\documentclass[11pt]{report}
\usepackage{graphics,graphicx}
\usepackage{indentfirst}
\usepackage{makeidx}
\usepackage{amssymb,amsmath}
\usepackage{wasysym}
\usepackage[usenames, dvipsnames]{color}
\usepackage[ pdftex, plainpages = false, pdfpagelabels,
                 pdfpagelayout = useoutlines,
                 bookmarks,
                 bookmarksopen = true,
                 bookmarksnumbered = true,
                 breaklinks = true,
                 linktocpage=all,
                 pagebackref=false,
                 colorlinks = true,
                 linkcolor = BrickRed,
                 urlcolor  = blue,
                 citecolor = BrickRed,
                 anchorcolor = green,
                 hyperindex = true,
                 hyperfigures
                 ]{hyperref}
\usepackage{supertabular}
\usepackage{xifthen} 

\usepackage{etex}

\makeindex

\newcommand{\Adam}{Adam\index{Adam (biblical)}}

\newcommand{\Alice}{Alice\index{Alice (cryptographic)}}

\newcommand{\Appleby}{Appleby\index{Appleby, D. Marcus}}
   \newcommand{\Marcus}{Marcus\index{Appleby, D. Marcus}}

\newcommand{\Barrett}{Barrett\index{Barrett, Jonathan}}

\newcommand{\Bayes}{Bayes\index{Bayes, Rev.\ Thomas}}

\newcommand{\Bell}{Bell\index{Bell, John S.}}

   \newcommand{\Ingemar}{Ingemar\index{Bengtsson, Ingemar}}

\newcommand{\Benka}{Benka\index{Benka, Stephen G.}}
\newcommand{\Bennett}{Bennett\index{Bennett, Charles H.}}
   \newcommand{\Charlie}{Charlie\index{Bennett, Charles H.}}
\newcommand{\Theo}{Theo\index{Bennett, Theodora M.}}
   \newcommand{\BennettT}{Bennett\index{Bennett, Theodora M.}}

\newcommand{\Bentsen}{Bentsen\index{Bentsen, Lloyd M. Jr.}}

\newcommand{\Bernstein}{Bernstein\index{Bernstein, Herbert J.}}
   \newcommand{\Herb}{Herb\index{Bernstein, Herbert J.}}

\newcommand{\Bob}{Bob\index{Bob (cryptographic)}}

\newcommand{\Bohr}{Bohr\index{Bohr, Niels}}

\newcommand{\Born}{Born\index{Born, Max}}

\newcommand{\Brassard}{Brassard\index{Brassard, Gilles}}
   \newcommand{\Gilles}{Gilles\index{Brassard, Gilles}}

\newcommand{\Brukner}{Bruk\-ner\index{Brukner, \v{C}aslav}}

\newcommand{\Bub}{Bub\index{Bub, Jeffrey}}
   \newcommand{\Jeff}{Jeff\index{Bub, Jeffrey}}

\newcommand{\Buck}{Buck\index{Buck, Joseph R.}}

\newcommand{\Burns}{Burns\index{Burns, Marshall}}

\newcommand{\Buzek}{Bu\v{z}ek\index{Bu\v{z}ek, Vladimir}}

   \newcommand{\Adan}{Ad\'an\index{Cabello, Ad\'an}}

\newcommand{\Caves}{Caves\index{Caves, Carlton M.}}
   \newcommand{\Carl}{Carl\index{Caves, Carlton M.}}

\newcommand{\Comer}{Comer\index{Comer, Gregory L.}}
   \newcommand{\Greg}{Greg\index{Comer, Gregory L.}}

\newcommand{\Cox}{Cox\index{Cox, Richard T.}}

\newcommand{\Daroczy}{Dar\'oczy\index{Dar\'oczy, Zoltan}}

\newcommand{\Davidson}{Davidson\index{Davidson, Donald}}

\newcommand{\Finetti}{de Finetti\index{de Finetti, Bruno}}

\newcommand{\Demopoulos}{Demopoulos\index{Demopoulos, William G.}}

\newcommand{\Derrida}{Derrida\index{Derrida, Jacques}}

\newcommand{\Deutsch}{Deutsch\index{Deutsch, David}}

\newcommand{\Dewey}{Dewey\index{Dewey, John}}

\newcommand{\Dirac}{Dirac\index{Dirac, Paul A. M.}}

\newcommand{\Einstein}{Einstein\index{Einstein, Albert}}

\newcommand{\Enk}{Enk\index{Enk, Steven J. van}}

\newcommand{\Ericsson}{Ericsson\index{Ericsson, {\AA}sa}}
   \newcommand{\Asa}{{\AA}sa\index{Ericsson, {\AA}sa}}
\newcommand{\Everett}{Everett\index{Everett, Hugh}}

\newcommand{\EveC}{Eve\index{Eve (cryptographic)}}

\newcommand{\Faye}{Faye\index{Faye, Jan}}

\newcommand{\Fierz}{Fierz\index{Fierz, Markus}}

\newcommand{\Folse}{Folse\index{Folse, Henry J.}}

\newcommand{\Kiki}{Kiki\index{Fuchs, Kristen Michele (Kiki)}}

\newcommand{\Gleason}{Gleason\index{Gleason, Andrew M.}}
\newcommand{\God}{God\index{God}}

\newcommand{\Good}{Good\index{Good, Irving John}}

\newcommand{\Gottesman}{Gottesman\index{Gottesman, Daniel}}

\newcommand{\Haldane}{Haldane\index{Haldane, J. B. S.}}

\newcommand{\Hayden}{Hayden\index{Hayden, Patrick}}

\newcommand{\Heidegger}{Heidegger\index{Heidegger, Martin}}

\newcommand{\Heisenberg}{Heisenberg\index{Heisenberg, Werner}}

\newcommand{\Hoffding}{H{\o}ffding\index{H{\o}ffding, Harald}}

\newcommand{\Honner}{Honner\index{Honner, John}}

\newcommand{\James}{James\index{James, William}}

\newcommand{\Jaynes}{Jaynes\index{Jaynes, Edwin T.}}
\newcommand{\Jeffrey}{Jeffrey\index{Jeffrey, Richard C.}}

\newcommand{\Jozsa}{Jozsa\index{Jozsa, Richard}}

\newcommand{\Kant}{Kant\index{Kant, Immanuel}}

\newcommand{\Kennedy}{Kennedy\index{Kennedy, John B.}}
\newcommand{\KennedyJF}{Kennedy\index{Kennedy, John F.}}

\newcommand{\Kyburg}{Kyburg\index{Kyburg, Henry E.}}

\newcommand{\Lahti}{Lahti\index{Lahti, Pekka J.}}

\newcommand{\Liouville}{Liouville\index{Liouville, Joseph}}

\newcommand{\Lorentz}{Lorentz\index{Lorentz, Hendrik A.}}

   \newcommand{\Hideo}{Hideo\index{Mabuchi, Hideo}}

\newcommand{\MacLane}{Mac Lane\index{Mac Lane, Saunders}}

\newcommand{\Mead}{Mead\index{Mead, George Herbert}}

\newcommand{\Mermin}{Mermin\index{Mermin, N. David}}

\newcommand{\Mittelstaedt}{Mittelstaedt\index{Mittelstaedt, Peter}}

\newcommand{\Moller}{M{\o}ller\index{M{\o}ller, Poul Martin}}
\newcommand{\Moelmer}{M{\o}lmer\index{M{\o}lmer, Klaus}}

\newcommand{\Montreal}{Montr\'eal\index{Montr\'eal, Canada}}

\newcommand{\vonNeumann}{von~Neumann\index{Neumann, John von}}

\newcommand{\Omnes}{Omn\`es\index{Omn\`es, Roland}}

\newcommand{\Palge}{Palge\index{Palge, Veiko}}

\newcommand{\Pauli}{Pauli\index{Pauli, Wolfgang}}

\newcommand{\Peierls}{Peierls\index{Peierls, Rudolf}}
\newcommand{\PeirceB}{Peirce\index{Peirce, Benjamin}}
\newcommand{\Peirce}{Peirce\index{Peirce, Charles Sanders}}

\newcommand{\Peres}{Peres\index{Peres, Asher}}
   \newcommand{\Asher}{Asher\index{Peres, Asher}}

\newcommand{\Plunk}{Plunk\index{Plunk, Gabriel}}
   \newcommand{\Gabe}{Gabe\index{Plunk, Gabriel}}

\newcommand{\Poincare}{Poincar\'e\index{Poincar\'e, Henri}}

\newcommand{\Preskill}{Preskill\index{Preskill, John}}

\newcommand{\Putnam}{Putnam\index{Putnam, Hilary}}

\newcommand{\Quayle}{Quayle\index{Quayle, James Danforth ``Dan''}}

\newcommand{\Renyi}{R\'enyi\index{R\'enyi, Alfr\'ed}}

\newcommand{\Rorty}{Rorty\index{Rorty, Richard}}
\newcommand{\Rosen}{Rosen\index{Rosen, Nathan}}
\newcommand{\Rosenfeld}{Rosenfeld\index{Rosenfeld, L\'eon}}

\newcommand{\Schack}{Schack\index{Schack, R\"udiger}}
   \newcommand{\Ruediger}{R\"udiger\index{Schack, R\"udiger}}

\newcommand{\Schiller}{Schiller\index{Schiller, Ferdinand Canning Scott}}

\newcommand{\Schroedinger}{Schr\"o\-ding\-er\index{Schr\"odinger, Erwin}}
\newcommand{\Schumacher}{Schumacher\index{Schumacher, Benjamin W.}}

\newcommand{\Shor}{Shor\index{Shor, Peter W.}}

\newcommand{\SimpsonJ}{Simpson\index{Simpson, John W.}}

\newcommand{\Sipe}{Sipe\index{Sipe, John E.}}

\newcommand{\Smokler}{Smokler\index{Smokler, Howard E.}}

\newcommand{\Spekkens}{Spekkens\index{Spekkens, Robert W.}}

\newcommand{\Stamp}{Stamp\index{Stamp, Philip C. E.}}
\newcommand{\Stapp}{Stapp\index{Stapp, Henry P.}}
\newcommand{\Stairs}{Stairs\index{Stairs, Allen}}

\newcommand{\Szilard}{Szil\'ard\index{Szil\'ard, Leo}}

\newcommand{\Timpson}{Timpson\index{Timpson, Christopher G.}}

\newcommand{\Ulam}{Ulam\index{Ulam, Stanis{\l}aw}}

\newcommand{\Vaxjo}{V\"axj\"o\index{V\"axj\"o, Sweden}}

\newcommand{\Weizsacker}{Weizs\"acker\index{Weizs\"acker, Carl F. von}}

\newcommand{\Wheeler}{Wheeler\index{Wheeler, John Archibald}}

\newcommand{\White}{White\index{White, Horace}}

\newcommand{\Wigner}{Wigner\index{Wigner, Eugene P.}}

\newcommand{\Wittgenstein}{Wittgenstein\index{Wittgenstein, Ludwig}}

\newcommand{\Wootters}{Wootters\index{Wootters, William K.}}

\newcommand{\Zeilinger}{Zeilinger\index{Zeilinger, Anton}}
   \newcommand{\Anton}{Anton\index{Zeilinger, Anton}}
\newcommand{\Zurek}{Zurek\index{Zurek, Wojciech H.}}

\newcommand{\bv}{\begin{verse}}
\newcommand{\ev}{\end{verse}}
\newcommand{\be}{\begin{equation}}
\newcommand{\ee}{\end{equation}}
\newcommand{\bea}{\begin{eqnarray}}
\newcommand{\eea}{\end{eqnarray}}
\newcommand{\bq}{\begin{quotation}}
\newcommand{\eq}{\end{quotation}}
\newcommand{\tr}{{\rm tr}}
\newcommand{\qbar}{q\hspace{-1.10ex}{\rule[-0.255ex]{0.45em}{0.09ex}}}
\newcommand{\veec}[3]{\left[\begin{array}{c}{\!\!#1\!\!}\\{\!\!#2\!\!}\\{\!\!#3\!\!}\end{array}\right]}

\newcommand{\boxedalign}[1]{%
  \[\fbox{%
      \addtolength{\linewidth}{-2\fboxsep}%
      \addtolength{\linewidth}{-2\fboxrule}%
      \begin{minipage}{\linewidth}%
      \begin{align}#1\end{align}%
      \end{minipage}%
    }\nonumber\]%
}

\newtheorem{conj}{Conjecture}

\newcommand{\bconj}{\begin{conj}\protect$\!\!${\em\bf :}$\;\;$} \newcommand{\econj}{\end{conj}}

\newcommand{\ket}[1]{|#1\rangle}
\newcommand{\proj}[1]{|#1\rangle\langle#1|}
\newcommand{\bra}[1]{{\langle#1|}}

\newcommand{\vc}[2]%
{\left(\begin{array}{c}{\!\!#1\!\!}\\{\!\!#2\!\!}\end{array}\right)}
\newcommand{\D}{\displaystyle}

\def\tr{{\rm tr}\,}
\def\drangle{\rangle\!\rangle}
\def\dlangle{\langle\!\langle}

\newtheorem{theorem}{Theorem}

\newcommand{\editornote}{\textbf{Editor's Note:\ }}
\newcommand{\indexbullet}[1]{{\large\textbf{#1}}}

\newtheorem{aaronson}{Aaronsonism}
\newcommand{\bsa}{\begin{aaronson}\protect$\!\!${\em\bf :}$\;\;$}
\newcommand{\esa}{\end{aaronson}}

\newtheorem{appleby}{Applebyism}
\newcommand{\bma}{\begin{appleby}\protect$\!\!${\em\bf :}$\;\;$}
\newcommand{\ema}{\end{appleby}}

\newtheorem{atmanspacher}{Atmanspacherism}
\newcommand{\bha}{\begin{atmanspacher}\protect$\!\!${\em\bf :}$\;\;$}
\newcommand{\eha}{\end{atmanspacher}}

\newtheorem{audenaert}{Audenaertism}
\newcommand{\bka}{\begin{audenaert}\protect$\!\!${\em\bf :}$\;\;$}
\newcommand{\eka}{\end{audenaert}}

\newtheorem{bacciagaluppi}{Bacciagaluppism}
\newcommand{\bgba}{\begin{bacciagaluppi}\protect$\!\!${\em\bf :}$\;\;$}
\newcommand{\egba}{\end{bacciagaluppi}}

\newtheorem{bacon}{Baconism}
\newcommand{\bbaco}{\begin{bacon}\protect$\!\!${\em\bf :}$\;\;$}
\newcommand{\ebaco}{\end{bacon}}

\newtheorem{baeyer}{von Baeyerism}
\newcommand{\bhcvb}{\begin{baeyer}\protect$\!\!${\em\bf :}$\;\;$}
\newcommand{\ehcvb}{\end{baeyer}}

\newtheorem{bahrami}{Bahrami-ism}
\newcommand{\bmob}{\begin{bahrami}\protect$\!\!${\em\bf :}$\;\;$}
\newcommand{\emob}{\end{bahrami}}

\newtheorem{baker}{Bakerism}
\newcommand{\bdb}{\begin{baker}\protect$\!\!${\em\bf :}$\;\;$}
\newcommand{\edb}{\end{baker}}

\newtheorem{balian}{Balianisme}
\newcommand{\brb}{\begin{balian}\protect$\!\!${\em\bf :}$\;\;$}
\newcommand{\erb}{\end{balian}}

\newtheorem{barnum}{Barnumism}
\newcommand{\bhb}{\begin{barnum}\protect$\!\!${\em\bf :}$\;\;$}
\newcommand{\ehb}{\end{barnum}}

\newtheorem{barrett}{Barrettism}
\newcommand{\bjoba}{\begin{barrett}\protect$\!\!${\em\bf :}$\;\;$}
\newcommand{\ejoba}{\end{barrett}}

\newtheorem{belfer}{Belferism}
\newcommand{\birb}{\begin{belfer}\protect$\!\!${\em\bf :}$\;\;$}
\newcommand{\eirb}{\end{belfer}}

\newtheorem{bengtsson}{Bengtssonism}
\newcommand{\bib}{\begin{bengtsson}\protect$\!\!${\em\bf :}$\;\;$}
\newcommand{\eib}{\end{bengtsson}}

\newtheorem{benioff}{Benioffism}
\newcommand{\bpb}{\begin{benioff}\protect$\!\!${\em\bf :}$\;\;$}
\newcommand{\epb}{\end{benioff}}

\newtheorem{charlie}{Bennettism}
\newcommand{\bcb}{\begin{charlie}\protect$\!\!${\em\bf :}$\;\;$}
\newcommand{\ecb}{\end{charlie}}

\newtheorem{herb}{Bernsteinism}
\newcommand{\bhbe}{\begin{herb}\protect$\!\!${\em\bf :}$\;\;$}
\newcommand{\ehbe}{\end{herb}}

\newtheorem{bitbol}{Bitbolism}
\newcommand{\bmb}{\begin{bitbol}\protect$\!\!${\em\bf :}$\;\;$}
\newcommand{\emb}{\end{bitbol}}

\newtheorem{brassard}{Brassardisme}
\newcommand{\bgb}{\begin{brassard}\protect$\!\!${\em\bf :}$\;\;$}
\newcommand{\egb}{\end{brassard}}

\newtheorem{braunstein}{Braunsteinism}
\newcommand{\bslb}{\begin{braunstein}\protect$\!\!${\em\bf :}$\;\;$}
\newcommand{\eslb}{\end{braunstein}}

\newtheorem{brown}{Brownism}
\newcommand{\bhrb}{\begin{brown}\protect$\!\!${\em\bf :}$\;\;$}
\newcommand{\ehrb}{\end{brown}}

\newtheorem{brun}{Brunism}
\newcommand{\btb}{\begin{brun}\protect$\!\!${\em\bf :}$\;\;$}
\newcommand{\etb}{\end{brun}}

\newtheorem{bub}{Bubism}
\newcommand{\bjb}{\begin{bub}\protect$\!\!${\em\bf :}$\;\;$}
\newcommand{\ejb}{\end{bub}}

\newtheorem{buchanan}{Buchananism}
\newcommand{\bmab}{\begin{buchanan}\protect$\!\!${\em\bf :}$\;\;$}
\newcommand{\emab}{\end{buchanan}}

\newtheorem{busch}{Buschism}
\newcommand{\bpbu}{\begin{busch}\protect$\!\!${\em\bf :}$\;\;$}
\newcommand{\epbu}{\end{busch}}

\newtheorem{butterfield}{Butterfieldism}
\newcommand{\bjnb}{\begin{butterfield}\protect$\!\!${\em\bf :}$\;\;$}
\newcommand{\ejnb}{\end{butterfield}}

\newtheorem{buzek}{Buzekism}
\newcommand{\bvb}{\begin{buzek}\protect$\!\!${\em\bf :}$\;\;$}
\newcommand{\evb}{\end{buzek}}

\newtheorem{cabello}{Cabello-ism}
\newcommand{\bac}{\begin{cabello}\protect$\!\!${\em\bf :}$\;\;$}
\newcommand{\eac}{\end{cabello}}

\newtheorem{capelin}{Capelinism}
\newcommand{\bsc}{\begin{capelin}\protect$\!\!${\em\bf :}$\;\;$}
\newcommand{\esc}{\end{capelin}}

\newtheorem{cavalcanti}{Cavalcanti-ism}
\newcommand{\bec}{\begin{cavalcanti}\protect$\!\!${\em\bf :}$\;\;$}
\newcommand{\eec}{\end{cavalcanti}}

\newtheorem{caves}{Cavesism}
\newcommand{\bcc}{\begin{caves}\protect$\!\!${\em\bf :}$\;\;$}
\newcommand{\ecc}{\end{caves}}

\newtheorem{cho}{Cho-ism}
\newcommand{\bacho}{\begin{cho}\protect$\!\!${\em\bf :}$\;\;$}
\newcommand{\eacho}{\end{cho}}

\newtheorem{cleve}{Cleve-ism}
\newcommand{\brc}{\begin{cleve}\protect$\!\!${\em\bf :}$\;\;$}
\newcommand{\erc}{\end{cleve}}

\newtheorem{cohen}{Cohenism}
\newcommand{\boc}{\begin{cohen}\protect$\!\!${\em\bf :}$\;\;$}
\newcommand{\eoc}{\end{cohen}}

\newtheorem{greg}{Comerism}
\newcommand{\bgc}{\begin{greg}\protect$\!\!${\em\bf :}$\;\;$}
\newcommand{\egc}{\end{greg}}

\newtheorem{conway}{Conwayism}
\newcommand{\bjhc}{\begin{conway}\protect$\!\!${\em\bf :}$\;\;$}
\newcommand{\ejhc}{\end{conway}}

\newtheorem{dang}{Dangism}
\newcommand{\bhbd}{\begin{dang}\protect$\!\!${\em\bf :}$\;\;$}
\newcommand{\ehbd}{\end{dang}}

\newtheorem{dariano}{D'Ariano-ism}
\newcommand{\bgmd}{\begin{dariano}\protect$\!\!${\em\bf :}$\;\;$}
\newcommand{\egmd}{\end{dariano}}

\newtheorem{demopoulos}{Demopoulosism}
\newcommand{\bwd}{\begin{demopoulos}\protect$\!\!${\em\bf :}$\;\;$}
\newcommand{\ewd}{\end{demopoulos}}

\newtheorem{donald}{Donaldism}
\newcommand{\bmjd}{\begin{donald}\protect$\!\!${\em\bf :}$\;\;$}
\newcommand{\emjd}{\end{donald}}

\newtheorem{duncan}{Duncanism}
\newcommand{\btd}{\begin{duncan}\protect$\!\!${\em\bf :}$\;\;$}
\newcommand{\etd}{\end{duncan}}

\newtheorem{durrani}{Durrani-ism}
\newcommand{\bmd}{\begin{durrani}\protect$\!\!${\em\bf :}$\;\;$}
\newcommand{\emd}{\end{durrani}}

\newtheorem{emerson}{Emersonia}
\newcommand{\bje}{\begin{emerson}\protect$\!\!${\em\bf :}$\;\;$}
\newcommand{\eje}{\end{emerson}}

\newtheorem{vanenk}{van Enkism}
\newcommand{\bsve}{\begin{vanenk}\protect$\!\!${\em\bf :}$\;\;$}
\newcommand{\esve}{\end{vanenk}}

\newtheorem{ericsson}{{\AA}sa-ism}
\newcommand{\bae}{\begin{ericsson}\protect$\!\!${\em\bf :}$\;\;$}
\newcommand{\eae}{\end{ericsson}}

\newtheorem{ferrie}{Ferrie-ism}
\newcommand{\bcf}{\begin{ferrie}\protect$\!\!${\em\bf :}$\;\;$}
\newcommand{\ecf}{\end{ferrie}}

\newtheorem{fine}{Fine-ism}
\newcommand{\baf}{\begin{fine}\protect$\!\!${\em\bf :}$\;\;$}
\newcommand{\eaf}{\end{fine}}

\newtheorem{finkelstein}{Finkelsteinism}
\newcommand{\bjf}{\begin{finkelstein}\protect$\!\!${\em\bf :}$\;\;$}
\newcommand{\ejf}{\end{finkelstein}}

\newtheorem{flammia}{Flammia-ism}
\newcommand{\bstf}{\begin{flammia}\protect$\!\!${\em\bf :}$\;\;$}
\newcommand{\estf}{\end{flammia}}

\newtheorem{folger}{Folgerism}
\newcommand{\btf}{\begin{folger}\protect$\!\!${\em\bf :}$\;\;$}
\newcommand{\etf}{\end{folger}}

\newtheorem{folse}{Folsesm}
\newcommand{\bhf}{\begin{folse}\protect$\!\!${\em\bf :}$\;\;$}
\newcommand{\ehf}{\end{folse}}

\newtheorem{ford}{Fordism}
\newcommand{\bkwf}{\begin{ford}\protect$\!\!${\em\bf :}$\;\;$}
\newcommand{\ekwf}{\end{ford}}

\newtheorem{fraassen}{van Fraassenism}
\newcommand{\bvf}{\begin{fraassen}\protect$\!\!${\em\bf :}$\;\;$}
\newcommand{\evf}{\end{fraassen}}

\newtheorem{garisto}{Garistoism}
\newcommand{\brg}{\begin{garisto}\protect$\!\!${\em\bf :}$\;\;$}
\newcommand{\erg}{\end{garisto}}

\newtheorem{geabanacloche}{Gea-Banaclochesm}
\newcommand{\bjgb}{\begin{geabanacloche}\protect$\!\!${\em\bf :}$\;\;$}
\newcommand{\ejgb}{\end{geabanacloche}}

\newtheorem{gill}{Gillism}
\newcommand{\brdg}{\begin{gill}\protect$\!\!${\em\bf :}$\;\;$}
\newcommand{\erdg}{\end{gill}}

\newtheorem{girelli}{Girelli-ism}
\newcommand{\bfg}{\begin{girelli}\protect$\!\!${\em\bf :}$\;\;$}
\newcommand{\efg}{\end{girelli}}

\newtheorem{goheen}{Goheenism}
\newcommand{\beg}{\begin{goheen}\protect$\!\!${\em\bf :}$\;\;$}
\newcommand{\eeg}{\end{goheen}}

\newtheorem{gottesman}{Gottesmanism}
\newcommand{\bdg}{\begin{gottesman}\protect$\!\!${\em\bf :}$\;\;$}
\newcommand{\edg}{\end{gottesman}}

\newtheorem{gottfried}{Gottfriedism}
\newcommand{\bkg}{\begin{gottfried}\protect$\!\!${\em\bf :}$\;\;$}
\newcommand{\ekg}{\end{gottfried}}

\newtheorem{goyal}{Goyalism}
\newcommand{\bphg}{\begin{goyal}\protect$\!\!${\em\bf :}$\;\;$}
\newcommand{\ephg}{\end{goyal}}

\newtheorem{grandy}{Grandyism}
\newcommand{\bwtg}{\begin{grandy}\protect$\!\!${\em\bf :}$\;\;$}
\newcommand{\ewtg}{\end{grandy}}

\newtheorem{grangier}{Grangierisme}
\newcommand{\bpg}{\begin{grangier}\protect$\!\!${\em\bf :}$\;\;$}
\newcommand{\epg}{\end{grangier}}

\newtheorem{graydon}{Graydonism}
\newcommand{\bmag}{\begin{graydon}\protect$\!\!${\em\bf :}$\;\;$}
\newcommand{\emag}{\end{graydon}}

\newtheorem{griffiths}{Griffithsism}
\newcommand{\brbg}{\begin{griffiths}\protect$\!\!${\em\bf :}$\;\;$}
\newcommand{\erbg}{\end{griffiths}}

\newtheorem{grinbaum}{Grinbaumism}
\newcommand{\bag}{\begin{grinbaum}\protect$\!\!${\em\bf :}$\;\;$}
\newcommand{\eag}{\end{grinbaum}}

\newtheorem{grover}{Groverism}
\newcommand{\blkg}{\begin{grover}\protect$\!\!${\em\bf :}$\;\;$}
\newcommand{\elkg}{\end{grover}}

\newtheorem{halvorson}{Halvorsonism}
\newcommand{\bhh}{\begin{halvorson}\protect$\!\!${\em\bf :}$\;\;$}
\newcommand{\ehh}{\end{halvorson}}

\newtheorem{hardy}{Hardyism}
\newcommand{\blh}{\begin{hardy}\protect$\!\!${\em\bf :}$\;\;$}
\newcommand{\elh}{\end{hardy}}

\newtheorem{hartmann}{Hartmannism}
\newcommand{\bsh}{\begin{hartmann}\protect$\!\!${\em\bf :}$\;\;$}
\newcommand{\esh}{\end{hartmann}}

\newtheorem{harper}{Harperism}
\newcommand{\bwlh}{\begin{harper}\protect$\!\!${\em\bf :}$\;\;$}
\newcommand{\ewlh}{\end{harper}}

\newtheorem{harremoes}{Harremo{\"e}sism}
\newcommand{\bph}{\begin{harremoes}\protect$\!\!${\em\bf :}$\;\;$}
\newcommand{\eph}{\end{harremoes}}

\newtheorem{hayden}{Haydenism}
\newcommand{\bpah}{\begin{hayden}\protect$\!\!${\em\bf :}$\;\;$}
\newcommand{\epah}{\end{hayden}}

\newtheorem{healey}{Healeyism}
\newcommand{\brh}{\begin{healey}\protect$\!\!${\em\bf :}$\;\;$}
\newcommand{\erh}{\end{healey}}

\newtheorem{henderson}{Hendersonism}
\newcommand{\bleh}{\begin{henderson}\protect$\!\!${\em\bf :}$\;\;$}
\newcommand{\eleh}{\end{henderson}}

\newtheorem{shenry}{Henryism}
\newcommand{\bshenry}{\begin{shenry}\protect$\!\!${\em\bf :}$\;\;$}
\newcommand{\eshenry}{\end{shenry}}

\newtheorem{herling}{Herlingism}
\newcommand{\bgh}{\begin{herling}\protect$\!\!${\em\bf :}$\;\;$}
\newcommand{\egh}{\end{herling}}

\newtheorem{hiley}{Hileyism}
\newcommand{\bbjh}{\begin{hiley}\protect$\!\!${\em\bf :}$\;\;$}
\newcommand{\ebjh}{\end{hiley}}

\newtheorem{holevo}{Holevo-ism}
\newcommand{\bash}{\begin{holevo}\protect$\!\!${\em\bf :}$\;\;$}
\newcommand{\eash}{\end{holevo}}

\newtheorem{honner}{Honnerific}
\newcommand{\bjh}{\begin{honner}\protect$\!\!${\em\bf :}$\;\;$}
\newcommand{\ejh}{\end{honner}}

\newtheorem{ismael}{Ismaelism}
\newcommand{\bji}{\begin{ismael}\protect$\!\!${\em\bf :}$\;\;$}
\newcommand{\eji}{\end{ismael}}

\newtheorem{jacobs}{Jacobsism}
\newcommand{\bkj}{\begin{jacobs}\protect$\!\!${\em\bf :}$\;\;$}
\newcommand{\ekj}{\end{jacobs}}

\newtheorem{jaffe}{Jaffe-ism}
\newcommand{\bahj}{\begin{jaffe}\protect$\!\!${\em\bf :}$\;\;$}
\newcommand{\eahj}{\end{jaffe}}

\newtheorem{janssen}{Janssenism}
\newcommand{\bmj}{\begin{janssen}\protect$\!\!${\em\bf :}$\;\;$}
\newcommand{\emj}{\end{janssen}}

\newtheorem{jozsa}{Jozsa-ism}
\newcommand{\brj}{\begin{jozsa}\protect$\!\!${\em\bf :}$\;\;$}
\newcommand{\erj}{\end{jozsa}}

\newtheorem{kargin}{Karginism}
\newcommand{\bvk}{\begin{kargin}\protect$\!\!${\em\bf :}$\;\;$}
\newcommand{\evk}{\end{kargin}}

\newtheorem{kent}{Kentism}
\newcommand{\bak}{\begin{kent}\protect$\!\!${\em\bf :}$\;\;$}
\newcommand{\eak}{\end{kent}}

\newtheorem{khrennikov}{Khrennikovism}
\newcommand{\bakh}{\begin{khrennikov}\protect$\!\!${\em\bf :}$\;\;$}
\newcommand{\eakh}{\end{khrennikov}}

\newtheorem{kimble}{Kimble-ism}
\newcommand{\bhjk}{\begin{kimble}\protect$\!\!${\em\bf :}$\;\;$}
\newcommand{\ehjk}{\end{kimble}}

\newtheorem{king}{Kingism}
\newcommand{\bck}{\begin{king}\protect$\!\!${\em\bf :}$\;\;$}
\newcommand{\eck}{\end{king}}

\newtheorem{knuth}{Knuthism}
\newcommand{\bkhk}{\begin{knuth}\protect$\!\!${\em\bf :}$\;\;$}
\newcommand{\ekhk}{\end{knuth}}

\newtheorem{kobak}{Kobakism}
\newcommand{\bdk}{\begin{kobak}\protect$\!\!${\em\bf :}$\;\;$}
\newcommand{\edk}{\end{kobak}}

\newtheorem{lacour}{La Courism}
\newcommand{\bbrlc}{\begin{lacour}\protect$\!\!${\em\bf :}$\;\;$}
\newcommand{\ebrlc}{\end{lacour}}

\newtheorem{landahl}{Landahlism}
\newcommand{\bal}{\begin{landahl}\protect$\!\!${\em\bf :}$\;\;$}
\newcommand{\eal}{\end{landahl}}

\newtheorem{larsson}{Larssony}
\newcommand{\bjal}{\begin{larsson}\protect$\!\!${\em\bf :}$\;\;$}
\newcommand{\ejal}{\end{larsson}}

\newtheorem{lawrence}{Lawrencesm}
\newcommand{\bwel}{\begin{lawrence}\protect$\!\!${\em\bf :}$\;\;$}
\newcommand{\ewel}{\end{lawrence}}

\newtheorem{leifer}{Leiferism}
\newcommand{\bml}{\begin{leifer}\protect$\!\!${\em\bf :}$\;\;$}
\newcommand{\eml}{\end{leifer}}

\newtheorem{leung}{Leungism}
\newcommand{\bdwl}{\begin{leung}\protect$\!\!${\em\bf :}$\;\;$}
\newcommand{\edwl}{\end{leung}}

\newtheorem{lieb}{Liebism}
\newcommand{\behl}{\begin{lieb}\protect$\!\!${\em\bf :}$\;\;$}
\newcommand{\eehl}{\end{lieb}}

\newtheorem{mabuchi}{Mabuchism}
\newcommand{\bhm}{\begin{mabuchi}\protect$\!\!${\em\bf :}$\;\;$}
\newcommand{\ehm}{\end{mabuchi}}

\newtheorem{mana}{Manalogue}
\newcommand{\bpglm}{\begin{mana}\protect$\!\!${\em\bf :}$\;\;$}
\newcommand{\epglm}{\end{mana}}

\newtheorem{maroney}{Maroneyism}
\newcommand{\bom}{\begin{maroney}\protect$\!\!${\em\bf :}$\;\;$}
\newcommand{\eom}{\end{maroney}}

\newtheorem{martin}{Martinism}
\newcommand{\bkma}{\begin{martin}\protect$\!\!${\em\bf :}$\;\;$}
\newcommand{\ekma}{\end{martin}}

\newtheorem{maudlin}{Maudlinism}
\newcommand{\btm}{\begin{maudlin}\protect$\!\!${\em\bf :}$\;\;$}
\newcommand{\etm}{\end{maudlin}}

\newtheorem{mcdonald}{McDonaldism}
\newcommand{\bkmcd}{\begin{mcdonald}\protect$\!\!${\em\bf :}$\;\;$}
\newcommand{\ekmcd}{\end{mcdonald}}

\newtheorem{menicucci}{Menicuccism}
\newcommand{\bnm}{\begin{menicucci}\protect$\!\!${\em\bf :}$\;\;$}
\newcommand{\enm}{\end{menicucci}}

\newtheorem{mermin}{Merminition}
\newcommand{\bdm}{\begin{mermin}\protect$\!\!${\em\bf :}$\;\;$}
\newcommand{\edm}{\end{mermin}}

\newtheorem{merzbacher}{Merzbacherism}
\newcommand{\bem}{\begin{merzbacher}\protect$\!\!${\em\bf :}$\;\;$}
\newcommand{\eem}{\end{merzbacher}}

\newtheorem{milburn}{Milburnism}
\newcommand{\bgjm}{\begin{milburn}\protect$\!\!${\em\bf :}$\;\;$}
\newcommand{\egjm}{\end{milburn}}

\newtheorem{misak}{Misakism}
\newcommand{\bcm}{\begin{misak}\protect$\!\!${\em\bf :}$\;\;$}
\newcommand{\ecm}{\end{misak}}

\newtheorem{mohrhoff}{Mohrhoffism}
\newcommand{\bum}{\begin{mohrhoff}\protect$\!\!${\em\bf :}$\;\;$}
\newcommand{\eum}{\end{mohrhoff}}

\newtheorem{moelmer}{M{\o}lmerism}
\newcommand{\bkm}{\begin{moelmer}\protect$\!\!${\em\bf :}$\;\;$}
\newcommand{\ekm}{\end{moelmer}}

\newtheorem{talmo}{Morism}
\newcommand{\btalmo}{\begin{talmo}\protect$\!\!${\em\bf :}$\;\;$}
\newcommand{\etalmo}{\end{talmo}}

\newtheorem{musser}{Musserism}
\newcommand{\bgm}{\begin{musser}\protect$\!\!${\em\bf :}$\;\;$}
\newcommand{\egm}{\end{musser}}

\newtheorem{myrvold}{Myrvoldism}
\newcommand{\bwm}{\begin{myrvold}\protect$\!\!${\em\bf :}$\;\;$}
\newcommand{\ewm}{\end{myrvold}}

\newtheorem{yjng}{Ng-ism}
\newcommand{\byjn}{\begin{yjng}\protect$\!\!${\em\bf :}$\;\;$}
\newcommand{\eyjn}{\end{yjng}}

\newtheorem{jeff}{Nicholsonism}
\newcommand{\bjn}{\begin{jeff}\protect$\!\!${\em\bf :}$\;\;$}
\newcommand{\ejn}{\end{jeff}}

\newtheorem{nielsen}{Nielsenism}
\newcommand{\bmn}{\begin{nielsen}\protect$\!\!${\em\bf :}$\;\;$}
\newcommand{\emn}{\end{nielsen}}

\newtheorem{norsen}{Norsenism}
\newcommand{\btn}{\begin{norsen}\protect$\!\!${\em\bf :}$\;\;$}
\newcommand{\etn}{\end{norsen}}

\newtheorem{norton}{Nortonism}
\newcommand{\bjdn}{\begin{norton}\protect$\!\!${\em\bf :}$\;\;$}
\newcommand{\ejdn}{\end{norton}}

\newtheorem{overbye}{Overbye-ism}
\newcommand{\bdo}{\begin{overbye}\protect$\!\!${\em\bf :}$\;\;$}
\newcommand{\edo}{\end{overbye}}

\newtheorem{palge}{Palge-ism}
\newcommand{\bvp}{\begin{palge}\protect$\!\!${\em\bf :}$\;\;$}
\newcommand{\evp}{\end{palge}}

\newtheorem{pearle}{Pearley Quote}
\newcommand{\bpp}{\begin{pearle}\protect$\!\!${\em\bf :}$\;\;$}
\newcommand{\epp}{\end{pearle}}

\newtheorem{asher}{Asherism}
\newcommand{\bap}{\begin{asher}\protect$\!\!${\em\bf :}$\;\;$}
\newcommand{\eap}{\end{asher}}

\newtheorem{perezsuarez}{P\'erez-Su\'arezism}
\newcommand{\bmps}{\begin{perezsuarez}\protect$\!\!${\em\bf :}$\;\;$}
\newcommand{\emps}{\end{perezsuarez}}

\newtheorem{petz}{Petzism}
\newcommand{\bdep}{\begin{petz}\protect$\!\!${\em\bf :}$\;\;$}
\newcommand{\edep}{\end{petz}}

\newtheorem{pike}{Pike-ism}
\newcommand{\brp}{\begin{pike}\protect$\!\!${\em\bf :}$\;\;$}
\newcommand{\erp}{\end{pike}}

\newtheorem{pitowsky}{Pitowskyism}
\newcommand{\bip}{\begin{pitowsky}\protect$\!\!${\em\bf :}$\;\;$}
\newcommand{\eip}{\end{pitowsky}}

\newtheorem{plaga}{Plaga-rism}
\newcommand{\brplaga}{\begin{plaga}\protect$\!\!${\em\bf :}$\;\;$}
\newcommand{\erplaga}{\end{plaga}}

\newtheorem{plotnitsky}{Plotnitskyism}
\newcommand{\barkp}{\begin{plotnitsky}\protect$\!\!${\em\bf :}$\;\;$}
\newcommand{\earkp}{\end{plotnitsky}}

\newtheorem{plunk}{Plunkism}
\newcommand{\bgp}{\begin{plunk}\protect$\!\!${\em\bf :}$\;\;$}
\newcommand{\egp}{\end{plunk}}

\newtheorem{poirier}{Poirierisme}
\newcommand{\bhpoirier}{\begin{poirier}\protect$\!\!${\em\bf :}$\;\;$}
\newcommand{\ehpoirier}{\end{poirier}}

\newtheorem{pope}{Pope-ism}
\newcommand{\bdtp}{\begin{pope}\protect$\!\!${\em\bf :}$\;\;$}
\newcommand{\edtp}{\end{pope}}

\newtheorem{poulin}{Poulinism}
\newcommand{\bdp}{\begin{poulin}\protect$\!\!${\em\bf :}$\;\;$}
\newcommand{\edp}{\end{poulin}}

\newtheorem{preskill}{Preskillism}
\newcommand{\bjp}{\begin{preskill}\protect$\!\!${\em\bf :}$\;\;$}
\newcommand{\ejp}{\end{preskill}}

\newtheorem{price}{Pricey Quote}
\newcommand{\bhp}{\begin{price}\protect$\!\!${\em\bf :}$\;\;$}
\newcommand{\ehp}{\end{price}}

\newtheorem{quznetsov}{Quznetsovism}
\newcommand{\bgq}{\begin{quznetsov}\protect$\!\!${\em\bf :}$\;\;$}
\newcommand{\egq}{\end{quznetsov}}

\newtheorem{raginsky}{Raginskyism}
\newcommand{\bmr}{\begin{raginsky}\protect$\!\!${\em\bf :}$\;\;$}
\newcommand{\emr}{\end{raginsky}}

\newtheorem{rasco}{Rasco-ism}
\newcommand{\bbcr}{\begin{rasco}\protect$\!\!${\em\bf :}$\;\;$}
\newcommand{\ebcr}{\end{rasco}}

\newtheorem{rau}{Rau-ism}
\newcommand{\bjr}{\begin{rau}\protect$\!\!${\em\bf :}$\;\;$}
\newcommand{\ejr}{\end{rau}}

\newtheorem{raymer}{Raymerism}
\newcommand{\bmgr}{\begin{raymer}\protect$\!\!${\em\bf :}$\;\;$}
\newcommand{\emgr}{\end{raymer}}

\newtheorem{renes}{Renesism}
\newcommand{\bjmr}{\begin{renes}\protect$\!\!${\em\bf :}$\;\;$}
\newcommand{\ejmr}{\end{renes}}

\newtheorem{renner}{Rennerism}
\newcommand{\brr}{\begin{renner}\protect$\!\!${\em\bf :}$\;\;$}
\newcommand{\err}{\end{renner}}

\newtheorem{reynolds}{Reynoldsism}
\newcommand{\bpjr}{\begin{reynolds}\protect$\!\!${\em\bf :}$\;\;$}
\newcommand{\epjr}{\end{reynolds}}

\newtheorem{rosado}{Rosado-ism}
\newcommand{\bjir}{\begin{rosado}\protect$\!\!${\em\bf :}$\;\;$}
\newcommand{\ejir}{\end{rosado}}

\newtheorem{rozema}{Rozema-ism}
\newcommand{\blar}{\begin{rozema}\protect$\!\!${\em\bf :}$\;\;$}
\newcommand{\elar}{\end{rozema}}

\newtheorem{rudolph}{Rudolphism}
\newcommand{\btr}{\begin{rudolph}\protect$\!\!${\em\bf :}$\;\;$}
\newcommand{\etr}{\end{rudolph}}

\newtheorem{sanders}{Sandersism}
\newcommand{\bmds}{\begin{sanders}\protect$\!\!${\em\bf :}$\;\;$}
\newcommand{\emds}{\end{sanders}}

\newtheorem{savitt}{Savittism}
\newcommand{\bss}{\begin{savitt}\protect$\!\!${\em\bf :}$\;\;$}
\newcommand{\ess}{\end{savitt}}

\newtheorem{schack}{Schackcosm}
\newcommand{\brs}{\begin{schack}\protect$\!\!${\em\bf :}$\;\;$}
\newcommand{\ers}{\end{schack}}

\newtheorem{schlosshauer}{Schlosshauerism}
\newcommand{\bmaxs}{\begin{schlosshauer}\protect$\!\!${\em\bf :}$\;\;$}
\newcommand{\emaxs}{\end{schlosshauer}}

\newtheorem{schroeck}{Schroeckism}
\newcommand{\bfes}{\begin{schroeck}\protect$\!\!${\em\bf :}$\;\;$}
\newcommand{\efes}{\end{schroeck}}

\newtheorem{schumacher}{Schumacherism}
\newcommand{\bbs}{\begin{schumacher}\protect$\!\!${\em\bf :}$\;\;$}
\newcommand{\ebs}{\end{schumacher}}

\newtheorem{scudo}{Scudoism}
\newcommand{\bps}{\begin{scudo}\protect$\!\!${\em\bf :}$\;\;$}
\newcommand{\eps}{\end{scudo}}

\newtheorem{scully}{Scullyism}
\newcommand{\bms}{\begin{scully}\protect$\!\!${\em\bf :}$\;\;$}
\newcommand{\ems}{\end{scully}}

\newtheorem{shapiro}{Shapiro-ism}
\newcommand{\bjhs}{\begin{shapiro}\protect$\!\!${\em\bf :}$\;\;$}
\newcommand{\ejhs}{\end{shapiro}}

\newtheorem{shimony}{Shimonyism}
\newcommand{\bas}{\begin{shimony}\protect$\!\!${\em\bf :}$\;\;$}
\newcommand{\eas}{\end{shimony}}

\newtheorem{shor}{Shor Thing}
\newcommand{\bpws}{\begin{shor}\protect$\!\!${\em\bf :}$\;\;$}
\newcommand{\epws}{\end{shor}}

\newtheorem{siegfried}{Siegfriedism}
\newcommand{\btoms}{\begin{siegfried}\protect$\!\!${\em\bf :}$\;\;$}
\newcommand{\etoms}{\end{siegfried}}

\newtheorem{simon}{Simonism}
\newcommand{\bshs}{\begin{simon}\protect$\!\!${\em\bf :}$\;\;$}
\newcommand{\eshs}{\end{simon}}

\newtheorem{sipe}{Sipesm}
\newcommand{\bjes}{\begin{sipe}\protect$\!\!${\em\bf :}$\;\;$}
\newcommand{\ejes}{\end{sipe}}

\newtheorem{skyrms}{Skyrmsism}
\newcommand{\bbsky}{\begin{skyrms}\protect$\!\!${\em\bf :}$\;\;$}
\newcommand{\ebsky}{\end{skyrms}}

\newtheorem{slee}{Slee-ism}
\newcommand{\btsl}{\begin{slee}\protect$\!\!${\em\bf :}$\;\;$}
\newcommand{\etsl}{\end{slee}}

\newtheorem{slusher}{Slusherism}
\newcommand{\bres}{\begin{slusher}\protect$\!\!${\em\bf :}$\;\;$}
\newcommand{\eres}{\end{slusher}}

\newtheorem{smolin}{John Smolinism}
\newcommand{\bjas}{\begin{smolin}\protect$\!\!${\em\bf :}$\;\;$}
\newcommand{\ejas}{\end{smolin}}

\newtheorem{lsmolin}{Lee Smolinism}
\newcommand{\bls}{\begin{lsmolin}\protect$\!\!${\em\bf :}$\;\;$}
\newcommand{\els}{\end{lsmolin}}

\newtheorem{snyder}{Snyderism}
\newcommand{\bcs}{\begin{snyder}\protect$\!\!${\em\bf :}$\;\;$}
\newcommand{\ecs}{\end{snyder}}

\newtheorem{spekkens}{Spekkensism}
\newcommand{\brws}{\begin{spekkens}\protect$\!\!${\em\bf :}$\;\;$}
\newcommand{\erws}{\end{spekkens}}

\newtheorem{stairs}{Stairsism}
\newcommand{\bAllS}{\begin{stairs}\protect$\!\!${\em\bf :}$\;\;$}
\newcommand{\eAllS}{\end{stairs}}

\newtheorem{stamp}{Stampede}
\newcommand{\bpces}{\begin{stamp}\protect$\!\!${\em\bf :}$\;\;$}
\newcommand{\epces}{\end{stamp}}

\newtheorem{steinberg}{Steinbergism}
\newcommand{\bams}{\begin{steinberg}\protect$\!\!${\em\bf :}$\;\;$}
\newcommand{\eams}{\end{steinberg}}

\newtheorem{stoll}{Stollicism}
\newcommand{\bsks}{\begin{stoll}\protect$\!\!${\em\bf :}$\;\;$}
\newcommand{\esks}{\end{stoll}}

\newtheorem{sudbery}{Sudberyism}
\newcommand{\bts}{\begin{sudbery}\protect$\!\!${\em\bf :}$\;\;$}
\newcommand{\ets}{\end{sudbery}}

\newtheorem{summhammer}{Summhammerism}
\newcommand{\bjs}{\begin{summhammer}\protect$\!\!${\em\bf :}$\;\;$}
\newcommand{\ejs}{\end{summhammer}}

\newtheorem{svozil}{Svozilism}
\newcommand{\bks}{\begin{svozil}\protect$\!\!${\em\bf :}$\;\;$}
\newcommand{\eks}{\end{svozil}}

\newtheorem{tait}{Taitism}
\newcommand{\bmt}{\begin{tait}\protect$\!\!${\em\bf :}$\;\;$}
\newcommand{\emt}{\end{tait}}

\newtheorem{terno}{Terno-ism}
\newcommand{\bdt}{\begin{terno}\protect$\!\!${\em\bf :}$\;\;$}
\newcommand{\edt}{\end{terno}}

\newtheorem{timpson}{Timpsonism}
\newcommand{\bcgt}{\begin{timpson}\protect$\!\!${\em\bf :}$\;\;$}
\newcommand{\ecgt}{\end{timpson}}

\newtheorem{tipler}{Tiplerism}
\newcommand{\bft}{\begin{tipler}\protect$\!\!${\em\bf :}$\;\;$}
\newcommand{\eft}{\end{tipler}}

\newtheorem{topsoe}{Tops{\o}e-ism}
\newcommand{\bflt}{\begin{topsoe}\protect$\!\!${\em\bf :}$\;\;$}
\newcommand{\eflt}{\end{topsoe}}

\newtheorem{ududec}{Cozminism}
\newcommand{\bcu}{\begin{ududec}\protect$\!\!${\em\bf :}$\;\;$}
\newcommand{\ecu}{\end{ududec}}

\newtheorem{uhlmann}{Uhlmannism}
\newcommand{\bau}{\begin{uhlmann}\protect$\!\!${\em\bf :}$\;\;$}
\newcommand{\eau}{\end{uhlmann}}

\newtheorem{unruh}{Unruhism}
\newcommand{\bwu}{\begin{unruh}\protect$\!\!${\em\bf :}$\;\;$}
\newcommand{\ewu}{\end{unruh}}

\newtheorem{valente}{Valente-ism}
\newcommand{\bgv}{\begin{valente}\protect$\!\!${\em\bf :}$\;\;$}
\newcommand{\egv}{\end{valente}}

\newtheorem{valentini}{Valentini-ism}
\newcommand{\bav}{\begin{valentini}\protect$\!\!${\em\bf :}$\;\;$}
\newcommand{\eav}{\end{valentini}}

\newtheorem{wachter}{Wachterism}
\newcommand{\brfw}{\begin{wachter}\protect$\!\!${\em\bf :}$\;\;$}
\newcommand{\erfw}{\end{wachter}}

\newtheorem{waskan}{Waskanism}
\newcommand{\bjaw}{\begin{waskan}\protect$\!\!${\em\bf :}$\;\;$}
\newcommand{\ejaw}{\end{waskan}}

\newtheorem{waxman}{Waxmania}
\newcommand{\bnw}{\begin{waxman}\protect$\!\!${\em\bf :}$\;\;$}
\newcommand{\enw}{\end{waxman}}

\newtheorem{weinstein}{Weinsteinism}
\newcommand{\bsw}{\begin{weinstein}\protect$\!\!${\em\bf :}$\;\;$}
\newcommand{\esw}{\end{weinstein}}

\newtheorem{westman}{Westmanism}
\newcommand{\bhwe}{\begin{westman}\protect$\!\!${\em\bf :}$\;\;$}
\newcommand{\ehwe}{\end{westman}}

\newtheorem{wilce}{Wilce-ism}
\newcommand{\baw}{\begin{wilce}\protect$\!\!${\em\bf :}$\;\;$}
\newcommand{\eaw}{\end{wilce}}

\newtheorem{wiseman}{Wisemanism}
\newcommand{\bhw}{\begin{wiseman}\protect$\!\!${\em\bf :}$\;\;$}
\newcommand{\ehw}{\end{wiseman}}

\newtheorem{wolpert}{Wolpertism}
\newcommand{\bdhw}{\begin{wolpert}\protect$\!\!${\em\bf :}$\;\;$}
\newcommand{\edhw}{\end{wolpert}}

\newtheorem{wootters}{Woottersism}
\newcommand{\bbw}{\begin{wootters}\protect$\!\!${\em\bf :}$\;\;$}
\newcommand{\ebw}{\end{wootters}}

\newtheorem{wright}{Wrightism}
\newcommand{\bjw}{\begin{wright}\protect$\!\!${\em\bf :}$\;\;$}
\newcommand{\ejw}{\end{wright}}

\newtheorem{hulya}{Hulya-ism}
\newcommand{\bhya}{\begin{hulya}\protect$\!\!${\em\bf :}$\;\;$}
\newcommand{\ehya}{\end{hulya}}

\newtheorem{zeilinger}{Zeilingerism}
\newcommand{\baz}{\begin{zeilinger}\protect$\!\!${\em\bf :}$\;\;$}
\newcommand{\eaz}{\end{zeilinger}}

\newcommand{\germanism}[4]{#1 \hyperref[#4]{\emph{#2}} (#3)\dotfill\pageref{#4}}
\newcommand{\fantome}{\leavevmode\hphantom{MM}}
\newcommand{\myref}[2]{\hyperref[#1]{#2}}
\newcommand{\myurl}[2][]{\ifthenelse{\isempty{#1}}{\url{#2}}{\href{#1}{\tt #2}}}
\newcommand{\quantph}[1]{\href{http://arxiv.org/abs/quant-ph/#1}{{\tt quant-\allowbreak{}ph/\allowbreak{}#1}}}
\newcommand{\arxiv}[2][]{\ifthenelse{\isempty{#1}}{\href{http://arxiv.org/abs/#2}{{\tt arXiv:\allowbreak{}#2}}} {\href{http://arxiv.org/abs/#2}{{\tt arXiv:\allowbreak{}#2 [#1]}}}}
\newcommand{\pirsa}[1]{\href{http://pirsa.org/#1/}{{\tt PIRSA:\allowbreak{}#1}}}

\makeatletter
\renewenvironment{thebibliography}[1]
     {\subsection*{\bibname}
      \@mkboth{\MakeUppercase\bibname}{\MakeUppercase\bibname}%
      \list{\@biblabel{\@arabic\c@enumiv}}%
           {\settowidth\labelwidth{\@biblabel{#1}}%
            \leftmargin\labelwidth
            \advance\leftmargin\labelsep
            \@openbib@code
            \usecounter{enumiv}%
            \let\p@enumiv\@empty
            \renewcommand\theenumiv{\@arabic\c@enumiv}}%
      \sloppy
      \clubpenalty4000
      \@clubpenalty \clubpenalty
      \widowpenalty4000%
      \sfcode`\.\@m}
     {\def\@noitemerr
       {\@latex@warning{Empty `thebibliography' environment}}%
      \endlist}
\makeatother

\begin{document}

\thispagestyle{empty}

\vspace*{3in}
\begin{center}
\Huge {\bf My Struggles with the Block Universe}
\bigskip \bigskip

\LARGE Christopher A. Fuchs

\end{center}

\pagebreak

\pagenumbering{roman}

\pagestyle{empty}

\begin{center}
\Huge \bigskip {\bf My Struggles with the \bigskip Block Universe
}
\\
\Large \rm Selected Correspondence, January 2001 -- May 2011\\

\vspace{0.7in}

\LARGE {\bf Christopher A. Fuchs} \\

\vspace{0.5in}

\normalsize Department of Physics
\\
\normalsize University of Massachusetts Boston
\\
\normalsize 100 Morrissey Boulevard
\\
\normalsize Boston, Massachusetts 02125, USA\medskip
\\
(current)
\bigskip\medskip\\
AND \bigskip
\\\normalsize Quantum Information Processing Group
\\
\normalsize Raytheon BBN Technologies
\\
\normalsize 10 Moulton Street
\\
Cambridge, Massachusetts 02138, USA\medskip
\\
(at the time this document was first posted)
\bigskip\medskip\\
AND \bigskip
\\\normalsize Perimeter Institute for Theoretical Physics
\\
\normalsize 31 Caroline Street North
\\
\normalsize Waterloo, Ontario N2L 2Y5,
Canada\medskip
\\
(at the time most of this material was compiled)
\bigskip\medskip\\
AND \bigskip\\
\normalsize Stellenbosch Institute for Advanced Study (STIAS)
\\
\normalsize Wallenberg Research Centre at Stellenbosch University
\\
\normalsize Marais Street
\\
\normalsize Stellenbosch 7600,
South Africa\medskip
\\
(the soil where QBism finally took root)
\bigskip\bigskip\bigskip\bigskip\bigskip


{\bf Foreword by Maximilian Schlosshauer}\bigskip\\
{\bf Edited by Blake C.\ Stacey}
\end{center}
\pagebreak


\vspace*{3.7in}

\begin{center}
In honor of 10 May 2011, the $10^{\rm th}$ anniversary of my last Cerro Grande posting, \\
a day that should have seen the most significant struggles passed.\bigskip\bigskip

\normalsize \copyright~2014--2015, Christopher A. Fuchs
\end{center}

\pagebreak

\thispagestyle{empty} \vspace*{3.7in}

\begin{center}
\begin{flushright}
\baselineskip=3pt
\parbox{3.6in}{
\bq
\noindent
The urge to archive all quantum thoughts is becoming the all-consuming fire of my soul.\medskip
\\
\hspace*{\fill} --- to N. D. Mermin, 26 April 2001
\eq
}
\end{flushright}
  \end{center}

\pagebreak
\pagestyle{plain}
\phantomsection

\begin{center}
\LARGE {\bf Foreword}
\end{center}

\addcontentsline{toc}{chapter}{Foreword}

\bq

My first glimpse of the universe that is Chris Fuchs was a little more than ten years ago, at the Rob Clifton Memorial Conference in College Park, Maryland. I spotted a bespectacled guy hanging around a table offering up a stack of book copies bearing the awkward title \emph{Notes on a Paulian Idea:\ Foundational, Historical, Anecdotal and Forward-Looking Thoughts on the Quantum}. The author, according to the cover, was one Christopher A.\ Fuchs, who I figured had to be the person manning the table. I opened to a random page and skimmed a few more. ``What is \emph{this}?'' I thought. ``He's offering up his email correspondence \emph{as a book}?'' It struck me as a little vain, and that was about it.

I have come a long way since then. I cannot precisely remember what got me hooked, but I have spent countless hours getting lost in the very book I had once so foolishly dismissed. Drafts of the present collection, which traces the evolution of that Paulian Idea to Quantum Bayesianism and finally QBism, have accompanied me for the past few years. Not unlike a weary traveler reaching for the Bible in a motel's bedside drawer, I turn to \emph{My Struggles with the Block Universe\/} whenever I need to reaffirm my belief that quantum mechanics is simply the most worthy and exciting cause to which to devote one's intellectual energy.

In his foreword to \emph{Notes on a Paulian Idea}, David Mermin observed that ``nobody today writing about quantum mechanics combines poetry and analysis to better effect than Chris Fuchs.'' His assessment holds just as true today: QBism appeals to both heart and mind, blending vivid imagery (prosthetic hands, philosopher's stone, my-\emph{my-my\/} experience) with technical bravura (SICs), all spiced up by Fuchs's inimitable voice and wit. But what stands out most to me is the fervor Fuchs has devoted to the intellectual journey shared with us here. There is an infectious intensity in Fuchs's writing and thinking, a relentless search fueled by profound curiosity. Fuchs's modus operandi---to try out his ideas on as many thoughtful people as he can find---is used to its full potential. Although Fuchs is never shy to declare his convictions, his correspondences are not an excuse for preaching; they are his way of discovery and reflection.

Einstein once said that Bohm's deterministic theory seemed ``too cheap'' to him. In a similar vein, I see Fuchs's development of QBism as an attempt to guard against any answers that feel too cheap. QBism proceeds from the pronouncement that ``a large part \dots\ of the structure of quantum theory has always concerned information.'' But Fuchs, of course, is all too aware of Bell's probing questions---\emph{whose\/} information? information \emph{about what}?---to let it go at that. And thus begins the arduous journey (see this volume) from the notions of information and knowledge, which still suggest something external and objective as their reference, to the notion of belief. Beliefs, cashed out as gambling commitments, strictly belong to an individual agent, and they are not beliefs about something objectively existing but about that agent's future experiences.

What Fuchs studiously aims to avoid here is falling into the trap of using the term \emph{information\/} as a mere sugar-coating, as something that tricks us into thinking that we have made more palatable the chewy bits quantum mechanics tosses our way. But Fuchs does not stop there. The professed goal is to strip away all those elements of quantum theory that can be interpreted in subjective, agent-dependent terms. The hope is that whatever remains will hint at something essential and objective about nature. With this hope, QBism has also breathed new life into instrumentalism (a term originally coined by John Dewey to describe his brand of pragmatism). Instrumentalism, it seems, has a bad reputation. It is perceived as unable to explain the successes of our best theories, and as unwilling to reach for nature's essence. QBism tells us that this assessment may well be premature. What QBism shows is that an \emph{enriched\/} instrumentalist attitude---in this case, seeing quantum theory as something that ``deals only with the object--subject relation,'' as Schr\"odinger once put it---may in fact be our best bet of arriving at explanations more profound and radical than anything we could have imagined. And so it is perhaps not surprising that recently, with David Mermin having officially hopped on board, QBism has been elevated from an interpretation of quantum mechanics to a conception of science.

``If this is going to be done right, we must go this far,'' Fuchs once wrote. ``Nothing less will do.'' To have dared to go this far, and with such great persistence, is one of Fuchs's most admirable accomplishments. This collection of correspondence shows us the kinds of places we can reach if we refuse to settle for cheap answers.  It makes clear that whatever we may think of those places, the journey has already paid off. Whether QBism is the correct way of thinking about quantum mechanics I do not know. But I find its approach too refreshing and its possibilities too exciting not to want to cheer from the sidelines.

There is no need for you to feel daunted by the well over two thousand pages that follow. Just do as I usually do: Skip to a random page and start reading, and I promise you will come away inspired and delighted. We must be grateful to Fuchs for sharing this marvelous volume with us. But we must also thank his close collaborators (Carl Caves and R\"udiger Schack, to name just two) and the many correspondents who have engaged with him so thoughtfully over the years. After all, QBism has always been, and will always be, a collective and self-consciously unfinished endeavor.

\vspace{.5cm}

\begin{flushright}
Maximilian Schlosshauer\\
Portland, Oregon\\
April 2014
\end{flushright}
\eq

\pagebreak

\vspace*{.5in}

\begin{center}
\baselineskip=12pt
\parbox{4.7in}{\baselineskip=12pt

\hspace*{.25in} What does determinism profess? It professes that those parts of the universe already laid down absolutely appoint and decree what the other parts shall be. The future has no ambiguous possibilities hidden in its womb; the part we call the present is compatible with only one totality. Any other future complement than the one fixed from eternity is impossible. The whole is in each and every part, and welds it with the rest into an absolute unity, an iron block, in which there can be no equivocation or shadow of \smallskip turning.

\hspace*{.25in} Indeterminism, on the contrary, says that the parts have a certain amount of loose play on one another, so that the laying down of one of them does not necessarily determine what the others shall be.  It admits that possibilities may be in excess of actualities, and that things not yet revealed to our knowledge may really in themselves be ambiguous.  Of two alternative futures which we conceive, both may now be really possible; and the one become impossible only at the very moment when the other excludes it by becoming real itself.  Indeterminism thus denies the world to be one unbending unit of fact. It says there is a certain ultimate pluralism in it.
\\
\hspace*{\fill} --- {\it William {\James}}}
\end{center}

\vspace{.2in}

\begin{center}
\baselineskip=12pt
\parbox{4.7in}{\baselineskip=12pt

\hspace*{.25in} Chance is a purely negative and relative term, giving us no information about that of which it is predicated, except that it
happens to be disconnected with something else---not controlled,
secured, or necessitated by other things in advance of its own actual
presence. What I say is that it tells us
nothing about what a thing may be in itself to call it ``chance.''
All you mean by calling it ``chance'' is that this is not guaranteed,
that it may also fall out otherwise. For the system of other things
has no positive hold on the chance-thing. Its origin is in a certain
fashion negative: it escapes, and says, Hands off!\ coming, when it
comes, as a free gift, or not at all.\smallskip

\hspace*{.25in} This negativeness, however, and this opacity of the chance-thing when thus considered {\it ab extra}, or from the point of view of previous things or distant things, do not preclude its having any amount of positiveness and luminosity from within, and at its own place and moment. All that its chance-character asserts about it is that there is something in it really of its own, something that is not the unconditional property of the whole. If the whole wants this
property, the whole must wait till it can get it, if it be a matter
of chance. That the universe may actually be a sort of joint-stock
society of this sort, in which the sharers have both limited
liabilities and limited powers, is of course a simple and conceivable
notion.
\\
\hspace*{\fill} --- {\it William {\James}}}
\end{center}

\pagebreak

\phantomsection

\begin{center}
\LARGE {\bf Introduction}
\end{center}

\addcontentsline{toc}{chapter}{Introduction}

\bq
This document is the second installment of three in the {\sl Cerro Grande Fire Series}.  It is a collection of emails compiled in the same spirit as my previous collection \arxiv{quant-ph/0105039}, {\sl Notes on a Paulian Idea:\ Foundational, Historical, Anecdotal and Forward-Looking Thoughts on the Quantum\/} (later printed by {\Vaxjo} University Press, and still later reissued with a lengthy introduction by Cambridge University Press as {\sl Coming of Age with Quantum Information}). One could joke that the present document represents the loquacious side of my quantum-foundations research program, but that would be a far understatement. Nearly every one of my emails from the period January 2001 to May 2011 having anything to do with the conceptual side of quantum theory---from my growing understanding of it, to fights over its meaning, to the advocacy of my views\footnote{And, my gosh, will one find advocacy in this volume.  Be prepared for repetition.  If the despicable years of George W. Bush's presidency had anything positive which I incorporated into my own palette, it was, ``Stay on message.''}, to dreams, ``business and funding opportunities,'' jokes, even moments of simple reverence and awe over the formalism---anything and everything on the subject has some representation in this volume.  To zeroth-order approximation, the compilation represents a grand exercise in ``intellectual waste not, want not.'' But, I bring it forward with a more serious goal in mind.

One of the correspondents in my earlier collection wrote me just before its release, ``I would not object to [your selection of our correspondence] in your volume \ldots\ but I wonder if our amicable banter would really be of value/interest to a wider readership!?''  The question of interest, of course, can only be answered by the reader.  But on the question of value for those who {\it are\/} interested in issues of quantum interpretation, my answer is a resounding yes. I am convinced that the email compositions collected here provide an essential ingredient, unavailable anywhere else, for turning the point of view of quantum theory put forth in technical detail here
\begin{center}
\arxiv[quant-ph]{1003.5209}\smallskip\\
and\smallskip\\
\arxiv[quant-ph]{1301.3274}
\end{center}
into a ``live option'' within the vast spectrum of quantum foundational thought.

The papers cited above comprise the most developed statements of the ``Quantum Bayesian'' research program, or QBism, to date, but they are no substitute for the kind of understanding---the living feeling---that can only be gotten by the actual process of fighting one's way day after day, sleepless night after sleepless night, through an inchoate and only slowly forming worldview.

No textbook on any subject can leave a student knowing the subject from a mere reading of its pages.  The student must actively work at understanding. He must live through the struggles the book's exercises mean to provide, always testing his understanding, always practicing how to respond if confronted on this or that issue.  Yet, all this is only the start of a process; it is just the glimmer of knowing.  Ultimately, the student must set the book's subject matter to work in his wider stream of activities.  But if this is so for ``knowing'' a textbook's subject, so much more so it must be for a worldview---a vision of the thrust and character of the world.

To get oneself into seeing or comprehending a worldview at all, one must gain a feel for its proponents' temperaments.  William James---the veritable mascot of this volume---makes the point most forcefully in his 1907 monograph {\sl Pragmatism}:
\bq
The history of philosophy is to a great extent that of a certain
clash of human temperaments.  Undignified as such a treatment may
seem to some of my colleagues, I shall have to take account of this
clash and explain a good many of the divergencies of philosophies by
it.  Of whatever temperament a professional philosopher is, he tries,
when philosophizing, to sink the fact of his temperament. Temperament
is no conventionally recognized reason, so he urges impersonal
reasons only for his conclusions.  Yet his temperament really gives
him a stronger bias than any of his more strictly objective premises.
It loads the evidence for him one way or the other \ldots\ just as this
fact or that principle would.  He {\it trusts\/} his temperament.
Wanting a universe that suits it, he believes in any representation
of the universe that does suit it. He feels men of opposite temper to
be out of key with the world's character, and in his heart considers
them incompetent and `not in it,' in the philosophic business, even
though they may far excel him in dialectical ability.

Yet in the forum he can make no claim, on the bare ground of his
temperament, to superior discernment or authority.  There arises thus
a certain insincerity in our philosophic discussions:  the potentest
of all our premises is never mentioned.
\eq
With a bit different emphasis, he adds to that from an 1879 essay: 
\bq\noindent
Why does Clifford fearlessly proclaim his belief in the conscious-automaton theory, although the `proofs' before him are the same which make Mr.\ Lewes reject it?  Why does he believe in primordial units of `mind-stuff' on evidence which would seem quite worthless to Professor Bain?  Simply because, like every human being of the slightest mental originality, he is peculiarly sensitive to evidence that bears in some one direction. It is utterly hopeless to try to exorcise such sensitiveness by calling it the disturbing subjective factor, and branding it as the root of all evil. `Subjective' be it called!\ and `disturbing' to those whom it foils!  But if it helps those who, as Cicero says, ``vim naturae magis sentiunt,''\footnote{``feel the force of nature more''} it is good and not evil. Pretend what we may, the whole man within us is at work when we form our philosophical opinions. Intellect, will, taste, and passion co-operate just as they do in practical affairs \ldots.
\eq
The year 2014 rings no differently.  David Deutsch calls the Everett interpretation of quantum theory an ``implication of science.''  Yet, where he, David Wallace, and Simon Saunders---to name some of Everett's most striving and eloquent disciples---can see no evidence against such a reading of quantum theory, I can see nothing particular about the theory's formalism that compels the view at all.  In fact, I see nothing that even suggests it.  In counterpoint:  The identities and analogies between quantum theory and Bayesian probability theory that come alive for me, seem to be for them but dead afterthoughts of a formalism handed down from on high.  There is something about their mindset that is as foreign to me as mine is to them.

The conceit of this document is that {\it QBism vis naturae magis sentit}.  QBism feels the force of nature more.  The 2,300 or so pages to follow are the only ways I know to convey the varieties of ``intellect, will, taste, and passion'' needed to make a view like QBism sensible and deemed worth pursuing.  Plenty in the pages herein is indeed banter, but taking note of the statement and style, both of the banter and the more seriously considered exegeses, is, as I see it, the most honest way to expose the potentest of QBism's premises.  In all, these pages taken together form an expression of one temperament as it confronts a world that will either accommodate it, bend it, or potentially smite it down.

If one were to look for a single most-encompassing statement of QBism's vision of the world---the crucial extract of this whole document---it might well be captured by this passage from page \pageref{ForIntroduction1}:
\bq
QBism says that every quantum measurement is a moment of creation, and the formal apparatus of quantum theory is an aid for each agent's thinking about those ``creatia'' she is involved with.  But surely a Copernican principle applies just as much to QBism as to any other science.  QBism's solution starts by saying the last point just that much more clearly:
``Quantum measurement represents those moments of creation an agent happens to seek out or notice.'' It does not at all mean that there aren't moments of creation going on all around, unnoticed, unparticipated in by the particular agent, all the time.  The larger world of QBism is something aligned with James's vision of a pluriverse where ``being comes in local spots and patches which add themselves or stay away at random, independently of the rest.''
\eq
or a little more fully by this one on page \pageref{ForIntroduction2}:
\bq
Three characteristics set QBism apart from other existing interpretations of quantum mechanics. First is its crucial reliance on the mathematical tools of quantum information theory to reshape the look and feel of quantum theory's formal structure. Second is its stance that two levels of radical ``personalism'' are required to break the interpretational conundrums plaguing the theory. Third is its recognition that with the solution of the theory's conundrums, quantum theory does not reach an end, but is the start of a great journey.

The two levels of personalism refer to how the ``probabilities'' and ``measurement events'' of quantum theory are to be interpreted. With regard to quantum probabilities, QBism asserts that they be interpreted as personal, Bayesian degrees of belief. This is the idea that probability is not something out in the world that can be right or wrong, but a personal accounting of what one expects. The implications of this are deep, for one can see with the help of quantum information theory that it means that quantum states, too, are not things out in the world. Quantum states rather represent personal accounting, and two agents speaking of the same quantum system may have distinct state assignments for it. The second level of personalism appears with the meaning of a quantum-measurement outcome. On this QBism holds with Pauli that a measurement apparatus must be understood as an extension of the agent herself, not something foreign and separate. A quantum measurement device is like a prosthetic hand, and the outcome of a measurement is an unpredictable, undetermined ``experience'' shared between the agent and external system. Quantum theory, thus, is no mirror image of what the world is, but rather a ``user's manual'' that any agent can adopt for better navigation in a world suffused with creation: The agent uses it for her little part and participation in this creation.
\eq
But these formulations are the fruit of a tree somewhere 10-years-up a very steep slope.  They could not have been said at the time of my {\sl Notes on a Paulian Idea}.

Much of what is recorded here concerns a fairly major transition in my thought since that 2001 posting. Once upon a time, I expressed my position with the slogan, ``Quantum States are States of Knowledge.''  But I ultimately realized this phrase is inconsistent with a thoroughgoing Bayesian conception of probabilities---the conception of Frank P. Ramsey, Bruno de Finetti, and Richard Jeffrey, where the actual event a probability is intended for never urges the probability assignment itself.  Worse, the slogan is inconsistent with the idea that physical systems have {\it no\/} spooky action-at-a-distance, a notion which I believed should be preserved at all cost.  I say the latter because to reject it is really to reject the idea that pieces of the world have any autonomy at all.  (Certainly an example of ``intellect, will, taste, and passion'' cooperating if there is any in this book.)

The remediation of these problems required steps more radical than I ever imagined I could or would take.  For instance, I came to recognize that even probability-1 assignments must be personalist Bayesian probabilities of a cloth with any other Bayesian probability.  Moreover, I started to understand that not only quantum states, but the operators used for describing measuring devices and all quantum time evolution maps as well, even Hamiltonian evolutions, must be viewed in strictly personalist Bayesian terms.  If these steps were not taken, the view of quantum theory we had been striving to construct would simply topple from inconsistency.  But it was no easy pill for even my most sympathetic colleagues and collaborators to swallow.  Indeed it fairly threatened to tear apart the Bayesian alliance Carlton Caves, {\Ruediger} Schack, and I had identified ourselves with for nearly a decade.  (See the notes between late June 2001 and late June 2002 for some of the most violent fireworks.)

In an early version of this samizdat covering that period---infelicitously titled ``{\sl Quantum States:\ What the Hell Are They? The Post-{\Vaxjo} Phase Transition:  Knowledge $\rightarrow$ Information $\rightarrow$ Belief $\rightarrow$ Pragmatic Commitment}''---I quoted an October 2001 letter to David Mermin as its introduction (embellishing the quote with one extra sentence):
\bq\noindent
Collecting it up, it's hard to believe I've written this much in the
little time since {\Vaxjo}. I guess it's been an active time for
me. I think there's no doubt that I've gone through a phase
transition. For all my Bayesian rhetoric in the last few years, I
simply had not realized the immense implications of holding fast to
the view that ``probabilities are subjective degrees of belief.'' Of
course, one way to look at this revelation is that it is a {\it
reductio ad absurdum\/} for the whole point of view---and that will
certainly be the first thing the critics pick up on. But---you
wouldn't have guessed less---I'm starting to view it as a godsend.
For with this simple train of logic, one can immediately stamp out
the potential reality/objectivity of any of a number of terms that
might have clouded our vision.  With so much dead weight removed, the little part left behind may finally have the strength to
support an ontology.
\eq
That's where it all got very tricky, or at least I should say very weird.  Ontology?  Nearly everyone I knew at the time was either calling me an instrumentalist (the seeming term of choice for philosophers of science) or a solipsist (the term of choice for physicists).  Both were insults to me, as both implied that I had given up on the love of my life---trying to figure out the character of the stuff of the world, as {\it indicated by\/} quantum theory.

The key came from contemplating the subjective nature of quantum measurement operators, which I had been forced to for the consistency of our view.  What could a quantum measurement be if one no longer had any impersonal handles to hold on to it with?  What are these $j$ indexing the operators $E_j$ in a positive-operator-valued measure if the $E_j$ are subjective ascriptions essentially of the same status as quantum states?  
They seemed more and more like the grin left behind from the Cheshire cat.  And that's when it hit me!  Quantum mechanics is not about {\it it}; it has never been about {\it it}.  It has always been about {\it me}!  What {\it I\/} am gambling on when {\it I\/} use quantum mechanics to calculate {\it my\/} probabilities are the consequences of {\it my\/} actions for {\it myself}.  Quantum mechanics is a calculus for gambling on {\it my\/} experiences, {\it whenever I use it}.  And it is a calculus for gambling on {\it your\/} experiences, {\it whenever you use it}.  But it is not a calculus for gambling on anything else.  That, I realized, was the great import of all the no-go theorems like Bell's and Kochen and Specker's (and more recently Pusey, Barrett, and Rudolph's, and Colbeck and Renner's, and many others).

And {\it that\/} was a statement of ontology!  Physics is telling us that all we have to gamble on are our own experiences {\it because\/} the world as a whole, or as a way in which things truly are, isn't even there yet!  It is not finished; it is being made.  There was a reason I had been falling in love with the thoughts of William James.  It all came together.  Maybe I'll quote him again (you will find that I do that a lot in this volume):
\bq
The import of the difference between pragmatism and rationalism is
now in sight throughout its whole extent. The essential contrast is
that {\it for rationalism reality is ready-made and complete from all
eternity, while for pragmatism it is still in the making, and awaits
part of its complexion from the future}. On the one side the
universe is absolutely secure, on the other it is still pursuing its
adventures. \ldots

The humanist view of `reality,' as something resisting, yet malleable, which controls our thinking as an
energy that must be taken `account' of incessantly is evidently a difficult one to
introduce to novices. \ldots

{\it The alternative between pragmatism and rationalism, in the shape
in which we now have it before us, is no longer a question in the
theory of knowledge, it concerns the structure of the universe
itself.}

On the pragmatist side we have only one edition of the universe,
unfinished, growing in all sorts of places, especially in the places
where thinking beings are at work.
On the rationalist side we have a universe in many editions, one real
one, the infinite folio, or {\it \'edition de luxe}, eternally
complete; and then the various finite editions, full of false
readings, distorted and mutilated each in its own way.
\eq

This collection of letters is ultimately about my struggles with the block-universe concept and my slow realization that the quantum theory pulverizes it.  I hope it will be an aid to the readers' own adventures in this forever unfinished and malleable world.

\vspace{.5cm}

\begin{flushright}
Christopher A. Fuchs\\
Cambridge, Massachusetts\\
2 May 2014
\end{flushright}

\eq


\pagebreak

\phantomsection

\addcontentsline{toc}{chapter}{Disclaimers and Acknowledgements}

\begin{center}
\baselineskip=12pt
\parbox{5.0in}{\baselineskip=12pt
DISCLAIMERS:  \medskip\\
{\bf I.} This document represents a rather unusual way to communicate my foundational thoughts on quantum theory:  It is a collection of letters spanning more than 10 years. For this reason it does represent a danger to my friends and correspondents.   There are two things that should not be mistaken: 1) The potential for my memory to fail when reporting the views of others, and 2) that the quotes taken from my correspondents were composed in any way other than casually for our {\it private\/} use. With regard to the latter, I assert the right of my correspondents to deny without apology that their quotes represent accurate accounts of their thoughts (then or now). Indeed, I know that some correspondents have changed their opinions since the time of the snippets recorded here.  I have tried to guard against misrepresentation by keeping the number of quotes and correspondent-replies to a minimum:  The ones that are used, are used mainly as springboards for my own tendentiousness or because I judge them to contain a deep thought in support of the general lines of this document.
\medskip\\
{\bf II.} Various deletions of text have been made to the original
letters.  Most of these have been done
for the sake of protecting the innocent, protecting the
correspondents, and protecting the illusion that I am good-natured.
The same holds as well for a small number of explicit changes of
phrase (in my own writings, {\it never\/} the correspondents). In
most cases, I have tried to make the process look as seamless as
possible, with no evidence that the text may have been otherwise. In
my own writings, bare ellipses should be interpreted as punctuation;
bracketed ellipses indicate true editorial changes.
\medskip\\
{\bf III.} There is no claim that all the ideas presented here are
coherent. The hope is that the incoherent ones will earn
their keep by their entertainment value.}
\end{center}

\medskip

\begin{center}
\baselineskip=12pt
\parbox{5.0in}{\baselineskip=12pt
ACKNOWLEDGEMENTS: \medskip\\
\hspace*{.25in} I thank Scott Aaronson, {\Ingemar} Bengtsson, Paul Benioff, Jeffrey Bub, {\Adan} Cabello, Richard Campos, {\Carl}ton {\Caves}, {\Asa} {\Ericsson}, Robert Garisto, Lucien Hardy, Osamu Hirota, Adrian Kent, the late Eugen Merzbacher, Klaus {\Moelmer}, Ulrich Mohrhoff, Michael Nielsen, Y. Jack Ng, Marcos P\'erez-Su\'arez, the late Itamar Pitowsky, Arkady Plotnitsky, Damian Pope, Steve Savitt, Ben Schumacher, and Bill Wootters for helping me believe a project like this might not be {\it too\/} crazy.
\smallskip
\\
\hspace*{.25in} I thank Marcus {\Appleby}, Hans Christian von Baeyer, Eric Cavalcanti, Greg Comer, Steven van {\Enk}, Andrei Khren\-nikov, Matthew Leifer, David {\Mermin}, John Preskill, Huw Price, {\Ruediger} {\Schack}, and Rob {\Spekkens} for helping me believe the earlier drafts have even been worthwhile.  To Maximilian Schlosshauer, who would write such a flattering foreword to this volume endorsing its ways of communication, I will be a friend for life.
\smallskip
\\
\hspace*{.25in} Beyond all others, I wish to thank Blake Stacey for maintaining a three-year enthusiasm for the idea of this ``ultimate'' volume, for being a living example that the words in it can make a difference to someone's thinking, and for the nearly inexhaustible editorial work and hyperlinking needed to make the text accessible to others.  I am deeply grateful Blake.
\smallskip
\\
\hspace*{.25in} There is a major difference between this collection and my last:  It is not so busy with letters and
replies to Asher Peres. He is sorely missed. In all, the tone of the book is a little lonelier without him.}
\end{center}


\pagebreak

\setcounter{tocdepth}{0}
\tableofcontents

\pagebreak

\twocolumn

\phantomsection

{\Large \bf Correspondents:}\bigskip

\addcontentsline{toc}{chapter}{Index of Correspondents}

\noindent {\bf Scott Aaronson --} \pageref{Aaronson1}, \pageref{Aaronson2}, \pageref{Aaronson3}, \pageref{Aaronson4}, \pageref{Aaronson5}, \pageref{Aaronson6}, \pageref{Aaronson7}, \pageref{Aaronson8}, \pageref{Aaronson9}, \pageref{Aaronson10} \medskip

\noindent {\bf Samson Abramsky --} \pageref{Abramsky1}, \pageref{Abramsky2}, \pageref{Abramsky3} \medskip

\noindent {\bf Guillaume Adenier --} \pageref{Adenier1}, \pageref{Adenier2} \medskip

\noindent {\bf Ran Adler --} \pageref{Adler1} \medskip

\noindent {\bf D. Marcus Appleby --} \pageref{Appleby0}, \pageref{Appleby1}, \pageref{Appleby2}, \pageref{Appleby3}, \pageref{Appleby4}, \pageref{Appleby5}, \pageref{Appleby6}, \pageref{Appleby7}, \pageref{Appleby8}, \pageref{Appleby9}, \pageref{Appleby10}, \pageref{Appleby11}, \pageref{Appleby12}, \pageref{Appleby13}, \pageref{Appleby14}, \pageref{Appleby14.1}, \pageref{Appleby14.2}, \pageref{Appleby14.3}, \pageref{Appleby14.4}, \pageref{Appleby14.5}, \pageref{Appleby14.6}, \pageref{Appleby15}, \pageref{Appleby16}, \pageref{Appleby17}, \pageref{Appleby18}, \pageref{Appleby19}, \pageref{Appleby20}, \pageref{Appleby21}, \pageref{Appleby22}, \pageref{Appleby23}, \pageref{Appleby23.1}, \pageref{Appleby24}, \pageref{Appleby25}, \pageref{Appleby26}, \pageref{Appleby27}, \pageref{Appleby28}, \pageref{Appleby29}, \pageref{Appleby30}, \pageref{Appleby31}, \pageref{Appleby32}, \pageref{Appleby33}, \pageref{Appleby34}, \pageref{Appleby35}, \pageref{Appleby36}, \pageref{Appleby37}, \pageref{Appleby38}, \pageref{Appleby39}, \pageref{Appleby40}, \pageref{Appleby41}, \pageref{Appleby42}, \pageref{Appleby43}, \pageref{Appleby44}, \pageref{Appleby45}, \pageref{Appleby46}, \pageref{Appleby47}, \pageref{Appleby48}, \pageref{Appleby49}, \pageref{Appleby50}, \pageref{Appleby51}, \pageref{Appleby52}, \pageref{Appleby53}, \pageref{Appleby54}, \pageref{Appleby55}, \pageref{Appleby56}, \pageref{Appleby57}, \pageref{Appleby58}, \pageref{Appleby59}, \pageref{Appleby60}, \pageref{Appleby61}, \pageref{Appleby62}, \pageref{Appleby63}, \pageref{Appleby64}, \pageref{Appleby65}, \pageref{Appleby66}, \pageref{Appleby67}, \pageref{Appleby68}, \pageref{Appleby69}, \pageref{Appleby70}, \pageref{Appleby71}, \pageref{Appleby72}, \pageref{Appleby73}, \pageref{Appleby74}, \pageref{Appleby75}, \pageref{Appleby76}, \pageref{Appleby77}, \pageref{Appleby78}, \pageref{Appleby79}, \pageref{Appleby80}, \pageref{Appleby81}, \pageref{Appleby82}, \pageref{Appleby83}, \pageref{Appleby84}, \pageref{Appleby84.1}, \pageref{Appleby85}, \pageref{Appleby86}, \pageref{Appleby87}, \pageref{Appleby88}, \pageref{Appleby89}, \pageref{Appleby90}, \pageref{Appleby91}, \pageref{Appleby92}, \pageref{Appleby93}, \pageref{Appleby94}, \pageref{Appleby95}, \pageref{Appleby96}, \pageref{Appleby97}, \pageref{Appleby98}, \pageref{Appleby99}, \pageref{Appleby100}, \pageref{Appleby101} \medskip

\noindent {\bf Nathan Argaman --} \pageref{Argaman1}, \pageref{Argaman2} \medskip

\noindent {\bf Harald Atmanspacher --} \pageref{Atmanspacher1}, \pageref{Atmanspacher2}, \pageref{Atmanspacher3}, \pageref{Atmanspacher4}, \pageref{Atmanspacher5}, \pageref{Atmanspacher6}, \pageref{Atmanspacher7}, \pageref{Atmanspacher8}, \pageref{Atmanspacher9}, \pageref{Atmanspacher10} \medskip

\noindent {\bf Guido Bacciagaluppi --} \pageref{Bacciagaluppi1}, \pageref{Bacciagaluppi2}, \pageref{Bacciagaluppi3}, \pageref{Bacciagaluppi4}, \pageref{Bacciagaluppi5}, \pageref{Bacciagaluppi6}, \pageref{Bacciagaluppi7}, \pageref{Bacciagaluppi8} \medskip

\noindent {\bf Manuel B\"achtold --} \pageref{Baechtold1}, \pageref{Baechtold2}, \pageref{Baechtold3} \medskip

\noindent {\bf Dave Bacon --} \pageref{Bacon1}, \pageref{Bacon2}, \pageref{Bacon3}, \pageref{Bacon4}, \pageref{Bacon5}, \pageref{Bacon6} \medskip

\noindent {\bf Hans Christian von Baeyer  --} \pageref{Baeyer1}, \pageref{Baeyer2}, \pageref{Baeyer3}, \pageref{Baeyer4}, \pageref{Baeyer5}, \pageref{Baeyer6}, \pageref{Baeyer7}, \pageref{Baeyer8}, \pageref{Baeyer9}, \pageref{Baeyer10}, \pageref{Baeyer11}, \pageref{Baeyer12}, \pageref{Baeyer13}, \pageref{Baeyer14}, \pageref{Baeyer15}, \pageref{Baeyer16}, \pageref{Baeyer17}, \pageref{Baeyer18}, \pageref{Baeyer19}, \pageref{Baeyer20}, \pageref{Baeyer21}, \pageref{Baeyer22}, \pageref{Baeyer23}, \pageref{Baeyer24}, \pageref{Baeyer25}, \pageref{Baeyer26}, \pageref{Baeyer27}, \pageref{Baeyer28}, \pageref{Baeyer29}, \pageref{Baeyer30}, \pageref{Baeyer31}, \pageref{Baeyer32}, \pageref{Baeyer33}, \pageref{Baeyer34}, \pageref{Baeyer35}, \pageref{Baeyer36}, \pageref{Baeyer37}, \pageref{Baeyer38}, \pageref{Baeyer39}, \pageref{Baeyer40}, \pageref{Baeyer41}, \pageref{Baeyer42}, \pageref{Baeyer43}, \pageref{Baeyer44}, \pageref{Baeyer45}, \pageref{Baeyer46}, \pageref{Baeyer47}, \pageref{Baeyer48}, \pageref{Baeyer49}, \pageref{Baeyer50}, \pageref{Baeyer51}, \pageref{Baeyer52}, \pageref{Baeyer53}, \pageref{Baeyer54}, \pageref{Baeyer55}, \pageref{Baeyer56}, \pageref{Baeyer57}, \pageref{Baeyer57.1}, \pageref{Baeyer57.2}, \pageref{Baeyer58}, \pageref{Baeyer59}, \pageref{Baeyer60}, \pageref{Baeyer61}, \pageref{Baeyer62}, \pageref{Baeyer63}, \pageref{Baeyer64}, \pageref{Baeyer65}, \pageref{Baeyer66}, \pageref{Baeyer67}, \pageref{Baeyer68}, \pageref{Baeyer69}, \pageref{Baeyer70}, \pageref{Baeyer71}, \pageref{Baeyer72}, \pageref{Baeyer73}, \pageref{Baeyer74}, \pageref{Baeyer75}, \pageref{Baeyer76}, \pageref{Baeyer77}, \pageref{Baeyer78}, \pageref{Baeyer79}, \pageref{Baeyer80}, \pageref{Baeyer81}, \pageref{Baeyer82}, \pageref{Baeyer83}, \pageref{Baeyer84}, \pageref{Baeyer85}, \pageref{Baeyer86}, \pageref{Baeyer87}, \pageref{Baeyer88}, \pageref{Baeyer89}, \pageref{Baeyer90}, \pageref{Baeyer91}, \pageref{Baeyer92}, \pageref{Baeyer93}, \pageref{Baeyer94}, \pageref{Baeyer95}, \pageref{Baeyer96}, \pageref{Baeyer97}, \pageref{Baeyer98}, \pageref{Baeyer99}, \pageref{Baeyer100}, \pageref{Baeyer101}, \pageref{Baeyer102}, \pageref{Baeyer103}, \pageref{Baeyer104}, \pageref{Baeyer105}, \pageref{Baeyer106}, \pageref{Baeyer107}, \pageref{Baeyer108}, \pageref{Baeyer109}, \pageref{Baeyer110}, \pageref{Baeyer111}, \pageref{Baeyer112}, \pageref{Baeyer113}, \pageref{Baeyer114}, \pageref{Baeyer115}, \pageref{Baeyer116}, \pageref{Baeyer117}, \pageref{Baeyer118}, \pageref{Baeyer119}, \pageref{Baeyer120}, \pageref{Baeyer121}, \pageref{Baeyer122}, \pageref{Baeyer123}, \pageref{Baeyer124}, \pageref{Baeyer125}, \pageref{Baeyer126}, \pageref{Baeyer127}, \pageref{Baeyer128}, \pageref{Baeyer129}, \pageref{Baeyer130}, \pageref{Baeyer131}, \pageref{Baeyer132}, \pageref{Baeyer133}, \pageref{Baeyer134}, \pageref{Baeyer134.1}, \pageref{Baeyer135}, \pageref{Baeyer136}, \pageref{Baeyer137}, \pageref{Baeyer138}, \pageref{Baeyer139}, \pageref{Baeyer140}, \pageref{Baeyer141}, \pageref{Baeyer142}, \pageref{Baeyer143}, \pageref{Baeyer144}, \pageref{Baeyer145}, \pageref{Baeyer146} \medskip

\noindent {\bf Mohammad Bahrami --} \pageref{Bahrami1} \medskip

\noindent {\bf David B. L. Baker --} \pageref{Baker1}, \pageref{Baker2}, \pageref{Baker3}, \pageref{Baker4}, \pageref{Baker5}, \pageref{Baker6}, \pageref{Baker7}, \pageref{Baker8}, \pageref{Baker9}, \pageref{Baker10}, \pageref{Baker11}, \pageref{Baker12}, \pageref{Baker13}, \pageref{Baker14}, \pageref{Baker15}, \pageref{Baker15.1}, \pageref{Baker16}, \pageref{Baker17}, \pageref{Baker18}, \pageref{Baker19}, \pageref{Baker20}, \pageref{Baker21}, \pageref{Baker22}, \pageref{Baker23}, \pageref{Baker24}, \pageref{Baker25}, \pageref{Baker26}, \pageref{Baker27}, \pageref{Baker28}, \pageref{Baker29}, \pageref{Baker30}, \pageref{Baker31} \medskip

\noindent {\bf Roger Balian --} \pageref{Balian1}, \pageref{Balian2}, \pageref{Balian3} \medskip

\noindent {\bf Leslie E. Ballentine --} \pageref{Ballentine1} \medskip

\noindent {\bf Ning Bao --} \pageref{Bao1} \medskip

\noindent {\bf Karen Barad --} \pageref{Barad1} \medskip

\noindent {\bf Howard Barnum --} \pageref{Barnum1}, \pageref{Barnum2}, \pageref{Barnum3}, \pageref{Barnum3.1}, \pageref{Barnum4}, \pageref{Barnum5}, \pageref{Barnum6}, \pageref{Barnum7}, \pageref{Barnum8}, \pageref{Barnum9}, \pageref{Barnum10}, \pageref{Barnum11}, \pageref{Barnum12}, \pageref{Barnum13}, \pageref{Barnum14}, \pageref{Barnum15}, \pageref{Barnum16}, \pageref{Barnum17}, \pageref{Barnum18}, \pageref{Barnum19}, \pageref{Barnum20}, \pageref{Barnum21}, \pageref{Barnum22}, \pageref{Barnum22.1}, \pageref{Barnum23}, \pageref{Barnum24}, \pageref{Barnum25}, \pageref{Barnum26}, \pageref{Barnum27}, \pageref{Barnum28}, \pageref{Barnum29}, \pageref{Barnum30}, \pageref{Barnum31}, \pageref{Barnum32}, \pageref{Barnum33} \medskip

\noindent {\bf Jonathan Barrett --} \pageref{Barrett1}, \pageref{Barrett2}, \pageref{Barrett3}, \pageref{Barrett4}, \medskip \pageref{Barrett5}

\noindent {\bf Claus Beisbart --} \pageref{Beisbart1} \medskip

\noindent {\bf Jacob D. Bekenstein --} \pageref{Bekenstein1} \medskip

\noindent {\bf Viacheslav P. Belavkin --} \pageref{Belavkin1} \medskip

\noindent {\bf Israel ``Roly'' Belfer --} \pageref{Belfer1}, \pageref{Belfer2}, \pageref{Belfer3}, \medskip \pageref{Belfer4}

\noindent {\bf Ingemar Bengtsson --} \pageref{Bengtsson1}, \pageref{Bengtsson2}, \pageref{Bengtsson2.1}, \pageref{Bengtsson3}, \pageref{Bengtsson4} \medskip

\noindent {\bf Paul A. Benioff --} \pageref{Benioff1} \medskip

\noindent {\bf Charles H. Bennett --} \pageref{Bennett1}, \pageref{Bennett2}, \pageref{Bennett3}, \pageref{Bennett4}, \pageref{Bennett5}, \pageref{Bennett6}, \pageref{Bennett7}, \pageref{Bennett8}, \pageref{Bennett9}, \pageref{Bennett10}, \pageref{Bennett11}, \pageref{Bennett12}, \pageref{Bennett13}, \pageref{Bennett14}, \pageref{Bennett15}, \pageref{Bennett16}, \pageref{Bennett17}, \pageref{Bennett18}, \pageref{Bennett19}, \pageref{Bennett20}, \pageref{Bennett21}, \pageref{Bennett22}, \pageref{Bennett23}, \pageref{Bennett24}, \pageref{Bennett25}, \pageref{Bennett26}, \pageref{Bennett27}, \pageref{Bennett28}, \pageref{Bennett29}, \pageref{Bennett30}, \pageref{Bennett31}, \pageref{Bennett32}, \pageref{Bennett33}, \pageref{Bennett34}, \pageref{Bennett35}, \pageref{Bennett36}, \pageref{Bennett37}, \pageref{Bennett38}, \pageref{Bennett39}, \pageref{Bennett40}, \pageref{Bennett41}, \pageref{Bennett42}, \pageref{Bennett43}, \pageref{Bennett44}, \pageref{Bennett45}, \pageref{Bennett46}, \pageref{Bennett47}, \pageref{Bennett48}, \pageref{Bennett49}, \pageref{Bennett50}, \pageref{Bennett51}, \pageref{Bennett52}, \pageref{Bennett53}, \pageref{Bennett54}, \pageref{Bennett55}, \pageref{Bennett56}, \pageref{Bennett57}, \pageref{Bennett58}, \pageref{Bennett59}, \pageref{Bennett60}, \pageref{Bennett61}, \pageref{Bennett62}, \pageref{Bennett63}, \pageref{Bennett63.1}, \pageref{Bennett64}, \pageref{Bennett65}, \pageref{Bennett66}, \pageref{Bennett67}, \pageref{Bennett68}, \pageref{Bennett69}, \pageref{Bennett70}, \pageref{Bennett71}, \pageref{Bennett72} \medskip

\noindent {\bf Theodora M. Bennett --} \pageref{TheoBennett} \medskip

\noindent {\bf Joseph Berkovitz --} \pageref{Berkovitz1}, \pageref{Berkovitz2} \medskip

\noindent {\bf Herbert J. Bernstein --} \pageref{Bernstein1}, \pageref{Bernstein2}, \pageref{Bernstein3}, \pageref{Bernstein4}, \pageref{Bernstein5}, \pageref{Bernstein6}, \pageref{Bernstein7}, \pageref{Bernstein8}, \pageref{Bernstein9}, \pageref{Bernstein10}, \pageref{Bernstein11}  \medskip

\noindent {\bf Douglas J. Bilodeau --} \pageref{Bilodeau1}, \pageref{Bilodeau2}, \pageref{Bilodeau3}, \pageref{Bilodeau4}, \pageref{Bilodeau5}, \pageref{Bilodeau6}, \pageref{Bilodeau7},  \pageref{Bilodeau8} \medskip

\noindent {\bf Michel Bitbol --} \pageref{Bitbol1}, \pageref{Bitbol2}, \pageref{Bitbol3} \medskip

\noindent {\bf Margaret E. Blume-Kohout --} \pageref{BlumeKohoutME1} \medskip

\noindent {\bf Robin Blume-Kohout --} \pageref{BlumeKohout1}, \pageref{BlumeKohout1.1}, \pageref{BlumeKohout2}, \pageref{BlumeKohout3}, \pageref{BlumeKohout4}, \pageref{BlumeKohout5}, \pageref{BlumeKohout6}, \pageref{BlumeKohout7}, \pageref{BlumeKohout8}, \pageref{BlumeKohout9}, \pageref{BlumeKohout10}, \pageref{BlumeKohout11} \medskip

\noindent {\bf Nora M. Boyd --} \pageref{Boyd1}, \pageref{Boyd2} \medskip

\noindent {\bf Katherine Brading --} \pageref{Brading1}, \pageref{Brading2} \medskip

\noindent {\bf Gilles Brassard --} \pageref{Brassard1}, \pageref{Brassard2}, \pageref{Brassard3}, \pageref{Brassard4}, \pageref{Brassard5}, \pageref{Brassard6}, \pageref{Brassard7}, \pageref{Brassard8}, \pageref{Brassard9}, \pageref{Brassard10}, \pageref{Brassard11}, \pageref{Brassard12}, \pageref{Brassard13}, \pageref{Brassard14}, \pageref{Brassard15}, \pageref{Brassard16}, \pageref{Brassard17}, \pageref{Brassard18}, \pageref{Brassard19}, \pageref{Brassard20}, \pageref{Brassard21}, \pageref{Brassard22}, \pageref{Brassard23}, \pageref{Brassard24}, \pageref{Brassard25}, \pageref{Brassard26}, \pageref{Brassard27}, \pageref{Brassard28}, \pageref{Brassard29}, \pageref{Brassard30}, \pageref{Brassard31}, \pageref{Brassard32}, \pageref{Brassard33}, \pageref{Brassard34}, \pageref{Brassard35}, \pageref{Brassard36}, \pageref{Brassard37}, \pageref{Brassard38}, \pageref{Brassard39}, \pageref{Brassard40}, \pageref{Brassard41}, \pageref{Brassard42}, \pageref{Brassard43}, \pageref{Brassard44}, \pageref{Brassard45}, \pageref{Brassard46}, \pageref{Brassard47}, \pageref{Brassard48}, \pageref{Brassard49}, \pageref{Brassard50}, \pageref{Brassard51}, \pageref{Brassard52}, \pageref{Brassard53}  \medskip

\noindent {\bf Samuel L. Braunstein --} \pageref{Braunstein1}, \pageref{Braunstein2}, \pageref{Braunstein3}, \pageref{Braunstein4}, \pageref{Braunstein5}, \pageref{Braunstein6}, \pageref{Braunstein7}, \pageref{Braunstein8}, \pageref{Braunstein9}, \pageref{Braunstein10}, \pageref{Braunstein11}, \pageref{Braunstein12}, \pageref{Braunstein13}, \pageref{Braunstein14}, \pageref{Braunstein15} \medskip

\noindent {\bf Hans J. Briegel --} \pageref{Briegel1}, \pageref{Briegel2}, \pageref{Briegel3}, \pageref{Briegel4}, \pageref{Briegel5}, \pageref{Briegel6} \medskip

\noindent {\bf Harvey R. Brown --} \pageref{BrownHR1}, \pageref{BrownHR2}, \pageref{BrownHR3}, \pageref{BrownHR4}, \pageref{BrownHR5}, \pageref{BrownHR6}, \pageref{BrownHR7} \medskip

\noindent {\bf James R. Brown --} \pageref{BrownJR1}, \pageref{BrownJR2}, \pageref{BrownJR3} \medskip

\noindent {\bf \v{C}aslav Brukner --} \pageref{Brukner1}, \pageref{Brukner2}, \pageref{Brukner3}, \pageref{Brukner4}, \pageref{Brukner5}, \pageref{Brukner6}, \pageref{Brukner7}, \pageref{Brukner8} \medskip

\noindent {\bf Todd A. Brun --} \pageref{Brun1}, \pageref{Brun2}, \pageref{Brun3}, \pageref{Brun4}, \pageref{Brun5}, \pageref{Brun6}, \pageref{Brun7}, \pageref{Brun8}, \pageref{Brun9} \medskip

\noindent {\bf Jeffrey Bub  --} \pageref{Bub1}, \pageref{Bub2}, \pageref{Bub3}, \pageref{Bub4}, \pageref{Bub5}, \pageref{Bub6}, \pageref{Bub7}, \pageref{Bub7.1}, \pageref{Bub8}, \pageref{Bub9}, \pageref{Bub10}, \pageref{Bub11}, \pageref{Bub12}, \pageref{Bub13}, \pageref{Bub14}, \pageref{Bub15}, \pageref{Bub16}, \pageref{Bub17}, \pageref{Bub18}, \pageref{Bub19}, \pageref{Bub20}, \pageref{Bub21}, \pageref{Bub21.1}, \pageref{Bub22} \medskip

\noindent {\bf Mark Buchanan --} \pageref{Buchanan1} \medskip

\noindent {\bf Paul Busch --} \pageref{Busch1}, \pageref{Busch2}, \pageref{Busch3}, \pageref{Busch4}, \pageref{Busch5}, \pageref{Busch6}, \pageref{Busch7}, \pageref{Busch8} \medskip

\noindent {\bf Jeremy N. Butterfield --} \pageref{Butterfield-1}, \pageref{Butterfield0}, \pageref{Butterfield1}, \pageref{Butterfield2}, \pageref{Butterfield3}, \pageref{Butterfield4}, \pageref{Butterfield4}, \pageref{Butterfield4.05}, \pageref{Butterfield5}, \pageref{Butterfield6}, \pageref{Butterfield7} \medskip

\noindent {\bf Vladimir Buzek --} \pageref{Buzek1} \medskip

\noindent {\bf Ad\'an Cabello --} \pageref{Cabello1}, \pageref{Cabello1.1}, \pageref{Cabello2}, \pageref{Cabello3}, \pageref{Cabello4}, \pageref{Cabello5}, \pageref{Cabello6} \medskip

\noindent {\bf A. Robert Calderbank --} \pageref{Calderbank1}, \pageref{Calderbank2}, \pageref{Calderbank3} \medskip

\noindent {\bf Richard A. Campos --} \pageref{Campos1} \medskip

\noindent {\bf Simon Capelin --} \pageref{Capelin1}, \pageref{Capelin2}, \pageref{Capelin3}, \pageref{Capelin4}, \pageref{Capelin5}, \pageref{Capelin6}, \pageref{Capelin7}, \pageref{Capelin8}, \pageref{Capelin9}, \pageref{Capelin10}, \pageref{Capelin11}  \medskip

\noindent {\bf Nancy Cartwright --} \pageref{Cartwright1} \medskip

\noindent {\bf William B. Case --} \pageref{Case1}, \pageref{Case2} \medskip

\noindent {\bf David C. Cassidy --} \pageref{Cassidy1} \medskip

\noindent {\bf Ariel Caticha --} \pageref{Caticha1} \medskip

\noindent {\bf Eric G. Cavalcanti --} \pageref{Cavalcanti1}, \pageref{Cavalcanti2}, \pageref{Cavalcanti3}, \pageref{Cavalcanti4}, \pageref{Cavalcanti5}, \pageref{Cavalcanti6}, \pageref{Cavalcanti7}, \pageref{Cavalcanti8}, \pageref{Cavalcanti9} \medskip

\noindent {\bf Carlton M. Caves --} \pageref{Caves0}, \pageref{Caves0.1}, \pageref{Caves0.2}, \pageref{Caves0.3}, \pageref{Caves1}, \pageref{Caves2}, \pageref{Caves3}, \pageref{Caves4}, \pageref{Caves5}, \pageref{Caves6}, \pageref{Caves7}, \pageref{Caves8}, \pageref{Caves8.1}, \pageref{Caves9}, \pageref{Caves10}, \pageref{Caves11}, \pageref{Caves12}, \pageref{Caves13}, \pageref{Caves14}, \pageref{Caves15}, \pageref{Caves16}, \pageref{Caves17}, \pageref{Caves18}, \pageref{Caves19}, \pageref{Caves20}, \pageref{Caves21}, \pageref{Caves22}, \pageref{Caves23}, \pageref{Caves24}, \pageref{Caves25}, \pageref{Caves26}, \pageref{Caves27}, \pageref{Caves28}, \pageref{Caves28.1}, \pageref{Caves29}, \pageref{Caves29.1}, \pageref{Caves30}, \pageref{Caves31}, \pageref{Caves32}, \pageref{Caves33}, \pageref{Caves34}, \pageref{Caves35}, \pageref{Caves35.1}, \pageref{Caves36}, \pageref{Caves37}, \pageref{Caves37.1}, \pageref{Caves37.2}, \pageref{Caves37.3}, \pageref{Caves38}, \pageref{Caves39}, \pageref{Caves39.1}, \pageref{Caves39.2}, \pageref{Caves39.3}, \pageref{Caves40}, \pageref{Caves40.1}, \pageref{Caves41}, \pageref{Caves42}, \pageref{Caves43}, \pageref{Caves43.1}, \pageref{Caves44}, \pageref{Caves45}, \pageref{Caves46}, \pageref{Caves47}, \pageref{Caves48}, \pageref{Caves49}, \pageref{Caves49.1}, \pageref{Caves50}, \pageref{Caves51}, \pageref{Caves52}, \pageref{Caves53}, \pageref{Caves54}, \pageref{Caves55}, \pageref{Caves56}, \pageref{Caves57}, \pageref{Caves57.1}, \pageref{Caves58}, \pageref{Caves59}, \pageref{Caves60}, \pageref{Caves60.1}, \pageref{Caves60.2}, \pageref{Caves61}, \pageref{Caves62}, \pageref{Caves63}, \pageref{Caves63.1}, \pageref{Caves63.2}, \pageref{Caves63.3}, \pageref{Caves63.4}, \pageref{Caves64}, \pageref{Caves65}, \pageref{Caves65.1}, \pageref{Caves66}, \pageref{Caves67}, \pageref{Caves68}, \pageref{Caves69}, \pageref{Caves70}, \pageref{Caves70.1}, \pageref{Caves70.2}, \pageref{Caves71}, \pageref{Caves72}, \pageref{Caves72.1}, \pageref{Caves73}, \pageref{Caves73.01}, \pageref{Caves73.02}, \pageref{Caves73.1}, \pageref{Caves73.2}, \pageref{Caves73.3}, \pageref{Caves74}, \pageref{Caves74.05}, \pageref{Caves74.1}, \pageref{Caves74.2}, \pageref{Caves74.3}, \pageref{Caves75}, \pageref{Caves75.1}, \pageref{Caves75.2}, \pageref{Caves75.3}, \pageref{Caves76}, \pageref{Caves76.1}, \pageref{Caves77}, \pageref{Caves77.01}, \pageref{Caves77.02}, \pageref{Caves77.03}, \pageref{Caves77.1}, \pageref{Caves77.2}, \pageref{Caves77.3}, \pageref{Caves77.4}, \pageref{Caves78}, \pageref{Caves79}, \pageref{Caves79.1}, \pageref{Caves79.1.1}, \pageref{Caves79.1.2}, \pageref{Caves79.1.3}, \pageref{Caves79.2}, \pageref{Caves79.3}, \pageref{Caves79.4}, \pageref{Caves79.5}, \pageref{Caves79.6}, \pageref{Caves80}, \pageref{Caves81}, \pageref{Caves81.1}, \pageref{Caves81.2}, \pageref{Caves81.3}, \pageref{Caves81.4}, \pageref{Caves81.5}, \pageref{Caves81.6}, \pageref{Caves82}, \pageref{Caves83}, \pageref{Caves84}, \pageref{Caves85}, \pageref{Caves86}, \pageref{Caves87}, \pageref{Caves87.1}, \pageref{Caves87.2}, \pageref{Caves87.3}, \pageref{Caves87.4}, \pageref{Caves88}, \pageref{Caves89}, \pageref{Caves89.1}, \pageref{Caves90}, \pageref{Caves91}, \pageref{Caves92}, \pageref{Caves93}, \pageref{Caves93.1}, \pageref{Caves94}, \pageref{Caves94.1}, \pageref{Caves94.2}, \pageref{Caves95}, \pageref{Caves95.1}, \pageref{Caves95.2}, \pageref{Caves96}, \pageref{Caves96.0.1}, \pageref{Caves96.0.2}, \pageref{Caves96.0.3}, \pageref{Caves96.0.4}, \pageref{Caves96.1}, \pageref{Caves96.2}, \pageref{Caves96.3}, \pageref{Caves96.4}, \pageref{Caves96.5}, \pageref{Caves97}, \pageref{Caves98}, \pageref{Caves99}, \pageref{Caves100}, \pageref{Caves100.1}, \pageref{Caves100.2}, \pageref{Caves100.3}, \pageref{Caves100.4}, \pageref{Caves100.5}, \pageref{Caves101}, \pageref{Caves101.1}, \pageref{Caves101.2}, \pageref{Caves102}, \pageref{Caves103}, \pageref{Caves104}, \pageref{Caves105}, \pageref{Caves106}, \pageref{Caves107}, \pageref{Caves108} \medskip

\noindent {\bf Gregory J. Chaitin --} \pageref{Chaitin1} \medskip

\noindent {\bf Dimiter G. Chakalov --} \pageref{Chakalov1} \medskip

\noindent {\bf Giulio Chiribella --} \pageref{Chiribella1}, \pageref{Chiribella2}, \pageref{Chiribella3}, \pageref{Chiribella4} \medskip

\noindent {\bf Eric Chisolm --} \pageref{Chisolm1} \medskip

\noindent {\bf Adrian Cho --} \pageref{Cho1}, \pageref{Cho2}, \pageref{Cho3} \medskip

\noindent {\bf Hyung S. Choi --} \pageref{Choi1} \medskip

\noindent {\bf Joy Christian --} \pageref{Christian1}, \pageref{Christian2} \medskip

\noindent {\bf Paul Cilliers --} \pageref{Cilliers1} \medskip

\noindent {\bf J. Ignacio Cirac --} \pageref{Cirac1}, \pageref{Cirac2} \medskip

\noindent {\bf Lisa O. Clark --} \pageref{Clark1} \medskip

\noindent {\bf Richard Cleve --} \pageref{Cleve1}, \pageref{Cleve2}, \pageref{Cleve3} \medskip

\noindent {\bf Rob Clifton --} \pageref{Clifton1} \medskip

\noindent {\bf Bob Coecke --} \pageref{Coecke1}, \pageref{Coecke2} \medskip

\noindent {\bf Oliver Cohen --} \pageref{Cohen1}, \pageref{Cohen2} \medskip

\noindent {\bf Roger Colbeck --} \pageref{Colbeck1} \medskip

\noindent {\bf Gregory L. Comer --} \pageref{Comer1}, \pageref{Comer2}, \pageref{Comer3}, \pageref{Comer4}, \pageref{Comer5}, \pageref{Comer6}, \pageref{Comer7}, \pageref{Comer8}, \pageref{Comer9}, \pageref{Comer10}, \pageref{Comer11}, \pageref{Comer12}, \pageref{Comer13}, \pageref{Comer14}, \pageref{Comer15}, \pageref{Comer16}, \pageref{Comer17}, \pageref{Comer18}, \pageref{Comer19}, \pageref{Comer20}, \pageref{Comer21}, \pageref{Comer22}, \pageref{Comer23}, \pageref{Comer24}, \pageref{Comer25}, \pageref{Comer26}, \pageref{Comer27}, \pageref{Comer28}, \pageref{Comer29}, \pageref{Comer30}, \pageref{Comer31}, \pageref{Comer32}, \pageref{Comer33}, \pageref{Comer34}, \pageref{Comer35}, \pageref{Comer36}, \pageref{Comer37}, \pageref{Comer38}, \pageref{Comer39}, \pageref{Comer40}, \pageref{Comer41}, \pageref{Comer42}, \pageref{Comer43}, \pageref{Comer44}, \pageref{Comer45}, \pageref{Comer46}, \pageref{Comer47}, \pageref{Comer48}, \pageref{Comer49}, \pageref{Comer50}, \pageref{Comer51}, \pageref{Comer52}, \pageref{Comer53}, \pageref{Comer54}, \pageref{Comer55}, \pageref{Comer56}, \pageref{Comer57}, \pageref{Comer58}, \pageref{Comer59}, \pageref{Comer60}, \pageref{Comer61}, \pageref{Comer62}, \pageref{Comer63}, \pageref{Comer64}, \pageref{Comer65}, \pageref{Comer66}, \pageref{Comer67}, \pageref{Comer68}, \pageref{Comer69}, \pageref{Comer70}, \pageref{Comer71}, \pageref{Comer72}, \pageref{Comer73}, \pageref{Comer74}, \pageref{Comer75}, \pageref{Comer76}, \pageref{Comer77}, \pageref{Comer78}, \pageref{Comer79}, \pageref{Comer80}, \pageref{Comer81}, \pageref{Comer82}, \pageref{Comer83}, \pageref{Comer84}, \pageref{Comer85}, \pageref{Comer86}, \pageref{Comer87}, \pageref{Comer88}, \pageref{Comer89}, \pageref{Comer90}, \pageref{Comer91}, \pageref{Comer92}, \pageref{Comer93}, \pageref{Comer94}, \pageref{Comer95}, \pageref{Comer96}, \pageref{Comer97}, \pageref{Comer98}, \pageref{Comer99}, \pageref{Comer101}, \pageref{Comer102}, \pageref{Comer103}, \pageref{Comer104}, \pageref{Comer105}, \pageref{Comer106}, \pageref{Comer107}, \pageref{Comer108}, \pageref{Comer109}, \pageref{Comer110}, \pageref{Comer111}, \pageref{Comer112}, \pageref{Comer113}, \pageref{Comer114}, \pageref{Comer115}, \pageref{Comer116}, \pageref{Comer117}, \pageref{Comer118}, \pageref{Comer119}, \pageref{Comer120}, \pageref{Comer121}, \pageref{Comer122}, \pageref{Comer123}, \pageref{Comer124}, \pageref{Comer125}, \pageref{Comer126}, \pageref{Comer127}, \pageref{Comer128}, \pageref{Comer129}, \pageref{Comer130}, \pageref{Comer131}, \pageref{Comer132}  \medskip

\noindent {\bf Cher Communards --} \pageref{Communards1}, \pageref{Communards2}, \pageref{Communards3}  \medskip

\noindent {\bf John H. Conway --} \pageref{Conway1}, \pageref{Conway1.1}, \pageref{Conway2}, \pageref{Conway3} \medskip

\noindent {\bf Correspondent X --} \pageref{CorrespondentX1} \medskip

\noindent {\bf Correspondent Y --} \pageref{CorrespondentY1} \medskip

\noindent {\bf Rachel Couban --} \pageref{Couban1} \medskip

\noindent {\bf Oscar C. O. Dahlsten --} \pageref{Dahlsten1}, \pageref{Dahlsten2}, \pageref{Dahlsten3} \medskip

\noindent {\bf Hoan Bui Dang --} \pageref{Dang1}, \pageref{Dang2}, \pageref{Dang3}, \pageref{Dang4}, \pageref{Dang5}, \pageref{Dang6}, \pageref{Dang7}, \pageref{Dang8}, \pageref{Dang9}, \pageref{Dang10}, \pageref{Dang10.1}, \pageref{Dang11}, \pageref{Dang12} \medskip

\noindent {\bf Giacomo Mauro D'Ariano --} \pageref{DAriano1}, \pageref{DAriano2}, \pageref{DAriano3}, \pageref{DAriano4}, \pageref{DAriano5}, \pageref{DAriano6}, \pageref{DAriano7}, \pageref{DAriano8}, \pageref{DAriano9} \medskip

\noindent {\bf Shannon Dea --} \pageref{Dea1}, \pageref{Dea2} \medskip

\noindent {\bf Francesco De Martini --} \pageref{DeMartini1}, \pageref{DeMartini2}, \pageref{DeMartini3} \medskip

\noindent {\bf William G. Demopoulos --} \pageref{Demopoulos1}, \pageref{Demopoulos2}, \pageref{Demopoulos3}, \pageref{Demopoulos4}, \pageref{Demopoulos5}, \pageref{Demopoulos6}, \pageref{Demopoulos7}, \pageref{Demopoulos7.1}, \pageref{Demopoulos8}, \pageref{Demopoulos9}, \pageref{Demopoulos10}, \pageref{Demopoulos11}, \pageref{Demopoulos12}, \pageref{Demopoulos13}, \pageref{Demopoulos14}, \pageref{Demopoulos15}, \pageref{Demopoulos16}, \pageref{Demopoulos17}, \pageref{Demopoulos18}, \pageref{Demopoulos19}, \pageref{Demopoulos20}, \pageref{Demopoulos21}, \pageref{Demopoulos22}, \pageref{Demopoulos23}, \pageref{Demopoulos24}, \pageref{Demopoulos25}, \pageref{Demopoulos26}, \pageref{Demopoulos27}, \pageref{Demopoulos28}, \pageref{Demopoulos29}, \pageref{Demopoulos30}, \pageref{Demopoulos31}, \pageref{Demopoulos32}, \pageref{Demopoulos33}, \pageref{Demopoulos34}, \pageref{Demopoulos35}, \pageref{Demopoulos36}, \pageref{Demopoulos37}, \pageref{Demopoulos38}, \pageref{Demopoulos39}, \pageref{Demopoulos40}, \pageref{Demopoulos41}, \pageref{Demopoulos42}, \pageref{Demopoulos43}, \pageref{Demopoulos44}, \pageref{Demopoulos45}, \pageref{Demopoulos46}, \pageref{Demopoulos47}, \pageref{Demopoulos48}, \pageref{Demopoulos49} \medskip

\noindent {\bf David Deutsch --} \pageref{DeutschD1} \medskip

\noindent {\bf Ivan H. Deutsch --} \pageref{DeutschI1}, \pageref{DeutschI1.1}, \pageref{DeutschI2} \medskip

\noindent {\bf Persi Diaconis --} \pageref{Diaconis1} \medskip

\noindent {\bf Sara Diamond --} \pageref{Diamond1} \medskip

\noindent {\bf Michael Dickson --} \pageref{Dickson1} \medskip

\noindent {\bf David P. DiVincenzo --} \pageref{DiVincenzo1}, \pageref{DiVincenzo2}, \pageref{DiVincenzo3}, \pageref{DiVincenzo4} \medskip

\noindent {\bf Matthew J. Donald --} \pageref{Donald1}, \pageref{Donald2}, \pageref{Donald3}, \pageref{Donald4} \medskip

\noindent {\bf Bill Dreiss --} \pageref{Dreiss1}, \pageref{Dreiss2} \medskip

\noindent {\bf Ian Duck --} \pageref{Duck1} \medskip

\noindent {\bf Ken R. Duffy --} \pageref{Duffy1}, \pageref{Duffy2}, \pageref{Duffy3}, \pageref{Duffy4}, \pageref{Duffy5} \medskip

\noindent {\bf Todd Duncan --} \pageref{Duncan1}, \pageref{Duncan2}, \pageref{Duncan3}, \pageref{Duncan4}, \pageref{Duncan5}, \pageref{Duncan6}, \pageref{Duncan7}, \pageref{Duncan8} \medskip

\noindent {\bf Ian T. Durham --} \pageref{Durham1}, \pageref{Durham2}, \pageref{Durham3}, \pageref{Durham4} \medskip

\noindent {\bf Matin Durrani --} \pageref{Durrani1} \medskip

\noindent {\bf Rocco Duvenhage --} \pageref{Duvenhage1} \medskip

\noindent {\bf Armond Duwell --} \pageref{Duwell1}, \pageref{Duwell2}, \pageref{Duwell3}, \pageref{Duwell4} \medskip

\noindent {\bf John Earman --} \pageref{Earman1}, \pageref{Earman2} \medskip

\noindent {\bf Joseph H. Eberly --} \pageref{Eberly1}, \pageref{Eberly2} \medskip

\noindent {\bf Artur K. Ekert --} \pageref{Ekert1} \medskip

\noindent {\bf Joseph Emerson --} \pageref{Emerson0}, \pageref{Emerson1}, \pageref{Emerson2}, \pageref{Emerson3}, \pageref{Emerson4}, \pageref{Emerson5}, \pageref{Emerson5.1}, \pageref{Emerson6}, \pageref{Emerson7} \medskip

\noindent {\bf Berthold-Georg ``Berge'' Englert --} \pageref{Englert1} \medskip

\noindent {\bf Steven J. van Enk --} \pageref{vanEnk16.1}, \pageref{vanEnk17}, \pageref{vanEnk18}, \pageref{vanEnk19}, \pageref{vanEnk20}, \pageref{vanEnk21}, \pageref{vanEnk22}, \pageref{vanEnk23}, \pageref{vanEnk24}, \pageref{vanEnk0}, \pageref{vanEnk25}, \pageref{vanEnk26}, \pageref{vanEnk27}, \pageref{vanEnk28}, \pageref{vanEnk29}, \pageref{vanEnk30}, \pageref{vanEnk31}, \pageref{vanEnk1}, \pageref{vanEnk1.1}, \pageref{vanEnk2}, \pageref{vanEnk3}, \pageref{vanEnk4}, \pageref{vanEnk5}, \pageref{vanEnk6}, \pageref{vanEnk7}, \pageref{vanEnk7.1}, \pageref{vanEnk8}, \pageref{vanEnk9}, \pageref{vanEnk10}, \pageref{vanEnk11}, \pageref{vanEnk12}, \pageref{vanEnk13}, \pageref{vanEnk14}, \pageref{vanEnk15}, \pageref{vanEnk16} \medskip

\noindent {\bf Charles P. Enz --} \pageref{Enz1} \medskip

\noindent {\bf {\AA}sa Ericsson --} \pageref{Ericsson1}, \pageref{Ericsson2}, \pageref{Ericsson3}, \pageref{Ericsson4}, \pageref{Ericsson5}, \pageref{Ericsson6}, \pageref{Ericsson7}, \pageref{Ericsson8}, \pageref{Ericsson9}, \pageref{Ericsson10}, \pageref{Ericsson10.1}, \pageref{Ericsson10.2}, \pageref{Ericsson11}, \pageref{Ericsson12}, \pageref{Ericsson13}, \pageref{Ericsson14}, \pageref{Ericsson15}, \pageref{Ericsson15.1}, \pageref{Ericsson16}, \pageref{Ericsson17}, \pageref{Ericsson17.1}, \pageref{Ericsson18} \medskip

\noindent {\bf Cecilia Eriksson --} \pageref{Eriksson1}, \pageref{Eriksson2} \medskip

\noindent {\bf Aaron Fenyes --} \pageref{Fenyes1}, \pageref{Fenyes2}, \pageref{Fenyes3} \medskip

\noindent {\bf Christopher Ferrie --} \pageref{Ferrie0}, \pageref{Ferrie1}, \pageref{Ferrie2}, \pageref{Ferrie3}, \pageref{Ferrie4}, \pageref{Ferrie5}, \pageref{Ferrie6}, \pageref{Ferrie7}, \pageref{Ferrie8}, \pageref{Ferrie9}, \pageref{Ferrie10}, \pageref{Ferrie11}, \pageref{Ferrie12}, \pageref{Ferrie13}, \pageref{Ferrie13.1}, \pageref{Ferrie14}, \pageref{Ferrie15}, \pageref{Ferrie16} \medskip

\noindent {\bf Arthur Fine --} \pageref{Fine1}, \pageref{Fine2}, \pageref{Fine3}, \pageref{Fine4}, \pageref{Fine5}, \pageref{Fine6} \medskip

\noindent {\bf Jerry Finkelstein --} \pageref{Finkelstein1}, \pageref{Finkelstein2}, \pageref{Finkelstein3}, \pageref{Finkelstein4}, \pageref{Finkelstein5}, \pageref{Finkelstein6}, \pageref{Finkelstein7}, \pageref{Finkelstein8}, \pageref{Finkelstein9} \medskip

\noindent {\bf Steven T. Flammia --} \pageref{Flammia0}, \pageref{Flammia1}, \pageref{Flammia2}, \pageref{Flammia3}, \pageref{Flammia4}, \pageref{Flammia5}, \pageref{Flammia6}, \pageref{Flammia7}, \pageref{Flammia8}, \pageref{Flammia9}, \pageref{Flammia10}, \pageref{Flammia11}, \pageref{Flammia11.1}, \pageref{Flammia12} \medskip

\noindent {\bf Tim Folger --} \pageref{Folger1}, \pageref{Folger2} \medskip

\noindent {\bf Henry J. Folse --} \pageref{Folse1}, \pageref{Folse2}, \pageref{Folse3}, \pageref{Folse4}, \pageref{Folse5}, \pageref{Folse6}, \pageref{Folse7}, \pageref{Folse8}, \pageref{Folse9}, \pageref{Folse9.1}, \pageref{Folse10}, \pageref{Folse11}, \pageref{Folse12}, \pageref{Folse13}, \pageref{Folse14}, \pageref{Folse15}, \pageref{Folse16}, \pageref{Folse17}, \pageref{Folse18}, \pageref{Folse19}, \pageref{Folse20}, \pageref{Folse21}, \pageref{Folse22}, \pageref{Folse23} \medskip

\noindent {\bf Kenneth W. Ford --} \pageref{Ford1}, \pageref{Ford2}, \pageref{Ford3}, \pageref{Ford4} \medskip

\noindent {\bf David J. Foulis --} \pageref{Foulis1}, \pageref{Foulis2}  \medskip

\noindent {\bf Bas C. van Fraassen --} \pageref{vanFraassen1}, \pageref{vanFraassen2}, \pageref{vanFraassen3}, \pageref{vanFraassen4}, \pageref{vanFraassen4.1}, \pageref{vanFraassen5}, \pageref{vanFraassen6}, \pageref{vanFraassen7}, \pageref{vanFraassen8}, \pageref{vanFraassen9}, \pageref{vanFraassen10}, \pageref{vanFraassen11}, \pageref{vanFraassen12}, \pageref{vanFraassen13}, \pageref{vanFraassen13.1}, \pageref{vanFraassen13.2}, \pageref{vanFraassen14}, \pageref{vanFraassen15}, \pageref{vanFraassen16}, \pageref{vanFraassen17}, \pageref{vanFraassen18}, \pageref{vanFraassen19}, \pageref{vanFraassen20}, \pageref{vanFraassen21}, \pageref{vanFraassen22} \medskip

\noindent {\bf Cheryl N. Franklin --} \pageref{FranklinCE1} \medskip

\noindent {\bf Doreen Fraser --} \pageref{Fraser1}, \pageref{Fraser2}, \pageref{Fraser3}, \pageref{Fraser4}  \medskip

\noindent {\bf Laurent Freidel --} \pageref{Freidel1}, \pageref{Freidel2}, \pageref{Freidel3}  \medskip

\noindent {\bf Michael Friedman --} \pageref{Friedman1} \medskip

\noindent {\bf Kristen M. ``Kiki'' Fuchs --} in every thought on every page \medskip

\noindent {\bf William M. Fuchs --} \pageref{FuchsW0}, \pageref{FuchsW1} \medskip

\noindent {\bf Akira Furusawa --} \pageref{Furusawa1} \medskip

\noindent {\bf Maria Carla Galavotti --} \pageref{Galavotti1}, \pageref{Galavotti2} \medskip

\noindent {\bf Robert Garisto --} \pageref{Garisto1}, \pageref{Garisto2}, \pageref{Garisto3}, \pageref{Garisto4} \medskip

\noindent {\bf Julio Gea-Banacloche --} \pageref{GeaBanacloche1} \medskip

\noindent {\bf Sevag Gharibian --} \pageref{Gharibian1}, \pageref{Gharibian2} \medskip

\noindent {\bf Hendrik B. Geyer --} \pageref{Geyer1}, \pageref{Geyer2}, \pageref{Geyer3}, \pageref{Geyer4}, \pageref{Geyer5} \medskip

\noindent {\bf Richard D. Gill --} \pageref{Gill1}, \pageref{Gill2}, \pageref{Gill3}, \pageref{Gill4}, \pageref{Gill5}, \pageref{Gill6}, \pageref{Gill7}, \pageref{Gill8} \medskip

\noindent {\bf Florian Girelli --} \pageref{Girelli0}, \pageref{Girelli1}, \pageref{Girelli2} \medskip

\noindent {\bf Nicolas Gisin --} \pageref{Gisin1} \medskip

\noindent {\bf James Gleick --} \pageref{Gleick1}, \pageref{Gleick2}  \medskip

\noindent {\bf Marcelo Gleiser --} \pageref{Gleiser1} \medskip

\noindent {\bf Elizabeth Goheen --} \pageref{Goheen1}, \pageref{Goheen2}, \pageref{Goheen3} \medskip

\noindent {\bf Jaume Gomis --} \pageref{Gomis1} \medskip

\noindent {\bf Daniel Gottesman --} \pageref{Gottesman1}, \pageref{Gottesman2}, \pageref{Gottesman3}, \pageref{Gottesman4}, \pageref{Gottesman5}, \pageref{Gottesman6}, \pageref{Gottesman7}, \pageref{Gottesman8}, \pageref{Gottesman8.1}, \pageref{Gottesman8.2}, \pageref{Gottesman9}, \pageref{Gottesman10}, \pageref{Gottesman11}, \pageref{Gottesman12}, \pageref{Gottesman13} \medskip

\noindent {\bf Kurt Gottfried --} \pageref{Gottfried1} \medskip

\noindent {\bf Philip Goyal --} \pageref{Goyal1}, \pageref{Goyal2}, \pageref{Goyal3}, \pageref{Goyal4}, \pageref{Goyal5} \medskip

\noindent {\bf Christopher E. Granade --} \pageref{Granade1}, \pageref{Granade2} \medskip

\noindent {\bf Walter T. Grandy, Jr.\  --} \pageref{Grandy1}, \pageref{Grandy2}, \pageref{Grandy3} \medskip

\noindent {\bf Philippe Grangier --} \pageref{Grangier1}, \pageref{Grangier2}, \pageref{Grangier3}, \pageref{Grangier4}, \pageref{Grangier5}, \pageref{Grangier6}, \pageref{Grangier7}, \pageref{Grangier8}, \pageref{Grangier9}, \pageref{Grangier10}, \pageref{Grangier11} \medskip

\noindent {\bf Helena Granstr\"om --} \pageref{Granstrom1}, \pageref{Granstrom2} \medskip

\noindent {\bf Matthew A. Graydon --} \pageref{Graydon1}, \pageref{Graydon2}, \pageref{Graydon3}, \pageref{Graydon4}, \pageref{Graydon5}, \pageref{Graydon6}, \pageref{Graydon7}, \pageref{Graydon8}, \pageref{Graydon9}, \pageref{Graydon10}, \pageref{Graydon11}, \pageref{Graydon12}, \pageref{Graydon13}, \pageref{Graydon14}, \pageref{Graydon15}, \pageref{Graydon16}, \pageref{Graydon16.1}, \pageref{Graydon17}, \pageref{Graydon18}, \pageref{Graydon19}, \pageref{Graydon20}, \pageref{Graydon21} \medskip

\noindent {\bf Hilary Greaves --} \pageref{Greaves1} \medskip

\noindent {\bf Richard J. Greechie --} \pageref{Greechie1} \medskip

\noindent {\bf Robert B. Griffiths --}  \pageref{Griffiths1}, \pageref{Griffiths2} \medskip

\noindent {\bf Alexei Grinbaum --} \pageref{Grinbaum1}, \pageref{Grinbaum2} \medskip

\noindent {\bf Daniel M. Greenberger --} \pageref{Greenberger1}, \pageref{Greenberger2}, \pageref{Greenberger3}, \medskip \pageref{Greenberger4}

\noindent {\bf Lov K. Grover --} \pageref{Grover1}, \pageref{Grover2}, \pageref{Grover3}, \pageref{Grover4}, \pageref{Grover5}, \medskip \pageref{Grover6}

\noindent {\bf John R. Gustafson --} \pageref{Gustafson1}, \pageref{Gustafson2}, \pageref{Gustafson3} \medskip

\noindent {\bf Gus Gutoski --} \pageref{Gutoski1} \medskip

\noindent {\bf Ian Hacking --} \pageref{Hacking1} \medskip

\noindent {\bf Nicolas Hadjisavvas --} \pageref{Hadjisavvas1} \medskip

\noindent {\bf Amit Hagar --} \pageref{Hagar1} \medskip

\noindent {\bf Christopher P. Hains --} \pageref{Hains1} \medskip

\noindent {\bf Alan H\'ajek --} \pageref{Hajek1} \medskip

\noindent {\bf Hans Halvorson  --} \pageref{Halvorson1}, \pageref{Halvorson2}, \pageref{Halvorson3}, \pageref{Halvorson4}, \pageref{Halvorson5}, \pageref{Halvorson6}, \pageref{Halvorson7}, \pageref{Halvorson8}, \pageref{Halvorson9}, \pageref{Halvorson9.05}, \pageref{Halvorson9.1}, \pageref{Halvorson10}, \pageref{Halvorson11}, \pageref{Halvorson12}, \pageref{Halvorson13}, \pageref{Halvorson14}, \pageref{Halvorson15}, \pageref{Halvorson16}, \pageref{Halvorson17} \medskip

\noindent {\bf Alioscia Hamma --} \pageref{Hamma1}, \pageref{Hamma2}, \pageref{Hamma3} \medskip

\noindent {\bf Lucien Hardy --} \pageref{Hardy1}, \pageref{Hardy2}, \pageref{Hardy3}, \pageref{Hardy4}, \pageref{Hardy5}, \pageref{Hardy6}, \pageref{Hardy7}, \pageref{Hardy8}, \pageref{Hardy9}, \pageref{Hardy10}, \pageref{Hardy11}, \pageref{Hardy12}, \pageref{Hardy13}, \pageref{Hardy14}, \pageref{Hardy15}, \pageref{Hardy16}, \pageref{Hardy17}, \pageref{Hardy18}, \pageref{Hardy19}, \pageref{Hardy20}, \pageref{Hardy21}, \pageref{Hardy22}, \pageref{Hardy22.1}, \pageref{Hardy23}, \pageref{Hardy23.5}, \pageref{Hardy24}, \pageref{Hardy25}, \pageref{Hardy26}, \pageref{Hardy27}, \pageref{Hardy28}, \pageref{Hardy29}, \pageref{Hardy30}, \pageref{Hardy31}, \pageref{Hardy32}, \pageref{Hardy33}, \pageref{Hardy34}, \pageref{Hardy35}, \pageref{Hardy36}, \pageref{Hardy37}, \pageref{Hardy38}, \pageref{Hardy39}, \pageref{Hardy40}, \pageref{Hardy41}, \pageref{Hardy42}, \pageref{Hardy42.1}, \pageref{Hardy43}, \pageref{Hardy44}  \medskip

\noindent {\bf Vanessa Hardy --} \pageref{HardyV1}, \pageref{HardyV2} \medskip

\noindent {\bf William L. Harper --} \pageref{Harper1}, \pageref{Harper2}, \pageref{Harper3} \medskip

\noindent {\bf Peter Harremo\"es --} \pageref{Harremoes1}, \pageref{Harremoes2}, \pageref{Harremoes3}, \pageref{Harremoes4} \medskip

\noindent {\bf Aram W. Harrow --} \pageref{Harrow1}, \pageref{Harrow2}, \pageref{Harrow3}, \pageref{Harrow4} \medskip

\noindent {\bf David J. Harris --} \pageref{Harris1} \medskip

\noindent {\bf James B. Hartle --} \pageref{Hartle1}, \pageref{Hartle2}, \pageref{Hartle3} \medskip

\noindent {\bf Stephan Hartmann --} \pageref{Hartmann1}, \pageref{Hartmann2}, \pageref{Hartmann3}, \pageref{Hartmann4}, \pageref{Hartmann5}, \pageref{Hartmann6}, \pageref{Hartmann7}, \pageref{Hartmann8}, \pageref{Hartmann9}, \pageref{Hartmann10}, \pageref{Hartmann11}, \pageref{Hartmann12}, \pageref{Hartmann13}, \pageref{Hartmann14}, \pageref{Hartmann15}, \pageref{Hartmann16} \medskip

\noindent {\bf Patrick Hayden --} \pageref{Hayden1}, \pageref{Hayden2}, \pageref{Hayden3}, \pageref{Hayden4} \medskip

\noindent {\bf Richard Healey --} \pageref{Healey1}, \pageref{Healey2}, \pageref{Healey3}, \pageref{Healey4}, \pageref{Healey5}, \pageref{Healey6}, \pageref{Healey7}, \pageref{Healey8}  \medskip

\noindent {\bf Geoffrey Hellman --} \pageref{Hellman1}  \medskip

\noindent {\bf Meir Hemmo  --} \pageref{Hemmo0}, \pageref{Hemmo1}, \pageref{Hemmo2}, \pageref{Hemmo4}  \medskip

\noindent {\bf Leah Henderson --} \pageref{Henderson1}, \pageref{Henderson2}, \pageref{Henderson3}  \medskip

\noindent {\bf Scott Henry --} \pageref{SHenry1} \medskip

\noindent {\bf Gary Herling --} \pageref{Herling1}, \pageref{Herling2}, \pageref{Herling3} \medskip

\noindent {\bf D. Micah Hester --} \pageref{Hester1}, \pageref{Hester2} \medskip

\noindent {\bf Carl Hewitt --} \pageref{Hewitt1} \medskip

\noindent {\bf Basil J. Hiley --} \pageref{Hiley1} \medskip

\noindent {\bf Ian N. Hincks --} \pageref{Hincks1} \medskip

\noindent {\bf Osamu Hirota --} \pageref{Hirota1}, \pageref{Hirota2}, \pageref{Hirota3}, \pageref{Hirota4}, \pageref{Hirota5}, \pageref{Hirota6}, \pageref{Hirota7}, \pageref{Hirota8} \medskip

\noindent {\bf Christopher R. Hitchcock --} \pageref{Hitchcock1} \medskip

\noindent {\bf Alexander S. Holevo --} \pageref{Holevo1}, \pageref{Holevo2}, \pageref{Holevo3}, \pageref{Holevo4}, \pageref{Holevo5}, \pageref{Holevo6}, \pageref{Holevo7}, \pageref{Holevo8}, \pageref{Holevo9}, \pageref{Holevo10}, \pageref{Holevo10.1}, \pageref{Holevo11} \medskip

\noindent {\bf Thomas Homer-Dixon --} \pageref{HomerDixon1} \medskip

\noindent {\bf John Honner --} \pageref{Honner1}, \pageref{Honner2} \medskip

\noindent {\bf Sabine Hossenfelder --} \pageref{Hossenfelder1} \medskip

\noindent {\bf Don Howard --} \pageref{Howard1}, \pageref{Howard2} \medskip

\noindent {\bf Lane P. Hughston --} \pageref{Hughston1} \medskip

\noindent {\bf Jenann Ismael --} \pageref{Ismael0}, \pageref{Ismael1}, \pageref{Ismael2} \medskip

\noindent {\bf Kurt Jacobs --} \pageref{Jacobs1}, \pageref{Jacobs2}, \pageref{Jacobs3} \medskip

\noindent {\bf Ted Jacobson --} \pageref{Jacobson1}, \pageref{Jacobson2} \medskip

\noindent {\bf Andrew H. Jaffe --} \pageref{Jaffe1}, \pageref{Jaffe2} \medskip

\noindent {\bf Kannan ``Jagu'' Jagannathan --} \pageref{Jagannathan1}, \pageref{Jagannathan2} \medskip

\noindent {\bf Michel Janssen --} \pageref{Janssen1}, \pageref{Janssen2}, \pageref{Janssen3}, \pageref{Janssen4}, \pageref{Janssen5} \medskip

\noindent {\bf Zhengfeng Ji --} \pageref{Ji1} \medskip

\noindent {\bf Nick S. Jones --} \pageref{Jones1} \medskip

\noindent {\bf Richard Jozsa --} \pageref{Jozsa1}, \pageref{Jozsa2}, \pageref{Jozsa3}, \pageref{Jozsa4}, \pageref{Jozsa5}, \pageref{Jozsa6}, \pageref{Jozsa7}, \pageref{Jozsa8}, \pageref{Jozsa9}, \pageref{Jozsa10}, \pageref{Jozsa11} \medskip

\noindent {\bf Vladislav Kargin --} \pageref{Kargin1} \medskip

\noindent {\bf Antti Karlsson --} \pageref{Karlsson1}, \pageref{Karlsson2} \medskip

\noindent {\bf Adrian Kent --} \pageref{Kent1}, \pageref{Kent2}, \pageref{Kent3}, \pageref{Kent4}, \pageref{Kent5}, \pageref{Kent6}, \pageref{Kent7}, \pageref{Kent8}, \pageref{Kent9}, \pageref{Kent10}, \pageref{Kent11}, \pageref{Kent12}, \pageref{Kent13}, \pageref{Kent14}, \pageref{Kent15}, \pageref{Kent16}, \pageref{Kent17}, \pageref{Kent18}, \pageref{Kent19}, \pageref{Kent20}, \pageref{Kent21}, \pageref{Kent22}, \pageref{Kent23}, \pageref{Kent24}, \pageref{Kent25} \medskip

\noindent {\bf Larry Ketchersid --} \pageref{Ketchersid1} \medskip

\noindent {\bf Andrei Y. Khrennikov --} \pageref{Khrennikov1}, \pageref{Khrennikov2}, \pageref{Khrennikov3}, \pageref{Khrennikov4}, \pageref{Khrennikov5}, \pageref{Khrennikov6}, \pageref{Khrennikov7}, \pageref{Khrennikov8}, \pageref{Khrennikov9}, \pageref{Khrennikov10}, \pageref{Khrennikov11}, \pageref{Khrennikov12}, \pageref{Khrennikov13}, \pageref{Khrennikov14}, \pageref{Khrennikov15}, \pageref{Khrennikov16}, \pageref{Khrennikov17}, \pageref{Khrennikov18}, \pageref{Khrennikov19}, \pageref{Khrennikov20}, \pageref{Khrennikov21}, \pageref{Khrennikov22}, \pageref{Khrennikov23}, \pageref{Khrennikov24}, \pageref{Khrennikov25}, \pageref{Khrennikov26} \medskip

\noindent {\bf H. Jeffrey Kimble --} \pageref{Kimble1}, \pageref{Kimble2}, \pageref{Kimble3}, \pageref{Kimble4}, \pageref{Kimble5} \medskip

\noindent {\bf Christopher King --} \pageref{King1}, \pageref{King2}, \pageref{King3}, \pageref{King4}, \pageref{King5}, \pageref{King6}, \pageref{King7}, \pageref{King8}, \pageref{King9} \medskip

\noindent {\bf Marguarite Knechtel --} \pageref{Knechtel1} \medskip

\noindent {\bf Kevin H. Knuth --} \pageref{Knuth1}, \pageref{Knuth2}, \pageref{Knuth3}, \pageref{Knuth4}, \pageref{Knuth5}, \pageref{Knuth6}, \pageref{Knuth7} \medskip

\noindent {\bf Dmitry Kobak --} \pageref{Kobak1}, \pageref{Kobak2}, \pageref{Kobak3} \medskip

\noindent {\bf Simon Kochen --} \pageref{Kochen0}, \pageref{Kochen1}, \pageref{Kochen2} \medskip

\noindent {\bf Hermann K\"onig --} \pageref{Koenig1} \medskip

\noindent {\bf Fred Kuttner --} \pageref{Kuttner1}, \pageref{Kuttner2} \medskip

\noindent {\bf Cathryn E. Laake --} \pageref{Laake1} \medskip

\noindent {\bf Brian R. La Cour --} \pageref{LaCour1}, \pageref{LaCour2}, \pageref{LaCour3} \medskip

\noindent {\bf Raymond Laflamme --} \pageref{Laflamme1}, \pageref{Laflamme2}, \pageref{Laflamme3}, \pageref{Laflamme4}, \pageref{Laflamme5}, \pageref{Laflamme6}, \pageref{Laflamme7}, \pageref{Laflamme8}, \pageref{Laflamme9} \medskip

\noindent {\bf Pekka J. Lahti --} \pageref{Lahti1} \medskip

\noindent {\bf David C. Lamberth --} \pageref{Lamberth1}, \pageref{Lamberth2}, \pageref{Lamberth3}, \pageref{Lamberth4}, \pageref{Lamberth5}, \pageref{Lamberth6} \medskip

\noindent {\bf Andrew J. Landahl --} \pageref{Landahl1}, \pageref{Landahl2}, \pageref{Landahl3}, \pageref{Landahl4}, \pageref{Landahl5} \medskip

\noindent {\bf Henry J. Landau --} \pageref{Landau1} \medskip

\noindent {\bf Jan-{\AA}ke Larsson --} \pageref{Larsson1}, \pageref{Larsson2}, \pageref{Larsson3}, \pageref{Larsson4}, \pageref{Larsson5}, \pageref{Larsson5.1}, \pageref{Larsson6}, \pageref{Larsson7}, \pageref{Larsson8}, \pageref{Larsson9}, \pageref{Larsson10} \medskip

\noindent {\bf Walter E. ``Jay'' Lawrence --} \pageref{Lawrence1}, \pageref{Lawrence2}, \pageref{Lawrence3}, \pageref{Lawrence4}, \pageref{Lawrence4.1}, \pageref{Lawrence5}, \pageref{Lawrence6} \medskip

\noindent {\bf Sir Anthony J. Leggett --} \pageref{Leggett1}, \pageref{Leggett2} \medskip

\noindent {\bf Matthew S. Leifer --} \pageref{Leifer1}, \pageref{Leifer2}, \pageref{Leifer3}, \pageref{Leifer4}, \pageref{Leifer4.1}, \pageref{Leifer5}, \pageref{Leifer6}, \pageref{Leifer7}, \pageref{Leifer8}, \pageref{Leifer9}, \pageref{Leifer9.1}, \pageref{Leifer10}, \pageref{Leifer11}, \pageref{Leifer12}, \pageref{Leifer13}, \pageref{Leifer14}, \pageref{Leifer15}  \medskip

\noindent {\bf J. Brad Lentz --} \pageref{LentzB1}, \pageref{LentzB2}, \pageref{LentzB3}, \pageref{LentzB4}, \pageref{LentzB5}, \pageref{LentzB6}, \pageref{LentzB6.1}, \pageref{LentzB7}, \pageref{LentzB8}, \pageref{LentzB9}, \pageref{LentzB10}, \pageref{LentzB11} \medskip

\noindent {\bf Susie J. Lentz --} \pageref{LentzS1}, \pageref{LentzS2}, \pageref{LentzS3}, \pageref{LentzS4}, \pageref{LentzS4.1}, \pageref{LentzS5}, \pageref{LentzS6}, \pageref{LentzS7}, \pageref{LentzS8}  \medskip

\noindent {\bf Debbie W. Leung --} \pageref{Leung1} \medskip

\noindent {\bf Elliott H. Lieb --} \pageref{Lieb1}, \pageref{Lieb2}, \pageref{Lieb3} \medskip

\noindent {\bf Etera R. Livine --} \pageref{Livine1} \medskip

\noindent {\bf Seth Lloyd --} \pageref{Lloyd1}, \pageref{Lloyd2} \medskip

\noindent {\bf Norbert L\"utkenhaus --} \pageref{Luetkenhaus1}, \pageref{Luetkenhaus2} \medskip

\noindent {\bf Dan Lynch --} \pageref{Lynch1} \medskip

\noindent {\bf Hideo Mabuchi --} \pageref{Mabuchi0}, \pageref{Mabuchi1}, \pageref{Mabuchi2}, \pageref{Mabuchi3}, \pageref{Mabuchi4}, \pageref{Mabuchi5}, \pageref{Mabuchi6}, \pageref{Mabuchi7}, \pageref{Mabuchi8}, \pageref{Mabuchi9}, \pageref{Mabuchi10}, \pageref{Mabuchi11}, \pageref{Mabuchi12}, \pageref{Mabuchi12.1}, \pageref{Mabuchi13}, \pageref{Mabuchi13.1}, \medskip \pageref{Mabuchi14}

\noindent {\bf James D. Malley --} \pageref{Malley1} \medskip

\noindent {\bf Piero G. Luca Mana --} \pageref{Mana1}, \pageref{Mana2}, \pageref{Mana3}, \pageref{Mana4}, \pageref{Mana5}, \pageref{Mana6}, \pageref{Mana7}, \pageref{Mana8}, \pageref{Mana9}, \pageref{Mana10}, \pageref{Mana11}, \pageref{Mana12}, \pageref{Mana13}, \pageref{Mana14}, \pageref{Mana15}, \pageref{Mana16}, \pageref{Mana17}, \pageref{Mana18}, \pageref{Mana19} \medskip

\noindent {\bf Kiran K. Manne --} \pageref{Manne1} \medskip

\noindent {\bf Owen J. E. Maroney --} \pageref{Maroney1}, \pageref{Maroney2}, \pageref{Maroney3}, \pageref{Maroney4}, \pageref{Maroney5}, \pageref{Maroney6} \medskip

\noindent {\bf Keye Martin --} \pageref{Martin1}, \pageref{Martin2}, \pageref{Martin3}, \pageref{Martin4}, \pageref{Martin5}, \pageref{Martin6}, \pageref{Martin7}, \pageref{Martin8}, \pageref{Martin9}, \pageref{Martin10}, \pageref{Martin11}, \pageref{Martin12} \medskip

\noindent {\bf Tim Maudlin --} \pageref{Maudlin1} \medskip

\noindent {\bf James E. Mazo --} \pageref{Mazo1} \medskip

\noindent {\bf Kirk T. McDonald --} \pageref{McDonald1}, \pageref{McDonald2}, \pageref{McDonald2.1}, \pageref{McDonald3}, \pageref{McDonald4}, \pageref{McDonald5}, \pageref{McDonald6}, \pageref{McDonald8} \medskip

\noindent {\bf Zachari E. D. Medendorp --} \pageref{Medendorp1}, \pageref{Medendorp2} \medskip

\noindent {\bf Nicolas C. Menicucci --} \pageref{Menicucci1}, \pageref{Menicucci2}, \pageref{Menicucci3}, \pageref{Menicucci4}, \pageref{Menicucci5} \medskip

\noindent {\bf N. David Mermin --} \pageref{Mermin-3}, \pageref{Mermin-2}, \pageref{Mermin-1}, \pageref{Mermin0}, \pageref{Mermin1}, \pageref{Mermin2}, \pageref{Mermin2.1}, \pageref{Mermin2.2}, \pageref{Mermin3}, \pageref{Mermin4}, \pageref{Mermin5}, \pageref{Mermin6}, \pageref{Mermin7}, \pageref{Mermin8}, \pageref{Mermin9}, \pageref{Mermin10}, \pageref{Mermin11}, \pageref{Mermin12}, \pageref{Mermin13}, \pageref{Mermin14}, \pageref{Mermin15}, \pageref{Mermin16}, \pageref{Mermin17}, \pageref{Mermin18}, \pageref{Mermin19}, \pageref{Mermin20}, \pageref{Mermin21}, \pageref{Mermin22}, \pageref{Mermin23}, \pageref{Mermin24}, \pageref{Mermin25}, \pageref{Mermin25.1}, \pageref{Mermin26}, \pageref{Mermin27}, \pageref{Mermin28}, \pageref{Mermin29}, \pageref{Mermin30}, \pageref{Mermin31}, \pageref{Mermin32}, \pageref{Mermin33}, \pageref{Mermin34}, \pageref{Mermin35}, \pageref{Mermin36}, \pageref{Mermin37}, \pageref{Mermin38}, \pageref{Mermin39}, \pageref{Mermin40}, \pageref{Mermin41}, \pageref{Mermin42}, \pageref{Mermin43}, \pageref{Mermin44}, \pageref{Mermin45}, \pageref{Mermin46}, \pageref{Mermin47}, \pageref{Mermin48}, \pageref{Mermin49}, \pageref{Mermin50}, \pageref{Mermin51}, \pageref{Mermin52}, \pageref{Mermin53}, \pageref{Mermin54}, \pageref{Mermin55}, \pageref{Mermin56}, \pageref{Mermin57}, \pageref{Mermin58}, \pageref{Mermin59}, \pageref{Mermin60}, \pageref{Mermin61}, \pageref{Mermin62}, \pageref{Mermin62.1}, \pageref{Mermin62.2}, \pageref{Mermin62.3}, \pageref{Mermin62.4}, \pageref{Mermin63}, \pageref{Mermin64}, \pageref{Mermin65}, \pageref{Mermin65.05}, \pageref{Mermin65.06}, \pageref{Mermin65.1}, \pageref{Mermin65.2}, \pageref{Mermin65.3}, \pageref{Mermin66}, \pageref{Mermin67}, \pageref{Mermin68}, \pageref{Mermin69}, \pageref{Mermin70}, \pageref{Mermin71}, \pageref{Mermin72}, \pageref{Mermin73}, \pageref{Mermin74}, \pageref{Mermin75}, \pageref{Mermin76}, \pageref{Mermin77}, \pageref{Mermin78}, \pageref{Mermin79}, \pageref{Mermin80}, \pageref{Mermin81}, \pageref{Mermin82}, \pageref{Mermin83}, \pageref{Mermin84}, \pageref{Mermin85}, \pageref{Mermin86}, \pageref{Mermin87}, \pageref{Mermin88}, \pageref{Mermin89}, \pageref{Mermin90}, \pageref{Mermin91}, \pageref{Mermin92}, \pageref{Mermin93}, \pageref{Mermin94}, \pageref{Mermin95}, \pageref{Mermin96}, \pageref{Mermin97}, \pageref{Mermin98}, \pageref{Mermin99}, \pageref{Mermin100}, \pageref{Mermin101}, \pageref{Mermin102}, \pageref{Mermin103}, \pageref{Mermin104}, \pageref{Mermin105}, \pageref{Mermin106}, \pageref{Mermin107}, \pageref{Mermin108}, \pageref{Mermin109}, \pageref{Mermin110}, \pageref{Mermin111}, \pageref{Mermin112}, \pageref{Mermin113}, \pageref{Mermin114}, \pageref{Mermin115}, \pageref{Mermin116}, \pageref{Mermin117}, \pageref{Mermin118}, \pageref{Mermin119}, \pageref{Mermin120}, \pageref{Mermin121}, \pageref{Mermin122}, \pageref{Mermin123}, \pageref{Mermin124}, \pageref{Mermin125}, \pageref{Mermin126}, \pageref{Mermin127}, \pageref{Mermin128}, \pageref{Mermin129}, \pageref{Mermin130}, \pageref{Mermin131}, \pageref{Mermin132}, \pageref{Mermin133}, \pageref{Mermin134}, \pageref{Mermin135}, \pageref{Mermin136}, \pageref{Mermin137}, \pageref{Mermin138}, \pageref{Mermin139}, \pageref{Mermin140}, \pageref{Mermin141}, \pageref{Mermin142}, \pageref{Mermin143}, \pageref{Mermin144}, \pageref{Mermin145}, \pageref{Mermin146}, \pageref{Mermin147}, \pageref{Mermin148}, \pageref{Mermin149}, \pageref{Mermin150}, \pageref{Mermin151}, \pageref{Mermin152}, \pageref{Mermin153}, \pageref{Mermin154}, \pageref{Mermin155}, \pageref{Mermin156}, \pageref{Mermin157}, \pageref{Mermin158}, \pageref{Mermin159}, \pageref{Mermin160}, \pageref{Mermin161}, \pageref{Mermin162}, \pageref{Mermin162.1}, \pageref{Mermin163}, \pageref{Mermin164}, \pageref{Mermin165}, \pageref{Mermin166}, \pageref{Mermin167}, \pageref{Mermin168}, \pageref{Mermin169}, \pageref{Mermin170}, \pageref{Mermin170.1}, \pageref{Mermin171}, \pageref{Mermin172}, \pageref{Mermin173}, \pageref{Mermin174}, \pageref{Mermin175}, \pageref{Mermin176}, \pageref{Mermin177}, \pageref{Mermin178}, \pageref{Mermin179}, \pageref{Mermin180}, \pageref{Mermin181}, \pageref{Mermin182}, \pageref{Mermin183}, \pageref{Mermin184}, \pageref{Mermin185}, \pageref{Mermin186}, \pageref{Mermin187}, \pageref{Mermin188}, \pageref{Mermin189}, \pageref{Mermin190}, \pageref{Mermin191}, \pageref{Mermin192}, \pageref{Mermin193},
\pageref{Mermin194}, \pageref{Mermin195}, \pageref{Mermin196}, \pageref{Mermin197}, \pageref{Mermin198}, \pageref{Mermin199}, \pageref{Mermin200}  \medskip

\noindent {\bf Eugen Merzbacher --} \pageref{Merzbacher1}, \pageref{Merzbacher2}, \pageref{Merzbacher3} \medskip

\noindent {\bf Gerard J. Milburn --} \pageref{Milburn1}, \pageref{Milburn2}, \pageref{Milburn3}, \pageref{Milburn4}, \pageref{Milburn5}, \pageref{Milburn6} \medskip

\noindent {\bf Cheryl Misak --} \pageref{Misak1}, \pageref{Misak2}, \pageref{Misak3}, \pageref{Misak4}, \pageref{Misak5}, \pageref{Misak6}, \pageref{Misak7} \medskip

\noindent {\bf John W. Moffat --} \pageref{Moffat1} \medskip

\noindent {\bf Ulrich Mohrhoff --} \pageref{Mohrhoff1}, \pageref{Mohrhoff2}, \pageref{Mohrhoff3}, \pageref{Mohrhoff4}, \pageref{Mohrhoff5}, \pageref{Mohrhoff6}, \pageref{Mohrhoff7} \medskip

\noindent {\bf Klaus M{\o}lmer --} \pageref{Moelmer1}, \pageref{Moelmer2} \medskip

\noindent {\bf Tal Mor --} \pageref{TalMor1} \medskip

\noindent {\bf Michele Mosca --} \pageref{Mosca1}, \pageref{Mosca2} \medskip

\noindent {\bf Wayne C. Myrvold --} \pageref{Myrvold1}, \pageref{Myrvold2}, \pageref{Myrvold3}, \pageref{Myrvold4}, \pageref{Myrvold5}, \pageref{Myrvold6}, \pageref{Myrvold7}, \pageref{Myrvold8}, \pageref{Myrvold9}, \pageref{Myrvold10}, \pageref{Myrvold10.1}, \pageref{Myrvold11}, \pageref{Myrvold12}, \pageref{Myrvold13}, \pageref{Myrvold14}\medskip

\noindent {\bf George Musser --} \pageref{Musser1}, \pageref{Musser2}, \pageref{Musser3}, \pageref{Musser4}, \pageref{Musser5}, \pageref{Musser6}, \pageref{Musser7}, \pageref{Musser8}, \pageref{Musser9}, \pageref{Musser10}, \pageref{Musser11}, \pageref{Musser12}, \pageref{Musser13}, \pageref{Musser14}, \pageref{Musser15}, \pageref{Musser16}, \pageref{Musser17}, \pageref{Musser18}, \pageref{Musser19}, \pageref{Musser20}, \pageref{Musser21}, \pageref{Musser22} \medskip

\noindent {\bf William T. Myers --} \pageref{Myers1}, \pageref{Myers2} \medskip

\noindent {\bf William R. Newman --} \pageref{Newman1} \medskip

\noindent {\bf Alyssa Ney --} \pageref{Ney1}, \pageref{Ney2}, \pageref{Ney3}, \pageref{Ney4} \medskip

\noindent {\bf Jeffrey W. Nicholson --} \pageref{Nicholson1}, \pageref{Nicholson2}, \pageref{Nicholson3}, \pageref{Nicholson4}, \pageref{Nicholson5}, \pageref{Nicholson6}, \pageref{Nicholson7}, \pageref{Nicholson8}, \pageref{Nicholson9}, \pageref{Nicholson10}, \pageref{Nicholson11}, \pageref{Nicholson12}, \pageref{Nicholson13}, \pageref{Nicholson14}, \pageref{Nicholson15}, \pageref{Nicholson16}, \pageref{Nicholson17}, \pageref{Nicholson18}, \pageref{Nicholson19}, \pageref{Nicholson20}, \pageref{Nicholson21}, \pageref{Nicholson22}, \pageref{Nicholson23}, \pageref{Nicholson24}, \pageref{Nicholson25}, \pageref{Nicholson26}, \pageref{Nicholson27}, \pageref{Nicholson28}, \pageref{Nicholson29}, \pageref{Nicholson30}, \pageref{Nicholson31} \medskip

\noindent {\bf Michael A. Nielsen --} \pageref{Nielsen1}, \pageref{Nielsen2}, \pageref{Nielsen3}, \pageref{Nielsen4}, \pageref{Nielsen5}, \pageref{Nielsen6}, \pageref{Nielsen7}, \pageref{Nielsen8}, \pageref{Nielsen9} \medskip

\noindent {\bf Theo M. Nieuwenhuizen --} \pageref{Nieuwenhuizen1} \medskip

\noindent {\bf Yee Jack Ng --} \pageref{Ng1}, \pageref{Ng2}, \pageref{Ng3}, \pageref{Ng4}, \pageref{Ng5}, \pageref{Ng6}, \pageref{Ng7} \medskip

\noindent {\bf Travis Norsen --} \pageref{Norsen0}, \pageref{Norsen1}, \pageref{Norsen2}, \pageref{Norsen3}, \pageref{Norsen4}, \pageref{Norsen5}, \pageref{Norsen6} \medskip

\noindent {\bf John D. Norton --} \pageref{Norton1}, \pageref{Norton2}, \pageref{Norton3}  \medskip

\noindent {\bf Melvin E. L. Oakes --} \pageref{Oakes1} \medskip

\noindent {\bf Rachel Obajtek --} \pageref{Obajtek1} \medskip

\noindent {\bf Izumi Ojima --} \pageref{Ojima1}, \pageref{Ojima2}, \pageref{Ojima3}  \medskip

\noindent {\bf Yasser Omar --} \pageref{Omar1}  \medskip

\noindent {\bf Roland Omn\`es --} \pageref{Omnes1}  \medskip

\noindent {\bf Dennis Overbye --} \pageref{Overbye1}, \pageref{Overbye2}, \pageref{Overbye3}, \pageref{Overbye4}, \pageref{Overbye5}, \pageref{Overbye6}, \pageref{Overbye7} \medskip

\noindent {\bf Constantine Pagonis  --} \pageref{Pagonis1} \medskip

\noindent {\bf Veiko Palge --} \pageref{Palge0}, \pageref{Palge1}, \pageref{Palge2}, \pageref{Palge3}, \pageref{Palge4} \medskip

\noindent {\bf Matteo G. A. Paris  --} \pageref{Paris1} \medskip

\noindent {\bf Andrew J. R. Parker  --} \pageref{Parker1} \medskip

\noindent {\bf Edward R. Patte  --} \pageref{Patte1} \medskip

\noindent {\bf Kent A. Peacock --} \pageref{Peacock1}, \pageref{Peacock2} \medskip

\noindent {\bf Philip Pearle --} \pageref{Pearle1}, \pageref{Pearle2}, \pageref{Pearle3} \medskip

\noindent {\bf Sir Roger Penrose  --} \pageref{Penrose1} \medskip

\noindent {\bf Asher Peres --} \pageref{Peres1}, \pageref{Peres2}, \pageref{Peres3}, \pageref{Peres4}, \pageref{Peres5}, \pageref{Peres6}, \pageref{Peres7}, \pageref{Peres8}, \pageref{Peres9}, \pageref{Peres10}, \pageref{Peres11}, \pageref{Peres12}, \pageref{Peres13}, \pageref{Peres14}, \pageref{Peres15}, \pageref{Peres16}, \pageref{Peres17}, \pageref{Peres18}, \pageref{Peres19}, \pageref{Peres20}, \pageref{Peres21}, \pageref{Peres22}, \pageref{Peres23}, \pageref{Peres24}, \pageref{Peres25}, \pageref{Peres26}, \pageref{Peres27}, \pageref{Peres28}, \pageref{Peres29}, \pageref{Peres30}, \pageref{Peres31}, \pageref{Peres32}, \pageref{Peres33}, \pageref{Peres34}, \pageref{Peres35}, \pageref{Peres36}, \pageref{Peres37}, \pageref{Peres38}, \pageref{Peres39}, \pageref{Peres40}, \pageref{Peres41}, \pageref{Peres42}, \pageref{Peres43}, \pageref{Peres44}, \pageref{Peres45}, \pageref{Peres46}, \pageref{Peres47}, \pageref{Peres48}, \pageref{Peres49}, \pageref{Peres50}, \pageref{Peres51}, \pageref{Peres52}, \pageref{Peres53}, \pageref{Peres54}, \pageref{Peres55}, \pageref{Peres56}, \pageref{Peres57}, \pageref{Peres58}, \pageref{Peres59}, \pageref{Peres60}, \pageref{Peres61}, \pageref{Peres62}, \pageref{Peres63}, \pageref{Peres64}, \pageref{Peres65}, \pageref{Peres66}, \pageref{Peres67}  \medskip

\noindent {\bf Aviva Peres --} \pageref{AvivaPeres} \medskip

\noindent {\bf Lydia Peres-Hari --} \pageref{LydiaPeres} \medskip

\noindent {\bf Marcos P\'erez-Su\'arez --} \pageref{PerezSuarez1}, \pageref{PerezSuarez2}, \pageref{PerezSuarez3}, \pageref{PerezSuarez4}, \pageref{PerezSuarez5}, \pageref{PerezSuarez6}, \pageref{PerezSuarez7}, \pageref{PerezSuarez8}, \pageref{PerezSuarez9}, \pageref{PerezSuarez10}, \pageref{PerezSuarez11}, \pageref{PerezSuarez12}, \pageref{PerezSuarez13}, \pageref{PerezSuarez14}, \pageref{PerezSuarez15}, \pageref{PerezSuarez16}, \pageref{PerezSuarez17}, \pageref{PerezSuarez18}, \pageref{PerezSuarez19}, \pageref{PerezSuarez20}, \pageref{PerezSuarez21}, \pageref{PerezSuarez22}, \pageref{PerezSuarez23}  \medskip

\noindent {\bf D\'enes Petz --} \pageref{Petz1}, \pageref{Petz2}, \pageref{Petz3} \medskip

\noindent {\bf Laura Piispanen --} \pageref{Piispanen1} \medskip

\noindent {\bf Rob Pike --} \pageref{Pike1}, \pageref{Pike2}, \pageref{Pike3}, \pageref{Pike4}, \pageref{Pike5}, \pageref{Pike6}, \pageref{Pike7}, \pageref{Pike8}, \pageref{Pike9}, \pageref{Pike10} \medskip

\noindent {\bf Itamar Pitowsky --} \pageref{Pitowsky1}, \pageref{Pitowsky2}, \pageref{Pitowsky3}, \pageref{Pitowsky4}, \pageref{Pitowsky5} \medskip

\noindent {\bf Rainer Plaga --} \pageref{Plaga1} \medskip

\noindent {\bf Arkady Plotnitsky --} \pageref{Plotnitsky1}, \pageref{Plotnitsky2}, \pageref{Plotnitsky3}, \pageref{Plotnitsky4}, \pageref{Plotnitsky5}, \pageref{Plotnitsky5.1}, \pageref{Plotnitsky5.2}, \pageref{Plotnitsky5.3}, \pageref{Plotnitsky5.4}, \pageref{Plotnitsky6}, \pageref{Plotnitsky7}, \pageref{Plotnitsky8}, \pageref{Plotnitsky9}, \pageref{Plotnitsky10}, \pageref{Plotnitsky11}, \pageref{Plotnitsky12}, \pageref{Plotnitsky13}, \pageref{Plotnitsky14}, \pageref{Plotnitsky15}, \pageref{Plotnitsky16}, \pageref{Plotnitsky17}, \pageref{Plotnitsky17}, \pageref{Plotnitsky19}, \pageref{Plotnitsky20}, \pageref{Plotnitsky21}, \pageref{Plotnitsky22}, \pageref{Plotnitsky22.1}, \pageref{Plotnitsky23}, \pageref{Plotnitsky24}, \pageref{Plotnitsky25}, \pageref{Plotnitsky26}  \medskip

\noindent {\bf Gabriel Plunk --} \pageref{Plunk1}, \pageref{Plunk2}, \pageref{Plunk3}, \pageref{Plunk4}, \pageref{Plunk5}, \pageref{Plunk6}, \pageref{Plunk7}, \pageref{Plunk8}, \pageref{Plunk9}, \pageref{Plunk10}, \pageref{Plunk11}, \pageref{Plunk12}, \pageref{Plunk13}, \pageref{Plunk14}, \pageref{Plunk15}, \medskip \pageref{Plunk16}

\noindent {\bf Herv\'e Poirier --} \pageref{Poirier1}, \pageref{Poirier2}, \pageref{Poirier3} \medskip

\noindent {\bf Damian T. Pope --} \pageref{Pope1} \medskip

\noindent {\bf Eugene S. Polzik --} \pageref{Polzik1} \medskip

\noindent {\bf David Poulin --} \pageref{Poulin1}, \pageref{Poulin2}, \pageref{Poulin3}, \pageref{Poulin4}, \pageref{Poulin5}, \pageref{Poulin6}, \pageref{Poulin7}, \pageref{Poulin8}, \pageref{Poulin9}, \pageref{Poulin10}, \pageref{Poulin11}, \pageref{Poulin12}, \pageref{Poulin13}, \pageref{Poulin14}, \pageref{Poulin15}, \pageref{Poulin16} \medskip

\noindent {\bf John Preskill --} \pageref{Preskill1}, \pageref{Preskill2}, \pageref{Preskill3}, \pageref{Preskill4}, \pageref{Preskill5}, \pageref{Preskill6}, \pageref{Preskill7}, \pageref{Preskill8}, \pageref{Preskill9}, \pageref{Preskill10}, \pageref{Preskill11}, \pageref{Preskill11.1}, \pageref{Preskill12}, \pageref{Preskill13}, \pageref{Preskill14}, \pageref{Preskill15}, \pageref{Preskill16}, \pageref{Preskill16.1}, \pageref{Preskill17}, \pageref{Preskill18}  \medskip

\noindent {\bf Huw Price --} \pageref{Price1}, \pageref{Price2}, \pageref{Price3}, \pageref{Price4}, \pageref{Price5}, \pageref{Price6}, \pageref{Price7}, \pageref{Price8}, \pageref{Price9}, \pageref{Price10}, \pageref{Price11}, \pageref{Price12}, \pageref{Price13}, \pageref{Price14}, \pageref{Price15}, \pageref{Price16}, \pageref{Price17}, \pageref{Price18}, \pageref{Price19}, \pageref{Price20}, \pageref{Price21} \medskip

\noindent {\bf The QBies} (comprising varied combinations of D. M. Appleby. H. B. Dang, {\AA}. Ericsson, A. Fenyes, E. S. Gould, M. A. Graydon, A. Karlsson, L. Piispanen, G. N. M. Tabia, \& H. Yadsan-Appleby) {\bf --} \pageref{QBies1}, \pageref{QBies2}, \pageref{QBies3}, \pageref{QBies4}, \pageref{QBies5}, \pageref{QBies6}, \pageref{QBies7}, \pageref{QBies8}, \pageref{QBies9}, \pageref{QBies10}, \pageref{QBies11}, \pageref{QBies12}, \pageref{QBies13}, \pageref{QBies14}, \pageref{QBies15}, \pageref{QBies16}, \pageref{QBies17}, \pageref{QBies18}, \pageref{QBies19}, \pageref{QBies20}, \pageref{QBies21}, \pageref{QBies22}, \pageref{QBies23}, \pageref{QBies24}, \pageref{QBies25}, \pageref{QBies26}, \pageref{QBies27}, \pageref{QBies28}, \pageref{QBies29}, \pageref{QBies30}, \pageref{QBies31}, \pageref{QBies32}, \pageref{QBies33}, \pageref{QBies34}, \pageref{QBies35}, \pageref{QBies35.1}, \pageref{QBies36}, \pageref{QBies37} \medskip

\noindent {\bf Gunn Quznetsov --} \pageref{Quznetsov1} \medskip

\noindent {\bf John Racette --} \pageref{Racette1} \medskip

\noindent {\bf Andrzej Radosz --} \pageref{Radosz1}, \pageref{Radosz2} \medskip

\noindent {\bf Maxim Raginsky --} \pageref{Raginsky1} \medskip

\noindent {\bf Arthur P. Ramirez --} \pageref{Ramirez1}, \pageref{Ramirez2} \medskip

\noindent {\bf Bertis C. Rasco --}  \pageref{Rasco1}, \pageref{Rasco2}, \pageref{Rasco3}, \pageref{Rasco4} \medskip

\noindent {\bf Jochen Rau --} \pageref{Rau1}, \pageref{Rau2}, \pageref{Rau3}, \pageref{Rau4}, \pageref{Rau5}  \medskip

\noindent {\bf Michael G. Raymer --} \pageref{Raymer1}, \pageref{Raymer2}, \pageref{Raymer3} \medskip

\noindent {\bf Eugenio Regazzini --} \pageref{Regazzini1} \medskip

\noindent {\bf Joseph M. Renes --} \pageref{Renes1}, \pageref{Renes2}, \pageref{Renes2.1}, \pageref{Renes2.2}, \pageref{Renes3}, \pageref{Renes4}, \pageref{Renes5}, \pageref{Renes6}, \pageref{Renes7}, \pageref{Renes8}, \pageref{Renes9}, \pageref{Renes10}, \pageref{Renes11}, \pageref{Renes12}, \pageref{Renes13}, \pageref{Renes14}, \pageref{Renes15}, \pageref{Renes16}, \pageref{Renes17}, \pageref{Renes18}, \pageref{Renes19}, \pageref{Renes20}, \pageref{Renes21}, \pageref{Renes22}, \pageref{Renes23}, \pageref{Renes24} \medskip

\noindent {\bf Renato Renner --} \pageref{Renner1}, \pageref{Renner2}, \pageref{Renner3}, \pageref{Renner4}, \medskip \pageref{Renner5}

\noindent {\bf Peter J. Reynolds --} \pageref{Reynolds1} \medskip

\noindent {\bf Dean Rickles --} \pageref{Rickles1} \medskip

\noindent {\bf David Rideout --} \pageref{Rideout1} \medskip

\noindent {\bf Jos\'e Ignacio Rosado --} \pageref{Rosado1}, \pageref{Rosado2}, \pageref{Rosado3}, \pageref{Rosado4} \medskip

\noindent {\bf Bruce Rosenblum --} \pageref{Rosenblum1}, \pageref{Rosenblum2} \medskip

\noindent {\bf Carlo Rovelli --} \pageref{Rovelli1}, \pageref{Rovelli2} \medskip

\noindent {\bf Lee A. Rozema --} \pageref{Rozema1} \medskip

\noindent {\bf Terry Rudolph --} \pageref{Rudolph1}, \pageref{Rudolph2}, \pageref{Rudolph3}, \pageref{Rudolph4}, \pageref{Rudolph5}, \pageref{Rudolph6}, \pageref{Rudolph7}, \pageref{Rudolph8}, \pageref{Rudolph9}, \pageref{Rudolph10}, \pageref{Rudolph11}, \pageref{Rudolph12}  \medskip

\noindent {\bf Mary Beth Ruskai --} \pageref{Ruskai1}, \pageref{Ruskai2}, \pageref{Ruskai3} \medskip

\noindent {\bf Joe D. Sanders --} \pageref{SandersJD1} \medskip

\noindent {\bf Laverne ``Gayle'' Sanders --} \pageref{SandersG1} \medskip

\noindent {\bf Michael D. Sanders --} \pageref{SandersMD1}, \pageref{SandersMD2}, \pageref{SandersMD3} \medskip

\noindent {\bf Masahide Sasaki --} \pageref{Sasaki1}, \pageref{Sasaki2}, \pageref{Sasaki3}, \pageref{Sasaki4}, \pageref{Sasaki5}, \pageref{Sasaki6}, \pageref{Sasaki7} \medskip

\noindent {\bf Steven Savitt --} \pageref{Savitt1}, \pageref{Savitt2}, \pageref{Savitt3}, \pageref{Savitt4}, \pageref{Savitt5}, \pageref{Savitt6}, \pageref{Savitt7}, \pageref{Savitt8} \medskip

\noindent {\bf Valerio Scarani --} \pageref{Scarani1} \medskip

\noindent {\bf R\"udiger Schack --} \pageref{Schack0}, \pageref{Schack0.1}, \pageref{Schack0.2}, \pageref{Schack0.3}, \pageref{Schack1}, \pageref{Schack2}, \pageref{Schack3}, \pageref{Schack4}, \pageref{Schack5}, \pageref{Schack6}, \pageref{Schack7}, \pageref{Schack8}, \pageref{Schack9}, \pageref{Schack10}, \pageref{Schack11}, \pageref{Schack12}, \pageref{Schack13}, \pageref{Schack14}, \pageref{Schack15}, \pageref{Schack16}, \pageref{Schack17}, \pageref{Schack18}, \pageref{Schack19}, \pageref{Schack20}, \pageref{Schack21}, \pageref{Schack22}, \pageref{Schack23}, \pageref{Schack23.1}, \pageref{Schack23.2}, \pageref{Schack24}, \pageref{Schack25}, \pageref{Schack26}, \pageref{Schack27}, \pageref{Schack28}, \pageref{Schack29}, \pageref{Schack30}, \pageref{Schack31}, \pageref{Schack32}, \pageref{Schack32.1}, \pageref{Schack33}, \pageref{Schack34}, \pageref{Schack35}, \pageref{Schack36}, \pageref{Schack37}, \pageref{Schack38}, \pageref{Schack39}, \pageref{Schack40}, \pageref{Schack41}, \pageref{Schack41.1}, \pageref{Schack42}, \pageref{Schack43}, \pageref{Schack44}, \pageref{Schack45}, \pageref{Schack46}, \pageref{Schack47}, \pageref{Schack48}, \pageref{Schack49}, \pageref{Schack50}, \pageref{Schack51}, \pageref{Schack52}, \pageref{Schack53}, \pageref{Schack54}, \pageref{Schack55}, \pageref{Schack56}, \pageref{Schack57}, \pageref{Schack57.1}, \pageref{Schack58}, \pageref{Schack59}, \pageref{Schack60}, \pageref{Schack60.01}, \pageref{Schack60.02}, \pageref{Schack60.1}, \pageref{Schack60.2}, \pageref{Schack61}, \pageref{Schack62}, \pageref{Schack63}, \pageref{Schack64}, \pageref{Schack65}, \pageref{Schack66}, \pageref{Schack67}, \pageref{Schack68}, \pageref{Schack69}, \pageref{Schack70}, \pageref{Schack71}, \pageref{Schack72}, \pageref{Schack73}, \pageref{Schack74}, \pageref{Schack75}, \pageref{Schack76}, \pageref{Schack77}, \pageref{Schack78}, \pageref{Schack79}, \pageref{Schack80}, \pageref{Schack81}, \pageref{Schack82}, \pageref{Schack83}, \pageref{Schack84}, \pageref{Schack84.1}, \pageref{Schack84.2}, \pageref{Schack84.3}, \pageref{Schack85}, \pageref{Schack86}, \pageref{Schack86.1}, \pageref{Schack87}, \pageref{Schack88}, \pageref{Schack89}, \pageref{Schack90}, \pageref{Schack91}, \pageref{Schack92}, \pageref{Schack93}, \pageref{Schack94}, \pageref{Schack95}, \pageref{Schack96}, \pageref{Schack97}, \pageref{Schack98}, \pageref{Schack99}, \pageref{Schack99.1}, \pageref{Schack100}, \pageref{Schack101}, \pageref{Schack102}, \pageref{Schack103}, \pageref{Schack104}, \pageref{Schack105}, \pageref{Schack106}, \pageref{Schack107}, \pageref{Schack107.1}, \pageref{Schack108}, \pageref{Schack109}, \pageref{Schack110}, \pageref{Schack111}, \pageref{Schack112}, \pageref{Schack113}, \pageref{Schack114}, \pageref{Schack115}, \pageref{Schack115.1}, \pageref{Schack116}, \pageref{Schack117}, \pageref{Schack117.1}, \pageref{Schack118}, \pageref{Schack119}, \pageref{Schack120}, \pageref{Schack121}, \pageref{Schack122}, \pageref{Schack122.1}, \pageref{Schack123}, \pageref{Schack124}, \pageref{Schack125}, \pageref{Schack126}, \pageref{Schack127}, \pageref{Schack128}, \pageref{Schack129}, \pageref{Schack130}, \pageref{Schack131}, \pageref{Schack132}, \pageref{Schack133}, \pageref{Schack134}, \pageref{Schack135}, \pageref{Schack136}, \pageref{Schack137}, \pageref{Schack138}, \pageref{Schack139}, \pageref{Schack140}, \pageref{Schack141}, \pageref{Schack142}, \pageref{Schack143}, \pageref{Schack144}, \pageref{Schack145}, \pageref{Schack146}, \pageref{Schack147}, \pageref{Schack148}, \pageref{Schack149}, \pageref{Schack150}, \pageref{Schack151}, \pageref{Schack152}, \pageref{Schack153}, \pageref{Schack154}, \pageref{Schack155}, \pageref{Schack156}, \pageref{Schack157}, \pageref{Schack158}, \pageref{Schack159}, \pageref{Schack160}, \pageref{Schack161}, \pageref{Schack162}, \pageref{Schack163}, \pageref{Schack164}, \pageref{Schack165}, \pageref{Schack166}, \pageref{Schack167}, \pageref{Schack168}, \pageref{Schack169}, \pageref{Schack170}, \pageref{Schack171}, \pageref{Schack172}, \pageref{Schack173}, \pageref{Schack174}, \pageref{Schack175}, \pageref{Schack176}, \pageref{Schack177}, \pageref{Schack178}, \pageref{Schack179}, \pageref{Schack180}, \pageref{Schack181}, \pageref{Schack182}, \pageref{Schack183}, \pageref{Schack184}, \pageref{Schack185}, \pageref{Schack186}, \pageref{Schack187}, \pageref{Schack188}, \pageref{Schack189}, \pageref{Schack190}, \pageref{Schack191}, \pageref{Schack192}, \pageref{Schack193}, \pageref{Schack194}, \pageref{Schack195}, \pageref{Schack196}, \pageref{Schack197}, \pageref{Schack197.1}, \pageref{Schack198}, \pageref{Schack199}, \pageref{Schack199.1}, \pageref{Schack200}, \pageref{Schack201}, \pageref{Schack202}, \pageref{Schack203}, \pageref{Schack204}, \pageref{Schack205}, \pageref{Schack206}, \pageref{Schack207}, \pageref{Schack208}, \pageref{Schack209}, \pageref{Schack210}, \pageref{Schack211}, \pageref{Schack212}, \pageref{Schack213}, \pageref{Schack214}, \pageref{Schack215}, \pageref{Schack216}, \pageref{Schack217}, \pageref{Schack218}, \pageref{Schack219}, \pageref{Schack220} \medskip

\noindent {\bf Christoph Schaeff --} \pageref{Schaeff1} \medskip

\noindent {\bf Wolfgang P. Schleich --} \pageref{Schleich1}, \pageref{Schleich2}, \pageref{Schleich3}, \medskip \pageref{Schleich4}

\noindent {\bf Maximilian Schlosshauer --} \pageref{Schlosshauer1}, \pageref{Schlosshauer2}, \pageref{Schlosshauer-newname}, \pageref{Schlosshauer3}, \pageref{Schlosshauer4}, \pageref{Schlosshauer5}, \pageref{Schlosshauer6}, \pageref{Schlosshauer7}, \pageref{Schlosshauer8}, \pageref{Schlosshauer9}, \pageref{Schlosshauer10}, \pageref{Schlosshauer11}, \pageref{Schlosshauer12}, \pageref{Schlosshauer13}, \pageref{Schlosshauer14}, \pageref{Schlosshauer15}, \pageref{Schlosshauer16}, \pageref{Schlosshauer17}, \pageref{Schlosshauer18}, \pageref{Schlosshauer19}, \pageref{Schlosshauer20}, \pageref{Schlosshauer21}, \pageref{Schlosshauer22}, \pageref{Schlosshauer23}, \pageref{Schlosshauer24}, \pageref{Schlosshauer25}, \pageref{Schlosshauer26}, \pageref{Schlosshauer27}, \pageref{Schlosshauer28}, \pageref{Schlosshauer29}, \pageref{Schlosshauer30}, \pageref{Schlosshauer30.1}, \pageref{Schlosshauer30.2}, \pageref{Schlosshauer31}, \pageref{Schlosshauer32}, \pageref{Schlosshauer33}, \pageref{Schlosshauer34}, \pageref{Schlosshauer35}, \pageref{Schlosshauer36}, \pageref{Schlosshauer37}, \pageref{Schlosshauer38}, \pageref{Schlosshauer39}, \pageref{Schlosshauer40}, \pageref{Schlosshauer41}, \pageref{Schlosshauer42}, \pageref{Schlosshauer43} \medskip

\noindent {\bf Frank E. Schroeck, Jr.\ --} \pageref{Schroeck1}, \pageref{Schroeck2}, \pageref{Schroeck3}, \pageref{Schroeck4}, \pageref{Schroeck5} \medskip

\noindent {\bf Benjamin W. Schumacher --} \pageref{Schumacher1}, \pageref{Schumacher2}, \pageref{Schumacher3}, \pageref{Schumacher4}, \pageref{Schumacher5}, \pageref{Schumacher6}, \pageref{Schumacher7}, \pageref{Schumacher8}, \pageref{Schumacher9}, \pageref{Schumacher10}, \pageref{Schumacher11}, \pageref{Schumacher12}, \pageref{Schumacher13}, \pageref{Schumacher14}, \pageref{Schumacher15}, \pageref{Schumacher16}, \pageref{Schumacher17}, \pageref{Schumacher18}, \pageref{Schumacher19}, \pageref{Schumacher19.1}, \pageref{Schumacher20} \medskip

\noindent {\bf Andrew J. Scott --} \pageref{Scott1} \medskip

\noindent {\bf Petra F. Scudo --} \pageref{Scudo1}, \pageref{Scudo2}, \pageref{Scudo3}  \medskip

\noindent {\bf Marlan O. Scully --} \pageref{Scully1}, \pageref{Scully2}, \pageref{Scully3}, \pageref{Scully4}, \pageref{Scully5}, \pageref{Scully6}, \pageref{Scully7}, \pageref{Scully8} \medskip

\noindent {\bf Michael P. Seevinck --} \pageref{Seevinck1} \medskip

\noindent {\bf Lynden K. ``Krister'' Shalm --} \pageref{Shalm1} \medskip

\noindent {\bf Jeffrey H. Shapiro --} \pageref{Shapiro1} \medskip

\noindent {\bf Lana Sheridan --} \pageref{Sheridan1}, \pageref{Sheridan2}, \pageref{Sheridan3} \medskip

\noindent {\bf Abner Shimony --} \pageref{Shimony1}, \pageref{Shimony2}, \pageref{Shimony3}, \pageref{Shimony4}, \pageref{Shimony5}, \pageref{Shimony6}, \pageref{Shimony7}, \pageref{Shimony8}, \pageref{Shimony9}, \pageref{Shimony10}, \pageref{Shimony11}, \pageref{Shimony12}, \pageref{Shimony13}, \pageref{Shimony14}, \medskip \pageref{Shimony15}

\noindent {\bf Peter W. Shor --} \pageref{Shor1}, \pageref{Shor2}, \pageref{Shor3}, \pageref{Shor4}, \pageref{Shor5}, \pageref{Shor6}, \pageref{Shor7}, \pageref{Shor8} \medskip

\noindent {\bf Tom Siegfried  --} \pageref{Siegfried1}, \pageref{Siegfried2} \medskip

\noindent {\bf Daniel R. Simon  --} \pageref{SimonD1} \medskip

\noindent {\bf Linda Simon  --} \pageref{SimonL1} \medskip

\noindent {\bf Steven H. Simon  --} \pageref{Simon1}, \pageref{Simon2} \medskip

\noindent {\bf John E. Sipe  --} \pageref{Sipe1}, \pageref{Sipe2}, \pageref{Sipe3}, \pageref{Sipe4}, \pageref{Sipe5}, \pageref{Sipe6}, \pageref{Sipe7}, \pageref{Sipe8}, \pageref{Sipe9}, \pageref{Sipe10}, \pageref{Sipe11}, \pageref{Sipe12}, \pageref{Sipe13}, \pageref{Sipe14}, \pageref{Sipe15}, \pageref{Sipe16}, \pageref{Sipe17}, \pageref{Sipe18}, \pageref{Sipe19}, \pageref{Sipe20}  \medskip

\noindent {\bf Brian Skyrms --} \pageref{Skyrms1}, \pageref{Skyrms2}, \pageref{Skyrms3}, \pageref{Skyrms4} \medskip

\noindent {\bf Tom Slee --} \pageref{Slee1}, \pageref{Slee2}, \pageref{Slee3}, \pageref{Slee4} \medskip

\noindent {\bf Richart E. Slusher --} \pageref{Slusher1}, \pageref{Slusher2}, \pageref{Slusher3}, \pageref{Slusher4}, \pageref{Slusher5}, \pageref{Slusher6}, \pageref{Slusher7}, \pageref{Slusher8}, \pageref{Slusher9}, \pageref{Slusher10}, \pageref{Slusher11}, \pageref{Slusher12}, \pageref{Slusher13}, \pageref{Slusher14}, \pageref{Slusher15}, \pageref{Slusher16}, \pageref{Slusher17}, \pageref{Slusher18}, \pageref{Slusher19}, \pageref{Slusher20}, \pageref{Slusher21}  \medskip

\noindent {\bf Christopher Smeenk --} \pageref{Smeenk1}, \pageref{Smeenk2}, \pageref{Smeenk3}, \pageref{Smeenk4}, \pageref{Smeenk5}, \pageref{Smeenk6} \medskip

\noindent {\bf Graeme Smith --} \pageref{Smith1} \medskip

\noindent {\bf John A. Smolin --} \pageref{SmolinJ1}, \pageref{SmolinJ1.1}, \pageref{SmolinJ2}, \pageref{SmolinJ2.05}, \pageref{SmolinJ2.1}, \pageref{SmolinJ3}, \pageref{SmolinJ4}, \pageref{SmolinJ5}, \pageref{SmolinJ6}, \pageref{SmolinJ7}, \pageref{SmolinJ7.1}, \pageref{SmolinJ7.2}, \pageref{SmolinJ8}, \pageref{SmolinJ9}, \pageref{SmolinJ9.1}, \pageref{SmolinJ10} \medskip

\noindent {\bf Lee Smolin --} \pageref{SmolinL1}, \pageref{SmolinL2}, \pageref{SmolinL3}, \pageref{SmolinL4}, \pageref{SmolinL5}, \pageref{SmolinL6}, \pageref{SmolinL7}, \pageref{SmolinL8}, \pageref{SmolinL9}, \pageref{SmolinL10}, \pageref{SmolinL11}, \pageref{SmolinL12}, \pageref{SmolinL13}, \pageref{SmolinL14}, \pageref{SmolinL15}, \pageref{SmolinL16}, \pageref{SmolinL17}, \pageref{SmolinL18}, \pageref{SmolinL19}, \pageref{SmolinL20}, \pageref{SmolinL21}, \pageref{SmolinL22}  \medskip

\noindent {\bf Christian Snyder --} \pageref{Snyder1}, \pageref{Snyder2}, \pageref{Snyder3}, \pageref{Snyder4}, \pageref{Snyder5}  \medskip

\noindent {\bf Stanley E. Sobottka --} \pageref{Sobottka1} \medskip

\noindent {\bf Rafael D. Sorkin --} \pageref{Sorkin1}, \pageref{Sorkin2} \medskip

\noindent {\bf Ernst Specker --} \pageref{Specker1}, \pageref{Specker2} \medskip

\noindent {\bf Robert W. Spekkens --} \pageref{Spekkens1}, \pageref{Spekkens2}, \pageref{Spekkens3}, \pageref{Spekkens4}, \pageref{Spekkens5}, \pageref{Spekkens6}, \pageref{Spekkens7}, \pageref{Spekkens8}, \pageref{Spekkens9}, \pageref{Spekkens10}, \pageref{Spekkens11}, \pageref{Spekkens12}, \pageref{Spekkens13}, \pageref{Spekkens14}, \pageref{Spekkens15}, \pageref{Spekkens16}, \pageref{Spekkens17}, \pageref{Spekkens18}, \pageref{Spekkens19}, \pageref{Spekkens20}, \pageref{Spekkens21}, \pageref{Spekkens22}, \pageref{Spekkens23}, \pageref{Spekkens24}, \pageref{Spekkens25}, \pageref{Spekkens26}, \pageref{Spekkens27}, \pageref{Spekkens28}, \pageref{Spekkens29}, \pageref{Spekkens30}, \pageref{Spekkens31}, \pageref{Spekkens32}, \pageref{Spekkens33}, \pageref{Spekkens34}, \pageref{Spekkens35}, \pageref{Spekkens36}, \pageref{Spekkens37}, \pageref{Spekkens37.05}, \pageref{Spekkens37.1}, \pageref{Spekkens38}, \pageref{Spekkens39}, \pageref{Spekkens40}, \pageref{Spekkens41}, \pageref{Spekkens42}, \pageref{Spekkens43}, \pageref{Spekkens44}, \pageref{Spekkens45}, \pageref{Spekkens46}, \pageref{Spekkens47}, \pageref{Spekkens48}, \pageref{Spekkens48.1}, \pageref{Spekkens49}, \pageref{Spekkens50}, \pageref{Spekkens51}, \pageref{Spekkens51.1}, \pageref{Spekkens52}, \pageref{Spekkens53}, \pageref{Spekkens54}, \pageref{Spekkens55}, \pageref{Spekkens56}, \pageref{Spekkens57}, \pageref{Spekkens58}, \pageref{Spekkens59}, \pageref{Spekkens60}, \pageref{Spekkens61}, \pageref{Spekkens62}, \pageref{Spekkens63}, \pageref{Spekkens63.1}, \pageref{Spekkens63.1.1}, \pageref{Spekkens63.2}, \pageref{Spekkens64}, \pageref{Spekkens65}, \pageref{Spekkens66}, \pageref{Spekkens67}, \pageref{Spekkens68}, \pageref{Spekkens69}, \pageref{Spekkens69.1}, \pageref{Spekkens70}, \pageref{Spekkens71}, \pageref{Spekkens72}, \pageref{Spekkens73}, \pageref{Spekkens74}, \pageref{Spekkens75}, \pageref{Spekkens76}, \pageref{Spekkens77}, \pageref{Spekkens78}, \pageref{Spekkens79}, \pageref{Spekkens80}, \pageref{Spekkens81}, \pageref{Spekkens82}, \pageref{Spekkens83}, \pageref{Spekkens84}, \pageref{Spekkens85}, \pageref{Spekkens86}, \pageref{Spekkens87}, \pageref{Spekkens88}, \pageref{Spekkens89}, \pageref{Spekkens90}, \pageref{Spekkens91}, \pageref{Spekkens92}, \pageref{Spekkens93}, \pageref{Spekkens94}, \pageref{Spekkens95}, \pageref{Spekkens96}, \pageref{Spekkens97}, \pageref{Spekkens98}, \pageref{Spekkens99}, \pageref{Spekkens100}, \pageref{Spekkens101}, \pageref{Spekkens102}, \pageref{Spekkens103}, \pageref{Spekkens104}, \pageref{Spekkens105}, \pageref{Spekkens106}, \pageref{Spekkens107} \medskip

\noindent {\bf Allen Stairs --} \pageref{Stairs1}, \pageref{Stairs2}, \pageref{Stairs3}, \pageref{Stairs4} \medskip

\noindent {\bf Philip C. E. Stamp --} \pageref{Stamp1}, \pageref{Stamp2}, \pageref{Stamp3}, \pageref{Stamp4} \medskip

\noindent {\bf Andrew M. Steane --} \pageref{Steane1}, \pageref{Steane2}, \pageref{Steane3}, \pageref{Steane4} \medskip

\noindent {\bf Aephraim M. Steinberg --} \pageref{Steinberg1}, \pageref{Steinberg2}, \pageref{Steinberg3}, \pageref{Steinberg4}, \pageref{Steinberg5}, \pageref{Steinberg6}, \pageref{Steinberg7} \medskip

\noindent {\bf Graeme Stemp-Morlock  --} \pageref{StempMorlock1} \medskip

\noindent {\bf Sheri K. Stoll  --} \pageref{Stoll1}, \pageref{Stoll2}, \pageref{Stoll3} \medskip

\noindent {\bf Carlos R. Stroud, Jr.\  --} \pageref{Stroud1}, \pageref{Stroud2} \medskip

\noindent {\bf Roger H. Stuewer  --} \pageref{Stuewer1} \medskip

\noindent {\bf Anthony Sudbery --} \pageref{Sudbery1}, \pageref{Sudbery2}, \pageref{Sudbery3}, \pageref{Sudbery4}, \pageref{Sudbery5}, \pageref{Sudbery6}, \pageref{Sudbery7}, \pageref{Sudbery8}, \pageref{Sudbery9}, \pageref{Sudbery10}, \pageref{Sudbery11} \medskip

\noindent {\bf Johann Summhammer --} \pageref{Summhammer1}, \pageref{Summhammer2}, \pageref{Summhammer3}, \pageref{Summhammer4} \medskip

\noindent {\bf Karl Svozil --} \pageref{Svozil1}, \pageref{Svozil2} \medskip

\noindent {\bf Morgan Tait --} \pageref{Tait1}, \pageref{Tait2}, \pageref{Tait3}, \pageref{Tait4}, \pageref{Tait5}, \pageref{Tait6}, \pageref{Tait7} \medskip

\noindent {\bf Max Tegmark --} \pageref{Tegmark1}, \pageref{Tegmark2} \medskip

\noindent {\bf Kip S. Thorne --} \pageref{Thorne1} \medskip

\noindent {\bf Daniel R. Terno --} \pageref{Terno1}, \pageref{Terno3}, \pageref{Terno3}, \pageref{Terno4}, \pageref{Terno5}, \pageref{Terno5.1}, \pageref{Terno5.2} \medskip

\noindent {\bf Christopher G. Timpson --} \pageref{Timpson1}, \pageref{Timpson2}, \pageref{Timpson3}, \pageref{Timpson4}, \pageref{Timpson5}, \pageref{Timpson6}, \pageref{Timpson7}, \pageref{Timpson8}, \pageref{Timpson9}, \pageref{Timpson10}, \pageref{Timpson11}, \pageref{Timpson12}, \pageref{Timpson13}, \pageref{Timpson14}, \pageref{Timpson15}, \pageref{Timpson16}, \pageref{Timpson17}, \pageref{Timpson18}, \pageref{Timpson19}, \pageref{Timpson20} \medskip

\noindent {\bf Frank J. Tipler --} \pageref{Tipler1} \medskip

\noindent {\bf Tommaso Toffoli --} \pageref{Toffoli1} \medskip

\noindent {\bf Flemming Tops{\o}e --} \pageref{Topsoe1}, \pageref{Topsoe2}, \pageref{Topsoe3}, \pageref{Topsoe4} \medskip

\noindent {\bf Cozmin Ududec --} \pageref{Ududec1}, \pageref{Ududec2} \medskip

\noindent {\bf Jos B. M. Uffink --} \pageref{Uffink1}, \pageref{Uffink2}, \pageref{Uffink3}, \pageref{Uffink4} \medskip

\noindent {\bf Letitia Wheeler Ufford --} \pageref{Ufford1}, \pageref{Ufford2} \medskip

\noindent {\bf Armin Uhlmann --} \pageref{Uhlmann1}, \pageref{Uhlmann2}, \pageref{Uhlmann3} \medskip

\noindent {\bf Ole C. Ulfbeck --} \pageref{Ulfbeck1}, \pageref{Ulfbeck2} \medskip

\noindent {\bf William G. Unruh --} \pageref{Unruh1}, \pageref{Unruh2}, \pageref{Unruh3}, \pageref{Unruh4}, \pageref{Unruh4.1}, \pageref{Unruh5}, \pageref{Unruh6}, \pageref{Unruh7}, \pageref{Unruh8}, \pageref{Unruh9}  \medskip

\noindent {\bf Joan A. Vaccaro --} \pageref{Vaccaro1}, \pageref{Vaccaro2}, \pageref{Vaccaro3}, \pageref{Vaccaro4} \medskip

\noindent {\bf Giovanni Valente --} \pageref{Valente1}, \pageref{Valente2}, \pageref{Valente3}, \pageref{Valente4}, \pageref{Valente5}, \pageref{Valente6}, \pageref{Valente7}, \pageref{Valente8}, \pageref{Valente9}, \pageref{Valente10}, \pageref{Valente11} \medskip

\noindent {\bf Antony Valentini --} \pageref{Valentini1}, \pageref{Valentini2} \medskip

\noindent {\bf Frank Verstraete --} \pageref{Verstraete1} \medskip

\noindent {\bf Ralph F. Wachter --} \pageref{Wachter1}, \pageref{Wachter2}, \pageref{Wachter3} \medskip

\noindent {\bf David Wallace --} \pageref{Wallace1}, \pageref{Wallace2}, \pageref{Wallace3}, \pageref{Wallace4} \medskip

\noindent {\bf Zachary D. Walton --} \pageref{Walton1} \medskip

\noindent {\bf Manfred K. Warmuth --} \pageref{Warmuth1} \medskip

\noindent {\bf Jonathan A. Waskan --} \pageref{Waskan1}, \pageref{Waskan2}, \pageref{Waskan3}, \pageref{Waskan4}, \pageref{Waskan5}, \pageref{Waskan6}, \pageref{Waskan7} \medskip

\noindent {\bf Natasha Waxman --} \pageref{Waxman1}, \pageref{Waxman2}, \pageref{Waxman3}, \pageref{Waxman4}, \pageref{Waxman5}, \pageref{Waxman6}, \pageref{Waxman7}, \pageref{Waxman8}, \pageref{Waxman9}, \pageref{Waxman10}, \pageref{Waxman11} \medskip

\noindent {\bf Steven Weinstein --} \pageref{Weinstein1}, \pageref{Weinstein2}, \pageref{Weinstein3}, \pageref{Weinstein4}, \pageref{Weinstein5}, \pageref{Weinstein6}  \medskip

\noindent {\bf Paul Wells --} \pageref{Wells1}, \pageref{Wells2}, \pageref{Wells3}, \pageref{Wells4}, \pageref{Wells5}, \medskip \pageref{Wells6}

\noindent {\bf Reinhard F. Werner --} \pageref{Werner1} \medskip

\noindent {\bf Brad Weslake --} \pageref{Weslake1}, \pageref{Weslake2}, \pageref{Weslake3}, \pageref{Weslake4} \medskip

\noindent {\bf Hans Westman --} \pageref{Westman1}, \pageref{Westman2}, \pageref{Westman3}, \pageref{Westman4} \medskip

\noindent {\bf Michael D. Westmoreland --} \pageref{Westmoreland1}, \pageref{Westmoreland2}, \medskip \pageref{Westmoreland3}

\noindent {\bf M. Andrew B. Whitaker --} \pageref{Whitaker1} \medskip

\noindent {\bf Alice E. White --} \pageref{White1} \medskip

\noindent {\bf Karoline Wiesner --} \pageref{Wiesner1} \medskip

\noindent {\bf Alexander Wilce  --} \pageref{Wilce1}, \pageref{Wilce2}, \pageref{Wilce3}, \pageref{Wilce4}, \pageref{Wilce5}, \pageref{Wilce6}, \pageref{Wilce7}, \pageref{Wilce8}, \pageref{Wilce9}, \pageref{Wilce10}, \pageref{Wilce11}, \pageref{Wilce12}, \pageref{Wilce13}, \pageref{Wilce14}, \pageref{Wilce15}, \pageref{Wilce16}, \pageref{Wilce17}, \pageref{Wilce18}, \pageref{Wilce19}, \pageref{Wilce20}, \pageref{Wilce21}, \pageref{Wilce22}, \pageref{Wilce23}, \pageref{Wilce24}, \pageref{Wilce24.1}, \pageref{Wilce25} \medskip

\noindent {\bf George Will --} \pageref{Will1} \medskip

\noindent {\bf Howard M. Wiseman --} \pageref{Wiseman1}, \pageref{Wiseman2}, \pageref{Wiseman3}, \pageref{Wiseman4}, \pageref{Wiseman5}, \pageref{Wiseman6}, \pageref{Wiseman7}, \pageref{Wiseman8}, \pageref{Wiseman9}, \pageref{Wiseman10}, \pageref{Wiseman11}, \pageref{Wiseman12}, \pageref{Wiseman13}, \pageref{Wiseman13.1}, \pageref{Wiseman14}, \pageref{Wiseman15}, \pageref{Wiseman16}, \pageref{Wiseman17}, \pageref{Wiseman18}, \pageref{Wiseman19}, \pageref{Wiseman20}, \pageref{Wiseman21}, \pageref{Wiseman22}, \pageref{Wiseman23}, \pageref{Wiseman24}, \pageref{Wiseman25}, \pageref{Wiseman26}, \pageref{Wiseman27}, \pageref{Wiseman28}, \pageref{Wiseman29}  \medskip

\noindent {\bf David H. Wolpert --} \pageref{Wolpert1}, \pageref{Wolpert2}, \pageref{Wolpert3}, \pageref{Wolpert4} \medskip

\noindent {\bf James Woodward --} \pageref{Woodward1}, \pageref{Woodward2}, \pageref{Woodward3}, \pageref{Woodward4} \medskip

\noindent {\bf William K. Wootters --} \pageref{Wootters1}, \pageref{Wootters2}, \pageref{Wootters3}, \pageref{Wootters4}, \pageref{Wootters5}, \pageref{Wootters6}, \pageref{Wootters7}, \pageref{Wootters7.1}, \pageref{Wootters8}, \pageref{Wootters9}, \pageref{Wootters10}, \pageref{Wootters11}, \pageref{Wootters12}, \pageref{Wootters13}, \pageref{Wootters14}, \pageref{Wootters15}, \pageref{Wootters16}, \pageref{Wootters17}, \pageref{Wootters18}, \pageref{Wootters19}, \pageref{Wootters20}, \pageref{Wootters21}, \pageref{Wootters22}, \pageref{Wootters23}, \pageref{Wootters23.1}, \pageref{Wootters24}, \pageref{Wootters25}, \pageref{Wootters25.1}, \pageref{Wootters26} \medskip

\noindent {\bf Jessey Wright --} \pageref{Wright1} \medskip

\noindent {\bf Hulya Yadsan-Appleby --} \pageref{Yadsan1} \medskip

\noindent {\bf Andrew C. Yao --} \pageref{Yao1} \medskip

\noindent {\bf Bernard Yurke --} \pageref{Yurke1} \medskip

\noindent {\bf Arthur G. Zajonc --} \pageref{Zajonc1}, \pageref{Zajonc2} \medskip

\noindent {\bf Gerhard Zauner --} \pageref{Zauner1} \medskip

\noindent {\bf Anton Zeilinger --} \pageref{Zeilinger1}, \pageref{Zeilinger2}, \pageref{Zeilinger3}, \pageref{Zeilinger4}, \pageref{Zeilinger5}, \pageref{Zeilinger6}, \pageref{Zeilinger7}, \pageref{Zeilinger8}, \pageref{Zeilinger9}, \pageref{Zeilinger10}, \pageref{Zeilinger11}, \pageref{Zeilinger12}, \pageref{Zeilinger13}, \pageref{Zeilinger14} \medskip

\noindent {\bf Marek \.{Z}ukowski --} \pageref{Zukowski1} \medskip

\noindent {\bf Wojciech H. Zurek --} \pageref{Zurek1}, \pageref{Zurek2}, \pageref{Zurek3}

\onecolumn

\pagebreak

\vspace*{1.5in}

\phantomsection

\begin{center}
\Large {\bf Chronicles for the History of Quantum Information}
\end{center}

\addcontentsline{toc}{chapter}{Chronicles for the History of Quantum Information}

\begin{itemize}
\item \indexbullet{Born, Max Born}\par
\germanism{21-04-10}{Book Review Notes}{to H. C. von Baeyer}{Baeyer118}\par
\germanism{27-04-10}{Midnight Reading}{to H. C. von Baeyer}{Baeyer119}\par
\germanism{08-01-11}{Strange Einstein Remark}{to R. W. Spekkens}{Spekkens91}\par

\item \indexbullet{Einstein's views on quantum theory}\par
\germanism{05-05-01}{Author's Reply}{to A. Peres}{Peres12}\par
\germanism{07-03-04}{All Kinds of Veils}{to A. Peres}{Peres62}\par
\germanism{01-07-07}{First Einstein Quote}{to R. W. Spekkens}{Spekkens46}\par
\germanism{22-12-08}{Nonlocal Boxes and Ehrenfest}{to N. D. Mermin}{Mermin144}\par
\germanism{21-04-10}{Book Review Notes}{to H. C. von Baeyer}{Baeyer118}\par
\germanism{08-01-11}{Strange Einstein Remark}{to R. W. Spekkens}{Spekkens91}\par

\item \indexbullet{Excess baggage in hidden-variable models}\par
\germanism{30-11-01}{NMR Stuff}{to C. M. Caves}{Caves40.1}\par
\germanism{21-01-08}{Talk and Diverse}{to P. G. L. Mana}{Mana10}\par

\item \indexbullet{Genesis of the word ``qubit''}\par
\germanism{28-04-04}{Quantum History}{to B. W. Schumacher \& W. K. Wootters}{Wootters19}\par
\germanism{12-12-10}{The Qubit Story}{to A. Stairs}{Stairs4}\par

\item \indexbullet{History of the GHZ gedankenexperiment}\par
\germanism{07-02-01}{GHZM}{to N. D. Mermin}{Mermin-3}\par
\germanism{10-02-01}{Historical Accuracy}{to N. D. Mermin}{Mermin-1}\par

\item \indexbullet{Inspiration for and development of the Simon, Shor and Deutsch--Jozsa algorithms}\par
\germanism{10-01-11}{Q3, Part 1}{to N. D. Mermin}{Mermin196}\par
\germanism{20-09-07}{Easy Questions}{to D. Deutsch}{DeutschD1}\par
\germanism{20-09-07}{Easy Questions}{to R. Jozsa}{Jozsa7}\par
\germanism{20-09-07}{Easy Questions}{to P. W. Shor}{Shor4}\par
\germanism{20-09-07}{Easy Questions}{to D. R. Simon}{SimonD1}\par
\germanism{22-09-07}{Easy Questions, RPWS2}{to P. W. Shor}{Shor6}\par

\item \indexbullet{Inspirations for communication complexity theory}\par
\germanism{20-09-07}{Easy Questions}{to A. C. Yao}{Yao1}\par
\germanism{20-09-07}{Easy Questions}{to R. Cleve}{Cleve3}\par

\item \indexbullet{Inspirations for quantum computation}\par
\germanism{20-09-07}{Easy Questions}{to H. J. Briegel}{Briegel5}\par
\germanism{20-09-07}{Easy Questions}{to P. A. Benioff}{Benioff1}\par

\item \indexbullet{Inspirations for quantum cryptography}\par
\germanism{20-09-07}{Easy Questions}{to J. Preskill}{Preskill16.1}\par
\germanism{20-09-07}{Easy Questions}{to A. Kent}{Kent14}\par
\germanism{20-09-07}{Easy Questions}{to A. K. Ekert}{Ekert1}\par

\item \indexbullet{Inspiration for the Kochen--Specker theorem}\par
\germanism{15-05-08}{1:30 AM Note -- Decompressing the Day}{to D. M. Appleby}{Appleby34}\par

\item \indexbullet{Jaynes' rejection letters for Landauer's landmark paper}\par
\germanism{26-10-04}{More ``More on Landauer''}{to W. T. Grandy, Jr.}{Grandy3}\par

\item \indexbullet{Pauli presaging the Heisenberg uncertainty principle}\par
\germanism{27-09-10}{Pauli's Eye Quote}{to H. Atmanspacher \& H. C. von Baeyer}{Baeyer134.1}\par

\item \indexbullet{SIC-POVMs: Origins and prehistory}\par
\germanism{11-01-10}{Analogy?}{to H. C. von Baeyer}{Baeyer94}\par
\germanism{19-01-10}{Translation}{to H. C. von Baeyer}{Baeyer100}\par
\germanism{27-04-10}{Midnight Reading}{to H. C. von Baeyer}{Baeyer119}\par
\germanism{07-04-11}{Huangjun Zhu Visiting at the Moment}{to R. W. Spekkens \& M. S. Leifer}{Spekkens107}\par

\item \indexbullet{SIC-POVMs: Pure and mixed states are all probability distributions}\par
\germanism{04-06-02}{Random Questions}{to G. Plunk}{Plunk6}\par
\germanism{18-06-02}{SIC POVMs}{to C. M. Caves}{Caves66}\par
\germanism{04-06-08}{Hammerhead}{to G. L. Comer}{Comer115}\par

\item \indexbullet{The Remarkable Theorem of Flammia, Jones and Linden}\par
\germanism{20-08-07}{R\'enyi Order-3 and the Weird Object}{to F. Tops{\o}e \& P. Harremo\"es}{Topsoe3}\par
\germanism{20-08-07}{R\'enyi Order-3 and the Weird Object, 2}{to F. Tops{\o}e \& P. Harremo\"es}{Topsoe4}\par
\germanism{13-01-09}{The Remarkable Theorem}{to A. Uhlmann}{Uhlmann2}\par
\germanism{04-09-09}{Little Lemma, Big Theorem}{to N. S. Jones}{Jones1}\par

\end{itemize}

\pagebreak

\vspace*{1.5in}

\phantomsection

\begin{center}
\Large {\bf Index to the More Significant Arguments and Ideas}
\end{center}

\addcontentsline{toc}{chapter}{Index to Some of the More Significant Arguments and Ideas}

\begin{itemize}
\item
\indexbullet{Certainty}\par
\germanism{04-07-01}{Context Dependent Probability, 2}{to A. Y. Khrennikov}{Khrennikov5}\par
\germanism{16-08-01}{Subject-Object}{to P. Grangier}{Grangier1}\par
\germanism{22-08-01}{Identity Crisis}{to C. M. Caves \& R. Schack}{Schack4}\par
\germanism{02-09-01}{Intersubjective Agreement}{to N. D. Mermin}{Mermin35}\par
\germanism{05-09-01}{Unique Assignment}{to C. M. Caves \& R. Schack}{Caves14}\par
\germanism{06-09-01}{Some Comments}{to C. M. Caves \& R. Schack}{Caves18}\par
\germanism{27-06-02}{Compatibility Never Ends}{to D. Poulin}{Poulin2}\par
\germanism{27-06-03}{Utter Rubbish and Internal Consistency, Part I}{to R. Schack, C. M. Caves \& N. D. Mermin}{Mermin86}\par
\germanism{28-06-03}{Utter Rubbish and Internal Consistency, Part II}{to R. Schack, C. M. Caves \& N. D. Mermin}{Mermin87}\par
\germanism{14-01-04}{Hello Kitty}{to H. Mabuchi}{Mabuchi4}\par
\germanism{15-01-04}{Idealism}{to R. W. Spekkens}{Spekkens27}\par
\germanism{23-01-04}{The Wittgenstein Bug}{to R. Schack}{Schack79}\par
\germanism{28-01-04}{Merminizing}{to N. D. Mermin}{Mermin107}\par
\germanism{31-01-04}{Just the Opposite}{to N. D. Mermin}{Mermin108}\par
\germanism{27-04-04}{Jeffrey Knew Certainty}{to R. Schack}{Schack82}\par
\germanism{30-06-05}{Wittgensteinian Arm Twisting}{to H. Price}{Price5}\par
\germanism{15-11-05}{Canned Answers}{to B. C. van Fraassen}{vanFraassen12}\par
\germanism{24-03-06}{On Certainty, Quantum Outcomes, Subjectivity, Objectivity, \\ \fantome and Expanding Universes, Part 1}{to R. Schack \& C. M. Caves}{Caves82}\par
\germanism{17-11-06}{Certain Comments}{to C. G. Timpson}{Timpson11}\par
\germanism{20-12-06}{Paper}{to R. Schack}{Schack116}\par
\germanism{27-12-06}{Residue of the Category Error}{to S. J. van Enk}{vanEnk8}\par
\germanism{29-06-09}{Off-the-Shelf Ontology vs.\ the Republican Banquet}{to C. G. Timpson}{Timpson13}\par
\germanism{11-08-09}{From Moore-ish Sentences to Bohrish Ones}{to C. G. Timpson}{Timpson15}\par
\germanism{12-08-09}{More on Moore}{to M. Schlosshauer}{Schlosshauer3}\par
\germanism{20-09-09}{One of the Essences of Things, Maybe?}{to R. Schack}{Schack177}\par
\germanism{20-09-09}{A Way To Salvage Much of the Old}{to R. Schack}{Schack179}\par
\germanism{07-04-10}{LNC Section 6.8}{to N. D. Mermin}{Mermin175}\par
\germanism{18-04-11}{If Words Have Meaning}{to N. D. Mermin}{Mermin198}\par
\germanism{18-04-11}{If Words Have Meaning, 2}{to N. D. Mermin}{Mermin199}\par

\item
\indexbullet{Dutch book arguments: coherence vs.\ strict coherence}\par
\germanism{28-05-01}{Card-Carrying Greens}{to C. H. Bennett}{Bennett2}\par
\germanism{04-07-01}{Invitation, 2}{to A. Y. Khrennikov}{Khrennikov4}\par
\germanism{07-08-01}{Knowledge, Only Knowledge}{to T. A. Brun, J. Finkelstein \& N. D. Mermin}{Mermin28}\par
\germanism{22-08-01}{Identity Crisis}{to C. M. Caves \& R. Schack}{Schack4}\par
\germanism{02-09-01}{Intersubjective Agreement}{to N. D. Mermin}{Mermin35}\par
\germanism{04-09-01}{Fourth and Fifth Reading}{to C. M. Caves \& R. Schack}{Caves12}\par
\germanism{05-09-01}{Unique Assignment}{to C. M. Caves \& R. Schack}{Caves14}\par
\germanism{06-09-01}{The Underappreciated Point}{to C. M. Caves \& R. Schack}{Caves16}\par
\germanism{06-09-01}{Some Comments}{to C. M. Caves \& R. Schack}{Caves16}\par
\germanism{06-09-01}{Weak Point}{to C. M. Caves \& R. Schack}{Caves19}\par
\germanism{07-09-01}{Email Not Received}{to C. M. Caves \& R. Schack}{Caves21}\par
\germanism{07-09-01}{(Backbreaking) Analysis}{to C. M. Caves \& R. Schack}{Caves22}\par
\germanism{08-09-01}{Negotiation and Compromise}{to C. M. Caves \& R. Schack}{Caves23}\par
\germanism{10-09-01}{Short Reply}{to C. M. Caves \& R. Schack}{Caves24}\par
\germanism{16-09-01}{Nerve Therapy}{to R. Schack}{Schack20}\par
\germanism{03-10-01}{Kid Sheleen}{to N. D. Mermin}{Mermin45}\par
\germanism{04-10-01}{Replies to a Conglomeration}{to C. M. Caves \& R. Schack}{Caves35}\par
\germanism{27-10-01}{Coming to Agreement}{to C. M. Caves \& R. Schack}{Caves37}\par
\germanism{26-11-01}{PRA Proofs}{to C. M. Caves \& R. Schack}{Caves40}\par
\germanism{30-11-01}{Some Thoughts on Your Paper(s)}{to K. Svozil}{Svozil1}\par
\germanism{05-12-01}{Dear Prudence}{to C. M. Caves \& R. Schack}{Caves41}\par
\germanism{14-12-01}{Strong Consistency}{to C. M. Caves \& R. Schack}{Caves45}\par
\germanism{15-12-01}{Bah, Humbug}{to C. M. Caves}{Caves46}\par
\germanism{04-01-02}{Once Again}{to C. M. Caves \& R. Schack}{Caves49}\par
\germanism{08-01-02}{Help with ET Quote}{to C. M. Caves \& R. Schack}{Caves50}\par
\germanism{08-01-02}{Term Origin}{to C. M. Caves \& R. Schack}{Caves51}\par
\germanism{06-02-02}{The Commitments}{to C. M. Caves \& R. Schack}{Caves54}\par
\germanism{06-02-02}{How Did?\ What Did?}{to N. D. Mermin}{Mermin58}\par
\germanism{06-02-02}{Actually, the Both of You}{to C. M. Caves \& R. Schack}{Caves55}\par
\germanism{08-02-02}{Samizdats and Dutch Book}{to B. C. van Fraassen}{vanFraassen1}\par
\germanism{03-03-02}{De Finetti and Strong Coherence}{to P. F. Scudo}{Scudo3}\par
\germanism{28-03-02}{Men in Power}{to P. Hayden}{Hayden1}\par
\germanism{10-05-02}{Compatibility}{to D. Poulin}{Poulin1}\par
\germanism{15-05-02}{Bayes, POVMs, Reality}{to A. Shimony}{Shimony1}\par
\germanism{17-05-02}{More Balking}{to R. Schack}{Schack56}\par
\germanism{28-05-02}{Strong Coherence?}{to M. C. Galavotti \& E. Regazzini}{Regazzini1}\par
\germanism{29-05-02}{Notes from the Web}{to C. M. Caves \& R. Schack}{Caves65}\par
\germanism{29-05-02}{Strict Coherence?}{to B. Skyrms}{Skyrms1}\par
\germanism{29-05-02}{More Strict Coherence}{to B. Skyrms}{Skyrms2}\par
\germanism{24-06-02}{The World is Under Construction}{to H. M. Wiseman}{Wiseman6}\par
\germanism{27-06-02}{Compatibility Never Ends}{to D. Poulin}{Poulin2}\par
\germanism{29-06-02}{Incompatible Beginnings}{to D. Poulin}{Poulin4}\par
\germanism{05-07-02}{The Physics of Floyd}{to M. J. Donald}{Donald4}\par
\germanism{17-08-03}{Tearing Off the Duct Tape}{to N. D. Mermin}{Mermin102}\par
\germanism{16-03-04}{Bayes or Bust}{to A. H. Jaffe}{Jaffe2}\par
\germanism{23-04-04}{Demonizing Bayesians}{to J. D. Norton \& J. Earman}{Earman1}\par
\germanism{15-09-04}{Certain Epiphanies}{to M. P\'erez-Su\'arez}{PerezSuarez15}\par
\germanism{30-01-06}{Island of Misfit Toys}{to K. T. McDonald}{McDonald6}\par
\germanism{21-02-06}{London Overnighter}{to R. Schack}{Schack99}\par
\germanism{27-02-06}{Dutch Book}{to D. M. Greenberger}{Greenberger3}\par
\germanism{08-03-06}{A Little Wheelerfest Report}{to H. C. von Baeyer}{Baeyer20}\par
\germanism{19-04-06}{Church of the Smaller Hilbert Space}{to M. S. Leifer}{Leifer2}\par
\germanism{24-06-06}{Notes on ``What are Quantum Probabilities''}{to J. Bub}{Bub21}\par
\germanism{07-07-06}{Markus Fierz, RIP}{to H. C. von Baeyer}{Baeyer22}\par
\germanism{07-12-06}{(Select) Replies to Referees}{to R. Schack \& C. M. Caves}{Caves93}\par
\germanism{21-01-08}{Potential Topic of Discussion}{to R. Schack}{Schack126}\par
\germanism{08-05-08}{Triple Products in Dimension 3}{to D. M. Appleby}{Appleby33}\par
\germanism{08-05-08}{Wigner Connectives}{to R. Schack}{Schack133}\par
\germanism{03-08-08}{European Tour}{to H. C. von Baeyer}{Baeyer40}\par
\germanism{27-06-09}{Acz\'el}{to P. G. L. Mana}{Mana14}\par
\germanism{29-06-09}{Disturbing the Solipsist}{to H. M. Wiseman \& E. G. Cavalcanti}{Wiseman22}\par
\germanism{09-07-09}{Trying to Make a New Start}{to C. Ferrie}{Ferrie1}\par
\germanism{13-07-09}{Articulation, 2}{to C. Ferrie}{Ferrie5}\par
\germanism{03-08-09}{Interpretation of Bayesian Probabilities}{to D. H. Wolpert}{Wolpert1}\par
\germanism{07-04-10}{Me, Me, Me Again!, 2}{to N. D. Mermin}{Mermin176}\par
\germanism{09-04-10}{QB Decoherence}{to M. Schlosshauer}{Schlosshauer33}\par
\germanism{23-08-10}{Probability}{to H. C. von Baeyer}{Baeyer126}\par

\item
\indexbullet{Indeterminism}\par
\germanism{03-10-01}{Replies on Practical Art}{to C. M. Caves \& R. Schack}{Caves30}\par
\germanism{17-05-02}{Dueling Banjos}{to W. K. Wootters}{Wootters10}\par
\germanism{17-05-02}{The Divinity that Breathes Life}{to W. K. Wootters}{Wootters11}\par
\germanism{24-06-02}{The World is Under Construction}{to H. M. Wiseman}{Wiseman6}\par
\germanism{27-06-03}{Utter Rubbish and Internal Consistency, Part I}{to R. Schack, C. M. Caves \& N. D. Mermin}{Mermin86}\par
\germanism{28-06-03}{Utter Rubbish and Internal Consistency, Part II}{to R. Schack, C. M. Caves \& N. D. Mermin}{Mermin87}\par
\germanism{01-07-03}{Objective Chance}{to W. C. Myrvold}{Myrvold1}\par
\germanism{28-07-03}{New Twenty Questions}{to R. Schack}{Schack66}\par
\germanism{12-08-03}{Me, Me, Me}{to N. D. Mermin \& R. Schack}{Mermin101}\par
\germanism{13-08-03}{Renouvier}{to R. Schack}{Schack72}\par
\germanism{23-09-03}{The Trivial Nontrivial}{to S. Savitt}{Savitt3}\par
\germanism{22-12-03}{Tail Tuckers}{to D. B. L. Baker}{Baker6}\par
\germanism{19-05-04}{The Barbecue Quest}{to G. Musser}{Musser2}\par
\germanism{10-11-04}{Even Better Than Your Talk Title}{to H. Halvorson}{Halvorson3}\par
\germanism{02-11-05}{Carry Cameras, not Guns}{to C. H. Bennett}{Bennett43}\par
\germanism{01-01-06}{Quantum Events and Propositions}{to W. G. Demopoulos}{Demopoulos4}\par
\germanism{11-01-06}{Quantum Information and KS}{to J. H. Conway}{Conway1}\par
\germanism{30-01-06}{Island of Misfit Toys}{to K. T. McDonald}{McDonald6}\par
\germanism{20-03-06}{The Rest of the Story?}{to J. E. Sipe}{Sipe9}\par
\germanism{24-06-06}{Notes on ``What are Quantum Probabilities''}{to J. Bub}{Bub21}\par
\germanism{11-09-06}{My Swerves and Yours}{to M. S. Leifer}{Leifer4}\par
\germanism{13-10-06}{Rabid Pragmatists (but in the sense of W. James, J. Dewey, \\ \fantome  and C. Fuchs)}{to R. D. Gill}{Gill5}\par
\germanism{18-10-06}{Real Possibility}{to A. Shimony}{Shimony11}\par
\germanism{16-11-06}{Challenges to the Kierkegaardian Bayesian}{to A. Shimony}{Shimony12}\par
\germanism{07-12-06}{Changes I Made To Certainty}{to R. Schack \& C. M. Caves}{Caves92}\par
\germanism{24-12-06}{Christmas Conversations in My Head}{to C. H. Bennett}{Bennett55}\par
\germanism{27-12-06}{New Year's Delirium}{to H. J. Bernstein}{Bernstein8}\par
\germanism{09-01-07}{Facts-in-Themselves}{to H. C. von Baeyer}{Baeyer27}\par
\germanism{04-02-07}{The Painful Ambiguity of Language}{to W. G. Demopoulos}{Demopoulos11}\par
\germanism{11-06-07}{Objective Indeterminism}{to \v{C}. Brukner}{Brukner2}\par
\germanism{27-09-07}{The Joint-Stock Society}{to H. R. Brown}{BrownHR3}\par
\germanism{22-12-07}{One More Comment!}{to W. C. Myrvold}{Myrvold9}\par
\germanism{15-08-08}{Free Will and Renouvier}{to R. Schack}{Schack137}\par
\germanism{24-06-09}{(again?)}{to S. Savitt}{Savitt7}\par
\germanism{07-07-09}{PI Kids Quote?}{to N. Waxman}{Waxman1}\par
\germanism{12-08-09}{More on Moore}{to M. Schlosshauer}{Schlosshauer3}\par
\germanism{15-08-09}{Roiling Mess}{to C. G. Timpson and R. W. Spekkens}{Timpson18}\par
\germanism{17-08-09}{Footprints}{to M. Schlosshauer}{Schlosshauer4}\par
\germanism{17-08-09}{Timezones (a conversation)}{to M. Schlosshauer}{Schlosshauer5}\par
\germanism{16-10-09}{The More and the Modest}{to L. Hardy}{Hardy38}\par
\germanism{17-10-09}{The More and the Modest, 3}{to N. D. Mermin}{Mermin165}\par
\germanism{25-10-09}{More Factoids}{to D. M. Appleby}{Appleby78}\par
\germanism{30-10-09}{My Interiority Complex}{to H. Barnum}{Barnum26}\par
\germanism{06-11-09}{Quotable Schopenhauer}{to M. Tait}{Tait7}\par
\germanism{09-11-09}{Vienna Indeterminism}{to A. Zeilinger}{Zeilinger4}\par
\germanism{05-01-10}{Three Attachments \ldots\ No, Four}{to M. Schlosshauer}{Schlosshauer13}\par
\germanism{22-01-10}{Is the Big Bang Here?}{to A. Zeilinger}{Zeilinger8}\par
\germanism{15-02-10}{Notes}{to M. B. Ruskai}{Ruskai3}\par
\germanism{22-02-10}{Schelling, Quantum, Creation}{to I. Ojima}{Ojima1}\par
\germanism{01-03-10}{Curiosity}{to C. Ferrie}{Ferrie13}\par
\germanism{05-03-10}{Torture}{to A. Kent}{Kent22}\par
\germanism{22-03-10}{A Line I Shall Use in Templeton}{to R. Schack}{Schack193}\par
\germanism{24-03-10}{Notes on the Open Future}{to L. Smolin}{SmolinL20}\par
\germanism{24-03-10}{Revealing}{to D. M. Appleby, R. Schack \& H. C. von Baeyer}{Appleby89}\par
\germanism{12-04-10}{Notes on a First and Second Reading}{to D. M. Appleby \& H. Barnum}{Barnum32}\par
\germanism{17-09-10}{Free Will Saved James's Life}{to the QBies}{QBies27}\par
\germanism{09-11-10}{The Universe on a Social Analogy}{to P. Cilliers}{Cilliers1}\par
\germanism{08-01-11}{Strange Einstein Remark}{to R. W. Spekkens}{Spekkens91}\par

\item
\indexbullet{The mark of external reality}\par
\germanism{03-10-01}{Replies on Practical Art}{to C. M. Caves \& R. Schack}{Caves30}\par
\germanism{10-03-06}{Isms}{to J. E. Sipe}{Sipe8}\par
\germanism{16-10-06}{Our Professor}{to R. Schack}{Schack107}\par
\germanism{09-09-08}{Panpsychism}{to D. M. Appleby}{Appleby38}

\item
\indexbullet{Pragmatist theories of truth}\par
\germanism{27-04-01}{Folse Quotes with ``Relata''}{to N. D. Mermin}{Mermin2.2}\par
\germanism{07-08-01}{Kiki, James, and Dewey}{to J. A. Waskan}{Waskan1}\par
\germanism{23-08-01}{My Own Version of a Short Note}{to C. M. Caves}{Caves8}\par
\germanism{02-09-01}{Truth and Beauty}{to N. D. Mermin}{Mermin36}\par
\germanism{04-10-01}{Replies on Pots and Kettles}{to C. M. Caves \& R. Schack}{Caves33}\par
\germanism{16-10-01}{Craters on the Moon}{to J. A. Waskan}{Waskan2}\par
\germanism{17-10-01}{Quick Single Point}{to J. A. Waskan}{Waskan3}\par
\germanism{17-10-01}{Quick Second Point}{to J. A. Waskan}{Waskan4}\par
\germanism{17-10-01}{Quick Third Point}{to J. A. Waskan}{Waskan5}\par
\germanism{04-11-01}{Dreams of an Ever-Evolving Theory}{to A. Peres}{Peres21}\par
\germanism{20-11-01}{James' Loose Lips}{to R. Schack}{Schack36}\par
\germanism{21-11-01}{Pragmatism versus Positivism}{to R. Schack}{Schack37}\par
\germanism{30-11-01}{Later in the Book}{to J. M. Renes}{Renes8}\par
\germanism{07-01-02}{Correlation without Correlata}{to N. D. Mermin}{Mermin52}\par
\germanism{08-01-02}{Information $\rightarrow$ Belief $\rightarrow$ Hope ??}{to N. D. Mermin}{Mermin54}\par
\germanism{08-01-02}{Rorty on Religion}{to J. W. Nicholson}{Nicholson3}\par
\germanism{30-01-02}{Sweet Talk}{to N. D. Mermin}{Mermin57}\par
\germanism{02-02-02}{Colleague}{to C. G. Timpson}{Timpson1}\par
\germanism{04-03-02}{Sliding Off the Deep}{to H. J. Folse}{Folse8}\par
\germanism{17-05-02}{Slide Show}{to N. D. Mermin \& C. H. Bennett}{Mermin66}\par
\germanism{24-06-02}{The World is Under Construction}{to H. M. Wiseman}{Wiseman6}\par
\germanism{27-06-02}{Probabilism All the Way Up}{to H. M. Wiseman}{Wiseman8}\par
\germanism{28-10-02}{Blather, Lather, and Rinse}{to J. Bub}{Bub8}\par
\germanism{16-01-03}{A Footnote}{to H. J. Folse}{Folse18}\par
\germanism{20-01-03}{More Pragmatism}{to H. J. Folse}{Folse19}\par
\germanism{18-03-03}{The Big IF}{to A. Sudbery \& H. Barnum}{Sudbery2}\par
\germanism{18-09-03}{Instrumentalism}{to A. Peres}{Peres55}\par
\germanism{07-11-03}{Quantum Pragmatology}{to G. Valente}{Valente3}\par
\germanism{12-02-04}{The House Philosopher}{to J. Preskill \& H. Mabuchi}{Preskill11.1}\par
\germanism{21-04-04}{Essential Incompleteness}{to W. G. Demopoulos}{Demopoulos1}\par
\germanism{17-05-04}{Jamesian Exchangeability}{to R. Schack}{Schack84}\par
\germanism{01-06-04}{Shielding the Mathematics}{to I. Pitowsky}{Pitowsky4}\par
\germanism{18-06-04}{Retracing Thoughts}{to J. Woodward}{Woodward3}\par
\germanism{06-10-04}{Incompleteness}{to H. Price}{Price1}\par
\germanism{15-11-05}{Canned Answers}{to B. C. van Fraassen}{vanFraassen12}\par
\germanism{21-11-05}{Notes to van Fraassen}{to R. Schack}{Schack88}\par
\germanism{23-11-05}{After-Shower Thought}{to R. Schack}{Schack97}\par
\germanism{24-06-06}{Notes on ``What are Quantum Probabilities''}{to J. Bub}{Bub21}\par
\germanism{25-11-06}{The Subjectivity of Convincing}{to R. W. Spekkens}{Spekkens39}\par
\germanism{10-12-07}{SICs, FQXi, \& the Way Chris Thinks}{to D. M. Appleby}{Appleby24}\par
\germanism{27-12-07}{A Little Christmas Pragmatism}{to W. G. Demopoulos, \\ \fantome  J. E. Sipe \& R. W. Spekkens}{Demopoulos20}\par
\germanism{02-09-09}{Inconsistencies}{to R. W. Spekkens}{Spekkens69}\par
\germanism{07-09-09}{Labor Day Recreations}{to R. W. Spekkens, D. M. Appleby, \\ \fantome  W. K. Wootters \& R. Blume-Kohout}{Appleby69}\par
\germanism{10-09-09}{Euler and Classical Mechanics}{to P. G. L. Mana}{Mana16}\par
\germanism{06-01-10}{13 Quotes, 2}{to M. Schlosshauer}{Schlosshauer19}\par
\germanism{24-03-10}{Notes on the Open Future}{to L. Smolin}{SmolinL20}\par
\germanism{30-03-10}{Quantum Bayesianism}{to J. Wright}{Wright1}\par
\germanism{31-03-10}{Truth and Happening}{to N. M. Boyd}{Boyd1}\par
\germanism{08-04-10}{What I Mean}{to P. G. L. Mana}{Mana17}\par
\germanism{12-04-10}{Notes on a First and Second Reading}{to D. M. Appleby \& H. Barnum}{Barnum32}\par
\germanism{04-08-10}{Help}{to C. M. Caves}{Caves105}\par
\germanism{14-09-10}{The Latest Cleanest Latest!}{to H. C. von Baeyer}{Baeyer133}\par
\germanism{21-02-11}{Einstein on Religion; Pragmatism on Spinoza}{to {\AA}. Ericsson}{Ericsson16}\par
\germanism{28-02-11}{Postmodern Smoke}{to C. M. Caves}{Caves107}\par
\germanism{07-03-11}{Einstein, Indeed!}{to R. W. Spekkens}{Spekkens105}\par
\germanism{18-04-11}{If Words Have Meaning}{to N. D. Mermin}{Mermin198}\par
\germanism{18-04-11}{If Words Have Meaning, 2}{to N. D. Mermin}{Mermin199}\par

\item
\indexbullet{Quantum measurement as metaphorically like ``working the  philosopher's stone''}\par
\germanism{07-03-01}{Pauli and Wheeler}{to H. Atmanspacher}{Atmanspacher2}\par
\germanism{05-07-01}{Standing Up and Saying YES}{to J. Finkelstein}{Finkelstein2}\par
\germanism{15-12-03}{Alchemy}{to G. Valente}{Valente5}\par
\germanism{09-03-04}{Alchemy Quote}{to M. P\'erez-Su\'arez}{PerezSuarez10}\par
\germanism{25-05-04}{Cathy and Erwin, the Movie}{to C. G. Timpson}{Timpson4}\par
\germanism{01-06-04}{Quantum System as Philosopher's Stone}{to M. P\'erez-Su\'arez}{PerezSuarez13}\par
\germanism{29-06-04}{Late, Sorry}{to G. Musser}{Musser4}\par
\germanism{19-06-05}{Philosopher's Stone}{to G. L. Comer}{Comer72}\par
\germanism{08-11-05}{Quibbles, Actions, and Reading}{to H. C. von Baeyer}{Baeyer8}\par
\germanism{10-11-05}{``Action'' instead of ``Measurement''}{to B. C. van Fraassen}{vanFraassen7}\par
\germanism{23-11-05}{The Skinny Answer}{to R. Schack}{Schack95}\par
\germanism{28-12-05}{And a Cartoon}{to H. C. von Baeyer}{Baeyer13}\par
\germanism{01-02-06}{Enabling Alchemy}{to R. E. Slusher}{Slusher12}\par
\germanism{27-02-06}{Wheelerfest}{to K. W. Ford}{Ford4}\par
\germanism{10-03-06}{Isms}{to J. E. Sipe}{Sipe8}\par
\germanism{24-06-06}{Notes on ``What are Quantum Probabilities''}{to J. Bub}{Bub21}\par
\germanism{12-07-06}{Notation for Inspiration}{to H. C. von Baeyer}{Baeyer24}\par
\germanism{20-07-06}{Capacity for Creation}{to J. E. Sipe}{Sipe10}\par
\germanism{17-11-06}{Certain Comments}{to C. G. Timpson}{Timpson11}\par
\germanism{03-01-07}{New Year's Alchemy}{to H. C. von Baeyer}{Baeyer26}\par
\germanism{24-08-07}{Are You Nuts?}{to G. L. Comer}{Comer107}\par
\germanism{08-11-07}{Uncle Fierz Needs You}{to H. C. von Baeyer}{Baeyer28}\par
\germanism{22-12-07}{One More Comment!}{to W. C. Myrvold}{Myrvold9}\par
\germanism{09-09-08}{Panpsychism}{to D. M. Appleby}{Appleby38}\par
\germanism{06-10-09}{New Orleans, RMP, and Capacities}{to N. D. Mermin}{Mermin163}\par
\germanism{12-08-09}{More on Moore}{to M. Schlosshauer}{Schlosshauer3}\par
\germanism{16-10-09}{The More and the Modest}{to L. Hardy}{Hardy38}\par
\germanism{24-10-09}{Dimension as Capacity}{to D. M. Appleby}{Appleby74}\par
\germanism{27-10-09}{Partially Sipe Inspired}{to J. E. Sipe}{Sipe19}\par
\germanism{10-12-09}{Lawless World / Malleable World / Pluriverse}{to N. Cartwright}{Cartwright1}\par
\germanism{14-12-09}{Invitation to Lecture}{to J. Emerson}{Emerson3}\par
\germanism{04-01-10}{Birth Notes}{to M. Schlosshauer}{Schlosshauer12}\par
\germanism{16-01-10}{The Tetragrammaton}{to H. C. von Baeyer and D. M. Appleby}{Appleby82}\par
\germanism{16-01-10}{Paulian Alchemy, Sunday Reading}{to A. Zeilinger}{Zeilinger7}\par
\germanism{19-01-10}{The Reason I Like Your Axiom 1}{to \v{C}. Brukner}{Brukner4}\par
\germanism{24-03-10}{Notes on the Open Future}{to L. Smolin}{SmolinL20}\par
\germanism{03-04-10}{The Quantum Bayesian Glossary -- Some Updates, 2}{to M. Schlosshauer}{Schlosshauer29}\par
\germanism{24-04-10}{Saturday Morning Alchemy}{to the QBies}{QBies17}\par
\germanism{12-06-10}{Pauli, Four, and Me}{to R. Renner}{Renner3}\par
\germanism{03-10-10}{Scaling, of a QBist Flavor}{to M. A. Graydon}{Graydon11}\par
\germanism{29-11-10}{Mermin's Cautionary Tale}{to D. M. Appleby}{Appleby97}\par
\germanism{22-01-11}{Creatia!}{to the QBies}{QBies32}\par
\germanism{29-03-11}{NY Anytime Soon?, 2}{to A. Plotnitsky}{Plotnitsky26}\par

\item
\indexbullet{Radical probabilism}\par
\germanism{04-07-01}{Invitation, 2}{to A. Y. Khrennikov}{Khrennikov4}\par
\germanism{05-07-01}{Invitation, 3}{to A. Y. Khrennikov}{Khrennikov4}\par
\germanism{07-08-01}{Knowledge, Only Knowledge}{to T. A. Brun, J. Finkelstein \& N. D. Mermin}{Mermin28}\par
\germanism{22-08-01}{Identity Crisis}{to C. M. Caves \& R. Schack}{Schack4}\par
\germanism{02-09-01}{Intersubjective Agreement}{to N. D. Mermin}{Mermin35}\par
\germanism{04-09-01}{Objective Probability}{to C. M. Caves \& R. Schack}{Schack6}\par
\germanism{04-01-02}{New Breach of Faith}{to C. M. Caves}{Caves48}\par
\germanism{08-01-02}{Help with ET Quote}{to C. M. Caves \& R. Schack}{Caves50}\par
\germanism{06-03-02}{Poetry on Concrete}{to L. Hardy}{Hardy7}\par
\germanism{29-05-02}{Notes from the Web}{to C. M. Caves \& R. Schack}{Caves65}\par
\germanism{26-06-02}{The One Boolean Algebra Approach}{to I. Pitowsky}{Pitowsky1}\par
\germanism{05-07-02}{The Physics of Floyd}{to M. J. Donald}{Donald4}\par
\germanism{19-06-03}{Guidelines}{to M. P\'erez-Su\'arez}{PerezSuarez1}\par
\germanism{20-06-03}{Logical Probability and That}{to J. M. Renes}{Renes17}\par
\germanism{28-06-03}{Utter Rubbish and Internal Consistency, Part II}{to R. Schack, C. M. Caves \& N. D. Mermin}{Mermin87}\par
\germanism{30-06-03}{Probabilistic Dialogue}{to R. W. Spekkens}{Spekkens16}\par
\germanism{25-07-03}{Relative Onticity}{to R. Schack}{Schack64}\par
\germanism{17-08-03}{Tearing Off the Duct Tape}{to N. D. Mermin}{Mermin102}\par
\germanism{02-12-03}{Paper and Visit}{to R. Schack}{Schack76}\par
\germanism{27-04-04}{Jeffrey Knew Certainty}{to R. Schack}{Schack82}\par
\germanism{29-04-04}{Ontic Elements for Quantum Systems}{to H. Atmanspacher}{Atmanspacher3}\par
\germanism{13-05-04}{10 Lines and MaxEnt}{to R. Schack}{Schack83}\par
\germanism{18-05-04}{Agents, Interventions, and Surgical Removal}{to J. Woodward}{Woodward2}\par
\germanism{01-12-04}{BBQW Responses --- Progress Report, as I near the Pacific}{to S. Hartmann, \\ \fantome  C. M. Caves \& R. Schack}{Schack86.1}\par
\germanism{30-06-05}{Wittgensteinian Arm Twisting}{to H. Price}{Price5}\par
\germanism{12-08-05}{Quantum Bayesians and Anti-Bayesians}{to W. C. Myrvold}{Myrvold3}\par
\germanism{01-09-05}{BBQW Report}{to S. Hartmann}{Hartmann12}\par
\germanism{23-09-05}{Bayesians at FPP?}{to A. Y. Khrennikov}{Khrennikov11}\par
\germanism{15-11-05}{Canned Answers}{to B. C. van Fraassen}{vanFraassen12}\par
\germanism{31-12-05}{If We Make It Through December \ldots}{to G. Musser}{Musser20}\par
\germanism{10-01-07}{Anti-Algebra, the Reprise}{to V. Palge}{Palge3}\par
\germanism{03-02-07}{Bill's Thoughts on QL and QI Frameworks}{to W. G. Demopoulos}{Demopoulos9}\par
\germanism{24-06-06}{Notes on ``What are Quantum Probabilities''}{to J. Bub}{Bub21}\par
\germanism{26-07-06}{Another Question}{to J. E. Sipe}{Sipe11}\par
\germanism{08-11-06}{November 8th}{to N. D. Mermin}{Mermin124}\par
\germanism{17-11-06}{Certain Comments}{to C. G. Timpson}{Timpson11}\par
\germanism{25-11-06}{The Subjectivity of Convincing}{to R. W. Spekkens}{Spekkens39}\par
\germanism{06-07-09}{Refeeding Quantum Mechanics}{to B. C. van Fraassen}{vanFraassen21}\par
\germanism{09-07-09}{Trying to Make a New Start}{to C. Ferrie}{Ferrie1}\par
\germanism{25-09-09}{Why Such Radical Moves?}{to R. W. Spekkens}{Spekkens73}\par
\germanism{02-11-09}{One Good Rant Deserves Another}{to M. Schlosshauer}{Schlosshauer8}\par

\item
\indexbullet{Subjectivity of quantum states}\par
\germanism{07-08-01}{Knowledge, Only Knowledge}{to T. A. Brun, J. Finkelstein \& N. D. Mermin}{Mermin28}\par
\germanism{22-08-01}{Identity Crisis}{to C. M. Caves \& R. Schack}{Schack4}\par
\germanism{02-09-01}{Intersubjective Agreement}{to N. D. Mermin}{Mermin35}\par
\germanism{03-10-01}{Kid Sheleen}{to N. D. Mermin}{Mermin45}\par
\germanism{03-10-01}{Replies on Practical Art}{to C. M. Caves \& R. Schack}{Caves30}\par
\germanism{29-04-04}{Ontic Elements for Quantum Systems}{to H. Atmanspacher}{Atmanspacher3}\par
\germanism{13-05-04}{Epistemic Pure States}{to H. Atmanspacher}{Atmanspacher4}\par
\germanism{01-09-05}{BBQW Report}{to S. Hartmann}{Hartmann12}\par
\germanism{10-03-06}{Isms}{to J. E. Sipe}{Sipe8}\par
\germanism{16-10-06}{Our Professor}{to R. Schack}{Schack107}\par
\germanism{09-09-08}{Panpsychism}{to D. M. Appleby}{Appleby38}

\item
\indexbullet{Subjectivity of quantum operations}\par
\germanism{27-06-01}{Gleasons, and Me at AT\&T}{to H. Barnum}{Barnum3}\par
\germanism{10-07-01}{My Recent Foundations Posting}{to J. D. Malley}{Malley1}\par
\germanism{23-07-01}{The Principal PrincipleS}{to C. M. Caves}{Caves2}\par
\germanism{08-08-01}{The First Eye}{to C. M. Caves}{Caves3}\par
\germanism{04-09-01}{Note on Terminology}{to C. M. Caves \& R. Schack}{Schack5}\par
\germanism{03-10-01}{Kid Sheleen}{to N. D. Mermin}{Mermin45}\par
\germanism{03-10-01}{Replies on Practical Art}{to C. M. Caves \& R. Schack}{Caves30}\par
\germanism{04-10-01}{Replies on Pots and Kettles}{to C. M. Caves \& R. Schack}{Caves33}\par
\germanism{20-11-01}{James's Loose Lips}{to R. Schack}{Schack36}\par
\germanism{29-11-01}{Community}{to W. K. Wootters}{Wootters3}\par
\germanism{29-01-02}{{\Vaxjo} Contributions}{to C. M. Caves}{Caves52}\par
\germanism{06-02-02}{The Great Quantum Well}{to C. M. Caves}{Caves57}\par
\germanism{19-06-03}{Lurch}{to J. M. Renes}{Renes16}\par
\germanism{25-02-04}{Promised Message}{to N. D. Mermin}{Mermin110}\par
\germanism{29-04-04}{Ontic Elements for Quantum Systems}{to H. Atmanspacher}{Atmanspacher3}\par
\germanism{21-11-05}{Pull!}{to R. Schack}{Schack87}\par
\germanism{21-11-06}{The Self-Preservation Society}{to M. S. Leifer}{Leifer6}\par
\germanism{21-11-06}{The Self-Preservation Society, 2}{to M. S. Leifer}{Leifer7}\par
\germanism{22-11-06}{The Self-Preservation Society, 3}{to M. S. Leifer}{Leifer8}\par
\germanism{22-11-06}{CFS Philosophy, Ontic and Epistemic States and Evolutions,\\ \fantome  Funes the Memorious}{to C. H. Bennett}{Bennett48}\par
\germanism{25-11-06}{Hoffa and the Bayesians}{to C. H. Bennett}{Bennett49}\par
\germanism{25-11-06}{The Subjectivity of Convincing}{to R. W. Spekkens}{Spekkens39}\par
\germanism{10-01-07}{Anti-Algebra, the Reprise}{to V. Palge}{Palge3}\par
\germanism{18-10-07}{Noncolorable Histories}{to N. D. Mermin}{Mermin134}\par
\germanism{18-03-08}{Needing a Little Jung Myself}{to D. M. Appleby}{Appleby29}\par
\germanism{04-11-09}{QK, QB, QR -- First Read}{to H. Barnum}{Barnum27}\par
\germanism{04-11-09}{QK, QB, QR -- First Read, 2}{to H. Barnum, D. M. Appleby \& M. Tait}{Barnum28}\par
\germanism{17-02-10}{Strong Radio Silence}{to M. Schlosshauer}{Schlosshauer23}\par
\germanism{02-03-10}{Fatherhood/Paperhood}{to M. Schlosshauer}{Schlosshauer25}\par
\germanism{06-05-10}{{\Vaxjo} Abstract, 2}{to M. Schlosshauer}{Schlosshauer37}\par
\germanism{08-01-11}{Strange Einstein Remark, 3}{to R. W. Spekkens}{Spekkens94}\par

\item
\indexbullet{Objectivity of the quantum formalism}\par
\germanism{02-09-01}{Intersubjective Agreement}{to N. D. Mermin}{Mermin35}\par
\germanism{03-10-01}{Kid Sheleen}{to N. D. Mermin}{Mermin45}\par
\germanism{03-10-01}{Replies on Practical Art}{to C. M. Caves \& R. Schack}{Caves30}\par
\germanism{27-10-05}{It Wasn't You}{to H. C. von Baeyer}{Baeyer3}\par
\germanism{01-01-06}{Quantum Events and Propositions}{to W. G. Demopoulos}{Demopoulos4}\par
\germanism{08-11-06}{November 8th}{to N. D. Mermin}{Mermin124}\par
\germanism{16-11-06}{Challenges to the Kierkegaardian Bayesian}{to A. Shimony}{Shimony12}\par
\germanism{10-01-07}{Anti-Algebra, the Reprise}{to V. Palge}{Palge3}\par
\germanism{09-09-08}{Panpsychism}{to D. M. Appleby}{Appleby38}

\item
\indexbullet{Malleability of reality}\par
\germanism{10-07-01}{Replies on a Preskillian Meeting}{to A. J. Landahl}{Landahl1}\par
\germanism{05-09-01}{Malleable World}{to G. L. Comer}{Comer5}\par
\germanism{20-09-01}{Praise, Folly, Enthusiasm}{to W. K. Wootters}{Wootters2}\par
\germanism{04-10-01}{Replies on Pots and Kettles}{to C. M. Caves \& R. Schack}{Caves33}\par
\germanism{17-05-02}{No Nasty}{to T. Rudolph}{Rudolph4}\par
\germanism{28-05-02}{Anti-Anti-V\"axjination}{to J. Summhammer}{Summhammer4}\par
\germanism{27-06-02}{Probabilism All the Way Up}{to H. M. Wiseman}{Wiseman8}\par
\germanism{13-11-03}{Elephants}{to L. Hardy \& F. Girelli}{Hardy13}\par
\germanism{14-06-04}{RMTTR,BOWEAL}{to H. Mabuchi}{Mabuchi10}\par
\germanism{21-03-05}{Changing the World}{to C. H. Bennett}{Bennett39}\par
\germanism{07-04-05}{Recovery of Philosophy}{to J. E. Sipe}{Sipe4}\par
\germanism{23-08-05}{The Article for \underline{Science \& Vie}}{to H. Poirier}{Poirier2}\par
\germanism{27-03-06}{Block of Ice-Nine}{to C. M. Caves}{Caves83}\par
\germanism{01-05-06}{Conversations with God}{to W. B. Case}{Case2}\par
\germanism{11-02-08}{Anti-Block Meeting Invitees (working list)}{to L. Smolin}{SmolinL11}\par
\germanism{13-05-08}{Misak Tuesday}{to J. E. Sipe}{Sipe18}\par
\germanism{22-08-08}{Possible Collaboration?, 2}{to T. Duncan}{Duncan5}\par
\germanism{05-01-09}{What I Really Want Out of a Pauli/Fierz-Correspondence \\ \fantome  Study}{to H. C. von Baeyer \& D. M. Appleby}{PauliFierzCorrespondence}\par
\germanism{06-04-09}{A History-of-Knowledge Thing}{to D. P. DiVincenzo \& C. H. Bennett}{DiVincenzo4}\par
\germanism{21-06-09}{Rainy Day Train}{to C. Eriksson}{Eriksson2}\par
\germanism{06-07-09}{The Verdict}{to H. M. Wiseman}{Wiseman27}\par
\germanism{07-09-09}{Labor Day Recreations}{to R. W. Spekkens, D. M. Appleby, \\ \fantome  W. K. Wootters \& R. Blume-Kohout}{Appleby69}\par
\germanism{22-10-09}{QBism House Draft}{to N. Waxman}{Waxman4}\par
\germanism{23-11-09}{My Visit, James, Lotze, and the Malleable World}{to A. Zeilinger}{Zeilinger6}\par
\germanism{19-01-10}{Translation}{to H. C. von Baeyer}{Baeyer100}\par
\germanism{24-03-10}{Notes on the Open Future}{to L. Smolin}{SmolinL20}\par
\germanism{02-03-11}{New Scientist? \ldots\ New Sensationalist, Maybe!}{to H. C. von Baeyer}{Baeyer142}\par

\item
\indexbullet{John Archibald {\Wheeler}}\par
\germanism{23-01-01}{It's Really Real}{to the {\Vaxjo} Invitees}{VaxjoIsReallyReal}\par
\germanism{19-01-01}{Industrial Boys}{to A. Kent}{Kent1}\par
\germanism{07-02-01}{Wow!}{to A. Plotnitsky}{Plotnitsky2}\par
\germanism{06-03-01}{Pauli-ish Compendium}{to H. Atmanspacher}{Atmanspacher1}\par
\germanism{07-03-01}{Pauli and Wheeler}{to H. Atmanspacher}{Atmanspacher2}\par
\germanism{13-03-01}{FYI -- The History Channel}{to S. K. Stoll}{Stoll1}\par
\germanism{28-05-01}{Letters: the Long and the Short}{to A. Plotnitsky}{Plotnitsky3}\par
\germanism{12-06-01}{Something I Wrote Once}{to R. Obajtek}{Obajtek1}\par
\germanism{28-06-01}{Context Dependent Probability}{to A. Y. Khrennikov}{Khrennikov3}\par
\germanism{10-07-01}{$V^2 = U$}{to S. Aaronson}{Aaronson2}\par
\germanism{23-07-01}{Law without Law}{to J. Summhammer}{Summhammer1}\par
\germanism{31-07-01}{Parachutes}{to C. H. Bennett}{Bennett3}\par
\germanism{07-08-01}{Parachutes}{to J. Preskill}{Preskill3}\par
\germanism{18-09-01}{Hi Back}{to A. Peres}{Peres17}\par
\germanism{29-11-01}{Community}{to W. K. Wootters}{Wootters3}\par
\germanism{05-12-01}{Lucky Seven}{to B. W. Schumacher}{Schumacher4}\par
\germanism{06-12-01}{Baudrillard Maybe?}{to B. W. Schumacher}{Schumacher5}\par
\germanism{23-01-02}{Book Review of Nielsen \& Chuang's Book for American \\ \fantome  Journal of Physics}{to L. K. Grover}{Grover1}\par
\germanism{29-01-02}{Quotes That Moved Me Once}{to J. W. Nicholson}{Nicholson6}\par
\germanism{30-01-02}{Sweet Talk}{to N. D. Mermin}{Mermin57}\par
\germanism{15-02-02}{Friendship Call}{to L. Hardy}{Hardy3}\par
\germanism{21-02-02}{Quantum Fest, Bell Labs Style?}{to H. Mabuchi}{Mabuchi3}\par
\germanism{12-03-02}{Wheeler Link}{to R. Pike}{Pike6}\par
\germanism{13-04-02}{An Address}{to L. Hardy}{Hardy8}\par
\germanism{21-04-02}{Walton's Mountain}{to G. L. Comer}{Comer11}\par
\germanism{07-05-02}{This Tape Will Self Destruct}{to G. Plunk \& N. D. Mermin}{Plunk2}\par
\germanism{15-05-02}{And Only a Little More}{to W. K. Wootters}{Wootters8}\par
\germanism{16-05-02}{Exterior/Interior}{to W. K. Wootters}{Wootters9}\par
\germanism{17-05-02}{Dueling Banjos}{to W. K. Wootters}{Wootters10}\par
\germanism{01-06-02}{Wheeler Compendium}{to K. W. Ford}{Ford1}\par
\germanism{04-06-02}{Big File Coming}{to K. W. Ford}{Ford2}\par
\germanism{05-01-03}{ForAsher.tex}{to A. Peres}{Peres47}\par
\germanism{17-05-03}{Good News, Bad News}{to J. W. Nicholson}{Nicholson17}\par
\germanism{18-07-03}{Solipsism Concerns}{to N. D. Mermin}{Mermin97}\par
\germanism{23-09-03}{The Trivial Nontrivial}{to S. Savitt}{Savitt3}\par
\germanism{15-11-03}{Elementary Quantum Phenomena}{to R. Laflamme}{Laflamme1}\par
\germanism{19-12-03}{Time to Think about Time}{to G. L. Comer}{Comer47}\par
\germanism{26-02-04}{Thanks}{to A. Peres}{Peres61}\par
\germanism{29-04-04}{Turgidity}{to H. M. Wiseman}{Wiseman14}\par
\germanism{06-07-04}{A Little More}{to G. Musser}{Musser7}\par
\germanism{07-07-04}{A}{to G. Musser}{Musser9}\par
\germanism{07-07-04}{B}{to G. Musser}{Musser11}\par
\germanism{12-08-04}{Background Noise}{to G. Musser}{Musser13}\par
\germanism{12-10-04}{Blowing in the Wind}{to G. L. Comer}{Comer55}\par
\germanism{11-01-05}{Wheeler, Bayes, Schleich}{to W. P. Schleich}{Schleich1}\par
\germanism{16-03-05}{Stein}{to A. Radosz}{Radosz1}\par
\germanism{01-07-05}{Bell, Buddha, Pauli}{to M. O. Scully}{Scully2}\par
\germanism{10-11-05}{Kaine, Corzine, Kochen, Krazy Kats and All That}{to H. C. von Baeyer}{Baeyer10}\par
\germanism{10-11-05}{Wheeler's 20 Questions and Nordheim}{to B. C. van Fraassen}{vanFraassen8}\par
\germanism{02-12-05}{Princeton Quantum Informatics Conf}{to M. O. Scully}{Scully7}\par
\germanism{28-12-05}{Wheeler and the Pleasures of Life}{to D. Overbye}{Overbye3}\par
\germanism{28-12-05}{Wheeler and the Pleasures of Life, 2}{to D. Overbye}{Overbye4}\par
\germanism{31-12-05}{Wheeler Meeting at Princeton}{to W. G. Unruh and several others}{Unruh6}\par
\germanism{17-01-06}{Wheeler Meeting at Princeton}{to several at Princeton University}{Lieb1}\par
\germanism{17-01-06}{Wheeler Meeting at Princeton}{to K. T. McDonald}{McDonald3}\par
\germanism{18-01-06}{Wheeler Meeting at Princeton}{to several at Princeton University}{Lieb2}\par
\germanism{19-01-06}{The Wheeler Meeting}{to D. B. L. Baker}{Baker13}\par
\germanism{27-01-06}{Wheeler Quantum Information Meeting, \\ \fantome  February 24--25}{to the Wheelerfest participants}{WheelerfestInvite}\par
\germanism{30-01-06}{Loose Ends and Pedagogy}{to K. T. McDonald}{McDonald4}\par
\germanism{30-01-06}{Wheeler Accuracy?}{to J. Preskill}{Preskill16}\par
\germanism{16-02-06}{Back from the Silence}{to H. C. von Baeyer}{Baeyer17}\par
\germanism{22-02-06}{Wheelerfest}{to D. Overbye}{Overbye5}\par
\germanism{22-02-06}{Wheelerfest, 2}{to D. Overbye}{Overbye6}\par
\germanism{23-02-06}{John Wheeler's Visit}{to L. Wheeler Ufford}{Ufford1}\par
\germanism{27-02-06}{Wheelerfest, Thanks}{to M. O. Scully}{Scully8}\par
\germanism{27-02-06}{Wheelerfest}{to K. W. Ford}{Ford4}\par
\germanism{27-02-06}{Wheeler}{to W. P. Schleich}{Schleich2}\par
\germanism{28-02-06}{Stupider Chris}{to T. Rudolph}{Rudolph11}\par
\germanism{08-03-06}{A Little Wheelerfest Report}{to H. C. von Baeyer}{Baeyer20}\par
\germanism{20-03-06}{The Rest of the Story?}{to J. E. Sipe}{Sipe9}\par
\germanism{07-12-06}{Changes I Made To Certainty}{to R. Schack \& C. M. Caves}{Caves92}\par
\germanism{27-12-06}{Bayes, Born, and Everett}{to F. J. Tipler}{Tipler1}\par
\germanism{31-01-07}{Nature Giving a Flip}{to C. A. Fuchs}{FuchsC16}\par
\germanism{15-03-07}{Attributed to Einstein, Incorrectly?}{to G. Will}{Will1}\par
\germanism{15-03-07}{Attributed to Einstein, Incorrectly?, 2}{to C. M. Caves}{Caves96}\par
\germanism{12-06-07}{Quantum Chris}{to M. Tegmark}{Tegmark2}\par
\germanism{07-09-07}{That Wheeler Quote}{to M. A. Nielsen}{Nielsen7}\par
\germanism{18-09-07}{Shakespeare in Sweden}{to P. G. L. Mana}{Mana8}\par
\germanism{01-10-07}{Shakespeare in Sweden, 2}{to P. G. L. Mana}{Mana9}\par
\germanism{12-12-07}{For Lane Hughston}{to C. A. Fuchs}{FuchsC18}\par
\germanism{14-04-08}{Your Dad}{to L. Wheeler Ufford}{Ufford2}\par
\germanism{14-04-08}{John Wheeler's Death}{to G. L. Comer}{Comer113}\par
\germanism{14-04-08}{Sad Day}{to D. Overbye}{Overbye7}\par
\germanism{14-04-08}{Times and the End of Time}{to W. P. Schleich}{Schleich3}\par
\germanism{24-04-08}{Topos Theory Today}{to L. Hardy}{Hardy27}\par
\germanism{17-05-08}{Springtime in the Air}{to G. Musser}{Musser21}\par
\germanism{21-05-08}{John Wheeler}{to S. K. Stoll}{Stoll3}\par
\germanism{14-07-08}{Wheeler Scholars}{to W. P. Schleich, B. W. Schumacher \& W. K. Wootters}{Schleich4}\par
\germanism{11-08-08}{The Two Freedoms}{to H. C. von Baeyer}{Baeyer41}\par
\germanism{10-10-08}{Block U Alternative}{to T. Duncan}{Duncan6}\par
\germanism{20-01-09}{From Pragmatism to Pure Experience}{to M. B\"achtold}{Baechtold2}\par
\germanism{22-02-09}{The Shape of Hilbert Space}{to G. L. Comer}{Comer122}\par
\germanism{06-04-09}{A History-of-Knowledge Thing}{to D. P. DiVincenzo \& C. H. Bennett}{DiVincenzo4}\par
\germanism{07-04-09}{Your Thoughts}{to N. Bao}{Bao1}\par
\germanism{08-04-09}{Media Interview for FQXi about Hugh Everett}{to G. Stemp-Morlock}{StempMorlock1}\par
\germanism{08-04-09}{AJP Resource Letter}{to R. W. Spekkens}{Spekkens63.1.1}\par
\germanism{29-05-09}{The Elusive Nature of the Quantum State}{to G. J. Milburn}{Milburn4}\par
\germanism{13-06-09}{Update}{to D. M. Appleby}{Appleby63}\par
\germanism{16-06-09}{Latter-Day Wheeler}{to W. H. Zurek}{Zurek1}\par
\germanism{22-06-09}{More Thoughts}{to D. M. Appleby, H. C. von Baeyer and R. Schack}{Appleby65}\par
\germanism{23-06-09}{Fragments}{to D. M. Appleby and H. C. von Baeyer}{Appleby66}\par
\germanism{25-06-09}{Down with Unitarity}{to T. Rudolph, cc J. Rau}{Rudolph12}\par
\germanism{31-07-09}{Shelter Island and Bennett, Preskill, etc.\ Stories}{to D. Gottesman}{Gottesman13}\par
\germanism{20-08-09}{More on the Cover Story}{to S. Capelin}{Capelin9}\par
\germanism{25-09-09}{More On Normativity and a Quantitative Measure on Degrees of\\ \fantome  Detached Observers}{to C. G. Timpson}{Timpson19}\par
\germanism{25-10-09}{Metaphysical Club}{to D. M. Appleby}{Appleby76}\par
\germanism{20-11-09}{Wheeler on Quine}{to R. Healey}{Healey8}\par
\germanism{22-01-10}{Is the Big Bang Here?}{to A. Zeilinger}{Zeilinger8}\par
\germanism{22-02-10}{Schelling, Quantum, Creation}{to I. Ojima}{Ojima1}\par
\germanism{02-03-10}{Wheelerism Full-Throttle!}{to B. W. Schumacher}{Schumacher17}\par
\germanism{18-03-10}{Very Final Version}{to R. W. Spekkens}{Spekkens80}\par
\germanism{24-03-10}{Notes on the Open Future}{to L. Smolin}{SmolinL20}\par
\germanism{01-04-10}{Explanatory Note}{to M. A. Graydon}{Graydon6}\par
\germanism{08-04-10}{UNC Visit}{to Y. J. Ng}{Ng6}\par
\germanism{09-04-10}{QB Decoherence}{to M. Schlosshauer}{Schlosshauer33}\par
\germanism{13-04-10}{Possibly Final Version of Section 2}{to H. C. von Baeyer}{Baeyer114}\par
\germanism{17-04-10}{World Elements}{to M. Schlosshauer}{Schlosshauer34}\par
\germanism{20-04-10}{My Itinerary for DC}{to R. W. Spekkens}{Spekkens84}\par
\germanism{21-04-10}{World Elements, 2}{to M. Schlosshauer}{Schlosshauer35}\par
\germanism{24-04-10}{Weyl's Book and John Wheeler}{to the QBies}{QBies16}\par
\germanism{24-04-10}{Weyl's Book and John Wheeler, 2}{to the QBies}{QBies18}\par
\germanism{05-06-10}{A Question on Informational Gravitation}{to H. Poirier}{Poirier3}\par
\germanism{05-06-10}{The Consequences of Tommy}{to the QBies}{QBies22}\par
\germanism{07-06-10}{Time Flies}{to J. D. Norton}{Norton2}\par
\germanism{15-07-10}{Wheeler's 20 Questions}{to P. Wells}{Wells4}\par
\germanism{03-09-10}{The Strip Tease}{to H. C. von Baeyer}{Baeyer129}\par
\germanism{14-09-10}{From Coleman to Cuero}{to D. C. Lamberth}{Lamberth6}\par
\germanism{14-09-10}{The James Meeting}{to R. Schack}{Schack205}\par
\germanism{16-09-10}{The Final Submission}{to H. C. von Baeyer}{Baeyer134}\par
\germanism{12-10-10}{The APS March Meeting}{to G. Chiribella}{Chiribella2}\par
\germanism{02-11-10}{Probable Delay}{to R. Schack}{Schack214}\par
\germanism{10-12-10}{Texan Roots}{to M. E. L. Oakes}{Oakes1}\par
\germanism{09-01-11}{One Endorsement}{to M. Schlosshauer}{Schlosshauer41}\par
\germanism{14-01-11}{Great Incendiary Fun}{to I. T. Durham}{Durham4}\par
\germanism{22-01-11}{Creatia!}{to the QBies}{QBies32}\par
\germanism{07-02-11}{More QB Silliness}{to the QBies}{QBies35}\par
\germanism{17-02-11}{Symmetries of the Magic Measurement}{to J. D. Bekenstein}{Bekenstein1}\par
\germanism{24-02-11}{The Manic and the Depressive}{to H. C. von Baeyer}{Baeyer140}\par
\germanism{26-02-11}{Feelings of Guilt}{to the QBies}{QBies35.1}\par
\germanism{28-03-11}{Statement of Research}{to Correspondent Y}{CorrespondentY1}\par
\germanism{02-05-11}{A JTF William James Center?}{to H. S. Choi}{Choi1}\par

\item
\indexbullet{Wigner's friend}\par
\germanism{31-12-04}{New Year's Eve}{to A. Peres}{Peres67}\par
\germanism{10-11-05}{Our Own Rovellian Analysis}{to B. C. van Fraassen}{vanFraassen9}\par
\germanism{16-11-05}{Your Phrase}{to B. C. van Fraassen}{vanFraassen13}\par
\germanism{21-11-05}{Pull!}{to R. Schack}{Schack87}\par
\germanism{21-11-05}{Notes to van Fraassen}{to R. Schack}{Schack88}\par
\germanism{21-11-05}{Seeing the Light}{to R. Schack}{Schack89}\par
\germanism{22-11-05}{From the Poetaster}{to R. Schack}{Schack91}\par
\germanism{23-11-05}{The Skinny Answer}{to R. Schack}{Schack95}\par
\germanism{23-11-05}{Wigner's Impotence}{to R. Schack}{Schack96}\par
\germanism{27-12-05}{White Christmas}{to W. G. Demopoulos}{Demopoulos3}\par
\germanism{24-03-06}{On Certainty, Quantum Outcomes, Subjectivity, Objectivity, \\ \fantome  and Expanding Universes, Part 1}{to R. Schack \& C. M. Caves}{Caves82}\par
\germanism{16-10-06}{Our Professor}{to R. Schack}{Schack107}\par
\germanism{06-12-06}{Fourth and Fifth}{to J. Barrett}{Barrett1}\par
\germanism{02-08-07}{The One-Belly Theory of the Universe}{to S. J. van Enk}{vanEnk12}\par
\germanism{22-12-07}{One More Comment!}{to W. C. Myrvold}{Myrvold9}\par
\germanism{03-08-08}{European Tour}{to H. C. von Baeyer}{Baeyer40}\par
\germanism{20-08-08}{Progress}{to E. G. Cavalcanti}{Cavalcanti2}\par
\germanism{03-09-08}{Two (Important!) Things}{to C. Rovelli}{Rovelli1}\par
\germanism{24-09-09}{Timpson's Talk on Sept.\ 25}{to the QBies}{QBies3}\par
\germanism{25-09-09}{Why Such Radical Moves?}{to R. W. Spekkens}{Spekkens73}\par
\germanism{25-09-09}{More On Normativity and a Quantitative Measure on Degrees of\\ \fantome  Detached Observers}{to C. G. Timpson}{Timpson19}\par
\germanism{05-01-10}{Importance of Humanism and Dogs}{to M. Schlosshauer}{Schlosshauer15}\par
\germanism{17-01-10}{Oh Translator, 2}{to H. C. von Baeyer}{Baeyer99}\par
\germanism{17-02-10}{Strong Radio Silence}{to M. Schlosshauer}{Schlosshauer23}\par
\germanism{17-02-10}{Oh, Repository}{to R. Schack}{Schack189}\par
\germanism{04-03-10}{Fuchsianism / Ones For Which}{to R. Blume-Kohout}{BlumeKohout11}\par
\germanism{22-03-10}{Pong}{to R. Schack}{Schack192}\par
\germanism{30-03-10}{Qbism Questions}{to C. Ududec}{Ududec2}\par
\germanism{08-04-10}{Me, Me, Me Again!, 3}{to N. D. Mermin}{Mermin177}\par
\germanism{12-04-10}{Notes on a First and Second Reading}{to D. M. Appleby and H. Barnum}{Barnum32}\par
\germanism{03-06-10}{Padmanabhan, 2}{to L. Smolin}{SmolinL22}\par
\germanism{12-09-10}{Interpreting the Universe after a Social Analogy}{to D. C. Lamberth}{Lamberth4}\par
\germanism{01-02-11}{How To Remove Yourself from Someone's Xmas Card List}{to M. S. Leifer}{Leifer13}\par
\germanism{03-03-11}{Normalizing Fiducial Stability Groups}{to M. A. Graydon}{Graydon16}\par
\germanism{04-04-11}{Les Papillons}{to H. C. von Baeyer}{Baeyer143}\par

\end{itemize}

\pagebreak

\vspace*{1.5in}

\phantomsection

\begin{center}
\Large {\bf Index of Technical Points and Questions}
\end{center}

\addcontentsline{toc}{chapter}{Index of Technical Points and Questions}

\begin{itemize}
\item \indexbullet{Axioms for quantum mechanics}\par
\noindent Traditional\par
\germanism{10-05-01}{The Easy Part}{to N. D. Mermin}{Mermin6}\par
\noindent Why complex Hilbert spaces?\par
\germanism{06-07-01}{Stamina!}{to S. Aaronson}{Aaronson1}\par
\germanism{10-07-01}{$V^2 = U$}{to S. Aaronson}{Aaronson2}\par
\noindent   Deriving the tensor-product rule\par
\germanism{28-08-01}{Introduction}{to A. Wilce}{Wilce1}\par
\noindent   A Bayesian's version of Hardy's convex-linearity\par
\germanism{06-03-02}{Poetry on Concrete}{to L. Hardy}{Hardy7}\par
\noindent To motivate von Neumann entropy\par
\germanism{28-03-02}{Men in Power}{to P. Hayden}{Hayden1}\par
\noindent   Why the von Neumann measurement is an arbitrary ideal\par
\germanism{29-05-02}{That Damned von Neumann}{to N. D. Mermin}{Mermin68}\par
\noindent   Clifton--Bub--Halvorson\par
\germanism{12-01-06}{Sounds of Silence and CBH, 2}{to G. Brassard}{Brassard49}\par
\germanism{15-08-06}{Drink the Kool-Aid}{to C. H. Bennett}{Bennett45}\par
\noindent   Different axioms which Hardy could have chosen\par
\germanism{27-01-06}{Monday or Tuesday Meeting}{to H. Halvorson}{Halvorson9.1}\par
\noindent   Axiomatic approaches to distinguishability measures\par
\germanism{18-09-06}{Another Topic}{to M. S. Leifer}{Leifer5}\par
\noindent   Operational axioms for SICs\par
\germanism{05-10-07}{Start of an Answer}{to P. Goyal}{Goyal2}\par
\noindent   SIC existence as an axiom\par
\germanism{13-03-10}{Conceptual Barrier!}{to the QBies}{QBies6}\par
\noindent Tomographic locality\par
\germanism{28-04-10}{Stueckelberg??}{to L. Hardy}{Hardy40}\par
\noindent   Hybrid assumptions for rederiving QM\par
\germanism{05-10-10}{Galoshes}{to M. A. Graydon}{Graydon12}\par
\germanism{19-10-10}{Intersections}{to L. Hardy}{Hardy42.1}

\item \indexbullet{Bell--Kochen--Specker theorems}\par
\noindent Origin of the GHZ gedankenexperiment\par
\germanism{07-02-01}{GHZM}{to N. D. Mermin}{Mermin-3}\par
\noindent Appleby's response to Meyer--Kent--Clifton ``nullification''\par
\germanism{04-10-01}{Replies to a Conglomeration}{to C. M. Caves \& R. Schack}{Caves35}\par
\noindent   And consistent histories\par
\germanism{19-09-02}{Unsurprising Fact}{to N. D. Mermin}{Mermin74}\par
\germanism{18-10-07}{Copyrights?}{to N. D. Mermin}{Mermin133}\par
\germanism{18-10-07}{Noncolorable Histories}{to N. D. Mermin}{Mermin134}\par
\noindent   And rational vector spaces\par
\germanism{13-12-02}{More Addenda}{to A. Peres}{Peres44}\par
\germanism{01-11-04}{Rational Hilbert Space}{to S. Aaronson}{Aaronson6}\par
\noindent   Contextuality and classicality in $d = 2$\par
\germanism{04-11-03}{Slow Draw McGraw}{to R. W. Spekkens}{Spekkens22}\par
\noindent And intermediate Gleason theorems\par
\germanism{02-12-05}{Gleason Frames}{to I. Bengtsson}{Bengtsson1}\par
\noindent Specker's motivation\par
\germanism{15-05-08}{1:30 AM Note -- Decompressing the Day}{to D. M. Appleby}{Appleby34}\par
\noindent And cryptographic protocols\par
\germanism{17-10-10}{QBlue's Fall Quantum Information Prize}{to the QBies}{QBies30}\par

\item \indexbullet{Born Rule}\par
\noindent   As an empirical addition to Dutch-book coherence\par
\germanism{21-01-08}{Potential Topic of Discussion}{to R. Schack}{Schack126}\par
\noindent   As a deformed Law of Total Probability\par
\germanism{25-01-08}{Bristol, 3 AM}{to D. Gottesman}{Gottesman8}\par
\noindent   Shortcomings of Colemanesque ``derivations''\par
\germanism{03-06-02}{I Think She'll Know, 2}{to N. D. Mermin}{Mermin70}\par
\germanism{16-08-10}{Whaddya Think?, 2}{to T. Jacobson}{Jacobson2}\par
\noindent Historical predecessors\par
\germanism{21-04-10}{Book Review Notes}{to H. C. von Baeyer}{Baeyer118}\par

\item \indexbullet{Category theory}\par
\noindent   Keith Rowe's work thereon\par
\germanism{24-09-08}{Friends of the Tensor Product}{to H. Barnum and several others}{Abramsky1}\par
\noindent   Philosophical demerits and technical merits of topos-theoretic quantum logic\par
\germanism{24-04-08}{Topos Theory Today}{to L. Hardy}{Hardy27}\par
\noindent And Carnap's definition of ``functor''\par
\germanism{12-05-08}{Fascination with Words}{to W. G. Demopoulos}{Demopoulos23}\par

\item \indexbullet{Compatibility criteria for quantum states}\par
\noindent   Bayesian compatibility of density matrices\par
\germanism{07-08-01}{Knowledge, Only Knowledge}{to T. A. Brun, J. Finkelstein \& N. D. Mermin}{Mermin28}
\germanism{27-06-02}{Compatibility Never Ends}{to D. Poulin}{Poulin2}\par
\germanism{27-06-03}{Utter Rubbish and Internal Consistency, Part I}{to R. Schack, C. M. Caves \& N. D. Mermin}{Mermin86}\par
\germanism{07-12-06}{(Select) Replies to Referees}{to R. Schack \& C. M. Caves}{Caves93}\par
\noindent   Mutual unbias versus orthogonality\par
\germanism{13-07-02}{Yestopher}{to A. Plotnitsky}{Plotnitsky9}\par
\noindent And state pooling\par
\germanism{28-12-06}{Goose and Gander}{to R. W. Spekkens}{Spekkens41}\par

\item \indexbullet{Consistent/decoherent histories}\par
\noindent  And Kochen--Specker\par
\germanism{19-09-02}{Unsurprising Fact}{to N. D. Mermin}{Mermin74}\par
\germanism{18-10-07}{Noncolorable Histories}{to N. D. Mermin}{Mermin134}\par
\germanism{13-07-10}{QBism, Installment 4}{to H. C. von Baeyer}{Baeyer122}\par
\noindent   Issues with the time-symmetric ABL formalism\par
\germanism{18-03-04}{ABL, Appleby, and Grue}{to S. Savitt}{Savitt6}\par
\noindent And eavesdropping in BB84\par
\germanism{10-01-11}{Q3, Part 1}{to N. D. Mermin}{Mermin196}\par

\item \indexbullet{Criticisms of QBism}\par
\noindent ``Solipsism!''\par
\germanism{02-01-02}{Solipsism Story}{to C. H. Bennett \& J. A. Smolin}{SmolinJ2}\par
\germanism{10-07-03}{Solipsism}{to N. D. Mermin}{Mermin92}\par
\germanism{11-07-03}{One Final Thing in the Wee Hours}{to N. D. Mermin}{Mermin95}\par
\germanism{18-07-03}{Solipsism Concerns}{to N. D. Mermin}{Mermin97}\par
\germanism{19-07-03}{Definitions from Britannica}{to N. D. Mermin}{Mermin99}\par
\germanism{12-08-03}{Me, Me, Me}{to N. D. Mermin \& R. Schack}{Mermin101}\par
\germanism{08-10-03}{Heart of the Matter}{to M. P\'erez-Su\'arez}{PerezSuarez4}\par
\germanism{09-08-05}{Slides}{to H. M. Wiseman}{Wiseman18}\par
\germanism{07-03-06}{April}{to N. C. Menicucci}{Menicucci1}\par
\germanism{26-11-06}{Jimmy Hoffa's Bones}{to C. H. Bennett}{Bennett50}\par
\germanism{07-12-06}{(Select) Replies to Referees}{to R. Schack \& C. M. Caves}{Caves93}\par
\germanism{27-12-06}{New Year's Delirium}{to H. J. Bernstein}{Bernstein8}\par
\germanism{03-01-07}{New Year's Alchemy}{to H. C. von Baeyer}{Baeyer26}\par
\germanism{09-01-07}{Facts-in-Themselves}{to H. C. von Baeyer}{Baeyer27}\par
\germanism{01-02-07}{Metaphysics of the Time Process}{to L. Smolin}{SmolinL7}\par
\germanism{05-12-07}{The Misak Book}{to R. W. Spekkens}{Spekkens48}\par
\germanism{13-05-08}{Misak Tuesday}{to J. E. Sipe}{Sipe18}\par
\germanism{20-08-08}{Progress}{to E. G. Cavalcanti}{Cavalcanti2}\par
\germanism{08-09-08}{Progress, 2}{to E. G. Cavalcanti}{Cavalcanti3}\par
\germanism{20-01-09}{From Pragmatism to Pure Experience}{to M. B\"achtold}{Baechtold2}\par
\germanism{10-02-09}{Instance of James}{to R. W. Spekkens}{Spekkens60}\par
\germanism{03-03-09}{Dates}{to R. Schack}{Schack154}\par
\germanism{11-04-09}{QBism, Certainty, and Norsen, 2}{to R. Schack}{Schack160}\par
\germanism{17-04-09}{Two Quotes Before I Forget}{to R. W. Spekkens}{Spekkens66}\par
\germanism{13-06-09}{Once a Solipsist, Always a \ldots}{to H. M. Wiseman}{Wiseman19}\par
\germanism{22-06-09}{More Thoughts}{to D. M. Appleby, H. C. von Baeyer \& R. Schack}{Appleby65}\par
\germanism{30-06-09}{A New Name for Some Old Ways of Thinking}{to M. Schlosshauer}{Schlosshauer-newname}\par
\germanism{06-07-09}{The Verdict}{to H. M. Wiseman}{Wiseman27}\par
\germanism{06-10-09}{New Orleans, RMP, and Capacities}{to N. D. Mermin}{Mermin163}\par
\germanism{16-10-09}{The More and the Modest}{to L. Hardy}{Hardy38}\par
\germanism{20-01-10}{Bohr was Bayesian?}{to C. Ferrie}{Ferrie10}\par
\germanism{24-03-10}{Notes on the Open Future}{to L. Smolin}{SmolinL20}\par
\germanism{30-03-10}{I Knew He'd Just Call Me Solipsist Again, and he did}{to N. D. Mermin}{Mermin168}\par
\germanism{31-03-10}{Erwin Schrolipsism}{to N. D. Mermin}{Mermin169}\par
\germanism{30-03-10}{I Knew He'd Just Call Me Solipsist Again, and he did, 2}{to N. D. Mermin}{Mermin170}\par
\germanism{31-03-10}{Bell, Locality, Etc., 2}{to T. Norsen}{Norsen2}\par
\germanism{02-04-10}{Congratulations!, 3}{to the QBies}{QBies12}\par
\germanism{07-04-10}{Mid Conversation Frustration}{to T. Norsen}{Norsen4}\par
\germanism{07-04-10}{And After a Deep Breath}{to T. Norsen}{Norsen5}\par
\germanism{22-01-11}{Creatia!}{to the QBies}{QBies32}\par
\noindent   Finite dimensionality of Hilbert spaces\par
\germanism{04-07-03}{Solid Ground, Maybe?}{to G. L. Comer}{Comer33}\par
\germanism{28-04-05}{Infinite Limits}{to G. Valente}{Valente6}\par
\germanism{22-02-10}{Hilbert-Space Dimension in Second-Quantized Theories}{to N. C. Menicucci \\ \fantome  \& E. G. Cavalcanti}{Cavalcanti9}\par
\germanism{26-03-10}{QBist Double Slit}{to M. A. Graydon}{Graydon4}\par
\noindent Wither Lagrangians?\par
\germanism{28-11-06}{Whose Lagrangian?}{to M. A. Nielsen}{Nielsen5}\par
\germanism{28-11-06}{Whose Lagrangian?, 2}{to M. A. Nielsen}{Nielsen6}\par
\germanism{21-04-11}{Quick Question}{to H. C. von Baeyer}{Baeyer144}\par
\noindent Moore sentences\par
\germanism{17-11-06}{Certain Comments}{to C. G. Timpson}{Timpson11}\par
\germanism{12-12-06}{Small Changes}{to R. Schack}{Schack112}\par
\germanism{20-12-06}{I and It, 1}{to R. Schack}{Schack117}\par
\germanism{05-10-08}{Titles}{to R. Schack}{Schack138}\par
\germanism{29-06-09}{Off-the-Shelf Ontology vs.\ the Republican Banquet}{to C. G. Timpson}{Timpson13}\par
\germanism{11-08-09}{From Moore-ish Sentences to Bohrish Ones}{to C. G. Timpson}{Timpson15}\par
\germanism{12-08-09}{More on Moore}{to M. Schlosshauer}{Schlosshauer3}\par
\germanism{17-08-09}{Timpson}{to M. Schlosshauer}{Schlosshauer6}\par
\noindent Relativity and the block universe\par
\germanism{09-01-11}{One Endorsement}{to M. Schlosshauer}{Schlosshauer41}\par
\noindent   Published in journals\par
\germanism{04-04-11}{Les Papillons}{to H. C. von Baeyer}{Baeyer143}

\item \indexbullet{Decoherence}\par
\noindent Lack of foundational import\par
\germanism{07-02-01}{Reports and Chutzpah}{to A. Kent}{Kent4}\par
\germanism{25-11-01}{quant-ph/0106166}{to R. Cleve}{Cleve2}\par
\germanism{03-10-01}{Replies on Practical Art}{to C. M. Caves \& R. Schack}{Caves30}\par
\germanism{04-10-01}{Replies on Pots and Kettles}{to C. M. Caves \& R. Schack}{Caves33}\par
\germanism{02-02-02}{Colleague}{to C. G. Timpson}{Timpson1}\par
\germanism{01-03-03}{Wither Entanglement?}{to T. Siegfried}{Siegfried2}\par
\germanism{01-07-04}{Friends and Enemies}{to S. Hartmann}{Hartmann3}\par
\germanism{13-07-10}{QBism, Installment 4}{to H. C. von Baeyer}{Baeyer122}\par
\noindent And the preferred basis problem\par
\germanism{24-04-02}{A Stapp in the Right Direction?}{to B. W. Schumacher}{Schumacher9}\par
\germanism{24-06-06}{Notes on ``What are Quantum Probabilities''}{to J. Bub}{Bub21}\par
\noindent Versus Kofler--Brukner--Zeilinger imprecise measurement theory\par
\germanism{03-02-08}{Decoherence}{to M. Schlosshauer}{Schlosshauer1}\par
\germanism{25-04-10}{The Importance of Gadflycity}{to N. D. Mermin}{Mermin187}\par
\germanism{18-01-11}{What Is the Scale at which Quantum Theory Applies?}{to R. W. Spekkens}{Spekkens102}\par
\noindent As a mere application of formalism\par
\germanism{30-06-09}{A New Name for Some Old Ways of Thinking}{to M. Schlosshauer}{Schlosshauer-newname}\par
\noindent And van-Fraassenian reflection\par
\germanism{03-04-10}{The Science of Philosophy -- 2}{to M. Schlosshauer}{Schlosshauer30.2}\par
\germanism{09-04-10}{QB Decoherence}{to M. Schlosshauer}{Schlosshauer33}\par
\germanism{06-05-10}{{\Vaxjo} Abstract}{to M. Schlosshauer}{Schlosshauer36}\par
\germanism{06-05-10}{{\Vaxjo} Abstract, 2}{to M. Schlosshauer}{Schlosshauer37}\par

\item \indexbullet{De Finetti representations}\par
\noindent For quantum states\par
\germanism{19-04-01}{Quantum de Finetti}{to D. Petz and Other Friends}{Petz2}\par
\germanism{10-05-01}{Your Questions on de Finetti and Diffusion}{to R. W. Spekkens}{Spekkens3}\par
\germanism{14-05-02}{Quick Answer about Quantum de Finetti}{to K. Jacobs}{Jacobs2}\par
\germanism{05-07-02}{Coherence Everlasting}{to H. M. Wiseman}{Wiseman9}\par
\germanism{07-07-02}{And Continuing Consternation}{to H. M. Wiseman}{Wiseman10}\par
\germanism{08-07-02}{Short First Reply}{to H. M. Wiseman}{Wiseman11}\par
\noindent And coming to agreement\par
\germanism{07-08-01}{Knowledge, Only Knowledge}{to T. A. Brun, J. Finkelstein \& N. D. Mermin}{Mermin28}\par
\noindent For unknown quantum models\par
\germanism{19-11-01}{A Lot of the Same}{to C. M. Caves \& R. Schack}{Caves38}\par
\germanism{20-11-01}{A Bathtub Moment}{to C. M. Caves \& R. Schack}{Caves39}\par
\noindent   Bernoulli trials and quantum communication\par
\germanism{20-04-02}{Yes}{to T. Rudolph}{Rudolph2}\par
\noindent And MaxEnt\par
\germanism{19-09-03}{de Finetti vs.\ Jaynes}{to D. Poulin}{Poulin9}\par
\noindent   Limits on obtaining states by partial traces\par
\germanism{08-01-04}{Partial Exchangeability}{to D. Poulin}{Poulin11}

\item \indexbullet{Entropy}\par
\noindent   Principle of minimum cross-entropy in classical Shannon theory\par
\germanism{08-08-01}{Cross Entropy Min}{to J. Finkelstein}{Finkelstein6}\par
\noindent Motivating von Neumann entropy\par
\germanism{28-03-02}{Men in Power}{to P. Hayden}{Hayden1}\par
\noindent   Multivariate generalisations of mutual information\par
\germanism{30-07-02}{Coordinate Systems, 2}{to W. K. Wootters}{Wootters15}\par
\noindent   Mutual information, Holevo bound, preparation information and von Neumann entropy\par
\germanism{06-12-02}{Quantum Information Does Not Exist}{to A. Duwell}{Duwell1}\par
\noindent   Entropy bounds in general relativity\par
\germanism{04-07-03}{Bekenstein Bound Status}{to W. G. Unruh}{Unruh1}\par
\germanism{07-07-03}{Bekenstein Bound Status, 2}{to W. G. Unruh}{Unruh2}\par
\noindent   Relation of von Neumann and thermodynamic entropies\par
\germanism{23-01-04}{Entropy?}{to P. Grangier}{Grangier8}\par
\noindent   Jaynes' rejection letters for Landauer's paper\par
\germanism{26-10-04}{More ``More on Landauer''}{to W. T. Grandy, Jr.}{Grandy3}\par
\noindent   von Neumann entropy and SICs\par
\germanism{05-10-07}{My Sick Questions}{to D. M. Appleby \& S. T. Flammia}{Appleby23.1}\par
\noindent   Historical references on nonextensive entropies\par
\germanism{03-02-08}{Tsallis and Dar\'oczy}{to O. J. E. Maroney}{Maroney1}\par
\noindent   SICs and R\'enyi entropy\par
\germanism{02-11-10}{Probable Delay}{to R. Schack}{Schack214}

\item \indexbullet{Gleason's Theorem}\par
\noindent  Gleason's Theorem for POVMs\par
\germanism{27-06-01}{Gleasons, and Me at AT\&T}{to H. Barnum}{Barnum3}\par
\germanism{27-08-01}{A Little Contextuality on Noncontextuality}{to C. M. Caves}{Caves9}\par
\noindent  Naturalness of noncontextuality in Gleason-type theorems\par
\germanism{22-07-01}{Noncontextual Sundays}{to N. D. Mermin}{Mermin26}\par
\noindent   Quantumness of a set of states bounded by quantumness of a Hilbert space\par
\noindent   Modified Gleason's theorem and transforming among POVMs\par
\germanism{26-04-02}{Transformation Rules}{to A. Peres}{Peres30}\par
\germanism{02-09-03}{``Gleason'' Paper}{to P. Busch}{Busch5}\par
\noindent   Reverse Gleason's Theorem\par
\germanism{17-01-05}{Need Answer Quickly}{to R. W. Spekkens}{Spekkens32}\par
\noindent   Intermediate Gleason theorems\par
\germanism{02-12-05}{Gleason Frames}{to I. Bengtsson}{Bengtsson1}\par
\noindent   Distinctions between Demopoulos and QBism\par
\germanism{17-02-06}{Incompletely Knowable vs ``Truth in the Making''}{to W. G. Demopoulos}{Demopoulos6}\par
\germanism{18-02-06}{Two Questions}{to W. G. Demopoulos}{Demopoulos7}\par
\noindent   Subjectivity of unitaries and the possibility of a dynamical Gleason theorem\par
\germanism{21-11-06}{The Self-Preservation Society}{to M. S. Leifer}{Leifer6}\par
\germanism{21-11-06}{The Self-Preservation Society, 2}{to M. S. Leifer}{Leifer7}\par
\germanism{22-11-06}{The Self-Preservation Society, 3}{to M. S. Leifer}{Leifer8}

\item \indexbullet{History of quantum physics}\par
\noindent Origin of the GHZ gedankenexperiment\par
\germanism{07-02-01}{GHZM}{to N. D. Mermin}{Mermin-3}\par
\noindent   Standard Quantum Limit on position measurements\par
\germanism{26-07-01}{In the Middle}{to Y. J. Ng}{Ng2}\par
\germanism{26-07-01}{In the Night}{to Y. J. Ng}{Ng3}\par
\noindent   Historical references on noncontextuality of probabilities\par
\germanism{18-04-02}{Urgent Reference}{to H. Barnum}{Barnum4}\par
\noindent   Holevo on POVMs\par
\germanism{13-08-02}{Variola}{to R. W. Spekkens}{Spekkens7}\par
\noindent   Historical references on quantum fidelity\par
\germanism{19-09-03}{Fidelity}{to A. Peres}{Peres56}\par
\noindent   Pauli--Luba\'naski vector\par
\germanism{11-12-03}{Thomas, Pauli, Luba\'naski, and Enz}{to A. Peres}{Peres58}\par
\germanism{19-12-03}{Prizes, Hippies, and Vectors}{to A. Peres}{Peres59}\par
\noindent   Early references to ``quantum information''\par
\germanism{28-04-04}{Darwin, Qubit, Butterfield}{to C. G. Timpson}{Timpson3}\par
\noindent   Origin of the word ``qubit''\par
\germanism{28-04-04}{Quantum History}{to B. W. Schumacher \& W. K. Wootters}{Wootters19}\par
\germanism{12-12-10}{The Qubit Story}{to A. Stairs}{Stairs4}\par
\noindent   Entanglement in quantum field theory\par
\germanism{07-09-05}{EPR}{to A. Plotnitsky}{Plotnitsky18}\par
\noindent   Mermin shifting away from Ithaca Interpretation\par
\germanism{11-01-06}{For the Record}{to N. D. Mermin}{Mermin122}\par
\noindent Shor's factoring algorithm inspired by Simon\par
\germanism{20-09-07}{Easy Questions}{to P. W. Shor}{Shor4}\par
\noindent Simon's inspiration\par
\germanism{20-09-07}{Easy Questions}{to D. R. Simon}{SimonD1}\par
\noindent Gottesman's development of the stabilizer formalism\par
\germanism{20-09-07}{Easy Questions}{to D. Gottesman}{Gottesman5}\par
\noindent Preskill on security proofs in quantum cryptography\par
\germanism{20-09-07}{Easy Questions}{to J. Preskill}{Preskill16.1}\par
\noindent Kent on security proofs in quantum cryptography\par
\germanism{20-09-07}{Easy Questions}{to A. Kent}{Kent14}\par
\noindent Ekert on quantum key distribution\par
\germanism{20-09-07}{Easy Questions}{to A. K. Ekert}{Ekert1}\par
\noindent Yao on quantum Turing machines and communication complexity\par
\germanism{20-09-07}{Easy Questions}{to A. C. Yao}{Yao1}\par
\noindent Briegel on cluster-state computation\par
\germanism{20-09-07}{Easy Questions}{to H. J. Briegel}{Briegel5}\par
\noindent Benioff on unitary quantum computation\par
\germanism{20-09-07}{Easy Questions}{to P. A. Benioff}{Benioff1}\par
\noindent DiVincenzo on quantum information\par
\germanism{20-09-07}{Easy Questions}{to D. P. DiVincenzo}{DiVincenzo3}\par
\noindent Cleve on communication complexity\par
\germanism{20-09-07}{Easy Questions}{to R. Cleve}{Cleve3}\par
\noindent Deutsch on the Deutsch--Jozsa algorithm\par
\germanism{20-09-07}{Easy Questions}{to D. Deutsch}{DeutschD1}\par
\noindent Jozsa on the Deutsch--Jozsa algorithm\par
\germanism{20-09-07}{Easy Questions}{to R. Jozsa}{Jozsa7}\par
\noindent   Pre-QBism advocacy of quantum probabilities as subjective\par
\germanism{01-07-03}{Your Papers}{to N. Hadjisavvas}{Hadjisavvas1}\par
\germanism{27-09-08}{The Old Jenann}{to R. Healey}{Healey3}\par
\noindent  Stueckelberg and real vector space QM\par
\germanism{28-04-10}{Stueckelberg??}{to L. Hardy}{Hardy40}

\item \indexbullet{Information gain and disturbance}\par
\noindent And lasers\par
\germanism{15-04-01}{$P_E$}{to A. Peres}{Peres4}\par
\germanism{18-04-01}{Op-Ed}{to A. Peres}{Peres5}\par
\germanism{10-05-01}{Your Questions on de Finetti and Diffusion}{to R. W. Spekkens}{Spekkens3}\par
\noindent   Homework problems concerning information disturbance\par
\germanism{10-09-02}{Don't Forget Your Problem Sets}{to the Communards}{Communards3}\par
\noindent   State disturbance and POVMs\par
\germanism{30-03-06}{The Sqrt Operation}{to W. K. Wootters}{Wootters23}\par
\noindent   Information gain and disturbance for pure and mixed states\par
\germanism{11-05-07}{Information Gain Disturbance}{to R. W. Spekkens}{Spekkens42}\par
\germanism{11-05-07}{Information Gain Disturbance, 2}{to R. W. Spekkens}{Spekkens43}\par
\germanism{14-05-07}{Information Gain Disturbance, 3}{to R. W. Spekkens}{Spekkens44}\par
\germanism{14-05-07}{More Info Gain Disturbance Tradeoff}{to R. W. Spekkens}{Spekkens45}

\item \indexbullet{Locality}\par
\noindent   Instantaneous state-toggling at a distance can't communicate\par
\germanism{16-08-01}{Subject-Object}{to P. Grangier}{Grangier1}\par
\noindent Einstein's argument\par
\germanism{01-07-07}{First Einstein Quote}{to R. W. Spekkens}{Spekkens46}\par
\noindent   Einstein's clock-in-a-box as an early EPR argument\par
\germanism{22-12-08}{Nonlocal Boxes and Ehrenfest}{to N. D. Mermin}{Mermin144}\par
\noindent Popescu--Rohrlich boxes\par
\germanism{12-06-09}{In Defense of a Nomenclature that is Psychologically \\ \fantome  Rather Mature}{to J. Barrett}{Barrett5}\par
\noindent  References in favor of viewing QM as local\par
\germanism{21-08-09}{Quantum Locality}{to R. B. Griffiths}{Griffiths1}\par
\germanism{31-05-10}{References}{to N. D. Mermin}{Mermin188}\par
\germanism{31-05-10}{References, 2}{to N. D. Mermin}{Mermin189}\par
\germanism{31-05-10}{Another on the Side of Locality}{to N. D. Mermin}{Mermin190}\par
\noindent Tomographic locality as an axiom\par
\germanism{28-04-10}{Stueckelberg??}{to L. Hardy}{Hardy40}\par

\item \indexbullet{No bit commitment}\par
\noindent Standard MLC attack\par
\germanism{19-04-01}{CVs and Samizdats}{to A. Peres}{Peres6}\par
\germanism{19-04-01}{CVs and Samizdats, 2}{to A. Peres}{Peres7}\par
\germanism{22-04-01}{An Apple and a Commitment}{to A. Peres}{Peres8}\par
\noindent And induced decompositions\par
\germanism{22-04-01}{An Apple and a Commitment}{to A. Peres}{Peres8}\par
\germanism{14-05-02}{No BC's Role}{to R. Schack}{Schack50}\par
\noindent In the Spekkens toy model\par
\germanism{15-01-03}{Memory Lane}{to O. Cohen}{Cohen1}\par
\noindent And $C^*$-algebras\par
\germanism{12-01-06}{Sounds of Silence and CBH, 2}{to G. Brassard}{Brassard49}\par

\item \indexbullet{No-cloning and no-broadcasting}\par
\noindent   Classical information as that which can be copied\par
\germanism{02-08-01}{The Montr\'eal Commune -- Ditto}{to B. W. Schumacher}{Schumacher3}\par
\noindent   References on no-broadcasting\par
\germanism{01-04-02}{Broadcasting and Petra}{to A. Peres}{Peres27}\par
\noindent   How Wigner missed the no-cloning theorem\par
\germanism{05-05-02}{Wigner and Clones}{to A. Peres}{Peres31}\par
\noindent   Epistemically restricted Liouville mechanics\par
\germanism{22-02-06}{The Way the Cookie Crumbles}{to H. C. von Baeyer}{Baeyer19}\par
\noindent   How a qubit is like two probabilistic bits\par
\germanism{09-01-07}{Politically Correct}{to D. Bacon}{Bacon5}\par
\noindent   Classical no-cloning\par
\germanism{03-02-07}{Bill's Thoughts on QL and QI Frameworks}{to W. G. Demopoulos}{Demopoulos9}\par
\germanism{27-02-07}{Ref Help}{to W. C. Myrvold}{Myrvold7}\par

\item \indexbullet{Quantum computation}\par
\noindent   References on measurement-based quantum computation\par
\germanism{12-08-01}{A Silver Lining in the Deutsch Cloud}{to R. Pike}{Pike2}\par
\germanism{17-08-01}{Unloading Bayes}{to C. M. Caves \& R. Schack}{Schack3}\par
\noindent   Nonbooleanity of quantum properties as source of computational speedup\par
\germanism{06-10-01}{Your Spelling Conscience}{to J. Bub}{Bub5}\par
\noindent   Measurement-based quantum computation and violation of Bell inequalities\par
\germanism{12-01-02}{Raussendorf--Briegel}{to S. J. van Enk}{vanEnk25}\par
\germanism{21-01-02}{R, B, and P}{to A. Peres}{Peres23}\par
\noindent   Quantum operations are quantum states in disguise\par
\germanism{25-02-04}{Promised Message}{to N. D. Mermin}{Mermin110}\par
\noindent   BQP in rational Hilbert spaces\par
\germanism{01-11-04}{Rational Hilbert Space}{to S. Aaronson}{Aaronson6}\par
\noindent   Representation-dependent minus signs\par
\germanism{26-07-05}{Am I Turning Into You?}{to S. Aaronson}{Aaronson8}\par
\noindent   Violating a temporal Bell inequality to gain quantum computational power\par
\germanism{12-06-06}{Grover's Alg and Temporal Bell Inequalities}{to L. K. Grover}{Grover3}\par
\noindent Shor's factoring algorithm inspired by Simon\par
\germanism{20-09-07}{Easy Questions}{to P. W. Shor}{Shor4}\par
\noindent Simon's inspiration\par
\germanism{20-09-07}{Easy Questions}{to D. R. Simon}{SimonD1}\par
\noindent Gottesman's development of the stabilizer formalism\par
\germanism{20-09-07}{Easy Questions}{to D. Gottesman}{Gottesman5}\par
\noindent Yao on quantum Turing machines and communication complexity\par
\germanism{20-09-07}{Easy Questions}{to A. C. Yao}{Yao1}\par
\noindent Briegel on cluster-state computation\par
\germanism{20-09-07}{Easy Questions}{to H. J. Briegel}{Briegel5}\par
\noindent Benioff on unitary quantum computation\par
\germanism{20-09-07}{Easy Questions}{to P. A. Benioff}{Benioff1}\par
\noindent Deutsch on the Deutsch--Jozsa algorithm\par
\germanism{20-09-07}{Easy Questions}{to D. Deutsch}{DeutschD1}\par
\noindent Jozsa on the Deutsch--Jozsa algorithm\par
\germanism{20-09-07}{Easy Questions}{to R. Jozsa}{Jozsa7}\par

\item \indexbullet{Quantumness measures and communication channels}\par
\noindent Nonorthogonal states and stability against noise\par
\germanism{22-03-01}{Nonorthogonal States}{to R. W. Spekkens and J. E. Sipe}{Sipe1}\par
\noindent   Accessible fidelity and tensor product ensembles\par
\germanism{27-05-03}{Got It: Thanks!}{to O. Hirota}{Hirota2}\par
\noindent   Quantumness of a set of states bounded by quantumness of a Hilbert space\par
\germanism{09-06-03}{Quantumness}{to P. Busch}{Busch4}\par
\noindent   Choice of sign convention\par
\germanism{22-10-03}{Upside-Down Quantumness}{to S. J. van Enk}{vanEnk28}\par

\item \indexbullet{SIC POVMs}\par
\noindent   How a density matrix is a probability distribution over SIC outcomes\par
\germanism{04-06-02}{Random Questions}{to G. Plunk}{Plunk6}\par
\germanism{18-06-02}{SIC POVMs}{to C. M. Caves}{Caves66}\par
\noindent   Constraints on pure states in SIC representation\par
\noindent   SICs and eavesdropping\par
\germanism{24-02-03}{Fall Calendar and Symmetry}{to W. E. Lawrence}{Lawrence1}\par
\germanism{06-10-08}{SIC States}{to S. Abramsky and Y. J. Ng}{Abramsky2}\par
\noindent   Why a ``Bureau of Standards'' measurement should have $d^2$ outcomes\par
\germanism{26-03-04}{So Slow Chris}{to A. Peres}{Peres64}\par
\germanism{01-09-05}{BBQW Report}{to S. Hartmann}{Hartmann12}\par
\noindent   Seemingly basic yet unexplored questions in linear algebra\par
\germanism{30-03-05}{Civil Disobedience}{to G. L. Comer}{Comer64}\par
\noindent   Nice properties of SICs\par
\germanism{05-10-06}{How I Sold Us}{to D. M. Appleby \& H. B. Dang}{Appleby15}\par
\noindent   The Remarkable Theorem and the shape of quantum state space\par
\germanism{20-08-07}{R\'enyi Order-3 and the Weird Object}{to F. Tops{\o}e \& P. Harremo\"es}{Topsoe3}\par
\germanism{20-08-07}{R\'enyi Order-3 and the Weird Object, 2}{to F. Tops{\o}e \& P. Harremo\"es}{Topsoe4}\par
\noindent   Schmidt-like decompositions;\\
\noindent   Orthonormal bases related to SICs;\\
\noindent   Von Neumann entropy;\\
\noindent   SIC-Gleason theorem\par
\germanism{05-10-07}{My Sick Questions}{to D. M. Appleby \& S. T. Flammia}{Appleby23.1}\par
\noindent   Multiplicities in the spectra of SIC triple products\par
\germanism{09-01-08}{Wednesday, just before Toronto}{to D. M. Appleby}{Appleby25}\par
\noindent   Role of the triple-product constraint on shaping quantum state space\par
\germanism{16-05-08}{2:35 Happy Blues}{to R. Schack}{Schack134}\par
\noindent   The 7 in the QBic Equation\par
\germanism{04-06-08}{Hammerhead}{to G. L. Comer}{Comer115}\par
\noindent   SICs are as close to an orthonormal basis as one can get on the positive-operator cone\par
\germanism{06-10-08}{SIC States}{to S. Abramsky \& Y. J. Ng}{Abramsky2}\par
\noindent   Inability to express the Born rule in real vector space QM\par
\germanism{13-02-09}{It Wasn't Real Complex}{to R. W. Spekkens, \\ \fantome L. Hardy \& R. Blume-Kohout}{Spekkens61}\par
\noindent   State space from urgleichung-related assumptions; \\
Quantifying the degree to which physics is alchemical\par
\germanism{16-09-09}{Till Tomorrow}{to R. Schack}{Schack175}\par
\germanism{19-09-09}{The Cabibbo Angle}{to R. Schack}{Schack176}\par
\germanism{20-09-09}{One of the Essences of Things, Maybe?}{to R. Schack}{Schack177}\par
\germanism{20-09-09}{All Ways of Expression}{to R. Schack}{Schack178}\par
\germanism{20-09-09}{A Way To Salvage Much of the Old}{to R. Schack}{Schack179}\par
\germanism{21-09-09}{Wavering}{to R. Schack}{Schack180}\par
\germanism{22-09-09}{The Great Hardy}{to R. Schack}{Schack181}\par
\germanism{22-09-09}{Our Diophantine Equation}{to R. Schack}{Schack182}\par
\germanism{22-09-09}{Further Point}{to R. Schack}{Schack183}\par
\germanism{23-09-09}{All QBbibbo Angles}{to R. Schack}{Schack184}\par
\germanism{25-09-09}{The Scale of Engagement}{to H. C. von Baeyer}{Baeyer79}\par
\germanism{25-09-09}{Beta is the Parameter to Rule them All}{to R. Schack}{Schack185}\par
\germanism{25-09-09}{More On Normativity and a Quantitative Measure on Degrees of\\ \fantome  Detached Observers}{to C. G. Timpson}{Timpson19}\par
\germanism{19-10-09}{Purple Haze}{to M. Tait}{Tait4}\par
\noindent   SIC experiments\par
\germanism{09-11-09}{Seeking SICs}{to A. Zeilinger}{Zeilinger5}\par
\noindent Invention by Zauner\par
\germanism{11-01-10}{Analogy?}{to H. C. von Baeyer}{Baeyer94}\par
\germanism{19-01-10}{Translation}{to H. C. von Baeyer}{Baeyer100}\par
\noindent   SIC decomposition as analogous to Fourier decomposition\par
\germanism{11-01-10}{Analogy?}{to H. C. von Baeyer}{Baeyer94}\par
\noindent   Coupling-to-ancilla not the essence of POVMs\par
\germanism{12-03-10}{Counterexample}{to J. Emerson \& R. Laflamme}{Emerson5}\par
\noindent   SIC existence as an axiom;\\
\noindent   Building quantum state space one simplex at a time\par
\germanism{13-03-10}{Conceptual Barrier!}{to the QBies}{QBies6}\par
\noindent  Weyl--Heisenberg group and historical origins of SICs\par
\germanism{27-04-10}{Midnight Reading}{to H. C. von Baeyer}{Baeyer119}\par
\germanism{07-04-11}{Huangjun Zhu Visiting at the Moment}{to R. W. Spekkens \& M. S. Leifer}{Spekkens107}\par
\noindent  ``Classical'' states in SIC representation\par
\germanism{06-07-10}{Comments}{to C. Ferrie}{Ferrie16}\par
\noindent   Defining feature of SICs\par
\germanism{21-08-10}{One Ring to Bind Them, One Ring \\ \fantome  to Rule Them All}{to H. C. von Baeyer}{Baeyer124}\par

\item \indexbullet{Teleportation}\par
\noindent Laser light\par
\germanism{10-04-01}{Rudolph/Sanders/Teleportation}{to H. J. Kimble \& H. Mabuchi}{Kimble1}\par
\noindent  Alice, Bob, Chris and a qubit\par
\germanism{05-07-01}{Standing Up and Saying YES}{to J. Finkelstein}{Finkelstein2}\par
\noindent As described by Asher Peres\par
\germanism{14-05-02}{Qubit and Teleportation are Words}{to C. H. Bennett, W. K. Wootters, \\ \fantome  N. D. Mermin, A. Peres, J. A. Smolin and others}{SmolinJ2.1}\par
\noindent   No-signalling built into the quantum formalism\par
\germanism{26-02-04}{Thanks}{to A. Peres}{Peres61}\par
\noindent   Why ``steering'' is not shocking\par
\germanism{25-10-05}{GOBs, Bobs, Steering \& Teleportation}{to H. Halvorson \\ \fantome  \& B. C. van Fraassen}{vanFraassen5}

\item \indexbullet{Tomography}\par
\noindent References on tomography\par
\germanism{31-01-01}{Tomography}{to S. J. van Enk}{vanEnk19}\par
\noindent Informative and informationally complete POVMs\par
\germanism{10-06-01}{Answers}{to N. D. Mermin}{Mermin20}\par
\noindent   Tomography using projective measurements\par
\germanism{27-10-02}{Technical Question Minus}{to H. M. Wiseman}{Wiseman12}\par
\noindent Motivating the von Neumann entropy\par
\germanism{05-10-05}{Too Long Draft}{to A. P. Ramirez}{Ramirez1}\par
\noindent Tomographic locality as an axiom\par
\germanism{28-04-10}{Stueckelberg??}{to L. Hardy}{Hardy40}\par

\item \indexbullet{Toy theories reproducing subsets of quantum mechanics}\par
\noindent   Overlap-preserving time evolutions for Liouville densities\par
\germanism{02-01-03}{Glory Days}{to A. Peres}{Peres46}\par
\germanism{05-01-03}{ForAsher.tex}{to A. Peres}{Peres47}\par
\noindent Spekkens toy model\par
\germanism{15-01-03}{Memory Lane}{to O. Cohen}{Cohen1}\par
\germanism{27-04-03}{Bull by the Horns}{to R. W. Spekkens}{Spekkens13}\par
\germanism{11-06-03}{Getting the Word Out}{to R. W. Spekkens}{Spekkens15}\par
\germanism{01-07-03}{Rob Spekkens}{to R. D. Gill, K. M{\o}lmer \& E. S. Polzik}{Gill2}\par
\germanism{11-07-03}{Wow!}{to J. A. Smolin}{SmolinJ6}\par
\germanism{01-08-03}{Spekkens and Letter}{to A. Peres}{Peres52}\par
\germanism{02-13-03}{You Really Must Hurry}{to R. W. Spekkens}{Spekkens25}\par
\germanism{10-01-04}{Cohen Again}{to S. J. van Enk}{vanEnk30}\par
\germanism{16-01-04}{Also Holevo}{to R. W. Spekkens}{Spekkens28}\par
\germanism{01-03-04}{Wither Entanglement!}{to T. Siegfried}{Siegfried2}\par
\germanism{21-04-04}{Can Your Toy Model Do This?}{to R. W. Spekkens}{Spekkens31}\par
\germanism{13-05-04}{Epistemic Pure States}{to H. Atmanspacher}{Atmanspacher4}\par
\germanism{01-09-05}{BBQW Report}{to S. Hartmann}{Hartmann12}\par
\germanism{25-10-05}{GOBs, Bobs, Steering \& Teleportation}{to H. Halvorson \\ \fantome \& B. C. van Fraassen}{vanFraassen5}\par
\germanism{08-11-05}{Quibbles, Actions, and Reading}{to H. C. von Baeyer}{Baeyer8}\par
\germanism{15-08-06}{Drink the Kool-Aid}{to C. H. Bennett}{Bennett45}\par
\germanism{16-11-06}{Challenges to the Kierkegaardian Bayesian}{to A. Shimony}{Shimony12}\par
\germanism{22-12-10}{Did I Do You Good?}{to R. W. Spekkens}{Spekkens89}\par
\germanism{10-01-11}{Q3, Part 2}{to N. D. Mermin}{Mermin197}\par
\noindent Van Enk toy model\par
\germanism{29-01-07}{Toy Model, 1}{to S. J. van Enk}{vanEnk9}\par
\germanism{10-02-07}{Toy Model, 2}{to S. J. van Enk}{vanEnk10}\par
\germanism{10-02-07}{Inside and Outside}{to S. J. van Enk}{vanEnk11}\par
\noindent   Epistemically restricted Liouville mechanics\par
\germanism{22-02-06}{The Way the Cookie Crumbles}{to H. C. von Baeyer}{Baeyer19}\par
\noindent   How a qubit is like two probabilistic bits\par
\germanism{09-01-07}{Politically Correct}{to D. Bacon}{Bacon5}\par
\noindent   Classical no-cloning\par
\germanism{27-02-07}{Ref Help}{to W. C. Myrvold}{Myrvold7}\par
\noindent   Correlation monogamy and unperformed experiments having no outcomes\par
\germanism{09-12-08}{Quantum Random Numbers}{to R. Schack \& C. M. Caves}{Caves99}\par

\end{itemize}

\pagebreak

\chapter{2001: A Bayes Odyssey}
\pagenumbering{arabic}

\section{05-01-01 \ \ {\it In Defense of Interactionism} \ \ (to J. M. Renes)} \label{Renes1}

\bjmr
I've been reading up on some quantum logic, and finally have some idea of
what it is and how it might be useful in this ``research direction'' -- but
in some ways cuts against the Bayesian grain, and that might be an
opportunity to do some interesting (further) research.  I say further
because I think our Gleason-like result is a piece in this puzzle; the
puzzle being something like ``what does quantum logic look like when
using POVMs instead of projectors?''  That's probably misleading a bit, so
``starting from where Hilbert-space-projector quantum logic does, where do
POVMs take you? What are the logical structures and rules which
differentiate classical and quantum mechanics? How does the weirdness of
quantum mechanics arise from these rules and structures?'' I'm sure a lot
of the answers to these questions are trivial replacements of PVMs with
POVMs, but not always.
\ejmr

I don't know what Carl has said, but I like the sound of your ``research direction.''  I doubt any of the references below will be overly useful, but I met a guy at the quantum information tutorials in Edinburgh who has spent a lot of time generalizing quantum logics to effect algebras.  His name is Roberto Giuntini.  I just looked up the following articles on Web of Science.

I say I doubt it will be overly useful, because I gathered that, in his heart, he too was looking for the holy grail of most quantum logicians:  To find a way of pinning observer-independent properties on the world via algebraic properties of the theory's surface terms (i.e., states, observables, Hamiltonians).  In opposition to Carl---I won't say it's heresy---I think that's the wrong way to proceed.  I think we'll ultimately find {\it a quantum reality}, but we'll have to dig deeper than that.

All that said, the reason I like your research direction is because I think understanding the purely algebraic properties of POVMs is an important link nevertheless.  I would like to think that quantum mechanics is more about our interface with the world than the world itself \ldots\ and that it's at that interface that we'll find our glimpse of a ``quantum reality.''  What is it about the interface that makes our best Bayesian predictions of the quantum mechanical form?  That can be an algebraic question just as much as trying to pin naive properties on the world (regardless of how contrived the logic).  In that sense, there is some chance of maybe learning something from the Giuntinis of the world.

On another topic---not completely unrelated---I talked to my boss about your summer position here.  It's not completely 100\% yet, but it's pretty damned close (maybe 99.9\%).  So, we should have a good summer together.

\begin{itemize}
\item
Dalla Chiara ML, Giuntini R\\
Paraconsistent ideas in quantum logic\\
SYNTHESE 125: (1--2) 55--68 2000

\item
Cattaneo G, Dalla Chiara ML, Giuntini R, et al.\\
Effect algebras and para-Boolean manifolds\\
INT J THEOR PHYS 39: (3) 551--564 MAR 2000

\item
Cattaneo G, Dalla Chiara ML, Giuntini R \\
How many notions of ``sharp''?\\
INT J THEOR PHYS 38: (12) 3153--3161 DEC 1999

\item
Giuntini R, Pulmannova S\\
Ideals and congruences in effect algebras and QMV-algebras\\
COMMUN ALGEBRA 28: (3) 1567--1592 2000

\item
Cattaneo G, Giuntini R, Pilla R\\
BZMV(dM) algebras and stonian MV-algebras (applications to fuzzy sets and rough approximations)\\
FUZZY SET SYST 108: (2) 201--222 DEC 1 1999

\item
Giuntini R\\
Quantum MV-algebras and commutativity\\
INT J THEOR PHYS 37: (1) 65--74 JAN 1998

\item
CATTANEO G, GIUNTINI R\\
Some Results on BZ Structures from Hilbertian Unsharp Quantum Physics\\
FOUND PHYS 25: (8) 1147--1183 AUG 1995

\item
GIUNTINI R\\
Quasi-Linear QMV Algebras\\
INT J THEOR PHYS 34: (8) 1397--1407 AUG 1995

\item
DALLACHIARA ML, GIUNTINI R\\
Partial and Unsharp Quantum-Logics\\
FOUND PHYS 24: (8) 1161--1177 AUG 1994

\item
GIUNTINI R\\
3-Valued Brouwer--Zadeh Logic\\
INT J THEOR PHYS 32: (10) 1875--1887 OCT 1993
\end{itemize}

\section{08-01-01 \ \ {\it Little Miracles} \ \ (to R. Jozsa)} \label{Jozsa1}

I ran across the funniest thing by accident a minute ago.  I'll place the announcement below.  I especially got a chuckle from the last speaker's talk title.  Maybe that's been the problem between you and me all along!
\bv
Conference Announcement \medskip\\
Numbers, Sets and Structures\\
University of Bristol\\
Saturday 18th November 2000\medskip\\
Speakers:\medskip\\
``The origin and status of our conception of number''\\
Bill Demopolous (Western Ontario)\medskip\\
``Can you be sure the number three isn't Julius Caesar?''\\
Fraser MacBride (St.\ Andrews)\medskip\\
``Is there a unique natural number system?''\\
John Mayberry (Bristol)\medskip\\
``Anti-realists and classical mathematicians cannot get along''\\
Stewart Shapiro (Ohio State)
\ev

\section{08-01-01 \ \ {\it Lunch?}\ \ \ (to S. J. van {\Enk})} \label{vanEnk16.1}

Sorry, but I'm going to try hard not to exist until observed by Norbert.

\bv
There was a young man who said, ``God\\
Must think it exceedingly odd\\
If he finds that this tree\\
Continues to be\\
When there's no one about in the Quad.''\bigskip\\
REPLY\\
Dear Sir:\\
Your astonishment's odd:\\
I am always about in the Quad.\\
And that's why the tree\\
Will continue to be,\\
Since observed by\\
Yours faithfully,\\
NORBERT.
\ev

\section{09-01-01 \ \ {\it Please Come} \ \ (to L. Hardy)} \label{Hardy1}

The little blurb below is self-explanatory, and I'm being too lazy to personalize it.  But you're in the first round of people I'm contacting.  The reference to {\Montreal} refers to the previous meeting I tried to invite you to.  I hope you can come to this one; I'm especially eager to talk about your new paper (and I think Mermin is too).  Unfortunately, Wootters can't make it, but I'm hoping that some of the other guys (from the previous meeting) can.  If you have some grant money to spend, that'd be great, and it will help me get some of the more needy folk there.  But most importantly, I hope you can come either way.  Please let me know your thoughts as soon as possible.  Once I get a better picture of who can make it, I'll start contacting more of the room-and-board crowd.

June 11--16, 2001, there's going to be a conference in {\Vaxjo}, Sweden titled, ``Quantum Theory: Reconsideration of Foundations,'' and I've been asked to organize a session on ``quantum foundations in the light of quantum information'' (or at least that's my take on it).  I've been given the go-ahead to invite 3--4 people, all expenses paid, and 5 people with local expenses taken care of (but no travel).  I'm hoping hard that you'll say yes to being one of the 3--4 (or one of the 5 if you have enough grant money and are feeling generous).  In any case, {\it I'm hoping you'll say yes}.  In flavor and constitution, I plan to make the session as much like our little meeting in {\Montreal} as I can (given the larger, more diverse audience of the conference).  The other people that I presently know to also be attending are Enrico Beltrametti, Jeff Bub, Arthur Fine, Ed Nelson, Pekka Lahti, and Asher Peres \ldots\ but as a I say there will be a further contingent of quantum information people and a load of others beside that.  Please let me know your thoughts at your earliest convenience:  I've been told to get back to the main organizer (Khrennikov) with my recommendations by next Sunday (Jan 14).\medskip

\noindent Your madly-optimistic-for-the-quantum friend,\medskip

\noindent Chris

\section{09-01-01 \ \ {\it To Capitalize or Not To Capitalize?}\ \ \ (to S. J. van {\Enk})} \label{vanEnk17}

That is the question.  If you were me, would you capitalize the Greek letter starting the sentence below?
\bq\noindent
The words ``quantum state'' are used here, just as in the previous
formulation:  One cannot get away from that.  However, there are no
unknown quantum states.  $\rho^{(N)}$ is known by the experimenter if
no one else.  More importantly, the experimenter must be in a \ldots
\eq
Don't you wish you had my troubles?

\section{10-01-01 \ \ {\it Foundation-Sensitive Quantum-Information People} \ \ (to A. M. Steane)} \label{Steane1}

I really, really enjoyed my conversation with you in Mykonos this summer: It prompts the invitation below.  The blurb should be self-explanatory, so I'm going to be a little lazy and not personalize it.  But you {\it are\/} in the first round of people I'm contacting; it's just that most of those guys were at a previous meeting Gilles Brassard and I organized this spring.  Anyway, I hope you'll come.  I think some of your ideas are quite deep and deserve a wider circulation with people who enjoy this sort of thing.  To give you a flavor of the sort of people I'm seeking out for this session, I'll also place the {\it {\Montreal}\/} meeting blurb further below. (So, don't get confused by there being two conference announcements in this note.)  I don't know who I can get yet, but if things turn out like at the previous meeting, I think you'll have quite a bit of fun.

Finally, if you would like to come and have some grant money to spend, that'd be great, and will help me get some of the more needy folk there too.  But most importantly, I hope you can come either way.  Please let me know your thoughts as soon as possible.  Once I get a better picture of who can make it, I'll start contacting more of the room-and-board crowd.

\section{14-01-01 \ \ {\it Title and Abstract for Quantum Tutorials}\ \ \ (to J. E. Mazo)} \label{Mazo1}

Title and abstract for the talks (all three) below.

\bq\noindent
Title:  Enough Quantum Mechanics To Get Up and Running\medskip

\noindent Abstract:  I have a single goal in mind, to get as many people involved and cranking out research in quantum information science as possible.  The motivation is self-serving:  I really do believe there is nothing more exotic, more interesting, more important than quantum information.  These talks (three or so) will be designed with that goal in mind---to get the communication theorist thinking about quantum stuff.  If you can come to the lectures with any small memory of linear algebra (what a vector space is, what an inner product is, etc.), then I'll try to get you out with a workable knowledge of what you might accomplish in the quantum domain.  The lectures will be predominantly about the rules of the game from the communication perspective, but they will also be peppered liberally with simple relevant examples---quantum key distribution, quantum teleportation, Bell inequalities (no pun!), etc.
\eq

\section{17-01-01 \ \ {\it How About This?}\ \ \ (to A. M. Steane)} \label{Steane2}

That's too bad you can't come.  My confirmed list of sessioners is now:  Bub, Caves, Hardy, Jozsa, Lahti, Mermin, Peres, Schack, and Schumacher.  (Still waiting on Preskill.)  Almost surely now, we'll change the dates to June 17--22.  You would have been a great addition!  I'd really like to get some representation of your paper ``A Quantum Computer Needs Only One Universe.''  So, how about this idea?  Do you have a cohort sufficiently versed in the ideas (and convinced of their utility) to carry the torch?  If you can't talk on your paper, maybe that person could do just as well.  Any recommendations?

\section{17-01-01 \ \ {\it Shannon meets Bohr} \ \ (to A. Y. Khrennikov)} \label{Khrennikov1}

Here's where it stands so far.  I've drawn up a significantly more detailed synopsis for you.  What do you think?  I think the handwriting is on the wall as far as this group is concerned:  the best dates for the meeting would be June 17--22.  Shall we make those dates firm?

As I've told you already, what I've been doing in inviting people is giving them the choice to pay for part of the way out of their own grants.  I figure we can get a lot more interesting people that way, and not sacrifice any quality.  Indeed, the list is starting to include some of the best of the best in quantum information and still a lot of them are willing to pay their own travel expenses (especially, they say, if it will make for a better meeting).

So, how do your finances look given this new information?  Would you like me to pursue some more of my (relevant) friends?  Rarely do we find a congenial opportunity to talk about quantum foundations with each other---so this opportunity is quite a nice thing for us.  The way I would proceed is roughly according to the list I laid out below.  I say, ``the more, the merrier!''  But {\it you\/} are the financier; so it's {\it your\/} decision.

I read today about your meeting last Nov--Dec on {\sl Probability in Physics}!  It looked fascinating; I wish I had been there.

Anyway, for the present meeting, it looks like Shannon meets Bohr!

\section{17-01-01 \ \ {\it Up Your \ldots}\ \ \ (to H. J. Bernstein)} \label{Bernstein1}

Alley.  Up your alley, Herb.

Have a look at this article:
\bq\noindent
The Environment and the Epistemological Lesson of Complementarity \\
Folse HJ \\
Environmental Ethics {\bf 15}(4) 345--353 WIN 1993\medskip\\
Document type: Article  \ \   Language: English  \ \   Cited References: 23 \ \    Times Cited: 1\medskip\\
Abstract:
Following discussions by Callicott and Zimmerman, I argue that much of deep ecology's critique of science is based on an outdated image of natural science. The significance of the quantum revolution for environmental issues does not lie in its alleged intrusion of the subjective consciousness into the physicists' description of nature. Arguing from the viewpoint of Niels Bohr's framework of complementarity, I conclude that Bohr's epistemological lesson teaches that the object of description in physical science must be interaction and that it is now mistaken to imagine that physical science aims to represent nature in terms of properties it possesses apart from interaction.\medskip\\
KeyWords Plus: normative naturalism, intrinsic value, quantum-theory, science\medskip\\
Addresses:
Folse HJ, Loyola Univ, Dept Philosophy, 6363 St Charles Ave, New Orleans, LA 70118.
\eq

\section{18-01-01 \ \ {\it Indeed I Have} \ \ (to A. Plotnitsky)} \label{Plotnitsky1}

\barkp
I hope you have been thinking about Pauli and all in general
(naturally, no rush with sending anything).
\earkp

Indeed I have, but not nearly to the extent that I had hoped to.  Getting established in this new environment has been quite something else.  Did I ever write you after Mykonos?  I can't find any record of it.  OK, in the next note I'll send you a {\it very\/} incomplete compendium on Paulian ideas.  Perhaps that will help you get better oriented.

In the present note, I'll also put some thoughts prompted by Mykonos.  They were in a letter to David Mermin.  And then still further below that, I'll place a passage that my note to Mermin refers to, ``Genesis and the Quantum''.  [See 10-09-02 note ``\myref{Communards3}{Don't Forget Your Problem Sets}'' to the Communards.]  I hope you can sort it all out.  It would be interesting to hear to your thoughts.  I think there is some connection between the ``efficacity'' you spoke of and my ``Zing!''.  (I think we even spoke about this in Mykonos.)

By the way, could you send me the full bibliography of your writings on quantum stuff?

\subsection{Excerpt from 20 July 2000 note titled ``Zing!''\ to N. David {\Mermin}}

\bq
Let me give you yet another suggestion for reading (that'll you'll
probably ignore 2/3 of \ldots\ you told me you only read about 1/3 of
what I suggest).  I really enjoyed this one on my flight home the
other day.
\bq\noindent
H.~J. {\Folse}, ``Niels {\Bohr}'s Concept of Reality,'' in {\sl
Symposium on the Foundations of Modern Physics 1987: The Copenhagen
Interpretation 60 Years after the Como Lecture}, edited by
P.~{\Lahti} and P.~{\Mittelstaedt} (World Scientific, Singapore,
1987), pp.~161--179.
\eq
As always, I almost surely liked it because it was saying something
I wanted to hear.

Somehow I feel that I had an epiphany in Mykonos.  Do you remember
the parable of ``Genesis and the Quantum'' from my Montr\'eal problem
set?  [See 10-09-02 note ``\myref{Communards3}{Don't Forget Your Problem Sets}'' to the Communards.]  And do you remember my slide of an empty black
box with two overlays.  The first overlay was of a big $|\psi\rangle$
(hand drawn in blue ink of course).  I put the slide of the box up
first, and said ``This is a quantum system; it's what's there in the
world independent of us.'' Then I put the first overlay on it and
say, ``And this symbol stands for nothing more than what we know of
it. Take us away and the symbol goes away too.''  I then remove the
$|\psi\rangle$.  ``But that doesn't mean that the system, this black
box, goes away.''  Finally I put back up the $|\psi\rangle$ over the
box, and the final overlay. This one says: ``Information/knowledge
about what?  The consequences of our experimental interventions into
the course of Nature.''

Well, now I've made another overlay for my black box slide.  At the
top it asks, ``So what is real about a quantum system?''  In the
center, so that it ends up actually in the box, is a very stylistic
version of the word ``Zing!''  And at the bottom it answers, ``The
locus of all information-disturbance tradeoff curves for the
system.''  In words, I (plan to) say, ``It is that zing of the
system, that sensitivity to the touch, that keeps us from ever
saying more than $|\psi\rangle$ of it.  This is the thing that is
real about the system.  It is our task to give better expression to
that idea, and start to appreciate the doors it opens for us to
shape and manipulate the world.''  What is it that makes quantum
cryptography go?  Very explicitly, the zing in the system.  What is
it that makes quantum computing go?  The zing in its components!

Anyway, I'm quite taken by this idea that's getting so close to being
a technical one---i.e., well formed enough that one might check
whether there is something to it.  What is real of the system is the
locus of information-disturbance (perhaps it would be better to say
``information-information'') tradeoff curves.  The thing to do now is
to show that Hilbert space comes about as a compact description of
that collection, and that it's not the other way around.  As I've
preached to you for over two years now, this idea (though it was in
less refined form before now) strikes me as a purely ontological one
\ldots\ even though it takes inserting an {\Alice}, {\Bob}, and
{\EveC} into the picture to give it adequate expression.  That is, it
takes a little epistemology before we can get to an ontological
statement.

I looked back at your original Ithaca Interpretation paper, and I'll
be bold enough to say that this idea satisfies all your most
important desiderata:  (1), (2 suitably modified), (3), and (5).

Part of this, by the way, is why I liked so much {\Folse}'s paper.
Also, believe it or not, for a moment while reading it I thought I
could finally ``SEE'' correlation without correlata.  (Not lying.)
But then I thought I liked the phrase ``Interaction without
Interactoids'' even more.  My wife just thought I was being silly.
Maybe you will too.
\eq

\subsection{Arkady's Reply, ``Bibliography''}

\bq
Here is my ``quantum-mechanical'' bibliography. It only lists books and
articles primarily on Bohr and the epistemology of quantum mechanics.  There
are another dozen or so articles dealing with the relationships among
quantum epistemology, philosophy, and literature, as well some reviews in
scholarly and popular press ({\sl Chicago Tribune}).\medskip

\noindent {\bf Books:}
\begin{itemize}
\item
{\sl The Knowable and the Unknowable:\ Modern Science and Nonclassical Thought},
Ann-Arbor, Mich.: University of Michigan Press, forthcoming in 2001.
Contains substantial chapters on Bohr and Heisenberg.

\item
{\sl Complementarity: Anti-Epistemology After Bohr and Derrida}, Durham, NC: Duke
Univeristy, 1994.		
\end{itemize}	
		
\noindent {\bf Articles:}
\begin{itemize}
\item
``Reading Bohr: Complementarity, Epistemology, Entanglement, and
De\-co\-her\-ence,'' forth\-coming in the {\sl Proceedings of the NATO Advanced Research
Workshop on ``Decoherence and its Implications for Quantum Computation,''} eds.\
Antonios Gonis and Patrice Turchy (Dordrecht:  Kluwer, 2001)

\item
``Quantum Physics and Philosophy 1900--2001,'' forthcoming in {\sl Disciplinarity at
the Fin-de-Si\`ecle}, eds.\ Amanda Anderson and Joseph Valente (Princeton, NJ:
Princeton University Press, 2001)

\item
``Landscapes of Sibylline Strangeness:\  Complementarity, Quantum Measurement
and Classical Physics,'' {\sl Metadebates}, eds.\ G. C. Cornelis, J. P. Van
Bendegem, and D. Aerts (Dordrecht: Kluwer, 1999)

\item
``Techno-atoms:\ The Ultimate Constituents of Matter and the Question
Concerning Technology in Quantum Mechanics,'' {\sl Tekhnema: Journal of Philosophy
and Technology}, Winter--Spring 1999

\item
``Complementarity, Idealization, and the Limits of the Classical Conceptions
of Reality,'' in {\sl Mathematics, Science and Postclassical Theory}, eds.\ Barbara
H. Smith and Arkady Plotnitsky (Durham, NC:  Duke University Press, 1997)
\end{itemize}
\eq

\section{18-01-01 \ \ {\it Berkeley's Clones}  \ \ (to A. Peres)} \label{Peres1}

\bap
BTW, can God clone an unknown quantum state?
\eap

Funny that you ask this.  I just addressed this question in the paper I'm writing with Caves and Schack:
\bq
There is hardly a paper in the field of quantum information that does not make use of the idea of an ``unknown quantum state.''
Unknown quantum states are teleported [Bennett1993, Experiments1998], protected with quantum error correcting codes [Shor1995, Steane1996], and used to check for quantum eavesdropping [Bennett1984, CryptoExperiments]. The list of uses grows each day.  But what can the term unknown state possibly mean? In an information-based interpretation of quantum mechanics the term is an overt oxymoron:  For there, quantum states in their essence are states of knowledge and not states of nature [Hartle1968]. If a quantum state is used to describe a system, then it is {\it known\/} by someone---at the very least, by the describer himself.

This message is the main point of our paper. If a phenomenon ostensibly invokes the concept of an unknown state in its formulation, then the unknown state had better be a shorthand for a more fundamental situation, even if that more fundamental situation still awaits a complete analysis.  This means that, for any phenomenon using the idea of an unknown quantum state in its description, a consistent information-based interpretation of quantum mechanics demands that either
\begin{enumerate}
\item
The owner of the unknown state---a further decision-making agent or observer---must be explicitly identified.  In this case, the unknown state is just a stand-in for the unknown {\it state of knowledge\/} of an essential player that was skipped over previously. Or,
\item
If there is clearly no further decision-making agent or observer on the scene, then a way must be found to reexpress the phenomenon with the term ``unknown state'' consistently banished throughout its formulation. In this case, the end-product of the effort will be a single quantum state used for describing the phenomenon---namely, the state that captures the describer's state of knowledge.
\end{enumerate}

The use of unknown states we actually analyze in depth has to do with the measurement technique known as quantum-state tomography [Vogel1989b, Smithey1993, Leonhardt1995]. The usual description of tomography is this.  A device of some sort, a laser say, repeatedly prepares many instances of a quantum system in a fixed quantum state $\rho$ (pure or mixed). An experimentalist who wishes to characterize the operation of the device or calibrate it for future use may be able to perform measurements on the systems it prepares even if he cannot get at the device itself.  This may be useful because he may have some prior knowledge of the device's operation that can be translated into a probability distribution over states.  Thus learning about the state will also be learning about the device. Most importantly, though, it is assumed that the precise state $\rho$ is unknown. The goal of the experimenter, therefore, is to perform enough measurements, and enough kinds of measurements, on a large enough sample, to estimate the identity of $\rho$.

This is clearly an example where there is no further player to pin the unknown state upon as a state of knowledge.  Any attempt to do so would be quite unnatural:  Where would the player be placed?  On the inside of the device the tomographer is trying to characterize?$\,$\footnote{This move would be little more respectable than George Berkeley's famous patch to his philosophical system, a difficulty captured engagingly by the limerick of Ronald Knox and an anonymous reply:
``There was a young man who said, `God must think it exceedingly odd if he finds that this tree continues to be when there's no one about in the Quad.'\,''
\begin{quotation}
\footnotesize
\noindent Dear Sir: \\
Your astonishment's odd: \\
I am always about in the Quad. \\
And that's why the tree \\
Will continue to be, \\
Since observed by \\
\indent \indent \indent Yours faithfully, \\ \indent \indent \indent God.
\end{quotation}
}
The only strategy open to us is the second one listed above, i.e., to banish the idea of the unknown quantum state completely from quantum-state tomography's formulation.
\eq

\section{19-01-01 \ \ {\it Industrial Boys} \ \ (to A. Kent)} \label{Kent1}

\bak
But might you want to circulate a criticism of the article itself,
independent of the value of its John-component?
\eak

Thanks for the offer, but I think I have to shy away from that.  John has always been among my physics heroes.  That's why I was so incensed when I found his name attached to (1) mathematical Platonism, (2) some form of an Everett interpretation, (3) a hope for a final theory, and (3) the idea that the observer plays no role in the measurement, ALL in opposition to everything he's written for 30 years.  My lashing was a lashing out at Max.  I don't want to drag John any further through the dirt.

\section{20-01-01 \ \ {\it Sleeping Around \ldots}\ \ \ (to G. Brassard)} \label{Brassard1}

Maybe I should have completed the title of this note with, ``with Shannon and Bohr.''  Well anyway, you'll understand the relevance of the title when you read the invitation below!  I really hope you can come.  You, more than anyone, are probably responsible for this session's existence:  Without your financing last year, I wouldn't have been propelled into this position of {\it infamy}!!  (Nor would I have these continued chances to see my dream come true.)  I think your talk on the communication cost of Bell inequality violations would be perfect for this crowd.  But on top of that, at least Bub and Hardy (and soon to be many more I think) are taking our ideas about ``QM from QKD and no QBC'' seriously.

You'll note that Bill isn't in the list below; I already contacted him and he couldn't make it this time.  You'll also note that Charlie's not in there either:  In this case, it's because he refused to talk about foundations last time!  (You did too, but I won't hold that against you.)

So, come!

\bq
\noindent Dear friends,\medskip

I recently found out that I've gotten another opportunity to gather some friends in an exotic place to thrash out the idea of ``Quantum Foundations in the Light of Quantum Information.''  (This is starting to become a habit for me.)  Anyway, this time I've been asked to organize a session at a larger conference in {\Vaxjo}, Sweden titled ``Quantum Theory:\ Reconsideration of Foundations.''  The main organizer is Andrei Khrennikov, and he is planning to have 40--50 people attending, with further sessions on Bohmian mechanics, GRW mechanics, and other issues in quantum foundations (Bell inequalities, Kochen--Specker theorems, etc.).  Also I've been allowed to ask a small contingent of philosophers.

In flavor and constitution, I plan to make our session much like the meeting Gilles Brassard and I held in {\Montreal} last spring.  There the theme was organized around the strong feeling that we'll find the greatest things and technologies to come out of quantum mechanics when we finally grasp the parts of it that make us feel the most uncomfortable.  The theory is begging us to ask something new and profound of nature.

This time the list of quantum information invitees (so far) includes: Gilles Brassard, Jeffrey Bub, Carl Caves, Lucien Hardy, Richard Jozsa, David Mermin, Asher Peres, John Preskill, {\Ruediger} {\Schack}, John Smolin, and Ben Schumacher.  The philosophical invitees I'm in charge of trying to get are:  Doug Bilodeau, Henry Folse, Itamar Pitowsky, Arkady Plotnitsky, and Abner Shimony.  Depending upon how the money holds out, I might be able to call up even a few more in the community.  (Also, as I say, there are other sessions and thus other invitees, but I don't know who they are yet.)

I think we'll have a double-edged opportunity at this meeting, so I'm looking quite forward to it.  First, we'll get a chance to continue the lines several of us started in \Montreal---at that time, the attendees were Bennett, Bernstein, Brassard, Bub, Hayden, Jozsa, Mermin, Schack, Schumacher, and Wootters.  But second, we may be able to play the role of educators to a larger community interested in dabbling in these same issues.  That is, we will have an opportunity to get them to think of quantum information as a TOOL for exploring quantum foundations.

The dates for the meeting are June 17 to 22.  We think this choice will help make for quite a pleasant time:  Midsummer's eve is June 21, and Mermin has pointed out that picking wild strawberries in the midnight twilight is huge fun.  Furthermore, there are significant local festivities planned around that time of year.

The information I'd like to get from you (as soon as possible!)\  is:
\bv
1)  Can you confirm positively that you would like to come? \\
and \\
2) What kind of financial resources will it take for us to get you
there?\\
\hspace*{.15in} a) Full expenses paid? \\
\hspace*{.15in} b) Local expenses paid? \\
\hspace*{.15in} c) Nothing? \\
\ev

Concerning question 2), you can count this letter as an invitation, so feel free to answer any of the three options a), b), or c).  However, the more money I can get the invitees to throw into the pot (IF they have it available), the more interesting people I can get to Sweden to keep us entertained.  I've been told that I can have 3-4 invitees all expenses paid and 5 invitees with local expenses paid from the quantum information group.  You can see that I'm {\it stretching\/} my limits!  How am I going to do this?  I'm {\it hoping\/} that a significant number of invitees will be able to volunteer to pay their nonlocal expenses, i.e.\ their travel expenses.  Money is flowing pretty well into quantum information in Canada, the US, and the UK right now; and I do know that at least some of you have a surplus.  On the other hand, the most important thing is that you attend, regardless of your funding status.

So please come:  I think we'll have a lot of fun.  Please let me know your thoughts as soon as you can!  The sooner Khrennikov and I can get this preliminary information, the sooner we can start throwing stones at all the other problems.\medskip

\noindent All the best,\medskip

\noindent Chris
\eq

\section{23-01-01 \ \ {\it It's Really Real}\ \ \ (to the {\Vaxjo} Invitees)} \label{VaxjoIsReallyReal}

The main organizer of the {\Vaxjo} meeting, Andrei Khrennikov, now has all your financial-need information, etc., in hand.  The word is, the total session below will go through---physicists, philosophers, and all.  You should be hearing from him with an official invitation and conference announcement next week (when he returns from a conference in Italy).  So, please mark the dates in your calendars!

Think John Wheeler thoughts!: ``I want you \ldots\ to jolt the world of physics into an understanding of the quantum because the quantum surely contains---when unraveled---the most wonderful insight we could ever hope to have on how this world operates, something equivalent in scope and power to the greatest discovery that science has ever yet yielded up \ldots''

Looking forward to seeing you all in Sweden!

\begin{center}
{\bf Quantum Theory: Reconsideration of Foundations {\Vaxjo}, Sweden, 17--22 June 2001} \smallskip \\

Session: {\it Shannon meets Bohr:\ Quantum Foundations in the Light of Quantum Information}\medskip
\end{center}

\pagebreak

\begin{supertabular}{lll}
The Physicists       &                   &   Confirmed         \\
(Shannon meets Bohr) &    Institution    &   Participation     \\
\hline               &                   &                     \\
Gilles Brassard      &    U. {\Montreal} &   ??                \\
Jeffrey Bub          &    U. Maryland    &   yes               \\
Carlton Caves        &    U. New Mexico  &   yes               \\
Christopher Fuchs    &    Bell Labs      &   yes               \\
Lucien Hardy         &    Oxford U.      &   yes               \\
Richard Jozsa        &    U. Bristol     &   yes               \\
Pekka Lahti          &    U. Turku       &   ??                \\
David Mermin         &    Cornell U.     &   yes               \\
Asher Peres          &    Technion       &   yes               \\
Itamar Pitowsky      &    Hebrew U.      &   yes               \\
John Preskill        &    Caltech        &   tentative         \\
Ruediger Schack      &    U. London      &   yes               \\
Ben Schumacher       &    Kenyon College &   yes, $-$1st day   \\
John Smolin          &    IBM Research   &   yes               \\
                     &                   &                     \\
The Philosophers     &                   &   Confirmed         \\
(Bohr meets Shannon) &   Institution     &   Participation     \\
\hline               &                   &                     \\
Doug Bilodeau        &   U. Indiana      &   yes               \\
Henry Folse          &   Loyola U.       &   yes               \\
Arkady Plotnitsky    &   Purdue U.       &   yes               \\
Abner Shimony        &   Boston U.       &   ??                \\
                     &                   &                     \\
Student              &                   &   Confirmed         \\
Contingent           &   Institution     &   Participation     \\
\hline               &                   &                     \\
Joe Renes            &   U. New Mexico   &   yes               \\
Daniel Terno         &   Technion        &   yes               \\
\end{supertabular}

\section{24-01-01 \ \ {\it Split Personality?}\ \ \ (to S. J. van {\Enk})} \label{vanEnk18}

\bq\noindent
December 12, 2000, Tuesday\\
Science Desk\\
{\sl New York Times}\medskip\\
ESSAY: `A Practical Tool,' But Puzzling, Too\\
By JOHN ARCHIBALD WHEELER\bigskip

What is the greatest mystery in physics today? Different physicists have different answers. My candidate for greatest mystery is a question now a century old, ``How come the quantum?''

What is this thing, the ``quantum''? It's a bundle of energy, an indivisible unit that can be sliced no more. Max Planck showed us a hundred years ago that light is emitted not in a smooth, steady flow, but in quanta. Then physicists found quantum jumps of energy, the quantum of electric charge and more. In the small-scale world, everything is lumpy.

And more than just lumpy. When events are examined closely enough, uncertainty prevails; cause and effect become disconnected. Change occurs in little explosions in which matter is created and destroyed, in which chance guides what happens, in which waves are particles and particles are waves.

Despite all this uncertainty, quantum physics is both a practical tool and the basis of our understanding of much of the physical world. It has explained the structure of atoms and molecules, the thermonuclear burning that lights the stars, the behavior of semiconductors and superconductors, the radioactivity that heats the earth, and the comings and goings of particles from neutrinos to quarks.

Successful, yes, but mysterious, too. Balancing the glory of quantum achievements, we have the shame of not knowing ``how come.'' Why does the quantum exist?

My mentor, the Danish physicist Niels Bohr, made his peace with the quantum. His ``Copenhagen interpretation'' promulgated in 1927 bridged the gap between the strangeness of the quantum world and the ordinariness of the world around us. It is the act of measurement, said Bohr, that transforms the indefiniteness of quantum events into the definiteness of everyday experience. And what one can measure, he said, is necessarily limited. According to his principle of complementarity, you can look at something in one way or in another way, but not in both ways at once. It may be, as one French physicist put it, ``the fog from the north,'' but the Copenhagen interpretation remains the best interpretation of the quantum that we have.

Albert Einstein, for one, could never accept this world view. In on-again, off-again debates over more than a dozen years, Bohr and Einstein argued the issues -- always in a spirit of great mutual admiration and respect. I made my own effort to convince Einstein, but without success. Once, around 1942, I went around to his house in Princeton to tell him of a new way of looking at the quantum world developed by my student Richard Feynman.

Feynman pictured an electron getting from point A to point B not by one or another possible path, but by taking all possible paths at once. Einstein, after listening patiently, said, as he had on other occasions, ``I still cannot believe God plays dice.'' Then he added, ``But maybe I have earned the right to make my mistakes.''

Feynman's superposed paths are eerie enough. In the 1970's, I got interested in another way to reveal the strangeness of the quantum world. I called it ``delayed choice.'' You send a quantum of light (a photon) into an apparatus that offers the photon two paths. If you measure the photon that leaves the apparatus in one way, you can tell which path it took.

If you measure the departing photon in a different way (a complementary way), you can tell if it took both paths at once. You can't make both kinds of measurements on the same photon, but you can decide, after the photon has entered the apparatus, which kind of measurement you want to make.

Is the photon already wending its way through the apparatus along the first path? Too bad. You decide to look to see if it took both paths at once, and you find that it did. Or is it progressing along both paths at once? Too bad. You decide to find out if it took just one path, and it did.

At the University of Maryland, Carroll Alley, with Oleg Jakubowicz and William Wickes, took up the challenge I offered them and confirmed that the outcome could be affected by delaying the choice of measurement technique -- the choice of question asked -- until the photon was well on its way. I like to think that we may one day conduct a delayed-choice experiment not just in a laboratory, but in the cosmos.

One hundred years is, after all, not so long a time for the underpinning of a wonderfully successful theory to remain murky. Consider gravity. Isaac Newton, when he published his monumental work on gravitation in the 17th century, knew he could not answer the question, ``How come gravity?'' He was wise enough not to try. ``I frame no hypotheses,'' he said.

It was 228 years later when Einstein, in his theory of general relativity, attributed gravity to the curvature of spacetime. The essence of Einstein's lesson can be summed up with the aphorism, ``Mass tells spacetime how to curve, and spacetime tells mass how to move.'' Even that may not be the final answer. After all, gravity and the quantum have yet to be joined harmoniously.

On the windowsill of my home on an island in Maine I keep a rock from the garden of Academe, a rock that heard the words of Plato and Aristotle as they walked and talked. Will there someday arise an equivalent to that garden where a few thoughtful colleagues will see how to put it all together and save us from the shame of not knowing ``how come the quantum''? Of course, in this century, that garden will be as large as the earth itself, a ``virtual'' garden where the members of my imagined academy will stroll and converse electronically.

Here, a hundred years after Planck, is quantum physics, the intellectual foundation for all of chemistry, for biology, for computer technology, for astronomy and cosmology. Yet, proud foundation for so much, it does not yet know the foundation for its own teachings. One can believe, and I do believe, that the answer to the question, ``How come the quantum?'' will prove to be also the answer to another question, ``How come existence?''
\eq

\section{26-01-01 \ \ {\it Report:\  Wallace, `Worlds in the Everett Interpretation'} \ \ (to J. N. Butterfield, C. Pagonis, and D. Wallace)}   \label{Butterfield-1} \label{Pagonis1} \label{Wallace1}

Anyone who knows me knows that I am rather down on attempts to interpret quantum mechanics along Everett-like lines.  I think the most funny and telling statement of this in the present context is that, whereas Mr.\ Wallace speaks of ``Everettians,'' I often speak of ``Everettistas.''  Thus, I am almost surprised that you sent me this paper to referee.

My difficulties come not so much from thinking that an Everett-like interpretation is inherently inconsistent or that parallel worlds tax the imagination too much.  It's more that this line of thought strikes me, at best, as a complete dead end in the physical sense.  At worst, I fear it requires us to tack on even more ad hoc structures to quantum theory than we already have.  (Here, I'm thinking of a preferred basis for the Hilbert space and a preferred tensor-producting of it into various factors.)  For these reasons, among umpteen others, I have always been inclined to an epistemic interpretation of the quantum state.  Doing this has helped me (personally) to focus the issue to asking, ``What is this {\it property\/} of the quantum world---i.e., reality---that keeps us from ever knowing more of it than can be captured by the quantum state?''  To that extent, I consider myself something of a realist who---just as David Deutsch---takes the wavefunction absolutely seriously.  {\it But\/} absolutely seriously as a state of knowledge, not a state of nature.  I do well believe we will one day shake a notion of reality from the existing theory (without adding hidden variables, etc.), but that reality won't be the most naive surface term floating to the top (i.e., the quantum state).  When we have it, we'll really have something; there'll be no turning back.  Physics won't be at an end, but just at the beginning.  For then, and only then, will we be able to recognize how we might extend the theory to something bigger and better than quantum mechanics itself.

All that said, you're going to be surprised by my evaluation of this paper.  Without hesitation, I recommend you publish it!  Of all the things I've read on Everett-like interpretations over the years, this paper has struck me as the most reasonable of the lot.  This is a nice paper.  It changed none of my views, but it caused me deep pause for thought.  What better honor can one have from a sparring partner?

I am not in a position to judge whether the paper is a significant step forward over the many Saunders papers it cites---I've never read any of them---but I do know this much:  I have walked away from three of Simon Saunders' talks and not had a clue what he was talking about.  So, even if my evaluation (as an outsider) of the paper's technical merit leaves something to be desired, I think Mr.\ Wallace's paper clearly serves a purpose within our community.  The analogies between the Everett-like structure he proposes and the spacetime of general relativity are indeed intriguing and worthy of thought.  For me personally---whatever their ultimate merit---they give a new stone on which to hone the arguments for an epistemic interpretation of the quantum state.

Let me just make a few minor comments to round out this report.

1)  I very much enjoyed the discussion:  ``So there are no theory-neutral observations:  rather, there is an existing theory in terms of which our observations are automatically interpreted, and which we must take as our starting point when interpreting the theories of physics.''  I've seen a discussion like this before, and it probably should be cited.  Here are two references from my personal archive:

\begin{itemize}
\item
W.~Heisenberg, {\sl Physics and Beyond:~Encounters and Conversations}, translated by A.~J. Pomerans (Harper \& Row, New York, 1971), pp.~63--64.  Heisenberg reports Einstein as having once said the following to him:

\begin{quote}
It is quite wrong to try founding a theory on observable magnitudes alone.  In reality the very opposite happens.  It is the theory which decides what we can observe.  You must appreciate that observation is a very complicated process.  The phenomenon under observation produces certain events in our measuring apparatus.  As a result, further processes take place in the apparatus, which eventually and by complicated paths produce sense impressions and help us to fix the effects in our consciousness.  Along this whole path---from the phenomenon to its fixation in our consciousness---we must be able to tell how nature functions, must know the natural laws at least in practical terms, before we can claim to have observed anything at all.  Only theory, that is, knowledge of natural laws, enables us to deduce the underlying phenomena from our sense impressions.  When we claim that we can observe something new, we ought really to be saying that, although we are about to formulate new natural laws that do not agree with the old ones, we nevertheless assume that the existing laws---covering the whole path from the phenomenon to our consciousness---function in such a way that we can rely upon them and hence speak of ``observation.''
\end{quote}

This is discussed further in:

\item
M.~Jammer, ``The Experiment in Classical and in Quantum Physics,'' in {\sl Proceedings of the International Symposium, Foundations of Quantum Mechanics in the Light of New Technology}, edited by S.~Kamefuchi, H.~Ezawa, Y.~Murayama, M.~Namiki, S.~Nomura, Y.~Ohnuki, and T.~Yajima (Physical Society of Japan, Tokyo, 1984), pp.~265--276.
\end{itemize}

2)  Equation 1:  There's a typo.  ``Atom decayed'' also appears in the second term of the superposition.

3)  Section 10 (the big table).  This may have been my favorite part of the paper.  But I would do this:  Put the relativity column on the left and the Everett column on the right.  Somehow, I found it much easier to read in that manner.  It just seemed more natural to recognize a feature in relativity first, and then look for the analogous feature in Everett.

4)  There's a typo in the fifth entry of that table.  The words ``choice of choice of'' appear in the right-hand column.

\section{27-01-01 \ \ {\it The Test Particle} \ \ (to D. Wallace)} \label{Wallace2}

I presume by now you've seen my thoughts on your Everett article.  I meant it all; it's a very nice paper.  I'd like to encourage you to place the article on the {\tt quant-ph\/} archive.  I'd like to refer some of my friends to it to the get a debate going.  Having the paper easily accessible will help get that off the ground.

One piece of analogy (or disanalogy) that you didn't explore very much is the geodesic.  Is there an analogue to the geodesic in your Everettian system?  One thing that strikes me is that you might start seeing a conceptual divergence here.  It's not clear to me that one can concoct a good notion of ``test particle'' for this game:  even adding the smallest system possible to an existing multiverse (I hate that term) doubles its Hilbert space dimension (acting as if the Hilbert space is finite to begin with).

\section{31-01-01 \ \ {\it Ah Midsummer \ldots} \ \ (to R. Jozsa)} \label{Jozsa2}

Thanks for the report on Butterfield.  I just refereed a paper by David Wallace (for Butterfield's journal) on a variant of many worlds.  It's a paper that Butterfield is very sympathetic to.  The impressive thing about it was that maybe it made more sense than anything else I've read on the idea.  So I accepted it for publication.  I made it very clear that I continue to think that MWI is a dead-end, contentless idea.  But I feel morally obligated to think that only of (what might be) their best shot!

\section{31-01-01 \ \ {\it Corrections FUCHS Manuscript 060103PRA}\ \ \ (to {\sl Physical Review A})}

Along with this note, I will fax back my corrections listed directly on the page proofs.  All page numbers below refer to the marked up proofs.  On the faxed copy, all modified pages are marked with a star in the bottom right corner.

The vast majority of your changes improved the readability of the paper.  I thank you for making them.  However, please note that I am fairly adamant about my remarks concerning pages 1 and 4, which have turned into something of a bone of contention because of a previous experience.  Please consider my point of view with respect. \medskip

\noindent \underline{Page 1)} \bigskip

b)  First sentence of abstract.  I wish Physical Review would review its policy of blindly deleting all uses of the word ``new'' in titles and abstracts.  Sometimes authors make use of this term for exactly the opposite of self-aggrandizing purposes.  In my original draft I started with the sentence:
\bq\noindent
``In this paper we give a new way to quantify the folklore notion that quantum measurements bring a disturbance to the system being measured.''
\eq
You changed it to:
\bq\noindent
``In this paper we provide a way to quantify the notion that quantum measurements bring a disturbance to the system being measured.''
\eq
Written in this manner (without any extra qualification) it makes the authors appear arrogant.  Any reader who knows anything whatsoever of quantum mechanics will think with disdain, ``As if that's never been quantified before?!?!''  Having said my piece, however, I will accept your deletion of the word ``new.''

What I cannot accept is your complete deletion of the word ``folklore.'' It plays an important role in that sentence, and I ask that you replace it in its adjectival form ``folkloric.''  The original word (barring only its mistaken grammatical form) was placed there quite carefully.  It conveys the idea that we are speaking about something that is taken to be common knowledge, even when it has never been quantified.  Quantifying it is the point of our paper.  ``FOLKLORE:  the traditional beliefs, myths, tales, and practices of a people.''  Physicists, being human, are as susceptible to folklore as anyone else.  You write, ``Author--Please use more literal, meaningful modifiers.''  It was meant to be taken literally:  there is no word that captures the concept better.

The fight over this word is a funny one with {\sl Physical Review}.  Asher Peres and I in our paper PRA {\bf 53}, 2038 (1996) had to fight for it in our opening sentence then.  Finally Bernd Crasemann (then editor) accepted our point.  Here is the original conversation, drawn from my email archive:\medskip

\noindent Peres:
\bq
        \noindent Dear Sirs: \smallskip

        I just received the proofs of this article, and I am shocked that the fourth word of the first paragraph, ``folklore'', was changed into ``convention'' by the copy-editor, who added a comment ``we favor more literal and accurate terminology''.  The authors have carefully chosen the word ``folklore'' because it has, in the present context, a slightly derogative meaning. This is exactly the message we want to convey to the readers of this article. I understand that the role of a copy-editor is to correct typos, style and grammar errors, and the like. There should be no distortion of the original meaning of the text. The authors, not APS or AIP, are responsible for the contents of their article.

        Please confirm that you are willing to accept the term ``folklore'' that was chosen by us. In case of a negative answer, we shall have to reconsider whether to publish our work in {\sl Physical Review}, or withdraw it and submit it to another journal (we would prefer {\sl Physical Review}).
\eq
Crasemann:
\bq
        You convincingly justify your use of the word ``folklore'' and we will certainly let it stand, even though it is not listed in the index of your admirable book, {\sl Quantum Theory:\ Concepts and Methods}, which sits on my desk and has been a source of great edification.

        Permit me, however, to rise in defense of our staff's unceasing struggle to guard the integrity of the journal's scientific language. It was not a copy editor, but a Senior Assistant to the Editor, endowed with enviable Sprachgefuehl, who suggested the change in an effort to expunge unduly colloquial terminology. While no one is perfect, our colleagues in the Editorial Offices do a truly outstanding job safeguarding the journal's style.
\eq
I see no reason why the decision should be different this time around. \medskip

\noindent \underline{Page 2)} \bigskip

a)  Second paragraph, third sentence.  I originally wrote, ``If there were a set of hidden variables underneath \ldots''  You changed it to ``If there is a set \ldots''  This question was meant to be a hypothetical, and as such the verb should be ``were.''  There is no question in my mind that there {\it are no\/} hidden variable underneath quantum mechanics.  Please replace ``were.'' \medskip

\noindent \underline{Page 3)} \bigskip

a)  Second paragraph, sixth sentence.  I wrote ``What is novel here is that the encoding \ldots''  You changed ``novel'' to ``interesting,'' again charging that I should use more literal wording.  ``NOVEL: strikingly new, unusual, or different.''  ``INTERESTING: arousing or holding the attention.''  Novel is the appropriate word for that sentence (describing a discovery of 1970); please reinstate it. \medskip

\noindent \underline{Page 4)} \bigskip

a)  Second paragraph, fifth sentence.  I accept your point that the phrase ``founding fathers'' is a colloquialism.  However, substituting ``original description'' does not capture the proper idea either.  Please replace ``founding fathers'' with ``common folklore'' (in line with my remarks above).

b)  Third paragraph, first sentence.  Again, please substitute ``folklore'' for ``the founding fathers.''\medskip

\noindent \underline{Page 12)} \bigskip

a)  Sentence just above Eq.\ (16).  You changed ``\ldots\ one can think of the interaction as causing the system to further unitarily evolve to'' to ``\ldots\ to evolve further unitarily to.''  I understand you don't want me to split my infinitive, but ``unitarily evolve'' is the proper atomic verb.  Please change the phrase to ``\ldots\ to unitarily evolve further to.''\medskip

\noindent \underline{Page 13)} \bigskip

a)  First paragraph, second to last sentence.  Indeed ``get a handle on'' is a colloquialism, almost a Texanism.  But ``comprehend'' doesn't capture the right idea either.  Please replace the sentence with, ``Finally, it stands to reason that if we can delineate the tradeoff \ldots''\medskip

\noindent \underline{Page 15)} \bigskip

a)  Second paragraph, third sentence.  I wrote, ``When this obtains, the Shannon \ldots''  You wrote, ``When this answer is obtained, the Shannon \ldots''  Completely different meaning.  Please replace it with, ``When this is the case, the Shannon \ldots''\medskip

\noindent \underline{Page 19)} \bigskip

a)  Very top of page.  Consistency is the hobgoblin of my small mind.  99.9\% of all ``non'' words in the English language are nonhyphenated.  Check a {\sl Webster's\/} dictionary; check an {\sl American Heritage\/} dictionary.  It seems only in the physics community that this standard rule is broken.  Please reinstate ``nonnegative'' in place of ``non-negative.''

\section{31-01-01 \ \ {\it Tomography}\ \ \ (to S. J. van {\Enk})} \label{vanEnk19}

The second reference below sounds a lot like it's doing what you say?

\begin{itemize}
\item
K. Vogel and H. Risken, ``Determination of Quasiprobability Distributions in Terms of Probability Distributions for the Rotated Quadrature Phase,'' Phys.\ Rev.\ A {\bf 40}, 2847 (1989).

\item
D.~T. Smithey, M. Beck, M.~G. Raymer, and A. Faridani, ``Measurement of the Wigner Distribution and the Density Matrix of a Light Mode Using Optical Homodyne Tomography:\ Application to Squeezed States and the Vacuum,'' Phys.\ Rev.\ Lett.\ {\bf 70},  1244  (1993).

\item
U. Leonhardt, ``Quantum-State Tomography and Discrete Wigner Function,'' Phys.\ Rev.\ Lett.\ {\bf 74}, 4101 (1995).
\end{itemize}

\section{01-02-01 \ \ {\it Oh Translator}\ \ \ (to S. J. van {\Enk})} \label{vanEnk20}

Can you translate the German in this [quote of E. T. Jaynes]?

\bq
For some sixty years it has appeared to many physicists that probability plays a fundamentally different role in quantum theory than it does in statistical mechanics and analysis of measurement errors.  It is a commonly heard statement that probabilities calculated within a pure state have a different character than the probabilities with which different pure states appear in a mixture, or density matrix.  As Pauli put it, the former represents `\,\dots\,eine prinzipielle {\it Unbestimmtheit}, nicht nur {\it Unbekanntheit}'.  But this viewpoint leads to so many paradoxes and mysteries that we explore the consequences of the unified view, that all probability signifies only human information.
\eq

\subsection{Steven's Reply}

\bq
Literally it says
\bq
a principle undeterminedness, not just unknownness.
\eq
where principle here is used as an adverb meaning something
like fundamental.

Does that help?
\eq

\section{01-02-01 \ \ {\it Zoom, Zoom, Zoom}\ \ \ (to G. Brassard)} \label{Brassard2}

Can you come, can you come, can you come??!?!?!?!

\section{01-02-01 \ \ {\it Zing, Zing, Zing}\ \ \ (to G. Brassard)} \label{Brassard3}

Please come, please come, please come!!

\section{02-02-01 \ \ {\it Zub, Zub, Zub}\ \ \ (to G. Brassard)} \label{Brassard4}

Please come, please come, please come!

\section{02-02-01 \ \ {\it Zurek Criticisms} \ \ (to A. Kent)} \label{Kent2}

Considering the careful critic that you are, you must have at some point written up your thoughts on Zurek's view of the quantum interpretation problem (to the extent that he has a well-defined view).  Where can I find that?  Alternatively, do you know of any other good critique papers by others along those lines?

\section{05-02-01 \ \ {\it The Eerie Parallel} \ \ (to A. Kent)} \label{Kent3}

\bak
[Functionalism is] a view which has little to recommend it except a pleasing sense
of answering a deep question with no work, and so naturally has become
very widely held and respected among philosophers
of mind.   In this, it eerily parallels the Everett interpretation.
It is only fitting that the two should be combined into a grander
exercise in question-begging.
\eak

At times in the past I've found myself wanting to be able to write like Mark Twain, and like William James.  Both had this way of putting things---very different ways---that made me shiver from seeing the truth in their thoughts.  Today I found myself wanting to be able to write like you!

\section{06-02-01 \ \ {\it Interesting Coincidence}\ \ \ (to A. S. Holevo)} \label{Holevo1}

I'm writing up a paper (finally!)\ on this quantum de Finetti stuff you saw in Cambridge two years ago.  And while building the bibliography, I had to place the following two contiguous entries:
\begin{itemize}
\item
A.~S. Holevo, ``Information-Theoretical Aspects of Quantum Measurement,'' Prob.\ Info.\ Trans.\ {\bf 9}, 110 (1973).

\item
E.~Prugove\v{c}ki, ``Information-Theoretical Aspects of Quantum Measurements,'' Int.\ J. Theo.\ Phys.\ {\bf 16}, 321 (1977).
\end{itemize}
I cited them for different reasons, and the two papers have quite distinct contents, but isn't it funny how the citations fell next to each other!

\section{06-02-01 \ \ {\it Zoink!, Zoink!, Zoink!}\ \ \ (to G. Brassard)} \label{Brassard5}

Will you come?  Will you come?  Will you come?

(Every time I do this I have visions of the video version of {\sl Horton Hears a Who}.)

\section{07-02-01 \ \ {\it GHZM} \ \ (to N. D. {\Mermin})} \label{Mermin-3}

Weren't you the first person to show that there's something weird about the state $000+111$ (even though everyone calls that a GHZ state)?  If so, can you give me the full reference for that paper:  including title and page numbers.  If no, can you still give me the full reference for {\it that\/} paper:  including title and page numbers.

\subsection{David's Reply}

\bq
Very scholarly of you to notice.  The GHZ state is not in the GHZ
paper.  But they deserve the glory. I only constructed my version
after hearing about their result and thinking that there ought to be a
simpler way to make the same point.  (Standing on the shoulders
of GHZnts.)

I wrote about GHZ in two places.  One is a {\sl Physics Today Reference
Frame\/} column (the only Reference Frame column I know of to receive
citations in the technical literature) where I launch the sloppy
scholarship by suggesting that GHZ invented the $000 + 111$ version of the
argument as well as the general idea; the other's an AJP article.
\begin{itemize}
\item         ``What's Wrong with These Elements of Reality?,''
                Phys.\ Tod., 9--11, June, 1990.

\item               ``Quantum Mysteries Revisited,''
                  Am.\ J. Phys.\ {\bf 58}, 731--734, 1990.
\end{itemize}

There was a reason for my sloppy scholarship, by the way. I was quite
enchanted with what I had boiled their argument down to and realized
that if I attributed it to them, I could praise it extravagantly.  But
if I presented it as mine, I'd have to be boringly objective.
\eq

\section{07-02-01 \ \ {\it The Scholar and the Cut} \ \ (to N. D. {\Mermin})} \label{Mermin-2}

\bdm
Very scholarly of you to notice.
\edm

I am nothing if not a scholar.  (Too bad I haven't yet been a good physicist too.  I keep waiting for the day \ldots)

\section{07-02-01 \ \ {\it Wow!}\ \ \ (to A. Plotnitsky)} \label{Plotnitsky2}

Yesterday I received the package you mailed to me!  I can't express how grateful I am.  I never imagined you would send me your book!!  I'll make you a promise:  I'll work very hard to digest all of it before meeting you again.  I feel that a lot of our thoughts run fairly parallel, but I know that I have a lot of room for growth.  (You, by the way, couldn't have gotten a better endorsement in my eyes than that John Wheeler was interested in your book.)

\section{07-02-01 \ \ {\it Reports and Chutzpah} \ \ (to A. Kent)} \label{Kent4}

\bak
I gather you'd already read David's article: maybe in fact that
prompted the Zurekian query?
\eak

Yeah, I was pretty open with my report.  Might as well share it with you too; I'll paste it below.  [See 26 January 2001 entry titled ``Worlds in
the {\Everett} Interpretation'' in {\sl Coming of Age with Quantum Information}.]

But, no, that's not what prompted my Zurekian query.  That was prompted by the paper I'm writing for the NATO meeting last summer.  (I'm a little late!)  The title will be ``Quantum Foundations in the Light of Quantum Information,'' and mostly it will be devoted to laying out my program (rather than reporting results).  But because the meeting was on decoherence and its connection to foundations, I thought it behooved me to say in a crisp way that it has {\it no\/} connection to foundations before going into my own spiel.

\section{09-02-01 \ \ {\it A Little GHZM with Your States, Sir?}\ \ \ (to S. L. Braunstein)} \label{Braunstein1}

I was thinking about a conversation we once had, and it dawned on me that you're probably the only one I dare tell this joke to.  You once asked me for confirmation that David Mermin was actually the inventor of the three-particle GHZ state:  GHZ, in their original paper, only talked about a four particle state.  (I'll place the full story of this in David's words below.) [See 07-02-01 note ``\myref{Mermin-3}{GHZM}'' to N. D. {\Mermin}.] Anyway, I was toying with the idea of calling the three-particle GHZ state the GHZM state---maybe you once did too?---but how do you think that would end up being pronounced?  ``\verb+!@?$+'' is the first thing that comes to mind \ldots\ but we can't have that at professional conferences!

\section{10-02-01 \ \ {\it Historical Accuracy} \ \ (to N. D. {\Mermin})} \label{Mermin-1}

While I'm digging for complete historical accuracy \ldots\  Can you give me the complete reference for the very first appearance of GHZ's four-party state.  Was it some AJP article including Shimony?  There must have been something before that.

In any case, can you give me the complete reference including the title of the paper.  Sorry to keep bugging you.

\subsection{David's Reply}

\bq
The very first GHZ paper was: ``Going Beyond Bell's Theorem'', in {\it Bell's Theorem, Quantum Theory,
and Conceptions of the Universe} (! --- conceptions of the laboratory
would have been good enough for me), ed.\ M. Kafatos (Kluwer Academic,
Dordrecht, The Netherlands, 1989) pp.\ 69--72.

The AJP article you mention is ``Bell's Theorem without Inequalities'',
Daniel M. Greenberger, Michael A. Horne, Abner Shimony, Anton
Zeilinger, Am.\ J. Phys.\ {\bf 58}, 1131--1143 (1990).
\eq

\section{10-02-01 \ \ {\it Our Time with Brick and Keech} \ \ (to D. B. L. Baker)} \label{Baker1}

You know, the years I spent with you during our youth make you partly to blame for this!  I just wrote the following two footnotes for a paper by Caves, Schack and me, slated to appear in the {\it prestigious\/} {\sl American Journal of Physics}.  Tell me if you get the joke.  The greatest mark of success will be if you (and readers like you) get it \ldots\ but the editor doesn't and therefore lets it slip past his red pen.

Let me know.

\begin{itemize}
\item
\verb+\bibitem{Mermin1990}+\smallskip\\
N.~D. Mermin, ``What's Wrong with These Elements of Reality?,'' Phys.\ Tod.\ {\bf 43}(6), 9 (1990); N.~D. Mermin, ``Quantum Mysteries Revisited,'' Am.\ J.\ Phys.\ {\bf 58}, 731 (1990).  Though Mermin was the first to point out the interesting properties of this three-system state (following the lead of Ref.~\verb+\cite{Greenberger1989}+ where a similar four-system state was proposed), we call attention to the pronunciational perils of calling the state a ``GHZM state'' and, thus, defer to the more common label GHZ.

\item
\verb+\bibitem{Greenberger1989}+\smallskip\\
D.~M. Greenberger, M.~Horne, and A.~Zeilinger, ``Going Beyond Bell's Theorem,'' in {\sl Bell's Theorem, Quantum Theory and Conceptions of the Universe}, edited by M.~Kafatos (Kluwer, Dordrecht, 1989), p.~69.
\end{itemize}

\section{10-02-01 \ \ {\it Smoke on the Water} \ \ (to C. M. {\Caves} \& R. {\Schack})} \label{Schack0} \label{Caves0}

Fire in the Sky!  I finally got that pimple off my butt.

In the next email I'll send the quantum de Finetti paper as it stands presently.  Per my promise to {\Ruediger}, I didn't change his 14 page draft all that much:  It only just barely creeps over 25 pages now.

I suspect that I need not remind Carl of the occupational hazards of working with me, but while I'm here I might as well hint at the main one.  Given the (literally) hundreds of times the words in the Introduction rolled over in my head, if you change anything in that section, you'd better have a darned good reason for it.  And moreover, be prepared to exhibit the reasons automatically when sending out the revised version.  I don't take well to editors who make arbitrary changes to my English or ideas:  My presentation was anything but arbitrary; likewise should be any changes you make to it.  (You can see I had a bad experience with PRA last week.)

You'll certainly notice some idiosyncrasies in the paper.  Perhaps the biggest is that I banned all usage of the term ``Copenhagen interpretation.''  One thing I've learned from the lecture circuit is that any usage of the term is just plain suicide.  On the other hand, I've also learned that if you wrap the same ideas in the banner of ``informational point of view'' or ``information-based interpretation'' people perk their ears up and act interested.  They think they're going to hear something new, and then don't even recognize it when I blurt out the same old same old.  Similarly, I was a little more careful with the word ``subjective.''

Looking back on the paper now, after building up a fuller annotation, etc., I'm struck that the content is not half bad.  It might as well have been the case that the proof of the theorem didn't exist for the last 25 years.  This is because the presentations it has been given so far conveyed no sense whatsoever of the importance of the theorem.  I can well believe that Hudson and Moody, Accardi, and Petz had no strong reason for going through the mathematical machinations they did (other than to do it just for the sake of doing it).  So, indeed, our paper will be a contribution after all.

\section{11-02-01 \ \ {\it Explanatory Note} \ \ (to R. {\Schack})} \label{Schack0.1}

I hope you are, on the whole, pleased with the content of the draft I sent you yesterday.  Looking back this morning at my note preceding it, I know that I owe you an apology.  Its tone and abruptness has become sort of a habit for me lately.  The last few months have seen me become more and more belligerent, and I'm not completely sure why.  It's almost as if a kind of paranoia is creeping over me.  In its taproot, it definitely has something to do with the fire.  But there's no doubt that the problem is being exacerbated by the many academic/intellectual/administrative debts I have accumulated with my friends and colleagues.  I feel like I'm constantly fighting against bankruptcy.  But it's something that's got to pass, or I really will be bankrupt and a has-been.

We've been friends for over seven years.  I don't want to ask you to have patience with me, but maybe to turn a blind eye to my last few indiscretions.  I'm trying to clean myself up.

I'm glad you're coming to Sweden.  You and Schumacher pulled off the two best and deepest presentations at the {\Montreal} meeting.  If we get a repeat performance of that, the community will not help but take note that the times are changing.\medskip

\noindent On an observant Sunday,

\section{12-02-01 \ \ {\it AMS Abstract} \ \ (to myself)} \label{FuchsC1}

\bq\noindent
Title: The Power of Generalized Measurements (POVMs)\medskip\\
Abstract: It has become customary in quantum information theory to think of ``generalized measurements'' or ``positive-operator valued measures'' (POVMs) as a derivative concept.  This is because it is known that a POVM can always be built up from a standard von Neumann measurement on an ancillary system that has previously interacted with the target system.  In this talk we point out the utility of taking the POVM as a basic notion.  This strategy greatly reduces the difficulty of proving many very basic theorems in quantum mechanics and gives a deeper insight into why ``entanglement'' has the exact structure that it does.
\eq

\section{14-02-01 \ \ {\it Shannon and Bohr}  \ \ (to E. Merzbacher)} \label{Merzbacher1}

Since I haven't heard back from you concerning the ``Shannon meets Bohr'' meeting, I'm going to assume that you cannot come.  In any case, our funds are fairly exhausted now.  But I would have loved to have had you there, especially to expand on this wonderful thing you wrote me in 1999:
\bem
For a long time I have thought about writing an essay on what is meant
by the term ``physical system'', including some history of that slippery
concept.  I'll probably never do it, but your opinion piece has
brought this matter back to my mind.
\eem
Do it some day!  You'd automatically have an audience of at least one.  (That's something I cannot say for some of my papers!)

\section{17-02-01 \ \ {\it Saturday with Some Jazz}\ \ \ (to A. S. Holevo)} \label{Holevo2}

\bash
   The conference was interesting, among people you know were Milburn,
Massar, van Dam, Werner, \ldots\@. Quite remarkable was great interest from
classical probabilists and mathematical statisticians to learn quantum
information. Eurandom is a good place and we should keep it in mind. \ldots\ \ The highlight was supposed to be the lecture of a
Nobel-price winner 't Hooft \ldots\@.  The point of his lecture was that on Planck
distances, where gravitational interaction becomes important, one may
convert to nonlocal hidden variables in place of quantum theory. It was
interesting to see how classical prejudices are strong even in the most
distinguished physicists' minds.
\eash

Thanks for the report of the Europhys conference.  Indeed your remark about 't Hooft was most interesting.  Today I am taking a leisurely day, thinking about philosophical things and compiling further my Bohrish--Paulian compendium (I now have 355 references), and thus thinking about such things.

We had a very busy week here at Bell Labs.  Chris King gave a wonderful talk, and much of Jim Mazo's information theory group was in attendance.  Mary Beth outlined a proof of the equality conditions for the strong subadditivity and this led to some good discussion about notions of Markov chains for quantum states.  Serap and Gerhard gave a talk about ``quantum coding'' for commutative alphabets.  And then while Peter Shor was visiting, Emina Soljanin presented what she thinks is a proof of the achievability of (your) $\chi$ in the noncommutative case.  (Recall that Barnum and several of us proved $\chi$ to be a lower bound; M. Horodecki also proved it independently.)  Unfortunately I had to miss that talk.  Finally, I gave my final tutorial talk on basic quantum mechanics (from the quantum information slant) to Mazo's group; when I return from Japan I will start a set of tutorials on quantum information proper (for the smaller, more prepared audience).

\section{17-02-01 \ \ {\it My Bohr--Pauli Compendium}\ \ \ (to K. Barad)} \label{Barad1}

Your quantum writings have been brought to my attention by Herb Bernstein and Mike Fortun, and I wanted to tell you that your one article that I did read was wonderful!  (I'll place the passages that took me the most below.)

Mainly I'm writing this letter, though, in hopes that I can get you to send me some more of your writings or at least fill in the missing information for the two references (far) below.  Any way you can help would be most appreciated.  I am compiling a bibliographical compendium of sensible writings about quantum mechanics, and I would like to include all your work there.

\begin{itemize}
\item
K.~Barad, ``A Feminist Approach to Teaching Quantum Physics,'' in {\sl Teaching the Majority:\ Breaking the Gender Barrier in Science, Mathematics, and Engineering}, edited by S.~V. Rosser (Teachers College Press, New York, 1995), pp.~43--75.
\end{itemize}

\bq
The Newtonian worldview is compatible with an objectivist epistemology, in which the well-prepared mind is able to produce a privileged mental mirroring of the world as it exists independently of us human beings.  That is, what is ``discovered'' is presumed to be unmarked by its ``discoverer.''  The claim is that the scientist can read the universal equations of nature that are inscribed in [God's]
blackboard:  Nature has spoken. Paradoxically, the objects being studied are given all the agency, even and most especially when they are seen as passive, inert objects moving aimlessly in the void.  That is, these cultureless agents, existing outside of human space-time, are thought to reveal their secrets to patient observers watching and listening through benignly obtrusive instruments. Notice that agency is not attributed to human beings; once all subjective contaminants have been removed by the scientific method, scientists simply collect the pure distillate of truth.

The Newtonian worldview is still so much a part of contemporary physics culture that it infects the teaching of post-Newtonian physics as well.  That is, the stakes are so high in maintaining the mirroring view of scientific knowledge that quantum physics is presented as mysticism.
\eq
and
\bq
Notice that particular experimental arrangements can be used to give more or less definite meaning to each of the complementary variables, but due to the lack of object-instrument distinction \ldots\ it is not possible to assign the value obtained to the object itself.  The ``property'' being measured in a particular experimental context is therefore not ``objective'' (that is, a property of the object as it exists independently of all human interventions), but neither is it created by the act of measurement (which would belie any sensible meaning of the word {\it measurement\/}).  Bohr speaks of this ``interaction'' between ``object'' and ``instrument'' as a ``phenomenon.''  The properties then are properties of phenomena. That is, within a given context, classical descriptive concepts can be used to describe phenomena, our intra-actions within nature.  (I use the term {\it intra-action\/} to emphasize the lack of a natural object-instrument distinction, in contrast to {\it interaction}, which implies that there are two separate entities; that is, the latter reinscribes the contested dichotomy.  \ldots\ That is, the ambiguity between object and instrument is only temporarily contextually decided; therefore, our characterizations do not signify properties of objects but rather describe the intra-action as it is marked by a particular constructed cut chosen by the experimenter (see Ref.\ [Barad95] for more details).

The notion of ``observation'' then takes on a whole new meaning according to Bohr:  ``[B]y an experiment we simply understand an event about which we are able in an unambiguous way to state the conditions necessary for the reproduction of the phenomena'' (quoted in Ref.\ [Folse85], p.~124)\@.  According to the analysis of the previous section, this is possible because, in performing each measurement, the experimenter intervenes by introducing a constructed distinction between the ``object'' and the ``measuring device'' (e.g., deciding whether the photon is part of the object or the instrument).
The claim is that unambiguous, reproducible measurements are possible through the introduction of constructed cuts.  Notice that ``[n]o explicit reference is made to any individual observer'':  Different observers will get the same data set in observing any given phenomenon.
Therefore, reproducibility, not some Newtonian notion of objectivity denoting observer independence, is the cornerstone of this new framework for understanding science.

For Bohr, the uncertainty principle is a matter of the inadequacy of classical description.  Unlike the ``mirroring'' representationalism inherent in the Newtonian-Cartesian-Enlightenment framework of science, scientific concepts are not to be understood as describing some independent reality.  A post-Newtonian framework sees these constructs as useful (i.e., potentially reproducible) descriptions of the entire intra-action process (the phenomenon, which is context dependent by definition), not of an isolated object.  The implications of this finding are profound.  In Bohr's own words:
\bq\noindent
The extension of physical experience in our own days has \ldots\ necessitated a radical revision of the foundation for the unambiguous use of elementary concepts, and has changed our attitude to the aim of physical science.  Indeed, from our present standpoint, physics is to be regarded not so much as the study of something a priori given, but rather as the development of methods for ordering and surveying human experience. (Bohr, Ref.\ [Bohr63c], p.~10)
\eq
In other words:
\bq\noindent
These facts not only set a limit to the extent of the information obtainable by measurements, but they also set a limit on the meaning which we may attribute to such information.  We meet here in a new light the old truth that in our description of nature the purpose is not to disclose the real essence of [physical objects] but only to track down, so far as it is possible, relations between the manifold aspects of our experience. (Bohr, Ref.\ [Bohr63a], p.~18)
\eq
\eq
and
\bq
Bohr's philosophy of physics involves a kind of realism in the sense that scientific knowledge is clearly constrained, although not determined, by ``what is out there,'' since it is not separate from us; and given a particular set of constructed cuts, certain descriptive concepts of science are well-defined and can be used to achieve reproducible results.  However, these results cannot be decontextualized.  Scientific theories do not tell us about objects as they exist independently of us human beings; they are partial and located knowledges. Scientific concepts are not simple namings of discoveries of objective attributes of an independent Nature with inherent demarcations.  Scientific concepts are not innocent and unique.  They are constructs that can be used to describe ``the between'' rather than some independent reality.  (Why would we be interested in such a thing as an independent reality anyway?  We don't live in such a world.)  Consideration of mutually exclusive sets of concepts produces crucial tensions and ironies, underlining a critical point about scientific knowledge:  It is the fact that scientific knowledge is socially constructed that leads to reliable knowledge about ``the between''---which is just what we are interested in.  This shifting of boundaries deconstructs the whole notion of identity:
Science can no longer be seen as the end result of a thorough distillation of culture.  There is an author who marks off the boundaries and who is similarly marked by the cultural specificities of race, history, gender, language, class, politics, and other important social variables.  Reproducibility is not a filter for shared biases.  In stark contrast to the objectivist representationalism that is usually transmitted to students, the new framework inspired by Bohr's philosophy of physics is robust and intricate.  In particular, there is an explicit sense of agency and therefore accountability.  And so I refer to this Bohr-inspired framework, which shares much in common with central concerns in contemporary feminist theories, as ``agential realism.''
\eq

\subsection{Karen's Reply}

\bq
Thank you for your interest in my work. As per your request, below is a list of some of my other articles further elaborating Bohr's philosophy-physics. I hope you find these useful.

\begin{itemize}
\item
K.~Barad, ``Meeting the Universe Halfway:\ Realism and Social
Constructivism without Contradiction,'' in {\sl Feminism, Science,
and the Philosophy of Science}, edited by L.~H. Nelson and J.~Nelson
(Kluwer, Dordrecht, 1996), pp.~161--194.

\item
K.~Barad, ``Getting Real:\ Technoscientific Practices and the
Materialization of Reality,'' in Differences:\ A Journal of Feminist
Cultural Studies {\bf 10}(2), 87--126 (1998).

\item
K.~Barad, ``Agential Realism:\ Feminist Interventions in
Understanding Scientific Practices,'' in {\sl The Science Studies
Reader}, edited by M.~Biagioli (Routledge, NY, 1998), pp.~??--??.

\item
K.~Barad, ``Reconceiving Scientific Literacy as Agential
Literacy:\ Or, Learning How to Intra-act Responsibly Within the
World,'' in {\sl Doing Science + Culture}, edited by R.~Reid and
S.~Traweek (Routledge, NY, 2000), pp.~221--258.

\item
K.~Barad, ``Agential Realism,'' in {\sl Routledge Encyclopedia
of Feminist Theories}, edited by L.~Code (Routledge, NY, 2000).

\item
K.~Barad, ``Re(con)figuring Space, Time, and Matter,'' in {\sl
Feminist Locations}, edited by M.~DeKoven (Rutgers University Press,
New Brunswick, NJ, 2001), pp.~??--??.

\item
K.~Barad, ``Scientific Literacy $\rightarrow$ Agential Literacy
= (Learning + Doing) Science Responsibly,'' in {\sl A New Generation
of Feminist Science Studies}, edited by M.~Mayberry, B.~Subramaniam,
and L.~Weasel (Routledge, NY, 2001), pp.~??--??.
\end{itemize}
\eq

\section{21-02-01 \ \ {\it There Are No Quantum States}  \ \ (to A. Peres)} \label{Peres2}

\bap
There are no quantum states (in a relativistic theory). Therefore it is
pointless to discuss collapse, there is no EPR paradox, etc. What the
theory is about is propagators (or rather superpropagators $=$ completely
positive maps) from initial preparation to final observation. However,
an explicit evaluation of these superpropagators is best done by
introducing fake notions (quantum states), just as we use a vector
potential in electrodynamics, or a spacetime metric in general
relativity, because it is cumbersome not to use these gauge dependent
objects. I don't know what these ideas will bring, but at least I have
an impression that now I understand better.
\eap

\begin{quotation}\noindent
My thesis, paradoxically, and a little provocatively, but nonetheless genuinely, is simply this:
\begin{center}
QUANTUM STATES DO NOT EXIST.
\end{center}
The abandonment of superstitious beliefs about the existence of Phlogiston, the Cosmic Ether, Absolute Space and Time, \ldots, or Fairies and Witches, was an essential step along the road to scientific thinking. The quantum state, too, if regarded as something endowed with some kind of objective existence, is no less a misleading conception, an illusory attempt to exteriorize or materialize the information we possess.\medskip
\\
\hspace*{\fill} --- {\it the ghost of Bruno de Finetti} \end{quotation}

\section{21-02-01 \ \ {\it There Are No Quantum States, 2}  \ \ (to A. Peres)} \label{Peres3}

\bap
Where did you write that?
\eap

I've never put it in real print yet, only in my philosophical samizdat.  Maybe I'll find a way to get it into real print soon.

What am I up to?  Too much and not enough at the same time!

\section{21-02-01 \ \ {\it A Reference}  \ \ (to P. Pearle)} \label{Pearle1}

If I were going to cite one (and only one) reference on spontaneous collapse ideas, which one should it be?  I just want something, even a popular exposition, that can serve as a point of departure for my readers.

To give you a flavor of what I'm looking for I cited the following in connection with the Einselectionists, the Bohmians, the Everettistas, and the Consistent Historians.
\begin{enumerate}

\item
W.~H. Zurek, ``Decoherence, Einselection and the Existential Interpretation (The Rough Guide),'' Phil.\ Trans.\ R.\ Soc.\ Lond.\ A {\bf 356}, 1793--1821 (1998).

\item
J.~T. Cushing, A.~Fine, and S.~Goldstein, editors, {\sl Bohmian Mechanics and Quantum Theory:\ An Appraisal}, (Kluwer, Dordrecht, 1996).

\item
D.~Deutsch, {\sl The Fabric of Reality: The Science of Parallel Universes---and its Implications}, (Allen Lane, New York, 1997).

\item
R.~B. Griffiths and R.~Omn\`es, ``Consistent Histories and Quantum Measurements,'' Phys.\ Today {\bf 52}(8), 26--31 (1999).
\end{enumerate}

\section{26-02-01 \ \ {\it THE Book} \ \ (to D. P. DiVincenzo)} \label{DiVincenzo1}

I just wrote to Barbara explaining what a heel I was for forgetting to ask how she made it home during the big snow storm (after her interview here).  But I'm also a heel for not writing earlier to thank you for the wonderful Jammer book you bought me.  I really, really appreciate it.  It is THE classic in quantum foundations.


\section{27-02-01 \ \ {\it Claude Shannon's Death} \ \ (to the ``Shannon meets Bohr'' Invitees)}\label{ClaudeShannonDeath}

\noindent To those friends I contacted for the ``Shannon meets Bohr'' session in {\Vaxjo}: \medskip

We at Bell Labs were all deeply moved yesterday to learn of Claude Shannon's death over the weekend.  Like for many Alzheimer's patients, however, we knew that it may have been a blessing in disguise.  You can read more about Shannon's life in the {\sl New York Times\/} obituary section today.

Some weeks ago, I found myself using the following words for a recommendation letter I was asked to write:
\bq\noindent
As it turns out, today January 16, we had a dedication ceremony at Bell Labs for a bronze bust of Claude Shannon, the founder of classical information theory.  That struck me as symbolic.  Since joining the laboratory, I have been asking myself over and over what role I might play in furthering the legacy of Shannon?
\eq
Those words hold just as true for me for this midsummer's meeting.  Let us use information theory as a sword and finally defeat this mystery of the quantum.

\section{28-02-01 \ \ {\it Claude Elwood Shannon} \ \ (to C. E. Laake, W. M. Fuchs, \& C. N. Franklin)} \label{Laake1} \label{FuchsW0} \label{FranklinCE1}

\noindent Dear Family, \medskip

There was one man who essentially single-handedly created the field of information theory---the field that has become the larger part of my life's work.  His name was Claude Shannon, and, though I never met him, his papers and his students influenced me greatly.  Shannon died at the age of 84 last Saturday after many years of a deep Alzheimer's disease, much like our Grannie Thigpen's.  I wrote a little epitaph in honor of him this morning and sent it to the attendees of a meeting I'm organizing in Sweden in June.  [See 27-02-01 note titled ``\myref{ClaudeShannonDeath}{Claude Shannon's Death}'' to the ``Shannon meets Bohr'' attendees.]  The title of the meeting will be ``Shannon meets Bohr:\ Quantum Foundations in the Light of Quantum Information,'' and so it seemed so appropriate.  But while I was writing, I think I was thinking as much of Grannie as I was of Shannon.  So, in a way, the piece is in memoriam to her too.

Could one of you please pass it (along with the NYTimes article on Shannon) on to Mom?\medskip

\noindent \medskip Love,

\noindent Chris

\section{28-02-01 \ \ {\it History's Mysteries --- A Real One!}\ \ \ (to family and old friends)}

\noindent Dear Family and Friends (mostly from my old home town \ldots\ I'm too embarrassed to tell my colleagues), \medskip

I likely made a good fool of myself, but if you have cable TV, you might be on the lookout for my ugly mug on the History Channel the week of March 12.  I got the announcement below from them.  The time 8pm may be referring to California time, I don't know.  They interviewed me for about an hour---stuff to do with quantum teleportation---but I suspect they're going to use only a minuscule amount of that footage.  Steven Weinberg and Jeff Kimble will also be making appearances in the capacity of ``official, knowledgeable scientists.''

\bq\noindent
We have an air date!!  The show will air on the History Channel on the
show, `History's Mysteries:\ The Philadelphia Experiment' Monday, March
12 at 8pm. For more show information, you can access the History Channel web site at
\bv
{\tt www.thehistorychannel.com}
\ev
The show will repeat throughout the week and you can get that air info
from the web site.
\eq

\section{28-02-01 \ \ {\it Merzbacher Thought} \ \ (to G. L. Comer)} \label{Comer1}

While driving home it dawned on me that you might be interested in seeing this too.  I hadn't thought to put Merzbacher in my epitaph distribution list.  Nevertheless, he sent me this.  He's right:  Shannon really was one of the major scientists of the last hundred years.  (You should look at his collected works, edited by Neill Sloane.)  But I bet hardly a physicist knows of the things he did.

\subsection{Eugen's Preply}

\bq
The morning paper carried an obituary for Claude Shannon.  I was amazed to learn that he was so ``young'' and still around.  Clearly, one of the major scientists of the past 100 years.
Heard an interesting symposium on quantum computing at the AAAS in San Francisco last week.  Is Dorit Aharonov, who gave a good talk, Yakir's relative?  Andrew Steane gave good introductory survey.  David DiVincenzo was okay.  Dave Wineland was stimulating, as usual (between his almost weekly contributions, alternatingly to {\sl Science\/} and {\sl Nature}).  Michael Freedman did not show.
\eq

\section{06-03-01 \ \ {\it Pauli-ish Compendium} \ \ (to H. Atmanspacher)} \label{Atmanspacher1}

I am compiling a rather large compendium to be titled, ``The Activating Observer:\  Resource Material for a Paulian--Wheelerish Conception of Nature,'' and will ultimately be submitting it to the journal {\sl Studies in History and Philosophy of Modern Physics}.  Perhaps the subject of the compendium needs no explanation given the title.  It presently has about 400 items listed in it.

In any case, I wonder if I can ask of your assistance.  In the compendium, I presently have you listed with three titles:
\begin{itemize}
\item
H.~Atmanspacher, ``Wolfgang Pauli und die Alchemie,'' Zeitschrift f\"ur Parapsychologie und Grenzgebiete der Psychologie {\bf 34}, 3--32 (1992).

\item
H.~Atmanspacher, H.~Primas, and E.~Wertenschlag-Birkh\"auser, eds., {\sl Der Pauli-Jung-Dialog und seine Bedeutung f\"ur die moderne Wissenchaft}, (Springer, Berlin, 1995).

\item
H.~Atmanspacher and H.~Primas, ``The Hidden Side of Wolfgang Pauli:\ An Eminent Physicist's Extraordinary Encounter with Depth Psychology,'' J. Consc.\ Stud.\ {\bf 3}, 112--126.
\end{itemize}
Unfortunately, I have not been able to obtain these materials.  Might I ask you to send me copies of these papers (in the case of the book, maybe you wrote an introduction)?  Also, if there is anything else you've written on the subject since then, I would be most appreciative if you could send that too.  I will place my professional address below.

I obtained your email address from Andrei Khrennikov who just placed me on the advisory board to his ICMM in {\Vaxjo}.  I am in Germany at least once a year to visit my wife's parents in Munich; perhaps I will get a chance to visit with you sometime during such a visit (the next will be in August or September).

\subsection{Harald's Reply}

\bq
I am happy to give you a list of publications of mine on Pauli,
see below. Copies of the articles will be sent to you.

Concerning the title of your project, it is somewhat unclear to me
why you connect Pauli so tightly with Wheeler. You will have your
reasons to join those names. Personally, I would not connect Pauli
and Wheeler on a more than superficial level. Maybe you have insights
which I am lacking.

Whenever you are in Germany, please let me know if you see a chance
to come to Freiburg. As I spend some part of my time in Munich,
it's possible we can meet there as well.

\begin{enumerate}
\item H.~Atmanspacher. \\
      Alchemie und moderne Physik bei Wolfgang Pauli.\\
      In {\it Unus Mundus.} Edited by T.~Arzt, M.~Hippius-Gr\"afin D\"urckheim,
and R. Dol\-lin\-ger.
      Peter Lang Verlag, Frankfurt, 1992, pp.~135--181.
\item H.~Atmanspacher.\\
      Wolfgang Pauli und die Alchemie, Teil I: Biographische und historische
Aspekte. \\
      {\it Z.~Grenzgeb.~Psych.} {\bf 34}, 3--32 (1992).
\item H.~Atmanspacher.\\
      Wolfgang Pauli und die Alchemie, Teil II: Das opus alchymicum.\\
      {\it Z.~Grenzgeb.~Psych.} {\bf 34}, 131--162 (1992).
\item H.~Atmanspacher.\\
      Wolfgang Pauli und die Alchemie, Teil III: Impulse f\"ur die modernen
Naturwissenschaften.\\
      {\it Z.~Grenzgeb.~Psych.} {\bf 35}, 1--27 (1993).
\item E.~Wertenschlag und H.~Atmanspacher. \\
      Das Irrationale in den Naturwissenschaften: Wolfgang Paulis Begegnung mit
dem Geist der Materie
       (Tagungsbericht \& Kommentar). \\
      {\it Gaia} {\bf 2}, 214--225 (1993).
      Reprinted in  {\it Z.~Grenzgeb.~Psych.} {\bf 35}, 221--230 (1993).
\item H.~Atmanspacher, H.~Primas, und E.~Wertenschlag (editors). \\
    {\it Der Pauli-Jung-Dialog und seine Bedeutung f\"ur die moderne
Wissenschaft}. \\
    Springer Verlag, Berlin, 1995.
\item H.~Atmanspacher.\\
      Raum, Zeit und psychische Funktionen.\\
      In {\it Der Pauli-Jung-Dialog und seine Bedeutung f\"ur die moderne
Wissenschaft}. Edited by
      H.~Atmanspacher, H.~Primas, und E.~Wertenschlag. Springer, Berlin, 1995,
pp.~239--274.
\item H.~Atmanspacher. \\
      Paulis wissenschaftlicher Briefwechsel 1940--1949 (Book Review). \\
      {\it Physik.~Bl\"atter} {\bf 51}, 1104--1105 (1995).
\item{H.~Atmanspacher and H.~Primas. \\
      The hidden side of Wolfgang Pauli. \\
      {\it Journal of Consciousness Studies} {\bf 3}, 112--126 (1996). \\
      Reprinted in {\it Journal of Scientific Exploration} {\bf 11},
      369--386 (1997).}
\item{H.~Atmanspacher. \\
     Paulis wissenschaftlicher Briefwechsel 1950--1952 (Book Review). \\
     {\it Physik.~Bl\"atter} {\bf 53}, 349 (1997).}
\item{H.~Atmanspacher. \\ Wolfgang Pauli -- nicht nur Physiker. \\
      In {\it Jenseits von Kunst}. Edited by P.~Weibel. Passagen, Wien, 1997,
      pp.~246--248.}
\item{H.~Atmanspacher. \\
     Paulis wissenschaftlicher Briefwechsel 1953--1954 (Book Review). \\
     {\it Physik.~Bl\"atter} {\bf 55}(9), 79 (1999).}
\end{enumerate}
\eq

\section{07-03-01 \ \ {\it Wonderful} \ \ (to M. A. B. Whitaker)} \label{Whitaker1}

Thank your for your kind invitation to speak at the Belfast meeting.  I am flattered, and of course I will make time on my calendar to be there.  Indeed ``Quantum Foundations in the Light of Quantum Information'' has been one of my pet peeves lately, and I have spent quite a bit of energy in promoting that direction in our community.  (First organizing a small meeting in {\Montreal} last year with Brassard, and then a small meeting in Sweden this year.)  I think there is probably no way quantum information will make a deeper contribution to physics than in finally bringing to a satisfactory conclusion the foundation problem.

So please mark me down for a talk---I have a pretty good one now devoted solely to this idea that not many in the foundations community have seen---and please keep me abreast of the further developments.

\section{07-03-01 \ \ {\it Pauli and Wheeler} \ \ (to H. Atmanspacher)} \label{Atmanspacher2}

What a wealth of information!  Thank you so much.  I didn't realize that you had written so extensively on Pauli!  If only I could read German!  But, still, your sending those things will be most useful:  with the help of some friends I will be able to get the abstracts and some key pieces of the text translated for inclusion into the compendium.

You ask why I connect Pauli and Wheeler so tightly?  Well, really the whole document is centered around a theme to do with the potential ``malleability'' of our world.  I use Pauli and Wheeler as two figureheads, but really the document is much more about the overall theme.  There is also much material on Bohr, quite a lot on Rosenfeld, stuff to do with ideas about {\it evolutionary\/} physical laws \`a la Peirce, Bergson, James, etc., a sprinkling of historical things on alchemy, and so forth.  I'll place the (present) abstract and two quotes (one by Pauli, one by Wheeler) below to show why I've chosen those two men as figureheads.

Thank you again for your help!

\section{09-03-01 \ \ {\it Banach Conference} \ \ (to D. Petz)} \label{Petz1}

Thank you for the invitation to speak at the Banach Center meeting.  I am flattered that you would think of me in this regard.  If I were to give a talk there I would try to make sure that the subject was more along the lines of ``Some Nontrivial Theorems in Quantum Mechanics'' rather than as in my previous visit!!  Beside visiting you again, I would be specially intrigued to meet Csiszar and discuss his ``secrecy capacity'' in the quantum context.

May 21--26 is just potentially possible for me.

\section{13-03-01 \ \ {\it FYI -- The History Channel}\ \ \ (to S. K. Stoll)} \label{Stoll1}

\bsks
On a related topic, is it true you gave an interview to Swedish media about teleportation of ``souls''?  I'm puzzled.
\esks

Why do you ask?  It wasn't to the Swedes, it was before that.  It may have been a Japanese crew.  I tried to explain that what is teleported is the quantum state, not a material object.  And of course, they said, ``What's that?''  And I said, ``It's more like the soul than the body.''  Or some such thing.\footnote{Actually it was a Brazilian television crew.  On 7 November 1998, I wrote to Asher Peres, ``Teleportation has surely grasped the media's attention.  Wednesday a Brazilian television crew came here and shot 1.5 hours of footage of the lab, Kimble and me.  All that to make a five minute TV segment on some prime-time show.  The interviewer assured me that 70 million people would be seeing my face and hearing my thoughts.  It's funny, but when they told me that, all I could think of was the poor sick woman who sent Charlie Bennett a police report of some minor accident she had had about ten years earlier.  She had read enough of Wheeler's thoughts on quantum mechanics---with phrases like, ``reality is a function of the community of communicators,'' etc.---to think that if not more people knew about the smallness of the accident, it might turn into something really big and horrible.  I thought, ``I'm not starting to feel any more real yet.  I wonder if it'll happen when the viewers actually see the film?''  Oh well, I guess I don't know what it means to feel real anyway!  The interviewer was fixated on whether the soul would also get carried over by a teleportation device.  I'm sure my remarks will go over well in a Catholic country!''}

Did I actually make an appearance on the show?  Or did they completely cut me?

I'm in Tokyo right now.  Akira came for Steven's and my talks.  That guy is amazing:  It took him a two hour trip each way, just to come sit through our talks.  Friday we're going to go visit his laboratory.  He's an associate professor at U. Tokyo now.

\section{13-03-01 \ \ {\it Is Nice nice?}\ \ \ (to S. L. Braunstein)} \label{Braunstein2}

What's the business about below?  It seems they put my name on their flyer before I had a chance to answer them.  Oh well.  A little extra advertisement can't hurt.  Now we'll just have to see whether I'm free during that time.  Thanks for promoting me.

Is Nice a nice place?  I'm in Tokyo right now.  Charlie Bennett and I are sharing a room, and both up with jetlag.  The lights are off and the laptops are glowing.  A little like two fireflies in the night.

\subsection{Sam's Preply}

\bq
\noindent Dear Dr.\ Touddaint and Dr.\ Plet,\medskip

Mr. Anders Hansson was correct about my interest in giving a presentation at DASIA 2001, unfortunately, other commitments have been made before I knew what dates were involved.

Might I suggest an alternative (whom I have approached):
\begin{itemize}
\item
Dr.\ Chris Fuchs at Lucent\\
cafuchs@research.bell-labs.com
\end{itemize}
His especial interests are quantum cryptography, which I think you would find very interesting; he is is also an excellent speaker and would be worth getting to come.\medskip

\noindent Sincerely, \medskip

\noindent Sam Braunstein
\eq

\section{16-03-01 \ \ {\it Jammin' Jammer} \ \ (to H. Barnum)} \label{Barnum1}

Thanks for thinking of me.  Actually, I was just given a copy of Jammer's 1974 book ({\sl The Philosophy of Quantum Mechanics}) by David DiVincenzo.  But I don't have Jammer's 1966 book yet ({\sl The Conceptual Development of Quantum Mechanics}).  I'm not completely sure which one you were thinking of.

And by the way, thanks for sending me a report on Jeff Bub's seminar.  I was dancing a little jig for a week after.

\section{17-03-01 \ \ {\it RCFoC}\ \ \ (to J. B. Lentz)} \label{LentzB1}

Thanks for the article.  I'm still in Japan.  I'll be here until Thursday.  The History Channel show was an episode of History's Mysteries.  The title was something about ``The Philadelphia Experiment,'' some silliness about a supposed teleportation of a navy ship from Philadelphia to Norfolk, VA sometime in WWII.  From what I understand, of the hour-long interview they gave me, they only used about a minute of the footage (at most).

\section{17-03-01 \ \ {\it Pounding Rice Cakes} \ \ (to H. Barnum)} \label{Barnum2}

The allusion is this:  One of my greatest eye-opening experiences was in learning that when the Japanese look at the moon they see a rabbit pounding rice cakes.  Ever since being told that, I've been able to see it too.  It's plain as day.  Thirty-three years of not seeing it once, and then, boom, someone tells me of it and I see it all the time.  Likewise, I'll bet the next big step in physics will only require that we see something right here in front of us.  It'll be something no big multi-billion dollar particle accelerator will be needed for.  We just have to figure out how to take note of it.  I'm banking that a hint is already written down in the quantum.

\section{18-03-01 \ \ {\it The Copenhagen Picnic} \ \ (to H. J. Folse)} \label{Folse1}

\bhf
What is it that you're looking for in the Archives?  I spent six
months going through them 26 years ago.  Of course there's more there
nowadays.  But I doubt that there is any undiscovered gold to find,
and if there is it's deeply buried.
\ehf

Ultimately, I plan to dig deeper into whatever can be found in the US repositories you mentioned, but for the present Ben and I thought it might make a nice picnic with which to end our time in Scandinavia.

What am I looking for?  Basically, just fun historical tidbits to do with this train of thought I've been telling you about.  You never know what you might find \ldots\ especially when seen through ``prepared'' eyes.  If I need to pull out an official reason for being there, I can use this large compendium I'm putting together---if it doesn't create too much copyright trouble, I plan to have it appear in {\sl Studies in History and Philosophy of Modern Physics}.  And I'm sure I can get any number of letters vouching for my historical integrity:  you perhaps?, Mermin, Butterfield (editor of SHPMP), or maybe some more appropriate people if I think harder.

Speaking of the compendium, would you like to see it as it stands?  It's now got about 400 references in it and a nonnegligible amount of annotation.  It stands presently at 80 pages.  I'm projecting about 500 references and 150 pages for the completed project.  There's a whole lot of editing left to be done, but you might find it fun even at this stage.  And maybe you can tell me about any glaring omissions that you know of.

For the present, let me attach a note I wrote you on my way to Japan.  You never answered my two questions, even though you replied to later emails.  So maybe you never got it.

\subsection{From a 09 March 2001 note to Henry Folse, ``The Right Choice''}

\bq
Sorry for my long delay in getting back to you.  I'm off to Japan in
about five hours for two weeks, and {\it trying\/} to get things tied
up here before leaving has made the week pretty hectic.

Anyway, the most important message I wanted to tell you is that I'm
gettin' damned happy I invited you to V\"axj\"o.  Reading your papers
has been a really pleasant experience.  I think I hit my first dozen
last night at about this time.  (But maybe the bigger question is,
why am I up at this time?!?)  Below, I'll place my {\Folse}
compendium as it stands---every word typed in lovingly with my own
little fingers!  Of course, I have a few quibbles with some things
I've read, but I think I'll save my comments until I've read the
complete body of work.  27 years is a long time, and you could well
have changed your mind about some things:  I'll give you the benefit
of the doubt for now.

\bhf
I should certainly warn you that very few in this community agree
much with my reading of {\Bohr}.
\ehf

I couldn't care less about that:  I like it (or most of it), and that's
all that counts for me.  Besides, I've read {\Bohr} myself---fairly
carefully I've always thought---and your view significantly
coincides with my memory of that.

\bhf
I'm more concerned about my lack of knowledge of anything about
information theory.
\ehf

You need not be too concerned, but of course it wouldn't hurt you to
do a little reading on the side if you've got some time.  Somewhere
below, you wrote:
\bq
\noindent
In describing the phenomena of observational interactions, quantum
theory describes them as being caused by the interactions between
the observing systems and microsystems.  The fact that we can form
no representation, no mechanical picture, of the atoms on which the
mechanistic description of the phenomenal world is based, hardly
reveals that we are ignorant of what these entities {\it are}.
Rather it testifies to what we have learned about them -- that they
cannot be so represented -- through explorations of the atomic
phenomenon in which their strange behavior is revealed to human
experience.
\eq

It is my strongest opinion that the great fruits of quantum
information and computing get at precisely this point \ldots\ and in
spades!  The point is that this ``nonrepresentability'' in actual
fact boils down to a positive statement rather than a negative one.
So it would do you well to learn a little about our field.  (And,
luckily, most of what we do is not abstruse stuff:  it's just basic
quantum mechanics, viewed mostly from a new point of view with a new
set of goals in mind.)  Where to start?  Maybe a good place would be
three ``recent'' {\sl Physics Today\/} articles:
\begin{enumerate}
\item  {\Gottesman} and Lo, ``From Quantum Cheating to Quantum Security,'' PT
November 2000 (don't have the page numbers)

\item  {\Preskill}, ``Battling Decoherence: The Fault-Tolerant Quantum
Computer,'' PT June 1999, p.~24

\item  {\Bennett}, ``Quantum Information and Computation,'' PT October 1995,
p.~24 (interesting coincidence)
\end{enumerate}
If you get that far, let me know, and I'll suggest a couple of
really {\it mild\/} technical articles that'll be worth their weight
in gold in insight.

But why do I think it would do you well?  Because I think you have
an honest heart.  And, because while I believe {\Bohr} and his gang
certainly started to point us in the right direction, I think we
have a long, long technical way to go before we can claim a
particularly deep understanding of the quantum structure.  Here's
how I put it in exasperation to David {\Mermin} once:
\bq
What's your take on this passage?  Can you make much sense of it?
What does he mean by ``providing room for new physical laws?''  What
``basic principles of science is he talking about?''  What five pages
of derivation are lying behind all this business?

It nags me that {\Bohr} often speaks as if it is clear that the
structure of quantum theory is derivable from something deeper, when
in fact all the while he is taking that structure as given.  When did
he ever approach an explanation of ``Why complex Hilbert spaces?''
Where did he ever lecture on why we are forced to tensor products for
composite systems?  It's a damned shame really:  I very much like a
lot of elements of what he said, but as far as I can tell all the
hard work is still waiting to be done.\footnote{\editornote Rudolf
  Haag reminisces about a debate with Bohr during the early 1950s:
  ``I tried to argue that we did not understand the status of the
  superposition principle.  Why are pure states described as [rays] in a complex linear space?
   Approximation or deep principle?  Niels Bohr did not understand why
   I should worry about this.  Aage Bohr tried to explain to his
   father that I hoped to get inspiration about the direction for the
   development of the theory by analyzing the existing formal
   structure.  Niels Bohr retorted: `But this is very foolish. There
   is no inspiration besides the results of the experiments.'  I guess
   he did not mean that so absolutely but he was just annoyed.''  From
  R.\ Haag, ``Some people and some problems met in half a century of
  commitment to mathematical physics,'' European Physical Journal H
  \textbf{35,} pp.~263--307 (2010).}
\eq

The issue in my mind is {\it not\/} to {\it start\/} with complex
Hilbert space, unitary evolution, the tensor product rule for
combining systems, the identification of measurements with Hermitian
operators, etc., etc., and {\it showing\/} that {\Bohr}'s point of
view is {\it consistent\/} with that.  Instead it is to start with
{\Bohr}'s point of view (or some variant thereof) and see that the
{\it precise\/} mathematical structure of quantum theory {\it must\/}
follow from it.  Why complex instead of real? Why unitary, rather
than simply linear? Indeed, why linear?  Why tensor product instead
of tensor sum?  And, so on. When we can answer {\it these\/}
questions, then we will really understand complementarity.

I'm banking my career on the idea that the tools and issues of
quantum information theory are the proper way to get at this program.

OK, I've got to get some sleep.  I have this dream that I'm going to
get work done all the way to Japan.  But if I don't get some sleep,
I'll certainly be kidding myself.
\eq

\subsection{Henry's Reply}

\bq
Yes, if you have some particular theme to look for beforehand, then you can perhaps improve your chances of finding something of interest.  Actually when I first started exploring the Archives my interest was totally different, and I probably didn't see many things that might strike me as important now.  My original target was Heisenberg, not Bohr, and my original interest was the influence of Kantianism on German scientists, not realism or interpreting quantum theory.  I thought of Bohr as a positivist, but when I started reading more stuff, I came to believe that Bohr had been seriously misrepresented by philosophers in general, and so I ended up going in this direction.  Also they're probably in a lot better order today than they were then.  At that time Bohr's last secretary, Fru Hellman, was still very much alive and running the show.  She had personal knowledge of much of the files, but beyond that there was just a list of file names and dates.  I'm sure at that time there were a lot of pages of manuscripts that were more or less misidentified or otherwise inscrutable. It was really an old-world European kind of thing. The last time I was there it seemed they had made a lot of progress, but still had a ways to go.  I'd be interested to see how modern they might be nowadays.  I imagine I'll go thru Copenhagen prior to Sweden; I'm afraid I have to get home ASAP after the conference is over.
\eq

\section{18-03-01 \ \ {\it Hello from Tokyo} \ \ (to J. D. Sanders \& L. Sanders)} \label{SandersJD1} \label{SandersG1}

Here's a funny story about Cuero.  The last time I was in town, Kiki and I dropped in to the little bookshop/coffeeshop on Main street.  I picked up an old copy of Arthur Eddington's {\sl Our Mysterious Universe\/} for 50 cents.  When I bought it, I told the lady---and this is a true story---``You know, this is the first book I ever bought in Cuero!''  I had bought comic books in Cuero as kid, but never a real book (even a paperback).  All my real books had come from Victoria and the like.  But there was another weird coincidence.  That night there was an author there having a book signing.  It turned out, she works in Morristown (where I live presently) and lives in Murray Hill (where I work presently).  A small world, and it keeps getting smaller!

\section{20-03-01 \ \ {\it The Transcript Archive}\ \ \ (to B. W. Schumacher)} \label{Schumacher1}

That is a really nice site!

Thanks again for coming to Japan.  Charlie, Steven, and John are all gone now.  Only I'm left.  Things are much quieter now, and I am starting to get back to the hard job of writing some things down.

\subsection{Ben's Preply}

\bq
This is pretty exciting stuff.

I've been looking into the Archives for the History of Quantum Physics
(AHQP) stuff.  There is a web site at which one can find out more:
\begin{center}
\myurl{http://www.amphilsoc.org/guides/ahqp/}
\end{center}
Included is a list of everyone about whom there is material in the AHQP,
including interviews.  There are copies of the transcripts in the U.S.
at various institutions, but it would of course be much cooler to visit
them at the Bohr Institute.
\eq

\section{21-03-01 \ \ {\it {\Schroedinger}'s Cat} \ \ (to G. L. Comer)} \label{Comer2}

I was just reminded of Stephen Hawking by my ``quote of the day'' program.  Do you think this means that Hawking desires to be a modern day Goering?

\bq\noindent
``Whenever I hear the word culture, I reach for my revolver.''\\
\indent --- Hermann Goering (attributed to) (1893--1946), German Nazi leader.\medskip

\noindent Whether or not Goering ever said it, the only recorded reference to this remark is from the play Schlageter (1933) by Hanns Johst (1890--1978), Nazi playwright and president of the Reich Chamber of Literature. The line was said by a stormtrooper in act 1, sc.~1: ``{\it Wenn ich Kultur hore \ldots\ entsichere ich meinen Browning}'' (literally, ``I cock my Browning'').
\eq

\section{22-03-01 \ \ {\it Nonorthogonal States} \ \ (to R. W. {\Spekkens} and J. E. {\Sipe})} \label{Sipe1} \label{Spekkens1}

I just pulled up your new paper, \quantph{0003092}\@.  It looks like it's going to be interesting even for nonmodal people!

It dawned on me when I saw your title this morning---I'm jetlagged greatly just having got back from Japan---that maybe I never told you about my paper: C.~A. Fuchs, ``Nonorthogonal Quantum States Maximize Classical Information Capacity,'' Phys.\ Rev.\ Lett. {\bf 79}(6), 1162--1165 (1997).

(You can also find it on {\tt quant-ph}.)  It shows that there is a certain sense (motivated in an operational way by information transmission problems) in which nonorthogonal states are sometimes more stable against noise than orthogonal ones.  Peter Shor, John Smolin, and I even have examples now where it is best to pull the states from a set of linearly dependent ones!

I don't know if these examples have any impact on your proposed ontology, but it might be worth thinking about.

By the way, you misspelled Wootters.

\section{24-03-01 \ \ {\it Annoyed}\ \ \ (to S. K. Stoll)} \label{Stoll2}

I'm back from Japan now, and I decided to finally watch that history channel thing this morning at 4:00 AM!  You're right, I certainly do wish I had never been on it.

\bsks
In the last 5 minutes of the program, the photography of the lab was
great, they show that first, along with a panning shot of the
``Unconditional Quantum Teleportation'' paper, with voice over and a
clip of you saying something like, ``this experiment is so fantastical
that it is almost in the realm of science fiction \ldots'' (I kid you
not).  Then it has a clip of Jeff explaining how technically it is
impossible to teleport a material object in the foreseeable future
because of the massive bits of information necessary to do so, or some
such. Then the narrator says something about how Einstein started all
this talk of teleportation and it flips backs to you explaining that
his '35 paper was an argument against quantum mechanics.
\esks

More precisely, I said:
\bq\noindent
``This is something that is so fantastic \ldots\ scientifically \ldots\ at this point in time that it almost is in the realm of science fiction.''
\eq
But that was targeted not at Akira's experiment; it was about that silly ship teleportation stuff.  They certainly gave a sense though that I was talking about Jeff's lab.  Then the other thing about Einstein really grates.  The larger part of my career is devoted to debunking the idea that entanglement is about ``spooky action at a distance.''  It is just about information: what information Alice and Victor can obtain about Bob's site.  The only thing that is teleported in quantum teleportation is Victor's {\it predictions\/} \ldots\ from being about one particle to being about another.  But that is not only the larger part of my career, it was the larger part of my interview!!  And they just dropped that completely.  They played up this ``spooky action'' stuff, and that is just the opposite of what I had intended.

I live and I learn.

\section{26-03-01 \ \ {\it Home Sweet Home} \ \ (to O. Hirota)} \label{Hirota1}

I apologize for my delay in sending you word of my safe arrival.  Since arriving in New Jersey, most of my time has been spent in trying to wake up, battling a small cold, and becoming reacquainted with my family.  I've hardly looked at email at all.

There is no need to thank {\it me\/} for {\it my\/} involvement in the Tokyo party.  I don't deserve that honor.  We all owe {\it you\/} many thanks for giving us this wonderful opportunity.  I had a tremendous time, and was struck once again by the inner strength of the Japanese culture.  I also particularly thank you for helping to get me out of trouble my final day there.  My daughter became so happy with her ``Japanese gifts''; it was an opportunity I would have been ashamed to miss.

To continued progress in quantum information!

\section{26-03-01 \ \ {\it Big Daddy on the Way} \ \ (to H. J. Folse)} \label{Folse2}

I arrived back in my office today to find your newest mail to me.  Thanks so much for going to that trouble.

In the next email, I'll send you a PostScript file of ``The Activating Observer''---the compendium I told you about.  Please keep in mind that it is a very tentative version \ldots\ full of typographical errors, gaps, and other inanities.  In particular, the Introduction is still very much under construction.  I may well take out some of the nasty things I said about Whitehead for instance.

As of yesterday, I believe I only have about three of your papers to go reading-wise!  It's been a very nice experience.  In particular, over the weekend, I enjoyed your 1978 paper ``Kantian Aspects of Complementarity.''  In this connection, I have quite a pressing question for you.  Do you happen to have the full text of all the letters between Bohr and Pauli on the phrase ``detached observer?''  (The ones I know of from your paper are dated 15 February, 2 March, and 25 March 1955.  But maybe there are more.)  If so, have you (or someone) translated those texts into English?  When you open my compendium, you'll see why I'm particularly interested in this.  Those texts would be an invaluable addition to it.

Thanks again for going to such trouble for me.

\section{26-03-01 \ \ {\it Clifton Proposition} \ \ (to N. D. {\Mermin})} \label{Mermin0}

Have you thought about this?  [See 02-04-01 note ``\myref{Clifton1}{Present State of Thought}'' to R. Clifton.] I've thought about combining contributions from {\Montreal} and {\Vaxjo} as a {\it small\/} possibility, but then I was thinking as the target journal our new one on quantum information.  But as I say, I was only thinking of this in a small way:  most of my thoughts were purely self-serving \ldots\ namely, using it as a way to extract Schumacher's ``Doubting Everett'' paper from his head.  (Bennett with his Everettista fatigues was really getting on my nerves in Japan last week:  there's only so many days I can room with such a person!  Ben was my only breath of fresh air.)

\subsection{David's Reply}

\bq
All my life I have conscientiously declined invitations to edit anything
and have never regretted it.  I'm inclined not to abandon
that successful policy at this late stage, but I told Rob Clifton
(who incidentally is one of the more solid philosophers
of quantum mechanics) that I'd chew it over with you
in {\Vaxjo} this June.

On the other hand a volume that was {\Montreal}$+${\Vaxjo} might be a useful
contribution to western civilization.  I like the idea of it being in
Hist Phil Sci, because it would reach a bunch of readers who ought to
be more interested in this subject than they currently are.  (Look at
the profound effect you had on Bub.)

I don't think there's any hurry with this.
\eq

\section{29-03-01 \ \ {\it French Philosophers} \ \ (to J. M. Renes)} \label{Renes2}

Do you remember you once told me about a French philosopher who said something like:  neither the subject is real on its own, nor the object is real on its own; only their interface is real.  Or some such thing.  Can you give me a reference to that?

\section{29-03-01 \ \ {\it There's At Least One More} \ \ (to H. J. Folse)} \label{Folse3}

Looking through your book (I picked up a copy from \myurl[http://www.powells.com]{Powells.com}), I found at least one more letter in that exchange on the ``detached observer'' that I'm interested in.  It's dated 11 March 1955, from Pauli to Bohr.

I sure hope you have copies of these letters in their entirety!  (And I keep my fingers crossed that you've translated them.)  This is quite exciting to me.

Unfortunately, I'm going to have to slow down on my metaphysical project for the time being.  The mundane matters in my life are starting to pile up for the month.  Too many papers that desperately need finishing and---more frighteningly for me---I've got to meet with the president of Bell Labs in three weeks.  He's asked for a pitch on where we as a company should be going with quantum information.  The philosopher meets the executive:  that ought to be pretty.

\subsection{Henry's Reply, ``Bohr--Pauli Exchange''}

\bq
You're quite right that this an interesting exchange between Bohr and
Pauli.  I suspect that many have ignored them because of the late date.

The letters are in English and typed, but at least on one Pauli inserted
several comments in handwriting.  Since it's about 15 pages or so, I've
photocopied and mailed all four of them to you at your Bell Labs
address.
The copies are too poor to scan very easily.  The Pauli letters are
photocopies of the originals; the Bohr letters are photocopies of Bohr's
carbon copies.

Hope you enjoy reading them. \medskip

\bq\noindent
{\bf W. Pauli, letter to Niels Bohr, dated 15 February 1955, photocopy
obtained from the Niels Bohr Institute via Henry Folse.}\medskip

\noindent
Dear Bohr,\smallskip

It is with great pleasure that I received your nice letter and above
all, the text of your lecture on ``Unity of Knowledge''. The general
outlook of it is of course the same as mine.  Under your great
influence it was indeed getting more and more difficult for me to
find something on which I have a different opinion than you. To a
certain extent I am therefore glad, that eventually I found
something: the definition and the use of the expression ``detached
observer'', which appears on page 10 above of your lecture and which
reappears on page 13 in connection with biology. According to my own
point of view the degree of this ``detachment'' is gradually
lessened in our theoretical explanation of nature and I am expecting
further steps in this direction.

1) As you will see in the reprint on my lecture on ``probability and
physics'', which I have sent to you, it seems to me quite appropriate
to call the conceptual description of nature in classical physics,
which Einstein so emphatically wishes to retain, ``the ideal of the
detached observer''. To put it drastically the observer has
according to this ideal to disappear entirely in a discrete manner
as hidden spectator, never as actor, nature being left alone in a
predetermined course of events, independent of the way in which the
phenomena are observed. ``Like the moon has a definite position''
Einstein said to me last winter, ``whether or not we look at the
moon, the same must also hold for the atomic objects, as there is no
sharp distinction possible between these and macroscopic objects.
Observation cannot {\it create\/} an element of reality like a
position, there must be something contained in the complete
description of physical reality which corresponds to the {\it
possibility\/} of observing a position, already before the
observation has been actually made.'' I hope, that I quoted Einstein
correctly; it is always difficult to quote somebody out of memory
with whom one does not agree. It is precisely this kind of postulate
which I call the ideal of the detached observer.

In quantum mechanics, on the contrary, an observation hic et nunc
changes in general the ``state'' of the observed system in a way not
contained in the mathematically formulated {\it laws}, which only
apply to the automatical time dependence of the state of a {\it
closed\/} system. I think here on the passage to a new phenomenon by
observation which is technically taken into account by the so called
``reduction of the wave packets.'' As it is allowed to consider the
instruments of observation as a kind of prolongation of the sense
organs of the observer, I consider the impredictable change of the
state by a single observation---in spite of the objective character
of the result of every observation and notwithstanding the
statistical laws for the frequencies of repeated observation under
equal conditions---to be {\it an abandonment of the idea of the
isolation (detachment) of the observer from the course of physical
events outside himself}.

To put it in nontechnical common language one can compare the role of
the observer in quantum theory with that of a person, who by its
freely chosen experimental arrangements and recordings brings forth a
considerable ``trouble'' in nature, without being able to influence
its unpredictable outcome and results which afterwards can be
objectively checked by everyone.

Probably you mean by ``our position as detached observers'' something
entirely different than I do, as for me this new relation of the
observer to the course of physical events is entirely {\it
identical\/} with the fact, that our situation as regards objective
description in ``this field of experience'' gave rise to the demand
of a renewed revision of the foundation for ``the unambiguous use of
our elementary concepts'', logically expressed by the notion of
complementarity.

2) Passing now from physics to other sciences like psychology and
particularly biology I am most interested in your approach, which
certainly seems to me to go in the right direction. Without entering
a discussion of the dependence of such concepts as ``[art?]'', not
only on the state of motion but also on the psychological attitude of
the observer, I am very much looking forward to your article on the
organic evolution which you announced in your letter.

In discussions with biologists I met large difficulties when they
apply the concept of ``natural selection'' in a rather wide field,
without being able to estimate the probability of the occurrence
{\it in a empirically given time\/} of just those events, which have
been important for the biological evolution. Treating the empirical
time scale of the evolution theoretically as infinity they have then
an easy game, apparently to avoid the concept of purposiveness. While
they pretend to stay in this way completely ``scientific'' and
``rational'', they become actually very irrational, particularly
because they use the word ``chance'', not any longer combined with
estimations of a mathematically defined probability, in its
application to very rare single events more or less synonymous with
the old word ``miracle''. I found for instance {\it H.~J. M\"uller\/}
very characteristic for this school of biologists (see also his
recent article ``Life'' in {\sl Science}, issue of January 7, 1955,
which certainly contains very interesting material), but also our
friend Max {\it Delbr\"uck}. With him this is combined with vehement
emotional affects and a permanent thread to run away which I
interpret as obvious signs of overcompensated doubts.

You can imagine how much better than ``natural selection'' sounds for
me ``natural evolution'' which I never heard before from you and
which you use now on page 19 of your lecture. I hope that your
announced article will tell us more about your use of the latter
concept.

Concluding this letter, I add some remarks about your sentence on
page 14 concerning the ``medical use of psycho-analytical treatment
in curing neurosis''. I am quite glad about this sentence, as logic
is always the weakest spot of all medical therapeuts, who never
learned the rigorous logical demands of mathematics.

Historically the word ``the unconscious'' was used by German
philosophers of the last century, particularly by {\it
E.~von~Hartmann\/} [also E.~G. Carus], developing further older
allusions of Leibniz and Kant. The Psycholamarckist {\it A.~Pauly},
on whom we spoke already, quoted von~Hartmann in 1905 (Freud was not
known to him), when he called processes of biological adaptation,
already in plants, an ``{\it unconscious\/} judgement of the psyche
of the organisms''. In this way however, only a new name was
introduced, which did not explain anything. Freud was the first who
made practical applications of the unconscious replacing hereby this
word by ``subconsciousness'', which you also apply.  With this
change of the word Freud wanted to emphasize that all ``contents of
the subconsciousness'' were earlier in the consciousness and had been
suppressed (``verdr\"angt'') afterwards. In this way Freud's
subconsciousness was like a bag containing a finite number of
objects. The purpose of the psychoanalytical treatment was therefore
to make this bag again empty by upheaval of the suppression.

To this restricted concept of subconsciousness among others C.~G.
Jung is in opposition since about 1913. He reestablished the older
word the unconscious of the philosophers emphasizing that every
change of consciousness for instance in a medical treatment,
backwards also changes the unconscious, which therefore can never be
made ``empty of contents'', only a small part of which has ever been
in consciousness. The aim of the medical treatment according to Jung
and his school is therefore the establishment of a correct and sound
``equilibrium between consciousness and the unconscious'', like an
equilibrium between two powers. This process in which this
equilibrium is reached and reestablished, they also call ``the
assimilation or interaction of the unconscious to the
consciousness''.

I only refer here historically a situation without identifying myself
with this kind of terminologies, which seem to me rather far from
logical clarity. The Jung school is more broad minded than Freud has
been, but correspondingly also less clear. Most unsatisfactory seems
to me the emotional and vague use of the concept of ``Psyche'' by
Jung, which is not even logically self consistent.

I am very glad about the prospect of a visit in Copenhagen in the
autumn of this year, when also your 70th birthday will be
celebrated. Francas treatment is not yet finished entirely, but she
is much better and there is much hope, that she also will be able to
go to Copenhagen this next time.

With all good wishes from both of us to yourself, Margrethe and the
whole family,
\begin{flushright}
\noindent yours old, \quad \smallskip

\noindent W.~Pauli \quad
\end{flushright}
\noindent [PS:]\ Where your lecture on ``Unity of
Know\-ledge'' will be printed in case I would like to quote it? \medskip
\eq

\bq\noindent
{\bf N.~Bohr, letter to Wolfgang Pauli, dated 2 March 1955, photocopy
obtained from the Niels Bohr Institute via Henry Folse.} \medskip

\noindent Dear Pauli,\smallskip

On my return from the CERN meeting in Geneva, I am writing to thank
you for your letter of February 15th. It was very good of you to
write me so carefully about your reaction to my article and, as
always, you touch upon a very central point. A phrase like
``detached observer'' has of course like all words different
linguistic and emotional aspects, but using it in connection with
the phrase ``objective description'', taken as theme of the
discussion on Unity of Knowledge, it had to me a very definite
meaning. In all unambiguous account it is indeed a primary demand
that the separation between the observing subject and the objective
content of communication is clearly defined and agreed upon.  The
aim of the article is just to stress that this condition is
indispensable in all scientific knowledge, including biology and
psychology, while in art as well as in religious belief one allows
oneself to neglect or rather tacitly to shift such separation. In
this connection, the historical information in your letter about the
use of terminology by psychologists was very valuable to me, and I
was glad that you on the whole sympathize with my approach. Indeed,
contrary to what some of our common friends seem to believe of me, I
have always sought scientific inspiration in epistemology rather than
mysticism, and how horrifying it may sound, I am at present
endeavoring by exactitude as regards logic to leave room for
emotions.

It is on this background that it seems to me very important that we
fully understand each other in questions of terminology. Of course,
one may say that the trend of modern physics is the attention to the
observational problem and that just in this respect a way is bridged
between physics and other fields of human knowledge and interest. But
it appears that what we have really learned in physics is how to
eliminate subjective elements in the account of experience, and it
is rather this recognition which in turn offers guidance as regards
objective description in other fields of science. To my mind, this
situation is well described by the phrase `detached observer', and it
seems to me that your reference to our controversy with Einstein is
hardly relevant in this connection. Just as Einstein himself has
shown how in relativity theory `the ideal of the detached observer'
may be retained by emphasizing that coincidences of events are common
to all observers, we have in quantum physics attained the same goal
by recognizing that we are always speaking of well defined
observations obtained under specified experimental conditions. These
conditions can be communicated to everyone who also can convince
himself of the factual character of the observations by looking on
the permanent marks on the photographic plates. In this respect, it
makes no difference that in quantum physics the relationship between
the experimental conditions and the observations are of a more
general type than in classical physics. I take it for granted that,
as regards the fundamental physical problems which fall within the
scope of the present quantum mechanical formalism, we have the same
view, but I am afraid that we sometimes use a different terminology.
Thus, when speaking of the physical interpretation of the formalism,
I consider such details of procedure like `reduction of the wave
packets' as integral parts of a consistent scheme conforming with
the indivisibility of the phenomenon and the essential
irreversibility involved in the very concept of observation. As
stressed in the article, it is also in my view very essential that
the formalism allows of well defined applications only to closed
phenomena, and that in particular the statistical description just
in this sense appears as a rational generalization of the strictly
deterministic description of classical physics.

I am eager to learn your reaction to these points as I feel that it
is essential, not least for the approach to the wider problems on
which we are working, to be as precise as possible in terminology,
and above all to avoid any vagueness as to the demands of objective
description. It was a great joy to learn that we can expect a visit
of you and Franca in the autumn and perhaps there is an opportunity
of meeting you even earlier, since I am invited to give a talk in
Basel at the end of March on the general epistemological problems.

With kindest regards and best wishes to you both from us all,
\begin{flushright}
Yours, in every way, old\medskip
\end{flushright}
\eq

\bq\noindent
{\bf W.~Pauli, letter to Niels Bohr, dated 11 March 1955, photocopy
obtained from the Niels Bohr Institute via Henry Folse.}\medskip

\noindent Dear Bohr,\smallskip

I find your letter of March 2nd very youthfull, which is just the
reason that it is not easy for me to answer. Although we have the
same view ``as regards the fundamental physical problems which fall
within the scope of the present quantum mechanical formalism'' and
although I agree with some parts of your letter, the situation is now
complicated by your use in a publication of a phrase like ``detached
observer'' (without comment!)\ which I used already in some
publications in a very different way.  I believe that this should be
better avoided to prevent a confusion of the readers\footnote{An
explaining remark about it in your {\it new\/} article would be most
welcome!} and I don't cling at all to particular words myself. I also
felt, already before your letter arrived, that my brief
characterisation of the observer in quantum theory as
``non-detached'' is in one important respect misleading. As is well
known to both of us, it is essential in quantum mechanics that the
apparatus can be described by classical concepts. Therefore the
observer is always entirely detached to the {\it results} of his
observations (marks on photographic plates etc.), just as he is in
classical physics. I called him, however in quantum physics
``non-detached'', when he chooses his experimental
arrangements.\footnote{I still believe today that this more
restricted use of my terminology is very good and that it has been
unhappily obscured in your article in a non-logical way!}

I shall try to make my point logically clear, by defining my
concepts, replacing hereby the disputed phrase by other words. As I
was mostly interested in the question, {\it how much informative
reference to the observer an objective description contains}, I am
emphasizing that a communication contains in general {\it
informations on the observing subject}.

Without particularly discussing the separation between a subject and
the informations about subjects (given by themself or by other
persons), which can occur as elements of an ``objective
description'', I introduced a concept ``degree of detachment of the
observer'' in a scientific theory to be judged on the kind and
measure of informative reference to the observer, which this
description contains.  For the objective character of this
description it is of course sufficient, that every individual
observer can be replaced by every other one which fullfills the same
conditions and obeys the same rules.  In this sense I call a
referency to experimental conditions an ``information on the
observer'' (though an impersonal one), and the establishment of an
experimental arrangement fulfilling specified conditions an ``action
of the observer''---of course not of an individual observer but of
``the observer'' in general.

In physics I speak of a detached observer in a general conceptual
description or explanation only then, {\it if it does not contain any
explicit reference to the actions or the knowledge of the observer}.
The ideal, that this should be so, I call now ``the ideal (E)'' in
honor of Einstein.  Historically it has its origin in celestial
mechanics.

There is an important {\it agreement\/} between us that we find
Einstein not consequent in this formulation of the ``ideal E''.
Indeed, there is no a priori reason whatsoever to introduce here a
difference between the {\it motion\/} of the observer on the one
hand, and the realization of specified experimental conditions by
the observer on the other hand.  If Einstein were consequent he had
to ``forbid'' also the word coordinate system in physics (as not
being objective).  That the situation in quantum mechanics has a
deep similarity with the situation in relativity is already shown by
the application of mathematical groups of transformation in the
physical laws in both cases.

In this way I reached the conclusion to distinguish sharply between
the ``ideal of an objective description'' (meaning science) on the
one hand (which I warmly supported just as you do) and the ``ideal
of the detached observer'' on the other hand (which I rejected as
much too narrow).

What really matters for me is not the word ``detached'', but the more
active role of the observer in quantum physics, which is already
implied in your [consideration?]\ of the ``indivisibility of the
phenomena and the essential irreversibility involved in the very
concept of observation''.  According to quantum physics the observer
has indeed a new relation to the physical events around him in
comparison with the classical observer, who is merely a spectator:
The experimental arrangement freely chosen by the observer lets
appear {\it single\/} events {\it not\/} determined by laws, the
ensembles of which are governed by {\it statistical\/}
laws.\footnote{In this way we obtain just the logical foundations of
an ``{\it objective\/} description'' of the incidents (Ein begriffe)
which the quantum mechanical observer makes within his surroundings
with his experimental arrangements.  Attention: there is {\it no\/}
logical contradiction between a word like ``trouble'' and a
possibility of its objective observation and description.}  It is not
relevant to me, if you say the same thing using {\it different\/}
terminologies (but please use [essentially?]\ different words than I).
They will only confirm my statements again as all these statements on
the observer are part of an ``objective description''.

I confess, that very different from you, I do find sometimes
scientific inspiration in mysticism\footnote{By the way: the
``Unity'' of everything has always been one of the most prominent
ideas of all mystics.} (if you believe that I am in danger, please
let me know), but this is counterbalanced by an {\it immediate\/}
sense for mathematics.  The result of both seems to be my kind of
physics, whilst I consider epistemology merely as a logical comment
to the application of mathematics in physics.\footnote{We are here
{\it both\/} in our letter in a realm of information on the writing
subject, which do {\it not\/} belong to the ``objective content of
the communications''.}  Thus when I read a sentence as ``how to
eliminate subjective elements in the account of experience'' my
immediate association is ``group theory'' which then determines my
whole reaction to your letter.  Although the first step to
``objectivity'' is sometimes a kind of ``separation'', this task
excites in myself the vivid picture of a superior common order to
which all subjects are subjected, mathematically represented by the
``laws of transformations'' as the key of the ``map'', of which all
subjects are ``elements''.

I hope that it will be possible to find a terminology which will
turn out to be satisfactory for both of us, but it is no hurry with
it.  I propose to resume this discussion only when your new article
will be ready, which I am eagerly awaiting.  It will show me your
terminologies in more general cases of objective descriptions, of
which I am most interested in the application to biology, in
connection with your new expression ``natural evolution''.

From March 16th till about 27th I am away in Germany and Holland and
when I come back I hope either to see you or to hear from you (I
wrote to Basel to get informations on your lecture
there).\footnote{Meanwhile I heard from P. Huber in Basel, [Fierz is
in the United States], that your lecture there is on March 30.  On
this date I am very glad, because I shall be back from my trip by
then. Paa Gensje!}

Hoping that you will in the future (just as I do myself) enjoy the
enrichment coming from the different kind of access to science by
different scientists, expressed in different, but not contradicting
terminologies, I am sending, also in the name of Franca, all good
wishes to yourself, to Margrethe and to the whole family,
\begin{flushright}
\noindent as yours complementary old \quad \smallskip

\noindent W.~Pauli \quad\medskip
\end{flushright}
\eq

\bq\noindent
{\bf N.~Bohr, letter to Wolfgang Pauli, dated 25 March 1955, photocopy
obtained from the Niels Bohr Institute via Henry Folse.}\medskip

\noindent Dear Pauli,\smallskip

Thanks for your letter of which I was glad to see that even if I am
old you do not feel that I am yet so petrified that we cannot have
such animated and fruitful discussions as in our younger days. You
are certainly right that, as regards many personal utterances, in my
letters like in yours, we are not detached on-lookers, although of
course we have so much in common that it is a pure discussional
accident which words, like mysticism or logical systematism, the one
or other of us uses for mutual educational purposes. I also read
with great pleasure your beautiful Columbia radio-lecture which I
had not seen before. Of course, I appreciate the background for your
use of the phrase ``detached observer'' on that occasion, but in my
article I was using the phrase in a more generalized (or, if you
prefer, more limited) sense suited to point to the characteristics
of our position in science and art.

To characterize scientific pursuit I did not know any better word
than detachment, especially in connection with psychological studies.
As regards quantum mechanics and biology, I wanted to stress the
difficulties which even in these fields have had to be overcome to
reach the detachment required for objective description or rather for
the recognition that in such field we meet with no special
observational problem beyond the situations of practical life to
cope with which the word ``observer'' has been originally introduced.
To my mind, the lesson was merely that continued exploration of the
regularities of nature only gradually should teach us of the
necessary caution in looking for unambiguously communicable
experience. As you, I am of course prepared to change terminology
when it is clear that this will promote common understanding, but
before any of us decides on such steps, I wish to call your attention
to the pure scientific aims I have perhaps not sufficiently clearly
presented in my article.

To make myself more clear, I may for a moment remind of the days of
so-called ``classical'' physics and ``critical'' philosophy, when in
the description of the course of events the role of the tools of
observation was disregarded and space-time coordination and
causality were considered a priori categories. You are certainly
right that Einstein is not consequent when speaking of the ideal of
detached observer and neglecting his own wisdom of relativity, which
Eddington poetically described by the picture of how long man traced
a footprint in the sand until he recognized that it was his own.
Seriously, I mean that you are yourself as inconsequent in stressing
the difference in such respect between classical and quantum
mechanics. It is true that, before the epistemological aspects of
the observational problem were so widely cleared up, a certain
confusion was prevalent, but after the thorough lesson which we have
received, the whole situation including that of classical mechanics
appears in a new light. Though in a vast field of experience one
could neglect the interaction between what was regarded as separate
objects of investigation and tools of observation, one often
overlooked our reach of interfering with the course of events
through our freedom of choosing the experimental arrangement. Indeed,
in those days, relying upon the deterministic and reversible
character of the mechanical description, one might rather have
thought that such influencing within a large scope was possible in
unlimited detail.

On the basis of the recognition of the limited divisibility of
elementary phenomena we have, however, obtained a more generalized
description embracing new fundamental regularities of nature, the
orderly comprehension of which in principle implies statistical
account even as regards reversibility, and which for the exhaustion
of knowledge demands mutually exclusive experimental arrangements.
The point which I especially wanted to stress in the article is
that, just by avoiding any such reference to a subjective
interference which would call for misleading comparison with
classical approach, we have within a large scope fulfilled all
requirements of an objective description of experience obtainable
under specified experimental conditions. The freedom of the choice
of the experimental arrangement is indeed common to classical and
quantum physics and, considering all aspects of the situation, we may
say that in both cases a sharp separation between the ``observer''
and the ``phenomena'' is retained. The difference is only that in
quantum phenomena we have for their definition to include the
description of the whole experimental arrangement and that we have
less possibility of influencing the course of events.

Still, if the study of natural phenomena were exhausted by simple
experience, one might not take questions of terminology too serious,
at any rate within the scope in which order is already obtained. I
want, however, to challenge you whether you really mean that, in the
description of proper biological phenomena, we have to do with an
even greater interference with events on the part of the observer
than that you want to stress in quantum mechanics. To my mind, the
situation is entirely opposite, since the characteristics of our
position in biological studies is just the impossibility without
excluding the display of life to arrange the experimental conditions
required for well defined mechanistic description. It is of course
true that physiological research just consists in studying the
reactions of the organisms to experimental conditions open to our
choice, but it appears to me to be practical as well as rational to
include such reactions under varied specified conditions in an
exhaustive account of organic life. A further point which in this
connection is on my mind is to stress that, in the description of
the characteristic properties of the organism, reversal of events is
logically excluded, and just this circumstance is of course of
fundamental importance for speaking of ``natural evolution''.

I do not know if I in any such respect was able to make the
essential points sufficiently clear in my article, but I am glad
that, quite apart from our present dispute, you were not
unsympathetic with my striving for a unified attitude to the
scientific description of that nature to which we belong and in the
exploration of which we step by step have been reminded of
fundamental general aspects of our position as observers, which only
in limited fields may be disregarded. I shall here not go further in
repeating things which are not new to any of us and leave the battle
about the word ``detached'' to our meeting in Basel in preparation of
which I only wanted to remind of our resources for defence as well
as attack, irrespective of the word ``old''.

\begin{flushright}
With kindest greetings from home to home,\\
and p\aa\ gensyn,\\
Yours ever
\end{flushright}
\eq
\eq

\section{31-03-01 \ \ {\it Saturday Hangover Time} \ \ (to G. L. Comer)} \label{Comer3}

\bgc
It would help my career tremendously if I were smart.
\egc

Don't you know I think the same thing all the time!  Here's a true story (that I usually tell in a different order \ldots\ for a different effect).  At the 1996 QCM meeting in Japan, the three ``quantum communication award'' recipients were Horace Yuen, Alexander Holevo, and Charles Bennett.  When it came time for the acceptance speeches, Holevo stood up and thanked the teachers who had had such a great influence on him (Naimark, Gelfand, etc.).  Yuen stood up and essentially thanked himself.  But, Charlie Bennett stood up and thanked all the ``smart people'' he'd been fortunate enough to gather around himself---the ones who take some of his questions seriously.

Charlie stirs me at times.

I'm a little hung over as I write this note to you, and it's not so unrelated to the story above.  Kiki and I had a little too much wine last night after we put Emma to bed.  We were having such a nice time dreaming, we got a little carried away.  (Well, mostly I got carried away.)  Kiki's been bandying about this idea that we get some land in Massachusetts, Vermont or New Hampshire with an old, big farmhouse and start up a kind of quantum information institute.  It'd be kind of like some mixture between the Niels Bohr Institute and Ghost Ranch (where we got married) \ldots\ with Kiki playing the part of Mrs.\ Bohr.  She's semi-serious in her own way, and I must say she's sort of sucked me into this pipedream.  She figures I'm starting to know enough people in prominent funding positions, and I've organized enough international meetings, etc., etc., that if I start taking the idea seriously now, something might come of it in five years or so.

So we dreamed and dreamed.  And I pulled out a book (that Eugen Merzbacher had given me) on the history of the first ten years of the Bohr Institute.  It had loads of pictures, and even a floor plan of the original building (two stories plus attic and basement).  It had a list of all the visitors, their funding sources, their lengths of stay, and even a count of the papers they wrote while staying there.

Institute for Quantum Information.  Institute for the Activating Observer.  I'm pipedreaming now.

\section{02-04-01 \ \ {\it Present State of Thought} \ \ (to R. Clifton)} \label{Clifton1}

Good to finally meet you.  (Though I think I've had some email with one of the students in your seminar.)  Sorry to reply so late to your note, but I wanted to hear {\Mermin}'s thoughts first.  I'll just paste in the exchange we had below:  The conjunction of the two notes captures my present state of thought pretty well.  {\Mermin}, I see, does have a good point.  [See 26-03-01 note ``\myref{Mermin0}{Clifton Proposition}'' to N. D. Mermin.]  And since I do already have promises for three papers (beside one that I could write), it might be worthwhile running in this direction.  We'll let you know something at the end of June, maybe just at the start of your new tenure.

\subsection{Rob's Preply}

\bq
My name is Rob Clifton and, as of July 1st, I will be the new chief editor of
the journal {\sl Studies in History and Philosophy of Modern Physics}.  Jeremy Butterfield, my
predecessor, has told me that you might be interested in guest editing a special issue of the
journal devoted to the conceptual implications of quantum information theory, computation theory or
both.  To my mind, this is an excellent idea.  Are you still interested?  I've also mentioned the
idea to David {\Mermin}, with the thought that you and he could perhaps guest edit the issue together.
Might that work?
\eq

\section{04-04-01 \ \ {\it Ode to the Monolith} \ \ (to C. M. {\Caves} \& R. {\Schack})} \label{Schack0.2} \label{Caves0.1}

Did I say in my last note on the subject that I think this paper's not too bad?  Well perhaps it's just my mood, but right now I've changed my mind.  Try as I might, I could hardly make it through the beast.  In penance, over and over I've been dropping the heavy thing onto my unprotected toes, chanting ``Forgive me father, for I know I am predominantly to blame for this monolith.''

Maybe tomorrow I'll be out of my Eeyore-like mood.  In the mean time it's your lucky day, because it means I just don't have the strength to put up too much of a fight.  You'll see the bigger part of the fight in the first three or so comments below, but then it peters out almost immediately.

I'm sure we've all learned some lessons in making this paper.

I hope that what I offer below will be viewed as a compromise, as I know you {\it both\/} have compromised dearly for me throughout this project.

Let's put this thing on {\tt quant-ph} as soon as possible.  I had hoped to draw two figures for the paper, but that can wait.\medskip

\noindent Eeyore

\bq\noindent
What is a quantum state?  Since the earliest days of quantum theory, the predominant answer has been that the quantum state is a representation of the observer's knowledge of a system, without any objective objective reality of its own.  This information-based view the authors hold quite firmly.  Despite the association of this view with the founders of quantum theory, holding it does not require a concomitant belief that there is nothing left to learn in quantum foundations.  Quite the opposite!
\eq
$\qquad\qquad\mathbf{\Downarrow}$
\bq\noindent
What is a quantum state?  Since the earliest days of quantum theory, the predominant answer has been that the quantum state is a representation of the observer's knowledge of a system.  In and of itself, the quantum state has no objective reality.  The authors hold this information-based view quite firmly.  Despite the association of it with the founders of quantum theory, however, holding the view does not require a concomitant belief that there is nothing left to learn in quantum foundations.  It is quite the opposite in fact:  Only by pursuing a promising but incomplete program can one hope to learn something of lasting value.
\eq
NOTE:  I know that Carl really wanted to emphasize emphasize objective objective, but I think there are some rules and rules of thumb that I just can't let go of in these opening sentences:
\begin{enumerate}
\item
I like shorter sentences.
\item
I like redundancy, because I want people to remember what I said and
    what I wanted to emphasize.
\item
I believe in complete sentences.
\end{enumerate}

\bq\noindent
Challenges to the information-based view arise with increasing frequency
\eq
$\qquad\qquad\mathbf{\Downarrow}$
\bq\noindent
Challenges to the information-based view arise every day
\eq
NOTE:  I don't want ``increasing frequency'' because I don't want anyone thinking that they're getting the better of us.  The only reason there is an increasing frequency is because there is a bandwagon effect of nonthinking.  We don't need to promote that.

\bq\noindent
A challenge bested or at least blunted, one walks away with a deeper sense of the physical content of quantum theory and a growing confidence for tackling questions of its interpretation and applicability.
\eq
$\qquad\qquad\mathbf{\Downarrow}$
\bq\noindent
With each challenge successfully resolved, one walks away with a deeper sense of the physical content of quantum theory and a growing confidence for tackling questions of its interpretation and applicability.
\eq
NOTE:  Carl called my style ``playfully lofty.''  There is a fine line between that and writing in too baroque a style for the working man.  Because I'm already afraid of my tippy position, I'm {\it often\/} happy when Carl tones me down a little (despite my kicking and screaming the whole way).  This time he toned me up, and, in {\Ruediger}'s words, I don't feel comfortable with that.

\bq\noindent
start to feel tractable (and even connected) from this heightened perspective.
\eq
$\qquad\qquad\mathbf{\Downarrow}$
\bq\noindent
start to feel tractable (and even connected) from this perspective.
\eq
NOTE:  Even I see that I go too far at times.

\bq\noindent
Unknown quantum states are teleported, they are protected with quantum error correcting codes, and they are used to check for quantum eavesdropping.
\eq
$\qquad\qquad\mathbf{\Downarrow}$
\bq\noindent
Unknown quantum states are teleported, protected with quantum error correcting codes, and used to check for quantum eavesdropping.
\eq
NOTE:  I see three complete sentences here.  Why are they separated only by commas?  Thus I return them to their former unital state.

\bq\noindent
If quantum states, by their very definition, are states of knowledge and not states of nature---if a quantum state is a description of knowledge about a system---then the state is {\it known\/} by someone---at the very least, by the describer himself.
\eq
$\qquad\qquad\mathbf{\Downarrow}$
\bq\noindent
If quantum states, by their very definition, are states of knowledge and not states of nature, then the state is {\it known\/} by someone---at the very least, by the describer himself.
\eq
NOTE:  Rule of thumb:  Too many dashes for Chris, and for any other writer I've ever seen outside of the {\sl New Yorker}.  Besides, there was even redundancy here that even I could find no use for.

\bq\noindent
In this case, the unknown state is merely a stand-in for the unknown {\it state of knowledge\/} of an essential player who was omitted from the original formulation.
\eq
$\qquad\qquad\mathbf{\Downarrow}$
\bq\noindent
In this case, the unknown state is merely a stand-in for the unknown {\it state of knowledge\/} of an essential player who was missed in the original formulation.
\eq
NOTE:  There is a meaningful difference between ``omitted'' and ``skipped over.''  The first conveys the idea that the player was actively excluded, while the second portrays something accidental.  I tried to rectify this with the word ``missed'', but if you like something else better---that carries the same sense---use it.

\bq\noindent
More importantly, because of the quantum de Finetti representation theorem, the experimenter is in a position to make an unambiguous statement about the structure of the whole sequence of states $\rho^{(N)}$.
\eq
$\qquad\qquad\mathbf{\Downarrow}$
\bq\noindent
More importantly, the experimenter is in a position to make an unambiguous statement about the structure of the whole sequence of states $\rho^{(N)}$:
\eq
NOTE:  I couldn't make any sense of the extra insertion, especially as it was meant to be part of the premise that gets us to the q de F.

\bq\noindent
Indirect though this might seem, it corresponds to laboratory measurement procedures, and it can be a very powerful theoretical technique, sometimes revealing information that could not have been revealed otherwise
\eq
$\qquad\qquad\mathbf{\Downarrow}$
\bq\noindent
Indirect though this might seem, it can be a very powerful technique, sometimes revealing information that could not have been revealed otherwise
\eq
NOTE:  I cannot say what I do not believe.  I had left it suitably ambiguous enough for my taste---without feeling the need to go into a deeper exegesis---saying only ``any POVM can be represented formally as \ldots''  But you pushed it further, implying that that ``representation'' is what actually happens in the laboratory.  You know I don't believe that.

\section{05-04-01 \ \ {\it 24 Little Hours} \ \ (to C. M. {\Caves} \& R. {\Schack})} \label{Schack0.3} \label{Caves0.2}

Ladies and Gentlemen, Miss Dinah Washington \ldots\ ``What a difference a day makes.  Twenty-four little hours.''

I think Dinah helped get me through the night.  Because, though my toes are still sore, I found myself thinking today that the paper's quite enjoyable after all (even if long).  I just finished reading the whole thing again.  (It took me much longer than the one hour I was estimating.)

I made very few changes outside of the things brought to my attention by you two (my typo yesterday, the thing about rank-oneness, etc.).  I only tweaked a few words here and there \ldots

I didn't completely follow Carl's advice on the Maxwell demon discussion.  The passage was much more dramatic and convincing in its original form---thank you Carl.  So I wanted to stay as close to the original as possible, but---of course---I didn't want to cause undo strain on the historical record.  (You see, Carl, that very sentence indicates that in my heart I believe in some kind of reality.  One of these days, you will get it.)  I think my present rendition hits that compromise; I'll place it below for convenience.

So that's pretty much it for me.  If Carl makes no further Fitzgeraldian changes, I'm ready to sign off on this.

\bq\noindent
Demon passage:\medskip

As a rejoinder, we advise caution to the objectivist:
Tempting though it is to grant objective status to all the mathematical objects in a physical theory, there is much to be gained by a careful delineation of the subjective and objective parts. A case in point is provided by E.~T. Jaynes' insistence that entropy is a subjective quantity, a measure of ignorance about a physical system.  One of the many fruits of this point of view can be found in the definitive solution to the long-standing Maxwell demon problem, where it was realized that the information collected by a demon and used by it to extract work from heat has a thermodynamic cost at least as large as the work extracted.
\eq

\section{07-04-01 \ \ {\it Operational Approaches} \ \ (to R. W. {\Spekkens})} \label{Spekkens2}

\brws
I've been trying to become better acquainted with quantum information theory.  For this and other reasons, I became interested in determining how the axioms of quantum mechanics appear in an {\bf entirely} operational language. I noticed that most attempts at axiomatization (for instance those found in the standard textbooks) commit themselves to some degree of realism, assigning dynamical variables to particles, etc.  Only Peres' book seems to be entirely operational in its approach. Are there other places where one can find a strictly operational
axiomatization of quantum mechanics?
\erws

You can also look at the books by Karl Kraus and G\"unther Ludwig.  But I don't think their efforts are very convincing (though complex they certainly are).  Still another source might be the book by Paul Busch and coauthors; I think it's called {\sl Operational Quantum Mechanics}.

\brws
The impression I now have is that the task of deriving the axioms of quantum mechanics from a few physical principles is a task that one can hope to carry out entirely within an operational approach to the theory, the physical principles being operational principles.
So it seems to me that this project can proceed pretty much
independently of whether one can find a satisfactory realist
interpretation. Is this in line with your thinking?
\erws
Somewhat.  You can find much more about my foundational thoughts (though in a somewhat nonorganized form) at my website.

\brws
Anyhow, it's fascinating stuff. Is there any chance you'd be willing
to come to Toronto to give a talk on this or other research?
\erws
I'm pretty much tied up until the Fall, but then I should have plenty of time to travel again.  So I'd love to come to Toronto possibly in that time frame.  Let's keep in touch on this.

\section{10-04-01 \ \ {\it Rudolph/Sanders/Teleportation} \ \ (to H. J. Kimble \& H. Mabuchi)} \label{Kimble1} \label{Mabuchi0}

If you look at {\tt quant-ph} today, you'll find a new little production by Steven and me---``The Quantum State of an Ideal Propagating Laser Field,'' \quantph{0104036}---that we think pretty much clears up the trouble brought about by Rudolph and Sanders.  More than that, we think it clears up quite a few other things as well:  namely, how laser light can be used most generally as a source for quantum information experiments.  The point is a simple one and a basic one:  There is a useful difference between thinking about the quantum state on the inside of the laser cavity and the quantum state on the outside.  So, in a way, we're hoping the paper also gets attention outside of the circle of teleportation defenders---it's almost the kind of thing that ought to be in the classroom (if we say so ourselves).

In the end, Rudolph and Sanders dissolved most easily:  We were still quite confused the last time we talked to you (though we didn't feel it then).  Laser light can be used to generate continuous-variable entanglement after all (and you certainly did it in the Furusawa et al.\ experiment), it's just in the form of {\it distillable\/} entanglement.  It's much better than ``entanglement of assistance'' in the sense that you don't even have to recover the laser cavity to recover the entanglement.  And even better than that, you don't even have to distill it to do continuous-variable quantum teleportation!  So everything appears to be safe with those aspects of the experiment.

I had previously mentioned to you the possibility of writing a comment to PRL, following an earlier suggestion by Hideo concerning the same but for {\sl Physics Today}.  In light of the almost trivial solution in the end, that possibility is starting to fade from my mind.  But maybe I'm still open to entertaining some thoughts---I'm not ready to shut the door completely yet.  Tell me what you think of our new paper, and I might revise my disposition accordingly.  We'll see, but right now, like I say, I'm starting to get disinclined to the idea.  (I'm not sure what Steven's thoughts are.)

Anyway, that's it.  Wishing you both the best in sunny Southern Cal,

\section{11-04-01 \ \ {\it Piggybacking Philosophy on Physics} \ \ (to J. M. Renes)} \label{Renes2.1}

\bjmr
Answering physics questions relative to a static background philosophy
will not, I believe, yield what one would want it to --- the grand
questions are necessarily partly philosophical and thus must be
confronted head on.
\ejmr

I do appreciate the spirit, and I am part of the choir.  John Wheeler once said something that made a big impression on me:  ``Philosophy is too important to be left to the philosophers!''

You can see the implication.  If we're going to hope to make real progress we've got to have our feet firmly planted in the practice of physics.  Really good physics is the very best of philosophy.  And I want to see you do that.  So when you get here this summer, let's work on developing the art of physics in your heart.

Have you had any luck securing a place to stay for the summer?

If you're working under a real ``grand vision deficit'' I can send you my latest wacky document, ``The Activating Observer''---a large compendium of annotated references.  It's at 92 pages now and is in dire need of a little proofreading.  But when you arrive here, it's gonna be solid physics all the way.  Life's too short, and we've got to make progress by the most-likely-to-succeed means possible.

In the words of George Castanza, ``Do me a solid buddy.''

\section{15-04-01 \ \ {\it $P_E$}  \ \ (to A. Peres)} \label{Peres4}

\bap
Please give me a reference for the formula for $P_E$ in terms of
$\tr(\rho_1-\rho_2)$.
\eap

You meant in terms of $\tr|\rho_1-\rho_2|$, of course.  The reference is:
\begin{itemize}
\item
C.~A. Fuchs, ``Information Gain vs.\ State Disturbance in Quantum Theory,'' {\sl Fortschritte der Physik\/} {\bf 46}(4,5), 535--565 (1998). [Reprinted in {\sl Quantum Computation: Where Do We Want to Go Tomorrow?}, edited by S.~L. Braunstein (Wiley--VCH Verlag, Weinheim, 1999), pages 229--259.]
\end{itemize}
The paper is listed on quant-ph as \quantph{9611010}.  I had originally submitted it for the proceedings of PhysComp96 in {\sl Physica D}.  After waiting a ridiculously long time with no sign of movement from the editors, I finally withdrew it and submitted it to the special issue of {\sl Fortschritte}.

Kiki, Emma and I are in Texas for the Easter holidays.  I believe I had forgotten the meaning of the words humid, muggy, and sultry!!  They've come back to me now with a force that won't be forgotten!

\section{18-04-01 \ \ {\it Op-Ed}  \ \ (to A. Peres)} \label{Peres5}

I'll paste the entry from my dictionary below on the word op-ed.  (It's a common term in America.)

I'm glad that all is well with your family.  Kiki, Emma, and I are just back from a week in Texas.  I finally submitted the old quantum de Finetti paper to {\tt quant-ph}---it only took me two years!  It will appear tomorrow morning.  You may have an interest in it; it is one of my better pieces of work (conceptual {\it and\/} technical).  Petra has already expressed an interest in it.  You might get her to give you a report on it.  van Enk and I also gave a technical application of the result in \quantph{0104036}, ``The Quantum State of a Laser Field.''  (It was nice to see that it could be used in the real world of quantum information bickers!)  That may interest you as well.

\section{19-04-01 \ \ {\it Warsaw Wait} \ \ (to R. Jozsa)} \label{Jozsa3}

\brj
Anyhow it'll be great to catch up in Warsaw. Have you seen the amazing
papers of Koashi/Imoto recently on {\tt quant-ph}? Certainly among the best
papers for some years!!\ldots
\erj

Yes, I have \ldots\ and I agree with you completely.  (I wrote Koashi telling him the same.)

I'll come to Warsaw if I can.  The main issue is funding.  Despite the pre-history of big bucks at Bell Labs, Lucent has fallen on hard times---you warned me!---and there are no travel funds for this year.  I've been put on an ``expenses-paid invitation only'' alert.  So I hope I can come.  I'll talk about my stuff with Sasaki if I do.  (Now I've just got to finished the damned paper for him!  I'm a week late.)

Charlie's talk on ``time travel'' was less interesting than you might think.  I took grave issue with it, and didn't like it.  (But I bet that makes you want to hear about it even more now.)

Have fun in Japan.  I'm glad you're finally getting to see the cherry blossoms.

\section{19-04-01 \ \ {\it CVs and Samizdats}  \ \ (to A. Peres)} \label{Peres6}

\bap
Meanwhile it seems to me that I found a method for secure quantum bit
commitment. It is so simple that I can't see what can go wrong.
\eap

That is always dangerous.  Have you tried very hard to apply the standard Mayers--Lo--Chau attack on it?  Steven van Enk and I were able to break the ``definitive'' QBC protocol of Horace Yuen within about 5 minutes of discussion in just that way.

\bap
Aviva and I plan to be at ITP UCSB from Aug 13 till Sept 29. Shall we
have the pleasure of seeing you?
\eap

No, I will not be there.  They didn't accept my application because I listed a time window of less than four weeks availability.

My mother-in-law leaves today to go back to Germany.  She has been at our house since April 4 visiting.  Also though, while we were away in Texas, she completed painting the walls downstairs and up the stairwell.  I'm very thankful for all that she's done, but having extra duties of conversation have been time consuming for me.  Now that she's gone, I intend to work like gangbusters to get at least one more paper on the web before April is out (a paper by Sasaki and me).  Also I intend to post my big quantum samizdat on May 10; Mermin is writing a foreword for it presently.  Getting it together is an exhausting and painstaking affair, making sure that there are no places where I put my foot in my mouth (where I don't want to, that is).  It is 300$+$ pages at this stage.  I will send a copy to you and a few other friends pretty soon asking for permission to quote you in a small number of places.  I think no one need worry though:  I have worked very hard to make it safe and tasteful.

\section{19-04-01 \ \ {\it CVs and Samizdats, 2}  \ \ (to A. Peres)} \label{Peres7}

\bap
I have a paper by Chau and Lo on my desk, another on the table near
it, and also one by Brassard et al. I have some difficulty
understanding them but eventually I'll find out.
\eap

Maybe the best place to look for an easy understanding of this idea is Jeff Bub's paper.  I'll put the abstract below.  He really tried to be very didactic in that paper.

I'll tell you more about the samizdat in a couple of days.
\bq\noindent
\quantph{0007090}\\
The quantum bit commitment theorem\\
Authors: Jeffrey Bub\\
Comments: \LaTeX, 25 pages. Forthcoming in {\sl Foundations of Physics}, May 2001\medskip\\
Unconditionally secure two-party bit commitment based solely on the principles of quantum mechanics (without exploiting special relativistic signalling constraints, or principles of general relativity or thermodynamics) has been shown to be impossible, but the claim is repeatedly challenged. The quantum bit commitment theorem is reviewed here and the central conceptual point, that an `Einstein--Podolsky--Rosen' attack or cheating strategy can always be applied, is clarified. The question of whether following such a cheating strategy can ever be disadvantageous to the cheater is considered and answered in the negative. There is, indeed, no loophole in the theorem.
\eq

\section{19-04-01 \ \ {\it Quantum de Finetti} \ \ (to D. Petz and Other Friends)} \label{Petz2}

\noindent Dear friends who I recall having shown some interest in quantum de Finetti issues: \medskip

My coauthors and I have finally gotten off our lazy duffs and written up our work of two years ago on a quantum analog of de Finetti's theorem on exchangeable probability assignments.  The paper appeared on the Los Alamos Archive today:
\bq
\myurl{http://xxx.lanl.gov/abs/quant-ph/0104088}.
\eq

\noindent Apologies to all for the long delay,

\bq
\noindent Title:     Unknown Quantum States: The Quantum de Finetti Representation\\
Authors:   Carlton M. Caves, Christopher A. Fuchs, {\Ruediger} {\Schack}\\
Comments:  30 pages, 2 figures \medskip

\noindent Abstract:\\
We present an elementary proof of the quantum de Finetti representation theorem, a quantum analogue of de Finetti's classical theorem on exchangeable probability assignments. This contrasts with the original proof of Hudson and Moody [Z. Wahrschein. verw. Geb. 33, 343 (1976)], which relies on advanced mathematics and does not share the same potential for generalization. The classical de Finetti theorem provides an operational definition of the concept of an unknown probability in Bayesian probability theory, where probabilities are taken to be degrees of belief instead of objective states of nature. The quantum de Finetti theorem, in a closely analogous fashion, deals with exchangeable density-operator assignments and provides an operational definition of the concept of an ``unknown quantum state'' in quantum-state tomography. This result is especially important for information-based interpretations of quantum mechanics, where quantum states, like probabilities, are taken to be states of knowledge rather than states of nature. We further demonstrate that the theorem fails for real Hilbert spaces and discuss the significance of this point.
\eq

\section{22-04-01 \ \ {\it An Apple and a Commitment}  \ \ (to A. Peres)} \label{Peres8}

I printed out your paper and quickly skimmed it while I was eating my apple for lunch.  Forget about the cryptic details of cryptology, your paper is in fundamental contradiction to the HJW result.  I'm sorry, but I think that squashes your protocol.  Read below, and then look up the paper, and I think you'll understand.

I am really sorry about this.

Back to my other project, the samizdat.  I'll write you more details about that in a few days.  And thanks for getting me an invitation to Israel Dec 17--21 (or at least I suspect you were behind it).

\bq
\begin{center}
Review of:\\ Lane P. Hughston, Richard Jozsa, and William K.
Wootters,\\ ``A Complete Classification of Quantum Ensembles Having a Given Density Matrix,'' \\ Physics Letters A, Vol. 183, pp.\ 14--18, 1993.\medskip
\end{center}

Abner Shimony likes to say that entanglement gives rise to ``passion at a distance.''  He does this because when Alice performs a measurement on A of an entangled system AB, {\it something\/} changes for B, BUT that change cannot be used for the purpose of communication with a Bob at B.  If you ask me, this is language that is just asking for trouble; it is language that is poised to confuse a generation of new physicists.  Something indeed does change for B, what Alice can {\it say\/} of it.  But it is nothing more than that; to think that it is truly a physical change with respect to B alone---especially one that is so contrived as to not lead to communicability---is to open up a sink hole.  We are dealing here with changes of states of knowledge.

Within this context, it is quite reasonable and quite interesting to ask how many different ways Alice's knowledge can change.  Depending upon which measurement Alice wishes to perform on A, there will be any of a number of different state assignments for B that follow from that.  What are they, and what are their probabilities?  This is the main question addressed in this little paper.  It has a very clean answer:  a state assignment can be created for B from a measurement on A if and only if that state falls within a pure-state decomposition of the marginal density operator of B.  Moreover, the probabilities of Alice's measurement outcomes correspond precisely to the probabilities of the ultimate state assignments.

This result, it turns out, is not of purely academic interest. It has had a wide range of application in several more applied problems in quantum information:  it is crucial for the proof that no quantum bit commitment schemes exist, it plays a crucial role in proving the exact expression for the entanglement of formation for two qubits, and it is crucial for defining the notion of the entanglement of assistance.  This theorem is one that all quantum information theorists should have incorporated into their tool bag.

Can more be said?  Actually, it turns out that this result even has some historical significance.  For, unknown to the authors above, the same question was raised and even partially answered by Erwin {\Schroedinger} in a 1936 paper!  [E. {Schroedinger}, ``Probability Relations between Separated Systems,'' Proc. Cam. Phil. Soc. 32,
446--452 (1936).]
\eq

\section{24-04-01 \ \ {\it Endo, Exo}  \ \ (to A. Peres)} \label{Peres9}

\bap
You are an inexhaustible source of references. The terms ``exophysical''
and ``endophysical'' were coined by Primas (I believe). Do you know when
and where? Or could you find Primas's email (he is at ETH in Zurich, I
know nobody there). I'll need these terms when my current research
with Danny will bear fruit. This may take a long time, at least
several days.
\eap

You flatter me.  Actually I think David Finkelstein first invented the words.  The story of that is written down somewhere.  I think it is probably in:
\begin{itemize}
\item
H.~Primas, ``Endo- and Exo-Theories of Matter,'' in {\sl Inside Versus Outside:  Endo- and Exo-Concepts of Observation and Knowledge in Physics, Philosophy and Cognitive Science}, edited by H.~Atmanspacher and G.~J. Dalenoort (Springer-Verlag, Berlin, 1994), pp.~163--193.
\end{itemize}
At times like this, I dearly miss my old paper collection.  I had thousands of papers copied, and I could just turn to my file cabinet and find the answer.

My memory is awfully weak on this, but I think there is even a letter of Finkelstein's published somewhere.  If it's not in the volume above, try
\begin{itemize}
\item
O.~E. R\"ossler, {\sl Endophysics:\ The World as an Interface}, (World Scientific, Singapore, 1998).
\end{itemize}
Of this reference, I had written a note to myself:  ``This is a strange little book, full of crazy ideas, but still perhaps some are worth further scrutiny.''

As I recall, though Primas uses the same words as Finkelstein, he had opted to do something odd with their meanings in relation to Finkelstein's.  (I think he may have even {\it reversed\/} their meanings!)

I hope that helps.

\section{26-04-01 \ \ {\it Short Replies}  \ \ (to A. Peres)} \label{Peres10}

\bap
Indeed I found in the proceedings of the 1990 Joensuu conference
Primas's talk, where he refers to a 1983 paper by the Finkelsteins. I
also found the talk of Schroeck, where he uses the acronym POVM.
\eap

I procrastinated during my lunch hour the other day (lately I've been taking no lunches at all), and dug up the Primas article I told you about.  He gives a full history of the terms there and their variations of usage:  In particular, his is different from Finkelstein's.  The references he gives are, first R\"ossler,
\begin{itemize}
\item
O.~E. R\"ossler, ``Endophysics,'' in {\sl Real Brains, Artificial Minds}, edited by J.~L. Casti and A.~Karlqvist (North-Holland, NY, 1987), pp.~25--46.
\end{itemize}
then Finkelstein,
\begin{itemize}
\item
D.~Finkelstein, ``Finite Physics,'' in {\sl The Universal Turing Machine.\ A Half-Century Survey}, edited by R.~Herken (Kammerer \& Unverzagt, Hamburg, 1988), pp.~349--376.
\item
D.~Finkelstein, ``The Universal Quantum,'' in {\sl The World of Contemporary Physics}, edited by R.~F. Kitchener (State University of New York Press, Albany, 1988), pp.~75--89.
\end{itemize}
Primas, in fact, writes:  ``A pertinent distinction referring to internal vs.\ external viewpoints has been introduced by Otto R\"ossler and David Finkelstein under the notions `endophysics' and `exophysics', respectively.''   And he gives the references above.

\section{26-04-01 \ \ {\it Mysticism and Logic} \ \ (to N. D. {\Mermin})} \label{Mermin1}

\bdm
You never replied to my comment on it: that you were getting close to
Schr\"o\-dinger\-ian mysticism.  I suspect the only difference (if it
is a difference) is that you say the observer is in the world while
{\Schroedinger} --- this is a cartoon version of what he says but is
hard to resist in the current context --- says that the world is in
the observer.  Both of you say (with {\Pauli}) that they cannot be
separated.
\edm

But there is a big difference:  {\Schroedinger}'s view was not contingent
on quantum mechanics.  He said the same things before and after its
construction.

\ldots\ Though I guess I've heard the same accused of Bohr.

\section{26-04-01 \ \ {\it The Anti-Deutsch} \ \ (to R. Pike)} \label{Pike1}

Remind me again of the PR guy I'm supposed to talk to before interviewing and such.  I keep forgetting his name.

Discover magazine wants me to insult the many-worlds interpretation of quantum mechanics next week, and I'm more than happy to do it.  (What that really means is that they'll want to make Deutsch the shining good guy, and me the bad guy.  I'm sure it'll turn out that way, but somehow, someway, someone has got to say that this view is nonsense.)

\section{26-04-01 \ \ {\it Your Interview Request} \ \ (to T. Folger)} \label{Folger1}

Yes, I should be able to talk to you Monday or Tuesday of next week.
(Tuesday is preferred, but even Wednesday is fine.)  As Asher
suggested, I too would suggest that you read our articles before we
talk.  I'll send them both to you now to make that more convenient. I
would especially encourage you to read the smaller ``reply to
critics'' article [Physics Today {\bf 53}(9), pp.\ 14 and 90 (2000)]
(after reading the original) in preparation for the thing you say
below:

\btf
More specifically, I hope to discuss how most physicists take a very
utilitarian attitude towards quantum theory, and tend to avoid
considering what the theory might have to tell us about the
fundamental nature of reality.
\etf

From your choice of words in this, you hint very much that you've
already talked to David Deutsch.  He seems to have it in his head that
there is only {\it one\/} way to extract a scientific point of view
of Nature out of quantum theory:  namely, his way (a many-worlds
view). But I find that view essentially contentless, whether he wants
to call it scientific or not.

You will not find me saying that quantum mechanics teaches us nothing about the nature of Nature.  Just the opposite:  I would say it has taught us a lot, and there is still a load more to learn (just by contemplating quantum mechanics alone).  But we will never see the greatest things the theory has to offer if we first shut our eyes to its greatest lesson:  That is, that the terms in the theory are not about a {\it free-standing\/} reality.  Rather they are concerned with our interface with the world.  We are part of Nature---an inextricable part when it comes to the constructions of our descriptions of it---and that has to be reckoned with.  That is the great lesson of the quantum.

If I had one thing to do over again in the original article with
Asher, I would have inserted the words ``free-standing'' in front of
every instance of the word ``reality'' to make that absolutely clear.
I would also have done it in an effort to keep people like Deutsch
from taking it out of context.

\section{26-04-01 \ \ {\it Tuesday Afternoon} \ \ (to T. Folger)} \label{Folger2}

Tuesday afternoon will be fine.

Rereading the last note I sent to you, I realized that you wouldn't be able to have a clue what the last sentence of it was about:
\bq\noindent
I would also have done it in an effort to keep people like Deutsch from taking it out of context.
\eq
The ``it'' in this refers the following exchange, which I recently inserted into a big thing I'm writing up on the quantum foundations.  Let me paste that in for you:
\bq\noindent
AUTHOR'S NOTE:  In my {\sl Physics Today\/} article ``Quantum Theory Needs No `Interpretation'\,'' with Asher Peres, there is a passage that says the following:
\bq
\noindent Contrary to those desires, quantum theory does {\it not\/} describe physical reality.  What it does is provide an algorithm for computing {\it probabilities\/} for the macroscopic events (``detector clicks'') that are the consequences of our experimental interventions. \ldots\
\eq
This passage (usually taken out of context) has been found distasteful by several readers.  At least one of those readers was Alvaro Carvalho, and at least one reader of Carvalho was the iconic many-worlds activist David Deutsch.\footnote{It probably goes without saying, but perhaps I should have said ``many-worlds activist and Star Trek fan.''  See \myurl{http://www.qubit.org/people/david/David.html} for details.}
I know this because Michael Nielsen forwarded to me Carvalho's posting of 12 July 2000 and Deutsch's reply of 16 July 2000 from the electronic bulletin board {\tt Fabric-of-Reality@egroups.com}. Carvalho writes:
\bq
For those who have not read the Letter: ``Quantum theory needs no 'interpretation'\,'' by C. Fuchs and Asher Peres (Physics Today -March 2000), here are some short excerpts (with comments\ldots):
[\ldots]

``Contrary to those desires, quantum theory does not describe reality \ldots'' I wonder what it can possibly describe. Is there anything else beyond reality?
\eq
While Deutsch responds:
\bq
No, but that's not what they think. They think it describes our observations, but that we are not entitled to regard this as telling us anything about a reality beyond our observations. Why? Just for the Bohring old reason that they don't like the look of the reality that it would describe, if it did describe reality. Why? -- I have many speculations, but basically I don't know. I don't understand why.

It's sad enough when cranks churn out this tawdry old excuse for refusing to contemplate the implications of science, but when highly competent physicists -- quantum physicists -- dust it off and proudly repeat it, it's a crying shame.
\eq
In bemusement, I forwarded these quotes to David {\Mermin}, who responded with:
\bdm
Funny, there was this English Bayesian, Tony Garrett, who said more or less the same thing about anybody who had given up the search for hidden variables.  And I suppose there's a sense in which many worlds, insofar as it can be made coherent, is the ultimate hidden variables theory.
\edm
This comment spurred the following note on my part \ldots
\eq

\section{26-04-01 \ \ {\it The Itch, the Burn, the Fire} \ \ (to N. D. {\Mermin})} \label{Mermin2}

Oh, how I itch to put these in the samizdat too \ldots\ \ But I won't do it!  A man is nothing if he can't control his itches.  (And besides, enough is enough.) [See 26-04-01 notes ``\myref{Folger1}{Your Interview Request}'' and ``\myref{Folger2}{Tuesday Afternoon}'' to T. Folger.]

\section{26-04-01 \ \ {\it Memory Lane}\ \ \ (to G. Brassard)} \label{Brassard6}

I just took a walk down Memory Lane and I'm wondering how you're doing:  I just skimmed all 214 notes I've written you since 8 May 1997!  ``Why?,'' you must be asking.  I'm putting together compendium of all my q-foundational emails, and I'm leaving no stone unturned.  You ought to see this book---it's really turning into something.  Mermin is writing a Foreword for it, and I'll go public with it pretty soon.  (But I'll be sending you a preliminary draft of it before then.)

I {\it do\/} hope you'll come to {\Vaxjo}.  People are really taking this idea seriously that you and I have promoted.  I think you'll be quite pleased with what you get to see, hear, think about, and take part in if you come.  The quantum is waiting to be explained (and extended)!  And I think our ideas really are on the right track.

So, come.  I'll bring a little single-malt for us.

\section{27-04-01 \ \ {\it You Say Relate-a, I Say Relata} \ \ (to N. D. {\Mermin})} \label{Mermin2.1}

Did you ever run across anyone else using the word ``correlata?''
The excerpt of a note below to Howard Barnum shows that we were once
worried about that.

In the mean time, I have seen old papers by Henry Folse where he
essentially says that Bohr's point of view is ``relation without
relata'' \ldots\ and he really does use the word relata.  I've been
meaning to tell you about that.

I'll dig up some relevant Folse quotes and send them to you in the
next email.

\section{27-04-01 \ \ {\it Folse Quotes with ``Relata''}  \ \ (to N. D. {\Mermin})} \label{Mermin2.2}

\noindent From: H.~J. Folse, ``The Copenhagen Interpretation of Quantum
Theory and Whitehead's Philosophy of Organism,'' Tulane Stud.\ Phil.\
{\bf 23}, 32--47 (1974)---

\bq
\indent
While Whitehead and classical physics are in sympathy with regard to
experience as the starting point for natural philosophy, there is a
dramatic difference with respect to the manner in which the two
define ``nature.''  In the materialistic world-view of classical
physics, nature is the {\it object\/} that {\it causes\/} our
experiences; what we experience are observables correlated with the
primary and secondary properties of material substances that exist
without the aid of experience.  For Whitehead, however, ``nature is
that which we observe in perception through the senses.'' Nature is
not the cause of experience; it is the field of experience itself. On
this point Whitehead's philosophy insists: we are not concerned with
discovering the concrete object which is the objective cause of
experience, the nature of which is known in abstract concepts; we are
instead concerned with explicating the origins of our abstractions by
reference to the concrete factors nature revealed an experience. Thus
Whitehead claims that
\bq
we have nothing to do with the ultimate character of reality.  It is
quite possible that in the true philosophy of reality there are only
individual substances with attributes, or that there are only
relations with pairs of relata \ldots\@.  Our theme is Nature \ldots\
we confine ourselves to the factors posited in the sense-awareness of
nature \ldots.
\eq
Against the traditional view of science, the Copenhagen
Interpretation finds complete agreement with Whitehead's intention:
\bq
As a final consequence, the natural laws formulated mathematically in
quantum theory no longer deal with the elementary particles
themselves but with our knowledge of them \ldots\ we can no longer
view ``in themselves'' the building blocks of matter which were
originally thought of as the last objective reality \ldots\ basically
we can only make our knowledge of these particles the object of
science.
\eq
It is important to note that when Heisenberg refers to ``knowledge''
he is not calling upon a conceptual abstraction but rather the
experiences of investigators in their experiments.  Since what is
involved in this agreement between Whitehead and quantum theory is a
redefinition of the concept of ``nature'' and since physics seeks to
``explain'' nature, it is not surprising that Bohr should call for
``a reconsideration of our attitude towards the problem of physical
explanation.''  Such a reconsideration is precisely the goal of
Whitehead's philosophy.
\eq
and
\bq
\indent
In undertaking a metaphysics which would take the professed
empiricism of science seriously and yet offer a conceptual
elaboration of nature, Whitehead offers us an ontology of ``events''
as the real termini of our experiences in sense-awareness.  Each
event is an ``actual entity'' which is a concresence of prehensions
of all previous events. In a ``prehension'' the prehending event
brings within itself with a certain determinate ``subjective form''
the prehended event as an objective datum entering into the process
that makes up the life of the prehending event.  In this process the
prehending event ``becomes'' while the prehended event achieves
``objective immortality'' and ``perishes.''  There is no hylomorphic
structure to the actual entities of this ontology; there is not
``something that endures,'' material substance, and something which
the enduring thing ``has,'' properties, that can change.  What is, is
events, and events do not endure, they happen.  In a prehension an
event reaches out to ``feel'' other events; thus an actual entity
``acts.''  Activity, not endurance, is the basic ontological status
of entities in the philosophy of organism.

Since the actual entity has no hylomorphic structure, what is
experienced is not properties standing for the object experienced,
but rather the actual entity itself.  Experience does not reveal
nature mediately, but immediately.  The task of science is not to
explain nature through an appeal to a conceptual representation of
it, but rather to explain a conceptual representation through an
appeal to experience.  In this way the philosophy of organism is a
``critique of abstractions'' and the fallacy of misplaced
concreteness has been avoided.
\eq
and
\bq
\indent
If we endorse the materialistic ontology, then Bohr's position is
totally unacceptable, but on the Whiteheadian view his ideas follow
quite naturally.  The common point, as has been stressed, is that
both positions are interested in a reformulation of the doctrine of
physical explanation by an appeal to direct experience.  However,
where Whitehead analyzes experience in general, Bohr, as a physicist,
is interested in only a small set of experiences, namely those
observations in which a measurement of a microsystem  takes place.
Whitehead protested against the idea of experience as experience of
an impassive object ``out there'' related at each precise instance in
time to a subject that in no way reached out to modify the object
experienced.  Since at each instance the subject is allegedly related
to the object in a perfectly determinate way, there is no becoming
{\it within\/} an experience; becoming can only be the succession of
experiences, one following, instant after instance in a continuum,
upon the other.  If nothing becomes within the instant, nothing
happens; there is no activity in experience. Experience, on this
view, is quite literally, a point instant in time in which nothing
happens.  Against this notion Whitehead's process view holds that the
essence of an experience is the {\it activity\/} of the prehending or
experiencing entity.

This crucial role of activity in experience is the cornerstone for
the compatibility between Bohr's Copenhagen position and Whitehead's
view.  While Whitehead appeals to the logical absurdity of the
materialistic doctrine, Bohr appeals to a physically confirmed
theoretical assumption, namely, the quantization of action.  The
introduction of the quantum of action into physical theory requires
that in a measurement the system being measured and the system
performing the measurement share for finite period of time at least
one indivisible quantum of action; in other words, that they are in
{\it interaction}.  In speaking of ``isolated systems'' classical
physics used an idealization or abstraction which was tolerable
within the limits of accuracy relevant to large objects, but this
abstraction is inapplicable and leads to ambiguities if used in
microphysics. Thus complementarity gives a critique of abstractions.

The direct fit between Bohr's interpretation of physics and
Whitehead's metaphysics can be seen best if we express the
measurement interaction in terms of Whitehead's vocabulary.  An
experimental situation is a prehending in which the measuring system
prehends or measures a certain other actual entity or society of such
entities.  The measuring apparatus itself is a society of actual
entities, i.e.\ a nexus of mutually prehending entities each having a
specific subjective form of prehending such that all share in common
a defining characteristic that is determined by the function and
application of the measuring instrument.  The measurement itself is
an event, i.e.\ an actual occasion in which the society forming the
measuring system prehends the measured system with a certain definite
subjective form determined by the whole of the experimental setup.
If the experimental setup is changed, that the society of events
constituting the measuring system will prehend the measured system
with a different subjective form and the objective immortality
achieved by the prehended entity will differ accordingly.

On the classical view the fact that microphysical entities which are
all theoretically represented as being in the same state will in one
observational setup appear as particles and in another appear as
waves seems highly perplexing.  On the Whiteheadian view, there is
nothing remarkable in this fact, for what is ``measured'' is not an
{\it isolated\/} system, but an entity that is the coming together of
its relations to everything else, including the entities of the
measuring system and the situation in which the two interact.  Thus
Bohr also emphasizes ``the impossibility of separating the behavior
of atomic objects from the interaction of these objects with the
measuring instruments which serve to specify the conditions under
which the phenomena appear.''

Bohr often speaks of the above conclusion as the impossibility of
separating subject from object, and he held that ``but this situation
in physics has so forcibly reminded us of the old truth that we're
both onlookers and actors in the great drama of existence.''  Such a
conclusion is precisely what Whitehead advocates in his account:
\bq
The fundamental concepts are activity and process \ldots.  The notion
of self-sufficient isolation is not exemplified in modern physics.
There are no essentially self-contained activities within limited
regions \ldots.  Nature is a theater for the interrelations of
activities.
\eq
The point intended these passages is not that ``subjectivity'' in the
usual sense has intruded into scientific explanation, but merely that
like the notion of an ``isolated system'' the idea of an object
distinctly separate from a subject is an abstraction or idealization
which omits the factor of the interaction between systems.
\eq

\noindent From:  H.~J. Folse, ``What Does Quantum Theory Tell Us About
the World?,'' Soundings {\bf 72}, 179--205 (1989)---

\bq
\indent
Thus we find a way to relate the {\it philosophers'\/} question about
realism to the {\it scientists'\/} concerns about the systems
described by quantum physics.  If a neutral observer were to follow
the discussion between realists and their opponents when it comes to
quantum theory, and then is asked to whom to award the palm, one
reasonable reply would be to say, ``No decision can be made until we
first know what is this `something more' that realists want me to
believe and anti-realists find so unacceptable?''  Answering this
question is the job of a contemporary philosophy of nature, and it is
precisely here that the realist interpretation of quantum theory
finds itself most embarrassed.
\eq
and
\bq
\indent
This classical account of how we know the nature of reality behind
the phenomena relies on the presupposition that knowledge requires
the ``truth'' of theoretical statements to reside in a {\it
correspondence\/} between at least some terms in these statements and
the properties of independently existing entities.  This
correspondence account of truth implied that the resulting
``spectator theory of knowledge'' stipulates the {\it objective\/}
knowledge must describe reality in terms of the properties actually
{\it possessed\/} by an independent reality.  (To be sure, even in a
classical account, observation involves an interaction between
observing and observed systems, thus what is recorded in an
observation is strictly speaking a {\it relation\/} between the
interactors, as even the Ancients well understood.  But insofar as
this interaction involves systems that can be defined as existing in
separate mechanical states, such relations entirely supervene on the
{\it possessed\/} properties of the relata, and thus can be
``reduced'' to them.)  According to this outlook, the ``objectivity''
essential to {\it scientific\/} knowledge is guaranteed by the fact
that classical mechanics makes it possible to provide a description
of the object which eliminates any reference to the observer as a
physical system interacting with the object to produce the
``observations'' on which that description is based. Thus the
``subject'' is ``detached'' from the object by treating the
``observer'' ({\it qua\/} physical system) as mechanically isolated
from the ``observed'' object.  In this way the observer is treated as
a ``ghost spectator'' and any physical effect of observation is
eliminated from the account in order that the description can be
considered as referring to a physical world existing apart from
observation.
\eq
and
\bq
This was justified on its view because ``observation'' in the
Cartesian framework refers to an event in the cognitive domain, {\it
i.e.}, the human ``mind,'' and thus even though the careers of
spatio-temporal substances are the ``cause'' of this observation, it
cannot be described as a {\it physical\/} interaction.  The Cartesian
ideal which pictures what the universe would look like even if no one
was there to look at it is the viewpoint of a ghost spectator who
pilots without physical effort his corporeal submarine through a
space-time sea.

In a quantum-era philosophy of nature this dualist approach to
observation must be thrown out.  The empirical starting point of
science is the description of a phenomenon through a very specialized
set of concepts in which that phenomenon is described as an
observation of a neutrino or a quasar.  Thus the description of
observation as interaction which was exactly what had to be {\it left
out\/} in the classical account now becomes exactly {\it what it
is\/} that is described in the quantum description of microphysical
reality.

For this reason the realistic interpretation of quantum physics
requires not only that we discard the spectator account of knowledge,
but it also denies the presupposition that ``truth'' refers to a
property of statements and exists in virtue of a reference
relationship between terms in these statements and the properties of
an independent reality to which these terms correspond.  Now we learn
that in physics we characterize an independent reality {\it not\/} by
attributing properties to some substantial entity which is imagined to
possess those properties apart from our interaction.  Instead we
characterize it through the phenomena which occur in our interactions
with it.  Truth, then, is not a property of statements but a property
of the whole theoretical structure which allows us to predict those
sorts of phenomena, and such a theoretical structure has that
``truth'' in virtue of its power to predict successfully precisely
those phenomena in which we are said to observe these objects.

The collapse of the hope for a hidden-variables theory which would
preserve the separability of the states of mechanically isolated
systems, and the dispelling of the myth that quantum physics was
conceived in an anti-realist spirit now make it necessary to take
seriously a philosophy of nature which represents real microentities
as the seats of objective potentials for interaction.  Such a
philosophy of nature will no longer characterize as ontologically
fundamental those primary properties which characterized the
classical body.  Material objects are not vast collections of tiny
extended bodies, Democritean atoms or Cartesian {\it rei extensae}.
In breaking the presumed link between the primary properties of the
classical mechanistic framework and those properties which are
conceived to be ontologically fundamental, the quantum revolution
point towards a philosophy of nature which ``atomizes'' not bits of
matter, but elementary processes of interaction.
\eq

\noindent From: H.~J. Folse, ``Ontological Constraints and
Understanding Quantum Phenomena,'' Dialectica {\bf 50}, 121--136
(1996)---

\bq
\indent
The discourse of physics speaks about physical systems by attributing
properties to them in two distinct ways.  Insofar as physics is an
empirical science it must be possible to describe physical systems as
objects of {\it observation}.  Insofar as the ``explanation'' of the
observable properties offered by physics entails deductions from
theory which are confirmed by observation, it must also be possible
to attribute properties to physical systems on the basis of theory.
Although the resulting distinction between ``observational'' and
``theoretical'' properties is of course deeply entrenched in the
philosophical literature, the relations presumed to hold between them
rest on often tacit ontological presuppositions which the quantum
description shows to be idealizations acceptable at the macrolevel
but not at the microlevel.
\eq
and
\bq
\indent
The quantum theoretical description of atomic systems was born in the
attempt to devise mechanical models of the chemical ``atom'' as a
complex physical system composed of subsystems of the (then held to
be) truly atomic ``elementary particles.''  The hope that mechanistic
atomism could provide an ontological basis for understanding a wide
variety of phenomena traced to the behavior of the chemical ``atom''
seemed a rationally attainable goal because of the (then) well
corroborated assumption that the classical state of an isolated
system obeys strictly deterministic laws.  However, the discovery
that initiated the quantum revolution was the theoretical parameter
of ``action'' could formally explain the relevant phenomena only by
being theoretically represented as ``quantized.''  This discovery in
turn implied that in the obsevational interaction there is a kind of
{\it wholeness\/} which precludes attributing a classical mechanical
state to the observed object as an isolated system immediately (or
any time) after the interaction.  Consequently while the properties
in terms of which the classical mechanical state is defined can be
predicated as {\it observable}---i.e.\ relational---properties of the
empirical object, these same properties cannot unambiguously be
predicated of the isolated system as {\it theoretical}---i.e.\
possessed---properties.  Thus there is a crucial difference between
the ways in which classical and quantum descriptions of the phenomena
allow us to attribute properties to systems in the discourse of
physics.
\eq
and
\bq
\indent
Assuming that there are a plurality of individuals in nature, an
ontology must provide a means for individuating them from each other.
This need for a principle of individuation entails that ontological
discourse requires that at least some ``fundamental'' properties must
be {\it possessed\/} absolutely by the entities of which they are
predicated.  To see why this is so, consider that if no properties
were possessed, then all properties would be relational.  But in this
case there would be no way to individuate the relata as distinct
entities between which the relation is said to hold.  Indeed, the
supposed relation could not be external between the relata, but
becomes a relation ``internal'' to the complex whole of the supposed
relata.  But to say that a relation is internal to a nonreducible
whole, is in effect to say that the whole, treated as an individual,
possesses a certain property.  So we are back to the need for
possessed properties.
\eq

\section{27-04-01 \ \ {\it Primrose Lane}\ \ \ (to G. Brassard)} \label{Brassard7}

\bgb
I'm not yet saying no, but the chances of my coming are slim.
Alas!  Please forgive me and think of me again in the future.
\egb

In other words, you {\it are\/} saying no.  OK, I understand.  But you will be sorely missed.

\bgb
By the way, I have about 2.5 times more money to spend than in the
``gobs and gobs'' era!  Any fresh ideas how to deal with this?
\egb

Having gobs and gobs, and 2.5 times that(!), is a good thing.  I'll think hard about a postdoc for you.  Would you be willing to take someone like me?  (I.e., someone with zero knowledge of computer science and cryptology?)  Or do you want to stick on a more traditional basis this time?

In the mean time---you knew I would say this!---what say you about ``QFILQI-2''?  (It may not even be such a bad thing in the eyes of the computer scientists:  Larry Schulman recently organized a small meeting like this at Caltech.  Watrous gave an excellent talk there.)  We could do it bigger, better, a fresh set of people (with old standbys too), if you wish.  Or we could keep it small and cozy.  The point is, the time is ripe, and people---good people---are starting to take these issues seriously.  I could lean hard on everybody for a conference proceedings this time.  Think about it.\medskip

\noindent Your old scheming friend,

\section{30-04-01 \ \ {\it Rudolph/Sanders/Teleportation, 2} \ \ (to H. J. Kimble)} \label{Kimble2}

I'm glad you're waiting to ask your scientific questions.  I'm running ragged trying to get a big symphony finished.  Stay tuned to {\tt quant-ph\/} May 10:  I think it's likely to be the biggest submission ever.  (Single spaced, 400+ pages, no figures!)

\section{01-05-01 \ \ {\it Samizdat Revealed}  \ \ (to A. Peres)} \label{Peres11}

I finally reveal the samizdat!  I'll do it mostly by tacking on a form letter that I wrote for some of the more sensitive cases of my quoting frenzy.  Perhaps if you read that first, what I am about to write will make more sense.

The samizdat is a collection of some of my deepest and most forward-looking thoughts on quantum theory:  it's the kind of stuff that cannot appear in a scientific journal because it is not science {\it yet}.  Nevertheless I do not think it is all worthless---that is the main reason for my wanting to make it public.  Life is short, and I want to do as much as I can in this lifetime to advance physics.  Inspiring others to think about deeper things is part of what I now view as my mandate.

What I would like to ask of you is for you to follow the instructions below for getting a preview of the thing.  In your case, actually, I believe I have quoted nothing sensitive:  You will find no surprises.  But I do very much crave your approval for the 71(!)\ times that I do quote you.  (They are all clearly denoted as Asherisms, followed by a number.)  You {\it by far}, had the greatest influence on this document and my letters to you alone take up about 90 pages!  No other correspondent comes close.  (There are even two separate chapters devoted to you.)  Also I guess what I am searching for is a kind nod of tolerance for the chapter titled ``Diary of a Carefully Worded Paper.''  Beside my point of general emphasis below about reading the Disclaimers, please do read the introduction to the DCWP chapter.  I think you will find it pleasing in regard to our mission as teachers.

This document, despite its nonstandard character, is quite a serious work, I believe.  And some quite serious people think that of it (Bub, Mermin, Caves, Nielsen, etc.).  Worried slightly that you might not approve of my inclusion of the DCWP chapter, I had hoped that David Mermin would first share his Foreword with you---I believe he will laud the chapter there.  Unfortunately he is holding on to it until the original deadline I set for him, but he did write me to tell you this:
\bdm
If you want you can tell him I've been working hard on a foreword
which there's no way I would have done if I didn't take your project
extremely seriously and think the Samizdat well worth making public.
\edm

So, that's about it from me.  I hope I don't crash your computer.  And I hope I hear some happy words from you tomorrow.

\subsection{Form Letter}

\bq
May 10 will be the first anniversary of the Cerro Grande fire entering the township of Los Alamos.  The 400+ families that lost their homes that night will all likely commemorate the event in their own way.  For myself, I plan to do what I usually do on special occasions:  share my sins with the wider physics community.  In this case, it will be in the form of a posting (to the Los Alamos archive) of a 400+ page book of my quantum foundational emails.  I am in cahoots in this with David Mermin, who is writing a foreword for the thing.

In two or three days, I plan to start requesting proper approvals from my correspondents for the quotes of theirs I would like to use.  To make it easier on everyone, I hope to have a name index in place before then (listing not only the quotes but my conversations with others about their thoughts).  But in a couple of cases, I wanted to get pre-approvals even before that.  Yours is one of them.  The main questionable thing in connection to you is that I would like to include \underline{\phantom{David Albert}}.  I hope you will say ``yes'' as it adds much to the theme.  [In your case Asher, there is nothing really questionable; the only outstanding feature is its sheer mass.]

In either case I hope you can tell me your thoughts on this as soon as possible.  I would really be very grateful.  Trying to meet the anniversary date makes it such that I (and Mermin) are working around the clock.

Here's what to do:
\begin{enumerate}
\item Go to my website 
and download the PostScript file titled, Wacky File \#1.  The file is very large (3.5 megabytes or so), so do not do this over a modem; use Ethernet only.  Furthermore, don't bother to print it out; when it is complete I will send you a bound copy.  Use only a screen viewer for your reading (unless you want to kill a lot of trees).

\item Read the Introduction page (so you get a deeper feel of what I'm aiming for), then read the Disclaimers and Acknowledgements on the following page.  The Disclaimers are {\it especially\/} important in connection to your contribution.

\item In the Table of Contents, find your section and skip over to it.  There you'll find the material I'm talking about.  [In your case Asher that is Chapters 14 and 15 in the present version.]
\end{enumerate}
If you say that I should not use any of the direct quotes, I will delete them before my larger approval request.  But as I say, it adds much to the theme.  [More importantly Asher, in your case you are the backbone of the book.]

Thanks for everything.
\eq

\section{02-05-01 \ \ {\it Commitments and Burdens} \ \ (to J. Bub)} \label{Bub1}

I dug up a good tidbit from '96, especially in connection to you.  I'll send you a pointer once I get the new (internal) posting out.  I think it hints at a new---maybe more interesting---twist to the foundational value of the bit commitment issue. \ldots

Well maybe I oversold it, but there is certainly a thought there that I had completely forgotten about.  The places to look are the second two notes in the chapter on Michael Nielsen.  [See chapter on Nielsen in {\sl Coming of Age with Quantum Information}.]  What I found interesting was that my feelings were a bit equivocal about the possibility of a no bit commitment theorem at the time.  What I really seemed to think important was that quantum mechanics have the possibility of providing a provably secure digital signature.  Things were muddled in my head at the time though, because I didn't seem to understand fully that the two protocols might not be equivalent.

Anyway, there it is.  The further thing that I find intriguing is that Dan Gottesman slipped me a preprint by himself and Ike Chuang the other day where they {\it think\/} they have a provably secure digital signature scheme (that does not rely on bit commitment, of course).  (One glitch remains in the proof.)  So, putting the two together, it's just food for thought.

Enjoy London!  (If you see Rob {\Spekkens} there, tell him to look at the samizdat:  all the answers to his teleportation questions are there.)

\section{03-05-01 \ \ {\it Lyme Disease Newsletter} \ \ (to J. W. Nicholson)} \label{Nicholson1}

God, even if I don't have the disease, I've got the symptom:  I had {\it not\/} noticed that they used ``entitled''!

Here's a good one that Mermin caught me at:  I used to flaunt when I really wanted to flout.  Apparently I'm not the only one:  it is labeled as a ``usage problem'' in the {\sl American Heritage\/} dictionary.

\section{03-05-01 \ \ {\it Snap Decision} \ \ (to D. J. Bilodeau)} \label{Bilodeau1}

Let me write you a very short note at this late hour.  I am working frantically to put together the definitive version of my samizdat.  It will make its appearance on {\tt quant-ph} next Thursday, with a Foreword by David Mermin.  The note below written originally to Bill Wootters explains much of what I need to explain to you (applied to his special case, of course, though).  Read that now, and then come back to the next paragraph.

I've just been compiling the Bilodeau chapter and it dawned on me that my side of the writings in this case are not {\it nearly\/} as interesting as my correspondent's, i.e., you.  Thus I had the following wacky idea:  Would you be interested in allowing me to quote you {\it very\/} extensively.  That is, I would do it much like I did in the case of the Benioff and Bub chapters (with extensive reply subsections)---have a look at them once you finish reading this.  The main difference would be that your writings are much more massive than theirs, and much more full of really juicy thought.  It would sort of be a ``Chris's picks of Bilodeau's thoughts''.

I think this document despite its length could receive some good airplay, and that might benefit you.  For instance I had an interview with a {\sl Discover Magazine\/} reporter today, and will again after the samizdat's appearance.

Of course, I would pass all quotes by you for final approval.  But I am working against a huge time limit problem here, so---in your case---I would like to know before getting too deeply into the construction process whether you think I have a safe go ahead.

If you don't feel comfortable with my doing this, I'll just insert a few choice quotes like I did with everyone else \ldots\ and then run those past you (like I did with everyone else).  The thing I'm looking for here is a nod as to how much time I should be spending on your chapter before getting too deeply into it.

There is no doubt that you are one of the existing backbones of the Paulian theme.

\section{03-05-01 \ \ {\it Snap Decision, 2} \ \ (to D. J. Bilodeau)} \label{Bilodeau2}

Wow, that's a long letter:  I haven't even attempted to read it yet, outside of the first couple of lines.  In any case, your history (whatever it is---I'll read it later), would play no role in this.  It's your thoughts on the quantum that I crave; I care not about the mysteries that brings them together.

But---just back from dinner---I was just going to write you in any case.  Since writing you last night I decided to significantly scale back my plans concerning your chapter:  Upon reading a fraction of what you've sent me, I realized that you just had too much material for the present project.  A better place for it to appear---I think---will be in my big ``Activating Observer'' compendium, which will be one of the three companion volumes in this series.  I will treat your letters just as if they were regular papers, and quote them at length there.

Give me ten minutes, and I'll have a new posting on the web which includes your chapter.  Please do have a look at the ``Bilodeau-isms'' and the ``Doug's Reply'' subsections, just to make sure you are OK with everything.  So, I'll post that, and then I'll come back to read your long note.

\section{04-05-01 \ \ {\it Fermat} \ \ (to J. Preskill)} \label{Preskill1}

\bjp
If you make some copies, I'll buy one from you. You can even mark it up \ldots
\ejp
``I have found a truly marvelous proof that quantum mechanics is a law of thought.  But this margin is too small to contain it \ldots''

\section{05-05-01 \ \ {\it Author's Reply}  \ \ (to A. Peres)} \label{Peres12}

Thank you for your ``referee's report''.

\bap
Circa line 20200, you write ``It is the theory which decides what we can
observe''. I remember hearing these words from von {\Weizsacker} in Joensuu
in 1987. What is the reference? Who said that to whom?
\eap

In the samizdat I write:
\bq
Here's an idea attributed to Einstein by Heisenberg that I love.  It captures an extremely important point (in this fuller version that is not quoted so often):
\bq
It is quite wrong to try founding a theory on observable magnitudes alone.  In reality the very opposite happens.  It is the theory which decides what we can observe.  You must appreciate that observation is a very complicated process.  The phenomenon under observation produces certain events in our measuring apparatus.  As a result, further processes take place in the apparatus, which eventually and by complicated paths produce sense impressions and help us to fix the effects in our consciousness.  Along this whole path---from the phenomenon to its fixation in our consciousness---we must be able to tell how nature functions, must know the natural laws at least in practical terms, before we can claim to have observed anything at all.  Only theory, that is, knowledge of natural laws, enables us to deduce the underlying phenomena from our sense impressions.  When we claim that we can observe something new, we ought really to be saying that, although we are about to formulate new natural laws that do not agree with the old ones, we nevertheless assume that the existing laws---covering the whole path from the phenomenon to our consciousness---function in such a way that we can rely upon them and hence speak of ``observation.''
\eq
\eq

The whole paragraph of thoughts is meant to be a paraphrase of Einstein.  He had told this to Heisenberg on a walk, after Heisenberg's talk in Berlin.  The time period was sometime strictly between Heisenberg's discovery of matrix mechanics and his discovery of the uncertainty principle.  He claims that that conversation was his main motivation for searching for ``something like'' an uncertainty principle.

\section{05-05-01 \ \ {\it Full Reference}  \ \ (to A. Peres)} \label{Peres13}

\begin{itemize}
\item
W.~Heisenberg, {\sl Physics and Beyond:~Encounters and Conversations}, translated by A.~J. Pomerans (Harper \& Row, New York, 1971), pp.~63--64.
\end{itemize}
Also quoted in:
\begin{itemize}
\item
M.~Jammer, ``The Experiment in Classical and in Quantum Physics,'' in {\sl Proceedings of the International Symposium, Foundations of Quantum Mechanics in the Light of New Technology}, edited by S.~Kamefuchi, H.~Ezawa, Y.~Murayama, M.~Namiki, S.~Nomura, Y.~Ohnuki, and T.~Yajima (Physical Society of Japan, Tokyo, 1984), pp.~265--276.
\end{itemize}
This paper contains some absolutely wonderful material about the experimenter's role in varying experimental initial conditions and about how that is an essential piece in the construction of scientific theories.

\section{08-05-01 \ \ {\it Depth}  \ \ (to N. D. {\Mermin})} \label{Mermin3}

By the way, I find the opening sentence of this paragraph extremely
deep sounding!  Is that what I've been talking about all this
time?!?!

\bdm
The real issue is nothing less than how you and I can each construct
a representation of the manifold aspects of our individual
experiences (loosely known as a world), and the constraints that my
representation imposes on yours, and vice-versa.  By focusing
explicitly on the strange information-processing capabilities
inherent in the quantum mechanical description of physical reality,
the new discipline of quantum information offers an opportunity to
put on a sound foundation what was only hinted at in the convoluted
prose of Bohr, the facile sensationalism of Heisenberg, the aphorisms
of {\Pauli}, and the poetic mysticism of {\Schroedinger}.  If it hasn't
occurred to you that this is the real justification for your quantum
information-theoretic pursuits, then you owe it to yourself to pause
and peruse these pages.
\edm

\section{08-05-01 \ \ {\it No Irony}  \ \ (to N. D. {\Mermin})} \label{Mermin4}

\bdm
\bq\noindent\rm
[CAF wrote:]  By the way, I find the opening sentence of this
paragraph extremely deep sounding!  Is that what I've been talking
about all this time?!?!
\eq
I detect irony.

No, that's my current take on how what you're doing will ultimately
bring coherence to the Ithaca Interpretation.  I actually thought it
might irritate you.  And perhaps it has.
\edm

No irony at all.  Go back and read ``Fuchsian Genesis'' in Comer's
chapter.  What the heck do you think it's about?  I just thought you
said it in a particularly masterful way.

\section{10-05-01 \ \ {\it Leibnizian Thoughts}  \ \ (to N. D. {\Mermin})} \label{Mermin5}

\bdm
I know Fuchsian Genesis and its variations almost by heart at this
point, but I didn't think that was what I was saying.  I wasn't
thinking of the constraints in terms of your information gathering
requiring my information loss.  I had in mind something more like a
static situation.  Not very well defined (or I'd be done).  But maybe
it's the same thing.  Will think about it.
\edm

I'm just rambling.  Here's a thought from my drive in to work a
minute ago.

Given what you say above, maybe you should go back and read about
Leibniz's monads again (just like someone else had once suggested to
you).

Thinking about that led me back to the phrase ``correlata without
correlation.''  And thinking about that lead me to a thought I had
never thought before:  how ridiculously Cartesian the relative state
interpretation is!  There is res extansa (maybe I should call it res
physica to be more careful), and there is res cogitans.  And the only
way the two ``interact'' is by the happenstance---you sold me on the
word---of appearing in a Schmidt decomposition of some wave function!
Deutsch's multiverse boils down to about the same thing as a
modern-day pineal gland.

\section{10-05-01 \ \ {\it The Easy Part}  \ \ (to N. D. {\Mermin})} \label{Mermin6}

I'll give you a bonus:  a scotch, a beer, and comments.
Unfortunately for the bulk of the comments, you will have to wait
until Monday or Tuesday.  I have pledged to my wife to have the
computer completely turned off tomorrow for Mother's Day.

A serious problem is that the PostScript you sent me will not print.
[\ldots]

In the meantime, let me comment on footnote 5, because that's easy.
Yeah, you're right---I think---that the usual textbooks don't
emphasize the Born rule in the setting of bipartite systems. However,
I myself have always been reluctant to take the ``collapse
postulate''---i.e., that after a measurement, a system is left in an
eigenstate of the observable---simply because it is not true
generally.  The state of the system after a measurement depends in a
detailed way on the precise form of the measurement interaction, and
even upon whether the observer kept all the information available to
him.  (Remember, for me, a quantum state is a state of knowledge; and
no one has the right to say that I can't throw away some of my
information, etc.)  There is one setting, however, where one can make
something close to the collapse postulate come alive:  and that is in
considering measurements on subsystems of a total whole.

Below I'll place the ``axioms'' for quantum theory that I used in a
tutorial on ``Basic Quantum Mechanics'' for a DIMACS meeting in 1997.
Anyway, axiom 5 in particular emphasizes the point above, and comes
close to the point you were trying to make.  Though notice, in
contrast to you, I was silent on what happens to the state on system
$S_1$.

About an overall assessment, you're going to have to wait until I can
print out the thing and view it properly.

\begin{center}
\large \bf The DIMACS Meeting Axioms
\end{center}

\subsubsection{Quantum Axiom 1}

A {\it quantum system\/} is any domain of physical inquiry or
manipulation.

{\it Maximal\/} states of knowledge about such a system are in a
bijective correspondence with operators
$$
\Pi=|\psi\rangle\langle\psi|\quad\mbox{with} \quad
\langle\psi|\psi\rangle=1
$$
for $|\psi\rangle$ in some Hilbert space $\cal H$.
$$
\Pi = \mbox{``the quantum state''}
$$

\subsubsection{Quantum Axiom 2}

The most precise manipulation of a quantum system that can arise
without learning anything new of it corresponds to a map of the form
$$
\Pi \longrightarrow U\Pi U^\dagger
$$
where $U$ is unitary.

\subsubsection{Quantum Axiom 3}

Each question that can be asked of a system corresponds to a set
$\{P_1,\ldots,P_k\}$, $k\le\mbox{dim}\cal H$, of operators with
\bea
P_i P_j &=& \delta_{ij} P_j \nonumber \\
\sum_j P_j &=& I\;. \nonumber
\eea
The index $j$ labels the outcomes.

When state is $\Pi=|\psi\rangle\langle\psi|$, outcome $j$ can be
expected to occur with probability
$$
p(j|\Pi)=\tr\Pi P_j = \langle\psi|P_j|\psi\rangle\;.
$$

\subsubsection{Quantum Axiom 4}

Maximal states of knowledge about a composite system ${\cal
S}_1\oplus{\cal S}_2$---the components with Hilbert spaces ${\cal
H}_1$ and ${\cal H}_2$---are in bijective correspondence with all
$$
\Pi=|\psi\rangle\langle\psi|\quad\mbox{with} \quad
\langle\psi|\psi\rangle=1
$$
{\it where}
$$
|\psi\rangle\in {\cal H}_1\otimes{\cal H}_2\;.
$$
When $|\psi\rangle\neq|\alpha\rangle|\beta\rangle$, the state
$|\psi\rangle$ is said to be {\it entangled}.

\subsubsection{Quantum Axiom 5}

Suppose ${\cal S}_1\oplus{\cal S}_2$ consists of two noninteracting
systems (e.g.\ spacelike separated systems) and $|\psi\rangle\in
{\cal H}_1\otimes{\cal H}_2$ is of the form
$$
|\psi\rangle=\sum \sqrt{p_i}|e_i\rangle|\psi_i\rangle\;,\qquad \sum_i
p_i=1
$$
for some orthonormal set $|e_i\rangle\in{\cal H}_1$, and
$|\psi_i\rangle\in{\cal H}_1$ with $\langle\psi_i|\psi_i\rangle=1$.
(The $|\psi_i\rangle$ need not necessarily be orthonormal).

Then a measurement of $\{|e_i\rangle\langle e_i|\}$ on ${\cal S}_1$
revealing outcome $k$ leaves the observer in a maximal state of
knowledge about ${\cal S}_2$:
$$
\mbox{state}({\cal S}_2|k)=|\psi_k\rangle\langle \psi_k|\;.
$$

\section{10-05-01 \ \ {\it Your Questions on de Finetti and Diffusion} \ \ (to R. W. {\Spekkens})} \label{Spekkens3}

I still don't have enough time to come out of the water.  But let me surface to get a breath of air: I've got NATO pounding on me for a paper, and I've got a meeting with the president of Bell Labs Monday.  That combined with getting the big samizdat on {\tt quant-ph} this morning has been killing me.  Literally, only 3-4 hours of sleep each night for the past two weeks.  (That kind of thing takes its toll on an old man; I'm not young like you and Terry any more.)

\brws
I enjoyed reading your paper on the quantum de Finetti theorem.  I'm quite interested in the mathematical result, but I'm not sure I agree with
the way you've interpreted it.  In particular, your view that quantum states are states of knowledge doesn't make sense to me.

Let me first see if I properly understand what your view is {\bf not}. It seems to me that it is not an operational view.  This actually surprises me
because I somehow got the impression that you were an operationalist (perhaps because I take Asher Peres to be an operationalist and you wrote that
{\bf Physics Today} article together). My understanding of operational approaches to quantum mechanics is that they take quantum states to represent
preparation procedures (for instance, the setting on a dial on some apparatus).  If an observer has only partial information about which of a set of preparation procedures was implemented, it seems to me that an operationalist should describe this observer's state of knowledge by a probability distribution over density operators. I get the feeling from reading your paper that you would agree with my characterization of operationalism as distinct from your view, given that you come down against interpreting the density operator as ``the description of the `true' preparation procedure''. Have I got that right?

You state that ``the quantum state is a representation of the observer's knowledge of a system''.  What I don't quite get is what you mean by `knowledge of a system'.
\erws

But you provide the first example of how the samizdat can serve me.  (By samizdat, I mean the paper cited below.)  Let me give you pointers for where to look in there.  Three places:  In the Brassard chapter, read the introduction to the problem set for the {\Montreal} meeting.  Also look at Proto-Problem \#3.  Jeffrey Bub has been big on precisely this question too.  So read his chapter.  Finally, there is plenty of material on your question in Chapter 23, ``Diary of a Carefully Worded Paper.''

\brws
This actually surprises me because I somehow got the impression that
you were an operationalist (perhaps because I take Asher Peres to be
an operationalist and you wrote that {\bf Physics Today} article together).
\erws

You were perceptive to notice a difference between Asher and me.  We had to fight that one out.  See Chapter 23.

\brws
Although I have not yet fully thought through your arguments in the
``quantum state of the laser'' paper, it isn't clear to me that the
quantum de Finetti theorem is applicable to the state of a laser, since
the finiteness of any laser's coherence time seems to imply a lack of
infinite exchangeability.
\erws

I know I still need to write Terry, but I've just going bonkers timewise.  People with bigger hammers have been hitting on me.  Phase diffusion has little to do with the issue:  in that paper, we {\it ignored\/} phase diffusion (for the propagating beam) just as Terry and Barry do in their paper.  That of course is an approximation.  Our argument is about a proof of principle:  namely without phase diffusion (as Terry and Barry), there is after-all entanglement in a Kimble kind of set up (in contrast to Terry and Barry):  the {\it distillable\/} entanglement is equal to what was believed to be the case before, but then it was mistakenly identified as {\it pure\/} entanglement.  Terry and Barry's great contribution was in getting us to recognize that.  Taking into account phase diffusion will---as you say---invalidate infinite exchangeability.  But I imagine (actually, that's too weak, I really should say {\it know}) that the flavor of our conclusion will remain intact:  there will still generally be distillable entanglement (on some time scales), it will just be degrading as time goes on.  An interesting, but probably hard calculation, would be to work that out explicitly.  There is already a lot known about partial exchangeability classically:  see the concluding remarks in our de Finetti paper.  Combining that with estimates of coherence times, etc., in the present case is the path that needs to be taken.

I guess I'll CC this to Terry and also Steven van Enk:  chances are strong that I'll be (completely) dropping out of the email scene for a couple of weeks again.  Have a look at the beast below.

\bq\noindent{\tt
\quantph{0105039} [abs, src, ps, other] :\\
Title: Notes on a Paulian Idea: Foundational, Historical, Anecdotal\\ \indent and Forward-Looking Thoughts on the Quantum\\
Authors: Christopher A. Fuchs, (with foreword by) N. David {\Mermin}\\
Comments: 504 pages including introduction, table of contents, and \\ \indent index of names; no figures\\

\noindent This document is the first installment of three in the Cerro Grande \\ Fire Series. It is a collection of letters written to various \\ colleagues, most of whom regularly circuit this archive, including \\ Howard Barnum, Paul Benioff, Charles Bennett, Herbert Bernstein, Doug \\ Bilodeau, Gilles Brassard, Jeffrey Bub, Carlton {\Caves}, Gregory Comer, \\ Robert Griffiths, Adrian Kent, Rolf Landauer, Hideo Mabuchi, David \\ {\Mermin}, David Meyer, Michael Nielsen, Asher Peres, John Preskill, \\ Mary Beth Ruskai, {\Ruediger} {\Schack}, Abner Shimony, William Wootters, \\ Anton Zeilinger,  and many others. The singular thread sewing all the \\ letters together is the quantum. Some of the pieces are my best \\ efforts to date to give  substance to an evanescent thought I see \\ rising from the field of quantum information---I call it the Paulian \\ idea. To the extent I have communicated its misty shadow to my \\ correspondents and seen a twinkle of enthusiasm, it seemed \\ worthwhile to expand the jury on this anniversary occasion. (612kb)}
\eq

\section{10-05-01 \ \ {\it The de Finetti Theorem and the Samizdat}\ \ \ (to A. G. Zajonc \& K. Jagannathan)} \label{Zajonc1} \label{Jagannathan1}

(Could one of you also forward this note to Bob Romer, I don't have his email address.  Could one of you give me his email address?)

I've been meaning to write you for some time to tell you that the last talk I gave at Amherst has finally made it onto paper.  You can find it on {\tt quant-ph} now; I'll put the citation below.  [See ``On Unknown Quantum States:\ The Quantum de Finetti Representation Theorem,'' \arxiv{quant-ph/0104088}.] We think it's a fundamental and pretty nice little result, and does clear up quite a bit for those who want to take the wave function as a state of knowledge, rather than a state of nature.  We tried to make the paper chock full of motivation, so even if you don't read the technical parts I think you'll still get something out of it.  (In fact, for that reason, we submitted it to AJP, so maybe Bob has already seen it.  I don't know whether he sees all the papers that come through there at an early stage of the game or not.)

Also, I put something else on the web today that you saw an extremely early version of (during my first interview trip, I think).  I've been calling it the samizdat.  I think you might enjoy browsing parts of it.  (Actually Arthur makes a couple of appearances in there.)  David Mermin wrote a very nice foreword for it.  I have this dream (hope) that it'll somehow have an impact on the community, but it {\it is\/} probably too big for that. [``Notes on a Paulian Idea:  Foundational, Historical, Anecdotal \& Forward-Looking Thoughts on the Quantum (Selected Correspondence),'' \arxiv{quant-ph/0105039}.]

My wife and I very much want to spend a few days at the Bennetts' house this spring.  I'll try to drop by the college when I'm in town and see what you fellows are up to.

\section{10-05-01 \ \ {\it No Tests:\  I'm Beat!}\ \ \ (to A. Cabello)} \label{Cabello1}

\bac
More \LaTeX\ typos for the next edition: Garc\'{\i}a, \'Alvaro (if it is a Spanish name).

I have sent about 10 e-mails recommending your work. It is great!!!
\eac

Thanks for the word of encouragement!  And also thanks for already helping me catch some typos:  I didn't realize that I should \TeX\ your name as {\Adan}.  I will fix that in the next edition.

Anyway, the main goal here is to try to Copenhagenize people, but with a modern twist.  I think accepting Copenhagen-like ideas is the first step in a great journey---not an end to the road---and that's what I'm trying to turn people on to.  Thanks for showing that people are really starting to read the thing!

\section{11-05-01 \ \ {\it Tot Dinsdag}\ \ \ (to S. J. van {\Enk})} \label{vanEnk21}

\bsve
Congratulations on your submission that surely will make you even more
famous!
\esve
You made a spelling mistake:  you misspelled infamous.

Robert Garisto, that assistant editor I know at PRL, wrote me that it is the ``Many Letters Interpretation of Quantum Mechanics.''  (Actually, with hindsight now, I wish I would have written him a letter before/during our submission.)

\section{11-05-01 \ \ {\it No Worry, Just No Time}\ \ \ (to A. S. Holevo)} \label{Holevo3}

I am sorry that my silence has alarmed you.  I did not recognize that your request for a fax was an urgent matter.  Things have just been very hectic for me in trying to get my large samizdat onto {\tt quant-ph} in time for the anniversary of the Los Alamos fire.  I have only been sleeping about 3--4 hours per night for the last two weeks, while trying to get it compiled.  (It is a very big manuscript, 504 pages, so you should wait until I can give you a bound copy before attempting to read very much of it.  However, you might enjoy reading David Mermin's foreword to it from your screen.  I have never been so flattered.  It is \quantph{0105039}.)

On top of that though, NATO has been putting much pressure on to complete my paper for them.  (And also withholding \$1,000 of my reimbursement unless I get it submitted.)  So, I have had to multiplex in that way too.

So, there is nothing to worry about, I have just been very strained.

\section{11-05-01 \ \ {\it Shannon Meets Bohr}  \ \ (to P. Pearle)} \label{Pearle2}

I put a large samizdat about quantum foundation things on {\tt quant-ph} yesterday.  (Mermin wrote the foreword to it.)  Parts of it, I think, you might enjoy.  Other parts, though, are certainly going to irritate you.  (So I apologize in advance.)  In connection to your own point of view about quantum mechanics, you might have a read of the ``Jeffrey Bub Chapter'' and branch out from there.

\section{11-05-01 \ \ {\it Sigh of Relief}\ \ \ (to B. W. Schumacher)} \label{Schumacher2}

Great to hear you're definitely coming.  (It seems the only definite dropout will be Brassard; Preskill is still in superposition.)  And great to hear your term is coming to an end:  Maybe you'll have some time for some leisure reading!  My big samizdat appeared on {\tt quant-ph} yesterday (with a foreword by Mermin).  You and Mike Westmoreland might enjoy perusing some of it.  I'll put the abstract below.

\subsection{Ben's Reply}

\bq
Mike and I did notice your samizdat, though I haven't had time to read it.
Sounds like great incendiery fun --- which is perhaps appropriate, given
the title \ldots \medskip

\noindent Your fellow bomb-thrower,
\eq

\section{12-05-01 \ \ {\it The Observer in the Quantum Experiment}  \ \ (to B. Rosenblum \& F. Kuttner)} \label{Rosenblum1} \label{Kuttner1}

Last night I read your recent paper on {\tt quant-ph} with the same title as above.  [See Rosenblum and Kuttner's \arxiv{quant-ph/0011086}.] It was enjoyable reading, and I am in substantial agreement with significant parts of it.

I did however get the impression (from some of your choices of words near my citation and at the end), that you would not have guessed that I would say such a thing.  Quite the opposite!  That is what my paper with Asher Peres was about, albeit it in a somewhat subdued way and despite its provocative title.  For, though we did write,
\bq\noindent
Contrary to those desires, quantum theory does not
describe physical reality.  What it does is provide an algorithm
for computing probabilities for the macroscopic events
(``detector clicks'') that are the consequences of our
experimental interventions.
\eq
the word ``interventions'' is prominent there.  It is there precisely to refer to the paragraph previous to it where we write:
\bq\noindent
To begin, let us examine the role of experiment in science. An
experiment is an active intervention into the course of nature:
We set up this or that experiment to see how nature reacts.
\eq
The word ``nature'' is a stand-in for all the ``stuff'' of the world.  The issue is whether the theory gets at a free-standing reality rather than only our interface with nature.  I opt for the latter:  i.e., that the theory is indicating to us that there simply is no free-standing reality (at least now that we are on the scene).

I believe we say this more clearly in our follow-on paper C.~A. Fuchs and A. Peres, ``Quantum Theory -- Interpretation, Formulation, Inspiration:\ Fuchs and Peres Reply,'' {\sl Physics Today\/} {\bf 53}(9), pp.\ 14, 90 (2000).

In any case, I say these kinds of things from many, many different angles in my recently posted samizdat.  Given your paper, I think you might enjoy perusing it a bit.  It emphasizes many of the same points as you.  Maybe the first places to look are Chapter 14, a letter to Bob Griffiths, and also the beginning of the {\Montreal} problem set starting on page 85.  I'll place the full reference below.

\section{13-05-01 \ \ {\it Early Morning Sneaking}  \ \ (to N. D. {\Mermin})} \label{Mermin7}

Kiki's still asleep, so I snuck in here for a while.  It looks like I
can view this version just fine.  I presume therefore I can print it
out too, but that will have to wait until tomorrow when I get it to a
printer.

\bdm
The way you state it would lead most readers to think that it holds
only for special states of the composite system.
\edm

Yes, you're probably right about that.  But I certainly meant your
version of it rather than the misreading of it.

\bdm
I've never seen an argument that it's implied by 3, and am glad to
see that you too state 5 as an independent assumption. Does anybody
else?  Or has anybody showed that it is implied by 3? Or produced a
counterexample that satisfies 3 but not 5?
\edm

It certainly cannot be derived from 3 as I have it stated.  That only
addresses probabilities, not state changes.  5 only addresses state
changes, not probabilities.  (In this version of Gleason's theorem
that I'm writing up, there is a slight connection between the
probability rule and the form of quantum state changes, but that's a
pretty advanced matter that need not enter this discussion:  in
particular because it views quantum mechanical measurement as {\it
nothing more\/} than a state-of-knowledge change.)

Does anybody else take 5 as an independent assumption?  Now, that you
mention it, I guess I never have seen anyone else state it that way.
I do remember Bill Wootters in the audience at that meeting, combing
his beard with his fingers, and looking intent as if the talk were
actually interesting (remember it was a tutorial on basic quantum
mechanics).  That's the only place I've ever presented that talk
where there was someone who knew quantum mechanics in the audience.
But I don't remember ever having this conversation with anyone else
but you:  so there may well be a similar development out there, but I
haven't seen it.

Kiki should be up any minute now:  I'd better get going.

\section{14-05-01 \ \ {\it Today's Meeting Canceled -- Future of Quantum Information Processing}  \ \ (to N. D. {\Mermin})} \label{Mermin8}

The easiest way to answer is below:  I am so relieved!  (Arun had to have his gall bladder taken out.  In his own way he's probably relieved too.)  BTW:  do you know any of the others in the list?

Among other things, I found out last night---middle of the night---that there's been a load of work done on quantum cryptography over optical fiber networks (by Paul Townsend and the like at BT, now owned by Corning, our competitor).  If I can swing it all, these are papers I need to digest (just a little at least) before meeting the big cheese.

Thanks for the developing interest in my life:  it certainly helped the samizdat out.

Whose Knowledge?  (I saw the title of your {\Vaxjo} talk.)  My answer:  anyone
old enough to have a driver's license and write down a density
operator. (Actually, one probably doesn't need a driver's license,
but it might be safer that way.)

\section{15-05-01 \ \ {\it Many Letters $+$ One More} \ \ (to R. Garisto)} \label{Garisto1}

\brg
I read a few pages from your intriguing e-book and thought I'd send you a
response.  Although not tied to it, as you know I have some fondness for
the Many-Worlds Interpretation (MWI).  Although we will probably never
agree about the MWI, I will say a few words in its defense anyway.

First of all, to be glib, and to make a point, I will refer to your
e-book as the ``Many Letters Interpretation''.  It is a beautiful and
complex mixture of information, but it is essentially impossible to follow
without a table of contents or index.  And like all compilations,
superpositions if you will, it makes sense when one uses the index to
apply culling conditions on it, say, all pages which mention me (I note I
am equal to Aristotle {\rm \smiley}).

My point is that the MWI says that the world is a beautiful mosaic which
makes no sense until one imposes the human viewpoint of a classical world.
Once one applies the culling condition ``all parts of the quantum which are
consistent with me existing and having had my last thought'', the mosaic
falls into focus and the world is conceivable to the human mind.

Is this a slight to art, poetry, or free will?  I say that it is not.
There are certainly exponentially small areas of the quantum (notice how I
am avoiding assuming that this quantum info is actually embodied (rather
than simply described) by ``the wave function''?  I should surely get
brownie points for that {\rm \smiley}) \ldots
\erg

You certainly do.  And thanks so much for the appellation ``Many Letters Interpretation'' --- I like it!  (I have a feeling this won't be the last I hear of it.)  If there is a Chris Fuchs in another world who sways more toward MWI, I hope you meet him eventually.  Then maybe the two of you could confront me in a way that might just make some headway!  (My wife tells me I only listen to myself.) \medskip

\noindent With many thoughts (to match my many letters),

\section{15-05-01 \ \ {\it Many Letters $+$ One More, 2} \ \ (to R. Garisto)} \label{Garisto2}

\brg
Nah, it would be no fun if you agreed with me on everything.
\erg

In fact, it's necessary for the very notion of reality itself.  Here's how I just wrote it for an article that I'm writing (in an attempt to distill the samizdat to a paper-sized statement).
\bq
\noindent\bq\noindent
[S]urely, the existence of [the] world is the primary experimental fact of all, without which there would be no point to physics or any other science; and for which we all receive new evidence every waking minute of our lives.  This direct evidence of our senses is vastly more cogent than are any of the deviously indirect experiments that are cited as evidence for the Copenhagen interpretation.

\hspace*{\fill} --- {\it E.~T. Jaynes, 1986}\medskip
\eq
\bq\noindent
The criticism of the Copenhagen interpretation of the quantum theory rests quite generally on the anxiety that, with this interpretation, the concept of ``objective reality'' which forms the basis of classical physics might be driven out of physics. \ldots\ [T]his anxiety is groundless \ldots\  At this point we realize the simple fact that natural science is not Nature itself but a part of the relation between Man and Nature, and therefore is dependent on Man.\medskip

\hspace*{\fill} --- {\it Werner Heisenberg, 1955}
\eq

There are great rewards in being a new parent.  Not least of all is the opportunity to have a close-up look at a mind in formation. I have been watching my two year old daughter learn things at a fantastic rate, and though there have been untold numbers of lessons for her, there have also been a smidgen for me.  For instance, I am just starting to see her come to grips with the idea that there is a world independent of her desires.  What strikes me is the contrast between this and the concomitant gain in confidence I see grow in her everyday that there are aspects of existence she actually {\it can\/} control.  The two go hand in hand.  She pushes on the world, and sometimes it gives in a way that she has learned to predict, and then sometimes it pushes back in a way she has not foreseen (and may never be able to). If she could manipulate the world to the complete desires of her will, I am quite sure, there would be little difference between wake and dream.

But the main point is that she learns from her forays into the world.  In my more cynical moments, I find myself thinking, ``How can she think that she's learned anything at all?  She has no theory of measurement.  She leaves measurement completely undefined.  How can she have any true stake to knowledge?''

Hideo Mabuchi once told me, ``The quantum measurement problem refers to a set of people.''  And though that is a bit harsh, maybe it also contains a bit of the truth.  With the physics community making use of theories that tend to last between 100 and 300 years, we are apt to forget that scientific views of the world are built from the top
down, not from the bottom up.   The experiment is the basis of all
that we know to be firm.  But an experiment is an active intervention into the course of nature on the part of the experimenter; it is not contemplation of nature from afar. We set up this or that experiment to see how nature reacts. It is the conjunction of myriads of such interventions and their consequences that we record in our data books.

We tell ourselves that we have learned something new when we can distill from the data a compact description of all that was seen, and, more tellingly, when we can dream up further experiments to corroborate that description. This much can never change or science ceases to be science. It is the minimal requirement of what we shoot for in science.  If, however, from such a description we can {\it further\/} distill a model of a free-standing ``reality''
independent of our interventions, then so much the better.  I have no bone to pick with reality.  It is the most solid thing we can expect from a theory.  Classical physics is the ultimate example in this regard.  It gives us a compact description, but it gives us more.

However, there is no logical necessity that this worldview always be obtainable. If the world is such that we can never identify a reality independent of our experimental interventions, then we must be prepared for that too.
\eq

\section{16-05-01 \ \ {\it Reference} \ \ (to P. Busch)} \label{Busch1}

Can you give me the complete reference for your paper, ``Resurrection of von Neumann's No-Hidden-Variables Theorem'' (including page numbers)?

I've been meaning to write you for some time to tell you that Joe Renes, Kiran Manne, Carlton Caves, and I came across essentially the same proof of a ``a Gleason-like theorem for POVMs'' as you did.  We were getting all set to write it up as a Phys.\ Rev.\ Lett., when I leaked some word of it to David Mermin.  He remembered your paper and pointed me toward it \ldots\  and sure enough!  So, we never published it.  (And I think it shattered poor Joe's research life.)  But now I'm going to review it in a longish expository piece, so I want to get the reference right.

I'll put below the paragraph I'm just working on where the citation makes its first appearance.  I hope you're OK with the wording of that.  I just wanted to make sure Joe recovers even some small something for his effort.

I hope I'll see you at the Belfast meeting in September.

\bq
The structure of the remainder of the paper is as follows.  In Section 3 ``Why Information?,'' I reiterate the cleanest argument I know of that the quantum state is solely an expression of information---the information one has about a quantum system.  It has no objective reality in and of itself.  The argument is then refined by considering the phenomenon of quantum teleportation.  In Section 4 ``Information About What?,'' I tackle this common question head on.  The answer is, ``nothing more than the potential consequences of our experimental interventions into nature.'' Once freed from the notion that quantum measurement ought to be about revealing traces of some preexisting property (or beable, one finds no particular reason to take the standard account of measurement (in terms of complete sets of orthogonal projection operators) as a basic notion. Indeed quantum information theory, with its emphasis on the utility of generalized measurements or positive operator-valued measures (POVMs), suggests one should take those entities as the basic notion instead.  The productivity of this point of view is demonstrated by the beautifully simple (Gleason-like) derivation of the quantum probability rule found recently by Paul Busch and, independently, by Joseph Renes and collaborators.  In Section 5 \ldots
\eq

\section{16-05-01 \ \ {\it Principle Theory} \ \ (to J. Bub)} \label{Bub2}

Can you give me the full citation of your paper ``Quantum Mechanics as a Principle Theory?''  (Including starting and ending pages.)

Thanks.  I'm citing that paper for the question ``Information About What?''\ in regard to my view that the quantum state is solely subjective information.  (I'm trying to distill the essence of the samizdat into about a 40 page paper:  I'm about 25 pages toward it now.  I'm really hoping it's finished long before {\Vaxjo}.  But I write so damned slowly!)

\section{16-05-01 \ \ {\it Amicable Banter II} \ \ (to R. Jozsa)} \label{Jozsa4}

\brj
Thanks for your message --- I would not object to that inclusion into
your volume of correspondence (but I wonder if our amicable banter
would really be of value/interest to a wider readership!?\ldots).
\erj

More seriously, I've found that the thing (in scaled-down preliminary versions) turned out to be a surprisingly effective means for getting my (forming) point-of-view across.  To see Bub, for instance, make a 180 degree turn even after writing his two books on quantum logic, was the most shocking thing.  Similarly to see the dents I've made in David Mermin, John Preskill, Lucien Hardy, and a few others went a long way toward giving me some courage.   Amicable banter is kind of like a sugar coating that helps the medicine go down.

I keep counting on the day that you'll see I'm not a complete fool, only a partial one (but maybe one not completely worth dismissing).  It may or may not happen:  but you remain my friend and sparring partner as always.

Looking forward to some beers and a little rest in Warsaw!

\section{17-05-01 \ \ {\it Voracious Readers} \ \ (to J. Bub)} \label{Bub3}

I found the information on the reference I needed.  So, no need to reply on that.  But in the process of finding it, I found a new paper of yours that I had never seen.

I'd like to see that!  Can you send it to me?

\bq\noindent
J. Bub, ``Indeterminacy and Entanglement: The Challenge of Quantum Mechanics,'' Brit.\ J. Phil.\ Sci.\ {\bf 51}: 597--615 (2000).\medskip

\noindent Abstract: I explore the nature of the problem generated by the transition from classical to quantum mechanics, and I survey some of the different responses to this problem. I show briefly how recent work on quantum information over the past ten years has led to a shift of focus, in which the puzzling features of quantum mechanics are seen as a resource to be developed rather than a problem to be solved.
\eq

\section{17-05-01 \ \ {\it Pre-Irritation Irritation!}\ \ \ (to P. Pearle)} \label{Pearle3}

\bpp
I shall now get it out and get irritated!  But perhaps not.  I was amused by
your article with Peres in {\bf Physics Today} about quantum theory needing no
interpretation---amused because I have over the years come across this point
of view promulgated in one guise or another many times and I always wonder,
if that is so obvious, why there is a continual necessity, starting with
Bohr (so you are in a great tradition, for polemics against any other point
of view).
\epp

No, that is a very good point, and I am trying to tackle that issue head on in the present paper I'm writing.  Let me send you the very much incomplete, preliminary draft right now.  The first four sections are particularly important for what you say above.  And I just want to show you I've thought about that!

So drop the samizdat for the moment!  And go to this pre-irritation irritation.  I think you'll find I'm not so flaming as you think I might be!

It's attached below as a PostScript file.  If you need it in a different format, let me know.

\section{18-05-01 \ \ {\it Finally!}\ \ \ (to P. W. Shor)} \label{Shor1}

Sasha and I just put together the stuff below for Fred Roberts' request.  Why don't you make any corrections/additions you deem appropriate and then forward it to Fred.  I'm sure we missed some people on the potential invitee list, but maybe this is enough to get started \ldots\ and maybe you can remember some others that ought to be on it.

Sasha and I will both be gone next week.  (I'll be in Warsaw.)

Please CC us whatever you send to Fred.\medskip

\noindent \underline{Title:}\\
Capacities and Coding for Quantum Channels\medskip

\noindent \underline{Abstract:}\\
In 1973 the information theorist J. R. Pierce wrote for the 25th anniversary of Claude Shannon's original paper, ``I think that I have never met a physicist who understood information theory.  I wish that physicists would stop talking about reformulating information theory and give us a general expression for the capacity of a channel with quantum effects taken into account.''  Today, 28 years later, Pierce might still be saying the same thing!  For though there has been an explosion of activity in understanding quantum channel capacities in recent years, Pierce had no feeling for the breadth of the question he asked:  associated with a quantum channel, one can speak of its capacity for sending classical information, its one-way capacity for sending truly quantum information, its two-way quantum capacity, its entanglement-assisted classical capacity, its secrecy capacity, and the list of distinctions is still growing.  This tutorial and workshop will be  devoted to all aspects of quantum channel capacity and coding, from code-construction techniques to bounds on capacities and error exponents.\bigskip

\pagebreak

\noindent \underline{{\it Potential\/} Invited Speakers:}  (NOT FOR PUBLIC VIEW YET)\medskip

\begin{supertabular}{ll}
Rudolf Ahlswede          &    (Bielefeld)                                   \\
Alexei Ashikhmin         &    (Bell Labs)                                   \\
Howard Barnum            &    (LANL)                                        \\
V. P. Belavkin           &    (Nottingham U.)                               \\
Charles H. Bennett       &    (IBM Research)                                \\
Toby Berger              &    (Cornell U.)                                  \\
Gilles Brassard          &    (U. {\Montreal})                              \\
Rob Calderbank           &    (AT\&T Research)                              \\
Nicolas Cerf             &    (Brussels)                                    \\
Thomas Cover             &    (Stanford)                                    \\
Imre Csiszar             &    (Budapest)                                    \\
G. David Forney          &    (MIT)                                         \\
Michael Freedman         &    (Microsoft Research)                          \\
Fujiwara                 &    (Japan)                                       \\
Peter Gacs               &    (Boston U.)                                   \\
Nicolas Gisin            &    (Geneva)                                      \\
Daniel Gottesman         &    (U. California, Berkeley)                     \\
Markus Grassl            &    (U. Karlsruhe)                                \\
Osamu Hirota's group     &    (Tamagawa U.)                                 \\
Alexander Holevo         &    (Steklov Mathematical Inst.\ / Bell Labs)     \\
Richard Hughes           &    (LANL)                                        \\
Richard Jozsa            &    (U. Bristol, UK)                              \\
Louis Kauffman           &    (U. Illinois at Chicago)                      \\
Christopher King         &    (Northeaster U.)                              \\
Alexei Kitaev            &    (Microsoft Research)                          \\
Koashi and/or Imoto      &    (Japan)                                       \\
Emmanuel Knill           &    (LANL)                                        \\
Lev Levitin              &    (Boston U.)                                   \\
Hoi-Kwong Lo             &    (MagiQ Technologies, NY)                      \\
Peter Loeber             &    (Bielefeld, Germany)                          \\
Dominic Mayers           &    (NEC, Princeton)                              \\
Ueli Maurer              &    (ETH, Zurich)                                 \\
Hiroshi Nagaoka          &    (Japan)                                       \\
Prakash Narayan          &    (U. Maryland, College Park)                   \\
Michael A. Nielsen       &    (Queensland)                                  \\
Denes Petz               &    (Budapest)                                    \\
John Preskill            &    (Caltech)                                     \\
Eric Rains               &    (AT\&T Research)                              \\
Mary Beth Ruskai         &    (U. Massachusetts)                            \\
Masahide Sasaki          &    (Communication Research Lab., Tokyo)          \\
Serap Savari / G. Kramer &    (Bell Labs)                                   \\
Benjamin W. Schumacher   &    (Kenyon College)                              \\
Neil Sloane              &    (AT\&T Research)                              \\
John A. Smolin           &    (IBM Research)                                \\
Emina Soljanin           &    (Bell Labs)                                   \\
Andrew Steane            &    (Oxford)                                      \\
Yossef Steinberg         &    (Technion)                                    \\
Barbara M. Terhal        &    (IBM Research)                                \\
Michael D. Westmoreland  &    (Dennison U., Ohio)                           \\
Reinhard Werner          &    (Germany)                                     \\
Andreas Winter           &    (U. Bristol, UK)                              \\
Stefan Wolf              &    (Switzerland)                                 \\
William K. Wootters      &    (Williams College)                            \\
Horace Yuen              &    (Northwestern U.)        \\
\end{supertabular}

\section{19-05-01 \ \ {\it Golden Ages of the Quantum}  \ \ (to E. Merzbacher)} \label{Merzbacher2}

I'm just off to an information theory meeting in Warsaw.  But I'm so happy---and I wanted to tell you---I also just finished the fine book you gave me about the early years of the Bohr Institute.\footnote{Peter Robertson, {\sl The Early Years:\ The Niels Bohr Institute 1921--1930}, foreword by Aage Bohr, (Akademisk Forlag, Copenhagen, 1979).}  (Now I can use the flight to work rather than read.)  Thanks once more for the nice book collection.

Let me also tell you about something I put on the Los Alamos archive last week.  You might be interested in it, as it also documents a little bit of the history of quantum mechanics (both of the old golden age and the new golden age).  David Mermin wrote a foreword for it.  I'll put the abstract below.  [See {\sl Notes on a Paulian Idea: Foundational, Historical, Anecdotal and Forward-Looking Thoughts on the Quantum}, \arxiv{quant-ph/0105039}.]

\section{22-05-01 \ \ {\it  Go Figure} \ \ (to D. J. Bilodeau)} \label{Bilodeau3}

I wouldn't bother with all that.  Use what you need; cite the source in your talk if he happens to be in the audience.  God knows I steal things left and right for my talks.

(In fact I've been using one of Seth Lloyd's slides for about three years just to make fun.  He shows an icy mountain with the words ``novel quantum states'' at the base and ``Shor's factoring algorithm'' at the top.  The idea is to depict the great challenge in front of us.  ``But,'' I say, ``that's not the way I see it.''  ``The reason we're climbing this slope is to get a gauge of how far it is from the top to the bottom.''  Then I switch slides to one of an iceberg.  ``Because it's not a mountain, but the tip of an iceberg.  We're doing this just to get a better feeling for the other 8/9 that's underneath.'')

Relax as much as you can:  I don't think you'll find any citation hungerer's (at least in our side of the meeting).

I'm in Warsaw.  Yesterday, I met two really interesting information theorists from Copenhagen.  I've encouraged them to join us in {\Vaxjo}.  It's just a small train ride for them.

\section{25-05-01 \ \ {\it Quantum Axiomatics}  \ \ (to C. H. {\Bennett})} \label{Bennett1}

\bcb
Have you read the new papers by Aerts et al.\ on repairing
``defective'' axioms of quantum mechanics?  Are they worth reading?
\ecb

That's a very strange question coming from you Charles {\Bennett}.  No,
I haven't looked at that:  I am in Warsaw right now, with very
limited email access and essentially no web access.  I'll look at it
when I'm back in the states next week and get back to you.

You know (or ought to know) that my only trouble with the axioms is
that we have to deal with them at all.  I.e., that we don't seem to
have clear-cut way to pose the theory without invoking them.  (As one
might say that we do when it comes to special relativity.)  My
clearest statement yet on that point is made in the paper attached (I
am in the middle of writing it, so it's not complete).  But read the
first two sections and tell me whether it stirs you at all.  Or does
it still leave you limp?

\section{26-05-01 \ \ {\it The Warsaw Cafe}  \ \ (to N. D. {\Mermin})} \label{Mermin9}

I'm on the last day of my trip to Poland, writing you from a little
place with blue tablecloths and yellow flowers.  Mostly I'm writing
you again to thank you again:  maybe one of these days I'll say
enough is enough, but not yet.  The week had its ups and downs.  The
biggest up was seeing an audience of classical information theorists
light up to the idea of quantum cryptography.  The biggest down was a
collection of conversations with [Professor X] about the samizdat.  It seems to
eat at him that I broke the rules of academia in this way, and that I
would put such rubbish on the net for ``self-glorification.'' One
night I got so depressed, I found myself opening up your foreword and
reading it again.  It did me a world of good, and got me through the
night.  I knew that without you I would not have had the courage to
carry out the project in the first place.  What I didn't know was
that I would need you even after it was completed.  Thanks again from
the bottom of my heart.

The NATO paper has---of course!---gotten delayed again by this little
outing.\footnote{This refers to C.~A. Fuchs, ``Quantum Foundations in
the Light of Quantum Information,'' in {\sl Decoherence and its
Implications in Quantum Computation and Information Transfer:
Proceedings of the NATO Advanced Research Workshop, Mykonos Greece,
June 25--30, 2000}, edited by A.~Gonis and P.~E.~A. Turchi (IOS
Press, Amsterdam, 2001), pp.~38--82. Also \arxiv{quant-ph/0106166}.}  I
know that Gonis views me as one of the most evil men alive.  But if
it's not too arrogant, I think it might turn out pretty good over
all. There is a fine line between trying to shock people into action
and going too far, and I know I'm just playing it by ear.  The
manuscript just reached the 32 page mark. Once I get to the stage of
having the ``collapse rule argument'' completely written down, I'll
send the preliminary version your way: you seemed the most interested
in that part of it.

My off hours here have been devoted to going through Rosenfeld with
the same thoroughness that I gave Folse (and will eventually give
Plotnitsky).  I copied down a little passage for you.  If I'm not
mistaken, it's Rosenfeld's way of {\it trying\/} to say what you were
{\it trying\/} to say in one of the sentences in the Foreword:

\bq
And now comes the last great surprise:  it turns out that in
describing atomic phenomena one particular picture cannot suffice:
according to the circumstances we are obliged to make use of several
quite different pictures; one can only describe the atomic world, as
it were, from one point of view at a time, and the prospects it
offers from various angles are so different that they cannot be fused
into one single picture in the usual sense.

This situation has given atomic physicists occasion for fascinating
reflections upon the very essence of human cognition, which I shall
not enter upon here.  I just wanted to mention it as a concluding
touch in the picture of the atomic physicist that I have attempted to
outline.  I scarcely venture to hope that I have succeeded in giving
an adequate insight into his methods of work, but perhaps I have
managed to persuade you that the secret weapons that have secured him
such brilliant triumphs are persevering toil and common sense.
\eq

Plotnitsky, by the way, wrote me the most wonderful letter comparing
his reading of Bohr to Folse's.  He also included a load of analysis
on {\Pauli} and {\Wheeler}.  You should see the size of it:  it almost
makes my samizdat look like small cakes.  Most surprising though:
it's one of his best pieces of writing I've seen yet!  (I actually
understand a thing or two:  that used to take an actual face-to-face
conversation.)

But now my coffee's gone, and I've got to start contemplating the
taxi ride to the airport.  I'll be in touch again soon.

\section{26-05-01 \ \ {\it Your Correspondence}  \ \ (to Y. J. Ng)} \label{Ng1}

Thanks for the warm reception of my samizdat.  I'm just flying back from a meeting on (predominantly classical) information theory in Poland.  The reception of quantum cryptography was wonderful:  I could really see the enthusiasm for the idea building as the audience started to understand what it is all about.

I hope to write you some comments on your black hole paper (early) this summer.  It is an important issue, and I'm glad you've turned your mind to it.  Sometimes I wonder (but not nearly so strongly as in the quantum case) if one might pose a similar program for general relativity as I did in the beginning page of the samizdat (i.e., the one immediately following Mermin's foreword, with Preskill's question).  That is to say, might one be able to find a natural information theoretic justification for the principle of equivalence, etc.  If one could, then one might immediately find a new point of connection between GR and QM \ldots\ one that may have been passed over before.

\section{26-05-01 \ \ {\it Sitting in Zurich}  \ \ (to E. Merzbacher)} \label{Merzbacher3}

Thank you for your encouraging letter on the samizdat.  The reactions of course have been varied, but on the whole those who are writing me seem to think they might get something out of it.  I truly do want to see all this commotion about quantum mechanics ``not making sense'' disappear before my daughter goes to college.  Life is too short to be confused for too very long.

I'm trying to condense the program of the samizdat into a single paper (of about 40 pages).  It is being written for a NATO meeting I attended last summer whose hidden undercurrent was quantum foundations.  With some luck I'll be able to get it on the web by the end of next week.  I'll put the abstract below (to see if I can whet any more of your interest).  (Beside that though, the paper will have a couple of very basic theorems that might interest you in their own right:  it's time for me to get a little longer on analysis and a little shorter on poetry again.)

It sounds like you've been going to the New York theater much more than I have!  That's a real shame for me, considering I'm only 32 miles away!

\bq
\noindent In this paper, I try to cause some good-natured trouble.  The issue at stake is when will we ever stop burdening the taxpayer with conferences and workshops devoted---explicitly or implicitly---to the quantum foundations?  The suspicion is expressed that no end will be in sight until a means is found to reduce quantum theory to two or three statements of crisp {\it physical\/} (rather than abstract, axiomatic) significance. In this regard, no tool appears to be better calibrated for a direct assault than quantum information theory. Far from being a strained application of the latest fad to a deep-seated problem, this method holds promise precisely because a large part (but not all) of the structure of quantum theory has always concerned information.  It is just that the physics community has somehow forgotten this.
\eq

\section{26-05-01 \ \ {\it The Wallpole of Wisconsin} \ \ (to G. L. Comer)} \label{Comer4}

Well the week came and went and I got no new letter off to you.  It's so very depressing how little I get done.  I've got 4 hours and 20 minutes left in this flight and I'm absolutely miserable.  I'm flying SwissAir this time, rather than my usual carrier American, and I feel like a complete sardine.  Plus they have no outlets for me to plug my laptop into.  Remind me to never fly anyone but American!

By the way, what the heck is a wallpole?  It seems like it's a word, but I can't find it in my dictionary.  [\ldots]

So, there, I think I've spilled all my secrets.  In the mean time, Bell Labs treats me very well.  My boss even endorsed the samizdat.  But I know that things of that flavor cannot last forever.  Philosophy is rarely profitable.

By the way, just coming back from Poland, I should tell you that I learned that the word ``zurek'' means ``white sausage.''  I had a little zurek soup one night in fact.  Poland was quite a surprise to me in several ways. [\ldots]

This meeting was a rather unique one.  It gathered researchers from biology, economics, quantum, and just pure information theory.  The theme, of course, was simply using information theory in one's work.  I gave two lectures.  One with transparencies, and one at the chalkboard.  They both went very, very well.  I don't think I've had a reception like that in a while.  (Maybe not since the colloquium at UNC, in fact.)

I kind of wish I had some deep thought to tell you.  But there's nothing in the old noggin right now.  One day I hope you'll tell me an information theoretic reason for the equivalence principle, or tell me why one cannot be given.  We have discussed this before, haven't we?  One of these days, I will come back to GR, you know.  It may be 20 years from now, but I really think I will.  Maybe when we meet up in the retirement home we'll finally start a collaboration!

OK, I'd better go.  My battery is getting awfully low.  And I've got a few ``official'' things that still need doing before the flight is over.

\section{26-05-01 \ \ {\it The Activating Observer on Hold}  \ \ (to B. Rosenblum \& F. Kuttner)} \label{Rosenblum2} \label{Kuttner2}

Thanks for your nice letter.  I hope you will be patient with my reply.  In fact I am putting a paper on the web probably at the end of the week (or at the end of the following week, latest) that I believe will answer your questions.  If it still does not, then we can pick it up from there.

Things have been immensely busy for me lately and I've had to cut my email drastically.  I'm writing to you as I'm flying back from Warsaw.  (Most of my email is in fact limited to what I can do on a plane now, for the time being at least.)

Thank you for the invitation to speak at your university.  I am in California from time to time (to visit Caltech etc.).  I'll try to give you a heads-up when the next trip goes into the planning stage.

\subsection{Bruce and Fred's Preply}

\bq
Thanks for your email and the friendly comments.  You're right, we
would not have guessed that you would be in substantial agreement
with significant parts of our argument.   But in rereading your
opinion piece with Peres and the other stuff you referred us to, we
can see how we are not---necessarily at least---far apart.

We too might agree that ``quantum theory needs no `interpretation'.''
After all, the interpretations of quantum theory arise because of the
measurement problem, which can be seen as the apparent intrusion of
the observer.  Our main thesis is that this apparent intrusion arises
in the quantum experiment, logically prior to the quantum theory.  If
so, no mere interpretation of the theory can ever resolve this
problem.

I'm puzzled though.  Is the position you advocate (a version of) the
Copenhagen interpretation?  Your opinion piece ends with implying
that all that quantum mechanics need be is a useful guide to the
phenomena around us.  That sounds much like Bohr's ``Science is not
about nature but only about what we can say about nature,'' which more
or less seems to capture the Copenhagen spirit.  In a sense, of
course, the Copenhagen interpretation is not really an
interpretation.  (Peierls objects to the term ``interpretation''
because, according to him, Copenhagen is the only way to understand
quantum mechanics.)  Might, however, what you advocate be seen as
close to Mermin's extreme rendering of Copenhagen: ``Shut up and
calculate!''?

You do say that you would be the last to claim that the foundations
of quantum theory are not worth further scrutiny.  But then the
example you give (a search for minimal sets of physical assumptions)
seems to advocate a rather narrow range of scrutiny.  Reading other
things you have written---including your comment on our piece---we'd
have thought that you would welcome more far-reaching investigations.
Are we wrong?

While we don't think mere interpretation can resolve the measurement
problem, we do believe there likely is a real problem---that Nature may
be trying to tell us something we can't yet understand.  And,
moreover, interpretations may have a role in addressing it.

As we see it, the role of interpretation can be two-fold:
Interpretations can suggest new physical phenomena to address the
measurement problem.  For example: Bohm's suggested deeper levels and
implicate order; interpreting collapse to be a physical phenomenon
motivates the GRW and Penrose speculations; many worlds has suggested
non-linear terms connecting worlds; and extending von Neumann, Stapp
has a physical theory of consciousness.

Interpretations, on a yet grander scale, may suggest new world-views.
The presently dominant Newtonian world-view, still the basis of much
of our thinking and attitudes, is, of course, fundamentally flawed.
Insignificant for all practical purposes, perhaps, but world-views go
beyond practical purposes.  This may all be well outside the realm of
physics.  But physics may be encountering something of great concern
beyond physics---as it has in the past.

You say your opinion piece was motivated by a concern that the airing
of various interpretations of quantum mechanics ``may lead some
readers to a distorted view of the validity of standard quantum
mechanics.''  If you mean the readers of articles in {\sl Physics Today},
AJP, and the physics research journals, that is not a concern that
would worry us.  Most of our colleagues are completely immunized
against that problem by their courses in quantum mechanics and by
their continued work with the discipline.

We'd have two different concerns.  The first is that by ignoring the
enigma that Nature presents us with, physics may just miss something
truly important---albeit, perhaps, something beyond what is normally
regarded as physics.  (And we, personally, have the feeling that by
concentrating solely on interpretations of the quantum theory,
physics may indeed be missing something.)

A second concern is that the secret that physics has a problem is
out.  There is no way to honestly present the measurement problem to
someone not brainwashed by a serious course in quantum mechanics in a
way that doesn't leave them feeling that some mysterious spookiness
is involved.  To the extent that physicists don't develop these ideas
among themselves and present them properly to the interested wider
public, others less knowledgeable---or, worse, badly motivated---will
become even more the major communicators than they now are.

After all this talk about the measurement problem, we feel like
noting that when we teach quantum mechanics we spend almost zero time
on the enigma.  Aside from some, likely missed, offhand comments,
nearly all of our own students are probably as brainwashed as most.
They concentrate on what will be on the exams (and I can't blame
them).

To some good extent we feel that you generally agree with much of
what we say here.  Reading some of the very interesting stuff you
directed us to, it seems you take the issue quite seriously.  (Your
``daytime'' publications also look very interesting.)

Which brings me (BR) to ask if you are likely to be in the California
area during the next academic year.  If so, would you be able to come
to Santa Cruz for a colloquium? (Unfortunately, our limited
colloquium budget permits few invitations from the east coast.)  We'd
love to hear from you.
\eq

\section{27-05-01 \ \ {\it Mohrhoff}  \ \ (to A. Peres)} \label{Peres14}

\bap
Have a look at \quantph{0105097}. You'll probably enjoy it.
\eap

I am home finally! And, I have now seen what you were talking about.  You were right, it looks like it might be interesting.  (I only read the first page quickly.)  I remember that Mohrhoff impressed David Mermin in the past, but the last time I saw one of his papers I couldn't get through it.  Have you read this one?

Do you think people will one day write such mean things about my samizdat?  (I'm starting to get frightened that they might!)  I've now, by the way, received 19 responses to the samizdat's posting.  Only two of them have been overtly negative.  Yesterday, on the other hand, I found a very flattering one from Eugen Merzbacher.  Things like that keep my spirits high, and help build the belief that I may have done something worthwhile.

I read your paper with Danny yesterday on the flight.  I will formulate a response to it tomorrow.  As for the rest of today, I am going to spend some time with my family; they will be waking up soon.  I think I will try to find an IHOP to take them out for a pancake breakfast.

\section{28-05-01 \ \ {\it The Morristown Coffee Tank}  \ \ (to N. D. {\Mermin})} \label{Mermin10}

Thanks for the cheering letter.  I'm back in Morristown now, tanking
up for the morning.  No, I haven't seen the 't Hooft thing.

\bdm
But if it happened to get two readers who appreciated its literary
merits, do you think its subsequent appearance in PRL would make me a
laughing stock in the quantum-info community?  This worries me.
Please advise.
\edm

Yes, I can see that that might be a worry.

Here's how I opened my first talk in Warsaw the other day:
\bq
Well I'm going to do something unexpected; I'm going to change the
category of my talk from a research talk to an expository talk.  I
hope the organizers will forgive me; it just seemed I could make the
talk more relevant that way. \ldots\ You know, I referee a lot of
papers. [Someone in the audience yelled, ``You're the one?!?'']  And
for a while, whenever I thought a paper's content was skimpy, I would
describe it as a ``didactic'' paper in my anonymous referee reports.
The implication was that the paper was merely instructive, rather
than of research value.  But that led me to a little trouble, because
I sometimes write my colleagues directly to insult their papers.  And
I found that I was using the word didactic in my emails too!  You can
guess I was eventually caught! \ldots\  Anyway, today's talk will be
a didactic one, and I have one goal in mind \ldots
\eq

Let me withhold judgment until I see the revamped version.  PRL has
been publishing so much crap lately, and turning down some rather
good pieces---like Paul Busch's Gleason-like derivation---that they
might do well to publish a good didactic one.  And also, you might
convince me that it's more than that with this new prose.

\section{28-05-01 \ \ {\it Card-Carrying Greens}  \ \ (to C. H. {\Bennett})} \label{Bennett2}

I'm back in Morristown now.  Thanks for the (unusually long) note.  I
can see that I continue to leave you limp!

\bcb
Were you a Green?  Did you really believe there was no difference
between Gore and Bush?
\ecb
No, I wasn't a Green.  And, I didn't believe that there would be no
difference in the {\it consequence\/} of having Gore versus Bush in
the White House.

But I do believe it of the quantum campaigns.  And I do believe the
reason there continues to be money in quantum foundations research is
because the arguments all center around almost-invisible variations
of the same thing.

\bcb
Nice introduction and motivation of the problem, especially the math
vs physics versions of special relativity.  However, in your table on
Quantum Axioms and Imperatives, the one with the redundant right
column, it seems to me you beg the question with ``give an
information theoretic reason''.
\ecb
If I'm on the right track, the two columns really will be redundant.
So, that's a good sign.

\bcb
Isn't information theory, no less than linear algebra, a branch of
mathematics, rather than a branch of physics?
\ecb
But rarely is a branch of mathematics laid down without a more fuzzy
kind of motivation working in the background.  The fuzziness comes
first, the formalization comes next.  But the meaning and the value
of the final construct---i.e., the final formal system---remains in
the fuzziness.  What I'm really asking us to do in that paper is to
lay open the motivation for why we have the particular structure we
do in quantum mechanics.  If I had to place a bet, it would be that
the answer will be found along information-theoretic or
decision-theoretic lines.  You undersell those fields in not
realizing that they have a significantly firmer grip on their origin
than quantum mechanics does.  (Remember {\Ruediger}'s nice presentation
of the ``dutch book'' argument for the probability calculus in
{\Montreal} last year?  As I recall, you appeared, at least mildly,
impressed. That is the flavor of things I'm hoping for for quantum
mechanics.)

\bcb
And when you ask for a physical explanation, what do you mean by
``physical''?  It's like pornography---you can't define it but you
know it when you see it.
\ecb
I know you meant to be sarcastic with this, but, roughly, yes.  Yes,
we will know it when we see it:  the nagging, nasty feeling that
something is missing from our worldview will disappear (at least in
this avenue of science).  Funding for quantum foundations meetings
will---by community choice---dry up.  One could argue that one's pet
interpretation---like the Everett program, for instance---is already
virtually there \ldots\ if the rest of the community would just stop
being stubborn!  But I just don't think the rest of the community
will stop being stubborn until the essence of the theory can be
taught to a junior high student.  (Your Everett interpretation is not
there, except in the trivial sense of saying ``all possibilities
equally exist'' \ldots\ which one could have just as well have said
of classical physics, asserting that every initial condition is
equally valid.)

\bcb
You could say that the special relativity example, despite its patent
physicality, is really a geometric axiom asserting that spacetime
behaves like a Minkowski space rather than a Euclidean one. So how
much worse is it to have a bunch of axioms saying that states behave
like rays in a Hilbert space?  To me quantum mechanics is a more
essentially mathematical and less physical part of physics than
special relativity is. In that regard it is more like thermodynamics,
which concerns the macroscopic consequences of microscopic
reversibility, and would apply in any world, for example a 2
dimensional classical-mechanical world, a discrete cellular automaton
world, or a 5-dimensional special-relativistic world, as long as the
underlying dynamics was reversible.
\ecb

You will find little disagreement with me about part of this:  i.e.,
that quantum theory shares a lot with thermodynamics in its range of
applicability.  But that is precisely why I would call quantum
mechanics more a ``law of thought'' than a ``law of nature.''  Just
as Boole did of probability theory.  And just as Jaynes did with
(classical) statistical mechanics.  (In fact, either of you
two---i.e., you or Jaynes---might just as well have written your
quote above.)

You might say, and I suspect you will say, the distinction I draw
between ``nature'' and ``thought'' is only a semantic one.  But I
don't believe that:  I think it is precisely in making that
distinction clear (and operational) that we have a chance of closing
the quantum foundations debate and moving on.

Physical theory is about two things:  what is and what we know (or
what we believe).  It's the process of putting the two things
together that gives a prediction in any practical setting.  Quantum
theory, interestingly, seems to be a nontrivial jumble of those two
things.  I think that is a rather deep statement about the world, and
one that we have not yet come to grips with.

I hope you'll read the whole paper from beginning to end when I
finally post it.  You know that I crave your respect.  But I know
that I have to earn it.

\section{28-05-01 \ \ {\it Letters:\ the Long and the Short} \ \ (to A. Plotnitsky)} \label{Plotnitsky3}

Thank you so much for the amazing letter!  I haven't read the whole thing yet, but I've read enough to know that definitely going to get something out of it.  Very impressive.  When I get in to the office tomorrow, the first thing I'm going to do is print it out so I can read it properly.

Let me also alert you to something I posted on quant-ph a couple of weeks ago.  The citation is below.  [See {\sl Notes on a Paulian Idea: Foundational, Historical, Anecdotal and Forward-Looking Thoughts on the Quantum}, \arxiv{quant-ph/0105039}.]  It's the thing I've been calling the samizdat:  I don't know whether I've mentioned it to you.  You might enjoy parts of it.

\subsection{Arkady's Preply}

\bq
First of all, I apologize for being out of touch since receiving the latest version of your ``Compendium,'' but a busy end-of-semester schedule and too many pending (and some past-all-deadline) commitments have intervened.  I have managed, however, to finish reading the expanded version of the ``Compendium'' in the meantime, which, a bit slow as it was (in part because I also used the occasion to visit some of the things you cite), actually provided a kind of refuge.  I have by now a rather extended set of notes that nearly amounts to a small ``compendium'' of its own.  I have, however, distilled a kind of commentary out of them over the last few days, which I thought I would pass on to you.  (At your request, I also append in the end a few typos and other small glitches that I have noticed, although, as I was paying much more attention to the content, I could have easily missed others.)

This commentary started as a much shorter set of remarks on some of Henry Folse's work, which your cite extensively and which especially attracted my attention, since of the works that you mention it deals with Bohr most extensively and since I am already familiar with.  (By contrast, I was not familiar with D. J. Bilodeau's work, which you cite extensively as well and which I found, I take it in agreement with you, to be among the better commentaries on most subjects he addresses.)  My thought, however, quickly moved to more general concerns, prompted by other things (many of them quite extraordinary) that you cite and by my own thinking about the differences among Bohr's, Pauli's, and Wheeler's views (with some of Heisenberg's ideas, my current special interest, added to the mix), and developed into the argument to follow.  I shall, nevertheless, start with some comments on Folse's work, as a convenient entrance to this argument.  For the sake of convenience, I shall mostly restrict myself to citing your ``Compendium'' (hereafter referred to as ``Fuchs'').

Folse is among the more assiduous readers of Bohr, and his more recent readings seem to me to improve on his 1985 book, {\sl The Philosophy of Niels Bohr}, which I read when working on {\sl Complementarity\/} and on which I comment there, along with Dugald Murdoch's and Jan Faye's readings of Bohr.  I must especially commend Folse on commenting (he is one of the very few who has done so) on the significance of Bohr's new conception of atomicity, specifically in his ``Niels Bohr's Concept of Reality,'' in Pekka Lahti and Peter Mittelstaedt's {\sl Symposium on the Foundations of Modern Physics 1987:\ The Copenhagen Interpretation 60 Years after the Como Lecture\/} (Singapore: World Scientific, 1987), which you mentioned in one of your emails and which you cite in your ``Compendium'' (``Fuchs,'' \#135, p.\ 27).   Folse's interpretation of this conception and, in part correlatively, of Bohr's epistemology as a whole in this article and elsewhere is, however, quite different from the one that I have tried to pursue in my own recent articles.  I have no quarrel with Folse's insistence on Bohr's ``realism''; and he has good reasons for this argument, given some arguments against Bohr.  The question is what form this ``realism'' takes and the areas and limits within which such concepts can apply to Bohr's view.

In my view, while there is a place for some realism in Bohr's interpretation of quantum mechanics, I cannot think of any concept of reality that could, in my interpretation of his interpretation, apply at the ultimate (quantum) level of description.  Nor, accordingly, can I think of any form of realism attributable to Bohr's view of nature at this level, except of course that he believed that what we call ``quantum objects'' exist.  How could one deny this, given that no concept of reality applies to such ``objects''?  To what would such concepts be inapplicable, then?  And what would be responsible for the effects on the basis of which one would argue for this inapplicability?

Folse does not quite see Bohr's epistemology in terms of classical (knowable) effects, manifest in measuring instruments (or in other macro-phenomena) of nonclassical (unknowable) efficacities, defined by the interaction between quantum objects and measuring instruments, even though he (rightly) stresses the significance of this interaction, specifically for Bohr's concept of phenomenon.  The lack of this or (one can think of different ways of handling the situation) a comparable epistemology sometimes prevents him from following Bohr's key concepts (such as ``phenomenon'' or ``atomicity'') and arguments to their limit.  It also leads him to certain critical arguments against Bohr that are in fact, or in effect, answered by Bohr, for example, those concerning the identity of atomic systems (``Fuchs,'' \#146, p.\  31; \#149, p.\  33) or ``properties'' (``Fuchs,'' \#149, p.\  33).

On the first point, concerning the identity of atomic systems, I would argue as follows.  One can have, and, in my view, Bohr does, a perfectly logically and data-wise consistent argument, whereby all individuality, identification (say, of electrons) and so forth are defined in terms of possible effects upon measuring instruments, without appealing to the properties or identity of quantum objects themselves.  Bohr, incidentally, never says and, in my view, would not say, as Folse does, in reading him, that ``complementary `wave' and `particle' phenomena are {\it complementary phenomenal appearances of the same atomic system}'' (``Fuchs,'' \#146, p.\  31; Folse's emphasis).  Both types of appearances are, of course, possible, but at least two experiments, and hence two different systems, rather than ``the same system,'' would have to be involved in order to obtain two complementary phenomena.  I leave aside, for the moment, further  qualifications as concerns the possibility, or, as the case may be, impossibility, of a ``wave'' appearance of any given atomic system (defined, again, in terms of effects or phenomena, by a given initial detection or/as preparation).  This may appear a bit too pedantic, but it may be shown that these nuances have major epistemological implications, for example, in the EPR case, implications sometimes missed by Folse.

On the second point concerning ``properties,'' Folse is not wrong in saying that, in interpreting quantum physics, one cannot avoid speaking of ``properties'' at least at some level.  One cannot, however, quite see this point, as Folse does, as indicating a problem in or even incompleteness of Bohr's argument.  It is clear that all quantum-mechanical phenomena in Bohr's sense do possess ``properties,'' classical-like properties, at the level of effects and indeed are defined by such properties, without---this is almost the whole point of Bohr's concept of phenomenon---the possibility of attributing any properties to quantum objects themselves or their behavior.

Let me note that David Mermin's argument concerning ``correlations without correlata'' may be generalized to ``relations without relata'' and related to Folse's remarks on properties.  I might be bringing David's view closer to Bohr's than he himself has in mind, but one can see Bohr's argument in these terms in the following sense.  While in classical physics observable phenomena (in the usual sense) can be properly {\it related\/} to and {\it correlated\/} with the observable properties of actual objects, in quantum mechanics, in Bohr's interpretation, observable phenomena can only be {\it correlated\/} with the behavior of quantum objects.  It is not possible to establish the correlata of such correlations at the quantum level as properties of quantum objects or of their behavior, whether the quantum objects under observation or the quantum stratum of measuring instruments, through which the latter interact with quantum objects.  This impossibility is intimated already in Bohr's original criticism of Heisenberg's ``microscope'' thought experiment in Heisenberg's original uncertainty relations paper of 1927, although it took the EPR argument and then a decade or so for this point to crystallize.  It is this impossibility that eventually led Bohr to his redefinition of the term phenomenon in terms of the effects of the interactions between quantum objects and measuring instruments upon the latter (and thus, as effects, configurable in terms of classical physics properties of measuring instruments), something, again, under-appreciated by Folse.

What I find most problematic, however, is that Folse, in the essay in question and elsewhere, appears to attribute to Bohr the suspension, at least a possible suspension, of locality and to contrast Bohr's and Einstein's positions on this point, with obvious implications for Bohr's interpretation of EPR-type experiments.  This argument is, I would argue, difficult to sustain, if possible at all, as opposed to the argument that Bohr and Einstein held sharply contrasting views concerning the independent (local) reality of physical description at the level of quantum objects.  It is true, that, in this essay, Folse, following Don Howard's argument (which you cite as well, ``Fuchs,'' \#189, p.\ 41), attempts to assimilate Einstein's position on reality to his position concerning locality.  This attempt is, in my view, not altogether successful, on either Howard's or Folse's part, although I do agree that the connection between reality and locality in Einstein's thought should indeed by explored further.  I shall not address this subject here, since it is not required for my argument concerning Bohr's interpretation and specifically his view of locality (which derives from the overall argument outlined below).  Bohr's interpretation is local and is in part designed to respond to Einstein's arguments concerning locality by offering a local interpretation of quantum mechanics.  I do not think that one could even argue that Bohr's view is open as concerns the possible nonlocality of quantum physics.  (I entertained such a possibility myself at some point, a few years ago, but a relatively quick ``reality check'' of Bohr's argument showed that even this weaker argument is difficult to sustain.)  From this standpoint, I find Folse's (or Howard's) parallel assimilation of Bohr's wholeness (indivisibility) of phenomena to ``nonlocality/nonseparability'' problematic.  This is a crucial, perhaps uniquely central point, since, in my view, it is Bohr's concept of wholeness of the phenomenon in his sense that enables the locality of his interpretation, rather than allows for even the possibility of nonlocality.

Indeed, if one reads the passage from Folse's essay that you cite, it may appear that by ``separability'' (of Einstein's concepts of reality) and, conversely, ``nonseparability'' (of Bohr's concept of reality), Folse refers, respectively, to the possibility of an unambiguous reference to the properties of quantum objects independently of---``separately'' from---their interactions with measuring instruments vs.\ the impossibility thereof.  The former is indeed desired by Einstein, while the latter is argued for by Bohr and defines his concept of the wholeness (indivisibility) of phenomenon in his sense, which is to say, his concept of phenomenon, since all phenomena are indivisible in this sense.  Of course, by invoking ``nonseparability'' Folse has this indivisibility in mind as well.  It is clear from the essay as a whole, however, that he also means by ``nonseparability'' more than merely ``indivisibility'' in this sense, in view of the two assimilations just mentioned, that of Einstein's view toward locality (which is possible, or at least is worth exploring) and Bohr's position toward a form of nonlocality, which is, I would argue, untenable.  He means by it primarily (a form of) nonlocality.  It is true that, in the passage you cite (and in general) Folse separates ``locality'' and ``separability.''  First, however, he clearly argues that Bohr, in contrast to Einstein, does not see ``locality'' (rather than only ``separability'') as the necessary condition of ``physical reality,'' which is incorrect in any event.  Secondly, ``separability'' is first established by Folse as a form of nonlocality and only then is linked to Bohr's wholeness/indivisibility of phenomena.  ``Separability'' appears to be seen by him as a physical independence of spatially separated and physically disconnected events, on which, I would argue, both Einstein and Bohr equally insist, and which Bohr also argues to be possible in view of his interpretation.  Locality and nonlocality are seen in terms of, respectively, the impossibility and the possibility of instantaneous physical connections between spatially separated physical objects.  It is clear, however, that nonseparability would make it, at least in principle, possible to use the resulting mutual dependence of spatially separated entities to create nonlocal connection, even if undetectable physically.  Hence, it may be seen as a form of nonlocality, which makes it, again, unacceptable to both Einstein and Bohr (in the present reading).

In other words, while applied to the standard version and specifically to Bohr's interpretation of quantum mechanics, and thus making {\it both\/} nonlocal, Folse's view is ultimately that of proponents of Bohm's theory, which of course uses a different formalism.  You will recall that in some versions of Bohmian mechanics, nonlocality, that is, violation of relativity, is never manifest, but is nevertheless always a strict mathematical consequence of theory.   Folse's view is not unique, of course; for example, Henry Stapp argues for the nonlocality of quantum mechanics, which argument is, at least, unpersuasive, as his recent exchange with David Mermin shows as well, and of Bohr's view, which argument is simply untenable.  The point is of considerable significance for the recent debates.  The nonlocality of complementarity and, especially, of quantum mechanics itself, whether at the level of the data (such as the EPR correlations) or of the formalism would erase arguably the most crucial difference between the standard and Bohmian mechanics.

Folse's reading of Bohr's argument as, at least, tolerant of nonlocality (rather than his argument concerning Bohr's realism) was for me the main problem when I originally read the article (and a couple of his related pieces) a few years back.  Perhaps I am becoming over-concerned with locality.  A Bohmian I know (a philosopher, not a physicist) once asked me why I give so much importance to locality (leaving aside my argument that Bohr insists on it just as much as Einstein does).  What if locality were not true at the ultimate level?  This is of course not inconceivable in principle, even though so many things currently in place would have to go wrong (which has, however, happened in physics).  I replied by repeating Einstein, when asked a similar question concerning the validity of general relativity, which of course safeguards locality as well:  ``Then, I would be sorry for the Lord.''  I guess it is all about the beauty of relativity.  In any event, your email and, then, your ``Compendium'' prompted me to revisit Folse's argumentation and my own, ultimately opposing, argument, which I shall now sketch.  It uses my argument in ``Techno-Atoms'' and in my Mykonos paper as background, but the main points should be clear, especially given the preceding argument.  I shall comment on the question of reality first.

Here is my view of Bohr's position, which (perhaps, as I shall explain, taken a notch further) is also my own.  Bohr does not and, for the reasons explained above, could not deny the existence---and in this, but only in this sense, ``reality''---of quantum objects, nor (in contrast to the positivist or, to be cautious, naively positivist view) the relevance of the question.  What he does deny is that any concept of reality, that is, of the properties and behavior of quantum objects, conceived on the model of classical physics is applicable to them, while such a concept is germane to classical physics, and is retained by Bohr at the level of measuring instruments.  Indeed, which is equally crucial for Bohr, within certain limits, such concepts are germane in quantum physics, especially insofar as classical physics plays a role in quantum theory:  the correspondence principle in its various incarnations; the classical description of measuring instruments and of the effects of their interactions with quantum objects upon them; and so forth.  Thus, reality is suspended only as concerning the ultimate objects of description but is retained at the level of their effects upon what can be described in the realist way, such as the macro-world around us and, specifically, the measuring instruments involved in quantum physics.  Heisenberg's ``new kinematics'' of his first paper introducing quantum mechanics already epistemologically reflects this argumentation and, while influenced by Bohr in turn, may well be the ultimate source of Bohr's conception.

This view appears to me to be close to Asher Peres's and your own position both in your article and your replies to letters in {\sl Physics Today}.  In truth, however, the best I can say about some of those letters would be to repeat Pauli's remarks on the commentary by the editors of {\sl Nature\/} prefacing Bohr's article introducing complementarity (a much reworked version of the Como lecture):  ``{\it sancta simplicitas}.''  (As I was perusing your ``Compendium,'' I have read and, in some cases, reread some of Peres's articles and his book on quantum mechanics, many of which I found quite helpful.)  Some among the key aspects of Folse's argument concerning Bohr's realism are consistent with this view as well.  Thus, he writes: ``According to Bohr's concept of reality, real entities are entities that have power to interact, and in interacting, to produce the phenomena that comprise the natural world'' (``Niels Bohr's Concept of Reality,'' ``Fuchs,'' \#135, p.\  28).  This is correct, although, again, it does not sufficiently discriminate, in the way Bohr does, between the classical and quantum levels of description (or, in the quantum case, undescribability), since this statement would apply to both, which may have been Folse's point.  The differences, however, are hardly less crucial.  In other words, here, too, Folse does not fully explore or take advantage of the finer structure or layering of the architecture of phenomena in Bohr's sense, which leads to a number of problems in his argumentation (such as those mentioned above).

Now, although Bohr does not present his argument in so general a way in his writings and perhaps does not take it that far (for most of his purposes he did not need to), it appears that one can take a step further here, to what may be called ``the strong Copenhagen interpretation.''  Rather than only anything conceived of by analogy with classical physics, more radically, no conceivable concept of reality, or even any concept of ``existence,'' or ``object'' and ``quantum'' (the latter originating in classical physics, to begin with) or, in dealing with multiple systems, possibly even ``local'' or ``nonlocal,'' is applicable to ``quantum objects.''  I would like to add that this impossibility of applying either determination, ``local'' or ``nonlocal,'' at the quantum level does not mean that anything is nonlocal at the quantum level, and of course allows for the locality of all possible effects.  By the same token, this view is different from those versions of hidden variables theories, in which the locality of all observable effects would be maintained as well, but which configure the underlying quantum reality as nonlocal.  In accordance with this view quantum objects and behavior, including the ultimate nature of the interactions between them and the classical world (the world amenable to description in terms of classical physics) may be beyond our perceptive and conceptualizing capacities.  It is through a refinement of these capacities that we constructed the conceptuality and mathematization of classical physics, even if (this is somewhat more complicated) not mathematics itself.

This last point, coupled to the strong epistemological position just outlined, ``the strong Copenhagen interpretation,'' also suggests a better approach to Bohr's famous and often misunderstood insistence on the inevitability and necessity of using classical concepts in quantum mechanics.  Bohr's argument to that effect is at the core of {\Schroedinger}'s letter to Bohr, which, as you say, ``continues to haunt [you]'' (``Fuchs Compendium,'' \#200, pp.\  43--44).  You are, of course, not alone.  This apparently excessive emphasis on the necessity of classical physical concepts appears to have been one of the most haunting aspects of Bohr's arguments for complementarity throughout the history of its reception.  I think, however, that, eloquent as his letter is, {\Schroedinger} misunderstands Bohr on this point, as do most other critics of Bohr.  I shall spare you a reading of the letter itself, although such a reading would be instructive on several counts.  Part of my argument concerning Bohr and quantum mechanics all along has been that the significance of reading in physics is often underestimated, and this concerns as much the works of Bohr's key critics as those of Bohr himself, notoriously demanding and inattentively read, by critics and admirers alike, as his writing is.

Indeed I am not certain whether {\Schroedinger} sufficiently examined Bohr's overall argument, which is, I think, imperative in order to understand Bohr's point in question.  In accord with his argumentation as here outlined, Bohr's argument is, more or less, as follows.  Measuring instruments and their behavior, and, accordingly, the effects of the interactions between quantum objects and these instruments upon the latter, must be described in the classical manner in both physical and, correlatively, epistemological terms, while quantum objects and their behavior cannot be described at all, by any means, classical or other.  This, let me reiterate, is not the same as saying that the behavior of measuring instruments under the impact of their interaction with quantum objects is described classically, while that of quantum objects themselves is described quantum-mechanically.  In Bohr's view, quantum mechanics describes nothing, but only predicts the outcome of the interactions in question; and, again, most crucially quantum objects and processes, including those involved in the interaction between a quantum object under investigation and the measuring instruments involved in that investigation, are not describable by any means.  Otherwise, among other things, it may not be possible to maintain locality, for example, in the case of the EPR correlations.  The presence of this circumstance in quantum physics, which Bohr sees as uniquely significant to the difference between classical and quantum theory, is crucial to the rigorous description of any experiment, which hardly makes its manner classical---within Bohr's interpretation.  In other words, Bohr (his reply to EPR is crucially at stake in this exchange) argues that we do have an interpretation of quantum mechanics, as complementarity, that is both complete and local, which, accordingly responds to the EPR argument (it also specifically accounts for the specific experiment in question, or for ``quantum correlations,'' as we say now).  I shall return to the question of interpretation presently.  For the moment, my main point is that {\Schroedinger} does not quite see Bohr's point that the distinction in question is also necessary for locality, or, it appears, even completeness, and hence he missed the essence of Bohr's reply.  He thought the question posed by the EPR unresolved and that new ``concepts'' (and perhaps even new mathematics accompanying them) were necessary.  Bohr, by contrast, saw that old concepts could be used to solve the problem and to provide an interpretation of quantum mechanics as both complete, within its limits, and local.  Einstein, by and large (there are some further nuances concerning locality), accepted this point, and rejected Bohr's interpretation on epistemological grounds.  The epistemology rejected by Einstein was defined by Bohr's view that in quantum physics the classical description or describability in general of measuring instruments co-exists with the irreducible indescribability of quantum objects (and hence, the impossibility of classical-like reality at the quantum level).  {\Schroedinger} liked none of this epistemology either, but it appears that in his letter, or perhaps elsewhere, he does not get that far.

At this further juncture, the question becomes whether it is possible to find an interpretation or possibly a new theory (both Einstein and {\Schroedinger}, and sometimes even Bohr, under-appreciate this crucial distinction) that is more epistemologically palatable, including by virtue of being local.  Nonlocality would be epistemologically as unpalatable to Einstein or {\Schroedinger} as the impossibility of a realist description at the quantum level.  Ultimately, both appear to be nearly ready to give up on causality, but not locality and realism, which moreover they see as linked, in contrast to Bohr, who, accordingly, saw no such need, and, by and large, regarded the debate to be over.  The question itself remains, I would argue, unanswered, whether by means of old or, as Einstein and {\Schroedinger} thought possible, new physical concepts.  Bohm's hidden variables offer an old-style alternative, but it is nonlocal, and Einstein did not like it on both grounds---the old conceptuality and the nonlocality of the theory.  Einstein and {\Schroedinger} looked for new (nonclassical) physical concepts to reinstate the epistemologically classical (realist and preferably causal) local theory.  Bohr, by contrast, while seeking to preserve locality found a way to use classical concepts, but at the cost of a loss of classical epistemology or, it depends how one sees this philosophically, as a way, as a fringe benefit, of discovering the nonclassical epistemology.  As part of this process he invents quite a few new philosophical concepts, such as complementarity.  For the moment, I refer by ``complementarity'' to complementary physical description or experimental arrangements---mutually exclusive and hence never applicable simultaneously, yet both necessary for a comprehensive theoretical account---rather than to the overall interpretation of quantum mechanics.

The fact that, within Bohr's interpretation, quantum objects and their behavior are not, and cannot be, described by either classical or quantum-mechanical means, or, again, ultimately by any conceivably means, is crucial to Bohr's appeal to classical concepts, which link is usually missed by Bohr's opponents, {\Schroedinger} among them, and indeed by most other commentators on Bohr, critical or sympathetic.  For it follows that, whatever physical concepts we can ever conceive of could not be applicable for describing or even imagining, providing a (visualizable) intuition, {\it Anschaulichkeit}, concerning the properties of quantum objects and of their behavior.  (``Pictorial visualization,'' often used, including by Bohr, in English is adequate and part of what is at stake, but is not quite accurate and strong here, as opposed to the German term, used by both Heisenberg and Bohr, while the Danish word, used by Bohr, is very close to German.)  From this perspective, we may even define as {\it classical\/} whatever is conceivable and can serve for the purposes of a realist (and especially causal) {\it physical\/} description.  Mathematical formalism (such as that of quantum mechanics, which cannot be seen as descriptive in this sense, already by virtue of its dependence on complex numbers, as Bohr observed on several occasions) is another question; and Bohr never speaks of any limits in this respect.  Naturally, there are limits to what degree this type of mathematics, even two-dimensional complex Hilbert spaces (used in the case of spin) or indeed the field of complex numbers (the complex plane), let alone infinite-dimensional spaces of quantum mechanics, are available to our spatial intuition.  (Even if they were, though, and I do not think they are, we still could not use them to describe the behavior of quantum objects, or indeed anything physical; hence Bohr's appeal to mathematical ``abstractions'' in quantum mechanics, in contrast to classical physics.)

Bohr does not perhaps entirely deny that new concepts could be and are being developed, but only that, no matter what new concepts or mathematical tools (``abstractions'') we can and must develop, they will not bring us closer to describing or imagining the behavior of quantum objects.  This argument would of course only remain valid unless the nature of the available experimental evidence drastically changes (say, uncertainty relations or locality would be no longer valid), or, I would add (Bohr might be a bit less cautious here), unless a different consistent and complete local interpretation of quantum mechanics or an alternative local theory is found.  As Bohr says:
\bq
Such argumentation does of course not imply that in atomic physics, we have no more to learn as regards experimental evidence and the mathematical tools appropriate for its comprehension.  Indeed it seems likely that the introduction of still further abstractions into formalism will be required to account for the novel features revealed by the exploration of atomic processes of very high energy.  The decisive point, however, is that in this connection there is no question of reverting to a mode of description which fulfills to a higher degree the accustomed demands regarding pictorial representations of the relationships between cause and effect [which obtains in classical physics]. ({\sl The Philosophical Writings of Niels Bohr [PWNB]}, 3 vols.\ [Woodbridge, Conn.: Ox Bow Press, 1988), vol.\ 3:6)
\eq

One might indeed speak here of any pictorial representation, assuming that we could intuit anything in noncausal terms, which Wittgenstein, for example, if not already Hume and Kant, and certainly, Nietzsche questioned.  In other words, there is a kind of (en)closure---enclosure and closure (not the same as ``end'')---of physical (or of course other) concepts, defined by our perceptual and conceptual capacities, and perhaps ultimately linked to biological and evolutionary nature.  We cannot reach beyond this (en)closure, and we must depend on and (as both Bohr and Heisenberg grasped) can greatly benefit from this (en)closure even in our thinking about quantum objects, but, at the same time, this---beyond this (en)closure---is also where, in this interpretation, the nature of quantum objects and their behavior lies, forever hidden from us.  The latter placement of the quantum is, again, in this interpretation, itself a consequence of the effects of the interaction between quantum objects and the classical world (including measuring instruments) upon the latter, effects whose understanding is, accordingly, within this (en)closure and is available to us.  Classical physics is a refinement of these capacities, and there is of course a question of how far it reached on that road, but it would not affect the argument, here presented, concerning placing quantum objects beyond this limit in Bohr's interpretation, no matter how much further such a refinement reaches.  Perhaps it defines, if not the end of this road, a certain limit, a ``closure,'' indicating that we can only go so far and only along certain lines, perhaps incrementally infinitely, but the limit and the closure are implied, even if not expressly determined.  Possibly, it is quantum physics that establishes this limit.  Bohr's position on this point is not altogether clear, except, as the above statement suggests, on the point that quantum mechanics is beyond this limit, beyond the ``cut.''  (I am of course not using this term in its strictly quantum-mechanical sense, but one can contemplate certain links between both usages.)  As explained above, the overall argument just outlined remains, perhaps against Bohr's view (one can read him either way), conditioned by the particular interpretation here adopted, the strong Copenhagen interpretation, that is, such are epistemological consequences of this interpretation, on interpretation necessitated in part by the locality requirement.

Thus, {\Schroedinger}, just as Einstein, wants new concepts to return to the epistemologically classical models in physics, which would mathematically represent, in however idealized a way, independent and, it appears, causal physical reality.  By contrast, according to Bohr's interpretation (in this strong form), the appeal to classical concepts, beyond the practical role, reflects the impossibility of having such models.  The abstractions (in many ways, and indeed, as Bohr anticipated, ever more radical) of the quantum-theoretical mathematical formalism, from the original quantum mechanics to the field theories of the current standard model, enable excellent, even if only statistical, predictions under these conditions.  This is a kind of miracle, and an epistemological enigma, since, in contrast to classical physics, there appears to be no physical justification why they do so, and in this interpretation there cannot be, in part because the interaction between quantum objects and measuring instruments is itself quantum and subject to the same nonclassical epistemology.

From the present perspective, Bohr's interpretation, which may be called ``reciprocal-interactive,'' appears as somewhat different from Wheeler's more or, at least, still more, ``participatory'' or ``observer dependent'' view of the quantum world, which may be ultimately closer to Pauli's, although, in my view, it is uneasily suspended between Bohr's and Pauli's views.  The question is rather subtle and I should only make a few ``background'' points against which it could be considered, rather than addressing it as such, since to do it justice would require an engagement with Pauli and Wheeler on a much greater scale.  One cannot, however, stop short of anything but admiration for the grandeur of Wheeler's vision, which is well conveyed in your selection of quotations, let alone the works themselves cited.

I would like to comment briefly on Wheeler's discussions of the delayed-choice experiment, such as that in ``Law without Law'' (in J. A. Wheeler and W. H. Zurek, {\sl Quantum Theory and Measurement} and elsewhere, for example, in ```A Practical Tool,' But Puzzling Too,'' {\sl New York Times}, 12 December 2000, which you cite, ``Fuchs,'' \#383, p.\  74).  His analysis does not appear to me (perhaps I miss something) sufficiently to take into account the architecture of phenomena in Bohr's sense.  I, again, refer most especially to the wholeness-indivisibility of phenomena, in view of which we cannot attribute any properties, such as wave-like or particle-like ones, to quantum objects themselves even when a phenomenon becomes, using Bohr's term favored by Wheeler, ``registered'' as the result of an observation.  This view would help to refine Wheeler's argument that our observation of certain quantum events that took place in the past, possibly a very distant past, in no way influences what had happened at the time of such events, even though we have a (delayed) choice in setting our equipment.  This choice allows us to ``observe'' a different ``past event'' depending on a given setting.  Wheeler accurately puts the same type of quotation marks throughout his discussion.  Or, one can put it more accurately, and it is, I think, here that Wheeler's analysis could be refined and, I think, it would be by Bohr, as appears from his statement on or anticipating the delayed choice setup in {\sl PWNB\/} (Vol. 2, p.\  57), cited by Wheeler and Zurek in {\sl Quantum Theory and Measurement} (p.\  778).  The choice in question allows us to create different phenomena in the sense of what will have classically manifest itself in our measuring instruments at the time of measurement (see, for example, Wheeler's design in ``Law without Law,'' {\sl Quantum Theory and Measurement}, p.\  193), without, however, claiming anything concerning the properties of quantum objects or of their behavior.  For example, we can make no claim as to how the photons in question in Wheeler's design actually traveled before leaving their traces in our measuring instruments.  By the way, Wheeler's article in {\sl The New York Times}, which you cite, contains, in my view, an inaccurate statement, unless Wheeler's whole interpretation is off, at least as against Bohr's.  Wheeler says: ``if you measure the departing photon in a different [and mutually exclusive] way (a complementary way), you can tell if it {\it took both paths at once}'' (``Fuchs,'' \#383, p.\  74; emphasis added).  My understanding (which follows Bohr's and most other versions of the Copenhagen Interpretation) is that in such cases we {\it cannot know\/} through which route the photon has passed.  This lack of knowledge is reflected in the effects of the experiment, such as, once we repeat the experiment with a large number of photons, the interference-like pattern on a silver screen in the double-slit experiment.  (In the other complementarity observation measurement case we can establish through which ``route'' the photon travels.)   In ``Law Without Law,'' Wheeler's puts ``routes'' in all cases in quotation marks ({\sl Quantum Theory and Measurement}, p.\ 192), which may imply only a symbolic assignment of routes.  Even so, I am not altogether happy with his usage of ``both routes'' in ever speaking of a single photon without qualification.

Nor could we argue, counter-factually, concerning what could or would have happened to the same photon, if we had set our equipment differently.  Indeed, it appears that we cannot use counterfactual logic of that type in quantum mechanics without reinstating nonlocality, even though we can, of course, speak of different possible outcomes of future experiments, or of possible, that is, possibly obtainable, information concerning future events without introducing nonlocality.  This delayed-choice experiment seems to invite the former (counterfactual) reasoning, and Wheeler needs all the help he can get from Bohr to escape its traps, nonlocality included, but, to his great credit, by and large he manages to do so.

In sum, we can never ``see'' what happened with quantum objects as such in the past, nor argue, it follows, always counterfactually, on the basis of what would have but did not happen.  We can only predict (with certain probability) what will happen in future experiments we can perform, future effects of the interactions between quantum objects and measuring instruments upon the latter, on the basis of previously performed experiments and previously observed or measured effects of the same type.

From this viewpoint, one can speak of the ``constitutively participatory measurement'' in, and defining, quantum physics, or indeed of ``the activating observer,'' as you do.  I am not sure, however, in what sense one can speak, as Wheeler does of a ``participatory universe,'' even if in quotation marks and even if designating that the sense in which the universe is ``participatory'' is ``strange'' (``Law without Law,'' {\sl Quantum Theory and Measurement}, p.\ 194).  Unless, of course, he merely means that it is the universe that participates in these interactions, but the latter would be equally true in classical physics.  ``The activating observer'' would indeed be more accurate, since, while there is some reciprocity to these relationships, they are not exactly symmetrical.

While I am on the subject, let me point out that Pauli, in my view, also misses certain key nuances of Bohr's concept of phenomenon and his argument concerning the indivisibility of phenomena.  Thus, I cannot agree (nor I think would Bohr), in any event not without further qualifications (not offered by Pauli and not apparent from the article as whole), with the following statement in ``Phenomenal and Physical Reality'': ``[E]very act of observation is an interference is of undeterminable extent, with the instruments of observation as well as with the system observed, and interrupts the causal connection between the phenomena preceding and succeeding it'' (``Fuchs,'' \#257, p.\  56).   There is no such ``causal connection,'' or there would not be in Bohr's view of the situation, and, in all rigor, there is never either a phenomena (in Bohr's sense) of either preceding or succeeding an act of measurements, but only such acts.  Each such act gives rise to a phenomenon (no phenomena could appear otherwise) and such acts rarely, if ever, can present any ``sequence.''  Instead they entail repetitions using different quantum systems.  It seems to me that Pauli here thought of ``phenomena'' in terms of reference to the properties of quantum objects, or in any event he does not sufficiently elucidate the point.  If, however, he did think in this way, he could not have properly or at least in full measure understood Bohr's counter-argument to EPR and related arguments by Einstein, since Bohr's argument does not work unless it is assumed that all observable quantities or references to physical attributes pertain to measuring instruments and only to them.  The immediately following sentence in Pauli's elaboration, ``The gain of knowledge by means of an observation has as a necessary and natural consequence the loss of some other knowledge,'' is not altogether accurate either.  At the very least, one should, I think, say ``the loss of other possible knowledge.''  The point might, again, seem minor, but everything here is in finer micrological details and nuances.

As concerns the ``participatory universe,'' I would say, in general, that such statements as nature ``responds'' (the quotation marks are imperative) to the kind of questions we ask  (no quotation marks are necessary), and hence is defined by the latter participation on our part, can only be used metaphorically and, even then, with much caution.  We ``participate'' in nature only within the limits of our interaction with it, for example, by means of scientific experiments, even though this ``participation'' can take us far, as our knowledge, or what we learn we cannot know, is concerned.  Of course this participation also defines any conceivable view of reality, or, again, the impossibility thereof, we can have, for example, the possibility of describing much (not all!) of the macroworld in terms of classical physics, or (this difference is crucial in the present context) using quantum physics to learn how the quantum micro objects affect the macro world.   (Some quantum effects, such as those manifested in Josephson's devices, can of course appear at the macro level, but they would still be due to the micro constitution of nature.)   The so-called ``unreasonable'' (we merely exclude other things as ineffective)  ``effectiveness of mathematics'' belongs only to this level, as do certain archetypal correspondences (which attracted Pauli).  Heisenberg, in his later view (although one finds anticipations of it in his earliest works), combined both of these ideas and refined their application closer to Bohr, in arguing as follows:
\bq
If we attempt to penetrate behind this reality [the spatio-temporal reality of classical physics, or of our perception, to begin with] into the detail of atomic events, the contours of this `objectively real' world dissolve---not in the mist of a new yet unclear idea or reality, but in transparent clarity of mathematics whose laws govern the possible [the outcome of experiments?] and not actual [what actually happens?].  It is of course not by chance that ``objective reality'' is limited to the realm of what Man can describe simply in terms of space and time.  At this point we realize the simple fact that natural science is not Nature itself, but a part of the relation between Man and Nature, and therefore depends on Man.'' (``Fuchs,'' \#176, p.\  38)
\eq

This is not quite the way I (or, I think, Bohr) would put it, as my interpolating parentheses indicate, but it is essentially correct and, in its own way, admirable and philosophically appealing (Heisenberg clearly speaks here with a broader audience in mind).  In any event, the ultimate ``workings'' of ``nature'' are in no way participatory in the present view, even though, and because, they make any participation on our part possible.  By contrast, quantum mechanics, as a mathematical-experimental science of nature, is irreducibly participatory.  It appears to me that (for perhaps different reasons) both Pauli and Wheeler blur this difference, or want to dispense with it.  I prefer, however, to refrain from any definitive claim, since on a further reading this ``appearance'' or suspicion may yet dissolve.  Some of Heisenberg's elaborations, such as the one cited above, may seem to be a form of the mathematical Platonism of nature, while in fact they are not.  Plato's own view is yet another story and may even be closer to Heisenberg's and Bohr's than it appears, as Folse, incidentally, suggests by juxtaposing Democritus's and Plato's atomism, although I am, again, not in agreement with his reading of Bohr's atomism itself (``Fuchs,'' \#129, pp.\  25--26).

A (in my view) related remark by Heisenberg that you cite is also important here:  ``It is quite wrong to try founding theory on observable magnitude alone.  In reality the very opposite happens.  It is the {\it theory\/} which {\it decides\/} what we can observe.  You must appreciate that observation is a very complicated process'' (``Fuchs,'' \#177, p.\ 38; emphasis added).   One must qualify Heisenberg's point here (actually it is Einstein's).    First, ``the theory'' ought to refer to much more than a given mathematical apparatus, but instead to the whole conceptual framework of which the latter is a part, and indeed one may need to speak of a broader ``plane of mental immanence'' where these concepts emerge (I am not sure one can call this plane ``theory'') that shapes ``the theory.''  Second, it would be more accurate to say that the theory {\it shapes\/} what we observe and {\it decides\/} what observations are included, unless, again, we refer by the ``theory'' to a broader plane of immanence, just mentioned, in which case the theory might indeed be seen as {\it deciding\/} here.  The remainder of the (important) elaboration that you cite might be shown to illustrate this point.

The main point that I want to make is as follows.  The view here expressed by Heisenberg may appear to be in conflict with Heisenberg's approach in his first paper on quantum mechanics, to which he clearly refers here, and stressed the ``magnitudes, which in principle are observable,'' in other words, more or less individual quantum effects in the above sense.  The effects, rather than properties of quantum objects and of their behavior, that became subject to his ``new kinematics.''  There is no contradiction, however, and, as is well known and well documented, Einstein's point concerning the theory deciding what can be observed, was indeed was guiding Heisenberg in his work.   For his theory was not founded on such magnitudes alone; nor of course was Bohr's complementarity.  Heisenberg's famous, but not always carefully read, opening statement (a kind of abstract, but much more than that) is worth citing: ``The present paper seeks to establish a basis for theoretical quantum mechanics founded exclusively upon {\it relationships\/} between quantities [magnitudes] which are {\it in principle\/} observable'' (emphasis added).  ``Relationships'' is the key word here, and the title of the paper was, we recall, ``On Quantum-Theoretical Re-Interpretation of Kinematic and Mechanical Relations.''   ``In principle'' is quite crucial too, for, no matter how theory-laden and how complicated the processes of observation, the magnitudes in question could, in principle, be observed and, as it were, ``kinematized,'' while the classical-like physical (and ultimately any) properties of quantum objects and of their behavior could not.  In other words, even leaving aside for the moment the theory-laden character of all conceivable data (equally including that of classical physics), dealing with such, ``in principle observable,'' magnitudes is not the same as founding the theory on them, and Heisenberg's paper was not doing the latter.  While working with the available data of quantum physics (such as the Rydberg-Ritz formulas and the Bohr frequency relations), his theory qua theory was founded above all on Bohr's correspondence principle.  The latter was used to argue both that for large quantum numbers the data becomes the same as it would be in a classical case, at least as far as predictions are concerned (the description could, in all rigor, no longer be the same).  The principle was also used to argue, in part correlatively, that the equation should be formally the same as those of classical mechanics, the Hamiltonian equations.  (For large quantum numbers these of course would give correct predictions classically.)

It is this combination, this ``lethal combination'' (as far as classical physics is concerned), that leads to most remarkable properties of quantum mechanics, such as Bohr's probability rules, uncertainty relations, and so forth, and mathematically (in part correlatively) the irreducibility of complex numbers and replacing functions with operators as the kinematical and dynamical variables of the theory.  Both Dirac's and von Neumann's schemes are more or less automatic translations of Heisenberg's matrix mechanism.  Indeed, Heisenberg's stroke of genius (not altogether unprepared by, among others, Bohr, but a stroke of genius nevertheless) was itself a founding theoretical move.  That is, this arrangement of the relationships between observable magnitudes, made moreover into complex, rather than real, numbers (never observable as such), into infinite matrices is already {\it theory}, not observation.   (You may also recall that these matrices must be infinite in order to derive uncertainty relations.)

I, once again, apologize for not having enough to say on Pauli.  To me, his statement, which you cite from his essay ``Matter'' (but the statement, with minor variations, recurs throughout his lectures):  ``Like an ultimate fact without any cause, the {\it individual\/} outcome of a measurement is, however, in general not comprehended by laws'' (``Fuchs,'' \#251, p.\  54), has always been his greatest and most significant statement on quantum theory.  Indeed, it may well have been also on much that is beyond it, perhaps ultimately on life, or life/death, the randomness of whose occurrences, including in his own life, Pauli (I could hardly doubt this), was also contemplating here.  But then, Pauli appears to have been reluctant to separate life and quantum theory, anymore than Kepler wanted to separate it from his music of the cosmic spheres, either of which would indeed be difficult to do, at least in their own cases.

These are a few more or less immediate thoughts prompted by reading your ``Compendium,'' and clearly the stakes continue to be as enormous as ever, after three quarters of a century of debates, which may never end, and the reasons (I can think of several) why they may not end are worth considering, but cannot be addressed here.  I am, once again, grateful to you for allowing me to read it, and of course for putting it together.  As I said, the work, while seemingly consisting of citations, has a life and argumentation of its own, not reducible to the sum of citations, if indeed we can ever sum them up.  Perhaps, my sense of it is best captured by my reaction to Wheeler's remarks in his letter to Carrol Alley:  ``Today, the physics community is bigger and knows more that it did in 1939, but it lacks the same feeling of {\bf desperate} puzzlement.  I want to recapture that feeling for us all, even if it is my last act on Earth'' (``Fuchs,'' \#380, p.\  73).  Whether with Wheeler's (and Pauli's) help or not, I think your ``Compendium'' does bring the right sense of deep puzzlement, or perhaps better of profound complexity, whose resources are far from exhausted, and we have yet barely touched the epistemology of quantum field theories.  I am not sure one needs, at least at this point (perhaps ever) to feel desperate, or even {\bf desperately puzzled}.  Perhaps one needs more resolve to move forward from where quantum mechanics has already brought us and, of course, use what it gave us, in physics and philosophy alike---{\it increscunt animi, virescit volnere virtus}, the spirit grows, strength is restored by the wounds it receives, as Niezsche would have it.  One can hardly doubt that quantum mechanics inflicted one of the deepest wounds upon the spirit of physics; but physics survived, and more than merely survived; and, this is part of Nietzsche's meaning, what does not kill us sometimes (not always) can make us stronger.  In my view (I know that not everyone shares it, but I think that Bohr would), in this case it did.  Naturally, one could hardly doubt this either, this growth will bring us, it has already, new, yet deeper puzzles and puzzlement, but they will be, and some already are, new.
\eq

\section{29-05-01 \ \ {\it `Typo in Note 6' and `Worst of All'}  \ \ (to N. D. {\Mermin})} \label{Mermin11}

\noindent Wootters! Wootters! Wootters! Wootters! Wootters! Wootters!
Wootters! Wootters! Wootters! Wootters! Wootters! Wootters! Wootters!
Wootters! Wootters! Wootters! Wootters! Wootters! Wootters! Wootters!
Wootters! Wootters! Wootters! Wootters! Wootters! Wootters! Wootters!
Wootters! Wootters! Wootters! Wootters! Wootters! Wootters! Wootters!
Wootters! Wootters! Wootters!\, Wootters!\, Wootters!\, Wootters!
Wootters!\, Wootters!\, Wootters!

There are two t's.

\section{29-05-01 \ \ {\it Bayes and Lorentz: Never the Twain Shall Meet}  \ \ (to A. Peres)} \label{Peres15}

I still owe you some comments on your paper.  I guess in the end, I
don't have much to say.  I agree with its content of course. Not
least of all, because you taught me long ago that the correct analogy
is between quantum states and Liouville distributions, and not
between quantum states and phase space points. For me, that means
that both objects---quantum state and Liouville distribution---are
subjective entities, and can therefore be changed by the whims and
the fancies of the observer.  Their valuations are not tied to
physics per se.  The learning of information always presents just
such an opportunity:  there is no law of physics that says that all
observers must agree in their Liouville distributions for a given
system.  Measurement can lead to very nice inequities in that regard:
an application of Bayes' rule is not Lorentz invariant, but then
again it is not even {\it observer\/} invariant.  We apply it when we
gather information; we don't apply it when we don't.

But the question, I must ask is:  what new point did you make in this paper?  What did you say here that was not clearly in your mind when you wrote the RQM paper, for instance?  (I can even find our discussion of similar points in the samizdat, pp.\ 270--276.)  The answer to those questions don't come out clearly in your present exposition.

\section{29-05-01 \ \ {\it No Miracles}  \ \ (to A. Peres)} \label{Peres16}

\bap
Exophysical agents are essential, and most people tend to sweep them
under the rug.
\eap
Yeah, you might say my {\sl Notes on a Paulian Idea\/} is about
looking under the rug.

\section{29-05-01 \ \ {\it Substantiation}  \ \ (to N. D. {\Mermin})} \label{Mermin12}

What particular piece in the paper directly substantiates the
sentence:  ``It is possible to eliminate all couplings between the
source and the destination because quantum qubits have a richer range
of logical capabilities than do classical bits.''

Have you tried explicitly walking through all your circuits with a
classical Liouville distribution rather than a quantum state?  That
is to say, in Fig. 2 for instance, let $|\psi\rangle$ stand for a
(Bayesian) probability distribution over $|0\rangle$ and $|1\rangle$,
rather than a pure quantum state.  Then you would have a classical
state-swapping circuit:  where the word state now means without doubt
``state of knowledge.''

What goes break and where goes break in getting from figure 2 to
figure 10 in such a classical setting?  Well, nothing of course,
because we know that teleportation also works for mixtures.  But it
does probably mean that some components in the circuit are probably
more ``quantum'' than they need to be to carry out such a classical
project.  (I.e., the project of teleporting a classical state of
knowledge, via purely classical resources.)

Can you pinpoint where that occurs?  It has now struck me that your
view of teleportation may be a good laboratory for sussing that out.
I'm warming up as a referee.

Today Renes arrives, and this evening {\Caves} arrives, and \ldots\ and
next Tuesday I have committed myself to lecturing on Holevo's channel
capacity result, and---between all of that---somehow, someway, some
pig (Charlotte's web), I must complete the NATO paper.  I keep hoping
for a miracle:  none occurs.

You've never read it, I know, but you should read it:  de Finetti's
paper ``Probabilismo,'' written in his youth while in a fascist
fervor.  Have the mental strength to divorce it from the fascism, and
you will find it is an absolutely wonderful piece: B.~de Finetti,
``Probabilism,'' Erkenntnis {\bf 31}, 169--223 (1989).  [See also
R.~Jeffrey, ``Reading {\it Probabilismo},'' Erkenntnis {\bf 31},
225--237 (1989).]

I've come to think of the NATO paper as my own Probabilismo.  I just
wish I had the youthful fervor.

\section{31-05-01 \ \ {\it Quotes and Barbells} \ \ (to H. J. Folse)} \label{Folse4}

No, there is never a worry in quoting me.  I say right things, and I say wrong things, but I never say anything without the intention of lifting the veil of the quantum.  I am more than happy to be quoted if my right/wrong thing contributes to that.

We don't have a program worked up yet:  Unfortunately, both Khrennikov and I have been hugely tied up for the past weeks.  But we started the process yesterday---in fact---so maybe something will come out by mid-next week.  In any case, I'm fairly sure at this stage you can count on a 30 minute talk.  I think we all can; there have been several additions to the conference (Bernstein, Greenberger, and several others) even though a few dropped off the list (Shimony, Brassard, Preskill).

I printed out my samizdat for the first time yesterday.  It's a frightening thing, seeing it bend the table.  I don't suppose you've had any luck in retrieving it.  Maybe I'll just break down and send you loose-leaf copy of it. Can you send me your mailing address so I make sure I send it to the right place.

\section{03-06-01 \ \ {\it Quotes} \ \ (to myself)} \label{FuchsC2}

\begin{itemize}
\item
``Only those who will risk going too far can possibly find out how far one can go.'' \\
\ -- \ T. S. Elliot

\item
``Opportunity is missed by most people because it is dressed in overalls and looks like work.''
\ -- \ Thomas Edison
\end{itemize}

\section{06-06-01 \ \ {\it Comments and Answers}  \ \ (to N. D. {\Mermin})} \label{Mermin13}

Wow.  Thanks again for the very thorough reading!  (And concern over
my welfare.)

I reply to all your comments below.  You've certainly left me hungry
for more.

\bdm
I didn't really get the epigraph (``imprimatur'') and the apparent
explanation of it on p.\ 2.  Is that a coy reference to the bad-boy
tone?
\edm
Which meaning of coy did you have in mind?
\bq
\noindent 1. Tending to avoid people and social situations;
reserved.\\
2. Affectedly and usually flirtatiously shy or modest.\\
3. Annoyingly unwilling to make a commitment.
\eq

\bdm
What is an Einselectionst? (p.\ 2.) Everything else is familiar, but,
irritatingly, it's the one school you don't explain in footnote 2. Is
this a (bad) joke you're playing on the reader?
\edm
I'm letting that one stand: it was meant to be a little insulting.  I
can't figure out what their views are.

\bdm
p.\ 2. You should think twice about sentences like ``If I am ever
going to get \ldots\ someone else's war.''  I know it's true, that
physics would be a lot more fun if people weren't trained to hide
such sentiments, etc.  But it does have a certain self-indulgent
quality.
\edm
Stripped away with surgical precision \ldots

\bdm
p.\ 4. Say the whole thing: 1) the speed of light IN EMPTY SPACE is
[constant,] independent of the speed of its source.  (Einstein did
not say ``is constant'' --- it's not necessary.)  Keep ``is
constant'' if you like, but why leave out his ``in empty space'' ---
that's crucial.
\edm
Oh, alright!

\bdm
Bottom of p.\ 4.  Can you explain classical E{\&}M to a junior
high-school student giving the essence, not the mathematics?
Classical mechanics?  Thermodynamics?  I've always thought relativity
was special that way.
\edm
No.  Not yet.  Maybe.  Me too.  I.e., I consider E{\&}M much more
like the setting of a particular Hamiltonian within QM:  And I
wouldn't expect a high-school student to be in a position to
understand why we have chosen the particular Hamiltonians that we
have in any practical situation.  Since the overarching belief here
is that ``quantum mechanics is a law of thought'' (with an almost
trivial side-input of honest-to-god physics), it strikes me as being
in a category much more like relativity.  In the words of J.~Bub, it
is a ``framework theory.''

\bdm
Also p.\ 5, I've been meaning to ask you: why is $H$ a vector space
if the scalars are real or complex, but a module if they are
quaternions?  As I remember ``module'' has a precise meaning ---
maybe having to do with the scalars lacking inverses --- which is not
satisfied if they're quaternions.  Isn't $H$ still an ordinary vector
space?
\edm
According to Adler's book, it is a module.  Why one makes such a
distinction, I don't know:  I just followed convention.

\bdm
Is footnote 5 on p.\ 7 a little too cute?  [Just asking.]
\edm
Let me think about this one.  I added the footnote first as a joke to
myself.  But then I decided I sort of liked it.  The motivation came
from {\Bennett} having no clue about what imagery I was trying to convey
with zing.

\bdm
p.\ 11.  I find your brief knee-fortifying rejoinder to Penrose
somewhat of an anticlimax.  If you had added explicitly that the
answer to Penrose is that it's all about the RELATION between Alice
and the qubit (even without invoking correlations without correlata)
I would have thrown away my Ace bandage.
\edm
Yeah, I thought it was a bit of an anti-climax too.  I continue to
search for ways to strengthen my rejoinder to Penrose and Jozsa (the
z comes first).

But indeed you and I differ on the climaxes we seek.  As far as I can
tell, you seem to seek a predominantly static worldview.  I want to
see true becoming \ldots\ I think.

Perhaps I said something closer to what you'd like me to say
presently in the Samizdat, pages 407--408 (in a note titled Penrose
Tiles).  Have a read of that, and tell me whether that meshes with
what you would have liked to hear.  Who knows, I may well reconsider
(based on the old text).

\bdm
p.\ 11.  I'd never seen so direct a statement of correlations
(between Man and Nature) without correlata in Heisenberg before. Nice
quote. Where's it from?
\edm
W.~Heisenberg, ``The Development of the Interpretation of the Quantum
Theory,'' in {\sl Niels Bohr and the Development of Physics:~Essays
Dedicated to Niels Bohr on the Occassion of His Seventieth Birthday},
edited by W.~{\Pauli} with the assistance of L.~Rosenfeld and
V.~Weisskopf (McGraw--Hill, New York, 1955), pp.~12--29.
\bq
\indent
The criticism of the Copenhagen interpretation of the quantum theory
rests quite generally on the anxiety that, with this interpretation,
the concept of ``objective reality'' which forms the basis of
classical physics might be driven out of physics.  As we have here
exhaustively shown, this anxiety is groundless, since the ``actual''
plays the same decisive part in quantum theory as it does in
classical physics.  The Copenhagen interpretation is indeed based
upon the existence of processes which can be simply described in
terms of space and time, i.e.~in terms of classical concepts, and
which thus compose our ``reality'' in the proper sense.  If we
attempt to penetrate behind this reality into the details of atomic
events, the contours of this ``objectively real'' world
dissolve---not in the mist of a new and yet unclear idea of reality,
but in the transparent clarity of mathematics whose laws govern the
possible and not the actual.  It is of course not by chance that
``objective reality'' is limited to the realm of what Man can
describe simply in terms of space and time.  At this point we realize
the simple fact that natural science is not Nature itself but a part
of the relation between Man and Nature, and therefore dependent on
Man.  The idealistic argument that certain ideas are {\it a priori\/}
ideas, i.e.~in particular come before all natural science, is here
correct.  The ontology of materialism rested upon the illusion that
the kind of existence, the direct ``actuality'' of the world around
us, can be extrapolated into the atomic range. This extrapolation,
however, is impossible.
\eq

\bdm
p.\ 11. ``wake and dream''?  waking and dreaming?
\edm
I left that alone.  The {\sl American Heritage Dictionary\/} has a
huge usage note associated with ``wake.''  I took that to give me
license, plus I like the flow of it.

In general, I think that paragraph is the deepest part of the whole
paper.  It is the Paulian idea.

\bdm
p.\ 13.  Gleason derives the state-space structure?  What do you
mean? He talks about projections.  What are they projecting on?  What
does ``tr'' mean?  Isn't the state-space structure already there?
You yourself begin by saying ``Let $P_d$ be the set of projectors
associated with a \ldots\ Hilbert space \ldots''
\edm
\begin{center}
state space structure (for me) $\;\;=\;\;$  convex set of density
operators
\end{center}

The Hilbert space you are talking about---again for me---only has
significance in that it defines the set of potential measurements.
The quantum state, and with it the state-space structure, is a
secondary notion.

But I will try to ward off confusion by adding the definition above
to the text.

\bdm
p.\ 16.  ``sum total OF ways''  I wrote in the margin around here
that noncontextuality for POVMS seems to be an enormously stronger
assumption than noncontextuality for projections \ldots
\edm
Yes.  But so what?  The point is---physically---the assumption is
equally deep for both notions of measurement.  Noncontextuality IS
Bayes' rule.  (Or just about so much.)  It's just that applying it to
POVMs actually gets you somewhere (without having the mathematical
skills required to solve one of Hilbert's problems).

\section{06-06-01 \ \ {\it The Haunting}  \ \ (to N. D. {\Mermin})} \label{Mermin14}

You have this habit of haunting me:  even when I say I'm not going to
listen to you, I end up listening to you.  Demon!
\bq
Einstein was the master of clear thought. I have already expressed my
reasons for thinking this in the arena of electromagnetic phenomena.
Likewise, I would say he possessed the same trait when it came to
analyzing the quantum. For even there, he was immaculately clear and
concise in his expression. In particular, he was the first person to
say in absolutely unambiguous terms why the quantum state should be
viewed as information (or, to say the same thing, as a representation
of one's knowledge).
\eq

\section{06-06-01 \ \ {\it Reading de Finetti}  \ \ (to H. M. Wiseman)} \label{Wiseman1}

{\Carl} {\Caves} is visiting Bell Labs, and yesterday he shared with me some correspondence he's had with you about the quantum de Finetti paper.  Thanks a million for the interest!

I just wanted to make one comment on one of your comments.  You asked about the deeper reasons for my mean-spirited, off-hand remark about ``for a stretch of the imagination \ldots''.

\bhw
But of course the Bohmian particles behave in a nonlocal way, so this
means that our experience must be capable of nonlocal influence. That
is, we cannot rule out faster-than-light information transfer, if
information is transferred when a being becomes conscious of it. (Is
this a reasonable concept of information transfer? What a tangled
question! I will assume it is.) So in fact Bohmian mechanics is not
even compatible with relativity, unless conscious systems are
constrained somehow so as not (or at least with vanishing
probability) to have nonlocal effects. This seems an ad-hoc hypothesis
and even if it can be shown to work for us, I don't see how it could
be proven in general.
\ehw

Actually, something like this has been one of my longer-standing pet peeves with Bohm theory.  In particular, I've always thought it absolutely contrived that the nonlocal components in the theory could not be made use of.  That they could not be turned into a technology, say.  In that capacity, the Bohm theory seems to me to be no deeper than saying ``God wills each and every event in our quantum world.''  It is true that the theory has a sheen of equations---which a religion normally doesn't have---but beyond that, I see no great difference.  Somehow the mere possibility of writing down an equation has been deemed to be an acceptable state of affairs in the Bohmian community.  But if the parameters in that equation cannot be set by the experimentalist (with a sufficient amount of effort), then, in my mind, it is no different than a burning bush proclaiming, ``I am that I am.''

Now, not all Bohmians, believe that the particle trajectories cannot be controlled at all.  Antony Valentini is a notable exception.  And he has also studied the non-uniqueness of the trajectory equation at length, hoping one day to put the various options to experimental test.  (I don't have any references for his papers, but I see he has at least one paper on {\tt quant-ph}:  So you can probably find references therein.)  But it has been my experience that the vast majority of the Bohmians (or the Goldstein, Duerr, Albert flavors thereof) see no need to go to the trouble of actually calculating a particle trajectory.  It is enough, for them, to believe it exists.

I had not realized before that you have an interest in quantum foundations.  Lately I've been trying to energize the community to think hard about ``quantum foundations in the light of quantum information'' and have organized two meetings to that effect.  (One in {\Montreal} last year, and one coming up in Sweden at the end of next week.)  There is a good chance that Brassard and I will be doing still another in {\Montreal} next summer.  If you have an interest in coming, I'll surely make sure that you get an invitation.

\section{06-06-01 \ \ {\it Bohm and the Burning Bush}  \ \ (to H. M. Wiseman)} \label{Wiseman2}

\bhw
\bq\noindent{\rm
[Chris said:] Somehow the mere possibility of writing down an equation has been deemed
to be an acceptable state of affairs in the Bohmian community.  But if
the parameters in that equation cannot be set by the experimentalist
(with a sufficient amount of effort), then, in my mind, it is no
different than a burning bush proclaiming, ``I am that I am.''}
\eq
But that is because you have such a strong instrumentalist philosophy (is
that a fair description?). I'm in favour of giving a theory a fighting
chance of producing its own interpretation. After all, wasn't gas theory
criticised on your sorts of grounds a century or so ago?
\ehw
No, most of the gas laws were found empirically at first.  Then theory was tested quite well, by working explicitly in new regimes of pressure, temperature, etc.  Quantum theory, too, has been tested empirically.  We have means of preparing quantum states and checking how they evolve.  We don't have the same means---so we are told by Goldstein and company---for (Bohmian) particle positions.

\bhw
But I'd certainly like to understand an informational view of QM (and
stat mech too, for example; do you think they are related?) better.
\ehw
Yes.  That was a good bit of the point of the de Finetti paper.

\section{07-06-01 \ \ {\it Frost}  \ \ (to N. D. {\Mermin})} \label{Mermin15}

Just found this in my quote of the day:
\bq
\noindent Poetry is a way of taking life by the throat. \medskip\\
Robert Frost (1874--1963), U.S. poet. Quoted in: Elizabeth S.
Sergeant, {\sl Robert Frost: the Trial by Existence}, ch.\ 18 (1960).
\eq

I see you wrote me a note last night at 10:17.  I'll have a look at
it.

\section{07-06-01 \ \ {\it The Mud of von Neumann}  \ \ (to N. D. {\Mermin})} \label{Mermin16}

\bdm
\bq
\noindent  \rm [CAF wrote:]  Forget about the larger Hilbert space.
It is artifice.  It is only historical accident that has confused us
for so long.
\eq
Now that I've added the discovery of POVMs to my discovery of the
moons of Jupiter I'm less inclined to object to this.  Indeed, I'd
like it to be so.  But if I give you a resolution of the (2
dimensional) identity for a single qubit into 17 positive operators,
can you always tell me how to set up a corresponding procedure with
17 distinct outcomes without enlarging the system to one described by
a larger Hilbert space?
\edm

The point is:  If I give you a resolution of the (2 dimensional)
identity for a single qubit into 2 projection operators, can {\it
YOU\/} always tell me how to set up a corresponding procedure with 2
distinct outcomes without enlarging the system to one described by a
larger Hilbert space?

Von Neumann didn't know how.  He introduced the notion of a
``measurement model,'' enlarged the Hilbert space, and got us into
the muddle most of us are still in today.  Now, people like Zurek are
going around trying to justify the ``pointer basis'' on von Neumann's
ancillary Hilbert space by enlarging it still further (and calling it
``the environment'').  And I'm sure you'll recall that von Neumann
himself did even more dastardly things.

You just have to get into a different mindset.  My attitude is it's
time to cut the Gordian knot.  That's what Section 4 and 5 are about.
To the extent that one admits a mystery to the standard (von Neumann,
orthogonal projection-valued) notion of measurement, one gains {\it
nothing\/} by holding tight to it.  One might as well transfer the
mystery to the POVMs and be done with it.  The advantage of the POVMs
is that they are a conceptually simpler structure:  the old von
Neumann notion is just a horribly contorted surface of constraint
within that beautifully smooth space.  What is a measurement?  A
refinement of one's state of knowledge, full stop.  Any refinement
whatsoever?  Yes, any refinement.

Let me give you a homework exercise.  (Now I'm going to sound like
Gottfried.)  Go back to a classical setting where you have a
probability distribution $p(h,d)$ over two hypotheses.  Marginalizing
over the possibilities for $d$, you obtain an initial state of
knowledge $p(h)$ for the hypothesis $h$.  If you gather an explicit
piece of data $d$, then, using Bayes' rule, you should update your
knowledge about $h$ to $p(h|d)$.  The question is this:  Do you not
find that transition $p(h) \longrightarrow p(h|d)$ a mystery you
should contend with?  Does it not bother you that if someone asked
you for a {\it physical description\/} of that transition, you would
be at a loss for words?  I mean, after all, one value for $h$ is true
and always remains true.  One value for $d$ is true and always
remains true.  There is no transition for those variables.  The
transition is in your knowledge (or belief, if you will).  Should we
not have a detailed theory of how the brain works before we can trust
in the validity of Bayes' rule?  (I ask that rhetorically of course.)
Should we close all the gambling houses in Nevada, on the suspicion
that they know better of Bayes' rule (and its limitations) than we do
and have been using that to their (nondisclosed) advantage all along?

In my view, recognizing the ridiculousness of the rhetorical question
is the {\it first\/} step to freedom.  Now we've got a long haul to
go, but at least we're out of the parking lot.

I think I'll CC this note to {\Caves} (across the hall) since parts of
it seem to be a point of contention between us too.

\section{07-06-01 \ \ {\it Modules Over a Division Ring}  \ \ (to N. D. {\Mermin})} \label{Mermin17}

Actually, I picked up the word ``module'' in a footnote of Adler's,
where he explains that he is going to flout convention and call the
object a ``quaternionic Hilbert space.''  (Or, maybe he said vector
space instead.)

Anyway, ``quaternionic module'' does square with the definitions in
{\MacLane} and Birkhoff's book {\sl Algebra}.

p.\ 134: ``A division ring is defined to be a non-trivial
ring (not necessarily commutative) in which every non-zero element
has a two-sided multiplicative inverse.  A commutative division ring
is thus the same thing as a field.  An example of a non-commutative
division ring is furnished by the quaternions \ldots''

p.\ 190:  ``A module is an additive abelian group whose elements can
be suitably multiplied by the elements from some ring $R$ of
`scalars.' \ldots\  This chapter will be concerned with general properties of modules over arbitrary rings; special properties of modules over fields (`vector
spaces') will be studied in the next chapter.''

p.\ 253:  ``We have already noted that most of the properties of
vector spaces (modules over a field) are shared by modules over a
division ring.  As the most important example of such a division ring
we now construct the ring of quaternions.''

I would like to know what cherished property of vector spaces goes
bust with quaternionic modules \ldots\ but I just don't have the time
for that now.

\section{07-06-01 \ \ {\it Two Go Bust}  \ \ (to N. D. {\Mermin})} \label{Mermin18}

Actually {\MacLane} and Birkhoff say that only two of their theorems
in Chapter 7 ``Vector Spaces'' fail for modules over a division ring.
But I haven't been able to locate the theorems, and I think I'm going
to have to give up for now.\footnote{\editornote The first one is
  Corollary 2 to Theorem 8: ``Let $V$ and $V'$ be vector spaces of the
  same finite dimension.  Then any epimorphism $t: V \to V'$, and also
  any monomorphism $t: V \to V'$, is necessarily an isomorphism.''
  They add that this does not always hold for all modules of finite
  type over an arbitrary ring.

The corollary to Proposition 10 also does not apply if the field used is
replaced with a division ring.  This corollary states that if $V$ and $V'$
are two vector spaces, then the group of linear transformations from~$V$
to~$V'$ is also a vector space, and its dimension is the product of the
dimensions of~$V$ and $V'$.}

\section{07-06-01 \ \ {\it The Mud of {\Mermin}}  \ \ (to N. D. {\Mermin})} \label{Mermin19}

\bdm
\bq\noindent \em
[CAF wrote:]  The point is:  If I give you a resolution of the (2
dimensional) identity for a single qubit into 2 projection operators,
can {\it YOU\/} always tell me how to set up a corresponding
procedure with 2 distinct outcomes without enlarging the system to
one described by a larger Hilbert space?
\eq
Yes, I can.  (That's why I picked that example.)  The two orthogonal
projections are necessarily of the form $(1 \pm n\cdot\sigma)/2$. So
all you do is find a Stern--Gerlach magnet and rotate it so its axis
is along $n$.  (I'm taking my qubit to be associated with the
magnetic degree of freedom of a spin 1/2 particle that I can shoot
between the poles of the magnet whenever I please.  Tricky to do, of
course, but not conceptually challenging.)
\edm

Your rejoinder didn't faze me a bit.  I just go to that same fine
machinist who built your Stern--Gerlach device, and ask him to cut me
two very good and very small mirrors, add the tippiest tip of glue to
one of them, pass it through a dilute solution of just the right
chemicals (so that just the right amount sticks to the glue).  Et
voila! In THEORY you CAN call such a thing an ``atom in a cavity,''
that allows a $J=0 \longrightarrow J=8$ multipole transition (in
principle) to be excited by a tenuous beam of light.  We then apply a
magnetic field and shine some auxiliary lasers in to ``check'' which
of the 17 sublevels was ``actually'' excited.

I ask the machinist, ``Did you feel particularly different when you
worked for me than when you worked for {\Mermin}?''  He says, ``Well you
do pay better!  But I promise you my prices were set by strictly
objective criteria: I had to roam the earth to find that exotic
chemical, and believe you me, that machining job was no easy task.''
I ask him, ``Well, did you at least feel particularly quantum?''
Just a guess, but I suspect he'll look at me with the same blank
stare of the students on page 15 of my draft.

\bdm
\bq\noindent \em
[CAF wrote:] von Neumann didn't know how.  He introduced the notion
of a ``measurement model,'' enlarged the Hilbert space, and got us
into the muddle most of us are still in today.  Now, people like
Zurek are going around trying to justify the ``pointer basis'' on von
Neumann's ancillary Hilbert space by enlarging it still further (and
calling it ``the environment'').  And I'm sure you'll recall that von
Neumann himself did even more dastardly things.
\eq
I basically agree with all of this but I don't think it has much to
do with the question I asked you, which was how to describe on the
down-to-earth unphilosophical laboratory (FAPP) level how to set up
an experiment whose distinct outcomes correspond to the distinct
positive operators.  I can think of ways to do it, but if I want to
describe them in the language of QM I need a larger Hilbert space to
do it.  I want to know how to do it in a way that refers only to the
original qubit, analogous to what I told you above.
\edm

You're just not getting the point.  There is no difference in
principle, between the machinist's fine work for you and his even
finer work for me.  But von Neumann, silly von Neumann, invoked extra
Hilbert space for both jobs.  At least he was consistent.  You're not
being consistent.  What do you think the Stern--Gerlach device is if
it's not extra Hilbert space (in the von Neumann view)?

\bdm
\bq\noindent \em
[CAF wrote:] Let me give you a homework exercise. \ldots
\eq
Again, very nice, but it has nothing to do with what I was asking
you.
\edm

No, it has everything to do with what you are asking.  You're just
causing me to drag you in kicking and screaming.  The point is:  You
don't need to invoke physics to make sense of Bayes' rule.  A part of
the quantum confusion has come about precisely because people have
wanted to invoke ``physics'' to make sense of quantum collapse.  It
hasn't happened in 75 years, and it's not going to happen now---or at
least that's my bet.  In your Stern--Gerlach example you explicitly
throw away the issue of ``where the outcome comes from.''  (Von
Neumann tried to answer that issue but botched it.)  But your example
is precisely on the right track:  you don't need to ask where the
outcome comes from for Bayes' rule, and you don't need to ask it for
the quantum.

Now I've got to go to lunch, and then try to recover from all the
work I did NOT put into the paper today.  I hope though that these
notes I'm writing you are clearing a little bit of the mud away.

\section{09-06-01 \ \ {\it Nope} \ \ (to H. J. Folse)} \label{Folse5}

\bhf
I'm less hopeful than you that conferences such as next week's in {\Vaxjo} will cease --- or at least if they do cease it will be because these questions have been answered {\rm\bf for once and for all}.  I
suspect we suffer from the historical delusion that back in the {\rm\bf old days} of ``classical physics'' there was a clear cut conception of what the universe was like and how we had knowledge of it.  And what we'd like is a quantum era substitute.  But of course a moment of historical realism tells us that back in the old days the philosophical significance of the ``{\rm\bf classical} mechanical conception of the universe'' was just as much controversial  as is the ``{\rm\bf quantum} mechanical conception of the universe'' today.    Such questions tend more to get {\rm\bf outdated} than answered.
\ehf

Nope.  I don't believe in ``for once and for all'' for anything.  I'm just looking to clarify quantum mechanics enough so that its foundation stops being a burden to thought.  Then we'll be in a position to move on to the next step in physics, with all of the wonderfully mysterious things it entails.  Sort of like laying down special relativity, so we'll finally have a fighting chance of discovering general relativity.

I'm still working on the silly NATO paper after all.  I've de-ego-ized the introduction quite a bit, and added more discussion on the analogy between Bayes' rule and quantum collapse than I thought I'd have to.  But {\Mermin} has been reading it carefully---that's why I made those changes---and though its slowing me down, it's certainly helping the paper.  I'll bring a copy of the final to Sweden.

I'm keeping my fingers crossed that the rain will subside on your end of the world.

\section{10-06-01 \ \ {\it Answers}  \ \ (to N. D. {\Mermin})} \label{Mermin20}

\bdm
Before I torture myself making sure that the converse really holds, I
need a little more education in the formalism.  [\ldots]

So I'm putting myself in the place of your student on p.\ 16 and I'm
looking for a simple example where the standard postulate you've
dropped doesn't hold, so I can see for myself whether it makes a
difference.  What could be simpler than the POVM
$$
      E_b = p(b) I
$$
where $I$ is the identity and the $p(b)$ are non negative
probabilities that sum to 1.  Now I look at your rule for $P(b)$ and
it tells me that $P(b) = p(b)$, independent of the density matrix of
the system. So I raise my hand and ask why you call that a
measurement when it doesn't tell you a damn thing about the system.

You will not say, ``Ah, that's because the POVM you picked
corresponds to the case where you turn on no interaction between the
system and the ancilla, so of course the measurement of the ancilla
tells you nothing about the system.''  You will not say that because
(a) the point is to eliminate the ancilla and (b) to say nothing more
about ``measurement'' than is in your table.

My guess is you will say that some POVMS are more informative than
others and I have picked a particularly bad one (just as picking $I$
for my observable is a particularly bad choice under the standard
rules).
\edm
Yes.

\bdm
So then I say, ``So there is some figure of merit for POVMs
associated with how well their outcomes discriminate among various
density matrices for the system?  Could it be that the postulate you
have dropped is there because it somehow maximizes this figure of
merit?''
\edm

There are two ways in which a POVM can be informative.  1)  Suppose
you have a density operator $\rho$ for a system, and you know that
someone else has a density operator $\rho_i$ for that same
system---where $\rho_i$ falls within some fixed decomposition of
$\rho$.  You just don't know which value of $i$ he happens to be
using (outside of some prior probability).  Therefore, you perform a
measurement in an effort to obtain information about his $i$:  this
is the scenario of classical communication.  2)  There is no extra
player, there is simply $\rho$.  However, you are dissatisfied with
how little you can predict of a random measurement in the future,
given that very mixed state $\rho$.  So you perform a measurement
now, in the hope that you can say more about a random measurement in
the future.  That is to say, by performing a measurement, you can
reduce the mixedness of a state \ldots\ and in that sense a
measurement can be informative.

Unfortunately, I do not say a lot about 1) in this paper outside of
the paragraph you cite below.  I do say a lot about 2) though.  It
sounds like you haven't read through that.

Of course some measurements will be better or worse for each of these
tasks.  In some examples of 1), you most certainly have cases where
NON-vNM POVMs are required.  As regards 2), I haven't thought about
it enough:  i.e.\ suppose you wanted to changed your mixedness from
amount $x$ to amount $y$.  Could you always do that with a
sufficiently well-chosen vNM?

Your example of $p(b)I$ is bad for both these tasks.  Call it a
``measurement'' if you will.  I keep thinking ``intervention'' or
``act'' is better, but I'm not going to change 75 years of ingrained
language.

\bdm
Could it be that the postulate you have dropped is there because it
somehow maximizes this figure of merit?
\edm

That would be interesting if it were the case.  But I don't know of
any ways in which it is true.

\bdm
Is there some way to measure the ability of povms to discriminate
among all the possible density matrices a system might have which
shows that pvms are the most sensitive?  Or, conversely, is there
some natural measure of discriminatory ability for which a povm that
is not a pvm does better.  (I seem to remember an example in Peres
--- possibly even in your paper --- where a three outcome povm tells
you more about something or other than any possible pvm.)
\edm

Systems don't have density operators; people ascribe density
operators.  But what you have in mind is 1) above.

Actually, there is at least one more interesting way to think about a
POVM as being informative of a density operator.  (This way is more
conceptual than the two above, however.)  The space of Hermitian
operators is a $d^2$ dimensional vector space.  It turns out to be
possible to find POVMs with precisely $d^2$ linearly independent
elements.  For such measurements, the probabilities $P(b)={\rm
tr}(\rho E_b)$ uniquely specify $\rho$.  You could of course do the
same task by (conceptually) measuring $d+1$ standard observables
\ldots\ like the three {\Pauli} operators in the $d=2$ case. But
counting the total number of outcomes for this case (and lumping the
whole thing into a single POVM if you will), you would have $d^2 + d$
outcomes.  Thus there are ``minimal informationally complete POVMs''
if you move outside the von Neumann paradigm.  And that too makes
POVMs special over von Neumann ones.  (You can read about these kinds
of POVMs in our quantum de Finetti paper.)

\bdm
A second question.  I'm on page 20, and a student again.  My povm has
only one outcome (which always occurs, again, independent of the
initial density matrix rho).  It is just $E_1 = I$.  How am I to
understand the enormous range of possible final density matrices?
This would appear to be a measurement from which I learn nothing. Yet
it has an enormous capacity for altering the density matrix, which
encapsulates my knowledge.  What's going on?
\edm

There are good interventions, and there are bad interventions.  For
your example $E_1=I$ there are no ameliorating ways to do it.  You
can never end up with an increase in the purity of the state
describing the system.

\bdm
Your ground rules forbid you to tell me that the enormous range of
outputs have to do with the fact that I've turned on an interaction
that has entangled the poor system with an arbitrarily chosen
ancilla, so of course, if I've paid attention to what I've done to
it, the poor density matrix will change even if I subsequently
perform no test on the ancilla.  But what do you tell me?
\edm

The word intervention really is better than measurement.  (Did you
read the ``Penrose Tiles'' section I sent you to?)  Some ways I can
act on the world now will help me predict the consequences of my
actions in the future.  Some ways will not.  Quantum theory gives us
the full range.

\section{12-06-01 \ \ {\it Something I Wrote Once} \ \ (to R. Obajtek)} \label{Obajtek1}

Below is something I wrote a couple of years ago.  I reread it this morning. You do understand the implication of the Wheeler quote?  It's solely about self-confidence.

Let's just get started on your research project this morning.

\bq
\begin{center}
Undergraduate Research and Quantum Information
\end{center} \medskip

When I was an undergraduate at The University of Texas I had the opportunity to be associated briefly with the research group of John Archibald Wheeler.  Two things about Wheeler's style made a great impression on me.  First, he viewed research for both graduates and undergraduates in precisely the same light:  It's a frying pan and you've just got to jump into it! It got him results, and it trained a generation of excellent theoretical physicists. The second thing came in a question-answer session at the end of a seminar.  Someone in the audience asked, ``Do you see a difference between the students at Princeton and the students at UT?''  His answer was just as clean and as simple as it should have been, ``Yes I do; the students at Princeton {\it know\/} they're smart.  Next question.''

If you want to know my philosophy of how to advise research, then you need go no further than the paragraph above.  Great discoveries are waiting to be made at all levels of science.  And if there ever was a frying pan to jump into for the undergraduate, it is quantum information.  Some of the greatest discoveries of our field have been very literally at the level of a third-year undergraduate quantum mechanics course.  There is not a student who has studied Vol.~III of {\sl The Feynman Lectures on Physics\/} who could not have discovered quantum teleportation for himself.  There is not a student at that level who could not have discovered the idea behind quantum cryptography.  Wonderfully, these are not isolated incidents: there is a sense in which they define what the field is about.  The field is about looking at quantum mechanics in a new way and wringing everything we can from it.  The only tool a student really needs for a start in quantum information is to {\it know\/} that he's smart.

The bulk of present-day research in quantum information is truly an interdisciplinary effort.  Take quantum mechanics, computer science, information theory, and linear algebra, put them in a bowl and mix.  Because the field is in its infancy, the use of undergraduate-level ingredients from each of those disciplines is far from exhausted.
There is just so much fun work to be done; one cannot help but be thankful for the army of eager, questioning undergraduates that will teach their professors so much.
\eq

\section{15-06-01 \ \ {\it New Flight, New Letter} \ \ (to D. B. L. Baker)} \label{Baker2}

A new flight, a new letter to you, and still none in return since my last.  Do you ever read your email anymore?  I'm on my way to {\Vaxjo}---pronounced Vexsher by the Danish but not the Swedes; no English speaker alive can pronounce it like the Swedes---via Stockholm via London.  Of course, the flight took off an hour late and now I'm going to have a tight squeeze on my connection.

Do you ever read your email anymore?  I'll ask you again:  How are things going?  How's the little one?  Did you get the play area built that you were planning?  How's your significant(?)\ other doing?  How's your love life?  How's your intellectual life?  How's your health?  How's the stroke recovery progressing?  How's the heat in Texas?  How's your mom?  How's your sister?  How's your niece and nephew?  How's the job going?  What's the general feeling in Texas since G.W.'s taking office?  Does it still upset your stomach?  (It does mine.)

Do you ever read your email anymore?  Have you tried a new wine lately?  Do you have any picks you should tell me about?  Have you done any barbecuing in a while?  (I wish I had.)  Do you ever get to Austin?  Have you heard any live music of late?  (I wish I had.)  Did you get your taxes in on time?

Do you ever write your email anymore?  (Did I get you with that one?)  Have you heard from Michael D. in a while?  (I haven't.)  Do you know whatever happened to Terry?  Do you know if he really moved to Florida?  Do you know why I'm asking you this?  (I don't.)  Any other gossip of Cuero?  Any dreams of ever going to Turkeyfest again?  (I don't suppose I have any.)

Do you ever read your email?  Are you still driving the same car?  (I guess not.)  Are you still leasing a car?  (I bet so.)  Were you hit very hard with the Allison rains?  (I suspect not.)  Do you ever read your email anymore?

Do you remember the time I backed out of your driveway on French St.\ and hit a car parked across the street?  Do you remember the time you and I went to Port A in your Fiat just after you got it?  Do you remember the allure of the Pat Magee shirt?  Do you remember the time I met Pat Magee?  Do you remember the time we went to Troll Bridge and you beat your head on the roof of your car?  (Or was it my car?)  Do you remember the time we left Linda Henderson's trunk open in Houston and it rained?  Do you remember the time we sat in your car, parked in front of my house, and listened to an interview with Rush?  (Or was it my car?  And was it Rush?)  Do you remember the time we tried to find our way to the beach by following the signal of C-101?  Do you remember the time we were going to see Simon and Garfunkel and the hurricane came?  Do you remember how eerie it was in Cuero with all the refugees hanging out in the high-school parking lot?  Do you remember the blonde-haired girl who kept a diary and whose father worked at Texas Eastern?  (I bet you do.)  Do you remember the blonde-haired girl at UT orientation that I never had the nerve to talk to?  Do you remember Donna Slack?  (Is my statute of limitations out yet?)  Do you know whatever happened to Wendy G?  Do you know whatever happened to the girl that married Warner Scott (briefly)?  Or was that Scott Warner?  Or Scott Hamilton?  Scott Henry!  (Didn't I make that mistake before?)

Do you ever read your email!?!  Do you remember the wonderful taste of Malt Duck on an early morning cruise?  Do you recall why we thought we were the deepest thinkers in Cuero, TX?  (Besides it being a small world?)  Do you remember the wonderful three day party we had at your house on French St?  Do you remember who came?  (Do you remember who went?)  Do you remember The Nails?  Do you remember Debbie D?  (Or was that two D's?)  Do you remember the time you helped me write a paper for Linda H?  Do you still scorn me for getting so many kudos on that?

Do you remember our graduation day?  Can you tell I'm going to a quantum foundations conference?  Do you remember Craig Calk?  (Some things never leave us.)  Do you know whether Grunt ever became a dentist?  Do you remember the Rudolf christmas special?  Do you ever think of the girl you dated at UT?  (I can't think of her name right now.)  You used to.  But do you anymore?  And while I'm at it:  Do you ever read your email anymore?

Do you miss our old days?  When we hadn't grown so far apart?  (Read your email every now and then.)  Do you ever think of the Cuero Country Club?  With Gaye K. and Sheri T?  Do you think they could possibly look the same still?  Do you ever get to Stubb's BBQ anymore?  Do you think life is finite?  Or unbounded?  Do you remember a time some policeman in Cuero told me I was glassy-eyed and I told him that that was my contact lenses?  Do you remember the dry hamburgers you grilled for Kiki and me once?  Do you remember the time you came to {\Montreal}?  Could you forget the Vampire Lounge?

Dinner time.

Three little wines, each by my doorstep.  Swinging sweet song.  A melody pure and true \ldots\  Do you ever read your email?

Do you remember our first week at UT?  Do you remember the week before that?  Do you remember how we drove through the middle of Gonzales the day we left Cuero?  Do you remember the night we slept in the lawn of the Methodist Church?  Do you remember the Hobo Pack?  Do you remember Don Billings?  Do you recall the time the cat shit behind your bed and you wouldn't remove the evidence?  Do you remember your trip to Alaska with your uncle?  Do you remember any Simon and Garfunkel songs from beginning to end?  Do you remember the worn circle on the back pocket of Art Garfunkel's jeans in {\sl Concert in the Park}?

Do you remember any Madonna videos?  Do you remember what you told me about Rumble Fish?  Do you ever read your email anymore?  Did you ever understand why I mailed you that menu from a restaurant in Italy?  Do you remember the good life?  Do you ever think about the Cuero Livestock Show?  Do you think Brett Wright's belly is bigger than mine?  (You know, he and I share the same birthday.)

Did you know that Laura Lee had no eyebrows?  Did you ever notice at the time?  Did you have a look at my book on the web?  Did you ever pay that outstanding ticket you owed in some county near Houston?  Do you still watch Ally McBeal?  Dharma and Greg?  M*A*S*H?  Saturday Night Live?

Do you ever think of Janet Reno's waddle?  (Richard Fish does.)  Do you still pull evening shifts?  Or is it a 9 to 5 job now?  Do you still read so much?  Do you think ``a mind is a terrible thing?''  Do you remember the Moulin Rouge?  Do you ever wish you could recover it all?  All of what?  Do you ever get scared when a jet starts bobbing up and down 30,000 feet over the ocean?  (I do, and I am now.)

Do you ever think of the Swedish Bikini Team?  (Not me.)  Do you remember Mad Dog and Beans?  Captain Quackenbush's?  The Fajita King?  Uma Thurman?  Conan's Pizza?  The GM Steakhouse?  The Doll House?  Bunk Brantley?  Ken Adams?  Lisa Adams?  The chubby and skinny sisters that lived nearby?  Vitalis hair tonic?  Caskey Realty?  Life on Long Island?

Do you ever download your email?  Do you ever play the state lottery?  Do you remember the ferry to Port A?  And, who could forget Greta?  (She was Swedish, wasn't she?  Tell me it {\it is\/} so.)  Do you think there is a god in heaven?  Do you think there is a heaven?  Do you ever go to church anymore?  Do you remember John Hinckley Jr.?

Have you prayed for my sins?  Have you ever confessed to yours?  Twilight Zone.  Brando.  Helicopters.  Napalm?  Frog?  American Legion?  Deputy Sheriff Bobby Roberts?  Charlie's Angels?  The Mighty Mightor?  Sealab 2020?  Moby Dick?  Buck Sralla?  Can you forget the Thanksgiving morning I knocked on your door?  (Or was it the Friday after?)  Can you forget the girl from Yorktown?  (I don't think you knocked on my door.)  Can you remember a life before children?  Can you remember Immanuel Kant?  Can you remember the Square One?  (I just remember Elvis.)  And speaking of Elvis, can you remember an Icehouse in Houston, TX\@?  Elvis the Pelvis.  You know he had a brother, don't you?  Enis.

Do you remember Octogafest?  The morning after?  The Butthole Surfers?  Do you remember the mystical side of Coleto Creek?  Do you remember a swim in Fox Crossing?  Could you forget the Meyersville Country Club?  The graduation party the year before ours?  Could you forget Spider Man?  Sandy Boehne?  (I've forgotten how to spell her name.)  Life on Long Island?  (Again?)  I must know:  Do you ever read your email?  Can you find the reply button?  Is it spiked with needles and pins?  Poison tips?  Deadly fungus?  Fly paper?  The scent of a woman?

Do you think there is life on Mars?  Do you think there is permanence?  Do you think Bill Clinton's hands are big?  Is there any connection between a Woody and a woody?  (Some things are better left unasked.)  Have you ever worn a trench coat in San Francisco?  Have you ever read a Ginsberg poem?  Can you remember the poutine in {\Montreal}?  Are you afraid of fires?  Could you forget the Vampire Lounge?  (Never!)

Yeah, I did that on purpose.  Benny Williams?  His mat headed, blonde headed, bicycle riding girlfriend?  My crush on her?  My obsessive idealism?  Your cynical realism?  My love of Marx?  Your belief in ``human nature''?  Her name was Kathy---it finally came to me.  Do you think the Gonzales girl ever graduated from college?  The one you did the skinny dipping with?  Do you think Ricky Bluntzer ever amounted to anything?

Do you ever eat chicken fried steak any more?  Do you have a clear vision?  Do you have clear vision?!?!  Do you see the difference?  (Pun intended.)  Can you answer all these questions?  Will you answer all these questions?

Good night old friend,

PS.  Do you ever read your email anymore?

\section{24-06-01 \ \ {\it  Things You Need (a nonexclusive list)} \ \ (to D. J. Bilodeau)} \label{Bilodeau4}

Wootters email:  {\tt William.K.Wootters@williams.edu}.

Things to look up on {\tt quant-ph} for the question of why linearity:  Svetlichny, Gisin $+$ Buzek $+$ someone, Bru{\ss} $+$ Macchiavello $+$ someone.

More general things along a better track:  Wigner's theorem, Kadison's theorem.  Look in their books.  But also see some of the references below.

\bq
\begin{center}
{\bf \large Resource Material for an Information--Disturbance Foundation Principle}\smallskip

Christopher A. Fuchs

Norman Bridge Laboratory of Physics, 12-33 \\ California Institute of Technology \\ Pasadena, California 91125, U.S.A.\smallskip
%
\end{center}

\begin{enumerate}
\item
K.~Banaszek, ``Fidelity Trade-Off in Quantum Operations,'' {\tt quant-ph/0003123}.

\item
V.~Bargmann, ``Note on Wigner's Theorem on Symmetry Operations,'' J.
Math.\ Phys.\ {\bf 5}, 862--868 (1964).

\item
H.~Barnum, C.~M. Caves, C.~A. Fuchs, R.~Jozsa, and B.~Schumacher, ``Noncommuting Mixed States Cannot Be Broadcast,'' Phys.\ Rev.\ Lett.\ {\bf 76}, 2818--2821 (1996).

\item
C.~H. Bennett, D.~P. DiVincenzo, C.~A. Fuchs, T.~Mor, E.~Rains, P.~W. Shor, J.~A. Smolin, and W.~K. Wootters, ``Quantum Nonlocality without Entanglement,'' Phys.\ Rev. A {\bf 59},
1070--1091 (1999).  (The essence of this problem is actually an information-disturbance tradeoff.)

\item
S.~L. Braunstein, C.~A. Fuchs, and H.~J. Kimble, ``Criteria for Continuous-Variable Quantum Teleportation,'' J.\ Mod.\ Opt.\ {\bf 47}, 267--278 (2000).  (Don't let the title fool you:~This is really a paper about information vs.\ disturbance for coherent
states.)

\item
D.~Bru\ss, D.~P. DiVincenzo, A.~Ekert, C.~A. Fuchs, C.~Macchiavello, and J.~A. Smolin, ``Optimal Universal and State-dependent Quantum Cloning,'' Phys.\ Rev. A {\bf 57},
2368--2378 (1998).

\item
D.~Bruss, ``Optimal Eavesdropping in Quantum Cryptography with Six States,'' Phys.\ Rev.\ Lett.\ {\bf 81}, 2598--2601 (1998).

\item
P.~Busch and J.~Singh, ``L\"uders Theorem for Unsharp Quantum Measurements,'' Phys.\ Lett.\ A {\bf 249}, 10--12 (1998).

\item
R.~Cooke, M.~Keane, and W.~Moran, ``An Elementary Proof of Gleason's Theorem,'' Math.\ Proc.\ Camb.\ Phil.\ Soc.\ {\bf 98}, 117--128 (1985).

\item
C.~A. Fuchs and A.~Peres, ``Quantum State Disturbance vs.\ Information Gain: Uncertainty Relations for Quantum Information,''
Phys.\ Rev.\ A {\bf 53}, 2038--2045 (1996).

\item
C.~A. Fuchs, N.~Gisin, R.~B. Griffiths, C.-S. Niu, and A.~Peres, ``Optimal Eavesdropping in Quantum Cryptography. I. Information Bound and Optimal Strategy,'' Phys.\ Rev. A {\bf 56}, 1163--1172 (1997).

\item
C.~A. Fuchs, ``Information Gain vs.\ State Disturbance in Quantum Theory,'' Fort\-schritte der Physik {\bf 46}, 535--565 (1998).
[Reprinted in {\sl Quantum Computation: Where Do We Want to Go Tomorrow?}, edited by S.~L. Braunstein (Wiley--VCH Verlag, Berlin, 1999).]

\item
C.~A. Fuchs, ``Just {\it Two\/} Nonorthogonal Quantum States,'' in {\sl Quantum Communication, Computing, and Measurement 2}, edited by P.~Kumar, G.~M. D'Ariano, and O.~Hirota (Kluwer, Dordrecht, 2000), pages~11--16.

\item
C.~A. Fuchs and A. Peres, ``Quantum Mechanics Needs No `Interpretation','' Phys.\ Today {\bf 53}, 70--71 (2000).

\item
A.~M. Gleason, ``Measures on the Closed Subspaces of a Hilbert Space,'' J. Math.\ Mech.\ {\bf 6}, 885--894 (1957).

\item
E.~T. Jaynes, ``Predictive Statistical Mechanics,'' in {\sl Frontiers of Nonequilibrium Statistical Physics}, edited by G.~T.
Moore and M.~O. Scully (Plenum Press, New York, 1986), pp.~33--55.

\bq
\indent
Now let's look at that mind-boggling problem from a different side.
A single mathematical quantity $\psi$ cannot, in our view, represent incomplete human knowledge and be at the same time a complete description of reality.  But it might be possible to accomplish Bohr's objective in a different way.  What he really wanted to do, we conjecture, is only to develop a theory which takes into account the fact that the necessary disturbance of a system by the act of measurement limits the information we can acquire, and therefore the predictions we can make. This was the point always stressed in his semipopular expositions.  Also, in his reply to EPR he noted that, while there is no physical influence on $S$, there is still an influence on the kinds of predictions we can make about $S$.

With all of this, we can agree entirely.  The fact of disturbance of one measurement by another was equally true in classical physics (for example, one cannot use a voltmeter and ammeter to measure the current and voltage of a resistor simultaneously, because of this
``complimentarity'': however you connect them, either the voltmeter reads the potential drop across the ammeter or the ammeter reads the current through the voltmeter).

But in classical physics such limitations on our knowledge could be recognized and taken into account in our predictions without losing our hold on reality; for the separation into what was ``objective''
and what was ``subjective'' was never in doubt.  The coordinates and velocities remained ``objective'', while the ``subjective'' human information resided entirely in the probability distributions over them.  The probabilities could vary in any way as our state of knowledge changed for whatever reason; while the coordinates and velocities continued to obey the equations of motion.

Furthermore, the limitations on our ability to make measurements at the microscopic level did not prevent us from discovering the microscopic equations of motion, or from checking them accurately enough to discover their failure in quantum effects.  There are lessons in this for the present.

Could we make a theory of microscopic phenomena more like this which, while keeping a firm hold on what is ``objective'', also recognizes and represents explicitly the role of limited human information in the predictions it can make?  Such a theory need not, we think, contradict the successful parts of Bohr's theory; rather it would remove the contradictions that still mar it, thus fully realizing Bohr's goal \ldots \eq

\item
G.~Lindblad, ``A General No-Cloning Theorem,'' Lett.\ Math.\ Phys.\ {\bf 47}, 189--196 (1999).

\item
I.~Pitowsky, ``George Boole's `Conditions of Possible Experience' and the Quantum Puzzle,'' Brit.\ J.\ Phil.\ Sci.\ {\bf 45}, 95--125 (1994).

\item
I.~Pitowsky, ``Infinite and Finite Gleason's Theorems and the Logic of Indeterminacy,'' J. Math.\ Phys.\ {\bf 39}, 218--228 (1998).

\item
F.~Richman and D.~Bridges, ``A Constructive Proof of Gleason's Theorem,'' J. Func.\ Anal.\ {\bf 162}, 287--312 (1999).

\item
C.~S. Sharma and D.~F. Almeida, ``A Direct Proof of Wigner's Theorem on Maps Which Preserve Transition Probabilities between Pure States of Quantum Systems,'' Ann.\ Phys.\ {\bf 197}, 300-309 (1990).

\item
U.~Uhlhorn, ``Representation of Symmetry Transformations in Quantum Mechanics,'' Arkiv F\"or Fysik {\bf 23}, 307--340 (1963).

\item
E.~P. Wigner, ``The Probability of the Existence of a Self-Reproducing Unit,'' in {\sl The Logic of Personal Knowledge:~Essays Presented to Michael Polanyi on his Seventieth Birthday}, (Routledge and Kegan Paul, London, 1961), pp.~231--238.

\end{enumerate}
\eq

\section{25-06-01 \ \ {\it Bohr and Bits} \ \ (to H. J. Folse)} \label{Folse6}

It was wonderful to meet you finally.  But, my deepest regret of the meeting was in not getting a chance to really talk to you!  I think I should dig up a way to drop by Loyola sometime.

I'm in London right now, slowly working my way back home, taking care of some delinquent tasks, and adding some more words to my compendium.  So, let me send you a little reminder.  When you get back from your vacation, can you send me a more complete copy of the paper:
\begin{itemize}
\item
H.~J. Folse, ``The Formal Objectivity of Quantum Mechanical Systems,'' Dialectica {\bf 29}, 127--136 (1975).
\end{itemize}
Also, if you could supply me with the full list of editors for the volume below (including a description of the needed diacritical marks), that'd be greatly appreciated.
\begin{itemize}
\item
H.~J. Folse, ``Realism, Idealism, and Representation in the
Description of Nature,'' in {\sl Avartuva
  ajatus:\ Julkaisutoimikunta}, edited by U.~Ketvel and ?? [$=$ A.~Airola,
T.~Kallio-Tamminen, S.~Laurema, K.~Rainio, and J.~Rastas]
(Luonnonfilosofian seuran, Espoo, 1999), pp.~73--77.
\end{itemize}
I'm shooting to get a completed version of the compendium out before the end of the year.

Also I'm getting much closer to giving you some feedback on all your papers.  That should arrive in the form of a longer letter sometime in the coming month.  My remark after your talk was a hint of the general difficulty I'm having with your version of Bohr.  If I can ask enough testing questions to prod you into trying describe ``the quantum postulate'' (and a few other things) from a new angle, then maybe I will be filling a useful task.

{\Caves} and I really enjoyed the format of your ``Bohr's Best Bits.''  In particular, the boldface to let us know what you thought was most important, while keeping the remaining text to build a context.

\section{25-06-01 \ \ {\it London Calling} \ \ (to D. J. Bilodeau)} \label{Bilodeau5}

I just arrived in London.  Now I've got two hours of time to recharge my battery before boarding the next plane.

I finally finished rereading all your old emails to me.  As before, I came away struck by your many deep insights.  Let me just make a few remarks to leave a record behind of the morning's efforts.

\begin{itemize}
\item
31 August 1999.
I'm still a little confused on the distinction you want to draw between ``object'' and ``system.''  As you have some time, would you still be willing to give it one more shot?
\item
14 September 1999.
Ditto.
\item
13 October 1999.
The other day I told you that I think you got the professor in the von Neumann story wrong.  But it was I who was wrong!  Looking back, I see you couldn't remember his name in 1999.
\item
9 November 1999.
I liked your point about Bohmian ``non-mechanics'' (as Hiley might insist I call it).  I'm wondering if you tried out the point on Hiley?  If you did, what was his reply?
\item
17 January 2000.
A most interesting letter.  Especially as regards the introductory remarks and the material from Toulmin.  Can you give me a full citation of Toulmin's book {\sl Cosmopolis}?  I may have appreciated your/his remarks about real machines being nonideal for the first time.  Your comments on my paper with Peres were certainly on the mark:  It was exactly on those points that I got the most flak from the readers.  But, on the other hand, there was little I could do:  The paper was a compromise.  Peres is far, far, far more positivistic than I am.
\item
8 March 2000.
I especially enjoyed your discussion of Axioms 0 and 1.
\item
15 March 2000.
Good food for thought again.  You said, ``Kierkegaard said subjectivity is truth.''  Can you find the source of that quote?  I especially enjoyed your remarks on time.
\item
22 March 2000.
   ``Impressive scores!  I'll have to co-author a paper with you some time.''
Yes, let's do it!
\item
14 April 2000.
``I agree that the Hamiltonian is not the `solid bedrock' of QM. [\ldots]\ Its solidity is just a reflection of its constancy for a certain class of problems.  It doesn't get spookily entangled -- by definition.''  Yes it can.  That was exactly one of my points to Caves.  Hamiltonians can be ``toggled'' from a distance (and thus must be subjective properties, just as state vectors).

Concerning Roger Newton, now I understand how he had an influence on your 1996 paper.  I didn't realize he was at IU.

``The above is rather vague, so I will attempt here and now to sketch out the bones of a more explicit account of what I'm getting at.''
I very much enjoyed the sketch, but now I want even more!  Want to give it still another shot?  I've always got an open ear for beautiful ideas.
\item
15 September 2000.
I still need to look up \arxiv{gr-qc/0009023} on ``Process Physics.''
\item
17 January 2001. Can you give me the full references for the Lakoff books?
\end{itemize}

It was great to finally meet you.  Keep those quantum thoughts flowing.  With the new day, I'm once again energized that the end of our quantum troubles may be just around the corner.

\subsection{Doug's Preply, 31 August 1999}
\bq By my records, I've had 12 email responses to this paper [``Why
  Quantum Mechanics is Hard to Understand,'' \quantph{9812050}],
including yours, of which 5 were definitely positive, one was a
neutral comment suggesting I look at another reference, one had no
comment on my own work but was trying to sell me on the other guy's
ideas (which I couldn't make sense of), one from the British science
writer John Gribbin, and four who wanted me to send them a hard copy
or needed help in downloading.

John Gribbin wrote to say, ``But quantum physics is \emph{not} hard to
understand!  It runs counter to commonsense, but so does the idea that
the world is round.''  I wrote back about a page worth explaining my
ideas in brief and why I didn't think his example applied here.  He
answered, ``Hmm.  Maybe I should read the whole paper, not just the
abstract!''

One of the positive responses was from Tom Siegfried, a science writer
for the {\sl Dallas Morning News.}  He wrote a column about the paper
which appeared Jan.\ 4 of this year.  If only half the people in the
Dallas metro area subscribe to the DMN and even only 1 in 1000 is a
regular reader of the science column, then at least 1000 people read
an account of my ideas, a few of whom may have been physicists!
\smiley\ So this paper has had a pretty good run already.  It would be
nice to get it published in a regular journal just for the record and
for citation.  But I'm much more focussed on the next paper, which
will completely supersede this one.

[\,\ldots]

OBJECT is a practical, functional concept, not ontological or
constitutive in a mechanical sense.  On the other hand, paradoxically,
our whole concept of ``objective'' reality is based on ``objects''
(naturally) and all we can know or understand of the ``real'' structure
of the physical world must come by way of them.  Hence the need for
Kantian subtlety.  I know that's not very clear, but I'm working on
it.  For now, I just want a general quantum mechanical method
applicable to all cases which spells out the relationship between
objects and systems.  So I'm in the process of going through a number
of simple but real-world applications of QM, trying to find an optimal
general system for describing what's going on.  What people usually do
is fall back on the concept of particle, and treat the particles as a
classical objects whenever possible, sliding back and forth between
quantum and classical properties as needed.  But that is neither clear
nor consistent nor applicable to all cases.

I think you're right that information theory is tied to quantum
foundations.  But not in the sense that Wheeler meant with his ``IT
FROM BIT'' slogan, if I understand him.  Information (like objects
themselves) is a feature of the way we experience and interact with
the world.  I like your simple summary of relativity.  We do have to
do the same with QM -- the opposite of complex formalizations like
quantum logic.  But even Axiom~0 is not without difficulties.  ``Exist''
is a loaded word.  There are phenomena correctly described by quantum
systems, but people tend to confuse a system with an object or a
component of a mechanical world.
\eq

\subsection{Doug's Preply, 9 November 1999}

\bq
Your email comes at a good time.  I had been distracted by other
things for 2 or 3 weeks and was just ready to get back into thinking
about QM, and your letter provides a strong stimulus to do so.

Very sorry to hear about the salmonella.  That can be quite serious.
I hope you have recovered completely by now.

I do think that our views are very close; at least we are motivated by
many of the same insights.  The main difference is that I am still
leery about ``information'' as a foundational concept.  But I have an
open mind about it.  I think most of the conceptual confusion in QM
today is related to the inadequacy of the idea of ``particle'', which
carries with it too much of the old Cartesian notion of the
geometrical basis of physical existence.  Feynman has greatly advanced
our understanding of quantum physics, but his attachment to the
particle idea has persisted and made it more difficult to advance
further conceptually.  Perhaps the concept of information can help to
clarify the dynamical rather than geometrical/ontological nature of
the quantum.  [I'm actually very indebted to Feynman.  I had a pretty
  lousy freshman course in physics, because I was an astrophysics
  major at the time and also taking an honors math sequence, and the
  only first year physics course compatible in the schedule with
  astrophysics and honors math was the non-calculus course taught to
  pre-med students.  In the spring, however, I bought vols.\ 1 and 2
  of Feynman's {\sl Lectures on Physics} (which had just been
  published) and spent the whole summer devouring them.  I consider
  that to be my real introduction to physics.]

To answer your first question: When I wrote the comment in the JCS
paper about freedom of will, I was saying that to establish causal
connections or verify patterns in phenomena, it is necessary that the
actions of the experimenter not be correlated with the contingent
details of the phenomena under investigation.  To know what ``free
will'' in an absolute sense means is a problem more subtle and
difficult even than quantum mechanics.  It is possible that a lack of
correlation or the idea of measurement in general will turn out to
imply some kind of property of ontological independence in the
observer, but for the purpose of doing physics, I think it is
sufficient to assume that the behavior of the observer/experimenter is
not determined by some external agency which also controls the
phenomena being observed -- e.g., that the brightness of a star does
not change just because I decide to look at it or vice versa.  In any
case, the observer can be that way and still be free or deterministic,
I think.  The whole question of free will is perhaps not well posed.
I have read some of the writings you mention, but will say more in a
couple of days when I have had a chance to digest it better.

Re: your prospective article for {\sl Physics Today.}  I have long thought it
would be possible to blow Bohmian mechanics and many-worlds out of the water
with a simple analysis of what physics does and why those viewpoints were
introduced and how they fail to accomplish what they set out to do.
I think they fail pretty drastically and cover up their failures with
obviously feeble rationalizations.  I'll try to say more about that in a
couple of days, too.  Consistent histories I'm not sure that I really
understand at all.  Accounts I've seen in several papers seem really opaque
and poorly motivated to me, and I don't see that it means anything except in
simple cases where it reduces to ordinary calculation of Feynman amplitudes.
So I don't know how to critique it.

One point I can make now about Bohmian mechanics.  A lot of people seem to
be impressed by the figures of the quantum potentials for the two-slit
experiment (as for example in Bohm and Hiley, pp.\ 33--34). The trouble is
that the quantum potential is not like a true potential at all, and not just
because of non-locality.  In electrodynamics, the electric potential
describes the environment of a charged particle which is influenced by the
resulting electric force.  The field/potential can be taken as an
independent local physical entity or as a representation of the effect of
surrounding charges, but the distinction between the particle and its
environment is crucial to the explanatory power of the idea of the
potential.  Bohm's quantum potential for the two-slit experiment depends not
just on the configuration of the slits, but also on the de Broglie
wavelength of the particle and hence on its momentum.  It is not just the
interaction between particle and potential or the resulting force which
depends on the particle momentum (as with the Lorentz force), but the
potential itself.  The potential thus does not represent the environment of
the particle -- it is just another description of the whole dynamical
situation and adds nothing new.  Typically, both Bohmian mechanics and
many-worlds confuse the system with its environment or context in a way that
shows no physical insight at all. Bohm is trying to defend the
meaningfullness of particle trajectories; but if the particle is bouncing
around inside his potential, what is it which carries as an attribute the
momentum which determines the separation between the peaks and troughs of
the potential?

I don't know if the clarification of QM will lead to new physics or not.
If I had to bet, I'd guess the odds are 50/50.  I hope it will.  But anyway,
I think we should proceed with the assumption that it will.  That helps to
keep clarification from degenerating to trivial and unresolvable
controversies.  And if it does -- the payoff will be really great.
\eq

\subsection{Doug's Preply, 8 March 2000}
\bq
\noindent{\bf ``0. Systems exist''}

I think one of the more interesting differences between quantum and
classical mechanics is that a classical system is a set of objects which are
pretty much open to our inspection all the time -- in principle, every value
of every coordinate or other parameter can be measured precisely at every
moment;  a quantum ``system'', however, (it's really a different kind of
animal altogether) is a part of a unified flow which includes ourselves and
our acts of observation and cannot be measured at all, really.  All we have
are expectations for a certain sort of phenomenon on the basis of
information available to us and the facts we can ascertain about these
phenomena when they do occur in particular instances.  Our relationship to
the underlying stuff of nature is completely different.
Not only do our ideas of measurement change; our understanding of the
categories of matter, space, time and cause change.  We need a language to
describe quantum phenomena which will make these changes clearer.

\medskip

\noindent{\bf ``1. States correspond to density operators over a Hilbert Space $H$.''}

Why a Hilbert space rather than a configuration or phase space more
obviously related to the space and time of common experience?  Because QM
does not describe events going on in space and time.  It describes the
relationship between expectations and outcomes.  Perhaps here is where
``information'' enters.  Expectations for a certain physical situation are
based on our prior knowledge (interestingly not an initial state at some
time $t$ but all prior knowledge).  Outcomes are categorized according to what
we can measure -- the sets and symmetries of possible results.  We impose
our mechanical way of thinking on nature -- mechanical in the sense of
techniques and devices for achieving goals.  None of the conceptual
apparatus of physics is given, inherent in nature.  We can understand it
only by adding a framework of knowledge and purpose -- what is it about this
situation which tends to produce a certain result?  What can we change in
the setup to change the results in a certain way?  Hence the ideas of cause,
force, inertia, etc.  Information is artificial; so is dynamical analysis.

\medskip

\noindent{\bf ``2. Measurements correspond to positive operator-valued measures on~$H$.''}

Feynman's book {\sl The Theory of Fundamental Processes} influenced me
quite a bit.  In chapter 2, he is talking about gamma ray emission
from a nucleus: ``If the nucleus could be characterized by a single
amplitude, say, its energy, then the gamma ray would have to be
emitted with equal likelihood in all diretions.  Why?  Because
otherwise we could set things up so that the gamma ray comes out in
the $x$ direction (for we can always rotate the apparatus, the working
system; and the laws of physics do not depend on the direction of the
axis).  This is a different condition because the subsequent
phenomenon (gamma emission) is predicted differently.  One amplitude
for our state cannot yield two predictions. The system must be
described by more amplitudes.  If the angular distribution is very
sharp we need a large number of amplitudes to characterize the state
of the nucleus.''  From symmetry, he finds a relationship between
information (number of amplitudes required to describe the state) and
outcome (degree of isometry).  Feynman amplitudes may be a more
natural way to explore these relationships than a traditional state
description.  BTW, I think it was reading the problem at the end of
chapter 1 of that book (about interference of light from two stars
with small angular separation) that started my investigation the
foundations of QM many years ago.  How could a process in which the
photon seen by detector $A$ originates in star 1 and the photon seen at
detector $B$ originates in star 2 interfere with a process in which the
origins are reversed?  It seemed to involve an uncertainty in photon
position over astronomical distances, which was counter to my
intuition at the time.

\medskip

\noindent{\bf ``3. H is a complex vector space \ldots''}

Is the relationship between amplitudes and probabilities just an
irreducible feature of quantum theory, or can we tie it to something
more fundamental?  I don't know.

\medskip

\noindent{\bf ``4. Systems combine by tensor producting their separate vector spaces\ldots''}

[Ouch!  My ear trained by old-fashioned school-marm grammarians allows
``multiplying'' but not ``producting''.  But then I also can't stand the word
``read'' as a noun; to say a book is a ``good read'' is like fingernails on
chalkboard to me.  I know the usage goes back to the 18th century or so, but
it has become common only in the last 10 or 15 years(?).  I hate it, but all
blurb writers seem to love it.]

\medskip

\noindent{\bf ``5. Between measurements, states evolve according to trace-preserving
completely positive linear maps.''}

I.e. simple rotations in Hilbert space?  A change in point-of-view?
Maybe both this and \#2.\ fall out from symmetry considerations once we
have the principle of amplitudes and (trickier) what constitutes a
process to which an amplitude may correspond.
\eq

\subsection{Doug's Preply, 14 April 2000}
\bq
The story about Wheeler bidding the equations to ``fly'' (p.\ 57) was great.
It touches the core of what I want to see in physics.  Some days I also feel
that ``the possibilities in our world will open up like a blooming flower.''
Other days I feel lazy and think, ``Who am I to remake physics?''
Something that radical is needed.  As the Pauli quote indicates (p.\ 61), our
physics is still ``semi-classical''.  The solution is, as you say, not another
set of equations -- flight, time, living awareness will never fall out of a
pure formalism.  We need to remake the physical presumptions which are prior
to the equations.  ``It flies'' has to be put in from the start.  The way to
do this, I think, is to formulate a clear, limited set of concepts which
accurately embody the fundamentals of what is going on in action,
observation and interaction, with no compromises -- no shifting meanings or
hidden reservations.  Then we build up everything from there.
We still can hardly help thinking of the world as a sequence of static
configurations (cf.\ your ``static portrait of `what is'\,'' on p.\ 63), like
movie film (except that there are $\aleph_1$ frames per foot).  Such moments
are not really accessible to us, and in QM become meaningless.
\eq

\subsection{Doug's Preply, 15 September 2000}
\bq
After a long delay (during which I've been sadly distracted from quantum
thoughts entirely for the most part), I return to comments on your
compilation of correspondence: ch.\ 6, Bob Griffiths.

Before I read your comments, I went back and read an early paper by
Griffiths on this topic:  the one from 1984 called ``Consistent Histories and
the Interpretation of Quantum Mechanics''.  The main thing I noticed was some
confusion about what constitutes ``a sequence of events in a closed system''.
There is first of all the usual naive or uncritical notion that ``events in a
closed system'' should be a fairly straightforward concept in physics (as
simultaneity was once?)  Then there is some ambiguity as to whether an
``event'' should be something phenomenal or classical on the one hand (a
particle is here, or a particle is found here at this time), or something
from the state description formalism (system $A$ is in the state $B$, etc.).
This ambiguity is especially prominent in the Hartle \& Gell-Mann version of
consistent histories, which is what makes their discussion seem so vacuous
to me.

Griffiths says on p.\ 232 of his paper, ``The collapse employed in
calculating the numerator in~(2.12)'' [he says this in defense of his
  method as immune from the paradoxes of the measurement problem]
``obviously has nothing to do with measurements (though it can be
connected with a sort of ideal `measurability;; see Section~5), since
it occurs at a time $t_1$ \emph{before} the particle has interacted with
either of the counters.''  But the argument is phony.  Any use of
projection operators or any of the state apparatus of quantum theory
implies measurement -- states and operators have physical meaning only
in a context of measurement or a set of possible measurements.  A
sequence of states as a ``history'' may seem like a good candidate for
an ``element of reality'', but only as long as we confuse states with
observed outcomes which can be associated with the states only in
certain contexts.

I found that I agreed almost entirely with your comments to Griffiths,
especially your criticism of the use of the conjunction relation in quantum
logic and of the status of propositions for which there are no experimental
means of discovering their truth values.

I'm curious where the quote from Heisenberg comes from -- ``\ldots The
representation \ldots\ so to speak, contains no physics at all.''

I also liked your reply (with Asher Peres) to Griffiths in the latest
{\sl Physics Today.}  You express very well, I think, the crucial point that
the lack of an objective \emph{description} of reality is not the same
as the lack of an objective reality itself, and that the appearance of
quantum non-locality arises from looking at quantum phenomena with
classical, realistic prejudices.

The most pervasive classical prejudice is the uncritical way of
understanding the concept of particle which most physicists still
suffer from.  These days my immediate goal is to analyze the
registration events which Bohr focussed on as the key to physical
reality; not to understand ``amplification processes'' in terms of
elementary quantum entities, as some might do, but rather going the
other way and determining what sort of meaning we can give to the
concept of particle on the basis of the kinds of phenomena we can
observe.

I was somewhat disappointed with my investigation of the Feynman path
integral apparatus.  It is certainly a useful and powerful method, but all
the conceptual problems are swept into the calculation of the amplitudes.
There seems to be no advantage to approaching quantum foundations in this
way.

An interesting paper appeared on the \texttt{gr-qc} arXiv recently
(\href{http://arxiv.org/abs/gr-qc/0009023}{\texttt{0009023}}) with the
title ``Process Physics: Modelling Reality as Self-Organising
Information''.  The first page raised a lot of the right questions, I
thought (and echoed a few of the speculations I was raving about back
in May), but unfortunately after that the authors drifted off into a
logical/mathematical fantasy with no grounding in physical phenomena.
\eq

\section{27-06-01 \ \ {\it Gleasons, and Me at AT\&T} \ \ (to H. Barnum)} \label{Barnum3}

\bhb
I must apologize for still having only skimmed the paper you gave
me to read.  I really liked the Gleason's theorem for POVMs.  I used to
think such a thing would be superfluous except for 2D, since we have it
for orthonormal frames in 3D and hence, by extension, for POVMs in 3D and
higher.  The lack of it for the rationals didn't bother me because I
always thought the difference between the rationals and the reals is
``unphysical.''  (I think we discussed this business a bit in Los Alamos in
the fall of '99, and in fact you guys may have had the theorem by then, and
it was part of the discussion.)  One would need to use some kind of
continuity of states (sets of probability distributions for each possible
measurement) in the way we represent them.  This would impose continuity
of frame functions as a requirement, which I think is reasonable despite
the fact that Gleason can do without it.  Now I never tried to think about
what that would do in the rational case (I guess maybe one could define
continuity there?).  But I think it might well require that frame
functions on the rationals extend continuously
when the rationals are embedded in the reals.

Now I think the POVM version is really quite important (just how
important, I am still trying to decide). It provides another, perhaps
better, way of getting at this sort of issue.  The thing is that the
unphysicality of the difference between the rationals and the reals is
closely related to the idea that we can never measure a true orthonormal
frame \ldots\ we always measure a fuzzed version of it.  All measurements are
measurements of POVMs.  So, in a sense, what should really be proved is a
Gleason's theorem {\bf whose proof uses only finite strength (in the
terminology of you \& Kurt Jacobs) POVMs}.  I.e., one which holds for the
physical model in which {\bf all} measurements are finite strength POVMs.  So
one should look at your proof and Busch's and see if they only use finite
strength POVMs.  (It seems like it ought to be possible to modify them to
do so, if not.)
\ehb

Yes, it will still work there.  The finite strength measurements simply corresponds to the set of POVMs but with the boundary removed.

\bhb
Here's an interesting idea to explore within the
Bayesian/operationalist view of quantum theory.  Okay, states are
essentially (whatever else they may correspond to) summaries of our
knowledge about a system insofar as it's relevant to predictions of future
measurement results.  Measurements (POVMs) similarly have a subjective
aspect:  what measurement we think we're performing depends on the
``state'' (subjective) of the apparatus we start out with, and {\bf our beliefs
about} the interaction dynamics, etc \ldots\ I think some of that
subjectivity is irreducible, in that we won't want to say ``well, we didn't
know what POVM we were doing, but now we do''.  We can sometimes {\bf view}
some POVMs as ``convex'' combinations of some others, in a sense such
that we learn both which of the sub-POVMs was done, and the result of the
sub-POVM.  But that, ex ante, is still the original POVM.

So, the fuzz in a fuzzy measurement results from our incomplete knowledge.
\ehb

Watch out:  you'll start to tweak Carl's greatest fears with this.

\section{27-06-01 \ \ {\it Invitation} \ \ (to A. Y. Khrennikov)} \label{Khrennikov2}

\bakh
I also plan to work in the direction of $p$-adic [Gleason] theorem, but
it is not so simple as we discussed \ldots
\eakh

I'm really glad to hear this.  More than anything, I would like to get down to the bottom of why this theorem works.  It works for the reals, but not the rationals.  What crucial new property comes into play there?  It also seems to work when the field is the algebraic numbers.  So, it is not just having a completion of the rationals that makes it fly.  The more examples/counterexamples we have, the more clear the picture will become.

\bakh
I plan to write a paper in that I shall criticize your Bayesian
approach to quantum probabilities.  But I think it would be not so
bad. You could reply and there could be a good exchange of viewpoints.
\eakh

In a way I'm glad to hear this too.  Every opportunity for a clarification of the crucial issues is a welcome opportunity.  One thing I hope you will keep in mind though in building your criticism is that:  Even though Caves, Schack and I see no escape from a subjectivist account of quantum probabilities, that does not mean we strive for a subjectivistic account of Nature.  The issue is, instead, that quantum mechanics seems to be a jumble of epistemic and ontological entities.  The task we see is to disentangle those two aspects of the theory, with the main goal being to lay bare the ontological content.  Getting straight the epistemological nature of quantum probabilities---we believe---is the first step toward a greater goal.

I will write you some notes on your context-dependent probabilities before the day is out tomorrow.

\section{28-06-01 \ \ {\it Amherst-Area Visit}\ \ \ (to A. G. Zajonc \& K. Jagannathan)} \label{Zajonc2} \label{Jagannathan2}

I will be in the Amherst area July 1--3.  Would either of you be interested in my dropping by the college for a short visit?  (July 2 would probably be the best date for that, but the morning of the third would be OK too.)  I could also give an impromptu talk if there's any interest:  I've recently been involved in three very foundational theorems that might titillate you.  (I'll put a trouble-making abstract below that I wrote recently for a quantum foundations conference.)

\bq\noindent
Title:\\
Quantum Foundations in the Light of Quantum Information Theory \medskip\\
\noindent Abstract:\\
In this talk, I hope to cause some good-natured trouble.  The issue at stake is when will we ever stop burdening the taxpayer with conferences and workshops devoted---explicitly or implicitly---to the quantum foundations?  The suspicion is expressed that no end will be in sight until a means is found to reduce quantum theory to two or three statements of crisp physical (rather than abstract, axiomatic) significance. In this regard, no tool appears to be better calibrated for a direct assault than quantum information theory. Far from being a strained application of the latest fad to a deep-seated problem, this method holds promise precisely because a large part (but not all) of the structure of quantum theory has always concerned information.  It is just that the physics community has somehow forgotten this.
\eq

\section{28-06-01 \ \ {\it Context Dependent Probability} \ \ (to A. Y. Khrennikov)} \label{Khrennikov3}

\bakh
It was nice to meet you in {\Vaxjo} and discuss fundamental problems
of quantum theory.  Unfortunately, I have the impression that my
presentation on Contextual Probabilistic Interpretation of quantum
theory was not so clear for participants (conversations during
lunches and dinners). I try to present my views as short and clear as
possible.
\eakh

Thank you for valuing my opinion on your ideas; I am flattered.  So I
treated the problem in a conscientious manner:  I downloaded and read
three of your papers (\quantph{0103065}, \quantph{0105059}, and \quantph{0106073}).

I am indeed quite intrigued by the possibility that quantum mechanics
may be nothing more than a calculus for comparing probabilities when
the experimental context cannot be deleted from the results it brings
about.  In vague philosophical terms, I think this is precisely the
kind of idea Bohr, Heisenberg, and {\Pauli} were bandying about in
constructing their interpretation of quantum mechanics.  It is
certainly the kind of notion Bohr was trying to get at with his
emphasis on ``complementarity.''  So I would welcome a more precise
way (a mathematical way) of expressing the essence of all this.  I
myself have been attracted to this sort of thing for a long time:  it
is a large part of the thread connecting my ``Notes on a Paulian
Idea''---that is, that the observer sets the context, and, in the
words of {\Pauli}, cannot be ``detached'' from what he finds. Also you
can find discussions of it in Sections 4 and 8 of the large paper I
was circulating at the conference, ``Quantum Foundations in the Light
of Quantum Information.''  I say all this to make it clear that I am
more than sympathetic to your program.

However, as much as I would like to tell you otherwise (because you
are my friend), I do not see that your present formulation of the
problem moves very far toward quantum mechanics in a convincing way.
There are problems on at least two levels.

Maybe the most devastating and immediate is your move between Eqs.\ (5) and (6) of \quantph{0106073}.  (I'll focus on that paper
for specificity since I did not see you make a stronger argument in
either of the other two papers.)  You write:
\bq
\noindent
The perturbation term $\delta({\cal S},{\cal S}^\prime)$ depends on
absolute magnitudes of probabilities.  It would be natural to
introduce normalized coefficient of the context transition \ldots\
\eq
The question anyone will ask is, ``Why is this natural?''  What
compels the precise form of the normalization other than that it
forces the equation to look of a more quantum mechanical form.  Why
did you choose the square root rather than the third root, say?
Indeed, why not divide by the absolute value of $\delta$, or the
exponential of $\delta$, or any other combination of functions one
could pull out of a hat?  To put it not so gently, it looks as if you
built the desired answer in at the outset, with little justification
otherwise.

The second level of my problem is that, even if you do get this far,
how do you make the further step to vector space representations of
quantum mechanics?  Why are observables POVMs and not other exotic
entities?  What leads us to the starting point of Gleason's theorem?
Etc., etc.?  I don't see that you have enough structure to do that.
But more importantly, until you have done that I would have to say
that your theory remains fairly empty in making a connection to
quantum mechanics.  Too empty.

The way I view the problem presently is that, indeed, quantum theory
is a theory of contextual probabilities.  This much we agree on:
within each context, quantum probabilities are nothing more than
standard Kolmogorovian probabilities.  But the contexts are set by
the structure of the Positive Operator-Valued Measures:  one
experimental context, one POVM.  The glue that pastes the POVMs
together into a unified Hilbert space is Gleason's ``noncontextuality
assumption'': where two POVMs overlap, the probability assignments
for those outcomes must not depend upon the context.  Putting those
two ideas together, one derives the structure of the quantum state.
The quantum state (uniquely) specifies a {\it compendium\/} of
probabilities, one for each context.  And thus there are
transformation rules for deriving probabilities in one context from
another.  This has the flavor of your program.  But getting to that
starting point from more general considerations---as you would like
to do (I think)---is the challenge I haven't yet seen fulfilled.

I very much hope that I have not offended you with these comments. I
greatly respect your program.  But because of that I want much from
it.  I want it to stretch our understanding.  John {\Wheeler} used to
say, ``We must make as many mistakes as we can, as fast as we can, or
we'll never have a hope of gaining a true understanding!''  I let
that philosophy rule my research life.  Thus I can only commend you
for your exploration, and hold the strongest hope that something firm
will come from it with a little more work and contemplation.

\section{28-06-01 \ \ {\it The Oblique Observer} \ \ (to N. D. {\Mermin})} \label{Mermin21}

You haunt me so that I wake up in the middle of the night just to
spar with you.  It's this damned noncontextuality of quantum
probabilities.  If one walks into the game with the {\it firm\/}
belief that a quantum state is a state of knowledge, then
noncontextuality is almost a given.  That, in part, is what my
section ``Whither Bayes' Rule?'' is about.  Moreover, all questions
about instantaneous signaling through quantum-state change just
become silly:  such questions spring solely from a wrong-headed view.

But, you do not walk into this game with the firm belief that a
quantum state is a state of knowledge.  So I am forced to work to win
your approval.  In the end, with clarity achieved, your demands will
have been a great gift.  Right now, they are just annoying.

In any case, this leads me down the path of the oblique observer.
There is a way in which a von Neumann-like view of measurement may be
a virtue.  That is, a view of measurement where, to make it go, I
first introduce an ancilla to interact with the system.  Then I
introduce a second ancilla to interact with the first ancilla because
I didn't know how to solve the measurement problem at the new level.
Then I introduce a third ancilla, and so forth.  Von Neumann went to
this extreme because he was chicken to let the mental update that is
a measurement fall outside of physics (as Thomas Bayes would have).
So, he piled up superobserver after superobserver, just so there
would always be an outside view.  But so be it:  Let us glean what we
can from this freedom of our descriptions.

The main point is this:  You pick any measurement (any POVM) you wish
for some system, and I can always think of a way to get at that
measurement in an indirect way.  That is to say, I can always delay
my cutting of the Gordian knot until I get to a system with no
residual causal link to the one I'm really interested in.  In
language I don't like, it means I can always induce the ``physical
collapse'' somewhere else.  I can always push the measurement to an
arena where you would take the noncontextuality assumption (on the
system of interest) as a given.  My direct measurement on an ancilla
serves only to refine my knowledge about the actual system:  the
actual system cannot care how I came to that refinement, or, indeed,
if I ever pursued refining my knowledge at all.

You get my point---with that much repetition, you'd better.  This
gives rise to a vague idea that perhaps you can help me elicit into
reality.  (And save me some torture that you are the root of.)
Forgetting about the precise structure of quantum mechanics, why
should we not view all observations as oblique observations?  Whose
philosophy ever dreamt that we had {\it direct\/} access to the
minutest details of the world in the first place?  (I'll paste in the
great Heisenberg/Einstein quote below so you'll have quick access in
case you wish to remind yourself of it---especially the fourth, fifth,
and sixth sentences.  [See 05-05-01 note ``\myref{Peres12}{Author's
    Reply}'' to A. Peres.])

Moreover, coming back to quantum mechanics in particular, what is to
keep me (in my derivation of the tensor product rule) from thinking
of the two separate observations as each concerning the opposite
system?  The question is, can one get some quantitative mileage from
this.  At least that's the question on my mind as the sun is rising.
Any thoughts?

It's a good thing you left the {\Vaxjo} before Friday's lunch.  The
boiled potatoes were ridden with rocks:  I lost an eighth of a molar
in the process!  What was your overall impression of the meeting
(good and bad)?

\section{28-06-01 \ \ {\it The Kettle Black} \ \ (to N. D. {\Mermin})} \label{Mermin22}

I'm just reading through a friend's comments on my NATO draft.  Just
after my equations 65--67, where I write, ``The resemblance between
the process in Eq.\ (66) and the classical Bayes' rule of Eq.\ (38)
is unmistakable,'' he writes:  ``Seems to me contrived --- you want
to find a resemblance and then you find it.''

Thinking back to one of his recent productions, I think ``Boy, that's
the pot calling the kettle black!''

But aside from that, I do seem to be impressing less people with this
result than I would have thought.  Is it my presentation, or is it
really the substance?  This result, in particular, is the one I'm
most pleased with in the paper, but it gets the coldest reception.
You've never yet told me your truest thoughts on that section.  I'd
like to hear them.

\section{29-06-01 \ \ {\it Alex Favor} \ \ (to D. R. Terno)} \label{Terno1}

I wonder if I might ask a favor of Alex.  Could you ask her if she would be willing to type in the linguistic analyses she made at the {\Vaxjo} meeting?  She thrilled me to no end by telling me how consistent my language patterns were in my talk.  (Consistency is what I'm striving for!)  Most importantly, I'd like to see how some of the other attendees fared in this regard.  (I'm always willing to learn by others' mistakes!)

\section{29-06-01 \ \ {\it Opinion} \ \ (to S. J. van {\Enk})} \label{vanEnk22}

Do you have an opinion on this as my abstract for the NATO paper on {\tt quant-ph}?  I'll drop by your office in a little bit.
\bq\noindent
This paper reports three almost trivial theorems that nevertheless appear to have significant import for quantum foundations studies.
1) A Gleason-like derivation of the quantum probability law, but based on the positive operator-valued measures as the basic notion of measurement (see also Busch, \quantph{9909073}).  This theorem works both for 2-dimensional vector spaces and for vector spaces over the rational numbers (where the standard Gleason theorem fails).
2) A way of rewriting the quantum collapse rule so that it looks almost precisely identical to Bayes' rule for updating probabilities in classical probability theory.  And 3) a derivation of the tensor-product rule for combining quantum systems (and with it the very notion of quantum entanglement) from Gleason-like considerations for local measurements and classical communication on bipartite systems.
\eq

\section{02-07-01 \ \ {\it Objective Properties} \ \ (to D. G. Chakalov)} \label{Chakalov1}

Thank you for all the interest you've shown in the papers I have been
involved with.  I commend you in your efforts to get to the bottom of
what's going on in our world.  But I cannot believe it very likely
that distinct new kinds of {\it physics\/} arise in our brain
processes.  Instead the road I have chosen to develop is making sense
of quantum mechanics (as a theory predominantly of inference) from
{\it within\/} quantum mechanics.  I understand that your road is
distinct:  but life is short, and one has to make a cut or one will
certainly never get anywhere.  My own direction may turn out to be
completely wrong, but I have decided to pursue it with dogged
determination and not to get derailed.

I wish you luck in your own pursuits.

\section{02-07-01 \ \ {\it Making Good Sense} \ \ (to J. Finkelstein)} \label{Finkelstein1}

\bjf
I enjoyed reading your latest ``quantum states are states of
knowledge'' manifesto, \quantph{0106133}.  I do have sympathy
for that point of view, but I would like to put my two-cents-worth in
by remarking that it is not quite fair to imply that experimental
results such as those of Scarini et al.\ which you cite furnish
ADDITIONAL support for it. \ldots

It is certainly important to confirm the standard quantum predictions
under as wide a set of circumstances as possible, but that
confirmation does not distinguish between alternative interpretations
all of which agree with the prediction.  For example, neither I nor
(I believe) you are advocates of the many-worlds interpretation.  But
some folks are, and those folks would have expected Scarini, Zbinden,
Gisin etc to have found exactly the results that they did find.
Therefore I would say that the many-worlds interpretation has the
same (small) degree of plausibility after these experiments as it did
before.

Would you agree?
\ejf

Thanks for the note!  Yes, I guess I would (though only to the small
extent that I think many-worlds is coherent in the first place). But
there are two things working in the background.  1)  Probably plain
sloppiness on our part in our wording.  And 2) the fact that
{\Ruediger} is (presently) more conciliatory to MWI than {\Carl} and I
are.  He sees Bayesian probability as holding a place even in their
interpretation.  (A rough cartoon is:  In their interpretation, the
universal wave function serves an ontologic role, while the relative
states in a Schmidt decomposition with respect to an observer's mind
serve the same epistemic role we ascribe to them.)  We should
probably either remedy 1) or make 2) more clear, or both.  We'll have
to huddle for that.

\bjf
(And also by pointing out that the experiment with detectors in
relative motion was reported in Zbinden et al.\ (\quantph{0007009}), rather than in your ref.\ 2.)
\ejf

That's probably my screw-up:  I just assumed (from a search through
SciSearch) that Ref.\ 2 was the published version of Zbinden {\it et al}. I
presume you're telling me it's not.  Is there some history here that
we should be aware of?  Or a different published reference?

\bjf
I enjoyed reading your latest ``quantum states are states of
knowledge'' manifesto \ldots
\ejf

And there's still another one coming:  it should have appeared on
{\tt quant-ph} today.  I hope you'll read it too.  This one's a solo
flight by me (titled ``Quantum Foundations in the Light of Quantum
Information'').  (BTW:  Don't let the sober sounding abstract on {\tt
quant-ph} fool you; I'm as loquacious and philosophical as usual on
the inside.)

\subsection{Jerry's Reply}

\bq
My understanding of the history is as follows:
\begin{enumerate}
\item
H. Zbinden, J. Brendel, W. Tittel, and N. Gisin, \quantph{0002031} and
   {\sl Pramana-Journal of Physics\/} {\bf 56}, 349 (2001) [two somewhat different papers]
   discuss the results with moving detectors, and also claim the lower limit
   on the velocity of $10^7c$.
\item
A more detailed version of the above is  H. Zbinden, J. Brendel,
   N. Gisin, and W. Tittel (note permuted authors), \quantph{0007009},
   published as {\sl Phys.\ Rev.\ A.} {\bf 63}, 022111 (2001).
\item
The paper you cite ({\sl Phys.\ Lett.\ A\/} {\bf 276}, 1 (2000)) is the published version
   of V. Scarini, W. Tittel, H. Zbinden, and N. Gisin, \quantph{0007008}.
   Here they consider the velocity in the frame of the cosmic microwave
   background, and claim a lower limit of (only!)\ 1.5$\times 10^4c$.
\end{enumerate}
\eq

\section{02-07-01 \ \ {\it quant-ph/0106133} \ \ (to R. {\Schack})} \label{Schack1}

I would only temper what you just said by making one addition:

\brs
Actually, I believe that the relative state an observer in some
branch of a multiverse {\bf [has no choice but to assign]} to, say, a
qubit has a very natural interpretation as a state of knowledge.
\ers

The lack of free choice is important there, and to that extent the
whole scheme is non-Bayesian.  Bayesian probabilities are never fixed
by edict.  In a way, this is just a fancy version of David Lewis's
principal principle.

Top of the mornin',

\section{03-07-01 \ \ {\it Epistemic Probabilities and Zing} \ \ (to U. Mohrhoff)} \label{Mohrhoff1}

I have just started to read your new paper.

\bum
Epistemic interpretations of quantum mechanics fail to address the
puzzle posed by the occurrence of probabilities in a fundamental
physical theory. This is a puzzle about the physical world, not a
puzzle about our relation to the physical world.
\eum

I would appreciate hearing any thoughts you might have on my own newest: \quantph{0106166}.  In particular, whether you think the ideas (or the research direction) supported there might temper your statement above?  In a sense, it was written with precisely that purpose in mind.

\section{04-07-01 \ \ {\it Carts and Horses} \ \ (to U. Mohrhoff)} \label{Mohrhoff2}

\bum
I hate to dash your hopes \ldots\ \

Your pet idea that quantum information theory holds the key to the
mother of all mysteries is understandable, you being a quantum
information theorist, but it reminds me of something someone wrote
about the father of all mysteries, consciousness. Everyone concerned
with it (neuroscientists, psychologists, philosophers, AI-ers, etc)
thinks that his/her particular field holds the key, so a baker would
think that the secret of consciousness lies in the \'eclair. What
would a baker think about the puzzle posed by the occurrence of
probabilities in a fundamental physical theory?
\eum

You confuse the cart with the horse.  I was attracted to a career in quantum information---and it can be documented---precisely because I had wanted to express quantum foundations problems in information theoretic terms all along.  This was a process that started long before your publicational record and, indeed, long before Shor's factoring algorithm.  But, you say what you did because it is the easy thing to say.

I will read and remark on your papers in due time, despite A) your arrogance, and B) the offensiveness of your letter.  If there is something in your ideas, then it will be worthy of note regardless of its source.\medskip

[But see 16-07-01 note ``\myref{vanEnk24}{Insults}'' to S. J. van {\Enk}.]

\section{04-07-01 \ \ {\it Invitation, 2} \ \ (to A. Y. Khrennikov)} \label{Khrennikov4}

\bakh
Yes, this is very well! However, for me, the only bridge between
``reality'' and our subjective description is given by relative
frequencies \ldots
\eakh

But there other ways to make the bridge:  this is what gambling
situations (like the Dutch-book argument that {\Schack} spoke about) are
about.  They give a NON-frequency {\it operational\/} definition to
probabilities. Subjective probabilities make their {\it objective\/}
mark on the world by specifying how an agent should act when
confronted with them.

\section{04-07-01 \ \ {\it Context Dependent Probability, 2} \ \ (to A. Y. Khrennikov)} \label{Khrennikov5}

\bakh
P.S. But! How can you unify contextuality with subjective
probability?
\eakh

I just don't see this as a problem.  In choosing one experiment over
another, I choose one context over another.  The experiment elicits
the world to do something.  To say that the world is indeterministic
means simply that I cannot predict with certainty what it will do in
response to my action.  Instead, I say what I can in the form of a
probability assignment.  My probability assignment comes about from
the information available to me (how the system reacted in other
contexts, etc., etc.).  Similarly for you, even though your
information may not be the same as mine.  The OBJECTIVE content of
the probability assignment comes from the fact that NO ONE can make
{\it tighter\/} predictions for the outcomes of experiments than
specified by the quantum mechanical laws.  Or to say it still another
way, it is the very existence of transformation RULES from one
context to another that expresses an objective content for the
theory.  Those rules apply to me as well as to you, even though our
probability assignments WITHIN each context may be completely
different (because they are subjective).  But, if one of us follows
the proper transformation rules---the quantum rules---for going to
one context from another, while the other of us does not, then one of
us will be able to take advantage of the other in a gambling match.
The one of us that ignores the structure of the world will be bitten
by it!

\section{04-07-01 \ \ {\it Gleason} \ \ (to A. Y. Khrennikov)} \label{Khrennikov6}

I still need to answer your question about Gleason.  You can read about the POVM version of it in the section ``Information About What?''\ of my new paper ``Quantum Foundations in the Light of Quantum Information''.  But then the better places to look are the papers by Cooke, Keane, and Moran cited there, and also the paper by Pitowsky.  It is those versions of the Gleason theorem that I am particularly interested in, in connection to the question of $p$-adic numbers.

\section{04-07-01 \ \ {\it Book Reminder} \ \ (to the Los Alamos Historical Society)}

Kiki and I thank the Historical Society for their kind offer.  Our time in Los Alamos and our loss of everything material is something the family will have to reckon with for another generation:  there will be no getting around explaining the fire to our two children.

Please send the books to our new home address below.

\begin{itemize}
\item[1)]
{\sl Quads, Shoeboxes, and Sunken Livingrooms, A History of Los Alamos Housing} \\
Craig Martin

\item[2)]
{\sl Robert Oppenheimer} \\
Robert Bacher (Judy Gursky, editor)

\item[3)]
{\sl Tales of Los Alamos} \\
Bernice Brode (Barbara Storms, editor)

\item[4)]
{\sl Los Alamos:\ Beginning of an Era, 1943--1945} \\
Los Alamos Scientific Laboratory Staff

\item[5)]
{\sl Standing By and Making Do, Women of Wartime Los Alamos} \\
Jane S. Wilson and Charlotte Serber
\end{itemize}

\section{05-07-01 \ \ {\it Standing Up and Saying YES} \ \ (to J. Finkelstein)} \label{Finkelstein2}

Thanks for the comments.  I welcome any that you send me!

\bjf
This is not really any objection to what you have written, but the
story you tell on page 10 might produce even WEAKER knees with the
following modification:  Suppose that Alice, instead of choosing ANY
state $|\psi\rangle$ for her qubit, makes her choice from a finite
and previously-agreed-upon set.  She broadcasts the result of her
measurement, but keeps her choice a secret, except that she reveals
her choice in a sealed envelope which she sends to Chris (who
initially leaves it sealed).  Bob performs the appropriate {\Pauli}
rotation, then he makes a guess as to which state Alice chose, and
performs a yes-no measurement with that guess; he communicates his
guess, as well as the yes-no result, to Chris.

Chris can now open the sealed envelope; if it happens that Bob's
guess was in fact correct, then the result must have been ``yes''.
So, if one wanted to be contrary (and of course I do not) one might
say that, although when the yes-no measurement was performed nobody
knew that the guess was correct, and although Alice did not ``take
the time to \ldots\ interact with it'', nevertheless the qubit had
``the power to stand up and say YES all by itself''.
\ejf

I agree, this does sound even more dramatic.  And maybe I will start
using it in my presentations.  But the point remains the same:  it is
Bob's action that elicits a consequence.

You can see, I keep dreaming (modern) alchemical thoughts.  Below.
From:  W.~Heisenberg, ``Wolfgang {\Pauli}'s Philosophical Outlook,'' in
his book {\sl Across the Frontiers}, translated by P.~Heath, (Harper
\& Row, New York, 1974), pp.~30--38.

\bq
\indent
The elaboration of Plato's thought had led, in neo-Platonism and
Christianity, to a position where matter was characterized as void of
Ideas.  Hence, since the intelligible was identical with the good,
matter was identified as evil.  But in the new science the world-soul
was finally replaced by the abstract mathematical law of nature.
Against this one-sidedly spiritualizing tendency the alchemistical
philosophy, championed here by Fludd, represents a certain
counterpoise.  In the alchemistic view ``there dwells in matter a
spirit awaiting release. The alchemist in his laboratory is
constantly involved in nature's course, in such wise that the real or
supposed chemical reactions in the retort are mystically identified
with the psychic processes in himself, and are called by the same
names.  The release of the substance by the man who transmutes it,
which culminates in the production of the philosopher's stone, is
seen by the alchemist, in light of the mystical correspondence of
macrocosmos and microcosmos, as identical with the saving
transformation of the man by the work, which succeeds only `Deo
concedente.'\,''  The governing symbol for this magical view of
nature is the quaternary number, the so-called ``tetractys'' of the
Pythagoreans, which is put together out of two polarities.  The
division is correlated with the dark side of the world (matter, the
Devil), and the magical view of nature also embraces this dark
region.
\eq
and
\bq
\indent
When, in the spring of 1927, opinions on the interpretation of
quantum mechanics were taking on rational shape and Bohr was forging
the concept of complementarity, {\Pauli} was one of the first physicists
to decide unreservedly for the new possibility of interpretation.
The characteristic feature of this interpretation---namely, that in
every experiment, every incursion into nature, we have the choice of
which aspect of nature we want to make visible, but that we
simultaneously must sacrifice, in that we must forego other such
aspects---this coupling of ``choice and sacrifice,'' proved
spontaneously congenial to {\Pauli}'s philosophical outlook.  In the
center of his philosophical thinking here there was always the wish
for a unitary understanding of the world, a unity incorporating the
tension of opposites, and he hailed the interpretation of quantum
theory as a new way of thinking, in which the unity can perhaps be
more easily expressed than before. In the alchemistic philosophy, he
had been captivated by the attempt to speak of material and psychical
processes in the same language. {\Pauli} came to think that in the
abstract territory traversed by modern atomic physics and modern
psychology such a language could once more be attempted \ldots
\eq

\section{05-07-01 \ \ {\it Invitation, 3} \ \ (to A. Y. Khrennikov)} \label{Khrennikov7}

\bakh
I think that there is some mystification
in such a direct use of subjective probabilities.
Do you really believe that you choose probabilities
in gambling situation by your personal belief?
I think you (and everybody) do in the following way: you have some
experience with gambling (frequency!)\ and use this experience to
introduce ``subjective'' probabilities.
\eakh

No, I think it is just the opposite:  people almost never use
frequency data as the determiners of their information in any common
situation.  Instead they use symmetry.  If someone presents me with a
coin that I have never seen before, then after a quick examination, I
will likely {\it ascribe\/} a 50/50 probability to its coming up
heads {\it simply\/} because I have no reason to believe otherwise.
But if Danny Greenberger is the tosser of it, I know that he has the
skill to make it look superficially as if it were being tossed in a
haphazard fashion but it will still come up heads every time.  The
50/50 ascription is not a property of the coin!  It is simply a
property of ignorance.

\section{05-07-01 \ \ {\it QIC010531 (fwd)}\ \ \ (to S. J. van {\Enk})} \label{vanEnk23}

\bsve
I really get irritated by report I, you?
\esve
Yes.  We should probably just tackle it head on, pointing out how it is irrelevant.

Let me share some of my recent insults to cheer you up.  Below is a letter I sent yesterday and for which I received profuse apologies today.  What really annoyed me was that he insulted me for 24 kilobytes worth of email (in front of Peres and {\Mermin}), and then at the end of it, had the audacity to say, ``I trust you will find that we are striking consonant chords.''  So, I decided to strike {\it his\/} chords.

\section{05-07-01 \ \ {\it Par Avion}\ \ \ (to H. J. Bernstein)} \label{Bernstein2}

I just received your phone reimbursement.  Thanks!

And thanks for coming by Charlie's the other day.  What I really need to do the next time I come up, is just spend a day visiting you and visiting ISIS.  Would you guys like to hear a talk from me?  In fact, Kiki and I didn't get nearly as much (book) shopping time in as we had hoped to (mostly because of my Hirota duties):  So, we might just come for another visit very soon.  We'd probably just stay in a hotel in Amherst or Northampton or something to maximize our free time.  What's your summer schedule like?  What might be a good time for us to visit?

Charlie dismays me at times, calling my efforts to clean up quantum mechanics ``theology.''  Strangely, it does hurt---I guess because I respect him so much.  For instance, I doubt he'll even look at my latest paper just because he doesn't like the goal I have in mind.  But there are some meaty theorems there that he might find useful \ldots\ if he wouldn't just shut himself off to my trains of thought.  Getting something like this from Holevo maybe makes it even harder:
\bq\noindent
I downloaded your recent \quantph{0106166}. It has several interesting
observations, but I like particularly the argument concerning
derivation of tensor product of Hilbert spaces from the measurement
statistics.
\eq
There are two great information theorists in my life, but only one of them will read my papers.

\section{05-07-01 \ \ {\it Bayesian Computation} \ \ (to C. M. {\Caves}, H. Barnum \& J. M. Renes)} \label{Caves0.3} \label{Barnum3.1} \label{Renes2.2}

\bcc
Here's a question for all of you.  What's the Bayesian description of
the state at the output of an ideal quantum algorithm that is supposed
to store the answer as a particular state in the computational basis?
One knows the input state, one knows the exact sequence of unitaries,
yet one doesn't know the output.  Otherwise there would be no point in
doing the computation.  So the Bayesian output state is presumably a
density operator diagonal in the computational basis.  Why? Upon making
a measurement in the computational basis, one discovers the answer,
which is implicit in the input, but unknown because it can't be
calculated efficiently in our heads or in any classical machine for
that matter.  What guarantees the veracity of this answer?
\ecc

I have nothing to say other than that I really, really, really like this question.  Asking it is certainly a step in the direction of squashing away any imagery of ``massive parallelism'' in quantum computation.  Perhaps the first thing to do is ask for a Bayesian description of classical computation.

\section{06-07-01 \ \ {\it Stamina!}\ \ \ (to S. Aaronson)} \label{Aaronson1}

Thanks for the note!  I'll give you a longer reply later. But in the mean time:
\bsa
Anyway, I finished (!!!)\ ``Notes on a Paulian Idea.''  Your shamelessness in mixing quantum physics, philosophy, and your personal life is an inspiration; maybe it will encourage other scientists to try something similar.
\esa
Am I to take this to mean that you actually read the whole thing?  If so, I'm shocked!  I never imagined anybody would read the whole thing.

\subsection{Scott's Preply, ``Notes on a Preskillian Meeting''}

\bq
Remember me from Bell Labs?  I'm at Caltech now, and tonight was at
Preskill's group meeting where Andrew Landahl spoke on your ``Quantum
Foundations'' paper.  I thought you might want to know some highlights.
\begin{itemize}
\item
We took a vote on whether humans have free will or are mere machines.
Machines won, 7-5.
\item
Preskill defended the view that one can be an Everettista and still
agree with most of your thesis that quantum states are states of
knowledge.  I think his argument was that wavefunctions are indeed
mathematical constructs, but we can use them to describe those ``branches
of the multiverse'' that interest us.
\item
Andrew presented an imaginary dialogue in which he and Einstein chide
you for holding that interventions in the world just tell us about the
likely results of further interventions (I don't know if that's your
position).  Preskill said to Andrew, ``you and Einstein certainly hit it
off!''
\item
Andrew expressed concern for your daughter, after quoting the comment
that she has no theory of measurement.
\end{itemize}

Anyway, I finished (!!!)\ ``Notes on a Paulian Idea.''  Your shamelessness
in mixing quantum physics, philosophy, and your personal life is an
inspiration; maybe it will encourage other scientists to try something
similar.

I was struck by the recurrent question, why complex Hilbert spaces and
not real (or quaternionic) ones?  Someone ought to write a semipopular
article about this issue.  It seems analogous to asking why space has 3
visible dimensions and not 2 or 4, as opposed to the dozens of more
nebulous questions one could ask about space.

Do you know of an argument for complex Hilbert spaces that doesn't rely
on dimension-counting under tensor products (i.e.\ the Goldilocks
Principle)?  What about this: if we think time is continuous, then for
all operations $U$, there should exist a $V$ such that $V^2=U$.  Let $U$ be a
phase flip; then $V$ must involve complexes.

The obvious objection is, why are phase flips possible?  For example, in
real quantum mechanics with arbitrary rotations but no phase flips,
every operation does have a real square root.  One response is that a
phase flip in $n$ dimensions can always be simulated by a rotation in $n+1$
dimensions.  But since this rotation maps the $n$-dimensional subspace
onto itself, it must (we can declare!)\ have a square root that also maps
the subspace onto itself.  That forces us to complex numbers.

It's 5AM and I'm going to sleep.
\eq

\subsection{Scott's Reply}

\bq
\noindent
{\bf [Chris said:]}\ ``Am I to take this to mean that you actually read the whole thing?''\medskip

Well, I skipped some repetitive parts and extended quotations, but
besides that yes.  I think if condensed to, say, the best 200 pages, it
would be worth publishing as a book.  The jacket could say, ``Just like
{\sl The Fabric of Reality\/} by David Deutsch \ldots\ except antithetical in its
philosophical stance, and funnier!''
\eq

\section{09-07-01 \ \ {\it Quantum Optics, San Feliu} \ \ (to K. {\Moelmer})} \label{Moelmer1}

I just got a quote for a flight from Newark to Barcelona.  It was \$1182 (including taxes, etc.).  That comes to about 1394 euros.

\bkm
P.S. I started reading some of your letters in the compilation that you put
on {\tt quant-ph}. At first I thought: what a nut case, thinking that
anybody should take an interest in such stuff, but after reading a
bit, I got really impressed, and now I
only regret that life is too short to read both your letters and
William James. As a comment to Mermin's preface: I also had the kind
of nice feeling long time after having read his paper that you describe.
\ekm

Thanks for the soothing words about the samizdat.  I've gotten a lot of nice reactions (from about 25 people).  My very best PRL in contrast maybe got 2!  So, in the end, it has seemed worth it.

\section{09-07-01 \ \ {\it The O'bleak Observer} \ \ (to N. D. {\Mermin})} \label{Mermin23}

\bdm
You should not dismiss my feeling that you've not adequately
justified your assumption about noncontextuality as merely a
manifestation of a regretable atavistic tendency to reify the quantum
state.
\edm

The ``oblique observer'' note was a concession, not a dismission.  It
is evidence that I am taking your point very seriously (even though
I'd rather being out playing with the other kids).

\bdm
The question you're evading is what it means for one and the same
positive operator $E$ to appear in many different POVMS.
\edm

No, I don't think I'm evading it.  It means that those various
interventions or ways of gathering data---those POVMs---physically
diverse though they may be---all lead to at least one common
possibility for what my knowledge can be updated to (modulo the
unitary readjustment).

\section{09-07-01 \ \ {\it More O'bleakness} \ \ (to N. D. {\Mermin})} \label{Mermin24}

\bdm
\bq
\noindent
\rm
CAF Said: Moreover, all questions about instantaneous signaling
through quantum-state change just become silly:  such questions
spring solely from a wrong-headed view.
\eq
No!  The signalling has nothing to do with quantum-state change.
(We've been through this before.)  If Bob and Alice share a large
number of identically prepared pairs, then a very reasonable
requirement is that the statistical distribution of outcomes Bob gets
from his members of the pairs cannot depend on what Alice chooses to
do to her members.  (If it did Alice could send useable unmediated
signals to Bob.)  Again, this has nothing to do with how you like to
think about probabilities or quantum states.  I offered this to you
as an example of a situation in which you can, in fact, justify the
non-contextuality of certain probabilities by appealing to an
independent physical requirement (no remote signalling).
\edm

I hold firm in my opinion.  It has {\it everything\/} to do with how
you like to think about (quantum) probabilities.  If you think the
probabilities are subjective expectations for the local consequences
of one's experimental interventions, then the question never arises.

However, granting you a little distrust for that, the point about
oblique observations is that one might always be able to think of a
quantum measurement as being enacted on a system other than the
intended one.  This would give your point above a natural means for
being used to justify noncontextuality for {\it all\/} quantum
measurements.

Again, I'm starting to feel awfully comfortable with noncontextuality
as the very simplest generalization of Bayes' noncontextuality.  It
is the very glue that puts measurement outcomes into Hilbert space in
the first place.  (Otherwise we might just draw out an exhaustive
list of one-outcomed, two-outcomed, three-outcomed measurements etc.,
etc., and never even suppose a connection between them.)  But I offer
the above as an effort to go in the direction you want me to.

\section{10-07-01 \ \ {\it $V^2=U$}\ \ \ (to S. Aaronson)} \label{Aaronson2}

\bsa
Do you know of an argument for complex Hilbert spaces that doesn't
rely on dimension-counting under tensor products (i.e.\ the Goldilocks
Principle)?  What about this: if we think time is continuous, then for
all operations $U$, there should exist a $V$ such that $V^2=U$.  Let $U$ be a
phase flip; then $V$ must involve complexes.
\esa

I have heard some arguments not based on dimension counting.  Go to Adler's book\footnote{S. L. Adler, {\sl Quaternionic Quantum Mechanics and Quantum Fields}, (Oxford University Press, 1995).} and Wheeler's paper\footnote{J. A. Wheeler, ``The Computer and the Universe'' {\sl International Journal of Theoretical Physics\/} {\bf 21}(6/7), 557--72 (1982).}, both referenced on page 125 of the samizdat.
\begin{enumerate}
\item
In the first case it is that the generators of unitary time evolution cannot then be observables.
\item
In the second case it is that uniform distributions of pure states only lead to uniform distributions on the probability simplex of a fixed observable in the complex $H$ space case.
\end{enumerate}
Your argument is new to me.  It might be worth writing it up properly if some reference in Adler's book hasn't yet discussed it.

Thanks again for the note on John's meeting.  (And reading the samizdat!)

\section{10-07-01 \ \ {\it My Recent Foundations Posting} \ \ (to J. D. Malley)} \label{Malley1}

Thank you for your pleasant letter.  Indeed, good hearing from you again:  I trace that my last email from you was in 1996 when you were reading my PhD thesis.

I've gotten a lot of good feedback on this paper.

I checked out the Geisser book as soon as I got your letter.  His attitude to parameter estimation is particularly relevant for a debate Carl Caves and I are embroiled in presently.  And he raises some points similar to ones I've heard from {\Ruediger} Schack.  So, that was a very useful reference.

Let me reply briefly to your comments.

With regard to (1), this can be posed in a purely Bayesian way through use of the ``quantum de Finetti representation theorem.''  Maybe my description of it was too brief in the present paper.  You can find a significantly longer elaboration of it in my paper with Caves and Schack, \quantph{0104088}.

With regard to (2), yes, true enough.  But there is definitely a quantum analogue---a noncommutative analogue---to it if one only uses conditional states of knowledge (never joint).  That is what my ``Whither Bayes' Rule?''\ section is about.

With regard to (3), agreed.  (This, in large part, is my present debate with Caves \ldots\ but in the specific context of ``unknown Hamiltonians.'')

With regard to (4), I'm not sure I understand this.  If you're suggesting that my point of view on the quantum foundations will be more interesting if it suggests some new experiments, then I most certainly agree.  Unfortunately, I'm not there yet.

\subsection{Jim's Preply}

\bq
Thanks for the thoroughly enjoyable and very readable account of your take on quantum foundations and its connections with information theory (e.g.\ \quantph{0106166}, 29 Jun 2001).

I thought you might find it interesting to learn that your essential ideas have, it would seem to me, been the subject of much research in the statistical community for many years, though not with quantum things in mind. Specifically, there is a school of thinking which argues that as parameters in a model are generally not observable, they therefore should not be the focus of statistical inference.  Instead, it is argued that making predictions should be the real purpose of inference, and to this end one should use Bayes' theorem to construct the ``predictive posterior distribution.'' I think this strategy accords quite well with the quantum foundations argument you're making under your imprimatur.

Prominent among those making this argument has been Seymour Geisser at the U. of Minnesota ({\tt geisser@stat.umn.edu}). See for example, his textbook
\begin{itemize}
\item
S. Geisser (1993), {\sl Predictive Inference:\ An Introduction\/} (Chapman \& Hall).
\end{itemize}
As background for predictive inference in a more general context you could consult
\begin{itemize}
\item
A. H. Welsh (1996), {\sl Aspects of Statistical Inference\/} (Wiley); see pp.\ 34--35.
\item
B. P. Carlin and T. A. Louis (2000), {\sl Bayes and Empirical Bayes Methods for Data Analysis\/} (2nd edition) (Chapman \& Hall); see pp.\ 20--21.
\end{itemize}

Finally I'd like to add four, small comments:
\begin{itemize}
\item[(1)] given an unlimited amount of data (or, take a limit as sample size increases) the density operator is known with arbitrary accuracy, so in this sense is, finally, itself observed;

\item[(2)] use of Bayes theorem in any quantum setting is often a slippery business, since its rigorous derivation requires joint distributions for all observables: the events $A$ and $B$ that are outcomes in an experiment, and from which one constructs conditional probability $Pr(A | B)$, does not lead to a valid Bayes theorem unless the observables commute.

\item[(3)] it is not clear to me that the predictive posterior ever leads to any inference that is unique to its formulation---any inference obtained about parameters in a model can be considered as useful intermediates to prediction.

\item[(4)] any change in emphasis or construal of the quantum primitives has greatest interest only when new experiments are suggested, and the predictive posterior does not in itself suggest new experiments.
\end{itemize}

Again, thanks for the well-written and nicely organized discussion of this important topic.
\eq

\section{10-07-01 \ \ {\it Replies on a Preskillian Meeting} \ \ (to A. J. Landahl)} \label{Landahl1}

Wow, what a set of notes!  Thank you all for the interest in my silly
efforts.  This is a little unexpected.

\bal
The talk was a smash.  It went much better than I was expecting,
causing much discussion.  (As you know, some of the people in our
group are rather reticent, so that's really saying something.)
\eal

I am so glad to hear that.  It is really very flattering.

\bal
At the end of the talk, Sumit decided to go up to the chalkboard and
take a poll, the topic and results of which I'll leave as a surprise,
as I imagine John will tell you about them himself.  (If he doesn't,
just e-mail me back and I'll let you know.)
\eal

I presume this is the poll Scott mentioned.  I'll say more about that
later.

\bal
I also mentioned your program to establish an information-theoretic
foundation for all the laws of quantum mechanics (and physics?) in
this section.  This proposal met with much skepticism from the
audience.  I'm somewhat sympathetic to your cause (certainly more so
than some of our denizens!), but I don't believe that {\it all\/} of
physics has an information theoretic description.  For example, where
would the (dimensionful) physical constants enter into this scheme
(like Planck's constant and the speed of light)?  I don't see how
they could enter unless they {\it define\/} what physical dimensions
are, which is rather peculiar.
\eal

I really am very flattered by all this attention, but I do get
dismayed when I can't seem to get the most important point across to
my readers.  Even sympathetic readers!  Because of this, I have spent
months and months trying to clarify and refine my presentation.  But
for some reason it is amazingly difficult to get the point across.
At the very least I need people to understand what I want {\it
before\/} they declare that they disagree with it. (Disagreeing with
it would {\it then\/} be fair enough.)  The sentences above seem to
convey that you haven't gotten to the level of understanding what I
want.  How can the following sentences be consistent with what you
say above?

\begin{enumerate}
\item
Abstract, penultimate sentence.

This method holds promise precisely because a large part (but not
all) of the structure of quantum theory has always concerned
information.

\item
Section 1, last paragraph.

Our foremost task should be to go to each and every axiom of quantum theory and give it an information theoretic justification if we can. Only when we are finished picking off all the terms (or combinations of terms) that can be interpreted as information---subjective information---will we be in a position to make real progress. The raw distillate that is left behind, miniscule though it may be, will be our first glimpse of what quantum mechanics is trying to tell us about nature itself.

\item
Section 2, last paragraph.

The world is sensitive to our touch. \ldots\ The whole structure of quantum mechanics---{\it\underline{it is speculated}}---may be
nothing more than the optimal method of reasoning and processing
information in the light of such a fundamental (wonderful)
sensitivity.

\item
Section 3, penultimate paragraph, page 9.

The complete disconnectedness of the quantum-state change rule from anything to do with spacetime considerations is telling us something deep: The quantum state is information. Subjective, incomplete information. Put in the right mindset, this is {\it not\/} so intolerable.  It is a statement about our world. There is something about the world that keeps us from ever getting more information than can be captured through the formal structure of quantum mechanics. Einstein had wanted us to look further---to find out how the incomplete information could be completed---but perhaps the real question is, ``Why can it {\it not\/} be completed?''

\item
Section 5, last two paragraphs.

Perhaps the structure of the theory denotes the optimal way to
reason and make decisions in light of {\it some\/} fundamental
situation, waiting to be ferreted out in a more satisfactory fashion.

This much we know:  That ``fundamental situation''---whatever it
is---must be an ingredient Bayesian probability theory does not have. There must be something to drive a wedge between the two theories. Probability theory alone is too general of a structure. Narrowing it will require input from the world about us.

\item
Section 7, last two paragraphs.

    The quantum de Finetti theorem shows that the essence of
    quantum-state tomography is not in revealing an ``element of
    reality'' but in deriving that various agents (who agree some
    minimal amount) can come to agreement in their ultimate
    quantum-state assignments. This is not the same thing as the
    stronger statement that ``reality does not exist.''  It is
    simply that one need not go to the extreme of taking the
    ``unknown quantum state'' as being objectively real to make
    sense of the experimental practice of tomography.

One is left with the feeling \ldots\ that perhaps this is the whole point to quantum mechanics. That is:  Perhaps the missing
ingredient for narrowing the structure of Bayesian probability
down to the structure of quantum mechanics has been in front of
us all along.  It finds no better expression than in the taking
account of the limitations the physical world poses to our
ability to come to agreement.
\end{enumerate}

I certainly believe there are some things within quantum mechanics
that are beyond our subjective description.  As in your example,
Planck's constant could well be one of them.  The dimensionality $d$
of a Hilbert space is another one I feel fairly confident of.  That
number characterizes something intrinsic to a system.  To that
extent, it is not something that can be information-theoretic in
origin.

\bal
To press your point more forcefully in the future, you might consider
rephrasing the special relativistic axioms themselves in a more
information-theoretic light.
\eal

But it doesn't seem to me that special relativity is overtly about
Bayesian or information theoretic concerns in the way that quantum
mechanics is.  So I wouldn't want to express those axioms in a more
information-theoretic light.

\bal
Conclusion: quantum information theorists need to get out more!
\eal

Yes!  (And I speak for myself too.)

\bal
At least I got a good groan from John Preskill when Alice flipped Bob
a quarter for the ``two bits'' of classical communication she sent
him.
\eal

I got a similar groan from {\Mermin} when he first read the end of
Section 3.  BTW, footnotes 8 and 9 are not typos; several people have
asked me about that.

\bal
In The Future section I talked about Gleason's theorem for POVMs and
expressed my concerns about Emma's future psychological counseling
given that you already are pressing her for a theory of measurement.
None of us understood what the quote you ascribed to Hideo actually
meant, which I suppose I should ask Hideo about.  Do you understand
it?  It sure {\it sounds\/} amusing.
\eal

The point is, Emma gets by without a theory of measurement, and we
should all learn something from that.  It is the people who think
that knowledge acquisition, or better, belief acquisition, must arise
from a detailed dynamical theory that are the problem.

The point of view taken here is that ``detailed dynamical theories''
are theories of {\it inference}, and therefore lie outside of the
process of knowledge acquisition.  This does not bar quantum theory
from making contact with the REAL world---the world that was here
long before man ever arose (see points about reality below)---it just
means that one is not going to find it in the dynamics.

\bal
In the Learning section, I went over your argument for quantum
collapse being a kind of Bayesian conditioning.  I understand the
analogy you drew here, but I hardly believe this makes quantum
collapse any more gentle of a process.  That's because I don't
believe Bayesian conditioning is ``gentle.''  While it's true that
one can express the classical process as ``plucking'' a term out of a
sum over conditional probabilities, the change in probabilities can
be quite dramatic: the change can be from nearly zero to one in a
single step!
\eal

Fair enough.  Perhaps I overplayed the imagery.  The point I really
wanted to emphasize is that quantum collapse can be thought of as
predominantly a refinement of one's knowledge.

\bal
I also don't understand the meaning of the ``mental readjustment''
step in the quantum process.  Is this just a change-of-basis for the
description of the state?  Is it something more or less than this? If
that's all it is, then I really don't like this phrase ``mental
readjustment'' at all.
\eal

The track I'm on is that quantum state change is essentially Bayesian
updating of knowledge, but with the proviso that the things we have
to do to update our knowledge are (generally) not without effect on
the world.  This must be taken into account in some way. That the
updating is Bayesian-like has a trace in quantum mechanics through
Eqs.\ (57) and (58).  That we still have to take into account our
knowledge of our invasiveness, this has a trace in Eq.\ (59). That is
the ``mental readjustment''---i.e., taking into account what we know
about our own invasiveness.  (I agree, I should have found a better
word for it.)  When we know that our knowledge acquisition could not
have physically affected the system it was concerned with, then we
need do nothing whatsoever beyond Bayesian updating.  Eq.\ (64) is an
example of that.

\bal
I finished the Learning section by sketching how ``typical von
Neumann entropy'' as an uncertainty measure increases after every
measurement.  A question I had in your argument was why the
integration is done over only von Neumann measurements.  I looked
over your original paper on this subject and didn't find an answer
there either.  Some mention is made of projective measurements being
``maximally predictive,'' but I'm not totally convinced by this
argument.  I suspect that the true reason for restricting attention
to these measurements is technical. A measure over POVMs doesn't
exist, so one can't integrate over them. Wouldn't life be so much
nicer if there were one!  Is this the true reason for the von Neumann
measurement restriction?
\eal

Your suspicion is correct.  My choice was no deeper than that.

\bal
I spent most of my time discussing the Correlations section.  I went
in detail over your proof of what I called ``Gleason's theorem for
Classically Semilocalizable Operations (CSOs)'' in deference to the
terminology introduced in a recent paper by Beckman {\it et al}.
Personally, I thought it was cool that the tensor product arose out
of noncontextuality and the measurement model.  John Preskill wasn't
so impressed --- he believes that the tensor product will arise out
of any reasonable model of measurement which has the property of
locality.  (I.e where neither Alice's nor Bob's local actions can
meaningfully impact the other.)  He may be right, but for me that
isn't the point.  The point for me is that the proper way to view
Gleason's theorem is as a machine.  The input to the machine is the
measurement model and the output of the machine is the state space
structure and the probability law.
\eal

Indeed you did get the point.  Thanks.  The point is, how much of the
structure of quantum mechanics can we shove into the simple choice:
``measurements $=$ POVMs.''  How much of quantum mechanics is really
independent of that choice?  There has been a hell of a lot of work
trying to reduce all of quantum mechanics to the assumption of
unitarity.  I'm trying to go the other way.

\bal
What especially excites me about this point-of-view its potential
impact on quantum field theory.  The main point of the Beckman {\it et al}.
paper is that causal measurements and localizable measurements are
not one in the same.  Wouldn't it be interesting to see what happens
when we impose only causality on our measurement model and send it
through the ``Gleason machine?''  What do you suppose the resultant
structure of the state space would be?
\eal

If I understand you correctly, {\Mermin} in his talk in {\Montreal} and
Sweden has been wondering something very similar.  In fact, he would
like to see the quantum probability rule AND the tensor product rule
arise out of the idea that measurement cannot give instantaneous
signaling.  He doesn't yet feel comfortable with my (Gleason's)
noncontextuality assumption.  Yours is a good question; I'll try to
have a look at that paper.

\bal
On the whole I portrayed your ``party platform'' as the statement
that ``Quantum states are states of knowledge about the consequences
of future interventions.''
\eal

That statement, as it stands, is true.

\bal
In particular, those consequences aren't consequences to reality, but
rather consequences to states of knowledge about even further future
interventions.
\eal

That statement, as it stands, is not.  (Do you not see the
difference?!?!)

\bal
In this worldview Bayesian agents don't work to align their
predictions with an underlying reality.
\eal

They would if they could, but they don't because they can't.
Realizing this---it seems to me---is the first step to understanding
what the quantum world is about.

\bal
Instead they work to align their predictions with {\it each other}.
It is as if reality in this picture is solely the agreement of
predictions!

I'd be interested to hear if you believe that this is a fair
characterization of your party's platform.  After reading this paper,
I came to the conclusion that you didn't believe in reality at all.
(Or at best I thought you believed reality = knowledge.)  John
Preskill tells me you believe otherwise, namely that there {\it is\/}
a reality, which surprised me.
\eal

Yeah, you botch it pretty badly there.  John is right.  See my
diatribe under C) above.  But, let me also add to that:

\begin{enumerate}
\item
Section 4, first paragraph.

I have been watching my two year old daughter learn things at a
fantastic rate, and though there have been untold numbers of lessons
for her, there have also been a sprinkling for me.  For instance, I
am just starting to see her come to grips with the idea that there is
a world independent of her desires.  What strikes me is the contrast
between this and the concomitant gain in confidence I see grow in her
everyday that there are aspects of existence she actually {\it can\/}
control.  The two go hand in hand.  She pushes on the world, and
sometimes it gives in a way that she has learned to predict, and
sometimes it pushes back in a way she has not foreseen (and may never
be able to). If she could manipulate the world to the complete
desires of her will, I am quite sure, there would be little
difference between wake and dream.
\end{enumerate}

This wispy little piece is the closest I've been able to come to
giving substance what I call ``the Paulian idea.''  See my ``Notes on
a \ldots'', page vii.  The world must have some impredictability
about it, otherwise we would never be able to say we have seen any
trustworthy trace of a reality.

\bal
I'm curious to hear what you believe reality is.
\eal

Me too.  The idea is not well formed yet.  Perhaps this accounts
somewhat for people not getting my point that the first part of
attempting to identify what is real in the quantum world is to
identify what is subjective and governed by ``laws of thought.''  We
should do that because that's the easier part of the program.
Contemplating what's left behind is when the real fun will begin.

People are too used to seeing gurus (like Deutsch or Mohrhoff) sit on
high and declare what reality {\it is}.  My goals are more modest,
even if my method of advertisement is not.  I don't have an answer
yet; I just feel a direction.  One should not confuse my method of
attack with my answer.

If you were to push me real hard on this
``what-you-believe-reality-is'' business, I might be inclined to say,
``Read Schopenhauer's {\sl The World as Will and Representation}.''
But since I haven't read it myself, I can hardly expect you to do
that! Anyway, as a very {\it provisional\/} answer, I might say it's
something like the ``will'' (the quotes around that word are very
important) that Schopenhauer attributes to every piece of the world,
animate and inanimate alike.  For want of a better term, I call it
zing.

\bal
As for the mechanical details of the paper itself, I enjoyed your
refreshingly casual writing style.  I found one of your section
titles to be either exceedingly clever or merely a typographical
error.  Either way, I'm the only one out of a dozen people who
noticed it, even after I pointed it out.  I'm hoping that you
intended the convey the clever interpretation.
\eal

{\Caves}, Bilodeau, and Schumacher also asked me if it is a typo.  It is not:  you can be relieved.

\bal
If so, I suggest you correct the grammar on the section title to
``Wither Entanglement!'' to make the homonym less subtle without
sacrificing any wit.
\eal

Too late.  But in any case, I wanted all the section headings to be
questions, except the beginning and end ones.

\bal
The only other typo that jumped out at me was on page 13: ``shear
difficulty'' should read ``sheer difficulty.''
\eal

Thanks, I hadn't noticed the difference before.

\bal
Once again, great paper.  I'm psyched that it stirred up so much
discussion in our group meeting.
\eal

Me too!

Now, I said I would make some comments on Sumit's poll.  But I'm too
tired for that after all this writing.  So I won't tell you what I
think in any great detail at the present.  I'll just cut and paste
what Hans Primas thinks.  It's below.  I will say, however, that I
don't see that there should be a qualitative distinction between my
description of you (Andrew Landahl) and my description of the coffee
maker sitting to my right.  You are both physical systems embedded in
this thing we call the world.
\bq
\noindent\bq
\noindent
From: H.~Primas, ``Beyond Baconian Quantum Physics,'' in {\sl Kohti
uutta todellisuusk\"asityst\"a. Juhlakirja professori Laurikaisen
75-vuotisp\"aiv\"an\"a} (Towards a New Conception of Reality.
Anniversary Publication to Professor Laurikainen's 75th Birthday),
edited by U.~Ketvel (Yliopistopaino, Helsinki, 1990), pp.~100--112.
\medskip
\eq
The methodology of experimental scientific research and engineering
science is to a large extent characterized by the regulative
principles emphasized by Francis Bacon.  It is a tacit assumption of
all engineering sciences that nature can be {\it manipulated\/} and
that the initial conditions required by experiments can be brought
about by interventions of the world external to the object under
investigation.  That is, {\it we assume that the experimenter has a
certain freedom of action which is not accounted for by first
principles of physics}.  Without this freedom of choice, experiments
would be impossible.  Man's free will implies the ability to carry
out actions, it constitutes his essence as an actor.  We act under
the idea of freedom, but the topic under discussion is neither man's
sense of personal freedom as a subjective experience, nor the
question whether this idea could be an illusion or not, nor any
questions of moral philosophy, but that {\it the framework of
experimental science requires the freedom of action as a constitutive
though tacit presupposition}.

The metaphysics of Baconian science is based on the confidence that
only the past is factual, that we are able to change the present
state of nature, and that nothing can be known about nature except
what can be proved by {\it experiments}.  Francis Bacon's motto {\it
dissecare naturam\/} led to a preferred way of dividing the world
into object and observing systems.  An experiment is an {\it
intervention\/} in nature, it requires artificially produced and
deliberately controlled, reproducible conditions.  In {\it
experiments\/} in contradistinction to {\it observations\/} -- one
{\it prepares\/} systems in initial states, {\it controls\/} some of
the variables, and finally {\it measures\/} a particular variable.
The regulative principles of Baconian science require {\it power to
create initial conditions}, stress {\it the facticity of the past\/}
and {\it the probabilistic predictability of the future}, and reject
{\it teleological considerations}.
\eq

\section{10-07-01 \ \ {\it Old McBleak's Ale House} \ \ (to N. D. {\Mermin})} \label{Mermin25}

\bdm
I can't believe we're talking past each other on something this
basic! [\ldots]
\bq\noindent\rm
[CAF wrote:]  I hold firm in my opinion.  It has {\it everything\/} to do
with how you like to think about (quantum) probabilities.  If you
think the probabilities are subjective expectations for the local
consequences of one's experimental interventions, then the question
never arises.
\eq
To say that the question never arises is to say that probabilities
can never have any bearing on frequencies of experimental outcomes.

Try it in the language you prefer: if Alice's subjective expectations
for the local consequences of her experimental interventions differ,
depending on what kind of an experimental intervention Bob chooses to
make over in the next county --- not, I stress, on what Bob learns
from his intervention but just on how (or whether) he decides to
intervene --- then Alice will be wiped out by any competent Dutch
bookie, unless unmediated action at a distance is an objective
feature of the world.
\edm

We are talking past each other.

But, my wording was careful enough to cover your reply (modulo the
confusing parenthesis I put around the word quantum, for which I
apologize).  If a physical action associated with a POVM---by
definition---only affects the system associated with the POVM's
Hilbert space, then by definition that is all it affects.  Standard
quantum mechanics has that feature.

The issue is whether we should question the reasonableness of that.
Or, indeed, as you would like, turn the tables and check whether the
physical requirement of no-signaling gives rise to the standard
probability rule full stop.  Your question is a well-posed question,
I do not deny that.  But, as I view it, its motivation is a
throw-back to the days when entanglement was thought to have some
connection to the spooky ghosts of nonlocality.

I am torn.  75\% of the time, I think your question is a regressive
turn to the Popescu--Rohrlich--Aharonov--Shimony--Gisin
``passion-at-a-distance'' mentality.  It seems to me acknowledging
that as an interesting paradigm (even one to be ultimately shot down)
is a wrong turn.  But 25\% of the time, I think, ``Why not?  It is a
valid question, so answer it if you can.''

Still no sympathy for me?  (Probably not.)  But, am I at least
coherent?

\section{11-07-01 \ \ {\it Quantum Information / Evolutionary Universes}\ \ \ (to L. Smolin)} \label{SmolinL1}

We've met before (when I was at Caltech), but you probably don't remember me.  My name is Chris Fuchs; I'm on the research staff at Bell Labs, and my specialization is quantum information theory.

I'm writing you because lately I've taken a shine to evolutionary-universe ideas, because I've enquired with the Perimeter Institute about the opportunities they might hold for my quantum foundations program, and because I've heard that you will be joining them.  (The first part of the sentence---the part about the shine---is the cause, not the effect!)  Thus I would like to get to know you better.

For one thing, please allow me to share with you two of my writings on the subject:
\begin{itemize}
\item
C.~A. Fuchs, ``Notes on a Paulian Idea:  Foundational, Historical, Anecdotal \& Forward-Looking Thoughts on the Quantum (Selected Correspondence).'' 504 pages. Foreword by N. David Mermin.
See \quantph{0105039}.

\item
C.~A. Fuchs, ``Quantum Foundations in the Light of Quantum Information,'' to appear in {\sl Proceedings of the NATO Advanced Research Workshop on Decoherence and its Implications in Quantum Computation and Information Transfer}, edited by A.~Gonis (Plenum Press, NY, 2001). (Until then, see \quantph{0106166}.)
\end{itemize}

In the case of the book (the samizdat), a good place to start within it is the chapter on Rolf Landauer.  But also look up the references to John Wheeler and furthermore the discussion on {\Poincare} (and Boutroux) on page 195.  Both manuscripts, by the way, are also posted at my website listed below.  (At the website, you can find my CV in case you worry that I may not be legitimate.)

Of a more important nature, I wonder if you might help me with a project I am currently working on.  It is in putting together a large compendium of quotes from and citations to materials concerned with the idea that the universe may be a ``malleable'' entity.  It will eventually be submitted to {\sl Studies in History and Philosophy of Modern Physics\/} (probably this Fall), and presently stands at 96 pages, with 422 citations.  The working title is, ``The Activating Observer: Resource Material for a Paulian--Wheelerish Conception of Nature.''

If you wouldn't mind, I'd like you to look over it and point out any citations I missed.  Especially as regards your own work, I know that I must have missed several things.  But also, you may be able to help me with some obscure materials that I've never seen before.  I'm shooting to make the document as complete as I can.

\section{16-07-01 \ \ {\it Insults}\ \ \ (to S. J. van {\Enk})} \label{vanEnk24}

\bq\noindent
``The only gracious way to accept an insult is to ignore it; if you can't ignore it, top it; if you can't top it, laugh at it; if you can't laugh at it, it's probably deserved.''\medskip

--- Russell Lynes (b.\ 1910), U.S. editor, critic. {\sl Reader's Digest\/} (Brit.\ ed., Dec.\ 1961)
\eq
[The recording of this was prompted by my 04-07-01 note ``Carts and Horses'' to U. Mohrhoff.  It hardly needs to be said, but on that day I was not laughing.]

\section{18-07-01 \ \ {\it Horizons} \ \ (to J. Bub)} \label{Bub4}

That is awful news about your visual problem and its possible causes.
Please do keep me up-to-date on your health.  For my own part I will
cross my fingers and think of you often.

Don't worry at all about leaving me up in the air concerning a visit
to Provence.  Because of certain of our own medical issues, I've been
lobbying my wife to postpone her European vacation until September or
early October anyway.  (So, you see, I would have to leave you up in
the air right now too.)  If it happens, it happens. The main thing is
that it sounded like a good opportunity to pound out the similarities
and distinctions between our points of view on quantum mechanics
without being interrupted every three minutes.

I know I suggested I would write a longer letter soon, but I'm going
to wimp out of it again for now.  It would concern the main point of
distinction I see between us (and also between myself and Pitowsky).
Namely, A) that I view a large part of quantum mechanics as merely
classical probability theory (which on my view may be an a priori
``law of thought'') PLUS an extra assumption narrowing down the
characteristics of the phenomena to which we happen to be applying it
to at the moment, while B) you are more tempted to view quantum
mechanics as a {\it generalization\/} of classical probability theory
(and with it information theory).  I know that my view is not fully
consistent yet, especially as I have always distrusted mathematical
Platonism---which you pointed out to me I am getting oh so close
to---but it still feels more right (to me, of course).  Ben
Schumacher, {\Ruediger} {\Schack}, and I had a long discussion on this (on
a long walk) the day after the round table, and I'd like to record
that too.  Ben took a stance quite similar to yours, and maybe even
{\Ruediger} did too (despite his overwhelming Bayesianism).  So, I may
be the lonely guy out on this.  And my view may be subject to change.

What I probably really need right now is more conversation than
writing.  So, I do hope I get to see you in an uninterrupted way
soon.  (By the way, would it be possible for me to get a copy of the
talk you gave in {\Vaxjo}?  Could you copy that and mail it to me?)
For now, let me post below parts of two notes I wrote Andrei
Khrennikov.  They touch on the discussion above, even if they are
somewhat out of context here.  The second note, in particular, struck
me as a clean way of stating my position (in a way that I hadn't
explored before).  Maybe that'll help to zoom us in on the relevant
issues.

There is some good news on the horizon.  Gilles and I will be holding
another foundations meeting in 2002.  (Purely quantum info people,
much like the original.)  All the details aren't clear yet, but we
think we may be able to have desk space for people, it may be for an
extended period---maybe a month---with a revolving set of
participants, etc., etc.  I hope you'll be able to join us (once I
get the details to you).  The main difficulty is that it may have to
be in the fall (after the school semester starts), which will cause
participation trouble for those with a teaching load.

\section{18-07-01 \ \ {\it Letters:\ the Long and the Short, 2} \ \ (to A. Plotnitsky)} \label{Plotnitsky4}

I agree.  We ought to get together (for lunch?)\ sometime while you're in NYC\@!  I'll probably be tied up until at least July 25, though.  Are you in email contact all the time?  So that we can try to set something up more spur-of-the-moment if need be?  I'd probably just take a train in and meet you somewhere convenient in Manhattan.

Do you know anything about Schopenhauer's ``will''?  Could you give me a small lecture on it when we meet.  Would you say it's anything at all like your ``efficacity''?

\subsection{Arkady's Reply}

\bq
I can easily explain to you Schopenhauer's ``will'' and its relation to
``efficacity.'' The short answer is no, they are not the same, but one could
interestingly qualify why not and thus illuminate better what I mean by
efficacity or how quantum objects should be seen according to informational,
rather than ontological, view.  If we do not manage to meet this time
around, I could still explain this to you by email, or we might arrange for
a phone date one day.
\eq

\section{18-07-01 \ \ {\it Page 270!!}\ \ \ (to L. Smolin)} \label{SmolinL2}

I'm excited to hear that you're reading some of my things. Especially
since I've come across pages 270--272 of your book (paperback
edition)!  Indeed, there appears to be a significant overlap between
some of our toy ideas.  The one I'm speaking about is (in a technical
way) the undercurrent of my paper ``Quantum Foundations in the Light
of Quantum Information.''  But, you can find broader-view statements
of it on pages 156 and 190 of my samizdat. Also, you can see a trace
of it in {\Mermin}'s foreword, page iii, last paragraph.  Do you see the
overlap that I do?

Anyway, I find this quite intriguing:  Somehow, I had gotten the
impression that you were a staunch many-worlder, and that our views
of quantum mechanics might be diametrically opposite.  I really
apologize for my previous misreading (based on reading your book for
an hour in a bookstore one day).  I will read your book more
carefully, and also look at the papers you recommended.

I also apologize for writing you back so late.  Late last week I had to give a presentation to the president and vice-president of Bell Labs, and that really had me on edge until it was over with.  (The VP, Bill Brinkman, by the way, will become the president of the APS in January.)  After that, my wife and I took a long weekend in the country, and then I returned to find a hundred fires that needed putting out at the office.

I had contacted Perimeter (both Laflamme and Burton) previous to my contacting you, and I talked to Howard Burton for a long time on the phone Monday.  So, they are aware of my existence.  You may have been confused:  I'm not looking for a postdoc position, but for (possibly) a long-term one.  I already have a permanent research position at Bell Labs.  The issue I'm toying with in my mind is in turning a more directed focus on the quantum foundations.  (I've recently organized two large meetings with that theme, one in {\Montreal} and one in Sweden---and am now working on a third bigger one in {\Montreal}---and it's given me a taste for the hope that the quantum information community may be the key to this long-annoying puzzle.)  If Perimeter can help me in making these dreams come true, then I may be all for it.  What are your thoughts on Perimeter?

You asked me, ``What is the Paulian idea?''  I wish I knew!  More
seriously, the best summaries I can give you are 1) the {\it
conjunction\/} of two {\Pauli} quotes on page vii of the samizdat, and
2) the wispy little piece I wrote in the first paragraph of Section 4
of my paper ``Quantum Foundations in the Light of Quantum Information''.  This is a
very deep idea I think, and I don't know that I've ever seen it
expressed anywhere except (very sketchily) in {\Pauli}'s writings.  It
is that, in a world where the experimental context cannot be deleted
from the consequences it brings about, there must be a kind of
randomness or impredictability.  Else there would be no way to
distinguish between wake and dream for any observer who makes use of
such contexts.  It is the ultimate impredictability of the
consequences of our interactions with the world that gives us firm
evidence that there is something beyond us.  By this view, the world
is not real because it can be mathematized completely, but because it
cannot.

\section{20-07-01 \ \ {\it The Matchmaker} \ \ (to A. J. Landahl \& N. D. Mermin)} \label{Landahl2} \label{Mermin25.1}

Please allow me to introduce you two to each other:
\begin{itemize}
\item[]
David meet Andrew:
Andrew Landahl is a promising young grad student in quantum information at Caltech, with a little bit of a taste for fundamental questions.
\item[]
Andrew meet David:
David Mermin is a promising recent retiree from Cornell with a long history in fundamental questions, and a little bit of a taste for quantum information.
\end{itemize}
I woke up this morning and said to myself, ``If you can't do good physics yourself, be a matchmaker!''  So, I introduce you to each other because you've both recently written me about---I think---the same question.

Andrew said,
\bal
I spent most of my time discussing the Correlations section.  I went
in detail over your proof of what I called ``Gleason's theorem for
Classically Semilocalizable Operations (CSOs)'' in deference to the
terminology introduced in a recent paper by Beckman et al.\
[i.e., \quantph{0102043}]. Personally, I thought it was cool that the tensor product arose out of
noncontextuality and the measurement model.  John Preskill wasn't so
impressed --- he believes that the tensor product will arise out of
any reasonable model of measurement which has the property of
locality.  (I.e., where neither Alice's nor Bob's local actions can
meaningfully impact the other.)  He may be right, but for me that
isn't the point.  The point for me is that the proper way to view
Gleason's theorem is as a machine.  The input to the machine is the
measurement model and the output of the machine is the state space
structure and the probability law.  I suppose I should have realized
this a long time ago (especially once the POVM Gleason's theorem was
proven), but it took this proof to open my eyes.

What especially excites me about this point-of-view is its
potential impact on quantum field theory.  The main point of the
Beckman et al.\ paper is that causal measurements and localizable
measurements are not one in the same.  Wouldn't it be interesting to
see what happens when we impose only causality on our measurement
model and send it through the ``Gleason machine?''  What do you suppose
the resultant structure of the state space would be?
\eal
David said,
\bdm
The question you're evading is what it means for one and the same
positive operator $E$ to appear in many different POVMS. A POVM is
defined by the the entire collection of $E$'s that sum to the identity.
What does it mean to associate a probability with a single one of them
without reference to any of the others?  Isn't this a question that
simply has to be addressed, independent of whether you think a quantum
state is a state of knowledge or a state of the world, or whether you
think a probability is a degree of belief or an objective propensity?
Different POVMs (with one $E$ in common) are associated with entirely
different interventions.  But somehow you're thinking of that one $E$
these utterly different procedures have in common as an independent
outcome with a likelihood all its own.   It sounds to me as if you're the one who's reifying things. [\ldots]

No!  The signalling has nothing to do with quantum-state change.
(We've been through this before.)  If Bob and Alice share a large
number of identically prepared pairs, then a very reasonable
requirement is that the statistical distribution of outcomes Bob gets
from his members of the pairs cannot depend on what Alice chooses to
do to her members.  (If it did Alice could send useable unmediated
signals to Bob.)  Again, this has nothing to do with how you like to
think about probabilities or quantum states.  I offered this to you as
an example of a situation in which you can, in fact, justify the
non-contextuality of certain probabilities by appealing to an
independent physical requirement (no remote signalling).
\edm

1) Am I indeed on the right track that you two are thinking about the same things?  2) Am I to take it from Andrew's note that there is some evidence that no-signalling (in a Gleason-like theorem) may not be enough to specify the standard quantum probability rule?

If any sparks fly, I hope you'll let me be a voyeur.

\section{21-07-01 \ \ {\it The Reality of Wives} \ \ (to A. J. Landahl \& J. Preskill)} \label{Preskill2} \label{Landahl3}

This morning one of the local hospitals had a fund-raising flea
market, and I picked up a copy of Martin Gardner's {\sl The Whys of a
Philosophical Scrivener\/} for \$0.50.  I haven't been able to put
the thing down all day; it's quite good, and the beginning parts are
especially relevant to my recent discussion with you.

This evening while sitting outside enjoying the end of the day, I
couldn't help but read Kiki a cute little story from it.  Gardner
writes:
\bq
When I was an undergraduate philosophy student at the University of
Chicago I attended a seminar given by Bertrand Russell.  Carnap, then
a professor at Chicago, went to these sessions and often engaged
Russell in spirited debates which I only partly comprehended.  On one
occasion they got into a tangled argument over whether science should
assert, as an ontological thesis, the reality of a world behind the
phaneron.  [Phaneron was {\Peirce}'s term for the world of our
experience, the phenomenal world.] Carnap struggled to keep the
argument technical, but Russell slyly turned it into a discussion of
whether their respective wives (Russell's new wife was knitting and
smiling in a back-row seat) existed in some ontologically real sense
or should be regarded as mere logical fictions based on regularities
in their husbands' phaneron.

The next day I happened to be in the campus post office, where
faculty members came to pick up mail.  Professor Charles Hartshorne,
a whimsical philosopher from whom I was then taking a stimulating
course, walked in, recognized me, and stopped to chat.

``Did you attend the Russell seminar yesterday?'' he asked.  ``I was
unable to go.''

``Yes,'' I said.  ``It was exciting.  Russell tried to persuade
Carnap that his wife existed, but Carnap wouldn't admit it.''

Hartshorne laughed.  Then, by a quirk of fate, in walked Carnap to
get his mail.  Hartshorne introduced us (it was the first time I had
met Carnap; years later we would collaborate on a book); then, to my
profound embarrassment, Hartshorne said:  ``Mr.\ Gardner tells me
that yesterday Russell tried to convince you your wife existed, but
you wouldn't admit it.''

Carnap did not smile.  He glowered down at me and said, ``But that
was not the point at all.''
\eq

I followed that by saying, ``You know some of my friends are afraid
that I don't believe in reality.  So there, you're just a figment of
my imagination!''  She reacted in shock.  ``Well, I know that can't
be true,'' she said.  ``Clearly you'd make some changes!''

\section{22-07-01 \ \ {\it Noncontextual Sundays} \ \ (to N. D. {\Mermin})} \label{Mermin26}

I know you're busy, but I'm going to try again.  (Don't feel the need
to write back until you get some time.)  The issue is still
noncontextuality in the Gleason-like theorems:  Is it a natural
assumption or not?

Here was the best answer I gave you before, but now I'm going to try
to improve on it.

\bq\bdm
The question you're evading is what it means for one and
the same positive operator $E$ to appear in many different POVMS.
\edm
No, I don't think I'm evading it.  It means that those various
interventions or ways of gathering data---those POVMs---physically
diverse though they may be---all lead to at least one common
possibility for what my knowledge can be updated to (modulo the
unitary readjustment).
\eq

The point I'm going to try to make is that not only am I finding
noncontextuality a natural assumption, but actually it may be the
most {\it basic\/} assumption of the whole game.  (I.e., it may even
be prior to the notion that measurements correspond to POVMs.)  The
idea is captured above, but---I can see now---it is in too
idiosyncratic of a language to convince you easily.

Here's the new shot at it (emphasizing a slightly different aspect
than previously).

1)  Here's the scenario.  Forget about quantum mechanics for the
moment.  Let me take a system $S$ and imagine acting on it with one
of two machines, $M$ and $N$.  For the case of machine $M$, let us
label the possible consequences of that action $\{ m_1, m_2, \ldots
\}$.  For the case of machine $N$, let us label them $\{ n_1, n_2,
\ldots \}$.

2)  If we are good Bayesians, nothing will stop us from using all the
information available to us to ascribe probabilities to the
consequences of these two potential actions.  Thus we naturally have
lying around two probability distributions, $p_M(m_k)$ and
$p_N(n_k)$.

That's well and good, but it's hardly a physical theory yet.  We need
more.  So, let us suppose the labels $m_k$ and $n_k$ are at least
drawn from the same master set (possibly even a set with further
structure, like a vector space or something).  But then we must ask,
under what conditions should we identify two particular labels $m_i$
and $n_j$ with the same element in the master set?

There's really only one thing lying around to do it with, and that's
the probability assignments.  If $p_M(m_i) \ne p_N(n_j)$, then surely
we would not imagine identifying $m_i$ with $n_j$.  If, on the other
hand, $p_M(m_i)=p_N(n_j)$ {\it regardless\/} of the initial state of
knowledge about $S$, then we might think there's some warrant for it.

And that's the whole story of noncontextuality.  It is nothing more
than:  The consequences ($m_i$ and $n_j$) of our disparate actions
($M$ and $N$) should be labeled the same when we would bet the same
on them in all possible circumstances (i.e., regardless of our
initial knowledge of $S$).

By this point of view, noncontextuality is a tautology---it is built
in from the start.  Asking why we have it is a waste of time.  Where
we do have a freedom is in asking why we make one particular choice
of a master set over another.  Why should the $m_i$'s be drawn from
the set of ``effects'' (i.e., the positive operators smaller than the
identity on some Hilbert space)?  Recall the problem on page 86 of
the samizdat.  Not all choices of the master set are equally
interesting once we've settled on noncontextuality for the
probability assignments.

You see, I really never do dismiss anything you say!  Now I'm off to
buy a new BBQ grill.  (My family is tugging on me, and refused to let
me try to hone this letter.  But I really hope by this point it does
make some sense.)

\section{23-07-01 \ \ {\it Law without Law} \ \ (to J. Summhammer)} \label{Summhammer1}

I very much enjoyed reading your letter to {\Carl} {\Caves} titled
``promoting the Bayesian view.''  It was quite thoughtful, and makes
me regret not having talked to you more while we were in Sweden.

Please allow me to ask a couple of questions based on what you wrote.

\bjs
Here I think that, even if there is absolutely no order in the
physical world, it will exhibit statistical order to a rational
observer. The existence of rationality is to be taken as outside the
physical world. It is a transcendental fact. (The term ``law of
thought'' in your paper circles around the same thing.) Analysis of
brain functions and molecules explains nothing, because that analysis
is done by means of rationality.
\ejs

There have been times in my life when I have been very attracted to
ideas like this.  In particular, right now might be one of them
(though I have a history of going up and going down).  My first
exposure to the idea came from John {\Wheeler}'s writings on what he
termed ``law without law.''  In fact presently, I'm putting together
a large compendium of quotes and citations titled ``The Activating
Observer: Resource Material for a Paulian--{\Wheeler}ish Conception of
Nature.''  At the moment, it consists of 423 annotated citations,
taking up 96 pages of print.  The manuscript is far from complete,
but will eventually be submitted to {\sl Studies in History and
Philosophy of Modern Physics}.

I wonder if you have any suggestions for things I should include in
it (based on your passage above).  If so, please give me as complete
of references as possible.  If I'm not mistaken, I detect a Kantian
tinge in your thought:  that's an area I haven't explored too deeply
in my compendium.

\bjs
p.21: \ldots\ the Bayesian interpretation places actualization
outside its provenance \ldots\

I agree. Defenders of interpretations which claim to handle
actualization should read selected articles on the mind-body problem,
written over the last three thousand years.
\ejs

Would you mind expanding on this, and also what are some of those
selected articles?  Can you provide references?

Anyway, it was very nice meeting you for the first time.

\subsection{Johann's Reply}

\bq
Now I come to your mail.

As to references on what is to be accepted as transcendentally given I would
first think of Kant, in the {\it Critique of Pure Reason}. Then Hume should have
something on it, but I read only passages of Hume. Then there is C.~F. von
{\Weizsacker}'s {\sl Aufbau der Physik\/} (1985), (perhaps {\sl Reconstruction of Physics}) and his more philosophical book {\sl Zeit und Wissen\/} (1992), ({\sl Time and
Knowledge}), both of which, unfortunately, are not translated into English,
as far as I know. But he has a number of articles in English. Lyre has put a
number of papers onto the {\tt quant-ph} server in recent years. He is expounding
on von {\Weizsacker}'s ideas. (There is one paper of Lyre which I like quite
much. It dissects Bell's ``Against Measurement''. Perhaps it fits into your
compendium. Exposing the naive parts of Bell's thinking might be wholesome
to the community.)

I have scanned through {\sl Aufbau der Physik\/} and a little through {\sl Critique of
Pure Reason\/} over the weekend to find something that would support the
following passage of my mail that you wanted me to find references to:
\bjs
Here I think that, even if there is absolutely no order in the
physical world, it will exhibit statistical order to a rational
observer. The existence of rationality is to be taken as outside the
physical word. It is a transcendental fact. (The term ``law of
thought'' in your paper circles around the same thing.) Analysis of
brain functions and molecules explains nothing, because that analysis
is done by means of rationality.
\ejs

I found a compact one (but certainly not the best one) in {\sl Aufbau der Physik\/}, where v. {\Weizsacker} freely outlines Kant's thinking and then says in
which way his way of thinking about QM has gone beyond Kant.

It is in a section on Geist und Form (Mind and Form, pages 580--587). First
he says there is no reason why abstract quantum theory should not also be
applicable to knowledge which consciousness can obtain about itself.
Then he remarks that one might have objections against this, and, following
Kant, one objection might go like this: [I translate a section from pages
582--583]
\bq
``\ldots\ Loosely leaning on Kant we could make the following injection: Science
is knowledge, and therefore --- insofar as we may use the notion of
consciousness --- it is the contents of consciousness. Matter is the object of
knowledge. In scientific usage forms are notions. Therefore it appears
meaningful to explain the matter of physics by means of forms, thus notions;
as such physics is what we can know of matter. Consciousness is a
precondition of knowledge; consequently it would be circular to explain it
by the means with which knowledge operates, the forms. Following Kant, it is
for this reason that the knowing subject cannot be described as substance;
for substance is itself a category, a notion.

This injection is based on the argument of circularity, thus on the
hierarchical approach  of traditional philosophy (also see {\sl Zeit und Wissen\/} 5.2.3). [But] we are moving on the circular walk around and through nature.
Physics describes that which is lawful in nature, which in classical physics
was called matter. Abstract quantum theory can attempt with the same right
to describe that which is lawful in consciousness. \ldots\ The claim to achieve
a complete description of reality in this manner is presumably unattainable;
but it is legitimate to expect a consistent description. \ldots''
\eq

So much the quotation from von {\Weizsacker}. An important idea of his is the
``Kreisgang der Natur'' ($=$ walking through nature in a circle). This does not
mean circularity in reasoning, but recognizing that no matter with which
notions you start, you must start with some, and you can try to see how far
they can be illuminated when applying this whole set of notions to analyze
each of them. In this von {\Weizsacker} seems to have been influenced by Bohr
(as he himself often admits, his other influences being Plato and Kant). For
instance, you could start with what we know of molecules and see how much it
explains of the workings of the brain and the formation of concepts the
brain arrives at and why these should in turn lead to the idea of molecules.
What I like about this is that the question of what exists primarily is
tacitly shown as ill-conceived. (If I am not wrong Bohr liked to ask ``what
do you mean by this?'', and this outraged so many people, because they could
not accept that you can think penetratingly and live happily without
believing that all is based on material points that interact with each
other. Here, von {\Weizsacker} has a poignant statement: Druck und Stoss sind
keine Grundbegriffe der Physik. (Hitting and tossing are not basic notions
of physics.) You couldn't dismiss classical physics more thoroughly.
Personally I have a less dignified view of classical physics: It is the ape's way of projecting his muscular self-experience onto the world.)

Perhaps you should get in contact with von {\Weizsacker} directly, before he
dies. The man is 89, but from what I hear still very much active in
seminars. I have met him only once at a conference. He was a close friend of
Heisenberg. Lyre would be able to give you his coordinates.
If I am not mistaken, there was a common effort of Wheeler, von {Weizsacker},
D. Finkelstein and others, in the late sixties, early seventies, of really
getting on with quantum theory. von {\Weizsacker}'s input was ``Ur-theory'' and
that the distinction of past and future is fundamental (one is fact, the
other is possibility, like data and wavefunction).  In Ur-theory knowledge
is built up from empirically decidable alternatives, but that these
decisions must have a material embodiment. That's what struck me most, when
I first encountered this. That the distinction of knowing, and that what you
know about need not be considered so fundamental. Knowing constitutes the
thing, but not in the sense of idealism. Very quantum mechanical.
A nice idea that shows how this can permit you to at least try to break out
of the old realism-idealism prison is that, also human consciousness, as
something that can be talked about by means of decidable alternatives, must
have a physical manifestation. I translate a passage from page 581 of {\sl Aufbau
der Physik}: ``\ldots\ When we do the step towards concrete quantum mechanics, we
will build up the alternatives encountered in obtaining the knowledge
consciousness can have of itself by means of ur-alternatives. From this we
must conclude that it must be possible to describe human consciousness as a
body in three-dimensional space. We might expect that this should be the
human body. \ldots'' It is perhaps necessary to say that for von {\Weizsacker} three-dimensional
space follows almost trivially from yes-no alternatives. (To me it is not
that trivial.)

I think the recent effort of Anton Zeilinger and \v{C}aslav Brukner, of deriving
quantum theory by saying an elementary system carries one bit is a revisit
to ur-theory. (But I don't think Anton would agree, and \v{C}aslav might be
hesitant. They are reading it off from their EPR and teleportation
experiments, where you can posit something and then ask yourself how many
bits of information do you have about it, rather than saying, I know so many
bits, what something does this constitute?)

As to actualization as essentially the mind-body problem, I touched on this
with Kant and Schopenhauer. My occupation with the latter was 20 years ago,
and with Kant 15 years ago. But from works of von {\Weizsacker} I learnt that
already Aristotle and Plato were aware of the problem, and the modern
distinction of mind and body as two substances stems from Descartes. That is
why I said the problem of actualization has three thousand years of history.
To me it boils down to why there is something rather than nothing. To say
there is something implies two concepts: The sensor and the sensed, or The
observer and the observed, or The contemplator and the contemplated (ordered
with increasing levels of complexity). Anyway, the question has a long
tradition. In \quantph{0107005} on p.8 Ulrich Mohrhoff has just cited it in
the following form: ``Accounting for the existence of facts is the same as
explaining why there is anything at all, rather than nothing --- an impossible
task.'' Perhaps he knows who pondered this first.

Finally, a quote from Kant, which I happened to open accidentally just now.
It deals with empirical information ({\sl Critique of Pure Reason}, in the second
analogy of transcendental analysis; A210, B255, my translation from the
German edition):
\bq\noindent
``\ldots\ All increase of empirical knowledge, and every progress in perception
is nothing than an extension of the determinations of the inner sense, that
is, a progression in time, the objects may be what they please, appearances
[to the senses], or pure intuition. \ldots''
\eq
(My addition in rectangular brackets.)

It must have been this or a similar statement that lead me to the idea of
the robot observer whose information about the world can only increase,
which I mentioned in the mail to Carl Caves.

In my next mail I hope to get down to asking you a few questions on your
paper ``Quantum Foundations in the Light of Quantum Information''. And the
prior probability distribution of probabilities in the classical de Finetti
theorem is causing me some trouble. Also, I would appreciate a copy of your
paper (PS, PDF, or address for downloading preferred) ``How much information
in a state vector?'' Unfortunately I had not read any of your papers before {\Vaxjo}. There would
have been a number of interesting points to discuss.
\eq

\section{23-07-01 \ \ {\it A Nonbayesian Bayesian?}\ \ \ (to C. M. {\Caves})} \label{Caves1}

I enjoyed reading your dialogue with Summhammer.

I have one question of my own.

\bcc
I do believe that natural selection only works in a world with at
least a statistical order, which leads to the quasi-determinism of
the macroscopic world.   That we and other creatures are exquisitely
attuned to this order, to the point of often finding it where it's
not really there, is not surprising.  The costs of finding order
where there is only chaos must be less than the cost of failing to
notice and take advantage of order when it is there.
\ecc

What on earth do you mean by this?  Similarly when you write,
``statistical order is the first element of the Bayesian reality,''
in Section 7 of your {\sl Resource Material\/}?

Your phrase ``world with statistical order'' seems to teeter awfully
close to an objectivist notion of probability.  And it frightens me,
of course, having fully made a conversion now.

\section{23-07-01 \ \ {\it The Principal PrincipleS} \ \ (to C. M. {\Caves})} \label{Caves2}

It dawned on me that I should chide you on your discussion on page
21, starting with ``Two further comments on Hamiltonians \ldots.''
You left out the all-important part about how to connect the two
notions of Hamiltonian, the subjective---or effective, as you call
it---with the objective.  It must be a {\Caves}ian version of the
principal principle:  When the objective determinant of the time
evolution of one's subjective states of knowledge is known, then the
subjective determinant of one's subjective states of knowledge should
coincide with it.

Another typo btw (I think). Page 19: ``If you have maximal
information about a quantum system and you want to retain it, you
must know the system is Hamiltonian.''  Don't you mean ``system's
Hamiltonian''?

\section{24-07-01 \ \ {\it Feynman Quote} \ \ (to myself)} \label{FuchsC3}

\bq
If, in some cataclysm, all of scientific knowledge were to be
destroyed, and only one sentence passed on to the next generation of
creatures, what statement would contain the most information in the
fewest words?  I believe it is the atomic hypothesis (or the atomic
fact) that all things are made of atoms---little particles that move
around in perpetual motion, attracting each other when they are a
little distance apart, but repelling upon being squeezed into one
another.

Everything is made of atoms.  That is the key hypothesis.
\eq

\section{26-07-01 \ \ {\it BZZ} \ \ (to N. D. {\Mermin})} \label{Mermin27}

Renes and I are meeting Plotnitsky for lunch in NYC tomorrow.  The
agenda is Schopenhauer's ``will'' and Plotnitsky's ``efficacity''
(not to be confused with ``effervescence'').

\section{26-07-01 \ \ {\it In the Middle}  \ \ (to Y. J. Ng)} \label{Ng2}

As you can tell from the note I just cc'd you, I'm finally taking to your paper(s) in a serious way.  I'm not sure if you're familiar with all the terms I used in the letter (effects, operations, etc.), but my main worry is that your Eq.\ (1) numerically coincides with what has come to be known as the ``standard quantum limit'' (albeit in a different context), which we in the quantum measurement community know can be broken.  Maybe I told you about some of those troubles during my visit to NC.  Your Eq.\ (3) seems to be on firmer ground, but once again when we get into the quantum regime, it is not clear to me that troubles don't creep in.

Also I have some lesser---perhaps semantic---worries.  Namely, identifying a single-component system with a computational device.  That does not fall within the standard model of computing (either quantum or classical), the Turing machine.  So, one might call the black hole a limit point of a sequence of computers, but I think I would be hard pressed to identify it with a computer itself.

Also I worry about the general nature of your ``simple computers'', and why these bounds should only apply to such things.  This does strike me as a weak point.  You really should build a stronger argument for it (i.e., one better than ``this is the only case in which the derivation seems to work'').

But I certainly like the general flow of the paper.  It has the right feel about how to start to build up quantum considerations about spacetime---these are things that I have maybe pushed to the back of my mind for too long.

Anyway, once I hear from Ozawa, I will probably react again.

\section{27-07-01 \ \ {\it In the Night}  \ \ (to Y. J. Ng)} \label{Ng3}

\byjn
But the first thing that I am interested in is what replaces my Eq.\
(1) according to you.
\eyjn

My fear is that there is simply no limit at all.  (But it really is a fear, and I hope I'm wrong for your sake.)  While we await Ozawa's reply, have a look at the article below.  As I recall, it is written in a style that you would appreciate.  It gives a nice physical motivation for the SQL that is very similar in flavor to the derivations in your article.  Tell me whether you find its arguments believable.  The problem is, they turned out to be wrong.  You'll also note the significant overlap between that problem and yours.  So, even if it is otherwise irrelevant---which I am afraid it's not---we're all bound to learn something from tabulating the distinctions.  I really need to look up the old Wigner article, but I haven't had a chance to get to the library and I'm traveling this weekend.

\bq\noindent
Phys.\ Rev.\ Lett.\ {\bf 54}, 2465--2468 (1985) \medskip\\
Defense of the standard quantum limit for free-mass position\medskip\\
Carlton M. Caves\\
Theoretical Astrophysics 130-33, California Institute of Technology, Pasadena, California 91125\medskip\\
Measurements of the position $x$ of a free mass $m$ are thought to be governed by the standard quantum limit (SQL): In two successive measurements of $x$ spaced a time $\tau$ apart, the result of the second measurement cannot be predicted with uncertainty smaller than $(\hbar\tau/m)^{1/2}$. Yuen has suggested there might be ways to beat the SQL.  Here I give an improved formulation of the SQL, and I argue for, but do not prove, its validity.
\eq

\section{29-07-01 \ \ {\it Sunday Morning, Thinking of You}\ \ \ (to H. J. Bernstein)} \label{Bernstein3}

\noindent Hey old philosophy friend,\smallskip

I haven't heard from you in a while.  And you never even replied to my last message.  So, I wonder whether you got it.

Lately I've taken to trying to mix Arthur Schopenhauer, John Dewey and William James into a good quantum stew.  I'm certainly getting the feeling that some of their ideas were before their time (and, indeed, a better fit to the quantum world than the classical world of their focus).

But do write me.  I am lonely for your company.

\section{30-07-01 \ \ {\it The Morning After} \ \ (to S. Henry)} \label{SHenry1}

\bshenry
I remember you dude!  I guess the most vivid memory is
being at a party at Marie Clark's house.  If I
remember correctly you called me Mr.\ Magician because
I did a little trick with a quarter or something.  Or
maybe it was the way I could persuade Lori to do what
I wanted while you had a harder time getting her to
bend to your will!  (hehe)

It seems you've really excelled in science.  And yes
I'm totally interested in ``quantum teleportation'' and
physics in general.  I haven't read any of your
writings yet but will do so.  I've read several books
on physics/science over the years.  Some of my
favorites are {\bf Hyperspace} by Michio Kaku, and {\bf A
Brief History of Time} by Stephen Hawking.  And, of
course, I did read {\bf The Tao Of Physics} by Fritjof
Capra that explores the parallels between physics and
Eastern mysticism.  It sounded so much like {\bf The Tao
of Pooh} I just had to read it.  If you know of any
other books/authors that you think I'd be interested
in please let me know!
\eshenry

I remember deciding I would devote most of my life to quantum mechanics my second week in college after reading Heinz Pagel's {\sl The Cosmic Code}.  It made a big splash in my mind at the time:  I don't know what it'd do now, maybe the same.  For my present mentality, a very good book for the layman is David Mermin's {\sl Boojums All the Way Through}.  (Most importantly, read the chapters on ``quantum mysteries for anyone.'')  Both these books are out of print, I think.  But you can find them easily enough in a library, or by going to the used book section at {\tt BarnesAndNoble.com} (there's hardly a book one can't find there).  Things getting closer to what I do in the field---quantum information theory and quantum computing---can be found in Tom Siegfried's {\sl The Bit and the Pendulum}.  I've heard it's pretty good, but I haven't read it myself.  Also, tangentially, Simon Singh's {\sl The Code Book}---in connection to quantum cryptography---but, again, I haven't read it either.  I've just heard good things about it.

\bshenry
So how'd you find my webpage?  Were you looking for the DJ Scott Henry
and found my page instead?
\eshenry

No, it was 4:00 AM, and I was going through the search engines looking for things on Arthur Schopenhauer and his ideas about ``the world as will and representation.''  An advertisement for {\tt classmates.com} kept flickering at me, and I finally had a weak moment.  So, I went to have a look (like I did a few months earlier when I put my own name there).  This time I saw your name and went to your website.

\section{30-07-01 \ \ {\it Britannica} \ \ (to J. M. Renes)} \label{Renes3}

This was the only passage in the {\sl Encyclopedia Britannica\/} that
I could find about Nietzsche that even remotely resembled quantum
mechanics.  In general, he looks like a lot to wade through for
little return.

\bq
Perspectivism is a concept which holds that knowledge is always
perspectival, that there are no immaculate perceptions, and that
knowledge from no point of view is as incoherent a notion as seeing
from no particular vantage point. Perspectivism also denies the
possibility of an all-inclusive perspective, which could contain all
others and, hence, make reality available as it is in itself. The
concept of such an all-inclusive perspective is as incoherent as the
concept of seeing an object from every possible vantage point
simultaneously.

Nietzsche's perspectivism has sometimes been mistakenly identified
with relativism and skepticism. Nonetheless, it raises the question
of how one is to understand Nietzsche's own theses, for example, that
the dominant values of the common heritage have been underwritten by
an ascetic ideal. Is this thesis true absolutely or only from a
certain perspective? It may also be asked whether perspectivism can
be asserted consistently without self-contradiction, since
perspectivism must presumably be true in an absolute, that is a
nonperspectival sense. Concerns such as these have generated much
fruitful Nietzsche commentary as well as useful work in the theory of
knowledge.
\eq

\section{31-07-01 \ \ {\it Parachutes}  \ \ (to C. H. {\Bennett})} \label{Bennett3}

\bcb
I'd be glad to recommend you to the Perimeter Institute, and I think it a prudent preparation given what I've heard about
Lucent.  But before you place too much faith in a place called perimeter, I would remind you of a famous saying in general
relativity (a field I understand hardly at all) to the effect that the boundary of a boundary is nothing.
\ecb

John Wheeler would say ``the boundary of a boundary is zero,'' so you were pretty close.  But, yes, you pinpoint a worry of mine.

Today, the bosses are gonna give another big general announcement.  I'm sitting on needles and pins.  I just want to do physics, and stop worrying about jobs!

\section{01-08-01 \ \ {\it The {\Montreal} Commune} \ \ (to W. K. Wootters)} \label{Wootters1}

Gilles and I are once again putting together plans for a quantum
information/foundations party in {\Montreal}, though this one may be a
little more like a commune than a party.  The {\Montreal} Commune.
(Probably more officially, ``Workshop on the Impact of Quantum
Information Theory on Quantum Foundations,'' or some such thing.)  We
hope you will join us as a communard.  [\ldots]

On another subject, the meeting in Sweden went quite well, but we did
miss you.  Maybe some of the most interesting discussions centered
around Andy Steane's paper ``Quantum Computation Only Needs One
World'' (which Richard Jozsa presented).  Doug Bilodeau had the
wonderful idea that perhaps some combination of it and your old Ph.D.
thesis could give us a deeper insight into where quantum computing
derives its power from:  Quantum computers are not powerful because
they perform so many calculations in parallel (as the many-worlds
pundits imagine), but rather because they do so FEW calculations!
I.e., Their power derives from not doing anything they don't have to
do for the final result.  (Much like in your thesis, the
photon---which can only express its preparation through a
probabilistic law---does better by explicitly NOT carrying around the
baggage of a local hidden variable theory.)  So, your input at these
meetings really would be very valuable.

Let us hear from you as soon as you can.

\section{01-08-01 \ \ {\it Missing Anyone?}\ \ \ (to L. Hardy)} \label{Hardy2}

Good to hear back from you so quickly.  Let me ask your advice:  would you mind looking over my list of potential attendees and telling me whether I missed anyone really interesting?  The criteria for getting there are:
\begin{enumerate}
\item
Have to be practicing quantum information or something sufficiently close,
\item
Should have a healthy attitude that there are still some mysteries in QM.  (By this criterion, Bennett, Deutsch, and Griffiths probably should be taken off the list.)
\end{enumerate}

\section{02-08-01 \ \ {\it Schopenhauer} \ \ (to U. Mohrhoff)} \label{Mohrhoff3}

Schopenhauer-like, not Kant-like.

\section{02-08-01 \ \ {\it Quantum Philosophy} \ \ (to C. H. {\Bennett})} \label{Bennett4}

That guy Mohrhoff.  By his classification, you are indeed more Platonic and I more Kantian.  (Even though I don't think I'm overly Kantian.)  But more importantly, also according to him \ldots\ HAH!\ \ldots\ I am merely ``unilluminating'' whereas you are ``inconsistent.''

\section{02-08-01 \ \ {\it The {\Montreal} Commune -- Ditto} \ \ (to B. W. Schumacher)} \label{Schumacher3}

I've marked you down in my spreadsheet, and we'll let you know what's
up in a couple of weeks.

\bbs
An interesting result that I happened on, based on two propositions:
(1) ``Information'' resides in the relation between systems, and (2)
``classical'' information is exactly that information which may be
copied.  So we have two systems, $Q$ and $R$, in a joint state
$\rho_{RQ}$. Think of $R$ as a ``record'' of $Q$.  Suppose we require
that there exists an operation on $R$ only such that, at the end of
the day, there are two systems $R1$ and $R2$ so that $\rho_{R1Q} =
\rho_{R2Q} = \rho_{RQ}$. (So both $R1$ and $R2$ can have just the same
relation with $Q$ that $R$ had.) This is possible if and only if the
state is of the form
$$
          \rho_{RQ} = \sum_k p_k |k\rangle\langle k| \otimes \rho_k
$$
where $|k\rangle$ is an orthonormal set of $R$-states.  The result is
a pleasant application of the no-broadcasting theorem.
\ebs

I like your result (which can't be questioned)!

No subject without an object.  No object without a subject.  No
information without both.  I like that---it seems like a good track
for ontologizing information, to the extent that it can be.  The only
thing that scares me is your secret desire to reify the quantum
state---namely by translating (1) into a statement about bipartite
quantum states (which seem to me to have no other good interpretation
than information to begin with)!  It takes information to get
information off the ground?

\section{03-08-01 \ \ {\it Making Quantum Look Like Bayes}\ \ \ (to M. C. Galavotti)} \label{Galavotti1}

I was reading Brian Skyrms' book {\sl Pragmatism and Empiricism\/} this morning, and I thought of you.  Perhaps you remember me from the old note below?  It dawned on me that I haven't told you about two of my latest papers (listed below).  They can be found on the Los Alamos archive \myurl{http://xxx.lanl.gov/find/quant-ph} by doing a search on my name, or, if you wish, I can mail them to you (provide an address).  I think you may enjoy them, as they go the greatest distance yet in my efforts to draw out the analogies between quantum mechanics and Bayesian probability theory.

I hope things are going well for you and that you are continuing your own Bayesian efforts.

\begin{itemize}
\item
C.~A. Fuchs, ``Quantum Foundations in the Light of Quantum Information,'' to appear in {\sl Proceedings of the NATO Advanced Research Workshop on Decoherence and its Implications in Quantum Computation and Information Transfer} edited by A.~Gonis (Plenum Press, NY, 2001). (Until then, see \quantph{0106166}.)

\item
C.~M. {\Caves}, C.~A. Fuchs and R.~{\Schack}, ``Making Good Sense of Quantum Probabilities,'' submitted to {\sl Physical Review A}.  (See \quantph{0106133}.)
\end{itemize}

\subsection{Letter to M. C. Galavotti dated 5 May 1998, ``If It's Not Too Late \ldots''}

\bq
I have just read your paper ``Probabilism and Beyond'' and am now even more excited than I was several days ago!  It looks as if there is much that I have missed out on by not having studied Ramsey and de Finetti before!  Unfortunately, I am also finding that there is even more of your work that I am not able to get my hands on here at this small library.  (Though Caltech has had literally 26 Nobel Prizes distributed to its faculty and alumni, it has a library about the size of a pea.)  If it is not too late---if you have not already sent the last one---can you also send me your article ``F. P. Ramsey and the Notion of Chance''?  I am now hoping that something along the lines of Ramsey's ``objective chance''---to the extent that I understand the notion from your last article---is just the sort of thing that I have been looking for.

My collection of your papers presently consists of:
\begin{enumerate}
\item
M.~C. Galavotti, ``Comments on Patrick Suppes `Propensity Interpretations of Probability'\,'' Erkenntnis {\bf 26}, 359--368 (1987).

\item
M.~C. Galavotti, ``Anti-Realism in the Philosophy of Probability:\
Bruno de Finetti's Subjectivism,'' Erkenntnis {\bf 31}, 239--261 (1989).

\item
M.~C. Galavotti, ``The Notion of Subjective Probability in the Work of Ramsey and de Finetti,'' Theoria {\bf 62}, 239--259 (1991).

\item
M.~C. Galavotti, ``Probabilism and Beyond,'' Erkenntnis {\bf 45},
113--125 (1996).
\end{enumerate}
If there is anything else that you've written that you think I should be aware of, please send that on also!
\eq

\section{05-08-01 \ \ {\it Sunday Morning, Thinking of You, 2}\ \ \ (to H. J. Bernstein)} \label{Bernstein4}

\bhbe
ISIS's fall series is supposed to be about scientific and medical
ethics: Can you connect your proposed talk and simplify its level to
be appropriate?
\ehbe

Does that mean there's an honorarium that comes with giving the talk?

I might be induced to give a William James kind of lecture \ldots\ with a title something like, ``Ethics in an Ultimately Lawless World.''  Or, ``Ethics in a Law-Without-Law World.''  Or, ``Timeless Ethics in an Evolving Universe?''  Use your imagination.

\section{07-08-01 \ \ {\it Knowledge, Only Knowledge} \  (to T. A. Brun, J. Finkelstein \& N. D. {\Mermin})} \label{Mermin28} \label{Brun1} \label{Finkelstein3}

Below is a note I started composing last Friday---but then had to
leave for a long weekend for my wife's birthday---and only finished
up today.  In the mean time, Todd and Jerry have skirted very close
to the point I wanted to make.  So, the note is not quite as relevant
as it might have been, but maybe some of it is still worth
contemplating.

\noindent ---------------------------

Allow me to start off in a fanciful way (like usual) with a couple of
quotes:

\bq
      The subjectivist, operationalist viewpoint has led us to the
   conclusion that, if we aspire to quantitative coherence, individual
   degrees of belief, expressed as probabilities, are inescapably the
   starting point for descriptions of uncertainty.  There can be no
   theories without theoreticians; no learning without learners; in
   general, no science without scientists.  It follows that learning
   processes, whatever their particular concerns and fashions at any
   given point in time, are necessarily reasoning processes which take
   place in the minds of individuals.  To be sure, the object of
   attention and interest may well be an assumed external, objective
   reality:  but the actuality of the learning process consists in the
   evolution of individual, subjective beliefs about that reality.
   However, it is important to emphasize, as in our earlier discussion
   in Section 2.8, that the primitive and fundamental notions of
   {\it individual\/} preference and belief will typically provide the
   starting point for {\it interpersonal\/} communication and reporting
   processes.  In what follows, both here, and more particularly in
   Chapter 5, we shall therefore often be concerned to identify and
   examine features of the individual learning process which relate to
   interpersonal issues, such as the conditions under which an
   approximate consensus of beliefs might occur in a population of
   individuals.
              --- pp.\ 165--166, Bernardo and Smith, {\sl Bayesian
              Theory}
\eq

\bq
      What is the nature and scope of Bayesian Statistics within this
   spectrum of activity?
      Bayesian Statistics offers a rationalist theory of personalistic
   beliefs in contexts of uncertainty, with the central aim of
   characterising how an individual should act in order to avoid certain
   kinds of undesirable behavioural inconsistencies.  The theory
   establishes that expected utility maximization provides the basis for
   rational decision making and that Bayes' theorem provides the key to
   the ways in which beliefs should fit together in the light of
   changing evidence.  The goal, in effect, is to establish rules and
   procedures for individuals concerned with disciplined uncertainty
   accounting.  The theory is not descriptive, in the sense of claiming
   to model actual behaviour.  Rather, it is prescriptive, in the sense
   of saying ``if you wish to avoid the possibility of these undesirable
   consequences you must act in the following way.''
              --- p.\ 4, Bernardo and Smith, {\sl Bayesian Theory}
\eq

Thanks again to all of you for letting me look in on your interesting
emails!  I've learned a lot from this exchange.  Last night I was up
between 1:30 and 5:00 reading them all one more time (and David's
original paper too), and thinking much harder than I had before on
these issues.  So, now I hope to say some things in that regard, but
not inane things (as I had done a couple of days ago).

The main thing that started striking me more deeply last night is
that now it is very obvious that you {\it have\/} an answer,
but---much more than ever before---I don't really understand what the
question ought to be.  In a sense, I'm only coming across the same
troubles (Samizdat, p.\ 236) I've had ever since David first wrote me
on this issue, and it relates to the comment I made after his talk in
{\Vaxjo} (in case he remembers).  Let me try to explain.

David, in his paper, quotes Peierls as saying,

\bdm
In my view the most fundamental statement of quantum mechanics is
that the wavefunction, or, more generally the density matrix,
represents our {\it knowledge\/} of the system we are trying to
describe. \ldots\ [Yet, density matrices] may differ, as the nature
and amount of knowledge may differ.  People may have observed the
system by different methods, with more or less accuracy; they may
have seen part of the results of another physicist.  However, there
are limitations to the extent to which their knowledge may differ.
\edm

And David himself says,

\bdm
I have the feeling that if quantum mechanics is really about
knowledge and only knowledge, then there ought to be further
elementary constraints on the possible density matrices describing
one and the same physical system that are stronger than the very weak
second condition of Peierls, but not as strong as his overly
restrictive first condition.
\edm

What is being called for here---perhaps unintentionally---is a way to
think about quantum mechanics from the bottom up (as {\Caves}, {\Schack},
and I might like it, in a Bayesian way), rather than from the top
down (as Everett, Deutsch, and {\Bennett} might like it).  That is, one
should view quantum mechanics as a conduit for stitching our
individual pictures/thoughts/beliefs into a pastiche we ultimately
call ``the world.''  This contrasts with imagining that we have
miraculously grasped the ultimate reality (the universal
wavefunction, say), and can somehow see our individual points of view
as being derived back out of that.

But if this is the case, then I cannot understand Peierls' command,
David's quest, or the {\it answer\/} all three of you ended up coming
up with:

\bq
\noindent
Two density operators $\rho_a$ and $\rho_b$ can describe the same
system if and only if the support of $\rho_a$ has a nontrivial
intersection with the support of $\rho_b$.
\eq

This theorem is certainly consistent with a top-down view of the
Everett sort---that, to be explained below, is really is what it
seems to me to demonstrate---but it is not consistent with the
bottom-up view.  For, from the bottom-up view, there should not---and
more importantly, there cannot---be any constraints whatsoever on
what an agent can believe.  I have every right to be as wrong-headed
as I want to be with respect to you:  The density operator belongs to
me, not to you, and not to the system.  I have every right to say
inane things and make inane predictions---I do it all the time.  What
I should not do, however, if I want to remain rational, is refuse to
listen to you when you point out my inanities or refuse to listen to
the detector clicks that contradict my previous predictions.  From
the Bayesian view, it is the process of updating and the general
structure of beliefs that is constrained by rationality---i.e., by
the physical world, or the Platonic ideal, depending upon your
orientation.  It is not the actual beliefs themselves.

This point, perhaps more than any other, is why I (and {\Caves} and
{\Schack}) should adopt the word ``belief''---rather than ``knowledge''
or ``information''---for describing the operational significance of
quantum states.  (I will try to be more consistent in the future, but
that is really an aside as far as this note is concerned.)

Of course, your theorem is a theorem---or at least I can see nothing
wrong with it---the issue here is how I, with my little Bayesian
mind, can put it into a context I am more happy with.  At first, I
was the most pleased with Jerry's way of motivating it:  From that
point of view, what the theorem seems to express is simply the
conditions under which a third agent Carol can consistently
incorporate Alice and Bob's disparate beliefs into her own belief
system.

But from the Bayesian view, why should we care about a Carol at all?
What if there's no Carol to be found?  What if neither Alice nor Bob
ever intend to share their thoughts about this poor physical system
with anyone else?  To say that there is always a Carol about, or that
there ought to be one, is to come dangerously close to endorsing the
Everettian (or, for that matter, Bishop Berkeleyian) program.  This,
of course, may not bother Jerry or Todd---I'm not completely sure
about their foundational dispositions---but it does bother me, and I
suspect it might bother David, with his newfound deconstructionist
tendencies.

Thus, it now seems to me that Todd's original way of posing the issue
may be the safer way after all.  BUT that is not because it gives us
{\it the\/} answer, but instead {\it an\/} answer.  (I.e., there
should {\it only\/} be sufficient conditions, rather than necessary
ones.)  Let me let Bernardo and Smith speak again (and again):

\bq
      [T]here is an interesting sense, even from our standpoint, in
   which the parametric model and the prior can be seen as having
   different roles.
   Instead of viewing these roles as corresponding to an
   objective/subjective dichotomy, we view them in terms of an
   intersubjective/subjective dichotomy.  To this end, consider a
   {\it group\/} of Bayesians,
   all concerned with their belief distributions for the same sequence
   of observables.  In the absence of any general agreement over
   assumptions of symmetry, invariance or sufficiency, the individuals
   are each simply left with their own subjective assessments.  However,
   given some set of common assumptions, the results of this chapter
   imply that the entire group will structure their beliefs using some
   common form of mixture representation.  Within the mixture, the
   parametric forms adopted will be the same (the {\it
   intersubjective\/}
   component), while the priors for the parameter will differ from
   individual to individual (the {\it subjective\/} component).  Such
   intersubjective agreement clearly facilitates communication within
   the group and reduces areas of potential disagreement to just that of
   different prior judgements for the parameter.  As we shall see in
   Chapter 5, judgements about the parameter will tend more towards a
   consensus as more data are acquired, so that such a group of
   Bayesians may eventually come to share very similar beliefs, even if
   their initial judgements about the parameter were markedly different.
   We emphasize again, however, that the key element here is
   intersubjective agreement or consensus.  We can find no real role for
   the idea of objectivity except, perhaps, as a possibly convenient,
   but potentially misleading, ``shorthand'' for intersubjective
   communality of beliefs.
              --- pp.\ 236--237, Bernardo and Smith, {\sl Bayesian
              Theory}
\eq

\bq
      In the approach we have adopted, the fundamental notion of a model
   is that of a predictive probability specification of observables.
   However, the forms of representation theorems we have been discussing
   provide, in typical cases, a basis for separating out, if required,
   two components; the parametric model, and the belief model for the
   parameters.  Indeed, we have drawn attention in Section 4.8.2 to the
   fact that shared structural belief assumptions among a group of
   individuals can imply the adoption of a common form of parametric
   model, while allowing the belief models for the parameters to vary
   from individual to individual.  One might go further and argue that
   without some element of agreement of this kind there would be great
   difficulty in obtaining any meaningful form of scientific discussion
   or possible consensus.
              --- p.\ 237, Bernardo and Smith, {\sl Bayesian Theory}
\eq

You'll find something similar to this theme infused throughout my
paper ``Quantum Foundations in the Light of
Quantum Information,'' \quantph{0106166}.  In the present context the question is, under
what conditions (on the density operators themselves), can Alice and
Bob move toward consensus in their future density operator
assignments for a system?  You guys have answered this for the {\it
case\/} when there are some extra systems beside the one of interest
(all noninteracting) lying around for Alice and Bob to make
measurements on.  If there is no overlap between the initial
supports, then no measurement-at-a-distance, which can only have the
effect of refining a density operator, can get Alice and Bob any
closer to agreement.

What about necessity?  As I've tried to say, there can be no
requirement of necessity from the Bayesian view I'd like to see
prevail.  Instead the question is {\it always\/} to identify those
situations where agents (with disparate beliefs) can move toward
consensus be it by indirect measurements, direct ones, or even by
communication with further members of the community.

A good and indicative example comes from the quantum de Finetti
theorem along with the points made by \quantph{0008113} ({\Schack},
Brun, and {\Caves}).  From it, we have a natural case where two agents
can start out with distinct density operator assignments on a large
collection of systems, but through updating via some commonly viewed
measurement outcomes, they can move toward complete agreement in
their estimates of the outcomes of all future measurements.  The only
thing the agents need to walk into the room with is this much INITIAL
AGREEMENT:  (1) that the systems in the collection are exchangeable,
and (2) that the parametric form given by the de Finetti theorem has
full support (in the sense defined there). Without that initial
agreement, the techniques of quantum-state tomography would lead to
no final agreement at all.  No one can require that two observers in
Mike Raymer's lab must walk into it with such an initial agreement,
but if they happen to, there will be a reward at the end of the day.

That's my spiel.  How would I modify David's quest in light of all
that I've just said?  Here's my shot:
\bq
\noindent
I have the feeling that if quantum mechanics is really about
knowledge and only knowledge---or better, belief convergence and {\it
only\/} belief convergence---then FOR ANY GIVEN METHOD OF GATHERING
INFORMATION, there {\it should\/} be a way to ferret out of quantum
mechanics the necessary and sufficient conditions on two observers'
initial state assignments, so that the gathered information leaves
them in a better agreement than they started out with.
\eq

Let me give you an example of where such an exercise can go.  This is
a question I posed to {\Ruediger} while we were in {\Vaxjo}, but also
it is a generalization of Todd's considerations (now letting Alice
and Bob's measurements being disturbing ones).

Let us agree on a distance measure on density operators.  A
convenient one (and my favorite) is
$$
d(\rho, \sigma) = \tr | \rho - \sigma | ,
$$
but that's not so important for the considerations here.  (Well, it
may be in the long-run, but I don't want to let such a technicality
detract from posing the question.)

Suppose Alice walks into a room and says to herself that a system is
described by $\rho$, while Bob says to himself that it is described
by $\sigma$.  We can gauge their amount of initial consensus by
$d(\rho,\sigma)$.  Now, suppose Carol (or Alice or Bob for that
matter) performs a measurement on the given system whose
action---i.e., whose associated completely positive map---is to take
any initial state $\tau$ to
$$
\tau \longrightarrow \tau_b = \frac{1}{\tr(\tau A_b^\dagger
A_b)} A_b \tau A_b^\dagger
$$
depending upon the particular outcome $b$.  Here, of course,
$$
\sum_b A_b^\dagger A_b = I .
$$
(This is nothing more than the general form of an ``efficient''
measurement as defined in my QFILQI paper cited above.)  Let us allow
Alice and Bob to be privy to this map, and indeed to the actual
outcome $b$.

Thus, the final consensus of Alice and Bob will be gauged by
$d(\rho_b,\sigma_b)$.  The question is:  When can Alice and Bob
expect to be in better agreement after the measurement than before?
That is, as far as Alice is concerned, she will {\it expect\/} their
final distance to be
$$
D_A = \sum \tr(\rho A_b^\dagger A_b) d(\rho_b,\sigma_b).
$$
As far as Bob is concerned, he will {\it expect\/} their final
distance to be
$$
D_B = \sum \tr(\sigma A_b^\dagger A_b) d(\rho_b,\sigma_b).
$$
For a given set of $A_b$'s, what are the necessary and sufficient
conditions on $\rho$ and $\sigma$ so that
$$
D_A \le d(\rho,\sigma)
$$
and
$$
D_B \le d(\rho,\sigma)\;?
$$

Suppose we can answer these questions.  Then we will be able to
identify the minimal fact Alice and Bob must reveal to each other
(even without explicitly revealing their full beliefs captured by
$\rho$ and $\sigma$) so that they can expect to walk out of the room
in better agreement \ldots\ even if they still can't say with
certainty what each other now believes.

Beliefs, only beliefs.  But sometimes we can say something about
their convergence.  And you guys have provided an example.

\section{07-08-01 \ \ {\it Knowledge, Only Knowledge:\ Amendment} \ \ (to T. A. Brun, J. Finkelstein \& N. D. {\Mermin})} \label{Mermin29} \label{Brun2} \label{Finkelstein4}

I just reread the thing after sending it.  Let me be more careful
before one of you accuses me of being a flaming positivist again. (I
suspect Todd, in particular, will be inclined to do so, and I want to
fend that off before it happens.)  In my closing sentence, I wrote:
\bq\noindent
Beliefs, only beliefs.  But sometimes we can say something about
their convergence.  And you guys have provided an example.
\eq

Let me temper that to:
\bq\noindent
Beliefs, only beliefs.  But sometimes we can say something about
their convergence, as they are steered by our interactions with the
world external to us.  And you guys have provided an example.
\eq

Now, I can sleep safely \ldots

Good wishes to all!

\section{07-08-01 \ \ {\it The First Amendment} \ \ (to J. Finkelstein)} \label{Finkelstein5}

Thanks for the note.  But nothing changes for me (yet, at least).
Sorry for my overemphasis on Carol.

\bjf
If that can be agreed to, then the rest is essentially just algebra.
$S[\rho_a]$ is the orthogonal complement of the zero-eigensubspace of
$\rho_a$, which is the set of $|\phi\rangle$ such that Alice knows
with certainty that a measurement of $|\phi\rangle\langle\phi|$ must
yield the result zero.  So the support of the updated version of
$\rho_a$ must be a subset of the support of the original $\rho_a$.
Likewise, Alice can see that the support of the updated $\rho$ must
be a subset of the support of $\rho_b$. Etc.
\ejf

The issue for me is not whether Alice {\it might\/} be willing to
incorporate Bob's beliefs/knowledge into her own knowledge
base---that, I'll grant her, in which case I agree that everything
you say is true.  Instead it is whether she {\it must\/} be willing
to do it.  I think you are (tacitly) trying to get me to agree that
she {\it must\/} be willing to accept Bob's quantum state as extra,
valid information.  That's something I can't do.  I reserve the right
for Alice to think that Bob's quantum state is complete nonsense,
something that she would never want to incorporate into her own
knowledge base.  (It's part of being an American.)

To say that two observers {\it must\/} be willing to incorporate
their separate states of knowledge into a single state is---I
think---to tacitly (there's that word again) accept an Everettian
kind of view.  For it never allows that quantum states are states of
belief in the normal sense.  Instead it makes them more like
``objective points of view'' (relative states) that {\it must\/} be
derivable from a larger, more encompassing picture.  Why else would
the states have to be ``consistent'' with each other (for instance,
in your and Todd's sense)?

Is that helping any to get my wacky point of view across?

\section{07-08-01 \ \ {\it Kiki, {\James}, and {\Dewey}} \ \ (to J. A. Waskan)} \label{Waskan1}

The last month or so, I've been logging quite a few hours in the
philosophical world.  I read about 100 pages from this little book:
A.~Schopenhauer, {\sl The Philosophy of Schopenhauer}, edited, with
an introduction by I.~Edman, (Modern Library, New York, 1928).

It's a small collection of pieces from Schopenhauer's big masterwork.
I read about 50 pages on the world as representation, and about 50 on
the world as will.  Even that little was not easy for me.  It's
fanciful stuff, but maybe I extracted an idea or two that I like.
The main one is simply the idea of a dichotomy between what things
look like from the inside of any phenomenon (when there is a view
from that perspective)---Schopenhauer called it will---and what
things look like from the outside of the phenomenon (when there is a
view from that perspective)---Schopenhauer called it idea.  Or, maybe
I should have more safely said ``is'' rather than ``look like.'' But,
in any case, that distinction (and a strict separation between the
concepts) strikes me as useful or at least worth contemplating. Just
about all the rest, though, I probably wouldn't be able to accept:
the strict Kantian categories, the principle of sufficient reason,
etc., etc.

On the other hand, I have gotten {\it completely\/} carried away with
William {\James} and John {\Dewey}.  Here's how I put it to my friend {\Carl}
{\Caves} the other day:
\bq
Today I focused on rounding up some more William {\James}, John {\Dewey},
Percy Bridgman material.  I think {\James} is taking me over like a new
lover.  I had read a little bit of him before, but I think I was more
impressed with his writing style than anything.  But I was drawn back
to him by accident, after reading Martin Gardner's {\sl Whys of a
Philosophical Scrivener}.  Gardner devoted a lot of time knocking
down {\James}' theory of truth, because it is just so much easier to
accept an underlying reality that signifies whether a proposition is
true or false, rather than saying that the knowing agent is involved
in eliciting the very proposition itself (along with its truth
value).  And something clicked!  I could see that what {\James} was
talking about might as well have been a debate about quantum
mechanics.  He was saying everything in just the right way. (Let me
translate that:  he was saying things in a way similar to the way I
did in my NATO ``appassionato.'')  And things have only gotten better
since.
\eq
And indeed, they have only gotten better since!  Since coming to
Munich, I have not been able to put {\James} and {\Dewey} down (when I'm
not writing emails trying to translate their ideas into the quantum
mechanical context, in particular for a technical problem {\Caves},
{\Schack} and I are disagreeing violently on).  I read {\James}' {\sl
Pragmatism, and four essays from the Meaning of Truth}, and now I'm
about halfway through {\sl John {\Dewey}:\ The Essential Writings}.  I'm
moved by this stuff like nothing else I've ever read.

You can't tell me philosophers don't have the good life!

\section{07-08-01 \ \ {\it Parachutes} \ \ (to J. Preskill)} \label{Preskill3}

\bjp
[T]hanks for teaching me the word ``communard''. \ldots  [I] guess I have a conflict in October 2002. I have agreed
to go to Leiden for that month to be the ``Lorentz chair,'' teaching a short course on quantum computing. It is an odd thing to do for someone who travels as little as I do, but they convinced me that it is too big an honor to turn down. The list of past
recipients includes 10 Nobel Prize winners.
\ejp

Congratulations on the Lorentz chair thing.  That is quite an honor.  I looked at the list:  Steam started rising from my computer screen!  Here's my challenge:  Use the fanaticism of Wheeler, the thoroughness of Wigner, the religion of Rosenfeld, the science fiction of Klein, the common sense of Peierls, the intuition of Yang, and the poetry of Mermin to tell your audience in a convincing way that quantum computing and information is just the beginning of quantum mechanics, not the end.

\section{08-08-01 \ \ {\it Cross Entropy Min} \ \ (to J. Finkelstein)} \label{Finkelstein6}

\bjf
Which brings up a slightly different question.  Suppose, again, that
Alice describes a system by $\rho_a$, but this time let's say for
simplicity that she considers what would happen if Bob were to tell
her that his $\rho_b$ corresponded to a pure state $|\phi\rangle$;
then, given her knowledge of the system (ie, given $\rho_a$) what
could Alice say about what $|\phi\rangle$ might be?

It follows from what we have been saying that Alice knows that for
$|\phi\rangle$ to be possible, it must be in $S[\rho_a]$, and that
(in finite dimensions, at least) she cannot, based on her own
knowledge, rule out any $|\phi\rangle$ that is in $S[ \rho_a]$.  But
can Alice say any more than that? Would it make sense for Alice to
put a probability distribution on the possible $|\phi\rangle$ that
Bob might announce to her?  (It would have to be a probability
density; eg, if $\rho_a$ were a multiple of the identity, then Alice
would surely judge all states to be equally-likely.) If that did make
sense, it would mean that Alice would be constructing a particular
(continuous) ensemble representation of $\rho_a$.  What could that
be?
\ejf

As I recall, this is quite similar to the classical problem
(``principle of minimum cross-entropy'') explored in the two
references below.  But I'm not going to have a chance to refresh my
memory for a while.  If you've got the time, you might see if it's
relevant.

\begin{enumerate}
\item
John E. Shore and Rodney W. Johnson, ``Axiomatic Derivation of the
Principle of Maximum Entropy and the Principle of Minimum
Cross-Entropy,'' IEEE Transactions on Information Theory {\bf
IT-26}(1), 26--37 (1980).
\item
John E. Shore and Rodney W. Johnson, ``Properties of Cross-Entropy
Minimization,'' IEEE Transactions on Information Theory {\bf
IT-27}(4), 472--482 (1981).
\end{enumerate}

\section{08-08-01 \ \ {\it The First Eye} \ \ (to C. M. {\Caves})} \label{Caves3}

I am just about to get down to some serious (political) scheming to
do with our Bayesian program:  I'll let you know what I'm talking
about if it turns out to be successful.  First though, I want to take
a moment to tell you about a point of similarity between our (both
far-from-completely worked out) flavors of quantum ontology. This one
just struck me a few days ago.

Some time ago, I tried to explain to you what I was hoping for for
[sic] an ontology behind quantum mechanics.  (See Samizdat, pp.\
127--129.)  I said it would have something to do with the
information-disturbance tradeoff in quantum eavesdropping.  You
replied that, try as you might, you could see no ontological content
in such a statement.  I think what was troubling you was that the
information-disturbance relations (as I am thinking of them), by
their very nature, require explicit reference to the {\it
subjective\/} points-of-view/opinions/beliefs (i.e., quantum states)
of various {\it agents}.

I, on the other hand, have no problem with that.  For the way to
think about it is that the world (independent of our existence) has
latent within it a property that simply has no way of being properly
expressed without inserting information-manipulating agents into the
picture.  (Or, at the very least, that this anthropocentric way of
stating things may be our first firm handle for getting at a better,
more objective-sounding, formulation of the latent property.) Taking
this tack does not mean, of course, that there is no world
independent of human existence, and that is my point.  It just means
that we may sometimes have to take into account our (presumably
contingent) existence for expressing some of the world's properties.

Here is something, however, that you should think about in connection
to your ``world $=$ Hamiltonian'' hopes.  Let me put the ball back
into your court.  Can you explain to me the role of Hamiltonians in
your ontology in a way that does not make use---even tacitly---of the
concept of a (subjective) quantum state?  What is it that
Hamiltonians do if their primary role is not in evolving (subjective,
agent-required) quantum states?  In the classical world, one could
give an answer to this question by saying, ``They evolve the
positions of phase-space points.''  Such a statement makes no use of
the concept of information-bearing agents for its formulation.  (I
view it as just nitpicking to argue whether the points or their
trajectories (from which we can derive the Hamiltonian) are the more
primary of the entities.  Or whether they are equally primary.)  But
in the quantum case, I haven't yet seen what you can say if I take
away the linguistic tool of ``the quantum state'' from your
explanatory repertoire.  What can you say?

I'll put a few passages of Schopenhauer below to inspire you. [From:
A.~Schopenhauer, {\sl The Philosophy of Schopenhauer}, edited, with
an introduction by I.~Edman, (Modern Library, New York, 1928).]

\bq
\indent
``No object without a subject,'' is the principle which renders all
materialism for ever impossible.  Suns and planets without an eye
that sees them, and an understanding that knows them, may indeed be
spoken of in words, but for the idea, these words are absolutely
meaningless.  On the other hand, the law of causality and the
treatment and investigation of nature which is based upon it, lead us
necessarily to the conclusion that, in time, each more highly
organised state of matter has succeeded a cruder state:  so that the
lower animals existed before men, fishes before land animals, plants
before fishes, and the unorganised before all that is organised;
that, consequently, the original mass had to pass through a long
series of changes before the first eye could be opened.  And yet, the
existence of this whole world remains ever dependent upon the first
eye that opened, even if it were that of an insect.  For such an eye
is necessary condition of the possibility of knowledge, and the whole
world exists only in and for knowledge, and without it is not even
thinkable.  The world is entirely idea, and as such demands the
knowing subject as the supporter of its existence.  This long course
of time itself, filled with innumerable changes, through which matter
rose from form to form till at last the first percipient creature
appeared---this whole time itself is only thinkable in the identity
of a consciousness whose succession of ideas, whose form of knowing
it is, and apart from which, it loses all meaning and is nothing at
all.  Thus we see, on the one hand, the existence of the whole world
necessarily dependent upon the first conscious being, however
undeveloped it may be; on the other hand, this conscious being just
as necessarily entirely dependent upon a long chain of causes and
effects which have preceded it, and in which it itself appears as a
small link.  These two contradictory points of view, to each of which
we are led with the same necessity, we might again call an {\it
antinomy\/} in our faculty of knowledge, and set it up as the
counterpart of that which we found in the first extreme of natural
science.  The objective world, the world as idea, is not the only
side of the world, but merely its outward side; and it has an
entirely different side---the side of its inmost nature---its
kernel---the thing-in-itself.  This we shall consider in the second
book, calling it after the most immediate of its objective
manifestations---will.  But the world as idea, with which alone we
are here concerned, only appears with the opening of the first eye.
Without this medium of knowledge it cannot be, and therefore it was
not before it.  But without that eye, that is to say, outside of
knowledge, there was also no before, no time.  Thus time has no
beginning, but all beginning is in time.  Since, however, it is the
most universal form of the knowable, in which all phenomena are
united together through causality, time, with its infinity of past
and future, is present in the beginning of knowledge.  The phenomenon
which fills the first present must at once be known as causally bound
up with and dependent upon a sequence of phenomena which stretches
infinitely into the past, and this past itself is just as truly
conditioned by this first present, as conversely the present is by
the past.  Accordingly the past out of which the first present
arises, is, like it, dependent upon the knowing subject, without
which it is nothing.  It necessarily happens, however, that this
first present does not manifest itself as the first, that is, as
having no past for its parent, but as being the beginning of time.
It manifests itself rather as the consequence of the past, according
to the principle of existence in time.  In the same way, the
phenomena which fill this first present appear as the effects of
earlier phenomena which filled the past, in accordance with the law
of causality.  Those who like mythological interpretations may take
the birth of Kronos ($\chi\varrho o \nu o \varsigma$), the youngest
of the Titans, as a symbol of the moment here referred to at which
time appears, though indeed it has no beginning; for with him, since
he ate his father, the crude productions of heaven and earth cease,
and the races of gods and men appear upon the scene.
\eq

\section{09-08-01 \ \ {\it Lock Box Reference}\ \ \ (to J. A. Smolin)} \label{SmolinJ1}

Page 101 in:
\bq\noindent
C.~A. Fuchs, {\sl Notes on a Paulian Idea:\  Foundational, Historical, Anecdotal \& Forward-Looking Thoughts on the Quantum (Selected Correspondence)}. 504 pages. Foreword by N. David {\Mermin}. See \quantph{0105039}.
\eq

\section{10-08-01 \ \ {\it The Fifth Amendment} \ \ (to T. A. Brun)} \label{Brun3}

Thanks for the note, which I thoroughly enjoyed.  You hit a lot of
nails on the head with it.  Let me try to expand on some of the
points that---I believe---show that at least you and I are coming to
a little consensus. (As for Jerry and David, I will put them in a
superposition for the time being, and see how this this interaction
shakes things up.)

\btb
That doesn't mean that I completely disagree with you, Chris, but I
think you are making a point which is pretty far from the spirit of
this problem.
\etb

This is the point.  What I am trying to get straight is:  {\it
What\/} is the spirit of this problem?

You write,
\btb
We have been describing a consistency criterion.  {\bf If\/} one
wishes to combine two state descriptions of a single system into a
{\bf single\/} state description, the criterion tells one {\em
when\/} it is consistent to do so (i.e., when the two descriptions
are not actually contradictory).

I agree that nobody is holding a gun to Alice's head and forcing her
to incorporate Bob's information.
\etb

Putting it like that, I can certainly accept the proposition.  I want
to emphasize that.  The attractive feature for me is that it is built
on a conditional at the outset.

But you speak of {\it the\/} spirit of {\it this\/} problem.  How is
your statement to be reconciled with the tone of David's {\tt
quant-ph}? In particular, say, David's quote of Peierls (which he
takes as his guiding light):

\bdm
In my view the most fundamental statement of quantum mechanics is
that the wavefunction, or, more generally the density matrix,
represents our {\it knowledge\/} of the system we are trying to
describe. \ldots\ [Yet, density matrices] may differ, as the nature
and amount of knowledge may differ.  People may have observed the
system by different methods, with more or less accuracy; they may
have seen part of the results of another physicist.  However, there
are limitations to the extent to which their knowledge may differ.
\edm

Peierls does not use the qualification that you did---nor do any of
you three in many of the emails I have seen---and that is what
bothers me.

What I see in Peierls's version of the spirit is that two quantum
states cannot co-exist (even in a Platonic sense) unless they are
consistent (in one manner or another, yet to be fleshed out).  It is
as if the universe has these little properties floating about, called
quantum states, that MUST be consistent in the BFM sense (or some
other sense).  I will agree that that might be fine from an
Everett-kind of point of view.  But if one insists on consistency (as
one should with ontological, physical properties), then---it seems to
me---one breaks away from the desire to give the quantum state a {\it
purely\/} epistemological role \ldots\ which, as I understood it, was
the goal of Peierls and {\Mermin} (though I am willing to accept that it
may not be the goal for you and Jerry).

\btb
The point of consistency is to determine {\bf if\/} two points of
view can be combined into a single description, not to require that
they must be.
\etb

This I also agree with:  It was meant to be the whole point of my
note ``\myref{Mermin28}{Knowledge, Only Knowledge}.''  My only point of disagreement,
was that it was seeming to me that the tendency in David (and Jerry?)\ was the assumption that the ``if'' must be satisfied---in some
sense---in the ``real'' physical world.

I just don't know how to say this more clearly.  I think it is a
valid worry about the intent, the very definition of what the problem
is about.  I don't think I'm being subtle:  that is certainly the
last thing I want to be.

\btb
I will make one additional comment.  I think that in science there is
usually a {\bf tacit\/} (to use your word) assumption that the
separate states of knowledge of different observers can be combined,
provided that they make no errors and reason logically.  This then
implies that there {\bf is\/} a kind of ``global state,'' in the
limited sense of a state including all available knowledge.  If two
observers' beliefs are so inconsistent that they cannot be combined
together with any amount of communication and experimental data, they
might as well be living in different worlds.  This is why we say that
insane people are ``out of touch with reality.''
\etb

I can see why one would say that, especially if one believes that the
process of science has an end, and that there is a sense in which the
universe is pre-formed.  (So maybe I will admit to {\it some\/}
postmodernist tendencies.)  But then \ldots
\btb
Also for this reason, a rational person may very well never assign a
perfectly pure state to a system, but always a mixed state of the
form
$$
   \rho = (1-\epsilon) |\psi\rangle\langle\psi| + \epsilon
   \rho^\prime
$$
where $\epsilon$ contains the unspoken acknowledgement ``But I might
be wrong.''
\etb
giving an agent the right to set his own density operator seems to me
to be a concession to the quantum state's (purely?) epistemological
content.

A note on the ``The Second Amendment'' to be sent to David soon after
lunch.

\section{12-08-01 \ \ {\it A Silver Lining in the Deutsch Cloud} \ \ (to R. Pike)} \label{Pike2}

I like to think that there are silver linings to the clouds of my
many diatribes!  (E.g.\ the note I just sent Patrick.)  Anyway,
thinking more about Deutsch this morning (and his silliness that
quantum computing only makes sense from a many-worlds view), see the
two papers below.

Brooding on the Raussendorf--Briegel one especially (since David
DiVincenzo brought it to Steven's\index{Enk, Steven J. van} and my attention Thursday), I think
these are really important ones.  And they may give us a much easier,
nicer, more insightful way to describe what quantum computing is
about.
\bv
\quantph{0010033} \\
Title: Quantum computing via measurements only \\
Authors: Robert Raussendorf, Hans J. Briegel\medskip\\
\quantph{0108020}  \\
Title: Universal quantum computation using only projective
measurement, quantum memory, and preparation of the 0 state\\
Authors: Michael A. Nielsen
\ev

\section{13-08-01 \ \ {\it Out of the Frying Pan \ldots} \ \ (to R. Garisto)} \label{Garisto3}

My thoughts are with you, and thanks for including me in the list of updatees.

I don't know if it'll work for you, but in times when I'm hanging on the edge, listening to John Coltrane and contemplating quantum mechanics seem to help more than anything.  The two activities are not unrelated.

\section{13-08-01 \ \ {\it Reference You Wanted} \ \ (to D. P. DiVincenzo)} \label{DiVincenzo2}

The formula can be found on pages 26 and 27 of:
\begin{itemize}
\item[]
C.~A. Fuchs, ``Quantum Foundations in the Light of Quantum Information,'' to appear in {\sl Proceedings of the NATO Advanced Research Workshop on Decoherence and its Implications in Quantum Computation and Information Transfer}, edited by A.~Gonis (Plenum Press, NY, 2001). (Until then, see \quantph{0106166}.)
\end{itemize}

See you tomorrow probably.

\section{13-08-01 \ \ {\it A Silver Lining in the Deutsch Cloud, 2} \ \ (to N. D. {\Mermin})} \label{Mermin30}

By the way, you might enjoy these too if you haven't already run
across them.  I think there is some parallel here to thinking of
quantum teleportation in the two ways:  i.e., the original 1993 way,
and the quantum circuit way (like you wrote about recently).  The
dichotomy goes much deeper and may pervade all of quantum
computation, or least that's what is starting to strike me.

``Pinging the sensitive substrate, that's what quantum computing is
about.''  Or, at least, that's what I think would build a pretty
picture.  The two papers below may give us the right language to view
it that way.

\section{14-08-01 \ \ {\it Law without Law, 2} \ \ (to J. Summhammer)} \label{Summhammer2}

Thanks for your note.  Give me a few days to digest it.  In the meantime,
\bjs
In my next mail I hope to get down to asking you a few questions on
your paper ``Quantum Foundations in the Light of Quantum Information''.
And the prior probability distribution of probabilities in the
classical de Finetti theorem is causing me some trouble. Also, I would
appreciate a copy of your paper (PS, PDF, or address for downloading
preferred) ``How much information in a state vector?''
\ejs
let me give you the reference you requested.  It can be found on the Los Alamos archive (as can about 75\% of my publications).  Here's the abstract:
\bq\noindent
\quantph{9601025}\\
Date: Wed, 24 Jan 96 21:21:01 MST   (31kb)\medskip\\
Quantum information: How much information in a state vector?\\
Authors: Carlton M. Caves, Christopher A. Fuchs\\
Comments: 32 pages in plain \TeX, to appear in {\sl Sixty Years of EPR}, edited by A. Mann and M. Revzen, Ann.\ Phys.\ Soc., Israel, 1996\medskip

Quantum information refers to the distinctive information-processing properties of quantum systems, which arise when information is stored in or retrieved from non\-orthogonal quantum states. More information is required to prepare an ensemble of nonorthogonal quantum states than can be recovered from the ensemble by measurements. Non\-orthogonal quantum states cannot be distinguished reliably, cannot be copied or cloned, and do not lead to exact predictions for the results of measurements. These properties contrast sharply with those of information stored in the microstates of a classical system.
\eq

\section{15-08-01 \ \ {\it Knowledge, Only Knowledge -- Reprise} \ \ (to T. A. Brun)} \label{Brun4}

\btb
Frankly, I find your negativity about this problem puzzling.
\etb

On the contrary, I'm quite taken with it:  otherwise, I wouldn't have
given it the time of day.  Reading the collective emails has been a
great learning experience for me.

\section{15-08-01 \ \ {\it Compatible States} \ \ (to R. {\Schack})} \label{Schack2}

\brs
I got a number of rather incoherent messages on compatible states
from you. Is there anything more recent on this?
\ers

Yes.  About three thousand more notes. The issue refers to David
{\Mermin}'s recent {\tt quant-ph} paper ``Whose Knowledge?''

Where it stands right now is with Brun, Finkelstein, and {\Mermin} all
ganged up against me (as being the unreasonable one).  The conclusion
they've come to is something like this:
\bq
\noindent
Two density operators $\rho_a$ and $\rho_b$ can describe the same
system iff the support of $\rho_a$ has a non-trivial intersection
with the support of $\rho_b$.
\eq

I, on the other hand, think such a statement is far too dictatorial
for a Bayesian's taste.  Try as I might, they just don't think my
points are relevant.

I'll paste (what I deem to be) my most lucid notes below.  You can
judge for yourself.  I'm soon going to drop out of the debate, I
believe:  it's now at a point of diminishing returns.  But, I would
like to hear your thoughts!

\section{16-08-01 \ \ {\it Subject-Object} \ \ (to P. Grangier)} \label{Grangier1}

Thanks for your note!  I'm always amazed when anyone reads or skims
my papers:  You have a friend for life!

Last night I read your paper \quantph{0012122}, which I had
never seen before.  Thanks for bringing it to my attention.

\bpg
You are probably aware that statements such as :  ``The quantum state
is information. Subjective, incomplete information.'' \ldots\
``Quantum states are states of knowledge, not states of nature. That
statement is the cornerstone of this paper.'' are unwarranted, since
just opposite statements can be made without changing any physical
predictions or even any technical development. My personal view is
that these statements are even wrong, as soon as ``quantum state'' is
understood as ``pure quantum state'' (see eg \quantph{0012122},
missing in your list on p.1).
\epg

In your words,
\bq\noindent
[C]ontrary to the copenhagian dogma, a central point in our approach
will be to give an ``objective reality'' to the quantum state of a
physical system, in a sense which is developed below.  \ldots\ The
quantum state of a physical system is defined by the values of a set
of physical quantities, which can be predicted with certainty and
measured repeatedly without perturbing in any way the system.
\eq

Here is a problem I have with this conception.  (It is a problem I am
quite sure you are aware of, but for some reason you did not address
it directly in your paper.)  Consider two electrons originally
prepared in a spin-singlet state---one electron in the possession of
Alice, one in the possession of Bob.  Let us imagine now two
alternative scenarios.  In one, Alice measures $\sigma_x$ on her
particle; in the other scenario, she measures $\sigma_z$.  By your
criterion, Bob's particle does not start out with a quantum state
(since the two electrons are in an entangled state)---which is fair
enough (I have no qualm with that)---but {\it immediately upon\/} the
measurement it will go into a definite quantum state, either an
eigenstate of $\sigma_x$ or an eigenstate of $\sigma_z$, depending
upon the scenario.  We know this because if Bob were to thereafter
repeatedly send his particle through a Stern--Gerlach device with the
proper orientation (for the given of the two scenarios), Alice would
be able predict with complete certainty which way Bob's particle
would go.  Moreover, if Bob is careful enough, these further
measurements on his part will not perturb his system (in the sense of
changing the spin quantum state of the electron).  So, we have just
what you had wanted:  complete certainty and no necessary
perturbation.

But if the quantum state is an objective feature of the electron,
then we see that it can be toggled one way or the other
instantaneously from a distance.  (Alice's measurements causes Bob's
electron to go into one or another quantum state instantaneously.)
Thus, if you accept the objectivity of the quantum state, then you
must also accept the objectivity of instantaneous action at a
distance (that in no way diminishes with distance or the particulars
of the medium filling the space between Alice and Bob).

This is something I'm not willing to do.  Not out of dogma, but
because it strikes me that the world would have to be horribly
contrived to have this property:  a little private (instantaneous)
telephone line between each and every physical system that will not
accept outside calls.  I.e., I can never make use of this
instantaneous action for real, live communication even though it
really, really, really is there.  It stretches my imagination too
far.

You, of course, may accept it as you wish; but the reasons above are
mine for not doing so.  Let me address your other point that I quoted
above:
\bq\noindent
[Your statements] are unwarranted, since just opposite statements can
be made without changing any physical predictions or even any
technical development.
\eq

I certainly disagree with the latter part of this sentence:  that was
the whole point of my paper (i.e., that it can change the technical
development of the theory).  Taking one or another point of view
about the objectivity of the quantum state motivates different
directions of theoretical exploration.  In my case, it motivated
trying to find the four theorems I presented in that paper.  One who
believed in the objectivity of the quantum state---I am quite
sure---would not have sought those theorems in the first place.

\bpg
By the way I also disagree with your point of view that ``Quantum
Theory Needs No `Interpretation','' Phys.\ Today {\bf 53}(3), 70
(2000). The fact that a physical theory ALWAYS needs an
interpretation is in my opinion a central difference between physics
and mathematics.
\epg

You won't find a disagreement with me here.  The title and closing
sentence of that paper were meant to be tongue-in-cheek plays on
something Rudolf Peierls once said:  ``The Copenhagen interpretation
{\it is\/} quantum mechanics.''  The whole paper is very definitely
about an interpretation, and why one does not need to go any further
than it to make sense of quantum mechanics as it stands.  My paper
\quantph{0106166} and the large (more personal) collection \quantph{0105039} are about going the next step, i.e., what to do once
we have established the belief that quantum states are states of
knowledge.

When we do finally dig up an ontology underneath quantum mechanics,
I'm quite sure it will be an interesting one!

\section{17-08-01 \ \ {\it Your Computational Model} \ \ (to H. J. Briegel)} \label{Briegel1}

A couple of weeks ago, David DiVincenzo brought my attention to your paper with Raussendorf on models of quantum computation that make use of measurement only (once the initial state is set).  I must say, this model has me absolutely enamored!  (Besides being intrinsically interesting for our field, it seems to mesh well with the quantum foundational direction I have been trying to develop.)

I wonder if you (or Raussendorf) might have some time to let me hear the word from the horse's mouth, so to speak?  I tried reading the first small paper the other night, but it was a bit tough going for me.  So, I'd really like more to be eased into the subject with a presentation or two.  I will be in Munich (Zorneding actually) visiting my parents-in-law from August 30 through September 9.  If you've got some free time in the middle of that, I'd surely like to drop by your university for a discussion.

\section{17-08-01 \ \ {\it Unloading Bayes} \ \ (to C. M. {\Caves} \& R. {\Schack})} \label{Schack3} \label{Caves4}

Let me unload a couple more Bayesian thoughts on you---i.e., some
things we will probably want to address in the RMP article.

1)  Attached below is a note I wrote to {\Mermin} giving what I think is
the cleanest justification for noncontextuality in any Gleason
theorem.  [See 22-07-01 note ``\myref{Mermin26}{Noncontextual Sundays}'' to N. D. {\Mermin}.] In fact, it shows that noncontextuality is more basic in
the hierarchy of theories than anything else we've dealt with yet.
I.e., it comes far before the particular details of quantum
mechanics.  (Maybe this is what {\Carl} has been saying all along, but I
had to work through it for myself before it stuck.)

2)  Let me bring your attention to a cluster of papers that I think
are really important.
\bv
\quantph{0010033}: \\
Title: Quantum computing via measurements only \\
Authors: Robert Raussendorf, Hans J. Briegel (LMU Munich) \medskip\\
\quantph{0108020}: \\
Title: Universal quantum computation using only projective
measurement, quantum memory, and preparation of the 0 state
\\
Authors: Michael A. Nielsen \medskip\\
\quantph{0108067}: \\
Title: Computational model underlying the one-way quantum computer
\\
Authors: Robert Raussendorf, Hans Briegel \\
\ev

I am especially taken with the Raussendorf--Briegel development.  On
one level, I think it might be the simplest avenue for addressing
quantum computation from a Bayesian point of view:  One just builds
up the proper initial (universally valid) entangled state, and then
does a measurement site by site, doing proper (Bayesian) updating of
the quantum state of the remaining sites at each step.  At the end of
the day, one's knowledge is updated to the sought after answer. On
another level (not for RMP), I think it starts to capture what I have
been hoping for as an explanation of the power of quantum
computation:  it is not quantum parallelism that is doing the trick,
but the ``zing'' of quantum systems that makes them sensitive to our
interventions.

3)  Let me put further below my replies to Khrennikov on his
``contextual probabilities'' business.  [See 28-06-01 note ``\myref{Khrennikov3}{Context Dependent Probability}'' and 04-07-01 note ``\myref{Khrennikov5}{Context Dependent Probability, 2}'' to A. Y. Khrennikov.] I've already shared this with
{\Carl}, but not {\Ruediger} yet.  What I like about my statement there is
that it starts to put Gleason's theorem in a more Dutch-book kind of
light.  Just as in classical theory, the setting of initial
probabilities is completely free (and therefore subjective).  What is
set by coherence/rationality is the transformation rules.  To that
extent, this is the objective content of probability theory.
Similarly with quantum mechanics in the light of Gleason's theorem:
The objective content of quantum mechanics (or at least part of it)
is that if we subjectively set our probabilities for the outcomes of
any informationally complete POVM, we are no longer free to set them
arbitrarily for any other observable.  The probabilities are now
fixed and can be calculated uniquely from our previous subjective
judgment.  (I hate the American way of spelling ``judgement.'')

That's about it for now.  I've certainly written loads more (Bayesian
oriented stuff) since we've last seen each other.  But nothing else
that might have been striking is coming to mind right now.  I'll
probably unload more as I think of it.

As I'm hearing about more and more of you arriving in Santa Barbara,
I'm starting to get a little envious that I'm missing all the fun. At
least Kiki and Emma will be off to Munich Monday evening.  I won't be
joining them until the following Wednesday.  I have this great dream
of idling away my hours in the mean time with Arthur Schopenhauer,
William {\James}, and the Reverend Bayes.  But I'm sure some reality
will set in to knock me off my course of purity.

\section{20-08-01 \ \ {\it Unloading Bayes, 2} \ \ (to C. M. {\Caves})} \label{Caves5}

\bcc
I'm not sure anyone is going to be convinced by the above, but I
think it is a start.  So here's the scenario.  There is a big set of
things that can be true or false.  The big set is determined by the
(physical) theory you are dealing with.  There are further rules that
say how the elements in this big set can be gathered into subsets
that correspond to questions whose outcomes are exclusive and
exhaustive. Now, if you make noncontextual probability assignments to
the questions, you have just ignored the structure that led to your
original set.  That being the input from your basic theory, if you
make noncontextual assignments, you are deciding not to pay attention
to your own theory.  Not too bright.  So you should make
noncontextual assignments.
\ecc

I know this won't come as a surprise to you, but (really) I found my
(operational) explanation much more convincing.  So, I guess I really
didn't take my cue from you after all.  Probably more from
Pitowsky/Renes/Hardy.

Can you pinpoint what you didn't find convincing about my argument?

My trouble with your argument is that I still don't find the
statement, ``if you make noncontextual probability assignments to the
questions, you have just ignored the structure that led to your
original set,'' all that compelling.  I guess I still don't quite
understand what you are saying.

\section{21-08-01 \ \ {\it Contextual Reality $=$ Information??}\ \ \ (to P. Grangier)} \label{Grangier2}

Thanks for the longer explanation.  It has indeed clarified things
for me.

What I think is funny though, is that for precisely everything you
say below (which I quote), I would call the quantum state
``information'' rather than an ``objective reality.''

Let me ask you this:  Would you reread Section 3, ``Why
Information?''~in my \quantph{0106166} and comment why and how
you would use a different language in rewriting that?  It seems to me
that to say ``a (pure) quantum state is objective, but it is
contextual, i.e., defined relatively to a particular set of
measurements results'' (as you do), can only be to teeter very close
to admitting that the state is information (and only information). If
the state is only defined relative to Alice's measurement results,
why not call those measurement results the actual reality and be done
with it?  What does having the quantum state being objectively real
add to the story?

\subsection{Philippe's Preply}

\bq
Alice may obviously choose between several measurements, but she must
eventually decide for one and perform it. My definition of the
``objective quantum state of a system'' {\it requires\/} that Alice's
measurement is completed, which does not ``toggle'' the state, but
simply define it (at Alice's location).

Once Alice's measurement is completed, and given her measurement
result, she will be able to make predictions about Bob's state. But
obviously the corresponding information (orientation of Alice's
polarizer and measurement result ) will have to travel
(non-instantaneously) from Alice to Bob. Thus there is nothing here
like ``objective instantaneous action at a distance'', that I dislike
as much as you do. On the other hand, there is an objective quantum
state, in the sense that once Alice's results have reached Bob (and
not before !),  Alice and Bob will know ``the values of a set of
physical quantities, which can be predicted with certainty and
measured repeatedly without perturbing in any way the system''. In
this view BI are violated as a consequence of the lack of ``separate
reality'' of the two particles, rather than as a consequence of
non-locality.

To put it another way:  a (pure) quantum state is objective, but it
is contextual, i.e., defined relatively to a particular set of
measurements results. If new measurements are done that are not in
the initial set, the definition of the new physical state {\it must\/}
include the results of the new measurements. In some sense, this is
simply a restatement of Bohr's 1935 answer to the EPR article (and
maybe this can be related to your ``Bayesian'' approach?).
\eq

\section{21-08-01 \ \ {\it Noncontextuality} \ \ (to C. M. {\Caves})} \label{Caves6}

\bcc
This is a weird notion of operational, since you rightly note
immediately afterward that it is really tautological.  How does it
justify assigning the same probability to an element in the big set,
no matter what the context, to note that you have a rule that assigns
the same probability to an element in the big set, no matter what the
context?  It seems to me the real point is why you would ever
identify two elements from different contexts, and the reason is that
that they are the SAME element in the underlying set, which is handed
to you by your underlying theory.  Moreover, you only have one thing
to go on, and that is the fact that the theory tells you that two
elements are actually the same element in the underlying set.  This
means that the theory wants them to have the same probability in all
contexts.
\ecc

I think what is creeping in here (anew) is a fundamental
philosophical difference that is starting to come between us.  Maybe
I can characterize it in the following way.  You want the theory to
come first, and then to (somehow) recover our activity as scientific
agents back out of that grander picture.  I, on the other hand, am
becoming more and more content to start with the scientific agents
and thereafter pluck out those terms in their discourse with various
common features to call a theory.

\bcc
How does it justify assigning the same probability to an element in
the big set, no matter what the context, to note that you have a rule
that assigns the same probability to an element in the big set, no
matter what the context?
\ecc

It doesn't justify assigning the same probability.  Assigning the
same probability is the very reason for assigning a common element in
our theoretical descriptions of two very distinct devices.

\section{22-08-01 \ \ {\it Contextual and Absolute Realities} \ \ (to P. Grangier)} \label{Grangier3}

\bpg
This may sound like rhetoric, but if a theory explains nothing less than
the stability of matter, is able to calculate $g=2$ (and many other
things) with an incredible accuracy, and nevertheless does not speak
about ``physical reality'', what is physics? I thus consider much
more useful to put forward a reasonable definition of reality, that
allows me to say to a journalist : ``QM speaks about the reality of
micro-objects. It is a weird reality, \ldots\ but it is REAL.''
\epg

You ought to know that I could not agree more with this (the part
that I quoted, not the part that I did not quote).  The issue on my
mind is whether it is productive to view the quantum state in
particular as THE term in the theory that corresponds to the
objectively real.  The way I view the issue is this:  Quantum theory
is a mixture of objective and subjective elements and we will only
make progress in quantum foundations when we have had the
intellectual strength to cleanly separate those two ingredients. With
this point of view, I can answer the journalist just as soberly as
you (i.e., without carrying him off to a postmodern fuzz fest).

There are certainly elements in quantum theory that I am immediately
willing to identify as objectively real.  A good example comes from
our Alice and Bob example, generalized ever so slightly.  Suppose
Alice's system has a Hilbert space of dimension $E$ and Bob's has a
Hilbert space of dimension $D$, and that again Alice and Bob start
off with an entangled state for the bipartite system.  I say that the
quantum state must be (freely) subjective information because,
depending upon what measurement Alice chooses to perform on her side,
she will ascribe one or another quantum state to the system in Bob's
possession.  However, there is one obvious thing that Alice cannot
change by the free choices she makes on her side of the world:  It is
the dimensionality of Bob's system's Hilbert space. Thus, I would say
the number $D$ is the objective reality in this situation.  The
number $D$ remains constant regardless of what Alice does.

I could put this in language more to your liking, saying something
like:
\bq\noindent
The quantum state that arises for Bob's system after Alice's
measurement is a ``contextual reality,'' whereas the Hilbert-space
dimension of Bob's system is an ``absolute reality.''
\eq
but I personally don't see that as a road to further progress. I.e.,
it distracts from what I view as the main point of our task in
clearing up the foundations:  namely, working hard to separate the
subjective from the objective.

What I want to know in the most physical of terms is what does that
number $D$ signify?  I want to find a way of describing its meaning
that never once refers to a quantum state.  When I can do that, then
I will say that some progress has been made in identifying the
objective part of quantum theory.  But that is just an example.

\bpg
Remember my quotation about the 2 electrons in He: in their singlet
state is ``subjective information'', how do the electrons know it?
\epg

It seems to me, this is just a varied form of the Penrose argument I
wrote at the end of my Section~3.  I dismiss it in the same way that
I did there.  From my point of view, to say that the electrons are in
a singlet state is to give nothing more than a compendium of things
we can say about how they will react to our experimental
interventions into their nonhuman bliss.  It is not they who can
predict the consequences of my invasions into their territory, it is
me.

\section{22-08-01 \ \ {\it Identity Crisis} \ \ (to C. M. {\Caves} \& R. {\Schack})} \label{Schack4} \label{Caves7}

I enter this note with trepidation, because I know that what I am
about to say will not be taken lightly by either of you, and chances
are you will just view me as a troublemaker (again).  I don't want to
be a troublemaker, but I do have some concerns that are starting to
eat more and more at me.

The problem is, I am starting to have serious misgivings about our
Sections~II and IV of ``Making Good Sense'' (PRA Version).  Most of
this new thinking has come about through my taking David {\Mermin}'s
quest in his paper ``Whose Knowledge?''~to task, and my ensuing
debate with Todd Brun, Jerry Finkelstein, and David himself.  But
some of it, I think, flows directly from the spirit of Bruno de
Finetti, which I now believe I had shut my eyes to for too long.

The issue is no less than whether we really believe probabilities are
subjective or not.  I think a failure on our part to take their
subjective character completely seriously is causing us to go down a
path I'd rather not take.

Let me try to explain as best I can.  The trouble is localized in our
claim:
\bq
\noindent
\ldots\ if two scientists have maximal information about a quantum
system, Dutch-book consistency forces them to assign the same pure
state.
\eq
In the mildest version of my troubles, I am starting to think this
statement is contentless.  In the stronger versions, however, I find
it misleading, and I almost want to say ``wrong'' (though maybe I
won't go that far).

To make sense of what I mean by this, let me start by taking a cue
from Bernardo and Smith:
\bq
\noindent
   Bayesian Statistics offers a rationalist theory of personalistic
   beliefs in contexts of uncertainty, with the central aim of
   characterising how an individual should act in order to avoid
   certain kinds of undesirable behavioural inconsistencies.  \ldots\
   The goal, in effect, is to establish rules and procedures for
   individuals concerned with disciplined uncertainty accounting.
   The theory is not descriptive, in the sense of claiming to model
   actual behaviour.  Rather, it is prescriptive, in the sense of
   saying ``if you wish to avoid the possibility of these undesirable
   consequences you must act in the following way.''
              --- p.\ 4, Bernardo and Smith, {\sl Bayesian Theory}
\eq

If we accept this, then I think there is a much better way to word
our ``$p=1$ when certainty'' addition to the Dutch book argument in
Section II.  It seems to me it should more properly be viewed as a
normalization condition, subordinate {\it only\/} to internal
consistency/rationality as all the rest of the Dutch-book argument
is.  I say the latter to contrast it with how we presently have the
argument worded in our paper---namely, by making the $p=1$ condition
subordinate to some objective feature of the world.  We write:
\bq
\noindent
   The only case in which consistency alone leads to a particular
   numerical probability is the case of certainty, or {\it maximal
   information}.  If the outcome $E$ is certain to occur, the
   probability assignment $p<1$ means the bettor is willing to take
   the side of the bookie in a bet on $E$, receiving an amount $px$
   up front and paying out $x$ if $E$ occurs, leading to a certain
   loss of $x(1-p)>0$.  Consistency thus requires that the bettor
   assign probability $p=1$.  More generally, consistency requires a
   particular probability assignment only in the case of maximal
   information, which classically always means $p=1$ or 0.
\eq

What does it mean for an outcome $E$ to be ``certain to occur''?  I
think that phrase is much more loaded than we have previously treated
it.  In the Dutch book argument there are three players, two of them
animate and conscious (the bettor and the bookie) and one of them
presumably inanimate and unconscious (the world).  To which player
does the certainty get attached?  I don't think we make this clear in
the way we ought to.

If the certainty is to be attached to the world, then what business
does it have to do with my subjective judgments (which by definition
cannot be in a bijective correspondence with the world's states)?
Instead, I would say the ``certainty'' can only be a subjective
judgment in and of itself.  The Dutch-book argument for requiring
$p=1$ in the case of certainty should then be more accurately
advertised as a call to be ``true to our hearts.''  I.e., the
argument is really that, {\it when\/} we believe an event will happen
with certainty (a nonnumerical judgment), then we should ascribe
$p=1$ (a numerical judgment) for booking purposes.  That is,
Dutch-book coherence gives us a way to translate a nonnumeric belief
into a numeric one.

The thing that is really at issue here is that I think we should
remind ourselves always that ``certainty'' itself is nothing more
than a belief.  It may be a belief that can ultimately be tested
against the world in a single shot, but nonetheless it is a belief. I
believe with all my heart that my mother loves me; Schopenhauer
believes with all his heart that she hates me.  The only thing
Dutch-book consistency can give us is that ``if we wish to avoid the
possibility of undesirable consequences'' (Bernardo/Smith), then I
should ascribe $p=1$ and Schopenhauer should ascribe $p=0$ to the
proposition ``love.''  The Dutch-book argument prescribes that we
each should be true to our hearts---that we should both act in
accordance with our beliefs.  But it does not have within it the
power to make us believe the same thing \ldots\ EVEN in the case of
``certainty.''

Now, we whitewash all that by introducing the phrase ``maximal
information,'' which somehow makes ``certainty'' seem more objective,
but now I'm starting to think that that phrase is pretty impotent in
this context.  What role does it really play in our argument?  I
can't find any, other than that it is a euphemism for declaring that
we {\it believe\/} we have nothing left to learn (in the sense that
we {\it believe\/} there is nothing left to learn from the remainder
of the world that will help us refine our predictions for the system
at hand).  That belief may be wrong in the sense that rationality AND
the world will not allow us to perpetuate the belief AFTER the
experimental trial, but until the trial, ``certainty''---from the
Bayesian view---can mean nothing more than a metered belief.

You should be able to tell where I'm going with this by now.  In
Section IV, we write:
\bq
\noindent
    Maximal information in quantum theory instead corresponds to
    knowing the answer to questions that share one particular
    projector.
\eq
I suppose what I am saying is that I just cannot accept this anymore.
At least not in its present form.  Instead, if I were to modify it to
bring it into alignment with everything I said above, I would have to
write something like this:
\bq
\noindent
    Maximal information in quantum theory instead corresponds to
    believing adamantly that one knows the answer to all the questions
    that share one particular projector.
\eq

You might think this is nitpicking, but it completely takes all the
steam out of Section IV.  For it gives Dutch-book consistency no
grounds for enforcing that two agents ``with certainty'' should
believe the same thing.  And consequently it gives no grounds for
enforcing that two agents with ``maximal information'' should make
the same quantum-state assignment.

The only way I see to reinstate our original role for maximal
information is to say that two observers can only have maximal
information {\it when\/} they are both right (in the sense that the
world MUST CONFORM to their probability one predictions).  But then,
using our argument for Gleason's theorem, we would have DERIVED that
quantum probabilities are objective probabilities!  (This will be my
only exclamation point in the whole note, so you should take it in
seriousness.)  That is, we would be saying that we have maximal
information only when we {\it know\/} an objective reality, and by
our derivation, that objective reality would then be equivalent to a
compendium of probabilities.

Instead I think the best we can say is:  If Alice and Bob both
believe adamantly that they know the answer to some potential
measurement AND that measurement happens to be the same for both of
them, then Dutch-book consistency and Gleason's theorem will enforce
that they make the same probability assignments for all other
measurements (i.e., that they assign the same quantum state).  But
said that way, I don't think any non-Bayesian will be particularly
impressed:  For they would say that all we have shown is ``if two
people know the same things, then they will know the same things.''
Woop-ti-do.

The Bayesians among us will still have some room to be impressed: For
it will not be a priori obvious to them that beliefs about one
observable should have anything to do with beliefs about another. In
particular, it might even surprise them that a common belief in
certainty (for two observers) for any fixed observable should lead to
equal probability assignments for all other observables.  But even
then, I think our the shock-value of our paper will be diminished.
For I think in no way have we shown that when two observers make two
pure-state assignments for a system, those pure states MUST be
identical.

For me, this is a liberating thing to understand, i.e., that there
are no dictatorial constraints on quantum state assignments.  But I
suspect you will feel otherwise, at least on the first reading of
this note.  So, let me beg your forgiveness in advance.

As I alluded to in the beginning, these thoughts of mine don't live
in a vacuum.  They have been spurred by my debates with Brun,
Finkelstein, and {\Mermin}.  Thus let me give you some more material to
chew upon:  I'll attach it below, in the form of a composite note
that I've already sent to {\Ruediger}.  Perhaps it will help clarify
the things that have brought me to this position.  Some of it, of
course, will require that you try to imagine the context, but I think
the notes are self-contained enough that you will be able to fall
into the line of thought and see its relevance. \medskip

\noindent Best wishes (in spite of my predictable trouble),

\section{23-08-01 \ \ {\it My Own Version of a Short Note} \ \ (to C. M. {\Caves})} \label{Caves8}

I'm just back from a very long day in NY City (bookshopping), and a
very long night before that (reading).  So I won't reply to your
notes until I'm a little more refreshed.  But one quick comment:
\bcc
I get the feeling, strengthened by your own confession that it is
true, that my e-mail doesn't make much of a dent, so why bother with
it.
\ecc

That's absolutely not true.  I read everything you send me many, many
times over.  When they are reasoned well, I accept your arguments.
And you know I much prefer this method of communication, just so I
can have the opportunity to fully understand what my correspondent
hopes me to absorb---I've never been a quick thinker, and this helps
me fill in for that inadequacy.

Today I focused on rounding up some more William {\James}, John {\Dewey},
Percy Bridgman material.  I think {\James} is taking me over like a new
lover.  I had read a little bit of him before, but I think I was more
impressed with his writing style than anything.  But I was drawn back
to him by accident, after reading Martin Gardner's {\sl Whys of a
Philosophical Scrivener}.  Gardner devoted a lot of time knocking
down {\James}' theory of truth, because it is just so much easier to
accept an underlying reality that signifies whether a proposition is
true or false, rather than saying that the knowing agent is involved
in eliciting the very proposition itself (along with its truth
value).  And something clicked!  I could see that what {\James} was
talking about might as well have been a debate about quantum
mechanics.  He was saying everything in just the right way. (Let me
translate that:  he was saying things in a way similar to the way I
did in my NATO ``appassionata.'')  And things have only gotten better
since.

Have safe trips to everywhere you need to go.  Kiki is due December
23, but I'll see what can be done about ITP in November.

\section{23-08-01 \ \ {\it Identity Crisis, 2} \ \ (to C. M. {\Caves})} \label{Caves8.1}

And to answer your other question:
\bcc
I see now that this ``brief'' reply is longer than most e-mail I write,
so I might as well tack on something else.  Frankly, after {\Ruediger}
and I browbeat you into submission in Sweden, I didn't expect you to
stay browbeaten.  That's why I was trying to call you about a week
ago.  Since I couldn't extract a PRA-submitted version of our paper
from you, I was guessing that doubts were growing in your mind and
that you hadn't resubmitted it.  Let me know if my suspicions had any
foundation.
\ecc

Yes, partially.  You're starting to be able to predict me like Geraldine can.  (Thank god, Kiki hasn't caught on to the method yet.)  But the paper\footnote{C. M. Caves, C. A. Fuchs and R. Schack, ``Quantum Probabilities as Bayesian Probabilities,'' Phys.\ Rev.\ A {\bf 65}(2), 022305 (2002).} is submitted nonetheless.

\section{26-08-01 \ \ {\it Quantum Commune} \ \ (to the invitees)}

As many of you know, Gilles Brassard and I have become enamored with the idea that the field of quantum information and computing stands to tell us something deep about quantum mechanics itself.  To that end, we organized a small party in May of 2000 to flesh out the issue; we titled it ``Quantum Foundations in the Light of Quantum Information.''  This coming year we will do it again, but on a grander scale---something more like a commune than a party.  We hope you will join us as a communard.

\bq\noindent{\bf
Quantum Foundations in the Light of Quantum Information II\\
{\Montreal}, Canada\\
October 13, 2002 -- November 2, 2002}
\eq

The purpose of the meeting is to gather a group from the quantum information circle who think there are some aspects of quantum foundations that are more mysterious than they ought to be and, importantly, are intrigued by the idea of applying quantum information to the task of cleaning things up.  The atmosphere of the meeting should be quite relaxed with plenty of time for discussion and/or private brooding.  The autumn leaves in {\Montreal} will likely be quite beautiful.  Keeping with the tradition of the last meeting, we only ask that all attendees compile and share a list of concrete problems whose solution, they believe, will tell us something novel about quantum foundations.

We can accommodate up to 15 attendees at any one time, with all attendees having a desk and a computer in an office of one or two mates.  (Of course, the longer you can stay, the better, subject only to our head-count constraint.)  At this time, we can promise to cover all local and living expenses.  There may also be some money for travel costs.

What we would like to know from you for the present are three things:
\begin{itemize}
\item[{1)}]  Would you like to attend?

\item[{2)}]  If so, will your ability to attend be contingent upon travel
    reimbursement?

\item[{3)}]  What is your desired length of stay?
\end{itemize}

It is fairly important that we get this information before September 1, so that we can put the attendees' names on a poster the Centre de Recherche Math\'ematiques (our sponsors) will use to advertise their theme year.

We look forward to hearing from you.  (Since Gilles and I will both be traveling this week, please carbon copy your reply to both of us.)\medskip

\noindent With warm wishes,\medskip

\noindent Chris (and Gilles)\medskip

\noindent PS.  If you can attend, your companions may be any of the colleagues listed below.  (A * beside a name denotes attendance at the first QFILQI meeting.)
\bv
Howard Barnum
\\Charles Bennett (*)
\\Gilles Brassard (*)
\\Jeffrey Bub (*)
\\{\Adan} Cabello
\\Carlton {\Caves}
\\Ignacio Cirac
\\Richard Cleve
\\Claude Crepeau
\\Christopher Fuchs (*)
\\Nicolas Gisin
\\Daniel Greenberger
\\Lucien Hardy
\\Patrick Hayden (*)
\\Alexander Holevo
\\Richard Jozsa (*)
\\Adrian Kent
\\Hideo Mabuchi
\\Dominic Mayers
\\David {\Mermin} (*)
\\Tal Mor
\\Michael Nielsen
\\Asher Peres
\\Itamar Pitowsky
\\Sandu Popescu
\\{\Ruediger} {\Schack} (*)
\\Ben Schumacher (*)
\\Abner Shimony
\\John Smolin
\\Robert {\Spekkens}
\\Andrew Steane
\\David Wallace
\\John Watrous
\\William Wootters (*)
\\Arthur Zajonc
\\Anton Zeilinger
\ev

\section{26-08-01 \ \ {\it Lock Box} \ \ (to R. W. {\Spekkens})} \label{Spekkens4}

\brws
Onto my question.  It seems to be implied in the above quote that
the statements ``none of us can ever completely hide the effects of our interactions with the world'' and ``There are no lock boxes for
information'' are roughly synonymous.  So it seems that your conception of a lock box refers only to the property of concealment.  As far as I
can tell, in the sense you use the term here, a lock box need not be
binding.  Later you state ``Item (1) sounds a lot like the nonexistence of bit commitment.''  However, it doesn't sound that way to me because
being able to ``hide the effects of my interactions with the world''
requires only a BC protocol that is concealing.  Indeed, it seems to
me that item (1), as I am understanding it, is in fact false for QM.
QM {\bf does} allow me to hide the effects of my interactions with the world.  Simply use a BC protocol that is perfectly concealing but not at all binding.  Indeed, I would argue that if one only seeks lock
boxes that are concealing, then one need only look as far as Wiesner's
proposal for quantum money to see that such lock boxes {\bf are} provided by QM (note that Wiesner's scheme can easily be modified so as not to rely upon the ability to prepare `absolutely' pure states, so the finite strength of preparatory measurements is not an issue).  So my question, in a nutshell, is where does bindingness enter the picture?
\erws

You are certainly right about this; I was not being careful enough in my formulation.  Thinking back on it, I guess I was tacitly wanting to reject concealingness and while retaining bindingness, but the thoughts were still pretty vague at the time \ldots\ and {\it are\/} still pretty vague today.

Let me record a few random thoughts:
\begin{enumerate}
\item
I certainly have no problem with an information-writing agent being able to reverse his symbol, so long as the physical system on which he writes it can be thought of as an extension of his body.  But there should come a time when the agent becomes detached from the system, and at that point he should lose such privileges.
\item
In the quantum mechanical case, when Alice makes it so that the system is perfectly concealing, then it has zero binding power.  In that case, one might say that she has not ``really'' recorded any symbol at all.
\item
What is the meaning of the existence of intermediate cases?  Do they have anything to do with the motif I was trying to imagine?  Or do they lean on a completely different motif?
\end{enumerate}

I'll be sending you an invitation to a quantum foundations workshop Gilles and I are organizing in {\Montreal} for next October, in a very short while.  I hope you can come.

\section{27-08-01 \ \ {\it Bayesian Pill Taking} \ \ (to N. D. {\Mermin})} \label{Mermin31}

I am finally writing to reply to your long note on ``knowledge.''  I
apologize for keeping you waiting so long, but even now---since
{\Ruediger} is still in the woods---I feel a little like I am writing
you prematurely.  Nonetheless, I have this overpowering desire to get
this issue out of my mailbox and be done with it.  So here, I am.  I
would ask you, however, to please keep these thoughts private, at
least until {\Ruediger} finally emerges.  For, what I am about to say
involves him (and {\Caves}) directly.

I know that you think your note is ultimately conciliatory, writing:
\bdm
It seems to me that these are all valid statements about the formal
structure of quantum mechanics, independent of what interpretation
you favor.
\edm

But after much soul-searching, I still cannot agree with your
language.  The soul-searching was required because this position of
mine flies in the face of some of my very own published words (\quantph{0106133}).

The difficulty hinges on your Proposition 1:

\bdm
\label{MerminitionOrthogonal}1.  A system that is known (by somebody) to be in a state psi cannot
be found (by anybody) to be in a state orthogonal to psi.
\edm

In my new view, to say this is to throw away all that we have been
striving so hard for in establishing that quantum states are
subjective entities (and purely subjective entities).  You admit so
much yourself at the end of your Comment (c).  [You can take this to
mean that I also disagree with the beginning part of Comment (c).]

In 1880, I suspect there was not a single educated physicist who
doubted one iota that the speed of light was set with respect to the
stationary ether.  Regardless of that, in 1881 the first evidence
came out against the common belief.  The {\it spirit\/} of your
Proposition 1 would outlaw the very happening of that wonderful
historical event.

I see no way (nor even want to anymore) to get around this.  If the
quantum state is a subjective entity, without rigid connection to the
world in itself,---as I think Einstein's original argument for its
subjectivity indisputably shows us---then this is something we simply
have to live with.  We must allow that experimental data can speak
against our predictions, no matter how set we are to believe that
they will not.  Yet, on top of this, we must also accept that there
is no objective sense in which a (pure) quantum state is simply
``wrong'' before the experiment is performed.

I put all my heart and soul into presenting this point of view in the
note below to my coauthors, and I will share it with you.  It took me
a whole day to write the thing; I would like to think it is worth
reading.  (Also I have slight evidence that it makes sense in that
{\Caves} ultimately concurred himself.)  I believe the note is just as
relevant to you and your note on ``knowledge'' as it is to {\Caves} and
{\Schack}.  I will also send you the PRA version of \quantph{0106133}, so you can track the sections.

There.  Now I will clean out my mailbox.

\section{27-08-01 \ \ {\it A Little Contextuality on Noncontextuality} \ \ (to C. M. {\Caves})} \label{Caves9}

I've finally got enough time to write a small reply to your old note.

\bcc
There are two distinct approaches here, and I don't know which is to
be preferred.

1.  The first point of view, which I have been pushing (as a way to
justify noncontextuality), is that there is an underlying theory that
sets up the structure of questions.  This theory is primary.
Noncontextuality emerges as the natural assumption that probability
assignments should recognize the structure provided by the underlying
theory.

2.  The second point of view, which you have been pushing, starts
with the role of scientific agents, as you put it, and uses the fact
that probabilities are the same in different contexts to say that the
elements with the same probabilities in all contexts explains why
they are actually the same element in different contexts.

I do want you to understand my position, which is that I appreciate
both these points of view.  I'm not sure which will be the most
fruitful in the long run.  But they are trying to do quite different
things.  Here are two points:

1.  The second point of view doesn't provide a justification for
assuming noncontextuality, as you understand.  Coming at things from
the back door, as it were, it uses the fact of noncontextual quantum
probability assignments to conclude that apparently different things
are, in fact, the same.
\ecc

I would not use language quite like that.  I would say it IS a
justification for noncontextuality.  And it relies on quantum
mechanics not one iota.  Noncontextuality should be a property of any
instrumentalistic theory (where Bayesian probability has been grafted
onto to the world as the best way for us to steer our actions within
it).  By an instrumentalistic theory, I mean one where we explicitly
have to talk about our various possibilities for experimental
intervention into nature---a theory where we cannot detach the
experiment from the phenomenon.

\bcc
2.  The first point of view appeals to me presently because it
manages to make a long straight run to the state-space structure and
the quantum probability rule given only the Hilbert-space structure
of questions and probabilities faithful to that structure (i.e.,
noncontextuality).  I think the second point of view needs to address
the following question: given a set of elements to which
noncontextual probabilities are assigned, what structure is forced
onto the set by the existence of these noncontextual probabilities?
This question seems hopelessly underconstrained to me, but Howard
described to Joe and me some math research on this sort of question.
\ecc

What you say ``needs'' to be done, seems hopelessly underconstrained
to me too.  But I think you shouldn't view the problem that way. The
point is, one simply has a way of clearing the air of the
noncontextuality issue BEFORE getting down to the nitty-gritty of
quantum mechanics.  Noncontextuality is the base assumption upon
which one plays a new game:  What {\it physical\/} assumption makes
it so that our instruments should correspond to POVMs and not some
other mathematical structure?

By the way, you know I really dislike your phrase ``given only the
Hilbert-space structure of questions.''  I've probably said this to
you before, but let me try to articulate why in more detail so that
maybe you'll remember it a little better.  In your own words, the
phrase is ``hopelessly unconstrained.''  What does it mean?  It seems
to me there are all kinds of possibilities one could imagine if one
didn't know quantum mechanics beforehand.  Here's a simple example:
An even more basic feature of Hilbert-space before orthogonality is
linear independence.  When God came down and said, ``You will use
Hilbert-space structure for the questions you can ask of nature,''
why did he not mean that any set of linearly-independent vectors
corresponds to a valid question?  Presumably there are good reasons.
But those good reasons need to be spelled out, and are not at all
implied by the simple phrase ``Hilbert-space structure.''

Let me send you to pages 86--88 and pages 361--362 of the samizdat.
There it is shown that linear independence does not mesh so well
noncontextuality.  It is a dumb exercise, I know, but it does
indicate that ones need to be careful with one's phraseology.

Oh, let me tell you another thing, of historical note.  I talked to
Howard Friday, and he tells me that this kind of justification for
noncontextuality goes all the way back to Mackey (but then he settles
on the setting of ODOPs thereafter).

\bcc
I'm also quite interested in another question: How general are the
theories where maximal information does not lead [to] certainty, yet
does lead to unique noncontextual probability assignments?  In other
words, for what classes of theories is there a Gleason theorem?
\ecc

I think this is a really good question.  I was talking to Eric Rains
the other day and he thinks that the appropriate generalized setting
might be the Jordan algebras.  This is because this is the largest
structure he knows where there is a notion of positive operator.  (I
had shown him the trivial POVM way of proving Gleason.)

If Gleason can be proved in such a wildly general setting, I think it
would be quite interesting.  For it would tell us that the quantum
probability rule is not very closely connected to physics at all.
Dreaming of the process of deriving quantum mechanics as successively
tucking up the more general structure of Bayesian probability theory,
one might say that the real physical assumptions don't come until
much later in the game.  That would be worthy knowledge!

Waiting for a stupid doctor's appointment,

\section{28-08-01 \ \ {\it Introduction} \ \ (to A. Wilce)} \label{Wilce1}

Please allow me to introduce myself:  We have a mutual friend in Howard Barnum.  Howard brought my attention to some of your work, and he has also told me that you are reading my paper ``Quantum Foundations in the Light of Quantum Information.''

As you surely know by now, my section ``Whither Entanglement?''\ is just plain wrong.  Positivity alone on local measurements does not narrow down the bi-linear map to one derived from a density operator.  I am so ashamed of this mistake.  And I am already planning how I can make amends for my insult to nature.

I think what I would like to do rather than amend the original paper, is instead write a short comment and post it on {\tt quant-ph}, with a title something like ``Wither `Wither Entanglement?'?''  I could use this as another opportunity to inspire people to this issue, but also (importantly) to draw attention and advertise your and your colleagues' work to an audience not previously aware of it.

I can see at least two routes by which to stitch up my failed attempt of a theorem.  The first is to simply say that local measurements (with one-way classical communication) alone are enough to get us that ``states of knowledge'' must correspond to linear operators on the tensor product of the Hilbert spaces.  But then to completely nail down the state-space structure we contemplate putting the systems back together and imagine performing completely general measurements subject to this weaker property we have just derived.  Positivity {\it then\/} forces us to the density operators on the tensor product space.  Another way to say this is that local measurements alone tell us how to derive a dimensionality for the composite system---they just don't get us to positive semi-definiteness.

In a way, though, that is a dull way to patch the problem.  Another way that comes to mind is to ask what further conditions should we add upon the local-measurement statistics so that we nail down the density operators solely.  What I mean by this is that, since local measurements will get us to a {\it unique\/} linear map, it must be the case that some of can be ruled out by the local-measurement statistics they give rise to.

So, let me ask you these questions:
\begin{enumerate}
\item
Can you give me pointers to all the relevant literature to do with deriving the tensor product structure along our kinds of lines?
\item
Do you know of any conditions like the ones I'm contemplating in the second strategy above that will do the trick?  Where are they published?
\end{enumerate}
I hope you will be patient with me in that I am not so familiar with the language quantum logicians use.

I'm glad to get to know you.

\section{28-08-01 \ \ {\it Some Questions Regarding Your Comments} \ \ (to N. D. {\Mermin})} \label{Mermin32}

\bdm
The trouble with our exchanges is that I'm always trying to zoom in
on the issues under dispute and you're always trying to zoom out.
\edm

Thanks for the note.  Your point is a good one.  (It's just that
I've had to write so much email lately, and always, that it does take
a toll.  For instance, the present debate has been particularly
taxing in that regard.  Yet the issues have been important enough to
not give up.  So certainly I was hoping to recycle some material.)

I will try to write you a (focused) reply tomorrow, after recovering
from today.  Tomorrow evening, I fly out to Munich for a week and a
half.  Then I go to the quantum foundations meeting in Ireland to
tote the wares.

Still tonight I've got to work on packing, etc.

\section{31-08-01 \ \ {\it Tina}\ \ \ (to A. S. Holevo)} \label{Holevo4}

Thanks for the good wishes for my travel.  They worked!  The flights went without a hitch, and I got a significant amount of reading done.  (I've been focusing on the writings of William James, John Dewey, and the other American pragmatists lately.  I now believe they had a set of thoughts significantly in sync with the quantum foundational program I sketched in my paper.)

\section{31-08-01 \ \ {\it MSRI Q. Info Workshop Dates?}\ \ \ (to R. Jozsa)} \label{Jozsa5}

I hope Umesh can easily changes his dates:  that certainly would be the easiest option for the {\Montreal} thing.  I will keep my fingers crossed.  If I hear anything from Umesh I'll let you know as soon as I do.  Likewise, if you hear anything before I do, please let me know.  It'll give me a sigh of relief.

I'm in dreary Munich for the week.  But I like dreary weather like that:  I find it especially conducive to philosophical thought.  I'm going to visit Hans Briegel next week, by the way.  I'm quite enamored by this ``measurement-only'' computational model of his and coworkers'.

\section{01-09-01 \ \ {\it Left Wing, Right Wing, Not a Wing to Fly With} \ \ (to N. D. {\Mermin})} \label{Mermin33}

I was going to use my day today to write you a long, thoughtful (but
focused) note on all the recent issues you raised with me on ``Whose
Knowledge?'', but now you've gone and angered me.  I mean that.

\bdm
I took my left-right terminology from the Science Wars.  It seems to
me that in arguing against anything objective other than knowledge
Chris is taking a decidedly post-modernist position and therefore
allying himself with the ``Academic Left'' attacked by Gross and
Levitt.
\edm

Is my point of view so very subtle that even my most sympathetic
patrons cannot decipher it?  Or have I finally caught on that you're
really not listening to me after all?

You can't stand this, but you deserve it:  I will excerpt part of a
note I recently wrote to Andrew Landahl.  What set me off in his case
was when he wrote me the following after having read my paper
``Quantum Foundations in the Light of Quantum Information'' and given
a Caltech journal club talk on it.
\bq
       On the whole I portrayed your ``party platform'' as the statement
     that ``Quantum states are states of knowledge about the consequences
     of future interventions.''  In particular, those consequences aren't
     consequences to reality, but rather consequences to states of
     knowledge about even further future interventions.  In this
     worldview Bayesian agents don't work to align their predictions
     with an underlying reality.  Instead they work to align their
     predictions with {\it each other}.  It is as if reality in this
     picture is solely the agreement of predictions!

       I'd be interested to hear if you believe that this is a fair
     characterization of your party's platform.  After reading this
     paper, I came to the conclusion that you didn't believe in reality
     at all. (Or at best I thought you believed reality = knowledge.)
     John Preskill tells me you believe otherwise, namely that there
     {\it is\/} a reality, which surprised me.  I'm curious to hear what you
     believe reality is.
\eq

The scheme below is that everything marked with a ``$>$'' is a direct
quote from Andrew Landahl's letter.  Everything else is either me, or
a quote from my paper.  I will put only the very most relevant part
of my reply.

Please do read it before you---{\it yes}, you, the most trusted of my
academic friends---slander me again.  It is nothing if not {\it exactly\/} relevant to what you wrote about me above.

With surprisingly kind regards,

\section{02-09-01 \ \ {\it Left and Right} \ \ (to N. D. {\Mermin})} \label{Mermin34}

\bdm
I thought you were safely away in Ireland or pre-Ireland.
\edm

Yes, I am in pre-Ireland mode (in Munich), but keep in mind that I am
never safely away.

\bdm
I'll read what you sent in a little while.  But note that I would not
have made such a remark (even in jest) before I got your comments on
my summary (and the cc of your letter to your own coauthors) which
struck me (and I thought you too) as going beyond your earlier
position.
\edm

I will take this remark into account for the longer reply I am
presently constructing for your earlier query.  (It is sitting at 6K
in length now, and will likely be finished tomorrow.  Right now, I'm
having my first beer of the evening.)  But preliminarily, let me say
that the only thing that letter to my coauthors did was clarify my
position on the subjective character of the state vector.  In that I
went further than before, having gotten weak in the knees briefly
about my position on two agents sharing differing quantum states.
But, I do believe that I have long held fast in my opinion that there
is something in the universe beyond human ken:  It has always been a
problem of finding the right language and right ideas for expressing
what that something is, AND how it is PARTIALLY reflected in the
structure of quantum mechanics.

If this does not make sense yet, I hope it will make more sense after
my long note to you tomorrow.

\section{02-09-01 \ \ {\it Intersubjective Agreement} \ \ (to N. D. {\Mermin})} \label{Mermin35}

Let me finally throw myself into the ring of
intersubjective-agreement to see if I can wrestle you down a little.
I will try to be every bit as focused for you as the issue will let
me be.

\bdm
Well maybe you're more radical than I thought.  It was to avoid
correlations floating in the void, unattached to anything whatever,
that I've been interested in trying to follow you down the path of
knowledge.  If all it led to were knowledge, floating in the void,
unattached to anything whatever --- even to other knowledge --- then
I'd be no better off [than] when I started down the trail.
\edm

I think you misunderstand something very deeply here.  The point of
separating the categories ``knowledge'' and ``reality'' (or
``subject'' and ``object'' for that matter) is not to make knowledge
an objective reality in its own right or, even worse, to make it the
sole reality.  Rather it is to say that there is a distinction and
that that distinction should be recognized.  I see nothing wrong with
allowing a physical theory (such as quantum mechanics) to contain
formal elements that correspond to {\it both\/} categories.  The
issue in my mind is which elements should be thrown into which
category? The answer is not completely clear to me, but I am fairly
convinced of one thing:  The state vector should not be thrown into
the reality side of the line.

What I have ultimately NOT been able to stomach about your wording of
the whose-knowledge ``answer'', and Jerry's wording of the
whose-knowledge ``answer''---some of Todd's versions would actually
survive---is that you say, under certain circumstances, two
scientific agents (observers, or what have you) MUST assign
``consistent'' quantum states to a given system.  In the case of pure
states, the two agents MUST assign the {\it same\/} pure state to the
system.

Let me get that through your head:  What I object to is the word
MUST.  Todd once wrote it this way,
\btb
We have been describing a consistency criterion.  {\bf If} one wishes
to combine two state descriptions of a single system into a {\em
single} state description, the criterion tells one {\bf when} it is
consistent to do so (i.e., when the two descriptions are not actually
contradictory).

I agree that nobody is holding a gun to Alice's head and forcing her
to incorporate Bob's information.
\etb
and to this way of speaking I can agree.  But if you take away Todd's
``{\it If}'', then everything collapses in my mind.  Enforcing that
two agents MUST make the same state assignment if they are going to
be ``right'' at all reinstates the very objectivity, the very
agent-independence of the quantum state that the
Mechanica-Quantica-Lex-Cogitationis-Est program has been working so
hard to exorcise.  [As you know, we made a serious misstep in our
\quantph{0106133}, but that will be rectified in the next
edition.]

It is much like the old debate.  Is materialism right?  Or is it
Berkeley's idealism that is right?  Who cares, I say.  Both
philosophies are just simple samples of realism:  They only disagree
on the precise concept which ought to be taken as real, mundane
matter or sublime consciousness.  The way you characterize it above,
one would think that the only fruit of the Mechanica Quantica program
would be the RENAMING of a material reality into an ideal one---a
shift more of emphasis, rather than anything of grit.

\bdm
Are you also unable to agree with the statement that a photon that is
known (by somebody) to have just passed through a horizontal polaroid
cannot immediately thereafter be found (by anybody) to pass through a
vertical polaroid.

I'm asking you about this concrete example of the general proposition
because I can't tell whether you're objecting to the language in
which I generalized it or whether you object to the statement about
photon polarizations too.  If it's only the former I'm happy to use
less provocative phrasing.  All I meant by ``be in a state psi'' was
``has been found to be'' in the sense I specified prior to making the
objectionable statement.  But I worry that you object to both
statements.  In that case you are walking a dangerous path, denying
that one of the most elementary applications of quantum mechanics has
a legitimate meaning.
\edm

Here is what you are losing sight of.  In the Bayesian world, two
agents must agree a little before they can agree a lot.  Agreeing a
lot is the currency they are seeking, but agreeing a little to begin
with is not the limitation of their existence.  I'll come back to
this from a more positive angle in a minute, but let me tackle your
particular question before that.

What does it mean for ``a photon that is known (by somebody) to have
just passed through a horizontal polaroid?''  Presumably it means
that a particular quantum mechanical test, a POVM, $\{E_b\}$ has been
performed and one of the outcomes of that test has obtained---in this
case, the label $b$ is ``photon passed through horizontal polaroid.''
Now, you ask, ``immediately thereafter [can it] be found (by anybody)
to pass through a vertical polaroid?'' Implicit in that, you are
thinking that the state transformation, subject to the measurement
outcome, is
$$
\rho \longrightarrow E_b^{1/2} \rho E_b^{1/2}
$$
up to normalization.

Suppose you are the somebody spoken of above; and let me be part of
the anybody.  Now, let us say that I stubbornly insist that the state
transformation is
$$
\rho \longrightarrow U_b E_b^{1/2} \rho E_b^{1/2} U_b^\dagger ,
$$
where the $U_b$ are some unitary operators, and in particular, when
$$
b = \mbox{photon passed through horizontal polaroid}
$$
we have
$$
U_b = \mbox{horizontal} \rightarrow \mbox{vertical} .
$$
There is nothing in quantum mechanics (as a theory) that can keep me
from believing that, so long as the ONLY thing specified is the
``measurement'' $\{E_b\}$.  The point is, let us suppose we disagree
on how our beliefs should get updated upon the incorporation of a
measurement outcome into our knowledge bank.  [As an aside, notice
the distinction here:  $b$ is given the lofty title of knowledge,
whereas $\rho$ is subjected to being a belief.  I allowed myself to
do that because I am assuming we agree on the $\{E_b\}$, even if not
the state-change rule.  You might say we need at least this much to
get the problem off the ground.]

So, it boils down to this in more common language,
\bdm
     Are you also unable to agree with the statement that a photon
     that is known (by somebody) to have just passed through a
     horizontal polaroid cannot immediately thereafter be found (by
     anybody) to pass through a vertical polaroid?
\edm
And I say no, I cannot agree:  I saw Hideo Mabuchi in his lab
yesterday, and I saw that he inserted a really fancy polarizer into
his lab bench, one with an intriguing optical coating that allows
horizontal photons in, but has them come out as vertical ones.  I
insist that I saw him do that:  There's not a doubt in my mind.  You
insist that he is an honest upright boy, and he would never do such a
thing to confuse us.  With equal tenacity, there is not a doubt in
your mind.  We disagree, and in the strongest of ways.

How do we put our disagreement to test in the context of
photo-detector clicks?  WE insert a vertical polaroid---this one, I
assume, we both do agree on---behind the ``horizontal'' one and see
what happens.  Aha!  You were right after all!  Mabuchi really did
use an honest-to-god von Neumann polaroid; the input photon never
made it to the final detector.  It wasn't the fancy-coated polaroid
after all, but I could swear I saw him put it in.

By now, I know that you are thinking I have gone through a
ridiculously long-winded and pedantic way of describing a triviality:
that one of us made a FALSE assumption.  Implicit in your question
was the reasonable starting point---indeed, the one we use in all
common discourse---that all the agents involved start from a TRUE
state of affairs.

But what can TRUE and FALSE mean in a world where our only handle for
getting at things are SUBJECTIVE quantum states?  We get at the world
through our beliefs and belief updates---that's the fundamental tenet
for me.  And in that light, the only thing a FALSE belief can
mean---as I put it to {\Caves} and {\Schack} in the infamous email---is
that rationality (i.e., the Bayesian laws of thought) PLUS the world
(i.e., the detector clicks, whose meanings in the end are also set by
subjective considerations) will not allow the believer to perpetuate
any remnant of his initial belief after the experimental trial.  Let
me say that sentence again for extra emphasis:

\bq
\noindent
     The only thing a FALSE belief can mean is that rationality
     (i.e., the Bayesian laws of thought) PLUS the world (i.e., the
     detector clicks, whose meanings in the end are also set by
     subjective considerations) will not allow the believer to
     perpetuate any remnant of his initial belief after the
     experimental trial.
\eq

But if that is the case, what is so overpoweringly evil about having
a ``false'' belief?  Why must two valid quantum scientists
necessarily be aligned in their beliefs, even in the case of ``true''
and ``false''? I remain hardened:  I see no compelling reason for
asserting that necessity.  Indeed, such an assertion is antithetical
to the idea that a quantum state is a compendium of subjective
degrees of belief.

If you think TRUE and FALSE mean something more substantial than I
just described, then you tell me what role they play in my life other
than as a kind of shorthand for some characteristics of my belief
updates.

In pointing out these deficiencies, I am not ``denying that one of
the most elementary applications of quantum mechanics has a
legitimate meaning.''  I am coming nowhere near that.  I am merely
asserting each scientific agent's constitutional right to believe
what he will---i.e., to carry about whatever quantum state he
wishes---SO LONG AS those beliefs do not contradict the constitution
itself.

It is the latter that, from my view, is the most essential point you
have been missing.  So, let me get to that directly.  With this I can
finally start to define a positive program.

\bdm
I don't see that limiting ``objectivity'' to mean ``complete and
necessary intersubjective agreement'' is abandoning your quest.
Indeed, your Bayesian authorities say as much,
\bq
\noindent
\rm
We can find no real role for the idea of objectivity except, perhaps,
as a possibly convenient, but potentially misleading, ``shorthand''
for intersubjective communality of beliefs.
\eq
It must be the ``necessary'' that raises your hackles [\ldots]
\edm

If anyone cannot see by now that it is almost solely the word
``necessary'' that raises my hackles, then they are not listening.
You wrote at the beginning of your note that:
\bdm
     If all it led to were knowledge, floating in the void,
     unattached to anything whatever --- even to other knowledge ---
     then I'd be no better off [than] when I started down the trail.
\edm
But that is just not true.  We have gained a serious amount of
positive knowledge from this exercise.  It has allowed us to see much
more clearly what is firm and what is squishy in quantum mechanics.
The state assignments I would say are always squishy; the rules for
updating them are not.  To the extent that these rules fulfill an
edict in the spirit of Bernardo and Smith,
\bq
   Bayesian Statistics offers a rationalist theory of personalistic
   beliefs in contexts of uncertainty, with the central aim of
   characterising how an individual should act in order to avoid
   certain kinds of undesirable behavioural inconsistencies.  \ldots\
   The goal, in effect, is to establish rules and procedures for
   individuals concerned with disciplined uncertainty accounting.
   The theory is not descriptive, in the sense of claiming to model
   actual behaviour.  Rather, it is prescriptive, in the sense of
   saying ``if you wish to avoid the possibility of these undesirable
   consequences you must act in the following way.''
\eq

I would say we have identified an objective piece of quantum
mechanics.

It is not that ``physics as intersubjective agreement'' requires that
agents always agree, or, at least, that they must agree in certain
limiting circumstances.  Instead it is that there is a procedure in
any given case for deciding whether two agents will move closer to
agreement than not after looking at the world.  Quantum mechanics
gives us such a framework.  It might have been otherwise:  One can
imagine a world so chaotic that any percipient beings which happen to
arise in it would forever be in their own individual dreamlike
states, never realizing that it is even possible to come to agreement
with their fellow quixoticoids.

Maybe a good (but limited) analogy is this.  Think of an electric
potential function from which, by taking a gradient, we can derive
the electric field.  The potential itself cannot be the real stuff
because of its ridiculous freedom that can be set freely from
observer to observer.  Instead it is the way the potential changes
spatially that is what is of interest.  That spatial change which is
the common denominator of all the disparate potential assignments is
the real, real stuff.  Now think of the quantum state in the role of
the potential, and the quantum structures of POVMs, the Gleason
theorem, and the state-change rule in the role of the potential's
gradient.

As long as you and I play according to the quantum rules for updating
our beliefs in your example---you with your
$$
\rho \longrightarrow E_b^{1/2} \rho E_b^{1/2} ,
$$
me with my
$$
\rho \longrightarrow U_b E_b^{1/2} \rho E_b^{1/2} U_b^\dagger
$$
---who is to fault one of us for being irrational?  We just have
different beliefs about how a state ought to get updated in the
particular situation.  Neither one of us is taking the constitution
to task; neither one of us are using a state-updating method that
does not fall into the quantum mold.

The analogy of this with classical probability theory is that we both
might agree on the probability for some hypothesis $p(h)$, but
disagree on the joint distribution $p(h,d)$ for hypothesis and data.
Learning the data and using Bayes' rule, we will generally then come
to two distinct posterior assignments $p(h|d)$.  That nevertheless
gives us no warrant for backtracking our opinion that $p(h,d)$ is
just a subjective belief (as are all probability assignments).
Instead it helps us see that the objectivity working in the
background is Bayes' rule; it is our common denominator.

Quantum states---or through Gleason's theorem, nothing more than
compendia of quantum probabilities---do not float in a void.  They
are tied together more tightly than any other probabilities hitherto
ever found.  I cannot assign probabilities for $\sigma_x$ outcomes,
$\sigma_y$ outcomes, and $\sigma_z$ outcomes at the same time as {\it
independently\/} assigning them to the outcomes of any more exotic
POVMs.  In changing my probabilities for the outcomes of some
potential new measurements (just after a previous measurement), I had
better tie all those probabilities together along the lines of the
general form of the quantum state-change rule.

In this, we see something like a much greater deepening of the Dutch
book argument.  In the classical case, we find that we will bring
havoc upon ourselves if we allow ourselves to freely assign $P(A)$,
$P(B)$, $P(A\wedge B)$, and $P(A \vee B)$ all independently.  All
compendia of quantum probability assignments must be tied to the
particular structure of quantum states and the quantum state change
rules.  You should be thinking of the firmament rather than the void.

I think that's all the really general remarks I had wanted to say.
Let me now {\it briefly\/} address the remaining specific points in
your notes.

\bdm
If you do indeed object to both, then the only reason I can see for
it (and I agree that this does raises non-trivial issues) is that
probability 1 and probability 0 statements are idealizations --- that
nothing in the actual world we inhabit can be said to be certain or
impossible.  In that case, of course, the support of any acceptable
density matrix is the whole Hilbert space and there is no content
left to the criterion.  But since the theory does allow you to talk
about certain or impossible measurement outcomes, I'm reluctant to
declare that its illegitimate to consider them in trying to develop a
better understanding of the theory.
\edm

I hope that you can see by now that ``probability 1 and 0 statements
being idealizations'' (i.e., states of belief that we none of us,
even Job, are really ever in possession of) has nothing to do with my
considerations.  A belief is a belief.  Rationality itself cannot
infringe on what numerical value a belief ought to be.  It is
therefore perfectly legitimate to think about these idealized
situations.

\bdm
\bq
\noindent
\rm
CAF Said: on top of this, we must also accept that there is no
objective sense in which a (pure) quantum state is simply ``wrong''
before the experiment is performed.
\eq
[Do] you also require me to accept there is no objective sense in
which a pure quantum state is simply correct, after the experiment is
performed?
\edm

Yes.  The ghost of Bruno de Finetti haunts us:
\begin{center}
                       QUANTUM STATES DO NOT EXIST
\end{center}
And I understand that oh so much better now than I did two months
ago.

\bdm
P.S.  I also asked for clarification of your views on objectivity as
nothing more than intersubjective agreement, which on the one hand
you seemed to reject in accusing me of going objective in comment (c)
but on the other hand you seemed to endorse in quoting approvingly
your Bayesian gurus.
\edm

Bernardo and Smith would have never held fast to a ``necessity
clause'' like you seem to be doing.  That puts a gulf of distance
between your two separate uses of the phrase ``intersubjective
agreement.''

\bdm
P.P.S.  Just got a blast from the Eastern Front (Mohrhoff --- cc'd to
you, I believe).  I have the funny feeling that you two, who are so
far apart in opposite directions (knowledge-without-facts vs
facts-without-knowledge), may yet turn out to be strangely similar in
some respects.
\edm

To the extent that I understand him, I myself don't believe this is
likely.  The direction I see for physics, and quantum mechanics in
particular, was perhaps no better put than by William {\James}:
\bq
Metaphysics has usually followed a very primitive kind of quest. You
know how men have always hankered after unlawful magic, and you know
what a great part in magic {\it words\/} have always played. If you
have his name, or the formula of incantation that binds him, you can
control the spirit, genie, afrite, or whatever the power may be.
Solomon knew the names of all the spirits, and having their names, he
held them subject to his will.  So the universe has always appeared
to the natural mind as a kind of enigma, of which the key must be
sought in the shape of some illuminating or power-bringing word or
name.  That word names the universe's {\it principle}, and to possess
it is after a fashion to possess the universe itself. `God,'
`Matter,' `Reason,' `the Absolute,' `Energy,' are so many solving
names.  You can rest when you have them.  You are at the end of your
metaphysical quest.

But if you follow the pragmatic method, you cannot look on any such
word as closing your quest.  You must bring out of each word its
practical cash-value, set it at work within the stream of your
experience.  It appears less as a solution, then, than as a program
for more work, and more particularly as an indication of the ways in
which existing realities may be {\it changed}.

{\it Theories thus become instruments, not answers to enigmas, in
which we can rest.}  We don't lie back upon them, we move forward,
and, on occasion, make nature over again by their aid.
\eq

Mohrhoff, from what I can tell, sees a ``block universe'' (to use
another piece of {\James}ian terminology).  It is a completed thought in
the cosmic consciousness.

Good wishes, and I hope this document answers more questions for you
than it raises.  Now I've got to run to the biergarten again for a
little oompah-pah.

\section{02-09-01 \ \ {\it Truth and Beauty} \ \ (to N. D. {\Mermin})} \label{Mermin36}

Here's another passage from William {\James}'s {\sl Pragmatism\/} that
may help reveal a little more of my mindset.

\bq
\noindent
The truth of an idea is not a stagnant property inherent in it. Truth
{\it happens\/} to an idea.  It {\it becomes\/} true, is {\it made\/}
true by events.  Its verity {\it is\/} in fact an event, a process:
the process namely of its verifying itself, its veri-{\it fication}.
Its validity is the process of its valid-{\it ation}.
\eq

\section{03-09-01 \ \ {\it Subject/Object} \ \ (to M. A. Nielsen)} \label{Nielsen1}

\bmn
You may also be interested to hear that I'm engaged to be married :-)
\emn

Excellent!  This is only a joke partially, but lately I've been so
taken with the idea that unions can give rise to things greater than
those contained in the parts---thinking of quantum measurement, in
particular, from this angle---I've thought about calling my view on
QM ``the sexual interpretation of quantum mechanics.''

Many congratulations!

\section{04-09-01 \ \ {\it Note on Terminology} \ \ (to C. M. {\Caves} \& R. {\Schack})} \label{Schack5} \label{Caves10}

Thinking about it more, I would like to emphasize a point that was
buried away as an ``aside'' in my recent note to {\Mermin} titled
``Intersubjective Agreement.''

I am becoming more and more dissatisfied with the slogan ``A quantum
state is a state of knowledge, not a state of nature.''  The reason
for this is that people tend to view the word ``knowledge'' as
something that can be right or wrong, depending upon whether it is in
direct correspondence or not with something in the external world.
For this reason---as brought out clearly in my debate with {\Mermin},
Brun, and Finkelstein---I think we should get more into the habit of
calling a quantum state a state of BELIEF.  This is more in line with
the language both de Finetti and Bernardo and Smith use for
probabilities anyway, and therefore gets us into a quicker connection
with the personalistic Bayesians.

I now think it is much better to reserve the word KNOWLEDGE solely
for the outcomes of quantum measurements once they become part of the
mental makeup of an agent interested in them.  I walk into Mabuchi's
lab, and to the extent that he and I agree that he is performing some
POVM (denoted by a set of positive operators $\{E_b\}$), it seems to
me valid to call the outcome $b$ we both witness to be an addition to
our knowledge.  Now, what either of us may do with that knowledge is
a different story.  One thing is for sure, it ought to cause both of
us to update our beliefs.

Thus knowledge (and information) bear on how we change our beliefs,
and in that way---in a sense---become incorporated into our beliefs,
but there is no rigid connection between the two concepts.
Knowledge/information, as it is encoded in measurement outcomes, is a
bridge to the external world that the quantum state has no right to
be.

You may also recall another strange phrase I used in my note to
{\Mermin}:  ``the world (i.e., the detector clicks, whose meanings in
the end are also set by subjective considerations).''  This oddity is
reflected in my definition of knowledge above:  that is, I make a
distinction between the raw stuff of the world that the measurement
intervention brings about and the registration $b$ in our noggins (as
a flag for further actions in our role as agents).  What I am
thinking here is something roughly like the following.  Take the
famous white-on-black or black-on-white visual illusion that can be
viewed either as a vase or as two faces facing each other.  The raw
stuff of the world may be compared to the ink and the paper giving
the image.  In order to say, however, that Mabuchi and I gain the
same knowledge in viewing this we need the deeper cultural agreement
that we will {\it both\/} call it a face or instead a vase when we
see it.

Below I will put a glossary that tries to summarize where I have come
so far in my attempts to make sense of quantum mechanics.
Essentially, I'm expecting only the two terms above to be relevant to
the fights we'll be  having in writing our RMP article.  But maybe it
is nevertheless useful for me to lay my full set of language oddities
on the table.

Lately, I've been jokingly calling my view (as it stands) the
``sexual interpretation of quantum mechanics.''  (Most people turn
red and become uncomfortable when I do that and explain why.  I
suspect the same will be true even in your reading of this note. So,
brace yourselves.)  The essential idea is that something new really
does come into the world when two of its pieces are united. We
capture the idea that something new really arises by saying that
physical law cannot go there---that the individual outcome of a
quantum measurement is random and lawless.  The very fact that the
consequence of the union is random signifies that there is more to
the sum than is contained in the parts.  But I promise you I won't
reflect the licentious details of this view in the glossary below.
I'll leave the missing terms to your imagination.

\begin{itemize}

\item
ACT  --  The actual carrying out of a quantum
measurement/INTERVENTION, after a DECISION has been made by an AGENT
to do so.

\item
AGENT  --  Any participant in the construction of a scientific
theory.  In older language, the observer.

\item
BELIEF  --  In the context of quantum discussions, a quantum state.
Or one might say the quantum state is a compendium of BELIEFS.

\item
CONDITIONALIZING BELIEF  --  The rule one uses to update one's BELIEF
consequent to the completion of a measurement INTERVENTION. In the
language of Kraus and Preskill this would be called the ``quantum
operation'' or ``superoperator'', respectively.

\item
CONSEQUENCE  --  Whatever it is that a measurement INTERVENTION
elicits out of the world.

\item
DECISION  --  An AGENT, within his power can decide to perform one
ACT or another upon the world.  Just as physical law cannot impinge
on what determines the random outcome of a quantum measurement,
neither can it impinge on the mechanism behind an AGENT's decision.

\item
FACT  --  This is a word I do not like.  One might have said that the
outcomes of quantum measurement could be called facts just as well as
CONSEQUENCES:  But the word fact, to me, contains the connotation of
a kind of permanence that I do not see in the quantum world.  Facts
are irreversible additions to the furniture of the world.  But
measurement INTERVENTIONS (and their CONSEQUENCES) can be reversed
through the agency of a further outside intervener.

\item
INTERVENTION  --  The physical act that we call in older language the
measurement of a POVM.

\item
KNOWLEDGE  --  One's mental representation of the obtained
CONSEQUENCE of a given INTERVENTION into the world.  Implicit in the
use of this word, is that all communicating parties agree to the
meaning of the given INTERVENTION, i.e., that it is this POVM rather
than another.

\item
PROPERTY  --  A property is something possessed by a FACT.  I don't
like the word FACT.
\end{itemize}

\section{04-09-01 \ \ {\it Brilliance} \ \ (to N. D. {\Mermin})} \label{Mermin37}

\bdm
\bq
\noindent\rm
CAF Said:  I am becoming more and more dissatisfied with the slogan
``A quantum state is a state of knowledge, not a state of nature.''
\ldots\ I think we should get more into the habit of calling a
quantum state a state of BELIEF.
\eq
Brilliant! All kinds of trouble would have been avoided.
\edm

You know, I'm not one to turn down a ``Brilliant!''  But your second
sentence does clash a little with what you wrote on August 8:
\bdm
It seems to me Chris is getting much too subtle about this.  I would
talk about knowledge, not belief.
\edm
All kinds of trouble WOULD HAVE BEEN avoided?

Speaking of brilliant---real brilliance this time---today I'm going
into Munich to talk to Hans Briegel about his papers with
Raussendorf.  (I told you I would be in Munich, but I'm actually in
the little village of Zorneding outside of Munich.)  I think there's
something very deep in them, if they hold up.  You may recall I
recommended them to you once.

I've got more things of a philosophic nature to write you, but I've
just got to find some time to do it.  I'll try to be back to the
waves tomorrow.

\section{04-09-01 \ \ {\it Objective Probability} \ \ (to C. M. {\Caves} \& R. {\Schack})} \label{Schack6} \label{Caves11}

\bcc
I expect you to have a really hard time with this---please skip the
lectures on my not being sufficiently Bayesian---but it is, in my
present view, a necessary feature that expresses the tension that
exists in the notion that maximal information is not complete.  The
state assignment can't be verified by examining the system, but it
can be verified by examining the trail of evidence from which I
acquired maximal information.  If someone else finding that trail of
evidence could say that he didn't have maximal information or that he
had different maximal information, the notion that the information is
maximal would be untenable, since apparently something further would
be required to make it so.  This seems like a natural for someone who
takes seriously those quotes about the process of intersubjective
coming to agreement.  It grants to maximal information in quantum
mechanics some, but not all of the properties of maximal information
in a realistic world.
\ecc

It is hard for me to understand what that ``trail of evidence'' is a
stand-in for if it is not a compendium of OBJECTIVE probabilities.
You follow that trail, and you have NO CHOICE but to assign all the
probabilities that the Gleason theorem gives (presumably if you are
rational).  So, pure quantum states give rise to ``propensities''
\ldots\ when those pure states are ``right''?  Is that what you are
saying? (I said all of this, of course, in my original longer note,
but it seems good to isolate it here.)  Can you give me an
operational definition of this notion of ``propensity''?

And why can we toggle these propensities from a distance?  Are you
giving up on spacetime after all?  Or is this a new way of applying
the principal principle?

Now I really do have to join the family.

\section{04-09-01 \ \ {\it Fourth and Fifth Reading} \ \ (to C. M. {\Caves} \& R. {\Schack})} \label{Caves12} \label{Schack7}

\bcc
The point of our conclusion is that the Dutch-book argument leads to
a unique probability or density-operator assignment in the case and
only in the case of maximal information.  This is just an entirely
different thing from using frequency data---or something else---to
specify every component of a density operator.
\ecc

I still don't get it (though I've had a lot of wine by now).  By hook
or crook, I use the information available to me to assign a
probability distribution over the outcomes of some informationally
complete measurement.  That assignment gives rise to a unique density
operator.

I'm still having trouble seeing what is special about a ``maximal
information'' assignment.  I'm not lying; I'm not trying to cause
trouble; I'm just not seeing it.  (Think of me as the second referee
of the paper.  Would that be ethical?)

Good night!

\section{05-09-01 \ \ {\it Noncontextuality Again (and Again)} \ \ (to C. M. {\Caves})} \label{Caves13}

\bcc
The underlying structure is a specification of alternatives that can
be grouped in various ways---these are the contexts---to make
exhaustive sets.  We are required to make noncontextual assignments;
otherwise we are ignoring the fact that this specification doesn't
distinguish an alternative in two different contexts.  If it did, we
would be dealing with a different specification.  This is the
perspective of my first point of view, which justifies noncontextual
probability assignments in quantum mechanics from the Hilbert-space
specification of alternatives.
\ecc

I still don't entirely get this.  You say we are required to make
noncontextual assignments, otherwise we would be ignoring the fact
that the original groupings do not distinguish an alternative in two
different contexts.  But why could we not ignore it?  Perhaps the
underlying structure is there for an entirely different reason than
something to do with probabilities?  For some reason, this point of
view is just not clicking for me.

\bcc
Your perspective is different.  As I understood it, you think of an
alternative in different contexts as a single alternative because it
has the same probability in all contexts.  But where did you get this
equal probabilities?  Surely they're not measured or determined or
anything like that, since they are states of knowledge.  In quantum
mechanics you get that they're the same because the standard quantum
rule says so, but this is using noncontextuality, not justifying it.
\ecc

I don't know what more to say on this.  It means that identifying
this consequence of this intervention with that consequence of that
intervention is a SUBJECTIVE judgment.  (I.e., that identifying this
outcome of this measurement with that outcome of that measurement is
a subjective judgment.)

\bcc
\bq
\noindent\rm
CAF Said:  What you say ``needs'' to be done, seems hopelessly
underconstrained to me too.  But I think you shouldn't view the
problem that way. The point is, one simply has a way of clearing the
air of the noncontextuality issue BEFORE getting down to the
nitty-gritty of quantum mechanics.  Noncontextuality is the base
assumption upon which one plays a new game: What {\it physical\/}
assumption makes it so that our instruments should correspond to
POVMs and not some other mathematical structure?
\eq
I'm going to adopt your strategy, and just flatly say I don't get
this.
\ecc

Let me try again.  Here is the game we should be playing.  In the
most general terms, a measurement is defined to be a group of
elements (satisfying some given property) drawn from a set with a
given structure.  The individual elements correspond to the outcomes
of the measurement.  The question is, what should that structure be?
What should that property be?  What are the reasons for those
choices?  This much we will safely assume (for the reasons given
above):  The probabilities of the outcomes should depend only upon
the individual elements, not the group.

That's all I'm saying.  Here is an example of dumb theory.

\bv
UNDERLYING STRUCTURE $=$ one-d projectors onto a complex vector
space.
\\
GROUPING PROPERTY $=$ choose any set of projectors that project onto
a complete set of linearly independent vectors.
\ev

Then the only probability assignment that can be given to the
outcomes of such a notion of ``measurement'' is the uniform
distribution.

So, we start over and say, ``Maybe the grouping property ought to be
that the projectors add up to the identity.''  Aha, that gives us
quantum mechanics.  But you see, there are any number of other
combinations of structures and properties one might have played with.
The question is, what is essential about the structure and grouping
properties that we do use?  By saying that we have cleared the air of
noncontextuality, I simply mean that the existence of
noncontextuality in the probability assignments should not be a
question.  It was settled before we ever started the game.

\bcc
Well, when I say ``Hilbert-space structure of questions,'' I clearly
don't mean only that there is a Hilbert space, but that the questions
correspond to one-d projectors.  That's why I add ``of questions'' to
the phrase.
\ecc

No, what you mean precisely is:  A ``question'' corresponds to a set
of one-d projectors that sum up to the identity.  So why don't you
just say it in a precise way rather than a vague way?  If I were
uninitiated to quantum mechanics, I might have thought that you meant
the dumb theory above.  I'm serious about this.

\section{05-09-01 \ \ {\it Unique Assignment} \ \ (to C. M. {\Caves} \& R. {\Schack})} \label{Caves14} \label{Schack8}

\bcc
The point of our conclusion is that the Dutch-book argument leads to
a unique probability or density-operator assignment in the case and
only in the case of maximal information.  This is just an entirely
different thing from using frequency data---or something else---to
specify every component of a density operator.
\ecc

I guess my trouble stems from one of the things I said in the long
note announcing my worries about ``Making Good Sense.''  There I
said:
\bq
I think the best we can say is:  If Alice and Bob both believe
adamantly that they know the answer to some potential measurement AND
that measurement happens to be the same for both of them, then
Dutch-book consistency and Gleason's theorem will enforce that they
make the same probability assignments for all other measurements
(i.e., that they assign the same quantum state).  But said that way,
I don't think any non-Bayesian will be particularly impressed:  For
they would say that all we have shown is ``if two people know the
same things, then they will know the same things.''  Woop-ti-do.

The Bayesians among us will still have some room to be impressed: For
it will not be a priori obvious to them that beliefs about one
observable should have anything to do with beliefs about another.  In
particular, it might even surprise them that a common belief in
certainty (for two observers) for any fixed observable should lead to
equal probability assignments for all other observables.
\eq

What I am wondering is:  What would impress a devout Bayesian (who is
just learning quantum mechanics) about our argument?  Thus, given
what I said above, I wonder whether he would not be equally impressed
by the following.  By hook or crook, Alice and Bob individually come
to their own subjective probability assignments for the various
outcomes of a single informationally complete POVM. Then, because of
Gleason, they will have to match in their subjective beliefs about
the outcomes of all other measurements they might perform.  That
matching made no use of the concept of maximal information.  What
does the maximal information case give us in shock value?

I think what you're going to say is that in the case of nonmaximal
information, Alice and Bob may have come to different probability
assignments for that informationally complete observable.  And that
they couldn't have done that if they had had ``maximal information''
in the first place.  But as all this debate has already shown, I
think I reject that position.

As my glossary from yesterday attests, I think what is going on with
me is that I am becoming ever more uncomfortable with identifying
quantum states with information, maximal or otherwise.  Thus, instead
of calling a pure state maximal information, I am becoming more
inclined to something like:
$$
\mbox{pure state} \; = \; \mbox{maximally tight belief (or judgment)}
$$
or
$$
\mbox{pure state} \; = \; \mbox{a nonrefinable belief (or judgment)}
$$
or anything else more along those lines.

Leave the word information for what we gain when we see the outcomes
of measurements.  This entity we know two disparate observers should
agree upon---by definition---if they are in free communication with
each other.  But the quantum state, on the inside of one's head, is a
more personal state of affairs.

I'm going into Munich for much of the day (to visit with Briegel), so
you may not hear from me again until tomorrow.

\section{05-09-01 \ \ {\it Identity Crisis, 2} \ \ (to C. M. {\Caves} \& R. {\Schack})} \label{Caves15} \label{Schack9}

I've now given your long note the fourth and fifth readings it
deserves.  (I'm holed away in an office near Briegel's.)  And I'm not
sure how to respond yet.  I think I will await your responses to my
other notes first.  I think it is clear in the time that has elapsed
since our first communication that we have moved further away from
each other's position.  Or I should say I've moved further away, from
our original position.

Briegel just came; I'll be back.

\section{05-09-01 \ \ {\it Malleable World} \ \ (to G. L. Comer)} \label{Comer5}

Just a little quote I liked:
\bq
\noindent
``Once you bring life into the world, you must protect it. We must
protect it by changing the world.''
\eq
From:  Elie Wiesel (b.\ 1928), Rumanian-born U.S. writer. Interview
in {\sl Writers at Work\/} (Eighth Series, ed.\ by George Plimpton, 1988).

\section{06-09-01 \ \ {\it The Underappreciated Point} \ \ (to C. M. {\Caves} \& R. {\Schack})} \label{Caves16} \label{Schack10}

My fit of insomnia is running out, so I'm going to have to go back to
bed soon.

\brs
Please give me some feedback on these thoughts. I know that you wrote
A LOT more, but I find it easier to go through your emails paragraph
by paragraph.
\ers

You know, of that LOT, most of it was written very carefully and very
purposefully, so I do hope you will try to read it all with that in
mind before trying to get me to readdress too much.  (You can ignore
the stuff on noncontextuality for now; that's not important for
present issues.)  Next week I won't have the leisure of writing too
many notes, as I'll be out on the election trail trying to stump this
Bayesian point of view \ldots\ and trying to make it LOOK consistent.
(I.e., I'll be at the quantum foundations meeting in Ireland.)

But, before crawling back into bed, let me address your greatest
FEAR:
\brs
The most important thing to remember is the limited scope of the
paper.  It tries to show that Bayesian probabilities do have a place
in qm, not more. Remember that most physicists would reject this. We
show that, contrary to conventional wisdom, subjective quantum
probabilities are not arbitrary. Let me remind you that you agreed to
a paper of this limited scope last year in {\Montreal}. I do not want a
paper that is significantly expanded.
\ers
This must be addressing my single remark of September 3,
\bq
\noindent
Of course, as you know, I view this as a good opportunity (with page
limitations no longer of great concern) to expand some points and
give some more references.  I think we can read the referee as
agreeing with that.
\eq
because, in all my voluminous letters, I don't think I ever mentioned
modifying the paper otherwise.  Of course I think it is unhealthy
that your first trip-up would be that point \ldots\ but we are all
moved by different things.  (I don't fathom your pressures, and you
don't fathom mine.)  In any case, let me say this for the record to
try to clear the air:
\bq
\noindent
If given free reign---which I do not actually want---I think I could
modify the paper to my own tolerance without changing its length at
all, or, at most by a paragraph.  I would be more than happy if you
and {\Carl} would just find a language I could agree to, and modify the
paper accordingly.  The phrase {\it more references} was a euphemism
for citing more of Chris's papers.
\eq

Deep in my heart, I believe you guys fool yourselves in thinking that
this paper will be more widely read simply by being short, but that
is not the issue (and it has never been the issue).  My passion is to
get quantum mechanics straight:  So, let's get it straight.

On the whole, in reading your two notes, I found your method of
expression better fit to my present mentality than {\Carl}'s.  Maybe
I'll give more specific examples tomorrow.  For the present, let me
just mention two things.

\brs
It does indeed not follow from our Dutch book argument alone that two
agents must agree on the maximal info they have. But suppose agent A
has maximal information and agent B insists on assigning a pure state
that is not consistent with A's information. A can then extract money
from B. I don't think the symmetry of the situation is a problem
here. From A's perspective, B is wrong, in the same sense of wrong as
if A had a piece of classical knowledge that B chooses to ignore in a
bet.
\ers

Perhaps the greatest life change I have had is that I no longer like
the phrase ``maximal information'' in this context.  That little
phrase carries with it an entire philosophy, and it is one that, to
me, does not seem consistent with its roots and, more importantly,
does not seem right.

The most one can say on Bayesian principles is that:
\bq
\noindent
From A's perspective, B is wrong.  And from B's perspective, A is
wrong.
\eq

If A and B can have two pure-state assignments, and the most one can
say is the item above, then pure states should not be called
``maximal information.''  They are maximal ``something else'', but it
is not information.  (In another note, I have outlined what I think
that ``something else'' is.)

{\Carl} thinks he can fix this by invoking a ``trail of evidence'' that uniquely fixes which of two pure states is actually the case.  But
let me juxtapose two of his paragraphs and then try to reemphasize
the underappreciated point.

\bq
\noindent Paragraph 1:
\\
In quantum theory maximal information also constitutes a belief, but
we resist the notion that it corresponds to some objective reality
out there.  Why this resistance?  Ultimately it's because the maximal
information leads to a pure-state assignment that gives probabilities
whose only reasonable interpretation is subjective.  It is very
important to remember that this is the primary motivation for much of
what we do.  Probabilities are the subjective language used to deal
with situations of uncertainty, so wherever we find them, they must
be subjective.  The subjective view of pure quantum states gains
additional support from the fact that a pure-state assignment can't
be verified by consulting the system---the same can be said for a
probability assignment---and the fact that a state assignment for a
distant system changes when we obtain information about it without
ever getting close to it---this also holds for correlated probability
assignments.
\eq
and
\bq
\noindent
Paragraph 2:
\\
The state assignment can't be verified by examining the system, but
it can be verified by examining the trail of evidence from which I
acquired maximal information.  If someone else finding that trail of
evidence could say that he didn't have maximal information or that he
had different maximal information, the notion that the information is
maximal would be untenable, since apparently something further would
be required to make it so.  This seems like a natural for someone who
takes seriously those quotes about the process of intersubjective
coming to agreement.  It grants to maximal information in quantum
mechanics some, but not all of the properties of maximal information
in a realistic world.
\eq

If you hold fast to the view that that trail of evidence must {\it exist},
then you hold fast to the view that quantum probabilities (in some
cases) must be objective after all \ldots\ {\it independently\/} of the issue of intersubjective agreement.  And that negates Paragraph 1.

I am now of the opinion that if we can just clear the air [I'm fond
of that phrase] of this nonBayesian trapping from bygone times, we
will finally be in a position for real progress.  It is in
Dutch-bookian type coherence (as a general principle) that one finds
an objective statement in quantum mechanics; it is never in the
quantum state itself, even when that state is pure.  The OBJECTIVE
statement is:  All of you, each and every one of you, should
manipulate your compendia of beliefs according to the rules of
quantum mechanics if you wish to maximally avoid  undesirable
consequences in your gambles.  The particular quantum states at any
one time are just thin films of subjectivity floating on that wider
sea of objectivity.

But, please, please do read the other notes carefully.  I can only
write a finite amount.  I'll comment more particularly on your
present notes tomorrow (i.e., today, after I get back up).

\section{06-09-01 \ \ {\it Another Way} \ \ (to C. M. {\Caves} \& R. {\Schack})} \label{Caves17} \label{Schack11}

\bq
\noindent
If you hold fast to the view that that trail of evidence must {\it exist}, then you hold fast to the view that quantum probabilities (in some cases) must be objective after all \ldots\ {\it independently\/} of the issue of intersubjective agreement.  And that negates Paragraph 1.
\eq

Let me put it another way.  By {\Carl}'s view, if trails of evidence
MUST exist, then quantum states MUST exist, and the ghost of Bruno de
Finetti should have stayed in the netherworld.  For the probabilities
derived from the quantum state will exist after all.

\section{06-09-01 \ \ {\it Some Comments} \ \ (to C. M. {\Caves} \& R. {\Schack})} \label{Caves18} \label{Schack12}

Now I return from an unrestful morning in bed.

\brs
I guess you are right that we should be more explicit about ``whose
certainty''. It is the bettor's certainty.
\ers

The deeper issue is not that we {\it should\/} be more explicit about
``whose certainty,'' but {\it why\/} we should.

\brs
You should leave Schopenhauer and your mother out of this discussion.
The distinction between the cases of certainty (classical logic) and
reasoning in the face of uncertainty (probability theory) is useful.
\ers

I didn't understand this comment.

\brs
As I said in my previous message, two agents having conflicting
certainties is a completely classical situation. If you accept
classically that in this case, one of them must misread or ignore
some of the available information, then the point of the paper is
that the same classical argument gives you unique state assignment,
even though states are Bayesian. This is a forceful conclusion.
\ers

It is safer to have the wrong metaphysics in the classical case. This
is because certainty (i.e., overpowering belief in the outcome) for
one question means certainty for all questions.  And that certainty
can be verified or falsified in a single shot.  So, one gets in the
habit of thinking that the proposition (or its material counterpart,
as instantiated in the world) was already there before the
verification.  One can accept that metaphysics or leave it, but it is
usually more convenient to accept it.  In the quantum case, however,
if you assert that the proposition was already there (say, as
uniquely specified by {\Carl}'s ``trail of evidence'') then you have to
assert that all the rest of the quantum probabilities were already
there too.  That sounds an awful lot like objective, agent
independent probabilities to me.

You can retreat to objective probabilities if you wish.  But I say it
is better to be creative with our metaphysics.  JAW said it like
this, ``No question, no answer.''  And that distinction is rearing
its head in this very problem.

\brs
Making a pure-state assignment is an extreme statement. It entails
the conviction that assigning a different state is equivalent to
handing over money. It entails the conviction that the agent
assigning the different state is wrong in this sense, in the sense of
irrational behavior, not in the sense of not conforming to reality.
\ers

It ``entails the conviction.''  That is language I can accept.  It is
language I like.  Trying to instate that way of saying things has
been the whole point of my writing such detailed notes, especially
the point about ``not in the sense of not conforming to reality.''
But though you use it so nonchalantly now, it had no representation
in our previous discussion, and it has no representation in our
paper. At least looking at myself personally, I feel as if I have
come through a phase transition.

\brs
\bq
\noindent\rm
CAF Said: But said that way, I don't think any non-Bayesian will be
particularly impressed:  For they would say that all we have shown is
``if two people know the same things, then they will know the same
things.''  Woop-ti-do.
\eq
Still quite impressive. A and B know the same certain thing. Hence
they must assign the same {\it subjective\/} probabilities to all
questions. Even subjective probabilities $0<p<1$ are prescribed by this knowledge. The non-bayesian should be quite surprised and impressed
by this.
\ers

I said non-Bayesian.  Non-Bayesians do not accept subjective
probabilities.

\brs
\bq
\noindent\rm
CAF Said: For I think in no way have we shown that when two observers
make two pure-state assignments for a system, those pure states MUST
be identical.
\eq
If they are not identical, each agent has perfect reason to assume
that the other one is unreasonable.
\ers

I accept that.  But the point has been, and remains, that that is the
ONLY conclusion we can draw.

\brs
There are dictatorial constraints only in the limiting case of
maximal information.
\ers

Unless all of this email has been a grave mistake on my part, I
continue to not be able to accept this.  The only argument we have at
our beck and call is that Dutch Book $+$ Gleason dictates what {\it
I\/} must do in my head and what you must do in your head.  It tells
us each how to translate a {\it nonnumeric\/} belief (certainty)
about the outcomes of a single question, to a {\it numeric\/} belief
about the outcomes of all possible questions. Indeed, I will lighten
up: For a raw BAYESIAN that must be quite an impressive conclusion.
There must a good way to say that in the paper.  [For the
non-Bayesian however---one with no qualms about objective
probabilities, one with no qualms about the objectivity of quantum
states---I remain in my belief:  it will strike him as little more
than a tautology.]

But all of this does not lessen my debate with Brun, Finkelstein, and
{\Mermin} which started this whole affair.  There is no a priori
principle in the universe that will tell us that two quantum states
OUGHT to have overlapping supports.  The best one can say is that IF
Alice and Bob have overlapping support, then (if they wish) they may
be able to communicate the reasons for their beliefs and come to a
more refined consensus.  If they do not have overlap in their
supports, then the only they can do to lessen their strife is consult
the world.

The objectivity is not in the states, but in the state-space
STRUCTURE and in the answers the world gives us upon our
consultations.  When one has gained the latter, one has gained
information.  But the quantum state before and after remains belief,
pure state or not.

\section{06-09-01  \ \ {\it Weak Point} \ \ (to C. M. {\Caves} \& R. {\Schack})} \label{Caves19} \label{Schack13}

\bcc
I think we all agree that if states are Bayesian , then anyone can
assign any state he pleases, including any pure state.  He can be
misled or tricked, or he can just be crazy, but this sort of freedom
to assign any state is not of much interest for our paper.  An
objectivist will have no trouble agreeing that someone who is misled
or irrational will use the ``wrong'' quantum state.
\ecc

Let me try to say it again.  The main point is, in the quantum
mechanical world, these ``trails of evidence'' you are thinking of in
the back of your mind are NEVER enough to uniquely specify a quantum
state.  It has NOTHING to do with being misled or being irrational.
Even the purest of states is thoroughly infused with belief from the
get-go.  That is what my note titled ``Fw:\ Intersubjective
Agreement'' from September 3 is essentially about.  So, this is not a
case of measure zero, where the players are irrational or dumb to
begin with \ldots\ so long as we take our own arguments about
subjectivity seriously.

\bcc
What he wants to know is whether scientists acting like scientists,
sharing all information in a spirit of genuine co-operation, mutual
respect, and dedication to truth, can assign different pure states.
And we show that scientists acting like scientists can't: sharing
maximal information, they must make a unique quantum state
assignment.
\ecc

If that is what he wants to know, then he is not going to find it
from anyone's Dutch-book argument:  our last one, or our slightly
modified new one.  Your point is a weak one.  The Dutch-book game is
an adversarial game.  Anyone whose intention is to make his opponent
go bankrupt is NOT going to share everything he knows with him.  He
will be silent and bet his money.

\bcc
What rescues this conclusion from trivia?  First, it answers the
question of why science doesn't go down the drain: subjective state
assignments are constrained in the case of common maximal
information.
\ecc

I don't believe the conclusion is trivial; I said this to {\Ruediger}
yesterday.  But I also don't believe it has anything to do with
rescuing science.  Playing by the quantum rules ought to be enough.

\bcc
Second, it answers the question without referring to real, verifiable
properties of the system in question.  In a realistic world one might
justify the agreement in the case of maximal information by saying
that any disagreement can be resolved simply by looking at the system
and seeing who's wrong.
\ecc

Disagreements in the quantum world are resolved also simply by
looking at the system.  Suppose you and I agree to everything in the
world {\it except\/} the quantum state for a given system.  How do we resolve
our dispute?  We perform a maximally refined quantum measurement (a
POVM with rank-one elements).  We agree on the system's state
thereafter.  That is all that has ever been important in science
anyway---that the world provides us with a way to {\it come into\/} agreement
for all {\it future\/} predictions.  For god's sake, Albert Michelson did not
believe that the speed of light could be a constant.  But his tenure
was not stripped away when he found a negative result.  He revised
his ``impossible'' belief and got over it.

{\it All\/} that one need to demand from a theory is that it provide a way
for two agents to come to agreement for all FUTURE predictions.
Quantum mechanics (surely) satisfies that.  It has nothing to do with
re-objectifying quantum probabilities, and I can't see that it has
anything to do with this stuff we got in the habit of calling
``maximal information.''

\section{06-09-01 \ \ {\it The Well Appreciated Point} \ (to C. M. {\Caves} \& R. {\Schack})} \label{Caves20} \label{Schack14}

Your notes are well-appreciated themselves; I am finding reading them
productive.  Unfortunately, I cannot reply in detail tonight, but I
hope more will be waiting for me tomorrow morning.

Let me do say though that I think I addressed some of your points in
the note I just sent to {\Carl} (and CC'd to you).  The main thing was
this:
\brs
What the Bayesian can say is: If A assigns a pure state, he knows
with certainty that any other pure-state assignment is foolish
(handing over lots of money).

It is not a situation that can be resolved within science, by
discussion or experiment or comparison of notes. Both A and B are
certain there is nothing that could change their belief. For A, B
could just as well reject all of quantum mechanics.
\ers

It is not a sin for A and B to disagree about the present.  What
would be a sin is if they could not come to agreement in the future.
And quantum mechanics provides just such a mechanism.  It is not true
that experiment cannot change (absolutely firm) beliefs in the
quantum world:  quantum measurements are invasive, and thankfully so.
Each measurement gives us the opportunity to throw away the past and
start afresh.

\section{07-09-01 \ \ {\it Email Not Received} \ \ (to C. M. {\Caves} \& R. {\Schack})} \label{Caves21} \label{Schack15}

\brs
But, as I argued in my last email, I think that a modified betting
argument, now having A and B as adversaries ({\Carl} thinks that this
modified argument should not be called a Dutch book argument), shows
that starting from two different pure states to come to a later
agreement is not what science is about. A must dismiss B as a
crackpot. This argument would be useful in the Peierls debate. But
maybe not for our paper.
\ers

I didn't receive such an email; can either of you send it again?

Indeed I would bet that science cannot be made in a (purely)
adversarial environment.  Science is about cooperation, trying our
best to come to a consensus.  (That is why I have not lost heart in
writing all these ridiculous emails!)  But, nevertheless, from time
to time I do talk and try to come to consensus with people I deem
crackpots.  The point is, though someone may be adamantly wrong about
ONE thing (say, a pure-state assignment from my perspective), it does
not mean that he is adamantly wrong about ALL things.  And therein
lies a backdoor for a discussion with such a person.

The only thing that one has to trust in the making of science is that
one's colleague is internally consistent.  It is OK if he got SOME of
the {\it facts\/} ({\it The Well Appreciated
Point}) wrong (from my perspective), and that I got SOME of the
``facts'' wrong (from his perspective).  It is enough that he is
willing to join in with me in letting the world pull us together.
I.e., that each of us is willing to participate in trying to convince
the other that he is wrong by consulting the ultimate arbiter.

But, I'll write more later (in the context of your last two notes).

\section{07-09-01 \ \ \ {\it (Backbreaking) Analysis} \ \ (to C. M. {\Caves} \& R. {\Schack})} \label{Caves22} \label{Schack16}

\brs
We need to find some common ground.
\ers

Yes, that is true.  And I think we already have some, maybe even a
lot.

But, as I see it, there still remains a significant amount of trouble
in the language we chose to use in the past \ldots\ and that is what
is putting stoppers on our progress in the present.

Of course, I feel like I am repeating myself over and over, but let
me go to your explicit ``common ground'' paragraph and try to lay out
what I like and what I don't like about it.  The thing that keeps me
going is the hope that maybe this whole debate is a lot like beer: On
their very first taste of it, most kids think it is a foul stuff.
But after more and more of their friends offer it to them over time,
it starts to become a pleasant diversion.

\brs
Assume A has information of the kind we call maximal, i.e., A knows
that a measurement of a POVM containing the 1D projector P will give
the outcome corresponding to P. Then assigning any state but P will
be Dutch-book inconsistent. This will be A's inconsistency with her
own belief. The beliefs of the bookie or of Nature do not matter. A
knows that assigning any other state would make her accept a bet in
which she (not Nature or the bookie) knows that she will lose for any
outcome she believes is possible.

Now assume B has access to the same piece of maximal information. B
then knows that a measurement of a POVM containing the 1D projector P
will give the outcome corresponding to P. Then assigning any state
but P will be Dutch-book inconsistent. This will be B's inconsistency
with his own belief.

Hence: Two agents having access to the same maximal information MUST
assign the same state.
\ers

Here is how I would reword it to suit my present tastes.
\bq
Assume A is absolutely sure that a measurement of a POVM containing
the 1D projector P will give the outcome corresponding to P.  Then
assigning any state but P will be Dutch-book inconsistent.  This will
be A's inconsistency with her own belief.  The beliefs of the bookie
or of Nature do not matter.  A knows that assigning any other state
would make her accept a bet in which she (not Nature or the bookie)
is absolutely sure she will lose for any outcome.

Now assume that B is absolutely sure of the same thing, i.e., that a
measurement of a POVM containing the 1D projector P will give the
outcome corresponding to P.  Then assigning any state but P will be
Dutch-book inconsistent.  This will be B's inconsistency with his own
belief.

Hence:
\bq
\noindent Two agents having the same absolutistic belief about the
outcome of a measurement containing the projector P {\it MUST\/}
assign the same quantum state.
\eq
Or equivalently (but, to me, more forcefully):
\bq
\noindent
Two agents having the same absolutistic belief about the outcome of a
measurement contain the projector P {\it MUST\/} assign the same
(subjective) probabilities to the outcomes of all measurements that
can be contemplated.
\eq
\eq

What were the main substitutions?  Essentially, they were simply:
\begin{center}
``maximal information'' $\longrightarrow$ ``absolutely sure''\\
and\\
``maximal information'' $\longrightarrow$ ``absolutistic belief''
\end{center}

And the same substitutions count for the word ``know.''

To me, those simple substitutions completely change the metaphysical
complexion of the statement.  The statement goes no further than it
has to go to make the quantum Dutch-book theorem stand its ground.
Why go further?

{\Carl} gave his reasons:  To save science.  But I do not see that is
necessary in any way, and I do not personally believe that that
method is on the right track.

What is wrong with the word ``know''?  To my Western-trained mind, it
conveys the idea that there is something in the external world (the
world outside of my head and beyond my control) and that my mind
contains a mirror image of it.  It conveys the idea that the outcome
to the contemplated measurement already exists ``out there'' in some
deterministic or fatalistic sense.  It conveys the idea that I really
need never have a look to see if the outcome is produced by my
measurement:  It's already there, and I know it; why waste time on a
measurement?  Notice that I let the word ``knows'' stand when it came
to describing the very logic of the Dutch-book argument.

What is right about ``absolutely sure'' and ``absolutistic belief''
for me?  They convey the feeling that what I have in hand is a
belief, an extreme belief to be sure, but nonetheless a belief. That
phrase never reaches out to the external world for its
justification---or, at least it seems so to me.

What is wrong with ``maximal information''?  I think it screams out
no more clearly than in your concluding statement above.  In words
that Gary Herling might use:  The very phrase ``SAME maximal
information'' is an abomination of the English language.  In my mind,
information is much like the word ``know'' (though a little looser in
constitution).  It too conveys the idea of a mind or a newspaper
mirroring aspects of a preexisting reality.  Besides that, the very
fact that we have to go to the trouble to use the word ``same'' in
conjunction with ``maximal'' conveys the feeling that the word
``maximal information'' was never appropriate in the first place.  If
information is some stuff we have gathered from the world AND it is
maximal---the very most one can get---AND two agents really should be
gathering up the same stuff, else one of them is wrong, THEN why do
we have to go to the trouble of using the word ``same''?

Well, we use it (i.e., ``same'') to keep ourselves from contradicting
the belief that probabilities are subjective after all.  Fine, so
that is a good reason to keep the word ``same''.  BUT, it is a bad
reason to keep the phrase ``maximal information'' to merely convey
the concept that one is ``absolutely sure'' of the consequence of
some action that one might take (i.e., the measurement being
contemplated here).

\brs
\bq\noindent\rm
CAF Said: Suppose you and I agree to everything in the world {\it except\/}
the quantum state for a given system.
\eq
It follows from what I just wrote that this situation cannot arise in
the case of maximal information.
\ers

To me, that statement is a {\it non sequitur}.  I cannot find any
{\it logic\/} to bring it about.  And I say that especially if you
can agree to the validity of my attempt at expressing a ``common
ground'' for us in the highlighted paragraphs above \ldots\ no matter
how pedantic you think my actual phrasing is.

If the quantum state is not uniquely declared by some reality, then
there is nothing to stop us from agreeing on some aspects of the
world and disagreeing on others.

Please read the note I wrote {\Mermin} titled ``Intersubjective
Agreement'' again.  If you and I (in the presence of each other)
perform a given POVM consisting of rank-one projectors on a system,
then you might say that we will agree on the system's state
thereafter.  But that requires the assumption that we BELIEVE the
same quantum operation (for updating our states) will be associated
with that measurement.  If we don't agree on that at the outset, then
we will come to conclusion of two different pure states for the
system after the measurement is completed.

You say, well a quantum operation is surely an element of reality: It
is either right or it is wrong.  ({\Carl} would say that in any case; I
wouldn't.)  But suppose it is so---I will relax my debate with {\Carl}
for the moment.  How would we know which quantum operation we had?
We would have to have prepared a load of quantum states beforehand to
map which quantum operation is ``really'' there.  But then we would
have had to agree on our cache of exploratory quantum states in the
first place.  How did we get to that stage of agreement, I ask you?
And, on I will do the same, ad infinitum.

The point is, in a world where our only exploratory tools are quantum
states and quantum measurement outcomes, we can never terminate the
chain.  This is one aspect of what I meant yesterday when I said that
quantum states are infused with beliefs from the get-go.

Quantum measurement outcomes alone will never, ever be enough to
uniquely determine a quantum state.  One has to have some further a
priori information or beliefs to do that.  You can play the game---as
{\Carl} wants to---that that a priori information is the world's
Hamiltonian.  But then you will be about as stuck as Kant was with
his transcendental idealism:  you will still have to start off with
agents of some initial common belief before they will ever be able to
come to agreement about the Hamiltonian's form.  And how are we poor
finite beings to ever get to such a starting stage?

I say simply:  throw out any trappings that a quantum state can ever
be objective.

\brs
Is there ever maximal information? Yes. You give us an example where
C and F both obtain the same maximal information about the system:
\bq
\noindent\rm
CAF Said: We perform a maximally refined quantum measurement (a POVM
with rank-one elements).  We agree on the system's state thereafter.
\eq
\ers

In the sense that you want maximal information, i.e., something
beyond absolutistic belief, I would therefore say that there is never
any.  My example required that the two agents share an almost
strangulating amount of common belief.  See discussion in previous
section.

\brs
As {\Carl} writes, if C and F have the same maximal information, they
must assign the same state. This is an important situation, as
scientists share the information they have.
\ers

If we can get past the language, I will (clearly) agree that this is
an important situation.  Scientists share the data AND the beliefs
(interpretations, machine designs, etc.) they have.

\brs
\bq
\noindent\rm
CAF Said: The Dutch-book game is an adversarial game.  Anyone whose
intention is to make his opponent go bankrupt is NOT going to share
everything he knows with him.  He will be silent and bet his money.
\eq
No. The Dutch book game is about ONE agent's consistency.
\ers

Yes, you are absolutely right.  And I apologize for throwing in extra
junk that is not relevant.  But the only point I really meant was
that there is nothing in the Dutch-book set-up that forces the bettor
and the bookie to share their information.  That is an extra
requirement if you want it.  But it is a requirement that seems to me
almost to give up the whole spirit of the Dutch-book situation: it
involves no communication beyond the numbers $p$ and $x$.

\brs
That agents having access to the same maximal information must assign
the same state is all we need for our paper.
\ers

It seems to me, the only thing we need for our paper is the ``common
ground'' statement I made above.  I would not think that I need to
say it again, but I'm getting pretty fearful of the phrase ``maximal
information.''  At the very least, I would like to start using it in
a more limited sense or in a more limited way.  Or perhaps in a vague
enough way (for the present project) that I can worm out of it when I
want to write future papers of my own without you two.  (But this
issue is likely to haunt us all the way through to the end of the RMP
deal.)

All this email is starting to exhaust me, and it has certainly kept
me from making this trip to Munich even resemble a vacation.  I would
like to draw it to a reasonable end soon.  (But I do understand that
that will require flexibility on all our parts, even me.)  This may
help my samizdat production, but it no longer feels like it is
helping the rest of my life.  I feel like I have hold of some
important points that we were just too much in the ``classical''
tradition to recognize before.  If we ultimately disagree, then we'll
just have to do that, but I would rather not end up in that state of
affairs.

If you have to make choices on what to do with your own time, please
comment on my note titled ``Note on Terminology'' sooner rather than
later.  I fly out for Ireland Sunday morning.

\section{08-09-01 \ \ {\it Negotiation and Compromise} \ \ (to C. M. {\Caves} \& R. {\Schack})} \label{Caves23} \label{Schack17}

Let me tell you a little story I dreamed up while driving through the
Austrian countryside today.  It is based on one of the most annoying
realities of my life:  there are times when Kiki and I just cannot
come to agreement.  If I can use {\Ruediger}'s words, there are times
when I just think:
\brs
A is a physicist who would bet his career on his state assignment. If
he says B is wrong he means this in an absolute, very strong sense.
He has examined all the evidence, and there is no doubt left. He is
certain that B is missing some evidence. The Dutch book argument
shows that A is certain that B's position is equivalent to handing
over B's entire fortune. Wrong implies foolish, deluded. For A, B is
a crackpot, and the circumstance that B is certain that A is wrong
reinforces this position.
\ers
But there are realities:  Kiki and I are married; we share a bank
account.  And here and there, Kiki consorts with the Dutch.

What I am leading up to is that I think there is a place for Ben
Schumacher's observation about a three-person Dutch book in our
ongoing debate.

You two want to believe that there are god-given constraints on how
much two people can disagree.  I say there aren't.  It'll be a
miracle if we ever come to some consensus on this.  But I have never
said that there ought not to be reasons that two people might want to
come to agreement \ldots\ EVEN when they have differing but,
nevertheless, ``maximal information.''  (I use the phrase ``maximal
information'' despite my dislike for it in order to be sarcastic and
to underline a further difficulty with the term in a moment.)

I am internally consistent; there's no Dutch bookie who can take me
to the cleaners.  And despite my feelings for Kiki's complete
foolishness, I feel that she is internally consistent; there's no
Dutch bookie who can take her to the cleaners (as far as she is
concerned).  But we will be in deep trouble if that Dutch bookie
approaches us separately.  (Being married, we report all our beliefs
to each other.)

I can see two outs to this problem.  The first one---which is less
interesting---is that we make an effort to come to agreement by
consulting the world.  We make a measurement, and thereby, through
its invasiveness, force the quantum system into a state we can agree
upon.  (Assuming, as I keep harping on, we can agree on the quantum
operation associated with our measuring device.)  But what if we have
no access to the system of interest?  What are we to do then?

I think we would have no choice but to, each of us, back off in the
firmness of our beliefs.  That is, we should agree upon a density
operator that contains in its support both of our earlier
ascriptions.  We both give up some of our certainty is this process,
but the upshot is that we no longer have probability one of becoming
bankrupt.

So think about this:  Two agents start out saying that they are
absolute in their convictions about a some quantum measurement.  But
then the reality of their partner's stubbornness hits them, and the
only thing they can do is back off.

You continue to want to call a quantum state information.  But for
the present case, again, it seems the term ``information'' is
stretched beyond common usage by the factors people must sometimes
take into account in coming to their assignments.  From my point of
view, Kiki's foolish quantum state assignment is not information
about the physical world at all.  It is completely wrong, with no
reflection in the world as far as I am concerned.  Nevertheless, I
had better take it into account in making my bets if I don't want to
lose our whole joint bank account.

The point I take home is that is sometimes better to negotiate and
compromise even when one has ``maximal information.''

\section{10-09-01 \ \ {\it Short Reply} \ \ (to C. M. {\Caves} \& R. {\Schack})} \label{Caves24} \label{Schack18}

Just a very short reply to your latest posting.  But first let me say
something about this:

\brs
I hope this email establishes a little more common ground. I would
like to start, next week, on revising the paper in the view of this
discussion and the referee's comments.
\ers

Yes, do it.  With the draft in hand, I'll be better able to see which
statements make me feel like a liar and which do not. (Hopefully most
of them won't.)  And then we can be done with this, and then you can
finally stop saying to me, ``But not for this paper.''

\brs
\bq\noindent\rm
CAF Said: What is wrong with the word ``know''?  To my
Western-trained mind, it conveys the idea that there is something in
the external world (the world outside of my head and beyond my
control) and that my mind contains a mirror image of it.  It conveys
the idea that the outcome to the contemplated measurement already
exists ``out there'' in some deterministic or fatalistic sense.  It
conveys the idea that I really need never have a look to see if the
outcome is produced by my measurement:  It's already there, and I
know it; why waste time on a measurement?  Notice that I let the word
``knows'' stand when it came to describing the very logic of the
Dutch-book argument.
\eq
I do not think any of this is clear. I ``know'' something means I
have a firm belief in it. I don't think more is implied. The question
of whether something corresponding to the knowledge (or the belief)
exists out there is entirely separate from these wording issues.
\ers

{\Mermin} once wrote me this:
\bq
\noindent
It seems to me Chris is getting much too subtle about this.  I would
talk about knowledge, not belief.  I take ``knowledge'' to mean
simply ``true belief'', a definition that as I remember goes all the
way back to Plato and can be made unproblematic even in the quantum
context.  [Only a postmodernist would sneer at my saying this.]
\eq
And then this:
\bq
\noindent
C'mon, don't drag your heels.  QM is sometimes capable of assigning
probability 0 to certain outcomes.  For those one doesn't have to
argue about whether probability has to do with ensembles or degrees
of belief or anything else.  ``True belief'' seemed a good term to
describe such outcomes, and then I remembered that Plato (I think)
had used the same term (in contrast, as I remember, to ``opinion''.)
\eq
William {\James} writes it like this:
\bq
\noindent
The popular notion is that a true idea must copy its reality.
\eq
The {\sl Encyclopedia Britannica\/} says this (mainly the last
sentences are relevant):
\bq
In general, the philosophical tradition from the Greeks to the
present has focused on the kind of knowledge expressed when it is
said that someone knows that such and such is the case, e.g., that A
knows that snow is white. This sort of knowledge, called
propositional knowledge, raises the classical epistemological
questions about the truth or falsity of the asserted claim, the
evidence for it, and a host of other problems. Among them is the much
debated issue of what kind of thing is known when one knows that p,
i.e., what counts as a substitution instance of p. The list of such
candidates includes beliefs, propositions, statements, sentences, and
utterances of sentences. Each has or has had its proponents, and the
arguments pro and con are too subtle to be explored here. Two things
should, however, be noted in this connection: first, that the issue
is closely related to the problem of universals (i.e., whether what
is known to be true is an abstract entity, such as a proposition, or
whether it is a linguistic expression, such as a sentence or a
sentence-token) and, {\it second, that it is agreed by all sides that
one cannot have knowledge, in this sense of ``knowledge,'' of that
which is not true. One of the necessary conditions for saying that A
knows that p is that p must be true, and this condition can therefore
be regarded as one of the main elements in any accurate
characterization of knowledge.}
\eq

I put stars beside the hot stuff.  [It is italicized in this samizdat
version.]  And I'd send you more quotes if I could, but I'm a
gazillion miles away from home in a piss-poor dormitory room (with no
library).

\brs
There must be a misunderstanding. If you and I agree about everything
in the world, we also agree on the certainty of a particular
measurement outcome. We can not then disagree on the state to assign,
because at least one of us would be internally inconsistent.

You must mean that we agree on everything except the state AND the
measurement.
\ers

Yes.

\brs
\bq
\noindent\rm
CAF Said:  I now think it is much better to reserve the word
KNOWLEDGE solely for the outcomes of quantum measurements once they
become part of the mental makeup of an agent interested in them.  I
walk into Mabuchi's lab, and to the extent that he and I agree that
he is performing some POVM (denoted by a set of positive operators
$\{E_b\}$), it seems to me valid to call the outcome $b$ we both
witness to be an addition to our knowledge.  Now, what either of us
may do with that knowledge is a different story.  One thing is for
sure, it ought to cause both of us to update our beliefs.
\eq
This is a very difficult debate. I am not sure I understand why you
draw the line between knowledge and belief exactly where you do. Why
give belief in b a special role? I need to think much longer about
this.
\ers

Because I had assumed that Mabuchi and I had agreed to a fixed
``random variable.''  With respect to that prior assumption the thing
we gain is ``information'' in the standard Shannon sense.  I'm
willing to call that knowledge that we did not have before.  There is
nothing personalistic about it; we both have gained the same thing.

\noindent ------------------

I went to dinner with Jon {\Barrett}, Harvey Brown, Matthew Donald, and
David Wallace.  Getting Bayesian ideas across to them is going to be
a hard sell.  But the most amazing thing is that each and every one
of them was familiar with the Dutch book argument!  (Now, how could
anyone know it, and yet it not change their life?)

\section{12-09-01 \ \ {\it Ireland}\ \ \ (to S. L. Braunstein)} \label{Braunstein3}

Thanks for the offer.  If it looks like I'm stuck here, I'll surely take you up on it.  (Mostly I want to get home, safe and sound, and then {\it not\/} get on a plane again for some time.)

It's amazing here, at this meeting:  It has practically stopped.  People just can't get the news off their minds.\footnote{The conference in question was the ``10th UK Conference on Foundations of Physics'' in Belfast, Northern Ireland, September 10--14, 2001.  It is the meeting where I met Marcus Appleby for the first time, at a dinner with Matthew Donald.  Matthew said, ``Would you mind if Marcus Appleby comes along?  He's done some interesting work on Bohmian mechanics.''}

\section{13-09-01 \ \ {\it Re:\ Hope You Are Fine} \ \ (to H. J. Briegel)} \label{Briegel2}

Thanks for the concern.  They have indeed been horrible events.  But my family and I are both safe:  Kiki and Emma are in the States, and I am still in Ireland (I may be here for a while, depending upon the airlines).

I certainly enjoyed getting the little private lecture.  The word will be spread:  You've had a great set of papers.

I'll write you with information about {\Montreal} as soon as Gilles and I get our heads together (maybe next week or the week after).

\section{15-09-01 \ \ {\it Question on the Manuscript} \ \ (to R. {\Schack})} \label{Schack19}

Just send me the draft when you're pleased with it.  And I'll tell
you then what I can tolerate.

\brs
But as your (ugly) term ``absolutistic belief'' shows, they should be
extremely careful before making pure-state assignments.
\ers

I am not committed to that term, like you guys are to the
(inaccurate) term ``maximal information.''  It was just the best I
could come with at the time.  Change it if you wish.  The most
important thing is that whatever you substitute for it should carry
no flavor of a ``reflection theory'' of knowledge.

You won't buy it, and you'll think I'm just saying this out of less
than pure reasons, but you would be surprised at how many people have
now encouraged me to call the quantum state a ``belief'' rather than
``knowledge.''  (Four.)

I'd like to come to SB, but we'll have to see how things play out
with Kiki.

\section{15-09-01 \ \ {\it Melancholy Molly} \ \ (to N. D. {\Mermin})} \label{Mermin38}

I'm sorry to be writing you back so late:  this week has taken a big
toll on me.

\bdm
Well for better or for worse Todd, Jerry, and I came up with
something we could all agree on.
\edm
I suspect for the worse (if Renes's report to me on it is accurate).
I haven't been able to download the paper myself yet, and won't until
I get back into the states.  My personal opinion is that it is likely
to muddy the already troubled waters even more.

\bdm
Am still enchanted by the knowledge-belief breakthrough.  How do your
coauthors like it?
\edm
The world has become a little lonely for me lately.  I'll show you
the lengths I've had to go to with all of you below.  You are a
troublemaker, getting me started down this route!  (The document is
in the format of the samizdat.)

\bdm
Are you now in Ireland?  What part?  Be careful.
\edm
To think, I was once scared to be in Ireland.  I'll instead take ``be
careful'' to mean ``get back into the US safely.''  I am in London as
I'm writing this to you.  My original return flight was cancelled;
now I'm on the backlog with a million other people.
\begin{verse}
Blow up your TV, \\
Throw away your paper.\\
Move to the country,\\
Build you a home.\\
Plant a little garden,\\
Eat a lot of peaches,\\
Try to find Jesus on your own.
\end{verse}

\section{16-09-01 \ \ {\it For Worse} \ \ (to N. D. {\Mermin})} \label{Mermin39}

\bdm
Well for better or for worse Todd, Jerry, and I came up with
something we could all agree on.
\edm

Well, I managed to download your paper this morning after all.  I
ended up in the airport for over four hours.  Now, I'm somewhere over
the Atlantic.

I hate it, of course.  It's built around the same stubborn disregard
for the issues involved in trying to give the quantum state a
NONontological status that all the BFM emails were.

To use a term I picked up from a philosopher this week in Ireland,
the proposed outcome of a single quantum measurement simply cannot be
a ``candidate for knowledge.''  Its source must be considered
ineffable, else the wrath of Gleason's theorem would strip away the
nonontological status that was being sought for the quantum state in
the first place.

I can guess from the format of the paper that it is being submitted
to PRL.  If I were being mean, I might say, ``Good.  That'll give it
a higher chance of being rejected.''  But I'm not mean.  If I were
the referee I might even accept it.  But, ONLY if the authors use the
extra column at their disposal to explain why their statement of the
problem is not a deep endorsement of the Everettista manifesto. (Or,
if it is, then just come out and say it.)  What else is Zeno but a
baroque name for the quantum state of the universe (mixed or not) or
a candidate for the one that really gets it ``right''?

On a personal note:  If you're going to cite a personal communication
that appears in the samizdat, why don't you cite the samizdat?  (I'm
always looking for further ways to draw people into it.)

Best regards (in continuing disagreement),

\section{16-09-01 \ \ {\it Persnickety Business} \ \ (to N. D. {\Mermin})} \label{Mermin40}

Since my last note, I've looked over your paper again.  I must say I
found the sentence,
\bq
\noindent
It is surely a significant feature of the theory that consideration
of impossible outcomes and very little else leads, without any
invocation of ``the uncertainty principle'' or ``maximal
information'', to the fact that pure state assignments must be
unique, as well as the more general constraint on mixed-state
assignments.
\eq
a bit snobbish.  Especially since you gave zero devotion to the issue
of whether a quantum state could be ``true'' or not.  If they can be
``true,'' then why don't you just take them to be a bit of material
reality and forget all this crap about knowledge?  If they can't be,
then what do you mean by the word ``impossible''?

I still hate the paper, probably more so now.

A little further over the Atlantic (and wishing I were closer to
Greenland),

\section{16-09-01 \ \ {\it Arlo and Arlo} \ \ (to D. B. L. Baker)} \label{Baker3}

As I just wrote another David---i.e., {\Mermin} instead of Baker---right
now I'm somewhere over the Atlantic, wishing I were a little closer
to Greenland.  I'm on my way back home from Northern Ireland.  What a
week it has been.  I've never been so homesick for America before.

\bv
Good mornin' America, how are you?
\\
Don't you know me, I'm your native son.
\ev

I think I might just kiss the ground if I really do land at JFK.

Once upon a time, we could actually see the twin towers from
Morristown on a clear day.  They pierced the horizon like nothing
else in Manhattan.  Hearing Emma's voice on the phone the other night
was one of the toughest things.  She said, ``Hi Daddy-o,'' and then
told me all about her first day in playschool.  It's not easy to
juxtapose that with the death of 5000 people and the many more deaths
that may come in the near future.

Life is an essence in this universe.  Creative, productive,
reproductive life:  I'm not one of those scientists who think it is
just an illusion, an epiphenomena rolling on top of a clockwork or a
dice-rolling world.  It is something in itself.  It was latent there
all along, long before the trilobites had found their place, and it
is our task to see that it does not return back to that latency.

Very sad times.

\section{16-09-01 \ \ {\it Registered Complaints} \ \ (to J. M. Renes)} \label{Renes4}

Good to hear from you; thanks for the note.  I'm flying back from
Ireland as I write this to you.  I think the conference was
successful in many ways.  While there, I got to know {\Marcus} {\Appleby},
and he is a really good guy.  He's very {\James}ian in his perspective,
very clear in thought, and a real seeker of the truth.  I think I'm
going to work to get him to {\Montreal} next year (even though he does
not practice quantum information).

I agree with your assessment of the BFM paper.  I've already
registered my complaint with {\Mermin}.  But you know, it's not like
there are no pansies even in the midst of our own little group: {\Carl}
and {\Ruediger} have had no great fortitude when it came to this issue,
no courage of their convictions.

I'll place my small samizdat on the subject below in case you're
curious.

I hope your transition back to New Mexico has gone well.  We'll miss
you at Bell Labs.

\section{16-09-01 \ \ {\it Nerve Therapy} \ \ (to R. {\Schack})} \label{Schack20}

I'm in the last 1.75 hours of my flight to New York, and I'm doing
everything I can to keep from getting too nervous.  Let me finally
turn my sights to your note from a week ago.

\brs
I was referring to the argument that, from A's perspective, if B
assigned a different pure state, that would force B to accept a bet
offered by A that amounts to handing over money to A. Therefore B is
a crackpot from A's perspective (but not from his own, of course).
This is interesting because it is the very fact that B has maximal
information that forces him to accept A's bet: B ``knows'' A cannot
know anything that would give her an unfair advantage in the bet. I
still think this is a strong argument. Scientists would try to
overcome this critical situation by questioning the world, i.e., by
making measurements, but NOT on the system in question, but on
everything else, e.g., Hideo's optical table. After all they both are
convinced that the world is disentangled from the system in question.
\ers

Personally, I only find this restriction to not touching the system
of interest (by you and BFM) ad hoc.  Since when in science do we
restrict the experimentalists to never touch the objects of their
interest?  Besides, I think you may be missing one of the deepest
facts of the quantum world:  it has the power to cause agents to
agree {\it henceforth}, no matter how adamant their previous beliefs
\ldots\ AND EVEN without the underlying reality of preexisting
measurement outcomes that the classical world had.  With each quantum
measurement, a part of the world starts afresh.

\brs
\bq\noindent\rm
CAF Said: But there are realities:  Kiki and I are married; we share
a bank account.  And here and there, Kiki consorts with the Dutch.

What I am leading up to is that I think there is a place for Ben
Schumacher's observation about a three-person Dutch book in our
ongoing debate.
\eq
I cannot quite see where this situation offers any specifically
quantum insight. In any case, it is completely different from the
``adversarial game'' I am talking about.
\ers

I'm not sure that it does lead to any specific quantum insight.  But
then again, I had never seen you and {\Carl} be so adamant about taking
probabilities to be objective features of nature before.  (Don't do
it!  I know you're going to say ``NO, NO, NO, we have never wanted
objective probabilities!''  But all the pieces of evidence---to
me---point to the contrary.)

What it does show---and that is maybe where quantum theory comes in a
little bit---is that there are times when no matter how adamantly I
believe something, I should bet according to odds different from my
belief.  I.e., there are internal beliefs, and external ``beliefs.''

To that extent, I am starting to wonder if ``belief'' is even an
appropriate word for capturing the essence of a quantum state.
Instead it is starting to seem to me that it may more appropriately
be described as a negotiated signifier to external action.  I.e., in
some cases, it signifies the betting strategies that a community of
scientists can agree upon, even in the case of more refined and
divergent beliefs.

Whose knowledge, {\Mermin} asked?  Sometimes, mine.  Sometimes, yours.
Sometimes it's more a matter of the policy we have been able to
negotiate.

I do get a little worried in saying all of the above, in that, maybe
I have not taken Savage's and Bernardo and Smith's views of
probability into adequate account:  namely, that probabilities and
utilities spring into existence at the same time, and are a little
inseparable.  But presently I don't see how that fits this problem.
Nor does it extinguish the problem that sometimes it really is in
people's best interest to lie about what they believe.

\brs
\bq\noindent\rm
CAF Said: I am internally consistent; there's no Dutch bookie who can
take me to the cleaners.  And despite my feelings for Kiki's complete
foolishness, I feel that she is internally consistent; there's no
Dutch bookie who can take her to the cleaners (as far as she is
concerned).
\eq
I strongly believe that here marriage is fundamentally different from
science.
\ers

Perhaps.  But the point is, sometimes we bet according to situations
that have nothing to do with our beliefs.  We bet so as to obtain the
best common good.  And with that we do return to some aspects of
science (and politics).

\brs
\bq\noindent\rm
CAF Said: But we will be in deep trouble if that Dutch bookie
approaches us separately. (Being married, we report all our beliefs
to each other.)
\eq
This deep trouble arises even if both you and Kiki assign mixed
states with the same support, I believe (but I need to think more
about this). I am trying to make the point that pure-state
assignments are different.
\ers

That is true.  I chose the particular example of pure states to be as
dramatic as possible \ldots\ in an attempt to hit you and {\Carl} in a
point where you were being the most hardheaded.

\brs
\bq\noindent\rm
CAF Said: I can see two outs to this problem.  The first one---which
is less interesting---is that we make an effort to come to agreement
by consulting the world.  We make a measurement, and thereby, through
its invasiveness, force the quantum system into a state we can agree
upon.  (Assuming, as I keep harping on, we can agree on the quantum
operation associated with our measuring device.)  But what if we have
no access to the system of interest?  What are we to do then?
\eq
Well, I gave my answer above. We question everything in the world BUT
the system of interest. Using quantum measurements, of course.
\ers

Ad hoc.

\brs
\bq\noindent\rm
CAF Said: I think we would have no choice but to, each of us, back
off in the firmness of our beliefs.  That is, we should agree upon a
density operator that contains in its support both of our earlier
ascriptions.
\eq
Yes, the same is true for two reasonable scientists. But as your
(ugly) term ``absolutistic belief'' shows, they should be extremely
careful before making pure-state assignments. Your way out is similar
to having second thoughts after placing the bet.
\ers

No.  My way out is to point out that there is, after all, a
distinction between an (internal) belief and its manifestation as an
outward bet.  The belief comes before the bet.

OK, I've just passed Boston, and in the mean time, got another meal
in my stomach.  For the obvious reason, getting past Boston carries
some symbolic significance.  Let us believe in symbols.

\section{18-09-01 \ \ {\it Hi Back} \ \ (to A. Peres)} \label{Peres17}

Good to hear from.  Yes, I am home.  I was delayed by a day, spent a
night in London, had to return to JFK rather than Newark, and had the
most tense flight of my life, but I am home.  I held and held and
held on to Emma when I saw her waiting for me.

I have been in a huge email debate with David {\Mermin} about his two
latest papers.  Consequently, I also got into a huge debate with
{\Caves} and {\Schack}, though they are (slightly) more sensible.  (The
whole thing has been quite taxing.)  John {\Wheeler} once told us a
story of a condemned man who, while waiting for the firing squad,
calmed his nerves by contemplating Hamilton's beautiful equations. I
learned a lesson from this:  When I was worried about the safety of
my flight, I calmed my nerves by continuing my email debate with
{\Mermin}, {\Caves}, and {\Schack}!

I hope you and Aviva are able to get home on schedule.  I suspect
your grandchildren are waiting for your return with excited eyes.

\section{18-09-01 \ \ {\it Goodbye} \ \ (to N. D. {\Mermin})} \label{Mermin41}

I just don't know what more to say.  Clearly there is something that
is keeping us from communicating.  I write more and more, and you
just don't get my points.  It's probably better if I just write less
and less.

I'll record the way {\Caves} put the best part of your paper the other
day and just leave it at that.  I've never said anything different,
but maybe if you hear it from another voice something will click.

\bq
It seems to me that there are various kinds of coming to agreement or
inability to do so.  Much of this is a recapitulation of what we have
already discussed.  It is phrased in a way that is supposed to avoid
the notion that beliefs incorporated in a state assignment require
the world to do things. [\ldots]

1.  BFM consistency (a multi-party condition): If the parties share
their information, it is not ruled out that they can come to
agreement on a common state assignment.

If they share information, they must rule out all vectors in the
subspace generated by the union of their null subspaces, and they
must assign nonzero probability to all vectors in the intersection of
their supports.  They have the possibility of making a joint state
assignment if and only if the intersection of their supports is of
dimension 1 or more.

Notice that we don't have to say here, as BFM do, that an outcome
assigned probability zero by any party definitely cannot occur.
\eq

But let me say one last thing before I give up on this conversation.
In your last note you wrote:
\bdm
We've sent the hateful paper to Phys Rev A as an ordinary low-grade
submission.  My concern is that a referee will condemn it as trivial
or well-known, not outrageous or Everettistan.
\edm

You have probably summed up the referee's response correctly, but it
doesn't make it so.  Nor does it make you paper non-Everettistan
after all.  The whole point can be summed up with the following part
of your paper.

\bq
\noindent{\bf A necessary condition for compatibility}

Suppose Alice, Bob, Carol,$\,\ldots\ $ describe a system with density
matrices $\rho_a, \rho_b, \rho_c,\,\ldots\ $.  Each of their
different density matrix assignments incorporates some subset of a
valid body of currently relevant information about the system, all of
which could, in principle, be known by a particularly well-informed
Zeno.
\eq

This is an edict you are placing ON TOP OF quantum mechanics.  It is
a construction of your own doing.  I find it in no axiom system I
have ever looked at.  It IS a good half of the Everettista starting
point.

You might say the other half of Everett is that he takes the quantum
state explicitly to be an ontological entity.  But you've got that
too.  By imagining that facts of the world (measurement outcomes)
uniquely determine a quantum state, what else could a quantum state
be but a stand-in for those real things, those FACTS (the stuff that
one absorbs into one's mind and calls information)?  You simply
cannot say that quantum states MUST be consistent, if you want them
to retain a nonontological status.  The second you say they MUST be,
then you have given up the game:  states are then properties of the
world, and this question of consistency was a waste of time from the
beginning.

With this, I end the conversation:  I give up.

\section{19-09-01 \ \ {\it Morning Coffee} \ \ (to C. M. {\Caves} \& R. {\Schack})} \label{Caves25} \label{Schack21}

\brs
Yes, and this is great progress!
\ers

Well, the note really is some progress.  Let me just accept the gifts
the Lord has given me.

I have two questions.

1)  What does Peierls' consistency have to do with coming to
agreement?  I don't see it yet.  It seems to me to be more a
technical definition for the word ``wrong'' than anything else. Using
it, I can say when you are ``wrong'' with respect to my beliefs, but
that seems about it.

2)  This thing you call fuchsian consistency.  You bring to the
surface, a point I've always been a little worried about.  Why refuse
to consider outcomes lying in the null subspace of a density
operator?  Let me give an example:

\bv
Chris:  I say the state is spin-up in the z-direction.
\\
{\Carl}:   I say it is spin-down.
\\
Chris:  Spin-up!
\\
{\Carl}:   Spin-down!!
\\
Chris:  OK, let's test it. Here is an ideal z-spin-measuring device.
Its Kraus operation is just a von Neumann collapse.  Do you agree?
\\
{\Carl}:   Yes, I agree.
\\
Chris:  Are you sure you agree?  We need some agreement or this won't
be a meaningful game.
\\
{\Carl}:   Yes, I agree and am growing impatient.
\\
Hideo:  Spin measurement done.  It's spin-down.
\\
Chris:  Damn, I honestly believed with all my heart that it'd be
        spin-up!
\\
{\Carl}:   Well, I wouldn't have said it if I didn't know it.
\\
Chris: OK, from here out, I say the spin is spin-down.  I was a fool
before, but at least we agree now.
\ev

Your worry seems to be that the formal apparatus of quantum mechanics
will not allow me to propagate my belief.  According to the Kraus
rules, my new state should be the zero operator renormalized by a
probability zero:  i.e., it is an undefined object.  Nevertheless,
life goes on, just as the story above indicates.  This, at least,
does have a formal counterpart in that, if I back off in my belief
even the slightest amount to any mixed state (not on the boundary),
then the Kraus operation will take us both to the same place, to the
same one-d projector.  Consequently, the limit exists even if the
actual point is undefined.

So, in saying this, I don't think one has to do anything fancy with a
measurement to get this thing you call fuchsian consistency off the
ground.  No intricate constructions of states seem to be needed.
Choose ANY (mutually agreed upon) von Neumann measurement you like,
and {\Carl} and Chris will have to agree at the end of the day.  One of
them will feel foolish for having been so wrong, but as long as he is
rational, agreement can still be had.  As {\Ruediger} said, the idea is
that this is a property of quantum mechanics (in its capacity as a
law of thought), not of the initial states.

Alternatively, one can turn the problem around, as I tried to do in
my August 7 note to BFM:  suppose {\Carl} and Chris agree to a {\it
particular\/} Krausian state change rule (one not even closely
resembling a von Neumann collapse).  Now one can ask, under what
conditions on the initial states will C and C be left with more
agreement at the end of the day than they started with?  But that is
a different problem than this f-consistency condition you speak of.

\section{19-09-01 \ \ {\it 2001 Japanese-American Frontiers -- Background Information Needed} \ \ (to E. R. Patte)} \label{Patte1}

\bq
\noindent Title: Quantum Information, Quantum Channels \medskip \\
Abstract:
Most physics students, with their first lesson on the Heisenberg uncertainty principle, are given a subliminal message:  quantum mechanics is a limitation.  The attitude is, ``Quantum mechanics is something we deal with because we have to, but wouldn't the world have been so much better if we could just measure a particle's position and momentum simultaneously?''  This talk is about the counterpoint to that attitude.  Recent advances in the fields of quantum computation, quantum cryptography, and quantum information theory show that the physical resources supplied by the quantum world are anything but a limitation.  With these new resources we can do things almost undreamt of before, from the secure distribution of one-time pads for use in cryptography to the factoring of large numbers with a polynomial number of steps.  The magic ingredient in all this is something called quantum information.  I will illustrate the subtle strangeness of this new kind of information and the nice effects it buys with several concrete examples drawn from my own work in quantum cryptography and quantum channel-capacity theory.
\eq
Review Articles:
\begin{enumerate}
\item
D. Gottesman and H.-K. Lo, ``From Quantum Cheating to Quantum Security,'' {\sl Physics Today}, November 2000, p.\ 22.
\item
J. Preskill, ``Battling Decoherence:\ The Fault-Tolerant Quantum Computer,'' {\sl Physics Today}, June 1999, p.\ 24.
\item
C. H. Bennett, ``Quantum Information and Computation,'' {\sl Physics Today}, October 1995, p.\ 24.
\end{enumerate}

\section{19-09-01 \ \ {\it Tentative Hello} \ \ (to N. D. {\Mermin})} \label{Mermin42}

\bdm
I may take a shot at rewriting our argument, since I'd like to
produce something for the {\Vaxjo} proceedings.  At that point I may
try it out on you, but if you think it's likely to induce nausea, you
don't have to respond.
\edm

I owe you many things.  If you want me to look at it, I will look at
it, and I will try to respond (i.e., give you constructive criticism)
in a civil tone.

\section{19-09-01 \ \ {\it Practical Art} \ \ (to C. M. {\Caves}, N. D. {\Mermin} \& R. {\Schack})} \label{Mermin43} \label{Schack22} \label{Caves26}

I suspect all of you have seen, in one form or another, the optical
illusion where, looking at the drawing in one way, it appears to be a
beautiful young woman.  But looking at it in another way, it looks to
be an old hag.  Here's a true story about that.  The first time I
ever saw such a thing was in Roger Penrose's book {\sl The Emperor's
New Mind\/} in 1989 or so.  Below the picture was a caption,
explaining just exactly what I ought to see:  alternatively, a
beautiful woman and a old hag.  But the oddity was, I could only see
the beautiful woman.  As much as I tried to find a hag in the
picture, I couldn't do it.  That fascinated me---so much so that,
from time to time throughout the next year, I would pull the book off
the shelf and try to find the old hag.  I never did see her until one
day I was searching through the book for something completely
different and happened to come across the page.  In fact, that may
have even been two years after my initial encounter with the picture.

I think it is safe to assume from this that if Penrose had never
pointed out to me that there ought to be an old hag there, I simply
may have never seen her.

Let me apply this piece of art to a question on my mind.  Let us
imagine that some perpetrator has committed a dastardly deed to my
family and left a note promising that before the end of her life she
would do it again.  Beyond this I have no clue of the identity of the
perpetrator (or of her cause) except a sketch made of her by an
eyewitness, who has now also disappeared from the scene.  For
intentional completeness, let us suppose there is simply no further
evidence that I can lay my hands on---there is no more ``currently
relevant information,'' no deeper ``trail of evidence.''  The carbon
atoms laid out on the two pieces of paper---the note and the
drawing---are the only links I have to the cause of the crime.  The
trouble is, in the case of the drawing, the carbon atoms sketch out
the shape of Penrose's beauty-hag.

Now suppose an insurance salesman appears on the scene and is willing
to sell me insurance against a recurrence of the crime precisely 30
years hence at such and such a premium.  Do I buy it? That, of
course, will depend upon the probability I ascribe to the crime's
recurrence in precisely 30 years.  What probability do I ascribe?

It will depend upon what I see in the picture.  If I ``see'' the old
hag, it will be one thing.  If I ``see'' the beautiful woman it will
be another.  If I ``see'' there is an ambiguity, it will be still a
third.  But one thing is sure from this example, what I ``see'' is
not completely dependent upon the pattern laid out by the carbon
atoms.  Part of what I ``see'' is dependent upon psychological
factors that I {\it myself\/} may simply never have access
to---things deep within my head, things determined when I was first
toilet trained, things to do with the self-referential nature of
consciousness, things that physics just ought to (and maybe even has
to) leave alone.  My beliefs (my probability assignments) after
seeing the sketch are determined in part by something objective in
the world external to me, but also in part by my previous, purely
subjective beliefs.  This example ought to make it clear that I
cannot put all the weight of my posterior beliefs on the external
world.

Here's my question:  Where do you draw the line?  More importantly,
how do you draw the line?  What is it in your view that gives the
clicks of a quantum mechanical measurement a status that goes deeper
than the example above?  What makes the phrase ``currently relevant
information'' more decisive in the quantum mechanical case?  How does
that ``trail of evidence'' have more power to eradicate my subjective
pre-beliefs than any other?  So much so, in fact, that one gets the
feeling that we might even view the quantum state as determined
solely by such a trail?  You say, ``Well, there the information
conforms to a physical theory.''  But in what way does that change
things?  Can you pinpoint it?  Can you make that declaration
meaningful?  Indeed, can you pinpoint what about the example above is
not quantum mechanical?  It had atoms, it had systems, it had
probabilities.

I said I would not talk to David anymore, and here I am talking to
him.  I guess that just goes to show that that {\Mermin} can steal your
heart \ldots\ as can all of you.

\section{20-09-01 \ \ {\it The Stopgap} \ \ (to C. M. {\Caves} \& R. {\Schack})} \label{Caves27} \label{Schack23}

Attached is the latest draft.  You will find two (and only two)
changes.

\begin{enumerate}
\item
I added the word ``the'' in front of ``records.''
\item
I added two citations to myself.  One at the beginning, one at the
end.  If you don't like where I put them, that's fine---feel free to
change their positions---but I would like to have them somewhere in
the paper.
\end{enumerate}

Of course I am still troubled by the use of the phrases ``maximal
information'' and ``with certainty''.  I continue to think that they
convey an image that is better left unconveyed, but I'm the odd man
out here so I'll shut up.  The main thing is that you left me enough
wiggle room that I can defend my view of the quantum state when I
need to.  E.g., ``Oh, I'm sorry, that was just a bad choice of
language.  The thing to keep in mind is \ldots''  Rather than my
having to say, for instance, ``Yes, I believe we were wrong about
that, but I couldn't convince my coauthors.''

Another thing I guess I really didn't like---and this is only a new
one for me, it's not something that had eaten away at me before---is
the slogan ``Gleason's theorem can be regarded as the greatest
triumph of Bayesian reasoning.''  To say that is to imply that
Bayesian reasoning IMPLIES Gleason's theorem.  I don't think you mean
that. I think what you mean is that it is one of the most valuable
additions to Bayesian reasoning ever.  But I won't cause trouble
here:  I know you both like the saying.

I think I read the paper very carefully again---more carefully than
ever actually.  I will live with its consequences, and I apologize
for dragging my feet for three or four years.

Now, I just wish we three could come to agreement in our views of
quantum mechanics!!!!!!  So that I could look back on that day and
say, ``We few, we lucky few, we band of brothers \ldots''

\section{20-09-01 \ \ {\it Praise, Folly, Enthusiasm} \ \ (to W. K. Wootters)} \label{Wootters2}

A million thanks for your encouraging letter of August 30!  But also,
please accept my apologies for not writing you back for more than 20
days.  I had wanted to write you a rather long, contemplative note in
response, and I kept waiting for the right mood.  Somehow it just
never came---first with my travels to Munich and Belfast getting in
the way, and then with all the horrible events in the world in the
last 10 days.

I certainly like aspects of this speculation of yours.  Indeed, I
wish you would write it down so we could all have a chance to think
about it a more deeply.  (Would you be willing to do that?  You do
have tenure now, and our individual lives are finite: we should never
lose sight of that.)  But of what little bit I understand of your
ideas presently, they don't seem to have as much reciprocality or as
much dynamism as I would like to think the world has.  What I mean by
this is that your ideas seem to carry a significant flavor of the
Cartesian cut:  the graphs take the place of res extensa, and the
identifications take the place of res cogitans.  The two realms---as
I understood your explanation---don't really interact: The graphs are
timeless and independent of the identifications we might make between
them.

But, as you say, we are both speculating.  And we both realize that.
Here's the way William {\James} put it:

\bq
The history of philosophy is to a great extent that of a certain
clash of human temperaments.  Undignified as such a treatment may
seem to some of my colleagues, I shall have to take account of this
clash and explain a good many of the divergencies of philosophies by
it.  Of whatever temperament a professional philosopher is, he tries,
when philosophizing, to sink the fact of his temperament. Temperament
is no conventionally recognized reason, so he urges impersonal
reasons only for his conclusions.  Yet his temperament really gives
him a stronger bias than any of his more strictly objective premises.
It loads the evidence for him one way or the other, making a more
sentimental or more hard-hearted view of the universe, just as this
fact or that principle would.  He {\it trusts\/} his temperament.
Wanting a universe that suits it, he believes in any representation
of the universe that does suit it. He feels men of opposite temper to
be out of key with the world's character, and in his heart considers
them incompetent and `not in it,' in the philosophic business, even
though they may far excel him in dialectical ability.

Yet in the forum he can make no claim, on the bare ground of his
temperament, to superior discernment or authority.  There arises thus
a certain insincerity in our philosophic discussions:  the potentest
of all our premises is never mentioned.  I am sure it would
contribute to clearness if in these lectures we should break this
rule and mention it, and I accordingly feel free to do so.
\eq

Since reading this a month ago, I've been wondering what my
temperament really is.  What is the ``potentest of all my premises''?
I haven't completely tied it down yet, but I think it has to do with
the idea that the world can be moved (from within), that it is
malleable, that it is still under construction.  That the future, for
better or for worse, is not yet determined.  And that this
malleability---like the turtles---goes all the way down:  there is no
ultimate level that grounds it.

This, if it is, my ``potentest of premises'' has so far taken
manifestation in the speculation that the components of the world are
``sensitive'' or ``irritable.''  But I suppose that is part of the
``impersonal reasons'' that {\James} speaks of---i.e., that they form
not quite the whole truth of my motivation.  The better truth is
perhaps that the components of the world---the things that come out
of the ways we carve the world up---are movable in a deep sense.
They reach out and affect us in ways that we cannot foresee, and we
reach out and affect them in ways that they cannot.  And through that
intercourse, birth arises in the world in a sense every bit as real
as biological birth.

But I don't know how to make any of that more precise.  It stands at
just a program and a direction, but one I take seriously.

Your letter gave me courage, and with such speculations, believe me,
one needs courage!  How I wish we could get together more often to
hash these things out, letting our disparate speculations refine each
other.  I feel deep in my heart that there is progress to be
made---technical progress---it's just a question of building a
community with a critical mass of ideas, constructive opinions and
techniques.

What are your plans for your sabbatical this year?  In which
direction do you plan to use your time?  Will you be visiting IBM
often?  If I could get a chance to talk to you more often about the
sublime side of physics, I'd surely take it.

By the way, let me draw your attention to a mistake I made in the
paper you read.  Howard Barnum brought it to my attention last week,
and next week I plan to write a short comment on it and post it on
{\tt quant-ph} (before someone else does).  It is in the derivation
of the tensor product rule.  For the most part the derivation holds,
but I got sloppy in the very last step.  I.e., by requiring the
existence of a noncontextual probability assignment to the outcomes
of local measurements (with one-way classical communication), one
does indeed get that the probabilities are controlled by a linear
operator on the tensor product of the two spaces.  But, these
assumptions alone don't get you that the linear operator ought to be
a positive semidefinite operator.  That requires more assumptions. In
principle, one should be able to state those assumptions as a
restriction on the correlations one can obtain by local measurements,
but I don't quite see how to do it yet.  (An easy way out would be to
require that the linear operator give rise to positive probabilities
for all measurements, local and nonlocal. But, that's kind of a dull
answer after relying on locality for so much.  I'm sure one can find
a more interesting answer.)  For instance, I'm not quite sure how to
tackle the question of ping-ponging measurements in this framework.
Or even whether that's the sort of thing that should be looked at for
a natural restriction.

There's so much work to be done!  But, it's our place to do it \ldots

\section{20-09-01 \ \ {\it Pots, Kettles, and Frying Pans} \ \ (to N. D. {\Mermin})} \label{Mermin44}

More seriously.  I think the line you are talking about is not the
line I was talking about.

{\Caves} has been writing me things like this:
\bcc
I do want to conclude with my obligatory diatribe against wholly
e-mail exchanges.  You think all your messages were perfectly clear,
I think all mine were perfectly clear, {\Ruediger} probably thinks the
same, but the evidence is that they were not.  It's just not possible
to come to agreement by e-mail, the reasons among others being that,
first, questions arise in reading something after which further
comments get devalued and, second, peripheral and main points often
have their roles exchanged when a message is read.
\ecc

And ever more I am having to come over to his view on this subject.
But, it is just so hard to give up my old email habits, especially
since I've seen them become so ineffective of late.  It's like a
captain who just can't tear himself away from a sinking ship.

So let me say a few words in response to you.  You should know I've
got too much invested in the phrase ``knowledge about the
consequences of our interventions'' to back out:  it's part of my
whole being.  If you think that is what is going on with me, you are
mistaken.

I do not deny:  (1) that trails of evidence exist, and (2) that
trails of evidence are created in part by our interventions into the
world.  What I deny is that those trails of evidence can ever
uniquely determine a quantum state, even a pure one.  That is what is
at issue (with me at least), and that is what yesterday's note was
referring to.  (Somehow I get the impression that you saw something
completely different in the note.)

A detector goes click.  You write down that it went click; I write
down that it went click.  To that extent, the click is part of the
objective world independent of us; it is part of the trail of
evidence that you and {\Caves} speak of.  (Now, maybe the measurement
apparatus itself isn't completely part of the agent-independent
world---somebody built it to begin with---but that is a different
story.)  The formal structure of quantum mechanics says that we must
identify the click with an element in some POVM.  Fine.  I agree with
that.  (Presumably you do too.)

The issue is, which POVM?  And which state-change rule associated
with that POVM?  Show me a place in quantum mechanics where you are
told how to do that.  I say it is a subjective judgment, just like
the quantum state is.  Or, more precisely, it is exactly {\it that\/}
subjectivity that keeps our view of the quantum state as being
subjective from being an inconsistent notion.  In any given
experiment, if there is a single POVM (and state-change rule) that is
correct and objective, then so must be the quantum state.

Do you just not see the mapping problem here?  If a POVM (and
state-change rule) is an objective property of the interaction of two
systems, then so will be the post-measurement quantum state ascribed
to one of those systems.  Now, if the quantum state is an objective
state of affairs, why quibble about calling it ``knowledge'' or
``information''?  That would be just using two words for it that
never needed to be invoked in the first place.  If the quantum state
is objective, then call it ``quantum state'' and be done with it.

If you can stomach it, try to read my note ``Practical Art'' again.
It was meant to be a call to do some soul-searching.  It was meant to
try to persuade you that the subjective element can never be
eliminated in a theory that makes fundamental use of probabilities.
It was not meant to convey the idea that one cannot draw a line
between systems and apparatuses.  It was not meant to be a call to
join the ranks of the Everettista.

Chris (on a rainy day---it always happens on a rainy day)

\section{20-09-01 \ \ {\it Comment on Practical Art} \ \ (to C. M. {\Caves})} \label{Caves28}

You still don't recognize that the difficulty is a logical one, do
you?  There is nothing squishy, postmodern, deconstructionist, or new
age about the issue:  if POVMs (von Neumann ones in particular) are
nonsubjective, then so are quantum states.  Period.  You can't get
around that.  If you want to claim that quantum states are subjective
judgments, then you have to accept that POVMs (and their associated
state-change rules) are subjective judgments too.  Else the
post-measurement states that they give rise to would be objective.

I say this slightly better in my reply to {\Mermin}.  It is not that I
am leaving the realm of science:  it is the strict practice of the
art that put me here.  Assumptions $\longrightarrow$ conclusions.

Don't worry though:  I'm not offended by your note.  I'm just
(continuingly) surprised by your immense rigidity.

\section{25-09-01 \ \ {\it A Tough Decision} \ \ (to C. M. {\Caves} \& R. {\Schack})} \label{Caves28.1} \label{Schack23.1}

I hope you will both believe me when I tell you that this has been an extremely tough decision to make \ldots\ but I think I had better not come to Santa Barbara this year.  Three things have conspired to keep me from coming there:  (1) the general timing in relation to Kiki's pregnancy, (2) Lucent's downturn and the consequent cash troubles it makes for traveling, and (3) a good dollop of (probably irrational) fear about getting on planes more than I have to.  You should know that there are no two people I would like to spend my time with presently than you:  Getting this stuff settled about the meaning of the quantum state has been foremost on my mind, and I know that I will not feel at ease until we three can regain some equilibrium.  Also I would like to think that my giving a talk at the December meeting would help further the Bayesian cause in a way that is maybe unique to me.  But I have to let those opportunities pass and hope that a greater good will come out of this.

Thanks, though, to both of you for pushing on me so hard to come:  I am nothing if not paranoid about my value to this society, and feeling wanted always helps me get through the days.

\section{25-09-01 \ \ {\it Further Comment on Practical Art} \ \ (to C. M. {\Caves})} \label{Caves29}

\bcc
I was initially mightily annoyed by the tone and content of your
message.  My comment said nothing about and intended nothing about
being ``squishy, postmodern, deconstructionist, or new age.''
Moreover, your reply didn't address any of the points I raised,
apparently because they were all just further examples of
``rigidity.''
\ecc

Let me reply to your last point first.  Here's what happened:  I
thought I'd reply with a short and to-the-point comment first (about
what I saw as the overriding issue), and then follow that up with a
detailed reply to your particular points.  But then time ran out
before having to leave for a weekend trip to the beach.  I certainly
did not mean to ignore your points, and I certainly won't (as time
permits this week).  It was just that I was banned from email for the
weekend:  I'm sorry I didn't warn you of that.

Now, as far as the postmodern business, I apologize for being
oversensitive on that point.  It had to do with this remark:
\bcc
It suggests to me that agonizing over the sorts of questions in your
hag-beauty-insurance story is not going to help us understand what
we're doing as scientists and, moreover, that it will lead our
program into a place where no other scientists want to go.
\ecc

This is (what I recall to be) the third time you have made such an
allusion, i.e., that I am going places where no scientist ought to
want to go.  The only conclusion I could draw from this was that you
were implying that the subject of this whole discussion is
nonscientific.  I hope you will at least understand that I might find
that insulting.  And my reply was an attempt to put a quick end to
that train of thought.  Over and over, I feel that I have been trying
to make a simple point, and most importantly, a logical point---it
just happened to be an unpalatable point.  So, in the end I guess it
seemed to boil down to both of us accusing the other of being
unscientific.

I will write a (calm) set of emails replying to all the new issues
you and {\Ruediger} have brought up in the next day or two.

\section{26-09-01 \ \ {\it Form 1} \ \ (to R. Pike)} \label{Pike3}

Below is what I wrote up this afternoon.  You'll probably find it too long, but you can tell me what you'd like to see trimmed out.
\bq\noindent
\begin{center}
{\bf My First Year at Bell Labs}
\end{center}

If I had to pinpoint the thing I discovered liking most at Bell Labs, it would be the incredible freedom it afforded me to pursue pure research, unchained from the bureaucratic duties of my previous institutions. This year has been a good one for science. My focus (as promised last year) has been in mapping out the very foundations of quantum information. The idea is a simple one:  Where the foundations are strongest, the tallest, most sweeping structures will be built.  The best example of this in my own work came from trying to make sense of the concept of an ``unknown quantum state'' from the point of view of Bayesian probability theory.  The solution to the conundrum required the proof of a representation theorem for the most general form a density operator can take for sequences of quantum systems with permutation symmetry.  As abstract as this problem sounds, the same representation theorem allowed Steven van Enk and I to give a (surprisingly) never-explored analysis of the quantum state of a propagating laser field.  This analysis assured the validity of all recent and future quantum-information experiments to do with laser light, from Caltech's quantum teleportation experiments to Richart Slusher's squeezed-light experiments here at Bell Labs---both of which had come under recent attack in the literature.

Along the same foundational lines, I deem my deepest physical insight of the year to be in a reexpression of the content of quantum entanglement, long considered the main ingredient of all quantum information processing.  Carried along by the same current of thought that led to the representation theorem above, I was able to show that quantum entanglement is a secondary effect within quantum mechanics: Entanglement, it turns out, is subordinate to the structure of quantum mechanical measurements.  This lays open a whole new point of view on quantum information processing: Its power may actually come from the invasiveness of quantum measurements, not from the smooth flow of unitary evolution. Indeed, a concrete expression of this can be found in the new quantum computational model of Raussendorf and Briegel that makes use solely of quantum measurement as its computational primitive.  This is now a field of research that is wide open for rapid progress.

In all, I submitted six papers for publication in professional journals, and one very large paper as an invited contribution to a NATO Advanced Research Workshop.  In step with this activity, and in perceiving one of my main roles in the physics community as an enthusiastic promoter of the view that the power of quantum information arises from something deep in the quantum measurement process, I also posted a 504 page (edited) collection of correspondence on this subject on the Los Alamos archive.  I believe---and joking only slightly---this work has received more attention than all of my more technical publications combined.  Also in the same role, I accepted 10 invitations to present my work at international meetings and colloquia around the world, including Hungary, Northern Ireland, Japan, Poland, and Sweden.  I turned down three other ones because of time constraints. Furthermore, I co-organized two meetings of my own, and am in the process of co-organizing two major meetings for next year: ``Workshop on Quantum Foundations in the Light of Quantum Information'' to be held in {\Montreal}, and ``DIMACS-CCR Tutorial and Workshop on Capacities and Coding for Quantum Channels'' to be held in Princeton.  As far as media relations are concerned, I garnered a little attention for Bell Labs in two small appearances: (1) in an episode of {\sl History's Mysteries\/} on the History Channel, and (2) the September 2001 issue of {\sl Discover Magazine}.  Finally, in the capacity of general service to the physics community, I served as an Associate Editor for the new journal {\sl Quantum Information and Computation} and accepted a position on the Advisory Board of the International Center for Mathematical Modeling, {\Vaxjo} University, Sweden.

As concerns internal service for Bell Labs, here are some items I recall. (1) I gave a presentation to Arun Netravali and Bill Brinkman in a private session on ``The Future of Quantum Information Processing.'' With Brinkman mentioning quantum cryptography and computing as long-term research areas for Bell Labs in his July 31 address, I like to think the presentation was somewhat effective. (2) I played an active part in the recruitment of Alexander Holevo to our group. (Holevo is to quantum channel capacities, what Peter Shor is to quantum computing.)  (3) I gave six tutorial lectures to the Mathematical Sciences Research Center on quantum information theory. (4) I mentored an undergraduate student for the SRP program.  (5) I presented a talk in Arnold Auditorium for the Lucent Global Science Scholars Program.  And (6) I have spent a lot of time in general trying to get people throughout the company practicing research in quantum information.  Indicative examples can be found in the recent work of Lorenz Huelsbergen, Gerhard Kramer, and Serap Savari, but also in my role in co-organizing the Wednesday quantum talks.
\eq

\section{26-09-01 \ \ {\it How Close Is He?}\ \ \ (to G. Brassard)} \label{Brassard8}

Not as close as I want him to be, but there are a couple of good theorems in the paper.  Charlie Bennett once asked me, ``How will you know when you've got a clean, crisp, PHYSICAL statement of the content of quantum theory?  Is it like pornography:  you'll know it when you see it?''  I replied, ``Yeah, actually.''  Hardy's porn is good for a starter, but it's not as hot as I want it to be.

\section{01-10-01 \ \ {\it Difficulties} \ \ (to A. Furusawa)} \label{Furusawa1}

I wonder if you might consider allowing me to lay a huge burden upon you?

I am an invited speaker at the 4th Annual Japanese-American Frontiers of Science Symposium, sponsored by the Japan Society for the Promotion of Science and the U. S. National Academy of Sciences, Wednesday, October 10 through Friday, October 12 in Tokyo at the Tokyo International Exchange Center.  I think it is a fairly prestigious talk.  However, because of the recent terrorist threats to Americans, and because our ``state department'' issued an announcement Friday warning particularly of potential trouble in Korea and Japan, I have decided that I will not come to it.  My wife and I will be having a new baby in December and this would have been my fifth trip overseas this year:  In all, it seemed like a good time to curtail my travels, before my ``number'' came up.  Tomorrow I will announce this to the conference organizers.

But, all that said, I am plagued by a great sense of guilt in turning down my invitation on such short notice:  This is why I am writing you.  Would you consider giving a talk in my place?  The format of the conference is to have two talks in each of eight subjects:  one speaker in each subject is Japanese, one speaker is American.  In my session, the Japanese organizer is Hiroshi Imai (from your university), and the other speaker is Tetsuro Nishino.  My plan had been to give an introductory talk on quantum cryptography and quantum teleportation.  Because the audience is mixed, from all areas of science, the talk should be a very introductory one.

I fully realize that you are not an American; however, your quantum teleportation experiment was performed in America, and I am hoping to use that as a bargaining chip.  I think it would be a good talk for you to give, to get you and quantum information recognized by a wider audience.

If you are willing to do this, then I will send your name to the organizers tomorrow and write a strong letter of recommendation.  Thanks for any help you can give.

\section{02-10-01 \ \ {\it Consistencies} \ \ (to C. M. {\Caves} \& R. Schack)} \label{Caves29.1} \label{Schack23.2}

I'm still trying to find the time to reply to all the emails you and {\Ruediger} wrote me last week.  Yesterday I finally had the courage to buy a new car and to drop my engagement in Japan next week.  With those two mental burdens off me, I may actually start thinking science again (soon).

But let me make a brief comment on your latest email.
\bcc
As [{\Ruediger} and I] discussed today, another kind of consistency is that there is no
measurement such that an outcome assigned nonzero probability by
any party is deemed impossible by any other party.   This means that the
supports of all parties must be the same (quantum mechanically or
classically).  As we discussed, this is in some ways not a very reasonable
consistency condition, because there are many situations where it is
violated.  But we should note that if a set of density operators is BFM
consistent, then after pooling the information, the simplest way for
each party to update its density operator is to project into the
intersection of the supports, thus automatically yielding a set of
density operators that have this kind of consistency.
\ecc

It may be the simplest way, but I don't think it is the most enlightened way.  This is because your procedures amounts to Alice accepting that Bob's belief about the raw possibilities for the outcomes is relevant, while ignoring all the rest of his beliefs.  Why would she do that?  Is that natural?  Maybe under some conditions it is, but it is certainly not the most general thing.  In general, it seems to me, if Alice is willing to take into account Bob's opinions any at all, she wouldn't then draw an arbitrary line, accepting some of them while rejecting others.

Your problem sounds reminiscent (but not identical) to the one explored below:
\begin{itemize}
\item
John E. Shore and Rodney W. Johnson, ``Axiomatic Derivation of the Principle of Maximum Entropy and the Principle of Minimum Cross-Entropy,'' IEEE Trans.\ Info.\ Theory {\bf IT-26}(1), 26--37 (1980).
\item
John E. Shore and Rodney W. Johnson, ``Properties of Cross-Entropy Minimization,'' IEEE Trans.\ Info.\ Theory {\bf IT-27}(4), 472--482 (1981).
\end{itemize}

\section{03-10-01 \ \ {\it Kid Sheleen} \ \ (to N. D. {\Mermin})} \label{Mermin45}

What a busy couple of weeks it's been for me!  I am sorry it has
taken me so long to reply to you.  But cars needed buying, lots of
newspapers needed reading, lots of soul-searching needed doing
(before canceling what would have been my fifth trip abroad this
year), and Bell Labs needed some tolerance (with the transition from
Brinkman to Jaffe as our Research VP).

\bdm
Sorry.  Didn't mean to go on for so long.  {\Caves} is right; this is a
rotten way to have a conversation.  No need to reply.
\edm

Yeah, {\Caves} is right in many ways.  I've never had more frustration
in an email run before.  But then, I don't think I've ever made such
a headlong transition in open view before either.  Regardless,
however, I have certainly benefited from this:  I've never had to
strive so hard to make a simple point clear, and I think that gave me
a load more perspective on the issue than I would have come to by
myself.  Strangely, in a way, the whole affair has hardened me and
made me confident in this new direction of thought.

\bdm
so I'm sympathetic to what you're saying, but I worry that you're
giving up ``objectivity'' on too many fronts.
\edm

You think you worry!  That reminds me of a dialogue in the old movie
{\sl Cat Ballou\/} (with Jane Fonda and Lee Marvin):
\bv
Jackson Two-Bears:  Kid, Kid, what a time to fall off the wagon.
                         Look at your eyes.\\
Kid Sheleen:        What's wrong with my eyes?\\
Jackson Two-Bears:  Well they're red, bloodshot.\\
Kid Sheleen:        You ought to see 'em from my side.\\
\ev
But I don't think I've given up objectivity on too many fronts:  If
there's one thing for sure, it's that I don't want to go too far. The
tenet I hold fast to is that there is something happening in the
world, something was happening before scientists appeared on the
scene, and something will continue to happen (in one form or another)
should we wipe ourselves off the planet next year.

The issue is, how does that something interface with us, to what
extent can we grasp it, and how do we modify it by our very attempt
to grasp it?

Let me try to reply to some of the points in your letter.

\bdm
On the other hand you did seem to be undermining the impact of the
trails of evidence when you gave a central role to the question of
whether your Lucent colleague really had the polaroid oriented the
way you had been led to believe it was oriented.
\edm

Hideo Mabuchi is a young professor at Caltech (probably the
youngest); he won a MacArthur Fellow (i.e., \$500K) last year too. He
started grad school at Caltech at about the same time that I started
up with {\Caves}.  You can see he's the smart one.  We've been friends
ever since we first met at the Torino meeting of 1994.  He keeps the
napkin where I first explained the Holevo bound to him; I mention him
whenever I can (like in the NATO paper) because of all the pearls of
wisdom he gave me.

\bdm
I'll grant you that that's something to worry about, but it seems to
me on a different level from the characteristic quantum ambiguities,
\edm

Three months ago, I would have thought it was on a different level
too.  But now it is clear to me that the two NECESSARILY feed in to
each other.  (Though the whole issue has been building in me for two
years:  See Samizdat, page 127, in a note to {\Caves} titled ``The
Dangers of Probabilismo.''  I'll talk about that more in a minute.)

\bdm
unless you want to deal with your colleague and his polaroid on the
same footing as the photons, whence my joke about your closet
Everretism.  (Please read on before concluding that I still haven't
got the point.)
\edm

I did read on---many times---and I think you've got some of the
point, but not the whole thing.  In particular, I still contend that
Everettism is the antithesis to my point of view.

There is a sense in which Hideo and his polaroid are on the same
footing as the photons (and I have always thought this).  It is that
in reasoning about them---i.e., in reasoning about what kinds of
traces they will leave from my interactions with them---I am obliged
to use the formal structure of quantum mechanics if I want to do the
best job I can in that reasoning.  But I don't think your remark in
particular has anything to do with this.

Your remark seems to be more of the flavor:  Chris says POVMs (even
simple von Neumann measurements) are ``subjective judgments,'' so he
must mean that their outcomes are every bit as dreamlike and
subjective too.  Measurements that have no concrete outcomes?  What
else could this be but Everettism in disguise?

But there is a non sequitur there (if indeed that is the flavor of
your reasoning).  It is that you (or my caricature of you) think I am
identifying---in a supremely steadfast way---the ``click'' that makes
its way to our senses when we perform a measurement and the very
being of a POVM.

Instead, the idea is that the ``click'' is REAL, as real as you could
want anything to be (for those concerned, for those who know of it).
But, the index $b$ (from some POVM $\{E_b\}$ that we associate with
it) is not the thing-in-itself.  It just gives the ``click'' a NAME
and a {\it context\/} through which we can draw inferences and through which
we can start to contemplate further reasoning.  It is that
identification which is an ultimately subjective element; it is not
the ``click'' itself.

I had hoped to draw your attention to this distinction with the
Penrose beauty-hag example, but I see I completely failed on that
count (not only with you, but also with {\Caves} and {\Schack}).  There is
the stuff of the world---the ``click'' (in part).  And then there is
our description of it.  The two things are not the same things; one
lives in my head, and one lives partially outside it.  To accept
quantum mechanics, is to accept a template for that description and
to accept a method of manipulating one's judgments thereafter.

Does any of this make sense?  I think it is the main point I have
been wanting to make to you, but as far as I can tell, you've ignored
it.

\bdm
I agree that figuring out what measurements are associated with what
operators (even at the von Neumann level) is something you have to
bring in from the outside.  So is knowing what the Hamiltonian is.
I'm not so sure these are the same kinds of subjective judgments as a
state assignment, as you maintain next\ldots.
\edm

The new thing I've been saying since August is that the
identification of a particular POVM (and one element therein) with a
measurement ``click'' is a subjective judgment.  I had not clearly
appreciated that before.  However your mention of the Hamiltonian is
apropos, because that one, at least, I had caught before.  (That is
what Samizdat, p.\ 127 is about.)  There are several ways to see that
``the Hamiltonian'' must be on the same subjective footing as the
quantum state.  Here's one; I'll just quote you:

\bdm
I worry that ``objective'' is taking on too many different meanings
here.  For example, EPR makes it clear (at least to sensible people)
that the polarization state of a photon is not an objective property
of that photon.
\edm

Will you accept that the existence of quantum teleportation carries
as much force as the EPR argument that the quantum state is not an
objective property of a photon?  If you will, then can you tell me
what the import of all the recent papers on ``teleporting a unitary
operator'' is?  (See for example \quantph{0005061}, but there
are a plethora more.)  But even if you don't, the argument is simple:
Hamiltonians can be toggled from afar by our measurements on
entangled states.  ({\Carl} gets himself out of this by saying that the
only thing being toggled is the ``effective'' Hamiltonian, not the
underlying quantum circuit, but I say where there is a tear in the
fabric there is a rip.)

\bdm
I think you're saying that if the vertical alignment of the polaroid
is an objective fact, then the state --- ``vertically polarized'' ---
of the emerging photon is also an objective fact.  But that's not the
same as saying it's an objective property of that photon.
\edm

That is right, there is a difference.  And you have gathered what I
was saying (almost) correctly.  (I wrote ``almost'' to help remind
you of the points above about identifying ``clicks'' with POVM
elements.)  The point is, the quantum state had better not be an
objective fact, or the point of view that {\Caves} and {\Schack} and I have
been trying to build up will be in deep trouble.  What is wrong with
taking a quantum state to be an objective fact, as long as one drops
the insistence that that fact be localized with the photon? (I.e., as
long as one does not make it a property of the photon where it
stands.)  At first sight, maybe nothing:  I think that is probably
the point of view you are trying to build; it is also the point of
view Philippe Grangier has been trying to build in his recent {\tt
quant-ph}'s.  But, on second sight, one cannot forget that the
quantum state uniquely specifies a set of probabilities. If the
quantum state is an objective fact, then so are those probabilities.
And now it is on your shoulders to tell me what objective probability
can possibly be.  I won't stand for anything short of an operational
definition.

\bdm
I worry that ``objective'' is taking on too many different meanings
here.  For example, EPR makes it clear (at least to sensible people)
that the polarization state of a photon is not an objective property
of that photon.  It appears from the above that you believe you can
only consistently take this position if you deny objective status to
the outcome of the actual polarization measurement which enables you
to predict the outcome of the measurement that has not yet been made.
\edm

No.  See the point above, where I used a little TeX notation.  I
accede to the objective status of something happening in a
measurement intervention.  I just don't accede to an objective status
for what we decide to call it, i.e., for which POVM we decide to
associate it with.

\bdm
(Forgive me, but this smells like a many-worlds strategy again.  Your
answer to the EPR paradox would seem to be to deny that the first
measurement had an objective outcome.  Recall Henry Stapp who has
been saying for decades that only for the Everretista is nonlocality
not a consequence of EPR.)
\edm

I forgive you.  But I hope you'll tell me that, with your newfound
enlightenment, it doesn't smell so much like Everett anymore.  (Your
imagery conjures up my own imagery of walking on a warm day near a
trash can full of lobster parts behind some coastal New England
restaurant.)

\bdm
I'd prefer a middle ground which allows me to talk about objective
facts but not objective properties. (It just now strikes me that one
might call this correlations without correlata.)  I've never been
sure I can do this coherently, so I'm sympathetic to what you're
saying, but I worry that you're giving up ``objectivity'' on too many
fronts.
\edm

I am in partial agreement with your first sentence, and I would like
to think I have hit a sweet spot for that part.  I'll give you
objective ``clicks'' (though I might not call them ``facts'' \ldots\
but that's a story I probably shouldn't get into right now); I just
stand fast against the objectivity of the quantum state.  What is
more middle ground than that?  However, I do not share your aversion
for objective properties.

It seems to me, quantum systems do have some properties that we can
get our hands on.  I usually preach the bundle of
information-disturbance curves associated with a system, but let me
try from a different angle to convince you of at least one property.
I say that the quantum state cannot be an objective property because
we can toggle it from a distance.  I say that the Hamiltonian cannot
be an objective property because we can toggle it from a distance.
But what about a quantum system's Hilbert-space dimension $d$?  Can
you think of a way to toggle that number from a distance?  I can't.
And so, to that extent, I'm willing to call the raw number $d$ an
objective property of a part of the world.  Now, what is the physical
meaning of $d$?  Well, that's why I struggle with all this
information-disturbance stuff, but that's not the issue at hand.  The
issue is that one need not give up on all objective properties.

There are things in the world beyond our control:  One them is the
outcome of a quantum measurement, and one of them may be the
dimensionality of a Hilbert space.  Objectivity means nothing to me
if it doesn't mean that some things are beyond my control, are beyond
my whim and fancy.  To the extent that I'm willing to say this, I
don't think I'm giving up on objectivity on too many fronts.

Does this strike any chords in you? \medskip

\noindent PS.  By now you should have received our modified version of
``Making Good Sense.''  I won some good battles there:  We no longer
claim that two observers must be compelled to the same unique state
via a Dutch-book argument.  But I lost some too.  I continue to think
the paper is misleading as hell, always talking about a ``unique''
state assignment and using the word ``certainty'' in a way that still
troubles me.  We were able to compromise only in that I thought
things were now worded in a sufficiently vague way that I could worm
out of them in my future talks and publications.  I don't think we
say anything factually against my beliefs, but the reader will have
to be on his toes to not get fooled about where I really
stand.\medskip

\noindent PPS.  Here's another thing I ought to tell you.  PRA made the
mistake of asking me to referee the BFM paper.  Despite what I wrote
you earlier about probably accepting the challenge if it came up, I
decided to decline the opportunity.  I like the math of the paper,
but I just could not agree with what you make of it.  It seemed more
appropriate to let some less tainted souls than mine tell you what
they think of it.

\section{03-10-01 \ \ {\it Replies on Practical Art} \ \ (to C. M. {\Caves} \& R. {\Schack})} \label{Caves30} \label{Schack24}

\bcc
The philosophers tend to proceed by telling a story---reasoning by
analogy, they call it.  The actual problem is too hard for them to
formulate, so they immediately introduce a simple analogy, reach a
conclusion they like within the analogy, and then transfer the
conclusion back to the actual problem, without ever justifying why
the analogy has anything to do with the actual problem.
\ecc

Except for omitting the final justification---which is more than
important---is this so different from what you teach in your physics
classes?  I.e., That one ought to try one's ideas out on a simple
example first?  One that may already contain the essence of the
problem, before embellishing it with too many details?

\bcc
I enjoy reading your stories, but perhaps you're falling into the
same sort of trap in a less obvious way.  The difficult and very
personalistic questions about assigning probabilities in your
hag-beauty-insurance story are important in thinking about Bayesian
probabilities.  These personalistic factors are well known to be
present in a subjective interpretation of probabilities, but do we
really have to worry so much about them in the context of
interpreting quantum probabilities?
\ecc

My point was to remind you guys and {\Mermin} that these personalistic
factors always must exist, else we would have no need to take such
pains to talk about a ``subjective interpretation of probabilities.''
If they are well known (as you say), then they should not be
forgotten and replaced with purely objective ``trails of evidence.''

The point is, yes we must worry about them in the context of
interpreting quantum probabilities.  We must recognize that that is
part of the very problem.  Once we have recognized it, then we can
move on and almost forget that the issue was ever there---just as one
can do in whole textbooks on orthodox probability theory---but that
first step is a supremely important step.

\bcc
You and others write papers every day where this party assigns this
state and that party assigns that state, and I don't see any of these
papers agonizing over difficult, personalistic questions of what
state to assign.  You're right to keep badgering us to pinpoint why
this is true, but the fact that it is true---we don't worry about
this kind of stuff when making quantum state assignments---leads me
to believe that there is an answer.
\ecc

Nor do you see any sophomore-level textbooks on probability theory
agonize over these personalistic questions when posing its exercises
at the end of each chapter.  On the one hand, you completely missed
the point I was trying to make, but on the other you also completely
answered it.

The point is one does not have to worry about these personalistic
questions to get quantum mechanics as a {\it calculational\/} tool
off the ground.  In that regard, the present issue is no different
than with classical probability theory.  Indeed, this is probably why
in both theories a large sect of the practitioners have turned to
``objective'' interpretations of their main terms (alternatively,
probabilities and quantum states) in such a misguided way.  It is in
the very recognition that personalistic questions exist, that one is
compelled to finally get the foundations of the two theories
straight.

In practice, what almost always happens?  In the case of classical
probabilities, when given a specific problem, one reduces and reduces
the problem until one has transformed it into an equivalent problem
for which one feels confident in making the uniform probability
assignment.  Thereafter one {\it derives\/} the probabilities for the
problem of real interest by transforming and grouping, etc., until
one rearrives back at the starting point. (This is a point I probably
first learned from {\Ruediger}.)  Think for instance, if I were to ask
you what is the probability of obtaining a 7 or an 11 in a roll of
two dice.  Your mind would probably first jump to the judgment that
all six sides of each die are equally likely, and then let the
calculations flow.

Now of course, being a Bayesian, you would leave open the possibility
for something else than a uniform assignment in that step above.  But
in practice, there are some things that most of us can usually agree
upon \ldots\ and those are usually the starting points for textbook
problems.

The issue is little different in quantum mechanics.  When presented
with a problem of calculating a quantum state for a given physical
system, what do we usually do but reduce and reduce (or expand and
expand) the problem until we come across an equivalent one for which
we are confident we can predict the outcome of some measurement with
certainty?  Thereafter we work our way back just as before.  Just
think of Scully and Lamb's derivation of ``the'' quantum state of a
laser.  Alternatively, think about {\Moelmer}'s justification of the
same state.  [Indeed one might say that this is what the whole
(worthless) decoherence program amounts to:  deriving one subjective
state from another and then thinking there is something deep about
it.  But that's an aside.]

Now just as before, being a Bayesian, one ought to leave open the
possibility for something else than the particular pure state in the
basic step of this derivation.  But in practice, there are some
things that most of us can usually agree upon \ldots\ and these are
usually the starting points for textbook problems.

I think the similarity is overpowering.  It is enough in both cases
to recognize that ultimate personalistic issues exist, but then the
homework assignment goes on---the student reduces the problem to a
judgment few people would dissent on.

\bcc
It suggests to me that agonizing over the sorts of questions in your
hag-beauty-insurance story is not going to help us understand what
we're doing as scientists and, moreover, that it will lead our
program into a place where no other scientists wants to go.
\ecc

Looking back over this note again, your language really was very
scolding throughout---``trap,'' ``badgering,'' ``agonizing,'' ``a
place where no other scientists want to go''---in spite of the fact
that you warned me it would be ``a constructive and gentle
criticism.''  I understand that I am guilty of no less:  There is no
doubt that I can be arrogant and abrupt (and paranoid) at times. But
in all this massive email, I feel that I have been providing a
service, sharing ideas that I might not have if I didn't feel we
should be brothers in arms.  It became a little hard to gulp that all
these notes might be viewed as little more than a nuisance.

\bcc
\label{CavesismGleason} The answer might be as simple as this: we can only do science in
situations where we scientists have agreed that such personalistic
factors can be essentially eliminated, and quantum mechanics is the
very pinnacle of this kind of situation.  I think that's the content
of our statement that ``Gleason's theorem is the greatest triumph of
Bayesian reasoning'' and of our ``principle of quantum
indeterminism.''
\ecc

I think the answer might just be as simple as that, but at a level
higher than the one you are contemplating.  The agreement we need is
in accepting quantum mechanics as a method and a restriction for
shuffling about our more mundane, everyday beliefs.  What that
entails is accepting POVMs as the structure of the questions we can
ask a system and the Kraus state-change rule as our method for
updating our beliefs.

Everyone keeps asking, what is the objective piece of quantum
mechanics?  I answered some of my beliefs on that issue in the letter
I just sent off to {\Mermin} (and then forwarded to you).  But I think
there is also quite a bit to be learned on the issue by first turning
the question toward Bayesian probability theory.  What is the
objective piece of Bayesian probability theory?  I think all three of
us are in agreement that it is not in the particular probability
assignments that one might make.  But is there {\it any\/} objective
piece at all?  I think there is.  Take Bayes' rule as an example.  I
would say that it is something objective in the theory: it is the
ideal of behavior.  If one doesn't use it, one can be taken advantage
of.  You agree that Bayes' rule is the ideal of behavior, and I agree
that Bayes' rule is the ideal of behavior:  it would remain the ideal
of behavior if all of us were wiped off the planet.

Likewise, it seems to me, Gleason's theorem plays a similar role.
There must be a sense in which accepting that the structure of our
questions to the world (or, alternatively, our interventions upon it)
conforms to the structure of the POVMs must be the ideal of
behavior---something not so far removed from Bayes' rule itself.  It
is the ideal of behavior in the light of some crisp, physical fact. I
don't know what that fact is yet (in any precise sense), but that
does not stop me from seeing the outline of how the various
structures in quantum theory should be classified:

\begin{center}
\begin{tabular}{|c||c|} \hline
& \\
measurements $=$ POVMs  & objective feature (physical fact)
\\
Born RULE (via Gleason)  & objective feature (an ideal of behavior)
\\
Kraus state-change RULE  & objective feature (an ideal of behavior)
\\
quantum state    & subjective judgment (always)
\\
time-evolution map    & subjective judgment (always)
\\
Hilbert-space dimension   & objective feature (physical fact)
\\
particular POVM assignment & subjective judgment (always)
\\
particular Krausian assignment & subjective judgment (always)
\\
& \\
\hline
\end{tabular}
\end{center}

The list is not exhaustive; but I think these are the ones I see
clearly at the moment.

The point is:  Agreement required for science?  Yes.  Compelling
interpersonal agreement as a (potential) statement about the
agent-independent world?  Yes.  Agreement necessary at the level of
quantum states?  No.

\bcc
I'm not sending this to {\Ruediger} and {\Mermin}, but you can send it to
them if you think it's worthwhile to do so.
\ecc

Well, clearly I thought it was worthwhile to share my answers \ldots\
but there's that arrogance again.  \smiley

With a smile and a conciliatory tone,

\section{03-10-01 \ \ {\it Further Replies on Practical Art} \ \ (to C. M. {\Caves} \& R. {\Schack})} \label{Caves31} \label{Schack25}

\bcc
I think we agree that there are things that are effectively facts in
the effective reality of ordinary experience.
\ecc

In the words of {\Bennett}'s father (in such a context), ``These are
very deep waters.''  Since becoming enamored with {\James}, {\Dewey}, and
{\Schiller}---and having read copious (by my standards) amounts of
them---I'm not completely sure how I should answer you.  The issue
is, I'm not completely sure in which sense you are using the word
``fact.''  I have a feeling it is a more loaded sense than you would
guess.  But I don't want to get into that now:  You suggested some
simmer time, so I will leave you some until it becomes absolutely
necessary.

There is, however, one thing I dislike about this sentence, and that
is the phrase ``the effective reality of ordinary experience.''  But
you touched upon that very point in your note ``More on Pots and
Kettles''; so I'll say more to the issue in detail when I reply to
that note.  The main thing, though, is that I would say our ordinary
experience is the rawest stuff around:  It's the very stuff from
which we build these super-smooth pictures by way of which we derive
our further expectations.  There is nothing effective about it:  It
is the stuff, it is the starting point.  To use the word
``effective'' makes it feel secondary and derived (which is what you
have been striving to get at, not me).

\bcc
The questions arise in what those facts tell us or, perhaps, in
whether and what they compel us to believe.  The argument is about
pure-state assignments, not about mixed-state assignments.  You
believe that the subjectivity of pure states requires that it be
possible for different agents to assign different pure states.
\ecc

Yes.

\bcc
To say something is subjective is to say that it exists only because
of us and does not have an independent existence out there in the
world.  It also implies that different agents can disagree, the
degree of possible disagreement being just the flip side of the
degree of intersubjective agreement.
\ecc

Yes.

\bcc
Suppose we had the idea that facts in the effective reality force one
to a particular pure-state assignment.  The resulting pure state is
then based on a trail of evidence in the effective reality and is
embedded in each agent's mind.  Is the pure state then out there in
the world, independent of us?
\ecc

I would say, yes it is.  The agent's state of belief is then an
unneeded complication in everything under discussion.  The fact is
that there is a one-to-one correspondence between (sets of) facts and
quantum states.  You can say the agent's mind is nevertheless needed
to ``house'' that state, but then, to me that looks to be nothing
deeper than invoking the luminiferous ether to support the
electromagnetic field.

\bcc
Do I have it right that this is the issue, or at least an issue?
\ecc

Yes.  I have always perceived this to be the main issue.  To the
extent that I have said words all around this, it has been---I
believe---to attempt to give different angles for viewing the same
thing.  Every time I saw you, {\Ruediger}, {\Mermin}, Brun, etc., be
reluctant to accept the point, I tried to present it from a different
angle so as to be more convincing.  I take it now that everyone only
found that to be confusing.  But what else could I do?  And I can't
complain too much, because I think the whole process has sharpened my
presentation of the point (which I maybe only dimly perceived at the
beginning).

\bcc
I don't necessarily see where the pure state is if it's thought to be
out there in the world.  The trail of evidence is not a pure state;
we construct the pure state from the trail of evidence, but dogs
don't and dinosaurs didn't.
\ecc

A one-to-one mapping is a one-to-one mapping.  I do not see how YOU
cannot see that making these statements is not a tacit acceptance
that the quantum state is an objective entity after all.  Maybe you
have thought this all along.  Namely, that when you said a quantum
state is not a state of nature, what you really meant was that it was
simply not localized on the physical system it is meant to describe.
It is a state of nature, i.e., it is a collection of facts within
nature, it is just not living on the system it's intended for.  From
this point of view, it's clear why dogs don't use them:  Dogs
aren't clever enough or advanced enough technologically to discover
the true states of nature.

But I surely never thought this when I used the slogan, ``a quantum
state is a state of knowledge, not a state of nature.''  If facts can
uniquely determine a quantum state, and facts live in nature, then a
quantum state is a state of nature after all.

\bcc
The pure state is not out there in the system for the reasons we have
long discussed: the system can't report its pure state, and a
system's state can be changed to any pure state drawn from
incompatible sets without ever getting close to the system.  It looks
to me like the pure state is purely in our minds.
\ecc

I'm not sure how this remark fits in.  You might be making a call for
me to consider putting the ``trails of evidence'' into the mind, but
I'm not sure.

\bcc
You are insisting, I believe, that in order for a pure state to be
subjective, it must be possible for different agents to disagree on a
pure-state assignment.  You say, I believe, that if we are forced to
a particular pure-state assignment by the facts in the effective
reality, then the pure state becomes objective.
\ecc

Yes.

\bcc
I don't know where to come down on this.  It is one aspect of the
question I always ask of which aspects of maximal belief get
translated from realism to quantum mechanics.  It also has to do with
the nature of the ``facts'' in the effective reality and thus how the
effective reality arises out of quantum mechanics (this is the
content of {\Mermin}'s Pots and Kettles).  My own take at present is
that the effective reality is a form of intersubjective agreement.
\ecc

Fair enough that you don't know where to come down on this:  I will
try not to lose my patience any more.

\bcc
You are right in principle that nothing compels us to a particular
pure-state assignment, but clearly wrong in practice.  All our
experience with quantum mechanics suggests that we have no trouble
agreeing about pure-state assignments, so no matter how the facts in
the effective reality arise, there is nearly total intersubjective
agreement on what they imply for pure-state assignments (this is the
content of my Comment on Practical Art and further comments on Pots
and Kettles).
\ecc

I hope my previous note addressed this adequately.  In contrast to
what you say, I believe that I am right in principle and right in
practice.  You might have said the ``all our experience'' sentence
about classical probability theory if your name were Richard von
Mises.  He would have said that all our experience with dice shows
that we have no trouble agreeing that its outcomes are all equally
probable.  But you're not von Mises, and you've had the advantage of
having had 75 years of good Bayesians clear the air for you.  The
issue you bring up has little to do with the structure of the
physical world, and little to do with the structure of the
Hamiltonians we feel compelled to describe it with.

\bcc
Let me know if I have got your position straight.
\ecc

I think you did.

\bcc
If I have, then it seems to me that we are not far apart, the only
gap being how much we are willing to ascribe to the apparent
agreement that exists in assigning pure states.  You prefer to
emphasize that nobody can be coerced into this agreement, and I
prefer to emphasize that in practice nobody has to be coerced into
it.
\ecc

A point of emphasis can make a huge difference in a philosophy.  And
a difference in a philosophy can make a huge difference in the
practical and applied questions one might ask of quantum mechanics.

And I'm off to Lup\'e's for the best Mexican food in New Jersey. (You
know that's not saying much.)

More tomorrow!

\section{03-10-01 \ \ {\it Next Week} \ \ (to J. N. Butterfield)}   \label{Butterfield0}

Thank you for reminding me that you'll be in town soon!  It was a hard decision, but I decided not to go to Japan after all.  (It would have been my fifth trip overseas this year, in any case, which is maybe too much.)

I would love to meet with you.  Probably the most fun thing for me would be to meet you in the city somewhere, if you're game to that.  I could spend the morning book shopping, meet you for lunch and afternoon discussions somewhere in the Village or East Village, and then return to book shopping.  How does that sound?  Monday would probably be best for me; but Wednesday (second most favorite) and Tuesday (least preferred) would be doable too.

I would especially like to hear more about the deeper desires that underlie the program you outlined in your talk in Belfast.

I'll put all my phone numbers (work and home) below.  Let me hear from you before your departure and we can start to work on a more precise plan.

\section{04-10-01 \ \ {\it Champagne and Roses} \ \ (to J. N. Butterfield)} \label{Butterfield1}

\bjnb
Thanks for yours. Im very Glad theres a chance well meet; but I don't
want you to come to NY unlesss (i.e.\ and only unless!\ A
Halmos-ism) youd like to come for book-shopping etc anyway.
\ejnb

Please realize, you are the imperfection on the smooth surface of my champagne flute.  You are the catalyst in my converter.  No, I wouldn't normally go to NY City on a Wednesday, no matter the strength of my constant obsession to hoard books.  But all it ever takes to seed the bubble is the promise of a good conversation!  Let's meet in NY City.

I know of a good Afghani restaurant (``Khyber Pass'') in the East Village.  Would you like to meet there if it's still open?

\section{04-10-01 \ \ {\it Finicky Sins} \ \ (to C. M. {\Caves})} \label{Caves32}

By the way, in saying this yesterday,
\bq
\noindent
There is, however, one thing I dislike about this sentence, and that
is the phrase ``the effective reality of ordinary experience.''  But
you touched upon that very point in your note ``More on Pots and
Kettles''; so I'll say more to the issue in detail when I reply to
that note.  The main thing, though, is that I would say our ordinary
experience is the rawest stuff around:  It's the very stuff from
which we build these super-smooth pictures by way of which we derive
our further expectations.  There is nothing effective about it:  It
is the stuff, it is the starting point.  To use the word
``effective'' makes it feel secondary and derived (which is what you
have been striving to get at, not me).
\eq
it dawned on me afterward (on my drive home) that I was committing a
sin:  There was a time when I liked the phrase.  As you know, I
commandeered it when writing my {\sl Physics Today\/} articles with
Asher.

So, maybe I was being overharsh, or maybe just finicky.  In any case,
even when I used it unreservedly, I know that I had a distinct flavor
of the phrase in mind from the way you had been using it. [See our
discussion starting on page 133 of the Samizdat.]

Maybe I'll say more about this later today.

\section{04-10-01 \ \ {\it Replies on Pots and Kettles} \ \ (to C. M. {\Caves} \& R. {\Schack})} \label{Caves33} \label{Schack26}

This letter is going to be a hard one to reply to, because I don't
quite see how much of what I have said in the past led to the points
you make here.  So, let me just plunge into the thing and see what
comes out.

\bcc
I thought you were in the camp that holds that the our
experiences---our actions or interventions and our perceptions of the
world's response---are primary
\ecc

I thought I was too.

\bcc
and that the function of science is to account for them.
\ecc

It's this part of the sentence that I'm not so sure of (though it's
not clear to me exactly what you have in mind).  If our experiences
are primary, then it does not seem to me to be within science's
purview to account for them.  I believe, instead, the view I have had
for quantum mechanics for some time is best mimicked by these words I
picked up from William {\James} last month:

\begin{quotation}
Metaphysics has usually followed a very primitive kind of quest. You
know how men have always hankered after unlawful magic, and you know
what a great part in magic {\it words\/} have always played. If you
have his name, or the formula of incantation that binds him, you can
control the spirit, genie, afrite, or whatever the power may be.
Solomon knew the names of all the spirits, and having their names, he
held them subject to his will.  So the universe has always appeared
to the natural mind as a kind of enigma, of which the key must be
sought in the shape of some illuminating or power-bringing word or
name.  That word names the universe's {\it principle}, and to possess
it is after a fashion to possess the universe itself. `God,'
`Matter,' `Reason,' `the Absolute,' `Energy,' are so many solving
names.  You can rest when you have them.  You are at the end of your
metaphysical quest.

But if you follow the pragmatic method, you cannot look on any such
word as closing your quest.  You must bring out of each word its
practical cash-value, set it at work within the stream of your
experience.  It appears less as a solution, then, than as a program
for more work, and more particularly as an indication of the ways in
which existing realities may be {\it changed}.

{\it Theories thus become instruments, not answers to enigmas, in
which we can rest.}  We don't lie back upon them, we move forward,
and, on occasion, make nature over again by their aid.
\end{quotation}

Science does not account for our experiences.  Science builds on them
and gives us a structure by which to imagine pushing them to a new
extreme.  This is why I have laid such emphasis on calling the world
``malleable'' (for instance in my ``Activating Observer'' document
that I shared with {\Ruediger} \ldots\ and maybe you too, I can't
remember).  It seems to me, science does not say so much about what
is, but what can be (subject to the limitations to our actions
captured in the very structure of the given theory).

\bcc
Trouble is that when our interventions proceed to too fine a level,
the world's response is not deterministic and, furthermore, cannot be
described within the realistic language of ordinary experience.
Surprisingly we find that we can use the strange, unrealistic
formalism of quantum mechanics to describe the intrinsic
indeterminism that intervenes between our actions and our perception
of the world's response.
\ecc

What do you mean by ``too fine a level?''

\bcc
I thought you were ascribing some sort of objective or
intersubjective reality to our primary experiences.
\ecc

Pretty much.  Or, at least that's what I thought too.

\bcc
I thought the difference between you and me was that I think that we
must somehow derive from quantum mechanics---or, at least, make
consistent with quantum mechanics---the apparently realistic features
of the emergent ``effective reality'' of ordinary experience, whereas
you think this is unimportant, thus accounting for our different
reactions to the decoherence program.  But you now seem to be
demanding much more than I do,
\ecc

I've always thought that I've demanded much less than you, and I
don't think I've changed my tune on this account for several years.
For the view I dream of constructing, what is the classical world? It
is a world for which the agents describing it are full of ignorance
and the best to which they can muster is a lot of imprecise control.

I look out at one of the trees outside my window, and I ask how I
might capture everything I'm willing to say about it into a single
density operator.  I can't say much about that, but I'm willing to
bet that if I would carry the project through, what I would end up
with will be so mixed, so thermal, that it'll be just about commuting
with anything else I might have come up with, even if I had stared at
the tree a little longer.  This little fact---it seems to me---has
nothing to do with the Hamiltonians of the world (as if they were
objective things).  It is a function of my pure ignorance and my
unwillingness to tear the tree apart and refine my beliefs.

The idea toyed with here is that it is just ignorance, no matter how
we each walk into the room with it, that leads to the classical
world.  If I am so ignorant as to use an almost commuting set of
density operators for a given object, then any (gentle) attempt
you---as another scientist---may make to refine those beliefs will
not cause my beliefs to be any less valid:  Your information
gathering, will not cause a disturbance to my description.  And
therein---the speculation goes---lies the essence of classical
mechanics.

\bcc
that before we even start, we be able to explain exactly how the
effective reality works and at what point it emerges.  As David
points out, this is exactly what the Everettistas demand.
\ecc

As best I can tell, this remark can only come from viewing my program
(more accurately, my dream) from your philosophical predispositions.
I would never demand that we ``explain exactly how the effective
reality works and at what point it emerges.''  The classical world
comes first.  Quantum mechanics (as a theory of inference) extends
beyond it, by taking into account new phenomena that simply can only
be seen when working in a regime of less ignorance.

A relevant ditty to read might be my essay ``Always One Theory
Behind'' on page 464 of the Samizdat.

\bcc
Of course, after making the demand and finding present responses
unsatisfactory, you and the Everettistas go in quite different
directions.  They, out of an anal need for naive realism, simply
concoct a naive realism to go with the state vector.  You certainly
aren't going in that direction, but being risk averse and already
burned, I'm not going to risk a description of your direction here.
\ecc

I loved the phrase ``anal need.''

\bcc
To your credit I think you won't claim to have gotten your ideas
worked out entirely (perhaps the rest of us can be allowed some
access to that defense).  Still you might want to think about the
road you're traveling on and how it relates to this question of
taking as given the apparently objective experiences of our daily
lives.
\ecc

Thanks for the credit.  It's refreshing to be complemented for not
acting like a guru with all the answers.  People at foundations
meetings will have none of that.  It's been my experience that they
demand you tell them exactly what reality {\it is\/} \ldots\ before
they quickly tell you you're wrong.  (Matthew Donald told me he
couldn't take me seriously as a foundations researcher because I keep
evading the question of what reality {\it is}.)

Anyway, in conclusion, give me some feedback:  Did I answer anything
that you wanted answered?  (I sure hope I did:  I'm trying.)

\section{04-10-01 \ \ {\it Replies to Morning Coffee} \ \ (to C. M. {\Caves} \& R. {\Schack})} \label{Caves34} \label{Schack27}

\brs
Oh no, I thought we had reached some agreement. One problem is that
Kraus operations (unless they are 1D projectors) will not bring you
to a unique place. Which means that qm does NOT provide a universal
rule of coming to agreement. You need some ad hoc assignment.
\ers

Yes, I had always understood that.  I am sorry if I was sloppy about
expressing it, but I thought I had always emphasized that there are
two things that one can contemplate:  1) if Alice and Bob have
complete freedom to choose what measurement they might perform, and
2) if instead they have at their disposal some fixed measurement
(perhaps not of their choosing).  In the later case, only certain
initial states for Alice and Bob will lead to further agreement after
the measurement.  (See, for instance, my note to {\Mermin} and company
dated August 7.)

\brs
The other problem is that your conversation is far too playful. State
assignments are compilations of betting odds. They are COMMITMENTS.
Chris in your dialogue should have been deeply shaken. He would have
betted his house in New Jersey on this outcome to be impossible.
\ers

Yes, perhaps.  But, on the other hand, there is a counter trend in
you that troubles me.  And that is the basic philosophy that comes
across as the message, ``Once a quack, always a quack.''  What I mean
by this is, suppose I ascribe a pure state $|\phi\rangle$ to some
system, whereas you ascribe $|\psi\rangle$.  As we have laboriously
teased out of this correspondence, from your perspective, I am simply
wrong.  My judgment is not to be trusted (from your perspective).
This much we agree upon.  But I sense that you want to conclude more:
Not only am I not to be trusted in my conclusions about the given
system, but that I am not to be trusted about anything.  You conclude
that I am truly insane just because I adamantly disagree with you
about one thing (as captured by our differing pure states).  I say
that goes too far.

I tell too many stories, but here is a true story.  In discussing the
cardinality of the natural numbers versus the even numbers, Kiki will
accede that there is a one-to-one and onto mapping between the two
sets.  Nevertheless she contends that there are more natural numbers
than even numbers.  I have never had more annoying conversations than
the one we revisit about once a year on this subject.  I simply
cannot convince her that she is not being logical on this issue.  But
still I do find that I trust her judgments on other issues.

A ``misstep'' on a quantum state (even a pure state), it seems to me,
is not the end of the world precisely because of this.

Granted, I am a bit confused on what I think the ascription of a pure
state actually does capture, but I think making it carry the weight
of an agent's rationality or irrationality goes too far.

\brs
Yet another comment: You have said nothing in all your notes (to my
best knowledge\ldots) that tells me why this situation is different
from Chris being certain that there are two chairs in this room, and
{\Carl} with Hideo's help convinces him that he was wrong. I'd say
either Chris was tricked, or he had hallucinations. Todd said: ``This
is why we say that insane people 'out of touch with reality'\,''. I
said the same thing in a different way.
\ers

Since you ask this question more pointedly in another note, I'll wait
on answering it until I get there.

\brs
\bq\noindent\rm
CAF Said: So, in saying this, I don't think one has to do anything
fancy with a measurement to get this thing you call fuchsian
consistency off the ground.  No intricate constructions of states
seem to be needed. Choose ANY (mutually agreed upon) von Neumann
measurement you like, and {\Carl} and Chris will have to agree at the
end of the day.  One of them will feel foolish for having been so
wrong, but as long as he is rational, agreement can still be had. As
{\Ruediger} said, the idea is that this is a property of quantum
mechanics (in its capacity as a law of thought), not of the initial
states.
\eq
I found the idea quite attractive that in a quantum world,
differences can be resolved (using a well-chosen measurement) that
would be impossible to resolve classically. I thought that this was
what you had in mind.
\ers

As far as I can tell, no physical statement (no ascription of a phase
space point) is impossible to resolve classically.  What is different
is that quantum mechanics can do that even without the preexistence
of a phase-space point \ldots\ and that surely is a property of
quantum theory.  But I said I'd come back to this in another note.

\section{04-10-01 \ \ {\it Replies to a Conglomeration} \ \ (to C. M. {\Caves} \& R. {\Schack})} \label{Caves35} \label{Schack28}

Now let me reply to a conglomeration of notes from you two.

\subsection{First to {\Ruediger}:}

\brs
I remember you writing something to the effect that the click in a
measurement is the closest thing to a fact one could come up with
(sorry for not looking it up, but you write TOO MUCH).
\ers

Did the reply I wrote to {\Mermin} yesterday make any sense to you?  I
am now in the habit of drawing a distinction between a ``fact'' and a
``proposition.''  The difference was not so important classically,
but I now think it is paramount quantum mechanically.  The fact (or
consequence of our intervention) is the raw, uninterpreted,
unclassified stuff of the world.  It is the real stuff that makes its
way to our senses and then to our brain to be pondered.  The
proposition, on the other hand, by its very nature attaches a meaning
to the fact and, as such, is a subjective judgment.  What this means
in the quantum case is that to say there is a ``click'' is one thing:
Presumably that is not a subjective judgment if I say it, and Steven\index{Enk, Steven J. van}
says it, and everyone else we talk to says it. However, to say that
that means the particular outcome $E_b$ occurred in the POVM
$\{E_b\}$ is to lay down a proposition, a subjective judgment.

The reason we could get confused in the classical case, and think
that a proposition was more than a subjective judgment is because in
the classical case, propositions don't entail probability
assignments.

This distinction I'm drawing is not so different than the one {\Marcus}
{\Appleby} draws in criticizing the Meyer--Kent--Clifton ``nullification
of the Kochen--Specker theorem.''  See his paper, \quantph{0109034}.

\brs
I am afraid that I still don't know precisely where you stand,
despite of your effort at explaining.

Let's start from the classical notion of certainty. Let's consider
the case where a physicist is certain that some outcome will not
occur. Dutch book consistency implies that the outcome will be in his
nullspace. That's the quantum part.

Do you agree?
\ers

Yes.

\brs
1.) {\Carl} is certain that up will occur and Chris is certain that down
will occur.

2.) {\Carl} is certain that there are three chairs in the room, and
Chris is certain that there are two chairs in the room.

In both cases their beliefs are contradictory in the same, classical
sense.

Do you agree?
\ers

No, I do not think the statements have the same meaning.  In the
first case, in order to find out which of us is ``right'' and which
of us is ``wrong'' we must elicit the world to produce something that
it didn't contain beforehand---namely, the result of the measurement.
In the second case, we can go blissfully along thinking that one of
us is ``right'' and one of us is ``wrong'' simply because the world
has something in it that one of the two of our brains is mirroring
correctly.

There is a difference.  In the classical world, reality is the
ultimate arbiter of truth.  In the quantum world, where we are fairly
convinced that ``unperformed measurements have no outcomes,'' we are
actually lucky in a way that there is still an arbiter of
agreement---we just can't identify it with a preexisting reality. It
seems to me this is a feature of quantum mechanics:  The theory can
still bring us to agreement even without a preexisting truth value
for our propositions.  One might have imagined a more malicious world
where we would not have even been able to rely upon that.

\brs
If we make claims, we are COMMITTED to those claims (e.g., via
betting behavior implied by the claims). I believe that starting from
the notion that different scientists (different agents in the same
linguistic community) should not have contradictory beliefs is
eminently reasonable. To throw this notion over board, one needs
excellent reasons. I enjoy playing with the idea, but I am far from
converted.
\ers

This hits upon what I wrote to {\Carl} yesterday.  The gulf that
separates us seems only to be in where we think this agreement must
be applied to get the engine running.  I say both of us accepting
QUANTUM MECHANICS as a structure is good enough.  (I.e., accepting
the theory is our common agreement.)

\subsection{Now to {\Carl}:}

\bcc
Your point, as we see, is that we can think about life going on after
finding a result deemed to be impossible.  As you point out, there is
a limit (add on the null subspace with epsilon eigenvectors, get a
result in the epsilon subspace, update, and then take the epsilon
goes to zero limit) in which we can think of updating a state
assignment based on outcomes in the null subspace.  But I think this
misses the point.  This isn't updating a prior belief.  It's
realizing that your prior belief is full of it and abandoning it in
favor of life going on, as you put it.

Moreover, the really nice distinction between classical and quantum
Fuchs consistency is lost if we adopt your point of view.  If we
adopt your way of formulating Fuchs consistency, then it has no
content either classically or quantum mechanically.
\ecc

I'm just repeating myself now, I but I don't see that as contentless
at all.  In fact, though the effect is the same in both theories, the
content is quite different across the two of them.  In the classical
case, we can always ``pick up the pieces'' as you say, by realizing
that there is something really there and just checking what it is.
In the quantum case, we can always bend the world into something we
{\it will\/} agree upon, even if we violently disagree upon the
meaning of some subset of our previous interventions.

Here's the way, Josiah Royce put it in a letter in 1888:
\bq\noindent
Thus called upon to explain amid the trade-winds, and under the
softly flapping canvas, the mysteries of [quantum mechanics], I put
the thing thus:  ``There was once a countryman,'' I say, ``from Cape
Cod, who went to Boston to hear Mark Twain lecture, and to delight
his soul with the most mirth-compelling of our humorists.  But, as I
have heard, when he was in Boston, he was misdirected, so that he
heard not Mark Twain, but one of Joseph Cook's Monday Lectures.  But
he steadfastly believed that he was hearing Mark.  So when he went
home to Cape Cod, they asked him of Mark Twain's lecture. `Was it
{\it very\/} funny?'  `Oh, it was {\it funny}, yes,---it was {\it
funny},' replies the countryman cautiously, `but then, you see, it
wasn't so {\it damned\/} funny.'  Even so, Captain,'' say I, ``I
teach at Harvard that the world and the heavens, and the stars are
all {\it real}, but not so {\it damned\/} real, you see.''
\eq

\bcc
The parties can always come to agreement, no matter what their state
assignments, simply by getting amnesia regarding their prior beliefs
and then picking up the pieces in the only way they know how.
\ecc

But there's really more to the story.  They can always come to
agreement, indeed---regardless of how disparate their initial
opinions---if they are willing to make an essentially infinite
expenditure toward laboratory technique.  That is to say, the only
thing that will give assured agreement in all cases is a set of Kraus
operators all of rank-one.  Jacobs and I called those infinite
strength measurements:  the idea being that they are hard to actually
do.  In more real-world measurements, where the operators are never
really rank-one, coming to final agreement will generally require
some initial agreement.  Whence the point in my August 7 letter to
{\Mermin}.

\bcc
I believe that Fuchs consistency is about coming to agreement in the
light of the outcome of an agreed-upon measurement where no party has
to abandon his prior beliefs (certainly one has to agree that this is
a legitimate case to consider).  It could be that all parties are
dumbfounded by the result, but let's put that case aside.  For all
other outcomes, the point is that all the parties be able to come to
agreement by updating their prior beliefs.  This imposes a strong
constraint classically---all parties must have the same support---but
appears to be no constraint quantum mechanically.  That's an
important distinction, it seems to me.
\ecc

I do agree that that is a legitimate case to consider.  What I am not
seeing presently is that its study will shed some foundational light.
But I think I'm open-minded on this one:  I might be convinced yet; I
just don't see it now.

\section{06-10-01 \ \ {\it Your Spelling Conscience} \ \ (to J. Bub)} \label{Bub5}

I was committing one of your notes to my computer archive, and I caught such an intriguing spelling variation in it that I had to write.  In describing Steane's article below, you write ``inciteful.''  Wonderful!  How apropos for the context.  Did you do that on purpose?

\subsection{Jeff's Preply, 08-08-01}

\bq
\noindent\bq\noindent
{\bf [Chris said:]}\ Maybe some of the most interesting discussions centered
around Andy Steane's paper ``Quantum Computation Only Needs One
World'' (which Richard Jozsa presented).  Doug Bilodeau had the
wonderful idea that perhaps some combination of it and [Wootters'] old Ph.D.
thesis could give us a deeper insight into where quantum computing
derives its power from:  Quantum computers are not powerful because
they perform so many calculations in parallel (as the many-worlds
pundits imagine), but rather because they do so {\it few\/} calculations!
I.e., Their power derives from not doing anything they don't have to
do for the final result.
\eq
I think this is exactly right. I've been thinking along these lines myself.  I just wrote an article on `Quantum Entanglement and Information' for the {\sl Stanford Encyclopedia of Philosophy\/} (an online encyclopedia at \myurl{http://plato.stanford.edu}). It's not posted yet because I have to make a few minor changes. The concluding section says:
\bq
The favoured explanation among Deutsch and others of how a quantum system processes information is the so-called `many worlds' interpretation. The idea, roughly, is that an entangled state of the sort that arises in the quantum computation of a function, which represents a linear superposition over all possible arguments and corresponding values of the function, should be understood as a manifestation of parallel computations in different worlds. The quantum circuit is designed to enable the computation of a global property of the function by achieving some sort of `interference' between these different worlds. (For an insightful critique of this idea of `quantum parallelism' as explanatory, see Steane.)

An alternative view, not much discussed in the literature in this connection, is the quantum logical interpretation, which emphasizes the non-Booleanity of the structure of properties of quantum systems.  (The properties of a classical system form a Boolean algebra, essentially the abstract characterization of a set-theoretic structure. This is reflected in the Boolean character of classical logic, and the Boolean gates in a classical computer.) A crucial difference between quantum and classical information is the possibility of computing the truth value of an exclusive disjunction -- for example, the `constant' disjunction asserting that the value of the function (for both arguments) is either 0 or 1, or the `balanced' disjunction asserting that the value of the function (for both arguments) is either the same as the argument or different from the argument -- without computing the truth values of the disjuncts.  Classically, an exclusive disjunction is true if and only if one of the disjuncts is true. In effect, Deutsch's quantum circuit achieves its speed-up by exploiting the non-Booleanity of quantum properties to compute the truth value of a disjunctive property, without computing the truth values of the disjuncts (representing the association of individual arguments with corresponding function values.)
\eq
\eq

\section{06-10-01 \ \ {\it Invitation to Conference} \ \ (to C. M. {\Caves})} \label{Caves35.1}

I hope you will go to Jeff's workshop.  Where there's an invitation, there's usually a purpose.  But, no, I'm not invited.  So I especially  hope you'll be aware of your potential role as a spokesman for the Bayesian clan.

Since I've got thankless insomnia again, let me frizzle away some time with a few character profiles.

\begin{itemize}
\item
Rob Clifton (Princeton):  Mermin says he's a wonderful guy, but I've never met him; present editor of {\sl Studies in History and Philosophy of Modern Physics}; has recently had a big bout with cancer.
\item
Laura Ruetsche (Princeton):  don't know her or know of her.
\item
Hans Halvorson (Princeton):  don't know him or know of him.
\item
Frank Arntzenius (Rutgers):  don't know him or know of him; likely to be a Bohmian.
\item
David Albert (Columbia):  he'll be very aggressive and tell you (and the world) that you don't understand the goals and meaning of science.
\item
Antony Valentini (University College, London):  the most reasonable and interesting Bohmian I've ever met.
\item
Guido Bacciagaluppi (UC-Berkeley):  he's firmly planted in the Zurek camp of thought.
\item
Roman Frigg (LSE):  don't know him or know of him.
\item
Fred Kronz (University of Texas, Austin):  don't know him or know of him.
\item
Harvey Brown (Oxford):  Likes naive realism.
\item
Lee Smolin (Perimeter Institute, Canada):  I leave that one to your judgment.  Good DiVincenzo quote:  ``Barbara and I just read his book; it was written in a different style than you might be used to.  Settling the foundations of quantum mechanics isn't a big enough problem for Lee.''
\item
Jeff himself:  he continues to become more interesting to me.  I think he's got some ideas on where the power of quantum computation comes from that are worth exploring:  [See 06-10-01 note ``\myref{Bub5}{Your Spelling Conscience}'' to J. Bub.]
\end{itemize}

Good night and good morning!

\section{08-10-01 \ \ {\it Larger, Smaller} \ \ (to J. Summhammer)} \label{Summhammer3}

Thank you for your wonderful, thoughtful, long letter!  I have now
read it several times, and each time I think I've gotten a little
more from it.  Thank you also for your concern over my family and
associates in light of the September 11 attack:  As far as I know,
all my friends, and my friends' friends were left unscathed
physically.  But it is all a very frightening affair, and it is
certainly weighing on everyone's life on this continent and the
world.

Concerning the content of your letter, let me especially thank you
for the large number of YES's you wrote into the margin of my paper!
Let me make a couple of small comments on your one NO.

\bjs
It appears to me that quantum theory is the correct way of reasoning,
and classical probability theory is a certain limit of it. But both
spring from the SAME way of reasoning. So far, physics has stood in
the way of clarifying this. The perennial talk of systems and
properties of systems one is forced to carry along when dealing with
quantum theory is a real hindrance to clear thought.  Remnants of
mass points, forces, fields, etc. always sneak in, and with it the
need to allude to an objective world out there. As if repeatably
detectable structures in the statistics of probabilistic events and
their efficient description weren't objective and ``out there''
enough (to me mathematical truths and the Himalayas are equally ``out
there''; the former are mastered by acts of mental climbing, the
latter by acts of physical climbing, but both require willful action
to be conquered. You call it free-standing reality.)
\ejs

This issue has now come up in my email a couple of times since the
Sweden meeting.  Here's the way I put it to Jeff Bub on the last
round:

\bq
\noindent
The main thing is that it sounded like a good opportunity to pound
out the similarities and distinctions between our points of view on
quantum mechanics without being interrupted every three minutes.

I know I suggested I would write a longer letter soon, but I'm going
to wimp out of it again for now.  It would concern the main point of
distinction I see between us (and also between myself and Pitowsky).
Namely, A) that I view a large part of quantum mechanics as merely
classical probability theory (which on my view may be an a priori
``law of thought'') PLUS an extra assumption narrowing down the
characteristics of the phenomena to which we happen to be applying it
to at the moment, while B) you are more tempted to view quantum
mechanics as a {\it generalization\/} of classical probability theory
(and with it information theory).  I know that my view is not fully
consistent yet, especially as I have always distrusted mathematical
Platonism---which you pointed out to me I am getting oh so close
to---but it still feels more right (to me, of course).  Ben
Schumacher, {\Ruediger} {\Schack}, and I had a long discussion on this (on
a long walk) the day after the round table, and I'd like to record
that too.  Ben took a stance quite similar to yours, and maybe even
{\Ruediger} did too (despite his overwhelming Bayesianism).  So, I may
be the lonely guy out on this.  And my view may be subject to change.
\eq

To some extent, I can understand both motivations, i.e., to see
quantum theory as the larger of the two structures, and alternatively
to see it as the smaller of the two.  My thoughts are not completely
set yet about which direction is the best direction, but let me try
to explain a little about what I mean by probability theory possibly
being a larger structure than quantum mechanics. (As evidenced in my
paper, this is certainly the direction I lean most toward presently.)

Consider some physical system, say my house.  And consider some set
of questions you might ask about it.  For instance, what color is it?
(The answers being R, O, Y, G, B, I, V.)  Or, what kind of flooring
does it have?  (The answers being wood, tile, vinyl, or carpet.)  On
so on: Consider every question you might ask about it. If you were a
Bayesian, you would not hesitate taking all the information you know
about me and applying it to the construction of a probability
function for the outcomes of each such question that could be asked.
For instance, if you had gained the impression in {\Vaxjo} that I am
a sentimentalist, you might place a higher probability on my floors
being wooden than otherwise.

However it is also part of the Bayesian creed that there is no such
thing as an invalid probability assignment.  If there are no logical
connections between a set of questions, then there are no constraints
on the probability assignments I might give for their potential
answers.  So, for instance, though you might put a peaked
distribution on the answer to the question about my floors, you might
put a flat distribution on the colors.  And so forth, for every
elementary question that might be asked about my house.

However, when we come to quantum mechanics something changes about
this.  Now, the elementary questions correspond to POVMs.  But, using
Gleason's theorem, we are no longer free to assign probabilities to
their outcomes willy-nilly.  All but a very few probabilities
assignments are tied together via the existence of a density
operator.  For instance, viewing quantum probabilities as Bayesian
probabilities, one is completely free to assign any probabilities one
deems relevant to the outcomes of a $\sigma_x$, $\sigma_y$, and
$\sigma_z$ measurement.  However, once that is done, one is no longer
free to specify an arbitrary assignment for spin in the $n$
direction, for any other $n$.

From this point of view then, quantum mechanics allows only a subset
of the vastly larger set of probability assignments one might make to
the answers of the physical questions one might ask.  And one might
think that restriction is accounted for by some physical fact---the
yet-to-be-discovered fact that is the essence of quantum mechanics.

\bjs
Still, it is in this connection that I wrote a NO into your paper. On
p.\ 28 you say ``Probability theory alone is too general of a
structure.'', and at some other place you say there must be an input
from nature. Based on my own games with these questions I doubt this.
I think quantum theory is already contained in the basic notions that
lead to probability theory. The sum rule of probability is no
obstacle, if you ponder what ``mutually exclusive'' means from an
operational point of view. For this reason I see a valuable
contribution in Lucien Hardy's attempt of starting from a few axioms,
although an axiomatic approach is unsatisfactory as long as the
axioms aren't simple truths instead of formal assumptions.
\ejs

So, indeed, what I said above is what I meant as an input from
nature:  It is whatever binds us to Gleason's theorem.  (Gleason's
theorem being the string that ties all the various distributions for
a physical system together.)

You'll note actually that Hardy is almost an antithesis to this idea.
He starts with structures that are larger than both classical
probability theory AND quantum mechanics.  By adding an extra
postulate he can narrow it down to either one or the other (or any of
a number of other structures).  What I want is start with classical
probability and then narrow it down to quantum mechanics.

It could be the wrong direction, but it is the one that feels
predominantly right to me (and the one that seems to me to have the
highest probability of leading to some interesting physics).  It is a
subjective judgment of course, but that's all that each of us has.

By the way, {\Caves} and {\Schack} and I have been thinking about applying
for a visit to the Erwin {\Schroedinger} Institute next spring or early
next summer.  The plan would be to write (the bulk of) an RMP article
on all this Bayesian business while we're there.  It'd be great to
have your ear to test it out on, if we do follow through with the
plan.

\subsection{Johann's Preply, ``Quantum Foundations in the Light of QI''}

\bq
Now I come to your  paper ``Quantum Foundations in the Light of Quantum
Information''.  I had gone through it in summer, then practical problems
intervened, and now I scanned through it again. I liked it very much,
although I tended to skip the formal parts, due to a dislike of density
matrices. (The unspoken opinion of the experimentalist is this: If you can't
produce a pure state, you are not in control of the experimental conditions.
And while it is true that a pure state can never be produced accurately, it
can be approached arbitrarily well.)

Let me write down a few of the qualitative statements which captured my
interest because I agree with them:
\begin{itemize}
\item
p.6 Information about what?  \ldots\ nothing more than the consequences of our
experimental interventions into nature \ldots
\item
p.7  The whole structure of quantum mechanics --- it is speculated --- may be
nothing more than the optimal method of reasoning and processing information
(in the light of the world's sensitivity to our touch)
\item
p.9 The quantum state is subjective incomplete information.
(Meanwhile I understand in what sense you, Carl Caves and {\Ruediger} {\Schack} use
`incomplete' in this context. A kind of respectful bow towards Einsteinian
expectations from physics.)
\item
p.19 Quantum states are states of knowledge, not states of nature
\item
p.22 The uncertainty that decreases in quantum measurement is the
uncertainty one expects for the results of any possible future measurements.
\item
p.26 on maximal state of knowledge. I like the clarification that `nothing
new' is learnt when already having maximal state of knowledge, but that
`mental readjustment' happens. It is in line with the everyday truth that
{\it each\/} observation tells you something, in other words, that by additional
observation your overall information about the world can only increase,
never just remain the same. A fundamental theory must reflect this. (I had
remarked on this in my letter to Carl Caves.)
\item
p.28 \ldots\ perhaps the better part of QM is just a law of thought
\item
p.28 \ldots $|\psi\rangle$ must be information about the potential consequences of our
interventions into the world.
\end{itemize}

From the paper I also got an appreciation of POVMs, which I had always
looked at as an unnecessary extension. The hint of getting more information
from a two-level system if measured via a three-level ancilla was important.

I did not delve much into what you wrote about Gleason's theorem, although
it is probably important for the paper. It appears that contextuality and
counterfactuality are surprises only, if one is rooted in classical thought.

I also appreciated the part on why the tensor product of Hilbert spaces
(p.29 ff).

The application of Bayesian theory to quantum theory. It appears to me that
quantum theory is the correct way of reasoning, and classical probability
theory is a certain limit of it. But both spring from the {\it same\/} way of
reasoning. So far, physics has stood in the way of clarifying this. The
perennial talk of systems and properties of systems one is forced to carry
along when dealing with quantum theory is a real hindrance to clear thought.
Remnants of mass points, forces, fields, etc., always sneak in, and with it
the need to allude to an objective world out there. As if repeatably
detectable structures in the statistics of probabilistic events and their
efficient description weren't objective and ``out there'' enough. (To me
mathematical truths and the Himalayas are equally ``out there''; the former
are mastered by acts of mental climbing, the latter by acts of physical
climbing, but both require willful action to be conquered. You call it
free-standing reality.)

Still, it is in this connection that I wrote a NO into your paper. On p.28
you say ``Probability theory alone is too general of a structure.'', and at
some other place you say there must be an input from nature. Based on my own
games with these questions I doubt this. I think quantum theory is already
contained in the basic notions that lead to probability theory. The sum rule
of probability is no obstacle, if you ponder what ``mutually exclusive'' means
from an operational point of view. For this reason I see a valuable
contribution in Lucien Hardy's attempt of starting from a few axioms,
although an axiomatic approach is unsatisfactory as long as the axioms aren't simple truths instead of formal assumptions.

Quite generally, you seek an understanding of QM on the basis of two or
three statements of a crisp physical nature. And these should be about
information. In line with what I just said, I share this wish
wholeheartedly. Having grown up collecting clicks of individual neutrons I
would start with ``clicks'' as the only point of contact between observer and
observed. (``clicks'' $=$ outcomes of trials of probabilistic experiments). On
the other hand, you want to distill out certain features of the established
theory, take them as nature's core message, and attempt an understanding of
the whole theory from these.

Here, I hesitate a little. Quantum theory itself is already too remote from
the experience from which it was distilled. We may continue marveling at
our extracts from nature, and perhaps overlook that they are trivial
consequences of the scientific method applied to data which we have decided
to view as intrinsically probabilistic. In other words, the theory may
indeed just be a law of thought, but straightforwardly so, and  not in the
sense that we first have to accept its results (nonlocality, interference,
etc.)\ and from these be pointed to laws of thought which make these results
self-evident.

In clear moments I wonder why we still have to discuss quantum theory, just
as you did in the beginning of the paper, but from a slightly different
vantage point. All we have is probabilistic events. Simple events, and
coincidence events of arbitrarily high level. And all we can do is calculate
random variables from them. What should be difficult about finding the
random variables with the highest degree of invariance and the possible,
statistically testable relations between such random variables? That will be
our structure, and quantum theory will be contained in it. And nature will
fit in, or the idea that probability is basic is wrong.

So much for today. The priors-question in Bayesian theory and ``How Much
Information in the State Vector'' I must postpone to another time. Perhaps I will think of a practical
question in interferometry.\medskip

\noindent P.S.: I hope that neither you, nor anybody close to you, have been affected
by the recent terrorist attacks.
\eq

\section{09-10-01 \ \ {\it Writing Physics} \ \ (to N. D. {\Mermin})} \label{Mermin46}

I'll place the new supplement to the Samizdat in the next email as
plain text.  Please let me know whether you're able to \TeX\ it up
fine; if that doesn't work out, I can post it on my web page (as a
PostScript file) as I did before.

Collecting it up, it's hard to believe I've written this much in the
little time since {\Vaxjo}.  I guess it's been an active time for
me. I think there's no doubt that I've gone through a phase
transition. For all my Bayesian rhetoric in the last few years, I
simply had not realized the immense implications of holding fast to
the view that ``probabilities are subjective degrees of belief.'' Of
course, one way to look at this revelation is that it is a {\it
reductio ad absurdum\/} for the whole point of view---and that will
certainly be the first thing the critics pick up on.  But---you
wouldn't have guessed less---I'm starting to view it as a godsend.
For with this simple train of logic, one can immediately stamp out
the potential reality/objectivity of any of a number of terms that
might have clouded our vision.

You'll find the most useful stuff in here starting at my last note to
you, i.e., page 78 onward.  In particular you might enjoy the chart
on page 84.  It shows, I think, that when this exercise of
epistemologizing so much is over and done with, there's still a fair
amount left that one might be willing to call concrete reality.

I do hope you get something out of this.  Two of your questions were
the sources for the vast majority of the pages in the document.  If
you hadn't pushed me, I may have never seen that so much was waiting
in the wings to be made sense of.

\section{10-10-01 \ \ {\it Lunch Thanks} \ \ (to J. N. Butterfield)} \label{Butterfield2}

Reviewing the day, it dawned on my that I hadn't thanked you for lunch!  Thanks so much.  I very much enjoyed our conversation, and I have to commend you for giving me a lead:  you helped me realize that I really should read the principal principle paper again, and come up with a firm opinion.

I hope you will have a gander at my QFILQI paper:  the more criticism you can heap on it the better.  I know that I'll certainly come out the stronger for it.

It was wonderful seeing you again.  Don't forget to remind me about your particular dates for Princeton:  by the then I should know something for real about pragmatism, pro or con.  (But knowing me, I'll probably lie somewhere in the middle.)

\section{11-10-01 \ \ {\it Returning Home and Future Plans} \ \ (to A. Peres)} \label{Peres18}

\bap
Can you imagine Swissair bankrupt? I remember flying Braniff in one of
their last flights, but Swissair???
\eap

Computers get exponentially faster at the same time that world change gets exponentially faster.  The two phenomena are probably not so untethered to each other!

\bap
I sent you my preliminary draft ``Imprecise quantum measurements'',
which nullifies the Meyer--Kent nullification of the KS theorem. I got
a few comments, from Appleby, Cabello and Mermin.
\eap

I apologize, but I didn't have a chance to look at your manuscript.  (I'll also apologize for the format I'm using to reply to your present letter!  When time slips away, so does my good composition sometimes!)  Also, I figured with your large distribution list, you would get some good comments in any case.  How were Appleby's comments?  I met him for the first time in Ireland, and he struck me as a very clear thinker there.  But I have not read any of his papers.

\bap
Soon there will be the deadline that Khrennikov gave for his book
about the {\Vaxjo} conference. I talked there about my work with Petra,
and this is well documented in two PRL and a forthcoming JMO. No need
of repeating that. So I thought that an extended version of my
``Imprecise quantum measurements'' could be a nice chapter in such a
book. What is your opinion?
\eap

I think that's an excellent idea.

\section{16-10-01 \ \ {\it Craters on the Moon} \ \ (to J. A. Waskan)} \label{Waskan2}

In a way you stole me away from the family tonight; the little ride
home was full of thoughts about craters on the moon.  You said
something like, ``That there are a thousand craters on the dark side
of the moon, is a true statement regardless of whether it's useful or
not.''

Below is a passage I took from David Darling.  It once made a good
impression on me (long before my {\James} days), and I can't help but
feel that it is relevant to tonight's conversation too.  If all the
world is but atom and void---or substitute your favorite metaphysic,
for that matter---then, it seems to me, there is no strict sense in
which there are ``craters'' on the moon at all.  To interpret the
coarse information I have about some aspect of my experience as a map
of the craters of the moon, seems to me an ultimately subjective
judgment---one that I make because it is more or less useful.

Of course, I'm not wedded to these ideas:  the game I play is to pick
and choose anything from any philosophy that will help me make sense
of the physics I'm doing and to promote it to a new level. Sometimes,
after a sufficient amount of play, I change my philosophical mind.
But I think the observation below is not a completely idiotic one,
and it sends me some way toward the pragmatic conception of truth.
No proposition I use in my daily life can be strictly true or false
in the sense of reflecting the world as it is independently of me.
And if not the case in my daily life, then why should it be the case
at some more ``ultimate'' level (i.e., fundamental physics)
that---after all is said and done---was intellectually derived from
that daily experience in the first place?

\bq
The interface between mathematics and everyday reality appears sharp
and immediate at this point: one sheep, one finger, one token;
another sheep, another finger, another token, and you can take away
tokens or add them, as you can with your fingers. The tokens---the
numbers---are just abstracted fingers; the operations for dealing
with the tokens are just the abstracted raising or lowering of the
fingers.  You make a one-to-one correspondence between the tokens and
whatever it is you want to reckon, and then forget about the fingers.

At first, it seems clear from this that mathematics must be somehow
already ``out there,'' waiting to be discovered, like the grain of
the stone.  One sheep add one sheep makes two sheep.  Two sheep add
two sheep makes four sheep.  That is certainly the practical end of
the matter as far as the shepherd and the merchant are concerned. But
already, even in this most simple mathematical maneuver, something
strange has happened.  In saying ``one sheep add one sheep'' we seem
to be implying that any two sheep will always be identical.  But that
is never the case.  Physically, the first sheep is never exactly
equal to the second: it may be a different size, have different
markings. It takes only one molecule to be out of place between the
two, and they are not identical.  Indeed, because they are in
different places they are inevitably not the same on that basis
alone. We have extracted a perceived quality to do with the
sheep---namely, their ``oneness,'' their apartness---and then merged
this quality by means of another abstraction---the process of
addition. What does it mean, physically, to ``add'' things?  To put
them together?  But then what is ``putting together'' two sheep?
Placing side by side, in the same field---what?

All this may seem like nit-picking. But on the contrary, it brings us
back to the central mystery---the relationship between the inner and
the outer, the world of the rational mind and the world ``out
there.'' In the physical world, no two sheep are alike. But, more
fundamentally, {\it there are no ``sheep.''}  There are only some
signals reaching the senses, which the left brain combines and then
projects as the illusion of a solid, relatively permanent thing we
call a sheep.

Like all objects, sheep are fictions: chimeras of the mind. It is our
left hemispheres, having through natural selection evolved this skill
for extracting survival-related pieces of the pattern, that trick us
into seeing sheep, trees, human beings, and all the rest of our
neatly compartmentalized world.  We seek out stability with our
reasoning consciousness, and ignore flux.  We shut our eyes to the
continuous succession of events if those events seem not to
substantially affect the integrity of what we see.  So, through this
classifying and simplifying approach we make sections through the
stream of change, and we call these sections ``things.''  And yet a
sheep is not a sheep.  It is a temporary aggregation of subatomic
particles in constant motion---particles which were once scattered
across an interstellar cloud, and each of which remains within the
process that is the sheep for only a brief period of time. That is
the actual, irrefutable case.
\eq

\section{17-10-01 \ \ {\it Quick Single Point} \ \ (to J. A. Waskan)} \label{Waskan3}

Thanks for the note:  I'm still digesting it.  But let me quickly
reply to the one thing that I can reply to.

\bjaw
Also, it seems strange for the fellow you quote to concede that there
is light, eyes, the left half of one's brain, a process known as
natural selection, and at the same time to deny the existence of
sheep.
\ejaw

Indeed, it seems strange to me too:  He's pretty clearly not being
consistent.  But the role of the quote for me was as a motivating
piece. (I read Darling a few years ago, and the main reason I used
the quote last night was because it was already in my computer and I
could, thus, send off a quick note to you.)  My sentences above the
quote were meant to show that I toy with the idea of going a further
(more consistent) step:
\bq\noindent
And if not the case in my daily life, then why should it be the case
at some more ``ultimate'' level (i.e., fundamental physics)
that---after all is said and done---was intellectually derived from
that daily experience in the first place?
\eq

It is going that extra step that seems to me to be heading down the
track to {\James}ianism.

Let me read your note again \ldots\

\section{17-10-01 \ \ {\it Quick Second Point} \ \ (to J. A. Waskan)} \label{Waskan4}

Sorry, I can't wait for the beer.

\bjaw
Here's a daily affirmation for you.  Look in the mirror (not {\Rorty}'s
mirror of nature, the one in your house), and say the following:

I am pretty darned sure that I exist.  I think there is other stuff
in the universe too.  I'm pretty darned sure that I have beliefs
about (i.e., representations of) the other stuff that might be out
there. My beliefs are true insofar as the world is how I represent it
to be. If the world outside of my mind is in no way how I represent
it as being (e.g., IF THERE ARE NO SHEEP, etc.), then all of my
beliefs are false.  Even if all of my beliefs about the things
outside of my mind are false, this should have no bearing on the
nature of truth itself (i.e., correspondence).
\ejaw

If you want to use true and false that way, then---in my present
state of mind---yes, I would say that, strictly speaking, most all my
beliefs are false.  Beliefs play the role of coordinating our
actions, and, in that way, can be more or less useful.  But (in the
too small of thought I've given this) I can't find a role for the
concept of belief outside its use.

Like you, I am pretty darned sure of the existence of a world outside
myself.  But I would say that that surety comes NOT from some
(transcendental?)\ knowledge that my beliefs mirror that world as it
is. It is just the opposite.  I believe in a real world outside
myself because, throughout my life, things continue to take me by
surprise.  Significant numbers of my beliefs are systematically
INvalidated with each new day.  There's my evidence of the real
world.  Below is the way I put this point in a recent paper.

The point of view is not completely worked out yet---and it may never
be---but my experience in quantum mechanics makes it feel more right
than the other options I've seen so far.

Knowing me, I'll probably give your note another read, and be back
again tomorrow.  I hope you don't lose patience with me.

\section{17-10-01 \ \ {\it Quick Third Point} \ \ (to J. A. Waskan)} \label{Waskan5}

OK, one more for the day.

\bjaw
Also, sure no two sheep are exactly alike.  Neither are any two
bachelors.  That doesn't entail that there are no bachelors.  X is a
bachelor if X is unmarried, and X is male, and X is eligible (e.g.,
not a priest).
\ejaw

Granted.  But what I thought was at issue is whether there are
``bachelors'' without the agents who make up (and use) all these
judgmental categories.  If all the world is BUT atom and void---to
use an allegory I like but which should not be taken literally as my
view---then where do all these extra distinctions come from if not
the judgmental agent?

I say I thought this was the issue because my reading of the
pragmatic conception of truth is more to the following point: Without
agents, there are no ``propositions.''  Therefore ``propositions''
cannot be true or false in any absolute sense. Without the agent,
there is the world, and it is just whatever it is.  A proposition
adds something to the world that it itself did not possess before the
agent's attention was drawn to it (via the act of dreaming it up,
writing it down, acting upon it, etc.).

There's probably nothing worse than to have an armchair philosopher
in your presence \ldots

\section{17-10-01 \ \ {\it Final Point} \ \ (to J. A. Waskan)} \label{Waskan6}

I just read my last note over again and, to my great fear, I discovered you might read my closing two ways!!
\bq\noindent
There's probably nothing worse than to have an armchair philosopher in
your presence \ldots
\eq
I hope you understood that I was calling myself the armchair philosopher and not being a complete jerk.

\section{18-10-01 \ \ {\it No Doobies Here?}\ \ \ (to J. A. Waskan)} \label{Waskan7}

\bjaw
Lots of folks want to say that there are joints in the natural order
(though, admittedly, bachelorhood probably isn't one of those
joints).
\ejaw

I'm not sure what you mean by the term ``joints.''  Can you define it
precisely?  (Not knowing what you mean, it leaves me unable to reply
to most of your message.)

\bjaw
If, however, you take a step back and look at the big picture,
\ldots\ I suspect that the same can be said for the relationship
between atoms and tables.
\ejaw

You know what my worry is (fueled by the 75 year debate on the
interpretation of quantum mechanics).  It's that we just can't step
far enough back.  We are immersed in this thing called existence, and
there's just no way to get a view from outside it.  We do the best we
can from the inside, and that's called science.  (For me, the phrase
``best we can'' means to eliminate impredictability---i.e., (only
half jokingly) to delete reality as much as possible.  Cf.\
yesterday's note about Emma.)

Tell me what a joint is, so I can think a little more about what you
said.

\section{21-10-01 \ \ {\it Danny, Dewey, David, Delight, and Dilemmas}  \ \ (to A. Peres)} \label{Peres19}

Thanks for the note; I'm glad to hear about Danny's opportunities.

I suppose because of the war and a couple of other details closer to home, I've gone into a bit of an email slump for the last couple of weeks.  I am so far behind with all my correspondents.  At least in my spare time I have been motivated to read more philosophy.  And indeed I have come across a very good strain of that for once:  I've been absolutely taken away by the writings of William James and John Dewey.

I have written David Mermin about our scheme for the ITP, but I have heard no reply yet.  But he is enjoying the beauties of Vienna, so in a way, I expected no less.

\section{22-10-01 \ \ {\it Sad and Happy} \ \ (to R. Cleve)} \label{Cleve1}

Of course, I'm sad to hear that we couldn't entice you to New Jersey \ldots\ but, really, I expected as much.  Please do keep us in mind though in case anything ever changes.

On the other hand, I was really happy to hear that you liked the little sound bite in {\sl Discover Mag}.  I expanded those vague thoughts into a (slightly less vague) paper:  \quantph{0106166}, ``Quantum Foundations in the Light of Quantum Information''.  There is a small technical mistake in there---that I'm presently writing a comment on---but on the whole, I think the paper is relatively sound, and maybe entertaining.  I'd enjoy hearing any reaction you might have, and more importantly, any ideas you might have about how to proceed on the program posed there.

\section{22-10-01 \ \ {\it The Dilemmas of Subjectivism} \ \ (to R. {\Schack} \& C. M. {\Caves})} \label{Caves36} \label{Schack29}

I apologize for holding off so long in a reply to {\Ruediger}'s letter
concerning the RMP article.  The difficulty has come in that I didn't
know how to reply.  (I guess I still don't.)

The point of some potential consternation is this:

\brs
What I think we should be doing is a paper on ``Interpretations of
probability in quantum mechanics (with special emphasis on the
Bayesian viewpoint)''. The paper would NOT be on the interpretation
of quantum mechanics.
\ers

The problem is, I don't see how to separate the two issues.  Where
does the interpretation of probability fall off and the
interpretation of quantum mechanics kick in?  How can one have an
interpretation of quantum mechanics wherein the wavefunction is
objective, but still think of probabilities as being strictly
subjective?  Similarly, vice versa?

Let me do this:  Let me ship to you both the mini-samizdat of my
thoughts that came out of my post-{\Vaxjo} broodings.  You tell me
which {\it sentiments\/} will be banned and which won't if we end up
skinny-dipping together.  I can foresee some being excluded---like
the stuff in my letter to Wootters---and I can accept that; but for
the greater majority of the writings, I can't see myself drawing a
line, and I'm wondering where you will draw it.

Looking back on the BFM debate, I think the most important thing to
come out of it for us three in particular is that it makes it
absolutely clear that we need to get our thoughts straight on the
``principal principle'' before we can embark on a consistent
statement of our position.  For I see no way to erase the dilemma:
Either we accept that the ascription of one Kraus rule over another
in a measurement intervention is a subjective judgment, or we accept
a quantumatized version of Lewis's principal principle.  Why did we
all reject the principle before if we find ourselves accepting it
now?  This is something we ought to reflect upon more deeply.

\section{23-10-01 \ \ {\it Increased Surveillance} \ \ (to N. D. {\Mermin})} \label{Mermin47}

My spies tell me you gave a talk on ``Whose Knowledge?''\ yesterday.
How did it go?  Did anyone in the audience (unintentionally) lobby
any shells on my behalf?  If so, hand over their names, and I'll put
out a recruiting effort for them.

\section{23-10-01 \ \ {\it United We Stand Airlines} \ \ (to R. {\Schack})} \label{Schack30}

\brs
Do you think you could give me a lift at such an early time?
\ers

I do whatever it takes for the greater good of quantum mechanics. Of
course I can give you a lift!  (If you can stand to listen to me
babble that early in the morning.)

\section{26-10-01 \ \ {\it The Feynman Cult} \ \ (to J. N. Butterfield)} \label{Butterfield3}

I don't remember railing against the Feynman cult in your presence,
but your letter gives some evidence that I must have.  (Doing such
things was a common pastime for me at Caltech, but I've mellowed a
little in my efforts since leaving, i.e., since the cult hasn't been
in my face on a daily basis.)

But, anyway, yes I told Brandt that I would come.  (I hope you will
come too.)  I've even thought about how I will open my talk:  with
the Feynman quote below.  Lord knows I'm no materialist, so you can
rest assured that I'll do my best to ``zing'' it up afterward.  (See
my Samizdat, page 237.)

\bq
If, in some cataclysm, all of scientific knowledge were to be
destroyed, and only one sentence passed on to the next generation of
creatures, what statement would contain the most information in the
fewest words?  I believe it is the atomic hypothesis (or the atomic
fact) that all things are made of atoms---little particles that move
around in perpetual motion, attracting each other when they are a
little distance apart, but repelling upon being squeezed into one
another.

Everything is made of atoms.  That is the key hypothesis.
\eq

\section{27-10-01 \ \ {\it Coming to Agreement} \ \ (to C. M. {\Caves} \& R. {\Schack})} \label{Caves37} \label{Schack31}

\brs
In any case, do you know a reference for strong Dutch-book
consistency?
\ers

Nothing comes to mind:  I think I first learned of it from you (was
it in Cambridge?).  Might it have been in either of those papers we
looked up in a trip to the university library one day?  This was one
of them (though I don't have it anymore):
\bq
\noindent J.~G. Kemeny, ``Fair Bets and Inductive Probabilities,'' J.\
Symb.\ Logic {\bf 20}, 263--273 (1955).
\eq
Ahh, here was the other one:
\bq
\noindent
R.~S. Lehman, ``On Confirmation and Rational Betting,'' J.\ Symb.\
Logic {\bf 20}, 251--262 (1955).
\eq
It might be a good idea to dig those up again.

I'll try to answer your other questions next week, after I get a
chance to get a better grasp on the paper.

\section{27-10-01 \ \ {\it Literature} \ \ (to R. {\Schack})} \label{Schack32}

\brs
A question to you: How much literature is there on interpretations of
probability in q.m.?  Would it be feasible to review it all?
\ers

Attached is everything I had collected up previous to the Cerro
Grande fire.  A skim of that document might answer your question in
the most direct way.

By the way, I've asked Maria Carla Galavotti for her criticisms (and
any others that she knows) of the principal principle.  But I haven't
gotten a reply from her yet.

Thinking back on it, I can't remember if we ever had any NONquantum
reasons that were substantially different than J.~S. Mill's argument
against a ``substance'' underlying the phenomenal world:

\bq
\noindent
If there be such a {\it substratum}, suppose it at this instant
miraculously annihilated, and let the sensations continue to occur in
the same order, and how would the {\it substratum\/} be missed? By
what signs should we be able to discover that its existence had
terminated? Should we not have as much reason to believe that it
still existed as we now have?  And if we should not then be warranted
in believing it, how can we be so now?
\eq

Similarly one could say of objective probability:  Bayesian
coming-to-agreement would work precisely the same whether the
objective probability is there or not.  But did we have any other
arguments than that (that did not depend on upon quantum mechanics,
for instance in the nonuniqueness of the density operator
decomposition)?  Did we have any examples where believing in
objective probability in Lewis's sense would be downright misleading
in how it might suggest tackling a practical problem?

I so wish I had my old file cabinet back again; writing you this
morning has made that feeling more acute.

\section{31-10-01 \ \ {\it For the Bayesian Club} \ \ (to C. M. {\Caves} \& R. {\Schack})} \label{Caves37.1} \label{Schack32.1}

Cross-listings:
\bq
\noindent \arxiv{hep-th/0110253} [abs, src, ps, other] :\\
Title: Physics with exotic probability theory\\
Authors: Saul Youssef \smallskip

\noindent Probability theory can be modified in essentially one way while maintaining consistency with the basic Bayesian framework. This modification results in copies of standard probability theory for real, complex or quaternion probabilities. These copies, in turn, allow one to derive quantum theory while restoring standard probability theory in the classical limit. This sequence is presented in some detail with emphasis on questions beyond basic quantum theory where new insights are needed. (24kb)
\eq

\section{01-11-01 \ \ {\it Your Limerick}  \ \ (to I. Duck)} \label{Duck1}

A couple of days ago I opened up a package of old mail that Los Alamos had finally forwarded to me.  To my great delight, I found your letter and your limerick from February!!!  Thank you so much; this is a very great honor.  After reading your limerick I think I smiled all day.

I wish so that I had gotten your mail before May.  I would have then
been able to include (with your permission) your limerick in my large
samizdat posted on the web at: \quantph{0105039}.  It would have been
a great addition.  And I think you are right: ``the best we can ever
surmise'' catches the situation quite precisely.

Thank you again.

\section{01-11-01 \ \ {\it Delayed Reactions}  \ \ (to A. Peres)} \label{Peres20}

Two days ago, I received a package of old mail that was finally forwarded from Los Alamos.  Some of it was very old indeed, going back to November 2000.  In any case, two pieces were relevant for our {\sl Physics Today\/} articles.  One was from someone named Fritz Froehner, and it looks like he sent an identical copy to you.  But the far more interesting one was from a professor at Rice University (where, incidentally, {\Carl} {\Caves} got his undergraduate degree).

I scanned the letter into my computer this morning, and will put the whole thing below.  I have just written him thanking him for the warm feelings he gave me.

\bq\noindent
Dear Dr Fuchs:\medskip

I am sending you a copy of a limerick which was prompted by your OPINION article with Asher Peres in MARCH 2000 PHYSICS TODAY. I hope it amuses you, and that it catches the gist of your `No Interpretation' thesis. It is a fascinating subject which I believe to be completely rationalized by your point of view. I'm sure you have thought about this much more deeply than I. For instance, I just now realized that one does not have to invoke anything more miraculous than uninterrupted Hamiltonian evolution for the system. I had been thinking that it was a mystery that we should have gone from the mechanics of an inanimate object to a theory of knowing; but that is not what is happening. I think `the best we can ever surmise' catches the situation quite precisely.

I tried to interest Alan Alda in {\it Scientific American\/} with this, but I missed the deadline for his limerick contest and I don't think they ever forwarded it to him.\medskip

\noindent With best regards, \smallskip

\noindent Ian Duck, Professor of Physics\bigskip

\bv
\underline{LIMERICK on the INTERPRETATION OF QUANTUM MECHANICS}\bigskip\\
ON SCHROEDINGER'S CAT\medskip\\
It comes as a total surprise\\
That what we learn from the $\psi$'s\\
\verb+  + Not the fate of the cat\\
\verb+  + But related to that:\\
The best we can ever surmise.\medskip\\
IAN DUCK, FEB 2001
\ev
\eq

\section{01-11-01 \ \ {\it Professor Duck} \ \ (to C. M. {\Caves})} \label{Caves37.2}

Do you remember a Professor Duck from the physics department at Rice?  Since he wrote me a paper letter, I'm guessing he's an older guy.  The full story is below.  [See 01-11-01 note ``\myref{Peres20}{Delayed Reactions}''  to A. Peres.]

\subsection{Carl's Reply}

\bq
Sure, I remember Ian Duck.  Probably now in his
mid-60s.  Here's his web page,
\bq
\myurl{http://report.rice.edu/sir/faculty.detail?p=48D19A4186C80228}
\eq
which has a picture of him.  I liked his limerick.
\eq

\section{02-11-01 \ \ {\it Positive Operators and Measurement} \ \ (to K. Jacobs)} \label{Jacobs1}

Thanks for including me in the correspondence with Keiji Matsumoto.  Did his question arise from having read our paper?  (That would mean we might have 3 or 4 readers!!)  I was supposed to visit his institute in Japan last month, but I wimped out of leaving the country.  (In any case, I had already flown overseas four times this year \ldots\ and five seemed a bit too much.)

Let me make one comment on your points about the ``identity'' (as the unitary in the polar decomposition) giving a natural analog to Bayes' rule.  Have a look at Section 5 in my paper:
\quantph{0106166}.  What I show there is that by appropriately adjusting the various $U_i$ (but not to the identity), one can make a precise analog to Bayes' rule even in the noncommuting case.  Of course, it's a trivial point that we probably both understood while I was at LANL.  What's new though, is that I've now come to think it is an important conceptual point.

\section{04-11-01 \ \ {\it Dreams of an Ever-Evolving Theory} \ \ (to A. Peres)} \label{Peres21}

\bap
I am reading a wonderful book {\bf Dreams of a Final Theory} by Steve
Weinberg. Chapter 7 is ``Against Philosophy'' and I highly recommend
it.  I got that book for \$7.00$+$tax in a used books shop in Santa
Barbara, where Aviva was looking for something else.
\eap

Thank you for the tip; I did read Weinberg's chapter 7.  He writes in
a very crisp and no nonsense way, and I like that.  (You know I
constantly fight tendencies in the opposite direction in my own
writing.)  His points are well taken, especially the ones about how a
preset philosophy can create immense blinders for the scientist:
\bq\noindent
     ``\ldots\ in rejecting it the [PHILOSOPHY-X]ists were making the
     worst sort of mistake a scientist can make:  not recognizing
     success when it happens.'' --- page 177, paperback version.
\eq
But, I think a deeper point is the one he makes near the beginning of
the essay:
\bq\noindent
     ``I do not want to draw the lesson here that physics is best
     done without preconceptions.  At any one moment there are so
     many things that might be done, so many accepted principles
     that might be challenged, that without some guidance from our
     preconceptions one could do nothing at all.'' --- page 167
\eq
And I agree with the numerical tally of the next sentence:
\bq\noindent
     ``It is just that philosophical principles have not generally
     provided us with the right preconceptions.'' --- page 167
\eq
However, I part company with him in thinking that that is a strong
argument against pursuing philosophy as a sideline to science.  That
is, I don't know what the preconceptions can be if they're not
philosophies.

When it comes to philosophies and, not unrelated to that, scientific
research directions, I tend to take a lot of stock in a Darwinian
kind of conception.  That is, we each should do precisely what we
feel compelled to do; we each should research precisely what we feel
compelled to research.  There's probably nothing we can do about it
anyway.  Indeed ninety-nine percent of the time we will be on the
wrong track:  The world supplies a selection pressure for our
thoughts, just as it does for the lifespan of the drosophila.  And
just as it is not possible for the drosophila to change its genetic
makeup before it meets its demise, I think the only thing we can do
as scientists is cultivate to the best of our ability the
philosophical preconceptions that led us down our own paths.
Ninety-nine percent of us will be forgotten from the history books,
but the ones of us that remain will do so because the world is such
that it is less likely that we should fall.

For myself, I have DISCOVERED that I have chosen a direction of
thought that is very closely aligned with the philosophical movement
of pragmatism from the early part of the 20th century---a movement
the details of which have been nearly forgotten in modern times.
Interestingly, the thing that set me on to this realization was
Martin Gardner's essay ``Why I Am Not a Pragmatist'' in his book {\sl
The Whys of a Philosophical Scrivener}.  (You probably remember
Gardner from his column in {\sl Scientific American}.)  This happened
about three months ago.  I really recommend you read the article if
you get a chance.  Maybe your library has a copy of the book.  I
think in reading it, you will discover that the essay might just as
well have been titled, ``Why I Am Not in Agreement with Fuchs and
Peres's Physics Today Article.''  For, with each reason Gardner used
to explain why he was {\it not\/} a pragmatist, I found myself
thinking of quantum mechanics and saying to myself, ``ahh, I guess
that means I {\it am\/} a pragmatist.''  Really, the analogy is {\it
that\/} close even though the article has nothing to do with quantum
mechanics per se.

The issue is no less than whether ``unperformed measurements have no
outcomes.''  The pragmatists, for various reasons, thought it was
{\it safer\/} to assume that they didn't.  The movement then spent
the greater part of its time developing the (liberating) consequences
of this supposition.  Gardner (and Bertrand Russell and G.~E. Moore
and gazillions of others) thought ``how silly'' and ``how contrived''
when it is so much easier to use standard realist language to
describe the outcomes of experiments---to assume the outcomes are
there before one has a look.  But you and I know better, of course.
And, I think it is quite useful to know that there was a set of
people carrying through the detailed consequences of this line of
thought for their broader worldview long before you and I were on the
scene.  The way I view it, these old thoughts can be a resource to
our explorations of quantum mechanics just as much as any other:
However, their use is in setting the directions for potentially
fruitful lines of thought \ldots\ but that should always be the use
of any philosophy for any scientist.

Below I'll attach a letter I wrote Bill Wootters a while ago on a
similar subject.  [See 20-09-01 note ``\myref{Wootters2}{Praise, Folly, Enthusiasm}'' to W. K. Wootters.]  What I write just after the quote of William {\James}
better explains why I chose the title that I did for this note.

\section{04-11-01 \ \ {\it Slim Chance?}\ \ \ (to C. M. {\Caves})} \label{Caves37.3}

I suspect there is an awfully slim chance that you will be interested in this, but let me put it on the table anyway.  After returning to Israel from his ITP stay, Asher asked me if I would be interested in proposing (with him and a larger team yet to be built) a program at the ITP on quantum foundations.  The idea would be to divorce the program from my usual tack of quantum information's role in quantum foundations and let it be more freeform.  In particular, one should be careful not to give the impression of too much repetition with the present program.

The question is, would you be interested in participating?  I told Asher that one of my requirements for working in such a project would be that the organizers {\it not\/} be a random list of persons with similar agendas, but that the ties between them tend closer to friendship than mere colleagueship.  Others that have come to mind are Bub, Mermin, and Wootters.  Bub gave a definite yes for interest; Mermin gave a {\it definite\/} no.  I haven't contacted Wootters yet, but I bet I can guess he'll say no.

Asher also thinks Jim Hartle should be approached---because he is an ITP heavyweight---but I'm not sure where he fits with respect to my criteria above and with respect to my fears of YADFog (i.e., Yet Another Decoherence Fest).

So,
\begin{enumerate}
\item
I would like to see your participation; it would certainly help tip
   the scales for my own.  But,
\item
Even if you're not interested participating directly, would you be
   willing to give me advice on all the above.  Also, since you've been
   through this before, you are probably aware of various considerations
   that I've never dreamed of.  Therefore, please do feel free to say
   anything you think that might be relevant.
\end{enumerate}
Thanks for any thoughts.

\section{06-11-01 \ \ {\it Seventy Virgins and a Mule} \ \ (to G. L. Comer)} \label{Comer6}

\bgc
I have this awful tendency to dwell on my mortality now.
\egc

Me too, but for not such good reasons---for me it's been mostly the anthrax and nuclear target in my back yard.  On the other hand, maybe the biggest mental note I've taken away from September 11 is the dangers of a belief in an afterlife.

Consequently, lately I've been on a binge of asking myself what role death plays in the construction of our physical world.  What greater good does it open up?

It's a good mantra to toy with.

\section{14-11-01 \ \ {\it Samizdat II} \ \ (to N. D. {\Mermin})} \label{Mermin48}

\bdm
Expanding slightly on my reaction to Samizdat II: (Please don't
conclude that I've missed the point yet again until we meet):

It seems to me that all interpretations of QM have to come up against
what is loosely called the ``measurement problem'' in one form or
another, which I would describe as how we can reconcile a world empty
of ``facts'' to the ``facts'' of our own experience. (I would not
describe it as having to do with how wave-functions ``collapse'' or
``decohere''.)

Various people (at least those who don't want to decree that there is
a ``cut'' between two domains) deal with the problem by extending the
facts of our experience to everything else (Bohmians, GRW
collapsists).  Others deal with it by extending the indefiniteness of
QM to our apparently definite experience (many worlders).  It seems
to me you're following the second strategy (which is all I meant by
my irritating jokes about your Everettism --- your version of the
second strategy is obviously different from theirs) by insisting that
the subjective character of quantum probabilities requires us to take
the same subjective approach to classical ``facts'' --- i.e.\ to
insist that they too are beliefs that can also be dealt with only
through a (subjective) probabilistic treatment.

This is intriguing and well worth exploring. I do worry (with {\Carl}, I
think) that it's getting rather far from how physicists do physics --
or at least from how they think they do physics.  But I wouldn't say
it means the end of science.  Whatever that means.
\edm

Thanks for the extended comments.  I won't say you've missed the
point:  I think you've got it.  But I don't quite understand the
Everett analogy yet.  I would say their world---Deutsch's
world---abounds with facts.  Facts far, far in excess of what any of
us ever see.  (I.e., all their worlds.)  But I'll think much harder
about your note.  There's no need to reply yet again.

\section{16-11-01 \ \ {\it Great Quote} \ \ (to J. W. Nicholson)} \label{Nicholson2}

Now this is one I love!  Which supersedes which?  Christianity or humanity?  I didn't realize the two notions could be considered so opposed.
\bq
``It was very dramatic, right until the end,'' Taubmann said at a news conference Thursday night. He was wearing a new pair of pants that still sported the store tags, and he had sheared the scraggly beard and long hair he had when he had arrived in Islamabad.

``I am a Christian --- I have forgiven them [the Taliban] for what they have done,'' he said. ``But as a human being, I hate what they did to us.''
\eq

\section{19-11-01 \ \ {\it A Lot of the Same} \ \ (to C. M. {\Caves} \& R. {\Schack})} \label{Caves38} \label{Schack33}

\brs
Before you have me burned alive, please tell me why I am wrong!
\ers

Come on, you know it's the Thanksgiving season.  I would never burn
you, only roast you.  (Though my brother-in-law once fried his turkey
in hot oil.)

Sorry for the hiatus, but I just got inundated with email last week,
and I didn't have the proper mentality for replying to any of it. So
I shut down for a while.  Now I'm stuck with trying to clean out an
even bigger pile of old mail.  But let me compliment you by letting
you know that I'm tackling your letter first!  (It's the only
interesting one in the lot.)

\brs
I started writing up a summary of our discussions, and hit upon a
difficulty when I tried to formulate exchangeability for models. Here
is the problem.

In the traditional formulation of exchangeability, we say that we
have $N$ identical systems (same Hilbert space). At this stage, it is
thinkable to assign a different state to each system.  We then make
the judgement that the {\it state} of the $N$ systems is
exchangeable.

For models, we say that we have $N$ identical apparatuses.
Alternatively, we say that we use the same apparatus to perform $N$
measurements, let's say on different, independently prepared systems.
At this stage, I can't think of a good reason to even consider
assigning different models to the apparatuses. It seems to me that
one is forced to say that each apparatus performs the same operation,
so should be described by the same model. That leads immediately to a
heresy: there exists a true model. If we don't know it, we assign
probabilities to models. What we wrote down on your whiteboard is
consistent with this viewpoint. The difficulty we encountered
formulating exchangeability could mean that it is an unnatural
concept in this case.  Writing down a mixture of N-fold products of
models is completely natural however.
\ers

Let me try to allay your fears.  I think the issues here are almost
precisely the SAME as they are in our old de Finetti considerations.
To say it in a way that maybe {\Carl} would endorse, ``It's really all
about learning.''  Or in a way that I'm more tempted to these days,
it's all about demonstrating a willingness to update one's
beliefs---one's commitments, one's pragmatic strategies for action,
one's betting behavior---in the light of factual data.

Let me start with an example that's essentially already well-worn for
us by now.  Suppose we have a rather complicated quantum measurement
device whose manufacturer purports it to be the best $\sigma_z$
measurement device ever built.  Furthermore, suppose we have a fresh
supply of $10^8$ calcium atoms, all meticulously prepared to have
spin-up in the $x$-direction.  What do we expect to happen if we
individually dump all the atoms into the measuring device?  We expect
about 50\% of them to get registered as spin down and about 50\% of
them to get registered as spin-up.  But what happens if one after
another, all the registrations are of the spin-up variety?  Well,
that outcome sequence would be no less likely or no more likely than
any other outcome sequence if we walked into the laboratory with such
a radically adamant prior belief.  In a real-life situation, however,
we would be shocked; we would update our beliefs accordingly---for we
would have allowed for the possibility of ``learning.''

But in this situation, notice that there are at least two extreme
cases to which we could attach the possibility of learning.  The
learning could be about the device or it could be about the
preparations.  Who's to say that the learning is about something more
objective in the one case than the other?  Prosaically, it takes both
ingredients (the preparations and the device) to certify the device,
and you can't get away from that.

Let me try to tighten this up by sketching how we ought to start
thinking about a de Finetti theorem for unknown quantum models.  I
run a measurement device on $N$ independently and identically
prepared quantum systems.  Suppose I am absolutely confident of these
preparations---i.e., with respect to them, there is nothing left to
learn in the technical sense of i.i.d. statistics for any repeated
and KNOWN measurement.  Then, what can it mean---from a Bayesian
point of view---that the measurement device works according to an
unknown model?  It means that after all the outcomes are gathered,
there's still something left to be learned from the posteriori
quantum state for the systems that were measured.

That is, more simply, the best judgment we can make about the systems
that passed through the measurement device is that they are
exchangeable CONDITIONED on the registered measurement outcomes. For
instance, suppose the device spits out an index $i$ at each round.
The issue is, what pragmatic meaning should we give to each such $i$?
Quantum mechanically, the predictive meaning of an index $i$ is
specified by the Kraus operator $A_i$ we associate with that outcome.
(Its retrodictive meaning is given by the positive part of
$A_i$---the POVM.)

If we think we don't exactly know what the device is doing to each
individual system, then we shouldn't yet dare to make an association
$i \rightarrow A_i$.  (To make an extreme point of it, for all we
know, the device might be entangling all our test systems.)  We
should just rest confident that no matter what order we send the
systems through the device, we will end up with the same subjective
beliefs in the end.  Thus, if we gather up all the systems for which
an outcome $i$ occurred (as opposed to some other outcome $j$), then
the subjective density operator we assign should be exchangeable.
Using the standard quantum de Finetti theorem, we then get that that
density operator must be of de Finetti form.  Writing each of the
final density operators as a linear map acting on the initial density
operator, we (should) get something like the desired theorem for
unknown quantum models.  If we believe that we can learn something
about the model, then the probability distribution that appears in
the de Finetti form is restricted to being something other than a
$\delta$-function.

In summary, our belief that the best we can say of the outputs is
that they ought to be exchangeable (conditioned on the factual
outcomes), leads directly to a notion of mixture of models---i.e.,
that the output density operator is controlled either by a Kraus
operator $A_i$, or a Kraus operator $B_i$, etc., etc., about which we
capture our ignorance through some subjective probability
distribution.

Now, just as the regular de Finetti theorem cannot put an end to the
principal principle, we cannot use this (proposed) theorem to put a
stake through the heart of the true believer of objective quantum
models.  That is, David Lewis might say of the regular de Finetti
theorem, ``That is a very nice theorem, but it doesn't change the
fact that there really is always a `man in the box.'  His name is
God.''  And so he would probably also say of our quantum models (if
he knew quantum mechanics).  Instead, all the (proposed) theorem can
do is show that it is {\it possible\/} to get by without a man in the
box. We don't need him; all we need is something like the judgment of
exchangeability for the outputs (conditioned on the outcomes) along
with i.i.d.\ on the inputs.

So that's the sketch.  Now, how to dot the $i$'s and cross the $t$'s?
I can foresee at least one difficulty that I'm not clear-headed about
right now.  That is, using the description above, for each index $i$,
we will generate a probability distribution over models.  But by what
regularity condition can we assure that $p_i({\cal A})=p_j({\cal
A})$?  What I mean by this notation is that $\cal A$ stands for the
model in total (i.e., all the Kraus operators in it) and $p_i({\cal
A})$ stands for the probability distribution in the de Finetti
theorem derived for each index $i$.  It is probably so simple as
this:  If we were to imagine doing tomography on the posterior states
for each index, then the states derived from that should always
average up to a valid density operator.  But I'm not exactly sure how
to put that idea into action.

Oh, and here's another intriguing point that ought to be explored.
Suppose we focus our attention on a given exchangeable density
operator.  There are many ways that operator could arise, but suppose
that it came about as the posterior state arising from many identical
measurements (in the sense above).  The question I have in mind is
how much freedom do we have for trading off between an unknown
preparation and an unknown model for getting to the final state?  Can
one always find a fixed initial preparation and a mixture of models
that will give rise to the final exchangeable state?  Can one always
find a mixed model and a mixture of initial preparations?  Probably
yes and yes, but I'm not completely sure.

I'm so glad to hear that you may have reversed your opinion on
William {\James} (at least a little).  By the way, I hope you notice how
these de Finetti considerations are drawing out a lot of the
considerations I was trying to express to you in Samizdat-II and our
subsequent discussions.  For it helps draw the distinction between
the amorphous index $i$ that arises in a quantum measurement and the
meaning $A_i$ that I ascribe to that event.  The symbol $A_i$ plays
the role of a proposition that I write about $i$:  It carries the
information about how I will react after having seen it, how I will
place my bets.

I'm willing to believe this whole debate about ``truth'' might be a
red herring---i.e., that we might easily be able to get away with
never uttering the word.  But I think the realization of the last
paragraph had a primitive expression in {\James}'s worries about
``truth'' nevertheless, and to that extent maybe he and his movement
of pragmatism are worth contemplating (though of course not
subscribed to in toto).

\section{19-11-01 \ \ {\it William {\James}} \ \ (to R. {\Schack})} \label{Schack34}

\brs
You will be pleased to learn that I bought a copy of ``Pragmatism''
and enjoy reading it a lot! His rhetoric is the best I have ever seen
from a philosopher. He is definitely not tender minded.

It's a curious mix, though. Sometimes he seems to be very close to
Kantian ideas on truth, then he seems to subscribe to a naive
correspondence theory of truth, at least for simple facts such as
``this detector has clicked.''
\ers

Yes, I'm more attuned to that now, and I'm trying to get it straight
through extended readings (such as A.~J. Ayer's book {\sl The Origins
of Pragmatism\/} that I picked up in New York City with you).

\brs
I am also a little disturbed by his praise for Ostwald.
\ers

I don't remember his praise for Ostwald.  Who was Ostwald?  And what
does {\James} praise him about?

\brs
I am half way through writing a much improved draft of our paper.
\ers

Sounds great.

\section{20-11-01 \ \ {\it A Bathtub Moment} \ \ (to C. M. {\Caves} \& R. {\Schack})} \label{Caves39} \label{Schack35}

I'm sure I've already told you both the story of the time I happened
to end up at a British pub with Caroline Thompson, the famous
Bell-inequality conspiracy theorist, but let me repeat it for the
purpose of having it in this box.  Somebody at the table was speaking
of the great importance of intuition, of being able to ``see beauty
in a theory.''  I, with my usual example, piped up that I thought
that was hogwash:  I always thought Mary Ann was the prettiest girl
on Gilligan's Island; my best friend thought it was Ginger.  Anyway,
I followed that comment with, ``I never use any intuition in my
calculations; I don't even know what intuition can be in that
context.''  Caroline Thompson harrumphed, ``Well we could see that
from your talk!''

OK, so it won't be an intuition, but here's a hunch that hit me while
I was taking a shower this morning.  It's connected to the long note
on de Finetti I sent you yesterday.

Among the thirty other reasons I have been thinking that
trace-preserving quantum operations (and now measurement models
explicitly) are subjective entities is because one can make a
one-to-one correspondence between them and the density operators on a
larger Hilbert space.  That is, they have the same structure as the
states of belief that we've already toyed with.\footnote{By the way,
{\Carl}, as I recall, was never happy with thinking this point had any
significance for our program.  I think he saw it as little more than
a coincidence.}  This has suggested to me that there ought to be a
Gleason-like theorem for quantum operations (which I pursued a little
bit but never could quite make things connect).  But now this idea is
rearing its head again in the context of a de Finetti theorem for
models.

Yesterday, I blithely said something to the following effect.  To get
at a concept of an unknown model, what you do is 1) to 4) :

\begin{enumerate}
\item
Start with many copies of a quantum system, for which you believe 1)
that they are exchangeable and 2) that there is nothing left to
learn.  The standard quantum de Finetti theorem then gives us that
the density operator we ascribe to the collective system will be a
tensor product of identical quantum states.

\item
Now drop each of those systems into a measurement device and note the
outcomes $i$.  Separate the post-measurement systems into bins
according to those outcomes.

\item
Finally suppose we believe that the quantum state we ascribe to each
bin ought to be an exchangeable state for which we {\it can\/} learn
something.  The standard de Finetti theorem gives that this state
must be of de Finetti form (with a nontrivial support).

\item
The hunch was that the conjunction of 1), 2), and 3)---or them along
with some minor additional regularity condition---would specify the
content of the phrase ``an {\it unknown\/} quantum measurement
model.'' The unknown model is simply given by making explicit the
form of the linear maps connecting 1) to 3) for all possible inputs
into A.
\end{enumerate}

Notice that nowhere in there did I say anything about these maps
being completely positive.  I just chose the word ``linear'' for some
reason.  But surely the assumption of complete positivity must come
into this too.  So now my question:  UNDER THE ASSUMPTION that a
measurement model is a state of belief, can one adequately explain
the notion of an unknown measurement model by de Finetti techniques
WITHOUT the technical assumption that these linear maps are
completely positive?  Or, instead, is complete positivity absolutely
crucial to the program?

It strikes me that {\Carl}'s superoperator calculus has got to be the
way to go for exploring this question.

\section{20-11-01 \ \ {\it Copenhagen} \ \ (to C. M. {\Caves})} \label{Caves39.1}

Here's the little bit that Folse once wrote me about the Niels Bohr Institute that I was telling you about:
\bhf
It's a place of great history.  In Bohr's office they keep photographs of all of his prot\'eg\'es who won Nobel prizes.  It's quite a set of the 20th century's greatest.  Did you happen to notice that {\bf Travel and Leisure} magazine (I was reading in my ophthalmologist's waiting room) awarded Copenhagen the status of the most ``in'' city on the planet for 2000?
\ehf

\section{20-11-01 \ \ {\it {\James}'s Loose Lips} \ \ (to R. {\Schack})} \label{Schack36}

\brs
In lecture 2, {\James} says ``I found a few years ago that Ostwald, the
illustrious Leipzig chemist, had been making perfectly distinct use
of the principle of pragmatism [\ldots]''

In my own words, Ostwald rejected as meaningless any statement that
did not have observable consequences. A very pragmatic attitude
indeed. The trouble was that Boltzmann's ideas about atoms fell into
this category, at least that was the public opinion, led by Ostwald,
at the time. No cash value in the ``atom hypothesis''. I imagine
somebody like Ignacio (I pick him only because of his obvious
scepticism at my talk in Benasque) to ask: where is the cash value in
the Bayesian approach to quantum mechanics?
\ers

I agree with you now that that is a troubling praise coming from
{\James}.  I noticed similarly somewhere else in the book that he also
classified Mach in the ranks of the pragmatists.

But there is a grave distinction between positivism (Mach, Ostwald,
etc.) and pragmatism ({\James}, {\Dewey}, etc.) as I see it.  The
positivists eschewed all metaphysical assumptions---thus the egg on
their faces for not coming up to speed on the statistical mechanics
an atomic hypothesis can afford.  The pragmatists, on the other hand,
are willing to glorify any metaphysics with a cash value. This
relates to the passage by {\James} on Mill that I read you while you
were visiting.  With metaphysics, the cash value is not in its
explanation of any previously discovered facts, but in the concrete
actions its BELIEF will give rise to in the agent believing it. Thus
{\James}'s argument, for instance, for everyone's right to believe in a
God, even if that God will never have the opportunity of being
confirmed or falsified in an objective fashion.  A God's validity in
an agent is in his cash value for the agent's ethics, morals, and
mode of action for his daily life.

To put this in concrete terms for {\Carl} versus me:  I would say that
the ontological hypothesis I'm shooting for will show some cash value
in the amount of interesting physics it leads to, to the opening up
of new quantum computing and quantum control and quantum cryptography
methods.  And I think (or, more accurately, BELIEVE) it will help us
make the leap to the next stage of physics.  Whereas I would
say---but it's just a gut feeling---that the ontology {\Carl} has been
shooting for (i.e., the Hamiltonian) has no such cash value. Only the
money flow in the banks will ultimately tell.  (And unfortunately,
that can only be done with hindsight.)

{\James}---I think in his essay ``The Sentiment of Rationality''---has a
beautifully worded passage on these considerations that I'll try to
get scanned in tonight and sent to you.  But I think {\James} himself is
either not consistent in his writings, or he's pretty sloppy in
reading the other writers he wants to praise.  (What I know of him
now, I think it's probably mostly the latter.)

Sorry for writing all this.  I got carried away.  I hope to get your
new draft printed out today, and studied partially tonight.

By the way, I think Ignacio is a good benchmark with all this.  If we
can't find a way to impact him in five or seven years, say, then
maybe indeed all this is for naught.

\section{20-11-01 \ \ {\it One Horse's Mouth} \ \ (to J. M. Renes)} \label{Renes5}

I finally get to the last note I owe you.

\bjmr
You make this point in the nato paper explicitly (that the unitary
taking the quantum Bayes rule post measurement state to the orthodox
post measurement state is a ``mental readjustment'' and does depend
on the input state).  There's a lot here, though, especially since
Carl initially balked pretty hard at the idea since ``we're
physicists so I don't know what he's (you) talking about.''
\ejmr

I'm intrigued by your phrase ``there's a lot here, though.''  I don't
quite understand what you're trying to get at---that you agree that
it's a difficulty, or that it's a good thing?  Or that if {\Carl}
balked, that might be an unintentional mark in its favor?  Can you
explain a little better?

\bjmr
Some days I don't feel like a physicist (happily coinciding much of
the time with the days I don't want to) so I'm not initially troubled
by this.  However, as I said, there's a lot of ``stones unturned''
here. Are you saying that there is no physical picture of what's
going on, there must be some subjective element ``uncaptureable'' by
a physical picture? This seems to fit with your rep as being an
``extreme subjectivist'' but I'd rather hear it from the horse's
mouth.
\ejmr

I'm not sure exactly how I should reply to this.  You probably know
my thoughts at this point better than I know them myself.  Maybe I
should say it like this:  My pet idea at the moment is that there was
a world here before humankind ever appeared on the scene; there'll be
a world here after we disappear.  But I would say the world is still
under construction; there is no sense and no ultimate level at which
it is  already complete.  To that extent, I believe our beliefs, our
passions, our actions, our inventions, and our dreams modify the
world and form part of its construction in a nonnegligible way.

And I think our greatest hint of that comes from quantum mechanics. I
would say that what we're learning in a precise way from it is that
there is something about the stuff of the world that makes it
uncaptureable with a purely physical picture.  We find that we cannot
even draw a picture of the world without including our beliefs and
belief changes as a crucial background in the sketch. (How could we
if the world's not completed yet?)

Does that make me an extreme subjectivist?  I don't know.  Whatever
it is though that I should be called, I think this willingness to
accept a substantial part of quantum mechanics as simply ``law of
thought'' will keep me from going down a misguided path.  I.e., the
path of trying to ascribe all the easiest terms in the theory a kind
of physical reality independent of our presence as active agents.

One horse's mouth.

\section{21-11-01 \ \ {\it Pragmatism versus Positivism} \ \ (to R. {\Schack})} \label{Schack37}

Both quotes are taken from William {\James}'s essay ``The Sentiment of
Rationality.''
\bigskip

\noindent Quote I:

\bq
\indent
The necessity of faith as an ingredient in our mental attitude is
strongly insisted on by the scientific philosophers of the present
day; but by a singularly arbitrary caprice they say that it is only
legitimate when used in the interests of one particular
proposition---the proposition, namely, that the course of nature is
uniform. That nature will follow tomorrow the same laws that she
follows today is, they all admit, a truth which no man can {\it
know}; but in the interests of cognition as well as of action we must
postulate or assume it. As Helmholtz says: ``{\it Hier gilt nur der
eine Rat: vertraue und handle!}'' And Professor Bain urges: ``Our
only error is in proposing to give any reason or justification of the
postulate, or to treat it as otherwise than begged at the very
outset.''

With regard to all other possible truths, however, a number of our
most influential contemporaries think that an attitude of faith is
not only illogical but shameful. Faith in a religious dogma for which
there is no outward proof, but which we are tempted to postulate for
our emotional interests, just as we postulate the uniformity of
nature for our intellectual interests, is branded by Professor Huxley
as ``the lowest depth of immorality.'' Citations of this kind from
leaders of the modern {\it Aufkl\"arung\/} might be multiplied almost
indefinitely. Take Professor Clifford's article on the ``Ethics of
Belief.'' He calls it ``guilt'' and ``sin'' to believe even the truth
without ``scientific evidence.'' But what is the use of being a
genius, unless {\it with the same scientific evidence\/} as other
men, one can reach more truth than they? Why does Clifford fearlessly
proclaim his belief in the conscious-automaton theory, although the
``proofs'' before him are the same which make Mr.\ Lewes reject it?
Why does he believe in primordial units of ``mind-stuff'' on evidence
which would seem quite worthless to Professor Bain? Simply because,
like every human being of the slightest mental originality, he is
peculiarly sensitive to evidence that bears in some one direction. It
is utterly hopeless to try to exorcise such sensitiveness by calling
it the disturbing subjective factor, and branding it as the root of
all evil. ``Subjective'' be it called!\ and ``disturbing'' to those
whom it foils! But if it helps those who, as Cicero says, ``{\it vim
naturae magis sentiunt},'' it is good and not evil. Pretend what we
may, the whole man within us is at work when we form our
philosophical opinions. Intellect, will, taste, and passion
co-operate just as they do in practical affairs; and lucky it is if
the passion be not something as petty as a love of personal conquest
over the philosopher across the way. The absurd abstraction of an
intellect verbally formulating all its evidence and carefully
estimating the probability thereof by a vulgar fraction by the size
of whose denominator and numerator alone it is swayed, is ideally as
inept as it is actually impossible. It is almost incredible that men
who are themselves working philosophers should pretend that any
philosophy can be, or ever has been, constructed without the help of
personal preference, belief, or divination. How have they succeeded
in so stultifying their sense for the living facts of human nature as
not to perceive that every philosopher, or man of science either,
whose initiative counts for anything in the evolution of thought, has
taken his stand on a sort of dumb conviction that the truth must lie
in one direction rather than another, and a sort of preliminary
assurance that his notion can be made to work; and has borne his best
fruit in trying to make it work? These mental instincts in different
men are the spontaneous variations upon which the intellectual
struggle for existence is based. The fittest conceptions survive, and
with them the names of their champions shining to all futurity.

The coil is about us, struggle as we may. The only escape from faith
is mental nullity. What we enjoy most in a Huxley or a Clifford is
not the professor with his learning, but the human personality ready
to go in for what it feels to be right, in spite of all appearances.
The concrete man has but one interest---to be right. That for him is
the art of all arts, and all means are fair which help him to it.
Naked he is flung into the world, and between him and nature there
are no rules of civilized warfare. The rules of the scientific game,
burdens of proof, presumptions, {\it experimenta crucis}, complete
inductions, and the like, are only binding on those who enter that
game. As a matter of fact we all more or less do enter it, because it
helps us to our end. But if the means presume to frustrate the end
and call us cheats for being right in advance of their slow aid, by
guesswork or by hook or crook, what shall we say of them? Were all of
Clifford's works, except the {\sl Ethics of Belief}, forgotten, he
might well figure in future treatises on psychology in place of the
somewhat threadbare instance of the miser who has been led by the
association of ideas to prefer his gold to all the goods he might buy
therewith.

In short, if I am born with such a superior general reaction to
evidence that I can guess right and act accordingly, and gain all
that comes of right action, while my less gifted neighbor (paralyzed
by his scruples and waiting for more evidence which he dares not
anticipate, much as he longs to) still stands shivering on the brink,
by what law shall I be forbidden to reap the advantages of my
superior native sensitiveness? Of course I yield to my belief in such
a case as this or distrust it, alike at my peril, just as I do in any
of the great practical decisions of life. If my inborn faculties are
good, I am a prophet; if poor, I am a failure: nature spews me out of
her mouth, and there is an end to me. In the total game of life we
stake our persons all the while; and if in its theoretic part our
persons will help us to a conclusion, surely we should also stake
them here, however inarticulate they may be.
\eq

\bigskip
\noindent Quote II:

\bq
\indent
Now, I wish to show what to my knowledge has never been clearly
pointed out, that belief (as measured by action) not only does and
must continually outstrip scientific evidence, but that there is a
certain class of truths of whose reality belief is a factor as well
as a confessor; and that as regards this class of truths faith is not
only licit and pertinent, but essential and indispensable. The truths
cannot become true till our faith has made them so.

Suppose, for example, that I am climbing in the Alps, and have had
the ill-luck to work myself into a position from which the only
escape is by a terrible leap. Being without similar experience, I
have no evidence of my ability to perform it successfully; but hope
and confidence in myself make me sure I shall not miss my aim, and
nerve my feet to execute what without those subjective emotions would
perhaps have been impossible. But suppose that, on the contrary, the
emotions of fear and mistrust preponderate; or suppose that, having
just read the {\sl Ethics of Belief}, I feel it would be sinful to
act upon an assumption unverified by previous experience---why, then
I shall hesitate so long that at last, exhausted and trembling, and
launching myself in a moment of despair, I miss my foothold and roll
into the abyss. In this case (and it is one of an immense class) the
part of wisdom clearly is to believe what one desires; for the belief
is one of the indispensable preliminary conditions of the realization
of its object. {\it There are then cases where faith creates its own
verification}. Believe, and you shall be right, for you shall save
yourself; doubt, and you shall again be right, for you shall perish.
The only difference is that to believe is greatly to your advantage.

The future movements of the stars or the facts of past history are
determined now once for all, whether I like them or not. They are
given irrespective of my wishes, and in all that concerns truths like
these subjective preference should have no part; it can only obscure
the judgment. But in every fact into which there enters an element of
personal contribution on my part, as soon as this personal
contribution demands a certain degree of subjective energy which, in
its turn, calls for a certain amount of faith in the result---so
that, after all, the future fact is conditioned by my present faith
in it---how trebly asinine would it be for me to deny myself the use
of the subjective method, the method of belief based on desire!

In every proposition whose bearing is universal (and such are all the
propositions of philosophy), the acts of the subject and their
consequences throughout eternity should be included in the formula.
If $M$ represent the entire world minus the reaction of the thinker
upon it, and if $M + x$ represent the absolutely total matter of
philosophic propositions ($x$ standing for the thinker's reaction and
its results)---what would be a universal truth if the term $x$ were
of one complexion, might become egregious error if $x$ altered its
character. Let it not be said that $x$ is too infinitesimal a
component to change the character of the immense whole in which it
lies imbedded. Everything depends on the point of view of the
philosophic proposition in question. If we have to define the
universe from the point of view of sensibility, the critical material
for our judgment lies in the animal kingdom, insignificant as that
is, quantitatively considered. The moral definition of the world may
depend on phenomena more restricted still in range. In short, many a
long phrase may have its sense reversed by the addition of three
letters, {\it n-o-t}; many a monstrous mass have its unstable
equilibrium discharged one way or the other by a feather weight that
falls.
\eq

\section{23-11-01 \ \ {\it Bossa Nova} \ \ (to C. M. {\Caves})} \label{Caves39.2}

Thanks for doing such extensive preliminary work with Jim Hartle.

\bcc
P.S.  Jim also mentioned that the Perimeter Institute is currently
still in its big-shot hiring mode as far as foundations is concerned.
\ecc

As far as I can tell the ``big-shots'' haven't done a useful thing in this area for 66 years.

But many, many, many, many thanks for the rest of the comments.  Won't you come join us in being an organizer?

\section{23-11-01 \ \ {\it Long Hartle Note} \ \ (to C. M. {\Caves})} \label{Caves39.3}

Thanks for such a thorough note on your meeting with Hartle!  The information was valuable, indeed.  Now the thing I've got to think harder about is whether I really want to go through with this.  And maybe most importantly, is it worth it to me and the community to have a set of people whom I don't inherently trust judge what is and what is not interesting quantum foundations work?  I have the feeling I'm an outlier from the establishment in this regard.  It would certainly make a difference to me if you would accept being one of the organizers; but without that, I think I might turn toward a retreat on this.  I guess I just need to think about it more.

By the way, let me show you the line-up in {\Montreal} so far.  At some point, we'll expand the list a little beyond this until we fill about 30 slots.\medskip

\bq
\begin{supertabular}{llll}
\underline{Who}                 &  \underline{Where} &   \underline{Attend?} &   \underline{Preferences} \medskip\\
Howard Barnum       &  .us  &   Yes    &    1+ weeks     \\
Hans Briegel        &  .de  &   Yes    &    2+ weeks     \\
Jeffrey Bub         &  .us  &   Yes    &    1+ weeks     \\
Adan Cabello        &  .es  &   Yes    &    3 weeks      \\
Carlton Caves       &  .us  &   Yes    &    2+ weeks     \\
Richard Cleve       &  .ca  &   Maybe  &    1 week       \\
Nicolas Gisin       &  .ch  &   Yes    &    ??           \\
Daniel Greenberger  &  .us  &   Maybe  &    ??           \\
Lucien Hardy        &  .uk  &   Yes    &    2+ weeks     \\
Patrick Hayden      &  .us  &   Yes    &    1+ weeks     \\
Alexander Holevo    &  .us  &   Maybe  &    ??           \\
Richard Jozsa       &  .uk  &   Yes    &    1+ weeks     \\
Adrian Kent         &  .uk  &   Yes    &    2+ weeks     \\
Dominic Mayers      &  .us  &   Yes    &    1+ weeks     \\
David Mermin        &  .us  &   Yes    &    1+ weeks     \\
Michael Nielsen     &  .au  &   Yes    &    1+ weeks     \\
Asher Peres         &  .il  &   Yes    &    First week   \\
Itamar Pitowsky     &  .il  &   Yes    &    7--10 days    \\
Ruediger Schack     &  .uk  &   Yes    &    1 week       \\
Ben Schumacher      &  .us  &   Yes    &    3 weeks      \\
John Smolin         &  .us  &   Yes    &    2 weeks      \\
Robert Spekkens     &  .ca  &   Yes    &    2 weeks      \\ David Wallace       &  .uk  &   Yes    &    3 weeks      \\
William Wootters    &  .us  &   Yes    &    Oct 14, 15   \\
Arthur Zajonc       &  .us  &   Yes    &    few days     \\
\end{supertabular}
\eq

\section{25-11-01 \ \ {\it quant-ph/0106166} \ \ (to R. Cleve)} \label{Cleve2}

Thanks for the thoughtful comments!  Let me fire back a couple of my own.

\brc
What would the classical information theory version of the table on
page 4 look like? Somehow the axioms of probability theory don't come
out as crisply in English as those of relativity, do they?
\erc

I think that's a superbly relevant point; it's one that Charlie Bennett keeps making to me too.  Indeed, I do get a little worried about the axioms of information theory.  But, I'm fairly optimistic on the axioms of probability at least.  While you're at the ITP, you might take the opportunity to talk to {\Ruediger} {\Schack} or Carl Caves about the ``dutch book argument'' for the classical probability axioms.  They can sketch the argument for you, which I think you will find really intriguing.  In essence, with respect to a fairly natural game, it does squeeze the structure of classical probability theory into a couple of English sentences.

This sort of thing gives me a lot of hope for both classical information theory and quantum theory too.

\brc
Your comment on page 15 that starts on line 6 made me think of
something I recently worked out. Consider the case of a one-qubit
system in three possible states with an angle of 120 degrees between each pair and the goal being to determine which
state as well as possible -- is this the example you had in mind?
\erc

If you're thinking 120 degrees in Bloch-sphere space, then yes that was the example I had in mind.  On the other hand,
\brc
I had thought this was a good simple example where one can illustrate
how using a POVM is better than using von Neumann measurements
(assuming one isn't willing to go into a larger Hilbert space). But it
isn't. Instead of using the natural three-element POVM, one can first
randomly pick one of three orthogonal bases (120 degrees rotated from
each other) and then perform that von Neumann measurement. It turns
out that the net result of this process will yield exactly the same
information about which of the three states it was as doing the POVM
would.

Well, do you agree? My point here isn't against POVMs; I think there
are other examples where one really can do more with a POVM (without
enlarging the Hilbert space)
\erc
if you were thinking of what I said above, then I'm confident you made a mistake in your calculation.  What is unique about the famous three-outcome POVM is that no matter what the outcome, one of the three possible preparations is eliminated, leaving the posterior probability of either of the other two preparations at 1/2.  Thus the mutual information gained about the preparation is $\log(3/2)$.  On the other hand consider the randomized measurement you speak of.  First flip a (three-sided) coin, and then do one of the von Neumann measurements.  If you get a ``no'' answer for that two-valued measurement, then you will have indeed eliminated one of the three possible preparations \ldots\ giving $\log(3/2)$ bits.  On the other hand, if you get a ``yes'' answer---which happens some fraction of the time---you will get less than that.  For a ``yes'' does not eliminate any of the three possible preparations.  It only makes one of them more likely than the other two.

Does that make sense?  (It's a proof that goes back to one of Holevo's 1973 papers.)  On the other hand, maybe I didn't really understand the example you had in mind.

\brc
I am going to stop writing here (being halfway through the paper is a
good point to pause) and hope to follow up with more comments (for
what they're worth) soon.
\erc

They're worth quite a lot, and I'm very grateful.  More important than anything to me is that part of our community start thinking along these lines and make some progress (more than I can make with my feeble mind).  If you've got the interest, I've got the enthusiasm!

\subsection{Richard's Preply}

\bq
I made it out to ITP where I'll be for the last three weeks.
I've been meaning to read your paper for a while and on this
quiet -- and rainy! -- day where I'm experiencing some frustration
with a technical problem I decided this was as good a time as any
to read and think about {\tt 0106166}. Of course, your engaging literary
style gives a certain pleasure to the process. Here are some
(admittedly brief) comments.

The NATO meeting sounded nice. I agree with you that decoherence
doesn't occupy a sacred role in the foundations of quantum information.
And -- probably not surprisingly for someone with my background -- the
aspect of quantum mechanics that I also find the most interesting is
its underlying information theory.

I recall the table on page 4 from one of your talks and I always
find it compelling to see it, seeing the contrast between the
two columns. I guess that special relativity is about geometry
(more precisely a ``generalization of geometry'' that includes space
and time). The fact that our brains and language are well equipped for
spatial concepts, may be a reason why column one can be expressed
so eloquently in words in column two.

An analogous table for quantum information would be nice. It is
not really about geometry (in spite of geometric structure of
Hilbert spaces) but about information theory. (Like many others,
I think of quantum information as a ``souped-up'' version of classical information
-- i.e., a generalization of classical.)
What would the classical information theory version of the table
on page 4 look like? Somehow the axioms of probability theory
don't come out as crisply in English as those of relativity, do they?
On the other hand \ldots\ people don't seem to have that much trouble
with probability theory. Card players, actuaries, casino managers
all seem to do okay, in spite of numerous subtleties and the deep
philosophical issues that arise when really thinks about what
happens when one, say, flips a coin.

Your comment on page 15 that starts on line 6 made me think
of something I recently worked out. Consider the case of
a one-qubit system in three possible states with an angle of
120 degrees between each pair and the goal being to determine
which state as well as possible -- is this the example you had
in mind? I had thought this was a good simple example where one can
illustrate how using a POVM is better than using von Neumann
measurements (assuming one isn't willing to go into a larger
Hilbert space). But it isn't. Instead of using the natural
three-element POVM, one can first randomly pick one of three
orthogonal bases (120 degrees rotated from each other) and
then perform that von Neumann measurement. It turns out that the
net result of this process will yield exactly the same information
about which of the three states it was as doing the POVM would.
Well, do you agree? My point here isn't against POVMs; I think
there are other examples where one really can do more with
a POVM (without enlarging the Hilbert space).

I'm just reading the Bayesian stuff on page 21 and really like the
idea -- does the analogy really hold in some sort of sense?
I am going to stop writing here (being halfway through the paper
is a good point to pause) and hope to follow up with more comments
(for what they're worth) soon.
\eq

\section{26-11-01 \ \ {\it PRA Proofs} \ \ (to C. M. {\Caves} \& R. {\Schack})} \label{Caves40} \label{Schack38}

I'm in the office again finally, and I've read over the PRA proofs.
Of course, as always, I can't see that any of their changes were for
the betterment of the paper \ldots\ but this time at least, none were
overly annoying to me.

Here are my notes (which one of you two might want to incorporate
into a reply).

\begin{enumerate}
\item
page 2, ``If one accepts this conclusion \ldots'' :  Note they take away the ``a'' in front of the kets.  That grammatical change takes away the impression that the final state is unspecified within the
set. I vote that we force them to reinstate the ``a''s unless you can figure out a smoother way to express the proper idea.

\item
page 2, ``The physical basis of Einstein's \ldots'' :  They changed the end of the sentence to ``amenable to experimental testing.'' That seems odd to me.  Should we protest?

\item
page 3, ``We then use a version of the so-called Dutch-book \ldots''
: They want us to explain ``Dutch-book argument.''  That's pretty
stupid, given that they didn't ask us to explain ``Bayesian
probability theory'' or even ``Gleason's theorem.''  We even used the
warning sign ``so-called.''  I don't know what more can be said
without inserting Section II into the middle of this paragraph.
Perhaps you guys have a nicer way to approach it than I would.

\item
page 5, ``The probability assignment is thus inconsistent \ldots'' :
You'll note in the proof that the equation at the end of the sentence
is broken at the end of the line in an awful way.  Can we ask them to
keep it together?  Or perhaps we can simply display the equation.

\item
page 5, ``For example, normalization of the probabilities \ldots'' :
This is not a problem of theirs but a question of mine.  We end the
sentence with ``so obvious that it needs no justification.''  Do we
really need to insert that phrase?  I find it a little distracting
from the main point of the sentence.

\item
page 6, ``The keys to these results are \ldots'' :  I would change
the very last word to ``paragraphs.''

\item
page 10, ``The data gathered from the measurements are said \ldots''
: What is this, England?  ``The House of Commons have voted \ldots''
Horrible.  I view ``data'' as a collective noun.
\end{enumerate}

That's it really.

But let me take this opportunity to give one last ramble from the
heart. It has nothing to do with changing the present paper:  I just
want to say it because it's on my mind again.

After reading the paper once more, I found myself feeling awful
again.  I don't think there is a reader out there besides {\Schack} and
{\Caves} that will come away with the feeling that the Dutch-book
argument is {\it purely\/} an internal consistency argument (or
rationality check).  The English in a sentence like,
\bq
\noindent
Given the assumptions of Gleason's theorem, if a scientist has
maximal information, any state assignment that is different from the
unique pure state derived in the last paragraph is inconsistent in
the Dutch-book sense; i.e., it leads to a sure loss for a bet on the
outcome of a measurement on a single system that includes the unique
pure state among the outcomes.
\eq
is just loaded with imagery.  Who out there will read ``sure loss''
as anything other than a factual state of affairs---something
dictated by the world independent of the agent?  Who out there will
read the word ``unique'' as really meaning ``unique with respect to
the agent's belief of certainty''?  Who out there will not interpret
the phrase ``maximal information'' in an objectivistic way---i.e.,
that there is one and only one way to have maximal information?  Or
here's a better acid test:  If a reader were to be confronted with
our paper in its present form at the same time as the BFM paper,
would he be able to see any philosophical differences in the
approaches of the two papers?  I can't imagine it.  All that troubles
me very deeply: A small change of language really could have made all
the difference in the world.

\section{29-11-01 \ \ {\it Observation} \ \ (to J. M. Renes)} \label{Renes6}

\bjmr
Here's some stuff relating to what I wrote the other day from ``The
Taboo of Subjectivity'' by B. Alan Wallace:
\bq
\noindent
{\rm The disdain of scientific materialism for subjectivity has also
shaped the very concept of scientific observation. While
nonscientific kinds of observation also detect phenomena---such as
our joys and sorrows, hopes and fears, ideas and inspirations---they
are thought to be tainted by human subjectivity and are therefore
suspect. From the perspective of scientific materialism, human
sensory perception may be deemed not only unreliable but irrelevant.
For a scientific observation to take place, all that is required is a
detector, or receptor. The human eye is one type of receptor, which
detects a certain range of electromagnetic frequencies, but other
instruments also measure this and other types of information, and
they are regarded as more reliable.}
\eq
\bq
\noindent
{\rm In common parlance, for an observation to take place, the received
information must be transformed into humanly accessible information
that is, sooner or later, perceived and understood by a human being.
But according to scientific materialism, {\em observation} is
assimilated into the general category of {\em interactions}, thereby
freeing it from the subjectivity of its normal associations. This
interpretation is said to be central to grasping what is involved in
scientific objectivity in the search for knowledge and the
justification of belief.}
\eq
[reference to Dudley Shapere ``The Concept of Observation in Science
and Philosophy,'' Philosophy of Science {\bf 49} No.\ 4, 1982, pg
485.]
\ejmr

Beautiful quotes!  Do you have the whole Shapere article?  Do you
know whether he is for or against that conception of observation
(i.e., as interaction)?  If you do have the article, could you make a
copy and send it to me?  And what about the Wallace thing---is that a
book or an article?  How is it?

\section{29-11-01 \ \ {\it Observation, 2} \ \ (to J. M. Renes)} \label{Renes7}

\bjmr
As for Wallace, it's a book --- pretty good so far. He's arguing that
``scientific materialism'' (i.e.\ the way someone like Feynman thinks of
the world) is a bit like its own religion, with a major piece of
dogma being the exclusion of subjectivity from consideration.
\ejmr

Aha.  Have a look at William James's essay ``The Sentiment of Rationality'' (in most any collection of his essays) if you get a chance.  He also argues (for what amounts to) what you said above.

I'll try to see if I can find the Wallace book up here.

\section{29-11-01 \ \ {\it Community} \ \ (to W. K. Wootters)} \label{Wootters3}

What a beautiful letter; thank you.

In a way, I've been going to my own seminar on science and religion
lately---but with me, through my reading choices.  I've gotten stuck
on the ``pragmatism'' movement, predominantly William {\James}'s
version. I had never realized before what a wealth of material was
there (for the kinds of thoughts I'd like to think, about quantum
mechanics in particular).  Nor had I realized how {\Wheeler}esque and
Woottersesque {\James} was in his outlook---hanging so much on the idea
that the universe is (in part) a product of our collective
experience.

In that regard, but with respect to religion, I have never been so
impressed by the possibility---necessity even---of ``faith'' than
when I read {\James}'s articles ``The Sentiment of Rationality'' and
``The Dilemma of Determinism.''  They have made a huge effect on me.
If you happen to read them and have any reaction, I would love to
hear your thoughts.  (For fun I just had a look at the Williams
College library; you guys have a collection of 63 items penned by
{\James}!  In contrast, not one hit in the Bell Labs library \ldots\ but
who would have guessed otherwise.)

I sympathize with your unease with telepathy.  Here's how Stephen
Brush put it:
\bq
\indent
`{\Wheeler}'s dilemma' is this:  how can one maintain a strong version
of the Copenhagen Interpretation, in which the observer is
inextricably entangled with that which is observed, while at the same
time denying that our consciousness affects that which we are
conscious of---and thus accepting the possibility of telekinesis and
other psychic effects?  For {\Wheeler} himself there is no dilemma at
all; one simply has to recognize `the clear distinction between (1)
the strange but well verified and repeatable features of quantum
mechanics and (2) the pseudo-scientific, non-repeatable and
non-verified so-called extra sensory perception.' But {\Wheeler}'s own
views are likely to strike a non-physicist as being just as bizarre
as those of the parapsychologists he deplores. Indeed, no one has yet
formulated a consistent worldview that incorporates the Copenhagen
Interpretation of Quantum Mechanics while excluding what most
scientists would call pseudo-sciences---astrology, parapsychology,
creationism, and thousands of other cults and doctrines.
\eq

The issue whether there is something besides unitarity---whether it
can ever ``breakdown'' as you put it---is acute, but I think a lot of
the problem hinges on how one views unitarity's status in a physical
theory.  I am inclined to believe that unitarity---or more generally
``trace-preserving complete positivity''---does not breakdown either.
But that is because I am inclined to view the time-evolution mapping
one ascribes to a system as an epistemological entity (rather than an
ontological one), much as I view the quantum state one assigns the
same system.  That is to say, I am struck that there is a deep reason
one can make a bijective correspondence between the completely
positive maps on one Hilbert space and the density operators in a
larger space, a reason that has nothing to do with imagining
ancillary systems:  A CP map {\it is\/} a density operator, it is a
state of knowledge in just the same way any other quantum state is.

There are other (peculiarly quantum) reasons for saying what I just
did above, but it is also part of a larger program I have in mind (and
one I think {\James} had in mind too)---namely, to find a little
slippage between the notion of a physical theory and the world itself.
Below I'll quote a little piece I wrote to {\Carl} {\Caves} on the
idea.  It gets at the sentiment, even if not at the technical details!
[See 04-10-01 note ``\myref{Caves3}{Replies on Pots and Kettles}'' to {\Caves} \&
  {\Schack}.]

\section{30-11-01 \ \ {\it Ramsey Theory} \ \ (to W. K. Wootters)} \label{Wootters4}

Just a small note.  I want to bring up a reference that may or may not be relevant to thinking about your toy graph model---it's something I just learned about last week, called Ramsey Theory.  Here's an example, ``In any collection of six people either three of them mutually know each other or three of them mutually do not know each other.''  And more generally, ``Every system of a certain class possesses a large subsystem with a higher degree of organization than the original.''  It's a subfield of graph theory, quite developed, with loads of examples like that.

Anyway, for what it's worth, I was told that the following is a good starter reference:
\bq\noindent
Ronald L. Graham, Bruce L. Rothschild, and Joel H. Spence, {\sl Ramsey Theory}, second edition (John Wiley and Sons, New York, 1990).
\eq

\section{30-11-01 \ \ {\it Some Thoughts on Your Paper(s)} \ \ (to K. Svozil)} \label{Svozil1}

\bks
I would kindly like to ask two questions:

(i) are the Bayesian consistency requirement in any way related to
Boole's  ``conditions of possible classical experience'' (which have
been interpreted by Pitowsky in terms of the faces of classical
correlation polytopes)?

(ii) have you ever attempted to apply the methods to ``exotic'' but
non-quantum event structures? It might at least be pedagogically
helpful to consider such cases.
\eks

Thanks for your letter.  You ask two good questions, and
unfortunately I don't know the answer to either one!  I will however
have a good look at your papers and start thinking about them.

I think the Dutch-book argument for probabilities is not so strongly
tied to the Boolean structure of propositions (as, say, Cox's
argument for the probability axioms is).  For instance, just look at
Eqs.\ (1) and (2) in our paper.  It seems to me that that part of the
argument does not care one iota whether there is a distributive law
for this event structure or not.  If that's the case throughout the
remainder of the Dutch-book mechanics, then maybe this is an
important point.

\section{30-11-01 \ \ {\it Later in the Book} \ \ (to J. M. Renes)} \label{Renes8}

More great quotes!  Keep 'em coming (as you like).

\subsection{Joe's Preply}

\bq
Wallace says
\bq
James's philosophy of radical empiricism rejects the absolute duality of
mind and matter in favor of a world of experience, in which consciousness
{\it as an entity}, in and of itself, does not exist; nor is it a function of
matter, for matter {\it as an entity\/} in and of itself, does not exist either.
According to this view, the postulation of mental and physical substances
is a conceptual construct, as is the metaphysical distinction between
subject and object.  Mind and matter are constructs, whereas pure
experience, which is neutral between the two, is primordial. One
implication of the hypothesis that we are directly acquainted with reality
is that the contents of consciousness can no longer be regarded as being
``in the mind'' (let alone in the brain). Reality just {\it is\/} the flux of
experience.
\eq

So my thoughts on this subject aren't very original!  But to be in such
company is far better.  Wallace goes on to laud James some more, but does
mention some work of Hilary Putnam:
\bq
If all valid statements concerning the world of human experience have
both a conventional and a factual element, it follows that the referents
of language are also inseparable fusions of convention and reality. Thus,
the existence of a concrete object like a tree is also a matter of
convention, and our observation of a tree is possible only in dependence
on a conceptual scheme. The reason for this, according to Putnam, is that
`elements of what we call ``language'' or ``mind'' penetrate so deeply into
what we call ``reality'' that the very project of representing ourselves as
being ``mappers'' of something ``language-independent'' is fatally compromised
from the very start.' [from Putnam, {\sl Realism with a Human Face\/} 1990]
\eq
This ending quote from Putnam reminds me a lot of the squabbles about the
wavefunction and density matrix.

Okay, can't stop quoting.
\bq
Once we have chosen a conceptual scheme, there are facts to be discovered
and not legislated by our language or concepts. Our conceptual scheme
restricts the range of descriptions available to us, but it does not
predetermine the answers to our questions. As Putnam comments,
\bq\noindent
	``The stars are indeed independent of our minds in the sense of
	being causally independent; we did not make the stars \ldots\ The fact
	that there is no one metaphysically privileged description of the
	universe does not mean that the universe depends on our minds.''
\eq

On the other hand, if there were no language users, there would not be
anything true or anything with sense or reference. Thus, the rich and
ever-growing collection of truths about the world is the product of the
experienced world, with language users playing a creative role in the
process of production.
\eq
\eq

\section{30-11-01 \ \ {\it NMR Stuff} \ \ (to C. M. {\Caves})} \label{Caves40.1}

I just read the introduction and conclusions of your paper with Menicucci.  It's a nice paper, very sober style.\footnote{N. C. Menicucci and C. M. Caves, ``Local Realistic Model for the Dynamics of Bulk-Ensemble NMR Information Processing,''  Phys.\ Rev.\ Lett.\ {\bf 88}, 167901 (2002), \quantph{0111152}.}  Interesting how this hidden-variable model requires an encoding of the full quantum state at each site.  It reminds me of a description Charlie Bennett once gave to the Deutsch/Hayden paper (you can find it on {\tt quant-ph}).\footnote{D. Deutsch and P. Hayden, ``Information flow in entangled quantum systems,'' Proc.\ Roy.\ Soc.\ Lond.\ A\ {\bf 456}, 1759--74 (2000), \quantph{9906007}.}  He said something like, ``They show that you can always think of quantum mechanics as a local hidden variable theory {\it if\/} all the localized systems carry the information of the full quantum state.''  (Or, at least that was Charlie's take on the paper after hearing a description of it from John Smolin \ldots\ who couldn't confirm that that was his own opinion of the paper!!  So, Charlie never actually read the paper.)  But anyway, after Charlie said this, we all---Smolin, van Enk, Leung, and some summer students---spent an hour or so playing at the board trying to see if one could get rid of such an ``embarrassment of riches of information.''  In particular, Charlie raised the question of whether one could get away with each site only carrying information about its previous interactions \ldots\ or perhaps some other exponentially smaller information than the full quantum state.  Of course, we came to no conclusions, but it didn't look to be possible.

Anyway, I'm not sure I have a well-posed question, just a wonder \ldots\ \ How is bulk-ensemble NMR more hidden-variable friendly than full-blown quantum mechanics if what Charlie said above is correct?  Or is there an easy way to see that Charlie's description of Deutsch/Hayden is just wrong?

\section{04-12-01 \ \ {\it The Power of Advertising} \ \ (to A. Peres \& D. R. Terno)} \label{Peres22} \label{Terno2}

Actually, reading over this again from your upcoming paper,
\bq\noindent
The root of the difficulty we have to transform quantum expressions
from one Lorentz frame to the other is that the process called
``quantum measurement'' is an intervention in the quantum dynamics by
an ``exosystem'' [finkel], namely by an apparatus which is not
completely described by the quantum formalism [interv].
\eq

I wonder if I can ask a favor of you?  In my own paper,
\begin{itemize}
\item
C.~A. Fuchs, ``Quantum Foundations in the Light of Quantum Information,'' to appear in {\sl Proceedings of the NATO Advanced Research Workshop on Decoherence and its Implications in Quantum Computation and Information Transfer}, edited by A.~Gonis (Plenum Press, NY, 2001 or 2002). (Until then, see \quantph{0106166}.)
\end{itemize}
I go to great lengths to try to explain why the term ``intervention'' is a better one than ``measurement,'' and I try hard to promote its more frequent use in future dialogues on the subject.  Of course, I cite Asher for the original introduction of the term in that.  Nevertheless, might I ask you guys to cite this also?  Certainly feel free to ignore this request as you wish:  It just struck me that I might increase my readership if I happened to get a citation from you!

\section{05-12-01 \ \ {\it Yep} \ \ (to R. Pike)} \label{Pike4}

Regarding ``Challenging Particle Physics as Path to Truth'' by George Johnson in yesterday's {\sl New York Times\/}:
\bq
\myurl{http://www.nytimes.com/2001/12/04/science/physical/04SQUA.html}.
\eq
Yep, you know the anti-unificationists have my sympathy.

\section{05-12-01 \ \ {\it Dear Prudence} \ \ (to C. M. {\Caves} \& R. {\Schack})} \label{Caves41} \label{Schack39}

\noindent Fellow Bayesians, \medskip

(Let's see if {\Carl}'s knowledge of 1960s music can help him guess the
origin of this note's title.)

Anyway, to the real subject.  Dutch-book coherence?  Dutch-book
consistency?  Neither of them seem to be an accurate account of
what's going on with the theorem.  I know neither one of you will
accept it, but I think ``prudence'' gets significantly closer to the
mark.

\begin{itemize}
\item
{\bf Consistent} -- 1.  In agreement; compatible: ``The testimony was
consistent with the known facts.'' 2.  Being in agreement with
itself; coherent and uniform: ``a consistent pattern of behavior.''
3. Reliable; steady: ``demonstrated a consistent ability to impress
the critics.''  4.  In mathematics, having at least one common
solution, as of two or more equations or inequalities.

\item
{\bf Coherence} -- 1.  The quality or state of cohering, especially a
logical, orderly, and aesthetically consistent relationship of parts.
2.  In physics, the property of being coherent, as of waves.

\item
{\bf Prudent} -- 1.  Wise in handling practical matters; exercising
good judgment or common sense. 2.  Careful in regard to one's own
interests; provident. 3.  Careful about one's conduct; circumspect.

\item
{\bf Prudence} -- 1.  The state, quality, or fact of being prudent.
2. Careful management; economy.
\end{itemize}

How did {\Mermin}'s talk go?  How did {\Ruediger}'s talk go?

\section{05-12-01 \ \ {\it Lucky Seven}\ \ \ (to B. W. Schumacher)} \label{Schumacher4}

5)  I was a little disappointed to learn from your prospectus that you'll be going to Caltech for your sabbatical, but on the other hand I can easily imagine several reasons why that's the best choice for you.  Still, I'm going to miss having you around \ldots\ as will all the other guys in Bell Labs and the Bell Labs metro area, Charlie Bennett, David DiVincenzo, Steven van Enk, Danny Greenberger, Lov Grover, Mark Hillery, Alexander Holevo, Debbie Leung, Eric Rains, Terry Rudolph, Peter Shor, Dick Slusher, John Smolin, Barbara Terhal, Bill Wootters, and so forth. \frownie

6)  You know I'm always scheming, and this is one of the latest ones.  Actually it's an older one, but revived with a new head of steam.  Jeff Bub and I have decided to edit a special issue of {\sl Studies in the History and Philosophy of Modern Physics\/} (SHPMP) on \ldots\ (you can guess) ``Quantum Foundations in the Light of Quantum Information.''  But there are two things going on with this:  1) We pretty much want to get it done and published {\it before\/} the {\Montreal} meeting, and 2) we really only want to go through with it if we can get commitments for some really good papers.  Jeff wrote this in one of our emails:
\bjb
We'd really have to think hard about how to ensure that the authors
all deal with some conceptual or foundational question relevant to
quantum information in a philosophically serious way. We want to avoid
the issue just being `intro to quantum information for
non-specialists,' with some anecdotal interpretative comments as a nod
to philosophy. And it also shouldn't just be a forum for one point of
view.

The difficulty, of course, is that very few philosophers have actually
worked in this field. So we have to rely mostly or perhaps entirely on
scientists for the papers. It will be hard getting papers from
physicists and computer scientists for a philosophy journal. On the
plus side, it does seem to be the case that some people working in
quantum information have thought hard about the relevance of their
research for foundational questions. Somehow, we have to motivate
people to take it as a worthwhile and exciting challenge to write a
paper that tackles a philosophical issue in an intellectually
disciplined way --- as intellectually disciplined as these authors
invariably are when they write a technical scientific paper. Too
often, all analytical controls seem to be switched off when scientists
venture into philosophical terrain.
\ejb

We've already gotten a commitment from Andrew Steane to beef up his ``quantum computing needs only one universe'' paper for it.  Moreover, I know that you're also capable to the task, and already somewhat semi-committed to a paper on ``doubting Everett.''  So I'm hoping I can get an official nod from you in the YES direction.  If I can secure that, then I think I can make a weighty case to Charlie Bennett to record (more importantly defend) his thoughts on why he thinks Everett provides an adequate and conceptually simpler picture of quantum foundations than some other interpretive attempts.

Then beyond that I'm hoping to get some papers from {\it possibly\/} Jozsa, Caves \& Schack, Mermin, Wootters, and maybe a couple of others (and Bub and myself of course).  That's already more than enough.

Once I hear from you and Charlie, then Jeff and I will kick into ``phase two'' of our baiting, i.e., getting the other guys to say yes to getting a paper in before next summer.

7)  I really loved the ``How does it know?''\ story in your prospectus!  Let me give you two stories from my Samizdat in connection to that below.  Knowing and flying:  the two questions might be a little bit of the same thing.

\bq
\subsubsection{From a 17 December 1997 note to G. L. Comer, ``It's a Wonderful Life''}

Good holidays to you.  This morning, as I was driving to work, it
dawned on me that roughly this day 10 years ago, I was conferred my
degrees at the University of Texas.  Time does fly.

It made me think of a little anecdote about John {\Wheeler} that I heard
from John Preskill a few days ago.  In 1972 he had {\Wheeler} for his
freshman classical mechanics course at Princeton.  One day {\Wheeler}
had each student write all the equations of physics s/he knew on a
single sheet of paper.  He gathered the papers up and placed them all
side-by-side on the stage at the front of the classroom.  Finally, he
looked out at the students and said, ``These pages likely contain all
the fundamental equations we know of physics.  They encapsulate all
that's known of the world.''  Then he looked at the papers and said,
``Now fly!'' Nothing happened.  He looked out at the audience, then
at the papers, raised his hands high, and commanded, ``Fly!''
Everyone was silent, thinking this guy had gone off his rocker.
{\Wheeler} said, ``You see, these equations can't fly.  But our universe
flies. We're still missing the single, simple ingredient that makes
it all fly.''

Merry Christmas.

\subsubsection{From a 02 December 1997 note to J. Preskill, \ ``Flying Equations''}

I couldn't help but think of the anecdote about John {\Wheeler}'s
(non-)flying equations you told the other day when I came across the
following little passage (presumably Biblical in origin):
\bq
\noindent ``I forbade any simulacrum in the temples because the
divinity that breathes life into nature cannot be represented.''
\eq
\eq

\section{06-12-01 \ \ {\it Baudrillard Maybe?}\ \ \ (to B. W. Schumacher)} \label{Schumacher5}

\bbs
I certainly loved the samizdat stories you included.  Where did the
second quotation, the one about ``the divinity that breathes life into
nature'', come from?  What a counterpoint to Wheeler's futile ``Fly!''
\ebs

That's a good question; I really don't know anymore.  Looking further through the Samizdat I see that I reported reading Jean Baudrillard on November 23.  And I sort of faintly remember learning the word ``simulacrum'' from whatever book that was.  So, I'd bet it was from there.

\bbs
Well, yes, dammit.  I owe you this paper.  At this moment Mike W. and
I are finishing off a paper on approximate error correction, but I will
try to get something down afterwards.  [\ldots]  We have
also promised a paper to Mary Beth Ruskai for JMP. [\ldots]
But I owe you longer; I will put together a very rough draft over the
vacation, and run it past you --- I'd much value your comments.
\ebs

That's great!!!  I'll inform Jeff Bub, and move to Phase 1.5 --- i.e., coercing Bennett.  Wish me some success there; it's gonna take a lot of work to loosen him up.

\section{06-12-01 \ \ {\it Investigative Reporting} \ \ (to B. W. Schumacher)} \label{Schumacher6}

It looks like Baudrillard was the right path.  I just got on {\tt altavista.com} and did a search on simulacrum, Baudrillard, and divinity.  And I found the following:
\bq
Baudrillard has interesting views on the nature of postmodern
nihilism. In Simulacra and Simulation, he begins his discussion on the
precession of simulacra by describing what he calls the ``simulacra of
divinity.'' He points out that ``divinity that animates nature can never be represented'' (p.\ 4). The idea of God embodied within religious icons and iconography is a complex system of human belief, in which the visible is supposed to evoke the religious all-being, the
divine-referential. Yet when God's omnipotence and presence can only
be felt through these icons, it is becomes probable that the existence
of God is questioned, if the simulacra of divinity becomes the only
visible image of God's presence.\footnote{From \myurl{http://www.cyberartsweb.org/cpace/theory/baudrillard/lee.html}.}
\eq
Then I found an exact replica of the quote in someone's class notes about Baudrillard.  So, I'm sure I got it from him.

So the question now is where did Baudrillard steal the quote from?\footnote{\editornote Voltaire.  In his \emph{Dictionnaire philosophique} (1764), he puts words into the mouth of the second king of Rome.  ``Je suis Numa Pompilius, me dit-il ; je succ\'edai \`a un brigand et j'avais des brigands \`a gouverner : je leur enseignai la vertu et le culte de Dieu ; ils oubli\`erent apr\`es moi plus d'une fois l'un et l'autre ; je d\'efendis qu'il y e\^ut dans les temples aucun simulacre, parce que la Divinit\'e qui anime la nature ne peut \^etre repr\'esent\'ee.'' [``I am Numa Pompilius,'' he told me. ``I succeeded a brigand and I had brigands to rule.  I taught them virtue and the worship of God; after me, they forgot both more than once.  I forbade that there be any images in the temples, for the Divinity which animates nature cannot be represented.'']  Baudrillard's quotation is verbatim, but unattributed.\label{Voltrillard}}

\section{07-12-01 \ \ {\it A Sinking Ship?}\ \ \ (to C. M. {\Caves} \& R. {\Schack})} \label{Caves42} \label{Schack40}

\noindent Fellow Bayesians,\medskip

Especially in light of what {\Carl} wrote me yesterday, perhaps you
should both have a serious look at Gavriel Segre's Ph.D. Thesis at
\quantph{0110018v5}.  Van {\Enk} pointed it
out to me this morning, and believe me it'll be worth your time!  In
particular, pay attention to Theorem 5.2.21 ``Impossibility of a
Subjectivistic Bayesian Foundation of Quantum Probability Theory'' on
page 193.  Maybe we ought to get off this ship before it sinks?

Also, make sure you don't miss reading his acknowledgments on page 5.
This is surely the best quote I've ever seen on {\tt quant-ph}!

\section{07-12-01 \ \ {\it A Sinking Ship?, 2} \ \ (to C. M. {\Caves} \& R. {\Schack})} \label{Caves43} \label{Schack41}

\brs
Are you serious? Of course, the thought has crossed my mind.
\ers

The sinking ship remark was a joke.  The reference to the
(unaccepted) thesis was a scoff (or at least a little).  You did read
the ``acknowledgment'' of the thesis, didn't you?  Also look at the
other contents of the ``thesis.''

I have never been more confident about anything scientifically than
that we are on the right track in our quantum Bayesianism.  No
Bayesian that I've ever read called a probability assignment a
``state of knowledge.''  What's different about quantum states?  Why
should they be above the ``degree of belief'' that every classical
assignment is?  Taking the ``state of belief'' appellation seriously
is simply the cross we have to bear \ldots\ at least as I see it.
(Sorry for the Christian motif, {\Carl}.)  But I'm also confident that
it's not that bad anyway; it's just a question of packaging.

\section{07-12-01 \ \ {\it Weekend Curiosity} \ \ (to C. M. {\Caves} \& R. Schack)} \label{Caves43.1} \label{Schack41.1}

\noindent {\Ruediger}, Carl, Carl, {\Ruediger},

\bcc
{\Ruediger}'s talk was more of a problem and illustrated why we are in
heaps of trouble with the Bayesian view in qm and why we are going to
get in heaps more trouble with the ``states of belief'' formulation in
place of ``states of knowledge.''  You will be pleased to know that John
Preskill immediately focused on the exactly the difference between
these two that you have harped on, i.e., that knowledge can be relied
on as something approaching a fact, whereas belief is much weaker.
\ecc

Can either of you expand on this?  I sure would like to hear a more full account.  Perhaps we could talk by telephone today.  Will either of you be in your office today, or is the conference still going on?  (I've got Carl's number, but I don't have {\Ruediger}'s.)

\subsection{{\Ruediger}'s Reply}

\bq
I talked about entanglement as a state of belief, and I don't think
the audience appreciated it. This has affected my confidence: a bad
moment for a joke about sinking ships! There is clearly a danger that
I am ruining my scientific credentials by presenting stuff like this.
What this means is that we need theorems!
\eq

\section{10-12-01 \ \ {\it Lost It} \ \ (to C. M. {\Caves})} \label{Caves44}

I'll try to get in touch with you later today.  Sorry we missed each
other over the weekend.

\bcc
I managed to lose your last e-mail, where you expressed the view that
no Bayesian views a probability as based on a state of knowledge.  In
the absence of your full statement, here's a thought on that point.
\ecc

I don't balk against ``facts'' having a say in determining
probability assignments.  I balk against ``facts'' {\it uniquely\/}
determining them.

\section{10-12-01 \ \ {\it The Spirit of Gandhi} \ \ (to N. D. {\Mermin})} \label{Mermin49}

Well, you took away some of my fun with the letter you just sent me!
Over the weekend, I ran across your talk at the ITP and listened and
watched the whole thing:  In fact, I was going to send you some
direct comments on it.  Now some of my steam is taken away.

Still, let me gingerly point out some things I had wanted to point
out earlier \ldots\ even if they may not be quite as relevant
anymore.

Somewhere around 36 minutes into the talk---actually it was in your
reply to a question (which sounded to come from Jeff Kimble)---you
said:
\bdm
Our presentation, at least in the paper we submitted on the web, can
be read as being tinged with a view that quantum states are more than
a reflection of knowledge \ldots
\edm
the implication seeming to be that in the present talk you strove to
get around that.

However, at 23 minutes into the talk you said this:
\bdm
Of course there's the question of what it means for the combined
knowledge of all observers to constitute a consistent body of
knowledge about $S$, which is an interesting question.  I'm taking a
kind of dumb-physicist view, which is that there should be---at least
in principle---there should be one observer who has a lot of data
about various measurements and mutually commuting observables made on.
The observer having access to all the data will realize that if
\ldots
\edm

I know I have said all this before, but let me ``focus in'' on these
two passages and try to say it again (per your warning about me
``focusing out'' too often).  I would say that what you said in the
second passage has nothing to do with it being a ``dumb-physicist
view'' of things.   It is, however, the sine qua non of a view
``tinged'' with making the quantum state more than a simple epistemic
entity.  For it in essence says, there {\it should be\/} (your words) a
``right'' quantum state, or a range of ``right'' quantum states, and
that has nothing to do with any {\it actually existent\/} observers.

Please do think about the similarities between your ``dumb-physicist
view'' and the old limerick:

\begin{verse}
There was a young man who said, ``God \\
Must think it exceedingly odd \\
If he finds that this tree \\
Continues to be \\
When there's no one about in the Quad.''\bigskip\\

REPLY \medskip\\
Dear Sir: \medskip\\
Your astonishment's odd: \\
I am always about in the Quad. \\
And that's why the tree \\
Will continue to be, \\
Since observed by \medskip\\
     Yours faithfully,
     God.
\end{verse}

I would say that in supposing THERE SHOULD BE a superobserver, you
are supposing that the quantum state should---in essence---already be
there without any observer at all.  When Bishop Berkeley ran into
trouble with the question of where the trees go when there's no
observer, he invoked the idea that God was there all along.  Your
superobserver is in essence a God, who---through his own
objectivity---re-endows the quantum state with an objectivity I
THOUGHT you were trying to get rid of in the first place.

But maybe that was never your goal.

Martin Gardner said something very clear in this regard in his essay
``Why I Am Not a Solipsist,'' and so let me quote it:
\begin{quotation}
\noindent
In this book I use the term ``realism'' in the broad sense of a
belief in the reality of something (the nature of which we leave in
limbo) that is behind the phaneron, and which generates the phaneron
and its weird regularities.  This something is independent of human
minds in the sense that it existed before there were human minds, and
would exist if the human race vanished.  I am not here concerned with
realism as a view opposed to idealism, or realism in the Platonic
sense of a view opposed to nominalism or conceptualism.  As I shall
use the word it is clear that even Berkeley and Royce were realists.
The term of contrast is not ``idealism'' but ``subjectivism.''
\end{quotation}
(The phaneron, by the way, was C.~S. {\Peirce}'s term for ``the world of
our experience---the totality of all we see, hear, taste, touch,
feel, and smell.'')

If I were to give the BFM paper and your quest to make sense of
Peierls (and Bohr and Heisenberg, as evidenced by your talk) a
reading, I would say that what you are trying to do is give the
quantum state an idealistic interpretation (via the word
``knowledge'') and thinking that that somehow contrasts with a
realistic interpretation (and in doing so might fix everything up
quantum interpretation-wise).  But---in analogy to Gardner---for me,
``realistic'' and ``idealistic'' interpretations of the quantum state
amount to the same thing.  What I'm worried about is whether one can
make any sense of the quantum state at all without simultaneously
positing the active agent who makes use of it---I would claim you
can't.  And thus I'm left with the kind of ``subjectivism'' (for the
quantum state) that frightens so many people. \medskip

\noindent PS.  In case you're wondering why I titled this note ``The
Spirit of Gandhi,'' it is because I am hoping you will think of it as
a form of nonviolent protest.  Too many times in your talk you
pointed out what a violent reaction I had to the BFM result, and I
just can't think that that helped my reputation as a rational thinker
or the seriousness with which a Bayesian-kind of quantum-foundation
attempt should be viewed.  (But I love you just as much as
ever---maybe more with all the advertising you gave me---and there
are no grudges.)

\section{10-12-01 \ \ {\it Trumps and Triumphs} \ \ (to N. D. {\Mermin})} \label{Mermin50}

Also,

\bdm
I did urge {\Carl} and {\Ruediger} to do something about the phrase (in your other paper) ``Gleason's theorem is the greatest triumph of
Bayesian reasoning'' which I read as a claim that Bayesian reasoning
is the only way (or at least far and away the best way) to derive and
understand Gleason's theorem.  They explained that what you really
meant was that Gleason's theorem, by providing stringent constraints
on possible prior distributions, provided a powerful tool for
Bayesian reasoning. But I think they did an inadequate job of
removing the ambiguity, which is too bad, since I am rather sure it
will put off most of the people you ought to be addressing.
\edm

I'm glad you urged them.  Here's what I had said to {\Carl} and
{\Ruediger} in a September 20 note:
\bq
\noindent
Another thing I guess I really didn't like---and this is only a new
one for me, it's not something that had eaten away at me before---is
the slogan ``Gleason's theorem can be regarded as the greatest
triumph of Bayesian reasoning.''  To say that is to imply that
Bayesian reasoning IMPLIES Gleason's theorem.  I don't think you mean
that. I think what you mean is that it is one of the most valuable
additions to Bayesian reasoning ever.  But I won't cause trouble
here:  I know you both like the saying.
\eq

But they generally just view me as a trouble-maker.

\section{10-12-01 \ \ {\it Reread (pronounced re-red)} \ \ (to N. D. {\Mermin})} \label{Mermin51}

I'm just reading through all my emails of the day again.

\bdm
P.S.  Spent a lot of time talking with {\Carl} and {\Ruediger} in SB,
which I enjoyed very much.  I think the first half of your draft
joint paper formulates the basis for the ``necessary'' condition of
BFM much more coherently than we did: i.e.\ it is necessary if there
is to be any density matrix that does not contradict anybody's
strongly held beliefs.  Much better than talking about ``subsets of a
body of data that anybody with access to all of which would agree
constitutes nothing but valid results of measurements, none of which
invalidate earlier measurement outcomes\ldots.''  It's hard even to put
into a grammatical sentence.  The other four cases strike me as
interesting extensions of the case we (and Peierls) were addressing,
which concerned existing knowledge prior to any subsequent
interventions. My feeling was that there was no serious disagreement
among the three of us, but maybe they will give you different
reports.
\edm

Thanks, that's a very nice compliment.  Here's another way to put it,
which might be my preferred way.  Suppose two strongly consistent
agents want to stand a chance of coming to agreement, just by talking
to each other.  Then BFM is the necessary condition.  If it is not
satisfied, then the only way they will ever get anywhere is by
``consulting'' the system with a (mutually agreed upon) measurement:
With it, they can bend the world into something more congenial to
both.

\section{10-12-01 \ \ {\it Peace Pipe} \ \ (to T. Rudolph)} \label{Rudolph1}

Thanks for the peace pipe.

I have seen the paper by Fivel before.  We discussed it at a Maryland conference once.  It somehow doesn't have the right flavor for what I've been seeking though:  In the end, I want to find a kind of quantum ontology I can be happy with, not just something that feels more like a propositional calculus.  But it's a little like the debate about art versus porn:  You kind of only know porn when you see it.  And I'm looking for porn.

\btr
But as I said its just an attempt to get something going, and I'm
completely open to any suggestions you have -- in particular I'd like
to read any papers that you're thinking a lot about at the moment \ldots\
\etr

Are you kidding, I never read any papers any more!  I'm becoming a bit solipsistic.  But let me substitute this instead.  Below I'll attach my notes on the big phase transition I've made from ``knowledge'' to ``belief'' for my view of the quantum state.  If you need more of a context, first read the Brun--Finkelstein--{\Mermin} paper on {\tt quant-ph}.  I start up with criticism on that at about page 19, and then it goes pretty thick into it thereafter.  After October 9, {\Caves}, {\Schack}, and I started to work to turn a lot of that into a more technical argument:  we're hoping to finish the paper by the end of the week and post it soon thereafter.  (It stands at 14 pages at the moment.)

Anyway, the more radical stuff in those notes leads very quickly to some meaty technical questions.  For instance, what does it mean to be an ``unknown'' quantum operation?  It suggests a de Finetti theorem for completely positive maps, for instance, which {\Schack} and I are working on in our spare time.

\section{10-12-01 \ \ {\it Big Dreams}\ \ \ (to G. Brassard)} \label{Brassard9}

\bgb
I've been talking around about our little dream of founding quantum
mechanics on information theory. The more I talk about it to various
people, the more I realize this is the only scientific topic that
really excites me at this time.
And the more I am sorry that I'm not doing anything about it.
\egb

By the way, thank you for the very touching note.  You are misplaced, however, in saying, ``I am sorry that I'm not doing anything about it.''  At the very least, you are doing loads about it by organizing these meetings in {\Montreal}!

The truth of the matter is, just like you, nothing excites me more.  We're gonna prevail eventually.  In fact, at the very moment, I'm writing another (pleading) paper titled ``Quantum Mechanics as Quantum Information'' (for the {\Vaxjo} proceedings).  For your private fun---but please do keep it private---I'll attach a new mini-samizdat I've recently put together with some of my latest musings.  (It's probably hard to tell from it, but it has given rise to some new good theorems, that will be coming out soon in a paper with Caves and Schack.)

\section{14-12-01 \ \ {\it Strong Consistency} \ \ (to C. M. {\Caves} \& R. {\Schack})} \label{Caves45} \label{Schack42}

I know that I'm terribly slow to be coming up to speed on this
project.  But now I have another possibly frivolous question.

What was the origin of strong consistency?  What motivated {\Ruediger}
to invent the concept?  And indeed, did either of you ever find it
appearing in the literature otherwise?  It, of course, has a very
different flavor than normal Dutch-book---I know that's no surprise
to you---but for me, for the moment anyway, it seems to me to be
introduced more for mathematical nicety than anything else.  (I.e.,
It seems more to be introduced just for the purpose of cleaning
things up a bit axiom-wise.)  It doesn't have the same
operational/pragmatic flair that the rest of Dutch book has.

Indeed, it might be useful to lay this sort of issue on the line in
the paper.

I can think of at least one case where one would never want to
enforce strong consistency, namely for infinite event spaces.
Consider an infinite number of draws from an i.i.d.\ distribution.
One has that the probability of a typical sequence is one; however
one should never then say that that implies the bettor believes a
typical sequence will occur with certainty.\medskip

\noindent Reading ever so slowly (but I hope thoroughly),

\section{15-12-01 \ \ {\it Bah, Humbug} \ \ (to C. M. {\Caves})} \label{Caves46}

Humbug.  Did I ever tell you I hate Christmas:  It's the most
intrusive holiday I know of, with people pushing me all over with
obligations that they've invented for me.  (Susie, Kiki's mom,
arrived yesterday, by the way.)  Now Thanksgiving, that's my favorite
holiday:  you cook, you eat, you watch a parade, and you end up
rubbing your belly in a deep soporific ecstasy.

By the way, speaking of rubbing bellies, Kiki went into false labor
last night.  At first, when it was over, I thought, ``Thank God it's
false.''  But now, in the light of the morning, I'm thinking, ``When
are we ever going to get this over with?''

\bcc
There is evidence that it's out there in the literature (Sklar
mentions it without any references), but we haven't dug up any
references.  In fact, {\Ruediger} and I had the impression that you
might know where to find references.
\ecc

This is one of my biggest gripes about not being in a university.  I
don't have easy access to anything outside of the physics and
mathematics journals.  (In fact strangely, last night I dreamed that
I was back at the University of Texas, surrounded by books, books,
books on all subjects, and I was telling myself how glad I was to be
back.)

Did you have a look at the Kemeny and Lehman references I recommended
in an October 27 note?  [See 27-10-01 note ``\myref{Caves37}{Coming to Agreement} to C. M. {\Caves} \& R. {\Schack}.]  That at least would be my starting point.
Also Bernardo and Smith, as I recall, had a lot of references on
Dutch-book---in particular a lot of references dissatisfied with the
argument (as Bernardo and Smith, in fact, are).

\bcc
I don't know about the flair point.  I'll grant that strong
consistency isn't as compelling as ordinary consistency, but it still
registers on the flair scale: in a single trial, you never win, but
definitely sometimes lose.  Moreover---and this is the crucial
point---strong consistency is absolutely essential so that
probability assignments and density operators incorporate sufficient
information about one's beliefs that one can use them as surrogates
for the beliefs in our arguments.
\ecc

I wouldn't say that it's not ``as compelling'', it's just of a
different flavor.  To say that it's not as compelling, means it's
directly comparable to the Dutch-book argument.  But it's not clear
to me that it is.  Presently it seems to me much more like a Cox kind
of thing.  It says (at least in one case) my outward commitment {\it
must\/} reflect my internal belief---not because I will be imprudent
if I don't do so, but because I don't want my friends to see any
discrepancy between my beliefs and my actions.

I believe I understand what you call the ``crucial point.''  But I'm
guessing---just a psychological point---you're more attracted to it,
because it gives rise to (a better version of) the BFM statement.
Thus it gives the impression of giving something very solid. Whereas
simple Dutch book lets everything fly in the wind---maybe that's the
very thing I'm attracted to it.

In any case, I don't see anything wrong with exploring what strong
consistency leads to.  I'm just trying to get it straight in my
head---and possibly also for the reader---what it's all about.

\bcc
Chris said, ``Indeed, it might be useful to lay this sort of issue on
the line in the paper.''

Perhaps you could be more specific.  It seems to me that it is pretty
well laid out in the paper, but you're looking at it with a fresh
eye, so suggestions are welcome.
\ecc

I'm sorry, I don't quite know how to be more specific right now.
Maybe I would just like the reader to see more of the debate about
it.  And it should come across that it is (possibly) an invention of
C and S, not de Finetti or Ramsey.

\bcc
You're absolutely right here: probability 1 cannot mean certainty and
probability 0 cannot mean impossibility in the limit of an infinite
number of draws from an i.i.d., but it's not clear to me that the
problem is strong consistency.  To apply probabilities to infinite
sets, you need countable additivity for exclusive events, which
doesn't follow from the finite additivity given by the Dutch book. To
apply probabilities to continuous sets requires measure theory. These
things are additional mathematical structure beyond what the Dutch
book can provide.  Perhaps the best philosophy is {\Ruediger}'s:
probabilities really apply only to finite sets of events; the
generalization to infinite sets is an idealization, where additional
mathematical structure is added to make things work nicely.
\ecc

Point well taken.  But on the other hand, I'm still not convinced
either.  I think all three of us are aligned that the only thing
really worth considering, conceptual-wise that is, are finite event
spaces.  However, it still strikes me as a chink in the armor that I
can think of a limiting case where I would not want probability
$1-\epsilon$ to have anything to do with internal certainty.  That is
to say, maybe there's a deep reason that Dutch-book consistency is
NOT strong consistency \ldots\ and maybe we're missing something by
plastering over its unsightliness so quickly.

\bcc
Let me know if you have specific recommendations.  {\Ruediger} and I
have agreed on a number of changes in presentation and a large number
of additions on the POVM front.  I will be attempting to incorporate
these this weekend.
\ecc

I'm trying my best in the light of the present events.  (But I'm sure
you're not holding your breath for me.)  The main thing I'm shooting
for now, is to make sure I understand every aspect of the paper---in
a defendable way, that is---before {\Mermin} arrives Tuesday.

Did, by the way, you understand my issues with the {\it
sufficiency\/} of the $\det = 0$ conditions?  Was I missing
something?

\section{15-12-01 \ \ {\it One So Far} \ \ (to C. M. {\Caves})} \label{Caves47}

Just looking at the end of Section III, I was reminded that BFM
posted their final version of the paper on {\tt quant-ph\/} this
week. So, they have made some changes.  It might be worthwhile to
check that they didn't sneak in too drastically different a
formulation of the question.

Also, there you (or {\Ruediger}) write, ``In contrast, our derivation
is couched wholly in terms of the beliefs of the parties and does not
appeal to a real state of affairs; it is therefore preferable in a
Bayesian approach to quantum mechanics.''  I'm not sure---at least as
hinted by the present structure of the manuscript---that any
nonBayesian will have a clue why it's a preferable statement of the
problem.  ``What's wrong with letting the quantum state reflect an
objective fact, that what cannot happen cannot happen?,'' the reader
might ask.  ``Aren't these CFS guys just paining over inconsequential
points?''

Personally, I would like to see some discussion, perhaps in the
introduction or in Section III, along the lines of those Bernardo and
Smith quotes that I like so much.  I'll place them below in case you
forgot.  [See quotes in 07-08-01 note ``\myref{Mermin28}{Knowledge, Only Knowledge}'' to T. A. Brun, J. Finkelstein \& N. D. {\Mermin}.] I think they give the whole motivation for the work.

Finally, the present manuscript gives pretty short shrift to the
Peierlsian ideas that got David's blood up in the first place.  Nor
do we say much about the apparent inconsistency between the goals and
the ``solution'' of BFM in relation to the Peierls quest.

\section{17-12-01 \ \ {\it Births and {\Mermin}s} \ \ (to R. Pike \& S. J. van Enk)} \label{Pike5} \label{vanEnk0}

Well, Kiki informed me a little while ago (it's 3:00 AM now) that she's been in labor since about 9:00 PM last night.  So, I'm guessing---it's just a guess---I won't be in tomorrow.  (That's a joke.)

I'm hoping you two can hold down the quantum fort while I'm away.  In particular, David {\Mermin} should be showing up Tuesday between 2:00 and 2:30.  If by wild chance I'm not there---though really I have every plan to be---can one of you escort him around, make sure that he gets to his talk, and then make sure he gets to his room at the Turkey Hill Inn?  (And criticize his paper a little bit while you're at it.)

But I think that'll be an utter emergency:  this baby should certainly be born before today is out.  In particular, I'm planning to be at Bell Labs by 1:00 on {\Mermin} day (Tuesday).

{\Mermin} will be coming to Bell Labs directly from a meeting at Rutgers in New Brunswick.  For the heck of it, let me give you his contact number through Tuesday morning:  New Brunswick Hyatt, 732-873-1234.  But as I say, I should be at Bell Labs before his arrival.

Oh, and I'll cc this note to {\Caves} and {\Schack} so they'll understand why I might be silent on their latest draft.

\section{19-12-01 \ \ {\it Katie Viola Fuchs} \ \ (to the world)}\label{KatieViolaFuchs}

\noindent Dear Family, Friends, and Colleagues,\medskip

Katherine Viola Fuchs came to the world with Monday's sunrise, 17
December 2001.  She was 8 pounds, 19.5 inches, and a delight to her
parents' eyes---she is beautiful in every way.  And like her big
sister Emma, Katie has already made it clear that the future will not
just unfold before her, but be made in crucial part by her presence
and will.

With the greatest of expectations (and a modicum of pride), we send
greetings to everyone on Katie's behalf.\medskip

\noindent Chris, Kiki, and Emma Fuchs

\section{20-12-01 \ \ {\it Even Less Important} \ \ (to B. W. Schumacher)} \label{Schumacher7}

\bbs
Mike W. and I just posted the paper I mentioned to you about
approximate error correction.  (It ought to appear tomorrow at
\quantph{0112106}.) This is not earth-shaking stuff, but it does mean
that I can start working on the ``Doubting Everett'' paper next.
\ebs

That's great news, both about your paper with Mike and your ``continuing doubt.''  Jeff Bub and I have just started to firm up the plans for the special issue.  I'll let him know that we can count on you.

\bbs
By the way, I did mention that we could bring you here to give a talk.
[\ldots]  It turns out we have some open dates in late January and early February, too.  Not
that you should drop everything and come amuse undergraduates [\ldots] but if you'd like to come, we'd be delighted to have you.
\ebs

Let's think of February as a possibility (but not too firm of one).  What would be the latest date I can give you a firm decision?  It would certainly be great to get a chance to talk a bit:  I've recently made a further, more radical, Bayesian transition (and believe it or not, have had to drag Carl and {\Ruediger} kicking and screaming into it with me), and I'd like to test it out on as many moderately-sympathetic people as possible before opening my mouth to the general public.

\section{23-12-01 \ \ {\it Princeton and the Bud of Bayesianism} \ \ (to J. N. Butterfield)} \label{Butterfield4}

Thanks so much for the Italian lesson!

The last time we met, you mentioned something about a meeting in Princeton that you were organizing with Bas van Fraassen.  Can you remind me of when it will be, so I can get it onto my calendar?  Will it be a general philosophy of science meeting, or just something devoted to quantum mechanics?  If it's the former, and you don't mind, I would like to bring a local friend-of-the-family with me while I attend.  He has a Ph.~D. in philosophy from Washington University in St.\ Louis and is a philosopher of mind and a cognitive scientist; he teaches at a small college nearby (can't remember the name).  So, if the meeting is a general one, he'd fit right in, and I would certainly enjoy his company both there and on the drive over, etc.

I sure wish we could get together more often and talk.  I very much enjoyed and appreciated your knowledge of {\it both\/} de Finetti's and Lewis's ideas.  I believe I've made substantial progress in my Bayesianism in the last few months, and it would certainly great to have another test-ear to try it out on (and see it stand or fall) before I foolishly put it in print.

\section{30-12-01 \ \ {\it The Two-Girl Rule} \ \ (to D. Petz)} \label{Petz3}

Thanks for the happy note!

\bdep
There is a saying in Hungarian, something like that ``A clever man has
daughters''. I do not know if it is valid in the US but you may try
further experimental checkings. (I have two daughters, too.)
\edep

Here are the people (including myself) in quantum information that I can think of with two girls:
\bv
Herb Bernstein\\
Gilles Brassard\\
Ignacio Cirac\\
Claude Crepeau (adopted children)\\
Chris Fuchs\\
Jeff Kimble\\
Seth Lloyd\\
Asher Peres\\
D\'enes Petz\\
John Preskill\\
{\Ruediger} {\Schack} (2 girls, 1 boy)\\
Ben Schumacher
\ev
It is an interesting coincidence.

\chapter{2002: A World Under Construction}

\section{02-01-02 \ \ {\it Food for Thought} \ \ (to C. H. {\Bennett} \& J. A. Smolin)} \label{SmolinJ1.1} \label{Bennett5}

The more I think about it today, the more I'm tickled about the idea of either Charlie or the two of you writing something on thermodynamics and quantum information (rather than Charlie writing a pro-many-worlds piece).  I really hope you'll take the bait and view this as an opportunity to get some thoughts straight (including a response to Earman as Charlie suggested) and also to suck the philosophical audience into the excitement of quantum information \ldots\ and to let them see that quantum information is {\it actually\/} relevant to their subject matter.

Below I'll put a compendium of quotes from some of Jeff Bub's letters to give you a little more orientation on the goal of the project.  As soon as I get off my lazy butt, we should be sending out a more formal set of invitations soon.

\bq\noindent
\begin{enumerate}
\item Rob Clifton has asked me whether I'd like to guest-edit a special issue of {\sl Studies in the History and Philosophy of Modern Physics\/} (SHPMP) on quantum information. Apparently, he first made the suggestion to you some time ago, but since he hasn't heard back from you, he assumes that you are not too keen to take on the job. [\ldots]

The difficulty will be getting non-philosophers to write a paper that is not just informative for philosophers, but also philosophically convincing and articulate.

\item  Well, Bub and Fuchs as editors would be fine with me. But we'd really have to think hard about how to ensure that the authors all deal with some conceptual or foundational question relevant to quantum information in a philosophically serious way. We want to avoid the issue just being `intro to quantum information for non-specialists,' with some anecdotal interpretative comments as a nod to philosophy. And it also shouldn't just be a forum for one point of view.

The difficulty, of course, is that very few philosophers have actually worked in this field. So we have to rely mostly or perhaps entirely on scientists for the papers. It will be hard getting papers from physicists and computer scientists for a philosophy journal. On the plus side, it does seem to be the case that some people working in quantum information have thought hard about the relevance of their research for foundational questions. Somehow, we have to motivate people to take it as a worthwhile and exciting challenge to write a paper that tackles a philosophical issue in an intellectually disciplined way -- as intellectually disciplined as these authors invariably are when they write a technical scientific paper.  Too often, all analytical controls seem to be switched off when scientists venture into philosophical terrain.  [\ldots]

I don't think we should tie this to the {\Montreal} meeting. It would probably be better to go with people who have been thinking about a particular foundational issue for a while and whom we can persuade to write a paper on that issue, with the emphasis on philosophical relevance rather than the technical details.

\item This is probably not going to be easy, but I think it could be a very worthwhile project if we can get these people fired up about the idea of writing articles for a special issue that will clearly articulate the significance of the new work on quantum information and quantum computation for foundational and interpretational questions about quantum mechanics.
SHPMP is right now the premier journal for foundational and philosophical issues in modern physics. So this is the right place to do this. As I said before, what we don't want is a bunch of articles on dumbed down quantum information for philosophers. Rather, we want a serious and analytically sophisticated attempt to explore foundational issues relevant to quantum information, and to show precisely how looking at quantum information sheds new light on the old questions that plagued Bohr and Einstein.
\end{enumerate}
\eq

\section{02-01-02 \ \ {\it Solipsism Story} \ \ (to C. H. {\Bennett} \& J. A. Smolin)} \label{SmolinJ2} \label{Bennett6}

Here's the other story from Martin Gardner's ``Why I Am Not a
Solipsist.''

\begin{quote}
Russell once spoke on solipsism at a meeting chaired by Whitehead. As
Russell tells it in his autobiography, he said he could not believe
he had written those parts of Whitehead's books which he (Russell)
could not understand, although he could find no way to prove he
hadn't.
\end{quote}

\section{02-01-02 \ \ {\it Greedy Pragmatist} \ \ (to C. H. {\Bennett})} \label{Bennett7}

Kiki and I just got through looking through your photo-book.  Here are probably the three that strike us the most (in that order):
\begin{enumerate}
\item
Boat on Dam 1990
\item
Surf, Bass Hocks 1967
\item
Plant Saucers 1975
\end{enumerate}
It'd be truly wonderful if we could hang any or all of these up in our new house.  How about this:  If you'd mat them, and label and sign the mats of course, we'd frame them and reimburse you for all the materials (except the love you put into them).  Then, soon after that---after we get the house prim and proper---we would invite you and Theo up for the first showing.

Is that greedy of me?  Anyway, despite what John says, I'm not a solipsist; I do believe in a reality external to me.

\section{03-01-02 \ \ {\it Your Note}\ \ \ (to R. Balian)} \label{Balian1}

Carl Caves recently shared a note with me that you had written concerning our paper \quantph{0106133}.  Thank you so much for all the kind words!  I am very flattered.  Moreover, I will certainly read the papers you recommended to us:  I wish we had known of them before sending off our final proofs to the publisher.  In any case, I will not make the same oversight twice.

Let me address one of the points in your note:
\brb
What I appreciated the most in your work is the use of Gleason's
theorem. In my paper of 1989 I had assumed additivity of expectation
values even for non-commuting observables, and it is an important
progress, both conceptually and technically, to restrict to commuting
ones. However, what do you do in the case of a 2-dimensional Hilbert space where Gleason's theorem fails?
\erb

Yes, the two-dimensional case is interesting.  There actually turns out to be a way to handle it that continues with the spirit of Gleason, though not by the exact letter of his original assumptions.  The trick is to expand one's notion of measurement from the standard von Neumann type to the class of positive operator-valued measures (POVMs).  If one does that while retaining the remainder of Gleason's assumptions, then the theorem works even for two-dimensional Hilbert spaces.  Moreover the theorem also starts to work for Hilbert spaces over the rational number field---strangely, the original Gleason theorem seems to require the full power of the continuum to be valid---and beyond that, the proof of the theorem becomes essentially trivial (the sort of thing one can sketch on the back of an envelope).

I give what I believe is a fairly convincing argument for such a set of assumptions and a proof of the result in \quantph{0106166}.  I think you will find the spirited discussion in that reference easy and enjoyable reading and also see that it promotes our common quantum-foundational position in spades.

\section{03-01-02 \ \ {\it Prenuptial}\ \ \ (to R. Jozsa)} \label{Jozsa6}

In a few days Jeff Bub and I will be sending out some more formal-looking invitations, but I wanted to discuss this with you a little privately first.  The plan is to put together an issue of SHPMP (Studies in History and Philosophy of Modern Physics) on the impact of quantum information on quantum foundations and various related philosophical issues.

I know that you've been thinking hard for a long time about writing an article of this nature, and you pretty much (verbally) committed to me once that you'd do it if I could find a decent home for it.  (OK, maybe it was during one of our late-night outings!)  But, I really think this will be a decent home, and I hope you'll consider doing it.

So, far we have firm commitments from Andy Steane (he's going to put an extension of his ``Quantum Computing Needs Only One Universe,'' which so far hadn't found a home) and Ben Schumacher (who has started to write up his ``Doubting Everett'' lecture).  Beyond that, Charlie Bennett has pretty much said ``yes'' verbally and even seemed to relish the idea a bit.  We'll also be going after Kent, Mermin, Pitowsky, Popescu, and a couple of others (yet to be decided), but we haven't contacted them yet.  So I think I can promise you some ``relatively established'' company \ldots\ in case you're worried about reputation and such (since it is a philosophy journal).

As with Charlie Bennett yesterday, I'll place below some quotes from some of Jeff Bub's letters so you can get a better flavor for what the project is about.

I think whatever you wanted to write, it would be great, and we can give you about six months to do it.  I hope you know that I honestly believe this stuff and this debate is important (and will be ever more so for the generation after us).  Indeed, I'm guessing that a paper from you---it's just a guess, being tinged with mathematical Platonism and a fairly common-sense philosophical realism---would be a good counteragent to anything you might see Mermin and I spurt out \ldots\ and I hope you'll see all the more need for your own efforts because of that!!

So I'll leave it at that.  I'll hope and pray you say ``yes'' and await your reply.

\section{04-01-02 \ \ {\it New Breach of Faith} \ \ (to C. M. {\Caves})} \label{Caves48}

\bcc
I agree wholly with your statement in the following:

\bq
\noindent Also, there you (or {\Ruediger}) write, ``In contrast, our
derivation is couched wholly in terms of the beliefs of the parties
and does not appeal to a real state of affairs; it is therefore
preferable in a Bayesian approach to quantum mechanics.''  I'm not
sure---at least as hinted by the present structure of the
manuscript---that any nonBayesian [would] have a clue why it's a
preferable statement of the problem.  ``What's wrong with letting the
quantum state reflect an objective fact, that what cannot happen
cannot happen?,'' the reader might ask.  ``Aren't these CFS guys just
paining over inconsequential points?''
\eq

But I don't quite know what to do about it.  Your B\& S quotes are
nice, but they will just get in the way, in my view, of getting
people even to grasp the setting for what we're doing.
\ecc

Perhaps you misunderstood:  I am not asking that the Bernardo and
Smith quotes actually be used in the text.  What I am asking is that
we find a way to convey to the reader why any of this is important.
One thing you've got to realize is that more than once, I've heard
{\Mermin} describe the reception of the BFM paper with words like these:

\bdm
Sorry, I had the feeling that those few who understood anything at
all thought it was pretty obvious, but they were polite anyway.
\edm

What---it seems to me---is our task, is to convey to any readers who
had such a reaction as above that they were actually snookered by
BFM.  The Bernardo and Smith report of what Bayesianism is about
builds a context for our efforts---they make it clear why one should
expect a hierarchy of conditions (like the ones we explore in the
paper) rather than an {\it absolute\/} answer.

It is in saying terse, ``only established clique''-interpretable
(i.e., only preformed-radical-Bayes\-ian interpretable) things like,
``In contrast, our derivation is couched wholly in terms of the
beliefs of the parties and does not appeal to a real state of
affairs; it is therefore preferable in a Bayesian approach to quantum
mechanics'' \ldots\ and just kind of leaving it at that, without any
further buffer \ldots\ that is going to get us in trouble. Or at
least, that is one of my continuing fears whenever I work with you
and {\Ruediger}.

What is being laid in this paper is the groundwork for viewing the
quantum state in a way that people---even the Bayesians among
us---are not at all accustomed to:  Namely, taking the quantum
state's subjectivity absolutely seriously and to the extreme.  I
can't understand how leaning heavily on the motivation for this work
can hurt the paper.  Indeed it seems far more crucial than the
technical results in Section V if you ask me:  if no one cares about
the results, no one will read them in the first place.  (It's not
like this stuff can, with almost a single word, be advertised as a
new quantum algorithm like Shor's.)

What's a little annoying is that I know that you (from among the
three of us) are the one most up to the task of writing a beautiful,
yet businesslike introduction.  The only thing it seems that you need
to be convinced of is that people will actually read this paper if it
is well-written (and won't read it otherwise).  Strangely, enough, my
wacky 45-page papers get read (or at least skimmed): Explain that.
It's not the substance a priori (and maybe not even a posteriori),
but it might be the style.

Finally, in a last-ditch effort to shore up this point, I'm going to
perform a new breach of faith (and just reconcile it with St. Peter
when the day comes).  [\ldots]

OK, I'll say no more on the subject.

\bcc
Example: Kimble sat through {\Ruediger}'s talk at Caltech, nodding his
head in agreement the whole time, but then it emerged in the
discussion afterward that he thought each party had his own copy of
the system.  He and his student, Andrew Landahl, really had a hard
time with understanding that there is any [difference] between
``different states'' and ``different systems.''  They really just
couldn't help using these interchangeability.
\ecc

This is an oddity I have encountered before with several people.  I
don't understand its origin, but it is weird.  Even taking an extreme
realist view that there are objective quantum systems and objective
quantum states, one should be able to detect the matter-property
dichotomy in that and not confuse the two.  If you have a theory of
the confusion's origin, I'd be interested to hear it.

\section{04-01-02 \ \ {\it Once Again} \ \ (to C. M. {\Caves} \& R. {\Schack})} \label{Caves49} \label{Schack43}

\bcc
It IS interesting that strong consistency is required so that the
firm part of your belief structure can be read off your probability
assignments.  I would have thought that a Bayesian would want to know
what assumptions are required to translate beliefs into probability
assignments.
\ecc

Let me put the relevant part of my take on that down again:

\bq
\noindent
It [strong consistency] says (at least in one case) my outward
commitment {\it must\/} reflect my internal belief---not because I
will be imprudent if I don't do so, but because I don't want my
friends to see any discrepancy between my beliefs and my actions.
\eq

``Outward commitment'' means acceptable odds for the bettor.  What
strong consistency says is that there are cases I should lay my
precise beliefs on the table (for public view), even if I didn't have
to in order to avoid a sure loss.

Nowhere else does Dutch book do this.  For instance---with standard
Dutch book---I could internally believe $p(X,Y)$, but nevertheless
only accept bets according to $q(X,Y)$ so long as both these are
coherent assignments.  Is it part of the Bayesian creed to also
require honest reporting of my internal beliefs?  Maybe it is:  but
it seems to me that that is something on top of Dutch book (and in
fact we know that it is).

I'm not ``against'' strong consistency, I just want to understand
what motivates it other than that it leads to more equations in our
paper \ldots\ and thus leads to a more scientific look.

I'm going to try to get to a real library this weekend and dig up
those {\sl J.\ Symb.\ Logic\/} papers by Shimony etc.  Maybe that'll
demystify things for me.

\section{04-01-02 \ \ {\it Strong Consistency, 2} \ \ (to C. M. {\Caves})} \label{Caves49.1}

\bcc
I think we already explain it pretty well.  The point is that you need
strong consistency so that a probability assignment reflects one's
beliefs sufficiently that you can draw sensible conclusions from the
probabilities.  It's like our ``you can't ask the system'' argument.
Without strong consistency, you couldn't ask the assigner either,
since he might assign a pure state even when he thought all sorts of
orthogonal things were possible.  We make this point.  It can be made
more prominent.  Other suggestions are welcome.
\ecc

When I come back to read the paper more carefully next week, I'll rethink whether I'm being too harsh on this point.

Thanks for putting up with me.

\section{07-01-02 \ \ {\it Correlation without Correlata} \ \ (to N. D. {\Mermin})} \label{Mermin52}

Let's see what kind of reaction the longer quote below gets out of
you.  ``There are, so to speak, relations all the way down, all the
way up, and all the way out in every direction: you never reach
something which is not just one more nexus of relations.''

\bq
In the rest of this essay I shall be trying to sketch how things look
when described in antiessentialist terms. I hope to show that such
terms are more useful than terminologies which presuppose what {\Dewey}
called `the whole brood and nest of dualisms' which we inherit from
the Greeks. The panrelationalism I advocate is summed up in the
suggestion that we think of everything as if it were a {\it number}.

The nice thing about numbers, from my point of view, is simply that
it is very hard to think of them as having intrinsic natures, as
having an essential core surrounded by a penumbra of accidental
relationships. Numbers are an admirable example of something which it
is difficult to describe in essentialist language.

To see my point, ask what the essence of the number 17 is---what it
is {\it in itself}, apart from its relationships to other numbers.
What is wanted is a description of 17 which is different {\it in
kind\/} from the following descriptions: less than 22, more than 8,
the sum of 6 and 11, the square root of 289, the square of 4.123105,
the difference between 1,678,922 and 1,678,905. The tiresome thing
about all {\it these\/} descriptions is that none of them seems to
get closer to the number 17 than do any of the others. Equally
tiresomely, there are obviously an infinite number of other
descriptions which you could offer of 17, all of which would be
equally `accidental' and `extrinsic'. None of these descriptions
seems to give you a clue to the intrinsic seventeenness of 17---the
unique feature which makes it the very number that it is. For your
choice among these descriptions is obviously a matter of what purpose
you have in mind---the particular situation which caused you to think
of the number 17 in the first place.

If we want to be essentialist about the number 17, we have to say, in
philosophical jargon, that {\it all\/} its infinitely many different
relations to infinitely many other numbers are {\it internal\/}
relations---that is, that none of these relations could be different
without the number 17 being different. So there seems to be no way to
define the essence of seventeenhood short of finding some mechanism
for generating {\it all\/} the true descriptions of 17, specifying
all its relations to {\it all\/} the other numbers. Mathematicians
can in fact produce such a mechanism by axiomatizing arithmetic, or
by reducing numbers to sets and axiomatizing set theory. But if the
mathematician then points to his neat little batch of axioms and
says, `Behold the essence of 17!' we feel gypped. There is nothing
very seventeenish about those axioms, for they are equally the
essence of 1, or 2, of 289, and of 1,678,922.

I conclude that, whatever sorts of things may have intrinsic natures,
numbers do not---that it simply does not pay to be an essentialist
about numbers. We antiessentialists would like to convince you that
it also does not pay to be essentialist about tables, stars,
electrons, human beings, academic disciplines, social institutions,
or anything else. We suggest that you think of all such objects as
resembling numbers in the following respect: there is nothing to be
known about them except an initially large, and forever expandable,
web of relations to other objects. Everything that can serve as the
term of a relation can be dissolved into another set of relations,
and so on for ever. There are, so to speak, relations all the way
down, all the way up, and all the way out in every direction: you
never reach something which is not just one more nexus of relations.
The system of natural numbers is a good model of the universe because
in that system it is obvious, and obviously harmless, that there are
no terms of relations which are not simply clusters of further
relations.

To say that relations go all the way down is a corollary of
psychological nominalism: of the doctrine that there is nothing to be
known about anything save what is stated in sentences describing it.
For every sentence about an object is an explicit or implicit
description of its relation to one or more other objects. So if there
is no knowledge by acquaintance, no knowledge which does not take the
form of a sentential attitude, then there is nothing to be known
about anything save its relations to other things. To insist that
there is a difference between a nonrelational {\it ordo essendi\/}
and a relational {\it ordo cognoscendi\/} is, inevitably, to recreate
the Kantian Thing-in-Itself. To make that move is to substitute a
nostalgia for immediacy, and a longing for a salvatory relation to a
nonhuman power, for the utopian hope which pragmatism recommends. It
is to reinvent what {\Heidegger} called `the ontotheological tradition'.

For psychological nominalists, no description of an object is more a
description of the `real', as opposed to the `apparent', object than
any other, nor are any of them descriptions of, so to speak, the
object's relation to itself---of its identity with its own essence.
Some of them are, to be sure, better descriptions than others. But
this betterness is a matter of being more useful tools---tools which
accomplish some human purpose better than do competing descriptions.
All these purposes are, from a philosophical as opposed to a
practical point of view, on a par. There is no over-riding purpose
called `discovering the truth' which takes precedence. As I have said
before, pragmatists do not think that truth is the aim of inquiry.
The aim of inquiry is utility, and there are as many different useful
tools as there are purposes to be served.

Common sense---or at least Western common sense---has trouble with
the claim that numbers are good models for objects in general because
it seems counterintuitive to say that physical, spatiotemporal
objects dissolve into webs of relations in the way that numbers do.
When numbers are analysed away into relations to other numbers,
nobody mourns the loss of their substantial, independent, autonomous
reality. But things are different with tables and stars and
electrons. Here common sense is inclined to stick in its toes and say
that you cannot have relations without things to be related. If there
were not a hard, substantial autonomous table to stand in relation
to, e.g., you and me and the chair, or to be constituted out of hard,
substantial, elementary particles, there would be nothing to get
related and so no relations. There is, common sense insists, a
difference between relations and the things that get related, and
philosophy cannot break that distinction down.

The antiessentialist reply to this bit of common sense \ldots\
\eq

\section{08-01-02 \ \ {\it Correlation without Correlata, 2} \ \ (to N. D. {\Mermin})} \label{Mermin53}

\bdm
I don't like the stuff about 17 --- very unconvincing.  But after
that it gets more interesting.  It starts to get most interesting
just as you cut it off.
\edm

Of course it's not convincing:  philosophical mumbles---it seems to
me---can't serve that purpose.  I was just curious whether it struck
any chords with you on how to convey (whatever it is you've been
trying to get at with) your ``correlation without correlata.''  Is it
a good metaphor?  Does it carry any Ithacan soul?

Personally, I thought the analogy was nice from the second I read it.
Even the primeness of a number, for instance, can only be defined by
invoking the existence of all the other numbers.  Numbers just don't
have any properties in and of themselves (or at least none that I
could think of).

You're not going to QIP at IBM next week, are you?

\section{08-01-02 \ \ {\it Information $\rightarrow$ Belief $\rightarrow$ Hope ??}\ \ \ (to N. D. {\Mermin})} \label{Mermin54}

I started reading through the essay again and realized that if I were
to fulfill your request, I might just have to copy the whole damned
article.  I give up!  But at least I was kind enough to check that
you have the source in your library at Cornell (call number and
location below).

The essay is titled ``A World without Substance or Essences.''  It
should probably be read in conjunction with the essay previous to it,
``Truth without Correspondence to Reality.''  The book as a whole is
entertaining, but it is a little skimpy on firm argument. (Though,
the author admits it is an attempt at popularization \ldots\ so that
kind of makes it OK in my mind.)  He's certainly not the devil that
Steven Weinberg labeled him.  But I will admit, his version of
pragmatism may go too far for my tastes:  I'll hold on to the final
verdict for a while.  (William {\James} is still the best bet in my
eyes.) \medskip

\noindent PS.  The title to this note is a joke, based on {\Rorty}'s own
title (and some of the discussions in his book).  However, I am
seriously toying with the idea of making a distinction between
``beliefs'' and ``commitments.''  I.e., saying that a quantum state
ascription is explicitly a ``commitment'' rather than a
``belief''---a commitment to behaving one way or the other in the
face of some experimental data (yet to be gathered).  Sometimes
commitments explicitly correspond to beliefs (as Dutch-book takes to
be a definition), but it seems to me not always the case.

But as I say, I'm just toying with the idea:  I know {\Caves} and {\Schack}
will have a cow and beat me up if I have enough nerve to say anything
about it.  So, I'd better be sure of myself.

\section{08-01-02 \ \ {\it Help with ET Quote} \ \ (to C. M. {\Caves} \& R. {\Schack})} \label{Caves50} \label{Schack44}

\bcc
I've been trying to remember without success where ET discusses his
ability to toss coins \`a la what we saw from Dan Greenberger in
{\Vaxjo}. Can you help?
\ecc

I don't know that I've ever actually read the example:  I've just
heard stories of it (probably through {\Ruediger}), and maybe words to
that effect as a section heading or something.  I have a faint memory
that it was somewhere in Jaynes's big book, but I couldn't find it in the
table of contents.\footnote{\editornote Perhaps ``How to Cheat at Coin and Die Tossing'' in chapter 10, ``Physics of `Random Experiments'.''  Jaynes argues from the classical mechanics of tumbling motion that ``anyone familiar with the law of conservation of angular momentum can, after some practice, cheat at the usual coin-toss game and call his shots with 100 percent accuracy.  You can obtain any frequency of heads you want; and \emph{the bias of the coin has no influence at all on the results!}''}  {\Ruediger}'s probably a better person to ask.

One interesting thing I did find in looking for the answer to your
question though, is a Jaynesian diatribe against a Dutch-book
foundation for probability theory.  It's in Appendix A of his book.
The fear he expresses there strikes me as not so different from what
I was reading into your worries about my fall into ``radical
Bayesianism.''

\section{08-01-02 \ \ {\it Term Origin} \ \ (to C. M. {\Caves} \& R. {\Schack})} \label{Caves51} \label{Schack45}

Here's a little (unfortunately inconclusive) discussion on the origin
of the ``Dutch book'' term:
\begin{center}
\myurl[http://web.archive.org/web/20060524181532/http://www1.fee.uva.nl/creed/wakker/miscella/Dutchbk.htm]{http://web.archive.org/web/20060524181532/http:// www1.fee.uva.nl/creed/wakker/miscella/Dutchbk.htm}.
\end{center}

\section{08-01-02 \ \ {\it {\Rorty} on Religion} \ \ (to J. W. Nicholson)} \label{Nicholson3}

I wish I had had these quotes in stock when we were having our
conversation the other night. But I hadn't gotten that far in the
book yet \ldots\ if I were only a quick reader like you. [Disclaimer:
My copying these quotes for your thought (and our continued
discussion) neither represents an endorsement for or against their
content.]

From the essay ``Religious Faith, Intellectual Responsibility and
Romance'' in the R. {\Rorty}, {\sl Philosophy and Social Hope}:\medskip

\noindent page 153:

\bq\noindent
If one accepts that claim, one will have reason to be as dubious as
{\James} was of the purportedly necessary antagonism between science and
religion. For, as I said earlier, these two areas of culture fulfill
two different sets of desires. Science enables us to predict and
control, whereas religion offers us a larger hope, and thereby
something to live for. To ask, `Which of their two accounts of the
universe is true?' may be as pointless as asking, `Is the carpenter's
or the particle physicist's account of tables the true one?' For
neither question needs to be answered if we can figure out a strategy
for keeping the two accounts out of each other's way.
\eq

\noindent page 156:

\bq\noindent
Pragmatists are not instrumentalists, in the sense of people who
believe that quarks are `mere heuristic fictions'. They think that
quarks are as real as tables, but that quark talk and table talk need
not get in each other's way, since they need not compete for the role
of What is There Anyway, apart from human needs and interests.
Similarly, pragmatist theists are not anthropocentrists, in the sense
of believing that God is a `mere posit'. They believe that God is as
real as sense impressions, tables, quarks and human rights. But, they
add, stories about our relations to God do not necessarily run
athwart the stories of our relations to these other things.

Pragmatist theists, however, do have to get along without personal
immortality, providential intervention, the efficacy of sacraments,
the Virgin Birth, the Risen Christ, the Covenant with Abraham, the
authority of the Koran, and a lot of other things which many theists
are loath to do without. Or, if they want them, they will have to
interpret them `symbolically' in a way which MacIntyre will regard as
disingenuous, for they must prevent them from providing premises for
practical reasoning. But demythologizing is, pragmatist theists
think, a small price to pay for insulating these doctrines from
`scientific' criticism. Demythologizing amounts to saying that,
whatever theism is good for, it is not a device for predicting or
controlling our environment.
\eq

\noindent page 157--158:

\bq\noindent
I said earlier that many readers of `The Will to Believe' feel let
down when they discover that the only sort of religion {\James} has been
discussing is something as wimpy as the belief that `perfection is
eternal'. They have a point. For when Clifford raged against the
intellectual irresponsibility of the thesis, what he really had in
mind was the moral irresponsibility of fundamentalists --- the people
who burnt people at the stake, forbade divorce and dancing, and found
various other ways of making their neighbours miserable for the
greater glory of God. Once `the religious hypothesis' is disengaged
from the opportunity to inflict humiliation and pain on people who do
not profess the correct creed, it loses interest for many people. It
loses interest for many more once it is disengaged from the promise
that we shall see our loved ones after death. Similarly, once science
is disengaged from the claim to know reality as it is in itself it
loses its appeal for the sort of person who sees pragmatism as a
frivolous, or treasonous, dereliction of our duty to Truth.
\eq

\section{09-01-02 \ \ {\it Doin' the Dutch-Book Zombie} \ \ (to H. Mabuchi)} \label{Mabuchi1}

\noindent Hey MacArthur boy, \medskip

It's a funny thing being a parent of a newborn child.  It leads to a
kind of zombie-like state most every night:  You're never neither
really awake nor asleep.  And then your mind gets hung up on some
little thing---tonight being the Dutch-book argument---and you just
repeat it over and over, as if in a trance.  Does that build any imagery
for you?

Anyway, in looking up Andrew's email address last night (or was it
this night? --- time stops), I ran across your web page.  The
flattery to be listed among your collaborators!  (Keep it there!) But
you need to get the affiliation right --- Bell Labs, Lucent
Technologies.

Philosophically, lately, I've been taken away by William {\James} and
John {\Dewey}.  I've been reading their stuff with a pretty voracious
appetite (and there is a heck of a lot of it).  But of course one
thing leads to another---just like that marijuana---and last week I
found myself picking up a copy of Richard {\Rorty}'s {\sl Philosophy and
Social Hope}.  His flavor of pragmatism goes maybe too far even for
me, but he writes well and I find it easy to read him.  I guess I
just write you all this to tell you he speaks highly of your friend
Derrida!  One of these days, I am going to get the nerve to approach
that man.  (By the way, {\Mermin} told me recently that your old
professor Bas vF has been reading my papers \ldots\ and, apparently,
disagreeing with them \ldots)

I'm going to slip back under the covers now; like Nosferatu I keep a
little fresh earth there for comfort.

Just a sleepy note to let you know I miss you sometimes.

\begin{verse}
He did the mash \\
He did the monster mash \\
The monster mash \\
It was a graveyard smash
\end{verse}

\section{12-01-02 \ \ {\it Raussendorf--Briegel}\ \ \ (to S. J. van {\Enk})} \label{vanEnk25}

Just thinking a little more deeply about the Raussendorf--Briegel model, here's another thing I like about it.  Note that in the Nielsen--Leung models, one has to make a strict assumption about the post-measurement states in order to make it work.  That is, one is not free to have any old quantum operation (allowed by the Kraus specification) associated with the POVMs that are measured:  The measurements must be of the class of ``nondestructive measurements'' that Debbie talked about yesterday.  On the other hand the RB model has no such restriction.  The reason for this is because as soon as each measurement is done, they can simply throw the measured qubit away; it's never touched again.

I know this was obvious, but it just had never quite stood out for me before.  The main point it brings out---for me anyway---is that the exponential speedup of quantum computations really must just be a property of the nonclassical correlations that one can generate from such an entangled state.  And in that way, the exponential speedup appears to be deeply connected (or at least deeply analogous) to the violation of Bell inequalities.

\section{13-01-02 \ \ {\it Princeton Envy} \ \ (to H. Mabuchi)} \label{Mabuchi2}

\bhm
Sounds like domestic life is treatin' you good!
\ehm

Wasn't it Freud who said, once a school girl sees your Princeton,
she'll know what's missing from her life and be envious thereafter?

Well I'm not a school girl, but I saw your Princeton the other day
(ahem), and it certainly did start a deep yearning in me.  What a
wonderful place!  This was my first real trip there since moving to
New Jersey, and I was enchanted all day.  I found myself thinking,
what I wouldn't give to live the rest of my life in this little
cloister.

The thing that really struck me was the immense resources at one's
finger tips.  I found myself copying a little over \$35 of articles
in Firestone Library!  I couldn't believe it:  They had the complete
collection of the {\sl Danish Yearbook of Philosophy\/} and
(shockingly) 35 years of the {\sl Transactions of the Charles Sanders
{\Peirce} Society}, and I knew that those were just the tip of an
iceberg.  I was in heaven.

Emma and I play this game:  I say, ``When you go to college, I hope
you'll go to Harvard.''  She says, ``I want to go to Princeton.''  Or
if I say Princeton, she says Harvard---it's always the opposite. (You
can see a trend in our relationship starting to form.)

But that's just an aside (to tell you that the grass is always
greener).  Keep up the good work with all those good students.  Get
them to read William {\James}' {\sl Pragmatism\/} and tell them that
quantum mechanics is a much better motivation for all that he said
there \ldots\ but to never lose sight that the real goal is to get to
where he wanted to go.

\section{13-01-02 \ \ {\it Reality in the Differential} \ \ (to N. D. {\Mermin})} \label{Mermin55}

\bdm
Actually I'm leaving it in the original form for now, but you're
giving me a very hard time here.
\edm

Let me say touch\'e, and then little more than that.  Strangely,
actually, I've been having a little conversation of my own with You
this week.  And I've been planning to write you---i.e., the one with
the little y---all about it.  But what happens?  Now that I've got a
little time this Sunday morning, I'm finding that the inclination is
leaving me.

So this note is going to come out far weaker than I had planned. The
main thing was to build a conversation around another {\Rorty} quote,
and to tell you how pleased I am with the pragmatism movement in
general.  I'm finally finding a philosophy so close to what I'm
looking for that I'm willing to advertise it to my friends.

Somehow they go a little too far for me, but there are so many
beautiful gems I keep finding in their stuff that I find it better
not to dismiss it outright.  The second paragraph below struck me
especially last week.  It might as well be about my latest quantum
interpretation thoughts.

The reality is in the differential.  The quantum state represents a
(gambling) commitment on the part of the agent; it never represents
anything external to that agent.  If you're looking for where the
``reality'' of the external world creeps into the formalism, you
should look to how these commitments change.  That's what I've been
trying to say for a few months now \ldots\ but I guess you already
understood that (as maybe witnessed by our last real conversation,
during your visit to Morristown).

I had a pretty happy-sad week this week you might say.  The happiness
was that I visited Princeton for the first time since moving up here.
I was like a kid in a candy store!!  I ended up copying over \${35}
of articles in their wonderful library (which even subscribes to the
{\sl Danish Yearbook of Philosophy}!).  The sadness was in that I had
to leave that environment and go home at the end of the day; it
wasn't home itself.

I'm going to write you that longer {\James}--{\Dewey} note eventually.  But
right now I'll just leave you with a reminder that the reality is in
the differential.  From: R.~{\Rorty}, ``The Pragmatist's Progress:\
Umberto Eco on Interpretation,'' in his book {\sl Philosophy and
Social Hope}, (Penguin Books, New York, 1999), pp.~131--147:

\bq
As I see it, the rocks and the quarks are just more grist for the
hermeneutic process of making objects by talking about them. Granted,
one of the things we say when we talk about rocks and quarks is that
they antedate us, but we often say that about marks on paper as well.
So `making' is not the right word either for rocks or for marks, any
more than is `finding'. We don't exactly make them, nor do we exactly
find them. What we do is to react to stimuli by emitting sentences
containing marks and noises such as `rock', `quark', `mark', `noise',
`sentence', `text', `metaphor' and so on.

We then infer other sentences from these, and others from those, and
soon---building up a potentially infinite labyrinthine encyclopedia
of assertions. These assertions are always at the mercy of being
changed by fresh stimuli, but they are never capable of being {\it
checked against\/} those stimuli, much less against the internal
coherence of something outside the encyclopedia. The encyclopedia can
get {\it changed\/} by things outside itself, but it can only be {\it
checked\/} by having bits of itself compared with other bits. You
cannot {\it check\/} a sentence against an object, although an object
can {\it cause\/} you to stop asserting a sentence. You can only
check a sentence against other sentences, sentences to which it is
connected by various labyrinthine inferential relationships.

This refusal to draw a philosophically interesting line between
nature and culture, language and fact, the universe of semiosis and
some other universe, is where you wind up when, with {\Dewey} and
{\Davidson}, you stop thinking of knowledge as accurate representation,
of getting the signs lined up in the right relations to the
non-signs. For you also stop thinking that you can separate the
object from what you say about it, the signified from the sign, or
the language from the metalanguage, except {\it ad hoc}, in aid of
some particular purpose.
\eq

\section{13-01-02 \ \ {\it Indulgence}\ \ \ (to G. Brassard)} \label{Brassard10}

By the way, did I ever tell what I ended up doing on the single malt end?
I took all your suggestions:  I bought a Glenmorangie 18, a Talisker 10, and a Laphroaig 15.  (The Laphroaig wasn't cheap: \$77.50.)  It made for quite a Christmas!

Too bad you're not going to QIP this week.  I'm pretty sick (some kind of ear infection along with conjuctivitis in my eyes), but I'm going to be there.  (Just to infect everyone!)  By the way, have you seen the Raussendorf and Briegel model of quantum computation?  I.e., the model where one starts with a fixed entangled state for all computations (of a given spacetime size), and then simply performs a sequence of single qubit measurements thereafter to enact the particular computation one cares about.  I think this is a really very deep achievement.  Briegel will be at the meeting; I'm really looking forward to seeing that.

\section{21-01-02 \ \ {\it R, B, and P} \ \ (to A. Peres)} \label{Peres23}

\bap
I am happy that QIP2002 went well. I saw the papers of Briegel and
Nielsen on {\tt quant-ph} (please remind me the numbers, if you have them
ready). I was not favorably impressed, maybe I misunderstood them,
and I should read them again.
\eap

Actually, the better work is the Raussendorf/Briegel stuff.  If you
were not favorably impressed, then I think you should give it another
chance.  I think it is a beautiful construction, and, in fact, the
deepest thing I've seen in quantum computing for 2 or 3 years now.
The first paper to start with is \quantph{0010033}. Then more
details can be found in \quantph{0108067}, \quantph{0004051}, and \quantph{0108118}.

What is deep about this work is that all computations start off with
the SAME given entangled state for the qubits.  That is to say, a
given entangled state is taken as a resource for the task. Thereafter
the particular computation one is interested in is enacted by making
single-qubit measurements alone: there are no further unitary
evolutions.  I think that is quite remarkable and quite lovely.

An interesting feature of the Raussendorf/Briegel model (as opposed
to the Nielsen and Leung models) is that one need to take into
account NO details of the post-measurement state for the measured
qubits:  they can be thrown away immediately after the measurement.
And because the measurements are localized, the post-measurement
state of the remainder of the qubits is fixed completely by the POVM
(rather than the operation).  That is to say, for the relevant qubits
in this model, ``an effect only has one operation.''

\section{23-01-02 \ \ {\it Book Review of Nielsen \& Chuang's Book for American Journal of Physics} \ \ (to L. K. Grover)} \label{Grover1}

Good review, except I disagree with the following statement violently.
\blkg
Another rule that classical systems satisfy is something called the
locality principle (anything that happens depends on local conditions,
not on something that exists or happens far away). This is something
that quantum systems frequently violate. This led to a number of
paradoxical situations such as the EPR paradox. When Einstein and
others first discovered this, they thought they had disproved quantum
mechanics.
\elkg

If one takes the view (as, say, Caves, Cleve, van Enk, Heisenberg, Mabuchi, Mermin (somewhat), Milburn, Pauli, Peierls, Peres, Rudolph, Schack, Wheeler, Zeilinger, and I do) that quantum states are not objective properties of systems, but rather subjective states of belief or knowledge about systems, then quantum mechanics presents no evidence that the locality principle---as you define it above---is violated.

Beyond that, I would say you have the Einstein argument incorrect.  Einstein held fast to the {\it assumption\/} of locality and never budged from it.  The conclusion he drew from EPR situations---where localized measurements on entangled systems cause quantum-state updates for far away systems---is that the quantum state must stand for one's information/knowledge/belief rather than for something objective and observer-independent.  In his own words, the quantum state can only amount to an ``incomplete description.''  (See typical Einstein quote below.)  Thus he never felt that he had ``disproved'' quantum mechanics:  he only thought that there was {\it more\/} to say about nature than quantum mechanics had to offer.

You can find wordier versions of this argument, along with historical references, in Section 3 of my paper \quantph{0106166}.

The difference between Caves, Peres, Schack and myself---these guys' opinions I know for sure---from Einstein, is not that we reject his EPR-type argument for concluding that the quantum state must be subjective (i.e., dependent on the information any particular observer has gathered, etc.), but that we find no indication in this that quantum mechanics is incomplete \ldots\ and, thus, that there is more to say about nature than quantum mechanics can offer.

\bq\noindent
Here is the way Einstein put it to Michele Besso in a 1952 letter [found in J.~Bernstein, {\sl Quantum Profiles}, (Princeton University Press, Princeton, NJ, 1991)]:
\begin{quotation}
What relation is there between the ``state'' (``quantum
state'') described by a function $\psi$ and a real deterministic situation (that we call the ``real state'')?  Does the quantum state characterize completely (1) or only incompletely (2) a real state?
\ldots

I reject (1) because it obliges us to admit that there is a rigid connection between parts of the system separated from each other in space in an arbitrary way (instantaneous action at a distance, which doesn't diminish when the distance increases).  Here is the
demonstration: \ldots

If one considers the method of the present quantum theory as being in principle definitive, that amounts to renouncing a complete description of real states.  One could justify this renunciation if one assumes that there is no law for real states---i.e., that their description would be useless.  Otherwise said, that would mean: laws don't apply to things, but only to what observation teaches us about them.  (The laws that relate to the temporal succession of this partial knowledge are however entirely deterministic.)

Now, I can't accept that.  I think that the statistical character of the present theory is simply conditioned by the choice of an incomplete description.
\end{quotation}
\eq

\section{27-01-02 \ \ {\it What Would William {\James} Say?, 1} \ \ (to J. W. Nicholson)} \label{Nicholson4}

Of course I'm reading email in Texas!  (Or at least that's what I'd
say.)  I'll send your regards to my Mom.

\bjn
I had just settled down to a lovely lunch with my new issue of {\bf Scientific American\/} and was perusing the various articles, when I came across a regular column by Michael Shermer titled ``Skeptic.''  Now normally I enjoy this column (it's all about debunking psuedo-science and public misconceptions) and this month's topic was evolution vs creationism.  It got off to a good start with a quote by Richard Dawkins --- ``the universe we observe has precisely the properties we should expect if there is, at bottom, no design,
no purpose, no evil and no good, nothing but blind, pitiless indifference.''  I rather liked that.
\ejn

That may have been true at one time, i.e., at some stage in the
development of the world.  That's how it got off to a start so to
speak.  But now, I go the experimentalist (who does the hard work in
helping the theorist construct his theories), and ask him, ``What ya
doin?''  He says, ``Twiddlin' knobs.''  I ask, ``How come?''  He
says, ``I'm tryin' to hep this wacky friend of mine.  He wants to get
a theory of how iodine reacts to this and that?  I'm chartin' it out,
givin' him some clues.''  I ask, ``How do you do that?''  He says,
``By twiddlin' these knobs.''

The present stage of the universe doesn't look like it's full of
blind indifference to me.

The initial condition is always left separate from the theory;
there's a reason for that.  How else would the experimentalist be
able to twiddle his knobs?

\section{27-01-02 \ \ {\it What Would William {\James} Say?, 2} \ \ (to J. W. Nicholson)} \label{Nicholson5}

\bjn
Anyway, I was far more interested in your reaction to the quote
requiring God's absence from standard scientific theories.
\ejn

The point was, it doesn't even require the scientific agent's absence
from the ultimate gears and pinions of the world, much less a god's.
The scientific world view consists essentially of two components:
theories and initial conditions.  And, it seems to me, it is a tacit
assumption of the whole scientific enterprise that the
experimentalist can freely set the initial conditions he wishes to.

\bjn
If the universe really wasn't blindly indifferent, people wouldn't
fly airplanes into buildings, convinced they were on their way to
heaven.
\ejn

There's something about this sentence I just don't like.  I'll try to
put my finger on it in the next couple of days.

\section{29-01-02 \ \ {\it Quotes That Moved Me Once} \ \ (to J. W. Nicholson)} \label{Nicholson6}

\bjn
Only in the best of cases does an experimentalist have access to initial conditions.  Consider astronomers or evolutionary biologist who are left to mine a data trail, long after the fact.  And, I imagine a developmental psychologist who tried to monkey around with initial conditions would quickly find himself slapped with a malpractice lawsuit.
\ejn

Concerning your last note to me, here are some quotes that moved me once.  Maybe they do a more convincing job than I did, maybe they don't.  I'd be interested to hear your reaction.

\begin{itemize}
\item
D.~J. Bilodeau, ``Physics, Machines, and the Hard Problem,'' J.
Consc.\ Stud.\ {\bf 3}, 386--401 (1996).  The parts of this paper on quantum mechanics and the practice of science in general are absolutely excellent.
\end{itemize}

\bq
\indent
Even apart from the limits to measurement revealed by QM, there have been good reasons to doubt the power of physical observation to penetrate to ontological foundations.  I will focus on one of these, which is fundamental but not widely understood---the idea of dynamics.

The geometric idea of the physical seems simpler than it really is---there are simple geometrical entities (perhaps Newtonian
particles) which inhabit physical space, which move and interact according to simple physical laws and form larger structures which are the objects of our ordinary experience.  It might seem at first that, with the tools of modern science, the structure and nature of such a world would be open to observation with no conceptual difficulties. But, of course, actual observations require that we {\it use\/} this world in order to observe it.  We must manipulate and modify certain parts of it in order to create situations
(experiments) in which information about the structure of other parts can be conveyed to us. Some measurements necessarily preclude others.  We cannot dissect a microscope at the same time we are using it to study a sample of brain tissue (much less use the same brain tissue to think about the microscope, etc.).  In pure geometry, the intellect can wander over every detail of a geometric structure, as one can gaze over a blueprint or electronic schematic diagram, going here and there and back again at will.  But the physical observer has no such freedom. It is impossible to examine every point in space and time.  We will never be able to obtain more than an exceedingly tiny fraction of the total information contained in the structure of the universe (to the extent that it can be considered a structure).
\eq
and
\bq
\indent
The physicist {\it acts\/} and intervenes in nature.  Typical situations are limited and determined in part by what the experimenter can manage to contrive.  The idea of causality in physics depends on the freedom of the experimentalist to alter the conditions of the experiments.  I first gained an appreciation of the importance of this idea from an argument by Roger Newton:
\bq
The most practical and only foolproof method of scientifically testing a causal connection between A and B is `wiggling' one of them and watching the response of the other.  We are not interested here in what might be called `historical causality' (establishing a causal connection in a single chain of events) but in `scientific causality'
(establishing such a connection in repeatable events) \ldots\  It is the external control of A together with the correlation with B that establishes, in a good Humean sense, the causal connection between them, as well as the fact that A is the cause and B, the effect.
\eq

This observation contains a great profundity.  The laws of physics are dynamical---they are not laws of being but laws of action.  They are human constructions based on our experience of nature and in form and concept derive from our role as creative agents. [Footnote:  When I write of the physicist's `freedom' of action in setting up experiments and controlling parameters, I am not taking a position on the philosophical question of `freedom of the will.' I mean here only a pragmatic freedom which is independent of the physical entitities being observed.] The importance of action can be seen in the advance of celestial mechanics from Kepler to Newton.  Kepler observed the patterns of the motion of the planets and distilled them into three laws which described the elliptical shape of orbits and the speed with which the planets move along them at each point.  These are purely geometrical and kinematical ideas. Newton took Galileo's work on the motion of physical objects developed in experiments and refined it and extrapolated the concepts of gravitational force and mass (both dynamical rather than purely geometrical notions) out into the solar system and so was able to derive Kepler's laws from his own dynamical laws of inertia and gravitation.

The dynamical description of cause, force, and law stands in contrast to the `historical' denoting of particular things and events in the course of time.  These are two `modes of description' which are equally essential to any account of the physical world.
\eq

\begin{itemize}
\item
D.~Bilodeau, ``Why Quantum Mechanics is Hard to Understand,'' LANL eprint \quantph{9812050} (1998).  This paper was submitted to and rejected from the journal {\sl Foundations of Physics}; its most likely permanent home will be the LANL eprint archive.
\end{itemize}

\bq
A thing is historical insofar as it is objective (can be observed and treated as an object).  It then enters into the realm of recordable objective occurrences which can be ordered in historical space and time. It is dynamical insofar as it is defined as an abstract element of the dynamical theory which explains causal relationships between objects. \ldots

Imagine that we could see the universe as omniscient external observers, all space and time at once, and that what we ``see'' is a tangle of intersecting particle world-lines (cf.\ Ch.~1 of Misner, Thorne, and {\Wheeler}).  We might detect some patterns which would constitute physical rules or laws in some sense, but it would be quite difficult or impossible to know whether we had found all the important patterns, or to distinguish significant relationships from accidental ones.  Even more difficult would be to translate this omniscient description into the kinds of relationships and laws which would be observed by the huge clumsy bunches of world-lines which constitute ourselves.

When we set out to investigate Nature, we are not like that external omniscient observer at all.  We look for relationships and patterns in the behavior of objects we know.  We want to find out---does this kind of object always behave this way under these circumstances?  The phrases ``this kind,'' ``this way,'' and ``these circumstances'' imply the ability to abstract relevant or significant features from what are really unique events.  They also imply that we can find or (even
better) set up many instances of these typical situations. The result is that the concepts we develop to describe physical phenomena depend not only on what we can observe, but also on what we can do.

To say that A affects or causes or influences or interacts with B implies a counterfactual:  If A had been different, B would have been different, too.  The most convincing way to establish a connection is to ``wiggle'' some parameter in A more or less randomly and then observe the same odd pattern showing up in some property of B.  If I want to know whether a wall switch controls a certain light, I can flip the switch on and off and observe whether the light follows my actions. There is always the possibility that the light is being controlled by someone else or goes on and off spontaneously; but if I put the switch through a very irregular and spontaneous sequence of changes and the light still follows along, then the probability of a causal connection is very high (barring a conspiracy to deceive the experimenter).

Physical theory is possible because we {\it are\/} immersed and included in the whole process---because we can act on objects around the us. Our ability to intervene in nature clarifies even the motion of the planets around the sun---masses so great and distances so vast that our roles as participants seems insignificant.  Newton was able to transform Kepler's kinematical description of the solar system into a far more powerful dynamical theory because he added concepts from Galileo's experimental methods---force, mass, momentum, and gravitation. The truly external observer will only get as far as Kepler. Dynamical concepts are formulated on the basis of what we can set up, control, and measure.
\eq

\begin{itemize}
\item
H.~Primas, ``Beyond Baconian Quantum Physics,'' in {\sl Kohti uutta todellisuusk\"asityst\"a. Juhlakirja professori Laurikaisen 75-vuotisp\"aiv\"an\"a} (Towards a New Conception of Reality.
Anniversary Publication to Professor Laurikainen's 75th Birthday), edited by U.~Ketvel (Ylio\-pisto\-paino, Helsinki, 1990), pp.~100--112.
\end{itemize}

\bq
The methodology of experimental scientific research and engineering science is to a large extent characterized by the regulative principles emphasized by Francis Bacon.  It is a tacit assumption of all engineering sciences that nature can be {\it manipulated\/} and that the initial conditions required by experiments can be brought about by interventions of the world external to the object under investigation.  That is, {\it we assume that the experimenter has a certain freedom of action which is not accounted for by first principles of physics}.  Without this freedom of choice, experiments would be impossible.  Man's free will implies the ability to carry out actions, it constitutes his essence as an actor.  We act under the idea of freedom, but the topic under discussion is neither man's sense of personal freedom as a subjective experience, nor the question whether this idea could be an illusion or not, nor any questions of moral philosophy, but that {\it the framework of experimental science requires the freedom of action as a constitutive though tacit presupposition}.

The metaphysics of Baconian science is based on the confidence that only the past is factual, that we are able to change the present state of nature, and that nothing can be known about nature except what can be proved by {\it experiments}.  Francis Bacon's motto {\it dissecare naturam\/} led to a preferred way of dividing the world into object and observing systems.  An experiment is an {\it intervention\/} in nature, it requires artificially produced and deliberately controlled, reproducible conditions.  In {\it experiments\/} in contradistinction to {\it observations\/} -- one {\it prepares\/} systems in initial states, {\it controls\/} some of the variables, and finally {\it measures\/} a particular variable.
The regulative principles of Baconian science require {\it power to create initial conditions}, stress {\it the facticity of the past\/} and {\it the probabilistic predictability of the future}, and reject {\it teleological considerations}.
\eq

\section{29-01-02 \ \ {\it Qunix} \ \ (to J. M. Renes)} \label{Renes9}

\bjmr
In other news, I discovered a really interesting paper (actually an
undergrad thesis) by a guy at Oxford, who argues against Deutsch's
view of the quantum world. One thing he takes to task is the notion
that information is physical (which I think I've now discarded).
\ejmr

Yeah, I met {\Timpson} in Ireland and really enjoyed his company.  I'll
definitely look at his thesis.  Right before Rolf died, I had wanted
to write him a letter telling how much I had come to disliking the
phrase ``information is physical.''  I wanted to tell him that I
think a far more appropriate phrase would be ``information carriers
are physical.''  In fact, I told Charlie {\Bennett} about this---at the
time---and he told me, ``Too late; Rolf just had some fraction of his
brain removed last week.''

Concerning myself, I've gotten further carried down the path of
pragmatism.  I've even read some of the {\Rorty} blend now.  Indeed my
latest little epiphany hit me last week (during an operating-systems
talk in our center) when I came up with the following slogan:  ``A
physical theory really amounts to little more than a programming
language.''  Its rules, its specifications are more about the ways
we've come to naturally manipulate the world than anything intrinsic
to that same world.  I tried to say this in my Oct 4 letter to {\Carl}
(``\myref{Caves33}{Replies on Pots and Kettles}'' in the new mini-samizdat), but I
think the new slogan says it better.

\section{29-01-02 \ \ {\it Gleason Proof} \ \ (to C. M. {\Caves})} \label{Caves51.1}

I'm presently haphazardly going through a ridiculous amount of email \ldots\ trying to take the easiest path in my replies:  So don't be surprised to see a lot of my replies to your notes out of order.

\bcc
Are you aware of the following paper on Gleason: {\rm Richman and Bridges (1999), ``A constructive proof of Gleason's
theorem,'' Journal of Functional Analysis {\bf 162}, 287--312}?
\ecc

Yes, I am.  My full Gleason and Wigner theorems collection is below.

Funny though, I discussed the paper you mention very briefly with Andrew Gleason.  He was aware of it and seemed pretty skeptical that it could be a meaningful result.  I'm not sure how to interpret that, other than that maybe he was thinking there was a flaw in it.  So (maybe) watch out.  I have it printed out, but I've never studied it myself.

\begin{itemize}

\item
V.~Bargmann, ``Note on Wigner's Theorem on Symmetry Operations,'' J.\ Math.\ Phys.\ {\bf 5}, 862--868 (1964).

\item
P.~Busch and J.~Singh, ``L\"uders Theorem for Unsharp Quantum Measurements,'' Phys.\ Lett.\ A {\bf 249}, 10--12 (1998).

\item
R.~Cooke, M.~Keane, and W.~Moran, ``An Elementary Proof of Gleason's Theorem,'' Math.\ Proc.\ Camb.\ Phil.\ Soc.\ {\bf 98}, 117--128 (1985).

\item
A.~M. Gleason, ``Measures on the Closed Subspaces of a Hilbert Space,'' J.\ Math.\ Mech.\ {\bf 6}, 885--894 (1957).

\item
S.~Goldstein and A.~Paszkiewicz, ``Orthogonally Additive Functions on $B(H)$,'' Quantum Probability Communications, Vol.~X, 223--227 (World Scientific, 1998).

\item
R.~I.~G. Hughes, {\sl Structure and Interpretation of Quantum Mechanics}, 1992, Appendix A: Gleason's Theorem, pp.~321--346.

\item
I.~Pitowsky, ``Infinite and Finite Gleason's Theorems and the Logic of Indeterminacy,'' J.\ Math.\ Phys.\ {\bf 39}, 218--228 (1998).

\item
I.~Pitowsky, ``Range Theorems for Quantum Probability and Entanglement,'' \quantph{0112068} (2001).

\item
F.~Richman and D.~Bridges, ``A Constructive Proof of Gleason's Theorem,'' J.\ Func.\ Anal.\ {\bf 162}, 287--312 (1999).

\item
C.~S. Sharma and D.~F. Almeida, ``A Direct Proof of Wigner's Theorem on Maps Which Preserve Transition Probabilities between Pure States of Quantum Systems,'' Ann.\ Phys.\ {\bf 197}, 300-309 (1990).

\item
U.~Uhlhorn, ``Representation of Symmetry Transformations in Quantum Mechanics,'' Arkiv F\"or Fysik {\bf 23}, 307--340 (1963).
\end{itemize}

\section{29-01-02 \ \ {\it {\Vaxjo} Contribution} \ \ (to C. M. {\Caves})} \label{Caves52}

You know, if my true love is philosophy---{\Bennett} calls it theology
in my case, actually---yours is certainly sports.  You're just not
going to let this go without one hell of a fight, are you?

\bcc
So it appears that I'm in a pickle with evolutions that are mixtures
of unitaries, since quantum operations don't have unique
decompositions. But of course, all the decompositions into things
other than unitaries aren't of interest, and I've been able to
``show'' (a number of half-baked steps here) that mixtures of unitary
EVOLUTIONS are unique, at least in a sense that's good enough for me.
To put it more precisely, I've shown it for qubits and think I can go
further.
\ecc

I'm wondering what the ``sense that's good enough for me'' is?  For
instance, I already think of the depolarizing channel, where it
doesn't matter which $x$, $y$, and $z$ axes I use for defining the
Kraus operators.  But you probably have something up your sleeve that
will be more instructive than that.

\bcc
For pure states the problem shows up as an inability to make a clean
distinction between objective and subjective probabilities,
\ecc

And for the channel maps, I would say the problem shows up as an
inability to make a clean distinction between objective and
subjective probabilities.  [Just like my double footnote in the NATO
paper, I meant this to be taken seriously.]  The only difference now
is that the probabilities are of a conditional type, $p(y|x)$.  But I
know that this is too cryptic for you to make any sense of it at the
moment, and I already write you too much preachy email:  I just need
to try to write a damned paper and be done with it.

\section{29-01-02 \ \ {\it Lock In}\ \ \ (to A. M. Steane)} \label{Steane3}

I'm sorry, I haven't kept you up to date on how this special issue project of SHPMP is going.  Anyway, it is now definitely going to happen, and I hope we can still count on your commitment of publishing the ``one universe'' paper there.

So far, we've got definite yes's from Ben Schumacher, Howard Barnum, Richard Jozsa, and Itamar Pitowsky.  We've got a high-probability yes from Charlie Bennett.  And we've just heard from Carl Caves clamoring to write a paper on ``Can Hamiltonians Be Taken as an Ontology for Quantum Mechanics?''\ if Jeff can be convinced that it's appropriate.  Also there's a good chance that David Mermin will be saying yes.

All papers will be due mid-June of this year.

Hopefully Jeff Bub and I will be making an official announcement to all concerned sometime next week.

\section{29-01-02 \ \ {\it A Summer Masterwork?}\ \ \ (to N. D. {\Mermin})} \label{Mermin56}

\bdm
Funny that you should find {\Rorty} appealing.  I've liked a lot of what
he says too, although he is Public Enemy Number 1 for most of the
scientists engaged in the Science Wars.  No Geneva convention for
him!
\edm

You emerge!  It's so good to hear from you again.  If you have read
{\Rorty}, then why in the hell did you never tell me, ``Chris, what
you're shooting for in quantum mechanics really sounds a lot like
pragmatism!''?  You know you really might have saved me---more
importantly this program I dream of---a lot of time!  For instance, I
might not have wasted years thinking that a pure-state assignment
ought to be unique, if I had just read a little {\James}, {\Dewey}, and
{\Rorty}.

Anyway, I continue to go further off the deep end.  Here's what I
wrote Renes this morning:
\bq
Concerning myself, I've gotten further carried down the path of
pragmatism.  I've even read some of the {\Rorty} blend now.  Indeed my
latest little epiphany hit me last week (during an operating-systems
talk in our center) when I came up with the following slogan:  ``A
physical theory really amounts to little more than a programming
language.''  Its rules, its specifications are more about the ways
we've come to naturally manipulate the world than anything intrinsic
to that same world.  I tried to say this in my Oct 4 letter to {\Carl}
(``\myref{Caves33}{Replies on Pots and Kettles}'' in the new mini-samizdat), but I
think the new slogan says it better.
\eq

However, I write this letter for another reason.  The way I view it,
you play a unique role in our community:  You really do ``write
physics'' every bit as much as you exhort your readers to in that
nice essay on your webpage.  I say this because I would dearly love
you to make a statement, a solid statement, of where you think
quantum information can have its greatest impact on settling quantum
foundations questions.  Our community needs this kind of incitement,
and you have the writing skills that might even convince someone to
do something about it.  Even if you wrote a paper of nothing but
questions, it would be great and a great service.

Would you think hard about doing that?  It could be your masterwork
for the year.  Jeff Bub and I have finally gotten to the point of
organizing a special issue of SHPMP, and your contribution could help
set a good tone for it.

Below, let me copy a couple of letters I've already written in this
regard.  They'll fill you in on all the details of the way Jeff and I
see the project. [\ldots]

What say you?

\section{30-01-02 \ \ {\it Sweet Talk} \ \ (to N. D. {\Mermin})} \label{Mermin57}

\bdm
Saying that I've read {\Rorty} is a gross exaggeration.  I've dabbled
around and found a curious mixture of interesting and outrageous
assertions.  Beware of becoming a ``postmodernist'' yourself.
\edm

Funny that so many are viewing this as a new kind of onslaught to
science.  By his own admission, {\Rorty} is not saying anything
particularly different from {\James} and {\Dewey}.  And my reading is
starting to confirm that.  So, it looks like the train of thought
started before the turn of the (last) century.  What is postmodern
about it?

That said, the only thing I have ever wanted is a sensible of view of
what quantum mechanics is about.  If it takes rearranging our
thoughts about what the classical world (and classical physics) is
about, then so be it.  To that extent---I think---I run the danger of
becoming a postmodern.  The other day Hans Briegel was visiting and
he saw {\James}' book {\sl The Meaning of Truth\/} sitting in the back
seat of my car.  He ended up asking what pragmatism is about.  I
said, ``You know, these guys didn't know anything about quantum
mechanics, but I might venture to say it can be summarized as a
Copenhagen interpretation of classical physics.''

\bdm
Actually it's interesting the way foundations of QM have come up
again and again in the science wars, from Shelly Goldstein's attack
on all of physics in the notorious NY academy volume, to Mara
Geller's attack on Bohr, to my occasional mutterings that nobody
whose thought hard about foundations of QM could possibly think
science is as simple as Gross and Levitt would like it to be.
\edm

1)  What is the ``notorious NY academy volume''?  You've piqued my
interest.

2)  I love it:  you made the same spelling slip in Mara Beller's name
in an email of November 28, 1999!

3)  Who are Gross and Levitt?

\bdm
As for your latest attempt to sweet talk me into doing something
reckless, if I thought I had anything useful to say on that subject I
would have said it, instead of beating ``Whose Knowledge'' into the
ground (and possibly many feet under the ground).  I will, however,
keep the invitation in mind as I work my way through the second
edition of my Qcomp course.
\edm

If I thought it might help, I'd even say ``Pretty please, with sugar
on top.''  More seriously, I think the only thing that would be
reckless would be for you to NOT use these post-65 years to act a
little like a more sober version of John {\Wheeler} and get the physics
community going to its next great stage.  I know that you believe
that quantum information and computing have something deep to
contribute to quantum foundations studies.  Why do you believe that?
Can you articulate it?  Why do you think that Peierls might have been
on the right track?  Or do you even think that?  Why have you changed
your stand somewhat on nonlocality?  Why have you privately backed
out a little on your initial statement of the IIQM?  What problems
did you foresee?  What problems were brought to your attention?  What
part of the interpretation still stands a chance of being useful?
What troubles you about Everett and Deutsch's interpretation?  What
troubles you about Griffiths' interpretation?  What troubles you
about Bohm and Goldstein's interpretation?  What troubles you about
Zurek's interpretation?  {\it All\/} of these people {\it claim\/} to
have long since solved all the greatest mysteries of quantum
mechanics. Why should we not be listening to them?  Have you ever
gathered all your thoughts on all these questions?

You can't tell me that you don't already have a LOT to say, even
while \ldots\ all the while \ldots\ you are telling yourself that you
have nothing useful to say.  Most importantly, you have the means at
your disposal for saying these things in a way that people will
listen and think about them.  Even a study in self-indulgence, i.e.,
a reflection on why you've come to the positions that you have and
why you remain perplexed would be {\it immensely\/} useful to the
community.

Have you ever analyzed why I send so much email to you?  I mean,
``you'' in particular?  If I had to put my finger on it, it would be
for two reasons:  1) Because I've always felt that your high standard
for your own writing has induced me to a higher standard for mine.
And 2) because you are the least dogmatic and most clear-thinking
devotee of quantum foundations I've ever met.  It has made it a
pleasure for me to discuss my ideas with you; it has induced me to
sharpen and present in a more convincing fashion everything I have
ever wanted to say.  What I imagine for your contribution in this
volume is that same charm working on a public (archival) scale.

I really want to be able to reserve you a spot in the volume.

\section{30-01-02 \ \ {\it Qunix Redux} \ \ (to J. M. Renes)} \label{Renes10}

\bjmr
OK i'm done ranting. I hope this makes some sense, or is at least
enjoyable to read!
\ejmr

Indeed it was, and I'll read it many times over!

\subsection{Joe's Preply}

\bq
\noindent\bq\noindent
[CAF said:] I came up with the following slogan:  ``A
physical theory really amounts to little more than a programming
language.''  Its rules, its specifications are more about the ways we've
come to naturally manipulate the world than anything intrinsic to that
same world.  I tried to say this in my Oct 4 letter to Carl (``Replies on
Pots and Kettles'' in the new mini-samizdat), but I think the new slogan
says it better.
\eq

Ok, I see the pragmatism coming through --- the concepts of the world
arise because we get them from our observations and they're useful in our
manipulations. Now that I've been learning computer science and complexity
theory for the last few days (who knows why), this reminds me of
Solomonoff's notion of science as data compression. He's got some notion
of universal prior probability that I don't really know what to do with.
somehow this subject seems interesting, important, and useful, but I can't
quite put my finger on how, or if it jives with Bayesian views. A lot of
that complexity stuff seems aimed at justifying ``random number'' so we can
use Kolmogorov probability. Ugh.

However, things are beginning to come together in my mind, organized, as
ever, in the logico-algebraic approach. A while back Carl and Howard were
interested in bayesian views of computation; so am I. My strategy is to
always link these things to some logical/algebraic framework and then make
up probability distributions on the framework. Maybe computability can be
thought of the same way; maybe it's possible to ``understand the power of
quantum computation'' via this method. (Of course, I'm not particularly
interested in quantum computation per se, but the idea of using the same
framework to apply to physical theory and computer science is certainly
interesting.)

I really like this overall approach; there's many levels to it
philosophically, and tons of practical things to work out. I'm very
amenable to the notion that physical theory isn't true because it
corresponds to reality, but rather makes the most sense and use of our
data and our models. Within this context, however, it will be useful to
think of things as objective --- there's a nice section in kant's critique
of pure reason (in the transcendental aesthetic, the part on 'space') in
which he goes to some pains to assert that space is empirically real, as
it's objectively present in our sensation of the world, but it's
transcendentally ideal since it's really/ultimately/truly a part of our
intuition.

Perhaps we don't have to be that aggressive about what is
really/truly/ultimately, and instead take the fact of our experience as
given. Now we're really in with the pragmatist flow: from here on out
what's true is what's useful. And part of that will be thinking that
things are objectively real because it makes the most sense. So, maybe
Hamiltonians can be `real.' as real as anything, I suppose; though
certainly not real in the ultimate sense. We're out of that game now.
there's really not a lot we can say about it. It exists, inasmuch as 'it'
is our experience. But not much more than that. Besides, we don't have to;
we can get along quite nicely using what works.

That's it, back to computation, because I just realized what my mind has
been trying to think all these days. I don't think quantum computation is
fundamental, or as fundamental, in the logical sense. Computing, logic,
math, all that is about true statements and can you prove them. Quantum
logic, in this sense, is like fuzzy logic --- you can't assign truth
values to all the propositions. OK, this is the logical angle. QC is not
in the class of things we think of when we think of computing because its
logical structure is all wrong.

Now to the physical! In the physical domain there is a truth of the
matter, of sorts. The result of experiment. In the physical realm we shift
our focus a bit --- our probability distribution is over what will happen,
not what is the underlying truth value of the proposition in question. So
we can use our physical manipulation rules and probability distributions
for computation, if we like. But we're still performing a computation in
the classical sense: the factoring algorithm will give the correct factors
or not; there's a truth of the matter. But the fully quantum view of
logical processing there's no such thing as the truth of the matter, as we
know. Thus, in my mind quantum computation is divorced from the logical
business of computation; it's merely concerned with the physical business.

Still not coming out well. Put it this way --- it's the same idea as
there's no such thing as quantum information. When we talk about using a
quantum computer, we want one that we can put classical propositions into,
and get classical propositions out of, albeit probabilistically. There's
just no sense to the idea that it's doing some logical processing on
quantum propositions. Physical processing to be sure, but not logical. We
can use physical systems to perform computation, just as we can use them
for information transmission, but we shouldn't conflate the physical
carrier with the logical or semantic content.

OK I'm done ranting. I hope this makes some sense, or is at least
enjoyable to read!
\eq

\section{31-01-02 \ \ {\it Clean Sweep} \ \ (to A. J. Landahl)} \label{Landahl4}

Let me just say again how much I really enjoyed our conversations yesterday.  Your visit prompted me to think, and it prompted me to re-look at your old emails to me.

Maybe it's time I moved at least a little on your last email, even if I don't attempt to give you an answer anywhere close to my own satisfaction.

\bal
That said, I'm still not sure what you believe reality ``is''.  It seems
as though you wish to go in two different directions:
\begin{itemize}
\item[]
reality $=$ that which we cannot control/predict [Emma analogy]
\item[]
reality $=$ free will [Bacon citation].
\end{itemize}
Are these supposed to be the same?  They don't appear that way on the
face of things.
\eal

The short answer is, 1) yes, I take both these things to be components of reality, and 2) they are not supposed to be the same things.  For each of us, one is an inward reality and one is an outward reality.

At heart, I think I am a ``pluralist.''  Reality is made of lots of ``stuff,'' samples of which---by definition---cannot be reduced to each other.

So, I say, randomness and free will appear to be two components of reality.  At least as of today, I'm willing to say there may well be a few more things that we can glean from the structure of quantum mechanics and hypothesize as elements in the external world.  Among these is the concept of systems (each of which with associated integer properties, commonly called the Hilbert space dimension) and certain ideals of behavior (usually going under the name of collapse rule, etc.).

I don't expect this last sentence to make too much sense to you on a first reading.  But maybe a couple of passages from some of my correspondence with others will help.  To that end, let me attach a small samizdat that I've put together.  The relevant things to read are:  A) a note to Mermin titled ``Kid Sheleen'' starting on page 78, and B) a note to {\Caves} and {\Schack} titled ``Replies on Practical Art'' starting on page 82.  [See 03-10-01 note ``\myref{Mermin45}{Kid Sheleen}'' to N. D. {\Mermin} and 03-10-01 note ``\myref{Caves30}{Replies on Practical Art}'' to C. M. {\Caves} \& R. {\Schack}.] In particular look at the table on page 84.  (However, please keep this samizdat private:  I haven't edited it for complete public distribution yet.)

\bal
The part of your paper which smells the most like a lack of belief
in reality is your portrayal of Bayesian agents.  You say that
Bayesian agents would like to align their predictions with some kind
of underlying reality, and ``they would if they could but they don't
because they can't.'' Of course that doesn't mean that an underlying
reality doesn't exist, only that it doesn't allow that kind of
alignment.  (The use of the word alignment here is unfortunate; it
sounds like some kind of astrological prediction by Miss Cleo.)  Do I
understand correctly that you believe Bayesian agents instead attempt
to align their predictions with each other as a sort of ``plan B''?  If
so, then it sounds like all Bayesian agents learn through this process
is what other Bayesian agents will have to say about things.  How can
it be that Bayesian agents learn anything about reality in such a
process?  Unless, of course, that is all reality is---correlations
with other Bayesian agents' predictions.
\eal

Maybe the most effective way to answer this is to include a note I wrote Mermin the other day.  It is pasted below.  [See 13-01-02 note ``\myref{Mermin55}{Reality in the Differential}'' to N. D. {\Mermin}.] A hint of the external world is not to be found in the quantum states themselves, but in their changes.  The world pings me when I ask a question of it, and I update from one (possibly nonsensical) belief to another (possibly nonsensical) belief, i.e., from one quantum state to another.  Reality creeps into this description in two ways:  1) through the pressure that caused that very change, and 2) through the formal structure of quantum mechanics, as it makes possible---through its belief-update rules, i.e., the Krausian collapse rules---a convergence of our disparate beliefs.

I hope that helps.  I understand that it is partially vague.  But I view that predominantly as an effect of the fact that it is a point of view still under construction.  I am not a prophet, just someone searching for a clear view of what QM is all about.

\section{31-01-02 \ \ {\it Words, Only Words} \ \ (to A. J. Landahl)} \label{Landahl5}

I did a search on ``information-information tradeoff'' in my big samizdat \quantph{0105039} and found two hits, one in a letter to Mabuchi and one in a letter to Mermin.  I'll place them below in that order. [See notes ``Destinkifiers'' and ``Zing!''\ on pages 216 and 265, respectively, of C. A. Fuchs, {\sl Coming of Age with Quantum Information}, (Cambridge University Press, 2011).]

Another reason I send you these, is that they also both explicitly tackle the question you keep asking me.  What is real?  What is real?!  What is real!?!

The tentative answer here is the locus of all possible information-disturbance (or information-information) tradeoff curves for a system.  The main idea that connects this to what I was saying to you yesterday is that that locus should be characterized by a single number.  We call it the dimensionality of the Hilbert space.  A little further explanation can also be found in the letter to Caves titled ``The First Eye'' on page 26 of the new mini-samizdat I just sent you.  [See 08-08-01 note ``\myref{Caves3}{The First Eye}'' to C. M. {\Caves}.]

\section{01-02-02 \ \ {\it Dublin Visit} \ \ (to C. King)} \label{King1}

Here's my itinerary: [\ldots]  I'll place my title and abstract below.

\bq\noindent
Title: Squeezing Quantum Information through a Classical Channel:\ Measuring the ``Quantumness'' of a Set of Quantum States \medskip\\
Abstract:
In this talk, I propose a general method to quantify how ``quantum'' a {\it set\/} of quantum states is.  The idea is to gauge the quantumness of the set by the worst-case difficulty of transmitting the states through a purely classical communication channel.  Potential applications of the notion arise in quantum cryptography, where one might like to use an alphabet of states that promises to be the most sensitive to quantum eavesdropping, and in laboratory demonstrations of quantum teleportation, where it is necessary to check that quantum entanglement has actually been used in the protocol.  A more devious purpose for the talk, however, is to introduce some new concepts into current quantum foundations discussions.  Time permitting, I'll tell you what I think is ``real'' about a quantum system.
\eq

\section{02-02-02 \ \ {\it Colleague} \ \ (to C. G. {\Timpson})} \label{Timpson1}

Well, I still haven't read your anti-Brukner/Zeilinger paper, but I'm
writing this small note to tell you that I just finished reading your
undergraduate thesis {\sl Information and the Turing Principle}. It's
quite a work, and I very, very much enjoyed reading it!

I have to tell you, one of my first reactions after reading about the
first third of the thesis was, ``Finally there's something sensible
coming out of Oxford!''  It really was such a relief:  I had been
thinking that Deutsch had essentially brainwashed everyone in the
quantum information community there (except for possibly Hardy and
Steane).

Of course, what I like most is what I see as a significant overlap
between our attitudes toward scientific theories, the Church--Turing
thesis, the homunculus fallacy, and the misleadingness of the slogan
``information is physical.''  But I learned a lot from you, and you
helped me sharpen several points.

Here were some of my favorite pages:  4, 6, 8, 16, 24!, 32, 35-36,
38!, 46!, 47-48!, 52!, 60, 65!, 66, and of course chapter 5.  (An
exclamation means that something especially intrigued me.)

I know there are a few places where I distanced myself from the
phrase ``information is physical'' in my large Samizdat (\quantph{0105039}), but I'm having trouble finding them right now.
One is in a letter to {\Bennett} starting at the bottom of page 34.
However, the most intriguing moment in my mind is one instance I have
not recorded in email before.  Let me record it here.

I was visiting {\Bennett} at his weekend home in Wendell, MA one weekend
and somewhere in the night I got on a roll about how much I had
started disliking the phrase ``information is physical.''  In place
of it, I was arguing that a much better, much more accurate
call-to-arms we ought to be sending the physics and
information-theory communities is that ``INFORMATION CARRIERS are
physical.''  Taking that into account is what is behind all the new
questions we are asking in quantum information theory.  Then I told
{\Bennett}, ``I am starting to plan a long email that I would like to
write to Rolf on this subject.  But I know that I'm going to have to
word it delicately if I'm going to stand any chance at all of
catching his ear and not getting an immediate dismissal as a fool.''
Charlie replied, ``Too late; Rolf just had about a quarter of his
brain removed last week.''  Then he explained that Rolf was just
found with cancer in the brain, and that chances were strong he would
be dead soon.  About a week later Rolf died.

The Shanker 1987 paper looks especially intriguing to me.  I'll try
to pick it up the next time I'm at Princeton.

By the way, the homunculus fallacy struck me on several levels.  I
think one might view William {\James}'s argument against the
correspondence theory of truth in his a little book {\sl Pragmatism\/} as a little bit along the same lines, for instance.  But it also got me to thinking about one of the things that has long bugged me in the
Zurek-style versions of quantum foundations.  There, the starting
point is how bad the word ``measurement'' is and how it should be
banished from the foundation of the theory.  Yet, inevitably (just
watch them), whenever push comes to shove in their explanation of the
true importance of decoherence, to get the idea across, they start
saying things like ``in essence the environment `measures' the
system.''  (Zurek always makes little quote motions with his fingers
when he says it.)  And that's supposed to lead us to a deeper
understanding of that tabooed subject?!?!?  (I thought I had put that
complaint in a footnote in some recent paper, but for the present I
couldn't find it either!  I must be losing my memory or losing my
mind \ldots\ or both!)

Anyway, again, I really enjoyed the thesis.  Keep up the good work.

P.S. I also had a look at your webpage.  From the schedule of your
discussion group, it looks like you're trying hard to make sense of
the notions of `objective chance' and `propensity' to yourself.  I
went down that path once---roughly from 1991 to 1996---and it was
instructive.  I found that I couldn't buy any of the theories, and
that's what ultimately pushed me down the long road to Bayesianism. I
keep my fingers crossed that you meet the same frustration!!

\section{04-02-02 \ \ {\it Reverse Shannon} \ \ (to C. H. {\Bennett} and P. W. Shor)} \label{Bennett8} \label{Shor2}

Thanks for sending me a preliminary version of the reverse Shannon write-up; I'm having fun reading it.  And I'm happy to see that my little contribution---Eq.\ (5) and the eqnarray after it---was a useful thing.

Besides the full theorem, Lemma 1 is just plain nice conceptually.  In fact, it seems pretty closely connected to the development between Eqs.\ (62) and (65) of my wacky paper \quantph{0106166}.  There I kind of had a ``teleportation'' of a POVM, but I hadn't thought about applying it in addition to an arbitrary ancillary state.  In my case, it was just half of a partially entangled pair.  Your additional piece tickles me, and is making me think about its implications for my devious (theological) purposes.

Strangely, Andreas's matrices $\hat\rho_j$ in Theorem 1 also played a significant role in the same wacky paper.  They played a crucial role in the rewrite of the Kraus rules that I was seeking---Eqs.\ (54) to (59)---in order to make quantum-state updating look more like Bayes' rule.  In this case, though, it's probably just a coincidence.  But it leaves me wondering.

I'll try to read the draft more thoroughly tomorrow, after I awake from this medicated fog.  (Another middle ear problem.)

\section{05-02-02 \ \ {\it Fighting Windmills} \ \ (to C. M. {\Caves} \& R. {\Schack})} \label{Caves53} \label{Schack46}

Thanks for the revealing note on Feller (which I've finally had a
chance to read).

\brs
Chris, do you know about any other book or paper that explains a
``modern understanding of probability''?
\ers

No I don't really.  In fact I don't even really have a strong
understanding of where most physicists have gotten their prejudices
toward probability.  Maybe a good place to start in such a quest is to
look at the discussions of probability in the main graduate and
undergraduate quantum mechanics textbooks used today (Cohen-Tannoudji
or however you spell it, Merzbacher, Liboff, Bohm, and whoever
else).\footnote{\editornote In Chapter III of Cohen-Tannoudji, ``The
  Postulates of Quantum Mechanics'', we find this on p.~220:
\begin{quotation}
\noindent Assume that we want to measure, at a given time, the
physical quantity $\mathcal{A}$.  If the ket $|\psi\rangle$, which
represents the state of the system immediately before the measurement,
is known, the fourth postulate allows us to predict the probabilities
of obtaining the various possible results.  But when the measurement
is actually performed, it is obvious that only one of these possible
results is obtained.  Immediately after this measurement, we cannot
speak of the ``probability of having obtained'' this or that value: we
know which one was actually obtained.  We therefore possess additional
information, and it is understandable that the state of the system
after the measurement, which must incorporate this information, should
be different from $|\psi\rangle$.
\end{quotation}
So: a ket will ``incorporate'' information, and when one gains ``additional
information'', one writes a new ket.

The Cohen-Tannoudji view of probability is\ldots well, maybe the only word is
``physicist-y''. On p.~227, we read,
\begin{quotation}
\noindent The predictions deduced from the fourth postulate are
expressed in terms of probabilities.  To verify them, it would be
necessary to perform a large number of measurements under identical
conditions.  That is, one would have to measure the same quantity in a
large number of systems which are all in the same quantum state.  If
these predictions are correct, the proportion of~$N$ identical
experiments resulting in a given event will approach, as~$N \to
\infty$, the theoretically predicted probability $\mathcal{P}$ of this
event.  Such a verification can only be carried out in the limit where
$N \to \infty$; in practice, $N$ is of course finite, and statistical
techniques must be used to interpret the results.
\end{quotation}
Contrast this with the Feynman lectures' chapter on probability:
\begin{quotation}
\noindent An experimental physicist usually says that an ``experimentally determined'' probability has an ``error,'' and writes
\begin{displaymath}
P(H) = \frac{N_H}{N} \pm \frac{1}{2\sqrt{N}}.
\end{displaymath}
There is an implication in such an expression that there is a ``true''
or ``correct'' probability which \emph{could} be computed if we knew
enough, and that the observation may be in ``error'' due to a
fluctuation. There is, however, no way to make such thinking logically
consistent. It is probably better to realize that the probability
concept is in a sense subjective, that it is always based on uncertain
knowledge, and that its quantitative evaluation is subject to change
as we obtain more information.
\end{quotation}
Mehran Kardar's {\sl Statistical Physics of Particles\/} states plainly,
\begin{quotation}
\noindent All assignments of probability in statistical mechanics are
subjectively based.
\end{quotation}
This should by implication apply to \emph{quantum} statistical
physics, the subject of the book's last chapter, but the book does not
dwell on that.}

All I remember from my undergrad statistics course was that the
professor told us that the Bayesian flavor of statistics was
nonsense.  Unfortunately, I don't remember the text we used.  In fact
I got very little out of the course that I remember at all.

\section{06-02-02 \ \ {\it The Commitments} \ \ (to C. M. {\Caves} \& R. {\Schack})} \label{Caves54} \label{Schack47}

\brs
\label{TheSchackcosm}
The other problem is that your conversation is far too playful. State
assignments are compilations of betting odds. They are COMMITMENTS.
\ers

\noindent Dear old friends,\medskip

I'm going to try to write a note today that essentially has been
sitting in my head for over a month.  I'm sorry that I've held on to
it for so long, but the great difficulty has been that time has just
been stolen from me left and right ever since a few days before
Katie's birth.  (Katie herself, by the way, is the least of my
problems; she's really a dream of a child.)  Anyway, I think there
were times in this period where the issues were so at the top of my
mind that this note could have (or would have) turned out far more
passionate and, thus, maybe clearer.  But that time has past.  Still
I'll give it my best shot with the time I can muster today, and also
hope that it's not so late as to clash with anything {\Carl} has written
for the newest version of the post-BFM paper.  Here goes.

In a nutshell, what I'm going to say is that I now take Schackcosm
\ref{TheSchackcosm} absolutely seriously---indeed probably far more
seriously than it was ever meant to be taken.

I used to say that quantum states are ``states of knowledge,'' or
``states of information.''  But, you both know that brooding over the
BFM paper caused me to disabuse myself of that.  Then I got into the
habit of calling quantum states ``states of belief'' in analogy with
what the more left-wing, de-Finetti-flavored Bayesians say of
classical probabilities.  Now, what I'm going to tell you is that in
contemplating the points you {\it two\/} think are important in our
reply to BFM, I've gotten into the habit of calling quantum states
``states of commitment.''  A quantum state should be viewed most
properly as a compendium of commitments, gambling commitments.

Indeed you guys almost nudged me directly there by your love affair
(or at least {\Carl}'s love affair) with strong coherence.  I'll come
back to say what I mean by this in much greater detail in a minute,
but for the moment let me bring the postulate into a clearing so that
you can go ahead and start aiming your arrows at it:

\bq
\noindent
      A quantum state corresponds to a compendium of gambling
      commitments (i.e., just like the gambling odds of a Dutch-book
      argument) one is willing to make in various given
      practical situations.  However, the key new point is that
      these commitments are with respect to ALL THINGS CONSIDERED.
      There are times when one's commitments correspond to one's
      (internal) beliefs---and because of that the quantum state
      remains just as subjective as ever in my mind---but that
      need not always be the case.  There are times (and these are
      probably the vast majority of all real cases) where the
      quantum state one ends up ascribing to a system is something
      less than a compendium of ANYONE's beliefs.
\eq

I know that I don't have to remind you that I tried pretty darned
hard to choose all my words very carefully in that definition.

Now let me tell you about the sorts of thoughts that lead to this by
explicitly replying to some of your old notes to me.

{\Ruediger} wrote:
\brs
I don't quite understand Chris's problem. Strong consistency DOES
have a motivation which is very similar to ordinary consistency, and
which DOES have the same flavor. It's just stronger. If you violate
strong consistency, you are imprudent in the sense that you accept a
bet in which, according to your own state belief, you never win, but
you lose for at least one outcome that you believe is possible. That
qualifies as imprudent, I think. Actually, I believe that the term
``imprudent'' fits better here than in the case of ordinary
consistency, where ``outright crazy'' seems more appropriate.
\ers
And then {\Carl} wrote:
\bcc
I agree completely with the above, especially with the distinction
between ``imprudent'' and ``outright crazy.''  I think a violation of
strong consistency is somewhere between imprudent and outright crazy,
but I haven't been able to think up the right [word].  People who
take imprudent actions expect to win big, I think; they are judged
imprudent because soberer people can see that the chance of [a] big
win is small, whereas the chance of serious losses is large.
\ecc

First, let me try to settle the issues of language in these two
remarks, or at least try to say more clearly why I was
concerned---actually in a rather offhand way in the beginning (see my
note of 12/5/01, titled ``\myref{Caves41}{Dear Prudence}'')---with the
appropriateness of the usual terms, ``coherence'' and ``consistency.''
Then I'll tackle the more substantial issue {\Ruediger} brings up
before that.

I once had an officemate who committed suicide.  That is an action I
would call (and did call) ``outright crazy.''  Seeing the pain and
the soul-searching it put everyone through who was near him, I might
even have called it ``moronic.''  And that is important.  For, the
point I was trying to make explicit to you in my earlier (shoddy)
note, is that that kind of craziness is of a much less absolute
character than the kind of craziness one would be committing by
asserting both $A$ and $\neg A$.  In the first case, the craziness is
conditioned by one's culture and one's customs, or you might say by
the instinct for one's survival.  In the second, the craziness is in
the breaking of a timeless, Platonic, a priori, ideal ``law of
thought.''

To me, the word ``inconsistency''---and therefore the word
``consistency''---seems far more to connect to such an ideal Platonic
stasis than one's willingness to be taken to the cleaners by a
Dutch-bookie.  Similarly with the word ``incoherent.'' However, being
Dutch-book coherent strikes me much more as an expression of the
survival-instinct type than anything else.  It is a formal expression
of ``thou shalt not commit suicide.''  Why should that be considered
an inherently ``logical'' commandment? Why should it have the right
to live in the Platonic realm of Boolean logic?  If you ask me, I
would say it is probably much more a manifestation of simple
Darwinian evolution.  We try to stay Dutch-book consistent---it is
our ideal of behavior---precisely because of the survival tool it
represents for our kind.  However, we know by recent experience, that
some cultures bask in the idea that, at times, there are reasons to
override one's personal aversion to suicide.  Do you know of a
culture that, at times, finds it useful to override Boolean logic in
its mathematical proofs?

When I wrote my original offhand note, I never imagined the backlash
I would get from you guys.  Nor did I intend to make another proposal
to change established nomenclature.  I thought I would just simply
get a reply of the sort, ``Yes, we understand your concerns, but in
this case it really is too late to change the nomenclature. A good
fraction of the Bayesian community has been using the terms
`coherence' and `consistency' for almost 70 years.  The best you
should hope for is an extra paragraph in the paper's appendix
assessing why one ought to consider Dutch-book coherence
compelling.''  And I'll still stand by that.

But I actually think there is a point of more substance in this---one
that starts to bite much harder when one debates the relative merits
of regular Dutch-book coherence versus its stronger cousin (i.e.,
strong consistency, strict fairness, or what have you).  And the new
problem is no longer just a problem of language. To say it again,
{\Ruediger} writes:

\brs
Strong consistency DOES have a motivation which is very similar to
ordinary consistency, and which DOES have the same flavor. It's just
stronger. If you violate strong consistency, you are imprudent in the
sense that you accept a bet in which, according to your own state
belief, you never win, but you lose for at least one outcome that you
believe is possible.
\ers

You don't know how I ALMOST agree with the first sentence of this! In
fact, I could agree with it completely if you would just let me
change the first instance of ``does'' to ``can.''
\bq
\noindent
      Strong consistency CAN have a motivation which is very
      similar to ordinary consistency, and which DOES have the
      same flavor.
\eq

I said before that strong consistency was of a different flavor than
standard consistency, but let me be more careful now---i.e., now that
you've helped me sharpen my point.  It is not that the flavor is a
priori different; it is that in paying the price of strong
consistency, you actually get two flavors for the price of one. That
is to say, I cannot agree that strong consistency is ``just
stronger''---it is that and something more.

Here is a very different (and ostensibly much less satisfying) way of
motivating strong consistency:
\bq
\noindent
      AXIOM:  Whatever an agent's personalistic beliefs, i.e.,
      whatever personalistic probabilities he has managed to write
      down in his head for some event, when placing a bet (with a
      Dutch-bookie or otherwise), he MUST place it precisely
      according to the gambling odds his beliefs afford.
\eq

Clearly a special case of this is that when a person believes $p=1$
for some event, then he MUST bet as if he is certain that it will
occur.  I.e., he must be strongly consistent.

But you probably ask, ``What is wrong with that?''

I will tell you.  In doing such a thing, you would throw away loads
of freedom, loads of wiggle room that the standard Dutch-book
argument generally leaves in your command.  In essence you throw away
the freedom to concatenate your Dutch-bookie-game commitments with
any of a number of other games.  Or still another way to say
it---though it's not so nice---you throw away your freedom to lie,
even when it is in your best, overall, total interest.

An easy case to see this in is the ``double agent'' Dutch book
argument that I wrote you about on 8 Sept.\ 2001 in a note titled
``\myref{Caves23}{Negotiation and Compromise}'' (starting on page 64
of the mini-samizdat).  Suppose that Kiki and I draw our money from
the same bank account, and that for some given event I ascribe $p=1$,
whereas Kiki ascribes $p=1/2$.  We've talked about our disparity in
assignments many times, but I simply cannot convince her to accept the
evidence that led me to $p=1$; she holds her ground, and I know she
would do that in any circumstance, come hell or high water.  Now
suppose I'm later approached by a Dutch-bookie that I am fairly sure
has already gambled with Kiki.  What should I do?  If I don't want to
lose my shirt, then I had better not declare my adamant belief in the
ultimate occurrence of the event we are betting.  Instead, I should
adjust my GAMBLING COMMITMENT to agree with Kiki's (sorely) less than
optimal one.  (Sorely less than optimal from my point of view, that
is.)

If on the other hand, you say that in order to play the game of
Bayesian probability, I must be strongly Dutch-book consistent, then
I no longer have the option to save my own fortunes in this new game.
The point is, standard Dutch-book consistency concatenates well with
such a further, ancillary game, whereas strong consistency does not.

But this is just a contrived example.  Similar, but maybe better,
examples can also be found by looking at the old ``Keeping the Expert
Honest'' game (see pages 20--21 in Ph.D. thesis).  And I'm sure there
are still other more realistic situations that we could come up with
if we just tried.

So, what I am saying is Dutch-book consistency is only the tiniest
check on one's gambling strategies, and that is a GOOD thing.  A very
good thing.  If one imposes too much structure---apparently almost
anything more than simple DB consistency---then one will be left in a
lurch in the real world, a world where negotiation and compromise are
the keys to survival.

I truly, truly, truly hope you will see the point of this, but I
guess I have gained enough experience in the last email war to be
prepared for the worst.  Here is what I am mainly afraid of.  By
holding fast to strong consistency as a reasonable addition to
standard DB consistency, one ends up in the quantum case with a nice,
tightly mathematical looking theorem (almost) in the BFM style.  You
cannot tell me that at least the {\Carl} among you is not far more
attracted to results like that, than the willy-nilly result one is
left with if only standard DB consistency is enforced.  At the very
least, it makes the paper look far more constructive than destructive
\ldots\ and that's got to be deemed a good thing, right?

What I guess I am saying is that, concerning what {\Carl} wrote:
\bcc
I, as you know, like strong consistency and hope that in discussions
of it, we can separate the a priori reasons for liking or disliking
it from reasons based on the conclusions it leads to.
\ecc

I believe that I have done that.  However, I hope that you will live
up to these hopes too!  Every bit of the discussion above (excepting
the last paragraph) had nothing to do with quantum mechanics.  The
argument is purely classical and divorced from the BFM issue.  In our
post-BFM paper, I certainly have no problem whatsoever delineating
the whole hierarchy of conditions, regular DB, strong DB, etc.  What
worries me though is how the present draft gives pretty short shrift
to plain old regular DB --- the very one that I myself find the most
reasonable (for all the reasons above).

Of course, suspecting that you guys will try to wring me out after
spouting this blasphemy, I had hoped to come armed with good
knowledge of the literature.  Unfortunately, as I already expressed
way above, time hasn't been on my side these last few weeks.  What I
was at least able to do though, was run to the Princeton library one
day and amass copies of a lot of papers.  I ran from one paper to the
next, following the citation trail that each gave.  Below is the
result:  My complete collection of Dutch-book papers.

I wish I could say that I had read these, but for me the skim of a
title and abstract doesn't count as ``reading'' a paper \ldots\ as it
does for XXX, say. Nevertheless, a couple of points did stand out for
me. The most important one is that maybe I'm not alone in thinking
that strong consistency goes (far) too far. Hacking (1968) below
writes, for instance,

\bq
Abner Shimony called it {\it coherence}; John Kemeny called it {\it
strict fairness}; today many people speak of {\it strict coherence}.
According to Shimony's definition, a set of betting rates on a series
of propositions $h_i$ and $e_i$ is strictly incoherent, when ``there
exists a choice of stakes $S_i$ such that, if $X$ accepts the series
of bets at these stakes, then no matter what the actual truth values
of $h_i$ and $e_i$ may be, $X$ can at best lose nothing, and in at
least one possible eventuality he will suffer a positive loss.'' De
Finetti had a less demanding concept which is called {\it coherence}.
A set of betting rates is incoherent if, no matter what the actual
truth values of $h_i$ and $e_i$ may be, $X$ will suffer a positive
loss in every eventuality.  Logicians usually think that Shimony has
improved on de Finetti's concept of coherence, but statisticians,
including de Finetti himself, have seldom been persuaded.
\eq

After that, he starts writing about nonBoolean algebras, and I
haven't had a chance to try to decipher it.  But what is really
important to me is that this is the first indication I've run across
that de Finetti himself thought about strong coherence and then
rejected it.  I ask, ``Why?''  I know that {\Carl} has little use for
relying on authority---I'm referring to his quick rejection of even
wanting to hear de Finetti's opinion---but that is not what is at
issue here.  de Finetti is someone who thought long and hard about
probability from the Bayesian view.  I cannot see how it would not be
worthwhile to at least hear his arguments.  We might save ourselves
time, and we might save ourselves from making mistakes that will make
us feel foolish.

Then there is a large set of papers debating whether Dutch-book
coherence really has to do with ``rationality'' or rather something
else (as I alluded to above).  The titles should give those papers
away.  I've hardly even skimmed those at all, but you can see from
above where my present opinion lies.  DB consistency defines an
expedient for our actions, but it is hardly more rational or logical
than that.  Skyrms87 in particular, I am told, argues the opposite
point of view.  If that is a position that is near and dear to your
heart, then maybe it would be worth understanding what he has to say.

OK, clearly I'm petering out and it's after midnight now.  Soon I
won't even be coherent myself; I'm already starting to see signs of
it.  Let me try to quickly summarize the whole argument in a few
sentences, rather than checking and editing all the above to make it
more eloquent, more complete, and more connected.

\begin{enumerate}
\item  DB coherence strikes me as much more a pragmatic requirement than
as any rule of rationality (as the law of the excluded middle is).

\item  Thus one is more compelled to consider the pragmatic consequences
of standard DB versus strong DB.

\item  From that, one sees that strong DB is not just more of the same,
but carries with it whole new flavors of behavior.  In particular it
forces us all to be little George Washingtons---``I cannot tell a
lie''---when we have a $p=1$ assignment in mind.

\item Eschewing that, I am forced to divorce our (pragmatic) gambling
commitments from our actual beliefs.  Our beliefs can be our
commitments, but our commitments need not be our beliefs.

\item Thus it is better to say that ``probabilities are our gambling
commitments, ALL THINGS CONSIDERED.''  (with apologies to NPR)

\item Quantum states being compendia of probabilities are thus ``states
of commitment'' full stop.
\end{enumerate}

That's the argument.  I'm sure this letter is riddled with typos, but
I don't want to hold on to it anymore.  France Telecom is coming
Thursday and I've been tapped to convince them that quantum
information is interesting and that our dabbling in it is a little
value-added perk they'll get if they stay our customer rather than
running away to Alcatel.  Can you believe that?

Anyway, maybe {\Ruediger} will be a little happy to see this when he
gets into the office tomorrow \ldots\ to see that I haven't really
abandoned you.

\begin{enumerate}
\item
Brad Armendt, ``Is There a Dutch Book Argument for Probability
Kinematics?,'' Philosophy of Science {\bf 47}, 583--588 (1980).

\item
Patricia Baillie, ``Confirmation and the Dutch Book Argument,''
British Journal for the Philosophy of Science {\bf 24}, 393--397
(1976).

\item
David Christensen, ``Clever Bookies and Coherent Beliefs,''
Philosophical Review {\bf C}(2), 229--247 (1991).

\item
Barbara Davidson and Robert Pargetter, ``In Defence of the Dutch Book
Argument,'' Canadian Journal of Philosophy {\bf 15}(3), 405--424
(185).

\item
Richard Foley, ``Being Knowingly Incoherent,'' No\^us {\bf 26}(2),
181--203 (1992).

\item
Bas C. van Fraassen, ``Belief and the Will,'' Journal of Philosophy
{\bf 81}(5), 235--256 (1984).

\item
Ian Hacking, ``On Falling Short of Strict Coherence,'' Philosophy of
Science {\bf 35}, 284--286 (1968).

\item
Frank Jackson and Robert Pargetter, ``A Modified Dutch Book
Argument,'' Philosophical Studies {\bf 29}, 403--407 (1976).

\item
John G. Kemeny, ``Fair Bets and Inductive Probabilities,'' Journal of
Symbolic Logic {\bf 20}(3), 263--273 (1955).

\item
Ralph Kennedy and Charles Chihara, ``The Dutch Book Argument: Its
Logical Flaws, Its Subjective Sources,'' Philosophical Studies {\bf
36}, 19--33 (1979).

\item
R. Sherman Lehman, ``On Confirmation and Rational Betting,'' Journal
of Symbolic Logic {\bf 20}(3), 251--262 (1955).

\item
Abner Shimony, ``Coherence and the Axioms of Confirmation,'' Journal
of Symbolic Logic {\bf 20}(1), 1--28 (1955).

\item
Brian Skyrms, ``Coherence,'' in {\sl Scientific Inquiry in
Philosophical Perspective}, edited by Nicholas Rescher (Center for
Philosophy of Science, Lanham, MD, 1987), pp.~225--242.

\item
Jordan Howard Sobel, ``Self-Doubts and Dutch Strategies,''
Australasian Journal of Philosophy {\bf 65}(1), 56--81.

\item
Lyle Zynda, ``Coherence As an Ideal of Rationality,'' Synthese {\bf
109}, 175--216 (1996).
\end{enumerate}

\section{06-02-02 \ \ {\it How Did?\  What Did?}\ \ \ (to N. D. {\Mermin})} \label{Mermin58}

\bdm
I don't remember the precise content of some of the issues you seem
to be addressing, notably what is ``strong consistency''?
\edm

Then how on earth did you word your slide (where you gave our version
of the theorem) when you gave your talk here in Murray Hill.  Now my
curiosity is piqued; maybe I didn't notice some inanity in your talk.
One only gets something that looks even remotely like the BFM
criterion if one assumes strong consistency.  If you stick with
standard Dutch book, you get precisely what I had been trying to tell
you since the very beginning:  namely, there are {\it no\/}
constraints on the density operators required at all.

\section{06-02-02 \ \ {\it Actually, the Both of You} \ \ (to C. M. {\Caves} \& R. {\Schack})} \label{Caves55} \label{Schack48}

This morning I woke up and re-read the long note I sent you last
night.  In doing that, and looking back at your other notes again, I
now think I was too harsh in singling out {\Carl} when I wrote:
\bq
\noindent
By holding fast to strong consistency as a reasonable addition to
standard DB consistency, one ends up in the quantum case with a nice,
tightly mathematical looking theorem (almost) in the BFM style.  You
cannot tell me that at least the {\Carl} among you is not far more
attracted to results like that, than the willy-nilly result one is
left with if only standard DB consistency is enforced.  At the very
least, it makes the paper look far more constructive than destructive
\ldots\ and that's got to be deemed a good thing, right?
\eq

That's not fair to him.  In particular, both {\Ruediger} and {\Carl}
wrote:

\brs
Of course it is nicer to base one's approach on the weaker concept.
What we are discussing in the paper, it seems, is reasons for why the
weaker concept does not quite give us what we would like. These
reasons are secondary, they have nothing to do at all with the Dutch
book justification of strong consistency itself.
\ers

\bcc
{\Ruediger}'s two paragraphs nicely illustrate what I was hoping for
when I wrote that I ``hope that in discussions of it, we can separate
the a priori reasons for liking or disliking it from reasons based on
the conclusions it leads to.''  The first paragraph is about the a
priori reasons, and the second is about why its conclusions are
important.
\ecc

Let me focus in particular on {\Ruediger}'s sentence, ``What we are
discussing in the paper, it seems, is reasons for why the weaker
concept does not quite give us what we would like.''  What is it that
{\it we\/} would like?  And, why would {\it we\/} like it?

Clearly, I am most happy with the willy-nilly result that standard DB
coherence alone gives.  I'm glad you guys made that nice and
rigorous.  For me, it says that B, F, and M were just way off the
mark in trying to dictate what various observers MUST ascribe for
their quantum states.  It tucks nicely with my very first note to
them where I registered some protest.

Now, what is nice about the strong coherence version of the argument
is that it does at least get us the ``if'' part of BFM, and thus
gives us something solid with which to compare to their campaign. But
I suppose, I have always viewed that as a secondary, rather than the
primary, point of our criticism.  Its role is simply in that it
tells us what we have to ADD to pure (de Finetti flavored)
Bayesianism, to get something that comes close to resembling BFM.

So, my apologies to {\Carl}, and my consternation to the both of you!

\section{06-02-02 \ \ {\it Definition} \ \ (to C. M. {\Caves} \& R. {\Schack})} \label{Caves56} \label{Schack49}

Let me go back to the NPR thing briefly while I've got a couple of
moments.  ``All things considered,'' what does it mean?

I know that your knee-jerk reaction is going to be that it is a
hopelessly vague term.  However, I'm going to suggest that it is no
more and no less vague (and mysterious in its origin), than
``belief'' was in the first place.  It is just broader in scope.  In
fact, I think it amounts to little more than the sum total beliefs
one possesses.  (What more could one consider?)

Thus in setting a quantum state, one sets it according not only to
what one believes about the system of particular interest \ldots\ but
also according to the situation one believes he will be encountering in
the laboratory, the purposes of the information he will gather and
how it will be used, who will share that knowledge so gained, etc.,
etc.  It may even depend upon how the financial markets are doing,
which political party happens to be in power, and so forth.

It just takes seriously the idea, that to any quantum system, what we
say about it---i.e., what we are openly willing to bet on
it---depends upon many things beside the system itself.

Maybe all of this goes back to a conversation I had with {\Marcus}
{\Appleby} while I was in Northern Ireland, just after 9/11.  I tried to
explain my new point of view---that a quantum state is a state of
personal belief---and he replied that he though that had the right
feeling, but that he had some kind of ``unpinpointable'' fear that
maybe the idea didn't go far enough.  He seemed to be saying
something to the effect that beliefs can never truly be considered in
isolation from other beliefs.  I didn't understand his worry at the
time, but I think everything I wrote you yesterday starts to pick up
on this line.  In this regard, by the way, {\Appleby} suggested I read a
book by Michael Polanyi, {\sl Personal Knowledge}. (Polanyi was
apparently also a chemist of some renown, I believe he said.) Though
I haven't had a chance to dig it up yet.

In any case, the COMMITMENTS one (potentially) makes in the sum total
of all gambling situations quantify the quantum state.  They give it
an operational meaning---in terms of its ``cash value''---which goes
beyond niggling over the details and merits of ``beliefs'' versus
``culture'' versus ``all relevant things considered,'' etc.

\section{06-02-02 \ \ {\it The Great Quantum Well} \ \ (to C. M. {\Caves})} \label{Caves57}

\bcc
Not willing to discuss the technical issues much at present because
I'm not going to have time to flesh things out for a while.  I get a
long way by not being interested in Kraus decompositions, which
aren't generally convex combinations of unitaries, and I get a good
deal farther by insisting on time evolutions, not just single-time
stuff. The rest is going to come from having available some operators
that rigidify (and thus provide a physical interpretation) for the
vectors in Hilbert space, without which the quantum questions (and
Hamiltonians) aren't anchored to anything at all.
\ecc

I think the last sentence really captures what sets our goals apart.
The image I carry around in my mind presently is that the quantum
system is very much like an oracle.  We ping it, and it provides us
with something that we did not have before, something that we could
not foresee \ldots\ and therein lies its reality. Therein lies its
independence from us.  Each quantum system is an instantiation of
your great quantum well.

But the ``interpretation,'' the ``meaning,'' of the gifts those
oracles give us is set from the outside.  Completely from the
outside.  Or at least that's the point of view I'm pursuing.

I just don't see how that line of thought lies anywhere along the
lines of your accusation that quantum questions will thus not be
``anchored to anything at all.''

\section{06-02-02 \ \ {\it Jacobs Paper} \ \ (to C. M. {\Caves})} \label{Caves57.1}

\bcc
The paper is clearly a reaction to having heard Mermin's talk at the
ITP conference.
\ecc

In all fairness to Kurt, he's been talking about ``combining quantum states of knowledge'' and some variation of the present paper ever since before the Cerro Grande fire.  I remember several conversations with him on the subject, one of which led me---in a rather circuitous route---to rearranging the quantum collapse rule so that it made use of operator geometric means.  (See pages 90--91 of my big Samizdat, \quantph{0105039}.)  He had showed me something with a geometric mean in connection to his present paper, but I don't remember what.  Anyway, Mermin's talk---if it had anything to do with him on this score---was only a catalyst for the present write-up.

\section{07-02-02 \ \ {\it The Heisenbergs}\ \ \ (to J. B. Lentz)} \label{LentzB2}

I'm just reading some new revelations about Heisenberg's involvement in the Nazi A-Bomb project in the {\sl New York Times\/} this morning, and I ran across the following:
\bq\noindent
But others say questions about the meeting remain. One of Heisenberg's
sons, Dr.\ Jochen Heisenberg, who is now a physicist at the University
of New Hampshire, and Mr.\ Powers, say the documents show that Bohr
never understood the message Heisenberg meant to convey in Copenhagen.
\eq
Didn't you tell me you met this guy?

Anyway, the whole article is at \myurl{http://www.nytimes.com/2002/02/07/science/07BOMB.html}.

\section{08-02-02 \ \ {\it Samizdats and Dutch Book} \ \ (to B. C. van Fraassen)} \label{vanFraassen1}

I was talking to David {\Mermin} a while ago and he mentioned that you
had brought my name up in some email and seemed to be aware of some
of my papers.  (I hope I got the story straight.)  Anyway, that
piqued my interest:  If true, then you are very likely the only
philosopher who has ever noticed my existence!  I would certainly
love to hear your opinion---both con and pro---of my quantum
foundational thoughts (forming as they are).  Having the critique of
a true-blue philosopher would be most useful for steering me to
clarity, or even steering me away from the abyss!

My two most close-to-philosophical pieces can be found on the Los
Alamos preprint archive: \quantph{0106166} and \quantph{0105039}. Or, they can both be found at my (almost empty)
website with a couple of other pieces of supporting material.

Also I'm in Princeton from time to time making use of your wonderful
library system.  Maybe I could drop by for a chat?  (Sometime in
March or later, actually, since I'll be in Japan until essentially
then.)

By the way, I was trudging through various Dutch-book arguments
recently, and I came across your paper ``Belief and the Will,''
[Journal of Philosophy {\bf 81}(5), 235--256 (1984)] in the process.
It looks good.  I'll let you know if I form an opinion.

\section{11-02-02 \ \ {\it Oh Modern {\Wittgenstein}, 1} \ \ (to N. D. {\Mermin})} \label{Mermin59}

Concerning points 1 and 2 in your {\sl Tractatus Quantico-Philosophicus},
read the anecdotes below.  [See 02-02-02 note ``\myref{Timpson1}{Colleague}'' to C. {\Timpson}.]

I exclusively use the word ``qubit'' for the physical system, i.e.,
the ``carrier'' of the information, i.e., the object of one's belief,
i.e., the oracle the receiver consults at his end of the game.  I
never use qubit to mean a quantum state (and I don't think I ever
have), but I know it's a serious problem in the community.

\section{11-02-02 \ \ {\it Oh Modern {\Wittgenstein}, 2} \ \ (to N. D. {\Mermin})} \label{Mermin60}

\bdm
It's the Cbits I'm more concerned with.  ``bit'' -- which Charles
wants to reserve for the classical physical system clearly has an
important abstract meaning as well, relevant to both Qbits and Cbits.
\edm

Aristotle called it matter, the receptacle for accidental properties
(presumably some of which could be binary valued).  When we turn our
attention to a subset of such properties, and are completely ignorant
of which of the two is inherent, then we say that we are missing a
bit of information.

\section{11-02-02 \ \ {\it A Wackier Idea} \ \ (to J. Preskill)} \label{Preskill4}

I wonder if you would indulge me with still another wacky idea (actually a wackier idea than usual) to distract you from your usual workaday duties.  The only compensation I can give is my promise that in the long run---the long, long run---the world will be a better place if you make this effort.  (As if I'm in a position to talk that way!)

Here's the wacky idea.  Jeff Bub and I are putting together a special issue of the journal {\sl Studies in the History and Philosophy of Modern Physics\/} on the impact (if any) of quantum information on quantum foundations and other philosophical issues.  Here's what we've managed to line up so for for the issue.  The {\it definite\/} commits are:
\begin{itemize}
\item
Andrew Steane:  ``A Quantum Computer Only Needs One Universe,'' (extended version of \quantph{0003084})
\item
Ben Schumacher:  ``Doubting Everett''
\item
Richard Jozsa:  ``On quantum information, and the sense in which it is a distinct notion from its classical counterpart''
\item
Itamar Pitowsky:  ``On Bayesian quantum probability (deriving the quantum rules from a gambling scheme in the spirit of de Finetti or Ramsey)''
\item
Howard Barnum:  ``On quantum logic and quantum information''
\end{itemize}
Beyond this we have ``probable yeses'' from Charlie Bennett and David {\Mermin}.  Charlie would write something on Maxwell demons and David would write something {\Mermin}esque.

You might be able to tell from this that the lineup is starting to get a little one-sided against Everett.  I'm writing you because I'd like to find a little counterbalance to that.  In particular, I'd like a scientist---a real scientist---to explain as best he can, what he finds pleasing in the Everett picture, in what ways he finds it useful, how it might make quantum computing's power easier to stomach, and to what extent he finds the view problem-free or, instead, to what extent he foresees it needing a patch.

I've already tried to get Charlie Bennett to take this bait (since in private he is such an avid fan of Everett and calls my own quantum foundations quest ``theology''), but in his words, he ``just can't get it up to write an article on that subject.''  (See the note below, along with further details about the goals of the project.)  But I really think someone needs to do it, or at least try to do it.  Honestly, you're the only person besides Bennett that I trust with such a task.

I want a scientist \ldots\ not a counter-theologian like Deutsch or Vaidman.  I think only in that way, will we really get at the root of what's really good in the Everett view or where it remains weak.

Would you do this?  Could you do this?  (If I were a doctor, I'd send a sample of Viagra along with the note!)  We'd be able give you until about mid-June to get the article finished.  And of course, I could volunteer as much (constructive) feedback as you'd like for your drafts, if you're worried about the oddball audience you'd be writing this for.

I hope you'll think hard about the proposal.  There are few people that have such a broad overview of the field as you, and I'm banking that'll give you the resources to make a first-class contribution along these lines.

As soon as we hear from you one way or the other, Jeff and I will send out a more formal invitation to all involved, including deadlines, etc.

\section{11-02-02 \ \ {\it A Wackier Idea, 2} \ \ (to J. Preskill)} \label{Preskill5}

\bjp
Seems to me that you need someone who is passionate about this, and
I'm not.
\ejp

No, you're hitting the nail on the head with the latter part of the sentence, and that's exactly why I would like for YOU to write an article.  To the extent that you seem to find Everett a pleasing or satisfying picture, you have {\it always\/} seemed dispassionate about it.  That's a perspective that this debate on the subject seems to sorely need---the working scientist's point of view, without a religious fervor.  (I tried to convey that to you in my last note.)  A calm, collected account of why you find it a useful working picture.

\bjp
Actually, Deutsch seems like a natural, since he ought to be able to
pitch his message to an audience of philosophers, and you seem to want
that.
\ejp

Deutsch really is only about religious fervor (probably much like I am, if Charlie's accusations are on the right track).  I think what needs pitching to an audience of philosophers---for this particular subject---is what I said above.  You, like Bennett, think that the Everett picture has a certain cash-value in your everyday work (or, you wouldn't have taken the time to write about it in your lecture notes).  I think it would be useful for us all to understand that better (in general, but also with some explicit examples).

Deutsch would say the same things he has said a thousand times before.  He would invoke Popper, and Plato, and whatnot, and say that anyone who does not believe as he does is ``refusing to contemplate the implications of science.''  You would say things that John Preskill would say, in John Preskill's way.  And they may turn out to be the same things that Deutsch would have said, but then that will be a useful datum too.

\section{11-02-02 \ \ {\it Viagra Sample} \ \ (to C. H. {\Bennett} and J. A. Smolin)} \label{Bennett9} \label{SmolinJ2.05}

Below I'm gonna send the note again that I sent you on January 2 concerning the special issue of SHPMP that Jeff Bub and I are putting together.  Plus let me also send you the lineup of other authors for the issue we've put together since then.  I'm hoping you'll find it as stimulative as I find it!

Beyond these guys, Mermin and Preskill are also giving it some serious thought.

I sure would like to hear a ``definite yes'' from you guys soon.  (The deadline on the paper would be mid-June.)  Jeff and I are hoping to send out a more formal ``invitation letter,'' closing the ranks and sort of tying the whole project together by the end of this week.

The {\it definite\/} commits are:
\begin{itemize}
\item
Andrew Steane:  ``A Quantum Computer Only Needs One Universe,'' (extended version of \quantph{0003084})
\item
Ben Schumacher:  ``Doubting Everett''
\item
Richard Jozsa:  ``On quantum information, and the sense in which it is a distinct notion from its classical counterpart''
\item
Itamar Pitowsky:  ``On Bayesian quantum probability (deriving the quantum rules from a gambling scheme in the spirit of de Finetti or Ramsey)''
\item
Howard Barnum:  ``On quantum logic and quantum information''
\end{itemize}

\section{11-02-02 \ \ {\it Viagra Sample, 2} \ \ (to B. W. Schumacher)} \label{Schumacher8}

I thought you might enjoy this too.  (And I send it as a subtle reminder.)  [See 11-02-02 note ``\myref{Bennett9}{Viagra Sample}'' to C. H. {\Bennett} and J. A. Smolin.]

But let me also ask you a technical question related to your project.  Have you ever thought about what (if anything) it would change of your conclusions if one didn't augment the universe's Hilbert space with a fixed (preferred) basis---which definitely would demolish your whole point, as you have made clear in your talks---but instead one augmented it with a fixed (preferred) tensor-product decomposition?  Indeed, does that even quite make sense?

\section{12-02-02 \ \ {\it The Will to Believe} \ \ (to B. C. van Fraassen)} \label{vanFraassen2}

Thanks for the note.  You warmed my heart with the sentence, ``We
were all very intrigued with this `Bayesian' approach to probability
in QM.''   We (i.e., {\Caves}, {\Schack}, and myself, and sometimes {\Mermin})
know that it's all very much at the beginning stages, but things are
starting to fall into place so exponentially fast it evermore gains
the air of an inevitability.  We've now got a load of material that
we haven't published yet (and are working very hard to write up), and
I in particular have made a strong phase transition in my attitude
toward quantum time evolutions (i.e., their level of subjectivity
versus objectivity).  So things are just flying.

Thanks very much for putting me on the meeting list.  But also I hope
to meet you before then.  In any case, know that I am always, always
available on email.  (And as one of the documents I advertised in the
last note will attest, that is even my preferred means of
communication!)  So, please feel free to get a dialog going if you
wish:  I would relish it.

\section{12-02-02 \ \ {\it Samizdats and Dutch Books} \ \ (to P. F. Scudo)} \label{Scudo1}

I'm glad to hear that you are not frightened by de Finetti's ghost!  My whole life is tied up in it now.  Sometimes when one hits the right path to a solution, one just knows it---the feeling is overpowering.  And that is the case with me on this round.  I dare say there is no deeper, no more direct path to the quantum foundations than through Bayesianism.

\section{13-02-02 \ \ {\it Another Kent Paper} \ \ (to G. Brassard)} \label{Brassard11}

\bgb
What do you think of this?
\egb

I think ``nonlocality'' and especially questions about the
restrictions enforced on physical the\-ory---in particular, the
description of quantum phenomena---due to no-signaling criteria are
red herrings. Our brain pulp is better left for deeper matters. You
see, in playing the sorts of games that people have been playing
along these lines, the very starting point is to take the idea
seriously that the quantum state is a physical property, rather than
a description of information, knowledge, belief, betting-odds or what
have you. And that is an idea I stopped taking seriously a good while
ago.

See Section 6 (and its closing paragraph) in my paper \quantph{0106166}. By the way, there is a (relatively minor)
technical mistake in that section; let's see if you can find it!

\section{13-02-02 \ \ {\it One More for {\James}} \ \ (to C. M. {\Caves})} \label{Caves58}

\bcc
Curt?  Emphatically so.  Snide?  I don't think so.  But I can see
that my habitual curtness has led to more than its usual negative
reaction, so let me go back to square one.
\ecc

William {\James} likes to say that all beliefs are ``numerically
additional'' to the reality they take as their target, even ``true''
beliefs.

Thanks for the note.

I similarly need to automate my Outlook better:  Maybe that'll be one
of the great triumphs of our time together in Brisbane.

\section{15-02-02 \ \ {\it Friendship Call} \ \ (to L. Hardy)} \label{Hardy3}

I discovered yesterday that you'll be in New Jersey in March for the Wheeler-meets-rich-man-desperately-wanting-to-live-forever thingy.  I'm not sure whether I'll be going to that yet, but I wonder if I could interest you in dropping by my neck of the woods while you're in the state.  I should be around March 11 through 19 (inclusive).  March 19 at 6:00 PM I fly out for the little meeting in Dublin (quantum channel capacity stuff).

I can't offer you any travel funds, but I could put you up in our guest room and offer you some of my wife's home cookin'.  Also I've got a decent collection of single malts that might lead us to some insight.  (There's a great line in a Tom T. Hall song, ```We must've drank 10 quarts of German beer; my conscience and my sinuses were clear.'')  Also we're a 1 hour and 10 minute straight-shot train ride to New York City; so you could use us as a base of operations for a couple of days if you wished.

If you've got the time and haven't settled your plans yet with respect to this trip, I'd certainly love to see you.  I've got a load of things quantum I'd like to talk to you about.  (I'll be the poetaster---that's actually a word---and you can be the concrete master.)

I'm off to Japan for 10 days tomorrow, but I'll be in email contact every day.

\section{16-02-02 \ \ {\it Some Things Should Not Pass} \ \ (to several friends)}\label{ShouldNotPass}

Some things should not pass without our best effort to make them
indelible.  Yesterday, February 15, Kiki and I had to put our golden
retriever Wizzy to sleep.  He was the most loving and faithful dog
either of us had ever had.

As things happen, Wizzy's last day of life marked exactly eight years
from the time he first entered my apartment in Albuquerque; it was my
second date with Kiki.  I had schemed all day about how I might meet
her that evening, and the solution was to cook a meal, a large meal.
Hopping around the corner from my apartment to hers, I said, ``Would
you be interested in dinner tonight?  I accidentally made too much.''
She said, ``Sure; I was only going to warm up some potatoes and
cheese anyway.  I'll be over in about 20 minutes.''  A couple of
minutes later, I got a phone call asking if she could bring her dog
with her.  I said, ``No problem.''  It was a sweet and touching
sight:  Wizzy was a dog so insecure at the time, he never left Kiki's
side---he didn't sniff around or explore like most any other dog
would have; he stayed in bodily contact with her from the moment he
entered until the moment he left.

About four months before that night, Kiki had rescued Wizzy from an
animal shelter.  He must have had a hard life, we surmise, judging
from the scar on his head and the fear he had of brooms at the
beginning.  We'll never really know how old he was, but comparing him
to our other golden, Albert, he was probably 12 or 13 years old when
he passed.

The day Kiki met him, she had the intention of looking at two dogs
before making a decision of which one to take home.  Wizzy was the
first.  When the attendant let him out of the cage, he so leaned his
whole body into Kiki and seduced her with his big, loving eyes, that
she knew she couldn't put him back.

Wizzy in fact played a predominant role in my meeting Kiki.  For some
time I had seen her walking him around the neighborhood, and it
dawned on me that since no one was ever with her, she had a chance of
being single.  I waited for my moment, and it came one morning as I
walked across a neighborhood park.  I introduced myself by going
directly for Wizzy.  I said, ``What a beautiful dog; what's his
name?''  She told me, and then I asked in a sort of quizzical way,
``Is he a purebred?''  She said, ``Yes.''  I said, ``I don't think
so. I have a golden myself, and they don't look very much alike.''
Why she accepted a date with me a few weeks later remains a mystery.

This morning I broke the news to Emma, and she became sad.  She
asked, ``Where did he go?''  I said, ``Back to nature.''  When she's
ready to think about it harder, I'll tell her my (presently) favorite
metaphor for what happened:  Our finite lives are like little drops
of water that have parted from the sea.  For a small time we have the
chance to move around and determine our courses as we please---to
leave a trail behind us.  But we all eventually run back into the
sea.  We never stop being; we just become part of something bigger.

Kiki and I put Wizzy into the ground at sunset yesterday, like
Egyptians.    We gave him his blanket of eight years so that he would
never be cold.  We gave him his leash so that he could have an
infinity of walks, his bone so that he could have an infinity of
chews, and his rubber ring so that he could retrieve it for eternity.
I told him that I had always known he was a purebred.

\section{17-02-02 \ \ {\it Long Letter, Way Longer Flight} \ \ (to D. B. L. Baker)} \label{Baker4}

\noindent ``Baker-san. Kon-nichiwa. Fuchs-desu.''  [\ldots] \medskip

I've got such a long day ahead of me still, I don't even know where to start.  This time I'm going to the city of Sendai.  I think it's maybe 200 miles north of Tokyo.  So, once I get to Narita airport, I have to take a train to Tokyo station.  That ride lasts a little over an hour.  But then, after a 45 minute wait, I have to take a train from Tokyo to Sendai.  Even though it's labeled a ``super-express,'' that will also take over two hours.

For a while I was pretty frightened about navigating Tokyo station:  You wouldn't believe the ant hill that place is!!  But luckily after enough whining, my hosts thought it best to contact someone in Tokyo to meet me at the incoming train and lead me to the outgoing one!  These people---the Japanese---are so sweet and polite.  I've never seen anything else like it anywhere in the world. [\ldots]

About myself, I'm just plodding on---trying to act the part of a scientist/philosopher.  And trying not to worry too hard about the downfall of Lucent (and hence Bell Labs).  Philosophically, I've become an absolute junkie for the ``pragmatism'' movement.  There's hardly a day when I can be parted from my Willy.  William {\James} that is.  But also, John {\Dewey}, F.~C.~S. {\Schiller}, George Herbert {\Mead}, Richard {\Rorty} and so on.  Do you know any of these names?

It's sort of interesting how the first letter in my big samizdat---it's one to you---opens up by describing what a great respect I have for William {\James}' writing style.  But the truth of the matter is I really didn't start reading the man in depth until last August.  So, my knowledge of him is really very young.  Here's how I finished that letter to Mabuchi:
\bq\noindent
But that's just an aside (to tell you that the grass is always
greener).  Keep up the good work with all those good students.  Get
them to read William {\James}' {\sl Pragmatism\/} and tell them that quantum mechanics is a much better motivation for all that he said there \ldots\ but to never lose sight that the real goal is to get to where he wanted to go.
\eq
I'll tell you the same thing.  Read him; it's a philosophy for living, and writing the future.

Concerning the quantum world---the real world of my every waking moment---in the last six months, I think I've made the greatest progress I've made in years.  Go have a look at my paper:
\quantph{0106166}.
That was just the beginning of the transition in me.  Since then, I really believe I've climbed a peak that no one else has even been able to see from the ground.  Presently that insight is captured in about a 120 page document of emails (much like the bigger samizdat I published on the web), but I'm hoping to distill it into a real paper within the next month or so.  (As you can already tell from the size of this letter, I'm hoping to use this Japan trip to get a whole lot of writing done.)  [\ldots]

I haven't been to New York City nearly as much as I would like to; but I keep hoping for the day when I'll be able to increase the frequency of my visits.  The last time I was there (about a month ago), I had a pretty amazing experience.  I went to see this philosopher friend of mine who was in town to visit his sister.  He sent me her address and suggested that we meet at her apartment and then go to lunch.  The address was essentially the corner of Broadway and Bleeker Street, right in the middle of the Village.  I thought, ``Wow, that's in the thick middle of things.  The apartment has got to be an absolute shoe box; how else would an academic afford it?''  So I got quite excited about going there:  I thought I'd be able to see first-hand what squalid lengths people will go to just to live in an intellectually stimulating environment.  Well, it was anything but squalid.  It turned out to be a penthouse apartment on the top of the building.  Three levels of its own!  In square feet, it was essentially the same size as my house in Morristown!  Many millions of dollars worth of Manhattan space.  Here's the rub:  Arkady's sister was not an academic as I had assumed.  She's an investment banker, and her husband is an architect.  Arkady's girlfriend told me, ``You ought to see her place in the middle of Paris; it's ten times this big.''

It was funny:  One of the windows was placed explicitly so as to ``frame'' the World Trade Center.  It's kind of an empty view now.  I visited the old WTC site, by the way, sometime in early December.  It was quite a stirring experience.  The whole affair certainly caused me to think much harder about our finitude.  And it caused me to think much harder about what we all, each of us, might hope to do in this life.  Hence also the extracurricular reading of James.  (But I'm not quite sure which came first, the chicken or the egg---i.e., my seeking the books or the books seeking me.)

\section{17-02-02 \ \ {\it The Process} \ \ (to C. H. {\Bennett})} \label{Bennett10}

\bcb {\bf [Thoughts on Wizzy]}
Sorry to hear about the big change in your lives, lives
that would have been unimaginably different without
Wizzy.  My favorite metaphor for death is being
dropped into a black hole.  The main worthwhile
thing left behind is not your physical remains, nor
even their information content (which presumably
reemerges as Hawking radiation) but rather the
relative state you leave behind in the Church
of the Larger Hilbert Space.
If I were you, though, I wouldn't beat about the bush with Emma, but
just tell her that Wizzy died.  She's plenty smart, and probably knows
about death already, so euphemizing will not make it any less painful.
\ecb

I'm sorry; I didn't mean to give a misimpression to you.  Emma's asking ``Where did he go?''\ came a while after we had already discussed that Wizzy died (literal phrase).  That was exactly how I introduced the subject in fact:  ``I have some bad news.  Wizzy died last night.''  (All this happened, by the way, as I was getting her dressed so that she and Kiki could take me to the airport at 6:30 yesterday morning.  I'm in Sendai, as I write to you.)

I knew that she was a little familiar with the concept since she had seen a Babar movie where Babar's mother gets shot.  So she understood what was going on a little, but she didn't have all the aspects of it down yet.  She started by saying, ``I don't want Wizzy to die.  Why did he die?''  I said, ``Because he was old and his time had come.''  She said, ``I don't want him to be old; I want him to be new.''  I explained that we had buried him in the back yard, and that she could go to the spot and talk to him---``pretend talk''---whenever she wanted.  It was when I finally took her outside to show her the grave that she asked ``Where did he go?''  She was moving into a little more contemplative mode by then, so I thought I'd run with it.

Kiki said that after they returned from the airport, Emma wanted to go to the grave and talk.  When they got there she asked, ``When's he going to come out, Mom?''

\bcb
My favorite metaphor for death is being dropped into a black hole.
The main worthwhile thing left behind is not your physical remains,
nor even their information content (which presumably reemerges as
Hawking radiation) but rather the relative state you leave behind in
the Church of the Larger Hilbert Space.
\ecb

I don't suppose it's ever struck you what an excessive sort of
universe that would be. In a way, everything appears twice over. Once
in the state and once in the relative state.  (If you accept the
existence of the universal wave function, the one determines the
other uniquely.) What so moved God that he should make two copies of
everything? (Redundancy for the purpose of error correction won't do
as a reply!)

\section{18-02-02 \ \ {\it Psychology 101} \ \ (to J. Preskill)} \label{Preskill6}

Let me reply to some of your points in a way that doesn't reflect
their original order.

\bjp
In the past I have sensed that you and I differ in how we regard
ourselves. I believe that I am just another physical system governed
by the same fundamental laws as any other system. You seem to think
there is a fundamental distinction between yourself and the system
you are observing. To me the Everett view is appealing because it
turns away from this egocentrism.
\ejp

It's funny, but when I read this, my reaction went in two rather
peculiar directions.  First I thought, ``I wonder if, in the end, the
only thing the great quantum foundations struggles will leave behind
is a few psychological observations?  If so, what a shame.''  But
secondly, I imagined Galileo hoisting me up to the top of the Leaning
Tower of Pisa and dropping me off it along with his two famous
stones.  Even though I cursed and screamed the whole way down, I went
``splat'' at the same time that they went ``thud.''

Here's the psychological thought in a little more detail.  One of the
things that bugs me about the Everett view is what {\it I\/} consider
{\it its\/}  extreme egocentrism!  Now, how can that be---both of us
accusing the other's view as {\it the\/} egocentric view? I'll tell
you what I think, trying to express the problem from both sides of
the fence.

My side gets to go first.  What I find egocentric about the Everett
point of view is the way it purports to be a means for us little
finite beings to get outside the universe and imagine what it is
doing as a whole.  And what is it doing as a whole?  Something
fantastic?  Something almost undreamable?!  Something inexpressible
in the words of man?!?!  Nope.  It's conforming to a scheme some guy
dreamed up in the 1950s.

This whole fantastic universe can be boiled down to something
representable within one of its most insignificant components---the
brain of man.  Even toying with that idea, strikes me as an
egocentrism beyond belief.  The universe makes use of no principle
that cannot already be stuffed into the head of an average PhD in
physics?  The chain of logic that leads to the truth of the
four-color theorem (apparently) can't be stuffed into our heads, but
the ultimate operating principle for all that ``is'' and ``can be''
can?

It's a funny thing:  I don't think I've met anyone who would imagine
that mathematics will ever come to an end.  Or even that it {\it
can\/}  come to an end.  There'll always be new axiom sets to play
with, new formal structures to write down.  But with physics it's a
completely different story.  People are always wanting to say, ``Well
we've finally gotten there.''  Or, ``Even though we're not there,
we're pretty damned close.''  It's OK, even condoned, to have Dreams
of a Final Theory.  From this point of view, all the mathematics yet
to come is worthless as far as the essence of the universe goes; the
wad was already shot.

You get the point.  It's a psychological one, but it's one that I
find overwhelmingly powerful.  It is that anytime any of us ever has
the chutzpah to say, ``Here's an ultimate statement about reality,''
or even a potentially ultimate one, what we're really doing is
painting the world in the image of man.  We're saying that the measly
concepts we've managed to develop up to this point in time fit the
world in a way that none of our previous concepts have, that none in
the future will ever do better, and, most importantly, we view this
not as a statement about ourselves and the situation set by our
present evolutionary and intellectual stage, but rather as a property
of the universe itself.

Now let me start moving toward the other side of the fence.  The
question someone like me---someone who has these kinds of blasphemous
thoughts---has to ask himself is, how can I ever hope to be a
scientist in spite of all this?  What can science and all the great
achievements it has given rise to in the last 400 years be about if
one chooses to suspend one's dreams of a final theory at the very
outset?  (Or, to tribute Johnny, how can one have law without law?)

I think the solution is in nothing other than holding
firmly---absolutely firmly---to the belief that we, the scientific
agents, are physical systems in essence and composition no different
than much of the rest of the world.  But if we do hold firmly to
that---in a way that I do not see the Everettistas holding to it---we
have to recognize that what we're doing in the game of science is
swimming in the thick middle of things.  We're swimming in this
undulant sea, and doing our best to keep our heads above the water:
All the concepts that arise in a physical theory must be interpreted
to do with points of view we can construct from {\it within\/} the
world.

That is to say, we have to loosen the idea that a physical law is a
mirror image of what ``is'' in the world, and replace it with
something that expresses instead how each of us can best cope with
and hope to take advantage of the world exterior to ourselves. This,
it seems to me, is something that by its very definition can be
stuffed into the human brain.  The current state of science is our
presently best known means for survival.  A scientific theory indeed,
from this point of view, is yet another expression of Darwinian
principles. Scientific theories evolve and survive because the
survivors have a kind of staying power that none of the rest of the
competition have.  Not because they are part of the blueprint of the
universe.

The situation of quantum mechanics---I become ever more
convinced---illustrates this immersion of the scientific agent in the
world more clearly than any physical theory contemplated to date.
That is because it tells you you have to strain really hard and strip
away most of the theory's operational content, most of its workaday
usefulness, to make sense of it as a reflection of ``what is''
(independent of the agent) and---importantly---you insist on doing
that for all the terms in the theory.

I know you're going to find the last sentence debatable, but that is
what I see as the danger in the Everett point of view:  You are
able---or at least purportedly so---to view the universal state as a
reflection of something, but at the cost of deleting all the concrete
things it was meant to reflect in the first place.  What I mean by
this is, if we take any concrete situation in quantum mechanics---a
system, a measuring device, and some kind of model for the beginning
stages of a measurement---we can indeed construct a
Church-of-the-Larger-Hilbert space description of it.  I'll grant you
that.  But try to go the other way around without any foreknowledge
of the ``measurement'':  Start with the Church, and try to derive
from it that a concrete measurement has taken place, and you
encounter an embarrassment of riches.  You don't know how to identify
the valid worlds, etc., etc.  (And, if you ask me, invoking
decoherence as a cure-all is little more than a statement of faith
that some guy from Los Alamos has the all the answers to all the
tough questions the rest of us are too lazy to work out.)

So, I myself am left with a view of quantum mechanics for which the
main terms in the theory---the quantum states---express nothing more
than the gambling commitments I'm willing to make at any moment. When
I encounter various other pieces of the world, if I am
rational---that is to say, Darwinian-optimal---I should use the
stimulations those pieces give me to reevaluate my commitments. This
is what quantum state change is about.  The REALITY of the world I am
dealing with is captured by two things in the present picture:  1) I
posit systems with which I find myself having encounters, and 2) I am
not able to see in a deterministic fashion the stimulations (call
them measurement outcomes, if you like) those systems will give
me---something comes into me from the outside that takes me by
surprise.

OK, now let me put myself squarely in your pasture.  You worry that
having those main terms in the theory refer to {\it my\/} (or {\it
your}, or Joe Buck's) gambling commitments, is committing a kind of
egocentrism. What respectable theory would refer to my particular
vices, my desires, my bank account in making its most important
statements?

This is going to surprise you now, but I agree with you
wholeheartedly.  Even enthusiastically so.  Where I seem to disagree
is that I do not find this a good reason to promote those vices,
those commitments to an unearthly realm and call them ``states of the
universe'' (or relative states therein).  Instead, it seems to me to
be a call to recognize them for what they are and to redouble our
efforts for getting at the real nub of the matter.

Let me try to give you a way of thinking about this that you might
respect.  What was Einstein's greatest achievement in getting at
general relativity?  For the purposes of the present exposition, I
would say it was in his recognizing that the ``gravitational field''
one feels in an accelerating elevator is just a coordinate
effect---it is something that is induced purely with respect to the
description of an observer.  In this light, the program of trying to
develop general relativity thus boiled down to trying to recognize
all the things within gravitational and motional phenomena that
should be viewed as consequences of our coordinate choices.  Or to
use a phrase I've come to like, it was in identifying all the things
that can be viewed as ``numerically additional'' to the observer-free
situation which come about purely by bringing the observer
(scientific agent, coordinate system, etc.) onto the scene.

Now the point is, that was a really useful process.  For in weeding
out all the things that can be viewed as ``merely'' coordinate
effects, the fruit left behind could be seen in a clear view for the
first time:  It was the Riemannian manifold that we call spacetime.

What I dream for in my foundational program for quantum mechanics is
something just about like that.  Weed out all the terms that have to
do with gambling commitments (I used to call it information,
knowledge, or belief), and what is left behind will play a role much
like Einstein's manifold.

This much of the program, I hope and suspect you will understand even
if you are not sympathetic to it.  But, I don't know, you might be
sympathetic to it.  (Especially if I've done a good job above.)
However, it is also true that you have rightly suspected some
tendencies in me that go further.  In particular, in opposition to
the picture of general relativity, where reintroducing the coordinate
system---i.e., reintroducing the observer---changes nothing about the
manifold (it only tells us what kind of sensations the observer will
pick up), I do not suspect the same of the quantum world.  This is
why I recommend to all my friends that they read William {\James}'s
little article ``The Sentiment of Rationality.''  It sort of sets the
right mindset, even though it has nothing to do with quantum
mechanics (other than in the efficacy of taking gambles) and goes
much further on religion than I myself would go.

Anyway, here I suspect that reintroducing the observer will be more
like introducing matter into pure spacetime, rather than simply
gridding it off with a coordinate system.  ``Matter tells spacetime
how to curve {\it when it is there}, and spacetime tells matter how
to move {\it when it is there}.''  Observers, scientific agents, a
necessary part of reality?  No.  But do they tend to change things
once they are on the scene.  Yes.  Or at least that's the idea.

Does that mean that the scientific agent is something outside of
physical law?  Well, to give this an answer, you've got to go back
and be very careful to use the picture of ``physical law'' that I
built up at the beginning of the essay.  What we are ``governed'' by,
God only knows.  He's the one, if anyone, who sits outside the
physical universe and has a chance to look back at it whenever he
pleases. Our task is to build up as good and solid a set of beliefs
as we can from within it.  In that way, we increase our survival
power, and use our spare time to try to bring forth a few progeny of
our own. (I used the word ``governed,'' by the way, because you had
used it above.)

If Galileo had dropped me from the tower, I feel pretty confident
that I would have gone splat.

Aye yi yi, I wrote a lot.  That's the dangers of being jetlagged in a
foreign country without one's wife and kids.  (I'm in Sendai visiting
Ozawa.)

I'm going to have to reply to the other points of your note later.

\section{18-02-02 \ \ {\it Still Thinking} \ \ (to W. K. Wootters)} \label{Wootters5}

I've been meaning to write you for some time about your inside/outside distinction for entangled systems, i.e., the stuff you wrote in your last philosophical note to me.  Well, I'm still not quite able to say what I want to say, but let me set the stage by forwarding you the note I just wrote to John Preskill this morning.  [See 18-02-02 note ``\myref{Preskill6}{Psychology 101}'' to J. Preskill.]  It's probably just my jetlag, but I felt like I was being particularly clear in it.  The note is almost completely self-contained and refers to John's embrace and my disembrace of the Everett stance.  A good part of it was greatly inspired by two discussions I had with you (one in Princeton and one in {\Montreal}).  Maybe it'll show you the monster you created!

I hope to get my reply on your inside/outside idea off before this Sendai visit is over.

\subsection{Bill's Preply}

\bq
You apologized for taking 20 days to respond to my note, and now I've taken a couple of months to respond to yours!  Sorry about that!

You quoted William James on the temperaments that underlie apparently
dispassionate philosophical arguments.  His observations are certainly
correct.  Maybe one reason I have not written up my little graph model is that it doesn't fully jibe with all my underlying prejudices.  It captures one idea---that the universe we experience gets its form partly from being a {\it collective\/} experience---but in other respects I'm not satisfied with it.

This semester I've been participating in a seminar at Williams on science and religion.  It's been very interesting, and it has forced me to articulate, as least to some extent, my interpretation of quantum
mechanics.  I certainly don't have a fully worked out interpretation, but I do seem to hold firmly to two principles: (i) there is no physical breakdown of unitarity, and (ii) mind is crucial to the establishment of definite ``facts'' (such as a definite pointer reading, as opposed to endless entanglements and superpositions).  In these respects I find myself in line with most versions of the many worlds interpretation, in which definite ``facts'' occur in one's subjective experience, not in the true reality.  (Of course there are facts about the wavefunction of the universe, but that's not what I mean by ``facts'' in quotation marks.)  But I deviate from the many worlds interpretation in that I would like to attach a greater ontological status to the world of our experience and a lesser ontological status to the wavefunction of the universe.  That is, I want to say that the world of experience is a closer approximation to {\it reality\/} than is the wavefunction of the universe.  (The latter would be more like a mathematical device, something that defines the range of possibilities.)  This is pretty clearly a metaphysical distinction that cannot be tested by any conceivable experiment.  But it makes a difference for the questions that arise in the science/religion dialogue.  For example, I am led to say that mind is a crucial component of the structure of reality; and this makes it easier to say that God has something to do with the world.  If the
ultimate reality is the wavefunction of the universe, then the ultimate reality seems much more mechanical and less God-friendly, like the world of classical physics.

Or maybe there {\it is\/} a possibility of testing the above distinction, at least in principle.  (It depends on how I finish my interpretation.)  Let's consider David Deutsch's thought experiment in which a sentient computer measures, in a reversible way, some property that doesn't have a definite value for the object being measured.  Later, the computer will report that he observed a definite outcome but did not record which one, and the various possible paths will have been brought together successfully to exhibit interference.  That's David's prediction.  I would agree that interference should be possible, but I may disagree about what the computer will report.  If it is true that one needs to have a {\it shared\/} world in order to have a sensible world, then when the computer was all alone in his private entanglement with the object he measured, who knows what he experienced?  Maybe he experienced nothing more about that measured object than what one electron in a singlet pair experiences about the other electron.  Maybe he cannot honestly report later that he saw one outcome or another.

But I have to admit that this idea of ``sharing'' a world with others is very spooky.  I wonder if, by pushing the idea to its logical conclusion, I will have to embrace the possibility of telepathy.  I don't automatically reject this possibility, but I think it's time to change the subject!

What I'm actually working on these days is the subject of entangled rings.  I'm looking for a possible analogy between the {\it entanglement\/} among particles arranged in a lattice and the {\it action\/} associated with a lattice gauge field, but so far my lattices are just one dimensional (rings) and I have very little to offer in the way of an analogy.  Nevertheless, I'm in the process of writing a short paper on the subject to be posted on {\tt quant-ph} in the next week or two; so you'll see it there.

I hope your work is going well.
\eq

\section{19-02-02 \ \ {\it Re-Tackle} \ \ (to L. Hardy)} \label{Hardy4}

I'm in Japan for a couple of weeks at the moment, and I'm finally
getting some time put in the 16-hour days again.  (Like I used to in
the good old days.)

Anyway, I thought I'd tell you, though it is long overdue, I am
finally tackling your 5-Axiom paper again.  I'm starting to
appreciate it much more for sure.  If you just weren't so damned
non-Bayesian!!!  There's a lot of good stuff in it.  My main
difficulty at the moment is that you have a couple of moves that I
know I don't want to allow into my porn:  1) taking mixtures of
states (i.e., allowing probabilities of probabilities) as a
fundamental step, and 2) invoking extensions to the Church of the
LHS.

But I definitely think you are on the right track.  And it's probably
just a matter of my searching harder for some Bayesian ways of
looking at what you've already done.  (I'm doing that as we speak.)
The most essential things that strike me are 1) the move to column
vectors and thinking of measurement as a decomposition of the state,
and 2) invoking a relation between $K$ and $N$.  I think those
ingredients are definitely here to stay in my mindset.  Also I'm
warming up to the continuity axiom.  I'll try to write you the
reasons why soon.  (But I've made promises before.)

Anyway, I'm super-looking forward to your stay.

\section{19-02-02 \ \ {\it Where to Stop?}\ \ \ (to J. Preskill)} \label{Preskill7}

You know I've got a million ways of saying why I don't like the
Everett interpretation---none of which you find very convincing---but
here's a new thought that dawned on me as I was writing my last note
to you, and I wonder what you think.  Let me try it out on you.

Everett says, ``You know Chris, all these silly things you do like
leaving measurement as an undefined primitive, etc., will disappear
and find a more satisfactory solution if you'll just lay back, relax,
and recognize that the quantum states you're working with are really
just relative states \ldots\ ones derived from the universal
wavefunction under one or another decomposition.''

I say, ``Aha, OK.  Then what is this wavefunction of the universe?''

He says, ``Well for that, we ought to consult the Hartle--Hawking
paper.  Here it is:  It's $|\psi\rangle$.''

Then it dawns on me.  How do I know that that state they wrote down
isn't just the relative state of our universe with respect to some
super-universe?  And how do I know that that state is not itself some
relative state with respect to some super-super-universe?  And so on
ad infinitum.

The point is, what principle of science tells you where to stop?
None, I'd guess.  Is that troubling?  I don't know.  But it seems a
little fishy to me.

Everett tells me, ``You've just got to recognize that the
wavefunctions you use on a daily basis simply don't have the same
ontological status as my universal wavefunction.  You might call them
`states of knowledge' in a way.  But my universal wavefunction, now
that's the real thing; it's here independently of every man, woman,
and child.''  But then I ask, ``Well, why does yours get that exalted
position?  I claim that it itself is a relative state and you can't
prove me wrong.''

Like I say, I don't know what I think of this yet, but it does strike
me as fishy.  Once you get into the game of building a Church of the
Larger Hilbert Space, who tells you how many pews to put there?
That's not something it seems to me you can ever discern from within
the universe.  It's an article of religion, it seems to me, much like
the imagery the appellation seems to provoke.

\section{19-02-02 \ \ {\it One More Before Lunch} \ \ (to J. Preskill)} \label{Preskill8}

Here goes.  Let's see if I can be brief enough to finish in time. No
easy task for me!

\bjp
Still, I'm flattered by your persistence. Or are you (as I can't help
but suspect) slyly recruiting an Everettite who will make a weak
case?
\ejp

No, I was honest in all that I wrote in the flattering note.

\ldots\ Damn!  Didn't make it.  It's after lunch now.

What I was going to say was that I was absolutely honest in why I
want you to write the sort of paper that you might for the special
issue.  The point being that if a physicist really, really does find
Everett completely adequate to his needs \ldots\ and can argue that
it's not a superfluous addition to what he's actually doing when he's
doing a calculation, then that would be an interesting datum for the
freaky types like me who see it as an ugly picture of the world.

I wasn't slyly recruiting you to make a weak case.  But, of course, I
actually did have an ulterior motive---something much bigger in my
mind than the needs of the special issue---and now that you've forced
my hand, I ought to be up front about it.  I was banking that if you
really did put your heart into making a convincing case for the
Everettistas---i.e., the sort of thing that having to write a paper
on the subject might draw out of you---then your intellectual honesty
would cause you to see how much of the point of view really hasn't
been worked out (yet?\ or maybe ever will be?).  I.e., that they have
no convincing/relevant argument for the probability rule, that they
seem to require a preferred basis, that they seem to require a
preferred tensor product structure, that to make sense of two
systems, they need to invoke a third, and so on.  And when you
started to add all those things up, you would also realize that the
Everett picture really wasn't much help after all in getting you to
the point understanding what measurement is.  That's how I was being
sly.

\section{19-02-02 \ \ {\it More Psychology 101} \ \ (to J. Preskill)} \label{Preskill9}

\bjp
Sure, scientists are arrogant. That our puny brains can grasp
anything about how the world works is a miracle, and I can't pretend
to be able to explain it. But I believe it is so.
\ejp

It's not the claims of ``anything'' that worry me so much.  It's the
claims of ``everything.''  I.e., that our puny brains can grasp
everything (in the sense of an ``ultimate physics'') is the thing
that seems implausible to me.  And if we can't have that, then
we---or more realistically, those who are inclined to do so---ought
to be asking what it is we're shooting for.

Sorry I hit a nerve.

\section{19-02-02 \ \ {\it Sendai Morning} \ \ (to N. D. {\Mermin})} \label{Mermin61}

Thanks for helping pick me back up from my preskillsplat.  (That
could almost be a real German word.)

I'm enjoying Sendai and Ozawa's company greatly.  And I'm once again
thinking harder about complete positivity.  The fact that the
trace-preserving completely positive maps are isomorphic to the
density operators on a larger space has got to be a truly deep point
(in my quest to shore up my argument that the time evolution map is
itself a subjective belief).  But I just can't quite figure out how
to put that into a convincing physical context.

Also, by the way, I'm going through Hardy's five axioms again.  It's
making a much bigger impression on me this time around.  It's got a
lot of good stuff in it.  If it just weren't so damned non-Bayesian!
The point is, I think it's got a hell of a lot of cleaning up to be
done on it, but it really does have potential.

\section{20-02-02 \ \ {\it Out Loud} \ \ (to W. K. Wootters)} \label{Wootters6}

Thanks for thinking out loud.  I'll just respond to one point.

\bbw
Let me think out loud for a minute here about your note to John.  I can
think of a {\bf pragmatic} reason for being an Everettista.  (At
those times when I am particularly attracted to the Everett view,
this is what attracts me.)  Even if we can't hope to {\bf know}
reality, if we can {\bf guess} a model of reality, this guessing
helps science progress.  What Everett gives us is a guess at ultimate
reality.  So let's guess that Everett is right, and then work to
falsify this guess.
\ebw

It's about the guessing part.  I had meant to cover that case with
the word ``potentially'' in this paragraph:
\bq
You get the point.  It's a psychological one, but it's one that I
find overwhelmingly powerful.  It is that anytime any of us ever has
the chutzpah to say, ``Here's an ultimate statement about reality,''
or even a potentially ultimate one, what we're really doing is
painting the world in the image of man.  We're saying that the measly
concepts we've managed to develop up to this point in time fit the
world in a way that none of our previous concepts have, that none in
the future will ever do better, and, most importantly, we view this
not as a statement about ourselves and the situation set by our
present evolutionary and intellectual stage, but rather as a property
of the universe itself.
\eq

That's actually the point you first inspired in me with your aphorism
about the dog.  There are some things a dog will never understand;
there are even questions he can never understand.  Why should we
expect the evolutionary chain to stop with us?  In a way, this
cluster of thoughts that I'm starting to think is a rather strong
kind of anti-Church--Turing thesis.  That is, I think we've gotten
into the habit---and Deutsch tried to codify it in his 1985
paper---of thinking that the Church--Turing thesis implies that once
you've got a universal machine (people like to say the human brain is
one), then you've reached the end of the line.  But one should not
forget that what Turing was up to in his 1936 paper was to formalize
the notion of what is ``humanly computable.''  This was a point
brought home to me by Chris {\Timpson}'s excellent undergraduate thesis
\begin{center}
\myurl[http://web.archive.org/web/20010307183244/http://users.ox.ac.uk/$\sim$quee0776/thesis.html]{http://web.archive.org/web/20010307183244/http:// users.ox.ac.uk/$\sim$quee0776/thesis.html}.
\end{center}

By the way, not that it matters too much, but I refined one of my
paragraphs in the note before I archived it away:
\bq
The situation of quantum mechanics---I become ever more
convinced---illustrates this immersion of the scientific agent in the
world more clearly than any physical theory contemplated to date.
That is because it tells you you have to strain really hard and strip
away most of the theory's operational content, most of its workaday
usefulness, to make sense of it as a reflection of ``what is''
(independent of the agent) and---importantly---you insist on doing
that for all the terms in the theory.
\eq

And that---by the way again---may have been a point also inspired by
someone else, namely {\Schroedinger}.  Though I haven't been able to
completely track its origin in my mind.  Somewhere---maybe his 1935
paper in {\Wheeler} and Zurek---he says something like, ``understanding
quantum mechanics may not require the addition of more variables, but
rather taking some of them away.''

\section{20-02-02 \ \ {\it WQRST Revision with Figure} \ \ (to C. H. {\Bennett})} \label{Bennett11}

Thanks for the new draft.  Apparently we're both in Asia.  I'm in Japan right now.  I'm visiting Ozawa.  (By the way, I think in the past you've confused him with Ohya.  It was Ohya who said the wacky things about quantum computers solving NP-complete problems \ldots\ or at least that's what Ozawa says.  Ozawa is quite a good guy and not so silly \ldots\ and he likes to laugh, so you'd probably like him.)

BTW, you never answered my last two questions:
\begin{enumerate}
\item
Are you going to that Wheeler thing?
\item
Can I count on you for a Maxwell-Demon and/or I-Love-Everett article?
\end{enumerate}
Wish you were here, instead of there.

\section{21-02-02 \ \ {\it Quantum Fest, Bell Labs Style?}\ \ \ (to H. Mabuchi)} \label{Mabuchi3}

I just read your proposal for the Wheeler book.  Sounds good.  (And by the way, you smooch \ldots\ adding that flattering remark.)

I've been having a good time with Ozawa.  He's got some awfully good stuff, though he's not Bayesian enough for my taste.  I suggested to him that he ought to visit Caltech.  (And I'll suggest to you that you ought to invite him.  He'll be in America in the July--August time frame, with a little time free between QCMC and the Feynman Fest.)  I told him about your experimental interest in going beyond the SQL.  But also he's got some interesting new stuff that even goes beyond the uncertainty principle!  (Well \ldots\ that is, if you mean by that roughly the thing Heisenberg meant with his microscope model of attempting to infer position and momentum simultaneously.)  Anyway, I think it could make an awfully good splash for you, and he tells me he's in the  process of distilling a potentially relevant / Q-optics realizable model that would do the trick.  I've promised to help him work a bit on the style of the paper, so he can send it to {\sl Nature}.  (I always get myself into trouble with my promises.)

BTW, Ozawa is contemplating visiting Bell Labs in March too.  So I might have a bang-bang-bang quantum fest:  Hirota comes on March 12, Hardy will come for maybe two days before the Wheeler meeting, though the dates aren't settled yet.  And you'll give us a talk on March 18.  When were you thinking of coming into town?  Would you be staying the night of March 17, or the night of March 18, or both, or neither.  If Ozawa does come, he said he might try to line up with you.

On another subject, let me send you a little essay of my own.  [See 18-02-02 note ``\myref{Preskill6}{Psychology 101}'' to J. Preskill.]  I got carried away with replying to one of John Preskill's comments in a note to me the other day, and the end result was something I was particularly proud of as a statement of my quantum foundational ideas (broad sweep).  John, of course, hated it and almost reacted violently.  But Mermin and Wootters loved it.  So, I'm probably doing something right.

\section{22-02-02 \ \ {\it Sendai Thoughts} \ \ (to H. J. Folse)} \label{Folse7}

Greetings from Sendai, Japan!  I'm here for a small time of about 11 days, visiting Masanao Ozawa.  (He's done some of the most interesting technical work on the uncertainty relations that's been done in a long, long time.)

Anyway, I just thought I would tell you I was thinking of you.  Just a few moments ago at the university bookstore, I bought two volumes of Bohr's philosophical writings, translated into Japanese.  The total came up to maybe about \$11, so it seemed like a good investment.  I think I will write a little something in each and give them to my daughters.

By the way, while I'm here, let me remind you again to send me that article of yours I still haven't gotten:
H. J. Folse, ``The Formal Objectivity of Quantum Mechanical Systems,'' Dialectica {\bf 29}, 127--136 (1975).

Also, if you could, I'd like to get your {\Vaxjo} paper.

Let me also send you a little food for thought.  It's a little essay I wrote in response to a remark John Preskill wrote me the other day. [See 18-02-02 note ``\myref{Preskill6}{Psychology 101}'' to J. Preskill.]  I'm a little proud of it.  I think it's maybe one of my best expressions yet of where I'd like to see quantum foundations studies go.

\section{22-02-02 \ \ {\it Getting the Mindset} \ \ (to P. F. Scudo)} \label{Scudo2}

You asked for some materials to help you get more familiar with the
problem I'd like you to work on.

OK, I'm ready to send you some now.  And I'll inundate you, but don't
let that frighten you.  Only try to understand things to the extent
that you've got some free time.  (And I well expect you may have none
at all!)

The first thing to do is read the Brun, Finkelstein, {\Mermin} paper,
``How Much State Assignments Can Differ,'' \quantph{0109041}.
This is really the thing that started my thoughts off in the present
direction, for I completely disagree with them.  Their statement is
in ultimate conflict with the Bayesian idea of what a quantum state
can be.  So, understand their argument.

After that, start reading my new samizdat (i.e., underground
publication), which I will send you in a separate email.  It's quite
large ($150+$ pages), with plenty of repetition, but it mounts an
attack on B, F, and M from just about every direction conceivable.
Also there is the fact that as time went on, all the issues became
clearer and clearer with me, and so I found crisper and crisper ways
of expressing myself.  Still reading it (and reading it carefully)
might help you get in the right mindset for any number of problems we
might be discussing.

The upshot of much of the samizdat is that, for consistency in one's
Bayesianism, one must accept that the {\it assignment\/} of a POVM
(living on a piece of paper) to a measuring device (living in a
laboratory) is a subjective judgment at exactly the same level of
subjectivity as the quantum-state assignment.  Thus one is presented
with a cross-roads.  Either one accepts pure Bayesianism and gives up
the idea that POVMs and quantum time evolutions have objective
ascriptions (i.e., gives up the idea that they are independent of the
agent assigning them), or one continues with the belief of objective
POVMs and time evolutions and adjusts oneself to the idea that
probability has to be objective too.

The direction I personally take is that probabilities are subjective,
always.  They are never objective.  Therefore one must make sense of
what one means when one speaks of an ``unknown POVM'' or an ``unknown
quantum operation.''  This is where a new kind of de Finetti
representation must come in \ldots\ and hence your summer work.

But the most important thing for the present is understanding all the
motivation leading up to that point.  When you are ready to see a
sketch of how the theorem ought to go, look at the note of 19
November 2001 to {\Caves} and {\Schack} titled, ``\myref{Caves38}{A Lot of the Same}.''

For completeness sake, I will also send you a draft of the paper that
{\Caves}, {\Schack} and I are presently constructing to make some of this
official.  The part that is maybe the most relevant for your
education is the appendix on Dutch-book arguments.  You might try to
understand that.  And I'll write you more about that later.

\section{23-02-02 \ \ {\it Look Into My Thousand Eyes} \ \ (to G. L. Comer)} \label{Comer7}

I'm sitting in bed writing email early Sunday morning and cursing jetlag.  I woke up at 5:00 (again!), which means I'm keeping a pretty steady pace of between 5 and 6 hours of sleep each night.

It's been a good trip for me.  This is my fifth time here.  Join the Quantum Information Corps and see the world.  Semper Fidelis.  One of these days I'll have to count the number of times I've been to Europe, but I think it's hovering around 25 times at the moment.  And just yesterday, I got an invitation to go to India next January; this is a first for me.  Knowing me, I won't be able to turn it down:
\bv
This is living, this is style, this is elegance by the mile. \\
O the posh posh traveling life, the traveling life for me. \\
First cabin and captain's table, regal company. \\
Whenever I'm bored I travel abroad but ever so properly. \\
Port out, starboard home, posh with a capital P.
\ev
(Do you remember {\sl Chitty Chitty Bang Bang}?)

I've been using my time in Japan to write up a paper that's long overdue, but I've also been calculating a little and doing a {\it lot\/} of philosophizing.  (All the philosophizing had better pay off in the end with some kind of tangible, physical result, or I don't know what the hell Bell Labs is going to do with me!)

So, are you going to interpret your newest poem for me?  Or are you going to leave me guessing?  This one stumped me, I'll say.

Yesterday, by the way, I ran across this surreal scene just outside a Buddhist temple.  It was a cliff where, over the course of the years, the monks---of some religion or other, not Buddhist I think---carved these (mostly cubical) caves into it.  Into solid rock in fact!  Some of the caves were really very deep, like small homes.  Here and there in the walls of them, there would be an image of some god carved into it, as a statue sitting on a windowsill.  One predominant image was of a goddess with 11 faces, a thousand hands, and a thousand eyes.  Have you ever seen this one (or something like it) in your jaunts through the world of mythology?  Do you know what it symbolizes?  (BTW, I never could discern from the statues where the eyes actually fell on her body \ldots\ so that only left me more intrigued.)

OK, I've got to run.  Every morning of this trip, I've had a ritual of taking a long, long, hot bath while either a1) reading a paper by Lucien Hardy on a new axiom system for quantum mechanics, a2) reading a book by Richard Rorty, {\sl The Consequences of Pragmatism}, or b) staring at the wall, trying to make Hardy's axioms more amenable to a Bayesian way of looking at things.  Today---it being Sunday---the option will be a2 for sure.

\section{24-02-02 \ \ {\it Consistency} \ \ (to J. W. Nicholson \& K. M. Fuchs)} \label{Nicholson7} \label{KikiFuchs}

Consistent coffee quality the world over!  I'm just writing to report to both of you that I saw a Starbucks about an hour ago and \ldots\ despite my significant need for coffee and my overbearing homesickness for all things American, I want you to know I resisted it.  More than that, I want you to know I shunned it!  I'm sitting in a Japanese coffee shop as I write to you---itself a Starbucks wannabee, but not a Starbucks---drinking the most godawful mud-thick coffee, BUT I SHUNNED IT!!!  And in the grand scheme, I know I'll come out the better for this. \medskip

\noindent A big hello from breezy cold Sendai,

\section{25-02-02 \ \ {\it A Wonderful Life} \ \ (to W. K. Wootters)} \label{Wootters7}

Thanks for the two notes, and wow, thanks for reading the {\James}
essay.  Your questions were anything but naive.  In fact, they were
much needed.  In trying to answer them, I think I significantly
clarified---to myself even!---what I'm hoping to get at.  Besides, I
certainly don't have a final stand yet; the whole point of view is in
the process of formation and questions like yours really help.

I'll do my best to reply to your questions below, and in the process
I think I'll finally compose what I've been wanting to say about your
``private-world-within-entanglement'' musings.  At the end of the
note, I'll list some of the open questions on my mind.  (These are
likely to be the naive ones!)

\bbw
Of course I'm very sympathetic to the perspective you express in this
paragraph \ldots\ but couldn't one still argue that as a matter of
methodology, the tactic of pretending that we can know the whole
story has served science well?  We make up a model of the world, and
this model gives us something to shoot at.  We hang on to the model
until we have found an explicit flaw in it (other than the flaw of
hubris).  And then we move on to a new model.

I find this an interesting question.  On the one hand, I think this
strategy does work well in advancing science.  On the other hand,
scientists (and others) are much too prone to accept as true the
pragmatic lie that says we can fully understand the world.

Your note to John P. goes some way toward laying out an alternative
methodology.  You speak of science in Darwinian terms: the most
successful theories survive.  How then do we proceed as scientists? I
suppose the answer is that we still make up theories and test them,
but the theories are not tentative descriptions of the world. Rather,
theories are schemes for making predictions.  But you obviously also
want to say that our theories tell us {\bf something\/} about
reality, even if they are not descriptions of reality. Moreover, our
theories will tell us more about reality if we identify and remove
from them those aspects that are subjective.  So your view of science
is not entirely operational.  There is realism in the background.

Have I understood you correctly?
\ebw

Yes there is certainly a kind of realism working in the back of my
mind, if what you mean by ``realism'' is that one can imagine a world
which never gives rise to man or sentience of any kind.  This, from
my view, would be a world without science, for there would be no
scientific agents theorizing within it.  This is what I mean by
realism:  That man is not a priori the be-all and end-all of the
world.  (The qualification ``a priori'' is important and I'll come
back to it later.)

A quick consequence of this view is that I believe I eschew all forms
of idealism.  Instead, I would say all our evidence for the reality
of the world comes from without us, i.e., not from within us.  We do
not hold evidence for an independent world by holding some kind of
transcendental knowledge.  Nor do we hold it from the practical and
technological successes of our past and present conceptions of the
world's essence.  It is just the opposite.  We believe in a world
external to ourselves precisely because we find ourselves getting
unpredictable kicks (from the world) all the time.  If we could
predict everything to the final T as Laplace had wanted us to, it
seems to me, we might as well be living a dream.

To maybe put it in an overly poetic and not completely accurate way,
the reality of the world is not in what we capture with our theories,
but rather in all the stuff we don't.  To make this concrete, take
quantum mechanics and consider setting up all the equipment necessary
to prepare a system in a state $\Pi$ and to measure some noncommuting
observable $H$.  (In a sense, all that equipment is just an extension
of ourselves and not so very different in character from a prosthetic
hand.)  Which eigenstate of $H$ we will end up getting as our
outcome, we cannot say.  We can draw up some subjective probabilities
for the occurrence of the various possibilities, but that's as far as
we can go.  (Or at least that's what quantum mechanics tells us.)
Thus, I would say, in such a quantum measurement we touch the reality
of the world in the most essential of ways.

With that said, I now want to be very careful to distance this
conception of reality, from what I'm seeking in the foundation game
of quantum mechanics.  Here's the way I originally put it to John the
other day.  Let me repeat a good bit of it so that it's at the top of
your mind:

\bq
OK, now let me put myself squarely in your pasture.  You worry that
having those main terms in the theory refer to {\it my\/} (or {\it
your}, or Joe Buck's) gambling commitments, is committing a kind of
egocentrism. What respectable theory would refer to my particular
vices, my desires, my bank account in making its most important
statements?

This is going to surprise you now, but I agree with you
wholeheartedly.  Even enthusiastically so.  Where I seem to disagree
is that I do not find this a good reason to promote those vices,
those commitments to an unearthly realm and call them ``states of the
universe'' (or relative states therein).  Instead, it seems to me to
be a call to recognize them for what they are and to redouble our
efforts for getting at the real nub of the matter.  \ldots

What I dream for in my foundational program for quantum mechanics is
something just about like that.  Weed out all the terms that have to
do with gambling commitments (I used to call it information,
knowledge, or belief), and what is left behind will play a role much
like Einstein's manifold.

This much of the program, I hope and suspect you will understand even
if you are not sympathetic to it.  \ldots\  However, it is also true
that you have rightly suspected some tendencies in me that go
further.  In particular, in opposition to the picture of general
relativity, where reintroducing the coordinate system---i.e.,
reintroducing the observer---changes nothing about the manifold (it
only tells us what kind of sensations the observer will pick up), I
do not suspect the same of the quantum world.  \ldots

Anyway, here I suspect that reintroducing the observer will be more
like introducing matter into pure spacetime, rather than simply
gridding it off with a coordinate system.  ``Matter tells spacetime
how to curve {\it when it is there}, and spacetime tells matter how
to move {\it when it is there}.''  Observers, scientific agents, a
necessary part of reality?  No.  But do they tend to change things
once they are on the scene.  Yes.  Or at least that's the idea.
\eq

From some of my choices of words, I think you probably got the
impression that this thing---this structure within quantum
mechanics---that I'm hoping to find at the end of the day is meant to
be a model of ``reality.''  Or at least our ``current best guess'' of
what reality is.  But no, that's not really what I want.  And your
questions helped make that much clearer to me.  Remember, for me, the
mark of reality is its indescribability.

What I'm asking for instead is something like what one finds in the
old movie, {\sl It's a Wonderful Life}.  That is to say, in our
scientific theories, we codify some fraction of what we know about
manipulating the world and conditionally predicting the phenomena
about us.  However, suppose we wanted to get at a measure of our
place in the world.  How would we quantify it, or at least qualify
it?  That is, how might we ask how important our lives and agential
actions are with respect to the theory we ourselves laid out?

Our only tool, of course, is the theory; for it defines the frame for
optimal thinking (and imagination) at any given moment.  We can only
gauge our measure by deleting the free variable that is ourselves and
seeing what is left behind.  You surely remember what George Bailey
found when his guardian angel granted his wish in {\sl It's a
Wonderful Life}.  He found that his life mattered.  So too is what I
suspect we will find in quantum mechanics.

But all of that is the sort of thing I won't be able to say in a
conference presentation for quite some time.  It's the sort of thing
that we discussed once before, in the context of some {\James}ian quote.
It's the underground reason for the philosophy.

At the level of convincing our peers, let me put it to you this way.
Within quantum mechanics, there is an invariant piece which is common
to all of us by the very fact of our accepting the theory. That is
what we are in search of because in some sense---which need not
pertain to a realistic conception of a theory's correspondence to
nature---it is the core of the theory.  It is the single part that we
agree upon, even when we agree upon nothing else.  In the direction I
am seeking to explore, the quantum state is ``numerically
additional'' to that core.  (That is, the quantum state is a
compendium of Bayesian ``beliefs'' or ``gambling commitments'' and is
thus susceptible to the type of analysis {\James} gives in his
``Sentiment of Rationality.''  Our particular choice of a quantum
state is something extra that we carry into the world.)

I hope that clears up some of the mystery of my thoughts for you---it
did for me.  Given John's implicit acceptance of the idea that ``a
true theory is a mirror image of nature,'' I should not have said in
my note that I agreed with him ``wholeheartedly.''  I do not intend
for {\it any\/} part of the formal structure of quantum mechanics to
be a mirror image of nature (in the sense of a proposed final
theory).  However, I do not intend to give up the reality of our
world either.

From my point of view, the only ``true'' reality that creeps into
quantum mechanics is ``in the differential''---i.e., in the changes
we induce upon our (personal) quantum states for this and that due to
any stimuli we give to or take from the outside world.  That,
however, is a pretty amorphous thing as theoretical entities go.  It
is little more than what might have been called in older language,
the measurement ``click.''

There is a temptation to go further---to say that the POVM element
$E_b$ associated with a measurement outcome $b$ is itself an element
of reality.  But I think that has to be resisted at all costs. There
are several arguments one can use to show that the {\it ascription\/}
of a particular POVM to a measurement phenomenon is a subjective
judgment at the same level of subjectivity as the quantum state
itself.  (In fact the two go hand in hand, one cannot support the
subjectivity of the quantum state without also taking the
subjectivity of the POVM.)  Instead, one should view the
(theoretical) ascription of a POVM to an actual measurement device as
an attempt to set the significance and meaning of the ``click'' it
elicits.  Similarly for the Krausian quantum operation associated
with the measurement:  It describes the subjective judgment we use
for updating our quantum-state assignment in the light of the
``click.''  (If you want more details about these arguments, I can
forward you some of my old write-ups on the subject.)

So, you probably ask by now, ``What does that leave for the core of
the theory?  Aren't you throwing away absolutely everything?''  And
the answer is, ``No, I don't think so.''  Let me give you an example
of something which I think is left behind.  Recall my favorite
argument for why the quantum state cannot be an element of
reality---it's the Einstein argument I wrote about in Section 3 of my
NATO paper.  Once I posit a state for a bipartite system, even though
by my own admission my actions are purely local, a measurement on one
of the systems can toggle the quantum state of the other to a large
range of possibilities.  Thus, I say that the quantum state of the
far-away system cannot be more than my information or the compendium
of subjective judgments I'm willing to ascribe to that system.

Notice, however, that in positing the original state, I had to also
implicitly posit a tensor-product space for the bipartite system. Let
me ask you this:  Once this tensor-product space is set, is there any
way to toggle one of the factors from afar just as with the quantum
state?  As far as I can tell there is not.  Thus I would say that the
Hilbert space of the far-away system is a candidate for part of the
theory's core.  Well, the Hilbert space---once the choice of a
particular quantum state within it is excluded---really carries no
substance beyond its dimensionality $d$.  Thus, in a more refined way
of speaking, what I really mean to say is that when I posit a quantum
system, I am allowed to also posit a characteristic property of it.
It is a property that can be captured by a single integer $d$.

There are some other things which I can argue will be ``left behind''
in such an analysis, but I don't want to clutter this note too much.
Mainly I presented the example above so that I could give you a
clearer sense of how I want to draw a distinction between the rawest
forms of ``reality'' (the surprises the world gives us) and the
``core of a theory.''

It is the core of the theory (along with the theory as a whole) that
I am starting to view in Darwinian terms.  But don't we have every
right to posit that core as a property of the world itself, at least
as long as that belief serves us well?  This, as you point out, has
been the predominant image of what science is about heretofore.

The only answer I can give you is ``yes, we can'' (just as indeed we
have heretofore).  So, your point is well-founded.  What I am worried
about is whether we {\it should\/} posit it so.  You say that this
view has guided science well in the past.  But how do you know?  In a
world with a view that there is no ultimate law, how do you know that
we would not be a thousand years more advanced if we had only better
appreciated our role as the substratum of our theories?  I think it
boils down to the difference between an active and passive view of
what existence is about.  Or maybe the difference between a positive
and a negative view.

To make this point, let me try to put things back into the context of
regular Darwinian evolution.  Consider the word ``elephant.'' Does it
denote anything that exists in a kind of timeless sense, in a way
that we usually think---or in my case, previously thought---of
physical theories as existing?  If the concept of an elephant is
worthy of treating as a candidate for an element of reality, then so
too will a theory's core.

Well, if we have bought into Darwinism in any serious way, then I
would say, no, there is nothing particularly timeless about the
concept of an elephant.  There was once a chance that it might not
even arise in the world.  The ``elephant'' is merely a function of
the selective pressures that cropped up in our world's particular
history.  And, ashes to ashes, dust to dust, the poor elephant may
eventually disappear from the face of the universe, just like so many
species that arose in the course of evolution only to be never
discovered by a single archeologist.

But now, contrast the evolution of the elephant with the possible
future evolution of the human species.  The elephant was an accident
pure and simple, from the strictly Darwinian view.  But I would be
hard pressed to apply pure Darwinism to the future of mankind.  The
birth of my oldest daughter, for instance was no accident.  Her
traits were selected based on personal visions that both her mother
and I had for the future.  Similarly, but not so excitingly, with the
golden retriever, and all our other domesticated species.  The key
point is that in the present stage of evolutionary development, we
have it within our power to move beyond strict Darwinism.  This is
what our industry of genetic engineering is all about.

However, we would have never gotten to this stage if we had not first
realized that the concept of a species is not immutable.  As
strange---and as crazy and as scary---as it may sound, this is where
my thoughts are starting to roam with physical theories.  This does
not mean, however, that we can have exactly what we want with our
physical theories---that they themselves are little more than dreams.
Just as the genetic engineer can make a million viruses that will
never have a chance of surviving on their own, there is more to the
story than our whims and fancies:  There is the ever-present
selective pressure from the outside.  But that does not delete the
genetic engineer's ability to make something that was never here
before.

But now, I go far, far, far beyond what I needed to say to answer all
your questions.  Mainly, I just wanted to emphasize why I
intentionally placed the words ``a priori'' in my definition of
reality way above.

I fear now slightly that you're going to realize I'm one of the
craziest people you've ever met!  And, trust me, I'm not sure I
really believe all that I said in the last three paragraphs.  But it
does strike me as a productive, or at least hopeful, train of thought
that someone ought to explore.  I guess I offer myself as the
sacrifice.

\noindent --- --- --- --- ---

There. I think that's enough of my going around your questions in a
rather wide way.  Let me now zoom back to the center of one of them
for purposes of a final emphasis.

\bbw
But you obviously also want to say that our theories tell us {\em
something} about reality, even if they are not descriptions of
reality.
\ebw

I hope you can glean from all the above that I do indeed believe our
theories tell us something about reality.  But that something is much
like what the elephant tells us about reality.  Its presence tells
us something about the accumulated selective pressures that have
arisen up to the present date.  A theory to some extent is a
statement of history.  It is also a statement of our limitations with
respect to all the pressures yet seen, or---more carefully---a
statement of our limitations with respect to our imaginations for
classifying all that we've yet seen.  (I for instance, cannot jump
off the leaning tower of Pisa unprotected and hope to live; you, for
instance, cannot get into your car and hope to push on the
accelerator until you are traveling beyond the speed of light.)
Finally, to the extent that we the theory users are part of nature,
the theory also tells us something about nature in that way.

But for any theory, there is always something outside of it.  Or at
least that's the idea I'm trying to build.  \medskip

\noindent PS.  Way above, I said I would finally say a few words about
your ``private-world-within-entangle\-ment'' musings.  But somehow it
didn't quite fit in with the flow of the rest of what I wanted to
say.  So, let me try to present the statement in isolation.  From my
point of view, the quantum state, and with it entanglement, never
pierces into the quantum system for which we posit a parameter $d$
(the ``dimension'').  Similarly for any bipartite system for which we
posit two parameters $d_1$ and $d_2$.  The quantum state is only
about what I'm willing to bet will be the consequences when I reach
out and touch a system.  Otherwise, indeed, a quantum system denotes
a private world unto itself.  And similarly with bipartite systems.
We have very little right to say much of anything about the goings-on
of their insides.  (This part of the picture is something I've held
firmly for a long time; it even shows up in my {\sl Physics Today\/} article with Asher.)  Thus, I guess what I'm saying is that I can find a way of agreeing with what you wrote me:
\bbw
Or maybe there {\bf is} a possibility of testing the above
distinction, at least in principle.  (It depends on how I finish my
interpretation.) Let's consider David Deutsch's thought experiment in
which a sentient computer measures, in a reversible way, some
property that doesn't have a definite value for the object being
measured.  Later, the computer will report that he observed a
definite outcome but did not record which one, and the various
possible paths will have been brought together successfully to
exhibit interference.  That's David's prediction.  I would agree that
interference should be possible, but I may disagree about what the
computer will report.  If it is true that one needs to have a {\em
shared} world in order to have a sensible world, then when the
computer was all alone in his private entanglement with the object he
measured, who knows what he experienced?  Maybe he experienced
nothing more about that measured object than what one electron in a
singlet pair experiences about the other electron. Maybe he cannot
honestly report later that he saw one outcome or another.
\ebw
But maybe I'm coming at it from a different point of view. \medskip

\noindent PPS.  I also promised to end with some open questions.  But
I'm petered out now.  And if you've gotten this far, you're probably
exhausted too.  So I'll just leave it for the future, depending upon
how interesting you find the ideas above, or how much you think
they're nonsense!

Only two and half days left in Japan.

\section{26-02-02 \ \ {\it New Jersey Trip} \ \ (to L. Hardy)} \label{Hardy5}

I'm about to start on the long journey back to NJ this morning.  (I'm in Japan.)  There's a little good news on my end.  Within your framework, I finally found an absolutely simple Bayesian justification for the linearity assumption of the probability rule.  I hope to write you an explanation of that on the flight.  I think there's no doubt in my mind now:  Your axioms (or maybe only a very small variation thereof) are certainly the most promising starting point for what I've been hoping to see in quantum mechanics.  For the first time in my life I've understood (roughly) where the vector-space structure comes from.  Sorry it took me so long to come around.  If I were only smarter \ldots\

So, again, hopefully I'll write you a longer explanation on my flight.  (If I don't snooze the whole way.)

\section{26-02-02 \ \ {\it A Tired Old Man} \ \ (to G. {\Plunk})} \label{Plunk1}

I wrote the note below to David {\Mermin} the other day, and he gave you a pretty good grade.  Besides that, I understand from my superiors that there is still a {\it chance\/} (nothing certain) of filling at least one more position in our summer program.

Thus what it boils down to is for me to decide whether I've got enough spunk in me to take on a summer student this year \ldots\ and/or whether this note will frighten you away.  You unfortunately have written me at a time when I have just had a bad experience with a summer student.  That is, though the summer went well for her (by what she tells me), I felt that my summer was used for little more than giving basic lectures in linear algebra and giving emotional support to a kid on her first excursion to summer camp.  I lost a lot of time during the summer when I could have and should have been writing papers.  I hope that builds some imagery.

Thus if I'm going to take on a student, it has to be a student who will:
\begin{itemize}
\item[a)] Ask {\it some\/} of his own questions of nature, and not rely upon me for every input beyond the bigger picture and a reasonable amount of direction.
\item[b)] Use the resources available at Bell Labs to seek out an answer, which
   includes knocking on other people's doors besides mine some fraction
   of the time.  I.e., I want a team player.
\item[c)] Handle my being away for a good piece of the summer, with only email contact.
\item[d)] Be emotionally secure in the sense of not needing me as a den mother.
   And,
\item[e)] Believe that issues to do with quantum foundations are the most
   exciting ones in the physics of our day.
\end{itemize}

The pay here is good.  The towns are dull.  But you'd live near a train station, with Manhattan about 50 minutes down the line.

I'll give you a shot at convincing me.  Feel free to talk with {\Mermin} before composing your reply (or deciding I'm not worth the bother). In fact I'll CC this note to him, so he knows what's up.  But I'll need a reply pretty quickly, due to Bell Labs' time constraints \ldots\ and my own time to let a decision percolate through my mind.

I've got a long trip from Sendai back to New Jersey tomorrow, so I may not be in email contact again for a couple of days.

\section{27-02-02 \ \ {\it An Even Tireder Old Man} \ \ (to N. D. {\Mermin})} \label{Mermin62}

I forgot to include you as a recipient of the note below as I had
promised.  Let me not tell a lie:  It's attached this time around.

I made good progress on several fronts in Japan.  Maybe the most
important---though most trivial---was that I finally found a rather
natural Bayesian motivation for the linearity of the probability rule
in Lucien's system.  (Lucien is a trickster; he sneaks axioms in left
and right that he doesn't call axioms.)  Anyway, at the very
beginning of the paper already, I couldn't accept his motivation for
linearity, falling back as it does on the idea that a state refers in
an objective way to (a class of) preparations.  That is, he didn't
take states to be Bayesian beliefs pure and true, and that's crucial
to me.

Here's the solution; it's as simple as can be (now that I see it).
The upshot of Lucien's formalism is that a measurement is any
``I-know-not-what'' that causes a refinement of one's initial beliefs
(consistent with the other axioms).  Thus a measurement is simply an
application of Bayes' rule by its very definition---moreover, any
application whatsoever within the allowed range.  OK, that's good
enough for the single observer and already thrills me (because it
expresses in a more rigorous way a claim I made in the NATO paper at
the level of density operators).

But when there are two observers and you view measurement as nothing
but an application of Bayes' rule, you have to have some method of
saying when they agree that it is the same measurement (so that you
don't fall into the infinite regress I tried to warn you of when you
were going full-steam-ahead with BFM).  Well, here again the solution
is simple:  Two observers should call a refinement of their beliefs
the same measurement, precisely when they draw the same meaning for
the data they obtain.  Thus as long as both observers ascribe the
same $P(\mbox{data}|\mbox{hypothesis})$, we will say they ``agree''
upon the measurement.

And that's it.  That does the trick.  That and that alone gives
linearity.  No Gleason's theorem; not even anything fancy.  I'll make
this prediction right now:  Quantum mechanics will turn out to be one
of the simplest structures we've ever seen in physics, and for 75
years we've just been too pigheaded to see it.\medskip

\noindent Somewhere over the Pacific,

\section{28-02-02 \ \ {\it Bell Labs!!!}\ \ \ (to G. Brassard)} \label{Brassard12}

I'm just back from 11 days in Japan, and on what fuel I'm running I'm not sure.  In fact I'm deeply asleep as I write this.  But in my slumber I sifted through my pile of mail and found the CRM theme year poster.  What a delight to see our meeting listed!  But what horror to see Los Alamos as my affiliation!!  I hope Kiki will wake me up from this nightmare \ldots

\section{28-02-02 \ \ {\it Which Day?}\ \ \ (to L. Hardy)} \label{Hardy6}

What title would you like me to give the talk?  And while I'm at it, do you have an abstract?  Certainly talking on your axioms would be great.  Though---for business appearances---you might want to throw in a word or two (but no more than that) about how getting the essence of quantum mechanics straight may shed some light on what it is that powers quantum computing \ldots\ or some such money-grubbing sort of thing.

\section{01-03-02 \ \ {\it Drunken Nights} \ \ (to J. W. Nicholson)} \label{Nicholson8}

I almost forgot to send you the two letters I told you about.  I'm very proud of these.  In fact, I think they're my best written and most provocative pieces in a year.  This is what we've been talking about those drunken nights.  [See 18-02-02 note ``\myref{Preskill6}{Psychology 101}'' to J. Preskill and 25-02-02 note ``\myref{Wootters7}{A Wonderful Life}'' to W. K. Wootters.]

In what follows, \verb+\bjp+ and \verb+\ejp+ mean the beginning and ends of a John Preskill quote.  \verb+\bbw+ and \verb+\ebw+ means the same for Bill Wootters.  \verb+\bq+ and \verb+\eq+ mean the beginning and ends of some general quote.  The Everett point of view refers to the many-worlds interpretation of quantum mechanics.  Otherwise I think the two notes are self-contained.

You can see by the time stamp of the present note, jetlag has me by the nads.  I kind of doubt I'm going to come in today, so, sorry, probably no usual-Friday-lunch thing.

\section{02-03-02 \ \ {\it {\Mermin} Festschrift, 2} \ \ (to C. M. {\Caves})} \label{Caves59}

\bcc
I'm inclined to think that the {\Mermin} festschrift would be just the
right place for my speculations about the reality of Hamiltonians,
especially since he has kind of encouraged this idea.  What do you
think?
\ecc

I think, ``I wonder why you ask {\it me\/} this!?!?''  Are you
contemplating that we might have dual submissions?  Abbott and
Costello, Crosby and Hope, Lewis and Martin, the Smothers Brothers?
The problem is, both of us would want to be the straight man.

OK, enough sarcasm.  To answer your question, yes, that probably
would be a good idea.

By the way, here's a little technical meat for the hungry wolf.
Suppose I were willing to go with you on the objectivity of {\it the\/} interaction Hamiltonian or unitary in a measurement interaction. (I'm
not willing in actual fact, but that's beside the point for this
question.)  If we write down the POVM such an interaction ultimately
leads to, then we will get something like this for its elements
$E_b$:
$$
E_b = \tr_{\rm\scriptscriptstyle A}
\Big((I\otimes\sigma)U(I\otimes\Pi_b)U^\dagger\Big)\;,
$$
where all the terms have the usual interpretation.  I believe,
according to you, there should be only one subjective piece to the
right hand side of this equation, namely the ancilla's initial
density operator $\sigma$.

Here's my (innocent, first) question.  The subjectivity in $\sigma$
will afford a range in our interpretation of which POVM was actually
measured in any such interaction.  What is that range?  How large is
it?

A corollary question---this one is more rhetorical and less
innocent---is this.  You know I have never liked the above
formulation of what a measurement is because one has to invoke a
second measurement (the $\Pi_b$) to explain the first (the $E_b$).
This leads to an infinite regress because one can ask (as von Neumann
did), where do the $\Pi_b$ come from.  Thus, just to make sure we're
on board, let me reconfirm that I got the setting in your mind right
in my elaboration above.  Namely, that the only truly subjective term
you would accept in the displayed equation is $\sigma$?  The range in
$E_b$ is consequent only to that?  If that's the case, I would also
like to understand how you invoke a stopper at the level of the
$\Pi_b$.  That is, why do the $\Pi_b$ not have some of their own
range, induced by a higher level measurement model?

You can tell I'm still deeply jetlagged.

\section{03-03-02 \ \ {\it De Finetti and Strong Coherence} \ \ (to P. F. Scudo)} \label{Scudo3}

Thanks for the ghost story.

\bps
Will tell you more as soon as I finish my calculations for this
paper.
\eps

Not to worry at all.  I have a hard enough time leading a rushed
life.  You should not expect that I expect that out of you when I
cannot live up to that standard myself!  When you arrive here you'll have
plenty of time to practice the macabre.

But still, if you've got the interest, who am I to get in your way.
Here's some historical investigative work that you might tackle if
you find that have nothing better to do.  In fact, your friend
Regazzini might be a wonderful resource in this regard.  In the draft
of the anti-BFM paper that {\Caves}, {\Schack}, and I are putting together,
we make a distinction between simple ``Dutch-book consistency or
coherence'' (i.e., the notion that de Finetti and Ramsey first
introduced) and ``strong Dutch-book consistency.''  The latter is a
notion that apparently Abner Shimony first introduced, though {\Caves}
and {\Schack} stumbled across it independently in one of our many wars.
(See reference 12 on page 139 of the samizdat I sent you for the
original citation.)  [See 06-02-02 note ``\myref{Caves54}{The Commitments}'' to C. M. {\Caves} \& R. {\Schack}.]

Starting on page 133 of the samizdat, I write a fairly strong polemic
against the ``requirement'' of strong-consistency---it seems to me
that regular Dutch-book consistency is the most we can demand.
Here's my question.  As you'll see from the quote of Ian Hacking on
page 137, de Finetti was aware of Shimony's addition to coherence and
apparently rejected it.  Do you think it might be possible to find
out where he wrote about this?  More importantly, what were his
particular reasons for rejecting the stronger notion?

\section{04-03-02 \ \ {\it Sliding Off the Deep} \ \ (to H. J. Folse)} \label{Folse8}

I was able to print out your paper without a hitch.  And thanks
moreover, for your detailed notes on my letter to John Preskill. Let
me try to answer some of your questions (in particular the one about
Popper and propensities) by sending you still more rubbish. It's in
the form of a follow-on letter (to the Preskill one) that I sent Bill
Wootters while still in Japan.  [See 25-02-02 note ``\myref{Wootters7}{A Wonderful Life}'' to W. K. Wootters.]  Of course, I would appreciate any
further comments like your last ones if you've got the time!

Let me quickly reply to two of your points explicitly.

\bhf
Do any physicists still think this way?  I realize cosmologists talk
about a TOE, but surely the conduct of contemporary research is not
animated by the thought that we're closing in on some ``final''
description of the universe?  That was true enough a century ago, but
I think that the weight of philosophy of science at least in the 20th
century has pulled against that sort of image of science -- certainly
in a post-Kuhnian era.  It is curious perhaps that this kind of
attitude might still persist in so-called ``foundational'' studies,
whereas I would suppose in something like biological research
everyone would have become much more historicized by now.
\ehf

Nope, the attitude runs pretty rampant in the quantum information and
computing community.  (I'm definitely one of the outsiders there.)
Have a look at David Deutsch's {\sl The Fabric of Reality}. I think
he expresses much of the majority opinion of our little clique there.
But I think you'd be surprised to know that the ``dreams of a final
theory'' attitude runs pretty rampant even in such workaday fields as
laser physics.  I just got the following message from an
experimentalist friend of mine whose work is predominantly used for
the design of better fiber optics:

\bjn
One of the thoughts that continues to strike me is your optimism in
the continual evolution in physical theory.  I will admit that on bad
days I feel like we are in the twilight of physics---only incremental
progress is left to be had.
\ejn

Of the physicists I have met who have even heard of Kuhn or
{\Rorty}---there aren't that many---almost all of them view what little
they know of their thoughts (i.e., Kuhn and {\Rorty}'s) with a little
contempt.  (By the way, John Preskill's reaction to my letter was
particularly violent.  So I think it is the ideas, and not the men,
that bug their sensibilities.)

By the way, I have indeed just discovered {\Rorty} and I find him
wonderful.  I've read the two collections of essays {\sl Philosophy
and Social Hope\/} and {\sl Consequences of Pragmatism}.  It won't be
long, and I'll move on to the book of his that you recommended. Also
since {\Vaxjo}, I've gotten a good feel for {\James} (I read {\sl
Pragmatism}, {\sl The Meaning of Truth}, and Perry's massive
biography of him), and also of {\Dewey} (though I've only read {\sl John
{\Dewey}:\ The Essential Writings}).

Thanks for telling me about Kitcher; I hadn't heard of him.  And
thanks for the pointer to Toulmin; I'll look more deeply into him
than I have in the past.

Concerning one of your other points,

\bhf
It is arrogant perhaps, but I don't see this as the egocentricity.
Every attempt to sketch a conception of the universe from our best
theories at any date in human history in effect commits such
arrogance. Were the Newtonians of the end of the seventeenth century
being ``egocentric'' to think that Sir Isaac had done nothing less
than peer into the mind of the Divine and discerned God's blueprints
for the universe?
\ehf

Yes.  (In my opinion.)  And you might interpret {\James} and pragmatism
in general as a reaction to that.  However, I think in our modern age
with quantum mechanics we have a motivation and opportunity in front
of us that {\James} did not have.  Try to give quantum mechanics a naive
realist interpretation---you can do it, or at least both Everett and
Bohm tell us we can---and you find yourself contorting yourself
beyond belief.  It's as if nature is telling us for the first time,
``Please don't interpret me in a naive realist fashion. I can't stop
you, but please don't.''

OK, the sun is rising and the vampire must return to his native soil.

\section{04-03-02 \ \ {\it The Good Questions of Nicholson} \ \ (to J. W. Nicholson)} \label{Nicholson9}

\bjn
So.  My question to you is two-fold.  Do you still feel the same way
about particle physics experiments that try to continually increase
the energy of interactions between particles?  And, if not, what
extrema in physical theory should we be poking at in order to
continue the evolutionary progress?
\ejn

You ask a good question old friend---the second one---and I shall
have to think a while before I reply.  Or at least compose myself
before I reply.

In the case of your first question the answer is easy.  Yes, I still
feel that way.  And the feeling is not that the SSC would not have
been worthwhile science per se; it might have been.  It was just the
cost versus insight-and-control-of-nature ratio that doomed it in my
mind.  Furthermore, the paradigm the high-energy types presuppose
certainly puts me off.  They think they are doing something deeper
than tabulating the Hamiltonian of an iodine atom, say, but I don't
think they are.  They are just tabulating the Hamiltonian of another
system, and that has the value it always has.

But let me think harder about how to answer your second question. For
my present project (i.e., quantum mechanics), it strikes me not so
much as poking at the extrema that we ought to be spending our time
on, but in poking at the consistency of the worldview we wish to
embed the theory in.  To revert to my analogy of the construction of
general relativity, it was only consistency that Einstein was seeking
when he first started contemplating how gravitation fit within the
framework of special relativity.  Seeing that it did not (and weeding
out all the coordinate effects as I expressed earlier) is what led to
the bending of the manifold.

But don't take that as my definitive (or exhaustive) reply.  I'm
still thinking.

\section{04-03-02 \ \ {\it Green Light} \ \ (to A. Y. Khrennikov)} \label{Khrennikov8}

\bakh
I am waiting for your papers for proceedings.
Please send me a signal that you are working!
If yes, I shall wait one-two weeks.
\eakh

Yes, please do wait for me.  I am working, and I am trying to find a way to respond to your ``{\Vaxjo} interpretation'' both directly and indirectly.

\section{05-03-02 \ \ {\it A Hardy Reminder} \ \ (to C. H. {\Bennett})} \label{Bennett12}

OK, it's set, Hardy's going to give his talk on Thursday March 14, from 2:00 to 3:00.  And as I told you, Osamu Hirota will also be around.  (Not to mention the pleasures of my goofy laugh and van Enk's dry wit.)  So, you've got every reason to come.  We'll look forward to seeing you Thursday.

\section{06-03-02 \ \ {\it Poetry on Concrete} \ \ (to L. Hardy)} \label{Hardy7}

I decided to write this letter in \LaTeX\ so that I would have a
small chance of being clearer.  A few months ago, you wrote me this:
\blh
[M]y basic picture of preparation/transformation/measurement is
rather similar, though considerably less poetic, than your picture of
humans interacting with a world. \ldots\ After the poetry you need to
make the journey from vision to concrete construction.
\elh
You're quite right.  I think your paper has finally started to mix
the right sort of concrete to support a building like the one I want
to see built.

In what follows, all I'm going to do is keep true to this word I
learned---poetaster---and try to write a little amateur poetry at
your construction site. There are three things that really intrigue
me in your paper, and one that really confuses me.  I'll take them
one at a time.

The first is simply the absolutely beautiful ease with which you
bring us to a vector space structure.  I had known about the concept
of ``informationally complete'' POVMs for years\footnote{I think the
idea can be traced back to E.~Prugove\v{c}ki,
``Information-Theoretical Aspects of Quantum Measurements,'' Int.\ J.
Theo.\ Phys.\ {\bf 16}, 321 (1977).}---they're even a crucial part of
the proof of our quantum de Finetti theorem\footnote{See Section IV
of \quantph{0104088}.}---but before your work I had never
appreciated so clearly that the probability distributions they
generate ought to be taken as the very most primordial representation
of a quantum state.
\bq
\noindent \it What is a quantum state?  Nothing more than one's
gambling commitments with respect to the outcomes of a fiducial
measurement.  Full stop.
\eq
Or at least, that's the way I want to view it.\footnote{And I say this
  even though I was saying things right after the {\Vaxjo} meeting
  like I said to Khrennikov in the letter of 4 July 2001 attached
  below. [See 04-07-01 note ``\myref{Khrennikov5}{Context Dependent
      Probability, 2}'' to A. Y. Khrennikov.]  Somehow, I just never
  took the idea so seriously as I do now, i.e., after reapproaching
  your paper in Japan.  ``Fiducial measurement'' $+$ ``a notion of
  transformation from one measurement to another that carries the
  probabilities with it'' is what tells the whole story!}

Now, what you write presently is not completely consistent with that
characterization.  I'm going to do my best to try to make that plain
to you.  You write, ``I don't believe that my approach needs to adopt
a particular approach to interpreting probability.''  There is a
sense in which that is certainly true for the mathematics you've
already performed.  However, if we're ever going to get past this
foundational impasse in quantum mechanics, I would say firmly that we
cannot leave the story there.

Here is the difficulty.  If you take an objective approach to
probability, then you are necessarily left with the conclusion that a
measurement ``here'' in general changes something objective (or
physical) over ``there.''  And you will never get around that.  Just
consider any entangled state.  Making a measurement on Alice's side
changes her ascription---without any considerations of time or
distance---of the state on Bob's side.  If a state is purely
equivalent to a set of probabilities and probabilities are not
subjective degrees of belief or gambling commitments, but rather
objective and independent of the agent, then something physical must
change at Bob's site with that far away measurement. And if you
believe that, then you end up with conundrums out the wazoo.

I know that may not be your stand yet, but it is mine.  Thus I'm left
with the task of putting your mathematics into a framework I can
accept.  The first and clearest example of a technical mismatch
between our views is what we each might consider as a valid
motivation for the convex-linearity rule in your
Eqs.~(41)--(43).\footnote{By the way, I have to take a little
exception with your accounting rules for your axioms.  In similar
treatments from the ``operational school,'' say of Kraus, Holevo, and
Ozawa, a convex-linear or affine assumption (of exactly the same
spirit as your Eqs.~(41)--(43)) is always listed as a distinct
axiom.}

I think if we were to trace the roots of our mismatch, we would find
it in that I do not subscribe to your ``basic picture of
preparation/transformation/measurement'' \ldots\ which I think you
think is common to both of us.  Here's the way I put it to Asher
Peres when we were constructing our Physics Today paper:
\bq
In general I have noticed in this manuscript that you lean more
heavily on the word ``preparation'' than we did in our letter to
Benka.  \ldots\  Unless I misunderstand your usage of the word, it
may actually be a little too anthropocentric even for my tastes.  The
problem is this:  consider what you wrote in the paragraph about the
wave function of the universe.  It seems hard to me to imagine the
wave function of those degrees of freedom which we describe quantum
mechanically as corresponding to a ``preparation.'' Who was the
preparer?

It is for this reason that {\Carl} {\Caves} and I prefer to associate a
quantum state (either pure or mixed) solely with the compendium of
probabilities it generates, via the Born rule, for the outcomes of
all potential measurements.  And then we leave it at that. Knowing
the preparation of a system (or the equivalence class to which it
belongs) is one way of getting at a set of such probabilities.  But
there are other ways which surely have almost nothing to do with a
preparation.  An example comes about in quantum statistical
mechanics:  when the expected energy of a system is the only thing
known, the principle of maximum entropy is invoked in order to assign
a density operator to the system.  There may be someone beside me in
the background who knows the precise preparation of the system, but
that does not matter as far as I am concerned---my compendium of
probabilities for the outcomes of all measurements are still
calculated from the MaxEnt density operator.

To help ensure that I was not jumping to conclusions on your usage of
the term, I reread today your paper ``What Is a State Vector?'' [AJP
{\bf 52} (1984) 644--650].  There was a time when I agreed with
everything you wrote there \ldots.  But as of today at least, I think
a more neutral language as in our letter to Benka is more
appropriate.
\eq

Thus what I seek is a picture that involves only measurement and
transformation.  (And if the truth be known, I'd like to get rid of
transformation {\it in a sense}, leaving only measurement behind from
your trinity.  But that's a longer story, and I'll only give hints of
it below.)  However, I do think your formalism is just about up to
the task, despite your Figure 1 on page 4.

The way I see it, this thing called the ``preparation'' is just a
token for the {\it right and true\/} quantum state you imagine for a
system.  But from a steadfast-Bayesian\footnote{Some would say
  ``radical Bayesian.''} point of view, there is no such thing as a
right and true probability assignment for anything. Consequently, if a
quantum state is to be solely a probability assignment for the
outcomes of a fiducial measurement, from the steadfast-Bayesian view,
there can also be no such thing as a right and true quantum
state.\footnote{I have argued this point at great length in a
  correspondence with {\Caves}, {\Schack}, and {\Mermin}
  predominantly.  I'm going to post that on my website at just about
  the same time I send this note off to you.  The file is called
  ``Quantum States:\ What the Hell Are They?,'' and the relevant part
  with respect to my remark starts on page 19 and winds up somewhere
  around page 97. However, the most relevant notes are the ones to
  \myref{Mermin28}{Brun--Finkelstein--{\Mermin} dated 7 August 2001},
  to \myref{Schack4}{{\Caves}--{\Schack} dated 22 August 2001}, and to
  \myref{Mermin35}{{\Mermin} dated 2 September 2001}.  I hope you will
  have a look at those at the same time as reading the present note.}

Let me emphasize this once more at the purely classical level before
going on. For the Bayesian there is no such thing as a {\it right\/}
probability assignment; there is no such thing as a {\it wrong\/}
probability assignment.  A probability assignment is a set of numbers
that signifies which gambles one is willing to make, period. How
those numbers are set in any given instance is an issue that has
nothing to do with nature. Instead it has everything to do with all
the ugly things we try to keep out of science:  one's emotions, one's
intelligence, one's hopes, the rumors one has heard recently, and so
on. In fact, a probability assignment has nothing to do with the
world itself---in a sense it floats above the physical world.

What this boils down to is that---{\it within a Bayesian
paradigm}---your argument for convex-linearity cannot make any sense:
\bq
\noindent
Assume that the preparation device is in the hands of Alice.  She can
decide randomly to prepare a state ${\bf p}_A$ with probability
$\lambda$ or a state ${\bf p}_B$ with probability $1-\lambda$. Assume
that she records this choice but does not tell the person, Bob say,
performing the measurement. Let the state corresponding to this
preparation be ${\bf p_C}$.  Then the probability Bob measures will
be the convex combination of the two cases, namely
\begin{equation}\label{convexf}
f({\bf p}_C) = \lambda f({\bf p}_A) +(1-\lambda) f({\bf p}_B)
\end{equation}
This is clear since Alice could subsequently reveal which state she
had prepared for each event in the ensemble providing two
sub-ensembles.  Bob could then check his data was consistent for each
subensemble.  By Axiom 1, the probability measured for each
subensemble must be the same as that which would have been measured
for any similarly prepared ensemble and hence (41) follows.
\eq
Here's the difficulty.  What does the distribution $(\lambda,
1-\lambda)$ refer to?  From the Bayesian view, it can only refer to a
belief (or better yet, a gambling commitment) that Bob possesses
concerning Alice's actions.  On the other hand, the numbers $f({\bf
p}_A)$ and $f({\bf p}_B)$ refer to a couple of alternate beliefs a
completely different person, Alice, might possess.  However---{\it
from the Bayesian standpoint}---there is nothing {\it a priori\/}
that can be used to rigidly identify Bob's beliefs as a function of
Alice's \ldots\ as Eq.~(\ref{convexf}) above would imply.  Alice has
every right to believe anything she wants; Bob has every right to
believe anything he wants.\footnote{Nor does it help to try to divert
the discussion from a two-player situation and stuff all of these
beliefs into a single head (Alice's say). For at any instance, Alice
believes what she believes. And there is nothing within the Bayesian
creed to force what she ought to believe in various other
hypothetical situations---for example, where the distribution
$(\lambda, 1-\lambda)$ might describe the amount of amnesia she has
concerning a previous belief that she's just forgotten.}

For the Bayesian, all the action---all the science---is not in the
probability assignments themselves that various individuals might
make, but in how those assignments change upon the acquisition of new
data (steered by the influence of rationality).\footnote{Read my note
  to {\Mermin} titled ``\myref{Mermin55}{Reality in the
    Differential}'' starting on page 128 of my previously mentioned
  posting.} In particular, the important things to ferret out are the
conditions in any situation where two observers will converge in their
beliefs when they agree that they are acquiring the same data.
Bernardo and Smith put it in a beautiful way: \bq The subjectivist,
operationalist viewpoint has led us to the conclusion that, if we
aspire to quantitative coherence, individual degrees of belief,
expressed as probabilities, are inescapably the starting point for
descriptions of uncertainty.  There can be no theories without
theoreticians; no learning without learners; in general, no science
without scientists.  It follows that learning processes, whatever
their particular concerns and fashions at any given point in time, are
necessarily reasoning processes which take place in the minds of
individuals.  To be sure, the object of attention and interest may
well be an assumed external, objective reality: but the actuality of
the learning process consists in the evolution of individual,
subjective beliefs about that reality.  However, it is important to
emphasize, as in our earlier discussion in Section 2.8, that the
primitive and fundamental notions of {\it individual\/} preference and
belief will typically provide the starting point for {\it
  interpersonal\/} communication and reporting processes.  In what
follows, both here, and more particularly in Chapter 5, we shall
therefore often be concerned to identify and examine features of the
individual learning process which relate to interpersonal issues, such
as the conditions under which an approximate consensus of beliefs
might occur in a population of individuals.

$\qquad\qquad\qquad\qquad$ --- pp.\ 165--166, Bernardo and Smith,
{\sl Bayesian Theory}
\eq
\bq
What is the nature and scope of Bayesian Statistics within this
spectrum of activity? Bayesian Statistics offers a rationalist theory
of personalistic beliefs in contexts of uncertainty, with the central
aim of characterising how an individual should act in order to avoid
certain kinds of undesirable behavioural inconsistencies. The theory
establishes that expected utility maximization provides the basis for
rational decision making and that Bayes' theorem provides the key to
the ways in which beliefs should fit together in the light of
changing evidence.  The goal, in effect, is to establish rules and
procedures for individuals concerned with disciplined uncertainty
accounting.  The theory is not descriptive, in the sense of claiming
to model actual behaviour.  Rather, it is prescriptive, in the sense
of saying ``if you wish to avoid the possibility of these undesirable
consequences you must act in the following way.''

$\qquad\qquad\qquad\qquad$ --- p.\ 4, Bernardo and Smith, {\sl
Bayesian Theory}
\eq

And, it is precisely this that I'm going to use to get to your
Eq.~(41) from a Bayesian starting point.  But before I can do that,
let me praise the second thing that I see as a deep contribution of
your paper. This is the very {\it definition\/} of what a measurement
is:  It is {\it any\/} application of Bayes' rule consistent with
your remaining axioms.\footnote{There is no doubt that I have been
predisposed to saying something like this for a long time.  For
instance, at the top of page 26 of my NATO paper \quantph{0106166}, I say, ``Quantum measurement is nothing more, and
nothing less, than a refinement and a readjustment of one's state of
knowledge.'' The difference is, your paper brings this beyond just
some metaphor. There I tried to capture the idea with my
Eqs.~(57)--(59), which are explicitly in terms of density operators.
You on the other hand, do it with the real thing.  I.e., you make it
absolutely airtight that it is Bayes' rule operating in the
background, and not just some noncommutative analog of it.}

To make some sense of this, let me put the problem into a notation
that is slightly better for my purposes.  Suppose the outcomes of a
fiducial measurement are labeled by $h$ and the outcomes of some
other measurement we might contemplate are labeled by $d$.  Then I
will dually write the quantum state (via your identification of
states with probabilities for the outcomes of a fiducial measurement)
as a function $P(h)$ or as a vector $\bf P$.  What I mean by an
application of Bayes' rule is the supposition of a set of probability
distributions $P(h|d)$ (or ${\bf P}_d$ in vector notation)---one for
each $d$---such that
\be
P(h)=\sum_d f_{\bf P}(d)\, P(h|d)\;,
\label{HappyFuchs}
\ee
or alternatively
\be
{\bf P} = \sum_d f_{\bf P}(d)\, {\bf P}_d\;,
\ee
for some other probability distribution of $f_{\bf P}(d)$ over the
variable $d$.  When a particular data $d$ is collected, one---at
least tentatively---enacts the transition
\be
{\bf P}\; \longrightarrow\; {\bf P}_d\;.
\label{MmtIs}
\ee
This is what I claim you have taught us is ``what a quantum
measurement is.''\footnote{Note carefully that I used the word
``tentatively'' in this description. That is because there is a
possibility of the further quantum phenomenon that measurement can be
more invasive than supposed classically.  What this means
operationally is that the ${\bf P}_d$ may not ultimately concern the
original fiducial measurement, but may instead be about a completely
different fiducial measurement at the end of the process.  See my
discussion centered around Eqs.~(57)--(59) of \quantph{0106166}.}

A measurement is any ``{\it I know not what}'' that enacts a
transition of the form Eq.~(\ref{MmtIs}).  What is the variable $d$?
How is it defined?  It simply is not, except as a place holder in a
particular instance of Bayes' rule.  And that's it:  That's the story
of measurement.

Now, where does one get the convex-linearity of the probability rule
$f_{\bf P}(d)$ in a Bayesian-happy way?  The technique is to consider
when it is that two observers will think they are performing the same
measurement.  That is to say, I could walk into the room and think to
myself that the measurement device in front of me gives me warrant
for the decomposition in Eq.~(\ref{HappyFuchs}).  You on the other
hand, with a completely different set of beliefs and experiences, may
think that the device warrants you to the decomposition
\be
Q(h)=\sum_d f_{\bf Q}(d)\, Q(h|d)\;.
\label{HappyHardy}
\ee
When shall we say that we actually agree upon the measurement?
Classically, the answer is when the statistical model made use of by
each of us is the same:
\be
Q(d|h)\equiv \frac{f_{\bf Q}(d)\, Q(h|d)}{Q(h)} = \frac{f_{\bf
P}(d)\, P(h|d)}{P(h)}\equiv P(d|h)\;.
\ee
That is, if we had hold of which one of the fiducial outcomes
actually obtained, then we would draw the same meaning from it. Here,
by ``meaning'' I mean how much we would feel warranted in revising
our beliefs that a $d$ would have popped up if we had instead
performed an appropriate measurement for it.

By the way, notice one thing.  If this account deviates from standard
Bayesianism any at all, it is only in this.  Never once did I invoke
the {\it necessity\/} of a joint probability distribution
\be
P(h,d) \equiv f_{\bf P}(d)\, P(h|d)
\ee
in my description of Bayes' rule.  Of course, such a function exists
as a mathematical artifice---i.e., it has all the properties of a
joint probability distribution---but one should not try to make any
meaning for it beyond that. In particular---and, especially with
regards to the quantum context---one should not feel it necessary to
interpret the function as a degree of belief of the simultaneous {\it
existence\/} of a true $h$ and a true $d$.

OK, let me use that now to start talking about convex-linearity
again.  Suppose there are at least three agents on the scene, Alice,
Bob, and Charlie.  And suppose Alice subscribes to
Eq.~(\ref{HappyFuchs}) for her description of what the device in
front of her is about, whereas Bob subscribes to
Eq.~(\ref{HappyHardy}), and Charlie still further subscribes to,
\be
R(h)=\sum_d f_{\bf R}(d)\, R(h|d)\;.
\label{HappyBennett}
\ee
If that is as far as it goes, then there is nothing whatsoever we can
say about the relation between $f_{\bf P}(d)$, $f_{\bf Q}(d)$, and
$f_{\bf R}(d)$.  Indeed there need be no relation whatsoever.

However, suppose Alice, Bob, and Charlie share the minimal belief
that actually the same measurement is being performed with respect to
each of their descriptions. Then by definition
\be
P(d|h)=Q(d|h)=R(d|h)\;.
\ee
Fine.  Now consider those cases where, as it turns out,
\be
P(h)=\lambda Q(h) + (1-\lambda) R(h)\;.
\ee
Letting
\be
G(d)=f_{\bf P}(d)-\lambda f_{\bf Q}(d) - (1-\lambda) f_{\bf R}(d)\;,
\ee
we see immediately that
\bea
G(d) &=& \sum_h \Big[P(h)P(d|h)-\lambda
Q(h)Q(d|h)-(1-\lambda)R(h)R(d|h)\Big]
\\
&=& \sum_h \Big[P(h)-\lambda Q(h)-(1-\lambda)R(h)\Big]P(d|h)
\\
&=& 0
\eea
since
\be
f_{\bf P}(d)=f_{\bf P}(d)\sum_h P(h|d)\equiv\sum_h P(h) P(d|h)\;,
\ee
and so on.

And that's it, from this perspective.
\be
P(h)=\lambda Q(h) + (1-\lambda) R(h) \qquad \Longrightarrow \qquad
f_{\bf P}(d)=\lambda f_{\bf Q}(d) + (1-\lambda) f_{\bf R}(d)
\label{HappyKiki}
\ee
when and only when the agents who hold the beliefs $\bf P$, $\bf Q$,
and $\bf R$ ``agree'' to the meaning of the $d$-``clicks'' the
measuring device will elicit.

Part of me fears that you're going to view all this as little more
than rhetoric.  It took me five pages of explanation to do what you
did in one paragraph.  So let me try to reiterate the point of this
exercise from my perspective one last time, before I throw in the
towel. Here's the point. If you believe that the quantum state is
rigidly (or even loosely) connected to reality, then---it seems to
me---you will never find a way out of the conundrum of
``unreasonableness'' associated with ``state-vector collapse at a
distance.''  I.e., our community will always be left with a search
for the {\it mechanism\/} that makes it go.  Our community will
always be left with the embarrassing questions to do with its clash
with Lorentz invariance.  And, maybe most importantly, we will be
left with the nagging question of why we can't harness this mechanism
for more useful purposes.\footnote{If we can't see angels, why posit
them? If we can't see superluminal communication, why posit it?
Alternatively, if we do posit angels, the natural question to ask is
how might they help save our souls.  I would be suspicious of a world
where the angels served little purpose other than rounding out a
theology and not aiding in our souls' deliverance.} I view these
questions as a distraction from the ultimate goal we all ought to
have, of building a more interesting, more fantastic physics.

In connection to the present discussion, I would claim that THE
preparation is little more than an anchor for such a rigid
connection.  It is, even if implicitly, a reversion back to the old
issues that led to all the trouble in the first place.  Thus I am
compelled to find a way to absolutely disconnect the quantum state
from the physical world.  Yet I am still required to make sense of
what it is that we're doing when we practice quantum mechanics.  For
this, the trail has already been blazed by (radical) Bayesian
probability theory, and thus I take that as my cue.  Getting at
Eq.~(\ref{HappyKiki}) from within a subjectivist framework is one of
those first steps you just have to take \ldots\ and then hopefully
never have to think about again.

OK, with that, I'm ready to praise you my third and final time in
this letter.  But I think you'll probably hardly feel it's a praise.
Let me tell you another goal of seeing how much of quantum mechanics
can actually be run through with complete airtight consistency from
the subjectivist standpoint.  And that is to pick the theory apart.
For, you see, I see no difficulty whatsoever in imagining that any
theory can consist of two basic components---the completely
subjective and agent-dependent, and the completely objective (or
intersubjective if you will) and thus, agent-independent.  What is
the distinction between these components (I hope you ask)?  It is
this:  Once an agent posits one of the objective components in any
particular instance (whether it's ``really'' there are not), there is
no move {\it within\/} the theory that will allow him to change that
supposition.

Let me give you an example of the latter.  Posit a bipartite system
with Hilbert spaces ${\cal H}_{d_1}$ and ${\cal H}_{d_2}$ (with
dimensions $d_1$ and $d_2$ respectively) and imagine an initial
quantum state for that bipartite system. Now, I would say that the
quantum state must be a subjective component in the theory because the
theory allows me localized measurements on the $d_1$ system that can
change my quantum state for the $d_2$ system.  In contrast, I would
say that the number $d_2$ that was posited for the second system is an
objective component in the theory. There is nothing I can do at the
$d_1$ site that will allow me to change the numerical value of $d_2$.
The only way I can change that number is to scrap my initial
supposition.  Thus, to that extent, {\bf (on first pass)} I have every
right to act as if the numbers $d_1$ and $d_2$ are potential
``elements of reality.''\footnote{Please note that I emphasized the
  qualifier ``on first pass.''  The reason for this emphasis will
  become clear if you read the letters to Preskill and Wootters that I
  have pasted into the present letter to you.  [See 18-02-02 note
    ``\myref{Preskill6}{Psychology 101}'' to J. Preskill and 25-02-02
    note ``\myref{Wootters7}{A Wonderful Life}'' to W. K. Wootters.]}

Here's where I really think you sell yourself short by advertising
your system as an extension or generalization of classical
probability theory (with classical probability theory as a special
case that's gotten by deleting one of the axioms).  For I would say
that your framework of ``states'' as vectors and ``measurements'' as
applications of Bayes' rule is {\it classical probability theory},
full stop. Or, I should just say ``{\it probability theory}, full
stop''---for, the word ``classical'' seems to imply that it is a
subject somehow within empirical science (rather than ``law of
thought'' that antecedes science).  In showing me that even quantum
``measurements'' can be viewed legitimately as nothing more than
applications of Bayes' rule, you have done me a great service. For
you demonstrate to me more clearly than ever that the concept of POVM
ought to be put onto the subjective side of the shelf when I tear
quantum mechanics into its two components.  But your other intriguing
axioms---like the simplicity and composite-system axioms---which you
think give the possibility of generalizing upon classical
probability, I would say are nothing of the sort. Instead, I would
say they express just the opposite.  These axioms seem to me to say
something about what we are positing of nature.  They express
something that is not subjective and is not ``law of thought.'' They
in fact form part of the restriction to probability theory that I
asked for over and over in my \quantph{0106166}. Thus drawing
those axioms out explicitly strikes me as great
progress.\footnote{And that is why this minor spanking counts as a
praise!}

Thus to come back to my example of Hilbert-space dimension.  As I
have already explained, I would say that quantity is a (potentially)
objective feature of nature.  But now, after understanding your paper
better, I would say the same of your composite-system axiom. In fact,
that axiom is a way of elucidating the very meaning of the dimension
$d$.  As such, it forms part of the ``manifold-structure analogy''
behind quantum mechanics which I tried very hard to explain to
Preskill and Wootters in two further letters I'll paste below. (I
hope you will read them along with this note, as I think they greatly
elucidate what I was hoping to convey in the previous four
paragraphs.)

There.  Three praises; take them for what they're worth.  They are
the three things that really intrigue me about your paper.  But I've
only started my digestion process.  I'll leave a discussion of the
confusions until we get some single-malt in front of us next week.

\section{08-03-02 \ \ {\it Your Webpage} \ \ (to H. J. Folse)} \label{Folse9}

Thanks a million for the link to your webpage.  I had a quick flip though it before my flight this morning, and it looks great.  (I'm on my way to Atlanta for an AMS meeting.)  I'll have a much deeper read through it upon my return; I can tell you that it will certainly help orient me with respect to the various schools.  Just a snap judgment---maybe not to be taken seriously, since I've only mildly skimmed the page---but it looks like I might be closer to an ``entity realism'' than anything else.

I'll tell you how my van Fraassen impressions go after I meet him.  There's no doubt I'm more Bayesian-optimistic than ever.  In particular, I've now become convinced (during my Japan stay) that a combination of my NATO paper (\quantph{0106166}) and Hardy's ``five-reasonable-axioms'' paper tells us unequivocally that collapse (or L\"uders' rule or Kraus's generalization of it, or whatever you wish to call it), is NOTHING OTHER THAN an application of Bayes' rule along with a possible redefinition of which ``observable'' (POVM generally) the information is relevant to.  In other words, collapse is after all an utterly trivial notion---it is refinement and readjustment (to a new context) of information and nothing more.

BTW, I met Plotnitsky and his girlfriend in Manhattan a few weeks ago.  You should have heard the praise and the descriptions of beauty they gave your home!  This report came, by the way, as were sitting in one of Plotnitsky's sister's homes:  a three level penthouse on the corner of Broadway and Bleeker Street in Manhattan!  Surely, a multi-million dollar affair.

\section{09-03-02 \ \ {\it Creation and the Equivalence Principle} \ \ (to G. L. Comer)} \label{Comer8}

I started composing the letter below on my flight to Atlanta, and now it's becoming clear that I'm not going to have a chance to finish it for a while.  Let me just send you what I've got, so that you don't spend some days thinking that your last message made no effect on me.

The difficulty on this end is that I'm sharing a room with Howard Barnum and we haven't been able to shut up about quantum foundations for little more than a minute.  So, it probably won't be until my return home that I'll have a quiet moment.

By the way, it is so good to be in the South again!!!  Last night I went to a restaurant and for \$13.50 (including beer, which was \$3.50) I had some of the most satisfying food I've had in a couple of years!  And the service was so nice!  They had pictures of Jimmy Carter here and there and told me that he comes in about once every two months.  I believe it!

\bq
\ldots\ I felt completely lost.  For the first time in my life as a speaker, my mouth became so dry that my lips stuck to my teeth!  (Without exaggeration.)

But let's get back to science!  I'm on my way to Atlanta for an AMS meeting where, I promise you, my lips will not become stuck to my teeth.  I am really very intrigued by your idea.  Did you catch the phrase with which I ended my letter to Bill Wootters?  ``But for any theory, there is always something outside of it.''  That can be taken at face value, but I think there is a corollary that starting to strike me even more than that.  And that is, you've got to respect any theory that tells you of itself, ``There are some well defined questions (within me), that have no answers (within me).''
\eq

\section{12-03-02 \ \ {\it Wheeler Link} \ \ (to R. Pike)} \label{Pike6}

\brp
Wheeler in {\bf The Times}:\medskip\\
\indent \rm $\bullet$ Dennis Overbye, ``Peering Through the Gates of Time,'' {\sl New York Times}, 12 March 2002, \smallskip \\
\indent \phantom{$\bullet$} \myurl{http://www.nytimes.com/2002/03/12/science/physical/12WHEE.html}.
\erp
Thanks.  That was good.  Typical Wheeler.

\section{15-03-02 \ \ {\it Ice Cream and Reciprocity} \ \ (to C. H. {\Bennett})} \label{Bennett13}

Thanks for taking care of Lucien last night.  After a night of
alternating between sweat and shivers, I seem to be on the road to
recovery.  A comment and a question.

Comment.  I think you said that I should make an overlay for my
quantum-axiom slide that says ``Give an ice-cream reason, if
possible,'' for each of the axioms.  By this, you were indicating
that my ``Give an information theoretic reason, if possible,'' is a
rather arbitrary thing to be asking.  But, I say that it is not
arbitrary precisely because the main object of our attention in the
theory, the quantum state, specifies {\it probabilities}.  It
specifies how we ought to be taking nature into account when we make
our mortal decisions.  The quantum state does not specify flavors of
ice cream.  Thus it seems to me like an entirely natural question to
ask:  If the main object of the theory is of an information theoretic
(or decision theoretic or call it what you will) character, then how
much of its support structure might also be of the same character?

I think I put the goal of this program in a particularly clear way in a note to John Preskill.  [See 18-02-02 note ``\myref{Preskill6}{Psychology 101}'' to J. Preskill.]  I'll paste it below.  Read it if you have the time.

Question.  On the other hand, I rather liked what you said about
wanting to base quantum mechanics on the idea that interactions are
more symmetric in this theory than in classical physics.  In
interaction, both observer and observed are changed in the process.
Could I get you to write your own version of that in a small
paragraph, so that I can have something solid to read and think
about?

\section{24-03-02 \ \ {\it Leaving Dublin} \ \ (to C. King)} \label{King2}

I just wanted to let you know, I absolutely enjoyed my little time alone with you today.  Even if there had been nothing else to the rest of the meeting---and there was much to the rest of the meeting---today's discussions would have made it worthwhile for me to make the trip!  I hope I can repay you one day likewise.

Concerning the part of today's conversation about Darwinist conceptions of scientific theories, I told you that I would give you an exact pointer to my write-up of that.  The place to look is at the file titled {\sl Quantum States:\ What the Hell Are They?}\ posted at my website (link below).  In particular, have a look at the letter to Preskill titled ``\myref{Preskill6}{Psychology 101}'' starting on page 143 and the letter to Wootters titled ``\myref{Wootters7}{A Wonderful Life}'' starting on page 149.

The references I owe you concerning characterizations of completely positive maps are these:
\begin{itemize}
\item
J.~A. Poluikis and R.~D. Hill, ``Completely positive and Hermitian-preserving linear transformations,'' {\em Linear Algebra and Its Applications}, vol.~35, pp.~1--10, 1981.

\item
G.~P. Barker, R.~D. Hill, and R.~D. Haertel, ``On the completely positive and positive-semidefinite-preserving cones,'' {\em Linear Algebra and Its Applications}, vol.~56, pp.~221--229, 1984.

\item
L.~J. Landau and R.~F. Streater, ``On Birkhoff's theorem for doubly stochastic completely positive maps of matrix algebras,'' {\em Linear Algebra and Its Applications}, vol.~193, pp.~107--127, 1993.
\end{itemize}

Take care and enjoy your visit with your wife and children.

\section{24-03-02 \ \ {\it Packing Your Suitcase} \ \ (to F. Verstraete)} \label{Verstraete1}

I'm just about to get on the road home from a week-long stay in Dublin.  Today I had the greatest conversation with Chris King where we mapped out a good load of mathematical questions that arise naturally from my musings about the Bayesian nature of quantum time evolution operators.  So, if you're amenable, I think I'll have you work on precisely those problems during your visit to Bell Labs.  (CK, in fact, thought these questions were exactly up your alley \ldots\ based on a paper you wrote recently characterizing CPMs.)

Anyway, here's the stuff to be reading to start to get ready for the project.  (That is, if you've got any spare time.  If you don't, then certainly don't worry about this.)  The easy stuff is my new collection of letters posted at my website (link below).  The title is ``Quantum States: What the Hell Are They?''\ and pay particular attention to the parts where I give arguments for the ``subjectivity'' of the Kraus operation (and POVM) that one associates with a measurement.  Anyway, that forms my personal background and motivation for the technical questions I've got for you.

But you don't have to worry:  The technical questions stand alone and will be worthwhile to you even if you buy in to none of my particular motivation.  As a little preparation for that, you might obtain the following papers:
\begin{itemize}
\item
J.~A. Poluikis and R.~D. Hill, ``Completely positive and Hermitian-preserving linear transformations,'' {\em Linear Algebra and Its Applications}, vol.~35, pp.~1--10, 1981.

\item
G.~P. Barker, R.~D. Hill, and R.~D. Haertel, ``On the completely positive and positive-semidefinite-preserving cones,'' {\em Linear Algebra and Its Applications}, vol.~56, pp.~221--229, 1984.

\item
L.~J. Landau and R.~F. Streater, ``On Birkhoff's theorem for doubly stochastic completely positive maps of matrix algebras,'' {\em Linear Algebra and Its Applications}, vol.~193, pp.~107--127, 1993.
\end{itemize}

Is everything settled now for your visit?  Can you send me the schedule of when you'll be arriving?  We're gonna have to find a place for you to live.

\section{25-03-02 \ \ {\it Company from Hybernia} \ \ (to A. Plotnitsky)} \label{Plotnitsky5}

This is just a small note to tell you I'll be traveling with you today.  I'm in Dublin right now, but soon to go home.  I was in a bookstore yesterday, and I ran across a copy of Paul Feyerabend's posthumous book {\sl Conquest of Abundance}.  Seeing your recommendation printed on the back pushed me over the edge and I bought it!

\section{26-03-02 \ \ {\it Popper and Detractors} \ \ (to R. Pike)} \label{Pike7}

\brp
From Adam Gopnik's article about Karl Popper in the latest {\bf New Yorker}:
\bq\noindent\rm
The reason science gave you sure knowledge you could count on was
that it wasn't sure and you couldn't count on it.  Science wasn't the
name for knowledge that had been proved true; it was the name for
guesses that could be proved false.
\eq
\erp

Thanks, I ought to have a look at the full article.

Here's the way I put my own take on exactly the same subject in a letter to Bill Wootters recently.  [See 25-02-02 note ``\myref{Wootters7}{A Wonderful Life}'' to W. K. Wootters.]

\section{27-03-02 \ \ {\it Still More Zing!}\ \ \ (to J. Gea-Banacloche)} \label{GeaBanacloche1}

Thank you for the long, beautiful note.  I loved it, and it makes me
so wish that I had had enough time to get to the rest of my talk in
Dublin---i.e., the part where I try to give some substance to the
word ``Zing!''~I introduced on one of the early slides.  (Recall,
``Zing!''~was meant to be a place holder for the answer to the
question ``What is real about a quantum system?'')

\bjgb
You seem to have a pretty good idea of how to make most of the
postulates (on probability, tensor space structure, and even
wavepacket reduction) flow in a more or less natural way from some
reasonable information-theoretic ideas, once you are given the basic
formal structure of operators and Hilbert spaces. The main question
would appear to be, where does this formal structure come from, and
what does it actually say about the physical universe?
\ejgb

That is the main question.  And---{\it in spirit}---I believe our
proposed answers appear to be essentially the same:
\bjgb
the basic physical fact at the heart of quantum mechanics is the
uncertainty principle, which one could formulate very generally as
follows--

(P1) The nature of quantum mechanical systems is such that, even when
we have all the possible information we can have about them, we
cannot, in general, predict the outcomes of all the possible
experiments we could carry on them.  Specifically,
\ejgb

The only issue in my mind is how to carry out the word
``specifically'' in a way that would satisfy the aesthetic I'm
seeking.  In particular, I would really like to pin down where the
noncommutativity comes from in a way that does not make a priori use
of the notions of ``compatibility'' and ``incompatibility.''  In
other words, I'd even like ``incompatibility'' to be a secondary
notion, rather than a primary one.  I think it is possible, and my
present thinking is that it can be made to come out in a pretty way
as a natural consequence of the mismatch between the number of bits
that can be reliably stored in a quantum system and the number of
measurement outcomes required for an ``informationally complete''
representation of a quantum state.  (That is to say, something along
the lines of the mismatch between the numbers $N$ and $K$ in Lucien
Hardy's treatment in \quantph{0101012} and \quantph{0111068}.)

In any case, I have a discussion of ``Zing!''~in several places in my
paper \quantph{0106166}.  Especially in the closing section.
Since then, I've come quite a way toward what I was trying to express
above, but you'd have to dig harder for that---it's not exactly
published properly yet.  The place to look at the moment is in the
file titled ``Quantum States:\ What the Hell are They?''~posted on my
webpage (link below).  The upshot is the following (working)
statement:  Each quantum system can be postulated to have an
intrinsic amount of ``sensitivity'' to our experimental interventions
upon it, and that sensitivity can be captured by a single parameter
$d$ (call it the dimension).  From that, everything about the algebra
of observables for a system follows from a basic statement about the
very meaning of Bayes' rule in that context.

I'm working hard to get some of this in a proper paper presently.

By the way, I loved your Teilhard de Chardin quote:
\bq\noindent
The history of the living world can be summarized as an elaboration
of ever more perfect eyes within a cosmos in which there is always
something more to be seen.
\eq
In my own way, I tried to express something similar in two pieces
that I've come to be pretty proud of.  Have a look at my letter to
John Preskill titled ``\myref{Preskill6}{Psychology 101}'' starting on page 143 and my
letter to Bill Wootters titled ``\myref{Wootters7}{A Wonderful Life}'' starting on page
149 of the file I mentioned above.

I hope that you yourself made it home safely and comfortably from
Madrid, and also that you found your family doing well there. Emma's
chicken pox are already clearing up:  So maybe I was gone just the
right amount of time!

\section{27-03-02 \ \ {\it Invited Submissions} \ \ (to J. H. Shapiro)} \label{Shapiro1}

Attached below are two pdf files.  One, titled {\tt QCMC02.pdf}, contains the
abstract to my invited submission.  The other, titled {\tt honest.pdf},
represents a joint (contributed) submission with Patrick Hayden.  I hope
that the organizing committing will still consider it a valid
contributed submission even though we have missed the deadline.  (The
missing of the deadline was entirely my fault, as I somehow got the
impression that all submissions were due April 1.)

In any case, of the second submission, Hayden would give the actual
presentation.  Just to let you know, Patrick Hayden is presently a Prize
Postdoctoral Fellow at Caltech and one of the young movers of quantum
information.  (He also had a Rhodes Scholarship for his graduate work at
Oxford, and has recently been offered a faculty position at McGill
University in {\Montreal}.)  So, having him there, would certainly do well
for the conference's reputation.

\section{28-03-02 \ \ {\it Men in Power} \ \ (to P. Hayden)} \label{Hayden1}

Sorry to get back to you so late.  (If you haven't discovered by now, it's a habit with me.)  The trouble was, I was in Ireland last week and part of this week, and, of course, I got carried away with all the conversations of the moment.

Anyway, I finally sent in the abstracts yesterday.  Here's the reply I got from Jeff Shapiro today:
\bjhs
Thanks for submitting your title and abstract in such a timely
fashion.  As for the ``\,{\tt honest.pdf}'' paper, I am sorry but I feel that I
must be absolutely bureaucratic and reject it.  We have 182
submissions for QCMC'02 as it is, and even with posters this number is
considerably more than we'll be able to accept.  I have already
rejected several late submissions, and it would be unfair of me to
create an exception for you and Patrick Hayden.
\ejhs

I am sorry about this.  Maybe just maybe if I had gotten off my duff and read your email before just leaving for Ireland \ldots

In any case, I do think we should pursue doing something for real this summer.  I'll try to write you soon about scoring function issues.

Below, let me attach the abstract I actually sent Shapiro.  Mainly I made only the most minor of changes so that it would fit in two pages:  I.e., I killed the last section and took out the associated reference.  Also I shortened the acknowledgements just a tad.  Stylistically, I changed all the ``non-''s to ``non''s (it's a pet peeve of mine) and changed the title to ``Keeping the Quantum Experimentalist Honest''.  Oh yeah, I also used a \verb+\Big+ in Eq.\ (\ref{Bundley}).  And I think that's it.

Like I say, I've got a lot more to write to you about scoring functions, but let me give you the short of it.  The main reason I'm attuned to the issue has to do with some of my recent debates with Caves and {\Schack} concerning the strength of axioms one should assume in a Dutch-book foundation for probability theory.  I say the axioms should not be so strong as to keep an agent from lying (no matter what the case).  This is important when one considers concatenating probabilities for various events.  Something of an account of where I last left the story can be found on my webpage in the file ``Quantum States:\ What the Hell Are They?''.  The most relevant part appears in a letter titled ``\myref{Caves54}{The Commitments}'' starting on page 133.  A link to my webpage is below.

Again, I'm sorry about the QCMC thing, and I'm sorry you spent the time writing the abstract for nothing.  But maybe still something good will come out of this in the end.

\bq
\begin{center}
{\bf \large Keeping the Quantum Experimentalist Honest}\smallskip

Christopher A. Fuchs and Patrick Hayden
\end{center}

\subsubsection{von Neumann entropy and the honest expert}

There are a number of compelling axiomatic characterizations
[AczelD75, OhyaP93] of the von Neumann entropy $S(\rho) = -{\rm
Tr} \rho \log \rho$ of a density operator $\rho$ and the quantum
source coding theorem provides what is certainly its most important
functional characterization
[Schumacher95, JozsaS94, BarnumFJS96].  Here we give what could
be the \emph{simplest} characterization of the von Neumann entropy,
generalizing an idea of Acz\'{e}l's for characterizing the Shannon
entropy [Aczel80].  The approach is to show that the entropy
arises as the optimal expected payoff in a type of game between a
cash-strapped experimentalist and her employer. Thus, it provides a
meaningful interpretation of the von Neumann entropy in a completely
nonasymptotic setting, when only one realization of the density
operator $\rho$ is available.

Alice, an ambitious scientist attempting to build a quantum
computer, manages to produce states that, to the best of her
knowledge, are described by the density operator $\sigma$.  She
then sends a note to her employer, Bob, saying that she has
succeeded in producing the state $\rho$.  Bob, as a conscientious
scientific administrator, would like to ensure that Alice does not
lie about her progress. In other words, he would like to guarantee
that $\rho = \sigma$. Therefore, from time to time Bob will visit
Alice's lab and perform a measurement, conditioned on the data
$\rho$ that she sent him.  Her future funding will depend on the
outcomes of these measurements. The question is, what measurement
should Bob perform and how should he structure his payments to
Alice such that she never has any incentive to deceive him?

We propose the following strategy for Bob.  He should perform a
measurement in the basis $\{\ket{e_i}\}_{i=1}^n$ that diagonalizes
$\rho = \sum_i r_i \proj{e_i}$ and, upon getting outcome $i$, pay
Alice $C +  D \log n r_i$ dollars, for nonnegative constants $C$ and
$D$. Notice that if $\rho$ described a maximally mixed state then
$r_i = 1/n$ and, regardless of the outcome, Bob pays Alice $C$
dollars, ensuring that she will be able to support herself even if
her lab produces completely random states.

Now, assume to start that Alice has prepared a state $\sigma$ such
that $[\sigma,\rho]=0$. We can then write $\sigma = \sum_{i=1}^n
s_i \proj{e_i}$ and the expected payment from Bob to Alice is
\begin{equation}
C + D \log n + D \sum_{i=1}^n s_i \log r_i,
\end{equation}
which is maximized if and only if $s_i = r_i$ for all $i$, giving
an expected payment of
\begin{equation}
C + D \log n - D S(\sigma).
\end{equation}
Thus, under the assumption that Alice reports a state $\rho$ that
commutes with her $\sigma$, she maximizes her payment by choosing
$\rho = \sigma$.  Moreover, there is a built-in incentive for her
to try and produce pure states since she is penalized by an
expected amount $D S(\sigma)$.

Now consider what happens in the general case.  The probability
that Bob will measure outcome $i$ is $p_i = {\rm Tr}( \sigma
\proj{e_i})$ and the expected payment is
\begin{equation}
C + D \log n + D \sum_{i=1}^n p_i \log r_i.
\end{equation}
Again we find that
\begin{eqnarray}
- \sum_{i=1}^n p_i \log r_i
&\geq& - \sum_{i=1}^n p_i \log p_i \\
&=& H(p_1,p_2,\ldots,p_n) \\
&\geq& S(\sigma),
\end{eqnarray}
where the second inequality holds because the output entropy of a
complete projective measurement is always at least as large as the
entropy of the input density operator.  (See, for example, Ref.\
[NielsenC].) Thus, the expected payment is
\begin{eqnarray}
&\,& C + D \log n + D \sum_{i=1}^n p_i \log r_i \\
&\leq& C + D \log n - D H(p_1,p_2,\ldots,p_n) \\
&\leq& C + D \log n - D S(\sigma).
\end{eqnarray}
Equality in the first line occurs if and only if $p_i = r_i$ for
all $i$ and in the second line if the measurement is a complete
projection in the eigenbasis of $\sigma$.  (Again, see Ref.\
[NielsenC].) Therefore, under the assumption that the payment
as a function of the outcome probabilities is unique, the proposed
measurement strategy is the unique solution to the problem of
keeping Alice honest.

The following theorem of Acz\'el's guarantees this last fact for
density operators with $n\geq 3$. It is notable for the extremely
weak assumptions made of the payment function.

\begin{theorem}[Aczel80]
\label{Ithm:Aczel}
Let $n\geq 3$.  The inequality
\begin{equation}
\sum_{i=1}^n p_i F_i(q_i) \leq \sum_{i=1}^n p_i F_i(p_i)
\end{equation}
is satisfied for all $n$-point probability distributions
$(p_1,\ldots,p_n)$ and $(q_1,\ldots,q_n)$ if and only if there
exist constants $C_1,\ldots,C_n$ and $D$ such that
\begin{equation}
F_i(p) = D \log p + C_i,
\end{equation}
for all $i = 1,\ldots,n$.
\end{theorem}

\subsubsection{The efficient gambler and accessible information}

The accessible information ${\rm Acc}({\cal E})$ in an ensemble of states
${\cal E} = \{ \rho_i; p_i \}$ is defined to be the maximum over
all possible POVM measurements $\{M_j\}$ of the mutual information
$I(i:j)$, where $p(j|i) = {\rm Tr}(\rho_i M_j)$.  Despite the enormous
effort that has been spent studying the accessible information [Fuchs95],
however, justification for the definition remains unclear.  One motivation,
which we won't describe in detail, is via asymptotic coding
[HausladenJSWW96, Holevo98, SchumacherW97].  The approach
yields the accessible information as the maximal rate at which bits
can be sent using quantum codewords whose marginal distribution is given
by ${\cal E}$ but only if severe restrictions are imposed on
the encoding and decoding. Namely, the codewords must be product states
and the decoding must be performed by product measurements.  If the second
restriction is relaxed, it is the Holevo $\chi$ quantity
\begin{equation}
\chi({\cal E}) = S\Bigl(\sum_i p_i \rho_i\Bigr) - \sum_i p_i
S(\rho_i)
\label{Bundley}
\end{equation}
which answers the coding question, not the accessible information.
Thus, the delicacy of the asymptotic problem provides yet another
incentive for finding a single--realization interpretation of the
accessible information.

Such a realization exists, as the quantum analog of an old idea of
Kelly's [Kelly, CoverT].  We imagine that Alice is a bookie who takes
bets on the mutually exclusive events $\{i\}$, which occur with
probabilities $p_i$.  For simplicity, we assume that Alice gives
fair odds and does not collect a fee, so that if the gambler Bob
wagers $S_i$ on event $i$, he is paid $S_i / p_i$ when $i$ actually
occurs.  Now suppose that Bob is given private information about which
outcome occurred prior to it becoming public knowledge, so that he
could still place bets at the original odds.  Moreover, imagine that
Bob's source of information is a \emph{noisy quantum channel} which
outputs $\rho_i$ when event $i$ occurred.  Bob's task, therefore, will
be to perform some POVM measurement $\{M_j\}$ and then use the
information gained from the measurement to make a wager, all done
in such a way as to
maximize the exponential rate of growth of his capital
\begin{equation}
G = \lim_{n \rightarrow \infty} \frac{1}{n} \log \frac{C_n}{C_0},
\end{equation}
where $C_0$ is his initial capital and $C_n$ his capital after $n$
rounds of betting.  We show that this maximal rate of growth is the
accessible information: $G_{max} = {\rm Acc}({\cal E})$.

\subsubsection{Acknowledgments}
We thank Howard Barnum and Simon Benjamin for helpful discussions.
P.H. was supported by a Sherman Fairchild Fellowship and US National
Science Foundation grant EIA--0086038.

\begin{enumerate}

\item
J.~Acz\'{e}l.
\newblock A mixed theory of information. {V}. {H}ow to keep the (inset) expert
  honest.
\newblock {\em Journal of Mathematical Analysis and Applications}, 75:447--453,
  1980.

\item
J.~Acz\'{e}l and Z.~Dar\'{o}czy.
\newblock {\em On measures of information and their characterizations}, volume
  115 of {\em Mathematics in science and engineering}.
\newblock Academic Press, New York, 1975.

\item
H. Barnum, C.~A. Fuchs, R. Jozsa and B. Schumacher.
\newblock General fidelity limits for quantum channels.
\newblock {\em Phys. Rev. A}, 54:4707--4711, 1996.

\item
T.~M. Cover and J.~A. Thomas.
\newblock {\em Elements of Information Theory}.
\newblock John Wiley \& Sons, New York, 1991.

\item
C.~A. Fuchs.
\newblock {\em Distinguishability and accessible information in quantum
    theory}.
\newblock PhD thesis, University of New Mexico, 1995.

\item
P. Hausladen, R. Jozsa, B. Schumacher, M. Westmoreland and W.~K. Wootters.
\newblock  Classical information capacity of a quantum channel.
\newblock {\em Phys. Rev. A}, 54:1869, 1996.

\item
A.~S. Holevo.
\newblock The capacity of a quantum channel with general signal states.
\newblock {\em IEEE Trans. Inf. Theory}, 44:269--273, 1998.


\item
R.~Jozsa and B.~Schumacher.
\newblock A new proof of the quantum noiseless coding theorem.
\newblock {\em J. Mod. Opt.}, 41:2343--2349, 1994.

\item
J.~L. Kelly, Jr.
\newblock A New Interpretation of Information Rate.
\newblock{\em Bell System Technical Journal}, 35:917--926, 1956.

\item
M.~A. Nielsen and I.~L. Chuang.
\newblock {\em Quantum computation and quantum information}.
\newblock Cambridge University Press, Cambridge, 2000.

\item
M.~Ohya and D.~Petz.
\newblock {\em Quantum entropy and its use}. Springer-Verlag,
Berlin, 1993.

\item
B.~Schumacher.
\newblock Quantum coding.
\newblock {\em Phys. Rev. A}, 51:2738--2747, 1995.

\item
B. Schumacher and M.~D. Westermoreland.
\newblock Sending classical information via noisy quantum channels.
\newblock {\em Phys. Rev. A}, 56:131--138, 1997.

\end{enumerate}
\eq

\section{28-03-02 \ \ {\it And a Short Reply} \ \ (to R. Campos)} \label{Campos1}

Thank you so much for your kind letter.  I just printed out your PLA paper, and I look forward to understanding it.  [R. A. Campos and C. C. Gerry, ``A Single-Photon Test of Gleason's Theorem,'' Phys.\ Lett.\ A {\bf 299}, 15--18 (2002).] Sorry it has taken me so long to reply, but I was tied up with a meeting in Dublin from the 19th through the 26th and then it took me still a couple days beyond that to recover from the travel.

Just a couple of very quick remarks on your manuscript.  1) In your second sentence, you say ``Starting from the description of physical states as $\hat\rho$ as vectors in a Hilbert space, and observables as projectors $\hat\Pi$ in that space, Gleason proved that the \ldots\@''  That however, is a little inaccurate and downplays Gleason's achievement.  Gleason does not start at all with states as vectors in a Hilbert space.  He only starts with the observables that you mention and then, in the process of his proof, {\it derives\/} that the states can be represented as density operators.  Thus, in a sense, he gets the states for free.  In any case, from the rest of the paper it looks that you do understand this.  I'm just pointing out that the wording of the introductory paragraph of the paper is misleading.  2) In reference 5, you write C. C. Caves instead of C. M. Caves.

As I say I really haven't had a chance to understand your paper yet, but I will indeed study it.  How could I but help to with so many flattering remarks in it about me!

You ask about simplifications to the standard Gleason theorem.  There are two places to look (one you've already known):
\begin{itemize}
\item
R.~Cooke, M.~Keane, and W.~Moran, ``An Elementary Proof of Gleason's Theorem,'' Math.\ Proc.\ Camb.\ Phil.\ Soc.\ {\bf 98}, 117--128 (1981)
\item
I.~Pitowsky, ``Infinite and Finite Gleason's Theorems and the Logic of Indeterminacy,'' J. Math.\ Phys.\ {\bf 39}, 218--228 (1998).
\end{itemize}
Unfortunately I don't have copies of either of these papers any more.

However, there is a far greater simplification than the above, which can be found if one is willing to embrace the notion that POVMs can be taken as a basic concept of measurement in quantum mechanics.  I explain this in detail in my paper \quantph{0106166} and prove a version of Gleason's theorem in an absolutely trivial way.  (Also you can pick it up off my webpage:  the link for it is below.) In particular the proof by this method works also for 2-D vector spaces and even vector spaces over the field of rationals, in both cases of which the standard Gleason theorem fails.

Certainly keep up the good work.  This quantum mechanics is a wonderful thing.

\section{28-03-02 \ \ {\it Quantum-Information Information} \ \ (to R. Duvenhage)} \label{Duvenhage1}

I just want to write you a very short note to tell you how much I
enjoyed your paper \quantph{0203070}, ``The Nature of
Information in Quantum Mechanics'' and to express how much similarity
I think I see between our points of view.  In particular, I think you
expressed some things so very clearly that I would love to co-opt
your phrases!

Here's where I think we agree the most:
\begin{enumerate}
\item
``A measurement is by definition the reception of information by the
observer.''
\item
``This renders many problems surrounding the measuring process in
quantum mechanics no more difficult than in classical physics.'' And
consequently,
\item
Your discussion of the Heisenberg cut.
\end{enumerate}

You can find some reflection of these ideas in my own paper \quantph{0106166}, though not put quite so succinctly as in yours. In
particular I'm thinking of my discussion on pages 27 and 28 of that
paper, following the earlier discussion on page 11. Also, I agree
with your point about B's information being ``invalidated'' in your
discussion on page 13 of your paper. Similar ideas make an appearance
on pages 39 through 41 of mine. Finally, I also enjoyed your
discussion of the linearity of time evolutions. That was the sort of
thing, I was trying to describe in my notes of 22, 26, and 27
September 1999 on pages 408, 284, and 285, respectively, of my
samizdat \quantph{0105039}.

However, I think I've come a long way since that paper and those
notes.  In particular, I think I'm no longer really willing to say
that ``quantum collapse is a noncommutative [generalization of]
conditional probability.''  I think there is a sense in which quantum
collapse is {\it precisely\/} an application of Bayes'
conditionalization rule, modulo only a final redefinition of what the
posterior probability is relevant to.  What I mean by this in more
detail can be found in \myref{Hardy7}{my letter to Lucien Hardy}
beginning on page 159 of my collection ``Quantum States:\ What the
Hell Are They?''\ (which can be found on my webpage, link below),
especially toward the end of the letter.  In fact, I'm presently
striving to write that up in an updated version of \quantph{0106166},
and I hope to place it on the server soon.

Anyway, I really want to point out the similarities in our thoughts
rather than the contrasts.  I think you've done the physics community
a good service with your paper.  It's very well written, and a lovely
piece actually.

I really have the greatest hope and enthusiasm that we, the quantum information community, are on the verge of something very big in our understanding of quantum mechanics.  I'm glad to see some good young minds joining into the enthusiasm as well!

\section{29-03-02 \ \ {\it Building with Bayes} \ \ (to B. C. van Fraassen)} \label{vanFraassen3}

Sorry to take so long to acknowledge your note:  I've been running
around Ireland with some bad phone connections and probably a few too
many pints.

But I'm back now, and I did find your flyer waiting in my mailbox. I
tacked it to the wall, but the subject's not likely to attract any of
the physicists in my immediate vicinity.  I would like to bring a
visiting student with me though.  Her name is Petra Scudo, and her
present email address is {\tt scudo@techunix.technion.ac.il}.  She,
as things turn out, did some undergraduate work in Pavia under a guy
named Regazzini, who in turn was a student of Bruno de Finetti. While
Petra is visiting (for a month and a week), I'm going to have her
work on trying to pin down a kind of quantum de Finetti
representation theorem for time-evolution maps.  I.e., a theorem in
answer to the question, ``What is an `unknown' quantum operation?''
(In this context, the term ``quantum operation'' refers to the
generalization of unitarity that is common in quantum channel theory
--- namely, the trace-preserving completely positive linear map.
Sometimes people call it ``open-system dynamics'' but, from the
Bayesian perspective, it is little more (nothing more?)\ than a
noncommutative generalization of a conditional probability,
connecting as it does the prior (quantum-state) assignment to the
posterior (quantum-state) assignment.)

Technically, I've made a breakthrough of sorts recently.  I now know
how to think of quantum collapse as {\it precisely\/} an application
of Bayes conditionalization (importantly, followed by a redefinition
of which measuring instrument the posterior probability assignment is
relevant to).  I call this a breakthrough because until recently (for
instance see page 25 of my NATO paper, \quantph{0106166}), I
continued to think of collapse as a noncommutative {\it analogue\/}
of conditionalization.  But now, using a representation of the
quantum state similar to the one Hardy harps on, I can see that what
is going on is the true-blue thing (i.e., simply Bayes in disguise).
At the moment I'm working hard to get this written up in a sensible
way \ldots\ or at least give my readers a hint of it, until I can do
it properly.

By the way, in coming to all this, I've taken a more radical Bayesian
turn than I had expected at the outset.  I.e., though I started my
career in an Ed Jaynes kind of ``objective Bayesian'' camp, I'm now
finding myself in the left of the de Finetti camp and maybe a little
beyond that.  In case it interests you, I've documented a lot of this
transition in a new samizdat which I've placed on my webpage. The title of the file is ``Quantum States:\ What the Hell Are They?'' and it contains a lot of
new correspondence with {\Mermin} and others along these lines. (As an
aside, I've significantly revamped my webpage; it's not so minimalist
anymore, and maybe thus a little more attractive.)

While I'm here and I've invested this much into a long letter to you,
let me make the thing even longer by tacking on two pieces from the
above-said collection.  I'll put them immediately below---one is a
letter to John Preskill and one a letter to Bill Wootters.  Both
letters should be self-contained.  [See 18-02-02 note
  ``\myref{Preskill6}{Psychology 101}'' to J. Preskill and 25-02-02
  note ``\myref{Wootters7}{A Wonderful Life}'' to W. K. Wootters.]
Anyway, I place them here because I had forwarded them to Henry Folse
a while ago, and he wrote me back a rather excited letter saying that
I'm starting to explore some philosophical ground not so dissimilar
from where Bas van Fraassen has gone.  I wish I were in a position to
judge the validity of that!  But I just haven't read enough yet.  (I'm
trying to, believe me.  But, being a physicist, I've got a lot of
material to catch up on.)  Anyway, until then, I'll keep my fingers
crossed that maybe I can get some reaction directly from the horse's
mouth. Are there similarities between our views?  And what other
pieces of your work should I be reading if there are?

Finally, concerning your seminar with Elga, I'd love to attend!  So,
please do keep me on the list.

\section{29-03-02 \ \ {\it Historical Rampage}  \ \ (to A. Peres, N. D. {\Mermin}, A. Cabello, H. J. Folse, and A. Plotnitsky)} \label{Mermin62.1} \label{Plotnitsky5.1} \label{Peres24}

I am finally sitting down to write an introduction to the {\Vaxjo} proceedings, and I am wondering if I can bother any of you to give me some help in this regard.  The help I would like is related to a paragraph I just wrote:
\bq
Quantum theory in its full-fledged form has been with us for 75 years.
Yet in a very real sense, the struggle for its formation remains.  Indeed, not a year has gone by since 19??\ when the world has not seen at least one meeting of physicists or philosophers devoted either explicitly or implicitly to the foundations of quantum theory.  Our very meeting in {\Vaxjo}, ``Quantum Theory:\ Reconsideration of Foundations,'' is just one of a long lineage of meetings with this tormented soul.
\eq
What I would like to do is back up that claim with some substance, and also make the year ``19??''\ look as dramatic as possible.

Here is the way you can help me if you have some time.  Please just send me a list of the meetings you are aware of.  What I would like is 1) the title of the meeting, 2) the location of the meeting, and 3) its date.  Even if you only have partial information for any of these things, I would still like you to send me what you have:  it may give me enough clues to piece together the rest.  (Furthermore, if you know of any good resources that may help me in this quest, please let me know.)

For instance, just in looking in my own CV I have been able to dig up the list below.

For whatever help you can give, I will thank you lavishly in the article's conclusion!

\begin{itemize}
\item
Workshop on Squeezed States and Uncertainty Relations, College Park, Maryland, March 1991
\item
Fundamental Problems in Quantum Theory: A Conference Held in Honor of Professor John A. Wheeler, Baltimore, Maryland, June 1994
\item
Sixth UK Conference on Conceptual and Mathematical Foundations of Modern Physics, Hull, England, October 1997
\item
Second Workshop on Fundamental Problems in Quantum Theory, Baltimore, Maryland, August 1999
\item
Chance in Physics: Foundations and Perspectives, Ischia, Italy, November 1999
\item
Quantum Foundations in the Light of Quantum Information and Cryptography, {\Montreal}, Canada, May 2000
\item
NATO Advanced Research Workshop on Decoherence and its Implications in Quantum Computation and Information Transfer, Mykonos, Greece, June 2000
\item
Quantum Theory: Reconsideration of Foundations, {\Vaxjo}, Sweden, June 2001
\item
10th UK Conference on the Foundations of Modern Physics, Belfast, Ireland, September 2001
\item
American Mathematical Society Meeting, Special Session on Quantum Logic, Atlanta, Georgia, March 2002
\end{itemize}

\section{31-03-02 \ \ {\it Progress Report}  \ \ (to A. Peres, N. D. {\Mermin}, A. Cabello, H. J. Folse, and A. Plotnitsky)} \label{Mermin62.2} \label{Plotnitsky5.2} \label{Peres25}

With the predominant help of {\Adan}, here's what I can show so far.  If any of you know of a way to fill in any of the gaps, that would be great.  (Also if you know the data for the double question marks ??, that would be wonderful.)  However, there's no need to waste your times sending me things from the years where I already have an entry.  (Of course, I only really want the list for dramatic effect, not for complete completeness.)

Tomorrow, when I have the resources of a library, I'll try to fill in as much as I can.  Of course it'd be great if I could get a continuous run ever since 1970 (or even earlier).

\begin{itemize}
\item
1970, International School of Physics Enrico Fermi.\ Course IL:\ Foundations of Quantum Mechanics, Varenna, Italy
\item
1971
\item
1972
\item
1973
\item
1974
\item
1975
\item
1976
\item
1977, International School of Physics Enrico Fermi.\ Course LXXII:\ Problems in the Foundations of Physics, Varenna, Italy
\item
1978
\item
1979
\item
1980
\item
1981
\item
1982
\item
1983, Foundations of Quantum Mechanics in the Light of New Technology, Tokyo, Japan
\item
1984
\item
1985, Symposium on the Foundations of Modern Physics:\ 50 Years of the Einstein--Podolsky--Rosen Gedankenexperiment, Joensuu, Finland
\item
1986, 2nd International Symposium on the Foundations of Quantum Mechanics in the Light of the New Technology, Tokyo, Japan
\item
1987, Symposium on the Foundations of Modern Physics 1987:\ The Copenhagen Interpretation 60 Years after the Como Lecture, ??
\item
1988, Bell's Theorem, Quantum Theory, and Conceptions of the Universe, George Mason University
\item
1989, Sixty-two Years of Uncertainty: Historical, Philosophical and Physical Inquiries into the Foundations of Quantum Mechanics, Erice, Italy
\item
1990, Symposium on the Foundations of Modern Physics 1990.\ Quantum Theory of Measurement and Related Philosophical Problems, Joensuu, Finland
\item
1991, Bell's Theorem and the Foundations of Modern Physics, Cesena, Italy
\item
1992, Symposia on the Foundations of Modern Physics 1992:\ The Copenhagen Interpretation and Wolfgang Pauli, ??
\item
1993, International Symposium on Fundamental Problems in Quantum Physics, Oviedo, Spain
\item
1994, Fundamental Problems in Quantum Theory:\ A Conference Held in Honor of Professor John A. Wheeler, Baltimore, Maryland
\item
1995, Fundamentos de la F\'{\i}sica Cu\'antica San Lorenzo de El Escorial, Spain
\item
1996, 2nd International Symposium on Fundamental Problems in Quantum Physics, Oviedo, Spain
\item
1997, Sixth UK Conference on Conceptual and Mathematical Foundations of Modern Physics, Hull, England
\item
1998
\item
1999, Second Workshop on Fundamental Problems in Quantum Theory, Baltimore, Maryland, August 1999
\item
2000, NATO Advanced Research Workshop on Decoherence and its Implications in Quantum Computation and Information Transfer, Mykonos, Greece
\item
2001, Tenth UK Conference on the Foundations of Modern Physics, Belfast, Ireland
\end{itemize}

\section{01-04-02 \ \ {\it Last Little Push}  \ \ (to A. Peres, N. D. {\Mermin}, A. Cabello, H. J. Folse, and A. Plotnitsky)} \label{Mermin62.3} \label{Plotnitsky5.3} \label{Peres26}

Thanks for all the help you've given me so far.  But now here's my goal:  to exhibit 30 full years of conferences.  I'm only missing entries for 1976, 1978, 1981, and 1982!!  So, if any of you have any faint memories about those years, I'd really appreciate any lead you can give me!

\begin{itemize}
\item 1972, Symposium on the Development of the Physicist's Conception of Nature in the Twentieth Century, Trieste, Italy
\item 1973, Foundations of Quantum Mechanics and Ordered Linear Spaces, Marbourg, Germany [See 15-05-02 note ``\myref{Peres34}{Marburg, Strasbourg, Blunderburg!}''\ to A. Peres.]
\item 1974, Quantum Mechanics, a Half Century Later, Strasbourg, Germany [See 15-05-02 note ``\myref{Peres34}{Marburg, Strasbourg, Blunderburg!}''\ to A. Peres.]
\item 1975, Foundational Problems in the Special Sciences, London, Canada
\item 1976
\item 1977, International School of Physics ``Enrico Fermi'', Course LXXII:\ Problems in the Foundations of Physics, Varenna, Italy
\item 1978
\item 1979, Symposium on Quantum Theory and Gravitation, New Orleans, Louisiana
\item 1980, Quantum Theory and the Structures of Time and Space, Tutzing, Germany
\item 1981
\item 1982
\item 1983, Foundations of Quantum Mechanics in the Light of New Technology, Tokyo, Japan
\item 1984, Fundamental Questions in Quantum Mechanics, Albany, New York
\item 1985, Symposium on the Foundations of Modern Physics:\ 50 Years of the Einstein--Podolsky--Rosen Gedankenexperiment, Joensuu, Finland
\item 1986, New Techniques and Ideas in Quantum Measurement Theory, New York, New York
\item 1987, Symposium on the Foundations of Modern Physics 1987:\ The Copenhagen Interpretation 60 Years after the Como Lecture, Joensuu, Finland
\item 1988, Bell's Theorem, Quantum Theory, and Conceptions of the Universe, Washington, DC
\item 1989, Sixty-two Years of Uncertainty: Historical, Philosophical and Physical Inquiries into the Foundations of Quantum Mechanics, Erice, Italy
\item 1990, Symposium on the Foundations of Modern Physics 1990.\ Quantum Theory of Measurement and Related Philosophical Problems, Joensuu, Finland
\item 1991, Bell's Theorem and the Foundations of Modern Physics, Cesena, Italy
\item 1992, Symposia on the Foundations of Modern Physics 1992:\ The Copenhagen Interpretation and Wolfgang Pauli, Helsinki, Finland
\item 1993, International Symposium on Fundamental Problems in Quantum Physics, Oviedo, Spain
\item 1994, Fundamental Problems in Quantum Theory:\ A Conference Held in Honor of Professor John A. Wheeler, Baltimore, Maryland
\item 1995, The Dilemma of Einstein, Podolsky and Rosen, 60 Years Later, Haifa, Israel
\item 1996, 2nd International Symposium on Fundamental Problems in Quantum Physics, Oviedo, Spain
\item 1997, Sixth UK Conference on Conceptual and Mathematical Foundations of Modern Physics, Hull, England
\item 1998, Mysteries, Puzzles, and Paradoxes in Quantum Mechanics, Lago di Garda, Italy
\item 1999, Second Workshop on Fundamental Problems in Quantum Theory, Baltimore, Maryland, August 1999
\item 2000, NATO Advanced Research Workshop on Decoherence and its Implications in Quantum Computation and Information Transfer, Mykonos, Greece
\item 2001, Tenth UK Conference on the Foundations of Modern Physics, Belfast, Ireland
\end{itemize}

\section{01-04-02 \ \ {\it Broadcasting and Petra} \ \ (to A. Peres)} \label{Peres27}

\bap
Only many years later there was the no broadcasting theorem, then some
better understanding of distinguishability, the proof that completely
positive maps did not improve distinguishability, etc.
Could {\bf you} please give me a brief survey with references?
\eap

That CPMs never create distinguishability in the sense of ``relative entropy'' between two states---this is perhaps the very strongest statement of the notion of ``distinguishability nonincrease'' ---was first proved by Armin Uhlmann.  Unfortunately, I cannot find that reference.\footnote{But 11 years later Blake Stacey could:  It is A. Uhlmann, ``Relative Entropy and the Wigner-Yanase-Dyson-Lieb Concavity in an Interpolation Theory,'' Commun.\ Math.\ Phys.\ {\bf 54}, 21--32 (1977).} However, I do know the paper is cited in Ohya and Petz's book {\sl Quantum Entropy and Its Use}.  The property to be looking for is called the monotonicity of relative entropy.

In terms of operationally measurable quantities (the relative entropy is not really one such thing), like the fidelity or the trace-norm distance, I think we, in our no-broadcasting paper were the first to prove the property in the former case.  Here's is the reference:
\begin{itemize}
\item
H.~Barnum, C.~M. Caves, C.~A. Fuchs, R.~Jozsa, and B.~Schumacher, ``Noncommuting Mixed States Cannot Be Broadcast,'' {\sl Physical Review Letters\/} {\bf 76}(15), 2818--2821 (1996).
\end{itemize}

With respect to the trace-norm distance, I don't know who was the first.  I've certainly done it independently of anyone else, but on the other hand, I've never published.  (And, actually, I don't know if anyone has ever published it.)

Better work on the no-broadcasting theorem, i.e., an algebraic proof, can be found in the work of G\"oran Lindblad:
\begin{itemize}
\item
``A General No-Cloning Theorem,'' Lett.\ Math.\ Phys.\ {\bf 47}, 189--196 (1999).
\end{itemize}
I am very proud of that reference because, in it, he called our proof of the no-broadcasting result ``ingenuous'' (though from the context he meant ``ingenious'')!  He set as his task to simplify the proof, though he ended up with a stronger theorem.

The very best work yet and the strongest (possible) result along these lines is due to Koashi and his boss Imoto, \quantph{010114}, ``What is Possible Without Disturbing Partially Known Quantum States?''.

I hope that helps you.  Incidentally, I knew that Ghirardi had rejected the Herbert paper.  He told me the story in Ischia in November of 1999.  Later, upon my request, he sent me all the documentation.  Unfortunately, I lost those papers in the Cerro Grande fire.

\bap
Please take good care of her. Her health is not very strong.
\eap

I will do my best to take care of her.  (And extract a good, lasting result in physics from her too!)

\subsection{Asher's Preply}

\bq
I found your ``rampage'' letter, and I'll look at home for all the
conference proceedings I have --- quite many! Now I'll need similar help
from you. When I was at the Trieste meeting, someone who was speaking
about the no cloning theorem said that the Herbert paper ``FLASH\ldots'' was
wrong. He repeated it so many times that this antagonized me. When time
came for questions, I said that I was a referee of that paper and I had
recommended its publication. Obviously it was wrong, because it violated
special relativity, but Herbert knew that very well. I wanted someone to
find what was actually wrong in the amplification mechanism that he
proposed, and I thought that a serious discussion would lead to progress.
Indeed it led!

Then I had a big surprise: Ghirardi came to the podium, with a
transparency that he had ready. He said that he was a referee of that
paper and had rejected it. His reason: the no cloning theorem that he had
proved first. The transparency was a letter from the editor of {\sl Found.\
Phys.}, certifying ``to whom it may concern'' that Ghirardi had proved the
theorem in his confidential referee's report. He never bothered to
publish it, until it was too late! I want to write all that in
my contribution to the proceedings.

However I also said that Ghirardi should not have rejected the paper,
because Herbert had no exact cloning. All he had in his laser was a mess,
perhaps described by a density matrix, and the no cloning theorem did not
apply. Only many years later there was the no broadcasting theorem, then
some better understanding of distinguishability, the proof that completely
positive maps did not improve distinguishability, etc. Could {\it you\/} please
give me a brief survey with references?

I was so excited after this exchange with Ghirardi that I had to take a
tranquilizer, something I do very rarely. Also, a young man came to me
saying: thank you for accepting that paper. The no-cloning theorem is my
main source of income.

The San Feliu conference was also very good. Petra was there, and soon
she will be going to you. Please take good care of her. Her health is not
very strong.
\eq

\section{02-04-02 \ \ {\it A Little Urgent} \ \ (to J. Bub)} \label{Bub6}

I'm trying to put together a dramatis personae for my latest lambasting of the existence of quantum foundations conferences---i.e., I'm trying to make a case that we should work toward the conference to end all conferences---and I would like to exhibit 30 full years of conferences in it.  And I'm only missing one!!  It's for the year 1978.  I wonder if you can help.

I discovered that you have a 1979 paper titled ``The Measurement Problem of Quantum Mechanics'' which appears in a book {\sl Problems in the Philosophy of Physics\/} (72d Corso), Bologna: \ldots\ Could this paper, by chance, have been due to a Conference Proceedings in 1978??  If so, can you give me the following things:  1) the title of the conference, and 2) where it was held.

Thanks, and if you can help me, I'd be ever so grateful.

\section{02-04-02 \ \ {\it Desist}  \ \ (to A. Peres, N. D. {\Mermin}, A. Cabello, H. J. Folse, and A. Plotnitsky)}  \label{Mermin62.4} \label{Plotnitsky5.4} \label{Peres28} \label{Cabello1.1} \label{Folse9.1}

Just in case any of you are still looking, I've now filled all the slots between 1972 and present.  So, you can wipe your brow and call it a good day's work.  Thanks to all of you for all the help.

\section{06-04-02 \ \ {\it The Early Morning} \ \ (to A. Plotnitsky)} \label{Plotnitsky6}

I laid in bed this morning, reading and thinking about your review of
Feyerabend's book.  What a pleasant way to wake up.  I think you are
right:  the most important issue is, what is the right balance of
construction?  It plays a role---I am convinced---but how big of a
role?

You flatter me by sending me your newest book!  How can you be so
productive, while all the rest of us just twiddle by?  I will cherish
the book and have it read before you know it.

\section{07-04-02 \ \ {\it Latour} \ \ (to A. Plotnitsky)} \label{Plotnitsky7}

\barkp
You got my point on Feyerabend exactly right.  Besides, it is, I
think, far more interesting, at least at this point, to think through
that which is unconstructible, or, if one prefers, could be ``constructed''
as irreducibly unconstructible.  (I am writing, among other things, a
review of Bruno Latour's book, where this is a central issue as well.)
\earkp
Please send me that review when you've gotten it written!

\section{07-04-02 \ \ {\it My (Hard-Earned) Pomposity} \ \ (to G. L. Comer)} \label{Comer9}

\bgc
Really?  There's nothing to be learned from it {\bf [Penrose's
book]}? I mean I read Hawking's thing, didn't agree with everything,
but did learn a few things.
\egc

Sure, there's plenty to be learned by way of mathematical ideas and
mathematical definitions and results.  It's his implicit philosophy
that I shun now, a philosophy that he's not even completely aware of
(I think).  And it's that philosophy (and the very troubles it
causes) that makes up the two books' whole reason to be. [\ldots]

Let me come back to this Penrose thing again.  It has to do with the
words ``rational,'' ``logical,'' and ``sane.''  Penrose is a believer
in a kind of other-worldly realm beyond the physical world called the
realm of mathematical truth.  In that realm logic and rationality are
defined in a timeless fashion, in a way independent of human frailty.

I've now come to think that that is hogwash.  In positing such a
thing, one misses the whole point of this wonderful world --- that it
is still under construction, and can be made {\it to some extent\/}
in the image we want to make it.  By this account, what is rational
and logical is not in what is timeless and ``right,'' but in what
gives us the most {\it long-term\/} survival value.  Survival value
so that we can give rise to the most progeny (genetically and
intellectually); survival value so that we can stand a chance to
shape things.  And that, as Darwin taught us, is not something set
intrinsically, but is a function of our cumulative environmental
pressures.

\section{08-04-02 \ \ {\it Continuing}\ \ \ (to G. L. Comer)} \label{Comer10}

\bgc
But it seems that we agree that there is something lacking in a world
that is based solely on Penrosian notions of rationality and logic.
I guess you are for extending their meaning, whereas I seem to be
keeping the standard understanding, but arguing that there is an
aspect of existence that transcends logic.
\egc

Yes, that is what it looks like to some extent.  But I guess I would say this:  In ``keeping the standard understanding'' (even in a restricted regime) you commit yourself to a usage of those terms that I reject in all regimes.  Being rational, when it works, is simply being ``Darwin-optimal'' by definition.  That is, rational for this day, for this age, for this set of circumstances, for the present set of selective pressures, and so forth.  It is having an ability to survive, spread, multiply, and {\it create\/}, that is the deeper, more all-encompassing concept.

\section{09-04-02 \ \ {\it Thanks} \ \ (to H. J. Folse)} \label{Folse10}

\bq
\noindent
{\underline{\bf IMPORTANT}:}  See the correction to the present note
in my note ``Doing It and Doing It Right,'' dated 15 April 2002, to
Henry Folse.
\eq

\bhf
\bq\rm\noindent
That is, the quantum state is a compendium of Bayesian ``beliefs'' or
``gambling commitments'' and is thus susceptible to the type of
analysis {\James} gives in his ``Sentiment of Rationality.''  Our
particular choice of a quantum state is something extra that we carry
into the world.
\eq
Whoa, this paragraph eludes my comprehension. I thought the
measurement chose the state of the outcome.
\ehf

That's a long story.  Yes, measurements can determine states, but my
latest greatest realization is that they determine states in a way
not so different from the way probability distributions $p(x)$ and
$p(y|x)$ determine $p(y)$.  $p(x)$ is the classical analog of the
initial quantum state.  $p(y|x)$ is the classical analog of the
projection operator in the collapse rule.  $p(y)$ is the classical
analog of the final state.  To the extent that in the Bayesian view
{\it all\/} probabilities---even conditional ones like $p(y|x)$---are
completely and utterly subjective, so are the meanings of the
outcomes of one's quantum measurement (i.e., so are the basis vectors
to which the collapses occur).

\section{10-04-02 \ \ {\it Evidence 2}\ \ \ (to G. Brassard)} \label{Brassard13}

I did say two pieces of evidence:  I almost forgot to go on to the second one.  The one below is an excerpt from a long(er) letter by Jeff Bub.  [See 10-04-02 note ``\myref{Bub7}{Bitbol-ization}'' to J. Bub.] I'll try to read up on Bitbol soon.  It sounds like he indeed embraces a position similar to ours.  (I.e., that QM may come out of the idea that the observer cannot be detached \ldots\ i.e., (hopefully) that QKD exists.)  So, the ``waiting list'' for our meeting grows.

But really, we need to give our more important attendants (i.e., the ones on the poster) some more concrete plans.

\section{10-04-02 \ \ {\it NSERC Funding} \ \ (to R. W. {\Spekkens})} \label{Spekkens5}

\brws
Incidentally, I was reading an old paper the other day which I thought
might resonate with you.  It is by Arthur Fine, ``Do correlations need
to be explained?'', in {\bf Philosophical Consequences of Quantum Theory},
edited by Jim Cushing.
\erws

I am aware of that paper, and indeed was mightily impressed with it for a while.  Have a look at my Samizdat, pages 373--379.  Ultimately, however, I shook myself of the hidden hand.  The only reason I'm a (subjectivistic) Bayesian is because of all the hard work I put into trying to be a propensitist.

\section{10-04-02 \ \ {\it Bitbol-ization} \ \ (to J. Bub)} \label{Bub7}

Sorry myself for taking so long to get back to you.  Bitbol sounds
interesting.  I've put in a good word for him to Gilles; we'll see
what happens.  I absolutely love that phrase ``blinding closeness.''

\bjb
Back to Bitbol and your earlier letter to me with replies to Preskill
and Wootters. The essential point there seems to me very close to the
neo-Kantian view about quantum mechanics that Bitbol has been
developing in several books and articles.
\ejb

At the risk of getting into deep waters with a philosopher, I would
say the point I'm pushing has a much deeper affinity to the
philosophical tradition of pragmatism ({\James} and {\Dewey}'s versions in
particular) than anything of a Kantian flavor.  But I've learned my
lesson about saying the words ``Copenhagen interpretation'' to
physicists.  (See the story on the bottom of page 70 in the big
Samizdat.)  And similarly I'm learning a lesson about saying the
words ``{\James},'' ``{\Dewey},'' and ``pragmatism'' in front of
philosophers. The reactions I've gotten from {\Timpson}, Brown, Donald,
and Butterfield!  Even when, in other contexts, I got such pleasant
reactions from them about the ideas I was talking about \ldots\
namely pragmatism without the explicit label.  Sometimes a few words
are worth a thousand (mental) fences.

\subsection{Jeff's Preply}

Apologies for being out of touch. I am now in my house in Quinson in Provence for another couple of weeks, after spending an interesting week in Paris, where I met a philosopher of quantum mechanics by the name of Michel Bitbol whom I think you should get to know (more of Bitbol later).

To answer your question about my paper \ldots

Back to Bitbol and your earlier letter to me with replies to Preskill and Wootters. The essential point there seems to me very close to the neo-Kantian view about quantum mechanics that Bitbol has been developing in several books and articles:
\bq
I think the solution is in nothing other than holding
firmly---absolutely firmly---to the belief that we, the scientific
agents, are physical systems in essence and composition no different
than much of the rest of the world.  But if we do hold firmly to
that---in a way that I do not see the Everettistas holding to it---we
have to recognize that what we're doing in the game of science is
swimming in the thick middle of things.  We're swimming in this
undulant sea, and doing our best to keep our heads above the
water:  All the concepts that arise in a physical theory must be
interpreted to do with points of view we can construct from {\it
within\/} the world.

That is to say, we have to loosen the idea that a physical law is a
mirror image of what ``is'' in the world, and replace it with something
that expresses instead how each of us can best cope with and hope to
take advantage of the world exterior to ourselves.

\ldots

The situation of quantum mechanics---I become ever more convinced---
illustrates this immersion of the scientific agent in the world more
clearly than any physical theory contemplated to date.
\eq

What seems to me closely related to Bitbol's position is your
emphasis on replacing the idea of a physical theory (you say `law')
as a mirror image of what is in the world (in what {\Pauli} called a
`detached observer' sense), with an opposing view that takes
seriously the fact that our science can only reflect points of view
we can construct from {\it within\/} the world. I agree with you that
this is the right way to look at quantum mechanics (although in my
book I argued for the `detached observer' view, which I identified
with an Einsteinian view).  Now I would say that quantum mechanics is
not so much a descriptive theory of new sorts of non-classical
objects (particles that are also wave-like, or particles with
properties that hang together in a non-Boolean way, for example), as
a theory of mechanics constrained by certain explicit limits to the
process of objectification. (So, as Bohr said, there are no quantum
objects. Quantum mechanics is not about how nature is, but about what
we can say about nature. In this sense, it's a mechanical theory at
the information-theoretic level, unlike classical mechanics -- and
the claim is that we are stuck at this level just because we are not
`detached observers.') Bitbol talks about the `blinding closeness' of
the world in the title to one of his books -- a much more apt image
than d'Espagnat's image of a `veiled reality.'

Here's a suggestion: Take a look at Bitbol's website (you'll find him just by looking up Michel Bitbol on Google). If you like what you see (most of his publications are in French, but there are some things in English you can download from his website), why not invite him to the {\Montreal} quantum commune meeting in the Fall? Note that he's more of a philosopher trying to make sense of quantum mechanics than someone who works with the formalism, but he does seem to know what he's doing.

\section{11-04-02 \ \ {\it Zing D} \ \ (to I. H. Deutsch \& C. M. {\Caves})} \label{Caves60} \label{DeutschI1}

I am finally writing up my contribution to the {\Vaxjo} conference
proceedings, and in it I plan to make a statement about which
elements of quantum theory I would be willing to call ``ontological''
if push came to shove.  As {\Carl} knows well, my favorite is the $D$.
That is, for each system, when I ascribe to it a Hilbert space of
some dimension $D$, what I am really doing is ascribing it an integer
parameter of some (potentially) ontological significance.  I do not
let myself, however, assume such a significance for the states {\it
in\/} the Hilbert space or the operators {\it on\/} it.

When I have that discussion, I would like to cite the stuff I saw
Ivan present at the ITP meeting.  (I saw it via the web.)  Do you
have paper written on that?  If so, what are its coordinates?  What
I'll say is something like:  If you're looking for the magic
ingredient that powers quantum computing, it's not going to be first
and foremost in the subjective elements of the theory.  It's going to
be in those things that stand a chance of being objective.

By the way, Ivan.  I remember seeing one slide where you talked about
the various points of view for just the question above---i.e., what
might give quantum computing its power.  In it you had one bullet
devoted to ``information-disturbance tradeoff.''  That was
flattering, but surely it must be such a minority opinion as to not
be an opinion at all!!  (However, you ought to know what I'm aiming
for:  that that and $D$ are the same thing, after all.  So thanks in
advance!)

\section{12-04-02 \ \ {\it Dogs and Bones} \ \ (to C. M. {\Caves} \& I. H. Deutsch)} \label{Caves60.1} \label{DeutschI1.1}

I'm just starting to read and understand your paper with Ivan and Blume-Kohout.  Already I'm struck at the philosophical level as I had guessed I would be.  (Of course, that's the easiest thing to do, especially on a first reading.)

Let me make a quick comment for the old dog inside me.
\bq\noindent
Whereas Hilbert space itself is an abstract construction, the number
of dimensions available to a system is a physical quantity that
requires physical resources.
\eq
Did I really have no influence on the chain of thought that led you to a phrase like this (and some of its sister phrases in the text of the paper)?

I know that I have touted the idea to you before that Hilbert-space dimension is a more solid kind of thing than the Hilbert space itself.  For instance I know that you've at least had the opportunity to read the letters:
\begin{itemize}
\item 22 Aug 01, to Grangier, ``\myref{Grangier3}{Contextual and Absolute Realities}''
\item 3 Oct 01, to Mermin, ``\myref{Mermin45}{Kid Sheleen}''
\item 3 Oct 01, to Caves--Schack, ``\myref{Caves30}{Replies on Practical Art}''
\end{itemize}
But, of course, having the opportunity to read something and reading it are two different things \ldots\ and I don't know which applies to you.  (All these things, by the way, are posted publicly at my website---link below---in the form of the file {\sl Quantum States:\ What the Hell Are They?}\ which I've advertised to a gazillion people by now.  Also there, you can find a letter to Hardy, of 6 March 02, ``\myref{Hardy7}{Poetry on Concrete},'' that gives the most complete statement of the idea yet.)

To my knowledge, I'm the only person I've ever seen argue for the reality of the dimensionality in opposition to the rest of the structure of Hilbert space.  (I put what I consider my prettiest formulation of my research program---which includes this argument---in the paper attached, and soon to be posted on {\tt quant-ph}.  [See ``The Anti-{\Vaxjo} Interpretation of Quantum Mechanics,'' \quantph{0204146}.]  There's a much bigger paper in the works, also for the {\Vaxjo} proceedings, which says it all more rigorously---that's the one I was referring to when I wrote you and Ivan yesterday for a reference---but it's a little further from complete.)

Anyway, I write all this not out of anger, but simply in hopes that you will give the material at my website and/or the paper below a little advertisement (for the explicit reasons above).  I.e., that you'll throw me a bone.  The point is a simple one, both for you and for me:  The Bayesian babblings we've been making and the thoughts they lead to can make a difference---a technical difference.  If by the grace of God, your paper makes it into {\sl Science\/} (or even only makes it into PRA), it will carry a kind of immediate weight that my more purely foundational efforts will take longer to establish.  And it might even attract a crowd that would not have been attracted otherwise.  On the other hand, if your readers dig into my website they will get a broader view of the support structure that underlies your application.

For the past six months, I've been preaching the reality of dimensionality to a variety of audiences.  And I think I've gotten exponentially more response on the Bayesian program we're aiming for than anything in the past.  (Sorry to say for your and {\Ruediger}'s sales skills.)  It seems that people find themselves much more willing to be attracted to the program if they can sink their teeth into a little reality at the same time.  I want the painting to be a minimalist one; you don't quite.  But these are details we can draw the community into thinking about if we do it right.

\section{13-04-02 \ \ {\it Quantum Locality} \ \ (to R. Garisto)} \label{Garisto4}

Thanks for sending me your draft; I will try to have a deeper look at
it after my present project lets me come up for a bit of air. But,
yeah, you're right, I can already tell I'm going to disagree with it.
``What is the speed of quantum information?''  It doesn't have a
speed.  It could only have a speed if you endow the state vector with
an objectivity it does not have.  QUANTUM STATES DO NOT EXIST (in and
of themselves).  They merely express the gambling commitments one is
willing to make when one encounters a quantum system.  But that does
not leave the world empty; it just means that the quantum state is
not part of its substance.  Why is it that I should choose my
gambling commitments to be in accordance with the structure of
quantum mechanics?  When we can answer that, we will have learned
something clean and simple about the substance of the world.  But
until then, encumbering the world with an idea---nonlocality---that
is clearly wrong-headed (without a heck of a lot of contrivance to
try to make it go) will only distract us from the straight course to
that great goal.  What is it that makes quantum mechanics go?  It is
something deeper and far more interesting than the quantum
state---that much I firmly believe.

I'm glad to hear that things are going well with you cancer-wise.
Every time I think of you, I cross my fingers mentally for your
health and happiness.

\section{13-04-02 \ \ {\it An Address} \ \ (to L. Hardy)} \label{Hardy8}

Can you send me the name and email address of the guy who invited you to the Wheeler conference?  I want to get in touch with him for a little help with regards that big reference document I'm putting together (``The Activating Observer'').

By the way, I think I've got a better Bayesian-styled argument for linearity than the one I sent you in the big note.  [See 06-03-02 note ``\myref{Hardy7}{Poetry on Concrete}'' to L. Hardy.] It allows me to conglomerate the three protagonists there into one.  I'll ultimately write that up and send it off to you, but probably not for a while.  By the way, I've got a summer student coming in from Cornell this summer, and I've decided on a technical project for him that has to do with connecting our (i.e., yours and mine) views of your paper.

\section{15-04-02 \ \ {\it Doing It and Doing It Right} \ \ (to H. J. Folse)} \label{Folse11}

I was just looking my last note to you over again, and I was appalled
at what I had written.  Namely, I really botched my description of
the classical analog to collapse.  How I could do that, I don't
know.  And I am ashamed of myself.

Here's the proper way to say it, and in fact the way I am just
writing it up for a paper:

\bq
In Section 8, ``{\bf What Else is Information?},'' I argue that, to
the extent that a quantum state is a subjective quantity, so must be
the assignment of a state-change rule $\rho\rightarrow\rho_d$ for
describing what happens to an initial quantum state upon the
completion of a measurement---generally some POVM---whose outcome is
$d$. In fact, the levels of subjectivity for the state and the
state-change rule must be precisely the same for consistency's sake.
To draw an analogy to Bayesian probability theory, the initial state
$\rho$ plays the role of an a priori probability distribution $p(h)$
for some hypothesis, the final state $\rho_d$ plays the role of a
posterior probability distribution $p(h|d)$, and the state-change
rule $\rho\rightarrow\rho_d$ plays the role of the ``statistical
model'' $p(d|h)$ enacting the transition $p(h)\rightarrow p(h|d)$. To
the extent that {\it all\/} Bayesian probabilities are
subjective---even the probabilities $p(d|h)$ of a statistical
model---so is the mapping $\rho\rightarrow\rho_d$. Specializing to
the case that no information is gathered, one finds that the
trace-preserving completely positive maps that describe quantum
time-evolution are themselves nothing more than subjective judgments.
\eq

\section{15-04-02 \ \ {\it Simple Darwinism} \ \ (to C. M. {\Caves})} \label{Caves60.2}

\bcc
Well, of course, you had some influence on this point---a considerable
influence, I would say, since I remember mentioning to you last fall that
you would like our paper because it takes this point of yours as a starting
point.
\ecc

Thanks for your nice letter.

\bcc
I would be happy to acknowledge your influence on---even your
ownership of---this point, but now we come to the sticking point.  I'm
not willing to reference your web site in an archival journal.  I
think references in archival journals should generally be reserved for
things that are permanently out there in the public domain, not just
available, but guaranteed to be available in perpetuity in the same
form.  Personal web pages don't do that for me.  So what are we left with.
\ecc

I will be pleased if you cite my anti-{\Vaxjo} paper.  I will give you the {\tt quant-ph} number (i.e., the impersonal webpage number), as soon as I get it posted.  Through it, anyone will be able to find my webpage, anyway.  Alternatively I might give you the {\tt quant-ph} number for my ``Quantum Mechanics as Quantum Information (and only a little more)'' if I get it posted in time.  Concerning ownership, people cannot own ideas any more than they can own children.  But of course, whatever will get the ideas across at the same time as helping me build a stabilized career, will not be shunned:  I'd love to have the security to raise more children.

I anticipated your sticking point with the ``and/or'' in my original note---and will live with it---but I think it is just silly.  It is a throwback to a time when the world looked less changing.  The predominant point of writing, as I see it, is to convey information, to give the readers something for their intellectual investment and to mold their thoughts to the extent one can with a rational argument.  Writing ``C. A. Fuchs, private communication'' might be timeless, but it gives the reader nothing; it only gives C. A. Fuchs something.  A webpage, however fleeting, at least gives the opportunity of an immediate information source.  (And it creates no burdens for the typesetter that he does not already have.)  Moreover, the more hits such a webpage gets, the greater the chance its contents will become archival in the sense of being disseminated to the hard disks and minds of the community.  An acid test is to ask yourself which kind of citation an investigative historian of a hundred years from now would prefer.

Looking at three of your recent papers, I wonder which of the websites will last the longest?
\begin{itemize}
\item
\myurl{http://www.nsf.gov/cgi-bin/getpub?nsf00101}
\item
\myurl{http://www.research.microsoft.com/dtas/}
\item
\myurl{http://bayes.wustl.edu}
\end{itemize}
The government agency's, the corporation's, or the personal webpage of the devotee?  It's hard to know these things in advance.\footnote{\editornote Prediction is difficult, particularly on the Internet.  As of March 2014, the government agency's page no longer exists, but it redirects to a working location; the corporation's page gives a blandly uninformative error message; and the devotee's personal webpage still endures.  The Microsoft Research group apparently changed from ``Decision Theory and Adaptive Systems'' to ``Adaptive Systems and Interaction''; see \myurl{http://research.microsoft.com/en-us/groups/adapt/}.}

\section{16-04-02 \ \ {\it Oh Grammarian} \ \ (to C. M. {\Caves})} \label{Caves61}

I don't think I ever got the rules for ``farther'' and ``further'' straight.  Which would you have put in the paragraph below (and why)?
\bq\noindent
The complete disconnectedness of the quantum-state change rule from anything to do with spacetime considerations is telling us something
deep: The quantum state is information. Subjective, incomplete information. Put in the right mindset, this is {\it not\/} so intolerable.  It is a statement about our world. There is something about the world that keeps us from ever getting more information than can be captured through the formal structure of quantum mechanics. Einstein had wanted us to look further---to find out how the incomplete information could be completed---but perhaps the real question is, ``Why can it {\it not\/} be completed?''
\eq
Thanks for (far?, fur?)\ your advice.

\subsection{Carl's Reply}

\bq
Get yourself a copy of Strunk and White, and read it cover to cover.  Here
they are on your question (p.\ 46--47):
\bq\noindent
Farther, further.  The two words are commonly interchanged, but there is a
distinction worth observing: {\it farther\/} serves best as a distance word,
{\it further\/} as a time or quantity word.  You chase a ball {\it farther\/} than the
other fellow; you pursue a subject {\it further}.
\eq
I'd say you want {\it further}, unless you are thinking of Einstein as literally
looking into the distance.
\eq

\section{16-04-02 \ \ {\it My Own Homeopathy} \\ (to C. H. {\Bennett}, T. A. Brun, C. M. {\Caves}, P.~Grangier \& N. D. {\Mermin})} \label{Mermin63} \label{Caves62} \label{Bennett14} \label{Brun5}

I thought of all of you with a smile as I was writing the footnote
below for an upcoming paper.  I'll send it to you now.  Maybe I
really do practice homeopathy.  (Wait till you see the undiluted
version.)

\bq
In the previous version of this paper, \quantph{0106166}, I
variously called quantum states ``information'' and ``states of
knowledge'' and did not emphasize so much the ``radical'' Bayesian
idea that the probability one {\it ascribes\/} to a phenomenon
amounts to {\it nothing\/} more than the gambling commitments one is
willing to make with regards to that phenomenon. To the ``radical''
Bayesian, probabilities are subjective all the way to the bone.  In
this paper, I start the long process of trying to turn my earlier
de-emphasis around (even though it is somewhat dangerous to attempt
this in a manuscript that is little more than a modification of an
already completed paper). In particular, because of the objective
overtones of the word ``knowledge''---i.e., that a particular piece
of knowledge is either ``right'' or ``wrong''---I try to steer clear
from the term as much as possible in the present version. The
conception working in the background of this paper is that there is
simply no such thing as a ``right and true'' quantum state.  In all
cases, a quantum state is specifically and only a mathematical symbol
for capturing a set of beliefs or gambling commitments.  Thus I
variously call quantum states ``beliefs,'' ``states of belief,''
``information'' (though, by this I mean ``information'' in a more
subjective sense than is becoming common), ``judgments,''
``opinions,'' and ``gambling commitments.'' Believe me, I already
understand full well the number of jaws that are going to drop by the
adoption of this terminology. However, if the reader finds that this
gives him a sense of butterflies in the stomach---or fears that I am
or will become a solipsist$^1$ or a crystal-toting New Age
practitioner of homeopathic medicine$^2$---I hope he will keep in
mind that this attempt to be absolutely frank about the subjectivity
of \underline{\bf some} of the terms in quantum theory is part of a
larger program to delimit the terms that actually can be interpreted
as objective in a fruitful way.
\eq

\begin{enumerate}
\item
P.~Grangier, private communication, 2001.
\item
C.~H. {\Bennett}, T.~A. Brun, C.~M. {\Caves}, and N.~D. {\Mermin}, various
vibes, 2001.
\end{enumerate}

\section{16-04-02 \ \ {\it Rushdie Quote} \ \ (to D. R. Terno)} \label{Terno3}

\bq
Scientists get angry when laymen misunderstand, for example, the
uncertainty principle. In an age of great uncertainties it is easy to
mistake science for banality, to believe that Heisenberg is merely
saying, gee, guys, we just can't be sure of anything, it's so darn {\it
uncertain}, but isn't that, like, {\it beautiful}? Whereas he's actually
telling us the exact opposite: that if you know what you're doing you
can pin down the exact quantum of uncertainty in any experiment, any
process. To knowledge and mystery we can now ascribe percentage points.
A principle of uncertainty is also a measure of certainty. It's not a
lament about shifting sands but a gauge of the solidity of the ground. \medskip

\hspace*{\fill} --- Salman Rushdie, {\sl The Ground Beneath Her Feet}
\eq

Thanks for the Rushdie quote.  It's quite nice, especially the last line:  ``It's not a lament about shifting sands but a gauge of the solidity of the ground.''

I think it's something a little in the spirit of what I wrote in my NATO paper:
\bq
The complete disconnectedness of the quantum-state change rule from anything to do with spacetime considerations is telling us something
deep: The quantum state is information. Subjective, incomplete information. Put in the right mindset, this is {\it not\/} so intolerable.  It is a statement about our world. There is something about the world that keeps us from ever getting more information than can be captured through the formal structure of quantum mechanics. Einstein had wanted us to look further---to find out how the incomplete information could be completed---but perhaps the real question is, ``Why can it {\it not\/} be completed?''
\eq

\section{18-04-02 \ \ {\it Urgent Reference} \ \ (to H. Barnum)} \label{Barnum4}

Can you send me some full references on where the idea of ``noncontextuality'' of the probability rule was introduced?  I think you told me it was Mackey.  Also, though, can you give me the reference, to the criticism of the idea:  I think you said something like Hilgevoord?

\subsection{Howard's Reply}

\bq
The easiest thing is probably to attach my draft for the
Cesena proceedings, which you may already have.
Refs.\ 6 and 7 are probably what you want.
\begin{itemize}
\item[{[6]}] G. W. Mackey, {\sl The mathematical foundations of quantum mechanics}, W. A. Benjamin, New York, 1963.
\item[{[7]}] R. M. Cooke and J. Hilgevoord, ``A new approach to equivalence in quantum logic,'' in Current issues in quantum logic, E. Beltrametti and B. van Fraassen, Eds., New York and London, 1981, Plenum.
\end{itemize}
I'll see if I can get the page nos.\ from my
copy of Cooke \& Hilgevoord (but they may be unnecessary since its
an edited vol, not a journal). They are 101--114.  Mackey's statement is
part of his Axiom II on page 63.  There is also an article which I haven't read,
``Quantum mechanics and Hilbert space'', {\sl American Mathematical Monthly}
{\bf 64}, pp 42--47 (1957), which was apparently ``expanded into a book'' (the
one I cited), so this is probably the original source for Mackey.  Note
that Cooke and Hilgevoord, while citing Mackey, attribute the idea of
probabilistic equivalence to Bohr, but I think they aren't specific about where (I can't find
the article right now).

As you probably know, I think there was also a lot of discussion of
noncontextuality in the papers around von Neumann's hidden variables
result, notably Bell's paper on it, but I can't remember the details.
\eq


\section{18-04-02 \ \ {\it Reference} \ \ (to F. E. Schroeck)} \label{Schroeck1}

Can you give me the full reference for the paper in which you introduced the idea of informationally complete POVMs?  (By full reference I mean title, journal, volume, beginning page number, ending page number, and year.)

\subsection{Frank's First Reply}

\bq
    The earliest reference I can find to informational completeness is in a
1977 paper by Prugove\v{c}ki. It is listed in my book {\sl Q.M. on Phase Space}.
The earliest reference in my own work is in a 1989 paper entitled ``An
Overview of Q. M. on Phase Space''. I'll have to look up a reference to see
where it was published! In this paper, I refer to it being present in two
papers of 1989 --- Busch and S., Found.\ of Phys.\ {\bf 19} (1989), 807--872; and
Brooke and S., Nuclear Phys.\ B (Proc.\ Suppl.)\ {\bf 6} (1989), 104--106.
I'll have to check these two references at home to see if ``Info.\ Compl.''\ is
even defined in them. I also refer to it being in a paper that was not
published, it being part of a package prepared for a grant; it was surpassed
by another paper. Anyway, I'll send you a copy of that paper (and all other
papers that I will mention here) in two days. The paper that surpassed this
is ``On Informational Completeness of Covariant Localization Observables and
Wigner Coefficients,'' D. M. Healy, Jr. and S., J. Math.\ Phys.\ {\bf 36} (1995), 453--507.

    The first(?)\ unambiguous reference of mine to it appeared in
``Unsharpness in Measurement Yields Informational Completeness,'' in
{\sl Symposium on the Foundations of Modern Physics 1990}, P. Lahti and P.
Mittelstaedt, eds., World Scientific, 1991. In this paper I talk about its
ramifications in quantum signal processing and natural biological systems.

    There were a whole series of papers that I wrote, with an occasional
collaborator, on representations of groups, pattern recognition, frames, \ldots

    And then there is my book.

    I'm going to send all this because the applications of info.\ compl.\ are
everywhere, and the references are sometimes hard to come by. You have a day
or two of reading ahead of you!

    I'll get back to you on the three early references.
\eq

\subsection{Frank's Second Reply}

\bq
    I found one that predates the one from 1990. It is in ``Coexistence of
Observables,'' F. E. S., Int.\ J. of Theor.\ Phys.\ {\bf 28} (1989), 247--262. I'll
send that one also. I presume you don't any longer need the page no's from
the paper mentioned below.

    I also found info.\ compl. mentioned in a later paper from 1989. (``On
the Reality of Spin and Helicity'') It is enclosed as well.

    So, there you have it!
\eq

\section{19-04-02 \ \ {\it Lovely Circuits} \ \ (to N. D. {\Mermin})} \label{Mermin64}

Is there a typo in your paper, third paragraph?  Two consecutive
``before''s.  If it's a clever construction, I didn't get it.

What do you use to draw your lovely circuits?  I'm contemplating
putting a figure of a circuit in my present paper, and I've never
taken that bold step before.

``What the story demonstrates is the ability of entangled states to
store interaction in a highly fungible form that need not be cashed
in until the need arises.''  In comparing entanglement to classical
correlation and making a point similar to yours, I once called
entanglement ``all-purpose correlation.''  This was great fun because
it set me up to mention ``Martha White's All-Purpose Flour,'' a
product I remembered from my youth.  What tickled me the most was
that I got to cite Lester Flatt and Earl Scruggs, who sang the
``Martha White Theme'' during commercial breaks at the Grand Ole
Opry. You might know them for the Beverly Hillbillies theme.  It's
the little things that keep me going, you know.

I'll tell you how if the paper stirs me.

\section{20-04-02 \ \ {\it Yes} \ \ (to T. Rudolph)} \label{Rudolph2}

It just turned 5:30 AM, and clearly I ain't got nothin' better to do.

This note is just to acknowledge that your question is a good one.  Can one give a primary quantum state assignment $\rho^{(n)}$ that captures the idea of $n$ draws from a fixed set $\{\rho_i\}$ of quantum states (distributed according to some probabilities $p_i$).  Let me call such a source, a Bernoulli source.

That is, the question is how to characterize the class of density operators that look like this
$$
\rho^{(n)}=\sum (p_{i_1} p_{i_2}\cdots p_{i_n}) \rho_{i_1}\rho_{i_2} \cdots\rho_{i_n}
$$
where the summation is over all strings $i_1 i_2 \cdots i_n$.  It would indeed be nice to have a characterization of the flavor of a de Finetti representation.

That is, in a Bayesian account, how can one take into acknowledge the receiver's state of knowledge is more than simply $\rho^{\otimes n}$, without inserting a man in the box?  Or is the only way to do it to explicitly insert a man in the box?  I'm not clearheaded enough right now.  But it might indeed be the latter.  See, for instance, in the case of quantum cryptography (B92, say), with respect to Eve there is always a ``man in the box''---it's Alice, the source of the states.  Similarly for all the other cases I've ever thought about:  They're always communication games, where there is explicitly a sender and receiver.  A sender sending messages (or a key), and a receiver (or eavesdropper) trying to decode them.

Concerning your $A$ matrix account yesterday, Howard Barnum has thought a little bit about that.  In particular he used the $A$ to define some notion of ``coherent superposition'' of density operators.  I'm not quite sure what he did with that, but it's written up as one of the chapters of his PhD thesis.

I'll cc this note to our friend Professor Schack, as he might have some interest in it too.

\section{20-04-02 \ \ {\it Wacky Paper \`a Moi} \ \ (to G. Brassard)} \label{Brassard14}

I'm writing another wacky foundations paper, and I want to give one of the section headings a title like, ``Le Bureau International des Poids et Mesures \`a Paris''  (to express the phrase ``International Bureau of Weights and Measures in Paris'').  My question to you is, should the opening ``Le'' be there in a title?  Also, am I safe assuming that the ``\`a'' translates to my ``in''?

My French is atrocious.  (In case you didn't know.)

\section{20-04-02 \ \ {\it Helping a Friend}\ \ \ (to A. Cabello)} \label{Cabello2}

Yesterday, it was brought to my attention that you had increased the size of your massive bibliography since its first posting.  I hadn't noticed.  What an amazing document!

But really, I write you to ask if you have a Bib\TeX\ version of that database?  One of our visiting grad students here, Frank Verstraete, had a hard disk failure on his laptop, and he very literally lost three years of work!  (Of course, he is an idiot for not backing it up---he had never done it, not even once.  But then again, I am an idiot for not grabbing my two-thousand pages of calculations as my wife and I fled Los Alamos.)  I'm just trying to think of ways to help him.  One thing he mentioned was that he had an extensive Bib\TeX\ file built up on quantum information things.

I noticed that your file is not written in Bib\TeX\ format.  But is there a version of it in which it is?  If so, it'd be nice if you could help him to this extent.

I am putting together a bibliographic project of my own---very different subject matter than yours.  But I'll certainly cite your lovely work in it, for the purpose of cross-connections.  I'm calling mine ``The Activating Observer:\  Resource Material for a Paulian--Wheelerish Conception of Nature.''  It's sitting at four hundred and something references right now, many of which I've got annotated extensively.  You can read the abstract of it at my new webpage (link below), but I don't have it posted yet.  I was hoping to get it complete for the second anniversary of the Cerro Grande Fire, but I'm almost surely going to miss that deadline.  Maybe I'll be able to do it for the 1.5 year mark.

\section{21-04-02 \ \ {\it No Bib\TeX\ Version, Sorry}\ \ \ (to A. Cabello)} \label{Cabello3}

Oh well; we tried.

\bac
By the way, I've recently realized that there was an international
symposium on ``Quantum Theory and Reality'' in Oberwolfach, in July
1966. The proceedings, edited by M. Bunge, were published by Springer-Verlag, NY, in 1967 (before I was born!).
\eac

I actually knew about that one.  We have its proceedings in our library here at Bell Labs.  I didn't work too hard on the things before 1972 because I decided that 30 years might be a good cut-off mark.

It would kind of be interesting to know when the earliest conference ever on quantum foundations was.  I know that there was at least one sometime in the 1950s.  Let me dig up the reference.  Here it is:  {\sl Observation and Interpretation: A Symposium of Philosophers and Physicists}, edited by S.~K\"orner (Academic Press, New York, 1957).

The earliest ever can probably be found in one of Jammer's books.

\section{21-04-02 \ \ {\it Walton's Mountain} \ \ (to G. L. Comer)} \label{Comer11}

I hope things are going smoother for you; I haven't heard from you in a while.  Myself, I'm writing an extension to my big NATO paper (for another conference), and that's been taking a heck of a lot of time.  Also, I've got two grad students visiting (one from Belgium, and one from Israel), and I've been having the most marvelous time watching them turn my little wacky ideas into solid calculations and theorems.  They're both exceptionally good, and a real pleasure.

Today is the gloomiest, most drab day as far as the weather is
concerned.  But I like that.  I always feel academic and reflective
on such days.  Kiki tells Emma, ``If nature doesn't supply Daddy with
a few clouds, he makes his own.''

Lucent has been doing awfully poorly again in the market.  [\ldots]
And the rumors are flying that about 10,000 more are going to get the
pink slip Monday (after this quarter's losses are announced). [But]
I'm not feeling too panicked at the present though. [\ldots]

[F]or the present, I'm just happy being a high-paid philosopher, and
I'm not worrying too much.

And the philosophy I've been doing!  I can't put the pragmatists
down; I read them in the morning, I read them in the night.  I am so
taken by the thoughts of {\James} {\it et al}.  I don't know how I could have
missed these guys for years!  It's a crying shame and certainly has
stunted my growth.  The pragmatists are all about the things John
{\Wheeler} was about (in his heyday), but oh so much deeper.  John was
an absolute amateur in comparison.

But enough about me.  Since I haven't heard from you in so long, it's hard for me to lob a directed comment or two your way.  I guess this letter shows that:  Leave a loud mouth to his own devices and he'll speak.

\section{23-04-02 \ \ {\it {\Mermin}'s Mysterious Ways} \ \ (to A. Peres)} \label{Peres29}

\bap
What happened to David {\Mermin}: {\tt 0204107} ``deconstructing dense coding'',
which follows deconstructing teleportation in\/ {\em PRA {\bf 65} (2002) 012320}?
Can you understand his motivation?
\eap

Did you read his acknowledgements in the first paper?  He writes, ``I thank \ldots\ Chris Fuchs for asking why I found it interesting.''  I don't know that he ever gave me an answer that I understood.  (He attempted to do it in the paper's present introduction.)  I think part of it is surely in the fact that he is presently teaching a quantum information course, and he is trying to be very pedagogical for the students.  But then, why he is not publishing in AJP, I don't know.

Petra went off to Amherst and then to Williamstown to visit Bill Wootters this weekend.  She returns by bus (to NY) and train (to Murray Hill) today.  She called me last night to tell me that everything is OK.  She is doing quite well in her research, and I am very pleased with her.

Tomorrow I post a rather philosophical paper on {\tt quant-ph}.  [See ``The Anti-{\Vaxjo} Interpretation of Quantum Mechanics,'' \quantph{0204146}.]  I think it will be my first document there with zero equations.  (But Seth Lloyd does almost the same in many of his {\sl Nature\/} papers!)  The reactions it's gotten in pre-circulation have been quite interesting.  For instance, Bill Wootters seemed to love it, but John Preskill seemed to hate it.  Julio Gea-Banacloche seemed to love it, but Charlie Bennett seemed to hate it.  And so on.

\section{23-04-02 \ \ {\it Music in the Musician} \ \ (to G. L. Comer)} \label{Comer12}

Thanks for the thoughtful note.  I found myself thinking about it on
and off all last night, in both my periods of wake and sleep.  I
think you expressed the issues to do with chemistry versus
consciousness especially vividly.

I think we just have to get rid of this imagery that we are ``made''
of atoms.  Or none of us are ever going to make any progress in our
emotional lives OR our physical understanding.  By my present
thinking, a much better imagery is this.  Take me and an old log: we
both float in water.  That is to say, we have that much in common.
But there are a heck of a lot more things that we do not have in
common.  For any two entities, we can always find some
characteristics they have in common, if {\it we\/} are willing to ignore all
the ways in which they are distinct.  And that, I think, is the story
of atoms.  The atomic picture has something to do with what we all
have in common.  (Or, maybe more potently, it has something to do
with what is common in the {\it part\/} of existence that we have chosen not
to ignore for the moment.)  But to see the atomic picture shine
through, we have to dim down all the things that are unique in us.
Who said the particular shape of that rock is not important?  Who
said the pain you are feeling is only epiphenomena?

Such a picture of what physics and chemistry is about is every bit as
consistent as the worldview Steven Weinberg, say, would have us
believe.  And I would say that it is more so; for it gives us a power
and a hope for control in our lives that his can't even imagine.

Let me do two things for you.  First, I'll paste in two old emails,
that have to do with your music-on-the-page versus
music-in-the-musician imagery (which I think it is so apt and so
beautiful).  [See stories from {\sl Notes on a Paulian Idea} pasted into my 05-12-01 note ``\myref{Schumacher4}{Lucky Seven}'' to B. W. Schumacher.]  Mostly I'm pasting them because your note caused me to
go back and read them this morning.  And I'm just reconfirming that
I'm on the same wavelength.

But then I want to quote William {\James}.  (That will come in a little
later note.)  I know you're not much in the mood to read any
philosophy right now.  But if you read the {\it right\/} stuff, I cannot see
how it cannot help.  My side of the conversations with you, in any
case, is just a poor reflection of what William {\James} already said
with such flare.

\section{23-04-02 \ \ {\it Installment 1} \ \ (to G. L. Comer)} \label{Comer13}

From {\sl Pragmatism}, pages 30--32:

\bq
And do not tell me that to show the shallowness of rationalist
philosophizing I have had to go back to a shallow wigpated age. The
optimism of present-day rationalism sounds just as shallow to the
fact-loving mind. The actual universe is a thing wide open, but
rationalism makes systems, and systems must be closed. For men in
practical life perfection is something far off and still in process
of achievement. This for rationalism is but the illusion of the
finite and relative: the absolute ground of things is a perfection
eternally complete.

I find a fine example of revolt against the airy and shallow optimism
of current religious philosophy in a publication of that valiant
anarchistic writer Morrison I. Swift. Mr.\ Swift's anarchism goes a
little farther than mine does, but I confess that I sympathize a good
deal, and some of you, I know, will sympathize heartily with his
dissatisfaction with the idealistic optimisms now in vogue. He begins
his pamphlet on `Human Submission' with a series of city reporter's
items from newspapers (suicides, deaths from starvation, and the
like) as specimens of our civilized regime. For instance:

\bq
``After trudging through the snow from one end of the city to the
other in the vain hope of securing employment, and with his wife and
six children without food and ordered to leave their home in an upper
east-side tenement-house because of non-payment of rent, John
Corcoran, a clerk, to-day ended his life by drinking carbolic acid.
Corcoran lost his position three weeks ago through illness, and
during the period of idleness his scanty savings disappeared.
Yesterday he obtained work with a gang of city snow-shovelers, but he
was too weak from illness, and was forced to quit after an hour's
trial with the shovel. Then the weary task of looking for employment
was again resumed. Thoroughly discouraged, Corcoran returned to his
home last night to find his wife and children without food and the
notice of dispossession on the door. On the following morning he
drank the poison.

``The records of many more such cases lie before me [Mr.\ Swift goes
on]; an encyclopedia might easily be filled with their kind. These
few I cite as an interpretation of the Universe. `We are aware of the
presence of God in his world,' says a writer in a recent English
review. (The very presence of ill in the temporal order is the
condition of the perfection of the eternal order, writes Professor
Royce ({\sl The World and the Individual}, II, 385).] `The Absolute is the
richer for every discord and for all the diversity which it
embraces,' says F.~H. Bradley ({\sl Appearance and Reality}, 204). He means
that these slain men make the universe richer, and that is
philosophy. But while Professors Royce and Bradley and a whole host
of guileless thoroughfed thinkers are unveiling Reality and the
Absolute and explaining away evil and pain, this is the condition of
the only beings known to us anywhere in the universe with a developed
consciousness of what the universe is. What these people experience
{\it is\/} Reality. It gives us an absolute phase of the universe. It
is the personal experience of those best qualified in our circle of
knowledge to {\it have\/} experience, to tell us {\it what is}.  Now
what does {\it thinking about\/} the experience of these persons come
to, compared to directly and personally feeling it as they feel it?
The philosophers are dealing in shades, while those who live and feel
know truth. And the mind of mankind---not yet the mind of
philosophers and of the proprietary class---but of the great mass of
the silently thinking men and feeling men, is coming to this view.
They are judging the universe as they have hitherto permitted the
hierophants of religion and learning to judge {\it them}. \ldots

``This Cleveland workingman, killing his children and himself
[another of the cited cases] is one of the elemental stupendous facts
of this modern world and of this universe. It cannot be glozed over
or minimized away by all the treatises on God, and Love, and Being,
helplessly existing in their monumental vacuity. This is one of the
simple irreducible elements of this world's life, after millions of
years of opportunity and twenty centuries of Christ. It is in the
mental world what atoms or sub-atoms are in the physical, primary,
indestructible. And what it blazons to man is the imposture of all
philosophy which does not see in such events the consummate factor of
all conscious experience. These facts invincibly prove religion a
nullity. Man will not give religion two thousand centuries or twenty
centuries more to try itself and waste human time. Its time is up;
its probation is ended; its own record ends it. Mankind has not aeons
and eternities to spare for trying out discredited systems.''
\eq
\eq

\section{24-04-02 \ \ {\it Leibniz on de Finetti} \ \ (to T. Rudolph)} \label{Rudolph3}

\btr
``Two things are identical if one can be substituted for the other
without affecting the truth.''
\etr

But neither de Finetti nor I believe in ``truth.''  How does that affect things?

\section{24-04-02 \ \ {\it A Stapp in the Right Direction?}\ \ \ (to B. W. Schumacher)} \label{Schumacher9}

I just ran across the abstract below on the archive.  Maybe it's relevant to you doubting project.

John Preskill told me you'll be at Caltech for 10 months, starting in September.  He was trying to use his siren song to lure me too for a little while.  I'm pretty sure I can't do it this year.  But I'm gonna try hard to do it next year while you're still there.

\bq\noindent
Quantum Physics, abstract\\
\quantph{0110148}\\
From: stapp@thsrv.lbl.gov\\
Date (v1): Thu, 25 Oct 2001 17:11:21 GMT   (13kb)\\
Date (revised v2): Tue, 23 Apr 2002 18:24:28 GMT   (15kb)\medskip

\noindent The basis problem in many-worlds theories\\
Author: Henry P. Stapp, (Lawrence Berkeley National Laboratory)\\
Comments: This extended version is to be published in The Canadian Journal of Physics\\
Report-no: LBNL-48917-Rev\medskip

\noindent It is emphasized that a many-worlds interpretation of quantum theory exists only to the extent that the associated basis problem is solved. The core basis problem is that the robust enduring states specified by environmental decoherence effects are essentially Gaussian wave packets that form continua of non-orthogonal states. Hence they are not a discrete set of orthogonal basis states to which finite probabilities can be assigned by the usual rules. The natural way to get an orthogonal basis without going outside the Schroedinger dynamics is to use the eigenstates of the reduced density matrix, and this idea is the basis of some recent attempts by many-worlds proponents to solve the basis problem. But these eigenstates do not enjoy the locality and quasi-classicality properties of the states defined by environmental decoherence effects, and hence are not satisfactory preferred basis states. The basis problem needs to be addressed and resolved before a many-worlds-type interpretation can be said to exist.
\eq

\section{24-04-02 \ \ {\it The Program}\ \ \ (to B. W. Schumacher)} \label{Schumacher10}

\bbs
Next-to-last week of classes, and things are really humming.  Hope all
is well with you and yours.
\ebs

By the way, I'm gonna post a slightly weird paper on {\tt quant-ph} tomorrow or the next day.  [See ``The Anti-{\Vaxjo} Interpretation of Quantum Mechanics,'' \quantph{0204146}.] If you've got a few moments between the bars of your hum, maybe have a look at it.  (It's easy reading; no equations.)  Especially the parts titled Preskill and Wootters.  Just to see the kind of strange friends you hang out with.  (Tell me whether you'll disown me if I go through with this!)

\section{25-04-02 \ \ {\it Two Nonorthogonal Pure States} \ \ (to P. Grangier)} \label{Grangier4}

I apologize for taking so very long to reply to your note.  It's just that I've had a million things tugging at me from all directions.

Your reply to me was thoughtful, and requires no less of a thoughtful reply in return.  And I'm finally writing you now because I think I've done that to my satisfaction.  It just so happens to take the form of a paper which will appear on {\tt quant-ph\/} tomorrow (\quantph{0204146}).  I will attach a copy.  I hope you will read it, especially the last two sections (concerned with Preskill and Wootters).  At heart, I think there is no doubt that you and I are both realists.  The only place we disagree is where we hope to see a hint of reality in quantum mechanics.  I think in that paper I paint the clearest picture yet of the direction I am seeking.  I sincerely hope you're reading it will delete some of the mystery of my ways to you.

\bpg
Just a question: you are strongly insisting on gambling, but aren't
you not surprised that a (pure) quantum state allows one to predict
many things with probability ONE?  This is a fairly strange behaviour
for a ``belief, state of belief, judgment, opinion'' etc, that is
usually far less efficient \ldots
\epg

Indeed this is quite an interesting question.  And it is one I have thought quite a lot about in the last year.  There was a time when I would have been surprised, but I am over that now.  I say there is no right and true quantum state for a system, even when two distinct observers ascribe two distinct pure states to it (whether they are commuting or not).  I talk about this in great detail in my web document ``Quantum States:\ What the Hell Are They?''\ and eventually I'll be putting a lot of those arguments into a proper paper.  Nevertheless, I think our discussion could benefit from your reading those pages.  The link to my webpage is below.  The places to look in particular are pages \myref{Mermin28}{19--23}, \myref{Schack4}{35--38}, \myref{Mermin35}{42--48}, \myref{Schack5}{49--50}, \myref{Caves14}{53--54}, \myref{Caves16}{55--64}, \myref{Caves25}{72--75}, \myref{Mermin45}{79--88}.  And then there are many more places beyond that in the same document on the same subject.  In general, I think the explanations get better and more convincing on the later rather than the earlier pages.  (So you might read it backwards.)  However, it is certainly true that I tried to be clear in my writing throughout.  So, if you end up not agreeing with me, I hope you will still find the effort entertaining.

\section{25-04-02 \ \ {\it Two Nonorthogonal Pure States, 2} \ \ (to P. Grangier)} \label{Grangier5}

\bpg
Exchanges of emails are hard to read, I prefer more concise formulations \ldots
\epg
You ought to know by now that my emails are more soliloquies than conversations!  I might play off a couple of lines from my correspondents, but then they're almost solely just me spouting my mouth off.  The technique is to read them as if they were independent papers.

\bpg
By the way, may be you have seen \quantph{0203131} which is as short as I could do (3 pages only \ldots). You will find \ldots
\epg
I have indeed read it, twice now.

\section{25-04-02 \ \ {\it Short Thoughtful Reply} \ \ (to C. H. {\Bennett})} \label{Bennett15}

Thanks for the picture of the skunk cabbage.  I've always wanted one.
I'm sorry I wasn't able to reply to you earlier, but with all the
students visiting, etc., I've had a gazillion things going on at
once.

Let me give you a very short reply, for what it's worth.

\bcb
My main wonder about your beliefs is why do you find it so important
to emphasize the subjectivity of quantum states, \ldots \ What
difference does it make in any case?
\ecb

I am just trying to do what scientists always try to do:  understand
how things ``hang together.''   I.e., build a (satisfying) picture of
the world that has nothing to do with my personal qualities.  It just
so happens, that my favorite problem happens to be different from
your favorite problem.

\bcb
Do you think Katie really exists, or is she just a mathematical
symbol for a set of bets you would [be] willing to make?
\ecb

Of course I think there is a sense in which Katie exists (i.e., some
large remnant of Katie as she is now would be here even if I were
killed tomorrow).  I just make a distinction between all the stories
I might write about her in my samizdats and whatever it is that she
{\it is\/} in herself.  What I don't understand about you is why you
find that such a foreign concept.

In particular, I don't think I could ever write a sentence like this:
\bcb
My main wonder about your beliefs is why do you find it so important
to emphasize the subjectivity of quantum states, but not other kinds
of information, such as the dinner you just ate, your shoes, or other
people?
\ecb

I don't think I've ever thought of the dinner I just ate as the
information I just ate.  Presumably there is something substantial to
broccoli independently of my subjective judgments about how it
tastes.  Information, as motivated by Shannon, has something to do
with the concept of surprise.  If I believe strongly that broccoli
tastes bad, then I will be surprised if I find that I actually like
it.  In that sense I will find that I have gained a lot of
information when my subjective judgment makes a transition from its
old value (Yuck!)\ to its new value (Mmm!).  But that has nothing to
do with broccoli as it is completely independently of me.

Even if I thought of a quantum state as an objective rather than a
subjective quantity, I still don't think that I could ever talk as
you did above.  There is a difference between ``systems'' and
``properties.''  And there is something in your language that seems
to blur the distinction.

To make this concrete, take a classical description of a pendulum's
motion in terms of phase space.  I would never call the phase space
point $(x,p)$ ``the particle.''  The particle is what {\it carries\/}
the property $(x,p)$.  Within classical physics, both the particle
and its property might as well be assumed to exist even when there
are no physicists about.  But let me ask the same thing about the
Liouville distribution for a {\it single\/} instance of the particle.
Can the single particle be said to ``carry'' a Liouville distribution
in the same way it ``carries'' the coordinates $(x,p)$?  I think you
would be hard pressed to say that it does.  For if I were to delete
the physicist who is ignorant of the phase-space point the particle
actually possesses, then I would delete the Liouville distribution.
But I would not delete the value $(x,p)$---whichever one it is---that
the particle can be safely assumed to have.

And that is all I am striving for in quantum physics.  To figure out
what properties we might safely ascribe quantum systems even when
there are no physicists about.  I think there are awfully good
reasons for thinking that ``the'' quantum state is not such an
``objective'' thing.  And thus the quantum state carries more of an
analogy to the Liouville distribution than to the phase space point.
However, that does not mean that I think our beloved theory gives us
no hint of what the properties are that I can safely treat as
objective.  It is just that, among them, I do not see the quantum
state.

\bcb
When you say a quantum state is just a set of bets you would be
willing to make, what is the ontological status of you the bettor?
Are you just a collection of bets some other bettor would be willing
to make?
\ecb

Don't blur the distinction between the system and the state!

There, that's my short reply.  As I say, I wish I could have replied
to you earlier, but I had so many things tying me up.  I wonder if I
could make a birthday wish of you?  Since my birthday was the 21st,
would you give me this much of a present?  Just read the parts of the
paper I posted for {\tt quant-ph} tomorrow, to do with Preskill and
Wootters.  I'll send it to you shortly.  It's not long reading, and
it's not hard reading.  (Certainly no harder than a New Yorker.) And
just give me two binary digits of satisfaction:  After reading those
passages, would you 1) say that you still do not understand my views
of Katie, and 2) does it still look like a tower of turtles to you?

With enduring friendship (and a picture of skunk cabbage hanging on
my wall),

\section{25-04-02 \ \ {\it King Broccoli} \ \ (to C. H. {\Bennett})} \label{Bennett16}

\bcb
I relish the taste of broccoli (especially broccoli rabe) and brussels
sprouts and cabbage, and I assume you don't.
\ecb

Actually, broccoli is my favorite vegetable of all time.  Well, more particularly, broccoli as my mother---Texan through and through---makes it:  overcooked, with a lot of butter and a lot of salt.

I loved your reply, but I don't think you answered either of my two questions.

\subsection{Charlie's Preply}

\bq
Happy birthday.  I read your last 3 sections, including the
Acknowledgement [of \quantph{0204146}].  I knew when I wrote you my passionate and
polemical words, in reply to your passionate and polemical words,
that we were talking past each other.  Basically I think the state
vector is more like the point in phase space than like the Liouville
probability density, and I think you think the reverse, but the big
lesson we should draw from this is that it doesn't matter enough to
get so passionate about.  Passion distracts from doing science which
I think is much more fun.  For example I was recently reading your
excellent paper with Caves and Schack about the quantum de Finetti
theorem, which I became very interested in lately in connection
with the quantum reverse Shannon conjecture, and after a while I
was able to filter out the polemical remarks as effortlessly as all the
ads for get-rich-quick I get in my email and enjoy the science almost
as much as if I they had been absent, or had been ones I agreed with.

To return briefly to unimportant matters, I think all three---quantum
state, classical state, and classical probability density---can be
viewed either as autonomous realities or as bets someone is willing
to make.  Despite the strong feelings one may have, it's a fool's errand
to try to prove that something is or isn't real.  The scientific content
of any quantum theorem or algorithm, such as de Finetti or teleportation,
can be mapped almost effortlessly from one interpretation to another,
by devices such as my favorite, the Church of the Larger Hilbert
Space---with its unexplained preferred basis---or your favorite, the humble but
undefined classical observer, and all we are left with to prefer one
interpretation over another is its good or bad flavor, which ultimately
is determined by our genetic makeup or early influences of our respective
pedophiles.  I relish the taste of broccoli (especially broccoli rabe)
and brussels sprouts and cabbage, and I assume you don't.  I
conjecture that there is a correlation between love of broccoli
and sympathy for the Everett interpretation.  David Deutsch doesn't
count, because he apparently doesn't much like any sort of food.  Unlike
the interpretations, this is a scientific question, susceptible to
experimental proof or refutation.  While we're at it, how do you feel
about the Korean pickle known as kim chee?
\eq

\section{25-04-02 \ \ {\it Quant-ph Number} \ \ (to C. M. {\Caves})} \label{Caves63}

\bcc
We'll be posting our paper tomorrow.  I'm planning to reference you
just at that point in the conclusion where we talk about Hilbert space
being a strange thing to need because it's an abstract object.
\ecc

The {\tt quant-ph} information is below.  [See ``The Anti-{\Vaxjo} Interpretation of Quantum Mechanics,'' \arxiv{quant-ph/0204146}.]  Of course I'd like to be cited at the second sentence in your abstract---``Whereas Hilbert space itself is an abstract construction, the number of dimensions available to a system is a physical quantity that requires physical resources.''---but you can't do that.  And, unfortunately, I don't find such a nice crisp statement with the same thought anywhere else in the paper.  You know that I still don't buy into this business that, ``A Hilbert space gets a physical interpretation \ldots\ through privileged observables.''  And I don't really want to be associated with that thought.  (From my way of thinking, I think it is safer to say that a Hilbert space gets its interpretation from the outside, not from the inside.  I.e., From the subjective judgments an agent uses to give meaning to the measurement ``clicks'' he finds.)

\section{25-04-02 \ \ {\it A Little Comment} \ \ (to C. M. {\Caves})} \label{Caves63.1}

\bcc
Just to cool a bit your ardor for that phrase in our abstract, ``The primary resource for quantum computation is Hilbert-space dimension. Whereas Hilbert space itself is an abstract construction, the number of dimensions available to a system is a physical quantity that requires physical resources.''  I have
a feeling that you mean that THE Hilbert-space dimension of a quantum
system is a physical quantity, whereas what we mean is that the
Hilbert-space dimension AVAILABLE to the system (which depends on the
physical resources available) is a physical quantity.
\ecc

Not sure I understood this.  (But I'm thinking about it.)

\section{25-04-02 \ \ {\it Guilt} \ \ (to U. Mohrhoff)} \label{Mohrhoff4}

I am writing at the present moment because I have not been able to shake the feeling of guilt.  To my shame, I still have your letter of 9/9/2001 sitting (with about seven others from other long-neglected correspondents) in the inbox of my email program.  It stares at me every day and calls out for a reply.  And one of these days, I'm going to do just that.

But in the mean time, as I say, I feel overwhelmingly guilty.  This is because tomorrow morning another one of my foundational papers will hit the airwaves ({\tt quant-ph}), and I know that I will be caught red-handed in having NOT neglected the subject.  Also, next week I am going to put a rather large extension of my NATO paper on {\tt quant-ph} too.

I can give you little in the way of compensation for my bad behavior, other than to let you know what I am feeling.  Also, maybe I can let you know that I acknowledge your input in the larger paper (which will appear in the paper next week).

\bq
\noindent {\bf Acknowledgments}\medskip\\
I thank {\Carl} {\Caves}, Greg Comer, David {\Mermin}, and {\Ruediger} {\Schack} for the years of correspondence that led to this view, {\Adan} Cabello, Asher Peres, and Arkady Plotnitsky for their help in compiling the dramatis personae of the Introduction, and Andrei Khrennikov for infinite patience. Special thanks go to Charlie Bennett, Steven van Enk, Jerry Finkelstein, Philippe Grangier, Andrew Landahl, Hideo Mabuchi, Masanao Ozawa, John Preskill, Terry Rudolph, Chris {\Timpson}, and Alex Wilce for their many comments on the previous version of this paper---all of which I tried to respond to---and particularly, to Howard Barnum, for pointing out my technical mistake in the ``Wither Entanglement?''~section. Finally, I thank Ulrich Mohrhoff for calling me a Kantian; it taught me that I should work a little harder to make myself look {\James}ian.
\eq

\section{26-04-02 \ \ {\it Objective QM?}\ \ \ (to P. Grangier)} \label{Grangier6}

Let me answer your easy questions first:
\bpg
PS In your paper you strongly ``recommend'' the work by Lucien Hardy. I
have two comments:\medskip\\
\indent {\rm 1)} {\it it surprises me that you adhere with the ``relative frequency''
approach to probabilities that is used by Lucien (his first axiom). I
certainly agree with it, but I thought you would not.}\smallskip\\
\indent {\rm 2)} {\it Lucien is trying to make QM look like a new probability theory. In
my opinion, this cannot work because I am convinced that there is an
inherent ``geometric'' (in Wigner's sense) content to QM. My views on
this are explained in the {\tt quant-ph} preprints.}
\epg

It is good that you picked up on that.  Because those are two things that I dislike the most about Lucien's paper.  I explain this in great detail on pages 147 and 159--166 in the document I wrote you about yesterday.  [See 19-02-02 note ``\myref{Hardy4}{Re-Tackle}'' and 06-03-02 note ``\myref{Hardy7}{Poetry on Concrete}'' to L. Hardy.] (These are both examples of letters that need absolutely no background---other than having read Hardy's paper itself---in order to be read.  So, do give it a shot.)  However, that does not take away from the mathematical structure Hardy has set his focus on.  And that is what I think is important, and worth wider recognition.

So yes:
\begin{enumerate}
\item
quantum probabilities, like all probabilities, I take to be gambling commitments, and
\item
I view quantum mechanics as plain old probability theory plus some further restrictions, NOT a generalization of classical theory.  (Those further restrictions are what---I claim---carry the ontological content of the theory.)
\end{enumerate}
And in both these opinions I contradict Hardy's paper.  But I would rather accentuate the positive in his efforts for the moment.

\section{26-04-02 \ \ {\it Transformation Rules} \ \ (to A. Peres)} \label{Peres30}

Let me quote the piece of the paper you were concerned about:
\bq\noindent
But the contexts are set by the structure of the Positive
Operator-Valued Measures:  one experimental context, one POVM.  The
glue that pastes the POVMs together into a unified Hilbert space is
Gleason's ``noncontextuality assumption'': where two POVMs overlap,
the probability assignments for those outcomes must not depend upon
the context.  Putting those two ideas together, one derives the
structure of the quantum state.
\eq

What I was referring to with the point about the overlap is that one
can derive the quantum probability law purely from the following
simple assumption:  there exists a function $f$ from positive
operators to the unit interval, such that the value of $f$ sums to
one over all positive-semidefinite resolutions of the identity.  With
this assumption (and nothing else, like continuity or
differentiability), one gets that there must exist a density operator
$\rho$ such that $f(E)=\tr(\rho E)$.  This is the modification
to Gleason's theorem that I describe in \quantph{0106166}
(section 4). So, it's not a completely trivial result of the standard
probability rule; instead one can take it as an assumption and get
the standard rule back out as a little gift.

Anyway, with this linear rule, comes for free the idea that
probabilities transform linearly when one changes from one
(sufficiently informative) POVM to another (sufficiently informative)
POVM.  And it is the transformations from ``context'' to ``context''
that Andrei has been making such a big deal about.

I hope that clarified a little bit.

\section{03-05-02 \ \ {\it Accepting Quantum Mechanics---The Short of It} \ \ (to B. C. van Fraassen)} \label{vanFraassen4}

I'm feeling horribly guilty because I promised you a note over two
weeks ago and everything---just everything---has conspired against me
getting it constructed.  I'm sorry.  But still I want to write you
something before I meet you next week.  Let me try to do what I can
in the next hour and then call it quits until Monday:  The details of
what I am about to say will be in a paper that I plan to have
finished and can bring with me then.

\bvf
Now when it comes to theories that give us probabilities, whether
absolute or conditional, I'll agree with scientific realists that
literally read they say that there are objective probabilities in
nature.  But accepting such a theory does not involve believing that.
Rather it involves appointing the theory as an `expert' for guidance
of our subjective probabilities concerning observable events.  The
metaphor of `expert' is cashed out (as by Haim Gaifman) as follows.
Suppose that I appoint Peter as my expert on snuffboxes.  That means
for my subjective probability $P$ and Peter's subjective probability
$q$ the constraint:
$$
   P\Big(A\,|\,q(A) = x \Big) = x
$$
with generalizations of this to intervals, odds, conditional
probabilities, for statements A that are about snuff boxes.

Thus the issue of whether there are objective probabilities in nature
or whether to believe in them is finessed:  there are only the
theory's probabilistic \underline{pronouncements} accepted as input and my
own subjective probabilities.

That is clearly not how you are approaching it overall.  But perhaps
there are connections?  I'd like to know how the QM probabilities are
fed into your subjective probability as a whole -- I wonder if it
will not be similar.  After all, even if a quantum state is read as a
compendium of probabilities, and you say something like ``this
material is in quantum state such and such'', your own subjective
probability function has a domain much larger than facts pertaining
to this material.
\evf

Yes, you are right, I don't like (ultimate) experts, and I don't
think quantum mechanics has any more need of them than weather
forecasting, say.  In fact, I think any attempt to hold on to
objective probabilities---even in the finessed form that you talk
about, where there is a higher authority in whose judgment we place
our faith---will only get in the way of our finding the deeper heart
of quantum mechanics. Thus, I hold fast to the idea that there is no
right and true quantum state EVER---just as de Finetti held fast to
the idea that there is no right and true probability distribution
EVER---and I take it as the very definition of my foundations program
to see what is left behind.  From my view, the theory does not
pronounce probabilities; it only pronounces what we ought to be doing
with them once we have set them just as subjectively as the next guy.
(And to make the issue as pointed as possible, I even mean this for
pure quantum states.)

What does it mean to accept quantum mechanics then?  The imagery I am
starting to build looks something like this.

If one generalizes the notion of quantum measurement to the one that
has become essential in quantum information theory---namely to the
positive operator valued measure (POVM)---then, for each quantum
system, one can contemplate a {\it single}, fiducial quantum
measurement for which the probabilities of its outcomes completely
specify the system's quantum state.  That is to say, whenever I write
down a subjective probability distribution $p(h)$, for the outcomes
$h$ of such a fiducial measurement, I completely specify a quantum
state $\rho$ (pure or mixed).  Imagine now that that fiducial
measurement device is tucked away safely in some vault in Paris at
the International Bureau of Weights and Measures.

A quantum state can be viewed as very literally nothing more than my
subjective judgment for what would happen if I were to ever bring my
quantum system up to that standard measurement device.  What now can
one say about a real-world measurement device, like one that we might
have here at Bell Labs?  Well, bringing my quantum system up to it
will generally evoke a click of some sort that I might label $d$.
Using all that I believe of the device, all that I might believe of
Lucent's technical prowess, etc., etc., I would be a bad subjectivist
if I didn't allow the click to update my beliefs about what would
happen if I were to approach the fiducial device.  Thus I end up with
some updated probability distribution $p(h|d)$.  (I.e., some updated
quantum state.)

In what sense is this subjectivism connected to Bayesianism?  This in
part is what this paper of mine is about that I'll give you a copy of
Monday.  One can show that the usual story of quantum collapse can be
viewed as the conjunction of two things:  1) Bayesian
conditionalization, and 2) a final rotation of the axes of the
probability simplex for the fiducial measurement.  That is to say,
quantum collapse in this description is only a pretty damned mild
generalization of the Reverend Bayes.

But, again, what does it mean to accept quantum mechanics?  It is
this.  If one studies the properties of these kinds of fiducial
measurements, one finds that for no initial quantum state (in the
usual Hilbert space picture) and no outcome is it ever the case that
$p(h)=1$.  That is, when one accepts quantum mechanics, one eschews
certainty for the outcomes of a fiducial quantum measurement.  In
fact, the set of allowed distributions $p(h)$ forms a convex set that
is strictly contained within the probability simplex (i.e., the set
of all imaginable probability distributions over an appropriate
number of outcomes).

Thus, accepting quantum mechanics is not accepting the existence of
an expert, but---in large part---accepting the two ingredients above:
\begin{enumerate}
\item
voluntarily accepting a restricted range for one's beliefs $p(h)$
\item
accepting a slightly modified form of Bayesian conditionalization
   for updating one's beliefs (i.e., standard Bayes $+$ rotation)
\end{enumerate}

The NEED for 1) a restricted range and 2) a minor modification of
Bayes, is where I say we should be looking if we want to be looking
for the ``meaning'' of quantum mechanics.  What is it about the world
as we view it that compels us to accept those two ingredients?  That
I see as the important question.  And the ever more convoluted moves
I see from some of our friends who want to hold on to a
nonsubjectivist view of the quantum state, I see as a waste of good
brain power.

That is to say, I agree with you in that, ``Be a realist if you want
to be.''  But I add to it with respect to the interpretation of
quantum mechanics, ``Don't do it for those parts of the theory where
it is not productive to do so.''  If you're looking for a little
realism in quantum theory, fine, but then look for it in a more
clever place than in the state vector.

Anyway, that's my present take.

And I lied:  that took me an hour and 35 minutes.  I hope it's a
little clearer at least for the extra time.  See you Monday!

PS. \ You wrote:
\bvf
The two articles of yours that we took up in our discussion group in
the fall were clearly only the beginning, and you have now taken the
program much farther.
\evf

Can you tell me which two articles you're referring to?  That would
give me a clearer vision of which views I've changed since your
reading and which views I need to be careful not to let be propagated
in your mind.

\section{05-05-02 \ \ {\it Wigner and Clones} \ \ (to A. Peres)} \label{Peres31}

I just skimmed your new paper.  Near your sentence, ``Why wasn't that
theorem discovered fifty years earlier?'' I think you ought to cite
Wigner as a case example.  Below is a little review of his paper I
published at {\tt quickreviews.org}.\footnote{\editornote According to Michael Nielsen, this site was established in 1997 and discontinued in 1998:  \myurl{http://michaelnielsen.org/blog/the-future-of-science-2/}.  Dave Bacon wrote that it ``was an interesting idea and first showed me the barrier to high quality online science'' [\myurl{https://twitter.com/dabacon/status/425498495840030720}].  The site appears to have gone entirely defunct sometime between March and August 2003.  See \myurl{https://web.archive.org/web/20030815000000*/http://quickreviews.org}.}

The original volume might be hard for you to get hold of, but the
paper can also be found in Wigner's later collection {\sl Symmetries
and Reflections}, page 200.  The spot where he just misses getting
the no-cloning theorem---i.e., the spot where he actually gets it
wrong---is in the second paragraph after the paragraph containing
Eq.\ (5).  (In the S\&R version of the paper, it's at the very bottom
of page 205.)  He writes, ``Let us denote the $n$ vectors which
represent living organisms by $v^k$ \ldots\  Then every linear
combination of the $v^k$ will also represent a living state.''

Here's the story Bill Wootters told me, when I first told him about the Wigner paper in 1998:
\bbw
Thanks a lot for your note.  I find it amazing that Wigner didn't quite realize that cloning was fundamentally impossible.  Frankly, I'm still not impressed with the cloning paper Wojtek and I wrote.  I remember asking someone for his opinion of a draft of that paper, and he said, in a very friendly way, that it would be a tough paper to referee, because on the one hand lots of people already knew the result, but on the other hand it may not have ever been written down.
\ebw

\begin{center}
Review of:\\ Eugene P. {\Wigner}, ``The Probability of the Existence of a Self-Reproducing Unit,'' \\ in {\sl The Logic of Personal
Knowledge: Essays Presented to Michael Polanyi on his \\ Seventieth Birthday\/} (Routledge \& Kegan Paul, London, 1961), pp.\ 231--238.
\end{center}
\bq
The no-cloning theorem first discussed by {\Wootters} and {\Zurek} and (independently) by Dieks is now understood to be a significant part of the foundation of quantum information theory.  But have you ever explained it to another physicist and received a reaction of the form, ``Is that it?  That's the big deal everyone is talking about?''
I have.  And it's no wonder:  the issue of no-cloning boils down to almost an immediate consequence of unitarity---inner products cannot decrease.  In fact, {\Wigner}'s famous theorem on symmetries even shows that the group of time-continuous, inner-product preserving maps on Hilbert space is strictly equivalent to the unitary group.
Therefore, it comes as quite a shock to see that {\Wigner} himself just missed the no-cloning theorem!  In this paper, {\Wigner} tackles the question, ``How probable is life?'' He does this by identifying the issue of self-reproduction with the existence of the types of maps required for the cloning of quantum states.  He doesn't tackle the question of cloning for a completely {\it unknown\/} quantum state head on, but instead analyses the ``fraction'' of unitary operators on a tensor-product Hilbert space that can lead to a cloning transformation for at least some states.  Nevertheless, he states quite clearly that an arbitrary linear superposition of clonable states ought also to be clonable. But this, of course, cannot be.

I think this paper is quite interesting from the historical point of view of our field:  it gives us an appreciation of the beauty and simplicity of that little theorem in a way that simply learning about it cannot provide.  It gets at the heart of something deep in very present physical terms, terms that even a great mind like {\Wigner}'s missed.
\eq

\section{07-05-02 \ \ {\it This Tape Will Self Destruct} \ \ (to G. {\Plunk} \& N. D. {\Mermin})} \label{Mermin65} \label{Plunk2}

I don't know that I've ever told either of you how I started down this slippery slope of quantum foundations studies.  It probably started officially in about 1985, when I took an undergrad research course with John Wheeler.  He assigned ME a problem in computational general relativity.  On the other hand, he assigned THE HONORS STUDENT who also took the course---I wasn't an honors student---the project of ``deriving quantum mechanics.''  I was so jealous!  I guess I never worked out that frustration.  (I think the honors student dropped out of physics.)

So, now it's my turn with {\Gabe}.  And guess what I'm thinking about assigning him?

The problem is about fixing up (in a outlandishly Fuchsian way) the death and destruction I have tried to heap upon the Brun, Finkelstein, {\Mermin} paper.  That is, if you can't say a state is right and true, what can you say?  The note below to Bas van Fraassen explains what I mean by all this in lay terms.  [See 03-05-02 note ``\myref{vanFraassen4}{Accepting Quantum Mechanics --- The Short of It}'' to B. C. van Fraassen.]

The particular research project I have in mind for {\Gabe} is to get a handle on the possible shapes for the convex set I describe below, as I range over all possible ``standard quantum measurements.''  We'll start with single qubits.  In particular, what I would like to know is what are the invariant properties of all these convex sets?  (Read below and this question will start to make sense.)  Could it be the volume?  Something more esoteric?  We can start with numeric work, and, if we get lucky, do something analytic.

Thus you see, what I want to say is, ``When an agent accepts quantum mechanics, what he is doing is accepting a convex region on the probability simplex with such and such characteristics.''  It would be nice if the characteristic had a simple characterization.

This tape will self destruct in five minutes.  Tell me whether you think this is a problem that could turn into a good senior thesis for a Cornell undergrad.  I think it would.

\section{07-05-02 \ \ {\it Failed Jokes} \ \ (to G. {\Plunk} \& N. D. {\Mermin})} \label{Mermin65.05} \label{Plunk3}

I wrote:
\bq\noindent
This tape will self destruct in five minutes.  Tell me whether you
think this is a problem that could turn into a good senior thesis for
a Cornell undergrad.  I think it would.
\eq
But of course, the title would have made more sense if I would have first said, ``This is your mission if you accept it.''

\section{08-05-02 \ \ {\it Updated Version}\ \ \ (to B. C. van Fraassen, J. Butterfield, J. Bub, T. A. Brun, and D. Wallace)} \label{vanFraassen4.1} \label{Butterfield4.05} \label{Bub7.1} \label{Brun6} \label{Wallace3}

\noindent Dear friends \ldots \medskip

to whom I gave an early version of my paper ``Quantum Mechanics as Quantum Information (and only a little more)''.  I astonished myself yesterday by having the time to clean it up to the point of publication.  In particular, I think the introduction goes more smoothly now and I've made some different choices of words throughout that I think clarify things.

So, if you haven't looked at the version I gave you yet---and I know you haven't, with all the activity of the meeting---I'd encourage you to go to the real thing (when you have some time for sleepy reading).  It'll be posted on quant-ph as of Thursday morning.  Here's the link: \quantph{0205039}.

(And for Bas, in particular, you'll be able to find a PDF version there too; so you shouldn't have trouble printing.)

\section{08-05-02 \ \ {\it The Big Eye}\ \ \ (to H. J. Bernstein)} \label{Bernstein5}

Check out my new paper.  It's titled ``Quantum Mechanics as Quantum Information (and only a little more)'' and you can find it at my webpage (link below).  It'll also appear on {\tt quant-ph} tomorrow.

I give you and your book with Mike Fortun a little advertisement at the beginning of Section 10.1 and in the Appendix, Point 21.

\section{08-05-02 \ \ {\it The QMP} \ \ (to H. M. Wiseman)} \label{Wiseman3}

Thanks for the letter, and particularly, thanks for skimming the
paper.  I chose a provocative title for the paper in an attempt to
draw in the crowd, but the content of the two letters to Preskill and
Wootters was pretty serious for me.  What I'm looking for, in
particular, is a way to make sense of science from a point of view
that says at the same time, ``Don't even think the terms in your
theory ARE or CAN BE a reflection of reality.  For the universe is
big, and your head is small.  And maybe reality is not static and
unchanging---and thus describable in any finite terms, like in terms
of GR and QM---anyway.''

That's a sweeping statement, but I think it may be one of the two
great lessons of quantum mechanics.  (I'll keep my opinions secret on
the other great lesson.)  Anyway in particular, the way I see it, we
as a community should be working as hard as we can to STOP trying to
see the wavefunction as anything of a reflection of nature.
Wavefunctions live in our heads.  They live and die in our heads. And
when they do their changing, they do that in our heads too. That's
the point of view I'm trying to run through with as much consistency
as possible.  I.e., how can I hold to it, and still have quantum
mechanical practice be what we all think it is?  How can I hold to
it, and not have everything that we say of nature be just a dream?

I'm sorry I've delayed so long in replying to you, but I wanted to do
it right, and that required that I finish up a paper so that I could
point you toward it.  The paper's finished now, and I posted it on
{\tt quant-ph} this morning; it'll appear tomorrow.  But if you wake
up before {\tt quant-ph} does, you can also get it at my webpage
(link below).  The title is, ``Quantum Mechanics as Quantum
Information (and only a little more).'' The parts that are
particularly relevant to our discussion are Section 4.2 and all of
Section 6.

\bhw
He implies that the same problems apply in classical theory. I
disagree. The problem is that noncommutative algebra does not apply
to the information in our brains. \ldots\ I'm not saying that it is
impossible to do that, but it is an extremely difficult problem which
he dismisses.
\ehw

Let me give an analogy that I think is apt.  I would say Bohr's great
genius in developing his model of the hydrogen atom lay solely in one
little move:  In dismissing the research program his predecessors
laid out before him.  That is, in dismissing the idea that the atom's
spectrum required a mechanical explanation.  Beyond that, the rest of
his work was just details.  Likewise, I think it is the case here.

I think that wavefunction collapse simply calls out for no
explanation at all.  And that is because I see it as nothing but a
variety (and, in a way, an extension) of Bayesian conditionalization.
To that extent, the problem---or, by my view, the nonproblem---was
already there long before quantum mechanics ever showed up on the
scene.  Imagine that physical theory really were like the Newtonians
wanted it to be:  nice and deterministic, Laplace's demon and all
that.  Now consider a weatherman immersed in the world.  He wakes up
in the morning, thinks about all the weather readings he has taken
the previous days, complains about the fact that his computer can't
do as much number-crunching as he would like it to do, scratches his
butt, decides whether he is feeling optimistic or pessimistic (that,
of course, might depend upon how his girlfriend has been treating him
lately), and so on and so on. He churns that all into a big mental
pot, and finally comes up with a set of numbers $p(h)$ to describe
his beliefs about what the weather will be doing tomorrow.  In fact,
at the same time, he'll probably come up with a whole set of numbers
$p(h,d)$ to describe not only the weather tomorrow, but also the
weather today.

Now, suppose he goes and has a look at the weather today and finds
that it is $d$.  Then, using Bayes' rule, he will update his belief
for tomorrow from $p(h)$ to $p(h|d)$.  It's a gut-wrenching, horribly
discontinuous transition.  But does it call out for an explanation?
And if it does, does it call out for an explanation that has anything
to do with the system the weatherman is modeling, i.e., the earth's
atmosphere?  For after all, in this world, when the weatherman looks
out the window, one value of $d$ was true and always remains true;
one value of $h$ was true and always remains true.  The only
discontinuous change is in his beliefs \ldots\ and those beliefs are
presumably a property of his head.

So, yes, I think Duvenhage is absolutely right on this point.  It's a
point I've been trying to make for years---it's what Asher and I were
up to in our {\sl Physics Today\/} article---but of course sometimes
it takes a long time to get the expression right.  And I think
Duvenhage did a particularly nice job on this particular score.

Now you and I both know that there's something more going on in
quantum mechanics than there is for the weatherman.  It's just that I
don't think that ``something more'' is localized in the issue of
``collapse.''  As far as the discontinuous change of belief goes---if
you ask me---that happens both quantumly and classically, and in fact
has nothing to do with any particular scientific theory.  It is
pre-science; it is simply a part of one's living and changing his
beliefs in response to stimuli from the world outside himself.

By my present thought, part of the ``something more'' that goes on in
quantum mechanics (as opposed to simple Bayesianism) is that we no
longer have the right to assume that the things we do in our
laboratories to change our beliefs (i.e., quantum measurements) leave
the world unscathed in the process.  I.e., we ought to be taking into
account that when the world stimulates us, we stimulate it back in
the process.  But how do we take it into account in our description?
My answer is that there's only one place to put it in a formalism,
and that is in a further change in BELIEF up and beyond that dictated
by Bayes' rule.

Thus, quantum collapse deviates from Bayes' rule, but not because it
has anything to with something going on outside our heads.  It
deviates from Bayes' rule because the subject matter we are talking
about when we are doing quantum mechanics (i.e., quantum systems)
have an implicit action-reaction principle that Bayes, Cox, Ramsey,
and de Finetti overlooked when they first worked out the calculus of
belief change---i.e., Bayes' rule.

Now, you say things like,
\bhw
The problem is that noncommutative algebra does not apply to the
information in our brains. \ldots\ If we could believe that
non-commutative algebra did apply to our knowledge then of course
that would solve the quantum measurement problem. But how can we
reconcile that with our experience?
\ehw
but, by my present view, using words like that is a red herring. (And
Duvenhage is much more guilty of it than you.)  The way I would say
it now---please read the sections in the new paper I told you
about---is that any kind of noncommutativity in quantum mechanics is
just an artifact of a certain representation.  The usual
representation is a useful one to be sure---I could hardly calculate
anything without it---but for the present issue I think it detracts
from the clearer understanding we can hope to obtain if we'll just
suppress it.  (The ``it'' meaning the usual representation.)  In
fact, I think we're only going to get that understanding by exploring
quantum mechanics as 1) a restriction on the space of probabilities
(the probability simplex), with 2) a conditionalization rule that
goes just a bit beyond Bayesian conditionalization.

And all this causes me to reject it when you say,
\bhw
The QMP is to find (i) a cut, and (ii) a way to bridge the cut,
between the quantum systems that do have non-commutative information
(in his terminology) and classical systems that do not.
\ehw
That is not the quantum measurement problem for me.  People,
experimentalists, scientific agents, have information or lack
information when they are trying to talk about something outside
themselves.  Systems are just systems, and when they are treated as
such, I would say the concept of information has nothing to do with
them.  Information is something I have or lack about a system.  For
instance, when I am speaking about you---thinking of you as a
physical system---I lack quite a bit of information in the sense that
your behavior could surprise me at any moment.  But that means
nothing about any kind of ``information'' intrinsic to you.

So, when you say what you said above IS THE QUANTUM MEASUREMENT
PROBLEM, I would say that that is a problem that comes from trying to
think of the quantum state, or the algebra of observables, etc., as
literal properties of the things you are describing \ldots\ and
failing to recognize that the quantum state is a property of the
observer and not the system.  The quantum state is the full
compendium of gambling commitments you would be willing to make about
what the system will cause a measuring device to do.

What I myself see as the quantum measurement problem is to give
compelling reasons for the two items I listed above.  By compelling,
I mean in the terms I lay out in the paper.  When we can finally do
that, then we will finally understand what properties we are really
assuming for the ``reality'' of a quantum system.

Anyway, I hope that helps explain my position.  I'll attach another
piece of correspondence below with Bas van Fraassen that takes a
different tack on some of the same issues.  Maybe that'll help
supplement the paper.  I hope you'll get a chance to take a look at
it.  By the way, I quote one of your papers in Footnote 33.

Thanks for giving me the opportunity for trying to say these things a
little more clearly.  And certainly feel free to question or comment
on anything that still doesn't make sense.  (Anything that doesn't
kill me might just make me stronger!)  Hey, I'm coming to Brisbane
May 25 to June 17.  Do you think we'll get chance to talk sometime
during then?

\section{09-05-02 \ \ {\it Physics 7 Months Ago, Today} \ \ (to A. Peres)} \label{Peres32}

\bap
I received {\bf Physics Today} of October 2001 (I have not received
September, because they took a cheap mailing service via India, and
almost nothing is being retransmitted --- a well known swindle). There
is a gloomy article on Lucent. What is really the situation, seven
months later?
\eap

I'm sorry I haven't had a chance to write you before now; I have a backlog of email that is a mile high.

When I joined Lucent there were about 120,000 employees.  Now, there are 50,000 \ldots\ with Wall Street rumors running that Lucent will have to cut 5,000 to 10,000 more beyond that to return to profitability.  So, things are a bit gloomy.  We are now encouraged to try to bring in military grant money, etc.  I doubt Bell Labs will ever be again what it was in the glory days, but so far it has been a safe haven for my nonsense.  We shall see how long it will last.

Today Petra left.  The group will miss her indeed.  Frank Verstraete left last Friday.  It was a very good month we all had together.

Petra told me today that she has decided not to return to Israel immediately.  I suspect that was sad news to you.

Tomorrow Kiki and the family and I make the long drive to Ithaca for David Mermin's retirement party.  Bill Wootters, Charlie Bennett, Gilles Brassard, Abner Shimony, and Lucien Hardy will be there among many others that I don't know.  (Well, I've heard of some like Michael Berry.)  In any case I'm sure looking forward to seeing the ones I do know!  I will lecture on Sections 4.2 and 6.1 of my new paper, \quantph{0205039}.

\section{10-05-02 \ \ {\it Go Ask Alice} \ \ (to G. L. Comer)} \label{Comer14}

Well the days have come and gone, and now the week is just about
over.  I finally finished up my extension of the NATO paper posted on
the LANL archive the other day.  (I thank you in it, by the way. Have
a look at \quantph{0205039}.) And the last of my visiting
students, left yesterday.  So I guess I'm getting ready for some kind
of denouement in the coming week.

This morning Kiki and I are packing up the kids and driving them to Ithaca, NY.  I'm one of the lucky ones who gets to give a talk at the Merminfest.  David has started the process of retirement this year.  I'm gonna try to make a new talk, if I can get my slides together before Sunday morning; my talk is at 10:30.  But wouldn't you know it, Sunday is Mother's Day.  Kiki said, ``You will not put work before me Sunday morning.  If we have to get up at the crack of dawn, you're taking the family to a pleasant breakfast.''  And so it goes \ldots\ I start to shiver with stress.

But I am really looking forward to this meeting, like none I have in a while.  And I'm doing something else I haven't done in a while:  I guess I'm proud enough of this new paper that I printed out 25 copies and I'll take them with me for distribution.  Isn't it funny:  I have to take what I consider some of my best work and stump it like a politician.  But when they were putting together the application for the Walton thing for me, I noticed that our teleportation paper has been cited something like 230 times.  That is such a crappy paper.

Your note has had me singing that old Jefferson Airplane tune in my
head all morning.  (I got up a little before 5:00.)  Torturous old
man.  I did really like that line of yours ``Dualism is a
Degeneracy.'' It strikes a chord in me with respect to my efforts to
overthrow the idea that a scientific theory is (potentially) a mirror
image of nature.  I.e., it is not even potentially.

But I guess I have a hard intellectual time with the idea you express
with:  ``[L]anguage is necessarily limited; beautiful, but just crude
enough that my BElongINGs can never be completely shared.''  You can
blame it on the Richard {\Rorty} in me.  You presume that there is a
``person on the inside'' that goes deeper than what can be built from
language.  That if we were able to conceptually strip all the rest of
the world surrounding Greg Comer away, there'd still be something
left.  It's a long story, but I guess I don't buy that.  The self is
just a local ``center of narrative gravity,'' {\Rorty} put it.  See his
``Ethics without Principles'' in his book {\sl Philosophy and Social
Hope}.

\section{10-05-02 \ \ {\it Compatibility} \ \ (to D. Poulin)} \label{Poulin1}

Thanks for pointing me to your paper; in fact, I had already noticed it.  {\Caves}, {\Schack}, and I are long overdue in our finishing up our criticism of the BFM paper.  The manuscript made it up to 12 pages, and was essentially complete.  But then I started to complain about ``strong coherence,'' as opposed to the normal Dutch book argument.  Anyway, hopefully we'll settle our differences and finish up the project in the late part of May when we all meet in Australia.  Regardless of that, though, I certainly do find your measure of compatibility interesting.

\section{13-05-02 \ \ {\it Upside Downside} \ \ (to N. D. {\Mermin})} \label{Mermin65.06}

Well, we finally made it home safely!  Here's my travel report.

{\bf Upside:}
Emma got so excited about the idea of a ``vacation.''  She loved the hotel and the swims and the ice creams she got every day.

{\bf Downside:}
Kiki wrecked the back end of the mini-van Saturday.  No one hurt except the pocket book and the insurance rate.

{\bf Upside:}
Michael Berry introduced himself by saying, ``Oh, you're Chris Fuchs.''  Michael Berry knows of Chris Fuchs?  Maybe better than that, he told me how he liked the ``Anti-{\Vaxjo} Interpretation'' paper and what he liked about it (anti-reductionism expressed in an interesting way).  (It was a good pick-me-up after Carl Caves called the same paper ``cloyingly affected.'')

{\bf Downside:}
I got the worst case of hay fever in quite some time during your banquet.  And your banquet lasted four hours!

{\bf Upside:}
I found myself telling Kiki all the stories from the banquet on the ride home.  The telling was as fun as the hearing.

{\bf Downside:}
I made the mistake of trying to draw up a new talk for this meeting.  Thus you got the first shot of it.  As is my experience, I know it'll get steadily better with practice.  But that means the Merminfest got the worst version it could get.  Only finished half the damned thing.  And I could see that the audience was getting more and more lost as I babbled on.  Beyond that, I gave you about the most unclear answer to your question that I could conjure up.

{\bf Upside:}
I enjoyed so many of the talks.  The sea squirts and the exploding heads especially took me.

{\bf Downside:}
I didn't get nearly as much time to talk to Hardy and Wootters as I had wanted to.  (And I certainly didn't get enough time to talk to you; but that one I expected.)

{\bf Upside:}
We discovered that you and Dorothy have an absolutely lovely house and surrounding lands.  Kiki and I found ourselves once again dreaming of living in a nice rural setting like that.

{\bf Downside:}
While Kiki and I were shopping in Ithaca today, somebody decided to even out the car for us.  It was a hit and run, and we didn't even get the pleasure of seeing it happen.  It tore off the front bumper.  Once calling the police, it took them 45 minutes come by to do the accident report.

{\bf Upside:}
I got some really good books!
\vspace{-12pt}
\bq\noindent
\begin{itemize}
\item
{\sl William James:\ His Life and Thought}, by Gerald E. Myers, \$12.00
\item
{\sl The Influence of Darwin on Philosophy}, by John Dewey, \$6.00
\item
{\sl Essays in Experimental Logic}, by John Dewey, \$6.00
\item
{\sl The American Evasion of Philosophy:\ A Genealogy of Pragmatism},
    by Cornel West, \$8.00
\item
{\sl Experience and Nature}, by John Dewey, \$6.00
\item
{\sl Values in a Universe of Chance}, by Charles S. Peirce, edited by
    Philip P. Wiener, \$6.00
\item
{\sl The Quest for Certainty}, by John Dewey, \$6.00
\item
{\sl A Pragmatist's Progress? Richard Rorty and American Intellectual History},
    edited by John Pettegrew, \$14.95
\item
{\sl The Philosophy of John Stuart Mill:\ Ethical, Political, and Religious,
    by John Stuart Mill}, edited by Marshall Cohen, \$5.00
\item
{\sl William James on Psychical Research}, by William James, edited by Gardner
    Murphy and Robert O. Ballou, \$14.50
\item
{\sl Schopenhauer}, by Christopher Janaway, \$5.00
\item
{\sl Some Problems of Philosophy:\ A Beginning of an Introduction to
    Philosophy}, by William James, \$20.00
\item
{\sl The Copernican Revolution:\ Planetary Astronomy in the Development of
    Western Thought}, by Thomas S. Kuhn, \$7.75
\item
{\sl Pragmatism and Other Essays}, by William James, \$3.00
\item
{\sl An Introduction to Metaphysics}, by Martin Heidegger, \$3.00
\item
{\sl Art as Experience}, by John Dewey, \$5.50
\item
{\sl Becoming William James}, by Howard M. Feinstein, \$25.00
\item
{\sl The Pragmatic Movement in American Philosophy}, by Charles Morris, \$25.00
\item
{\sl Robert Oppenheimer:\ Letters and Recollections}, by J. Robert Oppenheimer,
    edited by Alice Kimball Smith and Charles Weiner, \$9.50
\item
{\sl William James on the Courage to Believe}, by Robert, J. O'Connell, \$8.50
\item
{\sl William James:\ The Message of a Modern Mind}, by Lloyd Morris, \$9.00
\end{itemize}
\eq

\section{14-05-02 \ \ {\it No BC's Role} \ \ (to R. {\Schack})} \label{Schack50}

Thanks for the flurry of letters.  They provoke a lot of thoughts in
me and I am grateful.

I'm not sure how I'm going to reply to you, maybe just randomly (as
the details occur to me).

\brs
In my discussions with classical cryptographers, I am often forced to
concede that QKD is really quite limited in scope. You either assume
an unjammable channel (which I believe cannot be assumed if you use
the internet for communicating) or you use classical authentication,
which means you share some initial key. Given these limitations, I am
not convinced that QKD deserves such an exalted status as suggested
by the Bub quote above.

Has anybody taken the Brassard/Fuchs speculation any further?
\ers

Good point, of course.  But I never meant for those two ``axioms'' to
be read so literally.  Perhaps my best expression of the idea is
captured in a letter to Jeff Bub, 10 December 2000.  It starts at
page 100 of the old samizdat.  The main point is that I see
information-disturbance as the key idea, along with the commitment
that information can never be completely locked away.  Read those
passages; I think they'll clear some of this up.

That said, precise versions of the no-bit-commitment ideas are coming
to the top of my head again for their foundational value. Namely, as
part of the extra assumptions that might get us to proper density
operators in the ``Wither Entanglement?'' entanglement section
(rather than simply linear operators).  The thing that really powers
the no-bit-commitment theorem is the Hughston--Jozsa--Wootters result
that localized measurements on one system of two (described by a
bipartite pure state) can ``induce'' any decomposition one wants for
the other system's density operator.  It turns out that the ``pure
states'' of those other wacky operators in my bipartite-Gleason
construction don't necessarily have this property.  So, it looks like
an assumption of such a nature might get me a little closer to the
goal.  (Though---even if I found it completely acceptable, and I'm
not sure I do---it still wouldn't get me all the way to the goal.)
But all of this is a long story, and maybe it'd be easier to talk
about at a chalkboard.

\section{14-05-02 \ \ {\it More Toenails} \ \ (to R. {\Schack})} \label{Schack51}

\brs
Why are you so harsh on entanglement? In my quantum information
lectures, I postulated the tensor product structure, as a natural
formalism to deal with local operations on several particles.
Entanglement is then derived, an unexpected consequence of the tensor
product structure. You make this line of reasoning more compelling by
showing that there is really no alternative to the tensor product
structure. Far from being withered, entanglement emerges invigorated
from your analysis.
\ers

The main point is that entanglement can be thought of as secondary to
the structure of quantum observables on localized systems.  To that
extent, one realizes that one can focus on the structure of simple
observables in one's foundational efforts and forget about
entanglement.  In other words, it seems to me entanglement is not, as
{\Schroedinger} said, ``the characteristic trait of quantum mechanics,
the one that enforces its entire departure from classical lines of
thought.''  It is derived and secondary.

Now what is the structure of observables?  The thing I try to argue
in the paper is that a measurement is {\it anything\/} that gives
rise to {\it any\/} convex decomposition of a one's original density
operator. In that sense, measurement is nothing more than an
arbitrary application of Bayes' rule.

Entanglement thus arises from the more basic idea of Bayesian
conditionalization in conjunction with the idea that the allowed
probabilities for a standard quantum measurement device do not
explore the whole probability simplex.

Let me put two notes below that might clarify where I'm trying to go
with this.  I'd like to think that they add nothing beyond the paper,
but it has been my experience that I can just never say enough.  (I
know what you're thinking: ``You've got it backwards. If you'd just
say it all in fewer words!  That's what you really need.'')  [See
  03-05-02 note ``\myref{vanFraassen4}{Accepting Quantum
    Mechanics---The Short of It}'' to B. C. van Fraassen and 08-05-02
  note ``\myref{Wiseman3}{The QMP}'' to H. W. Wiseman.]

\section{14-05-02 \ \ {\it Emphasis De-emphasis} \ \ (to R. {\Schack})} \label{Schack52}

\brs
I am not sure I like your emphasis: You describe what I think is the
exciting part as a ``further readjustment of the posterior state''.
\ers

Yeah, I think I agree with that.  There's no doubt that that's the
most exciting part for me.  That extra adjustment strikes me as
capturing our beliefs about how we are stimulating the system (rather
than how it is stimulating us).  And to that extent, I would like to
emphasize it.  However, I guess I chose the tack I did because I
wanted people to stop thinking of quantum collapse as something so
different from ``mere conditionalization.''

\brs
Personally I like the concept of a compendium of probabilities better
than your bureau, but it is very interesting to see how far one can
get with the Bayes' rule alone.
\ers

Again, here, it is a point of emphasis.  The message I am trying to
get across is that the structure of observables comes from Bayes'
rule.  They (measurements) are not defined independently of it.  The
SQMD struck me as an effective way to badger that point. ``Compendium
of probabilities'' really stands for ``compendium of ways of applying
Bayes' rule.''  That is, I think it builds a stronger case for the
idea that it is Bayes' rule all the way down when it comes to quantum
measurement.  The theory gives us no overt means to identify the
objective thing that goes on behind a quantum measurement outcome.
The only grounding we have is to {\it declare\/} a point somewhere
off in the distance for which we will do no further updating.  And
that is the role the SQMD plays.

\section{14-05-02 \ \ {\it Deletions and Their Obverses} \ \ (to R. {\Schack})} \label{Schack53}

\brs
I think that taking William {\James} and Darwinism seriously means to
acknowledge that quantum mechanics is most likely to be superseded
one day by a theory with even more cash value. This new theory may
not have any of the features that we regard as the core of quantum
mechanics. How then is Darwinism or pragmatism going to tell us
anything about the foundations of quantum mechanics?
\ers

You're right, I think you are definitely missing what I was hoping to
express.  But that probably just means I didn't express it so
effectively.

No, I did not mean that Darwinism and pragmatism tell us something
about quantum foundations.  Just the opposite.  I see quantum
mechanics as giving us a great hint that there is still something
deeper.  Quantum mechanics is the first rip in the old fabric that
told us our place in the world is a nullity.  That is to say, I think
quantum mechanics only gives us even better arguments for pragmatism.

\brs
It looks to me as if your desire to find the objective core of
quantum mechanics is against the spirit of both Darwinism and
pragmatism.
\ers

I think the core of the theory---I don't think I ever used the words
``objective core''---is just our best guess of what we cannot change
with our present level of skills.  It is our best attempt to imagine
what it would be like if we were not here.  Obversely, when we have
ferreted the core out, we will have a quantitative indication of how
much of the world we can hope to control (given our present skills,
present evolutionary level, etc.)

My desire for delimiting the core is expressed particularly in the
passages below (taken from the paper).  Maybe this makes no sense
without  A) watching the movie {\sl It's a Wonderful Life}, and B)
reading the ``Sentiment of Rationality.''  Have you done either of
these things?  Probably not.

Your point certainly makes it clear to me that my paper is not
self-contained!  (The movie is an American classic and I think a
large fraction of Americans have probably seen it; but {\James} is
another story.)

\section{14-05-02 \ \ {\it Imaginations} \ \ (to R. {\Schack})} \label{Schack54}

Continuing again \ldots

\brs
Actually, I found the part where you mention the selection of traits
for your daughter outright disturbing, without any compelling
connection to the discussion on quantum mechanics.
\ers

Yes, me too.  Because read in the wrong way---and maybe there was no
other way to read it---it surely evoked images of Nietzsche (at the
least) and Nazi Germany (at the worst).

But on the other hand, I don't know how to draw a meaningful
distinction between our tools and ourselves.  (Remember my point
about the prosthetic hand in the paper?)  Should we stop pursuing
medical research because it goes against the grain of nature? Should
we stop pursuing genetic techniques for controlling AIDS? Should we
stop falling in love based on an attraction to our partners'
complementary (positive) traits to our own?  Traits that we would
like to see (even if subliminally) appear in our children?

I guess I say no, no, no.  Instead we need to cultivate at the same
time a respect for everything in its time.  Children of Down's
Syndrome, say, deserve respect, not deletion.  The Nazi experiment
with nature was an atrocity.  But, at the same time, I would say we
cannot stop pursuing progress in genetic engineering.

The point you make is a deep one.  I have no solution.  Only an
intuition of fear and promise, both at the same time.

If you get a chance, read Richard {\Rorty}'s book {\sl Philosophy and
Social Hope}.  It's all about fear and promise and quite easy to
read.

\section{14-05-02 \ \ {\it More Still} \ \ (to R. {\Schack})} \label{Schack55}

\brs
Everything you write suggests that you want more than just the theory
with the highest cash value: you are looking for some form of
absolute truth, something that transcends looking for the theory that
makes the most accurate predictions.
\ers

Yes, I guess so.  I am looking for an indication that the world can
be moved.  I am looking for an indication that the only law in
physics is the law that there are no laws.  I am looking for an
indication that the world is still writable.

I think if you want to call those things the pursuit of an absolute
truth, you can.  But they're only absolute in a pretty negative
sense.

See, I told you you provoked a lot of thoughts in me.

\section{14-05-02 \ \ {\it Qubit and Teleportation Are Words} \ \ (to C. H. {\Bennett}, W. K. {\Wootters}, N. D. {\Mermin}, A. Peres, J. A. Smolin and others)} \label{SmolinJ2.1} \label{Wootters7.1} \label{Mermin65.1} \label{Peres33} \label{Bennett17}

I doubt I will be of any use in constructing a short dictionary
definition for the word teleportation, but let me try to explain my
difficulty with the word ``property'' with regards to both quantum
states and quantum entanglement.  I do this with a little
trepidation, but on the other hand, you're the ones who brought me
into this discussion and I feel I ought to say something.

The trouble I have with the word ``property'' has to do with one of
the main points Charlie brought up in his talk this weekend.  What
instantaneously and physically changes about Bob's system when Alice
performs a measurement on hers?  Charlie told us {\it nothing\/} and
I agree with that.  But then, I look in the {\sl American Heritage
Dictionary\/} and find:
\bv
{\bf property}:
\\
1)  Something owned; a possession. \ldots\
\\
4)  a. A characteristic trait or peculiarity, especially one serving
to define or describe its \\ possessor. b. A characteristic attribute
possessed by all members of a class. See \\ synonyms at {\bf
quality}.
\\
5)  A special capability or power; a virtue.
\ev

If you think of a quantum state as a property owned by the system of
which it is about, then you---Charlie {\Bennett} in particular---are
obliged to continue propagating this thing you told us was a
misconception.  At the completion of Alice's measurement, there is a
new quantum state for Bob's system.  If the quantum state is
interpreted as a ``special capability or power'' for the system at
Bob's end, then you cannot get around the conception that Alice's
twiddle caused a change to something localized way over there.

And that, it seems to me is dangerous business.  What I am saying has
nothing to do with hidden variables.  It just has to do with the word
``property''.  The trouble only has to do with the idea of a quantum
state as a kind of feature possessed by a quantum system.

If you want to think of the quantum state as a property of something,
it seems to me the best you can do is speak of it as a property of
Alice's head (or Bob's head, or whoever's).  For, the quantum state
represents the {\it predictions\/} she can make about measurements
upon the system in question.  Similarly I could say all the same
things about entanglement.

To my view, toying with the idea that a quantum state is a property,
is to toy with a kind of pantheism or anthropomorphism that my
materialist mind won't tolerate.  Do all rocks have souls?  You'd
laugh at that, but it seems to me that's about what you're attempting
to do in thinking of the quantum state as something possessed by the
system itself.  ``That rock judges his chances of reacting to the
measurement device to be such and such!''  What could be more
anthropocentric than that?

\bcb
``Properties'' can reasonably be taken to be much broader than hidden
variables, and may include all sorts of conditional and post-selected
behaviors, eg ``how a system would behave if I measured its Z spin
component, after having watched my favorite horse lose at the race
track.''
\ecb

But, as I see it, conditional properties fare no better than the
nonrelational type in this regard.  At the completion of Alice's
measurement, there is a new quantum state for Bob's system, and
thinking of it as a conditional property in Charlie's sense still
means something physical changed at Bob's end.  This is because the
system did not have that particular (conditional) property before
Alice's action.

The main point is this:  Whatever a property for a quantum system is,
it should not be something changeable by someone's twiddles far, far
away.  For instance, take the dimensionality posited for a quantum
system by the ascription of a Hilbert space to it.  I think this is a
perfectly good candidate for a property of a quantum system; it's one
I would endorse.  For, once set, there is nothing Alice can do at a
far away location to change it.

So there, quantum systems do have properties---or at least I'm
willing to bet they do.  It is just that the properties do not
include among them ``the'' quantum state \ldots\ and it is the
quantum state that is transferred in the process of teleportation.

What I find miraculous about teleportation is that Victor (the guy
who ascribes the original, unknown state) can transfer his
predictions from one physical system to another at the cost of only
two bits of physical action on the target system.  In that sense, it
is Victor's description that is teleported from one system to the
other with almost nothing whatsoever traveling in between.  But
that'll never make it into a dictionary.

Of course, this is an ongoing debate between Charlie and me.  I'll
paste below one piece of our correspondence that has to do with the
present conversation---it itself focused on the issue of properties.
[See 25-04-02 note ``\myref{Bennett16}{King Broccoli}'' to C. H. {\Bennett}.]

By the way, I agree with Bill that the word ``object'' is better than
``system.''  {\sl American Heritage\/} writes this in its first two
definitions for the word:
\bv
{\bf object}:\\
1)  A material thing. \\
2)  A focus of attention, feeling, thought, or action.
\ev

\subsection{Asher's Reply 1, ``Quantum Teleportation Defined:\ Nothing Happens at Bob's End''}

\bq
When the quantum teleportation process was conceived, we had no clear
understanding of what was going on, and this may still be true today.
I learnt a lot from Chris when we wrote our ``Opinion'' essay in {\sl Physics
Today}, March 2000, pp.\ 70, 71. Here is the passage about quantum
teleportation, mostly due to Chris:
\bq
The peculiar nature of a quantum state as representing information
is strikingly illustrated by the quantum teleportation process.$^4$
In order to teleport a quantum state from one photon to another, the
sender (Alice) and the receiver (Bob) need to divide between them a
pair of photons in a standard entangled state. The experiment begins
when Alice receives another photon whose polarization state is
unknown to her but known to a third-party preparer. She performs a
measurement on her two photons-one from the original, entangled pair
and the other in a state unknown to her-and then sends Bob a
classical message of only two bits, instructing him how to reproduce
that unknown state on his photon. This economy of transmission
appears remarkable, because to completely specify the state of a
photon, namely one point in the Poincar\'e sphere, we need an
infinity of bits. However, this complete specification is not what is
transferred. The two bits of classical information serve only to
convert the preparer's information, from a description of the
original photon to a description of the one in Bob's possession. The
communication resource used up for doing that is the correlated pair
that was shared by Alice and Bob.
\eq

The crucial point is in the last five lines: Alice and Bob know nothing
of the states of their particles (the one that the preparer gave to
Alice, and the one that Bob finally has). To make the process simpler
let the two bits be sent by Alice to the preparer, next to her, not to
Bob, far away. Nothing happens at Bob's end. Only the preparer knows
instantaneously what is the state that Bob can verify, if and when he
receives the relevant information (same two bits).

How do you explain that to non-believers?
\eq

\subsection{Asher's Reply 2, ``Proposed Definitions:\ Qubit, Entanglement, Quantum Teleportation''}

\bq
I tried to find definitions that are scientifically correct and do not
depart too much from the style of the {\sl American Heritage Dictionary}, so
that they may be acceptable to them.\medskip

\noindent {\bf qubit -- quantum bit}

The smallest component in a computer designed to manipulate or store
information through effects predicted by quantum physics. Unlike bits
in classical systems, a quantum bit has more than two possible states
labeled 0 and 1. It can also be in a combination of 0 and 1 according
to the superposition principle, or even have no definite state at all,
if it is entangled with other quantum bits.\medskip

\noindent {\bf entanglement}

A quantum mechanical situation where a composite system has a definite
state, but none of its constituents has a definite state. Moreover, if
separate measurements are performed on these constituents, the results
show strong correlations.\medskip

\noindent {\bf quantum teleportation}

The instantaneous transference of our knowledge of the properties of
a quantum object to those of another distant object, without physical
contact with the latter.

This is not quite the same as our original paper, because here Bob (if
he exists---he is not needed) does not perform a Pauli rotation to have
an identical state. Anyway, if the state of his particle is publicly
known, then the required Pauli rotation will eventually be known in the
whereabouts of the second object.
\eq

\subsection{Asher's Reply 3, ``Instantaneous Remote State Determination''}

\bq
Since Charles protested my use of the term ``instantaneous'' here is a
paragraph of my paper ``Classical interventions in quantum systems.\
II. Relativistic invariance.'' PRA {\bf 61} (2000) 022117:
\bq
A seemingly paradoxical way of presenting these results is to ask the
following naive question: suppose that Alice finds that $\sigma_z=1$
while Bob does nothing. When does the state of Bob's particle, far away,
become the one for which $\sigma_z=-1$ with certainty? Though this
question is meaningless, it has a definite answer: Bob's particle state
changes instantaneously. In which Lorentz frame is this instantaneous?
In {\it any\/} frame!  Whatever frame is chosen for defining
simultaneity, the experimentally observable result is the same, owing to
\verb+\Eq{etcr}+. This does not violate relativity because relativity is built
in that equation, as will now be shown in a formal way.
\eq

Indeed when Alice finds her result she definitely knows Bob's
{\it potential\/} result if and when he performed or will perform his
measurement.
\eq

\section{14-05-02 \ \ {\it Chris's World} \ \ (to J. A. Smolin, C. H. {\Bennett}, N. D. {\Mermin} and others)} \label{SmolinJ3} \label{Mermin65.2} \label{Bennett18}

Good to see you in the morass.

Chris's world:  It's a funny place with all these fancy words like
``ascription'' and ``posit'' to remind us that there's a head in the
background of every quantum state, but there's no instantaneous
action at a distance there---no one would have ever thought there
might be.  I know the language drives Charlie bonkers, and it
probably drives you bonkers too.  But mostly the complaints just
remind me of what I used to hear in my hometown in Texas when the
seatbelt law was first enacted.  ``Why that Majatek boy was thrown
clean from the car!  Not a scratch on him.  If he'd have had a
seatbelt on, his head would have been nothin' but mush now; we'd be
at his funeral today.  Damned politicians puttin' their noses into
places where they ain't got no business.''

So you see I view the language as a safety measure---one that I think
will allow us to drive farther, longer, and ultimately faster.  But
first you've got to learn how to drive with a belt on.

\bjas
One is naturally forced into Chris's world of saying all that changes
is what people predict about the state, but that's a property
properly defined as above.
\ejas

Chris would never say this.  When I write down a quantum state, I
think of it as my judgment or prediction for which of one or another
measurement outcome will turn up.  I don't predict things ``about the
state.''  I don't know what the TRUE state is or could be, and thus
my ignorance cannot be about it.  In contrast, I would say the
ignorance is always about further measurement outcomes.  Or a
pleasing picture is that one can ground the ignorance with respect to
a single device sitting in the National Bureau of Standards if one
wishes.  See Sections 4.2 and 6.1 of the fat paper I put on {\tt
quant-ph} last week (\quantph{0205039}).

That said, I also see the suppression of another crucial issue
here---one that also flies in the face of the word property,
especially with your emphasis on DEFINE in definition 4.  If you will
allow me to call a mixed density operator a quantum state, then I
know even you will agree that there is no unique quantum state for a
system.  Thus it cannot be a property.  So, I think you'll be left
with being only willing to call a pure state a property.  So be it.
(I wouldn't do that, but I'll let you do that for the time being.)
But now, let's go back to Alice and Bob.  By your account, first
Bob's system has no property, then Alice measures her system,
and---Zing!---now Bob's system does have a property.  I.e., first it
has no quantum state, then it does.  How does the system know it
ought to have that property if there's no action at a distance?
Alternatively, if it doesn't know it, why call it a property of the
system?

It is because of that conundrum that it seems to me to be more
fruitful to just give up thinking of the quantum state as a property
of the system it targets.

I think a good analogy can be found in {\it classical\/} information
theory.  A homework problem in a textbook gives you a discrete
memoryless channel by  specifying the transition probabilities
$p(y|x)$.  Then it asks that you calculate the channel's capacity.
One goes through all the work and gets a number.  From that, one
starts to get the feeling that the capacity can be an objective
property for a real physical channel.  Why else would it take so much
work to calculate something if it weren't real?  But it can't be real
in any absolute sense.  For, with respect to a Laplacean demon, there
is never any noise in the channel at all; he can predict which bits
will be flipped and which won't.  The point is, the capacity is only
objective WITH RESPECT TO a {\it subjective\/} judgment $p(y|x)$.
Similarly, I would say with all quantum states. Just because a
textbook says calculate such and such a property---the entanglement
of formation, the distillable entanglement, or whatnot---of a quantum
state, one finds that one gets into the same habit of thinking those
properties have no subjective component.

You can do so, but then it's your burden to explain to young school
children and journalists---who think of a property as something like
the color red, a ball either has it or doesn't---what changes in
Bob's system when Alice performs a measurement.

\section{14-05-02 \ \ {\it Quick Answer about Quantum de Finetti} \ \ (to K. Jacobs)} \label{Jacobs2}

\bkj
Quick question about the Quantum de Finetti representation.
The expansion for an exchangeble $\rho^{(N)}$ is unique --- but does $N$ have
to be greater than or equal to 2 for this to be true?
\ekj

The de Finetti representation is something that applies to a {\sl sequence\/} of density operators $\rho^{(N)}$ living on larger and larger Hilbert spaces $H^{\otimes N}$.  The theorem says there is a unique decomposition of the form $\int P(\rho) \rho^{\otimes N} d\rho$.  But of course, that does not negate the fact that for each $N$ on can find many ways to decompose the mixed state.  It is just that if one finds an alternative way of expressing the density operator at level $N$, that alternative expression will not work for level $N+1$.  The case $N=1$ is especially relevant in that regard.  There are a continuum of ways to decompose $\rho^{(1)}$, but only some of those ways will work to get you up to $\rho^{(2)}$, and even fewer will get up to $\rho^{(3)}$, etc.  In the limit, only one of the original decompositions will work.

\section{15-05-02 \ \ {\it And Only a Little More} \ \ (to W. K. Wootters)} \label{Wootters8}

I'm just taking care of some loose ends left from the meeting in Ithaca.  While we were there, I had hoped to put a copy of my new paper into your hand.  But every time I thought about it, I thought it would be awkward for you to have to carry it since you weren't carrying a briefcase.  Then at our final encounter, just before your leaving, I didn't have a copy with me!

Anyway, here's the link in case you're interested:
\bq\noindent
Quantum Mechanics as Quantum Information (and only a little more)\\
\quantph{0205039}
\eq
Of course, feedback, both good and bad is always welcome.  What I am striving for is consistency, and it's pretty hard to see lapses in that without an outside eye.

By the way, mention that philosopher's name again that you've been taken with.  I'll look her/him up.

\subsection{Bill's Reply}

\bq
The book I mentioned is {\sl Nature Likes to Hide}, by Shimon Malin, a physicist
at Colgate University.  The book is in three parts: (i) an introduction to
the ideas of quantum mechanics (the book is for a very general audience),
(ii) application of some ideas of A. N. Whitehead to the interpretation of
quantum mechanics, (iii) presentation of a worldview based on quantum
mechanics and ideas of Plotinus.  I wouldn't follow Malin all the way, but
there's something about part (ii) that struck me as exciting and worth
exploring further.  I had already been reading Whitehead and other writers
inspired by him (especially Hartshorne), as well as C. S. Peirce, who
inspired both Whitehead and Hartshorne.

Whitehead argued that a materialistic philosophy could never adequately
account for mind.  So in his view, the basic elements of the world are not
material objects but are more like moments of awareness: they are more like
events than persisting objects, and they always have both an interior and
an exterior aspect.  There is some similarity with Wheeler's view that the
basic elements are quantum events, except that with Whitehead there is more
of a balance between the interior and exterior aspects of each basic
element.  Wheeler thinks of quantum events merely as objective events, seen
only from the outside, as it were.  Whitehead says that at every level of
being, there is always an interior view.  At the higher levels, this
interior view is what we call consciousness.  At lower levels, it's not
consciousness but it exists nonetheless.  So consider a photon encountering
a beam splitter, with a detector in each of the two possible paths.  The
firing of one of these detectors is an event that can be viewed
objectively, but according to Malin (interpreting Whitehead), it can also
be seen from the inside: the freedom to choose one detector or the other
can be thought of as a kind of proto-experience.  It's too primitive to be
called either experience or consciousness, but it has some quality in
common with consciousness (whereas a purely material object would not have
this quality).

I think the above paragraph should give you a sense of the direction Malin
is going.  There are a lot of things about Whitehead's view, and Malin's
view, that resonate with my own ideas.  Of course I like the Wheelerian
idea of taking quantum events as basic, and I like this question:  Where
does a quantum event reside?  For example, where was the decision made to
make one of those two detectors fire?  It wasn't made at the location of
the detector that fired, because then how would the other detector have
known not to fire?  And we know very well that the decision wasn't made at
the beam splitter.  Pretty clearly, the decision wasn't made at any
particular place in spacetime.  So I don't think a quantum event resides
primarily in spacetime.  A quantum event is more primitive than spacetime,
as Wheeler has said many times.  What I like about Whitehead and Malin is
that they emphasize a commitment to consciousness as something central to
existence and something that one has to have in mind even when talking
about the most basic elements of existence.
\eq

\section{15-05-02 \ \ {\it Not Even Vague Thoughts} \ \ (to R. W. {\Spekkens})} \label{Spekkens6}

\brws
I've also been thinking that the analogy between the possibility of
``steering'' and the impossibility of cloning suggests that there is
something, like the task of steering but different, that is analogous
to broadcasting and which would truly separate the quantum from the
classical. So far I've only got vague ideas.  Have you given any
thought to this?
\erws

I like this line of thought you're pursuing, but no I haven't thought about it any more than the little I did in Princeton.  I guess the first question is, what does entanglement buy you as far as steering is concerned?  For, even with unentangled bipartite (mixed) states, Alice can steer Bob into noncommuting possibilities.  So the line is not drawn at commutivity.

You might be able to prove something like this:  Suppose Alice and Bob possess an unentangled mixed state, but one for which no decomposition exists in terms of commuting states on Bob's side.  There can be no device on Bob's end that will make a tripartite state $AB^\prime B$ for which the $AB$ and $AB^\prime$ states are identical copies of the original.  But then, what does that have to do with steering?

In any case, you know I hate the word steering.  What one needs for No-QBC is simply that Alice be able to predict what Bob will find if he performs any valid measurement in the protocol.  It has nothing to do with steering anything.

\section{15-05-02 \ \ {\it Marburg, Strasbourg, Blunderburg!}\ \ \ (to A. Peres)} \label{Peres34}

\bap
I started to read your QM as QI, a few pages each evening, when I am too
tired to do anything productive, but not yet ready to go to sleep. Will
this appear in the {\Vaxjo} proceedings? Two misprints in the table page 2:
\bv{\rm
1973 Marbourg $\rightarrow$ Marburg\\
1974 Strasbourg is in France, not in Germany (it was in Germany 1940--44
     and before that 1871--1918, but it is in France since 1945).}
\ev
\eap

Thanks for catching those mistakes.  [See 01-04-02 note ``\myref{Peres26}{Last Little Push}''  to A. Peres, N. D. {\Mermin}, A. Cabello, H. J. Folse, and A. Plotnitsky.] I'm a little ashamed that I made them.  And I am flattered that you are reading the paper; thank you for that independently.

Yes, an older version of the paper will appear in the {\Vaxjo}
proceedings. The version on {\tt quant-ph} has 25 or 30 spots where
the language has been changed somewhat.  I have contemplated trying
to send the final, final version to Andrei.  But I don't know that it
is worth my while.  With Andrei's whirlwind methods, I suspect it is
too late for me to send him revisions in any case.  Of course, the
only version people will ever read---if they read that much---will be
the {\tt quant-ph} version.

I just downloaded your no-cloning history to see how you took into
account Wigner's paper.  I was a little surprised when I didn't see
you mention him.  I would guess that you made that decision because
you did not muddy Wigner's name.  Still, though, I think his blunder
only adds to the importance of the theorem (despite its mathematical
triviality), and I wish the readers could have seen that at the same
time that you defend your decision on FLASH.

Kiki and I arrived back home from Ithaca late Sunday evening, with a
hobbling car.  (It is our Honda mini-van, only one year old.)  First Kiki had a crash Saturday morning that tore deeply into the left-hand fender and bumper.  No one was hurt in the accident, only our pocketbook and insurance rate.  But then Monday, while we were bookshopping, someone decided to even out the car's aesthetic for us:  Somehow they hit the front of the car and tore the front bumper off.  It was a hit and run, and no one in the neighborhood saw it happen.

The meeting in Ithaca was quite nice, with quantum informationists,
quasi-crystalists, biologists, magazine editors, science policy
advisors, etc.  I think the best thing that happened to me was that I
met Michael Berry for the first time.  He introduced himself as I was
talking to Philip Pearle.  He said, ``Oh, you're Chris Fuchs?''  I
thought, ``Michael Berry knows of Chris Fuchs?''  It was quite a
surprise to me.  Anyway, he proceeded to tell me how much he liked
our {\sl Physics Today\/} piece, and how he quotes from it to his students.
During the meeting, I also gave him copies of my two {\Vaxjo} pieces
and he gave me some good feedback on both.  Sometimes, I need
confidence builders like that.

\section{15-05-02 \ \ {\it Berry, Etc.} \ \ (to A. Peres)} \label{Peres35}

\bap
Michael Berry must know of you because you were offered a position in
his department. Here is a frequent visitor to Israel because his wife
has her parents here.

I didn't quote Wigner. Why kick a dead horse? However I added a quote
to Jozsa's recent {\tt quant-ph}, which is really very nice. It's Bill who
called my attention to it.
\eap

I accept your decision.  But still I contend that you missed a good
opportunity.  The point is not to kick a dead horse, but to glorify a
living theorem.  Wigner was one of the greatest minds in physics this
century---nothing can take that away---but yet he came to within an
inch of the theorem and then missed it.  I think that signifies
something interesting psychologically.  Perhaps it also signifies
what a radical departure quantum mechanics is from classical lines of
thought.

\section{15-05-02 \ \ {\it Bayes, POVMs, Reality} \ \ (to A. Shimony)} \label{Shimony1}

It was good talking to you this weekend.  If you wouldn't mind
committing your story of meeting de Finetti to email, I will see to
it that it is archived forever in one of my samizdats.  (See my
webpage; link below.)

Also, if you could send me your mailing address, I will send you
copies of my two new papers on Bayesianism, POVMs, and good
candidates for quantum reality.  Alternatively, if you are accustomed
to downloading things from the {\tt quant-ph} archive, here are the links:
\begin{enumerate}
\item
``Quantum Mechanics as Quantum Information (and only a little
more),'' \\ \quantph{0205039}
\item
``The Anti-{\Vaxjo} Interpretation of Quantum Mechanics,'' \\ \quantph{0204146}
\end{enumerate}

I wrote them both in an attempt to be entertaining.  I hope you find
them so.

\subsection{Abner's Reply}

\bq
Here is what I recall of my conversation with de Finetti. In 1971 the
3rd International Congress of Logic, Methodology, and Philosophy of
Science was held in Bucharest (where I had the dubious pleasure of
meeting Ceau\c{s}escu in the receiving line at the Palace of Ministers;
what an ugly hard face he had.) De Finetti was there, and I believe
that I introduced myself, saying that I had some questions. The only
question I recall was why he didn't use the strong version of
coherence. He said that he was aware of the option of using it rather
than the weak version, but he didn't like the consequence of the
substitution:  namely that $C(h/e)=1$ only if $e$ entails $h$. This
consequence is part of my (2') in Sect.\ 5 of my ``Coherence and the
Axioms of Confirmation'', p.~136 in vol.~I of my {\sl Search for a
Naturalistic World View}. I don't recall the details of his objection
to this principle, but I think he said that it would cause trouble if
one had a nondenumerable set of mutually exclusive possible
hypothesis, as in the case of probability on continua. I vaguely
recall agreeing with him that there would be a problem, because I
discuss the problem on pp.\ 137--140, op.\ cit. I vaguely recall
saying (or maybe just thinking) that epistemic probability doesn't
apply well to nondenumerable sets of hypotheses, but thinking that
the propensity interpretation of probability, usable in stat.\
mechanics, could properly deal with nondenumerable sets of possible
outcomes. De Finetti surely would not have liked this discrimination,
since he believed that the only clear sense of probability was
epistemic, and in particular personalist. In my later paper,
``Scientific Inference'', op.\ cit.\ I suggest pragmatically
reasonable strategies for dividing the entire set of hypotheses into
as denumerable set, or even a finite set, by properly lumping subsets
of hypotheses. This strategy seems to me in the right direction,
partly because it is part of a program taking the Bayesian formalism
as only a framework, which has to be supplemented by pragmatic and by
a posteriori considerations.
\eq

\section{16-05-02 \ \ {\it King Broccoli, 2} \ \ (to J. A. Smolin, C. H. {\Bennett}, N. D. {\Mermin} and others)} \label{SmolinJ4} \label{Mermin65.3} \label{Bennett19}

I'll close with this statement.  But then after your rebuttal---if
you care to make one---we should probably take this offline.  I
suspect no one cares to explore the issue further (not even you and
not even me).  But you said something so nicely, I thought I should
emphasize it.

\bjas
\bq
\noindent \rm [CAF wrote:] [F]irst Bob's system has no property, then
Alice measures her system, and---Zing!---now Bob's system does have a
property. I.e., first it has no quantum state, then it does.  How
does the system know it ought to have that property if there's no
action at a distance? Alternatively, if it doesn't know it, why call
it a property of the system?
\eq
It always had the property that IF Alice measured one thing, then it
would behave as state $\phi$.  Conditional properties like that do,
of course, imply a sort of action at a distance, but what's so bad
about that?  It's not the sort of violate-the-speed-of-light action
at a distance that we should be concerned about.  In the end, it
means just what you want it to, except for the word property.  To use
a classical example, suppose I have a box that comes from the factory
with either a red ball or a blue ball in it.  Surely one can say that
a property of the box is that it has either a red ball or a blue ball
in it---that's the entire definition of the box.  Now if I call up
the factory and they tell me what color the ball actually is, never
touching the box, does the box change?  Was I wrong to call that
other thing a property? But everyone understands what's going on.
Entanglement doesn't really bring in anything new here.
(Alternatively you could formulate it like Charlie sometimes does and
say well, it was ALWAYS a red ball, but the measurement result
travels backwards though time and fixes things up so it is ok to say
``always,'' but I'm sure you won't care for that).
\ejas

You said that perfectly, and, of course, I especially liked the
concession that this choice of words does entail a kind of
uninteresting action-at-a-distance \ldots\ but action-at-a-distance
nevertheless.  Maybe my point is just, how is the
non-quantum-practitioner supposed to know where to draw the line
between the uninteresting and the interesting versions of the effect?
Between the science and the science-fiction?  How will he ever be
able to shake the nagging feeling that we ought to be able to harness
the uninteresting version and turn it into the interesting one?  (For
that matter, how will the quantum practitioner?  Nicolas Gisin comes
to mind.)

About the particulars of your box example, I think the common man
would be hard pressed to take the textbook definition of a
problem---like the one you describe above---as a property possessed
by a physical system.  How can the box containing the red ball know
I've embedded it in a problem where the possible colors are {\{}red,
blue\}, or instead in a problem where they are {\{}red, blue,
green\}? People think of physical properties as the things physical
systems carry around with them independently of the rest of the
world.  You may say that this is a limiting conception of the word
``property,'' but I'm pretty sure it's the conception most people use.
They would call the set {\{}red, blue\} a property of the problem
you've defined, not of the system.  Within classical physics, they
would say the ball has whatever properties it really has (say, red OR
blue) \ldots\ and it is the physicist's task to figure out which of
the two it is.

The standard retort you and Charlie give me is that I am doing
nothing more than encumbering the language by saying ``a quantum
state is ascribed to a system'' rather than saying ``a quantum state
is possessed by a system.''  But, come on, the word count is the
same. It is not that I am encumbering the language; it is just that I
am beating on a prejudice you don't want to let go of.  Or maybe to
be more conciliatory, it is that you cannot imagine that this kind of
language could ever be useful, whereas you think there are loads of
examples where your own language has led to triumph.  But I think
Charlie's talk the other day about the public's perception of quantum
teleportation as a kind of instantaneous action-at-a-distance (in the
science-fiction sense) is a case in point.

Things are only interesting or noninteresting with respect to a
context.  I think there is a sense in which quantum teleportation is
less interesting with respect to the conception that a quantum state
captures a state of knowledge rather than a property possessed. What
could be less interesting to say than that, ``Quantum teleportation
is the transference of one's predictions about one object onto
another object that has never interacted with the first''?  Maybe
it's only this that's keeping you in the old bounds. Is it that if
you keep a little bit of the science-fiction imagery alive, it'll
help fuel the physics?

``In any case, none of this matters for doing physics,'' you say. But
I think it does in the long run.  (Certainly, you've got to concede
that there's something that fuels me---and it hurts to think that you
might think it is nothing more than irrationality.)  When Charlie
sent me the picture of a skunk cabbage in reply to these very issues,
I found myself thinking of King Broccoli.  The story goes that one
day, by divine providence, it came to King Broccoli that broccoli,
the vegetable, his namesake, actually tastes good.  Good in a way
that hitherto only gods and angels had known.  Every child who had
ever said, ``Yuck, I don't like broccoli; it tastes awful!''\ was
simply wrong \ldots\ or at least that's what the king realized. King
Broccoli, being the head of state, decided to do something about it.
Henceforward, all gardens in the kingdom should have a patch devoted
solely to broccoli.  It really wasn't much of a burden on the
national product (except, perhaps, for the psychiatrists who had to
treat all the movie stars who had never felt fulfilled in their
broccoli experience).  But think of the diversity of vegetables the
kingdom might have raised if its citizens hadn't been encumbered with
the king's notion that broccoli had an objective, but never
verifiable, taste?

Moral?  Maybe there's none.  But it is a documented fact that the
Kingdom of Broccoli eventually fell and was replaced by a liberal
democracy (where the ideals rather than the particulars have an
objective status).

Everyone in this mailing list knows by now---though Charlie and John
seem to keep forgetting this, or maybe they've never let it sink
in---that I think quantum mechanics is just the hint of something
much deeper, some fantastic physics yet to come.  But I also don't
think we will ever stumble upon that physics until we truly get rid
of our classical prejudices:  seeing quantum states as
``properties''---it seems to me---is one of these.\smallskip

\noindent Signing off,\smallskip

\noindent Chris \smallskip

\noindent My own disclaimer:  Though I implied above that quantum
teleportation becomes less interesting within a subjective conception
of the quantum state, I think effects like quantum cryptography
become {\it more\/} interesting from this view.  So there is a
tradeoff.

\section{16-05-02 \ \ {\it Words} \ \ (to C. H. {\Bennett})} \label{Bennett20}

\bcb
Sorry.   Point well taken.  You are a perpetual stimulus to me, if not
always in the ways you hope, and I would miss it terribly if you
stopped.  I should be more grateful. Like the other day, when John was discussing your automotive metaphor,
you inspired me to think that all the cautions you would have us take
against quantum misconceptions are---for me, if I did them---like
driving a car with the parking brake on all the time.
\ecb

Then it seems to me you should at least be consistent in your behavior.  I interpreted your talk Sunday as a genuine concern for the perceptions the masses hold about some of our favorite quantum effects.  Do you have a concern, or do you not?  Is it that the world really has some kind of instantaneous action at a distance---like John's note yesterday supported---and we're just not allowed to say the phrase in polite company, or does it not?  If the world does not, then so be it.  But if it does, why should we try to so hard to delete the phrase from polite conversation?  The main point I always wonder is how well you really have these issues worked out in your own head.  If you care about misconceptions, then care about them---I say in this slightly grumbling state---and if you don't, then don't.  But I have trouble understanding your mix of halfheartisms.

\subsection{Charlie's Reply}

\bq
I think the world does not have instantaneous action at a distance, and it's important to find ways of speaking that do not encourage the
frequent misconception that entanglement provides a means of faster than light communication.  I am less interested, and think you should be less interested, in what seem to me to be hair-splitting arguments about ``properties'' or ``changes in one's state of knowledge'' that merely reflect different ways of talking about situations in which we entirely agree about the predictions of outcomes for any experiment. That seems a matter of aesthetics only, and the protections you find so reassuring I find merely annoying, like a parking brake, since they complicate my language and don't prevent me from making any wrong predictions, just what you would call wrong or fuzzy interpretations.  I would say that if a way of thinking such as my own does not lead to wrong predictions, then it is not fuzzy in any serious way.  Or to put in another way, the distinctions you would have me make because they seem so important to you are in fact less real than the wave function you don't want me to believe in.

Talking with you first thing in the morning is way better than coffee.
\eq

\section{16-05-02 \ \ {\it Exterior/Interior} \ \ (to W. K. Wootters)} \label{Wootters9}

\bbw
Whitehead argued that a materialistic philosophy could never
adequately account for mind.  So in his view, the basic elements of
the world are not material objects but are more like moments of
awareness: they are more like events than persisting objects, and they always have both an interior and an exterior aspect.  There is some
similarity with Wheeler's view that the basic elements are quantum
events, except that with Whitehead there is more of a balance between the interior and exterior aspects of each basic element.  Wheeler
thinks of quantum events merely as objective events, seen only from
the outside, as it were.  Whitehead says that at every level of being, there is always an interior view.  At the higher levels, this interior
view is what we call consciousness.  At lower levels, it's not
consciousness but it exists nonetheless.

Of course I like the Wheelerian idea of taking quantum events as
basic, and I like this question:  Where does a quantum event reside?
\ebw

There is also a similarity to Schopenhauer.  He called the interior stuff the ``will.'' Moreover, Schopenhauer also placed the will outside of space and time.  But maybe that's where the similarity ends; I don't know enough about either philosopher to say.  Let me place below two little paragraphs from my samizdat in which I had written about Schopenhauer.  [See 10-07-01 note ``\myref{Landahl1}{Replies on a Preskillian Meeting}'' to A. J. Landahl and 07-08-01 note ``\myref{Waskan1}{Kiki, {\James}, and {\Dewey}}'' to J. A. Waskan.] So I think there must be some significant overlap in what you and I are searching for.

But there are some differences:
\bq
What I like about Whitehead and Malin is that they emphasize a
commitment to consciousness as something central to existence and
something that one has to have in mind even when talking about the
most basic elements of existence.
\eq
though I'm not committedly against this.  So, now thinking back on the long letter I wrote you, ``It's a Wonderful Life'' it dawns on me that you might have had trouble---at the very least!---accepting the part where I said something like ``I eschew idealism in all forms.''

I will try to come up to speed on Whitehead not too far in the future.

\section{17-05-02 \ \ {\it {\Spekkens} Summary} \ \ (to R. Pike)} \label{Pike8}

Is this the sort of thing you're looking for?
\bq
Robert {\Spekkens}'s research plan is to determine what sort of information-theoretic security can be achieved for two-party cryptographic tasks using quantum protocols.  Two-party cryptographic tasks are those that are implemented between a pair of spatially separated parties that do not trust one another (like bit-commitment, coin-tossing, and remote gambling).  These are some of the so-called ``post-cold-war'' applications of cryptography that may have commercial significance.  Information-theoretic security is the security that can be achieved regardless of the technological capabilities of the adversary (including the capability to build a quantum computer).  In other words, it is the security that can be guaranteed by virtue of the laws of physics alone.  In addition to being of obvious foundational significance to cryptography, this research addresses the practical question of the extent to which quantum cryptographic protocols can take the place of classical protocols if the security of the latter are compromised (for instance, by quantum computers).

 {\Spekkens} is well-positioned to carry through with this research plan, since he has already demonstrated an ability to make original and significant contributions at the forefront of the field.  In fact, some of his papers have defined the very field.  The plan is to support {\Spekkens} through a DIMACS postdoctoral fellowship---assuming he obtains it---which would pay for both his salary and benefits.  The only burden to Lucent we foresee is the donation of some office space.  He would most likely stay here through the tenure of the fellowship, which may be up to two years.
\eq

\section{17-05-02 \ \ {\it No Nasty} \ \ (to T. Rudolph)} \label{Rudolph4}

I won't be nasty; I enjoyed your note.

Your thinking has a lot of the flavor of the paper:  A. Peres and W.
H. Zurek, ``Is Quantum Theory Universally Valid?,'' Am. J. Phys. {\bf
50}, 807 (1982).  I'm not sure if you're aware of it.  Asher once
told me that he himself still likes the paper a lot, but Zurek
basically disavows it now.

\btr
what I'm trying to understand is the physics analogue of Turing's
construction -- what is it that I, a regular physicist, am doing in
my interactions with the world and my construction of theories to
explain those interactions?
\etr

I like this question a lot, actually.  Below is some of my own
attempt to ask the same thing.  In this regard, I suggest you read
Richard {\Rorty}'s book, {\sl Philosophy and Social Hope}.  I think if I
were to read it again, I would realize how much of an influence it's
had on my own thoughts.  In fact, my Anti-{\Vaxjo} paper may be
little more than a condensation of it.

I guess if you were to ask me now---in my present state of mind---I
would say we are doing more than constructing theories.  We are
constructing the world (in part).  See other note below.  But beyond
that, I don't know how to say more presently.

Keep thinking about your question!  Don't listen to {\Spekkens}.

By the way, I won't be around for essentially a month:  Today, I
leave for Texas until Thursday.  Then the Saturday after that I'm off
to Brisbane for a 3 weeks.  I return to Bell Labs June 17.  (I'm
still looking for volunteers on the fence.)

\subsection{Terry's Preply, ``Musings on what the heck it is we're doing \ldots''}

\bq
I'm trying to get a handle on what it is we (human) physicists actually take axiomatically (although implicitly) to be true before constructing our theories. From this, if Chris' many pages of papers are even an epsilon amount toward the truth, I should be able to get some glimmering about why quantum mechanics is the way it is -- that is, if it really is only my way of describing my  `pushing on the world' to paraphrase Chris. [This latter statement is of course is NOT something that ALL physicists take to be true, so it can't be part of my list!]

So here's a brief synopsis of 3 things that I feel go, in some sense
axiomatically, into any (rational) physicist's theories:
\begin{enumerate}
\item
{\bf Spacetime Invariance}\smallskip\\
We all believe, after some fashion, that an experiment can be replicated.  Whenever we do so we implicitly neglect certain things that are not EXACTLY the same about the second experiment. The things we neglect generally vary from one situation to the next --  we use our intuition about whether they're important or not and this can be dangerous.  However, in all physics experiments, we presume that the fact the second experiment must take place at a different point in spacetime from the first (otherwise it WOULD be the first experiment) is not important with regards to its testing of our theory etc.

(Wild speculation aside: Since space/time translation invariance of a
Lagrangian give, by Noether's theorem, conservation of momentum/energy, it would be interesting to conjecture that conservation of momentum/energy is a necessary consequence of a fundamental philosophical belief we have about the scientific method, and not a ``truly objective'' physical principle whatever
the heck that is.)

\item
{\bf Universality of Physics}\smallskip\\
We often say blas\'e things along the lines of the universe is defined to be all we can observe. I think we all believe that physics, in principle, should be able to describe everything in the universe, that is, everything in the ``natural'' and not ``super-natural'' so to speak. If so then we must
believe that it describes US as physical objects too. Anthropocentricity has failed us as an axiom too many times for anyone to rationally hold to it.

\item
{\bf Free Will}\smallskip\\
The whole study of science is meaningless without free will. I personally believe I have free will. By point 2, this means I must accept the possibility that other objects in the universe have free will -- and since I assume you all claim that you do too I'll accept you as candidates for those other free will objects (you could be computers programmed to claim it, but not actually have it, for all I can really tell!)
\end{enumerate}

What do I gain from all this? At the moment the only thing I can say is this. If I accept these three points, then I must conclude that any truly universal (complete?)\ physical theory we construct must be
non-deterministic. Once there are two objects A, B in the universe that object A accepts have free will, then object A cannot rationally devise a deterministic theory that includes object B. So either we give up on Point 2 or we give up on 3, or we accept non-determinism as a inevitable consequence of performing physics as we currently know it.
[Note that I am NOT saying the oft stated converse, which I think is
completely pithy (and thus probably due to Penrose), namely that the free will we have will one day be found to be because of the uncertainty principle.]

Those are my thoughts in a nutshell. They have evolved from thinking a lot about the following:
\bq\noindent
{\bf A Physics Turing Machine}\smallskip\\
Turing was trying to capture the whole process of a mathematician's
thinking. A mathematicians pen and paper were simply an extension of his mind Turing said. The symbols he drew came from a countable set he argued, and from arguments along philosophical line about the process of performing mathematical calculations came the Turing machine. Today we relegate this machine to the status of a Pentium. However what I'm trying to understand is the physics analogue of Turing's construction---what is it that I, a regular physicist, am doing in my interactions with the world and my construction of theories to explain those interactions? What sort of theories can object A devise about object A? [To call something which performs arbitrary unitary transforms a ``universal quantum turing machine'' is repulsive.]
\eq

So, what should I be adding or subtracting from 1--3 above?

And if you dont agree with something I've said don't be mean and nasty in your replies! Yes, I mean you too Fuchs.
\eq

\section{17-05-02 \ \ {\it Dueling Banjos} \ \ (to W. K. Wootters)} \label{Wootters10}

\bbw
Well, it's true that there's a strong idealistic current running through
me.  Here's an interesting paragraph from Hartshorne about realism and
idealism:
\bq\noindent{\rm
It appears, then, that the idealistic interpretation of reality as
essentially relative to or consisting of mind, experience, awareness, that
is, psychicalistic idealism, is entirely compatible with a realistic view
of the independence of the particular object and the dependence of the
particular subject, in each subject-object situation. It may also be urged
that we need the word ``realism'' to refer to the mere thesis that every act
of knowledge must be derivative from a known which is not derivative from
that act. Thus the practice of contrasting ``idealism'' and ``realism'' as
though they were contradictories, is of doubtful convenience. ``Realistic
idealism,'' or ``realistic subjectivism,'' has a reasonable and consistent
meaning.}
\eq
\ebw

Thanks for the Hartshorne quote.  I had forgotten that I had sort of
agreed with that a bit.  Here's the way Martin Gardner put it in his
essay ``Why I Am Not a Solipsist'':
\begin{quotation}
In this book I use the term ``realism'' in the broad sense of a
belief in the reality of something (the nature of which we leave in
limbo) that is behind the phaneron, and which generates the phaneron
and its weird regularities.  This something is independent of human
minds in the sense that it existed before there were human minds, and
would exist if the human race vanished.  I am not here concerned with
realism as a view opposed to idealism, or realism in the Platonic
sense of a view opposed to nominalism or conceptualism.  As I shall
use the word it is clear that even Berkeley and Royce were realists.
The term of contrast is not ``idealism'' but ``subjectivism.''
\end{quotation}
(The phaneron, by the way, was {\Peirce}'s term for ``the world of our
experience---the totality of all we see, hear, taste, touch, feel,
and smell.'')

Thus, in making the transition from the first paragraph to the first
sentence of the second paragraph in the excerpt from my old letter
below, I was making a non sequitur.

Let me ask you this about your ``idealistic current.''  Does it run
counter to what my first paragraph below says and what Gardner says
above?  I guess that's the main point for me.

I suppose if I were to start to label things, then this thing I was
telling you about the other day---``the sexual interpretation of
quantum mechanics (SIQM)''---would be a kind of dualistic theory.  I
said it metaphorically this way:  When things bang together,
something is created that is greater than the sum of the parts.  Or
again:  When things---that's the materialistic aspect---bang
together, something is created---that's the mentalistic aspect, for
it is like an act of the will or a decision.  But that's just a
thought that's hitting me at 4:00 in the morning.  (So trust it less
than even the usual things that come out of my mouth.)  I hadn't
thought about it in this way before, and I'm not sure I want to
continue to thinking about it this way.  In general, I don't like
dualisms.  (Though even saying that is paradoxical; for I think I
like ``pluralisms'' in the sense of {\James}.)  The excerpt from an old
letter far below gives a slightly longer introduction to the idea of
the SIQM.

\subsection{Bill's Reply}

\bq
I agree at least with the letter of Martin Gardner's paragraph,
because I don't think there's anything particularly special about
{\it human minds}.

But I disagree with the spirit of your sentence:
\bq\noindent
Yes there is certainly a kind of realism working in the back of my
mind, if what you mean by ``realism'' is that one can imagine a world
which never gives rise to man or sentience of any kind.
\eq

Let me explain.  I think the sort of world envisioned by classical
physics is in fact impossible.  If we really understood what it
takes to make a world exist {\it really}, and not just on paper,
I think we would see that one needs the subject-object relation
in order to hold things together.  This is not to say that one needs
dualism.  In Whitehead's system, everything is both subject and
object, depending on the point of view.  There is no dualistic
separation into two kinds of entity.  But everything is related
to something else as subject, and everything is also related to
something else as object.  In classical physics, there is no such
relation.  As {\Schroedinger} points out, from the very beginning we
eliminate the very notion of an experiencing subject.  I agree
with {\Schroedinger} that this is too extreme an abstraction.  In
making this abstraction, we have removed something essential
from our view of the world.

I know you will say that your realism is not the same as that
of classical physics.  But in what way, ultimately, is it not
the same, other than by the absence of determinism?
I think indeterminism is crucial, but I think one needs the
further step of restoring the subject-object relation as
fundamental.

I've heard some people argue as follows.  ``There are many
conceivable physical worlds, because one can imagine all
sorts of different physical laws, expressed as mathematical
equations.  Now, the mathematics certainly exists on its
own.  It doesn't need physical realization.  But since the
mathematics exists, and since the mathematical laws are
the essence of the physical world that they would describe,
then all these physical worlds must {\it actually exist}.
They exist because the mathematics that describes them
exists.''

One is led to this conclusion (which I think is a wrong conclusion)
because there really is hardly any difference between (i) an
actually existing material world that follows definite laws, and
(ii) a mathematical description of such a world.  There's
nothing important in the ``actually existing'' world that's
not already present in its description.  What difference does
``existence'' make in this case?  What does existence mean
in this case?

I would like to think that my view avoids the line of reasoning
that takes the mathematical description as the essence of
the world.  As John Wheeler says, a set of mathematical
laws will not ``fly'' by itself.  I think the necessary added
ingredient is something like {\it experience}.  And that's what
I find in Whitehead's view.

\eq

\section{17-05-02 \ \ {\it Slide Show} \ \ (to N. D. {\Mermin} \& C. H. {\Bennett})} \label{Mermin66} \label{Bennett21}

Boy you got me into a stink, didn't you, by getting me into that word
debate!  If you hadn't done that, I might have had some time to
answer your other questions this week.

Below for your continued amusement, I'll include a side conversation
I had with Charlie.  Maybe this debate is at least edifying in some
ways.  Charlie sees me as hair-splitting; I see Charlie as being
half-hearted and inconsistent.

I just read these words in a Martin Gardner article:
\begin{quotation}
A third aspect of aesthetic theory that bores me even more are all
those tiresome disputes, in book after book, about whether aesthetic
values are subjective or objective. Here the situation is not quite
the same as that of truth. In previous chapters I have argued that
the least confusing way to talk about truth is to assume that the
world and its structure are not mind-dependent. But beauty, so far as
humanity is concerned (we will not consider what beauty may mean to
birds or apes, to creatures on other planets, or to gods), obviously
requires a human mind. Where is the red of an apple? As I have said,
it is in the mind if by red we mean the sensation of red. It is on
the apple if by red we mean the structure of a surface that reflects
a pattern of visible light which causes a mental sensation of red.

I see no difference between this antique quibble and the question of
whether beauty (however defined) is a property of an art object or a
sensation in a brain. If by beauty you mean the pleasure aroused by a
beautiful object, of course it is subjective. If by beauty you mean
the structure of an object capable of arousing aesthetic pleasure,
then the beauty is a part of the object. Or you may prefer a third
approach and ground beauty in the combined dynamic structure produced
by the interaction of an object and a mind. It is all such a weary
waste of words. The last approach is the one taken by John {\Dewey} in
his influential book Art as Experience. Although I found fault with
{\Dewey}'s attempt to redefine truth in pragmatic terms, I find his
approach to aesthetics (essentially the same, by the way, as
Aristotle's) a sensible way of speaking. Again, it is not a question
of {\Dewey} being right or wrong. It is a question of the most useful
way to talk about aesthetic values.
\end{quotation}

And I find that I agree with most every word of this (except for the
part about the {\James}ian-{\Dewey}an theory of truth).  But the issue at
stake with quantum mechanics goes much deeper than this, and it
annoys me to no end that our friend Charlie lumps me in with the art
critics.

If you have any words of wisdom that could take a little fire out of
our relations, I'd love to hear them.  Maybe I'll CC this note to
Charlie too.

By the way, this was not intended to be the subject of this note.  I
wrote instead to tell you that I have now posted the slides from my
talk at your party on my website.  Maybe you'll enjoy seeing the
second half, if not hearing it.

\bq
\noindent Charlie,
\bcb
Surely you must have something more sensible to say than some of us
who have spoken.
\ecb

It would have been more neutral to say, ``than those of us who have
spoken.''  If I were trying to read between the lines, I might be
tempted to write a note just like this one.  But I'll refrain from
reading between the lines.\medskip

\noindent Chris\medskip

\noindent ---------------------------------\medskip

\noindent Dear Chris, \medskip

Sorry.  Point well taken.  You are a perpetual stimulus to me, if not
always in the ways you hope, and I would miss it terribly if you
stopped.  I should be more grateful.  Like the other day, when John
was discussing your automotive metaphor, you inspired me to think
that all the cautions you would have us take against quantum
misconceptions are---for me, if I did them---like driving a car with
the parking brake on all the time.\medskip

\noindent -CHB\medskip

\noindent ---------------------------------\medskip

\noindent Charlie,\medskip

\bcb
Like the other day, when John was discussing your automotive
meta\-phor, you inspired me to think that all the cautions you would
have us take against quantum misconceptions are---for me, if I did
them---like driving a car with the parking brake on all the time.
\ecb

Then it seems to me you should at least be consistent in your
behavior.  I interpreted your talk Sunday as a genuine concern for
the perceptions the masses hold about some of our favorite quantum
effects.  Do you have a concern, or do you not?  Is it that the world
really has some kind of instantaneous action at a distance---like
John's note yesterday supported---and we're just not allowed to say
the phrase in polite company, or does it not?  If the world does not,
then so be it.  But if it does, why should we try to so hard to
delete the phrase from polite conversation?  The main point I always
wonder is how well you really have these issues worked out in your
own head.  If you care about misconceptions, then care about them---I
say in this slightly grumbling state---and if you don't, then don't.
But I have trouble understanding your mix of halfheartisms.\medskip

\noindent Chris\medskip

\noindent ---------------------------------\medskip

\noindent Dear Chris,\medskip

I think the world does not have instantaneous action at a distance,
and it's important to find ways of speaking that do not encourage the
frequent misconception that entanglement provides a means of faster
than light communication.  I am less interested, and think you should
be less interested, in what seem to me to be hair-splitting arguments
about ``properties'' or ``changes in one's state of knowledge'' that
merely reflect different ways of talking about situations in which we
entirely agree about the predictions of outcomes for any experiment.
That seems a matter of aesthetics only, and the protections you find
so reassuring I find merely annoying, like a parking brake, since
they complicate my language and don't prevent me from making any
wrong predictions, just what you would call wrong or fuzzy
interpretations.  I would say that if a way of thinking such as my
own does not lead to wrong predictions, then it is not fuzzy in any
serious way.  Or to put in another way, the distinctions you would
have me make because they seem so important to you are in fact less
real than the wave function you don't want me to believe in.

Talking with you first thing in the morning is way better than
coffee.\medskip

\noindent -CHB
\eq

\section{17-05-02 \ \ {\it More Balking} \ \ (to R. {\Schack})} \label{Schack56}

\brs
\bq
\rm\noindent [CAF wrote:]
\bq
\noindent {\Ruediger} said, ``I am still convinced, despite your severe
scolding, that a priori, without looking at the consequences, strong
and weak coherence are similarly compelling axioms.''
\eq
I don't see how you can make this distinction.  It seems to me,
axioms are only compelling or not compelling insofar as their
consequences.
\eq
Then why would there be a difference between postulating the
Kolmogorov axioms and postulating coherence? Why would we bother with
the Dutch book arguments? Why do you bother looking for an
information-theoretical reason for the quantum axioms?

It seems to me that the compelling reason to accept the axiom of
(ordinary) coherence is that ``whatever the consequences are, I
certainly do not want to violate coherence, because I don't want to
hand over money''. It seems to me that a discussion of whether strong
coherence is a compelling axiom has to be carried out at this level.
\ers

But that is not what is at issue.  Strong coherence and regular
coherence are {\it two different theories}, and a theory's worth can
only be assessed by looking at the whole thing.

It seems to me a very simple issue.  In the theory of strong
coherence, there is a kind of normative behavior for all agents that
regular coherence does not have.  The theory says, in effect, ``Thou
shalt not tell a lie.''  If an agent writes down $p=1$, then in the
rigid world of strong coherence, I can trust his statement to be a
reflection of his true inner belief.  The world of regular coherence
does not have that.

You can say that one theory is just more of the same with respect to
the old theory.  But that is to only look at one aspect of the
problem.  In another aspect of it, the two theories become
qualitatively different.

There is a sense in which my mother-in-law is just a faster version
of Kiki at times.  That is, she's just more of the same when it comes
to cooking, artistry, and a couple of other aspects.  But I fell in
love with Kiki, not my mother-in-law.  This is because when I move
past a few isolated features of my mother-in-law, I discover she is
qualitatively different from the woman I love.

How can you reject that as a reasonable line of thought?

If you thought at the outset that probabilities could be proper
properties of things---like Shimony does---then you might indeed
accept strong coherence.  For strong coherence grounds the very
meaning of $p=1$.  A ``$p=1$''--statement is a TRUE statement.  But
if you start out from a strongly subjectivist stance on probability,
then the only leg you have to stand on is the kind of argument you
and {\Carl} give:  It is just more of the same of regular coherence; it
is just slightly more sensible or cautious betting behavior.  But I
counter that by saying, carried through consistently and without
exception, strong coherence makes it nigh on impossible to embed
one's particular bets about particular events in the larger framework
of all bets and all events.

It is an issue worthy of debate, I agree, whether my reason is a
decent reason to reject strong coherence.  (I think it is, clearly.)
But it's not fair to say in distinguishing horses and zebras, we
should all close our eyes and focus on tactile differences, eschewing
all the visual information around us.

\section{17-05-02 \ \ {\it The Divinity that Breathes
Life} \ \ (to W. K. Wootters)} \label{Wootters11}

\bbw
I know you will say that your realism is not the same as that of
classical physics.  But in what way, ultimately, is it not the same,
other than by the absence of determinism?  I think indeterminism is
crucial, but I think one needs the further step of restoring the
subject-object relation as fundamental.
\ebw

To the extent that I could write something like the passages below, I
don't think so.  [See 20-11-01 note ``\myref{Renes5}{One Horse's Mouth}'' to
  J. M. Renes.]  Or, say, to the extent that I find myself liking this
little quote, ``I forbade any simulacrum in the temples because the
divinity that breathes life into nature cannot be represented,'' I
don't think so.  I think there's more than simple indeterminism in my
forming view; there is something lower level than determinism and
indeterminism both.

But what I am not settled on is whether the ``divinity that breathes
life'' is the subject-object relation.  I will think hard about your
point, and I will think harder about the impression my writings give.

\section{18-05-02 \ \ {\it QM as QI, cont'd} \ \ (to A. Peres)} \label{Peres36}

Thanks for the support.  As I wrote Lou Hand, an accelerator physicist at Cornell, who wrote me something nice after my talk there, ``My fragile self-confidence needs that sort of thing every once in a while.''

Kiki, Emma, Katie, and I arrived in Texas late last night.  We will visit my mother until Wednesday morning.  It is her first time to meet Katie.

I hope Aviva and the rest of the family are doing well.

\subsection{Asher's Preply}

\bq
I read a few more pages of QM as QI, and I really enjoy it very much. [See C. A. Fuchs, \quantph{0205039}.]
A few remarks:
\begin{itemize}
\item

[21] Rosen also told me that I was a solipsist (about 40 years ago).

\item

Bureau International des Poids et Mesures is in S\`evres (not far from
Paris), not in Paris itself.

\item

[27] is a bad reference for POVM. Their explanations and formulas are
plainly wrong. I wrote it to them and they told me this will be
corrected in a future edition.

\item

page 9: I loved ``one can immediately write down a new state for the
distant system''. See the circular I just sent to Bennett et al.

\item

I loved footnotes 14--16.
\end{itemize}
More later.
\eq

\section{20-05-02 \ \ {\it Denying Free Will} \ \ (to T. Rudolph)} \label{Rudolph5}

\btr
This concept of free will bugs the hell out of me -- in a sense when
you say ``I think we're constructing the world (in part)'' I think any
physicist would have to agree with you (in part) or else they'd end up
denying free will.
\etr

I know a lot of physicists who deny free will; Charlie {\Bennett} and John Preskill come to mind.

\btr
Do you think there are more than the three ``implicit assumptions'' I
mentioned?
\etr

Yes and No.

\section{20-05-02 \ \ {\it Dizzy in Texas} \ \ (to J. A. Smolin)} \label{SmolinJ5}

\bjas
Now you guys are trying too hard not to think like Chris.  Nothing is
magic about the ``instantaneous'' first stage of teleportation, only
knowledge is being changed.  If I have two boxes, one with a ball in
it and one without and send one far away, when I open the remaining
one I instantly know if the ball is in the other one.  Big deal.
When I measure something, I find something out.  Wow!
\ejas

I shouldn't let your note encourage me.  (The sober side of myself
won't.)  But there is a way to make the sum content of all quantum
measurements look exactly like your analogy above, even in a formal
way.  The mathematics may interest you, even if not the philosophy.
You can find it in Sections 4.2 and 6.1 of the paper I put on {\tt
quant-ph} last week (\quantph{0205039}).  Or you can see it
sketched quickly in the {\Mermin}fest talk posted at my website.

The trick is to represent the quantum state as a probability
distribution with respect to a fixed, fiducial informationally
complete POVM.  That is, one imagines a ``standard quantum
measurement device'' sitting in the National Bureau of Standards
beside the standard meter and the standard kilogram, and the quantum
state captures nothing more than one's judgment for how the device
will react if one were to throw one's quantum system into it.  With
that picture in mind, all a regular, everyday measurement does is
update one's judgment concerning the outcomes of the standard
measurement. The twist that comes with quantum mechanics (over
regular probability theory) is that the update rule is Bayes' rule
plus a little more \ldots\ it's not just Bayes' rule full stop.

\section{23-05-02 \ \ {\it Plans and Plans} \ \ (to C. M. {\Caves})} \label{Caves63.2}

You've probably noticed that I'm now scheduled to be in LAX from 6:00 PM to 10:30.  So, I'm trying to figure out where I'm going to hang out.  I've got use of both the Admiral's Club in Terminal 4 and the Qantas Club in the International Terminal.  If I'm going to meet with you for any of that time, there may be problems with both. \ldots

Right now, I'm feeling pretty negative on traveling in general, but will try to cheer up before I meet you.  In particular, I'm feeling a little negative on this whole trip \ldots\ given our diverging views on where we want to see quantum foundations taken and given our diverging styles.  As it stewed in my head the other night, your opinion of my attempts to sketch/communicate a Bayesian/pragmatic/Darwinian conception of nature as ``cloyingly affected'' just plain hurt.  Clearly I crave your respect, but then someone like Michael Berry comes up and tells me how he liked the same article.  It seems I can't win.  More and more I feel that my life is short and I've got to do something with it.  I don't compile lists of people's comments on my papers to ``build an ego''; I do it to build confidence.  And I do it to check that I'm not going too far off the deep end when I try to see past the edge.  Confidence is something I have far less of than you seem to understand.  Everyday I fear the world.

\section{24-05-02 \ \ {\it Better Moods} \ \ (to C. M. {\Caves})} \label{Caves63.3}

OK, I'm in a better mood now \ldots\ and will appear before you with a smiling face in either LA or Brisbane.  In any case, that's the sort of thing I do when confronted with a human face.  I looked back at the note I wrote you yesterday and noticed that I used the word confidence throughout.  But the issue is self-confidence.  Somehow it seems to make a difference in meaning.

Kiki and the kids took off for Munich yesterday, and I already miss them horribly.  It's difficult to imagine a month without my two little girls (even though when they're around I'd rather only see them less than two hours a day).  Moreover, experience tells me that I will be wiped from Katie's memory in less than three weeks; we'll have to start our relationship over again.  So, I really hope we can make this a productive time.

I'll come armed with de Finetti's book, Bernardo and Smith, and Kyburg and Smokler.  Can you think of anything else I ought to be bringing?  Could you bring de Finetti's paper ``Probabilismo''?  I don't have a copy of that anymore.  Also, do you have any papers by Dick Jeffrey?  Thinking of that, there was also that paper by Jenan Ismael where she suggested using subjective probabilities for quantum probabilities.

\section{24-05-02 \ \ {\it Airplane Reading} \ \ (to C. M. {\Caves})} \label{Caves63.4}

I just posted the latest version of the samizdat {\sl Quantum States:\  What the Hell Are They?}  202 pages now.  If you're looking for something to read on the plane, it might help you catch up on all I've been spouting off since our last face-to-face meeting.

A condensed version of the latest ideas can be found in the talk ``Where's a Good Weatherman When You Need One?''

Finally, the fat paper is posted at {\tt quant-ph}.  I can't remember if I told you:  \quantph{0205039}.

See you tomorrow.

\section{24-05-02 \ \ {\it Snail Mail}\ \ \ (to S. L. Braunstein)} \label{Braunstein4}

Tell me about some good places to go in Brisbane.  Carl, {\Ruediger}, and I are converging there to work on this Rev.\ Mod.\ Phys.\ paper on Bayesianism in QM we're trying to write up.  I have a small fear that the details of the paper may strain our relationship to the point of divorce, but let us hope for the best.

\section{25-05-02 \ \ {\it Fun with Feyerabend} \ \ (to M. A. Nielsen)} \label{Nielsen2}

I'm just putting some notes into my computer before I set off for
travel again.  At your recommendation, I've paid a little more
attention to Paul Feyerabend.  The quotations below summarize a
little of what I see as likable and usable in his philosophy.

See you tomorrow.

From:  P. Feyerabend, {\sl Conquest of Abundance:\ A Tale of
Abstraction versus the Richness of Being}, edited by B.~Terpstra (U.
Chicago Press, Chicago, 1999).
\bq
{\it Humans as Sculptors of Reality.}  According to the first
assumption, our ways of thinking and speaking are products of
idiosyncratic historical developments. Common sense and science both
conceal this situation. For example, they say (second assumption)
that atoms existed long before they were found. This explains why the
projection received a response, but overlooks that vastly different
projections did not remain unanswered.

A better way of telling the story is the following. Scientists, being
equipped with a complex organism and embedded in constantly changing
physical and social surroundings, used ideas and actions (and, much
later, equipment up to and including industrial complexes such as
CERN) to {\it manufacture}, first, metaphysical atoms, then, crude
physical atoms, and, finally, complex systems of elementary particles
out of a material that did not contain these elements but could be
shaped into them. Scientists, according to this account, are
sculptors of reality---but sculptors in a special sense. They not
merely {\it act causally\/} upon the world (though they do that, too,
and they have to if they want to ``discover'' new entities); they
also {\it create semantic conditions\/} engendering strong inferences
from known effects to novel projections and, conversely, from the
projections to testable effects.  We have here the same dichotomy of
descriptions which Bohr introduced in his analysis of the case of
Einstein, Podolsky and Rosen.  Every individual, group, and culture
tries to arrive at an equilibrium between the entities it posits and
leading beliefs, needs, expectations, and ways of arguing. The
separability assumption arises in special cases (traditions,
cultures); it is not a condition (to be) satisfied by all, and it
certainly is not a sound basis for epistemology. Altogether, the
dichotomy subjective/objective and the corresponding dichotomy
between descriptions and constructions are much too naive to guide
our ideas about the nature and the implications of knowledge claims.

I do not assert that any combined causal-semantic action will lead to
a well-articulat\-ed and livable world. The material humans (and, for
that matter, also dogs and monkeys) face must be approached in the
right way. It {\it offers resistance}; some constructions (some
incipient cultures---cargo cults, for example) find no point of
attack in it and simply collapse. On the other hand, {\it this
material is more pliable than is commonly assumed}.  Molding it in
one way (history of technology leading up to a technologically
streamlined environment and large research cities such as CERN), we
get elementary particles; proceeding in another, we get a nature that
is alive and full of Gods. Even the ``discovery'' of America, which I
used to support the separability assumption, allowed some leeway, as
is shown by Edmondo O'Gorman's fascinating study, {\sl The Invention
of America}.  Science certainly is not the only source of reliable
ontological information.

It is important to read these statements in the right way. They are
not the sketch of a new theory of knowledge which explains the
relation between humans and the world and provides a philosophical
grounding for whatever discoveries are being made. Taking the
historical character of knowledge seriously means rejecting any such
attempt. We can describe the results we have obtained (though the
description will always be fatally incomplete), we can comment on
some similarities and differences that have come to our attention,
and we can even try to explain what we found in the course of a
particular approach ``from the inside,'' i.e., using the practical
and conceptual means provided by the approach (the theory of
evolution, evolutionary epistemology, and modern cosmology belong in
this category). We can tell many interesting {\it stories}. We cannot
explain, however, how the chosen approach is related to the world and
why it is successful, in terms of the world. This would mean knowing
the results of all possible approaches or, what amounts to the same,
we would know the history of the world before the world has come to
an end.

And yet we cannot do without scientific know-how. Our world has been
transformed by the material, spiritual, and intellectual impact of
science and science-based technologies. Its reaction to the
transformation (and a strange reaction it is!)\ is that we are stuck
in a scientific environment. We need scientists, engineers,
scientifically inclined philosophers, sociologists, etc., to deal
with the consequences. My point is that these consequences are not
grounded in an ``objective'' nature, but come from a complicated
interplay between an unknown and relatively pliable material and
researchers who affect and are affected and changed by the material
which, after all, is the material from which they have been shaped.
It is not therefore easier to remove the results. The ``subjective''
side of knowledge, being inextricably intertwined with its material
manifestations, cannot be just blown away. Far from merely stating
what is already there, it created conditions of existence, a world
corresponding to these conditions and a life that is adapted to this
world; all three now support or ``establish'' the conjectures that
led to them. Still, a look at history shows that this world is not a
static world populated by thinking (and publishing) ants who,
crawling all over its crevices, gradually discover its features
without affecting them in anyway. It is a dynamical and multifaceted
Being which influences and reflects the activity of its explorers. It
was once full of Gods; it then became a drab material world; and it
can be changed again, if its inhabitants have the determination, the
intelligence, and the heart to take the necessary steps.
\eq
and
\bq
{\Pauli}'s views have much in common with the general picture that
emerged from Aristotle's principle. In this picture we start with a
world (which I shall call the primal world, or Being) which behaves
in its own way and not necessarily in accordance with any one of the
laws that have been discovered by scientists. (Here we still have an
element of realism.) Humans are part of the primal world, not
detached aliens, and they are subjected to its whims: Being can send
scientists on a wild-goose chase---for centuries. On the other hand,
it permits partial independence \ldots\ and it provides some of those
acting independently (not all of them!) with {\it manifest worlds\/}
they can expand, explore, and survive in (manifest worlds are in many
respects like ecological niches). Inhabitants of a particular
manifest world often identify it with Being. They thereby turn local
problems into cosmic disasters. But the manifest worlds themselves
demonstrate their fragmentary character; they harbor events which
should not be there and which are classified away with some
embarrassment (example: the separation of the arts and the sciences).
The transition from one manifest world to another cannot be described
in either except by excising large regions originally thought to be
real---a good case for applying the notion of complementarity. Bell's
request that a fundamental theory should not contain any reference to
observation is satisfied, but trivially so. Being as it is,
independently of any kind of approach, can never be known, which
means that really fundamental theories don't exist.
\eq

\section{25-05-02 \ \ {\it Becoming William {\James}} \ \ (to G. L. Comer)} \label{Comer15}

A thought struck me today as I was flying from Newark to LA reading
Lloyd Morris's book {\sl William {\James}: The Message of a Modern
Mind}. It is this:  I think there's nothing I presently want more
than to become a modern (quantum) incarnation of William {\James}.

Let me leave you with a lovely quote from the end of Chapter 2.

\begin{quotation}
He conceived the individual's life, and all social progress, as a
form of perpetual experiment.  But he did not preach reckless faith.
The ``will to believe'' for which he argued is best defined as
courage weighted with responsibility.  Contingency signifies that no
precaution can absolutely eliminate all hazard of shipwreck.  The
individual must take everything into account that may tell against
his success, and make every possible provision to minimize disaster
in the event of his failure.  But having done so, he must act.  And
in this circumstance, {\James} preached the right of the individual to
indulge his personal faith at his personal risk.  The part of wisdom
would always be to believe what is in the line of one's needs, for
only by such belief is the need fulfilled.  Over a wide area of
existence, possibilities and not finished facts are the realities
with which we have actively to deal.  So {\James} argued, and pointed
out that ``as the essence of courage is to stake one's life on a
possibility, so the essence of faith is to believe that the
possibility exists.''  But his doctrine subordinated faith to action,
for the real utility of faith is to make action genuinely dynamic.
``These, then, are my last words to you,'' he told a group of Harvard
students.  ``Be not afraid of life.  Believe that life {\it is\/}
worth living, and your belief will help create the fact.''
\end{quotation}

Now I'm off to Brisbane.  For my own self, I believe it is worth
becoming William {\James}.

\section{27-05-02 \ \ {\it Late, but Never Too Late?}\ \ \ (to K. Svozil)} \label{Svozil2}

I hope you will excuse my horribly late reply to your letter of
4/19/02.  I have just gone through a month of hell of traveling
almost constantly, and I have gotten pathetically behind in all my
correspondence.  (Just check with Johann Summhammer there; he is also
in my queue!  I'm hoping to get to him later in the day, if not
tomorrow.)

Thanks for the continued interest in having me around for the quantum
structures meeting.  Regrettably, I think I'm going to have to bow
out of the possibility of coming.  Presently I'm in Australia,
separated from my wife and children for a month, and this trip is
making me realize that I shouldn't take on any further travel than
the stuff I'm already committed to for the summer.  It's a shame
really, because I am getting ever more involved in IQSA kinds of
ideas, and it would be a great opportunity for me to learn a lot more
about what is already ``out there'' mathematically \ldots\ just
waiting for me to plaster over with some words of Copenhagenish
flavor.  Beyond that though, I would love to have a chance to express
some of my point of view to that audience.  I think my talk at the
quantum structures session at the AMS meeting in Atlanta two months
ago or so went exceedingly well in that regard, and I found myself
really enjoying conversations with Dave Foulis, for instance,
afterwards.

I finally had a chance to read your paper ``What Could Be More
Practical than a Good Interpretation?,'' by the way.  There are
certainly large parts of it I agree with---I don't know if that will
shock you or not.  However, there was more to my paper with Peres
than just the title!  I do get a little shocked when I find people
reading the title of the paper as its sum content.  Here's the way I
put it to Philippe Grangier a few months ago:
\bq
\vspace{-.3in}
\noindent
\bpg
By the way I also disagree with your point of view that ``Quantum
Theory Needs No `Interpretation','' Phys.\ Today {\bf 53}(3), 70
(2000). The fact that a physical theory ALWAYS needs an
interpretation is in my opinion a central difference between physics
and mathematics.
\epg

You won't find a disagreement with me here.  The title and closing
sentence of that paper were meant to be tongue-in-cheek plays on
something Rudolf Peierls once said:  ``The Copenhagen interpretation
{\it is\/} quantum mechanics.''  The whole paper is very definitely
about an interpretation, and why one does not need to go any further
than it to make sense of quantum mechanics as it stands.  My paper
\quantph{0106166} and the large (more personal) collection \quantph{105039} is about going the next step, i.e., what to do once
we have established the belief that quantum states are states of
knowledge.

When we do finally dig up an ontology underneath quantum mechanics,
I'm quite sure it will be an interesting one!
\eq
[See 16-08-01 note ``\myref{Grangier1}{Subject-Object}'' to P. Grangier.]  And here was the way I put it to Paul Benioff a year before that:
\bq
\vspace{-.3in}
\noindent
\bpb
To me that is an interpretation of QM.  Interpretations are what give
otherwise empty theories their meaning.
\epb

You're quite right about that.  What Asher and I wrote about is
indeed a kind of interpretation of the quantum mechanical formalism.
The title and the ending words of the article were more for
attention-getting than anything else.  Also, though, the words were
meant to be a small slap in the face to some of the extremes people
have gone to (like Everett worlds, Bohm trajectories, and
Ghirardi--Rimini--Weber stochastic collapses) just to hold on to a
philosophic view that came around long before quantum mechanics was
ever heard of.  (Talk about people being set in their ways!)
\eq

I think the best renditions of my present views can be found in my
new papers, \quantph{0205039} and \quantph{0204146}.  I
certainly would appreciate any comments you have on those. Especially
I would love for you to articulate the weak points you see in the
ideas.  It is an evolving point of view.  And what I want out of the
effort is just what you suggest in your paper.  I want new
calculations, new effects to go search for, new mathematics, and {\it
really\/} a new world view in total.

Will by chance you be at either the Oviedo, Spain meeting or the QCMC
meeting at MIT this summer?  Maybe we could talk then?

\section{28-05-02 \ \ {\it Anti-Anti-V\"axjination} \ \ (to J. Summhammer)} \label{Summhammer4}

Thank you for the beautiful long note of 4/29/02.  Please allow me to
apologize for taking so long to reply.  I've been almost completely
incapacitated in my email efforts for a few weeks now by travels,
family vacations, company duties, etc.  Certainly I appreciate the
efforts you take to read my things, and definitely your questions and
comments are the best products of that!

Let me comment on a couple of your points.

\bjs
But this `core of the theory' is always tentative. New information, a
wider frame of thought, can change it. And yet it is hard to deny
that it captures something about that which has been observed. It is
like clouds in the sky. For some time they do look like an animal, or
the face of a witch, and anyone with eyes will agree, and a few
minutes later they are gone. The immutable part here, as well as in
physical theories, does seem to be the rules of thought. By the term
`core of the theory' you seem to want to say that they contain a
timeless truth. But I have difficulties believing that a physical
theory could ever achieve the degree of timeless truth as exhibited,
for instance, by mathematical theorems, which are particular
expressions of the rules of thought.
\ejs

Not quite.  I would not say that this thing ``the core of a theory''
contains a timeless truth.  Indeed I tried to be careful to squash
that idea when I wrote:
\vspace{-14pt}
\bq
\noindent
\bbw
But you obviously also want to say that our theories tell us {\bf something} about reality, even if they are not
descriptions of reality.
\ebw

I hope you can glean from all the above that I do indeed believe our
theories tell us something about reality.  But that something is much
like what the elephant tells us about reality.  Its presence tells us
something about the accumulated selective pressures that have arisen
up to the present date.  A theory to some extent is a statement of
history.  It is also a statement of our limitations with respect to
all the pressures yet seen, or---more carefully---a statement of our
limitations with respect to our imaginations for classifying all that
we've yet seen.  (I for instance, cannot jump off the leaning tower
of Pisa unprotected and hope to live; you, for instance, cannot get
into your car and hope to push on the accelerator until you are
traveling beyond the speed of light.) Finally, to the extent that we
the theory users are part of nature, the theory also tells us
something about nature in that way.
\eq

Thus even the core of the theory is as historical and contingent as
the elephant.  And just as the elephant could disappear from all
historical record, so could the theory and, with it, its core.

\bjs
I did not understand what you meant with the last paragraph on page
16 (`However, we would never haven gotten to this stage \ldots.') Sounds
as if you could envision that we could set the laws according to our
wishes. Reminded me of an old view of evolution: The will (the basic
entity) wanted a hand for this particular species, and so it came
about \ldots\ Sometimes I like this idea, because it permits to look
for other patterns and correlations in the history of life than
`mindless' Darwinism.
\ejs

Nor did I understand it really.  But, yes, I suppose I'm imagining
that we might have some control in shaping the ``laws'' of physics.
However---and this is important---that control should be no stronger
than the kind of control we might have for shaping a species with
genetic engineering.  The species has to be able to survive on its
own after being produced in a trial.  If it can't survive on its own,
then one would be loath to call the monstrosity so produced a species
to begin with.  And so too with what I imagine for this malleable
universe.

Did you see my larger paper \quantph{0205039}?  There, I tried to incorporate some of these ideas in a more technical way.  Also, I've now got a lot of supporting material posted at my website (link below).

\subsection{Johann's Preply, ``Anti {\Vaxjo}$\,$''}

\bq
Just read your Anti-{\Vaxjo} paper (\quantph{0204146}). Despite its 17 pages,
the assurance that it is free of equations was incentive to download it.

I like your distinction of the two aspects of reality (p.15): Its rawest
form as the `surprise of the sensations', and the `core of the theory'. I
would also see it this way.

The `core of the theory' is an abstraction arrived at by applying rules of
thought (a kind of immutable Platonic reality). But this `core of the
theory' is always tentative. New information, a wider frame of thought, can
change it. And yet it is hard to deny that it captures something about that
which has been observed. It is like clouds in the sky. For some time they do
look like an animal, or the face of a witch, and anyone with eyes will
agree, and a few minutes later they are gone. The immutable part here, as
well as in physical theories, does seem to be the rules of thought. By the
term `core of the theory' you seem to want to say that they contain a
timeless truth. But I have difficulties believing that a physical theory
could ever achieve the degree of timeless truth as exhibited, for instance,
by mathematical theorems, which are particular expressions of the rules of
thought.

In physics, we are still haunted by the remnants of a naive picture of
reality. We have thrown out the classical view, but still tend to think that
our theories retain a kind of one-to-one link with `the world as such'. We
do no longer say we describe how something is, but that we describe what we
know about that something. Thereby we still stipulate the existence of that
something. But I think any statement about distinctions in the world, that
should exist independent of us, is premature. We can say ``The world is,'' but
we cannot say ``The world is such and such'' and mean it to apply independent
of our categorizations.

Take gravity, for instance. Invented by great minds, to be sure. But a
statement that the universe contains masses which interact by gravity,
better expressed as curvatures of a manifold in which all the masses are
supposed to be located, is so obviously a human way of categorization of
observations, that any alien would coil up in laughter, if we wanted to sell
this as a deep truth (my human way of categorizing his/her/its reaction,
further, that aliens would be individual lumps like us, further, that they
would be interested in science at all --- a consequence of that particular
brain disease, which leads to an extreme slow down between observing and
acting and manifests itself as language, etc.).

The human categorizations leading to gravity:
\begin{itemize}
\item
spatial representation of sensory input (3-space as our mode of
representation, the concept of distance as an unquestioned primitive)
\item
the separation of the sensory input into distinct `things' according to
some invariance principles
\item
mass (the `cause' of muscular strain when lifting and moving around an
everyday `thing')
\item
force (the muscular strain itself)
\end{itemize}
and probably some more.

Nevertheless I grant the following: We will never get out of our
categorizations. And as these are ultimately not our own inventions, the
fact that {\it these\/} categorizations lead to {\it these\/} laws may --- in combination --- be
seen as containing a higher `truth'.

I did not understand what you meant with the last paragraph on page 16 (`However, we would never have gotten to this stage \ldots.') Sounds as if you
could envision that we could set the laws according to our wishes. Reminded
me of an old view of evolution: The will (the basic entity) wanted a hand
for this particular species, and so it came about \ldots\ Sometimes I like this
idea, because it permits to look for other patterns and correlations in the
history of life than `mindless' Darwinism.
\eq

\section{28-05-02 \ \ {\it Little Toes and a World of Experience} \ \ (to W. K. Wootters)} \label{Wootters12}

I'm finally down in Australia, with a little time to think.  I'll be
down here for three weeks.  Kiki and the kids are in Munich, visiting
Kiki's mom and dad.

If you don't mind, I'd like to ask you a couple of questions about
your last letter.

\bbw
I think the sort of world envisioned by classical physics is in fact
impossible.  If we really understood what it takes to make a world
exist {\bf really}, and not just on paper, I think we would see that
one needs the subject-object relation in order to hold things
together.  This is not to say that one needs dualism.  In Whitehead's
system, everything is both subject and object, depending on the point
of view.  There is no dualistic separation into two kinds of entity.
But everything is related to something else as subject, and
everything is also related to something else as object.  In classical
physics, there is no such relation.  As {\Schroedinger} points out, from
the very beginning we eliminate the very notion of an experiencing
subject.  I agree with {\Schroedinger} that this is too extreme an
abstraction.  In making this abstraction, we have removed something
essential from our view of the world.
\ebw

I suppose I too have a gut feeling that your first sentence above is
on the right track, but I wonder if you have an actual argument in
mind for supporting the case?  Also let me ask you this.  Would you
say the same thing about a world governed by quantum mechanics?  Or
do you think the quantum world differs from the classical world in
this respect?  Maybe, in a refinement of this, let me ask the same
thing not about the quantum world in general, but the quantum world
in the vision of the many-worlders, say David Deutsch and Charlie
{\Bennett}.  The way I understand what they envision for quantum
mechanics, it has never struck me as so very different from what I
would call classical physics.  (It is just now that the world as a
whole has this thing called a ``state'' and it is it that goes along
according to some mathematically precise law.)

\bbw
I would like to think that my view avoids the line of reasoning that
takes the mathematical description as the essence of the world.  As
John {\Wheeler} says, a set of mathematical laws will not ``fly'' by
itself.  I think the necessary added ingredient is something like
{\bf experience}.  And that's what I find in Whitehead's view.
\ebw

Would you flesh out this thing you call the ``subject-object
relation'' a little more?  What do you mean by it?  Let us focus on a
simpler system than one usually thinks of as a sentient being.  Say a
rock or the little toe on my right foot.  When we think of these
systems in their capacities as subjects, rather than objects, what is
it that defines those capacities?  What are their characteristic
traits?

When I think of a person as a subject, I think of him carrying around
sets of probability distributions for this and that.  That is, I
think of a subject as something that can carry beliefs. However, I
guess, I have a hard time thinking of a rock or my toe as carrying
around beliefs.  I also think of a subject as something that can play
an active role in shaping other parts of the world because of those
beliefs.

However, maybe you mean something completely different when you're
thinking of a rock as a subject.

Beyond that, let me ask about this word ``experience.''  What do you
mean by that?  I hope to try out one of Whitehead's shorter books
soon but I haven't gotten to that yet.  However, I did read another
small book on {\James} on my flight over and it got me into some
territory that I haven't yet seen of his.  Namely, his stuff from
{\sl A Pluralistic Universe\/} and {\sl Radical Empiricism}, neither
of which I've read yet.  The book I read was {\sl William {\James}:\ The Message of a Modern Mind}, by Lloyd Morris.  It's only 91 pages, and I found it an engaging little thing (at least for giving me a quick overview of all of {\James}'s views, even if not the arguments he used
for getting there).  Anyway, Chapter 5 was titled ``A World of Pure
Experience'' and what he described there seemed to have some overlap
with what you expressed in your last note.  Morris wrote:
\begin{quotation}
\noindent ``My thesis,'' {\James} declared, ``is that if we start with
the supposition that there is only one primal stuff or material in
the world, a stuff of which everything is composed, and if we call
that stuff `pure experience,' then knowing can easily be explained as
a particular sort of relation towards one another into which portions
of pure experience may enter.  The relation itself is a part of pure
experience; one of its terms becomes the subject or bearer of the
knowledge, the `knower,' the other becomes the object known.''

This doctrine is essentially monistic.  But it is radically unlike
the monistic doctrines of either idealism or materialism, which
respectively affirm that mind and matter are the ultimate substance
of reality.  Pure experience is neither mind nor matter, but is the
ground of both.  In itself it is, as {\James} asserted, neutral.
\end{quotation}

Morris also said that this had some feed-in to Whitehead's later
thought, but he didn't elaborate.  So I'm guessing there will be some
overlap between this and what you're thinking about.  But still I'm
having trouble understanding what all this might mean, especially
since I'm having trouble envisioning the mental life of my toe.

Now the other day I said the idea of a random outcome in a quantum
measurement might be viewed as having a mentalistic aspect.  From the
outside, it looks like a random occurrence; however, from the inside
one might think it looks like a ``decision.''  But right now, I'm
wondering what even I really mean by this.

Thinking of you from way down under.

\section{28-05-02 \ \ {\it Australiocentrism} \ \ (to M. J. Donald)} \label{Donald1}

Thanks a million for the letter of 4/27/02.  Please let me apologize
for not replying before now.  I'm about a month behind in my email
due to excessive travels, company business, etc.  Just yesterday I
arrived in Australia for a three-week stay without the family.  So
I'm hoping to finally get caught up a little!

Anyway, you flatter me by reading my drivel and taking the time to
comment.  But I wish you didn't think the universe is doing nothing.

One question:
\bmjd
I'm an idealist because in the course of making that explanation I'm
prepared to throw away any ontological presuppositions.
\emjd

How do you define idealism?  And how would you contrast your flavor
of idealism with this little thing Martin Gardner says in his essay,
``Why I Am Not a Solipsist'':
\bq
In this book I use the term ``realism'' in the broad sense of a
belief in the reality of something (the nature of which we leave in
limbo) that is behind the phaneron, and which generates the phaneron
and its weird regularities.  This something is independent of human
minds in the sense that it existed before there were human minds, and
would exist if the human race vanished.  I am not here concerned with
realism as a view opposed to idealism, or realism in the Platonic
sense of a view opposed to nominalism or conceptualism.  As I shall
use the word it is clear that even Berkeley and Royce were realists.
The term of contrast is not ``idealism'' but ``subjectivism.''
\eq
(The phaneron, by the way, was {\Peirce}'s term for ``the world of our
experience---the totality of all we see, hear, taste, touch, feel,
and smell.'')

And how would you contrast it to this thing Charles Hartshorne says:
\bq
It appears, then, that the idealistic interpretation of reality as
essentially relative to or consisting of mind, experience, awareness,
that is, psychicalistic idealism, is entirely compatible with a
realistic view of the independence of the particular object and the
dependence of the particular subject, in each subject-object
situation. It may also be urged that we need the word ``realism'' to
refer to the mere thesis that every act of knowledge must be
derivative from a known which is not derivative from that act. Thus
the practice of contrasting ``idealism'' and ``realism'' as though
they were contradictories, is of doubtful convenience. ``Realistic
idealism,'' or ``realistic subjectivism,'' has a reasonable and
consistent meaning.
\eq

\section{28-05-02 \ \ {\it A Comment on One of Your Comments} \ \ (to C. M. {\Caves})} \label{Caves64}

\bcc
\bq\noindent\rm [CAF wrote:]
But there's really more to the story.  They can always come to
agreement, indeed---regardless of how disparate their initial
opinions---if they are willing to make an essentially infinite
expenditure toward laboratory technique.  That is to say, the only
thing that will give assured agreement in all cases is a set of Kraus
operators all of rank-one.  Jacobs and I called those infinite
strength measurements:  the idea being that they are hard to actually do.  In more real-world measurements, where the operators are never really rank-one, coming to final agreement will generally require some
initial agreement.  Whence the point in my August 7 letter to {\Mermin}.
\eq
My puzzlement here is related to that above.  How can it be
that rank-one's lead to ``assured agreement'' when we can never be sure what measurement was made nor what rule to use for the
post-measurement state?
\ecc

You're getting off track with this in the same way that you were yesterday in our walk near the lake.  The issue is solely this in all issues of ``coming to agreement.''  When can some amount of previous agreement between two agents lead to an increased amount of agreement after an act of consulting the world?

In the view I'm trying to promote, a measurement and a state-change rule (i.e., a POVM and an associated conditional quantum operation) is a subjective judgment.  When you say something like ``we can never be sure what a measurement was,'' in my way of thinking it is exactly like saying ``we can never be sure what the true quantum state was.''  To the extent that the latter is a non sequitur for me, your phrase is a non sequitur too.  When two experimentalists walk into the same laboratory, one carries in with him a subjective judgment about the POVM and quantum operation of the measurement device, and the other may carry in a completely different conception.  But both of them know what they themselves think.  For neither of them is there a ``true'' quantum measurement working in the background.

Suppose they lay their beliefs on the table.  The question is this, what must the characteristics of those beliefs be so that, upon the completion of a measurement, they will ascribe the same quantum state to the post-measurement system regardless of what each initially said about the system to be measured?  The only claim is that it is sufficient that their separate POVM + operations be 1) identical {\it and\/} 2) consist of rank-1 elements.

I found myself laughing last night, by the way.  After all your preaching to me about the ineffectiveness of email for conveying one's thoughts, you nevertheless wrote this note while we were in the same room!

{\Ruediger}, we can't wait until you get here!

\section{28-05-02 \ \ {\it Strong Coherence?}\ \ \ (to M. C. Galavotti \& E. Regazzini)} \label{Regazzini1} \label{Galavotti2}

My colleagues, {\Caves} and {\Schack}, and I are once again having an argument about the foundations of probability.  In particular, one of the things we are getting hung up on is the issue of requiring A) ``strong coherence'' as introduced by Shimony and Kemeny, or instead B) ``normal de Finetti coherence'' in Dutch-book arguments.  I hope that you are aware of these distinctions.

 {\Caves} and {\Schack} are attracted to strong consistency as a normative criterion because it leads to certain results that they would like to see when we apply the idea to quantum mechanical betting situations.  I, however, am opposed to the criterion for just the opposite reasons.  What I would like to ask of both of you is might you point me to some of the literature where the issue of ``strong coherence'' versus ``regular coherence'' is discussed?

I am aware of de Finetti's own rejection of strong coherence from his book.  However, in particular, I am wondering if there are any independent arguments (i.e., other than de Finetti's, which concerns an infinite limit and probabilities on a non-discrete space) for rejecting strong coherence?  Could you send me references if there are any?

Thank you so much for your help, and I hope to hear from you soon.

\section{29-05-02 \ \ {\it I Think She'll Know} \ \ (to N. D. {\Mermin})} \label{Mermin67}

Remember what the dormouse said; feed your head.

You can tell I'm pretty darned behind in my email.  I'm in Australia
now, finally with a little time to think.  Kiki and the kids are in
Munich.  {\Caves}, {\Schack}, and I are down here for three weeks doing
Bayesiany things.

\bdm
I don't see what your teleportation example (pages 11, 12) adds to
ordinary EPR.  Aren't all the issues exactly the same if Alice ``in
her laboratory prepares'' the single qubit in (1) that she possesses
by an appropriate measurement (to be sure, she can't control which
outcome she'll get, but that doesn't seem to be central to your
point, or is it?) after which she and only she knows what the outcome
of the corresponding yes-no measurement on Bob's qubit will be.
\edm

Yeah, I'd agree that it doesn't add a heck of a lot to the old
argument.  Mostly I wanted to say something about teleportation:
Namely that if Einstein had known about it, then he might have used
it to the same devious purposes he did with his old argument.

But still, I guess there were a couple of features of this version
that I thought made it a bit cleaner than the old Einstein thing. 1)
With the supplementation of {\it only\/} two bits of physical action
on the part of Bob (i.e., one of four possible unitaries), Alice can
put Bob's system into any state she wishes.  So, in essence (i.e.\ up
to two bits), there's nothing even random and uncontrollable about
the process.  2)  In the case of teleportation, even examining the
measurement device before and after the measurement will tell you
nothing about the posterior state for Bob's system that Alice ends up
with.  Very literally, only Alice ends up doing some updating. If
it's she and only she, why not call the state her knowledge?

\section{29-05-02 \ \ {\it That Damned von Neumann} \ \ (to N. D. {\Mermin})} \label{Mermin68}

Now let me try to answer more adequately the question you asked after
my talk in Ithaca.  The point is simply this:  Suppose I tell you
that I've got a device that measures a standard observable $H$. How
do you know that you should accept my claim?  Let's say you do this:
You simply give me a supply of a gazillion nonidentical states you've
prepared (anyway you wish) in your laboratory.  I'll perform my
measurement on each of them and report the results I found.  If you
find that I'm giving you back outcomes with the (conditional)
statistics you expect, then you'll have some warrant for believing
I'm really performing the measurement I claim.

Now let me ask you this.  Suppose you confirm my measurement to your
satisfaction by that method.  Do you now have warrant to say anything
about the post-measurement state for each of the systems you gave me?
The answer is ``no'' of course.  The point is, you need to know more
about the particulars of the device.

Now, von Neumann said that for an ``ideal'' device the
post-measurement states for the systems will be eigenstates of the
observable.  But I claim that is an arbitrary notion of ideal, and
Kraus's theory of ``effects and operations'' backs that up in a kind
of technical way.  The Kraus theory says that the state change can
always be thought of as a collapse in the von Neumann sense PLUS a
trace-preserving completely positive map.

Now, the standard quip that is made is that the CPM part of this is
just extra noise that didn't need to be there.  But again, I claim
this statement is arbitrary.  Here is a simple counter example.
Suppose I perform the measurement $H$ on half of an entangled pair.
Then (via the entanglement), I can always think of this measurement
as {\it really\/} a measurement of some sort on the other half.  In
fact I can think of it as simply the observable that is the transpose
of the original one \ldots\ only performed on the second system
rather than the first. What could be a more minimally disturbing way
of measuring $H^{\rm T}$ on the second system than that?  But does
von Neumann's collapse postulate hold for this kind of measurement?
Try it, and you'll find that it doesn't.  The only state changes this
kind of measurement can produce are of the pure-refinement kind.

Von Neumann brainwashed the generations in a needless way.  Even
something so simple and ``ideal'' as a standard photon detector does
not follow the postulate.  When there's a click, the photons are
absorbed from the mode and the field is now in the $|0\rangle$ state.

I hope that helps make up for my lack of lucidity at your
celebration.

\section{29-05-02 \ \ {\it Notes from the Web} \ \ (to C. M. {\Caves} \& R. {\Schack})} \label{Caves65} \label{Schack57}

[[{\bf Note:}  This is a quote I thought worthy of recording the time.  Unfortunately I did not record its source.]]

\bq
Note that traditional conditionalisation is a special case of Jeffrey conditionalisation. It is important that Jeffrey stressed that this rule of the kinematics of rational belief is all that is needed. It is never needed to give up propositions that are assigned probability one. The assumption is that only logical tautologies have probability one and only contradictions probability zero. This view is defended by means of the condition called strict coherence. A probability function obeys strict coherence if it is coherent and there is no set of bets consistent with it such that the better might lose, and cannot win. Jeffrey, among others, has argued that by assuming strict coherence it follows that no contingent proposition should have probability one, because no rational agent would bet his head on the truth of any contingent proposition.
\eq

\section{29-05-02 \ \ {\it Strict Coherence?}\ \ \ (to B. Skyrms)} \label{Skyrms1}

I doubt you remember me, but we met between 1996 and 1998, when I was a postdoc at Caltech and you had a visiting position there.

I dawned on me today that you might be able to help me in my plight described below.  [See 28-05-02 note ``\myref{Regazzini1}{Strong Coherence?}''\ to M. C. Galavotti and E. Regazzini.] Is there any literature that you could send me to that expresses a dissatisfaction with ``strong coherence''?  I'd be very grateful for any pointers you could give.

\section{29-05-02 \ \ {\it More Strict Coherence} \ \ (to B. Skyrms)} \label{Skyrms2}

\bbsky
Sure, I remember you. I think that the usual response to Shimony is just that
strict coherence isn't a plausible requirement in the usual way of
doing probability theory with continuous mathematics.  Don't
remember any good references, though.
\ebsky

That's too bad.  What I'm really looking for is anything that
expresses a dissatisfaction with strict coherence with respect to
discrete event spaces.  It turns out that enforcing strict versus
normal coherence can make a pretty drastic difference for some
interpretive problems in (even finite dimensional) quantum mechanics.
Unfortunately, with respect to the attractiveness of the conclusion,
my coauthors and I have opposite opinions.  Thus, I am inclined to
require only normal coherence for the agents in our game; whereas
they are inclined to require strict coherence.

The main point, even in the classical case, is that strict coherence
requires that an agent ascribe probability 1 to an event if and only
if he believes the event is a certainty.  Whereas under the
assumption of normal de Finettian coherence, a probability-1
assignment cannot be used to conclude a belief of certainty on the
agent's part.  He might be assigning probability 1, not because it
reflects his true beliefs, but because it is advantageous for other
purposes.  That is, with respect to certainty, strict coherence
compels an agent to never ``tell a lie.''

I think that goes too far as concerns a foundation for ``rational
behavior.''  Whereas normal coherence appears to me to strike a sweet
spot.

In any case, what I am looking for is some confirmation of my
troubles in the published literature \ldots\ to help me build a case
for the inevitable battles I foresee with my coauthors.

By the way, you might be interested to know that we have been putting a
substantial effort into interpreting quantum probabilities as
personalist probabilities.  Let me recommend four of our papers to
you (along with the web links to get them) in case you're interested.
[\quantph{0205039}, \quantph{0204146}, \quantph{0104088}, \quantph{0106133}]

As you climb from bottom to top in this list, you'll find us moving
closer and closer to a personalist position.  Also, there is a lot of
supporting information posted at my website, link below.

\section{31-05-02 \ \ {\it Poor Young Duvenhage} \ \ (to H. M. Wiseman)} \label{Wiseman4}

Thanks yourself for your long note in reply to my long note!  You can
see I'm working with a much larger lag time than you in my emails.
Anyway I'm in Australia now, and slowly working off the jetlag and
becoming more productive as the days go by.

Let me say a few words on your last note.

\bhw
Freedom taken, thank you. And I think you will be stronger if you
don't try to prop up your position using another paper which I think
goes fundamentally against your views in a number of places.
\ehw

If you believe that, you misconstrue my purposes for citing
Duvenhage.  My citation was to give attention to and encourage this
young researcher.  Of course, I don't agree with a lot of the paper;
but that is beside the point for me.  There is one thing I certainly
agree with and I think he said it particularly cleanly.

You can find where I said this outright to Duvenhage himself by
looking at pages 167 and 168 of my web samizdat, ``Quantum States:
What the Hell Are They?''  (It's in \myref{Duvenhage1}{a 28 March 2002
  letter}; I hope you'll read it.)  There you will find me expressing
some of the same misgivings as you, though with less detail.  As I
told him, I would rather encourage the similarities at the moment than
the contrasts.

Duvenhage's paper will appear in {\sl Foundations of Physics}, and I
think that is a good thing.  Having the paper out in publication-land
just encourages someone to write a comment on its deficiencies.
Indeed, I would love for that to go further and get a discussion
stirring. For, what might get more people to pay attention to the
potential of a Bayesian approach to QM than a good stirring
discussion? Moreover, if people start to pay attention to the
potential of the approach, they might just get to work on filling its
other (I would say, more real) deficiencies!

So, there.  I can even imagine an eloquent writer to start the ball
rolling \ldots\  (OK, a hint:  His initials are HMW.)

Now, let me go back to the one thing I said ``I certainly agree
with.''

\bhw
But here is the difference between classical and quantal.

In the classical theory the belief $p$ of the weatherman could be
given a precise formulation in terms of a set of classical variables
(presumably related to his [brain]), and that belief would actually
evolve (for a ``true $h$'' etc.) in a deterministic way from $p(h)$
to $p(h|d)$ as he looks out the window. That is, the belief of the
weatherman can be treated ON THE SAME BASIS as the objects of that
belief (i.e.\ the physical world). That is not to say it MUST be
treated on the same basis, but it CAN be, and there is no NEED to
have Bayes thm in the foundations of the theory.

In the quantum theory, our belief cannot be treated on the same basis
as the physical world. We use a state matrix to encapsulate our
belief about the world: our expectation for the results of a fiducial
measurement if you like. In the absence of information gathering,
this state matrix evolves in some well-defined way. But if we try to
treat our belief on the same basis, as a function of physical brain
variables, we run into the QMP.
\ehw

I think this pinpoints in a pretty terse way the root of our
disagreement.  To say what you said is to 1) accept a kind of
reductionism that I no longer think is healthy, and 2) to posit a
strong faith that something {\it can\/} be done that, in actual fact,
has {\it never\/} been done.

Concerning 1) I apologize for using the phrase ``those beliefs are
presumably a property of his head,'' for it evokes an imagery I would
rather not have in this discussion.  Perhaps it would have been
better for me to use the word ``possession'' rather than
``property.''

Here's a belief I presently possess:  With probability greater than
.9, there will be another suicide bombing in Israel in the next
month.  Try to put that into physical terms.  What can the word
``suicide'' mean in classical mechanical terms (or even quantum
mechanical terms)?  Or to make it look just a little more physical,
here is another belief:  With probability greater than .99, I will
see at least one car today.  But why didn't I say a lump of atoms
with this characteristic, that characteristic, and the other
characteristic?  The point is:  The world independent of man does
not, and cannot, know that that lump of atoms signifies a car.  My
beliefs---at their starting points---are always about things denoted
at such a level of {\it practical\/} existence.

Thus I would claim, concerning 2), that it is nothing but a religious
faith to suppose that one can derive the form of Bayes'
conditionalization rule from a mechanistic physics.  It has never
been done before, and I would venture to guess that it will never be
done.

But still, let me suppose that it could be done after all---it's just
that it hasn't been done yet.  What would that do?  In agreement with
(a small part of) Duvenhage's paper, I would say we're left with one
conclusion:  We should call the difficulties so found the ``classical
measurement problem.''  For in classical physics we wouldn't know
where the observer begins and the world ends \ldots\ just as we don't
presently know it with quantum mechanics. Nor as I said above, would
we know how to get a Bayesian collapse in the beliefs of the observer
(even once he has been identified).

So, with regards to this particular aspect of things, I hold firm
with my assessment of Duvenhage.  If there is a distinction between a
classical and quantum conception of nature, it is somewhere else than
in the updating of one's beliefs.  I claim the same difficulties are
either absent from both conceptions, or present in both conceptions.

But \ldots\ maybe I say this all for nothing.  For you also wrote
this:
\bhw
Duvenhage is quite specific about what he is saying. He says ``the
Heisenberg cut is therefore no more problematic in quantum mechanics
than in classical mechanics''. This is true if you are a committed
Bayesian, because then the cut is necessarily there from the
beginning of your conception of the world.
\ehw
and that seems to express that my previous quotation of you was
directed more toward Mr.\ D than me.

In any case,
\bhw
I accept that. This is your solution to the QMP. You say ``there is
no QMP'', but isn't that the same as their saying ``I've solved the
QMP''? You don't have to answer that.
\ehw
I would say it doesn't solve anything; it just shifts the terms. For,
where it dismisses old problems, it creates new problems.  And it's
the problems that keep us all young.

\section{31-05-02 \ \ {\it `Reality'} \ \ (to M. J. Donald)} \label{Donald2}

Thanks for the clarification on your form of idealism.  I also had a
nice time reading the FAQ on your webpage.  I'm whiling away a little
time in Australia at the moment, visiting Nielsen and trying to write
some papers with {\Caves} and {\Schack}.

\bmjd
If the ``current wavefunction'' is just the best description of our
knowledge of the system, then what are we made of?
\emjd

Something that is not the wavefunction, but for which, once we have
accepted its existence, compels us to a structure of reasoning and
belief revision whose form is identical to what we once thought of as
``quantum mechanics.''  My best shots so far at saying this in a
clear way can be found in Sections 4.2 and 6 of the new paper \quantph{0205039}.  Or you can see the same thing in pictorial form
by looking up the talk ``Where's a Good Weatherman When You Need
One?''\ at my webpage.

And so the sparring match goes on \ldots

\section{01-06-02 \ \ {\it Finally, My Slides} \ \ (to R. J. Greechie)} \label{Greechie1}

I've been meaning to write you for a while.  At the AMS meeting in Atlanta, you asked me if I could mail you a copy of my slides from the talk I gave there.  Well, I still can't do that easily (I'm in Australia for three weeks), but you can download my slides from my website now.  The link for it is below.  I was hoping to get them printed out in color to mail to you before I left for Australia, but time ran out on me.  I hope this will do as a substitute.

Also let me advertise some papers that outline what I would like to see happen in quantum foundations.  The top paper in the list below (\quantph{0205039}) makes my strongest statement of that.  All of them, I think you'll find to be pretty easy reading.
\begin{itemize}
\item
``Quantum Mechanics as Quantum Information (and only a little more)''\\
\quantph{0205039}
\item
``The Anti-Vaxjo Interpretation of Quantum Mechanics''\\
\quantph{0204146}
\item
``Unknown Quantum States: The Quantum de Finetti Representation''\\
\quantph{0104088}
\item
``Quantum Probabilities as Bayesian Probabilities''\\
\quantph{0106133}
\end{itemize}

Unfortunately, I can't come to the IQSA meeting in Vienna this summer; time just became too short for me.  But I get the feeling that there is a large mass of results I could use from the quantum logic community to further my program.  ``Results which are,'' as I wrote Karl Svozil recently, ``just waiting for me to plaster over them with some words of a Copenhagenish flavor.''

\section{01-06-02 \ \ {\it Vienna and Amherst} \ \ (to D. J. Foulis)} \label{Foulis1}

I just wrote a letter to Dick Greechie giving links to some of my papers and talks.  Rather than dreaming up some new things to say, let me just paste that letter into this one for your use.  [See 01-06-02 note ``\myref{Greechie1}{Finally, My Slides}'' to R. Greechie.] If you can download it, I especially hope you enjoy my paper ``Quantum Mechanics as Quantum Information (and only a little more)'' \ldots\ and don't find it too offensive!

I'm so happy to have met you in Georgia.  You really helped build my confidence that this method of attack I keep trying to push the community into might just go somewhere.  From time to time, I am in Amherst visiting Charlie Bennett (he's got a weekend home in Wendell) or Herb Bernstein at Hampshire College.  It'd be great if I could swing by and have some further talks with you.

\section{01-06-02 \ \ {\it Wheeler Compendium} \ \ (to K. W. Ford)} \label{Ford1}

Lucien Hardy gave me your name as a contact.  The reason I am writing you is that I am working to put together a rather large compendium of references and quotations from the literature with the title, ``The Activating Observer:\ Resource Material for a Paulian--Wheelerish Conception of Nature.''  The compendium presently stands at 439 references and fills 107 printed pages.  I expect it to be significantly larger than that when it is finished.  As you can guess from the title, the thoughts of John Wheeler play a significant role in the project.  In fact they started me down the course; John has long been one of my heroes.

The reason I am writing you is that I wonder if I could ask your help in making sure that I have the document as complete as possible.  Also I would like to ask you if you have any ideas of where I might publish the thing once it is completed.  I had originally planned to submit it to the journal {\sl Studies in History and Philosophy of Modern Physics\/} but now I am starting to fear that the document is becoming much too long for that.

If you would like me to email you a copy of the manuscript-under-construction, I can do that.  (Either as a raw REV\TeX\ file, a DVI file, or a PostScript file.)  I would certainly welcome any help you could give.

About myself---in case you worry about my legitimacy---I am a research physicist at Bell Labs working in the field of quantum information and computing, am 37 years old, and have had postdoctoral positions at Caltech and Los Alamos.  My full CV can be found at my website (there is a link to it below).  Also, you can read about me in the nice foreword David Mermin wrote for my samizdat {\sl Notes on a Paulian Idea\/} published on the Los Alamos archive.  Here is a link: \quantph{0105039}. In general, you can read my views on quantum mechanics and how it fits in with some of John Wheeler's larger questions by either consulting my website or perusing the Los Alamos {\tt quant-ph} archive.

Anyway, I put all this information so that you won't think I am just another crackpot who has become smitten with John's words.  It is a serious project.

I hope I hear from you soon.

\section{01-06-02 \ \ {\it Back from the Grave} \ \ (to G. Brassard)} \label{Brassard15}

Good to hear from you.  Our letters just crossed paths.

\bgb
But can you tell me what's the homunculus fallacy?
\egb
Sounds dirty, doesn't it?  I think it's the same fallacy as the camera obscura fallacy.

\section{01-06-02 \ \ {\it High Dispute} \ \ (to P. Grangier)} \label{Grangier7}

I hope you will forgive me for not replying to your letters for so
long.  The volume of email I have been getting lately has started to
become something I am not equipped for.  Sometimes I just crack under
the pressure, and thus my silence.  But at the moment, I've got a
little time away from home:  I am in Australia, visiting Nielsen and
trying to write some papers with {\Caves} and {\Schack}. Beyond that, my
wife, kids, and Bell Labs have all been left behind!  So I am hoping
for a productive three weeks.

\bpg
If we speak about ``objectivity'', we have first to agree about what
it is.  To keep it simple, I stick with the ``naive'' view that if I
do (or if a student does) a measurement in the lab, this is an
objective process: this ``did happen'', and the fact that you don't
know the result of the measurement will not change it.
\epg

As far as I can tell, we do NOT disagree on this.  So, it kind of
annoys me that you keep bringing it up.  Instead, I would say our
disagreement lies right here:  For some reason YOU think it is
NECESSARY to uphold the idea that a pure quantum state is an
objective property (of something in nature) in order for your
sentence above to come about.  Whereas I do not.

I know that I can function just fine with subjective quantum states,
even with subjective pure quantum states.  You will find no logical
inconsistency in me; and I doubt I can find any in you.  At the level
of our squabble, it is to a large extent a matter of taste. However,
you and I both know that matters of taste can lead to matters of fact
with regards to the questions we will seek of nature.  Your taste will
lead you one way; my taste will lead me another.  Only history will
tell which of us will have had the more productive view.  Only history
will tell which one of us ended up asking the most interesting
questions of nature.  (Recently I tried to capture this in a little
story to Charlie {\Bennett} and John Smolin about the pleasures of
broccoli.  I will paste the story at the end of this note.  [See
  16-05-02 note ``\myref{SmolinJ4}{King Broccoli, 2}'' to
  J. A. Smolin, C. H. Bennett, N. D. Mermin and others.])

Here is the way I put the whole point in a message to Matthew Donald
just a minute ago.  [See 31-05-02 note
  ``\,\myref{Donald2}{`Reality'}\,'' to M. J. Donald.]  At the present
time, I just do not know how to say it any more cleanly than this:
\vspace{-.2in}
\bq
\noindent
\bq
\noindent If the ``current wavefunction'' is just the best description
of our knowledge of the system, then what are we made of?
\eq
Something that is not the wavefunction, but for which, once we have
accepted its existence, compels us to a structure of reasoning and
belief revision whose form is identical to what we once thought of as
``quantum mechanics.''  My best shots so far at saying this in a
clear way can be found in Sections 4.2 and 6 of the new paper \quantph{0205039}.  Or you can see the same thing in pictorial form
by looking up the talk ``Where's a Good Weatherman When You Need
One?''\ at my webpage.
\eq

The sections I recommended to him, in their focus on the Bureau
International des Poids et Mesures, have direct implication on you.
So I have a secret dream that you will read them \ldots\ and finally,
finally something will click in your head, and you will say, ``You
know, Chris is not being so unreasonable after all.''

\bpg
it is quite enough that a complete ($=$self-defining) set of physical
properties can be predicted with certainty. The crucial point is that
there is no ``ignorance'' left (you must admit that each time you
write that a pure state has zero entropy).
\epg

It is fine to note that some observables can be predicted with
certainty when a state is a pure state, but that does not delete the
fact that there is plenty of ignorance left for other observables.
One can even quantify that ignorance nicely.  See equations (77)
through (82) in my \quantph{0205039}.  (But the result goes back
to Wootters.)  The von Neumann entropy---which is what you are
thinking of in your statement---simply captures the BEST CASE
predictability.  So what?

\bpg
PS In your paper you strongly ``recommend'' the work by Lucien Hardy.
I have two comments:
\bq
\indent $\bullet$ It surprises me that you adhere with the ``relative frequency''
approach to probabilities that is used by Lucien (his first axiom). I
certainly agree with it, but I thought you would not.\smallskip\\
\indent $\bullet$
Lucien is trying to make QM look like a new probability theory.
\eq
\epg

If you look at my paper, you will note that I said:
\bq
Beyond that, let me recommend four other articles. The first two are
the most technically important for the enterprise I promote in my
other contribution to this volume:  Namely, to secure a transfer from
our present abstract, axiomatic formulation of quantum mechanics to a
more physically meaningful one.  I think some elements in Lucien
Hardy's papers {\it almost\/} carry us to the brink of that.  In his
work, I think the right emphasis is finally being placed on the right
mathematical structures. \ldots

I should point out, however, that in all four of the above
references, I think significant improvements could be made by
adopting a sufficiently Bayesian stance toward the use and meaning of
probability.
\eq

I said what I meant.  (Just as the experimentalist should strive to
perform his measurements accurately; the good theorist should strive
to read his friends' papers carefully.)  The interesting part of
Hardy's papers, in my view, is the mathematics.  Saying that
constitutes an endorsement of neither 1) the relative-frequency
interpretation of probability, nor of 2) Hardy's desire to generalize
probability theory to a larger structure.  Hardy and I, indeed, have
had extensive discussions on this.

\bpg
Let us simply admit that local realism is dead, but that physical
realism can do quite well without it.  Why is that so difficult to
accept?
\epg

One more time:  I do not deny realism.  Moreover, just as you, I am
happy with the death of local realism (if what one means by that is
``hidden variable realism'').  However, I simply do not find your
proposed solution to the whole shebang of quantum interpretation
problems to be as compelling as you find it.

Measurement clicks alone do not specify post-measurement quantum
states.  Rather, measurement clicks PLUS prior quantum states (for
one's description of the measurement apparatus) do.  Thus, where you
think a measurement outcome prepares a unique quantum state, I say it
has a subjective component.  But you say, my measurement device is
calibrated.  And I say with respect to what?  And on and on we could
go ad infinitum.  The Gordian knot of the state's subjectivity simply
cannot be cut by your assuring me that I ought to think otherwise.

\bpg
admit that there is a ``reality'' attached to the pair of particles,
but that there is no ``reality'' attached to each particle.
\epg

This I see as an arbitrary move.  Just as one man's unipartite system
is  half of another man's bipartite system, one man's bipartite
system is half of another man's quadripartite system.  And so it
goes.

Instead, I would say {\it all\/} systems have a reality attached to them. It
is simply that that reality has {\it nothing to do\/} with the quantum states
one ascribes to those systems.  A system has a Hilbert space, and a
Hilbert space has a dimension that does not depend upon which state
is ``alive'' within it.  That dimension, I would offer you, can be
treated as a reality for a quantum system.  But there I stop, whereas
you want to go further (by supposing a reality to some nonlocal
quantum state).

\bpg
the ``reality'' attached to the pair makes no problem with Lorentz
invariance, because it was created when the two particles interacted,
and it simply follows them if they move very far away. The same
conclusion apply to more fancy schemes like entanglement swapping,
that requires classical communications to effectively prepare the
remote entangled state.
\epg

In the usual sense of what one means by the word ``interacted,''
this---I would say---is just wrong.  Suppose Alice and Bob possess
two particles that interacted in the past so that they are now
entangled (by your way of speaking).  Further suppose Carol and Ted
possess two particles that interacted in the past so that they are
entangled.  Maybe even make both states---the AB state and the CT
state---to be singlets.

Suppose finally that Alice and Bob have never before seen Carol and
Ted.  A conclusion one can draw from this is that the A and C
particles have never interacted in the past.  And the B and T
particles have never interacted in the past either.

Let now Alice and Carol meet by chance and perform a Bell measurement
on their two particles.  If Alice shares all her knowledge of the
original AB interaction with Carol, and Carol shares all her
knowledge of the original CT interaction with Alice, then they will
both be warranted to update their assessments of the BT system.  They
each will immediately ascribe a pure Bell state (one of four
possibilities to that system).  With {\it respect\/} to Alice and Carol, the
BT system will now be in a Bell state, even though B and T have never
interacted in the past.

And this has nothing to do with Alice and Carol sending the
information about their measurement outcomes to Bob and Ted.  I'll
say it one last time:  {\it with respect to\/} Alice and Carol, the BT system
will be in a Bell state after the measurement.

You see, you get hung up because you want to think of entanglement as
a real objective property of two systems, not merely a property of an
observer's judgment about those systems.  Beyond that, your desire
for Lorentz invariance makes you want to think of entanglement as
only a consequence of local interaction.  But that just goes too far,
as the above example shows.  Entanglement, just like quantum states,
can arise from measurements in the distance.

You can call that a ``contextually objective'' affair if you wish,
i.e., that the entanglement between B and T arises only with respect
to the context set by A and C.  But then I say---as I've said
before---why not just call the state ascribed to the BT system Alice
and Carol's information about it?  Moreover, by calling it
information, you will find that you will stop forgetting the prior
(subjective) information that was crucial for defining the posteriori
states in the first place.

By the way, I have read your FAQ posted on {\tt quant-ph}; I have not
ignored it.

\bpg
Bell's inequalities do not hold!!!! \ldots\ if a measurement if
performed on one side, ABSOLUTELY NOTHING happens on the other side.
\epg
Well, at least we can agree on this.  But look how many of your words
I had to delete from the paragraph below to get us there:
\bpg
- since there is no ``reality'' attached to each particle, Bell's
inequalities do not hold !!!! All the job is done by the fact that
the individual particles have no quantum state, or no other property
whatsoever that would decide on the result of the measurement. Then
``action at a distance'' simply vanishes: if a measurement if
performed on one side, {\it absolutely nothing\/} happens on the other side.
\epg

And we most certainly agree on the following:
\bpg
What next? QM is a fantastic theory that can only stimulate one
question: why is it working so well? Then we may notice that QM was
invented 75 years ago in a somehow anarchic way, as an attempt to
understand atomic spectra. But we may speculate the following: QM is
actually the answer to a question that was never clearly formulated.
We have the answer, what about finding the question?
\epg

Even if we do not see eye to eye in the secular world, we seem to
dream of the same heaven.

PS.  Don't forget the story below.

\bq
\noindent
King Broccoli\medskip

``In any case, none of this matters for doing physics,'' you say. But
I think it does in the long run.  (Certainly, you've got to concede
that there's something that fuels me---and it hurts to think that you
might think it is nothing more than irrationality.)  When Charlie
sent me the picture of a skunk cabbage in reply to these very issues,
I found myself thinking of King Broccoli.  The story goes that one
day, by divine providence, it came to King Broccoli that broccoli,
the vegetable, his namesake, actually tastes good.  Good in a way
that hitherto only gods and angels had known.  Every child who had
ever said, ``Yuck, I don't like broccoli; it tastes awful!'' was
simply wrong \ldots\ or at least that's what the king realized. King
Broccoli, being the head of state, decided to do something about it.
Henceforward, all gardens in the kingdom should have a patch devoted
solely to broccoli.  It really wasn't much of a burden on the
national product (except, perhaps, for the psychiatrists who had to
treat all the movie stars who had never felt fulfilled in their
broccoli experience).  But think of the diversity of vegetables the
kingdom might have raised if its citizens hadn't been encumbered with
the king's notion that broccoli had an objective, but never
verifiable, taste?

Moral?  Maybe there is none.  But it is a documented fact that the
Kingdom of Broccoli eventually fell and was replaced by a liberal
democracy (where the ideals set forth in the constitution, rather
than the particulars laws of any given day, have an objective
status).
\eq

\section{01-06-02 \ \ {\it Postmodernity} \ \ (to N. D. {\Mermin})} \label{Mermin69}

Ever since you wrote me this,
\bdm
I liked the first half (the anti-Final Theory) part of your sermon to
Preskill very much.  You really could become the darling of the
postmodernists if you put your mind to it.
\edm
I've been wondering what the heck a postmodernist really is.  Despite
my occasional mention of Derrida and Baudrillard and the like, I've
never really read any of their essays.  The language barrier was
always just too big.

In fact, I had never even read Sokal's parody.  Well anyway, I've
started the process of remedying that now.  Yesterday, I finished the
book {\sl The Sokal Hoax:  The Sham that Shook the Academy}.  It is a
collection of maybe 50 or 60 articles and news reports that arose in
the aftermath of Sokal's paper.  Here is the entry I put for Sokal's
article in my upcoming Cerro Grande II, the compendium ``The
Activating Observer: Resource Material for a Paulian--{\Wheeler}ish
Conception of Nature.''
\bq
\noindent A.~D. Sokal, ``Transgressing the Boundaries:\ Toward a
Transformative Hermeneutics of Quantum Gravity,'' Social Text {\bf
Spring/Summer 1996}, 217--252 (1996).  This parody article contains a
wealth of references.
\eq

There appear to be some good references there, some of which I have
read I {\it know\/} that I like!

Anyway, reading the book was an eye-opening experience.  Now that I
know the perils, it should be interesting to look back 20 years from
now and see how successfully or unsuccessfully I managed to navigate
the waters.

By the way, I just noticed a mystical coincidence:  The big samizdat
last year was assigned the number \quantph{0105039}; whereas my
fire and brimstone ``Quantum Mechanics as Quantum Information'' from
this year was assigned the number \quantph{0205039}.  Pretty
cool.

Did I tell you that I'm in Brisbane now?  I've been here for a week,
and will be here for two more.  Kiki and the kids are in Munich until
June 18.

\section{01-06-02 \ \ {\it Smaller than the Circle} \ \ (to G. {\Plunk})} \label{Plunk4}

The more I chew on what your graphs indicate, the more I like it.  ``Accepting quantum mechanics is, in part, merely accepting that one's judgments for an SQMD ought to live within an ellipsoidal volume smaller than X.  And X is the volume of the largest sphere that can fit within the simplex.''

The statement will get simpler and simpler as we think about it more, {\it if it is true}.

The question is, how can we ferret out its truth.  I hope you downloaded the thing from Caves's webpage.  I think there's one calculation you should be able to do pretty easily for the SICPOVMs.  Namely, it shouldn't be hard to calculate the {\it average\/} distance (or distance squared) of a pure state (thought of as a probability distribution) from the center of the simplex.  Just use the methods from:
\begin{itemize}
\item
R.~Jozsa, D.~Robb, and W.~K. Wootters, ``Lower Bound for Accessible Information in Quantum Mechanics,'' Phys.\ Rev.\ A {\bf 49}, 668--677 (1994).
\end{itemize}

However, maybe it'll even be easier than that.  Maybe you can just use the abstract properties that define a SICPOVM to show that the distance of a pure state from the origin is invariant.  In fact, thinking about this in my head, it might just be a piece of cake.  But I'm probably missing something.  You never can tell until you sit down with a piece of paper and a pencil.

\section{02-06-02 \ \ {\it Lamb} \ \ (to G. Brassard)} \label{Brassard16}

\bgb
Is this the Lamb of Lamb shift? [Willis E. Lamb, Jr.\ and Jagdish Mehra (ed.), {\sl The Interpretation of Quantum Mechanics}, ISBN 1-58949-005-3]  Do you think it should be interesting?
\egb

Yep, that's THE Lamb.  Willis Lamb.  But he is quite old (over 90) and a bit off his rocker now.  I hear that he comes out a bit against quantum teleportation there.  I was at a talk of his about 2.5 years ago when he said, quote, ``quantum teleportation is beneath contempt,'' lumping it in the same set of things he was already putting down, like Bohmian mechanics and such.  He didn't at all explain why he said that.

Buy it at your own risk.

\section{02-06-02 \ \ {\it Too Silly?}\ \ \ (to M. A. Nielsen)} \label{Nielsen3}

\bmn
Can you send me a title and abstract for your talk this Thursday?
\emn

I just plucked one off my webpage.  If it sounds too silly to you, I can write something a little more staid and proper.  Your choice.

\bq
\noindent {\bf The Oyster and the Quantum} \medskip

I say no interpretation of quantum mechanics is worth its salt unless it raises as many technical questions as it answers philosophical ones. In this talk, I hope to convey the essence of a salty, if not downright briny, point of view about quantum theory: The deepest truth of quantum information and computing is that our world is a world wildly sensitive to the touch. When we irritate it in the right way, the result is a pearl. The speculation is that this sensitivity alone gives rise to the whole show, with the quantum calculus portraying the best shot we can take at making predictions in such a world. True to form, I ask more questions than I know how to answer. However, along the way, I give a variant of Gleason's theorem that works even for rational and two-dimensional Hilbert spaces, give another variant of Gleason's theorem that gives rise to the tensor-product rule for combining quantum systems, and finally derive a new form for expressing how quantum states change upon the action of a measurement.
\eq

\section{02-06-02 \ \ {\it Spheres, Spheres, Spheres} \ \ (to G. {\Plunk})} \label{Plunk5}

Yeah, I did get your Saturday email, but it came after I wrote the note to you.  (That's the danger of writing a note before logging in to a server.)

So, I'd say, try again.  Today, I'm hoping to work on my project with Petra and {\Ruediger}.  In your expansion of the state $\rho$ in terms of the sicpovm, did you a) take into account that $\tr\,\rho=1$, and b) take into account that $\tr\,\rho^2=1$.  I think you mentioned one of them, but I don't know if you mentioned both.  Also, did you make sure to tabulate the distance from the center of the simplex.  I remember you mentioning something along those lines, but I got a little confused as to whether you thought you could ignore it.

It'll work out; I'm starting to have faith.

\subsection{Gabe's Preply, ``More on Spheres''}

\bq
So get this.  I woke up this morning to head off to the lab because I'm
itching to get back to this sphere problem (more on that later) and I
run into Mark the neighbor who reminds me about the party tonite.
Funny, I thought the party was tomorrow night.  Then I start driving and
there isn't any traffic---lucky break I think.  But when I find an empty
lot at Bell Labs it all finally clicks and I realize that it isn't
Friday but Saturday!  Anyway, the moral of this story is that Steven van
Enk isn't here to help me with my problem so I'm gonna ask you.

I feel it is almost definite that we should see spheres in the simplex
for SICPOVMs.  I have a simple plan to prove it too.  Start with Caves'
form: POVM $= \{(1/D) \Pi_\alpha\}$.  Where $\Pi_\alpha$ are projection
operators.  The symmetric part comes in by requiring the inner product
between these $\Pi_\alpha$ to be all equal to some number, $1/(D+1)$.  OK, so
now we have this set of $D^2$ lin.\ ind.\ operators which form this operator
basis as Caves says.  So the idea is that we want to decompose a general
pure state, $\rho$, in terms of this operator basis.  So I write
$$
\rho = \sum a_i \Pi_i\;,
$$
where the $a_i$ all add up to one since $\rho$ is a
density matrix.

But to make these rho pure we put an additional restriction on the $a_i$
by demanding $\tr\, \rho^2 = 1$.  This gives a little relationship for the
$a_i$.  So we take this arbitrary pure state $\rho$ and the POVM and
calculate the $D^2$ probabilities.  These give the components of vector
which reaches into the simplex to a point on the sphere.  Take the
magnitude of this probability vector and substitute the relationship for
the $a_i$ and I think what {\it should\/} happen is that the $a_i$ drop out and we
get a quantity that is just dependent on the dimension $D$.  (Now this
isn't the radius of the sphere since we didn't project the probability
vector into the simplex but this isn't a problem since the component
perpendicular to the simplex is constant so just adds in a constant when
we calculate the magnitude of the probability vector.  It's just as easy
to first subtract the point $(1/D^2, 1/D^2,\ldots,1/D^2)$ which is the center
of the sphere and work forward from there.)  So with a probability
vector with constant magnitude we have proved our sphere (see previous
parenthetical note).

So that's it.  But I think I goofed somewhere.  Because I worked it out
a hundred thousand times and it doesn't quite work.  I've got residual
$a_i$ that won't die.  So I tried it for the case that I've already done
(rebits) and know for certain it should work \ldots\ and it doesn't.  What's
wrong?  Something with my construction of $\rho$?  Some additional
restriction on the $a_i$?

I guess I'm gonna head home.
\eq

\subsection{Gabe's Reply, ``Soccer Balls in the Probability Simplex''}

\bq
Did you see Brazil play Turkey?  It was a fun match.  The world cup
games must be at reasonable times in Australia.  Here you have to stay
up till 4AM or settle for a replay the next day.

We finally found those spheres.  Steven showed me something that I'd
missed.  It becomes embarrassingly easy after that.  The SICPOVM gives
rise to a sphere in the probability simplex.  Its radius squared is:
$$
R^2 = \frac{D-1}{D(D+1)}
$$
Right.  So we'd like the volumes for all ICPOVMs to be ellipsoidal and
less than the volume of the sphere.  I'll see what can be done in that
direction.  I have some ideas.

Oh and I'll do out the qubit too, I've been neglecting it in favor of
looking for these spheres.
\eq

\section{02-06-02 \ \ {\it Postmodern Rags} \ \ (to A. W. Harrow)} \label{Harrow1}

Here are four representatives of what it is that Carl and {\Ruediger} and a I are trying to get at:
\begin{itemize}
\item
``Quantum Mechanics as Quantum Information (and only a little more)''\\
\quantph{0205039}
\item
``The Anti-{\Vaxjo} Interpretation of Quantum Mechanics''\\
\quantph{0204146}
\item
``Unknown Quantum States: The Quantum de Finetti Representation''\\
\quantph{0104088}
\item
``Quantum Probabilities as Bayesian Probabilities''\\
\quantph{0106133}
\end{itemize}

The thing that comes closest to a postmodern rag is my Anti-{\Vaxjo} paper.  (That's the one I thought you might enjoy reading for the literary value.)  The top paper, however, ``QM as QI,'' is my present pride and joy.

Below is a passage from my samizdat about Derrida and quantum mechanics.  The samizdat, by the way, is
\quantph{0105039}

By the time you see this note, I'll be sitting next to you in 161.

\subsection{From a 09 June 1997 note to G. L. Comer, ``Dictionaries and Their Problems''}

\bq
How are you my friend?  It's been so long since I've written you anything of substance, I almost wonder if I can still remember how!
Lately, I've once again taken to reading about {\Bohr}'s (and the other founding father's) thoughts on the epistemological and ontological lessons of quantum mechanics.  I suppose part of my reason for getting back to these things is just a general tiredness of looking at equations; maybe it's a form of procrastination---papers need writing, papers need revising, papers need refereeing, talks need preparing \ldots\ and I'm getting a little tired of it all.

In any case, the exercise is having its own payoff.  Maybe I'll share a little with you.  Remember I told you that {\Mermin} suggested that {\Derrida}'s mumblings shouldn't be written off?  I guess I'm starting to think he was right (though I have to admit that I haven't yet read any of {\Derrida}'s own writings, only commentaries). It seems that the focal point of {\Derrida}'s thought centers around none other than your ``problem of the dictionary''!
Let me try to give you something of a flavor of how these things might be connected to the quantum.  My starting point has been an excellent essay by John {\Honner} titled, ``Description and
Deconstruction: Niels {\Bohr} and Modern Philosophy'' (found in {\em Niels {\Bohr} and Contemporary Philosophy}, edited by Jan {\Faye} and Henry J. {\Folse} (Kluwer, Dordrecht, 1994), pp.\ 141--151). I hope you enjoy the quotes:

\bq
{\Derrida} undermines the notion that words and signs can capture present experience:  our tracing of experience always discloses a supplement, a `difference'.  This attack is equivalent to a subversion of the notion of strong objectivity and correspondence theories of truth.  For the deconstructionist, the foundations for knowledge are never securely laid: words do not correspond exactly to the world.  ``Presence'' can never present itself to a present consciousness, and hence experience is always and already constituted as a text.  [I use `text' loosely here, of course, meaning any collection of signs---discourse, mathematical equations, pictures, poems, prose, drama, hand-waving---used to trace and express insight and experience.]  A text is a collection of signs and any sign presumes a presence which it represents, but the sign is not the same as that which it represents. In signifying our awareness of a presence a move is made from the presence to sign.  By the word `presence' {\Derrida} is indicating something like substance, essence, or object, but he rejects such `totalising' categories as these, for such terms assume more about the presence than perhaps we are entitled to assume.  The term may `trace' the presence, but a remainder is always left over.
\eq

\bq
Speaking and writing are, according to {\Derrida}, `linear'
activities which lock us into space and time.  ``The great rationalisms of the seventeenth century'', as {\Derrida} describes them, fall into the trap of objectivity and neglect the timelessness of self-presence.  The linearity of the words limits the conditions for the use of language: ``If words and concepts receive meaning only in sequences of differences, one can justify one's language, and one's choice of terms, only within a topic [an orientation in space] and an historical strategy.''  Here we have a curious serendipity.  Our usage of words is tied, arguably, to the reidentifiability of particular objects, which itself implies those bastions of classical physics, the conservation of position and momentum and an absolute space-time framework.  And it was precisely these bastions that {\Bohr} attacked.  As I have argued elsewhere, {\Bohr}'s fundamental arguments entail a provocative hint at a link between the given character of ordinary language and a deterministic-mechanistic view of the workings of nature.  For {\Bohr}, classical physics is the inexorable result of the use of language based on the identification of experienced material particulars; or, vice versa, the use of language based on identification of experienced particulars will ultimately lead to a sense of the persisting presence and movement of material object in space and time, and hence to principles of conservation, causal change, and continuous space-time frameworks.
\eq
\eq

\section{03-06-02 \ \ {\it I Think She'll Know, 2} \ \ (to N. D. {\Mermin})} \label{Mermin70}

\bdm
Kurt Gottfried got me back to thinking about the old ``derivations of
probability'' dating back to Hartle in the 1960's and going through
the Sidney Coleman application to many worlds.  Turns out Jeffrey
Goldstone did something on it and there's a nice paper by Gutmann
(\quantph{9506016}).  To remind you, in the modern version one's
only probabilistic assumption is that if $\tr(\rho E) = 1$ then
$E$ must happen. Combining that with some highbrow analysis of the
nonseparable hilbert space formed by infinitely many copies of a
system with itself, one derives all the usual probabilistic rules.

As I remember you were quite scornful of this approach, saying that
they were sneaking probability into the story without admitting it.
Was this because as a good Bayesian you regard probability 1 as no
different from any other probability --- merely the current best
guess.  (Our conversation about Coleman took place before we had our
arguments about the difference between probability 1 and ``has to
happen''.)  Or did you have some other leakage of probability into
the argument in mind?
\edm

Yeah, I've still got loads of issues with that approach.  But maybe
just let me mention the simplest one again:  $\tr(\rho E) = 1$
certainly should not be taken to imply ``must happen'' in an infinite
setting.  Take the converse.  Consider an infinite sequence of coin
tosses.  Each individual outcome string has probability strictly
zero, yet one of them does happen.

Shifting the problem to sets of measure 1 doesn't help either.  For
lots of inequivalent sets have measure 1.  What principle of nature
sets one out as important?

It's all ad hockery that these guys are up to \ldots\ shined up with
some high-powered mathematics so that it looks important.

\section{04-06-02 \ \ {\it Random Questions} \ \ (to G. {\Plunk})} \label{Plunk6}

At least we can probably bank on this much (subject to proof).  The volume of the maximal quantum state space is bounded above by the volume of the sphere with radius given by the number you sent me earlier.  Here's a question.  How does that volume compare to the volume of the simplex?  It looks to me like it becomes a smaller fraction of the simplex volume, but I'm not sure.

\subsection{Gabe's Reply, ``Random Answers''}

\bq
I think we have easily that non-pure states will not touch the sphere.
See the attached files (the {\tt .txt} file is \TeX\ output from Mathematica).\footnote{A key thing to note here is that Gabe presents an expression for $\rho$ in terms of the SIC outcome probabilities.  Somehow I missed this, perhaps not being able to open the files he had sent me at the time.  Thus, the surprise in my 18-06-02 note ``SIC POVMs'' to C. M. {\Caves}; apparently I still did not understand Gabe in the previous day's discussion.  Just discovering this now, sadly I have never before given Gabe credit for this result.}

\noindent --- --- --- --- --- --- ---

\bq\noindent
The SICPOVM is defined using projection operators $\Pi_i$ which satisfy
$$
\frac{1}{D}\sum_i \Pi_i=I
$$
and
$$
\tr\big(\Pi_i\Pi_j\big)=\frac{1}{D+1}\;.
$$
The SICPOVM is thus $\{E_i\}$ with $E_i=\frac{1}{D}\Pi_i$.

Write an arbitrary state $\rho$ as in terms of the operator basis $\{\Pi_i\}$:
$$
\rho=\sum_i a_i \Pi_i \quad\mbox{with}\quad \tr\,\rho=1 \;\; \Longrightarrow \;\; \sum_i a_i = 1\;.
$$
The probability of outcome $\alpha$ is
\bea
p(\alpha)
&=&
\tr\,\rho E_\alpha \sum_i a_i
\nonumber\\
&=&
\frac{1}{D}\left(\frac{1}{D+1} \sum_i a_i + \frac{D}{D+1} a_\alpha\right) \; = \; \frac{1}{D}\left(\frac{1}{D+1} + \frac{D}{D+1} a_\alpha\right) \;.
\nonumber
\eea
We can find the center of the sphere by calculating the probabilities with $\rho=\frac{I}{D}$.  We get that all probabilities are $\frac{1}{D^2}$.  Subtract this from $p(\alpha)$ to get the radius vector:
$$
r(\alpha)=\frac{1}{D}\left(\frac{D}{D+1} a_\alpha-\frac{1}{D(D+1)}\right)
$$
$$
r^2(\alpha)=\frac{1}{D^2}\left(\frac{1}{D^2(D+1)^2}-\frac{2 a_\alpha}{(D+1)^2}+a_\alpha^2\frac{D^2}{(D+1)^2}\right)\;.
$$
Now sum over $\alpha$ to get the radius squared
$$
R^2=\sum_\alpha r^2(\alpha)=\frac{1}{D^2}\left(\frac{D^2}{(D+1)^2}\sum_\alpha a_\alpha^2 - \frac{1}{(D+1)^2}\right)\;.
$$
Now we can learn about the quantity $\sum_\alpha a_\alpha^2$ by taking the trace of $\rho^2$:
$$
\tr\, \rho^2 = \frac{1}{D+1}\sum_\alpha a_\alpha^2+\frac{1}{D+1}\;.
$$
Define $F=\sum_\alpha a_\alpha^2$ and note that $\tr\, \rho^2 \le 1$.  And we get $F\le 1$, with equality if and only if $\rho$ is a pure state.  Plugging into $R^2$ \ldots
$$
R^2=\frac{1}{D^2}\left(\frac{D^2}{(D+1)^2}F-\frac{1}{(D+1)^2}\right)\;.
$$
For pure states
$$
R^2=\frac{D-1}{D^2(D+1)}\;.
$$
Otherwise it's less.
\eq
\eq

\section{04-06-02 \ \ {\it Morning Headaches} \ \ (to G. {\Plunk})} \label{Plunk7}

I've got a huge headache this morning \ldots\ but you are my aspirin.

\bgp
It seems a reasonable hope that we could get that the surface created
by the pure states is closed.  Then we would have that the entire
sphere is generated by the pure states (right?).
\egp

The more I think about it, the less likely I think a miracle will happen here.  The problem is that normalized pure states have a dimensionality of $D-1$, whereas the full set of states has dimensionality $D^2$. \ldots\ Though in saying that I'm a little comparing apples and oranges.  (Because the pure states don't form a subspace in the space of operators.)  So, it's worth further thought, but I fear \ldots

Maybe a better way to say it is that to parameterize the set of pure states you only need $2D-2$ real parameters (i.e., $D$ complex numbers minus normalization), whereas to parameterize the full set of states you need $D^2-1$.  When $D=2$, that means you get 2 and 3, respectively.  So that means one can think of one set of parameters living in the sphere and the other on its surface.  But, when $D>2$ that clean connection goes away.

\bgp
I found the $n$-Sphere paper.  You said you'd like a copy--would you
like me to mail it?  Will Lucent do this for me?
\egp

No, I just meant to make me a copy while you were making yours and hold onto it.

\section{04-06-02 \ \ {\it Big File Coming} \ \ (to K. W. Ford)} \label{Ford2}

Thanks for the encouraging email.  In the next note, I'll send you the compendium as it stands.  I'll send it both as a \LaTeX\ file (REV\TeX\ actually) and as a PostScript file.  I'll attach both to the same email.  If you have any trouble opening or reading these files or getting them to compile, let me know and I'll try to find a solution.

As you'll be able to tell, I'm quite in mid-project with this.  A lot of material still needs to be added (essentially all the entries marked with a \P), but also the introduction still has a long ways to go.  In particular, with regard to you, you'll notice very few quotations in the Wheeler area.  But that is only because I haven't yet gotten them out of the papers and into my scanner.  Most of those papers I read more than 10 years ago, long before I started this project.  Anyway, I'll be fixing that shortly, i.e., as soon as I get back to New Jersey.

\bkwf
Or I could send you a copy of his bibliography (59 pages) for you to eyeball.
\ekwf

That would be great!\footnote{NOTE: I would have worked to include the bibliography Ken sent me, but using it as a seed, Terry M. Christensen built a significantly more thorough bibliography for his PhD thesis, {\sl John Archibald Wheeler:\ A Study of Mentoring in Modern Physics}. The bibliography section of the thesis has been conveniently posted by Baruch Garcia here: \myurl{http://jawarchive.files.wordpress.com/2012/03/bibliography.pdf}.}  I appreciate the help.  If you can send it electronically, just send it now.  If, however, you only have it on paper, please send it to my Bell Labs address below.  Beyond that, maybe one of the best ways you could help me would be in helping me obtain copies of some of the papers I haven't been able to find.  For instance,
\begin{itemize}
\item[]
J.~A. Wheeler, ``Delayed-Choice Experiments and the Bohr--Einstein Dialogue,'' in {\sl American Philosophical Society and the Royal Society:~Papers Read at a Meeting, June 5, 1980}, (American Philosophical Society, Philadelphia, 1980), pp.~9--40.
\end{itemize}
seems to be a crucial paper.  And there are some ones beyond that.

\bkwf
I really have no good ideas about a place to publish it. Maybe it will have to end up on the Web with references to it published in journals and newsletters.
\ekwf
Yeah, that's certainly an option.  I guess I was half hoping there might be some room in the Wheelerfest volume you're associated with, and if so, that'd give me the incentive to get this finished right away.  But, in truth, the paper is already pretty large, and I expect it to maybe double in size by the time I'm finished.  So, it's probably way too large for anything like that.

\bkwf
I'm sorry that John Wheeler himself has slowed down so much that he is
probably no longer able to absorb your work. There was a time when it
would have meant a lot to him and when he no doubt would have loved to
enter into a colloquy with you.
\ekwf

How is John's health?  Both physical and mental?  I haven't talked to him since 1994.  I had been thinking about visiting him this summer if was still going in to his office in Princeton.  I read an article in the {\sl New York Times\/} that said he was still going in a couple of times a week.  And I thought, if his health is that good, then it might be good to establish contact now that I am in the neighborhood.

OK, the files will come in the next mail, after I go have a cup of coffee.

\section{05-06-02 \ \ {\it Touching the Bound} \ \ (to G. {\Plunk})} \label{Plunk8}

I guess the simplest and main question on my mind at the moment is:  How does volume of the ``state sphere'' (i.e., the sphere within the simplex that {\it surrounds\/} the full set of quantum states) scale as a fraction of the volume of the full simplex?  So, calculate the volume of the sphere in arbitrary dimension, and calculate the volume of the simplex, and take their ratio.

I suspect the volume tends to zero, but it'd be nice to get a formal expression of that.

This would give a formal expression to the idea that, as the size of a system grows, one accepts in quantum mechanics that one can know less and less about it.

\section{06-06-02 \ \ {\it Good News and Not-So-Good News} \ \ (to G. {\Plunk})} \label{Plunk9}

\bgp
In higher dimensions my results say that the $n$-sphere's volume divided
by the $n$-simplex's volume becomes unbounded(!).  This means (a) the
$n$-sphere of required radius does not fit inside the $n$-simplex and
consequently (b) the states trace out only part of the sphere--at most
the part that intersects the $n$-simplex.
\egp

Yeah, that means something's got to be wrong.  The sphere, by definition, has to be contained within the (regular) simplex.  So, I guess there are two options.  Either your original calculation of the radius is wrong (make sure you calculate it from the center of the simplex).  Or the formulas you've found for the volumes are off or misapplied.

I presented this picture of quantum mechanics in a talk yesterday at U. Queensland.  It went over well, especially with some of the postdocs. I told them also about your explorations.  Next week I do the same at Griffiths University.

Main question:  Are you having fun with this problem?

\subsection{Gabe's Preply}

\bq
Some good news.  I was just thinking a little about the parametrization
of pure states in terms of the SICPOVM elements.  In the analysis I sent
you I decompose using $D^2$ coefficients.  The conditions on the
coefficients are that they sum to 1 and the squares sum to 1.  In
``coefficient space'' this is the intersection of a plane and a sphere
which gives a sphere of dimension $D^2 - 1$ which requires $D^2-2$
parameters to describe.  This is exactly the dimensionality we need for
describing the sphere in the simplex.  Pure states have dimensionality
$D^2-2$ and the full set of states has dimensionality $D^2-1$.\footnote{This sentence and the one previous to it are incorrect, as the pure states have dimensionality $2D-2$; however I preserve it for historical accuracy.}  It's even
simpler.  Creating the ``radius'' vector in the simplex amounts to a shift
and rescaling of the ``coefficient vector'' (look at $r(\alpha)$ in the proof
I sent).  The coefficient vectors form a sphere so the radius vectors do
also.  It looks like we have the {\it complete\/} sphere after all.  A similar
analysis fills the interior of the sphere using the mixed states.

Finding the volume ratio is rough going.  For the qubit, my results for
the ratio are exactly as you'd expect for a sphere inscribed in a
tetrahedron.  In higher dimensions my results say that the $n$-sphere's
volume divided by the $n$-simplex's volume becomes unbounded(!).  This
means (a) the $n$-sphere of required radius does not fit inside the
$n$-simplex and consequently (b) the states trace out only part of the
sphere---at most the part that intersects the $n$-simplex.  That can't be
right (especially given what I said above).  But I'm using textbook
formulae for the volume of regular $n$-simplices and $n$-spheres.  I'll
attach a Mathematica file of what I've done.
\eq

\section{07-06-02 \ \ {\it Spheres and Simplexes on a Saturday Morning} \ \ (to G. {\Plunk})} \label{Plunk10}

\bgp
I'm dumb.  I forgot completely about positivity.  We don't have the
whole sphere, you were right all along. At most we have the intersection of the sphere with the simplex.
Maybe that's exactly what we have.  Is it obvious?  Does the fact that
the probabilities are contained in the simplex guarantee that a given $\rho$ is a valid positive density matrix?
\egp

You're not dumb.  I know you're smarter than me.  But goddamit, start thinking about the sphere from the center of the simplex!  That is the one that is of interest.  Not the one centered on the origin of the coordinate system.  That sphere---the one from the center of the simplex---is automatically contained {\it within\/} the simplex.\footnote{Clearly, I was the ``dumb one'' because this is not true except in $D=2$.}

In other words, the set of valid quantum states will be a subset of that sphere.  A proper subset.  Nevertheless, even though the sphere's volume will only be an upper bound on the volume of the true set of quantum states, we can still glean some interesting information by studying the sphere.  If the ratio of sphere's volume to the simplex's volume goes to zero as the dimension grows---as I suspect---then it will follow that the more interesting ratio (namely that of the volume of the valid quantum states to the simplex) will go to zero too.  It'd just be a  nice fact to know for sure.

\bgp
My principle hope is to come out of this with a deeper understanding
of quantum mechanics.
\egp

I hope so too.  Have you given my paper \quantph{0205039} a shot?  If you have any questions, I'd be glad to answer.  Or why don't you pick up the book {\sl Pragmatism\/} by William James or the book {\sl Philosophy and Social Hope\/} by Richard Rorty off my desk?  I think, with respect to quantum mechanics, they contain a lot more physics than you might believe.

\section{11-06-02 \ \ {\it Who Is asdf?\ Who Is jkl?}\ \ \ (to G. {\Plunk})} \label{Plunk11}

\bgp
So what I'm trying to say is that I've been doing this \ldots\ and the
sphere is still too big.  Here's how I resolve the discrepancy:
\egp

Hmm.  This is intriguing.  But I have difficulty seeing how it could happen.  I guess strictly speaking, what you have really proved is that all the pure states are equidistant from the center of the simplex.  Period.  And indeed, there is no requirement that the sphere those points sit on need be strictly contained within the simplex.  However, I find it difficult to see how that could happen.  It seems to me that that would require that the pure states form a set of disconnected sets (one for each element of the SICPOVM).  But we know the pure states form a connected set.

I'll be glad when I'm back to there, so I can have a chance to see these things more directly.  I get back very late Sunday night.  I'll probably spend the night with my friend Jeff and then come back to the house Monday morning.  Would you mind hanging out there until then, so that I can get a ride to the repair shop to pick up my mini-van etc.?

\subsection{Gabe's Preply, ``asdf likes jkl;''}

\vspace{-12pt}\bq
\noindent\bq\noindent
[CAF said:] ``goddamit, start thinking about the sphere from the center of the simplex!''
\eq
So what I'm trying to say is that I've been doing
this \ldots\ and the sphere is still too big.  Here's how I
resolve the discrepancy:

By taking linear combinations of the POVM elements (the
construction of $\rho$ in the proof I sent) and requiring
trace and trace of the square to be one, we don't
necessarily get valid states--they aren't necessarily
positive operators though they are hermitian and
normalized by construction.  I've been banking on these
things being actual states.  It turns out that some of
them give probabilities outside the simplex (i.e.
negative probabilities).  So those ones definitely
can't be states.

The question (which I tried to pose in the last email)
is if we take the points on the sphere that are inside
of the simplex can we say that all of them correspond
to actual, valid states?  Most of them certainly.  If
{\it all}, then I can attempt calculate the volume of the
intersection of the sphere and the simplex and get your
bound \ldots.

This is a strange result--the sphere-simplex
intersection as the quantal probability space.  But
I've been so careful checking the math that I think
it's right.  Just in case you don't believe me (which
is understandable given my vacillations) I went ahead
and calculated the radius of the largest sphere which
could fit in the simplex (the inscribed sphere).  Its
volume divided by the simplex volume goes to zero as
expected.
\eq

\subsection{Gabe's Reply, ``Spheres with Holes''}

\vspace{-12pt}\bq
\noindent\bq\noindent
[CAF said:] ``It seems to me that that would require that the pure states form a set of disconnected sets (one for each element of the SICPOVM).  But we
know the pure states form a connected set.
\eq

The picture I have in my head is a connected region with holes.  The
vertices and edges of the simplex always are further from the center of
the simplex than the sphere's radius.  Here the sphere is contained in
the simplex.  It is close to the center of the faces of the simplex (set
one probability to zero to get a face) that the sphere emerges from the
simplex ($D>2$).  $D^2$ faces give $D^2$ holes.  That the pure states are a
connected set follows from this picture I think.
\eq

\section{11-06-02 \ \ {\it Radical Probabilism} \ \ (to A. W. Harrow)} \label{Harrow2}

\bv
\myurl[http://www.princeton.edu/~bayesway/]{http://www.princeton.edu/$\sim$bayesway/}
\ev

\section{11-06-02 \ \ {\it Act II -- Griffith University} \ \ (to A. W. Harrow)} \label{Harrow3}

1. Tomorrow at 11 seems fine for a seminar, although I am yet to get a room booked. Could you send me a title and abstract, please?

2. Another good question I meant to ask, is how would you translate the following into subjective language? In quantum control, optimal feedback control is based upon using one's knowledge of the system (i.e.\ its quantum state). However due to practical processing power limitations (shades of the quantum computer problem), one often cannot compute the true quantum state conditioned on the measurement results quickly enough. Instead, one computes an approximation to this, a best-estimate given the limited resources available, and uses that as a basis for feedback control. By simulation, one can determine on average how well this does compared to feedback based on the true state.

3. On the ``let's call the whole thing off'' theme, if you used some Greek letter other than $\rho$, it could work really well. For example, ``you say beeta, I say bayta, you say eeta and I say ayta \ldots'' Or for pure states it could work: ``you say psigh, I say psee, you say phigh, I say phee \ldots''.

\section{12-06-02 \ \ {\it Consilience} \ \ (to C. M. {\Caves} \& R. {\Schack})} \label{Caves65.1} \label{Schack57.1}

Here's my own present take on the matter.  [See 23-04-02 note ``\myref{Comer12}{Music in the Musician}'' to G. Comer.] \ldots\ I couldn't have said it better than I said it in the note below (I put that in for a {\Ruediger} smile).

\section{12-06-02 \ \ {\it Receipt} \ \ (to M. J. Donald)} \label{Donald3}

I got your long note; thanks.  I'm going to mull over it for a while
before replying.  But I will reply.

\bmjd
I also haven't commented much on the points I agree with in the
paper, although there are a few of these!
\emjd

It would be nice to see what these are.  In ways, I'm more interested
in where we agree than in where we disagree.  A good discussion needs
some solid ground somewhere.

Beyond that, it would please me to no end to learn that you might
have found a thought or two that you could use in the thing.  But if
you didn't, you didn't.  It's a horrible feeling to think I might be
writing nothing for nothing.

Each day, I tell my daughters that they can change the world.  I tell
them that they can change it to the core.  But I never tell them that
they can believe anything they want.  There is a difference.  And you
don't see it.

The summary at the end of your note troubles me to no end.
\bmjd
Your know-nothing ism, like de Finetti's irrationalism (Gillies,
``Philosophical Theories of Probability'', page 86), have the dangers
of Bohr's writings on which I would agree with Beller (Physics Today,
September 1998, pages 29--34).  In particular, by leaving far too
much in vagueness, incoherence, and pious hope, you give the
religously-minded the official endorsement of the physics
establishment that they may believe anything they want, instead of,
by example, instructing them that they can believe anything they want
as long as it is rational, coherent, tentative, revisable, and
compatible with the evidence (and therefore contrary to naive
expectations, because if quantum theory, or indeed science in
general, tells us anything it is that the world is not how we would
have imagined it before we investigated); and they accept that they
may be completely wrong.
\emjd
I will certainly return.

\section{12-06-02 \ \ {\it Bell Ineqs as Limitations on State Distinguishability} \ \ (to A. Peres)} \label{Peres37}

I'm in Australia at the moment, but it won't be too long before I'm able to go home!  I'm getting quite homesick.  Kiki and the girls are in Munich; they arrive in New Jersey one day after me.

\bap
I spent two days in Geneva. The first was mostly business, related to
Helle's final PhD exam (a public lecture) and the second was relaxing
and touring the city. My preceding visit was in 1979, for three weeks,
when I was the guest of John Bell.

Helle's work was on Bell ineqs and cryptography, and relations between
the two (you probably remember Gisin's insistence on this point).
In an informal discussion, Helle asked me whether Bell ineqs could be
reformulated in terms of limitations on state distinguishability. My
answer was ``QM as QI: there is no God but Chris Fuchs, and Danny Terno
is His prophet.'' Can you be more specific?
\eap

I wish I could say more about that!  But unfortunately my mind runs dry right now.

\section{16-06-02 \ \ {\it Weddings and Connecting Flights} \ \ (to A. Peres)} \label{Peres38}

Thanks for the pleasant note keeping me up-to-date on everyone.  We are indeed like one big family.  Sam actually invited me to his wedding!  But Israel is a bit out of my reach at the moment.

Actually, I'm not home yet.  At the moment I am in the American Airlines Lounge in Los Angeles.  Still six more hours of flight to go.  Kiki and the girls return to New Jersey from Munich on the 18th.

An interesting coincidence happened to me on the flight from Brisbane to Los Angeles.  It stopped for an hour in Auckland, New Zealand.  To my surprise, one of the passengers who entered there was Andrew Doherty, one of Hideo Mabuchi's postdocs.  Moreover, his seat was directly behind mine.  He had been in New Zealand on vacation, visiting his family, and was now going back to Caltech.  We had some pleasant conversations about the latest results in entanglement.

Take care of yourself.

\section{17-06-02 \ \ {\it Jones Stuff} \ \ (to G. {\Plunk})} \label{Plunk12}

\begin{enumerate}
\item
K.~R.~W. Jones, {\em Quantum Inference and the Optimal Determination of Quantum
  States}.
\newblock PhD thesis, University of Bristol, Bristol, England, 1989.

\item
K.~R.~W. Jones, ``Entropy of random quantum states,'' {\em Journal of Physics
  A}, vol.~23(23), pp.~L1247--L1250, 1990.

\item
K.~R.~W. Jones, ``Principles of quantum inference,'' {\em Annals of Physics},
  vol.~207(1), pp.~140--170, 1991.

\item
K.~R.~W. Jones, ``Quantum limits to information about states for finite
  dimensional {H}ilbert space,'' {\em Journal of Physics A}, vol.~24,
  pp.~121--130, 1991.

\item
K.~R.~W. Jones, ``Towards a proof of two conjectures from quantum inference
  concerning quantum limits to knowledge of states,'' {\em Journal of Physics
  A}, vol.~24(8), pp.~L415--L419, 1991.

\item
K.~R.~W. Jones, ``Riemann-{L}iouville fractional integration and reduced
  distributions on hyperspheres,'' {\em Journal of Physics A}, vol.~24,
  pp.~1237--1244, 1991.

\item
K.~R.~W. Jones, ``Fractional integration and uniform densities in quantum
  mechanics,'' in {\em Recent Advances in Fractional Calculus} (R.~N. Kalia,
  ed.), (Sauk Rapids, MN), pp.~203--218, Global Publishing, 1993.

\item
K.~R.~W. Jones, ``Fundamental limits upon the measurement of state vectors,''
  {\em Physical Review A}, vol.~50(5), pp.~3682--3699, 1994.
\end{enumerate}

\section{17-06-02 \ \ {\it The Dreams Stuff Is Made Of} \ \ (to T. Rudolph)} \label{Rudolph6}

\btr
This would be a good title for one of your papers: ``Quantum mechanics: The dreams stuff is made of''.
\etr
I like the title.  Now we just have to figure out a content for it.

\section{18-06-02 \ \ {\it SIC POVMs} \ \ (to C. M. {\Caves})} \label{Caves66}

That result is rather neat.  Thanks for sending it.  Funny coincidence: at the end of the day yesterday, {\Gabe} and I parted after asking whether there was a simple way to express a quantum state in terms of the SICPOVM probabilities.\footnote{See 04-06-02 note ``\myref{Plunk6}{Random Questions}'' to G. {\Plunk}. Gabe had actually already solved this problem, and I was somehow not aware of it.}  But I didn't work on the problem after getting home.  Instead, I woke up this morning and find that you've sent me the answer!  Maybe I'll try the experiment again tonight \ldots

\subsection{Carl's Preply}

\bq
\noindent Here's a nice property of SIC POVMs.  Let
$$
p_\alpha=\tr(\rho E_\alpha)
$$
be the measurement probabilities for a SIC POVM
$$\
E_\alpha=\Pi_\alpha/D\;.
$$
Then
$$
\rho=\sum_\alpha\left((D+1)p_\alpha-\frac{1}{D}\right)\Pi_\alpha
$$
The complete story is in the attached \TeX\ file updating my little document on SIC POVMs.
\eq

\section{18-06-02 \ \ {\it SIC POVMs} \ \ (to G. {\Plunk})} \label{Plunk13}

Have a look at the document below that I just got from Caves.  [See 18-06-02 note ``\myref{Caves66}{SIC POVMs}'' to C. M. {\Caves}.]\footnote{Again, see 04-06-02 note ``\myref{Plunk6}{Random Questions}'' to G. {\Plunk}.}  The result is rather neat and relates to our conversation at the end of the day yesterday.  I'll attach a PS file of the account in case you can't \TeX\ yet.  It might be a good exercise to see if you can find a more elementary derivation of the fact.

\section{18-06-02 \ \ {\it Horrible!}\ \ \ (to T. Rudolph)} \label{Rudolph7}

\noindent As I said, horrible (title, that is):
\bq
\noindent INSTITUTE FOR QUANTUM INFORMATION SEMINAR\medskip

\noindent Tuesday, June 18: IQI Seminar\\
Quantum steering\\
Frank Verstraete, KU Leuven, Belgium\\
3:00 p.m., 74 Jorgensen\medskip

\noindent ABSTRACT: \url{http://www.iqi.caltech.edu/seminar_abstracts.html#verstraete}\medskip

\noindent Immediately following the seminar, refreshments will be served in 156 Jorgensen.\medskip

\noindent Directions to Caltech:\\
\url{http://www.admissions.caltech.edu/visiting/directions.htm#drive}\\
Campus map: \myurl{http://www.caltech.edu/map/}  (Jorgensen is bldg.\ \#80.)
\eq

\section{19-06-02 \ \ {\it Classical Essence} \ \ (to E. Chisolm)} \label{Chisolm1}

Thanks for your letter.  You ask, ``What is the essence of classical physics?''  I agree that this is something worth thinking about.  I don't have an answer.

Maybe the closest thing I can give you at the moment is to send you to a few passages in my samizdat, \quantph{0105039}.  First have a look at the passage titled ``Genesis and the Quantum'' on pages 85--86, and then have a look at the note ``The Oyster and the Quantum'' on pages 100--103.  Finally, have a look at the note ``Reality and the Symbol'' on pages 233--234.

They will kind of lead you down the path of where I think classical mechanics fits into the scheme of things.  As I say, I agree with you that it is worthwhile to try to characterize the essence of classical physics.  However, in possible opposition to you, I think trying to do that in isolation, independently of the quantum problem, won't lead to the greatest insight.  The difficulty as I see it is that quantum mechanics strikes me much more as a radical departure from classical physics than just a shift in ``calculational recipes.''  For instance, take Newtonian gravity.  I would argue that finding its essence gives one little insight into the essence of general relativity \ldots\ even though Newton's gravity is one of the limiting cases of general relativity.  The underlying ontologies of the two theories are just so very different.  Similarly I am thinking of classical versus quantum physics.

\subsection{Eric's Preply}

\bq
I recently read your e-print ``Quantum Mechanics as Quantum Information
(and only a little more)'', and I like your suggestion that our
understanding of QM right now is in a state similar to that of physicists
trying to understand length contraction et al.\ before special relativity.
I also like your idea (hope?)\ that there are a few simple physical
statements from which the formalism of QM emerges as naturally as special
relativity does from Einstein's two postulates, and I want to ask you a
few questions related to that idea.

Let me describe how I think about the overall structure of theories in
physics.  I see classical mechanics and quantum mechanics, understood in
their broadest senses to include the corresponding field theories, as
basically recipe books:  If you want to make a theory of the
electromagnetic field, or projectile motion over the surface of the earth,
or of the gravitational field, then you come up with a Lagrangian or
Hamiltonian and then either classical or quantum mechanics provides you
with a basic set of rules to follow (use the Euler-Lagrange equations in
one case or {\Schroedinger}'s equation in the other, states are in either a
phase space or a Hilbert space, etc.)\ to get a fully predictive theory.
Each one is an outline, so to speak, that tells you how to use a
Lagrangian or Hamiltonian to describe the behavior of a system.  On the
other hand, another question most theories must answer is how they
incorporate the principle of relativity, and this can be done in two ways:
Galilean invariance or Lorentz invariance.  Thus, for example, we can
formulate a theory of free body motion classically or quantum mechanically
and with Galilei or special relativity.  (This is a huge
oversimplification, I know; many theories aren't strictly classical or
quantum, not every theory has a Lagrangian, for some theories the
principle of relativity isn't included at all, etc., but this is a good
scheme to think about.)

In this scheme, quantum mechanics and special relativity are not naturally
parallel; instead, special relativity is parallel to Galilean relativity
and quantum mechanics to classical mechanics.  That leads me to ask this
question:  We know the simple, clear physical statements that lead to
special relativity, and we know the same for Galilean relativity (I
believe replacing Einstein's postulate about the speed of light with
something like ``velocities add'' would do it); how about classical
mechanics?  Now an answer to this question would be interesting; surely it
would help us understand what to look for as the ``essence'' of quantum
mechanics if we could clearly state the essence of its classical partner.
I honestly have no idea what simple physical ideas could serve as the
foundation of classical mechanics (broadly understood, including theories
like classical electromagnetism and general relativity).  Do you?
\eq

\section{19-06-02 \ \ {\it Experience} \ \ (to W. K. Wootters)} \label{Wootters13}

\bbw
I'm writing just to let you know that I have not ignored or forgotten
the excellent questions you've asked.  I've been busy with other
things, and I still need to do some serious thinking about your
questions (i.e., about my earlier statements).  I promise a response
before long!
\ebw

That's no problem, but certainly I'd like to hear your answers.  In fact, I'm just back from Australia and Kiki and the kids are back from a month in Munich, so I'll probably be taking a couple of days off from the intellectual life.

Nevertheless, I've picked up my first Whitehead essays to read.  Maybe I'll have a chance of better understanding your terminology this way.  They're from a book titled {\sl Philosophers of Process: Bergson, Peirce, James, \ldots, Dewey, \ldots, Whitehead}.  Given my fascination with some of the ideas of James and Dewey, I think that indicates I may have some of the groundwork already within me.  We'll see.

\section{20-06-02 \ \ {\it Superadditivity}\ \ \ (to S. L. Braunstein)} \label{Braunstein5}

Kiki and I got back from almost a month away from home yesterday (in Munich and Brisbane, respectively), and we found the invitation to your wedding in our mail pile.  Many congratulations to you and Netta!  Kiki and I wish you a long and happy marriage.  I wrote this to another friend some time ago---in this hurry to write you something tonight, let me just repeat the words:  ``Marriage is about stability, happiness, and the kind of leisure that leads to true productivity (which only two people propped against each other can ever hope to negotiate).''

It's hard to imagine you being more productive, Sam, but I suspect we'll see a kind of brilliance from you that we've never seen before.

\section{20-06-02 \ \ {\it Physics without Math} \ \ (to U. Mohrhoff)} \label{Mohrhoff5}

You may not believe it, but I have been struggling on and off for several months to make some comments on your papers.  The writing in each of them is locally beautiful, and it is for that reason to no small extent that I have kept coming back to give them another shot.  But tonight, with too many burdens upon me, I have decided I must admit defeat and try to shake this feeling of guilt I have placed upon myself.

Life is short, and one has to make decisions on how to spend it.  It is a gamble, it is always a gamble, but if one wants to do something of lasting value, one just has to proceed with dogged determination.  I can see in your writings that you are doing that.  But you have got to understand, I am doing it too.  I may be damned for taking the route I am taking, but I feel I will only be able to know that with hindsight.  My game is one of consistency:  I want to push the subjective interpretation of the quantum state harder than anyone has ever pushed it before.  I do that because I have a gut feeling that this will rattle something loose from quantum mechanics.  Something that is wonderful and new.  That little nugget---whatever it turns out to be---strikes me as something our community can hope to attain in the near future, and something that may even be revolutionary in a Kuhnian sense.

From your view, this strategy certainly looks like a stupid move.  I understand that.  I guess what I am writing is that I am willing to live with what you must be thinking of me.  I really am sorry that I have not been able to say something sensible and useful about your papers.

I don't quite understand why I have had such difficulty reading your papers.  They draw me in and then they push me back out.  This may be the most flippant thing I can say, and may reveal a certain artlessness in me, but over and over I find myself asking, where's the meat?  Where's the mathematics?  Where's something explicit I can point to and either find a logical flaw in or---with greater desire---find some logical satisfaction.  In your 14 papers (289 pages) on {\tt quant-ph}, I count only 58 displayed equations.  I lose orientation very quickly with a nonmathematical structure like that.  Beyond this, in all honesty, for the same reason I find myself losing faith in your program.  The gap between the size of your claims and the number of your equations is enormous.  My intuition is that that is a symptom of a deeper trouble.

A couple of times, for instance, when you spoke of deriving the quantum probability rule, you wrote, ``To find the quantum counterpart to $\cal P$, all we have to do is make room for nontrivial probabilities, and the obvious way to do this is to replace the subsets of a phase space by the subspaces of a vector space, or by the corresponding projectors.''  Well, it is not obvious to me; none of these things are obvious to me.  (I struggle everyday to see where the vector-space structure of quantum mechanics might come from.)  But I could give countless instances beyond that.  One of my greatest complaints about Bohr is how he spoke over and over of how the quantum formalism is forced upon us by complementarity.  But I never could see that he gave the slightest shred of evidence for that.  Instead, I only saw him attempt to demonstrate that his point of view was consistent with the structure of quantum mechanics already given in its abstract axioms.  I am sorry to say, but I fear the same for you.

You may indeed have a novel view that allows you to fulfill all your Vedantic wishes, just as I may have a view that will allow me to fulfill all my {\James}ian (pluralistic) ones, but I would say we both need more.  We both need to know what is it that allows us to have no other view.  That is, to be convincing, we need a hard-core derivation of the quantum formalism.  I see myself and the community I am trying to stir up as moving in that direction.  On that count, however, your papers leave me cold, and I am making the judgment to leave the issue now.

I hope that explains somewhat my obstinacy.  I think it probable that I have hurt your feelings, and I apologize for that; I don't wish any harm upon you.  It really only boils down to, as I said, that life is short, and we all have to make decisions.

\section{21-06-02 \ \ {\it Semi-Classical Mail} \ \ (to K. W. Ford)} \label{Ford3}

I just mailed off my ``Activating Observer'' document to you.  Hopefully it'll reach you in a couple or few days.  You called the method ``classic mail'', but I hope the document is so thoroughly infused with the quantum it would be hard to think of the package as anything but semi-classical \ldots\ at the very least!

\section{21-06-02 \ \ {\it Holevo Paper, Etc.}\ \ \ (to G. {\Plunk})} \label{Plunk14}

Well, I guess I don't have the Holevo paper any more after all.  Here it is, in case you want to dig it up and make copies for us:

\begin{itemize}
\item
A.~S. Holevo, ``Statistical decision theory for quantum systems,'' {\em Journal of Multivariate Analysis}, vol.~3, pp.~337--394, 1973.
\end{itemize}
It contains the most general result.

However, I also remembered a derivation of the result for the more limited case of rank-1 POVMs, as I was tossing and turning in bed.  The paper below lays it out essentially, and is a classic in any case.  Would you mind copying that for us:
\begin{itemize}
\item
L.~P. Hughston, R.~Jozsa, and W.~K. Wootters, ``A complete classification of quantum ensembles having a given density matrix,'' {\em Physics Letters A}, vol.~183, pp.~14--18, 1993.
\end{itemize}
To get the result we need, we just consider the case where the density operator is proportional to the identity.

\section{23-06-02 \ \ {\it Voice Recognition} \ \ (to C. M. {\Caves} \& R. {\Schack})} \label{Caves67} \label{Schack58}

At the moment, I'm taking a few minutes from a frustrating morning to do a little web-surfing for fun.  Reading about new voice-recognition technologies, I ran across the following.  Apropos to our present situation, wouldn't you say?

\bq\noindent
To understand the problem of speech recognition, one must merely compare written and spoken conversations. In a spoken conversation, we have hundreds of inflections that allow us to ``read'' further meaning into the words that are actually spoken. Those inflections are instructed by the intelligences at both ends of the conversation, by culture, by environment, by timing, by context, and a myriad other factors. When we write, it often takes many more words to make sure that the meaning is clear. Many times, a two or three sentence conversation can clear up a month's correspondence over some difficult point.\footnote{\editornote From an anonymous commenter on a {\sl PC Magazine} discussion forum thread: \myurl{http://discuss.pcmag.com/forums/thread/37991036.aspx}.}
\eq

\section{24-06-02 \ \ {\it Points} \ \ (to H. M. Wiseman)} \label{Wiseman5}

You see, I'm catching up on my email finally.  Thanks for the Nash
pointer and the Aussie WWII statistics.

\bhw
4. re.\ the processing power limitations on quantum states. You could
have a look at my paper ``Bayesian feedback versus Markovian feedback
in a two-level atom'' (Phys Rev A 2002, or {\tt quant-ph}) to see
what language I actually did use (I can't remember). But the point is
that the state an experimenter computes on the fly to do feedback
control is not necessarily less pure than the ``true'' state which
the experimenter does not know. But the experimenter still wants to
know how much his state is likely to be ``in error'' from the ``true
state''. I guess so the experimenter armed with this knowledge would
have a better state estimate, by accounting for the error by making
the state more mixed. But in any case, how can you call the ``true
state'' a state of belief or whatever if there is no one who believes
it? It is what the experimenter's state of knowledge should be, given
what they know. That is, it is their state of knowledge, even if they
are too dumb to know it. This is another reason you may use to avoid
the phrase ``state of knowledge'', but the challenge is still how to
describe this ``true state'' in that case, without circumlocuting
intolerably.
\ehw

What you are asking for is the experimenter to compare his pragmatic
gambling commitments under the real world constraints he is living
under at a given moment, to the beliefs or commitments he would
possess under more ideal conditions (for example, if he had limitless
computing power).  Just as there is no ``true'' quantum state,
independent of the agent---i.e., for two agents, there might be two
quantum states---the same may be the case for a single agent.  I like
your examples because they help draw that out.  I can contemplate how
I should bet believing what I believe now and knowing that my feeble
mind cannot analyze the full implications of those beliefs, or I can
contemplate how I should bet believing what I believe now and
imagining a supercomputer that draws out the proper conclusions from
those beliefs.

In all cases, whatever the final products of all calculations and all
approximations one writes down, those are one's gambling commitments.
And {\it with respect to\/} that situation, that is one's ``quantum
state'' for a system.

\section{24-06-02 \ \ {\it The World is Under Construction} \ \ (to H. M. Wiseman)} \label{Wiseman6}

\bhw
Do you believe that events in the world really are random? Or do you
believe they only appear to be random? In the first case, doesn't
that mean that you have to believe in objective probabilities? \ldots

Or are you saying that the real world is unanalysable, unthinkable
even? Everything we say should be couched in terms of gambling
commitments. First, that seems to be a cop-out, giving up on any
understanding of the Universe. Second, it can't explain anything in
the Darwinian way you mentioned, except Dutch-book consistency. It
can't explain why it is ``bad'' to hold a gambling commitment based
on the idea that all world cup soccer balls contain bombs that have a
50\% chance to blow up every time a goal is scored \ldots\ \ You
cannot say anything about animals that would have been likely to have
gone extinct because of poor (but consistent) gambling commitments,
because that is a statement using the concept of objective
probabilities. You cannot {\em\underline{explain}} anything that is
not strictly deterministic without using objective probabilities, it
seems to me.

I trust you understand my motives. I wouldn't bother discussing this
with you if I didn't think your ideas were potentially revelational.
What does not kill you makes you stronger.
\ehw

For a couple of days I have been thinking about how to reply to the
questions in your 6/14/02 email `reality', but this morning I found
myself significantly revising the response I had started to build. In
particular, I decided to hardly reply to your questions at all! This
may be a little bit annoying to you, but I think it will benefit our
longer term discussion.

Of the three options you gave me for answering your questions (I only
quoted the last one of the three above), I suppose if I were forced
to choose one, I would align myself with the one you called a
``cop-out.''  However, from my point of view, the language you use
builds about the ugliest picture it can for where this effort is
going. Indeed, you miss the very point, the very beauty, of the
``cop-out.'' So, what I'd like to do is set that right---right here
and right now---before we go much further. As advertised, in that
way, I will not reply straightforwardly to your questions.

You see, the very starting point for most of my latest thoughts---the
thing I think quantum mechanics gives us the deepest and most
thorough hint of---is that there is no such thing as THE universe in
any completed and waiting-to-be-discovered sense.  The thought I am
\underline{\it testing out} is that the universe as a whole is still
under construction. And when I say this, I am not thinking of just
bits and pieces of it; I am thinking of the whole shebang, all the
way to the roots.  Nothing is completed.  Not just the playhouse Kiki
is building for Emma and Katie, or the evolutionary track of the
human species, but even the ``very laws'' of physics. The idea is
that they too are building up in precisely the way---and ever in the
same danger of falling down as---individual organic species. That is
to say, it's Darwinism all the way down.

So when you ask me if I am ``saying that the real world is
unanalysable, unthinkable even,'' the answer in a way is ``yes.'' For
it is blatantly impossible to analyze to the last detail the
characteristics of a world that has not even been dreamt up (even in
its own mind's eye).

But how can I impress this upon you, or even make it seem reasonable
as a direction for research?  That is a tough call.  For, like with
beer or single-malt Scotch, it is surely an acquired taste that
builds only slowly and with the right company.  Of course, I could
just send you back to my paper \quantph{0204146} and ask you to
take it very seriously.   But this morning it dawned on me to maybe
spend a little time with my scanner to try to IV some thoughts
straight into your bloodstream.

At the moment, I can think of no better introductions to the line of
thought I'd like to expose you to than three articles by Richard
{\Rorty}:  ``A World without Substances or Essences,'' ``Truth without
Correspondence to Reality,'' and ``Thomas Kuhn, Rocks, and the Laws
of Physics.''\footnote{WARNING: \ Just because I say I can think of
no better introductions to these ideas, it does not mean I endorse
every statement in these papers; I may not endorse even half of them.
However, I think these papers go in the right direction, even if they
go too far \ldots\ and even if their arguments are far too weak. But
I choose the papers I do because they are easy reading, with
beautiful writing, and I suspect these thoughts are so foreign to you
that if you can find any sense in {\it some of them}, then it may be
a good start for a dialogue.  Moreover, I continue to stress that the
best justification yet to pursue this direction of thought---and this
is something {\Rorty} does not know---is quantum mechanics itself.  So,
rather than being the final words on things, these are just the
beginning words on things.} (Read them in that order, if you read
them.) All three papers can be found in his collection of essays,
{\sl Philosophy and Social Hope}.  If you absorb these, I think
you'll understand completely what I'm up to, and why I so dislike the
negative connotations you associate with the radical-Bayesian way of
viewing the quantum state.  Of course, it may not turn your head the
way it turns mine, but at least you'll know where I'm coming from,
and from what pool of enthusiasm I derive my strength to eschew the
``golden nuggets'' of {\it mere\/} quantum cosmology, {\it mere\/}
Bohmianism, and {\it mere\/} ``dreams of a final theory.''  The world
as I see it is a much bigger place than those stories can tell.  And
the interpretational issues at the core of quantum mechanics strike
me as our first rigorous indication that there is something more to
this idea than simply the hopes and desires of an enthusiast.

For now, let me give you a flavor of the thoughts in these papers,
and then leave you on your own in the case that you would like to
pursue this further.  The following quotes come from ``Truth without
Correspondence to Reality.''

\bq
In this essay I shall focus on Whitman's phrase `counts \ldots\ for
her justification and success \ldots\ almost entirely upon the
future'. As I see it, the link between Whitmanesque Americanism and
pragmatist philosophy---both classical and `neo-'---is a willingness
to refer all questions of ultimate justification to the future, to
the substance of things hoped for. If there is anything distinctive
about pragmatism it is that it substitutes the notion of a better
human future for the notions of `reality', `reason' and `nature'. One
may say of pragmatism what Novalis said of Romanticism, that it is
`the apotheosis of the future'.

As I read {\Dewey}, what he somewhat awkwardly called `a new metaphysic
of man's relation to nature', was a generalization of the moral of
Darwinian biology. The only justification of a mutation, biological
or cultural, is its contribution to the existence of a more complex
and interesting species somewhere in the future.  Justification is
always justification from the point of view of the survivors, the
victors; there is no point of view more exalted than theirs to
assume. This is the truth in the ideas that might makes right and
that justice is the interest of the stronger. But these ideas are
misleading when they are construed metaphysically, as an assertion
that the present status quo, or the victorious side in some current
war, stand in some privileged relation to the way things really are.
So `metaphysic' was an unfortunate word to use in describing this
generalized Darwinism which is democracy. For that word is associated
with an attempt to replace appearance by reality.

Pragmatists---both classical and `neo-'---do not believe that there
is a way things really are. So they want to replace the
appearance-reality distinction by that between descriptions of the
world and of ourselves which are less useful and those which are more
useful. When the question `useful for what?' is pressed, they have
nothing to say except `useful to create a better future'. When they
are asked, `Better by what criterion?', they have no detailed answer,
any more than the first mammals could specify in what respects they
were better than the dying dinosaurs. Pragmatists can only say
something as vague as: Better in the sense of containing more of what
we consider good and less of what we consider bad. When asked, `And
what exactly do you consider good?', pragmatists can only say, with
Whitman, `variety and freedom', or, with {\Dewey}, `growth'. `Growth
itself,' {\Dewey} said, `is the only moral end.'

They are limited to such fuzzy and unhelpful answers because what
they hope is not that the future will conform to a plan, will fulfil
an immanent teleology, but rather that the future will astonish and
exhilarate. Just as fans of the avant garde go to art galleries
wanting to be astonished rather than hoping to have any particular
expectation fulfilled, so the finite and anthropomorphic deity
celebrated by {\James}, and later by A. N. Whitehead and Charles
Hartshorne, hopes to be surprised and delighted by the latest product
of evolution, both biological and cultural. Asking for pragmatism's
blueprint of the future is like asking Whitman to sketch what lies at
the end of that illimitable democratic vista. The vista, not the
endpoint, matters.

So if Whitman and {\Dewey} have anything interesting in common, it is
their principled and deliberate fuzziness. For principled fuzziness
is the American way of doing what {\Heidegger} called `getting beyond
metaphysics'. As {\Heidegger} uses it, `metaphysics' is the search for
something clear and distinct, something fully present. That means
something that does not trail off into an indefinite future \ldots
\eq
and
\bq
So far I have been trying to give an overview of {\Dewey}'s place in the
intellectual scheme of things by saying something about his relation
to Emerson, Whitman, Kant, Hegel and Marx. Now I want to become a bit
more technical, and to offer an interpretation of the most famous
pragmatist doctrine---the pragmatist theory of truth. I want to show
how this doctrine fits into a more general programme: that of
replacing Greek and Kantian dualisms between permanent structure and
transitory content with the distinction between the past and the
future. I shall try to show how the things which {\James} and {\Dewey} said
about truth were a way of replacing the task of justifying past
custom and tradition by reference to unchanging structure with the
task of replacing an unsatisfactory present with a more satisfactory
future, thus replacing certainty with hope. This replacement would,
they thought, amount to Americanizing philosophy. For they agreed
with Whitman that America is the country which counts for its `reason
and justification' upon the future, and {\it only\/} upon the future.

Truth is what is supposed to distinguish knowledge from well-grounded
opinion---from justified belief. But if the true is, as {\James} said,
`the name of whatever proves itself to be good in the way of belief,
and good, too, for definite, assignable, reasons', then it is not
clear in what respects a true belief is supposed to differ from one
which is merely justified. So pragmatists are often said to confuse
truth, which is absolute and eternal, with justification, which is
transitory because relative to an audience.

Pragmatists have responded to this criticism in two principal ways.
Some, like {\Peirce}, {\James} and {\Putnam}, have said that we can retain an
absolute sense of `true' by identifying it with `justification in the
ideal situation'---the situation which {\Peirce} called `the end of
inquiry'. Others, like {\Dewey} (and, I have argued, {\Davidson}), have
suggested that there is little to be said about truth, and that
philosophers should explicitly and self-consciously {\it confine\/}
themselves to justification, to what {\Dewey} called `warranted
assertibility'.

I prefer the latter strategy. Despite the efforts of {\Putnam} and
Habermas to clarify the notion of `ideal epistemic situation', that
notion seems to me no more useful than that of `correspondence to
reality', or any of the other notions which philosophers have used to
provide an interesting gloss on the word `true'. Furthermore, I think
that any `absoluteness' which is supposedly ensured by appeal to such
notions is equally well ensured if, with {\Davidson}, we insist that
human belief cannot swing free of the nonhuman environment and that,
as {\Davidson} insists, most of our beliefs (most of {\it anybody's\/}
beliefs) must be true. For this insistence gives us everything we
wanted to get from `realism' without invoking the slogan that `the
real and the true are ``independent of our beliefs''\,'---a slogan
which, {\Davidson} rightly says, it is futile either to accept or to
reject.

{\Davidson}'s claim that a truth theory for a natural language is
nothing more or less than an empirical explanation of the causal
relations which hold between features of the environment and the
holding true of sentences, seems to me all the guarantee we need that
we are, always and everywhere, `in touch with the world'. If we have
such a guarantee, then we have all the insurance we need against
`relativism' and `arbitrariness'. For {\Davidson} tells us that we can
never be more arbitrary than the world lets us be. So even if there
is no Way the World Is, even if there is no such thing as `the
intrinsic nature of reality', there are still causal pressures. These
pressures will be described in different ways at different times and
for different purposes, but they are pressures none the less.

The claim that `pragmatism is unable to account for the absoluteness
of truth' confuses two demands: the demand that we explain the
relation between the world and our claims to have true beliefs and
the specifically epistemological demand either for present certainty
or for a path guaranteed to lead to certainty, if only in the
infinitely distant future. The first demand is traditionally met by
saying that our beliefs are made true by the world, and that they
correspond to the way things are. {\Davidson} denies both claims. He and
{\Dewey} agree that we should give up the idea that knowledge is an
attempt to {\it represent\/} reality. Rather, we should view inquiry
as a way of using reality. So the relation between our truth claims
and the rest of the world is causal rather than representational. It
causes us to hold beliefs, and we continue to hold the beliefs which
prove to be reliable guides to getting what we want. Goodman is right
to say that there is no one Way the World Is, and so no one way it is
to be accurately represented. But there are lots of ways to act so as
to realize human hopes of happiness. The attainment of such happiness
is not something distinct from the attainment of justified belief;
rather, the latter is a special case of the former.

Pragmatists realize that this way of thinking about knowledge and
truth makes certainty unlikely. But they think that the quest for
certainty---even as a long-term goal---is an attempt to escape from
the world. So they interpret the usual hostile reactions to their
treatment of truth as an expression of resentment, resentment at
being deprived of something which earlier philosophers had mistakenly
promised. {\Dewey} urges that the quest for certainty be replaced with
the demand for imagination---that philosophy should stop trying to
provide reassurance and instead encourage what Emerson called
`self-reliance'. To encourage self-reliance, in this sense, is to
encourage the willingness to turn one's back both on the past and on
the attempt of `the classical philosophy of Europe' to ground the
past in the eternal. It is to attempt Emersonian self-creation on a
communal scale. To say that one should replace knowledge by hope is
to say much the same thing: that one should stop worrying about
whether what one believes is well grounded and start worrying about
whether one has been imaginative enough to think up interesting
alternatives to one's present beliefs. As West says, `For Emerson,
the goal of activity is not simply domination, but also provocation;
the telos of movement and flux is not solely mastery, but also
stimulation.'
\eq
and
\bq
It may seem strange to say that there is no connection between
justification and truth. This is because we are inclined to say that
truth is the aim of inquiry. But I think we pragmatists must grasp
the nettle and say that this claim is either empty or false. Inquiry
and justification have lots of mutual aims, but they do not have an
overarching aim called truth. Inquiry and justification are
activities we language-users cannot help engaging in; we do not need
a goal called `truth' to help us do so, any more than our digestive
organs need a goal called health to set them to work. Language-users
can no more help justifying their beliefs and desires to one another
than stomachs can help grinding up foodstuff. The agenda for our
digestive organs is set by the particular foodstuffs being processed,
and the agenda for our justifying activity is provided by the diverse
beliefs and desires we encounter in our fellow language-users. There
would only be a `higher' aim of inquiry called `truth' if there were
such a thing as {\it ultimate\/} justification---justification before
God, or before the tribunal of reason, as opposed to any merely
finite human audience.

But, given a Darwinian picture of the world, there can be no such
tribunal. For such a tribunal would have to envisage all the
alternatives to a given belief, and know everything that was relevant
to criticism of every such alternative. Such a tribunal would have to
have what {\Putnam} calls a `God's eye view'---a view which took in not
only every feature of the world as described in a given set of terms,
but that feature under every other possible description as well. For
if it did not, there would remain the possibility that it was as
fallible as the tribunal which sat in judgment on Galileo, a tribunal
which we condemn for having required justification of new beliefs in
old terms. If Darwin is right, we can no more make sense of the idea
of such a tribunal than we can make sense of the idea that biological
evolution has an aim. Biological evolution produces ever new species,
and cultural evolution produces ever new audiences, but there is no
such thing as the species which evolution has in view, nor any such
thing as the `aim of inquiry'.

To sum up, my reply to the claim that pragmatists confuse truth and
justification is to turn this charge against those who make it. They
are the ones who are confused, because they think of truth as
something towards which we are moving, something we get closer to the
more justification we have. By contrast, pragmatists think that there
are a lot of detailed things to be said about justification to any
given audience, but nothing to be said about justification in
general. That is why there is nothing general to be said about the
nature or limits of human knowledge, nor anything to be said about a
connection between justification and truth. There is nothing to be
said on the latter subject not because truth is atemporal and
justification temporal, but because {\it the {\bf only} point in
contrasting the true with the merely justified is to contrast a
possible future with the actual present}.
\eq

I don't have to tell you that I find these ideas tremendously
exciting.  It is not that nature is hidden from us.  It is that it is
not all there yet and never will be; `nature' is being hammered out
as we speak.  And just like with a good democracy, we all have a
nonnegligible input into giving it shape.  That is the idea I am
\underline{\it testing} for consistency and utility.  On the chance
that it will lead somewhere, it seems to me, worth the gamble.

\section{24-06-02 \ \ {\it Five Fuchsian Reasons for Rejecting a Line of Thought} \ \ (to H. M. Wiseman)} \label{Wiseman7}

\bhw
Yes, perhaps. But there is another challenge for you. If I understand
Chris' latest to me, his response is to dismiss as impoverished the HEP's view of the world, rather than to try to incorporate it.
\ehw

I thought you might enjoy this.  It is a list {\Ruediger} distributed during our last days in Brisbane.

\bq
\noindent {\bf Five Fuchsian Reasons for Rejecting a Line of Thought:}
\begin{enumerate}
\item This question is not part of the foundations of quantum mechanics \ldots
\item \ldots\ and has no answer within quantum mechanics.
\item The question is ill-defined anyway \ldots
\item \ldots\ and, if formulated properly, would have a trivial answer.
\item And should it have a nontrivial aspect, working on it is a waste of
   time that really should be spent on proper foundational issues.
\end{enumerate}
\eq

\section{25-06-02 \ \ {\it Of Interest}\ \ \ (to G. {\Plunk})} \label{Plunk15}

This may be of some interest to us; I don't know.
\bq\noindent
\quantph{0206169} [abs, ps, pdf, other]:\\
Title: How to mix a density matrix\\
Authors: Ingemar Bengtsson, {\Asa} Ericsson\\
Comments: 13 pages, 3 figures\smallskip

A given density matrix may be represented in many ways as a mixture of pure states. We show how any density matrix may be realized as a uniform ensemble. It has been conjectured that one may realize all probability distributions that are majorized by the vector of eigenvalues of the density matrix. We show that this is not true, but a marginally weaker statement may still be true. (70kb)
\eq

\section{26-06-02 \ \ {\it The One Boolean Algebra Approach} \ \ (to I. Pitowsky)} \label{Pitowsky1}

I might as well let you know:  I was one of the referees on your proposal to the ISF to study a Bayesian approach to quantum mechanics.  Of course I was flattered by the prominent position you gave some of my papers, but, no matter, it is a good proposal.  I wrote about the best report I could; now, all we can do is sit back and keep our fingers crossed that the other referees agree with me.

Let me make one further comment. I too used to think that the PBA
approach was the way to go if one wanted to build up a theory along
quantum logical lines. But now, I'm not so convinced of it. That is
because I am starting to think that quantum mechanics is more
analogous to the epistemological theory Richard Jeffrey calls
``radical probabilism'' than anything else. From that view, there are
``probabilities all the way down'' with one never getting hold of the
truth values of {\it any\/} propositions. {\Ruediger} {\Schack} and I just
discovered a wealth of material on Jeffrey's webpage \myurl[http://www.princeton.edu/~bayesway]{http://www.princeton.edu/$\sim$bayesway/}.

In any case, I think what this leads to is that we ought to be
focusing much more on characterizing quantum mechanics solely in
terms of the ``logic'' of POVMs than anything else---these being the
structures analogous to what crops up in Jeffrey's ``probability
kinematics.''  Thus, if one is looking to characterize PBAs, the best
task might be to focus on the kinds of PBAs that POVMs generate,
rather than the ones of Kochen and Specker based solely on standard
measurements.  (This may or may not have some connection to what
people are calling ``effect algebras'' but I don't know.)

Beyond that, I am now of the mind that all one really ever needs for
understanding quantum mechanics is a {\it single\/} Boolean algebra
that is kept safely in the background (solely) for reference.  The
rest of the theory (and indeed all real-world measurement) is about
probability kinematics with respect to that algebra.  See Sections
4.2, 6, and 6.1 of my paper \quantph{0205039} for details.

Take care, and I'm so happy you find this approach interesting.

\section{27-06-02 \ \ {\it Compatibility Never Ends} \ \ (to D. Poulin)} \label{Poulin2}

I apologize for taking so long to get back to you.  My email box has
just been running over lately, and on top of that I've been traveling
a lot (just back from Australia, actually).

\bdp
About your approach: as far as I can see, the main difference (aside
from the ``vocabulary'') between your derivation of the BFM
compatibility criterion and theirs is that you are aware that it
follows from {\bf strong consistency}.
\edp

No, I think the difference between us and them runs deeper than that.
Let me try to express the point in a way that maybe I haven't used
before.  Concerning the BFM criterion:  When BFM call it ``necessary
and sufficient'' and when CFS call it ``necessary and sufficient,''
we mean two very different things.  For BFM the ``necessary'' part is
enforced by the supposed existence of a super-observer, Zeno.  That
is, they mean that their criterion is ``necessary'' because they
suppose a super-observer, Zeno, must exist. (See the final section of
their paper.)  What we (i.e., CFS) do instead is define a notion of
``compatibility'' based on the classical notion of ``strong
consistency'' and show that the BFM condition is necessary and
sufficient with respect to that.  Beyond that, however, we leave the
issue open.

That is, with regard to the question, ``MUST the quantum states of
two observers be compatible?,'' we make no statement in the paper.
That is because, from our view---or at least mine in
particular---there is nothing in nature that enforces that states
MUST BE compatible. This is quite important if one wants to get a
consistent Bayesian view off the ground.  (I try to say all this in a
lengthier and maybe more complete way in my samizdat {\sl Quantum
States:\ What the Hell Are They?}, pages 116--118, in a note titled
``\myref{Mermin49}{The Spirit of Gandhi}.''  Actually, I think I say it better there;
so have a look at that.)

Now, if there is nothing to enforce compatibility in the BFM sense,
what would happen if two observers are incompatible with respect to
this criterion?  That's a fair and decent question, and I think where
all the excitement begins.  What the W criterion shows is that the
two observers can just shake off their incompatibility and go forward
(if they are in a congenial mood).  That is, they can come to future
agreement, not simply by pooling their prior beliefs, but by making
an active intervention upon the system their states are about (i.e.,
by making a measurement).  I view this as an extremely interesting
property of the quantum world:  it allows possibilities for going
forward that the classical world does not.

On the other hand, if two observers are not in a congenial mood
(i.e., will not perform a measurement of the W variety), then a
``crisis'' can ensue.  What the full implications of that are, we
don't know yet.  But it may be even more exciting still.

\bdp
Maybe honesty can be related to strong consistency?
\edp

Yes, you are right, and this is an important point.  I discuss this
at length in a note titled ``\myref{Caves54}{The Commitments}'' on pages 133--138, and
again on pages 192--193 in a note titled ``\myref{Schack56}{More Balking}.''  The
upshot is, yes, strong consistency is deeply tied to honesty.  The
next question is, must one enforce honesty in a Bayesian approach to
things?  Your note implies that you tend to think, ``yes.''  The
pages I refer you to show that I tend to think, ``no.''

\bdp
Finally, while you relate the BFM condition to your ES, PP, W, and
W', it is fundamentally different in spirit. This distinction is
related to what we refer to as ``type of knowledge''. In our
vocabulary, the BFM criterion is appropriate to compare knowledge of
the quantum while the other are suited for quantum knowledge (just
like quantum fidelity is). To me, the most convincing argument is the
one of two nonorthogonal pure states: the ``measurement criteria''
will grant that they are compatible while, as we expressed it ``if
two observers claim to have complete knowledge of a system, their
descriptions had better agree completely.'' Comparing states of
knowledge and measurement outcomes predicted from these states are
two different things. We have discussed a similar subtlety by
comparing knowledge described by a density matrix and one described
by a preparation: these are two different questions which deserve
different answers although physically, they cannot be distinguished.
\edp

It's a question of perspective.  If one goes the full Bayesian route
(which I am starting to do), then it is much better to call a quantum
state a ``state of belief or judgment,'' and not a ``state of
knowledge.''  For then, one finds that one is never inclined to say
``their descriptions had better agree completely.''  There is nothing
to enforce that, except possibly Darwinism.  (I'm serious about that
statement.)  That is to say, when the world rears its head, someone
with a firm belief might be wiped out, but in a Bayesian approach
(where probabilistic statements always depend upon {\it subjective\/}
priors), there is nothing to enforce compatibility beforehand.  A
misjudgment can only be declared a misjudgment after the fact, not
before the fact (which in the quantum setting is created by the
process of measurement).

\section{27-06-02 \ \ {\it Probabilism All the Way Up} \ \ (to H. M. Wiseman)} \label{Wiseman8}

\bhw
Second, my wife, Nadine, wants to know why Kiki was building a
playhouse while you indulge yourself in philosophy.
\ehw

I'd like to think we both do what we do best.  But I suspect there's
no philosopher out there who would say I'm doing good philosophy.
(Kiki's artistry on the other hand is always a good hit.)

\bhw
Now, more seriously, you say that my language ``builds about the
ugliest picture it can for where this effort is going''. As I keep
saying, I mean to be provocative. I hope it drives you to new heights
in building a beautiful picture in response. Honestly I do see the
beauty in your program. And I think the more extreme it becomes, the
more beautiful it becomes. I am very interested to see where it ends
up.
\ehw

Thanks for the compliment.  And, indeed, your correspondence does
drive me to new heights (of something).  But now I worry that I
offended you with my phrase ``ugliest picture.''  It probably came
off that way, but it wasn't meant to be an emotional statement or a
point about you personally.  If some emotion did slip into it, it
most likely refers to a conversation I had with Harvey Brown, circa
September 11 of last year.  Harvey kept saying that I wanted to
``doom'' nature to being ``ineffable.''  But that language carries
such a negative connotation.  It carries the idea that there is
something there that we can never, or should never, attempt to speak
of.  So, when you said something similar in print, it gave me the
opportunity to try to reply in print.  (As you know, I try to have my
thoughts recorded so I can refer people to them.  One of the original
ideas was that it would save me time that way; so far, that aspect of
it hasn't worked out.)  Anyway, as I made clear, I want to combat
that with all my strength.  In particular, the way that I am thinking
about it, it is not a bad thing that there are some things beyond
description in nature.  Instead, it is just a statement that there
are more things to come; it is a way of leaving room for something
new.

\bhw
As it happens, I don't have much of a taste for beer or single-malt
Scotch. Also as it happens, I was reading a critique of Richard {\Rorty}
the very morning before I got your letter. Otherwise I never would
have heard of him. It was a 1997 article by Alex Callinicos
``Postmodernism:\ a critical diagnosis''. The most interesting
criticism in there was to say that {\Rorty} ``presumes what he needs to
establish, namely that science and philosophy can be assimilated into
literature. \ldots\ It is \ldots\ very hard in practice when trying
to explain why one theory can be said to be more useful than another
to avoid at least tacitly appealing to the idea that it captures how
things are better than its rival does.''

Perhaps this is one aspect of {\Rorty} you disagree with. But I wonder
about your saying that quantum mechanics is the best justification
for {\Rorty}'s philosophy, as if quantum mechanics is something you
accept to be real, an ``intrinsic nature of reality'', the very idea
of which {\Rorty} explicitly rejects.
\ehw

First, just a technical point.  The philosophies I am most attracted
to at present are those of {\James} and {\Dewey} and what {\James} says about
F.~C.~S. {\Schiller} (but I haven't read {\Schiller} himself yet).  {\Rorty} has
donned himself to be the spokesman of those guys---and I don't mind
that because he writes so nicely---but his writings also have a good
admixture of the postmodernist ideas (of Foucault, Derrida, etc.)
thrown into them to boot.  This business about science not being more
trustworthy or real than literary criticism presently strikes me as
going too far.

But to Callinicos' point (give me the reference, by the way, and I'll
read it)---``It is \ldots\ very hard in practice when trying to
explain why one theory can be said to be more useful than another to
avoid at least tacitly appealing to the idea that it captures how
things are better than its rival does.''---I would just reply,
``Darwinism.''  And then, if that didn't sink in, I'd say,
``Darwinism.''  The point is, from this conception, there is very
little to say beyond that.  Were elephants written into the
blueprints of the universe?  From the Darwinistic conception, they
were not.  Yet, the species fills a niche and has had a stability of
at least a few million years worth.  There is a sense in which an
elephant, like a theory, is a ``true'' component in a description of
the world.  But that ``trueness'' only has a finite lifetime, and is
largely a result of a conspiracy of things beyond its command
(selection pressures).  To put it another way, in contrast to
Callinicos, the elephant doesn't ``capture how things are better than
its rival does'' in any absolute sense---only in a transitory
sense---but that doesn't take away from the functional value of the
elephant today.  So too, I am trying to imagine with theories.

Henry Folse, by the way, wrote me that there is something of a
tradition with this evolutionary idea (beyond {\Rorty}).  So I've got a
big reading list ahead of me:  He tells me Toulmin, Kuhn, Kitcher,
and van Fraassen.

Now, to quantum mechanics.  You find something contradictory about my
liking both quantum mechanics and {\Rorty}.  Here is the way I would put
it.  Presently at least, I am not inclined to accept quantum
mechanics ``to be real, an `intrinsic nature of reality','' except
insofar as, or to the extent that, it is a ``law of thought,'' much
like simple (Bayesian) probability theory.  Instead, I view quantum
mechanics to be the first {\it rigorous\/} hint we have that there
might actually be something to {\James}'s vision.

I've already told you the history of this, haven't I?  I gave a talk
in 1999 at Cambridge on the quantum de Finetti theorem, after which
Matthew Donald came up to me and bellowed, ``You're an American
pragmatist!''  I didn't know what that meant really, but I kept the
thought in the back of my head; I figured one day, I'd figure out
what he meant.  As it goes, that happened on July 21 of last year.  I
came across this book of Martin Gardner's of which one of the
chapters was titled, ``Why I Am Not a Pragmatist.''  (Part of the
story is recorded on page 15 of my little samizdat in a note titled
``\myref{Preskill2}{The Reality of Wives}.''  You might read it for a little laugh.) As
I read it, it was like a flash of enlightenment. For every reason
Gardner gave for not being a pragmatist, I thought about quantum
mechanics and realized that indeed I was one.  Donald was right after
all; I am an American pragmatist.  And my further study of pragmatism
has borne that out to a T.

My point of departure, unlike {\James}'s, was not abstract philosophy.
It was simply trying to make sense of quantum mechanics, where I
think the most reasonable and simplest conclusion one can draw from
the Kochen--Specker results and the Bell inequality violations is, as
Asher Peres says, ``unperformed measurements have no outcomes.'' The
measurement provokes the ``truth value'' into existence; it doesn't
exist beforehand.  Now, go off and read about {\James}'s and {\Dewey}'s
theory of truth and you'll find almost exactly the same idea (just
without the rigor of quantum mechanics).  And similarly with lots of
other pieces of the philosophy.

So, I view quantum mechanics as the hint of something much deeper.
But the full story is not yet told.  That is, quantum mechanics
strikes me as being to our community what the Galapagos Islands were
to Darwin---just a hint of something bigger.

\bhw
You and {\Rorty} I guess would agree that ``dreams of a final theory''
will never be more than dreams. I guess that idea does not worry me
as much as it would some physicists, but it does seem like a defeat.
But perhaps that just says something of my personality. How much of a
role does personality play in one's preferred philosophy?
\ehw

Your question is a good one, and one I worry about a lot.  Where your
knee-jerk reaction is defeat, mine is one of unlimited possibilities
and newfound freedom.  On a similar issue, {\James} put it like this:
\begin{quote}
The history of philosophy is to a great extent that of a certain
clash of human temperaments.  Undignified as such a treatment may
seem to some of my colleagues, I shall have to take account of this
clash and explain a good many of the divergencies of philosophies by
it.  Of whatever temperament a professional philosopher is, he tries,
when philosophizing, to sink the fact of his temperament. Temperament
is no conventionally recognized reason, so he urges impersonal
reasons only for his conclusions.  Yet his temperament really gives
him a stronger bias than any of his more strictly objective premises.
It loads the evidence for him one way or the other, making a more
sentimental or more hard-hearted view of the universe, just as this
fact or that principle would.  He {\it trusts\/} his temperament.
Wanting a universe that suits it, he believes in any representation
of the universe that does suit it. He feels men of opposite temper to
be out of key with the world's character, and in his heart considers
them incompetent and `not in it,' in the philosophic business, even
though they may far excel him in dialectical ability.

Yet in the forum he can make no claim, on the bare ground of his
temperament, to superior discernment or authority.  There arises thus
a certain insincerity in our philosophic discussions:  the potentest
of all our premises is never mentioned.  I am sure it would
contribute to clearness if in these lectures we should break this
rule and mention it, and I accordingly feel free to do so.
\end{quote}

But I think the disparity between our views is in better shape than
that.  I think you're only seeing the program ``physics is the
ability to win a bet'' as a defeat because---even if you don't know
it---you're working within a kind of Kantian mindset.  That the
universe is already formed and there; that there is an ``a priori.''
Anything that can't be said about the universe is then most surely a
loss or limitation.  But, I think once you see that what the
pragmatist is trying to get at is not that, maybe your heart will
change.  Physics as the ability to win a bet will strike you as
something immensely positive.  Physics is like that because reality
is still forming, and the Darwinistic component (along with the
``non-detachedness'' of the observer in quantum mechanics) indicates
that it may be somewhat malleable.  From that point of view, to have
``dreams of a final theory'' is almost like admitting defeat.

But given what you've said, maybe you're already starting to feel
some of this.  And that tickles me immensely.

\bhw
To conclude, I can (or rather could) accept a lot, or even all, of
what you and {\Rorty} are exploring. But I am still not emotionally or
intellectually compelled to do so. And I am really not sure whether I
want to be compelled in one direction, or whether I want to be able
to contain conflicting philosophies. I have this idea that there is
an incoherence at the heart of things.  Irreconcilable levels of
description. Profound truths being the opposite of other profound
truths. Incompleteness theorems. That sort of stuff.
\ehw

Understood.  I know that there's nothing worse than an evangelist
knocking at your door on a Saturday morning.  Feel free to not reply
to this note at all.  In the mean time, I'll try to do my best to do
what I really ought to be doing:  proving theorems, simplifying the
quantum axioms, trying to find real-world physics problems for which
this view is the most powerful way to tackle it, etc., etc.  One
thing physicists never deny is a better method for solving a problem.

\section{28-06-02 \ \ {\it To Believe to Know} \ \ (to D. Poulin)} \label{Poulin3}

\bdp
Otherwise, you get incompatible statements with which Bayesian theory
cannot deal (like computing $p(x|y)$ when $y$ is assigned probability
0).
\edp

It's starting to sound like we mean two different things by Bayesian
theory.  See, \myurl[http://www.princeton.edu/~bayesway/KC.tex.pdf]{http://www. princeton.edu/$\sim$bayesway/KC.tex.pdf}, slide \#12 (on page
13 actually).  More seriously, see: \myurl{http://cepa.newschool.edu/het/essays/uncert/subjective.htm},
especially the parts on de Finetti and Ramsey.

\bdp
I think that most of the discussion is based around this distinction:
``state of knowledge'' and ``state of belief''. I would say that
states of knowledge must be (BFM) compatible while states of belief
can be incompatible. How are these two things defined?  Well I would
say that a state of knowledge is built with the help of an initial
state on which everybody agrees, the postulates of quantum mechanics,
and ``public actions''.
\edp

I will agree to your definition of ``state of knowledge.''  But,
backtracking from that, an initial state upon which everyone agrees?
If one is taking a subjectivist approach (or what I had been calling
a ``Bayesian approach'') to interpreting the quantum state, there is
nothing in nature to enforce an initial prior agreement.  God does
not come down from on high and say to all the agents (i.e., all the
observers), ``Your starting point shall be the quantum state
$\Psi$.''  Everyone is left to fend for himself.

That is to say, in the language of the second webpage I sent you to
above, in a world where the quantum state is not presupposed to be
objective---as had been {\Mermin}'s goal when he started this
exercise---there is nothing to enforce the ``Harsanyi doctrine.''

\section{28-06-02 \ \ {\it The Harsanyi Doctrine} \ \ (to C. M. {\Caves} \& R. {\Schack})} \label{Caves68} \label{Schack59}

I just happened to run across the following discussion:

\bq
Finally we should mention that one aspect of Keynes's (1921)
propositions has re-emerged in modern economics via the so-called
``Harsanyi Doctrine''---also known as the ``common prior'' assumption
(e.g.\ Harsanyi, 1968). Effectively, this states that if agents all
have the same knowledge, then they ought to have the same subjective
probability assignments. This assertion, of course, is nowhere
implied in subjective probability theory of either the Ramsey / de
Finetti or intuitionist camps. The Harsanyi doctrine is largely an
outcome of information theory and lies in the background of rational
expectations theory---both of which have a rather ambiguous
relationship with uncertainty theory anyway. For obvious reasons,
information theory cannot embrace subjective probability too closely:
its entire purpose is, after all, to set out [an] objective,
deterministic relationship between ``information'' or ``knowledge''
and agents' choices. This makes it necessary to filter out the
personal peculiarities which are permitted in subjective probability
theory.\footnote{\editornote From \myurl{http://cruel.org/econthought/essays/uncert/subjective.html}.}
\eq

\section{28-06-02 \ \ {\it Who Would Have Thunk?}\ \ \ (to N. D. {\Mermin})} \label{Mermin71}

Who would have guessed that all we were debating in our ill-fated
correspondence on the BFM criterion was simply the validity (or maybe
the jurisdiction) of the ``Harsanyi Doctrine''?  Not me!  But that's
what I found out this morning quite by accident.

If you want to read the story leading up to this, download my
mini-samizdat ``Quantum States:\ What the Hell Are They?'' and look at
the notes to Poulin on pages 215 and 219.  [See the 27-06-02 note
  ``\myref{Poulin2}{Compatibility Never Ends}'' and the 28-06-02 note
  ``\myref{Poulin3}{To Believe to Know}'' to D. Poulin.]  I now find
this quite interesting actually.  Harsanyi won the Nobel prize with
Nash, and there seems to be a bit of literature on his doctrine.

By the way, you never responded to (the main content of) my December
10 note ``\myref{Mermin49}{The Spirit of Gandhi}.''  Here's your
chance before I close the doors on this chapter of my life.  (I.e.,
I'm going to bind up and close this mini-samizdat by the end of the
day.  The next one is going to be titled, ``Darwinism All the Way
Down.'')

\section{29-06-02 \ \ {\it Old Boole} \ \ (to I. Pitowsky)} \label{Pitowsky2}

\bip
Anyway, the Bayesian
Quantum Probability paper is much better than  my research proposal (which
was written with some haste, but this is a different story). I'll send you
a copy when it's in a reasonable shape. I use projection operators, but have no prejudice against POVM's, only
I'm not clear how to use the gambling language with them (we talked a
bit about this problem in the {\Vaxjo} conference).
\eip

You play exactly the same game you did in your proposal, except now that a complete set of mutually exclusive events are identified with the elements in a POVM.  I.e., {\it stop\/} thinking of POVM elements as non-mutually-exclusive events.  (Well, I don't know that you haven't already abandoned that line, but much of the community tends to call them ``unsharp'' measurements and such.  And that just creates the wrong imagery.  They are sharp as anything else from this view.)  These elements correspond to measurement ``clicks'' and are thus mutually exclusive.  From my standpoint, the ``clicks'' are the closest we can come to grasping a ``truth value.''  And thus that is the formal role they take on.

\bip
The slogan ``probability all the way down'' is great. One should even
make a sharp distinction between ``probability equals unity'' and truth.
\eip

Yes.  And that is what the better part of my samizdat ``Quantum States:\ What the Hell Are They?'' posted at my website (link below) is about.  You might enjoy thumbing through it.  In my own view, it leads one along the lines of something like a Jamesian theory of truth \ldots\ but hopefully with the opportunity to make it more rigorous now.

\section{29-06-02 \ \ {\it Incompatible Beginnings} \ \ (to D. Poulin)} \label{Poulin4}

\bdp
Maybe requiring a single description on which everyone agrees is a
bit too strong, but something has to be imposed.
\edp

The whole point of the research program I am building up, and the
point of most of the 220 pages in {\sl Quantum States:\ What the Hell
Are They?}, and much of the point of my paper \quantph{0205039},
is that I don't buy that.  Remember the line in the Pink Floyd song,
``Mother's going to put all her fears into you.''?  The point is,
none of us are born with ``rational'' views.  We start out in life
with whatever our community pumps into us.  The mark of
``rationality'' instead is how we change our views in the light of
evidence, and how we gamble based on what we believe.  That is what
Dutch-book coherence is about, for instance.

So, when I say ``agents are left to fend for themselves,'' I mean it.
Every now and then agents start out compatible in one sense or
another and then they stay that way, indeed \ldots\ perhaps with ever
more similarities between their distributions as data flows in.  But
that is not the norm.  Instead most of the time, there is some
incompatibility in our world views lurking in the background. We can
live with that as long as nothing comes to light to challenge our
views.  But when the opportunity arises, there will be a ``crisis,''
as {\Ruediger} {\Schack} likes to say.

At that point, as you say, Bayesian theory can't handle things.  But
the view I am shooting for is that Darwinism can.  That is, if you
have a firm belief (i.e., a probability 1 ascription) for something,
by Dutch-book rules you are willing to bet your whole bank account on
the event.  If the event doesn't occur, then you have lost your whole
bank account, and in a way your life.

The program, as I see it, is to see how rigorous and fruitful on can
make that line of thought.

\section{01-07-02 \ \ {\it Nose Jobs}\ \ \ (to T. Rudolph)} \label{Rudolph8}

\btr
Yah -- finished Rorty, I'm actually rereading a bit of the stuff near the
beginning and the section on Weinberg/Kuhn etc. I'm not entirely satisfied with the way he answered Weinberg \ldots\ not
that Weinberg's statements were that great, which is the reason I
think Rorty could have been stronger.
\etr
Agreed.  (Clearly there was some emotion behind it that got out of hand.)  But funny, on my own second reading, I liked the reply better.

I posted some stuff in that connection in the form of some letters to Howard Wiseman.  You might be interested.  They can be found near the end of my ``Quantum States:\ What the Hell Are They?''\ on my webpage.  [See 24-06-02 note ``\myref{Wiseman6}{The World is Under Construction}'' and 27-06-02 note ``\myref{Wiseman8}{Probabilism All the Way Up}'' to H. M. Wiseman.]

One thing I need to understand better (and not just trust Rorty on) is Donald Davidson's theory of truth.  But my reading list is a mile high.




\section{05-07-02 \ \ {\it The Physics of Floyd} \ \ (to M. J. Donald)} \label{Donald4}

I said I would be back; I apologize for the delay.

\bmjd
Your know-nothing ism, like de Finetti's irrationalism (Gillies,
``Philosophical Theories of Probability'', page 86), have the dangers
of Bohr's writings on which I would agree with Beller (Physics Today,
September 1998, pages 29--34).  In particular, by leaving far too
much in vagueness, incoherence, and pious hope, you give the
religiously-minded the official endorsement of the physics
establishment that they may believe anything they want, instead of,
by example, instructing them that they can believe anything they want
as long as it is rational, coherent, tentative, revisable, and
compatible with the evidence (and therefore contrary to naive
expectations, because if quantum theory, or indeed science in
general, tells us anything it is that the world is not how we would
have imagined it before we investigated); and they accept that they
may be completely wrong.
\emjd

In saying this, you demonstrate that you have read Gillies page 86.
I'm glad I know that.  But, as I tried to convey before, I think you
show a deep misconstrual of de Finetti's program (or radical
subjectivism, or radical probabilism, or whatever you want to call
it, even irrationalism).  I never tell my girls that they can believe
anything they want.  I always try to be as impartial as possible, but
in the end, I'm sure all I really teach them is to believe what {\em
I\/} believe and what the community around me believes.  That's where
they take their start in life.

And so it is with all of us.  Pink Floyd recorded a great lesson with
the words, ``Mother's going to put all her fears into you.''  De
Finetti's theory of probability is simply a stark recognition of this
fact and a stark recognition that there is no getting around it.  The
point is, none of us are born with ``rational'' views, and a
particular opinion at any one moment is {\it neither\/} rational or
irrational.  The mark of ``rationality'' is instead in how we change
our views in the light of evidence, and how we gamble based on what
we believe.  This is what ``coherence'' and Bayesian updating are
about.  When Mother has put such a bad opinion into us that
rationality cannot realign it enough for us to survive, then we won't
survive, and that opinion will not propagate.

It is not ``anything goes'' here any more than it is for the species
in the animal kingdom (Darwin).  If you see in my writings a hint
that anything goes, then it seems to me you are fighting a battle
with someone else and hoping I will fill his shoes.  The mathematical
side of ``radical subjectivism'' is an attempt to quantify the ideas
above and to show their consistency and utility.  It is what the 586
pages of Bernardo and Smith's book {\sl Bayesian Theory\/} is about;
example after example that rationality is not in the prior, but in
the conditionalization and in the coherence.

Where you think you need a law-driven reality to explain the success
of our scientific activity, I think all I need is (something along
the lines of) de Finetti and Darwinism.  The most important point of
distinction between our views is in how we are willing to mark the
``badness'' of an opinion/belief/judgment.  I say it is only possible
to mark a judgment as a misjudgment---in any objective way---after
the fact, after the event to which it was directed has either
occurred or not occurred.  Whereas you want to think there is
something that defines the badness or goodness of a judgment before
the event.

This is the great gulf between us.  I say, ``unperformed measurements
have no outcomes.''  ``The universe is in the making.''  You say, if
I'd just step back far enough---until I'm outside of the universe,
actually---I'd see that the universe ``is.''  It's already there.
It's already complete and waiting for me to draw a precise picture of
its smooth surface.  And in the dimples of that smooth surface, I
would see which of my actions are right and which of my actions are
wrong long before they take place.  Indeed, in that timeless ground
state, I would see that everything I'm talking about is folly.

There's no denying that there is something about the former picture
that piques the lascivious side of my character.  But the problem is
deeper and more openly communicable than that.  A couple of times in
your notes, you have said something to the effect,

\bmjd
I do hold as an article of faith that there is a true view, although
I accept that we may not ever know it, let alone ever know that we
know it.
\emjd

If we may never know it, or even know that we know it, then I cannot
see that it plays any normative role whatever in helping us get along
in life.  Its explanation of the phenomena about us is little more
than a hollow promise.  The argument is a minor modification of J. S.
Mill's argument against substance:
\bq
If there be such a {\it substratum}, suppose it at this instant
miraculously annihilated, and let the sensations continue to occur in
the same order, and how would the {\it substratum\/} be missed? By
what signs should we be able to discover that its existence had
terminated? Should we not have as much reason to believe that it
still existed as we now have?  And if we should not then be warranted
in believing it, how can we be so now?
\eq
Just change ``substratum'' to ``immutable law.''

There, I've said my piece.  And I've probably said too much in too
condescending a way:  For that I apologize (almost).  But I wanted to
make a point and make it forcefully.  My purpose is not to give fuel
to the religious fire-burners.  It is to change our conception of how
the law-like stability we see around us arises, in a way that does
justice to what I view as the great lesson of quantum mechanics.
Quantum mechanics, that is, as viewed from inside the world, where we
actually live, instead of from some Everettic fantasyland (where 1957
stands out as the most atypical year ever, the year man first saw the
universe in its entirety and confirmed what the rationalists had
suspected for so long: its image is so bleak as to fit within the
mind of a single man).  The mark of pious hope is in the belief of
miracles:  I don't believe in miracles or miraculous years.

Is all that I've said vague?  Of course it is.  But that is why this
is a research program rather than a completed product.  There is so
immensely much to do.  {\Carl} {\Caves} instilled in me long ago that the
good physicist is the one who poses a problem he can hope to just
solve in the span of his lifetime.  That's the way I see this
program.  With sustained determination, some creativity, and the help
of a lot of friends, I see it as actually going through (and carrying
us to a vista that is almost unimaginable now).

Some of the best help we could probably get would be the mind of
Matthew Donald.  But that might require a midlife crisis!  (And I
would hate to think of what that might do to your wife and children.)

Now let me address some of your other comments.

\bmjd
Actually, it was great fun writing:  I've always wanted to be allowed
to say ``Nope'', but I've never got away with it before.
\emjd

Are you saying that you find it easier to let your hair down around
me than with your Cambridgean/Oxfordian contemporaries?  If so, then
I will reinstate the image of the romantic and mustachioed
revolutionary just for you.  What the whole movement needs is a
little letting down of the hair; glad to be of service.

\bmjd
page 9: ``the final state itself for B cannot be viewed as more than
a reflection of some tricky combination of one's initial information
and the knowledge gained through the measurement.''  To what extent
then is the ``final state for B'' actually ``for B'' and to what
extent is it ``for'' the observer?

similarly, on page 12, in the last paragraph of section 3, can we say
that the system ``has'' the state $|\psi\rangle$, or should we say
that for Alice the system has the state $|\psi\rangle$?
\emjd

A later point in your note indicates that you know I would say ``for
the observer'' here.  Indeed that is better language.  Sometimes it's
just hard to make the right rhetorical judgment; sorry if I caused
you confusion.  My intention instead was to put one foot into the
enemy camp, and try to fight on their terms.  That is to say, my
intention was to draw an absurdity from associating a quantum state
with a system rather than with an observer's judgment.

\bmjd
footnote 12: which view gives the most coherent, consistent,
plausible explanation of the most phenomena? which view is true?????
\emjd

I exasperate you, don't I?  The point is, I have my reasons for
refraining from that word, and I tried to explain some of them above.

\bmjd
page 22: I'm not sure what to conclude from section 4.2.  This is all
that comes to mind:

Somewhere in the classical division of the BIPM is a die testing
machine.  It is used to throw a die sufficiently many times to
provide a reasonable estimate of the probability of each face.  It
shares with its quantum analogue the features that in general the
probability of any face is never zero, and that many trials are
needed to provide useful statistics.
\emjd

If it were a good classical die testing machine---that is, if it
really did the same thing each and every time---then the toss would
always come out exactly the same.  That is to say, there are no good
classical die testing machines.

\bmjd
page 24: Are POVM's really ``the structure of our potential
interventions''? All of our potential interventions?   Our most
important interventions?  The structure of a two-year old child's
interventions? Or are they just one particular ``very carefully
contrived way by which we can sometimes manage to see fairly directly
into part of [reality's] deep structure''.
\emjd

The way I would say it now is this.  Our interventions are whatever
they are.  Even a two-year old's.  The concept of our interventions
having a structure is something we lay on top of them.  To that
extent, the use of POVMs as a formal descriptive device is a
normative one.  I would say the same thing for our general use of
probability theory.  That is to say, we don't use the probability
calculus in making all our decisions, but we ought to.  Bernardo and
Smith put it this way:

\bq
What is the nature and scope of Bayesian Statistics within this
spectrum of activity?

      Bayesian Statistics offers a rationalist theory of personalistic
   beliefs in contexts of uncertainty, with the central aim of
   characterising how an individual should act in order to avoid certain
   kinds of undesirable behavioural inconsistencies.  The theory
   establishes that expected utility maximization provides the basis for
   rational decision making and that Bayes' theorem provides the key to
   the ways in which beliefs should fit together in the light of
   changing evidence.  The goal, in effect, is to establish rules and
   procedures for individuals concerned with disciplined uncertainty
   accounting.  The theory is not descriptive, in the sense of claiming
   to model actual behaviour.  Rather, it is prescriptive, in the sense
   of saying ``if you wish to avoid the possibility of these undesirable
   consequences you must act in the following way.''
\eq

Similarly I would say of POVMs.  If you want to do your best
reasoning concerning the consequences of what you've just done (or
what you are about to do) then you should use the calculus of quantum
mechanics.  And that says lay a template down over what you've done
(or what you are about to do) that happens to have the shape of a
POVM.

\bmjd
page 34: ``Quantum measurement is nothing more, and nothing less,
than a refinement and a readjustment of one's initial state of
belief.'' But it is something more.  It is also the result of a
physical process.
\emjd

No, I would say, the result of the physical process is something
else.  It is a creation of sorts.  The structure of quantum
mechanics, I would say, has something to do with what I can say about
that creation \ldots\ thus my refinement and readjustment.  But
concerning the real stuff of the world, the theory hardly goes there.
(And that's what I try to say, for better or for worse, in my \quantph{0204146}.)

\bmjd
The idea in many-worlds interpretations is to drop von Neumann's
process 1 (collapse of the wavefunction).  In section 6, it looks as
if you're intending to drop process 2 (unitary time propagation).
But, the $A_{di}$ in (63), or the $\Pi_d$ in (65), should be derived
from the physics of the measurement; no doubt using environmental
decoherence at some point.
\emjd

Yes, I do intend to drop process 2 as being something more
fundamental than process 1.  I keep looking for ways to do that that
I find completely convincing.  From my view, process 2 is just a
special case of process 1, where the POVM is a one-element set.  The
word ``should,'' you should realize, is little more than a cultural
statement \ldots\ from a cult that I do not buy into.  The phrase
``no doubt,'' on the other hand is a statement of faith.  The reasons
for your use of that phrase I will leave as an exercise in
self-reflection.

\bmjd
Probability theory is only a law of thought in as far as it describes
how one should deal with new information. (vN process 1).  But
physics should also be able to describe how one acquires new
information. (vN process 2).
\emjd

Ditto.

\bmjd
It is the physical laws and initial conditions which make an agent's
beliefs about the forms and the probabilities of possible future
experiences either right or wrong given his past experiences.
\emjd

I told you that was the gulf between us.

\bmjd
The rational assignment of a POVM to a given experimental device is
not a matter of free choice.
\emjd

I claim it is; at least initially so.  The rationality lies purely in
coherence and updating, not in particular assignments.  But clearly
it's going to take a lot of inculcation into my culture to convince
you of that.  I know it won't help, but I'll throw this out for
evidence's sake that I take your skepticism seriously:  Just as there
is a de Finetti theorem to make sense of ``unknown states,'' there is
a de Finetti theorem to make sense of ``unknown measurements.'' (This
is a result {\Schack}, Scudo, and I will be posting soon.  [See ``A de Finetti Representation Theorem for Quantum Process Tomography,'' Phys.\ Rev.\ A\ {\bf 69}, 062305 (2004), \quantph{0307198}.])

\bmjd
The strict quantum Bayesian of the Fuchsian persuasion, however,
faces the quagmire that physical laws are as subjective as
probabilities, so that there is nowhere to start (not even with
nothing!).
\emjd

Darwinism.  Darwinism.  Darwinism.  Darwinism.  It's presently the
best nothing I've ever known.

\bmjd
page 51: Hilbert space dimension, of the small integer type you are
referring to, is itself only an effective concept, which depends on
not looking too closely at the systems studied.  For example, D=2
arises when you consider a photon which may hit one of two detectors,
and ignore the full (infinite-dimensional) photon-detector space.
Therefore D also is state and agent and context dependent and
subjective.
\emjd

Absolutely.  But it is of the harmless type that fixing a classical
phase space is:  and that's the point.  Any classical (real-world)
pendulum has more than one degree of freedom, but once the context is
set, once the approximation is made, the human interests that set
that context and made that approximation can be safely forgotten.
This is what I envision for Hilbert-space dimension (but not the
quantum state).\medskip

\noindent ------------ \medskip

Let me thank you also for the other uplifting note you sent me.
Despite my own combative reactions (as witnessed above), I do find my
conversations with you valuable.  And in ways, you are partially
responsible for the creation of this monster.  (In fact, I recently
set to paper the story about your involvement.  In case you're
interested, I'll narrow the search for you: it's somewhere between
page 202 and 221 of the mini-samizdat ``Quantum States:\ What the Hell
are They?''\ posted on my webpage.  [See 01-06-02 note
  ``\myref{Grangier7}{High Dispute}'' to P. Grangier.]  If you wade
through that, you'll get the full impact of the story.  By the way,
I'll be shutting the doors to that samizdat soon.  I'll equip it with
an index, etc., like I did with the last one, but I won't post it on
{\tt quant-ph} this time, only my webpage.  After that, I intend to
start constructing a new one titled ``Darwinism All the Way Down.''
The present letter to you will be my first entry.)

Finally, let me end with a long quote that I pulled from Richard
Jeffrey's paper, ``Reading Probabilismo.''  You inspired me to copy
it into my machine.

\bq
There is a most instructive contrast between de Finetti's
``irrationalist'' probabilism and Carnap's rationalist positivism.

For Carnap, probabilism was a fallback position from what one might
call dogmatic rationalism, i.e., the view that scientific forecasts
and universal hypotheses ought to be logically deducible from {\it
Protokolls\"atze}. Probability theory was to replace deductive logic;
dogmatic rationalism would give way to probabilistic rationalism.
Rationalism itself was seen as essential to empiricism. The
probability $c(h, e)$ of a scientific forecast or hypothesis $h$
relative to a sentence $e$ that reports the totality of one's
empirical evidence needed to be independent of who one might be,
provided only that one were ideally rational: qua scientist, an
ideally rational being would be individuated only by $e$, the report
of that being's total individual experience to date. If $c$ as well
as $e$ could vary from scientist to scientist then $c(h, e)$ would
represent a subjective judgment. The only scientific basis on which
the $c(h, e)$ values might vary must appear in the second
argument-place, not in the function $c$ itself, i.e., such a basis
must be empirical. Logical empiricism was wedded to that sort of
rationalism. {\it That\/} sort of rationalism---not the rationalism
of the bogeyman who thinks he can predict the future by pure reason,
or can prove that space must be euclidian, a priori.

De Finetti's probabilism was ``irrationalist'' in denying the
possibility of an intelligible split between reason, represented by
the conditional probability function $c$, and experience, represented
by the total observation-report $e$. ``Anti-rationalist'' would have
been a less provocative term. But if de Finetti's probabilism was
anti-rationalist it was anti-empiricist as well on the same showing,
and ``anti-empiricist'' has the same ring of madness that
``irrationalist'' has. The suggestion is that an anti-empiricist
opposes observation and experiment, just as an irrationalist is
against being reasonable and thinking things through. What's
rejected, though, is neither experience nor reason, but the dichotomy
underlying logical empiricism, according to which the two can be
cleanly separated---say, into $e$ and $c$. de Finetti rejects as
untenable the basic dogma of logical empiricism: the dualism between
reason and experience.
\eq

\section{05-07-02 \ \ {\it Coherence Everlasting} \ \ (to H. M. Wiseman)} \label{Wiseman9}

\bhw
The comment by van Enk and Fuchs on Ref.\ [1] can be understood as an
explicit calculation showing (part of) how a laser beam, without an
absolute phase, can function as a clock; how the phase information can
be distributed and how there is no harm in regarding the phase as
real.  This is essentially the same point originally made by Molmer,
that laser phase is a ``convenient fiction''.
\ehw

No.  Our point was not even remotely the same as Klaus's.  Klaus was concerned with an intracavity mode, where I would say there is no reason whatsoever for preferring one decomposition of the density operator over another.  This contrasts with the multi-mode system we were concerned with.  With respect to that {\it tensor-product structure}, there is a preferred decomposition, and it involves coherent states.  If the modes are believed to be exchangeable, as they standardly are with continuous-wave laser light, then the de Finetti theorem dictates that there is a unique decomposition of the big multi-mode density operator into a probabilistic mixture of pure states, each of which preserves the exchangeability property.  Yes, it is a convenient fiction, just like when anytime one has a density-operator description of a system, it is PURELY a fiction to furthermore think of that state as composed of a deeper (more real) set of pure states.

The point of the expansions of a density operator---as I tried to make clear in my lecture at Griffith---is not to talk about ``what is there,'' but rather to tell something about the potential measurements one can perform on a system, and what those measurements will teach us.  I.e., what predictive value they will give.  Measurements correspond to decompositions.  Full stop.  There is nothing deeper than that.  What is significant about the de Finetti expansion of a laser beam is that it tells you that IF you can measure the phase of one mode, THEN the remainder of the modes will be describable via a simple, pure product state.  If one were to think of the multi-mode case as a mixture due to an original mixture of intracavity photon-number states, one would have to work like hell to see that.  If that is not a reason to prefer one valid decomposition over another, I don't know what is.  One thing is for sure though:  It does not endorse the {\it reality\/} of any decomposition, number-state, coherent-state, or otherwise.

\section{07-07-02 \ \ {\it And Continuing Consternation}\ \ \ (to H. M. Wiseman)} \label{Wiseman10}

\bhw
Actually I don't think that is so important, if you mean the multimode
versus single mode thing. I hope you don't mean they should have
written down a statistical mixture, because then I would really
disagree with you.
\ehw

Yes, that is exactly what I meant.  Sargent, Scully, and Lamb, in their Chapter 17, do a perfectly fine calculation, and the result is a mixed-state density operator.  It represents the laser operator's state of knowledge about his laser field, given some standardly agreed upon information.  To think there is a purer state of affairs---i.e., to think that there is an ``unknown state'' over and above his mundane description---is to go against the grain of everything I (and {\Caves} and {\Schack}, and Peres, etc.) have been striving for in the interpretational game.  Those pure states that lie underneath a density operator are pure artifice from my point of view.

\bhw
Well you certainly do not make that clear in your paper. On my
reading, you are accepting Rudolph and Sanders' idea that there is such
a thing as a real coherent state. That before you measure it, the
laser is in a statistical mixture, and after you measure it, it isn't.
I am saying there is no need to ever write down a statistical mixture
for the laser. The laser can be its own clock, and hence (ignoring
phase diffusion) it is in a coherent state.
\ehw

Then, you carry a lot of baggage into your reading.  (Perhaps as I carried a lot of baggage into my reading of your comment.)  A few notes ago, you wrote the following:

\bhw
After looking at it with {\Carl} and {\Ruediger}, I take back what I said to
Chris about the paper having sneaked in objective probabilities --- I
was misled by reading ``probability'' as ``objective probability''.
\ehw

I would suggest you are doing something similar with my paper with Steven.  The idea that ``that before you measure it, the laser is in a statistical mixture, and after you measure it, it isn't'' is now a foreign concept to me.  And it was already a foreign concept by the time Steven and I wrote those two papers.

I just re-read both papers, the PRL and the QIC version, and though they are written in a more neutral way than you usually see from me---Steven did the bulk of the writing---I still find none of the thoughts you thought you saw there.

\bhw
An electronic oscillator is no more real than an optical oscillator.
\ehw

Agreed.  The issue in my mind is one of a chain of inference.  Given the previous ascription of a multi-mode mixed state, what pinging of the quantum world (i.e., what measurements) will allow one to refine that ascription to a pure state for some of the remaining (unmeasured) modes?  We think we know how to ping the electronic oscillator, and we draw a chain of inference.  That's all.  But without the electronic oscillator as an intermediary, we are stuck with the mixed state ascription.  (Notice, I keep using the word ascription, not description.)

\section{08-07-02 \ \ {\it Short First Reply}\ \ \ (to H. M. Wiseman)} \label{Wiseman11}

\bhw
Hey, I think I do understand the quantum de Finetti theorem. If you
have an infinite sequence of systems you are happy to swap around in
any way then it is always possible to represent your state of
knowledge as being AS IF there is an unknown quantum state common to
all of them. It's not that complicated. That's not to say of course
that I could reproduce your proof, so to avoid being offended, I could
assume that being able to do that is what you mean by ``strong
understanding''. On the other hand, your statement could hardly have
been said unless you thought I had demonstrated in the past a weak
understanding of the issues (e.g.\ my not understanding what is so
terrible about the idea of an unknown preparation procedure) then my
defence would be that I believe I am in quite good company.
\ehw

Yes, it is the latter (i.e., being in ``quite good company'') that I meant.  So, I really only mean the points made in the introduction of that paper.

\bhw
For an optical experiment the laser is the clock, and therefore it is
in a pure coherent state.
\ehw

Our agreement lasts for the first part of this sentence.  Where we part, is in the ``therefore.''

\bhw
All quantum states are pure artifice, full stop. They are in the
observer's head --- what can be more artificial. I thought that's what
you taught me. If I say a laser is in a pure coherent state, how can
you say I'm wrong?
\ehw

True enough.  My point is only one about consistency.  If one {\it accepts\/} the standard derivation, say of Sargent, Scully, and Lamb, Chap.\ 17, then one has NO RIGHT TO GO FURTHER and say, ``well, there's really a coherent state there after all, and I just don't know it.''  One is simply not being internally consistent.  If, on the other hand, one wants to believe that the field should be described by a pure coherent state, that is fine by me, but then one should know what that state is (otherwise, to my post-1999 ears, they are speaking nonsense).

\bhw
What other possible meaning can be taken from this quote:
``This measurement would create an optical coherent state from a
standard laser source for the first time.''
\ehw
\bv
{\bf create} (kr\^e-\^at') verb, transitive\\
created, creating, creates\\
1.	To cause to exist; bring into being.\\
2.	To give rise to; produce.\\
3.	To invest with an office or title; appoint.\\
4.	To produce through artistic or imaginative effort.
\ev

\section{13-07-02 \ \ {\it The Fabric of Sexuality} \ \ (to A. Plotnitsky)} \label{Plotnitsky8}

\barkp
As I was reading your {\Vaxjo} piece, indeed already as I started to
read it at Cornell, one impression it made on me was a bit peculiar.  The very concept and terminology of ``quantum states'' appeared to me (in part because of an appeal to a ``state'') to be a remnant of a more conventional ontological view, and your argument seemed to have difficulty to adjust the concept to the view advanced by you in your article and elsewhere.  Almost immediately, however, I recalled the discussion in ``Quantum States,'' where you speak, I think rightly, of the quantum state as a state of belief rather than knowledge, although ``state of expectation'' or ``probabilistic estimates'' may even be more precise.  (I agree that ``knowledge'' is better suited to refer to the outcomes of measurements already performed, even though, at least in my view, both knowledge and expectation concerns only what is manifest in measuring instruments, never anything pertaining to quantum objects themselves.)
\earkp

Yes, I see your point and I agree with it.  It is a remnant.  But so far, I've had no stroke of inspiration for a better term.

\barkp
It seems to me, however, that, rather than of QUANTUM STATES, it
would be more accurate to speak of the STATE VECTORS (the ket-vectors
defined by the psi-function in a corresponding Hilbert space) as
representing or reflecting the state of expectation.  I do not think one could speak of states vectors themselves as states of expectation, but only as a mathematical device representing information related to our expectations concerning the outcomes of certain possible experiments or events.  One might in fact call them ``EXPECTATION VECTORS.''
\earkp

And actually this one doesn't do it for me either.  It's hard to pinpoint why, but I guess mostly I would be searching for a more philosophically evocative term.  Secondary to that, you use the word ``vector,'' and it is crucial to my point of view that there is nothing more fundamental about a pure state (i.e., a state vector) than a general density operator.  Thus, perhaps, ``expectation operator'' might be more to my taste, but along the point I just made, I really would like something more evocative than that.

Right now, ``belief'' still does the most for me---because I am so taken with William James's essay ``The Sentiment of Rationality''---quantum states are, in part, those things that he talks about there---but I agree with you that the terminology ought to go, once something better comes along.

\barkp
With these qualifications in mind, however, ``quantum state'' may be a
possible conception and perhaps even a necessary one (especially in the EPR-type situations), insofar as it indicates an independent existence of ``quantum objects.''  That includes their capacity to enter interactions with our measuring instruments or other classical and quantum entities with which they can interact, and various occasions of what you would call ``Zinc!''\ that may arise as a result.
\earkp

Here I think I disagree with you:  the {\it ascription\/} of a Hilbert space (and with it a dimension) already strikes me as up to the job.  Thus, in a sense, the Hilbert space---but none of the vectors in it---is the proper mode for expressing an individuality or independent existence.

And by the way, it's ``Zing!''\ dammit.  (Howard Barnum wrote a paper recently, where he made slight fun of the term, and called his version of it ``Voom!'')  ``It don't mean a thing, if it ain't got that zing.''

\barkp
Now (this is perhaps a tad less banal) it appears to me that, at
least on this view, it is in this and only in this that one could speak of an UNKNOWN quantum state (the locution one sometimes find in, among others, your writings).  This ``state,'' however, could, on this view, never be represented by a state vector, which indeed reflects the state of expectation, or by any other mathematics.
\earkp

I believe I understand where you are going here.  But precisely because of the last sentence, I would be reluctant to call what you are talking about a state.  The way I imagine it, quantum systems have little to no properties in and of themselves \ldots\ and thus no ``states.''  To the extent that they have any properties at all, they are born in a sexual act, so to speak.  What is most real of a quantum system is what comes out of two of them going bump in the night.

Right now, I'm about two hours into my flight toward Madrid.  As you can see, I've been filling my time writing a little email.  But also I am forcing myself during this trip to finally read David Deutsch's book {\sl The Fabric of Reality}.  I figure I ought to do this since I might get a chance to talk to him at the QCMC meeting at MIT.  Before setting off to write you, I finished the first chapter, and have been finding it just as detestable as I had suspected.  God save me this week if I get all the way through the thing.

\section{13-07-02 \ \ {\it Yestopher} \ \ (to A. Plotnitsky)} \label{Plotnitsky9}

\barkp
I would only reiterate that all our knowledge (already obtained in
measurements) or predictive estimates only concern classical effects of the interactions between quantum objects and measuring instruments upon the latter, but not quantum objects themselves, although these effects would not of course be possible without the existence of quantum objects.  I am not saying ``reality,'' since there is no concept of reality I know or can think of applicable to such ``objects,'' while it seems to be possible to think of existence as a capacity to produce effects, in a classical-like (as in classical physics) or in a nonclassical fashion (as in quantum mechanics).
\earkp

Yes, I definitely like this statement.  But let me think more about the rest of your letter (for maybe a couple of days) before commenting further.  It might prod me to finally write down some thoughts about a divergence that I think is starting to develop between Bohr and me.

BTW, you don't mean ``orthogonal'' in this passage:
\barkp
Indeed, once we perform a momentum measurement on the first system we
completely cut ourselves from any possibility of making prediction
concerning the [position] measurements on the second system. I suppose
this point is linked in some ways to Mermin's observation to the
effect that once a given system is [in] a given state (in the sense of
psi-function) it cannot be in the state orthogonal to it.  This
language does not altogether satisfy me, however, since, as I said in
my previous email, ket-vectors in a Hilbert space do not describe
actual physical states.  One could say (in a more cumbersome but more
accurate way) that the possibilities of predictions defined by
orthogonal state vectors (expectation vectors) are always mutually
exclusive, which is of course to say, rigorously complementarity in
Bohr's sense, since they also entail mutually exclusive experimental
arrangements.
\earkp
Position and momentum eigenstates are not really orthogonal in the sense that Mermin is speaking of here.  Rather they are ``complementary'' (as Zeilinger would say) or ``mutually unbiased'' (as Bill Wootters would say).  That is, two vectors in a $D$-dimensional space are ``mutually unbiased'' when their inner product is $1/\sqrt{D}$ in magnitude.  It just so happens with position and momentum eigenstates that $D=\infty$, (and thus they are strictly speaking orthogonal), but that is a secondary concept or property.  Mermin's trouble with orthogonality arises, even in a $D$-dimensional space, when the inner product is zero.

\section{29-07-02 \ \ {\it 20 More Minutes}\ \ \ (to B. W. Schumacher)} \label{Schumacher11}

I was intrigued by your comment of, ``If you would have just had about 20 more minutes, I might have been convinced by your talk.''  Seriously, could I get you to tabulate your most serious objections to a Bayesian approach to quantum mechanics?  It would certainly help to orient me and the other boys: sometimes it hard to see the biggest leaks in a ship from the inside.  We need this kind of criticism if the point of view is going to hope to survive.

On another subject, I'm afraid you haven't lifted your pen yet to write up your doubting-Everett argument.  The original deadline has just about slipped past, and I really would like to have your paper.  There are some stories that only you can tell.  What can I do to help you, or help keep you on course?  Seriously; I'll be at your service if you need me.  (And if that doesn't work, remember the guilt thing.  Once upon a time you said, ``I owe you this paper.'')

\subsection{Ben's Reply}

\bq
Oh, I guess my basic worry is that the wave function is a bit too peculiar
to be an ``a priori'' object --- probabilities seem inescapable, but who
ordered the quantum?  So a program of thinking of quantum mechanics as
a rational calculus of inferences seems to be lacking some ingredient.
Wish I knew what it was.  (Is it the Zing?)  But until I have an inkling,
the project seems a little doubtful.

I wish you could see just how much I'm hanging my head in abject shame.
(It makes it hard to type.)  I'm leaving for sabbatical tomorrow, and
will take two weeks to drive across the country.  I'll be in Los Alamos
for a day next week.  Anyway, I'm taking my Everett stuff, with the
intention of trying to write the danged thing on the way.  I'll be in
touch and let you know how it goes.  I will {\it not\/} be on e-mail every day.

I suppose my basic problem is that I feel such a paper should be couched
as a massive and authoritative review of the conceptual issues in the
Everett interpretation.  This of course is madness.  I will ignore this
instinct and write only what I really have to say.

It was amazingly good to see you at MIT and I am really looking forward
to visiting {\Montreal}.
\eq

\section{29-07-02 \ \ {\it Complimentarity} \ \ (to A. Plotnitsky)} \label{Plotnitsky10}

OK, the title is probably a bad pun.  But I thought I might tell you something nice I heard Asher Peres say of you while we were in Spain.  ``That Plotnitsky is a wise man.''  He was quite impressed by your understanding of Bohr.

Here's another piece of complimentarity:  While I was in Cambridge last week, I picked up a copy of your book {\sl Reconfigurations}.  It's in beautiful shape and I got it for a great bargain, only \$11.75.

How much longer will you be in NYC?  I might drop in one day soon if you've got some free time.

\section{29-07-02 \ \ {\it Coordinate Systems, 1} \ \ (to W. K. Wootters)} \label{Wootters14}

This is a small note to let you know that after a little thought I decided I really liked your question at the end of my talk Wednesday.  Setting the standard quantum measurement device is more like setting a coordinate system than merely setting a length or time scale.  Now I'm not sure how to take that any further, but I'll keep you up to date as the thoughts arise.

Did you and Robin make any progress with the SIC-POVM?  If there's still some work left to be done, then Gabe and I will likely plunge back into the water as the week goes by.

\section{30-07-02 \ \ {\it Coordinate Systems, 2} \ \ (to W. K. Wootters)} \label{Wootters15}

\bbw
PS.  Have you seen the following concept in classical probability
theory?  Consider a probability distribution $P$ over $n$ variables.  Now consider all the $(n-1)$-variable marginal distributions derived from $P$. Let $\tilde{P}$ be the $n$-variable distribution that agrees with $P$ in all its marginals but otherwise maximizes the entropy.  Then the difference in entropy between $\tilde{P}$ and $P$ can be taken as a measure of the amount of information in $P$ that is not in its
marginals.

Have you seen that idea anywhere?  Noah Linden, Sandu Popescu and I
are doing the same thing for quantum states, and I'm wondering what's
been done classically along these lines.  Thanks.
\ebw

I meant to tell you, by the way, that that's a really beautiful idea.  I don't think I have ever seen it before, so there's not much I can tell you.  Presumably---in analogy to what you showed for the quantum case in your talk---in the bipartite setting it reduces to simply the standard mutual information.  (Well of course it does; I started to see it as I wrote that sentence.)  I'll ask around to some of the information theorists here.

By the way---now this is a completely different subject, but the memory was jogged by your question---here's a historical tidbit that might interest you.  In the mid 1850s, George Boole was also interested in the marginals a posited joint probability distribution could produce.  In fact he studied the inverse problem, given a set of potential marginals, under what conditions can there exist a joint distribution giving rise to them?  Those conditions are now called Bell inequalities, but Boole had expressions for some of them even in the 1850s.  Here's a great little article to read along those lines:
\begin{itemize}
\item
I.~Pitowsky, ``George Boole's `Conditions of Possible Experience' and the Quantum Puzzle,'' Brit.\ J.\ Phil.\ Sci.\ {\bf 45}, 95--125 (1994).
\end{itemize}

\section{30-07-02 \ \ {\it Copenhagen Visit} \ \ (to F. Tops{\o}e \& P. Harremo\"es)} \label{Topsoe1} \label{Harremoes1}

A friend and I will be visiting Copenhagen September 12--17, partially as a vacation, but partially for some work.  (He will be there because of a conference Sept 8--12 on laser physics; and I will be there because of an obligation in {\Vaxjo}, Sweden during the same time.)  Also while in Copenhagen, I thought I might do a little work, either visiting you and/or doing a little historical digging at the Niels Bohr Institute.

Do either of you have any suggestions for how this friend and I might make the trip affordable but still pleasant?  For instance, do either of your universities have any visitor housing?  Or could you recommend any affordable, but pleasant, hotels in an interesting part of town (with caf\'es, bars, etc., nearby)?

We're just fielding suggestions at this point.  Please tell me anything you can?

\section{30-07-02 \ \ {\it 20 More Minutes, 2}\ \ \ (to B. W. Schumacher)} \label{Schumacher12}

\bbs
I suppose my basic problem is that I feel such a paper should be
couched as a massive and authoritative review of the conceptual issues
in the Everett interpretation.  This of course is madness.  I will
ignore this instinct and write only what I really have to say.
\ebs

Yes, that would be madness.  Just think over and over, ``Transcribe my talk.  Transcribe my talk.  Transcribe my talk.  Capture the atmosphere of my talk.''  And when you need some filler, say something about information being relative, say something about what quantum information holds for quantum foundations studies.  Say what you feel, and everything will be OK.  If you need references, I'll dig them up.

Have a safe trip \ldots\ but make it your best mental writing experience ever.

\section{01-08-02 \ \ {\it Research Problem} \ \ (to M. Raginsky)} \label{Raginsky1}

\bmr
I must say that I thoroughly enjoy reading your papers on the
foundations of quantum theory.  However, I came of late to take any
``Quantum Theory as Information'' manifesto with a grain of salt.  Is it
just me, or are people insisting on an information-theoretic revision
of quantum theory simply because the field is so, shall we say,
``trendy?''  Certainly we may demand an ``information-theoretic reason''
for the fact, say, that a pure state of a quantum-mechanical system is
a ray in a complex Hilbert space, and that is fine as far as that
goes, but what about the good ol' classical information theory?  Why
doesn't anyone clamor for an ``information-theoretic reason'' for the
mathematical formalism of classical mechanics, with its symplectic
manifolds and Hamiltonian flows and the like? Perhaps the answer is
this:  the modern mathematical machinery of classical mechanics is,
after all, a fancy way to express the belief in the ultimate
determinism of the classical world.  So, insofar as there is no
``spooky'' artifact of this formalism, no one could give a damn about
any ``information-theoretic reason.''  It is only when our macroscopic
intuition gets blown to smithereens by the microworld (entanglement,
anyone?), do we suddenly demand a paradigm overhaul.
What I mean to say, so loquaciously, is that the ``classical''
(Shannon) information theory is far from being a physical theory the
way quantum information theory is shaping out to be.  Any thoughts on
this?
\emr

I promised you that I'd tell you what I said to Ulrich Mohrhoff when he said what you said (but in a meaner way).  It's pasted below.  [See 04-07-01 note ``\myref{Mohrhoff2}{Carts and Horses}'' to U. Mohrhoff.]

But that aside, I think your question is a good one, and it leads immediately to a research problem that might interest you.  Indeed I do think it is worthwhile to try to give information theoretic reasons for some aspects of classical physics---Hamiltonian flow being precisely one of them.  Go to my samizdat, \quantph{0105039}, and have a look at the following notes:
\begin{itemize}
\item
page 408, ``Andrei''
\item
page 283, ``End of Day''
\item
page 284, ``Wigner''
\item
page 285, ``contra-Koopman''
\item
page 165, ``The Evolution of Thought''
\item
page 166, ``On the Mark''
\item
page 167, ``More Linearity''
\item
page 168, ``Three References''
\end{itemize}
I think they explain what I'm talking about in pretty good detail.

It was good meeting you at MIT, and I'm looking forward to studying your channel papers if I ever get an ounce of time.  If you have any good ideas along the lines above, I'd love to hear them.

\section{01-08-02 \ \ {\it Manuscript} \ \ (to C. King)} \label{King3}

\noindent Dear Math-Chris,

\bck
I have one observation about your multiplicativity problem: if you
replace the ``maximum eigenvalue'' operation (aka the operator norm) by
the entropy function (you have to normalize the states first, and put
the normalization factors outside, so you have a weighted average of
entropies) then the corresponding minimal entropy quantity is
additive. Any interest?
\eck

That sounds intriguing, but I can't quite visualize which expression you're talking about.  If you can, just \TeX\ in a quick equation.  What would the interpretation of that quantity be (within the cluster of ideas that the paper is about)?\medskip

\noindent All the best,\medskip

\noindent Philo-Chris

\section{01-08-02 \ \ {\it Notes} \ \ (to C. King)} \label{King4}

Aha!  Thanks for the notes.  I don't think I had ever thought of the set of states as defining a doubly-stochastic channel before.  So, at least I understand your notation.  Very good.  And neat result of yours.  Does it lead to any insight on how to solve the original problem?  And here's another question:  Does $G(\Phi)$ define an interesting channel characteristic in general?  Have you ever seen it before in that kind of context?

\section{02-08-02 \ \ {\it Baby} \ \ (to C. H. {\Bennett} \& T. M. {\BennettT})} \label{TheoBennett} \label{Bennett22}

Beautiful pictures, but more importantly beautiful babies.  Congratulations to everyone in the family.  Busy days ahead.  Charlie knows I get philosophical when I think about life and all the  possibility it opens up in the universe---possibility that wasn't even there before.  These babies are part of that.  Wonderful creations.  Please give our best to George\index{Bennett's son} and Martha\index{Bennett's daughter-in-law}.

\section{05-08-02 \ \ {\it The Spirit that Breathes Life} \ \ (to G. L. Comer)} \label{Comer16}

``The divinity that breathes life into nature cannot be
represented.''

Thanks for your note of August 1.  It gave me food for thought; free
will saved another life.  Indeed, I think the divinity that breathes
the whole of life into nature is just that:  free will, the essence
of making a choice.

I don't think I ever told you this, but 132 years ago, the
realization that free will is key saved the life of another great
man.  It was William {\James}, and the realization is attributed not
only to saving his life---he had been in deep depression for some
time, with the accounts saying he was on the brink of suicide---but
in turning his philosophy and all his thought around.  The full road
to recovery took several years, but that moment was the starting
point.

Let me paste in a passage from his diary for you.

\bq\noindent
April 30, 1870\medskip

I think that yesterday was a crisis in my life. I finished the first
part of Renouvier's second ``Essais'' and see no reason why his
definition of Free Will---``the sustaining of a thought because I
choose to when I might have other thoughts''---need be the definition
of an illusion. At any rate, I will assume for the present---until
next year---that it is no illusion. My first act of free will shall
be to believe in free will. For the remainder of the year, I will
abstain from the mere speculation and contemplative {\it Gr\"ublei\/}
in which my nature takes most delight, and voluntarily cultivate the
feeling of moral freedom, by reading books favorable to it, as well
as by acting. After the first of January, my callow skin being
somewhat fledged, I may perhaps return to metaphysical study and
skepticism without danger to my powers of action. For the present
then remember: care little for speculation; much for the form of my
action; recollect that only when habits of order are formed can we
advance to really interesting fields of action---and consequently
accumulate grain on grain of willful choice like a very miser; never
forgetting how one link dropped undoes an indefinite number. {\it
Principiis obsta}---Today has furnished the exceptionally passionate
initiative which Bain posits as needful for the acquisition of
habits. I will see to the sequel. Not in maxims, not in {\it
Anschauungen}, but in accumulated acts of thought lies salvation.
{\it Passer outre}. Hitherto, when I have felt like taking a free
initiative, like daring to act originally, without carefully waiting
for contemplation of the external world to determine all for me,
suicide seemed the most manly form to put my daring into; now, I will
go a step further with my will, not only act with it, but believe as
well; believe in my individual reality and creative power. My belief,
to be sure, can't be optimistic---but I will posit life (the real,
the good) in the self-governing resistance of the ego to the world.
Life shall [be built in] doing and suffering and creating.
\eq

\section{06-08-02 \ \ {\it SHPMP Quantum Information Issue} \ \ (to the SHPMP participants)}

\noindent Dear friends,\medskip

I have to open up this note with some sad news.  Rob Clifton, the editor-in-chief of {\sl Studies in History and Philosophy of Modern Physics\/} and the main instigator of our special issue, died sometime late last week, succumbing to cancer.  He was a young man in the prime of his life and leaves behind a wife and two young children.  You can find out a little about Rob, his life, and his love for quantum mechanics at his webpage \myurl{http://www.pitt.edu/~rclifton/index.html}.

With that, I should get to the formal task of this note, our special issue on quantum information.  The original deadline for all papers was to be August 1.  Jeff Bub and I have talked to many of you privately about your progress, but especially in light of this tragedy, we think it is time to get serious.  With that, we propose to extend the deadline to October 1, but to make that one hard and fast.  It is important that we get the papers to the journal with plenty of time for refereeing and copy editing before the special issue's appearance.

I hope that you will all reconfirm your participation, and that we can count on you for a quality issue.  Please let us know your status as soon as possible.  Below, I list the proposed papers (forgive me if I didn't quite get your title right), and further below that I paste in the original invitation with details.\medskip

\noindent Best regards,\medskip

\noindent Chris (and Jeff)\medskip

\noindent {\bf Contents:}
\begin{itemize}
\item
H. Barnum \\     ``Quantum Information and Quantum Logic: Toward Mutual
                 Illumination''

\item
C. H. Bennett and J. A. Smolin\\
                ``Maxwell Demons, Bit Commitment, Lock Boxes, and Quantum
                 Information''

\item
L. Hardy  \\     ``Quantum Mechanics from $N$ Reasonable Axioms''

\item
R. Jozsa    \\   ``Quantum Computation and Quantum Foundations''

\item
N. D. {\Mermin}  \\ ``Philosophical Ruminations on Deutsch's Problem''

\item
I. Pitowsky    \\ ``Betting on the Outcomes of Measurements: A Bayesian
                 Theory of Quantum Probability''

\item
J. Preskill,  \\  ``Defending Everett''

\item
B. Schumacher, \\  ``Doubting Everett''
\end{itemize}

\section{07-08-02 \ \ {\it Trouble Man} \ \ (to N. D. {\Mermin})} \label{Mermin72}

\bdm
I completely forgot about this and have no idea what I was thinking
of when I sent you that title.  Better count me out.  I have too many
other things I'm supposed to do in the next two months.
\edm

Is there none of that college spirit left in you?  No loyalty to the
editorial board (of which you are a member)?

Honestly, is there really nothing you would find worthy to write up?
I remember loving the last line in this paragraph from your
``teaching'' paper:
\bq\noindent
   There are nevertheless some who believe that all the amplitudes
   $\alpha_x$ have acquired the status of objective physical quantities,
   inaccessible though those quantities may be.  Such people then wonder
   how that vast number of high-precision calculations ($10^{30}$
   different amplitudes if you have 100 Qbits) could all have been
   physically implemented.  Those who ask such questions like to provide
   sensational but fundamentally silly answers involving vast numbers of
   parallel universes, invoking a point of view known as the {\it many
   worlds\/} interpretation of quantum mechanics.  My own opinion is
   that, imaginative as this vision may appear, it is symptomatic of a
   lack of a much more subtle kind of imagination, which can grasp
   the exquisite distinction between quantum states and objective
   physical properties that quantum physics has forced upon us.
\eq
Surely, that sentence could be turned into a whole paper, at least
under my pen.

I remember the first time I met you---in {\Montreal}---I heard you say,
``Quantum computation is the biggest sham in the industry.  The
government gives you all this money to crack codes, and what you're
really doing is quantum foundations.''  Couldn't you find some way to
flesh that out and say it in print?  It would benefit you and it
would benefit the community.

\section{07-08-02 \ \ {\it Books and Notes}\ \ \ (to A. S. Holevo)} \label{Holevo5}

Thank you for the Humboldt book; I very much enjoy reading things like that.  And I will surely distribute copies of your own book to Ruskai, Schumacher, Caves, etc.  I will be seeing all of them in the month of October.

In sorting through your things yesterday, I put all of your handwritten notes into a single pile.  I also placed all the papers by other people on which you had written some notes into the same pile.  It really would be no problem for me to mail that to you.  If you will just give me an appropriate address, I will do it.

By the way, it was a surprise to see that you had started a draft titled ``A Remark on Conjecture of Bennett, Fuchs and Smolin.''  Too bad you were never able to complete it!  Regardless of the result, with a title like that it would have certainly helped my career.  Few people today associate me with that question, even though to my knowledge, Bennett and I were the first to ever bring it up (at a meeting in Torino in 1996).\footnote{\editornote This refers to the conjecture that entangled encodings can help a message-sender beat the channel capacity limit established by Holevo, Schumacher and Westmoreland.  See C.\ H.\ Bennett, C.~A.\ Fuchs and J.\ A.\ Smolin, ``Entanglement-Enhanced Classical Communication on a Noisy Quantum Channel,'' in {\sl Quantum Communication, Computing and Measurement,} edited by O.\ Hirota, A.\ S.\ Holevo and C.\ M.\ Caves (Plenum Press, NY, 1997), pp.~79--88, \quantph{9611006}.  In section~5 of this paper, it was suggested that entangling successive transmissions sent through a quantum channel could increase the channel's capacity for carrying classical messages.  That is, channel capacity might not be simply additive over multiple uses of the same channel.  Additivity was finally disproved, by means of a nonconstructive counterexample, in M.\ B.\ Hastings, ``A Counterexample to Additivity of Minimum Output Entropy,'' Nature Phys.\ {\bf 5,} 255 (2009), \arxiv{0809.3972}.}

\section{08-08-02 \ \ {\it The Recalcitrant P's} \ \ (to A. Peres)} \label{Peres39}

I almost titled this note ``Two Recalcitrant Positivists,'' remembering something you had called us long ago.  But then thinking of how I am more closely aligned with the pragmatists now (James, Dewey, Rorty, etc.)\ than with the positivists, I decided to be more careful.  We're just two recalcitrant P's, and I'll leave it at that.

I'm glad to hear that you and Aviva are home safely.

\section{09-08-02 \ \ {\it Conspiracy Theory} \ \ (to J. W. Nicholson)} \label{Nicholson10}

By the way, have a look at my conspiracy theorist article:
\begin{center}
\myurl{http://www.pcmag.com/article2/0,4149,440168,00.asp}.
\end{center}

\section{09-08-02 \ \ {\it Conspiracy Theory, 2} \ \ (to J. W. Nicholson)} \label{Nicholson11}

\bjn
Did you really say, ``I'm sure that the government is already using quantum cryptography
systems for real applications.''??
\ejn

That line really scares me, because he put quotes around it.  I can't imagine that I would have said it.  But on the other hand, I'm sure reporters know how to cover their asses.  I've been thinking about what to write this guy to lodge a complaint.  Maybe it was really said in a context that took some of the edge off it?  I don't know.  I can imagine that I would have said ``developing'' \ldots\ but he bases the whole story on the word ``using''.

\section{13-08-02 \ \ {\it Variola} \ \ (to R. W. {\Spekkens})} \label{Spekkens7}

\brws
A quick question for you regarding POVMs.  It seems to me that there
is an interesting distinction between, on the one hand, POVMs that can
be obtained by convex combinations of coarse-grainings or convolutions
of PVMs, and, on the other, those that can only arise by restriction
of a PVM on a larger Hilbert space.  An example of a POVM in the
latter category would be one with rank 1 elements numbering greater
than the dimensionality of the Hilbert space.  Has this distinction
been noted and studied anywhere?
\erws

Yes, this distinction was first noted by Holevo:  A. S. Holevo, ``Information-Theoretical Aspects of Quantum Measurement,'' {\it Problemy Pere\-dachi Informatsii}, {\bf 9}(2), 31--42 (1973); also listed as A. S. Kholevo, {\it Problems of Information Transmission}, {\bf 9}, 110--118 (1973).

Beyond that, a most interesting distinction has to do with thinking of the POVMs as a convex set.  The extreme points of that set have the character you mention, except that all the operators in the POVM must be linearly independent.  You can read about that in:
A.~Fujiwara and H.~Nagaoka, IEEE Trans.\ Inf.\ Theory {\bf 44}, 1071--1086 (1998).

\section{14-08-02 \ \ {\it Empty Head} \ \ (to J. W. Nicholson)} \label{Nicholson12}

I loved the Maureen Dowd ``empty-headed'' comment today.  Just sayin.

By the way, I've thought a little bit about the ethics of killing cattle and the culture of the ``compassionate cattle raiser/killer.''  Interesting issues with, shall we say, quantum implications.  I'm honing the thoughts for a (thoughtful) conversation in the near future.

\section{20-08-02 \ \ {\it Old Note to Rolf} \ \ (to C. H. {\Bennett})} \label{Bennett23}

Your remark last night about ``Gilles's idea'' to derive quantum mechanics took me by surprise.  The surprise came from the way you continue to associate my efforts with a kind of religion, rather than with an honest effort to get at some different ways of looking at quantum mechanics (by deriving it from information theoretic structures, etc.)---i.e., you don't even see that this is what I am always talking about.  I'd like to think some of my latest papers represent progress in that direction, and I guess I'd also like to think the pressures I put on the people around me (Hardy is a good example) have been part of the reason for their own good work in that direction.  Religion without theorems, or a dream to prove them, is just religion.  But religion that gives rise to some solid results in return is usually called science.

Anyway, thinking of the history of this, I came across a little passage I wrote to Rolf Landauer in July of 1998 which I thought was cute.  I'll paste it below.  It puts a lot of weight on your shoulders Atlas.\smallskip

\noindent With my usual love, Chris

\bq
The [speculated] information-processing limitation has to do with a funny property of the world we happen to live in.  It is this:
{\it my\/} information gathering about a given physical system will generally disturb {\it your\/} description and predictions for that selfsame system. Nonetheless we, as communicating beings, must come to a consistent description of what we see and know about that physical system.  This, I ever more firmly believe, is the essence of the quantum mechanical formalism.  Quantum theory is nothing more than Bayesian-like reasoning in a world with such a funny property.
(This I see as a large chunk of my personal research program:
clarifying, delineating, and searching for holes in this point of
view.)  Perhaps \ldots\ to put it in a more amusing way \ldots\ what I am saying is that I wouldn't be surprised if the whole edifice of quantum theory couldn't be constructed from the singular fact that ``quantum cryptography exists.''  That is to say, Hilbert spaces, the inner product rule, entanglement, and unitary evolution all from that clean, simple idea {\Bennett} and Brassard were the first to make some currency of.
\eq

\section{21-08-02 \ \ {\it Way in the Back} \ \ (to D. Gottesman)} \label{Gottesman1}

\bdg
Yes, I was wondering if I should suggest you as a recruit.  Some of the Perimeter recruiting is in Foundations of QM rather than quantum
computing proper; I guess Lucien is the first real hire in that area.
\edg

Thank you; I'm flattered.  I guess I don't have to advertise to you the depth of my faith that a beautiful spark will eventually fly out of rubbing those two fields together.

Charlie Bennett and I are sitting at the back of this enormous room in the OpryLand Hotel [at the ARO conference in Nashville] acting like it's our office.  He's reading science fiction from his laptop, and I'm writing email.  There seems to be a speaker saying something from way up in the front.

\section{22-08-02 \ \ {\it Opryland USA} \ \ (to G. L. Comer)} \label{Comer17}

I'm sitting in the Nashville airport, thinking about you.  I hope things are clear skies up there in St. Louis.  I just had a country fried steak with gravy and mashed potatoes for lunch; so all is certainly OK with me.

The reason I write you from the Nashville airport is that I've been in Nashville since Sunday evening.  I dropped in on the ARO (Army Research Office) annual review on all the quantum computing efforts they fund.  It was a contract/subcontractor only thing, but I finagled my way in nevertheless.  Lucent has put me on a mission to bring in outside funds for anything I can.  Surely a sign that Bell Labs is getting pretty near the crumble stage.  The company has decided to decrease the budget for Bell Labs Research by \$12M this coming year.  Since all equipment, travel, etc., expenditures are already on hold at the moment, that can only translate into one thing.  Salary.  And the only method proposed for reducing that is the FMP, forced management program.  That's a fancy way of saying ``layoff.''  So, roughly speaking, for every \$300K we can bring in through outside contract, that's one person's job for the coming year.  Pretty serious stuff.

But enough of that, let me tell you the exciting news.  Not only did I stay in Nashville, but I stayed in the Opryland Hotel!  It's probably the closest I'll ever get to living in a biodome \ldots\ or one of Gerard O'Neill's big toroidal space colonies (remember those?).  What an amazing place.  Have you ever seen it?  The little boy left in me was absolutely fascinated \ldots\ while, unfortunately, the liberal democrat in me teetered closer to the Green Party than ever (just calculating the waste of resources required to power those waterfalls and the nine acres under a glass roof).

I got a kick out of telling people here and there a little story about my old Momma in Texas.  You see, since Jan 1 this year, I've been to Japan, Ireland, and Spain; and I'll be going to Sweden and Denmark next month.  Every time I tell my Momma about one of these trips, she says, in a dreadful voice, ``Oh son, I wish you didn't have to go.''  But last week when I told her I'd be staying in the Opryland hotel, she sounded like a schoolgirl, ``Oh really?!?!''  I told her I'd get her Mel Tillis's autograph if I saw him.  I could almost see her blush on the other end of the phone.

Boy I tell you, this field of quantum computing is growing up.  There were over 200 attendees at this meeting.  That is to say, that's 200 people the army is funding to map out the most hopeful potential implementations for quantum computing.  I tell you I was literally lost at this meeting; I bet I didn't understand five or ten talks in the whole thing.  Phosphorous doped silicon implementations, superconducting-loop and Josephson junction implementations, quantum Hall effect implementations, quantum dot implementations, carbon nanotube implementations, \ldots\ and so it went.  The crowd at this meeting were a far cry from the foundational crowds I usually hang out with:  these guys were serious.

It's been a while since I've written you.  I guess my first story above indicates a little of why---indeed, I haven't even been able to put an entry in my samizdat since July 5---but as usual, there's no proper excuse for silence.  We really ought to get together one of these days.  I guess if I do end up at the Perimeter Institute in Canada we'll be that much closer together, but I suspect there's still a hell of a lot of miles between Toronto and St.\ Louis.  Gotta work harder on that teleportation thing.

\section{27-08-02 \ \ {\it Dimensionality} \ \ (to P. J. Reynolds)} \label{Reynolds1}

\bpjr
Thanks for the e-mail.  I love your web page!  I'll have to make some
time and look at those PowerPoint talks.
\epjr

Sorry for the long delay in my reply:  Any man who says he loves my webpage certainly deserves a prompt, if not immediate, reply!!  But I finagled my way into the ARO review meeting last week, and that ended up commandeering all my attention \ldots\ I'm just recovering now.

\bpjr
I justify the small amount still in my program as benefiting
metrology, clocks, and time transfer.  Is the Hilbert space ``resource''
different (versus ``more fundamental than'') entanglement?
In other words, does it enable different things than the things we
attribute to entanglement?  If so, then (depending on what those
things might be) we might have an ``in'' for you.
\epjr

What I am thinking of here is seeking ways to quantify the extent to which the dimensionality $D$ of a quantum system's Hilbert space signifies that quantum system's ``sensitivity to the touch.''  If you have a look at some of my more foundational writings, in fact, you'll see I'm starting to view this ``sensitivity to the touch'' as the most fundamental property there is for a quantum system---i.e., the most significant feature that sets quantum stuff apart from classical stuff is its potential to be perturbed.  From this point of view, quantum entanglement is a secondary effect, and has more to do with how this sensitivity scales with the number of quantum systems than anything else.  The more important question is what does $D$ tell us, and what can we do with it?

With regard to your program, I can foresee at least a couple of directions with which to take this idea.

1) In an upcoming paper with Masahide Sasaki---I can supply the manuscript under construction if you'd like---we try to quantify the ``quantumness of a Hilbert space'' by a system's ability to sniff out whether it is being sent through a classical channel or a quantum channel.  In fact, we view this ability as a Bureau of Weights and Measures kind of thing:  Lucent Technologies says it is supplying you with a good quantum channel.  How can you tell that it really is?

2)  Good ``sensitivity to the touch'' signifies some potential for good antennas.  Here I'm thinking of things like LIGO in particular (though in a more abstract setting).  What does Hilbert-space dimensionality buy you with respect to the ability to distinguish weak signals?  I've got some of this captured in my Caltech lecture notes ``Viewing LIGO through Schwarz Colored Glasses,'' which I can send you.  Also you can find some hints of the ideas in a paper by Childs, Preskill, and Renes, ``Quantum Information and Precision Measurement'' (\quantph{9904021}), of which I played some part.

Anyway, that's the sort of stuff I'm thinking about.  If I were to put this down in a proposal I could beef it up pretty significantly, and hit the theme from still a couple more directions.

I hope that builds a little clearer picture for you.

\section{30-08-02 \ \ {\it Transformations of This and That} \ \ (to G. L. Comer)} \label{Comer18}

Thanks for the enjoyable note.

\bgc
Why did this passage interest me?  Because of the problem of the
dictionary: How can one ``understand'' language?  It seems to me that
even when we native English speakers speak English, all we are doing
is using rules of translation.  Into what? I don't know.
Interestingly enough, when I think of relativity, I see a big
similarity.  For me, relativity is not only about gravity as a
manifestation of the curvature of spacetime, but also about how two
observers can communicate the results of their experiments in such a
way that they can agree that they have witnessed the ``same'' event.
As an example, in order to determine the energy of the cosmic
microwave background, one observer with unit four-velocity
$u^{\mu}_1$ will say the energy is $u^{\mu}_1 p_{\mu}$, where
$p_{\mu}$ is the momentum of the radiation, but another observer with
four-velocity $u^{\mu}_2$ will say the energy is $u^{\mu}_2 p_{\mu}$.
Even if the observers are at the same place, at the same time, when
they make their measurement, they will in general not record the same
energy because their motions will in general be different.  So how
can they ever agree on anything?  Relativity tells them how to
``translate'' one measurement into the other so that a consistent
description results.  What is objective about the energy?  Nothing.
Its value depends on the motion of the observer.  Where is the
objectivity?  Only in the rules of translation.  The only objective
thing to me seems to be the how one object is to be compared with
another.
\egc

Did you ever read my ``Anti-{\Vaxjo} Interpretation of Quantum Mechanics''
paper?  (If not, you can download it from my webpage; link below.  It
should be super-easy reading; no equations at all.)  If I'm not
mistaken I play up a similar point there, in the context of
noncommuting observables.

But I kind of like something about your twist.  Spacetime is the
dictionary.  And, like the dictionary, the ``divinity that breathes
life'' into its connections is its user, the active agent.  Without
the agent it is a snake that bites its tail.

I'm in Los Alamos at the moment, being forced by the scenery to
remember some of my own transformations.  It has been two years since
I've been here; the first time I've been back since the fire.  The
mountainside is still full of these toothpicks that used to be trees;
I guess it'll be like that for at least 20 more years or so, fading
slowly.  Yesterday I drove to the street where our house (and all my
material possessions) used to be.  A new house is finally being
constructed in its place.  Funny, I found myself thinking mostly
about the ground underneath it; I guess like a burial place.

Wojciech Zurek invited me so that we could fight over quantum foundations.  And fight we have.  Maybe the greatest mystery to me---beside the very lack of consistency in all he says---is how he does not see how very {\it dull\/} this point of view he's trying to construct is.

All this travel is taking a toll on me again.  I get home Sunday, but then have to turn around and fly to Sweden and Denmark next Friday for 11 days.  My girls are growing up in the distance.

But you sound like a busy man yourself.  Welcome back to teaching!

\section{03-09-02 \ \ {\it One Emotion} \ \ (to G. L. Comer)} \label{Comer19}

\bgc
We are the freedom to choose.
\egc

God I loved that last line!

\section{05-09-02 \ \ {\it Copenhagen Visit} \ \ (to O. C. Ulfbeck)} \label{Ulfbeck1}

My name is Chris Fuchs.  We have a mutual friend in David Mermin, and he has suggested that I contact you.  For a quick introduction to me and my work, you can read David's foreword to my pseudo-book posted on the Los Alamos archive; here is a quick link to it: \quantph{0105039}.  Alternatively, for a technical introduction to the sorts of things on my mind, see my paper ``Quantum Mechanics as Quantum Information (and only a little more)'': \quantph{0205039}.  (It is a coincidence that the two papers have almost identical locator numbers.)

In any case, I will be visiting Copenhagen September 12 through 17, and I would very much like to talk to you and/or Prof.\ Bohr about your quantum-foundations paper on genuine fortuitousness.  I would also like to spend some time in the Niels Bohr Archive if it is possible and I can figure out how to make an appointment there, etc.  Would you like to get together?

If so, there is a chance I will be able to maintain email contact within Denmark; so perhaps I can be reached that way.  Also, however, I will be staying at a guest room of the university, and should be findable by some means there.  Here is the information I have on it: [\ldots]  Finally I should be reachable through my hosts Peter Harremo\"es and Flemming Tops{\o}e: [\ldots]

If you have the time, I would love to hear from you. \medskip

\noindent PS.  I will also be giving two talks while in residence:
\begin{itemize}
\item
13/9 13.15-14.00
On Unknown Quantum States: The Quantum de Finetti Representation Theorem
\item
16/9 9.15-10.00
Quantum Mechanics as Quantum Information (and only a little more)
\end{itemize}

I'm not sure where they will be yet and there is some possibility that the order will be reversed, but those are their dates and times.  I would be flattered if you could attend either one; I will place the abstracts below.

\bq\noindent
Quantum Mechanics as Quantum Information (and only a little more) \medskip
\\
I say no interpretation of quantum mechanics is worth its salt unless it raises as many technical questions as it answers philosophical ones. In this talk, I hope to convey the essence of a salty, if not downright briny, point of view about quantum theory: The deepest truth of quantum information and computing is that our world is wildly sensitive to the touch. When we irritate it in the right way, the result is a pearl. The speculation is that this sensitivity alone gives rise to the whole show, with the quantum calculus portraying the best shot we can take at making predictions in such a world. True to form, I ask more questions than I know how to answer. However, along the way, I give a variant of Gleason's theorem that works even for rational and two-dimensional Hilbert spaces, give another variant of Gleason's theorem that gives rise to the tensor-product rule for combining quantum systems, and finally derive a new form for expressing how quantum states change upon the action of a measurement.\medskip
\eq
\bq\noindent
On Unknown Quantum States: The Quantum de Finetti Representation Theorem\medskip
\\
There is hardly a paper in the field of quantum information theory that does not make use of the idea of an ``unknown quantum state.''  Unknown quantum states can be protected with quantum error correcting codes.  They can be teleported.  They can be used to check whether an eavesdropper is listening in on a communication channel.  But what does the term ``unknown state'' mean?  In this talk, I will make sense of the term in a way that breaks with the vernacular:  an unknown quantum state can always be viewed as a known state---albeit a mixed state---on a larger ``multi-trial'' Hilbert space.  The technical result is a quantum mechanical version of the de Finetti representation theorem for exchangeable sequences in probability theory.  Interestingly, this theorem fails for real and quaternionic Hilbert spaces.  The implications of this theorem for the point of view that quantum states represent states of knowledge, rather than states of nature, will be discussed.
\eq

\section{05-09-02 \ \ {\it Everett Quote} \ \ (to N. D. {\Mermin})} \label{Mermin73}

I just want to make sure I have this quote archived.  An easy way for
me to do that is to send it to you.  It comes from a footnote in Hugh
Everett's original relative-state paper.

\bq
Note added in proof. --- In reply to a preprint of this article some
correspondents have raised the question of the ``transition from
possible to actual,'' arguing that in ``reality'' there is---as our
experience testifies---no such splitting of observer states, so that
only one branch can ever actually exist.  Since this point may occur
to other readers the following is offered in explanation.

The whole issue of the transition from ``possible'' to ``actual'' is
taken care of in the theory in a very simple way---there is no such
transition, nor is such a transition necessary for the theory to be
in accord with our experience.  From the viewpoint of the theory {\it
all\/} elements of a superposition (all ``branches'') are ``actual,''
none any more ``real'' than the rest.  It is unnecessary to suppose
that all but one are somehow destroyed, since all the separate
elements of a superposition individually obey the wave equation with
complete indifference to the presence or absence (``actuality'' or
not) of any other elements.  This total lack of effect of one branch
on another also implies that no observer will ever be aware of any
``splitting'' process.
\eq

\section{08-09-02 \ \ {\it Commune '02} \ \ (to the Communards)} \label{Communards1}

\noindent {\bf Quantum Foundations in the Light of Quantum Information II {\Montreal}, Canada October 13, 2002 -- November 2, 2002}\medskip

\noindent Chers communards,\medskip

Gilles and I are making final plans for the October {\Montreal} quantum foundations commune.  I think the meeting promises to be a lot of fun and more importantly promises to be the site of some good physics.

Below is the latest tabulation of attendees.  If your dates are still marked with an X or a ?, please make a final confirmation as soon as possible.  Alternatively if you cannot ultimately attend the meeting, please let us know that too.  (In the coming days, we hope to populate the remaining spots with friends in the waiting list.  Recall, we have room for about 15 attendees at any one time, so that everyone will have a desk.)

Beyond that, the main thing we need from you now is to make your flight plans and living arrangements.  (Because of the late date of this letter, this is something that should be taken care of rather urgently.)  Concerning the former, please try to include a Saturday stay in your plans so as to keep the airline fees down.  Concerning the latter, the procedure is the following.  Have a look at the following webpage, and decide which option looks best for you: \myurl{http://www.crm.umontreal.ca/en/niveau2/index_vis_sug.html}.
After that, write an email to Luc St-Pierre ({\tt stpierre@CRM.Umontreal.Ca}) at CRM telling him which option you would like reserved and what your precise dates will be.  Gilles and I would suggest the Chateau Versailles Hotel as the best option for most of you:  it is the most pleasant hotel and only a short, straightforward bus ride from the university.  If you prefer no busses, Hotel Terrasse Royale is a small walk from the university, but---be warned---it's not nearly as pleasant a place to live.

The format of the meeting will be relaxed and mainly devoted to work, conversation, and collaboration time.  We will, however, schedule one to two talks per day; so everyone should come prepared for that eventuality.  Also---and this is a key component to the meeting---we ask that everyone compile and share a list of concrete problems whose solution, they believe, will tell us something novel about quantum foundations.\medskip

\noindent {\bf Key:}\medskip

\noindent
\verb+X+ -- dates proposed by Fuchs (based on request and/or guesswork) \\
\verb+O+ -- dates confirmed by participant \\
\verb+?+ -- additional dates participant is contemplating and/or requesting \\
\verb+-+ -- means person is not in attendance on that date\medskip

\noindent (Saturdays and Sundays are signified by an {\tt S}.)

\noindent
\verb+       \             |  S S           S S           S S           S S+\\
\verb+  Who   \   Date     |  1 1 1 1 1 1 1 1 2 2 2 2 2 2 2 2 2 2 3 3 0 0 0+\\
\verb+         \           |  2 3 4 5 6 7 8 9 0 1 2 3 4 5 6 7 8 9 0 1 1 2 3+\\
\verb+ ---------------------------------------------------------------------+\\
\verb+ Howard Barnum       |  - - - - - - - - - - O O O O O O O O O O O O O+\\
\verb+ Charles Bennett     |  ? ? ? ? ? ? ? ? ? ? ? ? ? ? ? ? ? ? ? ? ? ? ?+\\
\verb+ Gilles Brassard     |  - O O O O O O O O O O O O O O O O O O O O O O+\\
\verb+ Hans Briegel        |  - O O O O O O O O - - - - - - ? ? ? ? ? ? - -+\\
\verb+ Jeffrey Bub         |  - - - - - - O O O O O - - O O O O O - - - - -+\\
\verb+ Adan Cabello        |  - O O O O O O O O O O O O O O O O O O O O O O+\\
\verb+ Carlton Caves       |  - - - - - - - - O O O O O O O O O O O O O O O+\\
\verb+ Richard Cleve       |  - - - - - - - - - - - - - - - - X X X X X X X+\\
\verb+ Claude Crepeau      |  - O O O O O O O O O O O O O O O O O O O O O O+\\
\verb+ Chris Fuchs         |  - O O O O O O O O O O O O O O O O O O O O O O+\\
\verb+ Nicolas Gisin       |  - O O O O O O O O O O - - - - - - - - - - - -+\\
\verb+ Lucien Hardy        |  - - - - - - - - O O O O O O O O O O O O O O O+\\
\verb+ Patrick Hayden      |  - O O O O O O O - - - - - - - - - - - - - - -+\\
\verb+ Pawel Horodecki     |  - X X X X X X X X X X - - - - - - - - - - - -+\\
\verb+ Richard Jozsa       |  - X X X X X X X - - - - - - - - - - - - - - -+\\
\verb+ Adrian Kent         |  - - - - - - - - O O O O O O O O O O O O O O O+\\
\verb+ Dominic Mayers      |  - - - - - - - - O O O O O O O O - - - - - - -+\\
\verb+ David Mermin        |  - - - - - - - - - - - - - - - - - O O O O O O+\\
\verb+ Itamar Pitowsky     |  - O O O O O O O O O O - - - - - - - - - - - -+\\
\verb+ David Poulin        |  - O O O O O O O O O O O O O O O O O O O O O O+\\
\verb+ Ruediger Schack     |  - - - - - - - - - O O O O O O O O O O O O - -+\\
\verb+ Ben Schumacher      |  - O O O O O O O O O O O O O O O O O O O O O O+\\
\verb+ John Smolin         |  - - - - - - - - - O O O O O O O O O O O O O O+\\
\verb+ Robert Spekkens     |  - - - - - - - - - - O O O O O O O O O O O O O+\\
\verb+ David Wallace       |  - X X X X X X X X X X - - - - - - - - - - - -+\\
\verb+ William Wootters    |  O O O O - - - - - - - - - - - - - - - - - - -+

\section{08-09-02 \ \ {\it Y'all Come} \ \ (to C. H. {\Bennett})} \label{Bennett24}

You will have noticed that I included you in the latest announcement for the {\Montreal} Commune.  I do hope you'll come for some of it, and I mean that from the bottom of my heart.  I'm spending a week in Sweden presently, and I'm typing this letter from the hotel lobby where John Smolin, Lucien Hardy, and I had so many pleasant conversations at the last quantum foundations meeting I put together.  John, as I've told you before, turned out to be one of the most useful and interesting participants of the meeting---that took me by surprise because I never imagined he would have any interest in this subject.  Nothing could be better than for you to surprise me the same way!

Looking forward to your response.

\section{08-09-02 \ \ {\it Information is Physical} \ \ (to A. Peres)} \label{Peres40}

\bap
I am looking for references that will appear in my RMP with Danny.
Ben Schumacher confirmed that he invented ``qubit'', and Gilles found
the father of Alice and Bob (Blum, 1981) after I found Savitt 1982. I
remember that Rolf Landauer used to say ``information is physical'' but
I can't find a written record. Would you know one?
\eap

I'm in Sweden with a bad connection, so I'll be brief.

The following comes from page 192 of my samizdat, \quantph{0105039}.

\bq
\subsection{08 October 1998, to Rolf Landauer, ``Info is Physical''}

I have room in a very tight conference proceedings paper that I'm writing to make one citation to your phrase, ``Information is Physical.''  Do you have a favorite paper that I should cite? Should I make the citation to your earliest mention of the phrase? Or should I make the citation to what you think is the clearest statement of it?  Please make the decision for me and send me the correct reference.  My paper is going off to the editors tomorrow, so if you could send me a quick note that'd be great!

\subsection{Rolf's Reply}

\bq
The concept, but not that phrase, first appeared in R.~L. {\sl IEEE Spectrum\/} vol 4, issue \#9, pgs.\ 105--109 (1967). (Like for {\sl Physics Today}, the page numbering restarts with every issue.) The exact wording PROBABLY first showed up in my 1991 {\sl Physics Today\/} paper. But it seems best to cite an early or a very recent paper. Two are on their way into print. One that is likely to appear within a few weeks: R.~L. in {\sl Feynman on Computation 2}, ed.\ by A.~J.~G. Hey, Addison Wesley, Reading (1998?). The title of that one ``Information is Inevitably Physical''. About 11 days ago at a session in Helsinki I complained to my audience: I have gotten a fair amount of acceptance for that phrase, but not for the message attached to it. But thanks for checking.

P.S. I'll mail both papers, even though they will arrive too late.
But if you take an instant dislike to the one you used, you can scratch it at galley proof time.
\eq
\eq

\section{08-09-02 \ \ {\it Dreams from Sweden} \ \ (to D. J. Bilodeau)} \label{Bilodeau6}

I was walking around the streets of {\Vaxjo} this morning, and I found myself thinking of you.  I'm staying here for five days for no particular reason; just using the time to get some work and thought done at Khrennikov's expense.  Following that, I'll go to Copenhagen for five days for roughly the same reason.  I hope things are going well with you.

I picked up a copy of the Jung--Pauli letters (that have now been translated into English) just before coming to Sweden.  They're quite nice and giving me food for thought.  I really need to finish up my ``activating observer'' project (remember the large compendium of quotes, subtitled ``Resource Material for a Paulian--Wheelerish Conception of Nature''), but I've gotten so far behind.  With some luck I'll get it onto {\tt quant-ph} by May 10 of next year.  I've got just a load of material to get into it.  In fact, I've found a motherload of material in the writings of William James, John Dewey, and Richard Rorty that certainly qualifies for the project.

\section{09-09-02 \ \ {\it Information Carriers are Physical} \ \ (to A. Peres)} \label{Peres41}

\bap
I could not understand your answer, but there is nothing urgent.
What I would like is the exact wording, and a reference that ordinary
readers can easily access ({\bf Physics Today} is fine, or some conference).
\eap

I was just demonstrating that my samizdat can be a resource for historical questions.  The words below are Rolf's words himself:
\bq
The concept, but not that phrase, first appeared in R.~L. {\sl IEEE Spectrum\/} vol 4, issue \#9, pgs.\ 105--109 (1967). (Like for {\sl Physics Today}, the page numbering restarts with every issue.) The exact wording PROBABLY first showed up in my 1991 {\sl Physics Today\/} paper. But it seems best to cite an early or a very recent paper. Two are on their way into print. One that is likely to appear within a few weeks: R.~L. in {\sl Feynman on Computation 2}, ed.\ by A.~J.~G. Hey, Addison Wesley, Reading (1998?). The title of that one ``Information is Inevitably Physical''. About 11 days ago at a session in Helsinki I complained to my audience: I have gotten a fair amount of acceptance for that phrase, but not for the message attached to it. But thanks for checking.
\eq
Thus you should cite either the {\sl IEEE Spectrum\/} and/or the {\sl Physics Today\/} articles of his.  I don't know the exact sentences in those papers; for that you should probably get Danny to go to the library.

By the way, I really wish Rolf had said ``information carriers are physical'' instead of ``information is physical'' \ldots\ but I never got a chance to tell him that while he was alive.

\section{09-09-02 \ \ {\it Dates and Talks} \ \ (to W. K. Wootters)} \label{Wootters16}

I think, it is best to assume that people will be trickling in on the 13th, and the events will start on the morning of the 14th.  Let me make an executive decision:  Events will start on the morning of the 14th.  There.

I would like to reserve dinner with you, or you and Ben, for Oct. 13.  Do you think we could do that?  (I sure wish you had the time to stay longer.)  I'd really like to get a chance to corner you on some of these Whitehead things.

On top of that, though, would you be willing to give the first talk of the meeting?  Well, the first technical talk; Gilles or I will probably give some procedural talk/s before that.  I would be particularly pleased if you would talk about your ``private world within entanglement'' ideas.  And beyond that, I'd also be pleased if you would add a smattering of discussion about the Whitehead stuff.  Even if you don't feel that these things are science yet---maybe just ideas for ideas---I think it would still be immensely useful for setting a tone for the meeting.

Moreover, can you dream up any concrete problems to share that are motivated by these points of view?  That's your challenge.

\section{10-09-02 \ \ {\it Tentative Talk Schedule for Commune} \ \ (to the Communards)} \label{Communards2}

\noindent Chers Communards,\medskip

Below is a tentative schedule for talks at the commune.  Not having an official title from any of you yet---you can't be blamed, I haven't asked for one---I decided to make up some titles of my own.  Mostly the titles are a little whimsical, but they do reflect what I've found interesting in my conversations with you.  (Thus, I guess I have a secret wish that you'll talk about these things.)  But do feel free to talk about anything you want.  Please send me a better title as you get a chance.

Also, further below you'll find the latest tabulation of attendance.  If I've gotten anything wrong please let me know.

Soon I will send more detailed instructions for how to get from the hotels to the meeting site.

Looking forward to seeing you all soon. \medskip

\noindent {\bf Quantum Foundations in the Light of Quantum Information II} \\
{\Montreal}, Canada
\\
October 13, 2002 -- November 3, 2002
\medskip

\underline{Monday, Oct 14}

\begin{itemize}
\item[a)] Gilles Brassard and/or Chris Fuchs\\
``Opening Remarks''\\
``Moment of Silence for the Paris Commune 1871''\\
``The Spirit of the Commune''

\item[b)] Bill Wootters\\
``Speculative Physics from Speculative Philosophy''
\end{itemize}

\underline{Tuesday, Oct 15}

\begin{itemize}
\item[a)] Nicolas Gisin\\
``Why Correlation?''

\item[b)] Dominic Mayers\\
``The Foundational Significance of Bit-Commitment and Coin-Tossing
No-Go Theorems''
\end{itemize}

\underline{Wednesday, Oct 16}

\begin{itemize}
\item[a)] Charles Bennett\\
``Why I Feel It Important to Always Defend Many Worlds in Private,
but Never in Public''

\item[b)] Patrick Hayden\\
``A Paper I Wrote Long, Long Ago with David Deutsch -- Pro and Con''
\end{itemize}

\underline{Thursday, Oct 17}

\begin{itemize}
\item[a)] Hans Briegel\\
``The Quantum-Foundational Significance of Measurement-Based Models of Quantum Computation''

\item[b)] David Poulin\\
``What Do You Mean `Simulating a Quantum Computation?'''
\end{itemize}

\underline{Friday, Oct 18}

\begin{itemize}
\item[a)] Stefan Wolf\\
``Intrinsic Information and Classical Analogs to Entanglement''

\item[b)] Problem session, roundtable, festival of outlandish ideas.
\end{itemize}

\underline{Monday, Oct 21}

\begin{itemize}
\item[a)] Chris Fuchs, 10:30 AM\\
``The End of the World as We Know It:  Doing Quantum States and
Quantum Time Evolutions on a Simplex''

\item[b)] Gilles Brassard, 3:15 PM\\
``Quantum Computing without Entanglement''
\end{itemize}

\underline{Tuesday, Oct 22}

\begin{itemize}
\item[a)] Rob {\Spekkens}, 10:30 AM
``In Defense of the View that Quantum States Are States of
Knowledge''

\item[b)] {\Adan} Cabello, 3:30 PM\\
``All the Latest on Kochen--Specker, and Why It's Important''
\end{itemize}

\underline{Wednesday, Oct 23}

\begin{itemize}
\item[a)] {\Ruediger} {\Schack}, 10:30 AM\\
``Exorzismus der objektiven Quantenoperationen''

\item[b)] Howard Barnum, 3:30 PM\\
``Quantum Information Processing and Quantum Logic, Toward Mutual
Illumination, \ldots\ or How I Became a (Closed, Convex, Pointed)
Conehead''
\end{itemize}

\underline{Thursday, Oct 24}

\begin{itemize}
\item[a)] Jeffrey Bub, 10:00 AM\\
``Characterizing Quantum Theory in Terms of Information-Theoretic
Constraints''

\item[b)] John Smolin, 1:35 PM\\
``Another Excuse to Say `Lock Box'\,''

\item[c)] Ben Schumacher, 3:30 PM, departmental colloquium\\
``Reversible Computation and Demonic Thermodynamics''
\end{itemize}

\underline{Friday, Oct 25}

\begin{itemize}
\item[a)] Claude Crepeau, 10:30 AM\\
``Quantum Authentication and Codes Correcting more than N/4 Arbitrary Errors:  A Possible Key to Quantum Foundations''

\item[b)] Ernesto Galvao, 3:35 PM\\
``What a Single Qubit Gives Us''

\item[c)] Jos\'e, 4:20 PM\\
``A Result in Quaternionic Quantum Mechanics''
\end{itemize}

\underline{Monday, Oct 28}

\begin{itemize}
\item[a)] Lucien Hardy, 10:30 AM\\
``The Classical and Quantum State Change Rules are the Same''

\item[b)] Roberto Floreanini, 3:05 PM, just after coffee\\
``Entanglement and Complete Positivity in Open System Dynamics''
\end{itemize}

\underline{Tuesday, Oct 29}

\begin{itemize}
\item[a)] Carlton {\Caves}, 10:30 AM\\
``The Point Fuchs Keeps Failing to Appreciate''

\item[b)] Marcus {\Appleby}, 3:05 PM, just after coffee\\
``The Man Who Mistook His Wife for a Hat, or Nullification of the
Nullification''
\end{itemize}

\underline{Wednesday, Oct 30}

\begin{itemize}
\item[a)] Chris {\Timpson}, 10:30 AM\\
``Claude Shannon Was Smarter Than You Might Think:  Quantum Mechanics Can't Stop His Information Measure''

\item[b)] Fotini Markopoulou, 3:30 PM\\
``What Can Quantum Information Do for Quantum Gravity and Quantum
Cosmology, and What Can They Do for Quantum Information?''
\end{itemize}

\underline{Thursday, Oct 31}

\begin{itemize}
\item[a)] David {\Mermin}, 10:00 AM\\
``Poetry without Poetata''

\item[b)] Ben Schumacher, 3:05 PM\\
``Checklist of What I DO NOT Find Compelling in any Quantum
Interpretations To Date''
\end{itemize}

\underline{Friday, Nov 1}

\begin{itemize}
\item[a)] Problem session, roundtable, workshop of outlandish ideas.
\end{itemize}

\section{10-09-02 \ \ {\it Don't Forget Your Problem Sets} \ \ (to the Communards)} \label{Communards3}

\noindent Chers Communards,\medskip

This time I am writing briefly to remind you that Gilles and I would like you to dream up a set of concrete problems to share with everyone at the meeting.  It would be so nice to see everyone playing, having fun, and perhaps a few collaborations spark up in the process.  By my count, the problems from the last {\Montreal} meeting led to at least four published papers.

To give you a feel for the sorts of things we are thinking of---the level of difficulty, etc.---below I place a write-up of the problems I myself presented at the last meeting.  (Some of them are still unsolved.)

Let's make this a great meeting!  It's mostly in your hands.

\bq
\subsection{``Problems Based on Information-Disturbance Foundation Quest,'' 15 May 2000}

I hope you've had a chance to think about the request Gilles and I made in our invitation letter:  namely, to compile a list of concrete problems whose solutions might shed some light on the foundations of quantum theory.  What we were thinking in particular is that no point of view about quantum foundations is worth its salt if, at this stage, it doesn't raise as many questions as it answers.
Why should we buy into a point of view if it doesn't lead to more fun or, at the very least, something more concrete than a stale philosophical satisfaction?

With that in mind, I've decided to grease the gears a bit by giving you a preview of some of the problems motivated by my particular ish-ism. If you haven't yet created a set of your own problems (based on your ish-ism of course), I hope this will give you a flavor of what we were thinking when we made our request. Certainly the more varied the sets of problems everyone brings, the greater the chance we have for making some real progress!

The point of view I'm likely to represent at our meeting is, I think, best captured (though perhaps a little flamboyantly) by a manifesto I wrote a couple of years ago.  Let me reproduce that here as an introduction and motivation to the problems that follow.

\bq
\small
\begin{center}
\bf Genesis and the Quantum
\end{center}
\bq \noindent In the beginning God created the heaven and the earth.  And the earth was without form, and void; and darkness was upon the face of the deep. And the Spirit of God moved upon the face of the waters. And God said, Let there be light: and there was light.  And God saw the light, that it was good; and God divided the light from the darkness. And God called the light Day and the darkness he called Night.  And the evening and the morning were the first day. \ldots\ [And so on through the next five days until finally \ldots]\ \ And God saw everything that he had made, and behold, it was very good. And there was evening and there was morning, a sixth day.  Thus the heavens and the earth were finished, and all the host of them. \medskip\eq

But in all the host of them, there was no science.  The scientific world could not help but {\it still\/} be without form, and void.
For science is a creation of man, a project not yet finished (and perhaps never finishable)---it is the expression of man's attempt to be less surprised by this God-given world with each succeeding day.

So, upon creation, the society of man set out to discover and form physical laws.  Eventually an undeniable fact came to light:
information gathering about the world is not without a cost.  Our experimentation on the world is not without consequence.  When {\it I\/} learn something about an object, {\it you\/} are forced to revise (toward the direction of more ignorance) what you could have said of it.  It is a world so ``sensitive to the touch'' that---with that knowledge---one might have been tempted to turn the tables, to suspect a priori that there could be no science at all.  Yet undeniably, distilled from the process of our comparing our notes with those of the larger community---each expressing a give and take of someone's information gain and someone else's consequent loss---we have been able to construct a scientific theory of much that we see. The world is volatile to our information gathering, but not so volatile that we have not been able to construct a successful theory of it.  How else could we, ``Be fruitful, and multiply, and replenish the earth, and subdue it?''  The most basic, low-level piece of that understanding is quantum theory.

The {\sl speculation\/} is that quantum theory is the unique expression of this happy circumstance:  it is the best we can say in a world where {\it my\/} information gathering and {\it your\/} information loss go hand in hand.\footnote{Why is that a happy circumstance?  Because it implies in part that the book of Nature may not yet be a written product.  ``The world can be moved.''} It is an expression of the ``laws of thought'' best molded to our lot in life.  What we cannot do anymore is suppose a physical theory that is a direct reflection of the mechanism underneath it all: that mechanism is hidden to the point of our not even being able to speculate about it (in a scientific way). We must instead find comfort in a physical theory that gives us the means for describing what we can {\it know\/} and how that {\it knowledge\/} can change (quantum states and unitary evolution). The task of physics has changed from aspiring to be a static portrait of ``what is'' to being ``the ability to win a bet.''

This speculation defines the larger part of my present research program.
\eq

\subsubsection{A. Some Concrete Problems}

\subsubsection{\protect\hspace*{.2in} Problem {\#}1:~~Pre-Gleason, or Why Orthogonality?}

Andrew Gleason's 1957 theorem is an extremely powerful result for the foundations of quantum theory.  This is because it indicates the extent to which the Born probability rule and even the state-space structure of density operators are {\it dependent\/} upon the theory's other postulates.  Quantum mechanics is a tighter package than one might have first thought.

The formal statement of the theorem runs as follows.  Let ${\cal H}_d$ be a (complex or real) Hilbert space of dimension $d\ge3$, and let ${\cal S}({\cal H}_d)$ denote the set of one-dimensional projectors onto ${\cal H}_d$.  We shall suppose that whatever a ``quantum measurement'' is, it always corresponds to some complete orthogonal subset of ${\cal S}({\cal H}_d)$. Particularly, within each such orthogonal set, the individual projectors are the theoretical expressions for the possible outcomes of the measurement associated with it.

Assume now that it is the task of the theory to assign probabilities to the outcomes of all conceivable measurements. Suppose all that we know of the way it does this is the following: There exists a function \be
p: {\cal S}({\cal H}_d)\longrightarrow[0,1] \label{Heimlich} \ee such that \be \sum_{i=1}^d p(\Pi_i)=1 \label{Maneuver} \ee whenever the projectors $\Pi_i$ form a complete orthonormal set. It might seem a priori that there should be loads of functions $p$ satisfying such a minimal set of properties.  But there isn't.
Gleason's result is that for any such $p$, there exists a density operator $\rho$ such that \be p(\Pi)=\tr(\rho\Pi)\;.
\ee
In words, Gleason's theorem derives the standard Born probability rule {\it and}, in the process, identifies the quantum state-space structure to be the density operators over ${\cal H}_d$. Moreover, he gets this from assumptions that are ostensibly much weaker than either of the end results. This theorem is quite remarkable in that it requires no further conditions on the class of allowed functions $p$ beyond those already stated. In particular, there is not even an assumption of continuity on the functions $p$.

A question on my mind is to what extent, if any, does the structure of this theorem support an information-disturbance foundation for quantum mechanics?  I think this might be fruitfully explored by thinking in the following way.  The assumptions behind Gleason's theorem naturally split into two pieces. (A)~The questions that can be asked of a quantum system {\it only\/} correspond to orthogonal projectors onto ${\cal H}_d$. A consequence of this is that there is no good notion of measuring two distinct questions simultaneously---that is, there is no good notion of an AND operation for two measurements. And (B), it is the task of physical theory to give probabilities for the outcomes of these questions, and we can say at least this much about the probabilities:  They are {\it noncontextual\/} in the sense that, for a given outcome, it does not matter which physical question (i.e., which orthogonal set) we've associated it with. This is the content of the assumption that the probability rule is of the form of a ``frame function'' (a function satisfying Eqs.~(\ref{Heimlich}) and (\ref{Maneuver})).

It seems to me that the first assumption to some extent captures the idea that information gathering is invasive.  If you gather some information and I gather some other information, there is no guarantee that we can put the two pieces of information into a consistent picture: my information gathering has disturbed the relevance of the information you've already gathered. The second assumption, however, appears to be more of the flavor that nevertheless such information gathering is not {\it too\/} invasive.
For otherwise one might imagine the probabilities for a measurement's outcomes to depend upon the full specification of the orthogonal set used in its definition. The Born probability rule clearly has a much weaker dependence on the measurement than it might have had.

A question whose answer could bolster (or discourage) this point of view is the following.  Why is the invasiveness of quantum measurement specifically captured by identifying measurement outcomes with orthogonal sets of projectors?  Hilbert space has a lot of structure; why single out precisely the orthogonal projectors for defining the notion of measurement? To get a handle on this, we could try to see how it might have been otherwise.

As a wild example, consider an imaginary world where quantum measurements are not only associated with orthogonal projectors, but with the projectors onto {\it any\/} complete linearly independent set of vectors.  This would be a notion of measurement that made use solely of the linear structure of ${\cal H}_d$, eschewing any concern for its inner product.  What kinds of probability rules can arise for such a notion of measurement?  In particular, can one have an interesting ``noncontextual'' probability rule in the spirit of Gleason's theorem?  More precisely, what kind of functions $p$ can satisfy Eqs.~(\ref{Heimlich}) and (\ref{Maneuver}) but with the summation in the latter equation satisfied for any linearly independent set?

Well, it's not hard to see that the only noncontextual probability rule that works for all ``measurements'' of this kind would have to be the trivial probability assignment of $1/d$ for each outcome, no matter what the measurement.  To give an example of how to see this, visualize three linearly independent unit vectors $v_1$, $v_2$, and $v_3$ in $R^3$ and imagine assigning them probabilities $p_1$, $p_2$, and $p_3$.  Hold $v_1$ and $v_2$ fixed and rotate the third vector whichever way you wish. As long as it doesn't fall on the two lines spanned by $v_1$ and $v_2$, then the projector associated with it must always be assigned the same probability, namely $p_3$.  Now do the same thing with vector $v_2$, holding $v_1$ and $v_3$ fixed.
This will make almost all vectors on the unit sphere associated with projectors of probability $p_2$, proving that $p_2=p_3$.  Finally, one does the same trick with $v_1$, proving that $p_1=p_2=p_3=1/3$.

The lesson is simple:  if every linearly independent complete set of vectors in ${\cal H}_d$ constituted a measurement, one could not hope to retain a noncontextual probability assignment for measurement outcomes without making the world an awfully dull place!

But maybe this version of the game is just too dumb.  So, let's try to spice it up a bit by explicitly using the inner product structure of ${\cal H}_d$, but in a nonstandard way.  Again consider $R^3$, the smallest Hilbert space on which Gleason's standard theorem can be proved. Suppose now that a ``measurement'' corresponds to any three vectors with a fixed angle relation between themselves. What I'm thinking of here is to start with three vectors $v_1$, $v_2$, and $v_3$ whose angles (moving around them cyclically) are $\alpha$, $\beta$, and $\gamma$.  Now rigidly rotate that structure in all possible ways to generate all possible measurements.  Are there any interesting {\it noncontextual\/} probability rules---again in the spirit of Gleason---that one can associate with this notion of ``measurement?''

Here I don't know the answer.  But I do know of some special cases where one again gets only the trivial assignment $p_1=p_2=p_3=1/3$.
For instance, take the cases where $\alpha=\beta=\gamma$ and $\alpha$ is such that if we rotate around $v_1$, $v_2$ will fall back upon itself after an odd number of ``clicks''---what I mean by a click here is rotating $v_2$ into $v_3$ and so on \ldots\ click, click, click.  What happens if we run through such a ``clicking''
process?  Well, $v_1$ must be constantly associated with the same probability value $p_1$ by the assumption of noncontextuality.  But then by that same assumption, as $v_2$ rotates into the old $v_3$, it must pick up the probability $p_3$.  And so on it will fluctuate up and down: $p_2$, $p_3$, $p_2$, $p_3$ \ldots\ until it finally falls upon its original position.  If this happens in an odd number of clicks, then it will have to be the case that $p_2=p_3$ or the assumption of noncontextuality would be broken.  Similarly we can see that the whole circle generated by rotating $v_2$ and $v_3$ around $v_1$ must be ``colored'' with the same probability value.
Finally, run through the same process but by rotating about the vector $v_2$.  This will generate a second circle that intersects with the first.  From that and the assumption of noncontextuality, it follows that $p_1=p_2=p_3=1/3$ and this will be true of any triad by the very same argumentation.

Another case where one can see the same effect is in the single qubit Hilbert space $C^2$.  There the game would be that any two vectors with a fixed angle $\alpha$ between them would constitute a measurement.  Thinking about the Bloch-sphere representation of $C^2$, one can use the argument similar to the one above to see that whenever $\alpha\ne 90^\circ$ the only possible noncontextual probability assignment is $p_1=p_2=1/2$ for all possible measurements.

\bconj
In the case of $R^3$, whenever one of the angles $\alpha$, $\beta$, or $\gamma$ is not identically $90^\circ$, then the only possible noncontextual probability assignment for measurements outcomes will be the trivial one $p_1=p_2=p_3=1/3$.
\econj

How to tackle such a problem?  I think it may not be too hard actually, especially if one assumes that the noncontextual probability assignments, whatever they are, must be continuous functions. The starting point would be to try to trace through Asher Peres's derivation of the standard Gleason theorem in his textbook.
There, something will surely fail when one looks at the expansion of the proposed ``frame functions'' in terms of spherical harmonics. (I hope someone will bring  Asher's book with them to the meeting: I would bring mine, but I don't have it anymore.)

What's to be learned from this problem?  I'm not quite sure, but I think mostly it will help reinforce the idea that our standard notion of quantum measurement is not simply an arbitrary structure.
It's there for a reason, a reason we still need to ferret out.

\subsubsection{\protect\hspace*{.2in} Problem {\#}2:~~Wootters Revamped with POVMs.}

Bill Wootters in his Ph.D. thesis explored an alternative derivation of the quantum probability rule.  His work was based on the hope that it could be obtained via an extremization principle much in the spirit of the principle of least action in classical mechanics.  (I believe he may talk about this very problem at the meeting.)  The quantity extremized was the Shannon information a measurement reveals about a system's preparation, under the assumption that one has many copies of the system all with identical measurements.

Specifically the scenario was this.  Consider a {\it real\/} Hilbert spaced of dimension $d$ and a fixed orthogonal basis within that space.  One imagines that one has possession of $N$ copies of a quantum system with that Hilbert space, all with precisely the same quantum state $|\psi\rangle$.  Which quantum state?  One drawn randomly with respect to the unique unitarily invariant measure on the rays of ${\cal H}_d$, only one doesn't know which.  The fixed orthogonal basis represents a measurement that one will perform on the separate copies in an attempt to ascertain the unknown preparation.\bigskip

\noindent [{\bf NOTE:}
{\it Here, through the remainder of this problem set, the words were thrown together hurriedly after the fire.}]\bigskip

Anyway, Bill's attempt of a derivation didn't really work so nicely for complex Hilbert spaces.  The question here is, can we make it work after all, if we start thinking of POVMs as a primitive notion of measurement in its own right.

Specifically, the first thing we must ask is does there always exist an informationally complete set of rank-one POVM elements all of equal weighting on ${\cal H}_d$.  And it might be even nicer if that set could taken to have precisely $d^2$ elements. That is, for each $d$, does there exist a set of $d^2$ projectors $|b\rangle\langle
b|$ and a positive number $g$ such that
\be
g\sum_{b=1}^{d^2}|b\rangle\langle b|=I\;?
\ee
We know that there does exist such a set if $d=2$ or $d=3$. When $d=2$, just take any four states corresponding to the vertices of a regular tetrahedron on the Bloch sphere.  For the case $d=3$, Bill has explicitly worked out an example that perhaps he can remind us of. Also I remember some vague murmurings by Armin Uhlmann that ``of course'' they exist in all dimensions. But still, we should treat the existence in all $d$ as an open question---I'm not sure how much of Armin's talk was statement of known fact, how much was conjecture, and how much was faith.

If such a set exists always, then we can ask of it precisely the same question that Bill did in his thesis.  Assume we don't yet know the quantum probability law: we only know that there is some function $f$ for which the probability $p_b$ is given by \be
p_b=f(|\langle\psi|b\rangle|^2)
\ee
when the system's ``unknown'' preparation is $|\psi\rangle$. What function $f$ extremizes the information we gain about $|\psi\rangle$ when we have only one copy of the system available? What function $f$ extremizes the information when we have a very large number of copies available?

What does this have to do with my manifesto?  Perhaps nothing. But I have always felt that Bill's attempt at derivation was missing something in that nowhere in it did it make use of the idea that quantum measurements are invasive beasts:  it talked about information, but it didn't talk about disturbance. If Bill's derivation does turn out to work nicely by the addition of POVMs, then maybe that will be some motivation for me to rethink my ish-ism.

\subsubsection{\protect\hspace*{.2in} Problem {\#}3:~~Post-Gleason, or Should I Think von Neumann Is Special?}

For a long time, I have disliked the tyranny of thinking of von Neumann measurements as more fundamental than other POVMs. Here's a question that might break some of that orthodoxy.

{\bf Suppose} such a set of informationally complete POVMs as described in the last problem exists.  Let us think of the class of all ``primitive'' measurements on ${\cal H}_d$ as those that can be gotten from acting the unitary group on that set.  For instance, for a single qubit, the primitive measurements would correspond to all possible regular tetrahedra draw on the Bloch sphere.

Let us now imagine a notion of ``frame function'' as in Problem
{\#}1 for these kinds of measurements.

Question 1:  Is there a Gleason-like theorem for these structures?
And in, particular does the extra freedom of having $d^2$ outcomes to play with simply the proof of Gleason's result?

Question 2:  By use of the Church of the Larger Hilbert space, can we construct an arbitrary POVM with this notion of primitive measurement.  That is, is there a kind of Neumark extension theorem for this notion of measurement?

\subsubsection{\protect\hspace*{.2in} Problem {\#}4:~~Where Did Bayes Go?}

From my point of view, quantum states are best interpreted as states of knowledge, not states of nature.  Quantum mechanics is {\bf mostly} a ``law of thought'' in that it provides a firm method of reasoning and making probabilistic estimates in light of the fundamental {\bf physical} situation that the world is ``sensitive to our touch.''

With that in mind, I have to ask myself why doesn't wavefunction collapse look more like Bayes' rule for updating probabilities under the acquisition of new information.  Recall Bayes' rule for when we acquire some data $D$ about a hypothesis $H$:
\be
P(H|D)=\frac{P(H)P(D|H)}{P(D)}\;.
\ee
On the other hand, when we perform an efficient POVM $\{E_b\}$ and find outcome $b$, we should update our quantum state $\rho$ according to \be \rho\; \longrightarrow\; \tilde\rho_b= \frac{1}{p_b}U_b E_b^{1/2}\rho E_b^{1/2} U_b^\dagger\;, \ee where $U_b$ is some unitary operator and $p_b=\tr\rho E_b$.

Forgetting about the unitary $U_b$ for the present discussion, notice the difference in expression of these two ``collapse''
rules.  Bayes' rule involves states of knowledge alone:  it is constructed solely of probabilities.  Quantum collapse, on the other hand, appears to involve two distinct kinds of entities: density operators and POVMs.  Can we put it into a form more reminiscent of Bayes' rule and perhaps learn something in the process.

Here's one way that I think might be fruitful.  For each density operator $\rho$ and each POVM $\{E_b\}$, we can construct a canonical decomposition or refinement of $\rho$:  just multiply the equation $I=\sum E_b$ from the left and the right by $\rho^{1/2}$.
We get,
\be
\rho=\sum_b p_b \rho_b\;,
\ee
where
\be
\rho_b= \frac{1}{p_b} \rho^{1/2} E_b \rho^{1/2}\;.
\ee
Note that with this, and just a little bit of algebra, we can rewrite the collapse rule (again forgetting about the $U_b$) to be \be \tilde\rho_b=\rho^{-1/2}\left(\rho^{1/2}\sqrt{\rho^{-1/2}
\rho_b\rho^{-1/2}}\rho^{1/2} \right)^2 \rho^{-1/2} \ee

This expression is, I think, quite intriguing.  This is because it turns that the quantity in the large parentheses above, \be G(\rho_b,\rho)\equiv \rho^{1/2}\sqrt{\rho^{-1/2} \rho_b\rho^{-1/2}}\rho^{1/2} \ee has been characterized independently in the mathematical literature before.  It appears to be the most natural generalization of the notion of ``geometric mean'' from positive numbers to positive operators.  Here are some references:
\begin{enumerate}
\item W.~Pusz and S.~L. Woronowicz, ``Functional Calculus for Sesquilinear Forms and the Purification Map,'' Rep.\ Math.\ Phys.\ {\bf 8}, 159--170 (1975).

\item T.~Ando, ``Concavity of Certain Maps on Positive Definite Matrices and Applications to Hadamard Products,'' Lin.\ Alg.\ App.\ {\bf 26}, 203--241 (1979).

\item F.~Kubo and T.~Ando, ``Means of Positive Linear Operators,''
Mathematische Annalen {\bf 246}, 205--224 (1980).

\item M.~Fiedler and V. Pt\'ak, ``A New Positive Definite Geometric Mean of Two Positive Definite Matrices,'' Lin.\ Alg.\ App.\ {\bf 251}, 1--20 (1997).
\end{enumerate}

Can someone bring these references?  As you know, my copies don't exist anymore.  (If you only have time to copy one, perhaps the last one is best \ldots\ as it gives a large summary of all things known.  Jozsa may be most interested in the first reference by Pusz and Woronowicz.  Kubo and Ando would be the third most useful.)

Something fun to do is to rewrite the classical Bayes' rule in a similar form as this revamped quantum rule.  Indeed the standard geometric mean crops up in precisely this way.  Now try to further the analogy if you can.

One possible fact I seem to recall is that the operator geometric mean can be characterized in the following way.  Start with $\rho_b$ and $\rho$ and consider the following matrix, where $X$ is also a positive semidefinite matrix:
\be
\left(\begin{array}{cc}
\rho &  X\\
X & \rho_b\end{array}\right)
\ee
Then $X=G(\rho_b,\rho)$ is the matrix that maximizes the above matrix in the sense of making it as large as it can be in the standard matrix ordering sense.

Can we learn something about why the collapse rule takes the form it does from this exercise.  (Perhaps Jeff Bub can give us an introduction to his other characterization of the L\"uders collapse
rule.)

\subsubsection{B. Some Not-So Concrete Problems}

\subsubsection{\protect\hspace*{.2in} Proto-Problem {\#}1:~~Computing Power vs.\ Error Correctability.}

We know (suspect) that quantum computing gives us a speed up over classical computing.  But we also know that we have to strain slightly harder to get error correction and fault tolerance for it.
Imagine now the set of all computational models (whatever that might mean).  Within that set will be both classical computing and quantum computing, but also a lot of other things. Could it be the case that quantum computing hits some kind of happy medium, the one where the ratio of speed-up to error correction resources is best?  (You can see that this is directly motivated by my parable:  could the speed-up of quantum computing be due to this world's wonderful sensitivity to our touch?)

\subsubsection{\protect\hspace*{.2in} Proto-Problem {\#}2:~~Entanglement Monogamy and {\Schroedinger}'s Insight.}

Can we think of a way of viewing quantum entanglement as a secondary effect?  The primary effect being information disturbance tradeoff.
Let me just cut and paste an old email here.  [See the note to Todd Brun, titled ``Information Theoretic Entanglement,'' dated 8 June 1999, and the note to Howard Barnum, titled ``It's All About Schmoz,'' dated 30 August 1999, in {\sl Coming of Age with Quantum Information.}]

\subsubsection{\protect\hspace*{.2in} Proto-Problem {\#}3:~~Down with Beables, Up with Dingables.}

Quantum logicians (and presumably Jeff Bub), like to think that quantum mechanics is about ``what is,'' i.e., beables not observables (a phrase coined by Bell).  The only way they can do that is by introducing a kind of logic about the ``facts of the world'' that is different from our usual logic of AND, OR, and NOT.
I, on the other hand, like to think that that kind of change of logic is a strong indicator that we just can't get a notion of a ``free-standing reality'' within physics.

So the question on my mind is what kind of algebraic structure (if
any) does my parable indicate/motivate for a purely algebraic approach.  I've already said, I don't think AND makes any sense at all in such a world.  Do the other notions from standard logic still make sense though?

I realize, I've left too much vague in this question.  Talk to me at the meeting.

\subsubsection{\protect\hspace*{.2in} Proto-Problem {\#}4:~~Intrinsic Characterization of Complete Positivity.}

I really don't like the Church of the Larger Hilbert space. Let's see how far we can get toward CPMs and POVMs without ever having to assume it.  Is there a physically motivated criterion for CPMs that makes no reference to the Church.

\subsubsection{\protect\hspace*{.2in} Proto-Problem {\#}5:~~Doing Gleason with Algebraic Numbers.}

Problem motivated by Bennett remark.  Recently Pitowsky and Meyer have raised some interesting questions about Gleason's theorem when the number field of quantum mechanics is restricted to the rationals.  It might be useful for our understanding of QM to ferret out what is essential and what is not in its formulation. Cabello and Peres completely discount Pitowsky and Meyer because their world has no superposition principle.  But who cares? Well, maybe if Gleason still works in the minimal world with a superposition principle (i.e., algebraic number fields), we should think harder about the meaning of the ``superposition principle.''

\subsubsection{\protect\hspace*{.2in} Proto-Problem {\#}6:~~Challenge to Everettistas.}

\begin{itemize}
\item
Would your interpretation still work if the number field of ${\cal H}_d$ were the reals instead of the complexes?  If it were the rationals instead?
\item
Would your interpretation still work if the time evolution of the universe as a whole were nonlinear instead of unitary?
\item
Would your interpretation still work if the collapse rule of QM were anything different from the standard one?
\item
Would your interpretation still work if \underline{\hspace{.5in}}?
[You fill in the blank, challenge yourself.] \end{itemize}

I understand that I'm being belligerent, but I {\it suspect\/} your answers to each of these will be ``yes.''  Cf.\ Any of David Lewis philosophical works on the modal logics and the plurality of worlds, or Max Tegmark's paper \arxiv{gr-qc/9704009}, ``Is the `theory of everything' merely the ultimate ensemble theory?''  The tentative conclusion I draw from this is that the Everettista has a contentless interpretation.
\eq

\section{12-09-02 \ \ {\it The Copenhagen Interpretation} \ \ (to C. King)} \label{King5}

I'm in Copenhagen at the moment and will be here until next Tuesday.  I'm giving two talks in the Math department---one tomorrow and one Monday---and then I've got an appointment at the Niels Bohr Institute Monday afternoon.  With some luck, there's a chance I'll get to talk to Niels Bohr's son Aage (Nobel prize sometime in the 1950s or 60s).  He's recently turned his attention to quantum foundations.  In any case, I'm definitely talking with his collaborator Ole Ulfbeck.  If the Bohr thing comes through, I'll call this my best vacation ever!

\section{17-09-02 \ \ {\it Land of the Rising Sun} \ \ (to G. L. Comer)} \label{Comer20}

Sorry for the silence; I got the note you forwarded from your friend.  I'm presently over the Atlantic again; I'm making my way home from 11 days in Sweden and Copenhagen.  I gave four talks there, and with the discussions and other matters beyond that, I guess I'm fairly exhausted.

Let me tell you though, Bohr was with me on this trip.  I've never been to a more wonderful city.  It's its own land of the rising sun.  I'd make it my home if it would have me.  I spent a little time at the Bohr Institute and, wow, what a place.  I was so taken with just walking into the empty lecture hall where literally the first lectures on quantum mechanics were given by Heisenberg, Pauli, Dirac, and Bohr himself.  I also had long discussions with Ole Ulfbeck who, with Aage Bohr (Bohr's son, with a Nobel prize of his own in 1975), is now speaking rather badly of the ``Copenhagen Interpretation.''  They have an article in {\sl Foundations of Physics\/} a few months ago, where young Bohr essentially retracts from his father's position.  Sad really.  They've got one decent idea in that paper that I like a little, but mostly it is empty.

My friend Jeff took a pretty good picture of me standing in front of a bust of Bohr; I'll try to send that to you soon.

I've been reading a volume of the Jung--Pauli letters on this trip.  (They've finally been translated into English.)  They're quite impressive and even a little eerie.

\section{17-09-02 \ \ {\it Wow!}\ \ \ (to G. Brassard)} \label{Brassard17}

Wow, read this introduction to Appleby's paper!!!!  You've just got to tell me that we can have him at the meeting (and pay his way if need be)!  (I feel like a kid in a toy store, screaming, ``Daddy, daddy, I've just got to have that toy!'')

\bq
\noindent {\bf Introduction}\medskip

The stimulus for this talk came from a number of conversations with Chris Fuchs about his information-theoretic approach to the interpretation of quantum mechanics (include references).

Fuchs starts from the position that the quantum state simply is a probability distribution\footnote{Specifically:  it is the probability distribution for the outcomes of an informationally complete (generalized) measurement.}:  that, and nothing more whatever.  He also takes a radically Bayesian approach to the interpretation of probability\footnote{The interpretation of probability gives rise to almost as many conflicting views as does the interpretation of quantum mechanics.  So saying that the wave-function  simply is a probability distribution is not yet a complete answer.  It is also necessary to say what a probability distribution is---which is not so easy (at least, it does not seem easy to me).}.  According to him the quantum state has a \emph{purely} subjective significance:  it simply represents one's expectations (or beliefs, or gambling commitments) regarding the
outcome of a measurement.   On these assumptions the collapse of the
wave-function is no more remarkable than the collapse of a classical probability distribution when new information is acquired---no more remarkable than, for example, the probability of a coin landing heads being 1/2 before the toss, but either 1 or 0 afterwards.

Of course, this proposition, that the wave-function should not be interpreted  as an objectively real entity, is the starting point of
the Copenhagen Interpretation.   So it might appear from the above
that Fuchs is proposing nothing very new.  However, I think that that would be wrong.  It seems to me that Fuchs's programme is best described as an attempt to reinvigorate the Copenhagen
Interpretation.  To breathe new life into the old bones.   The
qualification is crucial.  If Fuchs's hopes were fulfilled---indeed, if his hopes were only half-fulfilled---the consequences would, I think, be revolutionary.

The term ``Copenhagen Interpretation'' ,  as it has previously been
understood, is not very sharply defined.   To begin with, it isn't
really a single interpretation at all.  Rather it is a family of interpretations, held together by certain common features. Moreover, even if one focuses on a particular variant, one finds numerous obscurities.  The writings of Bohr, in particular, are notoriously difficult to follow.

If I have understood him correctly Fuchs would argue that what is contained in the work of the Copenhagen school is, not so much a completed interpretation, more a project: a direction of thought which, if pursued with sufficient determination, might eventually \emph{lead} to a satisfactory understanding of what quantum mechanics is really telling us.  Unfortunately, the programme which they initiated, during the heroic age of the subject, has languished ever since.

Fuchs suggests that if we were really to follow through on the basic insights of the Copenhagen school it might have major implications, not just for our philosophical understanding, but on an empirical level.  As he puts it
\begin{quote}
I have this ``madly optimistic'' (Mermin called it) feeling that Bohrian-Paulian ideas will lead us to the next stage of physics.
That is, that thinking about quantum foundations from their point of view will be the beginning of a new path, not the end of an old one.
(samizdat, p.\ 173)
\end{quote}
I find these ideas deeply interesting.  I share Fuchs's feeling that the present situation in quantum foundations is one of ``impasse''.
His programme excites in me the hope that this could be a way to break out of the sterile circle.

Before becoming acquainted with Fuchs's ideas I was definitely not an adherent of the  Copenhagen philosophy\footnote{This is not to say that I preferred some other interpretation.  I belonged to the school of the frankly perplexed, which does not pretend to know how quantum mechanics should be interpreted.}.  It was not so much that I disagreed with the Copenhagen Interpretation.  I could not get as far as that.  I felt that it was not sufficiently clear for me to be able to tell whether I agreed or disagreed.  In other words, I took the same view as Bell:  the Copenhagen philosophy struck me as ``unprofessionally vague and ambiguous'' (\emph{Speakable}, p.\ 173).    The degree of vagueness and ambiguity is, in places, such
that agreeing to these propositions would be like signing a blank
cheque:  you might be committing yourself to almost anything (I think Beller scores some palpable hits in this respect).

However, Fuchs's ideas have caused me to revise this assessment.  It is not, let me hasten to say, that, where I formerly saw vagueness and ambiguity, I now see sharpness and rigorous exactitude.  On the
contrary:  the Copenhagen Interpretation remains, in my view, as obscure as ever it was.  And I continue to think that this means it cannot be seen, in the way Bohr and others wanted it to be seen, as the terminus of all intelligible thought on the subject.  If, however, one looks at it from the standpoint Fuchs suggests, not as a \emph{finish}, but as a \emph{start}---not as a completed interpretation, but as a project---then the ideas of the Copenhagen
School appear much more interesting.   What I had not properly
appreciated, before reading Fuchs, was the extent to which my rejection depended on the claim to complete finality.  If one subtracts this claim---if, instead of regarding Bohr as the voice of unimpeachable authority\footnote{Which, it would seem, is how he all too often was regarded, in the past.  For a discussion of the ``overpowering,  almost disabling, impact of Bohr's authority'', see Beller [add reference].}, one merely sees him as someone groping toward ideas which he himself did not adequately comprehend---then I think one can easily acknowledge, without doing any damage to one's logical conscience, that there is much in Bohr's writings which is deeply suggestive.  Seen in this light the obscurities in Bohr's account may appear, less as grounds for rejection, more as a challenge.  One may be stimulated to think that, concealed among the obscurities, there could be the germs of something cogent, which it might be worth one's while to try to ferret out.

Reading Fuchs has not persuaded me of the basic correctness of the Copenhagen philosophy. It has not even persuaded me of the correctness of a kind of neo- or revitalized Copenhagenism.  Such a statement of belief would, in the circumstances, be premature.  In any case, it seems to me that, if history shows anything, it shows that we ought to be extremely cautious about making unreserved commitments at the level of basic concepts.  Like Fuchs, I do not believe in final theories.

However, what I am persuaded of is that Fuchs has opened up an extremely interesting line of thought, which deserves to pursued very seriously.  I do not know where his programme will lead.  But even if it does not lead in precisely the direction he now envisages, I feel it promises to take us significantly further forward.
\eq

\section{17-09-02 \ \ {\it No-Cloning}\ \ \ (to S. L. Braunstein)} \label{Braunstein6}

Not only does he miss the full result, but there is a line in there where he actually gets it wrong.  For the full story, see page 451 of my samizdat, \quantph{0105039}.

I'm in Heathrow at the moment, heading home from my first trip to Copenhagen.  What a wonderful place!  Copenhagen, that is.

\section{17-09-02 \ \ {\it Wigner}\ \ \ (to S. L. Braunstein)} \label{Braunstein7}

You still don't get it:  Wigner actually gets it WRONG.  There is a line in the paper where he says, (essentially) ``all linear superpositions of living states must also be living states.''  That cannot be; that's the point of the no-cloning theorem.  But he didn't even realize that he made a mistake.

\section{19-09-02 \ \ {\it Chris's Blurb} \ \ (to R. Pike)} \label{Pike9}

Steven van Enk is indeed a physicist.  I don't know what I am:  physicists tend to call me a mathematician, while mathematicians generally call me a physicist.  Information theorists call me a philosopher.  But I practice quantum information.  Predominantly I have been interested in formalizing and quantifying the idea that information gathering causes a disturbance when it comes to quantum phenomena.  That idea has steered me to several problems in the field, including:  developing criteria for successful quantum teleportation, calculations in quantum channel capacity theory, work in quantifying quantum entanglement, and exploring novel notions of quantum nonlocality.

\section{19-09-02 \ \ {\it Unsurprising Fact} \ \ (to N. D. {\Mermin})} \label{Mermin74}

\bdm
Did you know that Bob Griffiths believes that within each and every
one of his frameworks there is a ``history that actually occurs''?
\edm

Yes, I did.  He says so quite explicitly in the {\sl Scientific
American\/} article (or was it {\sl Physics Today}?) he wrote with
{\Omnes}. I talked with {\Omnes} at length about it at the NATO meeting in
Greece you sent me to a couple of years ago, picking particularly on
this point.

I'm not sure whether I've told you this in the past, but I am
convinced that the consistent historians do nothing beyond what (some
interpretations) of Bohr already do.  Focus for the moment on a
SINGLE standard observable.  If one considers that observable in
isolation---i.e., without consideration of all other possible
observables---then there is absolutely nothing to stop one from
acting AS IF one of the values of the observable obtains and all the
others don't.  Where one runs into trouble (via Kochen--Specker,
etc.)\ is if one tries to hold that view for all observables
simultaneously. Now, what do the consistent historians do?  Instead
of playing the game above for a single standard observable they do it
for a single so-called ``consistent set of histories.''  But from my
point of view, all a consistent set of histories is, is an
arbitrarily singled-out kind of multi-indexed POVM.  The point being,
woop-ti-do! You give me {\it any\/} POVM, and there is nothing to
stop me from acting {\it as if\/} one of its values obtain and all
the others don't. I would only run into trouble (through KS, etc.)\
if I were to try to play this game for various noncommuting POVMs
simultaneously. When Griffiths and company command that one cannot
consider distinct sets of consistent histories simultaneously, all
they are doing is what I could already have done with {\it any\/}
POVM. There is nothing deep there.

\section{19-09-02 \ \ {\it Black Sheep} \ \ (to A. Kent)} \label{Kent5}

\bak
I think there's lots of interest to talk about in quantum foundations,
and of course in quantum information, but I'm not at all sure that
quantum information actually sheds any new light on the deeper
problems of quantum foundations.  Is this heresy tolerable?!?
\eak

Tolerable, and in fact, maybe welcome.  That can be your theme for the week and the subject of your talk.  If you can defend it well, that'll surely lead to some good discussions.

\section{20-09-02 \ \ {\it How's It Look?}\ \ \ (to R. Pike)} \label{Pike10}

I just threw thing below together.  I'm going to lunch now, but when I get back I'll put the finishing touches on (if there are finishing touches to put).  Tell me what you think.
\bq
\begin{center}
{\bf Employee Report on Activities and Accomplishments} \\  Oct.\ 2001 -- Sep.\ 2002\medskip
\end{center}

This year has been another good one for science.  I had eight articles going to press, all of them confirming the thread of thought that has become so important to me:  the predominant structure of quantum information is automatically and deeply connected to the structure of Bayesian probability theory.  Where the two structures differ is where one should look for the power of quantum information as an information-processing and communication resource.  On the practical side, this has implications for understanding quantum information experiments performed with standard laser light, such as the quantum teleportation experiment performed at Caltech and more recently at the Australian National University.

I consider two results my better ones of the year.  The first had to do with the settling of a long drawn-out debate in the literature concerning the compatibility of quantum-state assignments made by different experimentalists with differing amounts or kinds of information about a single quantum phenomena.  (This work is soon to appear in Physical Review A, co-authored with C. M. {\Caves} and R. {\Schack}.)  The second had to do with the precise form of the quantum-state change rule under the gathering of experimental data.  This phenomenon is normally mentioned in conjunction with the ``von Neumann collapse postulate.''  Using a little-observed fact from linear algebra, I was able to express the collapse in a way that makes it look {\it almost\/} identical to standard probabilistic Bayesian conditionalization.  This I believe sheds significant light on the process of quantum-state-change-through-measurement, and, consequently, as already alluded to, points toward the key ingredient in the power of quantum information processing.  The evidence that the quantum information community also sees these efforts as shedding light can be found in the number of invitations to speak I have been given this year.

I count having given 14 external invited talks this year. These talks have carried me to Japan, Ireland, Australia, Spain, Sweden, and Denmark, all at zero cost to Lucent.  On the official-recognition side of things, I was awarded an (international) E.~T.~S. Walton Visitor Award by the Science Foundation Ireland, which would translate into a 20\% increase in salary and substantial equipment and travel funds, if accepted.  (Presently the award is being renegotiated into a small series of small visits.)  I made three media appearances, via short interviews in {\sl PC Magazine}, {\sl PC Magazine Online}, and the book {\sl The Best American Science Writing 2002}.

Concerning general service to the physics community, I served as an Associate Editor for the Rinton Press journal {\sl Quantum Information and Computation}, as a member of the Advisory Board for the International Center for Mathematical Modeling in Physics, Engineering and Cognitive Sciences at {\Vaxjo} University, Sweden, and as a member of the Advisory and Award Committee for the Sixth International Conference on Quantum Communication, Measurement and Computing (QCMC'02). I co-organized an international meeting {\sl Workshop on Quantum Foundations in the Light of Quantum Information}, to be held at University of {\Montreal}, October 13 -- November 3 this year, co-edited a special issue of the journal {\sl Studies in History and Philosophy of Modern Physics\/} devoted to quantum information and quantum computation, and mentored two visiting graduate students for DIMACS for six weeks this year.

As internal service goes, I mentored a summer intern for the SRP program, I gave two talks for Lucent customers, France Telecom and SBC, and attended the Army Research Office annual review of funding for quantum computation and quantum information to strengthen the ties between Bell Labs and government funding sources.  Finally, I played a significant part in a lively visitor program of quantum information scientists of all walks to present their research to the Bell Labs community.
\eq

\section{23-09-02 \ \ {\it Genuine Fortuitousness} \ \ (to H. J. Folse)} \label{Folse12}

I haven't talked to you in a while.  I was in Copenhagen last week, and I got a chance to meet with Ole Ulfbeck, who has recently written a quantum foundational paper with Aage Bohr titled ``Genuine Fortuitousness: Where Did That Click Come From?''  Have you seen it?  If not, you can find it in Foundations of Physics {\bf 31}, 757--774 (2001).

In it, young Bohr makes a quite sharp turn around from his father's position.  It's kind of a shame, and as I understand it all, not a system worked out to any great extent.  Just wondering if you've got any opinions.

By the way, I noticed that you never submitted you {\Vaxjo} paper to the {\tt quant-ph} archive.  I think you ought to do that since it is already written in \LaTeX.  If the concept frightens you, I could submit it for you; it's a rather easy procedure and would only take about 10 minutes of my time.  (Though, of course it would be best if you did it, so as to retain control, etc.)

\subsection{Henry's Reply}

\bq
Right now Isidore is bearing down on New Orleans, so they've canceled classes and I have time to catch up on email.  Sorry I didn't know you were in Copenhagen, I would be glad to provide an introduction to Jan Faye at Kobenhavns Universitetet, whose work you know.  I've also been corresponding recently with Erik Rudinger who is teaching a course on Bohr, I presume at NBI, and who edited a lot of the Archives.  He is a mine of Bohr information and I'd be glad to provide an introduction there as well.

Yes, I did have that Ulfbeck and A. Bohr article called to my attention.  [\ldots]

It definitely appears to me that this view rejects a big portion of Father Bohr's commitment to the reality of atoms and embraces an extreme anti-realist attitude which I, for one, do not attribute to Niels Bohr, though of course many others have indeed taken to be his view.  But at the same time I do see a lot of bits of the father's ways of thinking poking through the mix.  Niels constantly reflected on the theme of the limits of ``space-time description'' and in the period from 1913 to 1925 he  wrote repeated bits and pieces with the title ``Space-time Description'' (``Rum-tids Beskrivelse,'' in Danish).  In the end, i.e.\ by Como, he concluded it stood in a complementary relation to ``the claims of causality.''  I think this period is very revealing precisely because there was then no formalism to reason from, and one had to think out experimental interactions (observations) physically, so to speak, without any mathematical algorithm to help (or worry about).  Already in his 1913 model Niels {\it knew\/} that the electrons in an atom couldn't be considered as ``real particles'' in the classical sense.  Talk about the electrons ``orbiting'' the nucleus was thus already then not to be taken in a classical realist sense.  The discontinuity in change of state of atomic systems in an interaction between, as he put it, ``radiation and matter'' implied it was meaningless to ask whether the electron in one orbit before the ``jump'' was the same as the electron in a new ``orbit'' after the jump, and of course the attempt to trace the electron's spatio-temporal career through this interaction is explicitly forbidden.  So Bohr concluded that talk about elementary particles as though they were the sorts of things that traced classical trajectories was used solely for its instrumental value in ``interpreting'' the interaction as a ``measurement'' determining some property of the ``object'' system.  The interpretation of elder Bohr that I have defended is that he rejects regarding the citizens of the microdomain as realistically represented by ``particle'' or ``wave'' ``pictures'' (the ``accustomed demands for visualizability''), but that he nonetheless regards such entities as {\it real\/} (not as instrumental fictions) because they are one of the interactors in the ``indivisible'' interaction with the observing instruments which provide the empirical evidence for the theory.  Electrons, in short, are {\it real}, but they're not {\it really\/} particles (or waves, either).

Now U\&B clearly reject the reality of particles in a more uncompromising way.  I think this is a version of the class of interpretations you have categorized as ``correlations without correlata,'' and they of course do refer to Mermin's article. (What did he think about U\&B?)  In philosophical jargon we would say it is a macrorealism combined with a micro anti-realism.  ``Clicks'' are real; but electrons are just, as it were, ``the set of all electron clicks.''  Of course they're able to dispense with any commitment to a microreality only by paying a very high price in insisting that the occurrence of such clicks is ``genuinely fortuitous.''  Immediately after defining this phrase, they write: ``What makes the click `irrational' is seen to be that the variable, which manifests itself by the click, never enters spacetime.'' (p.\ 762) (The reference `irrational' struck me because Father Bohr was fond of saying ``the quantum postulate with its inherent irrationality'' which a lot of people completely misunderstood in a positively perverse way.)  Now this seems to me to be a very strange way of speaking; indeed, I get the feeling that their ``genuine'' fortuitousness, isn't so genuine after all.

We are supposed to undercut any problems with crossing the bridge from the formalism to the physics because ``matrix variables'' never ``enter spacetime.''  However, when they ``manifest themselves'' in spacetime, they do so with good old fashioned classical single valued measured outcomes.  It all sounds quite ghostly: if I am told ``poltergeists never enter spacetime but they manifest themselves by making noises and making doors creak open'' then I understand this to mean they are non-physical (out of spacetime) {\it causes\/} for physical (empirically observable) effects.  Thus it seems to me that they're really just {\it saying\/} the ``click'' is ``genuinely fortuitous'' but at the same time talking about how it is ``caused'' by the ``manifestation'' of a ``matrix variable'' when it ``enters'' spacetime.  Talk of ``variables'' of course originally was a way of talking about {\it properties\/} of macrophysical real entities which were functions of time.  Now it seems that matrix variables are a way of talking about properties of outside-of-spacetime entities (aka ``atomic systems'' or ``systems in quantum mechanical states'').  If clicks are real, they are the ``manifest{\it ed},''  but if the manifest{\it ed\/} are real, that implies the reality of a manifest{\it or}, the thing being ``manifest{\it ed}'' when it ``enters'' spacetime.  So what began as an anti-realist ontological minimalism (only click events are real) blossoms into a whole 'nother ``reality'' which is characterized by ``matrix variables'' in a way somehow analogous to the way old time classical reality was characterized by single valued variables.  We see only the surface of the pond, but the bubbles and eddies on its surface are the manifestations of the doings of submerged beings beneath that spacetime surface.  I expect U\&B would reject that, and say that's only a way of speaking to say that ``clicks'' are such ``manifestations,'' but the point is that it is damn hard to do physics when you stop talking about ``causes,'' and the temptation always lurks to simply paper over the problem with an alternative locution. [Kant for example, forbad any application of the category of ``causality'' transcending the phenomenal, empirical world.  But at the same time, in order to cling to realism and not cave into idealism, he wanted to see these phenomena as the ``manifestations'' of ``things-in-themselves'' not experienced but necessarily ``posited'' by the cognizing subject.  He realized, of course, he couldn't say ``noumena {\it cause\/} phenomena'' so instead he started talking about the noumenon as the ``ground'' of the phenomenon.  But everyone saw this as a transparent ploy to circumvent the strictures on ``causality,'' and in point of fact in history philosophy did cave in to idealism.]

The same things happens when U\&B talk about ``associating'' one set of ``source clicks'' with another set of ``measurement clicks'' in the case of alpha decay of uranium. (They also use expressions like ``give rise to,'' etc.) The phenomenon of ``alpha decay'' manifested in our macro-surface-of-the-pond  world of spacetime refers to, we are told,  a certain distribution of clicks.  With that phenomenon we {\it associate\/} another set of macro clicks which we refer to as the phenomenon of ``a lump of pure uranium turning into half thorium after half-life $t$.''  But if ``causes'' is treated as ``constant conjunction in time'' we have in effect just said the uranium's transmutation into thorium is the cause of the alpha radiation.  Moreover, we must still make use of particle pictures in designing the ``clicker'' as a detector of the sort of ``particles'' we are looking for, a fact which the elder Bohr certainly would have stressed in his talk about interaction. We can know of causal connections quite independently of knowing anything about the mechanism by which such connections are affected, or indeed if there be any mechanisms at all. We knew sexual intercourse causes babies long before we knew any biology. All U\&B have really professed here seems to me to be that uranium decay causes emission of alpha particles, but we don't know the mechanism by which it is accomplished, and we do know whatever it might be it isn't a classical one capable of being predicted deterministically. Instead we speak of these processes using the language of ``manifesting,'' ``entering spacetime,'' ``giving rise to,'' etc. So it seems, when all is said and done, quantum mysteries abide.  If this be demystification, it seems more like demystification by {\it fiat}, rather than one earned by explaining the mysteries. Not that this does not happen in the history of physics.

You might be interested to learn that the biennial meeting of the Philosophy of Science Association this coming November features no less than two sections on Bohr and the Copenhagen Interpretation, one of which is chaired by Jeff Bub, whom you know, of course.  The other is devoted to Mara Beller's inflammatory {\sl Quantum Dialogues}.  So it seems that interest in Bohr is heating up again.  The program is at: \myurl{http://www.pitt.edu/~psa2002/}.
It would no doubt be as weird to you as {\Vaxjo} was to me, but it'd be fun if you could come.

To change subject abruptly: I obviously do not keep up with the {\tt quant-ph} archive and am only dimly aware of it through your references.  I would greatly appreciate your posting the paper there.  What do you need me to do?

BTW, in connection with the Proceedings of the {\Vaxjo} conference I got an email from Andrei in June saying that ``Soon we sill be able to send you copies of this volume.''  I have never received any copy.  Did they actually send them?  Did I miss my copy?

Oh, also, did you ever get together with van Fraassen, and what happened if and when you did?
\eq

\section{23-09-02 \ \ {\it Appleby Confirmed} \ \ (to G. Brassard)} \label{Brassard18}

So, Appleby is confirmed for the last week of the meeting.  Thus, basically, the end of the meeting is full now.  The issue now is what is going to happen at the front of it?  It looks like we've got loads of room there.  (Though I still wait to hear from you how many people we can really house.)

If I don't hear from you as the day passes, I'll probably be calling you!

\section{23-09-02 \ \ {\it Your Future} \ \ (to R. W. {\Spekkens})} \label{Spekkens8}

\brws
The more I think about quantum foundations, the more I become sure that you are on the right track.  For a variety of reasons, I am now absolutely convinced that the quantum state represents information, and that we are on the verge of major advances in our understanding of quantum mechanics.  So, what's the news on funding?
\erws

You know, even if it weren't obvious that you were pulling my strings, I'd still want you as a postdoc.  But reality is not completely shaped by wants.  Tell me about your back-up plans. [\ldots]

I'll be in Waterloo, Oct 2--8, giving talk at Perimeter Oct 4 and a talk at IQC Oct 7.  Is there a chance I could visit with you then?  Are you very far down the road?  I'll probably have a lot of time on the weekend (the 6th in particular).

\section{23-09-02 \ \ {\it Let Me Count the Ways \ldots} \ \ (to R. W. {\Spekkens})} \label{Spekkens9}

\brws
First of all, I am not simply pulling your strings.  I am absolutely
sincere about my conviction that the state vector is a representation
of information.  (I'm not about to follow you in your attitudes on
realism though.)
\erws

How many times (and how many ways) do I have to say I'm a realist?!?!  I'm just a more subtle realist than most.  (Read the anti-{\Vaxjo} interpretation paper again.)

\section{24-09-02 \ \ {\it Title and Abstract} \ \ (to D. Poulin)} \label{Poulin5}

Does this sound like the sort of thing you guys might want to hear?  If not, I could probably put together a new talk on an information function Sasaki and I have been studying.  But the present talk has the advantage for me that half of it has already been prepared!
\bq
\noindent {\bf Title:} Exchangeable Quantum States / Exchangeable Quantum Operations\smallskip\\
{\bf Abstract:}
There is hardly a paper in quantum information that does not make use of the idea of an unknown quantum state.  Unknown quantum states are protected with quantum error correction, teleported, and used to check for quantum eavesdropping.  But what does the term ``unknown state'' mean?  In this talk, I focus on quantum-state tomography and make sense of the term in a way that breaks with the vernacular:  An unknown quantum state can always be viewed as a known state---albeit a mixed state---on a larger multi-trial Hilbert space.  The technical result is a quantum mechanical version of the de Finetti representation theorem from classical probability.  Interestingly, the theorem fails for real and quaternionic Hilbert spaces, and this teaches us something new about entanglement.  Furthermore, a variation of the theorem applies to quantum operations, where it tells us something about the nature of complete positivity.  The implications of both theorems for the point of view that quantum states represent states of knowledge, rather than states of nature, may be discussed briefly, time permitting.
\eq

\section{24-09-02 \ \ {\it Carriers of the Torch} \ \ (to A. Peres)} \label{Peres42}

\bap
Have you seen my \quantph{0209114}? It has no relation to {\tt quant-ph}. I had
tried to put it in {\tt hep-th}, but the {\tt arXiv} manager thought it was a joke
and refused to publish it. After lengthy negotiations, with the help of
Mermin, I was told to put it in {\tt quant-ph}, whose readers know my name.
If I am not concerned about my good reputation with {\tt quant-ph}, I'll be
able to cross reference it in {\tt hep-th} (so I did, also {\tt gr-qc} and {\tt astro-ph}).
\eap
I just had a read of it; thank you for pointing it out.  I'll think about the point you make.

\section{25-09-02 \ \ {\it Ulfbeck and Bohr} \ \ (to N. D. {\Mermin})} \label{Mermin75}

\bdm
While I don't think they're [Ulfbeck and Bohr] foolish by any means,
I was not terribly excited by what they had to say.  If you were, by
all means invite them.
\edm

I'm finally getting a chance to clean up my mailbox.  Sorry for the
long absence.

After writing you the note referred to above, I got a chance to
actually read the Ulfbeck/Bohr paper.  Before that, I had only had
discussions with Ulfbeck.  Here's what really surprised me.  In
talking to him, I got the impression that he was giving me something
of a pr\'ecis of their views, and that I would find the heart and the
details in their paper.  But I didn't!  In fact, as far as I could
tell, I didn't find anything more in their paper that I didn't
already find in our short conversation \ldots\ only said six times
over in paper form.  ``The older view did not adequately account for
the genuine fortuitousness of the measurement click.''

I know I've complained about Father Bohr's lack of detail when
asserting the origin of the quantum formalism, but I think they force
my complaints to a new level.

There is, however, one idea in the paper that I am inclined to keep
or, at least to me, seems worth trying to develop.  I say this
predominantly because of its William {\James}ian feel.  Here it is,
deleting the words of theirs that I don't like or don't agree with,
\bq
\noindent The click with its onset is seen to be an event entirely
beyond law. \ldots\ [I]t is a unique event that never repeats \ldots\
The uniqueness of the click, as an integral part of genuine
fortuitousness, refers to the click in its entirety, with all the
complexity required for a break-through onto the spacetime scene.
\ldots\ [T]he very occurrence of laws governing the clicks is
contingent on a lowered resolution.
\eq

You see, from the {\James}ian viewpoint of ``radical pluralism,'' every
piece of the universe, every crumb of its existence, is a unique
entity unto itself.  Here's a little quote in that direction from his
essay ``Abstractionism and `Relativismus'\,'':
\bq
Let me give the name of `vicious abstractionism' to a way of using
concepts which may be thus described: We conceive a concrete
situation by singling out some salient or important feature in it,
and classing it under that; then, instead of adding to its previous
characters all the positive consequences which the new way of
conceiving it may bring, we proceed to use our concept privatively;
reducing the originally rich phenomenon to the naked suggestions of
that name abstractly taken, treating it as a case of `nothing but'
that, concept, and acting as if all the other characters from out of
which the concept is abstracted were expunged. Abstraction,
functioning in this way, becomes a means of arrest far more than a
means of advance in thought. It mutilates things; it creates
difficulties and finds impossibilities; and more than half the
trouble that metaphysicians and logicians give themselves over the
paradoxes and dialectic puzzles of the universe may, I am convinced,
be traced to this relatively simple source. {\it The viciously
privative employment of abstract characters and class names\/} is, I
am persuaded, one of the great original sins of the rationalistic
mind.
\eq

I wish I could find a better quote than that---I have memories of
reading the idea expressed in much greater detail and so much more
eloquently---but this morning, try as I might, I can't find it.

So I'll end this little note with another note I wrote a few months
ago---it carries the sentiment, if not the eloquence.  [See 23-04-02
  note ``\myref{Comer12}{Music in the Musician}'' to G. Comer.]  It's pasted
below. Maybe I should have titled the present article, ``A Click is
but a Click Not: it is so much more.'' For the same holds with
``clicks'' as with ``atoms.''

\section{26-09-02 \ \ {\it More Ulfbeck and Bohr} \ \ (to H. J. Folse)} \label{Folse13}

I hope things are going well for you in spite of Isidore.  I see from the weather channel that it made an early-morning landfall in Louisiana.

Thanks for your long assessment of the Ulfbeck/Bohr paper.  I enjoyed reading it, and think I agree with many (all, possibly) of the things you said.  You asked what was Mermin's reaction to it.  The only record I have (other than that he called their ideas ``very Mohrhoffian,'' referring to Ulrich Mohrhoff) is contained in the note below.  [See 25-09-02 note ``\myref{Mermin75}{Ulfbeck and Bohr}'' to N. D. {\Mermin} and 23-04-02 note ``\myref{Comer12}{Music in the Musician}'' to G. L. Comer.]

Unfortunately, I think it is going to be almost impossible for me to attend the Pittsburgh meeting.  I'll be in {\Montreal} Oct 13--Nov 3 running an extended workshop on ``Quantum Foundations in the Light of Quantum Information''---I've got about 30 people coming on and off this time---and there are limits to how far I can push my family.  But, boy, I sure would like to be at those Bohr sessions.

\subsection{Henry's Reply}

\bq
\bhf
Quoting Chris Fuchs:
\bq\noindent
{\rm There is, however, one idea in the paper that I am inclined to keep or,
at least to me, seems worth trying to develop.  I say this predominantly
because of its William Jamesian feel.  Here it is, deleting the words of
theirs that I don't like or don't agree with,
\bq\noindent
  ``The click with its onset is seen to be an event entirely beyond law.
   \ldots\ [I]t is a unique event that never repeats \ldots\ The uniqueness of
   the click, as an integral part of genuine fortuitousness, refers to
   the click in its entirety, with all the complexity required for a
   break-through onto the spacetime scene. \ldots\ [T]he very occurrence of
   laws governing the clicks is contingent on a lowered resolution.''
\eq
}
\eq
\ehf
This part of Aage's view is consistent with his father's.  In a recent discussion of this, Erik Rudinger gave me a bunch of quotes from Niels: As a collector of quotes, I thought you might like these:
\bq\noindent
``In this connection, it is also essential to remember that all unambiguous information concerning atomic objects is derived from permanent marks -such as a spot on a photographic plate, caused by the impact of an electron- left on the bodies which define the experimental conditions. Far from involving any special intricacy, the irreversible amplification effects on which the recording of the presence of atomic objects rests rather remind us of the essential irreversibility inherent in the very concept of observation. The description of atomic phenomena has in these respects a perfectly objective character, in the sense that no explicit reference is made to any individual observer and that therefore, with proper regard to relativistic exigencies, no ambiguity is involved in the communication of information.'' --- (Mid-Century, p.\ 310 -- BCW 7, p.\ [390])
\eq
\bq\noindent
``\ldots\ the emphasis on permanent recordings under well-defined experimental conditions'' --- (Mid-Century, p.\ 313 -- BCW 7, p.\ [393])
\eq
\bq\noindent
``Moreover, the circumstance that all such observations involve processes of
essentially irreversible character lends to each phenomenon just that inherent
feature of completion which is demanded for its well-defined interpretation within the framework of quantum mechanics.'' --- (Dialectica, p.\ 317 -- BCW 7, p.\ [335])
\eq
\bq\noindent
``It may also be added that it obviously can make no difference as regards observable effects obtainable by a definite experimental arrangement, whether our plans of constructing or handling the instruments are fixed beforehand or whether we prefer to postpone the completion of our planning until a later moment when the particle is already on its way from one instrument to another.'' --- (Einstein Paper, p.\ 230 -- BCW 7, p.\ [370])
\eq
\eq

\section{27-09-02 \ \ {\it Email} \ \ (to R. W. {\Spekkens})} \label{Spekkens10}

Concerning the epiphany, you tease me:  I don't want to see the research fruits at the moment; I just want to see a psychological description of their originator.  You've got to know by now, I find even the emotional aspects of quantum mechanics extremely interesting.  Concerning my publishing your words publicly, you can always ask that a given thing you write not be repeated:  Plenty of people do that with me, and I always abide.

\section{27-09-02 \ \ {\it A Pearl!}\ \ \ (to A. Peres)} \label{Peres43}

\bap
Here is part of a letter I received from an old lady who was a pretty
student in 1974 and forced me to learn Bell's theorem. She is now
unemployed and has plenty of time to work on physics. She even intends
to publish her first paper.
\bq\noindent{\rm
What makes me read papers is to make sure that I am original. I have to tell you that I covered
quite a material. There were three things that I liked, I don't have
enough words. One is a paper of Chris Fuchs, a PEARL, another
one is Mermin's ``Ithaca Interpretation'' and the third is Gordon
Baym's book.}
\eq
\eap

That made my day!

\section{27-09-02 \ \ {\it Commune of the Incorrigibles} \ \ (to C. H. {\Bennett})} \label{Bennett25}

\noindent Communard Charles,\medskip

You're incorrigible.  (And you should not read that as merely reflecting a fear of mine about your foundational fetish.)  I've marked you for October 15 through 17.  It's looking like we're going to have plenty of extra space the first week; I think I'm going to try to see if Reinhard Werner can come then.

If you'd like to stay in a hotel follow the instructions below, post haste.  Alternatively, if you want to save Gilles some money, you can stay at the Rockledge Apartments with me:  I've got an empty second room during that time.  (But whichever way, let me know for accounting purposes.)\medskip

\noindent Communard Chris

\section{28-09-02 \ \ {\it Better Editing} \ \ (to H. J. Folse)} \label{Folse14}

That's it, I think.  You should check over the result, making any further changes you want, and then send it back to me.  I'll send it to {\tt quant-ph} as time permits (though I'll be in Canada Oct 2--8, so it probably won't get done until after I'm back).  [The final result was \arxiv{quant-ph/0210075}.]

Concerning the content of the paper, I believe I absorbed that pretty well too.  This time, however,---you know I've read all but one of your papers---I was struck more than ever for the need of a translation table to mediate between some of your choices (or Bohr's choices) of words and the more modern, more common quantumspeak that present-day practitioners of the field use.  I hope to get a chance to write you about this in detail while I'm at the {\Montreal} commune (Oct 13 -- Nov 3).  There is the seed of something good in your ``ontological lesson'' section, but I am afraid it will be obscured to most in our field because of the point above.

\section{28-09-02 \ \ {\it Substances and Properties} \ \ (to H. J. Folse)} \label{Folse15}

I'm back again.  As I said, (hopefully) I'll write in more detail about the content of your paper when I'm in {\Montreal}.  In the meantime, let me forward you some letters I wrote contra taking the quantum state (in the usual sense of a $|\psi\rangle$) as a {\it property\/} of a quantum system.
[See 25-04-02 note ``\myref{Bennett15}{Short Thoughtful Reply}'' to C. H. {\Bennett}, 14-05-02 note ``\myref{Bennett17}{Qubit and Teleportation Are Words}'' to C. H. {\Bennett} and others, 14-05-02 note ``\myref{Bennett18}{Chris's World}'' to J. A. Smolin and others, and 16-05-02 note ``\myref{Bennett16}{King Broccoli}'' to J. A. Smolin and others.] You made me think of this because of your talk of substances and properties in the ``ontological lesson.''  The letters below were spurred by a discussion about the words ``qubit,'' ``entanglement,'' and ``quantum teleportation'' soon appearing in English dictionaries.  (Charlie Bennett, in particular, was seeking better definitions than had been proposed by some lexicographer.  He wanted something that didn't confuse the reader into thinking that quantum teleportation involves instantaneous action at a distance.  {\it Yet}, he still wanted to call a quantum state a property.  I found these two demands inconsistent.)

You might enjoy the letters, for what they're worth.

\section{06-10-02 \ \ {\it Ed's Book} \ \ (to A. Plotnitsky)} \label{Plotnitsky11}

\barkp
I have a question.  {\Ruediger} in his piece refers to Jaynes' 2003 book
on probability. Would you recommended it? I'd like to revisit
probability.
\earkp

In fact, I wouldn't be able to think of a better place for you to start.  Jaynes' book---never finished before his death---is really the very best Bayesian-tilted book to start off with.  (It doesn't go as radically Bayesian as I am now going, but that is a minor blemish in comparison to the good it does.)

I see, looking in {\Ruediger}'s paper, that he has cited the soon to be published posthumous version.  At the moment, you can nevertheless download it at the website: \myurl{http://bayes.wustl.edu/}.

Beyond that though, a good book to understand the more philosophical issues around Bayesianism, have a look at the Kyburg and Smokler book he cites in reference 1.

\section{08-10-02 \ \ {\it FYI Comments on Recent Papers} \ \ (to M. B. Ruskai \& B. J. Hiley)} \label{Ruskai1} \label{Hiley1}

Thank you both for sharing your correspondence with me.  I very much enjoyed reading them on my flight from Toronto to Boston this morning.  There are substantial issues here and it is pleasant to read about them in an evenhanded way.

I got a nice snicker in reading Basil's comment,
\bbjh
In my own study of quantum theory I have often reached the stage when I am
led to wonder why a particular author has come out so strongly for a
certain point of view and against all others when there is no observational
evidence one way or the other.  My own long experience of being associated
with Bohm's work has brought me directly and forcefully into contact with
such questions.  For example I have been greeted with the phrase ``You're a
Bohmian aren't you?''  I would like to feel flattered by such a comment but
it is often meant as a reference to someone who is regarded as belonging to
the complement of sane!

I certainly feel that social and cultural factors strongly influence these
choices.  As I have already remarked above they are very implicit and very
difficult to bring to the surface, partially because it is universally held
that such factors should not have any role to play.  Yet why does the
debate about the interpretations of quantum theory generate such deep
emotions if these questions do not touch the very depths of what we believe
lies at the heart of nature and even our very being? There are only two
other areas where I have experienced the same passion and energy and that
is in politics and religion.
\ebjh
thinking about the pleasant breakfast we had together in {\Vaxjo} last year and the talk he gave soon after that:  ``Chris Fuchs bounded up to me this morning and said, `Are you a Bohmian?!?'''  Well, it's good to know a true Bohmian, and to know the distinctions between him and a Goldsteinian.

\section{10-10-02 \ \ {\it Infinitesimal Comments} \ \ (to H. J. Folse)} \label{Folse16}

Let me just make some infinitesimal comments on your comments of my comments on van Fraassen's comments.

\bhf
I interpret this to mean that the justification for your holding that
subjective judgment is its success in predicting the POVM's which are
themselves objective.
\ehf
Predicting the {\it outcomes\/} of POVMs, yes.  (A POVM---i.e., a {\it set\/} of operators with a certain property---is the theoretical expression of a (physical) measurement.)  The operators within the POVM express the possible outcomes of the measurement.

\bhf
The updating is in a subjective belief, but the cause of the updating  is something presumably as objective as a ``click''?
\ehf
Yes.

\bhf
Here it seems to me that the ``about the world'' part is the objective
part.  Consider two odds-makers for horse races who give different
odds for the same races.  Clearly one is rational to ``accept'' the odds made by the guy who, betting on those odds, wins the most money.  So if somebody asks me why do I go by these odds, I answer because of the outcomes. It's all about what I'm justified in believing regarding how to bet. But then if I ask why does one guy succeed over the other, and I discover the successful one has a lot more information about the
horses, jockeys, conditions, etc., isn't this ``objective'' and has
nothing to do with what I believe?
\ehf
I'm shooting for a Darwinian conception of success here.  Success is only defined with hindsight.  There is no objective state of affairs that predetermines a ``degree of success'' for one's future gambles.  In that sense, there are no objective probabilities, past gambling success or not.

\section{10-10-02 \ \ {\it Your First Quant-ph} \ \ (to H. J. Folse)} \label{Folse17}

OK, the dirty deed is done:  You've now had your first {\tt quant-ph} submission.  I've checked it and everything appears to be OK; it downloads fine and looks just like it was supposed to.

You once wrote me this:
\bhf
Actually I've had at the back of my mind the project of collecting a
subset of these papers and trying to get them published as a book.  If
you do read thru them all, I'd appreciate your telling me what you
think of the worthiness of such a project.
\ehf
I do indeed think that would be a worthy project; your papers certainly opened my eyes.  I'll pledge this:  If you ever decide to do it, if you're interested I'd volunteer to write a foreword to the book --- I bet you couldn't find a more enthusiastic foreword-writer than me.  In particular, I'd love nothing more than to see the quantum information community get a good healthy dose of Bohr.  The extent to which I can contribute to making that happen is bound to help me get a little closer to heaven.

\section{13-10-02 \ \ {\it Quantum Foundations Commune} \ \ (to L. Hardy and Perimeter Institute generally)} \label{Hardy9}

Partially because of the size of the auditorium we ended up with and partially because of our desire to keep the sparks flying, Gilles Brassard and I decided to selectively open the doors for attendance to our quantum foundations commune.  Thus, if anyone at Perimeter is interested in audience participation and general discussions, they are more than welcome to come.  There are no more official talk slots open at the moment, but that shouldn't stop anyone from giving an impromptu talk on the subject if they'd like to.  (Unfortunately though, there are no more travel funds.  And if there's any desk space in the offices, we'll probably dole it out on a rotating basis.)

If you wouldn't mind, please help spread the word by forwarding this note to anyone at Perimeter who might be interested.  (Presently, I'm cc'ing the note to those of you at Perimeter who are in my address book.)

Below, I'll place four items of information:
\begin{itemize}
\item[a)] The tentative talk schedule and attendance table.
\item[b)] Part of the original meeting announcement.
\item[c)] Instructions for how to obtain a hotel.
\item[d)] Directions to the meeting.
\end{itemize}
Please join us if you can.

\section{26-10-02 \ \ {\it Classical Measurement} \ \ (to K. Jacobs)} \label{Jacobs3}

\bkj
You know that classical measurement can just be written as quantum
measurements, so long as all the POVM elements and the density matrix
commute. But do you know of a paper that actually states this (other
than my paper on pooling knowledge). I would like to reference
somebody other than myself.
\ekj

Sorry to take so long to reply to you.  I've been tied up for two weeks with a workshop in {\Montreal} that Gilles Brassard and I organized, and I've hardly gotten a spot of email done in that time.

I don't recall any papers that have discussed the issue explicitly, but I think it is implicit in my discussion of Section 6 of my paper:
\begin{itemize}
\item
C.~A. Fuchs, ``Quantum Mechanics as Quantum Information (and only a little more),'' in {\sl Quantum Theory:\ Reconsideration of Foundations}, edited by A.~Khrennikov ({\Vaxjo} University Press, {\Vaxjo}, Sweden, 2002), pp.~463--543. \quantph{0205039}.
\end{itemize}
Have a look at that and let me know if it's appropriate.

\section{27-10-02 \ \ {\it A Crowded Placetime} \ \ (to G. L. Comer)} \label{Comer21}

I hope you'll accept a million apologies from me.  I've been in {\Montreal} burning \$70,000 worth of Canadian funding since October 13 and have hardly had a chance to touch my email at all.  It's a three week workshop on ``Quantum Foundations in the Light of Quantum Information'' that Gilles Brassard and I organized, and I'm getting a breather at the moment by flying home for the weekend.  (We've made a family tradition of pumpkin picking each fall, and I couldn't let my big girl down.  However, I'm certainly going to be letting her down by not being home on Halloween.)  Actually, the family was with me last week, and we're flying back together---they're all asleep at the moment.

Thanks for sending your poem about the solitary place.  I especially
liked the penultimate paragraph.
\bv
My choices are radioactive remedies\\
That rot the roots, spin down the spirals,\\
Snip the snares, and dampen the dark.\\
Something much larger than the universe,\\
With all that space and time and matter.
\ev
You make me want to quote William {\James} again.  This one comes from a
letter in his collection (of letters):
\bq\noindent
    All I can tell you is the thought that with me outlasts all others,
    and onto which, like a rock, I find myself washed up when the waves
    of doubt are weltering over all the rest of the world; and that is
    the thought of my having a will, and of my belonging to a
    brotherhood of men possessed of a capacity for pleasure and pains of
    different kinds.  \ldots\ And if we have to give up all hope of seeing
    into the purposes of God, or to give up theoretically the idea of
    final causes, \ldots\ we can, by our will, make the enjoyment of our
    brothers stand us in the stead of a final cause \ldots
\eq

\ldots\  \ldots\  \ldots\  \ldots\  \ldots\  \ldots\  \ldots\  \ldots\

By now, the weekend has passed and I'm waiting in the airline lounge for my flight back to {\Montreal}.

\bgc
I wanted to ask your advice on something.  You put your samizdat under
public scrutiny, and even tried to get it published.  Was the experience worth it?
\egc

Yeah, I think it was worth it.  It's not my most cited work, but it has been cited a few times already, and there are two papers written by some philosophical types that take it as their starting point (pro and con!).  One of the guys is a grad student in philosophy in Vancouver, and one is a physics prof, but philosophical nonetheless, at University of London, Queen Mary.  Of course, the thing has convinced some that I'm a nutcase, but I ignore that, knowing that there are some out there who are actually benefiting from it.  For instance, Jeffrey Bub is a good example.  His view of quantum mechanics was turned around by it, enough so that he now disavows his own book on the subject, and I take a great deal of pride in that.

\section{27-10-02 \ \ {\it Technical Question Minus}\ \ \ (to H. M. Wiseman)} \label{Wiseman12}

Sorry to take so long to reply to you.  I've been inundated with the duties of organizing this meeting in {\Montreal} I'm involved in (``Quantum Foundations in the Light of Quantum Information II'').  Luckily, however, I had to fly home this weekend for pumpkin picking, and so I've got a little time on the flights to get some email done.

\bhw
He was preparing a section on q.\ tomography, but I was not happy with
it as it did not address the question of how many different projective
measurements are necessary to fully characterize a state. I thought
this would be well known, but the answer does not seem to be in any
tomography papers Gerard can find. My hypothesis is that the answer is
$N+1$, where $N$ is the Hilbert space dimension. Michael Nielsen suggested
you would be the person to ask about where to find this result (or
whatever the relevant result is), as there was a time when you cared
about things less general than POVMs. Can you help?
\ehw

To my knowledge, the first guys to give that answer were Band and Park.  Here are some references:
\begin{itemize}
\item
W.~Band and J.~L. Park, ``The empirical determination of quantum states,'' {\em Foundations of Physics}, vol.~1(2), pp.~133--144, 1970.

\item
J.~L. Park and W.~Band, ``A general method of empirical state determination in quantum physics: {P}art {I},'' {\em Foundations of Physics}, vol.~1(3), pp.~211--226, 1971.

\item
W.~Band and J.~L. Park, ``A general method of empirical state determination in quantum physics: {P}art {II},'' {\em Foundations of Physics}, vol.~1(4), pp.~339--357, 1971.

\item
W.~Band and J.~L. Park, ``Quantum state determination: {Q}uorum for a particle in one dimension,'' {\em American Journal of Physics}, vol.~47(2), pp.~188--191, 1979.
\end{itemize}

A paper showing that something goes wrong with tomography in real Hilbert-space quantum mechanics is:
\begin{itemize}
\item
W.~K. Wootters, ``Local accessibility of quantum states,'' in {\em Complexity, Entropy and the Physics of Information} (W.~H. Zurek, ed.), (Redwood City, CA), pp.~39--46, Addison-Wesley, 1990.
\end{itemize}
To my knowledge, the first people to talk about doing tomography with a single (minimal informationally complete POVM) were Caves, Schack, and I:
\begin{itemize}
\item
C.~M. Caves, C.~A. Fuchs and R.~Schack, ``Unknown Quantum States:\ The Quantum de Finetti Representation,'' {\em Journal of Mathematical Physics}, vol.~ 43(9), pp.~4537--4559 (2002). [Reprinted in {\sl Virtual Journal of Quantum Information\/} {\bf 2}(9).] \ \quantph{0104088}.
\end{itemize}

\bhw
Now, moving swiftly into the metaphysical plane, given that you
profess to believe in a real world (apart from your state of belief
about it, which is a quantum state) \ldots
\ehw

No, my states of belief are in general captured by probability distributions.  In some special cases, those probability distributions have the character of quantum states.  As I'm presently toying with the idea, the world is a much more varied place than I would ever use quantum mechanics to describe in toto.  It is just I think that quantum mechanics hints at something particularly deep about how we (as active agents) are wired to the world.

\bhw
Another question: if your state of belief is your odds for future
occurrences (on which you can either win or lose money), what is your
state of belief about events after you are dead?  You can't win or lose
any money on them, so how can rational behaviour help? To put it
another way, how do you justify having life assurance? I haven't read
your tracts lately, so my point of view is [wandering?]\ all over the
place again.
\ehw
I don't understand the question.  But maybe that's just because I'm giving this reply a rush job.

\bhw
A simpler question: have you read Carlo Rovelli's work ``Relational
Quantum Mechanics''. He says quantum states are about information, but
I'm not sure if you'd agree with his approach.
\ehw

Aha!  You haven't actually read either of my papers \quantph{0205039} or \quantph{0106166}.  I caught you.  In both those papers, I wrote:
\bq\noindent
    ``But what nonsense is this,'' you must be asking.  ``Where else
    could they start?''  The main issue is this, and no one has said it
    more clearly than Carlo Rovelli. Where present-day
    quantum-foundation studies have stagnated in the stream of history
    is not so unlike where the physics of length contraction and time
    dilation stood before Einstein's 1905 paper on special relativity.
\eq
But, yes, though I really liked Rovelli's motivational sermon \ldots\ and even something of the flavor of his two principles, I think he floundered pretty badly when it came time to put some meat on the bones.

\section{28-10-02 \ \ {\it Blather, Lather, and Rinse} \ \ (to J. Bub)} \label{Bub8}

The following is a note I started to construct for you several weeks
ago.  Unfortunately, I never got a chance to finish it, but it seems
appropriate to send you the note as it stands nevertheless \ldots\ if
for nothing else as a starting point for this week's discussions. The
note was to be titled ``Blather, Lather, and Rinse.''  I had wanted
to polish it better and add some more detail, but maybe this will get
the ball rolling.
\begin{flushright}
\baselineskip=3pt
\parbox{3.4in}{
\bq
\noindent Pilate was probably not the first to ask
 what truth is, and he was by no means the
 last.  Those who ask it seek something
 deeper than disquotation, which was the
 valid residue of the correspondence theory
 of truth.
\medskip
\\
\hspace*{\fill} --- W. V. Quine
\eq
}
\end{flushright}\medskip

I apologize for taking so long to reply to your several letters, but
I wanted to read Quine's little book {\sl Pursuit of Truth\/}
(revised edition) before doing so.

I want to defend my position that there is some good stuff in
{\Marcus}'s manuscript.  At the very least, there are things that I
needed to read about and be exposed to.  One man's blather is another
man's rinse, I suppose.  But I think it is more than that.  There
were three things that caught my attention.

a) The strategy of taking the Bohmian and Everettista's strong desire
to find a ``designatable reality term'' within quantum theory to take
as its anchor and turning that against them.  The point he was trying
to make could be more polished and intensified, but it was one that I
found to be a keeper to be developed.

b) I found the discussion of the ``primitive theories'' that underlie
our developed theories quite well done, and I think it is an
important point that has yet to sink in to most of our mentalities. I
just loved some of his lines in that section.  Here was one that
particularly struck me:
\bq
My second point was that the function of a formal physical theory is
to extend the primitive theory.  And, indeed, modern physical theories
extend it to a degree which may seem almost miraculous (which, to a
Palaeolithic hunter, and perhaps even to an Elizabethan savant, might
really have seemed miraculous).  But, however striking the advance,
we are never in a position actually to dispense with the primitive
theory.  No matter how sophisticated the fighter aircraft, the pilot
still controls it in the time-honoured Palaeolithic manner, using
hands and feet.  Similarly here:  if you want to confirm the theory,
or to apply the theory, then you must have recourse to your
Palaeolithic sensory organs, and your Palaeolithic sensory cortex.
No matter how marvellous the theory, if you want to relate it to the
world it is supposed to concern, you have to go through these
Palaeolithic channels.  Which means the theory can have no commerce
with the world it is about except through the mediacy---the good
offices, as it were---of the primitive theory.   There is no escaping
it:  it's the way we're wired.
\eq

I think this is the proper way to understand (or at the very least
build upon) what Bohr was talking about when he kept referring to the
necessity of a classical description for making sense of quantum
phenomena.  You can see this mimicked (but in much less developed
form) in my Samizdat: p.\ 260, ``Foods for Thought [sic],'' p.\ 464,
``Always One Theory Behind,'' and in footnote 18 of my \quantph{0205039}.

In particular I found {\Marcus}'s discussion of this point extremely
relevant to an ongoing debate I've been having with {\Mermin}, {\Caves} and
{\Schack} for some time.  From where comes the judgment that a
beamsplitter is a beamsplitter (and not some other kind of device)?
Those fellows want to think that it somehow---in a way yet to be
explained!---comes out of quantum theory itself.  But I don't buy it.
The path I'm traveling indicates that that judgment will never come
out of quantum theory.  The judgment that a beamsplitter is a
beamsplitter is, in proper Bayesian terms, the assumption of a
``prior''.  And where an ultimate prior comes from, Bayesian theory
is silent.  One must seek an answer from somewhere else, if from
anywhere at all.  In fact, I would say this is the ultimate meaning
of my slogan ``Quantum Mechanics is a Law of Thought'' (at those
times when I hold to it).

c)  Finally, I rather liked the discussion of ``reality as a logical
requirement'' at the very end of his draft.  I mimic some loosely
related thoughts at the beginning of Section 4 of my \quantph{0205039}, and flesh it out a little more fully in Section 5
of my \quantph{0204146}.  The reason to draw a distinction
between the ``state space'' and the ``sample space'' is not because
the ``sample space'' captures the idea that there is a representative
that is ``really there'' and we are just ignorant of it, but because
it captures the idea of a spur on which we revise our beliefs.

The things in {\Marcus}'s draft that seemed to capture your attention
were those (small) things that fall within the frame of usual
(nonpragmatic) philosophical debate.  I'm thinking here of your
remark on disquotation.  But I think he and I both have bigger bones
to pick.  The deeper issue is to do away altogether with a
correspondence notion of truth, and to put into its place something
akin to a pragmatic account (where ``truth'' is made by the process
of ``measurement intervention'').  The old pragmatists, as I see it,
had a glimpse of something that quantum mechanics gives us a much
stronger reason to take seriously.  [I was going to write much more
here, but now no time.]

\section{30-10-02 \ \ {\it Time to Make a Difference} \ \ (to R. W. {\Spekkens})} \label{Spekkens11}

I've been thinking about it on and off all night, and I wonder if I can convince you to do a rush job on the paper you proposed writing (when we were talking in Waterloo).  Namely, putting down in paper form the talk you gave here last week.  What I'm thinking is that I would really love to have it in the special issue of {\sl Studies in History and Philosophy of Modern Physics\/} that Jeff Bub and I are co-editing.  We've already got more than enough stuff for the issue, but I'd like to squeeze you in \ldots\ if I can convince you to take on the project.

Your talk struck a chord with me as being the right way to wham people over the head (if done right).  And I'm now pretty sure you can polish it to that point.  Moreover I think with this special issue we've got the right opportunity to be convincing to a lot of (the right) people, all at once.

So I hope you'll think seriously about this.  If I can get a relatively final draft from you in one month---but not a week over that---then I think I'll be able to squeeze you in.  But I've got to be absolutely hard about the deadline.  I know you'll very likely have to seriously rearrange your schedule to pull this off, but I think this could be a grand opportunity for you (and for me vicariously) to make a difference in the community.

Let me list below the papers we've now got in hand to show you that you'll be in some respectable company.
\begin{itemize}
\item
Howard Barnum, ``Quantum Information Processing, Operational Quantum Logic, Convexity, and the Foundations of Physics''
\item
Charles Bennett, ``Notes on Landauer's Principle, Reversible Computation, and Maxwell's Demon''
\item
Armond Duwell,
``Quantum Information Does Not Exist''
\item
Itamar Pitowsky,
``Betting on the Outcomes of Measurements: A Bayesian Theory of Quantum Probability''
\item
Chris {\Timpson},
``On the Supposed Conceptual Inadequacy of the Shannon Information''
\item
David Wallace,
``Quantum Probability and Decision Theory, Revisited''
\item
Andrew Steane,
``A Quantum Computer Only Needs One Universe''
\end{itemize}
Talk to you later today or tomorrow.

\section{02-11-02 \ \ {\it Technical Question Minus, 2}\ \ \ (to H. M. Wiseman)} \label{Wiseman13}

\bhw
Let me try again. It makes no sense to have a bet on an event that
will happen after you are dead, as you can never benefit from it. So
if your state of belief is just your betting odds for future events,
logically you can have no state of belief about events after you are
dead. So why have life assurance for the benefit of your family, for
instance?
\ehw

I think it was Henry VIII that first said, I am England.

\subsection{Howard's Reply}

\bq
I've never heard that about Henry VIII. Does ``I am England'' mean
something different from ``England is me''? (cf.\ Louis XIV
``L'\'etat? C'est moi.'')

I understand (I think) what you are saying, but I don't buy it.
\eq

\section{03-11-02 \ \ {\it Visionlessness} \ \ (to R. W. {\Spekkens})} \label{Spekkens12}

\brws
Admittedly, I am feeling these days that the hidden variable program
envisioned by Einstein was never really pursued seriously and thus may
have been abandoned too quickly.
\erws

Your note struck a fear in me that I'm having a hard time shaking.  In coming to Bell Labs, it looks like you might be wasting a year of your time.  But worse than that, I fear you're going to waste a year of my short life.

I have little to no tolerance for this.  The way I see it, the quantum conundrum---after 75 years of waste---is only going to be broken by a leap of faith and imagination.  It needs something creative.  And, most importantly, it needs a vision of simplicity.  The thought you express above is neither of those.

\subsection{Rob's Reply}

\bq
Perhaps you are reading more into my comments than was intended. Your
reaction reminded of the following Emo Philips joke:
\bq\noindent
I was walking across a bridge one day, and I saw a man standing on the
edge, about to jump off. So I ran over and said ``Stop! Don't do it!'' ``Why shouldn't I?''\ he said. I said, ``Well, there's so much to live for!'' He said, ``Like what?'' I said, ``Well \ldots\ are you religious or atheist?'' He said, ``Religious.'' I said, ``Me too! Are you christian or buddhist?'' He said, ``Christian.'' I said, ``Me too! Are you catholic or protestant?'' He said, ``Protestant.'' I said, ``Me too! Are you episcopalian or baptist?'' He said, ``Baptist!'' I said,``Wow! Me too! Are you baptist church of god or baptist church of the lord?'' He said, ``Baptist church of god!'' I said, ``Me too! Are
you original baptist church of god, or are you reformed baptist church of god?'' He said,``Reformed Baptist church of god!'' I said, ``Me too! Are you reformed baptist church of god, reformation of 1879, or reformed baptist church of god, reformation of 1915?'' He said, ``Reformed baptist church of god, reformation of 1915!'' I said, ``Die, heretic scum'', and pushed him off.\\
\hspace*{\fill} --- Emo Philips
\eq
In the spectrum of views on quantum foundations, ours are not so
dissimilar.
\eq

\section{04-11-02 \ \ {\it Yoo-Hoo} \ \ (to J. W. Nicholson)} \label{Nicholson13}

I've been spreading the word with van Enk and my management that I'm still in Canada \ldots\ and will be through Wednesday \ldots\ but the truth of the matter is I'm exhausted (at home) and need to get a load of housework done or I'll go insane.

Also I had an epiphany on the plane yesterday---something I thought to be exceedingly deep---and I struggled much of last night to get it written up.

Shall we have lunch Friday?

\section{04-11-02 \ \ {\it Got It!}\ \ \ (to R. {\Schack}, C. M. {\Caves} \& N. D. {\Mermin})} \label{Mermin76} \label{Schack60} \label{Caves69}

\bq
\noindent
{\bf WARNING! WARNING! WARNING! WARNING! WARNING!}\smallskip\\
Subsections 2 and 3 of the present note are UTTER RUBBISH, and I
entirely withdraw the ideas put forth there.  (I still agree with
subsections 1 and 4, however.)  See the notes ``Utter Rubbish and
Internal Consistency, \myref{Mermin86}{Parts I} and \myref{Mermin87}{II}'' of 27 and 28 June 2003 for an
explanation. I nevertheless leave the ill-fated sections in this
collection because, despite their embarrassment for me, I deem them
crucial for understanding the surer path we all explored later.
\smallskip\\
{\bf WARNING! WARNING! WARNING! WARNING! WARNING!}
\eq

It finally sank in completely and whammed me over the head!  It's
beautiful; it's compelling; it was implicit the whole time.  It's
trivial even, and I am so excited!  There is absolutely no mystery in
the Penrose example, and there is absolutely no mystery in the
Einstein--Podolsky--Rosen example once one has accepted that POVM
ascriptions have the same non-ontic character as the state vector
itself.  Since my first note to you on my ``\myref{Schack4}{Identity
  Crisis}'' in August of last year---(``Quantum States:\ WHAT?'', page
35)---the solution has been staring us in the face.  In fact, the
examples above now appear as hardly worthy of ever having been called
conundrums or paradoxes in the first place.

I guess this is what {\Ruediger} was trying to get across to me (and to
David?)\ on Friday before he left.  Let me put the story in my own
words.

I'll break the note below into four sections for clarity:
\begin{enumerate}
\item
A general exegesis on how to think about POVMs
\item
The Penrose example
\item
The EPR example
\item
Forward-looking thoughts on POVMs and a pluralistic universe
\end{enumerate}

\subsection{1) The POVM as a function from raw data to meaning}

We generally write a POVM as an indexed set of operators, $E_d$. Here
is how I would denote the referents of those symbols.  The index $d$
should be taken to stand for the raw data that can enter our
attention when a quantum measurement is performed.  The whole object
$E_d$ should be construed as the ``meaning'' we propose to ascribe to
that piece of data when/if it comes to our attention.  It is
important here to recognize the logical distinction between these two
roles.  The symbol $d$ stands for something beyond our control,
something that enters into us from the world outside our head.  The
ascription of a particular value $d$ is not up to us, by definition.
The {\it function\/} $E_d$, however, is of a completely different
flavor. It is set by our history, by our education, by whatever
incidental factors that have led us to believe whatever it is that we
believe when we walk into the laboratory to elicit some data.  That
is to say, $E_d$ has much the character of a subjective probability
assignment.  It is a judgment.

I have tried to say this in various ways before.  Maybe the first
place in ``Quantum States:\ WHAT?'' is in the note
``\myref{Schack5}{Note on Terminology},'' pages 49--50, or in more
detail in ``\myref{Caves35}{Replies to a Conglomeration},'' page 92.
Maybe there are still better shots at it, but I didn't look further.
(I guess I also give another variation on the matter on page 42 of
\quantph{0205039}).  You can have a look at those if you think it'll
help, but I think the paragraph above says it as well as anything.
And I know that at least {\Ruediger} is on board with all this.
{\Carl} probably is too, but I'm not as sure.

\subsection{2) The Penrose example}

Here's one of the passages I recorded [from Penrose's book] in the
Samizdat, pages 402--404,
\bq
One of the most powerful reasons for rejecting such a subjective
viewpoint concerning the reality of $|\psi\rangle$ comes from the
fact that whatever $|\psi\rangle$ might be, there is always---in
principle, at least---a {\it primitive measurement\/} whose {\bf YES}
space consists of the Hilbert-space ray determined by $|\psi\rangle$.
The point is that the physical state $|\psi\rangle$ (determined by
the ray of complex multiples of $|\psi\rangle$) is {\it uniquely\/}
determined by the fact that the outcome {\bf YES}, for this state, is
{\it certain}.  No other physical state has this property.  For any
other state, there would merely be some probability, short of
certainty, that the outcome will be {\bf YES}, and an outcome of {\bf
NO} might occur. Thus, although there is no measurement which will
tell us what $|\psi\rangle$ actually {\it is}, the physical state
$|\psi\rangle$ is uniquely determined by what it asserts must be the
result of a measurement that {\it might\/} be performed on it.
\eq

Let's think about this from our perspective.  From our point of view,
both the state $|\psi\rangle$ and the measurement Penrose speaks of
are subjective judgments.  They both count as priors.  In principle,
they are distinct subjective judgments, but in this case they happen
to coincide in a meaningful sense.  Here's the meaningful sense. When
accepting quantum mechanics as a theory for reasoning, we are, among
other things, accepting the consequences of Gleason's theorem. And
with that comes the coordinated states and measurements Penrose
speaks of.  WHEN the state, THEN the given measurement outcome with
probability one.

But what does that mean?  It means little more than that, no matter
what the objective character of the raw data we find, we {\it
ascribe\/} to it a meaning appropriately associated with the given
state.  That is, the {\bf YES} boils down to essentially a
convention.  The meaning we ascribe to the raw data has no choice but
to be labeled {\bf YES}.

And that, I think, is the whole story.  I'll add some more
metaphysics to this in a minute, but first let me move on to the EPR
situation before going further.

\subsection{3) The EPR example}

To illustrate this one, consider a bipartite set of qubits in a
maximally entangled state and suppose one measures the same von
Neumann measurement on each.  In this case, we can't predict the
outcome of either measurement, but we can predict that the two will
match identically.  Is there anything mysterious about this?  Nicolas
Gisin tells me there is.  For he says, ``I'll grant you that at least
one of the results is a sort of free creation or birth---i.e., it
does not arise because of a local hidden-variable theory---but then
how does the other creation in the pair know how to go the same
way?'' ``That simply can't happen unless there is some kind of
superluminal communication going on or, even more radically,
spacetime itself is meaningless.''

Well, what do we have here but little more than an extension of the
situation above?  Now, there are simply three priors rather than two.
And those priors command us to interpret the two pieces of raw data
(coming from the two separated parts of the experiment) as having the
same meaning.  That is to say, in analogy to our solution to the
Penrose conundrum, it is simply a convention that the clicks are the
same in the two wings of the experiment.

And that too is the end of the story.  It is just a triviality that
the measurements come out perfectly correlated.  They came out that
way because WE labeled them that way with the choice of our priors.

\subsection{4) POVMs and radical pluralism}

Now let me go into a bit of the metaphysics of this.  Here's a point
of view that I'm finding myself more and more attracted to lately.

I think it is safe to say that the following idea is pretty
commonplace in quantum mechanical practice.  Suppose I measure a
single POVM twice---maybe on the same system or two different
systems, I don't care---and just happen to get the same outcome in
both cases.  Namely, a single operator $E_d$.  The common idea, and
one I've held onto for years, is that there is an objective sense in
which those two events are identical copies of each other.  They are
like identical atoms \ldots\ or something like the spacetime
equivalent of atoms.  But now I think we have no warrant to think
that.  Rather, I would say the two outcomes are identical only
because we have (subjectively) chosen to ignore almost all of their
structure.

That is to say, I now count myself not so far from the opinion of
Ulfbeck and Bohr, when they write:
\bq
\noindent
   The click \ldots\ is seen to be an event entirely beyond law.  \ldots\ [I]t
   is a unique event that never repeats \ldots\ The uniqueness of the click,
   as an integral part of genuine fortuitousness, refers to the click in
   its entirety \ldots . [T]he very occurrence of laws governing the clicks
   is contingent on a lowered resolution.
\eq
For though I have made a logical distinction between the role of the
$d$'s and the $E_d$'s above, one should not forget the very
theory-ladenness of the set of possible $d$'s.  What I think is going
on here is that it takes (a lot of) theory to get us to even
recognize the raw data, much less ascribe it some meaning.  In {\Marcus}
{\Appleby}'s terms, all that stuff resides in the ``primitive theory''
(or perhaps some extension of it), which is a level well below
quantum mechanics.  What quantum mechanics is about is a little froth
on the top of a much deeper sea.  Once that deeper sea is set, then
it makes sense to make a distinction between the inside and the
outside of the agent---i.e., the subjective and the objective---as we
did above. For even in this froth on the top of a deeper sea, we
still find things we cannot control once our basic beliefs---i.e.,
our theory---are set.

Without the potential $d$'s we could not even speak of the
possibility of experiment.  Yet like the cardinality of the set of
colors in the rainbow---Newton said seven, Aristotle said three or
four---a subjective judgment had to be made (within the wide
community) before we could get to that level.  If this is so, then it
should not strike us as so strange that the raw data $d$ in our
quantum mechanical experience will ultimately be ascribed with a
meaning $E_d$ that is subjectively given.  (I expressed some of this
a little better in a note I wrote to David last month; I'll place it
below as a supplement.)  More particularly, with respect to the EPR
example above, it should not strike us as odd that the phenomenon
comes about solely because of an interpretive convention we set:  All
quantum measurement outcomes are unique and incomparable at the ontic
level.  At least that's the idea I'm toying with.

I think that's enough for tonight.  I now intend to sleep for at
least 14 hours!  (It was a long three weeks.)

\subsection{David's Reply}

\bq
Thanks cc'ing me the latest epiphany.  I've only had a chance to
glance at it and am still torn between whether my response is ``that
was how I understood your position to be all week'' or ``what utter
rubbish'' so I will take a little time before responding.  (I have to
go to Germany for a week in a week so it may be a couple of weeks.)
\eq

\subsection{{\Ruediger}'s Reply}

\bq
I am happy with your section 1) on the POVM as a function from raw
data to meaning. But I don't get the point you are making in section
2). I would say that the sentence ``a $\sigma_x$ measurement of a
particle in the state $|+\rangle$ gives the outcome YES with
certainty'' is saying exactly the following: If our states of belief
about the POVM and the state are expressed by $\sigma_x$ and
$|+\rangle$, then consistency requires us to assign probability 1 to
the outcome YES. It's a statement about consistent application of the
formalism. To say about a particular experiment that it performs a
$\sigma_x$ measurement of a particle in the state $|+\rangle$ is a
subjective judgment. So far we are in agreement.

But then you write
\bq
\noindent But what does that mean?  It means little more than that, no
matter what the objective character of the raw data we find, we {\it
ascribe\/} to it a meaning appropriately associated with the given
state.  That is, the {\bf YES} boils down to essentially a
convention.  The meaning we ascribe to the raw data has no choice but
to be labeled {\bf YES}.
\eq

I probably just misunderstand this paragraph. It seems to me that
once we have ascribed the state (the pair $\sigma_x$, $|+\rangle$),
we are making a very strong statement. We have identified two
possible outcomes, YES and NO, and we make a commitment to accepting
bets on YES with arbitrarily unfavorable odds. It is logically
possible that the outcome will be NO. If the outcome is NO, there
will be a crisis. Ruin. Your paragraph suggests that in this case we
just relabel the raw data. If that's what you mean, then I agree with
David's ``what utter rubbish''. But I probably misunderstand you
here. (In a year's time, you'll write ``I never said it better than
in my note headed `Got It!' from 3 November 2002'' \smiley

I agree that for the purpose of this discussion, the EPR scenario
does not add much. But you get me worried again in section 3):
\bq
\noindent That is to say, in analogy to our solution to the Penrose
conundrum, it is simply a convention that the clicks are the same in
the two wings of the experiment.
\eq

Either I don't get it, or it's utter rubbish.

Section 4), I like. It highlights the chasm that exists between our
approach and the many-worlders and decoherence people. We all agree
that a click is not an elementary phenomenon. They want to reduce it
to something more fundamental. We say it's irreducible (which does
not mean that we cannot analyze a measurement apparatus or
decoherence of a quantum register in as much depth and detail as
anybody else). Your metaphysical bit is nice in that it makes clear
that ``irreducible'' is not the same thing as ``elementary''. As you
say it,
\bq
\noindent
All quantum measurement outcomes are unique and incomparable at the
ontic level.
\eq

I hope you got your 14 hours of sleep. My best night so far has been
5 hours long.
\eq

\section{04-11-02 \ \ {\it Carts and Horses, 2} \ \ (to A. Kent)} \label{Kent6}

\bak
There's a long tradition of physicists becoming over-enthused with
current technology as a source of metaphors and even fundamental
explanations for physics. \ldots\  I wonder if there's maybe a danger of
your program falling into the same class.
\eak

Ulrich Mohrhoff once wrote something similar in sentiment.  It's a good thing you're one of my friends, and I would never treat you like I treated him.  [See 04-07-01 note ``\myref{Mohrhoff2}{Carts and Horses}'' to U. Mohrhoff.]

\section{04-11-02 \ \ {\it One Day} \ \ (to A. Kent)} \label{Kent7}

\bak
I take the point
that you were interested in information-theoretic aspects of quantum
theory before quantum information became popular, and this earns you
several cosmic brownie points for originality (which are not dispensed
lightly).  Still, for me, the fundamental worry stands: so much of
physics doesn't look reducible to information theory that I wonder if
any of it is.
\eak

You're right about GR not looking like anything to do with information.  Nor would any physics whose main task is to set a Lagrangian, like QCD say.  However, quantum theory (as a principle theory in Bub's sense) stands out like a sore thumb to me.  I see so many problems melt away with an epistemic view of the quantum state, with no new problems coming back to haunt in its stead.  The task, as I see it, is to see how many further terms and rules in the theory (quantum operations, the Born rule, the tensor-product rule, the Hamiltonian, etc.) can also be taken to be epistemic in a fruitful way.  The ones that can't, finally with a little breathing space around them, will be able to scream out their ontic significance loudly and clearly.  We'll hear it, and, I'll bank money, know what to do with it for that next step in physics beyond flat quantum mechanics.

\section{14-11-02 \ \ {\it Probabilismo!}\ \ \ (to D. M. {\Appleby})} \label{Appleby0}

I'm sorry, I forgot to send you those references, didn't I\@?  Let me dig them up.
\begin{itemize}
\item
B.~de Finetti, ``Probabilism,'' Erkenntnis {\bf 31}, 169--223 (1989).

\item
R.~Jeffrey, ``Reading {\it Probabilismo},'' Erkenntnis {\bf 31},
225--237 (1989).
\end{itemize}
Also, I think that whole issue of Erkenntnis is worth perusing.  I think there is an article on de Finetti's approach to the ``problem'' of induction in there that Dick Jeffrey recommends \ldots\ unfortunately, I can't remember the author at the moment.  It might have been von Plato or Zabell.

\bma
I hope my niggling doubts weren't too exasperating.
\ema

Not at all; I very much enjoy your company.  And I appreciate your worries, even if I am only now learning how to respond to them adequately \ldots\ and, indeed, only now recognizing some of the weaknesses in what I spout.  (Though the latter is an ongoing thing with me.)

I hope one day you'll post a version of your ``Elements of a Prolegomenon'' to {\tt quant-ph}.  {\Ruediger} and I, at least, found parts of it very, very helpful.  And, self-servingly, I can't imagine a more flattering appraisal of the foundations program I'm trying to define than the one in your introductory section.  Inspiring words can help recruit a workforce, and you know I'm all for that!

Let me end this note by pasting in part of another note that I wrote soon after the {\Montreal} meeting.  (I'll paste in the part I still trust; parts of it, I think need to be redone now.)  [See 04-11-02 note ``\myref{Mermin76}{Got It!}''\ to R. {\Schack}, C. M. {\Caves} \& N. D. {\Mermin}.] Anyway, it leans heavily on explanatory modes that I developed while talking to you.

\section{17-11-02 \ \ {\it \Vaxjo} \ \ (to J. Finkelstein)} \label{Finkelstein7}

Good to hear from you.  You haven't seen any details on the meeting because none have been worked out yet.  Thus, I was a little surprised to see Andrei's announcement of the meeting in Y. S. Kim's conference listing; he probably just wanted to get something announced fast (for administrative purposes).  The last time we had talked, we were thinking hard (and I thought it had already been agreed to) about giving the meeting the slant:  Quantum Information meets Quantum Logic.  The idea being to get predominantly quantum information people who had some interest in learning about the quantum logic community's results and quantum logic people who were interested in learning about the results of quantum information.  And so on.  That, by the way, is Howard Barnum's predominant research theme at the moment.

We're definitely going to have to get geared up over the holidays to get the planning settled.

In any case, I think it'd be great if you'd come.  It'd be nice to meet you finally.  The last meeting was pretty successful, and I think once we put in some work, this one will probably come out the same.  I think there's a good chance we'll have Hardy, John Smolin, and Jeff Bub there.  And from the quantum logic and convex structures side, we're going to try to attract Dave Foulis, Alex Wilce, Dick Greechie, Paul Busch, and Reinhard Werner.  Clearly some in that list have already taken part in a cross-culturation.  Also, I guess there'll be the people that Andrei has already listed in his announcement.  Some of those, too, like Holevo, have done the same thing.

Would you need any funding, or would you be able to make it on your own travel grants.  We'll have some small amount of money, but I want to use it as advantageously as possible so that we'll be able to get the maximum number of interesting people there.

\section{17-11-02 \ \ {\it Fertilization?}\ \ \ (to J. Finkelstein)} \label{Finkelstein8}

Well I just discovered that ``culturation'' doesn't seem to be a word.  Maybe you understood what I was trying to get at anyway.

\section{19-11-02 \ \ {\it Ridiculous Interpretation}\ \ \ (to S. L. Braunstein)} \label{Braunstein8}

\bslb
Hi! Do you recall an odd interpretation about how the quantum
information gets across in quantum teleportation? It suggested the
quantum info goes backwards in time from Alice's Bell measurement to
the point of creation of the entanglement and then forward in time to
Bob. Do you recall where this odd idea came from? Or would you have
any idea whether it actually appears in print somewhere?
\eslb

Yes, you can find that in the original quantum teleportation paper itself.  It was an idea of Ben Schumacher's.

Personally, I hate the idea and think it is misleading.  For it gives the image that a quantum state is an objective property of something, and that that objective property just travels along a wire (and through time) in the process of quantum teleportation.  Whereas I would say, the only thing that travels backwards then forwards in time is our {\it chain of inference}.  Then, once it is realized to be inference, you don't need to talk about traveling at all:  Quantum teleportation is just a case of taking some prior information, gathering some data, and updating to a posterior.  See my description of teleportation in Section 3 of \quantph{0205039}.  (And note footnotes 14 and 15.)

\section{20-11-02 \ \ {\it Poetry as the Zing} \ \ (to G. L. Comer)} \label{Comer22}

Attached I've placed a note from Marcus Appleby, one of the participants of my {\Montreal} meeting.  I thought you might be interested in the parts where he talks about poetry as the stuff that makes the world fly.  It looks like you, he, and I are all in this together.

\section{03-12-02 \ \ {\it Poetry, Subjectivity, Mirroring and Algorithms} \ \ (to D. M. {\Appleby})} \label{Appleby1}

This is just to let you know that I got your email a while ago, and even read it once, but I've had almost no time to get philosophical for a while.  (Rather serious business concerns.)  What I'm trying to do at the moment is just clean out my email box before I go crazy.

So, please accept my apologies, but I'm not going to be able to reply in detail for still a while further.

Let me just comment on one key piece, for which much seems to hinge for you:
\bma
It is worth noting, incidentally, that you do ascribe objective status
to the dimension $d$, and to the bound on the volume of that shape in
the probability simplex.  This means you are still playing the same
game as {\Caves}:  trying to find something in our heads which genuinely
is the reflection of something real out there.  It is just that you
are much harder to please than {\Caves}.
\ema
I don't think you have me right on that count.  Please reread my paper
\quantph{0204146}---especially the part written to Wootters---and tell me whether you still think what you said above syncs with me.

\section{06-12-02 \ \ {\it SHPMP Issue on Quantum Information} \ \ (to the SHPMP participants)}

We are now in the very final stages of collecting manuscripts from contributors to the special quantum information issue of {\sl Studies in the History and Philosophy of Modern Physics}.  The issue will contain the following papers:
\begin{enumerate}
\item Howard Barnum: `Quantum information processing, operational quantum
logic, convexity, and the foundations of physics'

\item Armond Duwell: `Quantum information does not exist'

\item Lucien Hardy: `Probability theories in general and quantum theory in
particular'

\item David {\Mermin}: `Teaching computer scientists quantum mechanics, or how
I discovered the Copenhagen interpretation'

\item Itamar Pitowsky: `Betting on the outcomes of measurements: a Bayesian
theory of quantum probability'

\item Andrew Steane: `A quantum computer only needs one universe'

\item Chris Timpson: `On the supposed conceptual inadequacy of the Shannon
information'

\item David Wallace: `Quantum probability and decision theory revisited'
\end{enumerate}

The special issue is scheduled for publication as the September, 2003 issue of SHPMP.  To avoid being bounced to a later issue, we absolutely \underline{\it must\/} have the final version of your manuscript before the end of December.  So please send (or re-send) us a \LaTeX\ file of your paper, prepared in accordance with the journal's guidelines below, by December 20.

We are excited about this project and we look forward to hearing from you shortly.\medskip

\noindent Jeffrey Bub and Chris Fuchs

\section{06-12-02 \ \ {\it Call Me, Let Me Call You} \ \ (to C. H. {\Bennett})} \label{Bennett26}

When we were in {\Montreal} together, I remember one sleepy 4:00 AM conversation when you said that Jeff Bub and I could use your paper \arxiv{physics/0210005}, ``Notes on Landauer's Principle, Reversible Computation and Maxwell's Demon,'' for our special issue of {\sl Studies in History and Philosophy of Modern Physics}.

Is the paper still up for grabs?  If it is, we would really, really, really love to have it.  It would be perfect for the issue.  I'll put the issue announcement below, so that you can see who you're potential neighbors would be.  They're all really good papers.

All you'd have to do is say {\it yes\/} \ldots\ and I'd even format the paper properly for you.  I'll just take it directly off the archive.  You couldn't find a better bargain in Filene's Basement.

Just let us know as soon as possible (i.e., as soon as you read this note).  Also, you could either call me, or I could call you tomorrow/today.  Where will you be?

\section{06-12-02 \ \ {\it From the Real Chris} \ \ (to C. H. {\Bennett})} \label{Bennett27}

\bcb The subject line of your email almost motivated me to throw it
into the junk bin.  I expected that someone had hijacked your
From:\ address and was about to continue with, ``Nice lady wants to
meet you.''  I would be glad to have you publish
\myurl[http://arxiv.org/abs/physics/02010005]{physics/02010005} but I
am currently working on version 2, which will be much better, and will
address more critics of Landauer's principle.  I prepared version 1 in
a hurry so as to get an archive number in time for Rex and Leff to
cite in their new edition of the Maxwell's Demon book.  \ecb

The paper is already great as it stands (by our standards).  But, if you could finish the new version before December 20---at the absolute, absolute, absolute latest (see explanation below)---we could still accept that.

Don't forget, by the way, that Earman and Norton, Shenker, and Bub have all published their papers on the same subject in the same journal.  So, this really is the appropriate place (and the appropriate time) for your paper.

You want me to give you a call?  Where are you?

You encourage me!

\section{06-12-02 \ \ {\it Enjoyed} \ \ (to D. M. Greenberger)} \label{Greenberger1}

I enjoyed listening to you talk about an evolutionary approach to
quantum mechanics the other day.  When we get together next time
please tell me more!  You've got my ear on this subject.

Below I'll place some notes I've written on things to do with
Darwinism and quantum mechanics \ldots\ and their connections.  The
notes come from my samizdat, {\sl Quantum States:\ What the Hell Are
They?} [See 18-02-02 note ``\myref{Preskill6}{Psychology 101}'' to J. Preskill, 25-02-02 note ``\myref{Wootters7}{A Wonderful Life}'' to W. K. Wootters, and 24-06-02 note ``\myref{Wiseman6}{The World is Under Construction}'' to H. M. Wiseman.]

\subsection{Danny's Reply, ``Tiny Answer to Large Questions''}

\bq
Your emails, like your papers, are quite prolix, lots to read (not
meant as a criticism, only as an excuse for having not had time to
digest it yet), and so I can only tell you where I am at on a few
issues.  I will have to reread most of the comments in the email and
ruminate upon them.

But I strongly believe in Darwinism as the mechanism that drives
science.  In fact I define science by saying that man is always
reacting to the environment presented to him, in order to better
adapt to it, and science is his conscious effort to maximize this
adaptation.  So I strongly believe that we frame the theories that we
do because we perceive the world the way we do.  Our senses and
conceptualizations are partly built in and are partly a response to
what is out there, and we build theories and instruments to help us
better interact with it.  But if our senses and brain structure were
different, we would have evolved a different response.

For example, if like a snake we could sense infrared, our whole world
view would be different.  The snake sticks out his forked tongue and
says, ``ah, a mouse passed this way a while ago, and it is hotter to
the left than to the right, so I can track him to his lair.''  The
mouse and his infra-red image aren't very separable.

Now we see objects in visible light, of short wavelength, and so we
see sharp images.  So we develop rules that give things sharp
boundaries, and we say two things can't occupy the same place at the
same time. This in turn leads us to think that the integers are
important, and that enumerating things is important.  Our whole
counting system and mathematics is based on it.  If we perceived
things like a snake, nothing would be sharp, and I suspect the
integers would be way down on our list of what's important in
mathematics.  We would think a continuum approach would be much more
intuitive.

So you see, I don't even think mathematics is god-given.  I think it
is part of our response to our environment, based on our senses.  And
when people tell me, as almost all mathematicians do, that
mathematics is objective, and corresponds to a Platonic something
``out there'', my answer is that ``The reason you believe that two
plus two equals four, is because those of your ancestors who also
believed that, ate the ones who believed that two plus two equals
five.'' There was survival value in it for us as we are. But that
might not have been the case if we were different.  It's not
necessary, it's Darwinism.

And so science evolves in a manner reminiscent of the man who looks
for things under the lamppost, on a dark night, because the light
there is better.  We do the experiments we can do, with the equipment
we have, because otherwise we couldn't do anything.  But we end up
with a very skewed view of the world.  One of the things Darwinism
gave us was an insatiable curiosity, and so it is good to dream of
final theories, but the dream is rationally a silly one.  Every time
we open a new window on nature we see all sorts of things we never
dreamed of.  We are under a lamppost somewhere in the middle of
nowhere, and most of creation is totally unknown to us. Topics like
consciousness and ESP, etc, we ban from science, because we can't
begin to get a handle on them.  But that doesn't make them
unimportant. They are actually the essence of things for us.  Our
lamppost just doesn't throw any light in those directions.

About a real theory of everything, I always tell people that I
wouldn't want to live in a universe where I could understand
everything.  Or even the important things.  And I don't think there's
any danger of that.

So you see, I think the search for knowledge is a wonderful dream,
but it's a romantic, even Quixotic one.  I think we'll learn a lot
about controlling the small part of nature that we see.  But truly
understanding anything deep, when we aren't even aware of most of
what's out there?  Not very likely.

I'm afraid I can't even take science seriously at some level,
although I thrill to its beauty, and am willing to dedicate my life
to the search.  But it's hubris to expect too much, because any
reality we can perceive now, can only pertain to where we are now on
the evolutionary ladder.  I think that when we die, and go to heaven,
and ask God how things really work,  he will look at us and smile,
and say, ``first you need a brain transplant.''

As for my views on quantum theory, I'll tell you sometime when we're
drunk.  Anyway, thanks for coming, I enjoyed it very much.  And I
very much would like to come down and see you soon.
\eq

\section{06-12-02 \ \ {\it Quantum Information Does Not Exist} \ \ (to A. Duwell)} \label{Duwell1}

Jeff Bub passed on a copy of Appendix B to your PhD thesis to me, and I've got to tell you, I just love it.  And I love the title:  I don't know if you know it, but it is a wonderful play on the words Bruno de Finetti wrote in boldface in the preface to his book on probability theory:
\bq\noindent
     My thesis, paradoxically, and a little provocatively, but
     nonetheless genuinely, is simply this:
\begin{center}
                    PROBABILITY DOES NOT EXIST.
\end{center}
     The abandonment of superstitious beliefs about the existence
     of Phlogiston, the Cosmic Ether, Absolute Space and Time, \ldots,
     or Fairies and Witches, was an essential step along the road
     to scientific thinking. Probability, too, if regarded as
     something endowed with some kind of objective existence, is
     no less a misleading conception, an illusory attempt to
     exteriorize or materialize our true probabilistic beliefs.
\eq

I think you make a sound point in your paper and an important point, and I am behind you 100\%.  Indeed I have tried to make the same point many times over to my friends (like Jozsa), but I've never done it so clearly or thoroughly as you have done it in this document.  (As points of reference, you can have a look at the note titled ``\myref{Timpson1}{Colleague}'', on page 132 of my samizdat {\sl Quantum States:\ What the Hell Are They?\/}\ posted on my webpage (link below), or look at the ``Swedish Bikini Team'' and ``More Swedish Bikini Team'' stories on page 34 of my \quantph{0105039}.)  The point is, Landauer has caused a lot of trouble with his slogan ``information is physical.''  The real issue is that ``information carriers are physical,'' and by studying the information carrying properties and capabilities of those carriers we may get a new, more insightful way of expressing their very essence.

Anyway because of this respect for your work, Jeff Bub and I would like to invite you to include a copy of your paper in a special issue of {\sl Studies in History and Philosophy of Modern Physics\/} devoted to issues to do with quantum information that we are co-editing.  This is extremely short notice to give you because, essentially, we will have to have a final version of the paper by the end of next week.  But still, we hope you will say yes.  Let me exhibit a list of the confirmed papers (imaginatively including yours), so that you can get a feel for the gist of the issue.
\begin{enumerate}
\item
Howard Barnum: `Quantum information processing, operational quantum
   logic, convexity, and the foundations of physics'
\item
Armond Duwell: `Quantum information does not exist'
\item
Lucien Hardy: `Probability theories in general and quantum theory in particular'
\item
David {\Mermin}: `Teaching computer scientists quantum mechanics, or how I discovered the Copenhagen interpretation'
\item
Itamar Pitowsky: `Betting on the outcomes of measurements: a Bayesian theory of quantum probability'
\item
Andrew Steane: `A quantum computer only needs one universe'
\item
Chris {\Timpson}: `On the supposed conceptual inadequacy of the Shannon information'
\item
David Wallace: `Quantum probability and decision theory revisited'
\end{enumerate}
As I said above, we realize we are asking this on very short notice.  But I have read your paper twice now, and I would say that we can accept it almost exactly as it presently is \ldots\ and hope that that will be extra incentive for you to say yes.

Below, I'll just list a few (very minor) points that caught my eye, and suggest that you fix those things up if you do submit it for the special issue.  The scheme will be the following:  In a separate email, I'll send you the original manuscript that Jeff had sent me (in the form of a pdf file) for the purpose of knowing which pages I am talking about.  (I figure you might have changed things, or used a different format, since the time of the draft that I read.)\medskip

\noindent Hoping to hear from you soon (almost immediately actually),\medskip

\noindent Chris Fuchs\medskip

PS.  You might also enjoy the notes ``\myref{Bennett15}{Short Thoughtful Reply},'' ``\myref{Peres33}{Qubit and Teleportation Are Words},'' ``\myref{Bennett18}{Chris's World},'' and ``\myref{Bennett16}{King Broccoli},'' on pages 175, 184, 185, and 187, respectively, of my {\sl Quantum States:\ What the Hell Are They?}  These notes also connect somewhat---though not directly---to the points you make in your paper.\medskip

\subsection{Notes on ``Appendix B:\ Quantum Information Does Not Exist''}

\bq\noindent
\begin{itemize}
\item[1)]
p.\ 3, just a comment, not a request for change.  You write, ``\ldots\ it seems that Jozsa thinks that quantum information is somehow real.''  I can confirm that.  We've argued this over far more times than I care to remember!

\item[2)]  p.\ 5, footnote 2.  You give Brukner and Zeilinger more credit than they deserve on this.  That property goes back to Faddeev in the 1950s.  I think you can find a reference in the Timpson paper listed above (it can be found on quant-ph).  If you can't find it there, I'll dig harder.

\item[7)]
p.\ 15, bottom sentence.  I'm not sure I get this sentence.  Even epistemic entities or, if you like, classical information, can have qualities.  $H(X:Y)$, for instance, is bounded above by $H(X)$---that is a quality mutual information possesses.  But you must mean something more specific to the context here; so maybe just strive to say it a little better.

\item[9)]
p.\ 20, bottom paragraph.  You call $H(X:Y)$ the ``accessible information,'' but that is not the common usage.  Accessible information, as the term was coined by Schumacher in his PhD thesis, is defined as $\max H(X:Y)$ where the maximization is taken over all POVMs that can be performed at the destination.

\item[10)]
p.\ 20, bottom paragraph.  Also, the Holevo bound actually tells you very little in detail about the accessible information, in distinction to what you seem to indicate.  For instance, take a pure state ensemble whose density matrix is $I/d$ (the identity operator divided by the dimension).  If the ensemble happens to consist of orthogonal states, then the accessible information will equal the Holevo bound and be given by $\log d$.  On the other hand if the ensemble is the ``Scrooge ensemble'' that Jozsa, Robb, and Wootters talk about [Phys Rev A, {\bf 49} (1994), p.\ 668], then the accessible information will never exceed 0.61 bits, regardless of $d$.

\item[11)]
p.\ 21, first sentence, second paragraph.  I didn't like your phrase, ``one member of the singlet state.''  It seems that all too often I find people in the field of quantum information---for some reason that is mysterious to me---using the words ``system'' and ``state'' synonymously.  You might think that slip doesn't cause much trouble, but I think it causes a lot of damage.  (Especially if you are someone like me who tries to think of quantum mechanics as consisting of both ontic and epistemic terms, and seeing the interpretive task as classifying which is which.  For me, the system is ontic in nature, whereas the state is epistemic.)  Anyway, I might have said something like, ``Alice and Bob each possess a system drawn from a pair of systems prepared in the singlet state.''

\item[12)]
p.\ 21, last sentence, second paragraph.  Just a remark.  What is remarkable is that local operations alone can transform the initial state to states that live with the full span of the Hilbert space.  But I think you make this point somewhere later in the paper.

\item[13)]
p.\ 22, second paragraph.  I didn't like at all the part where you write, ``[S]he can establish correlations nonlocally.  This is not a perplexing property of information, but rather a well-known fact about quantum states.''  But I won't bother you about that.

\item[14)]
p.\ 23, first sentence in Section 8.  Again the system/state distinction problem.  Actually read the note ``Short Thoughtful Reply'' that I mentioned above, and you'll see more clearly why I'm bothered by this.

\item[24)]  p.\ 33.  You write, ``It is possible that given some ensemble which describes the information source, the number of qubits required for q-reliable communication can be less than the number of cbits required for reliable communication.''  It is not only possible, it is always true.  In the language {\Caves} and I used in the 1996 paper of ours that you cite, ``The preparation information is always greater than or equal to the von Neumann entropy.''  I think we prove it there.

\item[25)]  p.\ 34, the part below the displayed equation.  On my first reading of this, I thought you had just got it wrong.  I thought you were suggesting that the encoder in a Schumacher compression scheme could just throw away the qubits that were given to him and generate some new qubits prepared in an eigenstate of $\rho$ (doing so with the appropriate probabilities) and send those on.  But on second pass I realized you were not saying that.  The thing that saved you was the phrase ``formally equivalent.''  Still, I think it might be useful for your readers to make the distinction between this formal equivalence, and what is really done in the physical process of Schumacher compression.

\item[28)]  I would appreciate it very much if you would cite my samizdats (maybe the specific parts I listed above) in connection to the main thesis of your paper \ldots\ rather than just citing {\Caves} and Fuchs, 1996.  A good place for that might be your sentence on page 3, ``In fact, I will defend the position that quantum information theory is properly about properties of classical information in quantum systems and that no new concept of information is needed.''  The more readers I can draw into those samizdats, the more I can hope to have an impact on the community.
\end{itemize}
\eq

\section{07-12-02 \ \ {\it The Holiday Season} \ \ (to T. A. Brun)} \label{Brun7}

\btb
If you do go to the Republic of Ireland for this QI-fest, were you
planning on following up on the sort of stuff you talked about in your
``Quantum Mechanics as Quantum Information'' paper?  I don't mean to
sound like a broken record (and how long before no one remembers what
that simile refers to?), but I really did enjoy that paper; it's one
of the most interesting I've read in a long time, and I'm keen to see
where you go with it.  For instance, can you relate your
information-based approach to the physical process of gathering
information in an intuitive way?  You made a start at that, I think,
with your Bayes-rule-plus-disturbance formulation, but it seemed like
more might be possible along those lines.  That sort of thing
interests me very much.
\etb

Thanks for the encouraging note; I'm flattered.  Let me apologize for taking so long to reply.  I haven't had a chance to write nearly so much email in the last few months as I'd like to (and need to).

Concerning your first question, in fact that's all I plan to do in Ireland!  There's just so much to be done before this Bayesian picture can have any real substance.

The main lines I'd like to develop (or see developed) fall into two categories.  The first has to do with the shape of the restricted region in Figure 1.  Depending upon which measurement device is taken to be the ``standard quantum measurement device,'' that region will have various different shapes.  In particular, the volume of the region can be arbitrarily small.  It cannot, however, be arbitrarily large.  There is a supremal volume, and I'd like to know it.  For a qubit, the supremal volume is just the shape of a sphere.  In higher dimensions, unfortunately, it gets more complicated.  But I think finding a bound on the biggest volume could already be interesting.  In particular, I'd like to know how the ratio of sup region volume to full simplex volume scales with dimension.  Does it go to zero?  I suspect it does, but that's based on philosophy.

Beyond that, though, just the more we can say about the region, the better.  For instance, for a supremal-volume region, how many edges of the simplex will the region touch?  How many faces?  Etc.  What is for sure is that if we can get a characterization of these regions without having first started with POVMs, then we will have a new characterization of the state spaces of quantum mechanics.

The second line I'd like to see developed is the business I started in Section 7 of the paper.  Namely, giving some substance to the slogan, ``A quantum operation is just a density operator in disguise.''  Or more accurately, just as a quantum state can be viewed as nothing more than a probability distribution $P(x)$ restricted to a certain region of the simplex, a quantum operation should be viewable as nothing more than a conditional probability distribution $P(y|x)$ restricted to a certain region of the appropriate enveloping space.  Then, all the questions above arise again, but within this new regime.

I've shored up the ideas (and the arguments for them) in that Section pretty significantly since first writing it \ldots\ and also since writing the stuff on the subject in my samizdat {\sl Quantum States:\ What the Hell Are They?}\ posted on my webpage.  I'd like to get all that into a paper; it's just finding the time to do it.  If God will grace me, maybe I'll get it done in January after the holiday season passes.  In any case, some statement of it will definitely come out in the near future in the introduction and conclusions to the paper {\Ruediger}, Petra Scudo and I are writing on a de Finetti theorem for quantum operations.  Since {\Ruediger} is in charge of completing the draft that'll probably get done in short order.

Anyway, as I said above, I'm flattered that you find some of this stuff interesting.  And I'll be even more flattered if you find some of the answers to these questions!  If you get any good ideas, do let me know.  And if you'd like to collaborate on anything, let me know that too:  I'm hoping that after mid-January, I'll be able to regain my identity as a scientist again.

\section{07-12-02 \ \ {\it Prolix Boy} \ \ (to D. M. Greenberger)} \label{Greenberger2}

Thanks for the great note!  I especially liked the snake example.  I didn't realize before that we are so in synch with each other.

For fun, let me send you a PART of one my stories.  This one comes from page 237 of my samizdat, \quantph{0105039}.  It was originally written for Mermin.  By ``Firing Line,'' I really meant ``Reference Frame''---that always annoys him.

Why don't you come to Bell Labs some time in January.  You ought to give us a talk.

\section{09-12-02 \ \ {\it No Jack {\KennedyJF}} \ \ (to C. M. {\Caves})} \label{Caves70}
i
``Senator, I served with Jack {\KennedyJF}, I knew Jack {\KennedyJF}, Jack {\KennedyJF} was a friend of mine. Senator, you are no Jack {\KennedyJF}.''

I did tell you that Dick Slusher is my supervisor now, didn't I?  Well, he started reading my fat paper ``Quantum Mechanics as Quantum Information,'' today, and he came into my office for a while this afternoon to make a few comments on it.  In the course of that he also told me how he was concurrently reading the paper you wrote for a 1991/92 Spain meeting.  Apparently he has some interest in quantum chaos now.  I said, ``You shouldn't start with that; I think some of his more mature works were much better.''  He insisted that he needed to start at the beginning, saying, ``{\Carl} writes so clearly and methodically.''  I said---referring back to the part of the conversation about my own paper---``You know that's what I strive for too; {\Carl}'s been my model since the beginning.''  Dick was just about to walk out, but before he went, he looked me straight in the eye and said, ``Yeah, and you've got some way to go.''

I couldn't help but think of that Lloyd {\Bentsen} debate with Dan {\Quayle}.

\section{13-12-02 \ \ {\it More Addenda}\ \ \ (to A. Peres)} \label{Peres44}

Concerning Cabello, I find him just wonderful.  He gave a talk at our meeting in {\Montreal} on various KS things, but there was a part of it on MKC in particular.  I learned a lot in that section.  The main thing being, that just because {\it a\/} KS coloring exists, it does not mean that an {\it arbitrary\/} coloring exists.  Of course, that logical point should have been obvious, but it didn't strike me so much until {\Adan}'s presentation.

I had a funny conversation with David Meyer in Nashville this summer.  He told me, ``I had known about the Godsil and Zaks result for years, but didn't realize it was useful for anything until you asked me about the existence of KS theorems for rational vector spaces at an AMS meeting.''  It dawned on me then that I had indeed first asked that question, to Itamar Pitowsky, David Meyer, and probably to a few others, but I didn't recall ever having been thanked in the literature.  (All this raced through my head in a flash as I was talking to him.)  I said somewhat smilingly, ``Well, did you thank me for the question in your paper?''  He must have caught on, because his reply was, ``I don't know, did I?''  Since leaving the lobby and going back to my room that night, the smile turned into a little bit of a frown.  See pages 448 and 449 in my samizdat.

\section{19-12-02 \ \ {\it Which One Really?}\ \ \ (to G. L. Comer)} \label{Comer23}

Great new little poem.  Here, the lines that took me were:  ``The
freedom to choose!  The source of chance.''  Often someone will say
to me, as Howard Barnum did,
\bhb
     Here's a caricature, so feel free to object:  Bell's worry
     about the foundations of QM has been:  that we have ``measurement''
     as an ``unanalyzed primitive'' of the theory.  Everett shows us
     how to get around that.  You don't like Everett's resolution
     because you {\bf want\/} to have an unanalyzed primitive around so
     it can be the locus of free will.
\ehb
And I say it is not that.  The universe has within its categories two
species, one is chance, and one is free will.  Free will does not
rely on chance as its source.  Instead, it's only through the
intercourse of the two that we get a real birth.

Gotta run to a big division meeting.  Such things are always scary to
me.

\section{20-12-02 \ \ {\it Oh Haight Ashbury} \ \ (to N. D. {\Mermin})} \label{Mermin77}

\bdm
Footnote 13 repeats without acknowledgment a point that you, {\Carl},
and {\Ruediger} love to make.  More disturbingly, it promulgates a joke
that arose (for me) in the course of an email exchange with you that
I can no longer find.  For the life of me I can't remember whether I
made the joke or you did.  Googling on +``go ask Alice'' +``entangled
state'' produces nothing.  Feel free to delete footnote 13 if you
think I'm stealing your line.
\edm

Funny you can't remember the origin of the ``Go Ask Alice'' line.  We
had a conversation about it the moment I arrived at your {\Mermin}Fest.
Since it didn't sink in then, I'll tell you the story again.  At 6:56
PM that morning, I sent off a note to my old friend Greg Comer titled
``Go Ask Alice''---I'll place the first part of it below.  After the
part I quote here, however, the letter becomes quite personal.

So, you can imagine the shock when at 12:15 PM I receive a note to
you titled ``Go Ask Alice''!!!  I have always had a fear of getting a
note written and then, by accident, sending it to the wrong person! I
thought, ``Oh my God, I finally did it.''  However, reading your
note, I quickly realized that there was no connection at all \ldots
(other than a Jungian synchronicity).

\bdm
I don't see what your teleportation example (pages 11, 12) adds to
ordinary EPR.  Aren't all the issues exactly the same if Alice ``in
her laboratory prepares'' the single qubit in (1) that she possesses
by an appropriate measurement (to be sure, she can't control which
outcome she'll get, but that doesn't seem to be central to your
point, or is it?) after which she and only she knows what the outcome
of the corresponding yes-no measurement on Bob's qubit will be.

But this, of course, has been debated in the EPR context for
generations, and I don't see the force of your argument that the
would-be informationist should not be weak in the knees.  Everybody
agrees that Alice is the only one who can do the trick.  There's no
problem if she only does it once.  Bob says the YES was random and
only Alice knows that it had to be YES. Bob can think she has
delusions of grandeur.  But if she does it right 10,000 times, then on
run 10,001 Bob would be a fool if a certain confidence that Alice
can call it every time should start to enter his mind too.

So to find out what the qubit will do you do indeed have to go ask
Alice.  (Isn't that an old Jefferson Airplane (pre-starship) song?)
You ask her; you don't ask the qubit (as you guys like to say). But
by run 10,001 everybody who has been paying attention (except
confirmed Humeans) will be fairly sure that ``the system [is]
prepared to reveal'' whatever answer Alice has sent over in a
lock-box.

I'm not saying I agree with Penrose on this.  Just that I don't
see that you've helped very much in relieving the queasiness
one is left with when one denies the psi-ness.
\edm

On 5/29/02 I finally replied at length in a note titled ``I Think
She'll Know.''  The opening lines of the note were, ``Remember what
the dormouse said; feed your head.''

But back to your question:  I think beyond a doubt, for the present
context, you were the inventor of the phrase ``Go Ask Alice.''  It's
a nice way to put it.  (But only people of our generation are gonna
get it.)

But the importance of the point in our little group, as far as I can
remember, was first brought out clearly by {\Carl}.  It's always been
{\Carl}'s favorite ``argument'' for the subjectivity of the wave
function. This piqued my own curiosity to see where we first put it
in print. I could find a trace of the idea in Footnote 44 of our
paper \quantph{9601025}.  Also I could see it shining through
just after equation 4.105 on page 120 of my \quantph{9601020}.

If you think \quantph{9601025} gets sufficiently close to the
mark, it might be nice if you'd cite it.

\section{24-12-02 \ \ {\it Give Us a Pluriverse} \ \ (to G. L. Comer)} \label{Comer24}

\bgc
I've been fiddling around with the idea, though, that chance is a
result of the intercourse between free-willed entities.
\egc

Yes, I think I like that better.  I had played with still a different
turn for a while.  One that I might sloganize like this:  Chance is
what you call ``it'' when viewed from the outside; free will is what
you call ``it'' when viewed from the inside.  What I wrote you in the
last note, was a small attempt to get away from the monism of that
slogan. But what you said above, I think, might now appeal to me even
more.

Merry Christmas.

\section{28-12-02 \ \ {\it Two Things} \ \ (to J. W. Nicholson)} \label{Nicholson14}

Just reading a NYTimes article on cloning in which I read:
\bq\noindent
     Senator Bill Frist, the Tennessee Republican who was just chosen
     Senate majority leader and who favors a ban on all forms of
     cloning, called Friday's announcement ``disturbing'' and added:
     ``While its validity is unclear, it should serve as a chilling
     reminder that individuals are still trying to clone human beings.
     These actions offend our sensibilities and undermine fundamental
     respect for the decency of human life.''
\eq
Now, tell me how could someone say that if he didn't believe that the genetic makeup IS the person \ldots\ rather than predominantly the watermark?  (Recall our conversation on the way to lunch one Friday.)

\section{29-12-02 \ \ {\it Erratum?}\ \ \ (to A. Peres)} \label{Peres45}

I just noticed that Conway (of the Conway-Kochen noncolorability result) makes no appearance in the Author Index of your book.  Maybe that is an erratum, or maybe I don't properly understand your criteria for admission into that index.  Either way, cheers!

\subsection{Asher's Reply}

\bq
I did not include Conway in the author index because there was no
reference to cite. Recently he and Kochen published a paper
\begin{itemize}
\item
J. H. Conway and S. Kochen, ``The geometry of the quantum
paradoxes'', in R. A. Bertlmann and A. Zeilinger (eds.), {\sl Quantum
[un]speakables: From Bell to quantum information (Vienna, 2000)},
Springer-Verlag, Berlin, 2002, pp.\ 257--269,
\end{itemize}
but these are mostly anecdotes. The only proof I know of their
construction is the one in my book.
\eq

\chapter{2003: Seeking SICs}

\section{02-01-03 \ \ {\it Foundations Site?}\ \ \ (to P. Busch)} \label{Busch2}

Thanks for the inclusion of a link on your webpage.  The more people I can draw into my ``quantum dreams'' page, the more people I can hope to take the bait of thinking about quantum mechanics as the hint of something deeper and more wonderful.  And the more people I can hope to start applying their brain pulp to the project!  So, thanks again.

Congratulations on your promotion.  When I look at your research record I guess I would have thought you to have become a full professor long, long ago!  Certainly this is overdue.

I will be on a sabbatical from Bell Labs in Dublin in the coming year, so maybe I'll get a chance to drop by Hull for a visit.

Speaking of seeing you, Howard Barnum and I are organizing a session for one of Andrei Khrennikov's meetings in Sweden this summer.  (I think it is in the early part of June.)  Our plan is to give it the theme ``quantum information meets quantum logic'' and we're trying to figure out how to stretch the money we've been given to get as many interesting people there as possible.  Some of the people we're thinking about trying to attract are you, Jeff Bub, Lucien Hardy, Dave Foulis, Alex Wilce, Dick Greechie, and maybe a few others.  a) Would you be interested, and b) if so, would you need travel funding to get there?

\section{02-01-03 \ \ {\it Glory Days} \ \ (to A. Peres)} \label{Peres46}

\bap
Yesterday the Dept Chairman informed me that it was impossible to
extend by one more year my professorial position, because of the
dismal financial situation of Technion (well, the whole country is
bankrupt). \ldots

I am not unhappy to ``retire'' this fall. I won't have to teach or
bother with administrative duties, and I'll have more time for
research.
\eap

Congratulations on your new career move!  You should think of
yourself as being a postdoc again, and what glorious days you will
have!

Here:  I'll give you your first postdoctoral research project!  I learned
predominantly from you that the proper analog to the quantum state in
classical physics is the Liouville distribution.  However, I learned
from Wigner that the only time evolutions for quantum states that are
overlap preserving are (up to phase equivalence) the unitary
evolutions:
\begin{itemize}
\item
V.~Bargmann, ``Note on Wigner's Theorem on Symmetry
Operations,'' J. Math.\ Phys.\ {\bf 5}, 862--868 (1964).

\item
C.~S. Sharma and D.~F. Almeida, ``A Direct Proof of Wigner's Theorem on Maps Which Preserve Transition Probabilities between Pure States of Quantum Systems,'' Ann.\ Phys.\ {\bf 197}, 300-309 (1990).

\item
U.~Uhlhorn, ``Representation of Symmetry Transformations in Quantum Mechanics,'' Arkiv F\"or Fysik {\bf 23}, 307--340 (1963).
\end{itemize}

My question is, what are the complete set of time evolutions on a
classical phase space that are overlap preserving for Liouville
distributions.  As far as I can tell, this question has never been
tackled in the literature.

\section{02-01-03 \ \ {\it The Quantum Shirt}\ \ \ (to A. Cabello)} \label{Cabello4}

Yesterday morning, New Year's Day, as I went to get my wife's newspaper, I discovered that the previous day's mail had been accidentally delivered to my front door, rather than to the usual mailbox in the back.  Anyway, in it, there was a package from Spain.  Thanks so much for the shirt!  I showed it off to my whole family and the friends that came to our afternoon celebration.

You are a great friend, and I am flattered.

\section{03-01-03 \ \ {\it Growing Old} \ \ (to L. Hardy)} \label{Hardy10}

The last few days I've been editing my big samizdat---mostly just working on completing the name index---and what a task it has been!  (The index is now almost 10 pages long all by itself.)  But maybe one interesting thing (to me anyway) has come from the project:  I'm starting to realize just how little the broad outline of my program has changed over the years.  I don't know if that's good or bad, but it seems true.  Apparently things started to gel in me somewhere around 1995/1996, and I've been trying to make the thought more precise ever since.

For instance, in my \quantph{0205039} this year, I write in a footnote:
\bq\noindent
It is at this point that the present account of quantum
mechanics differs most crucially from Refs.~[Hardy01a]
and [Hardy01b].  Hardy sees quantum mechanics as a
generalization and extension of classical probability
theory, whereas quantum mechanics is depicted here as a
restriction to probability theory.  It is a restriction
that takes into account how we ought to think and gamble in
light of a certain physical fact---a fact we are working
like crazy to identify.
\eq
and I had thought that this characterization of what I'm shooting for arose only after reading your papers.  But just a few minutes ago, I found this old note to Sam Braunstein:
\bq
\noindent
{\bf 19 July 1996, \ to Sam Braunstein, \ ``The Prior''}\medskip

While in Torino, you really got me interested in the old {\Cox} Box question again.  I noticed in this version of the book that {\Jaynes} makes some points about how there are still quite a few questions about how to set priors when you don't even know how many outcomes there are to a given experiment, i.e., you don't even know the cardinality of your sample space.  That, it seems to me, has something of the flavor of quantum mechanics \ldots\ where you have an extra freedom not even imagined in classical probability.  The states of knowledge are now quantum states instead of probability distributions; and one reason for this is that the sample space is not fixed---any POVM corresponds to a valid question of the system. The number of outcomes of the experiment can be as small as two or, instead, as large as you want.

However I don't think there's anything interesting to be gained from {\it simply\/} trying to redo the {\Cox}ian ``plausibility'' argument but with complex numbers. It seems to me that it'll more necessarily be something along the lines of: ``When you ask me, ``Where do all the quantum mechanical outcomes come from?''  I must reply, ``There is no where there.''  (with apologies to [Gertrude Stein] again!)  That is to say, my favorite ``happy'' thought is that when we know how to properly take into account the piece of prior information that ``there is no where there'' concerning the origin of quantum mechanical measurement outcomes, then we will be left with ``plausibility spaces'' that are so restricted as to be isomorphic to Hilbert spaces.  But that's just thinking my fantasies out loud.
\eq
Maybe it's just evidence of hardening of the arteries.

By the way, when are you going to send us the final version of your SHPMP paper, with references added?  I hope you'll read this sentence and say, ``Now.''

\section{03-01-03 \ \ {\it QM with Finite Fields} \ \ (to J. M. Renes and C. M. Caves)} \label{Renes11} \label{Caves70.1}

I just came across a paper title that might interest some combination of us three:
\begin{itemize}
\item
J. P. Eckmann and Ph. Ch.\ Zabey, ``Impossibility of quantum mechanics in a Hilbert space over a finite field,'' Helvetica Physica Acta {\bf 42}, 420--424 (1969).
\end{itemize}
I have no clue what the actual content of the paper is, but it might be worth a look one day.  In particular, does it connect any with the fact that the POVM-version of Gleason probably breaks down when the vector spaces are over a finite field?  Or does it?

\section{05-01-03 \ \ {\it ForAsher.tex} \ \ (to A. Peres)} \label{Peres47}

\bap
Thank you for telling me that I'll be like a postdoc. It's so true.
Now this will be my way of telling it to friends. Only one thing will
be missing: a good adviser for my research (my first adviser in 1961
was Wheeler, although I was formally the postdoc of Misner). Will you
be willing to play that role?
\eap

Those are awfully big shoes, and I wouldn't dare to fill them.  Nor would I even dare to the presumption of being your advisor!  You have been my teacher since before we met, and my teacher you remain.  (I think I first read one of your papers in 1989.)

\bap
You already asked:
\bq\noindent{\rm
    What are the complete set of time evolutions on a classical phase
    space that are overlap preserving for Liouville distributions?}
\eq
All I can think of is Koopman's theorem (my book, pp.\ 317--318). The
phase space evolution is also a unitary evolution. The overlap of two
Liouville functions is constant in time in any Hamiltonian evolution.
\eap

Yes, that's right.  And what I'm asking is whether there is a converse to this.  If the converse were true, then it would mean that the sum content of Hamiltonian evolution is overlap preservation for Liouville distributions.  I think that would be pretty if it's true.  Also, if it is true, I don't think it is a completely trivial consequence of the usual Wigner theorem.  I say this because I am asking for the preservation of Liouville overlaps only, not necessarily the preservation of overlaps between arbitrary functions on the phase space.

\bap
Have you seen {\tt 0301001}, the first {\tt quant-ph} of the year? It's by
Grangier, who participated in the first Aspect experiment, and now
writes some ``foundational'' papers. I like the expression ``quantum
holism'' instead of ``nonlocality.'' But there are many errors in that
paper.
\eap

No I haven't seen that one.  I'll try to have a look at it eventually.  Philippe and I have had several discussions on quantum foundations, to no great avail.

\bap
I searched {\tt quant-ph\/} for the string ``unknown'' and I found your {\tt 0104088}.
Where was this published? Now I also found Mermin's ``Whose knowledge?''\
{\tt 0107151}. Probably these two references are enough for my purpose.
\eap

Yes, I think that will do.  Here's my full reference on it:
\begin{itemize}
\item
C.~M. Caves, C.~A. Fuchs and R.~Schack, ``Unknown Quantum States:\ The Quantum de Finetti Representation,'' {\sl Journal of Mathematical Physics\/} {\bf 43}(9), 4537--4559 (2002). [Reprinted in {\sl Virtual Journal of Quantum Information\/} {\bf 2}(9).] \quantph{0104088}.
\end{itemize}

The package of things I put together for you is concerned with some discussions I had with Charlie Bennett, John Smolin, and Philippe Grangier about the perils of thinking of quantum states as ``properties'' of the systems they refer to.  I think that gets singularly in the way of Charlie's public expositions of what quantum teleportation is about.  Putting together a package seemed a little relevant since you said you'd be giving a lecture on the subject:  I don't think you will learn anything from the content, but you may enjoy some of the turns of phrase.

I'll place the code of my file ``ForAsher.tex'' below.  [See 25-04-02 note ``\myref{Bennett15}{Short Thoughtful Reply}'' to C. H. {\Bennett}, 14-05-02 note ``\myref{Bennett17}{Qubit and Teleportation Are Words}'' to C. H. {\Bennett} and others, 14-05-02 note ``\myref{Bennett18}{Chris's World}'' to J. A. Smolin and others, 16-05-02 note ``\myref{Bennett16}{King Broccoli}'' to J. A. Smolin and others, and 01-06-02 note ``\myref{Grangier7}{High Dispute}'' to P. Grangier.]

\section{06-01-03 \ \ {\it Pedagogy} \ \ (to N. D. {\Mermin})} \label{Mermin78}

Now I'd like to use your patient ear as an excuse to think out loud.
This is what I was referring to when I wrote you yesterday, ``Much
more soon.''

Today, I've got to spend part of the day preparing a talk to give at
Bell Labs tomorrow.  Here's the title and abstract I sent in for it:

\begin{verse}
Title:     Representing Quantum Mechanics on the Probability Simplex
\\
Abstract:  Classical information theory is about input probability
distributions, output probability distributions, and the transition functions that connect them.  Quantum mechanics and so far quantum information theory, on the other hand, have been traditionally formulated in terms of linear operators on a complex vector space and the linear superoperators that connect them.  To automate a comparison
between the two theories, a means for expressing the newer theory in a way that leans toward the older, more established one ought to be sought.  It can be done.  This talk is about a small part of that project and a couple of mathematical questions it poses.
\end{verse}

In substance, the talk will focus predominantly on the stuff I
presented to you (privately) in {\Montreal}, but will give a little more emphasis to the stuff about symmetric POVMs that {\Gabe} looked at this summer.

However, I'd like to give the beginning of the talk a little
different slant than I had previously.  Here it is.  (Here's the part
where I'm thinking out loud.)

Everybody has their favorite speculation about what powers quantum
information and computing.  Some say it is the superposition
principle, some say it is the parallel computation of many worlds,
some say it is the mysteries of quantum entanglement, some say it is
the exponential growth of computational space due to the tensor
product.  For my own part though, my favorite speculation is that it
is Newton's Third Law:  For every action, there is an equal and
opposite reaction.  Indeed I sometimes wonder if the very essence of
quantum mechanics isn't just this principle, only carried through far
more consistently than Newton could have envisioned.  That is to say,
absolutely NOTHING is exempt from it.

What do I mean by this?  What might have been exempt from the
principle in the first place?  To give an answer, let me note an
equivalent formulation of old Newton.  For every REACTION, there is
an equal and opposite ACTION.  Strange sounding, but there's nothing
wrong with it, and more importantly, this formulation allows for the
possibility of an immediate connection to information theory.  In
particular, we should not forget how information gathering is
represented in the Shannon theory.  An agent has gathered
information---by the very definition of the process---when something
in his environment has caused him to REACT by way of revising a prior
expectation $p(h)$ (for some phenomenon) to a posterior expectation
$p(h|d)$ (for the same phenomenon).

When information is gathered, it is because we are reacting to the
stimulation of something external to us.  The great lesson of quantum
mechanics may just be that information gathering is physical.  Even
something so seemingly unimportant to the rest of the universe as the
reactions that cause the revisions of our expectations are not exempt
from Newton's Third Law.  When we react to the world's stimulations
upon us, it too must react to our stimulations upon it.

The question is, how might we envision a world with this
property---i.e., with such a serious accounting of Newton's law---but
in a way that does not make a priori use of the information gathering
agent himself?  If the question can be answered at all, the task of
finding an answer will be some tall order.  For never before in
science have we encountered a situation where the theorizing
scientist is so inextricably bound up with what he is trying to
theorize about in the first place.

It's almost a paradoxical situation.  On the one hand we'd like to
step outside the world and get a clear view of what it looks like
without the scientist necessarily in the picture.  But on the other
hand, to even pose the question we have to imagine an information
gathering agent set in the middle of it all.  You see, neither
Shannon nor any of modern information theory has given us a way to
talk about the concept of information gain without first introducing
the agent-centered concept of an expectation $p(h)$.

So, how to make progress?  What we do know is that we actually are in
the middle of the world thinking about it.  Maybe our strategy ought
to be to use that very vantage point to get as close as we can to the
goal.  That is, though we may not know what the world looks like
without the information gathering agent in it, we certainly do know
something about what it looks like with him in:  We know, for
instance, that he ought to use the formal structure of quantum
mechanics when thinking about physical systems.  Beyond that, we know
of an imaginary world where Newton's Third Law was never taken so
seriously:  It is the standard world of classical physics and
Bayesian probability.

Thus, maybe the thing to do first is to look inward, before looking
outward.  About ourselves, at the very least, we can ask how has the
formal structure of our {\it behavior\/} changed since moving from
what we thought to be a classical Bayesian world to what we now
believe to be a quantum world?  In that DIFFERENTIAL---the
speculation is---we may just find the cleanest statement yet of what
the quantum world is all about.  For it is in that differential, that
the world without us surely rears its head.

To do this, we must first express quantum mechanics in a way that it
can be directly compared to classical Bayesian theory, where the
information-gathering agent was detached from the world.  That is
what this lecture is about \ldots\

As I say, just thinking out loud.  Thanks for the imaginary ear.

\section{07-01-03 \ \ {\it Newton's Third Law} \ \ (to R. E. Slusher)} \label{Slusher1}

By the way, here's the set of notes I wrote for myself before giving the talk Monday.  Since the business about quantum mechanics being an expression of something like Newton's third law, but much deeper, seems to strike a resonance with something you were groping for one day in one of our discussions---i.e., when you were saying things like ``you can't leave anything out in the quantum world''---I thought might enjoy reading this.  [See 06-01-03 note ``\myref{Mermin78}{Pedagogy}'' to N. D. {\Mermin}.]

\section{08-01-03 \ \ {\it Down with Vader, Up with Construction!}\ \ \ (to G. Herling)} \label{Herling1}

That's great news to hear that you're becoming less Vaderish.  I've got my fingers crossed for the full recovery.

\bgh
Thanks for thinking of me, but, I hope, not in the same breath as
Feyerabend!!!  You quote him in the QM notes that we downloaded.
\egh

I find the ``we'' in the second sentence encouraging.  Who does ``we'' consist of?

I don't know too much about Feyerabend.  I presume you're talking about the note I wrote to Michael Nielsen titled, ``Fun with Feyerabend''?  [See 25-05-02 note ``\myref{Nielsen2}{Fun with Feyerabend}'' to M. A. Nielsen.] I do like that quote, and think there's something to it.  However, I read his (posthumous) book {\sl Conquest of Abundance\/} and didn't get much more than that out of it.  Maybe his other books are more worthwhile \ldots\ or maybe they'll just expose him to be the crank he really is \ldots\ but I'm going to keep the jury out for a while.

If you want to see what I'm really shooting for in such directions, have a look at two of the notes I wrote to Howard Wiseman:  ``The World Is Under Construction'' and ``Probabilism All the Way Up.''  [See 24-06-02 note ``\myref{Wiseman6}{The World is Under Construction}'' and 27-06-02 note ``\myref{Wiseman8}{Probabilism All the Way Up},'' both to H. M. Wiseman.]  I think they're much better expressions of the program.

\section{08-01-03 \ \ {\it Your Note to Grangier} \ \ (to N. D. {\Mermin})} \label{Mermin79}

I haven't read the Grangier paper outside of the sentence he quoted
from me, but I did read your letter.  I think you do me pretty good
justice, and I rather liked your explanation.

\bdm
At the beginning of section II of \quantph{0301001}, I believe
you miss Chris Fuchs's point.  He is not talking about Bell's
theorem; he is talking about something rather like pre-Bell EPR.
After you have made a measurement on subsystem A, the information you
acquire permits you to assign a quantum state (as defined by your
second paragraph [in boldface type]) to subsystem B.  Before the
measurement on subsystem A, subsystem B had no quantum state.

Fuchs uses this to argue that the quantum state of a system (or
subsystem) --- when it has one --- cannot be an objective property of
that system since statehood can (under EPR conditions) be conferred
on a system from afar.

You want to have it both ways --- denying action at a distance, yet
maintaining the objectivity of the quantum state.  I believe you can
do it, but only if you acknowledge that the objective state of
subsystem B after subsystem A has been measured (so B does indeed
have a state) is not a local property of subsystem B.

I would say that the state of subsystem B is a compact way of
summarizing (1) the preparation of the A-B system that resulted in
the original EPR state, (2) the fact that nothing further was done to
B, (3) the fact that a measurement was performed on A, and (4) the
outcome of that measurement.  Because (1)-(4) are all objective facts
(Fuchs would disagree about this) the state of B can be said to be
objective.  But it would be dangerously misleading to call it an
objective property of B, because this suggests something residing in
B, whereas (1)-(4) are statements about both A and B and their
earlier history.
\edm

True enough.  But then here's my challenge to you.  It's something I
should have challenged to you too long ago.  Once you have that a
state can be said to be objective, and once you have that a
measurement specification can said to be objective, then through the
Born rule you have that the probabilities generated by the
measurement can be said to be objective too.  If so, then you must be
able to give me a definition of what it means to be a particular
probability value $q$ in a way that 1) is not circular, and 2) makes
no necessary use of a gambler.  (For after all, invoking a gambler is
just another way of invoking the agent/observer/experimentalist once
again.)

\section{10-01-03 \ \ {\it Filth Under the Rug} \ \ (to N. D. {\Mermin})} \label{Mermin80}

\bdm
What's wrong with the old frequentist definition?  If $N$ different
sets of qubits are subject to those same objective conditions the
fraction of final measurements giving the outcome $x$ gets very close
$p(x)$ when $N$ is large.

Granted you need more probabilistic statements to say what you mean
by ``gets very close'' (is that what you meant by circular?)\ but
surely that's an issue for any non-Bayesian view of statistics and
not peculiar to the interpretation of QM.
\edm

Now it is MY turn to be shocked by the triviality of YOUR reply.  I
almost feel like it is 1996 again, and we haven't made a bit of
progress in our discussions.  That bugs the hell out of me.  What
have I been wasting my time on all these years?  You might as well
still be writing papers that say, ``If all quantum puzzles can indeed
be reduced to the single puzzle of interpreting objective
probabilities, I would count that as progress.''

If what you're still shooting for is to sweep the issues of quantum
mechanics under THAT rug---and there's every indication you
are---you're just going to find more filth and dirt there.  What a
shame really.  I thought you had been slowly absorbing the Bayesian
point all this time.  I guess I had just not expected you to fail my
challenge in this facile way \ldots\ and I'm taken aback.

I'm fairly confident I understand (from your notes to Philippe and
{\Ruediger} and also your newest paper) your latest view of what the
quantum state is about.  And it might as well be the same view as
expressed in Asher's paper ``What is a State Vector?'' [AJP, 1984, p.
644].  What he calls a ``procedure'' or an ``instruction set'' you
call a ``history,'' but that's essentially where the difference ends.
And just as that paper led Asher to no deeper or more convincing
insight for 15 years, so too it will be with you if you continue down
this \ldots\ I almost said ``path,'' but maybe I should say ``dead
end.''

Let me pick up on your last sentence above:

\bdm
Granted you need more probabilistic statements to say what you mean
by ``gets very close'' (is that what you meant by circular?)\ but
surely that's an issue for any non-Bayesian view of statistics and
not peculiar to the interpretation of QM.
\edm

You get this completely backwards.  The great insight Ed Jaynes had,
and that {\Carl} and {\Ruediger} and I have slowly been reckoning with, is
that because quantum mechanics is so intimately tied up with
probability, one cannot hope to disentangle the troubles of quantum
mechanics without FIRST clearing up what the formal structure of
probability theory is actually about.  And on that first count, we
think the Bayesians are the winners.

Once that is accepted---clearly you haven't accepted it yet, but that
is no matter for the argument I want to make---then the task is to
ask, what are the IMPLICATIONS of that acceptance for our
understanding of quantum mechanics?  What my debates with you, and to
a lesser extent {\Carl}, and to a still lesser extent {\Ruediger}, have
been about since the beginning of your BFM murmurs is just this: What
are the implications?

My starting point has been the unbending acceptance that
probabilities are of the (de Finettian) subjective caste.  What are
the IMPLICATIONS of this?  Well, the first thing one gets is that the
quantum state is of the same subjective caste.  But then---and I
don't know why it was so hard to stumble across this, except possibly
for sheer prejudice---the next thing one gets is that at least some
quantum operations are also of the same subjective caste.  For
beauty's sake, I then go further than {\Carl} and {\Ruediger} are
presently willing to go, and say, ``If so be it for SOME quantum
operations, then so be it for ALL quantum operations.''  But the main
point is that the first three steps of this paragraph are pure
implication.

You called it poetry at the {\Montreal} meeting---yes, it did hurt a
little---but it is logic just as clean as you can get it.  And it's
of the most elementary sort.  (That's what made your remark hurt.)
The only thing that makes it appear to be poetry to you is some deep
resistance and, I suspect, fear of where it leads.

So, I have implications that run FROM interpretation of probability
TO quantum mechanics.  So what?  If it just stays at you saying
objective every time I say subjective, then this is a worthless
exercise and a waste of time.  There had better be more to it.  And I
claim there is.

For, I would say the implications above lead me down a mathematical
path.  Whereas your hope to retain the word objective for these
structures leads you nowhere.

\begin{enumerate}
\item
My point of view COMPELS me to ask whether there is a way to think of
a quantum state as a (single) probability distribution, plain and
simple.  With a little toil, I find there is.

\item
My point of view COMPELS me to seek out the analogies between Bayes'
rule and quantum collapse.  With a little toil, I find an analogy
that's never been found before.

\item
My point of view COMPELS me to ask why the notion of quantum
measurement is anything other than the refinement of one's belief,
i.e.,  exactly what classical (Bayesian) measurement is about.  With
a little toil, I find that it is precisely that after all \ldots\
just that in the quantum case there is an extra little kick given to
my final state of belief.

\item
Here, I think this is the most important one:  My point of view
COMPELS me to ask, if a quantum operation is as subjective as a
quantum state, then why are the two mathematical structures not
formally identical?  And we are led back to Jamio{\l}kowski's and
Choi's old insight:  A quantum operation IS a density operator.  With
a little reflection, one sees that that had to be \ldots\ in the same
way that {\em prior\/} probabilities $p(h)$ and {\em conditional\/}
probabilities $p(h|d)$ are both probabilities nevertheless.
\end{enumerate}

I would dare say that your point of view---where probability theory
is, at very best, secondary to, or at very worst, absolutely detached
from the deeper issues of quantum mechanics---would leave all of
these things as little more than coincidences.  ``There is a way to
map quantum operations and unitary operators to density operators?
Who cares?  It's just as mysterious as the structure of quantum
mechanics to begin with.''

But as long as there are coincidences in the structure of the theory,
that structure will always be a mystery.  What I think Bayesian
probability theory does for us is COMPEL us to view as natural the
connections we see within the axioms of quantum theory, rather than
as miracles plain and simple.

So you see, you have depressed me.  If I can't make any headway with
my best and most sympathetic friends---you're one of them---I don't
see how I'm going to make any headway in the wider world.  Even YOU
had not realized that all this talk about Bayesian stuff was meant to
LEAD us, and not just be an afterthought tacked on for NOTHING BUT
philosophical reasons.

Have you read {\Carl}'s document ``Resource material for promoting the
Bayesian view of everything'' posted at his website \myurl[http://info.phys.unm.edu/~caves/]{http://info.phys.unm.edu/$\sim$caves/}?  It would do you some good. It's
about time you took a course in Bayesian Ideas 101.  ``What's wrong
with the old frequentist definition?''---that about knocked me over!!

If after reading this note, you don't think it is too offensive, I
may forward it to {\Carl} and {\Ruediger}.  I'll bet they too will be
shocked---though much more polite in reaction than me---by this
dangerous frequentist tendency you're starting to reveal.

I thought I was going to write a little report on your Copenhagen
Computation today, but I knew I couldn't touch it until I got this
off my chest.  Sorry about that.  I won't be able to write you a
report until Sunday now.  (I'll be in NYC tomorrow.)

In friendship, disappointment, and enduring hope, \ldots

\section{13-01-03 \ \ {\it Ouch} \ \ (to N. D. {\Mermin})} \label{Mermin81}

I told Kiki the other night, ``Well, I probably lost a friend
today.'' In an ominous voice she replied, ``What'd you do?''  I said,
``I wrote David {\Mermin} a scathing note about some stuff in quantum
mechanics.'' In a scolding tone, she said, ``Why do you always do
that?'' I said, ``I just couldn't take it.  This guy hasn't hardly
absorbed a thing in our six years of discussion!''  She said, ``Why
do you get so upset? You know your work's never going to be done:
Churches never go out of business, do they?''

Ouch.

\section{13-01-03 \ \ {\it Give Us a Pluriverse, 2} \ \ (to G. L. Comer)} \label{Comer25}

I like that word pluriverse!  I don't think I've ever heard it before.  Did you make it up?

\section{13-01-03 \ \ {\it Give Us a Pluriverse, 3} \ \ (to G. L. Comer)} \label{Comer26}

\bgc
Nope.  I stole it from the subject of one of your e-mails.  Or are you
pulling my leg here?
\egc

No, I guess I'm just being an idiot.  The ``Re:''\ in your title might have been a hint?!?!  I guess that brain cell died.  The word sounds very Jamesian.  I wonder if I stole it from him?

Your poem kind of has a quality like that song on Sgt.\ Pepper, can't remember the name.  Something like ``For the Benefit of Mr.\ Kite.''

\section{13-01-03 \ \ {\it Der Kopenhagener Geist} \ \ (to N. D. {\Mermin})} \label{Mermin82}

I actually don't have much to say in the capacity of a referee.

First, a couple of typos: [\ldots]

Finally, a little technical point: [\ldots]

Now, let me tell you the thoughts you provoked as I was reading the
paper.

\bdm
     The state of $n$ Qbits has no meaning going beyond the abstract
     state vector itself, together with the rules for how it can be
     constructed and the computational uses to which it can be put.
     We return to this below, merely noting for now that although we
     shall speak often, as everybody does, of ``the state of $n$
     Qbits'' the terminology is potentially misleading.  It must not be
     taken to imply that the state characterizes a property possessed by
     and directly inferable from those Qbits, as it does for Cbits.  A
     better, but clumsier usage, would be always to say ``the state
     associated with $n$ Qbits''.
\edm

For my own part, I would say, ``the state ascribed to $n$ Qbits.''
``Ascribed to'' is not a hell of a lot clumsier than ``of,'' and it
has the advantage that it makes clear and serves as a constant
reminder that the origin of the ``state'' is not in the system
itself, but in a system external to it---namely, the agent.

\bdm
     The fact that the generic multi-Qbit state is incompatible with
     associating states with the individual Qbits is already an
     indication that Qbit states have a much more abstract character
     than the states of Cbits, which are always products of one-Cbit
     states.
\edm

Nice sentence.  I like it.  That's the way everyone ought to view the
issue.

\bdm
     The state associated with the Qbits is merely an extremely
     convenient way of recording the potential consequences of the past
     actions of the computer on those Qbits.  The consequences of
     those past actions can become accessible in only one way, and this
     way is the only way to extract information from $n$ Qbits: one can
     {\it measure\/} them.
\edm

There's something about this that I don't quite like, but I'm having
trouble putting my finger on it.  Probably has something to do with
your using the word ``consequences'' in a way that I wouldn't
endorse.  See glossary on page 49 of my {\sl Quantum States:
W.H.A.T.?}  [See 04-09-01 note ``\myref{Schack5}{Note on Terminology}'' to C. M. Caves and R. Schack.]

\bdm
     The Born rule contains, as a special case, a quantum imitation of
     the unproblematic (and therefore usually unremarked upon) process
     of extracting information from Cbits.  \ldots\
     The statistical, state-altering character of the outcome of a
     measurement of $n$ Qbits in a general state becomes the
     deterministic, state-preserving, unproblematic classical extraction
     of information when the state is one of the $2^n$ classical-basis
     states.  (page 7)
\edm
and
\bdm
     The view of quantum mechanics I gave my computer scientists relies
     on a primitive notion of measurement, without which the computation
     has neither a beginning nor an end.  A measurement gate is a black
     box whose interaction with the $n$ Qbits results in an unambiguous
     output on a display, whose reading is as unproblematic as reading
     the display of an ordinary classical computer.  Measurement is
     where the quantum-computational process starts, by permitting the
     association of an initial state with the Qbits, and finishes, by
     producing an unambiguous digital output.  Quantum computer science
     delves no more deeply into how information is actually extracted
     from a measurement than does classical computer science, where the
     preparation of the initial state of the Cbits and the reading of
     their final state are steps too trivial to warrant explicit
     theoretical attention, though they are certainly of concern to the
     engineers who design the computer.  (page 9)
\edm

I like these lines.  But ask a philosopher if this is unproblematic.
The reason the physicist finds it unproblematic is the same reason he
finds the independence or nonindependence of the continuum hypothesis
unproblematic in transfinite set theory:  He never thinks about it.
But philosophical lives have come and gone on the question.
(Probably, the question first came to life for me in reading William
{\James}'s little book, {\sl Pragmatism}.)

Anyway, I view your lines above as the greatest contribution of the
present paper.  That is, because they sort of soften up the western
front for a (one-day-in-the-future) full-fledged assault.  To the
extent that you can find yourself willing to do that, I'm happy.

The way I would emphasize the issue, though, is to say that quantum
measurement is either as {\em problematic\/} or as {\em
unproblematic\/} as classical measurement, take your pick.  The main
point is that it is not MORE problematic.

Here's the way I put it in my \quantph{0205039}:
\bq
     As far as Bayesian probability theory is concerned, a ``classical
     measurement'' is simply any {\it I-know-not-what\/} that induces an
     application of Bayes' rule.  It is not the task of probability
     theory (nor is it solvable within probability theory) to explain
     how the transition Bayes' rule signifies comes about within
     the mind of the agent.
\eq
And here's the way Rocco Duvenhage put it in another paper in {\tt
quant-ph}:
\bq
     In classical mechanics a measurement is nothing strange. It is
     merely an event where the observer obtains information about some
     physical system. A measurement therefore changes the observer's
     information regarding the system. One can then ask: What does the
     change in the observer's information mean? What causes it? And so
     on. These questions correspond to the questions above, but now they
     seem tautological rather than mysterious, since our intuitive idea
     of information tells us that the change in the observer's
     information simply means that he has received new information, and
     that the change is caused by the reception of the new information.
     We will see that the quantum case is no different.
\eq

The reason I say you are softening up the western front is because,
though you seem to admit this for a single, particular quantum
measurement---the computational basis---you haven't yet had the heart
to admit it for ALL quantum measurements.

The only things, it seems to me, that set the quantum case of
measurement apart from the classical case is A) what we do with the
information we gather, and B) what we concede the information is
about.  In the classical case, we enact ``Bayes' rule full stop'' with
the information we've just gathered.  In the quantum case, we
generally do something more (unless we are confident that the system
we are talking about was not touched physically by our measuring
device).  Concerning B), we had gotten in the habit classically of
thinking that the information we've just gathered is about ``what
is'' or ``what was.''  Quantum mechanics instead teaches us to look
to the future.  The information is about ``what will be.''  (I use
the measurement device LOCKED AWAY in the bureau of standards to make
it dramatic.)

So, the action, the excitement, of quantum mechanics is not in the
measurement, but in what it is that we're presupposing about the
world that causes us to process our data differently than we would
have classically.

\bdm
     A state can be associated with the Qbits only if their prior
     history is of a certain special form.  The state can then be
     constructed out of the particular features of that history: the
     outcome of the initial measurement and the particular sequence of
     unitary gates subsequently applied to the Qbits prior to the moment
     at which one associates the state with the Qbits.  But there is no
     way to determine that state if one is simply presented with the
     Qbits; only those who know this history know what state to assign
     to the Qbits.  The state does not reside on the Qbits; it is a
     concise encapsulation of those features of their history, back to
     the initial measurement, that are relevant to the outcome
     statistics of a subsequent measurement.
\edm
and
\bdm
     But to describe this as a collapse of the state of the Qbits and to
     regard it as a second kind of time dependence that Qbits can have
     in addition to their unitary evolution, is to ignore the whole
     point of the state vector and its unitary evolution, and, of
     course, to confuse the Qbits themselves with the state vector that
     compactly summarizes the statistical implications of their past
     history for future measurement outcomes.
\edm

Beside these two passages, I wrote in the margin, ``Asher's `What Is
a State Vector?'.''  It's interesting that you (in your note Friday
evening) called this accusation to be that of a ``regress.''  This is
because I think an infinite regress is all this view is gonna get
you.  That is to say, ultimately you're going to have to admit that
the quantum state ``compactly summarizes the statistical implications
of [the] past history'' of the entire universe.

Look at the discussion around Eq.\ (14) in my \quantph{0205039}.
What you would call a measurement is always determined by a further
quantum state for the apparatus.  And off to infinity (or the
boundaries of the universe) it goes.

\bdm
     Indeed, the generalized Born rule demonstrates that their state
     {\it cannot\/} be regarded as a property carried by the Qbits,
     since it provides an indirect method for associating a state with
     $n$ Qbits that share an entangled state with an additional Qbit,
     and therefore cannot initially be associated with any state of
     their own.  By measuring only the additional Qbit, one disentangles
     the $(n+1)$-Qbit state and is able to assign a state to the $n$
     Qbits, even though nothing interacts with them during or after the
     one-Qbit measurement.  This is only possible because the
     association of all $n+1$ Qbits with an entangled state prior to the
     measurement requires the additional Qbit to interact with the $n$
     of interest at some time {\it before\/} it is measured.  If one
     knows enough about that past interaction to determine the original
     $(n+1)$-Qbit entangled state assignment, then it is not surprising
     that the outcome of the measurement on the single Qbit can provide
     enough additional information to permit the assignment of a state
     to the $n$ Qbits.  This newly assigned state cannot be a property
     inherent to the $n$ Qbits, because nothing interacts with them
     during the process that takes them from stateless and externally
     entangled to having a state of their own.
\edm

Tell this to Philippe Grangier a thousand times and tell me whether
you've made any more progress than I have.  I think it is just silly
to say ``a state is a property'' or ``a state is a reality,'' and
then say ``some systems have no properties'' (when they are
entangled).  The realization he should have is rather, systems have
whatever properties they have, it is just that ``the'' quantum state
is not one of them.

Finally,
\bdm
     [I]n our description of nature the purpose is not to disclose the
     real essence of the phenomena but only to track down, so far as it
     is possible, relations between the manifold aspects of our
     experience.   --- Bohr (1934)
\edm

I don't like this quote, precisely because it is going to cause
people to accuse you of what they always (unjustly) accuse me of:
BANNING a set of questions that every physicist in his right mind
ought to be asking.  What is the real essence of the phenomena?

I have a pretty good reply to that now.  I'll attach it
({\tt ForSlusher.pdf}) for your midnight reading.

\section{14-01-03 \ \ {\it Impact Analysis} \ \ (to L. Hardy)} \label{Hardy11}

As I was driving in to the office yesterday, I heard that January 13 is officially designated ``Make Your Dreams Come True Day.''  That seemed apropos, as I was using the long drive to tumble some things over in my head about Perimeter.

I do have a dream, and it is simple and obvious and just about written on my forehead:  It is to get at the heart of quantum mechanics (or help the friends around me to do it), and then turn that understanding toward the next big phase of physics.  I do not think I am fooling myself by believing so strongly that we are within arm's reach of this goal.  It is just a question of concentrated, right-headed thinking with a smidgen of creativity \ldots\ and probably the main issue is simply amassing enough intellectual resources in one place for a final resonance.

\section{15-01-03 \ \ {\it Vanity, Sayeth the Preacher} \ \ (to J. W. Nicholson)} \label{Nicholson15}

\bjn
Is everyone in quantum foundations as verbose as you?
\ejn

No one else in science is as verbose as me.

\section{15-01-03 \ \ {\it The Ability to Write} \ \ (to N. D. {\Mermin})} \label{Mermin83}

I got a critique of one of my email expositions from a friend this
morning, and he said:
\bjn
\noindent \ldots\ the tone is probably appropriate.  [But] it's about twice as
long as (I thought) it needs to be.  Is everyone in quantum
foundations as verbose as you?
\ejn
He's got a grain of truth there.  But I find it so hard to rope
myself in sometimes.  I guess the best writer says what needs to be
said and little more.  I've got to work on the demons inside me.

\section{15-01-03 \ \ {\it Memory Lane} \ \ (to O. Cohen)} \label{Cohen1}

Pasted below is a letter I started to write to you in early December but never finished because of the holidays and some personal things that got in the way.  \ldots

It seems I have email contact with you only about every three years.
I hope things are going well.  Where are you now, and what are you
doing professionally?  The last I remember, I recommended you seek a
postdoc position with {\Carl} {\Caves} in New Mexico; I guess that never
materialized.

Anyway, I'm writing you because I've been reading your paper
``Classical Teleportation of Quantum States'' this week.  It's a nice
paper, and I very much like the simplicity of your scheme and the
point you make with it.  I am in complete agreement.

In fact it took me a little down memory lane.  You see, Asher Peres
and I had used teleportation as an example in our March 2000 {\sl
Physics Today\/} article, ``Quantum Theory Needs No
`Interpretation','' precisely to illustrate the sensibility of the
conception of a quantum state as a ``state of knowledge, rather than
a state of nature''.  When the paragraph peaked in clarity (i.e.,
before the editor's knife), it went like this:
\bq
The peculiar nature of a quantum state as representing information is
strikingly illustrated by the quantum teleportation process. In order
to teleport a quantum state from one photon to another, the sender
(Alice) and the receiver (Bob) need a pair of photons in a standard
entangled state. The experiment starts when Alice receives another
photon whose polarization state is unknown to her, though known to
some preparer in the background. She performs a measurement on her
two photons, and then sends Bob a classical message of only two bits,
instructing him how to reproduce the unknown state on his photon.
This economy of transmission appears remarkable because to completely
specify the state of a photon, namely one point in the {\Poincare}
sphere, we need an infinity of bits.  However, the disparity is
merely apparent.  The two bits of classical information serve only to
transfer the preparer's information, i.e., his {\it state}, to be
from describing the original photon to describing the one in Bob's
possession. This can happen precisely because of the previously
established correlation between Alice and Bob.
\eq

At the time I was basing my opinion predominantly on the result of
Cerf, Gisin, and Massar (\quantph{9906105}, ``Classical
Teleportation of a Quantum Bit'') along with the long heartfelt
conviction that the idea of a quantum state as a state of knowledge
gave the most sensible and constructive point of view about quantum
mechanics. Sometime after that though, Steven van {\Enk} and I---to the
best of my recollection---worked out a scheme pretty similar to your
own at the chalkboard at Bell Labs.  We never wrote it down however.

But that doesn't take away from your discovery.  It's a very clean
example, isn't it?  The conclusion you draw, I think, is particularly
important: the phenomenon of quantum teleportation only looks
surprising and remarkable if one takes an ontic view of the quantum
state.  In fact, in the past, I have accused some of my friends (some
of whom were authors on the original teleportation paper) of sticking
with an ontic interpretation of the quantum state precisely because it
is the only way to keep the phenomenon surprising and newsworthy.  You
might enjoy reading some of my correspondence with them on the
subject: Have a look at pages 175--176 and 184--189 of my samizdat
{\sl Quantum States:\ What the Hell Are They?}.  The said accusation
comes on page 189.  [See 14-05-02 note ``\myref{SmolinJ2.1}{Qubit and
    Teleportation are Words}'' to C. H. Bennett and others.]  That
correspondence occurred, by the way, as they were debating about how a
dictionary definition of ``quantum teleportation'' should be written.
(To get a copy of this samizdat, you can download the pdf file for it
at my webpage; there's a link to the webpage below in my `signature'.)

If you are interested in seeing the struggle Asher and I had in
constructing the paragraph above (and the reasoning and cues behind
it at the time), have a look at the discussions in my other samizdat,
{\sl Notes on a Paulian Idea}, \quantph{0105039} (or you can
download a better indexed version of it from my webpage).  The pages
to look at are 312, 316, 319--320, 322, 326, 327--330.  I have pages
326, 328, and 329 marked as the most interesting in my notebook
(can't quite remember exactly in what way though).  Maybe the main
lesson in those discussions is how difficult it is to give up an
objectivist language when using quantum states, even for a
recalcitrant positivist like Asher, and even in an example intended
to be illustrative of why quantum states should be viewed as states
of knowledge, rather than states of nature.  (The philosophy being:
if quantum mechanics looks too very mysterious, then you're probably
being wrong-headed about it.  Case in point: if teleportation looks
mysterious, then you're probably being wrong-headed about it too.)

The sense I get from your paper is that you are much more neutral
about the lesson than I am.  You say simply:  ``[O]ur classical
version of teleportation is just as impressive as the original
protocol, if we think of quantum states as representing states of
knowledge. \ldots\ If, on the other hand, we think of a quantum state
as having ontological content, \ldots, then our classical version of
teleportation is not equivalent to the quantum case,''  and leave it
at that. However, there is a spate of evidence starting to come out
that a significant fraction of some of the most `remarkable'
phenomena in quantum information theory can be mocked up with
classical toy models just as your own.  The only requirement for
seeing it is that one must focus on the epistemic states (i.e., the
states of knowledge) in such models rather than the ontic states
(like the actual H or T in your own model).  For instance, Rob
{\Spekkens} has a toy model which he has presented in several
conferences and which he is writing up presently as a paper, ``In
Defense of the Epistemic View of Quantum States: A Toy Theory,'' in
which he can reproduce the following quantum mechanical and quantum
information-theoretic type phenomena in a pretty NONremarkable way:
the noncommutativity of measurements, interference, a no-cloning
theorem, a no information-gain-without-disturbance principle, the
multiplicity of pure state decompositions of a mixed state, the
distinction between two-way and intrinsic three-way entanglement, the
monogamy of entanglement, superdense coding, mutually unbiased bases,
locally immeasurable product bases (i.e., what we originally called
`nonlocality without entanglement'), unextendible product bases, the
possibility of secure key distribution, the impossibility of bit
commitment, and many others.  (In particular, he gets teleportation
too, just like you do.)  As Rob\index{Spekkens, Robert W.} puts it in his abstract:
\bq
    Because the theory is, by construction, local and non-contextual, it
    does not reproduce quantum theory.  Nonetheless, a wide variety of
    quantum phenomena have analogues within the toy theory that admit
    simple and intuitive explanations. \ldots\ The diversity and quality of
    these analogies provides compelling evidence for the view that
    quantum states are states of knowledge rather than states of
    reality, and that maximal knowledge is incomplete knowledge.  A
    consideration of the phenomena that the toy theory fails to
    reproduce, notably, violations of Bell inequalities and the
    existence of a Kochen--Specker theorem, provides clues for how to
    proceed with a research program wherein the quantum state being a
    state of knowledge is the idea upon which one never compromises.
\eq

So, given that your paper is an independent and particularly notable
link in that, and as opposed to his paper, your result is not buried
within over 70 pages (and counting) of text, I very much endorse it.
I think the lesson is this:  A good lot of quantum information theory
is simply regular probability theory and information theory applied
in ways that had not been deemed interesting before.  What is
interesting and unique to the quantum itself, thus, must be something
else.

In my paper \quantph{0205039}, ``Quantum Mechanics as Quantum
Information (and only a little more),'' I tried to give the community
to a call to arms by saying this:
\bq
    This, I see as the line of attack we should pursue with relentless
    consistency:  The quantum system represents something real and
    independent of us; the quantum state represents a collection of
    subjective degrees of belief about {\it something\/} to do with
    that system (even if only in connection with our experimental kicks
    to it).  The structure called quantum mechanics is about the
    interplay of these two things---the subjective and the objective.
    The task before us is to separate the wheat from the chaff.  If the
    quantum state represents subjective information, then how much of
    its mathematical support structure might be of that same character?
    Some of it, maybe most of it, but surely not all of it.

    Our foremost task should be to go to each and every axiom of quantum
    theory and give it an information theoretic justification if we can.
    Only when we are finished picking off all the terms (or combinations
    of terms) that can be interpreted as subjective information will we
    be in a position to make real progress in quantum foundations.  The
    raw distillate left behind---minuscule though it may be with respect
    to the full-blown theory---will be our first glimpse of what quantum
    mechanics is trying to tell us about nature itself.
\eq

What your work and {\Spekkens}'s work does, from my perspective, is give
the best illumination yet of what I was hoping for when I was
speaking of ``combinations of terms'' in that passage.
Teleportation---being a certain combination of uses of the axioms of
quantum mechanics---is nevertheless a purely probabilistic or
information-theoretic effect.  As such, it tells us very little about
the ontology behind quantum mechanics.

My own view---and the thrust of my research program presently---is
that these examples help us to realize that what is unique in quantum
mechanics is not the probabilities (i.e., the quantum states) but
what the probabilities are applied to.  There, I think, lies the
essence of quantum mechanics:  It is localized in the Kochen--Specker
theorem.  ``Unperformed measurements have no outcomes,'' as Asher
Peres likes to say.  That is to say, where quantum mechanics gets its
uniqueness is from breaking with the old idea that a probability (as
a subjective state of knowledge) must be knowledge about a
pre-existent reality.  Instead, probabilities can just as fruitfully
be applied to capturing one's knowledge of ``what will come about due
to one's actions.''  The predominant issue becomes how to formalize
the difference between  probability theory as applied to pre-existent
facts and probability theory as applied to ``creatables'' (for want
of a better word).

There are some lines for tackling this idea buried within Sections
6.0 and 6.1 of my \quantph{0205039} \ldots

\section{16-01-03 \ \ {\it Smells} \ \ (to C. M. {\Caves})} \label{Caves70.2}

To me that smells of Lee Smolin.  On one occasion Lee told me (and everyone else at the table), ``When I was 16 I came to the conviction that there must be a hidden-variable explanation for quantum mechanics, and I haven't changed my mind since.''  I think I've got that almost verbatim.  Later in a personal conversation, Lee said something like, ``Wouldn't you think it just silly if a linear theory were the final story of things?  Quantum mechanics just has to be a linear approximation to something deeper.''  I said, ``I don't know, linearity's got a pretty good record.  Look at classical mechanics, the greatest example of a perfectly linear theory.''  Of course, he didn't get it, and when I explained the business about Liouville evolution being linear it made no impression.

\section{16-01-03 \ \ {\it A Footnote} \ \ (to H. J. Folse)} \label{Folse18}

I've been reading William James's book {\sl Essays in Radical Empiricism\/} and I came across a footnote in ``How Two Minds Can Know One Thing'' to do with something Harald {\Hoffding} said.  It just piqued my interest because I know from your papers (and some other things) that there is an interest in the extent to which William {\James} may have influenced Bohr via the conduit of {\Hoffding}.  What I have never seen any talk of is the extent to which {\Hoffding} may have influenced {\James}.  Do you know of any works in that regard?

On a related point, now that I've started to take such an interest in {\James}'s theory of truth and what I perceive as its similarity to the stance Asher Peres and I have taken on quantum mechanics, I've started to wonder why there has been so little scholarly work on this before.  In Henry Stapp's 1972 paper ``The Copenhagen Interpretation'' there is a decently extensive discussion on pragmatism, but I don't think I've seen anything really beyond that.  Could it be that there's a big gap in my canvassing of the literature?  Do you have any leads?  Or is it just, unfortunately, a subject that has been overlooked heretofore?

\subsection{Henry's Reply}

\bq
There's no doubt that {\Hoffding} was very excited by James' radical empiricism, and I think that it is reasonable to say something like they were ``on the same wavelength'' but that they had arrived at their views more or less independently.  Both men were psychologist-philosophers in a time when psychology was just beginning its break away from the womb of philosophy. Both were empiricists, and both took the ``enemy'' to be the then dominant force of idealism.  H's interest in the question of free will was no doubt stimulated by his interest in Kierkegaard, but surely resounded strongly with {\James} central interest in that same issue.  H's {\sl History of Modern Philosophy\/} which was published in 1895 makes no mention of James.  So I suspect his admiration for James only began sometime between 1895 and 1904.  He was one year younger than James, when they met in 1904 James would have been 62 and H 61, so I think that their outlooks were pretty well already formulated by that point.  In the spring of 1905, right after his visit to James in the fall of 04, H gave a series of lectures on ``the psychology of free will'' which definitely featured {\James} (he also lectured on Renouvier, whom James admired, and Boutroux, both of whom Jammer has pegged as expressing the same theme that experience can never be completely captured by any conceptual scheme). It was also in 1904--05 that {\James} wrote and put in an envelope marked ``essays in radical empiricism'' most of the articles which Perry later posthumously published as {\sl Essays in Radical Empiricism}. Also, right after his return from America in 1905 H published an English translation of his 1902 Danish book called {\sl Philosophical Problems\/} which included a preface by James himself. We're not certain, but there is a good bit of circumstantial evidence that Bohr attended at least some of H's 1905 lectures.  Even if he didn't, it seems to me that it is inconceivable that he never discussed {\James} with {\Hoffding}. So clearly J can be said to have ``influenced'' H, but perhaps it would be best put by saying J's influence was one of underscoring themes that were already present in H's thinking before he visited James.

As to whether or not  H ``influenced'' J, I am not sufficiently familiar with the {\James} materials to venture more than a speculation.  However, I do know who probably could answer that.  He is John J. McDermott, who's at Texas A{\&}M and has edited a lot of {\James} materials.  When he visited Loyola 15 years or so ago I talked with him about James and {\Hoffding}, but I was of course primarily interested in both of their influences on Bohr, not on each other.  James had only 6 years left after he met H, and I doubt if he had read any of H's works earlier.  So there would not have been a lot of time for any great amount of influence, though James's work continued to evolve right up to the end, so one can't rule it out.  Of course most of H's writings were in Danish, which James would not have known, but he did know German, and I believe H had published his Kierkegaard book in German.  It is just possible that J could have read that, but I think what brought the two men together was more likely to have been the common interest in psychology.  In 1906 H published an English translation of his book on {\sl Philosophy of Religion}; I would imagine he very probably sent James a copy, and that surely would have interested James, but I think that H was a good deal less tender minded when it came to religion than was James.  So I would think that if there were any influence from H to J, it would have most likely had to come through their conversations in the fall of 04 and the subsequent correspondence that must have been concerned with the publication of {\James}'s Preface to H's English translation of his book.  As I recall, McDermott had definitely read this correspondence (I imagine it's at Harvard?)\ but as I said, I didn't ask about any influence of H on J.   I also recall one of my old mentors, Andy Reck at Tulane, who wrote a book on {\James} for the French, knowing something about {\Hoffding} and {\James}, so I'll ask the next time I see him, but I would think McDermott would be a better bet.

\bq\noindent
On a related point, now that I've started to take such an interest in {\James}'s theory of truth and what I perceive as its similarity to the stance Asher Peres and I have taken on quantum mechanics, I've started to wonder why there has been so little scholarly work on this before.  In Henry Stapp's 1972 paper ``The Copenhagen Interpretation'' there is a decently extensive discussion on pragmatism, but I don't think I've seen anything really beyond that.  Could it be that there's a big gap in my canvassing of the literature?  Do you have any leads?  Or is it just, unfortunately, a subject that has been overlooked heretofore?
\eq

I think that you're right if by pragmatism you mean particularly William {\James}, and/or the so called ``classical'' pragmatists {\Peirce} and {\Dewey}.  But in a larger sense pragmatism has seeped into a great amount of American philosophy and can be said to be a major element in the thought of people like van Fraassen (BTW, I had a very brief conversation with him about you last November) and Arthur Fine.  Certainly I would say my own interpretation of Bohr reflects a lot of pragmatist themes, and you could reasonably say I characterize Bohr as a ``pragmatic realist.'' (There is a different anti-realist branch of pragmatism which is represented by vF and Fine.)  If I had to name a pragmatist to affiliate this view with, I'd say it is close to C.~I. Lewis.  In my years of discussion of Bohr with my colleague Jan Faye, I've always pushed Bohr closer to pragmatism, while Jan pushes him closer to positivism.  Indeed I would say that a lot of what people see as ``positivist'' about Bohr, are really the themes that positivism shares in common with pragmatism, namely a hard nosed empiricism.

Have you looked at Max Jammer's two books?  I'm quite certain he discusses Bohr and James, but I don't own copies of his books.  Gerald Holton also does the {\James}-Bohr issue in his well known article in {\sl Thematic Origins of Scientific Thought\/} but there's nothing you'll learn there that isn't in my book, and I think he caves in too easily to Rosenfeld's ``authority'' on the James/Bohr influence.  Jan doesn't discuss pragmatism per se but he does discuss James a fair bit in his book on Bohr and {\Hoffding}. You ought to just take a look at his book, {\sl Niels Bohr:\ His Heritage and Legacy}, for its exposition of {\Hoffding} in Chap IV, which is without doubt the most detailed exposition of H available in English.

Now that you bring it up, I think one thing is that a lot of those who've written about the history of QM are Europeans and therefore perhaps less likely to emphasize pragmatism or see pragmatism at work in the quantum revolution than would Americans.  Most of the younger generation of philosopher physicists writing these days are not so interested in historical kinds of questions, but just take off from the formalism (you're exceptional in that you seem to be going in the reverse direction).  Another thing is that the only classical pragmatist to live long enough to actually write on QM was {\Dewey}, and he got uncertainty all messed up as a disturbance caused by elephants trying to measure grains of sand.  So that would hardly be likely to enthuse people to look for insights into QM from {\Dewey}.  But apart from stupid things he might have said about QM, there are in {\Dewey}'s vast corpus a lot of interesting insights about the scientific description of nature which could possibly be useful to someone looking to make sense of quantum mysteries.  The same thing might be said about {\Peirce} who saw ``truth'' as what an ideal rational body of inquirers would reach given an infinite amount of time.  Rather than explain the process in terms of the end goal, he explained the goal in terms of the end of the process, and that's really a very profound notion.  But if you're looking for contemporary pragmatists, there's no doubt that the leading ``pragmatist'' expositor on QM these days is none other than van Fraassen in your own backyard.  He's of course a pragmatic anti-realist, but one who oddly retains a realist correspondence notion of truth that pragmatists would typically reject.  It's just that he thinks this realist notion of truth has nothing to do with ``acceptance'' of scientific theory which is done solely on the basis of its empirical adequacy and supplementing pragmatic virtues.

I'll let you know if I think of anything else.
\eq

\section{18-01-03 \ \ {\it The Quantum Principle, whatever it might be}\ \ \ (to L. E. Ballentine)} \label{Ballentine1}

Somehow I've been spared from the Volovich, Khrennikov, etc., distribution list on ``What is Quantum Mechanics?''\ \ldots\ and I'd like it to stay that way!  But this morning David Mermin forwarded me one of your notes (pasted below), which I did enjoy.  The point you make is an important one, and I think the whole nub of the matter.

I try to express something similar in my paper ``Quantum Mechanics as Quantum Information (and only a little more)'', \quantph{0205039}.  I'd like to think that paper makes a little progress toward your goal, but, if so, there is still a long, long way to go.  In any case, I think you might enjoy the paper because of some things I remember reading in your papers and your book several years ago and also because of some discussions I've had with your student Joe Emerson.  In particular, playing up the analogy between the quantum state and the Liouville distribution for all it is worth is, I think, the most important move for getting the whole project off the ground.

If you have any comments on the project set forth in that paper (and expanded at my website, link below), and in particular on how we might go further, I'd certainly love to hear them.

\subsection{Ballentine's Entry in the Discussion}

\bq
I think that this question is best interpreted as, ``What are the
principles of Quantum Mechanics?''.  (Indeed, Igor Volovich referred us to his Seven Principles \ldots.)

The first two principles are usually given as:
\begin{itemize}
\item[(a)]  Observables are represented by self-adjoint operators on a Hilbert
space;

\item[(b)]  Pure (mixed) states are represented by vectors (statistical
operators) in Hilbert space.
\end{itemize}
What, if any, is the physical content of these postulates? They contrast most unfavorably with Einstein's two postulates, from which
all of Special Relativity can be derived:
\begin{itemize}
\item[(1)] Equivalence of all uniformly moving frames of reference;

\item[(2)] Invariance of the speed of light, $c$.
\end{itemize}
These are physical postulates.  At least (1) has strong intuitive appeal.
Both can be directly tested by experiment.  And all of SR follows from
these two postulates.

Return now to postulates (a) and (b) of QM. They lack any intuitive appeal.
They cannot be directly tested.
They are insufficient to solve even one real problem in QM.

By itself, (a) seems to impose no physical restriction, since the spectrum
of an operator may contain any possible range of discrete and continuous
values.  The Indeterminacy Principle is permitted, but not implied, since
the observables might, at this stage, be commutative or not.  Perhaps (b)
together with (a) imposes some physical restrictions, but that is far from
obvious.  And Gleason's theorem seems to make (b) inevitable after (a).

The specific content of the theory (needed in order to solve any physical
problems) enters when we specify which particular operators correspond to
which particular observables.  Although this is sometimes done by means of
extra postulates, in fact the most important operators can be derived by
requiring invariance under the space-time symmetries (displacements,
rotation, and Galilean transformations).  [See my book, {\sl Quantum Mechanics --
a Modern Development}, World Scientific, 1998.]  In particular, the
commutation relation between position and momentum, and hence the
Indeterminacy Principle, now follows.

But the same space-time symmetries hold in Classical Mechanics, where
there is no Indeterminacy Principle.  So postulates (a) and (b) must have
imposed some restrictions on the physical contents of the theory, in spite
of the appearance that they only defined a general mathematical form
without specific physical content.

So my questions are:
\begin{itemize}
\item
What are the {\it physical\/} principles of Quantum Mechanics?
\item
Can they be expressed in a more transparent form?
\item
To what extent do the usual postulates (a) and (b) restrict the physical
content (not merely the form) of Quantum Mechanics?
\end{itemize}
\eq

\section{20-01-03 \ \ {\it Munificence}\ \ \ (to S. Aaronson)} \label{Aaronson3}

Thanks for teaching me the word munificence.  I'll try to incorporate it into my vocabulary!

\bsa
How far would you take your famed battle cry, ``Give an
information-theoretic justification if possible''?  In particular,
venturing beyond QM, would you want/expect an information theoretic
justification for why space has 3 visible dimensions, or why the
cosmological constant is positive?  Both things have been worrying me
a lot lately.
\esa

Unfortunately, I don't know how to answer that yet.  My feeling is that eventually the battle cry ought to stop.  But where, I don't have a strong feeling of yet.  I try to lay out the philosophical underpinning of this idea in my \quantph{0204146}, where I talk about the ``core'' of a theory---i.e., a part that at some level likely does not have an agent-centered information-theoretic reason.

I also try to say it somewhat differently in a recent essay I wrote; I'll paste it below.  [See 06-01-03 note ``\myref{Mermin78}{Pedagogy}'' to N. D. {\Mermin}.]

\bsa
I've gotten increasingly annoyed with people who whine, ``We need
more algorithms!  Shor's and Grover's are not enough!''  I'm starting
to suspect that large-scale quantum computers would be incredibly
useful, but for entirely different reasons than the ones that excite
us computer scientists.  Is that reasonable?  I.e.\ am I mistaken to
think that in whole areas of chemistry and high-energy physics, the
biggest bottleneck right now is the lack of a quantum computer (to do
things like perturbation sums)?
\esa

Yeah, I think that's probably true.  But now a question to you.  To what extent has it been proven {\it rigorously\/} that quantum computers are efficient at performing simulations to do with interesting questions about {\it other\/} quantum systems?

\section{20-01-03 \ \ {\it And Believing Simulations} \ \ (to D. Poulin)} \label{Poulin6}

You're going to be ashamed of me---or maybe you always expected it!---but I've only now read (in detail) the note you sent me December 20!

I liked your phrase ``objectivity distillation.''

\bdp
I came across a very simple problem lately which forced me to choose
an interpretation for the wave function. It is a little bit hard to
explain in an email but I'll gave it a shot.
\edp

I did gather that your note is a specific---and, as opposed to what my lazy butt was willing to do, actually worked out---instance of the sort of thing I talked about a lot in my file {\sl Quantum States:\ What the Hell Are They?}  I.e., of a program I spelled out to Brun--Finkelstein--Mermin in \myref{Mermin28}{a 7 August 2001 note} in this way:
\bq
I have the feeling that if quantum mechanics is really about knowledge and only knowledge---or better, belief convergence and {\it only\/} belief convergence---then FOR ANY GIVEN METHOD OF GATHERING INFORMATION, there {\it should\/} be a way to ferret out of quantum mechanics the necessary and sufficient conditions on two observers' initial state assignments, so that the gathered information leaves them in a better agreement than they started out with.
\eq

But I missed this:  Which interpretation were you forced to choose!?!?  (I'm keeping my fingers crossed we'll see each other in heaven!)  Seriously, does this example cause you to lean more toward the epistemic interpretation of the wavefunction or further away?  And why?  I didn't understand your reasoning on that.

\subsection{David's Reply}

\bq
Yes, my note was definitely in this spirit. When I tried to find a way of
simulating this system, it was screaming out for an epistemic
interpretation for the wave function. But I don't have a complete
understanding of it. In its current formulation (which I would call hybrid
epistemic-ontic), the monte-carlo simulation will inevitably be observer
dependent. What I would like to do is to formulate it in a purely
epistemic way, just like you guys did with the tomography problem. Thus,
my goal would be to run a monte-carlo which simulates the measurements
performed by the observers but whose branching probability is not
observer-dependent. This would allow them to dilute objectivity in the
sense that I have tried to explained in my previous message. In fact,
whether the observers come to an agreement or not (a sort of agreement
measure) will also be observer dependent in this simple model which is
quite strange!!!

I could tell you many strange things about this simple model but there are
probably things you have already thought about \ldots\ and they are quite easy
to discover when you think of how you would go and numerically simulate
this system.

By the way, one of the {\Montreal} talks has really influenced me \ldots\ my own!
I am taking this physics simulation business quite seriously. In an
attempt to show that there is no fundamental (i.e.\ sub-polynomial)
advantage at using a Q-computer to simulate a physical system, I am
looking into information processing at the Planck scale. That's right!
\eq

\section{20-01-03 \ \ {\it More Pragmatism} \ \ (to H. J. Folse)} \label{Folse19}

Thanks for the detailed note.  Let me go through it sequentially and make just a few comments.

\bhf
Also, right after his return from America in 1905 H published an
English translation of his 1902 Danish book called {\bf Philosophical
Problems} which included a preface by James himself.
\ehf
I'll try to dig up a copy of that.  That sounds interesting to me.

\bhf
He is John J. McDermott, who's at Texas A{\&}M and has edited a lot of
James materials.
\ehf
In fact, I have a massive collection that he put together---{\sl The Writings of William James:\ A Comprehensive Edition}.  It was one of the first things I bought when I took an interest in James.  However, I soon discovered that I preferred the original (little) books; they're much easier to carry around in a coat pocket.  I'll have a look in the index to that book when I get home.  Also, though, I had planned on visiting Marlan Scully and Ed Fry at Texas A{\&}M sometime this year; so maybe I'll try to work up a visit with him too.

\bhf
Well, I don't remember being much impressed by the exposition of
``pragmatism'' in Stapp's paper, though it was ages ago that I read it.
\ehf
Yep, don't get me wrong:  I wasn't particularly impressed with it either.  I was only trying to imply that that is the only thing I know of along those lines.

\bhf
I think that you're right if by pragmatism you mean particularly
William James, and/or the so called ``classical'' pragmatists Peirce and
Dewey.
\ehf
Yes, that is exactly what I meant.

\bhf
But in a larger sense pragmatism has seeped into a great amount of
American philosophy and can be said to be a major element in the
thought of people like van Fraassen (BTW, I had a very brief
conversation with him about you last November).
\ehf
Aha, that might explain why I got an email from him November 21 that started out like this:
\bq
     I know I have not been very communicative recently, but I thought
     in any case I'd let you know about this bit of Bayesianism \ldots
\eq
I have wondered a little bit about whether I've rubbed him raw, since I didn't hear too much from him after his fairly enthusiastic introduction.

\bhf
Have you looked at Max Jammer's two books?  I'm quite certain he
discusses Bohr and James, but I don't own copies of his books.
\ehf
Yes, I have read them both (years ago).  And more recently I have noted the relevant material in them on this subject.  But it's not deep enough for my tastes.

\bhf
Jan doesn't discuss pragmatism per se but he does discuss James a fair
bit in his book on Bohr and {\Hoffding}. You ought to just take a look at
his book.
\ehf
I read it too, about 10 years ago.  I remember there were parts in it that I absolutely loved.  But then I remember there were parts (like his discussion of Bell inequalities) that I absolutely hated---I thought he went way off the mark.  I should reread the book from start to finish, to see how it affects me now.  I have certainly changed a lot in my opinions about quantum mechanics since those days.  In fact, maybe I'll try to see if I can buy the book.

\bhf
Another thing is that the only classical pragmatist to live long
enough to actually write on QM was Dewey, and he got uncertainty all
messed up as a disturbance caused by elephants trying to measure
grains of sand.  So that would hardly be likely to enthuse people to
look for insights into QM from Dewey.
\ehf
Yep, that's sad.  The insights I'm looking for have to do with a) James's theory of truth, b) James's pluralism, c) and the lovely idea that the universe is still under construction.

Let me attach a couple of essays I've written in connection to this cluster of issues.

1) To contrast myself to the Deweyian thing you mention above, I'll attach a note titled ``Pedagogy.''  [See 06-01-03 note ``\myref{Mermin78}{Pedagogy}'' to N. D. {\Mermin}.]

2) To expand on the pluralism thing, I'll attach part of a note titled ``Probabilismo!''  It builds on some previous stuff that I had sent you about Ulfbeck and Bohr the younger.  [See 14-11-02 note ``\myref{Appleby0}{Probabilismo!}''\ to D. M. {\Appleby}.]

3) Finally, concerning c) above, now that you can download my files, why don't you have a look at two notes I wrote Howard Wiseman in the collection ``Quantum States:\ What the Hell Are They?''  The titles are ``\myref{Wiseman6}{The World is Under Construction}'' and ``\myref{Wiseman8}{Probabilism All the Way Up}'' and start on pages 210 and 217, respectively.  This is where pragmatism strikes me the most.

\section{21-01-03 \ \ {\it {\Hoffding} and {\James}} \ \ (to H. J. Folse)} \label{Folse20}

I did manage to find the following by looking through Ralph Barton Perry's {\sl The Thought and Character of William James}.  (I had previously read the condensed one volume version, but just last week in Manhattan I picked up the full two volume version for \$30.  {\Hoffding} was not mentioned in the condensed version.  Now I'm chompin' at the bit to read the full thing---it has loads more stuff about Boutroux, etc.)

The first mention of {\Hoffding} recorded there is in a letter from {\James} to John {\Dewey} dated 17 October 1903:
\bq
It rejoices me greatly that your School (I mean your philosophic school) at the University of Chicago is, after this long gestation, bringing its fruits to birth in a way that will demonstrate its great unity and vitality, and be a revelation to many people, of American scholarship.  I wish now that you would make a collection of your scattered articles, especially on ``ethical'' subjects.  It is only books that tell.  They seem to have a penetrating power which the same content in the shape of scattered articles wholly lacks.  But the articles prepare buyers for the books.  My own book, rather absurdly cackled about before it is hatched, is hardly begun, and with my slow rate of work will take long to finish.  A little thing by Harald {\Hoffding}, called {\sl Philosophische Probleme}, which I have just read \ldots\ is quite a {\it multum in parvo\/} and puts many things exactly as I should put them.  I am sure of a great affinity between your own ``monism,'' since you call it, and my ``pluralism.''  Ever gratefully and faithfully yours,
\eq

The second comes from a 1910 article in {\sl Nation\/} titled ``A Great French Philosopher at Harvard.''  James writes:
\bq
The great originality of M. Boutroux throughout all these years has been his firm grasp of the principle of interpreting the whole of nature in the light of that part of it with which we are most fully acquainted, namely, our own personal experience.  \ldots\  Those readers who know something of present-day philosophy will recognize in my account the same call to return to fullness of concrete experience, with which the names of {\Peirce}, {\Dewey}, {\Schiller}, {\Hoffding}, Bergson, and of many minor lights are associated.  It is the real empiricism, the real evolutionism, the real pluralism; and Boutroux (after Renouvier) was its earliest, as he is now its latest, prophet.
\eq

\section{21-01-03 \ \ {\it Choice Quotes} \ \ (to G. J. Milburn)} \label{Milburn1}

I just canvassed all the book reviews in QIC, in preparation for the one I've got to write on Holevo's new book. I just loved a couple of choice quotes from your article in QIC {\bf 1}, p.\ 89.

\bq\noindent
[N]o one has seen a `probability' in the same way we see a coin land heads up or a pointer deflection on an instrument.
\eq

That one was fun because I had just had a conversation with Steven van Enk a week ago, in which he asked, ``How would you explain that we never experience a superposition?''  I replied, ``Have you ever experienced a probability distribution?''  A few days later, he came in and told me that that single line had made a big effect on him.  He also suggested that I should write a paper of single-liners (maybe in question and answer format):  He thinks that my usual papers are far too long, and that I'd get the message out more effectively if I'd trim them so that people can see what the point is!  If I ever do follow through, I'll make sure to cite your book review.

\bq\noindent
There is no consensus view on why quantum mechanics offers a bonus, if any, in computational efficiency.  The allusions to the many-worlds-interpretation of quantum mechanics in the chapter introducing quantum computation, provide an indication of how desperate the situation is.  The many-worlds-interpretation provides most certainly not a consensus view of quantum mechanics and, while seductive, it seems to me little more than the last desperate refuge of the classically minded.  As far as I am aware photons do not interact with each other in this universe, let alone with photons in other putative universes.
\eq

That is great!!  It's somewhat like another line that I love, that you might enjoy too.  It comes from David {\Mermin}'s pedagogical paper, \quantph{0207118}.  He writes,
\bq\noindent
There are nevertheless some who believe that all the amplitudes \ldots\ have acquired the status of objective physical quantities, inaccessible though those quantities may be.  Such people then wonder how that vast number of high-precision calculations ($10^{30}$ different amplitudes if you have 100 Qbits) could all have been physically implemented.  Those who ask such questions like to provide sensational but fundamentally silly answers involving vast numbers of parallel universes, invoking a point of view known as the many worlds interpretation of quantum mechanics.  My own opinion is that, imaginative as this vision may appear, it is symptomatic of a lack of a much more subtle kind of imagination, which can grasp the exquisite distinction between quantum states and objective physical properties that quantum physics has forced upon us.
\eq

Anyway, as usual, I just wanted to get this stuff into my computer and I used you as the sounding board.  Thanks for the opportunity.  I'll repay you with a little essay I wrote the other day (on the subject matter of 2 above); I'll paste it below.  [See 06-01-03 note ``\myref{Mermin78}{Pedagogy}'' to N. D. {\Mermin}.]

\section{21-01-03 \ \ {\it The Thawing Heart of David Poulin} \ \ (to D. Poulin)} \label{Poulin7}

Your notes are warming my heart!  To see you embrace this distillation idea (and develop it from your own perspective) so keenly, and to see you actually do something with it, is just great.

\bdp
In fact, whether the observers come to an agreement or not (a sort of
agreement measure) will also be observer dependent in this simple
model which is quite strange!!!
\edp

No, I think that's quite marvelous and something that one shouldn't lose sight of.  I'm guessing it's a general feature of all models that do this problem correctly.  See my attempt to formalize the issue on pages 23 and 24 of my {\sl Quantum States:\ What the Hell Are They?}\ on my webpage.  You'll see I accounted for the potentiality of that feature too.  [See 07-08-01 note ``\myref{Mermin28}{Knowledge, Only Knowledge}'' to T. A. Brun, J. Finkelstein and N. D. Mermin.]

\bdp
By the way, one of the {\Montreal} talks has really influenced me \ldots\ my
own!
\edp
I'm glad you got something out of one talk!

\section{21-01-03 \ \ {\it Peeling Away the Agent}\ \ \ (to S. Aaronson)} \label{Aaronson4}

\bsa
I loved your diatribe to Preskill.  But I realized we have different ideas of what it means for a fact to have an ``information-theoretic justification.''  For you, it means the fact is observer-dependent, and therefore we should ``peel it away'' if we want to uncover a theory's true physical content (as in your general relativity example).

For me, the fact could be as observer-independent as anything we know -- e.g., the dimensionality of space.  Giving an ``information-theoretic justification'' means you show 3 things:
\begin{itemize}
\item[(1)] The fact in question has some implication for the computational complexity, communication cost, etc. of some problem.  (That is, anything that would interest a computer scientist.)

\item[(2)] If the fact were different, the implication would change.

\item[(3)] We (or at least I) have strong intuitions regarding computation: for example, NP-complete problems should not be solvable efficiently.  Changing the fact would violate these intuitions -- thus yielding an ``information-theoretic justification'' for why the fact is the way it is.
\end{itemize}
\esa

The ground for a disagreement with that, actually, is maybe the biggest thing I took away from Timpson's thesis.  And I guess that's why I liked it.  In particular, I had not appreciated so much before the very human-centeredness of the Turing-machine concept.  That is, a mathematical model for what is humanly computable.

Also, just focus on the phrase ``communication cost.''  When it comes time to quantify that, a probability distribution is going to be introduced somewhere.  At that point, I say, you may not realize it but an agent has been invoked.  For, probability ultimately only makes sense as a gambling commitment an agent would be willing to make---i.e., probability only makes sense in the Bayesian sense.

Below, I'll paste a (part of a) note I wrote to Ari Duwell and I'll attach a copy of the paper he's putting in the special issue of {\sl Studies in History and Philosophy of Modern Physics\/} that Jeff Bub and I are editing.  [See 06-12-02 note ``\myref{Duwell1}{Quantum Information Does Not Exist}'' to A. Duwell.]  Between the two of them (including the references I put in the note), I hope it better explains this viewpoint I'm building up.

\bsa
Cases where I think the program has already been carried out include
``why is quantum mechanics linear?''\ and ``why is there a minimum length
scale (i.e.\ the Planck scale)?''\  Also, your de Finetti paper goes some
way toward carrying it out for ``why are amplitudes complex rather than
real or quaternionic?''
\esa

And I'd like to think that Sections 4.2 and 6 of my paper \quantph{0205039\/} get at ``information-theoretic reasons'' for why the structure of quantum measurements takes on that of the POVMs and, also, why the state-change rule (for what happens upon the completion of a measurement) is what it is.

Finally, let me comment on something that I missed in your last note.
\bsa
PS. I remember you talked about William James in the samizdat \ldots\ Have
you read the {\bf Principles of Psychology}?  I'm working through it now.
I've decided to recommend it to people as ``the most up-to-date,
state-of-the-art book about consciousness'' (without telling them the
publication date).
\esa
Is that really true?  Do you really believe that?  In any case, the comment warms my heart, as I have turned into a huge fan of William James since my first samizdat.  I only sort of knew him cursorily when I wrote that stuff.  But now he is splattered all over the place in the new samizdat, ``Quantum States:\ What the Hell Are They?''

You might enjoy the stories of my real discovery of him recorded therein.  Take a look at ``\myref{Preskill2}{The Reality of Wives}'' on page 15 (sent to Landahl and Preskill) and the bottom of page 218 in the note ``\myref{Wiseman8}{Probabilism All the Way Up}.''  The document is posted at my website.

I haven't read {\sl Principles of Psychology\/} yet, but I have read {\sl Pragmatism}, {\sl The Meaning of Truth}, {\sl Some Problems in Philosophy}, {\sl Essays in Radical Empiricism}, and a few other scattered articles like ``The Sentiment of Rationality.''  Also---though not written by James---I've read {\sl The Cambridge Companion to William James\/} and a couple of other second source books.  Finally, Ralph Barton Perry's book, {\sl The Thought and Character of William James (Briefer Version)}, is just wonderful.  And it's chock full of great letters.  Just last week in Manhattan I finally found a copy of the full version, in two volumes, and I'm chompin' at the bit to get at it.

\section{22-01-03 \ \ {\it Another Footnote} \ \ (to H. J. Folse)} \label{Folse21}

This one I dug up from somewhere on the web.
\bq\noindent
The classic discussion of the ``stream of thought'' is, of course, to be found in {\James}, W. (1890). {\sl The Principles of Psychology\/} (2 vols.). New York:\ Henry Holt; for a discussion of {\James}'s view on the ``stream of thought,'' see the essay on James in this volume.  While there are only two references to {\Hoffding} in {\James}'s {\sl Principles}, one having to do with {\Hoffding}'s theory of recognition memory, a second with {\Hoffding}'s position on the role of bodily sensation vs.\ ``spiritual affection'' in emotion, and both are mildly critical, {\James} nonetheless goes out of his way to indicate a general approval of {\Hoffding}'s work, in one instance referring to ``his excellent treatise on Psychology'' (James, op.\ cit., Vol.\ 2, p.\ 455) and in the other professing his ``respect for him as a psychologist'' (ibid., Vol.\ 1, p.\ 674).\footnote{\editornote This is from R.\ H.\ Wozniak, ``Harald {\Hoffding}: {\sl Outlines of Pychology} (1882; English 1891),'' in {\sl Classics in Psychology, 1855--1914: Historical Essays} (1999).}
\eq

\section{29-01-03 \ \ {\it Elegance} \ \ (to A. Peres)} \label{Peres48}

Elegance is always in the eye of the beholder.

I just ran across \quantph{0212062}, where the last sentence of the abstract reads, ``Thus we recover Fuchs and Peres' formula in an elegant manner.''

Still, it would be interesting if they have developed a powerful formalism.

\section{05-02-03 \ \ {\it Accuracy (Urgent)} \ \ (to J. M. Renes)} \label{Renes12}

I'm hurriedly writing the last section of a long overdue paper on a measure of quantumness for ensembles of states, and I mention symmetric informationally complete POVMs there.  Could you take a look at my footnote on page 23 and check it for accuracy?  (I mention the business about the long overdue paper to hopefully give you a sense of my urgency!)  In particular, I have heard of some other people involved in the SICPOVM project---for instance, the guy who constructed the $d=4$ case---and I know I should mention them too if I'm going to mention anyone.  But I don't know who they are.  I'll attach the paper as a PDF file.

\section{06-02-03 \ \ {\it Midnight Oil} \ \ (to M. Sasaki)} \label{Sasaki1}

Do you want a house in America?  Ours just showed up in the real estate listings today: [\ldots]

I am going to miss the place.  Do you see the rightmost section of the house, with two windows toward the front, and three windows on the side?  There are also two windows on the backside that you cannot see.  That is the section of the house that I used as my study.  I surrounded myself with sunlight and the philosophy of William James in there.  Too bad I never got a chance to invite you over for a stay.

But the science will be good for me in Ireland.  And that is always what keeps my heart beating.  John Lewis (of Davies and Lewis) will be my boss.  Also I will have a significant overlap with Chris King and Mary Beth Ruskai, who will themselves be visiting there for something like 2 and 4 months respectively.  My plan is to greatly develop the ideas of Section 7 in my paper \quantph{0205039}.  This will require that I much better understand the structure of the set of completely positive maps.  I think I have a set of questions in that regard that no one has ever explored before.  Basically, I want to develop the slogan, ``A quantum operation is nothing but a quantum state in disguise.''  Lately I am taken with the idea that the Jamio{\l}kowski representation theorem---equation 116 in my paper---is one of the deepest statements in all of physics \ldots\ second only to Einstein's principle of equivalence!  (I am only joking a little!)

Anyway, I want to very much apologize to you again for causing such a great delay in getting our paper into the public eye.  Deep inside I am a very selfish and undisciplined person, and it can hurt my friends and associates.  It is a shame that you got caught up in that.

\section{06-02-03 \ \ {\it Looking at the List Again} \ \ (to N. D. {\Mermin})} \label{Mermin84}

\bdm
\bq\noindent\rm [CAF wrote:]  I'm particularly keen to see the reaction
{\Spekkens} gets.  (Did he give you a private version of his talk in
{\Montreal} as I had asked him to?)
\eq
He may have talked to me in {\Montreal}.  In my usual irresponsible
muddle-headed way, I can't remember who he is or who I talked with.
\edm

He has a talk on about 27 reasons why you ought to think that the
quantum state is epistemic in nature:
\begin{enumerate}
\item You can't clone a quantum state.  Guess what?  You can't clone a
Liouville distribution either.

\item You can change a quantum state from afar by gathering information
nearby.  Guess what?  You can do the same with classical joint
probability distributions.

\item  \ldots\ and so the list goes.
\end{enumerate}
Moral:  You can believe the quantum state to be ontic in character if
you want.  But then you've got to create an ad hoc reason for
justifying each and every one of the effects above.  Wouldn't it be
so much simpler and more natural to accept the epistemic hypothesis
and follow out its consequences?

\section{06-02-03 \ \ {\it The Minimum Maximally Sensitive Set of States} \ \ (to J. W. Nicholson)} \label{Nicholson16}

\ldots\ or something like that will be our title.  Now that we've got a title, we've got to write a paper.

\section{06-02-03 \ \ {\it No It Doesn't [AJP 53(3), 70 (2000)]} \ \ (to D. Poulin)} \label{Poulin8}

Now the last note of yours was even sweeter music to my ears!  Sorry I haven't replied, but I've been trying to get a paper ready to get on quant-ph next week, and I was quickly losing all face with my Japanese coauthor.  But maybe I've also gained face with you in the process:  In contrast to my paper \quantph{0204146}, this paper has 135 equations.  (Have a look at the abstract on quant-ph and you'll see what I mean.)

And speaking of {\tt quant-ph/0204146},
\bdp
Last night, I went through the first 20 pages or so and found a
recurrent theme: if reality was deterministic, there wouldn't be any
distinction between dreams and awareness. I am not sure I fully
understand the deep meaning of this but I sure like it. Basically, you
are saying that nondeterminism is a proof that there is something out
there, right?
\edp

I don't know if I'd go as far as saying ``proof'', but, yes, that is basically the idea.  I think the place where I give the fullest account of what's on my mind about that---though definitely my opinion is still forming on the subject---and say it all the most clearly is in \quantph{0204146}, actually.  The sections to read are the last two, devoted to Preskill and Wootters.  If you read them, please let me know if you get anything out of them.  I got a great compliment from Michael Berry, who told me he really liked the paper.  But {\Ruediger} told me he couldn't understand a thing of it; and Carl Caves said something like ``it's sickening trash.''  It's important that you first rent and watch the movie {\sl It's a Wonderful Life\/} if you haven't ever seen it before.  (Too bad it's not Christmastime.)

Anyway, you helped ease my depression with your note.  If confident, young David Poulin will invest the time to read 20 pages in my samizdats, maybe just maybe I'm not writing for nothing.  And maybe just maybe our community will pool its strength enough to conquer this quantum beast.  I'm serious:  It made me happy that you took me seriously, and I know I'm gonna sleep better for it tonight.

\subsection{David's Preply, ``Quantum Mechanics Needs an Interpretation''}

\bq
How are you? Have you had a chance to look at the problem I send you? And
if you did, were you able to understand anything? I believe that this
problem and the compatibility problem are both quite interesting because
they are well defined mathematical problems for which any solution forces
you to give the wave function a meaning. For me, it went the other way
around with compatibility. When I wrote the paper with Robin, it seemed
clear that the solution was in the lines set by BFM. It is only later that
I realized that underlying this solution was an interpretation of the wave
function: a mixed state ``can be'' a derived product due to lack of knowledge
while pure states are really out there. I have put ``can be'' in quotes
since they could also be fundamental when entanglement is present: the
adepts of the Church of the Enlarged Hilbert Space would see no difference
between these two cases. As you probably suspect, I am no longer so much
sympathetical with this point of view. Nevertheless, I find it exciting
that there are mathematical problems which require an interpretation to be
solved.

I think that the day I can find a satisfactory answer to both these
problems will be the day I have made up my mind on the correct
interpretation. Of course, this raises the problem of what is a
satisfactory answer. My only hope is that it will be like pornography:
I'll know it when I'll see it.

The last few emails you have send me contain something like ``of course,
this is in my notes\ldots''. I have glimpsed through your notes before but
never had the time (or motivation) to go through the entire thing. I am
attempting it now. It will be hard since I am so busy with my two classes
and learning spin foams \ldots\ but it is quite entertaining. Last night, I
went through the first 20 pages or so and found a recurrent theme: if
reality was deterministic, there wouldn't be any distinction between
dreams and awareness. I am not sure I fully understand the deep meaning of
this but I sure like it. Basically, you are saying that nondeterminism is
a proof that there is something out there, right?
\eq

\section{08-02-03 \ \ {\it Is Is} \ \ (to M. Sasaki)} \label{Sasaki2}

I was just reading the {\sl New York Times\/} (my usual newspaper on the web), and I thought of you when I read the following paragraph:
\bq\noindent
``The question is, is this about American power, or is it about democracy?''\ Mr. Asmus asks. ``If it's about democracy, we'll have a broader base of support at home and more friends abroad. The great presidents of the last century --- F.D.R., Wilson, Truman --- all tried to articulate America's purpose in a way that other parts of the world could buy into. Bush hasn't done that yet.'' Before long, we'll find out if he cares to.
\eq

Actually, you had me worried about the ``is, is'' construction last week. Your change caused me to reflect that, actually, I was not sure if the thing was proper after all.  I'm still not sure.  But I do see that others use it too.

We now have three offers on our house!  We have decided that we will accept the highest as of noon Monday.  I will probably be able to put our paper on {\tt quant-ph} Tuesday.

\section{12-02-03 \ \ {\it Your Recent {\tt quant-ph} Posting} \ \ (to C. M. {\Caves})} \label{Caves71}

\bcc
I think you're a bit unfair to the [quantum optics] notions of
classicality. \ldots\ This is a very experimentally based notion of classicality, and that's
perhaps both its strength and (from your point of view) its weakness.
\ecc

I guess I have always been a smart mouth.

But, let me emphasize the possible connections between this quantumness stuff and some of your more recent work.  Maybe the thing that intrigues me the most at present is the conjectured connection between A) symmetric informationally complete POVMs and B) the minimum-cardinality maximally-sensitive quantum alphabet.

What is the deeper meaning of that?  The main thing (maybe the only thing) that intrigues me about the SIC-POVMs is that they have got to give A) the prettiest region on the simplex (in the spirit of my National Bureau of Standards stuff that I keep drawing in my lectures) and B) the maximal-volume region on the simplex.  Both these are ingredients that one might think makes quantum mechanics look as close to simple Bayesianism as it can be.  Yet, that happens for a ``maximally quantum alphabet.''  Why?

I don't have a clue what the connection is, but my religion tells me there is one.

\section{13-02-03 \ \ {\it Your Schedule} \ \ (to K. R. Duffy)} \label{Duffy1}

Quantum theory---I predict it will ultimately turn out---is nothing {\it but\/} control theory \ldots\ when the components have a wonderful sensitivity to the touch (more slippery than the classical kind).  So, you've been doing the right thing.

\section{13-02-03 \ \ {\it Doubting Our Coherence} \ \ (to C. M. {\Caves})} \label{Caves72}

See page 5 of \quantph{0110107} by Halvorson and Clifton.

\section{13-02-03 \ \ {\it BayesianResource.com} \ \ (to C. M. {\Caves} \& R. {\Schack})} \label{Caves72.1} \label{Schack60.01}

And this has got to be the best use yet of the quantum de Finetti theorem:

\bq
\noindent \quantph{0212162}\medskip\\
Unconditionally Secure Key Distribution Based on Two Nonorthogonal States\medskip\\
Authors: Kiyoshi Tamaki, Masato Koashi, Nobuyuki Imoto\medskip\\
We prove the unconditional security of the Bennett 1992 protocol, by using a reduction to an entanglement distillation protocol initiated by a local filtering process. The bit errors and the phase errors are correlated after the filtering, and we can bound the amount of phase errors from the observed bit errors by an estimation method involving nonorthogonal measurements. The angle between the two states shows a trade-off between accuracy of the estimation and robustness to noises.
\eq

\section{19-02-03 \ \ {\it Irish Dates} \ \ (to C. King)} \label{King6}

My main research plan for my time in Dublin is to get a better handle on the structure of CPMs.  What I'd like to get are some characterization theorems in line with my foundational ideas.  In particular, I'm thinking of the main issues in Section 7 of my \quantph{0205039} and also the samizdat {\sl Quantum States:\ What the Hell Are They?}\ on my webpage.  That is, I want to put some mathematical meat on my slogan, ``A quantum operation is truly and only a density operator in disguise.''

\section{20-02-03 \ \ {\it OK} \ \ (to C. M. {\Caves}, J. M. Renes \& K. K. Manne)} \label{Renes13} \label{Caves73} \label{Manne1}

OK, I desist.  Each day and each evening, I tell myself, ``Tomorrow I'll really get to that paper,'' and each morning following something transpires against me.  I'll give you what I have:  It's only a minor variation of what you originally sent me all those months ago.  Do with it what you please:  I know that I should be shot for my behavior.

Let me just tell you what I wish were strengthened, in case you decide to be better citizens than I:

The Introduction and the Conclusions.  What bothers me most about the present introduction is that after seven of the eight paragraphs in it, the reader still does not know what this paper is about.  What really motivates our attention on POVMs?  What was wrong, or less than fruitful, in the old von Neumann approach?  Why isn't our approach cheating given the Neumark theorem?  Why should people care about this work?  What have we learned by making the theorem work for qubits?  What have we learned by making the theorem work for complex-rational Hilbert spaces?  Why did we even care to explore the restricted frame functions in Section IV?  What did we learn by the quantum rule only coming about for the trines?

Maybe all these things were said, but definitely not with enough emphasis for my tastes.

Technically, I had the most trouble with two things.  1) The discussion about noncontextuality, and 2) (what strikes me as) an imprecise use of ``linear'' in Section III.  In the case of the former, I just deleted the present discussion on the subject.  I say, let's try to skirt the issue unless the referee calls us on it, or do it a lot better than was done previously.  Concerning 2) I'm talking about phrases like ``The function {\it f\/} is thus established to be linear in the rationals.''  No, the nonnegative rationals.  And the instances are multiplied beyond this.  Just a note to be more careful (to fix the language that I didn't already fix).

One point of philosophy.  I don't buy for a minute statements like, ``Properties of physical systems, though useful if they help in this task, are ultimately irrelevant.''  That kind of positivism has never been my guide (though it does fit Asher Peres quite well).  My only point has been that one should not mistake the quantum state for a property.  The foundational task, as I view it, is precisely to uncover what can be called a property of a quantum system and what cannot.  To that task, I see this paper as making a sound contribution.

\section{20-02-03 \ \ {\it Intro Draft} \ \ (to J. Bub)} \label{Bub9}

The more times I read over your draft for the intro, the more I like
it.  I'll send you a very mildly revised version this afternoon.

Possibly the only delicate point will be how I might de-emphasize the
role of entanglement enough to suit me, while still pleasing you.  A
lot of quantum information does not depend upon entanglement at all
(most quantum key distribution schemes, for instance, but also
quantum nonlocality without entanglement, incompleteable product
bases, etc.).  Or look at the present debate on the power of
unentangled states for quantum computation (cf.\ Gilles' talk at the
commune).  Finally, you ought to know by now that my own opinion is
that entanglement is likely to be a red herring in the deeper vision
of things:  I see it as subordinate to the structure of measurements.
Derived and secondary.

\section{20-02-03 \ \ {\it New Draft, Only a Drop More of Poetry} \ \ (to J. Bub)} \label{Bub10}

Attached are my modifications to your draft.

I debated for a long time toning down the theme on entanglement, but
in the end gave up.  I know that I compromise my beliefs somewhat
with some of the phrasings, but, very probably, only {\it I\/} know
that \ldots\ and, in the end, I may be able to make an easier
connection to the entanglement stampede this way.  (My role has
always been that of a mole.)

Just some notes for your reference, so that you'll know what I did
and what I thought: [\ldots]

2) I inserted ``they thought'' between ``which'' and ``spelled'' in
``could exist in certain states which spelled trouble for the
Copenhagen interpretation'' (second sentence, first paragraph).  I
did this to leave a window open for the possibility that EPR were on
the wrong track with their argument.  From my view, the only thing
they demonstrated ultimately is that the quantum state is epistemic
in character.  If Bohr had known that word---epistemic---I think he
would have agreed with them up to that point.  He would have only
disagreed that a more complete epistemic characterization could be
given.

3) Second paragraph, fourth sentence: ``That is, depending on what
measurement Alice chooses to perform \ldots\ and the outcome of the
measurement, Bob's system will be left in one of the states of some
mixture \ldots''  I'll just note this as a sentence that I would not
normally write anymore, even though I left it intact.  The trouble
with it is that it conveys the image that the ``state'' is something
possessed and inherent in the system---i.e., that it is a
``property'' of the system.  Instead, I view the state in question as
a property of Alice's head---i.e., her information about the
system---and would normally use language appropriate to convey that
idea.  Since, however, we are contrasting things on Schr\"odinger's
understanding, I can be OK with usage in this instance. [\ldots]

10) Last paragraph.  Added one small drop of poetry.

\section{21-02-03 \ \ {\it Pseudo-Spring Day} \ \ (to G. L. Comer)} \label{Comer27}

I finally get a moment to write you.  (Though I have a nasty hang nail on my left index finger, which makes it partially unpleasant \ldots\ physically, that is!)

Refereeing papers?  It feels like that's all I ever friggin' do.  I hate it, but yet I always feel responsible to do it nevertheless.

You know what chaps my own ass?  Mathematicians or physicists who get haughty and act like, ``Well, if you're talking about a {\it finite\/} dimensional vector space, it must be trivial \ldots\ solved long ago \ldots\ not worth thinking about.  Everything is trivial in finite dimensions.''  (I remember such a scene vividly from a conversation with Dan Finley, a relativist at UNM, when I was a grad student there.)  OK genius---I want to say---take a complex vector space of $d$ dimensions.  Can you find precisely $d^2$ unit vectors such that the modulus of the inner product between any two of them is equal to a constant?  It's a trivially stated problem, but show me the trivial solution if it exists.  On the other hand, if there be no such $d^2$ vectors, show me that too.  Trivial, you say?

By the way, that's the big problem on my mind at the moment.  It's on several other people's minds too (I count about 10).  If you've got any insight, I'd love to hear it.  We still haven't written a joint paper, you know!

\section{22-02-03 \ \ {\it Lorentz Chair} \ \ (to J. Preskill)} \label{Preskill10}

By the way, what was it like to sit at Lorentz's desk?  If you give me a good anecdote, I'll record it for posterity.

\section{24-02-03 \ \ {\it Born and Toeplitz} \ \ (to A. Peres)} \label{Peres49}

\bap
Do you know of an autobiography of Max Born? Danny told me that he saw
it (in Russian translation?)\ and that Born wrote that when he was a
student, he was a baby-sitter for the daughter of Prof.\ Toeplitz in
Gottingen. The reason I ask is that the daughter later came to Israel,
and recently passed away. She was a cellist in Lydia's orchestra. Her
father was an amateur photographer, and she had pictures of all the
great men, Hilbert, etc. Lydia borrowed them to show to the Dean of
Maths, who had a shock, and asked permission to display some of the
pictures in the hallways of his building.
\eap

I think there are two books that could be reasonably counted as autobiographies of Born:  1) {\sl My Life and My Views}, and 2) {\sl Physics in My Generation}.  (I just looked them up in my list of burnt books.)  I don't remember if there was any overlap in the essays there---I seem to recall that there may have been---but I read them both 10 or more years ago.  I do remember enjoying them.  (I actually read the Born--Einstein letters twice, but that's a different subject.)

Interesting to hear about Toeplitz.  I hadn't noticed his name until last year.  I used facts about Toeplitz matrices to calculate Eq.\ (40) in my \quantph{0205039}.

\section{24-02-03 \ \ {\it Fall Calendar and Symmetry} \ \ (to W. E. Lawrence)} \label{Lawrence1}

\bwel
I personally have gotten interested in two directions --- unbiased basis sets and related operator sets (and the question of why nobody has found a proof that these exist other than for power-of-prime dimension).  Is this just number theory, or is it Physics that I don't
understand yet?
\ewel

I'll bet that's a number theory reason, rather than a physics reason \ldots\ but you never know.  Still, though, I'm glad you're interested in the question.  I've been interested in quite a similar one:  the one of the existence of ``symmetric informationally complete POVMs.''  You can find these things defined on page 23 of my new paper \quantph{0302092}.  They are much like a complete set of mutually unbiased bases, but even more symmetric.  In contrast to the MUB question, though, we do know (numerically) that they exist in, at least, all dimensions from $d=2$ to $d=14$.  Mostly I'm interested in them on three counts:
\begin{enumerate}
\item
I think they will give the most beautiful definition for the ``standard quantum measurement device'' in Section 4.2 of my \quantph{0205039}.  The prettier I can make the region in Figures 1 and 2---the latter on page 38---the more insight I think we'll be able to glean from such a representation of quantum mechanics.
\item
Also, in connection to that, I suspect that the symmetric POVMs define the largest such region possible.  Thus, thinking of quantum mechanics as a modification of usual Bayesian reasoning (which makes use of the full probability simplex, not just a restricted piece of it), such a measurement defines a venue in which our reasoning can be as close to Bayesian as possible.  That is to say, if you're looking for the minimal change quantum mechanics makes to our methods of probabilistic reasoning, it will be there.
\item
Finally, referring to page 23 of \quantph{0302092} again, there's something quite interesting about these POVMs that I've just come to realize.  There appears to be a sense in which the elements in that set are just as quantum as they can be with respect to quantum eavesdropping.  (Since posting that paper, by the way, I have proven that the accessible fidelity for such an ensemble really is $2/(d+1)$, after all.)
\end{enumerate}
So, now I wonder what is the deeper connection between all these statements.

\section{25-02-03 \ \ {\it Book Review}\ \ \ (to A. S. Holevo)} \label{Holevo6}

I don't know whether you are able to get the journal QIC ({\sl Quantum Information and Computation}) at your institute, so I will send the file attached directly to you.  It is a small (very small) review of your new book that I wrote for the journal.  I hope you enjoy it, and that it does not offend you too much (via my greed of always wanting a deeper and deeper explanation of quantum mechanics).

I have distributed copies of your book to Caves and Bennett so far.  Friday, when I see Peter Shor, I will give him a copy too.  Unfortunately, I forgot to give one to Schumacher the last time I saw him in November (even though I had packed one up for him).  I will, though, give him a copy when I see him in May at the Bennett 60th-Birthday Symposium.

\section{25-02-03 \ \ {\it 9909073} \ \ (to P. Busch)} \label{Busch3}

Carl Caves's student Joe Renes has finally put together our work on Gleason-like theorems for POVMs into a manuscript, with special emphasis on 2-D Hilbert spaces (where one can restrict the class of POVMs and still get the result).  We need to cite you.  Has \quantph{9909073} still not been published?!?!?!  I just looked at your webpage and see no update of it.  We're hoping to post the manuscript this week.

\section{25-02-03 \ \ {\it Halmos, Ireland, Serotonin} \ \ (to G. L. Comer)} \label{Comer28}

\bgc
I guess I know a bit about finite-dimensional vector spaces, being a
relativist and all, but I don't know much at all about complex spaces.
What books did you look at when learning about complex vector spaces?
\egc

Two good sources are:
\begin{itemize}
\item
Paul R. Halmos, {\sl Finite Dimensional Vector Spaces}
\item
Roger A. Horn and Charles R. Johnson, {\sl Matrix Analysis}
\end{itemize}

Just about everything you'd ever need to know to be a mathematical muscle in quantum information is already in those two books.  It's interesting just how much prettier, and better behaved, complex spaces are over real spaces.  The tensor product is a really screwed up construction for real vector spaces, but is immaculately clean for complex spaces.  That's something deep, and I'm sure, the ultimate reason why quantum mechanics makes use only of the complex variety.

Let me give you a relevant example in the present context.  The question I wrote you about actually has been considered by mathematicians before (in the early 1970s) \ldots\ but for real vector spaces.  There the natural question is whether one can find $d(d+1)/2$ equi-angular rays.  There, after a lot of work, it was finally proved that such sets do exist, {\it but\/} only for about five different (noncontiguous) values of~$d$.\footnote{This is not technically correct.  I believe no one presently knows if there may not be an infinite class of dimensions where the bound is achieved.}  Thus, the concept didn't seem to be so robust or interesting, and no one ever did any work on it again.  In contrast, for complex vector spaces, it's looking like there's going to be a solution for all values of~$d$.

The reason I'm particularly interested in the question is because, with such a set of vectors I could gedanken-construct an ``international standard quantum measurement device'' that is prettier than all the rest (in pretty intuitive sense).

\section{26-02-03 \ \ {\it Distribution List}\ \ \ (to A. S. Holevo)} \label{Holevo7}

By the way, I count that I have three copies of your book left, beside my own.  (QIC had also sent me a copy since I was a reviewer.)  Who would you like me most to distribute them to?  Below is a list of people that I will either see on or before the Bennett-fest.

\bv
M. Gell-Mann \\
E. Fredkin             \\
T. Toffolli            \\
S. Wiesner             \\
G. Brassard            \\
N. D. Mermin           \\
W. Wootters            \\
A. Peres               \\
R. Jozsa               \\
P. Shor                \\
B. Schumacher          \\
J. Smolin              \\
S. Popescu             \\
R. Horodecki           \\
A. Zeilinger           \\
M. B. Ruskai           \\
C. King                \\
J. Lewis               \\
O. Hirota              \\
\ev

\section{05-03-03 \ \ {\it Quantum Anti-War Petition:\ Need Exponential Growth} \ \ (to 58 people)}

\noindent Dear friends and quantum colleagues, (i.e., this really is Chris),\medskip

If you are as disturbed about the prospect of war in Iraq as I am, the world-wide destabilization it may cause, and the trashing of the United Nations it indicates, I hope you will consider signing the petition at:
\bv
\myurl{http://www.moveon.org/emergency/}
\ev

As far as I can tell, {\tt MoveOn.org} is a serious, significant organization and one to be taken seriously.  The petition seems to be at least 300,000 signatures strong presently, but I hope it will grow exponentially before its delivery to the United Nations tomorrow (March 6).
You can read about the organization's legitimacy by having a look at:
\bv
\myurl{http://www.moveon.org/moveonpress.html}
\ev

I apologize if this note is an intrusion on your politics, but I tried to be selective in my distribution list, based on conversations I've had with each of you in the past.  (I hope I have remembered correctly.)  Most importantly, I hope you will pass on this note, or a similar one, to your own friends before tomorrow's deadline.

\section{07-03-03 \ \ {\it Quantum Anti-War Petition:\ Need Exponential Growth} \ \ (to A. Plotnitsky)} \label{Plotnitsky12}

You're probably right about the fatalism---that things have progressed too far to be turned around---but there is still time to sign the petition today.  The number of signatures has now reached over 550,000.  It's one of the very largest petitions in history.  They're now aiming for 750,000 signatures.  So, I hope you will spread the word.

\section{24-03-03 \ \ {\it Har Har} \ \ (to J. M. Renes)} \label{Renes14}

\bjmr
Read any good papers on Gleason's theorem lately?
\ejmr

Point well taken, and I assure you I was already feeling guilty before I got your note.  It's just that this darned move and house sale has been pretty much all-consuming.  I will, however, be in Ireland away from the family (on my house-hunting trip) for the full week this week.  I am hoping to get quite a bit of work done with this opportunity.  Your paper is at the top of the list.

\section{27-03-03 \ \ {\it Crypto Scheme} \ \ (to B. Yurke)} \label{Yurke1}

Here's a simple crypto scheme for these states.\footnote{As stated here the crypto scheme is incorrect, but easily modified into something workable.}  It's a little BB84 and a little B92.  Alice sends one of the $d^2$ states drawn at random.  After Bob receives the signal, but before he performs a measurement, Alice reveals that the state she sent belongs to a set $X$ of $d$ linearly independent states ($X$ being a subset of the original set of $d^2$ states).  Bob then performs an unambiguous state-discrimination measurement appropriate for the set $X$.  In this way, through a pre-agreed numbering of the states within each set $X$, Alice and Bob should share a key on the successful outcomes.

The questions are a) how secure is this system, and b) how efficient is it?  Clearly it is more efficient in the limit of large $d$.  But that should also be the limit in which it is the most secure \ldots\ unless I'm missing something.  In any case, to do this right is going to take some analysis.

\section{20-04-03 \ \ {\it The Young James} \ \ (to G. L. Comer)} \label{Comer29}

\bgc
It proved to me that I have a free-will, unique to me; likewise, every
person has a free-will unique to them.  If I could, I would scrap all
religions, and replace them with the simple fact that we all have
free-will, by default.  The most important part of existence is shared
equally among all: we can choose.  We no longer have to be poisoned
with the drug of reassurance, purchased for us from the gods by some
elite priesthood, who bankroll it via an ``us and them''
partitioning of humanity.
\egc

This, in fact, was why I sent you the book {\sl Becoming William James}.  I don't know how well that biography captures the story---I haven't read it in particular yet---but what you express sounds very close to what one might take to be James's shaman experience.  In 1870 or 1872 (I can't remember which), nearing suicide, James had the same great realization.  It saved him and became the theme of the remainder of his life.

\section{20-04-03 \ \ {\it And Finally, My Smolin Number} \ \ (to G. L. Comer)} \label{Comer30}

\bgc
P.S. We had a talk at Parks, and Rachael Obajtek was there.  At some
point I was making some remarks, and Rachael said ``You sound like
Fuchs.''  I said, ``Oh no, you're wrong; Fuchs sounds like me!''
\egc

This reminds me of how I was at a meeting once bragging about my Einstein number:  It's 3 because I've written papers with Peres who's written papers with Rosen who'd written papers with Einstein.  John Smolin happened to be in the circle, and at some point piped up, ``Hmm, I wonder what Einstein's Smolin number is?''  He had me beat hands down!

\section{21-04-03 \ \ {\it A Question of Condensation and Time} \ \ (to J. I. Rosado)} \label{Rosado1}

Thank you for your interest in my paper.  Have we met before? Perhaps
in Oviedo?  Anyway, I apologize for taking so long to reply to you,
but for over a month now I have been scrambling with the details of
relocating my family to Dublin.  Finally that task is coming to an
end, and I hope to be able to turn my attention back to science soon.

\bjir
In section 3 you give an argument, EPR$+$teleportation, to
support the idea that a quantum state is a state of beliefs, so that
they have nothing to do with real properties of the real physical
state. At the same time, I think, your argument also implies that the
outcome of a measurement doesn't reveal a preexisting property of the
physical system, inclusive for pure states, inclusive of the case of
a measurement of an observable that has that pure state as
eigenstate. Do you think that this conclusion is true? If true why
don't you have stressed it?  I think it is very important to be explicit
in this conclusion because in the vast majority of the quantum
literature we can find phrases like: ``quantum measurement, with very
few exceptions, cannot be claimed to reveal a property of the
individual quantum system existing before the measurement is
performed.'' so that the vast majority of the scientific community
thinks that if we can predict with certainty the outcome of a
measurement then this is so because we know a real property of the
individual quantum system.
\ejir

Yes, that is true.  I have tried to stress it very significantly to
many of my friends, but unfortunately I may not have yet carried the
point through as clearly as possible in my papers.  It is a question
of time, finding the right phrases, and biting the bullet for a big
job.  What I mean by the latter is condensing the notes in my
document ``Quantum States:\ What the Hell Are They?''\ into a proper
paper.  If you are not familiar with that document, I would encourage
you to try to follow some of the arguments there, starting around
page 35.  The content of the samizdat follows the format of email
messages, but often they are self-contained.  (You can download the
samizdat from my webpage; there is a link for it below.)  I have it
as a goal to try to write a real paper on the subject before the end
of the year, but we shall have to see if that really materializes.

\bjir
My conclusion, after I have read your paper, is that if we know with
certainty that a measurement will obtain a particular outcome we
cannot conclude that this implies a preexisting property of the
individual quantum system, only that if we perform that measurement
in that physical system we will obtain that outcome with certainty.
\ejir

Yes, and I think this realization has the potential to lead us down a
profound path.  (But that is only a dream at the moment; it is not
science yet.)  Anyway, it has in part led to my newfound attraction
to the philosophy of William {\James}, which you can also read about in
the same document.  Maybe the best notes are the ones to Wiseman,
starting on page 210.

There is much to do to put some technical flesh on these little
observations and to follow their implications unflinchingly wherever
they lead.  I am flattered that you have taken an interest in these
ideas, and I hope you'll help carry the torch in your own work.

\section{23-04-03 \ \ {\it The Rumor Mill}\ \ \ (to S. J. van {\Enk})} \label{vanEnk26}

\bsve
But I read several chapters in Dennett's book {\bf Freedom Evolves}.
\esve

I have read parts of his (early) book {\sl Elbow Room\/} (also on free will), in which I suspect he makes the same argument.  He certainly tried to draw the same conclusion.  But I should read this other book; I've never heard of it before.  Dennett is always good to read.  And he's an impressive speaker too.

\section{25-04-03 \ \ {\it Infinity}\ \ \ (to S. J. van {\Enk})} \label{vanEnk27}

``Jesus love me, this I know, because the Bible tells me so.''

``Separable Hilbert space only so, because von Neumann told me so.''

There's probably a deeper answer, but I don't know it.  It probably just means that the mathematicians had too many counter-examples to what they wanted to be true in nonseparable Hilbert spaces.  Physicists, never worrying about such matters, then just fell in line.

And yes, the answer to 2) is the same as the answer to 1).

I know that's of little help, but I'm in an airport again.

\subsection{Steven's Preply}

\bq
I have 2 questions about infinite Hilbert spaces: I hope you find some
time to answer my no doubt simple questions:

1) I remember that you once said that one always assumes Hilbert spaces to
be separable, which if I remember correctly for infinite Hilbert spaces
means their dimension is countably infinite. Is that so? Why does one
assume that?

2) For our laser paper we made use of a trick from a paper by Blow and
Loudon et al.: starting from a continuum of modes, characterized by a
continuous frequency $\omega$, one goes to a discrete set of modes by
first defining a complete discrete set of functions of time (or
frequency), and subsequently constructing a discrete set of modes.
That always looked convincing to me, but why is that allowed? Why doesn't
it matter that we go from an uncountable to a countable set of modes?
[Maybe this has the same answer as 1)!]
\eq

\section{26-04-03 \ \ {\it Vancouver Books} \ \ (to P. C. E. {\Stamp})} \label{Stamp1}

Are you serious, ``Vancouver has no good used bookstores downtown''?!?!  I had a field day!  I'm still downtown, in fact, having a pint.  Let me tell you what I found:
\begin{itemize}
\item
Arcanum Books, 317A Cambie St.
\subitem
Tom Burke, {\sl Dewey's New Logic:\ A Reply to Russell}, \$12.00.
\item
Albion Books, 523 Richards St.
\subitem
Paul Feyerabend, {\sl Killing Time:\ The Autobiography of Paul Feyerabend}, \$10.00.
\item
MacLeod's Books, 455 West Pender St.
\subitem
W. V. Quine, {\sl Ontological Relativity and Other Essays}, \$8.00.
\subitem
Michel Foucault, {\sl The Archeology of Knowledge}, \$15.00.
\subitem
William James, {\sl The Letters of William James}, 1926 edition, \$25.00.
\subitem
and the real catch of the day
\subitem
William James, {\sl The Correspondence of William James, Volume 7: 1890--1894}, \$45.00.
\end{itemize}
That amounts the best book-buying day I've had in a while.  (You can see I like to record things.)

Thanks again for inviting me to this meeting.  I think much did come out of it for my quantum foundations program, and I am grateful.

\section{27-04-03 \ \ {\it Bull by the Horns} \ \ (to R. W. {\Spekkens})} \label{Spekkens13}

I just wanted to tell you again how excited your letter has gotten me.  This sounds like a real progress in our understanding!  Have a look at the note I wrote Sam Braunstein in 1996 [{\sl Samizdat}, page 434] \ldots\ or better the 24 April 1998 letter to {\Ruediger} {\Schack} on page 390.  [I think I also put some words like that in \quantph{9601025} \ldots\ Aha, yes I did; see page 25 of it.]  I may have sloganized the idea, but you took the bull by the horns!  I must say, {\it I am jealous}!! \ But I am also proud of you in this regard:  You should have rightly been my postdoc this year, not Perimeter's!

Let me go through a few points of your letter for clarification and a couple of suggestions.
\brws
I've been busily trying to finish up a bunch of projects, and get
settled in here. In my spare time, I've been working on a toy theory
(Lucien has also been looking at something very similar) that I think strengthens the argument for the epistemic view.  Basically, it's
classical probability theory with an additional constraint: that the
amount one knows about a system must always equal the amount one
doesn't know about the system when one is in a state of maximal
knowledge.  Basically, one demands a balance between certainty and
uncertainty.
\erws

Mostly I'm wondering how to interpret this precisely.  Presumably, you're starting off with a fixed probability simplex over some number, $K$, of outcomes.  Are you explicitly taking $K$ to be a perfect square as it would be in quantum mechanics?  Alternatively, if you're not:  I wonder what special properties a perfect square has in this regard?

More importantly though, it sounds like the arena for your considerations is that the ``states'' ought to correspond to a convex region within the simplex.  The states of maximal knowledge are then the region's extreme points.  Is that right, or do you mean something else?  For instance, I could imagine that you might be thinking of the boundary rather than the extreme points \ldots\ though that is less likely.

Now, what is the nature of the constraint that defines your states of maximal knowledge?  Are you saying that for an event space of size $K$, the Shannon entropy of a state of maximal knowledge ought to be $\log\frac{K}{2}$ bits?  Or do you have a different way to quantify the idea.

In either case, let me tell you about the move I would make at this point.  The tack is look back at quantum mechanics, see what it says about states of maximal information in this language, and then try to make sense of it as a very simple statement, like in your toy model.  What I'm thinking of in particular, is to think of the allowed region on the simplex as generated by a ``standard quantum measurement'' of the SIC-POVM variety, i.e., the symmetric informationally complete POVMs.  You recall these from {\Montreal}, right?  In case not, I'll attach a note from another collaboration of mine that gives some relevant definitions.

What is most interesting about that POVM in this context, is that it causes all the pure states to live on a sphere surrounding the center of the simplex (the maximally mixed state).  By this, I mean in terms of usual Euclidean distance.  You'll see this in Eq.\ (13) of the document I'm sending you.

In particular, it is not the case that all pure states have an equal Shannon entropy in this representation (if that indeed is the assumption of your toy model).  Rather they have constant, but not maximal, distance from the state of maximal ignorance.  What is the meaning of that?  I don't know in any detail, but clearly it is one possible statement that maximal information cannot be complete.

Anyway, question:  Making the assumption of ``constant distance from center'' rather than your assumption of ``balance between certainty and uncertainty'', how much closer still (if any) does one get to the full theory of quantum mechanics?

I know that it will not go the whole way to quantum mechanics, either.  This is simply because true pure quantum states do not make completely fill the surface of the sphere that they live on.  So, we're still missing something.

But mostly I'm intrigued in what you've already found with the following:
\brws
such as the impossibility of achieving disturbance-free information
gain about non-orthogonal states with certainty, [\ldots]
The toy theory includes all these phenomena.
\erws

My immediate reaction when reading your note was that this could not be right, but upon a little thought, I might be willing to buy it.  The point is, there can be no absolutely clean probes in such a theory.  Eve has to always have a kind of ignorance about her probe when it is quantum too.  Following the algebra of the assumption (i.e., that system$+$probe state must also be a state of maximal information) must always leave both states more mixed (with respect to the standard quantum measurement) than they started off being.  Is that the sort of thing you're finding?

\brws
I'm far from done, but I'm sending you a tiny section that addresses
Penrose's argument against the epistemic view. I suspect that you will
disagree with my response, but, if so, I'd like to hear your
objections.
\erws

Yeah, you're right:  I enjoyed half the argument.  I was trying to capture something similar at the meeting in {\Montreal}, in that I kept trying to dream of a way in which a von Neumann measurement could always be viewed as a ``coarse-grained measurement'' (the words I was using then) rather than the fine-grained variety everyone always assumes it to be.  But I could never quite pull the analogy together in a satisfactory way.  What you are talking about here is something of the same flavor \ldots\ right before you fall into the quagmire of ``ontic significance.''

Anyway, you're doing great work.  Can't wait to see you in Maryland.

\subsection{Rob's Preply}

\bq
I've been busily trying to finish up a bunch of projects, and
get settled in here. In my spare time, I've been working on a toy theory (Lucien has also been looking at something very similar) that I think strengthens the argument for the epistemic view.  Basically, it's classical probability theory with an additional constraint: that the
amount one knows about a system must always equal the amount one doesn't know about the system when one is in a state of maximal knowledge.  Basically, one demands a balance between certainty and uncertainty. As you know, classical probability theory includes a whole bunch of phenomena typically deemed quantum, such as no-cloning, steering, no exponential divergence of states under chaotic evolution, and all the others from my checklist in {\Montreal}, but excludes many other phenomena, such as the impossibility of achieving disturbance-free information gain about non-orthogonal states with certainty, the existence of sets of states that cannot be broadcast, the impossibility of multi-party steering (a.k.a. the monogamy of entanglement), the fact that there are many extremal convex decompositions of a state, the non-commutativity of measurements, the existence of interference, the impossibility of teleporting non-orthogonal states with certainty in the absence of perfect correlation, the existence of locally immeasurable product bases, the existence of unextendible product bases, and, of course, all the quantum information processing tasks that can be built on these. The
toy theory includes all these phenomena.  The phenomena that don't arise in the toy theory are things like contextuality, Bell correlations, the fact that there are POVMs that are not convex combinations of PVMs, the gap between the probability of discriminating non-orthogonal states using an optimal measurement and using an optimal unambiguous measurement, and the fact that 2 levels of a 3 level system behave like a 2 level system. (These lists aren't meant to be exhaustive -- they only include some of the things I've had a chance to look at) So, it seems to me that the toy theory makes about half of the mysteries of quantum mechanics intuitive. I'm hoping that there is an additional constraint that can be imposed that gets the rest of it. I have some ideas on that, but it's all very speculative.

I'll be talking about it in Washington.  Will you be there?  I'm
also trying to write it all up.  I'm far from done, but I'm sending you a tiny section that addresses Penrose's argument against the epistemic
view. I suspect that you will disagree with my response, but, if so, I'd like to hear your objections.
\eq

\section{27-04-03 \ \ {\it Delighted to Meet You} \ \ (to G. Valente)} \label{Valente1}

Thank you for your kind letter.  It is always very flattering to know that someone is reading your papers, but it's absolutely dazzling to know that someone is enjoying and, perhaps, getting something useful from them!

Concerning your query about whether I can enter into an extended discussion:  Unfortunately, you've hit me at a moment of almost complete correspondence blackout.  With all the stresses added by my recent move to Ireland, several of my collaborators beating down on me to finish our incomplete papers, and a very busy travel schedule at the moment, I have had to essentially shut down my email production.  (Believe me, that is a very painful thing!)  Of course, please feel free to write me anything you wish, but it may be quite some time before I am able to reply in detail.

However, I have two suggestions for a substitute:
\begin{itemize}
\item[1)]  Chris Timpson, a recent graduate of Harvey Brown, understands the quantum foundations program I'm pushing quite well.  If you don't know him already, he'd be excellent to talk to.

\item[2)]  I very much encourage you to attend a meeting on quantum foundations in Sweden that I am co-organizing this Spring.  The dates are June 1--6, I believe, and you can find information about it here:
\begin{center}
\myurl[http://web.archive.org/web/20030425003339/http://www.msi.vxu.se/aktuellt/konferens/]{http://web.archive.org/web/20030425003339/http:// www.msi.vxu.se/aktuellt/konferens/}.
\end{center}
\end{itemize}
Beyond the attendees listed on that page, Barnum and I are also organizing a separate session on ``Quantum Logic meets Quantum Information'' which will be attended by: \medskip

\begin{supertabular}{ll}
Scott Aaronson        &      (U. California at Berkeley, USA)      \\
Paul Busch            &      (U. Nottingham, UK)                   \\
Bob Coecke            &      (Oxford U., UK)                       \\
Richard Greechie      &      (Louisiana Tech U., USA)              \\
Hans Halvorson        &      (Princeton U., USA)                   \\
Lucien Hardy          &      (Perimeter Inst., Canada)             \\
{\Ruediger} {\Schack} &      (Royal Hollaway, UK)                  \\
John Smolin           &      (IBM Research, USA)                   \\
Alexander Wilce       &      (Susquehanna U., USA)                 \\
Guido Bacciagaluppi   &      (U. California at Berkeley, USA)      \\
Rob {\Spekkens} &      (Perimeter Inst., Canada)                   \\
\end{supertabular}  \medskip

Thus you will have plenty of interesting people to talk to, some of which have tendencies in our same direction, but also I will have plenty of time to talk to you when Barnum, Hardy, Smolin and I inevitably go to the pubs in the evening.  (It is a usual thing; and, within that, usually a very philosophical time.)

\subsection{Giovanni's Preply, ``Touchable and Untouchable in QM''}

\bq
I'm an Italian student from Padua University.  My field of work is philosophy of physics and nowadays I'm in Oxford to write my dissertation. In particular I'm dealing with the question (maybe a bad question, in a Wittgensteinian sense) of realism in Quantum Mechanics. I share a skeptical point of view about the possibility to directly touch or to understand an ontological level and I also consider the probabilistic structure of QM as impossible to disregard. Thus, after our first discussion, Dr.\ Harvey Brown, my supervisor here, asked me if I knew Chris Fuchs' works and I professed my ignorance of it (I'm sorry, sir\ldots): he really looked amused by my convergence at something unknown (for me). Following his advices, I started reading your papers (``Quantum Mechanics as Quantum Information (and only a little more)'') and ``eavesdropping'' your private correspondence (``Quantum States:\ What the Hell Are They?''). I found your approach very deep, consistent and, most importantly, completely {\it immanent\/} to QM. I think we can't forget we are {\it disentangling\/} a physical theory; our aim is a reflection over the foundations of quantum mechanics. If one adds further elements (and nothing prevents to do this!), one must be aware of the fact that one is no more doing QM or philosophy {\it of\/} QM (but something else!). Physics can be regarded as a linguistic game (Wittgenstein): a tangle of mathematical, experimental, syntactical and hypothetical ropes, where the empirically successful is the premise and the goal at the same time. Our task should be to see deeply {\it inside\/} the theory and clarify {\it its\/} nature. Philosophy is ``only'': to play this game with the consciousness that it is a game. Science could be just a working fiction, why not?  I deem your works are very eye-opening in this sense, especially for QM. I'd like to explore this track and I hope you'll agree to answer me about some specific doubts I have.

In any case I've been appreciating your perspective and your way of reasoning. Furthermore, I find quite amazing (and sometimes annoying) to discover such a similarity between my still raw view and your articulate and solid plan. I must reveal that sometimes I feel frightened. For example. I woke up thinking about our alchemistic touch to the nature yesterday morning. As soon as I started studying, I read the same idea (alchemy) in your first statement about Pauli. I closed my books and I run away to phone my mum in Vicenza \ldots

Thank you very much for your willingness

P.S. I love Arthur Schopenhauer, too. He is a very deep philosopher. I think his thought was splendidly understood and embodied by Ettore Majorana. In fact he mystically (more than mysteriously) disappeared \ldots
\eq

\section{30-04-03 \ \ {\it Screed}\ \ \ (to A. Wilce)} \label{Wilce2}

\bq
\noindent {\bf screed:}\\
noun,\\
1. long usually tiresome letter or harangue.\\
2. layer of cement etc.\ applied to level a surface.
\eq

Glad to hear you'll be in Sweden!  And can't wait to hear about test spaces! (But finite dimensional ones, please \ldots\ if that concept applies.)

\section{01-05-03 \ \ {\it In the Turbulence} \ \ (to C. M. {\Caves})} \label{Caves73.01}

Despite my nasty note yesterday, I'm making a little progress on the paper \ldots\ when the plane's not falling out of the sky.  I'm on my way to the Clifton meeting.  If it's any consolation:  Though I don't always work on papers, I do always work on selling the point of view.  The meeting in Vancouver was particularly pleasing (with the philosophy side of it) on that point.  People are seriously taking note of the ontic/epistemic split we're seeking.  And I'm discovering to my surprise that there are quite a few Bayesians out there.  As I told {\Ruediger} the other day, I think the time is becoming ripe for a conference on ``Being Bayesian in a Quantum World'' which I think we ought to put together in a year or so.

\section{01-05-03 \ \ {\it Screed Again} \ \ (to H. Barnum)} \label{Barnum5}

I'm finally reading Hacking's book {\sl Representing and Intervening}, per your, and Chris Timpson's, and Rob {\Spekkens}'s, and Steven Savitt's suggestions.  It reads like eating candy; very nice.  Just in the interstices of today's flight when I wasn't working on the present paper with Renes and Co., I managed to read 67 pages.  That's a phenomenal amount for me!  I generally read much slower than that.

\section{05-05-03 \ \ {\it Pitowsky} \ \ (to T. Rudolph)} \label{Rudolph9}

\btr
Did you ever ask Itamar about your ``shape of the q-states within the simplex''
problem?

I just remembered that when I first met him at ESI all those years ago (back
when we both were still young) he told me he was working on some aspects of a
problem to do with parameterising the shape of the convex set of q-states
(though not on the same simplex you are I don't think).

He also knows a lot of the math literature, so could probably guide you to
anything that has been done \ldots
\etr

No, I haven't.  That's a good idea, and I'll follow through with it.  I know which convex set you're talking about with earlier work, and, you're right, it's not the same one I'm thinking about.  But he'd definitely be a good source of techniques that may be relevant for the present problem.

I decided not to come to Bell Labs today.  I'm just too exhausted; maybe a day's recuperation is what I need most.  I'll probably be in tomorrow.  I very much enjoyed this weekend's discussions, by the way.

\section{05-05-03 \ \ {\it On Bohr and Realism-X} \ \ (to R. W. {\Spekkens})} \label{Spekkens14}

Attached is my compendium-under-construction, ``The Activating Observer: Resource Material for a Paulian--Wheelerish Conception of Nature.''  To find the best reading on Bohr's particular flavor of realism, look at the Barad and Folse areas of the compendium.

I have read Bohr thoroughly, and I have read Folse thoroughly, and I think no one does a better job of rationally explicating Bohr than Folse.  Tell me if you disagree.

\section{10-05-03 \ \ {\it Departing for Ireland} \ \ (to C. H. {\Bennett})} \label{Bennett28}

A small note to tell you that the information I ate for lunch was just great!  You ought to try the place; it's a nice little cafe outside the Aer Lingus lounge in JFK.  The lunchtime special was ``all you can eat, \$5.99.''  They brought out the NY Times for dessert, but I was absolutely stuffed by then.

Happy birthday again!

\section{17-05-03 \ \ {\it Good News, Bad News} \ \ (to J. W. Nicholson)} \label{Nicholson17}

Thanks for the good news; too bad about the bad news.  The good news got me percolating \ldots\ since I watched an episode of {\sl Scrubs\/} the other night, and they explained what {\it always\/} happens on a third date!  If you can't trust TV, who can you trust?

About the bad news, for some reason it took my mind to a story I once heard from John {\Wheeler} about a physicist who was to be executed---I can't remember---either sometime in WWII or during the Stalinist purge.  Anyway, the fellow was put in a death-row cell one evening, with the execution scheduled for the next morning.  When morning time came, for whatever reason he was set free.  Years later, {\Wheeler} asked him how he survived the night.  How could he manage with that horrible dread all night long?  The man told him, ``I found myself thinking over and over about Hamilton's beautiful equations.  It set my mind at ease.''

I suppose everyone has to have a Hail Mary in some form or another.

\section{21-05-03 \ \ {\it Pouring Dublin} \ \ (to B. C. Rasco)} \label{Rasco1}

I'm here for a year:  April 15 to April 15.

\bbcr
I had a couple of things to ask you about quantum interpretation and
the observer.
\ebcr

I try to say what I know I can say in good conscience in my paper ``Quantum Mechanics as Quantum Information (and only a little more)''.  It's posted at my website, and you should be able to read it since it's posted in PDF.  (I recall once upon a time you weren't set up to read PostScript or something.)

By the way, I actually quote you---or maybe I should say paraphrase you---in my samizdat {\sl Notes on a Paulian Idea\/} (also posted on my webpage).  It was just published by {\Vaxjo} University Press.  If you'd like a real (i.e., nonweb) copy, I'll have them send one to you.

\section{22-05-03 \ \ {\it Index} \ \ (to B. C. Rasco)} \label{Rasco2}

By the way, the way to find your name in {\sl Notes on a Paulian Idea\/} is to look in the index.

\bbcr
Do ``we'' discover science or do ``we'' create science? What do you think?
\ebcr

I'm starting to think that science is more an expression of Darwinism than anything else.  It is neither a full creation of our own, nor in any sense a reflection of the world as it is without us.  It is just the best survival technique to date, due to various random confluences of events.  I have an extended statement on this if you dare to tread:  It's in my paper ``The Anti-{\Vaxjo} Interpretation of Quantum Mechanics'' (also on my website).  Skip the first three and a half pages, and start at the part titled ``to Preskill''.

\section{22-05-03 \ \ {\it Your Paper, Hope You Can Come} \ \ (to P. G. L. Mana)} \label{Mana1}

Chris King pointed out your paper \quantph{0305117} to me, and I downloaded it today.  It looks quite nice.  I have a significant interest in characterizing quantum mechanical state spaces as portions of a probability simplex (or simplexes), and I'd like to see the prettiest way to do that.  It's my hope that your paper will shed some light on the problem.

As for my own efforts in that direction, please let me recommend Sections 4.2 and 6 in my paper, ``Quantum Mechanics as Quantum Information (and only a little more)'', \quantph{0205039}.  There I don't use a table, but a vector representation for the outcomes of a single, sufficiently rich measurement.  So, the representation is different, but some of the ideas are the same.  Ultimately what I would like to do is divorce the representation away from quantum mechanics, and then start over (in the manner of Hardy and you) and recover the quantum formalism as the result of further physically-significant requirements \ldots\ but that is a project still in the making.

Anyway, I write to you mostly because I would love to get the chance to talk to you in person and, reciprocally, give you an opportunity to work with several people interested in problems of this ilk.  I am one of the organizers of a meeting in {\Vaxjo}, Sweden, to be held June 1 through 6, on quantum foundations.  You can find out information about the meeting at this website:
\begin{center}
\myurl[http://web.archive.org/web/20050427164645/http://www.msi.vxu.se/aktuellt/konferens/]{http://web.archive.org/web/20050427164645/http:// www.msi.vxu.se/aktuellt/konferens/}.
\end{center}

As I say, there will be several people there that I know will be interested in your work.  I'm thinking in particular of Howard Barnum, Lucien Hardy, Joe Renes, {\Ruediger} {\Schack}, and Rob Spekkens.  It would be wonderful if you could come.  Seeing that you are stationed so close at the KTH, perhaps our chances are better than zero to meet you!

\section{24-05-03 \ \ {\it Mechanica Quantica Lex Cogitationis Est} \ \ (to H. Barnum)} \label{Barnum6}

\bhb
Do you know the original source of the words ``quantum mechanics is a
law of thought?''  Is it you, JAW, or someone else?
\ehb

It was me.  I guess I started using the phrase early in 1996.  You can find the story on pages 247 and 370--371 of my samizdat {\sl Notes on a Paulian Idea}.

\section{24-05-03 \ \ {\it Maybe Do Give a Talk \ldots}\ \ \ (to J. M. Renes)} \label{Renes15}

I just got the note below from Andrei.  It got me thinking about your talk again.  Maybe it would be a good idea for you to give a talk after all.  I told Andrei to go ahead and set aside the time.  I think it would be good to talk about the trine and the Platonic solids in relation to Gleason.  It could be good to refresh everyone's memory on this.  Also you could speculate on what might be the higher dimensional analogues of this.  Maybe your talk would prod others to think about these issues and --- even if we're not going to work on it --- induce them to do something interesting in that direction.

Have you ever thought about how a POVM-version of Gleason where the frame functions are only defined on the extreme points of the POVMs.  That strikes me as an interesting arena of investigation.  Since such generalized frame functions include the ones from the standard Gleason theorem, the project has to work.  But I wonder to what extent the proof might get simplified in this way?  Also, the skeptics would not be able to claim that we're overtly assuming the linear structure by going into the interior of the convex set.  You might also propose something like this in a talk.

\section{26-05-03 \ \ {\it Practice?}\ \ \ (to P. G. L. Mana)} \label{Mana2}

Thanks for the invitation to Stockholm.  However, at the present time my travel schedule is already over-crowded, and I will be flying, both literally and figuratively, in and out of the meeting.  Maybe I could visit sometime in the coming year, when I am making my way to {\Vaxjo} again.  I have also wanted to meet G\"oran Lindblad for many years.

It's great to hear that you'll be at our meeting next week.  You've probably already guessed that we won't be able to provide any further support for participants \ldots\ but I wonder if you would be interested in giving a talk on your paper nevertheless?  One of the speakers from Barnum's and my part of the meeting (Hans Halvorson) had to drop out because his wife had her baby early; so we've got a little extra space for a speaker.  If you're interested, let me know immediately.  I cannot promise that the time will not already be gobbled up, but the sooner you let me know, the better chance you'll have if you're interested.

If you're a first year graduate student, it could be good practice.  Just remember to keep it simple, focusing more on the ideas and motivation, and far, far less on any derivations.  Less is always more in a talk.  Also it would be quite interesting to see the contrast between your ideas and Hardy's.

\section{27-05-03 \ \ {\it Popper Seminar, with a Side of Oxford} \ \ (to S. Hartmann \& J. Butterfield)} \label{Butterfield4.1} \label{Hartmann1}

Thank you both for the invitations.  I would be delighted to come give talks at LSE and Oxford.  Presently my schedule is completely open in February.  So why don't you two decide when you want me, and for how long you want me, and I will put the dates on my calendar.

The papa Bayesian of our group---{\Ruediger} {\Schack}---by the way, was a student of Schenzle too.

\section{27-05-03 \ \ {\it Got It:\ Thanks!}\ \ \ (to O. Hirota)} \label{Hirota2}

Thanks so much for sending a copy of the book {\sl The Foundation of Quantum Information Science\/}!!  Are you the author of the whole book?  Or is it a compilation of many contributions?  In any case, it looks very impressive.

I found my picture on page 214, as you had suggested.  Now, please translate what it says!  Do the four characters below my picture signify the word ``Fuchs''?  (Or fox, since fuchs means fox in German?)

It was an interesting coincidence that I would get your book today.  Just this morning, the secretary in Sweden told me that she mailed off several copies of my samizdat {\sl Notes on a Paulian Idea}.  You were in the list of recipients.  So it should be arriving soon.

Things are going well for quantumness here in Ireland.  Chris King and Koenraad Audenaert were able to prove that the accessible fidelity is a multiplicative function over tensor product ensembles.  They are presently writing a paper on the subject.  This had been one of the open questions Sasaki and I listed at the end of our paper.  What is especially pleasing about this new result is that it shows a very deep connection between quantumness and capacity questions for entanglement breaking channels.

\section{29-05-03 \ \ {\it See You Soon}\ \ \ (to A. Wilce)} \label{Wilce3}

\bq
\noindent 10.35--11.15. Alexander Wilce (Susquehanna Univ., USA) \\ \hspace*{2.1cm} Compactness and Symmetry in Quantum Logic\\
plus up to ten minutes for questions.
\eq
See you soon.

\section{09-06-03 \ \ {\it Quantumness} \ \ (to P. Busch)} \label{Busch4}

Thanks for coming to the conference!  Your presence really made a difference, especially for {\Ruediger} and me.

Let me give you a pointer to the background of the problem I mentioned on the last day.  It is:
\begin{itemize}
\item
C.~A. Fuchs and M.~Sasaki, ``Squeezing Quantum Information through a Classical Channel:\ Measuring the `Quantumness' of a Set of Quantum States,'' {\sl Quantum Information and Computation\/} {\bf 3}(5), 377--404 (2003). \ \quantph{0302092}.
\end{itemize}

After some massaging, the quantity in Eq.\ (16) there can be turned into the quantity I showed you, i.e., the $\inf \tr X$.  Of the open questions in the back, numbers 1, 2, and 6 have now been solved.  Koenraad Audenaert, Chris King, Peter Shor, and I will be writing up all that soon.  Question 6 especially required some nontrivial mathematics, and led to the expression I showed you.

The biggest problem on my mind at the moment is to show (or find a counterexample) that an ensemble of at least of $d^2$ states is required to achieve the quantumness of a Hilbert space.  That is, if an ensemble consists of $d^2 - 1$ states, then the quantumness of that ensemble is always bounded away from the quantumness of the Hilbert space.  King and I mostly, are working on that at the moment, but if you have any insight I'd love to hear it.  With you being a B, and he and I being a K and an F, you'd be assured to be first author on the paper!

\section{09-06-05 \ \ {\it Funding Possibility} \ \ (to C. M. {\Caves} \& R. {\Schack})} \label{Caves73.02} \label{Schack60.02}

As I told {\Ruediger} at last week's meeting, I got the note below from Miklos Redei, which says:
\bq
Some of you know already that the European Science Foundation (ESF) ( $=$ roughly the counterpart of NSF) has approved a so-called ESF Scientific Network entitled ``Philosophical and Foundational Problems of Modern Physics'' (PMP). The Network will be funded for three years, ending by December 31, 2005, and its main activity is organising 3 conferences and a couple of small workshops in Europe on topics indicated by the Network's title. ESF has a general agreement with NSF that enables US scholars to apply for NSF support to be used to participate in ESF Networks' activities, in our case: conferences and workshops. On behalf of the Coordinating Committee of the PMP Network I would like to ask you (all who receive this email message) if you would be interested in applying for NSF funds.
\eq

We might consider using this as a way of partially funding our Bayesian meeting.  Two other relevant facts came up at last week's meeting.  1) Eugene Polzik is now at the Niels Bohr Institute; so that might reinstate some of Carl's hope for their involvement.  2) Robin Hudson told us last week that Smith (of Bernardo and Smith) was fond of saying something to the effect that ``if quantum mechanics makes use of probability, then it must be Bayesian probability but I don't know enough quantum mechanics to say more than that.''  So, as {\Ruediger} and I discussed, he may be vulnerable to our funding onslaughts.

\section{10-06-03 \ \ {\it Quantum Information}\ \ \ (to T. Siegfried)} \label{Siegfried1}

We met briefly years ago at a meeting on quantum information and computing, but I doubt you remember me particularly.  Anyway, I've been meaning to write you for some time to tell you about a neat little coincidence I ran across as I was proofreading my samizdat {\sl Notes on a Paulian Idea:\  Foundational, Historical, Anecdotal \& Forward-Looking Thoughts on the Quantum\/} for a slightly more official publication by {\Vaxjo} University Press.

In the foreword to the book David Mermin writes, ``If Chris Fuchs (rhymes with `books') did not exist \ldots''.  On the other hand, in your book {\sl The Bit and the Pendulum}, there is some point where you write, ``Chris Fuchs (rhymes with books) applied Landauer's principle \ldots''.  In fact I actually report your line in my samizdat, but I doubt David was aware of it, as even I had forgotten of its existence until the final editing.  (I guess I say ``Fuchs rhymes with books'' to a lot of people.)

In any case, all this brings to mind that maybe you'd like a copy of the book.  Here and there you might find a story in it that you'd enjoy.  If you'd like, I can send you a copy:  Just give me your snail-mail address.  You can also find the book in PDF format at my webpage (link below), but you wouldn't want to actually print it out.  (That version is about 500 pages.)  Have a browse of the document, and if you think you'd enjoy a hard copy, I'll have the VUP secretary send you one.

\section{10-06-03 \ \ {\it Too Late, but Sympathetic} \ \ (to R. D. Gill)} \label{Gill1}

I wanted to apologize to you for not getting my act together in time to write a few words about your paper ``On Quantum Statistical Inference'' with Barndorff-Nielsen and Jupp for the Royal Statistical Society.  (They had invited me to, but gave a deadline of June 4; now I am away to conferences again until June 14.)

I wanted to particularly endorse your statement in Section 7, ``As statisticians, we would like to argue (tongue in cheek) that quantum probability is merely a special case of classical statistics,'' and write a few comments thereabouts.  I think one shouldn't be so tongue in cheek about this.

Anyway, as I say, I'm sorry I passed the deadline.

Too bad you didn't come to the {\Vaxjo} meeting:  I think you would have found it quite a bit more productive than last year's.  Several of the talks, in fact, addressed your tongue-in-cheek comment above:  Guido Bacciagaluppi, Paul Busch, Lucien Hardy, Luca Mana, {\Ruediger} {\Schack}, Rob {\Spekkens}, and also my own---all in quite a direct way.

\section{10-06-03 \ \ {\it Orthomodular Lattices}\ \ \ (to S. L. Braunstein)} \label{Braunstein9}

Maybe suggest to your friend R. I. G. Hughes's book {\sl The Structure and Interpretation of Quantum Mechanics}.  It talks quite a bit about quantum logic from several perspectives, and it also will give him loads of references if he wants to dig deeper.

\section{10-06-03 \ \ {\it Notes on a Paulian Idea}\ \ \ (to C. P. Enz)} \label{Enz1}

I don't know if you remember me:  You had very kindly sent and then re-sent me copies of all your quantum foundational papers---the latter after my house burnt down in the Los Alamos fires three years ago.

In any case, I'm writing to tell you that my book {\sl Notes on a Paulian Idea:\ Foundational, Historical, Anecdotal \& Forward-Looking Thoughts on the Quantum\/} has just been published by {\Vaxjo} University Press.  The book is very strongly laced with the influence of Wolfgang Pauli's ideas (to which the title refers), and is legitimized a little with a foreword by N. David Mermin.  If you have an interest in reading (parts of) it, I will send you a copy.  All I need is your mailing address.

If you are dubious, before making a decision, you may also have a browse through the book by downloading a copy from my website (link below).  Also there, you can find various of my other writings along with a Curriculum Vitae.

Another project that I am working very hard to complete this year is a large document to be titled ``The Activating Observer:\ Resource Material for a Paulian/Wheelerish Conception of Nature''.  Basically it is a compilation of quite a large number of references, along with extensive quotations from most.  Presently it stands at 109 printed pages, with 443 references.  By the time the project is complete, I expect it to run about 250 pages:  I still have a significant amount of material to put into place and have decided to enlist a good scanner along with optical character recognition software to help me in the project.  That document is not ready to be downloaded from the website, but you can read its abstract there.  If you would like a preview of the project, I could mail that to you directly and perhaps you could give me a few pointers for things I've missed.  For instance, I just learned last week that you yourself have just published a biography of Pauli!  I look forward to reading that as soon as I can purchase a copy.

\section{11-06-03 \ \ {\it Getting the Word Out}\ \ \ (to R. W. {\Spekkens})} \label{Spekkens15}

You know sir, I continue to be excited by your toy model even if we differ in our ultimate goals.  I had a conversation with John Preskill about it today, showing him the main ideas, etc.  He seemed intrigued.  In fact, intrigued to the point that he wants to invite you to Caltech to talk about it.  I think this would be a great opportunity.  I told John to just try to pick a time that I could make it too.  So, a) I hope you'll go, and b) I'd like to be there so that I don't miss a thing.  Can we coordinate?  I've always gauged the value of the program by the caliber of the people who start out skeptical about it, but then ultimately take note.  Early on I singled out both {\Mermin} and Preskill on this count, i.e., as markers of progress.  {\Mermin} yielded quickly, but Preskill has been a bit of a holdout.  So I'm particularly keen to see if we, through our combined efforts, can finally tip the scales.

While I'm on it, let me tell you about another strategical point that's been on my mind lately.  When I first contacted Luca Mana about liking his paper, he wrote me back the following:
\bpglm
Thank you so much for your letter, and for the words of appreciation
for my paper. It is just a little more than a draft, and so its form
is far from being nice, and many are the missing references. I already
knew ``Quantum Mechanics as Quantum Information (and only a little
more)''; indeed, your idea, expressed there, that time evolution could
be a density operator, stimulated (via Rodriguez' ``Unreal
probabilities'', \arxiv{physics/9808010}) some of the research that led to my
paper. \ldots

(So I hope you understand how honoured, and a little embarrassed, I
feel for having received your letter!)
\epglm

Yet one doesn't find a citation to any of my papers there.  It could be that he was just being nice and not quite meaning it.  However, if there really was an influence on his work, then his lack of a citation is a bit harmful.  If I am going to be able to keep getting funding for conferences and drawing attention to our collective efforts, it'd certainly help if the larger community sees that my attempts to write and communicate and inspire are part of the active process leading to the (significantly more substantial) results of you younger fellows.

I guess all I am saying is, if it influenced you at all toward your toy model, I hope you will not forget to cite \quantph{9601025} and \quantph{0205039} in your write-up.  (Sample below.)

From {\tt quant-ph/9601025}:
\bq\noindent
Formally, one says that in classical physics, maximal information is complete, but in quantum physics, it is not. What should we demand of a physical theory in which maximal information is not complete?  Maximal information is a state of knowledge; the Bayesian view says that one must assign probabilities based on the maximal information.  Classical physics is an example of the special case in which all the resulting probabilities predict unique measurement results; i.e., maximal information is complete.  In a theory where maximal information is not complete, the probabilities one assigns on the basis of maximal information are probabilities for answers to questions one might address to the system, but whose outcomes are not necessarily predictable (some outcomes must be unpredictable, else the maximal information becomes complete).  This implies that the possible outcomes cannot correspond to actualities, existing objectively prior to asking the question; otherwise, how could one be said to have maximal information?  Furthermore, the theory must provide a rule for assigning probabilities to {\it all\/} such questions; otherwise, how could the theory itself be complete?  Quantum physics is consistent with these demands.  A more ambitious program would investigate whether the quantum rule is the {\it unique\/} rule for assigning probabilities in situations where maximal information is not complete.  You won't be surprised to learn that we don't know how to make progress on this ambitious program.
\eq

\section{16-06-03 \ \ {\it Exhaustion, Depression, Stagnation, Integration} \ \ (to G. L. Comer)} \label{Comer31}

Below is a letter I started to write to you on my latest flight (Friday night).  It never got finished.  In particular, the last lines are not going to make much sense to you without the further sentences that would have completed the thought.  But the mood is broken now, and I don't much feel like drinking some beers this morning to get it going again!!  We'll talk eventually.

I wasn't able to finish the letter because it turned out that one of my collaborators was on the same flight.  We switched some seats and got down to work for the rest of the flight.

In any case, being back at home finally, life feels a little better.

\bq
Thanks for your very flattering letter.  You put a smile on my face at a time when it is starting to look blanker and blanker.  I'm flying back to Dublin from a quick trip to Washington, DC, and if all goes according to plan, I'm hoping to have a few beers with you this evening.

Looking at my records, I see that I haven't written you in over a month.  It's been a tough time for me (and the family) ever since the move to Dublin got underway.  Let me show you what I mean.  Here's what my schedule has looked like.
\begin{itemize}
\item
March 23--30, househunting in Dublin (without family)
\item
April 1--4, frantic efforts to get the house sale finalized and all our
            things in storage
\item
April 4--8, drive to Texas
\item
April 14--15, flight moving the family to Dublin
\item
April 20--27, trip to Vancouver
\item
May 1--4, trip to DC, Clifton Memorial Conference
\item
May 4--7, visit to Bell Labs
\item
May 7--11, Bennett Festschrift in New York
\item
June 1--7, Sweden conference
\item
June 10--14, ARDA meeting, Washington, DC
\end{itemize}
Here's your exercise:  tell me a) how many days in all of that that I got to see my family, and b) how many days were actually relaxing?  Whatever the numbers, it's pretty miserable.  Top that off with all the troubles one normally encounters with trying to get a family settled in a new country, and a few other medical difficulties, \ldots\ and one has a general disaster.

I'm exhausted, with a tinge of depression.  Maybe more than a tinge.  And my mental life has come to an almost complete stagnation.  In the tradeoff, my quantum foundations program is achieving a kind of recognition and integration into mainstream circles that seems quite promising.  People talk of the ``Fuchsian interpretation'' or ``Fuchsian view of quantum mechanics.''  And the material benefits follow.  UBC appears to be taking me serious for one of their ``Canadian Research Chairs'' (nontenured, but otherwise has some very enticing benefits beyond a usual appointment), and---get this---a professor at U.\ Nottingham is nominating me for a true-blue ``chair'' in his department (i.e., full professorship with external funding).  So, I should be happy, right?  I should be smiling, right?  My face is blank.

I'm just too tired.  My children are growing up, and to a large extent they only know Dad from his evening calls.  My mom gets older and older (74) and so rarely do I get to Texas to see her.  My friends (Comer, Peres, Mermin, Schack, and 300 people beyond that) write me and write me, but never do I reply.  The editors of all the journals pester me and pester me, but I can't stomach the email:  I file it away without a reply.  (Maybe there is some logic in that though:  Maybe I want them to get so sick of me, they'll never come back.)  My coauthors abuse me:  they get no reply.  Why can't they write a friggin' decent draft, I ask myself?  If they can't care to actually communicate their thoughts, why should I?

The world is a big world.  And our lives are ever such a small part
of it.  And ever such a big part of it.  But regardless, there are
limits.  I feel I'm not prepared for this.  I tell Emma all the time
that she is destined for greatness; she will do important things.
\eq

\section{18-06-03 \ \ {\it Spiders, Ants, and Honey Bees}\ \ \ (to P. W. Shor)} \label{Shor3}

I'll match your Morris Kline with a Francis Bacon:
\bq\noindent
The men of experiment are like the ant; they only collect and use; the reasoners resemble spiders, who make cobwebs out of their own substance.  But the bee takes a middle course; it gathers material from the flowers of the garden and the field, but transforms and digests it by a power of its own.  Not unlike this is the true business of philosophy for it neither relies solely or chiefly on the powers of the mind, nor does it take the matter which it gathers from natural history and mechanical experiments and lay it up in the memory whole, as it finds it; but lays it up in the understanding altered and digested.
\eq
For better or for worse, this is the way I see this sensation I'm trying to instill into our community, ``quantum foundations in the light of quantum information'':  I see it as a honey bee.

\subsection{Peter's Preply}

I just googled it.
\bq\noindent
The developments in this century bearing on the foundations of mathematics are best summarized in a story.  On the banks of the Rhine, a beautiful castle has been standing for centuries.  In the cellar of the castle, an intricate network of webbing had been constructed by industrious spiders who lived there.  One day a strong wind sprung up and destroyed the web.  Frantically the spiders worked to repair the damage.  They thought it was their webbing that was holding up the castle.
\eq
Reference given: Morris Kline, {\sl Mathematics:\ the Laws of Certainty}, Oxford Univ.\ Press, NY, 1984.

It seems to me that this also applies to the philosophy of the foundations of physics.

\section{19-06-03 \ \ {\it Lurch} \ \ (to J. M. Renes)} \label{Renes16}

\bjmr
PS in regards to some quantum operations are subjective, is the
subjectivity simply the case of assigning a state to the measurement
apparatus? This is how {\Carl} put it, and it clears up a lot of the confusion in my mind, but perhaps this isn't precisely the case.
\ejmr

No.  That way of putting it trivializes the point.  For it makes it look as if {\it all\/} the subjectivity (in the Bayesian sense) is in the state assignment to the joint system of object $+$ measuring device --- a concept almost straight from the Tibetan Book of the Larger Hilbert Space.  Then one can go around blithely thinking of the interaction Hamiltonian ($+$ preferred basis, so that the Hamiltonian cannot be transformed away) as objective components of the theory, well protected from the nasty subjectivity of the state.  I.e., one can think of these things as classical matters of fact, independent of any agent's belief.

What I want is to go beyond that:  And {\Carl} knows that.  All operations are subjective, I say, including the unitary derived from the interaction Hamiltonian above.  I give fifteen arguments for this, none of which is conclusive, but in sum weighty enough for me to be willing to take the leap.  Quantum foundations needs more leaps than baby steps, it seems to me, if we're going to see any real progress in our lifetimes.

\section{19-06-03 \ \ {\it Guidelines} \ \ (to M. P\'erez-Su\'arez)} \label{PerezSuarez1}

From the last letter you know about my final decision with regard to Benasque.  Just a word of warning:  If you are allowed to attend, and you give a talk, do it on a technical subject and keep the philosophical remarks to a minimum.  We will---with these ideas that you and I are attracted to---change the world.  But one has to choose one's battles wisely, and there is something to be said for easing the community into the mindset at a rate slow enough for the resistance to be smaller rather than larger.  The community attending the Benasque conference, for the most part, only cares about technical meat:  If a foundational program makes no impact on that hunger, then it will be seen as no program at all.

\bmps
Well, I have been to the [Niels Bohr Archive] and, of course, I asked about the William
James story. I saw the transcripts of that last interview and I can
tell you that things went as I remembered from my readings some years
ago. There is no indication of any technical problem during the
interview and Bohr's answers seem to have been completely tape-recorded with no remark in the transcript of any disturbance. You can
find the content of the answers in several well-known books. Although
I haven't consulted my references yet, if you want me to tell you
where you can find it, I have no problem to do some research in my
library (it is not as big as Howard's, I guess). By the way, I think
that perhaps it can be found in Folse's book on Bohr's notion of
Complementarity.
\emps

Then the transcripts must be missing the nuance.  Here is what Henry Folse himself told me:
\bhf
As you know, Bohr died after the 6th interview; 20 were originally
projected.  Jordan's are in German, and a few others; de Broglie in
French, of course.  However, there are holes in the transcripts where
the transcriber couldn't make the tapes out clearly enough. (the tapes
in Copenhagen are old reel to reel tapes and require an old machine to
play; I don't know if the American centers have them on cassettes).
Actually one such occurs when Bohr was talking about his admiration
for William James.  Just at that moment the noon air raid siren blew,
more or less obliterating about 30 seconds.
\ehf
Of course I embellished a little in our pub discussion, but I bet they are nevertheless an interesting 30 seconds.

\bmps
I have downloaded Richard Jeffrey's book on probability. I have
browsed through it and found it seems to be very readable (and quite
funny, too). I also have the recently published CUP edition of Jaynes'
book, and {\Ruediger} told me that Savage's book on probability was quite
interesting, too, so I will be purchasing a copy. Could you give me
some advice on what I should pay more attention to first?
\emps

Savage and Jeffrey, I would say.

\bmps
P. S.: I have found some typos (to be true, of no real significance)
in your paper. If you want a list of them, just let me know.
\emps
Absolutely!  I'd greatly appreciate that.

\section{20-06-03 \ \ {\it Very Late Reply} \ \ (to S. Weinstein)} \label{Weinstein1}

I apologize for the very late reply:  It was just that your email hit me at a very bad time when I was in a complete email-shutdown mode.

I think your first question may refer to a problem that still has not been resolved to complete satisfaction.  Unfortunately I've never taken too much note of the issue, so I don't have a strong opinion myself.  I do know of Ozawa's work [M. Ozawa, ``Quantum Nondemolition Monitoring of Universal Quantum Computers,'' Phys.\ Rev.\ Lett.\ {\bf 80}, 631 (1998)], but I remember hearing from Richard Jozsa and ?? Fahmy last year, that Ozawa's paper had a flaw in it.  Perhaps it would be best for you to write Richard yourself; maybe he'll explain what he was talking about and give some references.  I wouldn't mind hearing the answer.  Richard's e-address is \verb+rjozsa@cs.bris.ac.uk+.

Concerning your second question:
\bsw
I think we talked briefly about the ``little more'' that must be added
to informa\-tion-theoretic postulates in order to get quantum mechanics.
It strikes me that one of the key elements of quantum theory which is
not captured (at least in any explicit way) by the information-theoretic approach is that Planck's constant, a new constant of nature
which has the dimension of action, or angular momentum, plays a role.
Why action?  Why not some other physical quantity?  Classically,
action is associated with dynamics via the principle of extremal
action.  In quantum mechanics, however, action plays a kinematical
role as well.  What is that telling us?
\esw

You're certainly right.  Looking solely at the {\it structure\/} of finite dimensional quantum mechanics, one never sees Planck's constant at all.  Where it comes from, I don't know.  Nor do I know if information-theoretic considerations will ever tell us much about its origin.  A dull answer, right?  I wish I could tell you more.

Would you, by the way, like a copy of my samizdat {\sl Notes on a Paulian Idea}; it's just published and I have a number of copies for giving away.  Mostly, though, I'm trying to target people who might be interested (and won't throw it away immediately).  If you'd like a copy send me your mailing address.

\section{20-06-03 \ \ {\it Hold the Presses} \ \ (to G. L. Comer)} \label{Comer32}

\bgc
Do you know, or know of, Stapp?
\egc

Yep.  Stapp stands for most of the things I stand against.  The only overlap, as far as I can tell, is that we both like to quote Pauli.

\section{20-06-03 \ \ {\it Logical Probability and That } \ \ (to J. M. Renes)} \label{Renes17}

I think I agreed with much of your note.  That was the upshot of last week's (month's?)\ discussions.  To answer your final question (about Cox), I guess I don't have a strong opinion \ldots\ at the moment anyway.  Do any of the others in the discussion list have a strong opinion?  Mainly I never found Cox nearly as compelling as Jaynes did.  (Well, I did for a short while, but then I became less than enamored.)  I think one could still accept Cox's derivation of the probability calculus---even though I'm not so inclined (already said that)---without accepting Jaynes' further dogma about some priors being compelled by ``THE information''.  That's because Cox's derivation has nothing to do with priors.

Did you forget an ``all'' in the title?

\subsection{Joe's Preply}

\bq
These arguments are already beginning to leak out of my mind, so I'd
better commit them to paper or email to help reinforce them.

Let me see if I have things straight. Let's call ``Bayesians'' those who
don't believe in objective probability, logical Bayesians those who
think from background information a unique prior logically follows, and
subjective Bayesians those who think that background knowledge does
not determine for each subject (person/observer), a prior probability. The two
approaches do share the normative aspect that Bayes' rule is the correct
way to update knowledge given data. Are these fair assessments? Of course,
it's not fair that the subjectivists are somehow defined as
illogical in opposition to the logical Bayesians, but you know what I
mean.

The argument against logical Bayesian probability I picked up at the
conference I think goes like this. Let's specialize to the case of picking
``ignorance priors'' and recall how this is meant to work in the logical
probability framework. The background knowledge $B$ gives no
information about one outcome over or under another. Thus we study
the symmetry group of the outcomes $a_j$ and our prior must be invariant
under this group. Presto!\ the symmetry group of the outcomes determines
the prior probability.

But this doesn't represent an implication from propositions $B$ to the
probabilities $p(a_j|B)$ because to say that the outcomes have some kind of
symmetry is to say that some function on them is invariant under the swaps
of labels (surely the swaps are always permitted, even if the case of
symmetry doesn't obtain) and this function is precisely the probability
function. I said ``is to say'' which is quite different than ``implies''
the logical structure of $B$
that was to imply the uniform prior turned out to be the uniform prior
itself. Put differently: You can't say ``$B$ doesn't favor any outcome over
another'' without it being a statement of probability. So there's no real
implication going on except to say that ``when you believe in the symmetry,
you believe in the symmetry''.

On some re-readings of the above, I've thought that it says nothing and I
haven't stated the problem properly. But consider the argument made by
Jaynes in chapter 2 of probability as logic: ``But now suppose that
information $B$ is indifferent between propositions $a_1$ and $a_2$; i.e.\ if it
says something about one, it says the same things about the other, and so
it contains nothing that would give the robot any reason to prefer either
one over the other.'' Later, he says, ``It shows --- in one particular case
which can be greatly generalized --- how the information given the robot
can lead to definite numerical values, so that a calculation can get
started.'' But the point of the above argument is that the ``information
given the robot'' must be the uniform prior: the statement of symmetry {\it is\/}
the uniform prior. You can't say that ``whatever $B$ says about one it says
about the other'' in a non-probabilistic way, which means there's no
logical implication to it. That last statement is the one I like best.

I think there's a couple of additional points to make. The background
knowledge $B$ doesn't pertain directly to the outcomes $a_j$ themselves: if
they do we're talking logical probability. $B$, in some sense, has
nothing logically to do with $a_j$; the point is that the
connection between the background knowledge $B$ and the $a_j$ {\it is\/} the
probability distribution, which is subjective, as it's made by a person
thinking about the relationship between the two. This is just a
clarification in light of the above argument, about what the background
knowledge is or what it means.

The second is that perhaps my understanding so far is only crudely
scratching the surface of the real argument beneath. Perhaps it's this.
One way to see that $B$ doesn't pertain
directly to $a_j$ is that it doesn't tell you how to carve up the outcome
space into the $a_j$ and only then make a judgement of indifference. In
fact, the judgement of indifference is precisely this ``resolution'' of the
outcome space. This
seems to be the point of Jeffrey's problems 2 and 3 (though curiously
labeled 1 and 2 in the sequence 1,1,2,4) in chapter one of
{\sl Subjective Probability}. The carving up of the outcome space into
equiprobable events is a probabilistic judgement: ``these particular
outcomes $a_1,a_2,a_3,\ldots$ are equiprobable''.

Greater question: Just because one isn't a logical Bayesian doesn't mean
that one doesn't subscribe to the Cox arguments about probability as
generalized logic. Jaynes as a logical Bayesian stresses the Cox view of
course, but when it
gets down to the logical aspects, he wants probability to contain the
syllogism: $(a\Rightarrow b\; \&\; a) \Rightarrow b$ and its inverse $(a \Rightarrow b \;\& \sim b)\Rightarrow\; \sim a$, which it does,
using Bayes' rule. So are there some pitfalls to being a subjective
($=$ illogical, remember!)\ Bayesian and thinking of probability as
generalized logic rather than coherent/consistent betting preferences that
I don't immediately see?
\eq

\subsection{Joe's Reply}

\bq
No I didn't really mean to put ``all'' in the title, but not for any
particular reason. Carl and I hashed out a lot of that argument (the main
problem was that I didn't state it very well). He had a good way of
putting it: First we have to decide if the symmetry is a belief or a fact.
The symmetry is a belief, of course, and this can be seen in the arguments
given. There are a few. First, the symmetry is obviously not in the
system, since there are distinct outcomes. They wouldn't be distinct if
there were really a symmetry. Second, the argument I was making, the
symmetry condition is only stated (and can only be stated) in terms of
probability. Ergo, it's a belief. Third, and think what Jeffrey is getting
at in his book, the facts of the matter don't actually bear directly on
the outcome space, so there can be no logical probability. You must decide
what the relation is, and of course this is the probability. He uses an
example of two coins. What's the probability that there's 2 heads showing?
Is it 1/4 because there's only one way in four for the coins to be
arranged this way, or 1/3 because there's three possibilities?  The
background info simply doesn't tell you (or as he says, there's no way to
answer this question without making a judgment as to what is important).
This is similar to the first point in that the background info doesn't
``carve up'' the outcome space for you.

Anyway, just wanted to get that down for the record, as it were. I find
these arguments completely devastating to logical probability, and I agree
that one can still follow the Cox approach (I've not done a heap of
thinking on this last point though).
\eq

\section{23-06-03 \ \ {\it Bayesianism, Yes, but then Something More} \ \ (to N. D. {\Mermin})} \label{Mermin85}

\bdm
Alice measures a Qbit, gets 0, and then sends it to Bob.  When Bob
measures it he always gets 0.  How can this be if the instruction to
give 0 is not carried by (``resides in'') the Qbit Alice sent Bob?
\edm
\brs
I believe that this formulation is part of the problem. ``He always
gets 0'' is a loaded phrase.  In my view, it can mean one of two
things. It can mean (i) that Bob makes $N$ measurements, and each
time he happens to get 0. Or it can mean (ii) that for a single
measurement, it is certain that he will get 0. The second meaning is
what I try to deal with in my previous email, where I claim that
certainty always refers to somebody's state of belief. ``It is
certain'' is then always short for ``it is my state of belief that it
is certain''.

As for meaning (i), if Bob happens to get 0 every time in $N$
measurements, he can use this result to update his belief about
future measurements. Or he can update his beliefs about where the
qubits came from. But I don't see why he would have to conclude that
each qubit carried the instruction ``0'' before the measurement.
\ers
\bdm
``Have to conclude'' is too strong, but wouldn't it be natural for
him then to wonder why these two 0 results were so nicely correlated
and entertain as an explanation the notion that Alice's measuring 0
imposed the instruction ``0'' on the qubit?  What is it in his
Bayesian training that prevents this thought from entering his head?
Why should he be content with correlations without correlata?
\edm

There is nothing in his Bayesian training to prevent the thought from
entering his head.  Instead, it is his training in quantum mechanics
that bears the burden:  Bell inequality violations and
Kochen--Specker.  It is a question of looking at the whole package.

Quantum mechanics, as I see it (and maybe {\Ruediger} too), is a layer
added to the top of pure Bayesian reasoning.  Bayesianism by itself
does not care about the precise character of the left-hand argument
in a probability function $P(h|b)$.  The $h$ could be a pre-existent
fact living in the cold, hard world, or it could be something yet to
come--- something in existential character completely dependent upon
the catalyzing intervention of the agent himself.

All we ask of the community is that it recognize the category
distinction between the function $P(h|b)$ and its argument $h$.  That
is the first step in clarification.  Once one can accept that into
one's heart, then---the hope is---great progress will follow.

In particular, I would say, look long and hard at Figures 1 and 2 in
my paper \quantph{0205039}.  That region and the extra
``rotation'' within it signifies, symbolically at least\footnote{I
said ``symbolically at least'' above because I don't want to commit
myself to the particular representation of quantum mechanics
discussed in \quantph{0205039} as in any sense the key to all
mysteries.  Instead I view it as a kind of scaffold for sharpening
the issue---i.e.\ as a way of giving direct comparison between quantum
mechanics and general Bayesianism.  But like any scaffold it has to
be taken away ultimately \ldots\ and maybe there is a way to shortcut
the process without ever introducing the scaffold in the first
place.}, the extra layer on top of Bayesianism that quantum mechanics
is asking us to contemplate. Within the shape of that region (or how
its volume scales with system size, or some other relevant feature of
it) and the style of the ``rotations'' that we add after pure
Bayesian conditionalization lays hidden that which we are all really
seeking: A precise statement about the (existential) character of the
$h$'s.

I bank on the idea that we will find that they are not correlata or
relata at all, but rather creatia.  (Simply trying to ontologize or
reify the correlations while giving up on the correlata, as you try
with your slogan, won't do.)  The world is in constant birth.  And to
the extent that we focus our actions on anything and ask ourselves,
``What will come of them?''---i.e., make note of quantum
phenomena---we too are involved in that birth.

But I draw the discussion outward, while you want to bring it inward.
I'll trust that {\Ruediger} will do all the hard work of bringing it
inward for the present.  You should know, though, that you yourself
have the whole answer already; you said it completely:  ``So of
course there is no mystery if you don't feel the need for an
explanation of the first feature.''  The best I can do (and the best
I will do, when I follow through with my promise) is fluff that
sentence into a whole paper.  Why give up on quantum measurement
outcomes as the revelations of pre-existent facts?  It is all and
only a case of once bitten twice shy:  the other route has been tried
and tried too many times.

But to spin the result this way,
\bdm
Why should he be content with correlations without correlata?
\edm
is to make the project look so very negative.  ``Be content
with''---I hate that.  It seems to forget that there is a much better
world in return for this piddling little loss.  To give you more
detail, I'll attach the best shot I've had so far to make that
convincing.  (I hear {\Ruediger} chuckling:  ``I've never said it
better than in my note to \ldots''.)  The file is titled
{\tt ForSlusher.pdf}. If for some reason you can't open it---you often seem
to have troubles---it comes from notes to Wiseman on pages 210 and
217 of my samizdat {\sl Quantum States:\ What the Hell Are They?}~on
my webpage.  [See 24-06-02 note ``\myref{Wiseman6}{The World is Under Construction}'' and 27-06-02 note ``\myref{Wiseman8}{Probabilism All the Way Up}'' to H. W. Wiseman.]

Still beaming from {\Ruediger}'s notes.

\section{27-06-03 \ \ {\it Utter Rubbish and Internal Consistency, Part I} \ \ (to R. Schack, C. M. Caves \& N. D. Mermin)} \label{Schack60.1} \label{Caves73.1} \label{Mermin86}

\begin{flushright}
\baselineskip=3pt
\parbox{4.6in}{
\bq
\noindent
What's the good of Mercator's North Poles and Equators,\\
\indent\indent     Tropics, Zones, and Meridian Lines?''\\
So the Bellman would cry: and the crew would reply\\
\indent\indent     ``They are merely conventional signs!
\eq
}
\end{flushright}\medskip

It's time for me to pay the piper.  On November 4 of last year, I had
a mystical experience during my flight home from the {\Montreal}
meeting:  I thought for the first time in my life I had seen with
complete clarity why there was no mystery, either to EPR correlations
or to the Penrose argument calling for the objectivity of the quantum
state.  The whole strange flight from {\Montreal} to Chicago to Newark
(i.e., American Airlines Advantage Platinum Number D7E5856) I typed
away.  To the result, David replied:

\bdm
Thanks for cc'ing me the latest epiphany.  I've only had a chance to
glance at it and am still torn between whether my response is ``that
was how I understood your position to be all week'' or ``what utter
rubbish'' so I will take a little time before responding.
\edm

David's second instinct was the best:  IT WAS UTTER RUBBISH.  The
great discussions I had with {\Ruediger} the last two days a) brought
me back to the subject, which I had shamelessly not thought about
again since writing the original note (a sign that there must have
been some inner skepticism or reluctance on my part, but I didn't
have the intellectual integrity to confront it), and b) showed me the
gross errors of my ways.  This note represents both a heartfelt
thanks to {\Ruediger} and an attempt to tell the Quantum Bayesian tale
in a way that does it better justice.

Let me start out by describing what I was thinking previously, both
to set the tone and because I think it adds a little clarity to the
recent discussions {\Ruediger} and I have been having with David.  What
my own trouble boiled down to previously was a moment of weakness---a
moment when even I lost faith in our Bayesian dream (and sadly did
not realize it as such).

Bell inequalities and EPR correlations, what are they good for?  I
think one thing only:  the conjunction of the two concepts provides
us with the most dramatic evidence yet that quantum systems do not
themselves carry ``instruction sets'' for specifying the outcomes of
``measurements'' we can make on them.  As {\Ruediger} said it two days
ago, never was there a better exposition of this point than in some
of David's writings.  Thus, it is a point David should appreciate and
take to heart.  I know that I did, and it is this that led me astray
in other directions.

The question is, how to live with this result?  It's one thing to say
there is a formal demonstration that there can be no instruction sets
inherent within a system.  But it is another thing to feel good about
it.  In particular, why doesn't it get absolutely under our skins
that there can be situations where quantum mechanics specifies that
we have certainty about the outcome of a measurement?  The million
dollar question.  You see, if there's no instruction set, where could
the certainty possibly come from?  Or again:  If there are no
instruction sets, measurement outcomes must come out of the thin air,
determined by nothing else.  However, if they do truly come out of
thin air, how could we ever be certain which outcome WILL occur
before it actually does?

The world does what it does, but yet I have certainty?  Where did
that come from?!?  It couldn't come from the world (as demonstrated
through the force of Mr.\ {\Mermin}'s writings); it must come from me.
Or---to be absolutely clear---that was the path of thought I found
myself traveling last November.

IF THE CERTAINTY CAN'T COME FROM THE WORLD, IT MUST COME FROM ME. The
interpretation we ascribe our clicks must be a convention.  (As
{\Ruediger} pointed out, it all starts to sound too much like Umberto
Eco, and I agree now, but I was blinded at the time.)  How could that
be?

The mystical insight---the false prophet!---was that the trail had
just been blazed!  It all had to do intimately with our newfound
realization that at least some quantum operations must have the same
subjective status as quantum states.  (I put the ``at least some''
solely so that {\Carl} will not stop reading at this point; if it were a
letter to {\Ruediger} alone, I would leave out the qualification
completely.)  A quantum operation is a density operator in disguise.
That was somehow to be the key.

In rough terms, I wanted to say that {\em ascribing\/} a pure state
$P$ to a system and {\em ascribing\/} a description $\{P,I-P\}$ to an
associated measurement device was just the convention I needed:  No
matter what the outcome of the ``true'' physical intervention
(measurement), the result would be interpreted as $P$.  But what
utter rubbish!  And it's hard to see now what I could have been
thinking, or even how I could have found it pleasing.  It goes
against everything I have ever viewed as the great insight of the
Paulian idea:

\bq
     Like an ultimate fact without any cause, the individual outcome
     of a measurement is, however, in general not comprehended by laws.
     This must necessarily be the case \ldots\

     In the new pattern of thought we do not assume any longer the
     detached observer, occurring in the idealizations of this classical
     type of theory, but an observer who by his indeterminable effects
     creates a new situation, theoretically described as a new state of
     the observed system.  In this way every observation is a singling
     out of a particular factual result, here and now, from the
     theoretical possibilities, thereby making obvious the discontinuous
     aspect of the physical phenomena.

     Nevertheless, there remains still in the new kind of theory an
     objective reality, inasmuch as these theories deny any possibility
     for the observer to influence the results of a measurement, once
     the experimental arrangement is chosen.
\eq

The last part of the idea is absolutely crucial.  The observer must
not be able to influence the results of a measurement, even by hiding
it in a convention.  I had lost my grip on reality during that
flight, and something deep inside me must have sensed it.

Let me try to make this absolutely stark by running down the
{\Ruediger}ian path.  Consider two agents, a quantum system, and a
quantum measuring device.  Suppose both agents agree that the
relevant feature of the measuring device is that it has two possible
clicks, and furthermore suppose they both agree that the clicks are
to be interpreted as a measurement of the POVM $\{P,I-P\}$.  In our
language, these are both subjective ascriptions (more on the word
subjective later), but they just so happen to agree.  However suppose
in contrast to this single belief about the measuring device, they
are in wild disagreement about what they believe of the quantum
system soon to be intruded upon:  One says the quantum system's
quantum state is $P$, the other says it is $I-P$.  From our
standpoint, there is in principle nothing wrong with this.  A quantum
state is a subjective ascription (more on the word subjective later),
period. It is not determined by the world external to the agent; it
is his personally.  So what happens to the measurement device when
the quantum system is dropped into it?  There are two agents, two
absolutely incompatible beliefs.  Does the measurement device
explode?

No.  A click occurs, and one agent is {\em made\/} wrong.  If that
agent had bet his life on his utterly extreme belief, he will now be
dead. Darwinian evolution will have stepped in to see that his
extreme belief is not propagated.  That is no convention.  The world
will smite one of the agents.

Now, let us join the {\Mermin}ian discussion presently in progress.  Let
us conceptually erase one of the agents.  What can possibly change
for the other?  Nothing.  Once again a firm belief exists and once
again there is nothing to keep the world from smiting the believer.

OK, take a restroom break, get some popcorn, and I'll be back with
the second half of this note after lunch.

\section{28-06-03 \ \ {\it Utter Rubbish and Internal Consistency, Part II} \ \ (to R. Schack, C. M. Caves \& N. D. Mermin)} \label{Schack60.2} \label{Caves73.2} \label{Mermin87}

It was a long lunch.  Let me pick up where I left off:

\bq
     A click occurs, and one agent is {\em made\/} wrong.  If that agent had
     bet his life on his utterly extreme belief, he will now be dead.
     Darwinian evolution will have stepped in to see that his extreme
     belief is not propagated.  That is no convention.  The world will
     smite one of the agents.

     Now, let us join the {\Mermin}ian discussion presently in progress.
     Let us conceptually erase one of the agents.  What can possibly
     change for the other?  Nothing.  Once again a firm belief exists
     and once again there is nothing to keep the world from smiting the
     believer.
\eq

The ultimate issue is, is there really any difficulty with the idea
of an utterly certain belief about an admittedly contingent fact?
And, I'll add for later discussion, is that something uniquely
quantum mechanical?

I'll tackle the first question by repeating something {\Ruediger} said
yesterday, but in a windier way:  I'll paste in two sections from de
Finetti's article Probabilismo.  (That, by the way, was the reason
for my long lunch:  I ended up reading Probabilismo from beginning to
end again---I think my fourth time since 1996.  If I would have only
absorbed the darned thing the first time!  Each time I read it, I am
struck that it is the best thing on probability I have ever read. Yet
each time, I miss something really important.  In fact I end up
feeling bad about myself, for it shows me just what an amateur I am.
The article's {\em got\/} to be made standard reading!)

\bq
\begin{center}
18.\medskip
\end{center}

``A gambler wants to make a bet; he asks my advice.  If I gave it to
him I would rely on the probability calculus, but I could not
guarantee success. That is what I would call {\em subjective
probability}.  But I suppose that an observer is there, who notes the
outcomes over a long period; when he reviews the record he will see
that the outcomes fall out in conformity with the probability
calculus. That is what I would call objective probability, and it is
this phenomenon that we must explicate.''  [A quote from {\Poincare}.]

That is a difficulty that leads many into error:  how can one not be
persuaded---one would ask---that the value of probability is not
simply subjective?

In all these cases, in all similar arguments, what is impressive is
only one fact:  that a practically certain event actually comes
about, or it is foreseeable that it will come about.  But should we
be impressed?  When I say that an event is practically certain I mean
that I should be amazed if it didn't occur:  am I then entitled to be
amazed at having guessed, to be amazed that an extraordinary fact
whose occurrence would have amazed me did not in fact occur?

{\Poincare} says that those who are present at the game ``will see
that the outcomes fall out in accordance with the probability
calculus''.  First of all they would be able to see that {\em some
remarkable and practically certain circumstances\/} occurred,
relative, e.g., to the frequencies, while it is impossible that {\em
all\/} the practically certain facts have occurred.  It suffices to
think that it was practically impossible that the particular sequence
of outcomes that has taken place would have taken place.  Then we
must limit attention to just one or a few remarkable and practically
certain circumstances.  {\Poincare} says that they will happen.  But
why does he say it?  Because he is certain of it, not absolutely, but
practically.  And didn't we already have to assume that he was
practically certain of it?  When I evaluate a probability as very
close to 1, I express this sensation:  that, almost without doubt,
the event will occur.  Do I add anything more when I repeat that,
almost without doubt, it will occur?  Do I have the right to think:
first I evaluate a probability, and then I ask myself if I can
actually anticipate the event with the corresponding state of mind?
This is what many do, and, when they can answer affirmatively, they
say that probability has an objective value.

But, when I evaluate a probability, I only express my state of mind,
and what does it mean to ask whether I can or cannot have a state of
mind which corresponds with the state of mind which is actually mine?
If such a doubt corresponds to something which is not meaningless and
is actually mine, it was already a part of my state of mind, and I
will already have used it in my evaluation of the probability.  But
once I have evaluated it (and as long as I suppose that my state of
mind will not change:  if it changes, then certainly I can modify my
earlier evaluation!) it is meaningless to think that my evaluation is
wrong, because it is meaningless apart from me, it has no other
function than to express my state of mind.

Why, when an event appears to me as practically certain (i.e., when I
evaluate its probability as close to 1) have I the right to be
practically certain that it will occur?  Because when I say that an
event is practically certain (when I evaluate its probability as
close to 1) I do not say nor can I want to say more or less than
this:  that I am practically certain it will occur.

If it is true that ``{\em opium facit dormire}'', can we think it
true that ``{\em opium habet virtutem dormitivam}''?  This is no less
difficult and no less deep a philosophical problem!  I leave it to
the reader's acumen to see whether the comparison is apt.\medskip

\begin{center}
19.\medskip
\end{center}

It seems strange that from a subjective concept there follow rules of
action that fit practice.  And {\Poincare} keeps explaining why the
subjective explanation seems insufficient to him, mentioning
practical applications in the field of insurance.  ``There are many
insurance companies that apply the rules of the probability calculus,
and they distribute to their shareholders dividends whose objective
reality is incontestable.''

Basically, this is only the preceding case, simplified by the fact
that here it is very clear what the ``remarkable circumstance'' is
that one must consider, and it has a very concrete importance:  the
dividend.  We make a budget in such a way that it is practically
certain that the gains will be such-and-such.  Naturally, it is
meaningless to say ``practically certain'' if I don't say {\em for
whom\/} they are so; in this case it will be the managers, the
actuaries, the shareholders.  When an enterprise is sound and has
little risk, it is easy to reach a universal or almost universal
consensus on this opinion, and there is nothing to be surprised
about, since it is exactly because of this that the enterprise is
said to he sound and have little risk.

But this is not sufficient: it is not just a feeling of the managers,
actuaries, and shareholders, someone will say.  You will see that the
dividend will prove them right.  What does this mean?  I mean that
this someone shares the feeling, the persuasion, the faith, which the
managers, the actuaries, and the shareholders already have.  At the
end of the year the dividend is regularly distributed.  See that, one
will say, that certainty was not just my feeling, it must have had an
objective ground.  But why?  If he---even on the basis of a totally
groundless conviction---thought it unlikely that there would be no
dividend, he would have to find it very natural that there is the
dividend, and it would be pointless, unnecessary, and useless to look
for an explanation.  Least of all for a purely verbal and abstract
explanation, like the one that consists of inventing ``chance'' and
``the laws of chance''.

But let us look into the function of the probability calculus in the
field of insurance.

Whatever enterprise one wants to undertake, whatever firm one wants
to manage, one always proceeds by consciously or unconsciously making
a budget, in which we equalize the hope of profits and the fear of
losses, the hope and the fear that the profits and the losses will be
more or less great.  We can love risk more or less, we can be prudent
or speculative, and our preferences will be different.  We could be
guided by the hope of a risky gain and risk everything, or we might
prefer the modest tranquility of those who feel safe from the tricks
of fortune.  We are perfectly free with regard to this choice;
everyone can do as he wishes.  The probability calculus cannot say we
are right or wrong:  it is true, the concept of mathematical
expectation is known, and it is very important, but its task is not
(as some seem to think) that of constraining our freedom of choice in
this case.  The notion of moral expectation has also been introduced,
which, besides not solving the problem, is also an artificial and
unimportant notion.  In any case, we must consider all the
alternatives together with their probabilities and their
consequences, and then act as we see fit.

In the case of an enterprise that must be secure and have little risk
we must act so that, as in the case of insurance, our profits may not
be fantastic, but they should be sufficient and practically safe.
That's all that non-speculative firms do, without using the
probability calculus, and nevertheless this certainty is not too
often belied by the facts.  And there is nothing strange in this, for
an obvious reason:  if these forecasts were always belied, we would
not make them, and we would act in some other way, and it would be
this other way that would inspire us to have greater or less
confidence in the various alternatives.

That a fact {\em is\/} or {\em is not\/} practically certain is an
{\em opinion}, not a {\em fact}; that {\em I judge it practically
certain\/} is a {\em fact}, not an {\em opinion}.  That I should act
according to this opinion is only apparently a corollary, because
this opinion only exists in that I think I must govern my action in
accordance with it.
\eq

What of any of this loses force when we come to quantum mechanics, or
when we come to {\Mermin}'s 1985 {\sl Physics Today\/} article?  I think
almost nothing, except possibly de Finetti's rhetorical question:
\bq
\noindent
     If it is true that ``{\it opium facit dormire}'', can we think it
     true that ``{\it opium habet virtutem dormitivam}''?  This is no
     less difficult and no less deep a philosophical problem!  I leave
     it to the reader's acumen to see whether the comparison is apt.
\eq
But even that's up to debate, depending upon which part of the
quantum confusion one wants to place alongside the analogy.  Thus, if
you will allow me to drop those three sentences from the passage, I
will boldly declare that nothing whatsoever of de Finetti's point
changes when we come to quantum mechanics.  In particular nothing
changes even when we come to EPR/Bell phenomena.

The issue is just one of mindset.  I think {\Ruediger} said it very
nicely yesterday:

\brs
     David approaches him and asks ``But aren't you dying to find out how
     the whole thing can possibly work?''  Bob is puzzled.  He doesn't
     understand the question.  When he finally thinks he understands, he
     is even more puzzled. ``Why would I be more confident about my
     predictions if the photons carried instruction sets?'', he asks.
     Clearly, Bob and David speak at cross-purposes.

     Apparently, for David the only acceptable way to understand ``how it
     can possibly work'' would be a mechanistic, clock-work type model
     for it.  Bob, Bayesian to the bone, thinks that David meant to ask:
     ``How can you possibly be (almost) certain that tomorrow's outcome
     will be 0?''  And Bob's answer is ``What difference would the
     existence of an instruction set make to my beliefs about tomorrow's
     outcome?''
\ers

So, it is not certainty that is a mystery ever.  Certainty is just
the expression of someone's state of mind, whether the world will
bear it out or not.  Certainty is, at best, about internal
consistency.  But, we would have no right to draw from this that
there is no deep lesson to be learned from EPR/Bell phenomena.  On
the contrary, EPR/Bell is just as important as it ever was---it's
only been the conclusion that has been misplaced.  David drew from it
a mystery that has nagged him (and us) for at least 20 years.
Instead, its lesson should be accepted at face value, regardless of
the feelings of mystery that motivated Bell's initial analysis: There
are no instruction sets.  That's the lesson.  It is telling us
something about (what we believe of) the world.

Wow!  What more could one ask for than a precise, well-formulated
statement of what one actually believes.  Sometimes I have a precise
statement of the implicit beliefs that motivate my actions, but not
very often.  Here's a case where I can actually nail one down.

So, the whole business of understanding quantum mechanics starts to
feel even better.  It is about understanding one's priors!  (I wish I
could boom that out in a Ben Schumacheresque fashion!  I hope you can
hear it.)  In accepting quantum mechanics, one is making an implicit
statement about one's priors.  It is the structure of those priors
and not their precise values that is telling us something about what
we believe of the world independent of our particular experiences.

In summary, let me just say this.  Letting go of the mystery in
EPR/Bell is not defeatism.  Instead, through it we accumulate a fact
for our Bayesian understanding of quantum mechanics, i.e., as a
statement about the structure of our priors.  There's far more work
to be done in that regard.  And if one wants metaphysics (or at least
a post-positivism, i.e., something more than raw experience), that's
where it is to be found.

Thus, let me end Part II.  There'll still be a Part III, but now I've
got to go to dinner.

\section{30-06-03 \ \ {\it Probabilistic Dialogue} \ \ (to R. W. {\Spekkens})} \label{Spekkens16}

\brws
I found myself thinking about probability theory last night.  I
realized I need to really go deeply into these issues. If radical
Bayesianism is the answer, then presumably the only thing preventing
me from adopting this view is my ignorance of the problems with the
alternative.

So, what should I read to learn about the arguments for and against
both the logical and subjective interpretations of probability?  If
you could tell me the classic texts for both camps, as well as any
good articles providing a modern perspective on the debate, that
would be ideal.  (I'm not interested in the propensity
interpretation.)
\erws

I'm glad you're taking a deeper interest in the foundational issues
of probability now.  I think, as your own toy model shows to some
extent, getting the idea of probability straight is a large part of
the task of getting quantum mechanics straight.  In any case, it is a
part of the problem that cannot be ignored.

I think the most devastating critique of the classical and/or logical
interpretations is the problem of defining the reference class for
any given event.  What is the set of possibilities that an event {\em
must\/} be considered an element of?  In the real world of
statistical practice and decision, there is never a unique answer.
Of course, in fundamental physics, one might think that one is in a
holier position:  One can always declare a reference class by fiat.
But then the onus is on the declarer to show why that reference class
and a uniform (or even any nonuniform but otherwise fixed)
probability measure over it is actually relevant to statistical
practice.

I was hoping to pin down some really good quotes over the weekend to
accompany this note.  You see, for instance, Richard Jeffrey, the
inventor of the idea of ``radical probabilism'' and a wholehearted
supporter of de Finettian ideas, started off in the school of Carnap,
where the hope was that logical probability would someday be the new
messiah.  As I understand it, it was years of watching the project
fail that ultimately led Jeffrey in the other direction.  But I
couldn't find anything quite forceful enough for my tastes.

In any case, here are the things I think you ought to read (in the
following order):
\begin{enumerate}
\item
Leonard J. Savage, {\sl The Foundations of Statistics}.  Read all of
Chapters 1 and 4, especially Section 4.5.

\item
Richard Jeffrey, {\sl Subjective Probability (The Real Thing)}. Read
pages 8 to 11.  The book can be found at: {\tt
www.princeton.edu/$\sim$bayesway/}

\item
Jos\'e M. Bernardo and Adrian F. M. Smith, {\em Bayesian Theory}.
Read all of Chapter 1 and pages 99 to 102.  And finally, the very
best thing that anyone can read is

\item
Bruno de Finetti, ``Probabilism,'' Erkenntnis {\bf 31}, 169--223
(1989). (I have never gotten so much out of any of my reading on
probability as I have from this article.  But I should warn you that I can already guess you will find Sections 1, 2, and 3 of it
unpalatable, at least on first reading.  Therefore I hope you will be a little lenient on the man until you get past them.  After that (and excepting the very last section on fascism) I think there is so much that is absolutely firm in the article, I don't see how anyone could disagree with it.)
\end{enumerate}

Let me also add three attachments to the present note that I think
are directly relevant to you.  One is a little thing Joe Renes wrote
recently.  Despite the free-formness of the note, I think it hits the
points related to your present query quite well.  The other two are
notes that I wrote recently:  They have to do with the ``mystery''
(or lack thereof) of EPR from the Bayesian standpoint, and lean
heavily on the concept of ``certainty'' in the Bayesian sense.  I do
hope that they'll somehow make something click in you.  The best
answers are going to be the most trivial, not the most contrived.

I hope you'll let me know what you think of all these things.  In
particular, it'd be nice if we could start up a dialogue.  Can you
articulate what troubles you most about the de Finettian version of
probability?  What looks less than scientific about it to you?

I suspect it all boils down to the fact that you feel that
Bayesianism is somehow too arbitrary:  ``For God's sake, they won't
declare {\em any\/} probability assignments right or wrong!  Science
is not just `anything goes.''' \ldots\ but I don't want to put words
into your mouth.  If however that is roughly the case, I hope you'll
think hard about a point {\Marcus} {\Appleby} likes to make:  ``It's really
hard to believe something you don't actually believe.''  I.e.,
Bayesians, from my point of view, think anything but `anything goes.'
A probability assignment, when it is made, is one's best shot at
articulating what one believes; and what one believes is not up to
one's whim.

I would have thought that the de Finetti version of Bayesianism
should have been quite attractive from the point of view of the cut
you want to introduce into physics:  the ontic and the epistemic. Try
your best to say what the world does or is; that's what the ontic is
about.  Fine.  I'm OK with that.  But then why try to make our
fallible beliefs rigidly connected to the ontic world?  If you're
going to try to make probability assignments adhere to the concepts
of right and wrong (just like those ontic states of yours), why not
make them part of the ontic world too?  Another way to put it: What's
so blasted wrong about allowing a perfectly rational agent to be
`wrong' about the actual state of the world?  You want a) that there
is a world (with its one true state), and b) that any agent worthy of
being called rational or ideal, must by right about it to within some
tolerance.  But why?

Or maybe it is just the fear of having an agent in the background as
the anchor for a probability assignment that bothers you.  Didn't
Copernicus teach us that our place in the universe is not the center
of the universe?  Aha!  Maybe that's it.  So, if we could just find a
sound notion of objective probability, we could imagine the calculus
of quantum mechanics hanging around even when there's no one about in
the quad to make use of it.  Is that the deeper of the issues for
you?

OK, enough blabbering.  I'll shut up for a little while until you
give me some guidance.  In the meantime I'll leave you with some of
my favorite Bernardo and Smith quotes.  They're pasted below.  [See
07-08-01 note ``\myref{Mermin28}{Knowledge, Only Knowledge}'' to T. A. Brun, J. Finkelstein and N. D. Mermin.]

\section{30-06-03 \ \ {\it Fear of the Anchorman} \ \ (to N. D. {\Mermin} \& R. {\Schack})} \label{Mermin88} \label{Schack61}

Did you notice the ending lines I put in the note to {\Spekkens}?

\bq
I suspect it all boils down to the fact that you feel that
Bayesianism is somehow too arbitrary:  ``For God's sake, they won't
declare {\em any\/} probability assignments right or wrong!  Science
is not just `anything goes.'\,'' \ldots\ but I don't want to put
words into your mouth.  If however that is roughly the case, I hope
you'll think hard about a point {\Marcus} {\Appleby} likes to make:  ``It's
really hard to believe something you don't actually believe.''  I.e.,
Bayesians, from my point of view, think anything but `anything goes.'
A probability assignment, when it is made, is one's best shot at
articulating what one believes; and what one believes is not up to
one's whim.

I would have thought that the de Finetti version of Bayesianism
should have been quite attractive from the point of view of the cut
you want to introduce into physics:  the ontic and the epistemic.
Try your best to say what the world does or is; that's what the ontic
is about. Fine.  I'm OK with that.  But then why try to make our
fallible beliefs rigidly connected to the ontic world?  If you're
going to try to make probability assignments adhere to the concepts
of right and wrong (just like those ontic states of yours), why not
make them part of the ontic world too?  Another way to put it:
What's so blasted wrong about allowing a perfectly rational agent to
be `wrong' about the actual state of the world?  You want a) that
there is a world (with its one true state), and b) that any agent
worthy of being called rational or ideal, must by right about it to
within some tolerance.  But why?

Or maybe it is just the fear of having an agent in the background as
the anchor for a probability assignment that bothers you.  Didn't
Copernicus teach us that our place in the universe is not the center
of the universe?  Aha!  Maybe that's it.  So, if we could just find a
sound notion of objective probability, we could imagine the calculus
of quantum mechanics hanging around even when there's no one about in
the quad to make use of it.  Is that the deeper of the issues for
you?
\eq

Now that I've written that, I wonder how much of a problem exactly
the latter might be for the larger community.  Take, for instance,
David {\Mermin}'s Desideratum 1 in his original Ithaca interpretation
paper:  ``The theory should describe an objective reality independent
of observers and their knowledge.''  How would he ever fulfill
Desideratum 1 if probability must be interpreted in the Bayesian way?
Maybe it's that that worries people so much about the Bayesian creed
when it comes to quantum mechanics \ldots\ even (and particularly)
Bayesians!

\section{01-07-03 \ \ {\it Your Papers} \ \ (to N. Hadjisavvas)} \label{Hadjisavvas1}

I have been meaning to write you for some time to thank you for sending me the collection of papers that you did:
\begin{itemize}
\item
N. Hadjisavvas, ``\'Etude de Certaines Consequences d'une Interpr\'etation Subjective de la Notion d'Etat,'' Ann.\ Fond.\ Louis de Broglie {\bf 3}(3), 155--175 (1978).
\item
N. Hadjisavvas, ``Distance between States and Statistical Inference in Quantum Theory,'' Ann.\ Inst.\ Henri Poincar\'e {\bf 35}(4), 287--309 (1981).
\item
N. Hadjisavvas, ``The Maximum Entropy Principle as a Consequence of the Principle of Laplace,'' J. Stat.\ Phys.\ {\bf 26}(4), 807--815 (1981).
\item
N. Hadjisavvas, ``Properties of Mixtures of Non-Orthogonal States,'' Lett.\ Math.\ Phys.\ {\bf 5}, 327--332 (1981).
\end{itemize}
Particularly, I was very pleased to learn that you had toyed with Bayesian ideas within quantum mechanics already in your Master's thesis, so many years ago.  I will study all of your papers in detail and cite them duly in my upcoming work.

I wish I knew of which of my papers you found some interest in.  In case you haven't found it previously, you can find a body of my writings on quantum foundational things on my personal website (link below).  Also, Caves, Schack, and I have an accumulating list of Bayesian-type quantum results posted on the quant-ph archive at \myurl{www.arXiv.org}.

\subsection{Nicolas's Preply}

\bq
I recently saw (by chance, given that my research interests changed completely since almost 20 years) a couple of your papers on quantum information.  Please find enclosed some very old papers of mine on related subjects.  Actually, the paper in French was my Master's (DEA) thesis.  In it, the idea of a state describing the knowledge of the observer, rather than the system itself, is exploited for (still another) derivation of the projection postulate.
\eq

\section{01-07-03 \ \ {\it Rob {\Spekkens}} \ \ (to R. D. Gill, K. M{\o}lmer, and E. S. Polzik)} \label{Gill2} \label{Polzik1} \label{Moelmer2}

I just wanted to alert you guys to Rob {\Spekkens}, who has just applied to give a talk at the MaPhySto/QUANTOP meeting in August.  I hope his talk will be accepted.  I think he has truly one of the most interesting quantum foundational / quantum informational results I've seen in quite a while.  What he does is explore a toy model that is admittedly not quantum mechanics, but is a certain local hidden-variable theory.  What's interesting about the model in spite of this nonsense is that it incorporates a principle that Carl Caves and I were once hoping to found quantum mechanics itself upon:  Namely, that in quantum mechanics maximal information is not complete information.  The truly surprising result about {\Spekkens}'s toy model is just how much of standard quantum-information looking stuff his model contains:
\begin{itemize}
\item[a)] a no cloning theorem
\item[b)] superdense coding (without entanglement)
\item[c)] a no broadcasting theorem
\item[d)] a no-go theorem for bit commitment
\item[e)] an information-disturbance principle that can be used for secure key distribution
\item[f)] something that looks like entanglement monogamy (without entanglement)
\end{itemize}
and the list goes far beyond this.  The point is, one does not (and cannot) recover all of quantum mechanics from his toy model, but one can get a hell of a lot of it even in a hidden variable model (with the principle that maximal information cannot be made complete).

I think we stand to learn a whole lot about real quantum mechanics from this exercise.  And it is just a marvelous construction.  I hope there is still room in the schedule for young {\Spekkens}.

\section{01-07-03 \ \ {\it Gasp or Shudder?}\ \ \ (to J. W. Nicholson)} \label{Nicholson18}

Actually was it gasp or shudder?  My memory is getting fuzzy on the whole affair.  I know that I need to write this story down before it is completely gone.  Could I interest you in sending me some notes on what you remember?  Of course, when I write it up, I'll probably embellish things a little to try build atmosphere, but I would like to be decently accurate to what actually took place (and how things were said).

\section{01-07-03 \ \ {\it Samizdat} \ \ (to A. Fine)} \label{Fine1}

It was good meeting you at the Clifton memorial meeting a few weeks ago; I've long been a fan of your writings.  I'm writing to tell you, though, that my quantum samizdat {\sl Notes on a Paulian Idea\/} has just been published by {\Vaxjo} University Press, and I have a number of copies to give away.  (If you don't know what I'm talking about, you can peruse the file on my webpage (link below).)  If you think you might get something out of it, I'll have the publisher send you a copy.  All I need is your mailing address.

\section{01-07-03 \ \ {\it My Samizdat} \ \ (to W. E. Lawrence)} \label{Lawrence2}

I've been meaning to write you for a while.  Would you be interested in a copy of my quantum samizdat {\sl Notes on a Paulian Idea}?  The book has just been published by {\Vaxjo} University Press, and I have a number of copies to give away.  If you don't know what I'm talking about, you can peruse the file in PDF format on my webpage, link below.  (Mermin wrote a foreword to the book.)

If you think you might get something out of it, please let me know and I'll have the publisher send you a copy.  All I need is your mailing address.

\section{01-07-03 \ \ {\it Objective Chance} \ \ (to W. C. Myrvold)} \label{Myrvold1}

\bwm
It's several projects down the queue, and we may not get to it this
summer, but Bill Harper and I plan to write an article on why
Bayesians should believe in objective chance.  I'll send you a draft
when one exists.
\ewm

I look forward to it.  It's about the biggest thing on my mind at the
moment.  Particularly, I'm fairly of the opinion that trying to force
objective chance (rather than objective indeterminism) onto the world
is going to be counterproductive for a good understanding of quantum
mechanics.  But I'll give you a chance to convince me!  (I made the
distinction above, by the way, because I rather like indeterminism of
a sort, but I would be reluctant to try to ascribe a numerical
measure to it.)

In the mean time, anyway, can you give me some pointers to any
literature that tries to argue the same point as you'd like.  I'll
bank on your paper with Harper being a better version of it all, but
still I'd like to see some its predecessors.

\section{01-07-03 \ \ {\it Gasp or Shudder?}\ \ \ (to Friend X)}

Actually was it gasp or shudder?  My memory is getting fuzzy on the
whole affair.  I know that I need to write this story down before it is completely gone.  Could I interest you in sending me some notes on
what you remember?

\section{01-07-03 \ \ {\it Making the World Gasp Your Name}\ \ \ (to R. W. {\Spekkens})} \label{Spekkens17}

\brws
I had not heard of the Aarhus workshop until now.  I am thinking
about asking the organizers whether it is too late to attend and whether
there are any free slots left. Do you think it will be a good workshop for someone, like myself, who
finds himself currently absorbed with foundational rather than
practical issues? [\ldots]
\erws

I had a feeling it was something like that.  Have a look at this article before it disappears from the NY Times:
\bq
\myurl{http://nytimes.com/2003/07/01/international/europe/01DENM.html}
\eq
``Free Spirits in Their Fortress, the Law at the Gate.''

\section{01-07-03 \ \ {\it Samizdat and Potentia} \ \ (to A. Shimony)} \label{Shimony2}

\bas
About two weeks ago I sent you a note thanking you for your book, but
I fear that the address was wrong. If you have already received my
note of thanks, no harm is done. I certainly appreciated receiving
your remarkable book. You gave good excerpts from emails sent to you,
and these together with your replies constituted interesting
dialogues. As you know, there are serious differences of opinion
between us on the ontology of the wave function, but your arguments
always seem to me thoughtful. I shall consult your book often and
find stimulation in it.
\eas

Thank you for the kind words.

I also reread the lovely letter you sent me on May 27 of last year. I
know that there are differences of opinion between us, but I hope
that in the end they will not be so serious.  In particular it would
be nice if physical theory itself would lead the way to minimizing
those differences:  I think there is a chance, with a suitable
rewriting of quantum mechanics into more Bayesian-like terms.  At the
very least it'll be a new way to see things, and that may sharpen our
real (rather than apparent) points of contention.

You write, ``[I]n my opinion the greatest philosophical innovation of
QM is the discovery and exploration of a new modality of reality --
something in between actuality and logical possibility -- which
Heisenberg named `potentiality' in his book {\sl Physics and
Philosophy}.''  But I would like to think that I am quite attracted
to the same idea.  Throughout {\sl Notes on a Paulian Idea\/} you'll
find image after image of the quantum measurement process as a
creative process, something that brings about a transition from the
possible to the actual.  Indeed it is captured by the Paulian idea
itself (the conjunction of the two quotes at the beginning).  Here's
a slightly longer version of the same:
\bq
\noindent
     In the new pattern of thought we do not assume any longer the
     detached observer, occurring in the idealizations of this classical
     type of theory, but an observer who by his indeterminable effects
     creates a new situation, theoretically described as a new state of
     the observed system.  In this way every observation is a singling
     out of a particular factual result, here and now, from the
     theoretical possibilities, thereby making obvious the discontinuous
     aspect of the physical phenomena.
\eq
\bq
\noindent
     Nevertheless, there remains still in the new kind of theory an
     objective reality, inasmuch as these theories deny any possibility
     for the observer to influence the results of a measurement, once
     the experimental arrangement is chosen.
\eq
\bq
\noindent
     Like an ultimate fact without any cause, the individual outcome
     of a measurement is, however, in general not comprehended by laws.
     This must necessarily be the case \ldots\
\eq

Where we part company, I think, is only in a) my resistance to
summing up this new category with a numerical measure (i.e.,
objective probability or objective chance), and b) my desire to make
it clear that the quantum formalism is a calculus for manipulating
agent-centered probabilities.  Sometimes I put the latter point this
way:  Can a dog collapse a wave function?  Dogs don't use wave
functions.  I myself didn't collapse a wave function until I was 23.
But that doesn't mean that the quantum world will disappear without
the agent!  It only means that (judgmental, personalistic,
subjective) quantum probabilities disappear without the agent.

Here's the way I put it (again) to David {\Mermin} the other day:
\bq
\noindent
I bank on the idea that we will find that they are not correlata or
relata at all, but rather creatia.  (Simply trying to ontologize or
reify the correlations while giving up on the correlata, as you try
with your slogan, won't do.)  The world is in constant birth.  And to
the extent that we focus our actions on anything and ask ourselves,
``What will come of them?''---i.e., make note of quantum
phenomena---we too are involved in that birth.
\eq

Thus, the transition from possible to actual I am toying with is of a
sort of William {\James}ian flavor (now that I know a little bit about
{\James}'s philosophy).

Does that leave us with an even more gaping trench between us?  Or is
there some room for finding a common ground?

\subsection{Abner's Reply}

\bq
There are some bridges between your Weltanschauung and mine, and
there are also some chasms. Among the first are these:

(1) I too am a Bayesian, but in my view of scientific methodology.
There are, however, many different versions of Bayesianism -- logical
probability theory, subjective probability theory, personalist
probability theory (similar to subjectivism except for the constraint
of consistency), and communalist probability theory. Mine is none of
these. The probability theorist whom I most admire is Harold
Jeffrey[s], who announces himself as a logical probabilist but
modifies that position with what he calls ``the simplicity
postulate'', which is really a strategy. Maybe I should call my
rewriting of Jeffreys ``strategic Bayesianism.'' The position is
presented at excessive length in ``Scientific inference'', reprinted
in vol.\ 1 of my {\sl Search for a Naturalistic World View}. An
improved and shorter version is in the paper that follows it, called
``Reconsiderations on Inductive inference.''

(2) I maintain that Bayesian inductive methodology can be employed at
the level of generality of metaphysics. It seems to me that {\Pauli}
thinks the same, but the metaphysical principles which he derives
don't convince me, because of his emphasis on the subject -- a
residue of Kant. My best expositions of my theses are in the same
volume, one called ``Search for a world view that can accommodate our
knowledge of microphysics,'' and the other ``Reality, causality, and
closing the circle.''

Another bridge is our common admiration of the pragmatic tradition in
American philosophy. You love William {\James}, and I have respect for
{\James} but reverence for Charles S. {\Peirce}. If you look at their names
in the indices of both volumes of ``Search\ldots'' you will see my
reasons for these attitudes. Something that gave me great
satisfaction when I was Wigner's student was pointing out to him an
affinity between some of his philosophical papers and those of
{\Peirce}, whose name he had never heard of. He then read some {\Peirce}
and was excited:  ``this man has brains and imagination!''. And later
he cited {\Peirce} several times. It could happen to you, too. I am not
a missionary, but I like to share my enthusiasms with my friends
[Maybe that's another way of describing a missionary!].

(3) Although Bayesian probability theory does not, in my opinion,
suffice to characterize the probabilities implicit in quantum states,
it can be a valuable adjunct. I am now in the middle of writing a
Bayesian treatment of the problem of probability in ensembles that
are both pre- and post- selected -- a problem opened in a famous
paper by Aharonov, Bergmann, and Lebowitz, Phys.\ Rev.\ {\bf 134B},
1410 (1964), anthologized by {\Wheeler} and Zurek, and treated in many
papers by Aharonov and his school. There are, in my opinion,
systematic errors in their work, easily resolved by carefully using
Bayes's theorem. I'd like to present this work if we have the
conference on Bayes and QM that we talked about, but in any case I
shall send you the paper when it is done. I've written on the topic
before, but am dissatisfied with earlier expositions.

This is a beginning. I hope there will be sequels.

Did I tell you how impressed I was by the beautiful photo of you with
your daughter? No philosophy can pretend to depth without the message
of that picture!
\eq

\section{01-07-03 \ \ {\it Endorsing Probabilismo} \ \ (to N. D. {\Mermin})} \label{Mermin89}

I do hope by the way, when you come up for air from the wedding (is
it one of your children's?) you'll take a shot at reading de
Finetti's article ``Probabilism.''  Here are the coordinates:
\\
\indent Bruno de Finetti, ``Probabilism,'' Erkenntnis {\bf 31},
169--223 (1989).

The article is immediately followed by one titled ``Reading
Probabilismo'' by Richard Jeffrey.  It'd be a good idea to take a
look at that one too.

I thought I would send you two things to help prod you into taking
this homework assignment seriously.  One is an endorsement of the
paper by {\Ruediger} that I noticed he hadn't cc'd to you.  The other
is a quote from within the paper that sounds ever so much like
`correlation without correlata'.  (I know that nothing else so whets
your appetite.)

First the quote:
\bq
Concerning Aliotta, I think it necessary to report the following
passage, to avoid what might be an easy misunderstanding.

``It is necessary to distinguish relativism from relativism.  There
is one of its forms (the one commonly pointed to when relativism is
accused [of skepticism]) that relegates our knowledge to the realm of
relativity, opposing to it an absolute reality that will always elude
knowledge.  In this form relativism has a skeptical and agnostic
flavor and often goes together with mysticism.  In the blinding light
of the absolute our relative world devaluates, degenerating into a
vain apparent shadow.  We are the dream, the absolute is reality. And
life becomes the painful chase of those shadows, vainly trying to
become light.

``But there is another form of relativism (and this is mine), in
which what is relative is itself the reality and leaves nothing
outside itself.  What we know is not the shadow, but the light, not a
copy, but the true and concrete original'' ({\sl Relativismo e
Idealismo}, Naples, 1922, p.\ 92).

This is exactly my opinion, and I wish to note, for more complete
rigor, that the sentence ``what is relative {\em leaves nothing
outside itself}'' must not be understood as saying that the sentence
``there exists something outside what is relative'' is FALSE, but
that it is meaningless, so that it is impossible even to pose the
question as to its truth and falsity.  This is, after all, the
interpretation that conforms to Aliotta's thought, as appears clearly
further along in the text, where ``{\em the being in itself and
outside any relation of things}'' is seen as ``one of the many verbal
statements to which there correspond no ideas, and which have become
true and proper puzzles of philosophy'' (ibid.).
\eq

\section{02-07-03 \ \ {\it Photographs and Memories} \ \ (to A. Shimony)} \label{Shimony3}

Thank you again.  I will read your note many times over (as is my
habit), and I do hope this is only the beginning.

\bas
Did I tell you how impressed I was by the beautiful photo of you with
your daughter?  No philosophy can pretend to depth without the
message of that picture!
\eas

You flatter me \ldots\ but you might also enjoy two letters in the
Charlie {\Bennett} chapter, titled ``Emma Jane.''  They start on pages
52 and 53. The second of the two---[ending with the line ``giving her dad more reason to suspect that the world is so much more
than a mechanical contraption clinking along'']---was certainly meant to be a
philosophical statement.

\subsection{Abner's Reply}

\bq
Before answering you I looked at pp.\ 52--53, as you suggested. The
list of eighteen accomplishments at age six months
(5 June 1999) is astonishing. I have been a Chomskyan since I first met
him forty years ago and listened to the evidence that he had compiled
for infantile intelligence, but Emma Jane exhibits muscular and emotional
intelligence which Chomsky hardly notices.

She is programmed to elicit the responses that you are programmed to
exhibit. Isn't that a wonderful gift. And yes, the last accomplishment
in the list is a philosophical lesson of the first order.

You and your wife are fortunate to have such a daughter, and she
is fortunate to have parents who properly appreciate her.
\eq

\section{03-07-03 \ \ {\it The Dangers of Analogy} \ \ (to R. {\Schack})} \label{Schack62}

\brs
I just discovered that von Mises was right after all!  Proof by
analogy:
\begin{verse}
Einstein: Gravitational mass and inertial mass do not just have the
same numerical value---they are the same thing.
\\
Von Mises: Probability and limiting frequency do not just have the
same numerical value---they are the same thing.
\end{verse}
\ers

True enough.  But here's one I'm letting guide me ever more often:

\begin{center}
Hilbert space dimension \quad  $\longleftrightarrow$ \quad
gravitational mass
\end{center}

(If the Bekenstein bound is right \ldots\ or at least contains a
grain of truth \ldots\ perhaps it should be mass $\times$ area.)

\section{04-07-03 \ \ {\it Solid Ground, Maybe?}\ \ \ (to G. L. Comer)} \label{Comer33}

\bgc
P.S. When we're together, I want to press you somewhat on how
spacetime concepts enter, appear, etc either implicitly or explicitly
in your information theoretic program.
\egc

Yep, this is what the note is about.  Once upon a time I promised to
write you something about the information theoretic roots or NONroots
of the principle of equivalence if I ever had any thoughts on it.  I
think I had one thought.  Let me try to get it onto paper.

I go up and I go down when it comes to speaking the words gravity and
quantum in the same sentence.  At times I find myself thinking that
general relativity and quantum mechanics express two absolutely
incompatible worldviews.  The general relativistic universe is a
``block universe'' in William {\James}'s sense:  It's just there.  One
can talk about foliations and dynamics, etc., WITHIN the 4-manifold,
but in the largest view---the view from nowhere---the world and all
its history is just there.  It is a universe without life (in the
creative sense).  In contrast, the quantum world strikes me as a
malleable world---one that is still in formation, and in particular,
one for which it is impossible to get such a ``view from nowhere''
(as Nagel would call it).

At other times, I find myself feeling more lenient:  Perhaps the two
worldviews are incompatible, but that does not mean we cannot gain
insight about one of the theories from the other.  And when we find
the analogy, maybe it is just at that place where we should start
hammering away those inconsistencies.

Suppose you take any two pieces of the universe that your mind is
willing to call `matter.'  What can {\em you\/} tell me with
assurance, even if no further word is said about their constitutions?
You would tell me that they attract each other, and the force of
attraction (in the Newtonian view) is determined in part by two
numbers, one intrinsic to each of the two pieces of matter---their
masses $m$ and $M$.

Suppose now that I take any piece of the universe that my mind is
willing to label `matter.'  What can {\em I\/} tell you with
assurance, even if no further word is said about its constitution?  I
can tell you that with my free will I can write some number of
messages into it. Moreover, I can choose to write them in such a way
that that piece of matter will reveal (with some probability) whether
anyone else has had a look at my stored message.  Both the number of
messages I can write into the matter and my best probability of
catching an eavesdropper is determined by a single number:  The
matter's Hilbert space dimension $D$.

In my paper \quantph{0205039\/} ``Quantum Mechanics as Quantum
Information (and only a little more)'' and in my web samizdat {\sl
Quantum States:\ What the Hell Are They?\/} (on my homepage), I have
argued strenuously that it is ONLY the Hilbert space dimension $D$
that can be taken as a property intrinsic to a quantum system (a
piece of matter). The quantum state, nothing to do with entanglement,
or even anything to do with a Hamiltonian is intrinsic to it:  Just
the single, lonely number --- the dimension.

It strikes me that we have here a phenomenon of the power and scope
of universal gravitation, or maybe the principle of equivalence.

Why did I name this note `solid ground'?  It has to do with something
I tried my best to express the last time I was at the Perimeter
Institute, where it looks to me like there are so many people
flailing about without a clue as to what they should be up to when
they speak of combining general relativity with quantum mechanics.
Lee Smolin asks me how I could possibly imagine that the linear
structure of quantum mechanics will remain when one moves into such a
nonlinear regime as that given by the laws of gravity?  I say I'm not
fazed at all:  Most of what he means when he speaks of quantum
mechanics as an expression of physics, is for me but a law of
thought.  A wave function and its evolution are not properties
intrinsic to the system for which they are about.  Rather, if they
are properties of anything at all, they are properties of their
user's head---for they capture all his judgments about what might
occur if he were to interact with the system of interest.

The quantum foundational task as I see it is to baldly accept that A
LARGE PART of the theory is simply not about a world without
observers:  It is only about our interface with the world.  But there
is a part that remains, and that part must be given a firm
identification.  For only once we know how to do that will we know
how to move forward when it comes to gravity.  Only then will we see
that almost all of the ways that have hitherto been considered for
combining quantum mechanics with general relativity were far too
unconstrained:  I am willing to bet that they all essentially boil
down to sheer speculation.

So I write this note because I am starting to get a sense that there
might be some solid ground in these considerations.  Hilbert space
dimension is the universal factor for quantum systems that mass is
for gravitating systems \ldots\ or something like that.  Could they
be the same thing? \ldots\ Or something like that?

``What nonsense is this!??!?  Any of the simplest `real world' quantum
systems---not these paltry, specialized things you study in quantum
information theory---has an infinite dimensional Hilbert space.  Just
think of the hydrogen atom!''  Maybe.  But it's caused me to wonder if
I should take the Bekenstein entropy bound more seriously than I have
in the past.  The trouble that I had had with it before---after making
my transition to the subjective/Bayesian/Gibbsian school of
entropy---was that a bound on the ignorance one can have just doesn't
seem to make much sense: entropy is not an objective property.  But we
have to be careful in these things: there are levels of subjectivity.
What I mean by this is that though the cardinality $N$ of a sample
space may be subjective in the same sense that a probability
assignment is, once one has set it, one is obliged to declare a
maximum ignorance, $\log N$, with respect to that setting. And that
may be what is really going on the Bekenstein bound.

In other words, maybe in my language, all and only the number $ER/h$
signifies is a Hilbert space dimension.  That is one thread of
thought, wildly speculative and departing from solid ground though it
may be.

But the other thread of thought is in the interpretation of this
previous wild speculation.  If it's even on the right track, where
might it go?  Let me give you an example of very bad language:  ``The
Hilbert space dimension signifies the number of potential states a
system can inhabit.''  The lesson of all my (and Bell's, and
{\Mermin}'s, and many other guys) quantum foundational work is that that
is nonsense.  Hilbert space dimension signifies something else.

When I get poetic about it, I like to say it signifies something
about a system's ``sensitivity to the touch'', and when I'm getting
downright Paulian, I like to say it signifies something about its
potential for taking part in creation.  With my light touches, I can
send it off in ways unknown (to me and to it) and, counter to
intuition, the larger it is, the more I can do that.  Wojciech Zurek
has spent his life making up stories about why big things are
``classical.''  I say he has it all wrong, the bigger the thing, the
more Hilbert space it has, the more quantum it is!  The more
sensitive it is to the touch.  And thus the more ignorance an outside
observer generally has about it (except with respect to a very small
number of features) --- not for any particular reason to do with very
particular properties of Hamiltonians, but simply because it is big.
It is not the system that is classical, but the poor observer's
description of it that is.

So, let me leave you with an image for your flight.  It's about the
fleshpots of creation.  Take the Bekenstein bound with more than a
grain of salt---I'm not sure that I do yet, but it is fun to
play---and take a given region of space.  Ask yourself what you
should imagine to be there if you want the region to have the most
potential for creation.  And if you stumble across the answer I'm
guessing you will, what on earth could it mean?

PS.  Let me append another note I wrote a few months ago to David
{\Mermin} [\myref{Mermin78}{6 January 2003}].  Maybe there's a
connection in there somewhere.

\section{04-07-03 \ \ {\it Bekenstein Bound Status} \ \ (to W. G. Unruh)} \label{Unruh1}

Can you fill me in on the latest to do with Bekenstein's entropy
bound?  Is it still controversial?  Have any loopholes been plugged?
Is it dead in the water?  Things like that.  Or can you give me a
pointer to something to read on the latest state in the debate?

The reason I ask has to do with a recent {\em analogy\/} I've been
drawing between (finite) Hilbert space dimension and mass.  Maybe
it's taking me down heretical lines \ldots

Thanks!

\section{07-07-03 \ \ {\it Your Letter} \ \ (to P. G. L. Mana)} \label{Mana3}

Thanks for the very nice letter of 12 June; it all sounds very interesting.  Have you made progress since then?  Has Hardy replied with anything of interest?

I apologize for taking so long to reply myself, but many, many administerial distractions have come to me since our nice time in Sweden.

Concerning your first issue, let me ask a naive question.  Does your discovery boil down to something other than saying, in the end, one can always embed quantum mechanics in a hidden variable model.  (Bohmian mechanics is usually claimed to be proof by explicit example.)  Of course, one has to give up some nice features of standard quantum mechanics---like locality---but nevertheless one can embed it into a hidden variable model.  If that is what your result is, it may nevertheless be quite a new way of saying it which will lead us to a better understanding of what criteria of aesthetics we are using when we accept the standard quantum model.

Concerning your second issue of generalizing the quantum Bayes rule, I shouldn't think it too hard.  Will it not just correspond to taking the matrices $V_d$ in Eq.\ (95) of my paper to be (nonsquare) isometries rather than full unitaries?

Finally let me address one thing that would be useful for my efforts to raise funds for meetings, postdoctoral positions, and, in particular, sell this direction in quantum foundations that you, Hardy, I, etc., are exploring.  In the first note you wrote me you said:
\bpglm
Thank you so much for your letter, and for the words of appreciation
for my paper. It is just a little more than a draft, and so its form
is far from being nice, and many are the missing references. I
already knew ``Quantum Mechanics as Quantum Information (and only a
little more)''; indeed, your idea, expressed there, that time evolution
could be a density operator, stimulated (via Rodriguez' ``Unreal
probabilities'', \arxiv{physics/9808010}) some of the research that led to my paper.
\epglm
If that is true, would you mind re-posting a version of your paper that makes some citation to my own.  The issue is, I have to make a case here and there that these ideas of mine have some influence, even if it is you young guys out there doing all the nontrivial work.  Having proper documentation at those times would definitely help.

How far is Stockholm from Aarhus?  If you have the time, you might consider coming to the following conference: \myurl{http://www.maphysto.dk/events2/QPFA2003/}.

It'd be great to talk to you again on the short timescale.  On the longer timescale, we'll have to look into when you, {\Schack}, and I can all get together in Dublin.

\section{07-07-03 \ \ {\it Bekenstein Bound Status, 2} \ \ (to W. G. Unruh)} \label{Unruh2}

\bwu
My take has not changed. It is possible that some such entropy to
energy bound exists in the real world. It is certainly not necessary
and I can imagine theoretical worlds in which it is not true (the
simplest is that the entropy goes up as the number of species of say
massless particles goes up, so one can always violate any bound by
assuming a large enough supply of species). But it is also not needed
for saving black hole thermodynamics, which was what he invented it
for.
\ewu

That is an almost trivial argument, and I was aware of it --- heard
it from you long ago.  If, however, one could argue that (by fiat) in
the GR setting, the assumption of a $ER/h$ value is the {\em
assumption\/} of a QM Hilbert space of the same dimension, then one
would have it. Not a logical necessity, but in essence a new
postulate.  Still, it would have to be motivated from some
considerations:  So I guess I was asking about something along the
lines of your last sentence above.  What is the best thing to read in
that regard?

\section{08-07-03 \ \ {\it Slogans, Slogans} \ \ (to G. L. Comer)} \label{Comer34}

\bgc
Thinking a little about your idea of black holes and the
dimensionality of the Hilbert space, how does the dimensionality of
the spacetime enter?  Thinking classically for the moment, the
degrees of freedom of a Schwarzschild  black hole can (at least
qualitatively) be understood using the quasinormal modes.  I imagine
that one must sum over more and more $l,m$ etc mode numbers the
higher the spacetime dimension gets.  If we could `quantize' those
modes, would not the dimension of the Hilbert space change as the
spacetime dimension changes?
\egc

Oh, I don't know about any of these things; I hadn't thought about
spacetime dimensionality at all.

Mostly it's just my habit of driving research with slogans---you
don't know where you're going (or even where you want to go) unless
you can make a slogan of it.  As I was walking in, I was thinking I
could have just as well compacted my last long note into the
following little play.  Basically I was thinking, wouldn't it be so
nice if \ldots
\begin{verse}
   Greg:   How much Hilbert space do you think this tin can has?
\\
   Chris:  I don't know; let's weigh it.
\end{verse}

Don't tell me about its construction, its composition, its history;
just weigh it.

It's because I've been having outlandish thoughts like this, that I
started wondering about this Bekenstein bound again.  Maybe it's all
crap.  (Overwhelmingly likely it's all crap!)

Does Mr.\ Bekenstein's bound depend upon spacetime dimension?  (I'll
dig up some papers when you get here.)

By the way, when do you get here?

\section{08-07-03 \ \ {\it The Common Fear?}\ \ \ (to R. W. {\Spekkens} \& N. D. {\Mermin})} \label{Mermin90} \label{Spekkens18}

The last three days as I've been walking to work, I've been reading a
little book on the thoughts of J\"urgen Habermas.  I thought of both
of you (or, actually, my perceptions of both of you) when I read the
following lines.  Let me record them.

\bq
Every undergaduate with a unit of behavioural psychology in his or
her academic record will have had some experience of the most vulgar
example of what Habermas means.  In this, as in every other field of
postivistic science, we are asked to approach the object world as a
disaggregated jumble of discrete objects of perception, as a jumble
of `its'.  We are set the task of uncovering the regularities in the
behaviour of these atoms of substance by means of an experimental
method.  The criterion of success lies in the predictive power of the
uncovered `laws' that must produce replicable results.  This means
results that are  {\it independent\/} of the author, the inventor, in
short of the {\it thinking subject\/} who in the first place
conceived the problem, the method and the experiment and who thereby
created the knowledge.  One half of the underlying assumption is that
knowledge is always reducible to the totality of discovered
properties of the object world.  The other half is that the {\it
subject}---the actor, the creator, the knower, the inventor, the
scientist---is at worst a pollutant in his own purely objective
world, or at best, a ghost in the machine of science and something
that must be methodologically controlled and, so far as is possible,
eliminated.
\eq

\section{09-07-03 \ \ {\it You and PI} \ \ (to D. R. Terno)} \label{Terno4}

Things are nifty here and my family---especially my four year old---are quite acclimated already.

It's good to hear that you're liking it at PI.  What sorts of things are you working on now?  I hope the unhealthy hidden-variablist ideas floating around there aren't infecting you \ldots

\section{10-07-03 \ \ {\it The Anointed Snark} \ \ (to N. D. {\Mermin})} \label{Mermin91}

\bdm
\bq
\noindent\rm [CAF wrote:] Importantly though---just trying to keep
you on the cutting edge---have you read the de Finetti paper
``Probabilismo'' in your preparation?
\eq
Raced through it, probably too quickly, on a very hard bench in the
library, after spending 30 minutes hunting it down.  Non circulating
journal.  Too long to Xerox.  Has it been reprinted at anybody's web
site?  It didn't help me much the first time through.  The stuff on
correlations without correlata seems to be just a footnote.
\edm

What has the four page limit of Physical Review Letters done to our
brains?  Raced through it?!  I tell you that it's the best
explication/defense of subjective Bayesianism I have ever seen, and
you raced through it?  I send you evidence that {\Ruediger} thinks the
same, and you raced through it?!?  {\Ruediger} and I both write
detailed letters to show how embracing some of the ideas in it may be
just the analgesic your 1985 paper needs, and you raced through
it?!?!

Tomorrow when I go in to the office, I will Xerox my copy for you and
send it across the ocean.  Please don't race through it.  (But feel
free to give honest, open criticism of anything that doesn't make
sense in it.  With that, we'll all certainly learn something.)

One question in the meantime concerning this:

\bdm
Glad it's helped you.  Hasn't helped me much yet, but I'm still
interested in keeping it up.
\edm

Fair enough.  But can you tell me this:  Did I, in my note ``Utter
Rubbish and Internal Consistency, Part I'' capture what you think is
the essential mystery of the subjective view of the quantum state? If
so, then I've at least got the right starting point, and I can try to
refine what needs to be said.

PS.  Yes, the stuff about correlations without correlata was only a
footnote with no further mention.  I was just trying to use
everything at my disposal to lure you into the paper.  One thing
really of significance though:  All those philosophers he mentions
near that footnote represent the Italian school of ({\James}ian)
pragmatism.  My love affair with {\James} only deepens and deepens. Some
mixture of {\James}, de Finetti, {\Pauli}, and modern quantum information
are gonna ultimately rule the day on these issues.

\section{10-07-03 \ \ {\it Solipsism} \ \ (to N. D. {\Mermin})} \label{Mermin92}

\bdm
Thanks, I'd enjoy having a copy.

I think my problem is not with the subjective view of probability.
Between what I've learned from you and {\Ruediger}, reading around in
Jaynes, Cox's famous paper [which I came upon ten years ago --- did I
ever tell you that I independently derived and solved precisely his
functional equation in my ``Relativity Without Light'' paper (in
boojums)], the stuff I read in the Business Library (!) after first
meeting {\Ruediger}, etc., it all makes a lot of sense.

My problem is, and remains, how it ties in with Quantum Mechanics,
which appears (among other things) to tell you how to derive precise
probabilities under (apparently) precisely defined --- if idealized
--- circumstances.  ``Quantum Mechanics Is a Law of Thought'' is
certainly a good slogan for getting started, and your poetry is very
soothing, but I can't help feeling there is still an enormous gap,
despite nice things like your quantum de Finetti theorem. Declaring
the circumstances also to be subjective judgments appears to be a
good (perhaps even necessary) move, but I don't find it convincing
and (at best) it seems to lead into some kind of infinite regress in
which everything becomes subjective and we're back to solipsism
(which is irrefutable and therefore trivial).  I haven't been able to
articulate what I feel is missing well enough to send it on to you
even under the Littlewood-Hardy rules, but I'm still thinking about
it.

Rule 1 definitely applies to the above paragraph.
\edm

A) What do you mean by solipsism?  And, B) how does the ``Paulian
idea'' I sent you still strike you as falling into that category?

\section{11-07-03 \ \ {\it Wow!}\ \ \ (to J. A. Smolin)} \label{SmolinJ6}

\bjas
I am writing up a paper on the lockboxes, and I need to figure out where to send it.  I'd prefer a real journal, but perhaps the conference proceedings of Sweden are the best place.  In a couple of days I'll send you the paper if you're interested.
\ejas

Wow!  Excellent.  I never imagined that you'd actually write this stuff up.  I'll have a look at the actual draft next week.  I'll try to give you detailed comments.

The latter part of your note strikes me as really interesting:
\bjas
My plan is to actually write two papers.  I want to put a paper on the basic idea out there soon, then perhaps later a paper including a discussion of Rob's toy model and some foundational problems.  I also have some things I want to work out about Rob's model if he hasn't done it already by the time he write it up.
In particular, his love of the so-called underlying ontic states is, I
believe, what leads to the disallowing of some mixed states--The rule
is that a measurement on one system cannot change the ontic state of
another, but the right way to do it is to say that what you can't
change is the marginal measurement outcomes of the remote system.
This is definitely different under his model and I think leads to a
prettier result.
\ejas
You know I'm all for nononticity when it comes to the ``stuff'' that quantum states are about.

\section{11-07-03 \ \ {\it THE Paulian Idea} \ \ (to N. D. {\Mermin})} \label{Mermin93}

\bdm
Which ``Paulian idea''?
\edm

The one for which there are some notes on:

\bq
     In the new pattern of thought we do not assume any longer the
     detached observer, occurring in the idealizations of this classical
     type of theory, but an observer who by his indeterminable effects
     creates a new situation, theoretically described as a new state of
     the observed system.  In this way every observation is a singling
     out of a particular factual result, here and now, from the
     theoretical possibilities, thereby making obvious the discontinuous
     aspect of the physical phenomena.
\eq
\bq
     Nevertheless, there remains still in the new kind of theory an
     objective reality, inasmuch as these theories deny any possibility
     for the observer to influence the results of a measurement, once
     the experimental arrangement is chosen.
\eq
\bq
     Like an ultimate fact without any cause, the individual outcome
     of a measurement is, however, in general not comprehended by laws.
     This must necessarily be the case \ldots\
\eq

Or in more modern language (from \myref{Mermin76}{11/4/02}):

\bq
\noindent 1)  The POVM as a function from raw data to meaning.\medskip

We generally write a POVM as an indexed set of operators, $E_d$. Here
is how I would denote the referents of those symbols.  The index $d$
should be taken to stand for the raw data that can enter our
attention when a quantum measurement is performed.  The whole object
$E_d$ should be construed as the ``meaning'' we propose to ascribe to
that piece of data when/if it comes to our attention.  It is
important here to recognize the logical distinction between these two
roles.  The symbol $d$ stands for something beyond our control,
something that enters into us from the world outside our head.  The
ascription of a particular value $d$ is not up to us, by definition.
The {\it function\/} $E_d$, however, is of a completely different
flavor. It is set by our history, by our education, by whatever
incidental factors that have led us to believe whatever it is that we
believe when we walk into the laboratory to elicit some data.  That
is to say, $E_d$ has much the character of a subjective probability
assignment.  It is a judgment.
\eq

\section{11-07-03 \ \ {\it More Solipsism} \ \ (to N. D. {\Mermin})} \label{Mermin94}

Still trying to get your worry straight.  By your definition, could a
solipsist make changes?  (The note below makes sense of this
question.)  [See 21-07-01 note ``\myref{Preskill2}{The Reality of Wives}'' to
A.~J. Landahl and J. Preskill.]

\section{11-07-03 \ \ {\it One Final Thing in the Wee Hours} \ \ (to N. D. {\Mermin})} \label{Mermin95}

\bdm
Solipsism because when I try to look at things your way I find that
whatever I try to condition my subjective probabilities on turns out
also to be subjective and conditional on further subjective
judgments, ad infinitum.
\edm

Then, for you, does Bayesianism in general equal solipsism
(independently of quantum mechanical issues)?  For every Bayesian has
that infinite regress:  They ball it all up into something called
``the prior'' and ask not where it comes from.  (As Savage said it
very nicely in the passage {\Ruediger} recommended you read, the prior
might simply be a product of Darwinian evolution.)

Good night!  (It's after 3:00 AM for me now.)

\section{11-07-03 \ \ {\it Friendship} \ \ (to N. D. {\Mermin})} \label{Mermin96}

Your absorbing this de Finetti paper is important for our
discussions.  Consider it a present from the slush fund Kiki allows
me to dip into once a year (for the really important things).  It
will arrive at your office in Clark Hall sometime Monday.  (FedEx
tracking number 8410-8544-4295)

Solipsism indeed!

\section{11-07-03 \ \ {\it One Project Done} \ \ (to G. L. Comer)} \label{Comer35}

Well I got one project done that I had promised myself to do before
your coming to Dublin (even if I finished no others):  I finished
reading my copy of {\sl Becoming William {\James}}.  I just couldn't
bear the thought that you might know something about William {\James}
that I didn't know!  But let me apologize to you:  I found that I
hated the book, and I'm sorry I burdened you with it.  The only thing
that comes across in it is that WJ and the whole {\James} family is
troubled. One gets no feel for the utter greatness that was
developing all along at the same time as these (overemphasized) other
bits.  And the guy was just too full of it with of his own
psychological theories, rather than giving us a rounded glimpse of
the lives involved. Anyway, you can see it's on the bottom of my list
of {\James} biographies now.  I wish I had dug you up a copy of Perry's
{\sl The Thought and Character of William {\James}\/} instead.  It
carries the message I had wanted you to get, where this other book
utterly fails.

See you tomorrow.

\section{13-07-03 \ \ {\it Thanks Again} \ \ (to M. P\'erez-Su\'arez)} \label{PerezSuarez2}

Thanks for the latest notes.  I'll forward them to Joe Renes, who's taking care of the final draft.  (The paper will appear in {\sl Foundations of Physics}.)

Concerning the Paulian--Wheelerish compendium, I have finally gotten my computer set up with a scanner and top of the line OCR package (optical character recognition).  It works like a charm!!  So the compendium will be greatly expanding very soon.  I'm hoping to get the final product on the web by Christmastime.

\section{14-07-03 \ \ {\it Unitary Equivalence} \ \ (to A. Peres)} \label{Peres50}

\bap
During the {\Vaxjo} conference (June 2001) someone (Ben Schumacher?)\ gave
a talk explaining that any dynamical system could be canonically
mapped onto a harmonic oscillator and discussed the corresponding
quantum property. My comment was that unitary equivalence is not
equivalence from the point of view of physics (and I gave a reference
to Fong and Sucher). That talk does not appear in the conference
proceedings. Do you remember that talk and who gave it?
\eap

It was definitely Ben.  He uses that as an example to argue for the inconsistency of the many-worlds interpretation.  He's never published the talk as a paper, much to my dismay (I've tried to get him to on a couple of occasions).

What is the Fong and Sucher paper?

\subsection{Asher's Reply}

\bq
Fong and Sucher, J. Math.\ Phys.\ {\bf 5} (1964) 456 --- see my book,
penultimate paragraph of page 227.
\eq

\section{17-07-03 \ \ {\it Utrecht November Conference} \ \ (to J. B. M. Uffink)} \label{Uffink1}

For your meeting in November, let's try this:\medskip

\noindent {\bf Title:}  Quantum Information Does Not Exist \smallskip

\noindent {\bf Abstract:}  It is information {\it carriers\/} that exist---conceptually both classical and quantum.  To confuse the epistemic category (the information) with the ontic (the carriers) is to cause any amount of trouble.  Nonetheless, one thing is true when it comes to applications of information theory to classical and quantum phenomena:  There is a difference.  And, in that difference---this talk will argue---lies quantum theory's most direct statement about properties of the world by itself (i.e., the world without the information processing agent). \medskip

The title of the talk is actually a phrase popularized by Ari Duwell; in the talk I will give him proper credit.

\section{17-07-03 \ \ {\it {\Mermin}ism} \ \ (to R. {\Schack})} \label{Schack63}

Just got the note below from David {\Mermin}.  Is this man being a smart
ass or what?  Or is he being serious?

\bdm
Thanks for sending me Probabilism.  It does indeed require careful
perusal.  I shall bring it along to the sea shore next week.

Meanwhile I note that near the beginning it offers what I would call
a dictionary definition of solipsism.  I'm also inclined to recommend
it to my constructivist sociologist friends who I suspect would find
its point of view highly congenial.
\edm

\section{17-07-03 \ \ {\it Seven Pines VIII} \ \ (to J. Preskill)} \label{Preskill11}

\bjp
Is this a good thing to do?
\ejp

I don't know.  Probably.  In my own case, I'd be interested in
talking to Earman, Howard, Milburn, Unruh, and Wald.  So it could be
worthwhile.

In particular, I'd like to have the chance to field some questions to
the GR people \ldots\ as the universal characteristics of
Hilbert-space dimension have been on my mind, and I find myself
wondering to what extent there might be a (conceptual) analogy to
gravitational mass and universal gravitation here.  Far-fetched in my
usual way \ldots\ but maybe not completely stupid.

I think it'd be great if you'd be there.

\section{17-07-03 \ \ {\it Big, Big Favor?}\ \ \ (to R. W. {\Spekkens})} \label{Spekkens19}

I wonder if you'd mind doing me a big, big favor?  I'd like to get my hands on some pretty wacky articles, and I'm a little embarrassed to use the interlibrary loan services of CNRI to help me in the quest.

All the articles appear in a journal called {\sl Journal of Analytical Psychology\/} and it looks like the library at Laurier (apparently associated with Waterloo) has the right volumes.  Could I ask you to bring copies of the articles with you when you come to the Aarhus meeting.  (Did the organizers, by the way, give you a talk slot or a poster session slot?)

Let me know if you can, and I won't pursue any other options I might dream up.  I'll put the references I need below.

\begin{itemize}

\item
B.~Zabriskie, ``Jung and Pauli:\ A Subtle Asymmetry,'' J.
Analyt.\ Psych.\ {\bf 40}, 531--553 (1995).

\item
D.~Lindorff, ``One Thousand Dreams:\ The Spiritual Awakening of Wolfgang Pauli,'' J. Analyt.\ Psych.\ {\bf 40}, 555--569 (1995).

\item
D.~Lindorff, ``Psyche, Matter and Synchronicity:\ A Collaboration between C.~G. Jung and Wolfgang Pauli,'' J. Analyt.\ Psych.\ {\bf 40}, 571--586 (1995).

\item
V.~Mansfield and J.~M. Spiegelman, ``Quantum Mechanics and Jungian Psychology:~Building a Bridge,'' J. Analyt.\ Psych.\ {\bf 34}, 3--31 (1989).

\item
V.~Mansfield and J.~M. Spiegelman, ``The Opposites in Quantum Physics and Jungian Psychology, Part~I:\ Theoretical Foundations,''
J. Analyt.\ Psych.\ {\bf 36}, 267--288 (1991).

\item
V.~Mansfield, ``The Opposites in Quantum Physics and Jungian Psychology, Part~II:\ Applications,'' J. Analyt.\ Psych.\ {\bf 36}, 289--306 (1991).
\end{itemize}

\section{18-07-03 \ \ {\it Solipsism Concerns} \ \ (to N. D. {\Mermin})} \label{Mermin97}

Referring to
\bdm
I think my problem is not with the subjective view of probability.
Between what I've learned from you and {\Ruediger}, reading around in
Jaynes, Cox's famous paper [which I came upon ten years ago --- did I
ever tell you that I independently derived and solved precisely his
functional equation in my ``Relativity Without Light'' paper (in
boojums)], the stuff I read in the Business Library (!)\ after first
meeting {\Ruediger}, etc., it all makes a lot of sense.

My problem is, and remains, how it ties in with Quantum Mechanics,
which appears (among other things) to tell you how to derive precise
probabilities under (apparently) precisely defined --- if idealized
--- circumstances.  ``Quantum Mechanics Is a Law of Thought'' is
certainly a good slogan for getting started, and your poetry is very
soothing, but I can't help feeling there is still an enormous gap,
despite nice things like your quantum de Finetti theorem. Declaring
the circumstances also to be subjective judgments appears to be a
good (perhaps even necessary) move, but I don't find it convincing
and (at best) it seems to lead into some kind of infinite regress in
which everything becomes subjective and we're back to solipsism
(which is irrefutable and therefore trivial).  I haven't been able to
articulate what I feel is missing well enough to send it on to you
even under the Littlewood-Hardy rules, but I'm still thinking about
it.
\edm
and
\bdm
Solipsism because when I try to look at things your way I find that
whatever I try to condition my subjective probabilities on turns out
also to be subjective and conditional on further subjective
judgments, ad infinitum.
\edm
and
\bdm
Thanks for sending me Probabilism.  It does indeed require careful
perusal.  I shall bring it along to the sea shore next week.

Meanwhile I note that near the beginning it offers what I would call
a dictionary definition of solipsism.  I'm also inclined to recommend
it to my constructivist sociologist friends who I suspect would find
its point of view highly congenial.
\edm

I guess I am seriously concerned by the charge of solipsism you have
made of our program.  It is a serious charge.  (All one has to do is
look into the daily news to see the dangers of it.  See the Salon
article pasted below about G. W. Bush's shenanigans for a
particularly moving example.)  Why is it that we fail to communicate
on this point?

The world is not what I will it to be.  And there is nothing in this
view of quantum mechanics that our group is trying to construct that
hints of this.

Solipsism would come about if from the quantum formalism one could
prove
\begin{enumerate}
\item
that there are no ``instruction sets,'' and
\item
that the outcomes of all interventions (measurements) could actually
   be controlled by the agent instigating the intervention.
\end{enumerate}

But that is not the case.  I was thinking harder about this last
night as I was doing a little editing on my CG Fire Series, Vol.\ II.
The frontispiece contains a quote from John {\Wheeler} that I'll paste
below because it emphasizes precisely the right point.  The radical
constructivism or solipsism that you fear is blocked for each and
every quantum agent by his own stark admission that the outcomes of
his interventions are beyond his control.  If they are beyond his
control, then there is no need to suppose that they are products of
his mind.  What more needs to be said?

Let me leave it at that for this round.  {\Ruediger} is hoping to
construct a note for you within the next three hours or
so---hopefully before you leave for the seashore---making explicit
reference to some of the dangers we are starting to perceive in the
opening sections of Probabilismo \ldots\ probably precisely the ones
that are worrying you.  So, stay tuned to your email before leaving
for vacation!  (You're our most valued customer.)

First the {\Wheeler} quote:

\bq
The Universe can't be Laplacean.  It may be higgledy-piggledy.  But
have hope.  Surely someday we will see the necessity of the quantum
in its construction.  Would you like a little story along this line?

Of course!  About what?

About the game of twenty questions.  You recall how it goes---one of
the after-dinner party sent out of the living room, the others
agreeing on a word, the one fated to be a questioner returning and
starting his questions.  ``Is it a living object?''  ``No.''  ``Is it
here on earth?''  ``Yes.''  So the questions go from respondent to
respondent around the room until at length the word emerges: victory
if in twenty tries or less; otherwise, defeat.

Then comes the moment when we are fourth to be sent from the room. We
are locked out unbelievably long.  On finally being readmitted, we
find a smile on everyone's face, sign of a joke or a plot.  We
innocently start our questions.  At first the answers come quickly.
Then each question begins to take longer in the answering---strange,
when the answer itself is only a simple ``yes'' or ``no.''  At
length, feeling hot on the trail, we ask, ``Is the word `cloud'?''
``Yes,'' comes the reply, and everyone bursts out laughing.  When we
were out of the room, they explain, they had agreed not to agree in
advance on any word at all.  Each one around the circle could respond
``yes'' or ``no'' as he pleased to whatever question we put to him.
But however he replied he had to have a word in mind compatible with
his own reply---and with all the replies that went before.  No wonder
some of those decisions between ``yes'' and ``no'' proved so hard!

And the point of your story?

Compare the game in its two versions with physics in its two
formulations, classical and quantum.  First, we thought the word
already existed ``out there'' as physics once thought that the
position and momentum of the electron existed ``out there,''
independent of any act of observation.  Second, in actuality the
information about the word was brought into being step by step
through the questions we raised, as the information about the
electron is brought into being, step by step, by the experiments that
the observer chooses to make. Third, if we had chosen to ask
different questions we would have ended up with a different word---as
the experimenter would have ended up with a different story for the
doings of the electron if he had measured different quantities or the
same quantities in a different order.  Fourth, whatever power we had
in bringing the particular word ``cloud'' into being was partial
only.  A major part of the selection---unknowing selection---lay in
the ``yes'' or ``no'' replies of the colleagues around the room.
Similarly, the experimenter has some substantial influence on what
will happen to the electron by the choice of experiments he will do
on it; but he knows there is much impredictability about what any
given one of his measurements will disclose.  Fifth, there was a
``rule of the game'' that required of every participator that his
choice of yes or no should be compatible with {\it some\/} word.
Similarly, there is a consistency about the observations made in
physics.  One person must be able to tell another in plain language
what he finds and the second person must be able to verify the
observation. \\
\hspace*{\fill} --- {\it John Archibald {\Wheeler}} \\
\hspace*{\fill} Frontiers of Time, 1979
\eq

And, now from Joe Conason's Journal in {\sl Salon}: ``President
Bush's Astonishing New Reason for the War with Iraq: Saddam Wouldn't
Let Weapons Inspectors In.''

\bq
July 15, 2003  |  A ``darn good'' quote that almost nobody quoted
``We gave him a chance to allow the inspectors in, and he wouldn't
let them in.''

George W. Bush uttered that amazing sentence yesterday to justify the
war in Iraq, according to the Washington Post.

What? Yes, I promise that's what the man said. (And by ``him,'' the
president clearly meant Saddam Hussein -- not Kim Jong Il, who
actually has refused to let international inspectors into North
Korea.)

Now a presidential statement so frontally at variance with the
universally acknowledged facts obviously presents a problem for the
White House press corps. He wasn't joking, and he didn't sound
disoriented or unwell. Although Dana Priest and Dana Milbank wrote
the story as delicately as they possibly could, they couldn't make it
seem less weird:

``The president's assertion that the war began because Iraq did not
admit inspectors appeared to contradict the events leading up to war
this spring: Hussein had, in fact, admitted the inspectors and Bush
had opposed extending their work because he did not believe them
effective.''

Appeared to contradict the events leading up to war? Indeed, that's
an exceedingly mild description of what Bush said. There's no
plausible explanation, unless the president suddenly flashed back to
his Yale sophomore philosophy seminar, grappling with the argument
that everything we perceive is mere illusion.

For the moment, however, let's just assume reality does exist. What
possessed the president to make an assertion that everyone on the
planet knows to be untrue?  \ldots
\eq

\section{18-07-03 \ \ {\it Most In One Place} \ \ (to N. D. {\Mermin} \& R. {\Schack})} \label{Mermin98}

As I've told both of you, I'm starting to put together another
samizdat---this one to contain the last year's discussions.  Let me
drop it off with you as it stands at the moment.  Even though it's
not complete you might find it a little useful.  I think it already
contains all of the emails from my side in the latest round of
discussions \ldots\ if that's useful.

Parts of it are certainly diatribic \ldots\ and needlessly so.  But on
the other hand, there are some notes---the note of 10 January 2003,
``\myref{Mermin80}{Filth Under the Rug},'' comes to mind---that I
think express a truly heartfelt frustration.

David is right:  ``\,`Quantum Mechanics Is a Law of Thought' is
certainly a good slogan for getting started, and your poetry is very
soothing, but I can't help feeling there is still an enormous gap,
despite nice things like your quantum de Finetti theorem.''  There is
so very much that needs to be done by way of technical work.  But
that should not stop anyone from seeing that we are moving absolutely
in the right direction.  Just precisely what is it that is blocking
the vision?

\section{19-07-03 \ \ {\it Definitions from Britannica} \ \ (to N. D. {\Mermin})} \label{Mermin99}

\bdm
As far as I know solipsism is the claim that there is nothing more
than my own sense impressions.  It does not imply that I can control
what those sense impressions are.  They could be like a film I am
doomed to keep watching --- or an uncontrollable dream.

I believe that's the correct use of the term, but if I'm wrong, give
my ``serious charge'' another name.  Perhaps this clarification makes
it less serious?  Hope not.
\edm

I looked up several definitions for the heck of it.  The first set
below comes from the 2001 {\sl Encyclopedia Britannica}.  Following
that, I place my search results from several dictionaries for the
word solipsism in particular.  I guess the definition from {\sl
Wikipedia\/} best captures what I thought the word meant.

In any case, regardless of the labels, David's biggest worry seems to
be this:
\bdm
A) Solipsism because when I try to look at things your way I find
that whatever I try to condition my subjective probabilities on turns
out also to be subjective and conditional on further subjective
judgments, ad infinitum.
\edm

Let me try to tackle this one directly, now that maybe I understand
your worries better and after having talked to {\Ruediger} yesterday.
Take a good solid physicist like Steven Weinberg who stakes his
career on the search for a grand unified field theory.  Suppose he
finds it.  To find it (I presume) is to declare:  The world's
Lagrangian is $L$.  Now suppose I were to ask Mr.\ Weinberg, ``Why
$L$? Why not $M$?''  I know for sure his answer will be of the form,
``$L$ just is. It is the starting point.  It is an ultimate fact of
nature; it calls for no explanation.  In any case, if it calls for an
explanation, its answer must come from outside the realm of
science---religion? theology?---but I see no reason to go to such
lengths.''

Would that make Weinberg a solipsist?  A sensationalist?  A
phenomenalist?  The point is Weinberg's stance has nothing to do with
any of these labels.

Similarly for the Bayesian (even of the de Finettian variety, despite
the mumbo jumbo in the opening sections of Probabilismo).  For him,
``the prior'' on any event space is treated as an ultimate fact---an
ultimate fact about the agent.  There is no infinite regress because,
just as with Weinberg, ultimate facts call for no further
explanation.

Bayesian practice---and {\Ruediger} and I would claim the formal
structure of quantum mechanics too---is all about what to do once a
prior is established.  It is not about what to do before the prior is
established.  In the quantum mechanical case, establishing ``a
prior'' is 1) to write down a quantum state for all systems
considered, and 2) to write down a (conditional) quantum operation
for all measuring devices considered.

If there are two agents, there may well be two priors in the sense
above ---i.e., two ultimate facts (with respect to this level of
inquiry).  In that sense, the priors are ``subjective'', but that
does not take away their status as ultimate facts in this treatment.
It only calls for a recognition that the facts are about the agents.

The role of the separate system and measuring device---now
specializing on quantum me\-chanics---is that when the two are
combined they give ``birth'' to a new ultimate fact:  The ``click.''
There is no sense, however, in which this new ultimate fact is {\em
about\/} either of the agents:  It has a life of its own.  (In fact,
it is because of these lines of thought that I sometimes call my view
``the sexual interpretation of quantum mechanics.'')

Does this clarify anything?  Does this in any way address your fears
of solipsism/sensationalism/ phenomenalism?

Definitions below.

\bq
\noindent {\bf Solipsism:}

in philosophy, formerly, moral egoism (as used in the writings of
Immanuel Kant), but now, in an epistemological sense, the extreme
form of subjective idealism that denies that the human mind has any
valid ground for believing in the existence of anything but itself.
The British idealist F. H. Bradley, in {\sl Appearance and Reality\/}
(1897), characterized the solipsistic view as follows:

``I cannot transcend experience, and experience is my experience.
From this it follows that nothing beyond myself exists; for what is
experience is its (the self's) states.''

Presented as a solution of the problem of explaining human knowledge
of the external world, it is generally regarded as a reductio ad
absurdum. The only scholar who seems to have been a coherent radical
solipsist is Claude Brunet, a 17th-century French physician.
\eq

\bq
\noindent {\bf Subjective Idealism:}

a philosophy based on the premise that nothing exists except minds
and spirits and their perceptions or ideas. A person experiences
material things, but their existence is not independent of the
perceiving mind; material things are thus mere perceptions. The
reality of the outside world is contingent on a knower. The
18th-century Anglo-Irish philosopher George Berkeley succinctly
formulated his fundamental proposition thus: Esse est percipi (``To
be is to be perceived''). In its more extreme forms, subjective
idealism tends toward solipsism, which holds that I alone exist.
\eq

\bq
\noindent {\bf Sensationalism:}

in epistemology and psychology, a form of Empiricism that limits
experience as a source of knowledge to sensation or sense
perceptions. Sensationalism is a consequence of the notion of the
mind as a tabula rasa, or ``clean slate.'' In ancient Greek
philosophy, the Cyrenaics, proponents of a pleasure ethic, subscribed
unreservedly to a sensationalist doctrine. The medieval Scholastics'
maxim that ``there is nothing in the mind but what was previously in
the senses'' must be understood with Aristotelian reservations that
sense data are converted into concepts. The Empiricism of the 17th
century, however---exemplified by Pierre Gassendi, a French
neo-Epicurean, and by the Englishmen Thomas Hobbes and John
Locke---put a greater emphasis on the role of the senses, in reaction
against the followers of Ren\'e Descartes who stressed the mind's
faculty of reasoning. Locke's influence on 18th-century French
philosophy produced the extreme sensationnisme (or, less often,
sensualisme) of \'Etienne Bonnot de Condillac, who contended that
``all our faculties come from the senses or more precisely, from
sensations''; that ``our sensations are not the very qualities of
objects [but] only modifications of our soul''; and that attention is
only the sensation's occupancy of the mind, memory the retention of
sensation, and comparison a twofold attention.
\eq

\bq
\noindent {\bf Phenomenalism:}

a philosophical theory of perception and the external world. Its
essential tenet is that propositions about material objects are
reducible to propositions about actual and possible sensations, or
sense data, or appearances. According to the phenomenalists, a
material object is not a mysterious something ``behind'' the
appearances that people experience in sensation. If it were, the
material world would be unknowable; indeed, the term matter itself
would be unintelligible unless it somehow could be defined by
reference to sense experiences. In speaking about a material object,
then, reference must be made to a very large group or system of many
different possibilities of sensation. Whether actualized or not,
these possibilities continue during a certain period of time. When
the object is observed, some of these possibilities are actualized,
though not all of them. So long as the material object is unobserved,
none of them is actualized. In this way, the phenomenalist claims, an
``empirical cash value'' can be given to the concept of matter by
analyzing it in terms of sensations.
\eq

\bq
\noindent {\bf Positivism:}

in philosophy, generally, any system that confines itself to the data
of experience and excludes a priori or metaphysical speculations.
More narrowly, the term designates the thought of the French
philosopher Auguste Comte (1798--1857). The basic affirmations of
Positivism are (1) that all knowledge regarding matters of fact is
based on the ``positive'' data of experience, and (2) that beyond the
realm of fact is that of pure logic and pure mathematics, which were
already recognized by the Scottish Empiricist and Skeptic David Hume
as concerned with the ``relations of ideas'' and, in a later phase of
Positivism, were classified as purely formal sciences. On the
negative and critical side, the Positivists became noted for their
repudiation of metaphysics; i.e., of speculation regarding the nature
of reality that radically goes beyond any possible evidence that
could either support or refute such ``transcendent'' knowledge
claims. In its basic ideological posture, Positivism is thus worldly,
secular, antitheological, and antimetaphysical. Strict adherence to
the testimony of observation and experience is the all-important
imperative of the Positivists. This imperative is reflected also in
their contributions to ethics and moral philosophy, and most
Positivists have been Utilitarians to the extent that something like
``the greatest happiness for the greatest number of people'' was
their ethical maxim. It is notable, in this connection, that Auguste
Comte was the founder of a short-lived religion, in which the object
of worship was not the deity of the monotheistic faiths but humanity.
\eq

Further definitions of Solipsism:
\begin{itemize}
\item
{\bf Merriam-Webster Online Dictionary:} a theory holding that the
self can know nothing but its own modifications and that the self is
the only existent thing.
\item
{\bf Cambridge International Dictionary:} the belief that only one's
own experiences and existence can be known with certainty.
\item
{\bf American Heritage Dictionary:} 1. The theory that the self is
the only thing that can be known and verified. 2. The theory or view
that the self is the only reality.
\item
{\bf Encyclopedia Britannica, 1911 edition:} a philosophical term,
applied to an extreme form of subjective idealism which denies that
the human mind has any valid ground for believing in the existence of
anything but itself. It may best be defined, perhaps, as the doctrine
that all ``existence is experience, and that there is only one
experient. The Solipsist thinks that he is the one!'' ({\Schiller}). It
is presented as a solution of the problem of explaining the nature of
our knowledge of the external world. We cannot know
things-in-themselves: they exist for us only in our cognition of
them, through the medium of sense-given data. In. F.~H. Bradley's
words (Appearance and Reality):

``I cannot transcend experience, and experience is my experience.
From this it follows that nothing beyond myself exists; for what is
experience is its (the self's) states.''
\item
{\bf Wikipedia:} Solipsism is a metaphysical belief that one is like
a God, creating the reality in which one exists. Solipsism is
logically coherent, but not falsifiable, so it cannot be established
by current modes of the scientific method.

The classic objection to solipsism is that people die. However, you
have not died, and therefore you have not disproved it.

A further objection is that life causes pain. Why would we create
pain for ourselves? There may be some reason which we have decided to
forget, such as the law of Karma, or a desire not to be bored.

Solipsism is a common theme in eastern philosophy. Various
interpretations of Buddhism, especially Zen, teach that the entire
universe exists only in one's mind.
\end{itemize}

\section{24-07-03 \ \ {\it Subjective and Objective, Precursor} \ \ (to A. Sudbery)} \label{Sudbery1}

I promised to say something a little more scientific this afternoon,
rather than administrative.  But time slipped away, and I ended up
doing some mathematics with Chris King at the chalkboard.
Consequently, I didn't get Nagel's article finished, which I started
on the train yesterday.

But let me build up to a discussion on it (and then after it, on your
paper `Why Am I Me?' in the coming weeks).  Mostly, I just want to
say that, so far, it's not looking so good for Mr.\ Nagel in my eyes:
I'm finding that I'm not liking the paper as much as I was hoping to.
Of course, I could have guessed that I would get tangled in a
sparring match with it \ldots\ but I didn't think I would get tangled
in disagreement even with the motivation it sets out in the first
couple of pages.

Let me try to show you why.  Attached is an excerpt, relevant to the
present subject, from my smaller samizdat ``Quantum States:\ What the
Hell Are They?'' posted on my webpage. [See 24-06-02 note ``\myref{Wiseman6}{The World is Under Construction}'' and 27-06-02 note ``\myref{Wiseman8}{Probabilism All the Way Up}'' to H. M. Wiseman.]  I hope it shows why I would be
resistant to an `external view' at the very outset.  I will certainly
read Nagel thoroughly before this is all over with, but already on
the very first pages it looks that we come from very different
worlds.

More eventually \ldots

\section{25-07-03 \ \ {\it Agendas and Rubrics} \ \ (to R. W. {\Spekkens})} \label{Spekkens20}

\brws
These observations led me to wonder whether it might be impossible
to achieve a HVT with non-contextuality for the preparations for any
dimensionality of Hilbert space.  I believe I have now proven such a
no-go theorem using nine states from a $2d$ Hilbert space.  I'll try to
write up the proof tonight.

If it stands, I hope that this result may serve as another clue in
the mystery, another aid in determining the nature of the conceptual
innovation that needs to occur before one can devise a proper realist
interpretation of quantum theory, if such an interpretation can be
had.
\erws

What a novel idea!  I like the technical part.

Question:  What does it take to be a proper realist interpretation of
quantum theory?

I never did finish that note on black holes to you; I'll paste the
point I got up to in it below.  The only point I wanted to make (but
I wanted to do it in style!)\ was that, in my thinking about quantum
mechanics, Hilbert-space dimension plays a conceptual role that is a
bit like black-hole mass.  I.e., Hilbert-space dimension is the ontic
state of a quantum system.

Now, why wouldn't that fall under the rubric of ``proper realist
interpretation'' in your eyes?  I'm guessing that it won't.  In
particular, I'm guessing that it has something to do with the
``Common Fear?''\ I wrote you and {\Mermin} about a couple of weeks ago.
But, I don't know, you tell me.\medskip

\noindent --- --- --- --- --- --- --- --- --- ---\medskip

\noindent From just before the IBM meeting:\medskip

I spent a little while digging around in my email archive this
morning and was finally able to pull out this piece from June 7,
1991.  It was in a letter to Greg Comer.
\bq
     The other night I had a dream in which I was trying to calculate
     how many black holes would fit in my ice chest.  I could already
     see one in there and I was trying to figure out how many more
     would fit.  They're really beautiful, you know---contrasting
     against the white of the ice chest and the glistening of the ice.
     So, I pulled out another beer and watched for a while.
\eq

Consider a Schwarzschild black hole, to the exclusion of all else in
the universe.  What is the black hole's ontic state?

There is only one thing to pin it on:  It is the black hole's mass,
$M$.  Do you balk at that?  Does the starkness of the black hole's
characterization specify a dearth or emptiness of phenomena
associated with this lovely physical object?  Not at all, once the
black hole is embedded within the wider context of general
relativity, the orbits of various test bodies, and so forth.

Especially in light of the immense physics that comes from such a
single, simple number $M$ when \ldots\ [Never finished.]

\section{25-07-03 \ \ {\it Short Replies} \ \ (to G. L. Comer)} \label{Comer36}

Many short replies, as once again I'm feeling extremely pressed for time.

But mostly I'm intrigued by the idea of the project you suggested:  So I can't let that slip away unanswered!  I'd like to see your thoughts more fleshed up!  Try to say in more detail what you think the analogies are.

\bgc
``Classical versus Quantum Interfaces: Linear Structures,''
G. L. Comer and C. A. Fuchs
\egc

G. L. Comer and C. A. Fuchs, of course.  Since 1995 I have kept to strict alphabetical ordering (like the mathematicians and computer scientists do), with only the exception of the experimental teleportation paper.

\bgc
I like that word ``interlocutor.''
\egc

I'm not so taken with it.  It's probably because I'm not so taken
with the ``asking questions of nature and getting answers back''
imagery any more.  I'm more taken with the imagery of ``push on
nature and see how it reacts.''  We never ask of nature, we only push
on it---or at least that's my take at the moment.  (It's the reason
I'm calling my CG Fire Series, Vol.\ II, project ``The Activating
Observer''.)  But like my philosophy, I can be malleable.

\section{25-07-03 \ \ {\it No, Thank You!}\ \ \ (to J. Honner)} \label{Honner1}

Thanks for the nice compliments/complements.  Why did you leave
academia?  Not only did I enjoy your Bohr/Derrida article, but many
years before that, I enjoyed your book on Bohr---reading it was one
of my formative experiences in the field.  I'd love to get hold of it
again, as now in working on my ``The Activating Observer: Resource
Material for a Paulian/{\Wheeler}ish Conception of Nature'' project
(partially posted at my webpage) I'm going through many old things
once again, but with my own fine-toothed comb.  (Though, I guess I
have to admit, I did make a bit of fun of your book and its genre in
my paper \quantph{0104088}.  You can
see what I mean in endnote 5 and the text leading to it.)

By the way, if you think you might get something out of {\tt Notes on a Paulian Idea}, I'll have a copy sent your way.  {\Vaxjo} University Press published 100 copies (in proper book form) for my use, and I've still got about 30 copies to give away.

Anyway, it was really great hearing from you, and you can bet I'll probably consult with you again when I'm in the stages of tidying up ``The Activating Observer'' project.

\section{25-07-03 \ \ {\it Question} \ \ (to V. Kargin)} \label{Kargin1}

\bvk
I am trying to find out who said: ``Probability theory is measure theory with a soul''. It appears that some attribute this quote to Mark Kac and some to Kolmogoroff. I noticed that you used a couple of times this quotation in your papers. Can you help me in locating the source?
\evk

Indeed it was Mark Kac; I'm sorry if I caused any confusion in my writings.  In my review of Holevo's book, I wrote:
\bq\noindent
It is a probabilistic model that differs radically in character from the classical model first laid down by A. N. Kolmogorov in 1933 as ``measure theory with a soul.''
\eq
but the way that sentence should be parsed is the following:
\bq\noindent
It is a probabilistic model that differs radically in character from [the classical model first laid down by A. N. Kolmogorov in 1933] as [``measure theory with a soul.'']
\eq
I didn't mean to imply that the phrase should be attributed to Kolmogorov.  Basically I was leaving the attribution of the phrase blank so as not to complicate the sentence too much.  Maybe that was a mistake.

Anyway, here is the exact quote attributed to Mark Kac that I found in Holevo's book:
\bv
``Probability theory is a measure theory -- with a soul.''
\ev
I wish I knew the exact source.  Perhaps Alexander Holevo would know.

\subsection{Slava's Reply}

\bq
Thank you for your reply. I didn't mean to imply that the phrase in your review is misleading. I have seen the attribution to Kolmogoroff somewhere else.

I actually located the source of this quote. It is from a preface by Mark Kac to a book by Luis Santalo {\sl Integral Geometry and Geometric Probability}. Here is how it goes:
\bq\noindent
Above all the book should remind all of us that Probability Theory is measure theory with a ``soul'', which in this case is provided not by Physics, or by games of chance, or by Economics but by the most ancient and noble of all mathematical disciplines, namely Geometry.
\eq
\eq

\section{25-07-03 \ \ {\it Relative Onticity} \ \ (to R. {\Schack})} \label{Schack64}

There are two points I like in your last note:
\brs
What makes all this exciting for me is the fact that there appears to
be a straightforward logical argument from your ``Is the Moon
There?''\ article to de Finetti's position. Particles don't carry
instruction sets. Hence $|x\rangle$ is subjective.
\ers
and
\brs
I believe that you get an infinite regress only if you try to find
the objective ground behind your probability assignments. This is
futile. Your infinite regress argument is an argument FOR the
subjective viewpoint. The latter is not solipsism, but simply the
realization that our description in terms of maths is not the same
thing as the external reality.
\ers
Let's see what kind of impact they'll have on Mr.\ {\Mermin} now.

I thought a little more about our phone conversation on my walk to
work this morning---in particular, your point about, ``Why not let
Mr.\ {\Caves} (under the appropriate conditions) act AS IF his
Hamiltonians are real?  What's to be lost by it?''  I think the thing
to be lost is our hard won category distinction between A) the
probability function $P(x)$ and B) the values $x$ of its argument.
I.e., the beliefs and the facts.  Even when the Bayesian gives an
``as if'' theorem, like with the de Finetti theorem, he never drops
this category distinction.  I.e., he does not confuse the data he is
gathering (which with the assumption of exchangeability will allow
him to asymptotically settle on a product probability distribution)
with the updated belief.

At best, the ``as if'' theorems can give us something like the new
semi-ontic category that Shimony seeks:  the potentia.  That is to
say, there are ``actualities'' represented by the CLICKS, but then
there are the ``potentia'' represented by the probabilities of which
we can act AS IF they are objective.

I continue to think that it is of ultimate importance for this
program to not let Mr.\ {\Caves} go only halfway.  Where we are heading,
I think, is toward a wholly new kind of ontology---despite what looks
to me like your present attraction to sensationalism.  For that task,
we just shouldn't stop halfway.  I see us discovering and finding
convincing explanation for the idea that the world is truly on the
make.  To stop with the ``as ifs'' that we can all agree upon
(presumably after inculcation by the right physics community)---let
us say they really are Hamiltonians---strikes me overpoweringly as an
effort to once again introduce an overriding stasis to the universe.
The real, unchanging ground of the universe is {\em the\/}
Hamiltonian, full stop.  And what is a Hamiltonian?  In effect, it is
the potentia, or even a propensity.  (I.e., it is the generator of a
conditional probability which has been given an objective status by
fiat.)

Here's maybe another way of putting what I fear.  If we stop with
identifying the conditions under which one can act AS IF, then at
best we will draw a new picture of existing quantum mechanics.  We
will give new names to old prejudices.  It is simply not radical
enough for my tastes; when we solve the old problem of quantum
interpretations, we must get something new.  At the very least a new
world view, but I suspect something much more substantial and
technical than that.

But, anyway, why am I any better off with my Hilbert-space dimension?
Surely there is a sense in which it too is a subjective judgment.
True enough.  But then, why do I not see myself as falling into the
same trap as Mr.\ {\Caves}?  I want to claim that in opposition to the
Hamiltonian which is on the A) side of the category distinction
above, Hilbert-space dimension should be more rightly thought of as
on the B) side.  Hilbert-space dimension should be thought of as
taking the role of cardinality of the sample space in classical
probability theory.  And as you and I have already discussed, the
setting of a sample space (classically) is certainly a subjective
judgment.  But nevertheless the elements within the sample space play
the role of (potential) FACTS, not BELIEFS, after the setting is
made.  So, it is a relative onticity of a sort, but that is all I
have ever been trying to capture with my ``objective with respect to
the theory'' (or whatever phrase I used to use).

Enough.  I'll say more when I can say more.  Whereof one cannot
speak, thereof one must be silent \ldots

\section{28-07-03 \ \ {\it Popescu--Rohrlich Correlations, Gleason, Our Stuff} \ \ (to H. Barnum)} \label{Barnum7}

\bhb
Actually, maybe Chris and/or Joe's work, if I recall correctly, has
to do with perverted Gleason's theorems in which nonorthogonal sets of
vectors (with specific angle-sets) are identified with measurements
having mutually exclusive outcomes.
\ehb

You're right, but that was mostly work with Nurit Baytch.  Sadly, I never got off my duff to publish it, and now it's not particularly timely (at least in my own head).  Also the CG Fire put an end to the records of the calculation I had, which involved a lot of nasty stuff about spherical harmonics.

I think the result was the following.  Consider qutrits.  Suppose for whatever weird reason you wanted to consider three-outcomed ``measurements'' whose outcomes are associated with vectors that are angles $\alpha$, $\beta$, and $\gamma$ (all fixed) away from each other.  In the usual notion of von Neumann measurement $\alpha=\beta=\gamma=90^\circ$.  Now, try to build a concept of frame function for this kind of measurement and try to prove a Gleason theorem.  The upshot was that the {\it only\/} frame function existing in such cases were the uniform ones (i.e., flat functions), except in the particular case that $\alpha=\beta=\gamma=90^\circ$ in which one recovers the usual Gleason theorem.  [The style of problem is captured in my big samizdat, \quantph{0105039}, pages 86 through 88.]

I guess I'm not as interested in the result as before.  It came from a time when POVMs looked like particularly arbitrary structures to me.  Now though that I think I understand the deeper connection between the structure of measurements and Bayes' rule, I guess I wouldn't be going down such lines.

\section{28-07-03 \ \ {\it Tickles and Wiggles} \ \ (to H. Barnum)} \label{Barnum8}

By the way, in the evenings before bed, I've been finally reading Thomas Nagel \ldots\ since both you and Tony Sudbery have pushed me toward it.  I hope to give you a report of my thoughts on it in the decently near future.

\section{28-07-03 \ \ {\it Understanding Buridan} \ \ (to R. {\Schack})} \label{Schack65}

\brs
I think this is usually quoted in discussions about free will versus
determinism. I like it because it illustrates the absurdity of
indeterminism.
\ers
Does this mean that deep underneath it all you're a determinist who
believes in instruction sets?  This worries me a bit.

\section{28-07-03 \ \ {\it New Twenty Questions} \ \ (to R. {\Schack})} \label{Schack66}

\brs
Presently I am very much under the antirealist spell of de Finetti.
The way I understand de Finetti, indeterminism is just as meaningless
as determinism. But maybe I am taking the first three paragraphs of
Probabilismo too seriously\ldots\
\ers

The lines below are from my paper \quantph{0204146}.  Part of
them, even I am in disagreement with now.  But the part that gives a
definition of indeterminism, I am still OK with.  Are you OK with
that, or do you still contend that it is meaningless?

\bq
\noindent
In choosing one experiment over another, I choose one context over
another.  The experiment elicits the world to do something.  To say
that the world is indeterministic means simply that I cannot predict
with certainty what it will do in response to my action.  Instead, I
say what I can in the form of a probability assignment.  My
probability assignment comes about from the information available to
me (how the system reacted in other contexts, etc., etc.).  Similarly
for you, even though your information may not be the same as mine.
The OBJECTIVE content of the probability assignment comes from the
fact that {\it no one\/} can make {\it tighter\/} predictions for the
outcomes of experiments than specified by the quantum mechanical
laws.  Or to say it still another way, it is the very existence of
transformation {\it rules\/} from one context to another that
expresses an objective content for the theory.  Those rules apply to
me as well as to you, even though our probability assignments {\it
within\/} each context may be completely different (because they are
subjective).  But, if one of us follows the proper transformation
rules---the quantum rules---for going to one context from another,
while the other of us does not, then one of us will be able to take
advantage of the other in a gambling match.  The one of us that
ignores the structure of the world will be bitten by it!
\eq

\section{28-07-03 \ \ {\it New Twenty Questions, 2} \ \ (to R. {\Schack})} \label{Schack67}

Different question, but again referring to this:
\brs
Presently I am very much under the antirealist spell of de Finetti.
The way I understand de Finetti, indeterminism is just as meaningless
as determinism. But maybe I am taking the first three paragraphs of
Probabilismo too seriously\ldots\
\ers
What do you say about {\Mermin}'s anti-instruction-set derivation?  Is
that now contentless (at worst) or superfluous (at best) from this
anti-realist view?

\section{28-07-03 \ \ {\it Your Newest Turn} \ \ (to R. {\Schack})} \label{Schack68}

I did truly get into quite a bit of trouble tonight, not getting home
until 7:30.  In general it's just been a very bad day for me.

Let me give you an example of something I don't think you can allow
yourself to say in your newest turn.

When I wrote my ill-fated note ``\myref{Mermin76}{Got It!},'' I wrote in Section 4 of
it:

\bq
\noindent 4) POVMs and radical pluralism.\smallskip

Now let me go into a bit of the metaphysics of this.  Here's a point
of view that I'm finding myself more and more attracted to lately.

I think it is safe to say that the following idea is pretty
commonplace in quantum mechanical practice.  Suppose I measure a
single POVM twice---maybe on the same system or two different
systems, I don't care---and just happen to get the same outcome in
both cases.  Namely, a single operator $E_d$.  The common idea, and
one I've held onto for years, is that there is an objective sense in
which those two events are identical copies of each other.  They are
like identical atoms \ldots\ or something like the spacetime
equivalent of atoms.  But now I think we have no warrant to think
that.  Rather, I would say the two outcomes are identical only
because we have (subjectively) chosen to ignore almost all of their
structure.

That is to say, I now count myself not so far from the opinion of
Ulfbeck and Bohr, when they write:
\bq
\noindent The click \ldots\ is seen to be an event entirely beyond law.
\ldots\ [I]t is a unique event that never repeats \ldots\ The
uniqueness of the click, as an integral part of genuine
fortuitousness, refers to the click in its entirety \ldots. [T]he
very occurrence of laws governing the clicks is contingent on a
lowered resolution.
\eq

For though I have made a logical distinction between the role of the
$d$'s and the $E_d$'s above, one should not forget the very
theory-ladenness of the set of possible $d$'s.  What I think is going
on here is that it takes (a lot of) theory to get us to even
recognize the raw data, much less ascribe it some meaning.  In {\Marcus}
{\Appleby}'s terms, all that stuff resides in the ``primitive theory''
(or perhaps some extension of it), which is a level well below
quantum mechanics.  What quantum mechanics is about is a little froth
on the top of a much deeper sea.  Once that deeper sea is set, then
it makes sense to make a distinction between the inside and the
outside of the agent---i.e., the subjective and the objective---as we
did above. For even in this froth on the top of a deeper sea, we
still find things we cannot control once our basic beliefs---i.e.,
our theory---are set.

Without the potential $d$'s we could not even speak of the
possibility of experiment.  Yet like the cardinality of the set of
colors in the rainbow---Newton said seven, Aristotle said three or
four---a subjective judgment had to be made (within the wide
community) before we could get to that level.  If this is so, then it
should not strike us as so strange that the raw data $d$ in our
quantum mechanical experience will ultimately be ascribed with a
meaning $E_d$ that is subjectively given.  (I expressed some of this
a little better in a note I wrote to David last month; I'll place it
below as a supplement.)  More particularly, with respect to the EPR
example above, it should not strike us as odd that the phenomenon
comes about solely because of an interpretive convention we set:  All
quantum measurement outcomes are unique and incomparable at the ontic
level.  At least that's the idea I'm toying with.
\eq
to which you replied at the time:
\brs
Section 4), I like. It highlights the chasm that exists between our
approach and the many-worlders and decoherence people. We all agree
that a click is not an elementary phenomenon. They want to reduce it
to something more fundamental. We say it's irreducible (which does
not mean that we cannot analyze a measurement apparatus or
decoherence of a quantum register in as much depth and detail as
anybody else). Your metaphysical bit is nice in that it makes clear
that ``irreducible'' is not the same thing as ``elementary''. As you
say it,
\bq
\noindent All quantum measurement outcomes are unique and
incomparable at the ontic level.
\eq
\ers

In your newest turn, such discussion would be
``meaningless''---correct? You would no longer allow yourself to
contemplate what the CLICK $d$ might be in its own essence.  You
would no longer say that ``successive clicks with a single value $d$
are truly the same,'' nor would you contemplate that ``successive
clicks with a single value $d$ may actually be truly different.''

That is to say, in this new philosophy you are toying with, if
something (i.e., an event, a fact, etc.)\ is not a ``hook'' upon
which a probability can be conditioned, you are not willing to speak
of it. It is meaningless---I believe you say.  You don't even let
yourself conjecture about the stuff that is out there independently
of us. (Or at least I don't see how you're going to be able to do
this with such a strongly positivistic line.)

As I tried to express today, and as I'm trying harder to articulate
now, I guess I don't like that.

Pragmatists are not positivists, but more opportunists.  {\James}
allowed personal religions to be parts of reality, and I guess I'm
inclined to that.

The part of de Finetti's introductory section that I do like is his
organicist take on scientific theories.  I'm inclined to view the
PRESENT scientific theory of any type as part of the specification of
our species.  Such theories are part of the account of our being at
the moment; they express our possibilities and our limitations.  By
this account, for instance, Hamilton and Lagrange belonged to ever
slightly different species than you and me.  Silly, huh?  I don't
think so:  For what's important about this is that one sees by it
that a theory (and the entities within it) are every bit as real or
unreal as biological species.  Whatever they are, they have a
temporary hold on our description of things that cannot be denied.
Are they all and only AS IF statements?

I guess I don't think so.  In other words, I guess I'm saying that I
think you're going too far with your ASIFism.  As far as I can see,
all a representation theorem (say of the de Finetti type) can give us
is the conditions under which we can act AS IF our impredictability
is coming from a true but unknown PROPENSITY.  In particular, what
the AS IF theorems do not give are the sample spaces.  They are
always set before these representation theorems can be posed at all.
And, in that way, I say sample spaces obtain a status---though
subjective in the character of how they are set---that is somehow
different from what a representation theorem can give.

My goal continues to be ultimately realist in tone.  All I have ever
wanted to do in these last couple of years is strip away the
objective character of all probability statements and any part of
quantum mechanics that smells of being a probability statement in
disguise.  I have never imagined (nor do I think we have any warrant
from anything technical within probability theory) that the whole
structure of QM will go up in smoke in the process.

By the way---on a slightly different subject---let me readdress one
of the points in your Buridan's Ass note again:
\brs
Well, Buridan's ass stands in front of two buckets of hay, neither of
which looks to him any better than the other. Hence the poor ass
remains standing there and eventually dies. I think this is usually
quoted in discussions about free will versus determinism. I like it
because it illustrates the absurdity of indeterminism. The state
$|R\rangle +|L\rangle$ is symmetric with respect to left and right.
If nature really IS in a symmetric state with respect to left and
right, then what breaks the symmetry? If it is our state of belief,
however, that is symmetric, there is no problem. The ass chooses one
or the other, it's me who has no clue which one it will be. Either
one or the other detector goes click, it's me who has no clue which
one it will be.
\ers

I would say the very lesson of Bell and Kochen--Specker is that we
cannot (or rather should not) act AS IF nature is not at a juncture
in the quantum measurement setting.  ``Unperformed Measurements Have
No Outcomes.''

\section{29-07-03 \ \ {\it Wiggles and Bait} \ \ (to H. Barnum)} \label{Barnum9}

More about Nagel coming soon.  Sudbery has prodded me further out of my slumber (yesterday) \ldots\ and I can feel those natural defense mechanisms kicking in!  The email was starting to form in my head as I was walking in to the office today!

\section{29-07-03 \ \ {\it Please Do} \ \ (to J. Honner)} \label{Honner2}

\bjh
In 1995-7 I started writing what I thought was a half interesting
book {\bf [Bohr, Einstein, Bell, and the Feminist Critique of Physics]} \ldots\ on physics and the unconscious basically \ldots\  but after three
rejections of the core article I pulled my head in and changed my
life.  If you can bear reading the feminist critique of physics and a
fairly devastating set of reflections on Einstein, I'd be happy to dig
out a file and send it to you.
\ejh

Yes, please do send it!  Chances are I won't read it immediately, but it'll be in the queue then (and it may get a citation in the ``Activating Observer'' compendium).  When I do get to it, chances are I'll give you some feedback.

Also, by the way, if you could give me a complete list of all your quantum publications that would be most helpful.

\section{30-07-03 \ \ {\it Title} \ \ (to C. King)} \label{King7}

\noindent ``Two Characterizations of Complete Positivity that Evoke No Imagery from the Everett Interpretation of Quantum Mechanics''

\section{30-07-03 \ \ {\it Convictions, Courage, Clear Thinking} \ \ (to R. {\Schack})} \label{Schack69}

Your note was so chock-full, I had quite a sleepless night.  How I
wish I had time to reply to you right now in great detail---it is
something I need internally, there is no better therapy for me than
composing a note---but it is going to have to wait most likely until
I am in Germany.

I think you were right on most (or at least many) counts.  In
particular, I think this discussion represents a lack of courage on
my part, along with a small lack of clear thinking:  The former
likely caused the latter.  Funny how many times I have accused {\Carl}
(and sometimes you) for not having the courage of your convictions.
Stones and glass houses.

Also there is a slight problem of emphasis and particular choices of
words that might be getting in our way---i.e., that may be one of the
things that helped shunt us from complete and immediate agreement.
These things need fleshing out.

Mostly I want to say for the moment that your note inspired me, and I
think (maybe unseen to you) it gives me a way to have my cake and eat
it too.

I will be back with a vengeance in a few days (for hopefully an
adequate mix of email and verbal communication) \ldots

In the meantime, though, would you think about the tension in these
three statements of yours:
\brs
Well, Buridan's ass stands in front of two buckets of hay, neither of
which looks to him any better than the other. Hence the poor ass
remains standing there and eventually dies. I think this is usually
quoted in discussions about free will versus determinism. I like it
because it illustrates the absurdity of indeterminism. The state
$|R\rangle +|L\rangle$ is symmetric with respect to left and right.
If nature really IS in a symmetric state with respect to left and
right, then what breaks the symmetry? If it is our state of belief,
however, that is symmetric, there is no problem. The ass chooses one
or the other, it's me who has no clue which one it will be. Either
one or the other detector goes click, it's me who has no clue which
one it will be.
\ers
and
\brs
NOOOOOOO! You are joking, aren't you? The direct lesson of Bell and
Kochen--Specker is indeed that we should not act AS IF unperformed
measurement had outcomes. To conclude that nature is at a juncture is
adding a lot of baggage, however. I am sure de Finetti would be
rotating in his grave if he read this!
\ers
and
\brs
What de Finetti's anti-realism attacks is the prejudice that to the
terms of our description literally corresponds some real stuff out
there in nature. Isn't it ridiculous, he asks, to conclude that
nature is in the situation of Buridan's ass, only because we aren't
certain about what will happen next? Does that mean de Finetti
believes in instruction sets? I think this would do him injustice.
\ers
I'm with you on the middle one at the moment.  But the three together
sure look like a tense mix to me.  What actually would do de Finetti
justice on this issue?  It is not clear to me at all.

There's a lovely passage in an 1872 diary entry of William {\James}:  I
wish I had it here to get the exact wording.  It was something of the
order, ``my first act of free will shall be to believe in free will
\ldots\ I will give it a shot for one year and see where that gets
me.''

\section{01-08-03 \ \ {\it Mortified, Yes} \ \ (to N. D. {\Mermin})} \label{Mermin100}

\bdm
SHPMP Sept 2003 just arrived.

Well I knew you hated {\em Copenhagen Computation}, but making it the
one article in the collection that you don't say a word about in the
Introduction seems a bit much, particularly since your theme is the
vindication of Bohr over Einstein.  Putting it last underlines the
omission.  It will be clear to readers that I foisted it on you.  (As
I remember I didn't want to send in anything, but you kept
insisting\ldots)

I love you anyway.  Actually I assume it was an accident ---
Freudian, of course --- and trust that you are properly mortified.
\edm

I very, very, very much apologize.  I feel like utter crap.  How I
could let that happen I do not know.  Jeff wrote the first draft, and
I jiggled some things in it, but that I did not notice the omission
is inexcusable.

All I can say now is that I apologize, and I've got to find a way to
make it up to you in some other aspect of life.

It is not true that I did not like your paper.  I liked it a lot, and
I especially liked your presentation at the {\Bennett}fest.  It seems to
me that you get to the essential point when you point out the
non-problematicity of the ultimate readoff:  it is no more or less
mysterious than reading off of a piece of information from a computer
screen.  Either they are both deep problems, or neither of them are.
Showing that that is the best way to teach quantum computing to CS
people makes a deep statement \ldots\ and you're in tune with it.

Please do forgive me.  It wasn't intentional, and it wasn't even
Freudian.

\section{01-08-03 \ \ {\it Done}  \ \ (to A. Peres)} \label{Peres51}

Thanks!  What will you speak on in Aarhus?  Do you know if Petra has prepared a poster for the poster session?  I would recommend to you, by the way, Rob {\Spekkens}'s talk.  I think it is particularly deep, and the work has not been posted on {\tt quant-ph} yet.  Here is what I wrote to the organizers in my efforts to get him a speaking slot (at a late date):  [See 01-07-03 note ``\myref{Gill2}{Rob {\Spekkens}}'' to R. D. Gill, K. M{\o}lmer, and E. S. Polzik.]

\section{01-08-03 \ \ {\it {\Spekkens} and Letter}  \ \ (to A. Peres)} \label{Peres52}

\bap
I believe I saw something like you told me about {\Spekkens} in {\tt quant-ph},
or in {\bf Found.\ Phys.\ Lett.}, or elsewhere. I can't remember. I regret I
have no patience for these games.
\eap

But you should.  The whole point of {\Spekkens}'s toy model is to bolster our point that a quantum state is a state of knowledge.  The tactic is to show the extreme similarities between quantum mechanics and Liouvillian mechanics---a point I first learned from you.  The thing that is new in {\Spekkens}'s game is that he not only uses Liouville mechanics, but he adds a further restriction that the distributions cannot be too peaked.  With that he gets a theory that looks ever so much more (but not completely) like quantum mechanics.

The point to be made for people like Gisin and Jozsa who don't find quantum mechanics as palatable as they should:  Quantum mechanics is not so very weird and different from plain old Liouville mechanics.  That community should quit beating their heads trying to find nonlinear evolution equations and such.

I doubt you saw something like this (or at least of the prettiness of {\Spekkens}'s result) in {\sl Found.\ Phys.\ Lett.}

Give it a chance.  It is really very good and very creative work.

\section{12-08-03 \ \ {\it Me, Me, Me} \ \ (to N. D. {\Mermin} \& R. {\Schack})} \label{Mermin101} \label{Schack70}

Me, me, me; it's always about me! ---Yes.  But nonetheless it is
simply not solipsism.  Let me explain.

I guess I was actually fortunate today:  For the second time in a
month, I was called a solipsist by one of my friends.  (This time the
accuser was Howard Wiseman.)  On top of that, Asher Peres gave a talk
this morning that made me cringe, saying things like, ``When no one
performs a measurement, nothing happens [in the world].''  The
combination of these two bad experiences caused me to wander the
streets of Aarhus this afternoon in spite of the horrible heat.  I
suppose I needed to find a way to sweat the poisons from my body.

The fortune in this is that it caused me once again to strive for a
clearer and more consistent form of expression.  I want to try to
capture some of that in this note.  Mostly it is about not allowing
oneself to get hung up in someone else's (inconsistent) expectations
for what quantum theory ought to be.

In our 2000 opinion piece in Physics Today, Asher and I wrote:
\bq
\noindent
   The thread common to all the nonstandard ``interpretations'' is the
   desire to create a new theory with features that correspond to some
   reality independent of our potential experiments. But, trying to
   fulfill a classical worldview by encumbering quantum mechanics with
   hidden variables, multiple worlds, consistency rules, or spontaneous
   collapse, without any improvement in its predictive power, only
   gives the illusion of a better understanding. Contrary to those
   desires, quantum theory does {\it not\/} describe physical reality.
   What it does is provide an algorithm for computing {\it
   probabilities\/} for the macroscopic events (``detector clicks'')
   that are the consequences of our experimental interventions. This
   strict definition of the scope of quantum theory is the only
   interpretation ever needed, whether by experimenters or theorists.
\eq
But that is misleading and trouble-making.  In the second to last
sentence---with the experience of three more years of thinking on
this subject---I so wish we had said something more to the tune:
\bq
\noindent
   What quantum theory does is provide a framework for structuring MY
   expectations for the consequences of MY interventions upon the
   external world.
\eq
At least that is what the formal structure is about.  There is no
``we,'' there is no ``our.''  At this level of consideration, quantum
theory has nothing to do with intersubjective agreement. (By the way,
I'm not fooling myself:  Of course we could not have said what I said
above without restructuring the whole article---it would have opened
a can of worms!  I just want to try to do the idea better justice
right now.)

Here it is:  Any single application of quantum THEORY is about ME,
only me.  It is about MY interventions, MY expectations for their
consequences, and MY reevaluations of MY old expectations in the
light of those consequences.  It is noncommittal beyond that.  This
is not solipsism; it is simply a statement of the subject matter.

Is there any contradiction in this?  I say no, but how do I get you
into a mindset so that you might say the same?  Maybe the best way to
do this is to run through a glossary of quantum terms as I did once
before \ldots\ but now with all the latest slant.

\begin{itemize}
\item
SYSTEM:  In talking about quantum measurement, I divide the world
into two parts---the part that is subject to (or an extension of) my
will, and the part that is beyond my control (at least in some
aspects).  The idea of a ``system'' pertains to a part beyond my
control.  It counts as the source of my surprises, and in that sense
obtains an existence of its own external to me.  (Point 1 against
solipsism, but I will return for another.)

\item
POVM:  In the theory, this counts as an extension of my will.  It
counts as a freely chosen action on my part.  The whole concept of a
``measuring device'' as something distinct from me---I am now
thinking---just gets in the way.  It is a point that {\Pauli} made, but
I am coming ever more to appreciate it.  A ``measuring device'' is
like a prosthetic hand; its conceptual role is for the purpose of
recovering from our natural incapacities and, thus, might as well be
thought of as part of ourselves proper.  I perform a POVM on a
system---captured mathematically by a set---and one of its elements
comes about as a consequence.

\item
QUANTUM STATE:  As usual, the catalog of MY expectations for the
consequences of MY actions (i.e., POVMs) \ldots\ but now with
absolute, utter emphasis on the MY.

\item
UNITARY READJUSTMENT:  I'm talking here about the readjustment
appearing in Eq.~(95) of my paper \quantph{0205039}.  This, like
a quantum state, also captures a belief or expectation.  Its purpose
is to quantify the extent to which I feel the need to deviate from
Bayes' rule after learning the consequence of my action.  This is
what takes account of the nonpassive nature of MY interventions.

\item
QUANTUM DYNAMICS:  This is the unitary readjustment (or mixture of
decompositions and unitary readjustments) that I judge I ought to
apply if my action on the system is passive, i.e., if my POVM is the
singleton set.  It is how I readjust my expectations when I am
learning nothing.
\end{itemize}

Summing up the glossary, I would say quantum theory in its single
user implementation is about ME.  I act on the world and it reacts in
a way unpredictable to me beyond the expectations I build from MY
quantum state (about the system).

Why is this not solipsism?  Because quantum theory is not a theory of
everything.  It is not a statement of all that is and all that
happens; it is not a mirror image of nature.  It is about me and the
little part I play in the world, as gambled upon from my perspective.
But just as I can use quantum theory for my purposes, you can use it
for yours.  Thus, if I had not been seeking dramatic effect above, I
should have more properly said, ``Any single application of quantum
THEORY is about the ME who applies it.''  (Don't correct my English.)
When David {\Mermin} is a practitioner of quantum theory, what the
theory does is provide a framework for structuring HIS expectations
for the consequences of HIS interventions upon HIS external world.
\ldots\ And that is Point 2 against solipsism.

Recall the definition of solipsism I dredged up from the Encyclopedia
Britannica:
\bq
\noindent
   in philosophy \ldots\ the extreme form of subjective idealism that
   denies that the human mind has any valid ground for believing in
   the existence of anything but itself.
\eq

It seems to me we have plenty of valid ground for believing in the
existence of something besides ourselves:  It comes from all the
things we cannot control.  Indeed, as already emphasized, for those
things we can control, we might as well think of them as extensions
of ourselves.  Thus, to my mind, quantum theory already gives a
karate chop to solipsism because of the indeterminism it entails:
With each quantum measurement there is immediately something beyond
my control.

Beyond Point 1, though, there is Point 2.  It is a question of
finally getting straight what should and should not be in the purview
of the theory.  In this account, quantum theory is a theory of
personal action (and reaction).  The law-of-thought aspect of it
comes out with respect to each individual who uses it.  The textbook
poses an exercise that starts out, ``Suppose a hydrogen atom is in
its ground state.  Calculate the expectation of \ldots\ blah, blah,
blah.'' One might think it is asking us to calculate some objective
feature of the world.  It is not.  It is only asking us to carry out
the logical consequences of a supposed state of belief and a supposed
action that one could take upon the system.  And here's the clincher
about Bayesianism.  Just as no student in his right mind would find
it worthy to ask why the textbook writer posed the problem with the
ground state rather than the first excited state, no quantum theorist
should make a big to-do about it either.  It is simply an assumed
starting point.  An agent in the thick middle of a quantum
application can no more ask where he got his initial beliefs from,
than a pendulum can ask where it got its initial conditions from. The
cause of bottom-level initial conditions is ALWAYS left unanalyzed.
If such was not a sin in Newtonian mechanics, it should not be a sin
in a Bayesian formulation of quantum mechanics.

So, it seems to me, if anything, the Bayesian account of quantum
theory is essentially the opposite of solipsism.  Rather than a unity
to nature, it suggests a plurality.  An image that might be useful
(but certainly flawed) comes from Escher's various paintings of
impossible objects.  The viewer would initially like to think of them
as two-D projections of a three dimensional object; but he cannot.
Now imagine how much worse it would get if we were to have two
viewers with two slightly different paintings, each purporting to be
a different perspective on ``the'' impossible object.  Since neither
viewer can lift from his own two-D object to a three-D one, there is
no way to unify the pictures into a single whole.

Yet we live in one world, you say.  Maybe.  But, you should remember
that these quantum states we speak of are not perspectives.  They are
personal possessions.  To paraphrase Tilgher's quote at the beginning
of de Finetti's Probabilismo,
\bq
\noindent
   A quantum state is not a mirror in which a reality external to us
   is faithfully reflected; it is simply a biological function, a
   means of orientation in life, of preserving and enriching it, of
   enabling and facilitating action, of taking account of reality
   and dominating it.
\eq

``Are there other minds beside your own?,'' Howard Wiseman asks.  If
a mind is what it takes to write down a quantum state, then why not?
``If you leave the origin of the quantum state unanalyzed, why would
two minds ever agree on anything?''  That is the issue of
intersubjective agreement---something thankfully we can study within
the context of quantum theory.  But the first thing to get straight
is why the single user of quantum theory uses the very structure.
What is it precisely that he is believing of the world and his place
in it that leads him to the choice of quantum theory as his law of
thought?

That is, it is about ME and what I believe.  But what do I believe?
That's the research program!

\section{13-08-03 \ \ {\it The Tense Mix} \ \ (to R. {\Schack})} \label{Schack71}

Now that I'm back in Munich, I've tried to become re-engaged with the
conversation we left on, but I'm having difficulty.  I've read your
30/7/03 note ``levels'' three times over since yesterday, and I'm not
sure what I can add at the moment.  Clearly I've absorbed some of it
into my mentality.  In particular, I like the scolding you gave me:
\brs
NOOOOOOO! You are joking, aren't you? The direct lesson of Bell and
Kochen--Specker is indeed that we should not act AS IF unperformed
measurement had outcomes. To conclude that nature is at a juncture is
adding a lot of baggage, however. I am sure de Finetti would be
rotating in his grave if he read this!
\ers

You are right; the best I can say is that the lesson of quantum
mechanics is that we should not act AS IF unperformed measurements
had outcomes.  And that is good enough for me.  Like with William
{\James}, I will act as if nature is undetermined and see where that
leads me.  I can't do more than that.  (In fact, I have said it many
times that one may never be able to disprove either many-worlds or
Bohmian mechanics.  I only bank that those trains of thought will not
lead in any productive directions.  I suppose I suspended such
carefulness as my passions started to flare in our discussion.)

But as you might guess from the paragraph I just wrote, I still see
something of an uncomfortable tension in what you've written.  For
instance, I don't even understand your very next sentences:
\brs
I think de Finetti was both anti-realist and anti-indeterminist.  The
Buridan's ass passage illustrates this beautifully.  Your position is
clearly not anti-indeterminist. Be careful who you recommend
Probabilismo to!
\ers
Also I guess I don't like this emphasis of yours on the
MEANINGLESSNESS of various statements.  It doesn't ring true to me,
especially with the determinist/indeterminist issue.  It strikes me
as perfectly useful to take a stance on the issue (of course in the
AS IF sense that you have emphasized); and in that way it is not
meaningless.  That, I think, is why I brought up religions in an
earlier note; religions are never verifiable or falsifiable, but
faith in them is not meaningless.

It seems to me clear---and I don't think your note has changed
this---that when we practice quantum mechanics, we are implicitly
stating some beliefs about the world (as it is independently of the
agent).  You are right that we cannot move from that step to saying
that that is actually the way the world is.  Nonetheless we are
expressing {\it beliefs\/} about the world without our presence.  The
issue is to make those beliefs explicit---and that is your program of
AS IF\@.  I am OK with that.  To put so much emphasize on the
meaningless of this or that, though, repels me slightly.  That may be
what caused me to swing the pendulum too far in the opposite
direction.

I'm really looking forward to your visit to Munich.  I hope we can
find a way to get together.  Aarhus was a useful meeting:  Klaus
M{\o}lmer, Richard Gill, Ray Streater, Vladimir {\Buzek}, Mauro
D'Ariano, Hideo Mabuchi, and Howard Wiseman showed some decent
interest in the program.  But I must say, the divergences between us
and our desires and Rob {\Spekkens} are becoming clearer \ldots\ and
more disappointing.  I see him as having the potential to cause real
damage.  It really would be useful to have a more thoughtful critique
of his ``classical interpretation of probability over a fundamental
event space'' position.  Bernardo and Smith's dismissal is simply too
curt.

Somehow over the course of this meeting, I've decided the time is
right to try to put together a paper explicitly and solely on the
Penrose argument.  The little title I've toyed with in my head has
been ``On Quantum Certainty.''  What would you think?  Might we throw
in on it together?  (I guess I was partially induced to this by
{\Spekkens}'s assertion that we still have no adequate reply to Penrose
without the picture he is trying to construct \ldots\ based on
ignorance of an ontic state.)  Anyway, I started thinking that that
might be riper and more ready to go than a critique of the
Horodeckis.  Also it strikes me as paving the ground for all the
other discussions, and worthy of its own thorough treatment.

This is the first of a couple of notes I'll be writing you today.
Another one will be coming down the tubes in an hour or so.

\section{13-08-03 \ \ {\it Renouvier} \ \ (to R. {\Schack})} \label{Schack72}

I'm back a little faster than I thought I would be.  In this note, I
just want to send you a passage that I scanned into my computer from
a book on Charles Renouvier---the philosopher who convinced William
{\James} to try out free will.  There is something about the argument in
the last two paragraphs of page 2 that takes me---but I want to see
it made more rigorous, or stated better.  Also I hope you are as
impressed by Footnote~\ref{RightOne} as I was.  If those passages are
not too far out of context, they may show that he is going down the
same lines as de Finetti and us.

I'd love to get my hands on anything more technical to do with
Renouvier, but apparently there is almost nothing written on him in
the English language.  The only article I have been able to unearth
that might possibly be in some library in your university is: S.
Hodgson, ``M. Renouvier's Philosophy,'' Mind {\bf 6}, 31--61,
173--211 (1881). If you have time, could you check on that before
Munich?

\begin{center}
From: William Logue, {\sl Charles Renouvier, Philosopher of Liberty},
\\ (Louisiana State University Press, Baton Rouge, 1993), pp.~86--92.
\end{center}
\bq
Under the influences of discussions with his friend Jules Lequier, he
[Renouvier] became convinced of the reality of human free will and
its central importance for the understanding of everything
else.\footnote{Renouvier's account of his ``conversion'' to free will
is in the last part of Vol.\ II of the {\sl Esquisse}. Lionel Dauriac
(``Les Moments de la philosophie de Charles Renouvier,'' {\sl
Bulletin de la soci\'et\'e francaise de philosophie}, IV [1904], 23)
defined the high point of Lequier's influence---the writing of the
{\sl Deuxi\`eme essai}---as one of four ``moments'' in Renouvier's
philosophical development.} This conviction came to Renouvier while
he was still deeply under the influence of his first contact with the
Saint-Simonians. He experienced, not an overnight liberation from
their deterministic viewpoint, but a more gradual readjustment of his
views, which became a complete detachment from them only after 1851.
Perhaps the failure of the socialist movements in 1848, rooted as
they were in the would-be scientific philosophies of the preceding
three decades, finally persuaded him of the dangers of rejecting free
will.\footnote{Mouy ({\sl Id\'ee de progr\`es}, 43) argues that the
disappointments of 1848 played a key role in shaping Renouvier's idea
of liberty. For Renouvier, free will came to be seen as the ultimate
basis of political liberty ({\sl Deuxi\`eme essai}, 551). See {\sl
Histoire}, IV, 431, and especially {\sl Esquisse}, II, 382, and {\sl
Deuxi\`eme essai}, 371{\it n}1.} Alienated from political life during
the Second Empire, he would spend nearly two decades in the
construction and elaboration of his philosophy of liberty,
establishing its foundations and exploring its consequences.

Renouvier was aware that for a long time the question had been of
mainly religious significance: whether man's salvation depended on
free will or on predestination.\footnote{Renouvier saw free will as
one of the basic concepts of both philosophy and Christian doctrine
({\sl Histoire}, IV, 277).} This debate had reached its peak, in both
vehemence and subtlety, in the famous exchange between Erasmus and
Luther in the sixteenth century. The emergence of a secular debate
over free will was a result of the rise of the scientific worldview
in the seventeenth century. The ascendancy of the idea that the world
was governed by invariable laws, taking the role previously occupied
by an all-powerful, all-knowing God, seemed to leave less and less
room for the view that man was somehow an exception to the general
rule. The most heroic task for the modern philosopher was to find a
means of validating science and free will simultaneously, and the
most heroic effort of the eighteenth century was that of Immanuel
Kant. But for many in the next century, it seemed that Kant had saved
free will only at the cost of making it irrelevant.\footnote{What
does it matter to man if he has freedom in the world of the noumena
if his world of phenomena is entirely determined? Renouvier later
felt that Kant held on to free will solely for the sake of morals
while not really believing in it ({\sl Quatri\`eme essai}, 35--36).}
Fichte tried to rescue Kantian philosophy from this unhappy outcome,
but in the general opinion (only recently challenged by Alexis
Philonenko, Luc Ferry, and Alain Renaut) his effort led to the
fairyland of absolute idealism, denying reality to the material
world.\footnote{Renouvier praised Fichte as a defender of freedom and
criticized him as a mystic ({\sl Quatri\`eme essai}, 46). See Alain
Renaut, {\sl Le Syst\`eme de droit: philosophie et droit dans la
pens\'ee de Fichte\/} (Paris, 1986); Luc Ferry, {\sl Le Syst\`eme des
philosophies de l'histoire\/} (Paris, 1984), Vol.\ II of Ferry and
Renaut, {\sl Philosophie politique}; Alexis Philonenko, {\sl La
Libert\'e humaine dans la philosophie de Fichte\/} (Paris, 1966).}

Against the rising tide of determinism, Renouvier would try to show
that Kant could be the launching pad for a defense of free will that
would maintain its practical relevance and demonstrate its
compatibility with natural science, properly understood. He did not
claim to be presenting any new arguments in favor of free will; he
felt they were in any case unnecessary.\footnote{Renouvier was
concerned to establish a rationalist and not an empiricist view of
science. He saw free will as perhaps the main issue dividing the
rationalists and empiricists ({\sl Histoire}, IV, 262). He indicated
that there had been no new arguments in favor of free will since Kant
and Rousseau ({\sl Esquisse}, I, 280).} Renouvier's reasons for
coming to the defense of free will were partly shared with Kant and
partly his own. As we have seen in the previous chapter, the shared
part was the most familiar: a concern for the connection between free
will and moral behavior. Free will was for Kant the essential basis
of {\it practical reason}; without it, the whole idea of moral
obligation ceased to have meaning. For Renouvier, this consideration
remained central. Without moral responsibility, man would not be
distinct from the rest of the animal kingdom, and the whole of
civilization would be meaningless. But this was not the sole basis
for his concern with free will, and this additional concern moved
Renouvier beyond Kant and Fichte, bringing him closer to our own
time.\footnote{Renouvier saw Kant's German disciples as having
abandoned liberty for determinism, optimism, and pantheism ({\sl
Histoire}, IV, 467).}

It is not just the moral aspect of civilization that hangs on the
reality of free will, in Renouvier's opinion, but the whole of our
intellectual life. Free will is also the foundation on which
philosophy and the natural sciences rest.\footnote{See {\sl
Deuxi\`eme essai}, 227.} Without free will, our ability to know
anything, whether about man or about nature, is fatally undermined.
Scientists do not need to believe in free will, and as he knew, they
prefer to avoid this sort of question. In practice, they can
legitimately do so because in their narrow spheres of inquiry they
have developed techniques of investigation that work even when the
scientist is unconscious of the fundamental assumptions on which his
method rests. But without free will, the certainty of scientific
truths becomes illusory; a consistent determinism must lead to a
profound skepticism.\footnote{{\sl Histoire}, IV, 399; {\sl
Deuxi\`eme essai}, 327.} Renouvier would never despair of convincing
the scientists that just as our concepts of right and wrong depend on
free will, so do our concepts of true and false. Indeed, without free
will, we could not even talk sensibly about things being true or
false.

If, as he pointed out, I hold such and such a view to be true and I
am determined by forces outside my control to hold this view, the
person who disagrees with me is equally determined by outside forces
in his position. If these mutually contradictory positions are
equally necessary, what grounds can we have for the certainty that
either view is the correct one?\footnote{{\sl Histoire}, IV, 399;
{\sl Deuxi\`eme essai}, 306--307 (according to Hamelin, {\sl
Syst\`eme de Renouvier}, 242). Necessity destroys truth: ``If
everything is necessary, error is necessary just as much as truth is,
and their claims to validity are comparable'' ({\sl Deuxi\`eme
essai}, 327). For a restatement of his argument that freedom is
essential to the certainty of our knowledge, see {\sl Esquisse}, II,
270--74; see also, {\sl Science de la morale}, II, 377.} If our
belief that our ideas are determined is itself determined, so is the
other person's belief in free will determined. Under these conditions
how could it make any sense to speak of one view as ``right'' and the
other as ``wrong''? If, on the other hand, our choices are free, I
may freely choose to believe in free will or in spite of the apparent
contradiction, to believe in universal determinism. Of the four
possible positions revealed by this analysis, the only one that can
serve as a foundation for a rational certainty in the truth of our
beliefs is to freely believe in freedom.\footnote{The four are (1) we
are determined to believe in freedom; (2) we are determined to
believe in determinism; (3) we freely believe in determinism; (4) we
freely believe in freedom. See {\sl Deuxi\`eme essai}, 478; {\sl
Histoire}, IV, 399; Hamelin, {\sl Syst\`eme de Renouvier}, 273--74.}
But as Renouvier insists, this means that we must give up any
pretension to the absolute certainty of our
beliefs.\footnote{\label{RightOne} ``Certitude is not and cannot be
an absolute. It is, as is too often forgotten, a condition and an
action of man: not an action or a condition where he grasps directly
that which cannot be directly grasped---that is to say, facts and
laws which are outside or higher than present experience---but rather
where he places his conscience such as it is and as he supports it.
Properly speaking, there is no certitude; there are only men who are
certain'' ({\sl Deuxi\`eme essai}, 390). For Renouvier's battle
against the idea of evident truths, see {\sl Histoire}, IV, 75, 261;
certitude is a sort of ``personal contract,'' ``a real contract that
a man makes with himself'' (Lacroix, {\sl Vocation personnelle},
114).} The truth of free will cannot be proved so that no rational
person can doubt it. It is a relative truth, like all our other
truths, but more important because it plants a relativism at the very
core of our thought.\footnote{{\sl Deuxi\`eme essai}, 309--10
(according to Hamelin, {\sl Syst\`eme de Renouvier}, 242). There
were, however, ``great probabilities in its [free will's] favor''
({\sl Deuxi\`eme essai}, 475). ``It ought to be a universally
accepted maxim that {\it everything that is in the mind is relative
to the mind}'' ({\sl Deuxi\`eme essai}, 390). Philosophy needs to
take into account the existence of disagreement among philosophers
({\it ibid}., 414). Renouvier's approach to the existence of these
disagreements is one of the distinctive features of his philosophy.}

Scientists, Renouvier thought, should have no difficulty
understanding and accepting this because science is built on an
awareness of the conditional character of our knowledge, an openness
to the discovery of new truths and the abandonment of old
ones.\footnote{On the use of hypotheses in science, see {\sl Premier
essai}, 200.} In fact, he had to admit, many scientists were still
under the sway of older metaphysical conceptions of truth, except in
the conduct of their personal research, and were unaware of any
inconsistency in their position.\footnote{Renouvier credited the
English empiricists, following Hume, with freeing science from the
metaphysical concept of cause ({\sl Histoire}, IV, 273).} Some who
were aware were evidently afraid that to admit that an act of belief
was at the base of scientific knowledge would risk undermining the
claim of science to objectivity and, even worse, open the way to the
proliferation of pseudoscientific beliefs.\footnote{This is the
concern of Parodi ({\sl Du positivisme \`a l'id\'ealisme}, 184--85),
who finds in Renouvier a dangerous fideism. So does Brunschwicg ({\sl
Progr\`es de la conscience}, 625). For Renouvier's praise of
Boutroux's argument that the contingency of the laws of nature is not
a threat to science, see {\sl Histoire}, IV, 673--74.} In reality,
pseudoscientific beliefs were already proliferating under the aegis
of the belief in determinism. Without a critical analysis of the
nature and limits of scientific knowledge, however, our intellectual
life is subject to a constant abuse of the name and prestige of
science.

The abuse of science takes many forms: the application of research
methods to fields where they do not apply, the application of
particular concepts to areas other than those where they originated,
the confusion of ``Science'' with the operations of particular
sciences. One of the main intellectual trends of the nineteenth
century, which Renouvier called {\it scientisme}, usually rendered as
``scientism'' in English, was the product of this abuse. Renouvier's
relativism does not justify believing in whatever we want to
believe.\footnote{Dauriac (``Moments de la philosophie,'' 30--32)
strongly makes this point. It would be interesting to compare
Renouvier's conclusion on this point with the similar view expressed
by Richard {\Rorty}, coming from a rather different direction.} It
insists on submitting our opinions to every possible test of logic,
experiment, and experience. But we have to admit that our logic, the
hypotheses on which our experiments are based, the schemas of thought
by which we interpret our experience, all rest ultimately on acts of
belief and not on absolute certainties.\footnote{{\sl Histoire}, IV,
692; on the need for faith in reason, even though such faith is in
itself not a rational act, see Popper, {\sl High Tide of Prophecy},
218--19.}

If free will is thus essential to both morals and science, just what
does he mean by it? Over the centuries, most of the debate over free
will has failed to advance our understanding because of the lack of
agreement about what is meant by the term.\footnote{Adler, {\sl Idea
of Freedom}.} I cannot solve that problem, but I think we will see
that Renouvier's view makes the issue more comprehensible.

Free will, for Renouvier, is a capacity possessed by human beings,
and only by human beings, that enables them to choose whether to
accept one idea or another, whether to perform one act or a different
one. It is thus a rejection of the doctrine that holds that all
events, mental or physical, are absolutely determined and cannot be
other than what they are.\footnote{See definition of free will in
{\sl Histoire}, IV, 337; on liberty as choice, see {\sl Deuxi\`eme
essai}, 466; on real alternatives, see {\it ibid}., 339. Renouvier is
rejecting a causal necessity, not analytic necessity, as in the
syllogisms of logical operations ({\sl Premier essai}, 232--36).}
Free will is also a rejection of the doctrine of chance, for it is an
active power and not the ``liberty of indifference'' so belabored by
determinists.\footnote{{\sl Deuxi\`eme essai}, 330--34, 336, 337;
Hamelin, {\sl Syst\`eme de Renouvier}, 242--43,249.} Chance is also
hostile to liberty, since it denies man a real power of decision.

The existence of free will requires a measure of indetermination in
the universe but could not exist if nature were essentially
indeterminate.\footnote{``Liberty does not require the complete
indetermination of particular future events, even of those that are
directly connected to it'' ({\sl Deuxi\`eme essai}, 459); see also
{\it ibid}., 357; Hamelin, {\sl Syst\`eme de Renouvier}, 244.} Our
acts of free will are the beginnings of chains of consequences and
would have no meaning if their consequences were not subject to cause
and effect. ``Free acts are not effects without causes; their cause
is man, the ensemble and fullness of his functions. They are not
isolated, but are always closely attached to the preceding condition
of the passions and of knowledge. {\it A posteriori\/} they seem
henceforth indissoluble parts of an order of facts, although a
different order was possible {\it a priori}.''\footnote{{\sl
Deuxi\`eme essai}, 359; see also {\sl Science de la morale}, II,
361--62.} The laws that permit us to say this is followed by that do
not admit of an infinite regression into the past, according to
Renouvier. Therefore, every series of phenomena---and indeed the
existence of any phenomena---must have a beginning that we cannot
explain in terms of antecedents.\footnote{{\sl Premier essai}, 237;
{\sl Science de la morale}, II, 360--61. Most scientists today reject
the idea that infinite regression is an absurdity; {\sl Esquisse},
II, 378--79. We cannot explain beginnings because they are by
definition at the limits of our possible knowledge.}

The act of creation of the universe is thus replicated (in a much
smaller way!)\ in every act of free will. Every act of free will is
the creation of a new series of phenomena, a series that would not
otherwise have existed.\footnote{{\sl Esquisse}, II, 196-97.} These
new chains of cause and effect are not simply the product of the
intersection of existing but independent series, as A.-A. Cournot
argued, for such intersections, though they appear random from the
point of view of any one of the colliding series, would be necessary
from a higher viewpoint.\footnote{A.-A. Cournot, {\sl
Consid\'erations sur la marche des id\'ees et des \'ev\'enements dans
le temps modernes\/} (Paris, 1973), 9--10, Vol.\ IV of Cournot, {\sl
Oeuvres compl\`etes}, ed.\ Andr\'e Robinet.} They must be new
beginnings, arising from a conjuncture in which, given the
antecedents, more than one consequence was possible: ``ambiguous
futures,'' Renouvier called them.\footnote{{\sl Deuxi\`eme essai},
210. ``The real indetermination of various phenomena envisaged in the
future'' ({\sl Premier essai}, 240). ``[A determinist] would renounce
everything called reflection and reason, for these functions do not
work without the consciousness of a {\it representative
self-motivation}, which is itself linked to an awareness of the {\it
real ambiguity of future conditions\/} before it takes action'' ({\sl
Histoire}, IV, 769). See also {\sl Troisi\`eme essai}, xlvii;
Hamelin, {\sl Syst\`eme de Renouvier}, 230.} Free will is the
capacity to opt for one or another of those futures.
\eq

\section{13-08-03 \ \ {\it Pragmatism and QM} \ \ (to G. Valente)} \label{Valente2}

\bgv
I hope [to be] finishing my dissertation in a couple of months and taking my degree in November. I think the title will read ``Probability and quantum meaning: Chris Fuchs' pragmatism in quantum foundations''. I've been trying to understand the development of your idea, especially in a philosophical light, but I'd also like to understand if I've been understanding \ldots

Very roughly, I recognized three instances (however connected) of pragmatism in your analysis:
\bq
\indent $\bullet$
one due to the mere acceptance of the theory; that implies, for example, the tensor product rule to represent a basic trail of the formalism and the quantum states to be pragmatic commitments (following a Bayesian approach to quantum probabilities and assuming Dutch-book as [the] criterion of consistency)\smallskip\\
\indent $\bullet$
one peculiar of the preceding itself in quantum foundations, which banishes any metaphysical interpretation and leads operatively to identify a core of ``reality'' as un-veiling (``reality is in the difference'') and to drop important but not fundamental axioms of the theory (for example the time evolution)\smallskip\\
\indent $\bullet$ one ``deeper'' inspired by William James, that suggests a notion of reality as construction.
\eq
\egv

Thank you for your note, and I am flattered about your dissertation.  I very much hope you will send me a copy when it is complete.

I think your three points are roughly on the mark, and I will enjoy seeing them developed.  Concerning the last two points, I will add a little material for your thought.  The first is an out-take from my samizdat {\sl Quantum States:\ What the Hell Are They?\/}\ posted at my website.  It delineates a little more directly than usual what I see as the great hope in melding quantum theory with Jamesian and Deweyan kinds of thoughts.  (It is a file I put together for several other people, called {\tt Construction.pdf}.)  The other piece is pasted below.  It is a letter written to David {\Mermin} and {\Ruediger} {\Schack} over the last three days (and sent off yesterday evening).  I think it best captures my newest way of explaining things, and I hope it clarifies some points.  You can feel free to use the thoughts in it in your development, even though it is not yet posted on my webpage.  [See 12-08-03 note ``\myref{Mermin101}{Me, Me, Me}'' to N. D. Mermin \& R. Schack.]

In general, I would say the most encompassing source of the thoughts I've already gone public with is the document {\sl Quantum States:\ What the Hell Are They?\/}\ (whose last entry is dated 29 June 2002).  Since then, some ideas have changed but not too many; mostly I've spent my time developing the Darwinism issue \ldots\ and that is somewhat uncoupled from the older material.  Anyway, I am hoping to finally compress the whole {\sl Quantum States:\ W.H.A.T.?\/}\ document into a proper paper before December, but we shall have to see.  Unfortunately it will almost certainly not happen before you write your dissertation.

Concerning your very last question:
\bgv
Please, can I ask you any bibliographical indication about this issue
or something new about Bayesian probabilities?
\egv

I would say the very best thing to read is Bruno de Finetti's paper, ``Probabilismo.''  A month ago, I read it for my fourth time in six years, and this time I have to say it really, really sunk in \ldots\ and I was ashamed that I had not so much absorbed it before.  It is a masterpiece (except for the fascist statements at the very end) and has played a very deep role in my latest collaborations with {\Ruediger} {\Schack}.  Here is the English translation:  Bruno de Finetti, ``Probabilism,'' Erkenntnis {\bf 31}, 169--223 (1989).  Immediately following that is an article by Richard Jeffrey titled ``Reading Probabilismo'' which it would also be good to have a look at, for it sets the context of the paper.  In case it is more accessible to you, here is the original reference in Italian:  Bruno de Finetti, ``Probabilismo,'' Logos {\bf 14}, 163--219 (1931).

Good luck with everything!

\section{14-08-03 \ \ {\it Uncertainty} \ \ (to W. E. Lawrence)} \label{Lawrence3}

I'm glad you got my book.  I hope you'll get some bits and pieces of insight from it.  The development of the idea (the Paulian one) still goes on and seems to be gaining some momentum in the wider community.  (I am a decently good salesman, even if not a decently good physicist.)  Have you seen the September issue of {\sl Studies in History and Philosophy of Modern Physics\/} devoted to quantum information?  A lot of good articles there; Jeff Bub and I were editors.  I'm excited that in some small number of years, we're really going to tie this up and move on to the next step of physics.

\section{14-08-03 \ \ {\it Behind Again} \ \ (to P. G. L. Mana)} \label{Mana4}

I'm in Munich at my in-laws' house, with some moments to finally catch up on old mails.  Thanks for your insightful letter of 16 July.

I wish I had more to say at the moment; I just give a few nods of ``yes'' to you.

\bpglm
With regard to my paper, I wonder if I can ask your advice on a
point. I'd like to make there the bold statement that the concept of
probability table and its decomposition offer a simple point of view
to look at Gleason's theorem (and also at the one with general POVMs).
The point of view is the following (sketched):

1) The concept of preparations and effects can very intuitively (and
precisely, I think) be expressed through the idea of a `probability
data table'.

2) The assumption that effects are to be represented by vectors
implies that the only simple way to pass from a probability table to
these vectors is through the table decomposition in the manner I do in
the paper. Note that the `effects $=$ vectors' assumption is implicit in
Gleason's premises, since the starting objects used as effects
(projectors or POVMs) are additive. Note also that the table
decomposition explains why the preparations are also to be vectors,
and why linearity comes apparently for free in Gleason's theorem.

3) The decomposition and the fact that we are dealing with
probabilities (bounded within $[0,1]$) also leads easily to the fact
that the `shape' of the set of preparation-vectors determines the
shape of the set of effect-vectors, and vice versa. Now, Gleason
doesn't just make the assumption `effects $=$ vectors', but the stronger
one `effects $=$ projectors' (or `effects = POVMs'), which corresponds
to assuming also a definite shape of the set of effects -- thus it is
also clear why the theorem gets almost for free the geometry of the
set of quantum mechanical preparations.

Do you think that these points make some sense?
\epglm

Yes, I like this line of thought very much.  Ever since first drawing the picture that became Fig.\ 2 in my \quantph{0205039}, I've been thinking that Gleason's theorem is almost superfluous.  If you could dot the eyes and cross the tees in that in an even more general setting, that would be great!

\bpglm
In connexion with point 3) above, there is a very interesting and
non-trivial thing about the sets of effects and states in quantum
mechanics: they are the same, modulo a renormalisation of effects.
This, for a generic imaginary data table, doesn't happen; in fact, in
general it is even meaningless to speak of a `renormalisation'. For
example, if the set of states is a simplex, the set of effects is a
cube, which projects onto an hexagon; if the set of states is a
semicircle, the set of effects is a bi-pyramid which projects onto a
pointed ellipse \ldots\ Only in special cases does one have that symmetry
(e.g., a circle, a sphere, etc.). This puzzles me, and I wonder if this
symmetry has some physical meaning or if it just hides some assumption
we make in our concepts of `preparation' and `outcome'.
Because it amounts to saying, more or less, that there is a
correspondence between a preparation and an outcome -- which, in
general, I cannot see.
\epglm

I think the deep reason there is precisely that quantum measurement corresponds to an application of Bayes' rule.  In a measurement, one simply ``refines'' one's knowledge:  i.e., one moves from the barycenter of a convex decomposition of a density operator to one of the outlying points in the decomposition.

\bpglm
In Bayesian updating, I update the probability distribution that I
assign to some (mutually exclusive and exhaustive) propositions -- call
them hypotheses -- in view of new data. The point is that both my old
and my new probability distributions refer to the same set of
hypotheses.

The quantum ``updating'', instead, can bring me from one kind of system
to another kind of system, which can be completely different (in
particular, of different dimensionality, eg, when the $V_d$ are
unitaries). From a Bayesian point of view, this looks more like going
from a set of hypotheses to another set (which need not have something
to do with the former) ie from a probability distribution $\{P(A_i|L)\}$
to another $\{P(B_j|DL)\}$.
\epglm

Yes, you are right there, and I don't quite have my head around the point yet.  But you might also note in the Fig.\ 2 that I mentioned above, one could interpret the extra unitary readjustment in exactly the same way:  i.e., as readjusting the hypothesis set rather than the final state.  The problem is in part that the hypothesis set (i.e., the standard quantum measurement device at the Bureau of Standards) is just a scaffold and not real.  The main point of importance for me is the invariant piece:  for {\it any\/} standard measurement chosen, quantum updating looks like an application of Bayes' rule $+$ linear readjustment.

\bpglm
Finally, I'd like to ask you something about hidden-variable
theories and non-locality (I wanted to write and ask Mermin long ago,
but doubted to receive a reply). From a Bayesian point of view, isn't
the concept of ``correlation'' in the mind of the agent, and is it
something non-physical, just like quantum states? In Bell's theorem,
the assumption is made that the settings of the measuring devices are
uncorrelated (between each other, and between different times for any
one device). But can this assumption really be proven or disproven by
experiment, for the Bayesian reason above? What happens if one drops
it? Can one get a local deterministic hidden-variable theory that says
that the devices' settings are pre-determined, together with the
states of the two particles (of course, since the theory is
deterministic), and that they are pre-determined in such a way to give
the observed outcomes -- whatever degree of ``correlation'' they have? In
order to determine the behavior of all the pieces (particles,
devices, experimenters, etc.) wouldn't one only need to set up
appropriate initial conditions in a space-like slice of the past
light-cone which contains all the pieces and all the events?
\epglm

Yes, and you are right:  Several people have made that point.  (Not the Bayesian point, but the latter point.)  Unfortunately I can't think of any references---though, it may have been Bell himself first.

\section{17-08-03 \ \ {\it Tearing Off the Duct Tape} \ \ (to N. D. {\Mermin})} \label{Mermin102}

\bdm
Please take the damn tape off your mouth.  If I'm going in the wrong
direction I don't wish to waste anybody's time.
\edm

I apologize for keeping you waiting for another two days.  Once I
signed off Friday at Beer:30, I wasn't able to get back to email
until this morning.  The whole family ended up in Munich proper
yesterday (rather than a leisurely day here in Zorneding), and then
by the evening I was exhausted.

So, let's get to it.

\bdm
Quantum mechanics says (a) If two gates are measurement gates and (b)
if no other unitary gates are in the circuit but those two then (c)
if a qubit is sent through the circuit then with probability 1 both
gates will give the same reading.

This appears to be an assertion that under certain conditions any
rational person will strongly believe that both gates will give the
same reading.  Furthermore, the conditions are not that in the past
million runs only 00 and 11 have been registered and never 01 and 10
(as in the story you tell me about how Bob comes to his strongly held
belief).  They are conditions about the structure of the circuit,
through which no qubit may ever have been sent.  If, under these
conditions, all rational people must agree that the gates will give
the same reading, some might worry that it was nitpicking (rather
than deeply insightful) to nevertheless insist that this must be
viewed as a subjective judgment rather than an objective property of
the circuit.
\edm

Yes, exactly; your last sentence would certainly be my own take if it
were to stop at that.  I doubt you remember this, but it was the BFM
paper that started pushing me down the radical Bayesian path
precisely because I perceived its message to be something like your
point above.  (Of course you didn't say ``subjective judgment'' but
rather ``knowledge,'' but for me it had the same effect.)

I just took a walk down memory lane rereading my correspondence to
you, Brun, Finkelstein, {\Caves}, and {\Schack} between 7 August and 2
September 2001 to see if I could find any good quotes along those
lines.  Kind of depressing really: The sad thing was that I seemed to
be far more lucid then than I was by the end of our {\Montreal}
meeting in November 2002 (capped off by my fatal mistake on 4
November).  For instance, when I look back at my 22 August 2001 note
to {\Caves} and {\Schack} ``\myref{Schack4}{Identity Crisis},'' I see
that it had a perfectly good discussion of CERTAINTY---the very
starting point of this latest email conversation, as rekindled by
{\Ruediger}'s 18 June note this year.  Traveling in circles and
circles.  Reread some of that stuff and tell me whether you don't
think it is dead on the mark for today's very discussion.  Let me give
you an outtake, from \myref{Mermin35}{a note to you 2 September 2001}:

\bq
The point of separating the categories ``knowledge'' and ``reality''
(or ``subject'' and ``object'' for that matter) is not to make
knowledge an objective reality in its own right or, even worse, to
make it the sole reality.  Rather it is to say that there is a
distinction and that that distinction should be recognized.  \ldots

What I have ultimately NOT been able to stomach about your wording of
the whose-knowledge ``answer'', and Jerry's wording of the
whose-knowledge ``answer''---some of Todd's versions would actually
survive---is that you say, under certain circumstances, two
scientific agents (observers, or what have you) MUST assign
``consistent'' quantum states to a given system.  In the case of pure
states, the two agents MUST assign the {\it same\/} pure state to the
system. \ldots

What I object to is the word MUST.  Todd once wrote it this way, \btb
We have been describing a consistency criterion.  {\em If} one wishes
to combine two state descriptions of a single system into a {\em
single} state description, the criterion tells one {\em when} it is
consistent to do so (i.e., when the two descriptions are not actually
contradictory).

I agree that nobody is holding a gun to Alice's head and forcing her
to incorporate Bob's information. \etb and to this way of speaking I
can agree.  But if you take away Todd's ``{\it If}'', then everything
collapses in my mind.  Enforcing that two agents MUST make the same
state assignment if they are going to be ``right'' at all reinstates
the very objectivity, the very agent-independence of the quantum
state that the Mechanica-Quantica-Lex-Cogitationis-Est program has
been working so hard to exorcise. \ldots

It is much like the old debate.  Is materialism right?  Or is it
Berkeley's idealism that is right?  Who cares, I say.  Both
philosophies are just simple samples of realism:  They only disagree
on the precise concept which ought to be taken as real, mundane
matter or sublime consciousness.  The way you characterize it above,
one would think that the only fruit of the Mechanica Quantica program
would be the RENAMING of a material reality into an ideal one---a
shift more of emphasis, rather than anything of grit.
\eq

So, let me get to your better formulation straight away:
\bdm
I believe your answer to such worriers would be something like this:
that $(a)$ and $(b)$ imply $(c)$ is not an objective fact about the
world of circuits, but a law of thought (though it remains a part of
the research program to show precisely how it emerges as a law of
thought).  In isolation it cannot affect anybody's degree of belief
--- for example it cannot guide betting behavior --- unless $(a)$ and
$(b)$ can be established.  But $(a)$ and $(b)$ are subjective
judgments. Bob requires a million runs not to establish the law of
thought $(a)+(b) \Rightarrow (c)$, but to build up his strongly held
belief in $(a)+(b)$.
\edm

Yep, you're roughly on track here.  The law-of-thought (or maybe
better, law-of-ideal-thought) character in the implication $(a)+(b)
\Rightarrow (c)$ is due to Gleason's theorem plus Dutch-book
coherence. It's the kind of point we tried to make in \quantph{0106133}.  The definition of your first measurement gate in
(a) is that it gives ``maximal information'' about a reapplication of
itself, i.e., it gives certainty for the click (as judged by any
agent for which it is actually a measurement gate).  DB coherence
turns that certainty into a $p=1$ assignment, and Gleason's theorem
further turns that into a unique pure state.  The judgment of (b)
allows this pure state to be reconverted into a second $p=1$
assignment for the outcome of the later measurement gate.

The goal of the research program is indeed to show that all of this
together is a kind of Dutch-book coherence or, at least, Dutch-book
coherence modulo some particular assumption about the properties of
the (contingent, physical) world.  ({\Ruediger} and I may still have
slightly different goals here; it's hard to tell at this point.)

\bdm
In isolation it cannot affect anybody's degree of belief --- for
example it cannot guide betting behavior --- unless $(a)$ and $(b)$
can be established.
\edm

Yes.  In fact, it is just as normal Dutch-book coherence sets no
probability assignments at all.  (Emphasizing, of course, that even
in the case of ``certainty'' we have had to recognize the
subjectivity of the judgment.)  If DB coherence is taken to be akin
to a dynamical law in classical physics, (a) and (b) should be taken
to be akin to the initial conditions.  (I don't know if that analogy
helps.)

Let me only take point with one issue:
\bdm
But $(a)$ and $(b)$ are subjective judgments.
\edm
You should stop your development right there.  Conceptually, the
origins of (a) and (b) should be left unanalyzed (just as with
initial conditions in classical mechanics).  You can of course do
what you do in the next sentence (to somehow make it all more
palatable),
\bdm
Bob requires a million runs not to establish the law of thought
$(a)+(b) \Rightarrow (c)$, but to build up his strongly held belief
in $(a)+(b)$.
\edm
but you should realize that in doing this you expand the context of
the problem without adding anything of deeper significance to it. The
data from those million runs are of no significance without still a
further prior judgment.  ({\Ruediger} introduced exchangeability to
bring the language to the turf of your 1985 paper, but
exchangeability is still nothing more than a judgment.)  To know that
a million runs (rather than a billion runs or a trillion runs) is of
significance requires still more to be said:  In particular one must
choose a particular exchangeable assignment rather than another.  So
at some level, one is always stuck with making a quantum state (or
quantum operation) assignment just to get the ball rolling.
Thereafter, everything is empirical data plus law of thought applied
to that initial judgment.

For $(a)$ and $(b)$ to be ``subjective judgments'' in our sense means
(1) that their origins need not be analyzed, and (2) there is nothing
in the world that {\it requires\/} two agents to have the same such
judgments. The value of the game foundation-wise is that it tells you
that if you are looking for the ``objective'' or ``intersubjective''
in quantum mechanics, you should look elsewhere.

I hope you'll send the second installment soon!  {\Ruediger} and I are
now set to meet late morning Tuesday, and it would be great to
discuss your note face to face.

\section{18-08-03 \ \ {\it The Big IF} \ \ (to A. Sudbery \& H. Barnum)} \label{Sudbery2} \label{Barnum10}

I'm running far, far behind in all the things I've wanted to do this
summer, but maybe there's a chance I'll catch up.  In particular, I
have been trying to give Mr.\ Nagel a concerted effort during my
vacation here in Munich.  I went out and bought {\sl The View from
Nowhere\/} and am a little ways into it.  When I finish that, I'll
re-approach your article ``Why Am I Me?'' and give you a detailed
appraisal.

It's probably too early in my reading to tell, but my troubles with
Nagel may all boil down to ``The Big IF.''  That is, they may boil
down to the religion that lies behind this passage plucked out of his
article ``Subjective and Objective.''  (I'll capitalize the big IF
and a couple of other appropriate words so that you'll know what I'm
talking about.)  Here goes:
\bq
Since a kind of intersubjective agreement characterizes even what is
most subjective, the transition to a more objective viewpoint is not
accomplished merely through intersubjective agreement. Nor does it
proceed by an increase of imaginative scope that provides access to
many subjective points of view other than one's own. Its essential
character, in all the examples cited, is externality or DETACHMENT.
The attempt is made to view the world not from a place within it, or
from the vantage point of a special type of life and awareness, but
from nowhere in particular and no form of life in particular at all.
The object is to discount for the features of our pre-reflective
outlook that make things appear to us as they do, and thereby to
reach an understanding of things as they really are. We flee the
subjective under the pressure of an assumption that everything must
be something not to any point of view, but in itself. To grasp this
by DETACHING more and more from our own point of view is the
unreachable ideal at which the pursuit of objectivity aims.

Some version of this polarity can be found in relation to most
subject matter---ethical, epistemological, metaphysical. The relative
subjectivity or objectivity of different appearances is a matter of
degree, but the same pressures toward a more external viewpoint are
to be found everywhere. It is recognized that one's own point of view
can be distorted as a result of contingencies of one's makeup or
situation. To compensate for these distortions it is necessary either
to reduce dependence on those forms of perception or judgment in
which they are most marked, or to analyze the mechanisms of
distortion and discount for them explicitly. The subjective comes to
be defined by contrast with this development of objectivity.

Problems arise because the same individual is the occupant of both
viewpoints. In trying to understand and discount for the distorting
influences of his specific nature he must rely on certain aspects of
his nature which he deems less prone to such influence. He examines
himself and his interactions with the world, using a specially
selected part of himself for the purpose. That part may subsequently
be scrutinized in turn, and there may be no end to the process. But
obviously the selection of trustworthy subparts presents a problem.

The selection of what to rely on is based partly on the idea that the
less an appearance depends on contingencies of this particular self,
the more it is capable of being arrived at from a variety of points
of view. IF THERE IS A WAY THINGS REALLY ARE, which explains their
diverse appearances to differently constituted and situated
observers, then it is most accurately apprehended by methods not
specific to particular types of observers. That is why scientific
measurement interposes between us and the world instruments whose
interactions with the world are of a kind that could be detected by a
creature not sharing the human senses. Objectivity requires not only
a departure from one's individual viewpoint, but also, so far as
possible, departure from a specifically human or even mammalian
viewpoint. The idea is that if one can still maintain some view when
one relies less and less on what is specific to one's position or
form, it will be truer to reality. The respects in which the results
of various viewpoints are incompatible with each other represent
distortions of the way matters really are. And if there is such a
thing as the correct view, it is certainly not going to be the
unedited view from wherever one happens to be in the world. It must
be a view that includes oneself, with all one's contingencies of
constitution and circumstance, among the things viewed, without
according it any special centrality. And it must accord the same
DETACHED treatment to the type of which one is an instance. The true
view of things can no more be the way they naturally appear to human
beings than the way they look from here.

The pursuit of objectivity therefore involves a transcendence of the
self, in two ways: a transcendence of particularity and a
transcendence of one's type. It must be distinguished from a
different kind of transcendence by which one enters imaginatively
into other subjective points of view, and tries to see how things
appear from other specific standpoints. Objective transcendence aims
at a representation of what is external to each specific point of
view: what is there or what is of value in itself, rather than {\it
for\/} anyone. Though it employs whatever point of view is available
as the representational vehicle---humans typically use visual
diagrams and notation in thinking about physics---the aim is to
represent how things are, not {\it for\/} anyone or any type of
being. And the enterprise assumes that what is represented is
DETACHABLE from the mode of representation, so that the same laws of
physics could be represented by creatures sharing none of our sensory
modalities.
\eq

The two key ideas in this passage that I think quantum mechanics
plays the most havoc with are:
\begin{enumerate}
\item
the DETACHED agent (observer, scientist, etc.), and
\item
IF there is a way things really are \ldots
\end{enumerate}

I honestly believe one can take the Nagel worldview seriously---I
suspect there is no logical flaw in it.  One can legitimately try to
make quantum mechanics fit that worldview with more or less success.
My only point is the strong personal suspicion that with such a
project one forces quantum mechanics into shoes it does not fit. And,
as I see it, what bunions that will cause in the future!

The whole subject matter of my {\sl Notes on a Paulian Idea\/} is in
toying with the idea that the cleanest expression of quantum
mechanics will come about once one realizes that its overwhelming
message is that the observer cannot be detached from the phenomena he
HELPS bring about.  I capitalize the word HELPS because I want you to
take it seriously; the world is not solely a social construction, or
at least I cannot imagine it so.  For my own part, I imagine the
world as a seething orgy of creation.  It was in that orgy before
there were any agents to practice quantum mechanics and will be in
the same orgy long after the Bush administration wipes the planet
clean.  Both of you have probably heard me joke of my view as the
``sexual interpretation of quantum mechanics.''  There is no one way
the world is because the world is still in creation, still being
hammered out. It is still in birth and always will be---that's the
idea.  What quantum mechanics is about---I toy with---is each agent's
little part in the creation (as gambled upon from his own
perspective).  It is a theory about a very small part of the world.
In fact, I see it as a theory that is trying to tell us that there is
much, much more to the world than it can say.  I hear it pleading,
``Please don't try to view me as a theory of everything; you take
away my creative power, my very promise, when you do that!  I am only
a little theory of how to gamble in the light of a far more
interesting world!  Don't shut your eyes to it.''

The question is, how to get one's head around this idea and make it
precise?  And then, once it is precise, what new, wonderful, wild
conclusions can we draw from it?  That is the research program I am
trying to define.

Is it a SCIENTIFIC research program?  I think so, and in the usual
sense.  There will be lemmas, theorems, and corollaries.  (I would
like to think that my work and the work of the fellows I've drawn
down this path already evidences this.)  Ultimately there will be
calls for experiments.  There will be technologies suggested and
money to be made from the program's fruits.  Failure of nerve?
Anything but!:
\bts
Maybe you and [{\Rorty}] can shift me from my instinctive reaction to
pragmatism, which is that for a scientist it represents a failure of
nerve, a failure of imagination, and most seriously a failure of
curiosity. Being useful cannot, for a scientist, be the end of the
story about a statement or a theory; we immediately want to know {\em
why} one theory is more useful than another. That ``why?'' leads us
to an external world of some kind, maybe very strange (the stranger
the better, i.e.\ the more interesting, I would say) and to refuse to
follow where it leads seems to me to be a scientific copout.
\ets

I see it as anything but a failure of curiosity or a copout!  What
you wrote me above reminds me of a conversation I had with Chris
{\Timpson} in a pub one night.  I made the mistake of mentioning William
{\James}, and Chris quickly intoned, ``ALL {\James} was about was the
nonsense that truth resides in what is useful.''  The word ALL just
boomed!  A man's whole life was dismissed in a single sentence.  I
cut him short, ``William {\James} was about many things, ONE OF WHICH
was that the correspondence theory of truth holds no water.''
Similarly I will say to you, there is far more explored by the
pragmatist thinkers than that which is delimited by their ideas on
truth and warranted belief.  Pragmatism is not positivism; it is not
that there is nothing to be sought in science beyond the connections
between sense perceptions.  I see the classical pragmatists (and
myself) as ultimately realists, but honest realists---ones who have
realized that our theories are not mirror images of the underlying
reality, but rather extensions of our biological brains.

But that is going in a direction I don't want to go down at the
moment.  In any case, don't read {\Rorty} first!  Read {\James}' little
book Pragmatism to start off with.  More immediately, with respect to
the present Nagelian discussion, read ``Genesis and the Quantum'' on
pages 122--123, the dialogue between Adam and God on pages 118--120,
``Evolution and Physics'' and ``Precision'' on pages 267--270, and
some of Jeff Bub's expressions on the idea in Chapter 9, most notably
pages 139--140 and 141--142---all these things in the samizdat I sent
you. The game of ASSUMING the possibility of a detached observer, as
Nagel does, is just that:  a game of assuming. Thereafter, Nagel
tries to make sense of our more personal worlds in spite of this. The
pages I've just referred to in my samizdat try to sketch what quantum
mechanics might be talking about if one does not make such an
assumption.  In fact, they try to justify NOT making the assumption
at all.  I hope from these readings you will get the impression that
though there may be a fundamental disagreement between Nagel and me
at the outset, such a disagreement does not necessarily amount to a
copout on my part.

Finally, let me paste in a note below that I sent to David {\Mermin}
and {\Ruediger} {\Schack} the other day.  It's titled
``\myref{Mermin101}{Me, Me, Me!}''\ and gives my very latest attempt
to express the content of quantum mechanics in these lights.  I think
it does an adequate job \ldots\ but experience tells me I am always
over-optimistic.  In any case, it is directly related to all that was
said above.

I'll be back again after completing Nagel and your paper. \medskip

\noindent PS.  I'll also attach the note on Rorty-ish stuff that I sent you previously.  That way Howard will have a better handle on why I find excitement in dismissing the starting point ``IF there is a way things really are \ldots''

\section{21-08-03 \ \ {\it Everywhere and Nowhere} \ \ (to H. Barnum)} \label{Barnum11}

Thanks for the note in reply to my pre-Nagel thoughts.  You've given me courage in a way:  I feel more strongly than before that reading Nagel is worth my time.  I will follow it through to completion.

A couple of very little points.

\bhb
I'm not sure if ``religion'' is supposed to be a bad thing in this instance, but couldn't we just as easily call it a ``research program''?  Even a SCIENTIFIC one?  I guess my point here is that I don't think quantum mechanics can yet be considered decisive against such a program\ldots\ and the theorems, lemmas, technologies you mention (below) are coming from quantum mechanics taking a wide variety of interpretational positions (or none at all)\dots\ sometimes very similar theorems from people having (I think) different points of view.
\ehb

Fair enough.  I suppose my choice of the word was conditioned on the number of times I've had my own research program depicted as ``ignoring the results of science'' or as ``antithetical to the very idea of science'' (D. Deutsch and D. Albert, two examples).  The point I wanted to get at with the term ``religion'' was only that the possibility of satisfying ``The Big IF'' is a {\it faith}, not a self-evident truth.

(By the way, while I'm thinking of Deutsch and Albert and pointing out the depths of their religiosity, let me go the rest of the way and tell you the remainder of the thought.  These are religions, yes, but worse than that, they are {\it easy\/} religions.  Evidence:  Deutsch and Albert both have already found {\it the\/} answers, whereas I still strive away.)

\bhb
Again, I tend to think this ``assuming the possibility'' may just amount to ``attempting to coordinate the various perspectives into a coherent whole''.  Quantum mechanics suggests difficulties with this, enough to make it worthwhile to pursue your vision of ``assuming the
impossibility of a detached observer'' \ldots\ but to me, it {\it would\/} be a failure of scientific nerve/imagination to think quantum mechanics justifies abandoning the attempt to coordinate.  Let both approaches bloom \ldots
\ehb

Life is short!

\section{22-08-03 \ \ {\it Chance and Paper Letters} \ \ (to W. C. Myrvold)} \label{Myrvold2}

Funny you mentioned getting a paper letter from Abner.  Just yesterday when I returned from Munich I found a letter from Abner that had been forwarded to me from Bell Labs:  It was dated June 3!!  I was so delighted with it I scanned it into my computer this morning for my permanent records.  If Abner had only emailed it, it would have been in my computer three and a half months ago!

Concerning a conversation I'd ultimately like to start up with you, though---namely, subjectivist or personalist probability in quantum mechanics---a better source is my samizdat {\sl Quantum States:\ What the Hell Are They?\/}\ posted at my webpage.  In particular, you might enjoy skimming though it in connection with your project with Bill Harper.  Most of the fun in the document starts up on pages 19 and 35 (and thereafter).  That's where I take the turn of giving up on ``objective certainty,'' having realized that through quantum mechanics its logical derivative is ineluctably ``objective chance.'' [See 07-08-01 note ``\myref{Mermin28}{Knowledge, Only Knowledge}'' to T. A. Brun, J. Finkelstein \& N. D. Mermin and 22-08-01 note ``\myref{Schack4}{Identity Crisis}'' to C. M. Caves \& R. Schack.]

{\Ruediger} {\Schack} and I will be condensing much of that subject matter into a proper paper starting in the middle part of this month.

I'm actually surprised to hear what you say in this:
\bwm
I don't know of a good source on why Bayesians should believe in
objective chance (or, better: at least regard the notion as
intelligible), and neither does Bill Harper.
\ewm

I recall the words in David Lewis's paper, `A Subjectivist's Guide to  Objective Chance':  ``As philosophers we may well find the concept of objective chance troublesome, but that is no excuse to deny its existence, its legitimacy, or its indispensability.  If we can't understand it, so much the worse for us.''  I would have thought a lively discussion would ensue in the subjectivists' camps thereafter.

Anyway, I'm very much looking forward to this paper of yours.

\section{25-08-03 \ \ {\it Coordinate Systems} \ \ (to G. L. Comer)} \label{Comer37}

I read two things this morning that put me in the mood of coordinate
systems.  The first was a remarkable letter from {\Pauli} to Bohr, dated
11 March 1955.  [See 29-03-01 note ``\myref{Folse3}{There's At Least One More}'' to H. J. Folse.]  The second was your latest draft of ``linear
structures.''  I'm just trying to get these things clear in my head
and put them into a common language.  (By the way, I'm sorry I've
been so absent lately.  Unfortunately, after this brief return I'll
probably be absent again for a while; I've got to get a book review
to {\sl Physics World\/} by September 4, and I am far, far behind on my ARDA
roadmap duties.  Also I've got to finally get my application off to
the Bohr Institute this week---a much more difficult process than you
might imagine.  In total I'm going to be inundated through
mid-September.)

Anyway, I want to attach a PDF file containing two letters on quantum
mechanics that I've written recently (one to {\Mermin} and {\Schack},
and one to Sudbery and Barnum), along with the {\Pauli} letter
mentioned above.  Strangely enough---you must get tired of my saying
this!---I feel like I've had another epiphany in expression.  My two
letters try to capture that.  [See 12-08-03 note
  ``\myref{Mermin101}{Me, Me, Me}'' to N. D. Mermin \& R. Schack and
  18-03-03 note ``\myref{Sudbery2}{The Big IF}'' to A. Sudbery \&
  H. Barnum.]

Now for the common language business.  What I found remarkable upon
rereading the {\Pauli} letter---I last read it maybe a year ago---was
his discussion of coordinate systems and the way he alludes to
calling them ``actions on the part of the observer.''  Or, at least
in my cherry-picking mood, that's what I would like to glean from the
letter (ignoring that he did also use the word ``knowledge''). You'll
see why that takes me when you read the letter I wrote to {\Mermin} and
{\Schack}.  There I explicitly identify the set of POVMs (i.e., the full
set of quantum measurements upon a system) as the set of ACTIONS
one's {\it will\/} can take upon the system.

Clearly, {\Pauli} sees the coordinate system as the classical analog of
the POVM (and vice versa, the POVM as the quantum analog of the
coordinate system).  They are both mathematizations of ``the
observer,'' or better ``the agent,'' in the respective theories.  But
there are differences.  For one, classical coordinate systems are
completely passive---this {\Pauli} recognizes.  Equipping a manifold
with a coordinate system does not change the manifold or any of its
contingent properties (i.e., any of the fields living on it).  On the
other hand, POVMs are anything but passive.  This was a major point
for {\Pauli}.  A POVM elicits the world to do something it would not
have otherwise done.  The combination of the action of one's will
(the POVM) with a system external to it gives rise to a birth of
sorts---the flash, the event which we usually call (in older
language) the measurement outcome.

What is funny, however, is {\Pauli}'s discussion of Einstein.  As far as
I can tell, what he intends by the word ``consequent'' would better
be expressed by ``consistent.''  (Actually, upon looking up the word
in my dictionary, I find that this is its very meaning!) Consequently
I find {\Pauli}'s contention that Einstein was being inconsistent
puzzling.  The very reason Einstein felt no qualms about introducing
coordinate systems when necessary was because in relativity theory
the underlying observer-free description---i.e., the manifold
equipped with its fields---had been achieved.  In contradistinction,
this is something quantum theory has yet to achieve (and may never be
able to).  In fact, it was this that troubled Einstein \ldots\ saving him
from inconsistency.

Let me belabor the similarity and distinction.  I think you are
certainly right to point out one of the similarities via the mutual
use of projections in the two theories.  In the case of relativity
theory, a projection is used to tell what one WILL ``see'' from one's
perspective---you gave energy as an example.  The projection is an
operation used for expressing an aspect of a preexisting reality,
e.g., a tensor field.  In the case of quantum mechanics, one projects
one's state of knowledge (i.e., one's density operator for a system)
onto the potential consequences of one's action (i.e., onto the POVM
elements) to obtain a probability distribution.  This is just the
Born rule in a generalized setting.  The probability distribution so
obtained expresses one's expectations for the consequences of one's
action, not one's expectations for a preexisting reality.

So, projections both, but they encode very different things.

Let me close with another distinction that I'm wondering what to make
of.  It's connected to a point in the Anti-{\Vaxjo} paper I sent you
and a point in your latest draft.  As far as I can tell, I believe
both:
\bq
\noindent {\bf Me:} The OBJECTIVE content of the probability
assignment comes from the fact that {\it no one\/} can make {\it
tighter\/} predictions for the outcomes of experiments than specified
by the quantum mechanical laws.  Or to say it still another way, it
is the very existence of transformation {\it rules\/} from one
context to another that expresses an objective content for the
theory.  Those rules apply to me as well as to you, even though our
probability assignments {\it within\/} each context may be completely
different (because they are subjective).  But, if one of us follows
the proper transformation rules---the quantum rules---for going to
one context from another, while the other of us does not, then one of
us will be able to take advantage of the other in a gambling match.
The one of us that ignores the structure of the world will be bitten
by it!
\eq
\bq
\noindent {\bf You:} Even if the observers are at the same place, at
the same time, when they make their measurement, they will in general
not record the same energy because their motions will in general be
different.  So how can they ever agree on anything?  Relativity tells
them how to ``translate'' one measurement into the other so that a
consistent description results.  What is objective about the energy?
Nothing. Its value depends on the motion of the observer.  Where is
the objectivity?  Only in the rules of translation.  The only
objective thing to me seems to be how one object is to be compared
with another.
\eq

I.e., both theories make use of transformation groups in a crucial
way.  But what different ways!  In the case of relativity, I guess I
would have said that the translations are for external purposes. They
tell me how to connect one observer's potential observations to
another's.  Or another way to say it, they tell me how to translate
from one aspect of the preexisting reality to another.  In the case
of quantum theory, however, the translation is all internal:  What
should I believe of this, given that I believe such-and-such of that?
The quantum transformation rules I am thinking of here say nothing of
how to translate one person's beliefs into another person's.  In
fact, the whole point of personalistic probability is that there is
IN PRINCIPLE no way to connect someone's beliefs to anyone else's:
Probabilistic statements are personal betting strategies.

Thus I guess I find no warrant in what {\Pauli} says when he tries to
make the point:  ``That the situation in quantum mechanics has a deep
similarity with the situation in relativity is already shown by the
application of mathematical groups of transformation in the physical
laws in both cases.''

But then again, I start to wonder that if I took {\Pauli}'s starting
point completely seriously---i.e., if I were to start to think of
coordinate systems as ACTIONS of an individual observer---all my
troubles would melt away and maybe I'd start to see a deeper analogy
as he saw it.  I guess I'm still very much in the process of
thinking.

Enough for now.  (Don't forget the attached files---the one already
mentioned plus one further one relevant to the Sudbery discussion.)

\section{31-08-03 \ \ {\it Acceleration is Intervention!}\ \ \ (to G. L. Comer)} \label{Comer38}

What is it about Sundays that sets off the thought muscles?  Anyway,
I hope my title says it all.  That was the thought that hit me as I
was washing the dishes this morning.  Maybe coordinate systems are
interventions after all, when considered in the quantum context. [Cf.
last Sunday's discussion.]  What happens when a ``particle detector''
is accelerated?  You've told me before that it goes click. The usual
explanation goes that it has something to do with the perceived
quantum state of the field---namely it becomes something like a
thermal density operator.  But maybe instead, we should think of the
process as a change in the character of the measurement device.  That
is, maybe an accelerated particle detector is a distinct POVM from
its inertial cousin.  I.e., it intervenes on the field in way
different than the inertial detector.  (In older language, that is to
say, it ``measures'' a different observable.)

At least that's the thought that hit me.

Look at me getting pulled down the path of GR:  Analogies between
Hilbert-space dimension and gravitational mass, analogies between
coordinate systems and measurement devices.  What's happening to
me?!?!

\section{28-08-03 \ \ {\it Update} \ \ (to C. King)} \label{King8}

\bck
What are you working on now? Any more neat additivity conjectures to
throw my way?? You had mentioned the possibility of coming to Boston
some time -- now that Peter Shor is here, perhaps you have even more
reason to pop over?
\eck

Mostly I'm trying get my part of the ARDA roadmap finished.  It's a big annoying project.  And I may have to go to DC Sept 14--16 for it.

Mathematically, mostly I'd like to see whether or not there is a set of quantum states with $d^2-1$ elements which achieves the quantumness of the Hilbert space (i.e., $2/(d+1)$).

Coming to Boston definitely sounds nice!  By the way, would you have any to get (or get me) into the Harvard library?  There's an old PhD thesis that I'd like to get a copy of:
\begin{itemize}
\item
W.~H. Long, {\sl The Philosophy of Charles Renouvier and Its Influence on William James}, Ph.~D. thesis, Harvard University, 1927.
\end{itemize}
and it is not available through University Microfilms, etc.

\section{01-09-03 \ \ {\it Answering Correspondence with Correspondence} \ \ (to S. Savitt)} \label{Savitt1}

\bss
I've always (though perhaps incorrectly) thought of solipsism as the
view that my mind was the only mind that exists (or that I knew
exists, or that I could know exists). This way of looking at it is,
of course, compatible with materialism.
\ess

The predominant view seems to be that it connotes an extreme form of
idealism in which (one) self is the only reality, or at the very
least, the ground of reality.  I have a correspondence with {\Mermin} in
which I compile several definitions for the word pulled from various
sources; I'll attach that rather than pull it out of context.

\bss
I didn't get your remarks about Unitary Readjustment.  Is the kind of
updating of your beliefs that you do as the result of a measurement
in QM supposed to be incompatible with standard Bayesian updating. (I
thought one theme of your ``Quantum Information'' paper is to show
that these processes were at root the same.) Alas, the details of
your massive paper have begin to blur in my mind.
\ess

Yes it is generally incompatible; the deviation represents part of
the ``only a little more'' in the paper's title.  In Sections 6 and
6.1 of my paper I deemphasized the incompatibility (or rather,
``deviation from Bayes'') because I wanted to first clinch it in the
minds of the readers that quantum collapse has a large component that
is simply Bayesian conditionalization.  The unitary readjustment
quantifies the ``deviation from.''

There is an interesting case where quantum collapse is nothing at all
more than Bayesian updating:  That is, when there is no physical
interaction with the system being considered---for instance in the
standard EPR scenario.  By making a measurement on one of an
entangled pair of particles, one updates one's description for the
other particle.  The interesting thing one sees by writing collapse
in the way I endorse is that, in this case, the unitary readjustment
is just the identity operation \ldots\ which makes collapse precisely
Bayesian conditionalization.

The idea is that, when there is a deviation from Bayes, what one is
seeing is a mild trace of the much deeper statement:  The observer
cannot be detached from the phenomena he helps bring about.

\bss
Finally, despite your protest and because many will read your
writing, I will make the odd remark about grammar. It's not the one
you forestalled, though. It's that `like' is a preposition,
introducing a clause. So on page one of the 12 August note, I think
you really should say ``Maybe the best way to do this is to run
through a glossary of quantum terms as I did once before\ldots'' And
there's a similar mis-use on p.\ 2. Obviously, I don't think that
Winstons carried the day on this one.
\ess

Much appreciated.  I only started learning grammar about ten years
ago, and it shows.  I will readjust the sentences as soon as I send
off this email.  Please explain the Winstons remark though.  What
does it refer to?

\bss
Well, maybe I have another thing to add. I have read only little of
Shimony's technical work in QM, but here's someone who worked on
personal or subjective Bayesianism and doubtless had some ideas as to
how it applied to QM. Looking at your brief email correspondence with
him in the samizdat, it occurs to me that {\bf perhaps} your
difference (if there really is a difference) as to whether an
interpretation of QM must be ``ontological'' or not obscures the fact
that his Bayesianism is like yours.
\ess

Oh, there's quite a difference between Shimony and me on this score.
But there is enough mutual understanding that progress may be had in
the future.  In the attachment, I'll also include some correspondence
to and from Shimony.  The way I would characterize Abner is that he
is Bayesian (of a flavor) {\it about\/} theories, but, within a
theory, he is objectivist about its uses of probability (or at least
this is so for him with respect to quantum theory).

To me, making that move---i.e.\ giving up on the Bayesian account {\it
within\/} quantum theory---is the greatest impediment to
understanding. The empirical evidence I would give for this is simply
the number of years Abner spent taking as serious and reasonable a
search for collapse theories (say of the GRW variety). Also the
lengthy discussions he's had about the ``surprising'' consistency
between special relativity and wave-function collapse---an effect he
dubbed ``passion at a distance'' to contrast it with ``action at a
distance'' (i.e., trying to make it clear that one could not use
wave-function collapse for signaling superluminally).

The thing Abner does not yet appreciate is the immense leading power
of taking Bayesianism seriously {\it within\/} quantum theory.  The
way I would characterize the present stage of my research program is,
``Seek and ye shall find.''  An awfully good example comes from the
thing I pointed out above:  The EPR scenario actually represents pure
Bayesian conditionalization if written in the right way.  Far from
being anything mysterious---like passion at a distance---it simply
represents the usual conception of conditionalization.  In fact, it
is only the method of updating one's probabilities for systems with
which one makes physical contact that has any mystery or calls for an
explanation.  (I.e., the situations where one must insert an extra
unitary update.)

So, seek and ye shall find:  Look for a Bayesian reason for this or
that within quantum mechanics, and one finds---my experience
tells---that the theory is a much tighter package than maybe first
imagined.  It's only in the slight deviation from Bayesianism that
things really get interesting.

\section{01-09-03 \ \ {\it One Further Thought} \ \ (to S. Savitt)} \label{Savitt2}

One further thought.  I retumbled the note I wrote you earlier this
morning in my head as I was walking to work, and it dawned on me that
maybe I went too far in the other direction in trying to clarify
things for you.  I wrote:
\bq
\noindent
     Yes it is generally incompatible; the deviation represents part
     of the ``only a little more'' in the paper's title.  In Sections 6
     and 6.1 of my paper I deemphasized the incompatibility (or
     rather, ``deviation from Bayes'') because I wanted to first clinch
     it in the minds of the readers that quantum collapse has a large
     component that is simply Bayesian conditionalization.  The unitary
     readjustment quantifies the ``deviation from.''
\eq

I want to emphasize that this deviation only concerns {\it updating}.
However, I show in the fat paper that there is another role Bayes'
rule plays in quantum measurement (just prior to the update), and in
that role it is indeed ``Bayes' rule full stop'' that is the crucial
idea.  Namely, I show that there is a one-to-one correspondence
between the full set of quantum measurements (i.e., the POVMs) and
applications of Bayes' rule.

So, when you wrote \bss Is the kind of updating of your beliefs that
you do as the result of a measurement in QM supposed to be
incompatible with standard Bayesian updating. (I thought one theme of
your ``Quantum Information'' paper is to show that these processes
were at root the same.)  \ess you were more right than I was giving
you credit for.  There is a one-to-one correspondence between possible
applications of Bayes' rule and measurement devices (or, as I
expressed it in ``\myref{Mermin101}{Me, Me, Me},'' actions that my
will can take).  But, whereas in classical measurement the updating
stops there, in the quantum case there is a little more.

\section{01-09-03 \ \ {\it Fragmentation and Wholeness} \ \ (to N. D. {\Mermin})} \label{Mermin103}

I hope you are enjoying all the wholeness in well-being Japan and the
bodhisattva can afford.

As for your fragment, I am having quite some difficulty understanding
it:  There was only one thing I could latch onto enough to dislike.
Namely, I object to this phraseology:
\bdm
It's terrible for teaching computer scientists quantum mechanics. The
subject can't get off the ground with warnings that you can't be sure
a measurement gate is a measurement gate, a cNOT is a cNOT, that no
gates act except the ones you set up, etc.
\edm

You still slip into a mode of language that acts as if there is some
magical fact out there that is going to confirm that a cNOT is really
a cNOT or not.  The point is, I would just never talk this way.  Let
me give an example.  Suppose I am talking about a single toss of a
coin, and I say, ``For me, personally, the probability that it will
land heads is 50/50.''  In making such a statement, should I worry
myself that maybe my personal probability is not really 50/50 after
all?  Or to put it another way, and borrow your sentence structure,
would one ever say of classical probability theory:
\bq
\noindent
Things I don't like. \\
1. It's terrible for teaching gamblers how to gamble. The subject
can't get off the ground with warnings that you can't be sure a 50/50
probability assignment is a 50/50 probability assignment, \ldots
\eq

In our language, a measurement gate is a prior.  A cNOT is a prior.
Neither are open to empirical verification beyond soul-searching. The
only thing that ever needs be said to the student is, ``IF your
judgment of this is `measurement gate', and your judgment of that is
`cNOT', and that is `phase gate', etc., etc., then for coherence you
must judge the outcomes of the computation with such-and-such a kind
of probability.''  Notice the single quotes; they are important.  I
did not say ``If your judgment of this is that IT is a measurement
gate, and your judgment of that is that IT is a cNOT, \ldots''  The
latter kind of phraseology would lead one to think that the words
`measurement gate' correspond to a FACT of which one's judgment can
either be correct or incorrect.

I'll leave it at that for now.  Honestly, I got quite confused on
everything else you said.  Perhaps {\Ruediger} will fair better than I
did.

As you once told me, don't eat the fugu!  (At least until after you
clean up your fragment a bit.)

\section{02-09-03 \ \ {\it Peres Festschrift} \ \ (to 48 colleagues)}

\noindent Dear Colleague, \medskip

Asher Peres is turning 70 years old next year!  On this important occasion we would like to honor him with a festschrift for his seminal contributions to quantum theory, quantum foundations, quantum information, and relativity theory.  The festschrift will be published as a number of reserved issues in the journal {\sl Foundations of Physics}.

It is our pleasure to invite you to submit a research paper of about 20 to 30 pages, or even longer if the subject matter requires.  The manuscript would be due at the end of February 2004, leaving a reasonable time for reviewing and revision purposes if necessary.

Please let us know at your earliest convenience whether you will be able to contribute.  Your contribution would certainly be valued, and we hope for a positive reply!\medskip

\noindent With best regards,\medskip

\noindent Chris Fuchs and Alwyn van der Merwe

\section{02-09-03 \ \ {\it `Gleason' Paper} \ \ (to P. Busch)} \label{Busch5}

\bpbu
Unfortunately I haven't made any progress on the problem you proposed.
Have you got anywhere with it in the meantime?
\epbu

We've made progress in better understanding the quantumness function---that's recorded in \quantph{0308120}---but I've made essentially no progress in the question I asked you \ldots\ if I recall correctly the question I asked you.  I think I do.  Namely, to prove (or find a counterexample) that if the cardinality of a set of states is $d^2 - 1$ or less, then the quantumness of the set is bounded away from the minimum value $2/(d+1)$.

For Holevo's festchrift, I'm going to write a little paper proving the minimum value just mentioned and make a few statements about sets of states that achieve it.  Unfortunately, I'm just going to have to conjecture the converse theorem above.

\section{02-09-03 \ \ {\it The Archivist} \ \  (to I. Pitowsky)} \label{Pitowsky3}

Thanks for sending those pieces.  The Maxwell one was especially nice in that {\Ruediger} and I are presently working up a paper on ``On Quantum Certainty.''  Some of the phrases used in the letter fit it well!  Anyway, it's kind of nice being the historian of the field.

\subsection{Itamar's Preply}

\bq
From: {\sl The Life of J. C. Maxwell\/} -- L. Campbell and W. Garnett, London: Macmilian and Co., 1882. \medskip

A letter from Maxwell to Campbell probably from June 1850:

``\ldots I was thinking to-day of the duties of (the) cognitive faculty. It is universally admitted that duties are voluntary, and that the will governs understanding by giving or withholding Attention. They say that Understanding ought to work by the rules of right reason. These rules are, or ought to be contained in Logic; but the actual science of Logic is conversant at present only with thing either certain, impossible of {\it entirely\/} doubted, none of which (fortunately) we have to reason on. Therefore the true Logic for this world is the Calculus of Probabilities, which takes account of the magnitude of the probability (which is, or which ought to be in a reasonable man's mind). This branch of Math., which is generally thought to favour gambling, dicing, and wagering, and therefore highly immoral, is the only ``Mathematics for Practical Men,'' as we ought to be. Now, as human knowledge comes by the senses in such a way that the existence of things external is only inferred from the harmonious (not similar) testimony of the different senses, Understanding, acting by the laws of the right reason, will assign to different truth (or facts, or testimonies, or what shall I call them) different degrees of probability. Now, as the sense give new testimonies continually, and as no man ever detected in them any real inconsistency, it follows that the probability and {\it credibility\/} of their testimony is increasing day by day, and the more a man uses them the more he believes them. He believes them. What is believing?  When the probability (there is no better word found) in a man's mind of a certain proposition being true is greater than that of its being false, he believes it with a proportion of faith corresponding to the probability, and this probability may be increased or diminished by new facts. This is faith in general. When a man thinks he has enough of evidence for some notion of his he sometimes refuses to listen to any additional evidence {\it pro\/} or {\it con}, saying, ``It is settled question, {\it probatis probata}; it needs no evidence; it is certain.'' This is knowledge as distinguished from faith. He says, ``I do not believe; I know''. ``If any man thinketh that he knoweth, he knoweth yet nothing as he ought to know''. This knowledge is a shutting of one's ears to all arguments, and is the same as ``Implicit faith'' in one of its meanings. ``Childlike faith'', confounded with it, is not credulity, for children are not credulous, but find out sooner than some think that many men are liars.''
\eq

\section{03-09-03 \ \ {\it Old Dreams} \ \ (to C. M. {\Caves})} \label{Caves73.3}

Did I ever tell about the time when, in high school, I dreamed that I would become a Catholic priest?  It really made me very, very depressed.  I even had to talk to my Mom about it.  I thought, what a stagnant and empty life.

I had a flashback of that dream this morning.  When I opened my email I found an invitation to speak at a philosophy conference at U. Western Ontario.  \ldots\ But it wasn't that that depressed me.  It was that when I looked at my calendar to see if I could make it, I found some conflicts:  Just before the Ontario meeting I'm invited to speak at a philosophy meeting in Maryland, and overlapping with the Ontario meeting I'm invited to speak at a philosophy meeting in Minnesota.

It does sort of trouble me.  What a stagnant and empty life, that of the philosopher.

\section{03-09-03 \ \ {\it Stuff} \ \ (to A. Plotnitsky)} \label{Plotnitsky13}

\barkp
It's good to hear from you--and not without some synchronicity, as I was in fact reading your ``correspondences,'' now in print,
of course.  [\ldots]  Although I have of course read most of it earlier, there are parts of it, to which I continue to return, since they deal with truly essential and as yet (for me at least) not quite resolved points, plus it's a real pleasure to read. So I keep a copy handy.
\earkp

Thanks for all the flattery.  Actually, you know something that would be really useful to me, and maybe more entertaining for you, would be if you'd have a look at my newer samizdat {\sl Quantum States:\ What the Hell Are They?\/}\ posted at my webpage.  It's probably much harder reading than the older one, but also---by my own reckoning---it's much deeper:  It can't quite be nibbled on in the same way.  Anyway, {\Ruediger} and I are going to try to condense a good piece of that document into a single paper this fall.  The working title is ``On Quantum Certainty.''  After we post our paper on {\tt quant-ph}, I plan to post the said samizdat too.  So it'd be useful to get any feedback I can on it.  For instance, if you notice any missing cross-connects that make it difficult to read, etc., it be great to know about that!

By now that's old stuff for me, but it might be new stuff for you.  {\Ruediger} counts it as where we make our most decisive break with Bohr.  Here's the way he once put it to me.  He said, ``Bohr would always draw measuring devices as these heavy pieces of equipment, firmly bolted to table tops to emphasize that they are classical matters of fact.  But you don't that.  You bolt your measurement devices to the tops of turtles!  And then it's turtles all the way down.''

The next stage of my development is captured in still another samizdat, titled ``Darwinism All the Way Down (and probabilism all the way back up)''---about 100 pages---but unfortunately Mermin hasn't given me permission to share that with anyone yet.  I guess I need to edit him out a bit more first.

\barkp
I feel, though, a bit intimidated by the list of potential
contributors and I am worried that whatever I'd write may not be
technical enough.
\earkp

You shouldn't be intimidated at all.  I know that Asher respects you; that's why I put you in the list.  Also you met Asher's approval:  Believe me, not everyone in my initial list survived!  So, just write us an interesting paper, and everything will be OK.

\barkp
P.S. I was thinking of you in yet another context the other day.  Are
you aware that both formulas $E=mc^2$ (no surprise here of course) and
$E=h\nu$ apparently occur for the first time in Einstein's articles, on
special relativity and on photoeffects, both in 1905?  Planck, it
appears, did not write $E=h\nu$, although it is of course implicit in his
articles on the black body radiation.  He must have pondered their
connections, especially given Lorentz's work, although photons are of
course not electrons in terms of their (quantum) mechanics or
electrodynamics.
\earkp

I thought $E=mc^2$ first came up in a 1907 paper.\footnote{\editornote Einstein's fourth Miracle Year paper, ``Ist die Tr\"agheit eines K\"orpers von seinem Energieinhalt abh\"angig?'' [``Does the Inertia of a Body Depend upon its Energy Content?''], does come very close:  ``Gibt ein K\"orper die Energie $L$ in Form von Strahlung ab, so verkleinert sich Masse um $L/V^2$'' [``If a body gives off the energy $L$ in the form of radiation, its mass diminishes by $L/V^2$''], where Einstein uses $V$ to denote the speed of light.  In this paper, Einstein treats the radiating object as stationary---it radiates $L/2$ in each of two opposing directions, and so remains motionless.  What Einstein calls \emph{Energieinhalt} is therefore analogous to the more modern \emph{rest energy.}  His 1907 paper ``\"Uber die von Relativit\"atsprinzip geforderte Tr\"agheit der Energie'' [``On the inertia of energy required by the relativity principle,'' Annalen der Physik \textbf{323}, pp.~639--41] uses $\epsilon_0$ for the energy of a collection of point masses, as measured in the center-of-mass reference frame.  Einstein deduces that this quantity, which is the rest energy, equals $\mu V^2$.  In modern notation, we would write $E_0 = mc^2$, where $m$ is the rest mass.} You might check into that.  As far as $E=hv$, I guess I wasn't aware of that.\footnote{\editornote Planck \emph{did} say that the size $\epsilon$ of an ``Energieelement'' should equal $h\nu$; see section~10 of his ``Ueber das Gesetz der Energieverteilung im Normalspectrum'' [``On the Law of Distribution of Energy in the Normal Spectrum''], Annalen der Physik \textbf{309,} pp.~553--63 (1901).} However, I do seem to recall that in Planck's original paper he had an incorrect {\it derivation\/} of the (correct) black-body radiation formula.  And I believe it was Einstein who fixed that up in his PhD thesis (in 1903 or 1904).  This is not a fact emphasized in the physics textbooks.\footnote{\editornote In one part of his derivation, Planck presumed that an oscillator's energy could vary continuously; in another, he took the energy as discretized.  For a readable treatment, see A.\ Pais, ``Einstein and the quantum theory,'' Rev.\ Mod.\ Phys.\ \textbf{51,} pp.~863--914 (1979).}  But it must have put Einstein on the map with respect to Planck's attention.  I think a good source of these stories might be Kuhn's book on black-body radiation, but I've never read it.

By the way, I finally met Paula's friend Jane Nordholt at a quantum cryptography meeting in Virginia in June.  She said that she hadn't heard from Paula in quite a while (and I think maybe didn't know how to get in touch with her since her move to NY).  I said I'd write you about this---but then promptly forgot to!  Tell Paula I apologize.  Is everything going well with her?  Has she settled into NY fine?

\section{03-09-03 \ \ {\it Pronouncing Fuchs} \ \  (to R. {\Schack})} \label{Schack73}

If you are around, could you send me a pronunciation guide (perhaps using \LaTeX) symbols for how to pronounce the German words for ``book'' and ``fox.''  I might want to capture how Gell-Mann mispronounced my name in this {\sl Physics World\/} book review I'm writing (when describing the author's own discussion on the subject).  If I do it, I'm going to have to do it with tact.  But still, I would like to know how to write these things anyway.

\subsection{{\Ruediger}'s Reply}

\bq
Here is your pronunciation guide, with explanations of the symbols. The phonetic alphabet used is that of the International Phonetic Association (IPA), the explanations come from the Langenscheid Gro{\ss}w\"orterbuch
Deutsch-Englisch.

\begin{description}
\item[[u:]] Buch [bu:x] long, resembles English oo in boot, but closer and more
retracted than this

\item[[x]] Buch [bu:x] similar to Scottish ch in loch

\item[[{\scriptsize U}]] Fuchs [f{\scriptsize U}ks] short, resembles English u in bull, but closer and more retracted than this
\end{description}

Good luck!
\eq

\section{04-09-03 \ \ {\it Letting You Down, but Bringing You With} \ \ (to G. Brassard)} \label{Brassard19}

I had to make a tough decision yesterday:  I decided not to come to the Wye River Meeting.  My travel schedule just got too hectic again, and Richard Hughes had written me, ``if you would be coming from Ireland especially for this I would recommend saving the wear and tear unless you have other business in the US.''  So, I took him up on it:  I didn't really have any other business in the US.  But you probably don't need me anyway; you'll have plenty of fun with Charlie if you go.

Anyway, that's the bad news.  But let me follow through with a proposal that we've discussed before---you might find some good news in it.  Let us DO write a paper together for Asher's festschrift.  ``Brassard and Fuchs, together for the first time,'' the show bill could say.  I'd like to call the paper ``Quantum Foundations in the Light of Quantum Cryptography'' (or some variation thereof) and I'll do almost all the work for it.  The only thing you'd have to do is decide whether you can agree with most of the outlandish statements I'll be making.  I'll have plenty of equations, though, so that it'll look respectable.  In this paper, I want to emphasize an information-disturbance principle (akin to the mechanism behind secure QKD \ldots\ thus the title and your involvement) more than I have in the past with my ``Quantum Foundations in the Light of Quantum Information'' papers.  It'll be a good quality paper.  Will you join me?

More immediately, after I finish the book review I'm writing for {\sl Physics World\/} today, I plan to immerse myself in the ARDA roadmap business starting tomorrow or Monday latest.  I sure hope we'll be able to put our heads together for this.  I'm gonna need a lot of guidance!!

\section{04-09-03 \ \ {\it Nietsnie dna Rhob} \ \  (to R. {\Schack})} \label{Schack74}

\brs
In Munich, I visited a small Einstein exhibition in the Deutsches Museum, where the appended small letter {\bf [[from Bohr to Einstein]]} (my translation) was displayed. Isn't that exactly the Buridan's ass point?
\bq
\noindent Dear Einstein,\medskip

Many thanks for your kind lines. It gave us all great joy to express
our feelings on the occasion of your birthday. To continue in the same
playful tone, I can't help saying about the unsettling questions that,
in my opinion, the issue is not whether or not we should stick to a
reality amenable to physical description, but to follow the path shown
by you and to identify the logical conditions for the description of
the realities. In my bold way I would even like to say that
nobody---not even the good Lord---can know the meaning of an
expression such as ``playing dice'' in this context.\medskip

\noindent With best wishes Niels Bohr
\eq
\ers

You and your ass!  As always, I have trouble deciphering Bohr.  There is something about the language pattern that never sinks in to me.  I know you want to say that he is saying that it is {\it meaningless\/} to discuss whether the world itself is either determined or undetermined.  But why?  I don't see why it is meaningless.  It may not be fruitful to ask.  It may not be answerable from within our inside or finite perspectives.  Both of those things I would be willing to grant you, and I regimen my life appropriately taking them into account.  But why is it meaningless?  I don't think Bohr ever gave an answer, and that is why it is pretty hard to find him convincing.

And don't forget to tell me how to pronounce Fuchs.

\section{07-09-03 \ \ {\it correspondentx.com} \ \ (to Correspondent X)} \label{CorrespondentX1}

I have gone to your website and looked at your open letter. You seem
like a sincere person---one who practices a healthy amount of
self-criticism, which is one of the most important assets of a
scientist.

Concerning the content of your letter however, I am not the best
person to write.  Unfortunately I have gathered the reputation of an
enthusiastic correspondent:  No doubt it comes about predominantly
from my having published reams of my own correspondence.  In the last
year alone, I count 14 unsolicited letters from people unknown to me
offering either a) to derive quantum mechanics on the cheap, or b) to
demonstrate that quantum mechanics is completely wrong.  This is not
an exaggerated number; I keep all these letters in a single folder
within my email program.

What I mean by ``deriving quantum mechanics on the cheap'' is a
magical combination of English words---very rarely an equation---that
makes the whole edifice (i.e., all those equations, relations, and
mathematical structures) of quantum mechanics make sense or even, all
of a sudden, become obvious.

You've got to realize it is a difficult call for me.  On the one
hand, you might be the one who really has hit that magical
combination---one that conveys an idea that has never before been
conveyed, or at least, of an old idea, says it in a way that finally
makes it become crystal clear and undeniable.  Who am I to stifle
that?  On the other hand, of your letter, I can say it didn't ``feel
right'' to me.

Now, I could give you reasons for that, but if I were to do so I
would have another correspondent on my list.  And that would mean
more Saturdays and Sundays away from my family as I plug away on
letters to another complete stranger who hasn't yet entertained me.
It's harsh, but I have to make a judgment call or I would be
inundated---I already am inundated!  You ask is there a difference
between POVMs and q-numbers somewhere in the middle of your letter,
and I think, ``If he couldn't get this straight from my paper, what
good is more correspondence going to do?''

All I can say is, if you are confident of your work, publish it or
post it on {\tt quant-ph}.  Someone will eventually take the bait and
give you feedback, and if no one does, that is feedback in itself.

As I say, and I am serious about this, you strike me as a sincere
person.  I apologize if I have hurt your feelings, but I really can't
take on another correspondent right now.

\section{08-09-03 \ \ {\it Einstein and Bohr:\ Back to Basics}\ \ \ (to G. L. Comer)} \label{Comer39}

\noindent Dear old Einstein,

\bgc
{\bf [Referring to the note title ``Acceleration is Intervention!'']}\ Ah, now you're talking!  I hadn't thought about the Fulling-Davies-Unruh
thing in a long time.  I'm also excited by the fact that this is essentially
Einstein's elevator, but in a completely new light.  But, what a can of
worms!
\egc

I think the only thing to do is to go completely back to basics \ldots\ to rejoin the Einstein of 1907, but now with the insights of Kochen--Specker theorems and Bell inequalities.  There's plenty of work ahead of us; that much I can see.

\bgc
Do you think you'll get that research center going?  I would really
love to be your token relativist.  My only condition is that a
hospital be nearby, so that Holly can work.
\egc

In the long term, it is a serious thing.  I was discussing it with Mabuchi---my young Caltech-professor MacArthur-fellow friend, who just got tenure---when we were in Denmark a few weeks ago.  He still doesn't think it's completely crazy and knows another MacArthur fellow who's done something similar.  Right now though, clearly, my focus is on some background stability for the family.  That'll give me a base to work from.

In fact, it's back to basics in all respects \ldots\

Hoping that you will in the future (just as I do myself) enjoy the enrichment coming from the different kind of access to science by different scientists, expressed in different, but not contradicting terminologies, I am sending, also in the name of Kiki, all good wishes to yourself, to Holly and to the whole family, \medskip

\noindent as yours complementary old, \medskip

\noindent Bohr

\section{09-09-03 \ \ {\it The Missing Word} \ \ (to A. Plotnitsky)} \label{Plotnitsky14}

The missing word was ``leads.''
\bq
\noindent
It is surely a significant feature of the theory that consideration of impossible outcomes and very little else leads, without any invocation of ``the uncertainty principle'' or ``maximal information'', to the fact that pure state assignments must be unique, as well as the more general constraint on mixed-state assignments.
\eq
Thanks for catching that!

\section{09-09-03 \ \ {\it Nonperiodic Penrose Tiles} \ \  (to R. {\Schack})} \label{Schack75}

Below is the transcription of the notes from our conversation.  They're pretty holey---the {\it e\/} is pretty important at this stage, even if it might not appear in the end product.

\begin{itemize}

\item
``On Quantum Certainty''

\item
Brief mention of Bayesian program.

\item
Penrose seems to rely on correspondence theory of certainty.  (unwarranted assumption)

\item
If we could act ``as if''---Martin Gardner---then why not the above?

\item
But why we can't act is because of Bell.

\item
Must reassess certainty.

\item
Draw attention to our own bad ways?  A footnote.  Simply say we were less noncommittal in our meaning of certainty previously.

\item
(emails)
Objective Certainty + QM  $\Longrightarrow$   Objective Probability

\item
({\Montreal})
Objective Operations  $\Longrightarrow$   Objective Quantum States

\item
Repeat key arguments for subjectivity of states.

\item
Argument for subjectivity of POVM and preparation operation.

\item
Very careful to state(ment) what we mean by subjective judgments.  It does not mean one can believe anything one wants to believe.  One issue is to decide when intersubjective agreement can be had.

\item
{\Spekkens}!

\item
What means quantum randomness?  It means that if I make a pure state assignment, I am also certain that no one else has more privileged information than me.  ({\Ruediger}'s point against Gisin crew + anti-{\Vaxjo} paper)

\item
Once one has got the concept of certainty straight, what would having an underlying instruction set add?  ({\Ruediger}'s point.)
\end{itemize}

Unfortunately, that's all I have written down.  It seems like we said so much more.  Also, I'm not even sure what some of the notations above mean.

\section{09-09-03 \ \ {\it Warm Hearts} \ \ (to H. J. Kimble)} \label{Kimble3}

Thanks!  I think there'll only be one more application going off soon:  [\ldots]

But now this
\bhjk
I remain a big fan of your work, including your recent manuscript
that explored the question of the ``quantumness'' of states (even if
you should be burned at the Quantum Optics stake!).
\ehjk
really warmed my heart!  Thanks.  I'll post another paper soon on the quantumness of a Hilbert space, which I can now prove is $2/(d+1)$ for dimension $d$.  In it I want to say some fantastical things about the E\"otv\"os experiment \ldots\ how quantumness plays the role of mass in these new considerations.  And I'll say some things along these lines in my talk at Preskill's group meeting Oct.\ 16.  Hope you can come.  (And don't miss Rob Spekkens's talk Oct.\ 14; it's really really really good work.)

\section{09-09-03 \ \ {\it The Big Man Speaks} \ \ (to M. Sasaki)} \label{Sasaki3}

We just got a little more positive feedback on our paper.  This sentence just came from Jeff Kimble:
\bq\noindent
   I remain a big fan of your work, including your recent manuscript
   that explored the question of the ``quantumness'' of states (even if
   you should be burned at the Quantum Optics stake!).
\eq

Do you understand the joke?  In any case, it is a compliment to us.  He is saying that we should be set to fire and burned alive as religious heretics have been in the past.  A stake is what people are tied to when they are burned this way.  The reason he is saying this is because we said in our paper (essentially):  ``[Coherent states] cannot be cloned, and this holds whether the quantum optics community calls such states `classical' or not.''  In other words, Fuchs and Sasaki are heretics!

\section{09-09-03 \ \ {\it Feynman's Rainbow} \ \ (to M. Durrani)} \label{Durrani1}

\bmd
In the final paragraph you say: ``I know that Hilbert space makes
my heart flutter, but does an atom?'' Do you mean that your current
work (on atoms?)\ doesn't really interest you and that you would rather
be working on Hilbert space? Or that you are glad of the field you're
working in.
\emd

Does an atom make my heart flutter?  Not really.  But the Hilbert spaces of quantum mechanics do.  The implication is the question:  Would Feynman think of me as a physicist or not?  Should I think of myself as a physicist or not?  In other words, this is just like the self-examination Feynman gave Mlodinow.  Also, the word ``self-examination'' refers back to the third paragraph in the review, which is the running theme of the essay.

I read the paragraph to three guys down the hall, and they all got it.  I say, let's leave it as written.  Or how about this:  Just to make it absolutely clear, let me change ``Hilbert space'' to ``quantum-mechanical Hilbert space'':
\bq
Still, the book contains a nub of wisdom here and there. Near the end
of it, Mlodinow tells how Feynman challenged him to think hard about
the emotions he finds upon viewing an electron-microscope image of an
atom.  ``Does it make your heart flutter?''  I have found myself
thinking about this all week.  I know that quantum-mechanical Hilbert
space makes my heart flutter, but does an atom? Sometimes the
self-examinations of Caltech life are hard to leave behind.
\eq

\section{10-09-03 \ \ {\it cond-mat/0309188} \ \ (to S. H. Simon)} \label{Simon1}

\bshs
I've heard some buzz about the quantum measurement paper {\bf [``The quantum measurement process: an exactly solvable model,'' \arxiv{cond-mat/0309188}]}.  Have you guys looked at it?  Any comments?
\eshs

Well, I see I'm depicted as one of the good guys in their reference
8.   Still I doubt they've absorbed what {\Caves}, {\Schack} and I have
been writing about.

The way I approach the issue of quantum measurement is to first
forget about quantum mechanics and talk about undergraduate
probability theory.  The textbook gives you a probability
distribution $P(h,d)$ over two variables $h$ and $d$.  From it you
can derive a marginal distribution over $h$ alone; namely $P(h)$. Now
suppose you gather an explicit piece of data $d$.  That may be some
information you can use for updating your expectations about $h$. The
mechanism is simply to use Bayes' rule:  You update from $P(h)$ to
$P(h|d)$.

No big deal, right?  So, let me ask you this:  Might I have pulled
the wool over your eyes any?  In particular, shouldn't I tell you
about the precise mechanism the brain uses for updating from $P(h)$
to $P(h|d)$?  Shouldn't I give a physical explanation for the
process? Isn't the undergraduate textbook treating the problem
incompletely by not requiring that I fill in those missing steps?

I think you'd be crazy if you said ``yes'' to any of those things.
The axiom of updating is primitive within probability theory.  It
calls for no physical mechanism behind it, and none should be sought.

With that out of the way, let us go back to quantum mechanics and
quantum measurement.  One can show formally that quantum measurement
is precisely the process above---i.e., Bayes' rule in action---modulo
an extra readjustment that takes into account the idea that in
quantum measurements one might actually have to touch the system one
is interested in (and hence cause some kind of disturbance).  This is
worked out in Sections 4.2, 6 and 6.1 of my paper \quantph{0205039}. (Have a look at it:  I'll bet money---real
money---that my paper is clearer than theirs.)

Anyway, from this point of view, collapse calls for no explanation in
the same way that the process of updating from $P(h)$ to $P(h|d)$
calls for no explanation.  In fact, you would have to give a physical
explanation for latter in order to give a physical explanation for
the former---i.e., you would have to give a physical mechanism for
Bayes' rule itself.  And we've already agreed (or at least we should
have) that that would be silly.

Now concerning the paper you've asked about:  They clearly think
they've done something.  But I'll bet, in the last analysis, they've
done essentially nothing \ldots\ except lead the reader down an
infinite regress that they haven't had the guts to analyze.

Anyway, that's my take after giving the paper not more than a
three-minute skim.  After thinking about these issues for years, one
realizes that papers like this aren't so different than the latest
proposal for a perpetuum mobile.  It's a question of finding the
flaw, and then---and this is always the hard part---convincing the
authors that it is a flaw.  It's almost never worth the time.

Now, my sarcasm aside:  Why has there been some buzz about this paper
in the community?  What's been found exciting in it?

\section{10-09-03 \ \ {\it Careful Reading Sir} \ \ (to D. M. {\Appleby})} \label{Appleby2}

\ldots\ will be rewarded.

Thanks for the long note and the self-esteem course.  I needed the
latter!

And I was happy for the former too.  However, I think most of the
fight you fight in those pages has to do with a Chris that was too
sloppy with his words in his youth.

A better Chris to fight with is the one represented in the
``\myref{Mermin101}{Me, Me, Me}'' note.  You might think you read it
carefully, but all the evidence shows that you didn't.

In particular, I would localize a good bit of your troubles in your
paragraph:
\bma
You say somewhere that Alice has the right to assign any state she
wants, just as Bob has the right to assign any state he wants. So
what is to stop Bush saying he has the right to assign any
probability he wants to the proposition ``every person held at
Guantanamo Bay is a terrorist murderer''?
\ema

But how does that mesh with the following lines from ``Me, Me, Me'':
\bq
\noindent
   The textbook poses an exercise that starts out, ``Suppose a hydrogen
   atom is in its ground state.  Calculate the expectation of \ldots\
   blah, blah, blah.'' One might think it is asking us to calculate
   some objective feature of the world.  It is not.  It is only asking
   us to carry out the logical consequences of a supposed state of
   belief and a supposed action that one could take upon the system.
   And here's the clincher about Bayesianism.  Just as no student in
   his right mind would find it worthy to ask why the textbook writer
   posed the problem with the ground state rather than the first
   excited state, no quantum theorist should make a big to-do about it
   either.  It is simply an assumed starting point.  An agent in the
   thick middle of a quantum application can no more ask where he got
   his initial beliefs from, than a pendulum can ask where it got its
   initial conditions from.  The cause of bottom-level initial
   conditions is ALWAYS left unanalyzed.  If such was not a sin in
   Newtonian mechanics, it should not be a sin in a Bayesian formulation
   of quantum mechanics.
\eq

In the words of the great {\Marcus} {\Appleby}, ``It's really hard to
believe something you don't actually believe.''  When---long ago!---I
said that Alice had the ``right'' to assign, I meant that there is
nothing about the system itself that determines her state for it.  I
did not mean that she had the right to believe something she does not
actually believe.

Here's another way that I've put it more recently, in a different
excerpt from the same samizdat:
\bq
Take a good solid physicist like Steven Weinberg who stakes his
career on the search for a grand unified field theory.  Suppose he
finds it.  To find it (I presume) is to declare:  The world's
Lagrangian is $L$.  Now suppose I were to ask Mr.\ Weinberg, ``Why
$L$? Why not $M$?''  I know for sure his answer will be of the form,
``$L$ just is.  It is the starting point.  It is an ultimate fact of
nature; it calls for no explanation.  In any case, if it calls for an
explanation, its answer must come from outside the realm of
science---religion? theology?---but I see no reason to go to such
lengths.''

Would that make Weinberg a solipsist?  A sensationalist?  A
phenomenalist?  The point is Weinberg's stance has nothing to do with
any of these labels.

Similarly for the Bayesian (even of the de Finettian variety, despite
the mumbo jumbo in the opening sections of Probabilismo).  For him,
``the prior'' on any event space is treated as an ultimate fact---an
ultimate fact about the agent.  There is no infinite regress because,
just as with Weinberg, ultimate facts call for no further
explanation.

Bayesian practice---and {\Ruediger} and I would claim the formal
structure of quantum mechanics too---is all about what to do once a
prior is established.  It is not about what to do before the prior is
established.  In the quantum mechanical case, establishing ``a
prior'' is 1) to write down a quantum state for all systems
considered, and 2) to write down a (conditional) quantum operation
for all measuring devices considered.

If there are two agents, there may well be two priors in the sense
above---i.e., two ultimate facts (with respect to this level of
inquiry).  In that sense, the priors are ``subjective'', but that
does not take away their status as ultimate facts in this treatment.
It only calls for a recognition that the facts are about the agents.

The role of the separate system and measuring device---now
specializing on quantum mechanics---is that when the two are combined
they give ``birth'' to a new ultimate fact:  The ``click.'' There is
no sense, however, in which this new ultimate fact is {\em about\/}
either of the agents:  It has a life of its own.  (In fact, it is
because of these lines of thought that I sometimes call my view ``the
sexual interpretation of quantum mechanics.'')
\eq

Now, I will make the bold claim (because I don't have three days to
spend on writing a new note), that if you look carefully at the words
above and carefully back at ``Me, Me, Me,'' you will see a lot of
your other troubles melt too.  For instance, look harder at the
glossary. Is there not something in there that corresponds to your
``a way things really are''?  (Though I would prefer, ``a way things
really come out.'')  And further, notice the explicit dualism in the
system/POVM definitions.  I'm going to claim that that dualism saves
me from being {\it either\/} an idealist or a materialist.

I believe in FACTS just as much as the next guy:  Quit accusing me of
not.  I only stress that facts are made, they do not preexist.  And
it is because of this that there is no one way the world is.  (The
capitalization in Nagel, by the way, was all my addition.)

Finally, here are two excerpts from your note that I really, really
liked.  I tend to agree with the first (whether you agree with it or
not) and I tend to like the sound of the second.
\bma
     One final point. You say that probability assignments float above
     the world (or words to that effect). This is true (on your
     principles) however much they are updated. Rationality resides
     purely in the updating procedure. The assignments themselves are
     never rational. Not only are they not rational at the outset,
     before any updating has been performed. They are no more rational
     at the finish, after all the evidence has come in. In short, they
     float. They tell us nothing at all about ``the way things really
     are''.
\ema
and
\bma
     I am tempted to say that the ordinary human world essentially
     CONSISTS of probability assignments. That probability, so far from
     floating, is actually constitutive of the world (what we normally
     and humanly take to be the world).
\ema

\section{10-09-03 \ \ {\it Facts} \ \ (to D. M. {\Appleby})} \label{Appleby3}

One final thought, to complete my last note.  Looking back at this
passage:
\bma
     One final point. You say that probability assignments float above
     the world (or words to that effect). This is true (on your
     principles) however much they are updated. Rationality resides
     purely in the updating procedure. The assignments themselves are
     never rational. Not only are they not rational at the outset,
     before any updating has been performed. They are no more rational
     at the finish, after all the evidence has come in. In short, they
     float. They tell us nothing at all about ``the way things really
     are''.
\ema
though I generally liked it, I would not use the word ``evidence.''
Facts and data are just facts and data.  To say that they are
``evidence'' already presupposes a (subjective) probability
assignment.

\section{11-09-03 \ \ {\it Peres Festschrift} \ \ (to R. Omn\`es)} \label{Omnes1}

That is too bad.  If you decide to reconsider at any time, please just let me know.  I think philosophy of science would be just fine.  And certainly you would not have to commit to 20 pages; that was listed only so that contributors would realize that they could use this as an opportunity for a long paper if they wished.  But it is not necessary.

Speaking of your philosophy of science, I was so impressed with some of the lines in one of your papers that I copied them into my computer.  They are beautiful.  (I'll paste them below so that you can see the ones I am speaking about.)

Things go well for me, though I could use some mathematical help!
\bq\noindent
R.~Omn\`es, ``Consistent Interpretations of Quantum Mechanics,''
Rev.\ Mod.\ Phys.\ {\bf 64}, 339--382 (1992).\medskip\\
Perhaps the best way to see what it is all about is to consider what
would happen if a theory were able to offer a detailed mechanism for
actualization.  This is, after all, what the advocates of hidden
variables are asking for.  It would mean that everything is deeply
determined.  The evolution of the universe would be nothing but a
long reading of its initial state.  Moreover, nothing would
distinguish reality from theory, the latter being an exact copy of
the former.  More properly, nothing would distinguish reality from
logos, the time-changing from the timeless.  Time itself would be an
illusion, just a convenient ordering index in the theory.  \dots\
Physics is not a complete explanation of reality, which would be its
insane reduction to pure unchanging mathematics. It is a {\it
representation\/} of reality that does not cross the threshold of
actuality. \dots\ It is wonderful how quantum mechanics succeeds in
giving such a precise and, as of now, such an encompassing
description of reality, while avoiding the risk of an
overdeterministic insanity.  It does it because it is probabilistic
in an essential way.  This is not an accident, nor a blemish to be
cured, since probability was found to be an intrinsic building block
of logic long before reappearing as an expression of ignorance, as
empirical probabilities.  Moreover, and this is peculiar to quantum
mechanics, theory ceases to be identical with reality at their
ultimate encounter, precisely when potentiality becomes actuality.
This is why one may legitimately consider that the inability of
quantum mechanics to account for actuality is not a problem nor a
flaw, but the best mark of its unprecedented success.
\eq

\section{11-09-03 \ \ {\it Bayesian Beginnings} \ \ (to G. M. D'Ariano)} \label{DAriano1}

\bgmd
I really enjoyed our discussions in Aahrus on Quantum Mechanics, and
when I got back to US, I careful read your long paper ``Quantum
Mechanics as Quantum Information (and only a little more)'' published
on the Proceedings of the {\Vaxjo} Conference, keeping also notes
for myself (something that I do with quite few papers). I should say
that I completely agree with your point of view, and I was happy to
see that in many considerations that I made recently, I'm not alone.
You know, my school was Masanao Ozawa, and in the last years I taught
quantum mechanics of measurements using the Bayes principle. However,
after reading your paper I really became a strenuous Bayesian
(including tomography).
\egmd

Thank you so much for your heartwarming letter.  I'm really glad you
got something out of my paper.  I've been really enjoying your new
ones on the structures of POVMs too:  But I already told you that.

Yes, it would be great to visit Pavia.  Maybe we could work out a
good time in the Spring.

\bgmd
Sorry if I bothered you. My problem is that now I became too much
involved in such re-considerations, that I can hardly think to
anything different.
\egmd

You didn't bother me at all.  This is what I live for!  They say
every parent wants a better world for their children---a better world
than they themselves grew up in.  This is true:  I want to see a
complete understanding of the structure of quantum mechanics
developed before my daughters go to college.

We're going to do it!

\section{15-09-03 \ \ {\it Coffee} \ \ (to B. C. Rasco)} \label{Rasco3}

\bbcr
You read hard core philosophy, I read the fun stuff. Popper, Kuhn,
etc.\ and some how to educate people on science. In half of the stuff
I read there is a lot of mixing up of what science is and how it is
created. They seem rather separate to me, but many arguments seem to
mix them up.

Nothing new in that last paragraph. Do you have any suggested readings
along the lines of what is science / what does it mean? As opposed to
how it is created.
\ebcr

Nah, I read the easier stuff too.  In particular, because the easier stuff is probably closer to being right.  Philosophers are a dangerous lot when they try to build more and more complicated arguments to shore up what they {\it want\/} to be true.

I think I've only read one book particularly devoted to philosophy of science:  Ian Hacking's {\sl Representing and Intervening}.  I did get a lot out of it, at least in terms of terminology.

But if you want a real recommendation from me, read William James's little book {\sl Pragmatism}.  It's been essentially a life changer for me, if you can imagine that.  And it is really, really easy reading.

\section{16-09-03 \ \ {\it Get Religion} \ \ (to B. C. Rasco)} \label{Rasco4}

\bbcr
Do send derogatory remarks, they are always fun.

That is how I sum up philosophers too. Many scientists are like that also.
There is one guy at the community college I am at who is basically a
classical physicist and is pretty adamant about it. I actually like him
because he thinks I know enough to argue with him, and he admitted to me his
limitations, Charles Stores. He is a nice guy once you get through his
hatred of religion, but I think his faith in science is exactly the same
thing.

I am going to write a paper for him about science being a creation of man
just like all other explanations of the world (a.k.a.\ religion) but that it
involves logic and hence its actual usefulness, power, and predictability.
He probably will not like the elevation of religion, but it is fun to get
people flustered, they tend to think less clearly. The other thing I want to
say about it is that whether science can describe everything is up for grabs
based on whether or not the entire world is logical, I think not but I must
leave that open to debate.

Anywho is there any philosophy out there that you know of that
discusses those kinds of ideas? I just hate repeating others, because
they always seem to do it better.
\ebcr

If that's the direction you want to go, then definitely read the book {\sl Pragmatism\/} I suggested to you before.  But maybe more to the point, read William James's essay ``The Sentiment of Rationality.''  As I understand what you're wanting to get at, that article is exactly in the right line.  That essay is not included in the book {\sl Pragmatism\/} but you can find it in almost any collection of his essays in any bookstore.  I just happen to have a couple of extended excerpts from the essay in my computer; I'll paste them below.  [See 21-11-01 note ``\myref{Schack37}{Pragmatism versus Positivism}'' to R. Schack.]

Another thing I'd suggest reading---really easy reading---is Richard Rorty's little book {\sl Philosophy and Social Hope}.  Especially look at essays 2, 3, 10, 11, and 12 \ldots\ but in backwards order.  12 is titled ``Thomas Kuhn, Rocks and the Laws of Physics.''  11 is titled, ``Religion as Conversation-stopper.''  10, ``Religious Faith, Intellectual Responsibility and Romance.''  3, ``A World without Substances or Essences.''  And 2, ``Truth without Correspondence to Reality.''

The last two essays had a tremendous influence on me, even if I don't agree with everything in them.

If you get through this reading list and it seems along the right lines for what you want to explore, write me and I can recommend more.

\section{16-09-03 \ \ {\it Two Typos, One Disagreement} \ \ (to A. Peres)} \label{Peres53}

Two typos:
\begin{itemize}
\item[1)] One day, I came into Rosen's office, and I found sorting out his papers.

\item[2)] The EPR article was not wrong, but it had be written too early.
\end{itemize}

One disagreement:
\bap
Information is not an abstract notion. It is a physical object which
requires a physical carrier, and in particular is {\it localized}.
\eap

Let me lodge my disagreement with this language, by pasting in two old notes.  [See 25-04-02 note ``\myref{Bennett15}{Short Thoughtful Reply}'' to C. H. {\Bennett} and 02-02-02 note ``\myref{Timpson1}{Colleague}'' to C. G. {\Timpson}.]  The first explains my disagreement I think fairly well.  The second is there mostly for the purpose of relating an anecdote about Landauer.

Where your language above and [Charles Bennett's] language below come closest to each other is when he talks about the ``information you just ate.''

\section{17-09-03 \ \ {\it More Than Semantics, I Think} \ \ (to A. Peres)} \label{Peres54}

\bap
Information is not an abstract notion. It is a physical object which
requires a physical carrier, and in particular is {\bf localized\/}.
\eap

Our disagreement is localized in your phrase (effectively)
``information is a physical object.''  I would say the thing that we
have learned from quantum information is that we cannot, even in
abstracto, ignore the physical properties of information CARRIERS.
That was something ignored in the original Shannon information
theory.  But recognizing that is a far cry from saying ``information
is a physical object'' \ldots\ which, to me at least, conveys the
idea that information is something that can exist independently of
the agent possessing it.

At the end of your article, you write:  ``Quantum states are not
physical objects: they exist only in our imagination.''  True enough.
But, by the view we expressed in our {\sl Physics Today\/} Opinion, a
quantum state is nothing beyond one's information about a system.
Here's one of the ways we said it then:
\bq
\noindent
      From this example, it is clear that a wavefunction is
      only a mathematical expression for evaluating probabilities
      and depends on the knowledge of whoever is doing the computing.
\eq
It seems to me that if you are going to claim that information is a
physical object you will be stuck in an inherent contradiction (at
least in phraseology).

\section{17-09-03 \ \ {\it Went on to Organize} \ \ (to W. M. Fuchs)} \label{FuchsW1}

\noindent Dear Mike,\medskip

I've been thinking about what I'd like to write you for your
birthday.  50 is quite an important one.  On the one hand, I wanted
to say, ``Well, you're half way there.''  But that would have been a
lie.  Life is much, much bigger than that:  Even a hundred years
isn't the half of it.

There's a song that's been tumbling in my head the last few days that
Joan Baez sings.  It's called {\sl Joe Hill}, and it's about a labor
organizer who was active in the 1910s.  It's an important song.  It
conveys the idea that an honest and true life never dies.  With every
smile you give your nieces, your brothers, your sisters, your mom,
your friends, your coworkers, you'll never die.  With every piece of
wisdom you pass on, you'll never die.  Every bit of science I do
stems from the starlit conversations we had in the 70s.  Life
propagates in a lot more ways than the material.

Keep up the good work!\medskip

\noindent Happy birthday brother,\medskip

\noindent Chris

\bv
\underline{Joe Hill}\bigskip\\
I dreamed I saw Joe Hill last night,\\
alive as you or me.\\
Says I ``But Joe, you're ten years dead''\\
``I never died'' said he,\\
``I never died'' said he.\medskip\\

``The Copper Bosses killed you Joe,\\
they shot you Joe'' says I.\\
``Takes more than guns to kill a man''\\
Says Joe ``I didn't die''\\
Says Joe ``I didn't die''\medskip\\

And standing there as big as life\\
and smiling with his eyes.\\
Says Joe ``What they can never kill\\
went on to organize,\\
went on to organize''\medskip\\

From San Diego up to Maine,\\
in every mine and mill,\\
where working-men defend their rights,\\
it's there you'll find Joe Hill,\\
it's there you'll find Joe Hill!\medskip\\

I dreamed I saw Joe Hill last night,\\
alive as you or me.\\
Says I ``But Joe, you're ten years dead''\\
``I never died'' said he,\\
``I never died'' said he.
\ev

\section{18-09-03 \ \ {\it Instrumentalism} \ \ (to A. Peres)} \label{Peres55}

After reading your new article, I looked up the word
``instrumentalism'' in my {\sl Encyclopedia Britannica}.  Here is
what I found.
\bq
\noindent {\bf Instrumentalism:}  also called Experimentalism, a
philosophy advanced by the American philosopher John {\Dewey} holding
that what is most important in a thing or idea is its value as an
instrument of action and that the truth of an idea lies in its
usefulness. {\Dewey} favoured these terms over the term pragmatism to
label the philosophy on which his views of education rested. His
school claimed that cognition has evolved not for speculative or
metaphysical purposes but for the practical purpose of successful
adjustment. Ideas are conceived as instruments for transforming the
uneasiness arising from facing a problem into the satisfaction of
solving it.
\eq

\section{19-09-03 \ \ {\it de Finetti vs.\ Jaynes} \ \ (to D. Poulin)} \label{Poulin9}

\bdp
In thermodynamics, the hypothesis of exchangeability seems well
motivated, for noninteracting systems at least. Therefore, if one was
to measure a complete set of observables, one would be left with a
product state of the form
($\rho\otimes\rho\otimes\rho\otimes\cdots$), i.e.\ the ``probability
distribution over states'' appearing in the de Finetti representation
would have converged to a delta. But if one measures less
observables, e.g.\ a single one, his state assignment should not be a
product space unless his ``prior distribution over state'' has a very
special form, which I wouldn't know how to justify. On the other
hand, Jaynes tells us that after having measured say the average
energy, our state assignment should be a product state as above with
$\rho = \exp(-\beta H)$. As you know very well, this is done by
maximizing the entropy given the constraint observed: the product
emerges as a consequence that the systems are not interacting.
\edp

I understand your problem, and it's a mistake a lot of people make.
The main point to keep in mind is that a max-ent assignment is a
SINGLE system assignment.  One cannot extend it to a multi-system
assignment unless one is confident that further measurements will
reveal no further information useful for updating.

There is a fairly thorough discussion of this issue in \quantph{0010038} by Brun, {\Caves}, and {\Schack}, though in a different
context than the one you bring up above.  The substance of the issue
is the same, however.  The Horodecki's had made the assumption that
max-ent gives a multi-system assignment and it caused them to get a
nonsense result.  But rather than thinking hard about where they
might have gone wrong, they just plowed ahead and declared the
max-ent program to be ``wrong'' in quantum mechanics.

A more important thing to read is Jaynes' own discussion of the issue.
You can find that in reference 13 of the paper
above.\footnote{\editornote Reference 13 of the Brun, Caves and Schack
  paper is E. T. Jaynes, ``Monkeys, kangaroos, and $N$,'' in {\sl
    Maximum Entropy and Bayesian Methods in Applied Statistics,}
  edited by J. H. Justice (Cambridge, 1986), pp.\ 26--58,
  \url{http://bayes.wustl.edu/etj/articles/cmonkeys.pdf}.  But this
  doesn't appear to be the correct paper.  Instead, see E. T. Jaynes,
  ``Prior probabilities,'' IEEE Transactions on Systems Science and
  Cybernetics {\bf SSC-4} (1968), pp.\ 227--41,
  \url{http://bayes.wustl.edu/etj/articles/prior.pdf}.}

\bdp
How are you? Here, things are going very well, many interesting
research projects on the way. As always, I am preaching for the
epistemic interpretation of quantum mechanics. Believe it or not, I
have almost convinced Valentini that this is the right way to think
about states.
\edp

Now this is very interesting!!  The last I remembered, you were still
fairly agnostic on the issue.  I'm really glad to hear of this.

I've been doing some fairly technical things lately.  In particular,
I've been trying to shore up my old idea that quantum mechanics looks
the way it does because it is predominantly about how we should
manipulate and update our information in light of the fact that
quantum systems have a kind of ``sensitivity to the touch.''  Thus
I've been playing with this measure of quantumness that Sasaki and I
posted on {\tt quant-ph} a while ago.  In the coming month I'll put
another paper on {\tt quant-ph} showing how the quantumness of a
Hilbert space (by this measure) connects up to doing quantum
mechanics on a simplex (remember the pictures I incessantly drew in
{\Montreal}?).

The deeper thing on my mind, though, has been how this ``sensitivity
to the touch'' is universal \ldots\ in a weird analogy with
gravitation. An 18 dimensional Hilbert space, for instance, does not
care whether it is embedded in copper or gold.  Just like a given
amount of gravitational mass does not care if it is embedded in
copper or gold. (I use copper and gold because I think those were the
metals used in the E\"otv\"os experiment; I should look it up.)  This
intrigues me to no end.  I'm going to give a talk at Caltech on it in
October; I wish you could be there to keep me on my toes.

Also {\Schack} and I are finally distilling my big thing ``Quantum
States:\ What the Hell Are They?''\ into a single paper.  The
tentative title is ``On Quantum Certainty.''  In it we'll tackle the
Penrose argument head on and also how quantum operations are
epistemic in exactly the same way as quantum states.  Your revelation
of non-agnosticism, by the way, was the second piece of good news I
read today. Concerning this new paper of Schack and mine, David Mermin wrote me this:
\bdm
I think we're closer than it might appear (but I doubt that it's
close enough to write a joint paper about it with you and {\Ruediger},
much as I would enjoy doing it (and I really would, in spite of my
fulminations against collaborating in ``writing physics''.)
\edm
I was absolutely floored!  Now that signaled progress!  That he would even contemplate joining in on the project signaled that something is really going right here.

Workers of the world unite!, I say.  Good hearing from
you, and I hope the max-ent references given above will make the
answer to your question clear---I think they will.

\section{22-09-03 \ \ {\it Cosmology} \ \ (to C. H. {\Bennett})} \label{Bennett29}

Here's something I've been wanting to tell you about for a while, but
I kept forgetting.  Do you recall the conversation we had after our
dog Wizzy died?  You thought briefly that I had beat around the bush
with Emma, not quite telling her the truth about the event.  (I'll
place the old note below in case you don't remember.)

Anyway, about six months to a year after that---I wish I could
remember precisely when---Emma was hanging out in my office one day,
kind of swinging back and forth with the door, and out of the blue
asked, ``Dad, will you die one day?''  That one was a tough question:
I knew I had to say yes, but I didn't want to say yes.  In the end, I
told her, ``Yes, but you shouldn't worry about it so much; it won't
happen until you're grown and have children of your own.''  Now, that
WAS beating around the bush---of course I couldn't know when I'm
going to die.  Funny though how I thought the promise of having one's
own children at the time would soften the blow.

The reason I'm telling you this story after all this time is because
Saturday seemed to mark another, perhaps not unconnected, event in
Emma's development.  For the first time she asked me a cosmological
question.  She's about 4 and 2/3 years old now.  She asked, ``What
was here before the dinosaurs?''  I explained a little about the
earth before life, and about how the earth was formed in the
formation of the sun.  Then she asked, ``What was here before the
sun?'' I said, ``Ah, you see the pattern:  You can always ask a
`before' question, can't you?''  She said, ``Wow!''  (Not a lie.)
Then I said, ``Well I don't really know to any extent what was here
before the sun,'' deciding to skip a discussion about the big bang
and all that. I guess she decided to have some fun with me because
then she came back with, ``You don't know what happened before the
sun??!'' I said, ``Nope, no one does really,'' deciding to be
metaphorical for the moment---i.e., switching the sun and the big
bang in my mind. She questioned in surprise, ``Not even William
{\James}?'' I said, ``Not even William {\James}.''

\section{22-09-03 \ \ {\it More General Relativity} \ \ (to C. H. {\Bennett})} \label{Bennett30}

\bcb
Reading your paper on quantumness of a set of states, I see that you
define quantumness of a set of states by pessimizing the QCQ channel's
I/O fidelity over all choices of probabilities of states in the
ensemble, while optimizing it over measurements and preparations of
the states resp.\ before and after they are forced through the C part
of the channel. In this sense you view a set of states as more
fundamental than an ensemble.  But isn't a mixed state $\rho$ even more
fundamental than either a set of states or an ensemble?  What would be
wrong with defining the quantumness a density operator $\rho$ as the
least expected I/O fidelity for any ensemble realizing $\rho$; or if that
diverged, for any ensemble consisting of $r$ linearly independent pure
states, where $r$ is the rank of $\rho$?
\ecb

I don't think it would diverge because the maximum eigenvalue function (which appears in a higher order expression for accessible fidelity) is a convex function.  So, the definition is mathematically sensible.  I just hadn't seen a motivation for a quantity like that.  It kind of reminds me of a quantity that Oliver Cohen once wrote about \ldots\ just digging it up, \quantph{9907110}.

Actually if you want to go down a route like that, I think it is better to think of accessible fidelity as a function of an entanglement-breaking channel---it is the channel that is the fundamental thing.  As Audenaert, King, Winter and I show in the other paper on quantumness, for each ensemble, one can make a channel of particular interest.  However then, for any ensemble that gives rise to the same channel, one gets the same accessible fidelity.

In any case, what I'm most interested in at the moment is the quantity I called ``quantumness of a Hilbert space.''  I can now prove that it is numerically equal to $2/(d+1)$, and that to achieve that value on a tensor-product Hilbert space one must use an ensemble of entangled states.  I'm writing that paper right now.

What intrigues me most is the universality of some of these concepts like quantumness (as a potential characterization of Hilbert space itself).  I guess you've been saying something similar for a lot of years, talking about the fungibility of quantum information, but it had never really taken me until I started thinking about analogies with general relativity.  Here's the way I put it to David Poulin the other day:
\bq
The deeper thing on my mind, though, has been how this ``sensitivity to
the touch'' is universal \ldots\ in a weird analogy with gravitation.  An
18 dimensional Hilbert space, for instance, does not care whether it
is embedded in copper or gold.  Just like a given amount of
gravitational mass does not care if it is embedded in copper or gold.
(I use copper and gold because I think those were the metals used in
the E\"otv\"os experiment; I should look it up.)  This intrigues me to no
end.  I'm going to give a talk at Caltech on it in October; I wish you
could be there to keep me on my toes.
\eq

If you were there, maybe I'd be dancing a ballet!

\newpage

\section{22-09-03 \ \ {\it Lockbox Louder} \ \ (to J. A. Smolin)} \label{SmolinJ7}

Attached are my remarks.  They're almost all trivial and about presentation.

I enjoyed the paper a lot.  In fact, once you touch the paper up, I wouldn't mind reading it again.  I was particularly taken with your real toy model:  Mainly by just how much it resembles an EPR pair \ldots\ or correlation without correlata as David {\Mermin} would say.  (Well, the correlata are there but completely hidden---they almost might as well not be there.)  That is intriguing.

I'm sure I did a lot of things to the draft that will annoy you.  Certainly feel free to ignore any and everything you wish.

\bjas
A {\bf lockbox}, the basic unit of matter in this theory, is an
object akin to a physical box locked with a combination lock, which
can contain a bit value $b$.  The value cannot be read out of the
lockbox except if a particular string of bits $C$---the
combination---is presented to it.  The bit $b$ and combination
$C$ are chosen by the lockbox's creator at the time of its creation.
If the lockbox is presented with an incorrect combination, the bit
value is destroyed. This box need not be allowed by physics, it is
instead the building block of the toy theory.

Such a box cannot exist in classical mechanics.  It is often said
that one way in which quantum mechanics differs from classical
mechanics is that it cannot be represented by a local hidden variable
theory.  This statement hides a common oversight about classical
mechanics. Classical mechanics also is not correctly represented by a
local hidden variable theory, but by a local {\bf unhidden\/}
variable theory---in principle every possible property of a
classical system can be measured perfectly.
\ejas

You might cite the ``nonlocality without entanglement'' paper here.  Apparently also,
Bell himself somewhere along the way made this distinction.  There's
a nice discussion of it in Marcus Appleby's paper \quantph{0308114},
pages 14--16.  Why don't you cite him too?  He, or Bell, called them
``exposed variables'' interpretations.

\bjas
Thus a lockbox explicitly mimics the quantum property that unknown non-orthogonal states cannot be
cloned (copied) or even measured without disturbance.
\ejas

Have I ever
told you how much I hate it that physicists have taken it upon
themselves to start a new rule in English:  Thou shalt hyphenate all
words beginning with the prefix non.  And they do this in spite the
fact that almost every dictionary writes nonabsorbent, nonabsorptive,
nonadaptive, nonaddictive, nonadhesive, nonadjacent, nonadsorbent,
nonadsorptive, nonaged, nonagenarian, nonaggressive, nonalcoholic,
nonaligned, nonappointive, nonarbitrable, nonarbitrary, nonarboreal,
nonassertive, nonassociative, nonastringent, nonautonomous,
noncitizen, noncombatant, noncommissioned, noncompliance,
nonconductor, nonconformance, nonconformist, nonconformity,
nondepository, nondescript, nondevelopment, nondirectional,
nondiscretionary, nondisjunction, nondrinker, nondriver, nonentity,
nonequivalence, nonessential, nonevent, nonexistence, nonfat,
nonfeasance, nonfiction, nonfictional, nonflowering, nonfluent,
nongonococcal, nonindulgence, noninterference, nonintervention,
nonlinear, nonmalignant, nonmember, nonmetal, nonobservance,
nonoccurrence, nonparametric, nonparticipant, nonparticipation,
nonparticulate, nonpartisanship, nonpayment, nonperformance,
nonrapid, nonreader, nonreligious, nonremittal, nonresistance, \ldots.  This is a habit that has got to be squashed.

\bjas
It is straightforward to implement secure key distribution using
lockboxes.  As in quantum key distribution, Alice and Bob are assumed
to share an ordinary classical channel, which is unjammable and
authenticated, and to have the ability to create and send states, in
this case lockboxes, to each other.
\ejas

We've had this discussion before:  One sends systems not states.  There is a
difference in my eyes and an important difference.  One never speaks
of sending the color red.  One never speaks of sending happy down the
channel.  Lockboxes are not states, but systems.

\section{22-09-03 \ \ {\it A Thought of You} \ \ (to A. Shimony)} \label{Shimony4}

I thought of you when I wrote a little story about my daughter Emma and cosmology to Charlie Bennett earlier this morning.  In particular, I was thinking about how you wrote me this:
\bas
Another bridge is our common admiration of the pragmatic tradition in
American philosophy. You love William James, and I have respect for
James but reverence for Charles S. Peirce.
\eas
Clearly, you and Emma have attuned to the same vibe!

I'll attach the story (for your enjoyment) and the previous obituary of our dog Wizzy (so you'll understand the complete context).  Everything below is copied in reverse chronological order.  [See 22-09-03 note titled ``\myref{Bennett29}{Cosmology}'' to C. H. {\Bennett}, 17-02-02 note titled ``\myref{Bennett10}{The Process}'' to C. H. {\Bennett}, and 16-02-02 announcement titled ``\myref{ShouldNotPass}{Some Things Should Not Pass}''.]

No need to reply.  I was just thinking of you.

\subsection{Abner's Reply}

\bq
Thank you for the beautiful letter concerning Wizzy and Emma. I was
very moved by your conversation with her, both with her questions and
remarks and with your psychological difficulties in telling her some
painful truths. You were certainly right to meet her fear of your death
by the supposition that she would have children of her own. Why not
share with your child the thoughts that give you the most consolation --
namely, that something you care infinitely for will live after you? Most
of all you care for your daughter living after you, but why not
generalize somewhat and tell her that you care very much that there will
be a better world following you. Even a small child can understand the
idea of a surrogate for immortality in the continuation of things one is
devoted to. I feel that my sons have internalized this idea, and I hope
that it helped them when they lost their mother eight years ago.

Her elevation of William James to the status of prophet is both
amusing and inspiring.

\eq

\section{23-09-03 \ \ {\it The Trivial Nontrivial} \ \ (to S. Savitt)} \label{Savitt3}

\bss
Secondly, I realized (as I had not before) that a good way to think
about the issue I'm currently struggling with (in philosophy of time,
of course) would be to see it as trying to reconcile subjective (or,
at least, perspectival) and objective views of time. So, I am using
that slender excuse to send you a copy of my paper-in-progress. If
you do look at it, you won't see the connection till you get to the
second half, the constructive argument.
\ess

Well, I tried to give your paper a shot over the weekend, but it was
pretty tough going for me.  At times like this, I really feel the
lack of a philosophical training!  It can be a real impediment.
Probably the best critique I could muster for you would be to point
out two typos (``Temporal Ontolgy'' and ``simpliciter, on might
say'') and tell you to take away the hyphens behind your nons.
(Except for possibly ``non-usual,'' which is a non-usual word
anyway.)

More seriously, I wonder if you could help me classify the view of
George Herbert {\Mead} presented in his little book {\sl The Philosophy
of the Present}?  For the moment it strikes me as neither a species of
presentism nor a species of eternalism (to the extent that either of
those terms are coherent, or more to the point, to the extent that
{\Mead} is coherent!).  I'll attach a copy of a passage I scanned out of
Arthur Murphy's introduction to the book; I just happen to have
it in my computer (for my ``Resource Material for a
Paulian--{\Wheeler}ish Conception of Nature'' project)---you've probably
long since forgotten about {\Mead}.

Weirdly I'm starting to take this view more seriously than I had
before---even though I have previously used a version of it via John
{\Wheeler}'s slogan ``the past exists only insofar as it is recorded in
the present'' for dramatic purposes.  (See the ``Postpartum''
chapter, pp.\ 680--681, at the end of my samizdat.)

Of course, the reason I'm taking it weirdly seriously has to do with
quantum measurement theory.  It has to do with the Bayesian updating
I wrote you about previously.  One of the things I emphasized there
is that quantum updating (upon the recognition of a measurement
result) can be viewed as Bayes' rule simpliciter---thanks for teaching
me that word---as long as the quantum system whose quantum state is
being updated is causally disconnected from the measurement action.
That is true.  One thing I did not emphasize, however, is the nature
of the argument, $x$, for such a probability function, $P(x)$---i.e.,
the one we are talking about using Bayesian conditionalization upon.

The value of the random variable $x$ cannot be viewed as a
pre-existent fact of which we are ignorant.  Rather it must be viewed
as a potential consequence of our actions (i.e., measurements, in
older language).  Our ignorance is about what will be the actual
consequence of our actions, not about what is pre-existent.  Well,
the same holds true when we are using the outcomes of our quantum
measurements here and now for updating our quantum states for
something there and then.  The updating does not update our ignorance
of what was existent there and then, but rather updates our
predictability of what would come about {\it were\/} we to interact
with that system there and then.  Unfortunately, the latter we cannot
do. The best we can do is to wait for the system to move into our
present and interact with it then.

I think this is an important point.  Let me try to put it another
way:  Our quantum measurements in the present never tell us about the
past.  They only tell us about the consequences of the past for any
of our other measurements in the present.

I wish I could put it more clearly than that, but that's the best I
can do for now.  Despite first appearances, I don't think this is a
species of presentism:  It does not say ``only things in the present
exist'' (unless it is going to be in your non-usual way).  It just
says, the quantum states I write down are always ultimately about the
present:  They gauge my expectations for what {\it would\/} happen as
a consequence of my reaching out and touching my systems {\it now}.

Maybe I'll let it go at that for the moment.

\begin{center}
From: ``Introduction'' to George Herbert {\Mead}'s {\sl The Philosophy
of the Present}, \\ by Arthur E. Murphy
\end{center}
\bq
The present is to be taken as the locus of reality. This means, I
take it, that to consider anything as real is to consider it as
existing in, or in relation to, a present. Now what, in relation to
any present, is the status of its past? This is not to ask what it
was when it was present, for then it was not past and did not stand
in that relation by virtue of which it acquires the status of
pastness. The past of an event is not just an antecedent present.
This is Mr.\ {\Mead}'s main thesis throughout, but it does not often get
as clearly expressed as in the following statement. ``When one
recalls his boyhood days he cannot get into them as he then was,
without their relationship to what he has become; and if he could,
that is, if he could reproduce the experience as it then took place,
he could not use it, for this would involve his not being in the
present within which that use must take place. A string of presents
conceivably existing as presents would not constitute a past.''

The distinctive character of the past in its relation to the present
is manifestly that of irrevocability. As conditioning the present, as
making its occurrence possible, the past must have been of a
determinate character. It expresses the settled condition to which
the present must conform and without which it could not have been
what it is. And this means not merely antecedent occurrence, it means
causal determination or, as Mr.\ {\Mead} tends to put it, the ``carrying
on of relations.'' The past is that out of which the present has
arisen and irreversibility---the appeal might here have been made to
Kant---has its critical value in terms of such conditioning.

Yet this carrying on of identical relations is never the whole story.
The doctrine of emergence asks us to believe that the present is
always in some sense novel, abrupt, something which is not completely
determined by the past out of which it arose. A present, if it is
really new at all, will have in it an element of temporal and causal
discontinuity. Recent quantum physics has taught us to believe that
such indetermination is quite consistent with rigorous physical
analysis. But how is it possible to reconcile this novelty with
scientific determinism?

The answer to this question supplies the basic principles of the
theory. Before the emergent has occurred, and at the moment of its
occurrence, it does not follow from the past. That past relative to
which it was novel cannot be made to contain it. But after it has
occurred we endeavor to reconstruct experience in terms of it, we
alter our interpretation and try to conceive a past from which the
recalcitrant element does follow and thus to eliminate the
discontinuous aspect of its present status. Its abruptness is then
removed by a new standpoint, a new set of laws, from which the
conditions of our new present can be understood. These laws could not
have been a part of any previous past, for in the presents with
relation to which those pasts existed there was no such emergent
element. To assume a single determinate past to which every present
must wholly conform is to deny emergence altogether. But at the same
time, to treat the emergent as a permanently alien and irrational
element is to leave it a sheer mystery. It can be rationalized after
the fact, in a new present, and in the past of that present it
follows from antecedent conditions, where previously it did not
follow at all. As the condition of the present, the past, then, will
vary as the present varies, and new pasts will ``arise behind us'' in
the course of evolution as each present ``marks out and in a sense
selects what has made its own peculiarity possible.''

Is there any contradiction between this novelty of the past and its
essential irrevocability? None at all, for the two apply in different
senses. The irrevocable past is the past of any given present, that
which accounts for its occurrence. Its determining conditions will be
ideally if not actually fully determinable in the present to which it
is relative. But when a new present has arisen, with emergent facts
which were really not contained in the former present, its
determining conditions, hence its past, will of necessity be
different. The determinism then holds of the past implied in any
present, the emergence in the relation of one such present, with its
past, to another.

This hypothesis, in Mr.\ {\Mead}'s opinion, has two main advantages. In
the first place it accounts for the attitude of the research
scientist toward the data he is describing, an attitude otherwise
highly paradoxical. The laws of any science do in a sense reconstruct
the past out of which its given elements have arisen. So much is
assumed in the establishment of determinate laws, and for the
scientist to suppose that the present did not follow from the past in
terms of the laws he had established would be to deny their adequacy
to the data they interpret. So far as it goes in any field science
tends to be deterministic. Yet this ``following'' of present from
past is wholly relative to the data on which the interpretation is
based, and the scientist looks forward with equanimity to a new
interpretation, and hence a new past, relative to the emergent data
which the future will supply. And this combination of relative
determinism and future reconstruction which holds for the research
scientist, holds also, on this theory, for the nature he is
describing.

Secondly, this view is in harmony with the emergence of novelty in
experience, and the reorganization of experience in terms of it. This
is the theme of the first Supplementary Essay. Even those who
``bifurcate'' nature most relentlessly must admit that in experience
data may appear as intrusive elements in a world which has, in its
present constitution, no place for them. They stand in contradiction
to that world as currently interpreted and set a problem for
reconstruction. To interpret the world exclusively in terms of the
conditioning objects which a given period has isolated as the
permanent background of becoming is to relegate novelty to a merely
subjective experience. But in the case of data relevant to his own
problems a scientist makes no such bifurcation. Rather does he treat
the data as provisionally isolated in a world that does not now
account for them, but as candidates for admission to a reconstituted
world which may make the facts previously rejected the very center of
its interpretation. So it was, for example, in the status of the
Michelson-Morley experiment, first in its relation to classical
mechanics, then in the theory of relativity. Within experience new
objects are continually arising and a new present reorients the
settled conditions of an older era in the light of its discoveries.
And if the past is this orientation of settled conditions with
respect to present data, the past does empirically change as
evolution proceeds. This empirical description has been a part of
Mr.\ {\Mead}'s philosophy for many years. The novelty of the present
account arises from its correlation with the structure of temporal
reality as such, in the relation of a determining past to an emergent
present.

At this point the reader will be all too likely to object that it is
clearly only our viewpoint or interpretation of the past that has
altered here. The past in itself has surely not been changed by the
new way in which we have come to look at it. This however is just the
distinction that Mr.\ {\Mead}'s whole analysis attempts to supersede.
For a temporalist philosophy the past ``in itself'' is not a past at
all---its relation to the present is the ground of its pastness. And
this relation is empirically a causal one. If becoming is real that
causal relation is never such as to exclude emergence. When emergence
occurs a new perspective of the past, a new relatedness, will
ensue---a relatedness which is a natural fact about the new
situation, though it could never have occurred in the old. And what
is here new is precisely the way in which what, in the older present,
was merely novel and abrupt has become a part of the world of causal
objects, hence a part of the past through which they are supposed to
operate. The relatedness is real, and the perspective past it
generates, the past of the new present, is the real past of that
present, and only for a present can the past be real at all.

Mr.\ {\Mead}'s most objective version of his thesis occurs in Chapter
Two, in the contrast between the past as relative to a present and
the past as absolute. He holds, especially in criticizing Alexander,
that the past which physics requires is simply the expression of
identical relations in nature, not an antecedent environment,
existing in itself and giving rise, in its isolated being, to all
subsequent reality. Space-Time in Alexander's metaphysic, seems to be
a mathematical structure taken out of relation to the physical data
it interprets and transformed, in all its abstract independence, into
a metaphysical matrix from which all the complexities of nature are
somehow to be derived. This, on {\Mead}'s view, is just what the past
``in itself'' would be, a conditioning phase of natural process
turned into a metaphysical substance. The search for such a substance
is not ruled out for those whom it may concern. But the research
scientist cares for none of these things.

We seem, then, to have discovered in temporal transition itself a
unique sort of relativity, and a set of what we are now to describe
as ``temporal perspectives'' or ``systems.'' Each such system is
distinguished by the temporal center from which its relation to past
events is organized, and they differ primarily in this, that what is
external, contingent, hence ``emergent'' for one such standpoint will
``follow from'' and hence be reflected in the past of another. How
are such perspectives related, and how does the transition from one
to another take place? The answer can be given only when we have
inquired into the nature of relativity, and into its social
implications.
\eq

and

\bq
The argument returns at the end, as it should, to its point of
departure. It is in a present that emergent sociality occurs. And we
can now see that such a present is no mere moment of time,
arbitrarily cut out from an otherwise uniform ``passage of nature.''
A present is a unit of natural becoming; it is the period within
which something temporally real can happen. What has been and what
may be have their focus and actualization in a present standpoint and
it is from such a standpoint that creative intelligence, transforming
the novelty of emergence and the fatality of mere repetition into a
measure at least of meaningful development, brings to articulate and
self-conscious expression the pervasive form of natural process. It
is as the scene of such process that the present is the locus of
reality.

So original a hypothesis will naturally raise doubts and generate
formidable problems.
\eq

\section{24-09-03 \ \ {\it The Helping Hand} \ \ (to R. W. {\Spekkens})} \label{Spekkens21}

\brws
This allegory is especially apt for the situation in a toy theory
universe, but it misses part of what is going on in quantum theory,
specifically, the contextuality of quantum theory: the involvement of
the observer in what is observed.  My sense is that contextuality is
telling us that we need to abandon part of the traditional conceptual
framework of the realist, specifically, the notion of a primary
quality.  This notion dates back at least to Galileo and was refined
after the advent of Newtonian mechanics by John Locke.  It is meant
to be a quality that is inherent to a system and independent of its
relation to an observer.  I think that many of Berkeley's criticisms
of this notion are valid, but whereas his idealism seems to me to
have little content or guidance for a theoretical physicist, the
relationalism espoused by Leibniz may be a useful conceptual
alternative.  My hope is that if one takes seriously the idea that
the paradigm of systems and properties should be replaced by some
sort of paradigm of relations and relations among relations, then the
quantum state will be found to have a natural interpretation as a
state of knowledge about these relations.

Quantum theory is whispering something important to us, and these
vague ideas are the best I have come up with so far in trying to make
out the subject of the conversation.
\erws

I was reading a book as I was walking home the other day, as is my
habit here in Dublin, trying to avoid running into street lamps,
etc., when I came across a passage that seemed perfect for this
desire of yours.  It was actually something I had read before but
never thought to bring up to you.  Anyway, I scanned the passage in
this morning and will attach it as a PDF file to this note.  Let me
know if it helps you articulate what you're trying to get at.

For myself, it still doesn't feel sexual enough, but I have to admit
it feels like a move in the right direction.  It'll be interesting to
hear your reaction.

\vspace{-.2in}
\bq
\noindent\bq
\noindent {\bf \underline{NOTE}}: The following passage was taken from
Richard {\Rorty}'s article ``A World without Substances or
Essences''---pp.~47--71 in R. {\Rorty}, {\sl Philosophy and Social
Hope\/} (Penguin, London, 1999)---but it is {\it not\/} copied
verbatim.  In particular, I have removed all instances of the words
``{\Dewey},'' ``{\James},'' ``{\Peirce},'' ``pragmatism,'' ``human purpose,''
and a few other words of the same ilk.  Also I deleted some whole
sentences and paragraphs and, at least once, substituted the word
``antiessentialist'' for ``pragmatist.''  The goal was to see how the
passage would be received with such an ever so slightly different
slant.  Of course, I didn't reveal this to {\Spekkens} at the time; I
revealed it only after the experiment was complete.
\bigskip
\eq

We need to break down the distinction between intrinsic and
extrinsic---between the inner core of X and a peripheral area of X
which is constituted by the fact that X stands in certain relations
to the other items which make up the universe.  The attempt to break
down this distinction is what I call antiessentialism.  For
antiessentialists, there is no such thing as a nonrelational feature
of X, any more than there is such a thing as the intrinsic nature,
the essence, of X.

In the rest of this essay I shall be trying to sketch how things look
when described in antiessentialist terms. I hope to show that such
terms are more useful than terminologies which presuppose `the whole
brood and nest of dualisms' which we inherit from the Greeks. The
panrelationalism I advocate is summed up in the suggestion that we
think of everything as if it were a {\it number}.

The nice thing about numbers, from my point of view, is simply that
it is very hard to think of them as having intrinsic natures, as
having an essential core surrounded by a penumbra of accidental
relationships. Numbers are an admirable example of something which it
is difficult to describe in essentialist language.

To see my point, ask what the essence of the number 17 is---what it
is {\it in itself}, apart from its relationships to other numbers.
What is wanted is a description of 17 which is different in kind from
the following descriptions: less than 22, more than 8, the sum of 6
and 11, the square root of 289, the square of 4.123105, the
difference between 1,678,922 and 1,678,905. The tiresome thing about
all {\it these\/} descriptions is that none of them seems to get
closer to the number 17 than do any of the others. Equally
tiresomely, there are obviously an infinite number of other
descriptions which you could offer of 17, all of which would be
equally `accidental' and `extrinsic'. None of these descriptions
seems to give you a clue to the intrinsic seventeenness of 17---the
unique feature which makes it the very number that it is. For your
choice among these descriptions is obviously a matter of what purpose
you have in mind---the particular situation which caused you to think
of the number 17 in the first place.

If we want to be essentialist about the number 17, we have to say, in
philosophical jargon, that {\it all\/} its infinitely many different
relations to infinitely many other numbers are {\it internal\/}
relations---that is, that none of these relations could be different
without the number 17 being different. So there seems to be no way to
define the essence of seventeenhood short of finding some mechanism
for generating {\it all\/} the true descriptions of 17, specifying
all its relations to {\it all\/} the other numbers. Mathematicians
can in fact produce such a mechanism by axiomatizing arithmetic, or
by reducing numbers to sets and axiomatizing set theory. But if the
mathematician then points to his neat little batch of axioms and
says, `Behold the essence of 17!'\ we feel gypped. There is nothing
very seventeenish about those axioms, for they are equally the
essence of 1, or 2, of 289, and of 1,678,922.

I conclude that, whatever sorts of things may have intrinsic natures,
numbers do not---that it simply does not pay to be an essentialist
about numbers. We antiessentialists would like to convince you that
it also does not pay to be essentialist about tables, stars,
electrons, human beings, or anything else. We suggest that you think
of all such objects as resembling numbers in the following respect:
there is nothing to be known about them except an initially large,
and forever expandable, web of relations to other objects. Everything
that can serve as the term of a relation can be dissolved into
another set of relations, and so on forever. There are, so to speak,
relations all the way down, all the way up, and all the way out in
every direction: you never reach something which is not just one more
nexus of relations. The system of natural numbers is a good model of
the universe because in that system it is obvious, and obviously
harmless, that there are no terms of relations which are not simply
clusters of further relations.

To say that relations go all the way down is a corollary of
psychological nominalism: of the doctrine that there is nothing to be
known about anything save what is stated in sentences describing it.
For every sentence about an object is an explicit or implicit
description of its relation to one or more other objects. So if there
is no knowledge by acquaintance, no knowledge which does not take the
form of a sentential attitude, then there is nothing to be known
about anything save its relations to other things. To insist that
there is a difference between a nonrelational {\it ordo essendi\/}
and a relational {\it ordo cognoscendi\/} is, inevitably, to recreate
the Kantian Thing-in-Itself.

For psychological nominalists, no description of an object is more a
description of the `real', as opposed to the `apparent', object than
any other, nor are any of them descriptions of, so to speak, the
object's relation to itself---of its identity with its own essence.
Some of them are, to be sure, better descriptions than others. But
this betterness is a matter of being more useful tools---tools which
accomplish some purpose better than do competing descriptions. All
these purposes are, from a philosophical as opposed to a practical
point of view, on a par. There is no over-riding purpose which takes
precedence.

Common sense---or at least Western common sense---has trouble with
the claim that numbers are good models for objects in general because
it seems counterintuitive to say that physical, spatiotemporal
objects dissolve into webs of relations in the way that numbers do.
When numbers are analyzed away into relations to other numbers,
nobody mourns the loss of their substantial, independent, autonomous
reality. But things are different with tables and stars and
electrons. Here common sense is inclined to stick in its toes and say
that you cannot have relations without things to be related. If there
were not a hard, substantial autonomous table to stand in relation
to, e.g., you and me and the chair, or to be constituted, out of
hard, substantial, elementary particles, there would be nothing to
get related and so no relations. There is, common sense insists, a
difference between relations and the things that get related, and
philosophy cannot break that distinction down.

The antiessentialist reply to this bit of common sense is pretty much
the one Berkeley made to Locke's attempt to distinguish primary from
secondary qualities. The contemporary, linguistified form of
Berkeley's reply is: All that we know about this hard, substantial
table---about the thing that gets related as opposed to its
relations---is that certain sentences are true of it. It is that of
which the following statements are true: It is rectangular, it is
brown, it is ugly, made out of a tree, smaller than a house, larger
than a mouse, less luminous than a star, and so on and on. There is
nothing to be known about an object except what sentences are true of
it. The antiessentialist's argument thus comes down to saying that
since all sentences can do is relate objects to one another, every
sentence which describes an object will, implicitly or explicitly,
attribute a relational property to it.\footnote{The properties
usually called `nonrelational' (e.g., `red', as opposed to `on the
left-hand side') are treated by psychological nominalists as
properties signified by predicates which are, for some purpose or
another, being treated as primitive. But the primitiveness of a
predicate is not intrinsic to the predicate; it is relative to a way
of teaching, or otherwise exhibiting, a use of the predicate. The
putative nonrelationality of a property signified by a predicate is
relative to a certain way of describing a certain range of objects
having the predicate. One way of putting the lessons taught by both
Saussure and {\Wittgenstein} is to say that no predicate is
intrinsically primitive.

For a firm statement, of the contrasting view, see John Searle, {\it
The Rediscovery of the Mind\/} (Cambridge, Mass.: MIT Press, 1992),
p.\ 211.} We antiessentialists try to substitute the picture of
language as a way of hooking objects up to one another for the
picture of language as a veil interposed between us and objects.

Essentialists typically rejoin, at this point, that psychological
nominalism is a mistake, that we should retrieve what was true in
empiricism, and not admit that language provides our only cognitive
access to objects. They suggest that we must have some prelinguistic
knowledge of objects, knowledge that cannot be caught in language.
This knowledge, they say, is what prevents the table or the number or
the human being from being what they call a `mere linguistic
construct'. To illustrate what he means by nonlinguistic knowledge,
the essentialist, at this point in the argument, usually bangs his
hand on the table and flinches. He thereby hopes to demonstrate that
he has acquired a bit of knowledge, and a kind of intimacy with the
table, which escapes the reach of language. He claims that that
knowledge of the table's {\it intrinsic causal powers}, its sheer
brute {\it thereness}, keeps him in touch with reality in a way in
which the antiessentialist is not.

Unfazed by this suggestion that he is out of touch, the
antiessentialist reiterates that if you want to know what the table
really, intrinsically, is, the best answer you are going to get is
`that of which the following statements are true: it is brown, ugly,
painful to banging heads, capable of being stumbled over, made of
atoms, and so on and on'. The painfulness, the solidity, and the
causal powers of the table are on all fours with its brownness and
its ugliness. Just as you do not get on more intimate terms with the
number 17 by discovering its square root, you do not get on more
intimate terms with the table, closer to its intrinsic nature, by
hitting it than by looking at it or talking about it. All that
hitting it, or decomposing it into atoms, does is to enable you to
relate it to a few more things. It does not take you out of language
into fact, or out of appearance into reality, or out of a remote and
disinterested relationship into more immediate and intense
relationship.

The point of this little exchange is, once again, that the
antiessentialist denies that there is a way to pick out an object
from the rest of the universe {\it except\/} as the object of which a
certain set of sentences are true. With {\Wittgenstein}, he says that
ostension only works against the backdrop of a linguistic practice,
and that the self-identity of the thing picked out is itself
description-relative.\footnote{On the fundamental importance of this
latter {\Wittgenstein}ian point, see Barry Allen, {\sl Truth in
Philosophy\/} (Cambridge, Mass.: Harvard University Press, 1993).}
Antiessentialists think that the distinction between things related
and relations is just an alternative way of making the distinction
between what we are talking about and what we say about it. The
latter distinction is, as Whitehead said, just a hypostatization of
the relation between linguistic subject and linguistic
predicate.\footnote{It is useful to think of this Whiteheadian
criticism of Aristotle (a criticism found in other early
twentieth-century philosophers---e.g., Russell---who tried to
formulate a non subject-predicate logic) as paralleling Derrida's
criticism of logocentrism. Derrida's picture of a word as a node in
an infinitely flexible web of relationships with other words is
obviously reminiscent of Whitehead's account, in {\sl Process and
Reality}, of every actual occasion as constituted by relations to all
other actual occasions. My hunch is that the twentieth century will
be seen by historians of philosophy as the period in which a kind of
neo-Leibnizian panrelationalism was developed in various different
idioms---a panrelationism which restates Leibniz's point that each
monad is nothing but all the other monads seen from a certain
perspective, each substance nothing but its relations to all the
other substances. }

Just as the utterance of a noun conveys no information to people who
are unfamiliar with adjectives and verbs, so there is no way to
convey information except by relating something to something else.
Only in the context of a sentence, as Frege told us, does a word have
meaning. But that means that there is no way of getting behind
language to some more immediate nonlinguistic form of acquaintance
with what we are talking about. Only when linked up with some other
parts of speech does a noun have a use, and only as the term of a
relation is an object an object of knowledge. There is no knowledge
of the subject without knowledge of what sentences referring to it
are true, just as there is no knowledge of a number without knowledge
of its relations to other numbers.

Our sense that we can know a thing without knowing its relations to
other things is explained away by antiessentialist philosophers as a
reflection of the difference between being certain about some
familiar, taken-for-granted, obvious relations in which the thing
stands and being uncertain about its other relations. Seventeen, for
example, starts out by being the sum of 17 ones, the number between
16 and 18, and so on. Enough such familiar statements, and we begin
to think of 17 as a thing waiting to get related to other things.
When we are told that 17 is also the difference between 1,678,922 and
1,678,905 we feel that we have learned about a rather remote,
inessential, connection between it and something else, rather than
more about 17 {\it itself}. But when pressed we have to admit that
the relation between 17 and 1,678,922 is no more or less intrinsic
than that between 16 and 17. For, in the case of numbers, there is no
clear sense to be given to term `intrinsic'. We do not really want to
say that 17, in the secret depths of its heart, {\it feels\/} closer
to 16 than to numbers further down the line.

Antiessentialists suggest that we also brush aside the question of
whether the hardness of the table is more intrinsic to the table than
its color, or whether the atomic constitution of the star Polaris is
more intrinsic to it than its location in a constellation. The
question of whether there really are such things as constellations,
or whether they are merely illusions produced by the fact that we
cannot visually distinguish the distance of stars, strikes
antiessentialists as being as bad as the question of whether there
really are such things as moral values, or whether they are merely
projections of human wishes. They suggest we brush aside all
questions about where the thing stops and its relations begin, all
questions about where its intrinsic nature starts and its external
relations begin, all questions about where its essential core ends
and its accidental periphery begins. Antiessentialists like to ask,
with {\Wittgenstein}, whether a chessboard is {\it really\/} one thing
or 64 things. To ask that question, they think, is to expose its
foolishness---its lack of any interesting point. Questions which have
a point are those which meet the requirement that any difference must
make a difference.

The residual essentialism of common sense may rejoin to all this that
antiessentialism is a sort of linguistic idealism: a way of
suggesting that there was really nothing there to be talked about
before people began talking---that objects are artifacts of language.
But this rejoinder is a confusion between the question, `How do we
pick out objects?' and, `Do objects antedate being picked out by us?'
The antiessentialist has no doubt that there were trees and stars
long before there were statements about trees and stars. But the fact
of antecedent existence is of no use in giving sense to the question,
`What are trees and stars apart from their relations to other
things---apart from our statements about them?' Nor is it of any help
in giving sense to the sceptic's claim that trees and stars have
non-relational, intrinsic, essences which may, alas, be beyond our
ken. If that claim is to have a clear meaning, we have to be able to
say something more about {\it what\/} is beyond our ken, what we are
deprived of. Otherwise, we are stuck with Kant's unknowable
Thing-in-Itself. From the antiessentialist's point of view, the
Kantian lament that we are for ever trapped behind the veil of
subjectivity is merely the pointless, because tautologous, claim that
something we define as being beyond our knowledge is, alas, beyond
our knowledge.

The essentialist's picture of the relation between language and world
drives him back on the claim that the world is identifiable
independently of language. This is why he has to insist that the
world is initially known to us through a kind of nonlinguistic
encounter---through banging into it, or letting it bounce some
photons off our retinas. This initial encounter is an encounter with
the very world itself---the world as it intrinsically is. When we try
to recapture what we learned in this encounter in language, however,
we are frustrated by the fact that the sentences of our language
merely relate things to other things. The sentences, `This is brown',
or `This is square', or `This is hard', tell us something about how
our nervous system deals with stimuli emanating from the neighborhood
of the object. Sentences like, `It is located at the following
space-time coordinates' are, even more obviously, sentences which
tell us about what the essentialist mournfully calls `merely
relational, merely accidental, properties'.
\eq

\subsection{Rob's Reply}

\bq
Sorry I didn't respond to your last email sooner.  I really enjoyed the excerpt on relationalism.  This is definitely along the lines I've been thinking.  We will have to discuss this at Caltech.

For now, I have a more mundane question. [\ldots]
\eq

\section{25-09-03 \ \ {\it The Immediate} \ \ (to M. P\'erez-Su\'arez)} \label{PerezSuarez3}

I am intrigued by your proposal, but I am also a little frightened by your proposal.  A student---especially one in a foreign land far away from his home for any length of time---can be an immense responsibility.  I'll try to indicate the sorts of thoughts on my mind, by attaching a letter I wrote to (an undergraduate) Gabe Plunk when he inquired about working with me one summer.  [See 26-02-02 note ``\myref{Plunk1}{A Tired Old Man}'' to G. {\Plunk}.]

Do you feel of yourself that you can fulfill the criteria I laid out for him?  Also, do you think of yourself that you can skirt the worries I expressed to David Mermin?  Beyond that, what I think {\it the\/} program (i.e., the Bayesian quantum information program) {\it needs\/} is tireless criticism from withinside---that is the only way it will ever achieve consistency.  Do you think you could fulfill that role?  The last thing the program needs is a ``yes man'' who will take much of it on faith.

Think a little bit about it and write me back.

\section{26-09-03 \ \ {\it The Austin Interpretation of Quantum Mechanics} \ \ (to A. Fine)} \label{Fine2}

Chris Timpson (a grad student in the philosophy of science at Oxford) brought up the idea of suggesting a `Quantum Information and Foundations' symposium at the Philosophy of Science Association meeting in Austin, Texas in November 2004.  I think that's an absolutely great idea.  Would you be interested in throwing in with us on this?  (Or maybe just him, if I myself don't have the time.)  In any case, do you have any advice on the matter?

I'm picking on you because I read somewhere that you had been a president of the PSA, and also I noticed you seemed to have a developing interest in quantum information.

Here's a link Timpson gave me to look at:
\begin{center}
\url{http://web.archive.org/web/20040511031253/http://www.temple.edu/psa2004/Call_for_Symposia_and_Workshops.htm}
\end{center}
There's evidence that someone on the program committee might already find this an interesting idea.

\section{28-09-03 \ \ {\it PI Workshop on QC-QI-QG} \ \ (to F. Girelli \& E. R. Livine)} \label{Girelli0} \label{Livine1}

The idea of the meeting sounds great.  You can count on my being there; just give me the exact dates as soon as you can.  In the meantime, I've marked my calendar tentatively for February 25 through March 3.  Strangely enough, lately my work in quantum cryptography has been showing a significant analogy between (finite) Hilbert-space dimension and gravitational mass; I'll be posting a paper on the subject on the LANL archive in the coming month.  I have this little desire that it's more than just an analogy, though that's probably just a wild thought.  Anyway, the meeting will be a good opportunity for me to get some feedback on this stuff (outside my usual circles).

\section{28-09-03 \ \ {\it Rainy Dublin} \ \ (to A. Fine)} \label{Fine3}

Thanks for the reply; I've forwarded it on to Chris Timpson.  It'll be nice if something gets organized:  I think the time is definitely ripe.

\baf
I see you are in Dublin. Never been there. Is it nice?
\eaf

I like Dublin a lot.  The people are about the friendliest in the world, and the population in general is quite educated (in comparison with America's).  For instance, it's quite easy to strike up an {\it interesting\/} conversation with a taxi driver.  Not a lie:  Three times now, I've met taxi drivers who knew something about quantum cryptography or quantum computing (at the level of {\sl Scientific American\/} sort of stuff).

But boy does it rain here:  Seattle ain't got nothin' on us!

\section{29-09-03 \ \ {\it Equiangular Lines} \ \ (to H. K\"onig)} \label{Koenig1}

I am a physicist at Bell Labs, and I have been working on a problem in quantum measurement theory that requires me to construct $d^2$ equiangular lines in a complex vector space of dimension $d$.  By this I mean, can one find $d^2$ unit vectors $x_k$ such that $|(x_k, x_l)|^2 = 1/(d+1)$ whenever $k\ne l$?  It seems to be a much more difficult problem than I had initially suspected.

Anyway, in searching on the subject, I came across your paper ``Aspects of the Isometric Theory of Banach Spaces'' with Alexander Koldobsky.  Just preceding your Proposition 18---if I understand you correctly---you seem to say that such sets have only been found in dimensions 2, 3, and 8.

Does that remain the state of the art?  Also, can you give me a reference for where to find that result?  (I could not find a direct reference to it in your paper.)  I have been able to work out the $d=2$ and 3 cases myself, but I would be interested to see the $d=8$ case.  Finally, do you know if this is a problem with a long history?  Is it a well-known difficult problem and maybe I shouldn't spend too much time on it?  (I.e., I am not a mathematician, and don't want fool myself if the problem will not be reasonably within my reach.)

Thank you for any help you can give.

\subsection{H. K\"onig's Reply}

\bq
It is only known for $n=2,3,8$ that there are $n^2$ equiangular complex
lines, probably there are no more cases of existence. In the real case the
situation is explained very much in Lemmens-Seidel's paper which is
referred to in my paper which I attach. Actually, in the real case, only
$O(n^{3/2})$ equiangular lines are known to exist. For $n=8$, I believe the construction is
mentioned in Hoggar's or Delsarte-Goethals-Seidel's paper.  It uses
quaternions for simpler representation, as far as I remember. In the paper
I attach [``Cubature Formulas on Spheres''] I constructed for $n = \mbox{prime power} +1$, $N$ equiangular vectors, where
$N=n^2-n+1$, so it is almost the maximal number $n^2$ but not quite (see page
6). I attach the \TeX\ as well as the dvi-file of the paper.
\eq

\section{29-09-03 \ \ {\it Quotes That Bugged Me} \ \ (to A. Sudbery)} \label{Sudbery3}

I give up:  My backlog of email that needs replying to has grown too
enormous, and I'm going to have to make some cuts.

Let me just lodge a complaint about two of your quotes that bugged
me.  These are:
\bts
So I should keep to myself my enthusiasm for ambitious all-embracing
theories of reality and not go around insulting people who don't need
them, accusing them of copping out and God knows what. But I'm afraid
it's still true for me that not to be curious about what makes QM
work would be to stop doing science, to switch off what made me want to
learn about QM in the first place.
\ets
and
\bts
I have to say that I don't find the existence of objective reality so
easy to dismiss as just a faith, which one could simply choose to
abandon. The reason that the tension between subjective and objective
is a real philosophical {\bf problem\/} is that there seem to be good
reasons for holding the objective view.
\ets
\pagebreak
\noindent {\bf Curt Replies:}\medskip

\noindent 1)\medskip

Let me try to reiterate the goal of the program {\Caves}, {\Schack} and I
are developing.  It is not to say that there is no world external to
us; it is only to say that there are no quantum states external to
us, and then to see where that leads us in understanding quantum
mechanics.  There is a difference; why can't you see it?

I think it would be hard to call my 59 page paper, \quantph{0205039}, a ``lack of curiosity about what makes QM work.''
The whole point of the paper was that we don't have a decent handle
on what makes quantum mechanics work.  It then tries to
systematically explore a particular line of thought---that the
quantum state represents information, rather than material property.
It does not eschew the very existence of material properties.  It
simply says that among them, the quantum state is a very bad
candidate.

Here is the way I would caricature where you seem to stand in
relation to me.  Imagine a young student who first learned classical
electrodynamics solely in terms of the vector potential.  Then one
day someone pointed out to him that all the physical phenomena he
could see actually depended only upon the fields, not the complete
vector potential after all---there is a gauge freedom.  Well,
flabbergasted, it felt to this student that he was stripped of
something he ought to have.  So he spent the rest of his life
doggedly trying to find a justification for the TRUE gauge.  Of
course he had to give umpteen ad hoc reasons for why the true gauge
could not be measured, but that was the price to pay to do science.
(For if you're not trying to see the vector potential as a real
property of the world, you're not doing science.)  What a pity.

Likewise I would characterize where you want to go and where I want
to go with quantum mechanics.  You want to see the quantum state
simpliciter as a representation of reality.  Whereas I think we are
more likely to find reality in the ``differential.''  I.e., in the
support structure in which quantum states live and in the rules for
changing those states in light of how we are stimulated (by the world
external to us).  The structure of those rules represent something we
are assuming of the world as it is independently of us.

And wouldn't we like to know what we are assuming of the world as it
is independently of us? \medskip

\noindent 2)\medskip

Why do I go to pains to say things like this:
\bq
\noindent
SYSTEM:  In talking about quantum measurement, I divide the world
into two parts---the part that is subject to (or an extension of) my
will, and the part that is beyond my control (at least in some
aspects).  The idea of a ``system'' pertains to a part beyond my
control.  It counts as the source of my surprises, and in that sense
obtains an existence of its own external to me.
\eq
as I did in my ``Me, Me, Me'' note, when you nevertheless respond
with things like this:
\bq
\noindent I have to say that I don't find the existence of objective
reality so easy to dismiss as just a faith, which one could simply
choose to abandon.
\eq

In any case, my issue with Nagel---the big IF---is not about whether
there is a world external to or beyond all agents---objective reality
if you will.  But whether it is sensible to assume that there can be
a REPRESENTATION of it.  Like the boy fixated on his vector
potentials, one can always act as if there is such a REPRESENTATION
from the outside---the view from nowhere---but also like the boy, one
might be wasting one's time in doing so.  It is my sense that quantum
mechanics hints at the latter.

\section{03-10-03 \ \ {\it IC POVMs} \ \ (to G. M. D'Ariano)} \label{DAriano2}

I just finished reading your new paper on informationally complete measurements.  It is fascinating and very nicely done; I learned a lot.

I wonder however if I could ask to cite my paper \quantph{0205039} (the one you referred to above) in a reposting of yours---only the {\tt quant-ph} number please, not the {\Vaxjo} version.  I ask this, not because anything in your paper depends in a technical way on my paper, but because I want to spread the word as far as possible.  As I see it, it is an issue of getting a large enough workforce to move in these directions if we're going to make long-lasting progress.  The picture of doing quantum mechanics on a simplex and the ``standard quantum measurement device'' that I try to make attractive in that paper depends crucially on the existence of minimal informationally complete POVMs, which I emphasize.  For our colleagues to see the deeper connection between your work and mine, I think, can only be healthy.

Also, by the way, the existence of minimal informationally complete POVMs was crucial for our proof of the quantum de Finetti theorem (nonminimal IC POVMs could not be amended to that proof technique): C.~M. Caves, C.~A. Fuchs and R.~Schack, ``Unknown Quantum States:\ The Quantum de Finetti Representation,'' {\sl Journal of Mathematical Physics\/} {\bf 43}(9), 4537--4559 (2002). [Reprinted in {\sl Virtual Journal of Quantum Information\/} {\bf 2}(9).] \quantph{0104088}.

It is interesting that your paper can be used to connect the de Finetti theorem to group theory.

Anyway, if you could make a citation or two, I'd much appreciate it.

\section{03-10-03 \ \ {\it A Plea} \ \ (to R. Blume-Kohout, J. M. Renes, \& C. M. {\Caves})} \label{BlumeKohout1} \label{Caves74} \label{Renes18}

I want to make a plea to you guys to write up what you and your collaborators have on SIC-POVMs and get it out on the web in a hurry.  What I would suggest is that you write it up in a style appropriate for the journal {\sl Linear Algebra and Its Applications\/} and then ultimately submit it there.  For that journal, all you need to do is state the raw mathematical problem---that is enough motivation---and then state what you've got (including numerics).  I estimate that it would not take more than a week to write such a paper, at least to the quality of posting it on {\tt quant-ph}.

The reason I am writing this plea now are fourfold.

\begin{enumerate}
\item
I have only recently come to appreciate how old and well respected this problem is, and it appears to me you guys actually know more than anyone else.  I'll put a list of references with annotations below.  The upshot is that the problem is about 30 years old and has been hit by some {\it real\/} mathematicians.  The most telling tale comes from Hermann K\"onig (a mathematics professor in Kiel) who wrote me:  ``It is only known for $n=2,3,8$ that there are $n^2$ equiangular complex lines, probably there are no more cases of existence.''
\item
Have a look at \quantph{0310013} posted this morning by D'Ariano and collaborators.  You'll see that they're already working with your beloved group $\mathbb{Z}_d \times \mathbb{Z}_d$ for generating IC-POVMs.  I deem that it can't be long before they realize the interest in adding an S to that.  And, look, if they write a paper posing this problem you guys will lose a lot of credit for a lot of work:  It is not worth it to hold on to it privately anymore.
\item
I myself would like to talk more openly about the sensitivity of SIC ensembles with respect to eavesdropping, and I don't feel completely comfortable doing this while you guys continue to sit on your paper.  In particular I am writing something for the Holevo festschrift that I'll call ``On the Quantumness of a Hilbert Space'' showing that these ensembles (if they exist) are maximally quantum and some related facts.  It would be nice to have something solid of yours to cite.
\item
The most important reason:  You're holding up science!  What I suspect is, if you put your paper in the right audience, someone will be able to run with your $\mathbb{Z}_d \times \mathbb{Z}_d$ conjecture and prove it to completion.  Simply be happy that you've contributed this much, and give someone else a turn.
\end{enumerate}

I hope I've made my case strongly enough.

\begin{enumerate}
\item
P.~W.~H. Lemmens and J.~J. Seidel, ``Equiangular Lines,'' Journal of
Algebra {\bf 24}, 494--512 (1973).  This seems to be the father paper of the field, even though they do give some references to earlier work studying equiangular lines in the terminology of ``elliptic geometry'' (whatever that means).  This paper, however, is devoted solely to real vector spaces.

\item
P.~Delsarte, J.~M. Goethels, and J.~J. Seidel, ``Bounds for Systems
of Lines and Jacobi Polynomials,'' Philips Research Reports {\bf 30},
91--105 (1975).  This paper studies both real and complex spaces.  They make it clear that they know of the existence of maximal equiangular sets (i.e., the ones that achieve the ``Gerzon bound,'' namely $n^2$ in the complex case) in dimensions 2 and 3.  Looking at their references, these cases might have been known as early as 1914 but I have not confirmed that.

\item
H.~K\"onig and N.~Tomczak-Jaegermann, ``Norms of Minimal
Projections,'' \\ \arxiv{math.FA/9211211}.  I don't think this paper proves anything of use to the SIC-POVM problem---though I don't know---but it does show that these questions are intimately related to other issues in mathematics that people take seriously.  I.e., this is another reason your paper should be out there.

\item
S.~G. Hoggar, ``64 Lines from a Quaternionic Polytope,'' Geometriae
Dedicata {\bf 69}, 287--289.  Apparently the problem considered in this paper proves the existence of a SIC-POVM in $d=8$.

\item
A.~Koldobsky and H.~K\"onig, ``Aspects of the Isometric Theory of
Banach Spaces,'' in {\sl Handbook of the Geometry of Banach Spaces},
Vol.~1, edited by W.~B. Johnson and J.~Lindenstrauss, (North Holland,
Dordrecht, 2001), pp.~899--939. Available online at: \myurl[http://www.math.missouri.edu/~koldobsk/]{http://www.math. missouri.edu/$\sim$koldobsk/}.  I'll repeat what I said above:  I don't think this paper proves anything of use to the SIC-POVM problem, but it does show that these questions are intimately related to other issues in mathematics that people take seriously.  You can find an indication of the SIC-POVM problem on page 19 of the web-based manuscript.

\item
T.~Strohmer and R.~Heath, ``Grassmanian Frames with Applications to
Coding and Communication,'' \arxiv{math.FA/0301135}.  Not much of interest here, except a rederivation of the Gerzon bound.

\item
H.~K\"onig, ``Cubature Formulas on Spheres.''  Available online at:
\myurl[http://analysis.math.uni-kiel.de/koenig/pub.shtml]{http://analysis.math. uni-kiel.de/koenig/pub.shtml}.  This paper is interesting and getting close to the real thing.  He derives the existence of $d^2-d+1$ equiangular lines when $d=p^m+1$, $p$ a prime number and $m$ a natural number.

\item
B.~Et-Taoui, ``Equiangular Lines in $C^r$,'' Indagationes
Mathematicae {\bf 11}, 201--207 (2000).  Despite the title, this paper is not particularly interesting (I think):  The problem he addresses is much more particular than the title lets on.  But, again, it does show that people are thinking about these things.

\item
B.~Et-Taoui, ``Equiangular Lines in $C^r$ (Part II),'' Indagationes
Mathematicae {\bf 13}, 483--486 (2002).  Ditto.

\item
G.~M. D'Ariano, P.~Perinotti, and M.~F. Sacchi, ``Informationally
Complete Measurements and Groups Representation,'' \quantph{0310013}.
\end{enumerate}

\section{06-10-03 \ \ {\it Keep It Up!}\ \ \ (to J. M. Renes and C. M. Caves)} \label{Renes19} \label{Caves74.05}

I skimmed over your draft this morning [of what eventually became J. M. Renes, R. Blume-Kohout, A. J. Scott and C. M. Caves, ``Symmetric Informationally Complete Quantum Measurements,'' J. Math.\ Phys.\ {\bf 45}, 2171 (2004)].  I think it is starting to look good.  (However, when I suggested {\sl Linear Algebra and Its Applications\/} as a target, I was imagining something even more skeletal than that.  The whole point of that journal is to leave the words behind.)

Unfortunately I'm starting to think there is even more need for speed than I had thought before.  For instance, Koenraad Audenaert wrote me the following last night (I opened up the note this morning):
\bka
I may have something interesting here! (and then again, maybe not {\rm \smiley}) In Dublin you told me about the problem of constructing a SIC-POVM
in any dimension. Now I have been thinking about this now and again,
but without further results. Last Friday, however, there was this
preprint by Mauro d'Ariano et al about the construction of IC-POVMs
(not necessarily symmetrical ones). Upon reading it I was struck by
their first example (in section 4), where they constructed an IC-POVM
starting from an initial pure state and applying what seems to be all
possible displacement operators in the associated phase space of the
Hilbert space. Well, my left toe immediately started to itch.
Can one come up with SIC-POVMs in this way, starting from an
appropriately chosen pure state? I wrote a MatLab program that did a
brute force numerical search (simplex search) and, believe it or not,
for all dimensions I tried (2, 3, 4, 5 and 6) it {\bf did} come up with a
SIC-POVM (the inner products being correct up to 12 decimals!).

But maybe this is precisely the ``numerical evidence'' you mentioned.
\eka
I asked Koenraad to make sure if he finds something, he would not post before you, but I am not going to be able to ask everyone that.

Now, a couple of notes on your draft.
\begin{enumerate}
\item
I'm not an author.  I wish I had contributed enough to the project to be in the list, but unfortunately I haven't.  (I would be very happy, however, if you would find a way to cite my papers \quantph{0205039} and \quantph{0302092}.)  In general for this paper, you should be careful not to let the author list get out of hand.  I would say that hours spent working privately on the project should not be a criterion of authorship.
\item
I didn't understand your argument for linear independence on page 1.  The Gram matrix for an SIC ensemble (i.e., the matrix of inner products), will not be a real matrix; there will be phases everywhere.  I don't see why you don't use Carl's simple straightforward argument from his web notes to establish linear independence.
\item
``However, it is known that for $t=2$, the minimum number is $n=d^2$, so the SICPOVM is the smallest spherical 2-design.''  Where is this established?
\item
For Section III, I think you should exhibit the form of the expansion of a fiducial projector in terms of the displacement operators $D_{jk}$.  Then comment that the problem of existence boils down to any satisfying assignment for all the phases, and exhibit how nonlinear of a problem that looks to be.
\end{enumerate}

I'll send any further comments as they come to me, but this will probably be my last note for today:  I've got to give a talk at Trinity College this afternoon.

\section{06-10-03 \ \ {\it Agreement/Disagreement}\ \ \ (to J. M. Renes and C. M. Caves)} \label{Renes20} \label{Caves74.1}

\bjmr
OK, I'll try to tighten it up.  But Carl also suggested Journal of
Mathematical Physics, which in the end is probably what I was going
for. (Actually, I don't know so much about that journal either, so
what I was going for was the archive. I treat POVM as if the reader
already knows what it is, for instance.)
\ejmr

I agree that {\tt arXiv.org\/} is the very most top priority.  However, I disagree with J Math Phys.  It's a pure maths problem, and I deem it will have a much better chance of being solved if the paper ends up in Lin Alg App.  In the end it is a problem about equiangular lines, full stop; and the LAA audience is an audience hungry for those kinds of problems.  The mathematical-physics community, I suspect, will quickly forget about it if it's not easily tractable \ldots\ whereas the LAA audience will plug away at it year after year.  They won't denigrate it just because it is a finite-dimensional vector-space problem, where the JMP crowd might---at least that's the sort of reaction I personally have had from them before.  Finally, especially if it does turn out not to be so easily tractable, I think you'll ultimately generate more citations (outside of the {\tt quant-ph} crowd) with LAA.

\section{07-10-03 \ \ {\it Unruh Effect and Entropy}\ \ \ (to G. L. Comer)} \label{Comer40}

I just started writing my contribution to Holevo's festschrift.  In its introduction (before the irrefutable calculations) I'm going to make a couple of speculative remarks along the lines of the note I sent you on July 4, ``Solid Ground, Maybe?,'' but without all the black hole stuff---namely I just want to point out the universality of the information-disturbance stuff, and how it only depends on Hilbert-space dimension.  It doesn't, for instance, care whether the Hilbert space is associated with platinum or magnalium (just as in E\"otv\"os's experiment).

Here's how the first two sentences start: ``Memorable experiences can sometimes happen in elevators.  I have had two in my life.''

I'll send you a draft as it comes together.

\section{07-10-03 \ \ {\it The Nobel Prize}\ \ \ (to G. L. Comer)} \label{Comer41}

Just looking at the newspaper a minute ago, I see a new experience in my life:  Someone I actually know won a Nobel prize (i.e., my having known them before their prize).  It's Tony Leggett!  Mostly I find myself feeling kind of funky for how much I have denigrated his thoughts on quantum foundations.  I saw him two times in the last year, and we have made no progress at all.  In my talk in Vancouver, he almost looked visibly angry.

Let me dig up a note I wrote Asher Peres after meeting him for the first time:
\bq
I'm just back from a talk by Tony Leggett.  Perhaps I shouldn't compare him to Faris---because all his thoughts were certainly coherent, even if misguided---but I got quite annoyed with his whole program.  What a waste of long hours of calculation and even speculation.  Some speculation, I think, can be productive \ldots\ but this strand of it (i.e., macroscopically realistic theories \`a la Pearle and company) just struck me as a waste.
\eq

\section{07-10-03 \ \ {\it The Historical Record}\ \ \ (to G. L. Comer)} \label{Comer42}

I said,
\bq\noindent
Let me dig up a note I wrote Asher Peres after meeting him for the
first time.  I'll paste it below.
\eq

Actually, thinking about it, I realized I first met Tony Leggett in the summer of 1992 at the Aspen Center for physics.  He was always in the library calculating with a bound volume or two of {\sl Physical Review\/} opened near him.

\section{07-10-03 \ \ {\it Congratulations!}\ \ \ (to A. J. Leggett)} \label{Leggett1}

Congratulations on the Nobel prize!!  Now, almost for sure, I'm not going to get you to write something for the Peres festschrift---who would have the time?!

Anyway congratulations; from what David Mermin tells me about your work, I know it is well-deserved.

\section{07-10-03 \ \ {\it The Nobel Prize, 2}\ \ \ (to G. L. Comer)} \label{Comer43}

\bgc
I don't know anything of Leggett.  But if his work is like that of
Abrikosov and Ginsberg, then I can understand his discomfort with
what you are preaching.  I've met condensed matter types that have
a very practical side that I can see is just not receptive to the
basic aspects of your ideas.  I know someone who knows Abrikosov,
and he says he is absolutely brutal when in the audience.
\egc

Yep, he's definitely a bread and butter physicist.  And clearly very smart and rigorous and---if you could see the equations in his talks--- you'd understand that he must be absolutely single-minded when it comes to calculating.  It is only that all his later work is predicated on the obstinate belief that the quantum state {\it must\/} correspond to an objective property.  The Nobel prize was for the sort of work you are talking about.  I've only seen him in his later incarnation, i.e., worrying about (and proposing experiments for) quantum foundations.  He always expresses surprise and confusion that quantum mechanics has not been seen to break down yet.

Anyway, I just sent him a note of congratulations.  It is a very impressive feat to so change the world as he (and the other two) have done.

\section{07-10-03 \ \ {\it More Covariance} \ \ (to G. M. D'Ariano)} \label{DAriano3}

Thanks for your long letter, and certainly thanks for the citations in your ``informationally complete'' paper.  I printed out your ``Extremal Covariant Quantum Operations'' paper today, but now that one I cannot finish off in a lunchtime reading (as I did the last)!  It will take me sometime to digest it all.

In fact it will take me some time to digest your long note.  Unfortunately, I don't have any answers to your questions.

I think a visit to Pavia would be great as I said before.  Two weeks, in fact, would probably be optimal.  Unfortunately though, I really can't make any new plans until after the new year.  (I will be in Dublin until April 15.)

Finally, let me ask you:  Would you like a copy of my book {\sl Notes on a Paulian Idea\/} (\quantph{0105039})?  It is a samizdat, as David Mermin calls it, more for perusing than systematic reading, but I think it does capture some of the philosophical flair of what we are mutually shooting for in quantum foundations.  I have some copies of the {\Vaxjo} University Press edition for giving away, if you think you or your students might get something out of it.  (Kluwer will be reprinting it in a more proper edition, in their {\sl Fundamental Theories of Physics\/} series, but that will probably be a year away; and I won't have many copies to give away then.)  Anyway, if you want a copy, just tell me which mailing address to send it to.

\section{08-10-03 \ \ {\it Heart of the Matter} \ \ (to M. P\'erez-Su\'arez)} \label{PerezSuarez4}

\bmps
And, finally, on your note about ``You, you, you \ldots''. I usually enjoy your papers and notes, and this is no exception. I have never thought of your approach as implying solipsism. I don't even understand how someone might have thought it did. The only point I haven't been able to come to terms with yet is that sort of  ``unitary readjustment''. I do still think that it would have been ``nice'' being able to get Bayes' rule without any further modifications. Is this thought an indication that I am missing or misunderstanding something which is fundamental?
\emps

No, I don't think it would have been nice at all.  I think the modification to Bayesian conditionalization that we see here is one of our clearest (and most quantitative) indications yet that quantum measurement should not be viewed as a passive process by which the agent's mind comes to mirror a pre-existent reality.  If the agent were not changing his external reality in the process of having a look, his update rule would have been the regular Bayesian one.  The mantra is:  The quantum rule only reduces completely to the Bayesian rule when the agent believes he has no causal influence upon the system he is updating his beliefs about.  A good case in point for this is to consider making a measurement on one half of an entangled pair.  The update rule that one uses on the other half (subsequent to the data gathered on the first half) is precisely the Bayesian rule.

Open question:  Why is the readjustment unitary, rather than some more general linear or nonlinear transformation?

\section{08-10-03 \ \ {\it EnNobelization} \ \ (to N. D. {\Mermin})} \label{Mermin104}

That was great news hearing about Tony Leggett winning the Nobel
prize yesterday.  Strangely though, it made me feel very nervous.  I
guess I had always seen Tony as ``wasting a perfectly good mind'' (I
dug almost exactly that phrase in a note I had written to Asher Peres
in July 2000).  As I wrote to my friend Greg Comer in reply to one of
his points:
\bq
Yep, he's definitely a bread and butter physicist.  And clearly very
smart and rigorous and---if you could see the equations in his
talks---you'd understand that he must be absolutely single-minded
when it comes to calculating.  It is only that all his later work is
predicated on the obstinate belief that the quantum state {\it
must\/} correspond to an objective property.  The Nobel prize was for
the sort of work you are talking about.  I've only seen him in his
later incarnation, i.e., worrying about (and proposing experiments
for) quantum foundations.  He always expresses surprise and confusion
that quantum mechanics has not been seen to break down yet.

Anyway, I just sent him a note of congratulations.  It is a very
impressive feat to so change the world as he (and the other two) have
done.
\eq
Well, I guess his Nobel prize taught me!  A man shouldn't be judged
by his interpretation of quantum mechanics alone!

\section{08-10-03 \ \ {\it Can't Resist} \ \ (to H. Barnum \& A. Sudbery)} \label{Sudbery4} \label{Barnum12}

I know that I said I would not write again until I had finished
reading Nagel's book, but I came across a passage yesterday in an
essay of Richard {\Rorty}'s that I could not resist scanning into my
computer.  It deals somewhat with something Howard wrote us:
\bhb
During the ``Bohmian dialogue'' years ago at Hampshire (organized by
Herb Bernstein), I came to a couple of important realizations mostly
in the process of defining my views against ``what we've all learned
over the last twenty years'' (which included things like ``you can
choose your own myth,'' as I recall).

One of them was that, though I am not explicitly religious, I value
``transcendence'' (the term was being used as a putdown, I think).
That is perhaps a nicer word for what Nagel is calling
``detachment''\ldots\ getting beyond your own limited point of view to an
expanded point of view.\ldots\ even though of course that expanded point
of view is still gotten to by you, interacting with others, using
more of the different modalities and apparatuses available to you. It
is {\bf part of} ``variety and freedom,'' (the truth will
set you free, dontcha know) and ``growth,'' for me.

What I dislike most in some strains of ``antirealist'' modern
thought, is their disdain for the value of transcendence, their
desire to make everything a useful, folksy, comforting tool for
humans.\ldots\ not that useful, folksy, comforting are not good, but so
is getting outside oneself and recognizing the vast unbelievableness
of what becomes apparent as one does so\ldots.

I don't necessarily think your approach will end up negating that
value, in end..... our views may be closer than it seems.  I'll read
on....
\ehb

In the present passage, {\Rorty} is talking about culture and politics,
but he might as well have been talking about views of physical
theory.  In fact, in much of the rest of the book {\sl Philosophy and
Social Hope\/} he was.

I will return on Nagel (sooner rather than later).

Here's the passage:

\bq
Insofar as `postmodern' philosophical thinking is identified with a
mindless and stupid cultural relativism---with the idea that any fool
thing that calls itself culture is worthy of respect---then I have no
use for such thinking.  But I do not see that what I have called
`philosophical pluralism' entails any such stupidity.  The reason to
try persuasion rather than force, to do our best to come to terms
with people whose convictions are archaic and ingenerate, is simply
that using force, or mockery, or insult, is likely to decrease human
happiness.

We do not need to supplement this wise utilitarian counsel with the
idea that every culture has some sort of intrinsic worth.  We have
learned the futility of trying to assign all cultures and persons
places on a hierarchical scale, but this realization does not impugn
the obvious fact that there are lots of cultures we would be better
off without, just as there are lots of people we would be better off
without.  To say that there is no such scale, and that we are simply
clever animals trying to increase our happiness by continually
reinventing ourselves, has no relativistic consequences.  The
difference between pluralism and cultural relativism is the
difference between pragmatically justified tolerance and mindless
irresponsibility.

So much for my suggestion that the popularity of the meaningless term
`postmodernism' is the result of an inability to resist the claims of
philosophical pluralism combined with a quite reasonable fear that
history is about to turn against us.  But I want to toss in a
concluding word about the {\it un\/}popularity of the term---about
the rhetoric of those who use this word as a term of abuse.

Many of my fellow philosophers use the term `postmodernist
relativism' as if it were a pleonasm, and as if utilitarians,
pragmatists and philosophical pluralists generally had committed a
sort of `treason of the clerks', as Julien Benda puts it.  They often
suggest that if philosophers had united behind the good old
theologicometaphysical verities---or if {\James} and Nietzsche had been
strangled in their cradles---the fate of mankind might have been
different.  Just as Christian fundamentalists tell us that tolerance
of homosexuality leads to the collapse of civilization, so those who
would have us return to Plato and Kant believe that utilitarianism
and pragmatism may weaken our intellectual and moral fibre.  The
triumph of European democratic ideals, they suggest, would have been
much more likely had we philosophical pluralists kept our mouths
shut.
\eq

\section{08-10-03 \ \ {\it Can't Resist, 2} \ \ (to A. Sudbery)} \label{Sudbery5}

\bts
Curiously enough, I am toying with the idea of writing in defence of
quantum-mechanical pluralism for my contribution to Asher's
festschrift. Where does the Rorty quote come from?
\ets

It comes from the last essay in the book, {\sl Philosophy and Social Hope}.  Unfortunately I don't have the book with me at the moment to see what the title of the essay is.  I suggest you run to the bookstore and get a copy just as fast as you can.  The best things to read are the introductory essay (spurning Platonism), along with the three essays:
\begin{itemize}
\item
``Thomas Kuhn, Rocks and the Laws of Physics''
\item
``A World without Substances or Essences''
\item
``Truth without Correspondence to Reality''
\end{itemize}

\section{08-10-03 \ \ {\it Jaynes Stuff} \ \ (to D. Poulin)} \label{Poulin10}

Thanks for the long letter, and thanks for the invitation to work on a project with you and Raymond.  I've printed out your draft, and I'll see what I can understand from it.  But I fear you might have to give me a lecture in person before much will sink in!  (At the moment I am preoccupied with trying to get off to Caltech this weekend and making a talk that won't make me a laughingstock.)

I liked your way of putting things in the Jaynes' principle debate:  ``One can use Jaynes' principle to assign probabilities at the price of treating it like it wasn't a probability distribution.''  It seems to be of the right flavor \ldots\ though I can't say for sure that I have completely made up my mind on the value of the principle.  Since I am now drawn more to de Finetti's personalistic Bayesianism than Jaynes's version of it all, I suspect I would not put nearly as much stock in the ``principle'' as I used to.

\bdp
I have also had a look at Jaynes' 1986 paper. I don't see what Brun et
al.\ refer to when they say that ``Jaynes [13] has given a thorough
discussion of this problem.'' Maybe he did, but not in this reference.
\edp

Yeah, 1986 would be much too late.  Unfortunately, I don't have
Jaynes' book to figure out which is the right reference.  It had to do
with an issue brought up by Friedman and Shimony in the
1970s.\footnote{\editornote E. T. Jaynes, ``Prior probabilities,''
  IEEE Transactions on Systems Science and Cybernetics {\bf SSC-4}
  (1968), pp.\ 227--41,
  \url{http://bayes.wustl.edu/etj/articles/prior.pdf}.}  I did do a
little web search and found this nice paper by Jos Uffink: \bq
\myurl{http://www.phys.uu.nl/~wwwgrnsl/jos/mep2def/mep2def.html}.  \eq
I think it is dead on the mark with respect to the issue you're
interested in.  Also, Jos is a very thoughtful person, so I think it
is probably a very important one to read.

I'll try to write you again during my California stay.

\section{08-10-03 \ \ {\it Renormalization} \ \ (to F. Girelli)} \label{Girelli1}

I apologize for the delay in responding to your note.  I didn't realize until this morning that it had been eight days since I received it.

\bfg
My question is about renormalization. It is well known that there is
an interpretation of renormalization in terms of information: eg when
renormalizing we coarse grain so we loose some local information.  The
ideas of renormalization started with some spin systems, so it seems
that the de Finetti representation is a right tool to use.

Are you aware if the interpretation in terms of information has been
pushed forward?  And is it possible?

When you do the successive measurements in the context of the de
Finetti representation you refine your knowledge, which can be seen
as the opposite operation of the coarse graining.  It would make sense
as well to see how the Bayesian approach is implemented in the
renormalization scheme. For example if you deal with only the bare
coupling constants (I am taking the Polchinski point of view, or
Wilson's), it is not working, you have to measure some of them (the
relevant ones) and then update your measurement and everything gets
fine.

Do you have any references, ideas, intuitions about those things? I
would be very glad to hear about it.
\efg

I wish I could say something intelligent on the subject, but I don't think I can at the moment.  I had never thought about renormalization in information theoretic terms before reading your note.  I did a google search on the words ``renormalization group information theory'' and the only thing that caught my eye was Robert Tucci's paper ``An Application of Renormalization Group Techniques to Classical Information Theory'' posted on {\tt quant-ph} \ldots\ but I doubt it can be relevant.

Maybe when I come visit Perimeter during your meeting, you can fill me in in more detail what sort of thing you have in mind.  I have a hard time formulating a precise question at this stage.  It's probably the sort of thing we need a chalk board for.  You'd probably have to give me a few lessons before I'd have any good input.

Maybe for this problem---seeing that you are talking about throwing away information or disregarding information---it is best to scrap considerations to do with the de Finetti theorem.  Keep in mind that in quantum mechanics, the general expression (or mechanism) for information loss is the trace preserving completely positive map.  Which brings me to this point of yours:

\bfg
I am working with David Poulin on the notion of time that we can
extract from information theory, in order to understand the notion of
time in quantum gravity.
\efg

On this track, I have had some thoughts on recovering the general expression for quantum time evolution (i.e., CP maps) from a Gleason type theorem for the probabilities of outcomes of sequential measurements upon single systems.  In a way, that would be an information theoretic accounting of it.  What I had hoped to do was something along the lines of what I did in Section 5 of my paper \quantph{0205039}, the difference being that the two ``systems'' should be separated in time rather than in space.  The hope was, enforcing causality on the joint probabilities of outcomes (referring to measurements at two different times) would be enough to recover the general form of a completely positive map.  I made some progress, but I could never quite get it to work out.  Maybe this is another thing we could talk about when I visit.

\section{08-10-03 \ \ {\it The Push}\ \ \ (to J. M. Renes and C. M. Caves)} \label{Renes21} \label{Caves74.2}

I'll try to have a look at the new draft tonight.  One thing I'm just noticing is that you should put the $d=2$ case in there.  It's the only damned case anyone can visualize; put it in Bloch sphere language.  You can crib it directly out of Carl's notes.

Another thing:  I think it would be slightly cleaner to write the unitary operators in your Conjecture 1 as D'Ariano does in Eq.\ (21) of \quantph{0310013}.  Finally (for the moment):  I did not understand your equation 18 nor the discussion right around it.

Why am I giving you more push?  Because Koenraad just wrote me that he has now numerically confirmed the conjecture out to $d=24$ (and counting).  He really has come up with all this independently, but I don't think you deserve to add another author and water down the list that much more.

What is a Gabor frame?  What is a Weyl--Heisenberg frame?

I've got to run home now.  I'll probably be online later.

\section{09-10-03 \ \ {\it Pauli Back} \ \ (to K. Gottfried)} \label{Gottfried1}

Thank you for warm letter; it was quite nice of you.

\bkg
First, from where are those great quotes from Pauli on p.\ v1? The
two words in italics -- are they yours -- are so central to the
enigmas posed by quantum mechanics.
\ekg

The italics are Pauli's.  The two passages on that page come from his article ``Matter'', which is included in {\sl Writings on Physics and Philosophy}.  There are more Pauli quotes that intrigued me recorded on pages 192 through 195 of the samizdat, in the ``Greg Comer'' chapter.  It was in February and March 1995 that the Paulian Idea really took hold of me, while reading the collection you mention.

\bkg
I am a life-long Pauli devotee, and as a student in the early 1950s,
taking advantage of my childish knowledge of German, devoted a large
effort to studying his superb Handbuch article, which except for
Dirac, was the only serious text then available. Last year David M
steered me to ``Writings on Physics and Philosophy'' but I have not
been able to find the quotes there in an admittedly cursory search.
\ekg

I understand that Charles Enz has just published a biography of Pauli, but I haven't read it yet.  I have a much more extensive compilation of Paulian quotes in a document I'm putting together---including some amazing letters between Pauli and Bohr---called ``Resource Material for a Paulian--Wheelerish Conception of Nature.''  It's sitting at 112 pages (and 452 references), but it needs to be twice that size before I'll call it complete and post it.  (I.e., a lot of the references don't have quotes inserted into them yet.)  I will let you know when it's ready for perusal if you wish.  (Or if you'd like to see a sneak preview, I'm always looking for people who might catch a typo.)

\bkg
This summer the completely new second edition of my text finally
appeared and I'd like to send you a copy if you'll tell me your
mailing address.
\ekg

I'd love to have it!  Thank you.  One of these days I'd even love to teach a quantum mechanics course.  As long as the post will arrive before April 15, 2004, the mailing address to use is the one below.

\section{09-10-03 \ \ {\it Mocking Bird} \ \ (to N. D. {\Mermin})} \label{Mermin105}

\bdm
So don't underestimate him {\bf [Tony Leggett]}.  He's one of the
most impressive and thoughtful theoretical physicists I've ever met.
Have you ever looked at his little book ``The Problems of Physics''?
\edm

I hope you understand that the note I wrote to you yesterday was
written self-mockingly.  That was its whole purpose.

I hadn't seen or heard of the book before.  I'll try to have a look
at it while I'm at Caltech next week.

\section{09-10-03 \ \ {\it Parsing for the Uninitiated}\ \ \ (to J. M. Renes and C. M. Caves)} \label{Renes22} \label{Caves74.3}

Now I've skimmed the latest version of your paper.  At first I thought I'd read it very, very carefully this time around and give you detailed comments.  But then I found myself pulling my hair out with your first paragraph and ultimately gave up:  It's your paper not mine.

Take your first two sentences for instance:  ``A quantum measurement is termed informationally-complete if its statistics completely determine the input quantum state. An aesthetically appealing and potentially useful measurement is one which also possesses the symmetry that all pairwise inner product magnitudes are equal.''

What are these inner products taken between?  States?  Positive operators?  Complex vectors?  If that's not bad enough already, in the next sentence one starts to feel schizophrenic:  ``In the full regalia of quantum information jargon, such a set of states is a `symmetric informationally-complete positive operator-valued measure' \ldots''  In one half of the sentence you talk about states, in the other half you talk about positive operators.  Clearly you're thinking of a POVM as a set of state vectors (rather than positive operators adding up to unity), as you make more explicit in the next paragraph, but that is quite a nonstandard definition.  Finally, you talk about ``the existence of it'' as if there is only one.

Contrast this with your use of language in Section IV:  There you tell the reader that are only 2 SIC-POVMs when there are clearly a continuous infinity.  Presumably what you mean is that there are only two possible fiducial vectors (for this particular method of generating SIC-POVMs).  But what can it hurt to be precise?

I could keep going on.  For instance, just after Theorem 1, you invoke a nonstandard Gram matrix---I say it is nonstandard because previously you had only talked about sets of state vectors, but now you are considering sets of operators (in their own linear space), and you haven't even told us about the inner product you're using.  (You don't introduce that inner product until Section III.)

The point is, you should be have mercy on your reader, who, by definition, will not be sitting in your head.  Try to anticipate his questions and potential frustrations.  It's a worthy exercise and will help keep people reading your papers.

Enough.  I guess I'm just voicing frustration because I had wanted to give you detailed comments and you thwarted me---it became clear that it wasn't worth my while.

Here is one thing, however, that is worth my while.  Would you change your sentence, ``In quantum information theory such measurements are relevant to foundational issues
\begin{center}
\verb+\cite{fuchssasaki03a,fuchssasaki03b}+,
\end{center}
and useful for creating classical representations of quantum states, doing quantum state tomography, and quantum cryptography,'' to something more along the lines of the following:
\bq\noindent
    In quantum information theory such measurements are relevant to
    quantum state tomography \verb+\cite{A}+, quantum cryptography \verb+\cite{B}+
    and to foundational studies \verb+\cite{C}+ where they would make for a
    particularly interesting ``standard quantum measurement''.
\eq
\begin{itemize}
\item[A)]
C.~M. Caves, C.~A. Fuchs and R.~Schack, ``Unknown Quantum States:\ The Quantum de Finetti Representation,'' J. Math.\ Phys.\ {\bf 43}, 4537--4559 (2002).

\item[B)]
C.~A. Fuchs and M.~Sasaki, ``Squeezing Quantum Information through a Classical Channel:\ Measuring the `Quantumness' of a Set of Quantum States,'' Quant.\ Info.\ Comp.\ {\bf 3}, 377--404 (2003).

\item[C)]
C.~A. Fuchs, ``Quantum Mechanics as Quantum Information (and only a little more),'' \quantph{0205039}, and private communication.
\end{itemize}
You can throw out the reference to the other paper by Sasaki and me.

While I'm here I might as well make a couple of other trivial points.
\begin{enumerate}
\item
Your title:  No journal is going to accept POVM in it.  You might just make it ``Symmetric Informationally-Complete Measurements''.
\item
I think it is worthwhile to mention in the introduction that one of the results of Lemmens/Seidel is that symmetric informationally-complete measurements do not generally exist for real Hilbert spaces.  Therefore it is somewhat surprising (and very intriguing) that they seem to for the complex case.
\item
I noticed that Theorem 1 has changed.  The upper bound is gone (maybe it shouldn't have been there in the first place?); and a  state vector is not a subset.  Also you haven't given a proper citation to Benedetto et al.
\item
Definition of $f_t$.  Missing a subscript $t$ on $\cal H$.  Also $f_t$ is NOT equivalent to a choice of $F_t$, only on the symmetric subspace.  Also equations 6 and 7 are bad notations, if you're going to first write equation 5.
\item
Why introduce terms like Gabor frame if you don't use them?
\item
I still didn't get equation 18 and its surroundings.
\end{enumerate}
I'll desist at that.

\section{10-10-03 \ \ {\it Letters and Quantum Canaries} \ \ (to P. Hayden)} \label{Hayden2}

Of course I'll send off a letter for you.  I'll even spruce up the old one a bit before doing so.  I seem to recall you've published some decent papers since the last time I wrote for you.  \smiley

And I'll take you up on a dinner at your house---it won't count as a bribe.  But here's what I'd really like in return for my efforts:  Solve this problem.  You're probably just the man to do it anyway.

Fix a given set of pure states $\Pi_i$ on a Hilbert space of dimension $d$ with some associated probabilities $\pi_i$.  Imagine drawing from that ensemble just once and handing off a state to a very reserved eavesdropper.  She's very reserved in that she's not interested in gathering all the (expected) information she can; rather she settles for $I=\epsilon$.  Finally she gives back the system you originally gave her, and you check the yes-no question associated with the original state you had prepared.  Tabulate the disturbance in terms of fidelity $F$, just as Sasaki and I did in \quantph{0302092}.

The preliminary question is what is the functional dependence between
$I$ and $F$ in that limit?  The answer should depend upon the ensemble---for instance one might get $F=1-I^\alpha$, where $\alpha$ depends upon the details of the ensemble. Thus, more to the point, I would like to know what is the most sensitive ensemble in this sense, and how does the ultimate sensitivity depend upon $d$?  I.e., what are the best canaries for sniffing out the fumes in these mines?

In the case of the earlier stuff considered by Sasaki and me, I can now show that (along with the uniform ensemble) a symmetric informationally complete ensemble consisting of $d^2$ elements is a maximally sensitive ensemble.  I.e., in the terminology there, it achieves the ``quantumness of the Hilbert space.''  (I only kind of semi-conjectured it in the earlier paper.)  [[BTW, we must have talked about SIC-POVMs at some point?  It's a set of $d^2$ rank-one positive operators for which all pairwise Hilbert-Schmidt inner products are identical.  We still don't know for sure whether they exist, but that shouldn't stop one from using their formal properties.  In any case, Renes and company will post a paper next week showing numerical evidence that they exist at least up to $d=39$.  So I think it's darned likely they really do exist for all $d$.  And they even have a very nice symmetry that's been uncovered (again numerically)---namely, $\mathbb{Z}_d \times \mathbb{Z}_d$.]]  So a natural guess is that the SIC ensembles will also be good canaries in this sense, but I wouldn't have a clue about whether it's really true or how to solve it.

You solve that or give me some good ideas about how to do it, and you can bribe me for a lot of things.

\section{12-10-03 \ \ {\it Torture and Sleepless Nights} \ \ (to C. King)} \label{King9}

Well, after a lot of self-torture and sleepless nights since I last wrote you, I decided to give up on the SIC-POVM problem (for the time being).  It had become an obsession.  Then I turned my attention to prodding Renes and company to quickly get their conjecture out on the web before someone else did.  (A paper by the D'Ariano group a couple of weeks ago made it clear that he might be only one step away from the conjecture himself.  So promptness seemed called for!)

The upshot is, their paper will appear on {\tt quant-ph} Tuesday.  Have a look at it if you get a chance.

\section{12-10-03 \ \ {\it Moving On} \ \ (to C. M. {\Caves})} \label{Caves75}

I read the final draft of the paper today as I was (and still am) flying from Dublin to LA.  It's a nice paper, and I would be proud of it.  Only caught two typos: \ldots

Unfortunately I've got a horrible flu as I make this flight, and I'm leaving {\it everyone\/} home behind in the same condition.  If I can muster the strength, I'm going to plunge ahead on my Holevo festschrift ``On the Quantumness of a Hilbert Space.''  Hopefully I'll have it finished by the end of the week.  There isn't much that it doesn't mimic from your paper, but at least I found some of it on my own (though maybe through a subliminal suggestion from you guys).

If I can figure out the words to say it with, I'd like to make a metaphysical point in its introduction.  It is that Hilbert-space dimension plays a role in our Bayesian considerations that has an eerie conceptual similarity to rest mass in relativity theory.  When an agent hypothesizes a Hilbert-space dimension for an object, he is hypothesizing something {\it it\/} possesses.  (In contrast, when he hypothesizes a quantum state for the object, he is hypothesizing something {\it he\/} possesses.)  Also I want to make the point that the fungibility---a word I really dislike---of quantum information should maybe more properly conjure up images of the E\"otv\"os experiment and the equivalence principle than monetary considerations.  I hope that if I can say it in the right way in the end you'll approve (at least that this is a worthy direction to explore), but I have a long history of such hopes.

\section{14-10-03 \ \ {\it quant-ph/0310075}\ \ \ (to J. M. Renes, R. Blume-Kohout, A. J. Scott, and C. M. Caves)} \label{Renes23} \label{BlumeKohout1.1} \label{Scott1} \label{Caves75.1}

Good to see its appearance.  I think this is going to be a very important paper.

\section{19-10-03 \ \ {\it Fidelity} \ \ (to A. Peres)} \label{Peres56}

Good to hear from you.  I am visiting Caltech at the moment.  Rob {\Spekkens} is also visiting.  Through our combined forces, we are trying to make a full frontal assault on John Preskill's quantum foundations sensibilities:  The goal is to make him see that the best way to think of the quantum state is as a state of knowledge.

Concerning fidelity, the first time I ever heard the term was in Ben Schumacher's talk in PhysComp '92 in Dallas, TX.  The term subsequently appeared in
\begin{itemize}
\item
Benjamin Schumacher, ``Quantum Coding,'' {\sl Physical Review A\/} {\bf 51}(4), 2738--2747 (1995)
\end{itemize}
(which was based on the PhysComp talk) but the article had actually been submitted about two years previous to that---you can check the submission date.  By the way, this is also the article where the word ``qubit'' made its first appearance.

Finally, let mention this article:
\begin{itemize}
\item
Richard Jozsa and Benjamin Schumacher, ``A New Proof of the Quantum Noiseless Coding Theorem,'' Journal of Modern Optics {\bf 41}(12), 2343--2349 (1994).
\end{itemize}
It rederives the result in Schumacher's original (and also used the terms fidelity and qubit), but it actually appeared in print earlier!

\section{19-10-03 \ \ {\it SIC POVMs} \ \ (to A. Sudbery)} \label{Sudbery6}

\bts
We had our first weekly QIT meeting on Thursday. I talked about
mutually unbiased bases, which Sam picked up and is running with. And
my new student is full of ideas (but too shy to talk to anyone but me
so far). It's looking good!
\ets

Good.  If you're interested in that, you might also be interested in \quantph{0310075} by Renes and company on ``symmetric informationally complete POVMs''.  I think it's a very important problem, and as opposed to the MUBs, it looks like a solution exists in all dimensions.

\section{19-10-03 \ \ {\it Question} \ \ (to A. W. Harrow)} \label{Harrow4}

Which postmodern writer was it that made some sense to you?  And what was the book?  I was telling what I could remember of the story to some people here at Caltech tonight, but I want to make sure I get my facts right in the future.

\subsection{Aram's Reply}

\bq
Believe it or not, there have been a few postmodernists who have
made some sense to me at some point or other.  In particular, I liked
Foucault and Zizek.

However, the story I told you was about Jacques Derrida's {\sl Of Grammatology},
specifically part II, chapter 1.  He was writing about {\sl On the Origin of
Language\/} by Rousseau and {\sl Tristes Tropiques\/}  by Levi-Strauss, I think.

Glad this is still making the rounds.
\eq

\section{22-10-03 \ \ {\it Upside-Down Quantumness}\ \ \ (to S. J. van {\Enk})} \label{vanEnk28}

\bsve
He thought your measure of quantumness had the wrong ``sign'', with
which I agreed. Namely, your measure of quantumness is smaller when
the set is more quantum \ldots\ You agree?
\esve

Yes, I agree.  But here's the justification I used in the paper, for whatever it's worth:
\bq\noindent
Finally the {\it quantumness\/} of the set $\cal S$ is defined by
\be
Q({\cal S}) = \inf_{\{\pi_i\}} F_{\cal P}\;.
\label{Eugene}
\ee
This definition has the slightly awkward property that the {\it
smaller\/} $Q({\cal S})$ is, the more quantum the set $\cal S$ is.
This, of course, could be remedied easily by subtracting the
present quantity from any constant.  However, if we wanted to
further normalize the quantumness so that, say, its value achieves
a maximum when no set of states has a higher quantumness, we would
have to make use of a (presently) unknown constant in our
definition. Thus, it seems easiest for the moment to simply remain
with Eq.~(\ref{Eugene}).  This does, however, raise an important point---indeed one of paramount concern for the ultimate use of quantumness.  Just how quantum can a set of states be in the most extreme case?
\eq

\section{30-10-03 \ \ {\it Minor Correction} \ \ (to C. G. {\Timpson})} \label{Timpson2}

I quickly skimmed your proposal.  It looks pretty good.

One minor point about my biography: [\ldots]

Now, no need for a change on this one---it's not so important at this stage---but I'm not very comfortable with the language that ``measurement induces an uncontrollable disturbance in the object system.''  The reason for this is that I would not characterize the process of giving birth as an ``uncontrollable disturbance'' to one's wife.  I'll attach two excerpts from the latest samizdat that I think best characterize my present broader thoughts on quantum mechanics---the more key issue is the nondetachableness of the observer.  BTW, you actually played a bit part in one of the notes.  (Looking back at it, I realize I probably characterized you too harshly, but I hope you'll forgive me.)

\section{04-11-03 \ \ {\it Slow Draw McGraw} \ \ (to R. W. {\Spekkens})} \label{Spekkens22}

\brws
I forgot to ask you while you were here: where did you quote van {\Enk} as saying that there is nothing inherently nonclassical about a $2d$
Hilbert space?  I'd like to throw that reference into the contextuality paper.
\erws

The discussion you're thinking of about van {\Enk} is in my paper \quantph{0205039}.  It went like this:
\bq\noindent
There are two things that are significant about this much of the
proof.  First, in contrast to Gleason's original theorem, there is
nothing to bar the same logic from working when $D=2$.  This is quite
nice because much of the community has gotten into the habit of
thinking that there is nothing particularly ``quantum mechanical''
about a single qubit.\,  Indeed, because orthogonal
projectors on ${\cal H}_2$ can be mapped onto antipodes of the Bloch
sphere, it is known that the measurement-outcome statistics for any
standard measurement can be mocked-up through a noncontextual
hidden-variable theory.  What this result shows is that that simply
is not the case when one considers the full set of POVMs as one's
potential measurements.
\eq

But actually that was a dirty trick on Steven.  (I thought I was being cute, but he probably hates me for it!)  The reason is because I was partly responsible for Steven's having put a statement like that at the end of his anti-Meyer paper---I remember advising him on it, it was definitely the sort of thing I was thinking at the time.  Anyway, I looked for my email to him on the subject but couldn't find it; maybe it was something that came out of our discussions.  I'll cc this note to Steven to see what he says.

\section{04-11-03 \ \ {\it Belle and Beau}\ \ \ (to S. L. Braunstein)} \label{Braunstein10}

\bslb
Do you have explicit forms of Wootters' ``belle'' and ``beau'' mutually
unbiased bases? I am trying to pin down an error in something I am
doing and I suspect it occurs here.
\eslb

Actually I don't even know what the belle and beau bases are.  They do sound interesting though.  Can I infer from your question that complete sets of mutually unbiased bases come in at least two flavors?  What is the criterion?

This intrigues me mostly, because it dawned on me the other day that symmetric informationally complete POVMs may come in many different flavors.  For instance, the character of the ``triple products''
$$
\tr
(\Pi_i\Pi_j\Pi_k)
$$
can make all the difference in the world for certain questions.

\section{07-11-03 \ \ {\it Airline Tickets} \ \ (to J. B. M. Uffink)} \label{Uffink2}

Since I see Jean Bricmont is among the speakers, I will have to do my homework:  I.e., I'll read his book with Sokal, {\sl Intellectual Impostures}, before coming.  I have a sinking feeling that he'd label any attempt to Bayesianize quantum probability---and thus me---in the same way.

By the way, a couple of weeks ago when I had a little leisure time at Caltech I discovered your paper ``The Constraint Rule of the Maximum Entropy Principle.''  It's great!

\section{07-11-03 \ \ {\it Quantum Pragmatology} \ \ (to G. Valente)} \label{Valente3}

Thank you for sending me your thesis ``Probability and Quantum Meaning:\ Chris Fuchs' Pragmatism in Quantum Foundations''.  Unfortunately it came through my emailer as a ``corrupted file,'' so I had to do a good bit of reconstruction on it before I could get the \LaTeX\ to compile properly.  But when I did finally get it to print this morning, what a nice thing to behold!  I am very flattered by it all.  I never imagined that you were embarking on a work of such scope (140 pages!).  Is this a kind of ``senior honors thesis'' or rather is it a ``master's thesis''?  Somehow I had gotten the impression before that you were an undergraduate, but maybe I was wrong?

Anyway, at this point, I've read the first 21 pages and also Chapter 5, and it seems to flow very nicely.  I especially enjoyed your discussion in Section 2.1.2.  I think you're the first person to ever show any appreciation of that idea.  Also your analysis on pages 124 and 125 were quite enjoyable.  I have had some connected thoughts myself recently (i.e., that dimensionality would be a pretty worthless `property' unless all Hilbert spaces are, in fact, finite) \ldots\ which leads me to some rather fantastic thoughts about the various entropy bounds cropping up in black hole theory and quantum gravity.  I'll report some of this in my upcoming paper ``On the Quantumness of a Hilbert Space'' (which I'm working on at the moment).  If you would like, I'll send you a copy when the crucial pieces are in place.

Already seeing this much of your thesis---I hope the rest of it won't make me reassess!---I hope you will consider posting the final version of it on the {\tt quant-ph} archive.  At the very least it may stir up some discussion and get more people involved.

By the way, I've caught a few typos.  Maybe the most important is in your Musil quote:  You write, ``It is reality that awakens possibilities, and nothing would more perverse than deny it.''  Is it missing a ``be'' and a ``to''?  Maybe you should check the whole passage very carefully:
\bq
To pass freely through open doors, it is necessary
to respect the fact that they have solid frames. This principle,
by which the old professor had lived, is simply a requisite of the
sense of reality. But if there is a sense of reality, and no one
will doubt that it has its justifications for existing, then there
must also be something we can call a sense of possibility.
Whoever has it doesn't say, for instance: Here this or that has
happened, will happen, must happen; but he invents: Here this or
that might, could, or ought to happen. If he is told that
something is the way it is, he will think: Well, it could probably
just as well be otherwise\ldots\ \ A possible experience or truth is not
the same as an actual experience or truth minus its ``reality
value'' but has---according to its partisans, at least---something
quite divine about it, a fire, a soaring, a readiness to build and
a conscious utopianism that does not shrink from reality but sees
it as a project, something yet to be invented. After all, the
earth is not that old, and was apparently never so ready as now to
give birth to its full potential\ldots\ \ It is reality that awakens
possibilities, and nothing would be more perverse than to deny it. Even
so, it will always be the same possibilities, in sum or on the
average, that go on repeating themselves until a man comes along
who does not value the actuality above idea. It is he who first
gives the new possibilities their meaning, their direction, and he
awakens them\ldots\ And since the possession of qualities assumes a
certain pleasure in their reality, we can see how a man who cannot
summon up a sense of reality even in relation to himself may
suddenly, one day, come to see himself as a man without qualities.
\begin{flushright}
\hspace*{\fill} --- Robert Musil, {\sl The Man without Qualities}
\end{flushright}
\eq
(I've never actually read Musil myself.  In fact, I had never heard of him until last year when Frank Verstraete told me that his book seems to express my views on quantum mechanics!  The quote you used has convinced me that I should buy the book and try to read it!)  Another important mistake:  Look closely at Eq.\ (2.8).  Also the explanation ``moreover, the passage in the fifth line \ldots'' following the equation array on page 20 is incorrect.  $G$ will not necessarily commute with the $\Pi_d$.  Rather it is the general property that for any two operators $A$ and $B$, the products $AB$ and $BA$ have the same eigenvalues.

I am glad to hear that you will take a PhD in the philosophy of physics, and I hope you will continue to pursue this pragmatic line of thought.  Where to go?  Most of all, I'd recommend to stay away from the UK!  Pragmatism is certainly not viewed particularly positively there!  A few places to consider in the US might be:  1) the University of Maryland, Jeff Bub is there and he is really getting taken with this stuff, 2) University of Washington, Arthur Fine is there and seems to have quite an interest in it too, and finally 3) Princeton University, Bas van Fraassen is there and he seems to have some affinity (or at least understanding) of Bayesian ideas.  Unfortunately Richard Jeffrey died a few months ago, he would have probably been an excellent advisor at Princeton.  Those three choices definitely follow my personal preferences, so you should compensate for that in your calculation.  The are plenty of very strong Philosophy of Science programs in the US---like the University of Pittsburgh---but I doubt those places are as prepared for your line of thought as the three places I listed above.

I'll leave you with that.  As I wind my way through your thesis, I may write you more.  Oh, you asked about a visit to Dublin.  Sounds great; but my calendar won't be free until at least mid-January.

\section{07-11-03 \ \ {\it Toy Model Boy} \ \ (to C. M. {\Caves})} \label{Caves75.2}

I've been meaning to ask you:  What did you think of Spekkens's toy model?  I think it makes explicit a lot of things we were hoping for at an intuitive level.  Thereafter, Spekkens and I diverge on where the program should go---he still deeply desires objective probabilities and nonlocal hidden variables (doomed to failure)---but I think the toy model is stand-alone as a wonderful illustration.  Just wondering about your thoughts.

\subsection{Carl's Reply}

\bq
I am inclined to agree with you.  I think the toy model provides a very nice illustration of how far you can get with ``maximal information about underlying objective quantities is not complete.''  Surprisingly far, but not, of course, the whole enchilada.  Spekkens seems to feel that some further principle added to the toy model will give all of quantum mechanics, but he is, as you say, somewhat addicted to underlying objectivity.  It might be that some further principle will do it, but only if the first principle that ``maximal information is not complete'' is cleansed of all reference to underlying objectivity, and then I'm inclined to think the further principle has to do with having a continuum of possibilities (and evolution among these), \`a la Hardy.  Could, of course, be that nothing along this line will work or that some other type of additional principle is what is necessary.
\eq

\section{10-11-03 \ \ {\it Ever Vigilant}\ \ \ (to S. Aaronson)} \label{Aaronson5}

Where are you sending that lovely paper \quantph{0311039}??

\section{10-11-03 \ \ {\it Peres Theme}\ \ \ (to V. Buzek)} \label{Buzek1}

\bvb
I am honoured to write a paper to the Festshcrift for Asher Peres. Do
you have any specific underlying theme in mind?
\evb

That's great!  The underlying theme is Asher \ldots\ and all that entails.  There will be some submissions on relativity theory, but most will be on quantum information and quantum foundations.  Take your pick.

\section{10-11-03 \ \ {\it Who Knows What I Am?!}\ \ \ (to G. Valente)} \label{Valente4}

\bgv
Finally, let me know something I have been wondering since I started reading your ideas. Isn't what I called ``un-veiling the reality'' with respect to your program a suggestion you interiorized from your interest for Schopenhauer (dropping Maya's veil)?
\egv

About Schopenhauer, I really haven't read much of him \ldots\ only a little (actually very little) about him from secondary sources.  Effectively the only thing I really know about him is the quote below that I had put into my computer.  [See 08-08-01 note titled ``\myref{Caves3}{The First Eye}'' to C. M. {\Caves}.] So, your ``un-veiling''---I think---is new to me.

\section{11-11-03 \ \ {\it Working Title} \ \ (to G. Brassard)} \label{Brassard20}

We can't have a paper without a prospective title.  We've got to settle first things first.  How about, ``Prospects for Quantum Cryptographic Interpretation of Quantum Mechanics''?

Flying to Canada on a 747 tomorrow.  I hate those things.

\section{11-11-03 \ \ {\it Budget Butter} \ \ (to R. E. Slusher)} \label{Slusher2}

I wish I could say I had some exciting news, but mostly I'm just plugging away.  These symmetric POVMs have turned into an interesting and long-lasting problem.  Have you looked at Caves and Co's paper \quantph{0310075}?  They had originally had me on the author list of that, but in the end I withdrew thinking I hadn't done enough to merit it.  I'll probably kick myself in the butt for years to come:  I'm starting to feel that the conjecture there may remain open for quite some time.  I discussed it a lot when I was visiting Caltech and even the great Kitaev thought it nontrivial.

Anyway, Patrick Hayden and I do have some results on the accessible information of such an ensemble (i.e., no longer thinking of the states as elements of a POVM, but instead a signaling ensemble).  I'm in the process of writing that part of the work up.  I hope I'll put it on {\tt quant-ph} soon.

How is Bernie's patent coming along?  And how about your experiment of the same?

\section{13-11-03 \ \ {\it Pearle, Hampton, Hampshire} \ \ (to R. W. {\Spekkens})} \label{Spekkens23}

While it's on my mind, I've confirmed that Phil Pearle was at Hamilton College in New York State.  He retired in 2000.  Last night I said it was Herb Bernstein who was at Hampton College, but that was a glitch:  Herb is at Hampshire College in Massachusetts (with a name like that, you might have thought it'd be in New Hampshire).  And finally, on further reflection I don't know that I know of a Hampton College.  The only thing I can think of is Hampton Inn, but that's a hotel chain.

\section{13-11-03 \ \ {\it Variety and Freedom} \ \ (to L. Hardy)} \label{Hardy12}

Attached is the Rorty passages I sent to Howard Wiseman.  [See 24-06-02 note ``\myref{Wiseman6}{The World is Under Construction}'' to H. M. Wiseman.]  The phrase was actually ``variety and freedom''---that's the only goal.  Reading the passages again I still like them.  I think they nicely capture what would make a lawless world a better world than a law-fixed one.

\section{13-11-03 \ \ {\it Elephants} \ \ (to L. Hardy \& F. Girelli)} \label{Hardy13} \label{Girelli2}

Below I place some excerpts I dug out of my paper \quantph{0204146}, ``The Anti-{\Vaxjo} Interpretation of Quantum
Mechanics.''  They certainly do a better job of what I was on about last night than I did \ldots\ though in my sober reading this morning
they only seem slightly less drunk!

To Lucien's remark that scientific theories are cumulative---which he
wanted to use as an indication that succeeding theories are better
and better representations of reality---my reply might be that, from
this point of view which I'm trying to develop, the more relevant
concept is that succeeding theories have better ``feedback
mechanisms.'' That is, one might say that the human species is more
developed than the elephant species---even though neither species was
foreordained by nature---simply because humans can better adapt to
the changing conditions around them.  The human species has more
feedback mechanisms for adjusting to the environment around it.
Similarly, let us say for general relativity and Newtonian gravity.
The former can survive more experimental onslaughts than the latter
because its structure is more malleable and less rigid than the
latter's.  But I'm just shooting from the hip at the moment.

\section{15-11-03 \ \ {\it TOE Jam} \ \ (to C. Snyder)} \label{Snyder1}

I very much enjoyed listening to your excitement last night about a TOE-less world.  I think we're kindred spirits in this project.  Let me give you a pointer to two of my writings that you might enjoy in this regard.
\begin{itemize}
\item[1)] Look at Sections 4 and 5 of my paper ``The Anti-{\Vaxjo} Interpretation of Quantum Mechanics''
\item[and]
\item[2)] the little essay ``\myref{Wiseman6}{The World is Under Construction}'' on pages 210--215 in my samizdat ``Quantum States:\ What the Hell Are They?''.
\end{itemize}
Both documents can be found on my webpage in PDF format.  If you have trouble viewing the files, let me know and I'll send you excerpts in plain text.

By the way, the samizdat that Kluwer Academic will be publishing for me is {\sl Notes on a Paulian Idea}.  It's posted on the same webpage.

Hope to see you again later this week.

\section{15-11-03 \ \ {\it The Artist} \ \ (to C. Snyder)} \label{Snyder2}

I was just flipping through ``Quantum States:\ WHAT?''\ after telling you about it, and came across the birth announcement I wrote for my second daughter.  It's on page 121.  [See 19-12-01 note ``\myref{KatieViolaFuchs}{Katie Viola Fuchs}'' to the world.]  Maybe it captures a little bit (in a single sentence) of what moves you as an artist---that your painting will change the world.

\section{15-11-03 \ \ {\it Conjugalation} \ \ (to G. L. Comer)} \label{Comer44}

I bet you've never used the words ``lascivious'' and ``conjugal'' in
a talk abstract before.  Call me crazy, but I did.  (See below.)
Somehow it seemed like crazy words were called for at a crazy place.
I'm visiting the Perimeter Institute for a week and a half [\ldots]

Maybe one fortunate thing has come from this visit to Perimeter: I've
been reading loads of review material on quantum gravity.   I have
particularly liked the stuff I have read by Jacob Bekenstein and
Raphael Bousso.  I'm definitely tipping toward the entropy-bound side
now, which I am convinced can only mean ``assign a finite dimensional
Hilbert space to what was once thought to be a continuous system.''
I think this stuff is very likely revolutionary, and I am so sorry it
took me 15 years to appreciate it.

Another thing that has hit me is that with this Bayesian view of
quantum operations developed in {\sl Quantum States:\ What the Hell
Are They?}\ is that there is simply no information-loss ``paradox''
associated with black holes.  From my point of view, as long as a
time-evolution map is linear, completely-positive there is nothing to
keep one from assigning it as long as it is actually one's firm
belief for a system.  In particular there is no requirement that the
map be derivable from a unitary evolution on some composite physical
system. Thus the paradox, like the black hole itself, evaporates. I'm
going to try to write this up a little more clearly and pass it by
John Preskill soon.  I know, of course, that he won't buy it---he
could only do that if he had bought the starting point, i.e., that
quantum states and quantum operations are not ontic, but
subjective---but it'll still be a good exercise, and I know I'll
learn from his reaction (as long as it is not silence).

\bq\noindent
Title:\medskip\\
What is the Difference between a Quantum Observer and a Weatherman?\medskip\\
Abstract:\medskip\\
Not much.  But where there is a difference, there lies quantum
theory's most direct statement about properties of the world by
itself (i.e., the world without observers or weathermen).  In this
talk, I will try to shore up this idea by writing quantum mechanics
in a way that references probability simplexes rather than Hilbert
spaces.  By doing so, the connection between quantum collapse and
Bayes' rule in classical probability theory becomes evident:  They
are actually the same thing up to a linear transformation depending
upon the details of the measurement method.  Looking at quantum
collapse this way turns the usual debate in quantum foundations on
its head:  only local state changes look to be a mystery.  State
changes at a distance (as after a measurement on one half of an EPR
pair) are completely innocent---they simply correspond to
applications of Bayes' rule itself, without the extra transformation;
that is, collapse-at-a-distance is nothing more than the usual method
of updating one's information after gathering data.  Thus the idea
develops that if a quantum reality is to be found in the quantum
formalism, it will be found only in the formalism's {\it
deviations\/} from classical probability theory:  Reality is in the
difference. Time permitting at the end of the talk, I will try to
sketch, without getting too lascivious, how such a reality may be
best thought of in conjugal terms.
\eq

\subsection{John's Reply --- Yes, Really, John's Reply}

\bq\noindent
I was disappointed to hear that you won't learn from my silence.  I
may need to rethink my method of communicating \ldots
\eq

\section{15-11-03 \ \ {\it Elementary Quantum Phenomena} \ \ (to R. Laflamme)} \label{Laflamme1}

I looked in my email for that dinner or beer you said I promised you, but I couldn't find it.  You know, with my weird beliefs about quantum mechanics, a quantum phenomenon may not be an actual phenomenon until it is written in an email somewhere!  (Or was it John Wheeler who said that?)

\section{17-11-03 \ \ {\it Tossing and Turning} \ \ (to L. Smolin)} \label{SmolinL3}

Well you caused me quite a sleepless night of tossing and turning with your ``needling'' question yesterday.  When I snapped that your question is like my asking you why you haven't quantized gravity yet, maybe I should have bitten all the way through the flesh:  Maybe our two projects are the same damned thing!  And if you can't do it by yourself, why should I be expected to?  More seriously, I wonder whether identifying the ontic piece of quantum theory---i.e., the research program I am trying to build---might not just amount to (a large part of) getting at a quantum theory of gravity.

You said, ``Come back to me when you've got the ontic piece.''  (Well, you didn't quite say ontic but something of the same flavor.)  The part I see most clearly at the moment is Hilbert space {\it dimension\/} \ldots\ and maybe that's all there is.  At present, I am toying with the idea that $D$ quantifies a kind of ``nervous creative energy'' on the part of a quantum system.  More speculatively, I find myself wondering if it is too far-fetched to think that that creative energy gravitates much like its less metaphorical cousin.

So I was really quite excited last night on the bus ride home when I came across your appendix on Relational Quantum Cosmology in the review article you gave me!  I had never heard of Crane before, and I didn't know about your work with Fotini on what you call the weak holographic principle.  I almost melted when I read the sentence, ``The only property a screen has beyond its place in the causal network is the dimension of its Hilbert space.''

Anyway, in good faith to this excitement, let me place below some excerpts from my newest email samizdat, {\sl Darwinism All the Way Down (and Probabilism All the Back Up)}.  The connections to your sentence above should be evident, and I hope the passages convey a little of why I am thinking in these directions.  Of course the mumblings are not science yet (or they would be appearing in papers, not samizdats), but rather dreams of a solid direction to follow.  The real science strikes me as just around the corner.

\section{18-11-03 \ \ {\it Don't Forget Rorty} \ \ (to R. W. {\Spekkens})} \label{Spekkens24}

Don't forget to bring my Rorty back before I fly out.

Oh bring back, oh bring back, oh bring back my Rorty to me, to me \ldots

\section{25-11-03 \ \ {\it Philosophy Sauce} \ \ (to A. Valentini)} \label{Valentini1}

\bav
Sorry about the testosterone-fueled rant last night. It did come out
with an aggressive edge. Hope you have a good return trip.
\eav

No bothers; it just made for a saucier night.

For the record though:

The anti-{\Vaxjo} paper was passed by (at least):  Jeff Bub, Henry Folse, Bas van Fraassen, Wayne Myrvold, and Chris Timpson.

The ``World is Under Construction'' note was passed by (at least):  Richard Healey, Wayne Myrvold, Chris Timpson, and Steven Savitt.

To my knowledge \quantph{0205039} ``Quantum Mechanics as Quantum Information (and only a little more)'' has been read at some level by:  Guido Bacciagaluppi, Jeff Bub, Bill Demopoulos, Armond Duwell, Arthur Fine, Bas van Fraassen, Stephan Hartmann, Richard Healey, Wayne Myrvold, Itamar Pitowsky, Steven Savitt, Abner Shimony, Chris Timpson, and Giovanni Valente.

The large samizdat {\sl Notes on a Paulian Idea\/} has been distributed to: Guido Bacciagaluppi, Jeffrey Bub, Jeremy Butterfield, Arthur Fine, Henry Folse, Bas van Fraassen, Alexei Grinbaum, Hans Halvorson, Stephan Hartmann, Richard Healey, David Malament, Wayne Myrvold, Itamar Pitowsky, Steven Savitt, Abner Shimony, Chris Timpson, Giovanni Valenti, and Steve Weinstein.

Of the people above:  Amit Hagar has written a critique of my stuff in the journal {\sl Philosophy of Science}, Giovanni Valente has just submitted a Master's thesis in philosophy at the University of Padua on it, and Stephan Hartmann is including it as a component in his course at the London School of Economics.

The point:  I don't know whether you would be willing to call these guys philosophers, but it can be verified that they are all employed by philosophy departments.  The last thing I see myself doing is hiding from critique---a better word would be ``test.''

The only real goal is in understanding quantum mechanics.  And on this, we are in it together.

\section{25-11-03 \ \ {\it The Unfortunate Phrase} \ \ (to A. Valentini)} \label{Valentini2}

Let me follow up on your question about my view of the phrase appearing in one of my papers with {\Caves} and {\Schack}:  ``Gleason's theorem can be regarded as the greatest triumph of Bayesian reasoning.''

The place to look is my web samizdat {\sl Quantum States:\ What the Hell Are They?}.  See:
\begin{itemize}
\item
page 75, ``\myref{Caves27}{The Stopgap}''
\item
page 84, {\Caves}ism \ref{CavesismGleason} and below [03-10-01 note ``\myref{Caves30}{Replies on Practical Art}'' to C. M. Caves and R. Schack]
\item
page 118, ``\myref{Mermin50}{Trumps and Triumphs}''.
\end{itemize}
My own view of what the paper accomplishes can be found in the letters
\begin{itemize}
\item
``\myref{Schack4}{Identity Crisis}'' on pages 35 to 38,
\item
``\myref{Caves14}{Unique Assignment}'' on pages 53 to 54.
\end{itemize}
I thought I had said it somewhere more clearly in there, but I don't have the time at the moment to keep digging for it.

\section{25-11-03 \ \ {\it Worthwhility}\ \ \ (to J. E. {\Sipe})} \label{Sipe2}

I just wanted to say I'm sorry I didn't get a chance to talk to you longer at PI.  Your reaction after my talk made my day; in fact, it alone was enough to make my giving the talk worthwhile.  You wouldn't believe how much strength and stamina to go on I derive from reactions like yours.

Anyway, there's so much to do in this program and so many subtleties that still need hammering out, I welcome your interest.

\section{25-11-03 \ \ {\it I Dream of Everything} \ \ (to C. Snyder)} \label{Snyder3}

I just reread the note you sent me and enjoyed it again indeed.  Thanks for the great conversations last week.

\bcs
As artists we are subjective in the extreme \ldots\ (there are no wrong descriptions (or models),
only personal ones.
\ecs

This you might say is the program of ``Quantum States:\ What the Hell Are They?'':  It is to make the physicist realize that he has more in common with the artist than he might have thought.  The quantum state, in particular, I say, represents a subjective ascription.  Mimicking your words, there are no wrong quantum states, only personal ones.  I did a quick search and found some discussions of this on pages 39 (right after Merminition \ref{MerminitionOrthogonal}), 49--50, 54--56 (the \myref{Caves16}{underappreciated point}), and 159--166 (parts of \myref{Hardy7}{the letter to Hardy}) of the document you printed for me.  You might enjoy reading about this overlap of our concerns---it is quite real.

I hope we'll have an infinite time in the future to build on this.

\subsection{Christian's Preply}

\bq
Thanks for directing me to the various articles \ldots\ it will take some time to digest (I'm afraid I did a little more reading then perhaps I should have.)

I slept last night (got to bed late, as per always) and was pushed into a rather awkward dream (or state that resembled a dream). In this dream I was able to observe many things: myself, the universe, the lines of the forever.

Now I must apologize because I just spent 30 minutes trying to explain what it was that I saw in this dream, and realized I couldn't do it without some sort of pen and paper \ldots\ so I've erased it. Let's just say there was an $x$, $y$ and $z$ axis, a reflected sine wave and a bad morning hangover where everything I observed on the inside of my eyelid kept changing from a particle to a wave. It is a bad way to wake up when you have a soccer game an hour later.

Anyway, I was thinking, artists and you physics guys aren't really so dissimilar you know. We are both observers of the world, we both try to describe the world, but our models of the world are so fundamentally different it makes me wonder about you guys sometimes. As artists we are subjective in the extreme \ldots\ (there are no wrong descriptions (or models), only personal ones). As physics guys (and gals of course) you always want to be objective and your models (or descriptions) are always wrong. You know this going in \ldots\ you might get closer to the truth (assuming of course there is an objective overriding truth, which you don't believe in) but you guys do it with {\it your\/} data, with {\it your\/} machines.

Why is it that we expect objective data from machines we build?\\
{*LONG BREAK*}

Sorry, the clock on the wall says 3 AM, I'm going to bed. I'll write more later.\\
{*LONGER BREAK*}

Ahem, looks like I forgot to send this last night. Well here's to the early morning forgetfulness. Now I must promise not to re-read this thing before I send it 'cause otherwise I'll never send it.
\eq

\section{26-11-03 \ \ {\it Bartleby and What I Really Mean} \ \ (to C. H. {\Bennett})} \label{Bennett31}

I'm off to the Utrecht meeting tomorrow morning to declare ``Quantum
Information Does Not Exist.''\smallskip

\noindent {\bf Title:}  Quantum Information Does Not Exist \smallskip

\noindent {\bf Abstract:}  It is information {\it carriers\/} that exist---conceptually both classical and quantum.  To confuse the epistemic category (the information) with the ontic (the carriers) is to cause any amount of trouble.  Nonetheless, one thing is true when it comes to applications of information theory to classical and quantum phenomena:  There is a difference.  And, in that difference---this talk will argue---lies quantum theory's most direct statement about properties of the world by itself (i.e., the world without the information processing agent).

\section{02-12-03 \ \ {\it Alive and Well, but back in Dublin} \ \ (to A. Peres)} \label{Peres57}

I am just back in Dublin from a few days of email blackout in Utrecht.  Kiki accompanied me to this conference, and we left the kids in the care of their grandmother (who flew to Dublin from Munich just for the purpose).

I gave a talk with the belligerent title ``Quantum Information Does Not Exist'', but it was a variant of the one you saw me give in Aarhus.  I chose the title that I did to give Ari Duwell's nice paper of the same title some advertisement, but also because I wanted to emphasize de Finetti's version of Bayesian probability more this time.  For most of the audience though, I think the talk was a disaster.  The main thing was that I did not enter the arena with enough forethought of the ``crowd control'' I would need.  Roger Balian and Theo Nieuwenhuizen, in particular, heckled me quite a bit.  They would not accept that a POVM could actually be measured (since it contains noncommuting operators).  It was silly and only demonstrated that they just were not listening.  In the tradeoff at least, I could see the lights flashing in the eyes of Jos Uffink and Micha{\l} Horodecki.  Micha{\l}, for one, seemed to finally understand how you and I view the wavefunction (``Quantum Theory Needs No Interpretation''), and how we are not ``inconsistent''---something that he apparently thought of us all these years.  So, that was very nice to see such a change in him.

\section{02-12-03 \ \ {\it Back from Holland} \ \ (to G. L. Comer)} \label{Comer45}

I'm just back in Dublin from four days in Utrecht.

Thanks for the Gottesman--Preskill article.  Actually Gottesman is a pal of mine.  I had a dream about him a few weeks ago:  I dreamed he got a Nobel prize for his quantum error correcting codes.  In the dream I asked him, ``How could you win a Nobel prize for this when Shor and Steane were first?  You based your work on theirs.''  He replied, ``My codes were longer.''  He didn't say ``more efficient'' or something like that; he said ``longer''!  When I woke up, I thought this has got to be something Freudian.

Anyway, about the article, I'll definitely have a look at it.  I've been wanting to write that note to Preskill about the ``information paradox'' for some time.

Did I tell you that Gerard 't Hooft and I were both invited speakers at this meeting in Utrecht?  Well, he's actually a local \ldots\ so maybe his invitation didn't mean so much as mine.  Yeah, right!  Anyway, though I did show up for his talk, he didn't show up for mine!  His title was something like ``Black Holes:\ The Triple Point of GR, QM, and SM''.  My title was ``Quantum Information Does Not Exist.''  His talk was pretty bad, and mine went pretty badly --- so maybe we achieved some kind of parity!  Honestly, you would have been pretty disappointed with the sorts of ideas he's thinking about now.  What he is doing is trying to seek a {\it local\/} hidden variable model that is consistent with quantum mechanics.  But how can that be done in light of the experimentally tested Bell inequality violations?  't Hooft points out that the derivation of the Bell inequality violations of quantum mechanics rests upon some assumptions:  Like, for instance, that the localized experimentalists at each end of the entangled particles have the {\it free will\/} to set their measuring devices as they wish.  If one stops assuming such free will, then that blocks the usual derivation.  And he takes this absolutely seriously!

Sorry, I have to go:  My mother-in-law (visiting at the moment) just called me for dinner.

\section{02-12-03 \ \ {\it You Really Must Hurry} \ \ (to R. W. {\Spekkens})} \label{Spekkens25}

I'm just back from the Utrecht meeting and pretty exhausted, but I've got to tell you about a conversation I had with Micha{\l} Horodecki.  He was telling me about how Robert Alicki has been pushing them to think of quantum mechanics as just a restriction on knowledge.  And thus they had developed some ``toy model'' that, for instance, exhibits a kind of no-broadcasting theorem on top of a no-cloning theorem.  Prompted by that, I figured I had better tell him (a very little) about your toy model.  You could actually see the panic in his face.  He said something like, ``We must post our paper soon.''  I told him that priority shouldn't be an issue because your work has already been cited three times on {\tt quant-ph} and gazillions of people have heard your talk on the subject.  But I doubt I fazed him much with that:  These Horodeckis are wham-bam-thank-you-ma'am kind of guys when it comes to getting their ideas out there.

I know that you understand the point:  You are walking on thin ice with this result.  It's a beautiful piece of work, however the time is starting to become ripe for its re-appearance out of sheer Darwinism.  The pressures are mounting for the quantum state to be viewed epistemically and science is responding.

\section{02-12-03 \ \ {\it Now, For the Important Stuff} \ \ (to L. Henderson)} \label{Henderson1}

Here's the PhD thesis I was talking about:
\begin{itemize}
\item
W.~H. Long, {\sl The Philosophy of Charles Renouvier and Its Influence on William James}, Ph.~D. thesis, Harvard University, 1927.
\end{itemize}
There's another one along the same lines that I also want to get my hands on:
\begin{itemize}
\item
H.~C. Sprinkle, {\sl Concerning the Philosophical Defensibility of a Limited Indeterminism:\ An Enquiry based on a Critical Study of the Indeterministic Theories of James, Renouvier, Boutroux, Eddington, Bergson and Whitehead}, Ph.~D. thesis, Yale University, 1929.
\end{itemize}
but that's beside the point.  (I just wanted to have a record of both in the same email; I store most of my information in emails.)

Anyway, it would be lovely if you could check out the Long thesis before coming to Maryland conference next April 30.  Or maybe I could figure out a way to get to Cambridge before then.

Thanks so much for this help!

\section{02-12-03 \ \ {\it Paper and Visit} \ \ (to R. {\Schack})} \label{Schack76}

I'm just back from Jos Uffink's meeting in Holland.  Among other things, I got to meet Stephan Hartmann there; he's quite a Bayesian.  Also I had a fruitful time talking to Jos himself.  I asked him his views on the interpretation of probability and got quite a surprising answer:  He has not made up his mind, but he leans 1) mostly toward radical von Mises type frequentism (with kollectivs, randomness and place selections and all that), but 2) to a smaller extent toward radical subjectivism.  He sees those as the only two consistent alternatives.  In particular, he rejects propensitism, logical interpretations, objective Bayesianism, etc.  Strangely he also knows (and admits) that frequentism has no empirical content, but that doesn't seem to sway him.

Have you ever read his paper on the MaxEnt principle and thought about whether it is valid?  It's actually quite nice, and I think an indication that he is a thorough thinker.  You can find it here:
\myurl[http://www.phys.uu.nl/~wwwgrnsl/jos/mep2def/mep2def.html]{http://www.phys.uu.nl/$\sim$wwwgrnsl/jos/mep2def/mep2def.html}.

\section{02-12-03 \ \ {\it Scrap Paper} \ \ (to L. Hardy)} \label{Hardy14}

I just sent the attached picture to my brother so that he could print it out for my mom.  It's a drawing of Emma's from last week.  Upon reviewing it before sending it, its relevance to Perimeter dawned on me:  You'll see what I mean.  Whenever I come home from a trip, Emma asks me if I have any scrap paper for her.  I usually give her whatever was in the conference packet, etc.  This time it was a Perimeter Institute pad.  Her drawing skills are getting better and better; I'm pretty proud of her.

\section{03-12-03 \ \ {\it Holism, Nonholism, or Nihilism?}\ \ \ (to M. P. Seevinck)} \label{Seevinck1}

I had a chance to look at your paper as I walked in to work today.  So, I absorbed about 45 minutes worth of it.  I don't think I have anything substantial to say at the moment, other than I think I like your approach based on LOCC and operationalism.  It seems to be moving in the right direction.

I am a little curious to know how this train of thought I've been trying to hammer out in quantum foundations fits into one or another holism/nonholism distinction.  Tentatively I call my (forming) view ``The Sexual Interpretation of Quantum Mechanics''.  Its beginnings are in \quantph{0205039} ``Quantum Mechanics as Quantum Information (and only a little more)'' and \quantph{0204146} ``The Anti-{\Vaxjo} Interpretation of QM'' (Sections 4 and 5), but it only really takes on a kind of force in my document ``Quantum States:\ What the Hell Are They?'' posted on my webpage.  For only there do I come to realize that I fairly explicitly drop what you call the eigenvalue-eigenstate link in these newest trains of thought.

Is it holism?  Or, is it not holism?  On the one hand I wouldn't think my view a kind of holism, as the only {\it property\/} I find myself willing to suppose of a quantum system is its Hilbert-space dimension.  But then on the other hand, I find myself wondering how to interpret the following statements that I find myself saying:

\bq
\noindent page 48 of ``QS:\ W.H.A.T.?'': \medskip

Excellent!  This is only a joke partially, but lately I've been so
taken with the idea that unions can give rise to things greater than
those contained in the parts---thinking of quantum measurement, in
particular, from this angle---I've thought about calling my view on
QM ``the sexual interpretation of quantum mechanics.''
\eq

\bq
\noindent page 49 of ``QS:\ W.H.A.T.?'':\medskip

Lately, I've been jokingly calling my view (as it stands) the
``sexual interpretation of quantum mechanics.''  (Most people turn
red and become uncomfortable when I do that and explain why.  I
suspect the same will be true even in your reading of this note. So,
brace yourselves.)  The essential idea is that something new really
does come into the world when two of its pieces are united. We
capture the idea that something new really arises by saying that
physical law cannot go there---that the individual outcome of a
quantum measurement is random and lawless.  The very fact that the
consequence of the union is random signifies that there is more to
the sum than is contained in the parts.  But I promise you I won't
reflect the licentious details of this view in the glossary below.
I'll leave the missing terms to your imagination.
\eq

\bq
\noindent page 190 of ``QS:\ W.H.A.T.?'':\medskip

I suppose if I were to start to label things, then this thing I was
telling you about the other day---``the sexual interpretation of
quantum mechanics (SIQM)''---would be a kind of dualistic theory.  I
said it metaphorically this way:  When things bang together,
something is created that is greater than the sum of the parts.  Or
again:  When things---that's the materialistic aspect---bang
together, something is created---that's the mentalistic aspect, for
it is like an act of the will or a decision.  But that's just a
thought that's hitting me at 4:00 in the morning.  (So trust it less
than even the usual things that come out of my mouth.)  I hadn't
thought about it in this way before, and I'm not sure I want to
continue to thinking about it this way.  In general, I don't like
dualisms.  (Though even saying that is paradoxical; for I think I
like ``pluralisms'' in the sense of James.)
\eq

If you need more of the surrounding conversation to make sense of these statements, please go to my webpage and look at ``QS:\ W.H.A.T.?'' directly.

Also here are some pieces from my more recent correspondences:  [See 19-07-03 note ``\myref{Mermin99}{Definitions from Britannica}'' to N. D. {\Mermin} and 18-08-03 note ``\myref{Barnum10}{The Big IF}'' to A. Sudbery \& H. Barnum.] (In fact, I'll just attach a pre-packaged thing related to the Sudbery/Barnum correspondence.  Maybe you'll find it of interest in its own right.)

Anyway, all these statements certainly feel like a kind of holism as you define it early in your paper, except for possibly the ``creative'' aspect of a quantum measurement and ``subjective'' aspect of the quantum state that I keep emphasizing.

Is it holism, nonholism, or nihilism?  The last part of that is a joke, but you have piqued my interest in the question.

So, this note was hardly a comment on your paper, but I hope you see that you've gotten me thinking.

\section{03-12-03 \ \ {\it Recovery} \ \ (to M. P\'erez-Su\'arez)} \label{PerezSuarez5}

\bmps
Well, the point is that I thought there was a one-to-one
correspondence between quantum tests (by which I mean, basically, a
certain number of possible outcomes, with a probability assignment for
each one, as well as an updating for any quantum state assignment
after having gathered the actual data) and POVMs (by which I mean a
set of nonnegative operators comprising a resolution of the identity).
But there is not. Due to the dependence of the quantum state updating
on the particular Kraus representation for the effects (up to a
unitary relation) in the POVM, it turns out that one and the same POVM
corresponds to different tests, all of which, nonetheless, share the
number of outcomes (of course) and the probability distribution for
them. Thus, the different tests corresponding to the same POVM differ
only in the updated quantum state assignment. The reasoning behind
this is so simple that I don't think I have made any mistake, but I
felt surprised at this, though I guess it is a well-known result I had
simply not read about.  I'd like to know what you think this is
telling us (if anything) about the relation between those tests and,
maybe, by the way, about the nature of a quantum state assignment.
\emps

I've already tried to tell you what I think about it.  It is in my explanation of the extra ``mental readjustment'' in Section 6 of \quantph{0205039}.  Namely, one's (subjective)
estimate of one's influence on a system in the process of measurement is something over and above simple Bayesian updating.  I think there is still a whole lot more that needs to be said to flesh this out properly, but maybe that should wait until you arrive.

\section{03-12-03 \ \ {\it {\Vaxjo} Contributions} \ \ (to H. Barnum)} \label{Barnum13}

By the way, here are the titles that have been promised on the {\Vaxjo} proceedings.  Also I'm going to try to write something small.  I don't recall ever seeing a wink from you on whether you'll submit your big paper.  Are you gonna do it?
\begin{itemize}
\item[]
Scott Aaronson::  short, recreational paper about what goes wrong if you replace $|\psi|^2$ by $|\psi|^3$ or change QM in various other ways.

\item[]
Guido Bacciagaluppi::  ``Classical extensions, classical representations and Bayesian updating in quantum mechanics''

\item[]
Paul Busch::  ``Less (Precision) is More (Information): Quantum Information in Classical Embeddings''

\item[]
Bob Coecke and Keye Martin::  ``Partiality in Physics''

\item[]
Piero Mana::  ``Why can states and measurement outcomes be represented as vectors?''

\item[]
{\Ruediger} {\Schack}::  Titleless at the moment.

\item[]
Alexander Wilce::  ``Probability: classical, quantum, and otherwise'', or ``Symmetry and compactness in quantum logic''
\end{itemize}

\section{03-12-03 \ \ {\it A Question of Condensation and Time, 2} \ \ (to J. I. Rosado)} \label{Rosado2}

Because of the new paper that {\Ruediger} {\Schack} and I are constructing, tentatively titled ``On Quantum Certainty'', I was thinking of you the other day.  At first it was quite hard to find our correspondence in my files because I could not remember your name, but after some work I found it. [See 21-04-03 note ``\myref{Rosado1}{A Question of Condensation and Time}'' to J. I. Rosado.]

I don't think you made a further reply on the note below, but I just wanted to make sure.  In any case, if you have ever written anything on the subject of your note, {\Schack} and I should cite you.  Please let me know.

Where are you located?  Are you a student, postdoc, professor, etc.?  Tell me a little about yourself, so that I can better remember your name in the future.

\section{04-12-03 \ \ {\it Incompleteness} \ \ (to R. Plaga)} \label{Plaga1}

\brplaga
I read with great interest you paper ``Quantum Mechanics as Quantum
Information'' (\quantph{0205039}). I really like your idea that QM is
in a similar state as special relativity before 1905.

One question about your discussion of Einstein's criticism on p.\ 11.
You write ``the world has seen much in the mean time'' and quote the
theorems Bell and Gleason. Why do these theorems refute Einstein's
unwillingness to accept incompleteness?
\erplaga

Thanks for your interest in my paper!  I do not think that Bell and Kochen--Specker {\it refute\/} Einstein's goal of finding a more complete theory.  In fact, I do not think it can be refuted.  Rather, I would say that the theorems I refer to better delimit the costs of such a venture.  And by my own scale, the costs are now high enough that I find it worthwhile pursuing the opposite extreme:  I.e., to search for a sound, palatable, and instructive reason for why such an Einsteinian-type completion need not be sought.  Can we imagine a model of the world in which such a completion is blocked by the world's very properties?  That's the sort of thing I am after.

\section{04-12-03 \ \ {\it Degrees of Belief} \ \ (to C. Snyder)} \label{Snyder4}

Wow, you lot from the Jane Bond (this is my new name for the residents of Waterloo that I did not meet explicitly at the Perimeter Institute) are some pretty serious scholars.

\bcs
So \ldots\ Why this word BELIEF? (Please don't ask me to read de Finetti,
Bernardo or Smith \ldots\ I'm more interested in your own words.)
\ecs

I don't think any of the Webster's definitions capture the particular meaning I had in mind.  What I meant was something particularly abstruse.  Let me give you an example, that'll maybe clarify my usage a little.

Will it rain here in Dublin tomorrow?  I don't know.  At best I might have some kind of internal degree of belief $p$ ($p$ is a number between 0 and 1) for whether it will rain.  What does that mean?  Its operational meaning is this:  If anyone were to offer me a wager of $S$ dollars, then I should have no qualms about laying down ($p$ times $S$) dollars on the condition that IF it rains, he will return $S$ dollars to me.  Otherwise, he will keep my ($p$ times $S$) dollars for his own profit.

In this sense my degree of belief $p$ is experimentally testable:  It is manifested in how many dollars I would put down on some uncertain event.  But there is no sense in which my degree of belief can be right or wrong.  It will either rain or not in Dublin tomorrow, and it will do that independently of my beliefs (i.e., which bets I will accept).  To mimic you words a last time:  there are no wrong degrees of belief, only personal ones.

When I say that a quantum state for a physical system is a belief, what I really mean is that it is a SET of degrees of belief for the outcomes of all the different kinds of experiments I might imagine performing on that system.

Hope that helps.

\bcs
I would expect most quantum guys (this is my new name for your lot) to
not be terribly keen on the word BELIEF (even if you do define it in
the context of quantum discussion.)
\ecs

Yep, but my gut and my logic both tell me that I'm pushing the community in the right direction.  Some of them are only coming kicking and screaming at the moment, and some of them are fighting with all their lives, but that's slowly changing.

\subsection{Christian's Preply, ``To BELIEVE or to KNOW''}

\bq
Sorry about the delay in responding \ldots\ it has been quite a busy spell here for me. Listen \ldots\ I was reading your last email, and reading the things you asked me to read in your last email about subjectivity, but I had trouble getting beyond your ``Note on Terminology.'' I found your term BELIEF a very interesting one. You, of course, define it in the context of quantum discussions. And in that context define BELIEF as a quantum state. Good enough \ldots\ however Mermin would rather talk about knowledge, not belief.

Granted the game being played is one of definition, not entirely unlike the the white on black illusion you refer to on the previous page (you're both talking about the same thing, just defining/seeing it differently.) So to my question \ldots\

I would expect most quantum guys (this is my new name for your lot) to not be terribly keen on the word BELIEF (even if you do define it in the context of quantum discussion).

The original form (definition) of the word (outside of your context) is an assent to a proposition or affirmation, or the acceptance of a fact, opinion, or assertion as real or true, without immediate personal knowledge; reliance upon word or testimony; partial or full assurance without positive knowledge or absolute certainty; persuasion; conviction; confidence; as, belief of a witness; the belief of our senses. (That would be Webster's not Oxford \ldots\ but whatever.)

So is it the last line you like? (The belief of our senses and therein the
subjective.) But there is almost a contradiction in the word isn't there \ldots\ the assertion as real or true, without immediate personal knowledge, and yet can also be the belief of our senses. It is interesting that Webster's also lists opinion as a synonym.

It is also interesting that there is also a definition for ultimate belief (last in a train of progression or consequences; tended toward by all that precedes; arrived at, as the last result; final.) I wonder if there is a definition for ultimate knowledge \ldots\ if there is, it would no doubt be identical.

So \ldots\ Why this word BELIEF? (Please don't ask me to read de Finetti, Bernardo or Smith \ldots\ I'm more interested in your own words.)
\eq

\section{10-12-03 \ \ {\it Girelli's Workshop} \ \ (to W. G. Unruh)} \label{Unruh3}

I've been meaning to write you for some time, but I kept forgetting to.  Seeing the line-up at the Seven Pines thing in my email today finally reminded me.  Anyway, in particular, I had said to Florian Girelli that I would write you to encourage your participation in his workshop Feb 23--29 on quantum information and gravity at the Perimeter Institute.  I hope you will come.  The time does seem ripe to put these two things together (and knowing that you'll be there will help me justify in my mind my own participation!).  I'm hoping he'll also be able to get Raphael Bousso to come; it'd be interesting what comes out of the mix of you two.  In any case, the bigger, the better the crowd, the more interesting it'll be all around.

\section{10-12-03 \ \ {\it Thanks} \ \ (to J. B. M. Uffink)} \label{Uffink3}

By the way, I've been wanting to thank you for inviting me to your meeting.  I learned some things, and it certainly helped me discover that the Netherlands is every bit as nice as my wife had been telling me!

I conveyed a report of your unique view on interpretations of probability to {\Ruediger} {\Schack}.  (Unique, because you could see the merit in {\it both\/} strict-von-Mises frequentism and radical subjectivism.)  I hope I got the details right!  Anyway, I think it would be good for us to talk this through at length one of these days.  If {\Ruediger} and I can just pull off our ``Being Bayesian in a Quantum World'' conference (to which you would be invited), that'd be the perfect place.

\section{10-12-03 \ \ {\it Slogans} \ \ (to A. Duwell)} \label{Duwell2}

I've been meaning to write you for a couple of weeks to tell you a little story.  Sorry I didn't get a chance before now.  Anyway, it is about Jos Uffink's and Dennis Dieks' conference in Utrecht a couple of weeks ago.  Since they wanted me to talk about the meaning of quantum information, rather than quantum states per se, I knew that there was nothing more accurate I could title my talk than with your slogan, ``Quantum Information Does Not Exist.''  I hope you don't mind!  In particular, I opened my talk with some lines that went something like this:
\bq\noindent
   Unfortunately, I have to start my talk with a horrible admission:
   You have a thief among you!  For I have stolen the title for my
   talk from a nice paper of Ari Duwell that I would encourage all of
   you to read.  [Then I put up a slide with the title, your name,
   and the paper's publication details.]  At the end of my talk, I
   hope you will see a connection between what I am about to say and
   this slogan. However, for most of my time I will be talking about
   a formalism that might help you see the interpretational problems
   of quantum mechanics in a new light.  Though I stole the title, I
   hope you won't think I stole the contents of my talk!
\eq
It got a good laugh all around, and I hope it did encourage some in the audience to pay attention to your paper.

Which reminds me of something else I've been wanting to say to you for some time:  You really ought to post your paper on the {\tt quant-ph} archive.  A lot of people in the quantum information community are going astray.  If you don't post your paper there, a lot of people in that community are never going to see it \ldots\ and never get a chance to get their thoughts straightened out from it.

\section{10-12-03 \ \ {\it Sneaky Ways to See You} \ \ (to C. H. {\Bennett})} \label{Bennett32}

I'm writing you to encourage you to go to Jeff Bub's meeting April 30 through May 2 in Maryland.  See this link for details:
\myurl{http://carnap.umd.edu/philphysics/newdirections04.html}.
If you're going to go, could you drop him a line to let him know that you'll be there?  (And if you're not going to go, could you let him know that too?)

But I do hope you'll go.  The argument that you didn't actually settle the Maxwell demon problem seems to be gaining some momentum in that crowd, and I think if you'd just face them one-on-one for once, it would do a world of good.  In particular at the Utrecht meeting last week, a fellow named Owen Maroney---who will most surely be at the upcoming meeting too---gave what appeared to be a decently reasonable talk trying to pinpoint where you had gone wrong.  (I'll put his abstract below.)  If you had just been in the audience to ask the right question at the right time---I hate to say it again---it would have made a world of difference.

Given that you never wrote your planned review article for RMP, this is probably the best thing you could do as a substitute for the time being.  And besides, you and David Mermin and I could all have some fun.  Certainly David and I will have a lot more fun if you're around than not.

Please write me to let know if you'll be coming.  But more importantly, do please write Jeff with your decision.

\bq\noindent
{\bf Owen Maroney:
``The (absence of a) relationship between logical and thermodynamic reversibility''\medskip}

Landauer erasure seems to provide a powerful link between thermodynamics and information processing (logical computation).  The only logical operations which require a dissipation of energy are logically irreversible ones, with the minimum dissipation being $kT\ln2$ per bit of information lost.  Nevertheless, it can be shown that logical reversibility neither implies nor is implied by thermodynamic reversibility.  By examining thermodynamically reversible operations which are logically irreversible, it is argued that information and entropy, while having the same mathematical form, have significant conceptual differences.
\eq

\section{10-12-03 \ \ {\it City of Light} \ \ (to D. B. L. Baker)} \label{Baker5}

Do you know it's been over a year since I've written you a proper
letter?  (You ought to know by now that a letter from me is not
proper unless it contains at least 5K of raw text.)

I'm sure you've guessed from the title of this note that I'm in
Paris.  What a charming city \ldots\ even if it is full of surly
waiters! I haven't been here since 1994, and then only for one day in
a not-particularly-interesting part of the city.  So really my memory
has to reach back to 1991 for anything that remotely compares to this
weekend.  However even then, I would say I've grown so much and
learned to appreciate Europe that it is altogether quite something
new.  Frankly I'm carried away in a romantic/bohemian mood.  It
probably also helps that I earn a lot more money now, and so don't
really mind dropping 70 euro on a meal, as I did last night.  It
opens up exponentially more of the city, and even takes some of the
edge off the waiters.  Last night, a colleague and I dined next to a
Picasso.  It didn't make the food or the ambiance any better, but
maybe {\it they\/} gave Picasso the extra something he needed to
finally (after all these years) draw my attention more than
fleetingly.

The brick streets, the cheese shops, the wine shops, the caf\'es, the
life of the bohemian.  I am having a great time.  And I'm reawakening
slightly.

\ldots\ The long weekend has come and gone, and I'm now at the
airport, waiting on a delayed flight.  I thought I'd have a chance to
write you much more from the Parisian caf\'es, but a proper flux of
writing juices never materialized.  Looking back, what can I tell
you?  What do I feel like telling you?  Maybe the thing that will
interest you most had to do with an Italian student, Giovanni
Valente, who just finished his master's degree in philosophy in
Padua, writing a thesis titled, ``Probability and quantum meaning:
Chris Fuchs' pragmatism in quantum foundations.''  He'd been wanting
to meet me for a while, so a month ago I sent him my travel schedule,
saying that he could drop by Dublin one of the times in between.  Out
of the blue, I got this from him Friday:
\bgv
What about meeting in Paris? You wrote me that you have planned to go
there (6-9 December) and I could be there. I've been taking a little
rest after my viva and the GRE and my parents own a little flat in
front of the Moulin Rouge, so it may be a good occasion to meet.
Since I haven't been there for a while, I feel I could spend a few
days in my favorite place. Anyway, if you think we may meet and drink
something, let me know quite soon: I need to book a flight very
quickly.
\egv

Well, he wasn't lying!  He showed up in Paris the next day, and he
actually did have a flat just across the street from the Moulin
Rouge!  He made me a nice spaghetti dinner and we had a couple of
bottles of wine there one evening.  The flat was in an old bordello.
I certainly never imagined on that infamous New Year's Eve at your
house so many years ago that I'd one day get an opportunity like
that.  Apparently the boy's father was a socialist politician in
Italy for some years.  Who said socialism was the equalizer of
mankind?\footnote{In the editorial stage of this samizdat, Giovanni pointed out that I ``still don't fully get the difference between socialism and communism.''}

Monday I lectured at \'Ecole Polytechnique.  It went well.  But, the
really nice thing was that they put me up in a hotel nearby.  So I
was in the middle of the Latin Quarter for the stay.  That put a few
thousand restaurants within my reach, even if only to walk by and
look at (quite an experience in its own right).  And I was just a
mild walk away from the Pantheon, Notre Dame, the Louvre, Mus\'ee
d'Orsay, Luxembourg Gardens, the Sorbonne, etc.  I walked into the
courtyard of the Louvre today around 4:00 PM, just after all the
bustle had died down and just as the sun was moving into twilight.
(The sun sets early this far north.)  One could hear just a tinge of
traffic noise coming over the top of the building.  What a profound
aloneness:  I savored every second of it.  I just kept saying over
and over to myself ``Le Grand Palais du Louvre.''  It is quite a
monument to the power and creative force of mankind!  Big buildings
always do me in.

\section{10-12-03 \ \ {\it First Meeting} \ \ (to M. Bitbol)} \label{Bitbol1}

It was great meeting you the other day.  As I told you then, I have
greatly enjoyed finally reading some of your papers.

Let me paste below the passage from my samizdat {\sl Quantum States:
  W.H.A.T.?\/} that refers to you.  The text between the {\tt bjb} and
{\tt ejb} symbols are quotes of Jeff Bub.  Now that I understand that
Jeff's ``neo-Kantian'' might just mean ``somewhat pragmatic,'' I am
much more intrigued than I was before.  Have you written anything in
English on the ``blinding closeness'' that Jeff mentions?  [See
  10-04-02 note ``\myref{Bub7}{Bitbol-ization}'' to J. Bub.]

If you get a chance to have a look at it, I would greatly appreciate
hearing your thoughts on sections 5 and 6 of my paper ``The
Anti-{\Vaxjo} Interpretation of Quantum Mechanics''.  The paper can
be found at my web page; there is a link to it below.

\subsection{Michel's Reply}

\bq
Better late than never (I hope this faithfully translates the French
``Mieux vaut tard que jamais'').

I thought I had to answer your mail one day, but many other tasks
piled up. However, I finally found a free sunday morning and I read
some of your stuff with pleasure. The ``Anti-{\Vaxjo}
interpretation'' sounded very relevant since Andrei Khrennikov is
coming to visit us in CREA tomorrow!

I must say I was delighted to discover that, in this paper written by a physicist (you!), there is more excellent philosophy of quantum mechanics than I had ever seen before.  As you already know, I am disposed to agree on most of what you say.
I was drawn independently to very similar conclusions by combining a
radical version of Copenhagen interpretation (plus some insights from
{\Schroedinger}) and a network of philosophical readings ranging from
neo-kantianism to pragmatism. Here is, to begin with, a short list of
points of agreement (with some nuances):

1) Quantum mechanics is a theory of contextual probabilities:
Kolmogorovian probabilities in each context, several such probability
spaces being ``glued'' in a unified Hilbert space structure. To
summarize this idea, I called QM a ``Meta-contextual theory of
probabilities''.

2) The main terms of the theory refer to certain gambling attitudes.
With some stringent constraints however: coherence and
intersubjectivity (or rather inter-situationality, invariance of the
formal tools with respect to changes in situations/contexts), etc.
When this constraint of inter-situationality or intersubjectivity is
emphasized, one lands into a central concept of Kant's transcendental
philosophy: constitution of objectivity.

3) I also agree that physical theories express the conditions for our
successful life within the world rather than a faithful picture of
the world. Your insistence on Darwinian evolution, on viability
rather than mirroring, is very close to the ideas I developed after a
long contact with Varela's autopoietic theory of cognition. Here is a
short statement of this theory I recently wrote:
\bq
In the autopoietic theory of cognition, the relevant concept is not
inputs provided by the external world, but only local environmental
conditions for maintaining an operationally closed unit. The
invariants of this type of unit do not represent any feature of the
world, but rather identify with steady aspects of its own internal
dynamical organisation. As for the appropriate changes of an
operationally closed unit, it does not prove that the unit possesses
a faithful picture of the world, but only that its internal working
is viable in relation to environmental disturbances. One must
redefine the ``cognitive domain'' of the operationally closed unit
accordingly. This domain  is no longer some fraction of a
pre-existing world liable to representation, but a region of the
environment which has co-evolved with the closed unit and in which
the latter's organization may persist, develop, and reproduce despite
the disturbances.
\eq

As you rightly pointed out, the structure of an elephant (a special
case of autopoietic unit) does not represent its environment, but is
the end product of a long history of environmental alterations and
adaptative changes: it is a summary of the adaptative moves the
species had to do. One interesting (and very relevant) features of
this example is that the adaptation to the environment is not purely
passive. The behavior of elephants, their search for better-suited
environments (by migration) or even their own transforming activity,
is also part of the adaptative process. The adaptation then becomes
somehow mutual. The organism and its environment can be said to
``co-emerge''. In the same way, our theories are likely to be an
elaborated and stabilized byproduct of a cognitive history (made of
``conjectures and refutations'', gambling rules and failures) which
massively involves our experimental activity (our ``interventions''
by means of apparatuses).

The latter general overview was partly expressed in my ``blinding
closeness of reality'', which unfortunately exists only as a book in
French. But the basic idea is easy to summarize: Bernard d'Espagnat
(my master in this field) told that reality is quite difficult to
picture (actually, in the micro-domain, even impossible to describe)
because it is somehow remotely distant from us, because it is
separated by a ``veil'' from us. This was only a metaphor as I soon
understood by discussing with B.~d'Espagnat, but the happy effect of
this was to trigger a diametrically opposite metaphor. My idea was
that the reason why reality is impossible to describe as such is that
we are so deeply and intricately immersed in it, that we do not have
the opportunity of creating the objectifying distance. My motto,
inspired from {\Wittgenstein}, can be translated thus: ``The subject is
not facing the world; it is so much committed in it that this does
not allow description.'' But this does allow orientation in it, or
anticipation of part of what may happen for us in it.

I thus warmly approve your ``oceanic'' picture!

4) Reality manifests itself by its ``unpredictable kicks''. This is
perfectly true. But, as you know, there exists another, very popular,
conception of how reality occurs to us. According to many people, we
reach some real structure when we have extracted an invariant from
phenomena. Since it is invariant, they say, it does not depend on
particular points of view, or on particular subjects, and
``therefore'' (I am very reluctant when I read this ``therefore''),
it faithfully describes some independent feature of the world ``in
itself''.

Scientists usually combine (in various proportions) these two
conceptions of reality in the very dialectic of elaboration of their
theories. But they are basically wrong. And, I believe, you are
completely right: the only manifestations of something like reality
beyond the narrow boundaries of our little persons are unpredictable
kicks. Why is it so? Because there is a logical fallacy in the usual
inference from invariance to ``independent'' reality. Any feature of
a pre-structured ``independent'' reality would take for us the form
of an invariant; but invariants are not bound to express features of
some ``independent'' reality. We are not even sure that
``independent'' reality, if this sequence of words is meaningful at
all, has already a structure in store, ready to be disclosed (Rom
Harre calls the micro-world ``the glub'' to express the idea that it
is likely to be a plastic and dispositional stuff rather than a
pre-structured network of actual properties). So, if the cognitive
invariants do not disclose the structure of ``independent'' reality,
what do they do? They disclose stable, viable, and intersubjective
forms of our anticipations (our gamblings).

Instead of opposing two conceptions of the manifestation of reality
(as unpredictable variations and as structural invariants), we should
only keep one of them: yours, namely unpredictable variations. The
other conception only deals with what Kant calls ``objectivity'' in
his consistently anti-metaphysical acceptation. Reaching objectivity
in Kant's sense is by no means identical to mirroring reality. This
is what Kant meant when he carefully distinguished objectivity from
faithfulness to the ``thing in itself''.

I applied this remark to the status of QM in several papers
(including ``Some steps towards a transcendental QM'', published by
Philosophia Naturalis in 1998 and available on my website). In a
paper in French about ``Laws of Nature'', I concluded thus:
\bq
The boundary between what is transcendentally necessary and what is
irreducibly contingent has moved beyond recognition. All the
structural features of QM, including its law of evolution, are
transcendentally necessary in so far as they express the conditions
of possibility of any coherent (probabilistic) anticipation of the
contextual phenomena obtained by means of a systematic experimental
investigation. And the only element which remains definitely
non-contingent is non-structural; it is the very occurrence of
isolated experimental outcomes. Hence the aphorism: in QM, nothing
structural is contingent, and nothing contingent is structural.
\eq

5) I also agree with you that the many-worlds picture of reality is
amazingly daring and much more anthropomorphic than its supporters
think. However, there are other possible meta-interpretations of
Everett's interpretation than the many-worlds, and some of them
proved very close to your (to our!)\ view. An example of them is the
indexical view supported by Simon Saunders (and by me in
``L'aveuglante proximit\'e du r\'eel''). Here, the universal state vector
is not bound to describe the world as a whole, but rather to display
the set of possible particular situations we may occupy within it,
obtaining this or that unpredictable isolated outcome as a result of
an experiment.

6) At this stage, I have to state my major point of disagreement with
you: I suspect the remark you made about the ``many-worlds
interpretation'' applies quite well to some of what you say. Let me
explain this.

You write: ``What I find egocentric about the Everett point of view
is the way it purports to be a means for us little finite beings to
get outside the universe and imagine what it is doing as a whole.'' I
deeply agree with you.

But, then, you begin to do essentially the same as the naive
Everettian. You try to describe our situation in the world from a
sort of vantage point. Here is the sentence in your paper that made
me suspicious:  ``I think the solution is in nothing other than
holding firmly---absolutely firmly---to the belief that we, the
scientific agents, are physical systems in essence and composition no
different than much of the rest of the world. But if we do hold
firmly to that---in a way that I do not see the Everettist as holding
to it---we have to recognize that what we're doing in the game of
science is swimming in the thick middle of things.''

Of course, once again, I am delighted by the Pascalian metaphor of
``swimming in the thick middle of things''. But you seem to take it
as more than a metaphor: a definite belief about (not to say a
faithful description of) the world and our position in it. You write
seriously that ``we, the scientific agents, are physical systems in
essence and composition no different than much of the rest of the
world''. Isn't this a way of extrapolating one of our pragmatic -
adaptative concepts, one of the concepts we need to swim with some
success in the midst of the ``glub'' (here, the concept of a
``physical system''), in order to describe everything including us as
if it were seen from outside? I hear you saying something like ``the
world as I see it from my cosmic exile is made of physical systems
and each one of us is one of these physical systems''. But if we are
``swimming'' in the deep ocean of whatever we call ``reality'', we
have absolutely no context-independent concept at our disposal, not
even the very general meta-concept of ``physical system''. We must
say that we ignore everything of the ``thing in itself'', including
whether it is organized or not in a plurality of ``physical systems''.
And we must therefore content ourselves with stating the formal
conditions of our cognitive aptitudes (within it). This latter
attitude is typical of the Kantian and neo-Kantian lineage of
philosophy (when one gets rid of the foundational aspect of Kant
himself and hold a pragmatic variety of Kantianism, as I do). I hope
you'll recognize it as a radical variety of your view\ldots\ I wonder
whether you'll become a member of our radical club or rather decide
to stick to your position as it stands.
\eq

\section{10-12-03 \ \ {\it Thoughts in France} \ \ (to J. Bub)} \label{Bub11}

I'm just back from a few days in Paris.  I had a great thoughtful time reading your paper ``Why the Quantum?''\ during a long lunch and during my morning and afternoon coffees in the Latin Quarter yesterday.  I read it as I threaded in and out of the Louvre and Notre Dame.  There is much I like about the paper, but there is also much I disagree with (especially in the last third).  I owe it to you to get those thoughts down solidly.  At the moment, I'm shooting to get them to you before the end of the weekend.  Just a heads-up so you'll have a chance to stock up the liniment oil!

\section{10-12-03 \ \ {\it Down with Frequentism} \ \ (to K. R. Duffy)} \label{Duffy2}

Sorry I forgot to contact {\Schack} for a title to his talk Monday.  (Are you out there {\Ruediger}?)  Let's title it this tentatively:
\bq
Defining (Carefully) a Quantum-Mechanical Relative Frequency Operator.
\eq
{\Schack}'s affiliation is below.

\section{11-12-03 \ \ {\it My Communication Skills}\ \ \ (to R. Balian)} \label{Balian2}

It was very good meeting you in Utrecht two weeks ago.  I was dismayed a little bit by your reaction to my talk, however.  Without a doubt, I must have presented something in a confusing manner---for in the end all the things I said were rather trivial and elementary.  I.e., they were not the sorts of things that could even be controversial, much less between two Bayesians.

In any case, I hope that the paper associated with my talk ``Quantum Mechanics as Quantum Information (and only a little more)'' will do a better job than I did in person.  For your reference, you can find it here: \quantph{0205039}.  If you find anything in it that is still unclear after your reading, I would much appreciate your feedback.

\section{11-12-03 \ \ {\it Thomas, Pauli, Luba\'naski, and Enz} \ \ (to A. Peres)} \label{Peres58}

\bap
You are an inexhaustible source for references: where is the
Pauli--Luba\'naski vector? I asked Charles Enz who wrote a scientific
biography of Pauli, but got no answer (Enz, who is not young, was a
student of Pauli). Thanks.
\eap
\begin{itemize}
\item
L. H. Thomas, ``The Kinematics of an Electron with an Axis,'' Phil.\ Mag.\ {\bf 3}, 1--22 (1927).
\item
J. K. Luba\'naski, ``Sur la th\'eorie des particules \'el\'ementaires de spin quelconque. I.,'' Physica {\bf 9}, 310 (1942).
\end{itemize}
I'm not sure how Pauli came into the game though.  If you find out, please let me know.

About Enz, I had some contact with him soon after the fires in Los Alamos:  He replenished my supply of historical papers on Pauli that he had written.  His last note came to me 6/20/2000.  Since then I have written him on two occasions, the last being 6/10/2003, and have not gotten a reply.  This year or so, his book {\sl No Time to be Brief: A Scientific Biography of Wolfgang Pauli\/} appeared, but I found no reference to Luba\'naski in it.  I understand the book to be good, but I have not read it myself.

\section{14-12-03 \ \ {\it Digging Out} \ \ (to G. L. Comer)} \label{Comer46}

Sometimes, I think I'm never going to get out of this hole I've dug for myself email-wise.  It keeps flooding in the gates, and I just keep getting further and further behind.  I'm sorry to take so long to write you.  Earlier this week, I was in Paris, and \ldots\ {\it of course\/} \ldots\ all my email sensibilities completely shut down.  A wonderful city Paris is!  My hotel was in the middle of the Latin Quarter, and I was in heaven briefly.  At least I was able to walk and think and think and walk.  And one of these days I'll have to write the residue down.

\section{15-12-03 \ \ {\it Alchemy} \ \ (to G. Valente)} \label{Valente5}

I have been meaning to write you since my departure from Paris.  I hope you will accept my apologies for taking so long.  Mostly, I just want to say that I had a fabulous time discussing philosophy/politics with you.  Thanks also for the nice pasta meal.

Let me share one of my more secret projects with you as a little recompense.  I'll attach a document I'm compiling called ``The Activating Observer: Resource Material for a Paulian/Wheelerish Conception of Nature''.  It's still very far from complete, but I think you'll enjoy it even at this stage.

At times, I have dreamt fleetingly that the ancient alchemists may not have been so very far off course.   You have rekindled that dream in me again.

\section{16-12-03 \ \ {\it BIPM}\ \ \ (to R. Balian)} \label{Balian3}

Thanks so much for your notes.  I am pleased and excited to hear that you have printed out some of our papers.  For myself, I think quantum mechanics is at a crucial time.  The issue as I see it is that if we all can just muster the stamina and courage to be absolutely consistent in following through with a Bayesian interpretation of quantum probabilities some of the greatest physics may be just around the corner.

\subsection{Roger's Preply}

\bq
When printing your paper on qm and information, I noticed that you locate the ``Bureau International des Poids et Mesures'' in Paris. It is actually in a suburb, close to the one where I live, in a nice park on a hill just above the famous china manufacture of S{\`e}vres. Its address, that is mythical since we learn it at school when we are taught the metric system,
is: ``Pavillon de Breteuil, S{\`e}vres''.

The kilogram of the SI system of units is indeed in deep caves below this small end of XVIIIth century castle, but the meter, which is also there, has lost its official status \ldots
\eq

\section{16-12-03 \ \ {\it Reading Assignment} \ \ (to M. P\'erez-Su\'arez)} \label{PerezSuarez6}

I've been thinking a little bit on what we might work on.  There's a cluster of problems I've been thinking about to do with symmetric informationally-complete sets of states.  We'll probably pick up there after you get settled in here.

But let me give you a reading assignment as you suggested.

Take a look at:
\begin{itemize}
\item
Symmetric Informationally Complete Quantum Measurements\\
\quantph{0310075}
\item
Quantum Key Distribution Using the Trine Ensemble\\
\quantph{0311106}
\end{itemize}
And the draft of a paper I am writing, that I will attach.  I won't have any time to correspond about the ideas in these, but that's the sort of thing you can familiarize yourself with.

\section{17-12-03 \ \ {\it Pragmatism for the Holidays} \ \ (to L. Henderson)} \label{Henderson2}

\bleh
And by the way, have you sent me a copy of your book yet? I would
like one.
\eleh
Yes ma'am.  \ldots\ Well, actually I've sent your address to VUP.  But I don't know if the Swedish holidays will get in the way of a prompt delivery.

\bleh
I have also been eyeing my roommate's books on pragmatism with a
certain curiosity since our train conversation---it's not something I
know much about. Maybe in the holidays I will get a chance to have a
look. What do you suggest of William James?  We have {\bf Pragmatism} and {\bf The
Meaning of Truth\/} on the shelf.
\eleh

Definitely {\sl Pragmatism}.  It's the book that first hooked me.  I think {\sl The Meaning of Truth\/} causes more trouble than it mends.  When or if you're ready for that stage, I'll try to dig up some selections of Putnam, Dewey, or Rorty for you.  I might also suggest some very pleasant leisure reading in Louis Menand's book {\sl The Metaphysical Club:\ A Story of Ideas in America}.  Here's the Amazon.com link in case you want to read about it.

\section{17-12-03 \ \ {\it State of the Union} \ \ (to R. E. Slusher)} \label{Slusher3}

The world of quantum information remains firm.  I'm trying to get three papers out before February.  One on my quantumness measure for the Holevo festschrift, one with Schack on quantum-state tomography from a Bayesian view, and one with Brassard on the fundamental nature (for specifying quantum mechanics itself) of quantum cryptographic tasks.  Also, Bennett, Brassard, Preskill, and I just turned in our section for the ARDA quantum crypto roadmap.  So that's finished for a while!

Was there any progress putting Bernie's tetrahedral measurement box together?  I've got a PhD student joining me in January from Spain, and my plan is to get him working on some theory to do with these measurements.  Also I've started up a collaboration with a statistician, Richard Gill from Utrecht, to continue the search for things such a measurement would be optimal for.  One of these days we're bound to find something it's good for!

\section{17-12-03 \ \ {\it Sinking Depression} \ \ (to C. M. {\Caves})} \label{Caves75.3}

I'm just off the phone with Howard Burton, learning about my fate with Perimeter.  The file is finally dead:  They couldn't achieve an ``overwhelming majority'' in favor of me, whatever that might mean.  The biggest concern on the table, he said, was my ``lack of breadth'' in foundations work.

It's a load of malarkey, I know, but it's about the biggest slap in the face I've had in some time.  Quantum foundations has been my life since my first week in college in 1983.  In fact, it is the only reason I sought to get a physics degree.  There is not one paper I have written in my career without quantum foundations in mind in some way.

I am severely depressed at the moment.  My stomach is churning to the point of nausea.

This morning, I had told Kiki I was going to write you about how impressed I was in reading your holiday letter this year.  Your whole family has a twinkle of greatness to it.  I wanted to say so much more, but I suspect that's the most detail I'll be able to muster for a while.

\section{18-12-03 \ \ {\it Finally Some Answers} \ \ (to A. Sudbery)} \label{Sudbery7}

\bts
I wanted to make some general points about the nature of our disagreement and also some particular (to the point of being finicky) points about {\tt 0205039}. I'll probably have to stop before I've finished, so I'll take the general points first.
\ets

\bts
I want to write a piece called ``Quantum interpretation needs no
mechanics'', but perhaps ``polemics'' would be a better last word.
\ets

I try to use every trick in the bag to motivate the community to think about the connections and contrast between quantum mechanics and general probabilistic reasoning.  Sometimes some of the tricks work---and when they do they sometimes work spectacularly---and sometimes they just grate on the reader's nerves.  If you wait long enough before giving up hope on me, I might just find a trick that'll work on you.  The reason I believe this is I see {\it truth\/} here, and the truth will ultimately prevail (no matter how bad a prior one's earlier education may have drilled into them).

\bts
I'm sure I'm not alone in seeing your dissolving of the ``violent''
state change (on p.\ 34) as a cheat. Calling it a ``mental
readjustment'' doesn't make it any less violent. Incidentally, there's
something wrong with eq.\ (97); these $U_i$ are not unitary.
\ets

Fair enough.  It is violent---every bit as violent, discontinuous, and irreversible as a usual application of Bayes' rule.  But the point is that it is on the inside of the agent's head, and no one cares what is going on in the agent's head.  What physicists should be worried about is the physics of the external world (and not the agent's head).  And who knows what really happens there \ldots\ especially if physics itself is only really about the interface.

About the $U_i$'s you are correct.  I'm sorry, I've prepared an updated version with a lot of the typos, etc., fixed, but I haven't reposted it yet.  Here's what the new edition says:
\bq
\noindent
That is to say, there is a sense in which the measurement is solely
disturbance. In particular, when the POVM is an orthogonal set of
projectors $\{\Pi_i=|i\rangle\langle i|\}$ and the state-change
mechanism is the von Neumann collapse postulate, this simply
corresponds to a readjustment according to unitary operators $U_i$
whose action in the subspace spanned by $|\psi\rangle$ is
$|i\rangle\langle\psi|\;.$
\eq

\bts
There's something I'm failing to understand on p.\ 20. You seem to be
arguing that equiprobable events must be identical. Is that what you
mean, or am I being obtuse? (Also, in para 3, why do you think the
word ``measurement'' belongs only to quantum mechanics?)
\ets

Looking back, I don't know how to say it any more clearly than I said it there:
\bq
\noindent
The consequences ($m_i$ and $n_j$) of our disparate actions
($M$ and $N$) should be labeled the same when we would bet the same
on them in all possible circumstances (i.e., regardless of our
initial knowledge of $S$).  To put this maybe a bit more baldly, the
label by which we identify a measurement outcome is a subjective
judgment just like a probability, and just like a quantum state.
\eq
You might contemplate that God can see the difference between $m_i$ and $n_j$, but we can't.  And to that extent, we should call them the same thing.

\bts
I also, last weekend, enjoyed some more skimming through your {\bf Paulian Idea}, particularly the glimpse of the smithy where you and Asher forged your {\bf Physics Today} article, but I haven't got my notes with me. I remember only your impatience with your commentators. I repeat, we all have to learn tolerance. But not of interfering APS editors; fight with David {\Mermin} under the boojum banner!
\ets

The worst example I have ever seen of this came from the {\sl Physics World\/} editor I had to deal with for the thing I wrote for the November issue.  Here is a sample of my exchange with him.  (Note what an annoying character I appear to be in the first part of this, but then read my second exchange to see why I was furious.)

\begin{itemize}
\item[ME:]
     Thank you; that was a good and well-reasoned change, and I appreciate it.  However for most of the rest I am annoyed.  Could your copy-editor not see that I was in complete control of my writing?  All my words were chosen for reasons, and it would have paid him to spend a little time trying think of what those reasons might be. There is a difference between ``mustered'' and ``muttered'', for God's sake.
     In any case, the article will be written in my style (as long as it is grammatically sound), or it won't be written at all.  Luckily I have not yet released the copyright to {\sl Physics World}.  I will spend much of my morning reinstating a lot of my old word choices---ones that reflect my writing style and personality, not his---and send you a revised draft before the day is out.

\item[HIM:]
     I am sorry you did not like the subediting, which was done entirely by me. Please rest assured that it was not done to annoy you, but --- I hope --- to make sure your article is as readable as possible. I'm sorry also for the mustered/muttered confusion, which we can of course change.

\item[ME:]
     Changing almost every sentence in the text---19 out of 23 in the first column alone---is not acceptable.
\end{itemize}
In the end, that fraction had carried through to the whole article:  He had literally changed 83\% of the sentences!  I spent almost a whole day at Caltech changing almost all of his changes back, and writing an explanation for why I preferred my wording over his.  The final product is almost exactly as I had originally written it.  Tell me, does it look so bad to you?  Did it look worth an 83\% change?

\bts
However, all the modern dictionaries I have give an overwhelming
majority of hyphens with ``non'' -- it must be a transatlantic thing.
\ets

I had a similar encounter with John Smolin on this issue.  Here is what he wrote me:
\bjas
Regarding the ``non'' prefix.  I agree with you for the most part.
At first I didn't.  I considered your extensive list of non words as
evidence that you DO need the hyphen, since if the dictionary lists
so many words, it must consider all the ones it doesn't list as
nonwords.  And far be it from me to make up a new word, so I
preferred the hyphen.  But after looking around the net I found
basically one style guide that agreed with me and many that more or
less tell you to never use the hyphen.  So I'm changing sides.  I
still find it hard to write ``nonsecret'' as I'd never accept that as
a scrabble word.  Of course, the solution there is to use ``public''
instead of making up the word ``nonsecret'' anyway.  Oh, and
``nonagenerian'' should not have been in your list.  Cut-and-pasting
from the dictionary is a nongood idea.
\ejas

\section{18-12-03 \ \ {\it Finding Gleason} \ \ (to R. {\Schack} \& C. M. {\Caves})} \label{Caves76} \label{Schack77}

Just wanted to say how much I'm starting to like your work on the frequency operator stuff.  At first---though I never said so aloud---I thought, why beat a dead horse?  Eventually the ugliness and mystery of the construction would make it disappear on its own right, ably assisted by the simplicity of the POVM-Gleason construction.  But after {\Ruediger}'s talk here, I started to reassess.  What I was particularly impressed with is how you were able to uncover a hidden application of Gleason's very starting point.  And that's nice.

This leads me to pose a challenge, whose solution I would relish that much more.  Find a hidden use of Gleason's assumptions or theorem in Wojciech's ``envariance'' argument for the quantum probability rule.

You see, what more could there be to the quantum probability rule than Gleason?  Anyone who finds it by some other supposed means has to be faking something.

\section{18-12-03 \ \ {\it General Summary} \ \ (to R. W. {\Spekkens})} \label{Spekkens26}

\brws
Howard mentioned that you got the news today about your candidacy at
PI. A disappointing outcome.  How are you feeling about the whole
affair?  What are your plans?
\erws

\brws
I am still quite bummed out about the outcome of the discussions on
your candidacy.  You should know that there were many strong advocates
for you here.
\erws

Thanks.  I feel quite dicked over by this, you know.  One night at the Jane Bond, Valentini said jokingly (not so jokingly), ``We need to bump you off before you influence any others.''  That must have been the consensus in the mind(s) of whoever it was that actually quashed me.

Well, I haven't been bumped off.  And this is too powerful of a turn in quantum foundations that it won't live by itself.  But my life has been made immensely inconvenient by this decision.  And I am so pissed off that I won't now be able to be fully engaged in the birth of this baby.  It does quite sicken me.

It's just not a good thing to be up at 4:00 AM \ldots

\section{18-12-03 \ \ {\it Invitation} \ \ (to H. J. Briegel)} \label{Briegel3}

Congratulations again on the big move.  I suspect you are quite enjoying your new life.  Or, in any case, the promise of directing great progress must be very satisfying.

I would love to visit Innsbruck and am flattered by your invitation.  However, at the moment, my travel schedule is so full, I dare not make any more commitments for the foreseeable months.  Perhaps I will take you up on your offer of a visiting position at the new institute in (late) Autumn 2004 or Spring 2005.  However, even then, I hesitate to make a firm commitment at this point.  The truth is, I am quietly seeking an academic position outside of Bell Labs.  And if that materializes somewhere, I know that I will have to learn new responsibilities with respect to my travels.

But oh, how I would enjoy a discussion with you on some deep quantum thoughts.  I saw a very good talk by Andrew Childs recently in which he showed a couple of distinct ways to systematically generate states of the cluster-state type or variety (though they appeared to be inequivalent to your own cluster states).  It was quite fascinating, and I was once again struck by the depth of the approach.  I really want all of the most fascinating stuff in quantum mechanics to reduce to the structure of measurement, and your computational model seems to be a move in that direction (modulo the initial state)---I've told you that before.  But the opportunities look great with just the right thinking.

\section{19-12-03 \ \ {\it Time to Think about Time} \ \ (to G. L. Comer)} \label{Comer47}

\bgc
What's that line of {\Wheeler}'s about the past? Is it something like
``\ldots\ the past exists insomuch as at is consistent with the present''?
I've had this crazy idea about closed timelike curves, information
theory, no-cloning theorems, and time travel. The craziness is this,
can I perform a whole suite of information-gathering exercises today,
that would in effect change the past so that it looks like I actually
travelled back in time and effected the changes directly? Can nature
be so cockeyed as to allow something like that? I mean, can I squeeze
and prod nature in such a way that I can extract a present that is
consistent with some other past? In effect, could I then ``travel''
back to the ``past'', and ``kill'' my grandparents? Surely nature is
not this screwy.
\egc

It was never quite clear what {\Wheeler} was saying on this subject,
i.e., whether we can change the past with our quantum measurements, or
rather just change what we can say about the past. I think he
flip-flopped from time to time. My own present take is that it is
neither. A quantum measurement here and now cannot actually change the
past there and then (whatever that might mean). That is just as a
quantum measurement here cannot actually change anything physical
there (i.e., a spacelike separation away). Rather with regard to time,
it is all and only about the present. I would say this is inspired by
{\Wheeler}'s take on the subject, but not quite the same.  (Though who
can say, maybe it is). The closer alliance is to the thought of the
pragmatist philosopher George {\Mead}. I'll attach a note that
hopefully makes some sense of this. [See 23-09-03 note ``\myref{Savitt3}{The Trivial Nontrivial}'' to S. Savitt.]  Reading back over it, I still like my
quantum discussion, but there's no doubt it needs fleshing out. Tell
me if my discussion or the {\Mead} passage provokes any thoughts.

\section{19-12-03 \ \ {\it Prizes, Hippies, and Vectors} \ \ (to A. Peres)} \label{Peres59}

Congratulations on the Rothschild prize.  What was the particular citation for?  Years of theoretical service?  Or was it for a particular paper or series of papers that you wrote?  Did you get any money with the prize?

Thanks for the cheer-up note.  It certainly helped!  I just couldn't stop laughing about the commune-of-hippies remark:
\bq\noindent
I must confess that I am not well impressed by the research being done
at PI. It looks more like a commune of hippies, very pleasant to be
there, but not fruitful for physics.
\eq
(I read the whole thing to Kiki, who had just woken up.)

\bap
You gave me a reference to a paper by Thomas, about the Pauli--Luba\'naski
vector (the paper is in Phil.\ Mag.\ 1927). There is in it no mention of
anything like the P-L vector. What I am looking for is a statement
that it actually is a classical vector, indep of quantum mechanics.
The reason is that Czachor has sent to PRL a Comment on a paper by
Danny, Petra and me, and I have now to prepare a rebuttal.
\eap

All I did was a google search on Pauli--Luba\'naski vector, and those were the references I found.  They appeared in a few papers, so I felt confident that they must be right.  But maybe one cited the incorrect reference, and the effect just snowballed as one author simply went to another's paper rather than the original reference.

Anyway here is the passage I find in Enz's biography of Pauli that refers to Thomas:
\bq\noindent
    Frenkel did the only sensible thing: that is, use the analogy with
    the electromagnetic field to enlarge the magnetic moment vector
    $\vec{\mu}$ into a covariant antisymmetric tensor by combining it
    with an electric dipole moment and to develop the associated
    equation of motion [24d].  This work, which confirmed a Thomas
    factor of 2, was submitted for publication on 2 May 1926.  In
    January of the following year a second paper by Thomas appeared in
    which essentially the same formalism was developed [24b].  But Pauli
    had been convinced of the correctness of Thomas's result already
    before.
\eq
Reference [24b] is the one I've already given you.  Reference [24d] is:
\begin{itemize}
\item
J. Frenkel, ``Die Electrodynamik des rotierenden Elektrons,'' Zeit.\ fur Phys.\ {\bf 37}, 243 (1926).
\end{itemize}

Maybe I am not quite understanding the vector you speak of.  I found this essay on ``Relativistic Angular Momentum'' on the web
\bv
\myurl{http://panda.unm.edu/Courses/finley/P495/TermPapers/relangmom.pdf}
\ev
that makes a distinction between the ``spin vector'' and the Pauli--Luba\'naski vector.  It is written by an undergraduate, Nick Menicucci, who was Carl Caves' student for a while.  I know that Carl has much respect for him, calling him the best undergrad physics student he had ever had.  (They have one or two papers together.)  Maybe he's the right person to contact with your question.  He is now a graduate student at Princeton. In any case, there is no quantum mechanics in Menicucci's essay, and it certainly defines something called the Pauli--Luba\'naski vector (in section 5).

\section{22-12-03 \ \ {\it Shut Up and Calculate} \ \ (to C. M. {\Caves})} \label{Caves76.1}

\bcc
BTW, are you regretting your {\bf Phys Today} article with Asher when \quantph{0312058} says it is the premier example of the ``shut up and calculate'' school of interpretations, by which they mean there is nothing further to think about.
\ecc

I still calculate from time to time, but I doubt I'll ever shut up.

The paper's ending paragraph reads like this:
\bq\noindent
    All this said, we would be the last to claim that the foundations of
    quantum theory are not worth further scrutiny. For instance, it is
    interesting to search for minimal sets of {\it physical\/}
    assumptions that give rise to the theory. Also, it is not yet
    understood how to combine quantum mechanics with gravitation, and
    there may well be important insight to be gleaned there. However, to
    make quantum mechanics a useful guide to the phenomena around us, we
    need nothing more than the fully consistent theory we already have.
    Quantum theory needs no ``interpretation.''
\eq
but that's almost universally ignored.

I still believe everything we said in that paper, but especially the last paragraph I would have loved to come off more positive.  The whole paper was quite a struggle, you know.  I had never realized just how much Asher and I disagreed on things until we tried to put the article together.  Concerning the last paragraph alone and the compromise between my positivity and Asher's negativity, see the discussion on page 457 of the VUP edition of {\sl Notes on a Paulian Idea}.  The main point I tried to get across to him at the time was this:
\bq\noindent
   {\bf Comment:} How about a compromise here?  The thing I didn't like
   about your examples is that both seemed to be more about
   difficulties in {\it application\/} of quantum theory, rather than
   its interpretation.  I think it is important that the reader
   understand that we are not doomsayers about the interpretation---
   that is to say, we are not completely rigid in the way that {\Bohr}
   and {\Rosenfeld} were. A substantial number of {\Peres}'s papers in
   the last 15 years have been precisely about
   clarifying the foundations.  It has been a fruitful pursuit
   precisely because the foundations are not yet a closed book. Flesh
   and bones still need to be added even to Copenhagen-like ideas. So I
   made what I was trying to get at a little more concrete by changing
   the wording and adding two references of serious works. Then I gave
   an even more concrete reason for pursuing that: namely precisely
   your point about relativity and gravitation. Now I like the ending
   even better; I hope you will agree.
\eq
But my coauthors rarely listen to me.

\section{22-12-03 \ \ {\it Tail Tuckers} \ \ (to D. B. L. Baker)} \label{Baker6}

Sorry to hear about the decline of your pool skills.  But, to mimic Einstein, ``Whatever your troubles with pool, I assure you that mine are worse.''

\bdb
I'm glad to hear about your latest book.  Do you get any money out of the deal?
\edb

With the VUP edition, I didn't make anything monetarily.  The gain was purely the satisfaction that comes with spreading the word.  (They printed 200 copies and financed it with a Swedish science grant.)  For the Kluwer edition (initial printing 450 copies), I'll get 6\% of the net receipts and 20 free copies.  That's effectively making nothing again, except the satisfaction of spreading the word even further.

\bdb
I suppose your title refers to Linus Pauling?  Can't see how any of my
thoughts could have been any help. Maybe my comment about wearing a
beret?  I thing the coolest pictures of any ``respectable'' person
wearing a beret would be him and Leonard Bernstein.
\edb

The title refers to Wolfgang Pauli.  The ``Paulian Idea'' is something I made up:  It's effectively the conjunction of these three quotes, which I took out of one of his papers.
\bq
     Like an ultimate fact without any cause, the individual outcome
     of a measurement is, however, in general not comprehended by laws.
     This must necessarily be the case. \medskip

     In the new pattern of thought we do not assume any longer the
     detached observer, occurring in the idealizations of this classical
     type of theory, but an observer who by his indeterminable effects
     creates a new situation, theoretically described as a new state of
     the observed system.  In this way every observation is a singling
     out of a particular factual result, here and now, from the
     theoretical possibilities, thereby making obvious the discontinuous
     aspect of the physical phenomena. \medskip

     Nevertheless, there remains still in the new kind of theory an
     objective reality, inasmuch as these theories deny any possibility
     for the observer to influence the results of a measurement, once
     the experimental arrangement is chosen.
\eq
The book as a whole is mostly a study in the sociology of science---it is quite literally only a collection of my emails---but strangely they have been emails with quite an effect on quantum foundations.  For instance, they've been responsible for at least two Bohmians leaving the fold!  (That alone probably made the writing worth it, I would say.)

What was your role?  Lending me an ear, pure and simple.  In particular, the first letter in the collection ``Noodles of Nothing''---wherein I tell you how much I love the writing style of William James and how I love to read letter collections---sets the tone for the whole book.

\bdb
Any thought to text books?
\edb

That's a much more serious project, and yes I have thought of it and even started it.  The title will be {\sl Quantum Foundations in the Light of Quantum Information}.  I've been courted by three publishers so far (Kluwer, Springer-Verlag, and World Scientific), and I've told them all ``yes''.  But I haven't signed any papers.  For this one, I'm going to work much harder to get the best deal I can.  Don't hold your breath:  It'll be two or three years in the making.

Thinking of William James, let me attach a little story I wrote up about Emma a couple months ago.  [See 22-09-03 note ``\myref{Bennett29}{Cosmology}'' to C. H. {\Bennett}.]  Actually maybe I'll put the whole context, since we've never had too much parenting talk between ourselves.  I bet you've had to go through similar things yourself.  (Put yourself in the place of Abner and read from bottom to top.)  [See 22-09-03 note ``\myref{Shimony4}{A Thought of You}'' to A. Shimony.]

\section{23-12-03 \ \ {\it Back in the Fight} \ \ (to H. J. Kimble)} \label{Kimble4}

You may have heard that the Perimeter Institute---or should I say a faction within the PI---finally gave me the axe (and showed me just what a dangerous thing it is to act like one cares about quantum foundations).  So I'm out there acting like an unemployed physicist again for all intents and purposes (working under the assumption that Bell Labs will eventually tank).

\section{23-12-03 \ \ {\it Great Americans} \ \ (to C. M. {\Caves})} \label{Caves77}

I was thinking of the above title, and I was thinking of you.  And I decided to send you a small gift for the holidays.  Amazon.com says the package will arrive in about a week.  It contains two books that I have enjoyed very much:  {\sl Philosophy and Social Hope}, by Richard {\Rorty}, and {\sl The Metaphysical Club: A Story of Ideas in America}, by Louis Menand.  The writers were my two favorites of this year's reading.  You might know Menand from his writing in the New Yorker; the book is a Pulitzer prizewinner in history.

Anyway, I learned a lot about quantum mechanics from these books.  And I learned a lot about true Americanism (i.e., the kind of Americanism we need to strive to get back to \ldots\ and actually mould into a more stable, long-lasting form).  Done right, I think the two subjects are probably the same thing.

I hope you get as much out of the books as I did.  When I read a book, I keep an index card in it for recording small notes and page numbers on the things that really interested me (usually things I deemed somehow connected to quantum mechanics).  Below is the tabulation for these books.

\bv
\underline{Louis Menand, {\sl The Metaphysical Club: A Story of Ideas in America}}\medskip\\
Holmes on immortality -- 60\\
Holmes's pragmatism -- 62, 63\\
Darwinism discussion -- 122, 123\\
Darwinism as nonscience -- 127\\
{\James}'s speech on Shaw -- 147, 148\\
Quetelet's view on crime -- 188\\
Schweber reference -- 473 \#31, 474 \#44\\
nature and open markets -- 195\\
Bayesian probability -- 197\\
higgledy-piggledy -- 199\\
no view from nowhere -- 199, 200\\
the universe is only weather -- 210, 211\\
Holmes's bettabilitarianism -- 217\\
Renouvier on certainty -- 218, 219\\
reality creation and democracy -- 219, 220\\
Maxwell on free will -- 222, 477 \#45\\
{\Peirce} redefining true -- 222, 223\\
the universe aware of itself -- 269\\
Deutsch's take on law without law -- 275, 279, 280\\
time as a habit -- 277, 482 \#35\\
evolution of physical laws -- 278, 279\\
{\Dewey}'s writing style -- 304\\
participatory universe -- 304, 305\\
knowing as doing -- 322\\
whether the universe responds to our desires -- 337\\
Comer's ``music on the page vs.\ music being played'' -- 341, 342\\
Darwinistic conception of ``true'' -- 351-353, 407\\
Bayesian-style thought -- 354, 356, 363\\
Bayesianism and Darwinism -- 357, 358\\
thought and action, same thing -- 360-362\\
weakness of pragmatism -- 375\\
freedom giving rise to limits -- 409-411, 431\\
thirteen pragmatisms -- 413\\
natural law and gravity -- 422, 423\\
truth, free speech, and democracy -- 431, 433
\ev

\bv
\underline{Richard {\Rorty}, {\sl Philosophy and Social Hope}}\medskip\\
pragmatism and unexpected futures -- 27, 28\\
pressures on belief from the outside -- 32, 33\\
leading up to {\Dewey} -- 49\\
correlations without correlata -- 52-55\\
objects along Dennett lines -- 89 \#15\\
good summary of {\Dewey}an pragmatism -- 88\\
paradigm change as POVM change -- 176\\
pragmatism clearing the underbrush -- 96\\
self-creation -- 265\\
motivation of {\James} -- 269\\
intersubjective agreement -- 119, 155\\
hope -- 120\\
connection to Bayes' rule -- 139, 140\\
{\Peirce} equating belief and habit -- 148\\
similarity between Nietzsche and {\James} -- 165 \#9\\
religion (for Nicholson) -- 153, 156, 157\\
{\Heidegger}'s pragmatism -- 191\\
belief -- xxiv, xxv\\
reality as like God -- xxix, 269, 270\\
view from nowhere -- 11\\
being contrarian (for {\Spekkens}) -- 5\\
Zen and the Art of Motorcycle Maintenance -- 8\\
anything goes (not!) -- 18\\
{\Appleby}'s point -- 37\\
some things beyond our control -- 51\\
truth as a nonepistemic notion ({\Davidson}) -- 164 \#5\\
What is evidence? -- 149-151
\ev

\section{23-12-03 \ \ {\it Little Comments} \ \ (to R. {\Schack})} \label{Schack78}

I've read your paper now.  I like it; it's fine. [\ldots]

By the way, I'd definitely like to swipe some material from this for
the Matteo Paris thing.  Your discussion on subjectivity has me
thinking about the proper definition again.  I wonder if I would now
go so far as your first sentence in your section 2?  Mind you, I just
don't know.  An alternative I'm playing with at least might go
something like this:  ``To say that a quantum state for a system is
subjective means that the state is not determined by any objective
facts about the system.  The quantum state, instead, takes its ground
in the agent, whose beliefs the state is a compact summary of.''  Or
something like that.  It is a little more careful in that it does not
say that the state is determined by no facts, but rather if it is,
those facts are not determined by the actual system itself.

Here is a lovely passage drawn from near the conclusion of {\sl The
Metaphysical Club\/}:

\begin{quote}
Academic freedom and the freedom of speech are quintessentially
modern principles.  Since the defining characteristic of modern life
is social change---not onward and upward, but forward, and toward a
future always in the making---the problem of legitimacy continually
arises.  In a premodern society, legitimacy rests with hereditary
authority and tradition; in a modernizing society \ldots\ legitimacy
tends to be transferred from leaders and customs to nature.  Agassiz
and the senior Holmes and Benjamin {\PeirceB} all assumed that social
arrangements are justified if they correspond with the design of the
natural world \ldots\  But in societies bent on transforming the past,
and on treating nature itself as a process of ceaseless
transformation, how do we trust the claim that a particular state of
affairs is legitimate?

The solution has been to shift the totem of legitimacy from premises
to procedures.  We know an outcome is right not because it was
derived from immutable principles, but because it was reached by
following the correct procedures.
\end{quote}

And our own point is that Dutch-book coherence---and more generally,
though yet to be worked out fully, quantum `coherence'---is the
correct procedure.

\chapter{2004: For Once, a Productive Use for Hidden Variables}

\section{07-01-04 \ \ {\it Probability Tables} \ \ (to P. G. L. Mana)} \label{Mana5}

Sorry for the long absence.  I think your paper is great; it'll make a good contribution.  I also encourage you to post it on {\tt quant-ph} once you have finalized it.

A couple of comments on the parts of your text related to me.

1) ``It is unknown to me how the photons produced in this process tend to respond to various interventions.''  You call this sentence ``unquestionably meaningful'' even for the (de Finettian) Bayesian.  I don't think it is nearly as meaningful as you think.  For instance, how would you precisely define ``tend'' without a notion of ignorance or probability?  On the other hand, if you try to define it through a notion of objective chance or propensity, then you'll give the de Finettian heartburn again.

I think what you're driving at is that you really want to think of your ``preparations'' as ``objective facts''.  So, if they're objective facts, why don't you just drop the fancy language of preparations and be done with it?  But then, show me an independent definition or meaning or description of these objective facts outside of their use in your probability tables.  For me, your $s_1$ \ldots\ $s_M$ are just prior states of belief, and I think that is the only consistent account a Bayesian can have.  The conceptual role of the $r_1$ \ldots\ $r_L$, however, are different.  They represent data upon which we update; they represent sense impressions.  (Better yet:  They represent the feedback the world gives to our prior beliefs, consequent to our stimulations to or interventions upon it.)

Concerning the preparation-vectors $s_j$ themselves it is just as bad.  For you yourself say, ``Note, in particular, that the numerical values of the vectors $s_j$ and $r_i$ \ldots\ depend upon the {\it whole\/} collection of probabilities $p_{ij}$: if one of these is changed, then \ldots\ all the vectors will change.''  For the de Finettian, all probabilities---no matter where and for what phenomenon they are used---are of the same subjective flavor.  Thus, just as there can be no such thing as an unknown probability, there can be no such thing as an unknown preparation-vector.

2) ``[I]n fact, the quantum state rather appears to be {\it part\/} of a state of knowledge (or belief) \ldots\ \  Perhaps the point is that CFS's notion of a `quantum' state of belief implicitly assumes the {\it existence\/} and the {\it particular structure\/} of the whole set of quantum POVMs (i.e., the interventions).''

Agreed.  The upshot of Gleason's theorem (and Gleason-like theorems with POVMs) is that GIVEN the structure of measurements, a quantum state is nothing more and nothing less than a compendium of probability distributions (and thus a state of belief for the de Finettian).  However, the upshot of my analysis in Section 7 in \quantph{0205039} and much of my web book {\sl Quantum States:\ What the Hell Are They?\/}\ (posted at my webpage) is that the assignment of a given POVM to an intervention is every bit as subjective as the quantum state itself.  I.e., a POVM is a kind of belief itself.  So, the GIVEN in Gleason's theorem is not to be taken lightly or ignored.  And you, with your probability tables, have identified this issue from another track.  That's excellent.

Anyway, as I say, I enjoyed your paper.  I think you're doing excellent work.  Send us the reformatted version as soon as you can.  We can argue about the business above at our leisure.

\section{07-01-04 \ \ {\it Deep Breath} \ \ (to G. Brassard)} \label{Brassard21}

\bgb
My talk at QIP is next week and I still have not a clue how I could possibly fill 50 minutes with musings on
\bq
\noindent\rm
{\bf Title:} Quantum Foundations in the Light of Quantum Information\smallskip
\\
{\bf Abstract:}
The late Rolf Landauer has once claimed that ``information is physical''.
The main point of this talk is to argue that ``physics is informational''!  At the moment, quantum information theory is replete with beautiful theorems on what is and is not possible according to quantum mechanics. For example, quantum key distribution is possible but quantum bit commitment is not.  But the axioms of quantum mechanics are strange and ad hoc, reflecting at best the history that led to discovering this new world order.  It's time to pause and reflect on what is really fundamental and what are merely consequences.  Could information be the answer?
\eq
Panic is high and steadily rising.
\egb

Take a deep breath.  There's a lot of good stuff on the theme out on the web now.  You could take the opportunity to give a summary of the various issues out there and how they all converge in the same direction:  Namely---as I see it---quantum states and operations are expressions of information, rather than expressions of a reality independent of information-using agents.  If we can just swallow that idea whole and learn to savor it, rather than fight it---your conclusion could be---may be the best thing we can do to prepare ourselves for the best, most outlandish physics yet to come.

Here's some papers to look at that come to mind:
\begin{enumerate}
\item
\quantph{0205039}, ``Quantum Mechanics as Quantum Information (and only a little more),''
Christopher A. Fuchs.

\item
\quantph{0310017}, ``Classical Teleportation of Classical States,''
Oliver Cohen.

\item
\quantph{0203070}, ``The Nature of Information in Quantum Mechanics,''
Rocco Duvenhage.

\item
\quantph{0101012}, ``Quantum Theory From Five Reasonable Axioms,'' Lucien Hardy.

\item
\quantph{0211089}, ``Characterizing quantum theory in terms of information-theoretic constraints,'' Rob Clifton, Jeffrey Bub, Hans Halvorson.

\item
\quantph{0310101}, ``A note on information theoretic characterizations of physical theories,'' Hans Halvorson.

\item
\quantph{0310067}, ``Can Quantum Cryptography Imply Quantum Mechanics?,'' John A. Smolin.

\item
\quantph{0311065}, ``Can quantum cryptography imply quantum mechanics? Reply to Smo\-lin,'' Hans Halvorson, Jeffrey Bub.

\item
\quantph{0306179}, ``Gleason-Type Derivations of the Quantum Probability Rule for Generalized Measurements,''
Carlton M. {\Caves}, Christopher A. Fuchs, Kiran Manne, Joseph M. Renes.

\item
\quantph{0304159}, ``Quantum information processing, operational quantum logic, convexity, and the foundations of physics,'' Howard Barnum.

\item
\quantph{0208121}, ``Betting on the Outcomes of Measurements: A Bayesian Theory of Quantum Probability,'' I. Pitowsky.

\item
\quantph{0305088}, ``Copenhagen Computation: How I Learned to Stop Worrying and Love Bohr,'' N. David {\Mermin}.

\item
\quantph{0305117}, ``Why can states and measurement outcomes be represented as vectors?,'' Piero G. L. Mana.

\item
\quantph{0307198}, ``A de Finetti Representation Theorem for Quantum Process Tomography,'' Christopher A. Fuchs, {\Ruediger} {\Schack}, Petra F. Scudo.

\item
\quantph{0210017}, ``Quantum theory from four of Hardy's axioms,'' {\Ruediger} {\Schack}.

\item
\quantph{0308114}, ``The Bell--Kochen--Specker Theorem,''
D. M. {\Appleby}.

\item
\quantph{0311039}, ``Multilinear Formulas and Skepticism of Quantum Computing,'' Scott Aaronson.

\item
\quantph{0107151}, ``Whose Knowledge?,''
N. David {\Mermin}.

\item
\quantph{0109041}, ``How much state assignments can differ,'' Todd A. Brun, J. Finkelstein, N. David {\Mermin}.

\item
\quantph{0206110}, ``Conditions for compatibility of quantum state assignments,'' Carlton M. {\Caves}, Christopher A. Fuchs, {\Ruediger} {\Schack}.
\end{enumerate}

I'm sure I've missed a ton of things.

Skim and summarize!

\section{07-01-04 \ \ {\it Four More} \ \ (to G. Brassard)} \label{Brassard22}

Let me give you four other papers that might be useful.  None of these have appeared on the web yet.

\begin{enumerate}
\item
Ari Duwell's ``Quantum Information Does Not Exist'' (appeared in the September 2003 issue of {\sl Studies in History and Philosophy of Modern Physics}---a special issue edited by Bub and me on quantum information and foundations).

\item
Jeff Bub's ``Why the Quantum?''.  I think it will be his contribution for the Clifton memorial.  Conceptually, it may do a better job than the earlier papers.

\item
Piero Mana's ``Probability Tables''.  This isn't quite the final version, but it must be quite close.  It'll appear in our {\Vaxjo} proceedings this year.  If you want to mention it, you should get his permission.  But, at the least, it may help you understand his previously posted paper.

\item
Scott Aaronson's ``Is Quantum Mechanics an Island in Theoryspace?''  It'll appear in our {\Vaxjo} proceedings this year.  If you want to mention it, you should get his permission.  It's a fun (and funny) paper.
\end{enumerate}
OK, that exhausts my thoughts for the moment.

\section{07-01-04 \ \ {\it A Really Important One} \ \ (to G. Brassard)} \label{Brassard23}

Forgot a really important one:  Rob {\Spekkens}'s paper-in-the-writing, ``In Defense of the Epistemic View of Quantum States: A Toy Theory.''  The last draft I saw of it was 65 pages.  It's amazing the number of quantum-information style phenomena he nails with his toy theory.

You might ask him if he'd send you a copy of the draft to peruse, if you'd like to say a word or two about it.

Below is the abstract he used for his Caltech talk on the subject.

\bq
\noindent Title: In defense of the view that quantum states are states of knowledge\medskip

\noindent Abstract:
I present a toy theory that is based on a simple information-theoretic principle, namely, that when one has maximal knowledge of a system, one's knowledge is quantitatively equal to one's ignorance. The object analogous to the quantum state in the theory is a uniform probability distribution over hidden states. Because the theory is, by construction, local and non-contextual, it does not reproduce quantum theory. Nonetheless, a wide variety of quantum phenomena have analogues within the toy theory that admit simple and intuitive explanations. Such phenomena include: the noncommutativity of measurements, interference, no information gain without disturbance, the multiplicity of pure state decompositions of a mixed state, the distinction between two-way and intrinsic three-way entanglement, the monogamy of pure entanglement, no cloning, teleportation, dense coding, mutually unbiased bases, locally immeasurable product bases, unextendible product bases, the possibility of key distribution, the impossibility of bit commitment, and many others. The diversity and quality of these analogies provide compelling evidence for the view that quantum states are states of knowledge rather than states of reality, and that maximal knowledge is incomplete knowledge. A consideration of the phenomena that the toy theory fails to reproduce, notably, violations of Bell inequalities and the existence of a Kochen--Specker theorem, provides clues for how to proceed with a research program wherein the quantum state being a state of knowledge is the idea upon which one never compromises.
\eq

\section{08-01-04 \ \ {\it Calls to Copenhagen} \ \ (to M. P\'erez-Su\'arez)} \label{PerezSuarez7}

It's too bad your plans had to change to a Jan 20 arrival.  It turns out I will be gone Jan 23--27, for my mother's 75th birthday party in Texas.  Sorry I didn't warn you of this before, but I just found out myself yesterday.  I had to make a mad dash to get the tickets.

Beyond that, 20 minutes ago, I just got a call to Copenhagen!  The Niels Bohr Institute has asked me to come for a job interview February 2.  Thus you can probably count on my being gone Feb 1--3.

I hope you'll understand that in neither of these cases could I say ``no''.  I realize that this will leave you a little guidanceless at the very beginning of your stay, but I'll try to think of ways to make things go as smoothly as possible in my absence.

\section{08-01-04 \ \ {\it Partial Exchangeability} \ \ (to D. Poulin)} \label{Poulin11}

I'm glad to hear there was enough material in the paper to make it worth three readings!

\bdp
Is it possible to find, for example, a 3-qubit state which
(i) CAN be obtained by tracing out the 4th qubit of a 4-qubit
symmetric state but (ii) CANNOT be obtained by tracing out the 4th and
5th qubits of a 5-qubit symmetric state?
\edp
Yes it is.  Jaynes gives some good examples in
\begin{itemize}
\item
E.~T. Jaynes, ``Some Applications and Extensions of the de Finetti Representation Theorem,'' in {\sl Bayesian Inference and Decision Techniques}, edited by P.~Goel and A.~Zellner (Elsevier, Amsterdam, 1986), pp.~31--42.
\end{itemize}
for the case of classical probability distributions (i.e., commuting density operators) and gives some good intuitive explanation for what goes wrong.

There are similar results for partially exchangeable (entangled) quantum states.  Joe Renes is the expert on that, and can probably give you some good examples in the noncommutative case.

\section{09-01-04 \ \ {\it Where Did You Come From, Where Did You Go?}\ \ \ (to S. J. van {\Enk})} \label{vanEnk29}

Where did you come from Cotton-Eyed Joe?

Where have you been?  I never got a reply from you after I answered the note you sent to Kiki's account.

Anyway, I have a question in case you're listening.  Have you seen Oliver Cohen's paper ``Classical Teleportation of Classical States'', \quantph{0310017}?  I remember one day (two or three years ago) you and I were talking at the chalkboard, and we were toying with trying to write down a classical coin-toss ``teleportation'' scheme.  Do you remember if we succeeded?  I remember we were playing with something quite similar to this.  (I can certainly dig up a lot of places in my samizdats where I've taken it for granted that such a kind of scheme would work.)

Anyway, I've got to decide whether this paper should be a PRL.  I'll attach below the considerations I sent Bob Garisto this morning to help you see what I'm thinking.

\section{09-01-04 \ \ {\it Christmas with The Pogues} \ \ (to D. B. L. Baker)} \label{Baker7}

I didn't write you the letter I had hoped to during the Christmas holidays.

Instead, let me send you a song I learned this Christmas---a happenstance by being in Ireland.  With your better knowledge of music than me you've probably already known it for a long time.  But I'll place the words below and attach an mp3 file with the song.  The words, especially near the end, have such a powerful effect on me.  I haven't been able to get the song out of my head all this morning, and consequently the present note.

\bv
{\bf Fairytale of New York} \medskip\\
by The Pogues \medskip\\
I could have been someone\\
Well so could anyone\\
You took my dreams from me\\
When I first found you\\
I kept them with me babe\\
I put them with my own\\
Can't make it all alone\\
I've built my dreams around you\medskip\\
The boys of the NYPD choir\\
Still singing ``Galway Bay''\\
And the bells are ringing out\\
For Christmas day
\ev

\section{10-01-04 \ \ {\it Disappointment} \ \ (to M. P\'erez-Su\'arez)} \label{PerezSuarez8}

About books, I brought Nielsen and Chuang with me and you are free to use it (and any of my books).  I don't think I brought Peres with me, however.  I will try to remember to check in my office Monday.  As far as linear algebra, matrix analysis books go, I probably have anything you could need.  For library use we have the Dublin Institute for Advanced Studies, about a 20 minute walk from our offices.  They have a decent sized math/physics library.  Also, I have use of the Trinity College mathematics library (quite large).  I have a library of about 290 philosophy books (mostly about pragmatism, Wittgenstein, postmodern stuff, etc.)\ here in my own house.  So, basically, just tell me about any books you were thinking of bringing, and I will tell you whether there is overlap with what you can borrow from any of these sources.

\section{10-01-04 \ \ {\it Cohen Again}\ \ \ (to S. J. van {\Enk})} \label{vanEnk30}

\bsve
I saw the Cohen paper you mention, I even read some part of it and
decided it was nothing I didn't know. Moreover, I thought he missed
something, but now I don't remember what that was. I'll check more
carefully. I also remember vaguely we talked about this, and I
thought we did have some essential difference between classical and
quantum teleportation. Was it secrecy??
\esve

I think maybe in the version we came up with, Alice could learn something about Victor's state (i.e., probability distribution).  But in his version---where the `Bell basis' measurement automatically randomizes first---he doesn't have that problem.  Of course there could be the issue of a third party correlated with the `entangled pair'---like in the version Collins and Popescu consider.

So, I'm kind of torn.  Maybe it is too late for this paper.  If it had just been written five years ago.  (It is the sort of thing you and I should have written; but who would have guessed it could be turned into a whole PRL?)  On the other hand, I like the way the guy makes a clean simple point of the lesson:  quantum teleportation is simply not exciting or remarkable if one takes the epistemic view of quantum states.  It's a point I want more people to appreciate---even if a few of us realized it long ago---and his paper could be a good vehicle for that.

Also there's the issue that {\Spekkens} will soon release his toy model to the world.  Buried within that far more ambitious work (it's a 65 page manuscript and counting) is a classical teleportation protocol.  Should Cohen lose all credit, or all attention, in the blinding light of {\Spekkens}'s paper?  Alternatively might I help bring attention to {\Spekkens}'s paper by letting Cohen into PRL at the same time as forcing him to cite {\Spekkens}?

Cohen is not a power hitter.  He's only had six papers since 1999.  So I don't think I'm creating undue competition for us by letting a cheap paper past PRL, and I can see that it might actually help him hold onto a career he might otherwise be forced to drop out of.

I tell you, I turn more social democrat every day \ldots

Anyway, all of this (especially {\Spekkens}'s toy model) is clarifying and food for thought.  A good lot of quantum information theory is simply regular probability/information theory applied in ways that had not been deemed interesting before.  It helps us realize that what is unique in quantum mechanics is not the probabilities but what the probabilities are applied to.  There, I think, lies the essence of quantum mechanics:  It is localized in the Kochen--Specker theorem.  Unperformed measurements have no outcomes.  It's just a question of putting that in a more useful form before something can really come of it.

Thanks for calling me a father.  It helped a lot.

\section{10-01-04 \ \ {\tt quant-ph/0106166} \ \ (to G. Brassard)} \label{Brassard24}

\bgb
Is \quantph{0106166} obsolete in the face of \quantph{0205039}?
\egb

I think it is, and I try to sell it that way.  But strangely, a lot of people choose to read it first nevertheless---thinking of it as a `lite' version of {\tt 0205039}.  Certainly the intros of the two papers have different jokes and allusions; maybe they're worth looking at separately because of that.

Still, it would be nice if you did mention {\tt quant-ph/0106166} along with {\tt quant-ph/0205039}, after all \ldots\ since its title coincides with the title of your talk.

More a little later this evening.

I realized I told you in one of my other notes this morning to keep breathing slowly.  Don't breathe too slowly, though.  That wouldn't be healthy \ldots

\section{10-01-04 \ \ {\it More Help Please} \ \ (to G. Brassard)} \label{Brassard25}

\bgb
But seriously, there is something else you can do to help.
Please send me as much information as you can gather easily
on the 3 meetings that took place so far, i.e.\ both here in
{\Montreal} and the one in {\Vaxjo}.
\egb

Four meetings so far, and at least two coming up.
Though the fourth was really just a component in the larger {\Vaxjo} meeting:  This time, it was titled ``Quantum Logic meets Quantum Information.''  And the idea was to get people together who had an interest in making a bridge between the two communities.  So, one of the pre-requisites was that the participants from each community were supposed to have a little knowledge and sympathy of the other at the outset.  (`Quantum Logic', in case you don't know, is a certain community in quantum foundations that thinks that what is going on behind quantum mechanics is literally a change of the rules of formal logic \ldots\ i.e., from Boolean algebra to something else.)

I'll compile the information for you, but I can't do it until about 8 hours from now.

\section{11-01-04 \ \ {\it The First {\Montreal} Meeting} \ \ (to G. Brassard)} \label{Brassard26}

\bv
{\bf Workshop on Quantum Foundations in the Light of Quantum
Information}
\smallskip
\\
Centre de Recherches Math\'ematiques Universit\'e de Montr\'al
\smallskip
\\
16 -- 19 May 2000 \smallskip
\\
{\Gilles} {\Brassard} and Christopher A. Fuchs, organizers
\bigskip \\

\underline{17 May}\medskip\\

{\Gilles} {\Brassard}, Universit\'e de {\Montreal}\\
\hspace*{.4in} {\it Quantum Foundations in the Light of Quantum
Cryptography} \smallskip\\

Christopher A. Fuchs, Los Alamos National Laboratory\\
\hspace*{.4in} {\it Quantum Foundations in the Light of Quantum
Information} \smallskip\\

William K. {\Wootters}, Williams College\\
\hspace*{.4in} {\it Quantum Mechanics from Distinguishability?}\smallskip\\

Benjamin W. {\Schumacher}, Kenyon College\\
\hspace*{.4in} {\it Doubting {\Everett}} \bigskip\\

\underline{18 May}\medskip\\

N. David {\Mermin}, Cornell University\\
\hspace*{.4in} {\it Pre-{\Gleason}, Post-{\Peierls} \& Compumentarity}\smallskip\\

{\Herb}ert J. {\Bernstein}, Hampshire College\\
\hspace*{.4in} {\it Why Quantum Mechanics?}\smallskip\\

Richard {\Jozsa}, University of Bristol\\
\hspace*{.4in} {\it Foundations of an Interpretation of Quantum Mechanics\\
\hspace*{.4in} in the Light of Quantum Computing} \smallskip\\

Charles H. {\Bennett}, IBM Research at Yorktown Heights\\
\hspace*{.4in} {\it Entanglement-Assisted Remote State Preparation} \bigskip\\

\underline{19 May}\medskip\\

{\Ruediger} {\Schack}, University of London -- Royal Holloway \\
\hspace*{.4in} {\it Quantum Gambling and Bayesian Probability in
Quantum Mechanics}\smallskip\\

Patrick M. {\Hayden}, Oxford University\\
\hspace*{.4in} {\it Two Lessons from the {\Heisenberg} Representation, or\\
\hspace*{.4in} How I Learned to Stop Worrying \& Love Non-Commutativity}\smallskip\\

{\Jeff}rey {\Bub}, University of Maryland\\
\hspace*{.4in} {\it Some Reflections on Quantum Logic}\\
\ev

\section{11-01-04 \ \ {\it The Second {\Montreal} Meeting} \ \ (to G. Brassard)} \label{Brassard27}

Be warned:  Some of these titles are my own constructions! [See 10-09-02 note ``\myref{Communards2}{Tentative Talk Schedule for Commune}'' to the Communards.]

\section{11-01-04 \ \ {\it The First Sweden Meeting} \ \ (to G. Brassard)} \label{Brassard28}

\noindent Officially titled:
\bv
{\bf Quantum Theory: Reconsideration of Foundations}\smallskip\\
{\Vaxjo} University, {\Vaxjo}, Sweden\\
17-21 June 2001\\
Organizers: Christopher A. Fuchs, Pekka Lahti, and Andrei Khrennikov.\\
\ev
however, a significant part of the meeting was a session titled:
\bv
``Shannon Meets Bohr: Quantum Foundations in the Light of Quantum Information''
\ev
The participants and talks in this session were:
\begin{enumerate}
\item
Herbert J. Bernstein,
``Observer Participancy and Law without Law (or something like that)''
\item
Doug Bilodeau,
``Real Time: How Quantum Mechanics Describes the World''
\item
Jeffrey Bub,
``Information and Objectivity in quantum Mechanics''
\item
Carlton M. {\Caves},
``Quantum Dynamics and Maximal Information''
\item
Henry J. Folse,
``Bohr's Conception of the Quantum  Mechanical State of a System and Its Role in the Framework of Complementarity''
\item
Christopher A. Fuchs,
``Quantum Foundations in the Light of Quantum Information''
\item
Daniel Greenberger,
``The Inconsistency of the Galilean Transformation in Quantum Theory''
\item
Lucien Hardy,
``Quantum theory from Five Reasonable Axioms''
\item
Peter Harremo\"es,
``Bell's Inequalities and Graphical Models of Independence''
\item
Richard Jozsa,
``Quantum Computation and Foundational issues''
\item
N. David {\Mermin},
``Whose Knowledge?''
\item
Asher Peres,
``Indiscrete Quantum Information''
\item
Itamar Pitowsky,
``Range Theorems for Quantum Probability''
\item
Arkady Plotnitsky,
``Spectro-Cryptography: Bohr, Heisenberg and Quantum Mechanics as an Information Theory''
\item
{\Ruediger} {\Schack},
``Bayesian Probability in Quantum Mechanics''
\item
Ben Schumacher,
``Doubting Everett''
\item
John Smolin,
``Quantum Nonlocality without Entanglement''
\item
Joseph Renes,
``(Bayesian) Probability in Quantum Mechanics''
\item
Daniel Terno,
``Quantum Information in Relativity Theory''
\end{enumerate}

\section{11-01-04 \ \ {\it The Second Sweden Meeting} \ \ (to G. Brassard)} \label{Brassard29}

\noindent Officially titled:
\bv
{\bf Quantum Theory: Reconsideration of Foundations -- 2}\smallskip\\
{\Vaxjo} University, {\Vaxjo}, Sweden\\
1--6 June 2003\\
Organizers: Howard Barnum, Christopher A. Fuchs, Robin Hudson, and Andrei Khrennikov
\ev
however, a significant part of the meeting was a session titled:
\bv
``Quantum Logic Meets Quantum Information''
\ev
organized by Barnum and Fuchs.
The participants and talks in this session were:
\begin{enumerate}
\item
Scott Aaronson,
``Quantum Computing and Dynamical Quantum Models''
\item
Guido Bacciagaluppi,
``Classical Extensions, Classical Representations and Bayesian Updating in Quantum Mechanics''
\item
``Howard Barnum,
Using Probabilistic Equivalence to Construct Algebraic Structures from
Operational Theories''
\item
``Stephen Bartlett,
Restrictions to Quantum Information Processing''
\item
Paul Busch,
``Less (Precision) Is More (Information): Quantum Information in Fuzzy Probability Theory''
\item
Bob Coecke,
``Probability from Logic''
\item
Christopher Fuchs,
``What is the Difference between a Quantum Observer and a Weatherman?''
\item
Alexei Grinbaum,
``C. Rovelli's `Relational Quantum Mechanics' and the Problem of
In\-for\-ma\-tion Theoretic Derivation of Quantum Theory''
\item
Hans Halvorson,
``Tomita's Theorem and its Role in the Information-Theoretic Characterization of Quantum Theory''
\item
Lucien Hardy,
``Quantum Theory as a Probability Theory''
\item
Piero Mana,
``Why Can States and Measurement Outcomes Be Represented as Vectors?''
\item
Marcos P\'erez-Su\'arez,
``Quantum Mechanics and Information Theory: Some Further Musings on a Fuchsian Proposal''
\item
{\Ruediger} {\Schack},
``On Unknown Quantum Operations''
\item
John Smolin,
``Can Quantum Cryptography Imply Quantum Mechanics?''
\item
Rob {\Spekkens},
``In Defense of the Epistemic View of Quantum States''
\item
Alexander Wilce,
``Compactness and Symmetry in Quantum Logic''
\end{enumerate}

\section{11-01-04 \ \ {\it Hardy Facts} \ \ (to G. Brassard)} \label{Brassard30}

By the way, a little factoid about Hardy.  You might recall that he was invited to our first {\Montreal} meeting but had to withdraw at the last minute because of an illness in his family or something.  Anyway, his paper ``Quantum Theory from Five Simple Axioms'' was motivated directly---he tells me---by his invitation to our meeting.

More still to come, but I've got to get the kids to bed at the moment.

\section{11-01-04 \ \ {\it Special Issue of SHPMP} \ \ (to G. Brassard)} \label{Brassard31}

Continuing.  The September 2003 issue of the journal {\sl Studies in History and Philosophy of Modern Physics\/} was devoted to ``Quantum Information and Computation''.  The editors of the special issue were Jeff Bub and me.  Below is a tabulation of the papers, and further below that is the introduction we wrote for the issue.  N.B. there was a horribly serious omission in that introduction; pathetically we forgot to mention David {\Mermin}'s paper.  I am still atoning for that sin!
\begin{enumerate}
\item
Howard Barnum: `Quantum information processing, operational quantum logic, convexity, and the foundations of physics'
\item
Charles Bennett: `Notes on Landauer's principle, reversible
   computation and Maxwell's demon'
\item
Armond Duwell: `Quantum information does not exist'
\item
Lucien Hardy: `Probability theories in general and quantum theory in particular'
\item
David {\Mermin}: `Copenhagen computation'
\item
Itamar Pitowsky: `Betting on the outcomes of measurements: a Bayesian theory of quantum probability'
\item
Andrew Steane: `A quantum computer only needs one universe'
\item
Chris {\Timpson}: `On the supposed conceptual inadequacy of the Shannon information'
\item
David Wallace: `Quantum probability and decision theory revisited'
\end{enumerate}

\section{12-01-04 \ \ {\it Upcoming Conferences} \ \ (to G. Brassard)} \label{Brassard32}

I'm sure this is only a partial list, but it includes all the ones that I am confident will carry our theme.
\begin{itemize}
\item
Probability in Quantum Mechanics,
Centre for Philosophy of Natural and Social Science London School of Economics and Political Science, Monday, February 16, 2004,
Organizers: Stephan Hartmann and Roman Frigg
\item
Seven Pines Symposium VIII: ``Quantum Mechanics, Quantum Information, and Quantum Computation,'' University of Minnesota,
May 5--9, 2004, Organizer: Roger H. Stuewer
\item
Philosophy of Science Association, annual meeting 2004 (this is the biggest philosophy of science meeting) Austin, Texas, November 18--21, 2004 within it there will be a workshop titled `Quantum Information Theory and the Foundations of Quantum Mechanics: Is Information the Way Forward?,' Organizer: Christopher G. {\Timpson}
\item
This meeting should also have a significant component on quantum foundations in the light of quantum information:
New Directions in the Foundations of Physics, American Institute of Physics, College Park, April 30 -- May 2, 2004,
Organizer: Jeff Bub
\end{itemize}

\section{12-01-04 \ \ {\it Miscellania} \ \ (to G. Brassard)} \label{Brassard33}

\bgb
Your polite silence on what you think of my slides worries me!
\egb

Silent only because there wasn't much to say yet, and believe it or not, I am a busy man too.  I like Michelangelo.  And though I can imagine that the burning bush may have laid out ``I am that I am'' as fundamental principle (in constructing the universe), I agree with you that it is hard to imagine it declaring $C^*$ algebras in the same breath.

\section{12-01-04 \ \ {\it Diode Lasers} \ \ (to P. G. L. Mana)} \label{Mana6}

Thanks for the final draft; I've forwarded it to Andrei Khrennikov.

\bpglm
But I realise now, that I have not understood completely your ``philosophy'': roughly asking, do you consider a diode laser a sense impression?  Do you prefer to speak of sense impressions rather than ``(macroscopic) reality''?
\epglm

I consider a ``diode laser'' as very much akin to (or maybe even precisely) a prior.  (As in, the kind of prior a Bayesian would use as the starting point of his probabilistic calculations.)

\bpglm
What do you mean by `the world'? Your position seems almost
Leibnizian.
\epglm

I don't know what Leibnizian means.  What I mean by `the world' is everything external to the agent:  the source of all the stimulations upon him.

\bpglm
Apart from some possible differences in basic ``philosophical'' points of view (which concern something broader than quantum mechanics), \ldots\ and the fact that you believe in the ultimate validity of the quantum-mechanical formalism [\ldots]
\epglm

I wouldn't say I believe the latter at all.

In any case, I'll attach a note that expresses aspects of my latest position, such as it is.

\section{13-01-04 \ \ {\it Decisions}\ \ \ (to S. J. van {\Enk})} \label{vanEnk31}

\bsve
I just read Cohen's paper, and I don't know what I found wrong about
it before: I now agree! The only difference with quantum teleportation
is, which I mentioned before, you could use quantum teleportation for
key distribution, and you cannot use classical teleportation for that
purpose. Of course, QKD without teleportation is perfectly possible
too, so that argument doesn't really discredit Cohen's protocol.
\esve

If you don't disagree harshly then, I guess I'm going to recommend that it be published (with changes), despite the existence of
\bv\tt
\quantph{9906105} [abs, ps, pdf, other] : \\
Title: Classical Teleportation of a Quantum Bit\\
Authors: N. J. Cerf (1), N. Gisin (2), S. Massar\\
Comments: 4 pages, RevTex\\
Journal-ref: Phys.\ Rev.\ Lett.\ 84 (2000) 2521\bigskip\\
\quantph{0107082} [abs, ps, pdf, other] :\\
Title: A classical analogue of entanglement\\
Authors: Daniel Collins, Sandu Popescu\\
Comments: 13 pages, references updated
\ev
and {\Spekkens}'s new paper.  My support, as you know, is chiefly motivated by the moral he draws, which neither of the first two papers above seem compelled to.

I suspect I'm the only referee, because it was referred to me by ``a Divisional Associate Editor'' (had to be Bennett?).  So it seems to hinge on me (and by proxy, if there is a lack of protest, you).

\section{14-01-04 \ \ {\it Hello Kitty} \ \ (to H. Mabuchi)} \label{Mabuchi4}

I've been meaning to write you for a couple of days.  Emma turned five on January 12, and in her pile of presents was a Hello Kitty watch, wrapped with a tag listing ``Uncle Hideo'' as the source.  Emma asked, ``Who's that?''  Later I showed her a picture of you, and she said ``Wow.''  (She says wow a lot.)  Anyway, Kiki and I thank you!  Emma was proud to sport her new watch along with her new dress, bought especially for the occasion of her date with her dad later that evening.

On less happy matters, you've probably heard by now that the Perimeter Institute decided I was too risky a proposition for their world fame and dominance.  It seems that hidden-variables theories (or even sillier things) will carry the day.

You know my secret dream (and you know I've got a thousand secret dreams):  That someone of unique talent at Caltech would convince their physics and humanities departments to throw in half and half on me.  The possibilities!  Courses on Bayesian foundations for probability theory, courses fleshing out comparisons and contrasts between Bayesianism and pragmatism, courses on Wittgenstein and the foundations of quantum mechanics, courses on James and Dewey and the foundations of science, courses on why we should take some of the postmodern thinkers seriously, courses connecting the interpretational issues for quantum states to the philosophy of language, courses on beginning quantum mechanics for undergraduates, courses on standard quantum information theory, courses on matrix analysis methods in quantum information, \ldots.  What possibilities!  And what better place for such possibilities? \ldots\ Pipe dreams.

Actually {\Wittgenstein} has become my latest addiction.  In preparation for writing my new paper with {\Schack}---one that we've been tentatively calling ``On Quantum Certainty''---I thought I should read {\Wittgenstein}'s book ``On Certainty'' to see if, on the off chance, there was some overlap of ideas.  I had some indication that there might be from things I had read in Richard {\Rorty}.  Well, what a pleasant surprise!  {\Wittgenstein} is much deeper than I ever imagined.  It was just silly of me to be put off for so long by the style in which his books are written.  Within the first five pages of that 90 page book, he already said all that {\Ruediger} and I ever wanted to say and just kept going.  Here is a man who, 52 years ago, could have tackled the Penrose question head on:  How a pure quantum state can be epistemic and yet still give probability one predictions for some measurements?  Anyway, reading {\Wittgenstein} is certainly taking me down a more solid and careful path on the issue than I thought I was prepared for.

And speaking of the fancy term ``epistemic'', I hope you've noticed that Rob {\Spekkens} finally put his toy-model paper on {\tt quant-ph}.  It is a masterwork.  You want to teach your first-year students what quantum mechanics is all about?  Make this paper part of the curriculum.  It is that simple, and it is that convincing.  (That is, at least it gets half the answer.  We wouldn't want them to think there's nothing left to do.)

OK, now I'm just procrastinating.  Too many more official things to write, and too little desire to do it; email and gossip are always easier.

I hope things are going well for you and your family of 17.  And many thanks again---this time from Emma---for the watch.

\section{15-01-04 \ \ {\it Idealism} \ \ (to R. W. {\Spekkens})} \label{Spekkens27}

I'm planning to write you two notes, one titled ``Pragmatism'' and one titled ``Idealism.''  The one titled ``Pragmatism'' has to do with what I think I need for self-preservation at the moment, and not to do with my philosophy.  The one titled ``Idealism'' has to do with my philosophy, which is pragmatism, but has nothing to do with my career.

On page 2 you write, ``There is one case wherein the distinction between an ontic state and an epistemic state breaks down \ldots'' \ You know I still disagree with you:  epistemic is always epistemic, period.  If I could just get the discipline to finish my paper with {\Schack}, ``On Quantum Certainty,'' maybe we could communicate better.  In the mean time---though I doubt it will help---let me try to entice your soul not with my words, but Wittgenstein's.  Here's something I wrote Mabuchi yesterday:
\bq\noindent
Actually Wittgenstein has become my latest addiction.  In preparation for writing my new paper with {\Schack}---one that we've been tentatively calling ``On Quantum Certainty''---I thought I should read Wittgenstein's book {\sl On Certainty\/} to see if, on the off chance, there was some overlap of ideas.  I had some indication that there might be from things I had read in Richard Rorty.  Well, what a pleasant surprise!  Wittgenstein is much deeper than I ever imagined.  It was just silly of me to be put off for so long by the style in which his books are written.  Within the first five pages of that 90 page book, he already said all that {\Ruediger} and I ever wanted to say and just kept going.  Here is a man who, 52 years ago, could have tackled the Penrose question head on:  How a pure quantum state can be epistemic and yet still give probability one predictions for some measurements?  Anyway, reading Wittgenstein is certainly taking me down a more solid and careful path on the issue than I thought I was prepared for.
\eq

Here are a couple of paragraphs from {\sl On Certainty} that I had the word ``{\Spekkens}'' written by:
\bq\noindent
21.  Moore's view really comes down to this: the concept `know' is analogous to the concepts `believe', `surmise', `doubt', `be convinced' in that the statement ``I know \ldots'' can't be a mistake.  And if that {\it is\/} so, then there can be an inference from such an utterance to the truth of an assertion.  And here the form ``I thought I knew'' is being overlooked.---But if this latter is inadmissible, then a mistake in the {\it assertion\/} must be logically impossible too.  And anyone who is acquainted with the language-game must realize this---an assurance from a reliable man that he {\it knows\/} cannot contribute anything.
\\
22. It would surely be remarkable if we had to believe the reliable person who says ``I can't be wrong''; or who says ``I am not wrong.''
\eq
There were many more than that, but I'm too lazy to copy them in at the moment!

I'd say that a Wittgenstein reading is hardly a substitute for reading Fuchs and
{\Schack}, but who knows, maybe his mortars will soften the front.

\section{16-01-04 \ \ {\it Also Holevo} \ \ (to R. W. {\Spekkens})} \label{Spekkens28}

I forget to mention in my previous notes the business about Chapter 1 in Holevo's book.  You didn't mention the connection between that and your toy model in your paper.  It's certainly of the same flavor, but it is hard (for me at least) to tell from his general theorem (7.1) whether one would get all the sorts of quantum behavior from it that you do.  Maybe it's obvious though \ldots\ (And I'm starting to think it is as I write this sentence to you.)

\section{16-01-04 \ \ {\it More Diodes} \ \ (to P. G. L. Mana)} \label{Mana7}

\bpglm
\bq\noindent\rm
[CAF wrote:] I consider a ``diode laser'' as very much akin to (or maybe even precisely) a prior.  (As in, the kind of prior a Bayesian would use as the starting point of his probabilistic calculations.)
\eq
Now I must really say that I'm confused.
\epglm

I mean that within the framework of quantum mechanics ``diode laser'' is a name for a particular quantum operation (trace-preserving completely positive map) on the electromagnetic field.  And from my point of view quantum operations are as sinfully subjective as quantum states.  Their giveness just happens to be the starting point of most quantum mechanical problems, but it's nothing more than that.  So, in that role, quantum operations play the role that the prior (or ``statistical model'') plays in most textbook problems in a probability course.

Most of my samizdat {\sl Quantum States:\ What the Hell Are They?\/}\  (posted on my webpage) is devoted to this issue, but maybe pick up at the 19 September 2001 note ``\myref{Mermin43}{Practical Art}'' and then go to the 20 September 2001 note ``\myref{Mermin44}{Pots, Kettles, and Frying Pans}.''  And then some of the subsequent notes.  Maybe you'll find some of your wanted examples there.  The phraseology I used with you came up when {\Caves} and {\Schack} and I were discussing those passages later in Australia.  {\Caves} said, ``I call a beamsplitter an objective fact.''  I said, ``I call it a prior.''

\bpglm
But on the following do we perhaps agree again: this
not-quite-identified-yet something that makes the restriction -- {\bf this} must be the physical point, the trace -- and hopefully trail -- left by something we may call `reality'!
\epglm

Yes, I suspect I agree with that.

\section{19-01-04 \ \ {\it Shipbuilding, Part III} \ \ (to G. Brassard)} \label{Brassard34}

\bgb
Wow!  Had you ever mentioned this wild idea to me? This is getting more exciting by the minute.
\egb

Actually, no I hadn't.  But here's a slogan to go with it:  I told it to Howard Burton when I was visiting PI.  (So if you ever hear it re-surface, you'll know where it came from.)
\bv
     About some piece of matter:\medskip\\

     Question:  How sensitive do you thing we can make this thing to
                eavesdropping in a quantum crypto protocol?  What is
                its ultimate limit?\medskip\\

     Answer:    I don't know; let's weigh it.
\ev
(To be taken somewhat poetically, not completely accurately.)

\section{19-01-04 \ \ {\it NBI} \ \ (to G. Brassard)} \label{Brassard35}

\bgb
But (I repeat) had you told me about an interview at the Bohr Institute? When?  What?  Why? (And please don't tell me that the B.I. is in fact at Caltech!)
\egb

The Niels Bohr Institute is in Copenhagen.  It was literally the birthplace of quantum mechanics:  Heisenberg and Pauli were Bohr's postdocs at the time.

And, yes I told you about the application long ago.  But I hadn't told you that it got to the stage of an interview.

There.  That exhausts all my known interview information!

\section{20-01-04 \ \ {\it Caltech and {\Peirce}} \ \ (to A. Shimony)} \label{Shimony5}

Maybe in partial payment I should tell you that in the last couple of
evenings I've been reading a fun piece by Susan Haack called ``\,`We
Pragmatists \ldots': {\Peirce} and {\Rorty} in Conversation''.  It's a
fictitious conversation between the two men built out of quotes from
their various works.  It makes it quite clear how sly {\Rorty} is being when he continually invokes the phrase `We pragmatists \ldots'.  It'd
be nice if someone were to put together a similar dialogue between
{\Rorty} and {\Dewey}.  I sometimes wonder if the two of them are as close to each other as {\Rorty} portrays.

\section{21-01-04 \ \ {\it The Ambiguity of Language} \ \ (to R. W. {\Spekkens})} \label{Spekkens29}

\brws
Here is my response {\bf [to 15-01-04 note ``\myref{Spekkens27}{Idealism}'']}. Physical theories involve idealizations, like
fields upon which there is no back-reaction, point particles, isolated
systems, etc.  Quantum states are idealizations in the same sense.
Quantum theory is a model of a world wherein observers do have
certainty, even though our world may not admit such an idealization.
You must admit something like this occurs when it comes to quantum
state assignments that would, ideally, be adopted by an agent, but in
practice are not because they involve solving some computationally
difficult problem.
\erws
We seem to be reading two very different things.  Let's hold off discussion until {\Ruediger} and I write something that'll wallop you over the head.

\section{21-01-04 \ \ {\it WJ for T?}\ \ \ (to H. Mabuchi)} \label{Mabuchi5}

\bhm
What's your home address these days?  I wanna send you a book!
\ehm

What could be an encore to a Hello Kitty watch?  Did you manage to get a copy of {\sl William James for Toddlers}?!?

More on other matters before the day is out \ldots

\section{22-01-04 \ \ {\it Generalized Measurements} \ \ (to many)}

\noindent Dear friends in the quantum information and foundations communities,\smallskip

Many of you may not know it, but the concept of a generalized quantum measurement or positive-operator-valued measure was introduced in
\begin{itemize}
\item
E. B. Davies and J. T. Lewis, ``An Operational Approach to Quantum Probability,'' Communications in Mathematical Physics {\bf 17}, 239--260 (1970).
\end{itemize}
Last night, John Lewis passed away here in Dublin, his home of 28 years, from complications due to a recent surgery.  He was a good man, honest and upright, and left us a deep legacy.  He will be missed.

\section{22-01-04 \ \ {\it Bekenstein, Bohr, and Bohr} \ \ (to A. Peres)} \label{Peres60}

Your meeting sounds like it is going to be a great one.  I would be particularly keen to get a report on Jacob Bekenstein's talk.

\bap
I am slowly browsing throught your book ``Paulian Idea'' in the evening, when I am too tired to do anything productive, and I have until now two minor remarks. The first use of entanglement was not by {\Schroedinger} in 1935, but by
Heitler and London, Zeits.\ Phys.\ {\bf 44} (1927) 455.
\eap

Thank you for letting me know that.  I hadn't heard of that paper before.

\bap
Bohr's key idea was that ``classical'' and ``quantum'' labeled ways of
description, not properties of systems. Is this a Bayesian idea?
\eap

Not properly, no.  But what it does have in common with Bayesianism is that the quantum state (much like a probability distribution) is not a property of a system.

\bap
BTW, don't forget ``my'' Niels Bohr Medal (it's a medal, not a prize).
\eap

I certainly will not!  I had already searched on the web for information about nominations, but was not able to find it.  I will be in Copenhagen Feb 2 however for my job interview, and I was planning to ask Ole Ulfbeck and Eugene Polzik anything they knew then.  Do you have any leads on what the official procedure is?  (If I can figure that much out, I can organize the behind the scenes thing myself.)

\section{23-01-04 \ \ {\it Entropy?}\ \ \ (to P. Grangier)} \label{Grangier8}

Sorry to take a while to reply to you.  I'm on my way to Texas as I am writing you this note.  Unfortunately (or maybe fortunately) I don't have their paper here to look at, so I have to focus on the parts you've given me.

\bpg
I wanted to know your opinion about a recent preprint
(\quantph{0401021}) where the authors write on p1:
\bq\noindent
{\rm For a pure quantum mechanical state $\Psi$ the Von-Neumann definition
gives $S_H[\Psi ] = 0$. This seems to imply that a pure quantum
mechanical state is lacking a statistical nature. This is of course
not correct. For a general measurement we have uncertainty.}
\eq
\epg

There are many ways to define the von Neumann entropy of a density operator, and not simply by the raw formula.  My favorite comes from Wehrl (I believe):  the von Neumann entropy is defined to be the minimum Shannon entropy for the measurement outcomes of a fine-grained von Neumann measurement, where the minimum is taken over all such measurements.  (See pages 31 and 32, starting at Eq.\ (77), of my \quantph{0205039} to see a little discussion of this and to see it formalized.)

The von Neumann entropy of a pure state being zero does not mean that the pure state has no statistical character.  It only means that for some measurement it will give an outcome with certainty.  For the vast majority of other measurements it will give uncertainty.  (For the uncertainty on average, see Eq.\ (80) in my paper; it was a number first calculated by Bill Wootters.)

\bpg
Then they define an ``uncertainty-related'' entropy, which is non-zero
for a pure state.

What do you think about such an approach? Clearly it has the major
disadvantage of being in plain contradiction with the so-called ``third
principle of thermodynamics'', stipulating that (in most cases) the
statistical entropy is zero at zero temperature. Even worse, I wonder
whether it might be self-contradictory in some sense (that I was not
able to pin down), or at least contradictory with various points in
statistical physics. Your opinion would be appreciated.
\epg

They can make any measure they want, say something like the Eq.\ (80) I've already referred to.  But then it won't correspond to thermodynamic entropy.  You're quite right.  That number controls the free energy of a system---i.e., the energy that can actually be extracted once it is stored there---and there are already plenty of good arguments that it is the von Neumann entropy (going all the way back to von Neumann's book itself, but also detailed in Lindblad's book, and some of Jaynes's papers, etc.).  If you have stored energy in a system, and you have maximal knowledge of it---i.e., a pure state for it---then you can use that knowledge to re-extract the energy you've stored there.  That's what the von Neumann entropy being zero is about.

Now if one were to make a restriction on what operations can be done on a system, or a restriction on what measurements can be done on a system, that would change the flavor of everything.  But I doubt they've done anything like that.  It sounds to me like they're just confused.

Suggest that they read Wehrl's paper.

\section{23-01-04 \ \ {\it Flying to Texas} \ \ (to N. D. {\Mermin})} \label{Mermin106}

I'm flying to Texas as I write this to you.  I'm popping in for the
weekend for my Mom's 75th birthday.  I start heading back Monday
morning.  The following Sunday I'm off for a visit to the Niels Bohr
Institute.

Mostly I'm working on a proposal in the background, but I'll stop for
a minute to have some conversation with you.

\bdm
Somewhere in Jaynes's book (but I haven't been able to find it again)
he says that to assign something probability zero is, by Bayes's
theorem, to commit yourself to never updating the assignment of zero
on the basis of any new information.  This, of course, would be bad
news for your position.
\edm

I don't understand why you are thinking it would be `bad news'.  In
fact for our understanding it is actually good news, for it tells you
how opinions actually change:  By Darwinism.  As de Finetti points
out pretty strongly in his Probabilismo, applications of Bayes' rule
are not changes of opinion at all, but rather following one's initial
opinion through to its consequences.

In terms of a Dutch-book scenario, if one has certainty for some
event, then one is willing to bet one's life savings on it.  If that
event then does not occur, one has lost everything:  The whims of
Darwinian evolution have taken one out.  And one's beliefs then do
not propagate.

\bdm
But I believe what he has in mind is this:
$$
p(a|bx) = \frac{p(ab|x)}{p(b|x)} \leq \frac{p(a|x)}{p(b|x)}.
$$
So if $p(a|x) = 0$, then $p(a|bx) = 0$ too, for any $b$.

What he appears to overlook is that this is only so if $p(b|x) \neq 0$.
The correct statement should to be that if you assign something
probability zero, then only the occurrence of something else to which
you assign probability zero can lead you to update the first
probability to a non-zero value (assuming you are behaving rationally
--- i.e.\ following the laws of thought.)

I was reminded of this when I read in paragraph 69 of ``On
Certainty'':
\bq
   \ldots\ If I am wrong about this, I have no guarantee
   that anything I say is true \ldots
\eq
\edm

In the context of the other paragraphs surrounding {\Wittgenstein}'s 69,
{\Wittgenstein}'s point granted.  But I don't understand your `passing
thought'.  You can try again, but I think my point above already
negates it:  two deaths don't make a life.

\bdm
{\bf [Regarding {\Spekkens}.]}  Doesn't the fact that {\bf [his toy
model]} has both epistemic AND ontic states make it uninteresting?
\edm

I think you're missing something big here.  It is interesting
precisely because he has ontic states.  What his work helps argue is
that so many of these effects we find interesting in quantum
information---like teleportation, superdense coding, `nonlocality
without entanglement', the no-broadcasting theorem, entanglement
monogamy, etc., etc.---come about {\it solely\/} from the epistemic
nature of quantum states (including the less than full use of the
probability simplex, i.e., his epistemic restriction, or
knowledge-balance principle).  And in that light they are not nearly
as interesting as we once thought they were.

It's a question of refocusing yet again.  What his work shows us is
that the interesting questions lie elsewhere.  In particular, it
teaches us that the crucial question must be, `Information About
What?'  A quantum state is epistemic alright---all these examples are
meant to help you believe that---but epistemic about what?  That is
the great unanswered question.  Of course, I myself say, epistemic
about the outcomes of other interventions.  But what is left to do is
to make that sentence precise enough that, from it, we can rederive
the structure of quantum theory.  It is about formalizing THE Paulian
Idea---that is where he has taught us to refocus.  (Though he would
likely disavow much that I've said, being one who deeply longs for
nonlocal hidden variables or some such.)

Here's another way to put it.  Does THE Paulian Idea (i.e., that the
agent cannot be detached from the phenomena he helps bring about)
find any crucial expression in the phenomena of teleportation,
superdense coding, `nonlocality without entanglement', the
no-broadcasting theorem, entanglement monogamy, etc., etc.?  Probably
not.  All those phenomena seem to have more to do with the epistemic
nature of quantum states.

Instead, the Paulian Idea may well be localized solely in
Kochen--Specker phenomena.  (Which is something itself longing to be
reexpressed in Bayesian terms, but I need to write you a separate
note about that one day.)

I'll paste some related thoughts from a letter to Oliver Cohen below.
[See 15-01-03 note ``\myref{Cohen1}{Memory Lane}'' to O. Cohen.]

Does any of this help you see why I am excited by
Rob's\index{Spekkens, Robert W.} work?

\section{23-01-04 \ \ {\it The {\Wittgenstein} Bug} \ \ (to R. {\Schack})} \label{Schack79}

What a lovely {\Wittgenstein}ian reply to my little (and of course exaggerated) email!  If you had only numbered the paragraphs \ldots\

I'm stuck in a snowy Chicago airport at the moment, trying to make my way to Texas.  And I'm not thinking particularly clearly any more.  I'll try to say something sensible of your email next week.

\subsection{{\Ruediger}'s Preply}

\bq
I own {\Wittgenstein}'s complete works (in German) and I had read a lot of it in my undergraduate days. Your email made me reread quite a sample of ``On Certainty''. Is your edition split in 676 paragraphs as mine is?

My summary of the book: There are many intelligible ways of using the phrase `{\it I know x\/}', but Moore's is not one of them.

The book may be relevant for our program by helping us clarify the differences between the following statements:

I know that $2+2=4$.

I know that $12\times 12=144$.

I know that this is a chair.

I know that the earth existed 100 years ago.

I know that the ground state of the hydrogen atom is $|\psi\rangle$.

I know that the state of this hydrogen atom is $|\psi\rangle$.

I know that the outcome of this measurement will be 0.

I know that this is a beam splitter.

But then, it might not be relevant for our program. See par.~8, which states that in most contexts there is no difference between `{\it I know that x\/}' and `{\it I am certain that x\/}'. Here is another one-sentence summary of his book: `{\it I know that x\/}' is either meaningless or synonymous with `{\it I am certain that x\/}'. For our own program, we have the following:
\bq\noindent
`{\it I know that the state is $|{\rm up}\rangle$ and the {\rm Z} component of spin is measured\/}' is a reformulation of `{\it My state of belief is $|{\rm up}\rangle$ and {\rm Z}\/}'.
This reformulation is philosophically fraught and meaningless, exactly as the sentence `{\it I know this is a tree\/}' in {\Wittgenstein}'s text.
\eq
\bq\noindent
`{\it I know the outcome will be up\/}'
can be viewed as a careless way of saying `{\it I am certain that the outcome will be up\/}'
which in turn is equivalent to saying
`{\it I am willing to bet a lot of money on the outcome up\/}'.
\eq

I am rather puzzled by your sentence
\bq
\noindent Within the first five pages of that 90 page book, he already said all
that {\Ruediger} and I ever wanted to say and just kept going.
\eq
and the follow-up
\bq
\noindent Here is a man who, 52 years ago, could have tackled the Penrose question head on: How a pure quantum state can be epistemic and yet still give probability one predictions for some measurements?
\eq
I am not sure he said all we ever wanted to say about certainty.  He might even have doubted that the Penrose question as stated by you has any meaning. {\Wittgenstein} says
\begin{itemize}
\item[(i)]
You can be certain about something (`{\it the earth was here 200 years ago}') without being able to prove it. To doubt the thing would mean leaving the boundaries of the Sprachspiel (language game?).
\item[(ii)]
 The sentence `{\it I know that this is a tree\/}' is usually meaningless. It is only meaningful if there could exist reasonable doubt about this being a tree (for instance, one biology student could use the sentence to try to convince another student that the plant in question is a tree and not a shrub).
\item[(iii)]
 The sentence `{\it I know that this is a tree}' does not imply that this is a tree. It is not a sentence about the tree, but about me.
\end{itemize}

Is (i) relevant to the Penrose question? Not directly, I think.

How about (ii)? Maybe (ii) is relevant in the sense that the distinction epistemic versus ontic  is meaningless. You know (unobjectionable usage!)\ that I want to replace that distinction by the pair subjective versus nonsubjective.

Is (iii) relevant? No. Nobody claims that `{\it Mabuchi is certain that the outcome is 0\/}' implies that the outcome is 0. Mabuchi could be wrong.

What I think is that {\Wittgenstein} is trying to tackle a different question.

Here is another attempt:
\begin{itemize}
\item[(A)]
 I know the outcome is 0.
\item[(B)]
 I know the outcome will be 0.
\end{itemize}
{\Wittgenstein} would have no objections at all to (B), I believe. But (A) is tricky. It could be that all I want to say is that the outcome is 0. In that case I should just say that. On the other hand, if the outcome is in plain view and everybody can see that it is 0, then uttering (A) should make people doubt my sanity.

For us, (B) is the tricky one.

Reading {\Wittgenstein} certainly stimulates thought. Here is yet another version of our quantum question: How can we assign a pure state and at the same time grant that somebody else may legitimately assign another pure state? Or the extreme case: How can we assign a pure state and at the same time grant that somebody else may legitimately assign an orthogonal state?
Translated into {\Wittgenstein}'s terms: How can I know $x$ and at the same time grant that somebody else may know the negation of $x$? Wouldn't that mean that we play in different Sprachspiele? Which brings us back to my old and yet unanswered challenge to you: Give a satisfactory account of a crisis in a quantum measurement scenario.

I hope that some of all this makes sense. In any case, I KNOW that we are on the right track!
\eq

\section{23-01-04 \ \ {\it I Married a Teenage de Finetti?}\ \ \ (to G. Bacciagaluppi)} \label{Bacciagaluppi1}

\bv
Attack of the Giant de Finetti\\
The Bride of de Finetti\\
The Fearless de Finetti Killers (or Excuse Me but Your Dutch Book Is In My Neck) --- produced and directed by Roman Polanski, introducing
Sharon Tate\\
Plan de Finetti from Outer Space\\
Abbott and Costello Meet de Finetti\\
The de Finetti Rises\\
Attack of the Killer de Finettis\\
The Bride of de Finetti Returns!\\
The de Finetti of London\\
de de de Finetti:  de Finetti 3D!\ (Just when you thought it was safe to read David Lewis again.)
\ev

\section{26-01-04 \ \ {\it Randomness} \ \ (to G. L. Comer)} \label{Comer48}

It's been a long time since I've written you a note about randomness.  Well, maybe this one won't be {\it about\/} randomness either, but rather {\it of\/} randomness.  If all works out, it'll be a note of randomness.

So many years have passed since those early obsessions.

At the moment, I'm making my way back to Dublin from a quick trip to Texas.  My mom turns 75 tomorrow, and we threw a party for her Saturday.  Consequently I flew down Friday and now I'm flying back Monday.

The captain just told us to have a view of the Aurora Borealis out of the left-side windows!  I've never seen the Northern Lights before!!  Very exciting.

\ldots\ And so it petered out.  Dinner came, and then I found that I promptly fell to sleep.

\section{28-01-04 \ \ {\it {\Mermin}izing} \ \ (to N. D. {\Mermin})} \label{Mermin107}

Just wondering how you're doing?  Did you get my note in reply to
your `passing thoughts'?  I seem to have had some email problems,
with people not getting my messages again lately.  I hope you weren't
one of them.

My own favorite paragraph from {\sl On Certainty\/} comes from \#30:
\bq\noindent
Certainty is as it were a tone of voice in which one declares how
things are, but one does not infer from the tone of voice that one is
justified.
\eq

\section{31-01-04 \ \ {\it Just the Opposite} \ \ (to N. D. {\Mermin})} \label{Mermin108}

\bdm
I read in today's NY Times:
\bq \noindent
Mr.\ Bush no longer declares, as he once did, that he is certain that
sooner or later unconventional weapons will be found in Iraq.
\eq
Can Ludwig get him honorably off the hook?
\edm

It's just the opposite!  That was the part of the point of my last
long note to you.  For the Bayesian, the only way to back away from
certainty is death.  I.e., it's Darwin to the rescue!

\section{01-02-04 \ \ {\it Wilczek, Einstein, and Bohr} \ \ (to G. Brassard)} \label{Brassard36}

\bgb
And do you feel that Copenhagen and Bohr's ghost are receptive?
\egb

I was once at a conference where someone asked Frank Wilczek what it
was like to live in Einstein's old house in Princeton.  The fellow
asked, ``Have you ever felt Einstein's ghost?''  Wilczek answered,
``No, but every morning I wake up a little more dubious of quantum
mechanics.''

I'm going to bed now.  I'll tell you if Bohr enters my dreams.

\section{02-02-04 \ \ {\it Many Thanks} \ \ (to A. Shimony)} \label{Shimony6}

At the moment, I'm in Copenhagen, just finishing a visit to the Niels Bohr Institute.  I think the talk only went so-so.  But I did get to see Niels Bohr's office and desk!  That was quite nice.

For my leisure reading, I brought Donald Davidson's book {\sl Subjective, Intersubjective, Objective}.  I'm not finding him nearly as hard to read as everyone has been warning me:  It makes me worry that I'm missing something.

\section{03-02-04 \ \ {\it The Land of Bohr} \ \ (to G. L. Comer)} \label{Comer49}

I'm writing to you from Kastrup airport, just as I'm leaving the land
of Bohr.  Yesterday I walked into Bohr's old office and had a look at
his desk.  Thinking back on the moment, how I wish I had sat at it,
even if only briefly.  I could have; there was nothing to stop me.
But it didn't dawn on me at the moment that there was an experience
passing me by.  It reminds me of a couple of lines in The Pogues'
song, ``Fairy Tale of New York.''  The man laments ``I could have
been someone,'' to which the woman cries out in reply, ``Well so
could anyone!''

When I visited you in Meudon that year [1992], I went into the
Cathedral of Notre Dame and tried to imagine Napoleon taking the
crown away from the Pope at the last minute of his coronation.  ``I
crown myself Emperor,'' he said.

I hope I won't forget the lesson so quickly this time around.

\section{03-02-04 \ \ {\it Napoleonic History} \ \ (to D. B. L. Baker)} \label{Baker8}

\noindent Hey historian,\medskip

Can you point me toward a book where there's a good, vivid description of Napoleon crowning himself emperor?\medskip

\noindent Thanks (if you can),

\section{05-02-04 \ \ {\it Another Passing} \ \ (to G. L. Comer)} \label{Comer50}

I just did a big search to find the first mention of my dog Albert in my emails.  I came up with the following paragraph, from a 20 June 1991 letter to you:
\bq
     Worse than usual, things have been very busy in my life the last
     couple of weeks.  This is my excuse for not having been as
     conversant as usual.  My girlfriend has had me involved in a truly
     awesome construction project---rebuilding (new floor, new interior
     walls, etc.) and painting the back porch.  I'll have to send you
     pictures of the post-tornado house at some point in the future.  I
     suppose I should also mention the fact that we've recently given
     birth---a 3-4 pound, 6 week old, full-blooded Golden Retriever.
     His official name is Albert Winston Fuchs; we just call him Albert.
     Of course, I had to name my dog after (who I consider) two of the
     most prime specimens of the human race this century:  Albert
     Einstein (so that we could have long discussions about the
     foundations of physics and the determinism/indeterminism,
     event/phenomena debate---I'll play the part of Bohr) and, of
     course, John Winston Lennon (so that we could discuss the
     humanistic roots of these questions).
\eq

Albert passed away at approximately 6:45 this morning, Texas time.  My mom is going to have him buried next to my favorite dog from childhood, Frisky.  Both of them lived nearly 13 years.

Albert was a good boy and a constant friend.  And we did discuss quantum mechanics much in the lonely year of 1992.

Since I wrote you first about his coming, I'll write you first about his going.\medskip

\noindent In some sadness,

\section{05-02-04 \ \ {\it Facts, Values and Quanta} \ \ (to D. M. {\Appleby})} \label{Appleby4}

Wow!  Thank you for the color-full and value-laden paper.  I'm just
printing it out now.  I'll try to respond to you as soon as possible.

Will you be at the LSE meeting Monday February 16?

Most of the stuff in my talk you will have heard before (over and
over and over), but there is one point that is new (and I think
noteworthy).  Plus there is a line-up of several other good-looking
talks.  It'd be great to see you there.

BTW, regardless of content, I love your title!

\section{05-02-04 \ \ {\it Explain} \ \ (to H. Mabuchi)} \label{Mabuchi6}

Home from Copenhagen, and found a package waiting for me:  It's the book by Richard Powers you sent me.  Thanks!  Now explain.  Why did you send it?  Because it's a good novel, and you know that I haven't read any fiction in a while?  Or did you have something more specific in mind?

\subsection{Hideo's Reply}

\bq
As for the book: it's a novel --- read it!  I suspect you'll sympathize with one of the characters.  I declare it to be the best book I've ever read.
\eq

\section{06-02-04 \ \ {\it Replies to Your Comments} \ \ (to N. D. {\Mermin})} \label{Mermin109}

Here goes something of a reply.

\bdm
\bq\noindent\rm
[CAF wrote:]
This is where quantum information (including the collateral fields
of quantum cryptography, computing, and communication theory) has a
unique role to play.  Its tasks and protocols naturally isolate the
parts of quantum theory that should be given the most foundational
scrutiny. ``Is such and such effect due simply to a quantum state
being a state of information rather than a state of nature, or is it
due to the deeper issue of what the information is about?''
\eq
Is superconductivity (a famous quantum effect) a state of nature or a
state of information or is it due to \ldots.   I'm not trying to be
nasty. Just suggesting thoughts that may occur to some readers.
\edm
That is precisely the sort of thing that I am asking that we
examine---though I don't quite know where superconductivity fits into
the quantum-informational classification scheme yet.  How can I be
faulted for asking the same question that your straw man is?

\bdm
\bq\noindent\rm
[CAF wrote:] Recent investigations by several workers are starting to
show that a plethora \ldots
\eq
Have even you succumbed to the use of ``plethora''  (which means
excess, too much, etc.)\ simply to mean lots and lots without any
negative connotations?  Then the battle has been lost.
\edm
You forget that I did not know English before I started writing email
in 1991, and even then I did not get a real boost in my grammar until
I started corresponding with you in 1996.  It is an ongoing process
with me, and you have taught me the meaning of plethora.  It's not a
lost battle; thanks.

\bdm \bq\noindent\rm [CAF wrote:] On the other hand, other phenomena,
such as the potential computational speed-up of quantum computing,
seem to come from a more physical source: In particular, the answer to
the question, ``Information about what?''  \eq Quantum computational
speedup comes from the answer to the question, ``Information about
what?''  Do you really mean that?  \edm Yep, I mean that.  What is the
answer to ``Information about what?''? I like to say, ``the
consequences of our interventions into nature.'' But that needs to be
tightened up into a more formal statement.  It has its base in
Kochen--Specker, but I don't know how to carry the thought any further
than that at the moment.  Let me reiterate the point I made to you
about {\Spekkens} the other day: [See 23-01-04 note
  ``\myref{Mermin106}{Flying to Texas}'' to N. D. Mermin.]

\bdm
\bq\noindent\rm
[CAF wrote:] When we finally delineate an answer to this physics will
reach a profound juncture.  We will for the first time see the exact
nature of `quantum reality' \ldots
\eq
By putting the term in quotes I assume you mean we will finally
understand what the term means.   But it can also be read as saying
that there is a quantum reality that we will understand.   Not sure
what you want the reader to take from this.
\edm
Maybe both ideas.

\bdm
\bq\noindent\rm
[CAF wrote:] Trickle-down effects could be the solution to the
black-hole information paradox and even the meshing together of
quantum theory and gravitational physics---some of this can already
be seen in broad outline.
\eq
Really!?  Would you care to expand on this?
\edm
I was a little more careful in my formulation this time around.  The
`broad outline' part really only attaches to the black-hole
information paradox business.  Who would have thought that it is a
QM-foundational paradox on the order of all the usual ones, but I
think that's all it is.  In non-Bayesian approaches to quantum
mechanics, when one has a mixed state, one usually gets all fuddled
up either, a) trying to identify the REAL pure state underneath the
mixed state, or b) trying to seek a  purification of the state that
is the REAL state of some bipartite system.  From the Bayesian point
of view that is a fruitless exercise:  There is no demand that a
mixed state be derivable from any REAL pure state.  All those
ancillas, environments, and purifications are generally factitious.

Similarly for the Bayesian point of view of quantum operations, \`a la
Fuchs and {\Schack} (but maybe not yet {\Caves}).  There is no demand that
a proper trace-preserving quantum operation be derivable from some
REAL unitary operation on a larger system.  And that is all I think
is going on with the black-hole information paradox.  They can't find
any natural bipartite system to pin a unitary dynamics on.  But so
what?  It, like the EPR paradox, is a pseudo-problem.  It can be good
for clarifying concepts, but it is not a feature of nature.

\bdm
\bq\noindent\rm
[CAF wrote:] Since the beginning of quantum theory, much of what the
enthusiasts have called `foundational work' has been pseudoscience
pure and simple.  But the field can be made as respectable as quantum
theory itself, if done right. Quantum information is the technique
for the task.
\eq
Why get on the defensive?  You're the one who disparaged foundational
work.  Why should these guys share this negative view?
\edm
Just pounding on the idea that my program is quantum foundations with
that little something extra (i.e., legitimacy).

Anyway, why draw attention to it?  Look at one question I got in
Copenhagen the other day.  A fellow said, ``This is all very nice for
`Sunday physics,' but what do you do for the rest of the week?''  I
looked him straight in the eye and told him that this was the same
kind of physics that went into my calculations for Kimble's
teleportation experiment.  Apparently he needed to be told that.

People have to get it in their heads that this is serious physics, as
serious as anything else in the now vast edifice of quantum
information and computing.  They're drawing an imaginary line.

\bdm
\bq\noindent\rm
[CAF wrote:]
With regard to quantum mechanics then, the Bayesian view of
probability combined with Gleason's theorem on Hilbert-space
measures leads ineluctably to the idea that a quantum state is a
collection of gambling commitments and nothing more.
\eq
Ouch!  You've been trying for 5 years to bring me around to this point
of view and I'm still not there.  Try to tell a solid state physicist
that the BCS ground state of a superconductor is a set of gambling
commitments.  She'll kick you out of her office.  Try to tell a
chemist that a chemical bond is a gambling commitment.  Who is your
audience?
\edm
Position statement.  But your point is well taken.  Unfortunately, I
can't help it that a probability distribution is not a solid object
(as we've all been led to believe from our early educations), but I
can't lie about it either.  That point is the core of my research
program.  What one sentence could I write that would either a) soften
the blow, or b) make it all seem more reasonable?  One sentence
alone?

\bdm
\bq\noindent\rm
[CAF wrote:] What is already clear enough, nonetheless, is that from
a Bayesian approach the formal structure of quantum theory represents
not so much physical reality itself, but rather a behavior change
from standard Bayesianism for gambling agents {\it immersed\/} within
a quantum world. The trace of a `quantum reality' (which we would so
dearly love to formalize) must be found in the difference.
\eq
I've missed the punch line.  The difference between what and what?
Between behavior as described by standard Bayesianism and behavior of
gambling agents immersed in a quantum world?  It sounds like you're
saying that there is no reality for standard Bayesians but a little
bit emerges when quantum phenomena enter the story. Is this really
what you mean?
\edm
Something like that.  Standard Bayesianism makes no reference to
anything about reality.  Quantum mechanics seems to:  It is a layer
on top of Bayesianism that has to do with setting priors, and
modifying the update rule when there is physical contact between the
agent and the system he is stimulating (in old language, measuring).

\bdm
\bq\noindent\rm
[CAF wrote:] Quantum mechanics holds the promise of drastically
changing our world view.
\eq
That happened a long time ago.  Before even I was born, if you can
believe it.
\edm
I don't think so.  It changed physical practice long ago, but it has
yet to change our world view in a widespread way.  That's why there
is still so much effort to make Everett, decoherence (einselection),
consistent histories, Bohmian mechanics, modal interpretations, etc.,
etc., work.

None of those wimps have had the nerve to embrace the Paulian idea
(whatever it is).

OK, that's my response to all your comments.

Now, all that said, I am a little disappointed in what I've been able
to muster for this proposal.  I tried to be inspiring and sober at
the same time; I suspect I failed.

Writing is not an easy profession, is it?

\section{06-02-04 \ \ {\it For Your Amusement} \ \ (to H. J. Kimble)} \label{Kimble5}

I just wrote a note to David Mermin regarding some quantum foundational stuff we've been talking about, and the passage below came out.  I thought you might find some amusement in it.  When I said, ``looked him straight in the eye,'' I was imagining what you would do in a similar situation!  Probably make my actions look like cheesecake in comparison.

\bq
Look at one question I got in Copenhagen the other day.  A fellow said, ``This is all very nice for `Sunday physics,' but what do you do for the rest of the week?''  I looked him straight in the eye and told him that this was the same kind of physics that went into my calculations for Kimble's teleportation experiment.  Apparently he needed to be told that.

People have to get it in their heads that this is serious physics, as serious as anything else in the now vast edifice of quantum information and computing.  They're drawing an imaginary line.
\eq

\section{12-02-04 \ \ {\it The House Philosopher} \ \ (to J. Preskill \& H. Mabuchi)} \label{Preskill11.1} \label{Mabuchi7}

This morning I finally completed all the requirements for the IST faculty-search application.  As you'll see, I followed John's suggestion full stop:
\bjp
I also think you are right that to have a chance of success you need
to emphasize what makes you unique: your passion for and productive
contributions to the foundations of quantum theory.
\ejp
It felt good.

Attached for the heck of it are my teaching and research proposals.  It may not come through in the documents, but I gave quite a bit of thought to how I might actually pull off these course suggestions.

Thanks for giving me the incentive to go down this path, whatever the outcome:  It was a very good exercise.

\begin{center}
{\bf \large The Structure of Quantum Mechanical Information}
\end{center}
\bq
\noindent {\bf Quantum Foundations in the Light of Quantum
Information.} \ Nothing has done more to revitalize the idea that
quantum mechanics can be understood at a deep intuitive level---with
the concomitant benefits to physics this could bring---than the
development of quantum information. There is a reason for this.
Quantum mechanics is predominantly about information---plain,
ordinary Shannon information or uncertainty. Embracing this idea is
the starting point of my research program.

But it is not the end point:  No physicist would be doing his job if
he did not strive to map reality itself---that is, reality as it is
independently of any information processing agents. The issue is one
of separating the wheat from the chaff: Quantum mechanics may be
predominantly about information, but it cannot {\it only\/} be about
information.  Which part is which?  The usual way of formulating the
theory is a thoroughly mixed soup of physical and informational
ingredients.

This is where quantum information (including the collateral fields of
quantum cryptography, computing, and communication theory) has a
unique role to play.  Its tasks and protocols naturally isolate the
parts of quantum theory that should be given the most foundational
scrutiny. ``Is such and such effect due simply to a quantum state
being a state of information rather than a state of nature, or is it
due to the deeper issue of what the information is about?'' Recent
investigations by several workers are starting to show that many of
the previously-thought `fantastic' phenomena of quantum
information---like quantum teleportation, the no-cloning theorem,
superdense coding, and nonlocality without entanglement---come about
simply because of the epistemic nature of the quantum state. On the
other hand, other phenomena, such as the potential computational
speed-up of quantum computing, seem to come from a more physical
source: In particular, the answer to the question, ``Information
about what?''

When we finally delineate a satisfying answer to this, physics will
reach a profound juncture.  We will for the first time see the exact
nature of `quantum reality' and know what to do with it to achieve
the next great stage of physics. Trickle-down effects could be the
solution to the black-hole information paradox---perhaps already seen
in broad outline---and even the meshing together of quantum theory
and gravitational physics. In the meantime the approach proposed here
is a conservative and careful one; the work to be done is large. The
effort aims not to say first what `quantum reality' is, but what it
is not and gather insights all along the way.

Since the beginning of quantum theory, much of what the enthusiasts
have called `foundational work' has been pseudoscience pure and
simple. But the field can be made as respectable as quantum theory
itself, if done right. Quantum information is the technique for the
task.\medskip

\noindent {\bf Being Bayesian in a Quantum World.} \ But what is
information?  The Bayesian approach to the idea is that it has to do
with our individual expectations, however they might come about. In
particular, a probability distribution is a property of one's head,
not of what is outside it:  Its existence is manifested only in the
betting behavior of the agent who uses it. With regard to quantum
mechanics then, the Bayesian view of probability combined with
Gleason's theorem on Hilbert-space measures leads ineluctably to the
idea that a quantum state is a collection of gambling commitments and
nothing more.

There are, however, technical differences between quantum theory and
straight up Bayesianism.  For one thing, when one updates a
probability after gathering data, one uses Bayes' rule for the
transition.  But, when one updates a quantum state, one uses the
formal apparatus of quantum collapse.  The two transition rules,
though similar, are not identical.  What this (as only one example)
means is that the foundations of the Bayesian approach to probability
need to be carefully rethought when probabilities are brought to bear
on quantum phenomena.  If Bayes' rule is not to be used directly in
the quantum regime, something about the usual arguments for it must
fail to apply there. And so too, with all of Bayesian decision
theory.  The work to be done will fill a book. In fact, the finished
project should compare in scope and magnitude to the content of
Bernardo and Smith's magisterial {\sl Bayesian Theory\/} and have the
same impact on practice, {\it e.g.}~quantum feedback and control.

What is already clear enough, nonetheless, is that from a Bayesian
approach the formal structure of quantum theory represents not so
much physical reality itself, but rather a behavior change (from
standard Bayesianism) for gambling agents {\it immersed\/} within a
quantum world. The trace of this `quantum reality' we are striving to
formalize must be found in the difference.\medskip

\noindent {\bf Quantum Mechanics as a Powerful Hint.} \ In my opinion,
the most profound statement yet to come out of quantum theory is the
Kochen--Specker theorem.  For it licenses the slogan, ``Unperformed
measurements have no outcomes.''  This is just a beginning.  If one
canvasses the philosophic traditions for one that has significantly
developed this slogan, one will find the now mostly-forgotten
tradition of pragmatism fathered by William James and John Dewey. As
a source of ideas for what quantum mechanics can more rigorously
justify, no block of literature is more relevant: The connections
between the two fields cry out for systematic study. Quantum
mechanics holds the promise of drastically changing our worldview on
the wide scale. It is time to let that happen.
\eq

\begin{center}
{\bf \large Teaching Proposal for Caltech}
\end{center}
\bq
A good summary of my last 11 years would be that I have lived and
breathed quantum mechanics.  I tried to let it and the worldview it
suggests infuse every aspect of my life.  What I find most attractive
about Caltech is that it is an institution with enough resource,
foresight, and intellectual talent to play the role of a seed for
something like that on the grand scale.

The course suggestions proposed here are based on such an idea.
Proposed are five courses I would like to develop {\it over the next
few years}.  They all mutually interlock, and the insights from the
students' reactions to each are meant to play a significant role in
the development of the others.\medskip

\noindent {\bf Quantum Foundations in the Light of Quantum Information.}
This course, for graduate students and high-level undergraduates,
would be the torch bearer for the curriculum. There are deep rumbles
in the foundations of quantum mechanics, and enough technical
material has amassed to make a full survey course viable and
certainly desirable. The lectures would highlight papers by Scott
Aaronson, Marcus Appleby, Howard Barnum, Hans Briegel, Jeffrey Bub,
Paul Busch, Ad\'an Cabello, Ignacio Cirac, Rob Clifton, Carlton
Caves, Giacomo D'Ariano, myself, Nicolas Gisin, Hans Halvorson,
Lucien Hardy, Alexander Holevo, Piero Mana, David Mermin, David
Meyer, Itamar Pitowsky, Sandu Popescu, Robert Raussendorf, R\"udiger
Schack, Robert {\Spekkens}, Guifre Vidal, David Wallace, William
Wootters, and a few others.  The continuous thread running through
the lectures would be how each paper implicates the others. The goals
are 1) to leave the students with a palpable sense that real progress
is finally being made on this seemingly timeless problem, and 2) to
promote a clear vision of how they might use the techniques and ideas
explored in the course to good effect in their everyday research---be
it quantum information and computing, condensed matter theory, or
theoretical astrophysics.

The lecture notes in this course will be a crucial component in its
construction.  The plan is to put together a seamless-enough and
thorough-enough document to avail of a recent proposition for a book
of the same name by Springer-Verlag. The point is to leave a lasting
and developable mark in the practice and applications of quantum
mechanics.  Quantum foundations work is worth little if it cannot
leave that.
\medskip

\noindent {\bf Innovative Undergraduate Quantum Mechanics.}
I once wrote these words as a battle cry in one of my quantum
foundational papers,
\begin{center}
\parbox{5.2in}{\small
The task is not to make sense of the quantum axioms by heaping more
structure, more definitions, more science-fiction imagery on top of
them, but to throw them away wholesale and start afresh.  We should
be relentless in asking ourselves:  From what deep
{\it physical\/} principles might we {\it derive\/} this exquisite
mathematical structure?  Those principles should be crisp; they
should be compelling. They should stir the soul. When I was in junior
high school, I sat down with Martin Gardner's book {\sl Relativity
for the Million\/} and came away with an understanding of the subject
that sustains me today: The concepts were strange, but they were
clear enough that I could get a grasp on them knowing little more
mathematics than simple arithmetic. One should expect no less for a
proper foundation to quantum theory. Until we can explain quantum
theory's {\it essence\/} to a junior-high-school or high-school
student and have them walk away with a deep, lasting memory, we will
have not understood a thing about the quantum foundations.}
\end{center}
We are certainly far from that dreamy ideal today, but a soberer
version of the lesson is the same: The only meter for progress in
quantum foundations is the ease of quantum practice and the ease with
which students absorb the subject.

To that end, quantum information has already taught us several
lessons at the methodological level that can be made use of today.
What I envision is developing a course that takes its base in Hideo
Mabuchi's previously successful Ph125 and trying to tweak it to
perfection. The great lesson of quantum information is how much of
the meaning and mystery of quantum mechanics can be isolated in two
and three-level systems and their interactions. Therefore it is
useful to spend a solid amount of time sussing out that regime as
thoroughly as possible.  Simple quantum information effects like the
no-cloning theorem, quantum cryptography, Bell inequalities,
entanglement monogamy, quantum teleportation, and the Kochen--Specker
theorem are not `Sunday physics', but the very heart of the matter.
More than anything, they give an intuition for the necessity of the
Hilbert-space formalism.

The trick is in pushing that intuition deep into the students'
psyche, while at the same time not losing sight of the need to get
through a semi-standard curriculum of harmonic oscillators,
perturbation theory, hydrogen atoms, the Zeeman effect, etc., as
efficiently as possible. How to do it?  Some of physics is just
roll-your-sleeves-up hard work, but it is a good gamble that as the
course develops much of the method will lead its own way.
Inconsistencies in thought and method have a way of declaring
themselves---it is that system of checks and balances which has
tugged research in `quantum mechanics as quantum information' in a
uniform direction ever since the beginning anyway.  In all
seriousness, this is a case where basic quantum theory stands to gain
as much from the reaction of the undergraduate as any amount of
cloistered introspection.
\medskip

\noindent {\bf Bayesian Theory.}
Bayesian Theory is an approach to probability theory and statistical
decision-making that is quickly gaining traction in all aspects of
science. Jos\'e Bernardo and Adrian Smith characterize it this way in
their masterwork on the subject:
\begin{center}
\parbox{5.2in}{\small
The theory and practice of Statistics span a range of diverse
activities \ldots. What is the nature and scope of Bayesian
Statistics within this spectrum of activity? Bayesian Statistics
offers a rationalist theory of personalistic beliefs in contexts of
uncertainty, with the central aim of characterizing how an individual
should act in order to avoid certain kinds of undesirable behavioral
inconsistencies. The theory establishes that expected utility
maximization provides the basis for rational decision making and that
Bayes' theorem provides the key to the ways in which beliefs should
fit together in the light of changing evidence.  The goal, in effect,
is to establish rules and procedures for individuals concerned with
disciplined uncertainty accounting. The theory is not descriptive, in
the sense of claiming to model actual behavior.  Rather, it is
prescriptive, in the sense of saying ``if you wish to avoid the
possibility of these undesirable consequences you must act in the
following way.''}
\end{center}
My own interest in Bayesianism sprouts from the sensibleness it helps
bring to quantum mechanics:  For, one might consider much of the
formal structure of quantum theory as a prescriptive set of rules for
uncertainty accounting in the light of some (yet to be fleshed out)
more basic features of the world.  In other words, quantum theory is
not so much about physics itself, but rather decision theory in the
{\it presence\/} of physics.  This is the research program of `Being
Bayesian in a Quantum World.'

It is crucial for this research program, however, that the physics
community gain a proficiency and comfort with the subject long
previous to that.  This motivates me to try to introduce and maintain
a top-notch, Bayesian theory course for graduate students and
high-level undergraduates.  The course, based initially on Bernardo
and Smith's text {\sl Bayesian Theory}, would have a university-wide
appeal, being applicable to all the sciences and some fields in the
humanities. Covering both foundational topics and rigorous
technicalities in statistical modelling, it should be particularly
relevant to those students associated with the IST Center for the
Mathematics of Information.
\medskip

\noindent {\bf Quantum Mechanics and Anti-representationalist
Philosophies.} There are various threads connecting the quantum
research program proposed here to a wider philosophical tradition,
which to my knowledge has never been greatly examined in this
context.  The tradition comes under the rubric of what Richard Rorty
calls `anti-representationalist philosophies.'  This tradition,
spearheaded by the pragmatism of William James and John Dewey, also
includes thoughts of (the later) Ludwig Wittgenstein, Martin
Heidegger, Donald Davidson, Hilary Putnam, Rorty himself, and several
others.  How else can one understand the implications of the
Kochen--Specker theorem than by realizing it hints at something like
James' analysis of the concept of `truth'?  How else can one make
sense of a Bayesian take on pure quantum states than to explore the
same paths as Wittgenstein in his book {\it On Certainty\/}?

Since becoming immersed in the subject, I have found nothing more
exciting than these trains of thought.  For they indicate the extent
to which quantum foundations research may be the tip of an
iceberg---indeed, something with the potential to drastically change
our worldview, even outside the realm of physical practice. William
James once said this of the method of pragmatism,
\begin{center}
\parbox{5.2in}{\small
[Within it] theories \ldots\ become instruments, not answers to
enigmas, in which we can rest.  We don't lie back upon them, we move
forward, and, on occasion, make nature over again by their aid.}
\end{center}
Two thousand years hence, this may be the greatest, most lasting
legacy of quantum theory.

Within the humanities division, or possibly shared between physics
and the humanities, I would like to develop a course for graduates
and undergraduates that would meditate on these threads and tie them
together as much as possible. The lectures and discussions would
focus on a selection of materials drawn from my personal library of
nearly 300 books on the subject. The goal with respect to interested
philosophy of science graduate students would be to introduce them to
enough material and provide them with enough guidance for a solid
thesis topic in the subject.
\medskip

\noindent {\bf Writing Physics.}
The greatest physics laboratory for me has always been my keyboard.
With the practice of almost 17,000 emails since my first notebook
computer in 1997 (a plethora of which are long and detailed), I
believe I have taught myself a skill worth passing on to physics
students in general:  It is in line with what David Mermin has called
{\it writing physics},
\begin{center}
\parbox{5.2in}{\small
Physicists traditionally replace talk about physics by a mathematical
formalism that gets it right by producing a state of compact
nonverbal comprehension. The most fascinating part of writing physics
is searching for ways to go directly to the necessary modifications
of ordinary language, without passing through the intermediate
nonverbal mathematical structure. This is essential if you want to
have any hope of explaining physics to nonspecialists. And my own
view is that it's essential if you want to understand the subject
yourself.}
\end{center}
My trick has been to never write for my own mind, but always for
someone else's.  That is why email has been essential to me. (Some of
my best papers, technical and nontechnical, have been cut and pastes
from old emails.)  Indeed, email provides a kind of immediate
feedback that no other kind of writing can.

Based on these ideas, and Mermin's own from his paper of the same
name, I would like to create an innovative technical writing course
for undergraduates within IST.  The goal is to have better papers,
and better scientists because of better papers.  The goal is to have
IST writing physics.
\eq

\section{13-02-04 \ \ {\it Anniversary Cheer} \ \ (to J. Preskill)} \label{Preskill12}

\noindent\underline{\bf NOTE}: With hindsight, this letter seems to represent the first time I had incorporated William James's ``will to believe'' doctrine into my palette of thinking.\medskip

What welcome news to wake up to this morning, and what a great day to receive it on!  Ten years ago today, I worked up the nerve to trot across a neighborhood park and asked Kiki to dinner.  It was our first date.  (You can read about the weeks leading up to it in the story of our dog Wizzy, in ``Quantum States: W.H.A.T.?''\ on my webpage, pages 142--143.)  [See 16-02-02 note ``\myref{ShouldNotPass}{Some Things Should Not Pass}.''] Lesson learned:  Without nerve first, there isn't hope second.

\section{15-02-04 \ \ {\it Step Forward} \ \ (to G. Brassard)} \label{Brassard37}

Let me attach the proposals I finally sent off to Caltech last week.  [See 12-02-04 note ``\myref{Mabuchi7}{The House Philosopher}'' to J. Preskill and H. Mabuchi.]  I hope you enjoy them.

Today I fly out for the LSE meeting on Probability and Quantum Mechanics, and to give the ``Popper Colloquium'' the next day.  I think I need to do a particularly good job here, as we stand to gain a lot of interest from the philosophers.

I've gotta, gotta, gotta get to work on the Asher paper for us.  Let's hope that starts up in about a week:  In any case, I'm sure the editor of the special issue will wait for us.  I know him personally!

\section{17-02-04 \ \ {\it From London} \ \ (to A. Fine)} \label{Fine4}

Sorry to take so long to reply:  I'm at the LSE now and going to give my first colloquium ever to a philosophy department.  (Look in the obituary column tomorrow.)

\baf
I think we have some serious disagreements about how to think about
QM, the role of pragmatic considerations, etc.  But, then, choose any two folks who work in this area and you'll find the same sort of thing!
\eaf

Interesting.  I hope you'll explain as you get a chance.  In particular, it would be useful to me if you can pinpoint where you already disagree with my \quantph{0205039} ``QM as QI (and only a little more)''.  Following that, I'd like to hear about the disagreements with my later thoughts (for instance, the ones in the proposals) after pragmatism sunk more deeply into my soul.

\section{21-02-04 \ \ {\it Philosophical Training} \ \ (to A. Shimony)} \label{Shimony7}

I am just back in Dublin from a (mentally) long week in London at the LSE and then at All Souls College in Oxford.  It was my first extended engagement with full-fledged philosophy departments.  I tell you, I felt like I was in the frying pan the whole time!  It was defense, defense, defense; they made me earn my travel expenses.  But---no doubt---I got some good training in the process.

\section{21-02-04 \ \ {\it My Visit} \ \ (to S. Savitt)} \label{Savitt4}

I am just back in Dublin from an exhaustingly long week at the LSE in London and All Souls College in Oxford.  It was my first real engagement with philosophy departments.  I tell you, I felt like I was in the frying pan the whole time!  It was defense, defense, defense.  What does not kill you makes you stronger?  \ldots\  Actually, except for the constant effort I had to put into keeping from being bushwhacked, I think I fared pretty well.  Even picked up a convert or two \ldots

Looking forward to seeing you.

\section{22-02-04 \ \ {\it Fidelities}\ \ \ (to S. L. Braunstein)} \label{Braunstein11}

\bslb
I have a question, you may know the answer to. What is the fidelity
one needs to surpass to demonstrate teleportation of a qubit from an
alphabet $|0\rangle$, $|1\rangle$, $|0\rangle \pm |1\rangle$, $|0\rangle \pm i |1\rangle$?
Presumably it's higher than 2/3. Do you know of any reference which
has calculated this?
\eslb

No, it's actually precisely 2/3.  In fact, in $d$ dimensions, if one uses the $d(d+1)$ vectors drawn from a complete set of mutually unbiased bases, the breaking point fidelity one needs is $2/(d+1)$.

I'll attach the draft of a paper that I sorely need to finish which contains these calculations.  (It's a paper for the Holevo festschrift, and now the editors are really bothering me for it.)  The bit about MUBs in particular can be found around equations 43 through 46.

I hope things are going well for you.  For myself, I've been traveling far, far, far too much to get any work done.  I doubt the viewers of {\tt quant-ph} even remember me anymore.

\section{24-02-04 \ \ {\it Weird Forms of Realism} \ \ (to K. Brading)} \label{Brading1}

Thanks for the note; I very much appreciate it.  I wish I had met you at Oxford.

I think the best answer I can give to your question at the moment is recorded in Sections 4 and 5 of my paper, ``The Anti-{\Vaxjo} Interpretation of Quantum Mechanics.''  (You can find the paper at my website; there is a link below.)  Since then, I have developed those thoughts somewhat, but I think the gist of what I'm striving for is already expressed in a fairly easy-to-read form in that older paper (or, I should say two emails).  Reading over the sections again, I think they answer you quite directly.

I did a Google search on your name a little while ago to try to figure out who I'm talking to.  It didn't help much.  But I did learn that you've just obtained a faculty position at U. Notre Dame.  \ldots\ Wait!  I did meet you:  It was at the tea before the talk, wasn't it?  Right.  Well anyway, congratulations on your new position!  Carry out some great work there.

\subsection{Katherine's Preply}

\bq
I was at your talk last Thurs at the Phil of Physics seminar in Oxford, and wanted to drop you a line to say how much I enjoyed it. I don't know whether you knew that you were coming into a group dominated by committed realists --- of various stripes, but realists nonetheless --- and I thought you did a great job. Personally, I found your approach very congenial indeed, but I wondered whether you wanted to put your position concerning scientific realism as strongly as it appeared. You seemed to resist any temptation to talk about `The World', and yet one motivation you were giving for your approach is that we may hope that it will lead us to a new route `back to the world'.

Would you find the following way of expressing things acceptable, or is there a reason why you feel you need to go further:
\bq\noindent
The world is such that if I update my beliefs according to these rules (QM\ldots) then I will be successful in getting around.  But there is no inference from this to realist claims about `how the world is in itself', and without relation to my activity of going around in it. So when I make some claim, such as `The photons in this beam are vertically polarized', that's shorthand for `The world is such that if I believe that the photons in this beam are vertically polarized (and if I act accordingly) then I am likely to be successful in my further related activities'. Usually this shorthand doesn't get us into trouble (`There's a tree over there'), but sometimes it does.
\eq
How I understood the message of your talk was that, rightly understood, QM forces us to move to the epistemically modest longhand, rather than the naively realist shorthand, BUT that there are clues as to how we might then go forwards into approaching `the world in itself' again. I very much liked the suggestive discussion of disturbance at the end of your talk.

Anyway, whatever you think of the above --- if anything --- all I really wanted to do was to write a brief line saying that there were at least two not-so-realist bodies in the audience --- myself and Peter Morgan --- and that I for one like very much what you're doing.
\eq

\section{25-02-04 \ \ {\it Promised Message} \ \ (to N. D. {\Mermin})} \label{Mermin110}

\bdm
The following true story will appeal to nobody but you and maybe not
to you either, but here it is.

Do you remember my funny way of solving Bernstein-Vazirani (that I
spoke on at the {\Bennett} symposium)? You want to determine an n-bit
number $a$, and are given a $U$ that acts on a $n$-bit input register
containing $x$ to shift the one bit output register by the bit-wise
modulo-2 inner product of $a$ and $x$.  My solution starts by
replacing $U$ by a bunch of cNOT gates --- one for each non-zero bit
of a controlled by the corresponding bit of $x$, all of them targeted
on the output register.

Anyway I showed this to my class of physicists and computer
scientists, and somebody remarked that it was pretty clumsy having to
reconfigure the hardware every time you wanted to do it for a
different value of $a$.

Instantly one of the CS students said no, that wasn't necessary. All
you needed was an additional $n$-bit register into which you put $a$.
The hardware was then fixed, consisting of $n$ doubly controlled NOTS
(i.e.\ Toffoli gates) all targeted on the output register and
controlled by pairs of corresponding bits of $x$ and $a$.

Thinking back on this a few days later it struck me that the way I
presented it the choice of Hamiltonian (in the form of $U$)
associated with the different possible $a$'s was objective ---
different arrangements of the classical hardware.   But the way the
sharp CS student suggested doing it changed that choice of
Hamiltonian to a specification of the state of the additional
register. The selection of the Hamiltonian from among all possible
Hamiltonians in his scheme was on exactly the same footing as (and in
fact was identical to) the specification of a state vector.

That's all.  Sounds less entertaining now that I've written it out.
\edm

Nope, a very deep point I would say \ldots \  Sounds worthy of a
slogan. How about, ``A quantum operation is just a quantum state in
disguise.'' ??

\section{26-02-04 \ \ {\it Thanks} \ \ (to A. Peres)} \label{Peres61}

Thanks for writing me back; I really had started to worry.

\bap
Please remind me where you have written ``there are no quantum states
(the ghost of Bruno de Finetti)''. My question is related to Netanel's
PhD work on quantum gravity. In some spacetimes, there are obviously
no quantum states.
\eap

The place where I made the most to-do of the phrase ``quantum states do not exist'' was in \quantph{0205039}, ``Quantum Mechanics as Quantum Information (and only a little more)''.  However the phrase made its first public appearance in \quantph{0105039}, {\sl Notes on a Paulian Idea}.  In fact, it comes from a letter I wrote to you dated 1 December 1998:
\bq\noindent
\subsection{Letter to Asher Peres, 1 December 1998, ``Here Comes the Judge''}

\bap
The measuring process is an external intervention in the dynamics of
a quantum system.
\eap

I know that I've already expressed what I'm about to say several
times, but let me just set things off to a good start again: I really
like this turn of phrase!  For years now, I have believed exactly
what you say, namely,
\bap
[I]t is preferable not to use the word ``measurement'' which suggests
that there exists in the real world some unknown property that we are
measuring.
\eap
but I had never invented such a clever phrase to express it.  For a
long time instead I made use of the word ``creation'' in my
tract-like emails on the subject.  In some ways, I still
think---taking my cue from John {\Wheeler}---that that word captures
a certain central truth of quantum theory, but your word is more
encompassing.  I think it captures better the idea that quantum
measurement is a double-edged sword, learning {\it and\/} creating.

But I wouldn't be doing my scholarly duty, if I didn't remind you of
some passages of Niels {\Bohr}, whose words you seem to have so much
respect for.  In {\Bohr}'s 1949 article ``Discussion with
{\Einstein} on Epistemological Problems in Atomic Physics,'' he
wrote:
\bq
[At the Solvay meeting in 1927] an interesting discussion arose about
how to speak of the appearance of phenomena for which only
predictions of statistical character can be made.  The question was
whether, as to the occurrence of individual effects, we should adopt
a terminology proposed by {\Dirac}, that we were concerned with a
choice on the part of `nature,' or, as suggested by {\Heisenberg},
we should say that we have to do with a choice on the part of the
`observer' constructing the measuring instruments and reading their
recording. Any such terminology would, however, appear dubious
since, on the one hand, it is hardly reasonable to endow nature with
volition in the ordinary sense, while, on the other hand, it is
certainly not possible for the observer to influence the events
which may appear under the conditions he has arranged.  To my mind,
there is no other alternative than to admit that, in this field of
experience, we are dealing with individual phenomena and that our
possibilities of handling the measuring instruments allow us only to
make a choice between the different complementary types of phenomena
we want to study.
\eq
And, more to the point, in his 1958 article ``Quantum Physics and
Philosophy: Causality and Complementarity,'' he wrote:
\bq
In the treatment of atomic problems, actual calculations are most
conveniently carried out with the help of a {\Schroedinger} state
function, from which the statistical laws governing observations
obtainable under specified conditions can be deduced by definite
mathematical operations.  It must be recognized, however, that we are
here dealing with a purely symbolic procedure, the unambiguous
physical interpretation of which in the last resort requires a
reference to a complete experimental arrangement.  Disregard of this
point has sometimes led to confusion, and in particular the use of
phrases like ``disturbance of phenomena by observation'' or
``creation of physical attributes of objects by measurements'' is
hardly compatible with common language and practical definition.

In this connection, the question has even been raised whether
recourse to multivalued logics is needed for a more appropriate
representation of the situation.  From the preceding argumentation it
will appear, however, that all departures from common language and
ordinary logic are entirely avoided by reserving the word
`phenomenon' solely for reference to unambiguously communicable
information, in the account of which the word `measurement' is used
in its plain meaning of standardized comparison.
\eq

Do you see these passages as meshing well with your new terminology?
I don't really, but then again I don't feel that {\Bohr} was all that
right when he came to these points.  I think it might be useful for
your ultimate reader to add some words about your opinion on this.

\bap
As a concrete example, consider the quantum teleportation scenario.
The first intervention is performed by {\Alice}: she has two
spin-$\frac{1}{2}$ particles and she tests in which {\Bell} state these
particles are.
\label{Naugahyde}
\eap
In which state they {\it are\/}?!?!  There's got to be a better way
of putting this \ldots\ especially one more consistent with your
whole view of the measurement process.

\bap
I do not want to use the word ``histories,'' which has acquired a
different meaning in the writings of some quantum theorists.
\eap
What happened to your nice word ``chronicle''?  I liked it a lot.
Can't you find some way to reinstate it?  I guess I liked it because
it always reminded me of a passage that I like to quote from
{\Pauli}. In a letter to Markus {\Fierz} in 1947, he wrote:
\bq
\noindent
Something only really happens when an observation is being made
\ldots\,.  Between the observations nothing at all happens, only time
has, `in the interval,' irreversibly progressed on the mathematical
papers!
\eq

\bap
Quantum mechanics is fundamentally statistical, and any experiment
has to be repeated many times (in a theoretical discussion, we shall
imagine many replicas of our gedanken\-experiment).
\eap
\bap
Each one of these records has a definite probability, which is
experimentally observed as its relative frequency among all the
records that were obtained.
\eap
\bap
Note that the ``detector clicks'' are the only real thing we have to
consider. Their probabilities are objective numbers and are
{\Lorentz} invariant.
\eap

As a disciple of the Reverend {\Bayes}, you should know that I
strongly dislike all these expressions.  A good Bayesian would say
that probability quantifies a state of knowledge.  It is that and
that alone, meeting an operational verification {\it only\/} through
a subject's betting behavior.  In particular, probability has no {\it
a priori\/} connection to the frequency of outcomes in a repeated
experiment.  It matters not whether that repetition is real or,
instead, imaginary and virtual.

Let me just try to drive this home with two of the simplest possible
examples.

(1)  Consider a coin for which you have no reason to believe that a
head will occur over a tail in a toss.  Now imagine that you flip
that coin an infinite number of times, tabulating the number of heads
and tails.  Do you really believe that a frequency of precisely 1/2
{\it must\/} occur in the infinite limit?  Answer this question to
yourself very honestly.  What is to bar you from getting a head as
the outcome in literally every single toss?  What is there in the
coin to favor a random-looking sequence to a nonrandom one? Nothing.
To say something like, ``Well the nonrandom-looking sequences have
probability zero'' is just to beg the question.  For one thing, you
have to invoke the concept of probability again to even say it.  For
another, even if you allow yourself that, it still carries no force:
just take the set of nonrandom-looking sequences and add to it any
sequence that you would consider a valid random one (i.e., one that
you believe could be generated by a repeated coin toss).  That set
still has probability-measure zero, but now you would have to say
that that means the random-looking sequence could not be generated by
your imaginary coin.  The point:  probability has no direct
connection to frequency.

(2)  Consider now the case where I toss a coin repeatedly for you. I
assure you strongly that the coin is weighted 80--20 heads vs.\ tails
{\it or\/} 20--80, but I steadfastly refuse to tell you which way it
goes.  What probability would you ascribe to the outcome of a head
upon my first toss?  Fifty-percent of course. The point is, my
probability ascription is not your probability ascription.  As I toss
the coin ever more, if you are rational, your probability ascription
will {\it very likely\/} approach mine, but there are no absolute
assurances.  The point:  again probability has no direct connection
to frequency.  But also, two perfectly rational people can make
different probability assignments to the same experiment without
being inconsistent with each other.  There is nothing objective about
a probability assignment per se.

If you'd like to understand better this point of view, then I'd
strongly suggest the wonderful collection of essays, ``Studies in
Subjective Probability,'' edited by Henry~E. {\Kyburg}, Jr., and
Howard~E. {\Smokler} (Wiley, NY, 1964).  Also there is a later
edition of the book with a few other essays.  For a much more
in-depth study I couldn't recommend anything more than Ed {\Jaynes}'
book, ``Probability Theory: The Logic of Science.''  It unfortunately
may never be published properly, but preprints of the parts that were
written can be found at the ``Probability Theory as Extended Logic''
web page, \myurl{http://bayes.wustl.edu/}.

\bap
[P]robabilities are objective numbers \ldots\,. \ On the other hand,
wave functions and operators are nothing more than abstract symbols.
They are convenient mathematical concepts, useful for computing
quantum probabilities, but they have no real existence in Nature.
\eap

I think it is really cute the way you cite {\Stapp} here!

But again I want to come back to the last point above.  A good
Bayesian already knows that probabilities cannot be taken to be
objective numbers, existent ``out there'' in nature.  And a good
quantum physicist similarly knows that the wave function is {\it
not\/} something ``out there'' existent in nature.  The similarity of
these two points of view is not an accident: it had to be the case!
For, from one point of view, the wave function is nothing more than a
compendium of probabilities. It may be the case that ``unperformed
measurements have no outcomes,'' but it is not the case that
unperformed measurements have no probabilities!  If we specify the
probabilities of the outcomes for every {\it potential\/}
measurement, then we have specified the wave function.  There cannot
be a difference in the objectivity levels of these two mathematical
objects:  they are both abstract symbols that summarize our states of
knowledge.  Their empirical meaning comes about precisely in how a
rational being would behave in the light of that knowledge.

There is a famous quote in Bayesian probability theory due to Bruno
{\Finetti}.  It starts out the two volumes of his book.
\bq
\small
\noindent
My thesis, paradoxically, and a little provocatively, but nonetheless
genuinely, is simply this:
\begin{center}
PROBABILITY DOES NOT EXIST.
\end{center}
The abandonment of superstitious beliefs about the existence of
Phlogiston, the Cosmic Ether, Absolute Space and Time, \ldots, or
Fairies and Witches, was an essential step along the road to
scientific thinking. Probability, too, if regarded as something
endowed with some kind of objective existence, is no less a
misleading conception, an illusory attempt to exteriorize or
materialize our true probabilistic beliefs. \normalsize
\eq
In contrast, my paper ``Bayesian Probability in Quantum Mechanics''
with {\Caves} and {\Schack}, opens up with the following lines:
\bq
\small
\noindent
My thesis, paradoxically, and a little provocatively, but nonetheless
genuinely, is simply this:
\begin{center}
QUANTUM STATES DO NOT EXIST.
\end{center}
The abandonment of superstitious beliefs about the existence of
Phlogiston, the Cosmic Ether, Absolute Space and Time, \ldots, or
Fairies and Witches, was an essential step along the road to
scientific thinking. The quantum state, too, if regarded as something
endowed with some kind of objective existence, is no less a
misleading conception, an illusory attempt to exteriorize or
materialize the information we possess.\\
\hspace*{\fill} --- {\it the ghost of Bruno de {\Finetti}} \normalsize
\eq
I don't know of any better way to express it than this.

\bap
The physical evolution that leads to Eq.~(1) is the following. The
intervener receives a quantum system in state $\rho$ and adjoins to
it an auxiliary system (an {\it ancilla\/}) in a known state
$\rho_a$. The composite system, in state $\rho\otimes\rho_a$, is
subjected to a unitary transformation \ldots\
\eap

In my present way of speaking, I like to make this sound much less
absolute.  I usually say that this is only one {\it representation\/}
of a superoperator.  It is the superoperator that is the only thing
that need be given; anything else is just one way or another of
thinking about it.  I guess the point is I see no reason to give
unitary evolutions and von Neumann measurements a special status in
the axiomatics of the theory.  POVMs and superoperators are perfectly
good (and much less specific) starting points for the theory. You can
certainly do as you please, but it seems to me that there is a lot to
be gained from this point of view.  For instance, it seems to me to
de-emphasize the point of view that Charlie {\Bennett} labels ``the
Church of the Larger Hilbert Space'' (which is, as he admits, a
euphemism for the many-worlds interpretation).

\bap
If we wish to consider only the states just before and after the
intervention, without entering into the detailed dynamics of the
intervention, the result appears as a discontinuous jump of the wave
function (often called a ``collapse'').  Clearly, this jump is not a
dynamical process. It results from the introduction of an ancilla,
followed by its deletion, or that of another subsystem. The jump is
solely due to abrupt changes in our way of delimiting what we
consider as the quantum system under study.
\eap

I agree with you to some extent here, but not the whole way.  The
abrupt change that comes about when we conceptually delete a
subsystem from a larger composite system corresponds to a {\it
partial trace}.  Where does the {\it random\/} jump come from that
corresponds to the measurement outcome label $k$?  I've heard you say
things like this many times, i.e., that the statistics in quantum
mechanics comes about because of a mismatch in two languages, the
quantum and the classical---your talk at QCM was just one
example---but how does one see that you're not just talking about a
partial trace here?  I am perfectly willing to take
measurement/intervention as a primitive of the theory---it's the very
thing that gives the theory meaning---but you seem to want to go
further, to have it fall out of something more primitive, namely the
act of drawing a conceptual line in an overall unitary dynamics.  I
guess I still don't see it.

\bap
Summing over them is like saying that peas and peanuts contain on the
average 42\% of water, instead of saying that peas have 78\% and
peanuts 6\% [25].
\eap

I loved the citation here!  You don't think there might be a
connection between {\Stapp} and the USDA, do you?

\bap
Between these two events, there is a ``free'' (that is,
deterministic) evolution of the state of the quantum system. What
distinguishes such a free evolution from an intervention is that the
latter has unpredictable output parameters, for example which one of
the detectors ``clicks,'' thereby starting a new chapter in the
history of the quantum system. As long as there is not such a
branching, the evolution will be called {\it free\/}, even though it
may depend on external classical fields, that are specified by the
classical parameters of the preceding interventions.
\eap

I like this distinction, but I cannot completely agree with it. Take
any trace-preserving completely positive map that is not unitary,
i.e., most any quantum channel like the amplitude damping channel or
the depolarizing channel will do.  It gives rise to no ``branching''
changes in the quantum state, but I think one would be hard pressed
to call it a ``free'' evolution.

I overheard {\Herb} {\Bernstein} once say something like, ``Of course
there are two kinds of evolution in quantum mechanics; there are the
times when we learn something and then there are the times when we
learn nothing.''  The line struck me.  I know that this is something
like what you're trying to get at, but the cut isn't between unitary
and nonunitary then.

\bap
Quantum mechanics asserts that during a free evolution the quantum
state undergoes a unitary transformation.
\eap

This needs cleaning up exactly because of the point above.  By the
way, note that there is a typo in the sentence following Eq.~(7).

\bap
Note that ${\rm Tr}(\rho_f)={\rm Tr}(\rho'_f)$ is the joint
probability of occurrence of the records $k$ and $\mu$. This is the
only observable quantity in this experiment. \ldots\ \ {\Einstein}'s
principle of relativity asserts that both descriptions given above
are equally valid. Formally, the states $\rho_f$ (at time $t_f$) and
the state $\rho'_f$ (at time $t'_f$) have to be {\Lorentz} transforms
of each other.
\eap

I am afraid that this might be construed to mean that the only way
the equality of the traces can be satisfied is if $\rho_f$ and
$\rho'_f$ are ``{\Lorentz} transforms'' of each other.  Surely that's
not true and that's not what you mean.  Maybe you could clarify this
a bit.

\bap
As a further simplification, let all the $U$ and $V$ operators be
unit matrices, so that the two particles are really free, except at
the intervention events.
\eap

What do you mean by the phrase ``really free'' here?  Is it a joke?

\bap
We thus have to accept that unit matrices of any order may also be
legitimate {\Lorentz} transforms of each other.
\eap

So I guess what you're meaning is that you want this problem to
partially define what is meant by the very term ``{\Lorentz}
transform?''  I'm looking forward to understanding the business about
Green's functions much better.  After that I'll come back for some
more substantial comments on everything in this section following
Eq.~(10).

\bap
Quantum nonlocality has led some authors to suggest the possibility
of superluminal communication.
\eap

Notice the correction to {\it your\/} spelling mistake \ldots\ ahem.
Grin.  Anyway, in this connection let me point you to a very nice
article by a philosopher:  J.~B. {\Kennedy}, ``On the Empirical
Foundations of the Quantum No-Signalling Proofs,'' Phil.\ Sci.\ {\bf
62}, 543--560 (1995).  The point he makes---and I think validly
so---is that any so-called ``proof'' that quantum mechanics cannot
support superluminal signalling by using entanglement is essentially
circular.  It's not that the ``proofs'' are invalid, but essentially
that there was no use in doing them in the first place: no-signalling
is much more of an {\it axiom\/} than a result of the standard
structure of quantum mechanics.  {\Kennedy} argued by way of digging
into the historical references, finding for instance {\vonNeumann}'s
original motivation for introducing the tensor product rule for
combining Hilbert spaces---it was essentially to block the
possibility of superluminal signalling!  You might enjoy reading the
paper.

\bap
The classical-quantum analogy becomes complete if we use statistical
mechanics for treating the classical case. The distribution of the
bomb fragments is given by a {\Liouville} function in phase space.
When {\Alice} measures ${\bf J}_1$, the {\Liouville} function for
${\bf J}_2$ is instantly altered, however far {\Bob} is from
{\Alice}. No one would find this surprising, since it is universally
agreed that a {\Liouville} function is only a mathematical tool
representing our statistical knowledge.  Likewise, the wave function
$\psi$, or the {\Wigner} function, which is the quantum analogue of a
{\Liouville} function, are only mathematical tools for computing
probabilities. It is only when they are considered as physical
objects that superluminal paradoxes arise.
\eap

Indeed you have captured the point here.  Part of this, i.e., that
the proper comparison is between state vectors and {\Liouville}
distributions, is what {\Carl} and I were striving to convey in our
paper for Nathan {\Rosen}'s festschrift---so I'm already very
sympathetic to this.  (Of course I know that you already made this
point long ago in your paper ``What is a State Vector?'')  More
particularly, though, this is one thing that Steven van {\Enk} and I
had planned to write in a silly little paper---titled ``Entanglement
is Super \ldots\ but not Superluminal!''---for our contribution to a
book on superluminal signaling!  I hope you won't mind the overlap.

The thing that holds so many up from immediately grasping this point
of view is the {\it difference\/} between the quantum and classical
states of knowledge (i.e., those things that are changing
instantaneously as you say).  Classically, the probability is
attached to an existent property.  ``The particle has a position; I
just don't know it.  I capture what little bit I do know with a
probability assignment.''  Quantumly, the probability is attached to
{\it potential\/} measurement outcomes and nothing more.  How is it
that my probability assignment for some potential measurement outcome
can change from 50-50 to complete certainty when there is no sense in
which that measurement outcome can already be said to be ``out
there'' without the performance of a measurement?  That's the thing
that has people perplexed and, I think, is the source of their
confusion about using entanglement for communication.

\bap
In our approach, there is no `delayed choice' paradox. It is indeed
quite surprising that JAW, who is a hard core positivist, introduced
this `delayed choice' idea.
\eap

Sometimes it's pretty hard to figure out what precisely it is that
John is trying to say.  Perhaps you shouldn't be overly harsh on him.
In some of his discussions of ``delayed choice'' he makes it very
clear that he is not talking about affecting the ``past'' itself
(whatever that might mean) but instead ``what we have the right to
say about it.''  Combine that with one of the phrases he is fond of
saying, i.e., ``The past exists only insofar as it is recorded in the
present,'' and I think you will have to position him closer than not
to the positivist you used to think of him as.
\eq

\section{01-03-04 \ \ {\it Wither Entanglement!}\ \ \ (to T. Siegfried)} \label{Siegfried2}

Thanks for the letter.  I'm glad you're enjoying parts of {\sl Notes on a Paulian Idea\/} (even if some other parts rankle you).  Kluwer has picked it up to make for a more official-looking (even if ridiculously expensive) second edition this summer.  But that means I've got to get off my duff and get the thing re-edited by June.  I couldn't be a journalist for sure:  I'm so bad with deadlines.  I envy you.

\btoms
At the moment, I'm trying to work out a way of doing something on
quantum entanglement as a foundation for 21st century technologies
(based on a symposium at the recent AAAS meeting). Any suggestions of
non-obvious angles to pursue, or particularly significant papers
you've seen lately?
\etoms

It'd be hard for me to suggest an angle to pursue, as I don't really think entanglement is ultimately at the base of what is interesting in quantum information and computing.  I think entanglement is an auxiliary effect that is sometimes useful, but there is something deeper going on.  What the deeper thing is, now that's harder to say, but my feeling is that it has something to do with the particular structure of quantum measurements (which in turn capture something about the zing, zip, verve, pizzazz of the world).

But seriously, there are plenty of things in quantum information that don't derive their power from entanglement.  Simple BB84 and B92 quantum cryptography are two examples.  Or the stuff we called ``quantum nonlocality without entanglement'' (an old paper by Bennett and several of us, \quantph{9804053}) which makes use of the tensor product structure for combining systems, but not quantum entanglement in particular.

Also, in fact, one can see how the tensor product structure (and with it entanglement) arises in a simple way as a byproduct of the structure of measurements and a simple locality condition.  The argument is in the ``Wither Entanglement?''\ section of my \quantph{0205039}.  So, there is some substance to my speculation above.

Finally, let me just mention my favorite paper of the year:  Rob {\Spekkens}', ``In defense of the epistemic view of quantum states: a toy theory,'' \quantph{0401052}.  In that paper, Rob shows how a load of (what were once thought to be) `fantastic' quantum information effects can come about---qualitatively at least---even in a local hidden variable theory, as long as one is talking about the epistemic states of the theory.  Among these are things like teleportation, superdense coding, an analogue of entanglement monogamy, secure key distribution, and the list is quite a bit larger than that.  The point is, entanglement in the proper sense powers none of those effects---for his toy theory can have no entanglement; it is a local hidden variable theory.  The paper is quite easy to understand; in fact it contains almost no physics at all, just elementary combinatorics.  You ought to have a look at it.  It's the sort of thing that leaves you with a warm, fuzzy feeling that you've understood something quite deep.

\btoms
Also, any comments on Bub's recent paper?:
\quantph{0402149}
\etoms

It's a pretty good paper until he flubs it up by still thinking that decoherence is of any importance for quantum foundational issues.  In general the work by Clifton, Bub, and Halvorson is some nice stuff and a suggestive hint of the kinds of efforts we should keep striving at.  I think taking $C^*$ algebras as a starting point is far too advanced a starting point to get at the real nub of the matter, but what we're all doing is chipping away at an iceberg \ldots\ and I think their work contains a legitimate insight.

\section{01-03-04 \ \ {\it Midnight Murmurs} \ \ (to R. W. {\Spekkens})} \label{Spekkens30}

Good to hear from you.  I enjoyed your story about the Bielefeld meeting.  It sounds like the same meeting where I first concocted the image of a Holy City.  It was on the island of Ischia, near Naples, and I'm pretty sure a lot of the same people were there.  (The story is recorded in the letter spanning pages 450 and 451 of {\sl Notes on a Paulian Idea}, VUP edition.)  You didn't describe how your talk went though.  Did it make some impact?

\brws
Thanks for promoting my paper.
\erws

It's hard for me to keep my mouth shut about the paper.  [\ldots]  For instance, I just wrote this to Tom Siegfried (who writes the science column for the {\sl Dallas Morning News}):
\bq
   Finally, let me just mention my favorite paper of the year:  Rob
 {\Spekkens}'s, ``In defense of the epistemic view of quantum states: a toy
   theory,'' \quantph{0401052}.  In that paper, Rob shows how a load of
   (what were once thought to be) `fantastic' quantum information
   effects can come about---qualitatively at least---even in a local
   hidden variable theory, as long as one is talking about the epistemic
   states of the theory.  Among these are things like teleportation,
   superdense coding, an analogue of entanglement monogamy, secure key
   distribution, and the list is quite a bit larger than that.  The
   point is, entanglement in the proper sense powers none of those
   effects---for his toy theory can have no entanglement; it is a local
   hidden variable theory.  The paper is quite easy to understand; in
   fact it contains almost no physics at all, just elementary
   combinatorics.  You ought to have a look at it.  It's the sort of
   thing that leaves you with a warm, fuzzy feeling that you've
   understood something quite deep.
\eq
It is hard for me to keep my mouth shut.

\brws
Most recently, I needed to prepare a talk for the quantum gravity
meets quantum information workshop at PI.  (It's too bad you were
unable to make it to the workshop.  We needed a talk on how mass is
nothing but Hilbert space dimension.  How is that coming along
anyway?)
\erws

Basically, I'm not thinking at all.  And I don't foresee that I will again until I have the job issue settled and have a place to plant the family.  Mentally I'm empty, outside of those basic urges.  (Before you ask:  Yes, I could go back to Bell Labs.  But I think it would signal the end of all that I really love.)

\brws
How was the response to your talk at Oxford?  I would expect there to
be animosity towards your views there.  Did you manage to  shake the
convictions of any of the Everettians?
\erws

It felt like it was quite successful in some ways actually:  At the very least I did some of the best sparring I've done in a while, and I felt like it made an effect, though it was quite exhausting.  I got, I think, four emails, maybe five, afterward from people I didn't know expressing various kinds of interest.  For instance, I got this from Katherine Brading (never met her before):
\bq\noindent
   I was at your talk last Thurs at the Phil of Physics seminar in
   Oxford, and wanted to drop you a line to say how much I enjoyed it.
   I don't know whether you knew that you were coming into a group
   dominated by committed realists -- of various stripes, but realists
   nonetheless -- and I thought you did a great job. Personally, I found your approach very congenial indeed \ldots\
\eq
That's what it's all about:  Chipping away at an iceberg, getting people to rethink.

\section{01-03-04 \ \ {\it Thinking Out Loud, As Usual} \ \ (to H. Halvorson)} \label{Halvorson1}

Thanks for the note.  I guess we {\it have\/} never met; I've
certainly seen you in the distance.

\bhh
In some of my current work, I'm picking up a question that has been
close to your heart over the past few years --- viz., how much of QM
is just ``laws of thought''?  I'm going to have occasion to talk to some
general philosophical audiences about the topic, and I'd like to give
them an accurate representation of your position.  So, I was wondering
if I could ask you a favor: Could you point me to what you take to be
the two or three (or more, if you have time!)\ most significant
passages or results in your corpus that discuss, or bear on, this
topic?  In other words, where should one look first if one wants to
quickly learn the correct answer to the question?
\ehh

It's a phrase I stole from Boole, you know.  (See pages 527--529 and
351 in {\sl Notes on a Paulian Idea}, {\Vaxjo} U Press edition.  I
did send a copy of that to you, didn't I?)

It would be hard to tell you the ``correct'' answer because all of
these thoughts are in constant transition.  It's just been something
I've been groping for, for the last eight or nine years.  (I think I
invented the phrase at a bar in Albuquerque, Jack's Liquor and
Lounge, in the Fall of 1995.)

Be warned that by the phrase I don't mean something like a Kantian a
priori category, i.e., a position like von {\Weizsacker}'s in his book
{\sl The Unity of Nature}.  I don't mean something like, ``an
understanding using the terms of quantum mechanics is the
precondition for possible experience.''  Rather I have started to toy
rather strongly with a Darwinian kind of idea:  Using the rules of
quantum mechanics for manipulating and updating our expectations
(i.e., as a ``law of thought'') is the presently best known means for
survival, given that we are immersed in the particular world we are.
That is, I want to view quantum theory as a branch of decision theory
that is contingent upon properties of the world we live in \ldots\
and it is something we locked into only in our most recent turn in
evolutionary development.

Also, I should try to make it clear that, in this light, I view
quantum mechanics as a normative theory in a sense akin to the one
Bernardo and Smith use to describe Bayesian probability theory:
\bq\noindent
Bayesian Statistics offers a rationalist theory of personalistic
beliefs in contexts of uncertainty, with the central aim of
characterising how an individual should act in order to avoid
certain kinds of undesirable behavioural inconsistencies.  The
theory establishes that expected utility maximization provides
the basis for rational decision making and that Bayes' theorem
provides the key to the ways in which beliefs should fit together
in the light of changing evidence.  The goal, in effect, is to
establish rules and procedures for individuals concerned with
disciplined uncertainty accounting.  The theory is not descriptive, in the sense of claiming to model actual behaviour.  Rather, it is
prescriptive, in the sense of saying ``if you wish to avoid the
possibility of these undesirable consequences you must act in the
following way.''
\eq
That's the short of it.

The best reference for the long story at the moment is my samizdat
{\sl Quantum States:\ What the Hell Are They?},---unfortunately the
whole of it---posted at my webpage in pdf format.  But that's too much
material (and too loosely organized) for you.  So within that, let me
point you more specifically to: pages 49--50 ``\myref{Schack5}{Note on
  Terminology},'' pages 83--85 ``\myref{Caves30}{Replies on Practical
  Art},'' pages 144--147 ``\myref{Preskill6}{Psychology 101},'' and
pages 150--155 ``\myref{Wootters7}{A Wonderful Life}.''  Maybe also
pages 35--38, ``\myref{Schack4}{Identity Crisis}.''  That might do for
first pass.

Ultimately, I'd like to synthesize (and ``consistify'') these 235
pages of email into a single paper of 20 or 30 pages, but
unfortunately that hasn't happened yet.  If anything looks like sheer
nonsense, or the writing is detractingly ambiguous, let me know, and
I'll try to clarify for you.

Good luck.  If the stuff provokes any thoughts in you, I'd love to
hear them.  Also, I'd love to know the sorts of things you're already
thinking (that your note above alluded to).

\section{03-03-04 \ \ {\it Title and Abstract} \ \ (to UBC)}

Below is the title and abstract for my talk next week.  If there's anything else I've neglected to send you, let me know.

\bq\noindent
Title:  Two Strangely Similar Problems in Quantum Information \medskip\\
Abstract:  The deepest question one can ask in quantum information and computing is what makes it tick, what gives it its power?  The field is still new enough that no one really knows the answer---speculations range across the board, from quantum coherence to quantum entanglement to parallel worlds, and to stranger things still.  My own hunch is that the source of the power comes from the particular structure of quantum mechanical measurements, with entanglement being a secondary effect that derives from this source.  In this talk, I will demonstrate a few tools for exploring that hunch.  One is a representation of quantum measurements from the standpoint of probability theory.  Another one is connected to a question that first arose in the context of developing criteria for certifying the success and failure of quantum teleportation experiments.  Strangely enough, two open questions with regard to these tools seem to have the same answer.
\eq

\section{04-03-04 \ \ {\it Curl Activator} \ \ (to N. D. {\Mermin})} \label{Mermin111}

\bdm
Don't know if you wanted comments on the opening serenade of your
paper with {\Ruediger}, but here are a few:
\begin{itemize}\rm

\item (a) The Bayesian view of quantum states is that it is not: The
quantum state is not something the system itself possesses.

\item (b) What distinguishes this view from a more traditional
``Copenhagen-interpretation style'' view is that there is no pretense
that a quantum state represents a physical fact.
\end{itemize}
``this view'' in (b) seems to refer to (a).  But it's a non-trivial
jump from not being something the system itself possesses, to not
representing a physical fact.  E.g. the state of a quantum computer
represents the initialize procedure (measure all the qubits and apply
NOT to those that register 1) and the sequence of gates that have
been applied.  While this history is not ``possessed by'' the qubits
(since it can't be recovered from them) it's a big step to say that
the history (and the quantum state it gives rise to) is not a
physical fact.  (Namely your denial that gates [hamiltonians] are
objective, or that what has been measured [as opposed to the
measurement outcome] is objective.)
\bq\rm\noindent
   It is the outcomes of quantum measurements that represent physical
   facts within quantum theory, not the quantum states.
\eq
Repeating myself, shouldn't you acknowledge here that although the
outcomes of measurements represent physical facts, what it is that
has been measured is not, in your view, a physical fact.
\bq\rm\noindent
   In particular, there is no fact of nature to
   prohibit two different agents from using distinct pure states
   for a single system. [Footnote: Contrast this to the treatment
   of Refs.\ {\Mermin}2001 and Brun2001b.]
\eq
Please cite {\Mermin}2002 (J. Math.\ Phys., {\bf 43}, 4560-66) where the
argument is (in response to you guys) at least explicitly made
contingent on the assumption that probability 0 means objective
impossibility (and, I believe, no other assumption).
\edm

So, my words did curl your hair!  Makes me proud.

Thanks for the comments.  {\Ruediger} and I had a nice time on the
phone discussing them this evening.  The main thing it helped us
realize is that we've gotta, gotta, gotta get that ``On Quantum
Certainty'' paper written for you.

Anyway, the first main comment didn't cause us to make a change in
the draft:  The introduction wasn't the place to defend the view; we
just wanted to state the view.

About {\Mermin}2002, I've included it in the citations.  I've never
actually read the paper though (the dangers of not putting something
on {\tt quant-ph}).  Could you send me the file?

Concerning this one:
\bdm
\bq\noindent\rm
   In any case, this does not imply that a single agent can believe
   willy-nilly in anything he wishes. To quote D.~M. {\Appleby}, ``You
   know, it is {\it really\/} hard to believe something you don't
   actually believe.'' Difficult though this may be to accept for
   someone trained in the traditional presentation of quantum
   mechanics, the only thing it demonstrates is a careful distinction
   between the terms {\it belief\/} and {\it fact}.
\eq
It's not clear what ``this'' refers to.  From the syntax alone it
would appear to be {\Appleby}'s remarks.  This is reinforced by the
``believe'' ({\Appleby}) and ``belief'' (you).  But I assume you have in
mind some or all of what you have to say before {\Appleby} appeared on
the scene.
\edm
I think you got confused by not noticing that the ``Difficult though
this \ldots'' sentence, was outside of the footnote.  If you ignore
the footnote it works fine.  Still though I did adjust the words
slightly, just in case the reader loses track after a diversion to
the footnote.

Just in case you're curious, in the next email, I'll send you the
completed paper.  I'll have to send it as a pdf file though, so that
you get the figures.  As I told you before, it's pretty much a
throw-away paper, using as it does nothing but old technical
material.

Still, you might enjoy the Introduction, the Concluding Remarks, and
Section VI on ``Subjectivity of Quantum Operations.''  They package
the story in a way you may not have seen before.

Of use to me would be to know how the ``stimulus-response'' imagery
at the end strikes you.
\bq
Is there something in nature even when there are no observers or
agents about?  At the practical level, it would seem hard to deny
this, and neither of the authors wish to be viewed as doing so. The
world persists without the observer---there is no doubt in either of
our minds about that.  But then, does that require that two of the
most celebrated elements (namely, quantum states and operations) in
quantum theory---our best, most all-encompassing scientific theory to
date---must be viewed as objective, agent-independent constructs?
There is no reason to do so, we say.  In fact, we think there is
everything to be gained from carefully delineating which part of the
structure of quantum theory is about the world and which part is
about the agent's interface with the world.

From this perspective, much---{\it but not all}---of quantum
mechanics is about disciplined uncertainty accounting, just as
Bayesian probability theory in general. Bernardo and Smith write this
of Bayesian theory,
\bq
What is the nature and scope of Bayesian Statistics \ldots ?

Bayesian Statistics offers a rationalist theory of personalistic
beliefs in contexts of uncertainty, with the central aim of
characterising how an individual should act in order to avoid certain
kinds of undesirable behavioural inconsistencies.  The theory
establishes that expected utility maximization provides the basis for
rational decision making and that Bayes' theorem provides the key to
the ways in which beliefs should fit together in the light of
changing evidence.  The goal, in effect, is to establish rules and
procedures for individuals concerned with disciplined uncertainty
accounting.  The theory is not descriptive, in the sense of claiming
to model actual behaviour.  Rather, it is prescriptive, in the sense
of saying ``if you wish to avoid the possibility of these undesirable
consequences you must act in the following way.''
\eq
In fact, one might go further and say of quantum theory, that in
those cases where it is not just Bayesian probability theory full
stop, it is a theory of stimulation and response. The agent, through
the process of quantum measurement stimulates the world external to
himself.  The world, in return, stimulates a response in the agent
that is quantified by a change in his beliefs---i.e., by a change
from a prior to a posterior quantum state.  Somewhere in the
structure of those belief changes lies quantum theory's most direct
statement about what we believe of the world as it is without agents.
\eq

I think it's a particularly crisp way of expressing the program.
Anyway, it's all perfectly tame as written down, but the phrases
arose out of a discussion during my recent visit to the Oxford
philosophy department, where I was extolling the virtues of the
``sexual interpretation of quantum mechanics.''  Caught up in the
moment, I said something like:  ``{\Mermin}'s got this thing about
`correlation without correlata', but what I'm looking for in my
quantum foundations program is `stimulation without stimulata'!''  It
was nonsense and didn't fit, but we had a good laugh \ldots\ and the
word `stimulate' stuck with me, waiting to rise again on a more
legitimate occasion.

\section{04-03-04 \ \ {\it Catalog of Probabilities} \ \ (to J. N. Butterfield)} \label{Butterfield5}

I can't quite remember if it was you that I was talking to after Michael Redhead's talk at LSE about this.  \ldots\ But in any case, I'll get these quotes into my computer this way.

In Michael's talk, he implied (or rather said outright) that he introduced the term ``catalog of probabilities'' for quantum states.  I thought it was a much older term, probably going back to Pauli, but I haven't been able to confirm that.\footnote{\editornote ``This interpretation of $\psi$ as a catalogue of probabilities, or propensities as Popper would prefer to express it, is a perfectly reasonable and natural one'' [M.\ Redhead, ``Propensities, Correlations, and Metaphysics,'' Found.\ Phys.\ {\bf 22} (1992), pp.\ 381--94].  But a few years earlier we find ``The wave function is the catalogue of those probabilities that are mathematically implied by the knowledge gained in an experiment'' [M.\ Drieschner, Th.\ G\"ornitz and C.\ F.\ von Weizs\"acker, ``Reconstruction of abstract quantum theory,'' Int.\ J.\ Theor.\ Phys.\ {\bf 27} (1988), pp.\ 289--306].}  Anyway, I just ran across these quotes of {\Schroedinger} in a Brukner/Zeilinger paper; they were taken from S's 1935 papers on entanglement.  They quote the German and offer these as translations.  I'm not sure if these are their translations or, rather, they are the translations in the Wheeler/Zurek book.

\bq
For each measurement one is required to ascribe to the $\psi$-function ($=$ the prediction catalog) a characteristic, quite sudden change, which depends on the measurement result obtained, and so cannot be foreseen; from which alone it is already quite clear that this second kind of change of the $\psi$-function has nothing whatever in common with its orderly development between two measurements.  The abrupt change by measurement \ldots\ is the most interesting point of the entire theory. It is precisely the point that demands the break with naive realism. For this reason one cannot put the $\psi$-function directly in place of the model or of the physical thing. And indeed not because one might never dare impute abrupt unforeseen changes to a physical thing or to a model, but because in the realism point of view observation is a natural process like any other and cannot per se bring about an interruption of the orderly flow of natural events.
\eq
and
\bq\noindent
Whenever one has a complete expectation-catalog --- a maximum total knowledge --- a psi-function --- for two completely separated bodies \ldots
\eq

These don't explicitly contradict Michael, as they don't use the word ``probability,'' but they certainly convey the idea that a quantum state is catalog of something.

Anyway, words, just words --- not something particularly important.  But I like to get the history of QM straight when I can.

\section{06-03-04 \ \ {\it Receipt}\ \ \ (to G. Bacciagaluppi)} \label{Bacciagaluppi2}

Thanks for the long detailed note!  Now that's the kind of note I like to see!!

I hope to get back to you with something of substance before my physical arrival in Freiburg.

Since you brought up de Finetti, let me attach our latest concoction on the man.  {\Ruediger} and I just sent it off yesterday for a review book on quantum-state estimation.  It's mostly a throw-away paper with no new technical work, but I did use some new forms of expression in the Introduction and Conclusion sections, and also Section VI on the subjectivity of quantum operations.  In particular, I managed to clean up ``stimulation without stimulata'' enough, as to make it usable in the public forum.  Hope you enjoy.

\section{06-03-04 \ \ {\it Lost Cousins of de Finetti}\ \ \ (to G. Bacciagaluppi)} \label{Bacciagaluppi3}

Actually, I need to cite your paper ``Classical Extensions, Classical Representations and Bayesian Updating in Quantum Mechanics'' in the paper I just sent you.  Can you send me the citation information on that?  I didn't find it posted anywhere on {\tt quant-ph}.  Did you post it somewhere else?  Anyway, I'll get a chance to fix things up when the proofs come back to me next week.

You really should post it on {\tt quant-ph}!  (If you can post it fast enough, I'll get a chance to cite it in a way that people can find it.)\footnote{\editornote See \quantph{0403055}, submitted 07-03-04.}

\section{06-03-04 \ \ {\it Cutting the Umbilical} \ \ (to A. Plotnitsky)} \label{Plotnitsky15}

Please allow me to attach a paper that {\Ruediger\/} and I have just finished.  [See C. A. Fuchs and R. {\Schack}, ``Unknown Quantum States and Operations, a Bayesian View,'' \quantph{0404156}.]  It's mostly a regurgitation of old material---it was put together for a review volume on quantum-state estimation---but there are some new means of expression that may strike your sensibilities, or at least put you in a better position to contrast us with Bohr, Heisenberg, and even Pauli.  I'm thinking in particular here about the Introduction and Conclusion sections, and also Section VI on the subjectivity of quantum operations.  You'll understand why I titled this note ``Cutting the Umbilical'' when you read the first sentence of the second paragraph in the paper.  I don't know whether Ref.\ 23 still expresses Asher's point of view today---he has probably backed off on that slightly---but given your interest in the term ``quantum state'', and how it should be best used, I thought it might be worthwhile to point you in the direction.

\section{06-03-04 \ \ {\it Darwinism Down} \ \ (to M. P\'erez-Su\'arez)} \label{PerezSuarez9}

Attached is the samizdat ``Darwinism All the Way Down'' to the extent that it's been put together---a lot of material is still missing.

Anyway, the great mistake is recorded on page 9.  The especially sad thing was how it contradicted what I was saying as early as the note ``\myref{Schack4}{Identity Crisis}'' of 22 August 2001 (in ``Phase Transition''), where I had already pretty much sorted everything out.  It was as if I had forgotten everything of the previous year.  Anyway, it was the solution to this problem which effectively forced me to a Wittgensteinian notion of certainty.  (Though I only started reading Wittgenstein this January.)

\section{07-03-04 \ \ {\it All Kinds of Veils} \ \ (to A. Peres)} \label{Peres62}

\bap
I am absolutely elated by ``my'' discovery that there are no quantum
states (and therefore no ``problem of time'' in quantum gravity,
etc.). Heisenberg {\it et al}.\ used only algebras of operators to compute
observable quantities. Then {\Schroedinger} came and ``stole the
show''---and completely messed it up. He did that after Einstein
called his attention to de Broglie's thesis, and wrote ``he has
partly lifted the great veil'' or something like that. You are an
inexhaustible source of references. Where is that ``great veil''
mentioned?
\eap

I will look for the ``great veil'' reference tomorrow at the office,
where I have a copy of Jammer's {\sl The Philosophy of Quantum
Mechanics}. It is a very thorough historical reference.  Another
place where you are likely to find it, and other interesting things,
is a little book titled {\sl Letters on Wave Mechanics}, edited by
Karl Przibram, with letters by Einstein, {\Schroedinger}, Lorentz, and
maybe a couple of others.  Your library might have it.  I remember
finding one letter particularly amazing within it.  It was a letter
written by Einstein to {\Schroedinger}, in great excitement, after
Einstein's first quick reading of {\Schroedinger}'s paper on the time
independent {\Schroedinger} equation.  Einstein says something like
``what a great insight!''\ but then he quotes {\Schroedinger}'s
equation {\it incorrectly\/} and complains something like, ``You will
note that this equation has this and this and this undesirable
property. On the other hand, if you had considered this equation
[where Einstein now writes the correct {\Schroedinger} equation as if
he had never seen it], all of these problems will be fixed.''  But
then the really lovely thing that Einstein says next is something
like, ``However, for the life of me, I can't think of a physical
interpretation for the wave function that appears in this equation.''

Unfortunately, my copy of the book was burned up in the fire, and I
haven't replaced it since.  It's a nice little
book.\footnote{\editornote \label{Einstein1926} The letter, dated 16 April 1926, can be
  read in a somewhat awkward format starting on p.~35 here:

\myurl{http://www.scribd.com/doc/75421769/Albert-Einstein-Letters-on-Wave-Mechanics}

Here's Einstein expressing his dissatisfaction with the misremembered
wave equation:
\bq
If I have two systems that are not coupled to each other at all, and
if $E_1$ is an allowed energy value of the first system and $E_2$ an
allowed energy value of the second, then $E = E_1 + E_2$ must be an
allowed energy value of the total system consisting of both of them.
\eq
And then, a little later:
\bq
It also seems to me that the equation ought to have such a structure
that the integration constant of the energy does not appear in it;
this also holds for the equation I have constructed [the correct
  Schr\"odinger equation], but despite that I have not been able to
assign a physical significance to it, a matter on which I have not
reflected sufficiently.
\eq
I like this part in Schr\"odinger's reply:
\bq I am, moreover, very grateful for this error in memory because it
was through your remark that I first became consciously aware of an
important property of the formal apparatus. Besides, one's confidence
in a formulation always increases if one---and especially if
\emph{you}---construct the same thing afresh from a few fundamental
requirements.
\eq
Later, Einstein and Schr\"odinger get into the question of whether
$\psi$ is an expression of ignorance about a pre-existing
reality. Einstein associates this position with Born, and draws a
distinction between this and Bohr's thinking.

A distinction which, it seems, has been lost on many during the years
since\ldots\ but never mind.}

\section{08-03-04 \ \ {\it The Great Veil} \ \ (to A. Peres)} \label{Peres63}

\bap
Then {\Schroedinger} came and ``stole the show''---and completely
messed it up. He did that after Einstein called his attention to
de Broglie's thesis, and wrote ``he has partly lifted the great veil''
or something like that. You are an inexhaustible source of references.
Where is that ``great veil'' mentioned?
\eap

Apparently the quote is ``lifted a corner of the great veil.''
However, I haven't been able to pin down the exact origin of it yet.
You can at least read it here:
\bq
\myurl{http://www.aip.org/history/einstein/quantum1.htm}.
\eq

Jammer's book wasn't as much help as I thought it would be.  For
instance, I'm discovering how bad the index is.  My method of looking
was to cross-reference a) Einstein and de Broglie, and b) Einstein
and {\Schroedinger}.  There are quite a few places where I found the
names within the text, but no cataloguing within the index.

Actually, I just found this story:\footnote{See H. A. Medicus, ``Fifty Years of Matter Waves,'' Phys.\ Tod.\ {\bf 27}(2), 38 (1974) for more details.}
\bq
     Working on his doctorate in 1909, a young aristocrat, Prince
     Louis-Victor de Broglie, discovered a mathematical relationship
     between Planck's Constant and a yet to be observed wavelike
     property of moving masses. His examiners were of a mind to reject
     the paper, and wanting an outside opinion, sent a copy to Einstein
     who replied: ``He has lifted the corner of a great veil.'' The
     dissertation was accepted---fifteen years later it earned de Broglie
     a Nobel Prize, the first ever awarded for an academic thesis.
\eq

\section{08-03-04 \ \ {\it IC-POVM Entropies} \ \ (to A. H. Jaffe)} \label{Jaffe1}

Thanks for the letter.

\bahj
I hope you'll remember me --- we met briefly in London and then Oxford
--- I asked you about the Shannon entropy of an informationally
complete POVM but then had to go.
\eahj

I do remember talking to you in London right after my talk; you were to the left of the seat I was sitting in and wanted to know something about one of my transparencies.  Unfortunately, though, I can't visualize your face.  And I don't remember talking to you again in Oxford:  Everything about that trip is a big blur!

\bahj
My ``Bayesian bona fides'' can be indicated (I hope) by the
fact that I reviewed Jaynes' book for {\bf Science} last year.
\eahj

I'd very much like to read that.  Could you send me an electronic copy of the article?

\bahj
Anyway, on to something more specific. I've been thinking about the entropy of the probabilities related to a set of informationally
complete POVMs. Obviously, it depends on the detailed set of POVMs
chosen. But it seems that there are `better' and `worse' choices here
--- related, I wouldn't be surprised, but have not yet worked out --- to
the shape/volume of the accessible area of the probability simplex.
\eahj

There are probably a load of interesting questions one can ask here.  For instance, given a density operator $\rho$, what is the minimum Shannon entropy a measurement can generate, subject only to the condition that the measurement be an IC-POVM.  I suspect the answer is $S(\rho)+\log D$, where $D$ is the Hilbert-space dimension, but I don't have a proof of this.  In other words, the optimal POVM is just an infinitesimal variation of the basis that diagonalizes $\rho$, but having $D^2$ linearly independent outcome operators, rather than $D$.

Maybe a more interesting question is:  What is the {\it largest\/} Shannon entropy, subject only to the constraint that the POVM be a minimal IC-POVM (i.e., it have precisely $D^2$ outcomes)?  Again I don't know the answer, but I would bet it has something to with using {\it symmetric\/} IC-POVMs (at least if $\rho$ is a pure state).  (Actually it's probably obvious in the pure state case that the thing to do is align one of the outcomes of the SIC-POVM with the pure state.)

In general, the SIC-POVMs are the most interesting ones of the lot I think.  You can read about them here:
\bq\noindent
\quantph{0310075}:\\
Title: Symmetric Informationally Complete Quantum Measurements\\
Authors: Joseph M. Renes, Robin Blume-Kohout, A. J. Scott, Carlton M. Caves
\eq
It'd be an interesting project to say everything that can be said about the entropies such things can generate.

\bahj
Of course, you can also use the entropy to assign distributions in the
face of less-minimal information --- i.e., known averages, as usual.
But in this case we would expect that the results should depend on
POVM, since that's what determines what the averages correspond to.
Thinking about it a bit further, I imagine there's a further subtlety:
there must be something subjective about the POVM itself. So if two
different agents use maxent to assign a density matrix, under what
circumstances will they get `compatible' assignments (to use the
language of your work).
\eahj

Over the last couple or three years, Caves, Schack and I have been steadily moving further away from our Jaynesian roots to a more Bernardo and Smith kind of Bayesianism.  One of the upshots of this is a recognition that POVMs, too, have a subjective component (if they are not wholly subjective in the same way as any probability assignment -- which is what I personally believe).

For your fun, let me attach the draft of a new paper by Schack and me.  [See ``Unknown Quantum States and Operations, a Bayesian View,'' \quantph{0404156}.] (It's effectively complete, except that we'll add a couple of footnotes and references at proof time.)  Mostly it's a throw-away paper in that it has almost no new technical results; it was put together for a review volume on quantum-state estimation.  However, the Introduction, Concluding Remarks, and Section VI on the subjectivity of quantum operations, may give you some insight on where this program is going.

Good luck in your work, and I'm very happy to meet you.  There are so many things that need to be done to complete this Bayesian view of QM.

\section{09-03-04 \ \ {\it Alchemy Quote} \ \ (to M. P\'erez-Su\'arez)} \label{PerezSuarez10}

Thanks for the quote.  See you later today.

\subsection{Marcos's Preply}

\bq
I should be writing equations, but I find myself writing only numbers.
I'll take this exhaustingly boring exercise as a excuse for making a remark on my not well articulated answer to your question about the alchemists' psycho-physical parallelism. One of the leading ideas (I'd say, that THE core of the subject) in alchemy, leading back to the
(mythological?)\ Hermes Trismegistus (three times magister), is precisely this (translated from the supposedly original):
\bq\noindent
It is true without lie, certain and most veritable, that what is below is like what is above and that what is above is like what is below, [\ldots]
\eq

I have fortunately included this as a quote in one of the chapters for those Lecture Notes I have been writing. I'm afraid that most alchemical texts are so full of symbolism and esoteric that only an ``initiate'' (that is, someone already acquainted with this imagery, like Jung, for instance) can make any sense of them. But this is ``the point'' here: The Macrocosm (the Universe, if you want) is ``in a likeness'' (well, more than that, I'd say) to the Microcosm (the psyche), and they are intimately related to each other so that processes in one of the two domains have their correspondences in the other.

Hope that this provides further insight in Pauli's involvement with alchemical notions.

\eq

\section{09-03-04 \ \ {\it Reality of Wives, Part 2} \ \ (to J. Preskill)} \label{Preskill13}

I remember how much you seemed to enjoy my story a couple of years ago about the reality of wives (in a note to you and Landahl).  [See 21-07-01 note ``\myref{Preskill2}{The Reality of Wives}'' to A. J. Landahl and J. Preskill.] Attached is a throw-away paper that {\Ruediger} and I just sent off for a collection of review papers on quantum-state estimation---they wanted one from a Bayesian too.  I say it's throw-away because there are no new technical results, outside of a little foundational argument; predominantly it's a cut and paste job.  (We probably won't post it on {\tt quant-ph}.)  Nevertheless I send it because you might still get something out of reading the Introduction, Conclusions, and Section VI on the subjectivity of quantum operations.  They indicate a little of what I've been on about when I previously mentioned the issue in connection to the black-hole information problem.  The main point is that, to us, a quantum operation is to a conditional probability distribution what a quantum state is to an unconditioned distribution.  To the extent that a Bayesian does not demand a mechanical or dynamical explanation for a quantum state assignment, he need not (should not) demand one for a quantum operation either.

\section{09-03-04 \ \ {\it Bitbol} \ \ (to A. Plotnitsky)} \label{Plotnitsky16}

\barkp
I know Bitbol's work---he has a book on {\Schroedinger}.  I doubt I'd have
a chance to see him this time (too tight and too late to arrange), but
I am back to France in May/June for a longer time, and perhaps I
should write him, mentioning that you suggested this (with your
permission, of course).
\earkp

Sure, no problem.  Bitbol's later work has turned, I would say, toward more Copenhagenian styles of interpreting quantum mechanics.  Here's his web page:
\begin{center}
\myurl{http://perso.wanadoo.fr/michel.bitbol/page.garde.liste.html}.
\end{center}
Most of his papers are in French, but a few are in English.  I particularly enjoyed his ``Non-representationalist theories of knowledge and quantum mechanics.''

\section{12-03-04 \ \ {\it Your Post-Partum} \ \ (to G. Brassard)} \label{Brassard38}

I had a chance to read your report on our meetings.  It was great and to the point.

Let me attach my own latest concoction:  It is a contribution (quickly) pasted together for a review volume on quantum state estimation.  They wanted one article from a Bayesian point of view.  So, there's not a lot of new technical material there, but I think you might enjoy reading the Introduction, Conclusions, and Section VI on the subjectivity of quantum operations.  Particularly, I hope you enjoy the concluding remarks.  It's the closest I've come yet to declaring the ``sexual interpretation of quantum mechanics'' in the public forum.  The choice of the word ``stimulation'' in that section shouldn't be lost on you!  (Key point to notice:  In opposition to {\Mermin}'s desire for `correlation without correlata', there's no use to talk about the `stimulation' unless you're also going to talk about the `stimulata'!)

So the queue to the construction of our joint paper is getting shorter.  Upon my return to Dublin from Vancouver, I've got five days to put something together for the Holevo festschrift (I'm 3/4 way there already).  And then I'll get to work on our project.  As I say, the editor seems to be particularly tolerant on us \ldots

\section{12-03-04 \ \ {\it Putting It Sharply} \ \ (to W. G. Unruh)} \label{Unruh4}

I'm afraid I wasn't very clear yesterday about why I consider measurement the crucial component of quantum theory.  Let me attach an article {\Ruediger} {\Schack} and I just put together for a review volume on quantum state estimation.  I think the first three or four paragraphs in the introduction, plus the `concluding remarks' section, make our crispest statement yet of the interpretation of quantum mechanics we're shooting for (and distinguishes it from a certain thread in all the various `Copenhagen interpretations' I know).

Stimulation and response, and maybe a little creation in the process.  Let me know if the paragraphs clarify anything.  (And one day soon, I'll tell you more about how Section VI on the subjectivity of quantum operations impinges on the black-hole information problem.)

\section{14-03-04 \ \ {\it The House Philosopher, 2} \ \ (to S. Savitt)} \label{Savitt5}

Thanks again for dinner, the nice visit to your home, and the pleasant conversation about Goodman.  (I'll definitely plunge more carefully into Goodman.)

For the heck of it, let me go ahead and attach the proposals I wrote for Caltech.  (I'll just forward an old letter.)  [See 12-02-04 note ``\myref{Mabuchi7}{The House Philosopher}'' to J. Preskill and H. Mabuchi.]  The proposals had to be short (one and two pages), but maybe they still give a glimpse of the kind of program I'd put together if I lived in the best of all possible worlds.  But I'm not fooling myself:  The dice have only a very, very slim chance of falling in that direction.

I'll keep in touch with you about things.  Have a great trip to merry old England.

\section{15-03-04 \ \ {\it Via Alaska Airlines} \ \ (to D. M. {\Appleby})} \label{Appleby5}

I apologize for the long delay, but I've finally read your paper on
probability.  It happened on my flight from Vancouver to Los Angeles
the last few hours.  Now I'm in the Aer Lingus lounge in LAX, waiting
to make the rest of my way home to Dublin.

It's a good paper!  What parts were you afraid I would disagree with?
I think you did the community a great service with the paper, and
I'll try to advertise it as best I can.  I particularly enjoyed your
argumentation in Section 6 on retrodictive inferences.

A few very minor remarks.
\begin{enumerate}

\item  Of the paragraph in Section 1 starting, ``Hume famously argued
that one \ldots'', I wrote, ``Good way of saying it.''

\item  Of the sentence in the same section, ``On the face of it, taking
an epistemic view of the state vector amounts to giving up on the
idea of physical reality altogether,'' I wrote, ``Yuck.''  Rather
than ``on the face of it'' you might have written ``ostensibly'' or
``at first pass''.  I guess ``on the face of it'' conveys the idea
somewhat, but I fear the reader may not immediately see that you're
being rhetorical. You do, in fact, recover with:

\item ``However, I feel it may be consistent with a much more subtle and
interesting kind of realism \ldots'' to which I wrote ``Better.''

\item  In Section 4, where you say, ``For instance half-lives are
typically tabulated next to masses, as if they were just one more
physical property,'' I wrote ``Interesting!''  That is a good point.
However,

\item from that point onward, all the way to the end of the section, I
could not understand what you were getting at.

\item At section 9, I wrote ``Utilities!!''  I think you could do a lot
better in general in that discussion.

\item Finally Section 12 was far too abrupt!  You had me licking my lips
and then left me hanging.  (It looks a little like you were exhausted
by the time you got to the last section and, so, wrapped it up pretty
quickly.)  I was tantalized by this comparison of probabilities to
qualia---I had never seen that before---and I wanted to see how you
developed the point.  Also, despite the mention of Bohr in the last
two paragraphs, you never did come back to the point of a ``much more
subtle and interesting kind of realism.''
\end{enumerate}
All very minor points.

In an earlier note, you wrote me,
\bma
It is all about classical probability---coins, and the like.  I do
appreciate that quantum probabilities are fascinatingly and
intriguingly different.  And that is what I am going to start
thinking about now.
\ema
I don't think you'll find any consideration in your paper that is
changed in the quantum setting.  (``Where did you make use of the
difference between classical and quantum?,'' I ask rhetorically.)  At
least as long as you are talking about actually performed
trials---rather than counterfactually performed trials (as in my
bureau of standards)---I think everything will be OK.  In particular
if we look at things in the right way, I suspect the only difference
between classical and quantum will be in their notions of event, not
in their notions of probability.  In the quantum case, the events are
direct consequences of the agent himself (via his conjugal relations
with the external world)---stimulandum, stimulation, stimulata.
(Correlation without correlata?  No!  Rather, no stimulation without
stimulata!  No stimulation in the agent without a corresponding
stimulation to the external world.)  Somehow, taking that into
account in the agent's reasoning leads to an apparent modification of
the probability calculus with respect to counterfactual measurements
(i.e., how he updates his beliefs about the bureau of standards).

By the way, concerning,
\bma
Matthew thinks that he cannot pick up a glass of wine. Only an
American pragmatist could be so simple-minded---so horribly
unsophisticated---as to imagine that it is possible to genuinely do
anything.
\ema
thank you!  It's been a long time since I've been proud to be an
American!

All the best, and it really is a fine paper.  Thanks for writing it.

PS.  Let me attach the latest concoction {\Schack} and I brewed; it's
something we haven't posted yet.  There's not much new in it---it was
a quickly pasted together article for a review volume on quantum
state estimation---however you might still enjoy reading the
introduction, conclusion, and Section 6 on the subjectivity of
quantum operations.  In particular, I'm kind of proud of the language
choices I made in the first three paragraphs or so, and also in the
concluding section.  I really want to rearrange how we think of
quantum theory---from a big theory of everything to a little theory
from the inside.  A theory of stimulation and response, and maybe a
little creation in the process.  It's the latter part of that
sentence that gives me the will to keep working at this.

\section{15-03-04 \ \ {\it Via Aer Lingus} \ \ (to D. M. {\Appleby})} \label{Appleby6}

I also reread your note ``Poetry, subjectivity, mirroring and algorithms'' of 11/19/2002, which has been sitting in the ``must reply to'' box in my email program since then.  It's a great note, and I loved its construction, but I still don't know how to reply.  Hence, I think I'm going to file it into the regular ``{\Appleby}, {\Marcus}'' box at this point.

In your ``Facts, Values and Quanta'' you say ``a probability statement cannot be identified with a fact about the world, as it exists independently of us.''  I think that is all I have ever meant by ``subjective.''  But, then again, maybe that's just hindsight playing a trick on me.

\section{16-03-04 \ \ {\it Bayes or Bust} \ \ (to A. H. Jaffe)} \label{Jaffe2}

Thanks for sending your review.  It went down well with my coffee this morning.  It's quite nice.

\bahj
Conversely I suppose I share some of Jaynes's suspicion of the
fundamental status of Dutch book-type arguments, although I do think
they tell you something --- but perhaps not everything --- about the
meaning of probability assignments once you've derived the rules by,
for example, the Cox derivation.
\eahj

I hope I haven't given the impression in my writings (or the writings with Caves and Schack) that I think the Dutch-book argument is the end-all and be-all of probability theory.  I'm quite a bit more liberal than that.  The better go at it, I think, is given by a decision theoretic derivation, say, like the one in Bernardo and Smith.  Cox, on the other hand, is nice, but I personally have never found it nearly as convincing as decision theoretic ideas.  \ldots\ Well, that's not true:  I did once, but somewhere in 1994 I got suspicious of it and started to think its assumptions aren't nearly as compelling as Jaynes finds them.  (If I can recall my particular arguments, eventually I'll write them up and send them to you.)  Maybe my opinion is more like Earman's in his book {\sl Bayes or Bust\/}:  No single Bayesian style argument for the probability calculus is completely compelling, but the whole package taken together lends quite some evidence that everything is on the right track.

\section{16-03-04 \ \ {\it The IQSA Meeting} \ \ (to F. E. Schroeck)} \label{Schroeck2}

Thanks for the invitation to the IQSA meeting.  I wish I could be there, but I think now I'm so over-committed that I ought to make a decision not to get myself in any deeper.  I'll have just gotten home from a meeting in Waterloo at the time your meeting starts up:  My semi-constant absence is pretty tough on the family.

But good luck to you; it sounds like a great meeting.  Beside Barnum, another few people you might try out of the quantum information community are Jeff Bub ({\tt jbub@carnap.umd.edu}), Lucien Hardy ({\tt lhardy@perimeterinstitute.ca}), and Rob Spekkens ({\tt rspekkens@perimeterinstitute. ca}).  Spekkens in particular has done some really exciting work lately, which you can find on {\tt quant-ph}.

\section{18-03-04 \ \ {\it Hubert Space} \ \ (to R. {\Schack})} \label{Schack80}

Damned editors.  They are so very worthless.  I'll look at the proofs in the next couple of days. Did you notice they changed one of our Hilbert spaces to a Hubert space!

\section{18-03-04 \ \ {\it ABL, Appleby, and Grue} \ \ (to S. Savitt)} \label{Savitt6}

\bss
Second, it raises a question in my mind. Bill (Unruh) insists that the
right way to think of QM, to say what the theory is, is the
time-symmetric ABL formalism. Might it not be interesting to figure
out whether Paul Humphrey's criticism of the propensity view would or
would not apply to the probabilities needed for ABL if you tried to
think of them as propensities?
\ess

Good question.  Off the cuff, though, I'd bet the same criticism applies:  the main issue, I think, is about the difference (or whether there should be no difference) between predictive probabilities and inferential probabilities.  One might think of the ABL rule as giving propensities, but then one has to account for what one might mean by an ``unknown ABL probability.''  At that point, one invokes inferential probabilities, which are clearly epistemic.  So, one either ends up with a mixed theory (propensities and epistemic probabilities, which by a miracle obey the same formalism), or one says, ``this is good evidence once again for a purely epistemic view of probabilities.''

But that's off the cuff, and I'll certainly think about the issue further.

By the way, interesting coincidence.  On my flight home from Vancouver, I fulfilled an obligation to read Marcus Appleby's paper ``Facts, Values and Quanta,'' which is really mostly devoted to the interpretation of probability (outside of quantum issues).  Here's the link:
\bq
\quantph{0402015}.
\eq
If you can get past the sickly sweet introduction (praising some guy named Fuchs), it's not a bad paper.  There are some novel arguments in there that I liked.  In particular---and the reason I say ``interesting coincidence''---at one point he uses a modification of Goodman's grue to discredit frequency interpretations.  Given a couple of your remarks in our conversation the other night about ``predicted frequencies being seen,'' you might find reading the paper useful, at least as some food for thought.

\section{23-03-04 \ \ {\it Ride the Wave} \ \ (to D. J. Bilodeau)} \label{Bilodeau7}

Thanks for the nice letter.  Glad to see you're excited.  I hope you will eventually channel your enthusiasm into a new paper of the same outstanding quality as your 1998 one.  There could be no better tribute to progress.

\section{24-03-04 \ \ {\it Give Me Fungible} \ \ (to C. H. {\Bennett})} \label{Bennett33}

Can you give me a citation to the first time where you called quantum information fungible?  Better yet, can you {\it also\/} send me the text of the paper?

\section{25-03-04 \ \ {\it Reining in My Life} \ \ (to G. Brassard)} \label{Brassard39}

I'm finally beginning to rein my life back in.  I finally finished the Holevo festschrift and sent it off.  [See ``On the Quantumness of a Hilbert Space,'' \quantph{0404122}.] If you can, let me know what you think about the Introduction, Conclusions, and Eqs.\ (13) and (14).  More importantly, can you think of anything in quantum computing that would bolster the vision (or the hope) expressed in Paragraph 5 of Section 1.

Now, I finally start the process of writing our joint paper.  It is brewing.  I'll send you the thoughts as they come next week.  Part of it will certainly have to do with the content of the Holevo paper:  It is quantum crypto that powers quantum computation!  (I like pipe dreams.)

By the way, when are you going to take me to that Greek seafood place in {\Montreal} again?

\section{25-03-04 \ \ {\it Hot Off the Presses} \ \ (to P. C. E. {\Stamp} \& W. G. Unruh)} \label{Stamp2} \label{Unruh4.1}

Attached is a paper I just sent off to the editors of the Holevo festschrift.  [See ``On the Quantumness of a Hilbert Space,'' \quantph{0404122}.]  It contains the material I was hoping to get to for the second part of my talk in Vancouver.  The fifth paragraph of the introduction, in particular, tries to say (in a sentence) what I think all of this has to do with the source of power of quantum computing---something I was hoping to play up even further with you guys, but I ran out of time.

Anyway, I hope you enjoy the verbiage, if not the equations.

\section{26-03-04 \ \ {\it So Slow Chris} \ \ (to A. Peres)} \label{Peres64}

I apologize for being so slow to reply.  I have been hurriedly trying
to finish my paper for the Holevo festschrift, and I pretty much
dropped all else to get it done.  I will attach the \LaTeX\ file to
the present note:  Perhaps the introduction and conclusion sections
will amuse you.  (I haven't posted it yet on {\tt quant-ph}; I'll
probably do so Monday before departing for Germany myself.)

\bap
I have read again your \quantph{0205039}, and figures 1 and 2
reminded me of my paper ``Convex probability domain \ldots'' with
Danny Terno, JPA {\bf 31} (1998) L671 [\quantph{9806024}]. Is
there any relation?
\eap

Yes, there absolutely is.  It was quite an oversight on my part not
to cite your paper there.  (In fact, I thought I had until I received
this from you.)  It was because of your paper with Danny that I
decided to make the ``bureau of standards'' measurement have
precisely $D^2$ outcomes.  If the measurement had any more than $D^2$
outcomes, say $N$, then the allowed region would have zero volume in
the $N$-simplex, as you and Danny show.  I have been meaning to
repost that paper with various fix-ups.  This will surely be one of
them. (There, I just fixed it actually.)

What I think is the next important step is to say more about the
geometric properties of these convex sets.  For instance, with regard
to qubits, the convex region is actually always an ellipsoid (no
matter what the bureau-of-standards measurement).  The ellipsoid
varies from measurement to measurement, but it is always an
ellipsoid.  One question is, what is the proper characterization of
the region in higher dimensions?  What are the invariant geometrical
features?  And I can think of several questions beyond that.  Would
you be interested in collaborating on this?

I am glad to hear you survived your voyage back to Germany and did
not come back with a renewed bitter taste.  I very much enjoyed
reading the (partial) autobiography you sent me.  You have had a
stirring and very impressive life.

\section{26-03-04 \ \ {\it Nonuniqueness from Nonexistence} \ \ (to A. Peres)} \label{Peres65}

By the way, I've been meaning to tell you how pleased I am of your
embracing the point that ``quantum states do not exist'' in your
quantum gravitational research.

To that end, let me attach another paper of mine---this one with
{\Schack}.  There is not a lot new in it by way of technicality.  It was
effectively pasted together from old publications for the purpose of
a review volume on ``quantum state estimation,'' however, we reworked
the language in the old articles significantly to take into account
the more consistent view that we now have.  The main parts that will
be relevant to you are the introduction, the conclusions, and Section
6 on ``The Subjectivity of Quantum Operations.''

If quantum states do not exist (even pure states), then they surely
cannot be unique (even pure states).

You will note how I used your 1984 paper ``What Is a State Vector?''
as a point of contrast for our present view.  I hope you will not
think that I was picking on you.  The reason I singled out your paper
is because it is the very clearest statement on the subject.  If you
think, however, that I should change any of the language before
posting it, please let me know.  For instance, I did not attempt to
contrast your 2004 view with your 1984 view---they may be
significantly different!!  (And I would guess they are.)
Unfortunately, it may be too late for the version appearing in the
volume:  I was in such a rush sending off the draft---and returning
the proofs---that the possible sensitivity of my statement didn't
dawn on me at the time.  But it will not be too late for {\tt
quant-ph}; please let me know.

\section{26-03-04 \ \ {\it Hot Off the Press; Closing My File} \ \ (to J. Preskill)} \label{Preskill14}

Attached is a paper I just sent off to the editor of the Holevo festschrift.  [See ``On the Quantumness of a Hilbert Space,'' \quantph{0404122}.] I thought I might send it to you too.  It's the last remaining thing on my CV that's not publicly available.  With it you can close my ITS file~\ldots

I remember your asking me during my visit to Caltech in October, ``Remind me why this symmetric POVM is interesting?''  The present paper is an attempt to answer that.  I.e., it's because I think these structures lie closer to the heart of what Hilbert space is than do orthogonal sets.  In particular, I'm banking that ``No'' will be the answer to the open question in Section 6, and that the SIC ensembles will be the only minimal sets giving $Q_d$.  But I could be wrong.

If you want to throw in on the open question, I'd be grateful for any ideas (or even the flat-out solution).  Chris King and I spent a solid couple of weeks on it last summer, but didn't make any great progress.

\section{26-03-04 \ \ {\it Thanks} \ \ (to M. P\'erez-Su\'arez)} \label{PerezSuarez11}

I just wanted to say quickly (before I start to enjoy my Friday night beer) that I enjoyed our session together today.  It was stimulating.  Great to see you asking questions like that that are on the edge:  I.e., what can we glean from this Bayesian program?  How does this piece link with that piece?  So forth.  Keep it up!  That's what a researcher is:  A big bundle of questions.

\section{05-04-04 \ \ {\it Elevator Stories} \ \ (to W. K. Wootters)} \label{Wootters17}

I'm glad you are following through with a contribution to Asher's festschrift.  I'm still waiting on three other people too, so you're not the last one.  \ldots\ But the faster you can get it to me, the better!  (There are always further uncertainties with referees, etc., etc.)

I'm in Munich at the moment, taking a little holiday with Kiki and the kids at Kiki's parents' place.  A couple of days before that, however, I was in Freiburg visiting Guido Bacciagaluppi and Harald Atmanspacher and was lucky enough to run across a recent issue of Foundations of Physics at their institute:  It had your paper ``Why Things Fall'' in it, which I had not heard of before!  Very nice.  I got a chance to read it this morning.

In that regard, maybe let me attach two of my own papers that haven't made their way onto {\tt quant-ph} yet.  The first, ``Unknown States and Operations, a Bayesian View,'' is a kind of throw-away in that it doesn't contain really any new technical results---it was pasted together for a review volume on quantum tomography---but you might get something out of the introduction, conclusions, and Section 6 on the ``subjectivity of operations'' (which are new) that I haven't yet been able to adequately express to you.  In particular, I hope the choice of words there will better help you see how I am starting to view quantum mechanics as a whole---i.e., as predominantly a theory ``from the inside.''  As I see it, it is a theory that helps us gamble on our little parts in the act of creation.  But I also hope the argument in Section 6 will help you appreciate why I am starting to take such an austere position about what is ``real of a quantum system''---little more (and maybe no more) than its dimensionality.

In the tradeoff though, I hope this austerity provokes an image of another austerity in physics for you:  The equivalence principle.  And that's where the elevator stories come in, and my pleasure this morning in having read your article.  With ``elevator stories,'' I'm referring to my second attached article, ``On the Quantumness of a Hilbert Space.''  You'll see what I'm talking about if you open it up.

What I didn't go so far as to mention there is that I am starting to wonder if Hilbert-space dimension is not so unlike gravitational mass as to actually BE gravitational mass.  It's a wacky thought---almost surely wrong in detail, but maybe evocative of a fruitful direction of thought.  At least I hope so.  Q: ``What's the ultimate limit to this system's sensitivity to quantum eavesdropping?''  A: ``I don't know, let's weigh it.''  Wouldn't something like that be cool?

Anyway, part of this thinking is motivated by Bekenstein-entropy-bound kinds of things.  From a Bayesian point of view of QM, I can't see what the entropy bound could possibly mean but that:  If one posits an $E$ and an $R$ for a system, at the same time as positing a Hilbert-space dimension above that allowed by the bound (assuming a completely mixed density operator), one is living in a state of sin---i.e., one is being incoherent in one way or the other.  But part of the thinking is just motivated by a desire for even more austerity, full stop.

If you have any comments on the quantumness paper, let me know:  It's still at a stage where it can be modified before posting.

\section{05-04-04 \ \ {\it Gambling on Creation} \ \ (to W. K. Wootters)} \label{Wootters18}

Looking back over the note I just wrote you, I doubt you could understand what I meant by, ``As I see it, it is a theory that helps us gamble on our little parts in the act of creation.''  Let me try to make some sense of that by attaching a couple of old emails.  Both are written in a polemical style, but if you can ignore that they might help draw a picture.

\section{06-04-04 \ \ {\it Participatory Universe} \ \ (to Z. D. Walton \& T. Toffoli)} \label{Walton1} \label{Toffoli1}

I'm just starting to understand your paper.  I hope you'll forgive me if it takes a while.  I am certainly sympathetic to your goal:  ``Derive quantum mechanics from the premise that we live in a participatory universe.''  What a great thing it would be!

In the meantime, maybe you guys would enjoy (as a little ``bathroom reading'') something I put together a while ago:  It's titled {\sl Notes on a Paulian Idea}, \quantph{0105039}.  The thing I call ``the Paulian idea'' is the idea that the (quantum) observer cannot be detached from the phenomena he {\it helps\/} bring about.  If you'd like the file bound as a (paperback) book, let me know, and I'll have a copy (or copies) of the {\Vaxjo} University Press edition sent to you.

I also have a semi-historical document I'm putting together:  ``The Activating Observer:\ Resource Material for a Paulian--Wheelerish Conception of Nature''.  Presently, it stands at 124 pages of quotes drawn from (some fraction of) the 459 references it cites.  I'm hoping to post it next spring, but I've still got quite a lot of work to do to get it into shape before then.  (It'll probably double, if not triple in size before then.)  In any case, though, if you're interested in seeing it as it stands, I'll send you a PDF file of the draft.  (As payment, I only request that you compile and point out to me any typos you find in it.)

\section{15-04-04 \ \ {\it Slowly Coming Back} \ \ (to S. Hartmann)} \label{Hartmann2}

Slowly I'm coming back to speed on my backlog of work.  I decided to stay in Munich until the 14th (yesterday) and after Freiburg, I almost didn't do a thing work-wise.  In an impromptu way, I decided to take a real vacation (not something I do very often)---I neither read nor wrote almost any email!

Tomorrow I'll write {\Caves} and {\Schack} about your proposal of pitching in on the ``Being Bayesian in a Quantum World'' conference.  I'm quite thrilled about the idea.  In the meantime, let me paste in the list I had drawn up of potential participants to give you a firmer idea of the sort of thing I had been having in mind.

By the way, I looked at Hagar's paper to the level of being able to find the part you told me about:  ``Motivated in this way Fuchs then goes on to `trash' (Fuchs' own description of his attempt to understand QM) about as much of QM as he can \ldots''  I thought he could only be talking about my lines in \quantph{0205039} where I say:
\bq
The task is not to make sense of the quantum axioms by heaping more
structure, more definitions, more science-fiction imagery on top of
them, but to throw them away wholesale and start afresh.  We should
be relentless in asking ourselves:  From what deep {\it physical\/}
principles might we {\it derive\/} this exquisite mathematical
structure?  Those principles should be crisp; they should be
compelling. They should stir the soul. When I was in junior high
school, I sat down with Martin Gardner's book {\sl Relativity for
the Million\/} and came away with an understanding of
the subject that sustains me today:  The concepts were strange, but
they were clear enough that I could get a grasp on them knowing
little more mathematics than simple arithmetic. One should expect no less for a proper foundation to quantum theory. Until we can explain quantum theory's {\it essence\/} to a junior-high-school or
high-school student and have them walk away with a deep, lasting
memory, we will have not understood a thing about the quantum
foundations.

So, throw the existing axioms of quantum mechanics away and start
afresh! But how to proceed? I myself see no alternative but to
contemplate deep and hard the tasks, the techniques, and the
implications of quantum information theory. The reason is simple,
and I think inescapable.  Quantum mechanics has always been about
information.  It is just that the physics community has somehow
forgotten this.
\eq
But indeed he was right.  At the beginning of the ``Intermission'' section, I write:
\bq
Let us take a deep breath.  Up until now I have tried to trash about as much quantum mechanics as I could, and I know that takes a
toll---it has taken one on me.  Section 3 argued that quantum
states---whatever they are---cannot be objective entities. Section 4 argued that there is nothing sacred about the quantum probability
rule and that the best way to think of a quantum state is as a state of belief about what {\it would\/} happen if one were to ever
approach a standard measurement device locked away in a vault in
Paris. Section 5 argued that even our hallowed quantum entanglement
is a secondary and subjective effect. \ldots

Subjective.  Subjective!  Subjective!!  It is a word that will not go away.  But subjectivity is not something to be worshipped for its
own sake. There are limits: The last thing we need is a bloodbath of deconstruction. At the end of the day, there had better be some
term, some element in quantum theory that stands for the objective,
or we might as well melt away and call this all a dream.

I turn now to a more constructive phase.
\eq
It's good to have people out there watching you:  they set you straight from time to time when you forget what you've written.

\section{16-04-04 \ \ {\it What is the Difference between a Quantum Observer and a Weatherman?}\ \ \ (to A. J. R. Parker)} \label{Parker1}

Thanks for your interest.  You can find most of what I've written on quantum foundations posted at my website (link below).  Further materials can be found at the {\tt quant-ph} archive.

With regard to your query about a link between theories of consciousness and quantum foundations, I don't have any answers.  The only tangential link I see between my own quantum foundational program and such things is that I see some affinity between the direction I'm turning toward and the American pragmatist tradition (particularly James, Dewey, and Schiller).  In that connection, I bring up something the quantum computer scientist Scott Aaronson wrote me a while ago:
\bq\noindent
     PS. I remember you talked about William James in the samizdat \ldots\
     Have you read the {\sl Principles of Psychology}?  I'm working through
     it now.  I've decided to recommend it to people as ``the most
     up-to-date, state-of-the-art book about consciousness'' (without
     telling them the publication date).
\eq
Scott is a serious scientist (much more serious than I am).  Thus it might be worth your while.

Good luck.

\subsection{Andrew's Preply}

\bq
I note that you recently gave the above paper at All Soul's College Oxford.

I am a psychiatrist, but also studying for an MA in philosophy of mind. My interest is the nature of consciousness and whether quantum mechanics has serious implications for current neuroscientific and philosophical theories of consciousness. My current position is that it does \ldots\ that is, it seems, if my understanding of what it is to be an ``observer'' is broadly correct.

The latter issue seems to be addressed head on by very few people in the field --- although I have to admit, I have to skip over the equations given my non-physics/maths background. Perhaps there is an understanding growing which I am not aware of?

I wonder if your paper above may help, and whether it is available to send by e-mail.
\eq

\section{19-04-04 \ \ {\it Part 2.1} \ \ (to D. M. {\Appleby})} \label{Appleby7}

I'm sorry; you've caught me at one of those times when I am in ``bad
correspondent'' mode.  (At the moment it looks like this is going to
continue for a while---at least until I get my post-sabbatical plans
settled.)

Presently, I guess I don't have much to say about your note except
that I still don't like propensity, even as a view from somewhere.
For it would mean---as far as I understand the term---that from my
(or your or his or her, etc.)\ point of view nature has a certain
tendency to this or that (quantifiable by the probability
distribution each of us happens to use).  I just think that
terminology is misleading, attempting as it does to once again
materialize probabilities, rather than let them stand for sheer
ignorance or opinion.  The only direction I see forward is for all of
us to recognize starkly that the world owes us nothing. Regardless of
our probability assignments, the world owes us nothing.

Beyond that, I also don't think I'm happy with the conception of
physics as a {\it view\/} from somewhere.  The term I protest is
``view.'' Presently, I think it's all about survival, period, with
the last bit of representationalism banished.  I agree with the
``somewhere-centeredness'' you are striving for---though in my terms,
for each of us, it is about MY survival (remember the note I sent you
titled ``\myref{Mermin101}{Me, Me, Me}''?)---but I would be hesitant to call that a
``view.'' Do have a read of {\Rorty}'s ``philosophical papers,'' Volume
1---I think he does a pretty good job of what I'm shooting for (what
I'm shooting for in QM particularly, where I think the evidence is
greater than what he's got to work with).  But that's about all I can
say at the moment.

Or maybe I can swipe a few (relevant) words from a recent note that
Michel Bitbol wrote me.  I'll place them below.  Maybe they add a
little gloss to what I'd like to say myself (if it were a better
world with fewer time constraints).

Anyway, really, thanks for your long note.  I hope I'm not offending
with this short reply, but you're not alone:  I've only been replying
to small fraction (the most urgent) of my emails lately.

\section{20-04-04 \ \ {\it Accumulating Compliments} \ \ (to C. M. {\Caves})} \label{Caves77.01}

Physicist or philosopher?  The one calls me the other \ldots\ which of course means I'm neither \ldots\ which explains why I accumulate compliments rather than positions.

\section{21-04-04 \ \ {\it From Glub to Snowflakes, Creation, and Construction} \ \ (to M. Bitbol)} \label{Bitbol2}

Thanks for your thoughtful letter.  I have very much enjoyed reading
it (maybe five times now).

\bmb
At this stage, I have to state my major point of disagreement with
you: I suspect the remark you made about the ``many-worlds
interpretation'' applies quite well to some of what you say. Let me
explain this.

You write: ``What I find egocentric about the Everett point of view
is the way it purports to be a means for us little finite beings to
get outside the universe and imagine what it is doing as a whole.'' I
deeply agree with you.

But, then, you begin to do essentially the same as the naive
Everettian. You try to describe our situation in the world from a
sort of vantage point. Here is the sentence in your paper that made
me suspicious:  ``I think the solution is in nothing other than
holding firmly---absolutely firmly---to the belief that we, the
scientific agents, are physical systems in essence and composition no
different than much of the rest of the world. But if we do hold
firmly to that---in a way that I do not see the Everettist as holding
to it---we have to recognize that what we're doing in the game of
science is swimming in the thick middle of things.''

Of course, once again, I am delighted by the Pascalian metaphor of
``swimming in the thick middle of things''. But you seem to take it
as more than a metaphor: a definite belief about (not to say a
faithful description of) the world and our position in it. You write
seriously that ``we, the scientific agents, {\em are} physical
systems in essence and composition no different than much of the rest
of the world''. Isn't this a way of extrapolating one of our
pragmatic-adaptative concepts, one of the concepts we need to swim
with some success in the midst of the ``glub'' (here, the concept of
a ``physical system''), in order to describe everything including us
as if it were seen from outside? I hear you saying something like
``the world as I see it from my cosmic exile is made of physical
systems and each one of us is one of these physical systems''.  But
if we are ``swimming'' in the deep ocean of whatever we call
``reality'', we have absolutely no context-independent concept at our
disposal, not even the very general meta-concept of ``physical
system''. We must say that we ignore {\em everything} of the ``thing
in itself'', including whether it is organized or not in a plurality
of ``physical systems''. And we must therefore content ourselves with
stating the formal conditions of our cognitive aptitudes (within it).
This latter attitude is typical of the Kantian and neo-Kantian
lineage of philosophy (when one gets rid of the foundational aspect
of Kant himself and holds a pragmatic variety of Kantianism, as I
do). I hope you'll recognize it as a radical variety of your
view\ldots\ I wonder whether you'll become a member of our radical
club or rather decide to stick to your position as it stands.
\emb

I don't know that I have an answer for you at the moment.  (I hope
you understand that all my efforts, all my writings, are of the
groping variety---I have no final answers to anything.  Nor do I hold
any pretense of being consistent from one email or paper to the next.
The only thing I can promise is that I do strive for consistency, and
I welcome exercises like the one you've presented me.)

To make a start of an answer though, I think my usage of the term
``physical system'' is considerably more nuanced than you probably
guess.  (Though, it may not have been so nuanced at the time of my
writing the anti-{\Vaxjo} paper.  Alternatively, I may have simply
lapsed while I was writing those notes.)  I say that it is nuanced,
because I often toy with the idea that ``physical systems'' are
agent-defined.  You will find this idea probably first appearing in
my writings in the chart on page 292 of {\sl Notes on a Paulian
Idea}. There I ask, ``What is a quantum system?''  And I reply to
myself, ``A line drawn in the sand.''  (It goes back to a 1995 letter
to Greg Comer.)

Maybe I can do a better job of what I am thinking, though, by
attaching a couple of other emails that I've been promoting recently.
They're attached to this letter in a file titled {\tt
  For{\Marcus}.pdf}. In the letter ``\myref{Mermin101}{Me, Me, Me}''
to {\Mermin} and {\Schack} within that collection, I give a definition
of what I mean by ``system.''  I hope that definition will make you
think twice about the characterization of my views you gave above.

In general lately, I'm not even sure what I can make of the idea of
``thing in itself''---it now sounds too static for what I'm trying to
get at.  To that end, let me attach another little compilation---this
one titled {\tt ForSlusher.pdf}.  In particular, I hope the note
within that titled ``\myref{Wiseman6}{The World Is Under
  Construction}'' will help you better see what I am talking about.

Finally, let me paste below still another note to help muddy the
waters.  It is a string of earlier notes culminating in a few remarks
to Jeff Bub; so in reading it linearly you will be traveling
backward in time.  Its relevance here is that I think you think when
I utter the words ``physical system'' I am doing it in the sense that
William {\James} describes below as a case of ``nothing but.''  But I
sincerely hope that is not what I am doing.

Anyway, all of this is probably not the sort of thing you were
expecting as a reply:  It is quite roundabout.  But I am trying to
put three lines of thought (as expressed in the three collections
mentioned above) into a consistent whole.  If I can do that, I think
it will count as something of a direct answer to your query.

Does any of this make sense to you?  Or does it all look more like
the ramblings of a crazy man?

\section{21-04-04 \ \ {\it Essential Incompleteness} \ \ (to W. G. {\Demopoulos})} \label{Demopoulos1}

Thanks for sending me your paper ``Some Remarks on Elementary
Propositions and Partial Boolean Algebras.''  I've taken a shot at
understanding it---i.e., I've read it by some measure---but I
probably didn't fare as well as I should have.

As you argued in your earlier letter to me (one from last year
sometime), our views---or maybe just our languages---may not be so
incompatible as one might think.  However, I am left with the feeling
that this is only a contingent feature of the particular stages of
the game we happen to be at, at the moment.  In particular, from my
own view, I think it is quite important that we strive to stop
thinking of quantum states as states of knowledge about the TRUTH
VALUE of this or that proposition (even if truth value is not
invariant with respect to `experimental arrangement'---the idea you
are toying with).  My feeling is that the imagery of measurement
outcomes mapping to truth values (in this context anyway) will only
cloud our vision for how to take the next big step.

What is the next big step?  I think it is a deeper understanding of
how---very literally---the world ``is in the making'' (to use a
{\James}ian phrase).  To try to make that idea at least graspable (if
not either clear or consistent yet), and to try to show you quantum
theory's role in all this, let me attach four letters I've written
recently. They're contained within the attached files {\tt
For{\Marcus}.pdf} and {\tt ForSlusher.pdf}.  I think they are my best
statements to date of what I am shooting for; and I think that goal
fundamentally conflicts with the idea of ``measurement'' propositions
having truth values in the conventional sense.

That is not to say, however, that I am yet ready to give up on the
idea of physical systems having autonomous properties.  The question
is, what can still be pinned down as a property in the conventional
sense?

In the letter you wrote me way back you said,
\bwd
   What I've tried to address in my paper is the question whether there
   is anything in the quantum theory's conceptual framework that plays
   a role analogous to that played by a classical state. My suggestion
   is that the closest analog of this is given by the elementary
   physical propositions that are true of the system. They, rather than
   the quantum state, expresses the underlying reality that the theory
   purports to describe.
\ewd
As I said above, that's where I don't want to go.  Instead, at least
at the moment, the only thing I am willing to think of in quantum
theory that ``plays a role analogous to that played by a classical
state'' is a system's dimensionality.  [Question: What is this
system's ontic state?  Answer: $D$, just $D$.]  Here's the way I put
it in the paper ``On the Quantumness of a Hilbert Space'' which will
be appearing on {\tt quant-ph} tomorrow:
\bq
   In this paper, I present some results that take their {\it
   motivation\/} (though not necessarily their interpretation) in a
   different point of view about the meaning of a system's
   dimensionality. From this view, dimensionality may be the raw,
   irreducible concept---the single {\it property\/} of a quantum
   system---from which other consequences are derived (for instance, the
   maximum number of distinguishable preparations which can be imparted
   to a system in a communication setting). The best I can put my finger
   on it is that dimensionality should have something to do with a
   quantum system's ``sensitivity to the touch,'' its ability to be
   modified with respect to the external world due to the interventions
   of that world upon its natural course.  Thus, for instance, in
   quantum computing each little push or computational step has the
   chance of counting for more than in the classical world.
\eq
This language is definitely not completely consistent with the vision
I outline in the attached letters but it is the best I can do at the
moment.  (``Modification without modificata?!?,'' I hear you asking.
``What could it mean?'')

In any case, I have looked at your new paper, and I have re-looked at
the paper you gave me last year, and I will be bringing both of them
with me to Maryland.  It would be a pleasure to talk further about
all this---how we differ, or even how I missed something and the gaps
may be erasable after all.  (The latter would be very nice if it is
the case.)

\section{21-04-04 \ \ {\it Can Your Toy Model Do This?}\ \ \ (to R. W. {\Spekkens})} \label{Spekkens31}

Attached is an old set of notes that I finally glossed up into a paper and sent off to quant-ph today.  [See \arxiv{quant-ph/0404122}.] It dawned on me that one ought to ask a similar question of your toy model:  I'm thinking of the part of my paper surrounding Eqs.\ (13) and (14).  There one sees that quantum systems' sensitivity to eavesdropping scales supermultiplicatively in the number of systems making up a composite system.  Does a similar effect arise for your toy model?  I don't have a strong feeling one way or the other which way the answer will go.

\section{22-04-04 \ \ {\it Language Games} \ \ (to N. D. {\Mermin})} \label{Mermin112}

Correlation without Correlata \ldots\ Modification without Modificata
\ldots\ No Stimulation without Stimulata!

I actually have found a way to make sense of the last one, ``no
stimulation without stimulata,'' within my budding point of view
about quantum mechanics.  Have I told you?  No stimulation to the
agent (in old language ``the observer''), without simultaneously
recognizing that the world must be stimulated in return.  No
stimulation to the agent without the external world being
stimulata---i.e., that which is stimulated---in its own right.

Language games.

\section{23-04-04 \ \ {\it Demonizing Bayesians} \ \ (to J. D. Norton \& J. Earman)} \label{Earman1} \label{Norton1}

\bjdn
As you may know, I've been commissioned to speak on Maxwell's demon at
this  year's offering of New Directions and give my response to
Charles Bennett's reply. Last year I had only a visceral sense that
something was wrong.  Since I've got to stand and speak, I tried to
make these concerns more precise and now I think they are very
precise. The result can be downloaded (in draft form) at: \ldots

Of course with any paper like this there is always a nagging worry
that I have missed something obvious or not so obvious. So, if you do
look at the paper, I'd be grateful for any reactions.
\ejdn

I've started to absorb a bit of your paper now and am enjoying it.  I'm looking forward to overhearing the discussions in Maryland.  (My only advice at this stage is that you might find a way to make the title a little less belligerent---but that's your business.  I've got no room to talk given some of the introductions to my own papers.  See for instance, \quantph{0106166} where I demonized the Zurekians.)

In the meantime, could one or the other of you send me electronic versions of your 1998 and 1999 exorcist papers?  I don't have that journal easily available where I am.

It is good for me to think about these demon problems again, as I haven't really thought about them since long before my transformation to (some flavor of) personalistic Bayesianism.  The justification for Landauer's principle that I had used in my own mind previously always involved considering the erasure from two points of view:  1) what happens from the demon's view, and 2) what happens from a ``view from nowhere'' that includes the system the demon is acting upon and the demon himself.  Matching inside-view to outside-view probability distributions obtains (or, at least, so I had thought at the time) the result.  But a personalistic Bayesian has no {\it a priori\/} reason to match those distributions---for they must be attached to some agents if they exist at all, and there can be nothing that automatically requires the agents to have compatible assignments.

Anyway, it's a good opportunity for me to rethink all these things.  Now I am inclined to wonder whether something like a Dutch-book argument might not banish the demon forever.  Namely, could one show that an agent (i.e., a demon) who claims 1) to be ignorant of the molecule's position after removing the partition in a {\Szilard} piston, and 2) to be able to extract work ad infinitum, is simply incoherent (in something like a Dutch-book sense)?  Or something like that \ldots

Looking forward to seeing John N. again and meeting John E. for the first time.

\section{23-04-04 \ \ {\it Fest Paper} \ \ (to C. M. {\Caves})} \label{Caves77.02}

Excellent paper!  I started reading it with breakfast this morning and have thought it great so far.\footnote{\editornote See S. T. Flammia, A. Silberfarb and C. M. Caves, ``Minimal Informationally Complete Measurements for Pure States,'' Found.\ Phys.\ {\bf 35} (2005), 1985--2006, \quantph{0404137}.}  Silly me, but I tend to think there is something deep here beyond the technical result---something that will help us understand the very meaning of pure states.  (Somehow it brought me back to the old days when, for instance, I had read your paper with Sam, ``Wringing Out Better Bell Inequalities,'' from afar.  I thought, ``Now there's some guys who are going places.'')  This is good stuff.

Are these two gentlemen your newest crop of students?  Does it look like you've got a good, productive crop again?

I think you can leave your epilogue.  What does ``parvo non ex nihilo'' mean precisely?\footnote{\editornote Roughly, ``a little not from nothing.''  \emph{Parvum,} in the nominative case, may be better than \emph{parvo,} which is dative or ablative.}  Is all the dialogue modeled on Shakespeare, or is it a conglomeration of stuff?  (Unfortunately, I'm not literary enough to know the answer immediately; and I'm too lazy to Google enough to find the answer.)

\subsection{Carl's Reply}

\bq
Steve Flammia is my newest student.  I think he is excellent.  Andrew Silberfarb is Ivan's student, within a year of finishing. He takes an interest in everything---quantum gravity is his main side interest---and is able to make real contributions.  So this is just an example of what can happen when he takes an interest; the proof that $2D-1$ doesn't work was worked out as a collaboration between Steve and Andrew.

The epilogue is a combination of Shakespeare, the King James Version, and the Wyler reference in the paper, which was a spoof of Wheeler's writing style.
Parvo non ex nihilo comes from Wyler.\footnote{\editornote Well, not verbatim.  See J. A. Wyler, ``Rasputin, Science, and the Transmogrification of Destiny,'' Gen.\ Rel.\ Grav.\ {\bf 5} (1974), 175--82, \myurl{http://www.nr.com/JohnArchibaldWyler.pdf}.}
\eq

\section{24-04-04 \ \ {\it Trees for the Woods} \ \ (to R. {\Schack})} \label{Schack81}

I read {\tt t.tex} with breakfast this morning, and I don't quite know what to tell you at this stage.  When we last talked about it, I thought you were going to make a clean simple point in the paper:  That when one generates a 50/50 distribution via a quantum measurement on a pure state, one has much more than simply the 50/50 state of belief about the outcome.  One also has the belief that no further part of the world can have any insider information about the outcome.  And it is that that makes a quantum random number generator interesting.  But at least in the present draft I don't yet see that idea cropping up (or even presented in the bold way that it ought to be).  I.e., at this stage it's hard to see the forest for the trees.

\section{26-04-04 \ \ {\it Batting Eyes and Smiling Faces} \ \ (to L. Henderson)} \label{Henderson3}

\bleh
There is no option to photocopy it, except if they do it for you, and
in this case that would cost \$113.75, since the thesis is
455 pages long (it is a Long thesis, ho ho). If you still want it I can
order it for you, but it will take about 5--7 days so I wouldn't be
able to bring it to the conference.
\eleh

To show you how weird I am (read devoted), I'll take it!  (They'll regret it though:  I'll just post the damned thing to the internet once I get it and then they'll never get their \$113.75 again.)  Can I pay you back in Maryland this weekend?  (Cash.)  Also, maybe if you'll bat your eyes and give a big smile, they'll be compelled to give it to you before Thursday.  I'll pay for your extra trouble there too:  beer, dinner, fantastic ideas, \ldots\ some combination of all three.

See you soon!  And thanks again \ldots

\section{27-04-04 \ \ {\it Jeffrey Knew Certainty} \ \ (to R. {\Schack})} \label{Schack82}

\begin{itemize}
\item R.~C. Jeffrey, ``De Finetti's Radical Probabilism,'' {\sl Probabilit\`a e Induzione---Induction and Probability}, edited by P.~Monari and D.~Cocchi, (Biblioteca di Statistica, CLUEB, Bologna, 1993), pp.~263--275.
\end{itemize}

\section{27-04-04 \ \ {\it At It Again:\ Delirium Quantum} \ \ (to H. M. Wiseman)} \label{Wiseman13.1}

I'm at it again with my `samizdatery':  I put together a little collection of emails that may be relevant for a meeting on the philosophy and history of science I'm attending next week.  And I am flirting with the idea of posting it on {\tt quant-ph} (though I haven't made a decision yet).  The document is attached:  {\tt DeliriumQuantum.pdf}.  [See ``Delirium Quantum,'' \arxiv{0906.1968v1}.]

The reason I'm writing is to make sure the document doesn't offend you.  The first two letters contain direct quotes of yours, and Section 4 mentions you both at the beginning and at the end.  Let me know if I have your approval to post as is.

There is one little bit of new stuff in the paper that you may not have seen before (or at least a combination of words that you may not have seen before):  Footnote \#2 on page 7.  I dream that I might have finally hit the sweet spot of clarity with those words, but I probably haven't.

\section{27-04-04 \ \ {\it At It Again:\ Delirium Quantum} \ \ (to A. Sudbery)} \label{Sudbery8}

I'm at it again with my `samizdatery':  I put together a little collection of emails that may be relevant for a meeting on the philosophy and history of science I'm attending next week.  And I am flirting with the idea of posting it on {\tt quant-ph} (though I haven't made a decision yet).  The document is attached:  {\tt DeliriumQuantum.pdf}.  [See ``Delirium Quantum,'' \arxiv{0906.1968v1}.]

The reason I'm writing is to make sure the document doesn't offend you at a personal level.  Section 5 in particular quotes you directly.  Let me know if I have your approval to post the document as is, or whether you would suggest some changes.

There is one little bit of new stuff in the paper that you may not have seen before (or at least a combination of words that you may not have seen before):  Footnote \#2 on page 7.  I dream that I might have finally hit the sweet spot of clarity with those words, but I probably haven't.

\section{28-04-04 \ \ {\it Darwin, Qubit, Butterfield} \ \ (to C. G. {\Timpson})} \label{Timpson3}

\bcgt
Here's a question for you: Do you have any idea of when the phrase
`quantum information' might have begun to appear in print? It's in
Peres and Wootters' 1991 PRL, anything earlier do you think?
\ecgt

Good question!  I wish I knew the answer.  Looking through my own database, the earliest paper I could find with ``Quantum Information'' in the title was:
\begin{itemize}
\item
Masanori Ohya, ``On Compound State and Mutual Information in Quantum Information Theory,'' {\sl IEEE Transactions on Information Theory}, {\bf IT-29}(5), 770--774 (1983).
\end{itemize}
But I also seemed to recall an early paper by Roman Ingarden titled ``Quantum Information Theory''.  So I Googled it and found a nice paper by Barbara Terhal that references it.  It was a 1976 paper.  (Barbara's paper is on {\tt quant-ph}, titled ``Is Entanglement Monogamous?'')  As Barbara stresses though, what those people (as well as Holevo) were concerned with in the early days was using quantum states and quantum channels for transmitting classical information.

The idea of transmitting quantum states for the sheer beauty of sending intact quantum states was something that only arose with the {\Bennett}, Schumacher, Wootters club.  I wonder if maybe the phrase appeared in Ben's PhD thesis:
\begin{itemize}
\item
Benjamin Wade Schumacher, {\sl Communication, Correlation and Complementarity}, Ph.D. thesis, The University of Texas at Austin, 1990.
\end{itemize}

The word ``qubit,'' by the way, was invented by Schumacher.  And, as opposed to what Artur Ekert's webpage says, I believe it was in a private conversation between Schumacher and Wootters somewhere in New England, not at the Broadway meeting in the UK.

Maybe I'll check with Schumacher and Wootters on these things.

\bcgt
Enjoy the Seven Pines, I believe Jeremy B and Olly Pooley are going.
\ecgt
I didn't know Jeremy would be there!  That's great.  Even more fun now. (Don't know the other fellow though.)

\section{28-04-04 \ \ {\it Quantum History} \ \ (to B. W. Schumacher \& W. K. Wootters)} \label{Wootters19} \label{Schumacher13}

Since I'm turning into the amateur historian of the field, could I get either or both of you to record your earliest recollections of the words ``qubit'' and ``quantum information''?

I know that Ben produced ``qubit,'' but where and when?  (Ekert has a webpage that says the word was coined at the Broadway meeting in 1993, but I thought it came earlier than that.)

Concerning the idea of calling a quantum state ``quantum information,'' where did that come from?  Do either of you know?

\subsection{Ben's Reply}

\bq
OK, here is the whole story as I remember it.  I'm not sure that I've ever written it down.

In the spring of 1990 at the Santa Fe Institute, Bill (Wootters) and I discussed the idea that, by using suitable code words and decoding observables, you could reach the Holevo bound for classical information transfer over a quantum channel.  I don't think I knew that Holevo had previously conjectured this; Bill will have to speak for himself.
We worked on this for a couple of years.  The problem proved to be horribly difficult.

In 1991--92 I had a student named Eric Nielsen who was doing a senior honors project on Maxwell's demon.  I arranged to have Bill come to Kenyon for a couple of days in May of 1992 as an ``outside examiner''
for Eric.  He did, and we also spent a day or two trying everything we could think of to find our way into the Holevo capacity problem.
We didn't get anywhere.

After a couple of years of struggling with the problem, I was pretty discouraged.  I drove Bill back to the airport in Columbus, which takes about an hour.  As often happens when nothing is working, the talk turned a little crazy.  Maybe, we said, we are asking the wrong questions.  Maybe in quantum mechanics, the old ideas of information are just not appropriate.  Maybe we need a new idea of ``quantum information''.  And we could measure it in ``qubits''!  The idea of measuring something in ``qubits'' (like Noah's ark) struck us as immensely funny, and we laughed a good deal.

(Who made the joke?  Honestly, I do not remember for sure.  I think that I was the one who made the ``quantum information'' remark, but the joke sounds like Bill.  Don't know, though.)

Then I dropped Bill off at the airport and drove back.  During the drive back, I started thinking about our joke.  It occurred to me that it was not a bad idea, actually.  I understood several things immediately -- such as the fact that ``quantum coding'' could not be a copying process because of the no-cloning theorem, and the fact that the qubit would be a generic two-level system.  When I got back to Gambier, I spent the summer cooking up the data compression theorem.

Interestingly, I later ran across one of my notebooks from graduate school, circa 1985.  I had proved the theorem about typical subspaces then, so this was undoubtedly in the back of my mind somewhere, though I did not remember it and actually proved it a second time.  (I think it is also implicit in Ohya and Petz, but their book did not come out till 1993, I think, and I did not read that till much later.)  What was new in 1992 was the physical picture.

This was first presented at the IEEE meeting on the Physics of Computation in Dallas in October of 1992.  This was about a 12-minute talk, as I recall, so it went by pretty quick.  But Charlie Bennett was in the audience and thought that ``qubits'' were cool.  (I knew Charlie from previous meetings in various locales.  He had explained superdense coding to me in the fall of 1991 in Spain.)

At the time, I definitely talked about quantum information as a distinct concept -- in fact, the paragraph at the end of my 1995 data compression paper (which was written in 1993) is pretty much a verbatim quotation of end of my 1992 talk -- a paragraph I had carefully composed in advance to put my idea across in the brief time I had in Dallas.  I was quite consciously suggesting that there was a new field to be explored.

The Broadway meeting in 1993 was undoubtedly when lots of people who cared heard about this, and when Richard Jozsa made his suggestion for improving the proof.  I think that my talk in Broadway tracked the Dallas talk closely, except for being longer.

BTW, I never considered another spelling for ``qubit''.  So what {\Mermin} says is ``orthographically preposterous'' seemed obvious to me.

Later, of course, we returned to the Holevo capacity problem and were more successful, thanks to Bill's persistence and a few tidbits from the data compression paper.  So the story has a very happy ending.

So the whole thing started for me more or less as a really funny joke.
There's a moral there somewhere.  I will be interested to hear if Bill's recollection matches mine.
\eq

\subsection{Bill's Reply}

\bq
Ben's recollections sound entirely right to me.  In the course of our discussion in the car, Ben suggested that maybe we needed to think about a new kind of information that is fundamentally quantum mechanical.  And I remember very clearly that he is the one who came up with ``qubit'' as a unit of this information.  ``And we could measure it in qubits!''\ were very likely his exact words, including the exclamation point!
\eq

\section{29-04-04 \ \ {\it More Reading} \ \ (to G. Musser)} \label{Musser1}

I'm at the Dublin airport waiting for my departure, but I've got a moment to write.

\bgm
I've been reading your ``only a little more'' paper, actually.  I do not
pretend to understand all, or much, of it, but it is fascinating, and
engagingly written. It almost seems that the role of information theory is not to explain
the quantum per se, but to peel away the subjective aspects of the
theory and see whether some ontological nugget remains.
\egm

Thanks, I'm flattered.  And, yep, that's it.  Here's the way I put it in a recent proposal:

\bq
\begin{center}
   Quantum Foundations in the Light of Quantum Information. \medskip
\end{center}

   Nothing has done more to revitalize the idea that quantum mechanics
   can be understood at a deep intuitive level---with the concomitant
   benefits to physics this could bring---than the development of
   quantum information. There is a reason for this. Quantum mechanics is
   predominantly about information---plain, ordinary Shannon information
   or uncertainty. Embracing this idea is the starting point of my
   research program.

   But it is not the end point:  No physicist would be doing his job if
   he did not strive to map reality itself---that is, reality as it is
   independently of any information processing agents. The issue is one
   of separating the wheat from the chaff: Quantum mechanics may be
   predominantly about information, but it cannot ONLY be about
   information.  Which part is which?  The usual way of formulating the
   theory is a thoroughly mixed soup of physical and informational
   ingredients.

   This is where quantum information (including the collateral fields of
   quantum cryptography, computing, and communication theory) has a
   unique role to play.  Its tasks and protocols naturally isolate the
   parts of quantum theory that should be given the most foundational
   scrutiny. ``Is such and such effect due simply to a quantum state
   being a state of information rather than a state of nature, or is it
   due to the deeper issue of what the information is about?'' Recent
   investigations by several workers are starting to show that many of
   the previously-thought `fantastic' phenomena of quantum
   information---like quantum teleportation, the no-cloning theorem,
   superdense coding, and nonlocality without entanglement---come about
   simply because of the epistemic nature of the quantum state. On the
   other hand, other phenomena, such as the potential computational
   speed-up of quantum computing, seem to come from a more physical
   source: In particular, the answer to the question, ``Information
   about what?''

   When we finally delineate a satisfying answer to this, physics will
   reach a profound juncture.  We will for the first time see the exact
   nature of `quantum reality' and know what to do with it to achieve
   the next great stage of physics. Trickle-down effects could be the
   solution to the black-hole information paradox---perhaps already seen
   in broad outline---and even the meshing together of quantum theory
   and gravitational physics. In the meantime the approach proposed here
   is a conservative and careful one; the work to be done is large. The
   effort aims not to say first what `quantum reality' is, but what it
   is not and gather insights all along the way.
\eq
Anyway, you know what you're talking about:  You got the central idea.

Maybe I could recommend another thing you might enjoy reading (at least parts of):  a paper of mine titled, ``The Anti-{\Vaxjo} Interpretation of Quantum Mechanics,'' \quantph{0204146}.  Particularly I'm thinking of the two ending sections (built from letters to Preskill and to Wootters); ignore the other parts.  They try to build a picture of quantum theory as a ``theory from the inside.''  In addition to that, let me give what I think is my present best compact description of the idea.  It comes from a paper Schack and I posted on {\tt quant-ph} a couple of days ago:
\bq
   Is there something in nature even when there are no observers or
   agents about?  At the practical level, it would seem hard to deny
   this, and neither of the authors wish to be viewed as doing so. The
   world persists without the observer---there is no doubt in either of
   our minds about that.  But then, does that require that two of the
   most celebrated elements (namely, quantum states and operations) in
   quantum theory---our best, most all-encompassing scientific theory to
   date---must be viewed as objective, agent-independent constructs?
   There is no reason to do so, we say.  In fact, we think there is
   everything to be gained from carefully delineating which part of the
   structure of quantum theory is about the world and which part is
   about the agent's interface with the world.

   From this perspective, much---{\it but not all}---of quantum
   mechanics is about disciplined uncertainty accounting, just as is
   Bayesian probability in general.  Bernardo and Smith
   write this of Bayesian theory,
   \bq
   What is the nature and scope of Bayesian Statistics  \ldots\  ?

   Bayesian Statistics offers a rationalist theory of personalistic
   beliefs in contexts of uncertainty, with the central aim of
   characterising how an individual should act in order to avoid certain
   kinds of undesirable behavioural inconsistencies.  The theory
   establishes that expected utility maximization provides the basis for
   rational decision making and that Bayes' theorem provides the key to
   the ways in which beliefs should fit together in the light of
   changing evidence.  The goal, in effect, is to establish rules and
   procedures for individuals concerned with disciplined uncertainty
   accounting.  The theory is not descriptive, in the sense of claiming
   to model actual behaviour.  Rather, it is prescriptive, in the sense
   of saying ``if you wish to avoid the possibility of these undesirable
   consequences you must act in the following way.
   \eq
   In fact, one might go further and say of quantum theory, that in
   those cases where it is not just Bayesian probability theory full
   stop, it is a theory of stimulation and
   response. The agent, through the
   process of quantum measurement stimulates the world external to
   himself.  The world, in return, stimulates a response in the agent
   that is quantified by a change in his beliefs---i.e., by a change
   from a prior to a posterior quantum state.  Somewhere in the
   structure of those belief changes lies quantum theory's most direct
   statement about what we believe of the world as it is without agents.
\eq
Finally, for more on the ``golden nuggets,'' have a look at another new paper of mine (just the intro and conclusions): ``On the Quantmness of a Hilbert Space,'' \quantph{0404122}.

I'll write more about arranging a discussion once I'm in the States.  Gotta run now.

\bgm
We should get you to write for us someday!
\egm

Never thought of that.  But now that you mention it, if {\sl Scientific American\/} would ever like an article on ``progress in information-based interpretations of quantum mechanics'' or an article on ``Being Bayesian in a Quantum World,'' let me know and I'll drop everything else (well, almost) to get it done.

I'll be back in less than 48 hours.

\section{29-04-04 \ \ {\it Turgidity}\ \ \ (to H. M. Wiseman)} \label{Wiseman14}

\bv
{\bf turgid} (t\^ur'j\^{\i}d) adjective\\
1. Excessively ornate or complex in style or language; grandiloquent.\\
2. Swollen or distended, as from a fluid; bloated.\\
\ev
Which definition?

\bhw
I read the footnote and it seemed clear enough. However the last
sentence was a bit turgid: Somewhere in the structure of those belief
changes lies quantum theory's most direct statement about what we
believe of the world as it is without agents.

Just to pick this sentence apart, you have a theory of our beliefs
about the world (quantum mechanics) making a statement about our
beliefs about the world. Is this circular, or are we talking about
different sorts of beliefs? Or maybe in the ``what we believe of the
world as it is'', ``we'' is the authors of this paper. In this case I
suggest it would be much punchier if you replaced the above phrase by
``the world''.
\ehw

Lovely that you say that!  After I wrote the said passage for our paper, I wrote to {\Ruediger}:
\bq
I'm pretty pliable with a lot of the writing.  But I've got to admit
I'm pretty proud of the first two and the last two pages (before
references, that is).  So I'll pray that you spare the knife on those,
especially on the concluding remarks---my pride-and-joys.  In the
words of John Wheeler at his 83rd birthday conference, ``Give the old
man his firecrackers!''
\eq
To which he replied:
\bq
I am surprisingly happy with your intro and conclusion.  Just one
thing. The paradox in the penultimate sentence in the Concluding
Remarks, starting ``Somewhere\ldots'', is clearly intended. I am not too
sure it makes sense, though. I am not sure at the moment if I really
believe this sentence.
\eq

Yes, I was intentionally slightly mystical with that sentence:  I like to end my papers that way---probably a bad habit, but I have this secret dream that it'll help draw people into the program if the whole thing doesn't look too dry.  Have a look at the introductory and concluding sections of my paper ``On the Quantumness of a Hilbert Space'' (just posted a few days ago).  They're maybe my most mystical passages yet \ldots\ or at least I hope so!!  (If they're not, I've really been fooling myself!)

However, you do point out an ambiguity that I had not intended.  By ``we'' I meant ``we physicists,'' or maybe more particularly the community of those physicists who wish to squeeze some statement about (our beliefs about) reality out of quantum mechanics---not ``we authors.''  Being something of a budding pragmatist (in the philosophical sense), I have to add for myself the reminder that {\it at best\/} a physical theory is always a statement of belief about reality.  In its own way it is a gambling commitment.  That's why I wrote ``about what we believe of the world as it is without agents'' rather than simply ``about the world as it is without agents'' (i.e., more along the lines of your suggestion).

Beyond that, however, you were right that I was talking of two different sorts of belief in the sentence.  One can talk about accepting a quantum state $\rho$ and one can talk about accepting the very structure of quantum theory.  In both cases, I would say, one is accepting a state of belief.  However the objects of those beliefs are very different.  In the first case, I would say one is not at all accepting a belief about reality as it exists independently of the agent.  The quantum state, for me, is only my state of belief about how nature will react back upon me consequent to one of my potential stimulations upon it.  On the other hand, accepting the very structure of quantum theory is---I am banking---a hidden statement of what we believe of nature as it is without agents.  There is a hierarchy here:  At one level we cannot get rid of the agent in any convincing way, at the other level we might be able to.  That's the idea.  Finally, the question is how to unlock the hidden statement.  I am banking that the best place to look is in the structure of how the former types of belief (i.e., quantum states) are changed consequent to the acquisition of a measurement outcome.  (I.e., in the deviation from Bayes' rule encoded by the Kraus collapse rules.)

Hope that was worth reading.  (Now I return back down to 35,000 feet where the airplane is situated.)

\section{29-04-04 \ \ {\it Ontic Elements for Quantum Systems} \ \ (to H. Atman\-spach\-er)} \label{Atmanspacher3}

Thanks once again for your hospitality during my visit to IGPP.  And more than that, thanks for sending me your paper with Primas, ``Epistemic and Ontic Quantum Realities.''  I finally got a chance to give it a go today.  (I'm flying to the States as I write this.  It's a 10 day visit, three separate meetings, so I plan to have a few good days to let your ideas rattle around in my head before I have to return to my more usual world.)

You know of course how committed I am to making the ontic/epistemic split absolutely clear within quantum theory.  And I am very happy that it looks like (at least in some instances) we would be compelled to make the split at the same place.  For instance, for me, as for you, a density operator should be classified as an epistemic element.  (A radical Bayesian about probability must go further---as {\Schack} and I argue in our latest paper on {\tt quant-ph}---and also classify quantum operations (i.e., quantum channels) as epistemic, but that is not so important to me at the moment.)

What I am having difficulty understanding on this first reading, however, is what you are taking to be the ontic elements within quantum theory?  What is a quantum system's ``mode of being at a given instant'' (as you say)?  In my own case, the only thing I am willing to relegate to the ontic realm (at this stage in my thought) is the quantum system's Hilbert-space dimension (in the finite-dimensional case)---it signifies something about the system's ``sensitivity to the touch''.  (See my new paper, ``On the Quantumness of a Hilbert Space,'' \quantph{0404122}, to make that remark maybe epsilon clearer.)  Maybe I would be willing to relegate more than stark dimensionality to the ontic realm upon further reflection (and maybe that's where your paper will lead me), but that's where I stand at the moment.

In the hope of making some intellectual progress on these issues during this week of freedom, could I ask you to drop the $C^*$ language for the moment and consider a very simple case so that I might see how your view fares with regard to my pre-established prejudices.  Consider a simple qubit or qutrit, i.e., a 2-state or 3-state quantum system.  What would you identify as the potential ontic states for those systems.  Can you identify them in Hilbert-space language for me?  What are such a system's properties?  I should be in email contact almost every day despite my travels; if you could answer these questions in a language that I can hope to understand quickly, that would be most helpful for getting the ball rolling.

After that, I can see there may be some disagreements between the two of us that may not be so easily erasable.  But I think that is at a further step down the line and we shouldn't quite go there yet.  (For instance, I am thinking of your invoking systems with infinite degrees of freedom for a ``solution'' to the measurement problem.  In my own mind, recognizing that quantum states are epistemic already DISsolves the measurement problem enough.  That was part of the picture I tried to present when I spoke at your institute.)

On the other hand, I was very much intrigued by your idea of relative onticity.  Something feels very right about that, and I can see myself plunging into the idea and trying to adapt it to my own purposes.

Thus I hope you can see that my respect for your work is very strong.  There are points to be hammered out, but I am definitely adding your thoughts to my toolbox.

\section{02-05-04 \ \ {\it Precision} \ \ (to A. Wilce)} \label{Wilce4}

Do you still remember the precise quote of what David Albert said at the end of my talk?  If you do, I'd like to get it recorded into my records.

\subsection{Alex's Reply}

\bq\noindent
``I think we can pretty exactly quantify the interpretational progress made here -- and it's zero.''
\eq

\section{06-05-04 \ \ {\it SIC-POVMs} \ \ (to F. E. Schroeck)} \label{Schroeck3}

\bfes
I was reading ``On the Quantumness of a Hilbert Space'' (given to me by
Stan Gudder) and I wanted to look up Carlton Caves' paper ``Symmetric
Informationally Complete POVMs''. It wasn't online. In particular, I
wanted to know why the factor $1/(d+1)$ occurs in your eqn 15. Can you
give me either a hint, or Caves' e-mail address (or both)?

By the way, the first half of the paper is great, and the other half
is to be read.
\efes

Wow, thanks!

It's a simple consequence of the definition of a SIC-POVM.  The definition is that the POVM consists of $d^2$ rank-1 elements, all of equal trace, and such that the pairwise inner products of nonindentical elements are constant.

Using the fact that the $d^2$ elements form a resolution of the identity, and taking the trace of the identity, you get that the trace of each element must be $1/d$.

Now to get the values of the pairwise inner products, you just square the identity (and consequently square the resolution of the identity) and pull the same trick:  Namely taking the trace of the consequent.

Hope that makes it clear enough.

I'm kind of surprised you couldn't find Carl's notes at his webpage, but maybe he removed them.

\section{06-05-04 \ \ {\it Section 4.1} \ \ (to W. G. {\Demopoulos})} \label{Demopoulos2}

I'm on the plane now.  I guess, at this stage, I don't have much to
add to what I wrote you earlier after all.  The ``snowflakes'' bit
captures most of what I had wanted to say.  The only other thing I
can recommend to have a look at is Section 4.1 (trying to justify
noncontextuality for our probability assignments in QM) in \quantph{0205039} (i.e., ``QM as QI (and only a little more)''). I'll
be curious to know how that argument meshes with your own.

What I would like to go into further eventually is the similarity and
differences between 1) your point of view of the underlying events in
QM as being preexistent (but maybe unique) realities, and 2) my point
of view that they are best supposed as (unique) creations.  I also
get the feeling that your view collides with another one of my
favorite doctrines:  Namely, that we should take POVMs as just as
basic as standard quantum measurements (because they so prettily
match Bayes' rule).  But I'll have to think about that.

\section{06-05-04 \ \ {\it Middle of the Meeting} \ \ (to A. Peres)} \label{Peres66}

\bap
I finished to read (many parts of) {\bf No time to be brief}, Pauli's
biography by Enz. The book is overcomplete, with plenty of irrelevant
gossip on other physicists. Yet, there are many interesting things in
it. I never imagined that Pauli had such a tortured soul and needed
help from Jung and similar crooks. Why did you call your future book
{\bf Notes on a Paulian idea}?
\eap
Because most of the notes in there were deeply influenced by the collection of Pauli's more-philosophical papers {\sl Writings on Physics and Philosophy\/} that I had read in 1995.  Also the letters from Pauli to Fierz (that I had found in one of Laurikainen's books) made a similar effect.

Anyway, in particular, the phrase ``Paulian idea'' is meant to capture the two explicit quotes of Pauli's that I placed at the very beginning of the book.

I agree with your remark about Jung.  I had a fun time recently reading the Jung/Pauli correspondence volume.  There was a lot of trash from both sides, but on the other hand, at least there were several letters on Pauli's side that were quite interesting---and said, I think, some deep physics.  But with regard to Jung's side of the conversation, I really could not make heads from tails---I think it was very likely total nonsense.

Now I should probably return to listening in on the present talk.

\section{11-05-04 \ \ {\it The Late, Late, Late Show} \ \ (to A. Sudbery)} \label{Sudbery9}

I'm at it again with my `samizdatery':  I put together a little collection of emails that may be relevant for a meeting on the philosophy and history of science I'm attending next week.

There is one little bit of new stuff in the paper that you may not have seen before (or at least a combination of words that you may not have seen before):  Footnote \#2 on page 7.  I dream that I might have finally hit the sweet spot of clarity with those words, but I probably haven't.

\bts
The footnote on p.\ 7 is very clear, though I personally have problems
with the notion that quantum mechanics is all about your (my, our)
{\bf interventions} in the world and its response. I know that's what the textbooks say -- you only get a prediction for the result of a measurement -- but I don't think that's an accurate description of the theory (as it actually exists in the practice of working scientists).
\ets

You mean after all this technical work (of mine and so many others), and all my attempts to find the best English yet to tell the story (of ``measurement'' \ldots\ better yet ``intervention'' \ldots\ or still better ``stimulation''), I still haven't gotten any further than what the textbooks say?

One more round, Delia's gone, one more round. \smiley

\bts
I was sorry not to get to the LSE meeting on probability. Was it good?
Lots of elevated blood pressure? You know about our Foundations
meeting in York in September, don't you? No invited speakers, just a
free-for-all. I'd like to incorporate a round table or two, maybe one
on Everettism and one on Bayesianism, or even just one on both, face
to face.
\ets

Yeah, the LSE meeting was pretty good.  I enjoyed it.  And even more than that I enjoyed sparring with Butterfield, Saunders, Brown, and a few others in Oxford afterward.

I didn't know about the September meeting.  Is there a website?  I ought to try to come to that if I can.

\section{12-05-04 \ \ {\it Pragmatism, Quantum, Bayes, and pragmatism (no capital)} \ \ (to J. Woodward)} \label{Woodward1}

I had a look at your profile on your Caltech webpage this morning, and then I worked my way to your Stanford Encyclopedia article.  It was a nice experience reading parts of it---I'll give it a more serious consideration over the next couple of days---for it looks like there is some real overlap in our concerns.  In particular, one might almost describe what I'm shooting for in rewriting the notion of quantum measurement is something like a manipulationist account of it.  In place of ``measurement'', I think it is firmer ground to think of a quantum measurement click as a system's {\it response\/} to an external intervention \ldots\ that is, rather than think of it as the {\it revelation\/} of some pre-existent value.  My latest ways of putting these things are detailed in ``Delirium Quantum'', particularly in the section ``\myref{Mermin101}{Me, Me, Me}''.  Also, there's plenty to read about it scattered throughout my book {\sl Notes on a Paulian Idea\/} (I gave Jed a copy that you can borrow, or I can have one sent to you), for instance on pages 394 or 509--510.  What I have written is surely cloudy thinking in comparison to your corpus, but maybe it's enough to show that my heart is in the right place and that our interactions could be mutually beneficial.

Anyway, that's just a taste of things, and an attempt to put all the issues on the table.

\section{12-05-04 \ \ {\it Does Hilbert-Space Dimension Gravitate?}\ \ \ (to J. Preskill)} \label{Preskill15}

\bjp
In my view, you were the most cogent of the ``philosophers'' I talked
to in Stillwater. 
\ejp
That's because I'm a physicist in disguise.
\begin{center}
  ``Philosophy is too important to leave to the philosophers.'' -- J.A.W.
\end{center}

\section{13-05-04 \ \ {\it Epistemic Pure States} \ \ (to H. Atmanspacher)} \label{Atmanspacher4}

Thanks for the note.  I apologize for not getting back to you sooner, but the conferences I was at knocked me for more of a loop than I had expected.  I'm still recovering, and now my backlog of tardy work is piled even higher!

Let me just say a very short word at the moment.  With your note and the attached paper, you confirmed what I feared:  That you take a pure quantum state to be an ontic state.  This contrasts with our Bayesian approach where all quantum states (pure or mixed) are epistemic.

Let me attach two papers and one pseudo-paper that say something about this approach.  One paper is one of my own with {\Schack} which talks about a Bayesian view of quantum states in general.  The other is due to Rob {\Spekkens}.  In it, he gives the most compelling case I've ever seen that pure quantum states should be viewed epistemically.  He does this by getting to the core of the matter in a toy model (that is not quantum mechanics).  His model lacks a lot---for instance, his observers are still detached in that it is a hidden variable model---but I still think it is a strong argument for the epistemicity of the quantum state.  Finally the pseudo-paper is something I threw together for the Seven Pines meeting in Minnesota.  The relevant sections are the last two, ``\myref{Mermin101}{Me, Me, Me}'' and ``\myref{Sudbery2}{The Big IF}''.  What they try to give a hint of is a view in which quantum theory is not taken so much as a ``theory of the world,'' but rather as a theory of decision-making for agents {\it immersed\/} in and {\it interactive\/} with the quantum world.  Making that change in point of view is a crucial component in what we are after.

\bha
It could be that this provides a bridge between your and our
approaches. In my view and Primas', there is no absolutely
ontic description anyway (one cannot even speak without a context).
In this sense, our ontic states are already relative to a context
which, however, is unspecified.
\eha
I hope so.  I will think about it.

\section{13-05-04 \ \ {\it 10 Lines and MaxEnt} \ \ (to R. {\Schack})} \label{Schack83}

One of the meetings I'm back from is the Seven Pines meeting in Minnesota.  A few philosophers were there including John Earman and a guy named Geoffrey Hellman (a student of Hilary {\Putnam}'s).  In Hellman's (short) talk, he briefly mentioned that the interpretation of probability was important for quantum foundational issues.  He mentioned frequency interpretations, propensity interpretations, and our approach.  He very nicely complimented us by calling it a ``radical move'' (though of course, in private, he said that he didn't believe that it could be the ``right move'').  Earman said in one of our discussion sessions:  ``You mean $\psi^*\psi$ as a Bayesian degree of belief?  Could not be.''

So it goes:  Even the Bayesians won't believe us!

\section{13-05-04 \ \ {\it LNP Volume} \ \ (to M. G. A. Paris)} \label{Paris1}

Just use the abstract we already wrote.  It's only six sentences:
\bq\noindent
The classical de Finetti theorem provides an operational definition
of the concept of an unknown probability in Bayesian probability
theory, where probabilities are taken to be degrees of belief instead
of objective states of nature.  In this paper, we motivate and review
two results that generalize de Finetti's theorem to the quantum
mechanical setting:  Namely a de Finetti theorem for quantum states
and a de Finetti theorem for quantum operations.  The quantum-state
theorem, in a closely analogous fashion to the original de Finetti
theorem, deals with exchangeable density-operator assignments and
provides an operational definition of the concept of an ``unknown
quantum state'' in quantum-state tomography.  Similarly, the
quantum-operation theorem gives an operational definition of an
``unknown quantum operation'' in quantum-process tomography.  These
results are especially important for a Bayesian interpretation of
quantum mechanics, where quantum states and (at least some) quantum
operations are taken to be states of belief rather than states of
nature.
\eq

\section{17-05-04 \ \ {\it Big Bang 1910} \ \ (to G. L. Comer)} \label{Comer51}

The following passage comes from William {\James}'s posthumous book,
{\sl Some Problems of Phil\-o\-sophy}---started in March 1909 and worked
upon until his death in August 1910.

\bq
It is a common belief that all particular beings have one origin and
source, either in God, or in atoms all equally old. There is no real
novelty, it is believed, in the universe, the new things that appear
having either been eternally prefigured in the absolute, or being
results of the same {\it primordia rerum}, atoms, or monads, getting
into new mixtures. But the question of being is so obscure anyhow,
that whether realities have burst into existence all at once, by a
single `bang,' as it were; or whether they came piecemeal, and have
different ages (so that real novelties may be leaking into our
universe all the time), may here be left an open question, though it
is undoubtedly intellectually economical to suppose that all things
are equally old, and that no novelties leak in.
\eq

\section{17-05-04 \ \ {\it {\James}ian Exchangeability} \ \ (to R. {\Schack})} \label{Schack84}

[[The quote below comes from William {\James}'s book {\sl Some Problems of Philosophy}.  There was no further text in this note beyond the quote.]]

\bq
A concept, it was said above, means always the same thing: Change means always change, white always white, a circle always a circle. On this self-sameness of conceptual objects the static and `eternal' character of our systems of ideal truth is based; for a relation, once perceived to obtain, must obtain always, between terms that do not alter. But many persons find difficulty in admitting that a concept used in different contexts can be intrinsically the same. When we call both snow and paper `white' it is supposed by these thinkers that there must be two predicates in the field. As James Mill says: `Every colour is an individual colour, every size is an individual size, every shape is an individual shape. But things have no individual colour in common, no individual shape in common; no individual size in common; that is to say, they have neither shape, colour, nor size in common. What, then, is it which they have in common which the mind can take into view? Those who affirmed that it was something, could by no means tell. They substituted words for things; using vague and mystical phrases, which, when examined, meant nothing.' The truth, according to this nominalist author, is that the only thing that can be possessed in common by two objects is the same {\it name}. Black in the coat and black in the shoe are the same in so far forth as both shoe and coat are called black---the fact that on this view the name can never twice be the `same' being quite overlooked. What now does the concept `same' signify? Applying, as usual, the pragmatic rule, we find that when we call two objects the same we mean either (a) that no difference can be found between them when compared, or (b) that we can substitute the one for the other in certain operations without changing the result. If we are to discuss sameness profitably we must bear these pragmatic meanings in mind.

Do then the snow and the paper show no difference in color? And can we use them indifferently in operations? They may certainly replace each other for reflecting light, or be used indifferently as backgrounds to set off anything dark, or serve as equally good samples of what the word `white' signifies. But the snow may be dirty, and the paper pinkish or yellowish without ceasing to be called `white'; or both snow and paper in one light may differ from their own selves in another and still be `white,'---so the no-difference criterion seems to be at fault. This physical difficulty (which all house painters know) of matching two tints so exactly as to show no difference seems to be the sort of fact that nominalists have in mind when they say that our ideal meanings are never twice the same. Must we therefore admit that such a concept as `white' can never keep exactly the same meaning?

It would be absurd to say so, for we know that under all the modifications wrought by changing light, dirt, impurity in pigment, etc., there is an element of color-quality, different from other color-qualities, which we mean that our word {\it shall\/} inalterably signify. The impossibility of isolating and fixing this quality physically is irrelevant, so long as we can isolate and fix it mentally, and decide that whenever we say `white,' that identical quality, whether applied rightly or wrongly, is what we shall be held to mean. Our meanings can be the same as often as we intend to have them so, quite irrespective of whether what is meant be a physical possibility or not. Half the ideas we make use of are of impossible or problematic things,---zeros, infinites, fourth dimensions, limits of ideal perfection, forces, relations sundered from their terms, or terms defined only conceptually, by their relations to other terms which may be equally fictitious. `White' means a color quality of which the mind appoints the standard, and which it can decree to be there under all physical disguises. {\it That\/} white is always the same white. What sense can there be in insisting that although we ourselves have fixed it as the same, it cannot be the same twice over? It works perfectly for us on the supposition that it is there self-identically; so the nominalist doctrine is false of things of that conceptual sort, and true only of things in the perceptual flux.

What I am affirming here is the platonic doctrine that concepts are singulars, that concept-stuff is inalterable, and that physical realities are constituted by the various concept-stuffs of which they `partake.' It is known as `logical realism' in the history of philosophy; and has usually been more favored by rationalistic than by empiricist minds. For rationalism, concept-stuff is primordial and perceptual things are secondary in nature. The present book, which treats concrete percepts as primordial and concepts as of secondary origin, may be regarded as somewhat eccentric in its attempt to combine logical realism with an otherwise empiricist mode of thought.
\eq

\section{18-05-04 \ \ {\it Agents, Interventions, and Surgical Removal} \ \ (to J. Woodward)} \label{Woodward2}

(I'll drop the ``Professor Woodward'' bit and act like I know you,
now that this is a second contact.)

I had time over the weekend to better digest your Stanford
Encyclopedia article, ``Causation and Manipulability.'' Per my first
guess, we do indeed use a lot of words in common---intervention,
manipulation, agent---though your concern is causation and my concern
is the elicitation of ``quantum measurement outcomes.'' As evidence
of this, see for instance Section 4 of my paper ``Quantum Mechanics
as Quantum Information (and only a little more)'' posted at my
website (link below).

But there are also more than words in common. There is the common
concern of exorcising the (anthropomorphic) agent if possible from
each of our domains. Can it be done? My suspicion about quantum
mechanics is that there is a level at which it can be done and a
level of which it can't. Both levels are needed, however, for an
explication of what the theory is about.

What I liked in learning from your paper is the similarity of our
strategies. ({\Ruediger} {\Schack} tells me I'm a cherry-picker when it
comes to extracting content from papers; sorry if I'm doing that
here.) Anyway, the way I see it is we both start from rather firm
starting points---though they necessarily involve the notion of an
agent---and only then, make a step toward de-anthropomorphizing the
picture. In your case, it's to start with a manipulationist account
of causation and then to move to an interventionist one. The ultimate
goal is to talk of causes in a way that makes no reference to human
action.

In my case, I try first to make sense of a quantum measurement
outcome as a birthy sort of event that happens when two pieces of the
world metaphorically bump into each other {\it in the case that\/}
one of the pieces is an active agent (who is equipped with subjective
probabilities, etc., etc.). Of course, after that I'd like to go the
next step---to imbed these anthropocentric ``births'' in a larger
account of a world in continuous creation, in analogy to your
strategy of imbedding humanly manipulations within a larger
(de-anthropocentrized) interventionist account. At present, I only
have a glimpse of how to do that. About the only thing I feel
confident of is that, if it's going to be done, the project is going
to have to be carried out in a relatively oblique manner---namely,
one that stops talking about the births themselves and transfers
attention to systems' capacities to take part in these kinds of
mini-creations. (This is the stuff I call ``Zing!''\ in footnote 10
of the paper mentioned above.)

Why must the strategy be oblique? Your paper gave me a nice metaphor
for that:
\bq
Human beings cannot at present alter the attractive force exerted by
the moon on the tides (e.g., by altering its orbit). More
interestingly, it may well be that there is no physically possible
process that will meet the conditions for an intervention on the
moon's position with respect to the tides -- all possible processes
that would alter the gravitational force exerted by the moon may be
insufficiently ``surgical''. For example, it may very well be that any
possible process that alters the position of the moon by altering the
position of some other massive object will have an independent impact
on the tides in violation of condition (M2) for an intervention.
\eq
You see, as opposed to classical gravitational phenomena, I believe
quantum mechanics puts us in a new situation. The world appears to be
wired together so intricately that every attempt to extricate or
``surgically remove'' the agent from the quantum measurement process
leaves a conceptually empty structure (I'm thinking of Everettian and
Bohmian quantum mechanics, etc., here). What quantum mechanics is
mostly about is making-do in light of that.

That's on the negative side of things. On the positive side of
things, knowing that the world is made of a ``stuff'' that causes us
to ``make-do'' in this way is something I find tremendously exciting.
I.e., it's not the making-do that's exciting, but the stuff that
gives rise to it that is. This is because I think it leads to a far
more interesting and creative world than we might have imagined
otherwise. In the next note, I'll attach a historical/philosophical
document I've been working on that tries to convey some of this. The
title is, ``The Activating Observer: Resource Material for a
Paulian--{\Wheeler}ish Conception of Nature,'' and it's fairly large
(922K) \ldots\ so don't get frightened that something has gone wrong with
your email. The thing is still far from complete, but I hope it
conveys a flavor of the kinds of thoughts I'm up to in my most
philosophical moments.

How do I live with such a crucial piece of quantum theory (and the
next stage of physics it hints at) being so anthropocentric? Here's
the way I put it in that document: ``Observers, a necessary part of
reality? No. But do they tend to change things once they are on the
scene? Yes. If quantum mechanics can tell us something deep about
nature, I think it is that.'' And it is that idea that I'm working my
tail off to formalize.

Anyway, a long letter just to say I got a lot out of reading your
paper and to reiterate that I think we have plenty of potential for
fruitful interaction. [For instance---on a different subject---my
radically Bayesian nose smells the agent remaining, despite your
efforts, in your account of nondeterministic causation (right after
equation CD). That is because, for such a Bayesian, the probabilities
in the difference you consider still make crucial reference to an
agent for their very existence.]

\section{18-05-04 \ \ {\it Striving for Consistency} \ \ (to A. Grinbaum)} \label{Grinbaum1}

\bag
I am still thinking what to respond to your \quantph{0404122}. Frankly,
I was slightly unhappy to read that ``sensitivity to the touch'' means
``ability to be modified with respect to the external world due to the
interventions of that world upon its natural course''. This sounds so
realist, and the sentence is so different from the rest of your
article. I know that you had acknowledged that the dimension of the
Hilbert space is for you the only remaining element of physical
reality, but I certainly see no need to come to realism from
``sensitivity to the touch''. Why put accent on touch rather than
sensitivity?
\eag

You are right about that, and the only thing I can do for the moment is plead guilty.  I strive for consistency, but I don't always achieve it---worse yet, sometimes my repertoire for proper expression runs limp.  (You'll see in the first letter below how I made fun of myself on just this issue.) [See 21-04-04 note ``\myref{Demopoulos1}{Essential Incompleteness}'' to W. G. {\Demopoulos}.]

Anyway, let me try to make up for it---or make the matters worse!, I don't know which---by sending you some further expressions of what I'm trying to get at.  For one thing, see the attached pseudo-paper (which I haven't yet decided whether I will submit it to the archive or not, your vote would count).  [See ``Delirium Quantum,'' \arxiv{0906.1968}.] For another, see the further letter below that I wrote to Jim Woodward this morning. [See 18-05-04 note ``\myref{Woodward2}{Agents, Interventions, and Surgical Removal}'' to J. Woodward.]

To answer your question, it's very unlikely that I'll be in Europe in the Autumn.  I'm now scheduled to return to Bell Labs in mid-August, and the Fall is likely to be a quite hectic time with us getting settled into our new house, etc.  Sorry; I would have liked to have been at your defense.  But I know you'll do well.

Glad you enjoyed Dublin.

\section{19-05-04 \ \ {\it The Barbecue Quest} \ \ (to G. Musser)} \label{Musser2}

Sorry for the slow reply.

\bgm
To follow up our conversation regarding entanglement, I looked at the
``Wither Entanglement'' section of \quantph{0205039}.  If I follow, the
idea that entanglement is a statement about our knowledge, rather than
about a system, is demonstrated by the derivation of the
tensor-product rule.  What is the physical meaning of that rule?
\egm

I'm not sure what to say beyond what I said in the paper.  The point is simply that the tensor-product rule for generating a Hilbert space to describe composite systems is a consequence of a) the structure of localized quantum measurements, augmented solely with b) classical, subluminal communication between observers at each site.

Maybe I should put the point this way:  If one is looking for a mystery in quantum mechanics, one need look no further the structure of (local) measurements on single systems.  That's where the real mystery is.  Why is the structure of quantum measurements such that it can power a Kochen--Specker no-coloring theorem?  If we had a stick-to-your-ribs answer for that, I think we'd finally be at the end of the quest.

\bgm
I was also told to think of QM as nondeterminstic.  More recently I
was told (cf.\ John Earman), no, it is completely deterministic,
because the {\Schroedinger} equation uniquely maps a wave function at $t_0$
to another at $t_1$.  Now Griffiths tells me to think of it as
nondeterministic again: ``The successive events of a history are, in
general, not related to one another through the {\Schroedinger}
equation.''  Help.
\egm

I'm sure all three of us have very different reasons for saying what we say.  Here's the way I would put it.  You cannot forget the distinction between a) states of knowledge/expectation/belief, on the one hand, and b) what {\it is\/} or what {\it is happening\/} in the world, or more particularly what an agent can provoke of it, on the other.  Neither has an implication on the other.  Quantum time evolution---which is smooth and deterministic---is about how an agent's state of knowledge changes over time.  The singular, unpredictable, event-like ``measurement outcomes'' are about the response an agent's stimulation to a system external to himself will provoke.  Those events or responses seem to come out of the blue.  It is they that are (among) the real things of the world.  The quantum state is just a fancy way of organizing the agent's expectations---thus it is not among the real things of the world.  The only time there is a nonsmooth (indeterministic) change in a quantum state is when the agent holding it incorporates a new piece of {\it data\/} into his expectations.

I thought I had said this much better in {\sl Notes on a Paulian Idea\/} somewhere, but if I did, I couldn't find the place with a quick perusal.  The closest thing I could find was in the notes surrounding ``The Evolution of Thought'', pages 163--168 (in my website version).  Still, you might get something useful out of reading the note titled ``A Non-Randy Non-Bugger'' on page 163 nonetheless.

By the way, I had the secretary in Sweden send you a copy of the {\Vaxjo} University Press cut of {\sl Notes on a Paulian Idea}.  (Kluwer is publishing it later this year too.)  I hope it doesn't take too long to arrive.  The way to treat the book is as bathroom reading (i.e., not systematically)---think of it as a kind of snippet book on information-based interpretations of quantum mechanics.  It's main purpose is to inspire.  If you want to find my insults on various people, use the name index in the back.

\bgm
If after returning NJ you ever teach a senior-level or first-year
graduate-level course in QM, please let me know.  I took such a course
in grad school but the knowledge evidently has a short exponential
decay timescale.  I need to work through the math again.
\egm

Flattery.  (Thanks.)  Better flattery would be to tell me that {\sl Scientific American\/} wants a full article on this stuff.

\section{19-05-04 \ \ {\it Bathroom Reading} \ \ (to P. C. E. {\Stamp})} \label{Stamp3}

Per your request, I just had the secretary in Sweden send you a copy of the {\Vaxjo} University Press cut of {\sl Notes on a Paulian Idea}.  I'm sure I don't need to say it, but the way to treat the book is as bathroom reading (i.e., not systematically, to put it mildly)---think of it as a kind of snippet book on information-based interpretations of quantum mechanics.  Its main purpose is to inspire.

Particularly, something relevant to our last conversation (in the wrap-up session) can be found in two notes in the Barnum chapter, titled ``It's All About Schmoz'' and ``New Schmoz Cola''.  The best part is after Barnumism 5.

The main point I was trying to make to you---when I thought I could do it better than Aspelmeyer---is that in an information-based interpretation of quantum mechanics (like the one {\Caves}, {\Schack}, {\Mermin}, Peres, Hardy, and a few others of us are trying to put together) is that:  Though we take the wavefunction to be information, we never take it to be information about ``what is existent'' in the quantum system, or ``what is happening in the quantum system.''  That it cannot be information about something like that is what I take to be the great lesson of the Bell--Kochen--Specker theorem.  Rather it is always information about how a system will {\it respond\/} to an external stimulation (i.e., in usual language, a measurement \ldots\ but I want to get away from the usual language because it carries the wrong imagery).  This is the way to think of it when the wave function is about a single system and the way to think of it when the wave function is about a composite system---it doesn't matter.

So, what I am asking---among other things---is that the notion of ``measurement'' be replaced with the idea of ``stimulation and response.''  We only use wave functions to calculate our expectations with regard to how a system will respond to one or another of our possible stimulations upon it.  In this light, there is no contradiction between an ``inside view'' of a measurement where an unpredictable click really happens and an ``outside view'' (encompassing both the original system and the first observer), where a wave function for the composite system undergoes a smooth and deterministic transition.  This is because the ``outside view'' wave function refers again to {\it nothing more than\/} the outside observer's expectations for how that composite system will respond to external stimulations upon it.

Hope that helps clarify at least what I was trying to say.

\section{24-05-04 \ \ {\it Admissions} \ \ (to H. Mabuchi)} \label{Mabuchi8}

\bhm
Did you find time to read the Richard Powers novel I sent you?
\ehm

I've got to admit that I haven't read it yet.  But because of my upcoming visit, it's hittin' high priority.  You should be doubly honored, it'll be the first fiction book I've read in over 10 years.  (Unless you think I ought to count David Deutsch's book {\sl The Fabric of Reality\/} \ldots)

\section{25-05-04 \ \ {\it Bathroom Reading} \ \ (to G. Hellman)} \label{Hellman1}

It was good meeting you at the Seven Pines meeting.  I was particularly impressed with your fairness, balance, and patience in taking in a wide range of viewpoints.  If you could send me some of your ``relation without relata'' philosophy-of-mathematics papers, I would enjoy reading them.  (The best format is to send them electronically if you can, but snail-mail will do as well.)

I just had the secretary in Sweden send you a copy of the {\Vaxjo} University Press cut of {\sl Notes on a Paulian Idea}.  I'm sure I don't need to say it, but the way to treat the book is as bathroom reading (i.e., not systematically, to put it mildly)---think of it as a kind of snippet book on information-based interpretations of quantum mechanics.  Its main purpose is to inspire; I hope it fits the bill at least a little.  The name index at the back is hopefully a handy way to look up one or another point of view.  You might be particularly interested in the early discussions with Jeff Bub.

I intend to write a lengthy note to John Earman to reply to his take, ``Psi-star-Psi, a Bayesian degree of belief?  Cannot be!'', from the wrap-up session.  If you're interested, I'll forward it to you too once it's constructed.

\section{25-05-04 \ \ {\it Bathroom Reading} \ \ (to A. Duwell)} \label{Duwell3}

It was good seeing you again a couple of weeks ago.  (Get that paper on {\tt quant-ph}, young man!)

I just had the secretary in Sweden send you a copy of the {\Vaxjo} University Press cut of {\sl Notes on a Paulian Idea}.  I'm sure I don't need to say it, but the way to treat the book is as bathroom reading (i.e., not systematically, to put it mildly)---think of it as a kind of snippet book on information-based interpretations of quantum mechanics.  Its main purpose is to inspire.  I hope it fits the bill at least a little.

Concerning the little discussion we had on noncontextuality for quantum probability assignments, the section I was talking about was 4.1 in \quantph{0205039}.  If you have any thoughts on the argument (except for the sentence where I used a little circularity for rhetorical purposes), I'd like to hear them.

Finally, the paper with a discussion on the classical analogue(s) to Schumacher compression that I was telling you about is: \quantph{0008024}.  Don't know if it's as clear as I remember it to be, but maybe it's a start.

\section{25-05-04 \ \ {\it Bathroom Reading} \ \ (to A. J. Leggett)} \label{Leggett2}

It was good seeing you again at the Seven Pines workshop.  I particularly enjoyed the question you asked me at the end of my talk because I thought it got at the heart of the matter.  It is the laboratory procedures that refine our predictability that come first in defining the very notion of the (bad word) ``measurement.''  It is only post facto that a set of operators from the quantum mechanical formalism is associated with such a procedure.  Without the idea that knowledge is being refined, somewhere, somehow, the word measurement should never be used.

Anyway, I know you didn't request it, but I decided to have the secretary in Sweden send you a copy of the {\Vaxjo} University Press cut of my {\sl Notes on a Paulian Idea}, since I was already sending it to several others anyway.  I guess if anything, all I really want out of this is to present a convincing case to you that the stance I am taking on the wave-function is not giving up on realism (which is roughly the way you put it).  It's the opposite, as I see it---it's just recognizing a spade for a spade and trying to divert our attention to the parts of quantum theory that stand a chance of being realistically construed.  I try to make the case in several places in the book, but the parts you might be particularly interested in are the early discussions with David {\Mermin}, Jeff Bub, and Rolf Landauer.  Maybe some of the stuff to John Preskill too.  ({\Mermin}, by the way, wrote the foreword, if that'll help you get past the first page.)

Regarding the book as a whole, I'm sure I don't need to say it, but the way to treat it is as bathroom reading (i.e., not systematically, to put it mildly)---think of it as a kind of snippet book on information-based interpretations of quantum mechanics.  Its main purpose is to inspire if it can.  The name index at the back is the best bet for looking up one or another point of view; unfortunately I didn't have a subject index made for the initial publication.

If---in the end---you're interested in any of the more technical results needed to prop up these goals (and there is a lot of technical work to be done), most of my papers can be found at {\tt arXiv.org} in the {\tt quant-ph} listings (\quantph{0205039} being the most comprehensive).  In the one titled, ``Quantum States and Operations, a Bayesian View,'' you'll find extensive references to everyone else's work along the same lines as well.

\section{25-05-04 \ \ {\it Bathroom Reading} \ \ (to M. Janssen)} \label{Janssen1}

It was good meeting you at the Seven Pines workshop.  I particularly enjoyed the little discussion we had near the end.  I learned a lot from what you saw as a difference between Jeff Bub's and my approaches/goals.

I know you didn't request it, but thinking about this, I decided to have the secretary in Sweden send you a copy of the {\Vaxjo} University Press cut of {\sl Notes on a Paulian Idea}.  Anyway, I'm sure I don't need to say it, but the way to treat the book is as bathroom reading (i.e., not systematically, to put it mildly)---think of it as a kind of snippet book on information-based interpretations of quantum mechanics.  Its main purpose is to inspire; I hope it fits the bill at least a little.  The name index at the back is maybe the best bet to look up one or another point of view; sorry I didn't have a subject index prepared at the time.  Anyway, you might be particularly interested in the early discussions with Jeff Bub (which in turn was mostly based on Jeff's reading of the notes to {\Mermin}).

If---in the end---you're interested in any of my more technical results toward these goals, most of my papers can be found at {\tt arXiv.org} in the {\tt quant-ph} listings.

\section{25-05-04 \ \ {\it Cathy and Erwin, the Movie} \ \ (to C. G. {\Timpson})} \label{Timpson4}

You're not being dim-witted at all.  Instead, thanks for giving me a new opportunity to clarify things still further.  I'll take the task seriously if you can wait a day or two for my reply:  I want to do it right (certainly righter than before).

The central issue is to never forget that a quantum state is a state of belief {\it about\/} how a system will {\bf respond} to an external stimulation. Or better yet---and maybe necessarily so---as I put it in \quantph{0404156}, ``The agent, through the process of quantum measurement, stimulates the world external to himself.  The world, in return, stimulates a response in the agent that is quantified by a change in his beliefs---i.e., by a change from a prior to a posterior quantum state.''  In particular, the state should never be construed as being about ``what is existent'' or ``what is happening'' in the quantum system.

Aside:  From this point of view, the quantum system is a ``philosopher's stone'' that brings about a transformation in the agent.  (I just wanted to record that while I was thinking it.)

But anyway, I'll try to say all this more clearly and at length---and particularize it to the context of your two questions---as I get a chance to compose something.

\section{26-05-04 \ \ {\it Rorty, Delirium} \ \ (to A. Fine)} \label{Fine5}

Hey, did you know that Richard Rorty calls you his favorite philosopher of science?  I read that a while ago, and I've been meaning to tell you.  What kind of reaction does it elicit from you?  I'm a great fan of some of Rorty's (more tempered) writings.  (But I'm a cherry-picker when it comes to papers:  I absorb the parts I think I can use, and try to hold no ill will for the parts I simply ignore.)

Anyway, I was thinking of that when I was sending the attached cheerleading paper to a friend yesterday.  [See ``Delirium Quantum,'' \arxiv{0906.1968}.]  (The first and second sections lean heavily on Rorty.)  So while you're on my mind, I'll send it to you too.  I haven't quite decided whether I'm going to post it yet, or whether it's a little too delirious for the moment.

\section{26-05-04 \ \ {\it Bathroom Reading and Subliminal Suggestions} \ \ (to D. Howard)} \label{Howard1}

It was good meeting you at the Seven Pines workshop.

Per your request, I had the secretary in Sweden send you a copy of the {\Vaxjo} University Press cut of my {\sl Notes on a Paulian Idea\/} a couple of days ago. It shouldn't take too long to get there.  I'm sure I don't need to say it, but the way to treat the book is as bathroom reading (i.e., not systematically, to put it mildly)---think of it more as a kind of snippet book on information-based interpretations of quantum mechanics.  Its main purpose is to inspire if it can.  The name index at the back is the best thing presently for looking up one or another point of view; unfortunately I didn't have a subject index made for the initial publication.  Thinking of our discussions in Minnesota, you might particularly enjoy reading the correspondence with {\Mermin}, Peres, Bub, and Landauer; they're maybe the most relevant things to start off with.

The quote I read aloud after your talk, showing that Pauli clearly understood the concept of entanglement as early as 1927, can be found on page 334 in a note titled ``Pauli Understood Entanglement.''  You'll find the Rosenfeld quote that you're fond of on pages 299--301 in a note titled ``How Do I Sleep?''

Concerning the other quotes I promised to send you (from the Bohr--Pauli correspondence), they'll come in the next email in a big attachment---a little over 1MB---titled {\tt PauliProject.pdf}.  Let me know if you have any trouble opening the file; if necessary I'll seek another means to send it to you.

Sorry for the delay, but I was tweaking the document and resisted sending it until it found a new plateau of semi-stability.  In general the thing is a mess and very, very far from complete.  But I haven't given up hope that I might get it posted on {\tt quant-ph} by December.  It's mainly a question of getting a load of (pre-marked) material out of my file cabinet and into the computer, and then getting a useful indexing system made.  The introduction has a long way to go too, but that pales in comparison to the former task.

Definitely, if you have any suggestions for further materials that I should include in the thing (from your own work or anyone else's), or you catch any typos, please let me know.

\section{26-05-04 \ \ {\it The Pauli Program} \ \ (to D. Howard)} \label{Howard2}

I forgot to mention in my last long note a couple of references that might interest you.  These concern what you called ``The Pauli Program'' (I think)---namely, the idea of taking how subsystems combine into composite systems as one of the basic pillars of quantum mechanics.

Two things to read:
\begin{itemize}
\item[1)] Lucien Hardy's axiom-system paper.  You can find it here:
\quantph{0101012}.
Hardy's paper is one of the few---maybe the only one---I can think of where an axiom for combining systems plays such a prominent role.  (I would say his system is too positivistic/instrumentalistic for my taste, but nevertheless he makes wonderful mathematical progress.)

\item[2)] Section 5, ``Wither Entanglement?,'' of my paper \quantph{0205039}.
There I show how the tensor-product structure can be gotten out of a Gleason-type theorem.  It's not clear to me how such a result would fit into a Pauli program---it may be antithetical to it---but it might provide some food for thought nevertheless.  I myself certainly remain sympathetic to some aspects of the Pauli program.  (You can see by looking at pages 548--554 of the book I'm sending you that I've long had an interest in the idea.)
\end{itemize}

By the way, don't forget to send me the papers you promised.

\section{27-05-04 \ \ {\it Conflict} \ \ (to K. R. Duffy)} \label{Duffy3}

The unrelenting external world smites me again:  Kiki just told me that Emma's school performance is next Tuesday night.  So the Pub is out.

\section{31-05-04 \ \ {\it Menand in Context} \ \ (to C. M. {\Caves})} \label{Caves77.03}

Since you told me you'd give me your thoughts on Menand's book in the near future, let me supply you with some related reading in case it's useful for framing them.  The file is attached.  Talking with you was impetus for me to finish up the Menand section of my compendium, ``The Activating Observer.''  I've now included all the Menand quotes in it that have something to do with an activating-observer/law-without-law view of quantum mechanics.  Menand is entry \#294.

Also I've been working to beef up the William James sections (231--243).  I've now got some of my favorite quotes in there.  Maybe they'll indicate to you a little of what Menand was on about.

Anyway, here's the file.  Stuff it into your laptop and take it on your trip with you to kill some hours in the great wasteland---nothing better to recondition your batteries with.

Thanks again for the long talk the other day.  Made me long for the old, much simpler days of 1994--95.  Don't suppose you'd take me for a second PhD if all else fails?

\bq
The world is filled with unique things. In order to deal with the
world, though, we have to make generalizations. On what should we
base our generalizations? One answer, and it seems the obvious
answer, is that we should base them on the characteristics things
have in common. No individual horse is completely identical to any
other horse; no poem is identical to any other poem. But all things
we call horses, and all things we call poems, share certain
properties, and if we make those properties the basis for
generalizations, we have one way of ``doing things'' with horses or
poems---of distinguishing a horse from a zebra, for example, or of
judging whether a particular poem is a good poem or a bad poem. These
common properties can be visible features or they can be invisible
qualities; in either case, we create an idea of a ``horse'' or a
``poem,'' or of ``horseness'' or ``poetry,'' by retaining the
characteristics found in all horses or poems and ignoring
characteristics that make one horse or poem different from another.
We even out, or bracket, the variations among individuals for the
sake of constructing a general type.

Darwin's fundamental insight as a biologist was that among groups of
sexually reproducing organisms, the variations are much more
important than the similarities. ``Natural selection,'' his name for
the mechanism of evolutionary development that he codiscovered with
Alfred Russel Wallace, is the process by which individual
characteristics that are more favorable to reproductive success are
``chosen,'' because they are passed on from one generation to the
next, over characteristics that are less favorable. Darwin regretted
that the word ``selection'' suggested an intention: natural selection
is a blind process, because the conditions to which the organism must
adapt in order to survive are never the same. In periods of drought,
when seeds are hard to find, finches that happen to have long narrow
beaks, good for foraging, will be favored over finches with broad
powerful beaks: more of their offspring will survive and reproduce.
In periods of abundance, when seeds are large and their shells are
hard, the broad-beaked finches will hold the adaptive advantage.
``Finchness'' is a variable, not a constant.

Darwin thought that variations do not arise because organisms need
them (which is essentially what Lamarck had argued). He thought that
variations occur by chance, and that chance determines their adaptive
utility. In all seasons it happens that some finches are born with
marginally longer and narrower beaks than others, just as children of
the same parents are not all exactly the same height. In certain
environmental conditions, a narrower beak may have positive or
negative survival value, but in other conditions---for example, when
seeds are plentiful and finches are few---it may make no difference.
The ``selection'' of favorable characteristics is therefore neither
designed nor progressive. No intelligence, divine or otherwise,
determines in advance the relative value of individual variations,
and there is no ideal type of ``finch,'' or essence of ``finchness,''
toward which adaptive changes are leading.

Natural selection is a law that explains {\it why\/} changes occur in
nature---because, as Darwin and Wallace both realized after reading,
independently, Thomas Malthus's {\sl Essay on the Principle of
Population\/} (1798), if all members of a group of sexually
reproducing organisms were equally well adapted, the population of
the group would quickly outgrow the resources available to sustain
it. Since some members of the group must die, the individuals whose
slight differences give them an adaptive edge are more likely to
survive. Evolution is simply the incidental by-product of material
struggle, not its goal. Organisms don't struggle because they must
evolve; they evolve because they must struggle. Natural selection
also explains {\it how\/} changes occur in nature---by the relative
reproductive success of the marginally better adapted. But natural
selection does not dictate {\it what\/} those changes shall be. It is
a process without mind.

A way of thinking that regards individual differences as inessential
departures from a general type is therefore not well suited for
dealing with the natural world. A general type is fixed, determinate,
and uniform; the world Darwin described is characterized by chance,
change, and difference---all the attributes general types are
designed to leave out. In emphasizing the particularity of individual
organisms, Darwin did not conclude that species do not exist. He only
concluded that species are what they appear to be: ideas, which are
provisionally useful for naming groups of interacting individuals.
``I look at the term species,'' he wrote, ``as one arbitrarily given
for the sake of convenience to a set of individuals closely
resembling each other \ldots. [I]t does not essentially differ from
the term variety, which is given to less distinct and more
fluctuating forms. The term variety, again, in comparison with mere
individual differences, is also applied arbitrarily, and for mere
convenience sake.'' Difference goes all the way down.
\eq
and
\bq
What statistics seemed to show, in short, was that the market was
not, as people like Matthew Arnold complained, an invitation to
anarchy. Markets operate just the way nature does: left to
themselves, they can be counted on to produce the optimum outcome
over the long run. The individual pursuit of self-interest conduces
to aggregate efficiency. Of course, like all appeals to natural laws
as a justification for human arrangements, the ``discovery'' of the
laws reflected the arrangements to be justified. Nineteenth-century
liberals believed that the market operated like nature because they
had already decided that nature operated like a market.
\eq
and
\bq
Darwinism was a scandal to many Laplaceans. In the Laplacean
worldview, randomness is only appearance; in the Darwinian, it is
closer to a fact of nature---in some respects, it is {\it the\/} fact
of nature. Herschel, the man who had helped introduce Quetelet to
British readers, wrote in 1850 that if all the literature of Europe
were to perish and only Laplace's {\sl Syst\`eme du monde\/} and {\sl
Essai sur les probabilit\'es\/} remained, ``they would suffice to
convey to the latest posterity an impression of the intellectual
greatness of the age which could produce them, surpassing that
afforded by all the monuments antiquity has left us.'' But when {\sl
On the Origin of Species\/} appeared, in 1859, he ridiculed Darwin's
theory as ``the law of higgledy-pigglety''\footnote{Charles Darwin to
Charles Lyell, December 10, 1859: ``I have heard by round about channel that Herschel says my Book `is the
law of higgledy-pigglety'\,''.}

What does it mean to say we ``know'' something in a world in which
things happen higgledy-pigglety? Virtually all of Charles Peirce's
work---an enormous body of writing on logic, semiotics, mathematics,
astronomy, metrology, physics, psychology, and philosophy, large
portions of it unpublished or unfinished---was devoted to this
question. His answer had many parts, and fitting them all
together---in a form consistent with his belief in the existence of a
personal God---became the burden of his life. But one part of his
answer was that in a universe in which events are uncertain and
perception is fallible, knowing cannot be a matter of an individual
mind ``mirroring'' reality. Each mind reflects differently---even the
same mind reflects differently at different moments---and in any case
reality doesn't stand still long enough to be accurately mirrored.
Peirce's conclusion was that knowledge must therefore be social. It
was his most important contribution to American thought, and when he
recalled, late in life, how he came to formulate it, he described
it---fittingly---as the product of a group. This was the conversation
society he formed with William James, Oliver Wendell Holmes, Jr., and
a few others in Cambridge in 1872, the group known as the
Metaphysical Club.
\eq
and
\bq
[Chauncey] Wright did not consider himself an evolutionist. To him
the term denoted a belief that the world was getting, on some
definition, ``better.'' His loyalty was only to the theory of natural
selection, which he thought corresponded perfectly to his notion of
life as weather. ``[T]he principle of the theory of Natural Selection
is taught in the discourse of Jesus with Nicodemus the Pharisee,'' he
explained in a letter to Charles Norton's sister Grace. The allusion
may be a little gnomic today. The discourse with Nicodemus is in the
Gospel of John, and the words of Jesus Wright was referring to are
these: ``The wind bloweth where it listeth, and thou hearest the
sound thereof, but canst not tell whence it cometh, and whither it
goeth: so is every one that is born of the Spirit.'' Wright was, in
short, one of the few nineteenth-century Darwinians who thought like
Darwin---one of the few evolutionists who did not associate
evolutionary change with progress. ``Never use the word[s] higher \&
lower,'' Darwin scribbled in the margins of his copy of the {\sl
Vestiges of the Natural History of Creation\/} in 1847. The advice
proved almost impossible to follow to the letter, even for Darwin,
but if anyone respected its spirit, it was Chauncey Wright.

Wright's particular b\^ete noire was the evolutionist Herbert
Spencer, whose work seemed to him a flagrant violation of the
separation of science and metaphysics. ``Mr.\ Spencer,'' as he
declared, ``is not a positivist.'' Spencer's mistake was to treat the
concepts of science, which are merely tools of inquiry, as though
they were realities of nature. The theory of natural selection, for
example, posits continuity in the sequence of natural phenomena
(evolution does not proceed by leaps). But ``continuity'' is simply a
verbal handle we attach to a bundle of empirical observations. It is
not something that actually exists in nature. Spencer failed to
understand this, and he therefore imputed cosmic reality to what are
just conceptual inferences---just words. He did with the word
``evolution'' what Agassiz did with the word ``creation'': he erected
an idol.

``Mr.\ Spencer's philosophy contemplates the universe in its totality
as having an intelligible order, a relation of beginning and end---a
development,'' Wright said. But the universe is only weather.
\bq\noindent
Everything out of the mind is a product, the result of some process.
Nothing is exempt from change. Worlds are formed and dissipated.
Races of organic beings grow up like their constituent individual
members, and disappear like these. Nothing shows a trace of an
original, immutable nature, except the unchangeable laws of change.
These point to no beginning and to no end in time, nor to any bounds
in space. All indications to the contrary in the results of physical
research are clearly traceable to imperfections in our present
knowledge of all the laws of change, and to that disposition to
cosmological speculations which still prevails even in science.
\eq
``No {\it real\/} fate or necessity is indeed manifested anywhere in
the universe,'' he wrote to a friend, ``---only a phenomenal
regularity.''
\eq
and
\bq
Holmes was never keen to acknowledge the influence of other people on
his views, but he never had trouble acknowledging Wright's. He
identified with Wright's positivism: it suited perfectly his own
disillusionment. He took satisfaction in the notion that values are
epiphenomenal---that beneath all the talk of principles and ideals,
what people do is just a fancy version of what amoebas do. And he
therefore agreed with Wright that philosophy and logic don't have
much to do with the practical choices people make. He certainly
thought this was true in the law. ``It is the merit of the common law
that it decides the case first and determines the principle
afterwards,'' is the first sentence of the first law review article
he ever wrote, in 1870, two years before the Metaphysical Club came
into existence; and he spent much of his career as a philosopher of
jurisprudence explaining how the fact that judges conclude before
they reason does not mean that legal decision making is arbitrary.

Holmes eventually lost sympathy with the views of his friend William
James, which he thought too hopeful and anthropocentric. He never had
much interest in Peirce; he thought Peirce's genius ``overrated.''
But he continued to admire Wright, and years later cited him as the
inspiration for what he liked to call his philosophy of
``bettabilitarianism.'' ``Chauncey Wright[,] a nearly forgotten
philosopher of real merit, taught me when young that I must not say
necessary about the universe, that we don't know whether anything is
necessary or not,'' he wrote to Frederick Pollock in 1929, when he
was in his eighties. ``So that I describe myself as a {\it
bet\/}tabilitarian. I believe that we can {\it bet\/} on the behavior
of the universe in its contract with us. We bet we can know what it
will be. That leaves a loophole for free will---in the miraculous
sense---the creation of a new atom of force, although I don't in the
least believe in it.''
\eq
and
\bq
In his own way, and despite Holmes's distaste, William James was a
bettabilitarian, too. But he did believe in free will---what would it
mean to bet, after all, if we were not free to choose the stakes? He
was repelled by Wright's reduction of the world to pure
phenomena---he thought Wright made the universe into a ``Nulliverse''
and he regarded the abyss Wright insisted on placing between facts
and values as a fiction. James thought that Wright's decision to
separate science from metaphysics was itself a metaphysical
choice---that Wright's disapproval of talk about values was just an
expression of Wright's own values. Wright was a positivist because
positivism suited his character: moral neutrality was his way of
dealing with the world---and that, in James's view, is what all
beliefs are anyway, ``scientific'' or otherwise.

Wright was a regular visitor to the James family home in Cambridge
long before 1872, and in any case William James did not need the
Metaphysical Club to reach his conclusion about the nature of
beliefs. He had already arrived there by experimentation on what was
always his favorite human subject, himself. When he was living in
Germany in the late 1860s, he had got caught up in the speculative
frenzy about free will and determinism inspired by Buckle's book. As
usual, he found merits on both sides. ``I'm swamped in an empirical
philosophy,'' he wrote to Tom Ward shortly after getting back to
Cambridge in 1869; ``---I feel that we are Nature through and
through, that we are {\it wholly\/} conditioned, that not a wiggle of
our will happens save as the result of physical laws, and yet
notwithstanding we are en rapport with reason \ldots. It is not that
we are all nature {\it but\/} some point which is reason, but that
all is Nature {\it and\/} all is reason too.''

After he took his M.D. from Harvard, in June 1869, James collapsed.
He descended into a deep depression, exacerbated by back pains, eye
trouble, and various other complaints. His diary for the winter of
1869--70 is a record of misery and self-loathing. Then in the spring,
after reading the second installment, published in 1859, of a
three-part work called the {\sl Essais de critique g\'en\'erale\/} by
the French philosopher Charles Renouvier, he had a breakthrough.

Renouvier was a French Protestant from a family active in liberal
politics, but he had quit political life after the rise of the Second
Empire, in 1848, to devote himself to the construction of a
philosophical defense of freedom. Renouvier's argument was that ``the
doctrine of necessity'' is incoherent, since if all beliefs are
determined, we have no way of knowing whether the belief that all
beliefs are determined is correct, and no way of explaining why one
person believes in determinism while another person does not. The
only noncontradictory position, Renouvier held, is to believe that we
freely believe, and therefore to believe in free will. Even so, we
cannot be absolutely certain of the truth of this belief, or of
anything else. ``Certainty is not and cannot be absolute,'' he wrote
in the second {\sl Essai}. ``It is \ldots\ a condition and an action
of human beings \ldots. Properly speaking, there is no certainty;
there are only people who are certain.''

This was, in effect, Wright without the nihilism, and it was entirely
appealing to James. ``I think that yesterday was a crisis in my
life,'' he wrote in his diary on April 30, 1870.

\bq\noindent
I finished the first part of Renouvier's 2nd Essays and see no reason
why his definition of free will---the sustaining of a thought {\it
because I choose to\/} when I might have other thoughts---need be the
definition of an illusion. At any rate I will assume for the
present---until next year---that it is no illusion. My first act of
free will shall be to believe in free will \ldots. Hitherto, when I
have felt like taking a free initiative, like daring to act
originally, without carefully waiting for contemplation of the
external world to determine all for me, suicide seemed the most manly
form to put my daring into; now, I will go a step further with my
will, not only act with it, but believe as well; believe in my
individual reality and creative power.
\eq

As bold as this resolution sounds, it did not release James from his
depression. He seems to have been incapacitated by psychosomatic
disorders---in particular, an inability to use his eyes for reading
or writing---for another eighteen months, and he suffered chronically
from depression, eyestrain, and insomnia all his life. Henry's
mention of the formation of the Metaphysical Club in January 1872 is
one of the first signs, after the diary entry about Renouvier,
written a year and a half earlier, that William was socially active
again.''

Still, James believed that Renouvier had cured him, and he sent him
thanks. ``I must not lose this opportunity of telling you of the
admiration and gratitude which have been excited in me by the reading
of your {\sl Essais},'' he wrote to Renouvier in the fall of 1872.
``Thanks to you I possess for the first time an intelligible and
reasonable conception of freedom \ldots. I can say that through that
philosophy I am beginning to experience a rebirth of the moral life;
and I assure you, sir, that this is no small thing.'' Renouvier had
taught James two things: first, that philosophy is not a path to
certainty, only a method of coping, and second, that what makes
beliefs true is not logic but results. To James, this meant that
human beings are active agents---that they get a vote---in the evolving
constitution of the universe: when we choose a belief and act on it,
we change the way things are.
\eq
and
\bq
One of the first things Peirce did after he arrived at Hopkins in the
fall of 1879 was to start a Metaphysical Club. It was open to faculty
and graduate students from any department, and it met once a month to
discuss papers presented, usually, by the members themselves. \ldots

At one meeting, presided over by Morris, Dewey heard Peirce read a
paper called ``Design and Chance,'' and joined in the discussion
afterward. The paper is the germ of Peirce's later cosmology, and it
sums up in a few pages what was probably the substance of the
yearlong class Dewey had chosen not to take. Peirce's subject was the
laws of nature---the laws that Newtonian physicists believed
explained the behavior of matter and that physiological psychologists
believed explained the behavior of minds---and he began with a simple
question: Does the principle that everything can be explained have an
explanation? Or, as he also put it: Does the law of causality (which
is another name for the principle that everything can be explained)
have a cause? \ldots

Summarizing Peirce's Metaphysical Club paper on ``Design and Chance''
is \ldots\ not quite the same thing as paraphrasing it. The argument
begins with the point James Clerk Maxwell had made with his imaginary
demon: that a scientific law is only a prediction of what will happen
most of the time. Even ``the axioms of geometry'' said Peirce, ``are
mere empirical laws whose perfect exactitude we have no reason
whatever to feel confident of.'' The decision to treat a particular
law as absolute is a pragmatic one: sometimes we feel that
questioning it will only lead to confusion, and sometimes we feel
that questioning it is necessary in order to try out a new
hypothesis. A law, in Peirce's pragmatic view (derived, of course,
from Wright), is essentially a path of inquiry. It helps us find
things out---as the law of gravitation, for example, helped us
discover Neptune---and Peirce's first rule as a philosopher of
science was that the path of inquiry should never be blocked, not
even by a hypothesis that has worked for us in the past.

Maxwell's view was that laws are fundamentally uncertain because
there is always a chance that the next time around things will behave
in an improbable (though not an impossible) way---a chance that all
the fast molecules will congregate on one side of the container.
Peirce's point was that a chance occurrence like this can change the
conditions of the universe. His illustration was drawn from classic
probability theory: in a game with fair dice, a player's wins and
losses will balance out in the long run; but if one die is shaped so
that there is an infinitesimally greater chance that after a winning
throw the next throw will be a losing throw, in the long run the
player will be ruined. A minute variation in what seemed a stable and
predictable system can have cosmic consequences. In the natural
world, Peirce said, such minute variations are happening all the
time. Their occurrence is always a matter of chance---``chance is the
one essential agency upon which the whole process depends''---and,
according to probability theory, ``everything that can happen by
chance, sometime or other will happen by chance. Chance will sometime
bring about a change in every condition.'' Peirce thought that even
the terrible second law of thermodynamics---the law of the
dissipation of energy---was subject to reversal by such means.

As Peirce acknowledged, this was a Darwinian argument: ``my opinion
is only Darwinism analyzed, generalized, and brought into the realm
of Ontology,'' he said. What he meant was that since nature evolves
by chance variation, then the laws of nature must evolve by chance
variation as well. Variations that are compatible with survival are
reproduced; variations that are incompatible are weeded out. A tiny
deviation from the norm in the outcome of a physical process can,
over the long run, produce a new physical law. Laws are adaptive.

Pragmatically defined, variations are habits. They constitute a
behavioral tenden\-cy---for if they had no behavioral consequences,
they would have no evolutionary significance. Bigness in beak size is
whatever big beaks do for you (if you are a finch), just as (to use
an example from ``How to Make Our Ideas Clear'') ``hardness'' is just
the sum total of what all hard things do. What Peirce proposed in
``Design and Chance'' was that natural laws are also habits. This was
not a new thought for him. There is a story, attributed to William
James, about a meeting of the original Metaphysical Club in
Cambridge, in which the members waited patiently for Peirce to arrive
and deliver a promised paper.
\bq\noindent
They assembled. Peirce did not come; they waited and waited; finally
a two-horse carriage came along and Peirce got out with a dark cloak
over him; he came in and began to read his paper. What was it about?
He set forth \ldots\ how the different moments of time got in the
habit of coming one after another.\footnote{See M.~H. Fisch, ``Was There a Metaphysical Club in Cambridge?,'' in
{\sl Studies in the Philosophy of Charles Sanders Peirce, Second
Series}, edited by E.~C. Moore and R.~S. Robin (U. Massachusetts
Press, Amherst, MA, 1964), p.~11. The words are attributed to Dickinson
Miller. [Fuchs: Compare this to John Wheeler's story of the Pecan
Street Cafe grafitto.]}
\eq
It sounds like a joke, but the story is probably true. Peirce's paper
must have been an extrapolation from the nebular hypothesis---the
theory that the universe evolves from a condition of relative
homogeneity, in which virtually no order exists, not even temporal
order, to a condition of relative heterogeneity, in which, among
other things, time has become linear. How did time get straightened
out in this way? By developing good habits. In ``Design and Chance,''
Peirce put it this way:
\bq
Systems or compounds which have bad habits are quickly destroyed,
those which have no habits follow the same course; only those which
have good habits tend to survive.

Why \ldots\ do the heavenly bodies tend to attract one another?
Because in the long run bodies that repel or do not attract will get
thrown out of the region of space leaving only the mutually
attracting bodies.
\eq
If you are a heavenly body, in other words, gravitational attraction
is a good habit to have, in the same way that if you are a
proto-giraffe, a long neck is a good attribute to have. It keeps you
in the system. When gravitational attraction becomes the habit of
{\it all\/} heavenly bodies, then we can speak of ``the law of
gravity,'' just as when all surviving proto-giraffes have long necks,
we can speak of a giraffe species, and (presumably) when all moments
of time have the habit of following one another, we can speak of
past, present, and future. But the law of gravity did not preexist
the formation of the universe, any more than the idea of a giraffe
did. It evolved into its present state while the universe was
evolving into {\it its\/} present state. Gravity was a chance
variation that got selected. Objects that didn't have the
gravitational habit didn't survive.
\eq
and
\bq
That natural laws themselves evolve was the argument of a well known
book by the French philosopher \'Emile Boutroux, {\sl De la
contingence des lois de la nature\/} (``The Contingency of the Laws
of Nature''), published in 1874---a defense of free will written very
much in the spirit of William James's own French inspiration, Charles
Renouvier. ``Scientific laws are the bed over which passes the
torrent of facts,'' Boutroux wrote; ``they shape it even as they
follow it \ldots. They do not precede things, they derive from them,
and they can vary, if the things themselves happen to vary.'' The
tendencies of living beings to follow predictable paths, he said,
although they ``can look, viewed from outside, like necessary laws,''
are only habits. Without variation, everything would be dead matter.

Still, none of these arguments answered the question Peirce had
started out with, which was whether the law of causation has a cause.
To say that causality ``evolved'' is not an answer, because it uses
another law---the law of evolution by chance variation---as an
explanation, and leaves us with the question, Did the law of
evolution evolve? The search for a primal cause seems to suck us into
an infinite regression. Darwin and Maxwell's conception of chance
can't help us, because that conception simply expresses the
statistical notion of causality---the notion that outcomes are
distributed along a curve of probabilities on which extremes are
always possible. Darwin did not think that variations were
spontaneous in the sense of being uncaused, only in the sense of
being unpredictable, and he was willing to leave it at that. {\sl On
the Origin of Species\/} is actually silent on the question of the
origin.

But the origin was the problem that interested Peirce. His conclusion
was that there must be something he called ``absolute chance'' in the
universe. This was another way of saying that the answer to the
question, Does the law of causation have a cause? is, no. Causation
came about not as a consequence of the operation of some other law,
but by pure chance, out of the blue. Peirce spelled this idea out
more clearly a few years later in an essay---one of his most
ambitiously speculative, also unfinished and unpublished---called ``A
Guess at the Riddle.'' ``We are brought, then, to this,'' he says
toward the end of it:
\bq\noindent
conformity to law exists only within a limited range of events and
even there is not perfect, for an element of pure spontaneity or
lawless originality mingles, or at least, must be supposed to mingle,
with law everywhere. Moreover, conformity with law is a fact
requiring to be explained; and since Law in general cannot be
explained by any law in particular, the explanation must consist in
showing how law is developed out of pure chance, irregularity, and
indeterminacy \ldots. According to this, three elements are active in
the world, first, chance; second, law; and third, habit-taking.

Such is our guess of the secret of the sphynx.
\eq

Peirce's faith in the existence of absolute chance seems a vote for
freedom and originality and against statistics and systems; but that
is not how Peirce thought of it. Chance, in his theory, changes the
system, but it does not make it any less systematic. On the contrary.
As he put it in the paper on ``Design and Chance'': ``Chance is
indeterminacy, is freedom. But the action of freedom issues in the
strictest law.'' Unlike Renouvier and Boutroux, Peirce did not think
that he had come up with a scientific justification for liberty and a
belief in free will. He thought that he had found the Holy Grail of
post-Kantian system builders. He thought he had located the uncaused
cause.
\eq
and
\bq
Pragmatism is an account of the way people think---the way they come
up with ideas, form beliefs, and reach decisions. What makes us
decide to do one thing when we might do another thing instead? The
question seems unanswerable, since life presents us with many types
of choices, and no single explanation can be expected to cover every
case. Deciding whether to order the lobster or the steak is not the
same sort of thing as deciding whether the defendant is guilty beyond
a reasonable doubt. In the first case (assuming price is not an
object) we consult our taste; in the second we consult our judgment,
and try to keep our taste out of it. But knowing more or less what
category a particular decision belongs to---knowing whether it is a
matter of personal preference or a matter of impersonal
judgment---doesn't make that decision any easier to make. ``Order
what you feel like eating,'' says your impatient dinner companion.
But the problem is that you don't {\it know\/} what you feel like
eating. What you feel like eating is precisely what you are trying to
figure out.

``Order what you feel like eating'' is just a piece of advice about
the criteria you should be using to guide your deliberations. It is
not a solution to your menu problem---just as ``Do the right thing''
and ``Tell the truth'' are only suggestions about criteria, not
answers to actual dilemmas. The actual dilemma is what, in the
particular case staring you in the face, the right thing to do or the
honest thing to say really is. And making those kinds of
decisions---about what is right or what is truthful---{\it is\/} like
deciding what to order in a restaurant, in the sense that getting a
handle on tastiness is no harder or easier (even though it is
generally less important) than getting a handle on justice or truth.

People reach decisions, most of the time, by thinking. This is a
pretty banal statement, but the process it names is inscrutable. An
acquaintance gives you a piece of information in strict confidence;
later on, a close friend, lacking that information, is about to make
a bad mistake. Do you betray the confidence? ``Do the right
thing''---but what is the right thing? Keeping your word, or helping
someone you care about avoid injury or embarrassment? Even in this
two-sentence hypothetical case, the choice between principles is
complicated---as it always is in life---by circumstances. If it had
been the close friend who gave you the information and the
acquaintance who was about to make the mistake, you would almost
certainly think about your choice differently---as you would if you
thought that the acquaintance was a nasty person, or that the friend
was a lucky person, or that the statute of limitations on the secret
had probably run out, or that you had acquired a terrible habit of
betraying confidences and really ought to break it. In the end, you
will do what you believe is ``right,'' but ``rightness'' will be, in
effect, the compliment you give to the outcome of your deliberations.
Though it is always in view while you are thinking, ``what is right''
is something that appears in its complete form at the end, not the
beginning, of your deliberation.

When we think, in other words, we do not simply consult principles,
or reasons, or sentiments, or tastes; for prior to thinking, all
those things are indeterminate. Thinking is what makes them real.
Deciding to order the lobster helps us determine that we have a taste
for lobster; deciding that the defendant is guilty helps us establish
the standard of justice that applies in this case; choosing to keep a
confidence helps make honesty a principle and choosing to betray it
helps to confirm the value we put on friendship.

Does this mean that our choices are arbitrary or self-serving---that
standards and principles are just whatever it is in our interest to
say they are, pretexts for satisfying selfish ends or gratifying
hidden impulses? There is no way to answer this question, except to
say that it rarely {\it feels\/} as though this is the case. We
usually don't end up deciding to do what seems pleasant or convenient
at the moment; experience teaches us that this is rarely a wise basis
for making a choice. (``If merely `feeling good' could decide,
drunkenness would be the supremely valid human experience,'' as James
once put it.) When we are happy with a decision, it doesn't feel
arbitrary; it feels like the decision we {\it had\/} to reach. And
this is because its inevitability is a function of its ``fit'' with
the whole inchoate set of assumptions of our self-understanding and
of the social world we inhabit, the assumptions that give the moral
weight---much greater moral weight than logic or taste could ever
give---to every judgment we make. This is why, so often, we know
we're right before we know {\it why\/} we're right. First we decide,
then we deduce.

It does not follow that it is meaningless to talk of beliefs being
true or untrue. It only means that there is no noncircular set of
criteria for knowing whether a particular belief is true, no appeal
to some standard outside the process of coming to the belief itself.
For thinking just is a circular process, in which some end, some
imagined outcome, is already present at the start of any train of
thought. ``Truth {\it happens\/} to an idea,'' James said in the
lectures he published in 1907 as {\sl Pragmatism}. ``It {\it
becomes\/} true, is {\it made\/} true by events. Its verity {\it
is\/} in fact an event, a process: the process namely of its
verifying itself.'' And, elsewhere in the same lectures: ``\,`the
true' is only the expedient in the way of our thinking, just as `the
right' is only the expedient in the way of our behaving.'' Thinking
is a free-form and boundless activity that nevertheless leads us to
outcomes we feel justified in calling true, or just, or moral.
\eq
and
\bq
Pragmatists think that the mistake most people make about beliefs is
to think that a belief is true, or justified, only if it mirrors
``the way things really are''---that (to use one of James's most
frequent targets, Huxley's argument for agnosticism) we are justified
in believing in God only if we are able to prove that God exists
apart from our personal belief in him. No belief, James thought, is
justified by its correspondence with reality, because mirroring
reality is not the purpose of having minds. His position on this
matter was his earliest announced position as a professional
psychologist. It appears in the first article he ever published,
``Remarks on Spencer's Definition of Mind as Correspondence,'' which
appeared in the {\sl Journal of Speculative Philosophy\/} in the same
month that ``How to Make Our Ideas Clear'' was appearing in the {\sl
Popular Science Monthly}---January 1878. ``I, for my part,'' James
wrote,
\bq\noindent
cannot escape the consideration \ldots\ that the knower is not simply
a mirror floating with no foot-hold anywhere, and passively
reflecting an order that he comes upon and finds simply existing. The
knower is an actor, and co-efficient of the truth \ldots. Mental
interests, hypotheses, postulates, so far as they are bases for human
action---action which to a great extent transforms the world---help
to {\it make\/} the truth which they declare. In other words, there
belongs to mind, from its birth upward, a spontaneity, a vote. It is
in the game.
\eq

James returned to the argument in the last chapter---the chapter of
which he was proudest---of {\sl The Principles of Psychology}, where
he proposed to answer scientifically the question that Locke and Kant
had tried to answer philosophically: How do we acquire our ideas
about the world outside ourselves? Locke, of course, had attributed
all our ideas to sensory experience; Kant had pointed out that some
ideas, such as the idea of causation, cannot be explained by sensory
experience, since we do not ``see'' causation, we only infer it, and
he had concluded that such ideas must be innate, wired in from birth.

James agreed with Kant that many of the ideas we more or less
instinctively have about the world do not derive from what we
experience through the senses, but he thought there was a Darwinian
explanation: ``innate'' ideas are fortuitous variations that have
been naturally selected. Minds that possessed them were preferred
over minds that did not. But why? It doesn't make sense to say,
Because those minds mirrored reality more accurately. From a
Darwinian point of view, ``mirroring accurately'' is a gratuitous
compliment, an ex post facto rationalization-like saying that long
necks have been selected in giraffes because long necks look better
on a giraffe. Traits are selected because they help the organism
adapt. There are no other criteria.

The reason human beings came to possess the idea of causation, James
concluded, is not because causation really exists and would exist
whether we were around to believe in it or not. We have no way of
knowing whether this is so, and no reason to care. ``The word
`cause,'\,'' as he remarked in {\sl The Principles of Psychology},
``is \ldots\ an altar to an unknown god.'' The reason we believe in
causation is because experience shows that it pays to believe in
causation. Causation is a cashable belief. It gets us pellets. ``The
whole notion of truth, which naturally and without reflexion we
assume to mean the simple duplication by the mind of a ready-made and
given reality, proves hard to understand clearly,'' James declared
seventeen years later in the lectures on {\sl Pragmatism}. ``[A]ll
our thoughts are {\it instrumental}, and mental modes of {\it
adaptation\/} to reality, rather than revelations or gnostic answers
to some divinely instituted world-enigma.''
\eq
and
\bq
Philosophers, Dewey argued, had mistakenly insisted on making a
problem of the relation between the mind and the world, an obsession
that had given rise to what he called ``the alleged discipline of
epistemology''---the attempt to answer the question, How do we know?
The pragmatist response to this question is to point out that nobody
has ever made a problem about the relation between, for example, the
{\it hand\/} and the world. The function of the hand is to help the
organism cope with the environment; in situations in which a hand
doesn't work, we try something else, such as a foot, or a fishhook,
or an editorial. Nobody worries in these situations about a lack of
some preordained ``fit''---about whether the physical world was or
was not made to be manipulated by hands. They just use a hand where a
hand will do.

Dewey thought that ideas and beliefs are the same as hands:
instruments for coping. An idea has no greater metaphysical stature
than, say, a fork. When your fork proves inadequate to the task of
eating soup, it makes little sense to argue about whether there is
something inherent in the nature of forks or something inherent in
the nature of soup that accounts for the failure. You just reach for
a spoon. But philosophers have worried about whether the mind is such
that the world can be known by it, and they have produced all sorts
of accounts of how the ``fit'' is supposed to work---how the mental
represents the real. Dewey's point was that ``mind'' and ``reality,''
like ``stimulus'' and ``response,'' name nonexistent entities: they
are abstractions from a single, indivisible process. It therefore
makes as little sense to talk about a ``split'' that needs to be
overcome between the mind and the world as it does to talk about a
``split'' between the hand and the environment, or the fork and the
soup. ``Things,'' he wrote, ``\ldots\ are what they are experienced
as.'' Knowledge is not a copy of something that exists independently
of its being known, ``{\it it is an instrument or organ of successful
action}.'' ``The chief service of pragmatism, as regards
epistemology,'' he wrote to a friend in 1905, ``will be \ldots\ to
give the {\it coup de grace\/} to {\it representationalism}.''
\eq
and
\bq
Coercion is natural; freedom is artificial. Freedoms are socially
engineered spaces where parties engaged in specified pursuits enjoy
protection from parties who would otherwise naturally seek to
interfere in those pursuits. One person's freedom is therefore always
another person's restriction: we would not have even the concept of
freedom if the reality of coercion were not already present. We think
of a freedom as a right, and therefore the opposite of a rule, but a
right is a rule. It is a prohibition against sanctions on certain
types of behavior. We also think of rights as privileges retained by
individuals against the rest of society, but rights are created not
for the good of individuals, but for the good of society. Individual
freedoms are manufactured to achieve group ends.

This way of thinking about freedoms helps to explain why the two
people most closely associated with the establishment of the modern
principles of freedoms of thought and expression in the United States
were indifferent to the notion of individual rights. John Dewey and
Oliver Wendell Holmes had no particular interest in providing a
benefit to persons at the expense of the group. Dewey took a benign
pleasure and Holmes took a cynical pleasure in the spectacle of
personal wishes being subordinated to community will. But they both
saw the social usefulness of creating a zone of protection for
individual thought and expression, and the freedoms they helped to
establish are responsible for much of what is distinctive about
American life in the twentieth century and after. \ldots

The constitutional law of free speech is the most important benefit
to come out of the way of thinking that emerged in Cambridge and
elsewhere in the decades after the Civil War. It makes the value of
an idea not its correspondence to a preexisting reality or a
metaphysical truth, but simply the difference it makes in the life of
the group. Holmes's conceit of a ``marketplace of ideas'' suffers
from the defect of all market theories: exogenous elements are always
in play to keep marketplaces from being truly competitive. Some ideas
just never make it to the public. But it is the metaphor of
probabilistic thinking: the more arrows you shoot at the target, the
better sense you will have of the bull's-eye. The more individual
variations, the greater the chances that the group will survive. We
do not (on Holmes's reasoning) permit the free expression of ideas
because some individual may have the right one. No individual alone
can have the right one. We permit free expression because we need the
resources of the whole group to get us the ideas we need. Thinking is
a social activity. I tolerate your thought because it is part of my
thought---even when my thought defines itself in opposition to yours.
\eq
and
\bq
Academic freedom and the freedom of speech are quintessentially
modern principles. Since the defining characteristic of modern life
is social change---not onward or upward, but forward, and toward a
future always in the making---the problem of legitimacy continually
arises. In a premodern society, legitimacy rests with hereditary
authority and tradition; in a modernizing society, the kind of
society in which Louis Agassiz and Oliver Wendell Holmes, Sr., and
Benjamin Peirce lived and wrote, legitimacy tends to be transferred
from leaders and customs to nature. Agassiz and the senior Holmes and
Benjamin Peirce all assumed that social arrangements are justified if
they correspond with the design of the natural world---and so did
Adolphe Quetelet, Henry Thomas Buckle, Thomas Huxley, and William
Graham Sumner. But in societies bent on transforming the past, and on
treating nature itself as a process of ceaseless transformation, how
do we trust the claim that a particular state of affairs is
legitimate?

The solution has been to shift the totem of legitimacy from premises
to procedures. We know an outcome is right not because it was derived
from immutable principles, but because it was reached by following
the correct procedures. Science became modern when it was conceived
not as an empirical confirmation of truths derived from an
independent source, divine revelation, but as simply whatever
followed from the pursuit of scientific methods of inquiry. If those
methods were scientific, the result must be science. The modem
conception of law is similar: if the legal process was adhered to,
the outcome is just. Justice does not preexist the case at hand;
justice is whatever result just procedures have led to. Even art
adopted the same standards in the modern period: it became defined as
the realization of the aesthetic potential of the artistic medium.
Poetry was talked about as an exploration of the resources of
language, painting as a manipulation of canvas and paint, figure and
ground. The argument of Hand and Holmes---which was also the argument
of Jane Addams and John Dewey---about democracy had the same logic.
It is that a decision can be called democratic only if everyone has
been permitted to participate in reaching it.
\eq

\section{01-06-04 \ \ {\it Bathroom Reading, 2} \ \ (to M. Janssen)} \label{Janssen2}

Thanks for the great Slater quote.  I love collecting things like that.

Thanks also for the encouraging words about our research program.  After writing you, I used the same letter as a template and extended it a bit more in writing to Tony Leggett.  Let me paste that in below, as it might give you a little more orientation for where to seek out further information (for instance papers by Bub, Hardy, {\Mermin}, {\Spekkens}, etc., that I think are great steps forward in the effort).  [See 25-05-04 note ``\myref{Leggett2}{Bathroom Reading}'' to A. J. Leggett.]

\subsection{Michel's Preply}

\bq
For your amusement I quote from a letter of July 27, 1924 from John C. Slater to Minnesota's own John H. Van Vleck with Slater's impression of Copenhagen from which he is just returning (the letter is written as he sails along the coast of New England):
\bq
Don't remember just how much I told you about my stay in Copenhagen. The paper with Bohr and Kramers was got out of the way the first six weeks or so---written entirely by Bohr and Kramers. That was very nearly the only paper that came from the institute at all the time I was there; there seemed to be very little doing. Bohr does very little and is chronically overworked by it. The paper that was all written last Fall still hasn't been revised and printed, and I don't know when it will be. Bohr had to go on several vacations in the spring, and came back worse from each one. Kramers hasn't got much done, either. You perhaps noticed his letter to {\sl Nature\/} on dispersion; the formulas that he had before I came, although he didn't see the exact application; and except for that he hasn't done anything, so far as I know. They seem to have too much administrative work to do. Even at that, I don't see what they do all the time. Bohr hasn't been teaching at all, Kramers has been giving one or two courses.

There were visits from both Pauli and Heisenberg in the course of the spring. I don't know whether you have met either one. Heisenberg is a very nice red haired unassuming young chap, talks a little English, and everybody likes him. Pauli is as different as he could be, a big fat Jew, with a very good opinion of himself and a great liking to hear himself talk. Still, he is a good natured and accommodating person, and well liked.
\eq
\eq

\section{01-06-04 \ \ {\it Dreams} \ \ (to A. Sudbery)} \label{Sudbery10}

Thanks for keeping me apprised of the Foundations meeting in September.  I really feel like I should be there---I want to be there---but I don't think I can get back so soon after my move back to the States.  Also there's the little issue that Bell Labs has dried up as far as travel funds are concerned now, and I don't have any other grants.

On a different matter, strangely I had a dream about you the other night.  We were, of course, debating quantum foundations.  (To set the scene, you had had some more dental work done, so your mouth was bleeding again; I think we were in an empty conference room.)  And for some reason in the dream, I thought it was really important for you to read some passages I had scanned into my computer the night before---that is, passages that I had really scanned in, and not just in the dream.  I'm not making this up.  Anyway, upon waking, I couldn't figure out why it was so all-fired important that you read the passages, but still I'll send them to you.  They're attached [from William James's essay, ``The Sentiment of Rationality''].

\bq
A completed theoretic philosophy can thus never be anything more than
a completed classification of the world's ingredients; and its
results must always be abstract, since the basis of every
classification is the abstract essence embedded in the living
fact---the rest of the living fact being for the time ignored by the
classifier. This means that none of our explanations are complete.
They subsume things under heads wider or more familiar; but the last
heads, whether of things or of their connections, are mere abstract
genera, data which we just find in things and write down.

When, for example, we think that we have rationally explained the
connection of the facts $A$ and $B$ by classing both under their
common attribute $x$, it is obvious that we have really explained
only so much of these items as {\it is x}. To explain the connection
of choke-damp and suffocation by the lack of oxygen is to leave
untouched all the other peculiarities both of choke-damp and of
suffocation---such as convulsions and agony on the one hand, density
and explosibility on the other. In a word, so far as $A$ and $B$
contain $l$, $m$, $n$, and $o$, $p$, $q$, respectively, in addition
to $x$, they are not explained by $x$. Each additional particularity
makes its distinct appeal. A single explanation of a fact only
explains it from a single point of view. The entire fact is not
accounted for until each and all of its characters have been classed
with their likes elsewhere. To apply this now to the case of the
universe, we see that the explanation of the world by molecular
movements explains it only so far as it actually {\it is\/} such
movements. To invoke the ``Unknowable'' explains only so much as is
unknowable, ``Thought'' only so much as is thought, ``God'' only so
much as is God. {\it Which\/} thought? {\it Which\/} God?---are
questions that have to be answered by bringing in again the residual
data from which the general term was abstracted. All those data that
cannot be analytically identified with the attribute invoked as
universal principle, remain as independent kinds or natures,
associated empirically with the said attribute but devoid of rational
kinship with it.

Hence the unsatisfactoriness of all our speculations. On the one
hand, so far as they retain any multiplicity in their terms, they
fail to get us out of the empirical sand-heap world; on the other, so
far as they eliminate multiplicity, the practical man despises their
empty barrenness. The most they can say is that the elements of the
world are such and such, and that each is identical with itself
wherever found; but the question Where is it found? the practical man
is left to answer by his own wit. Which, of all the essences, shall
here and now be held the essence of this concrete thing, the
fundamental philosophy never attempts to decide. We are thus led to
the conclusion that the simple classification of things is, on the
one hand, the best possible theoretic philosophy, but is, on the
other, a most miserable and inadequate substitute for the fulness of
the truth. It is a monstrous abridgment of life, which, like all
abridgments, is got by the absolute loss and casting out of real
matter. This is why so few human beings truly care for philosophy.
The particular determinations which she ignores are the real matter
exciting needs, quite as potent and authoritative as hers. What does
the moral enthusiast care for philosophical ethics? Why does the {\it
\AE sthetik\/} of every German philosopher appear to the artist an
abomination of desolation?
\bq
Grau, teurer Freund, ist alle Theorie \\
\indent Und gr\"un des Lebens goldner Baum.
\eq
The entire man, who feels all needs by turns, will take nothing as an
equivalent for life but the fulness of living itself. Since the
essences of things are as a matter of fact disseminated through the
whole extent of time and space, it is in their spread-outness and
alternation that he will enjoy them. When weary of the concrete clash
and dust and pettiness, he will refresh himself by a bath in the
eternal springs, or fortify himself by a look at the immutable
natures. But he will only be a visitor, not a dweller, in the region;
he will never carry the philosophic yoke upon his shoulders, and when
tired of the gray monotony of her problems and insipid spaciousness
of her results, will always escape gleefully into the teeming and
dramatic richness of the concrete world.

So our study turns back here to its beginning. Every way of
classifying a thing is but a way of handling it for some particular
purpose. Conceptions, ``kinds,'' are teleological instruments. No
abstract concept can be a valid substitute for a concrete reality
except with reference to a particular interest in the conceiver. The
interest of theoretic rationality, the relief of identification, is
but one of a thousand human purposes. When others rear their heads,
it must pack up its little bundle and retire till its turn recurs.
The exaggerated dignity and value that philosophers have claimed for
their solutions is thus greatly reduced. The only virtue their
theoretic conception need have is simplicity, and a simple conception
is an equivalent for the world only so far as the world is
simple---the world meanwhile, whatever simplicity it may harbor,
being also a mightily complex affair. Enough simplicity remains,
however, and enough urgency in our craving to reach it, to make the
theoretic function one of the most invincible of human impulses. The
quest of the fewest elements of things is an ideal that some will
follow, as long as there are men to think at all.

But suppose the goal attained. Suppose that at last we have a system
unified in the sense that has been explained. Our world can now be
conceived simply, and our mind enjoys the relief. Our universal
concept has made the concrete chaos rational. But now I ask, Can that
which is the ground of rationality in all else be itself properly
called rational? It would seem at first sight that it might. One is
tempted at any rate to say that, since the craving for rationality is
appeased by the identification of one thing with another, a datum
which left nothing else outstanding might quench that craving
definitively, or be rational {\it in se}. No otherness being left to
annoy us, we should sit down at peace. In other words, as the
theoretic tranquillity of the boor results from his spinning no
further considerations about his chaotic universe, so any datum
whatever (provided it were simple, clear, and ultimate) ought to
banish puzzle from the universe of the philosopher and confer peace,
inasmuch as there would then be for him absolutely no further
considerations to spin.

This in fact is what some persons think. Professor Bain says ---
\bq
\indent
A difficulty is solved, a mystery unriddled, when it can be shown to
resemble something else; to be an example of a fact already known.
Mystery is isolation, exception, or it may be apparent contradiction:
the resolution of the mystery is found in assimilation, identity,
fraternity. When all things are assimilated, so far as assimilation
can go, so far as likeness holds, there is an end to explanation;
there is an end to what the mind can do, or can intelligently desire
\ldots. The path of science as exhibited in modern ages is toward
generality, wider and wider, until we reach the highest, the widest
laws of every department of things; there explanation is finished,
mystery ends, perfect vision is gained.
\eq

But, unfortunately, this first answer will not hold. Our mind is so
wedded to the process of seeing an {\it other\/} beside every item of
its experience, that when the notion of an absolute datum is
presented to it, it goes through, its usual procedure and remains
pointing at the void beyond, as if in that lay further matter for
contemplation. In short, it spins for itself the further positive
consideration of a nonentity enveloping the being of its datum; and
as that leads nowhere, back recoils the thought toward its datum
again. But there is no natural bridge between nonentity and this
particular datum, and the thought stands oscillating to and fro,
wondering ``Why was there anything but nonentity; why just this
universal datum and not another?'' and finds no end, in wandering
mazes lost. Indeed, Bain's words are so untrue that in reflecting men
it is just when the attempt to fuse the manifold into a single
totality has been most successful, when the conception of the
universe as a unique fact is nearest its perfection, that the craving
for further explanation, the ontological wonder-sickness, arises in
its extremest form. As Schopenhauer says, ``The uneasiness which
keeps the never-resting clock of metaphysics in motion, is the
consciousness that the non-existence of this world is just as
possible as its existence.''

The notion of nonentity may thus be called the parent of the
philosophic craving in its subtilest and profoundest sense. Absolute
existence is absolute mystery, for its relations with the nothing
remain unmediated to our understanding. One philosopher only has
pretended to throw a logical bridge over this chasm. Hegel, by trying
to show that nonentity and concrete being are linked together by a
series of identities of a synthetic kind, binds everything
conceivable into a unity, with no outlying notion to disturb the free
rotary circulation of the mind within its bounds. Since such
unchecked movement gives the feeling of rationality, he must be held,
if he has succeeded, to have eternally and absolutely quenched all
rational demands.

But for those who deem Hegel's heroic effort to have failed, nought
remains but to confess that when all things have been unified to the
supreme degree, the notion of a possible other than the actual may
still haunt our imagination and prey upon our system. The bottom of
being is left logically opaque to us, as something which we simply
come upon and find, and about which (if we wish to act) we should
pause and wonder as little as possible. The philosopher's logical
tranquillity is thus in essence no other than the boor's. They differ
only as to the point at which each refuses to let further
considerations upset the absoluteness of the data he assumes. The
boor does so immediately, and is liable at any moment to the ravages
of many kinds of doubt. The philosopher does not do so till unity has
been reached, and is warranted against the inroads of those
considerations, but only practically, not essentially, secure from
the blighting breath of the ultimate Why? If he cannot exorcize this
question, he must ignore or blink it, and, assuming the data of his
system as something given, and the gift as ultimate, simply proceed
to a life of contemplation or of action based on it.
\eq

\section{01-06-04 \ \ {\it Shielding the Mathematics} \ \ (to I. Pitowsky)} \label{Pitowsky4}

I thought of you the other day as I was reading a passage in one of Quine's books.  In particular I thought of the conversation we had on the plane to New Jersey, where you told me your thoughts on how a frequentist account of probability might be salvaged.  It struck me that the move you're toying with is in no longer ``shielding the mathematics'' in Quine's sense.

Anyway, I'll place the passage below for your enjoyment.

\begin{itemize}
\item
W.~V. Quine, {\sl Pursuit of Truth}, revised edition (Harvard U.
Press, Cambridge, MA, 1992).
\end{itemize}

\bq
Let us recall that the hypothesis regarding the chemical composition
of litholite did not imply its observation categorical single-handed.
It implied it with the help of a backlog of accepted scientific
theory. In order to deduce an observation categorical from a given
hypothesis, we may have to enlist the aid of other theoretical
sentences and of many common-sense platitudes that go without saying,
and perhaps the aid even of arithmetic and other parts of
mathematics.

In that situation, the falsity of the observation categorical does
not conclusively refute the hypothesis. What it refutes is the
conjunction of sentences that was needed to imply the observation
categorical. In order to retract that conjunction we do not have to
retract the hypothesis in question; we could retract some other
sentence of the conjunction instead. This is the important insight
called {\it holism}. Pierre Duhem made much of it early in this
century, but not too much.

The scientist thinks of his experiment as a test specifically of his
new hypothesis, but only because this was the sentence he was
wondering about and is prepared to reject. Moreover, there are also
the situations where he has no preconceived hypothesis, but just
happens upon an anomalous phenomenon. It is a case of his happening
upon a counter-instance of an observation categorical which,
according to his current theory as a whole, ought to have been true.
So he looks to his theory with a critical eye.

Over-logicizing, we may picture the accommodation of a failed
observation categorical as follows. We have before us some set $S$ of
purported truths that was found jointly to imply the false
categorical. Implication may be taken here simply as deducibility by
the logic of truth functions, quantification, and identity. (We can
always provide for more substantial consequences by incorporating
appropriate premises explicitly into $S$.) Now some one or more of
the sentences in $S$ are going to have to be rescinded. We exempt
some members of $S$ from this threat on determining that the fateful
implication still holds without their help. Any purely logical truth
is thus exempted, since it adds nothing to what $S$ would logically
imply anyway; and sundry irrelevant sentences in $S$ will be exempted
as well. Of the remaining members of $S$, we rescind one that seems
most suspect, or least crucial to our overall theory. We heed a maxim
of minimum mutilation. If the remaining members of $S$ still conspire
to imply the false categorical, we try rescinding another and
restoring the first. If the false categorical is still implied, we
try rescinding both. We continue thus until the implication is
defused.

But this is only the beginning. We must also track down sets of
sentences elsewhere, in our overall theory, that imply these newly
rescinded beliefs; for those must be defused too. We continue thus
until consistency seems to be restored. Such is the mutilation that
the maxim of minimum mutilation is meant to minimize.

In particular the maxim constrains us, in our choice of what
sentences of $S$ to rescind, to safeguard any purely mathematical
truth; for mathematics infiltrates all branches of our system of the
world, and its disruption would reverberate intolerably. If asked why
he spares mathematics, the scientist will perhaps say that its laws
are necessarily true; but I think we have here an explanation,
rather, of mathematical necessity itself. It resides in our unstated
policy of shielding mathematics by exercising our freedom to reject
other beliefs instead.
\eq

\section{01-06-04 \ \ {\it ReAktivation} \ \ (to M. P\'erez-Su\'arez)} \label{PerezSuarez12}

\bmps
Great, my copy of `The Activating Observer' includes 113 pages, 455 references.
\emps

It's now at 187 pages, 513 references.  Mainly I've significantly beefed up the pragmatism references---James is quoted extensively now.  Also I've added a fair amount from the Jung--Pauli correspondence (Pauli's side).  If you want it at this stage, I can send you the pdf; I don't think I'll be working on it again for a while.

\bmps
Btw, I have found a very interesting letter (at the very least, its
opening paragraph) from Pauli to Bohm, which is item 1313 (a nice
number) in the PWB collection, vol.\ 1950--1952. I think that is one of
his clearest statements on the ``attached observer''. I'll take the
book and copy it for you (the paragraph I referred to, not the whole
thing) later on.
\emps
I should peruse those volumes.  Are many of the letters in English?  Thanks for the offer.  Please try to include as much of the letter as possible.

\section{01-06-04 \ \ {\it Quantum System as Philosopher's Stone} \ \ (to M. P\'erez-Su\'arez)} \label{PerezSuarez13}

First a joke:  Notice the similarity between the symbol drawn near the top of this page
\begin{center}
\myurl{http://www.starchamber.com/1997/09/the-rewards-of-the-infinite.html}
\end{center}
and a modern quantum circuit.  Archetypes everywhere!

(I'll cc this note to David {\Mermin}; he might enjoy the joke too.)

More seriously, the phrase in the title of this note is something I've toyed with as a title for a paper I'd like to write one day.  It's been on my mind ever since I wrote these words for \quantph{0404156}:
\bq\noindent
In fact, one might go further and say of quantum theory, that in
those cases where it is not just Bayesian probability theory full
stop, it is a theory of stimulation and response. The agent, through the process of quantum measurement stimulates the world external to
himself.  The world, in return, stimulates a response in the agent
that is quantified by a change in his beliefs---i.e., by a change
from a prior to a posterior quantum state.
\eq

With the aid of a quantum system---and only with its aid---a transmutation takes place in the agent.  Of course, these ideas flow directly from Pauli; I'm just trying to give them more rigorous substance.  (See for instance, the Heisenberg entry ``Wolfgang {\Pauli}'s Philosophical Outlook'' in my compendium.  Or more generally, any number of entries under {\Pauli} himself.)

\bmps
As I promised, here you have the paragraph I told you from Pauli's letter to Bohm, dated December 3, 1951:
\bq\noindent\rm
``Since Descartes, it was the ideal of natural philosophy to
conceive a system of laws in which an entirely loose and untied
observer is looking from outside at a part of the world completely
determined by these laws. For me, however, it is much more
satisfactory if the laws of nature themselves exclude in principle the possibility even to \emph{conceive\/} the disturbances in the observer's
own body and own brain connected with his own observations.''
\eq
Well, it didn't turn out to be either the opening paragraph of the
letter or his clearest statement on the rejection of the ideal of the
``detached observer''. Actually, I am not sure as to what he means by
his last sentence (I have my own guess, though).
\emps
I think he means ``quantum system as philosopher's stone.''

\section{02-06-04 \ \ {\it Dreams, 2} \ \ (to A. Sudbery)} \label{Sudbery11}

\bts
We Everettians must be getting to you. {\rm \smiley}
\ets

Keep in mind that I (like most people) also have dreams about snakes and rats too.

\section{02-06-04 \ \ {\it The Chemical Wedding} \ \ (to M. P\'erez-Su\'arez)} \label{PerezSuarez14}

I promise to write more tomorrow.  Thanks for the references.  I have to admit that I'm getting more interested in the alchemical aspects of the whole issue again.  Maybe it's your dangerous influence upon me!

\subsection{Marcos's Preply 1}

\bq
Thanks for your note. As soon as I read your words under the heading ``Subject'', I found myself thinking that it would make for a great title!

Yes, the similarity is really amazing, much more than I was expecting it to be while downloading the webpage. But let me suggest to you two books, namely,
\begin{itemize}
\item
    A. Roob: {\sl Alchemy and Mysticism\/} (Taschen) (unbelievably INexpensive given its contents),
\item
    S. Klossowski de Rola: {\sl Alchemy.\ The Secret Art\/} (Thames and Hudson) (A tougher nut than the previous one. Sometimes it is not even that easy to discern whether what you are reading is the main text or paragraphs quoted from alchemical writings.)
\end{itemize}
and a website: \myurl{http://www.levity.com/alchemy/}.

The three of them are quite interesting if only for the huge amount of (enrapturing, breathtaking, amazing \ldots) images from many ages (although Middle Ages and Renaissance take most of them, for sure) that they include. But to me they are interesting in more than that.

I'll think harder on Pauli's sentence in that letter, but he writes:
\bq\noindent
[\ldots] much more satisfactory if the laws of nature themselves exclude in principle the possibility even to \emph{conceive\/} the disturbances in the observer's own body and own brain connected with his own observations.
\eq
I mean, he says ``excludes''. This is where I get lost.
\eq

\subsection{Marcos's Preply 2}

\bq
Take a look at this (included in one of the books I told you about in a previous note). This is my own translation from the Spanish (the edition I own):
\bq
At the microphysical level there is a tight ontological [imbrication?] between the subject and the object under observation. It has been necessary to admit that subjectivity is an active agent in the development of natural processes, processes that some alchemists define as the permanent inversion of the inner and the outer.
\eq
And later on, he quotes Whitehead after making (the author) some reference to the ``mercurial'' character of physics: ``Exactness is a fake''.

I insist that this is a wonderful book, if only for its pictorial reproductions. If only it were larger \ldots\
\eq

\section{02-06-04 \ \ {\it Leggett} \ \ (to G. Brassard)} \label{Brassard40}

Thanks for the note.  I'm just about to go out for a ``good-bye'' pub meeting with my friend Ken Duffy.  (When John Lewis died, he got reassigned to Maynooth; today was his last day here.)

Your note is cherished.  I'll send you that longer reply tomorrow.

In the meantime:
\bgb
You may be interested to know that I gave our talk on QFLQI (with your
name explicitly as coauthor on the title slide) a few days ago in
front of 2003 Physics Nobel Prize laureate Tony Leggett.  He showed
interest with several questions, but in the end I don't think he
bought the project.
\egb
Pasted below is my own recent letter to Leggett.  [See 25-05-04 note ``\myref{Leggett2}{Bathroom Reading}'' to A. J. Leggett.] Like with you, I saw him have little flashes of understanding, but ultimately he didn't buy it.  But that's OK, with concerted efforts, we'll eventually get there.

I did get this recently from a Minnesota professor (Michel Janssen) who was at the same meeting as Leggett:
\bmj
Anyways, may this serve as a surrogate for a substantial response to
your work. I too am glad that we had that chat. Without it, I'm
afraid, I would have just dismissed your stuff as crackpot physics.
Now I see it as the one serious contender to my favorite scheme---Everett $+$ decoherence.
\emj
These kinds of reactions keep rolling in  \ldots\ just no job market for me.

\section{03-06-04 \ \ {\it Your Address and Six or So Books Coming} \ \ (to H. Mabuchi)} \label{Mabuchi9}

If you don't mind, I might have some books from Amazon or Powell's shipped to your house to await my arrival.  That'll save me quite a bit in shipping. \ldots

I think in total I ordered six books and had them shipped to your home address.  Thanks for letting me do this.  The books are all on Schiller, Boutroux, and pragmatism.  Feel free to dig into them if you want.

\section{04-06-04 \ \ {\it Earwax} \ \ (to P. C. E. {\Stamp})} \label{Stamp4}

\bpces
Thanks for sending the copy of your book -- and I will put it in the bathroom, where it will be a useful antidote to the fashion magazines that my wife leaves there.
\epces

Quoting you:

\bpces
Regarding the information theory programme -- I have indeed got the
message that you are trying to do something analogous to, e.g.,
Skinner and Watson, with your black box (ie., stimulus-response)
approach. However I have several remarks to make on this

(i) Like many people I find the black box approach scientifically sterile -- and I didn't need, e.g., Chomsky to tell me why, it has been
understood to be sterile for a long time now (from this perspective
Skinner-Watson was apparently just an ephemeral aberration). The
problem with a black box approach is that it ignores what may be
going on inside or outside the box, which exists without the stimuli
and may not be completely characterised by the responses. Of course
hidden variable theories have been toasted so far, most generally by
Bell--Kochen--Specker, but that is no reason to cease looking for
something more than a ``measurement calculus'' (which is essentially
what the information approach is doing as far as I can see). This is
what physicists do, and they have ended up with quite a dazzling
display of objects ranging from non-Abelian gauge fields to
macroscopic order parameters and `wave-functions' to topological
quantum fluids to complex spin nets, etc. The attempt to strip this
down to nothing but stimulus and response, in a measurement or
information theoretic calculus, is very much like the attempt by
Skinner and Watson to take all the colour out of psychology -- at least in my opinion. This is something of a gut reaction, but I suspect it
is shared by many physicists.
\epces

If you can say this, you haven't listened to me very carefully.  It seems to me that you're just rather offhandedly drawing associations with things you have a distaste for because of little more than a similarity of words.

For instance, how can you make your characterization above mesh with these words that I wrote in \quantph{0205039}?  (I hope you'll read the passage all the way to the end; the average listener doesn't seem to hear anything beyond the first paragraph.):
\bq
So, throw the existing axioms of quantum mechanics away and start
afresh! But how to proceed? I myself see no alternative but to
contemplate deep and hard the tasks, the techniques, and the
implications of quantum information theory. The reason is simple,
and I think inescapable.  Quantum mechanics has always been about
information.  It is just that the physics community has somehow
forgotten this.

This, I see as the line of attack we should pursue with relentless
consistency:  The quantum system represents something real and
independent of us; the quantum state represents a collection of
subjective degrees of belief about {\it something\/} to do with that system (even if only in connection with our experimental kicks to
it). The structure called quantum mechanics is about the interplay
of these two things---the subjective and the objective.  The task
before us is to separate the wheat from the chaff.  If the quantum
state represents subjective information, then how much of its
mathematical support structure might be of that same character?
Some of it, maybe most of it, but surely not all of it.

Our foremost task should be to go to each and every axiom of quantum
theory and give it an information theoretic justification if we can. Only when we are finished picking off all the terms (or combinations
of terms) that can be interpreted as subjective information will we
be in a position to make real progress in quantum foundations.  The
raw distillate left behind---minuscule though it may be with respect
to the full-blown theory---will be our first glimpse of what quantum
mechanics is trying to tell us about nature itself.

Let me try to give a better way to think about this by making use of
Einstein again. What might have been his greatest achievement in
building general relativity? I would say it was in his recognizing
that the ``gravitational field'' one feels in an accelerating
elevator is a coordinate effect. That is, the ``field'' in that case
is something induced purely with respect to the description of an
observer. In this light, the program of trying to develop general
relativity boiled down to recognizing all the things within
gravitational and motional phenomena that should be viewed as
consequences of our coordinate choices.  It was in identifying all
the things that are ``numerically additional'' to the observer-free
situation---i.e., those things that come about purely by bringing
the observer (scientific agent, coordinate system, etc.) back into
the picture.

This was a true breakthrough.  For in weeding out all the things that
can be interpreted as coordinate effects, the fruit left behind
finally becomes clear to sight: It is the Riemannian manifold we call
spacetime---a mathematical object, the study of which one can hope
will tell us something about nature itself, not merely about the
observer in nature.

The dream I see for quantum mechanics is just this. Weed out all the terms that have to do with gambling commitments, information,
knowledge, and belief, and what is left behind will play the role of Einstein's manifold. That is our goal.  When we find it, it may be
little more than a minuscule part of quantum {\it theory}. But being a clear window into nature, we may start to see sights through it we
could hardly imagine before.\footnote{I should point out to the
reader that in opposition to the picture of general relativity, where reintroducing the coordinate system---i.e., reintroducing the
observer---changes nothing about the manifold (it only tells us what
kind of sensations the observer will pick up), I do not suspect the
same for the quantum world. Here I suspect that reintroducing the
observer will be more like introducing matter into pure spacetime,
rather than simply gridding it off with a coordinate system. ``Matter
tells spacetime how to curve (when matter is there), and spacetime
tells matter how to move (when matter is there).''
Observers, scientific agents, a necessary part of reality?  No.  But
do they tend to change things once they are on the scene?  Yes.  If
quantum mechanics can tell us something truly deep about nature, I
think it is this.}
\eq

The very point of the whole program is to see quantum systems as MORE than just black boxes.  But to do that, a lot of crud to do with observers or agents has to first be excised from the theory---or, at least, that's what the program I'm working on is about:  Getting to the physical kernel of quantum mechanics.  Here's the way I put it in a recent proposal:
\bq
No physicist would be doing his job if he did not strive to map
reality itself---that is, reality as it is independently of any
information processing agents. The issue is one of separating the
wheat from the chaff: Quantum mechanics may be predominantly about
information, but it cannot {\it only\/} be about information.  Which part is which?  The usual way of formulating the theory is a
thoroughly mixed soup of physical and informational ingredients.

This is where quantum information (including the collateral fields of quantum cryptography, computing, and communication theory) has a
unique role to play.  Its tasks and protocols naturally isolate the
parts of quantum theory that should be given the most foundational
scrutiny. ``Is such and such effect due simply to a quantum state
being a state of information rather than a state of nature, or is it due to the deeper issue of what the information is about?'' Recent
investigations by several workers are starting to show that many of
the previously-thought `fantastic' phenomena of quantum
information---like quantum teleportation, the no-cloning theorem,
superdense coding, and nonlocality without entanglement---come about simply because of the epistemic nature of the quantum state. On the
other hand, other phenomena, such as the potential computational
speed-up of quantum computing, seem to come from a more physical
source: In particular, the answer to the question, ``Information
about what?''

When we finally delineate a satisfying answer to this, physics will
reach a profound juncture.  We will for the first time see the exact nature of `quantum reality' and know what to do with it to achieve
the next great stage of physics. Trickle-down effects could be the
solution to the black-hole information paradox---perhaps already seen in broad outline---and even the meshing together of quantum theory
and gravitational physics. In the meantime the approach proposed here is a conservative and careful one; the work to be done is large. The
effort aims not to say first what `quantum reality' is, but what it
is not and gather insights all along the way.
\eq

The last thing I want to do is view quantum theory as a species of a rather sterile positivism, or as you put it, as no more than a ``measurement calculus''.  It is clearly more than that.

Where you misread me is in thinking that taking an epistemic view of the quantum state---epistemic, in particular, about the potential results of measurements and nothing more---is necessarily throwing out the baby with the bathwater.  It is not.  It is instead a strategy for getting around all the nasty foundational problems of a realistically construed ``wave-function collapse'' while not going so far as the emptiness of Everett or Bohm (or the unchecked and little-motivated speculation of GRW).  The point is to salvage realism by looking for it at a {\it higher level\/} than the level of quantum states and their equations of motion.

For one can ask, and one should ask:  Why this ``measurement calculus'' rather than another?  There is some {\it property\/} of the stuff of the world, or of physical systems, that forces the particular calculus we use upon us---it is not arbitrary, and to that extent, it tells us something about the world as it is without the gambling agent.

Here's yet another way I've put the program recently, in a letter to Jim Woodward:  [See 18-05-04 note ``\myref{Woodward2}{Agents, Interventions, and Surgical Removal}'' to J. Woodward.]

There, I hope that's enough to defend myself from your accusation of being to quantum mechanics what Skinner and Watson were to psychology.  Or, at least, to show you that my heart is not where you think it is.

You want color?  Then let me tell you about the actual origin of my use of the phrase ``stimulation and response''---it was not Skinner or Watson or the like.  Instead, the words comes from the palette of imagery I've been building up ever since I started calling my view ``the sexual interpretation of quantum mechanics''.  (See {\sl Quantum States:\ What the Hell Are They?\/}\ posted on my webpage, last paragraph page 49.  See also the note to {\Mermin} below.)  Put yourself in a lascivious mindframe, and then once again approach the word ``stimulation.''  I hope it leaves an indelible mark on you.

\section{14-06-04 \ \ {\it RMTTR,BOWEAL} \ \ (to H. Mabuchi)} \label{Mabuchi10}

The title to the note would have been ``Reality Modifiable To the Roots, But Only with Effort and Luck,'' but it looked too long, and so I shortened it.

Our conversation this morning emboldened me; I rather enjoyed the part about the Upanishads, etc.  In that regard, let me attach two of my pseudo-papers, ``Delirium Quantum'' and ``The Activating Observer''.  You might enjoy perusing parts of both.  Discussions like the one with you this morning make me feel like maybe I'm not going too far after all.  Indeed, maybe I'm not going far enough!

OK, the last sentence was a little of a joke \ldots\ but maybe ultimately not.  Still, if I use quantum mechanics as the firmest place I can start, I don't think it yet countenances a move that far, i.e., that that which is without is actually within (Brahman $=$ Atman).  But I think QM does move somewhat in the direction, by making that which is without to be somewhat malleable.  And to that extent, that which is without looks to be somewhat within.  (It's kind of fun writing like this.)  Anyway, true enough, the effort does remain with the distinction without/within---which I am presently comfortable with.  It doesn't completely erase it.

If you enjoy anything in these writings, give me feedback.

\section{14-06-04 \ \ {\it The Toy Intro} \ \ (to H. Mabuchi)} \label{Mabuchi11}

Finally, I'll share with you the toy intro I wrote for the basic quantum mechanics talk.  Since I almost surely won't say it out loud (or at least in such detail), I might as well share it with somebody.  [See 17-06-04 note ``\myref{Mabuchi12}{Preamble}'' to H. Mabuchi.]

\section{14-06-04 \ \ {\it Concrescence?}\ \ \ (to W. K. Wootters)} \label{Wootters20}

BTW, on my flight yesterday, I sketched out the following little introduction for my summer school lectures here at Caltech.  I don't know if I'll actually say those things.  But does any of it sound like the Whiteheadian ideas---I think you called it concrescence---you told us about in {\Montreal} two Falls ago?  [See 17-06-04 note ``\myref{Mabuchi12}{Preamble}'' to H. Mabuchi.]

\subsection{Bill's Reply}

\bq
Yes, it seemed to me when I was reading Whitehead that the emergence of a definite fact, out of quantum potentialities, should count as an example (maybe the only example) of concrescence.  It is the process of becoming concrete.

I don't have my Whitehead with me here, but from a couple of webpages I got the following definitions:

concrescence -- a dipolar process involving an interplay between physical feelings and mental valuations (PR 108) in which prehensions of early phases are contrasted in later phases. In short, it is the production of novel togetherness (PR 21.)

Whitehead defines `concrescence' as a process in which prehensions are integrated into a fully determinate feeling or satisfaction. A `satisfaction' is a unity of physical or mental operation attained by an actual entity.
\eq

\section{17-06-04 \ \ {\it Preamble} \ \ (to H. Mabuchi)} \label{Mabuchi12}

I think I would like you to also post the little text file below
along with my other suggested readings for my ``Intro to QM''
lecture. You can give it the title ``Preamble''.  It was something I
sketched out on my flight over here, and reading over it again, I
kind of like it.

\bq
A lecturer faces a dilemma when teaching a course at a farsighted
summer school like this one.  This is because, when it comes to
research, there is often a fine line between what one thinks and what
is demonstrable fact.  More than that, conveying to the students what
one thinks---in other words, one's hopes, one's desires, the
potentest of one's premises---can be just as empowering to the
students' research lives (even if the ideas are not quite right) as
the bare tabulation of any amount of demonstrable fact.  So I want to
use one percent of this lecture to tell you what I think---the
potentest of all my premises---and use the remaining ninety-nine to
tell you about the mathematical structure from which that premise
arises.

I think the greatest lesson quantum theory holds for us is that when
two pieces of the world come together, they give birth.  [Bring two
fists together and then open them to imply an explosion.]  They give
birth to FACTS in a way not so unlike the romantic notion of
parenthood:  that a child is more than the sum total of her parents,
an entity unto herself with untold potential for reshaping the world.
Add a new piece to a puzzle---not to its beginning or end or edges,
but somewhere deep in its middle---and all the extant pieces must be
rejiggled or recut to make a new, but different, whole.  That is the
great lesson.

But quantum mechanics is only a glimpse into this profound feature of
nature; it is only a part of the story.  For its focus is exclusively
upon a very special case of this phenomenon:  The case where one
piece of the world is a highly-developed decision-making agent---an
experimentalist---and the other piece is some fraction of the world
that captures his attention or interest.

When an experimentalist reaches out and touches a quantum
system---the process usually called quantum `measurement'---that
process gives rise to a birth.  It gives rise to a little act of
creation.  And it is how those births or acts of creation impact the
agent's {\it expectations\/} for other such births that is the
subject matter of quantum theory.  That is to say, quantum theory is
a calculus for aiding us in our decisions and adjusting our
expectations in a QUANTUM WORLD\@.  Ultimately, as physicists, it is
the quantum world for which we would like to say as much as we can,
but that is not our starting point.  Quantum theory rests at a level
higher than that.

To put it starkly, quantum theory is just the start of our adventure.
The quantum world is still ahead of us.  So let us learn about
quantum theory.
\eq

\subsection{Hideo's Reply}

\bq
From {\sl Three Farmers on Their Way to a Dance}, Richard Powers' first novel:
\bq\noindent
``Sander,'' a German photographer of the early 1900's, ``at the same time as those working in physics, psychology, political science, and other disciplines, blundered against and inadvertently helped uncover the principle truth of this century: viewer and viewed are fused into an indivisible whole.  To see an object from a distance is already to act on it, to change it, to be changed.''
\eq
\eq

\section{18-06-04 \ \ {\it Retracing Thoughts} \ \ (to J. Woodward)} \label{Woodward3}

Retracing yesterday's conversation in my head (I've got nothing better to do; my lectures being over), I realized I didn't quite adequately respond to one of your questions near the end.  It was just before we arose from the table.  You asked something like, ``Then what are the random variables in these probability distributions?''  I replied, ``Measurement outcomes.''  But I don't want to leave you with the impression that such an empiricist or positivist take is the end of the story.  Indeed the bit below, which I wrote for the research proposal John Preskill gave you and others, addresses the very issue with the phrase, ``Information about what?''

\bq
No physicist would be doing his job if
he did not strive to map reality itself---that is, reality as it is
independently of any information processing agents. The issue is one of separating the wheat from the chaff: Quantum mechanics may be
predominantly about information, but it cannot {\it only\/} be about information.  Which part is which?  The usual way of formulating the theory is a thoroughly mixed soup of physical and informational
ingredients.

This is where quantum information (including the collateral fields of quantum cryptography, computing, and communication theory) has a
unique role to play.  Its tasks and protocols naturally isolate the
parts of quantum theory that should be given the most foundational
scrutiny. ``Is such and such effect due simply to a quantum state
being a state of information rather than a state of nature, or is it due to the deeper issue of what the information is about?'' Recent
investigations by several workers are starting to show that many of
the previously-thought `fantastic' phenomena of quantum
information---like quantum teleportation, the no-cloning theorem,
superdense coding, and nonlocality without entanglement---come about simply because of the epistemic nature of the quantum state. On the
other hand, other phenomena, such as the potential computational
speed-up of quantum computing, seem to come from a more physical
source: In particular, the answer to the question, ``Information
about what?''

When we finally delineate a satisfying answer to this, physics will
reach a profound juncture.  We will for the first time see the exact nature of `quantum reality' and know what to do with it to achieve
the next great stage of physics. Trickle-down effects could be the
solution to the black-hole information paradox---perhaps already seen in broad outline---and even the meshing together of quantum theory
and gravitational physics. In the meantime the approach proposed here is a conservative and careful one; the work to be done is large. The effort aims not to say first what `quantum reality' is, but what it
is not and gather insights all along the way.
\eq

What these paragraphs are not detailed enough to convey is that the issue boils down to the need for a much better understanding of the implications of the Kochen--Specker theorem:  That is, the theorem that blocks measurement outcomes from being preexistent realities in the usual quantum description (i.e., without the supplementation of nonlocal hidden variables, etc.).  My own feeling is that it is this part of quantum theory which gives rise to the computational speed-up in quantum computation.  So, these things are intimately related I believe.

Thus, as much as I endorse taking ``measurement'' (or ``intervention'' as Asher Peres and I have called it in our articles; see pages 99, 146, 336, 384, 394, etc., and most notably page 417) as the very foundation of what the theory is about---those little acts of creation I always talk about---there is still so much work to be done to get this clarified.  Part of it hinges on clarifying the particular form of ``noncontextuality'' for the quantum probability assignment, but also more philosophically I think it is an opportunity to rethink variations on James and Dewey's notion of truth.  (Maybe I wouldn't want to apply that notion to everything in the world, as a true pragmatist (say Richard Rorty) would, but it does strike me as an appropriate concept for these particular quantum mechanical issues.)

Anyway, I hope that tells you a little more.

I've also been thinking a little more about ``Making Things Happen'' in the context of quantum computing.  I think there's a lot of thought one could give that.  I haven't yet decided if I would count a quantum measurement as an example of ``making things happen''.  However, one most certainly should count the enacting of a quantum computational algorithm---let's say, factoring a large composite number---as an example:  The number goes in the quantum computer, the factors come out.  What is interesting is that there are some new models of quantum computation that involve solely the action of quantum measurement, and make no use of time evolution at all.  How do these ideas mesh together?  Don't know at the moment.

I'm sitting in Steel, Rm.\ 305, till the end of the stay, by the way.  The phone number on my desk is 4493.

For the fun of it, I'll paste below the intro to my first lecture for the CBSSS summer school that I told you about.  I didn't do it quite so William-Jamesy when I did it in real life, but it conveys the idea.  (These were the notes I wrote myself on the flight the other day.)  The second lecture---the one I gave after we parted---went fabulously.  You must've put me in the right mood.  The students here are like eager sponges!  [See 17-06-04 note ``\myref{Mabuchi12}{Preamble}'' to H. Mabuchi.]

\section{21-06-04 \ \ {\it Teleportation} \ \ (to G. Musser)} \label{Musser3}

\bgm
P.S. How do you interpret teleportation?
\egm

Below is an outtake from a letter I wrote to Oliver Cohen a while ago; I think it answers your question.  [See 15-01-03 note ``\myref{Cohen1}{Memory Lane}'' to O. Cohen.] I decided to be lazy and not change the page numbers for {\sl Notes On a Paulian Idea\/} listed below:  They refer to the web edition---which can be downloaded at my webpage---rather than the VUP edition which I had sent to you.  (Did you get it?)  If you have trouble downloading it, let me know and if worse comes to worst, I'll fix the page numbers up for you.

\section{28-06-04 \ \ {\it Decisions, Decisions} \ \ (to J. Woodward \& C. R. Hitchcock)} \label{Woodward4} \label{Hitchcock1}

There's no doubt I think the time is ripe to meld philosophy of science and honest-to-god quantum information physics---both physics and philosophy will greatly benefit from the marriage---but finding a place that'll share this belief with me is mostly beyond my control.  Of course I'll try to turn the tide by writing more papers (and hopefully ever more convincing papers), but that will take some time in the coming.  In any case, I hope I have sufficiently piqued your interest in the subject that you might follow it a little and use your accumulated body of work on causality to make a contribution to it.  Vice versa, I hope the field will give you something to think about from a fresh perspective and maybe lead to something new in your own work.

In case you're interested, you can read about the quantum de Finetti theorems I told you about in my paper with {\Schack}, ``Unknown Quantum States and Operations, a Bayesian View,'' posted here: \quantph{0404156}.  In the time-evolution version of the theorem (as it stands), we only imagine a dynamics imposed upon our quantum systems from the outside---say, by some ``device.''  If we were to try to prove a version of the theorem sufficient for an account of an internal dynamics---i.e., like the stuff I was telling you about at the Red Door---then I am sure we would have to take into account something like the conditions for your ``interventions'' to make it fly.  Anyway, food for thought and some work for the future.

\section{29-06-04 \ \ {\it Late, Sorry} \ \ (to G. Musser)} \label{Musser4}

I'm sorry I'm late.  The travel and flu took a big toll on me, and I forgot about email for a while \ldots\ despite my promises.  Bad behavior.

Anyway, let me zoom in on the paragraphs where you mention me.

\bgm
\ldots\ Many of those struggling to grok quantum mechanics say Einstein's
mistrust [of QM] was well-founded. ``This guy saw more deeply and more
quickly into the problems that plague us today,'' says Christopher
Fuchs of Bell Labs \ldots.
\egm

Did I really say that?  It's kind of a strange construction.  But maybe it's only a question of intonation.  Does something like this more capture the moment:  ``This guy saw more deeply---and actually more quickly---into \ldots'' (emphasizing quickly maybe)?  Also, I am a bit reluctant to say that he saw into the ``problems that plague us today,''  because I'd rather not think of them as such horrible problems.  But if I said it that way, then I said it, and it's your property now.  But, given the chance, I think I would say it this way if I could:  ``This guy saw more deeply---and actually more quickly---into the central issues of quantum mechanics than many give him credit for.''  I say this because he so quickly gave such a clean argument for the idea that the quantum mechanical wave function represents a ``state of knowledge'' rather than a ``state of nature.''  And I think that realization quashes over half of the mysteries already.  See, for instance, Section 3 of my ``only a little more'' paper.

The only thing that differentiates Einstein from our little effort at this juncture is that he thought physics should ultimately be about the ``states'' of nature, whereas the guys in my camp think that the great lesson of quantum mechanics is that physics must be about something else.

\bgm
If hidden variables have a fatal flaw, it is that any attempt to
reconstruct quantum mechanics from something deeper is limited by
physicists' imagination.
\egm

I'm not sure I understand this.  What do you mean?

\bgm
Rather than try to build the theory from the bottom up, the most
respected efforts to explain the quantum work from the top down.
Fuchs, a leader in this area, says that so much of quantum mechanics
pertains to our information about a system rather than the system
itself. As Einstein argued, it is a theory of hearsay rather than of
direct knowledge. Fuchs is trying to strip away these subjective
aspects so that the objective properties of nature can stand out.
\egm

Our information about a system.  More accurately, our (statistical) information about how the system will evoke a response in us---i.e., the thing usually called a ``result'' of a measurement.  (Asher Peres likes to say, ``Unperformed measurements have no results.'')

\bgm
Physicists had long assumed, for example, that measuring a quantum
system causes it to ``collapse'' from a range of possibilities into a
single actuality. Fuchs argues it is just our uncertainty about the
system that collapses. Uncertainty about a quantum system is very
different from uncertainty about a classical one, because the process
of measurement cannot help but disturb a quantum system. And this
sensitivity to disturbance is a clue to what a quantum system really
is. Quantum mechanics may prove to be more complex than either its
inventors or detractors ever envisioned \ldots.
\egm

This is the paragraph that particularly troubles me.  For it makes this position of mine look like that of a closet hidden-variablist (i.e., one that hasn't had the intellectual honesty to see what he really stands for).

There are two issues to try to separate, and I'm not sure what's the best way to do it so that the paragraph is both accurate and pithy at the same time.  In a quantum measurement, I would say two things happen:  1) something is created that was not there before, the measurement ``click''---this idea bars hidden variables---and 2) something changes, this is the observer's wave function for the system, i.e., his knowledge.  Physicists often get in the bad habit of muddling the two things, the creation (``the measurement result coming out of nowhere'') and the change due to the creation.  The ``collapse'' properly only refers to the change of the wave function and, thus, to knowledge.  But what is the knowledge about?  It is not really about the system alone, or what is there, i.e., what properties it has---if it were, then we would be talking about hidden variables ultimately---but rather about what will come about the next time the observer touches it.

Think of a quantum system as a kind of philosopher's stone.  Its power is in transmuting an observer's state of knowledge.  And presumably it is transmuted too in the process.  Interestingly, quantum theory is predominantly about the observer's side of this state of affairs.  Think about how much two parents are transformed by the process of giving birth.  One parent is the quantum system, one parent is the observer; the child is the click.  Quantum theory effectively only interrogates one of the parents---the observer---and, to the extent that we get some characterization of the other parent from this interrogation, it is only second-hand.

I doubt these words are helpful to you, but I'm trying.

On another point, I know that I've used the word ``disturb'' before, but I have been trying to get away from it because of all the incorrect imagery it evokes---it just calls out for one to think of hidden variables being disturbed.  Thus, I would like to keep the idea of ``sensitivity,'' but I hope I never say ``sensitivity to disturbance'' anymore (if I ever did).  In the past, rather than the parenting imagery I used above, I have used the imagery of an oyster, a grain of sand, and a pearl to try to capture the idea of system, observer, and click respectively.  And it was in that context that I started to speak of a ``quantum system's sensitivity to the touch.''  No oyster, no pearl.  No sand grain, no pearl.  The pearl is a manifestation of the oyster's sensitivity to the touch in the presence of a sand grain.

``And the shape of this sensitivity is a clue to what a quantum system really is.''  --- For some reason I like that sentence, but I don't know how you would fit it into what you write.

If you need more help, I should be pretty steady on my email now that I'm back in Dublin.  (Though I'm flying out to house-hunt in NJ as of July 6, and you can expect an almost total shutdown until that's over with July 16 or 17.)

Sorry if I'm causing you more trouble than I'm worth.

\section{30-06-04 \ \ {\it Fancy Words First} \ \ (to M. Janssen)} \label{Janssen3}

\bmj
[L]et me just ask you what I should read to get over the next hurdle I'll have to take before this information stuff has a prayer of dislodging Everett. The hurdle is simply this: how is your information approach different from the old ignorance interpretation of QM.
And I'm sure you won't mind me spelling this out a bit to make sure you know what I mean. I take the ignorance interpretation to be saying things like this: of course, an electron has a definite position (as well as definite whatever) at all times, it's just that typically we don't know what that is. That's why we can only describe the electron in terms of a superposition of it being in all places it could possibly be. When we perform a measurement we find out what its real position is. So QM is just like classical statistical mechanics basically: it gives us probability distributions over values of variables we don't know the exact value of. The only difference is that QM tells us that we can't ever know all values of all variables at the same time, because typically the measurement of the value of one variable disturbs the value of others. I buy (what I take to be) the standard line against this view: the probability distributions generated by state vectors in Hilbert space cannot be reduced to classical probability distributions.

From talking to Christoph Lehner, who witnessed your exchange with David Albert at this year's Ryno-Bub fest, I understand that Albert's criticism was along these same lines (but maybe Christoph was just reading that into the exchange: it's certainly Christoph's first line of defense when I told him my Everettian faith had been tested). I'll be seeing Christoph in a week and a half and I'd appreciate it if you provide me with some target-specific ammunition.
\emj

Thanks for the note, and thanks for the alert to the {\sl Science\/} article.  I hope I can retrieve it.

Thanks also for the question.  I'll do my best to clarify things for you; give me a couple of days to work something up specific to your question.  The short story is that I am about as far away from the ignorance interpretation you speak of as one can imagine:  Quantum measurement is---for me---not a case of revealing, but a case of making (making something that was explicitly not there before).  That I take to be the significance of the Bell and Kochen--Specker theorems.

Particularly, you've got to understand that I draw a distinction between the quantum world (i.e., the world presumably as it really is) and the quantum formalism.  The latter is not a reflection of the former in any exact-correspondence sense.  That is to say, quantum theory is not a story of the world as it is completely independent of decision-making agents.  Rather it is---from this point of view---a normative theory for the optimal survival of agents immersed in this kind of world.  To the extent that we are using {\it this\/} normative theory of survival (i.e., this theory of decision-making and gambling) rather than {\it that\/} normative theory, we are saying something about the world as it is independent of the agent.  But the strategy for saying something about the world is necessarily more oblique than one thought one could obtain in classical times.

All of this is much too subtle of stuff for David Albert to have absorbed in my talk \ldots\ predominantly because he does not listen.  His actions (at my talk and every other one I have ever seen him attend) make it clear that his purpose for attending talks is not to listen, but to have a good spit.

Anyway, I will try to write you (or direct you to) something of detail within a couple of days.  In the meantime, let me attach a little poetic piece I wrote for my summer-school lectures at Caltech last week.  It says a little more about the point of view above.  Feel free to share any of these notes with your friend. [See 17-06-04 note ``\myref{Mabuchi12}{Preamble}'' to H. Mabuchi.]

\section{30-06-04 \ \ {\it Quail and Pheasant?}\ \ \ (to J. B. Lentz \& S. J. Lentz)} \label{LentzB3} \label{LentzS1}

\bq\it
\noindent Leggett, Preskill, and 22 other physicists,
philosophers, and historians have gathered
here for the Seven Pines Symposium.
Packed into a slightly ramshackle lodge in a
wooded state park, the scholars---all of them
men---will share their insights, suites of
rooms without telephones, and meals of roast
quail and pheasant at a long communal table.
Perhaps not since the famous Solvay Conferences
of the early 20th century, at which
Niels Bohr and Albert Einstein debated the
meaning of quantum theory in their free
time, has physics seemed so genteel.
\eq

I thought you might enjoy this article from {\sl Science} [A.~Cho, ``Elite Retreat Takes the Measure of a Weirdly Ordinary World,'' Science {\bf 304}, 1896 (2004)], which one of my acquaintances from the same meeting just sent me.  I learned from it that I was eating quail and pheasant!  I myself never imagined I was eating anything but chicken.

Anyway, I was particularly pleased to see the napkin pictured in the article:  It means someone was taking notes at my lecture or explaining it to someone else.  The fellow in the blue shirt looking down at me is the Mercedes man.

The atmosphere the author builds in the story is basically the sort of thing Kiki envisions when we dream of opening an institute in the New Hampshire countryside.  Scientists eating good food because someone has put it in front of them, and happy for it \ldots\ even if they don't quite know what it is.

\section{30-06-04 \ \ {\it Memory Lane} \ \ (to D. B. L. Baker)} \label{Baker9}

Just ran across this on the web and it reminded me of the time we both happened to see Vampire Circus and reported it to each other one Monday morning in about 1979/80.

\bq
Things weren't as complex in 1972 when the Spanish film {\sl La Noche de Walpurgis\/} was released in the grindhouses and drive-ins of America as {\sl The Werewolf vs.\ The Vampire Woman}. A big hit in Europe, cementing Paul Naschy's position as Spain's preeminent horror film star, it came to America with all the fanfare and bluster of a dusty, ratty, moth-eaten circus. I saw it as a young teenager one gummy-eyed late night in the '70's on San Antonio's KENS-TV's {\sl Project Terror\/} program (``Where the scientific and the mystifying merge \ldots''). {\sl Project Terror}'s programming consisted mainly of European horror at the time, and although some were heavily edited, one could see a glimpse of a bare breast once in a while and some cool but cheap gross-out gore. Ah, adolescence! I don't remember either actively liking or disliking it, but I do remember thinking that the werewolf makeup looked cheap and that the vampire woman (played by Patty Shepard) looked pretty hot, so much that I rooted for her in the climactic battle with the good guy wolfman Waldemar Daninsky played by Paul Naschy. A silly diversion, nothing more, nothing less. I hadn't thought about the movie since I saw it over 25 years ago. I've seen some other Naschy horror titles on video since then ({\sl House of Psychotic Women\/} is one that immediately comes to mind), and they're certainly fun to watch if you don't mind the fast, cheap, and out of control nature of the movies, but they contain none of the cruel poetry and transgressive power of the best of '70's Euro-horror (de Ossorio's {\sl Blind Dead\/} series, Rollin's vampire films, Argento's Suspiria, even a couple of Franco's films). Naschy's work is a throwback to the Universal horrors of the '30's and '40's, albeit with nudity and gore, but unfortunately adds nothing new to the mix.
\eq

\section{30-06-04 \ \ {\it Writing Physics} \ \ (to G. Musser)} \label{Musser5}

\bgm
Another issue that I'm struggling with is this business of separating
states of knowledge from states of nature.  Isn't a state of knowledge
a state of nature where ``nature'' includes the observer?  The
wavefunction is a statement about me, but I'm part of nature, right?
\egm

Honestly, I am quite intrigued by your question.  It's perceptive and a good one---you articulated it well.  I hope to use this as an opportunity to ``write physics'' in Mermin's sense.  If you haven't looked at it yet, read David's article ``Writing Physics'':
\begin{center}
\myurl{http://www.lassp.cornell.edu/mermin/KnightLecture.html}.
\end{center}

\section{01-07-04 \ \ {\it Friends and Enemies} \ \ (to S. Hartmann)} \label{Hartmann3}

Wow!  Thanks for the letter. After reading your letter, I wanted to hire the guy you were talking about; I hope I get a chance to meet him some day.

Anyway, in reading it, it dawned on me that I ought to contrast my emotions with the old phrase, ``With a friend like that, who needs enemies?''  After reading your letter, I thought, ``Wow, with a friend like that, maybe I can afford even a few more enemies!''

Which brings me to Amit.  I'd be more than happy to look at your student's paper!  \ldots Ahh, don't get me wrong, Amit and I actually have a very friendly relationship---he is truly a nice guy, and we've had a lot of personal conversations.  It's just that he didn't read me nearly as thoroughly or carefully as he thought he had.  I found his paper pretty annoying.  For instance, all this talk about ``decoherence'' and ``for all practical purposes'' in one of the sections had nothing to do with me:  I've even gone out of my way to disavow that decoherence has anything whatsoever to do with quantum foundational issues.  (See for instance the introduction to my \quantph{0106166}.  That's probably the strongest statement, but there are many places beyond that, and in more technical detail.)  Also, in his paper it keeps coming through that he had never grasped what I take the quantum state to be knowledge (now belief) about.  Else, when talking about me, he would have never talked about the wave function being ``knowledge of the state.''  That's a phrase I would never use.  Here's the way I explained it most recently to Philip {\Stamp}.
\bq
The main point I was trying to make to you \ldots\ is that in an
information-based interpretation of quantum mechanics (like the one
 {\Caves}, {\Schack}, {\Mermin}, Peres, Hardy, and few others of us are trying to put together) is that:  Though we take the wavefunction to be
information, we never take it to be information about ``what is
existent'' in the quantum system, or ``what is happening in the quantum system.''  That it cannot be information about something like that is
what I take to be the great lesson of the Bell--Kochen--Specker
theorem.  Rather it is always information about how a system will
{\it respond\/} to an external stimulation (i.e., in usual language, a
measurement \ldots\ but I want to get away from the usual language
because it carries the wrong imagery).  This is the way to think of
it when the wave function is about a single system and the way to
think of it when the wave function is about a composite system---it
doesn't matter.

So, what I am asking---among other things---is that the notion of
``measurement'' be replaced with the idea of ``stimulation and
response.''  We only use wave functions to calculate our expectations
with regard to how a system will respond to one or another of our
possible stimulations upon it.  In this light, there is no
contradiction between an ``inside view'' of a measurement where an
unpredictable click really happens and an ``outside view''
(encompassing both the original system and the first observer), where a wave function for the composite system undergoes a smooth and
deterministic transition.  This is because the ``outside view'' wave
function refers again to NOTHING MORE THAN the outside observer's
expectations for how that composite system will respond to external
stimulations upon it.
\eq
If Amit had understood that properly, he would not have misapplied Hemmo's argument to this context.

But they say in the Hollywood circles, no advertisement is bad advertisement.  (And Caltech is close to Hollywood.)  So, I'm keen to hear about your student's paper.  Certainly, feel free to share any of the technical parts of my emails with him (like some of the stuff in here) that you wish.  And I can certainly give pointers to other relevant materials after reading the draft if that would be useful to you guys.

I suppose I'm hard on the philosopher, being such a moving target to attack or analyze---i.e., the fact that my position is evolving rather than being settled yet (maybe it'll never be settled).  But all I'm shooting for is airtight clarity, and that can be a tough process.  Reading other's perceptions of the program is quite useful in that regard.

\section{01-07-04 \ \ {\it Quail and Pheasant?, 2} \ \ (to S. Hartmann)} \label{Hartmann4}

You might enjoy this and the story below too.  [See 30-06-04 note ``\myref{LentzB3}{Quail and Pheasant?}''\ to J. B. Lentz \& S. J. Lentz.]  It was nice to see ({\it what I hope will eventually be\/}) a quantum-Bayesian trademark---i.e., a probability simplex with a restricted area---making an appearance in such a prominent place.

\section{05-07-04 \ \ {\it Your Friend Philip} \ \ (to M. Janssen)} \label{Janssen4}

Let me come back to your question now:
\bmj
[L]et me just ask you what I should read to get over the next hurdle I'll have to take before this information stuff has a prayer of dislodging Everett. The hurdle is simply this: how is your information approach different from the old ignorance interpretation of QM.
And I'm sure you won't mind me spelling this out a bit to make sure you know what I mean. I take the ignorance interpretation to be saying things like this: of course, an electron has a definite position (as well as definite whatever) at all times, it's just that typically we don't know what that is. That's why we can only describe the electron in terms of a superposition of it being in all places it could possibly be. When we perform a measurement we find out what its real position is. So QM is just like classical statistical mechanics basically: it gives us probability distributions over values of variables we don't know the exact value of. The only difference is that QM tells us that we can't ever know all values of all variables at the same time, because typically the measurement of the value of one variable disturbs the value of others. I buy (what I take to be) the standard line against this view: the probability distributions generated by state vectors in Hilbert space cannot be reduced to classical probability distributions.
\emj

As I've already told you, as I see it, my view is pretty diametrically opposed to the ``ignorance interpretation'' you describe.  You can find some early and very direct discussion of this in my paper with Kurt Jacobs, ``Information Tradeoff Relations for Finite-Strength Quantum Measurements.''  Here's a link to the version sitting on the archive: \quantph{0009101}.

The part relevant for you is the introduction.  Reading over this again for the first time in almost four years, it doesn't strike me as half bad (even though I've become more metaphysical since then).  Anyway, read that, and see how it moves you.  Don't forget footnote 27.

It's hard to tell you where the very ``best'' thing to read is, as my view has been evolving (and I think becoming much more consistent) over the last few years.  The view will never be completely consistent until we get a full-fledged derivation of quantum mechanics from it (and I hope you'll contribute to it eventually).  But I'm still relatively proud of my paper ``Quantum Mechanics as Quantum Information (and only a little more).''  I think I gave you a copy of that, but in case you threw it out, get the version on my web page.

(The version on {\tt quant-ph} has more typos.)  The relevant parts for a first, philosophical reading are:  Section 3, ``Why Information?'', and Section 4, ``Information about What?''.  I think they answer your question pretty directly.  You might want to also look at Section 10, ``The Oyster and the Quantum.''  I'm not completely sure how that section fits in with the stuff I've been thinking most recently, but I think there's still a significant overlap.

Finally, let me supplement all that with some recent words that I recorded for Philip Stamp.  They're placed below.  I think that's about as complete a picture I can give you at the moment without getting into the project of writing a new paper. [See 19-05-04 and 04-06-04 notes, ``\myref{Stamp3}{Bathroom Reading}'' and ``\myref{Stamp4}{Earwax},'' to P. C. E. Stamp.]

\section{05-07-04 \ \ {\it Flavors of Realism} \ \ (to M. Janssen)} \label{Janssen5}

Let me finally send you two more supplements to what I've already sent you before you face your friend.  These things address the ways in which I see this foundational program for quantum mechanics as a flavor of realism (and a result of Darwinism).

The most important among these is Sections 4 and 5 in my paper ``The Anti-{\Vaxjo} Interpretation of Quantum Mechanics'' posted at \quantph{0204146}.

Second to that, though, let me also attach the handout I put together for the Seven Pines meeting, ``Delirium Quantum.''  (There was no particular reason that it needed putting together for Seven Pines per se; I just needed some event as a catalyst, and Seven Pines was it.)  The most relevant sections for your particular question are Sections 4 and 5 (again).

\section{05-07-04 \ \ {\it Flying Out} \ \ (to G. L. Comer)} \label{Comer52}

I feel a constant need to apologize.  I'm just a crappy correspondent lately.  Despite my continually telling myself (and some correspondents like you) that I'm on the verge of returning to the old copious-writing Chris, I just haven't been able to cope with email lately.

Tomorrow Kiki and I fly to New Jersey for eight days of house hunting.  No little amount of stress over this.

Before going, though, I wanted to comment on a few of the things you've sent me in recent (and not so recent) emails.

\bgc
So maybe you will let me play the same game that Preskill and you
played; you know, give an information-theoretic reason for item a,
item b, item c, and so on.
\egc
Absolutely great questions!  I love them.  Particularly this one:
\bgc
(vi) Give an information-theoretic explanation for why there are
bosons and fermions and their associated Bose-Einstein and
Fermi-Dirac statistics.
\egc
I say that because this one is probably clean enough and simple enough that some progress can be made.  And I promise you, I will be thinking about this.

But let's get the history straight:  Those questions had nothing to do with Preskill.  You probably got that impression because I used that quote of his at the beginning of {\sl Notes on a Paulian Idea}.  I used it as a dramatic device to set up the table, but something like that list was on my mind, say, by the 27 January 1996 note I wrote you (in the same book).  John's remarks were in response to a research proposal or an abstract or some such thing that I had sent him to see what he thought about it.  In some frustration, he wrote me:
\bjp
   You are a graceful writer, an admirable scholar, a deep thinker.  But
   I knew that already.  Just the same, I am a tad disappointed, because
   I feel that what you wrote did not live up to the expectations
   created by your title.

   I want you to frame a question, as sharp and clear as possible -- one
   to which you do not yet know the answer, but desperately want to
   know, and expect someday to know (October 2000?).

   Pretend to be David Hilbert.  The Millennium is approaching.  Issue a
   challenge to the quantum theorists of the 21st century.  List the key
   questions they should seek to answer.  Hard questions, but not
   hopelessly hard, questions whose answers could transform our
   understanding of how the physical world works.

   I need to know what the question is.  Then, perhaps, I can be more
   engaged in the search for the answer \ldots

   All else failing, you and I should enter a high-stakes game of
   Wheelerian 20 questions, and so discover not just the answer, but (no
   so incidentally) the question!
\ejp
I think he thought me a charlatan with a not-so-well-defined research program.  He probably still thinks that of me.

\bgc
I was thinking of you this morning as I was looking at the preprint
archive.  There was a paper there by Hartle, talking about umpteen
different interpretations of quantum mechanics.  I glanced to see if
you were cited.  The closest was a paper by Caves.  I was very
irritated by that, because regardless of whether someone agrees
with the Fuchsian point-of-view or not, there is enough serious
Fuchs-stuff in the literature now that it can't be {\bf annoyed}.
\egc

I hope you meant \underline{ignored}!  Anyway, thanks for bringing me out of the dumps the day you wrote that.  It is true that the approach gets more and more attention and recognition.  I get a decent rate of emails like this one coming in now:
\bmj
     Anyways, may this serve as a surrogate for a substantial response
     to your work. I too am glad that we had that chat. Without it, I'm
     afraid, I would have just dismissed your stuff as crackpot physics.
     Now I see it as the one serious contender to my favorite scheme ---
     Everett $+$ decoherence.
\emj
(This one was from Michel Janssen, a young prof in history and philosophy of science at U. Minnesota.)  And David Mermin wrote me this recently to cheer me up:
\bdm
     You already have made an important difference in quantum
     foundations.  You've had more of an impact on more people than just
     about any person now in the field that I can think of.  People have
     worked out of more peculiar places than Bell [Labs], so as long as
     they're still available don't throw in the towel.  And keep working
     on more congenial options.  I didn't realize Fine, Bub, and Shimony
     were all writing for you in addition to all the others.  You must
     have the most spectacular collection of letters ever put into a
     filing cabinet.  One of these days it has to lead to something.
     Hang on.
\edm
Notes like that give me hope that I'm not just fooling myself.

One thing about Hartle:  He was actually nicer to me at this meeting in Minnesota a couple of months ago than he's ever been.  First he came up to me after my talk and shook my hand to express approval---kind of a strange thing to do.  Then, in the general wrap-up discussion at the end of the meeting, he publicly ``endorsed'' (in front of the whole meeting) the program as something that should be pursued (``I endorse \ldots'' is the phrase he actually used).  I have known him since the early 90s, but he has always kept his distance from me.

Oh, and talking about the Minnesota meeting twice now.  Let me send you the write-up about the meeting that appeared in {\sl Science}.  I'll paste it in from the letter I sent my in-laws. [See 30-06-04 note ``\myref{LentzB3}{Quail and Pheasant?}''\ to J. B. Lentz \& S. J. Lentz.] The probability simplex with a restricted region (from my paper \quantph{0205039}) has become something of a trademark.  (When Paul Busch, or maybe it was Guido Bacciagaluppi, used it in his talk in Sweden last year, he put my name and a TM symbol by it.  It was flattering.)  Anyway, the main point is that the ideas are starting to get into the community's mentality.  And that means that maybe one day, with me or without me, they'll go somewhere.

But still I go back to Bell Labs.  Despite all this patting myself on the back, mostly it feels like my days are over.

Maybe more from New Jersey, as I get a chance.

P.S.  I also met Bob Wald at that meeting in Minnesota.  Very nice guy; I liked him.  I didn't get to talk to him too deeply though, and I certainly didn't mention any of our quantum-GR ideas to him:  It was just clear that the gulf is too big between us at the moment.  The first hurdle one has to get over is that both the wave function and the very notion of measurement are anthropocentric concepts.  And it was only that little piece of the iceberg that I had a chance to chip away on with him.

\section{06-07-04 \ \ {\it World and Observer Intertwined} \ \ (to G. Musser)} \label{Musser6}

I'm planted in Munich now for the evening, and it's two and a half hours until beer-thirty with my in-laws.  So, let me see what I can work up for you between now and then.  As I've already expressed, your main question is a good one, and I had hopes of writing you a little essay in reply.  But as usual, I wasn't being realistic with myself about my duties and my distractions.  Thus, the stuff below will have to do for the moment.

First off:
\bgm
\ldots\ most physicists still regard hidden variables as a long shot.
Quantum mechanics is such a rainforest of a theory, filled with
indescribably weird animals and endlessly explorable byways, that
seeking to reduce it to classical physics seems like trying to grow
the Amazon from a rock garden. Instead of attempting to reconstruct
the theory from scratch, why not take it apart and find out what makes
it tick? That is the approach of Fuchs and others in the mainstream of
studying the foundations of quantum mechanics.

They have discovered that much of the theory is subjective. It does
not describe the objective properties of a physical system, but rather
the state of knowledge of the observer who probes it. Einstein argued
as much when he critiqued the concept of quantum entanglement--the
``spooky'' connection between two far-flung particles. [The following is
presumably the problematic sentence.] Fuchs argues that what looks
like a physical connection is actually an intertwining of the
observer's knowledge about the two particles.
After all, if there really were a connection, physicists should be
able to use it to send faster-than-light signals, and they can't.
\egm

I like all of that quite a bit.  The only thing I would change is your sentence, ``They have discovered that much of the theory is subjective.''  I'm not sure a reader would quite get what you mean.  If you came upon this sentence cold, not preconditioned by my idiosyncrasies, would you know what it means for a theory to be subjective (or even have a guess at it)?  Your sentence immediately following it wards off quite a bit of the trouble, but you might be a little more preemptive.  Maybe I'd say something like this:
\bq\noindent
   They have discovered that, though conditioned by the peculiarities of
   nature, much of the theory concerns not nature itself but the
   observer's interface with nature.  In other words, the theory does
   not primarily describe the objective properties of physical systems,
   but rather the possible states of knowledge of the observers who
   probe them.
\eq
That's probably a bit long for you, but I'm sure it's shortenable with a little work (and can made to sound more in your own style), but most importantly I think it's more on the right track.

Now to the next paragraph.  Even up to here, I'm OK with it:
\bgm
Physicists had long assumed that measuring a quantum system causes it
to ``collapse'' from a range of possibilities into a single actuality.
Fuchs argues it is just our uncertainty about the system that collapses.
\egm
The only way I might amend the first sentence if I were writing it myself would be to make it something like this:
\bq\noindent
   Physicists had long assumed that measuring a quantum system causes it
   to literally `collapse' from a range of possibilities to a single
   actuality.  Fuchs argues it is just our uncertainty about the system
   that collapses.
\eq

The trouble only really starts with the next sentences:
\bgm
Uncertainty about a quantum system is very different from uncertainty
about a classical one, because the process of measurement inevitably
disturbs a quantum system. And this sensitivity to disturbance is a
clue to what a quantum system really is.
\egm
So, let's focus on them.  Maybe I would change them to something like this:
\bq\noindent
   It is only that the uncertainty is about a very different beast in
   the quantum case than in the classical one.  Classically when one
   speaks of uncertainty, one assumes that it is always uncertainty
   about a some preexisting property that has nothing to do with the
   act of observation.  In the quantum case the uncertainty is
   about something at the interface between the system in the
   observer---the result that comes about through the very act of
   probing the system.  Like Einstein's entangled particles, the world
   and the observers within it are wired together so tightly that
   the theory cannot be about the one without also being about the
   other.
\eq
I know it probably doesn't sound like something you yourself would compose---I get truly pissed off any time someone tries to change my own writing---but maybe it gives you something to work with that starts off a little closer to what's inside my head.

Here's the whole lot pasted together:
\bq\noindent
   \ldots\ most physicists still regard hidden variables as a long shot.
   Quantum mechanics is such a rainforest of a theory, filled with
   indescribably weird animals and endlessly explorable byways, that
   seeking to reduce it to classical physics seems like trying to grow
   the Amazon from a rock garden. Instead of attempting to reconstruct
   the theory from scratch, why not take it apart and find out what
   makes it tick? That is the approach of Fuchs and others in the
   mainstream of studying the foundations of quantum mechanics.

   They have discovered that, though conditioned by the peculiarities of
   nature, much of the theory concerns not nature itself but the
   observer's interface with nature.  In other words, the theory does
   not primarily describe the objective properties of physical systems,
   but rather the possible states of knowledge of the observers who
   probe them. Einstein argued as much when he critiqued the concept of
   quantum entanglement---the ``spooky'' connection between two far-flung
   particles. Fuchs argues that what looks like a physical connection is
   actually an intertwining of the observer's knowledge about the two
   particles. After all, if there really were a connection, physicists
   should be able to use it to send faster-than-light signals, and they
   can't.

   Physicists had long assumed that measuring a quantum system causes it
   to literally `collapse' from a range of possibilities to a single
   actuality.  Fuchs argues it is just our uncertainty about the system
   that collapses. It is only that the uncertainty is about a very
   different beast in the quantum case than in the classical one.
   Classically when one speaks of uncertainty, one assumes that it is
   always uncertainty about a some preexisting property that has nothing
   to do with the act of observation.  In the quantum case the
   uncertainty is about something at the interface between the system
   and the observer---the result that comes about through the very act
   of probing the system.  Like Einstein's entangled particles, the
   world and the observers within it are wired together so tightly
   that the theory cannot be about the one without also being about the
   other.
\eq
I hope that's helpful and not just annoying.  I might (repeat {\it might}) get a chance to write a little more, but let's start with this.

My wife and I are just in Munich for the night to drop off the kids with the grandparents while we're house hunting.  We fly out for NJ at 7:00 tomorrow morning.

\section{06-07-04 \ \ {\it A Little More} \ \ (to G. Musser)} \label{Musser7}

Back again.  It's beer-thirty now, but the family's still shopping.  Still no proper essay, but let me come back to that set of questions of yours.

\bgm
{\rm [CAF wrote:]
\bq\noindent
The only thing that differentiates Einstein from our little effort
at this juncture is that he thought physics should ultimately be
about the ``states'' of nature, whereas the guys in my camp think that
the great lesson of quantum mechanics is that physics must be about
something else.
\eq
}
What is it about, then?  Also, are you agreeing with Bohr that physics
isn't about nature but what we can say about nature?
\egm

I hope the reworking of your old paragraphs that I just sent you, along with the opening of my Caltech lectures, already answers this question.  I don't think Aage Petersen's rendition of Bohr---``There is no quantum world.  There is only an abstract quantum physical description.  It is wrong to think that the task of physics is to find out how nature {\it is}. Physics concerns what we can say about nature.''---gets it quite right.  Instead, the model of QM is predominantly about something arising between the world external to the observer and the observer himself.  Don't you see the difference of tone?  It is not as if we are talking about something from afar (as Petersen makes it sound like), but something we are deep within the middle of and intimately involved with.  A quantum mechanical measurement result has its source as much in the system as it does with the observer---that is an ontological statement and not just an epistemological one.  After all these years, there's probably still no better picture of the process than in John Wheeler's story of the game of twenty questions (surprise version):
\bq
The Universe can't be Laplacean.  It may be higgledy-piggledy.  But have hope.  Surely someday we will see the necessity of the quantum in its construction.  Would you like a little story along this line?

Of course!  About what?

About the game of twenty questions.  You recall how it goes---one of the after-dinner party sent out of the living room, the others agreeing on a word, the one fated to be a questioner returning and starting his questions.  ``Is it a living object?''  ``No.''  ``Is it here on earth?''  ``Yes.''  So the questions go from respondent to respondent around the room until at length the word emerges: victory if in twenty tries or less; otherwise, defeat.

Then comes the moment when we are fourth to be sent from the room.  We are locked out unbelievably long.  On finally being readmitted, we find a smile on everyone's face, sign of a joke or a plot.  We innocently start our questions.  At first the answers come quickly.  Then each question begins to take longer in the answering---strange, when the answer itself is only a simple ``yes'' or ``no.''  At length, feeling hot on the trail, we ask, ``Is the word `cloud'?''  ``Yes,'' comes the reply, and everyone bursts out laughing.  When we were out of the room, they explain, they had agreed not to agree in advance on any word at all.  Each one around the circle could respond ``yes'' or ``no'' as he pleased to whatever question we put to him.  But however he replied he had to have a word in mind compatible with his own reply---and with all the replies that went before.  No wonder some of those decisions between ``yes'' and ``no''
proved so hard!

And the point of your story?

Compare the game in its two versions with physics in its two formulations, classical and quantum.  First, we thought the word already existed ``out there'' as physics once thought that the position and momentum of the electron existed ``out there,'' independent of any act of observation.  Second, in actuality the information about the word was brought into being step by step through the questions we raised, as the information about the electron is brought into being, step by step, by the experiments that the observer chooses to make. Third, if we had chosen to ask different questions we would have ended up with a different word---as the experimenter would have ended up with a different story for the doings of the electron if he had measured different quantities or the same quantities in a different order.  Fourth, whatever power we had in bringing the particular word ``cloud'' into being was partial only.  A major part of the selection---unknowing selection---lay in the ``yes'' or ``no'' replies of the colleagues around the room.  Similarly, the experimenter has some substantial influence on what will happen to the electron by the choice of experiments he will do on it; but he knows there is much impredictability about what any given one of his measurements will disclose.  Fifth, there was a ``rule of the game'' that required of every participator that his choice of yes or no should be compatible with {\it some\/} word. Similarly, there is a consistency about the observations made in physics.  One person must be able to tell another in plain language what he finds and the second person must be able to verify the observation.
\eq

\bgm
But ultimately don't you suggest that there is more to nature than is
captured in the quantum formalism?
\egm

Yes I do.  Here's how I put it in the ``Delirium Quantum'' thing I sent you:
\bq
The whole subject matter of my {\sl Notes on a Paulian Idea\/} is in toying with the idea that the cleanest expression of quantum mechanics will come about once one realizes that its overwhelming message is that the observer cannot be detached from the phenomena he {\it helps\/} bring about.  I italicize the word {\it helps\/} because I want you to take it seriously; the world is not solely a social construction, or at least I cannot imagine it so.  For my own part, I imagine the world as a seething orgy of creation.  It was in that orgy before there were any agents to practice quantum mechanics and will be in the same orgy long after the Bush administration wipes the planet clean.  Both of you have probably heard me joke of my view as the ``sexual interpretation of quantum mechanics.''  There is no one way the world is because the world is still in creation, still being hammered out. It is still in birth and always will be---that's the idea.  What quantum mechanics is about---I toy with---is each agent's little part in the creation (as gambled upon from his own perspective).  It is a theory about a very small part of the world. In fact, I see it as a theory that is trying to tell us that there is much, much more to the world than it can say.  I hear it pleading, ``Please don't try to view me as a theory of everything; you take away my creative power, my very promise, when you do that!  I am only a little theory of how to gamble in the light of a far more interesting world!  Don't shut your eyes to it.''

The question is, how to get one's head around this idea and make it precise?  And then, once it is precise, what new, wonderful, wild conclusions can we draw from it?  That is the research program I am trying to define.
\eq
And in another part:
\bq
But, you should remember that these quantum states we speak of are not perspectives.  They are personal possessions.  To modify Tilgher's quote at the beginning of de Finetti's ``Probabilismo'' for our own purposes,
\bq\noindent
   A quantum state is not a mirror in which a reality external to us
   is faithfully reflected; it is simply a biological function, a
   means of orientation in life, of preserving and enriching it, of
   enabling and facilitating action, of taking account of reality
   and dominating it.
\eq
\eq

\bgm
That there is a there, there?
\egm

Are you a fan of Gertrude Stein?  Carl Caves and I wrote this in a 1996 paper:
\bq\noindent
    The information gathered from repeated measurements on quantum
    systems is indeed drawn from an inexhaustible well, but it is a
    well of potentialities, not actualities.  Asked where all this
    information resides, we reply, with apologies to Gertrude Stein
   ``There is no where there.''
\eq
I wouldn't use that language anymore (i.e., talking about measurement results as information unqualified), but I still like the idea of using Stein when possible.

\bgm
Or at least, you seem to suggest that there are ``buried'' variables --
they're not hidden, because they do appear in the formalism, but
they're so entangled with subjective elements that they are hard to
tease out.
\egm

I still don't know quite how to respond to this one.  The language of ``variables'' is on the wrong track.  But this is why I wanted to write a little essay for you.  (These things are never for free by the way:  You're the muse and the laboratory.  If the explanation turned out good enough, I'd find a way to incorporate it into a paper.)

Similarly I'd love to give you an extended explanation of this:
\bgm
Another issue that I'm struggling with is this business of separating
states of knowledge from states of nature.  Isn't a state of knowledge
a state of nature where ``nature'' includes the observer?  The
wavefunction  is a statement about me, but I'm part of nature, right?
\egm
The root of this one goes even deeper, all the way back to Bayesian probability.  Quantum mechanics just adds an extra layer (or reinforcement) to it.  The best I can do is give you a hint of an answer that comes from a little piece of an exchange with Marcus Appleby after his paper \quantph{0402015} (definitely recommend reading it if you want to go a little more deeply into this).  You are right:  The observer is certainly part of nature.  Here's the exchange:
\bq\noindent
   In your ``Facts, Values and Quanta'' you say ``a probability statement
   cannot be identified with a fact about the world, as it exists
   independently of us.''  I think that is all I have ever meant by
   ``subjective.''  But, then again, maybe that's just hindsight playing a
   trick on me.
\eq
But I need to write so much more on the subject.

\section{06-07-04 \ \ {\it Preliminary Yes} \ \ (to G. Musser)} \label{Musser8}

I might use ``pertains to'' rather than ``describes'', but otherwise I think the paragraph is accurate.  The only other thing is whether ``sexual interpretation'' is too risqu\'e for {\sl Scientific American}:  If it's not, I guess I'm OK with it.

Gotta run for the flight.  I'll think about the rest of your note over the Atlantic, and may possibly reply.

\subsection{George's Preply}

\bq
This is all very helpful.  I've tinkered with the first two paragraphs of that passage, but as you said, the real issues concern the third paragraph.  Here is my latest attempt:
\bq
Physicists had long assumed that measuring a quantum system causes it to ``collapse'' from a range of possibilities into a single actuality. Fuchs argues that it is just our uncertainty about the system that collapses.  Uncertainty about a quantum system is very different from uncertainty about a classical one, and this difference is a clue to the objective world. In the classical case, a particle has some velocity; uncertainty means that observers do not know this velocity until they measure it. In the quantum case, the velocity does not even exist until observers go to look for it. The concept of velocity describes the interface between system and observer. Fuchs calls this idea the ``sexual interpretation of quantum mechanics.'' He has written: ``There is no one way the world is, because the world is still in creation, still being hammered out.'' The same thing can be said of our understanding of quantum reality.
\eq
Please let me know asap if this works or not.
\eq

\section{07-07-04 \ \ {\it A} \ \ (to G. Musser)} \label{Musser9}

Stuck on the runway at Heathrow with the threat of an equipment change!  Arrggh!

\bgm
How literally should I take Wheeler's game of 20 Questions?  There
{\it is\/} a rhyme and a reason behind the answers I get to my 20 questions
-- namely the (pre-existing!)\ agreement among the participants.
There is an objective reality.
\egm

You should read that passage much more carefully!  (Second Arrggh!!)  It is not about the ``game of 20 questions'' but the ``game of 20 questions (surprise version)''.  The whole point of the story is exactly the opposite of what you took from it.  [But, see 12-08-04 note ``\myref{Musser13}{Background Noise}'' to G. Musser.]

\section{07-07-04 \ \ {\it A.5} \ \ (to G. Musser)} \label{Musser10}

B and C still coming, but now I'm stuck in Heathrow for five hours (at least) and not in a particularly good mood.  I'll probably return to complete them in NJ somewhere in the middle of some night.

\section{07-07-04 \ \ {\it B} \ \ (to G. Musser)} \label{Musser11}

\bgm
{\bf [CAF wrote:]} ``It is a theory about a very small part of the world \ldots\ a theory
that is trying to tell us that there is much, much more to the world
than it can say.''  How is this not hidden variables?

Sure, they may not be hidden variables in the pre-existing sense --
i.e.\ in the sense that a properly designed experiment can come
asymptotically close to ascertaining their pre-experiment value.  But
does not ``more to the world'' imply something hidden?
\egm

Take a break from me for a moment and ask yourself how the Everett
interpretation is not a hidden-variable theory?  (It almost seems you
would have asked the Everettian the same thing you asked me.)  A
hidden-variable theory is a very specific thing:  If one were to know
the value (even if only hypothetically and not operationally) of all
the variables (including possibly the ones on the inside of the
observer), then one can predict the outcome of all measurements with
certainty.  It is a fancy way of saying measurement outcomes
pre-exist, even if nothing one would ever call a measurement is
actually performed.

The determination or setting of specific measurement outcomes (in any
quantum mechanical experiment) has always been outside of the quantum
mechanical formalism.  There is nothing in the formalism that
determines whether one will get this click or whether one will get
that click in some measurement device.  But that does not make it a
hidden-variable theory.  What is hidden?

Here is the way {\Pauli} put it:
\bq
     Like an ultimate fact without any cause, the individual outcome
     of a measurement is \ldots\ in general not comprehended by laws.
     This must necessarily be the case \ldots

     In the new pattern of thought we do not assume any longer the
     detached observer, occurring in the idealizations of this classical
     type of theory, but an observer who by his indeterminable effects
     creates a new situation, theoretically described as a new state of
     the observed system.  In this way every observation is a singling
     out of a particular factual result, here and now, from the
     theoretical possibilities, thereby making obvious the discontinuous
     aspect of the physical phenomena.

     Nevertheless, there remains still in the new kind of theory an
     objective reality, inasmuch as these theories deny any possibility
     for the observer to influence the results of a measurement, once
     the experimental arrangement is chosen.
\eq

(The conjunction of these thoughts is what I call ``the Paulian
idea''---hence the name of my book.)  ``Like an ultimate fact without
any cause, the individual outcome of a measurement is not
comprehended by laws.''

The way I see it, quantum measurement outcomes are ultimate facts
without specific call for further explanation.  And indeed the
quantum formalism supplies none.  Thus there is more to the world
than the quantum formalism can supply.  Nothing to do with hidden
variables.

But more specifically, regarding your point:
\bgm
``It is a theory about a very small part of the world\ldots\ a theory
that is trying to tell us that there is much, much more to the world
than it can say.''  How is this not hidden variables?
\egm
How does the theory tell us that there is much more to the world than
it can say?  It tells us that {\it facts\/} can be made to come into
existence, and not just at some time in the remote past called the
``big bang'' but here and now, all the time, whenever an observer
sets out to perform (in antiquated language) a quantum measurement. I
find that fantastic!  And it hints that facts are being created all
the time all around us.  But that now steps out of the domain of what
the quantum formalism is about, and so is the subject of future
research.  At the present---as a first step---I want rather to make
the interpretation of the quantum formalism along these lines
absolutely airtight.  And then from there we'll better know how to go
further.

Doesn't that just make you tingle?  That (metaphorically, or maybe
not so metaphorically) the big bang is, in part, right here all
around us?  And that the actions we take are \underline{\it part\/}
of that creation! At least for me, it makes my life count in a way
that I didn't dare dream before I stumbled upon {\Wheeler}, {\Pauli}, and
Bell--Kochen--Specker.

But let me get away from this speculation and rope myself back in on
your particular question:  How is this not some hidden variables
account?  Simple:  If there are any extra facts being created around
us, they nevertheless do not impinge on the individual quantum
measurement outcome.

When I say that QM is a theory about a very small part of the world,
you should literally think of a map of the United States in relation
to the rest of the globe.  The map of the US is certainly incomplete
in the sense that it is obviously not a map of the whole globe.  But
on the other hand it is as complete as it can be (by definition) as a
representation of the US.  There are no hidden variables that one can
add to the US map that will magically turn into a map of the whole
globe after all.  The US map is what it is and need be nothing more.

Does that help any?

I think a good bit of the problem comes from something that was beat
into most of us at an early age.  It is this idea:  Whatever else it
is, quantum theory should be construed as a theory of the world.  The
formalism and the terms within the formalism somehow reflect what is
out there in the world.  Thus, if there is more to the world than
quantum theory holds out for, the theory must be incomplete.  And we
should seek to find what will complete it.

But my tack has been to say that that is a false image or a false
expectation.  Quantum theory from my view is not so much a law of
nature (as the usual view takes), but rather a law of thought.  In a
slogan:  Quantum mechanics is a law of thought.  It is a way of
plagiarizing George Boole who called probability theory a law of
thought.  (Look at the first couple of entries in the {\Ruediger} {\Schack}
chapter of {\sl Notes on a Paulian Idea}.)  Try to think of it in
these terms, and let's see if this helps.

Let us take a simple term from probability theory, namely a
probability distribution over some hypothesis $P(h)$.  This function
represents a gambling agent's expectations about which value of $h$
will obtain in an observation or experiment.  Suppose now the agent
gathers a separate piece of data $d$ from some other observation or
experiment and uses it to conditionalize his expectations for $h$;
i.e., he readjusts his expectations for $h$ to some new function
$P(h|d)$ by using Bayes' rule.  Now here's a question for you.  Is
there anything within abstract probability theory that will allow the
agent to predict precisely which value of $d$ he will find when he
gathers his data?  Of course not.  It's almost silly to pose the
question. Abstract probability theory has nothing to do with the
actual facts of the world.  But then, doesn't that mean that
probability theory is an incomplete theory?  It can't, for instance,
explain its own transitions $P(h) \longrightarrow P(h|d)$ since
probability theory alone can't tell us why this $d$ rather than that
$d$. Moreover if probability is incomplete in this way, shouldn't we
be striving to complete it? Both silly questions, and I hope for
obvious reasons.

So:
\begin{enumerate}
\item
There is no particular mystery in the transition $P(h)
\longrightarrow P(h|d)$.
\item
We would never expect probability theory to provide a mechanism to
determine which value of $d$ is found or produced in the experiment.
The value $d$ represents a fact of the world, and probability theory
is {\it only\/} a theory about how to manipulate expectations once
facts are given.
\item
But also no one would be compelled to call probability theory
incomplete because of this.
\item
In particular, admitting this does not amount to having a
hidden-variable explanation of probability theory.
\end{enumerate}

So I say with quantum mechanics.  The story is almost one-to-one the
same:  You just replace probability distributions with quantum
states.  \ldots\ But then you reply, ``But there's a difference; quantum
theory is a theory of physics, it is not simply a calculus of
thought.''  And I say, ``That's where you err.''  Quantum theory
retains a trace of something about the real, physical world but
predominantly it is a law of thought that agents should use when
navigating in the (real, physical) world.  In particular, just like
with probability theory, we should not think of quantum theory as
incomplete in the usual sense.  If it is incomplete in any way, it is
only incomplete in the way that the US map is incomplete with respect
to the globe: There's a lot more land and ocean out there.

Teasing out (your words) the trace of the physical world in the
formalism---i.e., the part of the theory that compels the rest of it
as a useful law of thought---is the only way I see to get a solid
handle on what quantum mechanics is trying to tell us about nature
itself.

With this let me now go back to the US map for one final analogy.  I
said that there is a sense in which the US map is as complete as it
can be.  However there is also a sense in which it tells us something
about the wider world:  If we tabulate the distances between cities,
we can't help but notice that the map is probably best drawn on the
surface of a globe.  I.e., the US already reveals a good guess on the
curvature of the world as a whole---it hints that the world is not
flat.  And that's a great addition to our knowledge!  For it tells a
would-be Columbus that he can safely go out and explore new
territories.  Exploring those new territories won't make the US map
any more complete, but it still means that there is a great adventure
in front of him.

OK, with that, I'm going to call this discussion to an end.  If you
still think of me as a hidden variableist, I've either failed in my
explanatory powers or my view is simply inconsistent.  (I would bank
on the former rather than the latter.)  Either way I'm going to call
the discussion to an end.

Good luck finalizing your article (if you haven't already finalized
it).

\section{07-07-04 \ \ {\it C} \ \ (to G. Musser)} \label{Musser12}

\bgm
I hate to say it, but if you write for {\bf Scientific American}, as I hope
and expect you will, it will be very similar to the process we are now
undergoing.  You will come up with something, I or one of my
colleagues will critique and perhaps rewrite it, and around and around
we will go.

The trick is to ensure that the writing converges before the pissedoffedness diverges.
\egm

I tend to be gentler when someone gains my trust \ldots

\section{07-07-04 \ \ {\it Flying Out, 2} \ \ (to G. L. Comer)} \label{Comer53}

\bgc
A rhetorical question at this point, since I
can't properly articulate now my objection to
Hartle's objection.
\egc

The Hartle objection listed in the article is a result of his still being under the impression that wave functions must be unique.  I.e., there must one ``right'' wave function for a system, and all other potential wave functions must be ``wrong.''  And in particular, he is still under the impression that systems have wave functions even when there is no one gambling on them \ldots\ i.e., a wave function is an objective property.  Thus he asks, if a wave function is solely an expression of knowledge, whose knowledge is it?

You have the right intuition about quantum states of the universe.  See my note to Preskill dated 8 Sept 1999.

Kiki and I are stuck in Heathrow with over a five hour delay.  I'm not a pleasant person at the moment!  So much for getting our house hunting started today.

\section{28-07-04 \ \ {\it Three, Two, One} \ \ (to G. Brassard)} \label{Brassard41}

I'm sorry for the absence:  I apologize to everyone now, and constantly.  What I really need is three months off to clean the palette and get a fresh start.  \ldots\ But I don't have that privilege.

\bgb
I spent 4 out of 5 consecutive nights with 3 to 4 hour sleep per
night last week$+$ for the purpose of finishing my survey of pseudo-telepathy, since I wanted to have something in the Asher special issue
(having given up on you), and I was more than late with the submission.
I would appreciate {\bf at least\/} a word of acknowledgement-of-receipt!
By the way, I'll be away (mostly vacation) all of August, and
therefore unable to react quickly if you ask me for an urgent and/or a
version typeset according to {\bf Foundations of Physics} nonsense (such as
double-spaced).
\egb

Great paper, the part I've read.  Please make sure to add the authors' affiliations, fix any typos, and email me the original \TeX/\LaTeX/whatever.  Also, yes, please do make it double-spaced.

Little point of history:  Podolsky actually wrote the EPR paper.  Einstein was not particularly pleased with it, but reluctantly agreed to publish it as it was.  He thought it obscured the cleaner argument he had---the one he reports in several letters (to {\Schroedinger}, Born, Besso, and others).  A good/fun place to read about all this is in the first couple of chapters of Arthur Fine's book {\sl The Shaky Game}.

\bgb
My talk on QFLQI in Toronto last Friday went extremely well.
Well over one hundred people in attendance (something never seen in
summer time) and by far the liveliest crowd I've ever seen!
They kept me with questions for 90 MINUTES after the end of the talk.
Wonderful, except that I almost missed my flight back!
Of particular interest among the attendance was Marlan Scully from
Texas A \& M University.  Do you know him?
\egb

That is great news!  I wish I could have been there to hear it.  It's nice to see this program getting so much enthusiasm. Yes, I know Marlan and have some good stories to tell about him.  What has his reaction?  Did he say anything interesting?

\section{02-08-04 \ \ {\it My Own Apology} \ \ (to C. G. {\Timpson})} \label{Timpson5}

I enjoyed your envoi and am thankful you wrote it:  Your willingness to take the program seriously is already a great compliment---in fact, it's its only toehold in the Oxford community.

I liked the way you made it clear that even if one adopts a kind of instrumentalism for the quantum state, one can still ask, ``Why this instrumentalism, rather than that instrumentalism?''  And to the extent that one can provide reasons for the choice, one is going beyond instrumentalism.  While I'm here, let me paste in something I wrote up for a reporter at {\sl Scientific American\/} that addresses the same issue.  I used a metaphor I hadn't used before that maybe better captures the idea:  It's how an accurate map of an isolated country can already reveal that there's a larger world out there (through the curvature required to make sense of the distances between cities).  Yet it's no sin of the map that it doesn't cover the whole world.  Likewise, I think of quantum theory.  [See 07-07-04 note titled ``\myref{Musser11}{B}'' to G. Musser.]

\section{12-08-04 \ \ {\it Background Noise} \ \ (to G. Musser)} \label{Musser13}

I'm coming back intermittently.  My wife, kids and I are presently in Texas, visiting my mom, and soon to start driving our mini-van back to New Jersey.  (We had it stored here, because we had to drive down our ailing old dog before taking off for Ireland.)

You had a point about the pre-agreement in Wheeler's game of twenty questions.  That is one part of the image that I'd like to not take too seriously.  I'd rather have Darwinism all the way down, if the world will support it.

If you haven't read it yet, read Louis Menand's {\sl The Metaphysical Club}.  (Menand also writes for the {\sl New Yorker}, I'm told.)  It's an easy in into the idea that the world is Darwinism all the way down.  Very enjoyable book; I loved it.

Yeah, there's a lot about your note below that I like.  But I've got too much to do to parse it at the moment.

\subsection{George's Preply}

\bq
You write: ``If you still think of me as a hidden variableist, I've either failed in my explanatory powers or my view is simply inconsistent.''  I'm not seeking to label your ideas as an ``ism''. I realize that the term ``hidden variables'' is a loaded one within the foundations-of-QM community, but I have been using it simply as a foil to try to understand your work. I'm not an antagonist but a persistent student.

Let me see whether I understand your vision of a world in continuous creation. Ernst has bequeathed me his cat; it arrives in a box. I open it to find the cat is \ldots\ alive. Before I do, the cat is neither alive nor dead; it is not even in a purgatorial superposition. In fact, the categories of cat alive and cat dead don't even exist yet.  When I open the box, I create the fact of a live cat.  That fact had to come into existence {\it sometime}. In the classical deterministic universe, it came into existence when the universe did. All the arbitrariness was in the ICs. The quantum universe makes up facts as it goes along. The arbitrariness is spread out over time. It's the distinction between my wife, who makes plans and carries them out, and me, who decides step by step on the fly.

If we're talking about, say, particle spin, I play an even more active role in the creation, by virtue of how I conduct my experiment. I tell the particle: You are spinning up toward the ceiling of my lab or down toward the floor; now choose.

Is that the general idea?  There are lots of issues in there, but I want to get this down.
\eq

\section{12-08-04 \ \ {\it The Fruit} \ \ (to G. Musser)} \label{Musser14}

By the way, let me show you the fruit of my extended absence---it hasn't been an easy few weeks.  We got a nice house in Cranford, the second town closer to NYC on the Raritan Valley Line than Westfield. (I don't know if you're familiar with the towns around there.)  Now I'm 8 miles from Bell Labs, 10 miles from Newark Airport, and 19 from the WTC.  Anyway, an 1890 house in great shape, three levels, three fire places, 0.4 acre yard, a nice patio, etc.  I think it's going to be a good place for thought.  Drop by some time.

\section{03-09-04 \ \ {\it The Fruit of Someone Else's Imagination} \ \ (to D. Poulin)} \label{Poulin12}

\bdp
You seem to be comfortable with this situation, but I'm not. In
particular, I'm not very happy being the fruit of someone else's
imagination, specially knowing that this person is the fruit of my
imagination. What's your response to this? Did I take a wrong turn or
do you agree with my conclusion? Is there any way out of this?
\edp

Oh student!  Yes, you took a wrong turn:  You're not the fruit of anyone's imagination.  (You've always been beyond my imagination anyway.)

But there's no time for email this week.  Maybe next week.  (In any case, what could I say differently this time than I've said before?)  Instead, let me let another French speaker---Michel Bitbol---talk to you for the time being.  He captures much of my position in the letter below.  [See 10-12-03 note ``\myref{Bitbol1}{First Meeting}'' to M. Bitbol.]  Does it help any?

PS.  You should come visit us sometime:  We now have the best house we've ever had:  about 400 square meters (three levels plus basement) on a .43 acre park-like yard, quiet street, walking distance to train station (.5 mile), and only 45 from New York thereafter.  The guest room is the biggest in the house.

\subsection{David's Preply}

\bq
It appears to me that there is no way around the fact that quantum states are not real, they represent our knowledge about systems. At first, I thought that one could get away with a ``mild flavored'' notion of subjectivity: the type discussed by Rovelli in his ``Relational Quantum Mechanics''. But I have been re-reading this lately, and I think that deep inside he is a many-world guy, but doesn't want to admit it.  He talks about certain consistency conditions and other imposed relations between distinct observers. He never says so clearly, but in the back of his head there is a God's eye point of view from which all observers' ``subjective'' state assignments only reflect the bits and pieces of the ``real'' wave function of which they have only partial access (they are ``stuck in a branch''). (Note that he rejects this in his paper \ldots\ but there is no way around it.)

But once you recognize that states represent your knowledge, you get caught in a spiral that leads you to the unavoidable conclusion that there is nothing real. I guess that this was the point of your Phys.\ Today letter: if you believe that quantum theory is right (and believe in causality), then there is no room in it for an objective reality. Of course, there is room for inter-subjective agreement among several independent observers (emergent objectivity), but this is not fundamental. A more profound theory might eventually admit elements of reality, but not quantum theory the way it is today.

You seem to be comfortable with this situation, but I'm not. In particular, I'm not very happy being the fruit of someone else's imagination, specially knowing that this person is the fruit of my imagination. What's your response to this? Did I take a wrong turn or do you agree with my conclusion? Is there any way out of this?
\eq

\section{03-09-04 \ \ {\it Newer, Newish Thoughts} \ \ (to D. Poulin)} \label{Poulin13}

I'll also paste in one of my more recent smears on ``reality''.  It's a PDF file.

Have a good weekend.

\section{03-09-04 \ \ {\it Reach of the Black Cow} \ \ (to C. H. {\Bennett})} \label{Bennett34}

I thought fondly of you as I drove to work this morning:  The radio announcer on JAZZ 88.3 said that the music set he just played was sponsored by the Black Cow in Croton.

Anyway, I'm back at Bell Labs now.  My new addresses and phone numbers are listed below.  We ought to have a good work weekend in Wendel some time.  Or alternatively, you and Theo should come visit us:  We've got a great house now in a great area.  (If you've got a broadband connection, I'll send you some pictures.)

\section{07-09-04 \ \ {\it Targeted Answer} \ \ (to D. Poulin)} \label{Poulin14}

When I wrote you the other day, I had forgotten about the little piece of literature below.  It's much more of a direct answer to what you were asking:
\bdp
But once you recognize that states represent your knowledge, you get
caught in a spiral that leads you to the unavoidable conclusion that
there is nothing real. I guess that this was the point of your Phys.\
Today letter: if you believe that quantum theory is right (and believe
in causality), then there is no room in it for an objective reality.
Of course, there is room for an inter-subjective agreement among several
independent observers (emergent objectivity), but this is not
fundamental. A more profound theory might eventually admit elements of
reality, but not quantum theory the way it is today.

You seem to be comfortable with this situation, but I'm not.
\edp

The outcome of a quantum measurement is not the fruit of the measurer's imagination, for he cannot predict it.  It is first-class evidence of a reality (at least partly) independent of himself.  The ``identity of the outcome'' is something in the world that quantum theory cannot capture.  But I don't see that as a blemish on the theory that is waiting to be {\it fixed\/} by a more ``profound theory.''  The way I explain this is below; it is something I wrote for a {\sl Scientific American\/} reporter.  [See 07-07-04 note titled ``\myref{Musser11}{B}'' to G. Musser.]

\section{07-09-04 \ \ {\it Summer-School Words} \ \ (to D. Poulin)} \label{Poulin15}

\bdp
Now, I should probably read the papers ten times before asking these
dumb questions, (I don't have much time to do this these days) but
what gets to provoke reality? You say many times that it is our
interventions into the world (measurements) that forces reality into
existence. Does this require consciousness?
\edp

I'll only comment with the words I wrote for the intro to my summer-school lectures at Caltech this summer.  They're below.  [See 17-06-04 note ``\myref{Mabuchi12}{Preamble}'' to H. Mabuchi.]

\section{07-09-04 \ \ {\it Realism} \ \ (to H. Barnum)} \label{Barnum14}

\bhb
I would like to be realistic about, say, electrons,
if not about the wavefunction.  Can you think of a way to
do that?  The problem is, one is led to say things like
``Yeah, there really are electrons, and there are some
right here in this box''.   The latter statement seems
to be about their wavefunction, however.  Maybe that's
not a problem.
\ehb

For the heck of it, while we're on the subject of realism, let me place below three recent letters/compositions that plead again that I'm really a realist.  I think I used some new modes of expression that might be useful.  I particularly like the analogy that makes use of a U.S. map and a globe that I explain near the end.

\bhb
I don't think it's just the dimension of their Hilbert space, though.  Maybe it's a little bit more:  a Hilbert space with a particular group action on it\ldots? Wavefunctions aren't real, but maybe the place where they live, equipped with its observables and their group-theoretic structure, is\ldots?)

Oh well\ldots
\ehb

Yeah, it is something more like that---I think in my more democratic moments.  Bill Unruh spent a good amount of time at the Minnesota meeting this summer trying to suggest something like that to me too.

\section{08-09-04 \ \ {\it Fuchsian Pick List} \ \ (to S. Hartmann, C. M. {\Caves} \& R. {\Schack})} \label{Schack84.1} \label{Caves77.1} \label{Hartmann5}

I'm trying to make up for lost time before {\Ruediger} arrives in New Mexico.  Attached is my reworking of Stephan's last list of potential invitees.

You'll see I blithely rearranged according to my tastes.  I hope I won't cause any offense.  I did try to annotate pretty thoroughly.  And definitely, I'm not above bargaining on any of these positions:  Anyone I moved into the ``waiting list'' can certainly be moved back out of it and vice versa.  The main thing I am shooting for is a happy, productive, mutually-respectful crew.  With the exception of Price, Fitelson, and Ismael, I know everyone in the present invitee list personally, and I think they will work well together \ldots\ providing a little friction, but not too much friction.  You'll see, I was pretty negative on Bohmians and other exotic sorts.  The main thing I personally want to see at this meeting is a sense of progress, rather than a continuing sense of standstill (which is what I see every time I see a confirmed Bohmian in the room).

If we could open the floodgates, I would love to see everyone in the present list plus the waiting lists all together hashing it out.  It would be a grand get-together.  But there is no doubt that it'd be much too large for the present meeting.

\section{10-09-04 \ \ {\it Soul's Return} \ \ (to S. Hartmann, C. M. {\Caves} \& R. {\Schack})} \label{Schack84.2} \label{Caves77.2} \label{Hartmann6}

After 37 days of crossing the ocean and (mostly) sitting in US customs, my household goods finally made their way to Cranford today.  That includes my 331 books on James, pragmatism, Wittgenstein, and Rorty:  It's almost as if my soul was returned to me!

Perhaps I'll be a more creative writer next week.

\section{13-09-04 \ \ {\it Blurb} \ \ (to Princeton University Press)}

Here's the blurb.\footnote{For R. Omn\`es, {\sl Converging Realities: Toward a Common Philosophy of Physics
and Mathematics}, (Princeton University Press, Princeton, 2005).}  Two points:  1) If you decide to use it, you have to keep the bad with the good, i.e., use the whole thing.  I.e., you can't just cut the positive part from the review.  My integrity is at stake.  And 2) my affiliation ``Bell Labs, Lucent Technologies'' must be used along with my name.  Blurb below.

\bq\noindent
Roland Omn\`es has a knack for writing in at least three of the world's
major languages:  English, French, and Science.  I have at times
found a lifetime's worth of insight from a single paragraph of his.
The present work is no exception. For though I disagreed with whole
swaths of the book, I learned much from it and had to fight every
inch of the way to shore up my position against its onslaught. No
discussion of reality should ignore this powerful and novel
exposition.
\eq

\section{13-09-04 \ \ {\it Silence from the Podium Is Sometimes Golden} \ \ (to S. Hartmann, C. M. {\Caves} \& R. {\Schack})} \label{Schack84.3} \label{Caves77.3} \label{Hartmann7}

\bsh
I think that it is important that we come up with a structure of the
conference very soon. We need this before we write the speakers to
give them instructions.
\esh

I actually think it preferable if not all invitees are allotted a formal speaking spot.  I'm thinking particularly of the format of the Seven Pines meeting in Minnesota this spring.  Maybe only a little over half the participants actually spoke.  The ones that did would speak in pairs (with somewhat related subjects), then we would take a coffee break, and come back for discussion.  The two speakers would be placed in chairs in front of the audience and the discussion would go for about as long as the talks.  The discussions were often more riveting than the talks.

Also, myself, I'm all for longer talks generally.  It is just too hard to convey one's point and the flavor of one's thinking in 30 minutes \ldots\ especially when a subject is not mature.  Thus, I would suggest strict 45 minute talks (two at a time), then break, and readjourn for one hour of discussion/debate.  Or something like that kind of scheduling.

Like I say, I think it worked quite well at Seven Pines.

How to weed out speakers from pure discussants?  As a first round, I would suggest something like the call that was made for the Banff meeting next week.  Here are the exact words:
\bq
    It is time to begin the process of scheduling of talks for the
    workshop. If you would like to speak, please reply to this message
    with a short description of what you would like to present.

    Also feel free to suggest a topic that someone else from the list of
    participants might speak about.

    We hope to follow a relaxed model for the workshop that will allow
    us to have ample time to talk and work with one another \ldots

    For this reason, please do not volunteer to give a talk out of any
    sense of duty or obligation, or if you would give a talk that many
    people have already heard.  Ideally less than half of the
    participants should give scheduled talks.
\eq
After we see what kind of response we get from that, then we could cajole as many people as we need to fill the gaps.

In the end, at Seven Pines everyone participated in just about the right way.

Any thoughts or counterpoints?

I'll write that blurb before you wake up in the morning.  I guess {\Ruediger} should be in Albuquerque by now.

\section{13-09-04 \ \ {\it If I Had a Million Dollars} \ \ (to C. M. {\Caves})} \label{Caves77.4}

At the moment I can't really see how I'll survive at Bell Labs.  But on the other hand, I've never lived in a place that feels more like home than Cranford.  I love the house; I love the town; I love the location.  I've got almost every material thing I've ever wanted there.  The place is good for my family and good for Kiki's mental health---she has her career back.  How could I trade that until I'm forced to?  Part of me just says, ``Make it work at all costs.''  If I could go into management to insure my position here, I might just do it.  If I could afford the house payments with a high-school teacher's salary, I might just consider it.  If I had a million dollars, so that I could pay off the house (and buy myself 20 years of lean existence to boot), I might just say ``to hell with it all.''

But none of that is practical.  Nor is QSydney really.  I would arrive there and the same story would unfold as unfolded at Caltech:  To every physicist in the world, I may look like a philosopher, but to no philosopher do I.  I will always be viewed by them as the ``interesting physicist''---the guy to visit or have visit.  And that's not enough to get a position in a philosophy department.  Philosophers see me as having ``no breadth'' in the philosophy of science.

Truth is, I've lost heart and I've lost confidence.  My mind is so blank that most of the time I can't even remember why I believe what I do about quantum states.  (It's a good thing there's a written record of it in my computer; maybe one day I'll study it again.)

\section{14-09-04 \ \ {\it Assessment Letter} \ \ (to whom it may concern)} \label{Khrennikov9}

\noindent To whom it may concern: \medskip

I have been asked to assess the International Center for Mathematical
Modelling in Physics, Engineering, and Cognitive Sciences at the
University of {\Vaxjo}, Sweden.  I gladly take on this task, as I
have nothing but positive things to say about the Center.  It has
been a driving force in world-wide quantum mechanical research,
bringing researchers together from every corner of that specialty for
a kind of symbiosis I have seen nowhere else in the world (even in my
dealings with the Newton Mathematical Institute at Cambridge
University, the Perimeter Institute for Theoretical Physics in
Waterloo, Canada, or the Aspen Center for Theoretical Physics in the
United States).

I first became involved with the ICMM in December 2000 when Andrei
Khrennikov invited me to organize a conference with him for the
following year (apparently at the suggestion of Jeffrey Bub of the
University of Maryland).  The meeting was titled {\it Quantum
Theory:\ Reconsideration of Foundations}. What an experience and what
a meeting that was!  Andrei gave me a pleasant budget of ``seed
money,'' and with it---along with significant matching funds from the
various researcher's own grants---I was able to attract 15
world-class physicists from the field of quantum information and
computing, a few of whom were actually founding fathers of the field,
for a special session on ``Shannon Meets Bohr: Quantum Foundations in
the Light of Quantum Information.''  That meeting was a watershed
moment for the field:  For the first time it became internationally
recognized that quantum information and computing have very deep
things to say about quantum foundations, and more importantly, vice
versa.  Progress was seen all around, and a real sense of unity came
of it.  The conference proceedings emanating from the meeting,
published by {\Vaxjo} University Press, have come to be a standard
reference on the subject.

Since then I have had several encounters with Professor Khrennikov,
organized another conference with him in {\Vaxjo} (a conference just
as successful as the first one), joined the advisory board of the
ICMM, and published a book in {\Vaxjo} University Press's series on
Mathematical Modelling in Physics, Engineering, and Cognitive
Sciences---all wonderful experiences.  In particular, the VUP
publication of my book {\it Notes on a Paulian Idea\/} has led to an
outpouring of research in these directions and gained enough
recognition that Kluwer Academic Publishers has picked up the book
for a second-edition printing in their series Fundamental Theories of
Physics.  In all, I have had more email traffic and praise about this
book than all of my other publications combined (and that is not a
skimpy record, with over 1500 citations recorded in the ISI Web of
Science, Science Citation Index).  All of this goes to show the
important role that the ICMM has come to take in our community,
through the meetings it has funded, the conference proceedings and
books it has disseminated, and the collaborations it has spawned.

And this is just my personal experience with the ICMM.  Since its
founding, literally hundreds of researchers have been affected by or
participated in ICMM events.  Moreover, with the conference
proceedings being widely read, the prestige of the ICMM conferences
is taking on a life of its own.  For instance, I recently received
this request from a respected Italian professor in quantum optics and
quantum information: ``I am very interested in the {\Vaxjo}
conferences. I presume there will be one next year in June, right?
Can you put me in the mailing list? I'm currently working extensively
on the convex structure of quantum mechanics (states, POVM's,
channels, and their relative connections), on which with my group, we
are writing a set of 4--5 papers \ldots.''  Evidence like this cannot be
ignored. I can attest from my interactions with so many in the
community that you would hear a similar stories all around.

The ICMM is a gem in the world of fundamental quantum mechanical
research.  I hope similar levels of activity will be funded for years
to come.\medskip

\noindent Sincerely,\medskip

\noindent Christopher A. Fuchs

\section{14-09-04 \ \ {\it Barbecue Blurb, First Draft} \ \ (to S. Hartmann, C. M. {\Caves} \& R. {\Schack})} \label{Schack85} \label{Caves78} \label{Hartmann8}

\noindent\bv
\bf Being Bayesian in a Quantum World\\
1--6 August 2005, Konstanz, Germany
\ev

To be a Bayesian about probability theory is to accept that probabilities, whenever used, represent degrees of belief.  This is in distinction to the idea that probabilities represent long-term frequencies or intrinsic, chancy propensities ``out there'' in nature itself.  But, how well does this mesh with the existence of quantum mechanics?  To accept quantum mechanics as a physical theory is to accept the calculational apparatus of quantum states and the Born rule for determining probabilities in measurement situations:  ``These  probabilities are given by fundamental physical law!''\ one might eject.  Does this dilemma spell the death of Bayesianism?  Or does it rather give us an opportunity to rethink what quantum mechanics is actually about?

There is no doubt that we live in a quantum world.  From transistors to lasers to nuclear warheads, the evidence is all around us.  One might take from this that the quantum formalism ideally represents a mirror image of the world:  That is, the wave function is so successful as a calculational tool precisely because it represents an element of reality.  A more Bayesian (or, at least, Bayesian-like) perspective is that, if a wave function generates probabilities, then they too must be Bayesian degrees of belief, with all that that entails.  In particular, quantum probabilities have no firmer hold on reality than the word ``belief'' in ``degrees of belief'' already indicates.  From this perspective, the only sense in which the quantum formalism mirrors nature is through the normative constraints it places on gambling agents doing their best to navigate within such a quantum world.  To the extent that an agent should use {\it this\/} structure (i.e., quantum mechanics) for his uncertainty accounting rather than {\it that\/} structure (i.e., some foil theory) tells us something about the world itself---i.e., the world independent of the agent and his particular beliefs at any moment.  The task of the quantum Bayesian is to make this argument explicit and rigorous and to reap any benefit this can give to philosophy and physical practice.

Hogwash or deep idea?  The time is ripe for a debate.  At the present meeting, in the beautiful surroundings of Lake Konstanz, we envision a fifty-fifty mix of philosophers (who have thought long and hard about probability and quantum foundations) and quantum-information physicists (who have developed an impressive box of mathematical tools for prying apart the probabilistic structure of quantum mechanics) to set the tone.  The goal is make real progress on these issues through a unique complementation of talents.  All proposed participants have been hand picked for their registered interest in the debate and their known ability to ``talk to the other side.''

The format for the meeting will be \ldots

At present we can commit to the following funding \ldots

We look forward to seeing you in Konstanz!\medskip

\noindent --- --- --- --- --- --- ---

\bv
This day is called the feast of Being Bayesian in a Quantum World:\\
He that outlives this day, and comes safe home,\\
Will stand a tip-toe when the day is named,\\
And rouse him at the name of BBQW.\\
He that shall live this day, and see old age,\\
Will yearly on the vigil feast his neighbours,\\
And say `To-morrow is Saint de Finetti:'\\
Then will he strip his sleeve and show his scars.\\
And say `These wounds I had on de Finetti's day.'\\
Old men forget: yet all shall be forgot,\\
But he'll remember with advantages\\
What feats he did that day: then shall our names.\\
Familiar in his mouth as household words\\
{\Ruediger} the king, Hartmann and Fuchs,\\
 {\Caves} and Poulin, Barnum and {\Timpson},\\
Be in their flowing cups freshly remember'd.\\
This story shall the good man teach his son;\\
And BBQ BBQWian shall ne'er go by,\\
From this day to the ending of the world,\\
But we in it shall be remember'd;\\
We few, we happy few, we band of brothers;\\
For he to-day that sheds his blood with me\\
Shall be my brother; be he ne'er so vile,\\
This day shall gentle his condition:\\
And gentlemen in Germany now a-bed\\
Shall think themselves accursed they were not here,\\
And hold their manhoods cheap whiles any speaks\\
That fought with us upon Saint de Finetti's day.\\
\ev

\section{15-09-04 \ \ {\it Quanta of Rain} \ \ (to H. J. Folse)} \label{Folse22}

I'm quite concerned about you, your family, and your house.  I will keep my fingers crossed that things don't go too badly for you---at this point it's fairly clear that you're going to get at least some troubles from the storm.  In any case, you're in my thoughts.

\ldots\ for that and another reason actually.  Caves, Schack, Stephan Hartmann, and I are organizing a meeting to be held in Konstanz next August, titled ``Being Bayesian in a Quantum World'' --- we're hoping for roughly 17 (quantum-information) physicists and 17 philosophers.  The theme is about hashing out how well Bayesian probability fits with quantum mechanics.  And we figure we ought to have one resident expert on Bohr and/or ``the'' Copenhagen interpretation around for part of the discussion.  As it stands, you're top on the list if we do indeed go that route.  We'll let you know something soon.

\section{15-09-04 \ \ {\it Certain Epiphanies} \ \ (to M. P\'erez-Su\'arez)} \label{PerezSuarez15}

\bmps
On certainty: Now I am certain (and not {\bf\rm just} in the sense of
assigning probability one) that an assignment of probability one to
an event does not make of it a certain event and that, in much the
same way, probability zero for an event does not make of it an
impossible event. At least, this is definitely what follows from the
measure-theoretic approach to probability, and both Dutch-book
coherence and probability theory in the more general framework of
decision theory (as in Bernardo and Smith's book) may be seen as
``just'' providing the former (axiomatic) approach with an operational
meaning, thus keeping intact those conclusions.
\emps
I'm glad for this epiphany.  When I had it, it may have been the most important transition in my life.

\section{18-09-04 \ \ {\it Caves and Schack} \ \ (to M. P\'erez-Su\'arez)} \label{PerezSuarez16}

I'm writing you from a wireless connection for my laptop in the American Airlines Admirals Club in Chicago, as I'm making my way to the Banff meeting.  It is kind of exciting; I've never done this before.

I think either Caves or {\Schack} would be fine for you.  I don't think you can go wrong with either of them.  Carl has the better track record under his belt of nurturing many students, but {\Ruediger} has his own strengths and might have more personal time for you (though beware).   Caves' group would help fill in the gaps though, and probably prove far more stimulating.

Only one comment about something you wrote:
\bmps
The only problem as for me is that I'm not sure how much I could get
involved in the work he and his group are doing, mostly due to his not
being a ``full fledged'' subjectivist, which might cause me some trouble
I use the verb ``might'' for a good reason: I simply don't know if it
would). On the other hand, his work on issues related in one way or
another to QKD is of course pertinent to me. But \ldots
\emps
You shouldn't underestimate Carl's breadth and versatility.  And he is often willing to travel where his students lead him researchwise.  He will be a full fledged subjectivist one day, if he's not already (and he might be already) \ldots\ simply because logic will ultimately lead him there.

I will write both Caves and Schack now preparing the way for your proposal.  Thereafter it's up to you.

\section{18-09-04 \ \ {\it Quantum Foundations in Sci Am} \ \ (to G. Musser)} \label{Musser15}

\bgm
Now that I've shaken the monkey of our Einstein special issue off my
back, perhaps we can return to the idea of your writing an article
for us.  The first step would be to prepare a proposal that I could
circulate among my colleagues for their formal approval.  The
ulterior motive of such a proposal is to get you (and me and our
high-energy editor, to whom I would probably hand the baton) thinking
about what the article would say, how it might be structured, and how
the topic might be motivated for the nonphysicist.

I think the article would be a service not just to the lay person but
also to the foundations-of-QM community.  The preparation of a
nontechnical account is often an excellent excuse to think through
issues.  Then again, I don't need to convince you of that, since
you're the one who referred me to Mermin's essay.
\egm

Let me now apologize for my latest extended absence:  You wrote that to me over five weeks ago!  This relocation and getting resettled in Bell Labs has just been a bear.  Anyway, don't think that I've lost interest!

I've put a marker in my calendar to come back to the matter Nov 15.  Unfortunately, I've got a big presentation to give to the president of Bell Labs Nov 11 on our efforts in quantum info, and between that and the conferences I'm attending in the meantime, I know that I won't honestly be able to do anything for you before then.  But I'm hoping to hit the project with a vengeance soon after that.

BTW, I was prompted to finally reply to you by seeing your article at a newsstand in Chicago.  (I'm on the way to Banff.)  And I would have bought the issue too!  But there was a tear in the cover and it was the only one they had left.  Anyway, I'll get it soon, as I have a chance.  I look forward to reading what you've written.

\section{23-09-04 \ \ {\it Tritonic} \ \ (to G. L. Comer)} \label{Comer54}

How on earth you can call me a muse, I do not know.  Given the sparse amount I'm around anymore, I'd think it much more likely that you'd just forget about me.

I'm on a flight from Banff, Canada back to New Jersey.  It was a beautiful place and a great place for a conference.

I much enjoyed the words to your new song.  I've read them over each day since you sent them, and each day I get a little more from the reading.  Worship of chance, random romance.

Have a look in this month's {\sl Scientific American\/} if you get a chance.  There's an article on quantum foundations in which you'll get an ever so brief introduction to the ``sexual interpretation of quantum mechanics.''  I guess this means it's made its world premiere.  [Did I ever tell you the story of how it came about (or at least started to weigh on my mind) in Copenhagen's ``Free City''?  If not I'll send it to you.]

[See 13-10-04 note ``\myref{Baker11}{The Free City}'' to D. B. L. Baker.]

So many stories to tell you \ldots\ about Caltech and Australia and New Jersey.  Of dangling carrots and cosmic jokes.  Of blending into the woodwork and still being called upon and not forgotten.

But for now, I'll probably just return to hibernation for a while.

\section{04-10-04 \ \ {\it Bush and Reality} \ \ (to myself)} \label{FuchsC4}

From Bob Herbert, ``Bush and Reality,'' {\sl New York Times}, 4 October 2004:

\bq
For 90 minutes, at least, democracy seemed to be working. The two men in dark suits took their places at the lecterns. The analysts, the handlers, the spinmeisters and the hangers-on had been cleared out of the way. With no commercial interruptions, more than 60 million Americans got a rare, unedited, close-up look at the candidates in one of the most important presidential elections in the nation's history.

John Kerry got the better of President Bush in last Thursday's debate in Coral Gables, Fla. The president seemed listless, defensive and not particularly well prepared. His facial expressions and body language at times were odd. Some of his strongest supporters were dismayed by his performance, and polls are showing they had reason to be concerned.

There undoubtedly were many reasons for Mr.\ Bush's lackluster effort. But I think there was one factor, above all, that undermined the president in last week's debate, and will continue to plague him throughout the campaign. And that was his problematic relationship with reality.

Mr.\ Bush is a man who will frequently tell you -- and may even believe -- that up is down, or square is round, when logic and all the available evidence say otherwise. During the debate, this was most clearly displayed when, in response to a question about the war in Iraq, Mr.\ Bush told the moderator, Jim Lehrer, ``The enemy attacked us, Jim, and I have a solemn duty to protect the American people, to do everything I can to protect us.''

Moments later Senator Kerry clarified, for the audience and the president, just who had attacked the United States. ``Saddam Hussein didn't attack us,'' said Mr. Kerry. ``Osama bin Laden attacked us. Al Qaeda attacked us.''

Given a chance to respond, Mr.\ Bush flashed an unappreciative look at Senator Kerry and said, ``Of course I know Osama bin Laden attacked us -- I know that.''
\eq

\section{06-10-04 \ \ {\it Incompleteness} \ \ (to H. Price)} \label{Price1}

By the way,
\bhp
I'm fascinated to hear that you've been reading stuff written under
what I usually think of as my other hat, such as ``Truth as convenient
friction''. Somewhere I've got a rough piece I wrote about 20 years
ago, trying to argue that there were two attractive avenues to explore
in QM (in the light of the difficulties with either answer to the
question whether the wave function is a complete description):
first, backward causation HV approaches, and second, a view which
rejects the question, on the grounds that the very notion of complete
description is suspect. I was reminded of it when I read the
attachments you sent before.
\ehp
could you send me that draft as you get a chance (if it is in electronic form, that is).  I would enjoy reading it.

Regarding on my own take on the ``incompleteness'' of wave functions, let me paste in a note I wrote a little while ago to a reporter at {\sl Scientific American}.  He kept hammering on how I could say that the wave function is not a complete description of a quantum system (actually, I would say it's not even a description \ldots\ at least in a correspondence sense) and yet not be a (closet) supporter of hidden variables extensions.  In my frustration, the note below resulted.  In writing it, I became quite keen on the metaphor of the map (especially the bit about curvature at the end). [See 07-07-04 note ``\myref{Musser11}{B}'' to G. Musser.]

\section{08-10-04 \ \ {\it Krugman, Reality} \ \ (to myself)} \label{FuchsC5}

From Paul Krugman, ``Ignorance Isn't Strength,'' {\sl New York Times}, 8 October 2004:

\bq
I first used the word ``Orwellian'' to describe the Bush team in October 2000. Even then it was obvious that George W. Bush surrounds himself with people who insist that up is down, and ignorance is strength. But the full costs of his denial of reality are only now becoming clear.

President Bush and Vice President Dick Cheney have an unparalleled ability to insulate themselves from inconvenient facts. They lead a party that controls all three branches of government, and face news media that in some cases are partisan supporters, and in other cases are reluctant to state plainly that officials aren't telling the truth. They also still enjoy the residue of the faith placed in them after 9/11.

This has allowed them to engage in what Orwell called ``reality control.'' In the world according to the Bush administration, our leaders are infallible, and their policies always succeed. If the facts don't fit that assumption, they just deny the facts.

As a political strategy, reality control has worked very well. But as a strategy for governing, it has led to predictable disaster. When leaders live in an invented reality, they do a bad job of dealing with real reality.
\eq

\section{08-10-04 \ \ {\it Emerson} \ \ (to M. P\'erez-Su\'arez)} \label{PerezSuarez17}

Emerson.  I don't have much to tell you, because I don't know that much about Emerson yet.  But I do know that Cornell West has argued that pragmatism has its roots in Emerson.  You can read about that in his book, {\sl The American Evasion of Philosophy: A Genealogy of Pragmatism}.  I read a couple of chapters of this book---the parts on Emerson in fact---but then stopped reading because his writing style drove me to distraction.  Unfortunately, I don't remember much of what West said.

Here's another source that you might look at:  David Jacobson's book, {\sl Emerson's Pragmatic Vision:\ The Dance of the Eye}.  I've got it in my collection, but a) I haven't read it, and b) it's still packed up in a box somewhere (until I get some new bookshelves), so I can't look at it.  However, here's a blurb on the book: \myurl{http://www.psupress.org/books/titles/0-271-00896-2.html}.
Also, I can tell you that William James knew Emerson personally; Emerson was a friend of the James family, in fact, when William was a child.  Beyond that, though, you're left to your own research.

Don't ask me questions about LSE and Kluwer.  Some things I'm too ashamed to talk about \ldots

\section{08-10-04 \ \ {\it Aage Bohr} \ \ (to G. Musser)} \label{Musser16}

\bgm
What do you think of Aage Bohr et al.'s letter in the October {\bf Physics Today}?
\egm

I haven't been able to see that article, because a) unfortunately the Bell Labs electronic library does not carry the most recent three months of the magazine (that would cost the library more), and b) my own subscription seems to have ended in August (or at least the September and October issues were not in my mail pile).  Maybe I'm behind in my APS membership.

Anyway, if you've got the article in electronic form, please send it to me.  If you don't, can you tell me what it said?

\section{12-10-04 \ \ {\it Reality, Cohen, Christopher Reeve} \ \ (to myself)} \label{FuchsC6}

From Richard Cohen, ``The Dreams and Realities of a Superman,'' {\sl New York Times}, 12 October 2004:

\bq
As she aged, Marlene Dietrich became a recluse. She was seen by a few people, but never by the public nor in public. She was determined to stay as she had been -- a striking beauty. Her decline and death would remain her own business. The myth would endure.

Christopher Reeve took a different tack. After considering and rejecting suicide, he thrust himself back into public life. In the words of the writer Roger Rosenblatt, who got to know Reeve when he worked on the actor's biography, Reeve had to choose between a ``horizontal and a vertical life. He aggressively chose the vertical life.''

This does not mean that Reeve ever stood again after being thrown from a horse in 1995. It did mean that he avoided becoming a recluse. It did mean that he made public appearances, was interviewed several times by Barbara Walters, appeared before Congress and at the 1996 Oscar awards ceremony, returned to film and even became a director. He refused to become invisible.

Death often rebukes, and this is one of those occasions. Reeve's insistence that he would someday walk again was proved sadly false -- as some always said it would be. He was criticized by knowledgeable people for what they characterized as his ignorant or opportunistic optimism -- he appeared in a TV ad promoting spinal cord research -- that gave false hope to others in his condition. What these people needed above all, Reeve's critics said, was the determination to face reality -- not the bogus dream that the past could somehow become the future. [\ldots]

There always was a kind of lie to the life of Christopher Reeve. His faith that he would walk again, his optimism in the face of insurmountable bleakness -- all of this was belied by what we all sensed was a dismal truth. But he kept at it -- kept us at it -- and made us think about what had become of him and others like him. ``It was his way of standing up,'' Rosenblatt said.

R.I.P., Christopher Reeve. You always stood tall.
\eq

\section{12-10-04 \ \ {\it Three Children} \ \ (to D. B. L. Baker)} \label{Baker10}

Thinking of Child \#3, there is something else I had wanted to send you in addition to the little note I wrote commemorating Elizabeth's birth \ldots\ but until now had forgotten.  It's the introduction to my first lecture at a summer school at Caltech a couple months back.  I'll paste it in below.  It is along the lines of something I wrote for Madelynn's birth:  Some themes die hard with me.  Please keep it for all three of your girls.  I had them all in mind along with my two as I wrote it.  [See 17-06-04 note ``\myref{Mabuchi12}{Preamble}'' to H. Mabuchi.]

\section{12-10-04 \ \ {\it Being Bayesian in a Quantum World} \ \ (to W. T. Grandy, Jr.)} \label{Grandy1}

Thanks for the nice note; I'm flattered.  And most importantly, I'm glad you're enjoying my paper.  There's a lot of work to be done, but more and more people are taking note of the sensibility of (at least the broad outline of) the program, and that leaves me hope that we'll see some real progress in quantum foundations in the next few years.

In case you're not aware of some of the other papers melding elements of Bayesianism with quantum mechanics, let me put a list below for you.\medskip

\noindent First, some of my own (with colleagues):
\begin{itemize}
\item
Quantum information: How much information in a state vector?\\
\quantph{9601025}
\item
Quantum Probability from Decision Theory?\\
\quantph{9907024}
\item
Information Tradeoff Relations for Finite-Strength Quantum Measurements\\
\quantph{0009101}
\item
Unknown Quantum States: The Quantum de Finetti Representation\\
\quantph{0104088}
\item
Notes on a Paulian Idea: Foundational, Historical, Anecdotal and Forward-Looking Thoughts on the Quantum\\
\quantph{0105039}
\item
Quantum Probabilities as Bayesian probabilities\\
\quantph{0106133}
\item
Quantum Foundations in the Light of Quantum Information\\
\quantph{0106166}
\item
The Anti-{\Vaxjo} Interpretation of Quantum Mechanics\\
\quantph{0204146}
\item
Quantum Mechanics as Quantum Information (and only a little more)\\
\quantph{0205039}
\item
Conditions for compatibility of quantum state assignments \\
\quantph{0206110}
\item
Gleason-Type Derivations of the Quantum Probability Rule for Generalized Measurements\\
\quantph{0306179}
\item
A de Finetti Representation Theorem for Quantum Process Tomography\\
\quantph{0307198}
\item
On the Quantumness of a Hilbert Space\\
\quantph{0404122}
\item
Unknown Quantum States and Operations, a Bayesian View\\
\quantph{0404156}
\end{itemize}
Secondly, some from my colleagues:
\begin{itemize}
\item
C. M. {\Caves} and R. {\Schack}, ``Properties of the frequency operator do not imply the quantum probability postulate''\\
\quantph{0409144}
\item
R. {\Schack}, ``Quantum theory from four of Hardy's axioms''\\
\quantph{0210017}
\item
R. {\Schack}, T. A. Brun, and C. M. {\Caves}, ``Quantum Bayes Rule''\\
\quantph{0008113}
\item
R. W. {\Spekkens}, ``In defense of the epistemic view of quantum states: a toy theory.''  (This paper is not explicitly Bayesian in bent, but it is big a step in the right direction.  I think this is the most important paper in quantum foundations in quite some time.)\\
\quantph{0401052}
\item
R. W. {\Spekkens}, ``Contextuality for preparations, transformations, and unsharp measurements.''  (This paper is not explicitly Bayesian in bent, but it is big a step in the right direction.)\\
\quantph{0406166}
\item
D. M. {\Appleby}, ``Facts, Values and Quanta''\\
\quantph{0402015}
\item
D. M. {\Appleby}, ``Probabilities are single-case, or nothing''\\
\quantph{0408058}
\item
D. M. {\Appleby}, ``The Bell--Kochen--Specker Theorem''\\
\quantph{0308114}
\end{itemize}

There are several more that I could recommend, but since I suspect your main interest is in the Bayesian aspects of the program, I'll leave it at that for the moment.

Also, by the way, if you would like a copy of the {\Vaxjo} University Press edition of my 700 page samizdat {\sl Notes on a Paulian Idea\/} (which you can preview online in a 500 page format at the link above), let me know and I'll have a copy sent to you.  All I'll need is your mailing address.  You can find my eulogy to Ed Jaynes on page 49 of the VUP edition under the heading ``Probability Does Not Exist,'' and general references to him throughout the book (by looking in the name index).  Your name, in fact, is listed in the index five times.

Thanks again for writing.  It makes me very happy to learn of people out there sharing in the spirit!

\section{12-10-04 \ \ {\it Blowing in the Wind} \ \ (to G. L. Comer)} \label{Comer55}

I thought about both of us as I was reading the article below
yesterday.  [Bob Thompson, ``The Answer, My Friend?\ A Generation Mined Bob Dylan's Lyrics For Meaning.\ Now There's a Memoir to Dig,'' {\sl Washington Post}, 11 October 2004.] I particularly loved the line, ``Half the people you knew
believed that if only they could figure out what Bob Dylan was
saying, the secrets of the universe would be revealed.''  I've
certainly done that:  With John Lennon, then Paul Simon, and then Bob
Dylan, and before and after and in between with John {\Wheeler} and John
Coltrane, and now mostly with William {\James} and Richard {\Rorty}.  I
keep waiting for the secrets of the universe to be revealed, and it's
usually in poetry and music and stirring prose.  Or at least I used
to.  Now mostly I worry about getting a new roof on the house and
getting the place rigged with central air and heat \ldots\ and how
much that's going to cost my pocket book.

\section{12-10-04 \ \ {\it Endophysics Meeting} \ \ (to H. Atmanspacher)} \label{Atmanspacher5}

Will you be going to this meeting:
\bq\noindent
Endophysics, Time, Quantum and the Subjective\\ January 17--22, 2005, Bielefeld, Germany
\eq
I saw your name listed in the invited speakers list (and that the meeting is invitation only).  I think I would be interested in attending, and I think I could speak on aspects of quantum information that are probably little known to the audience.  Do you think there is any room left?  Do you know the organizers?

\section{12-10-04 \ \ {\it Equiangular Lines} \ \ (to C. H. {\Bennett})} \label{Bennett35}

\bcb
Today I just noticed a dissertation \quantph{0410071} on your favorite
subject.
\ecb

You mean on maximal sets of equiangular lines in Hilbert spaces of general finite dimension $D$?

Good to hear from you.  When you get back, I'm hoping to drop in on you guys and get you fired up on this pesky question.  I've told John we can call it the ``SIC Project.'' (SIC is really for Symmetric Informationally Complete, but the overtones are clear.)  I want to know everything I can about these sets of quantum states, abstract and applied.  (Renes, for instance, now has a nice QKD protocol in terms of them.)

A much more important thing for you to look at (if, for instance, you've only got the stomach to look at one such thing this year) is Rob {\Spekkens}'s paper \quantph{0401052}.  I like to think that's really a very important paper.

Tell Theo hello for me and send my congratulations on her retirement.

What was quantum about your cobwebs?

\section{13-10-04 \ \ {\it The Incentive Program}\ \ \ (to G. Bacciagaluppi)} \label{Bacciagaluppi4}

Thanks for the note, and it was great to hear your good news.  I had such a wonderful time in Paris when I last visited (Grinbaum and Bitbol), that I can well see you might be in paradise.  I stayed at some little hotel in the Latin Quarter and found, to my surprise, that my brain could still produce fresh thoughts:  There is just something about even a simple walk in the streets there that is so promotional of thought.  And Bitbol, particularly, was a great find:  I like his style of philosophy and certainly have much in common with him when it comes to quantum foundations.  I think there is something very deep in his phrase ``the blinding closeness of reality.''

But I write this note for another purpose.  Thinking of our tripartite interests last night (i.e., yours, mine, and Jenni's), I decided it was time to finally unpack the box of single malts and get the bottles organized [\dots]

\section{13-10-04 \ \ {\it The Free City} \ \ (to D. B. L. Baker)} \label{Baker11}

I know that over the years you've heard me talk about the Copenhagen interpretation of quantum mechanics, the Bohmian interpretation (in Ischia in particular), the many-worlds interpretation, and so forth.  There's quite a long list of such ``interpretations'' now; they come a dime a dozen.  Well, to the list I've been trying to add my own---I call it the ``sexual interpretation of quantum mechanics.''  I call it this because the name really captures the main idea I want to get across---that quantum mechanics hints of a world in constant creation.  Moreover, that such creation comes about in an almost sexual way whenever two conceptual parts of the world ``touch'' each other.  (The first written mention of this that I can find in my records is in \myref{Nielsen1}{a letter dated 3 September 2001 to Michael Nielsen}, now posted in my {\sl Quantum States:\ What the Hell Are They?\/}\ on the web; I'm not sure how long I had been talking about the idea before then.  On the other hand, the phrase made its national debut in a little description in {\sl Scientific American\/} last month, page 91.)

OK, that's the introduction; now for the story I wanted to record.  A couple of years ago, a friend (who shall remain nameless) and I were in Copenhagen for a few days.  As it goes, we found out that there is a little section in the city, called Christiania or ``the Free City,'' that's something like a big hippie commune.  There's over 900 people who live there, and marijuana and hash are sold freely in the streets.  As long as you stay within Christiania's boundaries, the police won't bother you about possession and such.  It's great:  In one of the beer gardens, there was a huge banner proclaiming ``JUST SAY NO to hard drugs.''  My friend, giddy like a school kid, wanted to go to the Free City for the purpose of buying a joint---something he hadn't done in years---and I thought, ``Why not?''  So, I accompanied him.

We went to the beer garden, and he lit it up.  But here and there throughout the day we had been talking about the ``sexual interpretation,'' and it was really getting under his skin.  He would say, ``You're only doing this for shock value, to bring attention to yourself.  The name adds nothing substantial to our understanding of quantum mechanics.''  And I would say, ``No, no, no; it really does capture an important idea.  It is one that needs to be made as plain as can be.''  We went round and round in argument.  He took his first hit,   exhaled, and pleaded in exasperation, ``What are you really hoping to get out of this?!''  In equal exasperation, I clinched my teeth and said, ``I guess I just want to believe that with every quantum measurement I perform, I {\it please\/} nature just a little.''  He took his second hit. ``So what? What good would that do!?''  I couldn't help myself!  ``If I do it right, maybe she'll gasp my name!''

You should have seen the guy's eyes!  They lit up!  He looked at me with complete conviction and said, ``I get it.  I get it.''  And for a moment you could tell he really did.  \ldots\ And for a moment, I really got it too!

\section{15-10-04 \ \ {\it Bohr, Mottelson, and Ulfbeck} \ \ (to G. Musser)} \label{Musser17}

If you ever sent me the lost email again that I had requested, I never got it.

Anyway, I've seen the letter in {\sl Physics Today\/} now:
\bgm
What do you think of Aage Bohr et al.'s letter in the October {\bf Physics Today}?
\egm

But, I'm not sure what you want to know.

Once upon a time, I wrote a rather mean-spirited note to Mermin about Ulbeck and Bohr's ``genuine fortuitousness.''  I'll dig that up and paste it below, but I doubt it answers whatever you had in mind.  [See 25-09-02 note ``\myref{Mermin75}{Ulfbeck and Bohr}'' to N. D. {\Mermin}.]  But who knows, maybe it does.  \verb+\bdm+ and \verb+\edm+ refer to the beginning and end of a David Mermin quote; \verb+\bq+ and \verb+\eq+ refer to the beginning and end of a more general quote.  I later cleaned that letter up and used it in Section 3 of my pseudo-paper ``Delirium Quantum'' which I sent you a while back.

If your concern has to do with the phrase ``quantum world,'' and contrasting their use of the phrase with mine (I say it exists, they say it doesn't)---they're quite likely very different uses of the phrase---then it would probably take me longer to answer you.

\section{15-10-04 \ \ {\it Genuine Fortuitousness} \ \ (to O. C. Ulfbeck)} \label{Ulfbeck2}

I noticed from reading your letter (with Bohr and Mottelson) in the October {\sl Physics Today\/} that you three have a new article:  {\sl Foundations of Physics\/} {\bf 34}, (2004) p.\ 405.  Would it be possible for you to send me an electronic version of the paper (\TeX, \LaTeX, PDF, MS Word, etc.)---our library at Bell Labs no longer carries that journal.

\section{21-10-04 \ \ {\it Dowd on Faith and Certainty} \ \ (to myself)} \label{FuchsC7}

From Maureen Dowd, ``Casualties of Faith,'' {\sl New York Times}, 21 October 2004:

\bq
When I was little, I was very good at leaps of faith.

A nun would tape up a picture of a snow-covered mountain peak on the blackboard and say that the first child to discern the face of Christ in the melting snow was the holiest. I was soon smugly showing the rest of the class the ``miraculous'' outline of that soulful, bearded face.

But I never thought I'd see the day when leaps of faith would be national policy, when the fortunes of America hung on the possibility of a miracle.

What does it tell you about a president that his grounds for war are so weak that the only way he can justify it is by believing God wants it? Or that his only Iraq policy now -- as our troops fight a vicious insurgency and the dream of a stable democracy falls apart -- is a belief in miracles? [\ldots]

W.'s willful blindness comes from mistakenly assuming that his desires are God's, as if he knows where God stands on everything from democracy in Iraq to capital-gains tax cuts.

As Lincoln noted in his Second Inaugural Address about the Civil War, one can't speak for God: ``The Almighty has His own purposes.''
\eq

\section{21-10-04 \ \ {\it Words Aiming to Inspire} \ \ (to R. E. Slusher)} \label{Slusher4}

Finally a note that has nothing to do with the performance review \ldots\ Spurred by our conversations yesterday, let me send you the little introduction I wrote for my Caltech summer school lectures a few months ago.  It's pasted below.  [See 17-06-04 note ``\myref{Mabuchi12}{Preamble}'' to H. Mabuchi.]  I think it touches on quite a few themes of our discussion yesterday---both on the points where we seem to agree and the points where we seem to disagree.

I hope it paints a fairly evocative picture of what I see quantum mechanics indicating at its core:  That the world is in continued (everlasting) creation.  And that this creation comes about in an almost ``sexual'' way---that is, it is something that happens whenever two pieces of matter get together.  (This is what was being talked about when {\sl Scientific American\/} said I call my view the ``sexual interpretation of quantum mechanics''.)  But I am careful to say quantum mechanics {\it indicates\/} this:  Because another part of the lesson is that quantum mechanics itself is not a low-level physical theory making direct statements about physical reality.  Rather it is a very high-level theory whose main concern is about an agent's (i.e., a man, a computer, whatever it is that is needed to make an agent) making enlightened decisions in response these little acts of birth that it itself is involved in.

Anyway, that's what the note below is about.  Read it and see if it stirs any sympathy.  I think you might see some resemblance between my idea of ``birth'' and your attention to the discreteness of quantum measurement outcomes that you were mentioning yesterday.

Just waiting for Russo to begin to speak.

\section{21-10-04 \ \ {\it More on the S Interpretation} \ \ (to R. E. Slusher)} \label{Slusher5}

If the Caltech intro did anything for you, you might also be interested in this little pseudo-paper ``Delirium Quantum'' that I distributed at the Minnesota meeting (but never posted on the {\tt arXiv}).  [See ``Delirium Quantum,'' \arxiv{0906.1968}.]

\section{21-10-04 \ \ {\it Books, Books, Books!}\ \ \ (to J. B. Lentz \& S. J. Lentz)} \label{LentzB4} \label{LentzS2}

Thank you both for the fabulous birthday present!!!  You certainly shouldn't have \ldots\ but I'll certainly take it!  Tomorrow I'm taking the whole day off to go book shopping, and I suspect I'll be in the Strand for about six hours of that.  This is really the best birthday present I've had in ages.

I've been joking lately about how there's no better insulation for thin walls than a good collection of American pragmatism!  (In case you want to know more about the philosophy, go here:  \myurl{http://www.pragmatism.org/}.  William James is my hero of heroes.)  Anyway, let's see how the library grows tomorrow.  At present, that section of the library has 331 books.  I'll give you an update on what it looks like by the end of the weekend.

\section{22-10-04 \ \ {\it More from NY Times about Reality} \ \ (to myself)} \label{FuchsC8}

From Bob Herbert, ``Bush's Blinkers,'' {\sl New York Times}, 22 October 2004:

\bq
Does President Bush even tip his hat to reality as he goes breezing by?

He often behaves as if he sees -- or is in touch with -- things that are inaccessible to those who are grounded in the reality most of us have come to know. For example, with more than 1,000 American troops and more than 10,000 Iraqi civilians dead, many people see the ongoing war in Iraq as a disaster, if not a catastrophe.  Mr.\ Bush sees freedom on the march.

Many thoughtful analysts see a fiscal disaster developing here at home, with the president's tax cuts being the primary contributor to the radical transformation of a \$236 billion budget surplus into a \$415 billion deficit. The president sees, incredibly, a need for still more tax cuts.

The United States was attacked on Sept.\ 11, 2001, by Osama bin Laden and Al Qaeda. The president responded by turning most of the nation's firepower on Saddam Hussein and Iraq. When Mr.\ Bush was asked by the journalist Bob Woodward if he had consulted with former President Bush about the decision to invade Iraq, the president replied: ``He is the wrong father to appeal to in terms of strength. There is a higher father that I appeal to.'' [\ldots]

There are consequences, often powerful consequences, to turning one's back on reality. The president may believe that freedom's on the march, and that freedom is God's gift to every man and woman in the world, and perhaps even that he is the vessel through which that gift is transmitted. But when he is crafting policy decisions that put people by the hundreds of thousands into harm's way, he needs to rely on more than the perceived good wishes of the Almighty. He needs to submit those policy decisions to a good hard reality check.
\eq

\section{25-10-04 \ \ {\it Son of Samizdat} \ \ (to H. Price)} \label{Price2}

I'm pleased to hear that you're enjoying my samizdat!  If that's whet your appetite, I hope you'll move on to Son of Samizdat as you get a chance.  More officially, it's titled {\sl Quantum States:\ What the Hell Are They?\/}\ and it's posted on my website.  I consider it the best work of my life (though it still needs distilling into a paper).  It's where I took my strong anti-representationalist turn.  It was a very hard turn to take, but I think if we stick with it the rewards could be very big.

\section{25-10-04 \ \ {\it Inventory} \ \ (to J. B. Lentz \& S. J. Lentz)} \label{LentzB5} \label{LentzS3}

OK, here's the monster you helped create!  The list of Friday purchases is pasted in below.  If you want to borrow any of the books you funded, let me know \ldots

More seriously, get this title:  Michael Magee, {\sl Emancipating Pragmatism: Emerson, Jazz, and Experimental Writing}.  I didn't buy it, but the blurb on Amazon says:
\bq\noindent
{\sl Emancipating Pragmatism\/} is a radical rereading of Emerson that
posits African-Ameri\-can culture, literature, and jazz as the very
continuation and embodiment of pragmatic thought and democratic
tradition. It traces Emerson's philosophical legacy through the
19th and 20th centuries to discover how Emersonian thought
continues to inform issues of race, aesthetics, and poetic
discourse. Emerson's pragmatism derives from his abolitionism,
Michael Magee argues, and any pragmatic thought that aspires toward democracy cannot ignore and must reckon with its racial roots.
Magee looks at the ties between pragmatism and African-American
culture as they manifest themselves in key texts and movements,
such as William Carlos Williams's poetry; Ralph Ellison's discourse in {\sl Invisible Man\/} and {\sl Juneteenth\/} and his essays on jazz; the poetic works of Robert Creeley, Amiri Baraka, and Frank O'Hara; as well as the ``new jazz'' being forged at clubs like The Five Spot in New York. Ultimately, Magee calls into question traditional maps of
pragmatist lineage and ties pragmatism to the avant-garde American tradition.
\eq
Maybe that explains why I've been drawn to jazz since meeting you guys!

Thanks again for the birthday present.

\vspace{-18pt}
\bq\noindent
\begin{enumerate}
\item
John Dewey, {\sl Experience and Nature}, Second Edition, (Open Court, La Salle, IL, 1971), HC.
\item
John Dewey, {\sl John Dewey---The Later Works, 1925-1953. Volume 12: 1938}, editor Jo Ann Boydston, textual editor Kathleen Poulos, introduction by Ernest Nagel, (Southern Illinois University Press, Carbondale, IL, 1986), HC.
\item
William Joseph Gavin, {\sl William James and the Reinstatement of the Vague}, (Temple University Press, Philadelphia, 1992), HC.
\item
Philip Kitcher, {\sl The Advancement of Science: Science without Legend, Objectivity without Illusions}, (Oxford University Press, NY, 1993), PB.
\item
Arthur O. Lovejoy, {\sl The Thirteen Pragmatisms and Other Essays}, (The Johns Hopkins Press, Baltimore, 1963), HC.
\item
John McCormick, {\sl George Santayana:\ A Biography}, (Alfred A. Knopf, NY, 1987), HC.
\item
Ralph Barton Perry, {\sl Philosophy of the Recent Past: An Outline of European and American Philosophy Since 1860}, (Charles Scribner's Sons, NY, 1926), HC.
\item
Ralph Barton Perry, {\sl Present Philosophical Tendencies: A Critical Survey of Naturalism, Idealism, Pragmatism and Realism, together with a Synopsis of the Philosophy of William James}, (George Braziller, Inc., NY, 1955), HC.
\item
Philip P. Wiener, {\sl Evolution and the Founders of Pragmatism}, with foreword by John Dewey, (Harper \& Row, NY, 1965), PB.
\item
Tulane Studies in Philosophy, Volume XII, 1963:  Studies in Recent Philosophy, PB.  (Bought it because of three interesting essays on George Herbert Mead and pragmatism.)
\end{enumerate}
\eq

\section{25-10-04 \ \ {\it Stages of Development} \ \ (to N. D. {\Mermin})} \label{Mermin113}

I went through a hernia operation in the winter of '94 and didn't much
enjoy it. (Though, {\Carl} {\Caves} used to pride himself that it was his
kids that likely caused the hernia---when I was playing with them one
evening---and, by way of that, he could trace the cause of my meeting
Kiki back to himself.  You see, because of the operation, I actually
stayed home for a while and stopped looking for a girl friend in the
coffee shops and bars.  And, lo and behold, in my boredom looking out
the window, I first noticed this beautiful neighbor walking her dog.
The rest is history \ldots\ and two kids.)

Which reminds me of something that I've been wanting to record.
Lately when Katie (who is getting close to three years old) misplaces
something, if I ask her where it is, she will reply, ``anywhere.''
She doesn't say, ``I don't know.''  Or ``I can't remember,'' or
indeed any other phrase that would put the blame on her incapacities.
She just replies with ``anywhere.''  What has struck me is how much
this reminds me of some of the early expositions of the uncertainty
principle that I was exposed to.  Maybe in both cases, it's all about
stages in our development.

\section{25-10-04 \ \ {\it The Quantum and the Politic} \ \ (to N. D. {\Mermin})} \label{Mermin114}

By the way, as you can guess, I've been watching the presidential
race with greater than usual interest this time around.  Of course
there's the issue of how our nation is in a quagmire now \ldots\ and
my profound confusion about how so many Americans can still support
Bush. But there's been an undercurrent that has interested me at the
intellectual level:  All the talk about belief, reality, and
certainty.  It's everywhere!

In the first presidential debate, I couldn't have gotten a better
slogan for our own discussions of the last couple of years, than from
a remark Kerry made to Bush:  ``But this issue of certainty: it's one
thing to be certain, but you can be certain and be wrong.''

Even in quantum mechanics.

\section{25-10-04 \ \ {\it Ever More Interpretations}\ \ \ (to M. A. Nielsen)} \label{Nielsen4}

Did you happen to read David Mermin's diatribe in the May issue of {\sl Physics Today\/} about how he invented the ``Shut Up and Calculate Interpretation'' of QM?  (I'm just getting caught up with my {\sl Physics Todays}---they were all piled up in my Bell Labs mail box when I returned from Ireland.)  Anyway, I faintly remember you and I having that very conversation a few years back at Caltech.  We were in your office as I recall, and one of us said Feynman and one of us said Mermin.  Or, at least I faintly remember that.  Problem is, I don't remember which of us took which side!  Do you remember this?  I've been trying to replay this in my mind, and half the time I think I thought it was Feynman and half the time I think I thought it was Mermin!  If you can help me out, I might be able to make up a good joke to play on Mermin.

\subsection{Michael's Reply}

\bq\noindent
Funny, I remember having this discussion with someone, but I don't remember it being with you!  Dave Bacon comes to mind.  I'm not sure this helps you out in quite the way you hoped, although it does add to the story.  So many cases of mistaken identity \ldots
\eq

\section{25-10-04 \ \ {\it More on Landauer} \ \ (to W. T. Grandy, Jr.)} \label{Grandy2}

\bwtg
I was interested in the exchanges with Landauer, but his replies didn't reveal much. For a number of years now I've been wrestling with the definition of information -- to paraphrase the title of one of your papers, ``What the Hell Is It?'' In contemplating Landauer's claim that ``Information Is Physical'' I've had the impression that he was not distinguishing between information and its representation, and was really talking about the latter, which is indeed physical.

I'm inclined at this point to think that the concept of information
-- like randomness, probability, and entropy -- is basically anthropomorphic, or at least brain dependent down to some level.
But, then, anthropomorphic is certainly physical, though not in the sense usually envisioned by physicists, so maybe his characterization has something to it after all. This is a bit like punching a pillow, or wrestling with jello!
\ewtg

I am pleased to hear you are enjoying my samizdat.  Better than the Landauer section, I think you might get more from the exchanges with {\Mermin} and Peres---at least a lot more Bayesian flesh was put on my Copenhagen bones there.

You are right about Rolf confusing information with information carriers---I think that has caused quite a bit of trouble in our quantum-information community.  I'll append a note below where I tell a little story about this.  (This one comes from my second email samizdat ``Quantum States:\ What the Hell Are They?'' posted on my webpage.)  [See 02-02-02 note ``\myref{Timpson1}{Colleague}'' to C. G. {\Timpson}.]

\section{26-10-04 \ \ {\it More ``More on Landauer''} \ \ (to W. T. Grandy, Jr.)} \label{Grandy3}

\bwtg
Thanks for your last note and the Bennett anecdote re:\ Landauer. Sometime, perhaps after a sufficient amount of wine, I'll have to share my copy of Ed Jaynes' rejection of Landauer's paper ``Minimum Energy Dissipation in Logic.''  Other correspondence indicates that the editor was not at all happy with Ed's recommendation, but I'd guess the original still exists in the editorial archives of the IBM J. Research. It was published anyway, of course.
\ewtg

I would very much like to see that, if you ever feel like divulging it!  As you can see, I'm fashioning myself somewhat as a historian of quantum information.  And your copy of the report sounds like it will ultimately be important historically.

\subsection{Tom's Reply}

\bq
Upon reflection, and in the interests of history, I see no reason not to pass this on to you now, particularly since both Ed and Rolf are gone. I've attached a pdf file with both the request and Ed's report. In connection with this you might also find interesting an excerpt from an email Ed sent to me on 7 July 1996: ``Landauer just throws his readers off by insisting that erasing a memory requires a dissipation of energy of about $kT$ per bit. It obviously requires NO dissipation of energy, because computer memories are never ``erased''; they are just overwritten. Resetting every bit to zero does not destroy any information either -- it merely moves it into correlations between the memory and its environment. It is when you fail to take note of the changed environment that information is lost; but still no energy is dissipated. I have been trying to explain this to Landauer for 35 years now (having refereed his first paper in which he said this and trying again whenever I meet him) without the slightest success.''

An interesting point here is the discrepancy in dates in the above review and the ``35 years'', so there must be an earlier Landauer paper that Jaynes also refereed -- unless Ed's memory here was faulty, but it usually was prodigious.
\eq

\subsection{Letter from IBM Journal\ Editor to E. T. Jaynes}
\medskip

\bq{\tt
\noindent November 10, 1969\\
\\
Professor E. T. Jaynes\\
Department of Physics\\
Washington University\\
St.\ Louis, Missouri \ \ 63130\\
\\
Dear Professor Jaynes:\\
\\
Thank you for agreeing to review the enclosed manuscript,\\
``Minimum Energy Dissipation in Logic'' by R. W. Keyes and\\
R. Landauer.  It has been submitted for publication in\\
the {\it IBM Journal of Research and Development\/} and I would\\
like to know your opinion of its originality, technical\\
accuracy and general merit.\\
\\
A list of suggested review criteria is enclosed, but these\\
need not be followed on a one-for-one basis.  I would\\
also appreciate any comments or suggestions you think\\
appropriate.  Your report will be anonymous where the\\
authors are concerned; may I hear from you by November 24?\\
\\
There will be an honorarium and I have entered a compli-\\
mentary subscription to the {\it Journal\/} in your name.\\
\\
Sincerely,
\\
\\
R. J. Joenk\\
Associate Editor\\
IBM JOURNAL OF\\
RESEARCH AND DEVELOPMENT\\
\\
RJJ/mgk\\
Enclosures
}\eq

\subsection{E. T. Jaynes' Referee Report}
\medskip
\bq
{\tt
\begin{center}
REPORT OF REFEREE\medskip\\
\end{center}
\noindent R. W. Keyes and R. Landauer, ``Minimum Energy Dissipation in Logic''\\
\\
\indent I feel that the weakest aspect of this article is its lack of\\
convincing power; after reading it over several times, I remain\\
unconvinced that computer operations have anything to do with entropy,\\
or that there is \underline{any} fundamental minimal energy dissipation in logic\\
steps.\\
\\
\indent This doubt stems from two causes, both of which should be correctible\\
if the authors' conclusions are correct.\\
\\
\indent Firstly, they place a great burden on the reader, requiring him to\\
have read about a dozen previous articles from which they merely quote\\
the conclusions without giving any supporting arguments for them.\\
Surely, if these matters have been discussed so much, there must exist\\
by now some simple argument, requiring only a paragraph or two, \\
indicating to the reader \underline{why} any fundamental limitation must exist.\\
Presentation of \underline{one} such argument would accomplish more for the reader \\
than citing any number of ``authorities.''\\
\\
\indent Secondly, their analysis of one particular model does not seem \\
entirely air-tight, and in any event it gives no reason to think that \\
the conclusions are general.  A different model might lead to entirely\\
different conclusions.  I illustrate this by two counter-examples.\\
\\
\indent On the next to last page of the text, the authors say that \\
removal of a thin barrier leads to ``an unavoidable non-equilibrium \\
process'' with an entropy increase of k ln 2.  But this does not follow\\
from the laws of physics; there must be some hidden assumption involved.\\
For example, before the barrier is removed the particle might be in a \\
definite quantum state $\psi(0)$. Removal of the barrier then leads to a \\
problem which can in principle be solved to give the state $\psi(t)$ at \\
any later time, in which the wave function will in general be non-zero\\
on  both sides.  But it was initially in a pure state with entropy\\
zero, and it remains, according to the {\Schroedinger} equation, in a\\
pure state\\  with entropy zero.  The supposed ``irreversibility'' must\\
be the result of some unstated limitation of how deeply we are permitted\\
to analyze the situation.\\
\\
\indent A switching operation does not have to take place via adiabatic\\
changes in a potential well.  We can have simply two well-separated\\
holes A and B, into which a particle may be put.  We switch by lifting\\
the particle bodily out of a hole A, and placing it in hole B.  Whatever\\
energy was required to lift it out of A is recovered when we lower it\\
back into B.  If one considers a thermal situation, with the energy in\\
holes A and B fluctuating by kT, this conclusion remains true on the\\
average, for we have just as much chance of gaining energy as losing it\\
when the transfer is made.\\
\\
\noindent
If I were a computer engineer, I think I would simply ignore articles\\
that try to tell me what I \underline{cannot} do; and just go ahead and\\
do them anyway.  Almost all progress in technology has been made in the\\
face of theoretical predictions that it can't be done; and unless some\\
limitation is derived rigorously from some undoubted law of physics,\\
such as energy conservation, the lessons of history lead one to place\\
his bets on the clever inventor, to circumvent imaginary problems.
}\eq

\subsection{Subsequent Letter from IBM Journal to E. T. Jaynes}
\medskip

\bq{\tt
\noindent February 10, 1970\\
\\
Professor E. T. Jaynes\\
Department of Physics\\
Washington University\\
St.\ Louis, Missouri \\
\\
Dear Professor Jaynes:\\
\\
Thank you again for your critique of the manuscript, ``Minimal\\
Energy Dissipation in Logic'' by R. W. Keyes and R. Landauer.\\
A copy of the current, revised manuscript is enclosed.  There\\
were two other reviews of the original manuscript in addition\\
to yours; one of these is included for your information.\\
\\
You will observe, of course, that the authors have not adopted\\
the reasoning of your counterexamples.  In their opinion, the\\
first example is not appropriate because it is based on a \\
conservative system whereas, in fact, fluctuating external\\
forces must be included in a relevant model of machine manipu-\\
lations.  The second example is thought to be incomplete due\\
to neglect of ``prior information,'' or testing of information\\
states, which is necessar for a realizable computer operation.\\
Both the requirement of energy dissipation and that of testing \\
states are described in Landauer's original paper, which is\\
also enclosed.\\
\\
If you would like to comment further on this manuscript, I\\
would appreciate receiving your reply by February 18.\\
\\
Sincerely,
\\
\\
R. J. Joenk\\
Associate Editor\\
IBM JOURNAL OF\\
RESEARCH AND DEVELOPMENT\\
\\
RJJ:mgk\\
Enclosure
}\eq

\section{27-10-04 \ \ {\it NY Times on Reality} \ \ (to myself)} \label{FuchsC9}

From Nicholas D. Kristoff, ``Pants on Fire?,''\ {\sl New York Times}, 27 October 2004:

\bq
Whenever I say that President Bush isn't a liar, Democrats hurl thunderbolts at me. And when I say Mr. Bush isn't truthful, Republicans erupt like Mount St.\ Helens.

So what do I mean?

Let me offer an example -- not from Iraq but from Mr.\ Bush's autobiography. In it, he tells a charming little story involving his daughters in 1988, on the eve of the presidential debate between his father and Michael Dukakis:
\bq\noindent
``One night, Laura and I were out of town campaigning, and Barbara and Jenna spent the night at the vice presidential mansion. Dad had spent the day preparing for a debate with Michael Dukakis. Unfortunately, Barbara lost her sleeping companion, Spikey, her favorite stuffed dog. She complained loudly that she could not sleep without Spikey, so `Gampy,' better known as Vice President Bush, spent much of the night before his debate searching the house and grounds of the vice presidential residence, flashlight in hand, on a mission to find Spikey. Finally, he did, and Barbara slept soundly. I don't know if my dad ever went to sleep that night.''
\eq

It's a heartwarming tale of family values. And while it's not malicious enough to count as a lie, it's laced with falsehoods.

We know that because Mr.\ Bush's mother wrote about the same incident much earlier, in 1990, in {\sl Millie's Book}, nominally written by her dog. For starters, the episode occurred when the girls were five and a half, in 1987, a year before the presidential debate.

What's more, {\sl Millie's Book\/} says that Spikey was a cat, not a dog. And instead of searching all night and finally finding Spikey, Vice President Bush gave up, grumbling: ``I have work to do. What am I doing searching for a stuffed animal outdoors in the dark?'' Anyway, little Barbara had already fallen asleep with another stuffed animal. Spikey turned up the next day behind the curtains. [\ldots]

The current president's hyped version of the incident reflects his casual relationship with truth. Like President Ronald Reagan, reality to him is not about facts, but about higher meta-truths: Mom and Dad are loving grandparents, Saddam Hussein is an evil man, and so on. To clarify those overarching realities, Mr.\ Bush harnesses ``facts,'' both true and false.
\eq

\section{28-10-04 \ \ {\it Enjoyed and No} \ \ (to M. P\'erez-Su\'arez)} \label{PerezSuarez18}

I read your paper and enjoyed it.  And I think it is a great effort in getting out ``the word.''  Also the discussion on probability is top-notch and makes a much better effort than I ever have personally to explain why we adopt the Bayesian notion of probability (even for quantum states).

But this paper really is your production, and you should de-list me as an author.  With regard to my {\it own\/} research it doesn't contain anything new (unless you count the second part of Eq.\ 7 \ldots\ which I want to lay out in much more detail when I try to put together our springtime discussions in my own way for SHPMP).  And stylistically, it is your presentation and personality.  Which---actually---is a very good thing for the dissemination of the program:  The more angles we all hit it from, the more chance I suspect it will stick in the public's craw.  Without my name on it, also, it helps give the impression of an independent construction/exposition.  And thus, I hope the technique will bring in a new set of readers.

I hope you won't be offended by my withdrawal:  I mean to make no negative connotation by it.  It's just that time is short, and when Fuchs gets involved with something, he tends to spend too much time trying to work it into a Fuchsian style (whenever he can get away with it, that is)---and that is a bad trait, especially with respect to the present project.

Now that my head is more clear, and I'm back in a more technical environment, I dearly want to get back into doing calculations.  But I also must clear up all these old projects that are still waiting (like SHPMP and Kluwer and such).  So, the clearer I can keep my plate the better \ldots\ the better for us all.

\section{28-10-04 \ \ {\it Mathematical Metascience} \ \ (to D. J. Foulis)} \label{Foulis2}

Are you out there somewhere?  Would it be possible to send me an electronic copy of your article ``Mathematical Metascience''?  If you can't, I can get Howard Barnum to mail me a xerox copy of it; he was already on the verge of sending it, but I thought this might be faster.

\section{28-10-04 \ \ {\it Reference, 2} \ \ (to F. E. Schroeck)} \label{Schroeck4}

Below are the references to all three of my quantumness papers.  I hope you can understand clearly enough the archaic code that I used.  Renes, by the way, in a couple of papers on {\tt quant-ph}, now has a full-fledged quantum cryptographic application for these SIC ensembles.

I'm curious to see what you've written about.  Please send me a copy of the paper when you can---by email if possible.

No need for an absentee ballot:  We're back in New Jersey now!

\begin{itemize}
\item
C.~A. Fuchs, ``On the Quantmness of a Hilbert Space,'' in {\sl Quantum Information, Statistics, Probability: Dedicated to Alexander~S.
Holevo on the Occasion of His 60th Birthday}, edited by O.~Hirota (Rinton Press, Princeton, NJ, 2004), pp.~65--77. \quantph{0404122}.

\item
K.~M.~R. Audenaert, C.~A. Fuchs, C.~King, and A.~Winter, ``Multiplicativity of Accessible Fidelity and Quantumness for Sets of Quantum States,'' {\sl Quantum Information and Computation\/} {\bf 4}(1), 1--11 (2004). \quantph{0308120}.

\item
C.~A. Fuchs and M.~Sasaki, ``Squeezing Quantum Information through a Classical Channel:\ Measuring the `Quantumness' of a Set of Quantum States,'' {\sl Quantum Information and Computation\/} {\bf 3}(5), 377--404 (2003). \quantph{0302092}.
\end{itemize}

\section{29-10-04 \ \ {\it URGENT:\ Exact Quote} \ \ (to G. Brassard)} \label{Brassard42}

You're in luck.  The following paragraph can be found in the JMO paper:
\bq
The task is not to make sense of the quantum axioms by heaping more
structure, more definitions, more science-fiction imagery on top of
them, but to throw them away wholesale and start afresh.  We should
be relentless in asking ourselves:  From what deep {\it physical\/}
principles might we {\it derive\/} this exquisite mathematical
structure?  Those principles should be crisp; they should be
compelling. They should stir the soul. When I was in junior high
school, I sat down with Martin Gardner's book {\sl Relativity for
the Million\/} and came away with an understanding of the subject
that sustains me today:  The concepts were strange, but they were
clear enough that I could get a grasp on them knowing little more
mathematics than simple arithmetic. One should expect no less for a
proper foundation to quantum theory. Until we can explain quantum
theory's {\it essence\/} to a junior-high-school or high-school
student and have them walk away with a deep, lasting memory, we will
have not understood a thing about the quantum foundations.
\eq
I've got to run to lunch in a second, so I may not be around for the next couple of hours.

\section{29-10-04 \ \ {\it Intrigue!}\ \ \ (to W. K. Wootters)} \label{Wootters21}

I was on the web trying to find the title I gave for a talk at Jeff Bub's ``New Directions'' conference last year, and I ran across a schedule already posted for next year's meeting.  Beside your name, I saw the title:  ``Measurement-loops as Elemental Quantum Processes.''  I've got to say, that really, really intrigues me!  What are you gonna talk about?  If you've got the time to commit a sketch of the idea to email, I'd love to read it!

\section{01-11-04 \ \ {\it Abstract Abstracts}\ \ \ (to C. Beisbart)} \label{Beisbart1}

OK, here they are, 50 days late:

\bq\noindent
   Title:  Some of the Phenomena of Quantum Information
   Theory\medskip\\
\noindent
   Abstract:  The no-cloning theorem.  Quantum entanglement.  Quantum
   teleportation.  Quantum key distribution.  Positive-operator-valued
   measures (POVMs).  Quantum nonlocality without entanglement.  The
   monogamy of quantum entanglement.  Bell inequalities.  The Kochen-
   Specker theorem.\medskip
\eq

\bq\noindent
   Title:  Bayesian-Like Ways to Think of Those Phenomena\medskip\\
\noindent
   Abstract:  A Bayesian understanding of no-cloning.  A Bayesian
   description of quantum teleportation.  The connection between
   Bayesian conditionalization and POVMs.  \ldots\ And so on down the line.
   Until, finally, the Kochen--Specker theorem:  This one may actually be
   a statement about ontology rather than a statement about degrees of
   belief!  And without some ontology, Bayesians would have very little
   reason to be.
\eq

\section{01-11-04 \ \ {\it Rational Hilbert Space}\ \ \ (to S. Aaronson)} \label{Aaronson6}

I'm sure you're already aware of this paper, but I just ran across it this morning:
\bq
L. Adleman, J. Demarrais, and M.-D. A. Huang, ``Quantum Computability,'' SIAM J. Comput.\ {\bf 26}(5), 1524-1540 (1997).
\eq
Anyway, it caught my attention because it shows that the answer to my bar-room question to you, Toner, and Childs may not have been completely obvious.  At least someone found some aspects of it worth writing a paper about.

So, despite the fact that there is no usual Kochen--Specker theorem for rational Hilbert spaces, BQP survives unscathed for them.  Just an interesting little factoid for me.  Don't know what to do with it, but it's an interesting factoid for me.

I need to come to Princeton sometime soon to get a new library card (probably within the next two or three weeks).  Maybe we could meet up for lunch or something.

\section{01-11-04 \ \ {\it No Reductive Physicalism!}\ \ \ (to H. Halvorson)} \label{Halvorson2}

I ran across your title for the College Park meeting next year, ``No Reductive Physicalism, No Measurement Problem,'' and have become quite intrigued!  It looks very, very interesting from my perspective.  What sorts of things are you thinking now?

I never heard back from you after you requested some reading suggestions from me.  Was any of that useful?  Or did it all just look like a load of muck?

Anyway, I'm back in New Jersey again \ldots\ and this time I'd like to do things right.  In particular, not pass up the opportunity of visiting or collaborating with interesting colleagues in the area.  How are things with you?  I presume you're at Princeton permanently now.  It'd be great if I could give you a visit soon.  I've been meaning to come to campus to renew my library card in any case, and I definitely would like to do that before Butterfield departs so I get a chance to see him a little.  Might you have any time in the next two or three weeks?

\section{02-11-04 \ \ {\it Election Day Blog} \ \ (to C. M. {\Caves}, R. {\Schack}, and N. D. {\Mermin})} \label{Mermin115} \label{Schack86} \label{Caves79}

The {\sl New York Times\/} this morning started one of their columns in this way, and I thought it was particularly relevant for our collaboration:
\bq\noindent
Every four years, by journalistic if not political tradition, the
presidential election must be accompanied by a ``revolution.''  So
what transformed politics this time around?  The rise of the Web
log, or blog.  The commentary of bloggers -- individuals or groups
posting daily, hourly or second-by-second observations of and
opinions on the campaign on their own Web sites -- helped shape the
2004 race.  The Op-Ed page asked bloggers from all points on the
political spectrum to say what they thought was the most important
event or moment of the campaign that, we hope, comes to an end
today.
\eq

So, let me do my duty as an avid reader of the {\sl New York Times\/} and tell you what I think was the most important event of the campaign.  It was near the end of the first presidential debate, 3 October 2004, when John Kerry said, ``\ldots\ this issue of certainty: it's one thing to be certain, but you can be certain and be wrong.''

True of politics, true of human relations, true of what can be derived from a quantum state.

Unraveled, it is quite a deep statement, as we know.  One might even say, a ``revolution.''  Quantum certainty is not a state of affairs, but a state of mind.  And we knew it even before the election year!

Maybe I'll cc this note to {\Mermin} too.  Even he, I bet, would agree that one of the key issues in the campaign has to do with the meaning of ``certainty.''

\section{05-11-04 \ \ {\it I Had To Ask!}\ \ \ (to W. K. Wootters)} \label{Wootters22}

Well, I had to ask, didn't I?  \ldots\

I got both your notes, and I've read them several times over.  But I have to confess that I'm confused about what you're up to.

Is a measurement loop this (at least in the quantum mechanical setting)?  Ingredients:  1) a bipartite system comprised of identical components (and hence isomorphic Hilbert spaces), 2) a {\it fixed\/} initial state for that bipartite system, 3) a (unknown?)\ unitary operator on one of the systems (expressing their relative evolution), and 4) the set of all imaginable complete orthogonal measurements on the bipartite system?

Is part of your point---again, in the quantum setting \ldots\ I know you ultimately want to abstract away from that---that these ingredients are sufficient for doing tomography on the unitary?  Just a question so I can get better oriented.

On a maybe related matter, do you know this phrase I like?  That, ``a quantum operation is just a quantum state in disguise.''  What I mean by this is:  If from the Bayesian view, a quantum state is the analogue of a (unconditioned) probability distribution, one might ought to think of a quantum operation as the analogue of a conditional probability distribution.  And just as the conceptual difference between an unconditioned and a conditioned probability distribution is not so very important, so one might imagine for the conceptual difference between quantum states and operations---namely that there isn't really one.

Anyway, this train of thought led me to try to get the structure of quantum time evolutions out of something like a Gleason theorem for measurements upon a ``single system'' posed at two points in time.  I never could quite make that work the way I wanted it to (though I never completely gave up hope either).

Just thinking out loud now, and wondering whether there might be any connections between the two problems \ldots\ since they are both about tomography of time evolutions in way.  But before I do more thinking, I should understand whether I've gotten your scenario right.  I suspect I didn't.

\section{05-11-04 \ \ {\it Linear Physics, Nonlinear Politics} \ \ (to G. L. Comer)} \label{Comer56}

Do you know John Stachel?  I noticed that he'll be talking at the 4th annual New Directions in the Foundations of Physics meeting, for which I'll be one of the ``discussants'' this year.  (I was a speaker last year, and the year before that a discussant again.)  I read Stachel way back and recall liking what I read then.  I have a faint memory that he had some nonstandard ideas on quantizing gravity that I liked at the time.  Of course, he's quite old now, and we know we should take anything said by someone over 39 with a grain of salt.  (But, have you seen his collection of essays, {\sl Einstein, from B to Z}?  I think I might try to buy it when I'm in NY City Monday.)

Anyway, if you've got nothin' better to do next April, why don't you come to the meeting (if you've got a little funding lying around)?  Maybe we could have fun talking to Stachel about some of these things.

\section{07-11-04 \ \ {\it {\Mermin} on Dirac Notation} \ \ (to A. Wilce, H. Barnum \& J. M. Renes)} \label{Wilce5} \label{Renes24} \label{Barnum15}

Taken from Chap.\ 1 of his lecture notes for his course on quantum computation:
\bq\noindent
     This is a good example of the primary point of Dirac notation: it
     has many built in ambiguities, but it is designed so that any way
     you chose to resolve those ambiguities is correct. In this way
     elementary little theorems become consequences of the notation.
     Mathematicians tend to loathe Dirac notation, because it prevents
     them from making distinctions they consider important. Physicists
     love Dirac notation, because they are always forgetting that such
     distinctions exist and the notation liberates them from having to
     remember.
\eq

\section{10-11-04 \ \ {\it Being Bayesian in a Quantum World --- Invitation} \ \ (to all BBQWers)} \label{BBQWInvite}

\noindent Dear Colleague,\medskip

We hope you will accept our invitation to the meeting detailed below.  We expect to obtain some amount of funding to help cover your travel costs and expenses.  Please let us know as soon as possible whether you think you are interested in the meeting and will be able to come.  (I, of course, have already fielded the idea of the meeting with many of you:  So, I expect we shall have a great turnout!)  Once we hear back from you, we will provide more details as the organization proceeds.

On behalf of all the organizers,\medskip

\noindent Best regards,\\
Chris Fuchs

------------------------------------------------------------

\noindent {\bf BEING BAYESIAN IN A QUANTUM WORLD}

\noindent 1--6 August 2005, Konstanz, Germany \bigskip

\begin{supertabular}{lll}
Organizers:  & Carlton M. {\Caves} & (University of New Mexico) \\
             & Christopher A. Fuchs & (Bell Labs, Lucent Technologies) \\
             & Stephan Hartmann & (London School of Economics and\\
             & &                  University of Konstanz)\\
             & {\Ruediger} {\Schack} & (Royal Holloway, University of London)\\
\end{supertabular}\bigskip

To be a Bayesian about probability theory is to accept that probabilities, whenever used, represent subjective degrees of belief and nothing else.  This is in distinction to the idea that probabilities represent long-term frequencies or intrinsic, chancy propensities ``out there'' in nature itself.  But how well does a subjective account of probabilities mesh with the existence of quantum mechanics?  To accept quantum mechanics is to accept the calculational apparatus of quantum states and the Born rule for determining probabilities in a quantum measurement.  If there were ever a place for probabilities to be objective, one would think it ought to be here!  This raises the question of whether Bayesianism and quantum mechanics are compatible at all.  For the Bayesian, it suggests that perhaps we should rethink what quantum mechanics is actually about!

There is no doubt that we live in a quantum world.  From transistors to lasers to nuclear warheads, the evidence is all around us.  One could take from this that the individual elements in the quantum formalism give a mirror image of nature:  That is, that the wave function is so successful as a calculational tool precisely because it represents an element of reality.  A more Bayesian (or, at least, Bayesian-like) perspective is that if a wave function generates probabilities, then they too must be Bayesian degrees of belief, with all that such a radical idea entails.  In particular, quantum probabilities have no firmer hold on reality than the word ``belief'' in ``degrees of belief'' already indicates.  From this perspective, the only sense in which the quantum formalism mirrors nature is through the normative constraints it places on gambling agents trying to navigate through the world.  To the extent that an agent should use quantum mechanics for his uncertainty accounting rather than some foil theory tells us something about the world itself---i.e., the world independent of the agent and his particular beliefs at any moment.  The task of the quantum Bayesian is to make this argument explicit and rigorous and to reap any benefit such a clarification can give to philosophy and physical practice.

Hogwash or deep idea?  The time seems ripe for a discussion, pro and con.  At this meeting, in the pleasant surroundings of Lake Konstanz, we envision a roughly fifty-fifty mix of philosophers (who have thought long and hard about probability and quantum foundations) and quantum-information physicists (who have developed an impressive box of mathematical tools for prying apart the probabilistic structure of quantum mechanics) to set the tone.  The goal is to make real progress on these issues through a complementarity of talents and some good-hearted debate.  Might we better understand the power of quantum computation through Bayesian techniques?  Does a Bayesian conception of the quantum state really make ``the measurement problem'' go away?  How secure is quantum cryptography really if a quantum state is ``just'' a state of belief?  Less secure?  Or maybe more (honestly) secure?  Does a Bayesian conception of the quantum state lead us closer to or further away from the Copenhagen interpretation?  What about many-worlds?  \ldots\ And so the list of topics goes on.  We hope the discussions will go long into the night and fill the lakeside walks.

The proposed set of participants is as follows:

\begin{tabbing}
Marcus {\Appleby}  \hspace{.7in}     \=   (physicist, London, USA)
\\
Guido Bacciagaluppi  \>   (philosopher, Paris, France)                \\
Howard Barnum        \>   (physicist, Los Alamos, USA)                \\
Hans Briegel         \>   (physicist, Innsbruck, Austria)             \\
Todd Brun            \>   (physicist, U. Southern California, USA)    \\
Jeffrey Bub          \>   (philosopher, Maryland, USA)                \\
Paul Busch           \>   (physicist, Hull, UK)                       \\
Jeremy Butterfield   \>   (philosopher, Oxford, UK)                   \\
Ignacio Cirac        \>   (physicist, Max Planck Inst., Germany)      \\
John Earman          \>   (philosopher, Pittsburgh, USA)              \\
Arthur Fine          \>   (philosopher, Washington, USA)              \\
Jerry Finkelstein    \>   (physicist, San Jose, USA)                  \\
Branden Fitelson     \>   (philosopher, Berkeley, USA)                \\
Henry Folse          \>   (historian, Loyola, USA)                    \\
Bas van Fraassen     \>   (philosopher, Princeton, USA)               \\
Nicolas Gisin        \>   (physicist, Geneva, Switzerland)            \\
Hans Halvorson       \>   (philosopher, Princeton, USA)               \\
Lucien Hardy         \>   (physicist, Perimeter Inst., Canada)        \\
Jim Hartle           \>   (physicist, UC Santa Barbara, USA)          \\
Jenann Ismael        \>   (philosopher, Arizona, USA)                 \\
Norbert L\"utkenhaus  \>   (physicist, Erlangen, Germany)              \\
David {\Mermin}         \>   (physicist, Cornell, USA)                   \\
Gerard Milburn       \>   (physicist, Queensland, Australia)          \\
Wayne Myrvold        \>   (philosopher, Western Ontario, Canada)      \\
Michael Nielsen      \>   (physicist, Queensland, Australia)          \\
Itamar Pitowsky      \>   (philosopher, Hebrew U., Israel)            \\
David Poulin         \>   (physicist, Queensland, Australia)          \\
Huw Price            \>   (philosopher, Sidney, Australia)            \\
Benjamin Schumacher  \>   (physicist, Kenyon College, USA)            \\
Abner Shimony        \>   (jack of all trades, Boston, USA)           \\
John Smolin          \>   (physicist, IBM Research, USA)              \\
Rob {\Spekkens}         \>   (physicist, Perimeter Inst., Canada)        \\
Christopher {\Timpson}  \>   (philosopher, Leeds, UK)                    \\
Jos Uffink           \>   (philosopher, Utrecht, Netherlands)         \\
Bill Unruh           \>   (physicist, British Columbia, Canada)       \\
David Wallace        \>   (philosopher, Oxford, UK)                   \\
Reinhard Werner      \>   (physicist, Braunschweig, Germany)          \\
Howard Wiseman       \>   (physicist, Griffith U., Australia)         \\
Anton Zeilinger      \>   (physicist, Vienna, Austria)
\end{tabbing}
We look forward to seeing you in Konstanz!\medskip

\noindent Best wishes,\medskip

\noindent {\Carl} {\Caves}, Chris Fuchs, Stephan Hartmann, and {\Ruediger} {\Schack}

\section{10-11-04 \ \ {\it Being NON-Bayesian in a Quantum World}\ \ \ (to R. Werner)} \label{Werner1}

By now you'll have seen the invitation I sent you for our meeting ``Being Bayesian in a Quantum World.''  I know that you're negative on Bayesian probability as a whole, but for precisely that reason I hope you'll still come to the meeting.  {\Ruediger} Schack and I agree that you have given us some of the best runs for the money we've ever had!  And we think you could help keep the meeting quite lively in that regard.  Your arguments are sharp and to the point.   Whatever the outcome (i.e., pro- or anti-Bayesian), I want the meeting to be a productive one, and you would be great in that regard.

So, I'm definitely hoping to hear back from you!

\section{10-11-04 \ \ {\it $\psi^*\psi$, Subjective Probability?}\ \ \ (to J. Earman)} \label{Earman2}

By now you'll have seen the invitation to our BBQW meeting that I sent you.  I definitely hope you'll come!  I remember at the end of the Minnesota meeting, you said:  ``Psi-star psi generating subjective probabilities?  Could not be!''  So, I know you already have an opinion!  But it'd be great to see you engaged on the subject, arguing your point.  You'd be a great addition to the meeting.  And particularly, I have so many things to discuss with you that I first learned from your book, it'd be a great honor if you were there.

\section{10-11-04 \ \ {\it Do Correlations Need To Be Explained?}\ \ \ (to A. Fine)} \label{Fine6}

By now you'll have seen the invitation to the BBQW meeting I sent you.  I so hope you'll come.  That old paper of yours was quite influential on me several years ago.  So I was thinking of that one of your many hats, when I argued for your inclusion in our list.  It would be great to get your feedback, pro or con, on this whole movement as a foundational approach to quantum mechanics.  Plus I think you'd just have a lot of fun!

\section{10-11-04 \ \ {\it Even Better Than Your Talk Title} \ \ (to H. Halvorson)} \label{Halvorson3}

\bhh
This project was motivated by your overall program (which, if I have
interpreted it correctly, rejects the idea that there is a
``measurement problem'').  But I've added a bit of gloss that connects
this idea with recent debates in philosophy of mind.  (Debates about
how to formulate ``physicalism'', etc..)

In short, all philosophers of physics I know -- with the possible
exception of Bub -- seem to think that the ``measurement problem'' is a
problem that nobody can avoid, no matter what their metaphysical
persuasion.  But the more I re-read the standard derivations of the
problem, the more I suspect that some bogus metaphysics gets invoked
at the last step -- viz., reduction of mental events to physical
events.

Now, I'm not committed to anti-reductionism, or to some sort of
dualism.  If pressed, I might say the whole physicalism versus dualism
debate is ill-founded.  (I think that was James' position; and it is
certainly van Fraassen's position.)  But I think now that only strong
reductionists (namely, those who don't think that the problem is
ill-founded, and who think that, in fact, mental events are identical
to physical events) have a measurement problem.  This is good news for
those of us who have a sane view, because we can stop worrying about
the supposed problem and get on to more interesting topics!
\ehh

That sounds great!  Now you have my mouth watering to hear more about it all.  (I guess I shouldn't say my ears are watering \ldots\ that sounds disgusting.)  Certainly you know I have read a lot of James by now; he's turned into one of my heroes.  Anyway, it sounds like you've got some good meaty thought.  Looking forward to your explanations.

\bhh
By the way, John Conway is giving a talk ``Free Will, Elementary
Particles, \& the Kochen--Specker Paradox'' on November 19th at 4pm.  The talk is based on a recent theorem by Conway and Kochen.  They tell me that they have decisively proven, once and for all, that QM is
indeterministic.  Hmm!
\ehh

Frankly, I think I'd almost give up all of PSA to see this one!  My
own feeling is that nothing in quantum mechanics gets closer to a
(trial) ontological statement (or ontological relative to the theory)
in quantum mechanics than the Kochen--Specker theorem.  Here's the way
I put it tongue-in-cheek in a couple of talk abstracts recently (for
the summer school that will follow our Konstanz meeting):

\bq\noindent
   Title:  Some of the Phenomena of Quantum Information
   Theory\medskip\\
\noindent
   Abstract:  The no-cloning theorem.  Quantum entanglement.  Quantum
   teleportation.  Quantum key distribution.  Positive-operator-valued
   measures (POVMs).  Quantum nonlocality without entanglement.  The
   monogamy of quantum entanglement.  Bell inequalities.  The Kochen-
   Specker theorem.\medskip
\eq

\bq\noindent
   Title:  Bayesian-Like Ways to Think of Those Phenomena\medskip\\
\noindent
   Abstract:  A Bayesian understanding of no-cloning.  A Bayesian
   description of quantum teleportation.  The connection between
   Bayesian conditionalization and POVMs.  \ldots\ And so on down the line.
   Until, finally, the Kochen--Specker theorem:  This one may actually be
   a statement about ontology rather than a statement about degrees of
   belief!  And without some ontology, Bayesians would have very little
   reason to be.
\eq
Anyway, I suspect Conway will be talking about another version of
Kochen--Specker (or something similar).  I'd like to see the mechanics
of that.

Hope to hear a positive response from you soon on the BBQW meeting!
I think this is going to be the best (most productive, most exciting,
most memorable) meeting I've ever been involved in organizing!

\section{10-11-04 \ \ {\it The BBQW Thing}\ \ \ (to N. L\"utkenhaus)} \label{Luetkenhaus1}

By now you'll have seen the invitation I sent you to the BBQW meeting.  I hope you'll seriously think about coming and will not feel out of place with so many philosophers.  Particularly, I think you could make an important contribution to this meeting, by helping us understand whether our attempt for a Bayesian-like formulation of QM has any implication on information security (or vice versa).  Or, even how to pose various protocols---{\Ruediger}, in particular, would like to get into this aspect of the discussion.  Joe Renes is likely to be there too.  (Also maybe you've seen the recent papers by the Maurer group on quantum de Finetti theorems and could comment, etc.)  Anyway, I'm just saying, your inclusion in the list of invitees was very intentional, and we hope you can come.

\section{11-11-04 \ \ {\it Free (Bayesian) Housecleaning!}\ \ \ (to A. Shimony)} \label{Shimony8}

\bas
The conference in Constanz is very tempting. I would learn a lot and
would be forced to sharpen the presentation of my objectivist
position. But I anticipate such a burden this coming summer that I
cannot make a commitment to attend. If there is a change I shall
inform you.
\eas

Oh, I do hope that burden will go away, because I dearly want you at our meeting.  I want to get so many of these quantum-probability issues settled in our community sooner rather later, and your input really is crucial for that.  You've been thinking deeply about probability for so many years, I think it would be great if our participants could get a sense how those thoughts apply to our particular context.  If it'd help you out, I could even come to Boston to clean your house this summer!

Seriously, if there's any way you can make the trip happen, I think our whole debate would be much better primed for progress with your presence.  Please let me know if there's any way you can rearrange.

\section{11-11-04 \ \ {\it BBQW and Unique Butterfields} \ \ (to J. N. Butterfield)} \label{Butterfield6}

I'm glad you're intrigued by the conference!  But I oh so hope you can come.  Your presence would fill a unique role there, the way you strive to find common ground between seemingly opposed points of views.  We need you actually.  Seriously.  So I do hope things will work out with your and your family's schedule.

\section{11-11-04 \ \ {\it Making Every Effort} \ \ (to D. Poulin)} \label{Poulin16}

It's great to hear that you want to come to BBQW!

On your comment, actually I think Smolin, Gisin, and Hartle are all pretty good to being ``open to discussion''---each in their own way, of course.  I've had very good (and productive) discussions with all of them in the last year or two.  But, in any case, rest assured that we're doing our best to round up some {\it productive skeptics\/} for the meeting.  The notes below show a sampling of the effort specifically in that direction.

The point is, I think you'll have plenty of opportunity to sharpen your viewpoint, rather than either just a) preaching to the choir, or b) being buried asunder.  It's gonna be a great meeting:  Definitely, you gotta come!

\section{12-11-04 \ \ {\it Philosopher Henry} \ \ (to H. J. Folse)} \label{Folse23}

That is great news to hear that you can come!  It's gonna be a good meeting, and I'm pretty excited about it.

Sorry to have called you a historian; for some reason, I thought that was your official job title.  I should have checked.

\bhf
What are you looking for from participants, a 20--30 minute paper?
I'll have to think up something to put on a power point.
\ehf

That's hard to say at this point.  The note below is our last exchange on the subject (with which at least Stephan agreed), but I can't say yet whether it represents the final word.

Be thinking about what Bohr would have thought about de Finetti!  Is there enough evidence in his work to give us any indication?  It would be wonderful if you could give us a report on that.  I'm thinking of say:  B. de Finetti, ``Probabilism,'' Erkenntnis {\bf 31}, 169--223 (1989).  Exempting the first couple of sections (where I can't quite figure out how far deF is really going), I think this is his definitive philosophical statement.

\section{16-11-04 \ \ {\it Copenhagen Hoping} \ \ (to A. Zeilinger)} \label{Zeilinger1}

I just sent you an invitation for our meeting ``Being Bayesian in a Quantum World'', and I'm writing again because I hope, hope, hope you can come.  The main thing is that I've had enough contact with the philosophers recently to see that we're starting to chip away at their resistance to the Copenhagen interpretation.  And I'm hoping with this meeting we'll finally have the strength to make a full frontal assault.  If you could be there that would be a great help, and I think you would find the experience very satisfying.

I hope you will give a reply as soon as possible.

\section{16-11-04 \ \ {\it Symmetry Magazine and Article Request} \ \ (to D. J. Harris)} \label{Harris1}

I apologize for keeping you waiting.  I thought a lot about your proposal, but ultimately, I don't think I'm going to be able to find a way to tie in what I'd like to say about Einstein and quantum foundations with his photoelectric effect paper.  If your call were to write a 650-word version of what I think was the main quantum foundational theme of his life (from 1932 to the end)---i.e., a story like I wrote up on pages 9--11 of \quantph{0205039}---then I think I could do it.  And I'd do it because that's how the man should be remembered, and the widest possible community should recognize that of him:  He had a very clear understanding---maybe more than anyone else---that the quantum state must represent incomplete information.  And where we have come to nowadays---particularly in quantum informational approaches to the foundations---is in trying to explain why the information cannot be completed.  But I just haven't found a clever way to tie that in with the photoelectric effect.  So, I better desist at this point; I understand your deadlines.

\section{17-11-04 \ \ {\it The Plasticity of Memory, and the Hardness of Reality} \ \ (to G. L. Comer)} \label{Comer57}

You write about the hardness of reality, and I keep hoping that there's some malleability to it.  Sorry for the absence, but I've been off the airwaves again---just making a quick pit stop---and now I'm flying off to Austin for the PSA 2004 meeting.

\section{17-11-04 \ \ {\it Bayes, Bennett, and Bull} \ \ (to M. P\'erez-Su\'arez)} \label{PerezSuarez19}

Sorry to hear you're not quite happy there.  Just keep in mind the crowd you're hanging out with and learn what you can with regard to their specialties.  There'll always be time to Bayesianize at other meetings.

On your Smolin query, it might well be John.  He told me that he's going back to Cambridge for a week in December.  Also it sounds like Charlie Bennett is back in town now.  You'll find that he's not very philosophical (in fact he tries to be anti-philosophical), nor is he very sophisticated (he tries to be anti-sophisticated), but he can be great fun.  And I venture to say he's probably the greatest genius (if there can be an overall mark for such things) in the whole city.

I hope things are on course for you to spend some time in Albuquerque.

\section{01-12-04 \ \ {\it BBQW Responses --- Progress Report, as I near the Pacific} \ \ (to S. Hartmann, C. M. {\Caves} \& R. {\Schack})} \label{Schack86.1} \label{Caves79.1} \label{Hartmann9}

\noindent Friends in Barbecue, \medskip

Below is how it stands, just as I am poised for my jump over the Pacific to Sydney.  We've had a couple more replies, and at least one E has turned into a Y.  I think Carl was going to write Nielsen again; and Stephan was going to write Uffink.  We can probably write off Fine and Werner at this stage, since I've written them secondary notes and still gotten no response.

Also, see the note from Milburn below.  It looks like he will contact Stephan with some more funding ideas.  Carl, is there something you're pursuing that I should help with?

To answer Stephan's question, I think the PSA session on quantum information and foundations went very well.  Or at least I think so.  The biggest bright eyes in the audience were Bas van Fraassen, Harald Atmanspacher, Steve Savitt, Jossi Berkovitz, someone named Antigone (prof in Minnesota?), John Norton, and maybe Michael Dickson. Berkovitz and Antigone were particularly interested in the quantum de Finetti theorem.  In addition to that, though he wasn't at the session, I met Alan H\'ajek at the conference.  He spoke at the Richard Jeffrey memorial session (along with Brian Skyrms, Lyle Zynda, and Persi Diaconis), and I got a pretty good impression of him.  Afterward, I fielded the question of whether he would be interested in BBQW, and he was actually very enthusiastic.  I think now I'm inclined to say he would be my top pick for the Fitelson spot.  But why don't we invite both H\'ajek and Howson?  (Also, does anyone know Urbach?  What's he like?)

Beyond that, I wonder whether we might consider Diaconis?  I got one heck of a good impression from his talk.  On top of that, it's clear that he's got technical skills like no other:  Skyrms may talk ``radical probabilism'', but Diaconis calculates ``radical probabilism''!  And I do think we could use some representation of that.  Diaconis gave the impression that he may be particularly interested in foundations at this point in his life.

Are things in good shape for Stephan to apply to the VW foundation?  Or should we send out the second round of invitations first.  I could do that from Sydney this week:  I'm hoping to get a lot of work done from there in the wee hours of the morning.  We'd just need to decide upon who's next.

That's all I can think of at the moment.

\section{04-12-04 \ \ {\it From Australia} \ \ (to J. Bub)} \label{Bub12}

I'm sorry I forgot to give you the recommended reading list before I got involved with my brother's visit to NJ and then had to scuttle off to Australia.  (I'm visiting Huw Price.)  Here's what I can give you at the moment as a ``must read'':
\bv
B.~de Finetti, ``Probabilism,'' Erkenntnis {\bf 31}, 169--223 (1989).
\ev
It seems way off the mark in the first two sections (seemingly portraying that de Finetti is almost a solipsist  \ldots\  maybe something got really screwed up in the translation), but it soon recovers after that to become what I would say is THE most powerful article on Bayesian probability I have ever read.  It was the article that finally convinced me of the nonfactivity of all probability distributions.

Now, it took me four separate readings of the article over six years, and a lot of rumination in between, to get to that point.  But with the hindsight of that experience, I hope I'm in a much better position to help you guys slog through it more effectively than I did.

You might also want to read Richard Jeffrey's accompanying article
\bv
R.~Jeffrey, ``Reading {\it Probabilismo},'' Erkenntnis {\bf 31},
225--237 (1989).
\ev
to help set the historical context.

I also want to suggest to you some pages from Bernardo and Smith's book,
\bv
J.~M. Bernardo and A.~F.~M. Smith, {\sl Bayesian Theory}, (Wiley, New York, 1994).
\ev
but it'll have to wait till I get back home Wednesday before I can suggest anything specific to you.

How does that sound?  Is that enough to get you started until I can get back to you Wednesday or Thursday.

\section{08-12-04 \ \ {\it BBQW Topics --- Just Some Ideas} \ \ (to the BBQW Boys)}

Here are some other topics for your VW Foundation proposal for our meeting, and more generally, for the conference itself.  I guess I'll also send this to Gerard Milburn in case he might need it for his own funding efforts.

-----------------------------------

\begin{enumerate}
\item
Could the Everett Interpretation of quantum mechanics really give rise to a Bayesian derivation and interpretation of the quantum probability rule?  (Wallace, Uffink, Barnum, \ldots\ and Greaves if we sneak her in as a discussant)
\item
Can or cannot one contemplate doing quantum cosmology with a Bayesian interpretation of quantum states?  Assuming one can, what does Bayesianism add for helping to solve existent technical problems?  (Hartle, Unruh, Brun, {\Caves}, \ldots\ and maybe Srednicki.)
\item
How does a Bayesian interpretation of the quantum state impact information security in the sense of quantum cryptography?  Is there a serious difference between subjective and objective interpretations of the quantum state in this regard?  Does one really need an objective notion of randomness to make the idea of a quantum random number generator fly?  Even if one will not buy into the idea of a Bayesian interpretation of the quantum state, will Bayesian techniques for standard statistical practice nevertheless make a difference for security analyses?  (L\"utkenhaus, Gisin, {\Schack}, \ldots\ and maybe Renner and Renes.)
\item
What does objective chance in the sense of David Lewis's ``principal principle'' or a propensity interpretation (as in Giere's early paper or as a special case of Popper's) add to the interpretational issues of quantum mechanics?  Or does an objective interpretation of probability only impede quantum foundational progress?  (Myrvold, Shimony, Butterfield, Gisin, Howson -- pro objective chance; {\Appleby}, Fuchs, {\Schack}, {\Caves} -- con objective chance)
\item
Quantum computing.  How might quantum computational algorithms be understood as Bayesian inference engines?  Do any of the known quantum computational models (unitary-gate implementations, measurement-based models, or adiabatic models) favor or disfavor a Bayesian interpretation of quantum states and operations?  How to pose some known quantum computational algorithms as Bayesian updating questions?  (Nielsen, Briegel, Cirac, Wiseman, {\Caves}, Milburn, Barnum, {\Mermin}, \ldots)
\item
What would Bohr think?  How close is a Bayesian interpretation of quantum states {\it really\/} to anything Copenhagen like?  Where is a Bayesian interpretation less radical and anthropocentric than the Copenhagen interpretations (in the Bohr, Pauli, and Heisenberg varieties)?  Where is it more?  (Folse, Fuchs, {\Mermin}, and Zeilinger)
\item
Bayesian interpretations of the quantum state tend to suggest novel quantum measurements to use in their representations (for instance, the symmetric informationally complete positive-operator-valued measures and complete sets of mutually unbiased measures).  They also suggest interesting quantum-state and operation tomography questions (like the pure-state informationally (PSI) complete measurements and the de Finetti representation theorems).  It would be nice to get some technical reports on these issues.  ({\Appleby}, {\Caves}, {\Schack}, \ldots\ and Diaconis, Renner, and Wootters.)
\item
To what extent does quantum mechanics come about solely through information-theoretic constraints?  To what extent is it not much more than a theory of missing information?  Bub-Clifton-Halvorson theorem; {\Spekkens} toy models; Smolin objections.  Is a further ontology underlying the information-theoretic constraints needed or not?  Can Smolin really give an Everett interpretation to {\Spekkens}'s toy model?  (Bub, Halvorson, {\Spekkens}, Smolin, Poulin \ldots)
\item
Is quantum mechanics an extension of (Bayesian) probability theory?  For instance, through $C^*$-algebras or effect algebras.  Or is it rather a particular application with some special assumptions?  For instance, might it be little more than a theory of priors?  Or is it some combination of those two ideas?  (Bacciagaluppi, Busch, Hardy, Fuchs, Barnum)
\item
How well does quantum mechanics mesh with the ``radical probabilism'' of de Finetti, Richard Jeffrey, and Brian Skyrms?  Does it require radical probabilism for its understanding?  Does the natural noncommutivity that arises in more general kinds of conditioning (like Jeffrey conditionalization) have anything to do with quantum mechanical noncommutivity?  (H\'ajek, Hartmann, {\Schack}, Fuchs, \ldots\ and Diaconis)
\item
Quantum Bayesianism and wider philosophical issues, for instance from the philosophy of language to antirepresentationalism (i.e., the sorts of things in later Wittgenstein, Rorty, Davidson, later Putnam, etc.)  How do the views of Fuchs and {\Schack} on quantum certainty mesh with Wittgenstein's book {\sl On Certainty}?  In what ways is it useful to recognize the ``nonfactivity'' ({\Timpson}'s term) of the subjective Bayesian's view of quantum states?  ({\Timpson}, Price, Fuchs, {\Appleby}, {\Schack})
\item
Quantum versions of foundational Bayesian arguments:  Like Dutch book (synchronic and diachronic), Cox, de Finetti representations, decision theoretic axioms, etc.  (Pitowsky, Uffink, {\Caves}, {\Schack}, Fuchs, Poulin, Brun, Barnum, Wallace)
\item
\ldots\ And I'm sure I forgot a ton of other interesting potential topics at this sitting.  But I think this is enough for now, and 13 is an unlucky number anyway.
\end{enumerate}

\section{09-12-04 \ \ {\it Looking Forward}\ \ \ (to G. Bacciagaluppi)} \label{Bacciagaluppi5}

Now for other, less urgent business.
\bgba
I have been thinking further about ``the idea'': I think it can make
precise in what sense the measurement problem becomes more benign. I
shall try to write down a few pages and send them to you.
\egba
I'll look forward to seeing that.

On a somewhat related matter, I went to Hans Halvorson's talk at Princeton last night and came away very disappointed and self-reflective.  The disappointment (this time) was not with myself but with Halvorson.  I guess I put myself up for a let down:  You see, Halvorson titled his talk ``No Reductive Physicalism, No Measurement Problem''---a title that I could easily imagine myself writing to try to convey some of my most beloved ideas about quantum mechanics.  To top it off, Hans had told me this in one of his emails:
\bhh
This project was motivated by your overall program (which, if I have
interpreted it correctly, rejects the idea that there is a
``measurement problem'').  But I've added a bit of gloss that connects
this idea with recent debates in philosophy of mind.  (Debates about
how to formulate ``physicalism'', etc.)
\ehh

So, I thought, ``Great!  He's going to make precise and much more convincing all this stuff I've been muttering about for the last couple of years.''  (For instance, I was quite pleased with the fine points of our program that Chris {\Timpson} made clearer at the end of his thesis, where he points out that it is crucial for us that quantum states be ``nonfactive.''  That seemed a worthwhile ``gloss'' that may help in several ways.)  Thus, it was quite a shock to me when I saw that what Hans was talking about is diametrically opposed to almost everything I think of QM.  His quantum states are still {\it ontic\/} states, and (physical!)\ collapse comes about from some mysterious interaction between (ontic) brain states and (epistemic?\ ephemeral?\ Platonic?\ in any case, not ontic!)\ mental states.  It was almost like seeing the old von Neumann solution all over again:  The measurement chain stops at consciousness.  I.e., it's all about that mysterious last step.  It was awful.

I motivated that?  That's what disturbed me.  One of the most hotshot of the hotshot young philosophers could read me no more carefully than that?  Am I that absolutely unclear?  It's easy for me to see how someone could disagree with me on quantum foundations if they've understood but reject the Bayesian point of view---and that's fair enough---but what I can't understand is how when I strive so hard to be clear, I can still be continually misread.  Especially from someone like Hans, whose recent work with Bub might have indicated that he would have at least a nominal interest in trying to get these ideas straight.

The only thing I can think is that a large class of philosophers really don't take me seriously, or at least beyond a little lip service.  Halvorson put no gloss on my program; he muddled it with some pretty bad (and certainly completely independent) ideas.

Thus I really am quite thankful for the clarity you help give me.  With you, I've never had the impression that you misunderstand what I'm aiming for.  On the contrary, I often feel quit dumb around you, feeling that your understanding (of my own program) far outstrips mine.

But all of this is just to repeat in a long-winded way what I said in the first line about the note you're going to send me:  ``I'll look forward to seeing that.''

In what sense does the measurement problem become more benign?  You know what I've been aiming for:  A way to make convincing that quantum-state updating is no more (or if you rather, no less) philosophically interesting that the process of conditionalizing a probability.  It is just technically different than usual conditionalization.  The extra unitary rotations are not mysterious physical actions but simply different rules for conditionalizing.  Thus the technical problem becomes trying to find a way (or modification to standard Bayesian arguments) to justify why {\it this\/} conditionalization rule rather than {\it that\/} one.  But the overall point:  One should {\it only\/} find quantum state change mysterious if one also finds standard Bayesian conditionalizing in need of some kind of further metaphysical explanation---i.e., if one deems it important to find a literal or mechanistic explanation for how facts change beliefs.

Also, you know that that's only half the coin for me.  What I think \underline{\it is} philosophically interesting is that the character of the event space for our probability functions changes in going to quantum mechanics:  I.e., we shouldn't---from my view---think of the $h$'s in a $P(h)$ as signifying unknown but pre-existent properties.  Rather they are values that come about from our actions upon a physical system.  All I'm trying to do is cleanly isolate what I view as two very distinct issues:  Change of the conditionalizing rule?  Ultimately not so interesting.  And, to make use of your phrase, a kind of benign change from our old view.  Change of the character of the event space from pre-existent properties to the results of an agent's actions?  That---my intuition tells me---is where the really good stuff is!

But I blabber, blabber.  At least it's made me feel better to write all this down.

\section{10-12-04 \ \ {\it The Further Reading} \ \ (to J. Bub)} \label{Bub13}

I had promised to send a further list of things to read upon my return from Australia.  Here they are, I hope it's not too late.

From Jose M. Bernardo and Adrian F. M. Smith's book {\sl Bayesian Theory\/} (Wiley, 1994):
\begin{itemize}
\item[1)] Sections 1.1, 1.2, and 1.3.
\item[2)] The subsection titled ``Subjectivity of Probability'' on pages 99--102.
\item[3)] Section 4.8 to page 237.
\end{itemize}
Also I think Marcus Appleby's paper ``Probabilities Are Single-Case, or Nothing'' (\quantph{0408058}) could be quite helpful.  Particularly, I like the argument of Section 3.

My plan for your seminar is mostly to have a free roaming discussion about the contents of these readings.  Is that OK with you?  Or did you have something else in mind?

\section{13-12-04 \ \ {\it The Beast}\ \ \ (to S. Aaronson)} \label{Aaronson7}

\bsa
Hope everything's well with you.  Yep, my own big 400$+$ page ``samizdat'' is lumbering toward completion over the next couple days \ldots\ {\bf [See Scott's thesis, ``Limits on Efficient Computation in the Physical World,'' \quantph{0412143}.]}

Would you mind reading it through by tomorrow night and sending me detailed comments?  Ha ha, just kidding.  But you're mentioned in the acknowledgments, and the prologue (``Aren't You Worried That Quantum Computing Won't Pan Out?'')\ is intended to amuse.  I'd love to hear your thoughts.
\esa

I loved it!  And you have helped me immensely with your Chapter 1:  Soon after the new year, I've got to construct a talk to give to Jeff Jaffe (the president of Bell Labs research) and Dave Bishop (the new VP over physical-science research) on why Bell Labs should imagine spending even a cent on quantum foundations research.  Among other things, I have wanted to argue that it has been the foundational mysteries all along that have motivated so much of technical quantum computing and quantum information.  And that these guys who want to suck off the governmental tit spurred by this field shouldn't forget its very roots!  I.e., I'm going to partly lash out at the guys who call what I'm interested in ``Sunday physics'' but nevertheless (have to) cite my papers in their (self-proclaimed) ``hard core'' physics papers.

Anyway, I was going to quote liberally from Feynman, Deutsch, early Bennett and Brassard, Wootters, and the like.  But now I can also see that I'll be quoting liberally from Aaronson!  Even if you never prove another great theorem again, you're turning into a great writer.

BTW, I hope you will post your thesis on {\tt quant-ph}.  But before you do, compactify it with single-space, etc.  Save the trees man!

I've still got to come visit you in Princeton.  I'll try to do that soon after the new year too.

\section{14-12-04 \ \ {\it Bush Torment} \ \ (to G. L. Comer)} \label{Comer58}

My post-election resolution has been to try to think about politics/Bush/state-of-the-material-world as little as possible.  The election burned me badly enough that I was finally reminded that I should focus my energies on where I can make a difference in the world:  Namely, quantum mechanics and pragmatic philosophy.  Besides, I was going insane with the torment of the idiocy of my fellow Americans.  So, after a week of grave depression right following the election, I have been mostly sticking my head in the sand with regard to that aspect of our times.  Locally (i.e., given the crisis of our days) it's probably the wrong thing to do, but globally (i.e., my potential effect on the ages) I think it's probably the right thing for ME to do.  But everyone has a different calling.

Concerning your particular remarks in the last note:  Yes, I do agree with you.  You needn't worry on that account.

And a little tidbit concerning my remark on ``the ages'' above, the other day I got my most flattering career accolade yet.  This one I am very proud of:  {\sl The Oxford Dictionary of American Quotations\/} wants to quote me in their newest revision!  That in my eyes beats the hell out of another PRL any day.

\section{14-12-04 \ \ {\it Thoughts on Non-nonlocality} \ \ (to H. Price)} \label{Price3}

Finally coming back to your request for what to read within my
writings on how a Bayesian conception of quantum states takes care of
(or should take care of) issues to do with nonlocality in QM.  Let me
suggest you read Section 3, ``Why Information?'', and the beginning
part of Section 4, ``Information About What?'', in my paper ``QM as
Quantum Information (and only a little more)'', \quantph{0205039}. In that regard, let me also paste in below some
correspondence I had with David {\Mermin} and Jerry Finkelstein on those
sections---I think they clarify things a bit further---and let me
also attach an abridged version of the same paper that I never posted
on the archive.  In that version, I changed the ``Why Information?''
section somewhat to better reflect what I told {\Mermin} below.

The key issue is, what do I mean when I say that nothing physically
changes on Bob's side when Alice performs a measurement on her half of
an entangled pair?  In that regard, I think you might also get
something out of my discussions on pages 175--176 and 184--189 of my
samizdat {\sl Quantum States:\ What the Hell Are They?}.  These
discussions were motivated by some standard dictionary (can't remember
which) wanting to include a definition of quantum teleportation in its
latest revision, and {\Bennett} supporting a definition that said
something like, ``The transference of one particle's properties onto
another particle via the assistance of \ldots'' The thing I objected
to strenuously was the idea that any properties at all were
transferred.  [See 25-04-02 note ``\myref{Bennett15}{Short Thoughtful
    Reply}'' to C. H. Bennett and 14-05-02 note
  ``\myref{SmolinJ2.1}{Qubit and Teleportation Are Words}'' to
  C. H. Bennett and others.]  Also, let me point you to some
discussion on the same point in {\sl Notes on a Paulian Idea}.  See
pages 465--466, starting at the words ``Quote from Draft,'' and see
pages 467--471 in the note titled, ``Detailed Commentary.''  Getting a
description of quantum teleportation that Asher and I could both agree
on for the paper we were writing was a point of serious contention for
us.  And I went at great lengths to try to get my point across to him.
It's exactly the same point that's relevant to our own discussion
(i.e., Fuchs and Price).

I guess you could say the overarching idea is ``no surprises, no
reality.''  That is an idea I flesh out in pretty good detail in my
paper ``The Anti-{\Vaxjo} Interpretation of Quantum Mechanics.'' See
sections 4 and 5.  (They are completely independent of the rest of
the paper; so don't read any of the rest of the paper.)  I think (I
hope) actually, that you'll find the discussion there interesting
independently of any issues to do with our discussion of nonlocality.
In particular, if you read those sections, I don't think you could
fail to give me the credentials of a ``subject naturalist.''  Am I
right?

Let's see, anything more?  One last thing:  Let me also suggest
Chapter 8 of Chris {\Timpson}'s PhD thesis {\sl Quantum Information
Theory and the Foundations of Quantum Mechanics}, \quantph{0412063}, and also the ``Envoi'' at the very end.  He does
quite a nice job of explaining how the ``nonfactivity'' of our
quantum states buys us the ability to stop talking about nonlocality.

There, that's one thing marked off my to-do list for you.  (Certainly
there's no rush to respond; I know you're traveling.  Just archive
this note until you have the time \ldots\ even if that's six months
from now.)

\section{14-12-04 \ \ {\it Backward Causation} \ \ (to C. H. {\Bennett} and B. W. Schumacher)} \label{Bennett36} \label{Schumacher14}

Did you guys (or either of you guys) ever write anything up on the stuff that got you so excited in Japan a couple of years ago?:  I.e., the stuff on time travel via quantum teleportation?  If you haven't written anything up, do you have some PowerPoint presentations (or such) that you wouldn't mind sharing with me?  (I know that some exist because I've seen you talk on this before.)  I've been discussing this stuff with Huw Price, an Australian philosopher who is quite interested in exploring ideas about backward causation in the context of QM, and I've told him about your stuff.  Price is quite a good guy, and I want to make sure I've got the story straight with all this stuff.  Would you mind if I also share your slides with Huw in the course of our discussion?  I don't see how it could hurt (since I promise he's a nice guy), but I understand you might have some objection if you haven't published this yet.

Anyway, the way I pick on all three of you (i.e., you two and Huw) is that I think you do something dangerous with a simple ``chain of inference'' by mystifying it with all this forward and backward time-travel talk.  But now to compliment you:  I think your Japan examples are great for illustrating that!

\section{14-12-04 \ \ {\it Directions, Etc.}\ \ \ (to J. Bub)} \label{Bub14}

\bjb
Unfortunately, the only copy of Bernardo and Smith is out of the
library, so probably no one will have read that. A free roaming
discussion is fine, I think, but you should probably be prepared to
talk a bit in general first (or perhaps second) about what specific
foundational issues in quantum mechanics push one towards
Bayesianism, and perhaps also about information in physics, since
this is the topic of the seminar. Does that sound OK?
\ejb
I'll try.

But most importantly, you gotta realize that, concerning
\bq\noindent
     ``what specific foundational issues in quantum mechanics push one
     towards Bayesianism,''
\eq
it's the other way around.  It's clear thinking about probability and
the use of it that leads to Bayesianism---nothing to do with quantum
mechanics.  {\it Then}, recognizing that a quantum state is
(mathematically and conceptually) {\it nothing more than\/} a
compendium of probabilities, one gets to the Bayesian conception of
quantum states. In other words, it is Bayesianism that carries a hope
of clarifying quantum foundations---not that something in quantum
mechanics specifically pushes toward Bayesianism.

I'll try to see if I can get Bernardo and Smith scanned into my
computer and turned into a text file.  There's some chance that that
might happen tonight.

\section{15-12-04 \ \ {\it Favorite Bernardo and Smith Quotes} \ \ (to J. Bub)} \label{Bub15}

Well, try as I might, I couldn't get my scanner connected up last
night.  It always takes much more time to install these things than
one expects!  Very frustrating.  It just confirms what I've been
thinking:  I never want to go through another move again.  {\it I
never ever want to move again!}  Moves just disrupt everything in
life, and never for the better.

In light of last night's failure, let me do this as a stopgap.  Here
are two of my favorite Bernardo and Smith quotes that I already have
in my computer.  [See 07-08-01 note ``\myref{Mermin28}{Knowledge, Only Knowledge}'' to T. A. Brun, J. Finkelstein and N. D. Mermin.]
  Chew upon them, because I think they especially
express the flavor of the reconception of quantum mechanics I'm
shooting for:  That quantum mechanics is a {\it normative\/} theory
of personal behavior (or uncertainty accounting) IN LIGHT OF the
peculiar, particular world in which we are immersed.  Different
world, different normative theory for agents immersed within it.
Moreover, because of the last point, though quantum theory is a
theory about agents (rather than of objective reality per se), one
can still hope to glean some hypotheses about objective reality
itself from it.  One just has to do so obliquely.  But that's our
task, and that's what we're setting ourselves up to do.

I'll try to emphasize these things in a preamble to our discussion
tomorrow night.  But if you don't mind, distribute this note to your
students first so that it'll have had at least a little time to
percolate in their minds.

\section{15-12-04 \ \ {\it And Just a Little More To Read} \ \ (to J. Bub)} \label{Bub16}

I know you guys have already read this stuff, but maybe it'd be worthwhile going over it again before my arrival:  Namely Chapter 8 of Timpson's thesis (paying attention to the issues surrounding Footnote 4) and his ``Envoi'' at the very end.  I think he does good justice to what (and why) we Bayesians are up to.  It's all about breaking the ``factivity'' of quantum states.

\section{15-12-04 \ \ {\it And Just a Little More To Read, 2} \ \ (to J. Bub)} \label{Bub17}

\bjb
Re Timpson, I'd be interested in your comments on his footnote on p.\ 181: `Although he does not himself put it in these terms, Fuchs'
awareness of the factivity of the terms ``knowledge'' and ``information'' and his related criticism of {\Mermin}, mark the change from the
objective Bayesianism of Fuchs (2001) to the more consistent subjectively Bayesian position of Fuchs (2002a).'
\ejb

Plan to.  But also, it might be useful for you to look over the pages he cites from my ``Quantum States: W.H.A.T.?''\ (namely, 19--25 and 42--51).  I think I might read 'em too.  [See 07-08-01 note ``\myref{Mermin28}{Knowledge, Only Knowledge}'' to T. A. Brun, J. Finkelstein \& N. D. Mermin; and 02-09-01 note ``\myref{Mermin35}{Intersubjective Agreement}'' to N. D. Mermin.]

\section{16-12-04 \ \ {\it Troubles of the Slusher Kind} \ \ (to R. E. Slusher)} \label{Slusher6}

I'm taking off to Maryland now to give a seminar to Jeff Bub's quantum foundations group at UMD.  But I've been troubled all night about how I can get us past our point of disagreement from yesterday:  I.e., to try to find a way to convince you that we Bayesians are not anthropomorphizing quantum mechanics in an unnecessary way.  And to try to find a way to convince you that quantum mechanics is not like Newtonian mechanics, even in the broadest of all senses.  I.e., to convince you that Newtonian mechanics is a theory of systems (as they are by themselves), and quantum mechanics is rightly a theory of observers (and only observers), not a theory of systems.

I intend to convince you of that---so you're good exercise for me.  That'll be my new year's present.  In the meantime, have you read Section 4 of my \quantph{0204146}, ``The Anti-{\Vaxjo} Interpretation of Quantum Mechanics''?  It addresses the issue at least in a way that turned Preskill's head.

\section{16-12-04 \ \ {\it The Doctrine of Preemption} \ \ (to C. H. {\Bennett})} \label{Bennett37}

\bcb
Reading the end of your letter I find the usual backhanded compliment:  Presumably the last word ``that'' refers to the dangerousness of forward and backward time travel talk.  I seem to have spooked you so much with my loose talk that now you are even afraid of forward time travel, which you never used to be.
\ecb
OK, you got me \ldots\ in fact, preempted me.  I loved the lines near the end of your presentation:
\begin{itemize}
    \item[Q.]  Is it time travel?
    \item[A.]  It depends on what your definition of ``is'' is.
\end{itemize}
It was as if you had already heard my snide remark of 2004 in 2002.  Had you?

I'm glad to hear the Newton meeting is almost over.  I'm looking forward to having you back in the neighborhood.  If you've got the time, I'd love to start dropping in from time to time sometime in January.  In particular, at least at the outset, if I could interest you guys in the SIC-POVM problem (i.e., do they always exist?, and if they do, how are they constructed?, what can you do with them?, etc.) it'd be great.  I need someone(s) to bounce ideas off of, and I'm awfully lonely here.  I keep figuring there must be some simple solution, and that one really doesn't need to go hog-wild with abstract mathematics to get at it.  It probably just needs some experimental mathematics like the Smolin minimizer could give, a little Bennettian playfulness with ideas, and a Fuchsian smile.  (You've long known that I'm attracted to quantum mechanics because deep inside I think I can charm things into existence \ldots\ and QM is the one scientific thing that gives me hope that there's no pre-existent.)

And thanks for the pictures!

\section{20-12-04 \ \ {\it The Doctrine of Preemption, 2} \ \ (to C. H. {\Bennett})} \label{Bennett38}

\bcb
What's a SIC-POVM?  Is it an actual POVM or a misspelled unwell POVM?
\ecb

It stands for ``symmetric informationally-complete'', and yes, the pun is intended.  They're annoying critters.  Take the space of operators over a Hilbert space of dimension $d$; it is a vector space of dimension $d^2$.  Now ask, can one find a complete {\it orthogonal\/} basis $A_i$ on that space consisting solely of positive, rank-one operators?  The answer is no.  So, ask the next best thing.  If one cannot satisfy $(A_i,A_j)=0$ for $i\ne j$, can one at least satisfy  $(A_i,A_j)=\mbox{constant}$ for $i\ne j$?  The answer {\it appears\/} to be yes, and sets of such operators that sum to the identity are dubbed SIC-POVMs.  So, it's a really very elementary question \ldots\ but it seems to require something clever for an actual proof.

\bcb
My ``It depends on what your definition of `is' is,'' was meant to be a
quote from Bill Clinton, not a time-traveled reaction to you.
\ecb

I understood that.  But in a world with backward causation, are you so sure you can draw such distinctions?

Have a safe trip home.

\section{20-12-04 \ \ {\it Magical Cameras} \ \ (to G. L. Comer)} \label{Comer59}

Awfully busy at the moment \ldots\ but I couldn't resist replying to you!

\bgc
Consider what I just found in {\bf Nature}, Dec.\ 16, 2004:
\bq\noindent{\rm
Electrons frozen in motion\smallskip\\
Henrik Stapelfeldt\smallskip\\
A single electron cloud in molecular nitrogen has been photographed. The snapshot is recorded so rapidly that it might become possible to image electron clouds as they change during fundamental molecular processes.}
\eq
Now when I see such things, I wonder ``What Would Fuchs Do?''  So,
here's your chance to enlighten me!  Note, that when I read ``electron
cloud'' I'm thinking ``wave function'', and since I can't put either one
on some kind of Einstein pedestal of objectivity, I'm really confused
as to what it {\bf is} that is being photographed.
\egc

Amazing magical cameras!  Not only do they photograph the facts, but they photograph the theories too!  Get one now, while supplies last \ldots

\section{21-12-04 \ \ {\it Know Any Young German Bayesians?}\ \ \ (to P. Busch and H. J. Briegel)} \label{Busch6}

I should apologize for my provocative title!  But I wanted to get your attention!

First off, let me give you both an update on how our meeting in Konstanz is going.  I'll do that by pasting two notes below.  One lists the responses we've gotten so far, and one describes various suggested topics for the meeting.  I know I don't need to clench your interest since you've both already said that you would come, but I figure it can't hurt to try to stoke the enthusiasm.

Now for the real purpose of my note.  Soon I'll be sending out a second round of invitations, and I would like to ask your advice.  In total, we're hoping for about 36 participants (excluding the students we'll probably slip in).  Thus we've got room for maybe 10 more invitees.  Here's the question:  Can either of you suggest any good young Germans who might be a good fit for the subject matter of the conference?  You can also suggest some older Germans too!

The issue is that it seems the VW Foundation has some interest in funding a substantial part (if not all) of the meeting, but they want an increased emphasis on 1) younger researchers (i.e., ones who have not yet obtained a permanent position) and 2) Germans.  So, since there's some room anyway in our ranks, I don't mind trying to accommodate their wishes if I can \ldots\ especially if I can do it while still having the meeting of my dreams.

Thus if you have some suggestions, they would be most useful.  Who am I overlooking from the German quantum information, quantum foundations, or philosophy communities who could really add something to our meeting?  There's no promise that we'll add their names to the roster, but we'll definitely take anything you say seriously.

Hope to hear back from you.

\section{31-12-04 \ \ {\it Dylan and the New Year} \ \ (to J. B. Lentz \& S. J. Lentz)} \label{LentzB6} \label{LentzS4}

Well, I got a little time this morning to myself while the girls were out shopping, and I finally got a chance to finish reading Bob Dylan's {\sl Chronicles}.  I had wanted to do that before the new year started up.  Thank you two so much for sending it.  Ever since reading these lines in the Washington Post,
\bq
     The hype seems enough to turn the most faithful acolytes
     into unbelievers. How can it possibly be true, as the magazine
     claims, that this week's celebrity memoirist is ``the most
     influential cultural figure now alive''?

     Oh, you thought he was once, back in the '60s. The image of that
     faraway decade has now been so magnified by myth and distorted by
     culture-war invective that you have to strain to recall what it
     felt like to be alive then, but you do remember this:  Half the
     people you knew believed that if only they could figure out what
     Bob Dylan was saying, the secrets of the universe would be
     revealed.
\eq
I had wanted to sink my teeth into the book.  That last line, especially, was something that caught me.  I've been through that so many times myself:  from listening to John Lennon to reading John Wheeler to listening to Paul Simon to reading William James to listening to John Coltrane \ldots.  And even now, to reading Bob Dylan:  I knew that's how I'd tackle the book if I got my hands on it.  The secrets of the universe are in the end, at the high level, not the low level:  Quarks and electrons are ways that we can focus our attention so that we can distill simple truths here and there, but they're not what we're ``made of''.  Sometimes I think we're made of poetry, if anything.  Or, at least, that's closer to the right idea.

Anyway, it was a great book.  I enjoyed whole swaths of it and took a lot of notes.

BTW, there's a little good news on my side with respect to these same musings (on poetry and such) that I don't think I've told you about before:  I learned a few weeks back that the {\sl American Heritage Dictionary of American Quotations\/} is going through a revision (it'll soon be the {\sl Oxford Dictionary of American Quotations}), and they want to include a couple of lines from some of my writing.  It's very flattering to think that my little poetry could be in the same book with William James's and Bob Dylan's very big poetry, but mostly it's a statement about raw luck---the good kind for me and the bad kind for them.  Still, with everything one writes, whether it is preserved by luck or not, as long as it is preserved, it's a way of living a little a longer.  And the idea of that always feels good.

Good luck to yourselves in this new year, and many happy returns to both of you.

\section{31-12-04 \ \ {\it New Year's Eve} \ \ (to A. Peres)} \label{Peres67}

\noindent Dear Asher,\medskip

As I write you these words, it is indeed New Year's Eve, but the title of the note is meant to be read more metaphorically.  It has been a very long time since I have written you.  I hope your recovery from your stroke is coming along as well or even much better than you had indicated in one of your notes.  Maybe even everything is back to normal by now?  Or at least I'd like to dream that.

I don't know where I left off in my conversation with you.  I guess you know that I did not end up getting an offer for a faculty position, either in Copenhagen or Vancouver.  Nor did the Mabuchi/Preskill ploy of trying to create an interdepartmental position (physics and philosophy) for me at Caltech work out either.  Thus, with no place to go, I had to come back to Bell Labs.  Of course, far worse things could have happened---I look at what happened to you, and I know how selfish I am---but I fell into a fairly deep, dark depression for over four months.  An amazingly illogical depression.  I've done enough to survive my wife (like buying us a house and getting the family moved into it), and I've done enough to survive Bell Labs (speaking when I needed to speak and doing the administrative things required of me), but I've done little beyond that.  The depression has been much worse than even after our fire:  At least after the fire I kept writing and doing science.  But even my writing had come to a standstill this time.

For instance, after those very nasty problems with our email server---you probably don't remember, but I told you how reams of my email had been randomly eaten up for some undetermined time on the order of two months---I just used the opportunity to effectively slip out of the email ether, and very few have heard from me in these months.

Anyway, I'm writing you now because something has been jarred loose, and my self-esteem seems to have started to come back.  Or maybe my body just got weary of being depressed.  I'm not sure of the source of the optimism, but I'm now finding myself looking forward to an intense new year---intense with work and intense with thought.  And it's New Year's Eve!

\ldots\ All of that just so I could say ``Happy New Year!''\ to you in clear conscience.  I wish you and Aviva and the whole family all the best.

On another note---perhaps some welcome news for you---I had never written you about the completion of your festschrift in my period of silence \ldots\ but at least the festschrift itself went on.  All the papers have been with {\sl Foundations of Physics\/} for some months now.  They're a fine lot of 26 papers (including your autobiography).  I'll put the full list below.  vdM has most recently told me that they're next in queue after the Cushing festschrift (which itself got delayed for some reason or other), and they should start appearing early this new year.

Expect more letters from me this year!  I'll be back soon.\medskip

\noindent All the best,\medskip

\noindent Chris \bigskip

{\bf Papers in the Peres Festschrift:}
\begin{enumerate}
\item
H. Bechmann-Pasquinucci, ``From Quantum State Targeting to Bell Inequalities''

\item
Jacob D. Bekenstein, ``How does the entropy/information bound work?''

\item
Paul Benioff, ``Towards a Coherent Theory of Physics and Mathematics: The Theory -- Exper\-i\-ment Connection''

\item
Robin Blume-Kohout and Wojciech H. Zurek, ``Redundancy of Information Storage in Multi-qubit Universes''

\item
Gilles Brassard, Anne Broadbent, and Alain Tapp, ``Quantum Pseudo-Telepathy''

\item
\v{C}aslav Brukner, Markus Aspelmeyer, and Anton Zeilinger, ``Complementarity and Information in `Delayed-choice for entanglement swapping'{\,}''

\item
Dagmar Bruss and Chiara Macchiavello, ``How the First Partial Transpose Was Written''

\item
{\Adan} Cabello, ``Bell's theorem without inequalities and without unspeakable information''

\item
S. Deser, ``A Note on Stress-Tensors, Conservation and Equations of Motion''

\item
Bernard d'Espagnat, ``Consciousness and the Wigner's Friend Problem''

\item
David P. DiVincenzo and Barbara M. Terhal, ``Fermionic Linear Optics Revisited''

\item
Steven T. Flammia, Andrew Silberfarb, and Carlton M. Caves, ``Minimal Informationally Complete Measurements for Pure States''

\item
Friedrich W. Hehl and Yuri N. Obukhov, ``To consider the electromagnetic field as fundamental, and the metric only as a subsidiary field''

\item
Karol Horodecki, Micha\l\ Horodecki, Pawel Horodecki, and Jonathan Oppenheim, ``Information theories with adversaries, intrinsic information, and entanglement''

\item
Micha\l\ Horodecki, Ryszard Horodecki, Aditi Sen(De), and Ujjwal Sen, ``Common origin of no-cloning and no-deleting principles -- Conservation of information''

\item
Elena R. Loubenets, ``{\,}`Local Realism', Bell's Theorem and Quantum `Locally Realistic' Inequalities''

\item
N. David Mermin, ``What's wrong with this criticism''

\item
Asher Peres, ``I am the cat who walks by himself''

\item
Murray Peshkin, ``Spin-Zero Particles Must Be Bosons: A New Proof
within Nonrelativistic Quantum Mechanics''

\item
Arkady Plotnitsky, ``On the Precise Definition of Quantum Variables
and the Relationships between Mathematics and Physics in Quantum Theory''

\item
Abner Shimony, ``An Analysis of Stapp's `A Bell-type theorem
without hidden variables'{\,}''

\item
Henry P. Stapp, ``Comments on Shimony's Analysis''

\item
Anthony Sudbery and Jason Szulc, ``Compatibility of subsystem states''

\item
Daniel R. Terno, ``Inconsistency of a quantum-classical dynamics, and what it implies''

\item
William K. Wootters, ``Quantum Measurements and Finite Geometry''

\item
Mario Ziman, Martin Plesch, and Vladimer Buzek, ``Reconstruction of superoperators from incomplete measurements''

\end{enumerate}

\bigskip
\noindent {\Large \bf Asher Peres died 1 January 2005.}
\bigskip

\chapter{2005: After Asher}

\section{01-01-05 \ \ {\it Our Dear Asher} \ \ (to D. R. Terno)} \label{Terno5}

That is very sad news indeed.  I feel knocked over.  Asher was a great man---among the greatest I have ever known and ever read.  I owe him my whole career in fact---from the things I learned from him to the way he supported me to the ways he lifted my life by taking me seriously like few others.  He was one of the few people who actually read my samizdat, for instance, and I cherished his opinion.  I feel that I betrayed him greatly with my comparatively petty problems of the last few months---they kept me silent when maybe he needed me most.  Below is the letter I wrote him just yesterday, explaining what was up---finally breaking my long silence.  I am sure he never saw it.  I am very sad.

I think that you should send an announcement to the quantum information community, much like the ones I sent when Jaynes, Landauer, and Lewis each passed away.  If it would be useful to you, I could send you my email address book with maybe 300 or 400 quantum-information/foundations names in it.

I will call Aviva tomorrow when the time is more proper in Haifa.  She is a gem, and I hope that she will keep shining.

Asher was a great man, a great friend, a great scientist.

\subsection{Announcement I Sent Out Broadly 3 January 2005 on Behalf of the Authors}

\bq
\noindent {\it Quantum information science lost one of its founding fathers.}\smallskip\\
{\it Asher Peres died on Sunday, January 1, 2005.  He was 70 years old.}\medskip

A distinguished professor at the Department of Physics, Technion -- Israel Institute of Technology, Asher described himself as ``the cat who walks by himself''.  His well-known independence in thought and research is the best demonstration of this attitude.  Asher will be missed by all of us not only as a great scientist but especially as a wonderful person.  He was a surprisingly warm and unpretentious man of stubborn integrity, with old-world grace and a pungent sense of humor.  He was a loving husband to his wife Aviva, a father to his two daughters Lydia and Naomi, and a proud grandfather of six.  Asher was a demanding but inspiring teacher.  Many physicists considered him not only a valued colleague but also a dear friend and a mentor.

Asher's scientific work is too vast to review, while its highlights are well-known.  One of the six fathers of quantum teleportation, he made fundamental contributions to the definition and characterization of quantum entanglement, helping to promote it from the realm of philosophy to the world of physics.  The importance of his contributions to other research areas cannot be overestimated.  Starting his career as a graduate student of Nathan Rosen, he established the physicality of gravitational waves and provided a textbook example of a strong gravitational wave with his PP-wave.  Asher was also able to point out some of the signatures of quantum chaos, paving the way to many more developments.  All of these contributions are marked by a surprising simplicity and unbeatable originality.

Of all his publications, Asher was most proud of his book {\sl Quantum Theory:\ Concepts and Methods}.  The book is an example of Asher's scientific style: an uncompromising and deep understanding of the fundamental issues expressed in a form which is as simple and accessible as possible.  It took Asher six years to carefully weave the threads of his book together.  The great quality of the work is acknowledged by anyone acquainted with the final result.

In a favorite anecdote, Asher told about a reporter who had interviewed him on quantum teleportation.  ``Can you teleport only the body, or also the spirit?''\ the reporter had asked.  ``Only the spirit,'' was Asher's reply.  Our community has been privileged to know him and have been touched by his spirit.

``I am the cat who walks by himself'' is the title of a charming twelve-page autobiography covering his life from his birth in the village Beaulieu-sur-Dordogne in France until his meeting with Aviva on a train to Haifa [\arxiv{physics/0404085}].  The rest of his story is in his formal CV. \medskip

\noindent Netanel Lindner, Petra Scudo, Danny Terno

\eq

\section{01-01-05 \ \ {\it A Backup for Us Both} \ \ (to J. W. Nicholson)} \label{Nicholson19}

\bq\noindent
\begin{center}
A Time to Mourn\\ (From {\sl New York Times}, 1 January 2005)\\ by David Brooks
\end{center}
\medskip

I have this week's front pages arrayed on the desk around me. There's a picture of dead children lined up on a floor while a mother wails. There's a picture of a man on the beach holding his dead son's hand to his forehead. There are others, each as wrenching as the last.

Human beings have always told stories to explain deluges such as this. Most cultures have deep at their core a flood myth in which the great bulk of humanity is destroyed and a few are left to repopulate and repurify the human race. In most of these stories, God is meting out retribution, punishing those who have strayed from his path. The flood starts a new history, which will be on a higher plane than the old.

Nowadays we find these kinds of explanations repugnant. It is repugnant to imply that the people who suffer from natural disasters somehow deserve their fate. And yet for all the callousness of those tales, they did at least put human beings at the center of history.

In those old flood myths, things happened because human beings behaved in certain ways; their morality was tied to their destiny. Stories of a wrathful God implied that at least there was an active God, who had some plan for the human race. At the end of the tribulations there would be salvation.

If you listen to the discussion of the tsunami this past week, you receive the clear impression that the meaning of this event is that there is no meaning. Humans are not the universe's main concern. We're just gnats on the crust of the earth. The earth shrugs and 140,000 gnats die, victims of forces far larger and more permanent than themselves.

Most of the stories that were told and repeated this week were melodramas. One person freakishly survives while another perishes, and there is really no cause for one's good fortune or the other's bad. A baby survives by sitting on a mattress. Others are washed out to sea and then wash back bloated and dead. There is no human agency in these stories, just nature's awful lottery.

The nature we saw this week is different from the nature we tell ourselves about in the natural history museum, at the organic grocery store and on a weekend outing to the national park. This week nature seems amoral and viciously cruel. This week we're reminded that the word ``wilderness'' derives from the word for willful and uncontrollable.

This catastrophic, genocidal nature is a long way from the benign and rhythmic circle of life in {\sl The Lion King}. It's a long way from the naturalist theology of Thoreau's {\sl Walden} or the writings of John Muir.

The naturalists hold up nature as the spiritual tonic to our vulgar modern world. They urge us to break down the barriers that alienate us from nature. Live simply and imbibe nature's wisdom. ``Probably if our lives were more conformed to nature, we should not need to defend ourselves against her heats and colds, but find her our constant nurse and friend, as do plants and quadrupeds,'' Thoreau wrote.

Nature doesn't seem much like a nurse or friend this week, and when Thoreau goes on to celebrate the savage wildness of nature, he sounds, this week, like a boy who has seen a war movie and thinks he has experienced the glory of combat.

In short, this week images of something dark and unmerciful were thrust onto a culture that is by temperament upbeat and romantic.

In the newspaper essays and television commentaries reflecting upon it all, there would often be some awkward passage as the author tried to conclude with some easy uplift---a little bromide about how wonderfully we all rallied together, and how we are all connected by our common humanity in times of crisis.

The world's generosity has indeed been amazing, but sometimes we use our compassion as a self-enveloping fog to obscure our view of the abyss. Somehow it's wrong to turn this event into a good-news story so we can all feel warm this holiday season. It's wrong to turn it into a story about us, who gave, rather than about them, whose lives were ruined. It's certainly wrong to turn this into yet another petty political spat, as many tried, disgustingly, to do.

This is a moment to feel deeply bad, for the dead and for those of us who have no explanation.
\eq

\section{02-01-05 \ \ {\it I Hope You Are OK} \ \ (to U. Mohrhoff)} \label{Mohrhoff6}

I was reading the news tonight and learned that Pondicherry had significant damage from the tsunami.  My knowledge of geography is very bad, and I did not know before that Pondicherry is near the sea.  I so hope you are safe.  I keep my fingers crossed.

\subsection{Ulrich's Reply}

\bq
I am really touched by your inquiry. We (that is my wife and I) felt the earthquake as we sat down for breakfast (at this distance only a gentle swaying of the ground). We then monitored the news to know the location of the epicenter, and when we learned of the magnitude of the quake, we waited for the tsunami. Pondy town is sufficiently above sea level to have remained undamaged, but fishing communities nearby and some Auroville beach settlements were hard hit. One office near the beach had 20 computers flushed out.

P.S. I still feel that our views on QM have so much in common that we should form a common front against the quantum state realists \ldots\ If you can spare 30 minutes, he a look at my PowerPoint presentation ``Beyond causality\ldots'', which you can be downloaded from my website. (You are even quoted there. Favorably!)
\eq

\section{02-01-05 \ \ {\it Omphaloskepsis} \ \ (to J. W. Nicholson)} \label{Nicholson20}

I think you and Sumitra might get a kick out of how this dictionary described my behavior tonight.

\bq
\noindent OMPHALOSKEPSIS: Contemplating one's navel as an aid to meditation.\medskip

This word seems to be relatively new, at least the Merriam-Webster ``Word of the Day'' column claims it to have been invented only in the 1920s.  It turns up in only a few dictionaries and seems to be a word that survives more for the chance to show off one's erudition than as a real aid to communication. If so, this article is a further perpetuation of its unreal status. It is formed from two Greek words, omphalos, ``navel, boss, hub'', and skepsis, ``the act of looking; enquiry''. The former turns up in words such as omphalotomy, ``cutting of the umbilical cord'', in the related omphalopsychic for one of a group of mystics who practised gazing at the navel as a means of inducing hypnotic reverie, and omphalomancy, an ancient form of divination in which the number of children a woman would bear was determined from counting the knots in her umbilical cord at birth.\footnote{\editornote From \myurl{http://www.worldwidewords.org/weirdwords/ww-omp1.htm} (though the text of the page changed between December 2011 and February 2012).  The Oxford English Dictionary does provide, however, an instance of \emph{omphaloskeptic} in an Aldous Huxley letter, circa 1915: ``You must admit that no omphaloskeptic, nay, not Plotinus, could have so utterly realized the Infinite as at moments one did to night.''}
\eq

\section{03-01-05 \ \ {\it Safety} \ \ (to U. Mohrhoff)} \label{Mohrhoff7}

I am so relieved to hear that you are safe.  I became very worried.  Some time good luck happens too (but I've gotten where I don't expect it much anymore).

Now that I know you are safe, let me tell you some sad news that I could have told you yesterday.  Asher Peres passed away New Year's Day.  It was either a stroke or a heart-attack; it happened very quickly.  He was buried yesterday.  I will distribute an obituary written by his last three students in a few hours; you are in the distribution list.  I'm going to miss him very much.

With regard to your ``P.S.'', thanks for sending me your webpage link.  Indeed for some reason I feel re-energized to try to sort our similarities and differences in point-of-view about QM (and metaphysics).  I hope that I can come to a better understanding.  I tried downloading the PowerPoint presentation you told me about, but had no luck.  Could perhaps you just send it to me as an email attachment?  Likely, I will be able to open that.  I will also enjoy trying to read some of your papers again.

\subsection{Ulrich's Reply}

\bq
I really appreciate your concern. Thanks again.

Although I did not have the good fortune to meet Asher Peres, I am saddened
by the news of his passing. I treasure his one (unsolicited) email in which
he expressed appreciation for my AJP paper ``What quantum mechanics is trying
to tell us'' (responding to Mermin's ``What is quantum mechanics trying to
tell us?''). My copy of his {\sl Concepts and Methods\/} has disintegrated from
overuse, and I loved him for his insistence that ``there is no interpolating
wave function giving the `state of the system' between measurements''. (It's
echoing through all my papers.) If I have recently taken to criticizing the
paper you coauthored with Asher Peres, it is because the closer one's views
are, the larger loom the remaining differences \ldots.

\bq\noindent
[Quoting CAF:] \ldots\ for some reason I feel re-energized to try to sort our
similarities and differences in point-of-view about QM (and
metaphysics).  I hope that I can come to a better understanding \ldots.
I will also enjoy trying to read some of your papers again.
\eq

I very much look forward to that. (Let's bear in mind the warning my
grandmother gave to my father long before email made matters worse: ``Letters
don't smile.'') My recent relatively short \quantph{0412182} (to appear in the
Indian journal of physics {\sl PRAMANA}) may be a good starting point. I also
would like you to have a look at my class notes, once they are online. The
trouble is that they keep changing all the time \ldots\ I can see why it took
Asher Peres six years to write his book.
\eq

\section{03-01-05 \ \ {\it Songwriting, Interrupted} \ \ (to G. L. Comer)} \label{Comer60}

I started the note below to you on New Year's Day:
\bq\noindent
I want to start this note on New Year's Day, even if I may not finish it today.  Happy New Year old friend!  I wish you and your family a strong and productive one.
\eq
It was to be titled ``Songwriting'', and I had planned to cover several things and try to tie them together:  the tsunami, ideas about the size of the human race and space exploration, some quotes from Bob Dylan's book, some stuff about your own songwriting, and some stuff about pragmatism and quantum mechanics.  It was going to be a note like I haven't written you in quite some time.  But just as I wrote you the third sentence below, I got a note saying that Asher Peres had died.  It was either a stroke or a heart-attack; it happened very quickly.  It hit me very hard.  He was a good friend and like a father to me.  (I'll be distributing an obituary written by his last students in a few hours to the quantum information and foundations communities; you are in the distribution list.)

Anyway, I'm exhausted now and the writing muse has been taken out of me again.  I'll just say this:  Thanks for sending your Christmas poem.  I collect all your writings; I never destroy them; and I appreciate every one of them.  I appreciate every one of them for the insights they give me into this big puzzle.  And about Dylan, let me just say this:  Read the book if you get a chance.  It's all about songwriting, and I would love to hear the reactions of an actual songwriter to it.  (The note further below is the little report I gave Brad and Susie on Dylan's book just before my ill-fated note to you.) [See 31-12-04 note ``\myref{LentzB6}{Dylan and the New Year}'' to J. B. Lentz \& S. J. Lentz.]

\section{03-01-05 \ \ {\it The Alchemical New Year} \ \ (to M. P\'erez-Su\'arez)} \label{PerezSuarez20}

I apologize for taking so long to write you back; I had wanted to send you a timely New Year's greeting myself.  But when I sat down at my computer New Year's Day, I found the news that my mentor, colleague, and friend Asher Peres had died.  That led to a lot of sadness and also a lot of email in its own right.  You'll be receiving an obituary by Asher's students that I'm helping distribute in a few hours.

Let me come back to our happy alchemical musings though.  Your note intrigued me.  I'm looking forward to whatever you're going to write up.

In the meantime, let me send you some of my own mumbo-jumbo.  This comes from something I had written to a {\sl Scientific American\/} reporter earlier in the Fall.  Tell me if you can see the alchemical strain running through it---I think there's one there.  [See 07-07-04 note ``\myref{Musser11}{B}'' to G. Musser.]  Didn't the alchemists too see the big bang as all around us?

Best wishes for the new year!  This year I'm planning to make it quite a productive one (to counterbalance last year).

\section{05-01-05 \ \ {\it Fingers Crossed} \ \ (to T. Mor)} \label{TalMor1}

\btalmo
I was there in the funeral. Did you know he is now buried right next
to Nathan Rosen?  He bought that area, and even planned carefully that
the height of the stone on his grave is a little {\bf below} Rosen's stone.
[This I learnt yesterday, when visiting Aviva and Asher's daughter
Lydia.]
\etalmo

That's a great story; I didn't know that.

\section{07-01-05 \ \ {\it Krugman on Reality (again)} \ \ (to myself)} \label{FuchsC10}

From Paul Krugman, ``Worse Than Fiction,''\ {\sl New York Times}, 7 January 2005:

\bq
I've been thinking of writing a political novel. It will be a bad novel because there won't be any nuance: the villains won't just espouse an ideology I disagree with -- they'll be hypocrites, cranks and scoundrels.

In my bad novel, a famous moralist who demanded national outrage over an affair and writes best-selling books about virtue will turn out to be hiding an expensive gambling habit. A talk radio host who advocates harsh penalties for drug violators will turn out to be hiding his own drug addiction.

In my bad novel, crusaders for moral values will be driven by strange obsessions. One senator's diatribe against gay marriage will link it to ``man on dog'' sex. Another will rant about the dangers of lesbians in high school bathrooms. [\ldots]

In my bad novel the administration will use the slogan ``support the troops'' to suppress criticism of its war policy. But it will ignore repeated complaints that the troops lack armor.

The secretary of defense -- another ``good man,'' according to the president -- won't even bother signing letters to the families of soldiers killed in action.

Last but not least, in my bad novel the president, who portrays himself as the defender of good against evil, will preside over the widespread use of torture.

How did we find ourselves living in a bad novel? It was not ever thus. Hypocrites, cranks and scoundrels have always been with us, on both sides of the aisle. But 9/11 created an environment some liberals summarize with the acronym Iokiyar: it's O.K. if you're a Republican.

The public became unwilling to believe bad things about those who claim to be defending the nation against terrorism. And the hypocrites, cranks and scoundrels of the right, empowered by the public's credulity, have come out in unprecedented force. [\ldots]

Either way, when the Senate confirms Mr.\ Gonzales, it will mean that Iokiyar remains in effect, that the basic rules of ethics don't apply to people aligned with the ruling party. And reality will continue to be worse than any fiction I could write.
\eq

\section{11-01-05 \ \ {\it Our Bayesian QM Meeting --- Why You} \ \ (to P. Diaconis)} \label{Diaconis1}

By now I hope you have seen the invitation I sent you last night for our meeting ``Being Bayesian in a Quantum World'' to be held in Konstanz later this year.  In case you are wondering why I sent it to you, I should say that I heard your talk at the PSA meeting this year, and I liked what I saw.  In particular, it struck me that you could be a very helpful participant in our meeting with regard to the technical side of things:  I.e., helping us distill various philosophical issues into some well-posed mathematical questions \ldots\ and to help us see them solved!

So, I very much hope that you can come and will want to come.  I think we've made great progress recently in inserting the Bayesian conception of probability into quantum mechanics and quantum information theory.  But our effort is still in its infancy, and we need all the help we can get.

To help convince you that there is some serious work to be done, please allow me to point you to a few papers of mine along with colleagues that I hope will give you a flavor of the sorts of technical issues that you might help address or learn about etc., etc., through your participation at our meeting.  These things are along the lines of quantum versions of various de Finetti representation theorems, new kinds of diachronic coherence, etc.:
\begin{itemize}
\item
C. A. Fuchs and R. Schack, ``Unknown Quantum States and Operations, a Bayesian View,''
\quantph{0404156}
\item
C. M. Caves, C. A. Fuchs, and R. Schack, ``Conditions for Compatibility of Quantum State Assignments,''
\quantph{0206110}
\item
C. A. Fuchs, ``Quantum Mechanics as Quantum Information (and only a little more),''
\quantph{0205039}
\item
R. Koenig and R. Renner, ``A de Finetti Representation for Finite Symmetric Quantum States,''
\quantph{0410229}
\end{itemize}

I hope that's enough to pique your interest, and enough to indicate that your presence at the meeting would be very valued!

\section{11-01-05 \ \ {\it Wheeler, Bayes, Schleich} \ \ (to W. P. Schleich)} \label{Schleich1}

Thank you for sharing your memories about Asher.  I have collected many words of remembrance in reply to our announcement, and I will pass them on to Asher's family after the collection looks complete.

I hope you will also note the conference invitation I sent you last night.  I very much hope you will attend.  Caves, {\Schack} and I---and I especially---regard these efforts to get a Bayesian view of quantum mechanics as a detailed carrying-out of the program John Wheeler first set us onto:  ``law without law'' and ``it from bit''.  John Wheeler was a great influence on me, as he was on you.  Your participation in this whole program would be a great addition to it, and our conference, I think, will really get you in the mood.  So, please do say {\it yes\/} (so that we can hold you a spot), even if your schedule is only tentative at the moment!

As it turns out, I looked at your webpage yesterday (to make sure that I had the correct email address for you), and I was reminded of your early work on phase-space representations of quantum mechanics.  Let me draw your attention to a couple of my own papers where I show how one can get something like a coordinate-space representation of the quantum state (on finite-dimensional Hilbert spaces) in terms of a {\it single\/} probability distribution.  And by ``probability distribution'' I really mean that:  It's not something that goes negative like the Wigner function.  Instead, rather than through negativity, the unique quantum features of a system are expressed by other properties of the distribution.  Anyway, here are the papers, and I'd enjoy any feedback you can give me:
\begin{itemize}
\item
``Quantum Mechanics as Quantum Information (and only a little more),'' \quantph{0205039}.
(See particularly Sections 4.2 and 6.1 for the representations I'm talking about.)

\item
``On the Quantumness of a Hilbert Space,'' \quantph{0404122}.
(See equations 28--32 for a particularly clean representation of the variety introduced in the other paper.)
\end{itemize}

I think we're at the beginning of something very big in physics; we're carrying the torch that John first lit.  I hope you'll write to say, ``Yes, I'll see you in Konstanz!''

\subsection{Wolfgang's Preply, ``Asher Peres''}

\bq
Many thanks for your message concerning Asher Peres. I am deeply saddened and shocked by the news. The last time I have met Asher was in Rome during the celebration of De Martini's 70th birthday. At that time he did not look very good and he also looked rather confused. I was worried about him and had a longer talk with him. He told me that he will not travel again without his wife.

I have known Asher since 1984 when he came to Austin to give a talk on the influence of initial state to problems in quantum chaos. It is interesting that recently we have returned to just his approach again in some common work with Mark Raizen and Thomas Seligman. Asher was a great scientist and a wonderful human being who taught us all a lot in science and humanity. He will be certainly missed by the whole community.
\eq

\section{13-01-05 \ \ {\it Another Chance \ldots} \ \ (to G. Brassard)} \label{Brassard43}

\bgb
Here's another chance for us to write the paper that we had planned
for Asher's 70th birthday \ldots
\egb

This is a good point.  I will try, but I can't make a promise at the moment:  Almost all desire to write (even my usual nonsense) has left me for a while.  I have to struggle with each and every word lately---and I've been dreaming of the day that I wouldn't have extant and unfulfilled writing projects hanging over me.  But Asher deserves much, and I will try to find some way to accommodate.  Still I don't venture a promise at the moment.

\section{17-01-05 \ \ {\it Need Answer Quickly} \ \ (to R. W. {\Spekkens})} \label{Spekkens32}

Could we get you to come out and give us a visit this Spring?  Ideally I'd like to have you around for a week or more and give us a seminar on your noncontextuality work (pointing out connections between it and issues in quantum computing).  I ask because Lov has some money to spend and we haven't been able to get rid of it in big chunks.  So we're resorting to a more vigorous short-term visitor program for the short term (and are trying to lure some people here for sabbatical in the longer term).  I'd very much like to have you out, and I've gotten Lov's approval.

The reason I'm asking for a quick answer is that DARPA will be out here on the 20th to review Lov, and he'd like to have as complete a list as possible of visitors to advertise.  So, could you give me a quick reply letting me know if it's possible.

\brws
I would like to hear your thoughts on a result I just came up with the
night before last (hot off the presses).  I like to call it the
``reverse Gleason theorem''.  The theorem states that if a functional $f$ on the space of density operators satisfies
$$
0 \le f(\rho) \le 1,
$$
and
$$
f\left(\sum_k p_k \rho_k\right) = \sum_k p_k f(\rho_k)
$$
for some probability distribution then it can be written as
$f(\rho) = Tr(\rho E)$ for some effect $E$.
\erws

Yeah, I knew that one.  One can prove it with essentially the same techniques that we proved the POVM Gleason theorem with, though you may have come across another way.  In fact it really is the same theorem---certainly technically (except for little issues about scaling), but also conceptually (if one has already accepted that POVMs are refinements of knowledge full stop).

With hindsight, I also know that Holevo knew it sometime in the 70s.  He told me about it as we were flying to Pasadena in 1997; he said I could read about it in his 1980s book.  He kept emphasizing how difficult it was to prove the theorem in the infinite dimensional case, which I didn't care about, and, particularly, which obscured the content of what he was saying to me:  I didn't realize any connection between this and POVM-Gleason until long after our/Busch's papers when I was writing a review on Holevo's second book.  It brought the conversation back to me.

Anyway, Holevo's starting point for the theorem is the operational approach to quantum mechanics (of Davies and Lewis and himself, etc.), and I never liked that motivation.  I.e., he starts by considering the probability of a preparation (identifying a preparation with a density operator), but for a Bayesian about quantum states, that's like talking about a probability of a probability.  At the time I'd have none of that.  It's nice as an expedient in speech, but for a Bayesian it shouldn't be a low-level way of talking.  So more analysis is needed to get at the nub of the formulation.  And that's where I left it at the time.

But who knows, maybe you'll have some spin on it that'll intrigue me afresh.  Come out for a visit, and we can talk about all these things.

\section{18-01-05 \ \ {\it Two More For You} \ \ (to L. K. Grover)} \label{Grover2}

OK, I've got two more visitors for you to mark down for your presentation.

1) Rob Spekkens said he could come April 25th through May 4th, and he would talk on how the source of quantum computational speedup may be the Kochen--Specker theorem (or contextuality as he likes to call it) rather than entanglement per se, etc.  So this is something you can properly call quantum {\it computation\/} research to the funding agents rather than quantum communication (as you had indicated to me earlier).

2) Walter ``Jay'' Lawrence, a professor at Dartmouth College, who works with Zeilinger from time to time, can come for two or three days in mid-March.  So he'll be a cheap visitor---that doesn't help you spend a lot of money, but I'm trying my best.  Anyway, he'll give us a talk on the mutually unbiased basis problem.  And, at least in the case of he and I, we'll be collaborating on the same problem that Wootters and I will be collaborating on when he comes for sabbatical:  representing quantum mechanics by way of these structures.  By the way, for your own edification, you might have a look at Paz and collaborators' stuff on such representations of quantum {\it computing}:
\begin{itemize}
\item
\quantph{0204149}
\item
\quantph{0204150}
\item
\quantph{0410117}
\end{itemize}
I plan to do something similar with my SIC-POVM representation of QM.

Also, Wootters reiterated in a note last night that he'll be coming for a couple of weeks in the Fall and a couple of weeks in the Spring.

I hope that helps you out.

\section{24-01-05 \ \ {\it Proposal Criteria} \ \ (to G. J. Milburn)} \label{Milburn2}

\bgjm
The other aspect that worries me a little are two rules regarding what
they support:
\bq\noindent
{\rm participation in international activities, such as meetings and
workshops, that are likely to move the research agenda of the
field(s) forward, and shape collaborative and complementary research
programs to address significant problems or capture new opportunities}
\eq
but do not support:
\bq\noindent
{\rm conference organisation or attendance where the purpose is the
reporting of research findings.}
\eq
It is my belief that our BBQW is a workshop of the kind that they do
support. Is that also your view?  I will need to make this clear in
the application.
\egjm

I {\it definitely\/} view the conference as a specimen of their first criterion:  That's effectively the whole point of the meeting.  However, I would have never thought that to be mutually exclusive to their second criterion (which seems to be what they are implying).  It is true that I'm hoping we will get some presentations like:
\bv
 {\Spekkens}\\
\quantph{0406166}\\
\quantph{0401052} \medskip\\
Appleby\\
\quantph{0412001} \medskip\\
Renner\\
\quantph{0410229} \medskip\\
Wallace\\
\quantph{0312157} \medskip\\
Srednicki\\
\quantph{0501009} \medskip\\
Wootters\\
\quantph{0401155}
\ev
etc., etc.  To the extent that these are all relatively new works, this conference will be partially devoted to ``the reporting of research findings.''  But the ulterior purpose of that is ``to move the research agenda of the field forward, and shape collaborative and complementary research programs to address significant problems [and] capture new opportunities.''  That is, I want to see our various diverse efforts turned into a single agenda---the BBQW agenda---but for that to happen we will necessarily have to share our recent results with each other.

Does that help any for how to spin your proposal?

\section{25-01-05 \ \ {\it MaxEnt 2005 Invitation} \ \ (to K. H. Knuth)} \label{Knuth1}

Thanks for the call yesterday.  Your invitation to MaxEnt is very flattering.  When I was a grad student, I used to pore over the old proceedings trying to get ideas---so a potential appearance at a MaxEnt would be a great honor.  Also, as you seem to know, I think there is plenty of room to a Bayesian about probabilities at the same time as rejecting the idea of a hidden-variable explanation underneath quantum mechanics.  I like to spread the word of that view anytime I can \ldots\ and I would particularly like to spread the word the attendees of MaxEnt, who would likely make the most progress with it.

But the trouble is, I've already committed myself to lecturing at the Konstanz summer school on ``Philosophy, Probability, and Physics'' and the dates of that are also August 7 to 13.  [An aside:  Also, in fact, the week before that, Aug 1--6, I will be in Konstanz running my own conference ``Being Bayesian in a Quantum World'' (with Carl Caves, {\Ruediger} Schack, and Stephan Hartmann) with about 20 philosophers and 20 quantum information theorists in attendance.]  After that I was to rejoin my family for a week in Munich, where I they will deposited with my in-laws while I'm in Konstanz.

So, I'm torn.  On the one hand I don't want to turn you down.  But on the other, I did already commit to lecture at the summer school.  Let me ask you a couple of questions.  1) Will your budget permit you to fly me to and from Germany, along with my local expenses?  And, 2) would it be acceptable for me to attend only part of MaxEnt?  If there's a yes answer to both of these questions, I could try to see if the Konstanz guys will schedule my lectures for one end or other of the time period, whichever is the more convenient for your meeting.  Just an idea; I don't know if I've got enough nerve to really pull it off (i.e., all the quick flights), but if you're game, I would think about it harder.\footnote{Indeed, my records show that I arrived in San Jose, California at 5:42 PM August 9, and departed 9:37 AM August 12 to head back to Munich, Germany and finish up the other trip!  My travel schedules in those days were like Ezekiel's ``wheels within wheels''.}

\section{30-01-05 \ \ {\it My Last Email to Asher} \ \ (to Lydia Peres-Hari \& Aviva Peres)} \label{LydiaPeres} \label{AvivaPeres}

Below is the very last email I sent Asher.  I don't know whether he
saw it before passing away.  I will share it with you, though it
contains some fairly private information.  Sadly, I did not write
Asher much in the last three months; the note below explains why.  I
regret that very much now.

Asher was a great man, and in many ways, far beyond a friend to
me---a father actually.  I don't know whether he knew that, but I
hope he did.  In physics he had a great deal to do with my learning
to stand on my own two feet.  Without his support in 1994, I very
likely would have stayed crawling.  He gave me confidence in myself
like no one else.

I looked back in my email records after talking to Aviva this
morning.  Since May 9, 1997, I saw that I had 1129 emails from Asher!
And I myself had sent him 789 in that same period.  We also had
several hundred emails beyond that, written between November 7, 1994
and May 1997; unfortunately I used a different filing system back
then, so it is not easy to figure out how many we actually exchanged.
Still, by any measure, that's a lot of writing!  (At least I have all
our notes recorded, so they can be retrieved with a sufficient
effort.)

I loved him, and love him.

\section{31-01-05 \ \ {\it Quote of the Day} \ \ (to N. D. {\Mermin})} \label{Mermin116}

I just ran across the following quote of Nietzsche in Jauch's book on
quantum foundations, and it reminded me of you:
\bq\noindent
    That things have a quality in themselves quite apart from
    interpretation and subjectivity, is an idle hypothesis:  It would
    presuppose that to interpret and to be a subject are not essential,
    that a thing detached from all relations is still a thing.\\
\hspace*{\fill} ---  from {\sl The Will to Power}
\eq
I don't know that you'll like the first part; but I'll bet that the
second part makes the phrase ``correlation without correlata'' pop
into your head (for whatever reason).

\section{31-01-05 \ \ {\it Made My Reservations, Incompleteness, and (Irrelevance of) Decoherence} \ \ (to J. Bub)} \label{Bub18}

OK, I made my reservations at the Jury's Normandy.  For my own records, this is the confirmation number:  94740.

I've been meaning to write you ever since my visit to UMD to try to frame some of the issues you helped me think through with your questions.  But my recent anathema to email has really gotten in the way of any writing bug I may have had.  So, as a stopgap, and perhaps a starting point for further discussion when we get together again, let me attach a file with two old notes in it.  The first concerns the sense in which I think it is fair to say that quantum mechanics is incomplete (even though I think hidden variable completions are the wrong way to go).  [See 07-07-04 note ``\myref{Musser11}{B}'' to G. Musser.]  The second comes out of my ``Notes on a Paulian Idea'' and concerns what I think is the more proper direction for study of ``classicality'' than notions to do with ``decoherence.''  The point of the latter is that the classical or macroscopic world is not a state of nature in need of an explanation, but rather an artifact of our common states of knowledge.  And so, as with much of quantum mechanics, take out the observer, and you take out whether some phenomenon can be classified as classical or not.  Anyway, that's my point of view presently.  Though admittedly it's only the hint of a program.

Luckily, I've been thinking very little about these things the last few weeks.  Instead, I've been calculating, calculating, calculating.  One finds a kind of calm in that that's found nowhere else.

\section{09-02-05 \ \ {\it Measures of Our Field} \ \ (to R. Laflamme)} \label{Laflamme2}

I am putting together a high-level presentation for the Bell Labs executives for next week to give some evidence of how ``big'' the field of quantum information and computing is, and I'm trying to do it by various measures.  I wonder if I can ask you can to help me a little with some numbers?  Is it possible to put an easy figure on how much money Canada is spending on quantum information and computing research each year lately?  Do you know how that breaks down between theory and experiment?  Also do you have an estimate on how many students and postdocs there are there working in the field?  And if you can't answer these questions, do you know where I might try to find the answers?

I apologize for bothering you with this, but it'd help my presentation look spiffier if I had some numbers like this, and I'm guessing you may be able to help (and Gilles Brassard suggested the same).  Thanks for any help you can give!

\subsection{Ray's Reply}

\bq
The only reasonable thing I can offer is a rough guess for the last 4 years:
\begin{itemize}
\item
around \$40M from private sector (\$33M at IQC, \$7M at D Wave)
\item
infrastructure $+$  personnel, about \$17M  (\$12M at IQC, the rest at {\Montreal}/Cal\-ga\-ry/ Toronto and CIAR)
\item
\$1.5M/year for personal grants in the area (NSERC/MITACS/CIPI)
\end{itemize}
Total \$63M in the last 4 years

Hope that helps and good luck with your talk.
\eq

\section{10-02-05 \ \ {\it Being Bayesian in a Quantum World -- for Qubit News} \ \ (to Qubit News)}

A previous post said, ``It looks like they, in fact, are amidst quantum computing scientists.''  Indeed!  And we're hosting our first international conference on the subject.  The meeting is titled ``Being Bayesian in a Quantum World,'' and will be held in Konstanz, Germany, August 1--6, 2005.  It is now closed to further participants---because of space limitations and a desire to keep the discussion intimate---but I'll attach the meeting's announcement and the confirmed-participant list to give some flavor of the subject and to indicate whose writings to turn to if you want to learn more.  [See 10-11-04 note ``\myref{BBQWInvite}{Being Bayesian in a Quantum World --- Invitation}'' to the invitees.] Not all participants are by any means Bayesian (and also, there are many flavors of Bayesianism), but it shows that there are several quantum information scientists who consider the subject serious enough to entertain its potential.

\section{14-02-05 \ \ {\it Boredom} \ \ (to G. L. Comer)} \label{Comer61}

Thanks for sending me that word!  At first I didn't recognize it, but after looking it up I remembered learning it from Bruno de Finetti himself!

Presently I'm putting together the most boring talk I've ever put together.  It's for the president of Bell Labs Research.  Part of the problem is that they're forcing me to write it in PowerPoint which I only started to learn yesterday.  Oh well, it will remain boring:  That is the price they will pay for depriving me of my freedom!

\section{14-02-05 \ \ {\it Fenomeno Aleatorio} \ \ (to G. L. Comer)} \label{Comer62}

The memories are slowly coming back to me.  See the 27 April 1998 letter to {\Ruediger} Schack in my {\sl Notes on a Paulian Idea}.  I hope I never forget aleatoric again!

\section{18-02-05 \ \ {\it Lost Opportunity} \ \ (to R. E. Slusher)} \label{Slusher7}

I just read this quote in Feynman's 1982 paper on quantum computing, and boy how I wish I had used it to end my talks to Jaffe and Bishop.  It might have also benefited Bernie when Bishop was being stubborn about quantum limits too.  Here it is:
\bq\noindent
     \ldots\ full attention and acceptance of the quantum mechanical
     problem---the challenge of explaining quantum mechanical
     phenomena---has to be put into the argument, and therefore
     these phenomena have to be understood very well in analyzing
     the situation.  And I'm not happy with all the analyses that
     go with just the classical theory, because nature isn't
     classical, dammit, and if you want to make a simulation of
     nature, you'd better make it quantum mechanical, and by golly
     it's a wonderful problem, because it doesn't look so easy.
     Thank you.
\eq

Let's make that road trip to Peirce's house sometime.

\section{22-02-05 \ \ {\it Quantum Lobster} \ \ (to J. B. Lentz \& S. J. Lentz)} \label{LentzB6.1} \label{LentzS4.1}

A friend brought this link to my attention:
\bv
\myurl{http://www.nature.com/news/2005/050214/full/050214-4.html}.
\ev
You see, quantum mechanics is indeed responsible for the most important things in life!

\section{02-03-05 \ \ {\it Will You Find It Interesting?}\ \ \ (to M. P\'erez-Su\'arez)} \label{PerezSuarez21}

Thanks for bringing that book to my attention!  It looks exciting.

I'm glad to hear that you got the approval for a very long stay in Albuquerque.  I think you'll find the time productive (much more productive than your time with me).

I'm on my way to Poland at the moment to talk on quantum foundations.  Why I'm going there, I'm not quite sure, \ldots\ I guess it's because I'm easily flattered.  But you're right:  aside from this anomaly, I've been spending most of my time trying to calculate something.  (Unfortunately to little success.)

\subsection{Marcos's Preply}

\bq
I apologize for my silence.  But I have found a moment to tell you that you should take a look at this book, which I have been browsing in the last days:
\begin{center}
\myurl{http://www.springer.com/physics/quantum+physics/book/978-3-540-20856-3}.
\end{center}
William James, the ``ideal of the detached observer'', subjectivism \ldots\  a lot of subjects of your interest (even though you have readjusted yourself ---or was it the world that made the readjustment on you?--- a ``true physicist'' once again) are considered in the book (surprisingly enough, Harald's work is not mentioned).

I'll return to you in a few days (at least I hope so) in order to make my point on the ``alchemical connection'' and other issues.

Good news (at least, they are for me): The application for my stay with Carl has been accepted for the whole period of six months. I'll tell Carl about this, and I'll try to begin arranging things concerning it as soon as possible.
\eq

\section{02-03-05 \ \ {\it And Yesterday} \ \ (to R. E. Slusher)} \label{Slusher8}

\bres
I finished ``The Sentiment of Rationality'' and some other James stuff. Now I need more counseling from you!
I am determined to understand the dynamics of measurement by an apparatus. I still don't see the human observer being involved (except to come by and readout the apparatus hard drive) --- especially since I've been reading a lot about how complex the ``SELF'' is.
\eres

You can't blithely say the parenthetical remark.  The very most important thing we have learned from quantum mechanics is that there is no such thing as readout.  Supplement your William James (particularly the essay {\it Pragmatism}) with:
\bv
N. D. Mermin, ``Hidden variables and the two theorems of John Bell,'' Rev.\ Mod.\ Phys.\ {\bf 65}, 803 (1993).
\ev
That's one of the best articles on the Kochen--Specker theorem you can read.

But you're right, human is not the essential concept \ldots\ unless to bear probabilities and make decisions in the face of uncertainty is to be human.  (But I don't think it is.)

I'll gladly talk about this when I get back.

\section{02-03-05 \ \ {\it That Blasted Fuchs} \ \ (to C. M. {\Caves})} \label{Caves79.1.1}

Why am I ``that blasted Fuchs''?  Am I not already paying enough for my sins?

The {\sl Foundations of Physics\/} issues are now scheduled for June, July, and August according to van der Merwe.

I'm flying to Poland at the moment to give some lectures on quantum foundations---why I accepted this invitation, I don't really know.  In any case, I hope they're ready for the blast you indicate.

\section{07-03-05 \ \ {\it I Won't Be There Tuesday}\ \ \ (to J. B. Lentz)} \label{LentzB7}

What am I doing in Wroc{\l}aw?  It's the usual evangelical mission.  They invited me to give some lectures on my point of view of quantum mechanics.  When confronted with an offer like that, it's hard to turn down the invitation \ldots\ no matter where it is.  It's the usual story we encounter in most aspects of life:  One tries one's best to increase his children's chances of survival.  Mostly I did it out of a sense of duty.  The trip however, has turned out to be a pleasant surprise.  They have treated me royally.  The vice-president of the city even attended my first (more public) lecture, and presented me with a nice coffee-table book about the city.  The food, the cognac, the jazz restaurants, have all been very nice too.

\section{07-03-05 \ \ {\it No Laws at All} \ \ (to C. G. {\Timpson})} \label{Timpson6}

Indeed, I love that quote!  Thanks for sending it.  I have been told by someone (can't remember who at the moment) that I should read Cartwright's book, but until now I've put it off.  However you have an awfully good track-record with your recommendations to me of books, and this quote definitely seals it.  Better than reading it:  I'll buy it and then read it!

I'm in Wroc{\l}aw in Poland at the moment, finally enjoying a beer after all the hard work of this week.  When I wasn't lecturing, I couldn't seem to stop my obsession with a certain calculation:  It kept me up day and night, and thus the jetlag was much worse than usual.  But at least I learned a lot by my random walk in Platoland.

\subsection{Chris's Preply}

\bq
I am re-reading Cartwright's {\sl How the Laws of Physics Lie} and am being struck how the position in that book seems in many ways sympathetic to the kind of view of science, and qm in particular, you'd like to adopt.

Here in particular is a little passage I thought would please you:
\bq\noindent
Covering-law theorists tend to think that nature is well-regulated; in the extreme, that there is a law to cover every case. I do not. I imagine that natural objects are much like people in societies. Their behaviour is constrained by some specific laws and by a handful of general principles, but it is not determined in detail, even statistically. What happens on most occasions is dictated by no law at all.  (OUP 1983, p.49)
\eq
\eq

\section{08-03-05 \ \ {\it Spelled Wroc{\l}aw, Pronounced Vrotswav, but Germans Still Say Breslau} \ \ (to D. B. L. Baker)} \label{Baker12}

``It's been a long time since I've written you from a far-away place.
I guess that's because far-away places have become little more than a
pain for me.  Trip after trip seems to drone on, and I've lost so
much purpose.  But tonight I feel like writing something.  I'm
captured by the charm of what I see around me.  The snow coming down,
the muffled sounds and the romance in this city square; a kiss I
spied in a dark corner of the cathedral.''

I wrote those lines in my head last night, thinking that I would go
back to my hotel room, have a beer or two, and write you a long note.
It didn't happen.  I had the beer, you see, but I got wrapped up in
music and the newspaper.  Hans Bethe died yesterday at the age of 98,
and I got sucked into stories about him.  He was a great physicist
and a great man.  The last time I saw him, about 6 years ago, he was
still reporting his research!  In slow motion, but he was still
reporting topical research!  Amazing.  Get on Google News if you can,
and read about him.  He was an important part of America and of all
time; he brought the cores of stars into the hands of man.  You read
about someone like that, and I challenge you to think that man is an
insignificant force in nature.

It's early morning, and I just arrived at Wroc{\l}aw airport.  I have a tortured route today:  First to Munich, then to Paris, then to Chicago, and finally to Newark.  Humbug, bah humbug.  But at least I found last night romantic in the extreme.  There were such big flakes of snow coming down, while a few people in the city square waited patiently for the Zamboni to smooth the ice in the skating rink; the couple I saw at the corner of the cathedral; the music trickling out of the basement clubs.  It was those things combined with all that I had seen and heard this week; it came to a sort of fester.  It's a strange world here.  One that is still getting over World War II (everyone still talks about it, even the twenty-somethings), and one that reminisces the ``funny'' days of communism.  It is a world a few years behind ours, at least emotionally.

I had the nicest experience and was treated royally by my hosts.  What's the phrase?  ``Famous in Peoria,'' or something like that?  The vice-president of the city even attended my first lecture and presented me with a big coffee-table book (of city things of course).  There was a set of people that took a four hour train-ride to hear my lecture.  And, I can't believe this one:  There was a student who wanted my autograph!  That's a first \ldots\ but then again it was Peoria \ldots\ I mean Wroc{\l}aw.  Ah, what's the difference.  Still, it was touching.  I wish I could live up to that young guy's dreams, but I don't have it in me anymore.  Most of the time I feel like such a has-been.

But last night, everything felt right just briefly.  In the city square I felt so small, but in just the right way.  You know what I mean?  I felt like \ldots

Well, once again, I didn't manage to finish my note to you before having to get on the plane.  Now after all the delays and the nearly missed flights, the mood is broken.  Munich has passed, Paris has passed, and I'm somewhere over the Atlantic.  I strongly suspect however that my luggage isn't.  Over the Atlantic, that is.  Sad how some bad flights can shatter a day \ldots\ which I guess gets me back to where I started:  ``It's been a long time since I've written you from a far-away place.  I guess that's because far-away places have become little more than a pain for me.''

Maybe I'll have better luck the next time I try to write you.

\section{08-03-05 \ \ {\it Firebrands, at Big, Small, and Small-minded Universities} \ \ (to G. L. Comer)} \label{Comer63}

I want to finally take some time and think about your life.  At the moment I'm somewhere between Paris and Chicago, and I think somewhat more than halfway.  (Aha, they just put the map back up and we're over Canada now.)  I started in Wroc{\l}aw, Poland this morning (pronounced Vrotswav!), then hopped to Munich, and then to Paris.  What a tortured flight schedule!

I'm really proud to hear that you got two ``excellent''s on your NSF proposal.  Now that I've served on a panel, I know that those marks don't come easily.  Of the 13 proposals I was assigned, I only gave one mark like that, and that was because it was just over and above good science with respect to the other proposals.  And at least on our own panel (mathematical physics), I saw everyone else very conscientiously doing similar things.  Is there any indication that you may get funding?

As an aside, I tell you, serving on that NSF panel was a real eye-opener for me on the inner workings of these things.  It was some of the hardest work I've done in ages, and in my own case, it sometimes felt like I was in the middle of the movie {\sl Sophie's Choice}.  The Nazis said one of your children will live and the other will die; you make the choice or they will both die.  It was really agonizing like that toward the end.  The good side of the process was to see how conscientious everyone on the panel was for trying to see science develop:  I hadn't quite expected that, but really I only saw the best brought out in everyone.  Another pleasant surprise was to meet up with Louise Dolan after all these years; we sat beside each other.  For some reason I remember being slightly frightened of her at UNC---there was some kind of impression that she was a firebrand---but she really, really impressed me this time around.  She was a very motherly figure for young scientists---trying to support them over bigger names---and she was quite the champion for proposals from ``this or that guy stuck in some small out-of-the-way college.''  You literally came to mind when she was defending that idea once.  She said (this quote is pretty much literal), ``He could probably use a break.  What he's doing there is a thankless task.''  But what do you do when there are 74 proposals and at most 15 will get funded.  You can only hope that there is someone on the panel who will fight like hell for your proposal, or at least sees yours as the very best of a field that wouldn't otherwise be represented.  (BTW, do you know that there was not one pure GR proposal in our ranks, unless possibly there was one that got thrown out before the second day when we all convened out of our smaller groups.  Any idea why I didn't see anything from GR?)

One final aside, I also gained an appreciation for the kind of glibness one sees in the final ``panel summaries'' that come from these things.  The thing that I've come to appreciate is that they're not really reflective of the amount of thought or debate that went on about a proposal.  The reason is that the person who writes that summary is assigned as a ``scribe'' for the debate that ensues about that proposal and is usually assigned that role because he is not an expert on that particular subject.  So, the summary can come off awfully weak because of that.  Also, those summaries are thrown together the last afternoon as everyone is frantically trying not to miss their flights and don't reflect the interminable debates that led up to that point.

OK, enough of that.  As I say, I'm proud to hear of your excellents.

I liked these passages in your notes very much:
\bgc
It's another aspect of the dictionary problem, which is once you
allow yourself to be labeled, then you get attached to yourself
everything that has been written and discussed with regard to that
particular label. If I have a free will, I guess I should also have
my own philosophical label, the philosophy of Greg, which is
everything that makes Greg Greg.  And whatever that philosophy is
changes every day, as Greg becomes more and more.

Now, substitute you for me, Chris for Greg, or Kiki, or Holly, or
Sarah, or Katie, or Sam, or Emma Jane, etc., until all permutations
are exhausted.

Maybe we can be allowed the same freedom as in music, and that is
that the music written on the page is not meant to be absolute, but
rather just a general guideline as to what the musician could pursue
as the music is being played.  In fact, I guess that's what the
musician necessarily has to do, since the music on the page can never
be equal to the music played.
\egc

The only thing I wonder is how do you take into account a case like mine, where the peak came somewhere around 2002:  Since then, I've become less and less.  (That's almost a joke, but unfortunately, not quite.)  I told the story of Wheeler's flying equations to the Poles the other night, and I was very pleased to see a couple of their eyes glow with a kind of real understanding.  The point:
\bgc
By now, somewhat old themes as far as human lives go, but nevertheless
worth repeating, and putting in a different light.
\egc
Amen.  In the lingo of Richard Rorty, we're antirepresentationalists.  But, I know, I know, that's a label!

\ldots\ \ldots\ \ldots\ \ldots\ \ldots\ \ldots\ \ldots\

As it turns out, at the moment of writing this sentence, I am now waiting for the remainder of the flight to board for Newark.  A lot of annoying things ensued since my last sentence above.  Of course, my luggage was lost.  What's particularly annoying about that is that the suitcase contains my full cache of quantum foundations transparencies, and 90\% of them have never been scanned into my computer.  So, it just puts me on edge.  The reason the transparencies were in my suitcase was because I had not expected to check my luggage---in fact I very carefully planned ahead this time not to check anything because of my ridiculous number of connections.  But wouldn't you know, these damned over-friendly Polish hosts of mine surprised me with a ``small gift'' for my children at the airport at the last minute:  A big bag full of Polish dolls, books, etc.  When I saw it, my heart sank:  I knew immediately that I'd have to check my luggage.  And just as immediately, I said to myself, ``It'll be lost.''

So, here's the philosophical question I'll leave you with.  In this indeterministic world of ours---this world still under construction---why can we foresee some things but not others?

\section{11-03-05 \ \ {\it A Quiet Fuchs \ldots\ for once} \ \ (to H. Price)} \label{Price4}

I'm back from Poland, and I'm trying to get caught up on my email and other things.  (Among them, emails from long before Poland!)

Question:  Have you sent off your ``Truth as Convenient Friction'' to the memorial volume?  Sorry I never sent you that list of typos.  Trouble was, I wanted to reread it, and reread the relevant Rorty, and make some idiosyncratic comments at the same time \ldots\ maybe to show you that I really do have a philosophical streak.  It was laid out beautifully in my mind:  I would title the note ``Value Added.''  But then a bad thing happened---I got obsessed with a certain calculation, and I haven't been able to let it go for over two months now!

Still, if you haven't sent your paper to the publishers, I might use that as an excuse to reread it this weekend (skipping Rorty, and making no promises to even attempt value-addedness, but at least refreshing myself of your arguments as a kind of payment for the editorial work).

\section{11-03-05 \ \ {\it Chance} \ \ (to J. Ismael)} \label{Ismael0}

I'm just coming back to email after a long hiatus.  (Even for a guy whose professional career is based on his knack for correspondence, sometimes a hiatus is overdue.)  Anyway, thanks for the paper.  I just printed it out and will {\it certainly\/} read it.  Too bad you're not coming to Konstanz to read it to us!

There's a good (subjective) chance I'll be in Sydney at your meeting, but I don't want to give it a certainty yet.

\bji
But be patient; after hearing your talk in October, and rereading some
of your work, I realized we share something perhaps more important
than the surface disagreement:  viz., the conviction that probability
is information.
\eji

Without doubt, I'll be patient.  But it's hard at this point to imagine that our disagreement will only be a surface one:  I have so much invested in taking a subjective interpretation of probability down into quantum theory's core---as a research program---that a reversal at this point might amount to schizophrenia.  In fact, I think my ideas on quantum foundations hang together {\it only\/} because of a de-Finetti style interpretation of probability.  But still, I know I'll learn a lot by studying your paper.  And we shall see.

I'm curious to know which of my things you've read, so that I can know which version of me you've seen.  I'm fairly fluid in some aspects.  But, in particular, if I think I might further our discussion on chance by pointing you to some particular passage, it would be useful to know which part of my writings you've digested.

\section{11-03-05 \ \ {\it Historical Notes on Your Basis} \ \ (to C. M. {\Caves})} \label{Caves79.1.2}

Question:  What is the origin of the orthonormal basis of Hermitian operators you review in Section 2 of your ``Qudit Entanglement'' paper with Rungta and others?  Is that standard lore from somewhere?  Who introduced it?  More importantly, are there any other interesting properties that it possesses that you haven't tabulated there?  Where do I look?

\subsection{Carl's Reply}

\bq
We learned about it in the first of the following, and it gets a more thorough treatment in the latter:
\begin{itemize}
\item
K.~Lendi, ``Evolution Matrix in a Coherence Vector Formulation for Quantum Markovian Master Equations of $N$-Level Systems,'' J. Phys.\ A {\bf 20}, 15--23 (1987).
\item
R.~Alicki and K.~Lendi, {\sl Quantum Dynamical Semigroups and Applications}, Lecture Notes in Physics, Vol.\ 286, (Springer-Verlag, Berlin, 1987).
\end{itemize}

We've recently been thinking it's a bad choice and have been working in general orthonormal operator bases.  I've appended a half-finished document by Menicucci and me on Bloch space.  Menicucci was at UNM doing his Princeton's Master's thesis with me, and he went crazy over my notation and extended it in ways that are almost embarrassing.  But he did end up proving in a straightforward way the conditions for a Bloch rotation to corresponding to unitary conjugation, something that I had proved only in a very indirect way.
\eq

\section{15-03-05 \ \ {\it Down with Properties!}\ \ \ (to S. J. van {\Enk})} \label{vanEnk1}

\begin{itemize}
\item
The wavefunction is not a property of the quantum system.
\item
Nor is a measurement outcome the signifier of some property (either one pre-existent or one created from the measurement process itself).
\item
States and POVMs play essentially the same roles:  Subjective judgments.
\item
Measurement outcomes (I would guess) do have a kind of reality---they do come into existence via the measurement process---but they don't signify properties.  Rather they are rawer than that:  Kind of like the nodes in your graphs.  Anyway, see ``Preamble'' at the end of the samizdat.  New outcome, new graph?  Just playing with ideas.
\end{itemize}

\subsection{Steven's Preply}

\bq
When I read your email to {\Spekkens} I saw why you're interested in the book about how Asians think.

I liked {\Rorty}'s description of the world and his example of numbers. That somehow made me think of dictionaries, where words are defined only in relation to other words. Whereas I used to think a dictionary, therefore, doesn't define anything, thanks to {\Rorty} I now think the
opposite: it determines everything! In fact, I think if you give me a Swahili dictionary I could in principle translate almost every Swahili word into English by just looking at how all Swahili words relate to one another.

That in turn made me think of graphs and the graph-isomorphism problem: a graph is defined completely by relations between all nodes, where the nodes only have completely meaningless labels. The graph-isomorphism problem is like finding the translation of a dictionary.

And that, I suppose, is also the essence of {\Mermin}'s correlations without correlata.
\eq

\section{16-03-05 \ \ {\it Moment of Insanity} \ \ (to C. M. {\Caves})} \label{Caves79.1.3}

Thanks for the notes on Hermitian bases.  I never expected such a detailed reply!  I wish I had it in me to be a real scientist like you again.

Let me tell you about my moment of insanity.  I just asked Jennifer Chayes at Microsoft Research if MS would be willing to hire the two of us as a package to start a Bayesian-quantum-info research group there.  She'll get a good laugh out of the note, probably.  Why did I write it?  I'm not sure:  It just happened to be on my mind.  And why was it on my mind even though I tell myself I could never leave New Jersey?  That's a mystery even to me.

\section{16-03-05 \ \ {\it Stein} \ \ (to A. Radosz)} \label{Radosz1}

I've been meaning to write to you since my return, but I've only now gotten your email address.

I did find the Gertrude Stein quote you mentioned in one of my papers!  It was in a review I wrote of Alexander Holevo's new book:
\bq\noindent
The clear-cut purpose of the monograph is not to go through the mechanics of each theorem, but to paint
a picture in modern art. Throughout, there is an emphasis on how each and every quantum
result differs from its analog in the classical probabilistic model. One does, however, fear
that a central illuminating thread is somehow missing from the exposition, though the author
cannot be blamed: It is a thread no one has yet seen. A radical departure from classical
probability theory, yes. But why? What is the all-important, undeniable physical fact that
forces this revised calculus upon us? Can we state it in a way that does not consist of a dozen
disparate theorems? One day we will get there. In the mean time, though, an exposition like
this is part of the necessary journey. It is likely to be as Gertrude Stein said of the more
classical kind of modern art, ``It looks strange and it looks strange and it looks very strange;
and then suddenly it doesn't look strange at all and you can't understand what made it look
strange in the first place.''
\eq
I'm sorry I forgot that.  I was first alerted to that particular quote of Stein's by John Wheeler.  To see another Gertrude Stein quote in action, see an old paper of mine with Carl Caves:  \quantph{9601025}, ``Quantum information: How much information in a state vector?''.  See page 26 and footnote 43.

It was great meeting you and dining with you, and thanks again for the lovely introduction you gave to my talk!

\section{16-03-05 \ \ {\it Reply from the Lusitania} \ \ (to K. H. Knuth)} \label{Knuth2}

\bkhk
I have been extremely interested in the logic of questions, and have
recently found that they follow the free distributive algebra and possess a
measure that is a natural generalization of information theory.  I can now
perform computations that determine the degree to which one question
answers another, and I have been applying some of this work here at NASA to
design intelligent instruments that decide which measurements to take next
given the data that they already possess.  You may be interested in the
three papers I am attaching for your perusal \ldots. I will be giving a tutorial on this work at MaxEnt 2005 on Sunday.
\ekhk

I have been mightily impressed by a skim of some of your papers.  I hope I get a chance to actually understand them over the course of this Spring and Summer.  In particular, the two that most rise to my attention at the moment are ``What is a Question?''\ and ``Lattice Duality.''  For about three years now I've been emphasizing the direct analogy between the formal structure of quantum measurements (aka questions) and quantum states (aka Bayesian degrees of belief in the quantum context).  If a quantum state is akin to a probability (and therefore subjective), a quantum measurement is akin to a conditional probability (and therefore subjective)---that's the kind of idea.  You can find some of these thoughts tabulated in \quantph{0205039}, particularly in Section 6, but I go into much greater depth in {\sl Quantum States:\ What the Hell Are They?\/} posted at my website.  (Unfortunately, the latter has not been distilled into a proper paper yet, but maybe it's entertaining nonetheless.)  As I understand it, these papers of yours (and the Cox work that started them off, which I did not know about) seem to indicate that this is a much larger phenomenon than I had contemplated.  I'll definitely have to understand this stuff that you're uncovering.

\section{17-03-05 \ \ {\it It's All About Gram Matrices} \ \ (to K. R. Duffy)} \label{Duffy4}

\noindent Dear Old Friend,\medskip

By the time you read this note, I know you'll have a horrible headache (assuming you read email at all the day after St.\ Patty's Day), but I thought I'd say hi anyway while the day is still here.  Ireland has been on the whole family's mind all week---in lots of ways we all miss it, and we've been talking about it a lot and reminiscing.  I even endured watching Bobby Flay (who I don't like) on the Food Channel last night so that we could see ``A Taste of Ireland.''   The scenery made it worth it (at least whenever Flay wasn't in the same picture).  Hey, we even put an Irish flag on the pole at the front of our house this week, finally retiring the University of Texas flag we put up when we first moved in.

How are things go with you?  For myself, I've been scrambling the last two months to get a good result to present at the JTL meeting---not a lie.  At the other meetings I've been going to recently, I've been presenting slight variations on the same-old-same-old, knowing that I could always get away with it.  But at John's meeting, I want to present some mathematics that would have maybe put a smile on his face.  I keep thinking I'm this close ``'', then it all vanishes.  Still, there's a running chance!  And I'm little if not determined.

Say hi to Jo and Dave and anyone else I'd remember if you see them.  And take some Nurofen!

\section{21-03-05 \ \ {\it Changing the World} \ \ (to C. H. {\Bennett})} \label{Bennett39}

Here's a little story I've got to tell you and {\Theo}.  Ever since you wrote me these lines soon after Katie's birth,
\bcb
{\Theo} enjoyed your birth announcement for Katie.  When I
read her that Katie was beautiful in every way and would shape
the world by her presence and will, she said she wasn't
surprised.  That is just what she would expect from another baby
that came from the same parents as  Emma.  {\Theo} told me that her grandfather, when told of the birth of a baby from someone he
admired, would typically say ``That baby will change the
world.'' (After the world changes it a few hundred times, I couldn't help thinking.) Congratulations to all four of you from the two of
us.
\ecb
I have from time to time told her that one day she will change the
world---that that's her destiny.  Last night I told her that as I was
tucking her into bed and asked, ``Do you know it?''  How did she
reply? ``Yeah. Today I changed the batteries in my toy.''

I loved it and had to record it.

\section{25-03-05 \ \ {\it [SPAM?]\ At It Again:\ Delirium Quantum}\ \ \ (to H. M. Wiseman)} \label{Wiseman15}

\bhw
I'm writing a foundational paper (partly to prepare myself for your workshop, but its been on the go for years)
 and want to quote parts of the following:
\bq\noindent
{\rm Any single application of quantum theory is about ME, only me. It is about MY
interventions, MY expectations for their consequences \ldots\ and MY reevaluations of MY old expectations in the light of those consequences.
It is noncommittal beyond that.  This is not solipsism; it is simply a statement of the subject matter. \ldots\
What quantum theory does is provide a framework for structuring MY expectations for the consequences of MY interventions upon the external world.
At least that is what the formal structure is about. There is no ``we,'' there is no ``our.''}
\eq
from your ``Delirium Quantum'' collection. Is this going to be made public any time soon so I can give
it a proper reference?
\ehw

Hmm, what do you think?  You think I should post that?  How much more foolish than usual would I look with this one?  [See ``Delirium Quantum,'' \arxiv{0906.1968v1}.]

Another question.  Guido Bacciagaluppi once told me that ``Delirium Quantum'' was an abomination of a title:  It should be ``Delirium Quanticum'' so that both words are Latin (after I explained to him that I was thinking of ``delirium tremens'' when making the title).  In a way he's right, but on the other hand I've got a certain redneck kind-of persona to preserve:  I'm not sure I want to get too fancy with my titles.  The present one seems just incorrect enough to fit me.

\section{30-03-05 \ \ {\it Civil Disobedience} \ \ (to G. L. Comer)} \label{Comer64}

Thanks for your notes bringing me up-to-date on your tripartite
career: poetry, music, and physics.  And I applaud your efforts to
find linearity in each of them!  (I.e., the place of the
observer/actor/agent in each of them.) [\ldots]

I look forward to reading your discussion of linearity in GR.
Definitely, send it by me as it comes along.

For myself, I've been obsessed by a basic linear algebra for two
months now, and I've thrown myself into it like I was a graduate
student again.  That is one of the great things about quantum
information:  One starts to realize how many very basic questions in
finite-dimensional linear algebra have simply never been explored.
And sometimes, they're really tough questions despite the finite
dimensions.  Take the linear vector space of $d \times d$ complex
matrices.  That's a $d^2$ dimensional vector space, and you can equip
it with an inner product, say the Frobenius one $\tr(A^\dagger
B)$ where $\dagger$ denotes Hermitian conjugation.  Question:  Can
one always construct a complete orthonormal basis for this vector
space consisting solely of unitary matrices?  It's not obvious, but
the answer is yes---it took Schwinger to note it.  New question:  Can
one always construct a complete orthonormal basis consisting of
Hermitian matrices?  Much easier, and the answer is yes.  New
question:  Can one always construct a complete orthonormal basis
consisting of positive semidefinite Hermitian matrices?  Much harder!
But the answer is no.  In fact, it looks like the closest one can
come is to get a normalized basis with pairwise inner products of
value $1/(d+1)$. But proving it, that has turned out to be really
hard. Even constructing a set of (positive semidefinite Hermitian)
matrices with those inner products has turned out to be elusive.
Constructions exist for $d=2, \ldots, 8$ and 19, but there's only
numerical evidence for the other dimensions up to 45 (and after 45
there's not even numerical evidence, the problem becomes too big for
easily accessible computers).  But dammit, it's an important
question.  For these are precisely the things we should have been
using all along to express quantum mechanics, rather than expansions
in terms of orthonormal bases.

But such is the life!  Interpreting and reinterpreting the scripture.

\section{01-04-05 \ \ {\it My Bloodline} \ \ (to G. L. Comer, G. Herling, and C. M. Caves)} \label{Comer65} \label{Herling2} \label{Caves79.2}

Hey look at this:
\begin{center}
\myurl{http://www.genealogy.math.ndsu.nodak.edu/}
\end{center}
and backtrack from there.  I just discovered that Arnold Sommerfeld is my academic great-great-great grandfather:  Sommerfeld $\longrightarrow$ Herzfeld $\longrightarrow$ Wheeler $\longrightarrow$ Thorne $\longrightarrow$ Caves $\longrightarrow$ Fuchs.

I think that's particularly funny because I once told Carl he was ``the Arnold Sommerfeld of quantum information.''  Maybe I could somehow mystically sense that long lost ancestry.

\section{01-04-05 \ \ {\it Contradiction, Disappointing} \ \ (to G. L. Comer, G. Herling, and C. M. Caves)} \label{Comer66} \label{Herling3} \label{Caves79.3}

On the other hand, I just found this on the web, in a page titled ``Karl Herzfeld Memorial Lecture.''  It indicates that Herzfeld got his PhD in Vienna and was {\it only\/} a collaborator of Sommerfeld's.  Too bad; the other was a better story!

\bq
Karl F.\ Herzfeld was born in Vienna, Austria in 1892. He completed his doctoral studies at the University of Vienna and studied with Einstein in Zurich in 1914. After serving in the Austrian army during World War I, he became a faculty member at the University of Munich, where he stayed from 1919 until 1926, collaborating with Sommerfeld. Among his students were Pauli, Heitler, and Heisenberg.

In 1926, Herzfeld came to the United States to accept a professorship at Johns Hopkins University. In 1936 he became Chairman of the Physics Department at The Catholic University of America, a position he held until 1961. He continued to teach until 1970, and to be active professionally until his death in 1978.

Among Herzfeld's many research interests were the mechanisms of sound absorption in gases, liquids, and solids, the theory of the binding of atoms in crystal lattices, and the foundations of statistical mechanics and thermodynamics. During his distinguished career he was the author of than 130 scientific articles and co-author of fourteen books. He received many honorary degrees, medals, and was awarded the Navy's highest civilian award. He was elected to membership in the National Academy of Science.

No description of Herzfeld that did not include his profound interest in philosophy and theology would be accurate or complete. He contributed many scholarly articles on the relationship between the humanities and science. Karl Herzfeld's impact on science derives not only from his own scholarship, but also from his influence on those who studied under him. As the beloved professor of hundreds of students, he had a profound impact on both the professional and spiritual aspects of their lives.\footnote{From \myurl{http://physics.cua.edu/KarlHerzfeld.cfm}, possibly.}
\eq

\section{02-04-05 \ \ {\it Entanglement, IVP, and QM} \ \ (to G. L. Comer)} \label{Comer67}

\bgc
Now, what about entanglement?  Honestly, there is something
about this that I've always found frustrating.  Invariably, when I
try to read about this, there is always some statement like ``\ldots\ and
because the electrons were paired-up in some orbital of an atom,
they `interact' and become entangled.''  This is so frustrating!
Because no one ever seems to explain what the ``interaction''
actually is!  Once this magic statement is made, then an amplitude
is written down that represents the entangled objects, and I'm
left with my jaw hanging down to the floor, in a Homer Simpson,
or Barney, drunken-stupor kind of way.  Drool all over my face.
\egc

Yeah, entanglement has nothing to do with interaction \ldots\ at least not necessarily so.  It has everything to do with states of knowledge---that's the overriding issue (and as you say, that's like an initial condition).  For a very simple demonstration of this look at: \quantph{9904042}.  It's a simple example that shows that entanglement is really only about how you sort your subsets.  No knowledge, no entanglement.  Knowledge that gives you the right perspective, and, boom, entanglement.

Sayeth the preacher:  At a later stage in physics, we'll forget about the ``boom'' part, and just talk about the knowledge that gives us the right perspective.  That's essentially the beginning and the end of the story.

\section{06-04-05 \ \ {\it Internal Realism?}\ \ \ (to J. E. {\Sipe})} \label{Sipe3}

Finally, let me point you to \quantph{0204146}.  The relevant sections for our {\Putnam} discussion last night are 4 and 5.  I'm curious to know whether that sounds like {\Putnam}'s internal realism to you (as I think Rob thought it did), or whether it sounds like a different beast?  Of course, I should read {\Putnam} myself \ldots\ but your opinion might tide me over until I get a chance to.

\section{06-04-05 \ \ {\it Being Bayesian in a Quantum World} \ \ (to G. Musser)} \label{Musser18}

I am sorry I've been out of contact for so long, but of course nothing has worked out on the schedule I had wanted to keep to this year.  In particular, I hope you and {\sl Scientific American\/} might still be interested in an article from me.  I have decided that if I were to write one, I would like to title it ``Being Bayesian in a Quantum World'' and with it convey some of the excitement a rather large community of us are starting to feel for this aspect of the ``Quantum Foundations in the Light of Quantum Information'' program I described to you previously.

Now, what I think I'd like to do is start putting some words together right after our big meeting on the subject in Konstanz in early August of this year.  That way I think I'd be better able to take into account all the things I see and hear around me there, which would probably quite benefit the article.  The particular subject of the article would pretty much be along the lines of a little abstract I just wrote up for the MaxEnt and Bayesian Methods meeting, which will be held in San Jose just after BBQW:  I'll place the abstract below so that you might get a feel for the subject matter.

Would {\sl Scientific American\/} be interested in an article from me along these lines?

The abstract below is actually a variation on some paragraphs I wrote to announce our Konstanz meeting.  Thus, in a separate email, I think I will send you the full (original) announcement.  More particularly, if you're interested you can take a look at our website to see the final list of participants, the program, the research questions, etc.  Here's the web address:
\begin{center}
\myurl[http://web.archive.org/web/20090511060206/http://www.uni-konstanz.de/ppm/events/bbqw2005/]{http://web.archive.org/web/20090511060206/http:// www.uni-konstanz.de/ppm/events/bbqw2005/}.
\end{center}
At least given your seeming excitement in our discussions last year, you might just be interested---I hope so!  Anyway, one idea that comes to mind is that if you think you might like to cover the meeting for a little description or blurb or whatever in {\sl Scientific American}, you are certainly welcome to come.  There will be plenty of good people to talk to there, and I doubt you'd be bored.  You'd just have to let me know fairly quickly so that we can make sure there is still a room available at the conference hotel.

\section{07-04-05 \ \ {\it Recovery of Philosophy} \ \ (to J. E. {\Sipe})} \label{Sipe4}

\bjes
Here's the quote by {\Dewey} that I mentioned:
\bq\noindent
{\rm Intellectual advance occurs in two ways.  At times increase of knowledge is organized about old conceptions, while these are expanded, elaborated and refined, but not seriously revised, much less abandoned.  At other times, the increase of knowledge demands qualitative rather than quantitative change; alteration, not addition.  Men's minds grow cold to their former intellectual concerns; ideas that were burning fade; interests that were urgent seem remote.  Men face in another direction; their older perplexities are unreal; considerations passed over as negligible loom up.  Former problems may not have been solved, but they no longer press for solution.}
\eq
I find this really very amazing, especially since it was written in
1917.  I don't think Thomas Kuhn ever knew about it (-- while I
certainly haven't read everything Kuhn wrote, I'm not aware of him
acknowledging any debt to {\Dewey} anywhere --) especially since the
quote occurs in an essay (``The Need for a Recovery of Philosophy'')
that {\Dewey} wrote about contemporary philosophy in America.
But the approach is pure Kuhn (or rather Kuhn-before-Kuhn), with a clear articulation of what would later be called ``normal science'' and ``paradigm shifts.''
\ejes

Hey, I've read that article---relatively thoroughly in fact.  But I didn't catch the similarities to Kuhn in it!  Below are the quotes I took from the article for my upcoming ``The Activating Observer: Resource Material for a Pauli'an-{\Wheeler}'ish Conception of Nature'' \ldots\ which pretty much reveals the parts of the article I was paying attention to at the time.

The quotes come from (and the page numbers refer to),
J.~{\Dewey}, {\sl John {\Dewey}: The Essential Writings}, edited with
introduction by D.~Sidorsky, (Harper \& Row, New York, 1977).\medskip

\noindent {\bf malleable world, pp.\ 70--71:}
\bq
[I] set out with a brief statement of some of the chief contrasts between the orthodox description of experience and that congenial to present conditions.

(i) In the orthodox view, experience is regarded primarily as a knowledge-affair. But to eyes not looking through ancient spectacles, it assuredly appears as an affair of the intercourse of a living being with its physical and social environment. (ii) According to tradition experience is (at least primarily) a psychical thing, infected throughout by ``subjectivity.'' What experience suggests about itself is a genuinely objective world which enters into the actions and sufferings of men and undergoes modifications through their responses. (iii) So far as anything beyond a bare present is recognized by the established doctrine, the past exclusively counts. Registration of what has taken place, reference to precedent, is believed to be the essence of experience. Empiricism is conceived of as tied up to what has been, or is, ``given.'' But experience in its vital form is experimental, an effort to change the given; it is characterized by projection, by reaching forward into the unknown; connection with a future is its salient trait. (iv) The empirical tradition is committed to particularism. Connections and continuities are supposed to be foreign to experience, to be by-products of dubious validity. An experience that is an undergoing of an environment and a striving for its control in new directions is pregnant with connections. (v) In the traditional notion experience and thought are antithetical terms. Inference, so far as it is other than a revival of what has been given in the past, goes beyond experience; hence it is either invalid, or else a measure of desperation by which, using experience as a springboard, we jump out to a world of stable things and other selves. But experience, taken free of the restrictions imposed by the older concept, is full of inference. There is, apparently, no conscious experience without inference; reflection is native and constant.

These contrasts, with a consideration of the effect of substituting the account of experience relevant to modern life for the inherited account, afford the subject matter of the following discussion.
\eq

\noindent {\bf time, pp.\ 73--74:}
\bq
The preoccupation of experience with things which are coming (are now coming, not just to come) is obvious to any one whose interest in experience is empirical. Since we live forward; since we live in a world where changes are going on whose issue means our weal or woe; since every act of ours modifies these changes and hence is fraught with promise, or charged with hostile energies---what should experience be but a future implicated in a present! Adjustment is no timeless state; it is a continuing process. To say that a change takes time may be to say something about the event which is external and uninstructive. But adjustment of organism to environment takes time in the pregnant sense; every step in the process is conditioned by reference to further changes which it effects. What is going on in the environment is the concern of the organism; not what is already ``there'' in accomplished and finished form. In so far as the issue of what is going on may be affected by intervention of the organism, the moving event is a challenge which stretches the agent-patient to meet what is coming. Experiencing exhibits things in their unterminated aspect moving toward determinate conclusions. The finished and done with is of import as affecting the future, not on its own account: in short, because it is not, really, done with.

Anticipation is therefore more primary than recollection; projection than summoning of the past; the prospective than the retrospective. Given a world like that in which we live, a world in which environing changes are partly favorable and partly callously indifferent, and experience is bound to be prospective in import; for any control attainable by the living creature depends upon what is done to alter the state of things. Success and failure are the primary ``categories'' of life; achieving of good and averting of ill are its supreme interests; hope and anxiety (which are not self-enclosed states of feeling, but active attitudes of welcome and wariness) are dominant qualities of experience. Imaginative forecast of the future is this forerunning quality of behavior rendered available for guidance in the present. [\ldots] Imaginative recovery of the bygone is indispensable to successful invasion of the future, but its status is that of an instrument. To ignore its import is the sign of an undisciplined agent; but to isolate the past, dwelling upon it for its own sake and giving it the eulogistic name of knowledge, is to substitute the reminiscence of old age for effective intelligence.
\eq

\noindent {\bf time and control, p.\ 78:}
\bq
Experience, to return to our positive conception, is primarily what is undergone in connection with activities whose import lies in their objective consequences---their bearing upon future experiences. Organic functions deal with things as things in course, in operation, in a state of affairs- not yet given or completed. What is done with, what is just ``there,'' is of concern only in the potentialities which it may indicate. As ended, as wholly given, it is of no account. But as a sign of what may come, it becomes an indispensable factor in behavior dealing with changes, the outcome of which is not yet determined.

The only power the organism possesses to control its own future depends upon the way its present responses modify changes which are taking place in its medium. A living being may be comparatively impotent, or comparatively free. It is all a matter of the way in which its present reactions to things influence the future reactions of things upon it. Without regard to its wish or intent every act it performs makes some difference in the environment. The change may be trivial as respects its own career and fortune. But it may also be of incalculable importance; it may import harm, destruction, or it may procure well-being.
\eq

\noindent {\bf control, p.\ 79:}
\bq
As we have already noted, the environment is rarely all of a kind in its bearing upon organic welfare; its most wholehearted support of life activities is precarious and temporary. Some environmental changes are auspicious; others are menacing. The secret of success that is, of the greatest attainable success---is for the organic response to cast in its lot with present auspicious changes to strengthen them and thus to avert the consequences flowing from occurrences of ill-omen. Any reaction is a venture; it involves risk. We always build better or worse than we can foretell. But the organism's fateful intervention in the course of events is blind, its choice is random, except as it can employ what happens to it as a basis of inferring what is likely to happen later. In the degree in which it can read future results in present on-goings, its responsive choice, its partiality to this condition or that, become intelligent. Its bias grows reasonable. It can deliberately, intentionally, participate in the direction of the course of affairs. Its foresight of different futures which result according as this or that present factor predominates in the shaping of affairs permits it to partake intelligently instead of blindly and fatally in the consequences its reactions give rise to. Participate it must, and to its own weal or woe. Inference, the use of what happens, to anticipate what will---or at least may---happen, makes the difference between directed and undirected participation. And this capacity for inferring is precisely the same as that use of natural occurrences for the discovery and determination of consequences---the formation of new dynamic connections---which constitutes knowledge.
\eq

\noindent {\bf existence of reality, p.\ 81:}
\bq
One of the curiosities of orthodox empiricism is that its outstanding speculative problem is the existence of an ``external world.'' For in accordance with the notion that experience is attached to a private subject as its exclusive possession, a world like the one in which we appear to live must be ``external'' to experience instead of being its subject matter. I call it a curiosity, for if anything seems adequately grounded empirically it is the existence of a world which resists the characteristic functions of the subject of experience; which goes its way, in some respects, independently of these functions, and which frustrates our hopes and intentions. Ignorance, which is fatal, disappointment, the need of adjusting means and ends to the course of nature would seem to be facts sufficiently characterizing empirical situations as to render the existence of an external world indubitable.
\eq

\noindent {\bf Bergson, pp.\ 86--87:}
\bq
What are the bearings of our discussion upon the conception of the present scope and office of philosophy? What do our conclusions indicate and demand with reference to philosophy itself? [\ldots]

It is easier to state the negative results of the changed idea of philosophy than the positive ones. The point that occurs to mind most readily is that philosophy will have to surrender all pretension to be peculiarly concerned with ultimate reality, or with reality as a complete (i.e., completed) whole: with {\it the\/} real object. The surrender is not easy of achievement. The philosophic tradition that comes to us from classic Greek thought and that was reinforced by Christian philosophy in the Middle Ages discriminates philosophical knowing from other modes of knowing by means of an alleged peculiarly intimate concern with supreme, ultimate, true reality. To deny this trait to philosophy seems to many to be the suicide of philosophy; to be a systematic adoption of skepticism or agnostic positivism.

The pervasiveness of the tradition is shown in the fact that so vitally a contemporary thinker as Bergson, who finds a philosophic revolution involved in abandonment of the traditional identification of the truly real with the fixed (an identification inherited from Greek thought), does not find it in his heart to abandon the counterpart identification of philosophy with search for the truly Real; and hence he finds it necessary to substitute an ultimate and absolute flux for an ultimate and absolute permanence. Thus his great empirical services in calling attention to the fundamental importance of considerations of time for problems of life and mind get compromised with a mystic, nonempirical ``Intuition''; and we find him preoccupied with solving, by means of his new idea of ultimate reality, the traditional problems of realities-in-themselves and phenomena, matter and mind, free will and determinism, God and the world. Is not that another evidence of the influence of the classic idea about philosophy?
\eq

\noindent {\bf reality, pp.\ 88--89:}
\bq
It is often said that pragmatism, unless it is content to be a contribution to mere methodology, must develop a theory of Reality. But the chief characteristic trait of the pragmatic notion of reality is precisely that no theory of Reality in general, {\it {\"u}berhaupt}, is possible or needed. It occupies the position of an emancipated empiricism or a thoroughgoing naive realism. It finds that ``reality'' is a denotative term, a word used to designate indifferently everything that happens. Lies, dreams, insanities, deceptions, myths,
theories are all of them just the events which they specifically are.
Pragmatism is content to take its stand with science; for science
finds all such events to be subject matter of description and inquiry---just like stars and fossils, mosquitoes and malaria, circulation and
vision. It also takes its stand with daily life, which finds that such
things really have to be reckoned with as they occur interwoven in
the texture of events.

The only way in which the term reality can ever become more than a blanket denotative term is through recourse to specific events in all their diversity and thatness. Speaking summarily, I find that the retention by philosophy of the notion of a Reality feudally superior to the events of everyday occurrence is the chief source of the increasing isolation of philosophy from common sense and science. For the latter do not operate in any such region. As with them of old, philosophy in dealing with real difficulties finds itself still hampered by reference to realities more real, more ultimate, than those which directly happen.

I have said that identifying the cause of philosophy with the notion of superior reality is the cause of an increasing isolation from science and practical life. The phrase reminds us that there was a time when the enterprise of science and the moral interests of men both moved in a universe invidiously distinguished from that of ordinary occurrence. While all that happens is equally real---since it really happens---happenings are not of equal worth. Their respective consequences, their import, varies tremendously. Counterfeit money, although real (or rather {\it because\/} real) is really different from a valid circulatory medium, just as disease is really different from health; different in specific structure and so different in consequences. In occidental thought, the Greeks were the first to draw the distinction between the genuine and the spurious in a generalized fashion and to formulate and enforce its tremendous significance for the conduct of life. But since they had at command no technique of experimental analysis and no adequate technique of mathematical analysis, they were compelled to treat the difference of the true and the false, the dependable and the deceptive, as signifying two kinds of existence, the truly real and the apparently real.
\eq

\noindent {\bf malleable world, p.\ 91:}
\bq
The epistemological universe of discourse is so highly technical that only those who have been trained in the history of thought think in terms of it. It did not occur accordingly, to nontechnical readers to interpret the doctrine that the meaning and validity of thought are fixed by differences made in consequences and in satisfactoriness, to mean consequences in personal feelings. Those who were professionally trained, however, took the statement to mean that consciousness or mind in the mere act of looking at things modifies them. It understood the doctrine of test of validity by consequences to mean that apprehensions and conceptions are true if the modifications affected by them were of an emotionally desirable tone.

Prior discussion should have made it reasonably clear that the source of this misunderstanding lies in the neglect of temporal considerations. The change made in things by the self in knowing is not immediate and, so to say, cross-sectional. It is longitudinal---in the redirection given to changes already going on. Its analogue is found in the changes which take place in the development of, say, iron ore into a watch spring, not in those of the miracle of transubstantiation. For the static, cross-sectional, nontemporal relation of subject and object, the pragmatic hypothesis substitutes apprehension of a thing in terms of the results in other things which it is tending to effect. For the unique epistemological relation, it substitutes a practical relation of a familiar type---responsive behavior which changes in time the subject matter to which it applies. The unique thing about the responsible behavior which constitutes knowing is the specific difference which marks it off from other modes of response, namely, the part played in it by anticipation and prediction. Knowing is the act, stimulated by this foresight, of securing and averting consequences. The success of the achievement measures the standing of the foresight by which response is directed. The popular impression that pragmatic philosophy means that philosophy shall develop ideas relevant to the actual cries of life, ideas influential in dealing with them and tested by the assistance they afford, is correct.
\eq

\section{11-04-05 \ \ {\it Another Query} \ \ (to A. Kent)} \label{Kent8}

\bak
Another query on [\quantph{9512023}].   I'm puzzled by the comment on p.~12 which says that A$_{rs00}$ implies $| Phi_{00} \rangle = 0$ and similarly for 11.  That doesn't immediately seem to fit with eq (12), which suggests that you'd need $A_{00rs} = 0$ to imply $| Phi_{00} \rangle = 0$.   Also, I'm struggling to see how the optimal Eve strategy in the paper defines a unitary evolution if $| Phi_{00} \rangle = | Phi_{11} \rangle = 0$.

Am I confused or going mad?   Or is it possible there's a glitch in the conventions somewhere?   (Is the {\tt arXiv} version isomorphic to the published version, btw?  I'm working from the {\tt arXiv} one.)   You probably won't remember after 10 years, I realize.   But if you happen to be able to shed any light, I would be grateful.
\eak

Wow, that was a long time ago!  I'd have to relearn the paper---like you seem to be doing at the moment.  All my notes (and there were substantially more calculations than made it into the paper) burned up in the Cerro Grande fire.  As a first pass, does my paper \quantph{9611010} help any?  I remember trying to tie up some loose ends there (though for some different measures).  Also, concerning the conjecture---I presume you mean the conjecture that a two-state probe is good enough---I don't know that anyone ever proved it.  I gave up trying after {\tt 9611010}, though I did make a little progress in the direction there (see the discussion in the right column of page 10).  I think maybe Bob Griffiths tried his hand at it too, but didn't make headway---though I'm not sure.

Anyway, let me know if any of that helps, and if it doesn't I'll try to think about it next week (after I get the seminars I've got to give this week off my back).

\section{12-04-05 \ \ {\it Chris Fuchs, Apr.\ 14, MH} \ \ (to H. J. Landau)} \label{Landau1}

Here's something of an abstract.  Sorry to be getting back to you so late.

\bq\noindent
Abstract:
Sets of quantum states good for quantum key distribution.  Discrete positive-operator-valued measures on finite-dimensional vector spaces.  Representing quantum states as probability distributions.  Gram matrices.  Schur product theorem.  Hadamard exponential.  Schwarz inequality.  If any of these words or phrases interest you, come to the seminar.
\eq

\section{12-04-05 \ \ {\it Your Pretty Appendix} \ \ (to C. M. {\Caves})} \label{Caves79.4}

It probably won't surprise you to learn that the part I enjoyed most in the preprint you sent me on Bloch-space transformations is the Appendix.  Simple things for simple minds.  The characterization of pure states in A.3 is really nice.  Question:  Is this a folklore theorem?  Or is it something you two found?  I don't recall seeing it before, but my memory has been definitely failing me lately.

\subsection{Carl's Reply}

\bq
It is indeed nice.  Here it's been discovered twice, once in the Bloch-sphere language by Steve Flammia (I didn't believe it at first) and again by Menicucci in the operator language in the appendix (and again I didn't believe it at first).  We realized they were the same thing at a group meeting.

I mentioned it to Howard, who said he thought that Blume-Kohout had once mentioned it to him.  That's the extent of my knowledge of antecedents.
\eq

\section{21-04-05 \ \ {\it Another Idea} \ \ (to R. W. {\Spekkens})} \label{Spekkens33}

Another thing that might make your visit fun.  Maybe we could drive to Princeton one day and have a visit with John Conway and get him to give us a presentation of this new KS construction he's rumored to have with Kochen.  (I did a quick google search and found this, for instance, but haven't read it yet: \myurl{http://www.cs.auckland.ac.nz/~jas/one/freewill-theorem.html}.)

Do you know if there's anything new in this?

\section{25-04-05 \ \ {\it Your Talk (or your talks)} \ \ (to R. W. {\Spekkens})} \label{Spekkens34}

\brws
I'm thinking that it might be useful for us to discuss what sorts of
information-theoretic axioms might get us to the tensor product
structure of Hilbert space.  Another interesting project would be to
try to reduce Lucien's axioms about subspaces and subsystems to a
single axiom.  [\ldots]  Then again, we could try to work out the number
of ``mutually unbiased bases'' in the toy theory.  This might be easier
to solve than the equivalent problem in quantum mechanics.  Or \ldots\
something entirely different.
\erws

I wouldn't mind talking about a range of things, as I've grown flaccid in my old age.  Maybe something new from you could perk me up.

Though probably this is what I'll be most interested in:  I'm presently in the process of working out what the qutrit state space looks like, solely in terms of a constraint on the probability distributions over nine possible events.  Think ``toy model.''  Similarly, I {\it think\/} that with enough coffee, I could explicitly work out the constraints for the two-qubit case.  Say that I can actually get that done.  I'd like to understand a) in what way or not those constraints look like your toy model's constraints, and b) if getting the quantum constraints exactly right (i.e., what I'm doing) adds any qualitative phenomena over the long list your toy model already recovers.  Surely, it'll give entanglement, whereas your model cannot, but what does that amount to?

\section{26-04-05 \ \ {\it Thanks!}\ \ \ (to W. E. Lawrence)} \label{Lawrence4}

\bwel
Thanks for the great visit and kind hospitality.    It was
invigorating and informative.
\ewel

No, thank you.  I got every bit as much out of it, if not more.  You reminded me of how important it is to just role up your sleeves and get to work.  For instance, to stop toiling with the abstract when you're stuck and see what comes out in the three-dimensional case---it's super-important to make that move every now and then, and I sometimes forget.

\section{26-04-05 \ \ {\it Eagle Eyes} \ \ (to C. G. {\Timpson})} \label{Timpson7}

Wow, with a friend like you, who needs \ldots\ any other friends at all!  Just one good one would do.  Thanks for the defense!  I read it and enjoyed it.  You always have the power to lift my spirits simply by understanding the (essentially trivial) things I've been saying --- you're a rare bird in the philosophical community.  Even Jeff Bub and Hans Halvorson surprise me in these matters.  Every time I see Jeff try to re-ontologize ``information'' I cringe---I sometimes wonder why he even bothered to go through the motions of the Clifton--Bub--Halvorson exercise, when information to him is just another ontic entity.  With that misreading of the very idea of information, it seems to me at best one makes no progress at all; at worst one just mystifies the problem further and erodes any confidence in the community that this is an interesting research direction.

I haven't seen the {\Palge}/Konrad paper.  I wonder why they didn't send it to me too?  Interesting.  Probably, it was because I was a bad boy about replying to an email of questions Veiko had when he first started the critical project.  He hit me at a time when I could hardly stand to lift a finger for email.  It takes a toll trying to say and resay old things in better and better ways just to fend off another critical paper.  Somehow I need to get re-energized to try to write a ``definitive'' paper on the view, and then let come what may in the aftermath.

BTW, yes Eq.\ (97) was a typo, just as you sensed.  Try as I might---and I tried all weekend---I can't remember who first pointed it out to me.  I was able to trace some correspondence about it back to Tony Sudbery, 18 December 2003 [see 18-12-03 note ``\myref{Sudbery7}{Finally Some Answers}''], which went:
\begin{itemize}
\item[] {\bf Tony:}
    I'm sure I'm not alone in seeing your dissolving of the ``violent''
    state change (on p.\ 34) as a cheat. Calling it a ``mental
    readjustment'' doesn't make it any less violent. Incidentally,
    there's something wrong with eq.\ (97); these $U_i$ are not unitary.

\item[] {\bf Chris:}
    Fair enough. It is violent---every bit as violent, discontinuous,
    and irreversible as a usual application of Bayes' rule. But the
    point is that it is on the inside of the agent's head, and no one
    cares what is going on in the agent's head. What physicists should
    be worried about is the physics of the external world (and not the
    agent's head). And who knows what really happens there \ldots\
    especially if physics itself is only really about the interface.

    About the $U_i$'s you are correct. I'm sorry, I've prepared an
    updated version with a lot of the typos, etc., fixed, but I haven't
    reposted it yet. Here's what the new edition says:

    That is to say, there is a sense in which the measurement is solely
    disturbance. In particular, when the POVM is an orthogonal set of
    projectors $\{\Pi_i=|i\rangle\langle i|\}$ and the state-change
    mechanism is the von Neumann collapse postulate, this simply
    corresponds to a readjustment according to unitary operators $U_i$
    whose action on the subspace spanned by $|\psi\rangle$ is
    $$
    |i\rangle\langle\psi|\;.
    $$
\end{itemize}
For the life of me, I can't remember who first brought up the point, and my high-tech desktop-search tools haven't been of help this time.  I do distinctly remember it was someone before Tony.  Anyway, little things like this bug me at times \ldots\ when I can't put my finger on them.  If the person is not already in the acknowledgement list, he should be.

Seeing that someone made a big deal of that one line of bad exposition in a paper, though, tells me that I really should get the thing re-posted.  Most of the other typos were even more minor---for instance, I originally placed Strasbourg in Germany rather than in France---but there is another one that's on the level of the one above---i.e., one with the potential to cause real confusion.  If you look carefully at Eq.\ (61), you'll see that there's a slip-up of the same order there.  I wrote the swap-operator on the wrong space.  To do it right, I need to add two or three more sentences and change the notation.  I should do that, and then re-post the thing.

All that said, I actually do look forward to the day that you'll write a critical analysis of the Bayesian program.  I feel in my heart that that analysis will be worthwhile, and the flaws you dig up will force us to a new level of consistency.  (Of course, your analysis could simply break us, but I'm betting---I can't do otherwise!---that the outcome will much more positive than that.)

\section{28-04-05 \ \ {\it Infinite Limits} \ \ (to G. Valente)} \label{Valente6}

Thanks for sending me your paper.  I've printed it out and will bring it to Maryland with me today.  I'll look forward to a personal explanation of its content when I see you.

I have to tell you right now though that I am skeptical of its conclusion.  This comes predominantly from a long-standing belief that was made forcefully upon me when I was a student of Carlton {\Caves} (who himself was influenced on this issue by Ed Jaynes)---namely, that the infinite cases {\it never\/} carry the {\it essence\/} of a problem.  In the present context what this means is:  If something works in all finite dimensional cases, but {\it apparently\/} not in the infinite dimensional case, my faith in the former won't be eroded at all.  It just means something has gone awry with one's analysis of the infinite dimensional case.  And this conviction comes about because, in my mind, the infinite is just a convenient approximation, of use in this or that calculation, but without any autonomy.

I dug up a great quote from Chapter 2 of Jaynes' book on probability that frames the issue, and I'll place it below.

You'll also have to tell me about how miserable College Park is.  I suspected it would be so.  But when I saw that your being there would be a grand opportunity for a fellow alchemist\footnote{I remember fondly what you wrote in your very first letter to me:  ``I wake up thinking about our alchimistic touch to the nature yesterday morning. As soon as I started studying, I read the same idea (alchemy) in your first statement about Pauli. I closed my books and I run away to phone my mum in Vicenza\ldots''} to make some influence on Jeff, I closed my eyes to it.  Perhaps I shouldn't have.  Still, I am very happy that you are there to influence Jeff.

Jaynes quote below.
\bq
It is very important to note that our consistency theorems have been
established only for probabilities assigned on finite sets of
propositions. In principle, every problem must start with such finite
set probabilities; extension to infinite sets is permitted only when
this is the result of a well-defined and well-behaved limiting
process from a finite set. More generally, in any mathematical
operations involving infinite sets the safe procedure is the finite
sets policy:
\bq\noindent
     Apply the ordinary processes of arithmetic and analysis only to
     expressions with a finite number of terms. Then after the
     calculation is done, observe how the resulting finite expressions
     behave as the number of terms increases indefinitely.
\eq
In laying down this rule of conduct, we are only following the policy
that mathematicians from Archimedes to Gauss have considered clearly
necessary for nonsense avoidance in all of mathematics. But more
recently, the popularity of infinite set theory and measure theory
have led some to disregard it and seek short-cuts which purport to
use measure theory directly. Note, however, that this rule of conduct
is consistent with the original Lebesgue definition of measure, and
when a well-behaved limit exists it leads us automatically to correct
``measure theoretic'' results. Indeed, this is how Lebesgue found his
first results.

The danger is that the present measure theory notation presupposes
the infinite limit already accomplished, but contains no symbol
indicating which limiting process was used. Yet as noted in our
Preface, different limiting processes---equally well-behaved---lead
in general to different results. When there is no well-behaved limit,
any attempt to go directly to the limit can result in nonsense, the
cause of which cannot be seen as long as one looks only at the limit,
and not at the limiting process.
\eq

\section{09-05-05 \ \ {\it Being Bayesian in a Quantum World --- Conference Invitation} \ \ (to J. B. Hartle)} \label{Hartle1}

I'm sorry to hear you can't come after all.  Particularly, with what would have been your talk plan, your participation seems even more urgent than ever.  Too bad!  Well, we will think of you---among other things, I will present my standard slide of the ``The Jim Hartle 1964, Sect.\ 4, Interpretation of Quantum Mechanics (suitably modified)'' in my introductory talk---and please pass on my best wishes to your daughter.

\section{09-05-05 \ \ {\it Becoming in the World (much more important than being)} \ \ (to H. J. Briegel)} \label{Briegel4}

I am sorry to take so long to respond to you.  There are a million things I need to get going on with the conference organization \ldots\ and there'll soon be a flurry of activity in that direction.

The format of the meeting is that we hope to have less than about half of the attendees speaking:  Two talks in the morning, two talks in the afternoon.  Then after each set of two, the idea is to have open floor discussion somewhat related to their topics.  It's based on the model of the Seven Pines meetings, where I've seen it work well.  At some point soon, we'll start soliciting ``volunteers''.

Would you be interested in giving a talk on one-way quantum computing, and how it might shake up the ``standard'' ideas about what powers quantum computing.  And---if possible---say something about how well that meshes (or does not mesh) with a Bayesian take on probability theory?  I might put you back-to-back with Nielsen for a session.

Is that enough detail to help you in your planning?  We actually have a conference web site that's very close to being posted.  But we've been keeping it secret (and you should too) until some issues with a funding agency get resolved.

\section{09-05-05 \ \ {\it Justifying Conditionalization} \ \ (to H. Greaves)} \label{Greaves1}

I'm always desperate; it's part of my character:  And so \ldots\ thanks a million for the paper!  Indeed I'm quite curious about this ``kick'' you mention.

Are you sure we can't convince you to come to Konstanz if we can secure funds for you?  In another act of desperation, I'll show you our secret website in case it might help tip the scales for you:
\begin{center}
\myurl[http://web.archive.org/web/20090511060206/http://www.uni-konstanz.de/ppm/events/bbqw2005/]{http://web.archive.org/web/20090511060206/http:// www.uni-konstanz.de/ppm/events/bbqw2005/}.
\end{center}
Please peruse it, and if you reconsider let me know.  (BTW, when I say it's a secret website, I'm relatively serious:  we can't go public with it until some issues with one of our funding agencies are resolved.  So please do keep the address to yourself until we go public with it ourselves.)

\subsection{Hilary's Preply}

\bq
Good to see you again this weekend. I mentioned a paper by myself and David Wallace, justifying conditionalization (QM aside) via an expected-epistemic-utility proof. In case you're interested, it's at
\bq
\myurl{http://philsci-archive.pitt.edu/archive/00002212/}
\eq
Incidentally, the one type of updating rule that is allowed in principle, as a consequence of our framework (it dropped out of the proof and refused to go away --- we didn't want it), consists of rules of the form conditionalization-followed-by-a-kick \ldots

\ldots\ I would be astounded if this stuff had any relevance whatsoever for your project, but if you're desperate and/or curious, you might like to take a look.
\eq

\section{10-05-05 \ \ {\it BBQW Things} \ \ (to S. Hartmann)} \label{Hartmann10}

I hope that bug has now left your body; sorry for the late reply to your note.  Indeed, I've been meaning to write you for quite a while now, so I'm running far later than you probably even perceive.

From looking at the conference website, it looks like you got the note from Hartle saying that he cannot attend after all.  It is kind of a shame because his subject did look quite good.  But in the light of this, let me cause some new trouble:  Looking back at our attendee list, 47 starts to look like such an unsymmetric number!  I fear to ask this, but might you consider upping it to 50?  My reasons for inquiring are both scientific and political.

First the science.  Two weeks ago, I discovered an {\it outstanding\/} new candidate.  His name is Matthew Leifer, and he is a postdoc at the Perimeter Institute.  If you look at his webpage, via this link \ldots\ or \ldots\
you'll see that among his interests are ``the Bayesian approach to quantum probabilities'' but that far understates his knowledge and ambition in the area.  Indeed, I've come to the opinion that he would be one of our most interesting participants (when we can get him to open his mouth --- he's a very quiet person in a crowd, though he definitely opens up in one-to-one talk and that too would be a good role for him).  He has some extremely interesting ideas about how quantum states better connect up with de Finetti's more general notion of ``previsions'' than with probabilities per se, and he has started to explore some very concrete proposals in this direction.  A very, very interesting guy; he spent a week here recently, and I was ashamed that I had not pushed for him at our meeting before.  (Well, I couldn't really be ashamed --- I didn't know him --- but I'd be ashamed now if we didn't get him; we really could get a good bang for the buck from him.)  And Matt would really like to be there.

Now, making the transition from science to politics, though still on the scientific end: [\ldots]

Now, all the way on the political end. [\ldots]

Thus my question is, would these new additions (or some portion thereof) be OK with you?  Can the budget still afford it?

As you suggest, it is about time to start soliciting talks and scheduling.  I think it would be most useful for the website to be accessible before I start making those requests large scale.

Concerning number of speakers, I think I would suggest that we have about 20 total, and indeed follow the Seven Pines model.  Thus we could have four talks and two discussions per day, with two talks and one discussion on the day of the excursion.  That would only leave us with two other talks to squeeze in somewhere else.  Maybe we could have one evening session.

If it is honorable to suggest it, I think I would like to give the opening talk.  What I would do is give some very general words about the Bayesian program for QM as a whole, and suggest discussion points and open questions.  I'd also give a quick overview about how the various talks in the schedule fit together.  In other words, I'd like my talk to serve the purpose of giving a direction to the meeting; it would not be particular to my own latest research efforts.  As you hoped for me when I came to the LSE, I will try to be provocative.

Concerning your website questions, I hope to have some detailed comments for you in the near future.  One thing I already know that I'd like to do is start a section that has various papers for download.  For instance, I intend to scan in de Finetti's ``Probabilism'' and a Richard Jeffrey article on radical probabilism and issue these as suggested readings.  Once the speakers have been decided upon, we can also ask them to post their own picks of relevant readings.

\section{11-05-05 \ \ {\it LOL} \ \ (to C. P. Hains)} \label{Hains1}

After all these years, I finally hear from you, and what do you send me??!  National Orgasm Day?  I like it.  But you probably knew I'd be particularly receptive to the idea.  It's appropriate.  Since we haven't talked in so long, you probably don't know that I finally lighted on a name for the view of QM I've been trying to develop over the years.  I now call it ``the sexual interpretation of quantum mechanics''---Copenhagen and many-worlds interpretations are so dull in comparison.  Anyway, it makes for a lot of good bawdy bar-room talk at conferences.  Do a Google search on the term if you want to learn a little about it.

I hope things are going well for you---I imagine they are.  The big excitement here is that Kiki decided to paint the house orange.  \ldots\ I guess I should have seen it coming 11 years ago!

\section{12-05-05 \ \ {\it Smelling Correlation without Correlata} \ \ (to N. D. {\Mermin})} \label{Mermin117}

I was listening to a talk of Robert Raussendorf yesterday, who is
visiting us here at Bell Labs, on cluster-state computation, and I'm
almost ashamed to admit it, but I just about blurted out
``Correlation without Correlata!''\ right in the middle of the talk!
There was something in all he said that made the idea come to the
surface in a way it hadn't (for me) in years.  Particularly, I was
moved by his point (actually, the particular way he presented it
because I had already known it) that in a direct simulation of a
quantum circuit in this model, one could perform the measurement that
represents the readout in the first step, in the last step, or
anywhere in between. It's only the correlation, not the correlata
(not the particular measurement outcomes themselves), that matter in
this model of quantum computation.

Anyway, I found myself wondering if this model of quantum computation
might not be a better home for your ``Copenhagen Computation'' ideas
than the usual unitary model.  Have you thought about that?

\section{12-05-05 \ \ {\it The Gauntlet} \ \ (to G. L. Comer)} \label{Comer68}

\bgc
You see, for me, general covariance is not a statement about what is or is not ``meaningful.''  What I perceive, God Dammit, is meaningful!  The fact that I can communicate that meaningfulness to you in such a way that you can confirm my measurements is what general covariance is about.

I just think the GR community is going wrong from the get-go when trying to construct a quantum theory.  They want so much to maintain this Einstein worldview that one only wants a theory that talks about the purely objective, instead of taking the point of view that we admit to the subjective, but then focus on how to communicate that subjectivity in as objective a way as possible.

In this case, I have to now say that I don't quite agree with a
statement of yours: you talk about Einstein's program for GR in
analogy to what you want for quantum mechanics.  That is, you talk
about how Einstein stripped away all the unnecessary bits, and at the
end of the day was left with the geometry.  I'm not sure that this
analogy is actually helpful for your program; it seems to me to be
the same hole that the GR types have fallen into, and can't manage to
climb out of.

OK, there's the gauntlet thrown down!
\egc

OK.  Not much of a gauntlet, though.  Probably half the time I agree with you.

At least in my defense, you shouldn't forget about Footnote 6 (I assume you're referring to my paper ``Quantum Mechanics as Quantum Information (and only a little more)'':
\bq
     I should point out to the reader that in opposition to the
     picture of general relativity, where reintroducing the coordinate
     system---i.e., reintroducing the observer---changes nothing about
     the manifold (it only tells us what kind of sensations the
     observer will pick up), I do not suspect the same for the quantum
     world.  Here I suspect that reintroducing the observer will be
     more like introducing matter into pure spacetime, rather than
     simply gridding it off with a coordinate system.  ``Matter
     tells spacetime how to curve (when matter is there), and
     spacetime tells matter how to move (when matter is there).''
     Observers, scientific agents, a necessary part of reality?  No.
     But do they tend to change things once they are on the scene?
     Yes.  If quantum mechanics can tell us something truly deep about
     nature, I think it is this.
\eq

\bgc
I just think the GR community is going wrong from the get-go when
trying to construct a quantum theory.  They want so much to maintain
this Einstein worldview that one only wants a theory that talks about
the purely objective, instead of taking the point of view that we
admit to the subjective, but then focus on how to communicate that
subjectivity in as objective a way as possible.
\egc

Rob Spekkens and I ended talking a lot about quantizing/non-quantizing GR when we were roomating at this last conference in Maryland.  I've gathered that you are not completely taken with all of Ted Jacobson's opinions, but his talk did provide some good food for thought.  (I guess he was talking about this paper: \arxiv{gr-qc/9504004} or some new extension of it.)  If spacetime geometry is an expression of a thermodynamic situation, it is not something to be quantized in the usual sense.  You don't quantize ``pressure''; you don't quantize ``temperature''; you don't quantize ``entropy''---from the Bayesian view, these are all subjective quantities to begin with.  Something about that feels right.  But then, I always ask myself, ``So then what is wrong with the old (simple and basic) Feynman argument for why gravity must be quantized?''  What is wrong with it?  Or, remember that old Unruh paper on the subject in some conference proceedings, where he tries to lay the argument out with a little more care than Feynman did?  What is wrong there?  That's your homework assignment!  I wish I could get my head around it.

By the way, one of the things that came to the surface again when Spekkens and I were talking is that there simply must be something wrong with the ``usual'' ``partial-trace'' ``EXPLANATION'' (quotes around everything) of the Unruh effect.

I hope you don't mind, but I gathered together some of the notes I've written you, and I shared them with Rob---not that he'll actually read them.  I'll attach the result.  BTW, referring to your throwing the gauntlet down, you will notice the following line in that collection:
\bq\noindent
     The very reason Einstein felt no qualms about introducing
     coordinate systems when necessary was because in relativity theory
     the underlying observer-free description---i.e., the manifold
     equipped with its fields---had been achieved.  In
     contradistinction, this is something quantum theory has yet to
     achieve (and may never be able to).
\eq
And a final BTW:  Did I ever advertise Rob's work to you?  I think these two papers of his represent the best quantum foundations work in the last two or three years:
\begin{itemize}
\item
\quantph{0401052}
\item
\quantph{0406166}.
\end{itemize}

\bgc
My question for you is this: do the philosophers talk about the
biologist's BIG problem?
\egc

You're right, it does seem to be a forgotten problem for the circles I roam around in.

I sometimes think that to be living or not living is a subjective statement.  Neither category really fits at the ontic level; i.e., they always require reference to someone else's state of knowledge or belief.

\section{17-05-05 \ \ {\it Spontaneous Syntax Breaking} \ \ (to G. L. Comer)} \label{Comer69}

\bgc
I refereed a paper last Friday.  In it there was a comment to the effect that ``\ldots\ we thereby break general covariance by introducing a separation of space and time with our choice of coordinates.''  AAARRRRGGGGHHHHH!\@!  Maybe I'm too stupid, closed-minded, or both, but that kind of thing just kills me.  General covariance is the process through which results in one coordinate system can be translated into results for another coordinate system.  How can it be broken, then, by introducing a coordinate system?  It seems that one is introducing it {\bf with} the choice of coordinates, because then the process of general covariance has a starting place, i.e. the original choice of coordinates.

It comes from this particle physics spontaneous symmetry breaking notion, I think.  That is, there is this big desire to understand general covariance as just one of the n$^{\rm th}$ symmetries that nature has, and so we can play a familiar game by ``breaking'' it.

There you go.  My soap-box ranting for today.
\egc

I feel your pain \ldots

On a related subject (to which your note brings me back, as did a
couple of sentences in this talk by Healey that I heard a couple
weeks ago:\  \myurl{http://carnap.umd.edu/philphysics/healey.htm}),
back in my early years at UNC I went through an obsession for a while
of looking up formal developments of the idea of a differentiable
manifold.  The thing that I was troubled by at the time was that in
all the developments I could find, the idea of a set {\it plus\/} a
set of coordinate charts (along with a requirement that it be
possible to paste them together) was {\it primary}.  I could find no
definition of differentiable manifold that didn't rely on a more
primitive definition of coordinate chart.  The reason that troubled
me---these are all vague memories now, so please don't press me to
try to defend any of the ideas anymore or even try to make them
precise---was that when it came time to describe what general
relativity was about {\it but in words}---and all books of course
strived to do that---the differential manifold was taken as a {\it
primary\/} construct with coordinate charts being something of a
secondary nature.  See the circularity in that? You can't get the
idea of a differentiable manifold off the ground without the idea of
a coordinate chart, but then you're told that the differentiable
manifold is the primary idea and coordinate systems are only
gratuitous additions.

Do you remember in discussions of ours along these lines (circa
1991/1992)?  And if so, can you dig up any of the old mails/emails?

Friday I leave for beautiful Vienna.  But my romanticism for the
place is marred by the sad fact that I have no idea yet what I'll say
in my talk!  There'll be a lot of sleeplessness between now and then.

P.S.  Mostly I wrote this just to get that phrase ``spontaneous syntax breaking'' in print.  It came to me this morning, while taking a shower.

\section{17-05-05 \ \ {\it Numbers and Deeper Things} \ \ (to S. H. Simon)} \label{Simon2}

Going back to our conversation yesterday, it dawned on me that it might be useful for you to have the numbers I compiled for my presentation to Jaffe too.

Now to more fun things also related to our conversation yesterday:  You might enjoy reading through the Andrew Steane paper I mentioned.  Here's a link:  \quantph{0003084}.  I read it again over breakfast this morning, and it's as good as I remembered it.  (The last version I had read of it was in 2002 when I edited it for a special issue of SHPMP.)  It's easy reading, but it's deep reading.  In particular, I think it contains the roots of a much better way to start thinking about quantum computation, and thus a better way to present the phenomena to anyone interested (like our executives and VPs).  Steven's way of presenting things yesterday might be viewed as a stopgap in that direction, but clearly there's a lot of room for honing.

In any case, if we get the ideas right, results can't help but flow faster---and that's what we're about.  I think this paper deserves to be better known.

\section{17-05-05 \ \ {\it Steane Paper} \ \ (to R. E. Slusher \& S. J. van Enk)} \label{Slusher9} \label{vanEnk1.1}

I read the Steane paper that I mentioned yesterday over breakfast this morning, and it's every bit as good as I remembered it!  You two should have a look at it too:  \quantph{0003084}.  It's easy reading, but it's deep reading.  And it touches on a lot of points that Dick was worried about yesterday.

Quoting Steane, and inserting my own snide remark,
\bq\noindent
     The answer to the question `where does a quantum computer manage
     to perform its amazing computations?' [this is Deutsch's standard
     rhetorical question showing that he has little imagination beyond
     his cooked-up many-worlds imagery] is, we conclude, `in the
     region of spacetime occupied by the quantum computer'.
\eq

The paper's great!  And looking back, it clearly needs to be better known!  Citebase shows that it has only accrued 9 citations in the last 5 years.

\section{18-05-05 \ \ {\it Being Bayesian, Pro or Con} \ \ (to T. Rudolph)} \label{Rudolph10}

Now that I've decided I wholeheartedly like you again (never again steal my badge young man!), I've also decided that I'd like to invite you to our meeting ``Being Bayesian in a Quantum World,'' August 1--5, in Konstanz.  If I haven't too offended you by this delayed invitation, I hope you'll consider and let me know as soon as possible.  On the off chance that you'll say yes, we've gone ahead and reserved a hotel room for you.  We can definitely pay your local expenses, and will most likely be able to pay your travel expenses too.

In a day or two I'm going to start soliciting volunteers for talks.  Though we plan to max out with a participation of exactly 50 people, I'm hoping for no more than 20--22 actual speakers---the rest will be ``discussants'' during the long discussion sessions after each pair of talks.  Thus, having to have a talk to actually give is by no means mandatory.

\section{18-05-05 \ \ {\it Coordinate Charts} \ \ (to G. L. Comer)} \label{Comer70}

The point I didn't emphasize in my last note---where at least it
looks like I was successful in emphasizing that, in all the formal
developments I've seen, ``coordinate chart'' is a more basic notion
than ``differentiable manifold'' (otherwise the definition of
manifold would not make use of coordinate charts in any way)---is that
this may mesh very well with what you wrote me:
\bgc
     General covariance is the process through which results in one
     coordinate system can be translated into results for another
     coordinate system.
\egc
From that point of view, coordinate chart should indeed be primary,
with differentiable manifold only secondary:  ``Differentiable
manifold'' is only the formal object that falls out of the {\it
physical\/} statement that all coordinate systems should be smoothly
translatable into one another (locally).

A little bit of these thoughts were on my mind when I wrote that old
draft/notes ``an information theoretic hole in general covariance''
(or with such title), circa '91.  You probably don't remember it
because you probably thought I was a quack then (maybe you still do).
Anyway, the wacky idea on my mind at the time was that {\it I
hoped\/} that if one restricted oneself to {\it only\/} the
coordinate charts that were computably (in the Turing sense)
generatable from one another, one might not be able to distill a
full-fledged manifold from the process but a different, more loosely
connected structure for spacetime.  In this way, I hoped to make
sense of EPR phenomena by explaining that the two particles in it are
not so far from each other after all.

With the years and the thoughts that have intervened, I know that
those ideas were indeed very misplaced:  They were indeed quackery.
But it laid the groundwork for me to better appreciate your point:
Coordinate systems are a more basic mathematical structure than
manifolds for a rather deep reason.  They are proxies for potential
observers (agents, in my favored language), and the key issue is
translatability.

\section{20-05-05 \ \ {\it Room for One Very Last One}\ \ \ (to A. M. Steane)} \label{Steane4}

Carl Caves, Stephan Hartmann, {\Ruediger} {\Schack} and I are organizing a rather large (but invitation only) conference in Konstanz this summer, August 1--5, titled ``Being Bayesian in a Quantum World.''  The reason I'm writing you is because we have one last spot and, as it turns out---literally by coincidence---I happened to read your paper ``Quantum Computing Only Needs One Universe'' again the other day.  I was struck once again by how much I think you're on the right track in that paper, and I was re-energized to think we might have a good discussion along those lines if you were to be in Konstanz.  I had already planned a discussion of measurement-based quantum computing, as Hans Briegel and Mike Nielsen will be there.  But your participation would round out things even better.

Below, I'll paste in the original announcement.  [See 10-11-04 note ``\myref{BBQWInvite}{Being Bayesian in a Quantum World --- Invitation}'' to the invitees.]  Also, let me refer you to the ``secret'' webpage for the meeting (we haven't gone public with it, and can't until a certain funding agency makes a decision):  There you can read more about the topics that will be explored at the meeting, and also see the almost-final participant list.  It's really---I think---going to be a fabulous meeting.  There will be no more than 20--22 talks---i.e., not every participant will be speaking---so you won't be overloaded with talks.  Plenty of discussion and thinking time.

If you're interested in coming, we can fund you completely.  Just let me as soon as possible what your decision is.

\section{23-05-05 \ \ {\it Whitehead and the Quantum} \ \ (to R. E. Slusher)} \label{Slusher10}

Too bad you're not here:  I think you would have enjoyed the talks and exchanges between Abner Shimony and Shimon Malin about Whitehead's philosophy and what it might have to do with quantum mechanics.

I didn't remember that Bill Wootters had once written me about Malin's book until I ran across his old note with Google Desktop by accident a while ago.  I'll paste it below as you might get something out of it too.

My talk went well today; an audience of 250 or so.  Not too many fainted when I said, ``Roughly speaking, a Bayesian is someone who believes there cannot be probabilities without gamblers.''

\section{27-05-05 \ \ {\it Guess Where?}\ \ \ (to G. L. Comer)} \label{Comer71}

\begin{tabular}{ll}
Height:                  &    35,000 feet \\
Air Temperature:         &    $-72$ F       \\
Ground Speed:            &    590 MPH     \\
Distance to Destination: &  690 Miles
\end{tabular}\bigskip

Guess where I'm writing and SENDING this note to you from?  This is so cool!  (But so expensive.  You might say curiosity is killing {\Schroedinger}'s cat.)

\section{02-06-05 \ \ {\it This Summer} \ \ (to C. M. {\Caves})} \label{Caves79.5}

Still no word from Zurek?  He only showed up in Vienna for about 1.5 days.  I talked to him very little, though he did ask me at the end of my talk, ``If you are the Bayesian in this quantum world that you talk about, what are you made of?''  I tried to answer, ``I'm made of the same stuff as you, and you're an ordinary physical system as far as I'm concerned,'' but it didn't come out so cleanly of course.  In any case, he walked out of the session as soon as I finished my reply to him---he didn't wait to hear any of the other questions.  (Easy to see since he was on the front row.)

\section{04-06-05 \ \ {\it Quantum Bayesians \& Anti-Bayesians} \ \ (to all BBQWers)}

Thank you all once again for planning to participate in our upcoming meeting Being Bayesian in a Quantum World, Konstanz, 1--5 August 2005.  It's going to be a great time, and I think we're going to get a lot done.

With that in mind, it's time to start planning the talks.  There is now a conference webpage set up at
\begin{center}
\myurl[http://web.archive.org/web/20090511060206/http://www.uni-konstanz.de/ppm/events/bbqw2005/]{http://web.archive.org/web/20090511060206/http:// www.uni-konstanz.de/ppm/events/bbqw2005/}
\end{center}
and there, along with various pieces of practical information, you can catch a glimpse of your fellow participants and some suggested research topics and questions.  (However, please do keep the webpage private at the moment, making no links to it, etc., as certain issues with our funding agencies must be settled before we can go public.  Nevertheless, the webpage is already there for your use.)

The final number of participants at the meeting is 47, and we think we have struck a good balance of Bayesian pro and con, physicist and philosopher, introvert and extrovert, and so on.  From this group, ideally we'd like to have about 20--22 speakers---enough to give some focus to the meeting, but not so many as to make all our heads swim.  In fact, the plan is to follow the method of the Seven Pines meetings that some of you have participated in before:  Each day there will be four talks, two in the morning and two in the afternoon.  The talks will be 45 minutes each, and following each pair, there will be a coffee break and then an extended roundtable discussion (45 minutes to 1 hour) to further explore the topic at hand.  There may be one evening session to help compensate for an afternoon off.  The organizers will try to pair up the talks so that they complement one another and make for an even discussion afterward.

With that in mind, we'd like to start asking for volunteer speakers.  If you would like to speak, please send me a tentative title and enough of an abstract that the organizers can get a feel for the content of your proposed talk.  If there are too few volunteers, you can count on us to do some arm-twisting and guilt-tripping until we get what we want.  If there are too many---the much preferred option---we will have to make some tough decisions, excising overlap, etc.  In any case, it is expected that all participants will have a significant impact on each other, either through their lively exchanges at the roundtables or via their good-hearted debate over a beer or three.

Please let us know your thoughts as soon as possible.  It'll be great to hear from each of you.

\section{04-06-05 \ \ {\it Being Bayesian in a Quantum World} \ \ (to A. Cho)} \label{Cho1}

I don't know if I ever told you how much I enjoyed the article you wrote for Science covering the discussions at the Seven Pines meeting last year.  I mailed it to all my relatives saying, ``Quail and pheasant?  I thought it was chicken.''  But, particularly, I loved the napkin you pictured that showed a figure from my lecture in it!

Anyway, I'm writing you to see if you might be interested in repeating the experience in Konstanz, Germany this summer?  I and a few others have organized what I think is a very nice meeting on information theoretic views of the quantum state---i.e., views centering around the view of the quantum state you depicted in that napkin.  If you were interested in attending to cover the meeting, you would certainly be welcome!

To try to whet your appetite, I'll place the original conference announcement below.  Also, you can have a look at our ``secret'' website (we haven't gone public yet) to see the final list of speakers, the conference program, the talk topics, etc.  I think you'll note that we have some of the best of the best in quantum information and computing there---you'll have to take my word for it if you don't know the field---but in any case you'll probably at least know some of the more famous names (like Milburn, Unruh, Mermin, Nielsen).

Anyway, if you're interested, please do come.  I think it is fair to say you would not be bored!  Just let me know as soon as possible if you would like to come, so that we can make sure that there is still room at the conference hotel if you're interested.

\section{04-06-05 \ \ {\it BBQW Sauce}\ \ \ (to H. M. Wiseman)} \label{Wiseman16}

No, I'm just going to have to drop out of the Sydney workshop:  I'm way overbooked, I've realized.  I intend to write Huw about this in the next few hours.  (I'm in Miami, making my way to Stockholm; plenty of email time.)

Anyway, to answer your question:  See the mass mailing I just sent out.

Concerning Delirium Quantum, I'll post it soon \ldots\ especially if you send me one more note nudging me in that direction.  I wish you were going to be at this meeting in Sweden:  I'm going to unveil the story of two islands (and its impact on the quantum Bayesian view):  The Island of Nurturing Wives and the Island of Bad Girlfriends.  (There is an alternative universe where the two islands are the Island of Supportive Husbands and the Island of Annoying Boyfriends.  I'll be sure to mention both universes.)  Anyway, I think you would find it efficacious, and I would value your feedback.  Too bad.

In a hurry \ldots

\section{05-06-05 \ \ {\it MaxEnt 2005 Invitation, 2} \ \ (to K. H. Knuth)} \label{Knuth3}

I am so sorry I am running almost a month behind on replying to your email!

Please title my talk ``Being Bayesian in a Quantum World'' and maybe use the two paragraphs below as my abstract.  I hope it is not too long, but now that I've spent a while working on it, I rather like it.

I hope to be able to drop this email off in London, on my way to Sweden.  At the moment, I'm somewhere over the Atlantic.

\bq\noindent
Abstract:\smallskip

To be Bayesian about probability theory is to accept that probabilities represent subjective degrees of belief and nothing more. This is in distinction to the idea that probabilities represent long-term frequencies or objective propensities.  But, how can a subjective account of probabilities coexist with the existence of quantum mechanics?  To accept quantum mechanics is to accept the calculational apparatus of quantum states and the Born rule for determining probabilities in a quantum measurement.  If there were ever a place for probabilities to be objective, one might think it precisely here!  (And many do.)  This raises the question of whether Bayesianism and quantum mechanics are compatible at all.  For the Bayesian, it only suggests that we should rethink what quantum mechanics is actually about.  Is it ``law of nature'' or really more ``law of thought,'' though ``law of thought'' conditioned by the particularities of our world?

From transistors to lasers, the evidence abounds that we live in a quantum world.  However, one should not confuse the quantum WORLD with quantum THEORY.  In particular, one should not jump to the conclusion that wave functions are as successful as calculational tools as they are because they mirror some kind of elements of reality.  A more Bayesian-like perspective is that if wave functions generate probabilities, then they too must be Bayesian degrees of belief, with all that such a radical idea entails.  In particular, quantum probabilities have no firmer hold on reality than the word ``belief'' in the phrase ``degrees of belief'' already indicates.  From this perspective, the only sense in which the quantum formalism mirrors nature is through the normative constraints it places on gambling agents who wish to better navigate through this (quantum) world in which they are immersed.  It might be thought that this is rather thin information about nature itself---and thus that the whole view collapses into a kind operationalism or positivism---but the information is not insubstantial!  To the extent that an agent should use quantum mechanics for his uncertainty accounting rather than some other theory tells us something about the world itself---i.e., the world independent of the agent and his particular beliefs at any moment.  In this talk, I will try to shore up these ideas by showing what quantum mechanics looks like when represented using probability simplexes rather than Hilbert spaces.  It can be done, and when done, one starts to get a feeling for how little quantum theory deviates from Bayesianism after all.
\eq

\section{08-06-05 \ \ {\it Classical No-Cloning} \ \ (to J.-{\AA} Larsson)} \label{Larsson1}

Two references:
\begin{enumerate}
\item
C. M. Caves and C. A. Fuchs,
``Quantum information: How much information in a state vector?'' in {\sl The Dilemma of Einstein, Podolsky and Rosen -- 60 Years Later (An International Symposium in Honour of Nathan Rosen -- Haifa, March 1995)}, edited by A. Mann and M. Revzen, Annals of The Israel Physical Society, vol.\ 12, pp.\ 226--257, 1996. \quantph{9601025}.

\item
H.~Barnum, C.~M. Caves, C.~A. Fuchs, R.~Jozsa, and B.~Schumacher, ``Noncommuting Mixed States Cannot Be Broadcast,'' Physical Review Letters {\bf 76}(15), 2818--2821 (1996).  [Reprinted in {\sl Quantum Information and Quantum Computation}, C.~Macchiavello, G.~M. Palma, and A.~Zeilinger eds.\ (World Scientific, Singapore, 2000), pp.~195--198.] \ \quantph{9511010}.
\end{enumerate}
The '96 paper with the basic point was actually written first, but posted later.

\section{09-06-05 \ \ {\it The Bell Quote} \ \ (to M. O. Scully)} \label{Scully1}

What was that God, Buddha, Bell quote you challenged me to find?  If you give it to me again, I'll bet decent odds I find it.

[See 08-07-05 note ``\myref{Scully3}{The Bell Quote, at your request}'' to M. O. Scully.]

\subsection{Marlan's Reply}

\bq
Bell said something like:
\bq\noindent
``What if all this led us to the existence of God or Buddha; wouldn't that be very, very interesting?''
\eq
If you can find anything like that I would be grateful (and surprised).

Concerning your book and Pauli---does the connection with Pauli have to do with his ideas on the objective and mystical as complementary (Bohr) opposites?
\eq

\section{10-06-05 \ \ {\it Film (!)\ from 1927 Solvay Conference, on Web} \ \ (to H. Barnum)} \label{Barnum16}

\bhb
A good friend of mine, Alex Wilce (Univ.\ of Susquehana,
Mathematics) shared the following link with me;  the site has short
home-movie film clips from the 1927 Solvay Conference:
\bq
\myurl{https://www.youtube.com/watch?v=8GZdZUouzBY}
\eq
Here's the description from the website:
\bq\noindent
Following is a ``home movie'' shot by Irving Langmuir, (the 1932 Nobel Prize winner in chemistry). It captures 2 minutes of an intermission in the proceedings. Twenty-one of the 29 attendees are on the film. The film opens with quick shots of Erwin {\Schroedinger} and Niels Bohr. Auguste Piccard of the University of Brussels follows and then the camera re-focuses on {\Schroedinger} and Bohr.
\eq
\ehb
That is really great!  It cheered me up in this annoyingly delayed-flight time while I twiddle my thumbs in the American Airlines club.

\section{13-06-05 \ \ {\it Wonderful Meeting} \ \ (to A. Y. Khrennikov \& C. Eriksson)} \label{Khrennikov10} \label{Eriksson1}

Now that I'm home (\ldots\ though I am just about to turn around to go to Europe again, for the J. T. Lewis Memorial Meeting), I wanted thank you both for a very fine meeting.  I got lots of ideas at this one, and think it's rejuvenated my research---I have not felt this good in a long time.  You guys are performing a valuable, valuable service to the physics community.

\section{19-06-05 \ \ {\it Philosopher's Stone} \ \ (to G. L. Comer)} \label{Comer72}

It looks like 17 days have gone by, and I haven't written you any of the note I had promised.  I thought it would be easy to get my head back together and put on my writing cap while I was away from home.  But first Sweden passed without a sparkle, then a short weekend back at home in exhaustion, and now even Ireland has come and gone.  I'm over the Atlantic again---it's the eighth time this year---and come what may, I'm going to write you something today.

I hope your recovery from surgery is coming along well and that anxiety is not getting the best of you.  The last I heard from you was the story of the barium shake.  I can absorb a little at an intellectual level of what your feelings and fears must have been like.  Not pretty.  I just hope things are going much better now.

It's going to be a long day:  I'm actually overshooting Newark on my way home---flying from Dublin to Chicago, and then turning around to finally arrive in Newark.  American Airlines was giving 7,500 bonus frequent-flyer miles for roundtrips between Chicago and Dublin this month, and I thought, ``What the hell?''  But what a way to spend Father's Day!  Did your kids get you anything cool for the big day?

I've been doing really silly things for miles lately.  For instance, when I flew to Sweden, I went:  Newark to Miami, Miami to London, London to Stockholm.  I thought, ``Well, I'll get a good day of uninterrupted work in that way, and at the same price as any other route, I'll replenish my mileage supply.''  Geez, what a bad start to the trip it was!  Things just didn't click, and I got stuck next to bad neighbors on the flights, etc.  And the icing on the cake was that I thought I would land in Stockholm, get a train, and be in {\Vaxjo} within an hour---how wrong I was!  Three train connections and six (6!)\ hours passed by before I finally landed in my room.  I was a mess, and I still had to make several transparencies for my talk.  Like I say, it was a bad start.

But I started to get a lift as soon as I gave my talk.  I was the second speaker after Marlan Scully---old laser guy (you probably don't know him, but he's a big name in quantum optics circles; he and Willis Lamb having written the main textbook on lasers that's been used for probably the last 30 years).  It was great seeing him sit down and get enthralled in my talk, and I guess that got my engine going again.  The week turned out to be quite productive, and ideas started flowing.

The main thing on my mind has been viewing the quantum measurement
process as a kind of modern version of alchemy.  I guess I've been
talking about this on and off for several years now, having been
exposed to the analogy through {\Pauli}'s letters to Jung and Fierz, but
the transformation in Sweden was that the stuff really got into my
way of thinking, and I could start to see how to formalize it.  I'm
toying with the idea of writing a paper titled, ``The Consequences of
Our Interventions,'' to try to redo my paper with Asher ``Quantum
Theory Needs No Interpretation'' from this new perspective.  In that
paper, we had written,
\bq
   The thread common to all the nonstandard ``interpretations'' is the
   desire to create a new theory with features that correspond to some
   reality independent of our potential experiments. But, trying to
   fulfill a classical worldview by encumbering quantum mechanics with
   hidden variables, multiple worlds, consistency rules, or spontaneous
   collapse, without any improvement in its predictive power, only
   gives the illusion of a better understanding. Contrary to those
   desires, quantum theory does {\it not\/} describe physical reality.
   What it does is provide an algorithm for computing {\it
   probabilities\/} for the macroscopic events (``detector clicks'')
   that are the consequences of our experimental interventions. This
   strict definition of the scope of quantum theory is the only
   interpretation ever needed, whether by experimenters or theorists.
\eq

First off, I wish I had never said, ``quantum theory does not
describe physical reality''---I really only meant ``the wave function
does not describe reality'' and should have stuck with that
formulation.  But more importantly, what precisely are these
``consequences of our interventions''?  From the wording we used, one
surely gets the impression that, whatever they are---we said
``detector clicks,'' but what a glib phrase!---they somehow live
outside of the agent performing the experiment.  And I guess that's
what I thought at the time.

Now I'm quite internal about it all; those ``consequences'' really
live in the agent himself, not in the quantum system or even in some
``detached'' measurement device.  And I think I now have a clear
enough formulation of the idea that I ought to go public with it.

Quantum THEORY (I emphasize theory to draw a distinction between
quantum theory and the quantum world, i.e., that stuff outside of us
which is independent of the agent) is predominantly about the changes
we bring about within ourselves by interacting with the external
world.  Very little of quantum theory is about the external world
itself.  Particularly, the clicks are not ``out there,'' but rather
``in here.''

I've been scouring the {\Pauli} quotes I have in my computer to see
exactly where I got these ideas from, and so far I haven't read
nearly as forceful of a version of it as I seem to remember---so I'm
pretty sure I still haven't quite found the right passage.  The only
thing that comes close, so far, is this passage from {\Pauli}'s article
on Kepler:
\bq
   Now, there is a basic difference between the observers, or
   instruments of observation, which must be taken into consideration by
   modern microphysics, and the detached observer of classical physics.
   By the latter I mean one who is not necessarily without effect on the
   system observed but whose influence can always be eliminated by
   determinable corrections.  In microphysics, however, the natural laws
   are of such a kind that every bit of knowledge gained from a
   measurement must be paid for by the loss of other, complementary
   items of knowledge. Every observation, therefore, interferes on an
   indeterminable scale both with the instruments of observation and
   with the system observed and interrupts the causal connection of the
   phenomena preceding it with those following it. This uncontrollable
   interaction between observer and system observed, taking place in
   every process of measurement, invalidates the deterministic
   conception of the phenomena assumed in classical physics:  the series
   of events taking place according to pre-determined rules is
   interrupted, after a free choice has been made by the beholder
   between mutually exclusive experimental arrangements, by the
   selective observation which, as an essentially non-automatic
   occurrence (Geschehen), may be compared to a creation in the
   microcosm or even to a transmutation (Wandlung) the results of which
   are, however, unpredictable and beyond human control.

   In this way the role of the observer in modern physics is
   satisfactorily accounted for.  The reaction of the knowledge gained
   on the gainer of that knowledge (Erkennenden) gives rise, however, to
   a situation transcending natural science, since it is necessary for
   the sake of the completeness of the experience connected therewith
   that it should have an obligatory force for the researcher (f\"ur den
   Erkennenden verbindlich).  We have seen how not only alchemy but the
   heliocentric idea furnishes an instructive example of the problem as
   to how the process of knowing is connected with the religious
   experience of transmutation undergone by him who acquires knowledge
   (Wandlungserlebnis des Erkennenden).  This connection can only be
   comprehended through symbols which both imaginatively express the
   emotional aspect of the experience and stand in vital relationship to
   the sum total of contemporary knowledge and the actual process of
   cognition.  Just because in our times the possibility of such
   symbolism has become an alien idea, it may be considered especially
   interesting to examine another age to which the concepts of what is
   now called classical scientific mechanics were foreign but which
   permits us to prove the existence of a symbol that had,
   simultaneously, a religious and a scientific function.
\eq

Or maybe I picked it up from this passage in Heisenberg's article,
``Wolfgang {\Pauli}'s Philosophical Outlook,''
\bq
   The elaboration of Plato's thought had led, in neo-Platonism and
   Christianity, to a position where matter was characterized as void of
   Ideas.  Hence, since the intelligible was identical with the good,
   matter was identified as evil.  But in the new science the world-soul
   was finally replaced by the abstract mathematical law of nature.
   Against this one-sidedly spiritualizing tendency the alchemistical
   philosophy, championed here by Fludd, represents a certain
   counterpoise.  In the alchemistic view ``there dwells in matter a
   spirit awaiting release. The alchemist in his laboratory is
   constantly involved in nature's course, in such wise that the real or
   supposed chemical reactions in the retort are mystically identified
   with the psychic processes in himself, and are called by the same
   names.  The release of the substance by the man who transmutes it,
   which culminates in the production of the philosopher's stone, is
   seen by the alchemist, in light of the mystical correspondence of
   macrocosmos and microcosmos, as identical with the saving
   transformation of the man by the work, which succeeds only `Deo
   concedente.'\,''  The governing symbol for this magical view of
   nature is the quaternary number, the so-called ``tetractys'' of the
   Pythagoreans, which is put together out of two polarities.  The
   division is correlated with the dark side of the world (matter, the
   Devil), and the magical view of nature also embraces this dark
   region.
\eq
Or finally, maybe from this letter from {\Pauli} to Fierz:
\bq
   All of this then led me onto further, somewhat more phantastic [{\it
   sic\/}] paths of thought.  It might very well be that we do not treat
   matter, for example viewed in the sense of {\it life}, ``properly''
   if we observe it as we do in quantum mechanics, {\it specifically
   when doing so in complete ignorance of the inner state of the
   ``observer.''}

   It appears to me to be the case that the ``after-effects'' of
   observation which were ignored would {\it still\/} enter into the
   picture (as atomic bombs, general anxiety, ``the Oppenheimer case''
   e.g.~etc.), but in an {\it unwanted form}.  The well-known
   ``incompleteness'' of quantum mechanics (Einstein) is certainly an
   existent fact somehow-somewhere, but certainly cannot be removed by
   reverting to classical field physics (that is only a ``neurotic
   misunderstanding'' of Einstein), it has much more to do with {\it
   integral relationships between ``inside'' and ``outside'' which the
   natural science of today does not contain\/} (but which alchemy had
   suspected and which can also be detected in the symbolics of my
   dreams, about which I believe them specifically to be characteristic
   for a contemporary physicist).

   With these vague courses of thought I have reached the border of that
   which is recognizable today, and I have even approached ``magic.''
   (From this standpoint observation in quantum mechanics might even
   appear to someone as a ``black mass'' after which the ``ill-treated''
   matter manipulates its counter-effect against the ``observer,''
   thereby ``taking its revenge,'' as a ``shot being released from
   behind'').  On this point I realize well that this amounts to the
   threatening danger of a regression into the most primitive
   superstition, that this would be much worse than Einstein's
   regressive remaining tied to classical field physics and that
   everything is a matter of holding onto the positive results and
   values of the {\it ratio}.
\eq
Still I have the nagging feeling that there is something much better
if I could just dig it up.

Anyway, the way I view quantum measurement now is this.  When one
performs a ``measurement'' on a system, all one is really doing is
taking an ACTION on that system.  From this view, time evolutions or
unitary operations etc., are not actions that one can take on a
system; only ``measurements'' are.  Thus the word measurement is
really a misnomer---it is only an action.  In contradistinction to
the old idea that a measurement is a query of nature, or a way of
gathering information or knowledge about nature, from this view it is
just an action on something external---it is a kick of sorts.  The
``measurement device'' should be thought of as being like a
prosthetic hand for the agent---it is merely an extension of him; in
this context, it should not be thought of as an independent entity
beyond the agent.  What quantum theory tells us is that the formal
structure of all our possible actions (perhaps via the help of these
prosthetic hands) is captured by the idea of a
Positive-Operator-Valued Measure (or POVM, or so-called ``generalized
measurement'').  We take our actions upon a system, and in return,
the system gives rise to a reaction---in older terms, that is the
``measurement outcome''---but the reaction is in the agent himself.
The role of the quantum system is thus more like that of the
philosopher's stone; it is the catalyst that brings about a
transformation (or transmutation) of the agent.

Reciprocally, there should be a transmutation of the system external
to the agent.  But the great trouble in quantum interpretation---I
now think---is that we have been too inclined to jump the gun all
these years:  We have been misidentifying where the transmutation
indicated by quantum mechanics (i.e., the one which quantum theory
actually talks about, the ``measurement outcome'') takes place.  It
should be the case that there are also transmutations in the external
world (transmutations in the system) in each quantum ``measurement'',
BUT that is not what quantum theory is about.  It is only a hint of
that more interesting transmutation.  And, as you know, somehow out
of all this I think of the agent and the observer as being, together,
involved in a little act of creation that ultimately has an autonomy
of its own---that's the sexual interpretation of quantum mechanics.
(However the ideas in that last sentence are a little murkier than
what I'm trying to get at now.)

Does ``measurement'' in this new sense explicitly require
consciousness (whatever that is)?  I don't think so.  But it does
require some kind of nonreductive element---some kind of higher-level
description that cannot be reduced to a lower-level one.  Here's the
way I've been putting it in my last three lectures, when talking more
particularly about quantum mechanics from the Bayesian perspective. I
point out that a Bayesian is, roughly speaking, someone who believes
that without gamblers, there cannot be probabilities. Probabilities
are not external to gamblers.  Then someone always asks, must you
have consciousness to have probabilities?  And I say, ``No.''  Take
as an example my laptop computer loaded with a Bayesian spam filter.
It is a perfectly good gambler in the Bayesian sense, but I think
most people would be hard-pressed to call it conscious. Similarly, I
think we're going to ultimately learn an analogous lesson about all
this transformation/transmutation/creation/measurement business.

On the other hand, I do find myself being tickled toward a more
Whiteheadian-like view that, whatever this higher-level description
is, every piece of nature has more or less of it, from people all the
way to stones and atoms.  It's just that you have to be on the inside
of the philosopher's stone to see it.

\section{30-06-05 \ \ {\it Wittgensteinian Arm Twisting} \ \ (to H. Price)} \label{Price5}

One of the people I was really hoping would volunteer to talk at Konstanz was you!  Since, however, I haven't heard anything from you, maybe it's time to do a little arm-twisting.  Particularly, I wonder if I could entice you to give a talk that shuffles around in some way or other, or gives an introduction to, Wittgenstein's book {\sl On Certainty}?  As I recall, you'll be going to the Wittgenstein symposium the very next week, so he'll certainly already be on your mind.  This aspect of his thought may not be on your mind for {\it that\/} meeting, but without doubt you'll be the closest we have to an expert on the subject at {\it this\/} meeting!

Still more particularly, I would love some sort of discussion to build up about the ties (or contradictions) between Wittgenstein's notion of certainty and the radical Bayesian notion of certainty (\`a la de Finetti and Richard Jeffrey)---i.e., that even ``certainty'' is a state of mind, just like a 50/50 assignment (or any other one, for that matter).  (Compare how Wittgenstein says certainty is a ``tone of voice''.)

If you'd take the task, I'd be forever grateful.  And I can supply you with relevant passages in Jeffrey's writings and even some of our debate with {\Mermin} to give you something of a foil to work against, if you'd like.

It'd be really great to have a discussion of this subject, and there's no doubt you'd be the right man to lead it.

\subsection{Huw's Reply}

\bq
I'm afraid there's no way I could talk about {\sl On Certainty}. I'm not much of a Wittgenstein scholar in any case, and what little I do know
(certainly) doesn't include anything about {\sl On Certainty\/} (I've been invited to the Kichberg meeting as an expert on the arrow of time, not Wittgenstein!) Sorry!
\eq

\section{30-06-05 \ \ {\it Flattered} \ \ (to A. Duwell)} \label{Duwell4}\

I'm flattered by your abstract!  A couple of typos though that maybe you can fix:
\bq\noindent
{\bf Sorting the objective from the subjective: remarks on Fuchs}\medskip

Chris Fuchs seeks to identify the features of quantum theory that reflect the structure of the world and are not artifacts of the quantum formalism. To this end, Fuchs advocates attempting to interpret as many elements of quantum mechanics as reflective of subjective beliefs about quantum systems [as possible??]. The elements of the theory that resist this subjective interpretation are thereby candidates for reflecting elements of reality. The obvious problem with this methodology is that just because one can interpret an element of a theory subjectively, doesn't mean one ought to interpret it subjectively. In this paper I advocate the use of eliminative or demonstrative induction as a partial solution to the problem of sorting subjective from objective elements of theories. I discuss the application of this method to quantum theory and the limitations of the [what comes here??]
\eq

\section{30-06-05 \ \ {\it Konstanz Talk?}\ \ \ (to J. B. M. Uffink)} \label{Uffink4}

You were one of the guys I was really hoping to hear from regarding a talk at the BBQW meeting in Konstanz, but I haven't heard anything from you!  If you'd take the task we'd certainly be delighted!  Particularly, even something like the talk you gave at LSE last year could be very useful.  Might we entice you to give your thoughts on anything around the subject matter---it could be pro or con, it doesn't matter---I just want to get a lively discussion going.  Or if you wanted to use this as an opportunity to formalize your thoughts on Everettian attempts to derive the quantum probability rule that could be very useful too.  Whatever.  The main thing is, it'd be great to see your clarity of thought in action with respect to this cluster of ideas.

Please let me know as soon as possible whether you think you could give us a talk, or whether you're definitely going to decline (despite my heartfelt plea!).

\section{30-06-05 \ \ {\it A BBQW Talk?}\ \ \ (to A. Hagar)} \label{Hagar1}

Since I haven't heard back from you following my call for ``volunteers'', I wonder if we might entice you to submit a talk proposal for the BBQW meeting?  Particularly, if you still think the BBQW way is ``doing quantum mechanics in the dark'' it'd be great if you'd organize your ideas on the subject afresh and give us a presentation.  It could be a lively discussion point.  If you have a tentative title and abstract, it'd be great to throw it in the pot of possibilities.  We'll probably be making our talk decisions by the end of next week.

\section{30-06-05 \ \ {\it New Delirium}\ \ \ (to H. M. Wiseman)} \label{Wiseman17}

I'm sorry:  I've been very slow to reply to your proposal for a talk at BBQW.  I think it sounds great, and almost surely we'll pick it from the pool of potential talks.  We'll let everyone know the final verdict next week (but as I say, you might as well start planning your talk now).

I like the phrase in your abstract, ``\ldots\ accepting that quantum mechanics is incomplete'' because if you nuance it enough, and take out the idea of hidden variables, I'd even agree with it.  Thus, thinking about that, I decided to extend Delirium Quantum a little with two more excerpts before posting it for you.  Particularly, the new Sections 6 and 7 might give you a way of thinking ``quantum mechanics is incomplete'' in a way that you haven't thought before.  I'll attach the new version.  [See ``Delirium Quantum,'' \arxiv{0906.1968v1}.]

Looking much forward to this meeting and sparring with you there.

\section{01-07-05 \ \ {\it Bell, Buddha, Pauli} \ \ (to M. O. Scully)} \label{Scully2}

Thanks for the Bell phrase to be on the lookout for.  I'll keep my eyes peeled, and if I find something I'll let you know.

\bms
Concerning your book and Pauli---does the connection with Pauli have to do with his ideas on the objective and mystical as complementary
(Bohr) opposites?
\ems

My original interest in Pauli had more to do with a point he made over and over about how A) the quantum mechanical observer takes {\it part\/} in a little act of creation with each quantum measurement (very John {\Wheeler}ian also), yet B) there remains a kind of objective reality within the quantum description in that the observer cannot control the outcome of that creation.  The observer is participatory, but not overpowering.  Here it is, in one version, in Pauli's own words:
\bq
In the new pattern of thought we do not assume any longer the
detached observer, occurring in the idealizations of this classical type of theory, but an observer who by his indeterminable effects
creates a new situation, theoretically described as a new state of the observed system.  In this way every observation is a singling
out of a particular factual result, here and now, from the
theoretical possibilities, thereby making obvious the discontinuous aspect of the physical phenomena.

Nevertheless, there remains still in the new kind of theory an
objective reality, inasmuch as these theories deny any possibility for the observer to influence the results of a measurement, once
the experimental arrangement is chosen.

Like an ultimate fact without any cause, the individual outcome
of a measurement is, however, in general not comprehended by laws. This must necessarily be the case  \ldots\
\eq
It's the conjunction of all that that I call ``the Paulian idea'' (and thus the title of my book, which is really an elaboration of those ideas).  The part after the ``Nevertheless'' is particularly important.

However, I am aware of the complementarity you mention, though with Pauli it was more precisely a complementarity between ``physical'' and ``psychic'' or ``psychological''.  He devoted a lot of discussion with his assistant Markus Fierz and with the psychologist Carl Jung to try to flesh out a ``neutral language'' that would somehow be able to drop the distinction between ``physical'' and ``psychic''.  I'll give you an example of the sort of stuff he was up to, by pasting in a letter that Pauli had written to Jung below.  I've got a lot of material like that in my upcoming posting to {\tt quant-ph}, ``The Activating Observer: Resource Material for a Paulian--{\Wheeler}ish Conception of Nature.''

Anyway, that sort of idea is starting to intrigue me more now than it used to.  Another interesting fellow to read along those lines is William {\James}, the great American pragmatist.

I'll look for the Bell quote.\medskip

\noindent {\bf Letter from Pauli to Jung, 31 March 1953:}
\bq
\indent
The labeling of ideas as either of spiritual origin or physical (or
physiological) origin and your corresponding definition of physics as a science of ideas of the second kind has revived memories of my youth.

Among my books, there is a somewhat dusty case containing a Jugendstil silver goblet, and in this goblet there is a card. A gentle, benevolent, and cheerful spirit from days of yore seems to be issuing forth from this goblet. I can see him shaking your hand in a friendly way, welcoming your definition of physics as a pleasing, albeit somewhat belated, indication of your insight and understanding; he goes on to add how suitable the labels are for his laboratory, and expresses his satisfaction at the fact that metaphysical judgments in general (as he was wont to say) ``have been relegated into the realm of the shadows of a primitive form of animism.'' This goblet is a baptism goblet, and on the card it says in an old-fashioned ornate script: ``Dr.\ E. Mach, Professor at the University of Vienna.'' It so happened that my father was very friendly with his family, and at the time totally under his influence mentally, and he (Mach) kindly agreed to take on the role of my godfather. He must have had a much stronger personality than the Catholic priest, with the apparent result that I was thus baptized in an antimetaphysical manner rather than in a Catholic one. Be that as it may, the card remains in the goblet, and despite all the great mental changes I went through later on, it remains a label that I myself bear---namely: ``of antimetaphysical origin.'' And in fact, to put it in a somewhat simplistic way, Mach regarded metaphysics as the root of all evil
in this world---in other words, in psychological terms, as the Devil himself---and that goblet with the card remained as a symbol of the {\it aqua permanens\/} that keeps evil metaphysical spirits at bay.

I do not need to describe Ernst Mach more closely, for if you look at your own description of the extroverted sensation type, then you will see E. Mach. He was a master at experimentation, and his apartment was crammed full of prisms, spectroscopes, stroboscopes, electrostatic machines, and the like. Whenever I visited him, he always showed me some neat experiment, already completed, partly so as to eliminate unreliable thinking, with the ensuing illusions and errors, and partly to support it and correct it. Working on the assumption that his psychology was a universal one, he recommended everyone to use that inferior auxiliary function as ``economically''
as possible (thought economy). His own thought processes closely followed the impressions of his senses, tools, and apparatus.

This letter is not meant to be a history of physics, nor the classical case of type opposites: E. Mach and L. Boltzmann, the thinking type.\footnote{I should like to recount an anecdote here because I am sure you will really enjoy it. Mach, who was by no means prudish and was most interested in all the intellectual trends of the day, once pronounced judgment on the psychoanalysis of Freud and his school. He said: ``These people try to use the vagina as if it were a telescope so that they can see the world through it. But that is not its natural function---it is too narrow for that.'' For a while this became a popular quotation at the University of Vienna. It is very typical of Mach's {\it instrumental\/} way of thinking. For him, psychoanalysis immediately conjures up the vividly concrete image of the wrongly applied instrument: namely, the female organ where it does not belong.} I last saw Mach just before the First World War, and he died in 1916 in a country house near Munich.

What is interesting in connection with your letter is Mach's attempt to fall back on psychic facts and circumstances (sensory data, ideas) within the realm of physics as well and especially to eliminate as far as possible the concept of ``matter.'' He regarded this ``auxiliary concept'' as grossly overrated by philosophers and physicists and viewed it as a source of ``pseudo problems.'' His definition of physics basically coincided with the one proposed by you, and he never failed to stress that physics, physiology, and psychology were ``only different in the lines of investigation they took, not in the actual object,'' the object in all cases being the constant psychic ``elements'' (he exaggerated their simplicity somewhat, for in reality they are always very complex). I was surprised that despite your sweeping criticism of what later came to be called {\it ``Positivism''\/} (Mach used this term a great deal), there are nevertheless also fundamental similarities between you and this line of thought: In both cases there is the {\it deliberate elimination of thought processes}. And of course there is nothing at all wrong with these labels for ideas and the corresponding definition of physics, especially as it accords perfectly with the idealistic philosophy of Schopenhauer, who consciously uses ``Idea''
and ``Object'' synonymously. But it all depends on {\it how one proceeds}. What Mach wanted, although it could not be carried out, was the total elimination of everything from the interpretation of nature that is ``{\it not\/} ascertainable {\it hic et nunc}.'' But then one soon sees that one does not understand anything---neither the fact that one has to assign a psyche to others (only one's own being ascertainable) nor the fact that different people are all talking about the same (physical) object (the ``windowless monads''
in Leibniz). Thus, in order to meet the requirements of both instinct and reason, one has to introduce some {\it structural elements of cosmic order}, which ``in themselves are not ascertainable.'' It seems to me that with you this role is mainly taken over by the archetypes.

It is right that what one does or does not call ``metaphysics'' is, to a certain extent, a matter of taste. And yet I agree with you totally that in practical terms, great value is to be attached to the demand that metaphysical judgments be avoided. What is meant by that is that the ``not in themselves ascertainable'' factors (concepts) that have been introduced do not completely escape the controlling, checking mechanism of experience, and that {\it no more of them may be introduced than is absolutely necessary\/}: They serve the purpose of {\it making statements about the possibility of ascertainments hic et nunc}. This was the sense in which the concept of ``possibility''
was meant, and it was in this sense that I called such concepts {\it ``symbolic things in themselves''\/} and the {\it ``rational aspect of reality.''} As you rightly point out, {\it there is absolutely no need to make statements of Being in the metaphysical sense about these ``things in themselves.''} In the natural sciences, one makes the {\it pragmatic statement of usefulness\/} about them (in order to understand the ordering system of the ascertainable); in mathematics there is just the {\it formal logical statement of consistency}. In psychology, those ``not in themselves ascertainable'' concepts include, the unconscious and the archetypes, and in atomic physics, they include the totality of the characteristics of an atomic system that are {\it not all simultaneously\/} ``ascertainable {\it hic et nunc}.''

In my last letter, I referred to that which is actually ``ascertained {\it hic et nunc\/}'' as ``concrete phenomenon'' and the ``irrational aspect of reality.'' It is always present in the psyche of an observer, whatever the ``label of origin'' might be. At this point, however, the question arises of whether the description ``psychic''
or the term ``psyche'' can go {\it further\/} than the ``ascertainable {\it hic et nunc}.'' I am inclined to reply to this question in the negative and to take the ``not in themselves ascertainable'' structures, which are introduced as conceptual indications of possibilities of the ascertainable, and give them the definition {\it ``neutral''\/} and {\it not the definition ``psychic.''}

To me, this view also seems to be supported by Plato's expressions {\it meson\/} (middle) and {\it tritoneidos\/} (third form), which both meet my requirements for ``neutrality'' ($=$middle position), nay, actually seem to {\it emphasize\/} it. Plato certainly had the word ``psyche'' at his disposal, and if he opts to use a different word instead, then it must be one with a deeper meaning, one that calls for careful consideration. For me, this deeper meaning lies in the need to make a clear distinction between the experience of the individual, which exists in his psyche as something ascertainable {\it hic et nunc}, and the general concepts, which, ``nonbeing ascertainable in themselves,'' are suitable for taking up a middle position. Your {\it identification of psyche $=$ tritoneidos\/} thus seems to me a {\it retrograde step}, a loss in terms of conceptual differentiation.

With my call for ``neutral'' general concepts, I find myself in agreement with your article ``The Spirit of Psychology,'' which struck me as fundamental, especially when you say: ``The archetypes have \ldots\ a nature which one cannot definitely describe as psychic. Although by the application of purely psychological considerations I, have come to question the solely psychic nature of the archetypes, etc.'' I feel that you should certainly {\it take these doubts seriously and not once again\/} make too much of the {\it psychic factor}. When you say that ``the psyche is partly of a material nature,''\footnote{Your letter, p.\ 6. This statement seems to me alarmingly close to the definitely {\it meaningless\/} statement that ``everything is psychic''! For ``psychic'' to acquire any meaning, there must also be something nonpsychic involved, and this is where the ``not in itself ascertainable'' seems to me suitably {\it neutral}.} then for me as a physicist this takes on the form of a metaphysical statement. I prefer to say that psyche and matter are governed by common, neutral, ``not in themselves ascertainable'' ordering principles. (Unlike the psychologist, the physicist has no problem, for example, with saying ``the U field''
instead of ``the unconscious,'' which would thus establish the ``neutrality'' of the concept.)

But I wish to make it quite clear that my hope that you might agree with this general point of view is based on the impression that some of the {\it pressure needs to be removed\/} from your analytical psychology. The impression I have is of a vehicle whose engine is running with overloaded valves (expansion tendency of the concept ``psyche''); that is why I should like to relieve some of the pressure and let off steam. (I shall come back to this later on p.\ 10 below).

I would also hope that a clarification of the scope of the concept of the psyche might include your {\it de iure\/} recognition of the fact that the heart is not just a psychological symbol but also a conception labeled ``of physical origin.'' Economy with the inferior function on the lines of E.~Mach often serves to fulfill a function, even if it is not actually that of thinking!
\eq

\section{07-07-05 \ \ {\it Hook or Crook} \ \ (to K. H. Knuth)} \label{Knuth4}

OK, I booked some flights now.  But I had to get tricky to trim the price down.  The way I did it was to book TWO separate tickets:  One roundtrip between Munich and Brussels, and one roundtrip between Brussels and San Jose. [\ldots]

Here's the schedule:  [\ldots]

Another request:  I know there are going to be a lot of pressures on you by being the organizer of this meeting, but might you be able to find some time August 10 to give me a private rendition of the tutorial ``Probability and Entropy:\ Statements and Questions,'' you're giving the first day.  Or at least a gentle introduction to your work on the logic of questions?  I'd like very much to use this opportunity to get those ideas pumped into my head.

I think that's it for now.  I'll try my best to give you guys a rousing talk.  At these travel prices, I certainly won't be giving you your money's worth, but I'll try my best to inspire the young ones that there is a wide-open research field out there for being Bayesian in a quantum world.  Just because you're Bayesian, it doesn't mean you have to believe in hidden variables underneath quantum mechanics:  The world is way more interesting than that.

\section{08-07-05 \ \ {\tt  Your Touch, My Touch} \ \ (to M. S. Leifer)} \label{Leifer1}

Concerning:
\bml
The main point is that it is not necessary to include ``mental
readjustment'' unitaries in a world that only includes a single agent.
They are only required if a measurement is made by a second agent who
has incompatible information.  To summarize, I would consider
replacing your motto ``quantum mechanics describes a world that is
sensitive to our touch'' by ``QM describes a world in which my
information is sensitive to your touch, your information is sensitive
to my touch, but my information isn't necessarily sensitive to my
touch and yours isn't necessarily sensitive to your touch.'' (less
catchy I know).
\eml

But as you say, more accurate.

Anyway, this sounds strongly akin to something I once wrote in the introduction to \quantph{0009101} (a paper with Kurt Jacobs).  Read that intro---as you get a chance!---and tell me whether that meshes with your motto.  Particularly, note the stuff that starts up at the passage:
\bq\noindent
   Note the two ingredients that appear in this formulation.
   First, the information gathering or measurement is
   grounded with respect to one observer (in this case, the
   eavesdropper), while the disturbance is grounded with
   respect to another (here, the sender). In particular,
   the disturbance is a disturbance to the sender's previous
   information-this is measured by his diminished ability
   to predict the outcomes of certain measurements the legitimate
   receiver might perform. No hint of any variable
   intrinsic to the system is made use of in this formulation.
\eq
Don't forget the footnotes.

Speaking of mottos, since I still haven't written that note about the {\Vaxjo} insight, let me try to put it in a single sentence.  The extra unitary captures the agent's {\it judgment\/} of how he should conditionalize believing what he does about how he interacts with the ``measured'' system.  The unitary is literally about the extent he judges he should reject Goldstein conditionalization.  We know from Richard Jeffrey's remarks that even the method of updating is a judgment (that's why he mentions in one of those papers even wilder things than Goldstein).  That's the idea anyway---that diachronic coherence relies itself on a prior judgment---and I'd like to see if that can be made precise (and applicable to the quantum case).

\section{08-07-05 \ \ {\it The Bell Quote, at your request} \ \ (to M. O. Scully)} \label{Scully3}

It took some devious ways, but I found it.  Here's the part you were most interested in:
\bq\noindent
     Suppose for example that quantum mechanics were found to resist
     precise formulation.  Suppose that when formulation beyond FAPP
     is attempted, we find an unmovable finger obstinately pointing
     outside the subject, to the mind of the observer, to the Hindu
     scriptures, to God, or even only Gravitation?  Would not that
     be very, very interesting?
\eq
It's probably best, however, to look at the extended quote below to get a more complete sense of what he was getting at.  I'll place the extended quote at the end of this note.

The source is Bell's article:  J. S. Bell, ``Against `Measurement','' Physics World 3, pp.\ 33--40 (1990).

I believe the advertised reward for finding this stuff was \$100?  Maybe I have the exact number wrong.  In any case, I'd be happier to take a faculty appointment at Princeton instead.

\bq \noindent
{\bf\large Extended Bell Quote:}\medskip

SURELY, after 62 years, we should have an exact formulation of some serious part of quantum mechanics?  By `exact' I do not of course mean `exactly true'.  I mean only that the theory should be fully formulated in mathematical terms, with nothing left to the discretion of the theoretical physicist \ldots\ until workable approximations are needed in applications.  By `serious' I mean that some substantial fragment of physics should be covered.  Nonrelativistic `particle' quantum mechanics, perhaps with the inclusion of the electromagnetic field and a cut-off interaction, is serious enough.  For it covers `a large part of physics and the whole of chemistry' (P A M Dirac 1929 Proc.\ R. Soc.\ A {\bf 123} 714).  I mean too, by `serious', that `apparatus' should not be separated off from the rest of the world into black boxes, as if it were not made of atoms and not ruled by quantum mechanics.  \ldots

[Dirac] divided the difficulties of quantum mechanics into two classes, those of the first class and those of the second.  The first-class difficulties concerned the role of the `observer', `measurement', and so on.  Dirac thought that these problems were not ripe for solution, and should be left for later.  He expected developments in the theory which would make these problems look quite different.  It would be a waste of effort to worry overmuch about them now, especially since we get along very well in practice without solving them.

Dirac gives at least this much comfort to those who are troubled by these questions:  he sees that they exist and are difficult.  Many other distinguished physicists do not.  \ldots

I agree with [``the most sure-footed of quantum physicists''] about [this]:  ORDINARY QUANTUM MECHANICS (as far as I know) IS JUST FINE FOR ALL PRACTICAL PURPOSES.

Even when I begin by insisting on this myself, and in capital letters, it is likely to be insisted on repeatedly in the course of the discussion.  So it is convenient to have an abbreviation for the last phrase: FOR ALL PRACTICAL PURPOSES $=$ FAPP.

I can imagine a practical geometer, say an architect, being impatient with Euclid's fifth postulate, or Playfair's axiom:  of course in a plane, through a given point, you can draw only one straight line parallel to a given straight line, at least FAPP.  The reasoning of such a natural geometer might not aim at pedantic precision, and new assertions, known in the bones to be right, even if neither among the originally stated assumptions nor derived from them as theorems, might come in at any stage.  Perhaps these particular lines in the argument should, in a systematic presentation, be distinguished by this label --- FAPP --- and the conclusions likewise:  QED FAPP.

I expect that mathematicians have classified such fuzzy logics.  Certainly they have been much used by physicists.  But is there not something to be said for the approach of Euclid?  Even now that we know that Euclidean geometry is (in some sense) not quite true?  Is it not good to know what follows from what, even if it is not really necessary FAPP?  Suppose for example that quantum mechanics were found to resist precise formulation.  Suppose that when formulation beyond FAPP is attempted, we find an unmovable finger obstinately pointing outside the subject, to the mind of the observer, to the Hindu scriptures, to God, or even only Gravitation?  Would not that be very, very interesting?
\eq

\subsection{Marlan's Reply}

\bq
Very impressive---well done!

A 200\$ check ``in the mail''. How about coming down to Princeton next week for lunch?
\eq

\section{14-07-05 \ \ {\tt  hput diracqphsub.tex} \ \ (to H. Barnum)} \label{Barnum17}

\bhb
Make sure you mention to your boss that the paper primarily concerns
states that as far as we know, aren't physical!
\ehb

I think he's already well aware that almost everything I do is unphysical \ldots\ by his standards of the word ``physical'' that is.  But he's good to me nonetheless.  (We're even planning a pilgrimage to the home of C. S. {\Peirce} together.)

\section{15-07-05 \ \ {\it Krugman on Reality, Again} \ \ (to myself)} \label{FuchsC11}

From Paul Krugman, ``Karl Rove's America,''\ {\sl New York Times}, 15 July 2005:

\bq
John Gibson of Fox News says that Karl Rove should be given a medal. I agree: Mr.\ Rove should receive a medal from the American Political Science Association for his pioneering discoveries about modern American politics. The medal can, if necessary, be delivered to his prison cell.

What Mr.\ Rove understood, long before the rest of us, is that we're not living in the America of the past, where even partisans sometimes changed their views when faced with the facts. Instead, we're living in a country in which there is no longer such a thing as nonpolitical truth. In particular, there are now few, if any, limits to what conservative politicians can get away with: the faithful will follow the twists and turns of the party line with a loyalty that would have pleased the Comintern.

I first realized that we were living in Karl Rove's America during the 2000 presidential campaign, when George W. Bush began saying things about Social Security privatization and tax cuts that were simply false. At first, I thought the Bush campaign was making a big mistake -- that these blatant falsehoods would be condemned by prominent Republican politicians and Republican economists, especially those who had spent years building reputations as advocates of fiscal responsibility. In fact, with hardly any exceptions they lined up to praise Mr.\ Bush's proposals.

But the real demonstration that Mr.\ Rove understands American politics better than any pundit came after 9/11.

Every time I read a lament for the post-9/11 era of national unity, I wonder what people are talking about. On the issues I was watching, the Republicans' exploitation of the atrocity began while ground zero was still smoldering.

Mr.\ Rove has been much criticized for saying that liberals responded to the attack by wanting to offer the terrorists therapy -- but what he said about conservatives, that they ``saw the savagery of 9/11 and the attacks and prepared for war,'' is equally false. What many of them actually saw was a domestic political opportunity -- and none more so than Mr.\ Rove.

A less insightful political strategist might have hesitated right after 9/11 before using it to cast the Democrats as weak on national security. After all, there were no facts to support that accusation.

But Mr. Rove understood that the facts were irrelevant. [\ldots]

[W]hat we're getting, instead, is yet another impressive demonstration that these days, truth is political.
\eq

\section{15-07-05 \ \ {\it Arm Twisting}\ \ \ (to N. L\"utkenhaus)} \label{Luetkenhaus2}

We never heard back from you when we made a call for speakers at BBQW in Konstanz.  Thus maybe it's time to do a little arm twisting now---particularly as we have been studying the mix of volunteered talks and it would be nice to have one or two more on ``applications.''  Could we get you to give a talk on something to do with quantum crypto (predominantly) and any thoughts you have (even tangential or wildly speculative) on how the Bayesian view of quantum states might make a difference (or not) in the planning, design, or implementation of quantum cryptosystems?  It would just make our program a little more solid and less philosophical to see some of these ideas in action.

If you can work up the nerve to take the task, it would be great!  Remember this is a very informal meeting---only about science, and not about egos---so really you have nothing to lose, and you might spur some great discussions or applications with your talk.

Whatever your decision, please let me know asap.

\section{17-07-05 \ \ {\it Take the Task!}\ \ \ (to D. M. {\Appleby})} \label{Appleby8}

I apologize for taking so long to get back to you.  As it turns out, we finally made all the final talk decisions today (since {\Caves}, Hartmann, {\Schack} and I all happen to be in Waterloo).  Anyway, the decision is Yes, please do give us a talk.  And particularly, yes please do give us your passioned defense of the subjectivist account of probability.  So, you've got a task ahead of you:  But remember you basically asked for the challenge.

Your slot is on the first day, the third talk down the line.  Speakers are in order:
\begin{itemize}
\item[]
\underline{Morning:}
\item
{\Spekkens} (toy model, and arguing that quantum states should be viewed
epistemically)
\item
{\Schack} (not only epistemically, but subjectively---because that's the only sensible account of probability and ``objective chance'' plays no role anyway---and the consequences of that view for quantum
operations)
\item[]
\underline{Afternoon:}
\item
{\Appleby} (you give your talk just as you had outlined; subjectivist ideas are so foreign to some it would be good to hear them twice even if there is any overlap with {\Schack}; so basically don't worry about
overlap)
\item
Myrvold (rebutting, tries to argue that objective chance really is needed, otherwise Bayesianism would collapse when confronted with a fundamentally probabilistic theory like QM)
\item
{\Timpson} (``if the quantum state is to be associated with any cognitive state, it must be belief not knowledge,'' \ldots\ if you want to solve the ``measurement problem'')
\end{itemize}

Thus ends our arguing-for-the-subjective-view day.

Anyway, I hope that gives you a feeling for how you'll fit.

We'll post the full schedule in a couple of days; I'll send out a general announcement soon.  In the meantime, could you please send us an official title and a short abstract ASAP (prefererably before Thursday).

Certainly thanks for doing this {\Marcus}!  As I've told you before, I think your papers on probabilities are gems.

\section{18-07-05 \ \ {\it Your BBQW Talk} \ \ (to T. A. Brun)} \label{Brun8}

Sorry to be getting back to you so late.  Sadly, we've only finally made our final plans for the talks in Konstanz (Carl, Stephan, {\Ruediger} and I are all together this week at a meeting in Waterloo).

Anyway, yes please do plan to give a talk on compatible state descriptions.  You'll be talking on a more technical, less philosophical day, along with Mermin, Briegel (both on quantum computing things), Renner, and Srednicki (on de Finetti representation stuff).  But don't let that deter you from making foundational statements too.  Since we don't have any other representation of compatibility questions, your talk will be ideal---much needed as an intro to the question(s)  and subject.  More than likely, you can expect that {\Caves}, {\Schack}, or I will make some comments along the lines of our paper in the discussion session:  so that could be a nice discussion.

Looking forward to seeing you soon!

\section{18-07-05 \ \ {\it Your BBQW Talk} \ \ (to D. Wallace)} \label{Wallace4}

Thanks again for offering to give a talk at BBQW.  I hope you don't mind my scratching my earlier approval of your ``global criticism'' suggestion:  We've finally gotten to making decisions on the talks, and believe me, there's more than enough criticism of the Bayesian approach!  (Myrvold, Mermin, Wiseman, Hagar, and probably a couple others).

Anyway, we think it'd be much more productive for you to talk about your decision theoretic derivation of the probability law in Everettian QM---actually that subject sorely needs some representation.  Your talk will be the Wednesday morning.  In the same session, we're hoping to also get Barnum to talk a little dually on your stuff and on Zurek's recent derivation.  Also Carl Caves may talk on the frequency operator approach of Hartle and Farhi--Goldstone--Gutman.

We'll send out an announcement with the full talk schedule soon.  However, in the meantime, if you could send an official title and a short abstract ASAP that'd be great!

I hope you'll find this meeting a productive one; in any case, I'm looking forward to seeing you again.

\section{25-07-05 \ \ {\it Sessions and Chairs} \ \ (to V. {\Palge})} \label{Palge0}

Below you'll find the session titles and chairs.  Also, please change the title of my opening talk to ``Fuchsian BBQ(w), a Sampler of Meats and Sauces''.  After lunch, I'll send you some extra abstracts from some of the speakers that I have accumulated.  I'll be back in about an hour.

\begin{itemize}
\item
Monday Morning\\
``Epistemic Quantum States and the Paths They Lead To''\\
Chair:  Stephen Bartlett

\item
Monday Afternoon\\
``Subjective vs.\ Objective Probabilities in QM''\\
Chair:  Jeffrey Bub

\item
Tuesday Morning\\
``Implications from Quantum Computing''\\
Chair:  Mauro D'Ariano

\item
Tuesday Afternoon\\
``Technical Bayesian Theorems''\\
Chair:  Norbert L\"utkenhaus

\item
Wednesday Morning\\
``Bayesian Probabilities in Everettian Worlds''\\
Chair:  Huw Price

\item
Thursday Morning\\
``Updating Beliefs and Diachronic Coherence''\\
Chair:  Jos Uffink

\item
Thursday Afternoon\\
``Updating Beliefs and a Touch of Copenhagen''\\
Chair:  Holger Lyre

\item
Friday Morning\\
``Separating Subjective from Objective in QM (if possible!)''\\
Chair:  Paul Busch

\item
Friday Afternoon\\
``Changing the Course of Physics, Some Ideas''\\
Chair:  Joseph Renes
\end{itemize}

\section{25-07-05 \ \ {\it Bad Jokes} \ \ (to N. D. {\Mermin})} \label{Mermin118}

Regarding:
\bdm
Please use the full title I sent you:
\bq\rm
\noindent Does being Bayesian illuminate the quantum world?\\
Or is the quantum world an embarrassment to the Bayesian?\medskip\\
Abstract: The speaker will meditate on what he likes and what he
dislikes about taking a Bayesian view of quantum theoretic
probabilities.   The talk will be very informal if only because, as
we all know, there is no quantum world.
\eq
\edm
As you should know, I think there is ONLY a quantum world.  Did you
forget about me when you wrote your abstract?

\bdm
Just an example of the kind of bad jokes I hoped I'd be able to
string together.  I had in mind the Bohr quote.  Nothing more.  But I
did think it was a Law of Thought for you.  Not the same as a world.
\edm
I knew you had the Bohr quote in mind; I was trying to make my own
bad joke.

Concerning, ``But I did think it was a Law of Thought for you.''
\ldots\ you never cease to shock me \ldots.  And you never cease to
cause me to strive to try to convey the very simple little idea more
effectively! I wonder when I'm gonna finally hit the sweet spot?
Quantum THEORY, a law of thought:  Yes.  Resoundingly yes.  But the
quantum WORLD---i.e., that situation, that world, that reality, which
conditions us to choose THIS law of thought rather than THAT law of
thought (in other words some alternative or imaginary law of
thought)---is something else entirely.  It's the stuff that's here
whether there are any law-of-thoughters around or not.  That's what I
really want to get at; that's what I've always really wanted to get
at.

\section{26-07-05 \ \ {\it Am I Turning Into You?}\ \ \ (to S. Aaronson)} \label{Aaronson8}

\bsa
Consider the terrifying evidence at
\begin{center}\myurl{http://www.scottaaronson.com/papers/are.ps}\end{center} and in particular, the following sentence:
\bq\noindent
{\rm ``If exponentially-long strings were rocket fuel, and probability distributions were grape juice, then quantum states would be wine -- the alcoholic `kick' in this analogy being the minus signs.''}
\eq
Headed for an {\tt arXiv} near you (though there's still time to revise, and thereby reduce the foolishness quotient \ldots).
\esa

For God's sake let's hope not!  I've seen what it looks like from the inside, and I can tell you it ain't pretty.

However, I very much like this turn I'm seeing in you of dispelling {\it some\/} of the mystery of quantum computing by comparing quantum states to probability distributions.  The exponentiality in the latter is something Caves and I have played up here and there over the years.  Have you ever read our old paper, \quantph{9601025}?  Caves probably has something better by himself by now, but we laid the groundwork there.

Anyway, you don't realize it, but you might just be starting to fall down the slippery slope of becoming a quantum Bayesian.

Incidentally, I'm starting to think there must be something wrong with the idea that it's the minus signs that give the kick.  For the minus signs are representation dependent.  Take the representation of QM that I keep playing up, where the quantum state is represented as a single probability distribution full stop.  No minus signs at that level.  Now it is true that the minus signs then reappear in the time evolution equations.  But if one talks about the one-way computational model of Raussendorf and Briegel, even those minus signs go away.  So a kind of probabilistic model for quantum computation with no minus signs ever appearing.  What appears instead is an interesting relation between the marginal and joint event spaces.  One that apparently can't be sustained classically.  I wish I had your brain so that I'd be better equipped to explore these kinds of questions.

I'm finally going to go to Princeton today.  I guess you won't be there!

\section{27-07-05 \ \ {\it Wow!}\ \ \ (to M. Dickson)} \label{Dickson1}

I just saw your talk proposal for BBQW,
\bq\noindent
Subjective and Objective in Quantum Theory\medskip\\
Chris Fuchs has proposed to understand the quantum state in terms of the ``Bayesian idea that the probability one ascribes to a phenomenon amounts to nothing other than the gambling commitments one is willing to make on it'' (Fuchs 2003). Correctly predicting that jaws will drop in reaction to such an idea, Fuchs goes on to emphasize that it is part of a larger project of disentangling the subjective from the objective in quantum theory. This talk will be a reflection on that project. First, drawing on philosophical movements from the late 19th and early 20th century, specifically surrounding the philosophy of geometry and the role of analytic truths in physical theories, I will suggest that the interplay between subjective and objective in physical theory is complex and subtle, and itself depends on certain choices made by us. (Fuchs drops hints in the same direction. I will try to flesh out my own understanding of these subtleties.) Second, I will illustrate this idea, drawing on themes from Bohr (and possibly other Copenhagen-type approaches to quantum theory, but I make no claims on that point). Specifically, I will consider, as (I would argue) Bohr did, the question of how we make `contact' between quantum state-ascriptions and actual experimental observations. The result is suggestive of a neo-Kantian (or neo-Carnapian) philosophical foundation for the `subjectivist' approach to quantum theory, and in any case it might contribute to Fuchs's project.\medskip\\
Reference: C. A. Fuchs, ``Quantum Mechanics as Quantum Information, Mostly,'' J. Mod.\ Opt.\ {\bf 50}, 987 (2003).
\eq
and wow!  I'm flattered that you'd take me seriously enough to talk about some of my stuff (even if only to speak to its difficulties).

I guess I've talked more about the specific subject of your abstract in several places (other than the paper you cite, that is), but maybe let me draw your attention to the most relevant one in case you haven't seen it yet---I think it's a substantial development of what you've seen (if that's all you've seen).  It's a weird pseudo-paper titled ``The Anti-{\Vaxjo} Interpretation of Quantum Mechanics''.  You can find it here \quantph{0204146}, and the most relevant sections for you would be Sections 4 and 5.

I've had only one significant response to it from a philosopher---it was from Michel Bitbol.  Since you will also talk of the neo-Kantian tradition, let me place it below. [See preply to 10-12-03 note ``\myref{Bitbol1}{First Meeting}'' with M. Bitbol.]

I don't know if these materials will help you in any way---particularly at this late date---or only paint a more inconsistent picture of myself, but in the interest of making as much progress as we can in our limited time, I thought I should point them out to you.

I'm definitely looking forward to meeting you; I've only seen you in the distance before.  And I'm looking forward to learning from you.

\section{28-07-05 \ \ {\it Asynchrony!}\ \ \ (to A. Plotnitsky)} \label{Plotnitsky17}

\barkp
I assume that you got an invitation from George Welch for the Snowbird
conference, since I just got one myself.  I did manage to get in touch
with David from Australia, who cannot come--too bad of course.
\earkp

No, I did not get anything from George Welch.  Could you have him send whatever he sent again?  (I don't know George Welch.)  Coincidentally, I just had lunch with Marlan Scully a couple of days ago.  We talked mostly about his son's book and about free will.  Marlan has quite a religious side too---I had not realized the extent of that before.

At the very least, looking forward to seeing you in Snowbird \ldots\ but I hope we hook up again before that.

Maybe I should give as a title for my talk, ``Not the Incompleteness of Quantum Mechanics, but the Incompleteness of Bohr's Reply to EPR''.  Who are you going to get as a replacement for David?

\section{28-07-05 \ \ {\it Calendar Marked} \ \ (to W. E. Lawrence)} \label{Lawrence4.1}

OK, I've marked out my calendar for Sept 29 through Oct 1.  I'll plan to drive in the Thursday morning (well before the afternoon talk), and take off Saturday morning after breakfast.  Maybe there's some chance I can get my family to come with me; if I do, I'll certainly pay any extra hotel costs that incurs.  I'll let you know their decision as the time draws nearer.

Have fun on your vacation.  We're just taking off for Germany ourselves today.  The family will be there for 3 weeks, as I shuttle about here and there for meetings.  I'm very proud of the meeting I organized on Bayesian probability in quantum mechanics.  You might snoop around the webpage if you get a chance:
\begin{center}
\myurl[http://web.archive.org/web/20090511060206/http://www.uni-konstanz.de/ppm/events/bbqw2005/]{http://web.archive.org/web/20090511060206/http:// www.uni-konstanz.de/ppm/events/bbqw2005/}.
\end{center}
I suspect there are several talks you would have enjoyed there.  So, anyway, that part of my shuttling about will be a labor of love.  The ridiculous part of my shuttling comes from my jumping back to San Jose for a two-day visit to the MaxEnt conference in the middle of all that!  Now that there's no way for me to get out of it, I'm really starting to regret it!  Thus much is now for sure though:  From August 19, when we return to New Jersey, to September 29, when I come to Hanover, I'm not traveling anywhere else!!

So I should be fresh and happy again by the time I give your seminars.

\section{28-07-05 \ \ {\it Bernardo and Smith and Savage} \ \ (to S. Hartmann \& V. {\Palge})} \label{Hartmann11} \label{Palge1}

Do either of you have a copy of Bernardo and Smith's book {\sl Bayesian Theory\/} and Savage's book {\sl Foundations of Statistics\/} or the book {\sl Studies in Subjective Probability\/} (can't remember the editors)?  Anyway, I'm bringing a few books, like Jeffrey's two books and de Finetti's collected papers, but I can't fit much more into my suitcase.  If you guys have the above-said books (and maybe Howson and Urbach too), could you bring them to the meeting?  We could put things like this out on a display table for perusing or if anyone needs something for a sleepless night.  Also it'll make it easier for us to dig up a quote if need be.

Also, we should have a display table for the participants to lay out copies of recent papers, etc., if possible.

\section{09-08-05 \ \ {\it Slides}\ \ \ (to H. M. Wiseman)} \label{Wiseman18}

Would you mind sending me a copy of the talk you gave at BBQW?  I'd like to study the slides and ultimately reply.

(You know, if I were a solipsist, I wouldn't have to ask you for the slides \ldots\ nor would I ever have to use the Born rule, as I'd know the outcomes of all measurements in advance.  Solipsists have no need to make requests or gamble.)

\section{09-08-05 \ \ {\it Figure Supplement}\ \ \ (to H. Poirier)} \label{Poirier1}

It dawned on me that this figure from my talks might be a useful supplement for you with respect to the paper I sent you last week.  Unfortunately I don't have the latest version of it scanned into my computer, but this might still help.  (The stick figure now says, ``Ouch, $d$!''\ rather than ``Aha, $d$!''  Small conceptual difference, as ``aha!''\ conveys a discovery, while ``ouch!''\ conveys the idea of a simple physiological reaction.)

Anyway, the figure is attached.  Looking forward to your article.

\section{10-08-05 \ \ {\it Targus and American} \ \ (to C. M. {\Caves})} \label{Caves79.6}

\bcc
Home now.  Sorry I didn't get to say good-bye to you.  Although
I was in a pretty deep funk on Friday morning, I recovered as the day
went on and left in a pretty good mood, thinking the whole thing [BBQW] went
pretty well.
\ecc

I go up and I go down about the meeting.  Mostly I go down.  Mostly I think it was an ``opportunity lost'' \ldots\ predominantly due to my bad planning.

What was your final impression of your discussions with Unruh?  Did anything fruitful come of that?  Did he nail us on any problems we haven't thought about?

I'm at MaxEnt at the moment and tremendously exhausted.  Whereas you told Kathryn Laskey that the BBQW meeting enlivened you, mostly it has just further endeadened me.

\section{11-08-05 \ \ {\it So Slow Fuchs} \ \ (to A. Wilce)} \label{Wilce6}

Sorry to take so long to reply to you; I've fallen by the wayside again with this exhausting schedule of travels and particularly the Konstanz meeting.  As I write to you, I'm returning to Munich from the MaxEnt meeting in San Jose.

\baw
The conference has left my head buzzing with ideas, some of which I'd
like to discuss with you at some point.
\eaw

Glad to hear that.  And I always enjoy my discussions with you; I'm always impressed by your simultaneous mix of precision and tolerance (particularly for the imprecise likes of me).

\baw
Also -- I've finally unearthed an ancient, but possibly very
interesting, doctoral dissertation on Bayesian methods in generalized
probability theory (read, quantum logic), by Marie Gaudard, a student
of C. H. Randall. Would you like to have a copy?
\eaw

Yes please.  I have this dream of traveling no more until the new year come August 19.  If I can follow that, there some chance I might actually a paper or two read this Fall!

\section{12-08-05 \ \ {\it Per Diem?}\ \ \ (to K. H. Knuth)} \label{Knuth5}

Thanks again for inviting me.  I got a lot out of interacting with you, Ariel, Carlos, John Skilling, Andy Charman, Mike West and others.  We definitely should have had you all at our QM-Bayesian meeting in Konstanz; that was a big oversight, and I won't make it again.

\section{12-08-05 \ \ {\it Thanks}\ \ \ (to R. Renner)} \label{Renner1}

\brr
I am writing you to thank you again for the invitation and also for the organization of the nice BBQW workshop. I really enjoyed it and, clearly, I learned a lot. (Unfortunately, I had to run for the train last Friday, so I missed to say you goodbye personally.)
\err

I'm glad you enjoyed it.

Homework Assignment:  Go to Bernardo and Smith's book {\sl Bayesian Theory}, turn to Chapter 4 on ``modelling.''  Prove quantum versions of all the representation theorems in there.

Take care, and I hope to see you soon.

\section{12-08-05 \ \ {\it Quantum Bayesians and Anti-Bayesians} \ \ (to W. C. Myrvold)} \label{Myrvold3}

Thanks for including me on the distribution list of your long note.  This is something I want to eventually come back to, but at the moment I'm going to have to plead ``organizer's exhaustion''.  As I write to you, I'm on my way back to Munich from the MaxEnt conference in San Jose.  Soon after completing this note, my plan is to simply fall into a deep, deep sleep---it may last a couple of weeks.  (First I plan to vacation a little in Munich, but then it'll probably take me a week to get my life established again in New Jersey upon my family's return home.)  Plus by the time I finally open my sleepy eyes, I'll have the benefit of being able to review everyone else's reply to you before composing my own thoughts.

Keep up the good work!

Hey, here's a challenge to you.  I think the single most significant reading for sending me toward the radical-probabilist end of Bayesianism was de Finetti's paper, ``Probabilism,'' in {\sl Erkenntnis\/} {\bf 31}, pp.\ 169--223 (1989).  I'm pretty sure the same is true for {\Ruediger}, and possibly even for {\Appleby}.  Here's my challenge:  Pinpoint the places where you think de Finetti errs in that paper.  If you can convince me of some significant errors in that paper, you might just turn me back toward the temple of objective chance (from whence I originally came into this game).  Maybe Unruh, Duwell, and Harper could take the challenge too.

But as I say, for now I close my sleepy eyes.

\subsection{Wayne's Preply}

\bq
I don't know about you guys, but I found the Konstanz conference very valuable, and I believe I have a better grasp now of the views on the `quantum Bayesians', and my conversations with people at the conference helped me to sharpen my own thoughts.  Following is meant to be an attempt to continue the discussion, and to get clearer in my own mind what I think.  Response (and disagreement) welcome!

We're all interested, I think, in the question: What do we accept when we accept qm?  And accepting qm means, at minimum accepting the quantum rules as the ones to adjust one's credences to.  And so I'm going to couch everything that follows in terms of what accepting certain credences commits one to.

Start with a case that has nothing distinctively quantum about it.  I'm thinking of a quantum experiment with a fixed experimental setup, but that won't be essential.   Suppose there is a sequence of events over which Bayesian Peter (a character I borrow from Bas van Fraassen) has an exchangeable credence function (and here I'm departing a bit from deF's terminology, which makes ``exchangeable'' modify ``sequence''; from deF's own point of view, this is misleading, because exchangeability has to do with a credence function and not with the events themselves).   What does such a credence function commit Peter to?

(Though I'm going to assume exchangeability, pretty much everything I'm going to say goes through if Peter's credences can be written as a mixture over a richer set of distributions, including, perhaps, correlations between elements of the sequence or between elements of the sequence and other events; invoke the appropriate generalization of the de Finetti representation theorem).

Suppose that Peter has an opportunity to bet on the 101st element of the sequence, and he is given a choice between betting according to his current credences, and learning the outcomes of the first 100 elements, and then betting on the 101st.  Unsurprisingly, he can't prefer the former; he must regard the more informed credences as being at least as good as the less informed, and, provided that he attaches nonzero credence to the proposition that his credence in the 101st will undergo some change as a result of conditionalizing on the first 100, he will strictly prefer the informed credences.

Of course, we could save Peter the trouble of conditionalizing and just tell him what his credences would be if he were to conditionalize on the first 100, and he would gladly swap these credences for his current ones.  Moreover, we can write his current credences regarding element \#101 as his current expectation value of what his credences will be after learning the first 100 elements of the sequence.

And so on; Peter will prefer credences conditional on more data to credences conditional on less, in the sense that he will pay some money to be told what they are, and will adopt them once he knows what they are. Consider the set of credence functions consisting of all the credence functions that would result from Peter's priors by conditionalizing on finite initial segments of the sequence.  Peter believes --- that is, he assigns credence one to it --- that this set has a limiting distribution.  Moreover, if he knew what this limiting distribution is --- if God could hand it to him on a platter --- he would prefer it to all the members of the set.

Of course, what Peter would prefer to all of these, would be for God to hand him on a platter the results of the whole infinite sequence, once and for all.  Moreover, if Peter is a determinist, he believes that these results are already determined by the current state of the world, which is of course imperfectly known to him.  Since Peter will, at any given time, only have a finite body of data, data about past events, all he will ever have, with regards to future events, is credences conditional on finite initial segments.  The limiting distribution is the least upper bound, in Peter's preference ordering over things that God could hand him on a platter, of the set of finitely attainable credences.

Peter has certain credences about what this limiting distribution is, and his current credences about future events are his epistemic expectation values of the probability they have on the limiting distribution.  (That is, though Peter doesn't know what the limiting distribution is, he has some opinions about it, and his current degree of belief in an event $E$ is his estimate of what probability is given to $E$ by the limiting distribution).

What status does this limiting distribution have, for Peter?  It is {\it not\/} epistemic, in the sense of being anyone's degree of belief.   It's also the sort of thing once can have degrees of belief about, and Peter regards his process of updating on data as becoming better and better informed about what the limiting distribution is.

Now suppose we have a situation like that envisaged by Lucien Hardy in his 5 axioms papers: we have a preparation device, a transformation device, and a measurement device, all with various knobs on it.  Peter starts fiddling with the apparatus, doing various experiments with the knobs at various settings, and conditionalizing on the results.  If his initial credences concerning the outcomes of the experiments are reasonable, he will, with enough experimentation, start to converge towards the quantum rules concerning correlations of outcomes with the settings of the various knobs.  For any setting of the knobs on the devices, Peter's credences will be a mixture over possibilities of an ideal limiting distribution associated with those settings.  This ideal limiting distribution has the same status as it did before.  It is not Peter's or anyone else's actual belief state, but an imperfectly known, ideally optimal belief state for a given configuration of knob settings.  (`Optimal' meaning judged by Peter to be optimal, in the sense that he prefers having it to any finitely attainable belief state.)

If Bayesian Alice starts with priors different from Peter's, no finite amount of data will bring them into exact agreement.  But, provided neither regards the other as excessively dogmatic, they both believe (attach probability one to) the proposition that experimentation indefinitely continued would lead  both of them to converge to the same limiting distribution.  And they can both regard their current differences in credences to be differences in opinion about what that ideally optimal limiting distribution is.  (And this last remark would be true even if the conditions for convergence were not satisfied.  Both could regard their differences in opinion as differences in opinion about the ideally optimal distribution, but each regard the other of incapable of properly learning about it)

So --- at what point in the above train of thought would the quantum Bayesians want to get off?  Because we've gotten close to how I characterized objective chances. They are associated with chance set-ups (knob settings), and are optimal credences in the sense that they cannot be bettered on the basis of information that is in principle available to agents making wagers on the events in question.  They are the sorts of things one can have degrees of belief about, have differences of opinion about, and regard ourselves as learning about.
\eq

\subsection{Wayne's Reply}

\bq
Chris suggested that I look at ``Probabilismo,'' and it so happened that, when I got that e-mail, I had  just come back from the library, where I was copying it.  I've since finished reading it, and I must say that last paragraph came as a nasty shock!  (I had previously known nothing of de Finetti's political convictions). It should come with a disclaimer: Warning: Political content may offend some reader!

But of course it's not de Finetti's politics that are at issue.  Here are a few preliminary comments on ``Probabilismo.''

1. It is simply presumed throughout that ``probability'' is univocal; the notion that (as Poisson pointed out) ``probability'' is used in two distinct senses is wholly absent.  Hence there is a presumption that we are looking for a single notion of probability to cover all its uses.  This is a royal road to subjectivism regarding probability. If the choice had to be between holding  (a) ``probability'' always refers to degrees of belief, and (b) ``probability'' always refers to objective chance, then (a) wins hands down and in fact (b) doesn't even get out of the gate.

This isn't a minor point; I think that a major source of confusion in the literature on the interpretations of probability has been the idea that we have to pick {\it one\/} interpretation as the sole legitimate notion.  It's as if someone were reading my e-mails over the past few months, unaware that there are two cities named ``London''; such a person could easily come to the conclusion that London is a strange and peculiar place or that at least I have bizarre or contradictory beliefs about it (E.g, I got on a plane in London for a 7-hour plane ride to Toronto, after which I took a 2-hour bus ride to London.)

Having decided that probability is always epistemic, de Finetti calls the notion of an unknown probability of heads that we are trying to estimate nonsensical and metaphysical.  As indeed it would be if we were talking about an unknown credence.  An unknown chance that we have credences about is less obviously nonsensical.

The closest the de Finetti comes to offering an argument against admitting chances is the assertion that the hypotheses that the chance of heads has a certain value is not directly verifiable.  Which is true; we test such hypotheses by constructing a sequence of events that we are reasonably certain all have close to the same chance, and update our credences on the results of that sequence.  I count this as a measurement of the value of the chance.  Rejecting this as excessively indirect would be to reject far too much, as the procedure is not essentially different from the measurement of {\it any\/} physical quantity using noisy data (and {\it all\/} data is at least somewhat noisy).  Any epistemological considerations that could be held against one could equally well be held against the other.  In particular, it does no good to point out that one must start the procedure with certain assumptions, or that no amount of data will dictate a unique value as our estimate of the parameter or a unique set of posterior credences regarding that value of the parameter; this holds for {\it any\/} parameter estimation via statistical techniques, not just to estimation of chances.

2.  De Finetti's subjectivism seems, at least in part, to be motivated by a fallacious inference from the fact that every judgment is subjective --- that is, made by and belonging to a subject --- to a conclusion about the content of the judgment, namely, that the subjective judgment cannot be a judgment {\it about\/} anything that is not subjective.  This inference, fortunately, is not a valid one, because, if it were, we would have to conclude, not merely that all probabilities are subjective, but that everything is, and we would land in idealism or even solipsism.  To his credit (?), de Finetti does indeed take this extra step.  In section 30 (p.\ 214) he denies the existence of ``external reality'' (his scare-quote), and in section  2 (p.\ 171) he quotes Tilgher, apparently with approval: ``All the objects, men, and things of which I speak are, in the last analysis, only the content of my present act of thought: the very statement that they exist outside and independently of me is an act of my thought \ldots.''  This is Tilgher's paraphrase of what Berkeley commentators call Berkeley's ``master argument'' for idealism.  And it's no better here than it is in Berkeley.

3. De Finetti frequently draws an analogy between judgments of probability and matters of taste.  This is not part of any argument but merely part of the rhetoric of his presentation.  But in case this rhetoric is having any persuasive effect, perhaps I can curb the appeal of this analogy by pointing out that, from the point of view of Bayesian decision theory, the analogy is not a very strong one.  Matters of taste come into decisions via their influence on utilities; and utilities and probabilities play essentially different roles in decisions.  One important difference is that we are obliged to regard new information as potentially improving our probability judgments; given the choice of choosing according to my present credences, or first updating on information relevant to my choice and then making my choice, an expected utility calculation tells me to choose the latter.  There is nothing comparable when it comes to utilities.  Suppose I am reasonably certain that Alice started out with priors similar to mine, and then gained new information, so that her current credences are close to what mine would be if I were to update on the information she has.  Then, judged by my own current credences, I would be better off following her choices than making the choice based on my own.  If, on the other hand, I know that Alice started out with tastes similar to my own and then underwent a process that made her prefer castor oil to cognac, I will still order cognac after dinner; I do not regard the fact that Alice chooses castor oil as relevant to my decision.  One way to put this is: we can regard change of credences as pure learning experiences.  A chance of tastes (utilities) is {\it never\/} a pure learning experience.
\eq

\section{12-08-05 \ \ {\it Your Summary} \ \ (to W. G. Unruh)} \label{Unruh5}

Thanks for sending me the summary of your concerns.  As you'll see from the other note that I just wrote to Wayne Myrvold (and cc'd you on) though, I'm going to drop out of discussion for a couple of weeks.  So I don't have a detailed reply for you at the moment.

I did however take ten minutes to compile a couple of emails that I wrote to Mermin a couple of years ago on the very subject you bring up.  [See 27-06-03 note ``\myref{Mermin86}{Utter Rubbish and Internal Consistency, Part I}'' and 28-06-03 note ``\myref{Mermin86}{Utter Rubbish and Internal Consistency, Part II}'' to R. Schack, C. M. Caves \& N. D. Mermin.] The notes are somewhat inadequate as a reply to you, but they're along the lines to a beginning of a reply, so I'll go ahead and attach them to the present email.  You'll see their relevance if you have a read, but I can predict in advance that I didn't say anything that will convince you.

More later, but it may be a couple or three weeks.

\subsection{Bill's Preply}

\bq
Hope you had a good trip to Munich, got there safely and had a good
birthday celebration with your wife.

Just to summarize my concerns.
You want the probabilities to be a state of belief because you want to
argue that when, in the Bell problem, $A$ measures her $S_x$ and finds $+1/2$,
then the change in the probability at $B$ of $S_x$ from 50-50 to ``certainty''
for $+1/2$ is simply a change in the state of belief of the person who who
has this additional fact. That change in belief, of the person, has nothing
to do with the world itself, and such a change in belief thus does not
(non-locally) affect the world itself.
I would say that the main problem with that line of argument is that then
there is an additional fact, namely if you bet on the basis of that belief,
then you keep winning. Now, the belief may be personal, but the fact that
that that belief keeps you winning is more than just personal, it also
tells you something about the world. There is some matter of fact of the
world which puts its behavior into compliance with your state of belief.
That compliance, that correlation is something that needs explanation, just
as in your opinion, in the Bell case, where the probabilities are
considered objective facts about the world, the change in the probabilities
for B in the face of the measurements by A is something that requires
explanation (and you explain by non-locality).
I.e., it is not at all clear to me that making the probabilities to be beliefs and
subjective buys you much (anything) except to change what it is that needs
explanation.
Saying that that correlation between your beliefs and outcomes at $A$
requires no explanation seems to me to be as problematic as saying that the
change in objective probabilities at $B$ due to the measurement at $A$ requires no
explanation.

On the commutation of the informations, I guess I would take the state of
two two level particles to be $\sqrt{.51} |11\rangle+ \sqrt{.49} |00\rangle$. This is all that $B$ knows and for him
the state which best describes the second system is diagonal in the above
$|0\rangle$,$|1\rangle$ basis, with .51 and .49 on the diagonal. However $A$ measures the
first particle in the $|+\rangle$,$|-\rangle$ basis (call it the $z$ basis
$|+\rangle=1/\sqrt{2}(|1\rangle+|0\rangle)$, and finds its value to be $+$. For $A$, the optimal
state of the second particle is in the state $\sqrt{.51}|1\rangle+\sqrt{.49}|0\rangle$
Whose projector does not commute with $B$'s density matrix.

The 0 probability case is more difficult (I think I misunderstood that
constraints on the problem when on the train), and I am not sure what to
think in that case. Have you actually come up with a case where it is zero
(i.e., where the density matrix for $B$ times that for $A$ is 0)?
\eq

\section{18-08-05 \ \ {\it The Big Bang Happened Here} \ \ (to T. Duncan)} \label{Duncan1}

Not much time for a long note at the moment:  My wife, children, and I are on the last day of our vacation in Munich (seeing the grandparents), and we've got a dinner date soon \ldots\ followed by packing up for an early morning flight tomorrow.  So I can't write much.  But I've been enjoying getting to know you a little today.

First off thanks for the note---which I'll study and reply to eventually---and thanks for the links at the bottom of it that led me to the SII pages and others.  I've got to say, I'm quite intrigued by your life and efforts---what a beautiful idea the SII is!

As I say, I'll try to address your note specifically in the near future; you have a point worth mulling over in more detail (about what one might mean more specifically by ``incomplete information'' in a world without hidden variables).  In the meantime though, let me attach a little something somewhat on the subject that I have been contemplating posting on {\tt quant-ph} (at least Howard Wiseman wants me to, so that he can reply to it officially).  It's attached as a PDF file; the title is ``Delirium Quantum''.  I titled the present email after your textbook's title because it reminded me of a line I use in Section 7 of this paper:  ``That (metaphorically, or maybe not so metaphorically) the big bang is, in part, right here all around us?''

Listen, I didn't give you a copy of my book {\sl Notes on a Paulian Idea\/} in Sweden, did I?  At least I don't have your name recorded in my list.  If you'd like a copy, I'll have one sent to you:  I still have loads of copies of the VUP edition to give away (before the Springer edition comes out), and you might enjoy perusing it given your interests.

\subsection{Todd's Preply}

\bq
It was good to meet you in {\Vaxjo} back in June --- thanks for the references you pointed me to. The {\Spekkens} paper helped clarify the vague question I was trying to articulate after your talk, about how interference effects arise in the epistemic view of quantum states. From the correspondence between spin states and epistemic states in the toy model, I see now how the incompleteness of maximum knowledge makes it possible to define a binary operation that behaves as a coherent superposition (and hence produces interference). This is appealing since it's a glimpse of a deeper principle that could underlie the coherent superposition of quantum states.

I'd still like to gain a more concrete understanding of why this happens and what it means. It's difficult to get direct intuition into what's going on because much is buried in whatever ``nature'' (in the toy model) must be doing in order to enforce the knowledge balance principle. I guess this is essentially the point you make for the real universe, in asking (of the fact that maximal knowledge of a system is incomplete knowledge), ``Why can it not be completed?''  The heart of the matter lies in that question.  As I think about it, I'm not even sure what it means for knowledge to be incomplete other than in relation to hidden variables in the spirit Einstein described.  But as you said, that cannot be the type of incompleteness that's really going on.

This leads to a point that is still muddy to me. It may be a naive question and I'm not quite sure how to phrase it, but let me try this to get started: Given that we must ask what the ontic states can be, such that they force this restriction of incomplete knowledge on the epistemic states, why not just place the entire restriction directly on the ontic states themselves?  Why not simply say there is a limit to what we can {\it know\/} about the state of the world because there is a fundamental limit to how much information can actually be ``stored'' about its state? As I understand it, this is something like the view expressed e.g.\ by Zeilinger (``A Foundational Principle for QM'', {\sl Found.\ Phys.}\ 1999). I realize this is not what you're saying. So, I'd like to better understand why you emphasize the distinction between epistemic and ontic states, when it seems possible that all of the restrictions on our knowledge could be absorbed as consequences of the restrictions on the ontic states.  Anyway I'd appreciate any thoughts you have to point me in the right direction here.

On a more general note, I've read and pondered more of your writing.  Your ideas are inspiring --- I found myself cheering you on as I read, even (perhaps especially) the asides and footnotes. (I'm thinking particularly of your replies to Preskill and to Wootters, in ``The Anti-{\Vaxjo} Interpretation of QM.'') I have no idea if you make these comments for the same underlying reasons that I make similar comments, but I'd very much like to learn more about your philosophical perspective behind them. So I'll highlight a few points that struck me, and if you have time/interest, I'd love to hear your additional thoughts.

First, thanks for leading me back into William James' writing. I remember reading several of his essays and finding encouragement when I was in grad school trying to articulate the deeper motives and questions guiding my work. I especially like the opening pages of ``The Present Dilemma in Philosophy'' because he articulates so well the practical importance of one's view of the universe:  ``The philosophy which is so important in each of us is not a technical matter; it is our more or less dumb sense of what life honestly and deeply means. It is only partly got from books; it is our individual way of just seeing and feeling the total push and pressure of the cosmos.'' ({\sl Pragmatism}, lecture 1 -- The present dilemma in philosophy)

Next, your reminder that the usefulness of theories does not mean they are part of the blueprint of the universe is well articulated --- and very important, I think. Have you read Edward Harrison's book, {\sl Masks of the Universe}?  My background is in cosmology so I'm used to thinking of the issue in that language, but it's very similar. Harrison expresses it by highlighting the distinction between the real Universe (capital U) and our model universes (lowercase u) which are masks that we try to fit to the real Universe. Here's a passage from Harrison which captures the essence of the point:
\bq
The universes are our models of the Universe. They are great schemes of intricate thought --- grand belief systems --- that rationalize the human experience. They harmonize and invest with meaning the rising and setting Sun, the waxing and waning Moon, the jeweled lights of the night sky, the landscape of rocks and trees, and the tumult of everyday life. Each determines what is perceived and what constitutes valid knowledge, and the members of a society believe what they perceive and perceive what they believe. A universe is a mask fitted on the face of the unknown Universe.
\eq

Of course it's one thing to acknowledge superficially that, yes, of course, our models are limited and don't really capture everything about reality, \ldots\ but then going about our day-to-day work acting as if we can capture everything. (And as Wootters mentioned, there probably is some practical value in pretending this as a matter of methodology.) But what particularly intrigues me about what you're doing is that you seem to be bringing this awareness directly into the practical methodology of doing science. I'm not sure I said that very well ---I guess what I mean is that there could be situations where there is practical value in NOT pretending that theories can perfectly mirror reality. Explicitly building that point into a theory (as you seem to be doing for QM, if I understand correctly) is tremendously important I think.

This email is already getting long, so I'll stop here and save comments about other points from your writing for another day.
\eq

\section{22-08-05 \ \ {\it Exercise 1.image} \ \ (to T. Duncan)} \label{Duncan2}

\btd
Thanks for the ``Delirium Quantum'' paper, which I've started browsing. I
love your idea that new facts are being created around us all the time.
{\sl The Big Bang Happened Here\/} title evolved from a somewhat weaker point I
like to make in lectures --- that when we learn about the history of the
cosmos, we are not just learning about distant things detached from us.
Rather, we are discovering the history even of the space we are
immersed in right here. A little bit of space we can hold in our hands
once glowed with the heat of the big bang. Your point is a stronger
one, that the process is ongoing even now. And yes, it is
spine-tingling to think about that, to feel that our actions may
``count'' in some truly fundamental way. (BTW the title of our textbook
has changed now to {\sl Your Cosmic Context}, so it's funny that you found
an old reference to it. I've attached a draft of the first chapter
since it has some comments on models of the universe that might be of
interest to you.)
\etd

OK, so you already know how I'd answer your Exercise 1.image:  ``I am a cosmic artist -- a contributor to a universal creative process.''  Thanks for sending Chapter 1 of your book, {\sl Your Cosmic Context}.  I just enjoyed reading the first 13 pages of it.

I sent a note the secretary at VUP to have a copy of {\sl Notes on a Paulian Idea\/} sent to you.  Per my promise, I will eventually get back to your more substantial points (from your first email), I just have to have a thought or two before embarking on the mission.  (And I've got to get caught up a lot, after having been absent from Bell Labs for almost a month!)

\section{23-08-05 \ \ {\it The Article for \underline{Science \& Vie}}\ \ \ (to H. Poirier)} \label{Poirier2}

\bhpoirier
Hardy proved that QM can be seen as a Generalised Probability Theory,
or in other words, as a tool to calculate the probability associated
with each outcome of any measurement that may be performed on a
system prepared by the associated preparation.

As far as I understand your position, this redefinition of QM is not
so far from yours (when you say that the quantum state is solely an
expression of subjective information about the potential consequences
of our experimental interventions into nature).

So my problem is: If all the characteristics of QM can be seen as
generalised probability tools, how can you think that there will
remain a piece of quantum theory with no information theoretic
significance?
\ehpoirier

The formal structure of quantum mechanics is like a sacred text, isn't it?  There are so many ways to read it, so many diverse and conflicting meanings to be found in its pages.  At weaker moments, one starts to wonder how one is going to reconcile all those meanings.

You ask a good question:  ``If all the characteristics of QM can be seen as generalized probability tools, how can you think that there will remain a piece of quantum theory with no information theoretic significance?''

This is a place where Hardy and I presently diverge, I think.  There is great stuff in his derivation, and a lot that we've all learned from it---particularly, for me, I'm thinking of the origin of the multiplicative structure of Hilbert space dimension---but by thinking of quantum mechanics as a kind of {\it generalized\/} probability theory, I think Lucien swings too far toward a purely operational interpretation of quantum mechanics.  It becomes a theory {\it purely\/} of knobs and transformations and clicks, and stops saying anything particular about reality itself.

What is the origin of the divergence between us?  I think it is in his move of thinking that probability theory and information theory are malleable or fluid or empirical to some extent.  In my approach, on the other hand, I think of probability theory as a kind of ``a priori,'' much like simple arithmetic is ``a priori''---its structure does not depend upon the particular details of the world.  In other words, probability theory is not something that can be generalized; it is what it is.

Now, from that point of view, if we find that the content of quantum mechanics is, say, 1) a restriction on the set of probability distributions used to describe one's expectations about measurement outcomes, and 2) a modification from Bayesian conditioning---these are the two ingredients I standardly advertise in my own rewrite of the sacred texts---then those two ingredients are saying something about empirical reality.  They are not part of the a priori structure of probability theory.

Another way I sometimes say it in my lectures is that, in the past there has been a lot of work in trying to view quantum mechanics as a kind of larger structure than probability theory, and one recovers standard probability theory in the commutative case.  (Hardy's work is along the lines of that tradition.)  What I am shooting for, however, is something different.  I want to view quantum mechanics as a proper subset of probability theory to the extent that I can.  To the extent that I can't, then that is saying something about reality.

I think Hardy's derivation has some element of that approach in it too, but in order to disentangle it all, I think there's still some work to be done.  In particular, we've still got to get a better handle on what part of his derivation is actually {\it generalizing\/} probability theory---perhaps some of what he was doing has actually been misidentified.  I.e., in the end, maybe he wasn't generalizing probability at all, but applying it to a specific physical context.

Anyway, that's my take on it at the moment.

I hope that helps.  Good luck with your writing.

\section{25-08-05 \ \ {\it De Finetti's Fascism} \ \ (to W. C. Myrvold)} \label{Myrvold4}

\bwm
I've since finished reading it, and I must say that last paragraph
came as a nasty shock!  (I had previously  known nothing of de
Finetti's political convictions). It should come with a disclaimer:
Warning: Political content may offend some readers!
\ewm
Sorry I didn't mention that detail---his (early-in-life) fascism is certainly an embarrassment.  In any case, in that regard, you should have a look at the disclaimer at the end of page 31 of Carl's ``Resource Material for Promoting the Bayesian View of Everything'' posted at
\bq\noindent
  \myurl[http://info.phys.unm.edu/~caves/thoughts2.2.pdf]{http://info.phys.unm.edu/$\sim$caves/thoughts2.2.pdf}.
\eq

\section{30-08-05 \ \ {\it From Mineralarians to Pastafarians} \ \ (to C. H. {\Bennett} and J. A. Smolin)} \label{Bennett40} \label{SmolinJ7.1}

I'm guessing you guys have already heard about the Pastafarians, but in case you haven't, let me forward on this link that I just got from my friend Jeff.

\bjn
\myurl{http://www.venganza.org/}

May you be touched by His noodly appendage.  RAmen.
\ejn

\section{01-09-05 \ \ {\it Jazz in Westfield} \ \ (to J. W. Nicholson)} \label{Nicholson21}

\bjn
Now where is my Geritol?
\ejn
That reminds me of a time when I was visiting Charlie Bennett and Herb Bernstein in Amherst.  Bernstein stopped the conversation to take a break when he remembered that he should take his Metamucil.  Then he came back into the room with the bottle and a spoon asking if anyone else would like some.

\section{01-09-05 \ \ {\it BBQW Report} \ \ (to S. Hartmann)} \label{Hartmann12}

How's this?  Hope it fulfills your needs pretty well.\medskip

\noindent {\bf Being Bayesian in a Quantum World:  A Report}\medskip

``We've got to change the course of physics; nothing less will do!''
That was the rallying cry for our meeting, {\sl Being Bayesian in a
Quantum World}, in Konstanz, August 1--5, and the VW Foundation can be
sure that we tried our best to live up to this call.  With a
gathering of nearly 50 of the world's best quantum information
theorists and philosophers of quantum mechanics, the sparks were
certain to fly and they did.  There arose a palpable sense that
something new and powerful is brewing in the foundations of quantum
mechanics with this Bayesian turn, and the posing of clear-cut
research problems became the task of the day. The city of Konstanz
provided the perfect setting for the meeting, with just the right
variety of restaurants, biergartens, caf\'es, and walks to keep
everyone stimulated from beginning to end. The discussions, often
heated, literally started every day in the hotel's breakfast room
with the first cups of coffee and often lasted late into the night,
moving with the participants from coffee break to lunch to dinner to
a caf\'e here or there.

The meeting was organized around sets of two or three formal talks at
a go---for instance, two talks in the morning, followed by three in
the afternoon, most days---with long audience-wide discussions
following each set.  In total, there were 24 formal talks, with the
remainder of the invitees playing the role of session chairs and
official discussants.  In the following pages, we attempt to give a
sense of the subject matter of these talks along with some of the
discussion that surrounded them.

For instance, the first three talks of the conference went as
follows.

\begin{itemize}

\item
Christopher Fuchs:  This talk tried to set the tone of the meeting by
demonstrating that much of the content of finite-dimensional quantum
mechanics reduces to two simple modifications of Bayesian
ideology---1) the setting of a theory of prior probabilities with
regard to the outcomes of a single special quantum measurement, and
2) a modification of the standard Bayesian conditionalization rule
for updating probabilities in the light of new information.  From
this perspective, the formal structure of quantum mechanics becomes
{\it mostly\/} a ``law of thought'' (in the same sense that George
Boole called probability theory a ``law of thought'') rather than a
``law of nature.''  Where nature still rears its head---i.e., makes
its contingently given empirical content known---is through the
higher-level set of reasons for why decision-making agents in this
world should use {\it this\/} law of thought (i.e., quantum
mechanics) rather than {\it that\/} law of thought (i.e., some foil
theory other than quantum mechanics). Much of this talk was based on
\quantph{0205039}, but there were also
many newer points that have not yet been published.  For instance, it
was shown that the theory of priors mentioned above could be written
as two simple restrictions on a probability distribution $p(i)$:
$$
\sum_i p(i)^2=\mbox{constant} \qquad\mbox{and}\qquad \sum_{i\ne j\ne
k} C_{ijk}\, p(i)p(j)p(k) =\mbox{constant}
$$
for a certain set of coefficients $C_{ijk}$.  A research problem
posed to the audience was to give an information theoretic reason for
such constraints, particularly for the second one.

\item
Robert {\Spekkens}:  This talk set itself the task of giving a very
strong argument for the view that quantum states should be viewed
epistemically (i.e., as having to do with states of mind) rather than
ontically (i.e., as having to do with states of nature, independently
of any agent or observer).  The structure of the argument was based
on delineating the properties of the epistemic states about a certain
toy-model universe and showing that (qualitatively at least) a large
number of the most significant results of quantum information theory
can be recovered in this model. The reason for introducing the toy
model, rather than argue for the epistemic nature of the quantum
state directly, is that the toy model allows a means for making a
clear-cut distinction between ontic and epistemic states at the
outset---that is, it leaves no room for confusion between the two
concepts by its very construction.  Much of this talk was based on
the paper \quantph{0401052}, but the
case was made even more convincing with several new examples
exploring the structure of time evolution laws for such toy models.

\item
{\Ruediger} {\Schack}:  This talk attempted to go still further than the
last by arguing that not only should quantum states be viewed
epistemically, but more particularly as personalistic Bayesian
degrees of belief.  That is, even the idea of ``a quantum state as a
state of knowledge,'' with knowledge interpreted as ``justified true
belief,'' does not go far enough for a satisfactory Bayesian
interpretation of quantum states.  The argument was made quite
thoroughly and generally by mapping the uses of quantum states to the
uses of probability itself and then arguing for a ``radical
probabilist'' interpretation of probabilities along the lines of
Bruno de Finetti and Richard Jeffrey.  Following that setting of the
stage, a list of more technical results was shown to the audience.
For instance, it was demonstrated that a nearly identical repertoire
of theorems are available to the ``quantum radical probabilist'' as
to the ``classical radical probabilist'' for responding to
frequentists and propensitists who feel that quantum experiments can
be doing nothing other than manifesting objective chancy
propensities---a good example is the quantum de Finetti
representation theorem for exchangeable density operators.  This talk
was in part based on \quantph{0404156},
but also contained much new material.

\end{itemize}

Thus went the first evening and then the first morning of the first
full day. And the participants saw that it was good!  In the
afternoon, they came back for more---this time, the subject turned to
refining the debate set forth by Professor {\Schack} in the morning:
Namely, does or does not the Bayesian project of interpreting quantum
states require the notion of ``objective chance'' in sense of David
Lewis or perhaps in the sense of something like Karl Popper's
``propensities''?

To that end:

\begin{itemize}

\item
Marcus {\Appleby}, in his talk, gave a detailed argument for why
probabilities should be viewed in the radical probabilist way and
significantly expanded a line of thought started in his paper
\quantph{0402015}, where he compares probabilities to qualia in
the old primary-secondary quality distinction.

\item
Wayne Myrvold, in his talk, argued strenuously that even Bayesians
need objective chance to make sense of what they are doing, along
with a guiding idea like Lewis's ``principal principle'' for
connecting objective chance to subjective credence.  This work is
presently being prepared as a paper with William L. Harper, ``Why
Bayesians Should Countenance Objective Chance.''

\item
Christopher {\Timpson}, in his talk, still further firmed up the debate
by using various definitions and techniques drawn from the philosophy
of language.  Part of his argument can be read at \quantph{0412063} where the main conclusion
is:  If one is going to get any interpretive traction in solving
issues like the quantum measurement problem or the issue of
nonlocality in quantum mechanics, ``the cognitive state with which
one must associate the quantum state is the state of belief, not that
of knowledge.''

\end{itemize}

The second day of the meeting, the presentations turned to far more
technical and less foundational material.  The first two talks
concerned the revolutionary subject of quantum computing.

\begin{itemize}

\item
N. David {\Mermin}, in his talk, considered the model universe of a
quantum computer in the way that he presents it in his Cornell course
on quantum computing for computer scientists.  The main conclusion he
draws from this is that one is very naturally led to something like
Bohr's view of quantum mechanics; see \quantph{0305088}.  In
the present context, he particularly focused on how a radically
probabilist Bayesian view of the quantum state could contend with what
he now considers the main mystery of quantum mechanics---those
situations where quantum mechanics can make predictions with
certainty.

\item
Hans Briegel, in his talk, gave a detailed description of a model of
quantum computation (different from {\Mermin}'s) where the computation
is enacted solely from quantum measurements (rather than unitary time
evolution) on an array of quantum systems initially prepared in a
fixed entangled quantum state; see \quantph{0301052} and
\quantph{0504097} for two reviews of this very deep idea.  In
the present context, it was explored how this model seems to
demonstrate that it is nonclassical {\it correlations\/} that lie
behind the power of quantum computation, rather than the
sensationalistic imagery used by the popular press and David
Deutsch---namely, that quantum computation attains its power from
parallel computations in parallel universes.  To the extent that the
idea of correlation can be given a purely Bayesian treatment, one can
start to search for deep connections between a Bayesian interpretation
of quantum states and the power of quantum computation.

\end{itemize}

The afternoon talks concerned various Bayesian-like representation
theorems for quantum states:

\begin{itemize}

\item
Mark Srednicki addressed the quantum de Finetti representation
theorem, exploring whether one could indeed make sense of the idea of a ``probability of a probability'' and, consequently, the probability of an unknown quantum state.

\item
Renato Renner gave a representation theorem for finitely exchangeable quantum states (along the lines of a similar classical theorem of Diaconis and Freedman) and then showed that the theorem was crucial for proving the unconditional security of certain quantum cryptosystems.

\item
Todd Brun displayed a criterion (and proved a uniqueness theorem) for when two separate agents with distinct quantum states for a single quantum system can pool their expectations to form a new and improved quantum state.  He also explored several notions for quantifying the amount of compatibility of distinct states, and posed various technical questions to the participants along those lines.  Some of those problems were solved by Michael Nielsen and will eventually lead to publication.

\end{itemize}

Thus the afternoon and the morning of the second day passed.  And the
participants saw that it was still good!  So too went so much of the
rest of the conference.  The third day saw in the talks of David
Wallace, Howard Barnum, and Carlton {\Caves} an eager, and very
technical, debate on whether the quantum probability rule can be
derived (or even made sense of!)\ in the context of a many-world
interpretation of quantum mechanics (which is so popular in the
quantum computing community). The fourth day brought attention back
to some of the ideas explored in the opening talk of Fuchs:  Namely,
the extent to which the quantum updating rule can be viewed as a
variation on Bayesian conditionalization.  The talks of Veiko {\Palge},
Matthew Leifer, Guido Bacciagaluppi, and Thomas Konrad all dealt with
various details and critiques of this idea.  On a more philosophical
note, the talk by Michael Dickson---editor of the prestigious journal
{\sl Philosophy of Science}---placed the discussion of a Bayesian
view of quantum mechanics within a neo-Kantian or neo-Carnapian
framework. On the fifth day of the meeting, Armond Duwell and Amit
Hagar opened the morning session by exploring the extent to which one
might really hope to make a clear-cut separation objective and
subjective aspects of quantum theory.  The afternoon talks by David
Poulin, Howard Wiseman, and Lucien Hardy explored the extent to which
Bayesian and operationalistic interpretations of quantum theory can
be motivating factors in finding new physics---one particular item of
concern was in aiding the formation of a quantum theory of gravity.

In general, the meeting was a significant success.  As already
stated, the debate was strong, but the sense of a certain momentum
forward was even stronger.  It is fair to say, for instance, that a
significant fraction of the participants will be having their
students, graduates students, and postdocs exploring subjects brought
to the fore in this meeting for a significant time to come.

Finally, we mention that a further sense of the excitement felt at
this meeting can be read about in the French magazine {\sl Science et
Vie}, which is devoting its October 2005 cover story to our meeting,
and also in an upcoming article in the American magazine {\sl Scientific American\/} by one of the organizers.

\section{06-09-05 \ \ {\it A Homer Simpson Drool} \ \ (to J. W. Nicholson)} \label{Nicholson22}

Look at this baby, \myurl{http://www.pcmag.com/article2/0,1895,1850249,00.asp}.  I'm toying with the idea of getting something like this.

I picked out a book for you, I think:  Michael Polanyi's {\sl Science, Faith, and Society}.  It's a small one.  I might try to read it again before I hand it off to you (to see if I'm able to believe it now, like I think I did before).

\section{07-09-05 \ \ {\it EPR} \ \ (to A. Plotnitsky)} \label{Plotnitsky18}

\barkp
I am not sure, however, that I am quite ready to make any serious
statements concerning entanglement in qft theory, and for the moment I
mostly doing some reading. There is not that much actually.
\earkp

Let me try remedy your last sentence, a little at least, by giving you a few more references.  First and foremost let me recommend a tutorial paper by my colleague here at Bell Labs, Steven van Enk.  You can find it at: \quantph{0403119}.  His discussion there is quite in line with the sorts of things I was telling you the last time we met (was it in Sweden?).

Steven also recommended the following two papers by Werner (and a student maybe)---S. J. Summers and R. Werner, J. Math.\ Phys.\ {\bf 28}, 2440 (1987); ibidem {\bf 28}, 2448 (1987)---though he warned that they are quite technical.

Finally, for the heck of it, I used the ISI index to see what papers have cited the Summers--Werner paper, and I paste in my findings below.  Articles 1, 3, and 4 may be particularly useful to you.  And within that you might get the most from the Peres--Terno paper.  That paper can also be found on {\tt quant-ph\/} at: \quantph{0212023}.  Have a look at the section titled ``Entanglement in quantum field theory'' starting on page 21; that may give you some useful pointers.

Good luck in your thinking.

\section{07-09-05 \ \ {\it Quantum Theology} \ \ (to M. O. Scully)} \label{Scully4}

Now that Labor Day has passed, it's time for me to get back to work in a serious way.  I apologize for taking so long to comment on your son's draft---and I'm still not going to do so in any detail today---but after my schedule of constant meetings this summer, I found myself so mentally drained that I seem to have lost the ability to speak for a while.  But my tongue is getting loose again.

I read Chap. 11---the one on quantum theology---of Rob's\index{Scully, Robert J.} draft on a plane somewhere between Brussels and San Jose a while ago.  First off, I caught a lot of typos [\ldots]

Second, before I comment in any detail, I want to read it again.  So, if there have been any major revisions, maybe it'd be a good idea to email me the new version before I plunge ahead.

Concerning one thing you seemed to ask about particularly at our lunch meeting---i.e., about the spaceship example, where the uncertainty principle is amplified up to a level where it may affect practical human decisions---I think I like it.  {\it But\/} I wouldn't call it a ``proof'' of free will as you seemed to want to do at lunch that day.  Instead I think its import is that it forces us to contemplate as ubiquitous (rather than rare) the kind of situation William {\James} talks about in two quotes that I'll attach as a PDF file.  Particularly, look at the second quote.  Its message is that free-willed agents can ``shape'' reality (in a way) when they have to, and what Rob's example shows is that it is a consequence of quantum mechanics that we have to do such a thing all the time.

But as I say, I'll try to comment in detail later.  In the meantime, you and Rob\index{Scully, Robert J.} might want to read (and start thinking about) these great {\James} quotes.  [See 21-11-01 note ``\myref{Schack37}{Pragmatism versus Positivism}'' to R. Schack.]

\section{07-09-05 \ \ {\it Heisenberg Quote} \ \ (to M. O. Scully)} \label{Scully5}

Right, I had also promised to send you a quote of von {\Weizsacker}'s about meeting Heisenberg.  One such quote is below, but it is not quite as I remembered it.  I thought I remembered von {\Weizsacker} as saying that Heisenberg told him that he had ``proven free will'' that day---a much more dramatic statement than what he says below.  So, either I was thinking of a different version of the same story---and I'll have to work harder to dig up that quote---or maybe my memory is simply failing me finally.  I'll let you know if I come to a conclusion.

From C.~F. von {\Weizsacker}, ``Physics and Philosophy,'' in {\sl The Physicist's Conception of Nature}, edited by J.~Mehra (D.~Reidel, Dordrecht, 1973), pp.~736--746:
\bq
I remember very well how I met Heisenberg for the first time when I was a boy of fourteen. We happened to be in Copenhagen at the same time. Soon afterwards, in a taxi in Berlin in April 1927, he told me about the uncertainty principle. I was fourteen years old, and I was greatly moved by this new idea. I got the impression that if this was physics, one must study physics. This was the first moment when I saw that there was a hope of bringing together the two different parts:
the objective world described by classical mechanics and the world of man. I didn't know how, but somehow it meant that there was a connection between the two. And this was the way in which Heisenberg himself was expressing it when he said that the sharp distinction between subject and object was no longer possible in quantum theory.
\eq

\section{07-09-05 \ \ {\it The Activating Observer} \ \ (to M. O. Scully)} \label{Scully6}

While I'm sending you multiple emails today, let me go ahead and send you another.   I call it ``The Activating Observer: Resource Material for a Paulian--{\Wheeler}ish Conception of Nature,'' and it's been the source of some of the quotes I've sent you (more accurately, it's where I've been storing them).  With regard to Rob's\index{Scully, Robert J.} book-writing project, he may find some useful material in there.

As you'll see, I've long had an interest in the idea that quantum mechanics may show that nature is a little more malleable to the presence of decision-making agents than had previously been conceived.  Part of Rob's\index{Scully, Robert J.} Chapter 11 seems to be concerned with that too, so that's why I say he may find some useful resources in the present draft.  Sorry John {\Wheeler} is not yet better represented in this draft, but the problem there is a historical one:  I had read all those papers of his that I cite a long time ago, before I started collecting quotes.  Now I need to go back and scan that material in with my scanner.  It'll happen eventually; it just hasn't been done yet.

\section{21-09-05 \ \ {\it Egregiously Ingenuous} \ \ (to G. Brassard)} \label{Brassard44}

\bgb
According to my dictionary,
\bv
{\rm {\bf egregious}: adjective\\
\hspace*{.2cm} 1. outstandingly bad; shocking: egregious abuses of copyright.\\
\hspace*{.2cm} 2. archaic: remarkably good.}
\ev
\egb

Here's another word I've always loved as having dual and contradictory meanings:
\bv
{\bf ingenuous} (\^{\i}n-j\`en$^\prime$yi-es) adjective\\
\hspace*{.2cm} 1.	lacking in sophistication or worldliness; artless.\\
\hspace*{.2cm} 2.	obsolete: ingenious.
\ev
And just in case you don't know:
\bv
{\bf ingenious} (\^{\i}n-j\^en$^\prime$yes) adjective\\
\hspace*{.2cm} 1.	Marked by inventive skill and imagination.\\
\hspace*{.2cm} 2.	Having or arising from an inventive or cunning mind; clever.
\ev
I learned the meanings of ingenuous (it wasn't a word I had known before) when G\"oran Lindblad wrote in his paper ``A General No-cloning Theorem''
\bq\noindent
     A number of different versions of no-cloning theorems have been
     published. In particular, the recent no-broadcasting theorem of
     Barnum et al.\ is a very general result. The proof is ingenuous,
     and the method used is far from obvious. Their result inspired me
     to try to find another approach which uses standard results on
     operator algebras and systematically exploits the structure of
     completely positive maps.
\eq
I was always very proud of Lindblad's saying that, but who knows!

Compost is finished now, and the new grass is already coming up.  {\Ruediger} {\Schack} is visiting and we're working on our latest Bayesian screed.  (You tell me which meaning.)  And in particular, I'm heartened again to get at the question, when it comes to quantum states, if they are nothing more than information, then ``Information about what?''

By the way, we're also dropping in on Hans Halvorson and Bas van Fraassen's course today at Princeton.  Have a look at:
\myurl{http://www.princeton.edu/~hhalvors/teaching/phi538_f2005/}.

\section{23-09-05 \ \ {\it Bayesians at FPP?}\ \ \ (to A. Y. Khrennikov)} \label{Khrennikov11}

\bakh
I would like to try to stimulate you to submit a paper to our
conference proceedings which will be published by American Institute
of Physics.
\eakh

I'm sorry, I would like to, but I just can't get a paper out at this time.  I will definitely make a promise to give you a paper for your next proceedings.

Speaking of that, as I understand, you want me to be one of the organizers of your next meeting on Foundations of Probability and Physics.  Do you know when that will be held yet?  Anyway, I have an idea:  What would you think of having a session devoted to the Bayesian or ``radical probabilist'' approach to quantum probabilities?  I can think of several very good people I could invite (including Persi Diaconis).  You'll note I organized a rather large international meeting on this subject just this year with a lot of relatively famous people there (look at
\begin{center}
\myurl[http://web.archive.org/web/20090511060206/http://www.uni-konstanz.de/ppm/events/bbqw2005/]{http://web.archive.org/web/20090511060206/http:// www.uni-konstanz.de/ppm/events/bbqw2005/}.
\end{center}
So, I hope to impress on you that this is a growing and important subject, and it would be nice if you had some representation of it beyond me at your FPP meeting.  As you can guess---as I did before for you---I'll try to bring the American entrepreneurial attitude to my organizing of this session to maximize its success.

Also, there is a very large Bayesian meeting in Valencia, Spain from June 1--6, 2006.  See \myurl{http://www.uv.es/valenciameeting}.  There will be over 600 attendees there, they say, and maybe that will help us draw in some of the bigger names as they will be flying to Europe during a time very similar to the time of your meeting anyway.

\section{28-09-05 \ \ {\it Visitor Exhaust} \ \ (to J. W. Nicholson)} \label{Nicholson23}

Did you have any luck with the Polanyi book?  Believe it or not there were couple of times this and last week when I wanted to be able to quote one of the passages in it to {\Ruediger}.  (Probably just a function of it being fresh on my mind.)

\section{05-10-05 \ \ {\it Nobel} \ \ (to J. W. Nicholson)} \label{Nicholson24}

BTW, you must be pleased with this year's picks for the Nobel physics prize \ldots

\section{05-10-05 \ \ {\it Too Long Draft} \ \ (to A. P. Ramirez)} \label{Ramirez1}

I'll be in tomorrow at 9:00, and I'll drop by your office.  Attached is my ``first draft'' of a Form 1.  Somehow I got confused and thought the rules had changed this year and that we were supposed to write 2 pages front and back.  Dick set me straight today, so I'm going to condense significantly \ldots\ but not tonight.  Anyway, maybe the present draft will give us something to talk about tomorrow, and I'll attach it forthwith.  Also forgive me for the too flowery style---I'll be fixing that too---or at least tolerate me for a while.

\bq
\begin{center}
EMPLOYEE REPORT ON ACTIVITIES AND ACCOMPLISHMENTS \medskip
\end{center}

This year, unfortunately, I talked far more than I wrote \ldots\ but
at least I calculated more than I talked, if that's any consolation!
Allow me to report my progress in that order:  1) talks and
organizational efforts, 2) completed and near-completed scientific
papers, and finally 3) the yet unpublished calculations that consumed
much of my attention.

1) Much of this year represented a whirlwind of travel for me.  I
made 10 (invited) international conference appearances and gave 6
external physics colloquia and seminars.  I had to turn down at least
6 colloquia and conference invitations for lack of time or scheduling
difficulties---one invitation even a Caltech colloquium. (In dollar
terms, I tabulated that my hosts spent over \$19,000 on my behalf; in
contrast, I spent only \$1,500 of my allotted \$3,000 of Lucent
travel funds.)

My overwhelming reason for taking on all this travel was to establish
and entrench a certain point of view about quantum states within the
quantum information community.  This point of view---sometimes called
the Bayesian approach to quantum probabilities---is my strongest
intellectual child and has finally started to command serious
attention around the world.  For myself, I truly believe it is an
engine that will take quantum information theory and computing to new
heights.  The key idea is that {\it almost\/} all of the formal
structure of quantum theory (complex Hilbert spaces, state vectors,
unitary time evolution, etc.)\ represents not so much a {\it
direct\/} description of the quantum world itself, but rather a
modification of probability theory induced upon the gambling agents
(a.k.a.\ experimentalists) immersed within that peculiar world. Thus
the research task becomes to find the residue of quantum theory that
{\it is not\/} merely a modification of probability theory---the
suspicion is only therein lies the real source of power in quantum
information and computing. In fact, only once that part of quantum
theory is uniquely identified will we have a firm grasp on the most
interesting things that can be done with quantum information.

In Andrew Steane's seminal paper ``A Quantum Computer Only Needs One
Universe'' [{\sl Studies in History and Philosophy of Modern Physics}
{\bf 34B}, 469--478 (2003)], he fairly well demolishes the idea that
the power of quantum computers comes from massively parallel
computation (in parallel universes).  In its place, he puts the
following conjecture:
\begin{quote}
A quantum computer can be more efficient than a classical one at
generating some specific computational results because quantum
entanglement offers a way to generate and manipulate a physical
representation of the correlations between logical entities without
the need to completely represent the logical entities themselves.
\end{quote}
Or, as he goes on, ``[T]he basic fact which quantum computers take
advantage of is that multi-partite entanglement offers a way to
produce some computational results without the need to calculate a
lot of `spectator' results.  For example, we can find the period of a
function without calculating all the evaluations of the function.''
This idea rings true within the Bayesian approach to quantum
information, with its emphasis on `correlation' as just another
species of the peculiar probability calculus the world forces upon
us.  Fleshing out ideas like this are indeed a prime area for
exploration in the Bayesian approach.

Particularly to get the community thinking in these directions, I,
along with Carlton Caves, Stephan Hartmann, and R\"udiger Schack,
organized one of the more significant quantum information meetings
this year to explore just such issues.  The meeting title was {\sl
Being Bayesian in a Quantum World\/} and it was held in Konstanz,
Germany, August 1--5, by way of generous funding of the Volkswagon
Foundation (67,000 euro).  We had 54 participants drawn from some of
the best of the best of quantum information theory---the names
Michael Nielsen, David Mermin, Gerard Milburn, John Smolin, William
Wootters, Carlton Caves, Lucien Hardy, Norbert L\"utkenhaus, William
Unruh, and Hans Briegel will hopefully ring a bell or two.  For five
days the talks and discussions centered on the issues raised above,
and the verdict is that this point of view is definitely on the map.
For instance, the French magazine {\sl Science et Vie\/} made a
report of the meeting as their cover story for this month's issue,
and {\sl Scientific American\/} has invited me to write a full
article on the subject in the coming couple of months.

In total, I am quite pleased about the extent to which my talks and
organizational efforts have gotten the quantum information community
thinking along these directions.

2) With regard to publications, let me mention the two most
significant pieces of {\it new\/} work---one recently posted on the
archive and submitted to {\it Physical Review A}, and one in draft
form, soon to be posted.

The first---titled ``Influence-Free States on Compound Quantum
Systems'' and co-authored with H.~Barnum, J.~M. Renes, and
A.~Wilce---explores the connection between a previously published
derivation of mine of the tensor-product rule for combining quantum
systems and the idea that measurements on one side of an entangled
quantum state cannot be used to communicate instantaneously to the
other.  In particular it is shown that a kind of Bayesian updating of
the remote system's quantum state after a localized measurement goes
hand in hand with the idea of no-signalling.  Beyond that, we further
elucidate a trouble that came up in the previous derivation of the
tensor product rule:  Even though a no-signalling assumption
specifies a kind of Bayes rule, and indeed even specifies the
tensor-product rule for combining quantum systems, it is not strong
enough to uniquely pin down that the joint probabilities for the
outcomes of localized measurements must be derived from a standard
quantum state (i.e., a bipartite positive-semidefinite operator of
unit trace).  It only specifies the weaker requirement that the
probabilities be derived from bipartite operators that are ``positive
on pure tensors.''  Thus an extra assumption is clearly needed to get
all the way back to quantum mechanics.  What that extra assumption
is, unfortunately, we do not yet know.  However, we were able to show
in this paper that this more general class of operators are
pathological in the sense that they are not well behaved when they
are incorporated into a quantum teleportation scheme.  In general,
this work is part of a larger effort to get a the root of the idea of
quantum entanglement.

The second paper --- tentatively titled ``Accurate Quantum State
Estimation via `Keeping the Experimentalist Honest'\,'' and
co-authored with R. Blume-Kohout and P. Hayden --- takes another look
at the process of quantum-state tomography, but via Bayesian ideas
and methods.  It is based on first defining an ``honesty'' function
that derives from the old (Bayesian) mathematical puzzle of ``keeping
the expert honest.''  In this puzzle, an expert is hired to give his
opinion of whether one or another event $i$ will occur, and he
expresses that opinion in terms of a probability assignment $p_i$,
where $p_i\ge0$ for all $i$ and $\sum_i p_i=1$.  But {\it opinions\/}
are not objectively testable---certainly not by simply looking at the
actual event that occurred.  So how would one ever know whether the
expert is lying about his opinions?  For instance, the expert might
actually believe $p_i$, but because he does not particularly like his
customer and ultimately wants to see him fail, he might offer the
distribution $q_i$ instead.  The trick is to choose a payment scheme
that gives an incentive for the expert to keep honest. For instance,
if the client agrees only to pay the expert after the actual event
$i$ happens, and only then an amount $f(q_i)$ for some function $f$
acting on the expert's offered opinion $q_i$, then one can see that a
good incentive scheme should have the property that
$$
\sum_i p_i f(q_i) \le \sum_i p_i f(p_i)\;,
$$
with equality if and only if $q_i=p_i$.  In other words, if the
expert's real opinion is $p_i$, then he should expect to lose money
any time he offers an opinion different from his true one.
Remarkably, it can be shown that this inequality alone uniquely
specifies the function $f$ up to a constant and it is $f(x)=-\log x$.
(I believe this was first shown by Acz\'el in the mid 1970s.)  A
consequence of this result is that when an expert is honest, he can
expect to be paid (up to a constant) an amount that is equal to the
Shannon entropy $H(p)=-\sum_i p_i \log\, p_i$ of his true opinion.

In any case, on these ideas, we build the concept of an honesty
function and generalize it to the quantum case, where an expert
offers as an opinion a density operator $\rho$ rather than a
probability distribution.  In particular, we show that Bayesian
updating in the usual sense (for an exchangeable density operator
assignment) is the most honest strategy a quantum-state tomographer
can use before passing his opinion on.

3) Now let me come to my most passionate work for this year---it is
work that is still not complete and certainly not published.  What I
have been seeking is a well-behaved and {\it insightful\/}
representation for quantum states (in finite dimensional Hilbert
spaces) that is purely in terms of a single probability distribution
for each state. It is easy enough to see the starting point for such
an endeavor:  One merely chooses a generalized measurement or
positive-operator-valued measure $\{E_h\}$, $h=1, \ldots, d^2$, for a
Hilbert space of dimension $d$, where all the operators are linearly
independent to be a kind of fiducial measurement. Then the
probability distribution $p(h)=\tr\,\rho E_h$ for the outcomes
$h$ of such a measurement will uniquely specify the density operator
$\rho$.  This can done easily enough, as there are many constructions
$\{E_h\}$ that will do the trick.  But which of these constructions
will lead to the most insight with regard to the source and ultimate
power of quantum information processing?  The hope here is to find a
transparent way to see the extent to which the structure of Hilbert
space arises from a kind of information theoretic constraint on the
expectations any experimentalist should adopt for the outcomes of his
quantum experiments.

One promising candidate for an insightful fiducial measurement comes
from my work in defining a measure of ``quantumness'' for sets of
quantum states that I reported in last year's Form 1 (and was
published in the journal {\sl Quantum Information and Computation\/}
this year).  There, it was shown that if any set of $d^2$
one-dimensional projectors $\Pi_i=|\psi_i\rangle\langle\psi_i|$ exist
with the property that $\tr\,\Pi_i\Pi_j$ is constant for all
$i\ne j$, then such a set would fulfill a criterion for being the
most quantum a set of states can be.  Concerning the present issues,
a nice corollary of the old work is that if such a set exists, then
the operators $E_h=\frac{1}{d}\Pi_h$ would be linearly independent
and thus the set $\{E_h\}$ would be adequate for giving rise to a
fiducial measurement.

Ostensibly another promising measurement comes about by simply trying
to choose a set of $E_h$ to be as orthogonal to each other as
possible---i.e., one would like to get as close as possible to an
orthonormal basis of operators while still preserving that the set of
operators describes a potential measurement.  One of the things I
proved this year is that this other class of candidate measurements
actually turns to be identical to the previous one!

Thus the more reason to try to work out quantum mechanics in the
language of this particular class of fiducial measurements.  But do
they exist?  Computational work from several groups seems to confirm
that they do (for dimensions up to $d=45$ at least).  But despite a
handful of papers on the subject now posted on the {\tt quant-ph}
archive by various groups and a couple hundred pages of my own
calculations, an actual proof---and more interestingly an analytic
construction---is still up for grabs. I have, however, made plenty of
progress \ldots\ and in directions that no one else seems to have
taken.  For instance, to record my strongest result yet, I have
reduced the problem to simply showing the existence of a $d^2\times
d^2$ Hermitian matrix $G$ with constant diagonal entries and
off-diagonal entries all of the form $e^{i\theta_{hk}}$ such that
$$
\tr\,G^3=\tr\,G^4=d\;.
$$
If one can prove the existence of any solution whatsoever to those
two equations, then the existence of such a measurement will be
ensured.

On top of that I have been able work out what the full set of quantum
states looks like in this language if these very nice measurements
always exist---that is, I am getting very close to my real goal.  For
instance, in this language a pure quantum state is any probability
distribution $p(h)$ over $d^2$ outcomes satisfying the following two
equations:
$$
\sum_h p(h)^2=\frac{2}{d(d+1)}\;,
$$
$$
\sum_{jkl}p(j)p(k)p(l)\,{\rm
tr}(\Pi_j\Pi_k\Pi_l)=\frac{d+7}{(d+1)^3}\;.
$$
The first of these equations is very clearly an information theoretic
constraint, in terms of a R\'enyi entropy.  But what of the second?
This strikes of the sort of ``residue'' mentioned of the first
section of this report. The numbers $\tr(\Pi_j\Pi_k\Pi_l)$
represent universal constants whose values are deeply intertwined
with what quantum mechanics is actually about.  At the moment, I
think there is no more exciting question than pinning down what
exactly this means.

SUMMARY:  I hope the three sections above give some flavor of my
research activities this year.  As it turns out, yesterday I was
heartened to hear that Roy Glauber won part of this year's Nobel
prize in physics \ldots\ not because I think I could ever aspire to
those heights, but because Glauber's work reminds us of the power
that can be had by finding just the right representation for certain
physical phenomena---in his case recognizing the importance of
coherent states for representing quantum states of the
electromagnetic field.

Quantum information is still in search of the right representation
for its most fundamental phenomena, and I hope the work described
above is a significant contribution in that direction.
\eq

\section{13-10-05 \ \ {\it Nature Physics Article} \ \ (to G. Brassard)} \label{Brassard45}

\bgb
Ever heard of Frieden, Soffer or Brown?
\egb

Frieden's stuff is pretty soft---i.e., not to be taken seriously.  The basic thing he gets is the time-independent {\Schroedinger} equation to come out of a kind of stochastic (hidden-variable) model.  But that's a far cry from quantum mechanics \ldots\ and not even nearly as complete as Bohm's hidden-variable theory (of which I wouldn't consider a real derivation of QM either).  The only minor similarity between Frieden's work and our program is that we both use the word ``information''---there the resemblance stops.

\section{17-10-05 \ \ {\it Stomach Bug}\ \ \ (to S. L. Braunstein)} \label{Braunstein12}

I remember ending our conversation last Friday by saying, ``I'll almost surely be in next Monday.''  Teaches you something about the concept of certainty (and its fallibility).  I've been up much of the night with a stomach bug and there's every indication this morning that it still hasn't abated.  So, I think in good conscience, I probably shouldn't come in today \ldots\ for myself and everyone around me.  Hope that doesn't screw up your day's plans too badly, and I hope you get this note before you get on the road.

\section{18-10-05 \ \ {\it Hello} \ \ (to C. Hewitt)} \label{Hewitt1}

Thanks for all that; I'm flattered.  (Though, too bad you---or, your admiration at least---weren't around when I interviewed with the MIT EECS dept in '97 or '98!  I might be living in Cambridge, MA rather than Cranford, NJ!)  I'll have a look at those links; I had never heard about actor theory before.  But then again, I know almost nothing about programming---I've got a lot to learn.

If there really is a similarity between the actor model and my thoughts on quantum mechanics, then I suspect there is an even larger similarity between it and the ``radical pluralism'' of William James (i.e., a particular flavor of American pragmatism).  If you haven't discovered James yet, you might enjoy reading him---it's a whole philosophy built on the very idea that there's no ``global state.''

\subsection{Carl's Preply}

\bq
I am an admirer of your work.  In fact I made use of it in the following Wikipedia article (hopefully not distorting your views too much):\medskip\\
\url{http://en.wikipedia.org/wiki/Actor_model_history#Relationship_to_physics}\medskip

Also you might be interested in the following article:\medskip\\
\myurl{http://en.wikipedia.org/wiki/Actor_model_theory}\medskip

Your comments, questions, and suggestions are greatly appreciated.
\eq

\subsection{Carl's Reply}

\bq
Sorry that I missed your job talk.  But since I am CS my admiration might not have been of much help!

Meanwhile, your ideas have proved too radical for some of the current physics moss backs at the Wikipedia so the ideas have migrated
to\footnote{Apparently they are not represented there anymore either; maybe they were migrated into oblivion, or this is what was being referred to: \myurl[https://en.wikipedia.org/w/index.php?title=History_of_the_Actor_model&oldid=29339394]{https://en.wikipedia.org/w/index.php?title=History\underline{ }of\underline{ }the\underline{ }Actor\underline{ }model{\&}oldid= 29339394}.}
\bv
\verb+http://en.wikipedia.org/wiki/Actor_model%2C_mathematical_logic%+\\
\verb+2C_and_quantum_physics+
\ev
You should be able to read most of the above article in conjunction with
\bv
\myurl{http://en.wikipedia.org/wiki/Actor_model_theory}
\ev
even without any background in software.

I would greatly appreciate your comments, questions, and suggestions.\medskip

\noindent PS.  Yes, James is wonderful.
\eq

\section{18-10-05 \ \ {\it Lazy Abstract} \ \ (to H. C. von Baeyer)} \label{Baeyer1}

With time running short today, I think I'll just send you an old abstract (with only some very minimal modifications), rather than writing you something new.  Hope you don't mind.

\bq\noindent
Title:\medskip\\
Drawing Quantum Mechanics on a Probability Simplex\medskip\\
Abstract:\medskip\\
What is the difference between a quantum observer and a weather forecaster who uses classical probability theory?  Not much.  But where there is a difference, there lies quantum theory's most direct statement about properties of the world by itself (i.e., the world without observers or weathermen).  In this talk, I will try to shore up this idea by writing quantum mechanics in a way that references probability simplexes rather than complex Hilbert spaces.  By doing so, the connection between quantum collapse and Bayes' rule in classical probability theory becomes evident:  They are actually the same thing up to a linear transformation depending upon the details of the measurement method.  Looking at quantum collapse this way turns the usual debate in quantum foundations on its head:  only local state changes look to be a mystery.  State changes at a distance (as after a measurement on one half of an EPR pair) are completely innocent---they simply correspond to applications of Bayes' rule itself, without the extra transformation; that is, collapse-at-a-distance is nothing more than the usual method of updating one's information after gathering data.  Thus the idea develops that if a quantum reality is to be found in the quantum formalism, it will be found only in the formalism's {\it deviations\/} from classical probability theory:  Reality is in the difference.
\eq

\section{18-10-05 \ \ {\it Flight Schedule?}\ \ \ (to H. C. von Baeyer)} \label{Baeyer2}

\bhcvb
The question I am most interested in concerns Zeilinger's principle:
\begin{center}
\myurl{http://www.univie.ac.at/qfp/publications3/pdffiles/1999-10.pdf}.
\end{center}
\ehcvb

I would say the most serious technical development of an idea along those lines is Rob {\Spekkens}'s ``toy model'' even though, by its very construction, it can only go part of the way toward quantum theory.  Download: \quantph{0401052}.  In any case, I think it is one of the best, most convincing arguments for the epistemicity of quantum states around.  It's worth the study, both for what it does and what it can't do.

Is there something to Zeilinger's principle?  Maybe, but at present I think it's still more of an ``idea for an idea'' than a full-blown idea itself.  Certainly it seems important to better understand how (and then find the reasons for why) our uncertainties scale with the amount of quantumstuff that they do---there's strong indications that this is indeed a fruitful direction for research.  It'll be fun discussing these things with you.

\section{24-10-05 \ \ {\it Books, Bookcases, and Information} \ \ (to J. B. Lentz \& S. J. Lentz)} \label{LentzB8} \label{LentzS5}

Thanks for the nice electronic birthday card.  The kids and I both loved it.  I had a nice and thoughtful day for my birthday.  I went in to New York City and acquired 10 more books for the ``pragmatism collection.''  If we could just find the right bookcases, the front room---i.e., the library---would look great.  Who would have thought the books (given the subject) would be easier to get than the bookcases?

Thursday I fly to Newport News to give the physics colloquium at William and Mary in Williamsburg.  In preparation for that, I've been reading Hans Christian von Baeyer's book, {\sl Information:\ The New Language of Science}, this week.  I figure it is a nice courtesy since von Baeyer is the fellow who invited me (never met him before).  The book has turned out to be quite a pleasure though---accurate and entertaining, both; so I'm glad I took on the project.  Brad, I think this is one you might enjoy if you're looking for some leisure reading---the kind of thing, at least, that I might imagine as leisure reading for a Chief INFORMATION Officer.  Von Baeyer is exceptional in that his research field as listed by his department is literally ``public understanding of science.''  Anyway, he does a very good job on everything from Bayesian probability to quantum computing in this book---you might find it an enjoyable introduction.

\section{25-10-05 \ \ {\it GOBs, Bobs, Steering \& Teleportation} \ \ (to H. Halvorson \& B. C. van Fraassen)} \label{vanFraassen5} \label{Halvorson4}

\ldots\ and if you're patient there's even a little bit about
perspectivalism at the end.

Thanks again for letting me attend your seminar last Wednesday.  I
had a good time, and I hope I will push myself to continue coming.
(If it only weren't for that darned long drive:  45--65 minutes,
depending on traffic.)  The main thing I get out of the deal, of
course, is not so much the quantum-information material, but that the
lectures and discussion points give me a window into how philosophers
are starting to think about this subject.  That's valuable for me.

Anyway, spurred by what I heard at the last seminar, I put together a
few notes, and they're pasted below.  Feel free to not read any
further than this sentence if you're getting tired of my emails:
These exercises of sentence construction are always useful for me,
even if for no one else; so certainly I won't be hurt if you send
this email to the recycle bin and don't reply.  However, since you
were the discussion leaders, I figure I might as well share the notes
with you, even if only out of courtesy---I only hope they won't cause
you to give me a failing grade for the semester!

\subsection{Maudlin}

First off, let me tackle something Maudlin said near the end of the
seminar.  As I recall, he basically asserted that the phenomenon of
``steering'' is a very real effect (I'm tempted to put the word
``ontic'' into his mouth here), that Bell had taught us such, and
that the phenomenon has very useful consequences---for instance he
tried to make the last point dramatic by putting a gun to Bob's head.
As I recall, the point that started all this drama was Hans'
statement that he was reluctant to accept the idea that a localized
measurement could change the global {\it quantum\/} state for a
bipartite system.

The main point I want to make here is that Maudlin's drama really
carries no force as far as saying something {\it unique\/} about the
quantum world.  Let me give a contrived example that is a) not
quantum mechanical, but b) serves the role Tim sought to fulfill via
quantum entanglement (and quantum entanglement only).

Suppose Bob has in front of him four buckets, labeled 1, 2, 3, 4,
and that there is a ball in precisely one of them---the rest of them
are empty---but Bob has no clue which one.  Furthermore unfortunately
for him, these buckets along with the one ball are all themselves
hermetically sealed in a GREAT OBFUSCATING BOX, or GOB.  The
definition of a great obfuscating box is that one can query it to ask
a yes-no question of the ball's position within the collection of
buckets, but it will never let the questioner know the position more
precisely than two buckets' worth.  For instance, one might ask
``1v2?'', meaning ``is the ball in bucket 1 or 2?'', to which the GOB
will answer yes or no.  If it answers ``yes'', then one is assured
that the ball is indeed in bucket 1 or bucket 2; if it answers
``no'', then the ball must be in either bucket 3 or 4.  Similarly one
could ask the GOB ``2v3?''  A ``yes'' answer narrows the ball's
position down to 2 or 3, and a ``no'' answer narrows it to 1 or 4.
And similarly still, one could ask the GOB ``1v3?''\ to get an
appropriate answer for that question.   The nasty trick is that only
one of these three questions can be asked and never another one.

So, with a GOB in place of a qubit, let's now go back to the rest of
Maudlin's scenario.  Charlie holds a gun to Bob's head and says,
``I'm going to ask the GOB this question ``XvY?'', and if you
correctly predict the outcome before I get the result from the GOB
itself, then I'll let you go free; but if you get it wrong, then
BAM!, a bullet in the head!  By the specification of the scenario,
without any further help, there's very little Bob can do but make a
guess and say his prayers.

But suppose Bob's friend Alice happens to overhear what's going on
and can surreptitiously talk to him.  If it turns out that she knows
she has an identically manufactured GOB---assured by the factory to
have a ball in precisely the same bucket as Bob's---then everything
changes.  She can help her friend by asking her GOB the question
Charlie is about to ask his GOB and then slipping the answer back to
Bob.  Bob's GOB, we know, will give the same answer if she asks the
same question.  Alice, of course, still won't know precisely what
bucket her ball is in (and consequently which bucket Bob's is in),
but the information she gained is enough to keep Bob from getting
shot.

So there:  All the drama of Maudlin's example, all the same
conclusions operationally, but where's the quantum mechanics? Where's
the nonlocality?  There is none.  The example is powered solely by
Bob's having a friend who is capable of relieving him of some of his
uncertainty.  Alice passes off her own incomplete information to Bob,
and that saves the day.  In the context of quantum mechanics,
{\Schroedinger} called that first action of Alice's (before passing off
the information) ``steering,'' but that seems like such a weird term
for an example like this:  Alice isn't steering anything at all;
she's just trimming her uncertainty one way or another by making the
choice to ask one question or another.

Having a hidden variable---i.e., the actual position of the ball in
the buckets---serves only as a dramatic device in this example:  For
it makes it absolutely clear that Alice's action of learning at her
site changes nothing about the reality at Bob's site.  The reason the
situation {\it seems\/} to look different in quantum mechanics is
because we have all become convinced (through Bell-like arguments)
that there can be no local hidden-variable theories underlying
quantum mechanics.  But from my own perspective, that's an over-hasty
conclusion.  In fact, it's a {\it non sequitur}.  One can have
uncertainties about many things, from localized variables to
relations between very distant systems to the consequences of one's
own actions.  And in all of those cases, one can invent ``passing off
privileged information'' scenarios like the one above to make
dramatic a kind of ``steering phenomena.''  Ruling out localized
hidden variables (which I definitely believe has been done in quantum
mechanics) in no way pushes the further consequence that ``steering''
is an ontic phenomenon rather an epistemic one.

In fact, I have always hated the word ``steering'' as a name for this
phenomenon in the quantum context because it already loads the dice
toward an ontic interpretation of quantum states.  Indeed, one can
see that from the very beginning in {\Schroedinger}'s early 1935 papers
and his correspondence with Einstein.  {\Schroedinger} {\it starts\/}
with the assumption that quantum states are ontic states (rather than
epistemic states), and is therefore led to introducing some ad hoc
rules for entanglement-decay so as to get out of a kind of action at
a distance.  On the other hand, Einstein (as exhibited in
correspondence with {\Schroedinger} reprinted in Fine's book {\sl The
Shaky Game}) starts with the rejection of action at a distance, and
concludes that ``steering'' must be an epistemic phenomenon.

I side with Einstein, of course.  Thus, to load the dice toward my
own interpretation, I much prefer the words ``conditioning'' or
``conditionalizing'' (as from simple probability theory) or even
``updating'' over ``steering.''  These words much more adequately
convey the passive nature of what is going on in updating the quantum
state of a far away system.

By the way, now that I've said all that, let me report that I'm doing
nothing with my GOBs that Rob {\Spekkens} isn't already doing with his
toy model:  \quantph{0401052}.  I think that is conceptually the
most important paper written concerning the interpretation of quantum
information written in long while, and well worth everyone's
understanding.

\subsection{Teleportation}

But isn't quantum teleportation the indication of something much
deeper going on with quantum steering than with epistemic updating?
No, it is just about the same thing fancied up into a more
complicated situation.

Let me try to make that clear by writing down an example that I told
Hans about in conversation at our second-to-last meeting.  In usual
teleportation, the cast of characters includes an Alice and a Bob who
share two systems in a maximally entangled state, and implicitly, a
Charlie who prepares a third system in the state of his choice and
then hands it off to Alice.  Alice then performs a measurement on the
two systems in her possession and announces the result of the
measurement to Bob.  The teleportation process is completed with Bob
performing an operation on his system conditioned upon his newly
acquired information.

In what sense is it completed?  Only in this:  If Charlie has the
promise that Alice and Bob went through all the actions described
above, then he can safely ascribe the same quantum state to Bob's
system that he had originally ascribed to the system he handed off to
Alice.

Here's a corresponding classical example.  In place of entanglement,
let us equip Alice and Bob each with a coin (oriented heads or tails)
encased in a magical opaque box.  These magical opaque boxes have the
following properties:  1) though one cannot see how a coin is
oriented within it, one can nevertheless reach inside a box and turn
the coin over if one wishes, and 2) if one touches two of these boxes
together, they will glow green if the coins within them have the same
orientations, and they will glow red if they have opposite
orientations---the glowing reveals nothing about the actual
orientation of either coin, only about their relationship.  Finally
let us stipulate the following for Alice and Bob:  That their opaque
boxes contain identically oriented coins, but Alice and Bob (or
anyone else for that matter) know nothing more about the coins beyond
that.  In other words, Alice and Bob possess HH or TT, but they do
not know which.

Now, as in quantum teleportation let us introduce a third character,
Charlie.  Charlie has an opaque box of his own.  But let us give him
some partial certainty about the orientation of his coin.
Particularly, let us suppose he ascribes a probability $p$ for his
coin to be heads.  This is a real number between 0 and 1, and in
principle it might take an infinite number of bits to specify.

Here's the protocol.  Charlie hands off his coin (encased in a
magical opaque box) to Alice.  Alice touches her newly acquired box
to her old box.  The two glow red or green, and she communicates the
result to Bob.  If the result was green, Bob leaves his coin alone.
If the result was red, he reaches into his opaque box and turns the
coin over.  At that point the ``teleportation'' process is completed.

In what sense is it completed?  Only in this:  If Charlie has the
promise that Alice and Bob went through all the actions described
above, then he can safely ascribe the same probability $p$ to the
coin in Bob's box (i.e., $p$ that it will be heads) that he had
originally ascribed to the coin in his own box.  In other words,
Charlie has everything it takes to update his epistemic state about
the orientation of the coin in Bob's box to what he had originally
thought of the coin in his own box.

Is this wildly exciting?  The stuff that would make headlines in
papers all around the world and be called ``teleportation''?  At the
material cost of transferring a single bit from Alice and Bob, an
infinite number of bits (in the form of the real number $p$) has been
transferred between the two sites instantaneously?  Not at all!  The
only thing that was materially transported from one site to the other
was a single bit (that the boxes glowed red or green).  The rest was
just ``conditionalizing'' or ``updating''.  And there is no mystery
whatsoever in that.  Dumb and dull, it would never make a headline.

And so too, I would say with quantum teleportation, even though it
has built my career.  (Our Caltech experiment of it has over 600
citations now.)

At the end of this note, I'll place some of the correspondence I had
with Asher Peres as we were writing our ill-fated paper ``Quantum
Theory Needs No `Interpretation'.''  (Of course the whole paper is
about a particular interpretation, but most critics never read beyond
the title.)  In it, we had a paragraph on quantum teleportation and
how it helps to illustrate the epistemic nature of quantum states.
And boy did we fight about how to write that paragraph, and even---at
that time---about the meaning of teleportation.  Seeing some of our
own wrangling may be of use to you, if you care to explore this idea
further.  The main points of the part attached below are 1) how there
is no THE quantum state for a system (there as many as there are
potential observers) and how teleportation makes use of that idea,
and 2) when Alice performs her actions, nothing physical changes on
Bob's side.

Finally, let me comment on:

\subsection{Hans's Seemingly Black-and-White-ism or
All-or-Nothing-ism}

Two or three times now, I have heard Hans equate the idea of
interpreting quantum states epistemically (i.e., as knowledge or like
a Bayesian degree of belief or whatever) with giving up on the
project of thinking that {\it physics\/} has something to say about
the real world (i.e., presumably the world as it is without any
knowers or even Bayesians).  That is altogether too glib of a
position, and I want to do whatever I can to ease him out of thinking
it.

The stuff that {\Caves} and {\Schack} and {\Spekkens} and {\Appleby} and {\Timpson}
(and sometimes {\Mermin} and Unruh and Milburn and whoever else) have
been talking about is much more subtle than that, and it should be
recognized as such---this field of thought is at a respectable enough
level now that it shouldn't be caricatured.  I for one, for instance,
pretty much characterize myself as a realist in the time honored
sense:  That there is a real world out there beyond our whim and
fancy, and it is the task of science to hypothesize about its
attributes and properties.  I am not a philosophical idealist and
certainly not a solipsist.  And I don't think I am a positivist or an
empiricist.  Predominantly, I think I lean toward a kind of
materialism (though tempered by a lot of pragmatic subtleties).

The main way I think Hans errs is that when he hears a phrase like
``quantum states are states of knowledge,'' he thinks that that
throws {\it everything\/} away to knowledge (or Bayesian degree of
belief)---that nothing is actually hypothesized about the world.  But
quantum theory contains so much more than simply quantum states:
There are Hilbert spaces; there are Hilbert-space dimensions; there
are Hamiltonians; there are eigenvalues; there are notions of {\it
separate systems\/} whose joint description is worked up through a
tensor product. The quantum state is only one lonely piece of quantum
theory.

When it comes to quantum states, our point is simply scientific
open-eyedness and conceptual clarity:  1) Open-eyedness.  We see so
very many analogies between quantum states and incomplete
knowledge---just look at {\Spekkens}'s paper for more than 20 such---that
it would be scientifically foolish to not take those analogies
absolutely seriously and explore them for all they're worth.  But 2)
Clarity.  We think we get even more in return, via easy solutions to
some of quantum foundations' main supposed conundrums.  From the
epistemic view, there is no measurement problem, there is no
nonlocality.

Loads of questions still remain, but one shouldn't be frightened of
the methodology \ldots\ which is, as {\Spekkens} put it,
\bq\noindent
     The diversity and quality of these analogies provides compelling
     evidence for the view that quantum states are states of knowledge
     rather than states of reality, and that maximal knowledge is
     incomplete knowledge. A consideration of the phenomena that the toy
     theory fails to reproduce, notably, violations of Bell inequalities
     and the existence of a Kochen--Specker theorem, provides clues for
     how to proceed with a research program wherein the quantum state
     being a state of knowledge is the idea upon which one never
     compromises.
\eq
For instance, if you would ask {\Carl} {\Caves}, he would classify quantum
states as epistemic, but he thinks that Hamiltonians represent a
decent chance of remaining ontic terms within the theory (i.e., as
representing an agent-independent reality).  If you ask {\Spekkens}, his
own gut feeling is that quantum states represent incomplete knowledge
of {\it relations without relata\/} (whatever that would mean), but
particularly those relations are to be viewed ontically.  As for
myself, I tend to think that the ontology lies somewhere in a quantum
system's receptivity or sensitivity to external interventions
(whatever that would mean), and that receptivity is to be viewed
ontically.

So, there is a range of ways of staying hard-headed about the idea
that {\it quantum states are states of mind\/} without becoming a
postmodern or a deconstructionist---without taking a walk with
Derrida.  And that is what our research program is about.  Here's the
way I put it in my \quantph{0205039}:
\bq
     This, I see as the line of attack we should pursue with relentless
     consistency:  The quantum system represents something real and
     independent of us; the quantum state represents a collection of
     subjective degrees of belief about {\it something\/} to do with
     that system (even if only in connection with our experimental
     kicks to it).  The structure called quantum mechanics is about
     the interplay of these two things---the subjective and the
     objective.  The task before us is to separate the wheat from the
     chaff.  If the quantum state represents subjective information,
     then how much of its mathematical support structure might be of
     that same character?  Some of it, maybe most of it, but surely not
     all of it.

     Our foremost task should be to go to each and every axiom of
     quantum theory and give it an information theoretic justification
     if we can.  Only when we are finished picking off all the terms
     (or combinations of terms) that can be interpreted as subjective
     information will we be in a position to make real progress in
     quantum foundations.  The raw distillate left behind---minuscule
     though it may be with respect to the full-blown theory---will be
     our first glimpse of what quantum mechanics is trying to tell us
     about nature itself.
\eq

Now, look in that, and tell me where is this view of yours that we
drop the idea that ``physics should be about reality''?  I'll just
pose that as a challenge to you.

Hopefully I'll see you both again tomorrow.

\section{27-10-05 \ \ {\it It Wasn't You} \ \ (to H. C. von Baeyer)} \label{Baeyer3}

I just looked it up again.  At the web site, 
\begin{center}
\myurl[http://web.archive.org/web/20060104094204/http://www.physics.byu.edu/research/theory/papers/]{http://web.archive.org/web/20060104094204/http:// www.physics.byu.edu/research/theory/papers/}
\end{center}
someone posted this comment:
\bq\noindent
     Our own Dr.\ Evenson recently had dinner with Anton Zeilinger, and asked him what he thinks about the radical ideas of Christopher
     Fuchs.  Professor Zeilinger replied that ``he is not radical
     enough.''
\eq
I don't know who Evenson is, but he's not von Baeyer.

\section{30-10-05 \ \ {\it Correlation without Correlata in the Strangest Places} \ \ (to N. D. {\Mermin})} \label{Mermin119}

I read these lines in Charles Krauthammer's column in the Washington
Post this morning,
\bq\noindent
     This coldbloodedness is a trademark of this nation's most
     doctrinaire foreign policy ``realist.'' Realism is the billiard ball
     theory of foreign policy: The only thing that counts is how
     countries interact, not what's happening inside. You care not a
     whit about who is running a country. Whether it is Mother Teresa or
     the Assad family gangsters in Syria, you care only about their
     external actions, not how they treat their own people.
\eq
and thought, ``Is that really the billiard ball theory or rather
correlation without correlata?''

\section{31-10-05 \ \ {\it One World} \ \ (to C. H. {\Bennett} and J. A. Smolin)} \label{Bennett41} \label{SmolinJ7.2}

On another subject, I'd be curious to know your opinion(s) on Andy Steane's article \quantph{0003084}, ``A Quantum Computer Only Needs One Universe.''  I found a fair number of his points in there insightful \ldots\ but I'd be willing to listen if your opinion is that I'm not being critical enough.  I talked to Sam Braunstein last week about it, and he thought it was ``utter tosh,'' but he didn't say much that put a dent in my opinion.  Just the opposite, in fact:  It strikes me that there are some ideas in that paper worth really trying hard to develop, if I could just get a handle on what the next step ought to be.

\subsection{Charlie's Reply}

\bq
I rather like Steane's paper, which I hadn't read before. How delightfully put when he says the amount of computation is not properly measured by the number of steps that would have been required to do a computation some other way.  I quite agree with him that it generates great confusion to imagine that quantum computers are doing many computations at once.  I say they are doing one computation, and can get from input to output faster because of the extra maneuvering room Hilbert space provides during the intermediate stages.  Toward the end of his piece, Steane speaks of  entanglement as ``representing correlations between logical entities without representing the entities themselves''.  I think this is a less felicitous way of speaking, rather like Mermin's ``correlations without correlata''.  The trouble with both of these, in my view, is that they treat entanglement as a kind of correlation.  Rather entanglement should be viewed as the primary thing, and correlation as one of its manifestations.  Speaking in the liturgical language of the Church of the Larger Hilbert Space, correlation is a sacrament, the outward and visible sign of entanglement.   Alas, your seduction at a tender age by the atheistic doctrine of Jaynes probably prevents you from thinking in this satisfying and clear-headed manner.  To retreat from sectarian proselytizing, let me recommend my attached talk on publicity and privacy, which you might enjoy.
\eq

\section{31-10-05 \ \ {\it QMech Interpretation \& a Web-Thing Against the War}\ \ \ (to H. J. Bernstein)} \label{Bernstein6}

Thanks for including me as a recipient on this note, though I'm not quite sure what it's all about.  Did it have something to do with your last lines?  To wit,
\bhbe
What is YOUR best argument for those who take comfort (despite the
clear-cut differences between statistical ensemble ``lack -- of --
complete knowledge'' kind of probabilities and ``complex square roots
that superpose'' kinds of probability present even in the purest of
PURE quantum states)?
\ehbe

Anyway, just to record one of my latest strivings for clarity:  I've gotten out of the habit of regarding probability as expressing ``lack of complete knowledge.''  For that phrase effectively already implies that there's something out there that the user of the probability just doesn't have hold of.  Instead now, along with de Finetti's usage, I just think of probability as quantifying ``uncertainty'' full stop.  Uncertainty for whatever reason.

\section{31-10-05 \ \ {\it Cover to Cover} \ \ (to N. D. {\Mermin})} \label{Mermin120}

By the way, do you know this quote of Bohr's?  I've been meaning to
send it to you for a while (but kept forgetting), in case it
complements the one I usually hear you quote (i.e., the one about
``track[ing] down \ldots\ relations between the manifold aspects of
our experience'').
\bq\noindent
   The extension of physical experience in our own days has \ldots\
   necessitated a radical revision of the foundation for the unambiguous
   use of elementary concepts, and has changed our attitude to the aim
   of physical science.  Indeed, from our present standpoint, physics is
   to be regarded not so much as the study of something a priori given,
   but rather as the development of methods for ordering and surveying
   human experience.
\eq

\section{01-11-05 \ \ {\it Martha White's All-Purpose Correlation} \ \ (to C. H. {\Bennett})} \label{Bennett42}

I'm glad you give me some evidence that I'm not crazy for liking Steane's paper.

\bcb
Toward the end of his piece, Steane speaks of entanglement as
``representing correlations between logical entities without
representing the entities themselves''.  I think this is a less
felicitous way of speaking, rather like {\Mermin}'s ``correlations without correlata''.  The trouble with both of these, in my view, is that they
treat entanglement as a kind of correlation.  Rather entanglement
should be viewed as the primary thing, and correlation as one of its
manifestations.  Speaking in the liturgical language of the Church of the Larger Hilbert Space, correlation is a sacrament, the outward and
visible sign of entanglement.
\ecb

I'm not in disagreement with that \ldots\ particularly as the Grand Orgy of the Sexual Interpretation (of QM) would say the same thing.  (Ask John.)  Entanglement is a property of a quantum state with respect to a given tensor-product structure on the quantum state's Hilbert space.  Correlation, on the other hand, is something that arises only through consideration of probabilities for the outcomes of (two-indexed) quantum measurements.  Entangled states, consideration of measurements with only one index, NO notion of correlation; simple as that.  So correlation can't be viewed as the primary concept.

Here's something I wrote along those lines in a now long forgotten document (of 1998):
\bq
   The quantum world brings with it a new resource that senders and
   receivers can share:\ quantum entanglement, the stuff
   Einstein-Podolsky-Rosen pairs and Bell-inequal\-ity violations are made
   of.  This new resource, of all the things mentioned so far, is the
   most truly ``quantum'' of quantum information.  It has no classical
   analog, nor might it have been imagined in a classical world.

   What is quantum entanglement?  It is {\it not\/} probabilistic
   correlation between two parts of a whole.  Rather it is the {\it
   potential\/} for such a correlation.

   In a quick portrayal:

   {\it classical correlation}--- Alice and Bob entered a lottery for
   which they were the only players. They have not opened their
   ``winnings'' envelopes yet, but the messages in them say that one is
   the winner and one is the loser. Which is which, they do not
   know---they only know the correlation---but the answer is there,
   objectively existent, without their looking.

   {\it quantum entanglement}--- Alice and Bob will eventually perform
   measurements on the EPR pair their envelopes contain and the outcomes
   {\em will\/} be correlated. However, before the measurements are
   performed, there are no objectively existent variables already there.
   Different measurements can and will lead to different correlations.

   In a certain sense, entanglement is a kind of {\it all-purpose
   correlation\/} just waiting to be baked into something real---a
   quantum ``Martha White's Flour'' \verb+\cite{Flatt}+.  The uses for this
   all-purpose correlation are manifold within Quantum Information
   Theory.
\eq
[The citation \verb+\cite{Flatt}+ goes to Lester Flatt and Earl Scruggs for their {\sl Martha White Theme\/} song.  What the heck, let me include the words to that too; it's at the end of the note.]

What I'm really interested in is whether we might dream up some technical problems to try to work out based on Steane's paper.  Any ideas?

By the way, let me know if you get this note; I'm a little worried that your spam filter might pick it in retribution for my indiscretions above.

\bv
{\bf Martha White Theme Song} \medskip
\\
Now you bake right,\\
(band response:) ah ha,\\
With Martha White, \\
(band response:) yes ma'am, \\
Goodness gracious, good and light, \\
With Martha White.\medskip
\\
Now you bake better biscuits, cakes, and pies,\\
With Martha White Self-Risin' Flour,\\
(band response:) that one all-purpose flour,\\
With Martha White Self-Risin' Flour \\
You've done all right!
\ev

\section{01-11-05 \ \ {\it Thanks, Shelves, and Expenses} \ \ (to H. C. von Baeyer)} \label{Baeyer4}

Now that Halloween is over, I have a little time to reply to some emails finally.  Thanks so much again for the invitation to W\&M.  I had a great time, and it was lovely to see so many eyes sparkle from the ideas Friday.  (Though I wish I had had more of a chance to talk to Gene Tracy---his eyes particularly sparkled.)  And, I'm pleased that my ``job interview'' seemed to go well at the crabcake place \ldots

Finally, a further thanks for the detailed description of your bookshelves.  My wife and I spent Sunday morning intermittently staring at the walls in our (to be) library, trying to figure out just what we're going to do.  With my ``pragmatism'' collection now just shy of 400 volumes (and growing at an even pace), we've really got to get going on a solution.  I want that room to be the intellectual center of the house, and what better theme for it than that ``reality is on the make'' (one of James' and FCS Schiller's themes, particularly).

\section{01-11-05 \ \ {\it 01$\;+\;$10} \ \ (to T. Duncan)} \label{Duncan3}

Thanks for the latest note, and thanks for keeping your eyes peeled.

(The simple existence of) Your note reminded me that I still had not replied to your original query:
\btd
This leads to a point that is still muddy to me. It may be a naive
question and I'm not quite sure how to phrase it, but let me try this
to get started: Given that we must ask what the ontic states can be,
such that they force this restriction of incomplete knowledge on the
epistemic states, why not just place the entire restriction directly
on the ontic states themselves? Why not simply say there is a limit to
what we can {\it know\/} about the state of the world because there is a
fundamental limit to how much information can actually be ``stored''
about its state?  As I understand it, this is something like the view
expressed e.g.\ by Zeilinger (``A Foundational Principle for QM'', {\bf Found.\
Phys.}\ 1999). I realize this is not what you're saying. So, I'd like to
better understand why you emphasize the distinction between epistemic
and ontic states, when it seems possible that all of the restrictions
on our knowledge could be absorbed as consequences of the restrictions
on the ontic states. Anyway I'd appreciate any thoughts you have to
point me in the right direction here.
\etd

Somehow I never got inspired to give you the answer you deserve---I guess it would require something of an essay on why this distinction should be made at all, and I never mustered the energy.  But let me do this by way of a partial answer:  I'm just returning from having given the physics colloquium at William and Mary, and because Hans von Baeyer is there, I have also just finished his book, {\sl Information:\ The New Language of Science}.  It's really very, very good, I think.  Particularly, Chapter 2, ``The Spell of Democritus,'' I think goes a little toward answering your question.  Let me recommend then that you give that a shot first.  If it helps, let me know.  And if it doesn't, maybe try re-posing your question, and I'll try my best (with a layer of guilt for pushing it off this long) to give you some feedback.

Thanks also for bringing Louise B. Young's book to my attention.  Believe it or not, I've had that book on my shelf for five years and never opened it once.  (I remember picking it up at a hospital fundraiser.)  I'll certainly open it now.

\subsection{Todd's Preply}

\bq\noindent
\bq\noindent
[CAF wrote:] OK, so you already know how I'd answer your Exercise 1.image:  ``I am a
cosmic artist -- a contributor to a universal creative process.''
\eq

You can probably guess that I would answer similarly.\ \smiley\ \ I remember reading a book in high school by Louise B. Young, {\sl The Unfinished Universe}. Some of the technical details in the book are questionable, but the spirit of it resonated deeply with me. That spirit is pretty well expressed by the final paragraph: ``The universe is unfinished, not just in the limited sense of an incompletely realized plan but in the much deeper sense of a creation that is a living reality of the present. A masterpiece of artistic unity and integrated Form, infused with meaning, is taking shape as time goes by. But its ultimate nature cannot be visualized, its total significance grasped, until the final lines are written.''
\eq

\section{02-11-05 \ \ {\it von Baeyer}\ \ \ (to A. Y. Khrennikov)} \label{Khrennikov12}

I had the good fortune to meet Hans C. von Baeyer the other day and to read his new book, {\sl Information:\ The New Language of Science}.  It's a very good book, and I recommend it if you haven't read it yet.  Anyway, he told me that you had invited him to your next meeting in {\Vaxjo}.  I'm very glad you did, and it is excellent to have another Bayesian among us there!!  Have you been able to give any thought to my idea of a special session of ``Bayesians'' attending the meeting?

\section{02-11-05 \ \ {\it Zeilinger, Me, and von Baeyer Makes Three?}\ \ \ (to H. C. von Baeyer)} \label{Baeyer5}

I'm back, though ``lunch'' turned out to be effectively a two-day
affair! To business now:
\bhcvb
I have been dipping into your marvellous book all weekend, and have a
question (or seventeen). \ldots

In January I will give an invited talk at the 75th anniversary of the
American Association of Physics Teachers, in Anchorage, Alaska.  The
occasion is my receiving the Gemant Award of the AIP. My title is:
``How I learned to stop worrying about {\Schroedinger}'s cat.'' I intend
to chronicle my gradual adoption of your point of view about QM.  Now
the question:

It seems to me that the search is on for the irreducible kernel of
QM, you call it the Zing, Maxwell, as a boy, would have called it
``The GO of it.'' You and Zeilinger approach it from opposite ends:
You by stripping away all ``merely'' information theoretic baggage,
and he by guessing the answer.  You have not yet succeeded in getting
down to the bottom, and he has a long way to go before he builds up
to the full theory.  Analysis v.\ synthesis, top-down v.\ bottom-up.

Is this a fair way to characterize your approach?
\ehcvb

Thanks by this; I'm certainly flattered.  And I hope you'll ask me
all seventeen eventually.  The way to do good physics is to write
good physics, and in my case, I need good questions before I can even
hope for the latter.

And, by the way, speaking of good books, I really did enjoy reading
yours last week.  I think, whether it was intended to be or not, it
is a great service to the quantum Bayesian community---sort of a
mortar to soften up the popular beach---and I'm grateful for that.
For the heck of it, below, I've decided to type in the little notes I
made on my bookmark as I read it---they record my agreements and
disagreements and a little of my study of the English language.
Particularly, you'll note a remark I made about something on page 38.
It refers mostly to your sentence,
\bq\noindent
     If the wave function is nothing but a storehouse of information
     needed to make correct predictions, then the stuff of the
     world is really, at bottom, information.
\eq
I hope my talk at W\&M emphasized that I would be in disagreement
with that.

And in fact, to start to answer your question, I think that
disagreement probably captures the greatest gulf between the views
Anton and I are each striving to develop.  It's not a difference of
technique (analysis vs.\ synthesis) per se, but of our {\it
guesses\/} of an ontology---I think they are opposed to each other.

If you can give me a few days, I plan to come back to you with a much
more complete answer.  But first I want to re-read all of Zeilinger
and Brukner's stuff.  This will be a good opportunity for me to flesh
out the similarities and differences of what we're hoping to get at.
In any case, I'll get back to you long, long before your January
talk.

By the way, congratulations on the teaching award.  But I hope
Anchorage isn't too cold in January!

\noindent --- --- --- --- --- --- --- --- --- --- --- --- --- ---

\noindent Notes on H. C. von Baeyer, {\sl Information: The New Language
of Science}\medskip

vB asks ``why information not like energy'' -- 9, 10, 33

special relativity as inserting the observer into physics -- 12

{\Wheeler}, subjectivity -- 13

split between objective and subjective -- 13, 14

{\Wheeler} quote -- 15

relations without relata -- 23, 24, 52--53

({\Schroedinger}) information as surprise -- 25

Gell-Mann sobriety -- 33

bad inference -- 38

non-Bayesian view of probability? -- 57, 58, 72

questionable interpretation of Bohr -- 64, 65

probability as informed guess -- 75

not ``avoid'' but ``masks'' -- 95

quantum random numbers -- 107

hearing loss -- 125, 126

{\Wheeler} grafitto from Pecan St. -- 127, 128

as it were -- 132, 134, 155, 166, 201, 231

Nobel prize meat -- 140, 141

probability does not exist -- 172

bad language for superposition -- 180, 181

interventions -- 183

measurement as creation -- 184

qubit as state rather than system -- 187, 188

the godfather of soul -- 191

fey neurasthaenic -- 193

analogy, maybe more accurate than you think -- 195, 196

quantum crypto distinct idea from quantum computing -- 197

physics as comparison -- 203

Boltzmann titles -- 223

correct phrase on teleportation! -- 225

Zeilinger or Rovelli first? -- 226

information about what? whose information? -- 227

entanglement not necessarily the key to quantum crypto -- 230

\section{02-11-05 \ \ {\it Don't Forget Fierz!}\ \ \ (to H. C. von Baeyer)} \label{Baeyer6}

In fact, I'm back again already.  Just as I was sending off the last note, I realized that I hadn't mentioned the Fierz project again---surely I don't want to forget about that.  I wrote my friend Greg Comer today that my mouth is already watering to see your results.

Did by chance you read the Fierz article reprinted in my 29 August 1999 letter to {\Ruediger} {\Schack} in {\sl Notes on a Paulian Idea\/}?  If so, what did you think of it?

After sending off this note to you, I'll write Harald Atmanspacher with a brief note of introduction about you and the story of our meeting and similar interests, and then I'll put you two in contact with each other.

\section{02-11-05 \ \ {\it And Holladay} \ \ (to H. C. von Baeyer)} \label{Baeyer7}

I just can't get my email right today; I keep remembering things that
I forgot to reply to.
\bhcvb
Full disclosure: Wendell Holladay was my thesis adviser, but he died
last year.  De mortuis nihil nisi bene.  (Of the dead, say nothing
but good things.)
\ehcvb
I am sorry to hear about that.  Were you close?  Probably.  In
reviewing what I wrote about him in {\sl Notes on a Paulian Idea}, at
least I notice that a reply to his letter seemed to be the hardest
part of the paper for me to construct---i.e., apparently he gave me a
run for the money. He was probably a great guy.  Again, I'm sorry.

\section{02-11-05 \ \ {\it Carry Cameras, not Guns} \ \ (to C. H. {\Bennett})} \label{Bennett43}

I really enjoyed thumbing through your talk.  Particularly, I enjoyed
this quote
\bq
     Could it be that every major past phenomenon, say Sappho's other
     poems, or Jimmy Hoffa's murder, can be recovered from physical
     evidence in principle, if not in practice?

     To believe otherwise is venturing dangerously close to the
     deconstructionist view, abhorred by most scientists, that history
     is not what ``actually'' happened, only what we think happened.
\eq
as food for thought.  It put the importance of the black-hole
information question in a new light for me.

But also it reminded me of some stuff I once read about the
interesting things that can happen when one tries to mesh the idea of
``physical indeterminism'' together with the ideas of ``history'' and
``archeology'' as autonomous sciences.  As it happens, I have an old
PDF file which includes a description of George Herbert {\Mead}'s
thoughts on the subject.  I'll attach it for your fun.  (Ignore the
stuff I wrote at the beginning of it, and just go straight to the
copied description of {\Mead} himself on page 2.)  By the way,
{\Mead}---one of the founders of American pragmatism---was born in South
Hadley, MA in 1863.  So I suspect his ghost is practically a neighbor
to you guys in Wendell on the weekends.

\section{02-11-05 \ \ {\it Air Guitar} \ \ (to G. L. Comer)} \label{Comer73}

Great to hear that things are going well for the band \ldots\ and for you as a rock star.  I wouldn't mind a middle-aged woman or two playing air guitar in front of me.

I'm just back from a wonderful couple of days at the College of William and Mary.  I gave their physics colloquium Friday and got a much better reaction than usual; so I'm pretty pleased.  Also, I got to know a (soon retiring) professor Hans Christian von Baeyer and was quite impressed with the guy.  As it turns out, his main interest is in accurately popularizing science; he was the editor of the magazine {\sl The Sciences\/} for many years.  Consequently he also has a few popular books---the latest being {\sl Information:\ The New Language of Science}, and I was extremely pleased with its presentation of our field, and even our Bayesian view of the quantum state.  (If you've got the time for some easy reading, I'd definitely recommend it; I picked it up at the Princeton book store for \$16, paperback.)

More importantly than all that, though, I found out that the guy's godfather was Markus Fierz's brother.  Thus, when I suggested to him that he translate Pauli and Fierz's correspondence on all these various alchemical analogies to quantum measurement, he very interestingly promised to take on the task!!  Particularly, he thought it would be a nice easy project to fill the cracks between the more major projects he plans to start in retirement.  So, I'm very tickled about all this, and my mouth is already watering.

For fun, let me send you a couple of things after this note---two things that I wrote Charlie Bennett this morning.  1) Some pictures of the kids and me in lederhosen.  And 2) some stuff about rewriting history (and a little thought to do with the black-hole information problem).  I'll include Charlie's original talk in the second of those for context; so these will be some pretty big files.

\section{07-11-05 \ \ {\it The Nauseating Terminus} \ \ (to G. L. Comer)} \label{Comer74}

I thought your note was on the mark.  What it shows is that even in
our state of relative ignorance about the goings on of the universe,
we should still expect life to arise somewhere/somehow simply due to
the vast size of this universe.

But also keep in mind that, from the Bayesian view, probabilities
(and the surprise or lack of surprise they entail when one sees the
outcome of a trial) represent nothing other than one's degrees of
beliefs.  Probabilities are not part of the furniture of the
universe.  In particular, I suspect if we understood chemistry a lot
better, then---conditioned on that knowledge---life wouldn't look so
surprising at all:  That is, {\it with respect to\/} a better state
of knowledge, one wouldn't have to shore up one's argument for the
inevitability of life by statements of the vastness of the universe.
It'd be a simple consequence of chemistry as such (and not rare
initial conditions).

So that in the end, as scientists we would say:  Because our
fundamental fields have these Hamiltonians, life is effectively
inevitable.  The religious-minded at that point should say, ``So who
ordered those Hamiltonians?  You yourself contemplate that they could
have been otherwise.  Hence, at just that point, there's the concrete
expression of an intelligent design.  From it, as you have just
proved to me dear scientist, everything inevitably flows.''  And how
do you combat that?  We each have our terminus of inquiry:  ours is
an equation, and theirs is an old man behind the equation.  By why
should either of us terminate?  We should both have a horrible sense
of dissatisfaction.

I'll try to dig up the Bennett story.

\subsection{Greg's Preply, ``A Bayesian Look at Miracles''}

\bq
I teach a course called Intro to Physics.  It is for entering freshmen physics majors.  We learn how to use Mathcad, have guest speakers come in to talk about careers in physics and recent developments in physics, and discuss various readings from a ``physics for poets'' book.  I purposefully try to pick chapters that discuss ``sexy'' topics in physics.

Last week we discussed the chapter on life throughout the universe.  You know, are we alone?

During that discussion, we were led, naturally enough, to the question of whether or not God exists.  We talked about how dangerous the universe is, and about all the ``miracles'' that must occur in order for advanced civilizations to develop.  You know, if our evolution is typical, then a solar system must be stable in all respects for several billion years (i.e.
stable orbits, protection from comet, asteroid, etc impacts, and so on).  While religious conservatives won't accept the billion-year timescale, they will no doubt accept that the universe is dangerous, and the emergence of highly developed species requires ``miracles.''  So, ironically, the scientific and religious communities are in something of an agreement.
The difference, of course, is that the religious community says this points to a universal intelligence in control.  The scientific community, more or less, looks to chance.

What was striking about the discussion was the sense of awe that we experienced.  It really is amazing that we exist, and can actually have a discussion about our origins.  I had a new insight on that awe.  In fact, it was inspired by that tale you told me once of how Charles Bennett bet you---in a bar of course!---that he could get a particular result for the flipping of a coin.  (I don't remember the details, but I'm sure you can recall them for me given these few clues.)  Maybe I'm just finally realizing the full extent of your tale.

Imagine each human is given an unbiased coin.  Each is told to flip that coin 10 times in a row.  None will be surprised to get essentially 5 heads and 5 tails.

But what about those few individuals who manage to flip 10 heads in a row, or ten tails?  For them, the result will be, well, miraculous!  For the collective, however, the result will be, well, normal!  With so many humans flipping coins, surely we can expect the unusual to occur?

Think now about our universe.  For each star in our galaxy, there is at least a galaxy in our visible universe.  If the universe is infinite in extent, and describable by one of the homogeneous and isotropic solutions of the Einstein equations, there will be an infinite number of galaxies.  Certainly, with so many possibilities for solar systems to develop and evolve, we should not be surprised that in certain of those the ``miracles'' required for an advanced civilization to appear will in fact transpire.  And just through sheer chance.

If you can poll some universal analog of humans, you will no doubt find that these analogs are not surprised by the result.  But when you poll humans, they should be, on average, amazed that the miracles for them have occurred.  Just like the one human that flips ten heads in a row, or the one who wins the lottery.

I guess a miracle is always a posteriori.  Events can only become ``miraculous'' after the data is collected and the information has been updated, through that human lens that is always expecting cause and effect.
\eq

\section{07-11-05 \ \ {\it Macaroni} \ \ (to G. L. Comer)} \label{Comer75}

Here it is, dated 28 July 1998:

\bq
\noindent Dear Gregaroni, \medskip

I've just been enjoying a Barcelona morning, taking care of some e-mail odds and ends.  Today is my last day in Spain; tonight I spend the night in London, finally to be on my way home tomorrow.  Thank you very much for your stories, especially the ones about your dad.

The last time I was in Europe (last month), I met up with Charlie Bennett as usual.  I don't know how it happened, but we got into a conversation about coin tossing (real coin tossing, not the philosophical issue).  Charlie said, ``You know, there's a really easy method for predicting someone else's tosses; I'll show you.''  I said, ``Oh bull.''  Then he pulled a quarter from his pocket and gave it to me.  I flipped; he called heads.  It was heads.  So I did it again.  He called heads.  It was heads.  I said, ``Oh you just got lucky again.''  ``Well, do it again,'' he says.  So I flip.  He calls heads.  It was heads!!  In astonishment, I say, ``OK what's the trick you have up your sleeve?  Did you give me a fixed coin?!?''  He bursts out laughing, ``No, not at all.  One time in eight you get lucky, and when you do, it's a really good joke!''

But still, synchronicity happens.  The evidence is our two stories.  I'll never forget mine; maybe you'll never forget yours.\medskip

\noindent Take care my friend, \medskip

\noindent Chris
\eq

\section{07-11-05 \ \ {\it Bar Stories} \ \ (to G. L. Comer)} \label{Comer76}

\bgc
Got it!  Thanks.  Although, it doesn't imply that you were in a bar,
drinking when you heard the story.  That's too bad.
\egc

Bennett doesn't really drink.  I remember one time in Japan, Alexander Holevo told us he would make dinner for us if we would eat a true Russian meal.  We agreed.  Then we found out that the meal consisted of vodka, caviar, boiled eggs, and borscht.  Well, I could hardly stomach more than a couple of bites of caviar, and Bennett took only a couple of sips of vodka.  On the other hand, Bennett gobbled down the caviar, and I helped Holevo with more than half the bottle of vodka.  Bennett joked that between the two of us, we made one good Russian.

\section{07-11-05 \ \ {\it The Second Terminus} \ \ (to G. L. Comer)} \label{Comer77}

\bgc
I'm also not sure that I've assumed that probabilities are real. I
readily admit that I introduced them via the question asked.
\egc

I didn't mean to imply that you weren't being a good Bayesian.  I was
just changing the subject a bit \ldots\ so that I could make the
point that the religious can always find a place for their old man in
the sky. For they can always pick up wherever we (the scientists) say
THIS is the fundamental equation or THAT is the necessary class of
initial conditions.  On the other hand, they think they're getting
somewhere deeper by positing their old man in the sky than we are
with our fundamental equation or initial condition (or both).  To
that I ask, why have they gotten any deeper:  It looks to me to be of
effectively the same level.  For I can ask of their god, who ordered
that?  In fact, one can turn their own argument upon themselves.  If
they ever say that the beauty or the complexity of the world around
us cannot be explained without a creator, then I say that the beauty
or complexity of that creator could not be explained without an even
more powerful and harmonious creator.  And so on ad infinitum.  The
only leg I've ever see them stand on is to reply that, ``But God
needs no explanation'' \ldots\ end of argument.  And then I retort,
``But then why do you demand it of my equations.''  And of course it
just goes round and round.

Anyway, sorry, I wasn't disagreeing with you \ldots\ just changing
the topic a little.  But also, I wasn't even sticking to one topic
really.  The point about my personal {\it belief\/} that life is
largely independent of initial conditions (just like planet formation
seems to be, for instance), was really a side point.

Or to say it another way, mostly in my last note I was just ranting
because the religious right have pissed me off too.

\section{07-11-05 \ \ {\it Another Story about Beginnings} \ \ (to G. L. Comer)} \label{Comer78}

Here's another story that I'm not sure I ever shared with you.  See mostly the last paragraph of the note below. [See 22-09-03 note ``\myref{Bennett29}{Cosmology}'' to C. H. Bennett.]

It's always like that you know:  No one can explain the origin of their own priors.  Not even William James.  In probability theory, the main difference between the Bayesians and the nonBayesians is that they're at least honest with themselves on this point.

\section{07-11-05 \ \ {\it Zeilinger, Orange House, and Halvorson} \ \ (to C. G. {\Timpson})} \label{Timpson8}

You've been on my mind recently, as I was forced to rethink ``Zeilinger's Foundational Principle'' for a visit with Hans C. von Baeyer at the College of William and Mary last week.  It still doesn't fare better in my mind (even though I know it helped inspire Rob {\Spekkens}'s ``toy model,'' for which I have great respect).  Thus I recommended your paper on the subject to von Baeyer.

Von Baeyer is an interesting guy---his profession now is in the {\it accurate\/} popularization of science---and I enjoyed his book {\sl Information:\ The New Language of Science\/} quite a bit as a really good attempt in that direction.

Anyway, I'm writing this note because I understand you'll be in New Jersey in mid-December.  Where will you be flying into and out of?  Newark, presumably?  If you have any extra time, it'd be great to have you in the Murray Hill / Cranford area for a day or two (or more).  We could put you up in our house and fatten you up with some of Kiki's home cookin' while you and I try to come up with something new in quantum mechanics.  (Perhaps I should also mention my single-malt collection?  Or would that defeat the purpose?)  Finally, I wouldn't mind showing off the best book stores in NY City if you happen to bring an empty case.  Just let me know your plans and if you have any time.  I can get you from Princeton to Cranford, and then from Cranford to the Newark airport when the time comes.

The reason I know you'll be in NJ is because I've been attending van Fraassen and Halvorson's seminar on quantum information for a few of the sessions. Connected with that, in the next note I'll forward you something I wrote to vF and H after listening to Halvorson's lecture on teleportation.  [See 25-10-05 note ``\myref{Halvorson4}{GOBs, Bobs, Steering \& Teleportation}'' to H. Halvorson and B. C. van Fraassen.] It's a bit long, but maybe you'll find some parts of it entertaining.  To two of the questions in your Budapest slides, ``How is so much information transmitted?''\  and ``Just how does the information get from Alice to Bob?,'' my own answer would be:  {\it No\/} information is transmitted in the process of teleportation (excepting the two bits that tell Bob which action to be performed).  The only nontrivial thing transferred in the process of teleportation is {\it reference}.  Charlie's information (in the sense of Bayesian degrees of belief) stops being {\it about\/} the qubit he just handed off to Alice and starts being {\it about\/} Bob's.  I don't know what you say in your paper ``The Grammar of Teleportation,'' but I just printed it out and hope to have it read in the next couple of days.  If we disagree substantially, you can guess I'll be writing you again!

\section{07-11-05 \ \ {\it Charlie Bennett Stories} \ \ (to C. H. {\Bennett})} \label{Bennett44}

My friend Greg Comer asked me to remind him of a story I had once told about coin tossing.  I dug up the note below and sent it off to him to fulfill his request.  [See 07-11-05 note ``\myref{Comer75}{Macaroni}'' to G. L. Comer.]  Remarkable thing dawned on me though:  I've been telling stories about you for at least five years.  I should replenish my supply.

\subsection{Charlie's Reply}

\bq
Your version is partly apocryphal.   I was demonstrating Martin Gardner's 3-cup trick, which he calls mathematical 3-card monte, to Merrick Furst at CMU.  He didn't see how the trick worked, and thought it amazing that I had ``guessed'' the correct cup--the one holding the little piece of crumpled paper--3 times running.  When he asked me how the trick worked, I noticed that, by chance, the correct cup had always been the leftmost one, so I said, ``I just always choose the one on the left.''  This amazed him even more.
\eq

\section{08-11-05 \ \ {\it Closet Haecceity} \ \ (to H. Barnum)} \label{Barnum18}

By the way, I was amused with your comment
\bhb
``Well, his theory's got haecceity, and that's been d\'eclass\'e
since the late middle ages.  You just can't have a theory like that.''
\ehb

Truth is, I guess I'm a closet haecceitist.  I've probably told you that before.  Indeed I love this quote by William James,
\bq
Let me give the name of `vicious abstractionism' to a way of using concepts which may be thus described: We conceive a concrete situation by singling out some salient or important feature in it, and classing it under that; then, instead of adding to its previous characters all the positive consequences which the new way of conceiving it may bring, we proceed to use our concept privatively; reducing the originally rich phenomenon to the naked suggestions of that name abstractly taken, treating it as a case of `nothing but'
that concept, and acting as if all the other characters from out of which the concept is abstracted were expunged. Abstraction, functioning in this way, becomes a means of arrest far more than a means of advance in thought. It mutilates things; it creates difficulties and finds impossibilities; and more than half the trouble that metaphysicians and logicians give themselves over the paradoxes and dialectic puzzles of the universe may, I am convinced, be traced to this relatively simple source. {\it The viciously privative employment of abstract characters and class names\/} is, I am persuaded, one of the great original sins of the rationalistic mind.
\eq

And you yourself also have some of the blame:  When I retrace the footsteps that led me to this fetish, the point of departure seems to have been a conversation with you---when you first convinced {\Ruediger} and me (through de Finetti's argument) that even ``exchangeability'' is a subjective judgment.  The root of the issue is all right there.

I know you've seen an earlier version of this ``Delirium Quantum'' thing of mine [now online at \arxiv{0906.1968}], but I'll send it again:  Section 3 on ``snowflakes'' expands a little on this point in the quantum context.  Is it so bad being a closet haecceitist?

\section{08-11-05 \ \ {\it Quibbles, Actions, and Reading} \ \ (to H. C. von Baeyer)} \label{Baeyer8}

\bhcvb
Re Bayesian:  I have a couple of the volumes of the {\Pauli} letters
here on my desk.  If you look up Bayes in the index, you come to a
letter from P. to F. dated 12 July 1952.  It is distinctive in
including the dream-drawing of quantum mechanics and deeper reality
that you are going to paste into your {\tt PauliProject.pdf}.  Where
did you see it?  It also appears in a letter to Jung a year later. Is
{\Pauli}'s letter available in English?

Fierz wrote at least two detailed drafts of a reply.  One bit that
you would love is this:  {\Pauli} brings up people who believe -- or
disbelieve -- a theory so much that they remain unconvinced by
statistical evidence.  They stubbornly continue to wait for a
measurement that disagrees significantly with the theory.  {\Pauli}
suggests that the state, or society, could then declare that these
people belong in an insane asylum.  But he himself doesn't dare to
make such rules.

Fierz replies:
\bq\rm
I am totally against the putative opponents, in the name of the state
or of society, to hold their views.  Rather, one should challenge
them to a {\it bet}.  If they accept it, they will {\it lose their
money}, which proves that they have not lost their reason, but
everything else, in particular, money.

On the other hand, if they reject the bet, it is demonstrated that
their conviction is without practical consequences, in other words,
that they they don't really believe what they claim.

I agree with your claim, and have always shared it, that purely
logically nothing follows from the laws of Bayes and Bernoulli about
a concrete single event.  This is the answer to your question 2.

However, the maxims of our practical actions can never be justified
logically.  If that were the case, empirical observations could be
reduced to logic.  Every action requires a decision, and that is an
act of will.
\eq

Betting good money, that's where it's at!!!
\ehcvb
I like it!  What a great quote.  Find me a thousand more like that,
and I'll be forever in your debt.

Concerning your earlier Zeilinger question, I did take the time to
read the Zeilinger and Brukner--Zeilinger papers.  Also I reread
Hall's and {\Timpson}'s commentary on them.  (I suspect Mana's
commentary is also good, but I didn't discover that one until this
morning; so I can't say that I read it, though I know him and think
he is a very promising student.)  {\Timpson}'s comment in particular, I
think is very relevant for what you were wanting to know.  So I would
suggest you have a look at it, if you haven't already.  Here are the
coordinates: \quantph{0112178}. The paper is written in too much
of an argumentative style for my tastes, but I do agree with the
substantive points.

Particularly, I agree that there's nothing to Zeilinger's principle
{\it as presently sketched} without already invoking quantum
mechanics itself.  See {\Timpson}'s Appendix A.  This is not to say that
we may not gain some insight from these ruminations.  As I said
before, I think {\Spekkens}'s toy model is a great example in that
regard.  So Zeilinger's ideas are certainly important ideas, even if
ultimately flawed.  Indeed, I think what {\Spekkens}'s paper really shows
is that the heart of quantum mechanics is something beyond the
Zeilinger principle.  For one might say that {\Spekkens}'s toy model
already effectively embodies the Z principle---formally through its
``knowledge balance principle''---and yet the model is {\it not\/}
quantum mechanics.  Rather it is a local hidden variable theory.  As
Rob\index{Spekkens, Robert W.} says in his abstract, ``A consideration of the phenomena that the
toy theory fails to reproduce, notably, violations of Bell
inequalities and the existence of a Kochen--Specker theorem, provides
clues for how to proceed with this research program.''

For my own money, I now think the real heart of the theory is in the
Kochen--Specker theorems; and the Z principle just doesn't capture
that.  In particular, I think the KS theorems are likely the most
direct statements we yet have that the ``observer cannot be
detached'' in an ultimate account of reality.  Particularly, that's
where I think our research needs to focus most:  Finding a way to
justify a Kochen--Specker-like theorem directly from the idea of
engaged/activating observers, and bypassing the details of quantum
mechanics.  But I just don't see how to go that route yet.

Finally, let me go back to what I view as the more fundamental
disagreement in outlooks for Zeilinger and me at the moment.  (The
disagreement may not be a permanent thing, but it is definitely here
for the moment.)  The point is made by {\Timpson} pretty clearly:
\bq
    Before stating the Foundational Principle, it is helpful to
    identify two philosophical assumptions that Zeilinger's position
    incorporates. The first is a form of phenomenalism: physical
    objects are taken not to exist in and of themselves, but to be
    mere constructs relating sense impressions [Footnote: Here I take
    phenomenalism to be the doctrine that the subject matter of all
    conceivable propositions are one's own actual or possible
    experiences, or the actual and possible experiences of another.];
    the second assumption is an explicit instrumentalism about the
    quantum state:
\bq\noindent
           The initial state \ldots\ represents all our information
           as obtained by earlier observation \ldots\ [the time
           evolved] state is just a short-hand way of
           representing the outcomes of all possible future
           observations.
\eq
\eq

Zeilinger and I are certainly in agreement on the second assumption,
but it is the first that drives a wedge between us presently.  In
fact, just last week, I wrote Bas van Fraassen and Hans Halvorson
these words:
\bq\noindent
    I for one, for instance, pretty much characterize myself as a
    realist in the time honored sense:  That there is a real world
    out there beyond our whim and fancy, and it is the task of science
    to hypothesize about its attributes and properties.  I am not a
    philosophical idealist and certainly not a solipsist.  And I don't
    think I am a positivist or an empiricist.  Predominantly, I think
    I lean toward a kind of materialism (though tempered by a lot of
    pragmatic subtleties).
\eq

That's the key point:  As far as I can tell, I think I almost lean
toward a kind of materialism.  Metaphorically:  Outside of the
alchemist is the philosopher's stone.  Without it, there is nothing
to enact the transmutation of the baser metals or of the alchemist
himself.

If you decide after reading a good portion of the stuff I've already
sent you that I'm not completely crazy, and you're interested, then
I'll go on a limb and tell you more about what I meant in the last
two lines above.  And I'll tell you why I'm really so interested in
this {\Pauli}-Fierz correspondence (and Fierz's own writings).  But if
you're already thinking I'm a little crazy, I don't want to fuel the
fire.

\section{08-11-05 \ \ {\it A Few More Things} \ \ (to H. C. von Baeyer)} \label{Baeyer9}

At your house, you asked why I called my compendium ``The Activating
Observer'' rather than ``The Participating Observer'' or some such.
Looking in the \LaTeX\ file, I see that I marked out at least this
many other tentative titles:
\bv
            The Undetached Observer:\\
            The Catalyzing Observer:\\
            The Malleable Reality:\\
            A Malleable Reality:\\
            The Malleable Substrate:
\ev

\bhcvb
While I am happy that you are reading my book carefully, I won't
spend much effort on defending it. I have learned a lot since I wrote
it, and am much more of a Bayesian now.
\ehcvb

If you want to go to the real extreme---the place where {\Caves},
{\Schack}, and I now sit in the spectrum of Bayesianisms---then let me
recommend you read these things:
\begin{enumerate}
\item
Bruno de Finetti, ``Probabilism: A Critical Essay on the Theory of
Probability and on the Value of Science,'' Erkenntnis {\bf 31},
169--223 (1989).

\item
Bruno de Finetti, {\sl Theory of Probability\/} (Wiley, New York,
1990), volume 1.

\item
Richard Jeffrey, {\sl Subjective Probability (The Real Thing)\/}
(Cambridge University Press, Cambridge, UK, 2004).

\item
Richard Jeffrey, {\sl Probability and the Art of Judgment\/}
(Cambridge University Press, Cambridge, UK, 1992).
\end{enumerate}

The two de Finetti pieces are phenomenal, and they were life-changing
for me (particularly the first reference, {\it after I read it the
fourth time}!).  (It's this life change that is recorded in my
samizdat, {\sl Quantum States:\ What the Hell Are They?})  My only
warnings are: 1) to take his hint of solipsism in Section 2 of the
first reference above with a grain of salt---he backs off on it by
the time he gets to his book---, and 2) and to disregard his support
for Mussolini and fascism in the last section of the same paper.
Those things are independent of the rest of the development, and one
should not get distracted by them---the rest of the paper really is
phenomenal.

\bhcvb
If you look up Bayes in the index, you come to a letter from P. to
F. dated 12 July 1952.  It is distinctive in including the
dream-drawing of quantum mechanics and deeper reality that you are
going to paste into your {\tt PauliProject.pdf}.  Where did you see it?  It
also appears in a letter to Jung a year later.  Is {\Pauli}'s letter
available in English?
\ehcvb

Those figures that I'm going to paste in refer to two letters from
{\Pauli} to Jung:  One from 27 May 1953 and one from 31 March 1953.  The
{\Pauli}-Jung correspondence has been translated into English; that's
where I got those long quotes from.  They start up on pages 146 and
144 of {\tt PauliProject.pdf}, respectively.

There.  Now I think I've answered everything you've ever asked me.

\section{10-11-05 \ \ {\it Kaine, Corzine, Kochen, Krazy Kats and All That} \ \ (to H. C. von Baeyer)} \label{Baeyer10}

\bhcvb
By the way, the reason I am no longer worried about {\Schroedinger}'s cat
is not that I think that q.m. is now understood, but that I feel, for
the first time in my career, that the problem is in good hands with
young folks like you and your friends.  I think that where I can make
a contribution is in explaining to other physicists and to the public
that y'all are NOT crazy, that progress HAS been made, that new ideas
ARE being brought to bear, and that there actually IS a hope of
realizing Wheeler's sibylline predictions.
\ehcvb
I am very happy to hear that.  I didn't get a chance to write you this morning, but you made my day.

\section{10-11-05 \ \ {\it Delirium Quantum} \ \ (to B. C. van Fraassen)} \label{vanFraassen6}

I've got to say, I really, really enjoyed today's discussion.  Now,
I'm getting my money's worth!  I thought you did a great job of
giving a kind of (perhaps momentary) suspension of disbelief about
Rovelli's ideas---i.e., making an honest attempt to get your head
around them---and I thought Maudlin did a great job as skeptical
counterpoint.  I was very pleased with the food for thought it all
gave me, and it made the drive back to Cranford go like a flash.

Particularly I'm struck that there really is more of a similarity
between Rovelli's outlook and our quantum-Bayesian outlook than I had
thought.  Of course, there are plenty of differences too (and likely
big ones).  But I think it is much worthwhile for me to rethink all
the various issues surrounding his view.

Anyway, let me attach a file that I hope will amuse you; it's titled
{\tt DeliriumQuantum.pdf}.  The parts I'm interested in your reading
connect mostly to your remarks today about Rovelli's view of the {\it
  domain\/} of quantum theory.  In particular, I shoot for some
explication of how I can at once say ``quantum mechanics is
incomplete''---I do that a lot now---and still emphatically deny that
it would be worthwhile to try to complete it with a hidden-variable
theory. Particularly, I like the analogy to do with a map vs.\ a globe
that you'll see at the end of this file.  In total, the only parts
that I really care for you to see are Section 4
``\myref{Mermin101}{Me, Me, Me},'' Section 6
``\myref{Mabuchi12}{Preamble},'' and Section 7
``\myref{Musser11}{B}.''  Of course, you can look at all the rest if
you got nothin' better to do, but I don't think the other parts are
connected to your discussion today.

I apologize in advance that the pieces I'm recommending were written
in a more overtly William {\James}ian style than usual (even by my own
standards).  Sadly, I find more clarity for myself when I do that,
but I suspect not all my readers do!  I remember I once gave some
lectures at Case Western Reserve U., and one of the professors wrote
to John Preskill, ``During Fuchs's lectures everyone became convinced
that quantum information theory was the most important thing ever.
But when he left, no one could remember why!''  You may have a
not-unconnected reaction to these writings.

I'm taking the family on a little ``road trip'' vacation to New
Hampshire for the next four days---starting in about six hours if I
ever get to sleep---but I suspect I'll be dropping in on you
electronically over the next few days.  I'm quite excited about this,
and I will probably spout things from time to time as they come to
mind.  I hope you don't mind:  As always, feel free to ignore these
notes and chalk them up to being my diary entries if you wish.

\section{10-11-05 \ \ {\it ``Action'' instead of ``Measurement''} \ \ (to B. C. van Fraassen)} \label{vanFraassen7}

Let me also elaborate a little on the spiel I gave today condoning
the word ``action'' more than the words ``measurement'' or
``question'' in the context of ``quantum measurement.''

The following explanation comes from an email to another friend last
Spring.  I do think the very word ``measurement'' is fraught with
trouble, particularly for what Rovelli and us quantum-Bayesians are
trying to get after.

\bq
Anyway, the way I view quantum measurement now is this.  When one
performs a ``measurement'' on a system, all one is really doing is
taking an ACTION on that system.  From this view, time evolutions or
unitary operations etc., are not actions that one can take on a
system; only ``measurements'' are.  Thus the word measurement is
really a misnomer---it is only an action.  In contradistinction to
the old idea that a measurement is a query of nature, or a way of
gathering information or knowledge about nature, from this view it is
just an action on something external---it is a kick of sorts.  The
``measurement device'' should be thought of as being like a
prosthetic hand for the agent---it is merely an extension of him; in
this context, it should not be thought of as an independent entity
beyond the agent.  What quantum theory tells us is that the formal
structure of all our possible actions (perhaps via the help of these
prosthetic hands) is captured by the idea of a
Positive-Operator-Valued Measure (or POVM, or so-called ``generalized
measurement'').  We take our actions upon a system, and in return,
the system gives rise to a reaction---in older terms, that is the
``measurement outcome''---but the reaction is in the agent himself.
The role of the quantum system is thus more like that of the
philosopher's stone; it is the catalyst that brings about a
transformation (or transmutation) of the agent.

Reciprocally, there should be a transmutation of the system external
to the agent.  But the great trouble in quantum interpretation---I
now think---is that we have been too inclined to jump the gun all
these years:  We have been misidentifying where the transmutation
indicated by quantum mechanics (i.e., the one which quantum theory
actually talks about, the ``measurement outcome'') takes place.  It
should be the case that there are also transmutations in the external
world (transmutations in the system) in each quantum ``measurement'',
BUT that is not what quantum theory is about.  It is only a hint of
that more interesting transmutation.  And, as you know, somehow out
of all this I think of the agent and the observer as being, together,
involved in a little act of creation that ultimately has an autonomy
of its own---that's the sexual interpretation of quantum mechanics.
(However the ideas in that last sentence are a little murkier than
what I'm trying to get at now.)

Does ``measurement'' in this new sense explicitly require
consciousness (whatever that is)?  I don't think so.  But it does
require some kind of nonreductive element---some kind of higher-level
description that cannot be reduced to a lower-level one.  Here's the
way I've been putting it in my last three lectures, when talking more
particularly about quantum mechanics from the Bayesian perspective. I
point out that a Bayesian is, roughly speaking, someone who believes
that without gamblers, there cannot be probabilities. Probabilities
are not external to gamblers.  Then someone always asks, must you
have consciousness to have probabilities?  And I say, ``No.''  Take
as an example my laptop computer loaded with a Bayesian spam filter.
It is a perfectly good gambler in the Bayesian sense, but I think
most people would be hard-pressed to call it conscious. Similarly, I
think we're going to ultimately learn an analogous lesson about all
this transformation/transmutation/ creation/measurement business.

On the other hand, I do find myself being tickled toward a more
Whiteheadian-like view that, whatever this higher-level description
is, every piece of nature has more or less of it, from people all the
way to stones and atoms.  It's just that you have to be on the inside
of the philosopher's stone to see it.
\eq

\section{10-11-05 \ \ {\it {\Wheeler}'s 20 Questions and Nordheim} \ \ (to B. C. van Fraassen)} \label{vanFraassen8}

Below is {\Wheeler}'s account of the original incident.  It comes from
\bq\noindent
J.~A. {\Wheeler}, ``Bohr, Einstein, and the Strange Lesson of the
Quantum,'' in {\sl Mind in Nature:~Nobel Conference XVII, Gustavus
Adolphus College, St.~Peter, Minnesota}, edited by R.~Q. Elvee
(Harper \& Row, San Francisco, CA, 1982), pp.~1--23, discussion
pp.~23--30, 88--89, 112--113, and 148--149.
\eq

I'll work on digging up that Nordheim / von Neumann reference I told
you about.  I don't think it is the Hilbert / Nordheim / von Neumann
paper that one easily finds on the web, but something later.  But
maybe I'm wrong (and in more than one way).

\bq
What is the difference between a ``participatory'' reality and a
reality that exists ``out there'' independent of the community of
perceivers? An example may illustrate a little of the difference.
Edward Teller and I, and a dozen other guests, were sitting in the
living room of Lothar Nordheim in Durham after dinner. From general
conversation we moved on to the game of twenty questions. One, chosen
as victim, was sent out of the room. The rest of us agreed on some
implausible word like ``brontosaurus.'' Then the victim was let back
into the room. To win, he had to discover the word with no more than
twenty yes/no questions. Otherwise, he lost.  After we had played
several rounds, my turn came and I was sent out. \ldots
\eq

\section{10-11-05 \ \ {\it Nordheim?}\ \ \ (to A. Wilce)} \label{Wilce7}

Could you give me the reference to the Nordheim / von Neumann paper you mentioned in your talk (at Perimeter?\ and maybe {\Vaxjo}?)?  Also can you refresh me on what you actually said about it?  I ask because I was telling Bas van Fraassen today about Nordheim's involvement in Wheeler's famous story about playing the game of twenty questions in reverse.  And I remarked to vF about how it was interesting that someone involved in quantum logic would have been the instigator of that.

Below is the beginning of Wheeler's account of what happened. [See 10-11-05 note ``\myref{vanFraassen8}{{\Wheeler}'s 20 Questions and Nordheim}'' to B. C. van Fraassen.]

\section{10-11-05 \ \ {\it Our Own Rovellian Analysis} \ \ (to B. C. van Fraassen)} \label{vanFraassen9}

I particularly got a lot out of your ``puzzle'' analysis at the end.  There were some points there, particularly Puzzle 1, that I don't think I've thought much about (or maybe any) before.

About the issue of consistency in Wigner's friend scenario, here's how Asher and I put it in our {\sl Physics Today\/} piece:
\bq
Does quantum mechanics apply to the observer? Why would it not? To be quantum mechanical is simply to be amenable to a quantum description. Nothing in principle prevents us from quantizing a colleague, say. Let us examine a concrete example: The observer is Cathy (an experimental physicist) who enters her laboratory and sends a photon through a beam splitter. If one of her detectors is activated, it opens a box containing a piece of cake; the other detector opens a box with a piece of fruit. Cathy's friend Erwin (a
theorist) stays outside the laboratory and computes Cathy's wavefunction. According to him, she is in a 50/50 superposition of states with some cake or some fruit in her stomach. There is nothing wrong with that; this only represents his knowledge of Cathy. She knows better. As soon as one detector was activated, her wavefunction collapsed. Of course, nothing dramatic happened to her.
She just acquired the knowledge of the kind of food she could eat.
Some time later, Erwin peeks into the laboratory: Thereby he acquires new knowledge, and the wavefunction he uses to describe Cathy changes. From this example, it is clear that a wavefunction is only a mathematical expression for evaluating probabilities and depends on the knowledge of whoever is doing the computing.

Cathy's story inevitably raises the issue of reversibility; after all, quantum dynamics is time-symmetric. Can Erwin undo the process if he has {\it not yet\/} observed Cathy? In principle he can, because the only information Erwin possesses is about the consequences of his potential experiments, not about what is ``really there.'' If Erwin has performed no observation, then there is no reason he cannot reverse Cathy's digestion and memories. Of course, for that he would need complete control of all the microscopic degrees of freedom of Cathy and her laboratory, but that is a practical problem, not a fundamental one.
\eq

Of course that's not very satisfactory, but it was a {\sl Physics Today\/} piece.  This is extended and done much better on pages 322, 324(bottom)--325, and 19--23 of my {\sl Notes on a Paulian Idea}.  I like those discussions much better.

Still the whole thing can/should be done much, much better and more formally, and your discussion today inspired me to try do it from my perspective with all i's dotted and t's crossed.  Then it'd be nice to compare whatever comes out to your analysis of Rovelli's.  Certainly the developments are going to be similar, but there may be differences here or there.

\section{14-11-05 \ \ {\it American Private Equity} \ \ (to H. Mabuchi)} \label{Mabuchi12.1}

\bq\noindent
My firm recently spoke with Hideo Mabuchi in the quantum theory group at Cal Tech and David Deutsch at Oxford University among other leading minds, and I thought I would drop you a quick note.  My firm is currently looking to fund companies within the quantum technology sector.  I spend a portion of my day speaking to experimentalists and theorists in the quantum theory looking for applications.
\eq

Is this is a whole new style of crackpot coming at me \ldots\  Or by wild chance---not a chance!---is there something to this?  Is it time to fund our quantum information retreat in the woods of New England?  (Don't forget, you pledged \$100K!!)

\section{14-11-05 \ \ {\it Desperately Seeking Postdoc (Funds)} \ \ (to G. Brassard)} \label{Brassard46}

I'm sure you can guess the pressure Bell Labs is putting on all the researchers here to obtain outside funds for this and that.  Well, they've stepped it up a notch again \ldots\ and this is partially why I'm writing.

I had an idea (they don't come that often).  Can you think of any way we might be able to fund a kind of ``joint postdoc'' between the two of us to work on our qm-foundational program?  If we could do that---in such a way that Lucent could pick up some overhead from the project---something nice might come out of this pressure they're putting on me.  Indeed, I'd love to have a postdoc focused on this sort of stuff, rather than the more mundane issues in quantum information, and could make a promise to you to steer the guy right.  But funding is not my forte!  (As you already know.)

Any ideas?

\section{14-11-05 \ \ {\it Questions, Actions, Answers, \& Consequences} \ \ (to B. C. van Fraassen)} \label{vanFraassen10}

\begin{flushright}
\baselineskip=3pt
\parbox{4.0in}{
\bq
\noindent
But physicists are, at bottom, a naive breed, forever trying to come
   to terms with the `world out there' by methods which, however
   imaginative and refined, involve in essence the same element of
   contact as a well-placed kick.\medskip
\\
\hspace*{\fill} --- Bryce DeWitt and Neill Graham, 1971
\eq
}
\end{flushright}
\begin{flushright}
\baselineskip=3pt
\parbox{4.0in}{
\bq
\noindent
I think that the sickliest notion of physics, even if a student gets
   it, is that it is `the science of masses, molecules, and the ether.'
   And I think that the healthiest notion, even if a student does not
   wholly get it, is that physics is the science of the ways of taking
   hold of bodies and pushing them!\medskip
\\
\hspace*{\fill} --- W. S. Franklin, 1903
\eq
}
\end{flushright}\medskip

Thanks for the note.  In my own case, I didn't get nearly as much
reading and thinking done over the weekend as I had hoped
to---indeed, the time turned out to be a great vacation for the kids,
but not so much for dad.

\bvf
I thought I would after all not follow you in replacing the term
``measurement'', despite all the bad effects old connotations have
had in various discussions.
\evf
Fine.  That's your business.  I just offered my two cents because
Maudlin made one of his impassioned spiels precisely on the point of
what ``measurement'' could mean in the Rovelli context---it seemed
that some clarification was in order, and particularly worthwhile
reiterating the point that if better language were used, quantum
``measurement'' might stop looking so unfamiliar.

\bvf
We need to bracket the old connotations such as that a measurement
result reveals a pre-existing value for the measured observable. But
I think we can do that because:
\evf
Not trying to sway you anymore, but let me comment on a couple of
these.

\bvf
there is a certain kind of retrodictive inference possible
also on the basis of qm measurements.  For a long time the paradigm
was a source preparing a stream of particles in a certain state --
measurements on samples taken from the stream give a good basis for
conclusions about just what state the source was preparing, and these
conclusions can then be used to predict the outcomes of further
measurements made on later samples of the stream
\evf

This is what we in quantum information call quantum-state tomography.
One can indeed think of a quantum measurement outcome as {\it giving
information\/} in the old standard sense in that case and not simply
being the ``unpredictable consequence of one's action.''  But then
``giving information'' is quantified by Shannon's ``mutual
information,'' $I(X,Y)$ and not simply by his entropy function
$H(Y)$. That is, one has two random variables in the game---one
treated classically, namely the ``unknown preparation'' $X$, and the
other one purely quantum mechanical, the result $Y$ of the
measurement interaction.  Those two variables have quite different
roles, and one indeed would not want to think of $X$ as the
``consequence of one's interaction.''  On the other hand, without
making explicit mention of $X$ one has no means for thinking of the
elicitation of $Y$ as giving information about anything at all.
Before seeing the value of $Y$, one can expect to be {\it
surprised\/} to the extent quantified by $H(Y)$, but that's where the
story stops.

[For a more detailed Bayesian-like development of this point, you
might have a look at our paper ``Unknown Quantum States and
Operations, a Bayesian View,'' \quantph{0404156} and some of the
references therein. Particularly the Introduction and Concluding
section might be of some interest to you with regard to the present
discussion.]

The only point I want to make to you with regard to your remark above
is that, for these reasons, I would say it has no bearing on the
issue at debate:  I.e., whether it is better to think of a ``quantum
measurement'' as simply an action with an unforeseeable consequence,
or rather as a kind of ``question-asking'' or
``information-gathering.''  It is tangential.

On a completely different subject,
\bvf
Writers on the subject have emphasized that the main form of
measurement in quantum mechanics has as result the value of the
observable at the end of the measurement -- and that this observable
may not even have had a definite value, let alone the same one,
before.
\evf
your phrase ``MAY NOT even have a definite value'' floated to my
attention.  I guess this floated to my attention because I had
recently read the following in one of the Brukner/Zeilinger papers,
\bq\noindent
     Only in the exceptional case of the qubit in an eigenstate of
     the measurement apparatus the bit value observed reveals a
     property already carried by the qubit.  Yet in general the value
     obtained by the measurement has an element of irreducible
     randomness and therefore cannot be assumed to reveal the bit
     value or even a hidden property of the system existing before
     the measurement is performed.
\eq
I wondered if your ``may not'' referred to effectively the same thing
as their disclaimer at the beginning of this quote.  Maybe it
doesn't. Anyway, the Brukner/Zeilinger disclaimer is a point that
{\Caves}, {\Schack}, and I now definitely reject:  From our view all
measurements are generative of a NON-preexisting property regardless
of the quantum state.  I.e., measurements never reveal ``a property
already carried by the qubit.''  For this, of course, we have to
adopt a Richard Jeffrey-like analysis of the notion of
``certainty''---i.e., that it too, like any probability assignment,
is a state of mind---or one along (my reading of)
{\Wittgenstein}'s---i.e., that ``certainty is a tone of voice''---to
make it all make sense, but so be it.

I'm curious to understand whether Rovelli's writings already specify
an opinion on the issue.

\section{14-11-05 \ \ {\it Nordheim?, 2}\ \ \ (to A. Wilce)} \label{Wilce8}

Thanks for the reference, and also for bringing Kalmbach's book to my attention.  Nicely enough, we have the latter here in our library, and I just had a cursory look at it.  From it, I could tell that von Neumann had at least one paper where he mentioned a logical approach to quantum mechanics (a 1936 paper) before the Birkhoff--vN paper, but do you know if that is the earliest mention?  Also I found similar information in this paper by Redei, ``Why John von Neumann did not Like the Hilbert Space Formalism \ldots'' which can be found on the web.  Redei, in fact, quotes a November 1935 letter from vN to Birkhoff which seems to even indicate that vN had some understanding of a Gleason-like result even at that time.  Do you know anything about this?

\section{14-11-05 \ \ {\it Nordheim Again} \ \ (to B. C. van Fraassen)} \label{vanFraassen11}

Following up.  The paper Alex Wilce talked about was apparently,
\bq\noindent
Hilbert, D., von Neumann, J., and Nordheim, L., ``Uber die Grundlagen
der Quantenmechanik,'' Math.\ Annalen {\bf 98} (1927) 1--30.
\eq
It looks now, however, like I was wrong about these guys talking
about the possibility of a quantum logical approach.  Instead what
this paper seems to be famous for is simply giving some guidelines
for what it would mean to axiomatize quantum theory.  Von Neumann, as
best I can tell, only first talked about lattices (in the quantum
context) in a 1935 letter to Birkhoff and first in print in a 1936
paper (by himself, not the one with Birkhoff).

Sorry for feeding you some misinformation.

\section{14-11-05 \ \ {\it Fierz Translator} \ \ (to H. Atmanspacher)} \label{Atmanspacher6}

I hope you are OK, and that your back is better.  We definitely missed you in Konstanz.

Recently I've met a very interesting man, and I want to tell you about him.  He is Prof.\ Hans C. von Baeyer at the College of William and Mary here in the United States.  The last several years his interest has been in writing popularizations of science, and he recently won a couple of prestigious awards in that regard; have a look at \myurl{http://www.aip.org/aip/awards/gemawd.html} and
\hspace*{1pt}\myurl[http://www.wm.edu/as/physics/news/von-baeyer-receives-american-institute-of-physics-andrew-gemant-award.php]{http://www.wm.edu/as/physics/news/von-baeyer-receives-american-institute-of-physi\\ cs-andrew-gemant-award.php}.
I myself have read his latest book {\sl Information:\ The New Language of Science\/} and was quite pleased with its presentation.

Anyway, the main reason I write you is that it turns out that his godfather was Markus Fierz's brother, and I have gotten him interested in translating the Pauli--Fierz foundational/philosophical correspondence into English.  I think he would be great for the task---as he has personal reasons because of his relationship, he is fluent in English, German, and French, he knows quite a bit about our ``informational turn'' in quantum foundations, and he is looking for an ``easy'' project to fill the cracks of his larger writing projects (as he goes into retirement this coming year).

The main questions on his mind are 1) whether there is anyone else taking on a project of this nature, 2) who might be a good publisher for this kind of material, and 3) what are the best, most complete sources of this correspondence.  I told him you that would be the one most likely to have answers, and I told him something of your institute.  I'd like to know the answers too.

If you are interested, I could put you two in contact.  Or I could pass on any information you tell me to.

\section{15-11-05 \ \ {\it Canned Answers} \ \ (to B. C. van Fraassen)} \label{vanFraassen12}

\begin{flushright}
\baselineskip=3pt
\parbox{4.5in}{
\bq
\noindent
One does not infer how things are from one's own certainty.

Certainty is as it were a tone of voice in which one declares
     how things are, but one does not infer from the tone of voice that
     one is justified.\medskip
\\
\hspace*{\fill} --- L. {\Wittgenstein}, {\sl On Certainty}, par.\ 30
\eq
}
\end{flushright}
\begin{flushright}
\baselineskip=3pt
\parbox{4.5in}{
\bq
\noindent
But de Finetti's philosophical view does see determinism as a
     state of mind masquerading as a state of nature, and sees
     causality as a fancied magical projection into nature of our own
     patterns of expectation.  Beneath the mask of determinism is a
     state of mind---certainty---that is intelligible enough \ldots

Certainties are not the only states of mind that are made to
     masquerade as states of nature. Probabilistic previsions are also
     magically projected into nature, to produce ``real'' probabilities,
     i.e.\ ``objective chances'' or probabilistic ``propensities'', which
     de Finetti rejects as firmly as he does deterministic causality.
\medskip
\\
\hspace*{\fill} --- R. Jeffrey, ``De Finetti's Radical Probabilism''
\eq
}
\end{flushright}\medskip

Addressing your questions:
\bvf
Suppose that an observer assigns eigenstate $|a\rangle$ of $A$ to a
system on the basis of a measurement, then predicts with certainty
that an immediate further measurement of $A$ will yield value $a$,
and then makes that second measurement and finds $a$.  Don't you even
want to say that the second measurement just showed to this observer,
as was expected, the value that $A$ already had?  He does not need to
change his subjective probabilities at all in response to the 2nd
measurement outcome, does he?
\evf

It is not going to be easy, because this in fact is what {\Schack} and I
are actually writing a whole paper about at the moment---this point
has been the most controversial thing (with the {\Mermin}, Unruh,
Wootters, {\Spekkens}, etc., crowd) that we've said in a while, and it
seems that it's going to require a whole paper to do the point
justice. But I'll still try to give you the skinny of it:
\begin{itemize}
\item
   Q) He does not need to change his subjective probabilities at all
      in response to the 2nd measurement outcome, does he?
\item
   A) No he doesn't.
\item
   Q) Don't you even want to say that the second measurement just
      showed to this observer, as was expected, the value that A
      already had?
\item
   A) No I don't.
\end{itemize}

The problem is one of the very consistency of the subjective point of
view of quantum states.  The task we set before ourselves is to
completely sever any supposed connections between quantum states and
the actual, existent physical properties of the quantum system.  It
is only from this---if it can be done, and of course we try to argue
it can be done---that we get any ``interpretive traction'' (as Chris
{\Timpson} likes to say) for the various problems that plague QM.  (In
that regard, you might look at {\Timpson}'s explanation of the point in
the Envoi at the end of his thesis, starting page 223, in connection
with his discussion of the problem of the ``factivity of knowledge''
on pages 176--182, particularly footnote 4.)

This may boil down to a difference between the Rovellian and the
Bayesian/Paulian approach; I'm not clear on that yet.  I'm looking at
the first box on page 3 of your last week's handout at the moment.
Rovelli relativizes the states to the observer, even the pure states,
and with that---through the eigenstate-eigenvalue link---, YOU SAY,
the values of the observables.  I'm not completely sure what that
means in Rovelli-world yet, however.

I, on the other hand, do know that I would say that a measurement
intervention is always generative of a new fact in the world,
whatever the measurer's quantum state for the system.  If the
measurer's state for the system HAPPENS to be an eigenstate of the
Hermitian operator describing the measurement intervention, then the
measurer will be confident, CERTAIN even, of the consequence of the
measurement intervention he is about to perform.  But that CERTAINTY
is in the sense of Jeffrey and {\Wittgenstein} above---it is a ``tone of
voice'' of utter confidence.  The world could still, as a point of
principle, smite the measurer down by giving him a consequence that
he predicted to be impossible.  In a traditional development---with
ties to a correspondence theory of truth---we would then say, ``Well,
that proves the measurer was wrong with his quantum state assignment.
He was wrong before he ever went through the motions of the
measurement.''  But as you've gathered, I'm not about traditional
developments.  Instead I would say, ``Even from my view there is a
sense in which the measurer's quantum state is WRONG.  But it is MADE
WRONG by the ACTUAL consequence of the intervention---it is made
wrong on the fly; its wrongness was not determined beforehand.'' And
that seems to be the main point of contention.

I think I say some of this better, and give better argumentation for
it, in the attached document, but I'll let you be the judge of that.
Particularly, I hope the long de Finetti quote helps here.  The file
is {\tt Certainty.pdf}.

I think I have more to say in a positive vein on Rovelli, but I'll
come back to that after lunch.

\section{15-11-05 \ \ {\it February}\ \ \ (to A. Wilce)} \label{Wilce9}

Actually, if the slot is available, why don't we see if we can do it sometime between Feb 8 and 15 (inclusive) if you have a speaking date available then.

I would talk about symmetric informationally complete POVMs---what's known about them and what's not.

For our own personal pleasure, I have some new ideas about how working with them may help us out on our tensor-product derivation problem.  Hopefully I'll have something worked on the qubit and qutrit cases by then, so that we might be able to think about what to do in the general-dimensional case.

For my talk, I'll try to follow Hans C. von Baeyer's ``sawtooth model'':  plotting difficulty vs time, it should look like a saw blade, lots of peaks, lots of troughs.  The idea is to give everyone a little something, whatever their level of sophistication, and in periodic doses.  (By the way, I just met von Baeyer at William and Mary, when I gave the physics colloquium there.)  I'll certainly keep in mind the undergraduate side of the audience, and I'll ask you more about what that really means as the time draws near!

\section{16-11-05 \ \ {\it Your Phrase} \ \ (to B. C. van Fraassen)} \label{vanFraassen13}

What was that thing you said about torture in your last lecture?  It was in the context of measurement and Rovelli.  I want to get it right.  If I'm recalling it right, it might serve a purpose even in my own account of Wigner's friend.

\subsection{Bas's Reply}

\bq\noindent
I said that we also should not expect that, if we put someone to the question under torture, we would get an answer that he already had beforehand.
\eq

\section{16-11-05 \ \ {\it Phone Message and Relationalism} \ \ (to H. Barnum)} \label{Barnum19}

On Wednesdays, I go to Princeton to a) partly to get away, but b) partly to attend Halvorson and van Fraassen's seminar.  The last couple of sessions have really turned me on to van Fraassen.  He's been discussing Rovelli and potential relational interpretations of QM.  The part that I'm finding useful is that he is somewhat convincing me that, if one puts the interpretation of probability on the side for a moment, then there may not be that large of a gulf between our views (at least in any nonmetaphysical way).  This has taken me a little by surprise, and presently I'm trying to digest it all.

\section{16-11-05 \ \ {\it Red} \ \ (to R. W. {\Spekkens})} \label{Spekkens35}

\brws
Is ``red'' ontic or epistemic?  On the one hand, it refers to
a certain sensation which is certainly subjective and on the other it
refers to a particular wavelength of light, which is certainly
objective.  What good does it do us to argue about this?  It's much
better to try and articulate the precise sense in which colour vision
involves an interplay between the objective and the subjective.
Similarly for quantum theory.
\erws

Well put.

\section{17-11-05 \ \ {\it Dutch Meeting}\ \ \ (to A. Y. Khrennikov)} \label{Khrennikov13}

By the way, concerning,
\bakh
This is also related to the conference ``Beyond QM'' that we organize
with G. 't Hooft and other people in Netherlands, 29 May -- 2 June, so
we plan that a group of people will move from Netherlands to Sweden.
\eakh
can you tell me more about the meeting?  I might be interested in coming there too if possible.  Where will it be?  Who else will be there?  Is there a website?  \ldots\ And most importantly, is there any chance I could get invited?  (Even the Bayesian program I promote hopes to one day go ``beyond QM''!!  In fact, that's the whole aim.)

\section{17-11-05 \ \ {\it Rovelli Again} \ \ (to R. W. {\Spekkens})} \label{Spekkens36}

\brws
I understood that the comparison was between you and Rovelli.  In fact, I may even have mentioned this to you myself at some point.  I have a standard overhead listing a variety of papers that defend the ``epistemic view of quantum states'', and Rovelli's paper is on it.  Of course, it's not clear to me that he would agree with this characterization of his work, but the techniques are certainly similar.  He even has something akin to the knowledge-balance principle.  Others have tried to leverage what he's done into an axiomatic derivation of the formalism of quantum theory, but they have yet to be free of a few highly mathematical axioms.  This work tends to be in the operational or quantum logic tradition.
\erws

Can you give me references to some of those?  My frank opinion is that Rovelli really didn't even get off the ground for that part of the project.  (Much of the rest of the paper I like though.)  I am starting to think there is a deeper connection between Rovelli's ``even measurement outcomes are personal, i.e., relational'' and my ``any time two things get together, something new happens---quantum measurement and quantum theory just happens to concern only the cases when one of those things is me'' (i.e., F-theory), as van Fraassen has been trying to convince me.  But we'll see.

\section{17-11-05 \ \ {\it Cash Value} \ \ (to H. Halvorson)} \label{Halvorson5}

I probably laid too many cards on the table yesterday when you
attempted to elicit my opinion on Bohmian mechanics.  For, one of the
arguments I gave was essentially one I had made in a referee report
some time ago. Still it's always hard to keep these kinds of things
secret---they generally so infect one's way of life that surely
everyone can already guess the perpetrator.

Anyway, in principle, I am not a priori against the introduction of
extra, unobservable entities into a theory (over and above, say, its
rawest statement).  But to make me care at all about them, they have
to have a ``cash value.''  This is why I gave the example of
electromagnetic potentials.  In classical electromagnetism, they too
are unobservable; only the fields themselves are observable.  But
often it is quite useful to explicitly introduce the potentials and
work in terms of them when solving a problem (if one wishes, one can
even believe they're as real as real can be)---their facilitating my
problem solving is their cash value and my reason for taking them
seriously.

My argument is in the same vein when it comes to Bohmian mechanics.

Who knows, the Bohmians might one day indeed win out with me by
giving me a good dollop of cash value.  But I'm not banking on it.
And life is too short to pursue every wacky theory {\it simply\/}
because it is consistent.  One has to follow one's instincts, and my
instinct is that our biggest confusions with quantum theory come
about because the only way we know how to word it presently is
through the use of {\it too many\/} structures, not too few.  That
is, the number of distinct concepts in the axioms needs trimming, not
fattening.

But I don't say anything that I haven't already said a thousand
times.  And in your case, I'm preaching to the choir anyway!

\bq
The Bohmians are nothing if not an honest and faithful bunch.  They are faithful to a theory that, though it has more structure than standard quantum mechanics (i.e., some extra elements over and above it), nevertheless---at least they strive to show rigorously---is empirically equivalent to quantum theory in every way.  I am reminded of the secularist who complained to the priest that this is a godless world.  The priest replied that the world couldn't possibly make sense without a god suffused through all its parts:  The world only {\it looks\/} godless, but it only makes {\it sense\/} with a god.  The priest recommended that the secularist supplement his daily study of the newspaper with a daily reading of the Bible.  The difference between the Bohmians and the priest is that they have an equation to supplement standard quantum mechanics, whereas the priest only has unverifiable, humanly constructed words to supplement the daily events.  As far as this reviewer can tell, that's the only difference.

Is that difference enough to make Bohmian mechanics a science?  It is a structure within mathematics, by construction, and maybe one more or less worthy of study.  What will decide its support there are the whims and tastes of mathematicians, whose culture it is to explore various logical structures for reasons more detached than those of the physicist.  But is Bohmian mechanics mathematical physics?  Or is it more akin to an equationified religion?  This reviewer cannot help but feel that the answer is ``no'' to the former and ``yes'' to the latter.

There are so many tough and important problems in basic quantum mechanics waiting to be solved.  For instance, for two independent quantum channels (i.e., the tensor product of two trace-preserving completely-positive linear maps, say on finite dimensional Hilbert spaces), is the quantum entropy of the output minimized by an entangled or unentangled input quantum state in general?  No one knows the answer, but people have been working on the problem for almost 10 years now.  The standing conjecture is that it is an unentangled input that minimizes the entropy, but no one has proved it yet, nor has anyone found a counterexample.  Work like that, I would say, is an example of mathematical physics with respect to basic quantum mechanics. If a problem like this can be posed without the burden of the extra structure of Bohmian mechanics, one has to wonder whether adding the extra structure could possibly help in its solution. Maybe it can. But my attitude is, ``Show me.'' If the Bohmians can be the first to the finish line for one or more problems like the one above---i.e., one posed independently of Bohmian considerations, but badly in need of its guidance---I will certainly rethink my stance. It would be a lesson for me, and I can already foresee that I would come to think of myself as misguided as Mach was when he refused to take atoms seriously because of their unobservability.  But, show me.  And I will be a very different reviewer the next time I am asked to judge a Bohmian research proposal.
\eq

\section{19-11-05 \ \ {\it The Pleasures of Checking Consistency} \ \ (to H. Halvorson \& B. C. van Fraassen)} \label{vanFraassen13.1} \label{Halvorson6}

I'm having my Saturday coffee, reviewing last week's seminar in my head, and every now and then looking up to this week's episode of {\sl Ask This Old House}.  Good combination!

Anyway, thinking about the point I tried to make to Hans about the utility of checking for consistency of epistemic or instrumentalist views of the quantum state, here's the way I put it in \quantph{0104088v1}:
\bq
    What is a quantum state?  Since the earliest days of quantum
    theory, the predominant answer has been that the quantum state
    is a representation of the observer's knowledge of a system.
    In and of itself, the quantum state has no objective reality.
    The authors hold this information-based view quite firmly.
    Despite its association with the founders of quantum theory,
    however, holding this view does not require a concomitant belief
    that there is nothing left to learn in quantum foundations. It
    is quite the opposite in fact: Only by pursuing a promising, but
    incomplete program can one hope to learn something of lasting
    value. Challenges to the information-based view arise regularly,
    and dealing with these challenges builds an understanding and a
    problem-solving agility that reading and rereading the founders
    can never engender. With each challenge
    successfully resolved, one walks away with a deeper sense of the
    physical content of quantum theory and a growing confidence for
    tackling questions of its interpretation and applicability.
    Questions as fundamental and distinct as ``Will a nonlinear
    extension of quantum mechanics be needed to quantize gravity?''\
    and ``Which physical resources actually make quantum computation
    efficient?''\  start to feel tractable (and even connected) from
    this perspective.

    In this paper, we tackle an understanding-building exercise very
    much in the spirit of these remarks.  It is motivated by an
    apparent conundrum arising from quantum information theory.  The
    issue is that of the {\it unknown\/} quantum state.
\eq
Indeed I'm getting quite a bit out of this exercise of going over Rovelli.

\section{21-11-05 \ \ {\it Dutch Meeting, 2}\ \ \ (to A. Y. Khrennikov)} \label{Khrennikov14}

\bakh
This is very preliminary information about coming conference ``Beyond
QM'', it will be one of ``Lorentz workshops.'' In fact, I do not know
yet where in the Netherlands this series of workshops will take place. The
main strategy for selection is that one should really go BEYOND QM.
So people who just work on foundations of QM, but think that QM is the
final theory are not welcome. I already tried to discuss a few names
from ``{\Vaxjo}-conferences team'', but they were rejected on the basis
that there is nothing ``beyond.'' So this is the situation. Among
organisers there are G. 't Hooft, Theo (he was in {\Vaxjo}), I, two
people from SED community (they also were in {\Vaxjo}, Ana and Lois) and
some people from France (Balian and one two more names).
\eakh

If ``going beyond quantum mechanics'' in their sense means the unimaginitive move of {\it going back\/} to a hidden variable theory, then you're right, the meeting is not for me.  If there were instead some room for exploring other ideas---say, the idea that quantum mechanics is {\it only our first hint\/} of a fantastic new kind of reality (one even whose laws of physics can evolve like Darwinian species)---then I might get excited.  But looking at the list of the other organizers outside of yourself, it looks to me to be a more closed-minded, less-creative lot.  I'd probably stick out like a sore thumb.

Attached is a pseudo-paper of mine, that I might polish into a contribution for this year's {\Vaxjo} proceedings if you would like.  Section 7 expands on what I was saying in the last paragraph.

\section{21-11-05 \ \ {\it Pull!}\ \ \ (to R. {\Schack})} \label{Schack87}

I remember when I was a kid in Texas, when one went skeet shooting, the command for sending the clay pigeon into the air was ``Pull!''

Pull!

It'll be a busy month at Orange House Hotel (first Valente next week, then you the following one, and Timpson the week following that), but so much the better for quantum mechanics.

I think I now see the Wigner-friend issue with complete clarity.  It is not so much the subjectivity of unitaries or the definition of a man, but {\bf a)} things-happen-ism, + {\bf b)} extreme personalism (i.e., not only personal probabilities, but personal outcomes for quantum measurements), + {\bf c)} the analogy between quantum measurements and torture.  Ingredients {\bf a)} and {\bf c)} are somewhat related to your ideas, so this is not completely independent work, but I think I finally see it all in a total package.

I know you're sick of reading ``\myref{Mermin101}{Me, Me, Me},'' ``\myref{Mabuchi12}{Preamble},'' and ``\myref{Musser11}{B}'' in {\sl Delirium Quantum}, but with hindsight, it really was already there.  I think that was what was trying to come out of me the day we were watering the lawn.  So, I'm going to send you that document again.  Also, along with it, I'm going to send several notes I sent Bas van Fraassen last week to help bolster where I'm coming from.  They give something of the flow.

Later in the week, I hope to write up a non-poetic version of the whole thing (particular to Wigner) for you.

\section{21-11-05 \ \ {\it Notes to van Fraassen} \ \ (to R. {\Schack})} \label{Schack88}

[[This refers to my 10-11-05 notes ``\,`Action' instead of `Measurement'\,'' and ``Our Own Rovellian Analysis'' to Bas van Fraassen.]]

One final point, not in these notes but crucial:  I had an epiphany of sorts last week.  Von Neumann's description of quantum measurement (and consequently the root of the Wigner conundrum) is nothing more than the ``correspondence theory of truth'' (that William {\James} always assaults) suitably formalized to the quantum situation.  What I learned from Rovelli's paper (or vF's discussion of it, I don't know which) was how to break the instinct to formalize things that way.

\section{21-11-05 \ \ {\it Seeing the Light} \ \ (to R. {\Schack})} \label{Schack89}

\brs
It's the subjectivity of the sample space, stupid!

I am not sure this is what you are getting at, but your brief lines
on Wigner's friend suddenly made me see what could be the light. The
friend's sample space is {\rm \{pointer to the left, pointer to the right\}}. Wigner's sample space is different. By contemplating the superposition
$|\mbox{\rm friend sees left}\rangle+|\mbox{\rm friend sees right}\rangle$, Wigner does not treat ``pointer
to the left'' and ``pointer to the right'' as measurement outcomes. In
order to give a proper analysis of the so-called paradox, we must
first know Wigner's sample space. My mistake was to identify the
``left'' in Wigner's superposition with the ``left'' in the friend's set of outcomes. They are not the same thing. The latter is what the
friend sees. The former is a label in an abstract Hilbert space.
Wigner and his friend are not talking about the same quantum
measurement. There is no paradox.
\ers

Yep, that's essentially it.  The only thing further is that with all this ``Me, Me, Me'' business, one gives a gloss for why the sample spaces are different:  The sample spaces are tied to the agents by definition.

It's really nothing that Asher and I didn't already say in our {\sl Physics Today\/} thing that I quoted for Bas (and sent to you in the compilation).  It's just that now I feel much more comfortable with what I would do if confronted aggressively with a question like, ``Do you or do you not think a human intervention is necessary for a quantum measurement to come about?  And related, do you or do you not think things happen independently of human intervention?''

Reversing the friend's wavefunction may change the story Wigner will coerce out of him (thus the bit about torture), but it will not change what has happened in the world.  They are two different things.  Quantum mechanics is not about what happens in the world; it is only about what might come about from a ME's interventions into it---and that is just a very small part of reality.

\section{21-11-05 \ \ {\it Update on {\Schack} and Wigner} \ \ (to C. G. {\Timpson})} \label{Timpson9}

Let me give you an update on what {\Schack} is up to.  He is, after all, coming to New Jersey, but the dates he finally settled on are December 3--8.  Thus, if during your trip, you come by the Orange House Hotel, I'll have you all to myself.

By the way, I'm now also completely prepared for your visit too.  That is:  I'm not completely sure which troubles you had with a quantum-Bayesian account of the Wigner's friend issue (though {\Ruediger} gave me something of a sketch), but I am now quite confident that I can handle anything you throw at me.  (See my note to {\Ruediger} below.)  [See 21-11-05 note ``\myref{Schack87}{Pull!}''\ to R. {\Schack}.] Sometimes you have to rework these things in your head a thousand times, but in the end it's definitely all worth it.  Thanks for sewing a seed of doubt in me:  It caused me to strive as I thought I would, and I'm fairly pleased with the result.

We have a lot to talk about when you come.

\section{21-11-05 \ \ {\it AJP Resource Letter} \ \ (to R. H. Stuewer)} \label{Stuewer1}

The Anchorage meeting you mention must be the same one that Hans Christian von Baeyer wrote me about recently, at which he'll be receiving the Gemant Award (I think it was called).  He wrote me this very nice note saying that he would title his talk, ``How I Learned To Stop Worrying About Schroedinger's Cat,'' and said,
\bhcvb
     I intend to chronicle my gradual adoption of your point of view
     about QM. \ldots\
     By the way, the reason I am no longer worried about {\Schroedinger}'s
     cat is not that I think that q.m.\ is now understood, but that I
     feel, for the first time in my career, that the problem is in good
     hands with young folks like you and your friends.  I think that
     where I can make a contribution is in explaining to other
     physicists and to the public that y'all are NOT crazy, that
     progress HAS been made, that new ideas ARE being brought to bear,
     and that there actually IS a hope of realizing Wheeler's sibylline
     predictions.
\ehcvb
I'm so very happy he's giving exposure like that to our quantum foundational program (essentially the stuff I talked about at Seven Pines when you invited me) to such a good audience.  With exposure, we might just get a workforce \ldots\ and then who knows how things would really fly!

\section{22-11-05 \ \ {\it MGM Curiosity} \ \ (to C. M. {\Caves})} \label{Caves80}

\bcc
But the real point is that he never misses an opportunity to badmouth
quantum information, so I'm not inclined to have him putting those
opinions in a letter.
\ecc

What is his trouble with quantum information?

This kind of reminds me of the time I heard Willis Lamb talking about quantum information in the same breath as Bohmian mechanics, GRW theories, and the like.  I recall he ended the diatribe by saying ``quantum teleportation is beneath contempt.''

\section{22-11-05 \ \ {\it Difference Between Classical \& Quantum} \ \ (to C. M. {\Caves} \& R. {\Schack})} \label{Caves81} \label{Schack90}

I enjoyed finding these words in Joel Achenbach's piece in the {\sl Washington Post\/} today:
\bq\noindent
Reality gets fuzzier under closer scrutiny. It must be some kind
of law of physics. (Actually, I think it's called quantum
mechanics.) You would think that abundant evidence, steadily
compiled, would make everything clearer, but the opposite is true. Partial knowledge gives us a simple picture of the world and its
phenomena -- an extension of the ignorance-is-bliss rule. Look
deeper and you wind up scratching your head.
\eq
That's about the only truth I've ever seen in decoherence.

\section{22-11-05 \ \ {\it From the Poetaster} \ \ (to R. {\Schack})} \label{Schack91}

\brs
If a Bayesian spam filter can be an agent, then a quantum computer can act as an agent performing a quantum measurement.

I see two possibilities now: You don't agree with this in some
essential way, or your poetry is in trouble (I am hoping for the
latter).
\ers

Poetry can always be made to work; that's the beauty of it.  If ostensibly it can't, then it must not have been poetry to begin with, but poetastry.

I haven't gotten my head around what you're writing yet.  I'll keep trying (as time permits today).

But come on, do you really believe that a man (and consequently a Bayesian agent) CANNOT be built in a laboratory by the hand of another man?  That is what is at issue.  I believe there is no difference in principle between a natural human body and one that can be put together in the laboratory.  And if there is no difference, then we need a de-anthropocentrized way of putting what an agent is.

Yes, we are all already potential quantum computers.  As the only difference between being a classical or quantum system---as treated in physical theory---is someone's state of mind about it (i.e., one's degrees of belief as captured by their quantum states for it).  I can't accept that that there's any intrinsic difference:  that would be the myth of Zurek's decoherence program.

But I have to go to an annoying meeting at the moment.

\subsection{{\Ruediger}'s Preply}

\bq
\noindent
{\bf Note on Wigner's friend, for Chris from RS.}\medskip

If a Bayesian spam filter can be an agent, then a quantum computer can act as an agent performing a quantum measurement. Consider a joint system of two qubits, and let $s_{\rm prior}$ be a bit string that uses some code to describe the prior quantum state of the joint system. Now perform a projective measurement on the first qubit and let $s_{\rm post}(j)$ ($j=0,1$) be bit strings that encode the post-measurement state of the second qubit, conditioned on the  measurement outcome, $j$. Consider a composite Hilbert space, \begin{equation} {\cal H} = {\cal H}_{\rm qubit1}
   \otimes {\cal H}_{\rm qubit2}
   \otimes {\cal H}_{\rm prior}
   \otimes {\cal H}_{\rm post}
   \otimes {\cal H}_{\rm agent}   \;,
 \end{equation}
 where ${\cal H}_{\rm qubit1}\otimes{\cal H}_{\rm qubit2}$ is the Hilbert  space of the two qubit-system above, ${\cal H}_{\rm prior}$ is a quantum register  large enough to hold the string $s_{\rm prior}$ in the form of a  computational basis state, similarly ${\cal H}_{\rm post}$ for the strings  $s_{\rm post}(j)$, and ${\cal H}_{\rm agent}$ houses the working  space/internal state of the quantum computer/agent.

Let $\ket\phi,\ket{\phi_0},\ket{\phi_1}\in{\cal H}_{\rm agent}$ and $U$ be such that \begin{equation} U\,\Big(\, \ket\psi\otimes\ket{j}\otimes\ket{s_{\rm prior}}\otimes\ket0\otimes
\ket{\phi}\,\Big)
= \ket\psi\otimes\ket{j}
\otimes\ket{s_{\rm prior}}\otimes\ket{s_{\rm post}(j)}\otimes\ket{\phi_j} \end{equation} for any one-qubit state $\ket\psi$ and $j=0,1$.

To repeat, if a spam filter is an agent, then this quantum computer is an agent. It starts from a prior judgement and a sample space, interacts with the system, brings about a measurement outcome and records the post-measurement state in a register.

Is this any different from Wigner's friend? No.
Wigner sets up this device in the initial state \begin{equation} \ket{\Psi_{\rm ini}} = {\frac{1}{\sqrt{2}}}(\ket{00}+\ket{11})\otimes\ket{s_{\rm prior}}\otimes\ket0\otimes \ket{\phi} \;.
\end{equation}
He applies $U$ and writes down a superposition state. Does this mean that the friend is in a funny state? Not at all. Ask him (by looking, e.g., at the register encoding the posterior) and he will give you a simple answer. Reverse him, then ask him, and he will show you a register in the erased state $\ket0$, ready for the measurement.

I see two possibilities now: You don't agree with this in some essential way, or your poetry is in trouble (I am hoping for the latter). Does your poetry (hands in the form of meters extending into the world, the sexual interpretation of quantum mechanics, etc) work for a Bayesian spam filter?
\eq

\section{22-11-05 \ \ {\it The Point:\ Inanimate Object, Wigner, his Friend, or A\-gent --- It Doesn't Matter} \ \ (to R. {\Schack})} \label{Schack92}

\bq
I think the greatest lesson quantum theory holds for us is that when two pieces of the world come together, they give birth.  [Bring two fists together and then open them to imply an explosion.]  They give birth to FACTS in a way not so unlike the romantic notion of
parenthood:  that a child is more than the sum total of her parents, an entity unto herself with untold potential for reshaping the world.
Add a new piece to a puzzle---not to its beginning or end or edges, but somewhere deep in its middle---and all the extant pieces must be rejiggled or recut to make a new, but different, whole.  That is the great lesson.

But quantum mechanics is only a glimpse into this profound feature of nature; it is only a part of the story.  For its focus is exclusively upon a very special case of this phenomenon:  The case where one piece of the world is a highly-developed decision-making agent---an experimentalist---and the other piece is some fraction of the world that captures his attention or interest.
\eq
It's things-happen-ism, man.  But the point is, that shouldn't be confused with what quantum mechanics is about.  QM is only about that other point of view, what-I-happen-to-cause-ism.

\section{22-11-05 \ \ {\it Spoke Too Soon?}\ \ \ (to R. {\Schack})} \label{Schack93}

OK, maybe I spoke too soon when I said we were in essential agreement yesterday after you said, ``It's the subjectivity of the sample space, stupid!''

I looked at your {\LaTeX}ed note over and over today, but couldn't figure out what you were trying to get at.  Do you want to try again?

Was your note from today intended to be in disagreement with what I had written you in reply yesterday?

\section{22-11-05 \ \ {\it Story of Bohr's Office} \ \ (to G. Brassard)} \label{Brassard47}

Below is a story I wrote about my own visit to Bohr's office.  [See 03-02-04 note titled ``\myref{Comer49}{The Land of Bohr}'' to G. L. Comer.]  There is a certain spirit there, isn't there?

\section{22-11-05 \ \ {\it An Amusement} \ \ (to T. Maudlin)} \label{Maudlin1}

\btm
I wrote this little piece earlier in the semester, after there had
been some discussion in Bas and Hans's class about using the classical
notion of information in quantum contexts. I thought you might find
it amusing. I somehow think you will object to something in it, but I
don't know quite what\ldots
\etm

Thanks for sending me your piece.  I gave it a once-over this afternoon, and as my six-year-old likes to say, ``I didn't like it \ldots\ (pause) \ldots\ I loved it!''  I thought it was great; quite creative.  And I was particularly pleased to learn that, ``Dr.\ Psi, like all evil geniuses, is a perfect Bayesian.''

But you were right, I don't agree with everything in the document.  {\it However}, at least at my level of thoroughness for this once-over, I do think I agree with everything in the first 7.5 pages.  That's gotta count for something!  Particularly, I agree with the very first paragraph of the paper, and then the way you laid everything out and unrolled it thereafter in the rest of those 7.5 pages:
\bq
Quantum teleportation is a very interesting physical phenomenon, predicted by quantum theory and confirmed in the lab. There has been some question about what the proper information theoretic analysis of this phenomenon is. In particular, there have been doubts about whether ``classical information theory'' (i.e.\ Shannon's theory) can be applied to such a quantum situation. These worries are misplaced. Shannon's theory is ``classical'' only in the sense that it is a well-established, canonical theory, not in the sense that it applies only to ``classical'' physics (whatever that is). I hope the following little playlet will be instructive.
\eq

Where we start to part company is where you, indirectly, invoke a comparison between pure quantum states and classical phase space points in your discussion of the no-cloning theorem.  I say that's a comparison that shouldn't be made; it's like comparing apples to oranges, Frank Sinatra to M\"otley Cr\"ue, beliefs to facts.  And then we diverge still further with the discussion of nonlocality.

But I really did like everything else!  I remember in high school once writing a paper on Thoreau.  When my teacher returned it with an A$+$, I was really humbled when she furthermore said, ``Chris, you seem to have a feeling for Thoreau!''  Well, from the devious insight and style of conversation you put into this document, I almost want to say, ``Tim, you seem to have a feeling for Dr.\ Psi, the evil genius!''  But maybe that goes too far \ldots

See you tomorrow in class.

\section{22-11-05 \ \ {\it On the Fly} \ \ (to G. Musser)} \label{Musser19}

By the way, I should also tell you about how I have consciously adopted some of your phraseology in my own explanations.  It comes from this thing you once wrote me:
\bgm
     Let me see whether I understand your vision of a world in
     continuous creation. Ernst has bequeathed me his cat; it arrives
     in a box. I open it to find the cat is \ldots\ alive. Before I do, the
     cat is neither alive nor dead; it is not even in a purgatorial
     superposition. In fact, the categories of cat alive and cat dead
     don't even exist yet. When I open the box, I create the fact of a
     live cat. That fact had to come into existence {\bf sometime}. In the
     classical deterministic universe, it came into existence when the
     universe did. All the arbitrariness was in the ICs. The quantum
     universe makes up facts as it goes along. The arbitrariness is
     spread out over time. It's the distinction between my wife, who
     makes plans and carries them out, and me, who decides step by step
     on the fly.
\egm

The main point is the one in the last sentence, that the quantum universe makes up facts on the fly.  I can't tell you how many times I've used that phrase since, but the last example I can think of came last week in a letter to Bas van Fraassen (a philosophy professor at Princeton).

I'll paste it below for your amusement.  [See 15-11-05 note ``\myref{vanFraassen12}{Canned Answers}'' to B. C. van Fraassen.]

\section{23-11-05 \ \ {\it Cathy, the Friend, and the Device-Agent} \ \ (to R. {\Schack})} \label{Schack94}

OK, I start to appreciate what you're trying to get at.

Among other things, you call into question how I think I can make the inference from quantum mechanics (which is about what-I-happen-to-cause-ism) to ``the greatest lesson quantum theory holds for us'' (which is about things-happen-ism).  Is that correct?

That indeed is a good question.  It is something I should unroll a little more and be a little less declarative about.

But I also think your question (or maybe it is simply an independent question) has to do with the consistency of our {\Wittgenstein}ian notion of certainty (i.e., that one can have certainty and still not have it fulfilled).

Let me read and reread your two notes a little more, now that the kids are out of the house!

\section{23-11-05 \ \ {\it The Skinny Answer} \ \ (to R. {\Schack})} \label{Schack95}

\brs
There is no big problem with the following bit of poetry:
\bq\noindent\rm
The role of the quantum system is thus more like that of the
philosopher's stone; it is the catalyst that brings about a
transformation (or transmutation) of the agent.
\eq
The qubit is a catalyst which brings about a transformation of the
device-agent.

But how about the part where the world external to the agent makes one
of the outcomes happen? How does Wigner's description of the device-
agent tie in with the poetic tale that nothing prevents the external
world from smiting the device-agent if it bets its life (let's say,
its power supply) on the outcome? Doesn't Wigner have to conclude that
the device-agent is wrong in thinking that the external world has made
one of the outcomes come about?
\ers

Answering the last question in particular:  No.

You see, if I answered ``yes'' it would be equivalent to saying that it is the external view that is always the right one.  And that leads one down the path of many-world-ism, or at least Everettism.  Instead, the move here is to say, {\it all\/} internal views are equally valid---the device-agent's (who sees an actual outcome following his interaction), Wigner's (who sees an actual outcome following his interaction), Wigner's friend's (who sees \ldots), etc.  When all is said and done, the ``Me, Me, Me'' facts actually are disembodied facts---that's the things-happen-ism part.  But there is no contradiction with quantum mechanics because Wigner's degrees of belief are not degrees of belief about the device-agent's internal view.  His degrees of belief are about what he might coerce out of the device-agent or even the system (or both) if he were to interact with them.

That probably obfuscated things even worse for you.  But give me a reaction, and that will be a useful datum.

\section{23-11-05 \ \ {\it Wigner's Impotence} \ \ (to R. {\Schack})} \label{Schack96}

\brs
To ask it in a yet different way, can Wigner undo the outcome that the
world has made come about?
\ers

No, he cannot.  That is the disconnect between reality and Wigner's degrees of belief (including his certainty).  Wigner can only undo {\it state-vector\/} time evolutions for systems external to himself.

\section{23-11-05 \ \ {\it After-Shower Thought} \ \ (to R. {\Schack})} \label{Schack97}

I'm going to have to pack up and get on the Princeton trek soon, but let me just record one phrase that came to mind while I was in the shower.

\brs
I don't feel, though, that
\bq\noindent\rm
When all is said and done, the ``Me, Me, Me'' facts actually are
disembodied facts---that's the things-happen-ism part.
\eq
gives a satisfactory connection to the things-happen-ism part. What
and who determines the moment at which ``all is said and done''? You
certainly don't want to say it's when Wigner has queried the
device-agent. To repeat, what connects the ``actual outcomes'' or facts that the players see to the things that happen?
\ers

Quantum theory has no more power within it to say when (quantum) events happen than classical probability theory has the power within {\it it\/} to say when its own events happen.  Maybe this is why some developments of classical probability theory are careful to say that the $h$'s in a probability assignment $P(h)$ are propositions.  Propositions don't ``happen'', they are simply either true or false.  And thus, by fiat, one never has to address the question of when $h$ ``happens''.  But we know that we shouldn't try to make that further move (i.e., viewing the $h$'s as propositions with timeless truth values) in the quantum setting.  Quantum events do ``happen.''  But that doesn't mean that quantum theory should have the burden on it of explaining or determining ``the moment at which `all is said and done'.''

This is ultimately, I suspect, why our view will be greeted with the same revulsion that William {\James}'s theory of truth was by Russell and Moore and the like:  Because for {\James}, too, truth is something that is made to happen; it is not something that is just there in a timeless way.
\begin{quote}
The truth of an idea is not a stagnant property inherent in it.
Truth {\it happens\/} to an idea.  It {\it becomes\/} true, is {\it made\/} true by events.  Its verity {\it is\/} in fact an event, a
process:  the process namely of its verifying itself, its veri-{\it fication}.  Its validity is the process of its valid-{\it ation}.
\end{quote}

Have we made any progress?

\section{23-11-05 \ \ {\it Basil}\ \ \ (to A. Y. Khrennikov)} \label{Khrennikov15}

\bakh
I shall read your paper. I hope you are not against if I also give it Basil
Hiley. He is visiting me this week.
\eakh

I meant I might polish into a real paper for you upcoming proceedings, by adding an introduction, etc.  Of course it is fine to share it with Basil.

\bakh
More or less you are right. It is more or less about HV, or more generally about realism as the basis of QM. I think it is the continuation of the debate between Mach and Boltzmann. You are definitely Machian, he would like your new idea about the evolution of laws of nature. But do not forget that Mach rejected Boltzmann's ideas on atoms. There was no experimental evidence of the existence of atoms. Mach considered them as unuseful elements of knowledge. See! But finally Boltzmann was right!
\eakh

It's not true that I'm Machian.  In fact, I just wrote some stuff to Hans Halvorson and Bas van Fraassen the other day saying the opposite.  I'll paste it in below; it clarifies the point (and note the reference to Mach).  [See 17-11-05 note ``\myref{Halvorson5}{Cash Value}'' to H. Halvorson.]  However, it is true that I think that toying with hidden-variable extensions of quantum mechanics is a waste of time.  Trying to make an analogy between Boltzmann's introduction of atoms and Bohm's introduction of trajectories is not a good one.  I think you were much closer to the mark when you were still talking about ``contextual probabilities.''  The story of quantum mechanics, I think, is much more like that.

Read the stuff below.  And don't forget to read Section 7 in the pseudo-paper I sent you; I think---but maybe I'm wrong---it's a position that both you and Basil could respect.  (For it says, in the end, quantum mechanics is incomplete.)

\section{24-11-05 \ \ {\it Howard's Scream!}\ \ \ (to H. Barnum)} \label{Barnum20}

Quoting you:
\bhb
I tend to think that Many-Worlds (despite my having spent a lot of effort in my life playing devil's advocate for it) gets things backwards: stuff happens, we do scientific (i.e.\ some variety of roughly Bayesian in a very general, not necessarily conscious, sense) inference about the stuff that happens, the definite results we experience for measurements, we come up with a theory that systematizes the resulting beliefs (as evidenced by our willingness to bet, in a very generalized sense, on the outcomes of experiments and such). This systematization of our betting behavior faced with experiments can be represented in terms of probabilities given by the Born rule. Rederiving the Born probabilities from a part of the formalism that was cooked up, and especially, further developed and held onto, in part to give a good representation of just these probabilities, seems somewhat backwards. Without the probabilities, and the terrific guidance they give to our actions, who would have bothered with quantum mechanics anyway? I guess one can say that the rederivation is a sophisticated attempt to keep the probabilities and solve other problems that came along with quantum mechanics. But it still raises, for me, a serious problem of: what then of the formal and informal scientific reasoning, based on measurements having definite results, that brought us to the QM formalism and Born rule in the first place? Must we reconstruct it all in terms of Everettian branchings, with never a definite result?
\ehb

Hear, hear!  I like that line of argument.  And I wonder if you can put some real flesh on it.  It'd be nice to derive an honest-to-god contradiction.

But here's maybe something to think about.  It's a report from Heisenberg of something Einstein said to him:
\bq
It is quite wrong to try founding a theory on observable magnitudes
alone.  In reality the very opposite happens.  It is the theory which
decides what we can observe.  You must appreciate that observation is
a very complicated process.  The phenomenon under observation
produces certain events in our measuring apparatus.  As a result,
further processes take place in the apparatus, which eventually and
by complicated paths produce sense impressions and help us to fix the
effects in our consciousness.  Along this whole path---from the
phenomenon to its fixation in our consciousness---we must be able to
tell how nature functions, must know the natural laws at least in
practical terms, before we can claim to have observed anything at
all.  Only theory, that is, knowledge of natural laws, enables us to
deduce the underlying phenomena from our sense impressions.  When we
claim that we can observe something new, we ought really to be saying
that, although we are about to formulate new natural laws that do not
agree with the old ones, we nevertheless assume that the existing
laws---covering the whole path from the phenomenon to our
consciousness---function in such a way that we can rely upon them and
hence speak of ``observation.''
\eq
Particularly, I'm thinking about the part, ``When we claim that we can observe something new, we ought really to be saying that, although we are about to formulate new natural laws that do not agree with the old ones, we nevertheless assume that the existing laws---covering the whole path from the phenomenon to our consciousness---function in such a way that we can rely upon them and hence speak of observation.''

Could it be that this sort of thing helps save the move from ``definite outcome empirical world'' to ``nothing happens at all Everett world'' that the Everettians would like to make?  I.e., that the idea of definite outcomes is just some old-theory scaffolding that eventually falls away after the shining tower of Everett is open for public service?

Happy Thanksgiving, by the way.

\section{28-11-05 \ \ {\it Your Visit} \ \ (to G. Valente)} \label{Valente7}

Thanks again for pushing me about Ramsey.  I've been reading on and off about him over the holiday weekend, and I've learned much in the process---it's given me a lot of food for thought.

Next project, your paper!

\section{28-11-05 \ \ {\it Snowbird Hotel?}\ \ \ (to C. M. {\Caves})} \label{Caves81.1}

{\Ruediger} shows up Saturday for another week of wrangling.  You'd hate the idea of it, but how I think it'd be wonderful if we could open up a shop up here, Institute for Bayesian Quantum Information.  (Anywhere, of course, would do, but I'm turning into a sedentary lot.)  While you're talking to David Gross, why don't you try to turn him on to an idea like that!

\section{29-11-05 \ \ {\it Abstract for APS March Meeting -- Trial 1} \ \ (to myself)} \label{FuchsC12}

\bq\noindent
Title: \\ Quantum Mechanics in Terms of Symmetric Measurements\medskip\\
Abstract: \\ In the neo-Bayesian view of quantum mechanics that Appleby, Caves, Pitowsky, Schack, the author, and others are developing, quantum states are taken to be compendia of partial beliefs about potential measurement outcomes, rather than objective properties of quantum systems.  Different observers may validly have different quantum states for a single system, and the ultimate origin of each individual state assignment is taken to be unanalyzable within physical theory---its origin, instead, ultimately comes from probability assignments made at stages of physical investigation or laboratory practice previous to quantum theory.  The objective content of quantum mechanics (i.e., the part making no reference to observers) thus resides somewhere else than in the quantum state, and various ideas for where that ``somewhere else'' is are presently under debate---there are adherents to the idea that it is purely in the ``measurement clicks,'' there are adherents to the idea that it is in intrinsic, observer-independent Hamiltonians, there are adherents to the idea that it is in the normative rules quantum theory supplies for updating quantum states, and so on.  This part of the program is an active area of investigation; what is overwhelmingly agreed upon is only the opening statement.  Still, quantum states are not simply Bayesian probability assignments themselves, and different representations of the theory (in terms of state vectors or Wigner functions or $C^*$-algebras and the like) can take one further from or closer to a Bayesian point of view.  It is thus worthwhile spending some time thinking about which representation might be the most propitious for the point of view and might, in turn, carry us the most quickly toward solutions of some of the open problems.  In this talk, I will present several results regarding a representation of quantum mechanics in terms of symmetric bases of positive-semidefinite operators.  I will also argue why this is probably the most natural representation for developing a Bayesian-style interpretation of quantum mechanics.
\eq

\section{30-11-05 \ \ {\it TGQI Nomination Votes \ldots} \ \ (to H. Mabuchi)} \label{Mabuchi13}

Sorry I dropped out of sight almost as soon as you last heard from me:  Mr.\ Giovanni Valente arrived for his visit and has kept me fascinated with all kinds of ``noncommutative probability'' ever since.  But I'll forward on my picks in a couple of minutes, while he takes a breather from our debate.

\section{01-12-05 \ \ {\it Focus Session Quantum Foundations} \ \ (to R. W. {\Spekkens})} \label{Spekkens37}

OK, I did it.  Thanks for keeping me abreast of this.

Below is the abstract I originally wrote up before the submissions page made me trim it.  (I include it here so that I have a record of it.) \medskip

\noindent {\bf Title:} Quantum Mechanics in Terms of Symmetric Measurements\medskip

\noindent {\bf Abstract:}  [See report of it in 01-12-05 note ``\myref{Appleby9}{The Swedish Bayesian Team},'' and add these words to its end:] In this talk, I will present several results regarding a representation of quantum mechanics in terms of symmetric bases of positive-semidefinite operators.  I will also argue why this is probably the most natural representation for developing a Bayesian-style interpretation of quantum mechanics.

\section{01-12-05 \ \ {\it The Swedish Bayesian Team?}\ \ \ (to D. M. {\Appleby} \& Others)} \label{Appleby9}

I'm writing this note to see if you might be available to come to a
meeting in {\Vaxjo}, Sweden, June 4--9, 2006?  The general meeting is
{\sl Foundations of Probability and Physics -- 4\/} organized
predominantly by Andrei Khrennikov along the lines of his previous
meetings, but he has given me a small budget to contribute toward
trying to attract a group of quantum-Bayesians.  If this meeting goes
like the ones I've been associated with before, it will turn out to
be a pleasant and productive time for the group I hope to attract
there.  (It is a nice village with a couple of bars and a nice lake
to walk around and discuss things.)

Ideally, I'd like to have you, Itamar Pitowsky, Richard Gill, Matt
Leifer, Ariel Caticha, and {\Ruediger} {\Schack} there to contribute
toward a discussion.  [\ldots]

The sort of idea I have in mind for a focused discussion can be read
in some of the words I used in a recent abstract I wrote for my talk
at the APS March meeting.  I'll place those words below to give you a
kind of feeling.

\bq
\noindent Opening from Fuchs's APS abstract:

In the neo-Bayesian view of quantum mechanics that {\Appleby}, {\Caves},
Pitowsky, {\Schack}, the author, and others are developing, quantum
states are taken to be compendia of partial beliefs about potential
measurement outcomes, rather than objective properties of quantum
systems. Different observers may validly have different quantum
states for a single system, and the ultimate origin of each
individual state assignment is taken to be unanalyzable within
physical theory---its origin, instead, ultimately comes from
probability assignments made at stages of physical investigation or
laboratory practice previous to quantum theory.  The objective
content of quantum mechanics (i.e., the part making no reference to
observers) thus resides somewhere else than in the quantum state, and
various ideas for where that ``somewhere else'' is are presently
under debate---there are adherents to the idea that it is purely in
the ``measurement clicks,'' there are adherents to the idea that it
is in intrinsic, observer-independent Hamiltonians, there are
adherents to the idea that it is in the normative rules quantum
theory supplies for updating quantum states, and so on.  This part of
the program is an active area of investigation; what is
overwhelmingly agreed upon is only the opening statement.  Still,
quantum states are not simply Bayesian probability assignments
themselves, and different representations of the theory (in terms of
state vectors or Wigner functions or $C^*$-algebras and the like) can
take one further from or closer to a Bayesian point of view.  It is
thus worthwhile spending some time thinking about which
representation might be the most propitious for the point of view and
might, in turn, carry us the most quickly toward solutions of some of
the open problems.
\eq

\section{01-12-05 \ \ {\it State Spaces} \ \ (to H. Halvorson)} \label{Halvorson7}

Luckily, our library had the right volume of Alfsen and Shultz, and I checked it out.  This new kind of characterization of complete positivity has me quite intrigued:  Now if I can only translate the stuff I find in this book to finite dimensional language so I can see what it means!

What were the additional references you were going to give me on the relation between convex (state-space) structures and algebraic properties?

By the way, in times past, I've had the hope of deriving the idea of complete positivity from a Gleason-like theorem for the joint probabilities over measurement outcomes for measurements on a single system at two separate times.  I've had some concrete ideas about how to articulate that (and a good load of calculations), but I never could quite get a theorem to work.  If you'd like to throw in on a project like this, maybe we could discuss this too in the coming weeks.

\section{01-12-05 \ \ {\it Finite Sets (then limits) Policy} \ \ (to H. Halvorson)} \label{Halvorson8}

Also, here's the quote from Ed Jaynes that I almost read aloud in
your seminar yesterday.  I suspect nothing deeper than this is going
on with Redei's criticism of {\it whatever he is imagining is\/} the
Bayesian approach.  (I.e., there are many different approaches that
one might think of as more or less Bayesian, and whichever one Redei
is attacking, I doubt his problems are of any origin other than what
Jaynes expresses below.)

\bq
It is very important to note that [the theorems we have just proven]
have been established only for probabilities assigned on finite sets
of propositions. In principle, every problem must start with such
finite sets of probabilities; extension to infinite sets is permitted
only when this is the result of a well-defined and well-behaved
limiting process from a finite set. More generally, in any
mathematical operations involving infinite sets the safe procedure is
the finite sets policy:
\bq\noindent
     Apply the ordinary processes of arithmetic and analysis only to
     expressions with a finite number of terms. Then after the
     calculation is done, observe how the resulting finite expressions
     behave as the number of terms increases indefinitely.
\eq

In laying down this rule of conduct, we are only following the policy
that mathematicians from Archimedes to Gauss have considered clearly
necessary for nonsense avoidance in all of mathematics. But more
recently, the popularity of infinite set theory and measure theory
have led some to disregard it and seek short-cuts which purport to
use measure theory directly. Note, however, that this rule of conduct
is consistent with the original Lebesgue definition of measure, and
when a well-behaved limit exists it leads us automatically to correct
``measure theoretic'' results. Indeed, this is how Lebesgue found his
first results.

The danger is that the present measure theory notation presupposes
the infinite limit already accomplished, but contains no symbol
indicating which limiting process was used. Yet as noted in our
Preface, different limiting processes---equally well-behaved---lead
in general to different results. When there is no well-behaved limit,
any attempt to go directly to the limit can result in nonsense, the
cause of which cannot be seen as long as one looks only at the limit,
and not at the limiting process.
\eq

\section{02-12-05 \ \ {\it Princeton Quantum Informatics Conf} \ \ (to M. O. Scully)} \label{Scully7}

I'd be happy to help you organize a Princeton Quantum Informatics conference, particularly if it will involve in some part a tribute to John Wheeler (as you had mentioned previously).  Anyway, whichever way, I have a lot of friends in that community, and I think we could make a good meeting \ldots\ if I didn't try too many of their patiences with my quantum Bayesian conference in Konstanz last year!

As it turns out, I'll be in Princeton December 14 to attend Chris Timpson's lecture at Bas van Fraassen's seminar.  (Then Timpson will come back with me to Bell Labs for a couple of day visit.)  How about we get together then for a little planning session:  You can tell me much more about what you're thinking and what your constraints are, etc.

\section{02-12-05 \ \ {\it Gleason Frames} \ \ (to I. Bengtsson)} \label{Bengtsson1}

\bib
I'm looking for a topic for a Master's Thesis, and I am wondering whether I can ask the undergraduate in question to study Gleason's theorem, more specifically to understand the ``general POVM'' version, to get some feeling for why the ``only projective'' version is hard, and then to ask what happens if one places oneself somewhere in between---say admit ONLY tomographically complete but otherwise minimal POVMs, or something like that (I might ask her to invent restrictions herself). As a way to start, first study Kochen--Specker for $N=2$, which has been done for cubes inscribed in a dodecahedron (Aravind, I believe), and then try to do it with tetrahedra inscribed in a dodecahedron (I don't think I have seen that).

Question (that the supervisor ought to ask himself, really): Is the answer trivial, or known, or obviously too hard?

Any advice that you care to give would be appreciated.
\eib

I think the research project you suggest for your student is a great one!  (I wish I were your student, in fact, for this one.)

The question you ask is by no means trivial. Have you looked at our paper
\bq
\quantph{0306179}?
\eq

See Section IV in particular.  My original expectation was that the set of symmetric informationally complete POVMs would be sufficient for deriving the probability rule, but it was not!  (In fact, I'm still disappointed that that's not true:  It still seems that it should be.)  So, one question I might ask is how much asymmetry must be introduced before a Gleason theorem becomes viable again?  For instance, is it enough to consider one very asymmetric frame and all unitarily equivalent ones?

Keep me up to about your progress.  I think this is an exciting question.

\section{02-12-05 \ \ {\it Notes to van Fraassen and Delirium Quantum (some nested)} \ \ (to G. Valente)} \label{Valente8}

Contained within the attachments are all the notes I had promised to send you.  [See the 10-11-05 notes ``\,\myref{vanFraassen7}{`Action' instead of `Measurement'}\,'' and ``\myref{vanFraassen9}{Our Own Rovellian Analysis}'' to B. C. van Fraassen.] You will see why I say they touch on our view of the ``stability condition'' with respect to quantum measurement, and why I sometimes say that the very existence of repeatable measurements is something of a mystery from our perspective.

Thanks again for spending some time at the Orange House.  I got a lot out of the visit; you gave me much food for thought.

\section{02-12-05 \ \ {\it Sounds of Silence and CBH} \ \ (to G. Brassard)} \label{Brassard48}

OK, I'll accept your sounds of silence for a while \ldots\ until my own stress levels start to go over the top about how I might fund this magically gifted postdoc who wants to work on these things \ldots

Too bad you couldn't have been around here the last three weeks or so.  Bas van Fraassen and Hans Halvorson have been going over the CBH theorem in their seminar.  And Jeff Bub has been making the three- and four-hour drives (each way) to attend the sessions!  A lively discussion has ensued, with the obstinate Tim Maudlin in the audience.  Anyway, you should have been there.  I've gotten a lot from the discussion.

\section{02-12-05 \ \ {\it Snowbird Schedule}\ \ \ (to C. M. {\Caves})} \label{Caves81.2}

Scully wanted me to talk on whether I think Bohr had an adequate reply to EPR.  Kind of a silly thing to talk on, particularly in only 15 minutes.  Oh well, at least it'll give me a chance to plug our Bayesian program (to the three or four people in the audience).

\section{03-12-05 \ \ {\it Subjective Quantum Information} \ \ (to M. Bahrami)} \label{Bahrami1}

\bmob
I have studied the articles as you told me. There is a core stone in your attitude which can be called Subjectivness of Information.

I pretended I am on your side and accept this vision. Then I tried to exploit this vision in a real experimental setup. So I choose Interaction-Free Measurement as proposed by Elitzur-Vaidman. To be honest I feel I can say nothing about why the existence of an object in one of the ways causes such a strange result. I have checked the journals to find other physicists interpretations on this subject, but I could find just tow different attitude:
1) many-world interpretation of this experiment, mainly by Vaidman, 2)
objective interpretation of information by Auletta ({\sl Foundations of Physics}, May 2005)
Let alone the first one, but the second is in great contrast with the common beliefs of physicists like you on subjectiveness of information.

I would be grateful if you let me know how you interpret (or explain) this experiment with a subjective attitude toward information.
\emob

To be honest, that is a thought experiment that I never gave too much attention to, because I always thought it was hype without too much substance.  The concept of ``interaction-free measurement'' uncovers what had in earlier days been called {\it interference\/} and gives it a new, more dramatic name, but that's qualitatively as deep as it goes.  (Quantitatively, there are some differences with previous techniques, and this may be thought of as a clever interferometric technique.)

But now that you push me on it, I think this would be an excellent project for you to take on.  Why not let this be the subject of your Master's thesis?   Give a detailed quantum information theoretic analysis of this phenomenon.  And particularly, test your understanding of the Bayesian approach to quantum states with respect to it.  I, for instance, might learn a lot from your thinking, and then you can call the analysis your own.

\section{12-12-05 \ \ {\it The Great Pumpkin} \ \ (to J. W. Nicholson)} \label{Nicholson25}

I think you'll love this thing Greg Comer is going to do in his physics class today.  Read below.

\bgc
I can't believe I'm going to give the following question on my
physics exam today:
\bq\noindent {\rm
Charlie Brown tells Linus that he has developed a theory of
gravity that is based on mass and energy causing curvature in the
spacetime continuum.  He also shows Linus his experimental data which
confirms that the theory works within the accuracy and precision of the
experiments.  Linus tells Charlie Brown that, regardless of the
agreement with the data, the new theory is wrong, because it is known
that the Great Pumpkin is responsible for gravity, and the new theory
makes no reference to the powers of the Great Pumpkin.  How should
Charlie Brown respond?  Assume a rational and respectful discourse
from both.}
\eq
\egc

\section{19-12-05 \ \ {\it Superbroadcasting}\ \ \ (to C. M. {\Caves})} \label{Caves81.3}

\bcc
One of my students has written his class project on broadcasting, and
he describes new results on so-called superbroadcasting.  What's the
scoop on this?
\ecc

Apparently the ``if and only if'' become more complicated if the sender supplies two independent copies to the potential broadcasting device and requires that the marginals be equal to the inputs on three output systems.  Sounds like a much more complicated affair than we dreamed up, and I'm not sure what insight's to be gained from the exercise.  But that's all I know about it.

\section{19-12-05 \ \ {\it The Detached Fuchs} \ \ (to H. C. von Baeyer)} \label{Baeyer11}

I'm finally replying to your nice note:  Several visitors and Montezuma's revenge have gotten in my way since your writing me.
\bhcvb
I came across a footnote specifically about the term ``der losgel\"oste Beobachter: the detached observer'' in Wolfgang Pauli,
Wissenschaftliche Briefwechsel, Vol.\ IV part I, page 343, footnote 13
to the commentary on letter [1263] from Bohm to Pauli, which is of
course in English.
\ehcvb
Yes, of course, I would like that quote translated very much; as far as I know I don't already have it.  All in the greater cause!

\section{19-12-05 \ \ {\it Thanks and Question} \ \ (to H. Halvorson)} \label{Halvorson9}

This is a belated thanks for sending me the Alfsen, Schultz, and Mielnik references.  I've got to get on this!!

Now a question.  I noticed that you're on the PSA 2006 program committee.  The reason I was snooping there is because I had a flash that maybe I'd like to organize a session or workshop at the meeting comparing and {\it contrasting\/} epistemic views of the quantum state to Bayesian views of probability.  It'd be nice if I could get, for instance, Persi Diaconis or Brian Skyrms and Abner Shimony to give talks, along with, say, Rob {\Spekkens} and/or Huw Price and/or Itamar Pitowsky and/or Bill Demopoulos, etc., etc.---I'm just starting to think about it all.  The trouble is, the deadline for proposals has already passed.  Do you know whether this deadline is really strict?  I noticed that the proposals won't start being reviewed until mid-February.  If the deadline is a little loose and I work at it, I could probably have something submitted in the break between Christmas and New Year's.  Thanks for any advice you can give.

\section{19-12-05 \ \ {\it A Small(?)\ Request} \ \ (to A. Radosz)} \label{Radosz2}

I have skimmed your student's paper.  I'll just reply by telling you a story from my first year of study with Carlton Caves, in 1993.  At the time, I had just read a paper by Braunstein and Caves on information-theoretic Bell inequalities.  One of the things they note in that paper is that whereas their information-theoretic inequalities have an advantage over CHSH-type inequalities because of the ease and automated nature of their derivations, they have a {\it dis}advantage in that they are not violated over as large of a range of measurements as the CHSH ones.  Well, in my digging up various references in information theory (as I was learning the ropes), I learned about a whole new class of information functions---the {\Daroczy} entropies---and as I played with them, I discovered that they too could be used in a Bell inequality derivation \ldots\ just along the same lines as Braunstein and Caves had discovered for Shannon's measure of information.  Excited, I ran to Caves, told him about my result---I think it was something like the {\Daroczy} 2.73 entropy was optimal for finding a result, and with it I could find a violation over 12.7 degrees more than the Braunstein and Caves result.  Then I asked him if I should publish it?!?!  He replied, ``Well, I wouldn't.''  I replied, ``But that's you; you've got a hundred papers, and I have none.''  He replied, ``I'm just saying, if it were me, I wouldn't.''  I got the hint, finally, and never published it.  Caves always wanted his students to stay away from incremental work or isolated results.  In my case, it helped; but different students have different requirements.  It'll be your call.\footnote{I suppose for the purposes of subtle persuasion I intentionally left off an addendum to the story that I {\it do usually\/} include with it.  Not fair of me really!  Some time in the mid to late 1990s, I was having a walk with Micha{\l} Horodecki, and the conversation turned to the Braunstein and Caves inequalities. Micha{\l} noted, ``Sadly, they cannot be generalized to the {\Renyi} information measures.  The derivation just doesn't work.''  I said, ``Oh, that's right; that's because the {\Renyi} entropies are additive, but not subadditive.  The same derivation does however go through if you use the {\Daroczy} entropies instead. For, though they are not additive for independent random variables, they are subadditive generally, and that's all you need.'' Micha{\l} became quite excited and asked, ``Where have you written this?''  I said, ``Oh, I never published it. I worked it out a long time ago, but it didn't seem very important.''  He responded, ``That is wrong, you should publish it even now.''}

\section{27-12-05 \ \ {\it Randomness and Quantum} \ \ (to D. Overbye)} \label{Overbye1}

While up this morning with a little insomnia (it's 4:00 AM as I write this), I skimmed over your {\sl NY Times\/} article on ``Quantum Trickery.''  I especially enjoyed some lines near the end:
\bq\noindent
     As a result of the finiteness of information, he explained, the
     universe is fundamentally unpredictable.

     ``I suggest that this randomness of the individual event is the
     strongest indication we have of a reality `out there' existing
     independently of us,'' Dr.\ Zeilinger wrote in {\sl Nature}.

     He added, ``Maybe Einstein would have liked this idea after all.''
\eq
That's an idea I'm very attracted to and have explored in a decent set of my writings.  (There's a sampling of some posted at my webpage, link below.)  With regard to entertainment value along those lines, you yourself might enjoy Sections 4 and 5 of this 2002 pseudo-paper:
\bq
\quantph{0204146}.
\eq

Here's a sample paragraph from that connected to the above:
\bq
   I would say all our evidence for the reality of the world comes from
   without us, i.e., not from within us.  We do not hold evidence for an
   independent world by holding some kind of transcendental knowledge.
   Nor do we hold it from the practical and technological successes of
   our past and present conceptions of the world's essence.  It is just
   the opposite.  We believe in a world external to ourselves precisely
   because we find ourselves getting unpredictable kicks (from the
   world) all the time.  If we could predict everything to the final T
   as Laplace had wanted us to, it seems to me, we might as well be
   living a dream.

   To maybe put it in an overly poetic and not completely accurate way,
   the reality of the world is not in what we capture with our theories,
   but rather in all the stuff we don't.  To make this concrete, take
   quantum mechanics and consider setting up all the equipment necessary
   to prepare a system in a state $|\Psi\rangle$ and to measure some noncommuting
   observable $H$.  (In a sense, all that equipment is just an extension
   of ourselves and not so very different in character from a prosthetic
   hand.)  Which eigenstate of $H$ we will end up getting as our
   outcome, we cannot say.  We can draw up some subjective probabilities
   for the occurrence of the various possibilities, but that's as far as
   we can go.  (Or at least that's what quantum mechanics tells us.)
   Thus, I would say, in such a quantum measurement we touch the reality
   of the world in the most essential of ways.
\eq

\section{27-12-05 \ \ {\it A Little Christmas Randomness} \ \ (to A. Zeilinger)} \label{Zeilinger2}

Merry Christmas!  I hope you and your family are doing well and are having a relaxing holiday.

Anyway, you are on my mind again this morning because I read Dennis Overbye's article in today's {\sl New York Times\/} on ``quantum trickery,'' and I was struck once again by the similarities in the essential parts of our views.  Overbye wrote this about you:
\bq\noindent
As a result of the finiteness of information, he explained, the
universe is fundamentally unpredictable.

``I suggest that this randomness of the individual event is the
strongest indication we have of a reality `out there' existing
independently of us,'' Dr.\ Zeilinger wrote in {\sl Nature}.

He added, ``Maybe Einstein would have liked this idea after all.''
\eq

Compare that with some things I wrote in several places in {\sl Notes on a Paulian Idea}, but particularly in the pseudo-paper \quantph{0204146.pdf} where in Sections 4 and 5, I said:
\bq
I would say all our evidence for the reality of the world comes from
without us, i.e., not from within us.  We do not hold evidence for an
independent world by holding some kind of transcendental knowledge.
Nor do we hold it from the practical and technological successes of
our past and present conceptions of the world's essence.  It is just the opposite.  We believe in a world external to ourselves precisely
because we find ourselves getting unpredictable kicks (from the
world) all the time.  If we could predict everything to the final T
as Laplace had wanted us to, it seems to me, we might as well be
living a dream.

To maybe put it in an overly poetic and not completely accurate way, the reality of the world is not in what we capture with our theories,
but rather in all the stuff we don't.  To make this concrete, take
quantum mechanics and consider setting up all the equipment necessary to prepare a system in a state $|\psi\rangle$ and to measure some noncommuting observable $H$.  (In a sense, all that equipment is just an extension of ourselves and not so very different in character from a prosthetic hand.)  Which eigenstate of $H$ we will end up getting as our outcome, we cannot say.  We can draw up some subjective probabilities for the occurrence of the various possibilities, but that's as far as we can go.  (Or at least that's what quantum mechanics tells us.)  Thus, I would say, in such a quantum measurement we touch the reality of the world in the most essential of ways.
\eq
AND
\bq
So, I myself am left with a view of quantum mechanics for which the
main terms in the theory---the quantum states---express nothing more than the gambling commitments I'm willing to make at any moment.
When I encounter various other pieces of the world, if I am
rational---that is to say, Darwinian-optimal---I should use the
stimulations those pieces give me to reevaluate my commitments. This is what quantum state change is about.  The REALITY of the world I
am dealing with is captured by two things in the present picture:
\eq
\begin{enumerate}
\item
I posit systems with which I find myself having encounters, and

\item
I am not able to see in a deterministic fashion the
stimulations (call them measurement outcomes, if you like)
those systems will give me---something comes into me from the
outside that takes me by surprise.
\end{enumerate}

Just two examples, but I think they indicate I'm going to enjoy your paper very much!  Could I ask you to send me a copy of it electronically, however?  (I can't access it directly from the journal until I am back at Bell Labs, and that won't happen until the holidays are over.)  It would give me some holiday cheer.

\section{27-12-05 \ \ {\it Indistinguishable Particles}\ \ \ (to Y. Omar)} \label{Omar1}

Thanks for the warm Christmas wish!  I enjoyed it and laughed and laughed.

I will try to take care of the bureaucratic things I owe you within the first week after New Year's Day.

By the way, I think I have discovered a reason why nature allows for indistinguishable particles.  It comes from observing my two children play with their new toys on Christmas.  After listening all day to {\it each\/} say of the {\it other's\/} toys, ``I wish I had that,'' I thought, wouldn't it be nice if there were a way for them to not know whose was whose?  And, indeed, a way for them not to see any distinguishing qualities for the toys at all, other than the amount of sheer pleasure they give?  Then the parents' job would be so simple:  Simply give them identical quantities of pleasure to play with at their leisure, and be done with it.

\section{27-12-05 \ \ {\it Regarding the Detached Observer Attached} \ \ (to H. C. von Baeyer)} \label{Baeyer12}

\bhcvb
Dear Chris, the detached observer is attached.   I thought this
footnote might interest you not so much for its text, which is less
than limpid, as for its citations of the phrase.
\ehcvb
\bq
In Wolfgang {\Pauli}, {\sl Scientific Correspondence with Bohr,
Einstein, Heisenberg, a.o.}\ (Springer, 1996) Volume IV, Part I:
1950--1952, page 340 there is a three-page editorial comment on letter
number [1263], from David Bohm to {\Pauli}.  The penultimate paragraph
of this comment, in German, begins:
\bq
{\Pauli}, on the other hand, under the influence of his psychological
point of view, had progressively moved away from the
classical-Cartesian assumption of the detached observer (cf.\ letters
[1313] and [1314])$^{14}$ who assumes a {\it fixed or pre-arranged
game\/} behind the stage {\it (abgekartetes Spiel)\/}  [letter 1388]:
(in English) `For me, however, it is much more satisfactory if the
laws of nature themselves exclude in principle the possibility even
to conceive the disturbances in the observers own body and own
brain connected with his own observations\ldots.'
\eq
\{I believe the editor may be mixing up two different notions here,
that of the detached observer, and that of the view of nature as a
pre-arranged game, but I wanted to give the context of footnote 14.
HvB\}
\bq\noindent
$^{14}$This term first appears in this context in {\Pauli}'s letter
[1197] of 31 January 1951 to M.-L. von Franz and then again in the
letter from M. Fierz [1288] of 10 October 1951. From now on {\Pauli}
used it more frequently in his lectures, publications, and letters.
In a letter of 5 May 1953 to C.~F. von {\Weizsacker} he mentioned this
question in connection with his Kepler study: ``As I hinted in my
essay, it seems to me that Fludd was much closer to the symbolic
formulation of the {\it unity of existence}, which in turn so paradoxically
divides into `observer' and `external world' (`the cut'), than was
Kepler, with his `detached observer' of classical physics.  The
`archaic' Fludd had the stronger feeling for the proposition that the
`position of the cut' is arbitrary (Heisenberg).''  Similarly {\Pauli}
wrote on 15 May 1953 to M.-L. von Franz: ``Holding on to these
assumptions requires one to restrict oneself to {\it statistical\/} laws and
to `sacrifice' the individual case. Einstein, on the other hand,
would like {\it both\/} to `have his cake and eat it.' He yells `incomplete',
regresses to the detached observer of classical physics, and places
`world formulae' into a blue fog (which does not contain the
observer).
\eq
\eq

Thanks so much for the quote!  I've just incorporated it into my
``activating observer'' compendium (and given you credit for the
translation).  But now you'll start to see my insatiable appetite!
For I'd dearly love to see the FULL letters 1197, 1288, 1313, 1314,
1388, as well as the 5 May 1953 letter to von {\Weizsacker} and 15 May
1953 letter to von Franz (for some reason there were no numbers
listed next to the latter two).

In my existing collection, a search on the year 1951 only gives this
quote from a Primas paper,
\bq
   Faced with the wholeness of the reality, scientists have been slow to
   accept the challenge of discussing the premises of Baconian science.
   They have even been reluctant to consider the simplest modification
   of a mechanistic world view, namely the inclusion of teleological
   considerations as an essential part of their discipline -- a
   relatively simple problem for which in the framework of modern
   quantum mechanics all necessary tools are available presently.  But
   {\Pauli} was looking for an incomparably deeper approach which goes far
   beyond the limits of current quantum theory, and which includes {\it
   physis\/} and {\it psyche\/} as complementary aspects of the same
   reality.  A reality containing both rational and irrational elements,
   and alchemic {\it conjunctio\/} of spirit and matter.  In psychology
   as well as in physics, quaternity is taken to be an expression of all
   concepts of unbroken totality.  In a letter {\Pauli} wrote: ``Ich bin ja
   auf Kepler als Trinitarier und Fludd als Quaternarier gestossen --
   und f\"uhlte bei mir selbst, mit deren Polemik, ein\-en inneren
   Konflikt mitschwingen.  Ich habe gewisse Z\"uge von beiden, sollte
   aber jetzt in der zweiten Lebensh\"alfte zur quartern\"aren
   Einstellung \"ubergehen.  Das Problem ist, dass dabei die positiven
   Werte der trinitarischen Einstellung nicht geopfert werden
   d\"urfen.'' ({\Pauli} to Fierz on October 3, 1951, quoted in
   Enz, p.~509).  {\Pauli} could not solve his
   dilemma of three and four which plays a great role in alchemy as
   ``the axiom of Maria Prophetissa'' (``Out of the Third comes the One
   as the Fourth''), and this shows that we are at the bare beginning to
   understand reality.  But we again reached a turning point, a way of
   thinking is developing which is very different from that which has
   been dominant in the past decades, and which recognizes the
   repression of the irrational as incongruous.
\eq
(of which I don't know what the German part is saying) and this point
of interest from a Lindorff paper
\bq
   This statement, taken together with the way {\Pauli} associated the
   radioactive atom with the lapis, moved Jung to question whether `the
   archetype and its effects and the effect of the active atom on its
   environment is not more than a metaphor.'

   With these thoughts in mind Jung said he would amplify his essay,
   `Synchronicity: an acausal connecting principle'. He presented the
   material to the Psychological Club in two parts on 20 January and 2
   March 1951. The essay, together with {\Pauli}'s work on Kepler, was
   subsequently published in book form.
\eq

The only materials I have from 1953 are a few letters from {\Pauli} to
Jung; nothing from {\Pauli} or von {\Weizsacker} or von Franz or Fierz.

At the moment, I'm particularly interested in better understanding
the thing {\Pauli} wrote to von Franz in what you sent me:  ``For me,
however, it is much more satisfactory if the laws of nature
themselves exclude in principle the possibility even to conceive the
disturbances in the observers' own body and own brain connected with
his own observations \ldots''  That really intrigues me!  What
exactly does he mean by that?

In that regard, let me send you a little compendium I sent to David
{\Mermin}, in fact just today.  It's pasted below.  As you'll see, I'm
starting to lean very heavily on an alchemical analogy for building
some imagery of what quantum measurement is all about.  (I hinted a
little about this to you before, but supplied no details.)  And the
quote you gave me above seems to touch on that in a way I hadn't seen
before.  This is something I really would love to explore.

Thanks so much again for all this!  You're bringing me back to life
in these lazy days between Christmas and the New Year.

I hope you and your wife are having a great holiday yourselves.

\subsection{Hans's Reply, ``Translations,'' 31-12-05}

\bq
I am happy to translate this and all the other letters,  but in order to avoid an infinite regression, we need to establish some ground rules.
\begin{enumerate}
\item
My primary source for the next year or two will be the Pauli correspondence, with the [....] numbering.  This means that unless it is absolutely essential, I will avoid other sources.
\item
I am not good at filing, so you are in charge of keeping order.
\item
Both of us need to be vigilant about the existence of translations -- no use repeating someone else's work.  Of course I will make sure to check any translations I find for accuracy.  (The art historian Panofsky, Pief's brother I think, considered some of von Franz's translations from Latin, made for Pauli, to be distorted to the point of gobbledygook.)
\item
I had decided to start translating Fierz letters, but will start, instead, with the list you sent -- i.e., with the selection criterion ``detached observer'' instead of ``Fierz.''
\item
There will be three types of footnotes:  by the writers themselves, by Karl von Meyenn, and by me.  I will endeavor to flag them clearly. I will probably skip some parts, and will say so.  Any time you want me to fill in the blanks, I'll be happy to do so.  Letters to von Franz, for example, often have appended dreams.  If these don't seem relevant to the subject matter of the letter, but are simply entries in an ongoing dream-book, I will skip them, but say so.
\end{enumerate}
\eq

\section{27-12-05 \ \ {\it Words of Yours I Liked and Didn't} \ \ (to N. D. {\Mermin})} \label{Mermin121}

Happy holidays!  I've been missing you a lot lately, particularly as
{\Ruediger} was visiting a couple of weeks back, and once again we
picked up on our old thread of trying to develop a paper on
``certainty''---you don't know how many times your image came to us
in the conversations!

I've been meaning to write you, also, to thank you for having a copy
of your book sent my way!  It is beautiful.  Per my promise, I will
read it cover to cover; just give me a little time.

I have at least read the preface as of now, and I think I have
discovered why you have been so nice to me all of these years (or at
least tolerant of me)!  Is it because my heart is in the right place,
i.e., wanting to see quantum theory reduced to a statement that can
be taught in an ordinary high school?  I hadn't realized that you had
made such an important point of this with Special Relativity.
Wonderful!

Now, why am I writing you today in particular?  It's because I had
insomnia last night and I stumbled across Dennis Overbye's article in
the {\sl NY Times}.  When I read this,
\bq\noindent
   In an essay in 1985, Dr.\ {\Mermin} said that ``if there is spooky action
   at a distance, then, like other spooks, it is absolutely useless
   except for its effect, benign or otherwise, on our state of
   mind.''
\eq
I thought it was sheer genius, and I was ashamed that I didn't
remember having read it before.  I so wish I could write with your
cleverness! Words I liked.

But when I read this,
\bq\noindent
   ``I would say we have to be careful saying what's real,'' Dr.\ {\Mermin}
   said. ``Properties cannot be said to be there until they are revealed
   by an actual experiment.''
\eq

I thought, ``Tsk.\ Tsk.''  What trouble I continue to think that way
of expressing things makes for our imaginations.  Thinking of quantum
measurement outcomes as ``revealing properties'' (whether they are
pre-existent properties or not) is, I think, one of the biggest
problems getting in our way of making a decent myth for the quantum
mechanical world.  Words of yours, this time, I didn't like.

By way of saying Happy New Year again, I'll send you three recent
emails I wrote to Bas van Fraassen (all pasted below) where once
again I'm groping to find the right words to say what quantum
measurement ``is'' and what the clicks it gives correspond to.  A
quantum measurement outcome (considered on its own, without relation
to other issues) is just a ``surprise,'' full stop; it doesn't reveal
anything.  It's just a primitive reaction within the agent.  Or at
least that's the myth I'm groping for!  (You'll see in the notes
below that even I'm guilty of sometimes using the same words that I
said ``tsk, tsk'' to you about.  Once again---as always!---I learn
from reading even the tiniest things of yours; somehow you cause me
to introspect like few others do.)

[See 14-11-05 note ``\myref{vanFraassen10}{Questions, Actions,
    Answers, \& Consequences},'' 14-11-05 note
  ``\myref{vanFraassen11}{Nordheim Again}'' and 15-11-05 note
  ``\myref{vanFraassen12}{Canned Answers}'' to B. C. van Fraassen.]

Happy, happy, happy New Year!

\section{27-12-05 \ \ {\it White Christmas} \ \ (to W. G. {\Demopoulos})} \label{Demopoulos3}

I've been thinking about you most of the holiday season, but am only now getting around to sending out holiday greetings.  Happy holidays!  The event that seemed to start it off was watching the old movie ``White Christmas'' (with Bing Crosby and Danny Kaye) sometime early in December.  For some reason that I can't put my finger on, I was struck by how the actor who played General Waverly reminded me of you.  His name was Dean Jagger in real life, and I don't know if it was his facial features or his voice, but something about him made me think of you throughout the movie.

But, who knows, maybe I was already thinking about you earlier than that!  This morning, when I was reviewing some of my old emails from my last discussions with {\Ruediger} {\Schack} (for a paper we're writing on Wigner's friend), I came across a note I wrote that had these lines in it:
\bq
Quantum theory has no more power within it to say when (quantum) events happen than classical probability theory has the power within IT to say when its own events happen.  Maybe this is why some developments of classical probability theory are careful to say that the $h$'s in a probability assignment $P(h)$ are propositions. Propositions don't ``happen'', they are simply either true or false. And thus, by fiat, one never has to address the question of when $h$ ``happens''.  But we know that we shouldn't try to make that further move (i.e., viewing the $h$'s as propositions with timeless truth values) in the quantum setting.  Quantum events do ``happen.''  But that doesn't mean that quantum theory should have the burden on it of explaining or determining ``the moment at which `all is said and done'.''

This is ultimately, I suspect, why our view will be greeted with the same revulsion that William {\James}'s theory of truth was by Russell
and Moore and the like:  Because for {\James}, too, truth is something
that is made to happen; it is not something that is just there in a
timeless way.
\begin{quote}
The truth of an idea is not a stagnant property inherent in it.
Truth {\it happens\/} to an idea.  It {\it becomes\/} true, is
{\it made\/} true by events. Its verity {\it is\/} in fact an event, a process:  the process namely of its verifying itself, its veri-{\it fication}.  Its validity is the process of its valid-{\it ation}.
\end{quote}
\eq

When I read this, I thought of the nice walk and talk we had in Maryland a couple of years ago.  What I wrote above seems to be the main issue that separates us at the moment in our views of quantum mechanics, but it may be the only one!

All the best, and I hope you have a great New Year.

\section{27-12-05 \ \ {\it Happy Troubles} \ \ (to D. Overbye)} \label{Overbye2}

I actually got a lot out of reflecting over your article today.  So I wanted to say thanks and that I enjoyed it.  Particularly, I got a lot out of thinking about David Mermin's sparse remarks that you recorded and then noting an inconsistency in my own ways.  I tried to capture the point in an email I wrote to David a couple hours ago.  I'll paste it below, as I feel I owe you a little bit too.  (David and I are old friends and have been having a discussion on some of these points for a long time.)  [See 27-12-05 note ``\myref{Mermin121}{Words of Yours I Liked and Didn't}'' to N. D. {\Mermin}.]

And since you're not a full-time physicist, I'll give you a special treat:  A cartoon version of the whole story!  It is attached as a {\tt .jpg} file.  (Quantum theory is about all the stuff to the left side of the sparks; the stuff to the right is `reality,' the partial source of the sparks.)

\begin{center}
\includegraphics[width=10cm]{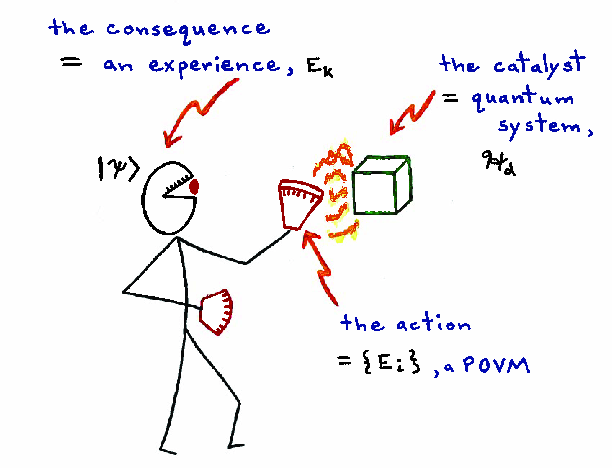}
\end{center}

\section{28-12-05 \ \ {\it Wheeler and the Pleasures of Life} \ \ (to D. Overbye)} \label{Overbye3}

I'm up again at this annoying hour, and, bringing my chance meeting of you last night to a closure, I read a little more about you on the web.  I enjoyed the interview posted on {\sl Edge}.  ``What else is there?\ Sex and physics.''  We might be kindred spirits \ldots\ \  When I first started calling my own view of quantum mechanics ``the sexual interpretation of quantum mechanics,'' one friend's reaction was that the title was of no real content---it was only a cheap way to draw attention to the view---but another friend's reaction was that I was looking for a way to combine my ONLY two pleasures in life \ldots\ \  (The latter friend doesn't think I'm particularly deep.)

More seriously, I also discovered that you seem to be a little bit of a fan of John Wheeler.  John has been a huge influence on my scientific life.  Before last night, I had not recalled reading anything of yours regarding him, but tonight, searching in this big document I'm putting together ``The Activating Observer: Resource Material for a Paulian--Wheelerish Conception of Nature'' (which is a compendium of a lot of quotations) I found two entries in your name.  I'll place them below for fun.  So, you've been writing about Wheelerish stuff since at least 1981!?  And apparently I've been reading you since about 1985 (when I started systematically digging up Wheeler stuff)?!  Good things to learn.

If you're interested in some funny Wheeler anecdotes, I've got a few compiled in my samizdat (which will soon be published as a real book by Springer) ``Notes on a Paulian Idea.''  (The version with the most complete index presently is posted on my webpage; it's slightly better to go there than {\tt quant-ph}.)  For instance, you might get a kick out of the story on page 149, which relates a little to my second quote of you below.

Anyway, as I say, bringing last night's meeting to a closure,

\begin{itemize}
\item
D.~Overbye, ``Messenger at the Gates of Time,'' Science81 {\bf 2}(5),
61--67 (1981).  This contains some biographical material on John Wheeler.  Also, there are some great quotes of John Wheeler:

\begin{quote}
The best way to find oneself outside the ranks of science is to find oneself inside the ranks of mysticism.
\end{quote}
and
\begin{quote}
If there's one thing in physics I feel more responsible for than any other, it's this perception of how everything fits together.  I like to think of myself as having a sense of judgement.  I'm willing to go anywhere, talk to anybody, ask any question that will make headway. I confess to being an optimist about things, especially this question about our hopes of being able to understand how things are put together.  So many young people are forced to specialize in one line or another that a younger fellow can't afford to try and cover this waterfront---only an old fogey who can afford to make a fool of himself.  If I don't, who will?
\end{quote}

\item
D.~Overbye, ``What Happened Before the Big Bang?,'' New York Times,
12 March 2002.

\bq
In their dreams, theorists of both stripes [string theory and loop quantum gravity] hope that they will discover that they have been exploring the Janus faces of a single idea, yet unknown, but which might explain how time, space and everything else can be built out of nothing. A prescription for, as the physicist Dr.\ John Archibald Wheeler of Princeton puts it, ``law without law.''

Dr.\ Wheeler himself, the pre-eminent poet-adventurer in physics, has put forth his own proposal. According to quantum theory's famous uncertainty principle, the properties of a subatomic particle like its momentum or position remain in abeyance, in a sort of fog of possibility until something measures it or hits it.

Likewise he has wondered out loud if the universe bootstraps itself into being by the accumulation of billions upon billions of quantum interactions---the universe stepping on its own feet, microscopically, and bumbling itself awake. It's a notion he once called ``genesis by observership,'' but now calls ``it from bit'' to emphasize a proposed connection between quantum mechanics and information theory.

One implication of quantum genesis, if it is correct, is that the notion of the creation of the universe as something far away and long ago must go. ``The past is theory,'' Dr.\ Wheeler once wrote. ``It has no existence except in the records of the present. We are all participators, at the microscopic level, in making that past as well as the present and the future.''

If the creation of the universe happens outside time, then it must happen all the time. The Big Bang is here and now, the foundation of every moment.

And you are there.
\eq
\end{itemize}

\section{28-12-05 \ \ {\it Wheeler and the Pleasures of Life, 2} \ \ (to D. Overbye)} \label{Overbye4}

\bdo
I am a huge fan of John Wheeler, actually. i wrote a long profile of
him in Science Times three years ago. Also he is featured in my book,
{\bf Lonely Hearts of the Cosmos}, currently a Little, Brown paperback,
1999.
\edo

More entries for my compendium!  I'm always hungry for more (and hopefully within the next year or so, I'll try to publish the thing).  I'll look those things up as soon as I can.

\section{28-12-05 \ \ {\it Grad Research and Quantum Information} \ \ (to C. Ududec)} \label{Ududec1}

I apologize for taking so very long to reply to you; I have gotten far, far, far behind on my email, and am only now trying to catch up during the holidays.

\bcu
I'm currently in my last year in a math/physics honours program at the University of Manitoba in Canada, and I'm interested in going on to graduate school and working on something in quantum mechanics and its foundations.  I've read a little about different approaches to the measurement and interpretation problems, and I've started reading your `Quantum Mechanics as Quantum Information' paper.  I'd be very interested in learning more about this kind of approach or working on something similar for my graduate studies.

I was wondering if you have any suggestions for other things I should
read, or if there is a good starting point for this topic.  I was also
wondering if you have any suggestions for schools I could look at, or
other people I should talk to.
\ecu

I would suggest reading anything posted on my website, and maybe supplementing that with
\begin{itemize}
\item
\quantph{0404156}
\item
\quantph{0404122}.
\end{itemize}
As for a good place for you to do graduate studies, if you are really interesting in this approach to quantum mechanics, I would suggest studying with Carl Caves at the University of New Mexico or {\Ruediger} {\Schack} at Royal Holloway University of London.

\bcu
Also, when you talk about the consequences of an experimental
intervention, are they just consequences for the one experimenter's
beliefs about the system and about future interventions?
\ecu
That is right.
\bcu
They aren't consequences for all other possible experimenters as well,
right?
\ecu
Correct.  A mantra that came up during {\Ruediger} {\Schack}'s last visit to NJ was, ``A quantum state represents one's personal beliefs about the personal consequences of one's personal interventions.''

\section{28-12-05 \ \ {\it And a Cartoon} \ \ (to H. C. von Baeyer)} \label{Baeyer13}

And I shouldn't forget to send you the cartoon that goes with the
last note I sent you.  It is attached as a {\tt .jpg} file.  Quantum
theory is about all the stuff to the left side of the sparks---i.e.,
the actions and transformations of the agent.  The stuff to the right
of the sparks is the `external reality,' the partial source of the
sparks---it is the `philosopher's stone,' without which the agent
would never get transformed.  The word `catalyst' in the cartoon is a
little misleading, as anyone with a little chemical training thinks
of a catalyst as something that remains unchanged in the reaction;
however, I'm thinking here of it in the broader sense of, say, the
{\sl American Heritage Dictionary\/} where a catalyst is simply ``an
agent that stimulates or precipitates a reaction, development, or
change,'' full stop.

The reason I'm intrigued by the {\Pauli} quote to von Franz is that he
says, ``For me \ldots\ it is much more satisfactory if the laws of
nature themselves exclude in principle the possibility even to
conceive the disturbances in the observer's own body and own brain
connected with his own observations \ldots''  That is, he seems to be
talking about something going on to the left of the sparks too, not
to the right of the sparks---i.e., just the part I claim quantum
theory is about.

\section{31-12-05 \ \ {\it If We Make It Through December \ldots} \ \ (to G. Musser)} \label{Musser20}

I don't guess you know the old Merle Haggard song, ``If we make it through December / / Everything's gonna be all right I know \ldots''  Well, that's where I am at the moment.  Thank you though for prodding me.

I apologize for the delay; it all started out innocently enough, before my interruptions snowballed.  The innocent part was that I started to worry some about your words:
\bgm
One yoke your article will have to carry is that talking of the
``subjective'' elements of quantum theory sets off alarm bells -- people
think of Fritjof Capra and get suspicious of a metaphysical agenda.
So you might want to state outright that interpretations of quantum
mechanics have a somewhat checkered history, but that this shouldn't
prevent us from plowing ahead and going where the physics leads us.
\egm

So, I took a little hiatus to learn more about Frank Ramsey's presentation of probability theory---the logic being that Ramsey's view of probability is the {\it same\/} as Bruno de Finetti's ``radically subjectivistic'' interpretation and yet, something about the way he presented the view wasn't nearly as incendiary as de Finetti's way.  I wanted to incorporate any presentational insight I could find from him.

But then the snowballing of delay came with 1) Thanksgiving, then 2) {\Ruediger} {\Schack}'s visit, then 3) Chris Timpson's visit, then 4) a long case of Montezuma's revenge, then 5) Christmas \ldots

Anyway, the new year is about to start and all these damned holidays will be over.  I have to go to a meeting in Snowbird, Utah, Jan 3--6; but I am hoping that that won't slow me down too much.  I very much hope to give you something the following week.  I've got to stop stalling; it's time to write something serious.

Thanks for pointing out the Zeilinger article too.  Somehow it didn't quite register when you wrote me about it (at that time, I was in the throes of the Montezuma thing I told you about), but my curiosity got piqued when I read Overbye's article in the {\sl Times\/} last week.  I ended up sending the note below to Zeilinger, to which I got a cordial response and a copy of the article to finally read.  [See 27-12-05 note ``\myref{Zeilinger2}{A Little Christmas Randomness}'' to A. Zeilinger.] We are definitely on the same wavelength about some things, as you'll see below.

\section{31-12-05 \ \ {\it {\Wheeler} Meeting at Princeton} \ \ (to W. G. Unruh and several others)} \label{Unruh6}

Marlan Scully and I are organizing a somewhat impromptu, but nonetheless serious, meeting in Princeton to give a little tribute to John Wheeler, law-without-law, it-from-bit, and quantum information.  It'll be a one-day meeting with maybe 10 short talks and a little socializing, held mid- to late-February (or possibly early March) on a Friday.  Wheeler has said that he will make an appearance, and we would like to invite you too.

Presently our preferred target date is February 17, but we'll also consider February 23 and March 3 if there are too many scheduling conflicts.  Please let me know as soon as possible if you would like to come and how your schedule fits with these possibilities.  There are likely to be funds available for your travel expenses too, if you need them.

On this first round of invitations, we're checking on the availability of the following colleagues to see if they can participate:  Yakir Aharonov, Carroll Alley, Janos Bergou, Michael Berry, Rob Calderbank, {\Carl} {\Caves}, Leon Cohen, John Conway, Ken Ford, Bas van Fraassen, Danny Greenberger, Hans Halvorson, Lucien Hardy, Mark Hillery, Simon Kochen, Seth Lloyd, Edward Nelson, Ben Schumacher, Yanhua Shih, Bill Unruh, Bill Wootters, Anton Zeilinger, Suhail Zubairy, and Wojciech Zurek.  More than likely, we've forgotten other colleagues who would also be very appropriate for this meeting; if you see some glaring absences, give us your suggestions, and we may be able to take them into account on a second round of invitations.\medskip

\noindent All the best, and a very happy New Year to you and your family,

\chapter{2006: Wheelerfest}

\section{01-01-06 \ \ {\it Speaking of 20-Questions Games} \ \ (to G. L. Comer)} \label{Comer79}

Speaking of 20-questions games, here's a little report I wrote up for my sister [Cathy Rother] with regard to the Christmas gift she gave my daughter Emma.  \ldots\ It's just elementary information theory!

\bq
[T]hanks so much for sending the girls gifts.  Katie just loved her tiara and earrings.  As soon as their presents were (mostly!)\ opened, both girls went upstairs to put on outfits for their rest of the day to play.  And Katie's first outfit was her Cinderella dress, so that she would have something to go with her tiara and earrings!  She looked beautiful.  And for Emma's game, that was the perfect thing:  When she first saw it, she said, ``I love this game!  I've played it at Nikki's!''  We assembled it and played it that afternoon.  I enjoyed it too, because it gave me a way to explain a little information theory (which, of course, was invented here at Bell Labs) to both Emma \ldots\ {\it and\/} Kiki!  (Emma happily absorbed the knowledge; Kiki poo-pooed the idea of getting a lesson on Christmas Day.  Shows you the difference between kids and adults!)  See, in a game like that, the best strategy (as Claude Shannon proved) is to try to ask questions that always eliminate about half of the possibilities, no matter what the answer.  It can always be done if you're clever with your questions.  For instance, I taught Emma to ask questions like, ``Does your person have facial hair?''\ \ldots\ instead of, on one go, asking, ``Does your person have a beard?''\ and, on another go, asking ``Does your person have a mustache?''  Things like that.  It so happened that the facial hair question, if you ask it in the first round eliminates about half of the possibilities no matter what the answer.  Then you adjust your strategy accordingly for the next round.  On average you'll always only need to ask the minimal number of questions that way.  (In the case of this game, only 4 to 5; because if you divide 24 by 2 you get 12, and if you divide 12 by 2 you get 6, etc.)  Well, Kiki hated it; thought it was cheating (somehow sneaking two questions in for the price of one).  But it wasn't cheating.  Emma and I would tell her, ``It's just elementary information theory Mom!''
\eq

\section{01-01-06 \ \ {\it Quantum Events and Propositions} \ \ (to W. G. {\Demopoulos})} \label{Demopoulos4}

\bwd
Maybe this is what separates us: I think an event's happening is
strictly analogous to a proposition's being true. Just as events
happen or fail to happen, propositions are true or false. So I'm not
sure what to make of the idea that props are true or false but events
just happen, seeing these as different ways of saying the same thing.
\ewd

Actually, I guess I fear that I agree with the ending part of this
statement, i.e., that ``I'm not sure what to make of the idea that
props are true or false but events just happen, seeing these as
different ways of saying the same thing.''  But, I think, the force
of it goes in the opposite direction for me than it does for you.
That is, I think I think [sic] one of the great lessons of quantum
mechanics is that it is a call to arms to rethink what is meant by
the truth value of a proposition.  Here's the way I put it in one of
the proposals I once made:
\bq\noindent
Quantum Mechanics as a Powerful Hint. In my opinion, the most
profound statement yet to come out of quantum theory is the
Kochen--Specker theorem.  For it licenses the slogan, ``Unperformed
measurements have no outcomes.''  This is just a beginning.  If one
canvasses the philosophic traditions for one that has significantly
developed this slogan, one will find the now mostly-forgotten
tradition of pragmatism fathered by William {\James} and John {\Dewey}. As a source of ideas for what quantum mechanics can more rigorously
justify, no block of literature is more relevant: The connections
between the two fields cry out for systematic study. Quantum
mechanics holds the promise of drastically changing our worldview on the wide scale. It is time to let that happen.
\eq

\bwd
On another matter, there is an idea in {\Pauli} that your remark about
quantum mechanics and classical probability theory reminded me of,
but I can't recall it exactly. Doesn't he say somewhere that what QM
does is to give a precise formulation of how the probabilities of
propositions all stand with respect to one another without, however,
specifying what actually occurs?

(I've quoted this: ``Just as in the theory of relativity a group of
mathematical transformations connects all possible coordinate
systems, so in quantum mechanics a group of mathematical
transformations connects all possible experimental arrangements.'')
\ewd

I don't recall {\Pauli} saying something like in your first version,
only your second, but he might well have:  I'm a horrible thief at
times.  Here's the way I put something similar in my anti-{\Vaxjo}
pseudo-paper \quantph{0204146}:
\bq
   In choosing one experiment over another, I choose one context over
   another.  The experiment elicits the world to do something.  To say
   that the world is indeterministic means simply that I cannot predict
   with certainty what it will do in response to my action.  Instead, I
   say what I can in the form of a probability assignment.  My
   probability assignment comes about from the information available to
   me (how the system reacted in other contexts, etc., etc.).  Similarly
   for you, even though your information may not be the same as mine.
   The OBJECTIVE content of the probability assignment comes from the
   fact that {\it no one\/} can make {\it tighter\/} predictions for the
   outcomes of experiments than specified by the quantum mechanical
   laws.  Or to say it still another way, it is the very existence of
   transformation {\it rules\/} from one context to another that
   expresses an objective content for the theory.  Those rules apply to
   me as well as to you, even though our probability assignments {\it
   within\/} each context may be completely different (because they are
   subjective).  But, if one of us follows the proper transformation
   rules---the quantum rules---for going to one context from another,
   while the other of us does not, then one of us will be able to take
   advantage of the other in a gambling match.  The one of us that
   ignores the structure of the world will be bitten by it!
\eq
That is, part of the rational part of quantum mechanics is much like
de Finettian or F. P. Ramseyan ``coherence'':  If you gamble this way
about this, and you gamble that way about that, etc., etc., then
you'd better gamble such way about the other, or you're not being
coherent with respect to your beliefs of the properties of the world.

\bwd
I'm back in Canada for the rest of the academic year and perhaps for
the foreseeable future. Jeff is supposed to visit PI for several
months beginning in February and Itamar should be here in May. Any
chance you might be through here? Will you be in Maryland this spring?
I plan to be there.
\ewd

Yes, I plan to be at the Maryland meeting in the Spring.  I don't think I'll be dropping by PI though anytime soon.  However, another way I might see you in the next year is that I've thought about organizing a symposium on ``subjective probabilities and quantum mechanics'' at the next PSA meeting in Vancouver.  When I wrote a couple of people in the organizing meeting to see if they'd give me a little time post-deadline to get a proposal in, I wrote this:
\bq\noindent
I noticed that you're on the PSA 2006 program committee.  The reason I was snooping there is because I had a flash that maybe I'd like to organize a session or workshop at the meeting comparing and {\it contrasting\/} epistemic views of the quantum state to
Bayesian views of probability.  It'd be nice if I could get, for
instance, Persi Diaconis or Brian Skyrms and Abner Shimony to give
talks, along with, say, maybe you and/or Rob {\Spekkens} and/or Huw
Price and/or Itamar Pitowsky and/or Bill {\Demopoulos}, etc., etc.---I'm just starting to think about it all.
\eq
They've given me an extension.  Now the question is what I might do.  What do you think of the idea?  Would you be willing to participate in the discussion or give a talk?  Do you think Brian Skyrms would be interested?  I don't really know him well yet.

\section{02-01-06 \ \ {\it Final Installment} \ \ (to W. G. {\Demopoulos})} \label{Demopoulos5}

Now let me tackle this point of yours
\bwd
I also don't see why we should need something as fundamental as KS to
sustain the notion that ``unperformed measurements don't have
outcomes.'' I'm being a devil's advocate here because I think what you
really mean is that without a measurement of whether the cat is
alive, the cat is neither alive nor not alive. But would you put it
this baldly? If not, why not?
\ewd
in a more longwinded way.  For that, I'll paste in an anthology of
emails I've had recently with David {\Mermin} and Bas van Fraassen.  I
think they refer to your point explicitly, but as always I'm still
groping to try to get the right language.  All the emails are
connected, and you can read them linearly from top to bottom.

To answer your question in the best way I know how at the moment, I
would say:  The transformation that quantum mechanics speaks about,
the transformation from a `superposition' to `aliveness' or
`deadness', is a transformation {\it within the agent}, and that
transformation cannot take place without some interaction with the
external physical system labeled by the word `cat'.  What happens to
`cat' itself (described in a way that makes no reference to the
agent)?  On that, I think quantum mechanics is silent.  With a
mantra:  Quantum mechanics is a theory for ascribing (and
intertwining) personal probabilities for the personal consequences of
one's personal interactions with the external world.

Does all this (and particularly the stuff below) go some way toward
answering your question?

\section{02-01-06 \ \ {\it The Oblique {\Pauli}} \ \ (to H. C. von Baeyer)} \label{Baeyer14}

Look at this little gem I discovered today.  I stumbled across it in
the Schilpp volume on Einstein as I was researching for the talk I
have to give in Utah Thursday:  The topic I got roped into is whether
I think Bohr gave an adequate reply to EPR.  (My talk's title is,
``Why I Never Understood Bohr's Reply to EPR, But Still Liked It.'')
The quote comes from page 683:
\bq
       It may appear as if all such considerations were just superfluous
   learned hairsplitting, which have nothing to do with physics proper.
   However, it depends precisely upon such considerations in which
   direction one believes one must look for the future conceptual basis
   of physics.

       I close these expositions, which have grown rather lengthy,
   concerning the interpretation of quantum theory with the reproduction
   of a brief conversation which I had with an important theoretical
   physicist.  He:  ``I am inclined to believe in telepathy.''  I:  ``This
   has probably more to do with physics than with psychology.''  He:
   ``Yes.'' ---
\eq

Who else could that ``important theoretical physicist'' be but {\Pauli}!
They were certainly discussing these sorts of things at length at
that time.  Einstein wrote his remarks in 1949 (I think), while {\Pauli}
had been visiting Einstein in 1948 (recall how {\Pauli} adjudicated the
quarrel between Born and Einstein on quantum foundations during that
time).

Also compare the similarity between what Einstein says above and
these words of {\Pauli} to Fierz, 10 August 1954:
\bq
   All of this then led me onto further, somewhat more phantastic [{\it
   sic\/}] paths of thought.  It might very well be that we do not treat
   matter, for example viewed in the sense of {\it life}, ``properly''
   if we observe it as we do in quantum mechanics, {\it specifically
   when doing so in complete ignorance of the inner state of the
   ``observer.''}

   It appears to me to be the case that the ``after-effects'' of
   observation which were ignored would {\it still\/} enter into the
   picture (as atomic bombs, general anxiety, ``the Oppenheimer case''
   e.g.~etc.), but in an {\it unwanted form}.  The well-known
   ``incompleteness'' of quantum mechanics (Einstein) is certainly an
   existent fact somehow-somewhere, but certainly cannot be removed by
   reverting to classical field physics (that is only a ``neurotic
   misunderstanding'' of Einstein), it has much more to do with {\it
   integral relationships between ``inside'' and ``outside'' which the
   natural science of today does not contain\/} (but which alchemy had
   suspected and which can also be detected in the symbolics of my
   dreams, about which I believe them specifically to be characteristic
   for a contemporary physicist).

   With these vague courses of thought I have reached the border of that
   which is recognizable today, and I have even approached ``magic.''
   (From this standpoint observation in quantum mechanics might even
   appear to someone as a ``black mass'' after which the ``ill-treated''
   matter manipulates its counter-effect against the ``observer,''
   thereby ``taking its revenge,'' as a ``shot being released from
   behind'').  On this point I realize well that this amounts to the
   threatening danger of a regression into the most primitive
   superstition, that this would be much worse than Einstein's
   regressive remaining tied to classical field physics and that
   everything is a matter of holding onto the positive results and
   values of the {\it ratio}.
\eq
So, I think it's just got to be {\Pauli} that Einstein is referring to!

On another matter, let me come back to something you wrote in your
last letter:
\bhcvb
I had decided to start translating Fierz letters, but will start,
instead, with the list you sent -- i.e., with the selection criterion
``detached observer'' instead of ``Fierz.''
\ehcvb

I apologize for causing trouble.  And the more even I think about it,
a well defined theme for you is probably called for.  How else would
you be able to turn your work into a book?  I fear a little, however,
that ``detached observer'' may be too narrow, as I think {\Pauli} and
Fierz must have discussed all kinds of ``mystical'' things
tangentially related to that topic, from the possibility of
physical/psychical neutral language to synchronicity to archetypes,
etc., etc.  And I think all that stuff is worthwhile to get into the
public eye.  On the other hand, as my getting carried away has
already demonstrated, some of what he wrote to von Franz and others
is probably quite interesting too.  So, where does that leave you?  I
hope not an infinite task!  Maybe outside of little side ventures of
gathering a little background material here and there, maybe indeed
it is better to stick with your original plan of tracking the
{\Pauli}--Fierz conversation.

Tomorrow I leave for Snowbird and may be out of email contact for a
little while.

Happy New Year again!

\subsection{Hans's Reply}

\bq
I think we will come across even more bizarre beliefs of Pauli!  For example, he and Fierz were apparently REALLY convinced that the ``Pauli effect'' was real. That's more than telepathy -- that's telekinesis!

Do not worry about my plans.  I am retiring, and don't ever have to write a book again.  The translations are very slow and arduous, and I'm going to just plunge in and let a theme find me, rather than the other way around.  The point, for me, is that translation forces me to try to understand these things very carefully.

B.T.W., did you know that the title of my Anchorage lecture is ``How I learned to stop worrying about {\Schroedinger}'s cat?''

Have fun in Utah!  Me, I'm going to the Bahamas for a few days.
\eq

\section{03-01-06 \ \ {\it The Oblique Pauli, 2} \ \ (to H. C. von Baeyer)} \label{Baeyer15}

\bhcvb
B.T.W., did you know that the title of my Anchorage lecture is ``How I
learned to stop worrying about {\Schroedinger}'s cat?''
\ehcvb
Actually, you did tell me already.  It is a great title, and more importantly a great transition!

As it turns out, I wrote a small something about {\Schroedinger}'s cat yesterday.  It was in a preamble to sending Bill {\Demopoulos} the same collection of letters I sent you in the note titled ``Regarding the Detached Observer Attached.''  I'll place it below for your amusement.  [See 02-01-06 note ``\myref{Demopoulos5}{Final Installment}'' to W. G. {\Demopoulos}.]

Enjoy the Bahamas.  I wish I were in your shoes rather than mine.

\section{03-01-06 \ \ {\it Who Is That Man in the Cowboy Hat?}\ \ \ (to G. L. Comer)} \label{Comer80}

From DFW, on my way to SLC \ldots

On the flight here, I've been reading Einstein's ``autobiographical notes'' and his ``reply to critics'' from the Schilpp volume.  Amazing stuff!  And I think something in it for all of us.  Have you ever read it?  Even if you have, read it again!

\section{06-01-06 \ \ {\it The Swedish Bayesian Team}\ \ \ (to A. Y. Khrennikov)} \label{Khrennikov16}

PS. I'm just returning from Marlan Scully's annual big meeting at some ski resort in Utah; there were 220 people there.  Roy Glauber, this year's Nobel Prize Winner, gave a talk, and in it he showed a picture of Willis Lamb from 1964.  What struck me was how much you and Lamb look alike!  (Maybe there's a Nobel Prize in your future?)

\section{09-01-06 \ \ {\it Old Brown Paper}\ \ \ (to C. M. {\Caves})} \label{Caves81.4}

I contacted Harvey Brown to confirm that he had a ``nonlocality-without-inequalities'' argument before GHZ, and he sent me this reference:
\begin{itemize}
\item
H. R. Brown and G. Svetlichny, Found.\ Phys.\ {\bf 20}, 1379--1387 (1990).
\end{itemize}
And, yep, as far as I can tell, it's basically the same style argument I gave at the Scully meeting.  Though, it seems they want to say it even excludes some NONlocal hidden-variables theories too:  I don't understand that bit at the moment.

Finally, they claim that their argument is a simplification of (but essentially the same as) an argument due to Stairs in 1983 in a quantum logical setting:
\begin{itemize}
\item
A. Stairs, Phil. Sci. 50, 587 (1983).
\end{itemize}
Anyway, you're right, I should reference this stuff if I ever present it again, particularly as Harvey wrote me this:
\bhrb
     I complained once to Dan Greenberger that no one cites the work of
     Heywood and Redhead and its progeny; he agreed that the
     Heywood--Redhead proof is the ``first of the GHZ proofs''. I thought
     his terminology was amusing!
\ehrb
Considering the fame of GHZ (and that Harvey is still not a full professor), he really has gotten the historical screw.  Kind of like poor old Dieks with no-cloning.

Heywood--Redhead's involvement was from a paper, also from 1983, that somehow used KS in a similar setting, but only led to a probabilistic result (along the lines of a Bell inequality).  So I don't know that Harvey should give him so much credit, but I wasn't able to get that paper in our library.

\section{09-01-06 \ \ {\it Malarkey and Me}\ \ \ (to C. M. {\Caves})} \label{Caves81.5}

\noindent PS. \  But you've got to stop saying I'm not a physicist.  It's destructive to what little self-respect I've got left.  If we really solve this quantum muddle like we think we can, we will be doing something that many a Nobel-prize winner in physics has tried to do and failed.  Who is to say that's not good physics?

\section{09-01-06 \ \ {\it Sunday Reading on Monday} \ \ (to A. Cho)} \label{Cho2}

\bacho
We met a couple of years ago at the Seven Pines symposium.
\eacho

Of course I remember you:  I tried to contact you last year to drum up some coverage for our meeting ``Being Bayesian in a Quantum World'' in Konstanz last August, but never heard back from you.  (I notice that your email address has changed; maybe that has something to do with it.)  Anyway, you missed a very good meeting with some of the most interesting people in the field of quantum information (Nielsen, Briegel, Mermin, Unruh, Milburn, Wootters, Hardy, etc.,)  Have a look at
\begin{center}
\myurl[http://web.archive.org/web/20090511060206/http://www.uni-konstanz.de/ppm/events/bbqw2005/]{http://web.archive.org/web/20090511060206/http:// www.uni-konstanz.de/ppm/events/bbqw2005/};
\end{center}
we would have loved to have you.

\bacho
I was hoping you might look at the attached paper (accepted by PRL) and give me your
take on its significance and importance in a phone interview.

In the paper, Francesco De Martini of the University of Rome, ``La
Sapienza,'' and colleagues report an experiment in which they've
performed a so-called minimal disturbance measurement on one photon by
entangling it with a second photon and measuring it. When the two
photons are maximally entangled, this effectively results in a von
Neumann measurement on the first photon; when the degree of entanglement
is lower, the measurement on the second photon produces a smaller
disturbance of the first, at the cost of some information about its
state. The researchers show that they can trace the optimal trade-off
between gaining information about the original photon and disturbing its
state, as was calculated in 2001 by Konrad Banaszek.

This strikes me as a very pretty experiment, and I like the fact that it
seems to put a quantitative experimental handle on a pretty fundamental
theoretical issue. But I'm just a journalist. I'd be most interested in
your opinion of the work.
\eacho

Yeah, I think it is a pretty fundamental theoretical issue; I have ever since Asher Peres and I wrote up the first calculation of this variety in 1995, ``Quantum State Disturbance vs.\ Information Gain:\ Uncertainty Relations for Quantum Information,'' \quantph{9512023} (published as PRA {\bf 53}, 2038--2045, 1996).
In fact, I have wondered on many occasions whether the effect they explore in this paper may be the single idea upon which all the formal structure of quantum mechanics can be built.

Let me give you some easy Sunday reading to skim before we talk; you may find some of it entertaining, particularly the imaginary conversation with God.  First, see the closing section, ``The Oyster and the Quantum,'' of my paper \quantph{0205039}.  Next, look at pages 83--84 and 156 of my book {\sl Notes on a Paulian Idea}, \quantph{0105039}.  It's purely coincidence that those two identifiers differ only by one number.  Also maybe look at the letter to Rolf Landauer on pages 190--191. (The book in its present edition is published by {\Vaxjo} University Press, but a more formal edition will be published this year by Springer.)

Anyway, that's a start for indicating what I think of the importance of the paper you cite.  I'm going to be pretty tied up tomorrow, but I should be able to talk Wednesday, any time between 11:00 AM, say, and 5:00 PM.  My office phone number is below.

\section{10-01-06 \ \ {\it Quantum Information at Princeton?}\ \ \ (to A. R. Calderbank)} \label{Calderbank1}

PS.  I was intrigued by Richard Jeffrey's remark in his book {\sl Subjective Probability:\ The Real Thing\/} that ``[Brian Skyrms] is my main Brother in Bayes and source of Bayesian joy.  And there are others---as, for example, \ldots\ Ingrid Daubechies, Our Lady of the Wavelets (alas!\ a closed book to me) but who does help me as a Sister in Bayes \ldots.''  I bring that up because a large component of my work is devoted to building a Bayesian-like way of viewing quantum information.  You or your wife might interested in looking at some of what we accomplished at the meeting I organized last year in Konstanz, ``Being Bayesian in a Quantum World''; here's a link:
\begin{center}
\myurl[http://web.archive.org/web/20090511060206/http://www.uni-konstanz.de/ppm/events/bbqw2005/]{http://web.archive.org/web/20090511060206/http:// www.uni-konstanz.de/ppm/events/bbqw2005/}.
\end{center}
Nicely, we had some of the most interesting people in quantum information there (Nielsen, Wootters, Briegel, Milburn, Mermin, Smolin, Hardy, etc.).  If you have a look at the picture
you'll see those guys and maybe some other familiar faces; I'm the guy in the middle with the coffee cup.

\section{10-01-06 \ \ {\it Stones and Glass Houses?}\ \ \ (to L. Smolin)} \label{SmolinL4}

Today I happened upon your letter in this month's {\sl Physics Today}, which in turn led me back to your June article which I had missed.  I found myself largely in agreement with most everything you said and, in particular, very much liked the original article.

However, I was struck by the conjunction of these words of yours in the June issue,
\bq\noindent
     Some other countries seem to be better at making room for the
     independent thinkers. \ldots\ Canada has opened the Perimeter
     Institute, whose specific mandate is to be a home for independent
     foundational thinkers, and other such projects are in planning
     stages around the world.
\eq
with these words of yours from this month's reply to letters,
\bq\noindent
     In fact, young people are contributing important new results and
     ideas to the foundations of quantum theory, but none are working at
     US research universities.  Let me name a few of them: Chris Fuchs,
     Lucien Hardy, Rob Spekkens, Antony Valentini, and David Wallace.
\eq
All I know from my own experience with Perimeter---to which I had dearly wanted to join---is that it was no more supportive of me than any ``US research university.''

\section{10-01-06 \ \ {\it Subject and Object} \ \ (to D. M. {\Appleby})} \label{Appleby10}

Thanks for the note, which I very much enjoyed reading tonight (while
listening to Abbey Lincoln and Hank Jones in the background).  It did
my soul good, and I'll certainly be reading it again.

Let me, however, give you a quick first reaction to this:
\bma
I know we agree that physics isn't any kind of mirror.  Well:  it
seems to me that if you start trying to identify some parts of
quantum mechanics as ``objective'' and other parts as ``subjective''
then you are going back on that.

In my talk at Konstanz I mentioned the idea of Galileo and Descartes
that properties like ``redness'' don't faithfully depict properties
actually in the object, whereas properties like ``cuboidal'' do.
They thought, in other words, that the property of ``redness'' is
subjective whereas the property of being ``cuboidal'' is objective.
And they thought that if one eliminates all the subjective features
of the visual field, one will be left with an accurate reflection of
things as they are in reality:  a mirror of reality.  Quite possibly
I have misunderstood you.  If I have please correct me.  But when you
talk about identifying the parts of quantum mechanics having
objective content it sounds, to my ears, as though you are thinking
in a similar way to Galileo and Descartes. Purifying the reflection.
Polishing the mirror.
\ema

Might I ask you to go back to Sections 4 and 5 (but particularly
Section 5) of my anti-{\Vaxjo} paper \quantph{0204146} and tell
me whether you think the words in there seem to alleviate any of my
sins?  I had been planning to use the ``It's a Wonderful Life''
allusion to open up my {\sl Scientific American\/} article, so I'd
like to know what you think of it in this context.  (If you don't
know the movie, find a way to rent it and give it a watch.)

Honestly, I'm getting confused on these issues even myself of late. I
have been meaning to read Donald {\Davidson}'s essay titled ``The Myth
of the Subjective'' and see if that helped me any, but I haven't had
the gumption yet.

You're definitely making me think about these things.

\section{11-01-06 \ \ {\it Quantum Information and KS} \ \ (to J. H. Conway)} \label{Conway1}

Peter Winkler, whom I was visiting at Dartmouth recently, told me a little (a very little) about your ``marvelous free will theorem,'' as he called it, with Simon Kochen.  He suggested that I might contact you to learn more about it.  I have a very deep interest in Kochen--Specker type theorems, and would like to learn if there is something new and interesting here that I and collaborators could use in our Bayesian-style approach to the foundations of quantum information.

Winkler also suggested that I send you a CV and a paper or two to demonstrate that I might be worth talking to.  I'll do that in the next email.  In fact, I'll also send you a little research proposal I wrote up for Caltech the last time I thought about jumping ship from Bell Labs -- it makes it clear that I was hoping to build a whole curriculum around the KS theorem!  So, I do take it seriously.  (Moreover, I take issues about ``free will'' seriously; thus my side interest in William James, Charles Peirce, and all of pragmatism, which that document also hints of.)  At the moment, whatever you have sounds to be awfully intriguing.

I need to come to Princeton next week, in any case, to dig up some old papers.  If you'll be around sometime, I'll try to tune my schedule to yours, and I would very much like to have a visit with you to get something of an explanation of all this.

\section{11-01-06 \ \ {\it For the Record} \ \ (to N. D. {\Mermin})} \label{Mermin122}

Well, I'm back at home, safely again at Bell Labs, after having spent
some time at a meeting in Snowbird, Utah.  I am afraid I angered our
friend, Prof.\ Plotnitsky, with my talk.  My original title for the
talk had been ``Why I Never Understood Bohr's Reply to EPR, But Still
Liked It''---but I wrote it on two overlapping transparencies so
that, at the appropriate moment, I could strip off the part that said
``But Still Liked It.''  (I hadn't originally intended to do that,
but it was the only thing I could do with honesty after rereading
Bohr.) And so the talk went.  I explained how I didn't see much in
Bohr's reply that EPR hadn't anticipated in their second- or
third-to-last sentence, ``\ldots\ one would not arrive at our
conclusion if one insisted that two or more physical quantities can
be regarded as simultaneous elements of reality only when they can be
simultaneously measured or predicted.''  (Which is a conception they
pretty much simply dismiss.) Then I showed how the EPR criterion of
reality can nevertheless be made to implode through a combination of
perfect predictability (through entanglement) and KS noncolorable
sets.  Then I read some long passages from Einstein's
autobiographical notes in the Schilpp volume, and claimed that his
own logic was flawless:  The conclusion being that a quantum state
cannot correspond to a ``real factual state of affairs''---when
Einstein was right, he was really right!  Finally I asked, so what is
the uncertainty given by a quantum state about? And concluded with a
picture that was meant to capture much of what I put in the emails to
you a couple of weeks ago.

There's just not a lot you can do by considering only {\it two\/}
noncommuting variables, and as far as I can tell, that's all Bohr
ever really did.

But that's not why I'm writing you---i.e., to make my own
record---but rather to get you to come down on the record.  I ask
because I don't know how many meetings in the last year or two where
I've heard people praise the Ithaca interpretation or ``correlation
without correlata''---this meeting in Snowbird was one of them (Ivan
Deutsch being the most recent admirer).  And in all cases, I've said,
``I don't think {\Mermin} subscribes to those ideas anymore, or at least
not fully.''  But when asked what your problems are with your earlier
ideas, I don't know that I've had adequate answers of your own point
of view.  So, could I ask you again to what extent you now disavow
the II and exactly why?  I'd like to get it on the record so I don't
screw up when trying to represent you.

On another issue, have you ever looked at the section in Max Jammer's
book {\sl The Philosophy of Quantum Mechanics\/} on ``relational
conceptions of the quantum state''?  It seems like there's probably a
lot of material in there that would interest you.  I myself don't
recall ever having noticed that section before last week (though
surely I read it before, as I read the whole book cover to cover in
the summer of 1984 or 1985), and I found it quite interesting.  I
also found his section on ``latency theories'' at the end, where he
reports Margenau's view, very useful:  In particular, I've started to
wonder if, DISREGARDING the part where Margenau thinks of the quantum
state as objective, the rest of his view might correspond quite
nicely with where {\Schack} and I stand at the moment.  Margenau,
apparently, views quantum measurement outcomes as secondary qualities
in Locke's hierarchy, not primary ones, much like the idea of
blueness (which I think you call qualia).  I found myself wondering
if, in the end, my own view might not just boil down to that.  (This,
of course, is not unconnected to some points you were trying to make in
your ``What Is QM Trying to Tell Us?''\ paper.)

See, I can write a paper on certainty, but I never can be certain
myself!

\subsection{David's Reply}

\bq
I wish I ran into people who liked the IIQM.  Maybe I should go to
more meetings.   I probably should write some sort of update before
the 10th anniversary of the AJP article, but from my perspective it
seems vaguely auto-erotic, since I haven't detected anything like the
amount of interest in the paper that you describe.

It seems to me that I touched on much of what bothers me (and
bothered me even then) in the final section of the AJP paper,
reproduced below for your convenience:

\bq
\noindent XII. A FEW FINAL REMARKS \medskip

At the risk of losing the interest of those who --- like myself ---
read only the first and last sections before deciding whether the
rest is worth perusing, I conclude with some brief comments about
loose ends.

As noted at the beginning, what I have been describing is more an
attitude toward quantum mechanics than a systematic interpretation.
The only proper subject of physics is how some parts of the world
relate to other parts. Correlations constitute its entire content.
The actual specific values of the correlated quantities in the actual
specific world we know, are beyond the powers of physics to
articulate. The answer to the question ``What has physical reality?''
depends on the nature of ``what.'' The answer is ``Everything!'' if
one is asking about correlations among subsystems, but ``Nothing!''
if one is asking about particular values for the subsystem correlata.

This alters the terms of the traditional debates. Traditionally
people have been asking what correlata have physical reality. The
many different schools of thought differ by answering with many
different versions of ``Some'' while the IIQM answers ``None!'' The
question of what correlations have physical reality, which the IIQM
answers with ``All!'' has not, to my knowledge, been asked in this
context. While I maintain that abandoning the ability of physics to
speak of correlata is a small price to pay for the recognition that
it can speak simultaneously and consistently of all possible
correlations, there remains the question of how to tie this wonderful
structure of relationships down to anything particular, if physics
admits of nothing particular.

At this stage I am not prepared to offer an answer, beyond noting
that this formulates the conceptual problem posed by quantum
mechanics in a somewhat different way, and suggesting that there may
be something to be learned by thinking about it along these lines. I
suspect our unfathomable conscious perceptions will have to enter the
picture, as a way of updating the correlations. To acknowledge this
is not to acknowledge that ``consciousness collapses the
wavepacket.'' But it is to admit that quantum mechanics does not
describe a world of eternally developing correlation, described by
``the wave-function of the universe''!, but a phenomenology for
investigating what kinds of correlations can coexist with each other,
and for updating current correlations and extrapolating them into the
future. This phenomenology applies to any system that can be well
approximated as completely isolated.

A skeptic might object that the problem of how to update correlations
is nothing more than the measurement problem, under a new name.
Perhaps it is, but at least the problem is posed in a new context:
How are we to understand the interplay between correlation as the
only objective feature of physical reality and the absolute
particularity of conscious reality? Is something missing from a
description of nature whose purpose is not to disclose the real
essence of the phenomena but only to track down relations between the
manifold aspects of our experience? Is this a shortcoming of our
description of nature or is it a deep problem about the nature of our
experience?

\ldots\ By acknowledging that in our description of nature the
purpose is not to disclose the real essence of the phenomena, we free
ourselves to construct from the manifold aspects of our experience
formal representations of the systems we want to talk about. We have
learned how to express their possible correlations by an appropriate
state space, and the evolution of those correlations by an
appropriate Hamiltonian. By setting aside ``the real essence of the
phenomena'' we also acquire the ability to replace the befuddling
spectre of an endlessly branching state of the universe---as
disturbing in the self-styled down-to-earth Bohmian interpretation as
it is in the wildest extravagances of the many worlds
interpretation---with a quantum mechanics that simply tells us how we
can expect some of the manifold aspects of our experience to be
correlated with others. While this may sound anthropocentric, it is
my expectation that anthropos can be kept out of everything but the
initial and final conditions, and often---but not always---even out
of those. But this remains to be explored.
\eq

What can I add to that, 8 years later?

1.  You persuaded me quite soon that ``objective probability'' was
problematic.   Until I met you I had never taken the notion of
subjective probability seriously, or even know very much about it.
While I'm still not convinced (sorry) that you've got it right
either, I'm much more aware that one of the pillars of the IIQM is
much more fragile than I thought.

2.  Not unrelated to 1, the notion of ``correlation'' is not well
defined, beyond my assertion that it means nothing more than ``joint
distribution''.   But what does it mean to say that joint
distributions are fundamental, while conditional distributions, which
can be constructed from joints, have no physical meaning?   And what
are these joint distributions describing?   How are they tied down to
experience?   Which brings us to

3.  Consciousness.   Although I say that the problem of consciousness
should be set aside, when I went around giving physics colloquia on
the IIQM in the late 90's, during questions everything kept coming
back to conscious perceptions.   It ended up being too big a rug to
sweep problems under.    This is a point in your favor.  If
probability is subjective, then there is a subject (with conscious
perceptions) built in at the beginning, and consciousness becomes the
starting point, rather than a completely mysterious add-on.

4.  Quantum computation (which I only started studying after writing
the paper) made me much more sympathetic to Copenhagen and a purely
instrumentalist (positivist?)\ view of the subject than I had ever
been.   (See ``How I stopped worrying and learned to love Bohr.'')

Doubtless there's more, and perhaps I should think about writing
something more careful and considered, but since my views have yet to
settle down, that still seems premature.

So if you want something to tell these people who you claim exist,  I
think it's too strong to say that I disavow the IIQM.   But I do
regard it as at best incomplete (as I think I made clear in the 1998
paper).   I guess I was hoping somebody would take up the cause and
complete it.
\eq

\section{11-01-06 \ \ {\it MY Anthropocentrism vs.\ YOUR Anthropocentrism} \ \ (to I. H. Deutsch)} \label{DeutschI2}

I've been meaning to write you ever since arriving back in New Jersey, but first I had to clear out a huge backlog of other things.  Anyway, I wanted to say I very much enjoyed the long walk and talk we had together.  I felt like I got to know you more in those few hours than I had in the last ten years.

Let me clear up a couple of loose ends.  Among other things, I wanted to come back to the issue of whether our Bayesian view is more overtly anthropocentric than it ought to be.  In that regard, let me attach one of my pseudo-papers that addresses the issue.  Whenever you get the time, have a look at Sections 4 and 5.  I think they better capture some of what I was trying to get across the other day, even if the whole story is not yet worked out completely.

In the next email, I'll send you a pseudo-paper more particularly devoted to F-theory.  In that one it's Sections 6 and 7 that you might want to look at, as your {\it leisure\/} time permits.

\section{12-01-06 \ \ {\it It's All Your Fault} \ \ (to H. C. von Baeyer)} \label{Baeyer16}

\bhcvb
Back from the Bahamas, I find three books from Princeton U. P. in
payment for a proposal I reviewed for them.  {\em [One I requested]}
on the strength of your blurb: {\Omnes}'s {\sl Converging Realities}.

Just a quick question:  Should I really read it, or were you just
being kind to him (them)?
\ehcvb

I'm not tellin'.  Your interests and taste will decide the first part
of your question within the first couple of chapters.

As for {\Omnes}'s ability to stir the soul, here's an example of {\Omnes} at
his best:
\bq
   Perhaps the best way to see what it is all about is to consider what
   would happen if a theory were able to offer a detailed mechanism for
   actualization.  This is, after all, what the advocates of hidden
   variables are asking for.  It would mean that everything is deeply
   determined.  The evolution of the universe would be nothing but a
   long reading of its initial state.  Moreover, nothing would
   distinguish reality from theory, the latter being an exact copy of
   the former.  More properly, nothing would distinguish reality from
   logos, the time-changing from the timeless.  Time itself would be an
   illusion, just a convenient ordering index in the theory.  \ldots\
   Physics is not a complete explanation of reality, which would be its
   insane reduction to pure unchanging mathematics. It is a
   {\it representation\/} of reality that does not cross the threshold of
   actuality. \ldots\ It is wonderful how quantum mechanics succeeds in
   giving such a precise and, as of now, such an encompassing
   description of reality, while avoiding the risk of an
   overdeterministic insanity. It does it because it is probabilistic in
   an essential way.  This is not an accident, nor a blemish to be
   cured, since probability was found to be an intrinsic building block
   of logic long before reappearing as an expression of ignorance, as
   empirical probabilities.  Moreover, and this is peculiar to quantum
   mechanics, theory ceases to be identical with reality at their
   ultimate encounter, precisely when potentiality becomes actuality.
   This is why one may legitimately consider that the inability of
   quantum mechanics to account for actuality is not a problem nor a
   flaw, but the best mark of its unprecedented success.
\eq

\section{12-01-06 \ \ {\it Sounds of Silence and CBH, 2} \ \ (to G. Brassard)} \label{Brassard49}

Good to hear from you.  I've been meaning to write to you for a while---thanks for reminding me.  What I had wanted to write you about is that, as a climax to the van Fraassen / Halvorson seminar, Chris Timpson gave a talk as an invited speaker on the CBH theorem.  It was a very, very good talk, and {\it sadly\/} very convincing.  Particularly, he argued that CBH really, really (despite what Bub says) {\it make no use of\/} the no-bit-commitment axiom after all.  In other words, they are really only getting the structure of quantum mechanics from 1) $C^*$-algebras, 2) no-signaling, and 3) no-broadcasting.  No-bit-commitment comes for free.  Foundationally, I don't see that that is devastating at all---it just means that ``no-bit-commitment'' doesn't get the airplay you had wanted of it.  But as Timpson points out, it does indicate that the $C^*$-algebraic starting point is doing much, much more work in the derivation than had been expected.  My own opinion is that {\it starting\/} with algebras like that (even weaker ones like Segal or JB algebras), is just the wrong way to go---one needs a pre-algebraic starting point.

You can read about all this in Timpson's PhD thesis, which can be found here:
\bq
\arxiv{quant-ph/0412063}.
\eq
The relevant section is 9.2, and particularly pages 206 onward might interest you.

So, what is the right way to go?  That's what our postdoc would work on!  It sure would be good if we could scheme something up here.

\bgb
I'm trying to emerge but at this point it is clear that I shall fail to do so before I leave for Paris (QIP06). Will I see you there?
\egb

Unfortunately, no.  But I just looked up the schedule, and I found that I was licking my lips over a large fraction of the talks.  I'm very envious of you!

\section{13-01-06 \ \ {\it Proxy for Dick Jeffrey?}\ \ \ (to B. Skyrms)} \label{Skyrms3}

We've had a little email correspondence in the past, but I don't know if you remember me.

The reason I am writing you now is that I am thinking very hard about organizing a symposium for the PSA 2006 meeting in Vancouver, November 2--5.  My tentative title for the symposium is ``Radical Probabilism WITHIN Quantum Mechanics,'' and I'm guessing that the title is clear enough for you to surmise what the subject will be.  I have run over deadline for getting a proposal in, but have had some correspondence with the program committee chairs, and they think it will still be OK if I get something in ASAP.  At the moment, I am shooting to get a proposal in by Monday or Tuesday.

I'm sorry to put pressure on you like this (i.e., to give me a very quick reply), but I would very much like you to give us a talk, at least partially as a proxy for Dick Jeffrey.  What I mean by this is that I would like you to give a talk with your thoughts on these two papers by Jeffrey:
\begin{itemize}
\item
R.~C. Jeffrey, ``De Finetti's Radical Probabilism,'' {\sl
Probabilit\`a e Induzione---Induction and Probability}, edited by
P.~Monari and D.~Cocchi, (Biblioteca di Statistica, CLUEB, Bologna,
1993), pp.~263--275.

\item
R.~Jeffrey, ``Unknown Probabilities: {\it In Memory of Annemarie
Anrod Shimony (1928--1995)},'' Erkenntnis {\bf 45}, 327--335 (1997).
\end{itemize}
or any thoughts you have that are an extension to what he writes about there.  Despite the titles, the two papers (particularly the first one) have very large parts devoted to quantum mechanics.

My own opinion is that he definitely gets the subjective interpretation of quantum states right, but when it comes to what those quantum states are effectively degrees of beliefs about, he shoots off the mark a bit.  Nevertheless, his contribution and historical role in the ``quantum Bayesian'' project deserves to be fleshed out a bit---and I can't think of a better person to do it than you.

What I mean by the quantum Bayesian project is a locus of several new results that, I think, have started to make this way of viewing quantum states very respectable (in fact a hot topic for research).  Let me give you a reference to a couple of my own papers that may help orient you:
\begin{itemize}
\item
``Unknown Quantum States and Operations, a Bayesian View''\\
\quantph{0404156}

\item
``Quantum Mechanics as Quantum Information (and only a little more)''\\
\quantph{0205039}
\end{itemize}
Also, you might look at the website (containing all the abstracts) for our meeting last year titled ``Being Bayesian in a Quantum World''
\begin{center}
\myurl[http://web.archive.org/web/20090511060206/http://www.uni-konstanz.de/ppm/events/bbqw2005/]{http://web.archive.org/web/20090511060206/http:// www.uni-konstanz.de/ppm/events/bbqw2005/}.
\end{center}

Finally you might look at Itamar Pitowsky's paper:
\begin{itemize}
\item
``Quantum Mechanics as a Theory of Probability''\\
\quantph{0510095}
\end{itemize}
for another variation on the same theme.

I hope you get the feeling (that I think you should!)\ that this is now a very large area of research with much promise.  And it is an area of research that philosophers of probability can give much to.  Thus I hope you will be able to give us a talk, and will let me know of your decision as soon as possible.

Beside you and myself, I'm also hoping to attract Pitowsky to give a talk, and it would be nice if Abner Shimony would serve as a counterpoint to all of us.

\section{13-01-06 \ \ {\it Counterpoint to the Proxy for Dick Jeffrey?}\ \ \ (to A. Shimony)} \label{Shimony9}

I just wrote the note below to Bryan Skyrms, asking him if he would participate in a symposium for PSA 2006 that I am hoping to organize.  [See 13-01-06 note titled ``\myref{Skyrms3}{Proxy for Dick Jeffrey?}''\ to B. Skyrms.] Now, I would like to ask you to participate, in the capacity of trying to keep us all honest:  i.e., as a counterpoint to Skyrms, Pitowsky, and me.  I know that you have thought very deeply about the subject described below---and I know that you disagree with us---so I hope that you will join us and use your eloquence to explain to the audience why they shouldn't agree with us either!

Please read the note below, and you'll get a feeling for what I have in mind.  And if I could hear back from you as soon as possible, I might stand a chance of actually getting this thing organized before the program committee loses patience with me.

I hope that all is well with you, and that your new year has gotten off to a great start!

\section{13-01-06 \ \ {\it Proxy for Dick Jeffrey?, 2}\ \ \ (to B. Skyrms)} \label{Skyrms4}

Thanks for your note.  Well, this is what I get for waiting a month past the last minute!  Neither you, nor Shimony, nor Pitowsky even, is in a position to be able to attend!  So, I'm not quite sure what I'm going to do---maybe I'll have to drop the project.

I can certainly find others to talk as a ``quantum Bayesian'' but what was attractive about Pitowsky was that he's a certified philosopher---all the rest I know are physicists by trade.  I'll have to think.

But who else but you could really fill the shoes of commenting on Jeffrey's radical probabilism papers?  Do you have any suggestions?

Anyway, thanks; I much appreciate your quick reply.

And congratulations by the way on your PSA presidentship.

\subsection{Brian's Reply}

\bq
I believe that Dick liked Stephen Leeds on subjective probabilities and QM. But I don't think that Dick's own views ever got completely developed---especially with regard to measurement.
\eq

\section{15-01-06 \ \ {\it Things That Are Outside of Time}\ \ \ (to A. Y. Khrennikov)} \label{Khrennikov17}

\bakh
P.S. My best greetings to your wife! (Did she finish the
reconstruction of your house?)
\eakh

She will never be finished reconstructing!  It would be against her nature \ldots

\section{16-01-06 \ \ {\it Quantum Information Coincidence?}\ \ \ (to K. T. McDonald)} \label{McDonald1}

This morning by chance, playing around, I told my daughter Emma, ``Come here; let's see if your name has ever appeared on the internet.''  So we typed ``Emma Jane Fuchs'' into Google and got the result in 0.10 seconds.  There weren't a whole lot of documents to come up, but what was surprising is that one of them is posted at your website:  It's a copy of my samizdat {\sl Notes on a Paulian Idea}!  What on earth?  I see that you have it stored in a folder called ``examples,'' so I was left wondering ``example of what?''  I'm almost afraid to hear the answer \ldots

If for whatever reason you'd like a copy of the paperback version (published by {\Vaxjo} University Press), let me know and give me a proper mailing address:  I have dozens to give away.

\section{16-01-06 \ \ {\it Quantum Information Coincidence?, 2}\ \ \ (to K. T. McDonald)} \label{McDonald2}

Thanks for allaying my fears!  I feared that you had used the thing as an example of nonsense.

\bkmcd
I'd be pleased to receive a copy of ``Notes on a Paulian Idea''.
\ekmcd

I sent a note to the secretary in Sweden to send you a copy.  It'll probably arrive in a couple of weeks.

Your lecture notes for your quantum computation course are very, very impressive!  If this is the norm for a course and I ever do wiggle my way into Princeton, clearly I'll have to adopt a new work ethic!!

\bkmcd
I gather that you stray into quantum metaphysics from time to time, so you might or might not be amused by my brief foray into this realm:
\bq
\myurl[http://puhep1.princeton.edu/~mcdonald/examples/evasion.pdf]{http://puhep1.princeton.edu/$\sim$mcdonald/examples/evasion.pdf.}
\eq
On the whole it's better to leave metaphysics alone, but its temptations are hard for many of us to resist.
\ekmcd

Concerning your question of ``metaphysics'', it depends on what you mean by a ``hidden-variable theory'' and what you mean by ``cloning.''  If the hidden-variable theory is like Goldstein's version of Bohmian mechanics, which (they say) is empirically equivalent to quantum mechanics for all experiments, then the existence of those hidden variables amounts to nothing beyond their inventors' warm, comfortable feelings---the comfort one usually finds in a religion.

On the other hand, I can think of a perfectly good ``unhidden variable theory'' that still has a no-cloning theorem:  It is simply classical Newtonian mechanics where the question is whether Liouville distributions can be cloned.  Since Hamiltonian mechanics is phase-space volume preserving, you can't clone overlapping Liouville distributions any more than you can clone quantum states.  (I.e., in my view, quantum no-cloning is not a particularly deep theorem.  Certainly not one to get metaphysical about.  Save that up for the Kochen--Specker theorem.)

\section{16-01-06 \ \ {\it Bushwhacking} \ \ (to H. R. Brown \& C.~G. {\Timpson})} \label{BrownHR1}

The title of this email refers {\it not\/} to what I think ought to be done with George Bush's policies (though that's what I think ought to be done), but to some of the content below.

I had a chance to read your new paper after all today, and I very much like it.  I even think I agree with most of it.  Particularly, I very much agree with this point you make:
\bq
[I]t is worth recalling \ldots\ that the dominant viewpoint in the philosophy of space-time physics over the last few decades puts a
very different gloss on Minkowski's contribution to SR.  Far from
being the basis of a mere algorithm for SR, the current orthodoxy
seems to be that Minkowskian geometry provides a constructive
dimension to SR (though it is not always put in these terms), and
thereby significantly enhances its explanatory power.  According to this view, it is the structure of the Minkowski space-time in which
they are immersed that ultimately explains why rods and clocks in
motion contract and dilate respectively.
\eq

As far as I know, that is the dominant viewpoint, and in my own opinion, it is rightly viewed as Minkowski's most important contribution.

This is a point I tried to emphasize in my PSA talk last year; I don't know if Chris remembers it.  In the talk, I gave my usual spiel about how we won't ever really clear up the muddle at the foundations of quantum mechanics until we can formulate the essential content of the theory in a couple of crisp, clear, physical statements (drawing on a principle-theory formulation of SR as an {\it example}).  But then, for that particular audience, to make it clear that I set myself apart from Jeff, I included a new transparency to make the point that one shouldn't think of the finding of the principles as the end of the line---rather, they're the real beginning, from whence one can hope to make greater progress.  The transparency is attached.  I pointed out that from the principles, one finally had a clear enough view to make the leap to Minkowski spacetime---which I see as a unification of the principles and a tentative {\it ontology\/} underlying SR.  Most importantly, it was only after that tentative ontology was in place, that one had a starting point to move physics forward to General Relativity---namely, to incorporate gravitational phenomena, you curve the manifold.

So I see it with quantum mechanics:  first, (convincing) principles rather than Hilbert space rubbish; then, tentative ontology (what I use the word ``Zing!'' as a placeholder for at the moment); and finally, a test that the ontology has some cash value over-and-above the principle-theory phenomenology:  can the ontology be used successfully as a starting point for a broader, more powerful theory?  Metaphorically, what happens if you curve the Zing!?

Here's the way I tried to express the sentiment in my old paper
\quantph{0205039}:
\bq
Our foremost task should be to go to each and every axiom of quantum theory and give it an information theoretic justification if we can. Only when we are finished picking off all the terms (or combinations of terms) that can be interpreted as subjective information will we
be in a position to make real progress in quantum foundations.  The
raw distillate left behind---minuscule though it may be with respect to the full-blown theory---will be our first glimpse of what quantum mechanics is trying to tell us about nature itself.

Let me try to give a better way to think about this by making use of Einstein again. What might have been his greatest achievement in
building general relativity? I would say it was in his recognizing
that the ``gravitational field'' one feels in an accelerating
elevator is a coordinate effect. That is, the ``field'' in that case is something induced purely with respect to the description of an
observer. In this light, the program of trying to develop general
relativity boiled down to recognizing all the things within
gravitational and motional phenomena that should be viewed as
consequences of our coordinate choices.  It was in identifying all
the things that are ``numerically additional'' to the observer-free
situation---i.e., those things that come about purely by bringing the observer (scientific agent, coordinate system, etc.) back into the picture.

This was a true breakthrough.  For in weeding out all the things that can be interpreted as coordinate effects, the fruit left behind
finally becomes clear to sight: It is the Riemannian manifold we call spacetime---a mathematical object, the study of which one can hope
will tell us something about nature itself, not merely about the
observer in nature.

The dream I see for quantum mechanics is just this. Weed out all the terms that have to do with gambling commitments, information,
knowledge, and belief, and what is left behind will play the role of Einstein's manifold. That is our goal.  When we find it, it may be
little more than a minuscule part of quantum theory. But being a
clear window into nature, we may start to see sights through it we
could hardly imagine before.
\eq

So, as I've already said, I think I agree with you much.  However, I wouldn't use the words that you used for Einstein's initial formulation of SR---i.e., as a ``strategic retreat''---to describe this methodology (at least in the case of quantum mechanics).  Rather, I think of it as an effective method for ``clearing out the underbrush''---i.e., making the real lay of the land visible.  It's not a strategic {\it retreat\/} at all, but rather simply good strategy.  I have big dreams for where we might take quantum mechanics, but first we've got to clear out the underbrush of extraneous observer-dependent detail in it.  And I see no sure-fire and nonspeculative way of doing that but to \underline{first} reduce the theory to something like a principle formulation.

Anyway, good paper.

\section{16-01-06 \ \ {\it Lovely Word} \ \ (to C. G. {\Timpson})} \label{Timpson10}

By the way, what was that lovely word you used in your lecture at Princeton to describe information's relation to reality?  As I recall, it struck me because the idea it conveyed was so similar to the {\James}ian phrase I often use in my writings, ``numerically additional''---there was an example of that phrase in the note I just sent you and Harvey.  Here are some other ways I've used it in various pieces:
\bq\noindent
   Suppose one wants to hold adamantly to the idea that the quantum
   state is purely subjective.  That is, that there is no right and
   true quantum state for a system---the quantum state is
   ``numerically additional'' to the quantum system.  It walks through
   the door when the agent who is interested in the system walks
   through the door.  Can one consistently uphold this point of view
   at the same time as supposing \ldots
\eq
or
\bq\noindent
   William {\James} likes to say that all beliefs are ``numerically
   additional'' to the reality they take as their target, even
   ``true'' beliefs.
\eq
or
\bq\noindent
   Within quantum mechanics, there is an invariant piece which is
   common to all of us by the very fact of our accepting the theory.
   That is what we are in search of because in some sense---which need
   not pertain to a realistic conception of a theory's correspondence
   to nature---it is the core of the theory.  It is the single part
   that we agree upon, even when we agree upon nothing else.  In the
   direction I am seeking to explore, the quantum state is ``numerically additional'' to that core.  (That is, the quantum state is a compendium of Bayesian ``beliefs'' or ``gambling commitments'' and is thus susceptible to the type of analysis {\James} gives in his  ``Sentiment of Rationality.''  Our particular choice of a quantum state is something extra that we carry into the world.)
\eq

Would that nice word of yours nicely substitute into these paragraphs?  If so, I think I might adopt it.

\subsection{Chris's Reply}

\bq
You know what: I think it might be even better fitted for these contexts than for the ends I had in mind. The term was `adventitious' adj.: of the nature of an addition from without; extrinsically added, not essentially inherent (OED).
\eq

\section{16-01-06 \ \ {\it Little Phrases} \ \ (to D. M. {\Appleby})} \label{Appleby11}

A subject and an object, from a physicalist perspective, are simply
two objects.  Suppose the subject has a belief about something to do
with an object (even if only a belief about how some interaction with
the object will lead to a certain sensation in the subject).  To the
extent that the belief is a possession of the subject, and the object
need not abide by it, the belief is ``subjective.''  I.e., it is a
statement of a property of the subject.  To the extent that the
belief encodes a statement of the subject's history and composition
(perhaps its genetic makeup, the culture in which it was raised, the
accidental things that happened to it throughout its life, how much
alcohol it had to drink just before the belief, etc.), it is
``objective.''  For, it is a settled and inalienable possession of
the actual historical record of that small piece of the world.

Does any of this wordplay have anything to do with settling the
issues you wrote me about last week?

\section{17-01-06 \ \ {\it {\Wheeler} Meeting at Princeton} \ \ (to several at Princeton University)} \label{Lieb1} \label{Calderbank2} \label{vanFraassen13.2} \label{Halvorson9.05} \label{Kochen0} \label{Conway1.1} \label{McDonald2.1}

\noindent Dear colleague in the Princeton community,\medskip

Prof.\ Marlan Scully and I are organizing a somewhat impromptu but nonetheless serious meeting in Princeton to give a little tribute to John Archibald Wheeler and his ``law without law'' and ``it from bit,'' and also to discuss some recent developments in quantum information.  It'll be a one-day meeting with maybe 10 short talks and a little socializing, held Friday, February 24.  Wheeler has said that he will make an appearance, and we would like to invite you too.

Please let me know as soon as possible if you would like to come and if your schedule will permit it.  Also please say if you would like to give a talk:  Tentative titles are encouraged, even if some will have to be eliminated in the end (with a lottery) due to time constraints.

On this round of invitations, we're checking on the availability of the following Princeton locals to see if they can participate:  Rob Calderbank, John Conway, Bas van Fraassen, Hans Halvorson, Simon Kochen, Kirk McDonald, and Edward Nelson.

Already from further quarters, we have commitments of participation from:  Carroll Alley, Ken Ford, Daniel Greenberger, Mark Hillery, Benjamin Schumacher, Yanhua Shih, William Unruh, William Wootters, and Suhail Zubairy.

It looks to be a promising meeting; we hope you can participate!

\section{17-01-06 \ \ {\it Bushwhacking, 2} \ \ (to H. R. Brown \& C.~G. {\Timpson})} \label{BrownHR2}

\bhrb
In my opinion, after the weeding, you end up with the wavefunction, or the vector in the Hilbert space. I confess I cannot see why that is `rubbish'. Maybe you think it is too complicated, too messy. Too many theoretical principles? I am reminded of Einstein's saying about how good it is to simplify as much as possible --- BUT NO MORE.
\ehrb

I'll try to get my hands on your book soon; that'll be fun.

As for your other questions, I'll let Chris {\Timpson} be my proxy if he'll take the task.  He'd probably do a better job than I, anyway:  What you trim, depends on the first count as to what you see as the more likely solution:  that the state vector as `inherent' to the quantum system or rather `adventitious' (a lovely word I learned from Chris).  Anyway, you two are probably closer to a common pub; and I know you enough already, Harvey, to know that your defenses will have to be down before any of this worldview will ever seep into you.

\section{17-01-06 \ \ {\it Wheeler Meeting at Princeton}\ \ \ (to K. T. McDonald)} \label{McDonald3}

\bkmcd
Bill [Brinkman] is very nuts and bolts about bits.  He is active in encouraging us to get involved with the implementation of quantum information processing, more than addressing theory issues.  But I'd hate to see quantum information go the way of high-energy physics in which there's too huge a gap between theory and experiment.

My own view is that quantum information formalism has gotten way out
ahead of lab reality, and that the next phase of hard work is more in
the lab than not \ldots
\ekmcd

True, but what else can a theorist do but follow his interests.  At least the gap in quantum information theory and experiment is not like the gap between string theory and experiment:  There's NO speculation in quantum information, no claims that anything must be right because of `beauty'---the results are either right or wrong, because after all quantum information is just elementary quantum mechanics pushed a few steps further.

\section{17-01-06 \ \ {\it No `Plunging Ahead' After All} \ \ (to S. Hartmann)} \label{Hartmann13}

Well, all my hopes for a Bayesian-QM PSA symposium got dashed this weekend:  1) Pitowsky couldn't make it because of teaching duties, 2) Skyrms couldn't give a talk because he'll be giving the presidential address, and 3) Shimony had too many commitments too.  I could easily find someone to replace Shimony---he wasn't that crucial---(maybe you would have liked to have spoken too, or maybe Elga as Hans had suggested, or maybe Bill Demopoulos), but without the two headliners of Pitowsky and Skyrms---both of which had ingredients that I couldn't figure out how to replace---there didn't seem much reason to go on.  Oh well, c'est la vie; maybe two years from now I'll have a more thought-out plan.

I'm sorry to have troubled you about this, to have it all come to naught.

Hope you are enjoying California.  (Outside of the professional part, is there anything to enjoy in Irvine?)

\section{17-01-06 \ \ {\it Diptera} \ \ (to D. M. {\Appleby})} \label{Appleby12}

I'm not sure what I did to do it, but it looks like I really angered
you.  All I guess is that you took very seriously my one (flippant?)\
use of the work ``physicalist,'' which probably has a lot of
connotations for you.  I'm not completely sure what it means (other
than the Webster definition), but I was trying to search for a word
that conveyed the idea that there is no true divide between subject
and object in a Kantian or Schopenhauerian sense.  What do you call
that?

Did you read the {\Davidson} essay ``The Myth of the Subjective''?  What
would you call that kind of idea?  Maybe not physicalism, but then
what?

You write, ``You are probably thinking that the physicalist
perspective you ask me to adopt \ldots''  But I didn't ask you to
adopt anything.  I ASKED, ``Does any of this wordplay have anything
to do with settling the issues you wrote me about last week?''  It
was my meager attempt to see if that was what was at issue; it was a
(lame?)\ attempt to clarify things to my mind.

\section{18-01-06 \ \ {\it New York Times Quote} \ \ (to myself)} \label{FuchsC13}

Betty Friedan in {\sl The Feminine Mystique}:
\bq\noindent
Down through the ages man has known that he was set apart from other animals by his mind's power to have an idea, a vision, and shape the future to it \ldots\ when he discovers and creates and shapes a future different from his past, he is a man, a human being.
\eq

\section{18-01-06 \ \ {\it What I Must Have Meant} \ \ (to D. M. {\Appleby})} \label{Appleby13}

I searched my memory on and off though my sleeping/sleepless hours
last night to see if I could find a clue to why I had used the word
``physicalist'' in my note to you.

Well, I found it:  It came to me through {\Rorty}.  ``Nonreductive
Physicalism'' is what he calls his view and what he calls {\Davidson}'s
in the ``The Myth of the Subjective'' and other places.  So, I
finally read {\Davidson}'s essay, and I RE-read {\Rorty}'s essay
``Non-reductive Physicalism'' in his book {\sl Objectivity,
Relativism, and Truth}.  And I think that view approximates to some
extent the direction I'd like to go.

Did you ever read {\Rorty}'s introduction to that volume, like I had
once suggested to you?  (I'll place my old note suggesting that
below.)  He states quite clearly there what he means by
antirepresentationalism---I think it is a view that approximates our
own two views (i.e., yours and mine).  Yet, later in the book {\Rorty}
has this essay ``Non-reductive Physicalism,'' and he certainly
doesn't see it at cross-purposes with his introduction.  So my
question is, what do you think of it?  Is it a meaning for the word
``physicalism'' that you could accept?

\section{18-01-06 \ \ {\it Integers and Real Numbers} \ \ (to D. M. {\Appleby})} \label{Appleby14}

Regarding this piece of your note:
\bma
I suppose one might argue, with some plausibility, that the integers
are a purely logical structure, devoid of physical content.  But I
have doubts even about that.  Not only does the concept of an integer originally arise in the context of counting sets of physical objects.
That is the still the way it is defined in  axiomatic set theory. And I would say that the concept of an object is a physical concept,
which embodies assumptions about the world.  Ditto the concept of a
set.  Furthermore, it seems to me that there is something very
classical about the concept of a set. The concepts of object and set
clearly have an approximate, everyday phenomenological validity.  But it seems to me that they have ``human'', in fact ``primitive human'',
stamped all over them.  I see no reason to assume that objects and
sets exist in ultimate reality.  Indeed, to my mind quantum mechanics rather suggests the opposite.
\ema
Below is a note that Danny Greenberger once wrote me, which I liked a lot.  Maybe you'd be interested in reading it in this connection.  (And maybe I'll cc this note to {\Mermin}, as he and I once had a discussion on the integers---I think it was in Capri---and I took the stubborn position that even they don't exist by themselves \ldots\ much like you seem to suspect.  I'm not sure I ever shared Danny's note with David.) [See 06-12-02 note titled ``\myref{Greenberger1}{Enjoyed}'' to D. M. Greenberger.]

\section{18-01-06 \ \ {\it {\Wheeler} Meeting at Princeton} \ \ (to E. H. Lieb)} \label{Lieb2}

Good to finally meet you; I've been a fan for many years.  (For a while, when I was working on my PhD, I thought the key to all the secrets of the universe might be found in one of your papers on operator-Schwarz inequalities.)

\behl
Marlan suggested a review of strong subbadditivity, but this
is well known stuff, or is it? Probably you have more up to
date material in mind.
\eehl

Well, the meeting is a bit of a hodgepodge of younger and older guys---the only common thread really being John Wheeler's influence on quantum foundational thought in the past.  Certainly, strong subadditivity is old stuff to some of the guys in my side of the invitation list---say, Schumacher, Wootters, maybe Unruh, etc.  But I doubt Greenberger, Alley, and others (including Marlan!)\ know anything about it.  So, I think it couldn't hurt to give a talk about it.  Maybe you could squeeze statements of some of your newer theorems in as asides \ldots\ but the talks will probably be pretty short if we try to get 10 of them in in one day.

\section{19-01-06 \ \ {\it The Wheeler Meeting} \ \ (to D. B. L. Baker)} \label{Baker13}

Well the holidays came and went and I didn't write you a thing.  I thought I would, and maybe send you some new pictures, but time just slipped away---it always seems to slip away now.

But I found myself thinking about you again yesterday, particularly that incident in 1984 when John Wheeler called your mom because you hadn't turned in a term paper for his class.  John was, I think, 73 then.

He's almost 95 now.  I've been involved in organizing a little conference for him at Princeton, and so all kinds of memories have been coming to me.  In this connection, I got the nice note (and attachment) below from Ken Ford, John's biographer and former student, that gives some sense of what the man is like now.  It was Ken's line about ``courtliness'' that made me think about John's calling your mother.  I hope you have MS Word installed, so that you can read the attachment; if you don't let me know and I'll put the words into a plain text file.

Mostly, Ken's closing lines got to me.  It kind of reminded me of that Joan Baez -- Kris Kristofferson duet, ``Hello in There''.  I think the song was written by John Prine.  Do you know it?  If you don't, some approximation to the lyrics can be found here:
\begin{center}
\myurl{http://www.cowboylyrics.com/lyrics/prine-john/hello-in-there-10843.html}.
\end{center}
I hope you and the family are all doing well.

\section{19-01-06 \ \ {\it Elevator Stories} \ \ (to A. Cho)} \label{Cho3}

Let me send you one last little thing to read while it's on my mind.  You might enjoy looking at the introduction and conclusions to this paper:  \quantph{0404122} (particularly the elevator stories).  Anyway, this has to do with the question you asked near the end of the phone conversation about what I would like to see experimentally explored next---this paper talks about that optimal alphabet I was telling you about.

Good luck with your writing.

\section{22-01-06 \ \ {\it QCMC, Hirota, and Giacometti}\ \ \ (to O. Hirota)} \label{Hirota3}

Thank you for the invitation to serve on the advisory committee for QCMC; I graciously accept.

It is a funny coincidence that I open my email to find a note from you this morning.  The reason is because, before turning on my computer, I had planned to write you a note myself this morning!!  This is really true.  Last week, I took my older daughter Emma to the New York Metropolitan Museum of Art to celebrate her 7th birthday.  All week, I have been meaning to tell you about how we saw two pieces of Alberto Giacometti's work there!  Did you see those pieces when you visited New York?  One was a sculpture, called ``Three Men Walking'' I think.  And the other was a painting of his mother, I believe.  They were striking!

I very much look forward to QCMC this year.  It has now been several years since I have been to Japan; I miss it very much.

I hope you will sometime come to visit us in our new home in New Jersey.    It is a big old Victorian home (from 1895) that my wife has been reconstructing wildly.  Most recently she painted it orange!!  (I will attach two pictures.)  It would be wonderful to talk of quantum information and noncausality within these orange walls with you!  And, Kiki will cook for you grandly!

\section{22-01-06 \ \ {\it Stranger Still!}\ \ \ (to O. Hirota)} \label{Hirota4}

Me again.  Reviewing some of my old emails to you, I discovered something stranger than the coincidence I wrote you about this morning!  I discovered a whole set of coincidences with you.  Note, for instance, what I wrote you on 1 June 2004:
\bq\noindent
   It was a coincidence that you wrote me today.  I had been meaning to
   write you again about the QCMC award---and I was going to do it
   today!
\eq
or as another example what I wrote you 27 May 2003:
\bq\noindent
   It was an interesting coincidence that I would get your book today.
   Just this morning, the secretary in Sweden told me that she mailed
   off several copies of my samizdat {\sl Notes on a Paulian Idea}.  You
   were in the list of recipients.  So it should be arriving soon.
\eq
Non-causality!!

\section{25-01-06 \ \ {\it Beyond QM and So On}\ \ \ (to A. Y. Khrennikov)} \label{Khrennikov18}

\bakh
And finally: I got again the proposal from AIP to publish Conf.\ Proc.
Would you like to be one of the Editors? It would be natural (then at the end you should write about people
from your team in the preface and make selection of papers related to
quantum information).
\eakh

As for my being an editor, that will be fine.  And yes, I will write an introduction for my team.

By the way, I met Simon Kochen for the first time Monday, and had an extended discussion with him (2.5 hours).  The main point I got from him is that he thinks the essence of quantum mechanics is that it is a {\it contextual probability theory}, and he thinks that he and John Conway have some strong mathematical results that support that claim.  I thought of you as he told me all that.

\section{26-01-06 \ \ {\it Some Words to Use, Maybe?}\ \ \ (to R. E. Slusher)} \label{Slusher11}

The great advantage of the controlled cold collision technique for optically trapped neutral atoms is the ease and simplicity with which it can generate Raussendorf--Briegel ``cluster states'' over any number of qubits [1]. These are entangled quantum states with the remarkable property that they are a {\it universal\/} resource for quantum computing [2]. If a technique can produce these states deterministically, then the remainder of any quantum computation amounts to simply performing single qubit measurements thereafter.  Moreover, the complexity class structure of this type of computation may be somewhat better than the older unitary gate style (or circuit model) of quantum computation because it allows for many of the measurements required of an algorithm to be performed in parallel---this is something the circuit model cannot support [3]. The discovery of the cluster-state method of quantum computation should not be underestimated:  It is among the top four theoretical discoveries in the history of the field (up there with the factoring algorithm, database search, and the existence of quantum error correction and fault tolerance).

However, beyond the technology-enabling universality property already discussed, cluster states are also of significant raw physical interest [4]. For instance, among the class of entangled states over $N$ qubits, there is a sense in which cluster states have the most persistent entanglement:  A user must perform measurements on at least $N/2$ qubits of the original $N$ before the entanglement in the remainder is completely depleted.  This is a property that Bell, GHZ, and error-correcting code states do not have.  Also, cluster states have the interesting property of being maximally connected:  By performing local measurements, one can teleport between any two sites in the array.  This physical phenomenon that has not been demonstrated in an array before.  Finally, the cluster-state model of quantum computation provides a test-bed for quantum foundational studies that take quantum measurement as the primary nonclassical feature of quantum mechanics [5], for, among other things, they show that even unitary evolution is ultimately representable in measurement and entanglement terms alone [6].

\begin{itemize}
\item[{[1]}] Recently, there have been a few experiments producing $N=4$ cluster states (and possibly $N=5$ in unpublished work)---see, for instance, P. Walther, et al., ``Experimental One-Way Quantum Computing,'' Nature {\bf 434}, 169 (2005); P. Walther, et al., ``Experimental Violation of a Cluster State Bell Inequality,'' Phys.\ Rev.\ Lett.\ {\bf 95}, 020403 (2005).  However the production of these states rely on down-conversion techniques that cannot produce cluster states {\it deterministically}.  Moreover, the experimental results reported are always of the post-selected variety:  I.e., the cluster state must be destroyed before it can be said to have {\it been\/} there.

\item[[{2]}] R. Raussendorf, D. E. Browne, and H. J. Briegel, ``Measurement-based Quantum Computation with Cluster States,'' Phys.\ Rev.\ A {\bf 68}, 022312 (2003).

\item[{[3]}] R. Jozsa, ``An Introduction to Measurement Based Quantum Computation,''\\ \quantph{0508124}.

\item[{[4]}] H. J. Briegel and R. Raussendorf, ``Persistent Entanglement in Arrays of Interacting Particles,'' Phys.\ Rev.\ Lett.\ {\bf 86}, 910 (2001).

\item[{[5]}] C. A. Fuchs, ``Quantum Mechanics as Quantum Information (and only a little more),'' \quantph{0205039}.

\item[{[6]}] M. A. Nielsen, ``Cluster-state Quantum Computation,'' \quantph{0504097}, to appear in Rep.\ Math.\ Phys.
\end{itemize}

\section{27-01-06 \ \ {\it Wheeler Quantum Information Meeting, February 24--25} \ \ (to the Wheelerfest participants)} \label{WheelerfestInvite}

\noindent Dear Colleague,\medskip

If you are receiving this email, it means that I have heard back from you with a firm or tentative yes that you will be able to participate in the ``Wheeler -- Quantum Information Meeting'' that Marlan Scully and I are organizing at Princeton.  The final dates for the meeting have now been set:  The meeting will run all day Friday February 24, reconvening Saturday morning February 25 for a further session (until noon).  Friday night, we will have a conference dinner.

If you have not done so yet, please send me a tentative title for your talk.

Presently we are either hoping for or expecting talks from:  Carroll Alley, Rob Calderbank, Ken Ford, Chris Fuchs, Danny Greenberger, Hans Halvorson, Mark Hillery, Simon Kochen, Elliot Lieb, Stephen Lyon, Kirk McDonald, Edward Nelson, Wolfgang Schleich, Ben Schumacher, Marlan Scully, Robert Seiringer, Yanhua Shih, Bill Unruh, William Wootters, Suhail Zubairy, Wojciech Zurek, and maybe John Conway and Anton Zeilinger.  Also, there will likely be some local students and postdocs giving talks.

Attached are instructions for making your hotel reservation and suggestions for how to get from the local airports to the hotel; we strongly suggest you fly into Newark airport if you are flying.  For the Friday of the meeting, all talks and the conference dinner will be held at a conference center near the Hyatt hotel; shuttle service from the hotel to the conference center will be provided by grad students.  The Saturday morning session will convene at a lecture hall on Princeton University campus; again shuttle service will be provided.

The present plan is that John Wheeler will be attending the Friday morning part of the meeting.  To give some sense of what John is (still) interested in, I'll also a attach a nice article that Ken Ford wrote recently for the Princeton Physics Department Newsletter.

Marlan and I look forward to seeing you all.  And I will be communicating with you all again before the date with further details.

\section{27-01-06 \ \ {\it Monday or Tuesday Meeting} \ \ (to H. Halvorson)} \label{Halvorson9.1}

\bhh
Thanks for coming to Princeton yesterday.  I really did appreciate the
fairly comprehensive overview of your current perspective.  By the
way, regarding upcoming meetings, I could also come up your way
sometime, or we could try to find a midpoint meeting place.  In any
case, I would like to try to maximize the chances that we will meet
again soon and often.
\ehh

I enjoyed it too, and it was nice to learn a little more about the things you're thinking and about our common ties through Davidson.  Thanks for your nice offer too; I'll certainly take you up on it sometime.

In the meantime, though, I was thinking about coming to Princeton Monday or Tuesday (if everything continues to go OK with my step-dad):  My in-laws are visiting, and I thought my father-in-law might like to explore a little of the town and campus.  So to kill two birds with one stone, maybe I could have a meeting with you while he's entertaining himself.  Would you have any time Monday or Tuesday?

By the way, I thought of something I should tell you with regard to Hardy's axioms.  You said you were particularly interested in Hardy's continuity axiom.  But that axiom, at least within the context of the other ones, is particularly weak.  In its place, Hardy could have inserted almost any distinction between classical probability theory and quantum mechanics, and it would have still done the trick.  For instance, instead of requiring the existence of a continuous transformation between pure states, he could have outlawed broadcasting.  That would have been just as effective for narrowing him down to quantum mechanics (as all one needs in that spot is a statement that works in the quantum case but not the classical one).  Thus, his axiom structure is not particularly tight.

\section{27-01-06 \ \ {\it Degrees of Freedom / Distinguishable States} \ \ (to H. Halvorson)} \label{Halvorson10}

Also, I spent a little time this morning trying to search my archive
of mumbles to see if I could find something that decently expressed
my distaste for taking ``the number of degrees of freedom'' or ``the
maximum number of distinguishable states'' as the foundation for
Hilbert-space dimension.  Unfortunately, I wasn't as successful as I
had hoped I would be:  Maybe it means my memory is failing, and I
haven't really yet written a clear statement of what I'm thinking.

So, let me give you the pointers that I could find in the meantime:
\begin{itemize}
\item
``On the Quantumness of a Hilbert Space,'' \quantph{0404122}
\item
pages 159--160 of {\sl Quantum States: W.H.A.T.?}
\end{itemize}

I want Hilbert-space dimension to denote not (a priori at least)
number of distinguishable states, but rather something to do with a
quantum system's sensitivity to the touch or its creative power.  And
that's partially what I am trying to get at with ideas like
``quantumness''.

\section{30-01-06 \ \ {\it Loose Ends and Pedagogy}\ \ \ (to K. T. McDonald)} \label{McDonald4}

It took me quite a while, but I finally found it.  I'm talking about the John Wheeler passage on $\hbar$ that I mentioned to you.  Let me paste it here:
   \bq
   The quantum, $\hbar$, in whatever correct physics formula it appears,
   thus serves as a lamp. \ldots

   Giving us its as bits, the quantum presents us with physics as
   information.

   How come a value for the quantum so small as $\hbar=2.612\times
   10^{-66}$ cm$^2$?  As well as ask why the speed of light is so great
   as $c=3\times 10^{10}$ cm/s!  No such constant as the speed of light
   ever makes an appearance in a truly fundamental account of special
   relativity or Einstein geometrodynamics, and for a simple reason:
   Time and space are both tools to measure interval.  We only then
   properly conceive them when we measure them in the same units.  The
   numerical value of the ratio between the second and the centimeter
   totally lacks teaching power.  It is an historical accident.  Its
   occurrence in equations obscured for decades one of nature's great
   simplicities.  Likewise with $\hbar$!  Every equation that contains
   an $\hbar$ floats a banner, ``It from bit''.  The formula displays a
   piece of physics that we have learned to translate into
   information-theoretic terms.  Tomorrow we will have learned to
   understand and express {\it all\/} of physics in the language of
   information. At that point we will revalue $\hbar=2.612\times
   10^{-66}$ cm$^2$---as we downgrade $c=3\times 10^{10}$ cm/s
   today---from constant of nature to artifact of history, and from
   foundation of truth to enemy of understanding.
   \eq

I certainly agree with significant pieces of that.  Particularly, through quantum information theory I think we are already seeing strong hints of how to formulate the essence of quantum mechanics without ever making mention of $\hbar$.  The idea is to let finite dimensional Hilbert spaces lead the way to telling what is really quantum and what is not.

It probably won't add much to that statement, but let me record that I really believe it by attaching a teaching proposal I wrote for Caltech the last time I thought about jumping ship from Bell Labs (a couple years ago).  At the time, there was a small chance of an interdisciplinary position between physics, philosophy, and engineering (through their Information Science and Technology initiative), and the proposal was written with that mix-match in mind.  The relevant part for you is the section on ``Innovative Undergraduate Quantum Mechanics.''  It took a little nerve to write that stuff, but I think it needs to be said.

I've been meaning to write you since returning from NIST, but I wanted to find the Wheeler quote first.  That meeting really fired me up for this DTO proposal we're writing.  Let me ask you this.  As you get a chance, have a look at:  O. Mandel, {\it et al}., ``Controlled Collisions for Multi-Particle Entanglement of Optically Trapped Atoms,'' {\sl Nature\/} {\bf 425}, 937--940 (2003).  If we (Lucent) could use our MEMS spatial light modulators to create 50,000 individually movable atom traps for enacting these controlled cold collisions---as we think we can in the near-term future---to set the scene for cluster-state quantum computation, would that count in your mind as ``doing physics'' within quantum information?  I'm just trying to get a feeling for the sorts of distinctions you seemed to be making at lunch the other day.

\section{30-01-06 \ \ {\it Loose Ends and Pedagogy, 2}\ \ \ (to K. T. McDonald)} \label{McDonald5}

\bkmcd
Your teaching proposal to Caltech looks like it might find better
reception in a philosophy department than in a physics department.
\ekmcd

Well, yes.  The potential position was explicitly to be interdepartmental if it materialized (it didn't materialize at the time).  I guess you're telling me my proposal looks to have left out the other half \ldots\ \ldots\ ouch.

\bkmcd
A clarification.  I have somehow (without being well informed) come to
associate ``Bayesian'' with the notion that there is no such thing as
intrinsic randomness in Nature.

I hope that this term means something different to you.
\ekmcd

\bkmcd
But I am a bit worried by the quotation:
``Bayesian Statistics offers a rationalist theory of personalistic beliefs in
contexts of uncertainty, with the central aim of characterizing how an
individual should act in order to avoid certain kinds of undesirable
behavioral inconsistencies.''

I interpret this as implying that Bayesian Statistics provides a kind of
deprogramming for those misguided souls who take too seriously the claims
that intrinsic probability plays a key role in Nature.
\ekmcd

I'll consider how to best respond to your other queries over the rest of the day.  In the meantime, let me ask if Sections 3, 4, and 5 of [\arxiv{quant-ph/0204146v1}] don't already answer your questions relatively directly?  And within that maybe pay particular note to the discussion near the ALL-CAPPED ``reality'' on page 6 and the last two paragraphs of page 8 for your question on ``intrinsic randomness.''

\section{30-01-06 \ \ {\it Wheeler Accuracy?}\ \ \ (to J. Preskill)} \label{Preskill16}

Marlan Scully and I are organizing a little impromptu meeting for John Wheeler and quantum information at Princeton Feb 24 \& 25, and I think I want to open my talk with that story you once told me about the flying equations.  It's recorded at the top of page 149 of my {\sl Notes on a Paulian Idea}, which you can get off my webpage below.  I wonder if I could ask you to check that write-up for accuracy, and tell me any other tidbits you remember that might be of use for setting the mood of the story.  I just want to make sure I get the dramatic effect right.

After that intro I will spend some time talking about our Bayesian view of QM:  That the formal structure of quantum mechanics should not be viewed as some kind of mathematical mirror image of the real world, but rather as a layer on top of Bayesian probability theory (molded by the real world).  It represents how an agent should tie together his gambles when the subject matter is the agent's ``kicking'' the external world and waiting for the reaction it produces within him (formerly known as quantum measurement).  Then, to end with making a connection back to John, I want claim that the disconnect between the ``real world'' and ``QM as normative theory of gambling'' is precisely the sort of space one needs to get the kind of flight John was talking about---no space, no flight.

I'm surely responsible for all the errors in the latter part of the talk, but I hope you can help me to be as historically accurate as possible in the first part of it.

By the way, if you're interested in coming (and/or talking), please do come.  I could send you the logistical information if you're interested.  John will make an appearance for two or three hours the Friday morning.  So far, outside of the Scully crew, I've only been inviting his old students and Princeton locals; but now that I think of it, you might possibly be interested too.  I'll paste in the tentative speaker list below.  I think everybody in the list will be coming except (probably) Zeilinger, Zurek, and Schleich---I just don't have talk titles yet from most (only previous confirmations).

\subsection{John's Reply}

\bq
Well, the moral of the story is correct. The course was actually called Honors Physics and it was for sophomores intending to major in physics. It started with Goldstein {\sl Classical Mechanics}, but actually we also did E{\&}M, relativity, statistical physics and quantum physics, all from a rather idiosyncratic perspective.

My recollection is that he didn't ask us for ``all the equations'' but for the ones we thought were most important and fundamental, the ones that succinctly sum up the known laws of physics. Otherwise, though, you have captured the moment pretty well, to the best of my recollection. The phrase ``But the universe flies'' had just the right tone of wonder, and there was a dramatic pause following, to let it sink in for a few seconds.

The meeting sounds like fun. Sorry I can't be there.
\eq

\section{30-01-06 \ \ {\it Locus of Free Will}\ \ \ (to H. Barnum)} \label{Barnum21}

In August 1999 you wrote me this:
\bhb
Ah!  I have come to understand (or remember, perhaps) the importance
of leaving a place for free will in your views on the foundations of
qm. \ldots

Here's a caricature, so feel free to object:
Bell's worry about the foundations of QM has been:  that we have
``measurement'' as an ``unanalyzed primitive'' of the theory.  Everett
shows us how to get around that.  You don't like Everett's resolution
because you {\bf want} to have an unanalyzed primitive around so it can be
the locus of free will.
\ehb
and I just thought of your writing it via our correspondence a few minutes ago in conjunction with a note I had written John Preskill a few minutes before that.  I'll place the note to Preskill below.  [See 30-01-06 note titled ``\myref{Preskill16}{Wheeler Accuracy?}'' to J. Preskill.]

Have you ever read the Wheeler story I refer to?  Anyway, it struck me that maybe I'm still talking about that same locus!

\section{30-01-06 \ \ {\it Island of Misfit Toys} \ \ (to K. T. McDonald)} \label{McDonald6}

\bkmcd
Looking over your anti-{\Vaxjo} note,  I infer that I am much less concerned with the relation of human beings to quantum theory than you.  So I doubt that I will be of much use to you as a sparring partner.
\ekmcd
Yeah, you're probably right.

\bkmcd
Instead, I'm led to infer that your view is that physics $=$
psychology.
\ekmcd
I must be a horrible expositor if that is what has come through to
you.

\bkmcd
Probability arises in situations involving people.
\ekmcd
No, it is that probability theory is as much prior to empirical
science and the peculiar properties of the objective, physical,
man-independent world (which I have no doubt is out there) as number
theory and mathematical logic are.  That is why we can practice
probability theory (just as we can mathematical logic) without
knowing anything necessarily of physics.
\bkmcd
People have lots of subjective impressions.
\ekmcd
They do, and I'm gaining one.
\bkmcd
Therefore, the notion of probability is probably subjective.
\ekmcd
Non sequitur.

The Bayesian view of probability, as developed for instance in
\begin{itemize}
     \item
     E.~T. Jaynes, {\em Probability Theory:\ The Logic of Science},
     edited by G.~L. Bretthorst (Cambridge University Press, Cambridge,
     2003).

     \item
     J.~M. Bernardo and A.~F.~M. Smith, {\sl Bayesian Theory} (Wiley,
     Chichester, 1994).

     \item
     L.~J. Savage, {\sl The Foundations of Statistics\/} (Dover, New
     York, 1972).
\end{itemize}
is that one's {\it initial\/} probabilities (for anything) are
largely beyond one's control.  For they depend upon one's particular
history, the accidents of one's learning, even the bad lessons one
might have received in college.  So, in that sense, probabilities are
``subjective.''  Initial probabilities are like the initial
conditions of a pendulum in a physics problem; their origins are left
unanalyzed as a matter of principle.  The best one can do is trace
back the initial condition of a pendulum to the initial condition of
some larger system (whose origin again goes unanalyzed), and so
similarly with probability---the best one can do in analyzing the
origin of some prior probability is to trace it back to some further
prior (whose own origin then goes unanalyzed).

What is {\it not\/} subjective and not dependent on history in
Bayesian probability theory, however, is how probabilities {\it
should\/} fit together.  As an example, if you gamble with odds
$P(A)$ for some event $A$, and $P(B)$ for some event $B$, and gamble
with odds $P(A\vee B)$ for the disjunction of $A$ and $B$, and with
$P(A\wedge B)$ for the conjunction of $A$ and $B$, then it had better
be the case that your numbers fit together according to this relation
$$
P(A \wedge B) = P(A) + P(B) - P(A \vee B)
$$
or you will lay yourself open to a sure loss.  (This is called the
Dutch book argument.)  Thus it is the transformation rules between
probabilities that are {\it objective}---the edict that one should
strive to make sure that one's probability assignments meet the
constraints of all Dutch book arguments is something that does not
depend on the peculiarities of one's actual history.  In that sense,
the transformation rules are not ``subjective,'' but rather
``objective.''

\bkmcd
\label{Hoompah!}
A clarification.  I have somehow (without being well informed) come
to associate ``Bayesian'' with the notion that there is no such thing
as intrinsic randomness in Nature.

I hope that this term means something different to you.
\ekmcd

Your association is the common one.  A.~J.~M. Garrett, for instance,
expresses it very eloquently,
\bq
   The nondeterministic character of quantum measurement can,
   and should, be taken to imply a deeper `hidden variable' description
   of a system, which reproduces quantum theory when the unknown values
   of the variables are marginalised over.  Differences in measurements
   on identically prepared systems then represent differences in the
   hidden variables of the systems.  Not to seek the hidden variables,
   as the Copenhagen interpretation of quantum mechanics arbitrarily
   instructs, is to give up all hope of improvement in advance, and is
   contrary to the purposes of science.
\eq
But {\Caves}, {\Schack}, {\Appleby}, Peres, Leifer, I (and I hope a few
others) are renegades from that.  If what you mean by ``intrinsic
randomness'' is that quantum measurement outcomes do not pre-exist
the measurement process---that they are made ``on demand''---and that
there is no fact of the matter in the universe that will determine
which way they will go, then {\Caves}, {\Schack}, {\Appleby}, Peres, Leifer,
and I all accept that.

But that does not contradict the main Bayesian point, which I
attempted to express in the section above:  That initial
probabilities are {\it subjective}, while the transformation rules
and the way probabilities tie together are not---they are {\it
objective}. Garrett-style hidden variables, regardless of what
Garrett might say, are thus not necessary for a Bayesian view of
quantum probabilities.

Which leads to this point:
\bkmcd
But I am a bit worried by the quotation: ``Bayesian Statistics offers
a rationalist theory of personalistic beliefs in contexts of
uncertainty, with the central aim of characterizing how an individual
should act in order to avoid certain kinds of undesirable behavioral
inconsistencies.''

I interpret this as implying that Bayesian Statistics provides a kind
of deprogramming for those misguided souls who take too seriously the
claims that intrinsic probability plays a key role in Nature.
\ekmcd
Quantum measurement is a ``context of uncertainty''---and it is {\it
because\/} of ``intrinsic randomness,'' as defined idiosyncratically
(but carefully) above.  As far as I can tell, that has nothing to do
with psychology---that is now a statement about our empirical world.
But that has no implication that quantum probabilities have to be any
more objective than Bayesian ones---they are still as much dependent
on the gambling agent's peculiar history as any agent's probabilities
in {\it any\/} context (because even the quantum states one ascribes
to a system depend on unanalyzed prior probabilities).

What Bernardo and Smith are talking about in that quote is the {\it
objective\/} content of Bayesian Statistics:  It is that one should
strive to be invincible to a Dutch book.  (Not being invincible is
what they mean by ``undesirable behavioral inconsistencies.'')  The
point that {\Caves}, {\Schack}, and I are attempting to develop is that the
{\it transformation rules\/} of quantum mechanics {\it too\/} are of
just such an {\it objective\/} flavor.  That is, if a gambling agent
bets this way on the outcomes of this measurement, and that way on
the outcomes of that measurement, and so on and so on, for a complete
or overcomplete set of quantum observables, and yet that gambling
agent does not relate his probabilities in the way quantum mechanics
prescribes (through the transformations between operators), then the
world will be able to smite him down as certainly as any Dutch bookie
can.

So, far from saying that physics is psychology, our Bayesian program
is a careful attempt to ferret out the objective content of quantum
mechanics at the same time as accepting ``intrinsic randomness'' (in
the sense above).  Only {\it part\/} of quantum mechanics is about
the gambler-independent world, and that part still calls for a full
identification.  This task of separating the wheat from the chaff is
the price one must pay if one accepts the idea that quantum {\it
states\/} are of essence information---luckily it is a task that
stands a chance of bearing fruit (and already has born fruit).  To
say `every physical system is a quantum computer' adds nothing to
existing physics---you would have been better to stick with the
tautology that a physical system is a physical system, full stop.
Using the terminology you do only anthropomorphizes things that
shouldn't have been anthropomorphized in the first place. It erases
the very distinction that one wants to make in order to get at an
objective statement.

OK, enough.

I have strived to use the word peculiar as many times as possible in
this note.

\section{30-01-06 \ \ {\it Not Quite Enough}\ \ \ (to K. T. McDonald)} \label{McDonald7}

And I still need to reply to this:
\bkmcd
As I sketch in my blurb, the issue as I see it is that the so-called
classical world (with its unexplained long-lived bound states) is
largely due to the quantum character of Nature.  \ldots\  The challenge is
to understand better how the classical view arises out of the quantum
world.

It may be that in your way you are addressing this challenge.  If so,
I need a little more help from you to be able to follow your thinking
on this.
\ekmcd
Indeed, I think that is a valid and important question (the part I quoted at least).  And, yeah, I do think that what {\Caves}, {\Schack}, the rest and I are doing is an important movement toward that goal.  Or, at least, I hope so.

But there's no sense in attempting to say anything about our efforts along these lines unless/until I can first instill a little respect in you for the substructure that supports our approach.  I.e., you have to have a little respect for the basics first---and I fear that we may not get there.

\section{30-01-06 \ \ {\it Krugman on Reality} \ \ (to myself)} \label{FuchsC14}

From Paul Krugman, ``A False Balance,''\ {\sl New York Times}, 30 January 2006:

\bq
``How does one report the facts,'' asked Rob Corddry on {\sl The Daily Show}, ``when the facts themselves are biased?''  He explained to Jon Stewart, who played straight man, that ``facts in Iraq have an anti-Bush agenda,'' and therefore can't be reported.

Mr.\ Corddry's parody of journalists who believe they must be ``balanced'' even when the truth isn't balanced continues, alas, to ring true. The most recent example is the peculiar determination of some news organizations to cast the scandal surrounding Jack Abramoff as ``bipartisan.''
\eq

\section{30-01-06 \ \ {\it Exponentiation}\ \ \ (to K. T. McDonald)} \label{McDonald8}

In the words of Scarlett O'Hara, ``Tomorrow is another day.''  A cooler head always prevails on the day after with me; maybe the process is already starting.  It was just---and you should know it---that I found some of your choices of words insulting.
I don't mind a skeptic raising doubt---being challenged to make our point of view consistent is what keeps the view moving along and making progress:  You can find any number of notes in my archives with polite and patient explanations, and plenty of evidence that I find the challenges I receive to be very useful.  But to use phrases like ``what little I can reconstruct from your note \ldots'' and ``I'm led to infer that your view is that physics $=$ psychology''---you've got to admit---had the feel of throwing down a gauntlet.  What physicist would not find it insulting that his work is seen as trying to say ``physics $=$ psychology''?

I'll let the issue rest, and actually thank you for firing me up:  I did rather like some of the formulations I used in the longer note I sent you---they will be useful in the future, in a calmer context.

I look forward to seeing you again at the Wheeler meeting.

\section{01-02-06 \ \ {\it Measurement Based Quantum Computation} \ \ (to H. Halvorson)} \label{Halvorson11}

While I'm thinking of it (because of the grant proposal I happen to
be working on), let me give you references to my three favorite
papers on measurement-based quantum computation:
\begin{itemize}
\item
Measurement-based quantum computation with cluster states \\
Authors: R. Raussendorf, D. E. Browne, H. J. Briegel \\
\quantph{0301052}

\item
Cluster-state quantum computation \\
Author: Michael A. Nielsen \\
\quantph{0504097}

\item
An introduction to measurement based quantum computation \\
Author: Richard Jozsa \\
\quantph{0508124}
\end{itemize}

That computational model strikes me as a veritable goldmine for
exploring how the notion of ``measurement'' I've been talking to you
about (i.e., action on a system, followed by unpredictable reaction
in the agent) is given {\it meaning\/} in the workaday sense.  Maybe
a way to put it is:  Quantum measurements don't ``inform,'' rather
they ``enable.''  And I think the Raussendorf--Briegel computational
model starts to give that slogan some precision.

\section{01-02-06 \ \ {\it Enabling Alchemy}\ \ \ (to R. E. Slusher)} \label{Slusher12}

Below, let me place three notes for you to look over.  The first one contains the slogan I just told you about.  The second one (somewhat about alchemy---don't tell anyone) provides some background for why I would say something like that.  The part that comes closest to a technical paragraph---and the part that you should take away in your memory---is the fourth paragraph from the bottom of it.  Finally, in the third note (to van Fraassen), I expand on how one should {\it not\/} think of quantum measurements as informative of anything.  [See 01-02-06 note ``\myref{Halvorson11}{Measurement Based Quantum Computation}'' to H. Halvorson, 19-06-05 note ``\myref{Comer72}{Philosopher's Stone}'' to G. L. Comer, and 14-11-05 note ``\myref{vanFraassen10}{Questions, Actions, Answers, \& Consequences}'' to B. C. van Fraassen.]

Let me also attach one illustration of the idea (hand drawn of course).

\section{03-02-06 \ \ {\it Red Spears} \ \ (to D. B. L. Baker)} \label{Baker14}

It was the strangest thing to my ear when you called him Red in your note to me the other day.  I hadn't heard him called that in so long---it kind of came back to me, ``That's right; people used to call him Red Spears.''\footnote{My curriculum vitae presently includes this passage in it:  ``To the present date, I have given over 175 invited lectures and seminars. Beyond traveling through or over most states in the United States, this has allowed me the opportunity to visit Australia, Austria, Belgium, Canada, China, Denmark, England, Finland, France, Germany, Greece, Hungary, Ireland, Israel, Italy, Japan, Mexico, The Netherlands, New Zealand, Northern Ireland, Poland, Portugal, Scotland, South Africa, Spain, Sweden, Switzerland, and Wales.  This is my tribute to my stepfather W. T. Spears, who would say, `Chris, travel is the best form of education.'\,''}

Red passed away this morning, about an hour ago.  It was during his sleep.  A pretty good way to go in the end, I suppose.

\section{08-02-06 \ \ {\it The Quotes} \ \ (to D. B. L. Baker)} \label{Baker15}

Here's the quote that'll be going in the revised edition of the {\sl American Heritage Dictionary of American Quotations\/} (though it has been bought by Oxford and will be renamed the {\sl Oxford Dictionary blah blah blah}):
\bq\noindent
   There is no one way the world is because the world is still in
    creation, still being hammered out.
\eq
It originally appeared in one of my samizdats, then made its way into a {\sl Scientific American\/} article by George Musser, and I guess this will be its final resting place.  Actually, I'm very proud of this little accolade (probably more than any of the other ones in my career); I'm a little surprised I had not already told you about it.  The fuller context of the quote is reproduced below.

It was so good seeing you again.  Your company is one of the very few reasons I would ever want to come back to Texas---otherwise, the place is a kind of desert to me.

At the moment, I'm finally on a plane bound for Newark and waiting for my two little bottles of red wine.  I was originally supposed to be home by 5:30 today; now I'm scheduled to arrive in Newark just at 10:00, where/when I'll have to catch a taxi home rather than be greeted by my family. [\ldots]

OK, well that exhausts all the quotes I could find tonight.  And it seems I'm not so good at holding my two little bottles of wine at 32,000 feet anymore.  Maybe I'll drift into one of those reveries like old John Wheeler.

\section{14-02-06 \ \ {\it Notwithstanding} \ \ (to D. Bacon)} \label{Bacon1}

Steven van {\Enk} pointed out to me the award you recently bestowed upon us on your blog.  [See \myurl{http://dabacon.org/pontiff/?p=1189}.]  Well, thanks---I'm honored.  (If I could only figure out a way to put it in my CV.)

Not completely unrelated (but mostly so), here's something that might tickle your fancy.  Once upon a time, Steven told me a story that he had heard from Klaus {\Moelmer}.  It went that Bohr needed to write a paper for a conference proceedings and insisted on starting it off with the word ``notwithstanding.''  Thereafter, for a day or two, Bohr was absolutely stumped and didn't write a word until he could compose his first sentence!

Well Steven and I, wanting to learn from the master, decided that we too would compose a paper around the word ``notwithstanding.''  The result was a little paper titled, ``Entanglement is Super \ldots\ but not Superluminal!,'' that appeared in a book of collected papers on spooky action-at-a-distance.  We were so embarrassed with the paper that we never posted it on {\tt quant-ph}!  But for you, since you've earned it with your kind accolades to us, here is its opening sentence:
\bq\noindent
     Notwithstanding its wonderful potential for enhancing and
     extending our capabilities within the realm of communication
     technology---through applications like quantum cryptography,
     quantum superdense coding, and quantum teleportation---quantum
     entanglement as a physical resource falls far short of being the
     feast the hungry seekers of superluminal communication would hope
     it to be.
\eq

\section{15-02-06 \ \ {\it Discussion} \ \ (to S. J. van {\Enk})} \label{vanEnk2}

Yeah, I still feel awful; I don't know why.  I've been in bed most of the day.  It's a constant mild nausea, but no other symptom.  I suppose it could be a stomach virus.

Anyway, you are right that in a certain sense, the only thing one ever {\it learns\/} (in the common sense of the word) from a ``quantum measurement'' are the consequences of one's actions.  Another way to put it is that, strictly speaking, measurement is nothing other than the changing of a prior state to a posterior state---and the only means the quantum formalism gives for enacting such a change is when one takes an action (i.e., a ``measurement'') on a system.

To say that one learns something in the common sense (i.e., of becoming aware of which of a set of alternatives is already there), one has to give up treating a system (or part of a bipartite system) quantum mechanically.  That is effectively what one is doing when one says that a sender ``prepares'' one state or another.  If one had treated everything quantum mechanically, then there would simply be one quantum state in the picture (circumscribing the preparer and the smaller system), as you point out.

Have another look at the ``clipping'' of the note to van Fraassen that I sent you a little while ago.  [See 14-11-05 note ``\myref{vanFraassen10}{Questions, Actions, Answers, \& Consequences}'' to B. C. van Fraassen.]

Does any of this help your worry?

\subsection{Steven's Preply}

\bq
If you're not too sick to read email, let me ask you a question about yesterday's discussion:

We discussed only very briefly the case where someone, Alice, prepares some qubit $A$ in a state $\rho$ in a particular way (say, tossing a coin, and preparing $|0\rangle$ or $|1\rangle$ depending on the result of that coin toss). It seemed you were going to say then that a measurement by you on system $A$ does teach you something, namely about the preparation.

But I don't see how that would be consistent: after all, you might as well consider Alice + the coin + qubit $A$ as one quantum system consisting of $N\gg1$ qubits, and assigning as state (in oversimplified but obvious $N\gg1$ form)
$$
|0\rangle^N + |1\rangle^N
$$
to Alice + coin + qubit $A$. But then you would say that you learn nothing from a measurement on $A$ \ldots

Isn't it consistent to say you never ever learn anything??

And in particular, you created this email by reading it!
\eq

\section{15-02-06 \ \ {\it Free Will} \ \ (to S. J. van {\Enk})} \label{vanEnk3}

\bsve
I also thought of another paradoxical statement about
counterfactuals: ``If a counterfactual statement could be experimentally tested, it wouldn't be a counterfactual statement, would it?''
\esve
I like that one.

On another subject, could I challenge you to write down in an email what you seemed to like in the discussion yesterday?  I'd like to be able to look over it and absorb it better.  You characterized how you thought we quantum Bayesians were saying something different than Bohr's position.  Could you repeat that?

I also liked the way you said something like we always include the observer in the counterfactual.  But I can't quite reconstruct what you really said.

Of myself, I also liked the little characterization I gave of how barring counterfactuals in the way we do is equivalent to the very idea that ``measurements'' are always generative of their outcomes.  But just as I can't reconstruct what you said, I can't seem to reconstruct what I said either!  Feel free to jump in if you've got a pithy way to say it.

It would be great if any of this clarifies the Hardy paradox for you.  Might it finally be time for another van {\Enk} -- Fuchs paper?

\subsection{Steven's Reply}

\bq\noindent
\bq\noindent
[CAF wrote:] {\it You characterized how you thought we
quantum Bayesians were saying something different than Bohr's position. Could you repeat that?}
\eq

I thought that Bayesians think they get out of paradoxes related to Bell inequalities etc.\ by the observation that
\bq\noindent
1) Alice's measurement changes something about {\it her\/} description of Bob's state, but nothing for Bob's description
\eq
True, that leaves nothing nonlocal but it avoids the ``real'' paradox which occurs when Alice and Bob get together, compare their measurement results and wonder what the heck they would have found had they chosen different measurements.  And
\bq\noindent
2) Both Bohr and Bayesians say: but quantum mechanics gives you all the probabilities you need for actually performed measurements, and you should not even ask the question what would have happened if something else had been measured.
\eq

So, again, what I like about yesterday's discussion is that now you {\it do\/} say something very clear about what would have happened if you had measured something else:  If you assign $\rho$ to some system, and measure $E_a$, with result $a_0$, what would you have found had you measured $F_b$ instead? Answer: You ignore the result $a_0$ and assign probabilities $p(b)=\tr(\rho F_b)$\,.

So, I guess that disproves my impression about what Bayesians say, but I still would guess that Bohr, if he were still alive [yep, another one!\/]\ would still insist on not giving an answer to counterfactual questions.

\bq\noindent
[CAF wrote:] {\it
I also liked the way you said something like we always include the
observer in the counterfactual.  But I can't quite reconstruct what
you really said.}
\eq
Darn, I tried many different ways, but I can't say clearly what I meant! I'll think about it more.

\bq\noindent
[CAF wrote:] {\it
Of myself, I also liked the little characterization I gave of how
barring counterfactuals in the way we do is equivalent to the very
idea that ``measurements'' are always generative of their outcomes.  But just as I can't reconstruct what you said, I can't seem to reconstruct what I said either!  Feel free to jump in if you've got a pithy way to say it.}
\eq

I agree you said that, and I think I agree with the content.  Isn't it
this:  Suppose a measurement does reveal some preexisting property of a system. Then my counterfactual statements about that system would be affected by that knowledge. In particular, had I done a very similar measurement on the system, then I would have found a very similar measurement result. And conversely, if a measurement generates the outcome, then if I had done a slightly different measurement, the outcome could have been very different, since it is as if you're doing a new measurement which will generate a new outcome from scratch. Anyway, obviously that's not the way you said it yesterday but it's how I understand it now.

\bq\noindent
[CAF wrote:] {\it
It would be great if any of this clarifies the Hardy paradox for you.}
\eq
I still think it does! The paradox arises there only if you consider 2
counterfactuals: if both Alice and Bob had done a different measurement then they would have found
\begin{enumerate}
\item
something that will never occur in an experiment, if you use the ``standard''  rule for counterfactuals
\item
exactly what will occur in an experiment if you use the new great rule for counterfactuals.
\end{enumerate}
And again, the only change I need to get rid of the paradox is change how I answer a counterfactual question!

\bq\noindent
[CAF wrote:] {\it
Might it finally be time for another van {\Enk} -- Fuchs paper?}
\eq
Yes! But only if it does not fall in the same category as ``Entanglement is super \ldots\ but not superluminal!''
\eq

\section{16-02-06 \ \ {\it Back from the Silence} \ \ (to H. C. von Baeyer)} \label{Baeyer17}

Thanks for the notes of 2/1 and 2/6.  I'm sorry to take so long to write back to you:  My stepdad became quite ill in late January and finally passed away February 3.  Between trips to Texas and all else, I got quite behind.

First off, I'm glad to hear that your AAPT talk went well!  I wish I could have seen it.  And I like the idea of this B3 diagram---it's good to have a trademark; it focuses people's attention.

Second off, thanks for showing me your first installment in the translation project!  I read it on the plane back from Texas the other day; so I should have just written you then, but I guess I was too exhausted.  I like your system for the endnotes.  One thing:  I detected a couple of typos.  How should we handle things like this in the future?  Would you like me to write a short note indicating them, or would you like me to simply insert the corrections directly into your Word file and send it back?  I could flag them by writing them in boldface or red or something.

Finally, let me apologize to you for a bad oversight on my part.  Marlan Scully and I have organized a little meeting at Princeton next Friday and Saturday in honor of John Wheeler.  In my invitations, I only approached old (quantum-oriented) students of John and Princeton locals for giving talks.  But I should have thought of you too, at least for participation!  It's probably too late for you now, but if you're interested, you're certainly welcome to come.  I'll attach the meeting program so you can get a feel for what it'll be like.  John himself, too, will make an appearance --- either for the first two or three hours Friday morning, or for the dinner Friday evening, we don't know which yet.  If I'm not mistaken, he'll soon turn 95.  John is very frail now, and almost completely deaf, but Ken Ford has said that he will be able to understand that the meeting is in honor of him, and he will recognize some faces.  If you're interested, let me know and I'll send you all the logistical information.

Within my ``Pauli Project'' folder on my computer, I have now started a subfolder titled ``von Baeyer'' to hold your translations and the like.  I can imagine my hard drive, if it could talk, saying, ``Now that's some comfortable real estate in my landscape.''

\subsection{Hans's Preply, ``AAPT Conference Talk,'' 01-02-06}

\bq
Here is an unanticipated fallout from my Anchorage lecture.  I am heartened to see the Bayesian view taking hold in Europe and Australia (is New Zealand part of that continent?), and also in industry and the elementary physics lab.

The ``B3 triad'' is a triangle with the words psi, information, and probability at its vertices, and the names Bayes, Born, and Bohr, respectively,  along their opposite sides.  My talk was well received by many, including my old friend Stuewer, who mentioned you.
\eq

\subsection{Hans's Preply, ``First Installment,'' 06-02-06}

\bq
Here is the first instalment of a translation.  You can regard it as a finger exercise of the kind piano students play.  I don't know yet what will develop.

Endnotes are labeled E, A, and T, for editor, author, and translator, respectively.  Letters are numbered in square brackets [\ldots].   References to the book's extensive bibliography are in curly brackets \{\ldots\}.

\subsubsection{[1286] Pauli to Fierz.}
\bq\noindent
Zurich, 3 October 1951\medskip

\noindent Dear Mr.\ Fierz!

	I have now read your essay\footnote{E1. Fierz had sent Pauli his essay ``The Development of the Science of Electricity as an Example of Physical Theory Making,'' which had been published as a brochure by the University of Basel {\normalsize\{}1951b{\normalsize\}}.} and thank you also for the written addendum.  Of course the psychology of the unitary versus dualist theories of electricity were very interesting to me, representing, as it does, a special case of  the psychology of the development of scientific theories (since the empirical facts to be explained were the same for both theories.)  Furthermore I was particularly interested in your written addendum about trinitarian and quaternarian scholars (the latter are sometimes non-thinkers), where you count Spinoza and Leibniz among the former, Voltaire and Kant among the latter.  (Your positive attitude toward Voltaire is very congenial to me.)

	I came upon Kepler as trinitarian, and upon {\it Fludd\/} as quaternarian -- and felt inside myself, with respect to their polemics, the resonance of an inner conflict.  I have certain traits from both, but now, in the second half of my life, I ought to change over to the quaternarian position.  The problem is that the positive values of the trinitarian position must not be sacrificed in the process.  (Mr.\ Panofsky in Princeton, who loves wordplay and bad puns, once wrote to me so amusingly: ``Today you can no longer pour out the Kepler with the Fluctibus\footnote{T1. Latin for Flood.}!'')\footnote{E2. Pauli's remark in letter [1278] to Panofsky.}

	By the way, I would like to remark that once upon a time (in Hamburg) my route to the exclusion principle was concerned with the difficult transition from 3 to 4: namely with the necessity to ascribe to the electron instead of {\it three\/} translations another {\it fourth\/} degree of freedom (which soon thereafter was explained as ``spin''.)  To wrestle my way to the realization that, contrary to the naive ``view'', the fourth quantum number is also a property of one and the same electron (just like the known {\it three\/} quantum numbers now called $n$, $l$, and $m$) --- this was really the {\it principal effort}.  (I had to fight so hard against the then current theories that had ascribed the fourth quantum number to the {\it rest of the atom\/} or core).

	When I began to work on Kepler's trinitarian point of view, I did not yet know anything about Fludd's polemic, and even less that the quaternity hat such an essential symbolic meaning for Fludd (the relevant longer passage is repeated and translated in my essay) --- I only knew that for Kepler the Pythagorean tetraktys, which he knew well, had {\it no\/} symbolic meaning.\footnote{E3. In his later letters [1383] and writings Pauli often worked on the special role of the tetraktys or quaternity which the Pythagoreans had already emphasized.  This interest had been awakened by Jung, who pointed in his 1937 {\sl Terry Lectures on Psychology and Religion\/} {\normalsize\{}1940, chapter 2{\normalsize\}} to the psychological meaning of the number four, or quaternity, which appeared also in Pauli's dreams.}  In this way, by following a psychological line, I happened {\it again\/} upon the problem of the transition from 3 to 4.  In both cases Mr.\ C. G. Jung had certainly not suggested this to me, nor did I have the intention from the start to wrestle with the problem 3 and 4, of all things.

	Therefore I am pretty sure that {\it objectively\/} an important psychological problem, and perhaps one of natural philosophy, is related to these numbers.

	In general I am completely in agreement with everything you said.  The question is, however, whether one should {\it add\/} several things --- perhaps on another occasion when you want to publish this matter.  (I have the feeling that in that case the work should be longer.)

	For example, one should note and explain that Dirac's hole theory is a not entirely successful new attempt to interpret the relativistic quantum mechanics of the electron in the sense of a unitarian theory of electricity.

	Then on to the problem of opposites:  in physics there are {\it compensatory pairs of opposites\/} (represented by positive and negative quantities like the two electrical charges) and {\it complementary pairs of opposites\/} (represented by non-commuting quantities like p and q).\footnote{E4. Cf.\ the preceding commentary to letter [1286].}  This is a very important distinction, because it my personal conviction that it has a counterpart in the realm of psychology:  A compensatory pair of opposites seems to me -- as it does to you -- to be good and evil, a complementary pair, on the other hand, conscious -- unconscious\footnote{T2. I use the word ``unconscious'' rather than ``subconscious'' to preserve the German distinction between ``Unbewusstsein'' and ``Unterbewusstsein''.} (in Chinese the corresponding pair is Yang and Yin.)  In my opinion analytical psychology is hitherto suffering severely from the absence of this distinction (which is related to the insufficient mathematical-scientific training of its representatives.)  (See below.)  I would be very glad if, in your essay, you insisted pedantically on this conceptual differentiation.

This matter brings me to another subject that is also treated in your essay:  The fundamental difficulty of the ``field'' concept.  On page 14 you very beautifully describe the {\it assumption\/} of the reality of the field: ``But Faraday thought that the field had to be there, whether we prove it or not, just as we believe that the moon is there, whether we look at it or not.''  One might add:  ``just as we assume that the motion of the moon is the same whether we look at it or not'' (which goes far beyond its mere existence.)\footnote{A1. I might call this ``the classicistic idea of objective reality in the universe.''}  This is of course the cloven hoof itself, both with respect to physics ({\it quantum\/} field theory) and to the psychological analogy.

	I have a slightly different opinion from you in that I do not ascribe the same significance to the impossibility of empty space in quantum field theory as you do.  In my opinion the cloven hoof remains {\it exactly the same\/} in quantum field theory as in the unquantised classical theory:  A field without the test bodies required for its measurement should not even be {\it thinkable\/} according to mathematics and logic.  Actually, however, in today's theory things stand as follows:   if you put $e = 0$, and thus describe light fields, then these are, regardless of whether they are classical fields or photons, mathematically possible {\it without\/} charges; if one puts $e \ne 0$ and describes electrons, positrons, or photons (Schwinger), then these are mathematically possible {\it without\/} the heavy masses in the measurement devices that are necessary in order to measure fields or charge densities in small spaces (of order $h/2\pi mc$, $m=$ electron mass).  As for the true complementarity relation between the possibility of regarding {\it the same\/} physical objects as fields or test bodies (measuring devices) (the former when {\it other\/} objects function as measuring devices), this choice is {\it not\/} expressed in today's formalism.  (N.B.  What good is it to me that {\it no\/} empty space is possible?)

	Now to switch to the psychological analogy of the physical field concept, it seems to me to reside in the concept of the {\it unconscious}.  The latter cropped up, and was applied practically, approximately simultaneously with the former.  The ``unconscious'' also ``assumes'' a reality --- in particular, just as in the physical field, an {\it invisible reality\/} (in the sense of everyday life) {\it which mediates a connection between spatially (and perhaps also temporally) distant events}.  For this reason it seems to me that there is here a more profound similarity than a mere analogy.  Consciousness corresponds to the test bodies.  And this conceptual correspondence between field and unconscious extends further to the fact that both exhibit the corresponding cloven hoof.  The concept of the unconscious was initially also used in analytical psychology as though one could observe it without changing it.  And although its representatives occasionally admit the opposite, many of their assertions about the unconscious are still much too close to what I called above ``the classicistic idea of objective reality in the universe.''  This seems to me to be especially the case when C. G. Jung tries to find law-like assertions about the sequence of ``archetypes'' (he calls this the ``dynamics of the self'')\footnote{E5. In the chapter XIV of his new book {\sl Aion}, which Pauli had mentioned in his preceding letter [1285] to M.-L. von Franz, Jung dealt with the {\sl Structure and Dynamics of the Self}.}  --- which are supposed to be valid regardless of the intervention of human consciousness.  These statements remind me in their type too much of Maxwell's classical equations.  An analogy to the quantum mechanical-complementary description does not exist yet, which seems to correspond to the absence of a satisfactory quantum field theory.

	This last paragraph is my personal opinion and probably not yet sufficiently demonstrable to be publishable.  But the correspondences between psychological and quantum mechanical concepts in general (I mention similarity, acausality, experimental set-up, correspondence, pairs of opposites, and holism -- these concepts are used in both disciplines, although they were formulated independently of each other) --- this correspondence is too striking not to believe in a more profound meaning of their {\it simultaneous\/} appearance in the history of ideas.

	And that is exactly what your essay wanted to show, using different examples from different times.

	I will be very interested in hearing your opinion of my ideas!

	Once more, many thanks for last Sunday at noon, and greetings to you and your wife.\footnote{E6. Pauli had evidently been invited to lunch by Fierz.}
\eq

\subsubsection{Editorial comment on letter [1286] from Pauli to Fierz, 3 October 1951}

\bq
The significance for the history of ideas of the problems of opposites, such as light-dark, warm-cold, one-many [1209, 1212], male-female [1391], good-evil [1291, 1364, 1373], idealism-materialism [1395], physical-psychic [1291, attachment], matter-form [1363, 1391], chance and necessity [1497], causal and acausal [1388], as well as their central role in Jung's psychology, henceforth begin to occupy Pauli more and more.\footnote{E1. The problem of pairs of opposites, and their resolution (by unification or conjunction), for which Pauli mentioned quantum mechanics as physical paradigm, also plays a central role in Pauli's later letters to Jung.  Cf., for example, Pauli's letter to Jung dated 27 February 1953 [1526].} Jung considered the {\it Self}, which included the conscious as well as the unconscious psyche, as the center of the spirit ({\it das seelische Zentrum}).  It was made manifest by a unification of opposites ({\it coniunctio}) which ``joins the temporal with the eternal and the specific with the most general.'' \footnote{E2. Jung, {\sl Alchemie und Psychologie}, 1972, p.~34.}

The parallel appearance of such pairs of opposites in modern physics was for Pauli another hint of the profound relationship which he assumed to exist between  physics and psychology [1179, 1180].  As Pauli said in his radio lecture on the occasion of the centenary celebration of Columbia University in 1953,\footnote{T1. Actually, the year was 1954, and the celebration was the bicentennial. } the biggest impression upon him was made precisely by the fact that ``in physics there are real pairs of opposites, such as particle v.\ wave, location v.\ momentum, energy v.\ time, whose contradiction can only be resolved in a symmetrical way.'' The special property of these novel pairs of opposites was that ``one partner is never eliminated in favor of the other, but both are taken over into a new kind of physical law, which expresses the complementary character of the pair in an appropriate way.''\footnote{E3. Cf.\ Pauli {\normalsize\{}1961/84, p.\ 8{\normalsize\}}.}

His interest as physicist was directed primarily toward the psychology of the creation of scientific concepts.\footnote{E4. Cf.\ Pauli's {\sl Bemerkungen zur Psychologie der naturwissenschaftlichen Begriffsbildung\/} in the biographical collection PLC Bi 67 in his literary estate. Cf.\ also von Meyenn {\normalsize\{}1994, pp.\ 18--23{\normalsize\}}. }  He paid particular attention to the development of the concept of matter, which was created in analogy with the doctrine of {\it privatio boni\/} in Platonism, which Pauli had aptly called the {\it Hole Theory of Evil\/} [1029,1278], after Dirac's hole theory.  Since Evil is defined here by the absence of the Good, matter, which is governed by a dark principle, is supposed to be explained by a deficiency of the spiritual [1283].\footnote{T2. In this 1951 letter to Panofsky, Pauli writes: ``The Christian theology of the `{\it summum bonum}' and the `{\it privatio boni}'  is both absurd and historically unique.  To call the entire {\it material\/} world `evil' is consistent; but to define the material as an `absence of the spiritual' would be absurd.'' }  A more positive expression was given to the concept of matter by Aristotle and his successors by calling matter, or {\it hyle}, something which actually existed because it was possible, and thus accessible to conceptual thinking.\footnote{E5. The Aristotelian concept of {\it hyle\/} was later translated into the Latin {\it materia}. Cf.\ also Happ {\normalsize\{}1971{\normalsize\}}.}

According to Pauli's opinion, the negative assessment of everything material, by its commingling with the ethical pair of opposites good-evil\footnote{E6. Cf.\ remarks in the letters [1354, 1357, 1362--1364, 1366, 1373, 1376, 1391, and 1396].}, thus entered into the European history of ideas, and through Fludd and the Platonism of the Renaissance permeates all Western thought.\footnote{E7. Cf.\ the attachment to letter [1328]. Pauli also presented his ideas on the development of the concept of matter in his ``{\it sermon}'' to the {\it International Conference of Scholars\/} in Mainz, March 1955. Cf.\ Pauli {\normalsize\{}1961/84{\normalsize\}}.}

However, compared to the doctrine of the {\it privatio boni}, Pauli was more impressed by the more symmetrical `conceptions of the whole' of the gnostics [1270] and by the ideas of Plotinus about the `problem of the one v.\ the many' [1236, 1278].  These ideas anticipated certain aspects of Bohr's notion of complementarity.

In the letter to Fierz below [1286] Pauli seeks to clarify even further the difference between  the two conceptions, by differentiating between {\it compensatory\/} (or polar) and {\it complementary\/} pairs of opposites.\footnote{E8. Pauli had already mentioned this subject in 1948 in his {\sl Background Physics}.  Cf.\ Meier {\normalsize\{}1992, p.~182{\normalsize\}}.}  As paradigms of complementary opposites he considers of course the non-commuting quantities $p$ and $q$ of quantum mechanics, which have already been mentioned.  Compensatory opposites, on the other hand, are quantities that cancel each other when they meet.  Pauli points out good-evil and conscious-unconscious as their psychological counterparts.

But what Pauli was striving for in these thoughts about pairs of opposites was the far more significant effort at a {\it comprehensive coniunctio (umfassende Coniunctio)}, as he put it in a commentary on a dream he had had on 20 December 1952 on the occasion of his visit to Bombay.\footnote{E9. Cf.\ also the comment on Letter [1489].}  ``I want to try,'' he explained there, ``to say somewhat more clearly what I really wanted to say in the last part of my Kepler essay\footnote{E10. Pauli {\normalsize\{}1952, p.~155{\normalsize\}}. }:   A firm grip on the {\it tail}, i.e.\ physics, provides me with unexpected means which may perhaps be useful in the larger enterprise, {\it to grasp the head}.  For it seems to me that the {\it complementarity in physics}, with its solution of {\it wave-particle duality}, is a kind of {\it model or paradigm for this other, more comprehensive coniunctio}.  The smaller {\it coniunctio\/} in terms of physics, the quantum or wave mechanics constructed by physicists, shows, quite without the intent of its inventors, certain characteristics which might turn out to be useful in the resolution of other pairs of opposites.''

To this, Pauli further added the following epistemological explanations: ``By admitting events and making use of possibilities which can no longer be considered pre-determined and existing independently of the observer, the quantum mechanical manner of explaining nature comes into conflict with the old ontology, which was able to state simply: {\it Physics is the description of reality\/} (words of {\it Einstein}), in contrast to, say, {\it the description of that which one merely imagines\/} (words of {\it Einstein}).  {\it Being\/} and {\it Non-being\/} are not unique characterizations of properties which can only be checked by statistical series of experiments with different set-ups, which may, in certain cases, exclude each other.''
\eq

\subsection{Hans's Reply, 16-02-06}

\bq
Thanks for the nice email and your kind invitation.   Your attachment is ominously labeled ``newsletter blurb\ldots doc.suspect'' and my computer haughtily rejects it.  But no matter --- I cannot get away on this short notice. I wish you and your friends, including John Wheeler, a wonderful weekend.  I happened to sit next to Ken Ford at the Anchorage banquet.

The comment and Pauli letter are background for Fierz's reply, which is so important that it comes in multiple drafts -- my next instalment.  I am grateful for all corrections, and would prefer a little email note about them, so I can make sure we are on the same page.  I am now in correspondence with Atmansbacher, and will probably visit him in the summer.  Furthermore, Roger Stuewer told me that one of his students had written a good thesis about Pauli's mysticism (my phrase), but has not emailed about it yet.  I must remind him.
\eq

\subsection{Hans's Reply, ``Trivium,'' 08-03-06}

\bq
I have been sick in bed and unable to translate, but things are better now.  I am reading a good thesis by John Gustafson, Roger Stuewer's student, claiming that contrary to conventional scholarship, Pauli became ``Jungian'' well before he met Jung. Here is a bit I thought you might
enjoy:
\bq\noindent
There would have been many good reasons for Pauli's father to leave Prague, to seek his fortune in nearby Vienna, to leave his family religion, and to embrace the Roman Catholic faith.  The Pascheles home at No.\ 7 on Prague's Old Town Square (Altst\"adter Ring ), had been a Paulan convent earlier, which may have been the reason he changed the family name to Pauli.
\eq
Did you learn anything, or teach anything, at the Wheelerfest?
\eq
\eq

\section{16-02-06 \ \ {\it Synchronicity} \ \ (to H. C. von Baeyer)} \label{Baeyer18}

\bhcvb
Cheers from sunny Williamsburg, where the total number of snowflakes
during the recent storm was 17.
\ehcvb

Interesting number, since it turns out that throughout my yard I
fairly evenly measured 17 inches.  First Fierz and {\Pauli}, and now the two of us!

\section{16-02-06 \ \ {\it Many Worlds Does Not Explain Much} \ \ (to B. C. van Fraassen)} \label{vanFraassen14}

Thanks for sending your notes.  I enjoyed reading them (twice
actually), and I think I get your point.  I don't know that I have
much to add at the moment by way of endorsement or, alternatively,
criticism, but your ideas are tumbling around in my head---maybe
something will eventually emerge.

Let me only point out a couple of connections that you evoked in me.
First, I liked the way you put this:
\bq\noindent
   In retort it will perhaps be suspected that I yearn for a classical
   understanding of the world, if I'm not willing to count the `many
   worlds' answer as explanatory.  Quite the contrary, I would say:  as
   I see it, quantum theory does leave unexplained why the actual
   outcome of a measurement is this rather than that one among the
   possible outcomes.  Yearning for a classical understanding means
   yearning for an explanation of all that this theory leaves -- and
   legitimately leaves -- unexplained.
\eq

Its tone reminded me mildly of something David {\Mermin} said in his
lecture notes for his quantum computing course:
\bq\noindent
   There are nevertheless some who believe that all the amplitudes $\alpha_x$ have
   acquired the status of objective physical quantities, inaccessible
   though those quantities may be.  Such people then wonder how that
   vast number of high-precision calculations ($10^{30}$ different
   amplitudes if you have 100 Qbits) could all have been physically
   implemented.  Those who ask such questions like to provide
   sensational but fundamentally silly answers involving vast numbers of
   parallel universes, invoking a point of view known as the many worlds
   interpretation of quantum mechanics.  My own opinion is that,
   imaginative as this vision may appear, it is symptomatic of a lack of
   a much more subtle kind of imagination, which can grasp the exquisite
   distinction between quantum states and objective physical properties
   that quantum physics has forced upon us.
\eq
[Actually the source of this quote is \quantph{0207118}.]

The other thing you evoked in me was a memory of an article by Markus
Fierz that I reprinted in my {\it Notes on a Paulian Idea\/} with the
title: ``Does a physical theory comprehend an `objective, real,
single process'?''  I read it again, between my two readings of your
note.  I think you too will enjoy the piece, and particularly near
the end of it, see some similarity to what you have written in your
paragraph above.  Tell me if I am on the money?  For your
convenience, I'll paste in the whole article below; I hope the
remnants of \LaTeX\ code in it won't bother you too much.

Finally let me remark on these words of yours:
\bq\noindent
   A beautiful program for the interpretation of quantum mechanics has
   lately received new life as new results, new techniques, and other
   approaches have been mined to aid in its elaboration.  The excitement
   in such work is its own reward.
\eq

My own feeling is that all their purported progress is illusory.  It
is a lot of technical-looking huffing and puffing (which builds a
shield of seeming protection), but in the end doesn't amount to much.
At a crucial point they make the identification ``weight $=$
probability'' for no other reason than that they know that's what
they have to do to get the answer they want.  Here's the way my
friend Howard Barnum put it recently:
\bq\noindent
The basic idea is that the Wallace argument is just some version of a Laplace's symmetry principle argument in another guise \ldots\ and that I think all these arguments have things backwards:  our belief that
there is a  ``physical symmetry'' just about IS the invariance of our
preferences under the relevant transformation \ldots
\eq
Basically, I say beware of giving the New Everettians too much
credit!

\section{16-02-06 \ \ {\it Wheelerfest Program, User-Friendly Version} \ \ (to the Wheelerfest participants)} \label{WheelerfestUserFriendly}

\noindent Dear Wheelerfesters,\medskip

Many of you have had trouble opening the MS Word document I sent out previously with the program for our upcoming meeting.  Thus, I have cut and paste the information directly into the present email.  The formatting is not quite as nice as it was before, but at least you can (without doubt) read the information this way.

Please note that there have been some changes to a couple of titles, and the final session has been shortened by one talk.

\bq
\noindent \underline{\bf Wheelerfest: Princeton University}\bigskip

\noindent Friday, February 24, 2006

\bq
\noindent 8:45-9:00 \ \ C. A. Fuchs and M. O. Scully \\ Welcoming Remarks\bigskip

\noindent \underline{Delayed Choice and Quantum Eraser}\medskip

\noindent 09:00-09:30  \ \   C. O. Alley \\
``Time For Choice Without Further Delay: `Deputy-General Relativity'
 (Professor Wheeler's name for the Yilmaz Theory) Must Now Take Command''\medskip

\noindent 09:30-10:00	 \ \   M. S. Zubairy	\\
``Time and Quantum: Quantum Eraser'' \medskip

\noindent 10:00-10:30	 \ \   Y. Shih	\\	
``A Random Delayed Choice Quantum Eraser''\bigskip

\noindent 10:30-11:00 Coffee \bigskip

\noindent \underline{Foundations of Quantum Mechanics, I}\medskip

\noindent 11:00-11:30	     L. Hardy	\\
``Quantum Foundations and Quantum Gravity'' \medskip

\noindent 11:30-12:00	     S. Kochen	\\
``It from Bit: Reconstructing the Quantum Formalism from Qubits''\bigskip

\noindent 12:00-13:30	Lunch Break\bigskip

\noindent \underline{Entanglement and Quantum Mechanics}\medskip

\noindent 13:30-14:00	     W. P. Schleich\\
``Entanglement and the Riemann Zeta Function''\medskip

\noindent 14:00-14:30	     M. Hillery	\\
``Programmable Quantum Circuits''\medskip

\noindent 14:30-15:00	     W. K. Wootters\\	
``How Come Phase Space?''\bigskip

\noindent 15:00-15:30	Coffee\bigskip

\noindent \underline{Foundations of Quantum Mechanics, II}\medskip

\noindent 15:30-16:00	     B. Schumacher\\
``The Physics of Impossible Things''\medskip

\noindent 16:00-16:30	     J. Conway\\
``The Free Will Theorem''\medskip

\noindent 16:30-17:00	     E. Nelson	\\
``Two Defects of Stochastic Mechanics''\medskip
\eq

\noindent \noindent Saturday, February 25, 2006\medskip

\bq
\noindent \underline{EPR-Bell}\medskip

\noindent 09:00-09:30	      M. O. Scully\\
``Do EPR-Bell Correlations Require a Nonlocal Interpretation of Quantum
 Mechanics?''\medskip

\noindent 09:30-10:00	      W. G. Unruh	\\
``Bell Inequalities and Nonlocality''\medskip

\noindent 10:00-10:30	      D. M. Greenberger\\
``A Bell Theorem for Two Particles - No Inequalities, Inefficient
 Detectors''\bigskip

\noindent 10:30-11:00	Coffee\bigskip

\noindent \underline{Understanding Quantum Mechanics}\medskip

\noindent 11:00-11:30	      C. A. Fuchs	\\
``The Equations of Quantum Mechanics Already Do Fly''\medskip

\noindent 11:30-12:00	      H. Halvorson	\\
``Deriving Quantum Mechanics from Information Theory:\ The CBH Theorem''\bigskip

\noindent 12:00-13:30	Lunch Break\bigskip

\noindent \underline{Quantum Information}\medskip

\noindent 13:30-14:00	      R. Seiringer	\\
``Some Refinements of Strong Subadditivity of Quantum Entropy and Their
 Applications to Statistical Mechanics''\medskip

\noindent 14:00-14:30	      S. A. Lyon	\\
``Low-Decoherence Electron Spin Systems for Quantum Computing''\medskip

\noindent 14:30-15:00       K. T. McDonald\\
``Reflections of a (Skeptical) Experimental High-Energy Physicist after Teaching a Course on Quantum Computation''\medskip
\eq

\noindent Posters: [Posters will be displayed throughout Feb.\ 24, 2006 outside the conference hall.]\bigskip
\begin{itemize}
\item T. Di:			
``Quantum teleportation of an arbitrary superposition of atomic Dicke
 states''

\item N. Erez:
``Bohm Trajectories: Realistic or Surrealistic?''					

\item A. Jordan:
``Fluctuations in Bose-Einstein Condensation, a path integral approach''			

\item M. Kim:
``Thermodynamic quantities in BEC''

\item A. Muthukrishnan:
``Precision phase measurement as a global quantum Fourier search''			

\item R. Ooi:
``An intense source of large non-classical two-photon correlation''			

\item A. Patnaik:		
``A new method to measure coherence dephasing via Raman photon
 correlation''		

\item Y. Rostovtsev:
``Bunching and anti-bunching of photons in coherently prepared media.''

\item A. Svidzinsky:
``Fluctuations in Bose-Einstein condensate''	

\item H. Xiong:
``From correlated spontaneous emission laser to an entanglement
 amplifier''		

\item L. Zhou:
``Generation of two-mode squeeze state and coherent-squeeze state in cavity QED system''\bigskip
\end{itemize}

\noindent Banquet will be held in the evening of Feb. 24, 2006 at the Conference Center at 6:00 pm.	
\eq

\section{17-02-06 \ \ {\it Free Will Again}\ \ \ (to S. Kochen)} \label{Kochen1}

Now that I've got a few bureaucratic burdens off me, I'm hoping to finally get a chance to skim this free-will manuscript you gave me over the weekend.  Free will at last!  Maybe we can talk about this a bit Monday.

I'm not sure you got this impression from me when we were talking a few weeks back, but I think this theorem may be deeply important for the Bayesian tack on quantum probability that {\Caves}, {\Schack} and I want to develop.  And your partial Boolean algebra view may not be so unrelated to our trains of thought either.  The paragraphs below taken from my pseudo-paper \quantph{0204146}, I think, give some decent indication of how seriously I take this ``measurement context dependence'' view.  Also, I'll excerpt from a recent email to Kirk McDonald on this subject.

Let me give you a couple of other pointers to our work.  I think these are the best things we've put together yet on our quantum foundational ideas:
\begin{itemize}
\item
``Quantum Mechanics as Quantum Information (and only a little more)''\\
\quantph{0205039}
\item
``Unknown Quantum States and Operations, a Bayesian View''\\
\quantph{0404156}
\end{itemize}

See you Monday.  Quotes below; maybe this will give us a little more to talk about.

\section{17-02-06 \ \ {\it Incompletely Knowable vs `Truth in the Making'} \ \ (to W. G. {\Demopoulos})} \label{Demopoulos6}

Now I have to apologize to you again for a long silence:  Soon after
the new year my stepfather became very ill, finally passing away a
couple of weeks ago.  It has been very tough on the family, and I am
only now catching myself back up.

However, don't think that throughout all that, your ideas have not
been on my mind.  Indeed I enjoyed another re-reading of your
`incompletely knowable domain' paper---it lifted me on my sad flight
back from Texas---and I tried to think hard about whether there
really is a substantial distinction between us or not.

Let me try to consider a situation and 1) try to imagine what you
would say of it (but probably in my idiosyncratic language), followed
by 2) what I think I would say of it \ldots\ and then see if there is a
substantial distinction.

Start with a finite dimensional Hilbert space, say of dimension 3,
and imagine it indicative of some real physical system within an
observer's concern.  From that Hilbert space, let us form all
possible sets of three mutually orthogonal one-dimensional projection
operators.  That is, let us consider all possible sets of the form
$\{P_1, P_2, P_3\}$.

What is it that you would say of those sets?  If I understand you
correctly, it is this.  Each such set $\{P_1, P_2, P_3\}$ corresponds
to a set of mutually exclusive properties that the system can
possess.  At any given time, one of those projectors will have a
truth value 1 and the other two will have values 0.  Now consider a
potentially different such set $\{Q_1, Q_2, Q_3\}$; again, at any
given time, one of those projectors will have a truth value 1 and the
other two will values 0.  What is interesting in your conception, if
I understand it, is that even if two elements happen to be identified
between those two sets---for instance, if $P_1=Q_3$---there is {\it
no requirement\/} that $P_1$ and $Q_3$ need have the same truth
value; $P_1$ might have the truth value 0, whereas $Q_3$ might have
the truth value 1.  Another way to say this is that the truth-value
assignments depend upon the whole set and not simply the individual
projection operators.  For you, all the identification $P_1=Q_3$
amounts to is that the {\it probability\/} for the truth value of
$P_1$ within the set $\{P_1, P_2, P_3\}$ is the {\it same\/} as the
{\it probability\/} for the truth value of $Q_3$ within the set
$\{Q_1, Q_2, Q_3\}$. (If you were a Bayesian about
probabilities---though I don't think you are---you would say, ``Well
$P_1$ has whatever truth value it does, and $Q_3$ has whatever truth
value it does (each within their appropriate set of mutually
exclusive triples), but my degree of belief about the truth value of
$P_1$ is the same as my degree of belief about the truth value of
$Q_3$.  That is the rule I am going to live by.'')  Then it follows
from Gleason's theorem that there exist no probability assignments
for the complete (i.e., continuously infinite) set of triples that are
not of the quantum mechanical form. In particular, one can never
sharpen one's knowledge to a delta function assignment for {\it
each\/} triple. This is how you cash out the idea of an `incompletely
knowable domain.'

That is a novel idea, and if I understand it correctly, I like it.

However, now let me contrast my characterization of you with what I
think has been my working conception.  I prefer not to think of the
triples $\{P_1, P_2, P_3\}$ as sets of mutually exclusive {\it
properties\/} inherent within the system all by itself, but rather
{\it actions\/} that can be taken upon the system by an external {\it
agent}. Each {\it set\/} of such projectors corresponds to a distinct
action; what the individual elements within each set represent are
the (generally unpredictable) {\it consequences\/} of that action.
What are the consequences in operational terms?  Distinct sensations
within the agent.  The reason I insist on calling them consequences,
rather than ``sensations'' full stop, is because I want to make it
clear that the domain of what we are talking about is sensations that
come about through the action of an agent {\it upon\/} the external
world.

The essential idea of the sexual interpretation of quantum mechanics
is that no element of a set $\{P_1, P_2, P_3\}$ has a truth value
before the action of the agent.  Rather the truth value---if you want
to call it that (maybe it is not the best terminology)---is generated
(or given birth to) in the process.  At the point, one of the $P_i$
stands in autonomous existence (within the agent), whereas the other
two fall.

I hope I have characterized both of us accurately!

Here is the question that has been troubling me.  Is there any real
distinction (one that makes an pragmatic difference) between our
views?  You say the truth value is there and revealed by the
measurement, and I say it's made by the measurement and wasn't there
beforehand.  So what?

If there is a pragmatic distinction, Steven van {\Enk} and I through
discussions this week have come to believe that it may show up most
clearly in how you and I would treat counterfactuals with regard to
measurement.  Let us take a situation where an agent ascribes a
quantum state $\rho$ to the system; contemplating the measurement
$\{P_1, P_2, P_3\}$, we know that he will ascribe probabilities
according to the Born rule $\tr(\rho P_i)$ for the various
outcomes. Suppose he now performs that measurement and actually gets
value $P_2$.

What does getting that outcome teach him about the quantum system?  I
think you would say it reveals which of the three mutually exclusive
properties the system actually had.  On the other hand, I would say
it teaches him nothing about the system per se; the outcome $P_2$ is
just the consequence of his action.  What is the implication of this
on counterfactuals?  Here's at least one.

Suppose after you get your outcome, you contemplate magically having
performed a distinct measurement $\{Q_1, Q_2, Q_3\}$ instead.  I
think you're careful to point out in your paper that the knowledge of
$P_2$ carries no implication for what you would have found with this
other imaginary measurement.  But what happens if you conceptually
transform this measurement $\{Q_1, Q_2, Q_3\}$ to one closer and
closer to the original, i.e., to $\{P_1, P_2, P_3\}$?  In the {\it
limit\/} when the two are identical again, I think you would say that
knowledge of the outcome $P_2$ in the original case implies that
$P_2$ will also be the outcome in the limiting counterfactual case.
But what would I say?  From my conception, there is no reason at all
to believe that the limiting counterfactual case will give rise to
the same outcome $P_2$.  The best one can do, either in the original
case or the counterfactual case, is to say that an outcome $i$ will
arise with probability $\tr(\rho P_i)$.  In fact, a
counterfactual analysis with this kind of result may be the very
meaning of the idea that quantum measurements are generative of their
outcomes.

At least that is a potential distinction Steven\index{Enk, Steven J. van} and I see at the
moment.  We are toying with the idea that this may have some
implications on the analysis of Hardy-type paradoxes, etc., where
counterfactuals abound, and if anything comes of that, we'll let you
know.  In the meantime, would you say that we have given you a fair
characterization?

\section{17-02-06 \ \ {\it Dutch Book Presentation!}\ \ \ (to R. {\Schack})} \label{Schack98}

In that talk, I'm going to try to say some of those things I told Kirk McDonald when I got pissed off.  I think our project when we get together in Sweden ought to be to try to make some of this business rigorous:  What is this extra ``coherence'' that quantum mechanics forces upon us significatory of?  The more I think about it, that has to be the content of the Born rule---that our probability assignments for seemingly distinct measurements are linearly related.

Before that of course, we have to finish the certainty paper \ldots

Let me hope I survive this week and next.  (Monday I give the Princeton applied mathematics colloquium; Friday and Saturday are the {\Wheeler}fest; and in the middle there is certainty, certainty, certainty.)

\section{18-02-06 \ \ {\it Two Questions} \ \ (to W. G. {\Demopoulos})} \label{Demopoulos7}

Here are two questions that came up in the discussions with Steven\index{Enk, Steven J. van}. I
probably won't have a chance to think about your answers too deeply
until the week after next (I've got to give the Applied Maths
Colloquium at Princeton and then run the {\Wheeler}fest there Thursday
and Friday \ldots\ and I've got a million things to do for both).
But let me throw the questions on the table anyway.

1)  Almost by definition, what you are proposing is a contextual
hidden variable theory.  But what is its status with regard to
locality?  At different times (while driving, taking a shower, etc.),
I've been able to convince myself that a constraint of {\it
locality\/} can be placed upon the truth values, but then I get
confused.  What can you say on the matter?

2)  Take two triads of one dimensional projectors, $\{P_1, P_2,
P_3\}$ and $\{Q_1, Q_2, Q_3\}$, as in my last note to you.  And as
before, suppose $P_1=Q_3$.  However this time, let us be careful to
assume that $P_1$ and $Q_3$ differ in truth value in their respective
sets.  What happens now when we consider nonelementary propositions
of the form $\{P_1, \neg P_1\}$ and $\{\neg Q_3, Q_3\}$ where by
$\neg P_1$ I mean the orthocomplement of $P_1$, etc.  Presumably you
still want to view these sets as representative of mutually exclusive
properties inherent within the quantum system.  However, by
construction the sets are identical: $\{P_1, \neg P_1\} = \{\neg Q_3,
Q_3\}$. How does one decide on a truth value assignment here, given
the previous truth value assignments for $\{P_1, P_2, P_3\}$ and
$\{Q_1, Q_2, Q_3\}$?

\section{21-02-06 \ \ {\it London Overnighter} \ \ (to R. {\Schack})} \label{Schack99}

I had a pathetic long talk with John Conway at Princeton about Dutch book yesterday (about three hours).  He buys none of it.  Can you believe this is the man who wrote the {\sl Atlasi of Finite Groups\/} (and actually has all that in his head \ldots\ as far as I can tell and as is rumored there).  Yet, a simple argument like Dutch book can't pierce his preconceptions.

John Nash was also at my talk:  I think he never blinked once, and he looked a million miles away.  I couldn't help but remember the passage in {\sl A Beautiful Mind\/} where he says that it was thinking about quantum foundations that pushed him over the edge.

\section{21-02-06 \ \ {\it Poetry} \ \ (to S. Kochen \& J. H. Conway)} \label{Conway2} \label{Kochen2}

I enjoyed the sparring yesterday, and I think just before we parted from the lounge Si had a good summary of a hefty aspect of what I'm shooting for.  So, thanks for the understanding, if not the agreement.  (Some of that stuff, Si, in case you haven't read it, is in the last note I sent you ``Free Will Again''---hopefully I'm a little clearer there than I was in person.)

While it's on my mind, let me come back to the part of the discussion where we had three different ways of talking about ``measurement'', going from left to right written on the board.  Recall John changed the word ``action'' to ``interaction.''  At about that time, I said something that made you both cringe.

With your indulgence, I want to say it again using the words I used at my Caltech summer school lectures last summer.  [See 17-06-04 note ``\myref{Mabuchi12}{Preamble}'' to H. Mabuchi.]  I'm sure it'll still make you cringe, but maybe you'll enjoy the pentameter of this version.  Like always when I talk, it emphasizes the distinction between ontology (``the quantum world'')---i.e., the exciting part---and epistemology (``{\it part\/} of quantum theory'')---i.e., the dull but necessary part.  Luckily, as I see it, in an oblique way, the epistemology is teaching us something about the ontology.  Maybe that's enough to leave us a little room for further discussion.

\section{21-02-06 \ \ {\it Heaping the Poetry} \ \ (to J. H. Conway)} \label{Conway3}

\bjhc
Well, it ended before it began.  Yeah, yeah.  JHC
\ejhc

It's a general problem; that's why I keep at it year after year.

I won't have the time to reply to you in depth until I get all the planning for this {\Wheeler}fest done, but don't you worry, eventually I'll be back in full force:  I, of course, think I have a reply to your last note that will run you in a circle.  But I want to spend some time writing it delicately.

In the meantime, let me give you a little something related to your present note that I happen to have in my archives.  (Menand's book was my very favorite of 2001, and I've recommended it to a lot of people.)  You should see the relevance to our discussion very quickly into the quote.  For me, a probability is like a beak:  That is all I am saying.  To that extent, probabilities are real in nature, but only to that extent.  Free decisions---as you call them---may be part of the fundamental furniture of nature (I'm willing to bet that)---that is, they are of a more respectable status than beaks---but it is going to be hard to convince me that probabilities are.

Still, I am a somewhat logical being, and as such I will give your arguments a hard go.

As I say, I'll be back eventually (sooner, rather than later).

From L.~Menand, {\sl The Metaphysical Club}, (HarperCollins, London, 2001):
\bq
The world is filled with unique things. In order to deal with the world, though, we have to make generalizations. On what should we base our generalizations? One answer, and it seems the obvious answer, is that we should base them on the characteristics things have in common. No individual horse is completely identical to any other horse; no poem is identical to any other poem. But all things we call horses, and all things we call poems, share certain properties, and if we make those properties the basis for generalizations, we have one way of ``doing things'' with horses or poems---of distinguishing a horse from a zebra, for example, or of judging whether a particular poem is a good poem or a bad poem. These common properties can be visible features or they can be invisible qualities; in either case, we create an idea of a ``horse'' or a ``poem,'' or of ``horseness'' or ``poetry,'' by retaining the characteristics found in all horses or poems and ignoring characteristics that make one horse or poem different from another.
We even out, or bracket, the variations among individuals for the sake of constructing a general type.

Darwin's fundamental insight as a biologist was that among groups of sexually reproducing organisms, the variations are much more important than the similarities. ``Natural selection,'' his name for the mechanism of evolutionary development that he codiscovered with Alfred Russel Wallace, is the process by which individual characteristics that are more favorable to reproductive success are ``chosen,'' because they are passed on from one generation to the next, over characteristics that are less favorable. Darwin regretted that the word ``selection'' suggested an intention: natural selection is a blind process, because the conditions to which the organism must adapt in order to survive are never the same. In periods of drought, when seeds are hard to find, finches that happen to have long narrow beaks, good for foraging, will be favored over finches with broad powerful beaks: more of their offspring will survive and reproduce.
In periods of abundance, when seeds are large and their shells are hard, the broad-beaked finches will hold the adaptive advantage.
``Finchness'' is a variable, not a constant.

Darwin thought that variations do not arise because organisms need them (which is essentially what Lamarck had argued). He thought that variations occur by chance, and that chance determines their adaptive utility. In all seasons it happens that some finches are born with marginally longer and narrower beaks than others, just as children of the same parents are not all exactly the same height. In certain environmental conditions, a narrower beak may have positive or negative survival value, but in other conditions---for example, when seeds are plentiful and finches are few---it may make no difference.
The ``selection'' of favorable characteristics is therefore neither designed nor progressive. No intelligence, divine or otherwise, determines in advance the relative value of individual variations, and there is no ideal type of ``finch,'' or essence of ``finchness,''
toward which adaptive changes are leading.

Natural selection is a law that explains {\it why\/} changes occur in nature---because, as Darwin and Wallace both realized after reading, independently, Thomas Malthus's {\sl Essay on the Principle of Population\/} (1798), if all members of a group of sexually reproducing organisms were equally well adapted, the population of the group would quickly outgrow the resources available to sustain it. Since some members of the group must die, the individuals whose slight differences give them an adaptive edge are more likely to survive. Evolution is simply the incidental by-product of material struggle, not its goal. Organisms don't struggle because they must evolve; they evolve because they must struggle. Natural selection also explains {\it how\/} changes occur in nature---by the relative reproductive success of the marginally better adapted. But natural selection does not dictate {\it what\/} those changes shall be. It is a process without mind.

A way of thinking that regards individual differences as inessential departures from a general type is therefore not well suited for dealing with the natural world. A general type is fixed, determinate, and uniform; the world Darwin described is characterized by chance, change, and difference---all the attributes general types are designed to leave out. In emphasizing the particularity of individual organisms, Darwin did not conclude that species do not exist. He only concluded that species are what they appear to be: ideas, which are provisionally useful for naming groups of interacting individuals.
``I look at the term species,'' he wrote, ``as one arbitrarily given for the sake of convenience to a set of individuals closely resembling each other \ldots. [I]t does not essentially differ from the term variety, which is given to less distinct and more fluctuating forms. The term variety, again, in comparison with mere individual differences, is also applied arbitrarily, and for mere convenience sake.'' Difference goes all the way down.
\eq

\section{21-02-06 \ \ {\it R\'enyi Distinguishability Measures} \ \ (to E. H. Lieb)} \label{Lieb3}

It was good meeting you yesterday.  I only wish I had had a chance to extend the discussion:  I've got loads of mathematical problems I'd love to get some expert input on!

Here's the reference where you can read a little about the quantum R\'enyi overlaps I was telling you about:
\quantph{9601020}.
The place to look is pages 60--64.  (Also pages 16--19 explain why this kind of quantity even comes up in the classical context.)  I had plenty more inequalities than that related to the quantity, but I never published them and they burned up in the Los Alamos fire.  In any case, none of them were probably very interesting, as they weren't tight and the thing I really wanted was the infimum.

\section{22-02-06 \ \ {\it Hollow Eyes} \ \ (to S. J. van {\Enk})} \label{vanEnk4}

Oh, let me tell you a funny little story, and then I should go.  During my Princeton talk, there was one of the oddest looking fellows in the audience.  His eyes just looked so hollow---I don't know how to describe it.  He hardly blinked the whole talk, and his face was completely expressionless.  After it was all over, I kept thinking, ``He looks familiar; who is that?''  Well the next morning in the shower, I got a hunch and checked it out on Google:  It was John Nash!  After discovering that, I amused myself by remembering how he claimed that it was his thinking about quantum foundations that pushed him over the edge to schizophrenia!  And I hoped I didn't push him over the edge again!

\section{22-02-06 \ \ {\it Tutorial} \ \ (to J. E. {\Sipe} \& R. W. {\Spekkens})} \label{Sipe5} \label{Spekkens37.05}

\bv
T3 Current Interpretations of Quantum Mechanics\\
Organizer: Rob Spekkens, Perimeter Institute for Theoretical Physics,\\
Room 302, Baltimore Convention Center  \ldots\ \\
Instructor: Professor John Sipe, University of Toronto
\ev
Too bad the pseudo-Bayesian view of Caves, Schack, Appleby, Fuchs and few others isn't a current interpretation of quantum mechanics \ldots

\ldots\ \smiley

P.S.  Below is a nasty note I wrote a while ago.  I kind of like the way I put some of the things in it (the explanatory things that is, not the nasty things)---I think they represent a little more mature way of putting things than I may have expressed to you two before.  Maybe you'll enjoy it for the content.  [See 30-01-06 note ``\myref{McDonald6}{Island of Misfit Toys}'' to K. T. McDonald.]

\section{22-02-06 \ \ {\it Probably More Than I Should Send} \ \ (to J. E. {\Sipe})} \label{Sipe6}

\bjes
Anyway, I haven't yet decided what (or where in the presentation) to
say something about your approach.  So this is a good chance for me to
ask a couple of questions.
\ejes

I hope you didn't take me too seriously; your tutorial title just looked like a good set-up to abuse you two, and I couldn't resist.

\bjes
As the saying goes, I didn't have time to write you a short
letter so I wrote you a long one \ldots\ apologies for the verbiage!
\ejes

You ought to know that {\it I\/} am more guilty of this than {\it anyone}!  I practically make a living of it.

If you can wait 'till Sunday or so, I'll write you some more delicate and specific answers to all your questions---at the moment, I'm struggling to get everything finished up for the Wheelerfest this weekend (including my own talk).

But let me send you two or three older emails in the meantime that I think fairly directly answer a couple of your questions.  Rob calls the ontology being spoken of (more accurately, dreamed of) here ``F-theory,'' where the F can be interpreted a couple of ways.

I'll come back to the issue of why I'm not very open-minded on hidden variables theories in the next note after the Wheelerfest, and I'll (probably) resculpt the partial answers that I'm sending you at the moment to be more in the shape of your present questions while I'm at it.

BTW, you and Rob have great gifts:  you both write very, very clearly, even in your hastily composed emails.  Your letter was a joy.

See notes below on ``F-theory'' and (what I don't view as) ``instrumentalism.'' [See 21-02-06 note ``\myref{Conway2}{Poetry}'' to S. Kochen \& J. H. Conway, 19-06-05 note ``\myref{Comer72}{Philosopher's Stone}'' to G. L. Comer, 14-11-05 note ``\myref{vanFraassen10}{Questions, Actions, Answers, \& Consequences}'' to B. C. van Fraassen, and 17-02-06 note ``\myref{Demopoulos6}{Incompletely Knowable vs `Truth in the Making'}\,'' to W. G. {\Demopoulos}.]

\subsection{John's Preply}

\bq
Well, I'm still writing the tutorial.  It's supposed to be four hours long.  As you can imagine, this is not an easy task.  In some ways (counting the slides that have to be made, for example!), four hours is dreadfully long.  But in another way it is too short to do much justice to anything.  My main goal in this is to ``first, do no harm'' in presenting the different views. We'll see how well I do \ldots

Anyway, I haven't yet decided what (or where in the presentation) to say something about your approach.  So this is a good chance for me to ask a couple of questions.  First, I am glad to see statements like:
\bq\noindent
To say that `everything is information,' as Zeilinger does, or to say `every physical system is a quantum computer,' as you and Lloyd do, adds nothing to existing physics---you would have been better to stick with the tautology that a physical system is a physical system, full stop.
\eq
I certainly feel that statements like Zeilinger's likely convey nothing, but only propagate buzz-words and probably, by their implicit over-hyping, do science no good in the long run. I also have a lot of sympathy for statements like:
\bq\noindent
Only {\it part\/} of quantum mechanics is about the gambler-independent world, and that part still calls for a full identification.  This task of
separating the wheat from the chaff is the price one must pay if one accepts the idea that quantum {\it states\/} are of essence
information---luckily it is a task that stands a chance of bearing fruit (and already has born fruit).
\eq
I gather from this that you see the ultimate task to be the identification of just what is the part of quantum mechanics that is ``about the gambler-independent world.''  The Bayesian analysis is a strategy, as I understand it, to identify that task and undertake it. It is a means, not an end.

But if I understand this correctly I am puzzled to see statements like:
\bq\noindent
If what you mean by ``intrinsic randomness'' is that quantum measurement
outcomes do not pre-exist the measurement process---that they are made
``on demand''---and that there is no fact of the matter in the universe
that will determine which way they will go, then Caves, Schack, Appleby, Peres, Leifer, and I all accept that.
\eq
Why could it not be that there is a ``fact of the matter in the universe'' that will determine the outcome of (maybe just some) measurements.  You say that
\bq\noindent
Garrett-style hidden variables,
regardless of what Garrett might say, are thus not necessary for a
Bayesian view of quantum probabilities.
\eq
Fair enough.  But might it not be that the ``gambler-independent world'' is in fact described by a hidden-variable theory of some sort?  For example, why would you (immediately!)\ rule out a Bohm--de Broglie universe (or a modified one, with the usual distribution over configuration space being a kind of `equilibrium distribution' that we might someday be able to overcome, such as Valentini suggests) simply because of your Bayesian strategy in trying to identify just what that ``gambler-independent world'' is?  Could it not be that Bohm and de Broglie have really identified the ``fact of the matter in the universe,'' or that some such approach would?  At least, must not this possibility be seriously considered?  How does a Bayesian strategy immediately rule it out? By rejecting even the possibility that some quantum measurement outcomes may pre-exist the measurement process, do you not prejudge the outcome of the larger quest, for which the Bayesian viewpoint is not a goal but a strategy?

Perhaps I simply have read too much between the lines of what you say, or perhaps I truly misunderstand your views.  In any case, (and to make sure in this tutorial I don't confuse somebody about what you really do hold!)\ could I get you to see if the way I describe your approach below is (a) basically right, (b) partly right but with major flaws, or (c) totally screwed up?  Below I am going to phrase things in the kind of language I'm going to use in the tutorial.  Please bear with me; I realize this is not the way you might want to frame the issues, and I really am not wedded to this as necessarily the `best way' to phrase the issues. But for better or worse it's the one I've adopted for the tutorial, and so it's the one in which I have to somehow (and hopefully without too much damage) fit the thoughts of Chris Fuchs.

Here we go.

I use the term ``interpretation'' of a modern physical theory to mean an identification of just what it is to which the abstract elements in the theory purportedly refer.  A ``type of interpretation'' identifies the type of things to which the abstract elements in the theory refer.

The abstract elements in quantum mechanics are density operators, POVMs, etc., etc. A ``realist'' type of interpretation identifies the abstract elements with stuff that purportedly really exists in the world, the ``furniture of the universe.''  An ``operationalist'' type of interpretation identifies the abstract elements with tasks in the laboratory, such as preparations, transformations, and measurements.  And so on.  An ``instrumentalist'' type of interpretation says that abstract elements don't refer to anything at all, except perhaps in the way a hammer refers to a nail.  That is, theories are just (only) tools for coping in the world.

Admittedly these categories are not as hard-and-fast as the language implies, but let's take them as a rough guide to thinking about these things.   Now you and your colleagues hold
\bq\noindent
``the idea that quantum {\it states\/} are of essence information''
\eq
and the obvious question a beginning student would ask---and undoubtedly someone will ask if I say something like this in the tutorial---is ``information about what?''  Perhaps one could argue that this question has no meaning, but I don't see how that view can be sustained.  In any case, from the talks I have heard you give I came away with the impression that, at least in the short term and from the point of view of our current work-a-day physics, you take this to be information, broadly speaking, about the impact and results of future tasks in the laboratory, such as measurements.  On the basis of this it seems to me you are in the ``operationalist'' camp.

Now within this camp there are various tribes. Some feel that any talk about underlying ``furniture of the universe'' is meaningless, or at least not worthwhile.  They are ``hard core operationalists.''  Gadgets and tasks for them are ultimate primitives in their theorizing, and for them an operationalist physics is all there can ever be, at least if it is to be what they would consider a proper physics.

I take you to be in a different tribe, whose members feel that ultimately we want a physics that does indeed talk about the furniture of the universe, the ``gambler-independent world.''  Indeed, you refer to that as the ``wheat,'' as opposed to the ``chaff.'' But that ``gambler-independent world'' has not yet been identified.  In the interim you argue that the abstract elements of quantum mechanics should be understood operationally, and the predictions and statements of quantum mechanics subjected to a kind of ``Bayesian rack'' to clarify exactly what beyond the objective rules of probability actually exists in them.  This analysis then provides a firm footing for the search for the actual ``furniture of the universe,'' whatever it is.  Ultimately you want a theory that does describe this furniture; I would guess you are uncertain as to whether quantum mechanics (sufficiently purified by Bayesian analysis) would provide that, or whether a successor theory would be required.  But in any case in your heart of hearts you hope someday for a realist science, or at least hope that that hope makes sense.

OK, that's my take.  How close to your thoughts is it?  It may be a while before I can respond to anything you say, because I am at a conference in Germany, but I really would appreciate some comments!

P.S.:  As the saying goes, I didn't have time to write you a short letter so I wrote you a long one \ldots\ apologies for the verbiage!
\eq

\section{22-02-06 \ \ {\it Wheelerfest} \ \ (to D. Overbye)} \label{Overbye5}

Here is something I should have told you about a good while back, but---I have to apologize---I didn't think about you until this morning.  Since you're such a fan of John Wheeler, you might be interested.

Marlan Scully and I have organized a little meeting in Princeton in honor of John Wheeler to discuss quantum foundations and quantum information.  It'll be held Friday Feb 24 and Saturday Feb 25; on the Friday it'll be at a certain conference center in Princeton, and on the Saturday it'll be on campus.  I'm sorry for the very, very short notice.

I'll place the meeting schedule below, so that you'll have some indication of what it's all about.  [See 16-02-06 note titled ``\myref{WheelerfestUserFriendly}{Wheelerfest Program, User-Friendly Version}''.] If you're interested in coming let me know.

John himself will join us Feb 24, the Friday, between 5:00 PM and 6:00 PM (and maybe a little longer, depending upon his stamina) to be greeted by former students, etc.  There'll be a conference dinner Friday evening at the same conference center (starting at 6:00), but we don't think John will stay all the way through that.  However, you're welcome to come---choice of prime rib or salmon.

\section{22-02-06 \ \ {\it Wheelerfest, 2} \ \ (to D. Overbye)} \label{Overbye6}

\bdo
I am a huge fan of Wheeler's and would love to have been at this meeting. Unfortunately I'm already committed to going to Pennsylvania for my mother-in-law's 80th birthday. I am especially intrigued by at least the title of Conway's talk about free will. Is that John Conway, the inventor of {\bf Life}? Sorry I can't be there but the weekend has long been taken.
\edo

Too bad!  I'm so sorry I didn't think about contacting you before today.  Yeah, that's John Conway---the inventor of ``life'' and a much more famous mathematician than that for his {\sl Atlas of Finite Groups\/} and many other things.  He's a very lively character.  The theorem he'll be talking about is one by Simon Kochen and him, that is a stronger, newer version of the sort of thing first proven by Kochen and Specker.  I would say it this way:  That quantum mechanical measurement outcomes do not pre-exist the process of measurement; nature makes a choice when forced to the point.  Or, another way I would put it is, that a quantum measurement forces a little act of creation (or birth).  They claim it's the strongest hidden-variable no-go theorem yet.  It's hasn't been published yet, but there has been plenty of press coverage due to Conway's talks.

\section{22-02-06 \ \ {\it The Way the Cookie Crumbles} \ \ (to H. C. von Baeyer)} \label{Baeyer19}

Thanks for the note.  As I just wrote to Marcus {\Appleby}, ``At the moment, I am scrambling to get something together for the Wheelerfest meeting I organized at Princeton this weekend.  I gave my talk the title `The Equations of Quantum Mechanics Already Do Fly,' but honestly I don't yet have a clue what I'm going to say.  I hope I get some inspiration before tomorrow morning, so that I'll have time enough to put it together before Friday morning!''  But I want to get caught up on some emails too!

Just a couple of comments on your note:
\bhcvb
Translating Fierz's sentence: ``As is often the case, when you don't understand anything, you think of all sorts of things that you feel are connected, but whose connection you cannot express clearly'' inspires me to speculate as follows.

How did the Greeks discover atoms?  By reasoning from the observation of steam that water could not possibly retain its characteristics under arbitrary dilution unless those characteristics were stored in tiny, finite lumps.

How did {\Schroedinger} invent molecular genetics?  By reasoning that living organisms could not possibly retain their characteristics over millions of years unless those characteristics were stored in tiny, material, finite lumps.

In both cases the argument was that the cookie would crumble into chaos unless its parts were quantized.

It seems to me that this is the basis of Zeilinger's principle.
Information would dissolve into ignorance if it were not quantized
into bits -- even though it is carried by qubits.  Conversely, if you
add analyticity to bits, you get qubits (Hardy). (This does not imply
that information is material.)
\ehcvb
With regard to the part about Hardy (which reminds me that I had wanted to invite him to my house this weekend---thanks), see the note below that I wrote to Hans Halvorson a few weeks ago. [See 27-01-06 note ``\myref{Halvorson9.1}{Monday or Tuesday Meeting}'' to H. Halvorson.]

\bhcvb
So if {\Spekkens} had a toy model in which the ontic content was
continuous, and the epistemic content countable, he would have quantum
mechanics.
\ehcvb
OK, as far as I know he doesn't have that.  But he and Terry Rudolph and Steve Bartlett do have one in which the ontic content is continuous.  See their talk at the APS March meeting; here is their abstract.  (Or at least I think that's the same toy model he told me about before, and as far as I understood, there was a continuous set of epistemic states allowed.)  [Though, now see S.\ D.\ Bartlett, T.\ Rudolph and R.\ W.\ Spekkens, ``Reconstruction of Gaussian quantum mechanics from Liouville mechanics with an epistemic restriction,'' Phys.\ Rev.\ A {\bf 86}, 012103 (2012), \arxiv{1111.5057}.]

\bq\noindent
{\bf 3:30PM D40.00004}\medskip\\ Liouville mechanics with an epistemic restriction and Bohr's response to EPR\medskip\\ TERRY RUDOLPH, Imperial College, STEPHEN BARTLETT, University of Sydney, ROBERT SPEKKENS, Perimeter Institute\medskip\\ We introduce a toy theory that reproduces a wide variety of qualitative features of quantum theory for degrees of freedom that are continuous. Specifically, we consider classical mechanics supplemented by a constraint on the amount of information an observer may have about the motional state (i.e.\ point in phase space) of a collection of classical particles -- Liouville mechanics with an epistemic restriction (This may well be how Heisenberg initially understood the Uncertainty Principle). We develop the formalism of the theory by deriving the consequences of this ``classical uncertainty principle'' on state preparations, measurements, and dynamics. The result is a theory of hidden variables, although it is not a hidden variable model of quantum theory because of its locality and noncontextuality. Despite admitting a simple classical interpretation, the theory also exhibits the operational features of Bohr's notion of complementarity. Indeed, it includes all of the
features of quantum mechanics to which Bohr appeals in his response to EPR. This theory demonstrates, therefore, that Bohr's arguments fail as a defense of the completeness of quantum mechanics.
\eq

In addition though, I'm pretty sure I don't agree with what you're hoping for here.  That is because, toy models, at least as conceived along the original {\Spekkens} line, have the property of noncontexuality (which gives rise to the fact that in these models, the measurement `outcomes' are already alive before the measurement actually takes place---i.e., they pre-exist).  In that sense, these are still detached-observer models.  One paper that I don't think I have recommended to you before, but that you might enjoy is this:
\bv
The Bell--Kochen--Specker Theorem\\
Authors: D. M. {\Appleby}\\
Comments: 22 pages\\
Journal-ref: Stud.\ Hist.\ Philos.\ Mod.\ Phys.\ {\bf 36} (2005) 1\\
\quantph{0308114}
\ev

\bhcvb
In any case, I feel (wish?)\ that we are very close to understanding
the quantum.  Maybe somebody will figure it out in Wheeler's lifetime.
\ehcvb
I can't disagree with that!

\section{23-02-06 \ \ {\it John Wheeler's Visit} \ \ (to L. Wheeler Ufford)} \label{Ufford1}

Thank you for your greeting.  It has been a pleasure to organize this.  John has long been one of my heroes, ever since I took a first-year college course with him in 1984.  Almost every paper I have written in physics has concerned one or more of the problems he brought to my attention way back then.  So, this is really a very small gesture in return.

I look forward to meeting you, your husband, Jackie and, of course, John tomorrow.  (It will be the first time I have seen him since 1994.)

\subsection{Letitia's Preply}

\bq
I am John Wheeler's older daughter and with Jackie Fuschini, his secretary, I and my husband will be bringing him to your Fest, hoping to arrive there a little before 5 p.m.  I look forward to meeting you and thanking you for arranging such a happy and encouraging event for him.
\eq

\section{27-02-06 \ \ {\it Wheelerfest, Thanks} \ \ (to M. O. Scully)} \label{Scully8}

I wanted to write you a note to try to express my heartfelt thanks for supplying the resources that made this meeting happen.  There is nothing that tickles me more than seeing our understanding of quantum mechanics advance, and meetings like this one are crucial in that regard.  In fact, I think on one of the things John Wheeler wrote in that letter to Carroll Alley that I read at the beginning of the meeting:
\bq\noindent
    Expecting something great when two great minds meet who have
    different outlooks, all of us in this Princeton community expected
    something great to come out from Bohr and Einstein arguing the great
    question day after day---the central purpose of Bohr's four-month,
    spring 1939 visit to Princeton---I, now, looking back on those days,
    have a terrible conscience because the day-after-day arguing of Bohr
    was not with Einstein about the quantum but with me about the
    fission of uranium. How recover, I ask myself over and over, the
    pent up promise of those long-past days?
\eq
I hope we fulfilled a little of John's dream:  Though the time was short, we did have a very impressive array of ``great minds with different outlooks'' arguing the great question.

Particularly, for me, I got a lot out of the interaction.  My favorite talks were yours and Unruh's (with which I heartily agreed) and Conway's and Kochen's (which stretched me to a new level).  Furthermore, Schumacher took my ``Dutch book'' challenge seriously, and, I think, may have already provided a very nice solution to it.  Finally, Hardy stayed at my house over the rest of the weekend and we got a good bit of conceptual work done with regard to his ``causaloid'' framework.  So, this conference turned out to be nothing but pleasure for me!

\section{27-02-06 \ \ {\it Wheeler} \ \ (to W. P. Schleich)} \label{Schleich2}

No, thank YOU very much for coming.  I was very much impressed by the humanity I saw in you; I had not got to know you so well before as I did this weekend.

John continues to be an inspiration for me --- as I hope my talk conveyed.

I hope that one day we'll get a chance to sit down and talk about this representation of (finite dimensional) quantum states in terms of true-blue probability distributions, which I mentioned briefly in my talk.  My own understanding of it is long in development, and I suspect there are some insights I could glean from Mr.\ Phase Space's view on it.  Schumacher, by the way, provided a nice ``Dutch book'' argument over the weekend, fulfilling the challenge I put at the end of the talk.  So, on the technical end even (i.e., even outside of the opportunity to see John honored), this meeting was quite useful to me.  It was a real pleasure to see it come together.

\section{27-02-06 \ \ {\it Convex Sets / Algebras Tutorial?}\ \ \ (to H. Halvorson)} \label{Halvorson12}

I very much enjoyed your talk at the Wheelerfest, the parts of it I understood at least---I'm terribly primitive when it comes to mathematics.

To continue our mutual quest, I wonder if the logical next step might be for you to help bring me up to speed on the connection between convex sets and algebras if you would?  That is definitely an important connection which I need to understand.  It would be great if I could pull a tutorial from you that a) concerned itself only with the finite dimensional case, and particularly b) used only the language of the finite dimensions in its presentation.  My mind gets unusually distracted when I hear terms like Hausdorff and compactification, and if you could help me get to the truly essential points (which I view always as the finite dimensional issues), I think it would help both of us exponentially (you from the teaching, me from the learning).

Might you have some time early next week?

\section{27-02-06 \ \ {\it Dutch Book} \ \ (to D. M. Greenberger)} \label{Greenberger3}

I'm sorry I never answered your question about Dutch book after my talk.  Maybe the simplest solution is for me to just point you to the nice notes Carl Caves put together on it:
\bq\noindent
\myurl{http://info.phys.unm.edu/~caves/reports/dutchbook.pdf}.
\eq
Another place to look is the appendix in our paper:
\bq\noindent
``Conditions for Compatibility of Quantum State Assignments''\\
Authors: {\Carl}ton M. {\Caves}, Christopher A. Fuchs, {\Ruediger} {\Schack}\\
\quantph{0206110}
\eq

It was good seeing you at the meeting; you had me laughing and laughing with your quips.

\section{27-02-06 \ \ {\it Wheelerfest} \ \ (to K. W. Ford)} \label{Ford4}

Thanks for the note.  I'm glad you enjoyed the meeting, and I'm particularly glad that you think John enjoyed it.  Seeing him at this late stage of life was just as inspiring for me as seeing him, and thinking about his thoughts, at every other stage.  He is an amazing man.

You're right about the gender balance issue, and your note helped promote it a little further to the top of my consciousness (and conscience).  I'll try to think harder next time when I work on an invitation list.

I hope you got something out of my talk.  I'm a little disappointed with it myself:  I tried to squeeze both too much history and too much technicality to get any live points across effectively.  (I was pleased though that Schumacher took my ``Dutch book'' challenge seriously over the weekend; I think he has provided a very nice solution to it!)  Your question at the end was a very good one.  It's one of the crucial issues---and I wish I had a much better answer for you by this stage.  As best I can put it at the moment, I think the issue is not consciousness at all; but the distinction between an ``inside description'' and an ``outside description''---analogous to one of the points Conway brought up in his talk about the ``free will theorem'' (which unfortunately you missed).  Maybe a way to put it is this:  What we call quantum theory has sadly been misidentified all these years as a ``description from the outside,'' when in fact it is almost completely a ``description from the inside.''  There is a hint of the outside in it---through these very powerful theorems of Kochen and Specker (and now Conway and Kochen)---and that is our best handle for getting at our next big step in physics, but the first thing we've really got to disabuse ourselves of is confusing the inside with the outside.  Quantum states, and their changes, live only on the inside; the ``acts of creation,'' the ``clicks,'' that trigger those changes live on the interface between the inside and outside---thus they should be viewed as distinct from the quantum-state changes themselves.  When we really get that cleared up (as I hope our Bayesian program is doing), then we'll be ready for some serious progress.

Let me paste in couple of notes below that go a little further along these lines.  [See 19-06-05 note ``\myref{Comer72}{Philosopher's Stone}'' to G. L. Comer and 17-06-04 note ``\myref{Mabuchi12}{Preamble}'' to H. Mabuchi.]  But you can see I've got a long way to go in my understanding before I'll be able to drop the poetry and put in its place a little clarity.

If you weren't able to recover the manuscript I gave you from the conference center, let me know and I'll send you another copy if you'd like one.

\section{28-02-06 \ \ {\it How Would You Reply?}\ \ \ (to C. M. {\Caves} \& R. {\Schack})} \label{Caves81.6} \label{Schack99.1}

How would you reply to someone who said this to you:
\bq
There are two main, and to my mind fatal, problems with the epistemic view [of probabilities in statistical mechanics].  The first stems from the role the uniform probability distribution plays in explanations of thermodynamic phenomena.  Consider the explanation of an ice cube's melting towards the future.  If we take an epistemic view of the uniform distribution that is placed over its current macrostate, then part of the reason for the ice's melting will be our ignorance of its initial microstate.  On the assumption that explanations of physical phenomena ought to be objective---and in any case not rely on our epistemic state---we should not maintain that part of the explanation that entropy increases (that ice melts, that coffee cools, that gases expand) is the extent of our knowledge.  How could our epistemic state have anything to do with the ice cube's melting?  This would be like saying that if we happened to be the kinds of beings who did have epistemic access to the initial microstate of the ice cube, then it might have behaved differently.  It is also a consequence of this view that no matter what kind of world we live in, we must assume that the ice is most likely to melt towards the future.  All of that seems crazy: we are after an objective, scientific explanation of thermodynamics.\footnote{\editornote From \myurl{http://aardvark.ucsd.edu/grad_conference/north.doc}.}
\eq

\subsection{{\Ruediger}'s Reply}

\vspace{-12pt}
\bq
\noindent
\bq
\noindent
$\bullet$ There are two main, and to my mind fatal, problems with the epistemic view
\eq

What is the second one?

\bq
\noindent
$\bullet$ Consider the explanation of an ice cube's melting towards the future.  If we take an epistemic view of the uniform distribution that is placed over its current macrostate, then part of the reason for the ice's melting will be our ignorance of its initial microstate.
\eq

No. Our particular state of knowledge is part of the reason for our prediction that the ice melts, not part of the reason for the ice's melting. The ice will either melt or not melt, and it couldn't care less about our ignorance.

Take a trillion systems, each consisting of a glass of water containing an ice cube. The following is then a proposition: ``Each of the systems has a microstate that leads to the ice melting.'' This proposition is either true or false. Its truth value is a fact about the world, independent of our knowledge.

But we don't know the truth value. No theory about the world gives us the truth value. We have to write down a probability, which reflects our ignorance.

\bq
\noindent
$\bullet$ On the assumption that explanations of physical phenomena  ought to be objective---and in any case not rely on our epistemic state---we should not maintain that part of the explanation that entropy increases (that ice melts, that coffee cools, that gases expand) is the extent of our knowledge.
\eq

``This ice cube will melt.'' A proposition. Its truth value: a fact about the world.

``Entropy increases.'' A true statement about how (our) probabilities evolve in time.

``Ice melts.'' An abbreviated version of the subjective judgement that the probability is close to one that ice cube, given the right circumstances, will melt.

\bq
\noindent
$\bullet$ How could our epistemic state have anything to do with the ice cube's melting?  This would be like saying that if we happened to be the kinds of beings who did have epistemic access to the initial microstate of the ice cube, then it might have behaved differently.
\eq

It won't behave differently (see above). But epistemic access to the initial microstate would mean that, in principle, you could extract more free energy than somebody who is ignorant about the microstate.

\bq
\noindent
$\bullet$ It is also a consequence of this view that no matter what kind of world we live in, we must assume that the ice is most likely to melt towards the future.
\eq

Why would this be true? If we lived in a different world, I'd think the first consequence would be that we had different beliefs.

\bq
\noindent
$\bullet$ All of that seems crazy: we are after an objective, scientific explanation of thermodynamics.
\eq

By disentangling facts from beliefs, we pave the way for an objective, scientific explanation of thermodynamics.
\eq

\section{28-02-06 \ \ {\it Our Special Sauce}\ \ \ (to L. Hardy)} \label{Hardy15}

I'm just back from a mandatory ``Bell Labs town meeting.''  It seems that the over-riding purpose of the meeting was to remind all the researchers here that they and their thoughts are just part of the food chain.  Here were some of the words and phrases that caught my attention:
\bq\noindent
``competition makes you stronger,'' ``value proposition,'' ``territory,'' ``what do we have to do to win,'' ``our special sauce,'' ``where do we want to compete to win,'' ``ways to attack certain verticals to win,'' ``attack,'' ``kickin' some butt,'' ``organic ways to win,'' ``where the puck is moving,'' ``we drive value,'' ``new competitive landscape,'' ``we need to develop stickiness,'' ``or else we don't eat,'' ``monetize the core product strategy,'' ``early alignment around our value proposition.''
\eq
This note seemed like a good opportunity to record them all.  Coming back to our conversation on the drive to the airport, I think it is this culture that is the main thing getting me down.

Your visit was a breath of fresh air.  Thanks for coming.  I've had the early Einstein arguments on my mind ever since, and I think you're on to something.

Concerning the Boston meeting, as I work on my schedule this afternoon, I think I probably ought to bow out:  I really can't afford it, both because of my travel budget and because of my time.  I leave for Lisbon March 27, and this would put me away from home immediately beforehand for still more days.  You go there, and remind them that decoherence, many worlds, and nonlocality really aren't all that interesting or creative.

Attached is a picture of Cap'n Hardy and ye maties.

\section{28-02-06 \ \ {\it Quantum Causal Structure}\ \ \ (to L. Hardy)} \label{Hardy16}

I'm back again:  I had forgotten that when Rob {\Spekkens} was here last, I put together a little collection of some of my wilder speculations on the connections between quantum and GR.  Let me share that file with you too.  I think it made little effect on Rob, but---given all the stuff you told me the other night---you might find more of use in it.  (Well, probably nothing of use, but maybe some food for thought.)

Particularly, for me, one of the phrases you used the other night has gotten me excited that it may be possible to give some true-blue substance to those old thoughts of Mead near the end of the document.  I.e., that taking account of ``this'' event may change the ``causal structure'' of all the rest:
\bq\noindent
   [T]he greatest lesson quantum theory holds for us is that when two
   pieces of the world come together, they give birth.  They give birth
   to FACTS in a way not so unlike the romantic notion of parenthood:
   that a child is more than the sum total of her parents, an entity
   unto herself with untold potential for reshaping the world.  Add a
   new piece to a puzzle---not to its beginning or end or edges, but
   somewhere deep in its middle---and all the extant pieces must be
   rejiggled or recut to make a new, but different, whole.  That is the
   great lesson.
\eq
I'm very attracted to that, and to some extent I took the imagery from Mead and his musings on time.

I will be studying your paper thoroughly, and rethinking old Einstein again.

\section{28-02-06 \ \ {\it Stupider Chris} \ \ (to T. Rudolph)} \label{Rudolph11}

\btr
I really like that second paragraph you quoted,
\bq\noindent
{\rm Is the entirety of existence \ldots\ built upon billions upon
billions of elementary quantum phenomena, those elementary acts of
``observer-participancy,'' those most ethereal of all the entities
that have been forced upon us by the progress of science? --- John Wheeler}
\eq
although to be honest I'm somewhat skeptical of the overriding sentiment (I suspect things are ultimately a little more mundane, the history of science is littered with the corpses of the Anthropocentricians).
\etr

Yeah, that's too bad:  Unfortunately, I agree with you.  However, I continue to wonder if one could de-anthropocentrify John's idea and still make something of it:  I.e., that the big bang is here, rather than way back there.  That part of it very much intrigues me.

Below is a daisy-chain of notes demonstrating my latest flailing for the right imagery.  [See 27-02-06 note ``\myref{Ford4}{Wheelerfest}'' to K. W. Ford.]

\section{28-02-06 \ \ {\it Foundational Cost Cutting?}\ \ \ (to A. Wilce)} \label{Wilce10}

Are you planning to go to the ``New Directions in the Foundations of Physics'' meeting, April 28--30?  If so, would you be interested in having a roommate to cut down on costs?  My travel budget is tiny this year, and I'm trying to do everything I can to stretch it.

Let me know what you think, or if you have any other suggestions.

\section{28-02-06 \ \ {\it I've Done It Now}\ \ \ (to G. L. Comer)} \label{Comer81}

Hey, here's something else I'm thinking.  I'm thinking about organizing a little meeting this Fall (or maybe next Spring) over the course of a weekend at my house.  Maybe I'd title it something like ``The Orange House Conference on Quantum Causal Structure.''  There'd be maybe 5 or 6 of the most creative guys I know, and the plan would be to ``get back to basics'' about putting the quantum and the equivalence principle together---notice I didn't say (or necessarily mean) ``quantize gravity'' in the standard sense.  Basically we'd meet on my back patio if the weather is good, or in our library otherwise.  The plan came up when Hardy was visiting me this weekend.  I haven't thought too deeply about this yet, but I'd probably invite you, Hardy, Schumacher, Spekkens, and a couple other guys; maybe Mermin could be there for fun.

I wouldn't have any funding, and with that many people, I'll have to ask everyone to stay at the Best Western in Westfield (2 miles down the road, charming town, lots of nice restaurants, etc).  Hotel runs \$115 a night there.  Anyway, something to think about and maybe plan your budget around.  Let me know what you think.

\section{01-03-06 \ \ {\it The Fuchs Award}\ \ \ (to G. L. Comer)} \label{Comer82}

\bgc
Absolutely!  If the poets and such are to show up, then I'm with you all the way.

I wouldn't want to come as the writer of refereed articles, referee of such articles, writer of NSF proposals, reviewer of such proposals, etc, if you see what I mean.
There are plenty of those to go around now anyway, and most of them, I
think, are completely misguided when it comes to ``tangent bundles over
pseudo-riemannian manifolds'' and the quantum.
\egc

No one who ever talks about such things would ever catch my attention.  Instead, I'm looking for the kind of guy who'd write a draft with a name like ``Classical versus Quantum Interfaces:\  Linear Structures.''  And it helps that he's a practicing general relativist (so he actually knows the classical version of the theory \ldots\ as opposed to the rest of the guys I'm thinking about inviting.)

Hardy and I see eye to eye on many things---for instance, I discovered an amazing overlap between his thoughts and some of the wilder speculations I've sent you before (I'll attach a file of them to remind you) during his visit this weekend---but neither of us really know GR.  Also Schumacher and Mermin both have texts on special relativity, but the same story may be true there too.

So, you see, you'd have a position of honor.

Literally the kind of idea I have in mind is that we'd all go back to 1907--1915 and think about elevators, rotating circles, and how it is only (spacetime) ``collisions'' that should count in our descriptions---but supplement that with all we've learned about ``quantum states as information'' and ``measurements as interventions with unpredictable consequences'' (i.e., a collision of sorts).  The idea is that the mix will be more interesting than the separate ingredients.

\section{01-03-06 \ \ {\it Foundational Cost Cutting?, 2}\ \ \ (to A. Wilce)} \label{Wilce11}

Too bad.  OK, either I'll try to look for another roommate, or I might even drop the plan of going myself.  I'm wishy-washy anyway:  Something deep inside me has been telling me that I've been going to too many foundational conferences (the payoff doesn't seem so big anymore).

\section{06-03-06 \ \ {\it Be Forgiving} \ \ (to G. L. Comer)} \label{Comer83}

\bgc
I'm very much convinced that the historical approach to teaching
special relativity is very much a disservice to the subject.  It
gives precisely the wrong foundational point-of-view.
\egc

I think you ought to be a little forgiving to something like a
historical process for thinking about the foundation of the
theory---or, at least admit that it may have its place.  It may not
be the best pedagogy for students, but it may have its place
elsewhere.

What I'm thinking of here particularly is how one might incorporate
quantum mechanical ideas into the scene.  If one starts the analysis
off---crucial words, ``starts the analysis off''---with spacetime
before rods and clocks (or clocks and maybe some stand-in for light
rays), then the only natural progression one can take may end up
being to try to ``quantize gravity'' in the usual sorts of ways (for
instance, wave functions over three-geometries and variations of that
program).

On the other hand, a more operationalistic approach---one that
doesn't assume spacetimes or wave functions of spacetimes at the
outset---may be just what is needed to free up the conceptual
playground enough to make good progress.  For instance, if a quantum
measurement event helps set the very notion of causal structure in
the first place, then maybe it is best to rethink where along the
lines of earlier thinking one could make the conceptual leap to
``metric is fundamental.''  In the quantum world, maybe one cannot
make that conceptual leap---and thus maybe that is not a good
starting point.

On a different subject:
\bgc
I mean not discussing the metric as fundamental is like presenting
quantum mechanics without mentioning the wave function.
\egc

You see, but that is what I would ultimately like to do!  For, for
me, wave functions are always about {\it someone's\/} degrees of
belief. And what we (quantum Bayesians) would like to get at is a
gambling-agent-independent account of what is up in the quantum
world.  As long as one invokes a wave function, one is always {\it
explicitly\/} talking about an observer.  See transparency \#12 in my
talk ``Being Bayesian in a Quantum World'' posted at the bottom of my
webpage.  It talks to that point directly!

\section{06-03-06 \ \ {\it Expanding the Mollusk} \ \ (to G. L. Comer)} \label{Comer84}

Expanding on my point in the last note a little, you might enjoy reading pages 53 and 54 of Hardy's paper:
\bv
Probability Theories with Dynamic Causal Structure:\ A New Framework for Quantum Gravity\\
Authors: Lucien Hardy\\
Comments: 68 pages, 9 figures\\
\arxiv{gr-qc/0509120}
\ev
It captures a little bit of the conversation we had the other evening which led to the idea of the Orange House Conference.  Maybe see pages 55--56 too.

\section{07-03-06 \ \ {\it Invitation to Grinnell and Talk Titles}\ \ \ (to W. B. Case)} \label{Case1}

Thanks for the invitation; I am flattered.  What is a Squire Lecturer?  It sounds like an honor (and even Bell Labs people think that \$800 sounds like good money).  I would very much like to come, but April is pretty much already filled up for me.  Would it be possible to come in May?  How long of a visit were you thinking?

Titles?  Maybe for the undergraduate talk:
\bq
              ``Drawing a Qubit in a Tetrahedron''
\eq
And how about this for the general talk:
\bq
              ``Quantum Information and the Malleable World''
\eq
They're just ideas.  If you don't like them, send me back to the drawing boards.

\section{07-03-06 \ \ {\it April}\ \ \ (to N. C. Menicucci)} \label{Menicucci1}

\bnm
I would really like to talk to you about Bayesianism in QM.  Howard Wiseman from Griffith University gave a talk recently where he called the Bayesian approach `solipsism' -- i.e., that there is no external world, only one's own mind.  We eventually convinced him that that was crap and not true of the interpretation, but he never really seemed to ``get it'':  How can there be an external world if all quantum states are states of my knowledge?  Well, I told him, they're states of your knowledge about the external world.  Somehow that didn't make sense.
This really got my goat because I think this interpretation might be one of the only ways of correctly looking at quantum theory, since it embraces the paradigm shift the theory inspires (although being a paradigm shift, it's hard for me to pin down in words), but Howard {\bf completely} missed the point and made it sound like all Bayesians are looney.  I tried to defend the position, but I was at a loss for words.
\enm

What you tell me is very sad.  A solipsist, it seems to me, would have no reason whatsoever to do physics, much less try to understand quantum mechanics.

I append a little story below from my ``Quantum States:\ What the Hell Are They?''\ about solipsism.  [See 21-07-01 note ``\myref{Preskill2}{The Reality of Wives}'' to A. J. Landahl \& J. Preskill.] (If you give me a mailing address, and you want a copy, I'll have a copy of the VUP edition of {\sl Notes on a Paulian Idea\/} sent to you.)

\section{07-03-06 \ \ {\it Some Reading}\ \ \ (to N. C. Menicucci)} \label{Menicucci2}

Going back to solipsism, I could point you to a lot of things I've written on this annoying issue, but let me start with this:
\quantph{0204146}.
Do Sections 4 and 5 help any already?

\section{08-03-06 \ \ {\it Counterfactual} \ \ (to S. J. van {\Enk})} \label{vanEnk5}

So, I've got half your counterfactual paper read.  I'll try to tend to the rest of it tomorrow morning and write you any comments that come to mind.

Speaking of counterfactuals, here are two: [\ldots]

2) I'd like to soon get back to the {\Demopoulos} issue.  Particularly, I would like to record better the change you pointed out in how we quantum Bayesians must be treating counterfactuals.  I wish there were a whole paper in that idea.  I guess I came back to the issue today as I was reading Thomas Marlow's paper, ``Relationalism vs.\ Bayesianism'' \arxiv{gr-qc/0603015}.  I couldn't really understand it, but he did drive one point home that you've been driving at with me for a while:
\bq\noindent
     It is often stated that EPR and Bell-like theorems are avoided
     (not disproved but just avoided) by using Bayesian probability
     theory (this is often the major reason people invoke Bayesian
     reasoning in quantum theory [6]), but rarely is it explicitly
     stated how the invocation of Bayesian probabilities overcomes
     arguments for causal nonlocalities. We shall attempt to argue why
     this is the case and why arguments that claim the opposite are
     lacking.
\eq
Ref.\ [6] refers to me.  Unfortunately, as I've already hinted, I didn't get his argument.  But his sentiment is good (and seems to second yours).  So, if by recording the conversation you and I have been having, we can provide a little service to the community, it might be worthwhile.

\section{08-03-06 \ \ {\it A Little {\Wheeler}fest Report} \ \ (to H. C. von Baeyer)} \label{Baeyer20}

\bhcvb
Did you learn anything, or teach anything, at the {\Wheeler}fest?
\ehcvb

Indeed I got a lot from the meeting.  And had many ridiculous fights
with John Conway, who is afraid of the Bayesian idea of probability.
Such a great mind; such a great prejudice!

Particularly useful for me in preparing my own talk was the {\it
strong\/} realization that the Born rule is not at all about setting
probabilities but transforming them!  Its role really is that of a
transformation rule.  There are hints of that in my writings before,
but finally I feel it with all my heart and soul---that's what I
meant by the ``strong realization.''  Anyway it indicates that maybe
one can get at the content of the Born rule through some
Dutch-book-like arguments. Thus I posed that as a question at the end
of my talk. Very nicely, Schumacher took the challenge and has
started to make progress in that direction.

I sort of like the talk I put together for this---even though it goes
out on the limb more than usual (even for me).  I will try to scan in
a copy soon (and send to you), and I will probably give an expanded
variation on it in Sweden.

\section{09-03-06 \ \ {\it Carl Again, a Few More Words}\ \ \ (to R. E. Slusher)} \label{Slusher13}

In the last three years, Carl has posted 13 papers to the quant-ph archive (with his students and postdocs).

The 5 papers I personally found the most interesting are:
\begin{itemize}
\item
\quantph{0304083} [abs, ps, pdf, other] :\\
Title: Physical-resource demands for scalable quantum computation\\
Authors: Carlton M. Caves, Ivan H. Deutsch, Robin Blume-Kohout
\item
\quantph{0306179} [abs, ps, pdf, other] :\\
Title: Gleason-Type Derivations of the Quantum Probability Rule for Generalized Measurements\\
Authors: Carlton M. Caves, Christopher A. Fuchs, Kiran Manne, Joseph M. Renes\\
Journal-ref: Found.\ Phys.\ 34, 193 (2004)
\item
\quantph{0310075} [abs, ps, pdf, other] :\\
Title: Symmetric Informationally Complete Quantum Measurements\\
Authors: Joseph M. Renes, Robin Blume-Kohout, A. J. Scott, Carlton M. Caves\\
Journal-ref: J. Math.\ Phys.\ 45, 2171 (2004)
\item
\quantph{0404137} [abs, ps, pdf, other] :\\
Title: Minimal Informationally Complete Measurements for Pure States\\
Authors: Steven T. Flammia, Andrew Silberfarb, Carlton M. Caves
\item
\quantph{0409144} [abs, ps, pdf, other] :\\
Title: Properties of the frequency operator do not imply the quantum probability postulate\\
Authors: Carlton M. Caves, Ruediger Schack\\
Journal-ref: C. M. Caves and R. Schack, Ann.\ Phys.\ 315, 123--146 (2005)
\end{itemize}

All of the papers are cut from the same cloth:  They are devoted to developing a technical apparatus for getting at what is quantum about quantum information.  Particularly, Carl's worldwide impact has been in his efforts to develop a view of quantum information along Bayesian lines---in fact he is the very father of that field.  To get some indication of the number of serious quantum information theorists willing to take this line of thought seriously, one can look at the participation list for the meeting ``Being Bayesian in a Quantum World'' for which Caves was a co-organizer:
\begin{center}
\myurl[http://web.archive.org/web/20090511060206/http://www.uni-konstanz.de/ppm/events/bbqw2005/]{http://web.archive.org/web/20090511060206/http:// www.uni-konstanz.de/ppm/events/bbqw2005/}
\end{center}
You'll see Hans Briegel, Lucien Hardy, David Mermin, Gerard Milburn, Michael Nielsen, John Smolin, Bill Wootters, and many others there.

Caves is a behind-the-scenes worker, a kind of theorist's theorist.  The press may not know him so much (because he's not a weird, journalist-attracting recluse like David Deutsch) nor may the funding agencies (because he's not a flamboyant stage actor like Seth Lloyd), but anyone in this field who is worth his salt knows of Caves's intellect and influence.

\section{10-03-06 \ \ {\it Opening of My APS Talk}\ \ \ (to R. E. Slusher)} \label{Slusher14}

Your story of how you opened your MIT talk inspired me for the opening of my own talk at the APS meeting Monday.

I'm going to start off by reading this quote of William James:
\bq\noindent
   Metaphysics has usually followed a very primitive kind of quest. You
   know how men have always hankered after unlawful magic, and you know
   what a great part in magic {\it words\/} have always played. If you
   have his name, or the formula of incantation that binds him, you can
   control the spirit, genie, afrite, or whatever the power may be.
   Solomon knew the names of all the spirits, and having their names, he
   held them subject to his will.  So the universe has always appeared
   to the natural mind as a kind of enigma, of which the key must be
   sought in the shape of some illuminating or power-bringing word or
   name.  That word names the universe's {\it principle}, and to possess
   it is after a fashion to possess the universe itself.
\eq
Then I'm going to show a big $|\psi\rangle$ and say we finally know its {\it name}:  It is ``expectation''---raw, unmitigated expectation with no further need for justification.  If we take that name and that idea seriously, we will ``possess the universe itself.''

I'm going to Princeton this morning to work with Halvorson a bit.  Depending upon when that finishes, I may or may not come in to Bell Labs.  If you really need me around, drop me a note.

\section{10-03-06 \ \ {\it Probably More Than I Should Send, 2} \ \ (to J. E. {\Sipe})} \label{Sipe7}

My post-Wheelerfest days have passed away, and it has just been one thing after another.  Now the APS meeting is upon us!  Time, time, time; what has become of me!

Still, I wouldn't mind writing you something today:  Never pass up an opportunity to expose a colleague to Bayesian ways of thought!  But I have to go to Princeton this morning to work with Halvorson.  May I ask, what is the last time you'll be checking email before leaving for Baltimore (if you haven't already left for Baltimore)?  You know, if there's a carrot in front of the horse, he'll pull the cart!

I'll plan my afternoon according to you.

\section{10-03-06 \ \ {\it Isms} \ \ (to J. E. {\Sipe})} \label{Sipe8}

\bjes
I'll be checking email until about 5PM today (Friday); I fly to
Baltimore tonight, and hope to be checking email there, but I'm not
sure that'll be possible until the conference starts, and my tutorial
is Sunday morning!
\ejes

Ouch, I've run out of time for all those delicate and carefully
constructed answers I wanted to give you!  My fault---you deserve
much better than me.  It's now almost 3PM, and I'm just finished
talking to Hans.  Now I'm in a Panera Bread in Princeton, having a
cup of coffee before my drive home.  Let me send you a very little
note answering some of your questions, and maybe we can talk more
next week.

I certainly liked the way you posed your questions.

\bjes
I gather from this that you see the ultimate task to be the
identification of just what is the part of quantum mechanics that is
``about the gambler-independent world.''  The Bayesian analysis is a
strategy, as I understand it, to identify that task and undertake it.
It is a means, not an end.

But if I understand this correctly I am puzzled to see statements
like: \ldots\ [blah blah blah measurements create outcomes]

Fair enough.  But might it not be that the ``gambler-independent
world'' is in fact described by a hidden-variable theory of some
sort? For example, why would you (immediately!)\ rule out a Bohm--de
Broglie universe (or a modified one, with the usual distribution over
configuration space being a kind of `equilibrium distribution' that
we might someday be able to overcome, such as Valentini suggests)
simply because of your Bayesian strategy in trying to identify just
what that ``gambler-independent world'' is? Could it not be that Bohm
and de Broglie have really identified the ``fact of the matter in the
universe,'' or that some such approach would?  At least, must not
this possibility be seriously considered? How does a Bayesian
strategy immediately rule it out? By rejecting even the possibility
that some quantum measurement outcomes may pre-exist the measurement
process, do you not prejudge the outcome of the larger quest, for
which the Bayesian viewpoint is not a goal but a strategy?
\ejes
Here is the way I once answered Jeff Bub when he asked a similar
question.  Jeff asked:
\bq
   \noindent
   You want to take quantum mechanics as a theory of the way information
   is represented and the limitations on the communication of
   information, and not a description of the behavior of particles, as
   in classical mechanics. Granted, the way the world is hard-wired
   might impose limitations on the gathering of information and the
   exchange of information -- limitations expressed precisely by the
   `limited sort of privacy' we have (i.e., secure kd, but no secure
   bc), hence by science necessarily taking the form of quantum
   mechanics. But how does it follow from this that we must interpret
   quantum mechanics as a theory of information, and not as a
   descriptive theory in the sense of classical mechanics?
\eq
My reply was:
\bq
   It doesn't.  You're completely on track there.  Do you remember my
   slide where I listed the axioms of quantum mechanics in {\Montreal}?
   In my presentation I said how I'm always struck by the stark contrast
   between that list of axioms and the ones we take for our other
   cornerstone theory of the world (referring to special relativity):
   (1) the speed of light is constant, and (2) physics is the same in
   all frames.  The debate over the foundations of quantum mechanics
   will not end until we can reduce the theory to such a set of crisp
   physical statements---I believe that with all my heart. However, just
   as special relativity will always be interpretable in Lorentz's
   way, quantum mechanics will likely always be interpretable in
   Bohm's way.  There's nothing we can do about that.

   What I'm really searching for is just a polite way to say, ``Ahh,
   blow it out your butt.  You can believe that Lorentzian way of
   looking at things if you want to, but why when have this absolutely
   simple alternative conceptual structure?''

   What I'm looking for is just a couple of crisp physical
   statements---contradictory appearing even, just as
   Einstein's---that can characterize what quantum mechanics is all
   about.  Something like (but more precise than):
\begin{enumerate}
   \item The effects of our interventions into the world are
   nondiminishable. And,

  \item But still we have science; all the world is not simply a dream of
   our own concoction.
\end{enumerate}
   Once we get that cleared up, we'll finally be ready to move to the
   next stage of physics, much like Einstein was ready to move on to
   general relativity once he had reduced the Lorentz contractions to
   the two statements above.
\eq
That's the best answer I think I can give you at the moment.  Even a
Bayesian approach to quantum mechanics cannot DISPROVE nonlocal,
noncontextual hidden variables theories.  It just emphasizes that
they don't smell right.

The ``proof'' of any approach will ultimately be in what new physics
it gives rise to.

Now for your most important question:
\bjes
In any case, (and to make sure in this tutorial I don't confuse
somebody about what you really do hold!) could I get you to see if
the way I describe your approach below is (a) basically right, (b)
partly right but with major flaws, or (c) totally screwed up? \ldots\

I use the term ``interpretation'' of a modern physical theory to mean
an identification of just what it is to which the abstract elements
in the theory purportedly refer.  A ``type of interpretation''
identifies the type of things to which the abstract elements in the
theory refer.

The abstract elements in quantum mechanics are density operators,
POVMs, etc., etc. A ``realist'' type of interpretation identifies the
abstract elements with stuff that purportedly really exists in the
world, the ``furniture of the universe.''  An ``operationalist'' type
of interpretation identifies the abstract elements with tasks in the
laboratory, such as preparations, transformations, and measurements.
And so on.  An ``instrumentalist'' type of interpretation says that
abstract elements don't refer to anything at all, except perhaps in
the way a hammer refers to a nail.  That is, theories are just (only)
tools for coping in the world. \ldots

Now you and your colleagues hold ``the idea that quantum STATES are of
essence information'' and the obvious question a beginning student
would ask -- and undoubtedly someone will ask if I say something like
this in the tutorial -- is ``information about what?''  Perhaps one
could argue that this question has no meaning, but I don't see how
that view can be sustained.

In any case, from the talks I have heard you give I came away with
the impression that, at least in the short term and from the point of
view of our current work-a-day physics, you take this to be
information, broadly speaking, about the impact and results of future
tasks in the laboratory, such as measurements.  On the basis of this
it seems to me you are in the ``operationalist'' camp.

Now within this camp there are various tribes. Some feel that any
talk about underlying ``furniture of the universe'' is meaningless,
or at least not worthwhile.  They are ``hard core operationalists.''
Gadgets and tasks for them are ultimate primitives in their
theorizing, and for them an operationalist physics is all there can
ever be, at least if it is to be what they would consider a proper
physics.

I take you to be in a different tribe, whose members feel that
ultimately we want a physics that does indeed talk about the
furniture of the universe, the ``gambler-independent world.'' Indeed,
you refer to that as the ``wheat,'' as opposed to the ``chaff.'' But
that ``gambler-independent world'' has not yet been identified.  In
the interim you argue that the abstract elements of quantum mechanics
should be understood operationally, and the predictions and
statements of quantum mechanics subjected to a kind of ``Bayesian
rack'' to clarify exactly what beyond the objective rules of
probability actually exists in them.  This analysis then provides a
firm footing for the search for the actual ``furniture of the
universe,'' whatever it is.  Ultimately you want a theory that does
describe this furniture; I would guess you are uncertain as to
whether quantum mechanics (sufficiently purified by Bayesian
analysis) would provide that, or whether a successor theory would be
required.  But in any case in your heart of hearts you hope someday
for a realist science, or at least hope that that hope makes sense.
\ejes

I think my way of looking at the terms {\it within\/} quantum
mechanics runs the whole gamut from instrumentalism to realism, and
there's no way to put it into one single category.
\begin{enumerate}
\item
When I posit a physical system about which I will speak (by assigning it a Hilbert space, etc.), I am doing that in an almost na\"{\i}ve realistic way.  I.e., the $\cal H$ I write down, represents a piece of the world that is out there independently of me.

\item
When I assign a dimension $d$ to that Hilbert space, I am
hypothesizing an inherent {\it property\/} of that system.  I might be right, or I might be wrong, but $d$ is something I hypothesize of it, even if only provisionally.  Realism again.

\item
When I draw a quantum state out of the space of operators defined by $\cal H$, however, I am expressing a bundle of my expectations. These are not properties inherent in the system.  They are subjective expectations that I bring into the picture (presumably because they have served me well in the past, or at least done me no harm).  If I were to conceptually delete myself from the picture, these expectations would disappear with me.  To that extent, the view might sound a little like---or at least be confused with---idealism (but that's only if one---as Howard Wiseman often does---forgets elements 1 and 2 above and one of the further elements below.)

\item
The subject matter of those expectations, i.e., what they are about, refers both to me and the system I posit.  They are MY expectations for the consequences (for ME) of MY interactions with the system.  That, you might think is a kind of operationalism:  For if I were to conceptually delete myself from the picture, those interactions would disappear too.

\item
On the other hand, the reason we use the formal structure of quantum mechanics to bundle our expectations, to manipulate and update them, to do all that we do with them, is to better cope with the world.  It is a means to help our species to survive and propagate.  That is a kind of instrumentalism.  That part of quantum mechanics is a tool like a hammer; it can be used to fix a lot of things, or simply as an aid to help defend ourselves.

\item
Still one can never forget the ultimately uncontrollable nature of
each quantum measurement outcome---and through Kochen--Specker, at
least the way I view it, the non-pre-existence of those ``outcomes''. That smacks of realism in the oldest, most time-honored way.  The world surprises us and is not a creation of our whims and fancies. Back to realism \ldots\ but the twist is quantum mechanics, as used by each individual user, only refers to the outcomes HE helps generate. (That smacks of alchemy \ldots\ dangerous to say so, but I call it like it is.)

\item
Nevertheless, having learned a little from Copernicus, it seems we
should ultimately try to abstract away from these personal encounters with the world (having learned what we could from the formalism concerned with gambling on them).  If a tiny little system and I create something new in the world when we get together---i.e., we give rise to a birth or new fact---so must it be likely, it seems to me, that any two things give rise to a birth when they get together. That is ``F-theory'' or the ``sexual interpretation of quantum mechanics'' \ldots\ but you'll have to wait for the movie if you question me any further on that \ldots
\end{enumerate}

I hope that answers you a little.  It took me a cup and a half of
coffee, and I think I just barely made deadline.

Have a safe trip.  It'll be fun talking to you next week.

\section{10-03-06 \ \ {\it Isms, 1} \ \ (to H. Halvorson)} \label{Halvorson13}

I didn't explain in the last note why I didn't leave Princeton until 4:00.  What happened was that I owed John Sipe a note before his leaving for the APS meeting to give a four-hour tutorial on quantum foundations Sunday---he wanted a little explanation of our Bayesian point of view so that he might incorporate it into the tutorial.  Well, after leaving Marx Hall, I started thinking, ``I don't quite know when he'll be leaving, so I'd better check my email.''  Thus I popped in to Panera's Bread, and sure enough I found out that he'd soon be checking his email for the last time before his flight.  So, I quickly composed the note below. [See 10-03-06 note ``\myref{Sipe8}{Isms}'' to J. E. {\Sipe}.]

The part that might interest you is the list at the end explaining the various isms that my view of the quantum mechanics seems to cross.  The list is incomplete:  I think I hit even more isms than that, but maybe this is a good starting point for categorization.

Another note on relevant isms coming after I get a chance to search the web.

Still a silent house!

\section{10-03-06 \ \ {\it Isms, 2} \ \ (to H. Halvorson)} \label{Halvorson14}

The other thing I wanted to write you about isms is this.  Here is
the paper by Huw Price that I was telling you about:
\bq\noindent
``Naturalism without Representation'' \\
\myurl{http://www.usyd.edu.au/time/price/preprints/naturalism-final.pdf}
\eq
The abstract for the paper is below.  I recall liking the paper very
much.  And to the extent that I understand the term, I believe I can
classify myself (presently) as what Huw calls a ``subject
naturalist.''

I wonder how you would characterize a) yourself, and b) your other
colleagues in your department, in these terms.

\bq\noindent
Abstract:  I begin with a distinction between two ways of taking
science to be relevant to philosophy. The first (``object
naturalism'') is a ontological thesis -- it holds that what exists,
what we should be realists about, is the world as revealed by
science. The second (``subject naturalism'') is a prescription for
philosophy, based on the belief that we humans (and in particular,
our thought and talk) are part of the natural world. What is the
relationship between these two kinds of naturalism? Contemporary
naturalists are apt to think that the latter view is a mere corollary
of the former. I argue that there is an important sense in which the
priority is the other way around: object naturalism depends on
``validation'' from a subject naturalist perspective -- in
particular, on confirmation of certain ``representationalist''
assumptions about the functions of human language. Moreover, I
maintain, there are good reasons for doubting whether object
naturalism deserves to be validated, in this sense. Thus, an adequate
naturalistic philosophy threatens to undermine what most contemporary
philosophers have in mind, when they call themselves philosophical
naturalists.
\eq

\section{10-03-06 \ \ {\it Singularities and Evolutionary Laws} \ \ (to H. Halvorson)} \label{Halvorson15}

Below are the three quotes I have by {\Poincare} concerning the
subject above.  The quotes seem horribly inadequate now (with respect
to what I remember of the niceness of the article); I wish I had more
extensive quotes in the computer.  Always too much to do \ldots\

From: H.~{\Poincare}, ``The Evolution of Laws,'' in his book {\sl
Mathematics and Science:\ Last Essays (Derni\`eres Pens\'ees)},
translated by J.~W. Bolduc, (Dover, New York, 1963), pp.~1--14.

\bq
Mr.\ Boutroux, in his writings on the contingency of the laws of
Nature, queried, whether natural laws are not susceptible to change
and if the world evolves continuously, whether the laws themselves
which govern this evolution are alone exempt from all variation.
\ldots\ I should like to consider a few of the aspects which the
problem can assume.
\eq
and
\bq
In summary, we can know nothing of the past unless we admit that the
laws have not changed; if we do admit this, the question of the
evolution of the laws is meaningless; if we do not admit this
condition, the question is impossible of solution, just as with all
questions which relate to the past. \ldots

But, it may be asked, is it not possible that the application of the
process just described may lead to a contradiction, or, if we wish,
that our differential equations admit of no solution?  Since the
hypothesis of the immutability of the laws, posited at the beginning
of our argument would lead to an absurd consequence, we would have
demonstrated {\it per absurdum\/} that laws have changed, while at
the same time we would be forever unable to know in what sense.

Since this process is reversible, what we have just said applies to
the future as well, and there would seem to be cases in which we
would be able to state that before a particular date the world would
have to come to an end or change its laws; if, for example, our
calculations indicate that on that date one of the quantities which
we have to consider is due to become infinite or to assume a value
which is physically impossible.  To perish or to change its laws is
just about the same thing; a world which would no longer have the
same laws as ours would no longer be our world but another one.
\eq
and
\bq
No doubt many readers will be dismayed to note that I seem constantly
to substitute for the world a system of simple symbols. This is not
due simply to a professional habit of a mathematician; the nature of
my subject made this approach absolutely necessary. The Bergsonian
world has no laws; what can have laws is simply the more or less
distorted image which the scientists make of it.  When we say that
nature is governed by laws, it is understood that this portrait is
still rather lifelike.  It is therefore according to this description
and this description only that we must reason, or else we run the
risk of losing the very idea of law which was the object of our
study.
\eq

\section{12-03-06 \ \ {\it The Spirit of Stevie Ray} \ \ (to G. L. Comer)} \label{Comer85}

Boy you sure send a lot of pictures of your guitar!  Handsome picture
of the singer though---you're right, you do look authentic.

I'm just off to the APS March meeting in Baltimore \ldots\ where our
little effort in physics has finally been recognized with a topical
group all of its own.  Sessions all week long.  I chair the one on
entanglement tomorrow morning, and then give my talk in the session
on quantum foundations in the afternoon.

Just made a transparency of this William {\James} quote:
\bq\noindent
Metaphysics has usually followed a very primitive kind of quest. You
know how men have always hankered after unlawful magic, and you know
what a great part in magic {\it words\/} have always played. If you
have his name, or the formula of incantation that binds him, you can
control the spirit, genie, afrite, or whatever the power may be.
Solomon knew the names of all the spirits, and having their names, he
held them subject to his will.  So the universe has always appeared
to the natural mind as a kind of enigma, of which the key must be
sought in the shape of some illuminating or power-bringing word or
name.  That word names the universe's {\it principle}, and to possess
it is after a fashion to possess the universe itself.
\eq
I'm gonna read that to the audience, and then put up a transparency
with a big $|\psi\rangle$ and nothing else on it.  Then I'll say that
since the beginning of quantum mechanics, the debate has been over
what this thing actually is.  But now we know its name:  it is
``expectation.''  Raw expectation.  (A better word than
``information'' I think.)  And now that we finally possess its name,
we will after a fashion possess the universe itself.

Then I'll bombard them with too many equations \ldots\ as one is
expected to do in a physics talk.

Good luck with your guitar!  Good luck with my sales pitch!

\section{20-03-06 \ \ {\it The Rest of the Story?}\ \ \ (to J. E. {\Sipe})} \label{Sipe9}

\bjes
Regardless of the strategy, the quantum world of this interpretation
is a fixed, static thing. It is a frozen, changeless place. Dynamics
refers not to the quantum world, but only to our actions, our
experiences, and our beliefs as agents. Or, more poetically (\`a la
Chris), life does not arise from our interventions; it is our
interventions.
\ejes

Just like in our conversation, I think I'm pretty sympathetic with your description \ldots\ until we get to here.

It is the substrate (and catalyst) that is changeless, but it is being written upon like a writing pad with the construction of a story.  And that part of the process seems to me to be as important as anything else.  In fact, I personally consider it a crucial part of the ontology.  I want to better understand your de-emphasis of it.

Let me give you a little material below to react to.  The first three passages represent what I have been calling the ``Paulian Idea''---they are almost literal quotes from {\Pauli} himself, but not quite, as I had to compress them somewhat to fit on the transparencies I was using for my presentation (at the {\Wheeler}fest).  I am not in complete agreement with these---for instance, sometimes he forgets his own emphasis that the measurement device should be considered as a prolongation of the agent---but maybe they can help get your gears turning, before I turn to more radical stuff.

Following {\Pauli}, I'll put one version of {\Wheeler}'s 20-question story that was influential on me.

Finally I'll end with one excerpt from my {\sl Notes on a Paulian Idea\/} and a couple of excerpts from my ``Anti-{\Vaxjo} Interpretation of Quantum Mechanics.''  These pieces emphasize, why I think the unpredictable outcome of a quantum measurement should be included in the ontology.

Do you see these things as consistent with your description above, or do you see them as contradictory?  I would like to get that straight, because as I say, I like much in your description previous to that.  (I'll come back to some more minor technical points in it later.)  And who knows, maybe I've been misunderstanding my own view!  It, like the world, is malleable after all.

\bq\noindent
{\bf Paulian Idea, 1}

[Einstein and I] often discussed these questions, and I invariably
profited very greatly even when I could not agree with Einstein's
views. ``Physics is after all the description of reality,'' he said
to me, continuing, with a sarcastic glance in my direction, ``or
should I perhaps say physics is the description of what one merely
imagines?'' This question clearly shows Einstein's concern that the
objective character of physics might be lost through a theory of the
type of quantum mechanics, in that as a consequence of its wider
conception of objectivity of an explanation of nature the difference
between physical reality and dream or hallucination become blurred.

The objectivity of physics is however fully ensured in quantum
mechanics in the following sense.  Although in principle, according
to the theory, it is in general only the statistics of series of
experiments that is determined by laws, the observer is unable, even
in the unpredictable single case, to influence the result of his
observation---as for example the response of a counter at a
particular instant of time.  Further, personal qualities of the observer do not come into the theory in any way---the observation
can be made by objective registering apparatus, the results of which
are objectively available for anyone's inspection. Just as in the
theory of relativity a group of mathematical transformations connects
all possible coordinate systems, so in quantum mechanics a group of
mathematical transformations connects the possible experimental
arrangements.

Einstein however advocated a narrower form of the reality concept \ldots\
\eq

\bq\noindent
{\bf Paulian Idea, 2}

It seems to me appropriate to call the conceptual description of
nature in classical physics, which Einstein so wishes to retain,
``the ideal of the detached observer''. To put it drastically, the
observer has according to this ideal to disappear entirely in a
discrete manner as hidden spectator, never as actor, nature being
left alone in a predetermined course of events, independent of the
way in which the phenomena are observed. ``Like the moon has a
definite position'' Einstein said to me last winter, ``whether or not
we look at the moon, the same must also hold for the atomic objects,
as there is no sharp distinction possible between these and
macroscopic objects. Observation cannot {\it create\/} an element of
reality like a position, there must be something contained in the
complete description of physical reality which corresponds to the
{\it possibility\/} of observing a position, already before the
observation has been actually made.'' It is this kind of postulate
which I call the ideal of the detached observer.

In quantum mechanics, on the contrary, an observation changes in
general the ``state'' of the observed system in a way not contained
in the mathematically formulated {\it laws}, which only apply to the
time dependence of the state of a {\it closed\/} system. I think here
on the passage to a new phenomenon by observation which is taken into
account by the so called ``reduction of the wave packets.'' As it is
allowed to consider the instruments of observation as a kind of
prolongation of the sense organs of the observer, I consider the
impredictable change of the state by a single observation to be {\it
an abandonment of the idea of the isolation (detachment) of the
observer from the course of physical events outside himself}.
\eq

\bq\noindent
{\bf Paulian Idea, 3}

Like an ultimate fact without any cause, the {\it individual\/}
outcome of a measurement is, however, in general not comprehended by
laws. This must necessarily be the case, if quantum mechanics is
interpreted as a natural generalization of classical physics, which
takes into account the finiteness of the quantum of action. \ldots\

The significance of this development is to give us insight into the
logical possibility of a new and wider pattern of thought.  This
takes into account the observer, including the apparatus used by him,
differently from the way it was done in classical physics. In the new
pattern of thought we do not assume any longer the {\it detached
observer}, occurring in the idealizations of this classical type of
theory, but an observer who by his indeterminable effects creates a
new situation, theoretically described as a new state of the observed
system.  In this way every observation is a singling out of a
particular factual result, here and now, from the theoretical
possibilities, thereby making obvious the discontinuous aspect of the
physical phenomena.

Nevertheless, there remains still in the new kind of theory an {\it
objective reality}, inasmuch as these theories deny any possibility
for the observer to influence the results of a measurement, once the
experimental arrangement is chosen.  Therefore particular qualities
of an individual observer do not enter the conceptual framework of
the theory.
\eq

\bq\noindent
{\bf {\Wheeler}ish Idea, 1}

But if I could have asked Bohr, how did he think the Universe came
into being, and what is its substance, what would he have said?

It is too late to ask.  The plan is up to us to find.

The Universe can't be Laplacean.  It may be higgledy-piggledy.  But
have hope.  Surely someday we will see the necessity of the quantum
in its construction.  Would you like a little story along this line?

Of course!  About what?

About the game of twenty questions.  You recall how it goes---one of
the after-dinner party sent out of the living room, the others
agreeing on a word, the one fated to be a questioner returning and
starting his questions.  ``Is it a living object?''  ``No.''  ``Is it
here on earth?''  ``Yes.''  So the questions go from respondent to
respondent around the room until at length the word emerges: victory
if in twenty tries or less; otherwise, defeat.

Then comes the moment when we are fourth to be sent from the room. We
are locked out unbelievably long.  On finally being readmitted, we
find a smile on everyone's face, sign of a joke or a plot.  We
innocently start our questions.  At first the answers come quickly.
Then each question begins to take longer in the answering---strange,
when the answer itself is only a simple ``yes'' or ``no.''  At
length, feeling hot on the trail, we ask, ``Is the word `cloud'?''
``Yes,'' comes the reply, and everyone bursts out laughing.  When we
were out of the room, they explain, they had agreed not to agree in
advance on any word at all.  Each one around the circle could respond
``yes'' or ``no'' as he pleased to whatever question we put to him.
But however he replied he had to have a word in mind compatible with
his own reply---and with all the replies that went before.  No wonder
some of those decisions between ``yes'' and ``no'' proved so hard!

And the point of your story?

Compare the game in its two versions with physics in its two
formulations, classical and quantum.  First, we thought the word
already existed ``out there'' as physics once thought that the
position and momentum of the electron existed ``out there,''
independent of any act of observation.  Second, in actuality the
information about the word was brought into being step by step
through the questions we raised, as the information about the
electron is brought into being, step by step, by the experiments that
the observer chooses to make. Third, if we had chosen to ask
different questions we would have ended up with a different word---as
the experimenter would have ended up with a different story for the
doings of the electron if he had measured different quantities or the
same quantities in a different order.  Fourth, whatever power we had
in bringing the particular word ``cloud'' into being was partial
only.  A major part of the selection---unknowing selection---lay in
the ``yes'' or ``no'' replies of the colleagues around the room.
Similarly, the experimenter has some substantial influence on what
will happen to the electron by the choice of experiments he will do
on it; but he knows there is much impredictability about what any
given one of his measurements will disclose.  Fifth, there was a
``rule of the game'' that required of every participator that his
choice of yes or no should be compatible with {\it some\/} word.
Similarly, there is a consistency about the observations made in
physics.  One person must be able to tell another in plain language
what he finds and the second person must be able to verify the
observation.
\eq

\bq\noindent
{\bf Flying Equations Story}

Good holidays to you.  This morning, as I was driving to work, it
dawned on me that roughly this day 10 years ago, I was conferred my
degrees at the University of Texas.  Time does fly.

It made me think of a little anecdote about John {\Wheeler} that I
heard from John Preskill a few days ago.  In 1972 he had
{\Wheeler} for his freshman classical mechanics course at
Princeton.  One day {\Wheeler} had each student write all the
equations of physics s/he knew on a single sheet of paper.  He
gathered the papers up and placed them all side-by-side on the stage
at the front of the classroom.  Finally, he looked out at the
students and said, ``These pages likely contain all the fundamental
equations we know of physics.  They encapsulate all that's known of
the world.''  Then he looked at the papers and said, ``Now fly!''
Nothing happened.  He looked out at the audience, then at the
papers, raised his hands high, and commanded, ``Fly!'' Everyone was
silent, thinking this guy had gone off his rocker.  {\Wheeler} said,
``You see, these equations can't fly.  But our universe flies.
We're still missing the single, simple ingredient that makes it all
fly.''

Merry Christmas.
\eq

\bq\noindent
{\bf V\"axjination, 1}

The situation of quantum mechanics---I become ever more
convinced---illustrates this immersion of the scientific agent in
the world more clearly than any physical theory contemplated to
date.  That is because it tells you you have to strain really hard
and strip away most of the theory's operational content, most of its
workaday usefulness, to make sense of it as a reflection of ``what
is'' (independent of the agent) and---importantly---you insist on
doing that for all the terms in the theory. \ldots\

So, I myself am left with a view of quantum mechanics for which the
main terms in the theory---the quantum states---express nothing more
than the gambling commitments I'm willing to make at any moment.
When I encounter various other pieces of the world, if I am
rational---that is to say, Darwinian-optimal---I should use the
stimulations those pieces give me to reevaluate my commitments. This
is what quantum state change is about.  The REALITY of the world I
am dealing with is captured by two things in the present picture:
\begin{enumerate}
\item
I posit systems with which I find myself having encounters, and
\item
I am not able to see in a deterministic fashion the stimulations
(call them measurement outcomes, if you like) those systems will
give me---something comes into me from the outside that takes me by
surprise.
\end{enumerate}
\eq

\bq\noindent
{\bf V\"axjination, 2}

Yes there is certainly a kind of realism working in the back of my
mind, if what you mean by ``realism'' is that one can imagine a world
which never gives rise to man or sentience of any kind.  This, from
my view, would be a world without science, for there would be no
scientific agents theorizing within it.  This is what I mean by
realism:  That man is not a priori the be-all and end-all of the
world.  (The qualification ``a priori'' is important and I'll come
back to it later.)

A quick consequence of this view is that I believe I eschew all
forms of idealism.  Instead, I would say all our evidence for the
reality of the world comes from without us, i.e., not from within
us.  We do not hold evidence for an independent world by holding
some kind of transcendental knowledge.  Nor do we hold it from the
practical and technological successes of our past and present
conceptions of the world's essence.  It is just the opposite.  We
believe in a world external to ourselves precisely because we find
ourselves getting unpredictable kicks (from the world) all the
time.  If we could predict everything to the final T as Laplace had
wanted us to, it seems to me, we might as well be living a dream.

To maybe put it in an overly poetic and not completely accurate way,
the reality of the world is not in what we capture with our
theories, but rather in all the stuff we don't.  To make this
concrete, take quantum mechanics and consider setting up all the
equipment necessary to prepare a system in a state $\Pi$ and to
measure some noncommuting observable $H$.  (In a sense, all that
equipment is just an extension of ourselves and not so very
different in character from a prosthetic hand.)  Which eigenstate of
$H$ we will end up getting as our outcome, we cannot say.  We can
draw up some subjective probabilities for the occurrence of the
various possibilities, but that's as far as we can go.  (Or at least
that's what quantum mechanics tells us.)  Thus, I would say, in such
a quantum measurement we touch the reality of the world in the most
essential of ways.
\eq

\section{20-03-06 \ \ {\it Replies to {\Sipe}} \ \ (to R. {\Schack})} \label{Schack100}

The history of the conversation with John {\Sipe} goes like this.  After reading the abstract for his tutorial at the APS meeting, I sent he and Rob {\Spekkens} this little quip (and asked Matt Leifer if I embarrassed him by using his name in some of the attached material):
\begin{verse}
T3 Current Interpretations of Quantum Mechanics\\
Organizer: Rob {\Spekkens}, Perimeter Institute for Theoretical
Physics
\\
Room 302, Baltimore Convention Center\\
\ldots\\
Instructor:
Professor John {\Sipe}, University of Toronto
\end{verse}
\bq\noindent
Too bad the pseudo-Bayesian view of {\Caves}, {\Schack}, {\Appleby}, Fuchs and few others isn't a current interpretation of quantum mechanics \ldots\
\eq
That got John\index{Sipe, John E.} to asking questions, but also Matt replied:
\bq
\noindent
I think that in Rob\index{Spekkens, Robert W.} and John's\index{Sipe, John E.} terminology the Bayesian approach is
not a ``current interpretation'' but an idea for an interpretation. They are careful to distinguish the two in their book. A full blown
interpretation is required to account for all of nonrelativistic
quantum mechanics and to make precise what the ontology of the
interpretation is. I don't think we have all agreed upon a consistent
answer to the latter question yet.
\eq

I tend to take Matt's point seriously, and I worry about losing the ground we've gained with {\Sipe}.  It was for this reason that I really tried to shy away from the issue of Hamiltonians when {\Carl} and I were talking to him the other day \ldots\ though {\Carl} really wanted to put it front and center.

Similarly I might say about a possible corrective to his description of ``measurements as [objectively described] tasks [via the formal assignment of a POVM].''  Which is what I think you were getting at with your words:
\brs
One point that surprises me is that he got the impression that there
is a fundamental difference in the way we look at preparations and
measurements. How on earth did that happen?
\ers
Partially it happened by my choosing to be silent on the issue.

Below I'll place the part's of John's\index{Sipe, John E.} description that I myself consider problematic, \ldots\ either because it doesn't convey properly our current unified thinking, or because it may pinpoint an area our current disagreement.

How to reply to him further, {\it if at all}, I am not yet clear on.

I've got to do a bunch of stupid company things today, but I really, really, really hope to start writing you two some detailed notes on the paper tomorrow.

Oh, also concerning {\Sipe}, did you already get from {\Carl} the one thing I did send him in reply a few days ago?  I'm somewhat timid to send it over the airwaves again, as {\Carl} has already pointed out its exorbitant length \ldots

--- --- --- --- --- --- --- --- --- --- --- --- --- --- --- --- ---

Things from {\Sipe}'s note and small comments.

\bq\noindent
Thus, while the abstract elements in the theory associated with
measurements are identified with tasks in the laboratory, as in
operationalism, the abstract elements in the theory associated with
preparations are identified with beliefs of the agent, signaling a
kind of empiricist perspective.
\eq
I understand {\Ruediger}'s worry, but there is some small truth in this and I think it worth nurturing at this stage in his thinking (without overwhelming him).  Yes, from our view, writing down a quantum state and writing down a set of effects is, in both cases, the writing down of beliefs.  But they are beliefs of different things.  And John\index{Sipe, John E.} carries the hint of that in his description.  A POVM represents the agent's belief about the action he will take on the external world; a quantum state represents his belief about which of the consequences will come about from his action.

\bq\noindent
In contrast, a usual Stern--Gerlach device oriented along the $z$
direction does not, in this interpretation of quantum mechanics,
reveal the $z$-component of angular momentum, or for that matter
anything else. The particular outcome of one experimental
run is simply a consequence of performing the experiment.
\eq
Agreed.

\bq\noindent
Nonetheless, repeated experimentation does reveal that the electron
associated with the atom passing through the device should be taken as a spin-1/2 particle. Here the attribute under consideration is taken
to be internal angular momentum, and the instance -- the irreducible
representation appropriate to the particle of interest -- spin-1/2.
\eq
This passage highlights the difference in ways that {\Carl} and I talk.  {\Carl} would say that a system's attributes of this nature---say the system's Hamiltonian---are set intrinsically.  The Hamiltonian (again, for instance) gives the system identity across time.  Whereas, I would say that such identity is set by a subjective judgment.

It is my judgment that I should write down a constant Hamiltonian through time for such and such system:  Measurement cannot reveal the TRUE Hamiltonian.  To the extent that it can, one is making a judgment analogous to exchangeability.  Because of the quantum de Finetti theorem, we know there is only the illusion that measurement can be used to infer the one true quantum state.  Similarly I would say of the one true Hamiltonian.

On the other hand, I appreciate {\Carl}'s point that measurements (and their outcomes) have to be given meaning by some means.  (This is what John\index{Sipe, John E.} was trying to get after with his phrase ``anchors of belief.'')  But I lean much more on the agent's initial involvement in that term than {\Carl}.  The bureau of standards measurement, however abstract, does the trick for me.  Its intention is to make clear that as long as an agent ascribes internal meaning to the outcomes of one measurement, then the meanings of all the rest are set too.  But it's no deeper than choosing a standard meter stick.  We have no right to say that the meter stick is the one true meter in nature's eyes, and similarly for the timeless quantum attribute.  We set it, and thereafter, the practice of quantum mechanics is in trying to being consistent (coherent) with that initial call.

Which leads us to this:
\bq\noindent
The role of an ``instance of an attribute'' in this interpretation is
not to specify one of a number of possible expressions of existence,
as it is in realist classical mechanics, but rather to specify one
class of possible beliefs -- the one that the theory recommends --
about the consequences of future interventions of a particular type.
\eq
which I can again accept, with emphasis on ``class of beliefs,'' before disagreeing with this for the reasons above:
\bq\noindent
The point of physics is to identify these nondynamical variables.
Repeated interventions by experimentalists, and the careful noting of the range of consequences that those interventions elicit, is how
these fixed instances are discovered.
\eq

Finally, I've already told John\index{Sipe, John E.} and {\Carl} know my strong dislike of this (and maybe it's been forwarded to {\Ruediger}):
\bq\noindent
Regardless of the strategy, the quantum world of this interpretation
is a fixed, static thing. It is a frozen, changeless place. Dynamics
refers not to the quantum world, but only to our actions, our
experiences, and our beliefs as agents.
\eq
but as I reread it, maybe I'm willing to find a grain of truth in its emphasis too.  John\index{Sipe, John E.} puts the formal structure on one side of a dividing line, and the extra-formal stuff (like the indices on a POVM) on the other side.  What we may be disagreeing over may only be whether the latter stuff is allowed to be called part of quantum mechanics.

--- --- --- --- --- --- --- --- --- --- --- --- --- --- --- --- ---

In all, after writing all this, I find myself wondering even more about how much of all this should be brought to John's\index{Sipe, John E.} attention.  Maybe it is best to leave John's\index{Sipe, John E.} temporary happiness as it is (he'll eventually come out of it anyway) and leave our internal arguments as a family matter for the moment.

\section{21-03-06 \ \ {\it The Unchanged Changer} \ \ (to R. {\Schack})} \label{Schack101}

\brs
\bjes
But in your interpretation some of the abstract elements of the
theory do describe something that ``actually exists''; it is these
``frozen'' elements, your Bureau of Standards measurement results,
which might be called the scaffold of our experiences.
\ejes
Am I right that he has misunderstood something?
\ers

I'm not sure at what level you're asking this question.  Is you question concerning the part about ``in your interpretation some of the abstract elements of the theory do describe something that `actually exists'\,''?  Or rather the part about ``your Bureau of Standards measurement results, which might be called the scaffold of our experiences''?

Since you know that I'm pragmatist about theories (and only willing to talk about ``objective with respect to a theory''), let me guess that your concern is with the second issue.  If that's the case, then let me rather focus on this passage of his:
\bjes
The results of your Bureau of Standards measurements, existing in
some Platonic heaven at some time in the future, are ``once and for
all.''  {\Carl} would say that the irreducible representation of a
spin-1/2 particle is the same today as it always was and always will
be.
\ejes

He's got the bit about the attributes being frozen in time right, for both {\Carl} and me (or, at least, as I understand {\Carl}).  That is one of the reasons I call the quantum system a ``catalyst'' in the attached diagram (which John\index{Sipe, John E.} would never look at in our conversation the other day)---it is the unchanged changer (at least at this level of our understanding).

Where he flubs things is where he attributes to me the idea that, ``The results of your Bureau of Standards measurements, existing in some Platonic heaven at some time in the future, are `once and for all.'\,''  I don't at all imagine the results as existing at some time in the future.  For no measurement results pre-exist an actual measurement process; or to say it better, without the actual action, there is no actual reaction.  Instead, to the extent that an agent judges that he is always thinking of the same system, he will fix a single B of S measurement with which to make reference (which in turn fixes an irreducible representation, as in {\Carl}'s version).  It is only that that is timeless in the way John\index{Sipe, John E.} is thinking, not any putative outcome for the Platonic measurement.

More generally about the note, he seems to have the flavor of the ``phenomenalism'' wrong that he attributes to Bohr and {\Pauli} (those two for sure), but also the one he attributes to me.  There is something about his description that seems to leave it disconnected from the quantum formalism.  Whereas I would say that I have come to this point precisely because of Kochen--Specker and Gleason, which surely rely on the quantum formalism.  The categories and their motivations just can't be separated so cleanly as he would like.

\section{22-03-06 \ \ {\it Web Page, Second Draft}\ \ \ (to R. E. Slusher)} \label{Slusher15}

\noindent 1.5.6 Fundamentally New Science and Technology\medskip

\noindent Quantum Computing, Communication, and Cryptography at Bell Labs\medskip

Quantum mechanics, as a branch of physics, has been with us since the late 1920s, and it can be safely said that without it our modern technological society would not exist.  Quantum effects power almost all that we know, from transistors, to lasers, to medical imaging, to nuclear weapons and much, much more.  However, only in the last 10--15 years has it been realized that the most arcane features of quantum mechanics---ones not presently used in any technology---can be harnessed for computing and communication tasks unimaginable within classical physics.  This is not so much because quantum devices can be made smaller and faster, but because the language of quantum mechanics allows the writing of new algorithms that simply cannot be written in classical terms.  The most famous algorithms in this respect are Peter Shor's prime factorization algorithm (which, if implemented, would make RSA public key cryptography insecure) and Lov Grover's quantum search algorithm, both of which were discovered here at Bell Labs.  Finding ever new uses for these arcane quantum effects and implementing them in real technologies is an active and lively field, known as Quantum Information.

Present work at Bell Labs ranges the field:  On the theoretical side, we specialize in algorithms, topological quantum computation, fundamental aspects of quantum information, quantum error-correcting codes, and supporting technologies calculations.  On the experimental side, we are participating in a large effort to implement quantum computation in scalable micro-ion-traps and scalable optical-lattices for cold trapped atoms.  As well, we have a strong effort to implement multi-channel gigahertz-rate quantum cryptography.  All our experimental efforts make crucial use of Lucent's expertise in MEMS device technologies, with fabrication of ion traps and spatial light modulators through the New Jersey Nanotechnology Consortium.

\section{23-03-06 \ \ {\it A Review of General Covariance} \ \ (to G. L. Comer)} \label{Comer86}

Waking up very slowly.  Silly, after an almost sleepless night.  All
night long I tossed and turned obsessing over the same
thought/question:
\bq\noindent
An `action' is defined by the set of its `consequences.'
However the only consequences I can see in a quantum measurement
are the refinement of one's expectations.  But some actions lead
to identical sets of refined expectations with differing
probabilities.  So they must be different actions after all.  But that contradicts the first part.  Repeat.  Repeat.  Repeat.
What's going on?  Repeat.  Repeat.  The only consequences of
quantum measurement should be refined expectations.  What's going on?
\eq
Somewhere around 6:00, I said, ``Aha, it's `likelihoods' that are the
consequences.''  I still don't really understand that, but somehow it
was enough to let me finally fall asleep.  But then it was almost
time to get up.

\section{24-03-06 \ \ {\it On Certainty, Quantum Outcomes, Subjectivity, Objectivity, and Expanding Universes, Part 1} \ \ (to R. {\Schack} \& C. M. {\Caves})} \label{Caves82} \label{Schack102}

Amazing, how hard it is to write this paper and even discuss these things internally between the three of us.  When we started this debate on certainty in August 2001, I had no grey hair; now I've got lots of it.  \ldots\ OK, maybe the real problem has been Lucent all along (which, as of this morning, may soon be Alcatel), but still it's been a tough paper to write.  Luckily---and despite all that---I do think {\Ruediger}'s outline/draft is pretty good.  Major chunks of it, to my mind, are certainly very usable.  It is a valiant attempt and certainly something I haven't been able to do.

But, there are some things in the present draft and in the latest discussion between you two that, I think, go in the wrong direction, or at least diverge from all the progress I thought we had made in confronting the Wigner's friend issue in December.  Maybe it was illusory, but I don't want to think so.

So, let me try to get this on paper before having a phone conversation about it.  (I know you two like to talk rather than write all this email; you find it more efficient.  But I don't most of the time.  I can pain over my sentences and pause as much as I want without getting ridiculed or sidetracked or rushed along, and that is pretty important for me.)

But even now, where to start?

Maybe a good place is these two sentences from {\Ruediger} (even if one or both are outdated by now \ldots\ and I think at least one of them is):
\brs
There is a {\bf quantum} reason for certainty being subjective, which
does not apply in the classical case.
\ers
and
\brs
I agree, reluctantly. The category distinction doesn't buy as much as
I hoped it would.
\ers

I don't believe I had ever thought of it that way.  I.e., that the category distinction between the elements of the event space and the probability function over the event space would ever 1) ``buy us'' something new---I only saw it as something that had to be emphasized to patch up a perceived inconsistency in our Bayesian position---, or 2) be such that there could be additional, specifically quantum reasons for invoking it.

The problem is, without a {\it philosophical\/} category distinction of \underline{some sort} between the $x$ and the $P()$ in $P(x)$---and not just the mathematical one---the Bayesian view of probability runs the risk of collapsing into a kind of solipsism or mysticism (as Howard Wiseman and Wayne Myrvold indeed, and sometimes Bill Unruh, seem to think it does).  That risk is already there without quantum mechanics.

One of our main contributions, I thought, was in emphasizing (like no one else ever had) that this category distinction {\it must} be maintained even in the limiting case of $P(x)=1$.  Then, after doing so, we would leverage that to say the Penrose argument carries no weight after all.  And partially this has been carried out in the draft.

But here's where I think things have gotten off track.  I'm seeing language of this order traveling between the two of you:  ``certainty never implies a fact,'' ``knowledge of a fact $h$ can, however, compel the assignment $P(h)=1$,'' ``some probabilities are objective after all.''  More colloquially, at some far off meeting, I think I even heard someone say this, ``this `category distinction' business is a red herring.''  These may not be precise quotes, but I think they capture the sorts of phrases that are going around.  And I think that is very dangerous business.

We had a clean point (or at least we were groping toward one); let's not lose it.

Part of the trouble, I think, is coming from derailed uses of the terms `subjective' and `objective' creeping back into this draft, and then starting a snowball effect.  Let me try to expand on that.  When I finish, I will come back to give specific replies to:

1) this thing that {\Carl} said:
\bcc
my concern is that the category distinction loses some of its force.
The original idea, I believe, was to say that facts never determine
beliefs (i.e., probabilities), either classically or in qm.  We can
still say that, but the important distinction between realism and qm
will be different: in a realistic world, knowledge of facts can
dictate delta probabilities, but in qm knowledge of facts cannot
dictate quantum states.  What do you think?
\ecc
2) this thing that {\Ruediger} said:
\brs
And you should make the point in this way when you talk to Chris \ldots\ \ Let's say $X$ is ``up'' in a spin measurement. Agents A and B agree about everything concerning the setup and the precise conditions under which you would call $X$ true. The measurement is performed, and they both agree that $X$ is true. Then $P(X)=1$ for both of them. This $P$ is objective.
\ers
and 3) this thing {\Carl} said:
\bcc
I am now thinking that it would be a mistake to get into
distinguishing ``facts'' from ``knowledge of facts,'' since that is not the distinction we want to make.  Rather we want to say that
``knowledge of facts'' can never compel a pure-state assignment, whereas it would seem that we are saying that ``knowledge of facts'' can compel a delta probability assignment.
\ecc

In all three cases I have a bone to pick.  In 1), it is in hinging too much on the nonoperational term `knowledge,' which I would like to banish in any case.  In 2), it is in a definition of `objective' that I do not like anymore, and I think runs into trouble or at least becomes confusing in the places where we need it most.  And ultimately in 1) and 3), I think this is just the wrong distinction, and, in any case, it is not true.

--- --- --- --- --- ---

I'll go ahead and send you this much now, to show you that I wasn't lying when I said that I've been thinking of these things, but before going on, I've got to go home to satisfy an urge to try to catch some squirrels.

Recommended reading:  Item \#7, ``Epistemology Probabilized,'' by Richard Jeffrey, at
\bq
 \myurl[http://www.princeton.edu/~bayesway/]{http://www.princeton.edu/$\sim$bayesway/}.
\eq

\section{27-03-06 \ \ {\it Block of Ice-Nine} \ \ (to C. M. {\Caves})} \label{Caves83}

\bcc
We should never have let Sipe get away with the ice-nine characterization of the Bayesian interpretation.  He is right that we posit a changeless world of attributes that physics is trying to get at, but this is less than half the story.  The rest of the story is the evolving universe in which we participate.  He would say that the only dynamics that occurs is the dynamics of our beliefs (or the descriptions based on our beliefs), but there is a world out there, which we are trying to describe as best we can, and the evolution of our beliefs is our best attempt to capture what's happening in the world.
\ecc
\bcc
It's a Vonnegut novel, in which all the water in the world follows a tiny seed into a solid thermodynamic state called ice-nine.
\ecc

Thanks for the definition.

I hate it when I can't pin down all my rantings.  I remember one day going off on someone (maybe it was Howard Barnum or maybe Jeremy Butterfield, but probably someone completely different) about how the many-worlds conception is of a universe that's a big block of ice.  And I became so tickled with that phrase that I just kept saying it over and over.

{\Ruediger}, if you remember, come to my rescue.  I remember us talking about this sort of thing when we were talking about that guy's book, {\sl The Paradox of Cause}.

Nothing more distasteful than a block-universe view.

Looking through my emails, I'm surprised by how little I've used the phrase ``block universe''!  And even more shocked that I couldn't find the phrase ``block of ice.''  I was going to send you some examples \ldots\ but I see I've now got to go catch my plane.

Ice-nine, I like it.

\bq
I go up and I go down when it comes to speaking the words gravity and quantum in the same sentence.  At times I find myself thinking that general relativity and quantum mechanics express two absolutely incompatible worldviews.  The general relativistic universe is a ``block universe'' in William {\James}'s sense:  It's just there.  One can talk about foliations and dynamics, etc., {\it within} the 4-manifold, but in the largest view---the view from nowhere---the world and all its history is just there.  It is a universe without life (in the creative sense).  In contrast, the quantum world strikes me as a malleable world---one that is still in formation, and in particular, one for which it is impossible to get such a ``view from nowhere'' (as Nagel would call it).
\eq

\section{30-03-06 \ \ {\it The Sqrt Operation}\ \ \ (to W. K. Wootters)} \label{Wootters23}

What are you doing up so early?!  I'm in Lisbon at the moment; so at least that's my excuse for reading your email at this time!

\bbw
You probably know the answer to this question.  Consider any POVM.  Depending on
how one implements the POVM, the system being measured could end up in many
different states, even for a specific outcome of the POVM.  (I'm assuming that
the system survives the operation.)  One very easy case to consider, at least
mathematically easy, is the case in which the final state of the system is
given simply by $\sqrt{\Pi_j}$ applied to the original state, where $\Pi_j$ is the
POVM element.  Is there any precise sense in which this particular
implementation of a POVM is the {\bf least disturbing} to the system's state?  (I
suppose there will have to be some sort of averaging over all the possible
initial states.  Some will be disturbed more than others.)
\ebw

Yeah, you've got it exactly right.  If the input is promised to be a random pure state (wrt unitarily invariant measure), and the measure of disturbance is 1-fidelity between input and output, then the operation you describe for a fixed measurement is the least disturbing on average.  This issue came up around 1996 and the proof was harder than one might think.  Howard Barnum has a large bit about it in his PhD thesis.  I think he finally published it here:
\quantph{0205155}.

\section{31-03-06 \ \ {\it Portuguese Beer} \ \ (to G. L. Comer)} \label{Comer87}

Thanks for sending me your abstract for your talk on connotation and denotation.  I'm not sure what your last sentence in the abstract means, but I sure wish I were there to hear the talk.

I'm writing this note to you from my hotel's bar in Lisbon, Friday night.  I'm just back from dinner.  Unfortunately, I don't have a connection here (like I did at the university where I retrieved your note), so it'll be a few days before you get this composition.

I once copied my friend Jeff Nicholson some quotes on religion from one of Richard Rorty's books, which I've just dug up from my files and I'll place below.  [See 08-01-02 note ``\myref{Nicholson3}{{\Rorty} on Religion}'' to J. W. Nicholson.] I'd be curious to know how Rorty's ideas grab you (or repel you).  Another essay of Rorty's that made an impression on me had a title like ``Religion as Conversation Stopper'' (which was meant to be descriptive of the role it usually serves in society); unfortunately I don't have any quotes from that article in my computer.

I wish I could write like Rorty.  Heck, I wish I could write like William James.  (And while I'm at it, I wish I could write like Steven Weinberg---I wouldn't write the same things, but I wish I could write like him.)

On other matters, I finally gave my talk here today (I've been here since Tuesday), and it was declared that I must be famous:  About 10--12 people came to the seminar, even one undergraduate, rather than the usual four.  Bet you didn't know you knew a famous man?

On the whole it's been pleasant here, but I've been working steady the whole time and eating ridiculously large meals.  Still it seemed I might need a few extra calories tonight \ldots

Have you ever been to Portugal?  The thing that I have found most shocking is the sound of the language:  Despite the written form of the word, it sounds nothing at all like Spanish!  It sounds like Russian!  It really does, and I've been studying it all week to see if that first reaction of mine would go away.  In fact, the way I first noticed the phenomenon was on the plane coming in:  When the announcements were made in multiple languages, I was so surprised that one of the languages was `Russian.'  I thought, ``Now how many Russians can they expect on this flight?  Must be a popular vacation spot for them.  I didn't know that.''

On another matter, I heard some very exciting things tonight about the teaching of science to the common folk and developing countries from one of my hosts.  But he swore me to secrecy:  So, when about six months have passed, remind me of this note and ask me what I was talking about.  I'd love to tell you in particular---I think you'd have a good ear for it.

Enough for now.  I'm gonna go get a few more calories before calling it an evening \ldots

\section{04-04-06 \ \ {\it Title} \ \ (to S. J. van {\Enk})} \label{vanEnk6}

``Had We Found a Better Title, We Would Have Chosen It''  [See 09-05-08 note ``\myref{vanEnk14}{Change of Plan}'' to S. J. van {\Enk}.]

\section{19-04-06 \ \ {\it Warning:\ A Manifesto Follows} \ \ (to G. L. Comer)} \label{Comer88}

\bgc
Would you ever get so mad at me that you would kick me out of your office?  That just happened here!  I was talking to a young faculty friend here, who is a religious conservative.  As always, the discussion leads to me saying unreasonable things like: ``I've read that the Church burned many of the early religious texts.''  I'm then asked ``Who said that?!?''  I then reply, ``I don't know, some historian.  I'm not a historian, but I've read this in several places, and seen it discussed in documentaries.''

Then the discussion degenerates into an argument about my sources vs his sources.  That's it.

The problem is my sources are human, and as far as I can tell not directly told by God what to write down.  If I quote a human source that disagrees with a particular interpretation of scripture, my source is obviously wrong; or if the number of people promoting the idea is huge---like the biology community---there is a vast conspiracy that keeps the lie going.  If I say something like ``Emperor Constantine had a commission that got together and put together what we now call the Bible,'' I'm asked, ``Who said that?''  I reply, ``historians.''    I'm then told ``show me your sources.''  I respond: ``Go to any library and look in the history section.''  Then I'm stuck.  The response is then, ``Well, you don't really know then, do you?  When you've got your story straight, come back to me.''  All of this in a discussion where I'm told ``99.9\% of the Bible is known to be consistent.''  Do you see the vicious circle?  It is a hangman's noose around all our necks.
\egc

Thanks for providing my coffee break!  I enjoyed your manifesto.

Do read Richard Rorty's ``Religion as Conversation-Stopper.''  It can be found in his book {\sl Philosophy and Social Hope}.  It's in your Pius Library and not checked out; here's the call number:   B945 .R52 1999   Go get it while you're still on a roll.

\section{19-04-06 \ \ {\it Warning:\ A Manifesto Follows, 2} \ \ (to G. L. Comer)} \label{Comer89}

\bgc
Is it legitimate, this Rorty person and his book?
\egc

He's presently my second biggest hero in philosophy.

\section{19-04-06 \ \ {\it Church of the Smaller Hilbert Space} \ \ (to M. S. Leifer)} \label{Leifer2}

I love it!

Almost semi-seriously, it'd be nice if you'd hone that and present it in {\Vaxjo}.

\subsection{Matt's Beautiful Preply: ``The Church of the Smaller Hilbert Space''}

\bq\noindent
I can't believe I wasted a whole hour writing this thing, and I can't even post it on my office door because some god-fearing member of the public might see it.
\bq\noindent
\begin{center}
TEN COMMANDMENTS OF THE CHURCH OF THE
SMALLER HILBERT SPACE\medskip\\
With apologies to Charlie Bennett and, of course, God.\medskip
\end{center}
\begin{enumerate}
\item
I am $\rho$, your state, who brought you out of wavefunction realism, the place of orthodox dogmatism.
\item
Do not have any other states except Me. Do not represent states by false purifications, conceived as ontological states of the Platonic forms above, of reality below, or of the space-time foam underlying reality. Do not bow down to such states or worship them. I am $\rho$ your state, a state that demands exclusive belief.
\item
Do not announce your state $\rho$ in vain. $\rho$ will not allow the one who announces it in vain to go unpunished by a Dutch bookie.
\item
Remember the CP-map $\cal E$ to keep dynamics meaningful. You can work things out using six different Kraus decompositions or Steinspring dilations and do all your tasks. But the CP-map $\cal E$ is an equivalence class to $\rho$ your state. Do not do anything that attaches meaning to the arbitrary tools you choose to work with. This includes your Hilbert Space basis $\{|j\rangle\}$, your purifications $|\Psi\rangle$, your Kraus decompositions $\{E_j\}$, ${\cal E}(\rho)=\sum_j E_j\rho E_j^\dagger$, your Steinspring dilations $U$, $U^\dagger U = UU^\dagger = I$, your
Naimark extensions $P_j$, $P_jP_k = \delta_{jk}P_j$, your path integral decompositions and your virtual particles. It
is for perturbation theory calculations in QED that $\rho$ is supplemented by the Feynman diagrams, the
path integrals, the Greens functions, and all that is in them, but the true evolution of $\rho$ rests on the
CP-map. $\rho$ therefore blessed the CP-map $E$ and made it meaningful.
\item
Honor your forefathers by using the Hilbert space algorithm they handed down to you to calculate your expected utilities. You will then live long on the land that $\rho$ your state describes your beliefs about.
\item
Do not commit murder, since there is no other ``branch of the wavefunction'' in which your victim will survive.
\item
Do not adulterate the {\Schroedinger} equation by adding nonlinear terms designed to cause collapse.
\item
Do not steal from classical physics by insisting that particle position or field configuration variables
must evolve deterministically.
\item
Do not testify as a false witness to the existence of histories of events that do not appear in the empirical records.
\item
Do not be envious of your neighbor's state $\sigma$. Do not be envious of your neighbor's dynamical CP-map
$\cal F$, his POVM elements $\{N_j\}$, his update CP-maps ${\cal F}_j$ , his Kraus operators $F_{jk}$, his donkey, or anything
else that is your neighbor's, for they only describe his beliefs (except for the donkey), which naturally differ from yours.
\end{enumerate}
\eq
\eq

\section{20-04-06 \ \ {\it Always Quantumness!}\ \ \ (to M. Sasaki)} \label{Sasaki4}

Thank you too for your other kind offer.  With friends like you, I am spiritually wealthy!  But do not break your back trying to realize an invitation for me if your budget is tight---certainly take care of important things first!

I hope things are going well for you scientifically.  A couple months ago, I talked about our quantumness again, for the first time in quite a while.  It was for the colloquium of the Applied Mathematics Dept.\ at Princeton University.  It was quite well received, in fact, as generating some interesting questions in mathematics.  In the audience there were Rob Calderbank (a coding theorist), Ingrid Daubechies (inventor of wavelets), Simon Kochen (a famous logician, particularly for the Kochen--Specker theorem in quantum mechanics), \ldots\ and even the super-famous John Conway ({\sl Atlas of Finite Groups\/} and the cellular automaton ``Life'') and John Nash (Nobel Prize in economics).  I'll place the abstract of the talk for you below.

Anyway, there are many things still to be done with the concept, and I hope that one day we will get a chance to work on it together again.  If the world of quantum information picks up here at Bell Labs in the next year (after the merger), I hope that I can reciprocate on your kind invitation.

\bq
\noindent Title:  Math Problems from the Far Side of Quantum Information \medskip

\noindent Abstract:  The field of Quantum Information has recently rightly attracted great interest for the technological fruits it may bear.  But there is a sect of its practitioners who think it stands a chance to bring us much more than that---namely, that its theoretical tools will give us a means for exploring what quantum mechanics is really all about and for settling some of the deepest problems in physics.  The roots of this optimism come from a very old thought:  that a quantum state has more to do with representing its user's information, than any inherent physical property of the system to which it is ascribed.  What is new and nice is that quantum information teaches us how to formulate this idea precisely and even check its consistency.  Nicer still for the mathematics community is the number of juicy mathematical problems the consistency-checking process poses.  In this talk, I will review some of the history of this and then quickly settle on a sample problem that has been annoying me a lot lately:  the question of the existence of symmetric informationally complete positive-operator-valued measures for finite dimensional Hilbert spaces.  I'm not alone---it turns out to be equivalent to a 30-year-old problem in coding theory---but I will say some things about it that you may not have heard before.
\eq

\section{20-04-06 \ \ {\it Church of the Big-Enough Hilbert Space, 1} \ \ (to M. S. Leifer \& D. Bacon)} \label{Leifer3} \label{Bacon2}

In contemplating the genius of Matt's commandments (and being especially motivated from watching the old Charlton Heston movie with my seven-year-old this weekend), I took a trip down memory lane tonight.  I found the passage below in my {\sl Notes on a Paulian Idea}:
\bq
I had a very nice conversation with {\Charlie} on the drive back to Wendell from your house, that, I think, has allowed me to sharpen what I think about things in quantum mechanics.  In particular, I would like to work on a point of view that substitutes in place of the ``Church of the Larger Hilbert Space,'' something along the lines of ``Church of the Big-Enough Hilbert Space.''  However, it'll require some writing for me to get it down coherently.
\eq
It led to the question is big-enough bigger than smaller---maybe, maybe not:  probably depends upon who you're kneeling with.  Anyway, I hope it's bigger than empty.

On another subject, I've also been contemplating dumping another 500 pages onto {\tt quant-ph} May 10 (for the fifth year anniversary of my last big dump).  Experienced bloggers that you both are, what do you think of the idea?  Do you think it goes too far, given my earlier foray?

\section{20-04-06 \ \ {\it Church of the Big-Enough Hilbert Space, 2} \ \ (to D. Bacon)} \label{Bacon3}

\bbaco
Whoever first framed the words ``Church of the Larger Hilbert Space''
was a genius.
\ebaco

Yes, John Smolin is a genius, and for many reasons.

\bbaco
I mean, it makes other interpretations just seem silly!
\ebaco

But this point worries me.  You do realize that Matt's\index{Leifer, Matt} manifesto, despite the humor is semi-serious?  It captures many of the essential points of what he, {\Carl} {\Caves}, {\Ruediger} {\Schack}, I and a few others, are trying to get at with our Bayesian view of quantum probabilities.  Particularly take head of his second commandment:
\bq\noindent
   Do not have any other states except Me.  Do not represent states by
   false purifications, conceived as ontological states of the Platonic
   forms above, of reality below, or of the space-time foam underlying
   reality.  Do not bow down to such states or worship them.  I am
   $\rho$ your state, a state that demands exclusive belief.
\eq
He means it.  And that statement is the antithesis of what Charlie Bennett and John Smolin are thinking when they invoke their church.

Thanks for endorsing another dump.  I'll take your vote into account as I make this painful decision \ldots\ (One day I'll have to tell you about the advice William {\James} once wrote to his brother Henry concerning the latter's hygienic habits.)

\section{25-04-06 \ \ {\it My Title and Abstract}\ \ \ (to A. Y. Khrennikov)} \label{Khrennikov19}

I'm not sure what title I sent you before, but let us now go with the one below.  It is a talk I gave at Princeton for their Applied Math Dept colloquium one week ago---I hope you don't mind if I recycle it, but I've run out of time for something new.

\bq
\noindent Title:  Math Problems from the Far Side of Quantum Information \medskip

\noindent Abstract:  The field of Quantum Information has recently rightly attracted great interest for the technological fruits it may bear.  But there is a sect of its practitioners who think it stands a chance to bring us much more than that---namely, that its theoretical tools will give us a means for exploring what quantum mechanics is really all about and for settling some of the deepest problems in physics.  The roots of this optimism come from a very old thought:  that a quantum state has more to do with representing its user's information, than any inherent physical property of the system to which it is ascribed.  What is new and nice is that quantum information teaches us how to formulate this idea precisely and even check its consistency.  Nicer still for the mathematics community is the number of juicy mathematical problems the consistency-checking process poses.  In this talk, I will review some of the history of this and then quickly settle on a sample problem that has been annoying me a lot lately:  the question of the existence of symmetric informationally complete positive-operator-valued measures for finite dimensional Hilbert spaces.  I'm not alone---it turns out to be equivalent to a 30-year-old problem in coding theory---but I will say some things about it that you may not have heard before.
\eq

\section{26-04-06 \ \ {\it Comerism P} \ \ (to G. L. Comer)} \label{Comer90}

\bgc
We went rafting this past weekend.  I broke my record for number of different states I had to pee in during a single day.  Before it was 4---established when I visited you in NM---and now it is 5.

Just for the record.  And, I expect this to go into some book of collected e-mail, to be labeled Comerism P.
\egc

\section{01-05-06 \ \ {\it Conversations with God}\ \ \ (to W. B. Case)} \label{Case2}

Attached are the two excerpts from my samizdat ``Notes on a Paulian Idea'' that I told you about last night.

\subsection{From a 23 January 2000 note to Gilles Brassard, ``I See Why Bit Commitment''}

\bq
Let me just say that I had a bit of an epiphany in the
shuttle bus at Dulles Airport:  for the first time I have understood
why you want to take {\it both\/} the EXISTENCE of secure key
distribution and the NONEXISTENCE of bit commitment as pillars in
your sought-after derivation of QM.  You have been thinking more
deeply than me since the beginning!

Let me place at the end of this note a little piece from my
samizdat.  [See letter to {\Greg} {\Comer}, 22 April 1999, titled
``Fuchsian Genesis.'']  It sort of presents what I've been trying to
get at in a dramatic way:  it may be my best presentation of the
idea of why quantum key distribution has something to do with the
foundations of quantum mechanics.  But more to the present point,
let me tell you about a second way I use to get the point across.
I've used this slide in a few talks. It consists of five frames with the
following little story.

\bq
\noindent In the first frame {\God} starts to speak to {\Adam} at a time just
before Genesis, ``{\Adam}, I am going to build you a world.  Do you
have any \medskip suggestions?''\\
{\bf {\Adam}}:  Mostly I don't want to be alone.  I want to have
friends \ldots\ and enemies to spice things up \ldots\ and generally
just plenty of people to talk to.
\medskip\\
{\bf {\God}}:  Done.  I'll give you a world populated with loads of
other people.  But you ask for a bit of an engineering feat when you
ask to be able to talk to them.  If you want to communicate, the
world can't be too rigid; it has to be a sort of malleable thing. It
has to have enough looseness so that you can write the messages of
your choice into its properties.  It will make the world a little
more unpredictable than it might have been for me---I may not be able
to warn you about impending dangers like droughts and hurricanes
anymore---but I can do that if you want.
\medskip\\
{\bf {\Adam}}:  Also {\God}, I would like there to be at least one
special someone---someone I can share all my innermost thoughts with,
the ones I'd like to keep secret from the rest of the world.
\medskip\\
{\bf {\God}}:  Now you ask for a tall order!  You want to be able to
communicate with one person, and make sure that no one else is
listening?  How could I possibly do that without having you two
bifurcate into a world of your own, one with no contact whatsoever
with the original?  How about we cut a compromise?  Since I'm already
making the world malleable so that you can write your messages into
it, I'll also make it sensitive to unwanted eavesdropping.  I'll give
you a means for checking whether someone is listening in on your
conversations:  whenever information is gathered from your
communication carriers, there'll be a reciprocal loss in what you
could have said about them otherwise.  There'll be a disturbance.
Good enough?  You should be able to do something clever enough with
that to get by.
\medskip\\
{\bf {\Adam}}:  Good enough!
\medskip\\
{\bf {\God}}:  Then now I'll put you in a deep sleep, and when you
awake you'll have your world.
\medskip\\
{\bf {\Adam}}:  Wait, wait!  I overlooked something!  I don't want an
unmanageable world, one that I'll never be able to get a scientific
theory of.  If whenever I gather information about some piece of the
world, my colleagues lose some of their information about it, how
will we ever come to agreement about what we see?  Maybe we'll never
be able to see eye to eye on anything.  What is science if it's not
seeing eye to eye after a sufficient amount of effort?  Have I doomed
myself to a world that is little more than chaos as far as my
description of it goes?
\medskip\\
{\bf {\God}}:  No, actually you haven't.  I can do this for you:
I'll turn the infor\-ma\-tion-disturbance tradeoff knob just to the point
where you'll still be able to do science.  What could be better?  You
have both privacy and science.
\medskip\\
So {\Adam} fell into a deep sleep, and {\God} set about making a
world consistent with his desires.  And, poof(!), there was QUANTUM
MECHANICS.
\eq

That's the tale.  But now I see the crucial spot of outlawing bit
commitment within it.  {\God} could have supplied {\Adam} with a set
of impenetrable boxes (and keys to open them) where he could place
his information whenever he wanted some secrecy.  A bit commitment
protocol could certainly be used in that secondary fashion.  But
{\God} chose to make all information open for all the world to see:
he just left the possibility of an imprint whenever someone has a
look.
\eq

\subsection{From a 22 April 1999 note to Greg Comer, ``Fuchsian Genesis''}

\vspace{-12pt}
\bq\noindent
\bq
\noindent
In the beginning {\God} created the heaven and the earth.  And the
earth was without form, and void; and darkness was upon the face of
the deep. And the Spirit of {\God} moved upon the face of the waters.
And {\God} said, Let there be light: and there was light.  And {\God}
saw the light, that it was good; and {\God} divided the light from
the darkness. And {\God} called the light Day and the darkness he
called Night.  And the evening and the morning were the first day.
\ldots\ [Day 2], [Day 3], [Day 4], [Day 5] \ldots\ And {\God} saw
everything that he had made, and behold, it was very good. And there
was evening and there was morning, a sixth day.  Thus the heavens and
the earth were finished, and all the host of them.\medskip
\eq

But in all the host of them, there was no science.  The scientific
world could not help but STILL be without form, and void.  For
science is a creation of man, a project not yet finished (and perhaps
never finishable)---it is the expression of man's attempt to be less
surprised by this {\God}-given world with each succeeding day.

So, upon creation, the society of man set out to discover and form
physical laws.  Eventually an undeniable fact came to light:
information gathering about the world is not without a cost.  Our
experimentation on the world is not without consequence.  When {\it
I\/} learn something about an object, {\it you\/} are forced to
revise (toward the direction of more ignorance) what you could have
said of it.  It is a world so ``sensitive to the touch'' that---with
that knowledge---one might have been tempted to turn the tables, to
suspect a priori that there could be no science at all.  Yet
undeniably, distilled from the process of our comparing our notes
with those of the larger community---each expressing a give and take
of someone's information gain and someone else's consequent loss---we
have been able to construct a scientific theory of much that we see.
The world is volatile to our information gathering, but not so
volatile that we have not been able to construct a successful theory
of it.  How else could we, ``Be fruitful, and multiply, and replenish
the earth, and subdue it?''  The most basic, low-level piece of that
understanding is quantum theory.

The {\sl speculation\/} is that quantum theory is the unique
expression of this happy circumstance:  it is the best we can say in
a world where {\it my\/} information gathering and {\it your\/}
information loss go hand in hand.  It is an expression of the ``laws
of thought'' best molded to our lot in life.  What we cannot do
anymore is suppose a physical theory that is a direct reflection of
the mechanism underneath it all:  that mechanism is hidden to the
point of our not even being able to speculate about it (in a
scientific way).  We must instead find comfort in a physical theory
that gives us the means for describing what we can {\it know\/} and
how that {\it knowledge\/} can change (quantum states and unitary
evolution).  The task of physics has changed from aspiring to be a
static portrait of ``what is'' to being ``the ability to win a
bet.''\footnote{The nice phrase ``physics is the ability to win a
bet'' is due to J.~R. {\Buck} (a grad student at Caltech) circa 19
February 1999.}

This speculation defines the large part of my present research
program.
\eq

\section{02-05-06 \ \ {\it Faster-than-Light} \ \ (to G. L. Comer)} \label{Comer91}

One lecture here down, one---public!---one to go.  The title tonight: ``Quantum Information and the Reactive World.''  I plan to get alchemical on them!

\section{09-05-06 \ \ {\it The Bayesian Big Bang} \ \ (to H. Barnum)} \label{Barnum22}

Wow, that's some report!  Thanks for sending me all that.  It sounds
like I really did miss a thought-provoking time.

I wouldn't know what to suggest to put in a note to Albert.

\bhb
Incidentally, the discussion with Albert ended with him saying ``so
you don't believe parts of the early universe do work on other parts
even though there is no subject who is using its knowledge to extract
work\ldots'' I was rather tired, so called a pause there and will
deal with that later. I think it can be dealt with, the main point
being that WE understand parts of the early universe as doing work on
other parts [\ldots], perhaps, because OUR KNOWLEDGE of those parts
is the same kind of knowledge (canonical ensemble at some
temperature, blah-de-blah) that would allow us, given large enough
apparatus etc\ldots, to extract that much work\ldots\ because some of
the processes that actually go on in the early universe are (I
guess\ldots) similar to various kinds of thermodynamic processes of
putting stuff in pistons and expanding, etc. [\ldots] [N]evertheless
I believe this is basically right\ldots\ our description of
early-universe processes in thermodynamic terms is because we have
the same kind of knowledge about it that we have about gases and
liquids being boiled and compressed etc\ldots\ ``The radiation
background was at XX degrees Kelvin when it decoupled from matter''
is a statement of what we think we know about it, not a property of
its microstate at the time.
\ehb

I've told you my wacky idea that for all Bayesians there must be a
big bang, haven't I?  In case not, the idea is this.  Consider my
beliefs about yesterday's events.  If we were to lump all of them
into a big joint probability distribution, it would have some
entropy.  Now consider my beliefs about the events the day before
that.  They too could be lumped into a distribution.  However, since
I'm probably less sure about the things that happened the day before
last than I am about the things that happened yesterday, the latter
distribution should have more entropy.  And so on we could go further
back in time.  Assuming I lose effectively all predictability as I
conceptually reach back to some finite time (based on the size of my
brain and my processing capability), I should end up with a
distribution of effectively infinite entropy.  Associating (somehow)
this distribution with one of canonical form---that's one of the hard
steps---I find an effectively infinite temperature in the finite
past.

Just another way of saying the universe came into existence (at least
with regard to my ability to extract work from it).

Homework Problem:  Given reasonable models of a typical brain's
inferential powers and the resolution of human senses, estimate the
number of years since the big bang.

\section{17-05-06 \ \ {\it Four Emotions} \ \ (to G. L. Comer)} \label{Comer92}

\bgc
``We are the freedom to choose.''
\egc

I still love that last line!

\section{17-05-06 \ \ {\it Beautiful Passage} \ \ (to G. L. Comer)} \label{Comer93}

Below is a beautiful passage from Danny Greenberger that I just pasted into my new samizdat; I don't know if I ever shared it with you before.  [See 06-12-02 note ``\myref{Greenberger1}{Enjoyed}'' to D. M. Greenberger.] But, since you're so much on my mind tonight, I'll go ahead and give you a sneak preview.  I hope you enjoy it as much as I did.  It's something I found as I was thinking about Bill Wootters making his trip to visit us next week.

Most of my work tonight though has been on going through your files---I'm up to December 2002 at the moment, so there's still a long way to go.  But it's been very enlightening to read our correspondence between each other again.  And so nice to read your poems again.  I feel like there's so much I want to tell you---about all the clarifications that have come to me about the transformation rules in quantum mechanics in the last year, about the disanalogy I start to see with relativity (relativity connects disparate observers, the transforms in quantum mechanics connect the gambles one single observer is willing to make, much like Dutch-book consistency), and about how so much of these thoughts probably find their origins in the things you were bringing to my attention in 2002.  But all of that is going to require new notes, not old!  So you have to wait old friend.

\section{17-05-06 \ \ {\it Transformation Rules} \ \ (to G. L. Comer)} \label{Comer94}

The note below comes from when I really got pissed off at this prof at Princeton who rubbed me the wrong way.  [See 30-01-06 note ``\myref{McDonald6}{Island of Misfit Toys}'' to K. T. McDonald.]  Anyway, nonetheless, I think there is a decent explanation in the stuff below on how to think of the significance of the transformation rules in QM.  The significance in QM is different from in relativity:  In QM it is about one single agent, in SR/GR it is about two or more.  That's an important difference that has only become clear to me in the last year or so.

\section{19-05-06 \ \ {\it Spring Is in the Air \ldots\ MUBs Are in the Air} \ \ (to W. E. Lawrence)} \label{Lawrence5}

It dawned on me today that it was about this time last year that you visited us and gave us a talk about MUBs and phase space.  There must be something about the time of year!  As it turns out Bill Wootters will visit us next week, and give the talk below.  What is it about the Spring?

Hope all is well with you.

\bq
\noindent RESEARCH TALK --\medskip

\noindent Phase Space for Qubits\\
Prof.\ William K. Wootters\\
Williams College\medskip

\noindent Wednesday -- May 24, 2006\\
11:00 AM\\
MH 1D-224\medskip

\noindent Abstract:\\
For a single particle moving in one dimension, the particle's quantum state can be represented as a real function of position and momentum, the Wigner function, which has certain properties that make it useful for state reconstruction. In this talk I show how a system of binary quantum objects (qubits) can be represented by a closely analogous function, a discrete Wigner function, on a discrete phase space. The two axis-variables of the discrete phase space, that is, the analogs of position and momentum, are associated with two ``conjugate'' bases of the state space, and are labeled by the elements of a finite field (in the algebraic sense of ``field''). The use of a finite field is by no means
arbitrary: one finds that the structure of the field accords remarkably well with certain features of the complex-vector-space structure of quantum mechanics.
\eq

\section{01-06-06 \ \ {\it Cirac Has Been Deservedly Awarded}\ \ \ (to M. P\'erez-Su\'arez)} \label{PerezSuarez22}

Thanks for the letter.  I'm sorry to hear that staying in the US put you in a funk, though.  Ignacio is a great guy; it is great to hear he has won such an award.  I talked to him for almost three hours at the APS March meeting, mostly about de Finetti theorems and SIC-POVMs, and it was such an exhilarating experience.  He is lightning fast and simply understands everything almost immediately.

Congratulations too on your book.  Thanks so much for sending me a copy.  I'll put it with pride on my shelf (even if I cannot read it).

We will miss you in Sweden.  I hope it is a productive meeting and actually expect it will be.  I will talk on recent things Bill Wootters and I have discovered on SIC-POVMs---that they are not all unitarily or anti-unitarily equivalent---also, I will frame better what I mean by saying that the Born rule for calculating quantum probabilities is really a kind of transformation rule (along the lines of a de Finettian coherence principle).  It would have been nice to have your critical input at the talk.

Send me your thesis as it comes together.

\section{01-06-06 \ \ {\it Critical Paper} \ \ (to J. Bub)} \label{Bub19}

Thanks for sending the paper!  It looks interesting.  I've just printed it out and will be taking it with me to Sweden tomorrow.  I'll send you my comments hopefully early in the week.

I didn't think there was any real difference between the Ramseyan and de Finettian views of probability (other than perhaps emphasis).  There may, however, be a difference between how you and I think of ``truth''---particularly in the quantum context.  I'll flesh out my thoughts to you just as soon as I can and give you feedback on your paper.

BTW, I've also printed out your most recent paper on quantum computation from quant-ph to take with me.  That looks very exciting.

\section{02-06-06 \ \ {\it Enjoyable Paper} \ \ (to J. Bub)} \label{Bub20}

As I wrote you earlier, I'm on my way to Sweden today (though I didn't tell you about how roundabout of a way I'm taking).  Anyway, I'm in Chicago at the moment and will board the next plane soon.

But I wanted to tell you that I used the first flight to read, take notes on, and contemplate your paper.  It was quite enjoyable.  And certainly I'm glad to see you making a movement in the Bayesian direction!

If I had known about your thinking along these lines, I would have certainly invited you to this meeting.  With the funding I had, I put together what I've been calling the ``Swedish Bayesian Team'' ({\Schack}, Leifer, Appleby, Pitowsky, and Caticha), so that hopefully we'll have a very focused discussion on some of the technical aspects of this subject (as opposed to the impasse that I mostly viewed the Konstanz meeting as).  You would have been a great contributor.

Anyway, as a proxy for not having you physically present, I'll be writing you this week.  I have plenty of comments and queries about your paper.

\section{13-06-06 \ \ {\it Little Chocorua} \ \ (to L. Simon)} \label{SimonL1}

I am a research physicist at Bell Labs doing quantum information theory and quantum foundations for a living, but have a great side interest in William James and pragmatism in general.  I even pride myself---probably incorrectly---on having the largest personal library on pragmatism in New Jersey!  (About 420 volumes.)

Anyway, I write to you because it dawned on me that you may be able to help me in a silly little quest.  My wife is constructing a playhouse for our daughters from 12 doors she recovered from an 1896 house being demolished near our home.  (Our own house is an 1896 Victorian, and the coincidence of the years gave her this clever idea.)  Of course it's one more door than James' home in Chocorua had, but nonetheless we've decided to call the playhouse ``Little Chocorua''---the kids seem to like the name and we've run with it.

This is where you come in:  Do you know of any good sources of pictures of the original Chocorua home?  Or do you perhaps have any of your own that you have scanned into your computer?  We're hoping to get ideas on how to give the playhouse some final touches so it fits its name even better.

The only picture of the Chocorua home that I have ever seen comes from your book.  (Wonderful book, by the way; I very much enjoyed it.)  But I also know from an interview with you on the web that I read a few months back, that you have physically visited the home.  Perhaps you took some pictures then.

Sorry to bother you with such a strange request \ldots\ but I thought it might be novel enough that you would be interested in helping us out.\medskip

\noindent  P.S\@.  If you are interested in seeing how some of James's idea strike me as having an intimate connection with quantum foundations issues, download the big document posted at the bottom of my webpage titled ``Cerro Grande II''.  Then simply do a search on ``William James'' within it, and maybe you'll find an interesting idea or two.  There is a link to my webpage below.

\subsection{Linda's Reply}

\bq
Dear Christopher Fuchs, your playhouse sounds wonderful, but I don't have photos. I was able to see the house because it was on the market at the time (Janice Hamel was the realtor) and I think it has since gone on the market again. Through the years, owners made lots of updates. It may be that the historical society, which I think is housed in the Chocorua Public Library, might be able to come up with some historical photos. Or maybe the realtor has some from when she handled it.

Thank you so much for your interest in my book and, of course, in James. I DO think you have a record number of books on pragmatism.

Good luck with the playhouse!
\eq

\section{19-06-06 \ \ {\it Report} \ \ (to Fellowships)}

This is an excellent proposal and I wholeheartedly recommend that it be funded.  Recent advances in quantum information theory have made it quite respectable within the physics community to view quantum states as representing agent-centered information about quantum systems, rather than as intrinsic physical properties for those systems.  Concomitant with these advances has been a reawakening in the philosophy-of-physics community for the need to study and develop philosophical systems that can support such an idea.  Pragmatism in one guise or another (Jamesian, Deweyian, Putnamian) seems to fit the bill, and there has been a flurry of activity in this direction.  Particularly nice so far has been the work of Prof.\ Michel Bitbol.  I see the present proposal [of a younger researcher] as a continuation down this line.  Novel to this approach is a melding of pragmatic and Kantian (or transcendental) lines of thought---this sort of work is long overdue, and I give it my highest recommendation.  It will have direct impact on the philosophical discussion of quantum foundations and direct impact on theoretical methods in quantum information science.

\section{10-06-06 \ \ {\it Title and Abstract---Special Treat} \ \ (to G. M. D'Ariano)} \label{DAriano4}

I apologize for not getting the title and abstract to you yesterday as I had promised:  From my hotel in Copenhagen, I could not get a connection to the internet (something to do with my wireless ``location profiles list'' being full, and I could not figure out how to fix it).  Now, though, I am in the Chicago airport, and can send you mail by my usual routes.

I know it was a further delay, but at least it allowed me the time to come up with something of a new abstract for you.  The result is below.  If you feel that your name ought properly to be in the list with the other guys, feel free to take out the part ``(maybe D'Ariano?)''\ and replace it with your name in the earlier list right after Caves (alphabetical order).

I'm looking forward to much better understanding your axiom system.  It looks indeed very promising---I am quite excited.  This is going to be a great trip.

I will take care of the whatever needs to be done in the other emails you wrote me, right after the weekend.

\bq
\noindent {\bf Title:} Where Is the Reality in a Bayesian View of Quantum Mechanics?\medskip

\noindent {\bf Abstract:} In the neo-Bayesian view of quantum mechanics that Appleby, Caves, Pitowsky, Schack, the author and others (maybe D'Ariano?)\ are developing, quantum states are taken to be compendia of partial beliefs about potential measurement outcomes, rather than objective properties of quantum systems.  Different observers may validly have different quantum states for a single system, and the ultimate origin of each individual state assignment is taken to be unanalyzable within physical theory---its origin, instead, ultimately comes from probability assignments made at stages of physical investigation or laboratory practice previous to quantum theory.  The objective content of quantum mechanics (i.e., the part making no reference to observers) thus resides somewhere else than in the quantum state, and various ideas for where that ``somewhere else'' is are presently under debate---there are adherents to the idea that it is purely in the ``measurement clicks,'' there are adherents to the idea that it is in intrinsic, observer-independent Hamiltonians, there are adherents to the idea that it is in the normative rules quantum theory supplies for updating quantum states, and so on.  This part of the program is an active area of investigation; what is overwhelmingly agreed upon is only the opening statement of this abstract---that quantum states are compendia of beliefs.  Still, quantum states are not simply Bayesian probability assignments themselves, and different representations of the theory (in terms of state vectors or Wigner functions or $C^*$-algebras and the like) can take one further from or closer to a Bayesian point of view.  It is thus worthwhile spending some time thinking about which representation might be the most propitious for the point of view and might, in turn, carry us the most quickly toward solutions of some of the open problems.  In this talk, I will explore various issues to do with the above and explain why I prefer a representation of quantum mechanics that makes crucial use of a single probability simplex.
\eq

\section{12-06-06 \ \ {\it Grover's Alg and Temporal Bell Inequalities} \ \ (to L. K. Grover)} \label{Grover3}

By the way, I met a really interesting young man in Sweden, who gave an excellent talk on some ideas surrounding your algorithm.  Here is the paper he presented:\medskip\\
\begin{tabular}{ll}
Title:   &    Information-theoretic temporal Bell inequality and quantum computation\\
Author:  &    Fumiaki Morikoshi \\
Journal-ref: & Phys.\ Rev.\ A {\bf 73}, 052308 (2006)
\end{tabular}\medskip\\
It's on the archive at \quantph{0602011}.

The basic idea is that he shows that your algorithm violates a certain kind of temporal Bell inequality.  Further he speculates that something like this might be at the heart of the speedup in all quantum algorithms.  It was certainly a thought provoking thought.

Anyway, I think he is very interesting and amiable, and he was very articulate in English too for a Japanese.  Also, he was very keen to talk and work with you.  If you want to see his full list of publications, go here: \myurl{http://arxiv.org/find/quant-ph/1/au:+Morikoshi/0/1/0/all/0/1}.

If you have any money left for visitors, I think this guy could be a winner for you.  His excitement was so high for visiting you and Bell Labs, that I suspect he would jump at the chance, even if you could only give him partial support.

\section{21-06-06 \ \ {\it Hoan Dang} \ \ (to A. P. Ramirez)} \label{Ramirez2}

Thanks.  At the moment, I'm planning to have Hoan work on one of these Wigner-function type representation of quantum mechanics, to see where the power of quantum computing comes from.  There have recently been some encouraging results along these lines.  See for instance,
\begin{itemize}
\item
\quantph{0506222}
\item
\quantph{0405070}.
\end{itemize}
Also, I'd like to see if we can translate those results into the particular $q$-function-style representation that I've been working on the last year.  There, the key issue wouldn't be negativity, but something else.

It all goes back to a mostly forgotten point made in Feynman's original paper on quantum computation---I think there's a lot of life in the idea and everyone has gotten carried away talking about entanglement and multiple universes.

Anyway, some meaty mathematical problems, and I'll probably try to use his programming skills to simulate a lot of things.  (Mabuchi at Caltech wrote me some very good things about his abilities.)

\section{22-06-06 \ \ {\it Triangles} \ \ (to H. C. von Baeyer)} \label{Baeyer21}

I was just taking care of some administrative things to do with the
{\Vaxjo} conference, and it dawned on me that I haven't yet written
you since the conference's end.  I was surprised to find you gone
already by the Wednesday.  I hope you didn't become ill or something.
It was also too bad to see you gone especially as I think you missed
three of the most interesting talks of the meeting:  Bengtsson,
Leifer, and D'Ariano.  For your reference, here are the two papers
associated with Leifer's and D'Ariano's talks:
\begin{itemize}
\item
\quantph{0606022} \\ Title: Quantum Dynamics as an Analog of
Conditional Probability \\ Author: M. S. Leifer

\item
\quantph{0603011}\\
    Title: How to Derive the Hilbert-Space Formulation of Quantum
           Mechanics From Purely Operational Axioms\\
    Author: Giacomo Mauro D'Ariano
\end{itemize}

The D'Ariano construction, to me, really starts to smell right from
the Paulian perspective---so, I'm quite pleased with it.  It still
could use a good bit of Bayesianization, but the mathematics, I
think, is starting to get in place.  Saturday, I'm off to Italy for a
little meeting that D'Ariano put together; it'll be another good
opportunity to delve into these things.

Below, I'll paste in the fragment of the note I started to write you
in March.  One of these days, I should finish that up.

I hope all is well with you, and that you got something out of your
{\Vaxjo} experience.

\noindent --- --- --- --- --- --- --- --- --- --- --- --- --- --- ---
---

\bq
\noindent Dear Hans,

I hope you're better by now.  I'm between meetings once again:  Last
Friday I got back from the APS March meeting in Baltimore, and this
coming Monday I zip off to Portugal.

While in Baltimore at dinner one evening, as we were waiting for our
crabs, somehow your name came up, and I ended up drawing your B3
diagram:
\bhcvb
   The ``B3 triad'' is a triangle with the words psi, information, and
   probability at its vertices, and the names Bayes, Born, and Bohr,
   respectively, along their opposite sides.
\ehcvb
(though I accidentally inverted the vertices and sides).

I was trying to explain why I thought the B3 idea was pretty catchy.
However, as we---that is, Rob {\Spekkens}, Matt Leifer, {\Carl} {\Caves},
Hideo Mabuchi, and I---discussed it, I found myself slowly morphing
the content of your diagram to better fit my predilections.  Here was
the result, which was left on a sheet of butcher paper in Obrycki's
Crab House:
\bq\noindent
   A triangle with the names Bayes, {\Pauli}, and Gleason at the vertices,
   and $p(x)$, ``contextuality'', and $|\psi\rangle$ along the opposite
   sides.
\eq

It won't be as catchy and memorable, I know (what small fraction of
physicists have ever heard of Gleason, and what even smaller fraction
have ever heard of ``contexuality''?).  But let me explain the
reasons.

At first it was just that I was dissatisfied with \ldots
\eq

\section{24-06-06 \ \ {\it Notes on ``What are Quantum Probabilities''} \ \ (to J. Bub)} \label{Bub21}

ABSTRACT: Jeff says, ``Chris Fuchs presents a different analysis of
the status of the projection postulate as Bayesian updating,
associated with a very different account of quantum probabilities as
degrees of belief than the view I want to argue for here. \ldots\ But
this rather complicated analysis misses the essential point.'' Chris
says, ``No.  It may not be perfect, but it captures the essential
point it was meant to capture.  It just so happens that that
essential point is different from the one Jeff wants to make.''

\subsection{Abstract}

\bjb
I distinguish between the measurement problem of quantum mechanics,
which is a problem about truth, and a problem about probabilities.
\ejb

I like that strategy and think it is an important step forward.  It
has been my strategy since 1995, though I didn't know it in these
terms then.  In July 2001, I learned a little about {\James}, {\Dewey}, and
pragmatist theories of truth, and realized that that's what I had
been on about with respect to quantum measurement.   But I started to
take the Bayesian step toward quantum probabilities long before that.

At this stage, I doubt very much that you'd be happy to think that
you might ultimately be led to a pragmatist notion of truth (in the
quantum measurement setting at least), but the distinction you make
is---in my eyes---the first step toward it.  In any case, there is
certainly nothing to be lost by starting to make a distinction
between the ``issue of quantum measurement'' and the ``problem of
quantum-state collapse.''  I will certainly admit that the former is
not refined or understood well enough for my own tastes at this
stage, but I think worrying about the latter anymore is just a waste
of time---I think the evidence is simply overwhelming.

Here's the way I put the distinction of the concepts to Oliver Cohen
in December 2003.  I will quote the passage at decent length because
it helps set the stage for the particular points I want to make with
regard to your own paper.

\bq
I'm writing you because I've been reading your paper ``Classical
Teleportation of Quantum States'' this week.  It's a nice paper, and
I very much like the simplicity of your scheme and the point you make
with it.  I am in complete agreement.

In fact it took me a little down memory lane.  You see, Asher Peres
and I had used teleportation as an example in our March 2000 {\sl
Physics Today\/} article, ``Quantum Theory Needs No
`Interpretation','' precisely to illustrate the sensibility of the
conception of a quantum state as a ``state of knowledge, rather than
a state of nature''.  When the paragraph peaked in clarity (i.e.,
before the editor's knife), it went like this:
\bq
The peculiar nature of a quantum state as representing information is
strikingly illustrated by the quantum teleportation process. In order
to teleport a quantum state from one photon to another, the sender
(Alice) and the receiver (Bob) need a pair of photons in a standard
entangled state. The experiment starts when Alice receives another
photon whose polarization state is unknown to her, though known to
some preparer in the background. She performs a measurement on her
two photons, and then sends Bob a classical message of only two bits,
instructing him how to reproduce the unknown state on his photon.
This economy of transmission appears remarkable because to completely
specify the state of a photon, namely one point in the {\Poincare}
sphere, we need an infinity of bits.  However, the disparity is
merely apparent.  The two bits of classical information serve only to
transfer the preparer's information, i.e., his {\it state}, to be
from describing the original photon to describing the one in Bob's
possession. This can happen precisely because of the previously
established correlation between Alice and Bob.
\eq

\ldots\ The conclusion you draw, I think, is particularly important:
the phenomenon of quantum teleportation only looks surprising and
remarkable if one takes an ontic view of the quantum state.  In fact,
in the past, I have accused some of my friends (some of whom were
authors on the original teleportation paper) of sticking with an
ontic interpretation of the quantum state precisely because it is the
only way to keep the phenomenon surprising and newsworthy. \ldots

If you are interested in seeing the struggle Asher and I had in
constructing the paragraph above see \ldots\ \  Maybe the main lesson
in those discussions is how difficult it is to give up an objectivist
language when using quantum states, even for a recalcitrant
positivist like Asher, and even in an example intended to be
illustrative of why quantum states should be viewed as states of
knowledge, rather than states of nature.  (The philosophy being:  if
quantum mechanics looks too very mysterious, then you're probably
being wrong-headed about it.  Case in point: if teleportation looks
mysterious, then you're probably being wrong-headed about it too.)

The sense I get from your paper is that you are much more neutral
about the lesson than I am.  You say simply:  ``[O]ur classical
version of teleportation is just as impressive as the original
protocol, if we think of quantum states as representing states of
knowledge. \ldots\ If, on the other hand, we think of a quantum state
as having ontological content, \ldots, then our classical version of
teleportation is not equivalent to the quantum case,''  and leave it
at that.  However, there is a spate of evidence starting to come out
that a significant fraction of some of the most `remarkable'
phenomena in quantum information theory can be mocked up with
classical toy models just as your own.  The only requirement for
seeing it is that one must focus on the epistemic states (i.e., the
states of knowledge) in such models rather than the ontic states
(like the actual $H$ or $T$ in your own model).  For instance, Rob
{\Spekkens} has a toy model which he has presented in several
conferences and which he is writing up presently as a paper, ``In
Defense of the Epistemic View of Quantum States: A Toy Theory,'' in
which he can reproduce the following quantum mechanical and quantum
information-theoretic type phenomena in a pretty NON-REMARKABLE way:
the noncommutativity of measurements, interference, a no-cloning
theorem, \ldots\ and many others.  (In particular, he gets
teleportation too, just like you do.)  As Rob\index{Spekkens, Robert W.} puts it in his
abstract:
\bq
    Because the theory is, by construction, local and non-contextual, it
    does not reproduce quantum theory.  Nonetheless, a wide variety of
    quantum phenomena have analogues within the toy theory that admit
    simple and intuitive explanations. \ldots\ The diversity and quality of
    these analogies provides compelling evidence for the view that
    quantum states are states of knowledge rather than states of
    reality, and that maximal knowledge is incomplete knowledge.  A
    consideration of the phenomena that the toy theory fails to
    reproduce, notably, violations of Bell inequalities and the
    existence of a Kochen--Specker theorem, provides clues for how to
    proceed with a research program wherein the quantum state being a
    state of knowledge is the idea upon which one never compromises.
\eq
So, given that your paper is an independent and particularly notable
link in that, and as opposed to his paper, your result is not buried
within over 70 pages (and counting) of text, I very much endorse it.
I think the lesson is this:  A good lot of quantum information theory
is simply regular probability theory and information theory applied
in ways that had not been deemed interesting before.  {\it What is
interesting and unique to the quantum itself, thus, must be something
else.}

In my paper \quantph{0205039}, ``Quantum Mechanics as Quantum
Information (and only a little more),'' I tried to give the community
to a call to arms by saying this:
\bq
    This, I see as the line of attack we should pursue with relentless
    consistency:  The quantum system represents something real and
    independent of us; the quantum state represents a collection of
    subjective degrees of belief about {\it something\/} to do with
    that system (even if only in connection with our experimental kicks
    to it).  The structure called quantum mechanics is about the
    interplay of these two things---the subjective and the objective.
    The task before us is to separate the wheat from the chaff.  If the
    quantum state represents subjective information, then how much of
    its mathematical support structure might be of that same character?
    Some of it, maybe most of it, but surely not all of it.

    Our foremost task should be to go to each and every axiom of quantum
    theory and give it an information theoretic justification if we can.
    Only when we are finished picking off all the terms (or combinations
    of terms) that can be interpreted as subjective information will we
    be in a position to make real progress in quantum foundations.  The
    raw distillate left behind---minuscule though it may be with respect
    to the full-blown theory---will be our first glimpse of what quantum
    mechanics is trying to tell us about nature itself.
\eq

What your work and {\Spekkens}'s work does, from my perspective, is give
the best illumination yet of what I was hoping for when I was
speaking of ``combinations of terms'' in that passage.
Teleportation---being a certain combination of uses of the axioms of
quantum mechanics---is nevertheless a purely probabilistic or
information-theoretic effect.  As such, it tells us very little about
the ontology behind quantum mechanics.

My own view---and the thrust of my research program presently---is
that these examples help us to realize that what is unique in quantum
mechanics is not the probabilities (i.e., the quantum states) but
what the probabilities are applied to.  There, I think, lies the
essence of quantum mechanics:  It is localized in the Kochen--Specker
theorem.  ``Unperformed measurements have no outcomes,'' as Asher
Peres likes to say.  That is to say, where quantum mechanics gets its
uniqueness is from {\it breaking\/} with the old idea that a
probability (as a subjective state of knowledge) must be knowledge
about a pre-existent reality.  Instead, probabilities can just as
fruitfully be applied to capturing one's knowledge of ``what will
come about due to one's actions.''  The predominant issue becomes how
to formalize the difference between  probability theory as applied to
pre-existent facts and probability theory as applied to
``creatables'' (for want of a better word). \ldots
\eq

\bjb
I show that the projection postulate can be interpreted as a
probability updating rule, and I argue for a subjective Bayesian
interpretation of quantum probabilities as rational degrees of
belief, in the sense of Ramsey rather than de Finetti.
\ejb

I would certainly call it a probability update rule too---so, if I
have missed an ``essential point'' it is certainly not that.  On the
other hand, I think you're hoping that there is an essential
distinction between Ramsey and de Finetti, where as far as I can tell
there is none.  However I'll have much more to say about both of
these subjects later.

\subsection{Introduction}

\subsection{From Classical to Quantum Mechanics}

\bjb
Faced with the conceptual puzzles of quantum mechanics, there is a
temptation to begin with a blank slate. It seems, then, that if one
were only careful enough with implicit assumptions about physical
theory and measurement, the characteristic features of the theory
could have been foreseen before classical mechanics.

I think this view is entirely mistaken.
\ejb

I am in agreement with your last sentence.  Attempts like the former
effectively erase the empirical content (or contingency) of quantum
mechanics, and I just don't see that.  When I myself invoke the
slogan ``Quantum Mechanics is a Law of Thought,'' I only do it
partially tongue in cheek, quickly correcting myself.  For if quantum
mechanics were {\it only\/} law of thought, it would be like one of
Kant's a priori categories of the understanding. On the other hand, I
come dangerously close to viewing ``probability theory'' (i.e., the
theory of coherent gambling) in such a Kantian kind of way.  And,
indeed, it is partially because of this that I would not want to view
quantum mechanics as a {\it generalized\/} probability theory.

\bjb
The transition from classical to quantum mechanics involves replacing
the representation of properties as a Boolean lattice, i.e., as the
subsets of a set, with the representation of properties as a certain
sort of non-Boolean lattice.
\ejb

The?  I would rather say {\it one possible way of looking at\/} the
transition from classical to quantum mechanics involves blah, blah,
blah. And, you partially recover from this a few paragraphs later
where you write:
\bjb \label{Bubism5}
Of course, other ways of associating propositions with features of a
Hilbert space are possible, and other ways of assigning truth values,
including multi-valued truth value assignments and contextual truth
value assignments. Ultimately, the issue here concerns what we take
as the salient structural change involved in the transition from
classical to quantum mechanics, and this depends on identifying
quantum propositions that take the same probabilities for all quantum
states.
\ejb
But let me hang on this point for a moment despite your partial
recovery. For when you say things like, ``Fuchs misses the essential
point,'' you should realize that that judgement (at most) comes from
within a context very different from the one I am working in.

I would, for instance, never say ``the representation of properties
in quantum mechanics involves as a certain sort of non-Boolean
lattice.''  That is just not the context I'm working in.  Similarly,
I would not say, as you say in the next section, ``Somehow, a
measurement process enables an indeterminate property, that is
neither instantiated nor not instantiated by a system in a given
quantum state, to either instantiate itself or not with a certain
probability.''---i.e., I would not say that a measurement process
instantiates any {\it property\/} at all for a quantum system.

Instead, the setting for our quantum Bayesian program (i.e., the
particular one of {\Caves}, {\Schack}, and me), is one where all the  {\it
properties\/} \underline{intrinsic} to a quantum system are timeless
and have no dynamical character whatsoever---moreover, those
properties have nothing to do with particular quantum state
assignments or particular quantum measurement outcomes. In that way,
the idea of a non-Boolean lattice simply doesn't apply to them.

John {\Sipe} recently made a nice write-up of our view for his book
that, I think, brings this one difference between you and me into
pretty stark relief.  Maybe it's worthwhile to quote it at length, as
it may lay the groundwork for a good bit of our later discussion:

\bq
This interpretation shares some features with operationalism. {\it
Measurements}, for example, are understood in a manner close to that
adopted by an operationalist. They are characterized by POVMs, and
those abstract elements are associated with tasks in the laboratory
undertaken with gadgets that are part of the primitives of the
theory. The result of any such measurement is simply one of a
possible number of outcomes, and there is no talk of these
measurements ``revealing'' the value of any variable, in the sense
that an arbitrarily precise position measurement in classical
mechanics is often described as revealing the position of a particle.
Yet, compared to the operationalist's quiet, unassuming terminology
of ``tasks'' and ``outcomes,'' advocates of this interpretation adopt
a more active manner of speaking, referring to ``actions'' (or even
``interventions'') undertaken by an agent, and the ``consequences''
that those actions elicit.

This indicates a role for the observer (or agent) in this
interpretation that is more significant than the role played by such
a person in operational quantum mechanics. The significance of that
role becomes clear when we consider the reference of density
operators in this interpretation. Density operators do not refer to
sets of tasks that define {\it preparations}, as they do in
operational quantum mechanics. Rather, a density operator is taken to
encode the beliefs of an agent concerning the probabilities of
different consequences of possible future actions. While these
beliefs may be {\it informed\/} by knowledge of the tasks involved in
setting up the particular gadgetry associated with a preparation,
they are not {\it determined\/} by it. Hence there is not a unique,
``correct'' density operator necessarily associated with each
preparation procedure, as there is in operational quantum mechanics.
In the present view two different researchers, one more skilled in
quantum mechanics than the other, could adopt different density
operators after being identically informed of the details of a
particular preparation procedure. One density operator might be more
successful than the other in predicting the possible consequences of
future actions, but each would be the correct density operator {\it
for that agent\/} insofar as it correctly encoded that agent's
beliefs.

Thus, while the abstract elements in the theory associated with
measurements are identified with tasks in the laboratory, as in
operationalism, the abstract elements in the theory associated with
preparations are identified with beliefs of the agent, signaling a
kind of empiricist perspective.

So in contrast to operational quantum mechanics, where density
operators are necessarily updated following a measurement --- since
the combination of the previous preparation and the measurement
constitutes a new preparation, and an operationalist associates the
new density operator with {\it that} --- in this view there is no
necessary updating of a density operator in the light of measurement
outcomes, since there is no {\it necessary\/} connection between the
consequences of an agent's action (more prosaically, ``measurement
outcomes'') and his or her beliefs. After all, foolish researchers,
like foolish men and women more generally, could choose not to modify
their beliefs concerning the consequences of future actions despite
their knowledge of the consequences of recent ones. And note that
even wise researchers will not update their beliefs concerning future
actions until they {\it know\/} the consequences of recent ones;
hence a wise researcher's ``personal density operator'' (the only
kind of density operator there is in this view!)\ will not change
until that researcher is actually aware of a measurement outcome.

Other abstract elements in the theory, such as the dimension of the
Hilbert space, and the dimensions of various factor spaces, are
actually associated with instances of attributes of physical objects.
Hence with respect to the reference of {\it these\/} abstract
elements this interpretation is realist. The manner in which this
works can best be seen by first reviewing the role measurement
outcomes play in revealing aspects of the universe in realist
classical mechanics, and then comparing that with the role such
outcomes play in this interpretation of quantum mechanics.

An arbitrarily precise position measurement of a bead moving along a
wire, in realist classical mechanics, reveals the position of the
particle, the instance (say, $x = 10$ cm) of a particular attribute
(bead position) of a physical object (bead) that actually exists in
Nature. In contrast, a usual Stern--Gerlach device oriented along the
z direction {\it does not}, in this interpretation of quantum
mechanics, reveal the z-component of angular momentum, or for that
matter anything else. The particular outcome of one experimental run
is simply a consequence of performing the experiment. Nonetheless,
repeated experimentation {\it does\/} reveal that the electron
associated with the atom passing through the device should be taken
as a spin-1/2 particle. Here the attribute under consideration is
taken to be {\it internal angular momentum}, and the instance -- the
irreducible representation appropriate to the particle of interest --
{\it spin-1/2}. The role of an ``instance of an attribute'' in this
interpretation is not to specify one of a number of possible
expressions of existence, as it is in realist classical mechanics,
but rather to specify one class of possible beliefs -- the one that
the theory recommends -- about the consequences of future
interventions of a particular type.

Note that, at least within nonrelativistic physics, the instances of
the attributes in this interpretation are fixed. A spin-1/2 particle
remains a spin-1/2 particle. Thus there are no dynamical variables in
this theory, only nondynamical variables analogous to the mass of a
particle in nonrelativistic classical mechanics. The point of physics
is to identify these nondynamical variables. Repeated interventions
by experimentalists, and the careful noting of the range of
consequences that those interventions elicit, is how these fixed
instances are discovered.

In this interpretation of quantum mechanics, with its mix of
operationalist, empiricist, and realist identification of abstract
elements in the theory, these fixed instances specify the [[agent
independent features]] of the ``quantum world,'' and it is the
business of physics to figure them out. This is done by
experimentation, and the theoretical linking of basis vectors in the
appropriate Hilbert space with various measurements, providing an
``anchor'' for those basis kets to our experience, the consequences
of our actions. Particularly significant is the Hamiltonian operator
and its basis kets [[in {\Caves}' particular version of all this]]. As
time evolves during what is colloquially described as ``unitary
evolution,'' we have the option to modify our beliefs or to modify
the anchors of those beliefs; the first strategy corresponds to the
usual {\Schroedinger} picture, the second to the Heisenberg picture.

Regardless of the strategy, the [[properties intrinsic to the]]
quantum world of this interpretation [[are]] a fixed, static thing.
[[This aspect of the quantum world]] is a frozen, changeless place.
Dynamics refers not to the quantum world, but only to our actions,
our experiences, and our beliefs as agents. Or, more poetically (\`a la
Chris), life does not arise from our interventions; it is our
interventions.
\eq

John\index{Sipe, John E.} doesn't represent us correctly in every detail of this
presentation---for the purpose at hand, it only seemed essential to
modify him in a few instances, which I have have marked with double
brackets [[$\bf \cdot$]]---but I would say he is roughly on track,
and he certainly gets it that we are not concerned with the usual way
of ascribing properties to quantum systems via the values of
measurement outcomes or probability-1 predictions (i.e., the
eigenvector-eigenvalue link).

Which brings me back again to your paper:
\bjb \label{Bubism6}
For a quantum state, the properties represented by Hilbert space
subspaces are not partitioned into two such mutually exclusive and
collectively exhaustive sets: some propositions are assigned no truth
value. Only propositions represented by subspaces that contain the
state are assigned the value `true,' and only propositions
represented by subspaces orthogonal to the state are assigned the
value `false.' This means that propositions represented by subspaces
that are at some non-zero or non-orthogonal angle to the ray
representing the quantum state are not assigned any truth value in
the state, and the corresponding properties must be regarded as
indeterminate or indefinite: according to the theory, there can be no
fact of the matter about whether these properties are instantiated or
not.
\ejb
You see, my way of looking at things wouldn't even allow me to say
what you say here.  It is just a very different world that I am
working in.

To try to make this point, let me quote a couple of emails I wrote to
Bas van Fraassen a few months ago.  It started with my saying this:
\bq
The way I view quantum measurement now is this.  When one performs a
``measurement'' on a system, all one is really doing is taking an
ACTION on that system.  From this view, time evolutions or unitary
operations etc., are not actions that one can take on a system; only
``measurements'' are.  Thus the word measurement is really a
misnomer---it is only an action.  In contradistinction to the old
idea that a measurement is a query of nature, or a way of gathering
information or knowledge about nature, from this view it is just an
action on something external---it is a kick of sorts.  The
``measurement device'' should be thought of as being like a
prosthetic hand for the agent---it is merely an extension of him; in
this context, it should not be thought of as an independent entity
beyond the agent.  What quantum theory tells us is that the formal
structure of all our possible actions (perhaps via the help of these
prosthetic hands) is captured by the idea of a
Positive-Operator-Valued Measure (or POVM, or so-called ``generalized
measurement'').  We take our actions upon a system, and in return,
the system gives rise to a reaction---in older terms, that is the
``measurement outcome''---but the reaction is in the agent himself.
The role of the quantum system is thus more like that of the
philosopher's stone; it is the catalyst that brings about a
transformation (or transmutation) of the agent.

Reciprocally, there [[may]] be a transmutation of the system external
to the agent.  But the great trouble in quantum interpretation---I
now think---is that we have been too inclined to jump the gun all
these years:  We have been misidentifying where the transmutation
indicated by quantum mechanics (i.e., the one which quantum theory
actually talks about, the ``measurement outcome'') takes place.  It
[[may]] be the case that there are also transmutations in the
external world (transmutations in the system) in each quantum
``measurement'', BUT that is not what quantum theory is about.
[[Quantum mechanics]] is only a hint of that more interesting
transmutation.   [[Instead, the main part of quantum mechanics is
about how]] the agent and the system [[together bring about]] a
little act of creation that ultimately has an autonomy of its
own---that's the sexual interpretation of quantum mechanics.
\eq
which led to the following dialogue:

\bq
\bvf
Writers on the subject have emphasized that the main form of
measurement in quantum mechanics has as result the value of the
observable at the end of the measurement -- and that this observable
may not even have had a definite value, let alone the same one,
before.
\evf

Your phrase ``MAY NOT even have a definite value'' floated to my
attention.  I guess this floated to my attention because I had
recently read the following in one of the Brukner/Zeilinger papers,
\bq\noindent
     Only in the exceptional case of the qubit in an eigenstate of
     the measurement apparatus the bit value observed reveals a
     property already carried by the qubit.  Yet in general the value
     obtained by the measurement has an element of irreducible
     randomness and therefore cannot be assumed to reveal the bit
     value or even a hidden property of the system existing before
     the measurement is performed.
\eq
I wondered if your ``may not'' referred to effectively the same thing
as their disclaimer at the beginning of this quote.  Maybe it
doesn't. Anyway, the Brukner/Zeilinger disclaimer is a point that
{\Caves}, {\Schack}, and I now definitely reject:  From our view all
measurements are generative of a NON-preexisting property regardless
of the quantum state.  I.e., measurements never reveal ``a property
already carried by the qubit.''  For this, of course, we have to
adopt a Richard Jeffrey-like analysis of the notion of
``certainty''---i.e., that it too, like any probability assignment,
is a state of mind---or one along (my reading of)
{\Wittgenstein}'s---i.e., that ``certainty is a tone of voice''---to
make it all make sense, but so be it.
\eq
and
\bq
\bvf
Suppose that an observer assigns eigenstate $|a\rangle$ of $A$ to a
system on the basis of a measurement, then predicts with certainty
that an immediate further measurement of $A$ will yield value $a$,
and then makes that second measurement and finds $a$.  Don't you even
want to say that the second measurement just showed to this observer,
as was expected, the value that $A$ already had?  He does not need to
change his subjective probabilities at all in response to the 2nd
measurement outcome, does he?
\evf

It is not going to be easy, because this in fact is what {\Schack} and I
are actually writing a whole paper about at the moment---this point
has been the most controversial thing (with the {\Mermin}, Unruh,
Wootters, {\Spekkens}, etc., crowd) that we've said in a while, and it
seems that it's going to require a whole paper to do the point
justice. But I'll still try to give you the skinny of it:
\begin{itemize}
\item
   Q: \ \ He does not need to change his subjective probabilities at all
      in response to the 2nd measurement outcome, does he?
\item
   A: \ \ No he doesn't.
\item
   Q: \ \ Don't you even want to say that the second measurement just
      showed to this observer, as was expected, the value that A
      already had?
\item
   A: \ \ No I don't.
\end{itemize}

The problem is one of the very consistency of the subjective point of
view of quantum states.  The task we set before ourselves is to
completely sever any supposed connections between quantum states and
the actual, existent physical properties of the quantum system.  It
is only from this---if it can be done, and of course we try to argue
it can be done---that we get any ``interpretive traction'' (as Chris
{\Timpson} likes to say) for the various problems that plague QM.
[[\ldots]]

This may boil down to a difference between the Rovellian and the
Bayesian/Paulian approach; I'm not clear on that yet. [[\ldots]]
Rovelli relativizes the states to the observer, even the pure states,
and with that---through the eigenstate-eigenvalue link---the values
of the observables.  I'm not completely sure what that means in
Rovelli-world yet, however.

I, on the other hand, do know that I would say that a measurement
intervention is always generative of a new fact in the world,
whatever the measurer's quantum state for the system.  If the
measurer's state for the system HAPPENS to be an eigenstate of the
Hermitian operator describing the measurement intervention, then the
measurer will be confident, CERTAIN even, of the consequence of the
measurement intervention he is about to perform.  But that CERTAINTY
is in the sense of Jeffrey and {\Wittgenstein} above---it is a ``tone of
voice'' of utter confidence.  The world could still, as a point of
principle, smite the measurer down by giving him a consequence that
he predicted to be impossible.  In a traditional development---with
ties to a correspondence theory of truth---we would then say, ``Well,
that proves the measurer was wrong with his quantum state assignment.
He was wrong before he ever went through the motions of the
measurement.''  But as you've gathered, I'm not about traditional
developments.  Instead I would say, ``Even from my view there is a
sense in which the measurer's quantum state is WRONG.  But it is MADE
WRONG by the ACTUAL consequence of the intervention---it is made
wrong on the fly; its wrongness was not determined beforehand.'' And
that seems to be the main point of contention.
\eq

Particularly this is going to be a key point when I finally come to
the analysis in Section 7 of your paper.

\subsection{The Probability Problem}

\bjb
The orthodox answer is that the probability assigned to a property of
a system by a quantum state is to be understood as the probability of
finding the property in a measurement process designed to ascertain
whether or not that property is instantiated. I will defend this
proposal later in the paper, but a little thought will reveal that it
is rather problematic. When the system is represented by a quantum
state that assigns a certain property the probability 1/2, say, this
property is indeterminate. Physicists would say that ascribing the
property to the system in that state is `meaningless.' But somehow it
makes sense to design an experiment to ascertain whether or not the
property is instantiated by the system. And in such a measurement,
the probability is asserted to be 1/2 that the experiment will yield
the answer `yes,' and 1/2 that the experiment will yield the answer
`no.' Clearly, a measurement process in quantum mechanics is not
simply a procedure for ascertaining whether or not a property is
instantiated in any straightforward sense. Somehow, a measurement
process enables an indeterminate property, that is neither
instantiated nor not instantiated by a system in a given quantum
state, to either instantiate itself or not with a certain
probability; or equivalently, a proposition that is neither true nor
false can become true or false with a certain probability in a
suitable measurement process.
\ejb

I found this paragraph interesting.  Particularly as I could both
agree and disagree with the last sentence!  The difference comes at
the semi-colon.  That is, I disagree with this:  ``Somehow, a
measurement process enables an indeterminate property, that is
neither instantiated nor not instantiated by a system in a given
quantum state, to either instantiate itself or not with a certain
probability.''  But I agree with ``A proposition that is neither true
nor false can become true or false with a certain probability in a
suitable measurement process.''  The trouble comes in at the
connective ``equivalently.''  I don't believe those separate thoughts
are equivalent at all.  The saving grace of the second of the two
clauses for me is that you don't explicitly mention what the
proposition is about.  In the first clause, on the other hand, you
are talking about properties of the system.

\subsection{The Measurement Problem}

\bjb
In classical theories, we measure to find out what we don't know, but
in principle a measurement does not change what is (and even if it
does change what is, this is simply a change or disturbance from one
state of being to another that can be calculated on the basis of the
classical theory itself). In quantum mechanics, measurements
apparently bring into being something that was indeterminate, not
merely unknown, before, i.e., a proposition that was neither true nor
false becomes true in a measurement process, and the way in which
this happens according to the theory is puzzling.
\ejb

Here again, I can agree.  You nicely did not say anything about what
the proposition refers to.  For instance, with your wording here---as
long as I am careful to take the paragraph out of context!---I am
free to think that the proposition that gains a truth value with the
process of measurement refers to MY sensations, not a property of the
system at all.

Indeed, I rather like the very next sentence:
\bjb \label{Bubism9}
The standard measurement problem of quantum mechanics is
fundamentally {\bf a problem about truth} \ldots, distinct from the
probability problem.
\ejb
as long as I am not forced to refill the ellipses with your
parenthetical ``(or the instantiation of properties).''  I think this
is the fundamental problem---it is about {\it truth}---and I am very
happy that you're saying that much.  When you say things in the
philosophy-of-science community people listen.  It is quite important
to make a careful distinction between the problem of probability and
the problem of truth, and not many people are doing that presently.

My own conviction that the measurement problem is fundamentally a
problem about truth explains my fascination with {\James}, {\Dewey},
{\Schiller}, {\Rorty}, and {\Putnam}, and the whole pragmatist framework for
truth. Here is something I wrote {\Carl} {\Caves} in 2001:
\bq
Today I focused on rounding up some more William {\James}, John {\Dewey},
Percy Bridgman material.  I think {\James} is taking me over like a new
lover.  I had read a little bit of him before, but I think I was more
impressed with his writing style than anything.  But I was drawn back
to him by accident, after reading Martin Gardner's {\sl Whys of a
Philosophical Scrivener}.  Gardner devoted a lot of time knocking
down {\James}' theory of truth, because it is just so much easier to
accept an underlying reality that signifies whether a proposition is
true or false, rather than saying that the knowing agent is involved
in eliciting the very proposition itself (along with its truth
value).  And something clicked!  I could see that what {\James} was
talking about might as well have been a debate about quantum
mechanics.  He was saying everything in just the right way. (Let me
translate that:  he was saying things in a way similar to the way I
did in my NATO ``appassionata.'')  And things have only gotten better
since.
\eq
My recommendation---tongue in cheek---I like the direction your paper
starts to move in, but I think you could take a good dose of {\James}!

Here's a little dialogue I had with Bill {\Demopoulos} earlier this
year:
\bq
\bwd
I also don't see why we should need something as fundamental as KS to
sustain the notion that ``unperformed measurements don't have
outcomes.'' I'm being a devil's advocate here because I think what
you really mean is that without a measurement of whether the cat is
alive, the cat is neither alive nor not alive. But would you put it
this baldly? If not, why not?
\ewd

To answer your question in the best way I know how at the moment, I
would say:  The transformation that quantum mechanics speaks about,
the transformation from a `superposition' to `aliveness' or
`deadness', is a transformation {\it within the agent}, and that
transformation cannot take place without some interaction with the
external physical system labeled by the word `cat'.  What happens to
`cat' itself (described in a way that makes no reference to the
agent)?  On that, I think quantum mechanics is silent.  With a
mantra:  Quantum mechanics is a theory for ascribing (and
intertwining) personal probabilities for the personal consequences of
one's personal interactions with the external world.
\eq

\bjb
The most sophisticated formulation of Everett's interpretation is
probably the Saun\-ders-Wallace version (Saunders, 1998; Wallace,
2003). Here the preferred basis is selected by decoherence, and
probabilities are introduced as rational degrees of belief in the
Bayesian sense via a decision-theoretic argument originally due to
Deutsch (Deutsch, 1999).
\ejb
Sophisticated in the sense of being complicated, yes.  Sophisticated
in the sense of hiding things in layer upon layer.  See our original
criticism of Deutsch---Barnum {\it et al}., ``Quantum Probability from
Decision Theory?,'' Proc. Royal Society London A {\bf 456},
1175--1182 (2000), and \quantph{9907024}---which still stands,
as well as some new nice stuff by Huw Price, \quantph{0604191}.
Or talk to Jos Uffink, who's got some very good replies.

\bjb
Of course, quantum mechanics could be false, but it seems wildly
implausible that a modification of quantum mechanics whose sole
motivation is to solve the measurement problem will survive
fundamental advances in physics driven by other theoretical or
experimental questions.
\ejb

I like this point very much.

\subsection{Solving the Probability Problem}

\bjb
[T]he physical world is {\bf nonlocal}, in that spacelike separated
systems can occupy entangled states that persist as the systems
separate.
\ejb
This language hints of an ultimately ontic view of quantum states.

\bjb
Solving the probability problem {\bf without reducing the problem to
a solution of the measurement problem (the truth problem)}, amounts
to treating quantum mechanics as a theory of information, in which no
measurement outcomes are certified as determinate {\bf by the
theory}. Rather, measuring instruments are sources of classical
information in Shannon's sense, where the individual occurrence of a
particular distinguishable event produced stochastically by the
information source lies outside the theory. In this sense, a
measuring instrument, insofar as it functions as a classical
information source, is ultimately a `black box' in the theory. So a
quantum description will have to introduce a `cut' between what we
take to be the ultimate measuring instrument in a given measurement
process and the quantum phenomenon revealed by the instrument. The
`cut' is just a reflection of the fact that quantum mechanics is a
{\bf theory about the representation and manipulation of
information\/} constrained by the possibilities and impossibilities
of information-transfer in our world, rather than a theory about the
ways in which nonclassical waves and particles move.
\ejb
I like much of this of course, if not not literally all.

\bjb
If we set aside the measurement problem, the Gleason probabilities
cannot be intelligibly interpreted as the objective chances or
relative frequencies that dynamical variables take determinate
values, and the only viable option is a subjective Bayesian
interpretation of the quantum probabilities as rational degrees of
belief.
\ejb
Why?  Of course, I like the sentiment of this, but it'd be nice to
see in your words an extended discussion of this point.  As it is
written right now, I think this represents a weak point of the paper.

On the other hand, I don't agree with any of this:
\bjb
To show that quantum mechanics can `stand on its own feet' as a
theory of probability, i.e., as theory of information, we need to
take account of the phenomenon of decoherence: an extremely fast
process that occurs in the spontaneous interaction between a
macrosystem and its environment that leads to the virtually
instantaneous suppression of quantum interference. What happens,
roughly, is that a macrosystem like a measuring instrument or
{\Schroedinger}'s cat typically becomes correlated with the
environment---an enormous number of stray dust particles, air
molecules, photons, background radiation, etc.---in an entangled
state that takes a certain form with respect to a preferred set of
basis states, which remain stable as the interaction develops and
includes more and more particles. It is as if the environment is
`monitoring' the macrosystem via a measurement of properties
associated with the preferred states, in such a way that information
about these properties is stored redundantly in the environment. This
stability, or robustness, of the preferred basis, and the redundancy
of the information in the environment, allows one to identify certain
emergent structures in the overall pattern of correlations---such as
macroscopic pointers and cats and information-gatherers in
general---as classical-like: the correlational information required
to reveal quantum interference for these structures is effectively
lost in the environment. So it appears that the information theoretic
constraints are consistent with both (i) the conditions for the
existence of measuring instruments as sources of classical
information, and (ii) the existence of information-gatherers with the
ability to use measuring instruments to apply and test quantum
mechanics, given a characterization of part of the overall system as
the environment. That is, by selecting a preferred basis, decoherence
provides an explanation for the emergence of classical information in
a quantum correlational structure.
\ejb
This is just Zurekian obfuscation.  You have already invoked the
black-box concept for measurement, explaining the need for this if
quantum mechanics is to be viewed as a theory of information.  You
did that nicely.  Why back off and re-ontologize all the ideas to try
to {\it explain\/} the existence of classical observers?  WHO is
writing down these decohering wavefunctions?  In an information
theoretic approach to quantum mechanics (by that I mean a non-ontic
approach to wave functions), it is not fair to give in to temptation
at the last moment and invoke God in the quad.

\bjb
If something like the above account of decoherence is acceptable,
then the probability problem reduces to showing that the
probabilities assigned to measurement outcomes by these
information-gatherers, in the subjective Bayesian sense, are just the
Gleason probabilities.
\ejb
Again, why?  It seems you're being very short on the key points you
want to make here (i.e., this point and the last point above where I
said, ``Why?'').

\subsection{The Projection Postulate as Bayesian Updating}

\bjb
An analysis of quantum probabilities as measures of ignorance in the
Bayesian sense, i.e., as degrees of belief measured by rational
betting behaviour, has been developed by {\Schack} {\it et al}. (2001), and
{\Caves} {\it et al}. (2002). In Pitowsky's formulation (Pitowsky, 2002,
2005), the structure of `quantum gambles' encoded in the subspace
structure of Hilbert space imposes nonclassical probabilistic
constraints that define a logic of partial belief in the sense of
Ramsey.
\ejb
What is this explicit distinction between Ramsey and de Finetti that
you keep invoking?  Can you articulate it?

Here is a quote by Keynes on Ramsey.  How do you react to it?
\bq
The application of these ideas [regarding formal logic] to the logic
of probability is very fruitful.  Ramsey argues, as against the view
which I had put forward, that probability is concerned not with
objective relations between propositions but (in some sense) with
degrees of belief, and he succeeds in showing that the calculus of
probabilities simply amounts to a set of rules for ensuring that the
system of degrees of belief which we hold shall be a {\it
consistent\/} system.  Thus the calculus of probabilities belongs to
formal logic.  But the basis of our degrees of belief---or the {\it a
priori}, as they used to be called---is part of our human outfit,
perhaps given us merely by natural selection, analogous to our
perceptions and our memories rather than to formal logic.
\eq
And here is a long quote by Sandy Zabell, ``Ramsey, Truth, and
Probability.'' How do you react to it, along with my parenthetical
comment?
\bq
The key point is that previous attempts to explain induction had
attempted to model the process by a unique description of prior
beliefs [[references]], or by a very narrow range of possibilities
[[references]].  De Finetti realized that because probability is a
logic of consistency, one can never---{\it at a given instance of
time}---uniquely dictate the partial beliefs of an individual; at
most one can demand consistency.  The essence of inductive behavior,
in contrast, lies not in the specific beliefs that an individual
entertains at any given point in time, but the manner in which those
beliefs evolve over time.  [[In this way it is exactly like classical
logic:  One is not judged as irrational for starting with the
incorrect truth value for some proposition in one's considerations;
one is judged irrational only if one makes an incorrect inference in
the proof process.---CAF]] \ Let us pause briefly over this point.

I change my mind slowly; you do so with rapidity; you think I am
pigheaded, I think you are rash.  But neither of us is of necessity
irrational.  Disagreement is possible even if we share the same
information; we may simply be viewing it in a different light.  This
is what happens every time the members of a jury disagree on a
verdict.  Of course it can be argued that the members of the jury do
not share the same body of facts: each brings to the trial the sum
total of his life experiences, and one juror tries to persuade another
in part by drawing upon those experiences and thus enlarging the
background information of their fellow jurors.  It is the credibilist
view of probability that if you knew what I knew, and I knew what you
knew, then you and I would---or at least should---agree.

Such a metaphysical stance may well be, as I. J. Good says,
``mentally healthy''.  But it is an article of faith of no real
practical importance.  None of us can fully grasp the totality of our
own past history, experience, and information, let alone anyone
else's.  The goal is impossible:  our information cannot be so
encapsulated.
\eq

\bjb
Here I want to show that the quantum rule is in fact just a
noncommutative version of the classical rule.
\ejb
The next three pages give a very nice treatment! (And I'm serious
about that.)  Why would I---Bayesian to the core---ever want to find
to find a different analogy between quantum collapse and Bayesian
conditionalization than the one you present here?  In a quick word,
the idea of a POVM (even a standard von Neumann measurement) as an
analog of an indicator function goes in the wrong direction for what
I want to squeeze out as the essential structure of quantum
mechanics.  Indicator functions carry the baggage of thinking about
quantum measurement as a process of revealing pre-existent values.
But I'll come back to this in a longwinded way, as you can guess.

\bjb
The prior probability assignment of an observer about to make a
measurement on a quantum system S is given by an initial density
operator $\rho_S\in {\cal H}_S$. What should an observer take this
density operator to be?
\ejb
Here we go again \ldots\  You sure have some dangerous remaining
objectivist tendencies when it comes to probabilities and quantum
states.

\bjb
Suppose the universe $U = S+E$ is in an initial pure state
$|\psi\rangle\in{\cal H}_{R}$.
\ejb
Whose pure state?  Is the state information (in the sense of Bayesian
probability theory), or is it not?

\subsection{Quantum Probabilities as Rational Degrees of Belief}

\bjb
Now, Fuchs points out (Fuchs, 2002b, p.\ 34) that the state change
following a quantum measurement of a POVM $\{E_d\}$ can be presented
as a 2-state process:
\ejb
I think you meant ``2-stage'' here.

\bjb
In the case of a projective measurement $\{E_d\} = \{E_d =
|d\rangle\langle d|\}$, where the $E_d$ are projection operators, the
state change on measurement with outcome $d$ is a collapse
corresponding to a readjustment by the unitary operator $U_d =
|d\rangle|\psi\rangle$.
\ejb
There's a typo at the end of the sentence here; an operator is not a
state.  Here is the way I said it in the final version of the paper
(after I had fixed my own typo at this very spot!):
\bq
\noindent
In particular, when the POVM is an orthogonal set of projectors
$\{\Pi_i=|i\rangle\langle i|\}$ and the state-change mechanism is the
von Neumann collapse postulate, this simply corresponds to a
readjustment according to unitary operators $U_i$ whose action on the
subspace spanned by $|\psi\rangle$ is
\be
|i\rangle\langle\psi|\;.
\ee
\eq

Finally, we come to the real point of contention:
\bjb
Fuchs concludes from this analysis that quantum collapse can be
regarded as a noncommutative version of Bayes's rule. But this rather
complicated analysis misses the essential point. It is precisely the
`violent' collapse transition (17), where measurement is only
disturbance, that has to be explained as a noncommutative variant of
Bayes' rule and not merely the `gentle' selection from an initial
density operator of a term corresponding to the outcome of a
measurement.
\ejb

``When a distinction in concepts can be made, a distinction in
concepts should be made.''  I thought that methodology was the
working bread and butter of the philosopher?  Let me try to better
explain what I did, and why I did it.  I made a distinction; I
followed through with the implications; I tried to learn a lesson.

The distinction, when it was found, was made because of this:  I am
working in a quantum foundational context that is trying very hard to
dispel that idea that quantum ``measurement'' has anything {\it
\underline{a priori}\/} to do with information gathering about things
intrinsic to the quantum system. I tried to give you ample evidence
and explanation of that above. Still, though, let me give one last
point of reference in that regard.  It comes from the same email
discussion with van Fraassen that I referenced before.  Then I'll
return to giving an account of the ``distinction'' I've alluded to.

\bq
\bvf
I thought I would after all not follow you in replacing the term
``measurement'', despite all the bad effects old connotations have
had in various discussions.  We need to bracket the old connotations
such as that a measurement result reveals a pre-existing value for
the measured observable. But I think we can do that because: \ldots

[T]here is a certain kind of retrodictive inference possible also on
the basis of qm measurements.  For a long time the paradigm was a
source preparing a stream of particles in a certain state --
measurements on samples taken from the stream give a good basis for
conclusions about just what state the source was preparing, and these
conclusions can then be used to predict the outcomes of further
measurements made on later samples of the stream.
\evf

This is what we in quantum information call quantum-state tomography.
One can indeed think of a quantum measurement outcome as {\it giving
information\/} in the old standard sense in that case and not simply
being the ``unpredictable consequence of one's action.''  But then
``giving information'' is quantified by Shannon's ``mutual
information,'' $I(X,Y)$ and not simply by his entropy function
$H(Y)$. That is, one has two random variables in the game---one
treated classically, namely the ``unknown preparation'' $X$, and the
other one purely quantum mechanical, the result $Y$ of the
measurement interaction.  Those two variables have quite different
roles, and one indeed would not want to think of $X$ as the
``consequence of one's interaction.''  On the other hand, without
making explicit mention of $X$ one has no means for thinking of the
elicitation of $Y$ as giving information about anything at all.
Before seeing the value of $Y$, one can expect to be {\it
surprised\/} to the extent quantified by $H(Y)$, but that's where the
story stops.

[For a more detailed Bayesian-like development of this point, you
might have a look at our paper ``Unknown Quantum States and
Operations, a Bayesian View,'' \quantph{0404156} and some of the
references therein. Particularly the Introduction and Concluding
section might be of some interest to you with regard to the present
discussion.]

The only point I want to make to you with regard to your remark above
is that, for these reasons, I would say it has no bearing on the
issue at debate:  I.e., the debate of whether it is better to think
of a ``quantum measurement'' as simply an action with an
unforeseeable consequence, or rather as a kind of ``question-asking''
or ``information-gathering.''  It is tangential.
\eq

So, what is this distinction that I was trying to capture in my
formalism that it seems your way of viewing quantum
conditionalization does not take into account?  Take a single quantum
system, for which some quantum-state assignment $\rho$ has been made.
Now imagine performing a quantum ``measurement'' on the system and
updating the quantum state to some new value $\rho_d$, consequent to
the measurement's outcome.  One can write that update in the usual
Kraus way or in my idiosyncratic way.  Let's even take the special
case where the update is just the L\"uders rule---noting that it too
can be viewed in both ways.

You have a state change; I have a state change.  In the end, they're
identical of course.  But whereas you have a change that looks {\it
solely\/} like an application of Bayesian conditionalization, I
don't. I have a handle for making a distinction---and in fact it is a
distinction I want to make.  You write, ``It is precisely the
`violent' collapse transition (17), where measurement is only
disturbance, that has to be explained as a noncommutative variant of
Bayes' rule and not merely the `gentle' selection from an initial
density operator of a term corresponding to the outcome of a
measurement.''

No.  It doesn't need to be, and I don't want it to be.  I want it to
remain a distant, maybe unrecognizable addition to Bayes, not a
variant at all. It quantifies the extent to which my previous opinion
changes radically (as compared to the change it would have made
solely through Bayes) upon the receipt of the data.  It gives me a
handle for exploring a new issue.  If a quantum ``measurement'' is
not merely the receipt of data with regard something pre-existent,
why should it look like a learning process at all?  Well, in truth,
the consequence of the measurement may allow me a refinement in
predictions for the next round of measurement.  But to some extent it
is simply new input into my beliefs that my previous opinion could
not take into account.

In the end, where I think I want to go with this kind of thing is
this.  Sandy Zabell puts it nicely in his essay on Ramsey.
\bq
Ramsey did consider the question of the dynamic evolution of belief.
Conditional probability is defined in terms of conditional bets; it
states the odds that someone ``would now bet on $p$, the bet only to be
valid if $q$ is true.'' This approach has since been adopted as the
basis of the commonly accepted subjectivist definition (see, e.g., de
Finetti, 1972, p.\ 193); but of course it does not address the
relation that conditional probabilities---thus defined---may have to
the actual degrees of belief one holds after the observation of an
event.  And here we run up against an apparent inconsistency in
Ramsey's views.  Initially, Ramsey notes there is no reason to
automatically identify the two quantities:
\bq
\noindent
the degree of belief in $p$ given $q$ is not the same as the degree to
which a person would believe $p$, if he believed $q$ for certain; for
knowledge of $q$ might for psychological reasons profoundly alter his
whole system of beliefs.
\eq
But further on in the last section of his essay Ramsey writes: \ldots

Clearly the emphasized portion of the quotation completely ignores
the profound insight of the preceding quotation; perhaps this second
passage represents a portion of the text written at an earlier stage.
\eq
(OK, I'm too lazy to copy the second passage of Ramsey, which has no
relevance to my argument.)  Anyway, this idea of Ramsey's has been
developed at great length in the work of van Fraassen, Skyrms, and
Richard Jeffrey---in fact it is the standard fare of what Richard
Jeffrey calls ``radical probabilism.''  Standard conditioning in the
probabilistic setting is NOT always the way to go.  (The precise
statement is that diachronic Dutch book arguments {\it always\/} need
supplementary assumptions that the synchronic ones don't need.)

I want to fit quantum conditioning into that framework, if possible,
and this distinction I draw---i.e., splitting the quantum state
change into two components---gives me some handle on that.  I regret
some of the phrases in my original description of this now---all this
business about ``gut-wrenching violence''---(remember, that was 4
years ago!), and I am sorry it may have caused confusion about my
motivations, but the technical result is still exactly where I want
to go.

\bjb
The above analysis simply illustrates the way the projection
postulate works in quantum mechanics. It does not explain it, in the
sense of showing how quantum collapse can be understood as a
noncommutative version of Bayes' rule for updating states of belief.
What Fuchs' analysis in terms of POVM's shows is that the relevant
features of my example are quite general. But, again, this in itself
does not demystify quantum collapse as a noncommutative version of
Bayes' rule.
\ejb
Well, I didn't think I had claimed to demystify the particular form
of the quantum collapse rule so much as quantify it in a new way (one
more in line with the point of view I am trying to build).  If my bad
writing style confused you, again I apologize.

\bjb
Fuchs' Bayesian interpretation of quantum probabilities follows de
Finetti rather than Ramsey and reflects de Finetti's instrumentalism.
\ejb
When I get back home, I'll send you a reading list on Ramsey.  The
divergence is not as drastic as you think.  The only difference I
have ever made note of is that Ramsey was willing to call
``intersubjective agreement'' on a probability assignment ``objective
chance.''  But mixing the logical and empirical realms, Ramsey did
not do or advocate.  I'll bet money on that.

\bjb
For de Finetti, science is just an extension of common sense and
cannot inform the `logical' aspect of probability formalized in the
probability calculus through the notion of coherence. So it would not
make sense to regard quantum mechanics as a nonclassical probability
theory or theory of information, where the formal features of this
new theory encode objective structural relations about the physical
world. Physics can only be relevant to the extra-logical and
context-dependent evaluation of probabilities.
\ejb
That is pretty much correct.  And it strikes me as exactly what one
would want if one were a {\it realist}!  There is our reasoning about
the world, and then there is the world itself.  When one starts to
mix those two ingredients, that's when one is becoming an
instrumentalist!

Yes, I definitely want to see the essence of quantum mechanics being
born out as an extra-logical and context-dependent (let me say
empirical) addition to probability theory.  It is a layer on top of
probability (for instance the restrictive region that I always draw
on the simplex), and in that way quantum reality (whatever that is)
keeps its autonomy from my thought.

\bjb
What Fuchs takes himself as establishing is that quantum states
represent subjective degrees of belief \ldots
\ejb
Yes, my paper was meant to be a foray in that direction.  But, the
remainder of your sentence
\bjb
\ldots and that quantum collapse is Bayesian conditionalization {\em
in the standard sense}.
\ejb
not at all.  Besides this blatantly contradicts the remaining
sentences in your paragraph:
\bjb
What the {\bf physics} tells us is summed up in his statement \ldots:
`The world is sensitive to our touch.' That is, there is an
irreducible nonclassical disturbance that occurs whenever we probe
the world. But this enters into the {\bf readjustment} of the
observer's probabilistic beliefs {\bf after} the application of
Bayesian updating, which is a straightforward refinement of prior
degrees of belief in the usual sense.
\ejb

And finally,
\bjb
The difference, ultimately, is between an instrumentalist approach to
quantum mechanics with the application of a strictly classical
Bayesian theory of probability, and an interpretation of quantum
mechanics as a nonclassical theory of information, in which the
structure of quantum gambles, considered as an objective feature of
reality, informs the correlational structure of quantum
probabilities.
\ejb
No, no, a million times, no!  My point of view is in no form an
instrumentalism.  With each of our theories we are making direct or
indirect statements of what we believe of the objective, external
world.  The quantum Bayesian view of {\Caves}, {\Schack}, {\Appleby}, me, and
whoever else to quantum mechanics is no less so.  On the other hand,
imagining melding empirical statements into the very structure of the
laws of thought---as you and Itamar seem to do---represents a real
danger with regard to the realist-instrumentalist divide.

Maybe more later, but I'll stop now and print this out.

\section{06-07-06 \ \ {\it Try This} \ \ (to R. E. Slusher)} \label{Slusher16}

\noindent {\bf Supercomputers of the Future} \medskip

These quantum computers, for some problems, would be unimaginably faster.  For, increasing a problem in size by a factor of 2, can often make it 4 times as hard to solve on a regular computer.  But not so with a quantum computer---there the problem only becomes 2 times as hard to solve. \medskip

\noindent {\bf Quantum Computers Vs.\ Current Computers} \medskip

Instead of classical bits, a quantum computer would work with qubits.  A qubit is a single atom, or a smaller subatomic particle, whose spin works like an oracle for a bit.  The oracle doesn't make up its mind until it is asked a question.

Depending on how it is {\it found\/} spinning, a qubit signifies a ``1'' or a ``0.''  It does not make up its mind before the process of observation.\medskip

\noindent {\bf Why Quantum Computers Could Be So Powerful} \medskip

Pairs of qubits can be much more powerful as computational components than pairs of bits because of a strange quantum feature called entanglement.  In a regular computer, if one were to stop it in its tracks and learn the values of some of its bit, this often is of only limited value for guessing the remaining bits.  But in a quantum computer, knowing the answers of some oracle calls can give a seemingly unnatural amount of information about other oracle calls.  This makes the logical operations of a quantum computer more tightly connected than in a classical computer, effectively allowing a quantum computer to skip computational steps that a regular computer has to meticulously strain to go through one by one.

\section{06-07-06 \ \ {\it College Keys!}\ \ \ (to G. M. D'Ariano)} \label{DAriano5}

\bgmd
You didn't give back the keys of the college room at S. Caterina.
\egmd

I'm so sorry about doing that!  When I read your note, I looked into my backpack and there they were!  In any case, I have mailed them straight off; I hope it won't take the keys too long to get there.

I have been meaning to write you for several days now, but ever since returning to the US I have had constant things to do because of Bell Labs (and this large nanotech meeting I am organizing for next week).  Anyway, I want to thank you again for inviting me; I very much enjoyed my time there and got a lot out of the interaction.  My discussions with you in particular will certainly make me rethink terminology, but more than that.  I'm not lying when I say I like the smell of your axiom system---it starts to seem right in a way that the others have not.  Ultimately, though, my desire is to get out of pure operationalism---i.e., to take a good look at the quantum formalism and distill from it some statement about the world that makes no direct appeal to the notion of an experiment taking place in the world.  I view your axiom system (what I understand of it) as a crucial step in that trek.

\section{07-07-06 \ \ {\it Markus Fierz, RIP} \ \ (to H. C. von Baeyer)} \label{Baeyer22}

\bhcvb
Today I asked Klaus Hepp (ETH Zurich), who happens to be a distant
cousin, whether Fierz is still alive, and he emailed back that he died
one week ago at 94.
\ehcvb

I'm always annoyed to hear news like this.  I think, ``Why on earth did I not seek out a way to talk to the person when I had a chance.''  I've had an interest in Fierz for 11 years.  If I had sought him out in 1995, when I was already traveling the world, I might have caught him at a fresh, relatively young, 83, and might have learned so many things.

\bhcvb
I looked at D'Ariano's paper, but found it pretty hard to understand.
However, if some of his postulates about ``informationally complete
observables'' and ``symmetric faithful states'' can be rendered into
English, they might indeed be promising.  Is the limitation of
information that nature allows us to collect about a quantum system
explicit or implied in his approach?
\ehcvb

I go up and I go down about the work, particularly after all the fights I had with him last week about terminology.  Might sound silly, but I think it reveals some serious underlying philosophical differences.  Certainly, I see ``purely operational axioms'' as a stopgap measure (though a necessary first step), whereas he may (probably) view it as the end of the line.  Still, I think leaning heavily on informationally complete measurements (and their properties) in an axiomatization is the way to go.

To answer your question, yes it has to be.  It is somehow already buried in the idea of an informationally complete measurement.  But ferreting the direct connection out may be a big problem.

In any case, this has been a very exciting summer for me foundationwise.  Itamar Pitowsky presented a very nice result at the Pavia meeting that is worth serious study, to do with a kind of pseudo-Dutch-book argument for the quantum probability rule.  (I say pseudo-Dutch-book because it had empirical elements in it---to do with Kochen--Specker things---that a straight-up Dutch book argument cannot have.)  He told us that the lecture is posted at his homepage, but I could not find it.  When I learn the exact coordinates, I will send them your way.

Attached are some notes I put together while in Pavia to comment on a draft of a paper that Jeff Bub is writing.  [See 24-06-06 note ``\myref{Bub21}{Notes on `What are Quantum Probabilities'}\,'' to J. Bub.]  The notes are certainly more about my point of view than what Jeff is actually writing about.  But I felt I had to do that, since part of his paper was strictly intended to be contra-me.  Still, for my discussions with you---on Pauli, Fierz, and alchemy---even though you've already seen various pieces of what I tell Jeff, it may be useful for you to see these various pieces of the description all tied together in one place.

\section{08-07-06 \ \ {\it Cerro Grande II} \ \ (to H. C. von Baeyer)} \label{Baeyer23}

\bhcvb
In passing I noted that ``Hans'' accused you of the view that the world does not exist. (If, surprisingly, I happen to be the only Hans in this context, then I am that Hans.  If there's another, my comment is moot.) The trouble is, I don't remember accusing you of such a drastic view.  I always thought of information as the go-between that mediates between the thing and the mind, and that we have no direct evidence of the material world.  Which doesn't mean it isn't there.

If I really said those things, fine.  Even if I didn't, I don't mind helping you out as a straw-man. But did I say them?
Only one person comes to mind who really did say them: the editor Herv\'e Poirier\index{Poirier, Herv\'e} in {\bf Science et Vie}. I recall that I was uncomfortable when he wrote that, but didn't object because I didn't know what your real views were.

Anyway, I am not in the least offended.  I'm just following your lead in trying to be as accurate as I can -- which is difficult with a lousy memory.
\ehcvb

Indeed it wasn't you.  But you should know that people accuse me of that ``drastic view'' all the time.  Literally all the time.  Here's a partial list that comes to mind:  Wayne Myrvold, Michael Nielsen, Andrew Landahl, Abner Shimony, Philippe Grangier, Matthew Donald, Hans Halvorson of course (though he doesn't any more), Simon Saunders, Todd Brun, Bob Griffiths, Wojciech Zurek, and untold numbers of people in the audiences of so many of these meetings I've gone to over the years.

Apparently it is a very thin line we are walking---thin at least with respect to the coarse-grained perceptions and views of the people who pipe up with said accusation.

\section{12-07-06 \ \ {\it Notation for Inspiration} \ \ (to H. C. von Baeyer)} \label{Baeyer24}

\bhcvb
And what of the future?  What notational revolutions can we expect, and what new understanding will they bring about?  We have to await a future Leibniz, Dalton, or Feynman to tell us -- but I can suggest a couple of areas in which a notational innovation would be welcome. [\ldots]

Quantum mechanics itself presents another opportunity for better notation.  The original version of the theory developed in 1925/26 includes two different ways in which atomic systems develop in time -- a smooth, predictable evolution according to well established rules, interspersed with abrupt random changes called quantum jumps or quantum leaps.  The two processes might be represented respectively by the steps and risers of a staircase, and both are well understood.  To describe an atomic process, physicists usually have to deal with two or three steps up and down the stairs.  In recent years, much effort has gone into trying to understand the puzzling interface between the quantum world of the atom, and our own, macroscopic experience. [\ldots]  Eventually I hope that an evocative notation will suggest a better way to understand how the world as we experience it, and as it is described by histories, arises from its quantum mechanical substrate.
\ehcvb

Thanks for sending me your article on notation.  I enjoyed it very much.  Do you know I've had Yates' book\footnote{Frances A. Yates, {\sl The Art of Memory}, (U. of Chicago, 1966).} on my shelf for several years now---ever since picking it up at a charity flea market---but have never read it.

The idea is simple and good.  Particularly, you are right in that we need a better notation for making sense of what quantum measurement is all about.

My own favorite image of what quantum measurement is all about is the one attached, which you have seen many times now.  But how to turn that image into an effective notation?  And notation to do what?  I don't really know.

Here's maybe an actual notational innovation in our field of quantum computation:  \quantph{0504097}.  If you haven't read about the Raussendorf--Briegel model of quantum computation yet, this may be a good place to start in any case.  Quantum computation enacted solely by quantum measurement and no unitary time evolution at all.  That thrills me to no end.  For myself, it makes me take the idea of ``measurement'' as ``action'' even more seriously.

It's all about philosopher's stones.  But how to put that into notation?  Your article did get me thinking.

\section{16-07-06 \ \ {\it Edward Lear?}\ \ \ (to A. Shimony)} \label{Shimony10}

\noindent\underline{\bf NOTE}: For an amusing story of one of the unintended consequences of this note, see the Abner Shimony story in the introduction, ``How to Stuff a Wild Samizdat,'' of my Cambridge University Press book {\sl Coming of Age with Quantum Information}.\medskip

I wonder if you're not in transit to Waterloo yet?

I am looking forward to your celebration and have been gearing up mentally for it in several ways.  One is I have just read Brent's biography of C. S. Peirce, so that I might get to know the man a little better.  (You know that my talk is titled ``Peirce, James, and the Quantum Bayesians''?)  What a tough thing to do, to read that book!  And actually what a depressing thing to do:  I couldn't help but note some of the parallels between his life and mine.  Though I lack his genius, I have certainly brought some of the same troubles upon myself.

Particularly now though, I'm writing you because I have just spent a frustrating hour on the web trying to hunt down the words of the poem you recited a little in Vienna last year.  The key line as I recall was ``going to sea in a sieve''.  It wasn't Edward Lear's ``The Jumblies'', was it?  What you read seemed so much more serious.  Anyway, I'd like to use the lines you actually did recite in my own talk in Waterloo.  (I'll use them in the obvious way \ldots\ to make fun of myself.)  If you're out there in email-land still, could you let me know?  If I don't hear from you, I'll use some of the lines from Edward Lear---they're certainly appropriate!

Can't wait to talk with you again.

\subsection{Abner's Reply}

\bq
It was very good hearing from you and hearing that you would be at the conference in Waterloo. Also that you are reading Peirce and about him, though your dark statements about parallels between his life and yours were disturbing.

The poem you enquired about is by Lear, and the first line is ``They all went to sea in a sieve, they did / In a sieve they went to sea.'' I don't know whether the poem is called ``The Jumblies'', though that name occurs in the poem.\footnote{\editornote It is ``The Jumblies''; the last stanza is as follows.
\begin{verse}
And in twenty years they all came back,\\
\quad   In twenty years or more,\\
And every one said, ``How tall they've grown!''\\
For they've been to the Lakes, and the Torrible Zone,\\
\quad   And the hills of the Chankly Bore;\\
And they drank their health, and gave them a feast\\
Of dumplings made of beautiful yeast;\\
And everyone said, ``If we only live,\\
We too will go to sea in a Sieve,---\\
\quad   To the hills of the Chankly Bore!''\\
\quad\quad      Far and few, far and few,\\
\quad\quad\quad         Are the lands where the Jumblies live;\\
\quad\quad      Their heads are green, and their hands are blue,\\
\quad\quad\quad         And they went to sea in a Sieve.
\end{verse}}  Most poetry books have indices not only by title but by first line.
\eq

\section{18-07-06 \ \ {\it Story about Peirce and Habits} \ \ (to J. Christian \& L. Hardy)} \label{Christian1} \label{Hardy17}

Here's the story about Peirce and linear time from Louis Menand's {\sl The Metaphysical Club}:

\bq
One of the first things Peirce did after he arrived at Hopkins in the
fall of 1879 was to start a Metaphysical Club. It was open to faculty
and graduate students from any department, and it met once a month to
discuss papers presented, usually, by the members themselves. \ldots

At one meeting, presided over by Morris, Dewey heard Peirce read a
paper called ``Design and Chance,'' and joined in the discussion
afterward. The paper is the germ of Peirce's later cosmology, and it
sums up in a few pages what was probably the substance of the
yearlong class Dewey had chosen not to take. Peirce's subject was the
laws of nature---the laws that Newtonian physicists believed
explained the behavior of matter and that physiological psychologists
believed explained the behavior of minds---and he began with a simple
question: Does the principle that everything can be explained have an
explanation? Or, as he also put it: Does the law of causality (which
is another name for the principle that everything can be explained)
have a cause? \ldots

Summarizing Peirce's Metaphysical Club paper on ``Design and Chance''
is \ldots\ not quite the same thing as paraphrasing it. The argument
begins with the point James Clerk Maxwell had made with his imaginary
demon: that a scientific law is only a prediction of what will happen
most of the time. Even ``the axioms of geometry'' said Peirce, ``are
mere empirical laws whose perfect exactitude we have no reason
whatever to feel confident of.'' The decision to treat a particular
law as absolute is a pragmatic one: sometimes we feel that
questioning it will only lead to confusion, and sometimes we feel
that questioning it is necessary in order to try out a new
hypothesis. A law, in Peirce's pragmatic view (derived, of course,
from Wright), is essentially a path of inquiry. It helps us find
things out---as the law of gravitation, for example, helped us
discover Neptune---and Peirce's first rule as a philosopher of
science was that the path of inquiry should never be blocked, not
even by a hypothesis that has worked for us in the past.

Maxwell's view was that laws are fundamentally uncertain because
there is always a chance that the next time around things will behave
in an improbable (though not an impossible) way---a chance that all
the fast molecules will congregate on one side of the container.
Peirce's point was that a chance occurrence like this can change the
conditions of the universe. His illustration was drawn from classic
probability theory: in a game with fair dice, a player's wins and
losses will balance out in the long run; but if one die is shaped so
that there is an infinitesimally greater chance that after a winning
throw the next throw will be a losing throw, in the long run the
player will be ruined. A minute variation in what seemed a stable and
predictable system can have cosmic consequences. In the natural
world, Peirce said, such minute variations are happening all the
time. Their occurrence is always a matter of chance---``chance is the
one essential agency upon which the whole process depends''---and,
according to probability theory, ``everything that can happen by
chance, sometime or other will happen by chance. Chance will sometime
bring about a change in every condition.'' Peirce thought that even
the terrible second law of thermodynamics---the law of the
dissipation of energy---was subject to reversal by such means.

As Peirce acknowledged, this was a Darwinian argument: ``my opinion
is only Darwinism analyzed, generalized, and brought into the realm
of Ontology,'' he said. What he meant was that since nature evolves
by chance variation, then the laws of nature must evolve by chance
variation as well. Variations that are compatible with survival are
reproduced; variations that are incompatible are weeded out. A tiny
deviation from the norm in the outcome of a physical process can,
over the long run, produce a new physical law. Laws are adaptive.

Pragmatically defined, variations are habits. They constitute a
behavioral tendency ---for if they had no behavioral consequences,
they would have no evolutionary significance. Bigness in beak size is
whatever big beaks do for you (if you are a finch), just as (to use
an example from ``How to Make Our Ideas Clear'') ``hardness'' is just
the sum total of what all hard things do. What Peirce proposed in
``Design and Chance'' was that natural laws are also habits. This was
not a new thought for him. There is a story, attributed to William
James, about a meeting of the original Metaphysical Club in
Cambridge, in which the members waited patiently for Peirce to arrive
and deliver a promised paper.
\bq
They assembled. Peirce did not come; they waited and waited; finally
a two-horse carriage came along and Peirce got out with a dark cloak
over him; he came in and began to read his paper. What was it about?
He set forth \ldots\ how the different moments of time got in the
habit of coming one after another.\footnote{See M.~H. Fisch, ``Was There a Metaphysical Club in Cambridge?,'' in
{\sl Studies in the Philosophy of Charles Sanders Peirce, Second
Series}, edited by E.~C. Moore and R.~S. Robin (U. Massachusetts
Press, Amherst, MA, 1964), p.~11. The words are attributed to Dickinson
Miller. [Fuchs: Compare this to John Wheeler's story of the Pecan
Street Cafe grafitto.]}
\eq
It sounds like a joke, but the story is probably true. Peirce's paper
must have been an extrapolation from the nebular hypothesis---the
theory that the universe evolves from a condition of relative
homogeneity, in which virtually no order exists, not even temporal
order, to a condition of relative heterogeneity, in which, among
other things, time has become linear. How did time get straightened
out in this way? By developing good habits. In ``Design and Chance,''
Peirce put it this way:
\bq\noindent
Systems or compounds which have bad habits are quickly destroyed,
those which have no habits follow the same course; only those which
have good habits tend to survive.

Why \ldots\ do the heavenly bodies tend to attract one another?
Because in the long run bodies that repel or do not attract will get
thrown out of the region of space leaving only the mutually
attracting bodies.
\eq
If you are a heavenly body, in other words, gravitational attraction
is a good habit to have, in the same way that if you are a
proto-giraffe, a long neck is a good attribute to have. It keeps you
in the system. When gravitational attraction becomes the habit of
{\it all\/} heavenly bodies, then we can speak of ``the law of
gravity,'' just as when all surviving proto-giraffes have long necks,
we can speak of a giraffe species, and (presumably) when all moments
of time have the habit of following one another, we can speak of
past, present, and future. But the law of gravity did not preexist
the formation of the universe, any more than the idea of a giraffe
did. It evolved into its present state while the universe was
evolving into {\it its\/} present state. Gravity was a chance
variation that got selected. Objects that didn't have the
gravitational habit didn't survive.
\eq

\section{18-07-06 \ \ {\it The Real Peres Number} \ \ (to J.-{\AA} Larsson)} \label{Larsson2}

For the Peres 33-ray example, you once told me how many rays and bases it would take if the jumble were to be completed into full sets of bases (i.e., no basis elements missing).  Could you tell me those numbers again?  (I guess the number of bases is 16---if I have that right---but what is the total number of rays when everything is counted properly?)

As before, I'm asking because I'm guessing you have these numbers at the top of your head.  If you don't, don't kill yourself---I just thought it might be nice to mention them in passing when I mention the KS theorem (in passing) in this week's talk.  (Actually, I'm going to put a lot of different material than usual in this one; wish you were here at the Abnerfest to discuss these things.)

\subsection{Jan-{\AA}ke's Reply}

\bq
The 33 vectors form 16 triads. But to do the proof for these you need to modify the ``two ones and a zero'' to ``maximum one zero in a pair'', sometimes you go via a rotation from one triad to a pair of vectors and then to another triad. Completing to triads only I counted to a total of 57 vectors, forming 40 triads interconnected by 96 rotations.

A table for different KS sets is in \quantph{0006134}.
\eq

\section{19-07-06 \ \ {\it My Draft} \ \ (to C. M. {\Caves} \& R. {\Schack})} \label{Caves84} \label{Schack103}

OK, I have read it all again, and now at least I agree with everything that is actually said in the paper.  At this stage, in preparation for printing it out tomorrow morning to hand off to {\Mermin}, the only changes I have made to {\Carl}'s latest draft are [\ldots]

As I said, now I at least agree with everything that is actually said in the draft.  That is progress.  That is not to say, however, I still wouldn't like to make some small additions.  There are the ones already marked for me to say something---and, I'm sorry again, but I haven't taken care of that yet.  But also,

1) I am a little worried that we have not defined the word ``objective'' all that well.  ``Subjective'' is OK, but ``objective'' is left a little dangling.  I worry about how to do that in a way we will all agree with and so as not to stir up a new hornet's nest.

2) The present paragraph on the ``fundamental conceptual difficulty'' in the Principal Principle in the coin-toss case is good, but we really didn't give any argument against it in the quantum case.  We simply said ``the Bayesian contends \ldots''  I don't think that is adequate, as much of the point of our paper is premised on the idea that ``objective probability'' is worthless.  The Lewisian says, ``Why worry about this category distinction business when both terms---the probability function and its argument---are simply objective in the quantum case?  It is a different thing than the classical Bayesian case.  There is the objective event and its objective propensity; no category distinction called for.''

3) I think it could be useful to end on a more rounded out note on the meaning of the Born rule.  What is the meaning of the Born rule if it is not specifying a probability by fundamental physical law?  Why do we all use the darned thing?  We give a negative answer to the first question---i.e., that despite appearances, quantum probabilities are not specified by law itself---but we never give a positive answer.  My own contention is that the sum total meaning of the Born rule is that it is a rule of transformation.  It tells us how to take our probability assignment for the outcomes of {\it this\/} informationally complete measurement (or set of measurements) and transform it into a probability assignment for the outcomes of {\it that\/} measurement.  Thus it is the transformation rule that is specified by ``fundamental physical law'' (a kind of empirical addition to Dutch-book coherence), not the probability assignment itself. [\ldots]

5) An equation or two might be called for at the discussion ``The quantum operation depends, at least partly, on an agent's beliefs about the device that executes \ldots''  For instance, something like what {\Ruediger} and I put in our tomography-volume paper:
\bq\noindent
For any trace preserving completely positive map $\Phi$ on a
system, one can always imagine an ancillary system $A$, a quantum
state $\sigma$ for that ancillary system, and unitary interaction
$U$ between the system and the ancilla, such that
$$
\Phi(\rho)=\tr_A \!\left( U(\rho\otimes\sigma)U^\dagger \right)\,,
$$
where $\tr_A$ represents a partial trace over the ancilla's Hilbert
space.  The Bayesian should ask, ``Whose state of belief is
$\sigma$?''
\eq

6) This one may not be for here---i.e., another can of worms thing that we probably don't want to get into for the present paper---but I want to record that I'm also a little worried about our dual use of ``facts'' and ``propositions''.  {\Appleby} has already called me on it during his reading in Pavia.  It could be a serious point, and it has come up in our conversation before.  For instance, the present dual use probably contradicts the usage I proposed in several of those old notes to you and {\Mermin}.  See for instance, 04-09-2001 ``\myref{Schack5}{Note on Terminology},'' in my letter collection.

{\Appleby}'s particular point is that our category distinction is really a logical distinction, and one should therefore stray away from any presentation that doesn't make it seem as such.  (But I'd probably have to work more than I have the time for now, and quote {\Wittgenstein} and such, to make this mean much to you.  So, I won't go into it any further than this.)

If you tell me which, if any of these, you agree with, and you yourself think is important, I will try to oblige.  Realistically though (even without any of the changes above) I don't see how we can post anything before {\Carl} returns from vacation.  I can promise to do my very best to stop my sinning of trashing deadlines, and see what I can work up to have waiting on your desks when you return August 6.

\section{19-07-06 \ \ {\it Quick Early Reaction} \ \ (to C. M. {\Caves} \& R. {\Schack})} \label{Caves85} \label{Schack104}

\bcc
God save us from {\Wittgenstein}.  We're writing this for physicists.  I
do think we want to be addressing things other than propositions,
because most physicists would think physics has to do with things
other than propositions.
\ecc

Believe it or not, I am not making this up.  (I may be vindictive, but I am not {\it that\/} vindictive.)  I just had lunch with {\Mermin} and the first thing he said was that he could not get past our first page.  He got hung up on the words ``proposition,'' ``data,'' and ``fact''---how they were potentially meant to be the same thing, or how they might possibly be different.  He didn't know which.  The word he was the most comfortable with was ``proposition.''  When I told him that ``acquiring data'' was meant to be the becoming aware of the truth value of a proposition, he was OK with that too and happier.  However, with the word ``fact'' we made no great progress.  He said he might be happier if we didn't have to use the word in our presentation.  At the very least, we should be more careful about the distinctions and identities.

 \ldots\ just quoting him to the best of my ability.  I hope he'll move on to page 2 now.

I know there have been plenty of times when {\Carl} has said that I am not a physicist.  (Though I certainly can't blame him:  Clearly it's an opinion shared by all the physics departments I've interviewed with.)  But do you really want to say that David {\Mermin} is not a physicist?

I think we're going to need some more precision.  At least {\Appleby} and {\Mermin} are a sympathetic readership; what's it going to be like when we hit someone who is nasty?

\section{20-07-06 \ \ {\it Capacity for Creation} \ \ (to J. E. {\Sipe})} \label{Sipe10}

That's the phrase I should have used.  That is what I would like to think of as permanent and unchanging for a quantum system.  (As far as I understand, it is not crucial for my ontology at this stage, but it is an idea I am testing out and intrigued by.)

The essay collection ``Delirium Quantum'' is attached.  [See \arxiv{0906.1968v1}.]

For the bit I read you tonight about how
\bq\noindent
Our quantum measurements in the present never tell us about the
past. They only tell us about the consequences of the past for any
of our other measurements in the present.
\eq
you can go to the present (incomplete) compilation of the new samizdat. [See 23-09-03 note ``\myref{Savitt3}{The Trivial Nontrivial}'' to S. Savitt.]

Thanks for the conversation tonight!  It was very useful to me.

\section{26-07-06 \ \ {\it Another Question} \ \ (to J. E. {\Sipe})} \label{Sipe11}

Thanks for the note.  I think it really goes to the heart of the matter (though, I recognize it is what you already brought up in our till-midnight conversation the other day).  In fact, it goes to the heart of the matter so much so that I had better search my soul before writing you more.  Your question is a really important one, and---I cannot lie to you---I am learning a lot by your forcing me on this point.

At the moment, I am inclined to say that I lean toward being a ``dualist Bayesian,'' though I don't think that definition (i.e., falling within a mere trichotomy) captures the right sense of distinction that should be made.  There is a whole spectrum of things that I think I would allow, and ``classical Bayesians'' and ``consequence Bayesians'' are just two points in that space (not necessarily extreme points).

Certainly I am sometimes in the habit of thinking of Hilbert-space dimension as akin to a classical property (something that one hypothesizes of a system).  Thus if one were to write down a probability distribution for the dimensionality of a system, one would be committing a type of ``classical Bayesianism,'' as opposed to ``consequence Bayesianism.''  What else could I mean by it?  On the other hand, I sometimes think of the distinction between dimensionality and a particular state vector as a {\it logical\/} distinction (one of the {\it level\/} of subjectivity).

See for instance the note ``\myref{Comer33}{04-07-03 Solid Ground, Maybe? (to G. L. Comer)}'' starting on page 209 of Cerro Grande II.  Particularly the remark on page 301 about levels of subjectivity.  Also have a look at the note ``\myref{Schack64}{25-07-03 Relative Onticity (to R. {\Schack})}'' starting on page 314.  Finally---though this may confuse things more than help things (because it is spiraling outward from your particular point)---if you have some extra time, {\it maybe\/} look at ``\myref{Schack68}{28-07-03 Your Newest Turn (to R. {\Schack})}'' starting on page 316.

Anyway, as I say, you deserve a thorough answer to this one, and it will be a good exercise to me to figure out my own opinion!

Let me now, if you don't mind, turn this note to a different kind of soul searching that I've been going through, and I hope you will give me an honest, straight-up opinion in reaction \ldots\ no matter what the reaction is.  One of my friends at the PI meeting, whose opinion I respect very, very much, absolutely hated my talk---so much so that he took me aside on a private walk to let me know.  The kinds of words that came out as a description of the talk were, e.g.:  ``too vague,'' ``not focused,'' ``irrelevant cartoons,'' ``doing a disservice to this otherwise interesting research program,'' [i.e., the quantum Bayesian program], overhearing people say that I'm a nut, things like that.  It was pretty devastating, and maybe I needed to hear it.  On the other hand, from another group of people at the conference, I thought---but maybe it was just wishful thinking---I got the opposite impression.  If I'm not mistaken Wayne Myrvold, for instance, said it was the clearest talk he had ever heard from me, and I remember some similar (though maybe not as extreme of) reactions from some of the other philosophers (and Bernstein, Greenberger, and Zeilinger).  When I encounter things like this, my head spins and I feel like a lost little sheep; in some ways I am very fragile.  All I care to do is get these ideas conveyed, and I have had a tendency to keep simplifying and simplifying to try to get this (what I view as a) rather simple message across:  In going from classical to quantum, the notion of probability and its import stays the same, but what changes is the notion of event (``consequence Bayesian'').  At the same time, one does not ``renounce the enterprise of physics'' (as Geoffrey Hellman said explicitly in his talk).  Our program, in fact, is exactly what is needed to get back on track in physics.  That's the message.  But maybe my talks are of absolutely the wrong strategy, and I am doing the program a great disservice after all.

Here's where I'd like your straight-up opinion.  Please be brutally honest.  I've been devastated once; it is bound to be much easier to be devastated a second time around, and I am prepared for it.  Do you think my talks on this subject would benefit by going heavier on technical details---quantum de Finetti theorems, senses of compatibility of quantum state assignments, POVM Gleason theorems, that kind of stuff?  I.e., that my method of speaking is way off key, even for a foundational meeting like the Abnerfest?  Taking Paul Busch's very clear talk---which I myself enjoyed immensely---as an example, do you think speaking more along the lines of that style would be of greater benefit to the quantum Bayesian program?

\subsection{John's Preply}

\bq
Let me try to me a bit more articulate about one of the questions I was trying to ask in our discussion at Perimeter.

Bayesians, as I understand them, take probability statements to be statements of beliefs.  The obvious follow-up is the question ``beliefs about what?'' As I understand your view, in quantum mechanics you take this to be an agent's belief about the consequences of future actions, and you have shown how the ket can be understood as encapsulating those beliefs through, for example, your ``bureau of standards'' approach.

Now suppose a Bayesian is not doing quantum mechanics, but rather paleontology, and is discussing the dinosaurs. If probabilities entered in his or her paleontology --- and presumably they would because of the kind of central significance that Bayesians give to probabilities --- those probabilities would again be taken to be statements of belief. Again one can ask ``beliefs about what?'' The man-in-the-street Bayesian (if such a person exists, but by the term I mean one not terribly familiar with quantum mechanics) might take those beliefs to be about when particular dinosaurs lived, what their physiology was
like, what they ate, what caused their extinction, etc., etc. That is, the ``$x$'' in $P(x)$ would be the usual kind of proposition about ``the external world'' --- in this case, ``the world of dinosaurs'' --- that is common in, say, usual readings of classical mechanics.

This would be different than the ``$x$'' in $P(x)$ that appears in your quantum mechanics, for there ``$x$'' is a proposition concerning the consequences of your actions, and not concerning ``the quantum world.''

In contrast to this, one could also imagine a Bayesian who insisted on using the same kind of ``$x$'' in paleontology as in quantum mechanics. That is, the ``$x$'' would refer to consequences of actions, such (as what we would colloquially describe) as ``seeing a fossil of such-and-such a type upon digging down one more layer at such-and-such a site.'' Any talk about dinosaurs existing, fighting, eating, etc., would be taken just to be heuristic devices for helping one think about the actual consequences of doing such fieldwork in paleontology, and the work of the discipline itself would understood to be properly concerned not with someone's heuristic thoughts of `dinosaurs' but rather with the fieldwork.

This leads to the point that, even if one is a fully-committed Bayesian, it seems to me that there are a number of types of which one might be a representative. Here are three:
\begin{itemize}
\item
A thorough ``consequence Bayesian,'' for whom the ``$x$'' in {\it any\/} $P(x)$ refers to the consequence of an agent's action.
\item
A ``dualist Bayesian,'' for whom the ``$x$'' in $P(x)$ {\it sometimes\/} refers to statements about the external world of the type common in the usual readings of classical mechanics, and {\it sometimes\/} refers to the consequence of an agent's action.
\item
A ``classical Bayesian,'' for whom the ``$x$'' in $P(x)$ {\it always\/} refers to statements about the external world of the type common in the usual readings of classical mechanics.
\end{itemize}

That is, simply committing to a ``belief understanding'' of probabilities does not in itself commit you to a particular view of to what kind of entity the ``$x$'' in $P(x)$ refers.

You would help me a lot if you would answer the question: In your view, what is the nature of the $x$'s in {\it all\/} the $P(x)$ that arise in science? Do they {\it all\/} just identify consequences of actions of the agent? Or can some of them identify propositions about the external world? And if some of them identify propositions about the
external world, what is the rule for establishing when that is possible? And if none of them identify propositions about the external world, paleontology is not about dinosaurs!
\eq

\section{27-07-06 \ \ {\it A Probability Calculus for density Matrices}\ \ \ (to M. Warmuth)} \label{Warmuth1}

Thank you for drawing my attention to your paper with Kuzmin.  To the moment, I have only given it a cursory look, but it looks very interesting!\footnote{\editornote See M.\ K.\ Warmuth and D.\ Kuzmin, ``Online variance minimization,'' in {\sl Learning Theory: Lecture Notes in Computer Science Volume 4005,} pp.~514--28 (Springer, 2006); M.\ K.\ Warmuth, ``A Bayes rule for density matrices,'' In
{\sl Advances in neural information processing systems 18 (NIPS'05)} (MIT Press, 2005); M.\ K.\ Warmuth and D.\ Kuzmin, ``Bayesian generalized probability calculus for density matrices,'' Machine Learning \textbf{78}, pp.~63--101 (2010).}

In fact, I hope that you will post both this paper and the other one on the {\tt quant-ph} archive, \myurl{http://www.arxiv.org/archive/quant-ph}.  It is effectively the main ``journal of quantum information and computing'' with most of us in the field looking to see what's new there daily.  It would be good for your results to be posted there for discussion and debate and building upon---this is certainly a community interested in such things.

Anyway, as I say, I look forward to understanding your papers.  There are indeed now several distinct things that one might be tempted to call a ``quantum Bayes rule.''  Exploring the realm of possibilities is good exercise and will lead to a better understanding of all that quantum mechanics has to offer.

My own preferred direction {\it at the moment\/} is actually {\it not\/} to think of density operators as ``generalized probabilities'' as you do in your papers, but rather as {\it single\/} (normal, usual, classical, whatever you want to call them) probability distributions for the outcomes of a fixed, fiducial measurement (an informationally complete POVM).  Then, the probability distributions for the outcomes of all other measurements can be thought of as coming about through a kind of coherence condition (analogous to Dutch book, but empirical in nature).  That is to say, the import of the Gleason theorem is that it tells us how to interrelate (or transform) probability distributions, not how to set them.

Caves, Schack, and I will make some remarks on this in our newest paper, which we hope to post soon.  However, let me give you a couple of existing pointers that at least broach the subject:
\begin{itemize}
\item
``Quantum Mechanics as Quantum Information (and only a little more)''\\
\quantph{0205039} \\
(See particularly, Subsection 4.2 and Section 6.)
\item
``Unknown Quantum States and Operations, a Bayesian View''\\
\quantph{0404156}
\end{itemize}

By the way, I already like the prettiness of this matrix operation you define in your Eq.\ (1).  (Though I was certainly aware of a similar operation used by Cerf and Adami---maybe though they used it in a more restrictive way---once upon a time, you helped bring it from the recesses of my memory.)  For your enjoyment (and contrast!), you might have a look at the monstrosity I once constructed to draw out some perceived similarity between quantum conditioning and Bayes' rule.  (The idea at the time was that the initial (pre-measurement) quantum state $\rho$ fulfills both the role of $P(h)$ and $P(d)$ in an application of Bayes' rule $P(h|d)=P(h)P(d|h)/P(d)$.)  You can find the thing at page 90 of \quantph{0105039} in the subsection titled ``Where Did Bayes Go?''

\section{27-07-06 \ \ {\it Course Adjustment?}\ \ \ (to W. L. Harper)} \label{Harper1}

It was good to finally meet you last week!  I greatly enjoyed our discussions, and I hope we'll pick them up again sometime.

For the moment though, I wonder if I might ask you a (probably difficult) question.  I hope you will give me your honest, straight-up opinion \ldots\ no matter what the opinion might be.  The reason I'm picking on you in particular is because it seems you got something out of my talk (I'm basing this on a couple of comments you made at the end) and because I don't think you've ever heard me speak before.  Thus perhaps you're a little more blank-slate for the question at hand than someone else I might ask.

OK, here goes.

One of my friends at the PI meeting, whose opinion I respect very, very much, absolutely hated the talk---so much so that he took me aside afterward to let me know.  The kinds of words that came out in his description of the talk were, e.g.:  ``too vague,'' ``not focused,'' ``irrelevant cartoons,'' ``doing a disservice to this otherwise interesting research program,'' [i.e., the quantum Bayesian program], overhearing people say that I'm a nut, things like that.  It was pretty devastating, and maybe I needed to hear it.  On the other hand, from another group of people, I thought---but maybe it was just wishful thinking---I got the opposite impression.  For instance, if I'm not mistaken Wayne Myrvold said it was the clearest talk he had ever heard from me (though maybe he was just being facetious).  Anyway, I certainly got some positive reactions from a few other people.

When I encounter opposite impressions like this, my head spins and I feel like a little lost sheep.  All I generally care to do in these quantum foundational talks is get the main ``quantum Bayesian'' message across, and I have had a tendency to keep simplifying and simplifying to try to get that done.  But maybe my talks are of absolutely the wrong strategy, and I am doing the program a great disservice after all, as my friend thought.

Here's where I'd like your straight-up opinion.  Please be brutally honest.  I've been devastated once; it is bound to be much easier to be devastated a second time around.  Do you think my talks on this subject would benefit by my going heavier on technical details---quantum de Finetti theorems, senses of compatibility of quantum state assignments, POVM Gleason theorems, that kind of stuff?  I.e., that my method of speaking is way off key, even for a foundational meeting like the Abnerfest?  Taking Paul Busch's very clear talk as an example, for instance, do you think my speaking more along the lines of that style would be of greater benefit to the quantum Bayesian program?  Or do you think it would be just the opposite, at least for someone's (like your own) first encounter with me?

I am truly perplexed, and if you could give me another data point to work with, that would be most useful for how I adjust my course.  Maybe I'll cc this note to Wayne, so he'll know what kinds of things I'm saying to his colleague (particularly if I'm misquoting him!).

\section{28-07-06 \ \ {\it Bad Reality Creation} \ \ (to G. L. Comer)} \label{Comer95}

I don't know if you read Paul Krugman in the {\sl NY Times}, but I wanted to have a recorded of the closing lines of today's column.  So, I'm going to send the whole column to you if you don't mind (in case you wanted to understand the context).  This is a very perplexing situation indeed---I've been scratching my head over it for six years!\medskip

From Paul Krugman, ``Reign of Error,'' {\sl New York Times}, July 28, 2006:

\bq
The climate of media intimidation that prevailed for several years after 9/11, which made news organizations very cautious about reporting facts that put the administration in a bad light, has abated. But it's not entirely gone. Just a few months ago major news organizations were under fierce attack from the right over their supposed failure to report the ``good news'' from Iraq --- and my sense is that this attack did lead to a temporary softening of news coverage, until the extent of the carnage became undeniable. And the conventions of he-said-she-said reporting, under which lies and truth get equal billing, continue to work in the administration's favor.

Whatever the reason, the fact is that the Bush administration continues to be remarkably successful at rewriting history. For example, Mr.\ Bush has repeatedly suggested that the United States had to invade Iraq because Saddam wouldn't let U.N. inspectors in. His most recent statement to that effect was only a few weeks ago. And he gets away with it. If there have been reports by major news organizations pointing out that that's not at all what happened, I've missed them.

It's all very Orwellian, of course. But when Orwell wrote of ``a nightmare world in which the Leader, or some ruling clique, controls not only the future but the past,'' he was thinking of totalitarian states. Who would have imagined that history would prove so easy to rewrite in a democratic nation with a free press?
\eq

\section{31-07-06 \ \ {\it To the Memory of Walter Philipp}\ \ \ (to A. Y. Khrennikov)} \label{Khrennikov20}

That is very sad news.  Do you know any details of how Walter passed away?  Was it a hiking accident or something?  Amazing how a person can be with you one day and gone the next.  There should be an object lesson in that which stays in the backs of all our minds.

\section{31-07-06 \ \ {\it Hi} \ \ (to W. E. Lawrence)} \label{Lawrence6}

\bwel
The Conference {\rm [``Quantum Reality, Relativistic Causality, and Closing the Epistemic Circle:\ An International Conference in Honour of Abner Shimony,'' July 18--21, 2006, Perimeter Institute, Waterloo, Canada]} looked as if it would be wonderful. What's your
impression of it? --- actually, before you answer --- I've got a mind to call you up and ask ``in person,'' as it were. But if you still want to answer \ldots
\ewel

Good to hear from you.  It was a fun conference.  I got quite a bit from the talks, and the ``dialog'' between Lee Smolin and Abner Shimony one night in the Black Hole Bistro was particularly fun.

For myself, I gave a sadly polarizing talk:  At the end of it, one of the philosophers came up to me and said it was the best talk he had ever heard from me.  On the other hand, one of the physicists (one whose opinion I respect very, very much) came up and said it was the worst talk I had ever given.  Unfortunately, I suspect the disparity in those comments indicates something deep!  Something I'd really rather live in denial about.

Anyway, you shouldn't feel that you completely missed the meeting.  You can still view the talks by going to this website:
\pirsa{C06009}. At the very least, you should enjoy the ``Bistro Banter'' that I mentioned above.

\section{02-08-06 \ \ {\it Collaboration?}\ \ \ (to L. K. Grover)} \label{Grover4}

Seeing your note to Bill coming by reminded me that I'd like to eventually discuss a possible collaboration if you're interested.  I was intrigued by some of your remarks at the July 11 meeting in the auditorium here, about how the $\pi/3$-phase-shift problem had a more understandable picture in terms of the evolution of probabilities rather than amplitudes.  That made me wonder how the Grover algorithm and all its variants would look if written in terms of this representation of quantum mechanics I've been developing --- where state vectors are replaced by probabilities for the outcomes of a certain informationally complete measurement.

Would you be interested in seeing if we could translate your algorithms into those terms?

I'm working rather steadily with my student Hoan on this representation until he leaves Sept 1.  But after Sept 1, I should be free to start another big project like the one above.  My hope is that it could give us some real insight into another way of looking at your algorithm.

\section{06-08-06 \ \ {\it The Story of Little Chocorua} \ \ (to J. B. Lentz \& S. J. Lentz)} \label{LentzB9} \label{LentzS6}

\bq\noindent
[[Flashback to three months before the present note:]] Kiki's gone freakier than usual on me.  A couple of weeks ago, she
found a hundred year old Victorian that's going to be torn down,
and she convinced the owner to let her have all the doors.  So,
I've got 16 doors (solid, mind you, and historical), but still 16
doors, stacked up in my garage.  Final destination:  She's going to
make the walls for the girls' playhouse from them. Too much time
on her hands and certainly too much imagination for her good (or at
least for the good of our relationship).  I told her she's got a
month to get them out of my garage.
\eq

I guess I have to admit that Kiki is a true blue carpenter now, and that I'm quite proud of her product.  The playhouse is fantastic.  And she's a pretty good salesman too, calling it Little Chocorua (after William {\James}'s weekend home in Chocorua, NH).  Did she tell you the story?  {\James} was so proud of the fact that there were 11 doors leading outside at his home; he could walk out and look in any direction on his land.  Well, the count of doors on Little Chocorua is 12.  I tell everyone we've done William {\James} one better.

By the way, here was another ploy of Kiki's for softening me up.  The big green patch on the right front door is chalkboard paint; I don't know if you knew.  Kiki says to the kids, ``You should think about how much you'd charge Dad for renting out Little Chocorua as a conference center.  He could bring his colleagues over and they could have small seminars around the chalkboard, if you'd rent your space out.''  That certainly did get me to start thinking, ``Hmm.  This might really be a good place for a small quantum information conference.''  She's a professional I tell you!

You two made quite a daughter.

\section{06-08-06 \ \ {\it Kiki's Wacky House} \ \ (to M. D. Sanders \& D. B. L. Baker)} \label{SandersMD1} \label{Baker15.1}

Let me give you a follow-up on the hair-brained scheme of Kiki's that I told you about a couple months ago.

I wrote the note below to Kiki's parents today since it's Kiki's birthday---it tells a little more of the story of the house---and included some pictures.  [See 06-08-06 note ``\myref{LentzS6}{The Story of Little Chocorua}'' to J. B. Lentz \& S. J. Lentz.]  Before stripping the pictures out of Outlook (so the {\tt .pst} file doesn't get too big), I thought I might send them to you too.  In the end, I'm pretty proud of the damned thing, and Kiki too.  She built every inch of the place all by herself---from the foundation to the frame to the roof---not one ounce of help from me or anyone else.  And even with the old ship's porthole window and the fire pole on the inside, the total cost was less than \$350.

Anyway, admitting that maybe she's not so crazy after all (to all the people I had told she was quite crazy) is part of my birthday present to her.

\section{15-08-06 \ \ {\it Drink the Kool-Aid} \ \ (to C. H. {\Bennett})} \label{Bennett45}

I have to apologize:  After our horse play the other day, I got carried away with some bureaucratic duties, and then simply {\it forgot\/} to come back to your note.  Pathetic.  Then Kiki, the kids, and I took a long weekend of sailing and amusement parks, etc.

I hope I'm not writing you back so late that the reply will be irrelevant.

\bcb
I thought first of Jonestown, but according to Wikipedia the Jonestowners drank a less well-known brand of grape drink.  ``Drink the Kool-Aid'' is said to have originated earlier, in the 1960's, from Ken Kesey's use of it as a vehicle for LSD.   Thus I can be said to have drunk the Kool-Aid of Many Worlds.  I thought that Lucent, like IBM, had drunk the Kool-Aid of DTO funding.

But what about basing QM on Yes QKD and No BC?   I am trying to write
a piece on conceptual foundations of quantum crypto.  As I recall
there was some difficulty with this program, and at one point it had
evolved to no-broadcasting and no superluminal communication, without
mentioning bit commitment.
\ecb

Yeah, the Bub, Clifton, Halvorson stuff wasn't too very convincing after my initial excitement wore off.  First, by starting off in a framework of $C^*$-algebras, they were effectively invoking quantum mechanics to begin with---or at least I always felt that way.  Then, it was pointed out that ``No BC'' wasn't doing any work for them at all.  And the last report I heard from Halvorson was that ``no superluminal'' was mostly redundant too.  A good explanation/critique comes in Chris {\Timpson}'s PhD thesis: \myurl{http://philsci-archive.pitt.edu/archive/00002344/}.
See pages 196--225; it's easy reading.

In the end, I don't think those early ideas of Gilles' and mine go very far toward getting at the essence of quantum mechanics.  The best counterexample, I think, is in the form of Rob {\Spekkens}'s toy model:
\quantph{0401052}.
It appears to have ``Yes QKD'' and ``No BC'', and much more (like a no-broadcasting principle and correlation monogamy) but it sure ain't quantum mechanics.  Instead it relies ``merely'' on an information limitation with regard to some hidden variables (i.e.\ pre-existing values).

I find that work of {\Spekkens} tremendously exciting because I learned so much from it.  It didn't lessen my conviction that the essence of quantum mechanics has to do with the intricate connection between observer and observed in the quantum world---in fact I think it is the best demonstration yet that quantum states are expressions of information, rather than stand-alone properties of the systems (or the universe) themselves---but it did help me to understand how much more interesting the story has to be.  {\Spekkens}'s toy model is based on the very pre-existence of measurement values (for that is what drives his combinatorics), but in the quantum world (because of Bell--Kochen--Specker), measurement values don't pre-exist the measurement process.  I now think it is that that is primary in quantum mechanics, and things like NO BC, YES QKD, no-broadcasting, teleportation, etc., are all secondary.

If you have any more questions, I promise to be more attentive this time.

\subsection{Charlie's Reply}

\bq
We miss you too.  Why don't you come up and visit us more often, so we can pretend to throw your computer away like in the old days?  Or we'll come down and visit you.   I thought of you when I was visiting my late mother's best friend Cynthia, 89 years old, the other day, along with her daughter Sally, my sister's best childhood friend, and her grandson Will (in his 20's).  Sally said ``quantum mechanics -- that must be hard to understand''.  Her mother said, ``We'll I'm taking a course in it for seniors, emphasizing the history and concepts, without the math''.  I said, ``Good for you.  The history is rather lively.  For example one of the greats, {\Schroedinger}, was a notorious womanizer, and historians are trying to figure out which mistress he was with when he came up with the {\Schroedinger} equation''.  Sally said, ``for a while they even thought of calling it condom mechanics.''
\eq

\section{18-08-06 \ \ {\it Aqualung} \ \ (to C. M. {\Caves} \& R. {\Schack})} \label{Caves86} \label{Schack105}

\begin{flushright}
\baselineskip=3pt
\parbox{3.3in}{
\bv
Feeling like a dead duck --\\
Spitting out pieces of his broken luck.\\
Sun streaking cold --\\
An old man wandering lonely.\\
Taking time\\
The only way he knows.\medskip\\
\hspace{\fill} --- Jethro Tull, ``Aqualung''
\ev
}
\end{flushright}

Attached is a draft where I made most of the changes I had suggested before.  I hope I didn't add anything contentious or take away anything that I wasn't supposed to either.  I tried not to cause trouble.  I did follow what {\Ruediger} earlier called the ``panacea''---I guess I'll find out how that flies.

Mostly I niggled a lot with some of the phrases to do with ``facts'' and ``the world'', in an attempt to be a little more careful, and so as not to give an impression that we do not think the agent is part of the world.  If you would like to get a better feeling for all the places I made changes (even trivially small ones), I could fax you my marked up manuscript.

Anyway---believe it or not---this time around I'm not so dead set on the particular way of expressing anything I added.  So, feel free to change expressions to your liking.  I tried my best to stay within the style that was already established.

You'll note that I added titles to all the references; this is because the paper is going to SHPMP rather than Phys Rev.  Some references still need to be added, but that can come next week.

The main thing that I regret not having touched very much is the discussion of objective chance and the weakness of the Principal Principle and such.  That really is the weakest point of the paper, and I wish we could do something better.  I suspect it is because of the weakness of that section that {\Carl}'s students came out swinging.  (Will they like this version better?  I did try to keep them in mind, particularly in the closing statements.)  And I also suspect that it is the weakness of the chance section that is going to lead to our most trouble in the future (particularly with the philosophical set, which is an annoying, but undeniable, fraction of our audience---I hate to say it, but they're the ones who at least take us more seriously than anyone else).

\section{23-08-06 \ \ {\it My Revision} \ \ (to C. M. {\Caves} \& R. {\Schack})} \label{Caves87} \label{Schack106}

\bcc
I think I understand why ``ascertaining'' was replaced by ``ascertaining or prompting'' in the second paragraph of the Intro: ``Gathering new data---i.e., ascertaining or
prompting the truth values of various propositions---allows the agent
to update his probability assignments, generally by using Bayes's
rule.''  It is probably meant to deal with the impression that ascertaining gives of finding out something that pre-exists.  I'm sympathetic to that, but suspect that the prompting will puzzle many readers.  Still, I don't want to change it.
\ecc
Yes, that was the motivation.  And, yes, you are probably right that a reader will be puzzled.  And finally, I'm happy that you don't want to change it.  But, I don't see that a small footnote with one sentence of explanation could hurt.  Might that be an effective analgesic?\footnote{Afterward, {\Carl} added this footnote to the paper: ``We introduce `prompting' here because `ascertaining' has the connotation of determining a pre\"existing property, which is in conflict with the central point of our paper.  Having made the point, however, we banish this connotation from the use of `ascertaining' and use it for the remainder of the paper.'' Later, he further changed `prompting' to `eliciting'; see 16-10-06 note ``\myref{Caves88}{Paper}'' to {\Caves} below.}

I think we're good to go on all else.  I'll certainly accept {\Ruediger}'s kind offer of making all submissions.  With all my computer problems, I barely have a functional office (still can't pin the problems down).

\bcc
In other words, here independent does not modify the verb demands,
which would call for an adverb, but is a predicate adjective in a
clause that has been abbreviated.
\ecc
In my professional career, I have always felt intellectually second-class whenever I'm in the presence of people who invoke terms like ``algebraic geometry'' or ``diffeomorphism invariance'' when speaking of a problem.  (Howard Barnum still frightens me to no end when I compare notes with him on any subject that I've been thinking about for some time.)  But ever since meeting Charles {\Bennett} and {\Carl}ton {\Caves}, I have realized that my greatest inadequacy is actually English grammar.

\bcc
It's always a privilege to write a paper with Chris (frustrating, but
a privilege), partly because the references make one look so much
more learned than one really is.
\ecc
The intent of the ``one'' can be read in at least two ways.  It's probably psychologically useful to me to believe that it is the more innocent of the two.

\section{23-08-06 \ \ {\it NATO Workshop Gdansk 10-13.09.2006} \ \ (to M. \.{Z}ukowski)} \label{Zukowski1}

Well, I've had no luck finding a speaker from the U.S. who could carry the torch of ``quantum states are information, not physical properties.''  But I have pinpointed a colleague from the U.K. who would do the point some serious justice, and in a technically important way.

It is Prof.\ {\Ruediger} Schack at U. London, Royal Holloway.  He would talk about recent work of he and Carlton Caves, that is some of the most important I've seen in our quantum-Bayesian effort in a while:  It is an analysis of ``quantum random number generators'' from our perspective.  The work ultimately connects to the precise meaning of how quantum cryptography can be more secure than classical cryptography even in this Bayesian view of quantum probabilities.  The differences from the naive view---which amounts to little more than blithely saying, ``it's secure because of physical law,'' which is an almost empty statement---are pretty striking.

So, I hope that piques your interest.  He would be a very good representative of our effort, and I saw the talk earlier this summer and know that it's excellent.

I know that he is from the U.K., but he would be talking about collaborations with his U.S. colleagues (Caves and me).  Do you think you can do it?  Would NATO allow it?

I apologize again for not being able to attend myself.  It looks like I am going to miss a really good and really important meeting.

\section{31-08-06 \ \ {\it My Two Cents} \ \ (to J. Finkelstein)} \label{Finkelstein9}

\bjf
It seems to me that to say that a detector has clicked is to make a statement about the state of that detector (or about the state of the device which records the clicks, which comes to the same thing). Since you take seriously the fact that experimental devices are necessarily quantum systems, I would have expected you to say that a statement about the state of a detector is a statement of subjective belief, not of objective fact.

To put it somewhat differently, in your paper you argue that since a preparation device is itself a quantum system, objective facts about that device do not determine the prepared quantum state.  But since the preparation device is a quantum system, how, in your view, could there be any objective facts about it?
\ejf

The way I would put it is to say that the cut is not three-way as you outline---that is, between the quantum system, the detector, and the agent---but rather two-way:  In our scenario there is only the agent and the external world.  The way one should think of the measurement device is the same way one should think of a prosthetic hand---it is an extension of the agent.

``Detector click'' is shorthand for {\it sensation\/} in the agent, and in my personal view, that is all it is.  The value of the ``detector click'' is beyond the control of the agent; it is not a function of his beliefs or his desires.  His interaction with the external world gives the class of potential sensations, but it does not set the value.

Thus, the value of the detector click---call it $d$---is objective for the agent.  (There is one sentence in the paper where I adopted almost that phraseology ``an event is a fact for the agent,'' and it might have been wise to adopt it more consistently.)  What is subjective from our view---i.e., a function of the agent and not the data or external world alone---is the POVM associated with those potential values.  That is, the set of operators $E_d$.

I expand on this in the essay ``\myref{Mermin101}{Me, Me, Me},'' [in this collection],
which I still think is worthwhile reading.   Let me also attach a file I wrote a couple of months back in response to some issues of Jeff Bub's (it's maybe a little more up to date).  [See 24-06-06 note ``\myref{Bub21}{Notes on `What are Quantum Probabilities'}\,'' to J. Bub.] Particularly relevant to the present discussion are the points I make right after Bubisms \ref{Bubism5}, \ref{Bubism6}, and \ref{Bubism9}. I'll also attach my favorite way of illustration of the point (in the form of a {\tt .jpg} file.)  There is an agent, there is an external world, there is an interaction between the two---the agent acts on the external world, it causes a reaction back in him (sensation).  As far as he is concerned that reaction is not subjective; it's as real as anything he's ever seen.

To make it clear:  I'm not speaking for {\Carl} or {\Ruediger} with these paragraphs.  They may disagree with one or another nuance of what I've said.  We tried to write the paper in a neutral way that didn't do a disservice to any of our particular views (though I think they are mostly aligned, except for particular choices of words).

\section{11-09-06 \ \ {\it My Swerves and Yours} \ \ (to M. S. Leifer)} \label{Leifer4}

\bml
For now, I'll just leave you with one comment.  I think a better name
for ``Zing'' would be ``the swerve'', in deference to Epicurus.
That's the earliest reference I can think of to any similar sort of
notion.
\eml

I've thought about ``swerve'' before (in the context of the old atomists and within our own), but I don't think the word really captures whatever it is I'm trying to get at.  For instance, somewhere in my first samizdat, you'll find this sentence:
\bq\noindent
The quantum mechanical indeterminism doesn't come about from an
indiscriminate swerve in the path of an atom; it comes from the point
of contact between the theory and the world---the measurement.
\eq
In my quantumness of a Hilbert space paper, I also remember writing this:
\bq
Associated with each quantum system is a Hilbert space.  In the case of finite dimensional ones, it is commonly said that the dimension
corresponds to the number of distinguishable states a system can
``have.''  But what are these distinguishable states?  Are they
potential properties a system can possess in and of itself, much like a cat's possessing the binary value of whether it is alive or dead?
If the Bell--Kochen--Specker theorem has taught us anything, it has
taught us that these distinguishable states should not be thought of in that way.

In this paper, I present some results that take their {\it
motivation\/} (though not necessarily their interpretation) in a
different point of view about the meaning of a system's
dimensionality. From this view, dimensionality may be the raw,
irreducible concept---the single {\it property\/} of a quantum
system---from which other consequences are derived (for instance, the maximum number of distinguishable preparations which can be imparted
to a system in a communication setting). The
best I can put my finger on it is that dimensionality should have
something to do with a quantum system's ``sensitivity to the
touch,'' its ability to be modified with
respect to the external world due to the interventions of that world upon its natural course.  Thus, for instance, in quantum computing
each little push or computational step has the chance of counting for more than in the classical world.
\eq

Zing is meant to capture something intrinsic to the system.  The bit of indeterminism that comes about in a quantum measurement I don't think of as intrinsic to any one thing.  Rather, it is more along the lines of the relational stuff you PI people always talk about---it is a function of two objects, not one.

\section{18-09-06 \ \ {\it Seeking Advice} \ \ (to L. Hardy, R. W. {\Spekkens}, G. M. D'Ariano, R. {\Schack}, D. M. {\Appleby}, and M. S. Leifer)} \label{Spekkens37.1} \label{Appleby14.1} \label{Leifer4.1} \label{Hardy18} \label{DAriano6}

\noindent Dear Friends and Acquaintances of the epistemic quantum state,\medskip

I wonder if I can ask your advice on the following subject---I'd like to get your opinions before making a decision myself.  I let the distribution list in sight, so you could see who else I was consulting, but please write back to me privately---I will keep your opinion in confidence.

Marlan Scully has given me the opportunity to have a plenary talk at his annual ``Physics of Quantum Electronics'' Conference in Snowbird, Utah (January 2--6), and more importantly, associated with that, the chance to organize a session around the subject with 4 or so other invited talks.  The thing I'm wondering is whether it would be worthwhile for our community for me to follow through with this?  If I were to organize a session it would be to highlight recent work (and potential applications) of epistemic/operational/Bayesian views of the quantum state.  Thus, this is why I'm writing to you.

It is decent sized meeting; I think there were over 300 people there last year, including a couple Nobel prize winners (Glauber, and I can't remember the name of the other one).  You can go to this website if you want to get a feel for what the line-ups the last few years have looked like:  \myurl{http://www.pqeconference.com/}.  Old programs since 2001 are posted there.

Anyway, the good I can see coming from this is that we would have a captive audience from a set of people who normally have far more to do with the applied side of physics than the ones we are usually around.  Maybe that would be useful for raising recognition of the promise of recent research in quantum foundations and its uses for the wider physics community.  The downside is that this is a very expensive conference, and there is no financial aid for it:  Registration is \$350 USD, hotel comes to \$180/night (though that could be halved by doubling up in rooms for anyone who cares to), meals are {\it not\/} included, and the only restaurants available are the ones of this resort hotel.  Not cheap.  They say the skiing is great (and the conference shuts down every afternoon for it), but for me personally, that is of very little consequence.

I would like to know your opinion of the value of such a get together before following through with this or not following through with this.  Also, would you yourself come?

\section{18-09-06 \ \ {\it Big Expensive Conferences} \ \ (to J. E. {\Sipe})} \label{Sipe12}

Sorry, I still owe you that note on the extent of my ``consequence Bayesianism.''  These are heady waters and I took a couple months off from foundational chit-chat to do some hard calculating while I had a good summer student.  Plus, after the Shimony meeting, I think my system needed a good purge.  However, I will come back, and probably soon (as we're starting to get referee reports on \quantph{0608190}, and some of their points concern the one you're also getting at).  But I don't feel too terribly guilty---you yourself never called me, after bringing up the possibility twice.

\section{18-09-06 \ \ {\it Another Topic} \ \ (to M. S. Leifer)} \label{Leifer5}

\bml
On a completely different topic, I wanted to ask you a question in
your role as chief expert on distinguishability measures in quantum
theory.  Do you know if anyone has done any work on axiomatic
approaches to distinguishability measures, along the lines of the
axiomatic approach to entanglement measures?  Specifically, I have in
mind that a distinguishability measure would be defined as a
functional of two states that is monotonically decreasing under CP
maps, with perhaps a few other regularity conditions, and then that
two such functionals could be found that provide upper and lower
bounds on all such measures.
\eml

You can look at Jozsa's {\sl J. Mod.\ Optics\/} article for a first pass.  But the real powerhouses in this regard are Uhlmann and Petz.  They have plenty written on the axiomatics of the subject.  Particularly, monotonicity has always been an essential ingredient.  A good place to get a start in the literature is probably the reference list in Ohya and Petz's book; or maybe flipping through Petz's web page:  \myurl{http://www.math.bme.hu/~petz/}.

\section{18-09-06 \ \ {\it New Course:\ Einstein's Universe} \ \ (to G. L. Comer)} \label{Comer96}

Word of warning on other matters:  Never ever get involved with 30 year old mathematical problems if you can help it.  They'll destroy your career just as surely as not doing any research at all will.  It's ``performance review'' time for us at Bell Labs this week, and I've got nothing at all to show for this last year.  God pity my mortgage.

\bgc
I think I have you beat: I'm working on dissipation in general
relativity, which has now about a 100 year history.  Maybe more
physics than math, but still unresolved.
\egc
Ah, but do you have something to show for your last year of work?  I'll bet I've got less than you! (Steven van {\Enk} once told me about a competition he saw where the contestants were each trying to see who could blow a noodle from his nose the farthest.)

\section{18-09-06 \ \ {\it Nonbayesian Musings} \ \ (to R. D. Gill)} \label{Gill3}

\brdg
I've made a first rough write up of some zen/frequentist quantum philosophy thoughts:
\bv
\myurl[http://www.math.leidenuniv.nl/~gill/waveparticle.html]{http://www.math.leidenuniv.nl/$\sim$gill/waveparticle.html}\\
\myurl[http://www.math.leidenuniv.nl/~gill/fifthposition.html]{http://www.math.leidenuniv.nl/$\sim$gill/fifthposition.html}\\
\myurl[http://www.math.leidenuniv.nl/~gill/BetterBell.pdf]{http://www.math.leidenuniv.nl/$\sim$gill/BetterBell.pdf} \ (Gdansk talk)
\ev
\erdg

Thanks for the coordinates (all of them).  Apparently though, I already had your new email address in my book---don't know how it got there.  Congratulations on your move to Leiden; I hope you fill Lorentz's shoes well!

Taking a look at your Gdansk talk, I noticed isn't there a contradiction in these two sentences in your closing transparency:
\begin{itemize}
\item[1)] ``I think that only detector clicks in the past are real.''

\item[2)] ``The probabilities are for real, the past is real, the wave function is objective \ldots''
\end{itemize}
Specifically, if ``{\it only\/} detector clicks in the past are real'' how can there be room for the probabilities to be real TOO?  I'll say (tongue in cheek):  Make up your mind!

\section{19-09-06 \ \ {\it Werner and Misrepresentation} \ \ (to R. D. Gill)} \label{Gill4}

I read this line in your Gdansk talk:  ``Recall the quantum information open problems site of Reinhard Werner (and let's wish him a speedy recovery too).''  What has happened to Reinhard?

And I read this line in one of your other pages:

\brdg
Contrary to the currently popular Bayesian interpretations of quantum physics, there is an objective reality (again:\ the past).
\erdg

That is simply a gross misrepresentation.  You grant reality to detector clicks; we grant reality to detector clicks.  And I certainly grant reality to the past.  I go further:  I grant reality to quantum {\it systems\/} (even when they're not in interaction with detectors) and their associated Hilbert-space dimensionality.  And {\Carl} {\Caves} among us, for instance, grants reality to the ``intrinsic'' Hamiltonians associated with quantum systems.  There are lots of things within the theoretical structure of quantum theory that we quantum Bayesians are willing to grant reality to.

The main things we {\it are not\/} willing to grant an intrinsic reality to (i.e., a reality outside a gambling agent that makes use of them) are {\it probabilities\/} and {\it quantum states}.  But that is a far cry from a position that ``there is no objective reality,'' as you effectively attribute to us above.

It's almost insulting that you cannot seem to get this simple, little point.

But you're not all bad, Richard.  I've always liked you.  And in the present instance, I very much like this sentence of yours:  ``As the present moves relentlessly forward the `past' crystallizes out of it, randomly.''  I am very much attracted to that idea, and have spent some time thinking about it myself.  If I understand your sentence correctly, the ideas are along the lines of those of George Herbert {\Mead}'s at the turn of the last century.  I'll paste in a little review of his ideas below.  [See quote in 23-09-03 note ``\myref{Savitt3}{The Trivial Nontrivial}'' to Steve Savitt.]

\section{22-09-06 \ \ {\it Chutes and Ladders} \ \ (to D. M. {\Appleby} \& H. B. Dang)} \label{Appleby14.2} \label{Dang1}

Thanks for the sympathetic note.

I spent a tossy-turny night having all kinds of wacky thoughts \ldots\ and trying to suppress them so that I might get some sleep.  The main thing I kept coming back to is:  Might we find some way of combining the first phase of this summer's work with the second?  Could we get any traction from that?

What I mean by the first phase is our general study of type-E matrices.\footnote{Let us call a $d^2 \times d^2$ matrix ``type-E'' if it is Hermitian and has $1/d$ on the diagonal and $1/d\sqrt{d+1}$ times various $e^{i\phi}$s for all the off-diagonal elements.  There is no restriction that a type-E matrix be positive semi-definite.  SIC Gram matrices (up to a scale factor) are special cases of the type-E matrices---they are simply type-E matrices that are rank-$d$ projectors.  For all type-E matrices $E$ it automatically holds that $\tr E = \tr E^2 = d$.  Thus if $E$ is a type-E matrix and $\tr E^3 = \tr E^4 = d$, then $E$ is a ``SIC-projector''.}  Particularly our reduction (under the assumption of Appleby form) to finding a set of $d^2$ phases so that a certain $d \times d$ matrix is rank-1 (the matrix given by Eqs.\ (11) and (12) in SIC-Notes C, 10 August).  Phase 2 was when we started thinking about minimizing the frame potential and/or explicitly solving the equations for a SIC-generating fiducial vector.

How to combine these two phases?

Here's one, almost surely useless idea.  (But it gives an indication of the sorts of things my mind was getting hung up on.)  Start with a Weyl--Heisenberg Gram matrix generated from some arbitrary fiducial state.  Next, find the nearest (in matrix norm) type-E matrix.  It won't be positive semi-definite or rank-1, but it will probably be closer to being such than an arbitrary choice.  So, next, find the nearest WH Gram matrix to this type-E matrix.  It generally won't be the original WH Gram matrix; it'll be a different one \ldots\ presumably being nearer to type-E form than the original.  And on we go iteratively.  A far-fetched question is:  Might one prove that this process generically converges to a sought-for SIC Gram matrix?  I.e., we just climb a ladder to get to it.

Sounds tough to me; the idea's probably a chute, rather than a ladder.  But mostly I'm interested in how we might combine our two separate trains of thought.

Here was another question that crossed my mind.  Might anything shake loose if we stopped thinking about the Gram matrices themselves (in the WH case), but the difference between the Gram matrices and our target.  Might this give us the tools to start considering small deviations from the target \ldots\ possibly getting a set of quadratic equations (rather than quartic)?

This is off the beaten track, but I remember Holevo's technique for proving his bound on accessible information.  He didn't try to prove the bound directly, but rather he started with an arbitrary ensemble of states and considered a one-parameter deviation from them.  Then he found the second derivative of the information with respect to that parameter.  The second derivative was relatively easily bounded, and then he just integrated back up.  I think you can read about the technique in my old paper ``Mathematical Techniques for Quantum Communication Theory'' (\quantph{9604001}).

Just thinking out loud \ldots

\section{26-09-06 \ \ {\it Something Significant} \ \ (to D. M. {\Appleby} \& H. B. Dang)} \label{Appleby14.3} \label{Dang2}

I just wrote you a very quick and dirty note.  The result is below.  I have no idea why these simplifications seem work, and have not made any attempt to prove them analytically yet.  But I am very, very pleased that they appear to be true.  You will see below.

I hope all is well with you.  I haven't had a chance to read your note from yesterday.  I've been sick in bed most of the day with a stomach virus.  I'll look at your note soon after this.

Recall that the frame potential in the Weyl--Heisenberg case is given by
\be
F=d^3\sum_{jn}\left|\sum_s a^*_s a_{s+j} a_{s-n} a^*_{s-n+j}\,\right|^2
\label{equinox1}
\ee
and the quantity will achieve its lower bound of $2d^3/(d+1)$ if and only if the following $d^2$ equations are satisfied:
\be
\sum_s a^*_s a_{s+j} a_{s-n} a^*_{s-n+j}=\frac{1}{d+1}\big(\delta_{n0}+\delta_{j0}\big)\;.
\label{StreetVendor1}
\ee

Defining the (unnormalized) vectors $|\psi_j\rangle$ by
\be
|\psi_j\rangle=\sum_s a_s^* a_{s+j} |s\rangle
\ee
another way to write Eq.~(\ref{StreetVendor1}) is as \be \langle\psi_j|X^n|\psi_j\rangle=\frac{1}{d+1}\big(\delta_{n0}+\delta_{j0}\big)\;.
\label{Somalia1}
\ee
(I'm only going to the trouble of introducing this notation so that one can identify the pieces of the little Mathematica routine I'm about to record.  It's not essential for what is about to be said.)

\subsection{Reducing the Number of Equations Drastically}

OK, with this set-up, here's the exciting thing.  Numerical work---at least from $d=4$ to 9 (I've only checked this little bit because I'm hoping Marcus will see this note before he goes to sleep, and so I'm rushed for time)---seems to indicate that we can effectively get rid of the $n$-index in the equations all together, being left with a little over $3d$ equations.

That is, it seems we can trim the full set of equations to (at least) only the following.
\noindent 1) $n=j=0$,
\be
\sum_s |a_s|^4 =\frac{2}{d+1}\;.
\label{lapis}
\ee
2) $n=0$, $j\ne0$, (and similarly the case $j=0$, $n\ne 0$), \be \sum_s |a_s|^2 |a_{s+j}|^2=\frac{1}{d+1}\;.
\label{lazuli}
\ee
3) $n=j$, $j\ne0$,
\be
\sum_s (a^*_s)^2 a_{s+j} a_{s-j}=0\;.
\ee
And 4) $n=j+1$, $j\ne0$
\be
\sum_s a_s^* a_{s+j} a_{s-1-j} a_{s-1}^*=0 \;.
\ee

There may be further simplification possible, but I haven't had a chance to look into it yet.  My first instinct was to hope that condition 3) could be taken away too---the roots of this were in one of Hoan's early hopes (one equation per diagonal in the outer product matrix, remember Hoan?)\ but at the moment this doesn't appear to be the case. (Or it may simply be that Mathematica's numerics are just having trouble settling down when I throw that constraint away. I'm just not sure at the moment.)

Anyway, you should probably check this independently as you have a chance.  Below is the Mathematica instruction set I used:

\noindent ------------------------------------------------------------------
\begin{verbatim}
a[i_] := b[i]+ I c[i]

NN = Sum[Abs[a[Mod[s,d]]]^2,{s,0,d-1}]

NN2 = Sum[Abs[a[Mod[s,d]]]^4,{s,0,d-1}]

CC[j_,n_] := Abs[Sum[Conjugate[a[Mod[s,d]]]*a[Mod[s+j,d]]*
             a[Mod[s-n,d]]*Conjugate[a[Mod[s-n+j,d]]],{s,0,d-1}]]

EC := Abs[CC[0,0]-2/(d+1)] + Sum[Abs[CC[j,0]-1/(d+1)],{j,1,d-1}] +
      Sum[Abs[CC[0,n]-1/(d+1)],{n,1,d-1}] +
      Sum[CC[j,j],{j,1,d-1}] + Sum[CC[j,j+1],{j,1,d-2}]

d = 7

nn = NMinimize[{EC, NN == 1, NN2 == 2/(d+1)},
     Join[Table[b[i], {i, 0, d-1}],Table[c[i], {i, 0, d-1}]],
     Method -> {"DifferentialEvolution", "RandomSeed" -> 3}]

bb = Table[b[i] /. nn[[2, i + 1]], {i, 0, d-1}]

cc = Table[c[i] /. nn[[2, i + d + 1]], {i, 0, d-1}]

aa = bb + I cc

psi[j_] := Conjugate[aa] RotateRight[aa, j]

Table[Conjugate[psi[j]].RotateRight[psi[j], n], {j,1,d},{n, 1, d}]

MatrixForm[%]
\end{verbatim}
------------------------------------------------------------------

To look at other dimensions, simply change the value of $d$ in there.
Also you have to jiggle a bit with the random seeds in each case, to get a correct answer.  Here are the random seeds that worked best for me (at least on my machine):
\begin{verbatim}
dimension            random seed
    4                    2
    5                    3
    6                    0
    7                    3
    8                    5
    9                    11
\end{verbatim}
The last line of the code above generates a view of how
Eq.~(\ref{Somalia1}) is satisfied.  $EC$ is minimized to zero as it should be with the constraints advertised above.  But what is nice is then that {\it all of\/} Eq.~(\ref{Somalia1}) is satisfied automatically.

I am very excited by this.

\section{28-09-06 \ \ {\it Simple, Getting Simpler} \ \ (to D. M. {\Appleby} \& H. B. Dang)} \label{Appleby14.4} \label{Dang3}

\subsection{Notation}

Recall that the frame potential in the Weyl--Heisenberg case is given by
\be
F=d^3\sum_{jn}\left|\sum_s a^*_s a_{s+j} a_{s-n} a^*_{s-n+j}\,\right|^2
\label{equinox}
\ee
and the quantity will achieve its lower bound of $2d^3/(d+1)$ if and only if the following $d^2$ equations are satisfied:
\be
G_{jn}\equiv\sum_s a^*_s a_{s+j} a_{s+n} a^*_{s+n+j}=\frac{1}{d+1}\big(\delta_{n0}+\delta_{j0}\big)\;.
\label{StreetVendor}
\ee
Defining the unnormalized vectors $|\psi_j\rangle$ by
\be
|\psi_j\rangle=\sum_s a_s^* a_{s+j} |s\rangle
\ee
we can also express the matrix elements $G_{jn}$ by
\be
G_{jn}= \langle\psi_j|X^n|\psi_j\rangle\;.
\label{Somalia}
\ee

\subsection{Understanding the Problem Better}

What it seems we were finding yesterday is this:  That the full set of $d^2$ equations
\be
G_{jn} = \frac{1}{d+1}\big(\delta_{n0}+\delta_{j0}\big)
\ee
is actually implied by the smaller set of $3(d-1)$ equations
\bea
G_{00} &=& \frac{2}{d+1}
\\
G_{j0} &=& \frac{1}{d+1} \quad\quad j=1,\ldots,d-1
\\
G_{jj} &=& 0 \quad\quad \qquad j=1,\ldots,d-1
\\
G_{j,j+1} &=& 0 \quad\quad \qquad j=1,\ldots,d-2
\eea
I still don't know how to prove this implication, but it is starting to seem much less surprising than when I first stumbled across it a couple days ago.  It might well be semi-trivial to prove it, if we just think about the problem in the right way.

The reason has to do with the thing Marcus pointed out about how the frame function has roughly a factor of 8 redundancy in it.  He worked with a more symmetric matrix $H_{jk}$ in the case of odd $d$ than the matrix $G_{jn}$ here, but the conclusion is surely almost exactly the same in both even and odd $d$ and can be seen directly in terms of the $G_{jn}$. I think it must simply come from careful index play/redefinition in the definition~(\ref{StreetVendor}) above---i.e., I haven't checked it directly yet, but I have loads of faith.  For, one can clearly see this kind of structure by taking a random $|\psi\rangle$ and looking at the magnitudes $|G_{jn}|$ numerically. They have a pretty and simple layout even without any special conditions on $|\psi\rangle$.  For instance, in $d=8$ one obtains a structure that looks like this
\begin{verbatim}
Z | Y X W V W X Y
-   - - - - - - -
Y | A B C D C B A
X | B E F G F E B
W | C F H I H F C
V | D G I J I G D
W | C F H I H F C
X | B E F G F E B
Y | A B C D C B A
\end{verbatim}
And in $d=9$, this
\begin{verbatim}
Z | Y X W V V W X Y
-   - - - - - - - -
Y | A B C D D C B A
X | B E F G G F E B
W | C F H I I H F C
V | D G I J J I G D
V | D G I J J I G D
W | C F H I I H F C
X | B E F G G F E B
Y | A B C D D C B A
\end{verbatim}

Anyway, in general one must have:
$$
|G_{j,n}|=|G_{-j,n}|=|G_{j,-n}|=|G_{-j,-n}|=|G_{n,j}|=
|G_{n,-j}|=|G_{-n,j}|=|G_{-n,-j}|\;.
$$
Whenever $j=-j$ or $n=-n$ or $n=j$, etc., some of the distinctions collapse.

What can be seen from this is that specifying the diagonal term to zero for $|G_{jn}|$, and similarly for the first off-diagonal, wipes out a whole lot of the terms!  For instance, in the examples above, only $C$, $D$, and $G$ remain {\it ostensibly\/} free.  But of course, we know they they are not completely free.  And it really shouldn't be that hard to see that they must actually vanish.

But how to show it, I don't know.  I tried several sorts of things with expansions of some vectors in terms of others, and with Schwarz inequalities, but so far I've failed.

In any case, looking at these structures promotes an even simpler looking conjecture.  And I propose this conjecture get the bigger proportion of study anyway.

\subsection{A Prettier Conjecture}

One can note that if one specifies the first three columns of these matrices, one is specifying just as many variables as if one specified according to the earlier scheme.  Thus, it seems that an even better conjecture would be:  That the following $3(d-1)$ equations imply the full set of $d^2$
\bea
G_{00}  = \langle\psi_0|X^0|\psi_0\rangle &=& \frac{2}{d+1}
\\
G_{j0} = \langle\psi_j|X^0|\psi_j\rangle &=& \frac{1}{d+1} \quad\quad j=1,\ldots,d-1
\\
G_{j1} = \langle\psi_j|X^1|\psi_j\rangle &=& 0 \quad\quad \qquad j=1,\ldots,d-1
\\
G_{j2} = \langle\psi_j|X^2|\psi_j\rangle &=& 0 \quad\quad \qquad j=1,\ldots,d-1
\eea
(In fact, by looking at the general structure of $|G_{jn}|$, one would really only need to specify about $3(d-1)/2$ equations.)

This is starting to strike me as looking extremely clean.  And I think it is as likely to be true as the previous conjecture.  I have checked it numerically for various dimensions here and there up to $d=13$---not as thoroughly and for as many seeds as previously, but enough to give me a sense that it must be true.

Supposing that it is relatively easy to show the equivalence: Can it really be so hard to satisfy the remaining equations?  Surely it is not even remotely on the same playing field with Fermat's last theorem or the Poincar\'e conjecture.

\section{28-09-06 \ \ {\it Simple, Getting Simpler, 2} \ \ (to H. B. Dang \& D. M. {\Appleby})} \label{Appleby14.5} \label{Dang4}

\bhbd
By the way, I'm curious how you found out about that set of $3(d-1)$
equations.  It's so amazing that it's true numerically! And do you know
that if you don't have all those $3(d-1)$ equations then it won't work,
i.e.\ can we further eliminate some of them?
\ehbd

Well, I can definitely see that they can be reduced to (roughly, i.e., plus or minus 1 or 2) a total of $3(d-1)/2$ equations.  That comes from the automatic symmetry of the $G_{jn}$ matrix.  But can they be reduced still further?  I don't know.  Possibly.  If you have a concrete proposal, I can test it out with my newfound Mathematica skills.  I had tried simply the equations for the main diagonal and the edges and that didn't work.  And I think I've tried other variations on this, but no great success.

How did I find the first set?  Well, I was enamored with the $n=j$ equations and had played with them for a day, trying to take their Fourier transform and such.  (I was hoping the magnitude equations---the ones on the edge---supplemented by a one-parameter set of equations that said something about the phases would do the trick.)  Then, once I got up and running numerically, I thought I would see if they (along with the edge equations) implied anything numerically.  The thing that surprised me was that they came pretty close to specifying the full set.  So, I thought, maybe if I take one of the off diagonals too, that would add enough.  And nicely, it did.  That's the basic story.

Now that you've got me thinking about this, I think I was able to successfully drop the $(0,0)$ equation in one of my earlier trials.  Let me see if I can still do that.  I don't know why I didn't think of this again this morning.\medskip

\noindent \ldots\ [time goes by]\medskip

Wow!  I quickly checked $d=7$ and $d=8$, and it seems indeed that that normalization equation---the $(0,0)$ one---can be dropped.  So, we're down to $3(d-1)-1$ equations!  (Or half that, as your taste may take you.)  That is a little more symmetric and prettier set.

\bhbd
My feeling is that it is some inequality (probably no
longer the Schwarz) that would do the proof, but I still can't prove
anything yet.
\ehbd

Yes, it would be good to use those known zeroes as upper bounds on the remaining terms.  That's why I was trying to use Schwarz yesterday, but no luck.  Maybe something to do with concavity?  Jensen's inequality?  Arithmetic -- geometric mean inequality?  It sure would be nice to get some sleep tonight!

Let me shuttle this note off to you and Marcus before checking further on dimensions.

\section{28-09-06 \ \ {\it Simpler, Simpler, Simplest?}\ \ \ (to H. B. Dang \& D. M. {\Appleby})} \label{Appleby14.6} \label{Dang5}

Furthering on what I wrote below.  I have now further checked that the conjecture is numerically so in $d=9$, 10, 11, and 12.  So, I think we're good to go on the new conjecture of $3d-4$ equations.  I also checked that some further deletions would not help.  For instance, keeping the $(0,0)$ normalization, but getting rid of the 0-column equations otherwise, is no good.  Nor does it help to reinstate one of the 0-column terms.

This may indeed be the minimal set.  Or if you really want to be picky in the accounting (eliminating all known redundancies):

\bq
\noindent If $d$ is odd.  The unadorned total comes to: \medskip

\begin{supertabular}{ll}
$(d-1)/2$    &  from the 0-column \\
$(d-1)/2$    &  from the 1-column \\
$(d-1)/2 - 1$ &  from the 2-column \\
\end{supertabular}\medskip

\noindent for a grand total of $3(d-1)/2 - 1$. \medskip

\noindent If $d$ is even.  The unadorned total comes to: \medskip

\begin{supertabular}{ll}
$d/2$        &  from the 0-column \\
$d/2$        &  from the 1-column \\
$(d-2)/2$    &  from the 2-column \\
\end{supertabular}\medskip

\noindent for a grand total of $3d/2 - 1$.
\eq

If you have Mathematica 5.0 (Marcus I know you don't), and you want to do your own experimentation, here is the code I used:

\noindent ---------------------------------------------------------------
\begin{verbatim}
a[i_] := b[i]+ I c[i]

NN := Sum[Abs[a[Mod[s,d]]]^2,{s,0,d-1}]

CC[j_,n_] := Abs[Sum[Conjugate[a[Mod[s,d]]]*a[Mod[s+j,d]]*
             a[Mod[s-n,d]]*Conjugate[a[Mod[s-n+j,d]]],{s,0,d-1}]]

EC := Abs[NN-1] + Sum[Abs[CC[j,0]-1/(d+1)],{j,1,d-1}] +
      Sum[CC[j,1],{j,1,d-1}] + Sum[CC[j,2],{j,1,d-1}]

d = 7

Do[Print[k] && Print[NMinimize[EC,
   Join[Table[b[i], {i, 0, d-1}],Table[c[i], {i, 0, d-1}]],
   Method -> {"DifferentialEvolution", "RandomSeed" -> k}]],{k,0,24}]

nn = NMinimize[EC,
     Join[Table[b[i], {i, 0, d-1}],Table[c[i], {i, 0, d-1}]],
     Method -> {"DifferentialEvolution", "RandomSeed" -> 3}]

bb = Table[b[i] /. nn[[2, i + 1]], {i, 0, d-1}]

cc = Table[c[i] /. nn[[2, i + d + 1]], {i, 0, d-1}]

aa = bb + I cc

psi[j_] := Conjugate[aa] RotateRight[aa, j]

Table[Conjugate[psi[j]].RotateRight[psi[j], n], {j,1,d},{n, 1, d}]

MatrixForm[%]
\end{verbatim}
---------------------------------------------------------------

The ``DifferentialEvolution'' method works pretty well up to about $d=11$, then it's better to use ``RandomSearch'' as the method.  Do the preliminary part first to find an adequate random seed, then do the rest if you want to check that you're getting a full set of equations.  Or if you want to test properties of the $|\psi_j\rangle$, etc.

\section{28-09-06 \ \ {\it Focus Session Invited Speaker} \ \ (to C. M. {\Caves})} \label{Caves87.1}

\bcc
It really can't be me this year, because I'm in charge of the whole
thing, but when Lorenza runs it next year, I (and maybe you) can hint
around to her that I would be keen to do it.
\ecc

But I really would like it to be you this year.  I took the license to ask you because you yourself had remarked to us:  ``Please don't rule out yourself if you think you're the best choice.''  What is good for the goose is good for the gander.

A) I really do believe the Bayesian point of view of QM is by far the best thing around, and the most poised for progress in all of quantum foundations.  I do not want yet another big mainstream talk endorsing nonlocality---those guys get far too much airplay.  Gisin?  Valentini?  Goldstein?  Even Spekkens?  Come on?

B) I don't want the Bayesian view to be forgotten by the community because of silence over the course of a year.  It's time for me to start giving talks on quantum information proper again; I'm out of the game---not that I would have been a good choice anyway.

C) The whole field needs some gravitas (see, even Howard Burton is not beyond being an influence on my thoughts), and particularly the Bayesian view.  I have failed to give it that.  {\Ruediger} is too terse in his talks to give it that.  And you, my teacher, are the best thing around at the moment.  ``The president-elect endorses foundations so much so, that he has allowed himself to be wrangled into giving a talk on it.''  I would make it clear to everyone that I forced you into this and take the heat.

You may think ``a year's delay is nothin','' but it's everything.  Out of sight, out of mind.  As {\Ruediger} wrote me about Poland:
\brs
    Also with Zeilinger, Gill, Grangier, Christandl.  Overall it
    seems to me that we have made little impact. People still get hung
    up on the point that you can check what the state is by doing
    tomography. So how could it not be objective?

    The experimenters didn't ask or comment. In particular, Zbinden
    (from Gisin's group) didn't seem to be interested. [\ldots]

    It's an uphill struggle!
\ers

I don't want to die having accomplished nothing with regard to the one thing that's important to me.  You give the view gravitas, and because of that people will think about it (if ever so slightly at first).

I would not be bugging you like this if you weren't truly my first choice for the invited speaker of this session.

\section{29-09-06 \ \ {\it Know These Guys?}\ \ \ (to G. L. Comer)} \label{Comer97}

\bgc
I looked at their reference to you.  I tried to figure out if they are
among the two classes of people you've corresponded with: a) those you
haven't pissed off and b) those you have pissed off.  It looks like
b), but I'm not sure.
\egc

Antony sent out a mass mailing to several of us, letting us know about the existence of the new book.  What is their reference to me?  Is it easy to find in there?  Guido and I are on very good terms; I've stayed with him in Freiburg (and had single malt with he and his wife), and he thinks he's contributed to the Bayesian view of QM (though his stuff was in the diametrically opposite direction, and I couldn't help but show my dissatisfaction).  Antony may have wanted to get a dig in at me.  We did have an exchange one night in the Jane Bond Bar where he got awfully aggressive with me.  [See 25-11-03 note ``\myref{Valentini1}{Philosophy Sauce}'' to A. Valentini.]  It led to the following email exchange below.  He had closed the discussion by saying something like, ``The problem with physicists is that they think they can play philosopher.  You---Fuchs---should pass your work by some real philosophers so that they can explain to you how bad it really is.''  That incensed me, as you'll see in the notes below.

\section{29-09-06 \ \ {\it The Lists We're In}\ \ \ (to G. L. Comer)} \label{Comer98}

A while ago you showed me a huge list of acknowledgements you were in because of a conversation with Don Page.  Let me show you a list I just found myself in:
\bq\noindent
   The article quotes the published work of highly reputable physicists
   (e.g.\ Einstein, Hawking, Heisenberg, and Fuchs) on quantum
   indeterminacy. Should the article be suppressed because their views
   are controversial?--Carl Hewitt 21:03, 17 October 2005 (UTC)
\eq

\section{29-09-06 \ \ {\it Lunch Break}\ \ \ (to C. M. {\Caves})} \label{Caves87.2}

Greg Comer brought this to my attention:
\bgc
It's in footnote d on page 166 of
\egc
\bv
\quantph{0609184} [abs, ps, pdf, other] :\\
Title: Quantum Theory at the Crossroads: Reconsidering the 1927 Solvay Conference\\
Authors: Guido Bacciagaluppi, Antony Valentini\\
Comments: 553 pages, 33 figures. Draft of a book, to be published by Cambridge University Press
\ev
(The book contains a full translation of the 1927 Solvay discussions; so it will definitely be widely read.)

Have a look at the footnote and the surrounding discussion on the previous page.  It gives the reason I had wanted you as the focus-session invited speaker {\it just that much more weight}.  It's amazing what kind of logically-loose crap is out there.

I wrote Greg this:
\bq\noindent
   I looked at it quickly, even though I shouldn't be taking the time.
   That argument is just crap.  And people know better about
   Bell---certainly Guido should.  The argument gives that EITHER a)
   locality goes, OR b) that measurement outcomes signify pre-existent
   properties goes (this is usually called the ``assumption of reality'',
   but is not so nearly inclusive as that phrase would indicate), OR c)
   both go.
\eq

The thing that impressed me about your talk at Scully-ville was that it was an attempt to set that right.  Also, I very much like the look of the talk you have posted.  Taking elements from both would make a real winner.

\section{05-10-06 \ \ {\it So Whiskey Tango Foxtrot?}\ \ \ (to M. D. Sanders)} \label{SandersMD2}

\bmds
\begin{center}
\myurl[http://web.archive.org/web/20061005192320/http://www.cnn.com/2006/TECH/science/10/04/teleportation.reut/index.html]{http://web.archive.org/web/20061005192320/ \\ http://www.cnn.com/2006/TECH/science/10/04/teleportation.reut/index.html}
\end{center}
Is this really going on?  This is kind of freaking me out.......but it is way cool.....
\emds

Yes, it's real.  But it's not nearly as exciting as the newspaper makes it sound.  ``Teleportation'' in the quantum information sense isn't so very much like the Star Trek version as the press always wants to portray it.  It's not about getting things from here to there without going in between, but about making my information stop being about this and start being about that without being about anything else in between.

Let me tell you a funny story about Polzik.  He's a guy who sought asylum in the US after escaping from Russia during the Cold War.  In 1998 we were roommates at a conference at Northwestern University (Evanston, IL).  We stayed in a dorm room.  The last night of the meeting around bed time---I had an early flight the next morning---Polzik said, ``I'm gonna go have a swim in the lake,'' while putting on his swim suit.  ``Do you want to come with me?''  I told him about my early flight and declined.  He grabbed a bottle of vodka and went out the door.  I got in bed, turned out the lights, etc.  About a half hour later, Polzik was back in the room pulling pants over his still dry swimming suit and muttering, ``No policeman is going to tell me I can't have a swim.''  I said, ``What's up?''  He said, ``I'm gonna get to the lake one way or other.''  I have no idea what he meant by that or what his plan was.  But I do know the next morning I found a wet swimsuit in the bathroom and the room reeked of vodka.

That was the same year we had the teleportation paper at Caltech, teleporting from light field to light field; Polzik was one of the authors with us.  (He had leant us a student who did much of the actual work.)

\section{05-10-06 \ \ {\it How I Sold Us} \ \ (to D. M. {\Appleby} \& H. B. Dang)} \label{Appleby15} \label{Dang6}

FYI, here are the words I used to try to sell our project with my superiors.  Here's the section on the subject I wrote for this year's performance report.  \ldots\ I deposit it here, just in case you too would like to know why you're working on this blasted problem!

\bq
The main portion of my research this year has been on using new techniques to tackle a 33-year-old open mathematical problem, and collaborators and I have made significant headway toward its solution---so much so that we think we are far ahead of the game with respect to anyone else thinking about it.  Most importantly, by way of this work, I believe I have gained my deepest understanding yet of the structure of quantum information.  The final goal is still a little out of reach (and we have felt compelled to hold back on publication of intermediate results), but it is a very exciting time.

The motivation for this particular problem is to find a useful way of formulating finite-dimensional quantum mechanics so that density operators are at the center of its considerations, rather than state vectors.  A technical reason for this is for efficiency in quantum information calculations, where density operators always play a prominent role in noisy, real-world situations; a fundamental reason for it has to do with its easier connection to the idea that quantum states represent information, rather than intrinsic properties of quantum systems---a prime component in my research for many years now.  For a $d$-level system (with a $d$-dimensional Hilbert space, a so-called {\it qudit\/}), its density operators live in a $d^2$-dimensional vector space of Hermitian operators.  In fact, they live in the positive cone of that vector space, i.e., the set of operators with non-negative eigenvalues.  An example of what I mean by bringing density operators to the center of considerations, is, for instance, that one might like to know how to formulate a superposition principle using elements solely within the cone.  For such a project, one should have an analog of an orthonormal basis of state vectors within the cone.

This summer, with my intern Hoan Dang, we were able to formulate this question in a more interesting way than had been done previously, and the answer was surprising.  Since there cannot be an orthonormal basis of $d^2$ operators in the cone, one might ask for a set of operators that, on average, are as close to being orthonormal as possible.  We found a rigorous bound on the average orthonormality and also the necessary and sufficient conditions for achieving that bound.  To my great pleasure, the conditions were ones I had seen before, but in fairly disparate work---particularly, my work on optimal alphabets of quantum states for sensitivity to eavesdropping in quantum cryptography.  This connection is something I had suspected, or more accurately hoped for, for some time, but I did not know how to formulate it very well or make it rigorous.  (In fact, I mentioned a preliminary result along these lines in last year's Form 1, but it was significantly weaker than the present one.  In the older version, the symmetry was built in by hand, rather than a result of the calculation.)  In the usual formulation of the superposition principle, the one at the state-vector level, basis vectors represent the {\it most classical\/} properties---ones that can, for instance, be cloned or eavesdropped upon without disturbance.  But in the density-operator formulation of a superposition principle, we are led to using bases of states that are as {\it quantum\/} as they can be---the ones that are most easily perturbed by their environment.  The reason I am very pleased about this is that it gives me a tool for exploring my pet idea of what powers quantum information and computing:  It is nothing to do with parallel universes or action-at-a-distance, or anything as science-fictiony as that, but simply quantum systems' greater sensitivity to external stimulus than classical systems.

The condition that must be satisfied for such a beautiful symmetry to be true is that one must be able to find a set of $d^2$ pure density operators $\Pi_i=|\psi_i\rangle\langle\psi_i|$ such that $|\langle\psi_i|\psi_i\rangle|^2=1/(d+1)$ when $i\ne j$.  But does such a set of states exist?  It is not obvious, and in fact the question is equivalent to the 33-year-old mathematical problem alluded to above---the problem of the ``maximal number of equiangular lines''---which had its origin in classical coding theory.  Furthermore, independently of the things reported here, the question of the existence of these states had already captured the interest of the quantum information community because of the set's intrinsic symmetry.  The working conjecture of several groups, and the one we are on the tail of, is that the operators $\Pi_i$ can always be generated by taking certain orbits of the standard set of unitary operators used in quantum error correction, the generalized Pauli operators.   One thing that is remarkable about representing quantum states in terms of the $\Pi_i$ is that then a general quantum error would be nothing more than an unknown permutation of indices---there are no longer any distinctions to be drawn between phase-flip vs.\ bit-flip errors, etc.   Thus, viewing quantum errors in this way may open a further book of ways in which powerful classical error correction techniques can be adapted to the quantum domain.  For all these reasons, proving the existence of these operators and giving a technique for actually constructing them would be a great achievement.

What has made my collaboration with Hoan Dang (now at Princeton) and {\Marcus} {\Appleby} (at Queen Mary) special in comparison to the rest of the pack considering the existence problem, is that we have opted for a basic---but steady---linear algebraic approach, rather than invoking abstract group theory and abstract number theory, as has been the fashion in most of the literature so far.  We are banking that the problem does not really call for all that, and that these high-level approaches have clouded more than they revealed.  For instance, until our work, the best numerical technique for checking for the existence of such a symmetric set scaled as a sum of $d^{12}$ terms; we have been able to reduce that to a sum that scales as $d^2$.  Concomitant with this, and more interesting actually, we have been able to reduce the exact problem to finding a single state vector, whose components must satisfy (roughly) $3d/2$ simultaneous quartic equations (of a very particular structure).  We feel that we are not all that far from a complete solution to this problem.  Personally, I feel this is some of my best work yet in quantum information theory.

What will come next is to ask in much greater detail what insight this kind of representation can give to quantum computing.  On two fronts I hope to tackle this directly; one through a collaboration with Lov Grover in our own department, and one through a collaboration with Al Aho, a Bell Labs alumnus who continues to come into the lab often.  Both of these collaborations are just starting up.
\eq

\section{10-10-06 \ \ {\it Phone Tag} \ \ (to J. E. {\Sipe})} \label{Sipe13}

PS. Carl Caves is visiting me for a couple of days at the moment, and we're being forced to confront all our sins and indiscretions in this quantum Bayesian business.  Not a pleasant way to spend one's time.

\section{13-10-06 \ \ {\it Small Thanks}\ \ \ (to C. M. {\Caves})} \label{Caves87.3}

Reviewing the week, I kind of doubt I expressed adequately how much I appreciated your visit here.  Particularly, it was so good working with you again.  The tension makes it a real challenge, and a real pleasure when there is finally some resolution; it was good to feel progress in the Bayesian byways again.  And, mostly, it gave me hope that the best in the subject is still waiting to happen.

\section{13-10-06 \ \ {\it Rabid Pragmatists (but in the sense of W. {\James}, J. {\Dewey}, and C. Fuchs)} \ \ (to R. D. Gill)} \label{Gill5}

Thanks for calling my attention to your updates.  I am impressed by
how much we both seem to be attracted to the same ideas.  (Of course,
at this level of description---like the one found in your
presentation---it is always easy to agree or to disagree.  But I
genuinely think there are similarities in our ontologies.)

\brdg
Detector clicks in the past are real.
\erdg
Yes.
\brdg
The future is a wave of possibilities.
\erdg
Yes.
\brdg
As now moves relentlessly forward the past crystallizes out of it
randomly.
\erdg
Yes.
\brdg
There is no nonlocality problem, no {\Schroedinger} cat problem, no
measurement problem (time is measurement).
\erdg
Yes.

The only place where we significantly disagree is on the issue of
probability in quantum mechanics.  I too---i.e., like you---believe
that quantum mechanics gives every indication of an {\it ontic\/}
indeterminism or randomness in our world.  I just part company with
you in thinking that the concept of an {\it ontic probability
function\/} sheds any light on this wonderful feature/fact of nature.
In fact, I think it does just the opposite: Holding onto the idea of
ontic probabilities---i.e., trying to make it make sense---only
clouds our vision and gets in the way of proper description of this
feature of nature.

In the note below (which I had originally written to a Princeton
physics professor in one of my own more rabid moments), there are a
few more details about what I actually mean when I speak of {\it
ontic indeterminism\/} and {\it subjective quantum probabilities\/}
in the same breath.  The two ideas can coexist.  (By the way, that
guy really pissed me off.  You never piss me off, so I wouldn't write
you a rude note like that.  But I forward the note to you because I
do like the formulation I gave to the issue you and I happen to be
talking about now.)

Tell me ultimately (i.e., as you get a chance), what you thought of
the details of G.~H. {\Mead}'s ideas.

\section{14-10-06 \ \ {\it Pragmatism and the Weather, 1} \ \ (to R. D. Gill)} \label{Gill6}

\brdg
or do you mean pragmaticist?
\myurl{http://en.wikipedia.org/wiki/Pragmaticism}
\erdg

No, I really meant the strain of pragmatism from William {\James} and John {\Dewey} (and an admixture of F.~C.~S. {\Schiller}).  {\Peirce}'s pragmaticism is quite a different thing, and not to my taste.

Talk about striking deep chords, I really liked this remark of yours:
\brdg
I have this idea that if intelligent life would arise on Jupiter and
develop mathematics then the most fundamental and deep mathematics
would be probability theory (since Jupiter is basically weather and
nothing else). The pure mathematicians would do statistics. The
applied mathematicians would do number theory. It is just a cultural
accident that probability is an embarrassingly tricky add-on to
mathematics. It could have been the other way round.
\erdg

That is a beautiful idea!  Danny Greenberger once wrote me an eloquent little passage about how it might have been the case that the real numbers would have been taken as mathematically primary (over the integers) if evolutionary pressures had been different---I'll include that passage below [see 06-12-02 note ``\myref{Greenberger1}{Enjoyed}'' to D. M. Greenberger]---but I like your idea even better!  And, in any case, you might not need Jupiter to pull it off.  It may already be the case with us here on earth; we've just strayed a little from the natural progression of ideas in the last few centuries.

I'll be back in a couple of minutes after I dig up another passage.

\section{14-10-06 \ \ {\it Pragmatism and the Weather, 2} \ \ (to R. D. Gill)} \label{Gill7}

The other thing I had wanted to send you was a passage in my computer about the philosopher Chauncey Wright.  He was a friend of James and Peirce and had been deeply influential on them (before ending his life early).  The reason I send you this is because of your remark that ``Jupiter is basically weather and nothing else.''  Wright was famous for saying on many opportunities, ``The universe is only weather.''  The passage below gives some flavor of that.

\bq
[Chauncey] Wright did not consider himself an evolutionist. To him the term denoted a belief that the world was getting, on some definition, ``better.'' His loyalty was only to the theory of natural selection, which he thought corresponded perfectly to his notion of life as weather. ``[T]he principle of the theory of Natural Selection is taught in the discourse of Jesus with Nicodemus the Pharisee,'' he explained in a letter to Charles Norton's sister Grace. The allusion may be a little gnomic today. The discourse with Nicodemus is in the Gospel of John, and the words of Jesus Wright was referring to are these: ``The wind bloweth where it listeth, and thou hearest the sound thereof, but canst not tell whence it cometh, and whither it
goeth: so is every one that is born of the Spirit.'' Wright was, in short, one of the few nineteenth-century Darwinians who thought like Darwin---one of the few evolutionists who did not associate evolutionary change with progress. ``Never use the word[s] higher \& lower,'' Darwin scribbled in the margins of his copy of the {\sl Vestiges of the Natural History of Creation\/} in 1847. The advice proved almost impossible to follow to the letter, even for Darwin, but if anyone respected its spirit, it was Chauncey Wright.

Wright's particular b\^ete noire was the evolutionist Herbert Spencer, whose work seemed to him a flagrant violation of the separation of science and metaphysics. ``Mr.\ Spencer,'' as he declared, ``is not a positivist.'' Spencer's mistake was to treat the concepts of science, which are merely tools of inquiry, as though they were realities of nature. The theory of natural selection, for example, posits continuity in the sequence of natural phenomena (evolution does not proceed by leaps). But ``continuity'' is simply a verbal handle we attach to a bundle of empirical observations. It is not something that actually exists in nature. Spencer failed to understand this, and he therefore imputed cosmic reality to what are just conceptual inferences---just words. He did with the word ``evolution'' what Agassiz did with the word ``creation'': he erected an idol.

``Mr.\ Spencer's philosophy contemplates the universe in its totality as having an intelligible order, a relation of beginning and end---a development,'' Wright said. But the universe is only weather.
\bq\noindent
Everything out of the mind is a product, the result of some process.
Nothing is exempt from change. Worlds are formed and dissipated.
Races of organic beings grow up like their constituent individual members, and disappear like these. Nothing shows a trace of an original, immutable nature, except the unchangeable laws of change.
These point to no beginning and to no end in time, nor to any bounds in space. All indications to the contrary in the results of physical research are clearly traceable to imperfections in our present knowledge of all the laws of change, and to that disposition to cosmological speculations which still prevails even in science.
\eq
``No {\it real\/} fate or necessity is indeed manifested anywhere in the universe,'' he wrote to a friend, ``---only a phenomenal regularity.''
\eq

\section{14-10-06 \ \ {\it That Pragmatism Crap}\ \ \ (to C. M. {\Caves})} \label{Caves87.4}

I just sent the note below to Richard Gill for other purposes, but it brought me back to your remark the other day.  [See 14-10-06 note ``\myref{Gill6}{Pragmatism and the Weather, 1}'' to R. D. Gill.] Particularly, I would say that what Greenberger describes as his vision of what science is (or is about) pretty much captures what I think of the subject now.  I guess I've always been intrigued by the idea that our world has ``Darwinism all the way to its roots''---that's why I was first attracted to John Wheeler's ``law without law.''  Your being my graduate advisor was a follow-on to that.

But, as I understood you the other day, you seemed to say that science is ``more than that'' or, at least, something else than that.  If you ever feel like wasting a little time composing a few words to get it on the record, I would like to better understand why you say that.  Particularly, what do you think I'm missing with this view of science?

\section{16-10-06 \ \ {\it Paper} \ \ (to C. M. {\Caves})} \label{Caves88}

\bcc
Chris, I have changed ``prompting'' to ``eliciting''.  I think it
captures much better what you have in mind, but won't mind going back
to ``prompting''.
\ecc
Just noticed this PS, which I hadn't noticed before.  Neither word, I think, in the end is completely adequate, but I don't know what's better at the moment.  Elicit is fine.  And you're probably right, it is closer to the mark at the moment.

\section{16-10-06 \ \ {\it Our Professor} \ \ (to R. {\Schack})} \label{Schack107}

\bcc
I think we ought to be thinking about a paper that lays out our
entire approach, particularly, the idea Chris and I discussed of a
three-pronged approach to subjectivity and objectivity: (i) quantum
states and probabilities are wholly subjective; (ii) system
attributes are wholly objective; and (iii) measurement outcomes are
where the rubber meets the road, i.e., where subjective and
objective meet to produce something that is not under the control of
the agent, but is also not out there in the world.
\ecc

This paragraph of {\Carl}'s is one that took me absolutely by shock.  I wish you could have been here for our first day of interaction, when {\Carl} was still in his mode of saying how I've been ``muddying things beyond belief.''  In fact, it all started with him going into a soliloquy his first day at the office.  ``I want to get one thing straight:  With this business about `fact for the agent', I'm not going there.  Even if my position is not consistent and that one is, {\it I'm not going there}.''  That, I'm fairly sure, is an almost verbatim record of what was said.  Then he outlined his story of ``facts'' being those things that are out there for all to see, and that {\it that's\/} what we've been talking about in this paper all along \ldots\ yada, yada, yada.

It was hard to get past it, but I held my ground.  I told him that you and I had both moved beyond that kind of understanding of ``a fact.'' (I used some of your emails to bolster the point.) I explained how he would always be forced into a corner by Finkelstein with his present view, and similarly be absolutely stuck by the time he ever got to Wigner's friend.

It was a priceless scene actually, him rubbing his temples, going silent, looking extremely annoyed, rubbing his forehead, and then cutting off all conversation.  The next day he woke up a fairly different man, saying, ``We're going to make some progress on this paper today.''  I said, ``Oh no,'' with complete dread.  He said, ``No, you'll be pleased.''  So, all of that was pretty cool.  But then to see the proposal above was a complete shock.

I hope these new duties at your university aren't killing you too much:  We need you man!

\section{17-10-06 \ \ {\it The Book Proposal} \ \ (to D. M. {\Appleby})} \label{Appleby16}

I've finally read your book proposal.  The words in it are nice, but I'm not sure that I got any insight toward answering these questions of yours: [\ldots]

With regard to the content, there were a couple of parts that I marked to look at a second time.

1) I enjoyed the two mid paragraphs on page 8, the ones about how ``It may seem \ldots\ as though the epistemic interpretation must lead \ldots\ to a depressingly positivistic view in which physics begins and ends with the task of predicting detector clicks.  For a long time I had the same fear.''  That is a point, as you know, that is near and dear to my heart.  It is a perception that I try to fend off at every point, and at every talk I give, to only small success.  I'm out of inspiration for how to make the point better, but it looks like you're not.  Therefore, I think it'd be great if you were to expand those paragraphs even further.  Find a way to really drive the point home.

2) The lines on page 9 that read, ``Calling the statement epistemic only means that there is not, in addition to such features as the fact that one of the wings is on the verge of falling off, an additional mind-independent physical property called `propensity for crashing'.''  This is because it reminded me of a conversation I had with Jos Uffink at our Konstanz meeting.  I remember saying something like, ``Why is it not good enough that the realm of the real world is already clearly marked out in a subjectivistic conception of probability?  It is marked out in the facts, the clicks, the outcomes, which are distinct from the probabilities.  The facts do not depend upon the agent.  Someone once said, the world is everything that is the case---and there it is; it's in the form of the facts that the subjective degrees of belief happen to be about.  But it seems that you, and all objectivists for probability, want more than that.  It's as if you want a soul that lives behind the facts.  The facts as they occur in nature are too secular, they're not good enough; they would be nothing if they didn't have a spiritual aspect that gives rise to them.  It's as if you want a soul to be responsible for them.''  Jos simply replied, ``That's right.''

\section{18-10-06 \ \ {\it Real Possibility} \ \ (to A. Shimony)} \label{Shimony11}

I was thinking about you the other day as I was writing Richard Gill
for another purpose.  I'll include that note below, which daisy
chains to still another one (written to Kirk McDonald).  [See 13-10-06 note ``\myref{Gill5}{Rabid Pragmatists (but in the sense of
W. {\James}, J. {\Dewey}, and C. Fuchs)}'' to R. D. Gill and 30-01-06 note ``\myref{McDonald6}{Island of Misfit Toys}'' to K. T. McDonald.] The subject
matter is, in essence, the idea of ``real possibility'' and quantum
mechanics' indication of it as a feature of our universe.  I found
myself thinking back on things you've said on the subject over the
years and, in particular, during the Perimeter Institute meeting this
summer.

In some ways, I feel there's really not that large a gap in our
opinions about this pretty idea that ``the universe is on the make''
and in the directions we think physics should turn to to get a better
grasp on it.  We are both intrigued by the category of potentia.  I
know that I think (some flavor of) the conception is the greatest
discovery in physics, certainly in the last 90 years, and maybe ever.
But there is this one thing that keeps getting in our way of
effectively communicating to each other and sharing our reverence for
it---it is our feeling about quantum mechanical probabilities.
Bayesian of a stripe though we both are, we certainly have different
feelings for the meaning of probabilities that come from quantum
mechanical pure states.

Thus I started to think that writing a careful statement of my view,
and the arguments I use to support it, might be a worthy contribution
to the festschrift Wayne Myrvold is organizing for you.  In fact, the
more I think about the idea, the more I am excited about it.

How can one have ``chance'' without ``objective probabilities''?
That's what the note to Gill (actually much more so the note to
McDonald) starts to address.  For me, chance is the primary ontic
category and, as far as I can tell, it cannot be quantified.  Quantum
mechanics, and the probability relations it suggests, from my view,
is a tool---a wonderful tool---for helping {\it physical\/} agents
swim in this world of chance.  The statement that we live in a world
of chance or real possibility is indeed represented in the very
structure of the quantum formalism, but as a kind empirical addition
to de Finettian / Ramseyan coherence.  Not in the specification of
some kind of objective probabilities.

That's the idea, anyway, that I think I would like to expand upon.
The groundwork has already been laid in some other things I've
written---1) the note below, 2) a paper with {\Caves} and {\Schack} that I'll be able to send you a draft of in a couple of days, and 3) a
sort of pseudo-paper (samizdat actually) that I'll attach to the
present email.  Particularly the last item goes into much more detail
about what I mean by ``objective indeterminism.''

I thought it might be fun to start to get some feedback from you before I draw up the plans for the new paper in too great of detail.  Your objections and queries could be part of the very structure and help me tailor my contribution to the things you find confusing or untenable.  Anyway, that's my thought and I'd dearly love to get some feedback from you.

It was great seeing you again this year, and for me, the Perimeter
Meeting was quite the most interesting conference I went to in some
time.  I'm very happy it happened (as it might not have in this world
of real possibility).

\section{18-10-06 \ \ {\it Back From My Long Silence} \ \ (to W. C. Myrvold and W. L. Harper)} \label{Myrvold5} \label{Harper2}

This note is way, way, way overdue, but I have not forgotten about your kindness to me, and I wanted to thank you both for giving me feedback when I was stressed over my presentations and presentation style just after the AbnerFest.  Both of your comments were very useful and helped me ``make it through the night,'' so to speak, during that tough time.  The reason this note is coming to you so late is that I got another comment from the fellow who had taken me aside at Perimeter, just after Wayne's writing and a little before Bill's leaving a message on my answering machine.  Basically it was just an email version of what he had already said, but it caused me to feel even worse than I had before, and I decided to simply shut down for a while on the issue.  (He is a physicist whose opinion I respect more than about anyone else's, and I got very confused about myself for a while.)  Thus my philosophizing, my usual samizdating, and my emailing fell by the wayside for the summer, as I became a hermit and plunged myself deep into calculation like I had not done in quite some time.  It was a form of crisis management.  Anyway the fruits of the tactic were definitely good---I'll present them at the QCMC meeting Tokyo this Fall (in a different style!\ \ldots\ I say facetiously)---and it allowed me to distance myself from the hurt.  But it's time now for me to face the issues I left behind, and more importantly, to become a little more balanced in my research and thought again.

You two helped me see that my work and the choices I've made for trying to convey it were not made in vain, even though at the same time there is probably a lot of wisdom in what my old friend said \ldots\ at least for particular audiences.  All I can do is keep trying, making sure to try to take into account each datum that comes my way.

Concerning the points where we left off:
\bwm
I think that there is less disagreement between my view and yours
than might appear.  (I can explain this more if you'd like).
\ewm

Yes, I would like you to explain this more as you get a chance.  I am curious (now that I've come back to being thoughtful again!).  Surely, I know that we agree on some things.  For instance, I think we would both say that quantum mechanics is a story of a world with inherent chance.  But certainly there are two sticking points that always get in the way of our communication:  1) objective, {\it quantified\/} chance (i.e., the Lewisian stuff), and 2) nonlocality.  As I understand you, you accept both.  As I understand myself, I reject both.  So there is some nontrivial distance between our views.  But if an open universe is the more basic of the things we are both hoping to get a real handle on---i.e., the bigger goal---it is certainly worth the attempt to open up a dialogue, and I'd welcome better understanding what you mean.

It's probably not a coincidence that I happened to write all the stuff below to Abner this morning:  These particular issues are on my mind again.  [See 11-08-06 note titled ``\myref{Shimony11}{Real Possibility}'' to A. Shimony.]  And, actually, all this concerns you in two ways.  As you'll see from my proposal to Abner, I think this subject will be the topic of my contribution to the festschrift you're editing.

Bill said,
\bwlh
At our discussions I suggested that you might find my introduction to
and many of the papers in the volume Stalnaker, Pearce and I edited.
Here is the reference: Harper, Stalnaker, Pearce, eds.\ {\bf IFS: conditionals, belief, decision,
chance and time}, D. Reidell Publishing Company, 1980.
I think we have many informative things to talk about.
\ewlh

As luck would have it, I just found a copy at Barnes and Noble online for only \$42.00!  The receipt says I should have it in my mailbox in 3 to 6 days.

Best wishes to both of you, and I apologize again for my long silence.

\section{20-10-06 \ \ {\it Slammed by the Closet Door} \ \ (to N. D. {\Mermin})} \label{Mermin123}

\label{Phish1}

\bdm
My point was simply that you badly undersell yourself and what you're doing in your talks \ldots\ \ But jokes and cartoons and metaphors and lack of focus on specific
controversial points (is ``quantum nonlocality'' nonsense?\ is there a
``measurement problem''?)\ won't convey this to anybody.  I mean you knew Leggett and Pearle, both of whom think the measurement problem is the most important unsolved problem in physics, would be there.
\edm

I knew that Abner would be there.

And more than anything I wanted to make a talk that somehow touched base with Jamesian pragmatism for him.  In the end, I failed my own standards (and I was aware of that).  But I tried in the off hours of that interesting meeting to get it all into a self-contained package.  You can't see it, but I actually worked very hard at that talk---much more than I have for any talk in quite a while.  You had seen everything before?  I should burn the transparencies and start completely over?  Maybe.  But, in fact there were 26 transparencies in there that were new; I just counted them again.  You had never seen them before.

Leggett and Pearle?  They don't mean as much to me as they mean to you.  I know the protest you will surely give me over what I am about to say, but they are ideologues.  They will carry the prejudice against epistemic quantum states with them to the grave.  Bill Clinton said it very nicely in a speech a couple of days ago, ``If you've got an ideology, you've already got your mind made up.  You know all the answers and that makes evidence irrelevant and arguments a waste of time.''  They may be great scientists when it comes to other subjects, but with regard to this one subject, they are ideologues.  The evidence is overwhelming that quantum states live in the mind rather than in the physical world (leaving aside, for the moment, the {\it detail\/} of whether they are knowledge, credence, information, belief, or whatever variation of such a kind of thing).  The {\it only\/} thing those guys have motivating their research in the other direction---the direction that gives rise to a ``measurement problem'' and ``nonlocality''---is the religion that a fundamental physical theory should not be IN PART about things to do with minds in the sense above.  That is their one motivator.  But the scale is completely imbalanced toward the other direction:  It screams out in almost every piece of the phenomena of quantum information---the no-cloning theorem, teleportation, the ``go ask Alice'' point, and so on.  And yet they don't see it.  (And while I'm here, you, with all your admirable open-mindedness and nonprejudice, don't see it either.)  The only thing that is really an open issue is getting a better answer to your question, ``Information about what?''  This was not the first time Leggett or Pearle were at one of my talks, or instead, one of Peres's, or one of {\Schack}'s, or Pitowsky's, or probably in the long past, one of John Wheeler's.  Tony has been to at least five or six of my talks, and in particular was at the Seven Pines meeting in Minnesota for a week where the epistemic view of quantum states got the vast majority of the discussion.  These guys are not ones that can be reached out to on this issue, my involvement or not.  Where I can hope to make an impact is, as Zeilinger said, with the young ones.  I may not be doing it successfully, but that's the only place I can hope to do it.

That aside, though, I did {\it try\/} to address directly those very questions you raise, and I did it with concrete examples too---think of the classical no-cloning theorem example I gave and the example of teleporting a probability distribution for the outcomes of a coin-toss.  Even YOU don't see the force of those examples (I could hear it in your question that day), and because of that, they don't seem like concrete examples to you at all.  But they carry the whole story.  There is absolutely nothing inherently quantum mechanical about quantum teleportation.  The only thing that is quantum mechanical is the answer to the question, ``Information about what?''  And that question is well-posed without anything to do quantum teleportation.  That was the point.

It was that IF one can just get one's head around the idea that quantum states are epistemic, then there is no measurement problem and there is no need to talk of nonlocality.  Peirce wrote this to James in 1904, ``[P]ragmatism solves no real problem.  It only shows that supposed problems are not real problems.''  And so too, this is the reason for adopting an epistemic conception of quantum states AS A STARTING POINT.  It is to say, there never was a measurement problem.

You tell me I should have addressed these things.  But I tried.  I used the wrong language for some of you, but I tried.

\section{22-10-06 \ \ {\it Wishful Thinking} \ \ (to C. M. {\Caves})} \label{Caves89}

Well it was wishful thinking on my part.  It's Sunday now, and I haven't done any tinkering on the paper; just looked at the draft and the comments.  Birthday and NY City took more of my attention away than I thought it would.  And now {\Appleby} is arriving today.  So, let me unlock the paper, so that you can return to tinkering on it if you wish.  I much more liked our method of working on it together; too bad we didn't have just one or two more days of that.

I had a very thoughtful day in NY (they don't come very often to me any more), and I drew up all kinds of grand plans for how best to present the program \ldots\ to philosophers at least.  I think the way I'll tackle it now is to start the presentation with all ontological sounding stuff---objects, events, indeterminism---get that established, and then throw in agents and subjective probabilities.  Traditionally I always started with subjective quantum states first, how that alleviates the measurement problem, etc., and then worked backward toward an ontology.  I sketched out a few pages of notes on the train and in coffee shops.  At least right now, I'm decently excited about this, {\it dreaming\/} that maybe it will stave off the usual sorts of fears and confusions, like Finkelstein's and everyone else's.  One thing I might try to do before Christmas with this is give a seminar at Princeton and see how it flies; another thing is I might try is to make it the format of my AbnerFest contribution.  If it looks after those two preliminaries to be on the right track, I may lobby for that with you and {\Ruediger} as the right way to go for our next joint venture (where surely all the language will be toned down from my preliminaries).

Thinking how 42 is supposed to be the answer to life, the universe, and everything, I've decided I want to try to have a different kind of 42nd year than Lewis Carroll and {\Ruediger} {\Schack}.  That's probably the most important result of my last three days of thought.

\section{25-10-06 \ \ {\it Maximum Uncertainty States}\ \ \ (to H. B. Dang)} \label{Dang7}

It seems the price of SICs just went up again today!  (Remember how you would say that every time we find a new question for which SICs are the answer, their price goes up.)  Well, remember how I had speculated how the SICs shared some features with coherent states, and therefore might in some sense be ``minimal uncertainty states'' for measurements with respect to a complete set of mutually unbiased bases?  Marcus and I finally drudged through the calculation today, and we found that I was just about right:  They're not the minimal-uncertainty but the maximal-uncertainty states!\footnote{Ha!  We soon learned from Bill Wootters that we had made a minus sign mistake!  They were minimal uncertainty states after all.  See D. M. Appleby, H. B. Dang, and C. A. Fuchs, ``Symmetric Informationally-Complete Quantum States as Analogues to Orthonormal Bases and Minimum-Uncertainty States,'' \arxiv{0707.2071v2}$\,$.}  (At least when $d =$ prime, this is true.)  It's been a very exciting day.  At the moment, we're trying to see if the SICs are the {\it only\/} maximal uncertainty states; for that we don't have an answer yet.

We still have not been able to figure out why the $d^2$ equations can apparently be reduced to only $\sim 3d/2$.  That would probably be our main topic of conversation, to see if we can make any progress there.

\section{28-10-06 \ \ {\it Please Email Abstract}\ \ \ (to K. H. Knuth)} \label{Knuth6}

This is a minor modification of the abstract I used for the Applied Maths colloquium at Princeton---making it look a little less mathematical and a little more physicsy, but I hope it will do.

\bq\noindent
Title: Math Problems from the Far Side of Quantum Information \medskip\\
Abstract: The field of Quantum Information has recently attracted great interest for the technological fruits it may bear.  But there is a sect of its practitioners who think it stands a chance to have an even greater legacy than that---namely, that its theoretical tools will give us a means for exploring what quantum mechanics is really all about and for settling some of the deepest problems in physics.  The roots of this optimism come from a very old thought:  that quantum states have more to do with representing their users' information than any inherent physical properties of the systems they describe.  What is new and nice is that quantum information teaches us how to formulate this idea precisely and even check its consistency.  Nicer still is the number of juicy mathematical problems the consistency-checking process poses.  In this talk, I will review some of the history of this and then quickly settle on a sample problem that has been annoying me a lot lately (the existence of a very symmetric, informative quantum measurement) and discuss the insights its existence would give for the structure of quantum mechanics as a whole.
\eq

\section{01-11-06 \ \ {\it Plunging Ahead} \ \ (to W. C. Myrvold)} \label{Myrvold6}

I never heard back from you concerning my proposal for an Abnerfest contribution.  Anyway, I'm plunging ahead.  Please let me know the last date you will be willing to accept the manuscript.

You know I always start my thinking with a slogan.  Here's the slogan for this paper (which I'll also send to Bill H.\index{Harper, Bill} in another note; I want to let him know I just got a copy of ``Ifs'').  The flavor of quantum Bayesianism that {\Caves}, {\Schack} and I are trying to establish concerns in great part the following thought:
\bq\noindent
    The {\it formal structure\/} of quantum mechanics represents an
    \underline{empirical addition} to Ramsey / de Finetti coherence, not an empirical addition to the setting of priors.
\eq
That, in a single sentence, is an attempt to capture the technical part of what I sent you last time.

I hope you're having fun at the PSA meeting.  I really regret having to miss it this year.

\section{01-11-06 \ \ {\it SBT, the Reprise} \ \ (to H. C. von Baeyer)} \label{Baeyer25}

\bhcvb
So, to cover up my guilt, I'll go on the attack. I'm reading {\Caves},
Fuchs, and {\Schack} on quantum certainty, and thinking that I'm getting
the drift, when I come up to an example -- which is my way of
understanding -- and whammo, on page 10 you hit me with 33 states,
which are (the cute little ``of course'' rubs salt in the wound)
connected by 96 rotations and one shoulder separation.  I can't
envision that.  Can't you use a simpler manifestation of KS like
{\Mermin}'s very cogent $3\times3$ matrix, or maybe it was $4\times4$?  As popularizer I'm always on the lookout for simpler versions.  (I remember that when GHZ first came out in about 1991 it was such an improvement over EPR that I wrote an article about it which came out so well that even Anton requested a reprint just this Fall.)
\ehcvb

Well, that's just the tip of the iceberg of my own troubles with that paper.  {\Ruediger} and {\Carl} had convinced me that we should try to write it in a ``metaphysically neutral'' manner---saying the things the authors could all agree on, without too greatly exploring the meaning of the words ``fact'' and ``truth'' in the quantum domain (a place where we ourselves we had significant differences).  But that was definitely the wrong thing to do:  For it is precisely the place where the {\it four\/} referees picked on us the most.  Clearly some clarity and commitment was called for!  (By the way, of the reports, two thought it was an outstanding paper (though needing plenty of revisions), one was somewhat neutral but generally positive and also asked for revisions, and one thought it was so hopeless we should simply scrap it at the outset.  The thing that tickled me most about the last report is that I had never before imagined that I could be wrong in so distinctly many ways.)

Anyway, I say this to warn you that I think a much better version of the paper is coming together now, and you shouldn't think too deeply about the present version.  We intend to have the new version posted before December.

About the section you mention, one of our referees even said this:
\bq\noindent
There is of course nothing wrong with the proof in section V, but
at this point in the history of the philosophy of quantum theory,
I confess that I don't see why it is there. Isn't a quick reference to the 10,000 other proofs of nonlocality sufficient? If not, can
you say what is special or new about this one? I would personally
rather see that space used for a more subtle discussion of the
newer issues, perhaps addressing some of the points that I've made
above.
\eq
In a way, he has a point; that section may bring us more trouble than it's worth.  Its main point was really to say only that neither 1) Kochen--Specker, nor 2) Bell inequalities alone, lead to conclusive results with regard to the question of the pre-existence of ``with-certainty'' measurement outcomes.  However, the conjunction of locality + KS (in the way of {\Stairs}) does.  To make that point, not all (or almost any) of the construction needed to be shown.  References alone might well have done the trick.  In any case, I'll share your concerns with my coauthors and treat your remark as a fifth referee report.

\section{08-11-06 \ \ {\it November 8th} \ \ (to N. D. {\Mermin})} \label{Mermin124}

\label{Phish2}

I apologize for my long delay in replying to you; this time it was not intentional.  First I had {\Appleby} here and we were completely single-minded on our simultaneous-quartic-equations problem to the exclusion of all else.  Then I had the misfortune of getting a very bad case of food poisoning (cold left-overs from Taco Bell) which sent me to Emergency Care and knocked me out for several days.  And, finally, yesterday was Election Day and I could not tear myself away from the television!

But like America as a whole on this glorious November 8th, I am recovering.

\bdm
\label{MerminitionForLuca}
Don't be so quick to dismiss Leggett as an ideologue, just because he
takes quantum states to be objective.  It's hard to spend decades
worrying about the phase diagram of helium-3 without believing that
quantum states are objective.  It's hard to spend a decade thinking
about information theory without believing that quantum states are
states of knowledge.  Does that make you an ideologue?  It seems to
me (probably too superficial a thought) that neither of you has
enough to say about the background that underlies the other's
ideology.
\edm

Your last sentence is fair enough:  There is such a divide between me and the nuts-and-bolts, roll-up-your-sleaves-and-actually-do-something-with-quantum-mechanics physicists that there is certainly a communication barrier.  (By the way, Tony happened to be visiting UBC in Vancouver during my interview there two and a half years ago.  And just coincidental with that and the remark above, I remember after the interview Philip Stamp telling me that someone on the committee said, ``Well, if he loves quantum mechanics so much, why doesn't he do anything with it!''  I doubt it could have been Tony (not being on any UBC committees), but these separate strands bring it all together in my mind.)  I don't know how to cross the divide; it's been a problem that has plagued me all of my career.  I am just a very narrow specialist in physics (who, like Ringo Starr, was always just lucky to be there), and to bridge the divide would require a real generalist \ldots\ one like you or David DiVincenzo.  Particularly someone far more capable than me.

With regard to your phase-diagram remark in particular, I think it's got to be instructive to try to imagine what a turn-of-the-century physical chemist (trained in the worldview of classical physics) should have thought of the tables of heat capacities and whatnot that he had spent much of his life compiling.  What can those numbers represent in the classical worldview?  Why would two distinct chemists compile the same tables of numbers if those numbers didn't signify something objective and intrinsic to the chemicals themselves?  Well, a Maxwell demon wouldn't compile the same tables.  And there is surely a continuum of positions between us and him.  What those tables represent is something about the relation between us, the coarseness of our senses, our inabilities to manipulate bulk materials, all kinds of things like that, and the materials themselves.  Those numbers are just as much about us (``anyman'') as about ``it.''

Anyway, forgetting about Leggett, how would I convince the classical physical chemist of such a point of view?  It would be damned hard (as history has already shown).  Instead of skirting so close to such an anti-Copernican idea---that knowledge gained in science is as much about ``us'' as about ``it''---he'd probably spend his life trying to poke holes in the conceptually simple Maxwell demon idea and try to return his beloved thermodynamical quantities to their pristine objectivity (a place where they'll last forever).

I just think it'd be instructive to carry that out in more detail before tackling anything more specific to quantum mechanics.

On your other point, no I don't consider myself an ideologue (clearly).  The reason is, I came to this position kicking and screaming \ldots\ as {\Carl} and {\Ruediger}, for instance, can attest to you.  It has been a very slow transition and has really only gained momentum in the years that I've happened to know you.  After the kicking and screaming phase---after {\Caves} and {\Peres} had laid the seed---``consistency'' kicked in and carried me the rest of the way to the ``radically subjective Bayesian'' where I am now.  And that's all that's happened.  (Can you really imagine that Tony went in the reverse direction, kicking and screaming all the way himself?)  In the years before I knew you, I had all kinds of crazy ideas about objective wave functions and objective probabilities and passion-at-a-distance and giving up on the notion of spacetime and \ldots.  And they all only expressed {\it one\/} thing:  Fear of taking the observer seriously as a component of what quantum theory is about.

To let that ``fear'' or its equivalent be the one over-riding piece of ``evidence'' for shaping one's foundational program, I call an ideology.  I've always liked the way {\Spekkens} put it in his toy-theory paper:
\bq
   We shall argue for the superiority of the epistemic view over the
   ontic view by demonstrating how a great number of quantum phenomena
   that are mysterious from the ontic viewpoint, appear natural from
   the epistemic viewpoint.  These phenomena include [about a million
   things]. \ldots\  The greater the number of phenomena that appear
   mysterious from an ontic perspective but natural from an epistemic
   perspective, the more convincing the latter viewpoint becomes. \ldots

   Of course, a proponent of the ontic view might argue that the
   phenomena in question are not mysterious \emph{if} one abandons
   certain preconceived notions about physical reality.  The challenge
   we offer to such a person is to present a few simple \emph{physical}
   principles by the light of which all of these phenomena become
   conceptually intuitive (and not merely mathematical consequences of
   the formalism) within a framework wherein the quantum state is an
   ontic state.  Our impression is that this challenge cannot be met.
   By contrast, a single information-theoretic principle, which
   imposes a constraint on the amount of knowledge one can have about
   any system, is sufficient to derive all of these phenomena in the
   context of a simple toy theory, as we shall demonstrate.
\eq
So, my statement is one about the mass of evidence.  It's one about good scientific practice \ldots\ even in the context of quantum interpretations.  To ignore a mass of evidence (in some context) and stick with one's gut is to be an ideologue (in that context).

\bdm
Like my referee's report?  You're right.  I still don't get it.  But
I want to get it.  The tension between what is known and what is a matter of judgment is somehow crucial to the whole business (and as
far as I'm concerned, is what the ``measurement problem'' translates
into in your way of looking at things.  It's still there.)  I guess
there was also tension among the three coauthors which did not promote
complete clarity of expression.
\edm

Your referee report was fabulous.  And much needed.  It was by far the best of the lot (there were four of them, along with detailed comments from Jerry Finkelstein and Chris {\Timpson}).  Actually they were all needed.  I blame the paper's lack of clarity on myself---it came about because I lost soul and didn't want to fight.  The real sin of the presentation is that we tried to write the discussion in a ``metaphysically neutral'' way, if you will, without making too much commitment about these hard words like ``fact'' and ``truth'' (in the quantum context or out of it).  In my gut, I didn't feel that was the right way, but I lost soul.  And also {\Ruediger} assured me that in all the ``unsaid things'' we three really didn't differ.  Well, the referee reports called us out on that:  We had set ourselves up to confuse everybody.  And when push came to shove I found out that Mr.\ {\Caves} hadn't agreed with {\Ruediger} and me at all in the fine metaphysical details.  So, no wonder the paper was confusing at a very deep level.

In fact, I actually had made a feeble attempt to say things a little more clearly when Jerry first commented.  (I'll place the note below that I had written to Jerry at the time; it is relevant to some of your own points.)  {\Carl} responded with an angry, ``I admit that my initial responses were too curt to be very useful, but now the waters are muddied beyond belief.''  At which point, I just curled up into a ball and went into hibernation.

Your report and the others at least helped dispel Carl's overwhelming obstinacy that nothing more needed saying.  And that in turn---like a good marriage counselor---led us to start speaking to each other again.
I think you will find a significant improvement in the paper when I can finally show it to you.  At least I think it reached much better clarity when we were working on it last \ldots\ even though it is still written in the sparse style my two coauthors prefer.  And I even gained ground:  1) it is not written in such a very ``metaphysically neutral'' way as it had been before (it attempts to make some statements of substance), and 2) {\Ruediger} and I even shockingly got this remark from {\Carl} soon after our meeting:
\bq
   I think we ought to be thinking about a paper that lays out our
   entire approach, particularly, the idea Chris and I discussed of a
   three-pronged approach to subjectivity and objectivity: (i) quantum
   states and probabilities are wholly subjective; (ii) system
   attributes are wholly objective; and (iii) measurement outcomes are
   where the rubber meets the road, i.e., where subjective and objective meet to produce something that is not under the control of the agent, but is also not out there in the world.
\eq
Point (iii) represents a very deep shift in {\Carl}'s thinking indeed---I honestly thought I had long since pounded the opposite conception out of him---and I really don't think we would have gotten there if your referee report hadn't forced him to confront these issues that I've been babbling about for the last couple years.

\ldots\ I just read that Donald Rumsfeld resigned while I was writing this note to you.  I didn't know that a third of a president could resign?  Quite a shock to me.  Anyway, more evidence that it is a glorious day!

So let me leave you with one remark on my favorite line in your report.  You say, ``your real point, which you then expand on nicely, is that you have a rather nice, somewhat off-beat interpretation of the Born rule.''  Thank you.  Because this really is the essence of where this research effort is now going.  I tried to put it in a slogan the other day and this is what came out:
\bq\noindent
    The {\it formal structure\/} of quantum mechanics represents an
    EMPIRICAL ADDITION to Ramsey / de Finetti coherence, not an
    empirical addition to the setting of priors.
\eq

What survival in the quantum world is imposing on us agents is not some kind of ``principal principle'' (i.e., that we should adopt this or that objective probability), but rather an addition to ``coherence'' (in de Finetti's sense).  When I am gambling on the sensations I will receive from interacting with a quantum system, my gambles should not only be coherent in de Finetti's sense, but more.  For, considering all the ways I can imagine interacting with the system, all my gambles for each of those individual ways should also be related to each other.  The precise way in which they should be related is an empirical statement:  And in our contingent world it happens to be via the Born rule.  (I.e., it could have been different if our universe were different.)  And that is where we have our clearest window on quantum reality.

But I tried to say all that in my Waterloo talk (never to be mentioned again!); maybe I said it more clearly this time.

\section{09-11-06 \ \ {\it Quantum Foundations at the APS March Meeting}\ \ \ (to a large distribution list)}

\noindent Dear quantum foundations friends,\medskip

Please note that there will be a ``special focus session'' on Foundations of Quantum Theory at the APS March Meeting in Denver, Colorado, March 5--9, 2007.  This session is organized by the TGQI---the Topical Group on Quantum Information, Concepts, and Computation---which hopes to promote the field of quantum foundations within the American Physical Society and the broader physics community.

We hope you will help make this session successful by submitting abstracts (to sorting category 23.8.1 ``Foundations of Quantum Theory (GQI)'').  The deadline for submissions is Monday, November 20.  More information can be found at the very bottom of the page at
\bv
\myurl{http://www.aps.org/meet/MAR07/abs.cfm}.
\ev

Please spread the word to others you may know who are interested in quantum foundations.

The meeting announcement is at \myurl{http://meetings.aps.org/Meeting/MAR07/}. 
You can join the TGQI at \myurl{http://www.aps.org/units/gqi/}.\medskip

\noindent Best wishes,\medskip

\noindent Chris Fuchs\\
for the TGQI Program Committee\medskip

\bq
\noindent 23.8.1    GQI    Foundations of Quantum Theory \medskip

\noindent Advances in both theory and experimental technique have made possible a golden age of investigations in the foundations of quantum mechanics. This session encourages both theoretical contributions that illuminate our understanding of foundational aspects of quantum physics and reports on the sophisticated new experiments that are probing the foundations of quantum theory.
\eq

\section{13-11-06 \ \ {\it Zing!\ Went the Strings of My Heart}\ \ \ (to H. J. Bernstein)} \label{Bernstein7}

\bhbe
I had some thoughts about the last (co-authored) paper of yours
that you shared; in fact I thought you were letting the pressure of
admitting to a classical world get you a bit far over from what we
really know we can say.  Which is the beauty of Bayesian approaches:
sure there is ``zing'' out there but updating our tables of
probabilities and conditionals is about all we know if it.

Rather than dig it out and try to get a recollection about the point
I'd make, I know you must have at least almost finished the paper if
not the negotiations with Carl \& whomever, so could you send a newer
version to me and my student Zac?
\ehbe

I'm not sure what you're saying, but it's always good to hear from you!

Actually, I'm working all this week to re-edit our paper once again.  So any input you might have would be really good.  Attached is the version of the paper that arose after David Mermin's detailed comments.  Next, I've got to tackle the other three(!)\ referee reports, and maybe take into account Jerry Finkelstein's and Chris Timpson's comments.  Why not join the club!  It'd be great to hear which parts you think could be made clearer.

\section{13-11-06 \ \ {\it `Many Worlds at 50' Invitation} \ \ (to A. Kent)} \label{Kent9}

Thank you for the invitation to the Many Worlds meeting.  I've marked my calendar in all worlds (i.e., one).  Please put me down for a talk titled, ``13 Direct Quotes from Everettian Papers and Why I Find Them Unsettling.''  I'm looking forward to this.

\section{13-11-06 \ \ {\it Quantum Foundations at the APS March Meeting} \ \ (to F. E. Schroeck)} \label{Schroeck5}

\bfes
I got your message about the Q Foundations focus group of the APS to
be held HERE, in Denver. I didn't know you had started this group.
Anyway, congratulations; I hope it gets off the ground.
\efes

I'm just a worker bee; the thanks for starting the group go to Danny Greenberger and Anton Zeilinger, and a big petition signed by over 400 people.  You should join the group; we definitely need larger representation from the quantum foundations community.

\bfes
I don't know what I could say that is new at this stage, but I'll
think about it. Maybe something about informational completeness in
measuring by placing holes in a ``screen'' at points in phase space in
the mean. I have talked about q.m.\ on phase space until I'm blue in
the face in Europe, but it still is mostly unknown here in the U.S.
\efes

That sounds good.

\bfes
What are you doing, researchwise now?
\efes

Well, there's the things I have to do for Bell Labs and (dreaming of) building quantum computers.

But I'm also taken with informationally complete observables, particularly very symmetric ones.  I'm writing a new paper on that at the moment, and I'll send you a copy when it is done in early December.

\section{13-11-06 \ \ {\it Dove \& Hudson Old Books}\ \ \ (to A. Caticha \& K. H. Knuth)} \label{Caticha1} \label{Knuth7}

By the way, Ariel, Kevin, I should have told you about the outcome of my book-buying excursion.  I found this place called Dove \& Hudson Old Books (on the corner of Dove and Hudson).  It was a wonderful place, and I walked away with about 15 new books (on pragmatism and Heidegger).  Anyway, on the small chance that you didn't know about it, I thought I should bring the place to your attention.

\section{14-11-06 \ \ {\it No Stand-Alone Event Space at All} \ \ (to V. {\Palge} \& W. G. {\Demopoulos})} \label{Demopoulos7.1} \label{Palge2}

I'm going to cc this note to Bill {\Demopoulos} since my reply to your technical question somewhat involves him.  I don't think the papers of his that I refer to are posted on the web; so he would probably have to send them to you if you want to dig into this more deeply.

\bvp
In the classical probability theory one assumes that the event space
has the structure of a sigma-algebra. In their quest to formulate the probability theory for the quantum realm in an analogous manner,
various approaches in the quantum logic tradition postulate
non-commutative structures, e.g., partial boolean algebras and the
like.

The question then arises: what is the structure of the quantum event
space according to your subjective Bayesian approach? Is it a
sigma-algebra, like in classical probability theory, or is it perhaps some non-commutative structure? If the latter, perhaps you can specify in more detail what it is?
\evp

The best answer I can give you, I think, is that neither of these kinds of structures map onto what I'm thinking.  The main reason for this is that I think it is incorrect to think of the process of ``quantum measurement'' either 1) as the {\it revelation\/} of a property inherent in the system under observation, or 2) as the {\it production\/} of such a property in the system.  And without that, I don't think there is enough glue to bind the events occurring in quantum measurements together into an algebraic structure (say, a lattice or a Boolean algebra, or even a partial Boolean algebra, where there are Boolean algebras tied together at the edges)---at least not in any useful sense that intrigues me as a physicist.

Unfortunately, the only reading recommendations I can give you at the moment (where I try to make this kind of idea of measurement clear) are in the form of little personal essays.  I am hoping to put this all together into a paper before the end of January (for the Abnerfest), but it is not there yet.  So, let me recommend the following pieces:
\begin{enumerate}
\item
``\,\myref{vanFraassen7}{`Action' instead of `Measurement'}\,'' (to van Fraassen)
\item
``Snowflakes'' (to Bub and others)
\item
``\myref{Mermin101}{Me, Me, Me}'' (to {\Mermin} and {\Schack})
\item
``\myref{vanFraassen10}{Questions, Actions, Answers, and Consequences}'' (to van Fraassen)
\item
``\myref{vanFraassen12}{Canned Answers}'' (to van Fraassen)
\item
``\myref{Demopoulos6}{Incompletely Knowable vs `Truth in the Making'}\,'' (to Bill {\Demopoulos})
\end{enumerate}
Maybe it is best to read the pieces in that order.

Particularly in light of your question, maybe it is important for you to understand the type of point of view that Bill {\Demopoulos} is trying to develop.  (I give my report of it in Item 6 above, but you should probably get Bill's original papers.)  I think there is a fruitful similarity between what he and I are seeking, even if we ultimately diverge.  It is this:  In both our views---and they are the only places I've ever seen this style of idea---even when two measurements share a common element (say a given projector $P_i$), there is no implication of a common truth value being imposed on $P_i$ across the measurements.  The reason for this for Bill is that the system's properties are bound up with the whole orthogonal family of projectors the individual $P_i$ happens to be embedded within.  In contrast, the reason for this for me is that I don't think of quantum measurement outcomes as signifying properties intrinsic to the system---they are simply consequences of actions for me.  What makes the element $P_i$ identified across measurements---for both of us---is not truth value, but that the {\it probabilities\/} for $P_i$ are identical in both cases.  What this means particularly from my Bayesian way of thinking is that a judgment is being made:  I, the agent, am identifying {\it this\/} potential outcome of this measurement with {\it that\/} potential outcome of that measurement because I judge their probabilities equal under all imaginable circumstances.  (See Section 4.1 of my \quantph{0205039}.)  It is not that they are identified in Nature itself.  Thus, for me, I think, there is no good sense in which they lie in the same event space at all.

I hope all of that helps.  That's the best I can do at the moment.

\section{14-11-06 \ \ {\it Visit?}\ \ \ (to A. Wilce)} \label{Wilce12}

\baw
Any chance we can lure you out here to give a talk this spring? (The
schedule's pretty open so far.)
\eaw

As a matter of fact, I had just started to accumulate my meeting schedule into a centralized view so I could contemplate the upcoming year.  When I put it all together, it started to look pretty frightening!  But I still owe you!  And it would be very nice to see you again.

Not listed in there is that my in-laws will probably be in and out of our house from Jan 1--13.

Thus, what would you think of sometime in the week of Jan 17?  Or maybe better, sometime in the week of Feb 7 or the week of Feb 14?

\begin{itemize}
\item
``Quantum Communication, Measurement, and Computing,'' Tokyo, Japan, 27 November -- 4 December 2006.
\item
``APS Sorters Meeting,'' Washington, DC, 7--9 December 2007.
\item
``Visit to Geraldine,'' Cuero, TX, circa 21--30 January 2007.
\item
``Pragmatism and Quantum Mechanics,'' Paris, France, 22--23 February 2007.
\item
``APS March Meeting,'' Denver, CO, 5--9 March 2007.
\item
``15th UK and European Meeting on the Foundations of Physics,'' Leeds, UK,
29--31 March 2007.
\item
``6th annual New Directions in the Foundations of Physics Conference,'' College Park, MD, 13--15 April 2007.
\item
``William James and Josiah Royce a Century Later,'' Harvard U., Cambridge, MA, 25--27 May 2007.
\item
``Vienna Symposium on the Foundations of Modern Physics,'' Vienna, Austria, 7--10 June 2007.
\item
``Workshop on Operational Probabilistic Theories as Foils to Quantum Theory,'' Cambridge, UK, 1--13 July 2007.
\item
``Many Worlds at 50,'' Perimeter Institute, Waterloo, Canada, 21--24 September 2007.
\end{itemize}

\section{15-11-06 \ \ {\it Rethinking and Rethinking} \ \ (to W. G. {\Demopoulos})} \label{Demopoulos8}

This line of yours intrigues me:
\bwd
I've been slowly rethinking things of mutual interest to us but am not
quite where I want to be about them; but maybe I'm getting close. I'll
write soon in any case
\ewd
For some reason, ``rethinking'' always has a positive connotation with me.  I myself am hoping to rethink a lot as I put together this Abnerfest contribution.  Particularly, I want to get straight the extent to which I want to ascribe any ontological statements to the ``quantum Bayesian'' view---I think I may end up with several (or at least events and objects, all that Donald {\Davidson} says I need anyway).

\section{15-11-06 \ \ {\it Erratum} \ \ (to C. M. {\Caves} \& R. Schack)} \label{Caves89.1} \label{Schack107.1}

OK, I've finally written the damned thing.  Does this cover everything?  Or was there some further concern?

If both of you find the document acceptable (or after any changes you want to make to it), {\Ruediger} would you communicate it to Matthias?

\bq
\begin{center}
{\bf Erratum: Unknown Quantum States: The Quantum de Finetti Representation}

[J.~Math.~Phys. {\bf 43}, 4537 (2002)] \bigskip
\end{center}

In our formal definitions of {\it exchangeability\/} for probability
distributions and quantum states, we wrote:

{\bf Definition 1:} A probability distribution
$p(x_1,x_2,\ldots,x_N)$ is said to be {\it symmetric\/} (or finitely
exchangeable) if it is invariant under permutations of its arguments,
i.e., if
$$
p\bigl(x_{\pi(1)},x_{\pi(2)},\ldots,x_{\pi(N)}\bigr) =
p(x_1,x_2,\ldots,x_N)
$$
for any permutation $\pi$ of the set $\{1,\ldots,N\}$. The
distribution $p(x_1,x_2,\ldots,x_N)$ is called {\it exchangeable\/}
(or infinitely exchangeable) if it is symmetric and if for any
integer $M>0$, there is a symmetric distribution
$p_{N+M}(x_1,x_2,\ldots,x_{N+M})$ such that
$$
p(x_1,x_2,\ldots,x_N)\; = \sum_{x_{N+1},\ldots,x_{N+M}}
p_{N+M}(x_1,\ldots,x_N,x_{N+1},\ldots,x_{N+M}) \;.
$$

{\bf Definition 2:} A joint state $\rho^{(N)}$ of $N$ systems is said
to be {\it symmetric\/} (or finitely exchangeable) if it is invariant
under any permutation of the systems. The state $\rho^{(N)}$ is said
to be {\it exchangeable\/} (or infinitely exchangeable) if it is
symmetric and if, for any $M>0$, there is a symmetric state
$\rho^{(N+M)}$ of $N+M$ systems such that the marginal density
operator for $N$ systems is $\rho^{(N)}$, i.e.,
$$
\rho^{(N)} = \tr_M\,\rho^{(N+M)} \;,
$$
where the trace is taken over the additional $M$ systems.

However, we should have made it explicit that the distributions
$p(x_1,x_2,\ldots,x_N)$ and the quantum states $\rho^{(N)}$ are to be
considered elements of infinite sequences, $p_N(x_1,x_2,$ $\ldots,x_N)$
and $\rho^{(N)}$ for $N=1,2,\ldots,\infty$, whose elements satisfy
the consistency conditions that $p(x_1,x_2,\ldots,x_N)$ is the
marginal of $p(x_1,x_2,\ldots,x_N,x_{N+1})$ and $\rho^{(N)} =
\tr_1\,\rho^{(N+1)}$ for all $N$.  Taking the incomplete wording in
our original definitions at face value, one can easily find
counterexamples to both the classical and quantum de Finetti
representation theorems. The discussion in Section 1 of our paper
makes it clear that the statements of the theorems concerns infinite
sequences, but we unfortunately left that out of the formal
definitions of the later sections.

We thank Matthias Christandl for bringing this issue to our
attention.
\eq

\section{16-11-06 \ \ {\it Quantum Foundations at the APS March Meeting} \ \ (to J.-{\AA} Larsson)} \label{Larsson3}

\bjal
Will you be very angry with me if I submit to Quantum Cryptography instead? {\rm \smiley}
\ejal

That's funny; I just got a note this morning from Rob Spekkens that said, ``Have you got a sense of whether many good foundations types will be submitting abstracts?''  You would have counted as a ``good foundation type''!

(You see a big worry of ours is that we get a sufficient number of serious people submitting \ldots\ since we want foundations to start looking legitimate in the APS's eyes.  Last year we were lucky enough to have one of the two sessions completely full of interesting people, with only one filled with fringe people.  If we can at least continue that kind of trend, we'll probably be OK.)

\section{16-11-06 \ \ {\it Challenges to the Kierkegaardian Bayesian} \ \ (to A. Shimony)} \label{Shimony12}

Thank you for the continued kind words.  [\ldots]

Anyway, on to more fun things:
\bas
I can't possibly say in one letter all that I have to say in response
to your ideas, just as you haven't said all you have to say. But I'll
start, and in due time we shall carry the discussion further.
\eas

Let me try to do that (at least a little) now.
\bas
Top of p.\ 3, point 4 of your list of similarities and differences
with Gill:  You don't give arguments that there is no measurement
problem and no nonlocality problem. As to the first, if the entire
dynamics of qm is linear, including the dynamics of macroscopic
systems, then it is hard to see how definite events occur at the
conclusion of a measurement -- hence a problem. As to the second, I
think quantum non-locality is a fact, and it is also a fact that it
cannot be used to send signals superluminally. Does this show that
there is tension between qm and relativity theory and nevertheless
``peaceful coexistence''? I used to think so, but Bell convinced me
otherwise -- hence there seems to be a problem.
\eas

It is true that I don't give any arguments there, but they have been peppered throughout my writings.  The key to much of it is to realize just how {\it thoroughgoing\/} an epistemic view my closest collaborators and I take on QUANTUM STATES.  Take for instance, your point that ``if the dynamics of qm is linear, including the dynamics of macroscopic systems, then it is hard to see how definite events occur at the conclusion of measurement.''

My reply to that is that the dynamics of quantum mechanics has nothing to do with how definite events occur.  It is silent on the matter.  It is silent on the matter in the same way that the formal structure of Bayesian probability theory is silent on the matter of where a datum $d$ comes from by which a gambling agent updates from a prior probability $P(h)$ to a posterior probability $P(h|d)$.  It is simply not within the power of the formal structure of probability theory itself to say either 1) how a truth value for $d$ comes about or from where it finds its origin, or 2) how the agent has become aware of $d$ so that he may conditionalize.

Similarly, I will say of quantum states.  For this clique of friends that I call ``the quantum Bayesians'' a quantum state represents a catalogue of Bayesian probabilities and has no independent existence beyond that.  A quantum state is nothing other than the complete catalogue of probabilities.  Probabilities for what?  States of the world?  No.  They are probabilities for things concerning the very agent himself who has ascribed those probabilities.  They are probabilities for something interactive:  Namely, the agent interacts with a piece of the world external to himself and there are consequences of that interaction for him---at the very least there are various sensations that may come about.  What the probabilities derived from a quantum state signify are the agent's expectations for the consequences of his actions.

What about the dichotomy of linear evolution and collapse?  For the quantum Bayesians it is no conceptual problem---they are simply two modes of the same thing, namely updating probabilities for whatever reason.  In the case of collapse it is of the cloth of conditionalizing.  In the case of linear evolution it is more of the flavor of Jeffrey conditionalization (and some of the updating processes van Fraassen has described).  But these are distinctions of detail, rather than conceptual distinctions.

I expand on some of this in the essays ``Raining Down in Cambridge,'' ``The Evolution of Thought,'' and ``More Linearity,'' between pages 222 and 239 of my book {\sl Notes on a Paulian Idea\/} (which I sent you a copy of).  You might find some of that amusing, if not instructive.

Similarly, I could go on about your statement, ``I think quantum non-locality is a fact.''  I would say it is a confusion that takes its origin in not properly appreciating the disconnection between quantum states (epistemic) and quantum events (ontic, but partial consequences of the agent himself).  I tried to say this as best I could in Section 3 of ``Quantum Mechanics as Quantum Information (and only a little more)'' (\quantph{0205039}).  I would be honored if you'd read that section.

Anyway, to sum up, none of these things are PROOF that a purely epistemic view of quantum states is the way to go, but instead it is a question of the mass of evidence.  And you know that I am a pragmatist.  I would think that any good Kierkegaardian Bayesian, too, would only adopt priors flexible enough that they could eventually be swayed by the mass of evidence if there is such.  In that regard, I think one of the most outstanding arguments (which you should be aware of) is Rob {\Spekkens}'s paper ``In Defense of the Epistemic View of Quantum States: A Toy Theory,'' which can be found at \quantph{0401052}.
I've always thought Rob put the main point so nicely with these words:
\bq
We shall argue for the superiority of the epistemic view over the
ontic view by demonstrating how a great number of quantum phenomena
that are mysterious from the ontic viewpoint, appear natural from
the epistemic viewpoint.  These phenomena include [about a million
things]. \ldots\  The greater the number of phenomena that appear
mysterious from an ontic perspective but natural from an epistemic
perspective, the more convincing the latter viewpoint becomes. \ldots

Of course, a proponent of the ontic view might argue that the
phenomena in question are not mysterious \emph{if} one abandons
certain preconceived notions about physical reality.  The challenge
we offer to such a person is to present a few simple \emph{physical}
principles by the light of which all of these phenomena become
conceptually intuitive (and not merely mathematical consequences of
the formalism) within a framework wherein the quantum state is an
ontic state.  Our impression is that this challenge cannot be met.
By contrast, a single information-theoretic principle, which
imposes a constraint on the amount of knowledge one can have about
any system, is sufficient to derive all of these phenomena in the
context of a simple toy theory, as we shall demonstrate.
\eq

That's the way I see it in our own debate---the one between you and me.  From my point of view, you are stuck with all these mysteries---the dichotomy of linear and nonlinear evolutions (when does one kick in and the other leave off?), passion-at-a-distance but no action-at-a-distance, and so on and so on.  Whereas I would say a thoroughgoing epistemic view of quantum states alleviates so many of these difficulties automatically that it is already worth exploring for that reason alone.  Here's the way Rob put it more eloquently in an earlier version of the paper already mentioned:
\bq\noindent
Because the [toy] theory is, by construction, local and non-contextual, it does not reproduce quantum theory. Nonetheless, a wide variety of quantum phenomena have analogues within the toy theory that admit simple and intuitive explanations. \ldots\ The diversity and quality of these analogies provides compelling evidence for the view that quantum states are states of knowledge rather than states of reality, and that maximal knowledge is incomplete knowledge.  A consideration of the phenomena that the toy theory fails to reproduce, notably, violations of Bell inequalities and the existence of a Kochen--Specker theorem, provides clues for how to proceed with a research program wherein the quantum state being a state of knowledge is the idea upon which one never compromises.
\eq

Here's what I would love to get a better sense of from you:  Does an argument like Rob's particular one (the paper is very easy reading, so I encourage you to give it a shot) \ldots\ does an argument like Rob's put any dent in your armor?  And if not, why not?

\bas
I agree, of course, that there is ontic indeterminism, for that is at
the heart of quantum potentiality, but I don't agree that ``ontic
probability function'' throws no light on this indeterminism.
\eas

Well, certainly my point is that it hinders our progress by presenting a false god---one that stands in our way of seeing a purer, more direct path to salvation!  More seriously, in toned down language:  That is my trouble.  The examples I pointed you to above give I think overwhelming technical evidence that quantum states should be viewed epistemically.  And that then has to trickle into my view of quantum indeterminism.  There is something about this world that keeps me, the agent, from having too much certainty about too many things (particularly the consequences of the various actions I might take on a given physical system).  This is expressed through the quantum probability rule---the Born rule (via Gleason's theorem).  The more subjectively certain I am about the outcome of spin-$x$ measurement, the less subjectively certain I should be about the outcome of spin-$y$ measurement, say.  This something---the thing that bars me from too much certainty about too many things---is, I would say, a direct handle on ``quantum potentiality.''  But I don't need to go so far as to assume that particular quantum-probability values are ontic to express this idea.  My very actions have uncertain consequences (this is undeniable under the assumption of quantum mechanics); no matter what I do I cannot gain a god-like foresight.  That seems to be a property of the world, but it is a property that finds its expression in epistemic terms.

Here's one of the ways I put it in the past (in my \quantph{0205039}).  The language is not quite as refined as what I would use now, but it carries a good piece of the point:
\bq
The last seventeen years have given confirmation after confirmation
that the Bell inequality (and several variations of it) are indeed
violated by the physical world.  The Kochen--Specker no-go theorems
have been meticulously clarified to the point where simple textbook
pictures can be drawn of them.  Incompleteness, it seems, is here to stay:  The theory prescribes that no matter how much we know about a
quantum system---even when we have {\it maximal\/} information about
it---there will always be a statistical residue. There will always be questions that we can ask of a system for which we cannot predict the
outcomes.  In quantum theory, maximal information is simply not
complete information.  But neither can it be completed. As Wolfgang
Pauli once wrote to Markus Fierz, ``The well-known `incompleteness'
of quantum mechanics (Einstein) is certainly an existent fact
somehow-somewhere, but certainly cannot be removed by reverting to
classical field physics.''  Nor, I would add, will the mystery of
that ``existent fact'' be removed by attempting to give the quantum
state anything resembling an ontological status.

The complete disconnectedness of the quantum-state change rule from
anything to do with spacetime considerations is telling us something deep: The quantum state is information. Subjective, incomplete
information. Put in the right mindset, this is {\it not\/} so intolerable. It is a statement about our world. There is something about the world that keeps us from ever getting more information than can be captured through the formal structure of quantum mechanics. Einstein had wanted us to look further---to find out how the incomplete information could be completed---but perhaps the real question is,
``Why can it {\it not\/} be completed?''
\eq

\bas
Certainly when one has systems simple enough to be characterized by
pure quantum states, and those states govern the transition
probabilities into other states -- as in emissions and absorptions --
then a quantitative character has been added to merely qualitative
indeterminism. Doesn't that throw light on the character of quantum
potentiality? And even more light is shed when one considers how these
quantitative transition probabilities are related to the geometric
structure of the Hilbert space of states, particularly to the inner
product of vectors in the Hilbert space.
\eas

Again there is a divergence between us because you are, at the outset, thinking of quantum states in ontic terms.  For me, if a system has a pure state, it is because I have a very strong, particular belief about what might arise as a consequence from my interactions with it.  If I could imaginatively ask the quantum system what its quantum state is, it would say, ``I don't know; I don't have a quantum state.  It's up to you to answer that question yourself---you're the only one who has a quantum state for me.''

Still, I think there is much to be gleaned about the actual structure of our world from the Hilbert-space structure of the quantum probability rule---that's predominantly what my research program is about.  It seems to me that the content of the Born rule is a statement about how to be ``coherent'' (in something like a Ramsey / de Finetti sense) when gambling on the consequences of our physical interactions.  The Born rule, as I see it, is an empirical addition to probabilistic coherence \ldots\ and that's where we stand a chance of learning something lasting about the precise nature of quantum indeterminism.

This idea is fleshed out in greater detail in the closing section of the attached paper (which {\Caves}, {\Schack}, and I are presently in the process of revising, so it is not ready for wide distribution yet).  But I figure I will give you a copy now as this point will be the core of my planned festschrift contribution for you.  If you have any particular criticism to this concept, I would welcome it.

\bas
There remains a difficult question of whether the world view of qm
accommodates indeterminisms that are not quantitative, such as the
outcome of a war.
\eas

Now, this point definitely intrigues me.  For, as you may have already gathered of me, I would say ALL indeterminisms are of this variety {\it including\/} the quantum ones.  In this regard I was deeply influenced by Pauli and Markus Fierz.  If you go to pages 577--582 of my book {\sl Notes on a Paulian Idea}, you'll find an essay of Fierz's on the subject, reprinted in full.  I'd be curious to know what you think of it?

\bas
But there are many varieties of Bayesianism, including logical
probability theory, personalism, tempered personalism, and a peculiar
variety that I call ``Kierkegaardian Bayesianism.'' I.J. Good has a
taxonomy of Bayesianisms somewhere that is very diverse and still not
complete. The important point is that some of the strategies for being
rational are not prior to empirical science but depend upon the
structure of the world and human adaptations to that structure:  we
not only learn by experience (using probability as a tool) but we
learn how to learn by experience (and therefore how to use
probability). As you see, I am not radically in disagreement with you,
but I think there are nuances.
\eas

And you are right.  On this point, I think we have an issue of potential convergence, rather than divergence.  I hope you'll see everything above in the light of this very point!

\section{17-11-06 \ \ {\it Certain Comments} \ \ (to C. G. {\Timpson})} \label{Timpson11}

Like I said yesterday, thanks so much for the comments.  Don't you mind sending me ``excessively long emails.''  When long emails are thoughtful like yours, I love them!

\bcgt
I enjoyed reading your certainty paper. I'm glad I'll now be able to
direct people to that now rather than encouraging them to read through
all of `Quantum States: WHAT'! I had a few comments, mainly about
presentation really.
\ecgt

Yes, it is long since overdue that {\Caves}, {\Schack}, and I have an official statement on this subject.  I'm not very pleased with the paper because we made too much of an attempt to stay ``metaphysically neutral'' on what the $h$'s in a probability assignment $P(h)$ stand for in the case of the quantum context.  But still the paper has improved significantly since the first posted version:  We had extremely detailed referee reports from two of our four(!) referees, and taking their points into account has improved the paper significantly \ldots\ and helped me win back a little territory of what I would have liked to have seen emphasized more in the paper.

However, now that we will soon have a stabilized triumvirate statement out, I plan to start embellishing it into something more to my taste for the Abnerfestschrift (where I'll be going it alone).  I plan to have that finished by end of January.

\bcgt
On the distinction between probability assignments and propositions.
It seems to me that, while pedagogically useful, this distinction
cannot be the whole story. To make a probability assignment may not be
to express a proposition (or at least not one concerning objective
facts in the world), but one can have truths concerning probability
assignments: ``He thinks that the probability is $p$'', for example, could express a truth. De Finetti allows this, I notice (having just checked the sections of Probabilisimo you refer to towards the end): ``That a fact is or is not a practically certain is an opinion, not a fact; that I judge it practically certain is a fact, not an opinion'' (end section 19). Thus one does need to add a claim to the effect that
judgements of probability aren't responsive to any facts about chances
in the world, as there are no such facts as the latter, on this view.
\ecgt
Yes, you are correct about that, and it probably should be reflected in better wording on our part.  I'll try to do the point better justice in the revised paper.

\bcgt
On that reading it would be ok to say that stating that the
measurement outcome is 1 with certainty {\bf does not express} a
proposition. But one could question whether it is necessary to go so
far anyway; all that is required for the position is that any
proposition expressed when making a probability assignment doesn't
concern how things are in the world external to the agent.
\ecgt
That, of course, is what we were shooting to express.  So, ditto my last remark above:  I'll work harder to make our language a little less clumsy.

\bcgt
p.2 ``Certainty is a function of the agent, not of the system.'' That
puts it nicely. The philosopher Alan White distinguishes between the
certainty of people and the certainty of things, arguing that these
are quite distinct things. Subjective probability concerns certainty
or otherwise of people. (Did I send you a copy of his paper
`Certainty'? I have a vague feeling that I might have done, but I'm
not sure.)
\ecgt
No, I don't believe you ever told me about Alan White.  I think I would have remembered that.  What of his should I read?  I didn't find anything in quick look on the web.

\bcgt
Against objective chance (Section 2). A quibble: I'm not sure I was
100\% persuaded by the argument you give in the final few paragraphs
here. Some accounts of objective chance -- I am thinking particularly
of Mellor's in {\sl The Matter of Chance} -- can deal with the problem of specifying initial conditions precisely. Mellor argues that a
propensity is a dispositional property of a chance set-up, so that if
the conditions allow the disposition to be manifest, {\bf then} a certain probability distribution will be displayed. There would presumably be a range of detailed initial conditions that would allow the display of this disposition, and other ranges which would not.
\ecgt

I think that is the weakest part of the whole paper.  In fact, so weak, as to be almost irrelevant \ldots\ or actually damaging.  What you read was the residue that remained after we scrapped our attempt at a direct blow to the principal principle \ldots\ one that was going to expand on Richard Jeffrey's against it.  But we got lazy and gave up on the project, particularly because {\Carl} kept saying, ``We're writing this paper for physicists,'' and it started to look too hard to get any consensus between us.

Maybe I'll try to do this better for the Abnerfest thing.  I think Jeffrey is right:  The principal principle simply cannot be properly formalized, and to the extent that it can be, it just captures judgmental probabilities after all.

\bcgt
But there's another challenge close by -- Howard's -- which is: can
{\bf one} be certain that q when one \underline{knows} that \underline{it} is not certain that q (when one knows that there is no fact of the matter whether q)?  It seems not; yet this is apparently the situation that a self-conscious Bayesian must admit. (There's a scope distinction in the negation here: one can still be certain that q when it's not the case that one knows that it is certain that q: here the form is ``not (Knows( Certain q))''; but one can't be certain when one knows it's not certain that q: ``knows (not (Certain q)''.)

So looking at the clause (p.\ 12 third para from bottom) `an agent could
not be certain about the outcome ``Yes'' without an objectively real
state of affairs guaranteeing this outcome \ldots, I think you are right to maintain that this is false. The assertion that there is such a state of affairs isn't needed for one to be certain. However, it seems we do require, in order to be certain, the absence of the assertion that there is {\bf no fact} guaranteeing the outcome.
\ecgt

This discussion of yours is very helpful, and it helps emphasize the pitfalls of our sparse representation.  I think it's more evidence that, despite {\Caves}' objections, we really did need to get more metaphysical.

Here's something relevant that I wrote to David {\Mermin} the other day:
\bq
With regard to your phase-diagram remark in particular [I.e.\ ``Don't
be so quick to dismiss Leggett as an ideologue, just because he takes quantum states to be objective.  It's hard to spend decades worrying
about the phase diagram of helium-3 without believing that quantum
states are objective.''], I think it's got to be instructive to try to
imagine what a turn-of-the-century physical chemist (trained in the
worldview of classical physics) should have thought of the tables of heat capacities and whatnot that he had spent much of his life
compiling. What can those numbers represent in the classical
worldview?  Why would two distinct chemists compile the same tables
of numbers if those numbers didn't signify something objective and
intrinsic to the chemicals themselves?  Well, a Maxwell demon
wouldn't compile the same tables. And there is surely a continuum of positions between us and him. What those tables represent is
something about the relation between us, the coarseness of our
senses, our inabilities to manipulate bulk materials, all kinds of
things like that, and the materials themselves.  Those numbers are
just as much about us (``anyman'') as about ``it.''

Anyway, forgetting about Leggett, how would I convince the classical physical chemist of such a point of view?  It would be damned hard
(as history has already shown).  Instead of skirting so close to such an anti-Copernican idea---that knowledge gained in science is as much
about ``us'' as about ``it''---he'd probably spend his life trying to
poke holes in the conceptually simple Maxwell demon idea and trying
to return his beloved thermodynamical quantities to their pristine
objectivity (a place where they'll last forever).
\eq
and something more particular to our paper that I had written to Jerry Finkelstein a good while back:
\bq
In our scenario there is only the agent and the external world.  The way one should think of the measurement device is the same way one
should think of a prosthetic hand---it is an extension of the agent.

``Detector click'' is shorthand for {\it sensation\/} in the agent, and in my personal view, that is all it is.  The value of the ``detector click'' is beyond the control of the agent; it is not a function of his beliefs or his desires.  His interaction with the external world
gives the class of potential sensations, but it does not set the
value.

Thus, the value of the detector click---call it $d$---is objective
for the agent.  (There is one sentence in the paper where I adopted
almost that phraseology ``an event is a fact for the agent,'' and it
might have been wise to adopt it more consistently.)  What is
subjective from our view---i.e., a function of the agent and not the data or external world alone---is the POVM associated with those
potential values.  That is, the set of operators $E_d$. [\ldots]

There is an agent, there is an external world, there is an
interaction between the two---the agent acts on the external world,
it causes a reaction back in him (sensation).  As far as he is
concerned that reaction is not subjective; it's as real as anything
he's ever seen.
\eq

I think the solution to your conundrum lies in extending these points.  Particularly this.  The reason the two statements
\begin{enumerate}
\item
   \underline{One} is certain
\item
   \underline{One} knows that ``\underline{It} cannot be certain''
\end{enumerate}
are not inconsistent is because of the ``metaphysical piece'' that's missing from most of our language in our paper.  The outcome of a quantum measurement is a joint product of the agent and the quantum system.  [I remember once writing on my notepad that the idea of quantum measurement---even in our conception of measurement (or, should I say, {\it particularly\/} in our conception!)---without a system to measure upon is as much of a koan as the sound of one hand clapping.]

When an agent expresses certainty about the outcome of some measurement, he is judging that, in effect, the quantum system's contribution to the outcome will be negligible.  (See the note titled ``Philosopher's Stone'' for an expansion on this conception of measurement. Also, in a separate email, I'll supplement this with a note I wrote to Veiko {\Palge} recently; there are several references in it that still further expand on this point.)  So the agent can self-consciously say ``\underline{I} am certain which will occur'' at the same time as saying ``\underline{it} is not certain,'' because ``neither \underline{it} nor \underline{I} is the whole story.''  The outcome comes out of a marriage of the two of us.  ``When \underline{I} interact with \underline{it} I am quite certain which sensation that interaction will lead to; \underline{it} won't surprise me.''

If I were to kick the cat after a hard day at the office, I am certain that it will lead to my satisfaction.  (For whatever crazy reason, I am also certain that it cannot react quickly enough to spoil the moment by scratching me.)  On the other hand, because of my fundamentalist Baptist upbringing, I might also believe that the cat cannot foresee how it will react to my kick, by running or fighting---``it's just a soulless dumb animal.''  Even more relevantly, it cannot foresee whether I will be satisfied or not.  In other words, performing a quantum measurement on a system for which I am certain of the outcome is like kicking a cat.

Does that make any sense to you?

Going back to the last point of yours quoted above,
\bcgt
However, it seems we do require, in order to be certain, the absence
of the assertion that there is {\bf no fact} guaranteeing the outcome.
\ecgt
what I said above, I think, may not contradict this.  It just means that we should be a little more careful to say, ``there is no fact {\it in the object\/} guaranteeing the outcome.''  Of course, there is no fact in the agent guaranteeing the outcome either; otherwise his certainty wouldn't be subjective certainty.

\bcgt
On a quite different matter: have you ever come across Diederik and
Sven Aerts' work?
\ecgt
I've seen a very little bit of it, but even that I haven't absorbed all that well.

\bcgt
They want to explain the appearance of quantum probabilities as
arising from what they call a hidden measurement model, involving
ignorance of some processes going on in measurements, rather than
ignorance of some underlying properties of the object systems. [\ldots] This seems to be close to what I think you have on occasion said might be one way -- one of the less interesting ways -- your and {\Ruediger} and {\Carl}'s programme might turn out.
\ecgt

I think you're probably right in this assessment. ``Hidden variables in the agent,'' (maybe a more honest way of saying ``contextual hidden variable theory'') is something that has crossed my mind from time to time as {\it possibly\/} consistent with my forming point of view of quantum mechanics.  But as you say, it seems for me that it would be ``one of the less interesting'' potential endings to our research program.  I clearly am immensely attracted to the idea that a quantum system and an agent, when put in conjunction with each other give rise to something that was in neither physical system alone.  That a quantum measurement outcome is a birth of sorts.  Even a ``relational'' viewpoint on measurements (say like {\Mermin} or Rovelli, or what {\Spekkens} desires) doesn't seem to be enough for me.  There's something in my gut that says that anything weaker than the radical, Paulian direction (the thing that in a bar room I call ``the sexual interpretation of QM'') is ultimately inconsistent.  But indeed I don't have a good argument for that right now.

On the other hand, I do think there is a pearl of wisdom in this quote you sent me:
\bq
If one wants to interpret our hidden measurements as hidden variables,
then they are hidden variables of the measuring apparatus and not of
the entity under study. In this sense they are highly contextual,
since each experiment introduces a different set of hidden variables.
They differ from the variables of a classical hidden variable theory
because they do not provide an `additional deeper' description of the
reality of the physical entity. Their presence as variables of the
experimental apparatus has a well-defined philosophical meaning, and
expresses the fact that we, human beings, want to construct a model of reality independent of our experience of this reality. The reason is
that we look for `properties' or `relations between properties', and
these are defined by our ability to make predictions independent of
our experience. We want to model the structure of the world
independently of our observing and experimenting with this world.
Since we do not control these variables in the experimental apparatus, we do not allow them in our model of reality, and the probability
introduced by them cannot be eliminated from a predictive theoretical model.
\eq
so maybe I should look at these guys more carefully.

\bcgt
My aim in the next couple of months is finally to write something more
substantial about your approach (which will end up as the final
chapter of my book), so I shall be sitting down to try and think hard
about it. I'll try not to pepper you with emails, but you may get one
or two. I see that there is a new collection of emails on your webpage
which I imagine I ought to take a look at to see how your thoughts
have been progressing.
\ecgt

It would be very useful for us to have your commentary.  So, please do pepper me any time you'd like.  If I don't answer right away, it'll only mean that I'm busy or traveling or on holiday; it won't mean at all that I'm bothered by your queries or objections.  I get more out of your emails than I've probably ever adequately expressed to you.  I want to understand quantum mechanics before I die; that is first and foremost the goal \ldots\ and whatever tension will lead me there is precisely what I need.

I hope {\Ruediger}'s inaugural lecture goes well.  I wish I could be there too.  I think what he's talking about (a quantum Bayesian definition of randomness) is very deep stuff, and it will be a while before it is appreciated.  I myself have yet to incorporate it into my metaphysics, but I think it will be a crucial piece of it ultimately.  Have a good time.

\section{17-11-06 \ \ {\it March Meeting} \ \ (to R. W. {\Spekkens})} \label{Spekkens38}

\brws
I was about to submit my abstract for the focus session on
foundations, and then I reread the email I received from APS.  It
says: ``the proposed title of your talk is `Interpretations of Quantum
Mechanics'.''  My best guess is that this wasn't meant to be a real
proposal for my title, and that I should speak on a topic of my choice
(I'm thinking of speaking on the toy theory), but I just wanted to
make sure that this was the case.
\erws

I think you should definitely talk on the toy model.  Put it in the wider context of the interpretation debate of course, and I would hope that you {\it not\/} spend much time on the latest and greatest variant or new achievements.  Instead, I hope you spend plenty of time getting the basic points across.

Was the paper ever published, by the way?

It's funny, I was just advertising the paper to Abner yesterday and rethinking that lovely closing line in the abstract of an earlier draft:
\bq\noindent
A consideration of the phenomena that the toy theory
fails to reproduce, notably, violations of Bell inequalities and the
existence of a Kochen--Specker theorem, provides clues for how to
proceed with a research program wherein the quantum state being a
state of knowledge is the idea upon which one never compromises.
\eq
Yes, Bell and Bell--Kochen--Specker are clues---the most important ones.  But not clues toward the epistemic nature of quantum states.  They are clues about the event space, and that is a much deeper issue.

\section{17-11-06 \ \ {\it Birds and Stones}\ \ \ (to S. L. Braunstein)} \label{Braunstein13}

\bslb
I must be getting old. I no longer believe in mixed states --- except as a convenient language/tool.
\eslb

Well, you know you shouldn't believe in pure states either.  It's funny, just yesterday I wrote this to Abner Shimony:
\bq\noindent
   Again there is a divergence between us because you are, at the
   outset, thinking of quantum states in ontic terms.  For me, if a
   system has a pure state, it is because I have a very strong,
   particular belief about what might arise as a consequence from my
   interactions with it.  If I could imaginatively ask the quantum
   system what its quantum state is, it would say, ``I don't know; I
   don't have a quantum state.  It's up to you to answer that question
   yourself---you're the only one who has a quantum state for me.''
\eq

\section{21-11-06 \ \ {\it The Self-Preservation Society} \ \ (to M. S. Leifer)} \label{Leifer6}

I read a good bit of your paper last night, and I really, really enjoyed it.  It's no-nonsense like your talks, and the whole thing is put together in a very compelling way.  This pretty construction of yours makes my mouth water again for a kind of Gleason theorem behind trace-preserving completely positive (TPCP) maps---it's got to be there ultimately.  We really ought to get together and think about this.  (There is some chance I'll be visiting PI in late Jan or early Feb; will you be around?)

I checked with Andrei and he says it's OK to post things on {\tt quant-ph}.  So there you go.

But before posting, let me bring up one political point.  I thought it was a bit unfair to me how you introduce the motivation for the paper:
\bq\noindent
   The situation is less clear when considering a unitary operation on
   the system of interest alone, since in this case the environmental
   state is irrelevant to the action of the operation on the system, and
   unitary operations do not cause convergence of distinct states. Thus,
   unlike the previously discussed cases, the subjectivity of unitary
   operations cannot be argued directly from the subjectivity of quantum
   states.
\eq
particularly
\bq\noindent
   This offers the opportunity for CFS to hold onto the objectivity of
   unitary operations, which might be tempting, since the specification
   of a Hamiltonian seems to encode the objective content of our most
   successful physical laws.
\eq
To be sure, your technical contribution is great and was needed, and I'll be advertising it to everybody as the best little piece of quantum foundations in a while.  But your wording gives no indication that the subjectivity of unitaries had ever crossed my mind or the preliminary and very public arguments I've given for it.  See, for instance,
\begin{itemize}
\item
Section 7 in \quantph{0205039},
\item
or Section 6 in \quantph{0404156}
             where particularly footnote 9 makes it clear,
\item
or Section 4 in \quantph{0307198}
             where I consciously did not single out unitaries as
             anything special among TPCPs,
\item
and there are certainly significant pieces in ``Quantum States: W.H.A.T.?''\ that I could pinpoint where I think the point is strictly argued.
\end{itemize}
The arguments may not have been conclusive, or even, you might think, right (though I think they are), but they started the whole subject.  The idea of the subjectivity of operations I consider my best work ever in physics---it was no easy wall to climb over.  Your work is a nice turning point, giving us something far more solid and productive to work with, but don't forget the old hint-giver who is still himself struggling for a stable career.  If that's my only role---and it's not too bad of a role---I still need to get credit for it, like the rest of you, to survive.

\subsection{Matt's Reply}

\bq
You are right.  Originally, I had wanted to say that this gives the C of CFS an opportunity to hold onto the objectivity of unitaries, since he's clearly the one who wants to, but I thought that was unfair because C has not yet written any paper to that effect.  I didn't mean to imply that $C \cup F \cup S$ actually wanted to take this option, only that it would not be a logical contradiction for any of them to do so.  Obviously, I am aware of your earlier discussions of the subjectivity of unitaries and these inspired my argument, but in my haste to finish the paper in time I may not have emphasized this.

If you can give me a couple of extra days I should be able to fix this.  It's not a big job, but unfortunately I have more than one deadline to deal with at the moment.

As regards visiting PI, I should be there in Jan and Feb, apart from the week of the 11th Feb.  I can see that a Gleason's theorem for TPP maps potentially follows from your tensor product argument, but I am still not completely sure what the motivation for studying it is.  I mean, any conceivable dynamics must map density operators to density operators, so the only real question is linearity.  For an operationalist, this would be argued from the possibility of forming mixtures.  I am guessing you don't like this because it uses a notion that is analogous to a ``probability of a probability''.  The more general issue of how to argue for convexity in the Bayesian approach without making use of the mixing argument is also worth discussing.  I know you have an email to Lucien on this in your Samizdat, and {\Ruediger} also discusses it in his response to Lucien's papers, but I still think there is something more to be fleshed out here.
\eq

\section{21-11-06 \ \ {\it The Self-Preservation Society, 2} \ \ (to M. S. Leifer)} \label{Leifer7}

\bml
I can see that a Gleason's theorem for {\bf TPCP} maps potentially follows from your tensor product argument, but I am still not completely sure what the motivation for studying it is.
\eml

The motivation is the usual one:  The search for a direct argument for something, rather than a roundabout or secondary one for it.  It would just be nice to get a straight-out argument for CPMs without {\it first\/} invoking the idea of a single-time density operator.  Moreover, it would be like a development of probability theory that defines conditional probabilities at the outset, without reliance on marginals and joints---it would be more radical probabilist at its core.

\bml
The more general issue of how to argue for convexity in the Bayesian
approach without making use of the mixing argument is also worth
discussing.
\eml

That would be good too.

\section{22-11-06 \ \ {\it The Self-Preservation Society, 3} \ \ (to M. S. Leifer)} \label{Leifer8}

You made the old man happy!  I think the paper's great.  And you really do ``significantly strengthen the case for the subjectivity of all TPCP,'' so we should uncork a bottle of bubbles when we're together.

My PI visit is now settled.  I should be there roughly Jan 29 to Feb 11.  It'll be good to see you.

This morning I dug up the notes I had written on trying to find a dynamical Gleason theorem.  Almost to my shock, I saw that they were dated ``1 December 2000, Vienna'' \ldots\ but it's not really to my shock:  I'm pathetic about ever completing anything.  (My biggest character flaw is that I seem to need ridiculous amounts of interaction to get anything done \ldots\ and Bell Labs can be a lonely place.)  At least I was pleased to see the way the notes start off:
\bq\noindent
Let us take the following idea as fundamental:  Hilbert spaces
need not be associated with systems per se, but with measurements.
Thus if I make a measurement on a `single system' at two distinct
times, I should consider that a POVM on $H \otimes H$.  If, later,
I might wish to consider $H$ as the only relevant Hilbert space, I
might do that --- but that is a derivative step, at least from this
conception.
\eq

If you give me a fax number where you can be reached, I'll fax you the notes (there's about 20 pages).  Maybe you can start thinking about all my wrong turns---and how to fix them---in the background of your mind before we meet up.

Tell me as soon as your paper is complete to the point of posting.  I want to bug Charlie Bennett about it before seeing him in person Monday.  Have you discussed the closing idea in your paper with Lee Smolin?  I should think if nothing else interested him about it (and all this subjectivity gobbledygook actually turned him off), he might still be intrigued by the formalism.

\section{22-11-06 \ \ {\it A Place in the World, 1} \ \ (to C. H. {\Bennett})} \label{Bennett46}

I just read these words of yours:
\bcb
Fortunately public interest is now such that lay people are
overcoming their fears and asking us questions like ``What is a
quantum computer?'', ``How can a particle be in two places at once?'', or ``How can observing one particle affect another particle?'' By
coming up with intelligible and respectful answers to these
questions, we will be doing our profession, and our fellow humans,
a great service.
\ecb
Made me feel like maybe I have a place in the world, after all.

Really looking forward to joking around with you in Japan next week.  I hope you'll be there the whole time.

\section{22-11-06 \ \ {\it A Place in the World, 2} \ \ (to C. H. {\Bennett})} \label{Bennett47}

\bcb
I'll be at QCMC there starting the late afternoon or evening of 28
November.  I look forward to joking around with you and trying on some
ideas about quantum Darwinism and the ontological status of Jimmy
Hoffa, a little further developed from the ones you heard before.
\ecb

Then here's a good way to start preparing for the joust.  I just got this paper from Matt Leifer [see \quantph{0611233}], which he'll be posting today.  I think it's really very good.  (And, of course, completely relevant to the ontological status of Jimmy Hoffa.)

See you the 28th.  I'll be waiting for you in the hotel lobby!

\section{22-11-06 \ \ {\it CFS Philosophy, Ontic and Epistemic States and Evolutions, Funes the Memorious} \ \ (to C. H. {\Bennett})} \label{Bennett48}

\bcb
Matt's paper deliciously deduces from your execrably anthropocentric CFS philosophy that if quantum states have no ontic existence, merely epistemic, then so do quantum evolutions.  This would include the identity ``evolution'' which leaves all ``states'' the same.    So even the process of staying the same is not a process, but a correlation between states of knowledge at two timelike separated events.   To thus problematize the process of merely existing, of remaining the same, is to begin to think like Borges' character Funes in {\bf Funes the Memorious}, who ``was disturbed by the fact that a dog at three-fourteen (seen in profile) should have the same name as the dog at three-fifteen (seen from the front). His own face in the mirror, his own hands, surprised him on every occasion.''
\ecb

I'm glad to hear you liked Matt's paper!!

Hey, tell {\Theo} Happy Thanksgiving from all of us in Cranford.  (I'm trying to get a big roaring fire going for tonight.)

\section{22-11-06 \ \ {\it The Usual Behavior of Chris} \ \ (to C. M. {\Caves} \& R. {\Schack})} \label{Caves90} \label{Schack108}

Let me give you a progress report as of 2:00 PM Wednesday, Nov 22, just before I take off until Friday morning (for Thanksgiving), whereupon I will start up on the project again.

Truth is, I've been mulling over and working on our paper since late last week, and corresponding privately with some of our critics.  Over the weekend I hit an extreme low point, and contemplated writing both of you that I would simply drop out of the project.  Upon carefully reading the reports and carefully reading again the paper as a whole, I just started to feel icky about the whole thing.  It dawned on me that the whole paper wasn't nearly as clear as I had pipe-dreamed it was and I didn't really feel like it represented my view on anything.  So why propagate the trouble I'm going to have to try to dig myself out of?  1) The paper doesn't convey any sense of the {\it privacy\/} of the single agent's description of his beliefs and the outcomes of his measurements, yet to be consistent requires it.  2) The paper is not consistent about the complete disconnect between the epistemic and the ontic.  (``Logical implication'' should be viewed is simply an instance of ``subjective certainty'' again---to talk of ``exceptions'' kills our whole point.)  3) It bandies about this term---this enemy---``the Copenhagen interpretation'' to complete distraction, when I myself think there are so many subtleties here that it would be best to never mention Copenhagen.

Anyway, that was my low point.  Since then I've started to climb the hill of optimism again, and at the moment at least, I am feeling like with enough {\it subtle\/} editing work I can pull it into a convincing package that has everything that {\it each\/} and {\it every one\/} of us wants.  I am trying.

{\Mermin} wrote me this:
\bdm
Eager to see your revised version.  Your problems confirm my
long-time practice of never having any collaborators if I can
possibly help it.  Particularly on the kind of paper you guys
are trying to write.  Reporting calculations is one thing, but what you're doing is just too subtle for a committee to get right.
\edm
He's right, you know.  I think we should think much harder about this in the future.

Below is a sample of some the changes made in Section 1.  Through them I am thinking that I am addressing the issues of three of the referees, but also {\Timpson}, Finkelstein, and von Baeyer.  I think I only have two days more work now that I'm on a positive, can-do note again.  (And my attitude really is positive again, even if the above does not convey it.)

I have every intention of shipping the thing to you before leaving for Japan Monday morning.

\subsection{Changes in Section 1}

\bq
\noindent Abstract.  Why did I change the abstract?  Predominantly because I was bothered by this sentence:  ``In this paper we investigate the concept of certainty in quantum mechanics, BECAUSE it is the with-certainty predictions of quantum mechanics that highlight the fundamental differences between our Bayesian approach on the one hand and Copenhagen \ldots''  This paper has slowly been morphing into a referendum on us against Copenhagen, and I don't like that.  For one thing, I don't know what the hell the Copenhagen interpretation is and no one else does either, and that is just setting ourselves up for years of pain.  I wish I hadn't allowed it to get this far, but I did and I can't back out now.  All I can try to do now is temper the damage, and then run for the hills as soon as I'm finished with this project.  In {\Ruediger}'s very first draft the word Copenhagen appeared 17 times; by the time we submitted to {\tt quant-ph}, the paper contained 28 instances.  Now, with Carl's latest modifications, we have 33.  We would have been better to never mention it once.\medskip

\noindent par.\ 2.  Too repetitive, so I trimmed it down.  Also made the logic flow a little better, and tried to tone down the ``fact for an agent'' which sticks out like a sore thumb.  Got rid of the parenthetical ``(or results)'' because we just don't use that terminology so much---doesn't need to be introduced.  Modified footnote on ``eliciting''---I don't like power-packed long sentences.\medskip

\noindent par.\ 3.  Got rid of the ``generally.''  Didn't see any need for is, as we make no admission to other kinds of conditionalizing, like Jeffrey, etc., in this paper.\medskip

\noindent par.\ 4.  Changed ``functions of'' of to ``depend upon'' because Referee 4 raised a flag.  Also changed ``probabilities must obey'' to ``gambling commitments should obey'' to make my Darwinian side feel better, also to head off the reading of the statement as a tautology.\medskip

\noindent par.\ 5, 6.  ``theory'' $\rightarrow$ ``world''.  Added ``purview'' and all kinds of stuff like that.  I took out the parenthetical ``(the Born rule)'' because people are not always so specific---they just like to flippantly throw out the phrase ``specified by physical law'' and act as if that means something obvious.  I know that you added that, Carl, to help introduce the next paragraph and to help it not seem so out of place (which it indeed is).  In general, I tried to round a lot of edges, connect things better, and build up to the next paragraph.\medskip

\noindent par.\ 7.  Fundamentally reworked.\medskip

\noindent par.\ 8,9.  In line with what I said above in the abstract, I changed descriptions to ``our main aim'' and ``along with this.''  The reason I added ``more carefully address'' is because previously we had already said, ``We have shown in a series of previous publications that \ldots\ all probabilities in quantum mechanics can be interpreted as Bayesian degrees of belief \ldots.  A consequence of the Bayesian approach is that all quantum states, even pure states, must be regarded as subjective.''  I also got rid of a little bit of repetition.  Also I backed off some instances of ``the Copenhagen interpretation'', and in places where the phrase ``and similar interpretations'' were invoked, I replaced it with the more compact ``Copenhagen-like interpretations.''  Note also that here and there (this paragraph and others) I peppered the phrase ``agent-independent'' to get across the idea of what is different in the Copenhagen-like interpretations.\medskip

\noindent (what had been) par.\ 10.  I've always thought this comment about not commenting on other interpretations had no natural place.  So, I've decided to place it in a footnote.  I think it much less breaks the flow this way.\medskip

\noindent 4th par.\ from last.  Added, ``(even when thought of as preparation devices simpliciter, rather than as quantum systems in a further analysis)'' and added a footnote to address Referee 3's concern about how Bohr dually thought of preparation devices as quantum and classical (just in different---complementary---modes of description).\medskip

\noindent Sect.\ 1, 3rd and 2nd par.\ from last.  Tightened up greatly to reduce repetition of previous paragraphs and also modified some language to try to alleviate confusion (of the sorts Timpson had).
\eq

\section{25-11-06 \ \ {\it Your Bravery} \ \ (to M. S. Leifer)} \label{Leifer9}

It dawned on me yesterday when I saw your posting that you are a brave man!  What a brave thing to get involved in any way with these loons, the subjectivist Bayesians!  I commend you.

By the way, I already got a little feedback from Charlie Bennett on your paper.  (Very dangerous to put ``deliciously'' and ``execrably'' in the same sentence.)  [See Bennettism in 22-11-06 note ``\myref{Bennett48}{CFS Philosophy, Ontic and Epistemic States and Evolutions, Funes the Memorious}'' to C. H. {\Bennett}.]

Guard your flank.

\section{25-11-06 \ \ {\it Hoffa and the Bayesians} \ \ (to C. H. {\Bennett})} \label{Bennett49}

I had a lovely email exchange with Matt Leifer this morning.  Below is Matt's reply, of which I asked him if I could forward it to you---I knew you'd enjoy the story at the end.  (When I see you, I'll tell you the one about the Jewish mystic I met on a plane.)  The ontological status of Jimmy Hoffa and subjective unitaries --- what is Osamu\index{Hirota, Osamu} going to think of us now?  (I just hope he won't slap me on the rear this time around.)

\subsection{Leifer Excerpt}

\bq
Finally, I think I should relate the story of what happened to me
last night.  Around 7pm I decided to go to a little Italian caf\'e for
dinner.  I had been writing an application for a faculty position
(which I should be finishing now), and was thinking about what I
should write about Bayesian stuff in my research statement.  Now, in this caf\'e the tables are quite close together and you often meet ``interesting'' characters.  It turns out
that the guy on the table next to me had some weird variant of
Tourettes that caused him to mumble incoherently to himself the whole
time.  At one stage, the conversation became coherent for just a few
moments and went as follows:
\begin{verse}
Him:	Are you studying here?\\
Me:		I'm a postdoc.\\
Him:	What subject?\\
Me:		Maths.\\
Him:	Do you like your......Bayesian.........Analysis?\\
Me:		Yes, I like it very much thanks.\\
\end{verse}
Of course, I tried to ascertain how he had known to say that, given
that I had not been talking out loud about this stuff anywhere
nearby, but the responses were all incoherent, mainly involving
``Bayesian arseholes'', ``lesbians'' and comments to the effect that he was more handsome than me.  In some ways this reminds me of your
story about the guy you sat next to on the airplane.  Is there some
sort of curse associated with writing papers on quantum Bayesianism?
Perhaps the savants of the world are trying to tell us something.
\eq

\section{25-11-06 \ \ {\it The Subjectivity of Convincing} \ \ (to R. W. {\Spekkens})} \label{Spekkens39}

\brws
I also wanted to mention that I am finally convinced that, within a Bayesian approach, even unitary maps must be considered to be subjective.  (Of course, I myself am not so sure that the Bayesian approach is the correct one, but that's irrelevant to the question.)  I remained unconvinced by your original argument, the recent CFS paper on certainty (actually it wasn't clear that it was arguing for this conclusion), and Matt's first draft of \quantph{0611233} (which he gave me a copy of).  But, after discussing it with Matt, he finally came up with an argument that did convince me (well, I haven't made a concerted effort to find fault with the argument, but intuitively, it seems right).  This is the one that is now found in sec.\ 5 of \quantph{0611233}.   Perhaps this argument was implicit in a previous argument made by one of you, but if so, it wasn't spelled out in enough detail for me to really get it until now.  If you haven't had a look at Matt's paper, you really should.
\erws

Yeah, I'm very proud of Matt's paper.  That's very nice work, and I am very impressed by him.

The CFS certainty paper made no mention of the issue.  1) It didn't need to for the point at hand (that objective facts don't determine quantum states), but 2) because {\Caves} is still an obstinate hold out on the issue of unitaries.  I'm hoping Matt's paper will finally make some dent in his armor.

Regarding your ``unconvinced'' point, that's fine.  But you shouldn't forget that ``being convinced'' depends upon a prior.  All argument of this variety is a function of 1) one's prior, and 2) the mass of evidence in comparison to the prior.  It is, just as you argue in your toy theory paper.  I became ``convinced'' a long time ago, not through one killer argument, but through several inconclusive arguments---programmable quantum circuits, teleportation of unitaries, the non-uniqueness of ancilla representations of quantum operations, the Choi-Jamio{\l}kowski isomorphism (though without a strong operational interpretation), various analogies to stochastic evolution equations (not as strong as Matt's), and so on.  But then the real cap was beauty and a sense of adventure.  The Bayesian edifice just looked sturdier and simpler without the {\Caves}ian cycles and epicycles (i.e., that ``true, there are some subjective, `effective' unitaries, but there are also the objective ones too'').  I understood that a conceptual leap might need to be made---i.e., that maybe we would never be able to make a strict logical argument {\it from\/} the subjectivity of states {\it to\/} the subjectivity of unitaries, i.e., that maybe it was an independent assumption.  But if anything, that thrilled me a little.

It gave me a small chance to play pretend to be William {\James}, and I could let W. K. Clifford be the proxy for my {\Carl} {\Caves}:
\bq\noindent
The talk of believing by our volition seems, then, from one point of view, simply silly.  From another point of view it is worse than
silly, it is vile.  [[As that]] delicious enfant terrible Clifford
writes: ``Belief is desecrated when given to unproved and unquestioned
statements for the solace and private pleasure of the believer\ldots.
Whoso would deserve well of his fellows in this matter will guard the purity of his belief with a very fanaticism of jealous care, lest at
any time it should rest on an unworthy object, and catch a stain
which can never be wiped away\ldots. If [a] belief has been accepted on insufficient evidence [even though the belief be true, as Clifford on
the same page explains] the pleasure is a stolen one\ldots. It is sinful because it is stolen in defiance of our duty to mankind. That duty is to guard ourselves from such beliefs as from a pestilence which may shortly master our own body and then spread to the rest of the
town\ldots. It is wrong always, everywhere, and for every one, to
believe anything upon insufficient evidence.''
\eq
Whereupon I could clench that, by his fears and lack of foresight, he was giving up the chance to make real progress in quantum foundations:  A REAL leap is needed to understand quantum mechanics---not just for us, but for any foundational direction in QM.  Pure logic won't do it; not for any of us (not for me, not for David Deutsch).  And to the extent that we quantum Bayesians really believe we smell the right direction, we should be willing to take a leap in it:
\bq\noindent
Believe truth! Shun error! --- these, we see, are two materially
different laws; and by choosing between them we may end by coloring
differently our whole intellectual life. We may regard the chase for truth as paramount, and the avoidance of error as secondary; or we
may, on the other hand, treat the avoidance of error as more
imperative, and let truth take its chance. Clifford, in the
instructive passage which I have quoted, exhorts us to the latter
course. Believe nothing, he tells us, keep your mind in suspense
forever, rather than by closing it on insufficient evidence incur the awful risk of believing lies. You, on the other hand, may think that
the risk of being in error is a very small matter when compared with the blessings of real knowledge, and be ready to be duped many times
in your investigation rather than postpone indefinitely the chance of guessing true. I myself find it impossible to go with Clifford. We
must remember that these feelings of our duty about either truth or
error are in any case only expressions of our passional life.
Biologically considered, our minds are as ready to grind out
falsehood as veracity, and he who says, ``Better go without belief
forever than believe a lie!''\ merely shows his own preponderant
private horror of becoming a dupe. He may be critical of many of his desires and fears, but this fear he slavishly obeys. He cannot
imagine any one questioning its binding force. For my own part, I
have also a horror of being duped; but I can believe that worse
things than being duped may happen to a man in this world: so
Clifford's exhortation has to my ears a thoroughly fantastic sound.
It is like a general informing his soldiers that it is better to keep out of battle forever than to risk a single wound. Not so are
victories either over enemies or over nature gained. Our errors are
surely not such awfully solemn things. In a world where we are so
certain to incur them in spite of all our caution, a certain
lightness of heart seems healthier than this excessive nervousness on their behalf. At any rate, it seems the fittest thing for the
empiricist philosopher.
\eq
Anyway, that was my little sense of adventure.  And Matt's work is a great cushion at the bottom of the leap.  Actually it's much more than that:  For it gives us a proper formal structure to start building upon.

By the way, with regard to this whole method of reasoning---i.e., beating down the opponent with a BIG mass of inconclusive evidence---I was reminded of your example (i.e., the toy theory paper) by a passage in Marcus {\Appleby}'s submission for the {\Vaxjo} proceedings with regard to interpreting probability.  Let me dig that up and paste it in for you.

\bq\noindent
The complaint of the Bayesians about the orthodox statistical
methodology has always been that it is (in the words of de Finetti)
``ad hoc'' and ``arbitrary''. Jeffreys makes the point with
characteristic irony when he says of Fisher (one of the founding
fathers of the orthodox methodology)
\bq\noindent
I have in fact been struck repeatedly in my own work,
after being led on general principles to the solution
of a problem, to find that Fisher had already grasped
the essentials by some brilliant piece of common sense.
[Jeffreys, p.\ 393]
\eq
This is, in a way, a compliment. However, the compliment is
distinctly back-handed:  for what Jeffreys is really saying is that
Fisher, notwithstanding his confusions and inconsistencies, often
contrives to get the right answer owing to the power of his
intuition.  It is rather as if a physicist were to congratulate a
snooker player on his ability to pot a ball notwithstanding his
ignorance of Newtonian mechanics; or to congratulate a fish on its
ability to swim notwithstanding its ignorance of the principles of
hydrodynamics.  I have argued elsewhere that that criticism is amply justified.  Generally speaking what drives the Bayesian school of
thought is a desire for clarity and logical cogency. By contrast the orthodox statistical methodology is driven by what Jaynes describes
as an ideological conviction that, if statistics is to be scientific, then probability distributions must be conceived as objectively real
entities. To attain that ideological end orthodox statisticians are
willing to make whatever sacrifice of logical coherence seems
necessary.
\eq

Happy Thanksgiving.

\section{26-11-06 \ \ {\it Jimmy Hoffa's Bones} \ \ (to C. H. {\Bennett})} \label{Bennett50}

\bq\noindent
   ``The hypothesis that there is an external world, not dependent on
    human minds, made of {\em something}, is so obviously useful and
    so strongly confirmed by experience down through the ages that we
    can say without exaggerating that it is better confirmed than any
    other empirical hypothesis.''\medskip\\
\hspace*{\fill}           --- Martin Gardner [and Chris Fuchs]
\eq
Just packing up and saw this on one of my transparencies, and thought of you.

\section{26-11-06 \ \ {\it Sample Talk} \ \ (to R. E. Slusher)} \label{Slusher17}

By the way, here's a sample of my present talks, just in case you would need it for anything.

\bq\noindent
Quasi-Orthonormal Bases for the Space of Density Operators \medskip

\noindent Recently there has been much interest in the quantum information community to prove (or find a counterexample to) the existence of so-called symmetric informationally complete measurements (SICs).  In this talk we show that there should be even more interest.  For, under a robust measure of orthonormality for operator bases (one that does not build in any symmetry at the outset), one can show that SICs, if they exist, come as close as possible to being orthonormal bases for the space of density operators.  Moreover, in contrast to the usual expression of the superposition principle (where bases are taken to be orthogonal sets of state vectors), writing a superposition principle in terms of SICs leads to a more intrinsically-quantum representation for quantum states.  This is because the basis states, rather than being the easiest to eavesdrop upon (as the usual ones are), are actually the hardest.  Furthermore, such states fulfill a few other extreme non-classical properties that make them very interesting.  Because of all this, writing the quantum-state space in these terms gives hope for a direct derivation of it from a plausible information-theoretic constraint principle.  Various aspects of this problem will be discussed.
\eq

\section{28-11-06 \ \ {\it Funes $+$ Borges} \ \ (to C. H. {\Bennett})} \label{Bennett51}

On the flight over, I read the Borges story you sent me.  It was great and much more meaningful than I had expected it to be.  That guy really builds atmosphere.  Anyway, you were right:  there is certainly a sympathy between my thoughts and Funes's.  But in another aspect of it, there is also a significant sympathy with Borges's.  Particularly the line ``To think is to forget a difference, to generalize, to abstract.  In the overly replete world of Funes there were nothing but details, almost contiguous details,'' struck me.  Though, for me, the importance is not quite in ``forgetting a difference'' but rather in paying no attention to it.

You'll see the similarity between what I think of quantum measurements and this point of Borges's if you dare enter the attached file [my paper ``Delirium Quantum,'' \arxiv{0906.1968}].  The appropriate part is Section 3, titled ``Snowflakes.''

\section{28-11-06 \ \ {\it Whose Lagrangian?}\ \ \ (to M. A. Nielsen)} \label{Nielsen5}

I'm shamefully ignoring a talk at QCMC in Japan at the moment.  I'm writing because I just looked at the scheduled and discovered that you're not here.  I don't know why, but I had it in my head that you were going to be here (maybe you were on an earlier schedule?).

Anyway, I was looking forward to talking to you.  I had wanted to discuss a paper with you posted by Matt Leifer last week---its number is \quantph{0611233}.  Particularly, I guess I wanted to know whether it alleviates any of your troubles about this whole Bayesian quantum state and operation business that {\Carl}, {\Ruediger} and I are trying to develop.  I remember your asking in Konstanz what good is the `search' for a fundamental Lagrangian if they're subjective (in the Bayesian sense) anyway.  And I remember your not looking too pleased with my answer (I also talked to Menicucci about this later)---fair enough, it wasn't a very good answer.  But now we've got a little more to work with, so it seemed like a good time for me to {\it try\/} to develop some answers.

(Hideo's sitting beside me doing his email too.  Two times shameful between the two of us.)

\subsection{Michael's Reply}

\bq
I haven't seen Matt's paper.  I'll be curious to have a look, although it
may take a while [\ldots]

On the general issue: as I recall, I essentially asked whether or not the
people who'd been searching for a single universal Lagrangian had been in
search of a chimera.  Your answer was, I think it's fair to say, rather
equivocal --- you didn't say ``yes'', but you didn't give a clearcut ``no'',
either.  Given the astounding success the program of searching for a
Lagrangian has had, I think any answer short of a resounding ``No'' spells
big trouble for the Bayesian program.
\eq

\section{29-11-06 \ \ {\it Whose Lagrangian?, 2}\ \ \ (to M. A. Nielsen)} \label{Nielsen6}

Quick question, upon which I'll then chew on the answer more slowly \ldots\ being careful to write you again only after you have some time to breathe.  Suppose a classical world, as was supposed, for instance in the days of Gibbs and Boltzmann.  Given the astounding success (at that time) of the canonical distribution for a wide range of statistical mechanical and thermodynamical calculations and predictions, would that have spelt big trouble for a Bayesian view of statistical mechanics?

\subsection{Michael's Reply}

\bq
I don't think the situation is at all analogous, for at least two reasons.

First, much of the importance of the ``quest for a Lagrangian'' is as a
guide to how we as physicists should work, and not so much as a theory of
how the world is (unlike stat mech).  The Lagrangian viewpoint might be
quite wrong, yet that wouldn't change the fact that it's been hugely
useful as a guide to making progress.  Indeed, even if it does turn out to
be wrong, it's reasonably likely to be replaced by some similar guiding
principle.  If the Bayesian point of view says we shouldn't look for a
Lagrangian, I therefore view that primarily as a strike against
Bayesianism, not Lagrangians.

Second, the Bayesian point of view explicitly addresses some shortcomings
of the old stat mech formalism.  I doubt very much that you're claiming
that this point of view is a serious contender for explaining some
shortcoming in the modern approach to quantum field theory.
\eq

\section{29-11-06 \ \ {\it The Funesian Path} \ \ (to C. H. {\Bennett})} \label{Bennett52}

Now that you've got me thinking about this issue more than I had been before, it seems I keep running across things that bring it back to the top of my attention.  Reading the book {\sl The Pragmatic Humanism of F.~C.~S. {\Schiller}\/} in this lovely Japanese bath tub, I came across this passage:
\bq\noindent
It is Aristotle's logic which is usually taken as the archetype of formal logic.  However, the `logical analysis of judgment' in anyone's hands is bound to come to failure if it discounts the personal intention and circumstances of the maker of the judgment.  Thus Hegel's formula of judgment as identity in difference disregards the fact that identities are always made or postulated by us through conscious disregard of those differences which we take for the moment to be irrelevant to the purpose of judgment.
\eq

\section{29-11-06 \ \ {\it Thinking of Brad}\ \ \ (to S. J. Lentz)} \label{LentzS7}

Hi from Japan.  I'm so glad to hear that Brad seems to have made it through the operation OK; I told so many people about it here yesterday, it became a topic of conversation.  Tell Brad that there were a load of quantum information physicists in Japan worried about his health!

Thinking of Brad's interest with the {\sl Eats Shoots And Leaves\/} thing, as I was lying here with jet lag, it dawned on me that you might take him this class of examples to think about.  (It was one of the topics of the dinner conversation last night.)
\begin{itemize}
\item
``Buffalo buffalo buffalo buffalo.''
\end{itemize}
If you parse it right, you'll see that it's a proper English sentence!  Not a lie.  (Don't tell him right away, but one reading of it can go like this:  ``The buffaloes from the city Buffalo bamboozle their fellow buffaloes everywhere.''  The reason this can be made to work is because one spelling of the plural of the noun ``buffalo'' is ``buffalo''.  When buffalo is used as a verb, it means bamboozle or baffle.)

But it gets worse, buffalo can also be an adjective.  My Webster dictionary gives it this definition as an adjective, ``of the kind of style prevalent in Buffalo, NY.''  So,
\begin{itemize}
\item
``Buffalo buffalo buffalo buffalo buffalo.''
\end{itemize}
is also a proper English sentence.  (Those ``Buffalo buffalo'' are a very self-deceptive lot.)  And it can actually be made still worse, with even more repetitions of buffalo, if you work at it.

And now here's a different kind of example---this one invented by Charlie Bennett many years ago.  If you cycle through the words ``toll, house, cookie, delivery, and truck'' you can make an infinite number of meaningful, distinct concepts.\footnote{Charlie also has a method of modifying this example so as to produce a {\it continuous infinity\/} of meaningful, distinct concepts!}  For instance,
\begin{itemize}
\item
Toll                               (that's meaningful enough)
\item
Toll House                         (meaningful too)
\item
Toll House Cookie                  (you know what those are)
\item
Toll House Cookie Delivery         (if we only had a company like that
                                    in Cranford!)
\item
Toll House Cookie Delivery Truck   (a truck that delivers those
                                    delicious cookies)
\end{itemize}
Now here's where it starts to get fun.

\begin{itemize}
\item
Toll House Cookie Delivery Truck Toll \\
     (it's silly but it's meaningful; it's a special charge required
      for toll house cookie delivery trucks to use the road)

\item
Toll House Cookie Delivery Truck Toll House \\
     (ok, those trucks have to take a special lane when they pay their
      tolls)

\item
Toll House Cookie Delivery Truck Toll House Cookie \\
     (a toll house cookie that happens to be in the toll house cookie
      delivery truck)

\item
Toll House Cookie Delivery Truck Toll House Cookie Delivery \\
     (not much market for it, but a very specialized company that
      delivers toll house cookies from the original toll house cookie
      delivery truck)

\item
Toll House Cookie Delivery Truck Toll House Cookie Delivery Truck \\
     (and that's the truck employed by that weird delivery service)
\end{itemize}
You get the pattern now.  And it can be made to go on forever, if you think about it.  For instance, those weird delivery trucks from the last example could also be tolled.  And there must be a place from where that toll is taken.  And so on, and so on, and so on.

When Brad's ready to think a little, see if he enjoys these.

\section{30-11-06 \ \ {\it Early Morning Rabbits} \ \ (to D. Gottesman)} \label{Gottesman2}

It's not very conclusive evidence, but it is along the lines of what I thought I remembered hearing:
\bq\noindent
   There is another myth that says that there are rabbits on the moon
   making mochi, pounding away with their mochi maker, since there is a
   rabbit shaped series of dark areas on the moon, according to
   Japanese myth \ldots\ then there is another myth that Tsuyoiko's rabbit is
   descendant from the rabbit on the moon \ldots
\eq
The more important outcome of this, though, is that I learned that the rabbit I have been ``seeing'' all these years is quite different from the ``official'' image:  See:
\bq
\myurl{http://en.wikipedia.org/wiki/Moon_rabbit}.
\eq

\section{01-12-06 \ \ {\it Chris's House from Space} \ \ (to K. R. Duffy)} \label{Duffy5}

Ahh Ken, I miss you!  I myself wouldn't say it's burnt Lamborghini orange---a little more subtle than that---but if having Lamborghini in the name will help me sell the place when the time (eventually, surely) comes, I'll use it liberally.  Attached is a colorful picture for your further pub stories.

And while I'm here, let me advertise the opening of the Orange House Quantum Information-Foundations Seminar Center.  That's what's in the second picture.  My girls have agreed to rent the space out from time to time, as the needs of science require.  Since that picture was taken, the two front columns have been rigged with fasteners for holding a large (green) chalkboard across the front, when meetings are in session.  (If you look carefully, you'll notice that the main walls are constructed completely of doors; there are 12 in total.  Kiki ripped them out of a nearby 1890's house.  And bless her soul, she built the whole place by herself, from floor to roof, without an ounce of help from her worthless husband or anyone else.)

I should send you a picture of the newest member of the family too, but I won't clog the airwaves.  Kiki decided that our retriever Murphy shouldn't be alone.  So, he now has a spry little border collie companion, Beamish.  Now, don't you think Ireland made some impression on us?

Finally to answer your question:  Yes, the house can be seen on Google Earth \ldots\ but not because of the color!  It's an old photo, most likely from a time before we purchased the house (the old roof can still be seen, and that is one of the first pieces of construction we had done).  Here are the coordinates, which I had written up for a relative some time ago: [\ldots]

Too bad you can't see the orange.

Greetings from Tsukuba Science City!  Wish you guys were here.  I'll try to get Charlie Bennett to send you pictures of the (before-cooked) menu selections he, Bill Wootters, and Lev Levitin had the other night.  I don't recall the animal, but three of the items were ``womb'', ``rectum'', and ``kidneys''---I nearly fainted at the sight.

\section{03-12-06 \ \ {\it Our Stuff}\ \ \ (to M. Sasaki)} \label{Sasaki5}

I found the Bartlett paper on the web, i.e., the one he presented to me at the poster session.  You might enjoy it too; it is:
\quantph{0608037}.

As you find some free time, let us develop this theory of ours more!

Thank you again for providing me with a wonderful week.  I truly enjoyed this meeting.  I think it was a great success.

\section{05-12-06 \ \ {\it The Future of Quantum Information}\ \ \ (to O. Hirota)} \label{Hirota5}

I just wanted to write you a short note of thanks for making it possible for me to come to Japan.  The meeting was wonderful; I learned so much from it.  Particularly I was happy for the venue it gave me to continue my debate with Charlie and the progress I made therein.  The issue is not just one of philosophy, I think, but rather something that will give rise to deep technical issues, and through them, a new means to move quantum information forward.  So, I should thank you too in proxy for ``the future of quantum information''!

In that regard, let me point you to two excellent papers that demonstrate the more technical side of the debate:
\begin{itemize}
\item
\quantph{0401052}
\item
\quantph{0611233}
\end{itemize}
If you have any students interested in such matters, I hope you will put them in the service of this army!  More seriously, not everything between Charlie and me is a joke.  Some of it leads to good equations!

\section{06-12-06 \ \ {\it Fourth and Fifth} \ \ (to J. {\Barrett})} \label{Barrett1}

I'll address your points ``Fourth'' and ``Fifth'' right away, and I'll come back with a longer, more specific reply to your note as a whole maybe by the end of next week.  (I'm just in the middle of significantly revising \quantph{0608190}, and that has to be done before I go to the APS sorter's session, which will block out my time through the end of this week.)

Funny that you bring these things up, because I've provisionally titled my Shimony festschrift contribution, ``Even Quantum Bayesians Like a Little Ontology from Time to Time.''  Anyway, for a provisional (though overly poetic) answer, see
\begin{enumerate}
\item
Sections 6 and 7 of my paper ``Delirium Quantum''
\item
Maybe also Section 4 in it, picking up at the paragraph that starts  with ``Why is this not solipsism?'', and
\item
the paragraphs below that were written in response to one of the points of one of the 4(!) referees for the present paper.
\end{enumerate}

If these passages provoke any other thoughts or other questions in you, feel free to write me with them, and I'll try to incorporate the issues into my larger response next week.  (I.e., feel free to mail me any further thoughts, just be prepared for my not replying for a little while.)

\bq\noindent
\underline{Remarks to a Referee of \quantph{0608190}}\medskip

This paper does not emphasize it, but no, we do not mean ``the facts
we are talking about here are facts for everybody.''  We mean that
the facts too are personal, though in a very careful sense.

We have tried to meld the phrase ``fact for the agent'' a little
better than previously into its surrounding text---so that it is
there at least as some shadow of our presently agreed upon view---but
it will have to remain somewhat awkward in the present context.

A better exposition would have emphasized the full setting of our
view, including some discussion of the ontological---i.e.,
noninstrumental---pieces of it.  1) There are two physical systems in
the story of this paper.  One takes the role of the agent---the
possessor of subjective degrees of belief and the activator of the
measurement process.  The other takes the role of the object system.
2) What is being discussed when one speaks of gathering data in the
quantum context is an interaction or transaction between the agent
and the object.  3) Without the agent, there would be no quantum
``measurement'' to speak of, but without the object, there would be
no means for the agent to obtain the data, the spikes upon which he
pivots his probabilities (his subjective degrees of belief).  4)
There are no other agents in this story.  5) We do not go any further
than points 1-5 warrant by giving the data obtained in a quantum
measurement an autonomous existence---for instance, as something
beyond the agent's sensations. That would run into inconsistencies in
a ``Wigner's friend'' scenario.  Nonetheless, quantum measurement
outcomes are beyond the control of the agent---they are only born in
the interaction---and thus are not functions of the agent in the way
that his degrees of belief are. The degrees of belief find their
source in the agent (the subject); the outcomes find their source in
the external quantum system (the object). But the outcomes lead back
to the agent in that they are personal to him.

This does not mean that the agent-object interaction and its fuller
consequences are not autonomous events in spacetime, but it does mean
that if there are any such things, quantum theory is not directly
concerned with them. In CAF's view, quantum theory takes its whole
definition as a normative theory for organizing an agent's {\it
personal\/} probabilities for the {\it personal\/} consequences of
his interactions with external physical systems.  The structure of
the agent-independent world that lies behind quantum mechanics is, in
the end, still codified by the theory, but only in a higher-order,
more sophisticated fashion than had been explored previously---it is
through the normative rules, rather than quantum states and
Hamiltonians.

Unfortunately a more detailed account like this will have to wait for
another paper.
\eq

\subsection{Jon's Preply}

\bq
Here's my own take. Parts at least are different from yours, so you might not like it.

First, both probabilities and quantum states are descriptions of degrees of belief and are thus subjective.

Second, quantum theory doesn't need to specify what these beliefs are
about, or What is a measurement?\ any more than classical probability
theory needs to specify a sample set, or What is a measurement? Classical probability theory has such broad application just because it does not specify the sample set, it can be applied to any betting scenario. Classical probability theory may be derived with Dutch book arguments. The hope is that quantum theory can be derived from similar arguments supplemented with a small number of physical postulates, which get us to Hilbert space, whence Gleason's theorem can take over. Quantum theory could then be applied in any betting scenario in which those postulates are true.

Third, classical probability theory on its own is not a theory of physics, it is a tool that can be applied in physics. A theory of physics needs to specify an ontology of events that agents can, at least in principle, bet on. Thus Newtonian mechanics gives an ontology of events (system at particular point in phase space at time t) that one could in principle bet on. Newtonian mechanics plus classical probability theory gives Liouville mechanics. By the same lights, quantum probability theory needs to be supplemented with a theory of physics that provides an ontology of events to bet on.

Fourth, at present we have only a provisional, half-baked theory in which these events are ``outcomes of interventions''. But a fundamental formulation of a complete theory must not use such vague terminology, and will eventually supply an ontology of primitive events that agents can in principle bet on.

Fifth, the relationship between this fundamental theory and current theory need not be anything so simple minded as a hidden variable interpretation of quantum theory, in which the fundamental theory supplies instruction sets, and quantum states are distributions over instruction sets. In this light, worries along the lines of, ``but hidden variables must be contextual'' are too shallow. In a fundamental theory, space and time themselves may well be emergent, in which case worries along the lines of ``but hidden variables must be nonlocal'' are also too shallow.

As I understand you would resist the fourth and fifth here, insisting that quantum theory is complete and fundamental, that the objective events I speak of are not supplied by further physical theory, but are simply ``outcomes of interventions'', and that pace Bell, ``Against Measurement'' etc, there is nothing wrong with this. But might I ask why? The view I outlined seems supported by many of your arguments, in so far as quantum states are subjective. It is relatively humble (physics is not there yet, it's not even close). It also avoids mysticism (we will never understand the universe, it's too ineffable, thus don't aim for a god's eye view, just predict measurement outcomes).
\eq

\section{06-12-06 \ \ {\it Made Wrong} \ \ (to J. {\Barrett})} \label{Barrett2}

Another quick point before I fade out of contact.  With regard to:
\bjoba
What is my point? Just that beliefs, described by quantum states, can
be objectively right or wrong, and we are forced to this point by the
need for objective outcomes of bets. This need not imply a complete
instruction set for every quantum system. But it does imply that a
theory must specify an ontology of events that give objective wrong
status to (at least some) quantum states.
\ejoba

Go to the file titled ``Cerro Grande II'' at the very bottom of my webpage, and in it do a search on the phrase ``made wrong''.  The essays surrounding that phrase may go a little way toward explaining my take on what you just said.  This very point is why I keep stressing the resemblance between quantum mechanics and pragmatic theories of truth---{\James}, {\Schiller}, {\Dewey}---rather than correspondence theories, which, I think, lies in the background of your remark.

\section{07-12-06 \ \ {\it 0611283} \ \ (to C. H. {\Bennett})} \label{Bennett53}

\bcb
I never liked GRW much, but this paper, rather rudely refuting the infamous ``Free Will'' paper, mentions our differences in its introduction, e.g.\ my seeking for a quantum theory in which the observer plays no role.
\ecb

Well, I'm all for getting rid of the observer.  But my strategy has always been to try to isolate which parts of quantum theory are explicitly about observers, and which parts aren't.  I don't believe the latter is an empty set.  But I don't believe the former is an empty set either---thus it needs to be carefully excised, and that is what I've been shooting for for a long time now.  Perhaps I'm wrong, but I view it as a very careful approach, whereas things like GRW strike me as sheer speculation, based on rather fantastic ideas.

I'll look at the paper you recommend later tonight (I'm in DC, gonna go see some monuments before it gets dark).

In the meantime, do send me a copy of your Tsukuba talk.  I want to think about it again.

\section{07-12-06 \ \ {\it Adding and Subtracting} \ \ (to C. H. {\Bennett})} \label{Bennett54}

Here's another way to put what I wrote to you in the last note that strikes me as a possibly useful formulation.  Traditionally the tack that's been taken in trying to recover an observer-free quantum mechanics (i.e., the thing you say you're seeking) has been in trying to {\it add\/} something to the raw formulation:
\begin{verse}
Bohm    \  $\longrightarrow$  \ hidden variables\\
GRW     \  $\longrightarrow$  \ a collapse mechanism\\
Everett \  $\longrightarrow$  \ a plethora of unseen worlds
\end{verse}

What I say of myself is that I too am seeking an observer-free formulation of the content of quantum theory.  But where I differ from the pack above is that, rather than adding anything to it to get the job done, I think it's just the opposite:  Some elements need to be taken away.  There's too much stuff in quantum theory as it stands---for, some of it is explicitly about observers.  (Otherwise there wouldn't be such a resemblance to good {\Spekkens}'s toy theory \ldots\ which is, by construction, about states of knowledge.)

\section{07-12-06 \ \ {\it The Cantankerous Colleague} \ \ (to R. {\Schack} \& C. M. {\Caves})} \label{Caves91} \label{Schack109}

It's a little after 12:30 AM, December 7, and I think I'm going to call it quits for now.  I'll ship you all the changes I've made, along with replies to the referees, and a list of most changes (with explanations for why I made them).  It's my hope that {\Carl} will be able to download these things and read them before we meet this afternoon.  (Be careful to take your blood pressure pills first.)

I think I have pretty much covered everything, except:
\begin{enumerate}
\item
Referee 2's point about extending the discussion of the distinction between ``truth'' and ``probability 1''.  That's probably an important point, but given the contention between us, I didn't feel like going into this now.  And maybe we shouldn't touch it at all \ldots\ ever.
\item
The needed extension of text around the sentence, ``The main issue here is whether there is any difficulty with the idea of an
utterly certain belief about an admittedly contingent fact.''  I think we can profitably build something based on a reply to {\Timpson}'s criticism (pasted in the referee replies document for your convenience).  It's quite an important point I think, but I've just lost steam now.
\end{enumerate}

Finally, I think it would be useful to give {\Mermin} some more extended replies.  Particularly we should say something about why this point he keeps bringing up---the one about how we should simply focus on never making probability-1 statements---is irrelevant, and just plain off the point.

In total, at this late stage, I don't like the paper very much.  {\Mermin}'s report, but maybe more so Referee 4's, made a big impression on me.  I became ashamed of myself, knowing that I knew better and that if I had just been more engaged this year, maybe everything would have been OK.  {\Ruediger} is right:
\brs
But without the joint effort it would not have been written at all.
And are you really confident that any of us would have written a
better paper by himself?
\ers

It likely would not have been written, but I don't know---the main delay on it, I think, was the issue of consensus.  Certainly, I am not confident that I would have written a better paper by myself, but it would have been a very different paper.  I've come to detest this business about the Copenhagen interpretation that runs throughout it.  I've tried my best to cover us against the community, but in my heart I know it's basically a lie (there's plenty of stuff out there on Bohr and {\Pauli} that slaps us right in the face).  I don't like that we never got ``metaphysical'' in the paper anywhere---without the imagery in the background, most readers will think that there's little but another version of positivism in it (the world is sense impressions) and/or they'll bring in their prejudices about what ``facts'' and ``truth'' are (correspondence and coherence theories), ones that I certainly don't share and I know that I could guard myself against in a paper that didn't demand consensus.  You think you guard against those misimpressions by being quiet about them; I think it's just the opposite.  {\Mermin}'s frustration was acute.  And the issue of the exception to the subjectivity of probability---I find it so dangerous and so distasteful.

The trouble is my mother didn't raise a very self-confident son.  For some reason, on this project---this now 5-year-old project---I urgently felt the need for consensus between the three of us.  Consensus was never reached in any but a superficial way.  We would have been better off to all write our own papers on our own thoughts.  The papers would have been finished long ago, and we could think in a civilized way how we agree or disagree with each other and maybe make some progress that way.

\section{07-12-06 \ \ {\it Changes I Made To Certainty} \ \ (to R. {\Schack} \& C. M. {\Caves})} \label{Caves92} \label{Schack110}

\subsection{Changes in Section 1}

\begin{itemize}
\item
Abstract.  Why did I change the abstract?  Predominantly because I was bothered by this sentence:  ``In this paper we investigate the concept of certainty in quantum mechanics, {\it because\/} it is the with-certainty predictions of quantum mechanics that highlight the fundamental differences between our Bayesian approach on the one hand and Copenhagen \ldots''  This paper has slowly been morphing into a referendum on us against Copenhagen and I don't like that.  For one thing, I don't know what the hell the Copenhagen interpretation is and no one else does either, and that is just setting ourselves up for years of pain (as effectively Referee 4's report attests to).  I wish I hadn't allowed it to get this far, but I did and I can't back out now.  All I can try to do now is temper the damage, and then run for the hills as soon as I'm finished with this project.  In {\Ruediger}'s very first draft the word Copenhagen appeared 17 times; by the time we submitted to {\tt quant-ph}, the paper contained 28 instances.  Now, with {\Carl}'s latest modifications, we have 33.  We would have been better to never mention it once.

\item
Abstract and throughout paper.  I replaced all instances of ``with-certainty'' with ``probability-1''.  I'm not adamant about this, but the new phrase did strike me as unneeded, and that the reader would more easily have a sense of what ``probability-1'' means and where the paper is going than taking a time to ponder over what we might mean with this never-before-seen adjective ``with-certainty.''  There were only four instances of ``with-certainty'' in the whole paper anyway.

\item
Abstract.  Left {\Stairs} reference, but took out phrase ``originally due to'' and replaced it with ``of.''  Don't want our abstract to seed a priority battle (even though {\Stairs} says that there is a version of the argument in his 1978 thesis, predating Heywood--Redhead significantly).  Anyway, in this way, our statement is certainly true:  I don't understand the Heywood--Redhead paper (so we can't be giving a version ``of'' their argument).

\item
par.\ 2.  Too repetitive, so I trimmed it down.  Also made the logic flow a little better, and tried to tone down the ``fact for an agent'' which sticks out like a sore thumb.  See discussion with Referee 4 for this point.  Got rid of the parenthetical ``(or results)'' because we just don't use that terminology so much---doesn't need to be introduced.  Modified footnote on ``eliciting''---I don't like power-packed long sentences.

\item
par.\ 3.  Got rid of the ``generally.''  Didn't see any need for it, as we make no admission to other kinds of conditionalizing, like Jeffrey, etc., in this paper.  Changed ``probabilistic argument'' to make Referee 4 happy.  Also changed ``functions of'' to ``depend upon'' because Referee 4 raised a flag.

\item
par.\ 3.  Also changed ``probabilities must obey'' to ``gambling commitments should obey'' to make my Darwinian side feel better, also to head off the reading of the statement as a tautology.

\item
par.\ 5, 6.  ``theory'' $\longrightarrow$ ``world''.  Added ``purview'' and all kinds of stuff like that.  I took out the parenthetical ``(the Born rule)'' because people are not always so specific---they just like to flippantly throw out the phrase ``specified by physical law'' and act as if that means something obvious.  I know that you added that, {\Carl}, to help introduce the next paragraph and to help it not seem so out of place (which it indeed is).  In general, I tried to round a lot of edges, connect things better, and build up to the next paragraph.

\item
par.\ 7.  Fundamentally reworked.

\item
par.\ 8, 9.  In line with what I said above in the abstract, I changed descriptions to ``our main aim'' and ``along with this.''  The reason I added ``more carefully address'' is because previously we had already said, ``We have shown in a series of previous publications that \ldots\ all probabilities in quantum mechanics can be interpreted as Bayesian degrees of belief \ldots.  A consequence of the Bayesian approach is that all quantum states, even pure states, must be regarded as subjective.''  I also got rid of a little bit of repetition.  Also I backed off some instances of ``the Copenhagen interpretation'', and in places where the phrase ``and similar interpretations'' were invoked, I replaced it with the more compact ``Copenhagen-like interpretations.''  Note also that here and there (this paragraph and others) I peppered the phrase ``agent-independent'' to get across the idea of what is different in the Copenhagen-like interpretations.

\item
(what had been) par.\ 10.  I've always thought this comment about not commenting on other interpretations had no natural place.  So, I've decided to place it in a footnote.  I think it much less breaks the flow this way.

\item
4th par.\ from last.  Added, ``even when derived from the ``ultimate measuring instruments'' of Ref.~\footnote{Letter from Niels Bohr to Wolfgang Pauli, 2 March 1955 (provided to us by H.~J. Folse).},'' again to address Referee 4's concerns.

\item
Sect.\ 1, 3rd and 2nd par.\ from last.  Tightened up greatly to reduce repetition of previous paragraphs and also modified some language to try to alleviate confusion (of the sorts {\Timpson} had).  Changed ``originally due to {\Stairs}'' to ``of {\Stairs}'' and also gave credit to Heywood--Redhead and Brown--Svetlichny as ``independent'' variations on the theme.  Also inserted ``noncolorability'' to better set apart the KS result from itself being a full no-preexisting-property proof.
\end{itemize}

\subsection{Changes in Section 2}

\begin{itemize}
\item
par.\ 1, last sentence.  Changed to ``we shall say that that facts are objective in the sense that \ldots'' to make it a little clearer that this is our definition of objective, rather than a categorical statement (potentially carrying the reader's own connotations, rather than ours).

\item
par.\ 2, last sentence.  Removed `in probability theory' because a) we don't elaborate on what that means, and b) because it created too many questions in Ref.\ 2.

\item
par.\ 3.  `takes this category distinction seriously' $\longrightarrow$ `takes this category distinction as its foundation'.

\item
par.\ 4.  Added ``predominantly'' because I modified (extended) the quantum discussion at the end of the section.

\item
2nd par.\ of PP discussion.  Changed `typical' to `exchangeable' because typical conveys the sense the there is a correct class of priors.  Why not be specific?  Stripped away the closing sentence, ``The whole idea, however, is plagued with a fundamental conceptual difficulty'' because it is not at all justified by our nonquantum, coin toss example.  It can't be a ``fundamental conceptual difficulty'' if it potentially works in the quantum world.  So, I played up the point that instead that chance simply serves no useful role.

\item
3rd par.\ of PP discussion.  Built a new transition into this paragraph.

\item
Last par.  Drastically reconstructed to take into account Referee 4's point about how we have the logic of the issue backward.  Also humbled the presentation somewhat.
\end{itemize}

\subsection{Changes in Section 3}

\begin{itemize}
\item
par.\ 4.  I removed the parenthetical phrase ``(often called the {\it model\/})'' (and the single further instance of ``model'' later in the paper), because I saw no use for the terminology.  No need to introduce an essentially unused term.  I also italicized ``exception'' and introduced an endnote to distance myself from the ``exception pack''\footnote{This was the footnote (it did not ultimately appear in print): ``One of the authors---CAF---strongly dislikes exceptions on matters of principle, and the present exception is no exception.  The problem, {\it as he sees it}, with the formulation that there is a case whereupon a fact strictly determines a probability is that it puts the horse before the cart:  It acts as if abstract `logical implication'---a statement of the form ``$d\Rightarrow h_0$''---has a meaning independent of the {\it subjective\/} judgment ``$\Pr(d|h)=0$ for $h\ne h_0$.''  (See for instance, F.~C.~S. {\Schiller}, {\sl Formal Logic: A Scientific and Social Problem}, (Macmillan, London, 1912), and F.~C.~S. {\Schiller}, {\sl Logic for Use: An Introduction to the
Voluntarist Theory of Knowledge}, (G.~Bell, London, 1929), for a development of this point.) Therefore the reader should associate neither the sentence of text cited by this endnote, nor the three sentences following it, with the views of CAF\@.  Since it is CAF's belief that these sentences undermine the very consistency of the quantum Bayesian point, it is a lucky thing that, {\it from his point of view}, they are not actually used anywhere in the paper's argumentation. Instead, they should be treated as dangling appendages that will be surgically removed in any subsequent CAF publications.''}.  By the way, I think the paper would be distinctly better if we could turn off this annoying feature of the REV\TeX\ 4 style file that turns footnotes into endnotes.  Independent of my definitely wanting this note of mine to be seen, it would help ensure that {\it all\/} notes aren't simply overlooked as references, and the reading would flow easier.  Is there anything in the paper that distinctly depends on the REV\TeX\ 4 style?  We are almost surely going to have to change style for submission to SHPMP anyway.

Just to be snide, by the way, I will point out that the following sentences of {\Carl}'s latest draft (i.e., the one just before my present modifications)
\begin{enumerate}
\item
``We first review the main arguments for the general claim that probabilities always represent degrees of belief.''
[in the abstract]
\item
``The updated probabilities always depend on the agent's prior
probabilities as well as on the data and thus can be different
for agents in possession of the same data.'' [in the introduction]
\item
``We argue, following de Finetti \footnote{B. de~Finetti, ``Probabilismo,'' Logos {\bf 14}, 163 (1931);
translation: ``Probabilism,'' Erkenntnis {\bf 31}, 169--223 (1989).}, that in
the last analysis probability assignments are always subjective in the sense defined earlier.''
[again in the introduction]
\item
``We emphasize that certainty is always an agent's certainty \ldots'' [once more in the introduction]
\item
``Probability assignments are not arbitrary, but they always have an irreducibly subjective component.''
[in section 2]
\end{enumerate}
are in direct conflict with the {\it exception}.  Similarly, without a more careful statement, this
\begin{enumerate}
\setcounter{enumi}{5}
\item
``Probabilities for outcomes or data are not facts.''
       [back in the introduction again]
\end{enumerate}
is in conflict too:  For one would have to be careful that the probabilities being spoken of here are distinctly not conditional probabilities, or at the very least, when they are conditional probabilities, a further analysis of their subjective status is required.

{\it Now}, that said, {\it I do not advocate changing statements\/} 1 {\it to\/} 6.  All I'm trying to do in my snideness is give you a feel for how {\it ugly\/} a formulation the paper would take on if you were consistent with yourselves.  For instance,
\begin{itemize}
\item[1'.]
``We first review the main arguments for the general claim that probabilities sometimes represent degrees of belief.''
\item[2'.]
``The updated probabilities mostly depend on the agent's prior probabilities as well as on the data and thus can be different for agents in possession of the same data.''
\item[3'.]
``We argue, following de Finetti, that in the last analysis probability assignments are fairly often but not always subjective in the sense defined earlier.''
\item[4'.]
``We emphasize that certainty is {\it always\/} an agent's certainty \ldots\ except in those cases when it is not''
\item[5'.]
``Probability assignments are not arbitrary, but here and there they have an irreducibly subjective component.''
\end{itemize}
Remind me:  Just what ``possible misunderstanding'' is this {\it exception\/} supposed to be saving the reader from?  I've never seen that it leads anywhere other than to weakening our argument.

\item
Fourth par.\ from last.  I instated a parenthesis around the `nontrivial'.  I figured if you allowed the parenthesis once a few paragraphs above, you'd allow it again.

\item
Par.\ where state preparation is given an equation.  Deleted ``essentially,'' because it struck me as apt to confuse---implying that there are exceptions.  Also took out ``thus the posterior
state is independent of the prior state'' for reasons of repetitiveness.

\item
Par.\ starting with ``The quantum operation depends, at least partly \ldots''  I have changed the logic of this argument because I feel that Referee 4 is right---it's just the problem of trying to say all this crap about the ``Copenhagen interpretation.''  It is dangerous business.  For instance, it's really hard---probably impossible---to reconcile these things we say with a passage like this one (drawn from Plotnitsky's book {\sl Complementarity\/}):
\bq\noindent
Bohr insists that indeterminacy affects the interaction, and thus
the possibility of sharp distinction, between the measured object
and the measuring instrument, since both must be treated as quantum systems (PWNB 1:11; PWNB 2:25-26, 72-74).
\eq

And anyway, I have always felt uncomfortable putting so much weight on a ``church of the larger Hilbert space'' view of state preparation.  As far as I am concerned, the really only conclusive argument (for the subjectivity of operations) is the one we present in the two paragraphs preceding this one:
\bq\noindent
It is tempting to conclude that objective facts, consisting of the measurement outcome~$d$ and a classical description of the preparation device, determine the prepared quantum state~$\sigma$. This would violate the category distinction by allowing facts to fully determine (nontrivial) probabilities derived from $\sigma$. What this can only mean for a thoroughgoing Bayesian interpretation of quantum probabilities is that the posterior quantum state $\sigma$ {\it must\/} depend on prior beliefs through the quantum operation.
\eq

It was fine to have the church argument lying around as a heuristic to help convince people (like {\Carl}) if needed, but when it is touted as 1) the ultimate reason for the subjectivity of operations and 2) an explicit contradiction to Bohr and the like, then it gets dangerous.  Because then every amateur reader of Bohr out there can be in a position to pounce.

We say in the last draft,
\bq\noindent
   Classical facts cannot suffice to specify a preparation device
   completely because, among other things, a complete description must
   ascribe to the device an initial quantum state, which inevitably
   represents prior beliefs of the agent who is attempting to describe
   the device.
\eq
Now, it was I who inserted the phrase ``among other things,'' but that really didn't go nearly far enough to save my conscience from agony.  Delete it, as almost surely no readers will take note of it anyway, being such a slight deviation from the rest of the flow, and what do you get?

\bq\noindent
Classical facts cannot suffice to specify a preparation device
completely because a complete description must ascribe to the device an initial quantum state, which inevitably represents prior beliefs of the agent who is attempting to describe the device.
\eq

Would Bohr himself buy that?  ({\Carl} will be quick to say that he doesn't give a damn what Bohr would say.  But without some firm foundation for this great enemy that we're supposed to be fighting, `the Copenhagen interpretation', we're just erecting a straw man that we're taking pleasure in kicking down.  {\Mermin} alluded to this, and so did Referee 4, and, honestly, in my own head I've always thought it.)

Here's a nice passage from Asher Peres's paper ``Karl Popper and the Copenhagen Interpretation'':
\bq
Quantum mechanics provides statistical predictions for the results of measurements performed on physical systems that have been
prepared in specified ways (Peres, 1995).  (I hope that everyone agrees at least with that statement. The only question here is
whether there is more than that to say about quantum mechanics.) The preparation of quantum systems and their measurement are performed by using laboratory hardware which is described in classical terms. If you have doubts about that, just have a look at any paper on experimental physics. The necessity of using a classical terminology was emphasized by Bohr (1949) whose insistence on this point was very strict:
\bq\noindent
       However far the [quantum] phenomena transcend the scope of
       classical physical explanation, the account of all evidence must
       be expressed in classical terms. The argument is simply that by
       the word `experiment' we refer to a situation where we can tell
       others what we have done and what we have learned and that,
       therefore, the account of the experimental arrangement and the
       results of the observations must be expressed in unambiguous
       language with suitable application of the terminology of
       classical physics.
\eq

The keywords in that excerpt are: classical terms \ldots\ unambiguous language \ldots\ terminology of classical physics.  Bohr did not say that there are in nature classical systems and quantum systems.  There are physical systems for which we may use a classical description or a quantum description, according to circumstances, and with various degrees of approximation. It is according to our assessment of the physical circumstances that we decide whether the q-language or the c-language is appropriate. Physics is not an exact science, it is a science of approximations. Unfortunately, Bohr was misunderstood by some (perhaps most) physicists who were unable to make the distinction between language and substance, and he was also misunderstood by philosophers who disliked his positivism.

It is remarkable that Bohr never considered the measuring process
as a dynamical interaction between an apparatus and the system under observation. Measurement had to be understood as a primitive notion. Bohr thereby eluded questions which caused considerable controversy among other authors ({\Wheeler} and Zurek, 1983). Bohr willingly admitted that any intermediate systems used in the measuring process could be treated quantum mechanically, but the final instrument always had a purely classical description (Bohr, 1939):
\bq\noindent
In the system to which the quantum mechanical formalism is
applied, it is of course possible to include any intermediate
auxiliary agency employed in the measuring process [but] some
ultimate measuring instruments must always be described entirely
on classical lines, and consequently kept outside the system
subject to quantum mechanical treatment.
\eq
\eq

So, I think, Bohr would reject our very starting point.  How can we really argue that he'd be wrong in rejecting it?  We'd say, ``But you're not really giving a complete description of the ultimate measuring device because a complete description must ascribe to the device an initial quantum state.''  He'd say, ``Young men, you just don't understand the practice and scope of quantum mechanics.''

OR, if I wanted to play like Bohr anticipated us all along, I think I could actually pull that off too.  For I can think of a completely different reading of Bohr where he might have been in essential agreement with us and even gone on to refine the point.  For instance, he might say,
\bq
Right you are young men that there is an essential subjective
   character in state preparation---that's what Plotnitsky was
   reporting on my behalf---but what do you think I meant by
   saying, ``The argument is simply that by the word `experiment' we
   refer to a situation where we can tell others what we have done and
   what we have learned and that, therefore, the account of the
   experimental arrangement and the results of the observations must
   be expressed in unambiguous language with suitable application of
   the terminology of classical physics.''?  I meant that two agents
   have to have common enough priors that they can actually speak to
   each other---there has to be enough intersubjective agreement that
   they're in the same `language game.'  Didn't you read Bernardo and
   Smith, who say:
\bq
       There is an interesting sense, even from our standpoint, in
       which the parametric model and the prior can be seen as having
       different roles. Instead of viewing these roles as corresponding
       to an objective/subjective dichotomy, we view them in terms of
       an intersubjective/subjective dichotomy.  To this end, consider
       a {\it group\/} of Bayesians, all concerned with their belief
       distributions for the same sequence of observables.  In the
       absence of any general agreement over assumptions of symmetry,
       invariance or sufficiency, the individuals are each simply left
       with their own subjective assessments.  However, given some set
       of common assumptions, the results of this chapter imply that
       the entire group will structure their beliefs using some common
       form of mixture representation.  Within the mixture, the
       parametric forms adopted will be the same (the
       {\it intersubjective\/} component), while the priors for the
       parameter will differ from individual to individual (the {\it
       subjective\/} component).  Such intersubjective agreement
       clearly facilitates communication within the group and reduces
       areas of potential disagreement to just that of different prior
       judgements for the parameter.  As we shall see in Chapter 5,
       judgements about the parameter will tend more towards a
       consensus as more data are acquired, so that such a group of
       Bayesians may eventually come to share very similar beliefs,
       even if their initial judgements about the parameter were
       markedly different. We emphasize again, however, that the key
       element here is intersubjective agreement or consensus.  We can
       find no real role for the idea of objectivity except, perhaps,
       as a possibly convenient, but potentially misleading,
       ``shorthand'' for intersubjective communality of beliefs.
\eq

   When I am talking about the ultimate measuring instrument, I am
   talking about precisely the situation where there is intersubjective
   agreement between agents.  So, you can purify the description of
   measurement if you wish (i.e., ascribe a quantum state to the
   preparation device), but then there is no further subjective freedom
   in those quantum states, for those hypothetical agents would be so
   in disagreement that they will not be able to communicate with each
   other.  That is to say, I say it's time for a classical description
   of a measurement device just at the moment where your argument about
   the subjectivity of states no longer holds for two disparate agents.

   A complete description of a preparation device is a fact?  Who said
   anything about facts?  I didn't.
\eq

This is why I think we have really worked ourselves into a corner---and done our relatively clear initial thoughts a disservice---by making the phrase `the Copenhagen interpretation' such a focal point.

But there's not much that can be done about this now, without scrapping the whole project.  Thus rather than attempting to delete the argument (which I'm sure would create an outcry like never heard before), what I propose is that we change the logic of the argument (and then only slightly \ldots\ enough to get us through the night).  The flow used to be ``a complete description of a preparation device {\it requires\/} that a quantum state be ascribed to the device.''  To at least stave off the Bohrians-in-the-know a little bit at this point, and to frame the argument more in terms that I can agree with, here's what I changed it to.  1) Take any purported agent-independent, complete classical description of the preparation device.  2) That description can always be `quantized' into one in the church of the larger Hilbert space, i.e., one of the variety we discuss.  But then, 3) the latter description of the process clearly has a subjective element in it in the form of a prior quantum state for the device.  4) Since there is no operational distinction between the descriptions, either the initial quantum state in the latter description is in fact objective (contrary to our assumption), or the initial purported classical description had a subjective element in it that was unacknowledged previously.

By the way, let me just put it down for the record why I'm not all that attracted to going to the church of the larger Hilbert space to justify the subjectivity of preparation operations.  The reason is that it is effectively like saying that to prove the subjectivity of conditional probabilities you have to start from the subjectivities of the joints and the marginals.  But the conditional should not require the existence of a joint distribution before its own existence---it is a stand-alone entity, as {\Ruediger} emphasizes in his classes.  And so, its subjectivity is stand-alone too.  Similarly, I would say, for quantum operations.
\end{itemize}

\subsection{Changes in Section 4}

\begin{itemize}
\item
par.\ 2.  Changed tense in this paragraph.  Deleted the use of ``with certainty'' that confused Referee 2.

\item
par.\ 3.  Changed `incomprehensible' to `inconsistent' for Referee 4.

\item
last par.\  Deleted `the quantum state of' because 1) my previous railing on church-of-the-larger-Hilbert-space justifications, and 2) because it gives the impression one can have beliefs about the (true) quantum state, rather than admitting forthrightly that quantum states are beliefs.  Also added ``that is true or false of the system'' to ``The statement that the measurement outcome is 1 with certainty is thus not a proposition, but an agent's belief---and another agent might make a different prediction.'' for the purposes of Referee 2.
\end{itemize}

\subsection{Changes in Section 5}

\begin{itemize}
\item
par.\ 3.   Added ``external to the agent'' to guard against those who would ask, ``What, the agent is not part of the world?''  I think that's as far as we want to go anyway.

\item
last par.\  ``slippery slope'' $\longrightarrow$ ``path''  The former seemed too alliterative for the {\Ruediger}ian style we had settled on.  Also, it seemed too judgmental to be stated without back-up.
\end{itemize}

\subsection{Changes in Section 6}

\begin{itemize}
\item
par.\ 6.  Added reference to Garrett.

\item
par.\ starting ``It might still be \ldots''  Add ``if he is bound by a classical worldview'' to the last sentence.

\item
par.\ starting ``For instance, take a complete \ldots''  Changed ``raw'' to ``bare''---seemed less harsh that way (I'm sensitive because of {\Carl}'s ``meaty'' jokes).  Also added ``law-of-thought'' as an adjective to help emphasize the point.  ``shows its head'' $\longrightarrow$ ``shows itself''

\item
Last paragraph.  Deleted ``of the world''---didn't seem needed in that context.  ``dogged'' $\longrightarrow$ ``plagued'' (I had gotten carried away previously).  ``finally'' $\longrightarrow$ ``most importantly'' (didn't want an overuse of finals).  Changed ``what features of the quantum formalism'' to ``what features of the quantum formalism beyond the ones discussed here'' just to make it clear that we had already discussed some.
\end{itemize}

\section{07-12-06 \ \ {\it (Select) Replies to Referees} \ \ (to R. {\Schack} \& C. M. {\Caves})} \label{Caves93} \label{Schack111}

\subsection{Replies to Referee 1}

\bq\noindent
{\bf Referee:}
In fairness to Copenhagen shouldn't you add that though
Copenhagen holds the quantum state to be objective in the sense you describe, this (by itself) does not imply that the state is an objective property of the system to which it is assigned.  Surely Bohr would never have said such a thing.  It is an objective property of the ``whole experimental setup'' which includes both the system and
the preparation device.  You do indeed say this a little further
down the page, but only reluctantly (``perhaps supplemented
by\ldots'').  You seem to be trying to make Copenhagen sound worse
than it is.
\eq

Agreed.  We softened the language and made the enemy more specific
throughout the paper. You are correct that Bohr would have likely
called the quantum state (though he rarely spoke of quantum states)
as an objective property of the ``whole experimental setup''.  What
he failed to realize though---in CAF's reading of Bohr---was the
dependence of the phrase ``whole experimental setup'' upon an agent's
judgment. (An interesting exchange between Bohr and Pauli from 1955
bears this out to some extent.)

\bq\noindent
{\bf Referee:}
``The main issue here is whether there is any difficulty with the
idea of an utterly certain belief about an admittedly contingent
fact.''  Yes indeed.  And this is the issue that underlies most of my
concerns about your not paying enough attention to what constitutes a
fact, and not giving any example of the certain beliefs of two agents
that underly different state assignments. This is central to your
argument.  Is a reference to de Finetti all you're going to give the
reader?  (This reminds me of a friend who was receiving Catholicism
lessons from a priest so he could be married in a church.  He kept
asking the priest why he should believe in the immaculate conception,
and then why he should believe that an angel had announced it, etc.,
until he got to a point where the priest said triumphantly, because
German scholarship has established it.)  Surely you owe your readers
some gloss on de Finetti, or an explanation of why he is not central
to your case.
\eq

You are quite right about this, and initially we tried to fill in the
gap here.  But then we realized it was too large a task to try to
bite off in a ``committee report'' and, we backed off. So we removed
the paragraph completely.  Further elucidation will have to await
another publication.

\bq\noindent
{\bf Referee:}
I once read somewhere in Jaynes that no Bayesian would take
$p(h)=1$
for any $h$ other than a logical tautology, because no acquired data
whatever could then permit updating the probability to anything less
than 1.  I assume that what he meant was that since $p(d)$ in (2) is
the sum on $h$ of $p(d|h)p(h)$, then $p(d)$ is necessarily equal to
$p(d|h)$
when $p(h)=1$.

Given this, why isn't your entire paper about a situation (certainty)
that no self-respecting Bayesian would ever find himself in?
Presumably the reason it interests you is that quantum theory does
assign $p=1$ to certain data given certain state assignments, and this
requires some interpretation.  But then shouldn't the position of the
Bayesian be that pure state assignments are ideal limiting cases that
one can never encounter in practice?  Analogous to initial conditions
specified to infinite precision in chaotic classical dynamical
systems. (You have seduced me into making dubious classical
analogies.)  All physically achievable states are mixed states.  A
pure-state assignment should not be viewed as subjective, but as
impossible. A quantum operation may well be subjective but this has
nothing to do with pure states being subjective, because there is no
physical quantum operation that prepares a pure state. The best you
can ever do is prepare a mixed state of very low entropy.
\eq

One of us---CAF---has never seen the relevance of this point (which
the referee has made previously in other contexts) for what is at
issue here.  CMC, on the other hand, says, ``This is an interesting
point that we are well aware of, but it is not a point we agree with,
nor is it one we wish to deal with in this paper.''

But, back to CAF.  Your concern seems to be this.  If one should,
methodologically, never assign a pure state, then there are no
handles by which to address pure states' subjectivity or objectivity.
(In fact it seems, one should just write them out of the theory.) But
that's not true. The Born rule tells us how to calculate
probabilities from pure states---so the issue has to be addressed, and
in fact there's something to be learned from it. If one has accepted
all the arguments for (or simply the beauty and simplicity of)
interpreting \underline{\it all} probabilities as subjective,
Bayesian degrees of belief, then one must conclude that the pure
state has a subjective element or the measurement description has a
subjective element or both have a subjective element.

\subsection{Replies to Referee 2}

\bq\noindent
{\bf Referee:}
The category distinction between facts and probabilities is explained
quite nicely. Nonetheless, the authors might wish to emphasize this
point further. The confusion I have in mind (which might be my own)
is the following one. It is typical to assert that assignments of
probabilities 0 and 1 to a proposition are equivalent to assignments
of truth and falsity to that proposition.  If this is so, then one
cannot say that ``the truth value of a proposition is a fact'' and
simultaneously that no probability assignment is a fact.  Presumably,
the mistake is in identifying probability 0 and 1 with truth and
falsity.  Or, perhaps there is a more subtle distinction to be made
between the actual truth value of a proposition and the truth value
that an agent assigns to it (which might be different from the actual
value).  In any case, treating propositional logic as a special case
of probability theory is sufficiently ubiquitous that more
explanation seems in order here.
\eq

Indeed we all agreed with the referee on this point, but we could
never come to great agreement among ourselves on how to handle the
issue in the context of this paper---where much of the groundwork
required to address it properly had not been laid. Thus, we added a
small---perhaps confusing---footnote to nonetheless mildly address
the issue. We will try to do your point better in further
publications.

\bq\noindent
{\bf Referee:}
It should be emphasized by the authors that they are making the
strong claim that the Bayesian approach allows one to hold on to the
assumption of locality.  This is likely to be a very controversial
claim and consequently should not be left implicit.  Wiseman and
others have argued that maintaining locality is only possible if one
is willing to deny the reality of distant observers, that is, it is
only possible if one is willing to adopt a solipsist view.  The
authors may choose not to address such issues in detail in this
paper, but should certainly say more than has been said about
locality in the Bayesian approach.
\eq

Wiseman (personal conversations with CAF) has never understood the
category distinction we are making---and that is the root of his
further misunderstandings. Since we have laid out the category
distinction fairly carefully in this paper, we think we have said
enough for the time being.  There's only so much that can be done in
one paper.

\subsection{Replies to Referee 3}

\bq\noindent
{\bf Referee:}
The authors attempt to defend the following general thesis. There
are no facts about probabilities in the world (see e.g.\ Section II).
Attempts (e.g.\ Lewis's PP) to connect between probabilities and facts
are highly problematic. Moreover, in principle the gap between
probabilities and facts is unbridgeable. Probability assignments
always have an irreducibly subjective component, essentially because
they involve priors. This holds for both deterministic (e.g.\
classical statistical mechanics) and indeterministic theories (where
the probabilities are given by dynamical laws, e.g.\ some versions of
quantum mechanics).
\eq

This captures part of our paper relatively accurately.

\bq\noindent
{\bf Referee:}
I agree that PP-style and similar approaches have problems. But
this does not mean that probabilities are not grounded by facts. Here
is a counter-argument to the above. Under quite general
circumstances, it is provable that if an agent updates probabilities
in accordance with Bayes' rule the probabilities will converge to the
long term relative frequencies (for large enough samples), no matter
what the prior probabilities are (except for priors equal to 0 and
1). The dependence on the priors (for large enough samples) washes
out in the long run, and so the priors, however subjective, don't
matter. The authors seem to be aware of this point (see p.~4, last
paragraph), but they don't directly address it. The same argument
applies also to the quantum mechanical probabilities and state
assignments (Section III).
\eq

The counter-argument holds no water for a subjective Bayesian. `Quite
general circumstances'?  For whom?  For sufficiently distinct priors,
there need be no convergence.  What you point out is only true if two
agents for whatever reason---perhaps simply because of their common
genome---happen to adopt `compatible' priors of one variety or other.
We explore this issue in detail (particularly in the quantum case) in
C.~M. {\Caves}, C.~A. Fuchs, and R. {\Schack}, ``Conditions for
Compatibility of Quantum-State Assignments,'' Phys.\ Rev.\ A {\bf
66}, 062111 (2002)---so, yes, we are aware of it.

The book by Bernardo and Smith that we cite in the present paper puts
the issue very nicely:
\bq
[T]here is an interesting sense, even from our standpoint, in
   which the parametric model and the prior can be seen as having
   different roles.
   Instead of viewing these roles as corresponding to an
   objective/subjective dichotomy, we view them in terms of an
   intersubjective/subjective dichotomy.  To this end, consider a
   {\it group\/} of Bayesians,
   all concerned with their belief distributions for the same sequence
   of observables.  In the absence of any general agreement over
   assumptions of symmetry, invariance or sufficiency, the individuals
   are each simply left with their own subjective assessments.  However,
   given some set of common assumptions, the results of this chapter
   imply that the entire group will structure their beliefs using some
   common form of mixture representation.  Within the mixture, the
   parametric forms adopted will be the same (the {\it
   intersubjective\/}
   component), while the priors for the parameter will differ from
   individual to individual (the {\it subjective\/} component).  Such
   intersubjective agreement clearly facilitates communication within
   the group and reduces areas of potential disagreement to just that of
   different prior judgements for the parameter.  As we shall see in
   Chapter 5, judgements about the parameter will tend more towards a
   consensus as more data are acquired, so that such a group of
   Bayesians may eventually come to share very similar beliefs, even if
   their initial judgements about the parameter were markedly different.
   We emphasize again, however, that the key element here is
   intersubjective agreement or consensus.  We can find no real role for
   the idea of objectivity except, perhaps, as a possibly convenient,
   but potentially misleading, ``shorthand'' for intersubjective
   communality of beliefs.
\eq
Particularly, there is no requirement of nature that there must be
intersubjective agreement of the type discussed.  If it happens, it
is often nice---it leads to less violent worlds for instance---but
nature cannot decree it.

\bq\noindent
{\bf Referee:}
p.\ 4, last par -- I think that the authors don't address the
real problem faced by any purely subjectivist interpretation of
probabilities. Take for example classical statistical mechanics. The
problem isn't about whether the uniform probability measure brings
about the melting of the ice cube. Of course it doesn't.
\eq

Of course it doesn't?  Like you, we would have said so.  But you
should not mock.  To quote the reference [35] you bring up in your
Point (8) below (i.e., J.~North's paper) at length:
\begin{quote}
There are two main, and to my mind fatal, problems with the epistemic
view [of probabilities in statistical mechanics].  The first stems
from the role the uniform probability distribution plays in
explanations of thermodynamic phenomena.  Consider the explanation of
an ice cube's melting towards the future.  If we take an epistemic
view of the uniform distribution that is placed over its current
macrostate, then part of the reason for the ice's melting will be our
ignorance of its initial microstate.  On the assumption that
explanations of physical phenomena ought to be objective---and in any
case not rely on our epistemic state---we should not maintain that
part of the explanation that entropy increases (that ice melts, that
coffee cools, that gases expand) is the extent of our knowledge.  How
could our epistemic state have anything to do with the ice cube's
melting?  This would be like saying that if we happened to be the
kinds of beings who did have epistemic access to the initial
microstate of the ice cube, then it might have behaved differently.
It is also a consequence of this view that no matter what kind of
world we live in, we must assume that the ice is most likely to melt
towards the future.  All of that seems crazy: we are after an
objective, scientific explanation of thermodynamics.
\end{quote}
We have added a more explicit reference to this paper, so that
perhaps a reader will be led to it.

\bq\noindent
{\bf Referee:}
The authors seem to imply (Sections V and VI) that the Bayesian
approach to QM is local, since it doesn't assign determinate
instruction sets (even in eigenstates). This, however, is completely
mistaken. Assuming that measurements have outcomes, any local theory
is committed to Bell's inequality. So any theory that produces the
statistical predictions of quantum mechanics must be nonlocal, no
matter whether or not it admits instruction sets.
\eq

`Completely mistaken' is a rather strong expression for a referee who
hasn't shown many indications of having understood much of the point
of view of the paper.

\bq\noindent
{\bf Referee:}
The authors claim that there is no fact of the matter about
probability assignment (including quantum states; Section VI). But
they accept the Dutch-book argument (Section I). This is a
contradiction. The Dutch-book argument presupposes that there are
factual constraints (sure loss) on probability assignments.
\eq

It is not a contradiction.  We view the Dutch-book argument as a
normative principle---not a statement of fact or a property of the
physical world.  Our view is along the lines of these words from our
Bernardo and Smith reference:
\begin{quote}
      What is the nature and scope of Bayesian Statistics \ldots?

Bayesian Statistics offers a rationalist theory of personalistic
   beliefs in contexts of uncertainty, with the central aim of
   characterising how an individual should act in order to avoid certain
   kinds of undesirable behavioural inconsistencies.  The theory
   establishes that expected utility maximization provides the basis for
   rational decision making and that Bayes' theorem provides the key to
   the ways in which beliefs should fit together in the light of
   changing evidence.  The goal, in effect, is to establish rules and
   procedures for individuals concerned with disciplined uncertainty
   accounting.  The theory is not descriptive, in the sense of claiming
   to model actual behaviour.  Rather, it is prescriptive, in the sense
   of saying ``if you wish to avoid the possibility of these undesirable
   consequences you must act in the following way.''
\end{quote}
Dutch-book coherence is something to strive for.  For if not
satisfied, an agent with any logical powers will see that his
gambling commitments could lead, in the appropriate circumstances, to
sure destruction.

To help abate misunderstandings like yours, we have changed the
sentence in our Introduction reading
\begin{quote}
The Dutch-book argument shows that, to avoid sure loss, an agent's
probabilities must obey the usual probability axioms.
\end{quote}
to
\begin{quote}
The Dutch-book argument shows that, to avoid sure loss, an agent's
gambling commitments should obey the usual probability axioms.
\end{quote}

\bq\noindent
{\bf Referee:}
Some items in the references list don't appear in the text, e.g.
[10], [35].
\eq

We thank the referee for pointing out his likely identity.

\subsection{Replies to Referee 4} \label{RepliesToReferee4}

\bq\noindent
{\bf Referee:}
A final point about this passage: you write
\begin{quote}
The occurrence or nonoccurrence of an event is a fact for the agent.
Similarly, the truth or falsehood of a proposition is a fact. Facts
are objective; they are not functions of the agent's beliefs.
\end{quote}
What work, precisely, is the modifier `for the agent' supposed to be
doing here? It comes across as slightly odd to say that occurrence is
a fact `for the agent' and then go on to emphasize the non-dependence
of these fact[s] on the agent. (Of course, there is no inconsistency
here: apparently, facts are {\it facts for everybody\/} on your view,
and so {\it a fortiori\/} they are facts `for the agent'. But still
[the] way you put things here strikes me as misleading somehow.)
\eq

This paper does not emphasize it, but no, we do not mean ``the facts
we are talking about here are facts for everybody.''  We mean that
the facts too are personal, though in a very careful sense.  There is
only so much work one can put into a paper as a view is evolving (and
as the authors are {\it slowly}, ever so very {\it slowly}, coming to
agreement), without scrapping the whole project and starting over.

We have tried to meld the phrase ``fact for the agent'' a little
better than previously into its surrounding text---so that it is
there at least as some shadow of our presently agreed upon view---but
it will have to remain somewhat awkward in the present context.

A better exposition would have emphasized the full setting of our
view, including some discussion of the ontological---i.e.,
noninstrumental---pieces of it.  1) There are two physical systems in
the story of this paper.  One takes the role of the agent---the
possessor of subjective degrees of belief and the activator of the
measurement process.  The other takes the role of the object system.
2) What is being discussed when one speaks of gathering data in the
quantum context is an interaction or transaction between the agent
and the object.  3) Without the agent, there would be no quantum
``measurement'' to speak of, but without the object, there would be
no means for the agent to obtain the data, the spikes upon which he
pivots his probabilities (his subjective degrees of belief).  4)
There are no other agents in this story.  5) We do not go any further
than points 1-5 warrant by giving the data obtained in a quantum
measurement an autonomous existence---for instance, as something
beyond the agent's sensations. That would run into inconsistencies in
a ``Wigner's friend'' scenario.  Nonetheless, quantum measurement
outcomes are beyond the control of the agent---they are only born in
the interaction---and thus are not functions of the agent in the way
that his degrees of belief are. The degrees of belief find their
source in the agent (the subject); the outcomes find their source in
the external quantum system (the object). But the outcomes lead back
to the agent in that they are personal to him.

This does not mean that the agent-object interaction and its fuller
consequences are not autonomous events in spacetime, but it does mean
that if there are any such things, quantum theory is not directly
concerned with them. In CAF's view, quantum theory takes its whole
definition as a normative theory for organizing an agent's {\it
personal\/} probabilities for the {\it personal\/} consequences of
his interactions with external physical systems.  The structure of
the agent-independent world that lies behind quantum mechanics is, in
the end, still codified by the theory, but only in a higher-order,
more sophisticated fashion than had been explored previously---it is
through the normative rules, rather than quantum states and
Hamiltonians.

Unfortunately a more detailed account like this will have to wait for
another paper.

\bq\noindent
{\bf Referee:}
You write:
\begin{quote}
The Copenhagen assumption that a preparation device can be given a
complete classical description neglects that any such device is
quantum mechanical and thus cannot be specified completely in terms
of classical facts.
\end{quote}
I said above that I wouldn't quibble about what the `Copenhagen
Interpretation' says, and I won't but I will argue that Bohr, at
least, is not fairly described in this way. For one thing, Bohr does
not deny that the physical instruments that play the role of
measuring devices can be described quantum-mechanically. (Bohr says
this in several places. I was quickly able to locate one (1949
Schilpp volume): ``\ldots\ of course, the existence of the quantum of
action is ultimately responsible for the properties of the materials
of which the measuring instruments are built and on which the
functioning of the recording devices depends''.) Indeed, it must be
possible to do so, on Bohr's view, since they could themselves become
the object of measurement. (The quotation just above even indicates
that Bohr will agree that in order to analyze the functioning of the
measuring device, one must use quantum theory.) What Bohr will insist
on is: (1) that the only `facts' about the physical world to which we
ultimately have access are `phenomena', by which he means
``observations obtained under specified circumstances, including an
account of the whole experimental arrangement'' (also from the
Schilpp volume); and (2) the `specification of circumstances' must be
given in terms of concepts from classical physics. Now, one could
well dispute either of these claims, but if we accept them for the
moment, then Bohr has an easy reply to the point that you are making
here: for the purposes of describing the phenomena that a given
measuring device is capable of eliciting, the device must be
described classically, but if we are to investigate how or why it
functions as it does, we describe it quantum-mechanically. (Of
course, an {\it empirical\/} investigation of the device will itself
involve observations of the device, and these observations will be
made by means of yet other devices, which will be described
classically, for the purposes of revealing phenomena that involve the
original device.) Given Bohr's (highly debatable) assumptions, it
seems to me that he has a reasonable response to the point that you
make here.
\eq

Actually we take explicit issue with the point you make near the end
of this paragraph:
\bq
Bohr has an easy reply to the point that you are making here: for the
purposes of describing the phenomena that a given measuring device is
capable of eliciting, the device must be described classically, but
if we are to investigate how or why it functions as it does, we
describe it quantum-mechanically.
\eq
The logic goes like this.  1) Take any purported agent-independent,
complete classical description of the preparation device.  2) That
description can always be `quantized' into a quantum mechanical one
on a larger Hilbert space, i.e., one of the variety we discuss, where
there is only unitary evolution. But then, 3) the latter description
of the process clearly has a subjective element in it in the form of
a prior quantum state for the device. 4) Since there is no
operational distinction between the descriptions, either the initial
quantum state in the latter description is in fact objective
(contrary to our assumption), or the initial purported classical
description has a subjective element in it that was unacknowledged
previously.

We have tried to emphasize the logic of this argument by rewriting
this section somewhat.  We have also bolstered the point by quoting
Bohr directly, where he says things similar to your summary above,
and saying a little on how we differ.

\bq\noindent
{\bf Referee:}
It is important not to give the impression that in `the
Copenhagen' view, there is {\it no room\/} for subjective
probabilities.
\eq

That is probably true, and we hope that we have taken this into
account by being more careful to say that our concern in this paper
are ``realist readings'' of Copenhagen.  Finally, to defuse the issue
of what is or what is not ``the Copenhagen interpretation'' a little,
we have replaced most instances of that phrase with ``the
objective-preparations view'' which better identifies the issue at
hand in any case.

\bq\noindent
{\bf Referee:}
After all, a Copenhagen-ist can still agree that there is some
measure of uncertainty about whether a device really prepares the
state that we generally think it prepares. (Perhaps it is true for
them that the {\it true\/} classical description of the device fixes
the state, but they are not committed to the claim that we know what
the true description is.)
\eq

But we would never say something like this.  Remember, for us, a
``preparation'' is a synonym for an initial quantum state (pure or
mixed).  There is never a question about whether a device {\it
really\/} prepares the state that we generally think it
prepares---such a statement ontologizes states in a way that we're
trying to get away from.

\bq\noindent
{\bf Referee:}
I'm more concerned about the presumption that there must be facts
that guarantee the outcome in the first place, {\it even\/} for the
Copenhagen view. Indeed, in at least one common form, the Copenhagen
view is more or less an anti-realist, more specifically
instrumentalist, view. Now, anti-realists will exactly {\it deny\/}
that there must be `facts' that underwrite the accuracy of the
predictions made by theory. Copenhagen-ists of this stripe will not
have any problem with your conclusion that ``we must abandon
explanations in terms of pre-existing properties'' -- they couldn't
agree more! See, for example, van Fraassen's and Fine's essays in the
volume ``Philosophical Consequences of Quantum Theory'' (ed.\ Cushing
and McMullin). Both of them argue explicitly for the view that even
predictions with certainty do not have to be `underwritten' by facts
about the world, of whatever sort. As van Fraassen puts it,
succinctly, ``There does not have to be a reason for everything.''
\eq

This is a good point.  We have tried to take into account your
concern by defining a little better what we are calling the
``Copenhagen-like interpretations.''

On the other hand, we wonder whether you have van Fraassen right, at
least in this particular context.  We say this because one of
us---CAF---had the following email exchange with him on 14 November
2005.
\begin{quote}
{\bf CAF:} [With regard to van Fraassen's sentence]
\bq\noindent
Writers on the subject have emphasized that the main form of
measurement in quantum mechanics has as result the value of the
observable at the end of the measurement -- and that this observable
may not even have had a definite value, let alone the same one,
before.
\eq
your phrase ``MAY NOT even have a definite value'' floated to my
attention.  I guess this floated to my attention because I had
recently read the following in one of the Brukner/Zeilinger papers,
\bq\noindent
     Only in the exceptional case of the qubit in an eigenstate of
     the measurement apparatus the bit value observed reveals a
     property already carried by the qubit.  Yet in general the value
     obtained by the measurement has an element of irreducible
     randomness and therefore cannot be assumed to reveal the bit
     value or even a hidden property of the system existing before
     the measurement is performed.
\eq
I wondered if your ``may not'' referred to effectively the same thing
as their disclaimer at the beginning of this quote.  Maybe it
doesn't. Anyway, the Brukner/Zeilinger disclaimer is a point that
{\Caves}, {\Schack}, and I now definitely reject:  From our view all
measurements are generative of a NON-preexisting property regardless
of the quantum state.  I.e., measurements never reveal ``a property
already carried by the qubit.''  For this, of course, we have to
adopt a Richard Jeffrey-like analysis of the notion of
``certainty''---i.e., that it too, like any probability assignment,
is a state of mind---or one along (my reading of)
{\Wittgenstein}'s---i.e., that ``certainty is a tone of voice''---to
make it all make sense, but so be it.

To which he responded:
\bvf
Suppose that an observer assigns eigenstate $|a\rangle$ of $A$ to a
system on the basis of a measurement, then predicts with certainty
that an immediate further measurement of $A$ will yield value $a$,
and then makes that second measurement and finds $a$.  Don't you even
want to say that the second measurement just showed to this observer,
as was expected, the value that $A$ already had?
\evf
\end{quote}
Thus at least on a surface reading of this, it seems we are saying
something in this paper on quantum certainty that even takes van
Fraassen by surprise.

\bq\noindent
{\bf Referee:}
There is of course nothing wrong with the proof in section V, but
at this point in the history of the philosophy of quantum theory, I
confess that I don't see why it is there. Isn't a quick reference to
the 10,000 other proofs of nonlocality sufficient? If not, can you
say what is special or new about this one? I would personally rather
see that space used for a more subtle discussion of the newer issues,
perhaps addressing some of the points that I've made above.
\eq

There is a folklore in the physics community that a paper without
equations will not get read.  We wanted physicists to read our paper
too.  More seriously, though, the context of this use of KS is quite
different from previous uses of it.  You yourself call it a ``proof
of nonlocality.''  That's not what it is to us:  It is a proof that,
under the assumption of locality, probability-1 predictions in
quantum mechanics correspond to subjective certainty rather than
objective certainty.

\section{07-12-06 \ \ {\it Getting It Right} \ \ (to S. J. van {\Enk})} \label{vanEnk7}

Attached is a draft of a revised version of my ``certainty'' paper with {\Caves} and {\Schack}.  I remember you thinking I didn't capture your point about the Tsirelson bound correctly at the end of it.  Could you give me a recommendation for a better more accurate statement?

\subsection{Steven's Reply}

\bq
What I meant was this: a Bayesian is not surprised as much as a
realist is by violations of the Bell inequality because he considers
the correlations are (at least partly) in his head, not really in the
quantum systems (oversimplifying it a bit). That's good, you solved
part of the mystery. But what I say then is why are the correlations
in your head still restricted by Tsirelson's inequalities? That should
 be much harder to explain.
 \eq

 \section{07-12-06 \ \ {\it Catching Up on a Southbound Train} \ \ (to D. M. {\Appleby})} \label{Appleby17}

 I'm sorry I've been silent.  Things have just been a little overwhelming, and since I didn't have much to tell you, I didn't write.  At the moment, I'm on a train making my way to Washington, DC.  I've got to work for the APS tomorrow to get the quantum information topical group sessions organized.  (At last count, there were 214 abstracts.)

 The talk in Tskuba went fine.  I checked the essential part of your minimum-uncertainty state derivation, and I found no mistakes.  So, it was part of the overall presentation.  [\ldots]

 Going back to science a little, the discussions last week with {\Bennett} were the most fruitful I've ever had with him (foundation-wise, that is).  Lately he's been worried about information erasure again, and what quantum mechanics adds to the idea.  Taking a strong many-worlds kind of stance as his starting point, he asks, if ``facts'' are just correlations between branches, what happens when those correlations move away from the original systems to which they were concerned?  Are the facts as ``real'' as they used to seem to be?  Well, of course this gets him into all kinds of trouble---the sort of trouble he's not used to since he normally shuns all discussions of foundations.  So it was an absolute surprise to me one morning when he calls down to say that our discussions had caused him to think a lot, and, quote,  ``\ldots\ it almost pains me to say this \ldots\ but I can see why you might use the language of calling a quantum state a state of knowledge.''  OK, he's coming up a little toward the Chris of about 10 years ago (not the radical Bayesian), but given what I know of him, it was almost a miracle to hear this!  It was an essentially infinite step really!  Here's what I wrote Kiki that night:
\bq\noindent
   But over all, it was a good day.  Probably one of my proudest
   moments ever:  Charlie took me aside early this morning and said
   that because of our discussions yesterday, he was starting to adopt
   a view of quantum mechanics that comes closer to mine---he wouldn't
   go so far as me, mind you, or at least he said so---but it was
   ``closer.'' [\ldots] I was proud and humbled at
   the same time.
\eq
Now, to nurture this seed!

\section{11-12-06 \ \ {\it Chris's House from Space}\ \ \ (to J. Racette)} \label{Racette1}

Thanks for the holiday letter.  I enjoyed it, and it was great to learn where you guys are at in your lives now.  Please give Kim a big Merry Christmas from Kiki and me.  (Include yourself too.)

Of course, you were always more literary than me:  In all these years I've never put together a holiday letter.  Maybe the closest I can come at the moment, is to forward you a letter (with the associated photos attached) that I had written to an Irish friend recently.  [See 01-12-06 note ``\myref{Duffy5}{Chris's House from Space}'' to K. R. Duffy.] (It was spurred by him telling another colleague in a pub that the orange of my house could be seen from space.)

Our two kids are growing too.  Emma is setting herself to conquer the world.  She reads like a scanner, and works out like a gymnast.  And Katie's coming along quite nicely too, though her enjoyments are pulling weeds with the old man and cooking.  Kiki is constructing and reconstructing the house constantly---partially restoring it to 1896 and partially just Kikifying it.  Me, I'm just trying to understand quantum mechanics as usual, and traveling the world while I do it.  This year I went to Portugal, Sweden, Italy, Canada and Japan.  (Last year was busier; I went to Australia, Poland, Austria, Sweden, Ireland, Canada, and Germany (twice).)

Sadly I don't have satellite TV yet (still with trudgy old cable):  Too bad I'm not in central New Mexico!

\section{11-12-06 \ \ {\it Quantum Brunch}\ \ \ (to P. Grangier)} \label{Grangier9}

\bpg
Following our ``quantum breakfast'' in Tsukuba, here is the talk which
includes the arguments with Gleason and Wigner theorems, given at
ENS in Paris in November 2004 (see
\myurl[http://www.spectro.jussieu.fr/Seminaires/Resumes_2004/Resu041116.html]{http://www.spectro.jussieu.fr/Seminaires/Resumes\underline{ }2004/Resu041116.html}).

I have various improvements as drafts in my
files, but I did not find time to write them down.
\epg

I enjoyed the quantum breakfast indeed.  Thanks for sending your presentation.  I shall study it, and maybe we can get together for a quantum brunch in February.

I'm just back from the APS sorters meeting, preparing for the March meeting.  That was a lot of hard work.  But it was also a lot of hard work in the many fights Carlton Caves and I had about our own interpretive stuff.  I wonder why I get involved in these things sometimes.

As I say, I will study your presentation.  It would be nice if it turned out that there is more common ground between us than was initially apparent.

\section{12-12-06 \ \ {\it More Stable Version of Manuscript on Operational Axiomatization of QM}\ \ \ (to G. M. D'Ariano)} \label{DAriano7}

Thanks for the note.  I too much enjoyed our discussion.  I'm sorry about the notes; when I saw you at breakfast, I was in such a hurry everything spilt out of my mind.  Anyway, I am working on the paper with Appleby now, and hope to get it posted early next week.

Thanks for the coordinates of your latest posting.  You are right that there is probably much latitude for joining our ideas.  I admire your axiomatic system---of all that I have seen, it is the one that smells the most right for leading to a deeper understanding of the essence of quantum phenomena.  On the other hand, my Bureau-of-Standards measurement idea is only a small piece of the puzzle, and I certainly need to wed it to a larger structure.

I will inform you of the coordinates of the thing with Appleby just as soon as I can.

\section{12-12-06 \ \ {\it Small Changes} \ \ (to R. {\Schack})} \label{Schack112}

Since we haven't heard from you, I went ahead and made some small changes to {\Carl}'s last draft.  I'm guessing that they won't interfere with anything that you may be doing to it.

1)  I modified the last paragraph to better capture van {\Enk}'s point (he had remarked that I didn't quite get the point across that he thought was important).  On top of that, with this modification, the flow of the logic is a better than it was before anyway.

2)  I wimped out on a longer exposition of the Principal Principle troubles again, which I was going to put in a footnote.  So, I just deleted the flagged footnote.

I'd probably like to do a few more {\it small\/} (literally) changes, but I'll go ahead and send you this much in case you do happen to be working on the draft.

The biggest issue probably---the one that will require the most work---is about de Finetti's previous statements on certainty.  As it stands, we say this:
\bq\noindent
   The main issue here is whether there is any difficulty with the idea
   of an utterly certain belief about an admittedly contingent fact.
   This issue arises in the classical setting, where it has been dealt
   with masterfully in Bruno de Finetti's ``Probabilismo''
   (Sect.~18-19).  Nothing of de Finetti's argument loses force when
   we come to quantum mechanics; indeed, the quantum denial of
   instruction sets seems to give it yet more force.
\eq
I don't think that quite frames it right.  (Plus---but this is slightly an aside---Bruno de Finetti's world was neither `classical' nor `quantum' in the usual physics sense.)  I think the real issue is closer to the one that {\Timpson} brings up:  The agent is certain of the outcome, even though he is also certain that the system {\it cannot\/} be (objectively) `certain'.  That is the apparent conundrum.  I'm not now sure that de Finetti considered such a stronger situation.  Or is my memory/understanding now failing?  (You can read what I had written previously on the issue---when I thought de Finetti had indeed done the job---at [now defunct pages; see 28-06-03 note ``\myref{Mermin87}{Utter Rubbish and Internal Consistency, Part II}'' to R. Schack, C. M. Caves \& N. D. Mermin].)  So the question is what to write.  Do you want to tackle this one?

My present feeling is that the solution---or at least one solution---is implicit in our ontology (though I'm not all that happy with what I had written {\Timpson} previously).  All that's being recognized with a Bell--KS argument is that quantum outcomes cannot be a function of the world external to the agent alone.  They come about, in part, by the participation of the agent himself.

On the previous point of contention---this phrase ``Bayesian updating is, as it should be, consistent with logical deduction of facts from other facts''---I'm now willing to leave {\Carl}'s draft alone with regard to it.  But I don't think it is nearly as meaningful as you two think it is.  In fact, I still think our disagreement is more than linguistic.  I'll explain in more detail in a later note.

\section{13-12-06 \ \ {\it Seeking an Objective Substructure, Yes; Theory of Fundamental Events, Maybe No} \ \ (to J. {\Barrett})} \label{Barrett3}

\noindent\underline{\bf NOTE}: This letter was apparently never finished or sent off.  I suspect it was composed roughly on the date listed here.\medskip

I've finally gotten back to reading your note again and thinking about it for a day.  I'm sorry for the delay.  I have enjoyed your questions and your thoughts.  Let me reply to a few of them, though I don't think I have as much to say as I had originally thought I would.

First let's go back to the issue of ``fundamental events,'' since I've already referred you to some of my ruminations on the subject.  You wrote this in your note:
\bjoba
at present we have only a provisional, half-baked theory in which
these events are ``outcomes of interventions''.  But a fundamental
formulation of a complete theory must not use such vague terminology,
and will eventually supply an ontology of primitive events that
agents can in principle bet on.
\ejoba

And let me recall the thing I wrote for the referee of our ``certainty'' paper:
\bq
   [Q]uantum measurement outcomes are beyond the control of the
   agent---they are only born in the interaction---and thus are not
   functions of the agent in the way that his degrees of belief are. The
   degrees of belief find their source in the agent (the subject); the
   outcomes find their source in the external quantum system (the
   object). But the outcomes lead back to the agent in that they are
   personal to him.

   This does not mean that the agent-object interaction and its fuller
   consequences are not autonomous events in spacetime, but it does mean
   that if there are any such things, quantum theory is not directly
   concerned with them. In CAF's view, quantum theory takes its whole
   definition as a normative theory for organizing an agent's {\it
   personal\/} probabilities for the {\it personal\/} consequences of
   his interactions with external physical systems.  The structure of
   the agent-independent world that lies behind quantum mechanics is, in
   the end, still codified by the theory, but only in a higher-order,
   more sophisticated fashion than had been explored previously---it is
   through the normative rules, rather than quantum states and
   Hamiltonians.
\eq
Also a piece of the Preamble section of ``Delirium Quantum'':
\bq
   I think the greatest lesson quantum theory holds for us is that when
   two pieces of the world come together, they give birth.  [Bring two
   fists together and then open them to imply an explosion.]  They give
   birth to FACTS in a way not so unlike the romantic notion of
   parenthood:  that a child is more than the sum total of her parents,
   an entity unto herself with untold potential for reshaping the world.
   Add a new piece to a puzzle---not to its beginning or end or edges,
   but somewhere deep in its middle---and all the extant pieces must be
   rejiggled or recut to make a new, but different, whole.  That is the
   great lesson.

   But quantum mechanics is only a glimpse into this profound feature of
   nature; it is only a part of the story.  For its focus is exclusively
   upon a very special case of this phenomenon:  The case where one
   piece of the world is a highly-developed decision-making agent---an
   experimentalist---and the other piece is some fraction of the world
   that captures his attention or interest.
\eq

So, clearly, I have sympathy for {\it something\/} that goes beyond quantum theory.  But what is that something?  The idea behind the Preamble passage has its origin in a kind of Copernican principle.

\section{13-12-06 \ \ {\it Offline Discussion} \ \ (to R. {\Schack})} \label{Schack113}

Here are the passages you and {\Carl} wrote:

\brs
I am, however, totally puzzled by your problem with what you call the
``exception''. I have said over and over again that from our
perspective this is a trivial issue. I think you have some kind of
mental block here. {\bf All} probabilities are degrees of belief. {\bf All} probabilities reside in the agent, not in the world. But the agent has to be coherent. The agent's probabilities generally depend on facts (these are facts for the agent) and his prior. Given a prior and some facts, the agent is forced by coherence to a particular
probability assignment. There are trivial cases, however, where the
prior doesn't come in. If $x=5$ is a fact for the agent, then the
agent is forced to the probability assignment $P(x=5)=1$ by coherence alone. There is nothing deep here at all.
\ers
and
\bcc
The point of the exception is to make clear that we are not saying
that probability theory is inconsistent with logic.  It is consistent with logic. Now in logic, you could say that some facts imply other
facts.  When thinking in terms of probabilities, you would say the
same thing as some unit probabilities imply other unit probabilities
via Bayes's rule. Now you can always maintain a distinction---and we
would want to maintain one---between the facts themselves and
certainty about facts, and I personally would be happy to change
things so that we're saying the right thing.

The main point is that sometimes in the case of certainties, facts do imply some other unit probabilities by logic, and there is no
escaping that.  For example, suppose we observe that one thing has
the value two and another thing has the value three.  These are now
facts (for the agents), but the agent cannot escape assigning unit
probability to the sum of the two having value five.  That's the only thing I want to guard against.
\ecc
And here's what I had written in the now deleted footnote:
\bq
\noindent
The problem, {\it as [Fuchs] sees it}, with the formulation that there is a case whereupon a fact strictly determines a probability is that it
puts the horse before the cart:  It acts as if abstract `logical
implication'---a statement of the form ``$d\Rightarrow h_0$''---has a meaning independent of the {\it subjective\/} judgment ``$\Pr(d|h)=0$
for $h\ne h_0$.''  (See for instance, F.~C.~S. {\Schiller}, {\sl Formal Logic: A Scientific and Social Problem}, (Macmillan, London, 1912), and F.~C.~S. {\Schiller}, {\sl Logic for Use: An Introduction to the
Voluntarist Theory of Knowledge}, (G.~Bell, London, 1929), for a development of this point.)
\eq
Let me try to say the same thing, but in a little more detail.

In my weakest criticism of this exception business, I would point out that you two seem to be drawing no distinction between analytic and synthetic truths (facts).  It seems you're wanting to apply a blanket rule to both, and that's partially what's getting under my skin:
\bq
\noindent
We may \ldots\ adopt the unsophisticated definition that a synthetic
statement has meaning only in terms of matter of fact, whereas an
analytic statement has meaning independently of matter of fact
(roughly, the Leibnizian distinction between {\it v\'erit\'es de
fait\/} and {\it v\'erit\'es de raison}).  An analytic statement, as has been charged, may still be considered as relative to the language
or symbolism in which it is stated \ldots\  However, these considerations should not affect the essence of the distinction, which is that, in
theory, an analytic statement, since it is concerned only with the
relations of meanings, cannot be verified or falsified by any
conceivable human experience, whereas a synthetic statement,
regardless of its ``certainty,'' may always, in theory, be upset by
experience.  A typical analytic statement is, ``two plus two equals
four.'' If the meaning of the words ``two,'' ``plus,'' ``equals,''
and ``four'' is made clear, it is impossible for this statement not
to be valid.  Synthetic statements are the material of science, and
their degree of certainty may approach any assignable limit, with the qualification that they are always subject to verification in
experience, i.e., may conceivably be falsified.
\eq
Propositions corresponding to quantum measurement outcomes are not analytic statements (nor are classical measurement outcomes for that matter).  And so, so-called ``logical implication'' between any two such statements can be nothing other than a subjective judgment of certainty, $\Pr(d|h)=1$.  Logical implication in this kind of situation has no independent definition.

As previously written in the paper, {\Carl}'s choice of words seemed to indicate a contradiction to this point.  (Actually, it went further than that---and may still in {\Carl}'s mind---as evidenced by all the earlier attempts of {\Carl} to draw a distinction between classical and quantum based on such ideas.)

My heels dug in when I looked at the passage on Bayes' rule as it had been written:
\bq
\noindent
Classically, Bayes's rule,
\begin{equation}
\Pr(h|d) = \frac{ \Pr(d|h) \Pr(h) }{ \Pr(d) }\;,
\end{equation}
is used to update probabilities for hypotheses $h$ after acquiring
facts in the form of data $d$.  The posterior probability,
$\Pr(h|d)$, depends on the observed data $d$ and on prior beliefs
through the prior probabilities $\Pr(h)$ and the conditional
probabilities $\Pr(d|h)$. The only exception to the dependence on
prior probabilities occurs when the observed data $d$ logically imply a particular hypothesis $h_0$, i.e., when $\Pr(d|h)=0$ for
$h\ne h_0$, thus making $\Pr(h_0|d)=1$.
\eq
and saw nothing but subjective judgments in every slot on the right-hand side, yet talk of {\it exceptions}.  To me, it sent the wrong message about our whole enterprise.  Once one exception, why not two?  Particularly if some subjective judgments are forced by facts (without prior judgment), why not others?  Why not, in fact, allow that some nontrivial probabilities---quantum probabilities---can even be forced by facts (without prior judgment)?

Since the real subject matter of our paper was for $h$ and $d$ that are  measurement outcomes, I saw no reason to sew a seed of doubt.

But---you might say---``That was just a little trouble in the way {\Carl} wrote up the point.  What we really meant to convey is that
$\Pr(h|h)=1$, and that holds true without subjective judgment even for synthetic statements.  It's just coherence.''

And I say, ``NO, even that is wrong---or at least irrelevant---as long as $h$ is synthetic.''
\bq
\noindent
It is Aristotle's logic which is usually taken as the archetype of
formal logic.  However, the `logical analysis of judgment' in
anyone's hands is bound to come to failure if it discounts the
personal intention and circumstances of the maker of the judgment.
Thus Hegel's formula of judgment as identity in difference disregards the fact that identities are always made or postulated by us through
conscious disregard of those differences which we take for the moment to be irrelevant to the purpose of judgment.
\eq
If the right-hand entry and the left-hand entry of $\Pr(h|h)$ are meant to be two distinct empirical instances of ``$h$'' (the quotes important), then it is a subjective judgment.  If the right-hand entry is synthetic, while the left-hand is treated as (now) analytic, I don't know what the very expression $\Pr(h|h)$ means.

Take your particular point:
\brs
If $x=5$ is a fact for the agent, then the agent is forced to the
probability assignment $P(x=5)=1$ by coherence alone.
\ers
Why would the agent write down $P(x=5)=1$ unless he were contemplating gambling on whether $x=5$ (and is quite certain of the outcome)?  The only sense I can make of that situation is that what is really happening is that the agent is contemplating gambling on a NEW instance of $x=5$, and because of the previous observation he is certain he will find it again.  But that is a subjective judgment.  If you take away my right to think of it as a new instance, then I will say he cannot be gambling on it at all---the statement $x=5$ is a dead fact; it has passed away.  And what you must mean by $P(x=5)=1$ is something other than a gambling commitment.

The same can be said for {\Carl}'s example above (though he does slip in a little ``2+2=4'' issue on the side, which as far as I can see is irrelevant for the observation he really wants to make \ldots\ which is the same as the one you made, or at least he told me so).

So, now we get to the point.  My present opinion is that probability statements are really only ever meant to be made about ``live options'' (to use a phrase of William {\James}, though I may not be using it exactly the same way as he did).  One can indeed write $P(x=5)=1$, but then
$x=5$ had better a live option, rather than a ``fact for the agent.''  If it is a fact for the agent, then we are reinstating the category error.  The only role for the ``facts for the agent''---despite their very real consequences, like `life' or `death'---are in pivoting the agent's subjective probabilities.  They are never used for the setting of probabilities unconditionally.

Could I make the criticism stronger?  Yeah, I think so, if I were to argue along the lines of {\Schiller} that there simply are no such things as analytic judgments anyway.  But that's probably further than I need to go.

Do you see my concern any more clearly now?

\subsection{{\Ruediger}'s Reply}

\bq
Let me try to tell you what I understand about facts and certainty.  You say that once {\it A\/} is a fact for the agent, the agent can't be certain of it in the sense of assigning probability 1 to it. When something is a fact for the agent, it doesn't normally make sense for him to ask ``Am I certain of the fact?''. This is one of {\Wittgenstein}'s points. And I fully appreciate it. In the quantum context, this has to do with the status of ``clicks'', which are directly given to the agent. Facts are facts for the agent and don't have an independent existence out there. This is why the agent should not assign probabilities, not even 0 or 1, to facts.

But where does leave the analytic/synthetic distinction? Isn't that a pre-Witt\-gen\-stein\-ian distinction? I think the analytic/synthetic distinction moves the whole discussion in the wrong direction.
\eq

\section{13-12-06 \ \ {\it More on Identity} \ \ (to R. {\Schack})} \label{Schack114}

Let me supplement what I just sent you with a little email conversation I had with Charlie {\Bennett} on the very point of `identity'.  [See 28-11-06 note ``\myref{Bennett51}{Funes $+$ Borges}'' to C. H. {\Bennett}.] The messages below are in reverse time order.  What Charlie is referring to is Matt Leifer's paper, \quantph{0611233}; and CFS refers to you, {\Carl}, and me.

There's a funny anecdote that goes with it too.  One evening while walking back from dinner in Tsukuba, Charlie said to Daniel Gottesman while pointing up to a traffic light, ``For instance, Chris wants to say that that traffic light right now is not the same the traffic light as \ldots''  And, by wonderful fate, just as Charlie gestured up to it, the light changed color!  We all had a very good laugh, and it's clearly already become one of Charlie's palette of stories:  I heard it at least three times over again at the meeting.

\section{13-12-06 \ \ {\it Analytic/Synthetic} \ \ (to R. {\Schack})} \label{Schack115}

\brs
But where does leave the analytic/synthetic distinction? Isn't that a
pre-{\Wittgenstein}ian distinction? I think the analytic/synthetic
distinction moves the whole discussion in the wrong direction.
\ers

And a pre-{\James}ian and pre-{\Schiller}ian distinction too.  Yes, it does move the discussion in the wrong direction.  But I left my exposition couched in terms of it because I thought it might strike a chord more quickly in you that way.  (It was a first assault.)  That's why I said, ``In my weakest criticism of this exception business, I would point out \ldots''  Later I hinted at the corrective with this:
\bq\noindent
Could I make the criticism stronger?  Yeah, I think so, if I were to
   argue along the lines of {\Schiller} that there simply are no such
   things as analytic judgments anyway.  But that's probably further
   than I need to go.
\eq
{\Schiller}, too, rejects the distinction.

Blending that point into the mix, the simplest solution to the dilemma caused by the ``exception'' was to have never imagined it to exist in the first place.  No mention of analytic/synthetic, but no mention of exceptions either.  (We had tried that solution once in an earlier draft, but it only made Prof.\ {\Caves} that much more bitter.)

\section{14-12-06 \ \ {\it What a Friend!}\ \ \ (to J. A. Smolin)} \label{SmolinJ8}

I just sent these words off to Mr.\ Khrennikov for part of the Preface to his latest volume.  Now, read that and tell me I'm not a good friend!  At some stage relatively soon, we've got to get back to working that out.  \ldots\ At the moment though SIC-existence is more important.

\bq
{\bf Swedish Bayesian Team.} Eight of the meeting's participants came
to {\Vaxjo} as a task group to further develop an understanding of
quantum probabilities as Bayesian degrees of belief and to examine
various technical issues surrounding the idea.  These were D. Marcus
{\Appleby}, Hans C. von Baeyer, Ariel Caticha, G. Mauro D'Ariano, C.~A.
Fuchs, Matthew S. Leifer, {\Ruediger} {\Schack}, and John A. Smolin.
Quantum Bayesianism is a view of quantum mechanics that has been
making much progress and is becoming widely recognized; it was thus
good to be able to come back to {\Vaxjo}, which played a pivotal
role in its development already with the ``Shannon meets Bohr''
session of the 2001 meeting. As concerns the talks, D.~M. {\Appleby}
gave a nice introduction to the Bayesian conception of quantum-state
tomography and explored the extent to which complete sets of mutually
unbiased bases (so-called MUBs) and symmetric informationally
complete measurements (so-called SICs) are optimal measurements for
tomography.  A.~Caticha gave a potential derivation of the quantum
formalism along lines inspired by R.~T. Cox's derivation of the
axioms of probability theory.  G.~M. D'Ariano presented some very
fresh work on an axiomatic system for quantum mechanics inspired by
the practice of quantum tomography and many of the work-a-day
features of quantum information theory.  C.~A. Fuchs gave a
presentation aiming to clarify some the ontological aspects of the
quantum Bayesian position, arguing that, though quantum states are
subjective from this conception, the Born rule for transforming
probabilities from one measurement context to another represents an
objective feature of the world.  M.~S. Leifer presented a new and
very useful isomorphism between bipartite quantum states and quantum
operations and discussed its implications:  Particularly, he argued
that from a Bayesian conception, to be consistent, quantum operations
should be interpreted as of the same subjective character as quantum
states. R.~{\Schack} discussed the issue of what it could possibly mean
for a quantum random number generator to be better than a classical
one and gave a rather deep answer:  When an observer writes down a
pure state for a quantum system, he is not only expressing his
certainty for the outcome of some measurement, but he is also
expressing his certainty that no one else can have what he would deem
as ``inside information.''  Finally J.~A. Smolin posed what was
perhaps the deepest question of the conference (for the Bayesians at
least!):  Forget about better and better Bell inequality violations;
to what extent has the basic Born rule for calculating probabilities
ever been independently tested, and indeed, can it be experimentally
tested?

We might add that three other participants of the meeting count as
{\it honorary\/} members of the Swedish Bayesian Team (whether they
would like such an association or not!).  First, there is Arkady
Plotnitsky, whose riveting talk on Bohr's reply to EPR, helped make
clear the single-case conception of all quantum phenomena (thus
seemingly requiring a Bayesian notion of probability). Then there is
Ingemar Bengtsson, who discussed the still-open (for $>30$ years!)\
problem of the existence or nonexistence of MUBs for non-prime-power
dimensional Hilbert spaces.  This kind of work and related
constructions are likely to be crucial for better Bayesian
reconstructions of the quantum formalism. And finally Jan-{\AA}ke
Larsson, for his role as the conscience for Bell-violation
experimentalists; we fear that one day he will be the conscience of
the quantum Bayesians!
\eq

\section{15-12-06 \ \ {\it Background Independent Bayesians} \ \ (to L. Smolin)} \label{SmolinL5}

I know he's already in your neighborhood, and you probably know him better than I do, but I just thought I'd tell you that you guys have an excellent postdoc in Matt Leifer.  I've been editing the volume for the {\Vaxjo} proceedings and discovered I really like Matt's submission:
\quantph{0611233}.
(Well, of course I would like it---and you'll easily catch why if you read the Introduction.)  But because of the closing remarks in the paper it struck me that you might have some interest in the paper too.  I think it is a good point, and if this quantum Bayesian---i.e., me---were to try to dabble in quantum gravity, it might be the first observation I'd try to get some traction from.

Also, in spite of the motivation for the formalism (i.e., if you can't stomach those Bayesians), you might find the formalism independently useful.

\section{15-12-06 \ \ {\it Swedish Delirium} \ \ (to D. Gottesman)} \label{Gottesman3}

I don't know if you could see it in my face, but I was taken quite aback when you said at the Tsukuba banquet (something like), ``The trouble with {\it all\/} the interpretations is that, by hook or crook, they always try to describe quantum mechanics in some classical way.''  Well \ldots\ ahem \ldots\ of course I thought that about every {\it other\/} interpretation, but I had never imagined such a criticism being leveled at me!

Anyway, I thought of you again last night as I was hurriedly annotating the attached ``pseudo-paper'' for the wacky Swedish proceedings.  And I wondered whether after reading it you'd still drop me in the same pot with all the rest.  (Maybe I just want to think I'm ``more anti-classical'' than the rest so that some one will call me MAC-daddy some day.)  But seriously \ldots\ \  You probably will (drop me in the same pot, that is).  But I'd like to better understand why, particularly in light of the peculiarities of the attached document.

On a different subject, might I ask you to send me a copy of your Tsukuba presentation?  I gave your paper a perusal, but didn't find your nice little example in it with the transition matrices written explicitly.  I think I'll need that kind of gentle leading to absorb your point.

\section{18-12-06 \ \ {\it I and It} \ \ (to N. D. {\Mermin} \& C. G. {\Timpson})} \label{Mermin125} \label{Timpson12}

I'm going to cc this note to Chris Timpson, since it is directly relevant to him.  (I.e., I will treat it as out of the context of my other notes to you today.)

\bdm
Footnote 25.  I can't imagine how a successful gambler can be
certain that red will turn up next without also feeling that it is
certain that red will turn up next.  So this footnote just confuses
me.
\edm

I think your confusion here comes from not parsing the sentence the right way.  (We probably need to add something more to the text to make sure no one else has the same trouble as you.)

It comes from some things that Chris {\Timpson} brought to our attention.  Let me paste in {\Timpson}'s explanation. [See 17-11-06 note ``\myref{Timpson11}{Certain Comments}'' to C. G. {\Timpson}.]

Or maybe that's not the trouble at all---as I keep staring and staring at your sentence, I start to wonder.  Maybe it is actually the same trouble that {\Timpson} expressed below?

Here's a far-fetched example of how it could possibly be.  {\Schroedinger} knew---nay, was certain---that with a fresh lover he would be able to make a great scientific discovery.  The lover is the catalyst for the discovery.  No lover, no discovery.  Yet, ask the lover, and she will have no clue about what her presence does for his science.  He is certain, but she is not.

From our view---let me be careful, from my view---a quantum system is a catalyst.  It brings about a transformation in the agent who interacts with it.  Sometimes the agent is certain (in the subjective sense) about the transformation that will be the result; sometimes he is not.  But even in those times in which he is certain, there is nothing in the system itself that is informative of his certainty.

Buy it?

\section{18-12-06 \ \ {\it Bennett's Hoffa Talk and Certainty} \ \ (to R. E. Slusher)} \label{Slusher18}

By the way, here's a copy of Bennett's painful (to him) Tsukuba talk.

Also attached is a much better version of our certainty paper than you had seen before. [C. M. {\Caves}, C. A. Fuchs, and R. {\Schack}, ``Subjective Probability and Quantum Certainty,'' Stud.\ Hist.\ Phil.\ Mod.\ Phys.\ {\bf 38}, 255--274 (2007).] It's what gave rise to the {\Schroedinger} story below.

\bq
Here's a far-fetched example of how it could possibly be.  {\Schroedinger} knew---nay, was certain---that with a fresh lover he would be able to make a great scientific discovery.  The lover is the catalyst for the discovery.  No lover, no discovery.  Yet, ask the lover, and she will have no clue about what her presence does for his science.  He is certain, but she is not.

From our view---let me be careful, from my view---a quantum system is a catalyst.  It brings about a transformation in the agent who interacts with it.  Sometimes the agent is certain (in the subjective sense) about the transformation that will be the result; sometimes he is not.  But even in those times in which he is certain, there is nothing in the system itself that is informative of his certainty.
\eq

\section{19-12-06 \ \ {\it Big Time Debt} \ \ (to C. M. {\Caves} \& R. {\Schack})} \label{Caves93.1} \label{Schack115.1}

\bcc
I made changes to the relevant footnote.  I'm sure you can do better.
He seems to want the quantum state to have a sort of disembodied
objectivity, like an objective concept, say, the probability rules.
That doesn't seem very reasonable.  I wouldn't be opposed to dropping
this footnote, as Mermin suggests below.
\ecc

No, the footnote Mermin suggested dropping was this one:
\bq\noindent
   And if you added Footnote 56 just to make me happy, please feel
   free to drop it.  Readers will wonder why it's there.
\eq
That footnote was this:
\bq
\noindent \verb+\bibitem{LogicalDeduction}+\\
Bayesian updating is consistent, as it should be, with logical deduction
of facts from other facts, as when the observed data $d$ logically imply a particular hypothesis $h_0$, i.e., when $\Pr(d|h)=0$ for $h\ne h_0$, thus making $\Pr(h_0|d)=1$.  Since the authors disagree on the implications of this consistency, it is fortunate that it is irrelevant to the point of this paper.  The point of this paper is the status of quantum measurement outcomes and their probabilities, and quantum measurement outcomes are not related by logical implication.  Thus we do not discuss further this consistency, or its implications or lack thereof.
\eq

\section{20-12-06 \ \ {\it All Unambiguous Account} \ \ (to C. M. {\Caves})} \label{Caves94}

\bcc
One thing I discovered as I was doing this.  Shouldn't that quote from
Bohr to Pauli start with ``In all unambiguous accounts'' instead of
``account''?
\ecc

Nope, that is just as Bohr said it.  I was very careful in entering those words into my computer, which come from a fascinating set of letters between Bohr and Pauli (written in English!)\ that Henry Folse supplied me with.

Bohr's letters can be found on pages 14--16 of the attached document.  (And a quote of Folse quoting the phrase ``in all unambiguous account'' too can be found on page 39.)  Pauli's more interesting (by my estimate) replies can be found on pages 132--135.

We could add an ``[sic]'' to the Bohr quote.

\section{20-12-06 \ \ {\it New Jersey Atoms} \ \ (to D. Gottesman)} \label{Gottesman4}

Thanks for sending me your talk.

\bdg
When you talk about an ``event'', that is a classical concept.  Less
obviously, ``information'', ``knowledge'', ``belief'' are also all classical concepts.  Of course, the actual classical concepts don't quite fit the quantum objects you'd like to assign them to, so you have to start to modify them, to give them more quantum properties.
\edg
I pretty much expected you'd say that, at least with ``event''---you said about as much in Tsukuba (though I think you used the word ``data'').  But I guess I was hoping you'd expand on what you mean by all this.  For instance, I don't know what you mean when you say ``belief is a classical concept.''  Classical in the sense of classical physics?  For myself, I would think terms like ``belief'' have essentially nothing to do with any particular conception of the world.  For instance, one could speak of beliefs (and even rudimentary probability theory) at a time when the world was thought to be animated with little spirits or when rocks were thought to fall according to Aristotelian rules.  Or in Hinduism where the world is but the dream of four brothers dreaming Vishnu dreaming the world containing four brothers dreaming Vishnu dreaming \ldots

If I see you in early February, maybe I can get you to tighten up what you mean.

\bdg
This, I think,
is not at its root different from the way most interpretations work:
They take some classical concept and try to modify it to allow it
to describe quantum mechanics.  I don't know that there's anything
{\bf wrong} with that, but I also don't know why one would expect there to be a uniquely ``best'' way of doing it.
\edg
True enough.  I understand your point on this.  But I do think the quantum formalism is trying to tell us something unique about the world, and despite our using the formalism every day, we don't yet have a good grasp of what ``it'' is trying to get at.  Whatever it takes to move forward on that project is, in the end, the important thing.  (And so: I'm ready to be Darwin'd out if it so happens, but at the moment the Bayesian way strikes me as the best way to proceed.)

\bdg
While there are certainly many ways in which atoms can be distinct,
they also can, in fact, be exactly the same.  The very notion of
``separateness'' of atoms breaks down in a BEC.
\edg
There's something in the way you pose this that makes me think you don't quite get what I was shooting for in that rumination.  (And it's no wonder, because I don't like the way I wrote it now either.)  From my {\it toyed-with\/} conception, atoms (and events) are not something larger bodies (and histories) are built up from.  But rather atoms are residues of in-common judgments about the bodies.  So, even from this point of view atoms can be exactly the same (as you would have them).  But now the statement almost turns into a tautology, rather than being empirical.  That is to say, if one has adopted an epistemic view of quantum states as a working hypothesis (as {\Caves}, {\Schack}, and I have), a symmetrized or anti-symmetrized quantum state is still an epistemic state, rather than a state of the world.

In connection with that point, I'll attach the latest {\Caves}--Fuchs--{\Schack}ism in case there's an evening when you find you can't fall asleep.  (This version is significantly different from the one we put on the web earlier under duress.)

\section{20-12-06 \ \ {\it Paper} \ \ (to R. {\Schack})} \label{Schack116}

\brs
Not that I disagreed with what it says, but I thought it would disturb
some readers unnecessarily. So I have some sympathy with the idea of
dropping it.
\ers

Disturb some reader unnecessarily in what sense?

The whole point of our paper is that we say that quantum mechanics leads us to think that there can (validly) be subjective certainty without an underlying objective certainty.  What is wrong with citing\footnote{The citation in question was this one:
More philosophical, quantum independent, precedents for this notion
of `certainty' can be found in A. White, ``Certainty,'' Proc.\
Aristotelian Soc. {\bf 46}, 1 (1972).  For example, White puts it
this way,
\begin{quote}
The certainty of persons and the certainty of things are logically
independent of each other.  Somebody can be (or feel) certain of
something which is not itself certain, while something can be certain
without anybody's being (or feeling) certain of it.  `He is certain
that $p$' neither implies nor is implied by the impersonal `It is
certain that $p$'.  The same thing cannot be both certain and not
certain, though one person can be certain of it and another of its
opposite \ldots\  People can become more or less certain of something
which itself has not become any more or less certain.  \ldots\ A gambler
need not feel that {\it it\/} [our italics] is certain that red will
turn up next in order to feel certain that it will.
\end{quote}
See also, L.~{\Wittgenstein}, {\sl On Certainty}, (Blackwell, Malden,
MA, 1975).
} someone who backs up a similar idea---or at least an idea that moves in a similar direction---and, in a clearer way than we do, in another context?

{\Timpson} may be wrong that this is relevant, but he is not dismissive and is thinking hard about the issue.  We need people like him, and if it helps to put our thoughts a little into the context of his world (and people's like his), I don't see the harm.  Saying the truth---as we understand it at any moment---is the only thing that's ever going to make us free (almost a JC\index{Christ, Jesus} quote).

{\Carl} says this paper is ``written for physicists.''   What physicist is going to look at a journal like SHPMP?  The only people who will look are the ones who really want some clarity, or have a philosophical bone to pick.  Either way, it seems to me, we ought to be providing them with as much material as we can, when we can.

It strikes me that you guys always work from a starting point of the ``the less said the better''.  But I think we're tearing a huge whole in these guys' (our reader's) worldview and if we don't put something in its place---some kind of metaphysics, some kind of world picture---we're just asking for trouble after trouble upon ourselves.  Wisemans calling us solipsists; Bubs calling us instrumentalists; {\Stamp}s calling us behaviorists; Pereses calling us positivists.  Everyone thinking we're making the world smaller, rather than bigger.  It is true that we may never get rid of all those biting, deaf dogs, but if we don't try, we've really only got ourselves to blame.

That said, I am willing to remove the quote of White (rather than adding a little more to it to help set the context) and leave only a citation of he and {\Wittgenstein} if you really wish.  But I'd hope you'd do it out of a reason deeper than fear.  (Fear, that is, that some jerk out there who says, ``This is not physics; I know what physics is.  This is philosophy with no impact on physics, and I am therefore warranted to close my mind.''  It is that kind of arrogance that keeps such people from making the really big discoveries, and I want a community around us that will make big, really big, discoveries, even if we never do.)

\section{20-12-06 \ \ {\it I and It, 1} \ \ (to R. {\Schack})} \label{Schack117}

{\Mermin}'s replies on I and It.  The first was to my note (about {\Schroedinger} and his lovers).  The second was to {\Timpson}'s long linguistic analysis on the Moore-type sentences.  (Which---I disagree with you---are relevant if one thinks of quantum measurement in the common ``question-answer'' motif, {\it rather than\/} in the alchemical motif I keep trying to refine.  In fact, the questions he raises give extra impetus, give extra reason for the alchemical/Paulian motif.  For that view, I think, sidesteps the issue he fears.  {\Timpson}'s distrust, I believe, has its source in what I railed about earlier:  Our tearing too much out of his worldview without replacing it with a new image.)

\section{22-12-06 \ \ {\it Done} \ \ (to C. M. {\Caves} \& R. {\Schack})} \label{Caves94.1} \label{Schack117.1}

Just to let you know, I'll be getting the paper out the door by tomorrow morning---unfortunately I'm tied up now (the leaves in the rain thing took me longer than expected).  I'll add one sentence introducing White better.  And one other minor, minor change:  I reordered two of the Bohr footnotes, and introduced the Murdoch quote with: ``Even more extreme is this reading of Bohr, from p.\ 107 of Murdoch (1987):'' (since Murdoch explicitly says the values are `preexistent' once the measurement is set):
\bq\noindent
Bohr, then, held what I shall call the {\it objective-values
theory\/} of measurement, according to which successful observation
or measurement reveals the objective, pre-existing value of an
observable. \ldots\ It is important here to note that the
objective-values theory should not be confused with what I shall call
the {\it intrinsic-values theory\/} of properties \ldots\ according
to which all the observables of an object have, at any moment,
definite values.  The latter theory, which Bohr rejected, is
logically independent of the former.
\eq
Other than that I'll leave everything the same.  I'll also write a swaying note to Hartmann saying (nicely) that it would be a disaster if he tried to modify anything.

\section{23-12-06 \ \ {\it Real Data!}\ \ \ (to G. L. Comer)} \label{Comer99}

\bgc
If you take a look at \arxiv{astro-ph/0605007} you'll find a paper of Nils, Reinhard Prix, and me referenced.  This is quite unique for me, because our model for neutron star glitches is being put to the test via real data analysis.  And \ldots\ we're still in the hunt!  It turns out that glitch data and our model share a certain linearity.
\egc

That's exciting!  What a way to go into the Christmas season.

I've never had anything comparable in my life---to the extent I have made any positive predictions at all, they've always been of use for engineering, never explaining any naturally occurring phenomenon.  I can tell you one funny story in that regard that happened recently.  A couple of weeks ago in Japan I was walking through the poster session at QCMC, and ran across two guys from MIT with a very professional looking poster titled, ``Experimental Realization of the Fuchs--Peres--Brandt Probe.''  I thought, ``what's that?''\ and stopped to talk.  Well, they tell me all about my original paper with Asher---or so I thought that's what they were telling me at first---and about how they used the calculations there to design a real-world optimal eavesdropping setup.  I ask them how much it cost to perform the experiment and that kind of stuff---about \$10K in parts plus a very, very expensive laser that was already in the lab, etc., they tell me.  Then, at some point, it all starts to sink in:  Whatever they've done, they've used it for eavesdropping on BB84.  Whereas my paper with Asher had solely to do with B92!  So, I explain this, and they say, ``No, there's a part on BB84.''  I say, ``No, no, there's nothing on BB84 in it.''  They say, ``Yes there is, Professor Shapiro told us so!''  Well, like a big Emily Litella, I think the whole experiment---or at least the title of it!---was a big ``Never mind.''

To think, there's a neutron star out there that knows the name of Greg Comer.  (Newton's Third Law ultimately says knowledge is reciprocal, you know.)

\section{24-12-06 \ \ {\it {\Spekkens} and Kant} \ \ (to R. W. {\Spekkens})} \label{Spekkens40}

Happy holidays!\medskip

\noindent{\bf Rob {\Spekkens} on wave functions:}
\bq
We shall argue for the superiority of the epistemic view over the
ontic view by demonstrating how a great number of quantum phenomena
that are mysterious from the ontic viewpoint, appear natural from
the epistemic viewpoint.  These phenomena include [about a million
things]. \ldots\ The greater the number of phenomena that appear
mysterious from an ontic perspective but natural from an epistemic
perspective, the more convincing the latter viewpoint becomes. \ldots

Of course, a proponent of the ontic view might argue that the
phenomena in question are not mysterious \emph{if} one abandons
certain preconceived notions about physical reality.  The challenge
we offer to such a person is to present a few simple \emph{physical}
principles by the light of which all of these phenomena become
conceptually intuitive (and not merely mathematical consequences of
the formalism) within a framework wherein the quantum state is an
ontic state.  Our impression is that this challenge cannot be met.
\eq

\noindent{\bf Immanuel Kant on ghosts:}
\bq\noindent
I do not dare wholly to deny all truth to the various ghost stories,
but with the curious reservation that I doubt each of them singly,
but have some belief in them all taken together.
\eq

\section{24-12-06 \ \ {\it Christmas Conversations in My Head} \ \ (to C. H. {\Bennett})} \label{Bennett55}

I read the lines below (which I'd like stored in my computer anyway, from Reuben Abel's book {\sl The Pragmatic Humanism of F. C. S. Schiller}), and I thought again of our conversations over the years.  Particularly I wondered whether the quote of {\Schiller} (the very last sentence in the paragraph below) might characterize our relationship.

\bq
This amount of indeterminism does not upset the structure of science. Alternative modes of conduct are, at least partially, calculable.
Continuity of character is a limitation on complete freedom.  But
this is not equivalent to saying that, since character and habits
affect choice, choice is therefore not free.  It is commonplace that people frequently do things on impulse, or without reason, or
``unexpected of them.''  Moreover, character is not a fixed entity, but a growing, evolving, changing concept.  Is it ever completely formed?
And besides, there exist situations in which the motives to contrary actions balance each other.  A choice can frequently be shown to
have, in retrospect, rational connections with the antecedent
circumstances.  But this does not imply that another choice in the
same circumstances would not also have, in retrospect, such rational connections.  Alternative courses may be equally free, yet they need
not be haphazard.  Determinists have argued that if the course of
events were not completely and rigidly determined, it must be
indeterminable.  This, {\Schiller} shows, is not necessary.  The two
postulates do not clash.  There is a certain amount of
indeterminism in the world which is determinable in alternative ways. The choice between these is a real freedom which does not involve the
dire consequences of either complete determinism or complete
indeterminism.  Scientific calculability and moral freedom are not in conflict.  Determinism as a scientific postulate is not endangered;
as an ontological dogma, like all other statements of metaphysics,
one may freely take it or leave it.  ``As we cannot vindicate our
freedom unless we are determined to be free, so we cannot compel
those to be free who are free to be determined, and prefer to think
it so.''
\eq
Just my weird way of saying:  Merry Christmas; you're never far from my thoughts.

\section{27-12-06 \ \ {\it My Christmas Present?}\ \ \ (to G. L. Comer, S. J. van Enk, and J. W. Nicholson)} \label{Comer100} \label{vanEnk7.1} \label{Nicholson26}

\noindent To my three most open-minded friends,\medskip

I've just got to share this thing I discovered with someone.  Go to Google and do a search on the three terms:  ``Chris Fuchs'', ``crazy'', and ``quantum''.  On the results page, click on the link at the bottom that'll take you to page 3(!)\ of the results, and look at the very last hit.  It's to something titled, ``Captain Stabbin -- Captain Stabbin Interpretation of Quantum Mechanics.''  If you're not on a company machine, and you're feeling brave, click on the link.  You'll be quite surprised!\medskip

\noindent Happy holidays!

\section{27-12-06 \ \ {\it New Year's Delirium}\ \ \ (to H. J. Bernstein)} \label{Bernstein8}

\noindent Hey Herb, you old reality maker,\medskip

I hope you're having a nice holiday, despite the meanness Charlie and I heaped on you the other day.  I'm having a nice time, thinking a lot about what quantum mechanics is trying to tell us, and reading up on F.~C.~S. Schiller's flavor of pragmatism.  (I'm giving a talk in Paris, Feb 23 at a meeting {\sl Pragmatism and Quantum Meeting}, titled ``William James, F.~C.~S. Schiller, and the Quantum Bayesians''.)  Schiller's ideas are fascinating and I think, perhaps, deeper than James's even.

Anyway, I was thinking I should send you a holiday letter and I came across a quote today that's giving me an opportunity.  Ultimately, it's about the ``big bang'' being right here, right now, all around us.  I'll give you the idea in three flavors, first in the form of a transcribed conversation between John Wheeler and R.~Q. Elvee, second in an excerpt from my pseudo-paper ``Delirium Quantum'' (which is attached in its final form, just sent to the publisher), and third in my newfound quote from Schiller.  It comes from his 1924 book, {\sl Problems of Belief}.

Wishing the best to you and your family for the coming year.  It sure would be nice to see you again sometime soon.  (Will you be going to the Zeilinger foundations meeting in Vienna in June?)

\bq\noindent
{\bf Flavor 1:}\medskip

ELVEE: Dr.\ Wheeler, who was there to observe the universe when it
started? Were we there? Or does it only start with our observation?
Is the big bang here?

WHEELER: A lovely way to put it---``Is the big bang here?'' I can
imagine that we will someday have to answer your question with a
``yes.'' If there is any conclusion that follows more strongly than
another about the nature of time from the study of the quantum nature
of space and time, it is the circumstance that the very idea of
``before'' and ``after'' is in some sense transcended.

There are two aspects of this idea. First, Einstein's theory of space
and time tells us that in order to predict all of space and time for
time to come, we have to know what the conditions of space are now
and how fast they're changing. Only then do we have enough
information to predict all the future. The uncertainty principle of
quantum theory tells us that if we know the condition of space now,
we cannot know how fast it's changing. Or if we know how fast it's
changing now, we cannot know what the geometry is now. Nature is so
built with this complementary feature that we cannot have the
information we need to give a deterministic account of space geometry
evolving with time.

That deterministic account of space evolving in time is what we mean
by spacetime. Everything that we say in everyday language, about time
is directly built on that concept. And with determinism out, the very
ideas of before and after are also out. For practical everyday
matters, this indeterminism, this indefinability of spacetime is of
no concern. The uncertainties only show up effectively at distances
of the order of $10^{-33}$ cm. Nobody at present has equipment fine
enough to reach down to a distance so small.

What does all this have to do with the big bang? At the very
beginning of time we know that---according to Einstein's
account---the universe was indefinitely small. Things were
indefinitely compact. When we talk about time when the universe
itself is so fantastically small, we deal with a state of affairs
where the very words ``before'' and ``after'' lose all meaning. This
circumstance puts one heavy restriction on the usefulness of the word
``time.'' There is another.

When we do our observations in the here and the now on photons,
quanta of light, hunks of energy coming from distant astrophysical
sources, we ourselves have an irretrievable part in bringing about
that which appears to be happening. We can put it this way: that
reality is, in a certain sense, made up of a few iron posts of
definite observation between which we fill in, by an elaborate work
of imagination and theory, all the rest of the construction that we
call reality. In other words, we are wrong to think of the past as
having a definite existence ``out there.'' The past only exists
insofar as it is present in the records of today. And what those
records are is determined by what questions we ask. There is no other
history than that. This is the sense in which we ourselves are
involved in defining the conditions of individual elementary quantum
phenomena way back at the beginning of the big bang.

Each elementary quantum phenomenon is an elementary act of ``fact
creation.'' That is incontestable. But is that the only mechanism
needed to create all that is? Is what took place at the big bang the
consequence of billions upon billions of these elementary processes,
these elementary ``acts of observer-participancy,'' these quantum
phenomena? Have we had the mechanism of creation before our eyes all
this time without recognizing the truth? That is the larger question
implicit in your comment. Of all the deep questions of our time, I do
not know one that is deeper, more exciting, more clearly pregnant
with a great advance in our understanding.
\eq

\bq\noindent
{\bf Flavor 2:}\medskip

The way I see it, quantum measurement outcomes are ultimate facts
without specific call for further explanation.  And indeed the
quantum formalism supplies none.  Thus there is more to the world
than the quantum formalism can supply.  Nothing to do with hidden
variables. \ldots

How does the theory tell us that there is much more to the world than
it can say?  It tells us that {\it facts\/} can be made to come into
existence, and not just at some time in the remote past called the
``big bang'' but here and now, all the time, whenever an observer
sets out to perform (in antiquated language) a quantum measurement. I
find that fantastic!  And it hints that facts are being created all
the time all around us.  But that now steps out of the domain of what
the quantum formalism is about, and so is the subject of future
research.  At the present---as a first step---I want rather to make
the interpretation of the quantum formalism along these lines
absolutely airtight.  And then from there we'll better know how to go
further.

Doesn't that just make you tingle?  That (metaphorically, or maybe
not so metaphorically) the big bang is, in part, right here all
around us?  And that the actions we take are \underline{\it part\/}
of that creation! At least for me, it makes my life count in a way
that I didn't dare dream before I stumbled upon Wheeler, Pauli, and
Bell--Kochen--Specker.
\eq

\bq\noindent
{\bf Flavor 3:}\medskip

On a very minute scale, but in a very real sense, our preferences and our acts are contributing to the shaping of the world, and sharing in the unceasing process of creation, which did not come to an end 5,928 years ago, but is continuously manifested in the all-pervasive creativeness which engenders \ldots\ novelties in every region of the universe.

If Humanism then is right, human agency is {\em not\/} the illusion it is so tempting to make it.  Truth may be, like the other values, like our moral and aesthetic ideals, a real contribution to reality, which the real might not possess unless we had made it.
\eq

\section{27-12-06 \ \ {\it Bayes, Born, and Everett} \ \ (to F. J. Tipler)} \label{Tipler1}

\bft
You probably have never heard of me, but I was a post-doc of John
{\Wheeler}'s from 1979 to 1981.
\eft

Of course I've heard of you.  I read your book with Barrow on anthropic principles about 20 years ago and enjoyed it immensely.  (My copy burned up in the Cerro Grande fire that came through Los Alamos six years ago, but I used to keep one of Sam Hurt's Eyebeam cartoons in it as a bookmark.  It was a scene of Eyebeam explaining anthropism to Ratliff; when each would turn his back on the other, the other would disappear from the scene.)

Your sentence indicates though, that you thought John {\Wheeler} might be a hook for my interest.  Indeed I have the greatest respect for some of John's more outlandish ideas on quantum mechanics.  (You might enjoy the {\Wheeler} story on page 149 of my \quantph{0105039}.  And you can find more {\Wheeler} stories by looking in the index; there's a better indexed version on my website.)  I wish I had known you were one of John's postdocs; I would have invited you to the little meeting Marlan Scully and I organized at Princeton last year for John on quantum information and quantum foundations.  Schumacher, Wootters, Unruh, and various of John's other quantum-interested associates came.  John even made an appearance for about 30 minutes.  It was quite nice to see him even at this stage in his life.  You could tell he was understanding the emotion of the occasion, even if not much else.  His smile and the thrusting of his fist into the air made it all worth it.

\bft
In reference to your ``Being Bayesian in a Quantum World,'' conference
last year,  I've just put a paper on the {\tt arXiv} (\quantph{0611245}), wherein I derive the Born interpretation using the Bayesian approach to probability, namely that probabilities are precise quantitative measures of human ignorance.
\eft

I have downloaded your paper and will study it carefully.  As it turns out, I'll be speaking at the ``Many Worlds at 50'' conference in Waterloo in September, and the title I gave the organizers for my talk is, ``13 Direct Quotes from Everettian Papers and Why I Find Them Unsettling.''  Not a lie.  I hope your paper doesn't make my list!

You certainly have a different use of the Bayesian idea of probability in mind with regard to quantum mechanics, than my closest colleagues and I do.  For fun, I'll attach two pieces of writing that aren't posted on the web yet---one's a paper, and one's a pseudo-paper---so you might see what I mean by this remark.  However, I might also send you to \quantph{0205039} which is the paper I'm most pleased with at the moment.  (It is not a mistake that the number above is {\tt 0105039}, while this one is {\tt 0205039}; it's just an interesting coincidence.)  Anyway, of the two papers attached, the pseudo-one tries to express what I don't like about block-universe (in the sense of William {\James}, say) conceptions of the universe---the Everettian idea is a species of this.  I'd rather think the universe is not complete in any sense, and I think quantum mechanics gives us plenty reason for exploring this turn of thought.  The other paper, particularly in its concluding section, tries to express what my colleagues and I see as the import of the Born rule:  The idea we are trying to develop is that it is not a rule for {\it setting\/} probabilities, but rather a rule for {\it transforming\/} Bayesian/personal/subjective probabilities.

\section{27-12-06 \ \ {\it Residue of the Category Error} \ \ (to S. J. van {\Enk})} \label{vanEnk8}

\bsve
Do I get this right, at least with high probability?
\esve

So, it's a live option for you?!  Not a fact?

Anyway, indeed you are on the right track (probably just using mildly different language than I use).  $h$ cannot both be a fact for the agent and a live option.  One makes the category error (i.e., thinking a probability assignment can be implied by a fact alone, and therefore itself is a fact) if one thinks $h\rightarrow h$ (a statement about a fact) enforces $P(h|h)=1$.  The latter is a judgment because, in order for $P(h|h)$ to be meaningful at all, the left-hand $h$ must be a live option.  That is, one judges that the live option will be the same as the dead fact.

\subsection{Steven's Preply}

\bq
What you say here sounds right to me:
\bq
So, now we get to the point.  My present opinion is that probability
statements are really only ever meant to be made about ``live options'' (to use a phrase of William {\James}, though I may not be using it
exactly the same way as he did).  One can indeed write $P(x=5)=1$, but
then $x=5$ had better a live option, rather than a ``fact for the agent.''
If it is a fact for the agent, then we are reinstating the category
error.  The only role for the ``facts for the agent''---despite their
very real consequences, like `life' or `death'---are in pivoting the
agent's subjective probabilities.  They are never used for the setting
of probabilities unconditionally.
\eq
In particular, I needed the part about ``reinstating the category error'' to be convinced.

``$x=5$'' could be either in the part labeled ``$d$'', or in the part labeled ``$h$'' in probabilities like $P(h|d)$, but they have different meanings then, one is a fact, the other is a live option.
So, writing $P(h|h)$ or $P(d|d)$ are both category errors.

Do I get this right, at least with high probability?
\eq

\section{28-12-06 \ \ {\it I and It, 2} \ \ (to R. {\Schack})} \label{Schack118}

\brs
I am just not sure what the problem is with simultaneously (a) knowing
there is nothing objective in the world guaranteeing an outcome and
(b) being certain of the outcome.
\ers

I'm finally replying to this.  Though, not much to say actually.  It is just that I think analyses of the {\Timpson} type are valuable for showing how twisted things can become under the ``question-answer'' motif for quantum measurement---that is, viewing a quantum measurement as asking a question of a system, and the system coughing up an answer.  With that imagery, how could the agent be certain of the outcome if he acknowledges that the system is not certain?  I can see his quandary.  But if one takes the alchemical view, as far as I can tell, it doesn't seem to be any big deal at all---it is not a great quandary.  The quantum system's role is that of a catalyst, not an answerer of questions---there is no reason for it to be ``certain'' of the outcome, i.e., there is no reason to have a fact residing in the system that indicates the outcome of the measurement.  And so my desire for {\Timpson} to run through a pretty thorough analysis of the opposing conception is a bit self-serving.  It's a useful exercise and someone should do it.

\section{28-12-06 \ \ {\it The Scandal of Bayesianism} \ \ (to A. H\'ajek)} \label{Hajek1}

Once upon a time, you showed me a draft of a paper that contained these words:
\bq
The scandal of Bayesianism is that no rational constraints on
subjective probabilities beyond probabilistic coherence have been
widely accepted.  \ldots\  Bayesians maintain and even celebrate a
remarkably anarchistic attitude to priors, (and derivatively to
posteriors): as long as they are coherent (or appropriately derived
from a coherent prior), anything goes.

\ldots\ Just as our beliefs aim for more than consistency, our
probabilistic judgments surely aim for more than coherence.  What,
then, do they aim for?  What does such a judgment's `fitting' the
world amount to?  What plays the role for subjective probability
analogous to the role that truth plays for all-or-nothing belief?
Presumably something plays this role, for probabilistic judgments
seem also to be governed by a norm of veracity.  How else can we
explain the fact that we regard some weather forecasters as better
than others, even when they couch their predictions
probabilistically?  And when different people assign different
probabilities to the same event---as the managers and engineers did
the day before the ill-fated space shuttle launch in 1986---we think that they are {\it disagreeing\/} about something.  There is surely
some fact about the world that a subjective probability assignment
strives to track.  Beliefs strive for truth; probabilistic judgments strive for \underline{\quad(fill in the blank)\quad}?
\eq
Could I ask you to email me a copy of that paper if you still have it around?

Thanks!

\subsection{Alan's Reply}

\bq
\bq\noindent
[CAF said:] ``Thanks for the paper.  Was it never published?''\medskip
\eq
Not yet. I did send it to a journal. The referee's report wrote that I ``didn't even discuss the obvious objection that nothing fills in the blank''. Not only DID I discuss this objection, but it got a section to itself, with an italicized heading. Maybe next time I should put it in bold face as well.  When I complained to the editors, they were very good about it --- apologetic, and they invited me to resubmit the paper. I eventually will, when I've revised it further.
\eq

\section{28-12-06 \ \ {\it Prepare Yourself} \ \ (to N. D. {\Mermin})} \label{Mermin126}

Sorry for my holiday absence (not that you probably minded!).  I started gearing up this morning to reply to your 12/18 note, particularly what I deem to be your main trouble expressed in it:
\bdm
All I'm saying is that this is rotten pedagogy, because it gives the
student no basis whatever for getting the whole process started.
Part (b) of the figure makes this dramatically clear.  All the circuit
does is transfer the state from the ancilla to the Qbit.  It doesn't
begin to answer the student's question of how you make an initial
state assignment independent of any prior state assignment.
Whether the initial state assignment is objective or subjective, the
grounds for making one remain utterly obscure.
\edm
and
\bdm
With reference to what you say below, the prior, as I understand
Figure 1  is that the damned ancilla on the left has the state
assignment $|0\rangle$, so you haven't told the [poor] student anything about what it means to make such a state assignment, either objectively or subjectively.  It's all circular.
\edm
So, I did a few Google searches.  For instance, I did one on ``Bayesian'' and ``prior come from.''  Well, there's some tasty quotes out there:
\bq
Bayes' Rule is central to modern economics and modern psychology.
According to Bayes' Rule, a rational person starts with some beliefs
about probabilities (his ``priors'') and changes them in a particular
way as new information arrives, in order to reach new beliefs (his
``posteriors''). Psychologists usually emphasize that people should use Bayes' Rule; economists are more likely to assume that people do use
Bayes' Rule.

The main problem with Bayes' Rule is that it doesn't say where priors
come from, or which ones you should have. It is tempting to say that
every prior is just your last posterior. But where did your first
prior come from? If you picked the wrong one, then everything based
on it could be wrong as well.
\eq
Maybe more relevantly I dug into my personal archives and found, for instance, this nice thing by Alan H\'ajek (from a 1998 draft of something):
\bq
The scandal of Bayesianism is that no rational constraints on
subjective probabilities beyond probabilistic coherence have been
widely accepted.  Bayesians maintain and even celebrate a
remarkably anarchistic attitude to priors, (and derivatively to
posteriors): as long as they are coherent (or appropriately derived
from a coherent prior), anything goes.

\ldots\ Just as our beliefs aim for more than consistency, our
probabilistic judgments surely aim for more than coherence.  What,
then, do they aim for?  What does such a judgment's `fitting' the
world amount to?  What plays the role for subjective probability
analogous to the role that truth plays for all-or-nothing belief?
Presumably something plays this role, for probabilistic judgments
seem also to be governed by a norm of veracity.  How else can we
explain the fact that we regard some weather forecasters as better
than others, even when they couch their predictions
probabilistically?  And when different people assign different
probabilities to the same event---as the managers and engineers did
the day before the ill-fated space shuttle launch in 1986---we think that they are {\it disagreeing\/} about something.  There is surely
some fact about the world that a subjective probability assignment
strives to track.  Beliefs strive for truth; probabilistic judgments strive for \underline{\quad(fill in the blank)\quad}?
\eq
But then I went back to bed \ldots\ and so now this is a restart.  (I hate it when I do that, for it forces me to change the planned titles of my notes.  The present title was meant to be the second title in the series.  But I've decided that maybe I'll just go for broke and put my complete reply in the present note.)

What I read into your note is simply a species of this general worry about Bayesianism.  ``Where does the prior come from?  Without telling me where a prior comes from, how can you have given me a complete solution to the meaning of probability?''

Compare:  Where does the initial quantum state come from?  Without telling me where the initial quantum state for a quantum computer comes from, how can you have given me a solution to the meaning of quantum states?  Particularly, how can you make any sense of what a quantum computer is doing if you don't give an answer for where its initial quantum state comes from other than saying that it's conditioned from (effectively) a previous quantum state?

Remember, for me, a quantum state is {\it nothing but\/} a single probability distribution (for the outcomes of a singled-out informationally complete observable).  Thus, to the extent that a subjective Bayesian leaves the ultimate origin of {\it any\/} probability assignment unanalyzed, precisely the same point holds for quantum states.

You call that circular.  The subjective Bayesian calls it necessary:  It is the cutting of a Gordian knot.  Without the stark recognition that all (ultimate) priors are personal to the agent, one would have an infinite regress.  Not circularity, but an infinite regress.  So, one says, ``There's a starting point, I can't get past that.''

For myself, I think, ``Why is that admission any more mysterious than what one finds in Newtonian mechanics?''  A pendulum, for instance, has an equation of motion and an initial condition.  Why doesn't anyone in a classical physics class complain, ``But where does the initial condition come from?  Without telling me the origin of the initial condition, you have not told me the complete story of the pendulum!''  Or why don't the instructors complain to the textbook writers, ``You think this is good pedagogy?  How am I going to explain to the students that they must just take initial conditions for granted \ldots\ and that it is beyond the bounds of the problem to ask where the initial conditions come from?''

As far as I can tell, in the two classes of problems the {\it only\/} difference of any note is in what the initial conditions are about.  There's no great conceptual or philosophical difference between the two.  The ordered pair $(x_0, p_0)$ is the initial condition of the pendulum.  The prior $P(h)$ is the initial condition of the agent.  Dynamics leads to $(x_1, p_1)$; conditioning leads to $P(h|d)$.  Why does the initial condition of the agent call for any more explanation or justification than the initial condition of the pendulum?

Or maybe let's put the two examples together.  $(x_0, p_0)$ is the initial condition of the pendulum, but $P_0(x,p)$ is the initial condition of the classical observer using Liouvillian mechanics (the agent).  Both systems, the pendulum and the agent, have unexplained initial conditions --- what's wrong with that?

Above I said ``only difference of any note,'' but of course there is a difference of detail here and there.  When we come to the quantum case particularly, there is this difference between the classical Liouvillian observer and the quantum observer.  When the classical observer writes down $P_0(x,p)$, it is his initial uncertainty of a preexistent reality for the pendulum; when the quantum observer writes down a quantum state $|\psi_0\rangle$, it is his initial uncertainty for {\it what will come about for him\/} in the case that he takes a certain fiducial action upon the system (i.e., makes a certain measurement)---it is not an uncertainty for a preexistent value.

Is that difference of detail enough to require that $|\psi_0\rangle$ be given more explanation or justification than $P_0(x,p)$?  We say (over and over), no.  What I should better understand about you is why you think it needs more.  My guess at this point is that it is nothing other than the usual fear of Bayesianism (even in the classical context of $P_0(x,p)$):  ``Surely there is a `right' prior, and whatever makes it `right' is what should be delineated.''  That's what you are asking for.

Now, before moving on to another point, let me come back to part of your second quote above:
\bdm
With reference to what you say below, the prior, as I understand
Figure 1 is that the damned ancilla on the left has the state
assignment $|0\rangle$, so you haven't told the [poor] student anything about what it means to make such a state assignment, either objectively or subjectively.
\edm

That is just incorrect.  We have told the poor student over and over what it means to make the state assignment $|0\rangle$.  It means that for his subjective assignment that IS the measurement gate (that particular quantum operation), he is quite certain what the consequence of its action upon the quantum system will be for him---he will get the datum 0.  He can predict the outcome with (subjective) certainty.  That is the {\it meaning\/} of $|0\rangle$:  It is the statement of certainty.  It is the statement that the student is prepared to gamble his life on the outcome.

You don't have to like our answer, but that is the answer we give.  It is clearly a meaning for the state $|0\rangle$---a very personal one, with operational consequences.  What we don't do is give a larger story that explains the origin of the agent's state of mind, $|0\rangle$.  Where did he get that from?  I don't know.  It might be a good question to ask, but it is not one that the calculus of probability can answer.  For, the calculus of probability requires as its very input a probability assignment---just like the calculus of Newtonian mechanics requires as its very input an initial state for the system under discussion.

So, why did I title this note ``Prepare Yourself'' (other than for its expected length)?

Just what is happening, you ask, for the quantum computer scientist with the initial preparation gate in your depiction of a quantum computer?  What is he doing when we {\it say\/} he is preparing the qubit?  From the Bayesian point of view, he is preparing himself.  He is tightening his state of belief for what will come about in future interactions with the qubit.

I guess that's why I showed such dismay at your talk in Konstanz:
\bdm
As I understand you it is unacceptable because that would give the
state an objective character. (The basis for the remark I made early
in my talk in Konstanz that made you say I still hadn't understood a
goddam thing.  Probably I still don't. If the reading of the
measurement gate is a fact, then the only room for subjectivity is
whether the measurement gate is indeed a measurement gate --- e.g.\ is
it properly aligned, or, in Figure 1, the subjectivity of making the
state assignment $|0\rangle$ for the ancilla.)
\edm

I remember you had a big question mark as the output of the preparation gate---seemingly implying that the quantum computer scientist would not know what the output of the gate was.  But he knows exactly what it will be.  The ``preparation gate'' is an action whose consequence will prepare his state of mind to something he wants it to be.  The subjectivity of the measurement gate (or preparation gate, I can't remember your exact terminology) means that it is {\it his\/} description of his action upon the qubit.  Making it {\it his\/} description---rather than a fact about ultimately remote parts of the universe external to him, as the conception you are shooting for does (see my previous note)---makes it no less useful to him.  (Maybe it makes it more so.)

At the moment, I fail to see how this would be so very mysterious to a computer scientist.  To perform a (quantum) computation, you build your confidence about some things to do with a machine (what you will see if you press this button).  Then you bank on that to let the machine help you build up various inverse probabilities for something you are not very sure about (say the factors of some large number).

Does any of this answer anything for you?

P.S. In return for all this labor this morning, I would love a reply from you on what I'm getting wrong with my critique of your ``disembodied facts'' account of quantum states.

\section{28-12-06 \ \ {\it Goose and Gander} \ \ (to R. W. {\Spekkens})} \label{Spekkens41}

\brws
On the subject of the ontic/epistemic status of the quantum state,
at the end of a talk to the history and philosophy of physics group at
Leeds, Ian Lawrie (a physicist) had a comment which, after I had finally understood it, impressed upon me (more strongly than before) that there's a sense in which the distinction breaks down when one tries to treat agents as physical systems.  I'm curious what you think about the following argument. Suppose one devises a physical theory for the ontic states of the world -- no external observer is required for the theory to be well-defined; it is a ``god's eye view'' type of theory.  And suppose that this theory does in fact provide a useful model of known physics and, in particular, of observers and
the process of observation. In other words, all observers are treated
internally to the theory.  Assuming a materialist position on the mind-body problem, everything that is known to any observer is encoded in the ontic state of the world according to this theory (either in the intrinsic properties of their brains, or, more likely, in the relational properties of their brains and the rest of the world).  According to the epistemic view of quantum states, different quantum states represent different beliefs. However, in such a physical theory, different beliefs correspond to different ontic states of the world, and so different quantum states correspond to different ontic states of the world.

Personally, I don't see much of a problem with granting that quantum
states are ontic in this sense.  They are simply not ontic in the
traditional sense of describing the reality of the system to which the state is assigned.  Furthermore, even though they may be ontic in this sense, they certainly will not correspond to a complete description of reality in such a theory.  In other words, quantum theory can't be this ``\"uber''-theory. What's your take on it?
\erws

Well, I don't think much of such hoped-for god's-eye-view theories, because I doubt the concept is even consistent at all.  They don't have enough fire in the loins, so to speak, to power a real universe.  (That's part of what I think quantum theory is trying to tell us.)  But I'll grant you the possibility, for the purpose of getting this discussion off the ground.

If you limit me to that context, then I'd say I probably pretty much agree with your second paragraph above.  In my presentations, I know that I often use language like this (the presentational Chris Fuchs is usually not as radical as the Chris Fuchs of correspondence---I tend to think it's better to be a little less than true-to-myself in such a forum for the purpose of more progress and less confrontation; Lord knows I already have enough confrontation).  I find myself saying things like I said to Greg Comer in a 4 July 2003 note:
\bq\noindent
A wave function and its evolution are not properties intrinsic to the
system for which they are about.  Rather, if they are properties of
anything at all, they are properties of their user's head---for they
capture all his judgments about what might occur if he were to
interact with the system of interest.
\eq

Indeed, that's not so very different from calling a quantum state an ontic state of the agent (when treated as an agent per se).  [Just for completeness sake, let me point out the difference between that and a situation where {\it I\/} treat {\it you\/} as a physical system (rather than an agent) of which I can contemplate performing measurements on.  For instance, I might ask you what you believe of that electron for all measurements you could perform on it.  Your answer to me will be in the form of a density operator---that is to say, the outcome of my quantum measurement will be something that symbolizes a quantum state in your possession.  Since I generally won't know what your answer to me will be---treating you as a physical system---I encode my uncertainty about the answer you'll give my questions in a density operator.  That density operator---which is an epistemic state for me---has nothing to do with the ontic state we are talking about above.  Nor is the outcome of my quantum measurement directly so either---for the outcomes of quantum measurements are not revelational of properties intrinsic to the system (they are a joint product of the observer and the system).  This, I think, is another way of saying what you said above, that quantum theory will not be its own \"uber-theory.]

Or to give a more recent example of my using this kind of imagery, I might paste in the note I composed for David {\Mermin} this morning.  Throughout it, I lean on the idea that a quantum state is effectively an ontic state for the agent (along the lines of how $x$ and $p$ are ontic states for a pendulum).  I think the note is relatively self-contained and you'll see its relevance for the present issue.  The figure {\Mermin} refers to is in the attached paper---a revised version of the CFS paper ``Subj Prob and Q Certainty''.  The issue that seems to bother {\Mermin} is that Bayesians give no account of the origin of a prior; I say that is no deeper a problem than Newtonian mechanics giving no account for the origin of any initial condition.  What's good enough for the goose, should be good enough for the gander.  [See 28-12-06 note ``\myref{Mermin126}{Prepare Yourself}'' to N. D. Mermin.]

\brws
On a different topic, I recently finished a paper on pooling quantum
states.  The central result is something that I worked out during the
BBQW workshop after Todd Brun's talk and have since developed with
Howard Wiseman.  It appeared on the {\tt arxiv} on Christmas day (I'm
attaching a copy).  I'm sure that you will be unhappy with it for the
same reasons that you're unhappy with the Brun-Finklestein-{\Mermin}
compatibility criterion.  Still, as someone who likes to think about
quantum analogues of features of classical probability theory, you
might be interested in the problem.
\erws

Thanks for sending the paper.  I'm quite OK with exercises like this---I think they're essential---as long as they are advertised for what they are:  Examples of conditions under which disparate agents will move closer to agreement.  One would like to identify those cases!  If Agent X believes this, and Agent Y believes that, then they will be willing to pool their beliefs for a sharper belief.  But to demand that beliefs always be poolable---I say---effectively transforms them back into ontic states (defeating the purpose of trying to draw an ontic/epistemic distinction in the first place)---they become incomplete perspectives on the ``true'' ontic state and regain an ontic character in that way.  BFM could just never get that into their heads (it's really hard to give up the idea of an ontic quantum state).

I'll certainly study your paper.  I hope you advertise the result the right way!!

\brws
Matt Leifer and I are writing a follow-up paper which makes extensive
use of Matt's ``conditional density operators'', and goes much further
on the problem than my paper with Howard does.
\erws

Glad to hear about that.  Reflecting on Bayesianism, one knows that the foundation is really in the idea of conditional probability $Pr(h|d)$ rather than unconditioned probabilities $Pr(h)$.  Similarly I'm starting to think perhaps quantum mechanics is focusing on the wrong mathematical object for a smooth development along Bayesian lines.  It is not the quantum state, but something {\it like\/} (if not exactly) Matt's conditional density operators.  So, it'll be good to get a better feel for these objects in as many ways as possible.

Happy New Year to you too!

\section{29-12-06 \ \ {\it Your Amazing Efficacity} \ \ (to A. Plotnitsky)} \label{Plotnitsky19}

I came into Bell Labs today and found to my great surprise the package containing your new book!  You really are quite amazing.  Prolific.  I wish I had a fraction of your productivity!  I'll certainly study your new material.

When will you be in New York again?  I'd love to get together with you one day for a long lunch.

Attached you will find my own latest, with {\Caves} and {\Schack}: \quantph{0608190}.  [\ldots]

For the moment though, I hope you will hang on the words ``fact for the agent'' which I introduced into this article.  I think it is the key for much of what I want to say in the next paper.  Here's a little discussion I had with one of the referees of the present paper.

\bq
This paper does not emphasize it, but no, we do not mean ``the facts
we are talking about here are facts for everybody.''  We mean that
the facts too are personal, though in a very careful sense.

We have tried to meld the phrase ``fact for the agent'' a little
better than previously into its surrounding text, but
a better exposition would have emphasized the full setting of our
view, including some discussion of the ontological---i.e.,
noninstrumental---pieces of it.  1) There are two physical systems in
the story of this paper.  One takes the role of the agent---the
possessor of subjective degrees of belief and the activator of the
measurement process.  The other takes the role of the object system.
2) What is being discussed when one speaks of gathering data in the
quantum context is an interaction or transaction between the agent
and the object.  3) Without the agent, there would be no quantum
``measurement'' to speak of, but without the object, there would be
no means for the agent to obtain the data, the spikes upon which he
pivots his probabilities (his subjective degrees of belief).  4)
There are no other agents in this story.  5) We do not go any further
than points 1-5 warrant by giving the data obtained in a quantum
measurement an autonomous existence---for instance, as something
beyond the agent's sensations. That would run into inconsistencies in
a ``Wigner's friend'' scenario.  Nonetheless, quantum measurement
outcomes are beyond the control of the agent---they are only born in
the interaction---and thus are not functions of the agent in the way
that his degrees of belief are. The degrees of belief find their
source in the agent (the subject); the outcomes find their source in
the external quantum system (the object). But the outcomes lead back
to the agent in that they are personal to him.

This does not mean that the agent-object interaction and its fuller
consequences are not autonomous events in spacetime, but it does mean
that if there are any such things, quantum theory is not directly
concerned with them. In CAF's view, quantum theory takes its whole
definition as a normative theory for organizing an agent's {\it
personal\/} probabilities for the {\it personal\/} consequences of
his interactions with external physical systems.  The structure of
the agent-independent world that lies behind quantum mechanics is, in
the end, still codified by the theory, but only in a higher-order,
more sophisticated fashion than had been explored previously---it is
through the normative rules, rather than quantum states and
Hamiltonians.
\eq

When we get together, I'd like to come back to your old `efficacity' concept.  Ever more I feel it's relevant to what I want to say \ldots\ but I wonder if that is one of the things you have changed your views on (as I don't see the word appearing in your index).

\section{29-12-06 \ \ {\it The Finish Line} \ \ (to S. Hartmann)} \label{Hartmann14}

\bsh
Could you send me the final version by the end of December? This
would be great.
\esh

OK, we just barely made your deadline!  Attached, please find our paper (both in \LaTeX\ form and PDF) and our replies to the referees (they are all in one single PDF file).

I can tell you, you have no idea how much (and how deeply) we fought in constructing this paper.  The whole thing was built out of negotiation---and you'll get some sense of that if you read our replies to the referees (particularly our remarks to Ref 1---i.e., {\Mermin}---and Ref. 4).  And every single word was pained over.  Thus, we pretty much beg of you not to ask for any further changes to the text.  Very seriously.  My word processor shows our word count to be 10,582---it is over the 10,200 word limit you and Roman set, but only by a little.

I hope you enjoy the new version of the article.  I think you got much more than you bargained for:  You asked for a review, but we gave you a new research article.  Thanks for giving us this opportunity.

By the way, I found Referee 4's comments excellent (as I hope my replies to him show).  If he is amenable, I would like to further the conversation with him, as I think I would get much out of the exercise and because I have many points to add to what I already said to him (that I couldn't express when speaking for {\Carl} and {\Ruediger}).  Thus, if you will, please give him my card (i.e., this paragraph), and tell him to contact me if he has any further interest.

Best wishes to you and your wife for the coming year!

\section{31-12-06 \ \ {\it Tipler Speaks} \ \ (to R. {\Schack})} \label{Schack119}

I don't think I'm going to reply to Tipler's reply to my reply to Tipler, but I am curious in what pithy way you would answer his sentence, ``Anyone with the same information would assign exactly the same probabilities.''  I'm afraid we have Ed Jaynes to blame for this one, and so we'll keep hearing it.  It might be good to develop a good canned answer.

\bq\noindent
{\bf Chris:} The other paper, particularly in its concluding section, tries to express what my colleagues and I see as the import of the Born rule: The idea we are trying to develop is that it is not a rule for
{\it setting\/} probabilities, but rather a rule for {\it transforming\/} Bayesian/personal/subjective probabilities.
\eq

\bq\noindent
{\bf Tipler:} Here I think you are indeed taking the wrong approach to probability theory.  There is nothing personal or subjective about assigning probabilities.  Anyone with the same information would assign exactly the same probabilities.  This was also Laplace's view, and he was a determinist like me.  Probabilities are not in nature, they are only precise measures of human ignorance.  Which is why I prefer to call my approach ``Laplacean probability'' rather than ``Bayesian probability.'' And I think that you will have difficulty improving the priors that come from quantum indistinguishability.
\eq

\chapter{2007: ``Who's Everett, and What's His Interpretation?''}

\section{03-01-07 \ \ {\it David Signals} \ \ (to N. D. {\Mermin})} \label{Mermin127}

Judging from your present note, it doesn't look like you'll like the latest incarnation of the footnote.  (I'm sorry; I just realized that I had not sent you the finalized manuscript.  It is attached.)  In this version we had said:

\bq\noindent
   N. D. {\Mermin} (private communications, 2003 and 2006) characterizes as
   ``dangerously misleading'' the idea that the post-measurement quantum
   state $|\psi\rangle$ of $B$ is an objective property of system $B$
   alone. He ``reject(s) the notion that objective properties (must)
   reside in objects or have physical locations.''  Yet the quantum
   state of $B$, if objective, is a property of the world, external to
   the agent, and since it changes as the world changes, it is hard to
   see how it can have only the disembodied objectivity {\Mermin} is
   describing.  If the objectivity of $|\psi\rangle$ {\em does\/} reside
   somewhere, say, in the entire experimental setup, including the
   device that prepares $A$ and $B$ and the measurement on $A$ and its
   outcome, then consider a measurement of an observable of $B$ for
   which $|\psi\rangle$ is an eigenstate with eigenvalue $\lambda$. If
   $\lambda$ is an objective property, how can it fail to reside in $B$
   (under the assumption of locality), thus making $|\psi\rangle$ a
   property of $B$ after all?
\eq
For, clearly here we use that bad word ``property'' again.

But, while we're on this subject, you say:
\bdm
The naughty things I'm searching for are facts --- in this case the
facts of the ``preparation procedure''.  There's nothing wrong with some
of the facts being facts about us. (Can't have ``correlations between
the manifold aspects of our experience'' without invoking us.)
\edm
In what way do you distinguish the phrase ``facts about us'' from ``properties of us''?  I would normally think of these two phrases, ``facts about'' and ``properties of,'' as synonyms, and I think everything I have written can be interpreted that way.  But you seem to reject that usage.

However, there's a still more worrisome issue here.  Previously I would have {\it never\/} thought you would allow the facts for which you take quantum states to be short-hand of to refer to the agent in any way (or the preparer, if you will).  But now it seems that is part of your very formulation.  Has it always been so?

Because the quantum state---from our view---depends on the AGENT as well as the data, we have insisted on calling it subjective.  As I understood you, you wanted to think that quantum states could be made to refer to facts alone---agent-independent facts particularly---and you therefore wanted to call (pure) quantum states objective.  If you are willing to explicitly renounce the requirement of agent-independentness (which I had thought you had always been making), in what substantial way does your view differ from ours?

Well, I would guess it still differs deeply in that you say:
\bdm
     I would say that the ``something in the world \ldots\ that guarantees
     the outcome YES'' has to be the whole state preparation --- all
     the relevant history of the system prior to the measurement.
\edm
For even if you were to substitute into that:
\bq\noindent
     I would say that the ``something in the world \ldots\ that guarantees
     the outcome YES'' has to be the whole state preparation --- all
     the relevant history of the system AND THE AGENT prior to the
     measurement.
\eq

CFS would still get hung up on your idea of ``guarantees.''

Let me paste in a note that I wrote to Rob {\Spekkens} (just after writing you last) that I think is directly relevant to the issue.  [See 28-12-06 note ``\myref{Spekkens41}{Goose and Gander}'' to R. W. Spekkens.]  In that note to Rob I say one might validly call a quantum state an ontic state of the agent (though maybe I should have called it `ontic aspect').  Is that all that your way of phrasing things is trying to capture?

If it is, then in a later note I'll try to argue that our way of phrasing things in better, but first I've got to understand this NEW (I think) twist I'm hearing from you.

\section{03-01-07 \ \ {\it New Year's Alchemy} \ \ (to H. C. von Baeyer)} \label{Baeyer26}

Thank you for the sweet New Year's note.  Though I've only known you for a short while now, I've learned that when I need some reassurance that I'm not wasting my life away, I can at least look to your emails.  That's more resources than I had before!  The idea of the Republica Literaria sounds charming.

So what should I call my upcoming lump of collected emails if not samizdat?  Your description was inspiring, but I need to distill the idea into a name that won't make me appear to be self-aggrandizing.  At the moment, I run dry.

Have you had a chance to read through (or translate) any more of the Pauli--Fierz correspondence?  I've been thinking about modern alchemy again---aka quantum measurement---as I've been thinking about how to put together my Shimonyfest contribution and also reading F.~C.~S. {\Schiller} in preparation for my lecture in Paris.  (Will you still be in Paris Feb 22 and 23?  If so, you should come; the meeting is titled ``Pragmatism and Quantum Mechanics,'' and my talk is titled ``William {\James}, F.~C.~S. {\Schiller}, and the Quantum Bayesians.''  {\Schack} will be there too.)  Maybe I'll be able to get a quote of {\Schiller} into my computer in the next few days and send it to you; it'll help reveal a little of what I see as the connection between {\Schiller}'s metaphysics (of a plastic reality) and Pauli's alchemy.

Attached is a copy of the finalized version of my recent paper with {\Caves} and {\Schack} on certainty.  (We haven't re-posted it yet; that'll have to wait until {\Ruediger} is back from vacation.)  It is much changed since the version you last saw; though not too much with regard to the particular point you made.  Sorry about that.  Still I hope the paper makes a little more sense this round.  I'm not all that pleased with the paper, but it'll have to do.  Particularly, the next stage of development really has to be an explicit and open discussion of the ontology that underlies our view of quantum mechanics.  That is what my Shimonyfest contribution will attempt to be about; I want it to be a much more detailed development of the point I made to one of our referees, which I'll paste in below.  [See 29-12-06 note ``\myref{Plotnitsky19}{Your Amazing Efficacity}'' to A. Plotnitsky.]  Measurement involves {\it two\/} systems, so it is not solipsism (or positivism) even if the outcomes make reference only to the observer's sensations.  The quantum system is the philosopher's stone that transforms the agent.  (In that way, there is a bit of Pauli in me; but the passage also shows where I differ from Pauli---his and Bohr's `outcomes' were automatically public, for the quantum Bayesians, they remain private \ldots\ at least till communicated.)

\subsection{Hans Christian's Preply, ``Happy New Year,'' 30 December 2006}

\bq
Dear Chris, for the New Year I wish you new insights, renewed certainty in your employment, and much joy.  Also good wine.

I am in gray, cold, wet Paris, which is nevertheless brightly illuminated, snug, and exhilarating indoors.  In the course of reviewing two new biographies of Emilie du Ch\^atelet, Voltaire's lover and Newton's translator, I am reading a book that reminds me of you.  In {\sl The Republic of Letters\/} the historian Dena Goodman describes the 18th century forerunner of today's ``scientific community'', which is distinct from both the university world and the international system of professional societies. Both of the latter already existed during the Enlightenment, but the Republic of Letters was broader in both subject matter and membership.

The Republic of Letters was a network (you see where this is going?)\ of people interested in the arts, humanities, and sciences.  As early as 1699 it was described as follows: ``The Republic of Letters is of very ancient origin. It embraces the whole world and is composed of all nationalities, all social classes, all ages, and both sexes [!].''

Goodman singles out three features: In the 18th century the RoL depended on two fundamental inventions of the modern world -- the printing press and the postal system -- which together made possible the cosmopolitan polity of the RoL. Further on she quotes another historian: ``\ldots\ it was the strict duty of each citizen of the Republica Literaria to establish, maintain, and encourage communication, primarily by personal correspondence or contact.''  Citizens considered it their duty to bring others into the republic through the expansion of their correspondence.  And finally: ``Reciprocity is the distinctive feature of correspondence as a mode of communication \ldots. Reciprocity was the fundamental virtue of the republic.''

It was this last sentence that struck me -- the difference between your style of research and the conventional style of publication. The RoL was a social construct that co-existed but contrasted with the monarchy, and, according to Goodman, eventually helped to bring about the revolution.  This is certainly a more cogent reading of history than the fairy tale of the masses rising up spontaneously against the evil king, but it's not what interests me here.

I once told you that I didn't think samizdat was a good description of your work, but I didn't have a better suggestion.  Now I realize that you are restoring the Republic of Letters by means of the latest technology, which replaces both the printing press and the postal system.  This insight puts your work into historical context and, in my mind at least, enhances its value and importance. I doubt that you intend to or will start a revolution, but I do think that your style will flourish.
\eq

\section{06-01-07 \ \ {\it Free Will, William James, NY Times} \ \ (to myself)} \label{FuchsC15}

From Maureen Dowd, ``Monkey on a Tiger,''\ {\sl New York Times}, 6 January 2007:

\bq
There was a touch of parody to the giddy Democrat takeover this week: Nancy Pelosi indulging her inner Haight-Ashbury and dipping the Capitol in tie-dye, sashaying around with the Grateful Dead, Wyclef Jean, Carole King, Richard Gere, feminists and a swarm of well-connected urchins.

The first act of House Democrats who promised to govern with bipartisan comity was imperiously banishing Republicans from participating in the initial round of lawmaking. Even if Republicans were brutes during their reign, Democrats should have shown more class, letting the whiny minority party offer some stupid amendments that would lose.

Perhaps the Democrats' power-shift into overdrive is a neurological disorder, or neuropolitical disorder.

If free will is an illusion -- if we are, as one philosopher put it, ``nothing more than sophisticated meat machines,'' doomed to repeat the same mistakes over and over -- that would explain a lot about the latest trend in which everyone is reverting to type.

William James wrote in 1890 that the whole ``sting and excitement'' of life comes from ``our sense that in it things are really being decided from one moment to another, and that it is not the dull rattling off of a chain that was forged innumerable ages ago.''

But in Science Times this week, Dennis Overbye advised Dr.\ James to ``get over it,'' observing that ``a bevy of experiments in recent years suggest that the conscious mind is like a monkey riding a tiger of subconscious decisions and actions in progress, frantically making up stories about being in control.''

As Mark Hallett of the National Institute of Neurological Disorders and Stroke told Mr.\ Overbye, ``Free will does exist, but it's a perception, not a power or a driving force. \ldots\ The more you scrutinize it, the more you realize you don't have it.''

That would explain why, after voters insisted that the president wrap it up in Iraq, he made a big show of pretending to listen, then decided to do a war do-over.

Is this just the baked-in stubbornness of one man, or is W.'s behavior evidence that he has no free will? Is the Decider freely choosing another huge blunder or is he taking instructions from his genetic and political coding, fearing that if he admits what a foul hash he's made of Iraq, he'll be labeled a wimp, as his dad was?
\eq

\section{09-01-07 \ \ {\it Harsanyi Principle}\ \ \ (to P. Grangier)} \label{Grangier10}

Thank you for your note; let me think about how I will reply.  In the meantime, you can read one of my mumbles on the principle Jaynes is pushing here:  \myurl[http://netlib.bell-labs.com/who/cafuchs/nSamizdat-2.pdf]{http://netlib.bell-labs.com/who/ cafuchs/nSamizdat-2.pdf}.  See pages 251 (note to David Poulin) to 253.  [See 19-09-03 note ``\myref{Poulin9}{de Finetti vs.\ Jaynes}'' to D. Poulin.]

\subsection{Philippe's Preply}

\bq
I stepped down on Jaynes's book on (strongly Bayesian, actually he tells Laplacian!)\ probabilities, and it does contain a lot of funny and interesting things (plus a few weird ones).  Here is one (up to you to decide in which category it fits):
\bq\noindent
``Subjective'' vs.\ ``Objective''\medskip\\
These words are abused so much in probability theory that we try to clarify our use of them. In the theory we are developing, any probability assignment is necessarily ``subjective'' in the sense that it describes only a state of knowledge, and not anything that could be measured in a physical experiment. Inevitably, someone will demand to know:  ``Whose state of knowledge?'' The answer is always: ``The robot---or anyone else who is given the same information and reasons according to the desiderata used in our derivations in this Chapter.''
Anyone who has the same information but comes to a different conclusion than our robot, is necessarily violating one of those desiderata. While nobody has the authority to forbid such violations, it appears to us that a rational person, should he discover that he was violating one of them, would wish to revise his thinking (in any event, he would surely have difficulty in persuading anyone else, who was aware of that violation, to accept his conclusions).
Now it was just the function of our interface desiderata (IIIb),
(IIIc) to make these probability
assignments completely ``objective'' in the sense that they are independent of the personality of the user. They are a means of describing (or what is the same thing, of encoding) the information given in the statement of a problem, independently of whatever personal feelings (hopes, fears, value judgments, etc.) you or I might have about the propositions involved. It is ``objectivity'' in this sense that is needed for a scientifically respectable theory of inference.
\eq

So: a pure quantum state (my ``modality'') is ``subjective'' in the sense that you certainly have to know something (and in particular the ``context'') to speak about it.  But it is ``objective'' in the sense that every ``robot'' (or whatever ``thinking observer'') which has that information can predict results with certainty, repeatedly, and without changing the system.

So is that Jaynes enough to convince a Bayesian that a pure quantum state is ``contextually objective''?\medskip

\noindent With Best Wishes for 2007
\eq

\section{09-01-07 \ \ {\it Harsanyi Principle, 2}\ \ \ (to P. Grangier)} \label{Grangier11}

\bpg
Here the mumbles do not shake me too much. You write:
\bq\noindent
{\rm I will agree to your definition of ``state of knowledge.'' But, backtracking from that, an initial state upon which everyone agrees? If one is taking a subjectivist approach (or what I had been calling a ``Bayesian approach'') to interpreting the quantum state, there is nothing in nature to enforce an initial prior agreement. God does not come down from on high and say to all the agents (i.e., all the observers), ``Your starting point shall be the quantum state $|\psi\rangle$.'' Everyone is left to fend for himself.}
\eq
But for QM, don't you think that there IS something to ``enforce an initial prior agreement'', it is simply what I call the ``context'', which is more or less what most people would call ``reality''? (in the classical sense $=$ if you don't believe in the wall, try to kick in it \ldots)

So again I am tempted to think that your ``extreme subjectivism'' denies ALL KINDS of ``realities'', quantum as much as classical.

Actually, I start liking this idea of ``prior knowledge'', but in Jaynes' sense, which apparently is not exactly yours (I hoped so!), but is closer to this ``Harsanyi doctrine'', which has the infamous goal of trying to be ``objective'' \ldots

So we have to try again!
\epg

Maybe not, we shall see.

Also let me send you the latest version of my paper with Caves and Schack.  (It's quite different than the posted version; can't update it until Schack is back from vacation.)  The closing section of it again comes back to the issue of what---in our view---is objective in the quantum world.  It is more than nothing, but it is probably less than you are hoping for with your contextual objectivity.

But that's only a short reply.  If one will allow ``levels of objectivity'' maybe the differences between us will disappear after all.

\section{09-01-07 \ \ {\it Talk about Anthropocentric!}\ \ \ (to D. Bacon)} \label{Bacon4}

I just enjoyed your anti-Wolfram [\myurl{http://dabacon.org/pontiff/?p=1402}] and ``Why I'm Not a Bohmist'' [\myurl{http://dabacon.org/pontiff/?p=1339}] columns with my lunch.  Nice.  I agree with both points.  Particularly your ``I do not believe in computers'' put a big smile on my face.  It has some of the flavor of the de Finetti phrase I co-opted for our quantum Bayesian program.  De Finetti said, ``Probability does not exist,''  and I keep advertising ``Quantum states do not exist.''  The point is, for {\Caves}, {\Schack}, a few others, and me, (all) probabilities and quantum states are anthropocentric constructions.  To say that is not to give up on the project of figuring out what physics is telling us about the world in itself (i.e., the world without us), but simply to recognize those {\it particular\/} mathematical objects for what they are (and what they are has something intrinsically to do with us).  Thus, when we say something like that, we're simply saying, ``Don't look here'' if your question is ``What is the universe made of?''  Look somewhere else.

Similarly I would say of a computer:  It's as anthropocentric of a concept as a hammer or screw driver.  A computer is a tool that satisfies a need; it is a {\it piece\/} of nature that has been harnessed for an external purpose.  There is no more indication that computers are any more likely to be parts of the fundamental constituents of the world than are hammers.  You didn't emphasize the anthropocentric aspect of it, but I think your point is pretty much the same.

Let me paste in a couple of pieces of my own writing that I hope will similarly put a smile on your face.  The first is a nasty referee report that I wrote on a Bohmian funding proposal.  I think the last paragraph of it gets at what really bugs me about what they're doing, and I think it makes much the same point as you were shooting for.  The other is a letter I wrote to a professor at Princeton when he pissed me off by saying that what {\Caves}, {\Schack} and I am up to is trying to turn physics into a branch of psychology.  I came out swinging!  Anyway, he too is one of those people who says ``the universe is a computer'' and thinks that's saying something significant.  Since I view that contrarily as an attempt to anthropocentrize the universe, I was doubly offended when he made the remark about psychology!  [See 17-11-05 note ``\myref{Halvorson5}{Cash Value}'' to H. Halvorson and 30-01-06 note ``\myref{McDonald6}{Island of Misfit Toys}'' to K. T. McDonald.]

\section{09-01-07 \ \ {\it Politically Correct}\ \ \ (to D. Bacon)} \label{Bacon5}

\bbaco
So every few months I try to come up with some new way of phrasing
foundational problems, so that I have something to write about on my
blog :)  Today I've decided my next one will be ``why is a qubit so
much like a probabilistic bit?''
\ebaco

Sounds like a good blog.  When you write it send me an announcement, as I don't visit your blog on a regular basis.  (Today's visit occurred because I wanted to see what your website name is; I'm thinking I need to get one for myself \ldots\ for when Bell Labs eventually lays me off.)

On a technical point, maybe your question should be ``why is a qubit so much like two probabilistic bits?''  For, if one wanted to think of a qubit's quantum state as a simple probability distribution (for the outcomes of an informationally complete measurement), the minimal event space has to have four outcomes.  Similarly, a qudit requires an event space worth $2\log d$ classical bits.  It's probable that you've seen the paper, but in case not, have a look at Rob {\Spekkens}'s ``toy theory'' \quantph{0401052}.  It's just a nice list of similarities between qubits and partially known cbits (maybe I'll call them pcbits).  And he, like you, suggests that the place where we're really going to learn something about quantum mechanics is in its deviation, not from cbits (i.e., the event space), but from pcbits (the space of allowed probability distributions over the event space).  So there:  quantum theory's real value comes in precisely where it is not pc!  (Those bad boys of quantum mechanics.)  Who would have expected anything else?

\section{09-01-07 \ \ {\it Facts-in-Themselves} \ \ (to H. C. von Baeyer)} \label{Baeyer27}

\bhcvb
My Fierz translation is on hold.  When I planned it, I thought that in
retirement I would have nothing else to do.  Then retirement turned
into an avalanche of other things to do, and besides, Fierz died
(though that should really spur me on all the more).
\ehcvb

That is too bad; I hope your conscience will send you back to the project eventually.  That's because it's not Fierz, the man, that is important (and so the project should have nothing to do with his death, one way or other), but the set of ideas he and Pauli were exploring.  The issue is getting those ideas into the consideration of a wider audience.  I'm always about building workforce; what can I do to get you to think that way too?

Thanks for the Russian analysis:
\bhcvb
The big difference between samizdat and your technique, besides the
issue of secrecy, is that the former obliges you to copy, whereas
letters oblige you to reply.  This principle of reciprocity is really
fundamental, and should be reflected in the name of your work.  But I
have not coined a sexy slogan yet.
\ehcvb
You have me intrigued.  I hope one of us can dream up a term.

You're right that reciprocity is really fundamental.  It is in letter writing and in the world.  Below is the F.~C.~S. {\Schiller} quote I had promised you (originally from a 1903 Int.\ J. Ethics paper).  I don't agree with it completely, but I like it much for the things I do agree with---particularly the part about reciprocity.

In fact I think I can localize the precise transition in the words that I don't like.  He writes, ``The simple fact is that we know the Real {\it as it is when we know it}; we know nothing whatever about what it is apart from that process.  It is meaningless therefore to inquire into its nature as it is in itself.''  Particularly, I agree with the first sentence, while disagreeing with the second.  I will agree that quantum measurements reveal nothing of the thing-in-itself, nor do they even generate facts-in-themselves---they are ``facts for the agent.''  I think {\Schiller}'s description is quite nice in getting that kind of idea across.  It's another take on the Paulian alchemical idea.  But what of the Born rule for transforming probabilities from measurement to measurement?  Of that, I would say, when we postulate it, we are taking a guess about some property of the world---a property of the world as it is in itself.  It's an indirect statement, but it's a statement nonetheless.  What's left is figuring out the precise nature of that guess (and its rather tight connection to some of the other things in {\Schiller}'s paragraphs!).

\bq
That the Real has a determinate nature which the knowing reveals but
does not affect, so that our knowing makes no difference to it, is
one of those sheer assumptions which are incapable, not only of
proof, but even of rational defence.  It is a survival of a crude
realism which can be defended only, {\it in a pragmatist manner}, on
the score of its practical convenience, as an avowed fiction.  In
this sense and as a mode of speech, we need not quarrel with it.  But
as an ultimate analysis of the fact of knowing it is an utterly
gratuitous interpretation.  The plain fact is that we can come into
contact with any sort of reality only in the act of `knowing' or
experiencing it.  As {\it unknowable}, therefore, the Real is {\it
nil}, as {\it unknown}, it is only potentially real.  What is there
in this situation to sanction the assumption that what the Real {\it
is\/} in the act of knowing, it is also outside that relation?  One
might as well argue that because an orator is eloquent in the
presence of an audience, he is no less voluble in addressing himself.
The simple fact is that we know the Real {\it as it is when we know
it}; we know nothing whatever about what it is apart from that
process.  It is meaningless therefore to inquire into its nature as
it is in itself.  And I can see no reason why the view that reality
exhibits a rigid nature unaffected by our treatment should be deemed
theoretically more justifiable than its converse, that it is utterly
plastic to our every demand---a travesty of Pragmatism which has
attained much popularity with its critics.  The actual situation is
of course a case of interaction, a process of cognition in which the
`subject' and the `object' determine each other, and both `we' and
`reality' are involved, and, we might add, {\it evolved}.  There is
no warrant therefore for the assumption that either of the poles
between which the current passes could be suppressed without
detriment.  What we ought to say is that when the mind `knows'
reality both are affected, just as we say that when a stone falls to
the ground both it and the earth are attracted.

We are driven, then, to the conviction that the `determinate nature
of reality' does {\it not\/} subsist `outside' or `beyond' the
process of knowing it.  It is merely a half-understood lesson of
experience that we have enshrined in the belief that it does so
subsist.  Things behave in similar ways in their reaction to modes of
treatment, the differences between which seem to us important.  From
this we have chosen to infer that things have a rigid and unalterable
nature.  It might have been better to infer that therefore the
differences between our various manipulations must seem unimportant
to the things.

The truth is rather that the nature of things is not {\it
determinate\/} but {\it determinable}, like that of our fellow-men.
Previous to trial it is indeterminate, not merely for our ignorance,
but really and from every point of view, within limits which it is
our business to discover.  It grows determinate by our experiments,
like human character.  We all know that in our social relations we
frequently put questions which are potent in determining their own
answers and without the putting would leave their subjects
undetermined. `Will you love me, hate me, trust me, help me?'\ are
conspicuous examples, and we should consider it absurd to argue that
because a man had begun social intercourse with another by knocking
him down, the hatred he had thus provoked must have been a
pre-existent reality which the blow had merely elicited.  All that
the result entitles us to assume is a capacity for social feeling
variously responsive to various modes of stimulation.  Why, then,
should we not transfer this conception of a determinable
indetermination to nature at large, why should we antedate the
results of our manipulation and regard as unalterable facts the
reactions which our ignorance and blundering provoke?  To the
objection that even in our social dealings not all the responses are
indeterminate, the reply is that it is easy to regard them as having
been determined by earlier experiments.
\eq

\section{10-01-07 \ \ {\it Hyle} \ \ (to A. Shimony)} \label{Shimony13}

I'm very happy to hear that you are OK.  I did start to get worried because the silence seemed a little unlike you.

\bas
I suggest that you read certain of my epistemological papers which
are relevant but perhaps not familiar to you: In my collection {\bf Search for a Naturalistic World View} vol.\ 1 (Cambridge U. Press 1993), papers 1, 2, 3, 9 (sections 3, 4, 5), and 10; and ``Some Intellectual Obligations of Epistemological Naturalism'' in {\bf Reading Natural Philosophy}, ed.\ by David Malament.
\eas

I will do.  In fact, I just ordered your book from Amazon.com.  They say it will ship by tomorrow.

The last two weeks I've been reading F.~C.~S. {\Schiller} in preparation for the talk I'm giving at the ``Pragmatism and Quantum Mechanics'' meeting in Paris Feb 22--23.  (I hope I do much better than I did in my talk in Waterloo for you this year.)  Now he is an obscure philosopher!---I wonder if you've ever read him at all---but I find that I'm even more attuned to him than {\James} or {\Dewey} \ldots\ probably why he is a forgotten philosopher.  He too was a fan of Aristotle's hyle, but in a much different way than {\Peirce}.

\section{10-01-07 \ \ {\it Anti-Algebra, the Reprise} \ \ (to V. {\Palge})} \label{Palge3}

I'm sorry to take so very long to reply to your email.  I'm behind on everybody's email.

\bvp
It seems that one of the main reasons you reject a well-structured
event space is that it assumes a realistic interpretation: its
elements would correspond to intrinsic properties of quantum systems
(either possessed by systems before a measurement or being generated
by a measurement). However, I wonder if one is forced to such
realistic commitment. Can't one understand the e.g.\ partial Boolean
event space in an instrumentalistic and hence more Bayesian spirit?
Can't one just take the elements as corresponding to clicks and blips in the measurement apparatus? What I have in mind is Pitowsky's 2003
(``Betting on the outcomes of measurements: a Bayesian theory of
quantum probability,'' SHPMP, {\bf 34}). As you probably know, he assumes that quantum event space is a partial Boolean algebra and then lays down the rules a Bayesian must follow in a world whose event space ultimately has this structure.

In short, I am wondering if your argument applies to the Pitowskian
Bayesian picture which is free from a realistic understanding of
properties?

The reason this issue seems important to me is that I believe
eventually a Bayesian should say something about the structure of the event space---this seems like a minimal condition for calculating
probabilities. And what this structure is has interesting
implications on further issues in QM.
\evp

Let me start with your first sentence:  ``It seems that one of the main reasons you reject a well-structured event space is that it assumes a realistic interpretation: its elements would correspond to intrinsic properties of quantum systems.''  That is the wrong direction of reasoning---though I am probably the cause of this misimpression through the restrictive choices of readings I recommended to you.  It is not that I reject a well-structured event space because it {\it assumes\/} its elements would correspond to intrinsic properties of quantum systems, but rather this is the {\it result\/} of a thoroughgoing subjective interpretation of probabilities within the quantum context.  What cannot be forgotten is that quantum-measurement outcomes, by the usual rules, {\it determine\/} posterior quantum states.  And those posterior quantum states in turn determine further probabilities.

Thus, if one takes the timid, partial move that Itamar and Jeff Bub, say, advocate---i.e., simply substituting one or another nonBoolean algebra for the space of events, and leaving the rest of Bayesian probability theory seemingly intact---then one ultimately ends up re-objectifying what had been initially supposed to be subjective probabilities.  That is:  When I look at the click, and note that it is value $i$, and value $i$ is rigidly---or I should say, {\it factually}---associated with the projector $\Pi_i$ in some nonBoolean algebra, then I have no choice (through L\"uders rule) but to assign the posterior quantum state $\Pi_i$ to the system.  This means the new quantum state $\Pi_i$ will be as factual as the click.  And any new probabilities (for the outcomes of further measurements) determined from this new quantum state $\Pi_i$ will also be factual.

So, the starting point of the reasoning is to {\it assume\/} that there is a category distinction between probabilities and facts (this is the subjectivist move of de Finetti and Ramsey).  Adding the ingredient of the usual rules of quantum mechanics, one derives a dilemma:  If there is a rigid, factual connection between the clicks $i$ and elements $\Pi_i$ of an algebraic structure, then probabilities are factual after all.  Holding tight to my assumption of a category distinction between facts and probabilities, I end up rejecting the idea that there is a unique, factual mapping between $i$ and $\Pi_i$.

Let me point you to two papers of mine that emphasize (and make a little more formal) this kind of reasoning.  The first is \quantph{0404156}, and the particular place to look in it is Section VI, starting on page 22.  The second is attached to this note.  It is a fairly drastic revision of \quantph{0608190} that we haven't posted yet (can't re-post it until {\Ruediger} gets back from vacation with the passwords).  In large part, this paper is on the very point we are discussing here.  Finally, let me recommend an excellent recent paper by Matt Leifer that extends these points: \quantph{0611233}.

But let me return to your paragraph before signing off.  If I read this question of yours in isolation:  ``Can't one just take the elements as corresponding to clicks and blips in the measurement apparatus?''  Then at one level my answer is, ``Of course; I've never said otherwise.''  What is at issue here is whether the events---the clicks and blips themselves---fall within the kind of algebraic structure you speak of, or whether it is something else (something a conceptual layer above the events) that falls within it.  From the CFS point of view, for a single device with clicks $i$, one agent might associate the clicks with a set of orthogonal projectors $\Pi_i$, and another agent might associate them with a set of noncommuting effects (i.e., POVM elements) $E_i$.  This was the sense in which I meant there is no stand-alone event space at all:  For us, the algebraic structure of the events (the clicks and blips), their level of commutivity or noncommutivity and whatnot, is just as subjective as the quantum state.  The clicks $i$ themselves are objective (in the sense of not being functions of the subject's beliefs or degrees of belief), but their association with a particular set of operators $E_i$ is a subjective judgment.

I hope this completely answers your question now.

But let me extend the discussion a little to try to give you a more positive vision of what we're up to.  The starting point is the category distinction between facts and probabilities applied to the quantum measurement context.  From this we glean that quantum operations and quantum states are of the same level of subjectivity.  But that is not our ending point.  Because, implicit in everything we have said there are these autonomous, realistically-interpreted quantum systems:  The agent has to interact with the quantum system to receive his quantum measurement outcome.  No quantum system, no measurement outcome.  Thus the CFS position is more than a kind of positivism or operationalism.  The objects of the external world with which we interact have a certain kind of active power, and we become aware of the presence of that active power particularly in the course of quantum measurement.  When we kick on a quantum system it surprises us with a kick back.

Can we say anything more explicit about the active power?  Can we give it some mathematical shape?  Yes, I think we can, but that is a research project.  Still, I think the hints for it are already in place \ldots\ and indeed they are in the quantum formalism, as we would expect them to be.  One of the hints is this:  The Born transformation rule.  When we make probability assignments in quantum mechanics, we are assuming more than de Finettian / Ramseyian coherence.  We assume that if we set the probabilities for the outcomes of this measurement this way, then we should set the probabilities for the outcomes of that measurement that way and the rule of transformation is a linear one.  This, from the CFS view, is the content of the Born rule (see the last section in the attached paper).  And it is empirical, contingent:  A different world than the one we live in might have had a different transformation rule.  So, if we're looking for something beyond personal probabilities in quantum mechanics, that is a point to take seriously.  It hints of some deep property of our world, and I'd like to know what that property is.

\section{10-01-07 \ \ {\it Anti-Algebra, the Moral} \ \ (to V. {\Palge})} \label{Palge4}

I walked into the hall to get a glass of water, and immediately felt that I had not properly clenched off the last two paragraphs of my last note.  There were several points made there, but among them is this, and I just want to be explicit about it:

Though CFS banish the algebraic structure of Hilbert space from having anything to do with a fundamental event space (and in this way their quantum Bayesianism differs from the cluster of ideas Pitowsky and Bub are playing with), they do not banish the algebraic structure from playing any role whatsoever in quantum mechanics.  It is just that the algebraic structure rears its head at the conceptual level of coherence rather than in a fundamental event space.  It is not that potential events are objectively tied to together in an algebraic way, but that our gambling commitments (normatively) {\it should be}.  This is another point of contact between my view and Bill's\index{Demopoulos, William G.}.

\section{10-01-07 \ \ {\it Mulling} \ \ (to N. D. {\Mermin})} \label{Mermin128}

I've been mulling over your letters ``State Preparation'' and ``The Grangier Letter'' from 1/2/07 for several days now, but the more I think about it, I don't know how to respond in any detail until I hear your thoughts on my note ``\myref{Mermin127}{David Signals}'' from 1/3/07.  Particularly, much will hinge on you respond to the part that starts off, ``However, there's a still more worrisome issue here.''

I know it's getting very close to your leaving for vacation; so I am not betting strongly to hear back from you on the subject for quite a while (probably not till you're visiting Princeton).  Basically, you just give me the signal when you're ready (by answering those questions), and we'll pick back up where we left off.

At the moment I'll only observe this much, with regard to these two things you said:
\bdm
You have to understand that much of my attitude towards quantum
foundations these days really does grow out of the past six years of
teaching the subject to computer scientists.  From this narrow
perspective quantum mechanics is a set of rules enabling us to
describe important aspects of the operation of a particular bounded,
isolated, piece of machinery.
\edm
and
\bdm
And it wipes out the beautiful symmetry between measurement gates
being crucial at both the start and the end of the computation, which
I can't help think is telling us something (about correlations between
aspects of our experience.)
\edm

In some ways, you seem to be roughly where I was when I started writing that ill-fated piece for Physics Today with Asher in 1999---basically the position of Asher in his 1980s paper, ``What Is a State Vector?''  (Does this mean I've devolved since then?)  Anyway, I paste in a bit of the correspondence I had with Asher at the point when I started to turn away from the master.

Best wishes, and stay clear of the lava,

\subsection{Excerpt of: 15-11-1999, to A. Peres, ``Have a Good Trip''}

\bq
One point that we may need some private side discussion on before we set it in stone, is captured among other places in your sentence:

\bap
The notion ``state'' refers to a {\it method of preparation,\/} it is not an intrinsic property of a physical system.
\eap

In general I have noticed in this manuscript that you lean more heavily on the word ``preparation'' than we did in our letter to {\Benka}.  (In fact, I can't find any mention at all of the word ``preparation'' in that letter.)  Unless I misunderstand your usage of the word, it may actually be a little too anthropocentric even for my tastes.  The problem is this:  consider what you wrote in the paragraph about the wave function of the universe.  It seems hard to me to imagine the wave function of those degrees of freedom which we describe quantum mechanically as corresponding to a ``preparation.''
Who was the preparer?

It is for this reason that {\Carl} {\Caves} and I prefer to associate a quantum state (either pure or mixed) solely with the compendium of probabilities it generates, via the {\Born} rule, for the outcomes of all potential measurements.  And then we leave it at that.  Knowing the preparation of a system (or the equivalence class to which it belongs) is one way of getting at a set of such probabilities.  But there are other ways which surely have almost nothing to do with a preparation.  An example comes about in quantum statistical mechanics:  when the expected energy of a system is the only thing known, the principle of maximum entropy is invoked in order to assign a density operator to the system.  There may be someone beside me in the background who knows the precise preparation of the system, but that does not matter as far as I am concerned---my compendium of probabilities for the outcomes of all measurements are still calculated from the MaxEnt density operator.

To help ensure that I was not jumping to conclusions on your usage of the term, I reread today your paper ``What is a state vector?''
[AJP {\bf 52} (1984) 644--650].  There was a time when I agreed with everything you wrote there (in fact, I think it was the first paper with which I got to know you).  But as of today at least, I think a more neutral language as in our letter to {\Benka} is more appropriate.  [\ldots]
\eq

\subsection{Excerpt of: 20-10-2000, to P. Benioff, ``Not Instructions, but Information''}

\bpb
If I recall correctly you wrote a paper with {\Peres} on QM without interpretation but stated that a quantum state corresponds to an algorithm for preparing it.
\epb

In this, though, you're confusing two things.   {\Asher} wrote a
paper in the early 1980s for AJP titled ``What is a State Vector?''
and in that paper he took the point of view you mention.  That is, that a quantum state corresponds to nothing more than the (equivalence class of) instructions for preparing a system one way or the other. I've never felt completely comfortable with that point of view, so in the paper that  {\Asher} and I co-wrote for {\sl Physics Today\/} last year we worked around that.  I think also that I may have even swayed  {\Asher}'s opinion on this issue, but you would have to check with him directly.  (Perhaps I'll just carbon copy this letter to  {\Asher}.) Here's how I put it in a note to  {\Asher} 15 November 1999: [\ldots]

I believe this point of view is adequately expressed in the article  {\Asher} and I wrote for {\sl Physics Today\/}:  a quantum state is nothing more and nothing less than one's best (probabilistic) information on how a system will react to our experimental interactions with it.
How we may have come by that information---be it through a preparation, through a sheer guess based on all available evidence, or the principle of maximum entropy---is something I view as largely outside quantum theory proper.  The structure of quantum theory instead codifies how we should manipulate our information (this is what time evolution and the collapse rule is about) and enumerates the varied ways with which we may gather new information (this is what the structure of observables or POVMs is about).

In this sense, I would call what we are talking about an ``information interpretation'' or ``Bayesian interpretation'' of quantum theory, rather than an ``algorithmic interpretation.''  For the most part, however, I would like to avoid the wording of an interpretation.

I will go further---and this is one point where I may diverge from  {\Asher}---and say that I do suspect that we will one day be able to point to some ontological content within quantum theory.  But that ontological statement will have more to do with our interface with the world---namely that in learning about it, we change it---than with the world itself (whatever that might mean).

\section{10-01-07 \ \ {\it Pragmatism in Paris} \ \ (to A. Grinbaum)} \label{Grinbaum2}

\bag
First, all the best wishes for 2007! It's been long time since we
talked--hope all is well. I was told that you'll be in Paris around
February 22 for a seminar: what are the exact dates? It would be
wonderful if you had time to talk and perhaps visit our new
foundations of physics lab at the CEA, LARSIM, where I now work.
\eag

Good to hear from you.  That's quite interesting:  A nuclear physics research center investing in quantum foundations?!  That's very nice.  And congratulations on your new position.  Is it permanent?  Or is it a limited appointment?  I hope it is the former and affords you much leisure to do good work.

I will arrive in Paris just in time to make it to the start of the meeting, Feb 22.  I'll be coming from London with {\Schack} on a Eurostar.  Then I fly directly out of Paris on the 25th.  Thank you on your kind invitation to visit your lab, but I think I will take a rain check on it this time.  It is true that I'll be in Paris an extra day beyond the meeting, but I had already drawn up grand plans for the day:  To walk around the city all alone and use its lonely sounds to try to think deeply and possibly provoke some new thoughts in my head.

Happy new year to you too, and see you in February.

\section{11-01-07 \ \ {\it Pyroclastic Surge} \ \ (to N. D. {\Mermin})} \label{Mermin129}

\bdm
The hope is that quantum foundations will be clarified by a
spectacular sunset and not by a pyroclastic surge.  But my main
reading project is {\bf Origin of Species}, which I somehow missed growing up.
\edm
Worthy book.  Read it, and then take all those ideas to physics.  Might I recommend as supplementary material:\medskip

\begin{supertabular}{ll}
Database: 	 &     Cornell University Library\\
Author/Creator: &	Wiener, Philip Paul, 1905--\\
Title: 	    &  Evolution and the founders of pragmatism /\\
            &      by Philip P. Wiener ; with a foreword by John Dewey.\\
Published: 	 &     Cambridge, MA : Harvard University Press, 1949.\\
Description: &	xiv, 288 p.\ ; 25 cm.\medskip\\
Location: 	 &     Library Annex\\
Call Number: &	B818 .W64\\
Status: 	 &     Not Charged\\
Location: 	 &     Uris Library\\
Call Number: &	B818 .W64\\
Status: 	 &     Not Charged\\
\end{supertabular}

\section{11-01-07 \ \ {\it Matter, Money, and Change} \ \ (to A. Shimony)} \label{Shimony14}

\bas
I am pleased that you have ordered my papers. I felt frustrated
engaging in a written debate with you when my full position cannot be
presented briefly, and when I have already spent years trying to
present it adequately. After you have read the papers you may be
convinced by me, or you may have cogent criticisms which will persuade
me.
\eas

Yes, you are probably right that that is a more efficient method.  By the way, your book only cost me \$13.29 (and is supposed to be in ``essentially new'' condition).  Since my Mom sent me \$20 for Christmas, I still have almost \$7 left!  Thanks for providing the form to the hyle, and thanks for the change!

\section{12-01-07 \ \ {\it Out of the Blue} \ \ (to H. Halvorson)} \label{Halvorson16}

I wonder:  1) if I can ask a favor of you that is surely beneath a professor, and 2) if I might have a conversation of deeper value with you as I pick up on favor 1 (if indeed you take up that task)? (I'll explain below.)

First the favor to ask:  I'm giving a talk at the ``Pragmatism and Quantum Mechanics'' meeting in Paris, Feb 22--23, titled, ``William James, F.~C.~S. Schiller, and the Quantum Bayesians,'' and partly because of that I'm doing everything I can to strengthen my knowledge of Schiller.  It's not easy, as he's a pretty obscure philosopher---though you can guess he stirs me like no one else!---and particularly I haven't been able to get my hands on one of his most important papers:  ``Axioms as Postulates.''  This is the favor that's beneath a professor, but I wonder if I can ask it of you anyway (for a friend)?  The Princeton Library has a copy of the volume that contains the paper, but it's archived at some off-site facility.  Could I ask you to recall the volume from there?

If you wouldn't mind doing that, I could drop by and copy the paper one day next week.

\section{13-01-07 \ \ {\it Conference ``Pragmatisme et M\'ecanique Quantique''} \ \ (to M. B\"achtold)} \label{Baechtold1}

Thanks for making the reservation for me.  I'll forward your note to {\Schack}; I'm sure the accommodation will be fine with him.

One thing about the program:  The title of my talk should be ``William James, F.~C.~S. Schiller, and the Quantum Bayesians'' rather than the original title you have listed.  (I sent you a note earlier changing the title, but maybe you didn't get it.)  The talk will be on an updated, quantum mechanical rendering of this quote of Schiller:
\bq
That the Real has a determinate nature which the knowing reveals but
does not affect, so that our knowing makes no difference to it, is
one of those sheer assumptions which are incapable, not only of
proof, but even of rational defence.  It is a survival of a crude
realism which can be defended only, {\it in a pragmatist manner}, on
the score of its practical convenience, as an avowed fiction.  In
this sense and as a mode of speech, we need not quarrel with it.  But
as an ultimate analysis of the fact of knowing it is an utterly
gratuitous interpretation.  The plain fact is that we can come into
contact with any sort of reality only in the act of `knowing' or
experiencing it.  As {\it unknowable}, therefore, the Real is {\it
nil}, as {\it unknown}, it is only potentially real.  What is there
in this situation to sanction the assumption that what the Real {\it
is\/} in the act of knowing, it is also outside that relation?  One
might as well argue that because an orator is eloquent in the
presence of an audience, he is no less voluble in addressing himself.
The simple fact is that we know the Real {\it as it is when we know
it}; we know nothing whatever about what it is apart from that
process.  It is meaningless therefore to inquire into its nature as
it is in itself.  And I can see no reason why the view that reality
exhibits a rigid nature unaffected by our treatment should be deemed
theoretically more justifiable than its converse, that it is utterly
plastic to our every demand---a travesty of Pragmatism which has
attained much popularity with its critics.  The actual situation is
of course a case of interaction, a process of cognition in which the
`subject' and the `object' determine each other, and both `we' and
`reality' are involved, and, we might add, {\it evolved}.  There is
no warrant therefore for the assumption that either of the poles
between which the current passes could be suppressed without
detriment.  What we ought to say is that when the mind `knows'
reality both are affected, just as we say that when a stone falls to
the ground both it and the earth are attracted.

We are driven, then, to the conviction that the `determinate nature
of reality' does {\it not\/} subsist `outside' or `beyond' the
process of knowing it.  It is merely a half-understood lesson of
experience that we have enshrined in the belief that it does so
subsist.  Things behave in similar ways in their reaction to modes of
treatment, the differences between which seem to us important.  From
this we have chosen to infer that things have a rigid and unalterable
nature.  It might have been better to infer that therefore the
differences between our various manipulations must seem unimportant
to the things.

The truth is rather that the nature of things is not {\it
determinate\/} but {\it determinable}, like that of our fellow-men.
Previous to trial it is indeterminate, not merely for our ignorance,
but really and from every point of view, within limits which it is
our business to discover.  It grows determinate by our experiments,
like human character.  We all know that in our social relations we
frequently put questions which are potent in determining their own
answers and without the putting would leave their subjects
undetermined. `Will you love me, hate me, trust me, help me?'\ are
conspicuous examples, and we should consider it absurd to argue that
because a man had begun social intercourse with another by knocking
him down, the hatred he had thus provoked must have been a
pre-existent reality which the blow had merely elicited.  All that
the result entitles us to assume is a capacity for social feeling
variously responsive to various modes of stimulation.  Why, then,
should we not transfer this conception of a determinable
indetermination to nature at large, why should we antedate the
results of our manipulation and regard as unalterable facts the
reactions which our ignorance and blundering provoke?  To the
objection that even in our social dealings not all the responses are
indeterminate, the reply is that it is easy to regard them as having
been determined by earlier experiments.

In this way, then, the notion of a `fact-in-itself' might become as
much of a philosophic anachronism as that of a `thing-in-itself,' and
we should conceive the process of knowledge as extending from
absolute chaos at the one end (before a determinate response had been
established) to absolute satisfaction at the other, which would have
no motive to question the absolutely factual nature of its objects.
But in the intermediate condition of our present experience all
recognition of `fact' would be provisional and relative to our
purposes and inquiries.
\eq

It captures a good piece of what we quantum Bayesians are up to.  But it also contrasts some---for we would say that the quantum formalism is telling us {\it something\/} of the world in itself.  Below is how I wrote Veiko {\Palge} on the subject recently. [See 10-01-07 note ``\myref{Palge3}{Anti-Algebra, the Reprise}'' to V. {\Palge}.] Anyway, that's what the talk will be about, and I'll tie it all in with a) the quantum formalism, and b) with William James too.

\section{16-01-07 \ \ {\it Axioms as Postulates} \ \ (to H. Halvorson)} \label{Halvorson17}

I don't know if you looked any at the document you copied, but already in the outline, Section 1, I love it!  Matter is the resisting medium.  That is just the right conception, I think, for a quantum system.

Read it as a novel (not for literal truth, but for inspiration) as you get a chance!

\section{16-01-07 \ \ {\it Don't Forget Schiller!}\ \ \ (to H. Price)} \label{Price6}

I've only read the first page of your Federation Fellow proposal, but in that little bit I already like it much.  And said to myself, ``I've got to print this out.''  In the meantime, I'll fly this note off to you.

When I read your listing of Peirce, James, and Dewey, I thought, ``Don't forget Schiller!''  My recent fascination with his thinking is the partial source of the title of my last note.  I'm now starting to surmise that his take on pragmatism perhaps better tacks with what I'm aiming for in quantum mechanics than even James.  We shall see.

Anyway, two notes below on my plans with Schiller that you may enjoy.  Particularly the long Schiller quote itself at the bottom.  [See 03-01-07 note ``\myref{Baeyer26}{New Year's Alchemy}'' and 09-01-07 note ``\myref{Baeyer27}{Facts-in-Themselves}'' to H. C. von Baeyer.]

I'll head toward the printer now.

\section{16-01-07 \ \ {\it The Information *Was* in the Warning}\ \ \ (to H. Price)} \label{Price7}

I just finished the ``Aims'' section, and I'm still with you.  Very nice.  Now, I'm off to the ``Background'' section.

I must have shown you this montage I put together once, playing on the words of Tilgher (some early Italian pragmatist, I think):
\bq\noindent
   A quantum state is not a mirror in which a reality external to us
   is faithfully reflected; it is simply a biological function, a
   means of orientation in life, of preserving and enriching it, of
   enabling and facilitating action, of taking account of reality
   and dominating it.
\eq

\section{17-01-07 \ \ {\it Ego}\ \ \ (to A. R. Calderbank)} \label{Calderbank3}

It's of no consequence now, but let me just record this for my own satisfaction.  As I was driving home this evening, reviewing the day in my head, I thought:  ``Wow did I ever undersell myself with that line about how `my main interest in a cross appointment arose because I'm neither a great mathematician, nor a great philosopher, but a little good at each'.''  Rest assured, I think much more of myself than that!  Particularly, I'm quite confident that the school of thought in quantum foundations I am building up around the world is deeply significant and paving the way for the next big step in physics.  All I really meant to imply is that I fit no department per se, and these are the closest fits.  But I don't doubt for a minute that I would inspire the students around me to do some dazzling work in these fields (by those departments' own standards).

Thanks, by the way, for talking to me.  Your clarity was much appreciated.  As we can schedule it (with my fairly full travel schedule this Spring), I'd be more than happy to give a seminar for you guys.

\section{18-01-07 \ \ {\it Pragmatism Old and New} \ \ (to R. E. Slusher)} \label{Slusher19}

Keeping with the idea of `something old, something new \ldots', I thought you might like to take this old book with you down to Georgia.  Certainly you'll need pragmatism (with a little p) to make big things happen in your lab.  But perhaps the book will also give you a chance to reflect on how {\sl Pragmatism\/} (with a big P) finds its best argument in our quantum world.

Good luck, and many more years of good science from you.

\section{29-01-07 \ \ {\it Quick Answer to Your ``Advice while I'm Thinking of It''}\ \ \ (to C. M. {\Caves})} \label{Caves94.2}

I'll be giving a talk Feb 8 to the philosophers (and some from the applied math dept) at the University of Western Ontario titled ``Subjective Quantum States in an Objective Quantum World''.  I'm hoping I can muster a full frontal attack that will get these guys to finally recognize that our Bayesian program is neither:
\bv
1) solipsism\\
2) instrumentalism\\
3) operationalism\\
nor\\
4) positivism
\ev
It is simply physics in the old sense of the word --- trying to figure out the character of the world.

\section{29-01-07 \ \ {\it Toy Model, 1} \ \ (to S. J. van {\Enk})} \label{vanEnk9}

Well, I've finally gotten your paper printed out.  I'll be taking it to Toy Model Central (i.e., PI) with me on my flight tomorrow.  I want to understand this. \ldots\ [some days]

Now, I've started putting work into understanding your paper.  I still think it's nice (very nice actually), but I'm not sure to what extent it has a claim to being called a toy model for epistemic states.  Once the ``probabilities'' come out of the ``information theoretic principle'' (both quotes have to be taken seriously now, because those words cannot be interpreted literally any more) as having to be negative, they lose their epistemic meaning.  They're something else now.

I'm thinking, though \ldots

\section{30-01-07 \ \ {\it Polchinski Story} \ \ (to L. Smolin)} \label{SmolinL6}

As it turns out, I just finished reading your new book as I landed in Toronto today.  (Literally so!  I was reading the last paragraph of the last page as the airplane wheels hit the pavement.)  I enjoyed it and got a lot out of it, particularly the chapters on ``the anthropic solution'' and  ``what is science.''  I also thought your last chapters made a powerful argument for the place and style of foundational work---nice.

Let me tell you a story about Joe Polchinski that one of your pages brought back to mind.  On page 137, you write:
\bq
\noindent
\ldots\ from 1984 to 1995, we were like amateurs \ldots\ [missing] most of what was necessary to make the system work until Polchinski discovered the missing essentials.  \ldots\ In the fall of 1995, Polchinski showed that a string theory, to be consistent, must include not only strings but surfaces of higher dimensions moving in the background space.  \ldots\  If a string, which is a one-dimensional object, can be fundamental, why can't a two-dimensional surface, be fundamental?  In higher dimensions, where there is a lot of room, why not a three-, four-, or even five-dimensional surface?
\eq

Well, sometime around 1985 or 1986 ('87 at the latest), I was at a physics colloquium at the University of Texas and Joe Polchinski was the speaker.  Steven Weinberg was in the first or second row of the auditorium in his usual seat.  Polchinski talked about strings.  What was funny was that early in the talk, Weinberg interrupted to ask, why stop with strings?  Why not two- or higher dimensionsal objects?  And, in particular, he ended, ``Once you move away from the particle concept, why not just entertain a universe filled with gummy bears?''  Polchinski at first just paced back and forth on the stage, looking down \ldots\ a bit of a silence.  Then Weinberg said, ``I'm not being facetious!  Why not gummy bears?  What in the theory constrains you to strings?''  The audience, of course, laughed.  And I remember Polchinski saying, ``I know you're not being facetious.''  However, I don't remember his answer beyond that.  Just a cute story you might like (and I might record).

\section{31-01-07 \ \ {\it Nature Giving a Flip} \ \ (to myself)} \label{FuchsC16}

\bv
``Perhaps nature doesn't give a flip about such questions, because $c$ is finite.''
\ev

Good point.  John Wheeler used to say of counterfactual games like this (at least with quantum mechanics):  ``Why talk of them?  They're worlds that never were and never could be.''

\section{01-02-07 \ \ {\it Metaphysics of the Time Process} \ \ (to L. Smolin)} \label{SmolinL7}

Here's some words from the paper I was telling you about, F.~C.~S. {\Schiller}, ``The Metaphysics of the Time-Process,'' originally published in Mind, 1895.

The abstract to the paper reads:
\bq
Significance of Dr.\ McTaggart's admission that the Hegelian Dialectic cannot explain the reality of succession `in Time.'  The reason of its failure, viz.\ that Time, Change, and Individuality are features of Reality we abstract from in our formation of Concepts.  Hence abstract metaphysics always fail to account for Reality.  Must we then either accept scepticism or reject a procedure on which all science rests?  No; for to admit the defects of our thought-symbols for reality need merely stimulate us to improve them.  As for science, it uses abstractions in a radically different way, to test and to predict experience.  Thus `law' is a methodological device for practical purposes.  Science practical both in its origin and in its criterion, and ethics as the science of ends conditions metaphysics.  Such an ethical metaphysic accepts and implies the reality of the Time-process.  And {\it therefore\/} it has a right to look forward to the realization of its ends in time, and forms the true Evolutionism.
\eq

More to the point with regard to yesterday's considerations after Tegmark's talk are these.  (In the long quote Schiller says about the same thing five times over, but I think I'll record it all in case there are some subtle distinctions I'm missing at the moment.)

\bq\noindent
[T]he incompatibility between the assertion of the reality of the Time-process and its comprehension by any system of `eternal' logical truth (whether Hegel's or any one else's) has its origin in very simple and obvious considerations.

Dr.\ McTaggart cannot find room for the reality of the Time-process, i.e.\ of the world's changes in time and space, within the limits of Hegel's Dialectic.  But is this an exclusive peculiarity or difficulty of Hegel's position?  Is the Time-process any more intelligible on the assumptions of any other purely logical\footnote{I.e. intellectualist.} system, as, for instance, on those of Plato or Spinoza?  I think the difficulty will be found to recur in all these systems.  And this shows that it is not accidental, but intrinsic to the {\it modus operandi\/} of all systems of abstract metaphysics.

They cannot account for the time-factor in Reality, because they have {\it ab initio incapacitated\/} themselves from accounting for Time as for change, imperfection and particularity---for all indeed that differentiates the realities of our experience from the ideals of our thought.  And their whole method of procedure rendered this result inevitable.  They were systems of abstract truth, and based on the assumption on which the truth of abstraction rests.  They aimed at emancipating philosophy from the flux to which all human experience is subject, at interpreting the world in terms of conceptions, which should be true not here and now, but `eternally' and independently of Time and Change.  Such conceptions, naturally, could not be based upon probable inferences from the actual condition of the world at, or during, any time, but had to be derived from logical necessities arising out of the eternal nature of the human mind as such.  Hence those conceptions were necessarily {\it abstract}, and {\it among the things they abstracted from was the time-aspect of Reality}.

Once abstracted from, the reference to Time could not, of course, be recovered, any more than the individuality of Reality can be deduced, when once ignored.  The assumption is made that, in order to express the `truth' about Reality, its `thisness,' individuality, change and its immersion in a certain temporal and spatial environment may be neglected, and the timeless validity of a conception is thus substituted for the living, changing and perishing existence we contemplate.  \ldots

The true reason, then, why Hegelism can give no reason for the Time-process, i.e.\ for the fact that the world is `in time,' and changes continuously, is that it was constructed to give an account of the world irrespective of Time and Change.  If you insist on having a system of eternal and immutable `truth,' you can get it only by abstracting from those characteristics of Reality, which we try to express by the terms individuality, time, and change.  But you must pay the price for a formula that will enable you to make assertions that hold good far beyond the limits of your experience.  And it is part of the price that you will in the end be unable to give a rational explanation of those very characteristics, which you dismissed at the outset as irrelevant to a rational explanation.  Thus the whole contradiction arises from a desperate attempt to eat one's cake and yet have it, to secure the eternal possession of absolute truth and yet to profit by its development in time!  \ldots\

If these considerations are valid, the idea of accounting for the time-process of the world on any system of abstract metaphysics is a conceptual jugglery foredoomed to failure, and must be declared mistaken in principle.
\eq

\section{02-02-07 \ \ {\it Something Fun} \ \ (to C. H. {\Bennett})} \label{Bennett56}

I wonder if you'd do something fun for me.  I'm visiting the Perimeter Institute, and I've been bragging about how their cafeteria could take some lessons from IBM Research's cafeteria.  Particularly the little paper and pencil boxes you have sitting on the tables.  (It came to me as I was having great difficulty writing on a napkin.)  Would you take a nice picture of one of them, or show how they're situated on the tables, so I can make a case to the people here that they should do the same thing.

Wish you were up here; I'm having a great time, just hanging out and being the old Chris again.

\section{03-02-07 \ \ {\it Bill's Thoughts on QL and QI Frameworks} \ \ (to W. G. {\Demopoulos})} \label{Demopoulos9}

\bwd
I've had an idea which I would like to run by you. It was prompted by
several things: your exchange with Veiko {\Palge}, a re-reading of some of our previous correspondence, and some more general reflections
suggested by work of Lucien and others. (Among ``others,'' I ran
across two very suggestive papers of Colin Howson on Ramsey on
probability and logic which I can send you the details of if you're
interested.)
\ewd

Yes, I would like to know what you're talking about here.  I remember
being extremely disappointed when I first met Howson at the LSE a
couple of years ago.  For despite his Bayesian credentials---I had
cut my eyeteeth on his book with Urbach---he seemed to hold fast to
the idea that subjective probability {\it had to be\/} supplemented
with objective chance \ldots\ and for the silliest of reasons:  To
explain how predictable frequencies could come about. I was taken
aback!  In case, just in case, you harbor your own reservations about
this issue, let me recommend Marcus {\Appleby}'s papers ``Facts, Values
and Quanta,'' and ``Probabilities Are Single-Case, or Nothing,'' \quantph{0402015} and \quantph{0408058}, respectively.  He makes
the argument there about as decisive as I've ever seen it.

\bwd
What I have to say is pretty abstract and at the level of the
``framework'' which I believe is implicit in your approach, and I
think that for just this reason it may clarify a little further the
connection between  us.
\ewd

You definitely get several of the elements right, I would say. Though
I do have some quibbles, as you'll see below.  Of course, these may
just correspond to parts of the framework you reject.  We'll see.

\bwd
I want to contrast what I'll call the quantum logical (QL) and
quantum informational frameworks (QI), where by a framework I mean
something very general, having to do with the language or conceptual
structure within which one seeks to formulate things, rather than any
particular thesis. One thing that I think emerges from looking at
things in this way is that it makes it clear that realism vs.\
anti-realism is not an issue, and I think it also shows that some of
the classical foundational issues, like whether values of observables
are created on measurement, are also not to the point.
\ewd

Well, I certainly agree with the first part of this sentence, i.e.,
``that realism vs.\ anti-realism is not an issue,'' but the jury for
me remains out on the second issue \ldots\ at least until I better
understand precisely what you mean by ``values of observables are
created on measurement.''

\bwd
It may raise the question of instrumentalism, but it does so in a
sense that doesn't contrast with realism.
\ewd

I've always been impressed by this passage of Richard {\Rorty}'s (from
the introduction to his volume, {\sl Objectivity, Relativism, and
Truth: Philosophical Papers Volume 1}):
\bq
The six papers that form Part I of this volume offer an
antirepresentationalist account of the relation between natural
science and the rest of culture.  By an antirepresentationalist
account I mean one which does not view knowledge as a matter of
getting reality right, but rather as a matter of acquiring habits of
action for coping with reality.

Philosophers in the English-speaking world seem fated to end the
century discussing the same topic---realism---which they were
discussing in 1900.  In that year, the opposite of realism was still
idealism.  But by now language has replaced mind as that which,
supposedly, stands over and against ``reality.'' So discussion has
shifted from whether material reality is ``mind-dependent'' to
questions about which sorts of true statements, if any, stand in
representational relations to nonlinguistic items.  Discussion of
realism now revolves around whether only the statements of physics
can correspond to ``facts of the matter'' or whether those of
mathematics and ethics might also.  Nowadays the opposite of realism
is called, simply, ``antirealism.''

This term, however, is ambiguous.  It is standardly used to mean the
claim about some particular true statements, that there is no
``matter of fact'' which they represent.  But, more recently, it has
been used to mean the claim that {\it no\/} linguistic items
represent {\it any\/} nonlinguistic items.  In the former sense it
refers to an issue within the community of
representationalists---those philosophers who find it fruitful to
think of mind or language as containing representations of reality.
In the latter sense, it refers to antirepresentationalism---to the
attempt to eschew discussion of realism by denying that the notion of
``representation,'' or that of ``fact of the matter,'' has any useful
role in philosophy. Representationalists typically think that
controversies between idealists and realists were, and controversies
between skeptics and antiskeptics are, fruitful and interesting.
Antirepresentationalists typically think both sets of controversies
pointless.  They diagnose both as the results of being held captive
by a picture, a picture from which we should by now have wriggled
free.
\eq
At least when it comes to quantum states, I know that I am thinking
of them in an antirepresentationalist way.  Thus, the issue of
realism vs.\ antirealism with regard to the quantum-Bayesian view
(which asserts outright that quantum states are subjective) is just
off the mark.

\bwd
I'm imposing this on you as a way of getting my head around some part
of what I will say in Maryland in April, so I hope I'm catching you
at a time when you are able to send me your reactions.

The idea of QI that I want to focus on is its rejection of the QL
framework of propositions and their structure in favor of an emphasis
on what  I'll call ``effects,'' a term whose significance will I hope
become clear.

I include with QI's rejection of the propositional framework all
related and to my mind equivalent ideas. I know that you sometimes
distinguish between propositions which are true or false vs.\ events
which happen or not, just as one also distinguishes both of these
categories from properties which hold or fail to hold, but I can't
really see any basis for marking such distinctions and see them all
as of one piece.
\ewd

I don't know if this will help, but let me at this point include
something I wrote to {\Ruediger} {\Schack}, when we were having a fiery debate on what to include in our ``certainty'' paper,
\bq \quantph{0608190}.
\eq
[See 12-12-06 note ``\myref{Schack113}{Offline Discussion}'' to R. {\Schack}.]  The distinction I keep drawing and that you don't like---I think---has to do with all that.

\bwd
QI rejects them {\bf all}.  By contrast, the QL tradition takes as its starting point the classical Kolmogorov theory of probability
which is naturally regarded as formulated within a framework based on
an algebraic structure of {\bf propositions}, \underline{events} or \underline{properties}; these latter notions are respectively {\it about}, \underline{associated with}, or \underline{belong to} the objects whose probabilistic behavior is described in a classical probability framework. QI doesn't deny that elementary particles sustain relations to propositions, events and properties, it simply does not take the propositions, events or properties associated with them to be the proper objects of quantum probability assignments.
\ewd

I like this last sentence very much.

\bwd
What is conventionally described as a property of a particle is, for
the purposes of the probability theory, replaced by the notion of an
{\bf effect} which the particle produces at the observable level of experimental results; and what is conventionally understood as the
probability that a particle has a particular property is re-conceived
as the probability that a particular effect will be produced under
certain experimental conditions. The particle contributes to this
effect and is not equated with the totality of its effects, but it is
effects---rather than properties of particles---that are the proper
subject-matter of quantum probability assignments.
\ewd

Yes.

\bwd
Effects are all describable at the {\bf macroscopic} level of
experimental results; they are not properties of the particles that
produce them, although their occurrence is a function of particle
behavior. Hence, in one sense, there is no question of realism, since
the framework of QI assumes the reality of the micro-level. I think
QI can even assume the {\bf observer-independent} reality of the
micro level without in any way compromising what is novel and
important about its framework, but I'll leave this to another letter.
\ewd

Yes, I agree with all this too.  But I would put it differently.  I
don't like drawing the distinction between the micro and macro
levels.  It sends one on the goose chase (like Zurek's decoherence
program) of thinking that the distinction is a physical one, rather
than a categorical one. The distinction to be drawn---as far as I am
concerned---is only between the {\it object\/} and the {\it agent}.

\bwd
One reason why Bayesianism is so attractive as an approach to
probability within the QI framework is because of the naturalness of
the idea of a fair betting quotient as an account of the meaning of
probability and the fact that this is readily paired with the primacy
which QI gives to the notion of an effect which is {\bf conditional on} the performance of an experiment:  probabilities are regarded as the fair odds associated with conditional bets, in the present case with bets which are conditional on an effect occurring should a particular experiment be performed.
\ewd

You're right about it being {\it one\/} reason.  As a matter of
history, I came across it from the other end.  I started out as a
strong objectivist about quantum probabilities, thinking that that
conception could eventually be made well-defined.  But in reading
everything I could on objective chances and propensities (from '91 to
'96), I slowly came to the belief that those notions were
contentless.  Eventually the Ramseyan notion struck me as the only
one with any operational content.

\bwd
As with conditional bets in general, if the experiment is not
performed, the bet is off. This is the basis for the claim, often
made by some proponents of the QI framework, that unperformed
experiments have no outcomes. So understood, the claim is neither
contentious nor surprising, and has no obvious implications for
anti-realism about the micro level. It does suggest an element of
instrumentalism, however.
\ewd

Instrumentalism for the quantum state, indeed.  But such is true for
{\it all\/} Bayesian probabilities:  they are never representational
of ``matters of fact'' about their objects.  Thus, along the lines of
{\Rorty}, it seems to me the wrong distinction to make.  Bayesian
probabilities never had a chance to represent anything about their
objects.

\bwd
A useful way of putting the difference between the two frameworks is
the centrality which one gives to {\bf conditional probability} and the other to {\bf conditional bets}. Within the QL framework of propositions or events we can show that the [L{\"u}ders] rule
generalizes the notion of conditionalization from a classical or
Boolean algebraic context. This is a formally elegant aspect of the
QL framework, but its interpretation is elusive for reasons connected
with the Kochen and Specker theorem. (Part of what I am thinking is
that perhaps this theorem should make us re-evaluate the QL
framework. I'm unsure whether the QI framework relieves the tensions
that the theorem poses for QL.) We can perhaps retain some of the
formal elegance of the QL account of conditionalization without
incurring its interpretational difficulties by focusing on the
probabilities of effects. The idea is that conditional bets have this
advantage: whenever a conditional bet takes place, there is no
question of the determinacy of the effects bet upon; their
determinacy raises no puzzles and is beyond question.
\ewd

I think I need you to explain what you mean by ``determinacy.''  I've
heard you say this word many times, but I've never pushed on you to
explain what you mean.  What do you mean?

And does what you mean have anything to do with {\Rorty}'s use of the
word in the same Introduction that I already quoted.  (I don't know
what {\Rorty} means either.)  Here's his usage of it:
\bq
The later {\Wittgenstein}, {\Heidegger}, and {\Dewey}, for example, would all
be as dubious about the notion of ``truth-makers''---nonlinguistic
items which ``render'' statements determinately true or false---as
they are about that of ``representation.''  For representationalists,
``making true'' and ``representing'' are reciprocal relations: the
nonlinguistic item which makes $S$ true is the one represented by
$S$.  But antirepresentationalists see both notions as equally
unfortunate and dispensable---not just in regard to statement of some
disputed class, but in regard to all statements.

Representationalists often think of antirepresentationalism as simply
transcendental idealism in linguistic disguise---as one more version
of the Kantian attempt to derive the object's determinacy and
structure from that of the subject.  This suspicion is well stated in
Bernard Williams's essay ``{\Wittgenstein} and Idealism.'' Williams says
there that a {\Wittgenstein}ian view of language seems committed to the
following chain of inference:\medskip
\begin{verse}
(i) `$S$' has the meaning we give it.\\
(ii) A necessary condition of our giving `$S$' a meaning is $Q$.\smallskip\\
{\it Ergo}\smallskip\\
(iii) Unless $Q$, `$S$' would not have a meaning.\\
(iv) If `$S$' did not have a meaning, `$S$' would not be true.\smallskip\\
{\it Ergo}\smallskip\\
(v) Unless $Q$, `$S$' would not be true.\medskip
\end{verse}

Since the values of $Q$ will typically include human social
practices, the conclusion of this set of inferences is, indeed,
reminiscent of transcendental idealism.  But the
antirepresentationalist will reply that (v) merely says that unless
certain social practices are engaged in, there will be no statements
to call ``true'' or ``false.'' [{\it Though, this part starts to
sound a little bit like what you are saying.---CAF\/}]  Williams,
however, rejoins that ``it is not obvious that for the later
{\Wittgenstein}ian view \ldots\ we can so easily drive a line between the
sentence `$S$' expressing the truth, and what is the case if $S$.''
His point is that antirepresentationalists typically do not think
that, behind the true sentence $S$, there is a sentence-shaped piece
of nonlinguistic reality called ``the fact that $S$''---a set of
relations between objects which hold independently of
language---which makes `$S$' true. So, Williams concludes,
antirepresentationalists, and in particular the later {\Wittgenstein},
are committed to the idea that ``the determinacy of reality comes
from what we have decided or are prepared to count as determinate.''

The trouble with this conclusion is that ``comes from'' suggests
causal dependence.  The picture called up by Williams's terminology
is some mighty immaterial force called ``mind'' or ``language'' or
``social practice''---a force which shapes facts out of indeterminate
goo, constructs reality out of something not yet determinate enough
to count as real.  The problem for antirepresentationalists is to
find a way of putting their point which carries no such suggestion.
Antirepresentationalists need to insist that ``determinacy'' is not
what is in question---that neither does thought determine reality
nor, in the sense intended by the realist, does reality determine the
thought.  More precisely, it is no truer that ``atoms are what they
are because we use `atom' as we do'' than that ``we use `atom' as we
do because atoms are as they are.''  {\it Both\/} of these claims,
the antirepresentationalist says, are entirely empty.
\eq

\bwd
A principal difference between QI and QL is that QL seeks to explain
the quantum probability assignments by appealing to the algebraic or
logical structure of quantum propositions. (Here I'm thinking of Jeff
and Itamar; my recent work is different in just this respect, since I
don't credit the logical structure (i.e., the ``basic'' rather than
the ``derived structure'' of my incompletely knowable domain paper)
with that kind of explanatory power.) This idea is a reasonable one
in view of Gleason's theorem. For it to be wholly successful, it
would be necessary to show how the structure of Hilbert space can be
recovered from logical or algebraic constraints on the family of
physical propositions, which is the central goal of the quantum
logical axiomatic program. It would also be necessary to show that
this is the correct direction of explanation---i.e.\ we should go
from logic to probability, rather than in the opposite direction or
in neither.

By contrast, I take it that QI envisages {\bf no} such explanation of the probabilities. For QI, the situation is the same as with classical probability theory: classical {\bf logic} is insufficient for explaining the coherence of classical probabilistic reasoning. Its explanation may plausibly be said to depend on the interpretation of probabilities as fair betting quotients and the theorem of de Finetti. So also in the context of quantum probability theory, where one also seeks a general strategy for all possible gambles---but where significantly, not all gambles can be taken simultaneously---if one's betting quotients do not obey the quantum probabilities, one is sure to lose.
\ewd

That is the idea.  Now, Itamar claims to prove just this.  But he
takes the quantum logical structure of propositions as given.  The
quantum Bayesians of the CFS variety, however, do not want to take
that as given.  And in that regard, what you say---``if one's betting
quotients do not obey the quantum probabilities, one is sure to
lose''---is only an idea for an idea:  It is a research program.

\bwd
The remarkable and striking fact here is not the suggestion that
logical relations explain probability assignments, but that an
empirical theory of the probability of outcomes of experiments should
inform the consistent distribution of probabilities over a family of
hypothetical gambles, {\rm [\ldots]}
\ewd

I agree to this point.

\bwd
{\rm [\ldots]}  a distribution which contravenes the classical constraints on their distribution.
\ewd

But I disagree here.  I don't think there's any sense in which the
Born rule specifying probabilities for all measurements contravenes
the usual Ramsey / de Finetti coherence.  As I see it, the extra quantum
restrictions are an addition on top of usual coherence.  (Remember,
as I see it, quantum measurement outcomes for distinct measurements
are not within a single algebra of any interesting variety.)

\bwd
Following tradition, I call the puzzling phenomena which comprise the
first grouping, quantum paradoxes, and the second, the measurement
problem. I depart from tradition by calling the problem of hidden
variables the theorem which is often invoked as a solution to what
usually passes for this problem; by the problem of hidden variables I
mean the problem of understanding the significance of this result in
the formulation which has become canonical since the work of Kochen
and Specker. {\rm [\ldots]}

I claim that the classification is nevertheless a useful one because
it isolates a conceptual issue that is more purely philosophical from
those that are peculiar to quantum theory, and which are therefore
more properly regarded as foundational issues specific to that
theory; the more purely philosophical issue is the problem of hidden
variables. Even though the theorem of Kochen and Specker might never
have suggested itself were it not for the development of quantum
mechanics and the presumption that it might bear on the foundational
problems mentioned earlier, the conceptual issues the theorem poses
are rightly distinguished as philosophical rather than foundational
because the notions on which their formulation depends are of such
generality that they are held in common by virtually every theory.
\ewd

Aha!  You may not agree with me in detail, but you seem to agree with
me in sentiment.  Recall what I wrote in the old Caltech proposals:
\bq
\noindent {\bf Quantum Mechanics as a Powerful Hint.} \ In my opinion,
the most profound statement yet to come out of quantum theory is the
Kochen--Specker theorem.  For it licenses the slogan, ``Unperformed
measurements have no outcomes.''  This is just a beginning.  If one
canvasses the philosophic traditions for one that has significantly
developed this slogan, one will find the now mostly-forgotten
tradition of pragmatism fathered by William {\James} and John {\Dewey}. As
a source of ideas for what quantum mechanics can more rigorously
justify, no block of literature is more relevant: The connections
between the two fields cry out for systematic study. Quantum
mechanics holds the promise of drastically changing our worldview on
the wide scale. It is time to let that happen.
\eq
and
\bq
\noindent
{\bf Quantum Mechanics and Anti-representationalist Philosophies.}
There are various threads connecting the quantum research program
proposed here to a wider philosophical tradition, which to my
knowledge has never been greatly examined in this context.  The
tradition comes under the rubric of what Richard {\Rorty} calls
`anti-representationalist philosophies.'  This tradition, spearheaded
by the pragmatism of William {\James} and John {\Dewey}, also includes
thoughts of (the later) Ludwig {\Wittgenstein}, Martin {\Heidegger}, Donald
{\Davidson}, Hilary {\Putnam}, {\Rorty} himself, and several others.  How else
can one understand the implications of the Kochen--Specker theorem
than by realizing it hints at something like {\James}' analysis of the
concept of `truth'?  How else can one make sense of a Bayesian take
on pure quantum states than to explore the same paths as {\Wittgenstein}
in his book {\it On Certainty\/}?

Since becoming immersed in the subject, I have found nothing more
exciting than these trains of thought.  For they indicate the extent
to which quantum foundations research may be the tip of an
iceberg---indeed, something with the potential to drastically change
our worldview, even outside the realm of physical practice.
\eq

You may become a pragmatist yet!

\bwd
The idea I propose to develop is based on the assumption that no
satisfactory interpretation of quantum probabilities is possible so
long as the probabilities are understood to be defined over
propositions ``belonging to'' a particle. I say that a proposition
belongs to a particle if its constituent property is a possible
property of the particle. I claim that such an interpretation is
compelling only when the propositions form a Boolean algebra.
\ewd

That sounds good.

\bwd
The essential idea of my positive proposal is that the probabilities
of ``quantum probability theory'' are not defined over the
totality---or even a sub-totality---of the propositions belonging to
a particle, but over the totality of effects which are induced by an
interaction with an experimental set-up and that are registered in
the experimental apparatus.
\ewd

I hope you'll give some credit to old Asher Peres here,
``Quantum Theory Needs No Interpretation.''  This is a formulation he
would have accepted:
\bq
\noindent
Our purpose here is to explain the internal consistency of an
``interpretation without interpretation'' for quantum mechanics.
Nothing more is needed for using the theory and understanding its
nature. To begin, let us examine the role of experiment in science.
An experiment is an active intervention into the course of Nature: We
set up this or that experiment to see how Nature reacts. \ldots\ What
[quantum theory] does is provide an algorithm for computing {\it
probabilities\/} for the macroscopic events (``detector clicks'')
that are the consequences of our experimental interventions. This
strict definition of the scope of quantum theory is the only
interpretation ever needed, whether by experimenters or theorists.
\eq
and
\bq
\noindent
We surely agree with Brun and Griffiths that ``in science, one cannot
rule out alternatives by fiat; one must evaluate them on their
merits.'' We do not find any merit in the various alternatives that
were proposed to the straightforward interpretation of quantum
theory: It is a set of rules for calculating probabilities for
macroscopic detection events, upon taking into account any previous
experimental information. Brun and Griffiths may think this a
``straitjacket,'' but it prevents the endless conundrums that arise
solely from shunning quantum theory's greatest lesson---that the
notion of experiment plays an irreducible role in the world we are
trying to describe.
\eq

\bwd
Effects, which comprise the macroscopic historical record of such
interactions, are the proper objects of quantum probability
assignments. My claim is that it is this shift in conception that is
mandated by the classical hidden variable theorems. Because these
theorems are formulated in a propositional framework, they possess a
kind of ``dialectical'' significance by showing the necessity of
replacing the propositional framework on which they are based. As a
point about the conceptual framework of quantum theory, the claim
lends itself to the following formulation: The logical form of the
representation of an elementary particle is not that of a class of
propositions whose constituent properties are possible properties of
the particle, but that of a function which, when subjected to a
specified class of operational procedures, produces a family of
effects.
\ewd

I tentatively like the sound of this formulation, but I'm not
completely sure yet that I know what you mean.

\bwd
The representation as a function implicates the particle in an
essential way in the production of the effects associated with its
interactions with a class of experimental procedures. But it is not
committed to the idea that an effect would have been the same had the
particle been presented with a different experimental set-up than the
one with which it was in fact presented. By contrast, the view of the
particle as a class of propositions is committal on this
counterfactual claim in just the way that the functional
representation is not.

Within this framework the problem of determinism is posed as follows:
Given a class of experimental procedures, to predict their effects
with perfect knowledge, i.e.\ to predict uniformly, and without
foreknowledge of the experimental procedure to which the system will
be subjected, the answer to every question regarding the occurrence
of a possible effect. The no hidden variable theorems show that the
quantum probabilities of such effects are not compatible with the
existence of a function which predicts every possible effect with
probability 0 or 1; hence, neither is it possible to construct a
reconstruction of the state of a particle on which to base such
predictions: the probabilities of the effects are logically
inconsistent with the existence of such a state. The representation
of an elementary particle as a function which, when presented with an
experimental configuration, yields an effect, is interchangeable with
its representation as a class of propositions only when the effects
are predictable with 0-1 probability. The fact that the
particle-effects described by quantum probability theory are not so
interchangeable with propositions belonging to them should be implied
by, and perhaps itself implies, the impossibility of cloning.
\ewd

You're getting off track with this last sentence:  The no-cloning
theorem has no power to imply the Kochen--Specker result.  No-cloning
is always true for probability distributions, no matter what the
character of arguments of the probability functions (i.e., whether
they are ``effects'' or classical (Kolmogorovian) propositions
belonging to the particle).  Furthermore, no-broadcasting is
generally true too, as long as the probability functions are not
allowed to become too peaked.  Simply see {\Spekkens}'s ``toy model''
paper.

Kochen--Specker is far deeper than no-cloning \ldots\ and far deeper than many of the standard phenomena of quantum information theory
(teleportation, superdense coding, entanglement monogamy, etc.).

\section{03-02-07 \ \ {\it My First Fortune Cookie in Waterloo} \ \ (to L. Hardy)} \label{Hardy19}

``You are going to pass a difficult test.''

\section{04-02-07 \ \ {\it QT Needs No Interpretation} \ \ (to W. G. {\Demopoulos})} \label{Demopoulos10}

\noindent Boy my ideas have significantly moved on since then!!  (Or at least gotten more subtle/resilient \ldots\ if you can imagine such a combination.)

\section{04-02-07 \ \ {\it The Painful Ambiguity of Language} \ \ (to W. G. {\Demopoulos})} \label{Demopoulos11}

\noindent\underline{\bf NOTE}: This letter was never finished or sent, though the part below was \medskip composed on this date.

Reading your notes, now I am VERY confused.  I thought I had understood your point of view, but now I wonder.  And when you had said in your earlier note, ``What I have to say is pretty abstract and at the level of the `framework' which I believe is implicit in your approach,'' I thought you were roughly right, but now I wonder about that too.

\bwd
The idea behind determinacy: Nothing can be colored without being
some specific color. The terminology is perhaps derivative from W. E.
Johnson (whom I've never read, but likely should have) who was
apparently an important influence on Keynes and Ramsey. Johnson
introduced the terminology of {\bf determinates} falling under a common \underline{determinable}, from which my example evidently derives.
\ewd

This must be the terminology {\Schiller} was using in that quote of his from 1903 that I like so much [F.~C.~S. {\Schiller}, ``The Ethical Basis of Metaphysics,'' in his {\sl Humanism: Philosophical Essays}, second edition, (Greenwood Press, Westport, CT, 1970)]:

\bq
That the Real has a determinate nature which the knowing reveals but
does not affect, so that our knowing makes no difference to it, is
one of those sheer assumptions which are incapable, not only of
proof, but even of rational defence.  It is a survival of a crude
realism which can be defended only, {\it in a pragmatist manner}, on
the score of its practical convenience, as an avowed fiction.  In
this sense and as a mode of speech, we need not quarrel with it.  But
as an ultimate analysis of the fact of knowing it is an utterly
gratuitous interpretation.  The plain fact is that we can come into
contact with any sort of reality only in the act of `knowing' or
experiencing it.  As {\it unknowable}, therefore, the Real is {\it
nil}, as {\it unknown}, it is only potentially real.  What is there
in this situation to sanction the assumption that what the Real {\it
is\/} in the act of knowing, it is also outside that relation?  One
might as well argue that because an orator is eloquent in the
presence of an audience, he is no less voluble in addressing himself.
The simple fact is that we know the Real {\it as it is when we know
it}; we know nothing whatever about what it is apart from that
process.  It is meaningless therefore to inquire into its nature as
it is in itself.  And I can see no reason why the view that reality
exhibits a rigid nature unaffected by our treatment should be deemed
theoretically more justifiable than its converse, that it is utterly
plastic to our every demand---a travesty of Pragmatism which has
attained much popularity with its critics.  The actual situation is
of course a case of interaction, a process of cognition in which the
`subject' and the `object' determine each other, and both `we' and
`reality' are involved, and, we might add, {\it evolved}.  There is
no warrant therefore for the assumption that either of the poles
between which the current passes could be suppressed without
detriment.  What we ought to say is that when the mind `knows'
reality both are affected, just as we say that when a stone falls to
the ground both it and the earth are attracted.

We are driven, then, to the conviction that the `determinate nature
of reality' does {\it not\/} subsist `outside' or `beyond' the
process of knowing it.  It is merely a half-understood lesson of
experience that we have enshrined in the belief that it does so
subsist.  Things behave in similar ways in their reaction to modes of
treatment, the differences between which seem to us important.  From
this we have chosen to infer that things have a rigid and unalterable
nature.  It might have been better to infer that therefore the
differences between our various manipulations must seem unimportant
to the things.

The truth is rather that the nature of things is not {\it
determinate\/} but {\it determinable}, like that of our fellow-men.
Previous to trial it is indeterminate, not merely for our ignorance,
but really and from every point of view, within limits which it is
our business to discover.  It grows determinate by our experiments,
like human character.  We all know that in our social relations we
frequently put questions which are potent in determining their own
answers and without the putting would leave their subjects
undetermined. `Will you love me, hate me, trust me, help me?'\ are
conspicuous examples, and we should consider it absurd to argue that
because a man had begun social intercourse with another by knocking
him down, the hatred he had thus provoked must have been a
pre-existent reality which the blow had merely elicited.  All that
the result entitles us to assume is a capacity for social feeling
variously responsive to various modes of stimulation.  Why, then,
should we not transfer this conception of a determinable
indetermination to nature at large, why should we antedate the
results of our manipulation and regard as unalterable facts the
reactions which our ignorance and blundering provoke?  To the
objection that even in our social dealings not all the responses are
indeterminate, the reply is that it is easy to regard them as having
been determined by earlier experiments.
\eq

I don't know that I would go as far as {\Schiller} and say that every aspect of matter is ``determinable indetermination,'' but I certainly think the idea corresponds decently with the situation in quantum measurement.  Take a complete orthonormal basis that corresponds to some quantum measurement that can be performed on a system.  The basis as a whole (i.e., not the subspace spanned by it, but the collection)---I would say---corresponds to a determinable belonging to the object.  (This is a variant of your terminology of ``belonging to.'')  That is to say, the basis as a whole is among the hypothesized properties of the object---for it has something to do with how any agent can interact with the object.  But what do the individual elements of the basis correspond to?  Truth values of propositions belonging to the object?  I would say no (just as apparently you would say no).  Thus I would say they are ``indeterminate of the object.''

But I think that's probably a very different use of the words ``determinate'' and ``determinable'' than here:
\bwd
In philosophy of QM it's thought to be the lesson of K-S that some
propositions are not determinately true or false. Since however their
disjunction may always have probability $= 1$, it is said that the
disjunction is true, but without any [disjunct?]\ being true or
false. Thus the determinable property holds without any determinate
which falls under it holding.
\ewd
For I think what you have in mind is that some quantum logicians would take the disjunction (i.e., the subspace spanned by the vectors) to be true OF THE OBJECT.  That the disjunction's truth value is a property of the object (a property that belongs to the object) even if \ldots

\section{04-02-07 \ \ {\it Giving Up for the Night} \ \ (to W. G. {\Demopoulos})} \label{Demopoulos12}

I started to construct a reply to your notes from today, and then it became long and unwieldy and mostly confused.  (I don't even want to show you the partial construction---it's pretty bad.)  So, I think I'm just going to give up tonight \ldots\ and indeed just give up until I see you in person.  I think at this stage, it might be more fruitful to have a conversation at a blackboard rather than write.

\section{06-02-07 \ \ {\it Physics, Economics, and Death by Interpretation} \ \ (to D. M. {\Appleby}, C. M. {\Caves}, and R. {\Schack})} \label{Appleby18} \label{Schack120} \label{Caves95}

How would you respond to someone who wrote this:
\bq
Economics and physics fundamentally diverge, however, on an entirely different front: people. To a physicist, our existence is profoundly irrelevant to the structure of the universe: we are made of the same basic building blocks as both quinces and quasars and the universe would unfold according to precisely the same structural laws if we were to suddenly disappear.$^1$ Economics, on the other hand, is entirely different. Without people, economics is not merely ``abstract''; it is simply non-existent, nonsensical. One obviously can't study the wealth of nations or the consumption of goods and services without tacitly assuming that there are humans$^2$ to produce and consume the goods.

$^1$I imagine there will be some astute reader who will point out that our present understanding of quantum mechanics argues for some special role of human observers. Without in any way wading into this particular miasma, the point is that the current ambiguity which makes such an argument defensible is nearly universally regarded by physicists as a problem with quantum theory, or at least our current interpretation of it.

$^2$Or aliens sufficiently similar to humans as to be, to all intents and purposes, the same thing.
\eq
Just wondering out loud within the earshot of some potentially like minds.

\section{07-02-07 \ \ {\it Research Program} \ \ (to P. Busch)} \label{Busch7}

\bpbu
I have read Marcus's papers but as I told him; I have more or less
come round to seeing probabilities as a ``logical'' tool, for making
uncertain or incomplete inferences (paraphrasing a quote Marcus gave
from Maxwell's writings). But then this still has to be brought
together with the ontology of the world out there.
\epbu

Yes, absolutely.  That's the nub of the research program, the part that has the greatest potential.  But to get this far---to this starting point really---we have had to have a hard, hard fight with the community to get the epistemic view of quantum states even to be taken seriously.

\section{07-02-07 \ \ {\it Epistemic/Ontic -- Subjective/Objective \ldots\ and All That} \ \ (to P. Busch)} \label{Busch8}

\bpbu
Well, well \ldots\ it did take you guys a while to realize that you were in danger of ``throwing the baby out with the bath water'', didn't it? {\rm \smiley}
\epbu

I'm not sure I'm understanding the correct intonation I should be reading this with (e.g., I can see a smile on your face), but under the assumption that I am:

I would say, NO, it was the {\it community\/} that took the long time to {\it recognize\/} (or finally listen) that we had no intention of ``throwing the baby out with the bath water''.  See, for instance, the bottom of the chart on page 5 and particularly the larger discussion on pages 5 and 6 of \quantph{0205039}.  That was five years ago, and the goal was already articulated quite clearly there.  What more can a poor worker do?

\section{10-02-07 \ \ {\it Toy Model, 2} \ \ (to S. J. van {\Enk})} \label{vanEnk10}

\bsve
If you want to violate Bell inequalities you either give up locality
or realism: I think if you give up realism then you can't keep your
underlying ontic state anymore \ldots\ don't you agree?
\esve

Yes \ldots\  But what does that have to do with your paper?  The reason you call it a ``toy model'' is 1) because there are underlying ontic states, and 2) there are ``epistemic'' states (for the ontic states) that satisfy some information theoretic requirement.  No?  What you find is that the requirement cannot be satisfied with ``epistemic'' states that are true-blue probability distributions---they must go negative.  The trouble is then I don't know how to really think of these things as epistemic states any more.  Thus, I'm not really sure in what sense this is an instructive ``toy model'' for QM after all.

\section{10-02-07 \ \ {\it Inside and Outside} \ \ (to S. J. van {\Enk})} \label{vanEnk11}

\bsve
I agree it's no longer knowledge about the ontic states. But I can view it as knowledge about the observables (i.e.\ the $Q$'s): there is a constraint on how much one can know about those. In the case that the $P$'s are all positive, the two constraints are equivalent; if not, then you have to use a trick: namely use Zeilinger/Brukner's measure of information.

But you're right, going to negative probabilities changes the character. On the other hand, I don't see how you can ever violate Bell inequalities, if you don't change the character of the {\Spekkens} model!
\esve

Have you ever thought about putting ``half of the hidden variable'' inside the observer?  I'm thinking something like this:  1) The system has an ontic state (say like in the {\Spekkens} model), but 2) the observer also has an ontic state (say, perhaps buried deep in his belly).  And though the value of the observer's belly state is within him, it is completely inaccessible to his knowledge.  That's what I mean by ``half of the hidden variable'' being inside the observer.

Now what about a measurement on the system?  Maybe {\it instead\/} of a measurement giving partial information about the system's ontic state, like in the {\Spekkens} model, the kind of measurement of interest tells something about the {\it relation\/} between the observer's belly state and the system's ontic state.  For instance, a measurement might reveal partial information about the parity of the inside bits and outside bits.

Anyway, the crucial idea here is that when an observer makes a measurement on the left member of a bipartite system, there is no good sense in which there is a local hidden variable at the right-hand member signifying the truth value of the measurement.  The outcome of a measurement only has some existence with respect to the hidden variable within the observer.  In other words, the observer has to take his belly variable over to the right-hand system before a {\it local\/} hidden variable account of the right-hand situation is complete.

Maybe another way to put it is that there is a sense in which this is a nonlocal hidden variable theory (the observer always carries half of the hidden variable)---and he is always with either the left-hand system or the right-hand system, but never localized at both.  So, though the hidden variable theory is nonlocal, it is nonlocal in an innocent kind of way.

Do you think one might get a Bell inequality violation out of toying with a kind of idea like that?

Just an idea; probably crazy.

\section{12-02-07 \ \ {\it Lord Zanzibar} \ \ (to J. Christian \& L. Hardy)} \label{Christian2} \label{Hardy20}

\noindent Dear Lord Zanzibar and Sir Lucien,\medskip

Here's that lovely quote I was telling you two about.  But I misremembered it a little; it's from G.~K. Chesterton's book {\sl Heretics\/} and only quoted by {\James}.
\bq\noindent
There are some people---and I am one of them---who think that the
most practical and important thing about a man is still his view of
the universe.  We think that for a landlady considering a lodger it
is important to know his income, but still more important to know his philosophy.  We think that for a general about to fight an enemy it
is important to know the enemy's numbers, but still more important to know the enemy's philosophy.  We think the question is not whether
the theory of the cosmos affects matters, but whether in the long run
anything else affects them.
\eq

My view of the universe is that it is many---that it ultimately cannot be unified, for it is alive and changing and creative in a very deep sense.  Moreover, that reality is, to some not-yet well-understood extent, plastic:  It can be molded by our actions.  Thus, though humanity is quite well a Darwinian accident, now that it is here, it is a significant component of the universe that must be reckoned with.

The question is, with that in the open would a landlady let me through the door!

\section{13-02-07 \ \ {\it Tegmark, 101 Years Ago} \ \ (to L. Smolin)} \label{SmolinL8}

I'm sorry I missed your talk last Thursday; I had to give my own talk in London that day.  It sounds like I quite missed something!  But we will have a chance to discuss these things many times, as I have made a firm decision now that I will be coming to PI for an extended stay as soon as possible.  More than discussing, though, I hope we will find a way to put some of these pragmatist ideas into solid, lasting physics---physics made of fantastic new equations and not just words.

At the moment though, let me send you some old words again.  You know I'm preparing for this ``pragmatism and qm'' conference in Paris next week, and because of that I'm rereading some {\James} and {\Schiller}.  Well, look at this nice tidbit I found in {\James} today.  It was as if he had been listening to Tegmark's talk himself.  Really, not a lot has changed in 101 years for a certain strain of mind.  ({\James}' lecture was in 1906.)  A modern like Tegmark uses a phrase like ``logical system,'' ``mathematical system,'' or ``multiverse,'' but it boils down to nothing essentially different from the ``Absolute'' and the ``mind of God'' that had taken hold of the ``rationalists'' at the time.  You'll see what I mean:
\bq\noindent
[I]f you are the lovers of facts I have supposed you to be, you find
the trail of the serpent of rationalism, of intellectualism, over
everything that lies on that side of the line.  You escape indeed the
materialism that goes with the reigning empiricism; but you pay for
your escape by losing contact with the concrete parts of life.  The
more absolutistic philosophers dwell on so high a level of
abstraction that they never even try to come down.  The absolute mind
which they offer us, the mind that makes our universe by thinking it,
might, for aught they show us to the contrary, have made any one of a
million other universes just as well as this.  You can deduce no
single actual particular from the notion of it.  It is compatible
with any state of things whatever being true here below.  And the
theistic God is almost as sterile a principle.  You have to go to the
world which he has created to get any inkling of his actual
character:  he is the kind of god that has once for all made that
kind of a world.  The God of the theistic writers lives on as purely
abstract heights as does the Absolute.  Absolutism has a certain
sweep and dash about it, while the usual theism is more insipid, but
both are equally remote and vacuous.
\eq

\section{14-02-07 \ \ {\it Rivers of Nows} \ \ (to G. L. Comer)} \label{Comer101}

\begin{flushright}
\baselineskip=3pt
\parbox{2.5in}{
\bv
{\bf Hollow Flow}\smallskip\\
In Einstein rapids,\\
Instants, rivulets,\\
Shadows shake\\
Beneath the foam.\smallskip\\
In carved canyons,\\
Sifted sandbars,\\
Imperfect memories\\
Attend the self-aware.\smallskip\\
In quantum pools,\\
Non-waves crash,\\
For milky ways\\
And silt-filled streams.\smallskip\\
In recollections,\\
Our reflections,\\
Are rivers of nows\\
And melting glaciers.\medskip\\
\indent Gregory Lee Comer \\ 02/13/07
\ev}
\end{flushright}

Thanks for the poem.  I'm glad I checked gmail!  (I usually only use gmail when I'm away from home, at a really backwoods place.)

Rivers of nows.  I love that phrase.  Rivers made of droplets?  So are there droplets of now?  Carriers of now?  Carriers of now.  Whose now?  My now.  I am the carrier of my now.  Maybe I should say, I'm the maker of my now.  Makers of now.  Whereupon synchrony, then?  Communication.  No communication, no global now; not even a quasi-local now.

Oh, you've carried me off today!  And this snow-locked day at home probably isn't helping either.  But I must resist.  Too much work needs to be done.

\section{15-02-07 \ \ {\it The Kochen--Specker Spectre} \ \ (to G. L. Comer)} \label{Comer102}

Thanks for the article.  Gregor Weihs will be one of my colleagues in Waterloo (though, he's at the university).  He's very reasonable and thorough kind of guy.

Couple of tiny points about the article.  I really disagree with blanket statements like this:
\bq\noindent
Thus, unless one allows the existence of contextual hidden variables
with very strange mutual influences, one has to abandon them --- and,
by extension, `realism' in quantum physics --- altogether.
\eq
It's the `by extension' part that bugs me.  For, I don't think that we quantum Bayesians have given up on realism at all.  Quantum mechanics allows for quantum systems---things out there independent of the experimentalist.  If they weren't out there for the experimentalist to perform experiments on, his actions in the laboratory would be as much of a koan as the sound of one hand clapping.  And yet, there are no hidden variables (from the quantum Bayesian view).  So, there is no `by extension'---it's just a non sequitur.

I think John Sipe does a good job of summarizing how we qB's can let go of hidden variables at the same time as holding on to a deeper, more interesting realism.  I'll paste in the section he wrote for his book below (wherein I've toyed with a couple of phrases to bring it further in line with my own thoughts).  [See Sipe passage reported in 24-06-06 note ``\myref{Bub21}{Notes on `What are Quantum Probabilities'}\,'' to J. Bub.]

\bq\noindent
We know this for photons already, but the corroboration in a
different system should help to convince doubting Thomases, as well
as assure the rest of us.
\eq

A quibble with regard to the need for funding these kinds of things:  For the skeptics who have not been convinced by experiment yet, they never will be.  Nor, does a positive result really assure anyone the likes of me---if one is already certain of something, one doesn't go out of his way to test it.  If there is a reason to do these experiments, it lies elsewhere:  And that is to develop laboratory techniques for quantum information processing.  Some QI protocols rely on KS for their very existence, and that's the real argument for using taxpayer money to do these things.

\section{15-02-07 \ \ {\it Bayesian Strategies} \ \ (to R. {\Schack})} \label{Schack121}

I'm looking forward to spending some time with you next week.  I feel like I'm bubbling over with ideas, but of course can't articulate a darned thing.

I'm writing this note to list a few things that I'd like to think about while together with you.

First an operational note:  I hope you know Paris well enough that we'll be able to negotiate the trip between the train station, the hotel (to check in and put away our suitcases), and the conference site in the allotted time.  We arrive in Paris Nord at 12:53, whereas the conference starts up at 3:00.

Now, for things to think about.

\begin{enumerate}
\item
I'd like to definitely further explore the conversation we started in completing the certainty paper, about ``live'' versus ``dead'' propositions.  Also, to discuss what better to say to {\Timpson}.
\item
The issue that John Sipe wants to draw out of us:  to what extent are we ``consequence Bayesians'' exclusively, or dualists between that position and ``classical Bayesianism'' (in his terminology).
\item
I'd like to figure what to say to Lucien Hardy who repeatedly challenges me on his paper with Galvao, ``Substituting a Qubit for an Arbitrarily Large Number of Classical Bits'' (\quantph{0110166}).  He somehow thinks this is a challenge to a Bayesian view of quantum states, and I just don't know what to say in defense.
\end{enumerate}

\section{27-02-07 \ \ {\it Ref Help} \ \ (to W. C. Myrvold)} \label{Myrvold7}

\bwm
While you were here I asked you for a reference for the classical no-
cloning theorem that goes via the invariance under Liouville evolution
of the fidelity of two classical densities---that is, integral of
$\sqrt{f(x) g(x)}$.  You referred me to {\Caves} and Fuchs, ``How Much
Information \ldots?'', \quantph{9601025},  but it's not in there.
\ewm

Well, it's mentioned, even if it's not spelled out in detail.  See the last paragraph of the paper, at the end of page 31 and the start of page 32.  That paper was started in summer of 1995, just at the time that I had first realized the point.  (I remember the moment perfectly; I was sitting in front of ``tangelo,'' my computer, writing on my thesis, and my jaw dropped.)  But I hadn't recalled that we waited so long before putting the paper on {\tt quant-ph}.  ({\Carl} was really dragging his feet on that one, and I remember saying, ``If we don't get this paper out quickly, Nathan Rosen may well never see it.''  Well, Nathan never saw it.)

I guess two publicly earlier references on the point are:
\begin{enumerate}
\item \quantph{9511010} (the no-broadcasting paper, pages 2 and 3),
\item \quantph{9601020} (my thesis, submitted Fall '95, pages 113 and 114)
\end{enumerate}
There's probably longer discussions buried somewhere in my {\sl Notes on a Paulian Idea}---if you want me to dig in there and find the appropriate stuff, let me know.

I was in Paris this week at a very nice meeting on pragmatism and QM.  I met Hacking there for the first time; a wonderful guy.  More importantly though, {\Ruediger} and I had many long discussions on the points you brought up at dinner in London.  I will try my best to put those thoughts down for you, along with a reply to your last note, soon.

\section{28-02-07 \ \ {\it Chris}\ \ \ (to A. Y. Khrennikov)} \label{Khrennikov21}

\bakh
Yes, it would be great! Thus I shall put you in the list of invited
speakers that we shall print today. Then you will contact with me
about whom will come?
\eakh

I'm just off the phone with {\Ruediger} {\Schack}, and I asked him if he would be available in my stead.  I told him that you would be able to pay his local expenses, but not the travel expenses.  He became very intrigued about the meeting, and is working to rearrange his schedule---he does not know for sure whether he can yet.  I think {\Ruediger} would be particularly good for the subject chosen for this year's meeting in that we have just prepared a paper making a direct comparison between our Bayesian view and older Copenhagen style views, and {\Ruediger} will have a presentation prepared on just that subject.

\section{01-03-07 \ \ {\it Soul Tired} \ \ (to D. M. {\Appleby})} \label{Appleby19}

You are a powerhouse of living scientists:  Your stamina and creativity amazes me.  You have reached so far further on this problem (and allied problems) that I wonder if I will ever catch back up.  I like your ``navigation'' analogy, which is the part of the note that I understood.

See what a horrible state I am in?  Actually, I am quite excited---extremely excited---about the move to PI.  It is absolutely clear that it will give me a new lease on life, and effectively set back the clock to the days when I first knew you (say, at the Oviedo meeting):  Life, and particularly quantum foundations, was full of possibility then.  But it is now getting there, and all the uncertainty in between, that is taxing my soul.

On other fronts, Paris was an extremely intense meeting for me.  I was surrounded by pragmatists, real pragmatists, for once and it was a completely new experience.  There were {\Dewey}ians, {\Peirce}ans, and even one {\James}ian.  There was a fellow who could recall detailed points in {\Wittgenstein}'s ``On Certainty''.  That, plus the general intensity of Paris (along with the general hope of PI), too made me feel that the future will be bright again.  Particularly useful were all the discussions with {\Schack} about an alchemical world and the ``potentest of all my premises.''

In that regard, let me ask you this:  Can you articulate why you feel that the study of SICs is profoundly important?  I would like to hear your answer without tainting it with any contemporary statements of my own, i.e., why I think SICs are profoundly important.  I reported all of those to {\Schack} and for the most part I think they left him cold.  So, I'm wondering how your version of things will come out.  Your stamina on this problem, I suspect, arises from a deep understanding of something---something I myself have not yet grasped---and I would like to pinpoint that source in you.

Besides the various tidbits above, my working life has been filled mostly with the politician's work:  Writing letters of recommendation, working behind the scenes to secure positions for good Bayesians (you're not the only one), getting the student-award program set up for the APS March meeting, working with PI to put together the foundations summer school, catching up on my editorial duties for QIC, crap like that.  And like a good American politician, I even got stuck in a snowstorm overnight during a recent flight:  Though in my case, I really did get stuck in a snowstorm, rather than slipping off with a mistress.

Kiki has been scrambling to complete all her uncompleted projects on the house, so that we might command a higher price.  And daily we look through the real estate adverts for the Waterloo area, tabulating distances and whatnot.

On my flight from Paris, I read a small book {\sl {\Wittgenstein}:\ A Memoir\/} that contained a small biography by G. H. von Wright, a memoir by Norman Malcolm, and a complete set of {\Wittgenstein}'s letters to Malcolm.  One thing struck me from the ebb and flow of W's letter-writing style:  It had a certain something in common with yours.  I never could put my finger on it, but at times (briefly), I felt like I was reading one of your letters.  It made me wonder if {\Wittgenstein} had influenced you deeply indeed.

\section{01-03-07 \ \ {\it More Progress on New Directions} \ \ (to W. G. {\Demopoulos})} \label{Demopoulos13}

Thanks for the latest draft.  I read a little of it during lunch---up to page 9---just before my lunchtime nap.  Let me make some completely superficial remarks at this stage.  I'll come back later with any more substantial remarks, after I get a chance to complete the document.  [\ldots]

I'm just recovering from my trip to Paris, where I had a thoroughly enjoyable, but intense time.  {\Ruediger} {\Schack} and I spent much time discussing what I keep calling ``the potentest of my premises.''  I also met several very knowledgeable living pragmatists, which was useful for me, and finally had several conversations with Ian Hacking, whom I got to know for the first time at the meeting.

\section{11-03-07 \ \ {\it Quantum Certainty} \ \ (to G. Quznetsov)} \label{Quznetsov1}

\bgq
On your ``Subjective probability and quantum certainty'' in LANL:  You already can be not worried about these topics because all these your problems are eliminated by my book {\bf Logical Foundation of Theoretical Physics}, Nova Science Publishers, N.Y. (2006).
\egq

Thank you for saving us a lot of work.  I will look at your book right away.

\section{12-03-07 \ \ {\it Remarks on Ludwig} \ \ (to D. M. {\Appleby})} \label{Appleby20}

I've finally sat down to read your long {\Wittgenstein} note seriously.  I am really, really sorry about this:
\bma
I {\bf think} that after all this time (more than 20 years) I have got the poison safely neutralised, but with something like this one can never be totally sure.  Which is why your comment, about similarities between my style and Ludwig's, worried me.  I feared that the toxins were breaking out.
\ema

It was only an offhand remark of mine!!  Really.  I certainly didn't mean to cause such a fear in you.  You never speak like an oracle.  If either of the two of us is guilty of it, it is I.  So, please rest your soul:  You are not {\it that\/} much like {\Wittgenstein}.  As long as you are a repository of some of his better thoughts, and can bring them to bear on physics, everything will be OK.

\bma
I am not like your friend at Paris:  I can't quote from the man
verbatim (a pretty pointless achievement, it seems to me).
\ema
Well, I didn't go quite that far.  I had only said, `There were {\Dewey}ians, {\Peirce}ans, and even one {\James}ian.  There was a fellow who could recall detailed points in {\Wittgenstein}'s ``On Certainty''.'  I don't see anything wrong in being a {\Wittgenstein}ian in that sense.

\bma
{\Wittgenstein} was a ball-crusher:  he tended to destroy the
intellectual virility of those around him (I use the word
``virility'' deliberately:  I get the impression that it was
specifically the men around him who were chiefly at risk (Anscombe, I
get the impression, was pretty weird, but I suspect she was like that
before she met Ludwig)).  And the effect wasn't dependent on personal
contact.  I mentioned the logical positivists, and the ordinary
language philosophers.  But aside from these there is a third
school:  the {\Wittgenstein}ians as they are often called.  These are
people who have fallen for the man completely.  They're a pretty sad
bunch of nonentities.  But they are not just sad in the way that
epigones are usually sad.  They have lost something vital, located in
the region of the crotch, and they all sing in the same kind of
distinctively precious voice.  Ludwig's castrati.
\ema

I wonder if I can infer from this that you'll think I wasted \$15 in buying G. H. von Wright's {\sl Philosophic Logic\/} in a Denver bookstore the other day?  I bought it for his chapter ``The Epistemology of Subjective Probability'', but of course have not read it yet.  The squirrel keeps gathering nuts for the long Canadian winters.

\section{12-03-07 \ \ {\it More Serendipity} \ \ (to G. L. Comer)} \label{Comer103}

\bgc
Do you know what I am now studying?  Reichl's {\bf Statistical Physics} book, in an attempt to understand better things like the Boltzmann
equation.  I used to hate that book.  But now I really like it.
Anyway, I've been reading about Liouville equation, etc.

So the book was unexpected and serendipitous!
\egc

Actually more serendipitous than you might think.  I've been thinking a little about what sorts of research pushes I might start to make after arriving at PI.  And one thought that has occurred to me is that I think it is time to start working on a ``quantum Bayesian'' understanding of Bose-Einstein and Fermi-Dirac statistics.  You see from a q-Bayesian point of view, symmetrization or anti-symmetrization is a {\it judgment}, nothing less, nothing more.  I.e., it is nothing less or nothing more than a property of a proposed quantum state, and a quantum state is a judgment.  So, the question then must be, if one {\it judges\/} either BE or FD, what is one judging?  I want to get a much deeper understanding of that.  (BE certainly looks as if it can be made to come out of an strengthier version of a finite quantum de Finetti theorem; but FD---maybe to be expected---is a completely wild mystery.)  These considerations have been hitting me from several directions lately.  One was discussions with the (eminent) philosopher Ian Hacking in Paris, who has recently taken a philosophical interest in the ``making'' of BECs.  (If you don't recognize the name, see \myurl[http://en.wikipedia.org/wiki/Ian_Hacking]{http://en.wikipedia.org/wiki/Ian\underline{ }Hacking}.)  The other is discussions with Daniel Gottesman at PI; see this paper of his:  \arxiv{cond-mat/0511207}.  But going back to Hacking, he asks the reasonable question:  Surely the stability of cold, far away stars does not depend on anyone's judgment.  The q-Bayesian needs a good, detailed answer to that question.

Anyway, I wonder would be interested in getting involved in this question in a serious way?  If you were to, I could justify bringing you to PI for a couple or three weeks in the coming year with my visitor budget.  PI would pay all your expenses (travel and accommodation), give you a \$40 CAD per diem for eating, and of course give you a desk.

Something to think about.  As you know, it'll be a while before I'm there physically---and because of that, it'll be until then that I can start thinking about these questions more seriously myself.  But maybe this is finally the right project to get us working together more closely.

\section{12-03-07 \ \ {\it A Little Hacking} \ \ (to G. L. Comer)} \label{Comer104}

I should have also mentioned that Hacking is partially at U.\ Toronto too---he's usually there for at least a semester each year.  Maybe it'd be good to schedule you at PI for a time when Hacking will be around so that we could go have discussions with him.

Also, at the moment, you might enjoy Section A.1 of this:
\begin{center}
\myurl{http://plato.stanford.edu/entries/physics-experiment/}.
\end{center}
And a couple of Hacking tidbits pasted and attached.

\bq
\begin{center}
Why Physics is Easy and People are Hard\medskip

      Ian Hacking\\
Coll\`ege de France, Paris, France\medskip
\end{center}

Of course physics is not easy. But it has evolved ways to turn the wonderful ``blooming, buzzing'' confusion around us, into problems simple enough for us to solve. ``Give me a laboratory and I will move the world.'' A current example will serve: the ultracold, the amazing domain of almost absolute zero. In the past few years it has been turned from the unattainably complex into something with which we can interact. It is teaching us new things about the universe almost every week.

Contrast the enigma of autism. It shows up early in the life of a child. Something is wrong neurologically, which may have genetic origins. Autism is devastating for parents. Despite optimistic announcements, we have no idea what causes it. We have only a little practical knowledge about how best to help autistic people. It is a deep psychological and biological problem that may teach us something about the human brain, the human mind, the human being. But only when we have overcome its complexity.

Both autism and the ultracold are fascinating. Between them they show that there is not just one kind of complexity, but at least two. The complexity of climatic modelling points to a third. Degrees of complexity in the theory of information and computation are different again. We should not address ``complexity'' as if it were one thing. We need to understand its many faces.
\eq

\section{13-03-07 \ \ {\it Recent Book Review?}\ \ \ (to N. D. {\Mermin})} \label{Mermin130}

Someone at the APS March meeting (I can't remember who) told me that you had written a book review that had something tangentially to do with quantum Bayesianism.  I thought they told me the most recent issue of Physics Today, but I can't seem it.  (It could be that it appeared in January, in which case, I can't even find that issue of {\sl Physics Today}!)  Curiosity is killing the cat.  Could you just send me the file if such a thing exists?

\section{13-03-07 \ \ {\it Recent Book Review?, 2} \ \ (to N. D. {\Mermin})} \label{Mermin131}

That was a very nice review.\footnote{\editornote The book being reviewed was Rosenblum and Kuttner's {\sl Quantum Enigma: Physics Encounters Consciousness} (Oxford University Press, 2006); see Am.\ J.\ Phys.\ {\bf 75}, 3 (2007), pp.\ 287--88.}  Makes me want to read the book (and scoff at part of it).  I liked several of your turns of phrase this time around.

I was also struck by this part:
\bq\noindent
    On the contrary, the reader is assured that quantum states are real
    states of affairs. ``In some very real sense, the wavefunction of an
    object is the object.'' Or, less guardedly, ``{\/}`The wavefunction of
    the atom' is a synonym for `the atom'.''
\eq
You know Charlie Bennett has been guilty of almost identical sentences before (and certainly remains guilty in spirit, as evidenced in Japan in November).  I've told you the story before, haven't I?

\section{14-03-07 \ \ {\it Aiming to Provide}\ \ \ (to C. M. {\Caves})} \label{Caves95.1}

\bcc
I do think the sentence about providing a firm basis is pretty bad.
What do you think of the following: Foundational studies aim to
provide a firm basis for quantum physics.\footnote{The conversation concerned the mission statement of the American Physical Society Topical Group on Quantum Information, which I was assisting Carl in writing at the time.}
\ecc

That one is a little better.  It's more honest.  The further ones became a little contorted.  Really, I think, the sentiment is this:  ``Foundational studies aim to provide a firm basis for quantum physics, and whether they get there or not, historically there has been a fruitful exchange of ideas between Quantum Foundations and Quantum Information Science.  We think that should continue.  Some even say that it should not be forgotten that QIS is little more than `applied QF'.''  That's the sentiment.  But getting it recorded in a non-hokey way is the challenge.

\section{15-03-07 \ \ {\it Last Try for a While}\ \ \ (to C. M. {\Caves})} \label{Caves95.2}

\bcc
The Topical Group recognizes a special role for foundational studies
of quantum mechanics and is committed to serving as the home within
the American Physical Society for researchers in the foundations of
quantum mechanics.  Foundational studies aim to provide a firm basis
for quantum physics.  Whether or not that succeeds, the Topical Group
encourages a continuation of the active and beneficial exchange of
ideas between quantum foundations and quantum information science.
\ecc

Well, I rather like this version.  Now my main fear is what the rest of the foundationsies will think of it.  I.e., whether they will think the phrase ``whether or not that succeeds'' will be conceding too much.  (But in my own mind it is true.  And this formulation might help take some of the heat off from naysayers like Bennett, Chuang, and Bill Philipps.)

\section{15-03-07 \ \ {\it Attributed to Einstein, Incorrectly?}\ \ \ (to G. Will)} \label{Will1}

I am a regular reader of your column and enjoy it very much, even if I disagree with it often.  This morning's column, however, left me in a position in which I could {\it finally\/} disagree with some authority!  Thus I thought I would write you:  One takes one's merit badges where one can.

Today's column started with the slogan ``The only reason for time is so that everything doesn't happen at once,'' which you ``attributed to Albert Einstein.''  However, I don't think that attribution is correct.  I have been familiar with a variation of the phrase for many years, one promoted by the Princeton physicist (and Einstein associate from Einstein's later days) John Archibald {\Wheeler}.  ``Time is nature's way to keep everything from happening at once,'' he would say.  {\Wheeler} was very fond of the phrase and put it in several publications, however he always attributed it to a graffito found on a wall in the Pecan Street Cafe in Austin, Texas.  For instance, in {\Wheeler}'s 1989 paper, ``Information, Physics, Quantum:\ the Search for Links,'' he cites it this way, ``Discovered among graffiti in the men's room of the Pecan Street Cafe, Austin, Texas.''

Of course, it could be the case that the phrase made it from Einstein to {\Wheeler} indirectly, via a men's room, but I find it unlikely given {\Wheeler}'s involvement in the development of general relativity in the late 1950s onward.

With regard to the history of ideas, let me point out one other variation of the theme that comes from Louis Menand's book, {\sl The Metaphysical Club}, a wonderful quadruple biography of William {\James}, Charles Sanders {\Peirce}, John {\Dewey}, and Oliver Wendell Holmes Jr.\ and a very good introduction to American pragmatism.  With regard to one of the meetings, Menand has this to say about {\Peirce},
\bq\noindent
They assembled. {\Peirce} did not come; they waited and waited; finally a two-horse carriage came along and {\Peirce} got out with a dark cloak
over him; he came in and began to read his paper. What was it about? He set forth \ldots\ how the different moments of time got in the habit
of coming one after another.
\eq

\section{15-03-07 \ \ {\it Attributed to Einstein, Incorrectly?, 2} \ \ (to C. M. {\Caves})} \label{Caves96}

I sent the note below to George Will this morning, only to stupidly do a Google search immediately thereafter.  I.e., I should have done it immediately before!  Anyway, the web reveals the quote variously attributed to Einstein, {\Wheeler}, and Woody Allen.  My guess is that Woody Allen really is the originator, now that I've seen it suggested.  Trouble is, no one seems to pin down when/where he said it.  Do you or Karen have any friends who are big Woody Allen fans that might recall the source of the phrase?\footnote{\editornote Ray Cummings' science-fiction novel {\sl The Girl in the Gold Atom} (1922) has the following in chapter five: ``The Big Business Man smiled. `Time,' he said, `is what keeps everything from happening at once.'\,''  As is the case with much science fiction of the period, this novel was first published in magazine serial form.  See {\sl All-Story Weekly,} 15 March 1919, which also contained a portion of Abraham Merrit's {\sl The Moon Pool,} another story with some science-history interest: \myurl{http://skullsinthestars.com/2009/01/21/a-merritts-the-moon-pool/}. }

\section{16-03-07 \ \ {\it Sorry, Sorry, Sorry, Sorry} \ \ (to N. D. {\Mermin})} \label{Mermin132}

I was very impressed with your book review.  I think it is because it expressed better than ever before what you see as the lesson of quantum mechanics.  Or at least I felt like it got to the nub.  Most particularly for me, I think it helped me pinpoint what I feel uncomfortable with in your view.  If you can give me some time, I'll try to articulate that.  (And tell you the Bennett story, etc.)

You are a good man David {\Mermin}, and I feel sorry about being such a sorry disciple.  I'll try to write more on more interesting matters soon.

\section{17-03-07 \ \ {\it Attributed to Einstein, Incorrectly?, 3} \ \ (to C. M. {\Caves})} \label{Caves96.0.1}

\bcc
I'm not so good at attributions, as you know.  I did enjoy reading
your message to Will.  Has he paid any attention to it?
\ecc

Well, I sent it to you not because of your own knowledge on such matters.  (For all I know, {\it you\/} don't even know who Woody Allen is.  When I first met you, you claimed not to know who Meryl Streep was.)  But you are tied to a culture that I would guess has statistically much better knowledge of Woody Allen than I.  Particularly, I thought Karen might know someone who's a real Woody Allen expert.  I think it would be quite funny if John had found great physical insight in one of Woody Allen's phrases.

You probably don't remember it, but I used a Woody Allen quote (which learned from {\sl The New Yorker}) in one of my papers, \quantph{0106166}: ``I hate reality, but, you know, where else can you get a good steak dinner?''

George Will, of course, didn't reply, but the Washington Post staff has a nice reply worked up:
\bq
   This is a note to let you know that George Will received your e-mail.
   Even if you do not get a personal response, please know that Mr.\ Will
   or one of his editors will read your e-mail. Thank you for taking the
   time to write.
\eq

\section{19-03-07 \ \ {\it Quantum Computing} \ \ (to L. Ketchersid)} \label{Ketchersid1}

I don't mind your quoting me (or anything in my writings), but I don't see how the present quote ties in with the subject of your article.

I read your draft.  I'll just make a couple of trivial comments that may be useful to you, and then sign off.  (Sorry, but you've caught me at a bad time these last couple weeks, with some rather big life changes in the works.  I will start to be more sociable again in the Fall, if you want to revisit these issues then.)

1)  Your sentence, ``A qubit is represented as a quantum object's nuclear spin (rotation),'' directly contradicts your later sentences, ``\ldots\ qubits are of course represented by quantum particles. Photons, electrons, quantum dots and the current flow across weakly coupled superconductors (Josephine Junctions) are a few of the quantum materials tried as qubits.''  The second of the two is truer:  A qubit can be made from just about anything, so long as it has at least two distinguishable states.  {\Schroedinger}'s cat is a perfectly good qubit, as it can be found alive or dead under the appropriate measurement procedure.  A qubit does not have to be a nuclear spin.

2)  In the same sentence already quoted, those are Josephson junctions, not Josephine junctions.  Josephine Junction is a town in Ontario.

3)  I don't much like the imagery of ``quantum parallelism'' and think we have been mis-steering the public (and ourselves) a good long time with it.  It's just a cheap way out; that's why everyone uses it.  Some models of quantum computation---for instance measurement-based QC---clearly have nothing to do with parallelism \ldots\ so that already gives a proof-of-principle contradiction to the whole imagery.  Have a look at these articles for your general education:
\quantph{0003084} ``A Quantum Computer Only Needs One Universe'', and
\quantph{0504097} ``Cluster-State Quantum Computation''.

4)
\bq\noindent
``The roadmap presented by D-Wave Systems, Inc., at their
    demonstration evidences that quantum computing may follow a similar
    pattern, with an `Online Quantum Computing Service' projected to be
    available in Q1 2008 and `Enterprise Deployable System' available
    one quarter after that.''
\eq
To propagate something clearly so farcical is dangerous, I think.  A more reliable timeline, I would say, should you wish to quote something, can be found in the ARDA document, ```Quantum Computation Roadmap,'' which can be found here:  \myurl{http://qist.lanl.gov/qcomp_map.shtml}.  Look at page 4 of the Overview chapter, to see where we will probably really be by 2012.\footnote{\editornote The high-level goal for 2012 was to ``implement a concatenated quantum error-correcting code [\ldots] which requires on the order of 50 physical qubits''.}  That's a far cry from an Online Quantum Computing Service.

5)  On a different subject---i.e., I don't see how it's related to your present article---in your second note to me, you wrote, ``I'd still be interested in putting some thoughts in this article from your paper on quantum mechanics as quantum information, specifically around the fact that some non-mathematical concepts are going to be required to describe what ``it'' is in order to get the computing community to take on quantum computing.''  I'm not quite sure what you're talking about, but if you are interested in what I see as the ``essential (and ineradicable) incompleteness'' of a quantum description of the world, let me recommend the following two (almost lay-) pieces to you.  Maybe they'll tell you a little more on the subject, or at least give you some thoughts.
\begin{enumerate}
\item
See Sections 4 and 5 of \quantph{0204146}.
\item
See Sections 1, 5, 6, and 7 of the attached paper (which I don't have posted, but appeared in this conference proceedings: \\
\myurl{http://proceedings.aip.org/proceedings/confproceed/889.jsp}).\footnote{\editornote See \arxiv[quant-ph]{0906.1968}.}
\end{enumerate}

But that's about all I can do for you now.  If you want to ``pelt'' me with questions (as you said in your first email), it'll have to wait until sometime in the Fall.

Good luck on your article.

\subsection{Larry's Preply}

\bq
My name is Larry Ketchersid, I am an old (in duration and age) friend of your father in law, Brad Lentz (from our Compaq days together). I have an interest in quantum mechanics and he has forwarded me one of your papers in the past.

I run a security software company ({\tt www.mediasourcery.com}), but in my spare time (usually between 1am and 2am) I write. I've written a book, {\sl Dusk Before the Dawn}, and I am writing technology articles for a newsletter called {\sl The Global Intelligencer\/} ({\tt www.theglobalintelligencer.com}).

I am working on my next article for the {\sl Intelligencer}, and would like to do one on quantum computing. With D-Wave Systems demonstrating a quantum computer recently, it is a timely topic and one I have some personal interest in.

The audience is not the most technical in the world, but I like to cite good sources (as I cited Mr.\ Schneier in my last article).

Would you mind being pelted with questions and potentially quoted?
\eq

\section{28-03-07 \ \ {\it An Anti-Fodor?}\ \ \ (to W. G. {\Demopoulos})} \label{Demopoulos14}

Thanks for the Fodor review of Frayn.  I got a lot out of it, even though I read it in the wee hours (as I couldn't sleep---something that always happens when I really need to).  Some remarks regarding our New York discussion later, but for the moment let me tell you about how his sentence, ``For better or worse (I think, in fact, it's much for the better), almost nobody has `a philosophy' any more,'' reminded me of a note I wrote to Joy Christian and Lucien Hardy recently.  I'll paste it below.  [See 12-02-07 note ``\myref{Christian2}{Lord Zanzibar}'' to J. Christian and L. Hardy.]

It sounds like I don't go as far as Frayn goes, but on the other hand, it does seem pretty clear that I'm an anti-Fodor nevertheless.

More piffle later (his words), when I have some light to see by.

\section{30-03-07 \ \ {\it Symmetry Considerations} \ \ (to C. H. {\Bennett})} \label{Bennett57}

Let me record a story while it's on my mind.  You seem like a good repository for this one.  The story's on my mind because the family and I are at Niagara Falls making our way to Waterloo for a short visit, and we saw the most brilliant rainbow through the Falls yesterday afternoon.  They're just waking up as I write this.  The Horseshoe Falls is 32 floors and a little hill's worth below me, just out the window to my left.

St.\ Patrick's Day was a couple of weeks ago, and Emma and I were having a discussion about leprechauns.  I asked her if she thought they were real, and she said yes, though it seems for no strong reason.  ``And they'll take you to their gold if you catch them?''  ``Yep,'' she replied.  ``But, I don't believe there's a pot of gold at the end of the rainbow.''  ``Why's that,'' I asked.  ``Because both sides of the rainbow are exactly the same.  How could you make a distinction between the beginning and the end?''

I was a very proud father.

Say hello to {\Theo} and the family for me.

\section{01-04-07 \ \ {\it Bohm and Cash Value} \ \ (to W. G. {\Demopoulos})} \label{Demopoulos15}

\bwd
Any thoughts on this? It's from a recent email from Jeff.

I think the criticism from the Bohmians and the Everettians will be something like this: ``You've given your analysis of the significance of the quantum revolution. That's interesting, but there is a less radical alternative: we can continue to adopt the standard view of systems characterized in terms of their properties on Bohm's interpretation if we take position, and properties that supervene on position, as the only properties; alternatively, we can hold onto the standard view if we interpret superposition in terms of multiplicity on the Everett interpretation. So why should we accept your view?''

I had pretty much this reaction from Simon Saunders, David Wallace, and Harvey Brown to my talk at Oxford.  And I have to say that I am still struggling with this -- i.e., why I don't accept the Everett interpretation (it seems to me if you are going to be a Bohmian, you might as well be an Everettian, since you don't really need the trajectories -- assuming you buy their decision-theoretic analysis of probability).
\ewd

I won't answer you on Everett at the moment, but I've got a quick temporary solution with regard to Bohm.  It's in the form of a note originally written to Hans Halvorson.  Pasted below.  [See 17-11-05 note ``\myref{Halvorson5}{Cash Value}'' to H. Halvorson.]

The main point I try to make is that one can make up {\it unobservable\/} ontologies underneath quantum mechanics ad infinitum.  Bohmian mechanics is not unique in that regard:  I could just as well say there is an individual angel guiding each quantum system, and that would be an ontology.  But the Bohmians pretend to be doing science rather than something so fanciful because they have an equation.  But so what?, I ask.  If the equation is unobservable, it is no better than the angel if it doesn't have some further cash value in the sense of facilitating physical calculations.  If there were some evidence that it did do that---facilitate calculations---I'd be all for it.  (If the angel ontology facilitated calculations, I'd be all for it too.)  But I've never seen any evidence of that sort.

The psychological origin of this criterion comes from my early interactions with Jeff Kimble.  When I was first a postdoc at Caltech, he would never speak to me.  I mentioned this to Mabuchi once and he said, ``Oh you just have to beat him at something, and then he'll take you seriously.''  And as it happened, one day during one of his group meetings, I disagreed with Kimble on some physical point.  When I was able to defend myself, he became a different person to me.  And that's my point about Bohmianism.  Let me take a standard problem of quantum mechanics---one still unsolved by all known means.  If the addition of Bohmian trajectories were to facilitate the solution, I would take the trajectories seriously.  They would remain, of course, unobservable, but their cash value would be clear enough to me and, hence, worth contemplating.

You see, for the physicist---as opposed to what I take for the average
philosopher of science---consistency of a theory is not enough to give it credibility.  It has to feel like it's going to give something new in the {\it next\/} step, not just tie up a loose end in an {\it old\/} one.  I like the way Scott Aaronson put it in his blog a while ago.  See the entry ``Mistake of the Week:\ Belief is King'' at \myurl{http://scottaaronson.com/blog/?p=200}.  Two
relevant paragraphs in there are:
\bq\noindent
The reason I'm harping on this is that, in my experience,
laypeople consistently overestimate the role of belief in
science. Thus the questions I constantly get asked: do I {\it believe\/} the many-worlds interpretation? Do I {\it believe\/} the
anthropic principle? Do I {\it believe\/} string theory? Do I
{\it believe\/} useful quantum computers will be built? Never
what are the arguments for and against: always what do I
{\it believe\/}? \ldots

In my view, science is fundamentally not about beliefs: it's
about results. Beliefs are relevant mostly as the heuristics
that {\it lead\/} to results. So for example, it matters that David
Deutsch believes the many-worlds interpretation because
that's what led him to quantum computing. It matters that
Ed Witten believes string theory because that's what led
him to \ldots\ well, all the mindblowing stuff it led him to. My
beef with quantum computing skeptics has never been
that their beliefs are false; rather, it's that their beliefs
almost never seem to lead them to new results.
\eq
My beef with Bohmian mechanics is where does it lead us further than the angel theory?

In the case of an ontology of objects and effects, or whatever you want to call it---i.e., the stuff you and I have been talking about---I feel it will lead to the NEXT step.  And its cash value has already long been apparent to me.  The conception told me that I should look for counterparts to almost all quantum information phenomena (e.g.\ no cloning theorem, teleportation, superdense coding, etc.) in Liouville mechanics.  I and several others looked, and we have found.  (See for instance the {\Spekkens} paper I recommended to you.)  One would like to see the conception lead to many more things still, but that's what should be required of any potential ontology.  Bohmianism seems to have been stillborn in that regard.  It's had a 50-year shot at it; it's time to move on.

That's my answer.

\section{10-04-07 \ \ {\it Title and Abstract}\ \ \ (to \v{C}. Brukner)} \label{Brukner1}

Let me apologize again for being so late on this.  Title and abstract below.  I hope they will work for you.

\bq\noindent
{\bf Good Coordinate Systems for Quantum Foundational Questions?}\medskip

Recently there has been much interest in the quantum information community to prove (or find a counterexample to) the existence of so-called symmetric informationally complete measurements (SICs).  In this talk we show that there should be even more interest.  For, under a robust measure of orthonormality for operator bases (one that does not build in any symmetry at the outset), one can show that SICs, if they exist, come as close as possible to being orthonormal bases for the space of density operators.  Moreover, in contrast to the usual expression of the superposition principle (where bases are taken to be orthogonal sets of state vectors), writing a superposition principle in terms of SICs leads to a more intrinsically-quantum representation for quantum states.  This is because the basis states, rather than being the easiest to eavesdrop upon (as the usual ones are), are actually the hardest.  Furthermore, such states fulfill a few other extremal properties that make them very interesting.  Because of all this, writing the quantum-state space in these terms gives hope for a direct derivation of it from a plausible information-theoretic constraint principle---a principle along the lines of the Brukner--Zeilinger proposal of information limitation, but applied solely at the single system level.  Various aspects of this problem will be discussed.
\eq

\section{10-04-07 \ \ {\it The Story of the No-Cloning Theorem} \ \ (to A. Wilce)} \label{Wilce13}

\baw
During our talk last week, you mentioned that you and Carl Caves were
responsible for the law-of-large numbers strategy for proving the
no-cloning theorem. I'd like to include a reference to this in our
paper -- where does it appear?
\eaw

The paper is this one: \quantph{9601025}. Its final resting place is:
\begin{itemize}
\item
C.~M. Caves and C.~A. Fuchs, ``Quantum Information: How Much
Information in a State Vector?,'' in {\sl The Dilemma of Einstein,
Podolsky and Rosen -- 60 Years Later (An International Symposium in
Honour of Nathan Rosen -- Haifa, March 1995),} edited by A.~Mann and
M.~Revzen, Annals of The Israel Physical Society {\bf 12}, 226--257
(1996).
\end{itemize}
It looks like Section 7 is the place, though I now wonder how much of the argument was us and how much of the argument was already folklore already long before then.

\section{10-04-07 \ \ {\it Squeamish on Everett} \ \ (to W. G. {\Demopoulos})} \label{Demopoulos16}

\bwd
I'll look forward to what you say about Everett. Regarding it, I would
have thought that no one would want to be stuck with such an absurd
view unless they were hammered into it. But that appears not to be the
case. I recall Harvey and Jeremy once trying to argue that it was no less absurd than the thought that there is a continuum of events, but I've never understood why they see this to be the same. If the issue were one of the continuum, then perhaps so. But that isn't what makes one balk at Everett.
\ewd

That is because it is a {\it temperament\/} that has essentially nothing to do with the details of quantum mechanics.  And that temperament has only found a recent home in quantum mechanics.  Read the {\James} quote below to see what I mean.  If you listen to one of Max Tegmark's talks on the beauty of many-worlds, you'll see little difference between his take on it and {\James}' description of the Hegelians.

I've written a couple of things contra-Everett.  They're not as tight of formulations as they should be, but they reveal my essential squeamishness with the idea.  See Section 4, ``Psychology 101,'' of
\quantph{0204146}.  Another place to look is in my {\sl Notes on a Paulian Idea}; see notes to Howard Barnum dated 30 August 1999 and 5 September 1999, ``It's All About Schmoz'' and ``New Schmoz Cola.''

The problem, in the end, is that Everettism is simply empty.

\section{12-04-07 \ \ {\it Effects of Effects} \ \ (to W. G. {\Demopoulos})} \label{Demopoulos17}

I'm sure this note will get to you too late for you to see it before the Bubfest, but let me wish you the best of luck with your talk.  I'm biased of course, but I think it'll be one of the most important at the meeting.  Let's hope this will be the beginning of a long list of effects arising from the recognition that effects are central for our understanding what QM is about.

I write this because I won't be able to come to the meeting after all. What's going on is that it's become clear that just too many things have to be done before we can put our house on the market [\ldots]  So, though the Bubfest represents probably the most important meeting for me all year long, I couldn't come to it in good conscience.  I hope you will understand.

I hope you'll give me a full report of your talk and the reaction to it as you get a chance.  And when I get to Waterloo, let's conquer the quantum world!

\section{13-04-07 \ \ {\it Happy Bubfest} \ \ (to J. Bub)} \label{Bub21.1}

It's with some sadness I have to write you this note:  Happy Bubfest!  Paradoxical, but necessary.  The reason is I won't be able to come to the New Directions meeting this year.  I really, really apologize and am kicking myself that this had to happen.  What's happening is that I'm in the middle of buying/selling houses, in preparation for my move to the Perimeter Institute, and everything hit me at once and at last minute.  My wife and I were up until after midnight last night negotiating the Waterloo house (our third offer finally got accepted), and our house here in Cranford now has to go on the market tomorrow.  And with the latter it's been a mad dash to the finish line in getting parts of the house freshly painted and decluttered, etc., for showing tomorrow morning.  Finally, yesterday afternoon it just became completely clear that I'd have to make a decision to stay at home this weekend.  That's the story.

I know I'm missing the best foundations meeting this year, and it hurts (hurts me, hurts the program, everything).  Still, at least I am confident you'll enjoy this well deserved honor of a meeting in your name.  Thanks for all the inspiration over the years.

\section{13-04-07 \ \ {\it Happy Bubfest} \ \ (to H. Barnum, A. Wilce, \& M. Leifer)} \label{Barnum22.1} \label{Wilce14} \label{Leifer9.1}

I'm sorry I'm going to miss all you guys today and this weekend.  (The story of my absence below.)  [See 13-04-07 note ``\myref{Bub21.1}{Happy Bubfest}'' to J. Bub.] The line-up of talks at this meeting looks fantastic, and on the top of my list is Howard's.  Good luck with it.  It has a chance of being a wake-up to the community, and I hope it will be.  The philosophers don't seem to be very good at seeking counterexamples:  All this talk about $C^*$-algebras ``covering a {\it vast\/} range'' of theories really grates, and I hope your talk will be a good splash of cold water for them.  There's so much to be done, and it'd be great if they'd start using their thoughts more wisely.  Lead the way!

\section{13-04-07 \ \ {\it Relations and Information} \ \ (to B. C. van Fraassen)} \label{vanFraassen15}

I am so sorry I'm going to miss your talk at the Bubfest.  I got caught up in all the stresses of a house sale/purchase, and it became clear that I could not afford to be away from home this weekend.  Many of the talks look fantastic, but particularly the remark in your abstract, ``It is a fascinating world in part because of Rovelli's reliance on the information-theory approach to the foundations of quantum mechanics \ldots'' has me intrigued.  If you have a paper version of the talk, I'd very much like to see it.

\section{11-05-07 \ \ {\it Information Gain Disturbance} \ \ (to R. W. {\Spekkens})} \label{Spekkens42}

\brws
What is a good reference for the following well-known result?
The only set of pure states about which one can get information without creating a disturbance is a set of orthogonal states.
\erws

That would be the Bennett, Brassard, {\Mermin} PRL that came out soon after the Ekert protocol:
\bq\noindent
C.~H. Bennett, G.~Brassard, and N.~D. {\Mermin}, ``Quantum Cryptography without Bell's Theorem,'' Phys.\ Rev.\ Lett.\ {\bf 68}, 557--559 (1992).
\eq

\section{11-05-07 \ \ {\it Information Gain Disturbance, 2} \ \ (to R. W. {\Spekkens})} \label{Spekkens43}

\brws
I just had a look at it.  They do show (around eqs.\ 3 and 4) that if two nonorthogonal states are left undisturbed, then no information is gained about them.  It's obvious that the result doesn't apply if the states are orthogonal, but unfortunately, they don't say so explicitly.  It would be great to find a reference that did say this explicitly.  Can you think of one?

Now that I think about it, for mixed states, the condition is no doubt that the set be a commuting set.  Surely, somebody has pointed this out somewhere.
\erws

Just a short note:  I've come in for a small break between mowing the left side of the yard and the right.

For mixed states one can gain information without disturbance even if they are noncommuting.  So long as they are block diagonal with the same blocks.  See Section 5 of \quantph{9611010}.  Koashi later proved if and only if somewhere.\footnote{\editornote See M.\ Koashi and N.\ Imoto, ``Operations that do not disturb partially known quantum states,'' Phys.\ Rev.\ A.\ \textbf{66}, 022318 (2002), section~VIII.B.}

I'll be able to think about the fine details of your question and trawl my memory tomorrow morning.  We've got to show the house to a potential customer tonight.  That's why I'm scrambling for all to look beautiful.

\section{11-05-07 \ \ {\it Big Drags and Little Drags} \ \ (to W. G. {\Demopoulos})} \label{Demopoulos18}

\bwd
I extracted a significant part of the paper and read it; I don't often do that, but the nature of the argument seemed to warrant this style of presentation. {\rm [\ldots]} I found the question period a little disappointing; Paul Teller said that he couldn't see how I was saying anything different from Bohr.
\ewd

Maybe he's right \ldots\ but the holy scriptures can be interpreted many ways.  More importantly that doesn't bar us from trying to say things more clearly than Bohr ever could \ldots\ and we are definitely saying them more clearly.  Anyway, seriously, I think I would say Paul is probably close to the right judgment.  In fact, it strikes me that in our own discussions---the discussions between you and me---you are to Bohr what I am to {\Pauli}.  I.e., their divide is our (present) divide.  For Bohr, the ``device'' was the repository of the effect; for Pauli the ``device'' should be considered a prosthesis of the agent.

\section{14-05-07 \ \ {\it Information Gain Disturbance, 3} \ \ (to R. W. {\Spekkens})} \label{Spekkens44}

\brws
I just had a look at [BBM].  They do show (around eqs.\ 3 and 4) that
if two nonorthogonal states are left undisturbed, then no information
is gained about them.  It's obvious that the result doesn't apply if
the states are orthogonal, but unfortunately, they don't say so explicitly.  It would be great to find a reference that did say this explicitly.  Can you think of one?
\erws

Is this more like the statement you were looking for:
\bq\noindent
On the other hand, if Eve definitely knows that the initial $|\psi\rangle$ is one of the orthonormal vectors
$|e_n\rangle$, but she does not know which one of them it is, she can unambiguously settle this point
by a non-demolition measurement [4], which leaves the state of the system unchanged.
\eq
It is from the top of page 3 of my old paper with Asher:  \quantph{9512023}.  I don't know how reference 4 had said things.

\ldots\ just catching up.

\section{14-05-07 \ \ {\it More Info Gain Disturbance Tradeoff} \ \ (to R. W. {\Spekkens})} \label{Spekkens45}

\brws
All of this talk has jarred my memory.  I now remember how at the end of one of your papers on state disturbance you show that the condition for information gain without disturbance is not commutativity, but rather commutativity on a subspace.   I also remember that I had a problem with this argument, so I was led to dig up some old notes on this question of infogain-disturbance for mixed states which I wrote way back when I was working on cryptography with Terry.  Here is what I said:

At the end of his paper, Fuchs suggests that all that is required for a measurement to be non-disturbing is for the state of the system to be unchanged. However, this seems to me to be necessary, but not sufficient. A sufficient condition is that the state of any larger system of which the system is a part is unchanged. The reasoning is as follows: if the correlations between A and B are disturbed due to a measurement on B, then clearly one's predictions about the outcomes of measurements on B suffer, specifically, one has reduced predictability for measurements on the composite of A and B.

A natural measure of this revised notion of disturbance is the probability of failing a test for a purification of the state.

So, whereas Fuchs argues that one can gain information without
disturbance for commuting density operators, and even for density
operators that fail to commute (but do commute on some subspace), I
would say that the notion of disturbance considered by him is an
unnatural one. Under the revised notion of disturbance, I suspect that
the only sets of density operators about which one can hope to gain
information without disturbance are orthogonal density operators.
\erws

I think your question is an interesting one, and I don't know the answer.  But I don't think---for a q-Bayesian, at least---that there can be a most-natural / least-natural distinction.  These problems are driven by the assumption of what the agent knows and which systems he has in his possession and control.  You imagine a situation where the sending agent has a reference system lying about and that he has a declared quantum state for the totality, even though he only sends part in one shot.  I imagined a situation where the agent sent everything he has per shot.  Just a different situation; neither is more fundamental or natural as far as I can see.  It would be very nice to find info-disturbance questions for which the answers are 1) ``iff orthogonal density operators'' and 2) ``iff commuting density operators.''  I always liked the achievement of the Holevo bound on accessible information in that it was ``iff commuting.''

\section{15-05-07 \ \ {\it Couldn't Slice It} \ \ (to R. {\Schack})} \label{Schack122}

Unfortunately, no matter how I tried to slice it, I couldn't make a small stay in London viable.  So, I guess I won't be seeing you this trip after all.  The trouble of course is my damned insistence on restricting myself to American Airlines.  Bypassing any time in London and arriving in Newark at 10:00 PM, made the cost of the trip within ethical bounds.  (Basically the lowest price I could have found anywhere, though of course other airlines would have given much better schedules.)  Arriving for a small visit to you in London at 4:00 PM would have upped my price \$120, and arriving at 1:00 PM would have upped it \$220---and besides, I would have still gotten into Newark relatively late in the day.  So, I just dropped the idea, and swallowed the idea of arriving at 10:00 PM (but a day earlier).

My existential state is complete misery at the moment.  It is the true evil of money---not living the life of an Erd\H{o}s, or a Wittgenstein, or a Tolstoy.  Hardly a couple has even looked at our house, and it is already mid-May. And as you can guess, monetarily, I've got everything tied up in this house.  I have no fluidity at all.  So, between the Waterloo house pending (July 31 latest) and all the expenses of the flood damage (waiting even to file insurance claims, much less collecting on them), I am up in arms.  And there's no great sign of relief in the real estate market here.  So my behavior has become one of OCD for this dirty subject; not at all the life of an academic.  Pray I survive to do science once again one of these days!

On better things, have a look at John Baez's blog ``This Week's Finds in Mathematical Physics'' (week 251, the latest edition).  He gives a small explanation (very small, but not infinitesimal) of the quantum de Finetti theorem.  [See \myurl{http://math.ucr.edu/home/baez/week251.html}.]

\section{29-05-07 \ \ {\it Foils Is a Very Broad Term} \ \ (to D. M. {\Appleby})} \label{Appleby21}

\bma
Rob did get in touch and I said I did want to go.  However, I am feeling a bit worried.  When you originally asked me I didn't pay much attention to the title {\rm [``Operational Probabilistic Theories as Foils to Quantum Theory'']}.  It is a workshop on quantum mechanics, you are going to be there, along with several other people I find interesting, so I automatically said ``yes''.  But Rob\index{Spekkens, Robert W.} has now sent me this:\ldots\ \ At the end he speaks of bringing ``everyone up to speed with what others are working on''.  But I am not working on operational theories.
\ema

Anyway, I don't think you'd be an imposter at that meeting at all.  And, looking carefully at Rob's\index{Spekkens, Robert W.} call for abstracts, I couldn't find the word ``operational'' or ``operationalism'' at all.  Probably it shows in the official title of the meeting \ldots\ but then that should have tweaked your fears long before the abstract request.

That aside, I view the issue of foils to quantum theory a very general question---a game that can be played whatever one's interpretational stance.  And in the list of attendees that I saw, I would say at least Wootters, Kent, {\Caves}, {\Barrett}, and Leifer could not be counted as ``operationalists.''  And in the case of {\Spekkens}, and maybe Barnum, I'm pretty sure they only see it as a means to an end anyway---a kind of stopgap method for bringing them ultimately to some kind of realism.

In my own case, what I want to know, and what I want to focus thinking on during the meeting is what is the significance of the damned $\tr\, A^2 = 1$ and $\tr\, A^3 = 1$ conditions for defining the quantum state space.  Who ordered that?  Why?  Particularly, when written in terms of an SIC expansion, what do they mean?  For, I want to see the two operator equations as information theoretic constraints on sets of probabilities, and---I think---the best way to translate them into something of that flavor is to use SICs as the intermediary.  Then the first condition is a constraint on a {\Renyi} information.  But what is the second equation?

What would be a foil theory to QM in this case?  Drop the one condition while retaining the other.  What would the resulting theory entail?  At least, that was the sort of thing I envisioned as good busy work while waiting for a better or more natural question.

And because I was thinking I wanted to think about those kinds of things, much more so than ``nonlocal boxes'' and the like, I thought it'd be great if you were there---that maybe you'd share some of my interest in these questions.  Moreover, I thought that you might get something out of thinking about all your SIC-existence work within this kind of context.  So, scientifically I think the meeting would be fruitful for you.

\section{31-05-07 \ \ {\it Vienna Traffic Lights} \ \ (to C. H. {\Bennett})} \label{Bennett58}

Noticed you're on the list for the Vienna meeting next week.  When do you get there; when do you leave?  We'll have to have some conversations near a traffic light and see if we can reproduce the phenomenon we noted in Tsukuba.  (Contradictory to the point of debate though that sentence be.)  [See 13-12-06 note ``\myref{Schack114}{More on Identity}'' to R. {\Schack} for an explanation of this.]

\section{11-06-07 \ \ {\it Unruh's Tenacity} \ \ (to C. M. {\Caves} \& R. {\Schack})} \label{Caves96.0.2} \label{Schack122.1}

One thing you got to give Bill Unruh is his tenacity.  {\it By his choice}, we ended up having a discussion to do with the objectivity/subjectivity of probability at the banquet in Vienna the other night.  I'm on my flight home now, but loaded this email just before leaving.

I sure would like to say just the right thing to Bill when I reply.  Do either of you have any suggestions?  In essence, Bill is saying that in the Alice--Eve scenario, Alice is better ``calibrated''---{\Ruediger}, recall our discussion on this in NJ---where calibration for him is defined in terms of her winning the bets.  Furthermore, he says, that calibration is reflective of an objective situation in the world.

If we can't win Bill Unruh over, how are we ever going to succeed?  In any case, it'll be good practice to formulate something.  I hope you have suggestions; I'd like to take them into account, and---this time actually---make sure we're all on board with each other.

Bill gave an excellent talk on sonic horizons in fluids, which was a lot of fun, and he tried to come to terms with whether one might actually see sonic Hawking radiation in the lab.  He said that Mark Raizen seems to think he might just be able to see it.  At the end, Bill came back to real black holes, and asked whether their entropy might come about from ignorance of an unknown state (either an internal state or a state of the horizon or the history of how the black hole was made).  He opted that it could not be any of those and that ``black hole entropy is like nothing we've ever seen before.''  That somehow it is more intrinsic than what we're taught in stat mech, where the entropy is the log of a number of potential states.  But I caught myself thinking, how is this than any different from any quantum mechanical entropy?  Hilbert space may come equipped with a dimensionality, but that dimensionality hardly signifies the number of states of the system.  As we know now, the states are not ``of'' the system, but ``of'' the observer in any case.  Just food for thought.

\subsection{Bill's Preply (apparently I never replied)}

\bq
Just to clarify. I agree with you that the probabilities are probably best
understood in terms of betting. The question is whether they represent (not are, but represent) real objective aspects of the world as well? I was
arguing that they do.  In your example where A sends a stream of photons
always chosen out of a set of up-down left-right, and never from 45-degree
possibilities. This fact which is a fact about the objective world alters
the probability description which A uses to describe the system in QM (i.e.,
alters the effective density matrix). That alteration of the probability
represents an objective fact. The probabilities themselves are as you
emphasise just betting odds. They themselves do not live in the world. They
live in the minds of the physicists. But they represent objective facts---in this case the biased choice of the experimentalist in sending the
particles. It is Alices true knowledge of that bias that results in her
changed distribution. Eve without that knowledge has a different
distribution, different set of betting. And the fact that her lack of
knowledge costs her because when she and Alice bet on the outcome of the
various experiments, she keeps losing.
Alice's true knowledge of the world reflected in her distribution, lets
here win the bet. That winning of bets is an objective fact, and reflects
the objective fact about the world represented by Alice's knowledge's change
in the betting probabilities.

I.e., probabilities while they themselves are just betting, are also
representations of objective features of the world.
\eq

\section{11-06-07 \ \ {\it Objective Indeterminism} \ \ (to \v{C}. Brukner)} \label{Brukner2}

I enjoyed talking to you the other night; I hope it is the sign of much more to come.  At the moment I write this, I am over the Atlantic working my way back home.  Particularly, I've been thinking about our little exchange on how one might have objective indeterminism at the same time as subjective probabilities---my view of course, that is, the one you took issue with.  Anyway, it led me to dig through my old correspondence to see what I had actually written on the subject.  Apparently not as much as I thought I had.  Still, I thought I'd compile for you what I found, even as incomplete as the thoughts are at this stage.  The file is attached.

The main letter to read is the second one, to Kirk McDonald.  I wrote it in a fit of passion, as he had written an insulting letter to me (saying that my view boiled down to ``physics $=$ psychology''), but nonetheless, I think that letter best captures the point.  [See 30-01-06 note ``\myref{McDonald6}{Island of Misfit Toys}'' to K. T. McDonald.]  Bayesian probabilities need not solely be ``ignorance probabilities'' (as I think you said) where the ignorance is of some actual existent, but rather probability in any context of uncertainty \ldots\ whatever the cause of the uncertainty.  In the case of quantum measurement, the uncertainty is there because the outcomes do not pre-exist the measurement.  Measurement is an act of creation, bringing something new into the world:  That is the ultimate reason for the uncertainty.  And what more can a poor quantum gambler do but express his beliefs about what will come into existence?  Alice has her beliefs about an event; Eve has hers.  They are not in conflict because the beliefs are not objective properties of the event.

Actually, while I'm on this subject, I think I'll also attach a little pseudo-paper of mine, ``Delirium Quantum'' (it appeared in an APS conference proceedings, but I never posted it on {\tt quant-ph}).  It describes in more detail the sort of sense in which I think nature is ``on the make''---that is, being hammered out, being created, in quantum measurement interactions.  [See \arxiv{0906.1968}.]

Thanks for organizing a great conference.  I thoroughly enjoyed it and got so much out of the discussions with so many.  Despite your weary words the evening of the banquet, I hope you'll organize another one soon!

\section{11-06-07 \ \ {\it Sonic Horizons} \ \ (to G. L. Comer)} \label{Comer105}

I haven't written you in a long time.  At the moment, I'm over the Atlantic again---this time coming back from a few days in Vienna.  It was a good meeting, a good admixture of people.  Of more interest to you probably, though, was a subset of them.  I had plenty of chance to talk with Bill Unruh, Paul Davies, and Ted Jacobson.  G.~F.~R. Ellis was there too, and I heard his talk (promoting the end of the spacetime concept, which of course sounded cool), but I didn't get a chance to talk to him.  Unruh and I, like usual, ended up fighting about probability.  Davies was glib mostly.  I had some very nice discussions with Ted Jacobson, who apparently just ``discovered'' me---it was kind of nice to go from being completely ignored to being an object of interest.  Have you followed at all his purported derivation of the Einstein equations as thermodynamic equations of state?  I'm curious to know what you think of it.

Mostly I really wish you had been there for Bill Unruh's talk on ``deaf and dumb holes''---sonic horizons.  Remembering that early stuff you did in Israel, I was wondering how the state of the art has progressed and whether you're getting due credit in the history of this stuff.  What was particularly exciting was Unruh saying that Mark Raizen at UT Austin was taking seriously the idea of doing an experiment to try to see the effect (minuscule though it be).  That would really be wonderful.

\section{12-06-07 \ \ {\it The Sweep and Dash of Mathematical Structures} \ \ (to M. Tegmark \& C. H. {\Bennett})} \label{Tegmark1} \label{Bennett59}

I'm flying over the Atlantic, slowly making my way home, and I'm using the time to take care of various loose ends that arose at the conference.  Below is the quote by William {\James} with the highfalutin' words I told you both about.  I thought it captured the point Charlie brought up with his question, but maybe not.  Charlie can be the judge of that.

What {\James} was talking about directly was a certain strain of philosophy at the time---a belief that the world arose from something called the Absolute.  The name to look up in an encyclopedia if you're interested is ``Absolute Idealism.''  Hegel's thought was an example of it, as was F. H. Bradley's.

{\James} wrote these words in 101 years ago in 1906:
\bq\noindent
[I]f you are the lovers of facts I have supposed you to be, you find the trail of the serpent of rationalism, of intellectualism, over
everything that lies on that side of the line.  You escape indeed the materialism that goes with the reigning empiricism; but you pay for
your escape by losing contact with the concrete parts of life.  The
more absolutistic philosophers dwell on so high a level of
abstraction that they never even try to come down.  The absolute mind which they offer us, the mind that makes our universe by thinking it, might, for aught they show us to the contrary, have made any one of a million other universes just as well as this.  You can deduce no
single actual particular from the notion of it.  It is compatible
with any state of things whatever being true here below.  And the
theistic God is almost as sterile a principle.  You have to go to the world which he has created to get any inkling of his actual
character:  he is the kind of god that has once for all made that
kind of a world.  The God of the theistic writers lives on as purely abstract heights as does the Absolute.  Absolutism has a certain
sweep and dash about it, while the usual theism is more insipid, but both are equally remote and vacuous.
\eq
In place of ``absolute mind'', as far as I can tell, you could insert Max's ``mathematical structures'' and little of the meaning would change.  Now, of course, ``mathematical structures'' makes reference to equations, symbolic logic, and the like, whereas the Absolute does not.  So it would seem to have a certain sweep and dash about it, far beyond anything the Absolute could muster.  More accurately, it gives the appearance of being more impersonal and scientific than the theistic God.  But, like {\James}, I cannot help but feel that this is only an appearance.  When a plan of the multiverse becomes so disconnected from what is seen in our actual universe, one might as well invoke a bearded old God---the two conceptions have no operational distinction, precisely because, as you pointed out (celebrated even), the ultimate, all-inclusive mathematical structure has zero information content.

{\James}, by the way, in that same lecture makes the first use of the term ``multiverse'' that I know of---though he meant it in quite a different way than you guys.

Max, I did enjoy your talk (and even more so the PI version of it) and many of your comments, which gave me lots of giggles.  Charlie says that's the best way to do science, and listening to you helped me clarify several of my own thoughts.  I also liked your comments to the students at the round table too; they were on the mark.

\section{12-06-07 \ \ {\it Quantum Chris} \ \ (to M. Tegmark)} \label{Tegmark2}

You also asked me about my own view of quantum mechanics.  For some reason, at that moment, I felt a little reluctant to talk about it.  Let me try to make up for that, in case you're still interested in the answer.

Maybe the single best place to look for a starter is this paper: \quantph{0205039}.  But that can maybe be supplemented with: \quantph{0404156} (particularly the introductory and concluding sections), and \quantph{0608190} (particularly the concluding section).  A couple of other significant pieces of the story are told in these papers by Marcus {\Appleby} and Matt Leifer: \quantph{0402015} and \quantph{0611233}.  (FQXi, by the way, partially funds Leifer \ldots\ and thus, I would guess this kind of thought?  Maybe you should watch where your dollars are going!!)

The view is a kind of realism about quantum objects, but one that strikes reality from the wave function.  The significant, stand-alone property of quantum systems is instead that they are catalysts.  They are catalysts that take part in the creation of events.  It is a view in me that owes much to the John {\Wheeler} of the 1970s and 1980s.

I hope that helps some.

\section{16-06-07 \ \ {\it Article for Nature} \ \ (to M. Buchanan)} \label{Buchanan1}

Funny you should ask for my thoughts on many-worlds; I'm not exactly a sympathizer.  For instance, my talk at the {\sl Fifty Years of Everett\/} conference coming up in September is to be titled ``13 Direct Quotes from Everettian Papers and Why I Find Them Unsettling'' (or at least that's the title I sent the organizers a few months ago).  But, yes, we can talk if you like---though Tuesday or Wednesday would be better dates for me.

\bmab
I've read a few things that suggest to me that the Everett view (and I
suppose there are many variations of it) has been gaining adherents in
recent years. Is this true, and if so why?
\emab

It is true that the Everett view is gaining adherents.  But I think it's powered mostly by a lack of imagination and a usual historical cycle working its way through.  Putting it bluntly, it's just the newest species of the ``serpent of rationalism.''  Read this little piece William James wrote 101 years ago for his Lowell Lectures:
\bq
   [I]f you are the lovers of facts I have supposed you to be, you find
   the trail of the serpent of rationalism, of intellectualism, over
   everything that lies on that side of the line.  You escape indeed the
   materialism that goes with the reigning empiricism; but you pay for
   your escape by losing contact with the concrete parts of life.  The
   more absolutistic philosophers dwell on so high a level of
   abstraction that they never even try to come down.  The absolute mind
   which they offer us, the mind that makes our universe by thinking it,
   might, for aught they show us to the contrary, have made any one of a
   million other universes just as well as this.  You can deduce no
   single actual particular from the notion of it.  It is compatible
   with any state of things whatever being true here below.  And the
   theistic God is almost as sterile a principle.  You have to go to the
   world which he has created to get any inkling of his actual
   character:  he is the kind of god that has once for all made that
   kind of a world.  The God of the theistic writers lives on as purely
   abstract heights as does the Absolute.  Absolutism has a certain
   sweep and dash about it, while the usual theism is more insipid, but
   both are equally remote and vacuous.
\eq
``The multiverse which they offer us, the mathematical structure that makes our universe by containing it within, might, for aught they show us to the contrary, have made any one of a million other universes just as well as this.  You can deduce no single actual particular from the notion of it.  It is compatible with any state of things whatever being true here below.''  My own feeling is that the MWI is simply empty of content---thus a dead end for physics.

Let me give you some links to pieces I've written to shore that feeling up a bit---ones written in a less formal, (hopefully) more entertaining fashion.  See:
\begin{itemize}
\item
My pseudo-paper \quantph{0204146}, particularly Section 4 of it (originally a letter to John Preskill), for my most sweeping discontent with the notion.
\item
My samizdat \quantph{0105039}.  Just look in the index for ``Everett'' and then flip to the pages.  Maybe pages 19--22, 92--93 (``Challenge to the Everettistas''), 356--358 and other letters to Preskill, 461--462 (letter to Spekkens), 472--473.
\item
And for something a little more serious, maybe the introduction and first couple of sections of this paper \quantph{0205039}.
\end{itemize}
The main point I try to get at there is that Everett (like most other interpretations) takes Hilbert space, state vectors, unitary evolution, inner products, tensor products, and all that mathematical machinery as the {\it starting point\/} of the interpretation, and then {\it contrives\/} a story around them.  (And in the Everett case, it's the most trivial story---one that could have been written, and as the quote above shows, was in fact written, by 1906.)  It should not be that way.  It should be the opposite:  The story should be written first, and one should be forced to the mathematical formalism from the story alone.

To give you some hint that this is not simply a pipe dream, have a look at the introduction to Rob Spekkens's ``toy theory'' paper:  \quantph{0401052}. It's a beautiful piece, even though it's not quantum theory directly.  What Rob does is start with a simple principle that ``maximal information is not complete information'' for a toy system and then looks at the properties of these states of information.  The lovely thing is that they have {\it so many} properties that one might have thought were unique to quantum mechanics.  And yet, there is no hint of parallel worlds in the theory at all.  ``Is that a coincidence?,'' one should ask.  No, of course not---the toy theory is explicitly constructed with the idea that there is one true world and what one is speaking of is one's ignorance of it.  But yet, as John Smolin points out, one could make a many-worlds interpretation of the toy theory!  And he's right.  But notice the order of things.  The mathematical structure was forced by the information principle alone; the many worlds interpretation of it is a grotesque appendage.  (Though John, being a sympathizer of MWI, would not call it grotesque as I did.)

Maybe a better person to ask this of is Howard Barnum.  He can give you a detailed account of what he sees as positive in the newest papers, but also, he can pinpoint exactly where he thinks they ultimately fail in their goal (particularly, of deriving the Born rule for quantum probabilities).

That's enough for now, enough to get you started.  Let's correspond again Monday or Tuesday about a good time to talk; I'm not quite sure what my schedule is going to be yet.

\section{01-07-07 \ \ {\it First Einstein Quote} \ \ (to R. W. {\Spekkens})} \label{Spekkens46}

Attached is part of the Einstein quote I was telling you about, from his ``Reply to Critics'' in the Schilpp volume.  Unfortunately, I do not have the whole quote with me here.  Where I write ``granted'' and quote him, that is the part where he describes what he means by ``physicist B.''  What he is grasping to express there is, I'm pretty confident, a physicist like myself or Asher or {\Carl} or {\Ruediger}, one for whom ``unperformed measurements have no outcomes.''

The main reason I brought this up tonight is I think you are right that Einstein didn't make a clear distinction between psi-epistemic and psi-supplemented:  I would guess for him, that he was simply using the word ``incomplete'' in the restricted fashion of meaning psi-epistemic only.  However, he did at least make one distinction that's pretty important to me:  between ``psi-epistemic about pre-existent properties'' and ``psi-epistemic about the consequences of interventions''.  I think he had that distinction pretty firmly in mind (probably from talking to Pauli), and in the present passages attempted to show that, upon the assumption of ``the real factual situation at 2 being independent of what is done at 1,'' wave functions could not be ``psi-ontic about the consequences of interventions.''  I.e., they could not be objective propensities or some such.

Had a nice long walk with Howard and Marcus tonight, talking about SICs and convex structures.  This is already shaping up to be a really good meeting.

\begin{center}
{\bf The More Pure Einstein}
\end{center}

\bq
\noindent {\bf Granted:}  ``The individual system (before the
measurement) has no definite value of $q$ (or $p$).  The value of the
measurement only arises in cooporation with the unique probability
which is given to it in view of the $\psi$-function only through the
act of measurement itself.''
\eq
\bq
\noindent {\bf Consider spatially separated systems $S_1$ and $S_2$
initially attributed with an entangled quantum state $\psi_{12}$.}
\eq
\bq
``Now it appears to me that one may speak of the real factual
situation at $S_2$. \ldots\ [O]n one supposition we should, in my
opinion, absolutely hold fast: the real factual situation of the
system $S_2$ is independent of what is done with $S_1$. \ldots\
According to the type of measurement which I make of $S_1$, I get,
however, a very different $\psi_2$ for [$S_2$]. \ldots\ For the same
real situation of $S_2$ it is possible therefore to find, according
to one's choice, different types of $\psi$-function.

If now [physicist B] accepts this consideration as valid, then [he]
will have to give up his position that the $\psi$-function
constitutes a complete description of a real factual situation.  For
in this case it would be impossible that two different types of
$\psi$-functions could be coordinated with the identical factual
situation of $S_2$.''
\eq

\section{01-07-07 \ \ {\it Second Einstein Quote} \ \ (to R. W. {\Spekkens})} \label{Spekkens47}

And below is the second Einstein quote I mentioned tonight, along with a Pauli quote.  It's a section from my new samizdat (a letter to Hans Christian von Baeyer).  [See 02-01-06 note ``\myref{Baeyer14}{The Oblique Pauli}'' to H. C. von Baeyer.] I can't believe I so botched my reporting of it tonight---my brain wasn't working.  Anyway, this little quote came soon after the earlier, more technical quote I just sent you.

\section{06-07-07 \ \ {\it Renner Monday} \ \ (to R. {\Schack})} \label{Schack123}

I hope you'll arrive here early Monday morning.  The schedule has been revised and Renato Renner will be giving the 10:00 talk before Ben Toner's on de Finetti theorems in the afternoon.  The subject of Renato's is supposed to be something like ``the de Finetti theorem as a precondition to doing science.''  Sounds fishy to me.  It'd be nice if you'd be in the audience to help me out at those tough moments.

\section{16-07-07 \ \ {\it Question, Possibly Addled} \ \ (to A. Wilce)} \label{Wilce15}

It was great to see you too.  Your clever comments throughout the week were enjoyed as ever.  And I'm glad too of your story about Harvey Brown.  These guys should stop and think, in the way you suggested, about what actually does compel them {\it if anything\/} to the MWI---they rarely do.  I think the class of examples in general convex set theories you bring up is great for cornering them on the subject.

About SIC-POVMs, yes, the definition does require that the elements of the POVM be rank-1.  You ask, ``if so, what advantage does that yield?''  Aha, I caught you!  The word is that that point was explained at a certain talk at Susquehanna University \ldots\  (Or at least I hope it was.)  I don't think there was originally any motivation for that particular restriction, other than to keep the problem from being too easy.  (Your construction, for instance, is just fine.)  But now things are a little more compelling.  My favorite reason for the restriction is that, by way of it, one comes geometrically as close to an orthonormal basis as one can on the positive cone.  See Section 2 of the attached paper.  So a SIC-set is a kind of stand-in for the idea of an orthonormal basis.  That's got to be good for something, I think, though I've yet to be able to express completely why.

I think it'll also be the case that if one generalized the idea of minimum uncertainty states we talk about in Section 3 to generally mixed states, one will find again the rank-1 density operators are the only ones that will fulfill the criterion.

Maybe there are other reasons to prefer the symmetric sets of rank-1 states over constructions (like your own) made of mixed states.  I should try to think of whether there are other reasons.  For instance, would mixed-state constructions like yours wreak any havoc on their equations corresponding to (5) and (6) in the attached paper for defining pure states?  Maybe they wouldn't?  Maybe they'd actually make for cleaner equations.  That would intrigue me greatly if it could possibly be so.

Can you send me references, by the way, for your papers on topologizing test spaces?  If you have any interest in exploring it further, I'd like to come back to your suggestions for generalizing Gleason:  I.e., identifying or classifying or finding some interesting characteristic of those test spaces for which the frame functions must be continuous (so that one has no dispersion free states).  Once at PI, I'd like to think much harder about that.

\section{16-07-07 \ \ {\it Trim URL?}\ \ \ (to C. M. {\Caves})} \label{Caves96.0.3}

I wish we had solved the SIC-POVM problem, but it's just a small piece of incremental progress.  The stuff you already heard about in fact; I just finally got around to writing it up.  It appeared on the {\tt arXiv} today.  Unfortunately, in the end, I could not attribute the bringing of SICs to physics solely to you.  We got {\Ruediger} to look at Zauner's PhD thesis, and sure enough, in a small part of it, he suggested constructing a measurement from these minimal two-designs.  Too bad.  So, we cited you two equally as introducing the idea.

{\Ruediger} and I had nice discussions on clarifying what one might mean about the locality-nonlocality issue from a Bayesian point of view.  So saying we ``solved everything'' was perhaps an overstatement \ldots\ carried away by the fact that we had our discussion in the same room of The Eagle pub where Watson and Crick announced that they had discovered ``the secret of life.''

\section{17-07-07 \ \ {\it Your Mom}\ \ \ (to C. M. {\Caves})} \label{Caves96.0.4}

I've been searching all day for something appropriate to say, but I think now that the day is gone I just give up.  I'm sorry to hear about your mom.  I'm sorry that one of the consequences of this time concept which so strikes me---that the world is not a frozen block, but a flux of creation---is that there is also passing.  That our greatest loves necessarily pass into memories, and memories which fade, until they are as the nothingness from which they came.  Why I see the one side of the equation as happy, but the other as sad, I don't know.  But I do.  I'm glad, as you said, there was a quick blessing at the end.

\section{31-07-07 \ \ {\it New Address} \ \ (to M. G. Raymer)} \label{Raymer1}

\bmgr
Thanks for including me on your list. I really enjoy reading your various papers and long, recorded email diatribes on QM.

In my quantum state, my belief is that quantum states are not simply matters of belief, but I think the idea goes a long way toward dramatizing the matter in the right direction.
\emgr

Thanks for the kind words.  And you shock me that have read any of my samizdats!  Very dangerous business \ldots

But one thing to keep in mind about Caves, Schack, and me in our terminology:  Try to resist the temptation to think that just because we categorize quantum states as ``states of belief'' that it means any unique individual can believe anything he wants to believe.  I.e., something like, ``I'll make up any quantum state I damned well please.''  If that's part of your aversion to the idea, then try to resist it.  In the words of Marcus Appleby, ``It's really hard to believe something you don't actually believe'' and the same is true of the conception of quantum states as states of belief.  When an experimentalist writes down a particular quantum state, he does it for good reason; it isn't made up on the fly, out of nothing.  All we are saying is that no part of that ``good reason'' has anything to with some property intrinsic to the quantum system.

Looking forward to more discussions another day.  Steven van Enk had invited me to give a colloquium at U. Oregon this year, but I had to drop out because of too many pressures.  Maybe next year you guys will give me a second chance, and we can pick back up then.

\section{01-08-07 \ \ {\it Your Conference} \ \ (to F. Tops{\o}e \& P. Harremo\"es)} \label{Topsoe2} \label{Harremoes2}

\bflt
Does this mean that you (and \ldots?)\ may be interested in our workshop
``facets of entropy'', cf.\ \myurl{http://facetsofentropy.fys.ku.dk}?
\eflt

Well, I didn't know about your workshop, but now that I do, indeed I am very interested.  In fact, I had been meaning to write you and Peter (to whom I'll cc this note) for some time to ask you a question on a rather funky kind of ``entropy'' (though I dare use that word for this) that has been fascinating me lately.  Truth is, I don't actually know what the object is, or whether it has any information theoretic interpretation at all.  And I was going to ask you if you've ever seen anything like it {\it within\/} standard, classical information theory, i.e., outside of quantum considerations.

In particular, I was going to write up the note a little more formally in \LaTeX\ and send a PDF to you once arriving at PI.  But now that I'm writing you anyway, let me pose the question in a little sloppier way.  I'll just refer you to the quantity on the left-hand side of Eq.\ (6) in \arxiv{0707.2071}.  Now, divorce it from its original motivating problem, and just imagine the coefficients $c_{ijk}$ are fixed constants handed down from heaven.  The question is, does that kind of quantity have any interesting information theoretic (or maybe game theoretic) meaning?  Are there any constraints on the $c_{ijk}$ needed in order to make the expression make sense information theoretically?

It strikes me that if I could get a handle on that {\it style\/} of quantity (i.e., somewhat independent of the particulars of the $c_{ijk}$) in a way completely independent of quantum mechanics, then I would have learned something deep.

But a potentially independent question.  If you read the paragraph surrounding Eq.\ (6), you will find that it arises from a consideration to do with a quantum mechanical version of the {\Daroczy} order-3 entropy.  But the {\Daroczy} order-3 entropy??  Who ordered that?  Have you ever run across any actual use of that particular entropy before?

Now, that I've posed the question, I'll probably be somewhat silent for a while \ldots\ but don't let that stop you from writing me even if I don't reply.  It's just that I've got a lot to do to get my house in New Jersey sold and get moved up to Waterloo in the next 9 days.  I'll be back in touch for sure, once I arrive at Perimeter.

\section{02-08-07 \ \ {\it Perimeter Visit} \ \ (to H. Mabuchi)} \label{Mabuchi13.1}

\bhm
I've been invited for a Perimeter colloquium --- when would be a good
time for me to come, in terms of your schedules?
\ehm

God knows with me.  Particularly, until I fulfill my 50,000 mile quota for the year with American Airlines (so that I keep my platinum status), I'm likely to be a wild card \ldots\ selling my soul to travel whenever the opportunity arises.  And I'm running behind this year, only 32,000 so far.  Pathetic, huh?  Anyway, it means at the least I may set up a trip to Beijing before the year is out.  Not quite sure when yet.

Here's a couple of blackout dates: [\ldots]

At the moment, Spring is wide open for me.

Clearly, the strategy ought to be that you just give me your dates, and I try my best to be at PI while you're around.

Congratulations on inheriting Ed Jaynes' office.  Now here's a real task for you:  Find the desk that Ed Jaynes actually sat at.  If you can dig it up, you'll have to let me sit at it, at least for just a minute.  Once upon a time, I had a chance to sit at Niels Bohr's desk while no one was looking, and I didn't take the opportunity.  I've been kicking myself since.

But more congratulations than for the office, congratulations for the new position!  Is it a fancy named professorship?  If so, what's your name?


\section{02-08-07 \ \ {\it The One-Belly Theory of the Universe} \ \ (to S. J. van {\Enk})} \label{vanEnk12}

By the way, I started to compose a note to you several months ago, and then never came back to it.  It was titled ``The One-Belly Theory of the Universe.''  When I get to Waterloo, I'll finish it I promise---I think it's actually an important idea.  Here's the upshot of it.  Alice and Bob are at two ends of a lab, checking for Bell inequality violations.  The question arises, are the detectors at both ends of the lab going ``click'' in this process.  My answer is, it depends upon one's perspective.  From Alice's perspective, her detector is certainly going click, as she collects the outcomes into her awareness.  But also from her perspective, Bob's is NOT going click at all.  As far as Alice is concerned, Bob has simply become entangled with the particles.  The reason I make these remarks has to do with the last technical discussion we had---I think it may be the key to understanding \ldots\  well, everything.  In the next note, I'll at least send you the beginning of the note I had started.  Then you pressure me some in two weeks and I'll finally finish it.  But you're so clever, I'm sure you'll see the idea right away \ldots\ invalidating my needing to finish it after all.  Still, I will.\medskip

``\underline{The One-Belly Theory of the Universe}'', first paragraphs, to be completed

\bsve
I'm still not sure how I would violate Bell inequalities with only positive probabilities and extra hidden variables in my belly.
After all, the hidden variables of the system on Alice's end and in her belly, and the variables of the system on Bob's end plus his belly are still local hidden variables, aren't they?
\esve

That's the wrong way to think about it.  I don't know whether thinking about it in the right way will give rise to a Bell inequality EITHER, but at the very least I can point out that the above is the wrong way to think about it.  Here goes.

What I'm going to say is prompted by this thought about the actual quantum situation.  Go to any exposition of the Bell set up, and one will see the whole affair laid out in terms of {\it two\/} devices going click-click-click, one on Alice's side and one on Bob's side.  Then one asks if the data can be ``explained'' by a local hidden-variable model.  What that is taken to mean is whether the actually-observed frequencies are the ones predicted with near certainty by various i.i.d.\ distributions over the sequence of experimental runs (conditioned by the $A$ and $B$ devices' settings).  Then one focusses on the single-shot distributions (for the various settings) and asks whether they can be derived as the marginals of a certain variety from a single deeper probability distribution.  Thus arises the Bell inequality, and the conditional probabilities given by quantum mechanics violate it.

Then a sweeping claim is almost always made:  That the Bell inequality was derived in a {\it theory-independent\/} fashion (aside from the assumptions of locality and the existence of a grand joint distribution that could be marginalized), and quantum mechanics---a particular theory---violates it.

But I'm now starting to think that the premises of the Bell derivation don't even fall within the framework of a quantum description of the whole thing.  Thus, perhaps, it is no wonder that quantum theory violates it.  Think Wigner's friend.

\section{07-08-07 \ \ {\it Last Email from Bell Labs} \ \ (to A. E. White)} \label{White1}

I'm just signing out, and I wanted to send you a short note to say thanks for ``tolerating me'' at the Lab for so long:  I know that my research didn't exactly fit in with the present incarnation of Bell Labs and the other departmental interests.  But the company nurtured me nevertheless, and I am grateful for my years there.  I wish you all the best of luck in the future.  If you're ever near Waterloo, Ontario, please visit us at the Perimeter Institute for Theoretical Physics.

\section{15-08-07 \ \ {\it Title} \ \ (to L. Hardy)} \label{Hardy21}

\noindent ``Quantum States as Uncertainty, pure and simple.  But, Uncertainty about What?'' \medskip

Probably the longest title you have, but I like it.  Subjects include:  Einstein's pre-EPR argument,  contrast between frequency and Bayesian interpretations of probability, Dutch book argument, no-cloning and no-broadcasting theorem, example of classical no-cloning theorem and yes-broadcasting, standard (non-Spekkens) Kochen--Specker theorem, words on the Paulian ontology in light of KS, ending with fiducial measurements (SICs and MUBs) and the shape of quantum state space.

Does that fit well (or contrast well) with the other talks?  I hope so.

\section{20-08-07 \ \ {\it {\Renyi} Order-3 and the Weird Object} \ \ (to F. Tops{\o}e \& P. Harremo\"es)} \label{Topsoe3} \label{Harremoes3}

I never heard back from either of you concerning the note below.  So, now that I am physically at Perimeter, I'll take the opportunity to send it again, just in case you didn't receive it.  It is pasted below.  [See 01-08-07 note ``\myref{Topsoe2}{Your Conference}'' to F. Tops{\o}e \& P. Harremo\"es.] (Maybe you didn't get it because I sent it out too close to my departure time on the Bell Labs server.)

If you don't understand the question, let me know, and I'll write it up more formally for you.  But also, while I'm here, maybe I can also ask the following.  Do either of you know any properties of the {\Renyi} and {\Daroczy} order-3 entropies that set them apart from the pack of other {\Renyi} and {\Daroczy} entropies?  Do the order-3 entropies have any {\it interesting\/} special properties that the others don't have?

\section{21-08-07 \ \ {\it {\Renyi} Order-3 and the Weird Object, 2} \ \ (to P. Harremo\"es \& F. Tops{\o}e)} \label{Topsoe4} \label{Harremoes4}

Thanks for the replies.  In case it helps to understand my main question, I attach a slightly more detailed description than my previous.  Maybe it helps to set the question outside the context of my previous paper.

Peter wrote:
\bph
The result that $\tr M^2 =1$ and $\tr M^3 = 1$ implies that $M$ is a 1-dimensional projection
can be derived as follows: \ldots\

For any odd $n>3$ one can replace the condition $\tr M^3 = 1$ by $\tr M^n = 1$
and get the same conclusion. Therefore I think that there is nothing
special about the power 3 in the setup.
\eph

Yep, I had understood that derivation and your caveat at the end for quite some time.  Still, I call the Jones/Linden result a ``remarkable theorem'' because it strikes me as being like the no-cloning theorem:  namely, trivial in a mathematical sense, but---I think---deep in physical meaning.  Plus you would be surprised at the caliber of some people who were not aware of it until it was brought to their attention.  For instance, neither Carl Caves, nor Dan Gottesman, nor Rob Calderbank, nor Beth Ruskai, nor Elliot Lieb!!!\ knew of that simple result---I checked with each one of them.  Just one of those simple things overlooked by history (much like the no-cloning theorem until 1981).

Anyway, about the caveat, I'm not sure it makes such an impact on me.
It may be the case that there is nothing special about the power 3 in
the setup where ``quantum states'' are already assumed to be operators $M$ whose extreme points are characterized as above.  But part of the goal for me is to jettison the starting point of operators and instead start at the level of a probability simplex.  Then the question is to add interesting (or information theoretically motivated) constraints so that the extreme points of some convex set within the simplex are isomorphic to 1-dimensional projectors.  Thus my focus on a function of the form $W$.

Of course, that question may have nothing to do with the (classical) {\Renyi} order-3 entropy.  But I had just wondered whether there's anything special about the order-3 one.  Certainly the order-2 entropy has been studied to pieces.  But why has no one ever moved up the chain to the next higher power?  Why is interest lost at power-2?

\bph
I have attached a paper discussing the use of entropies of integer
order in a quantum setting.
\eph

Actually, you didn't attach it.  Could I ask you to send it this time around?

Finally Flemming wrote:
\bflt
As for my part, I will have it in mind, but guess it is the kind of thing you will have a greater chance of getting good comments to at precisely a workshop as the one we plan (VERY strong persons expressed an interest but I hesitate to say who as one never knows, except for the keynote speakers who already accepted to come, if they will really come. Sorry to disappoint you, hoping to see you for the workshop.
\eflt

Now that brings up a problem.  It now looks like I won't actually be able to come \ldots\ so what you are saying is torture.  I wish I could be there.  The trouble is my colleague Appleby will have his Fall break October 20--28, and we had planned on his visiting PI during that time, for a rather intense exchange on SIC-states existence.  Thus, I hope I can ask you and Peter, as friends, to pose the attached question to some of your knowledgeable colleagues and send me any feedback you get!

\bq
We will consider probability vectors $\vec{p}=[p(1),p(2),\ldots,p(d^2)]$
on an event space of $d^2$ events, $d\ge 2$ a positive integer.  Let
$c_{ijk}$ represent a given 3-index ``table'' of $d^6$ real
numbers---the numbers need not all be positive, some may be negative
in fact. We will assume that $c_{ijk}$ is symmetric under
interchanges of pairs of indices, and also invariant under cyclic
permutations.  (Eventually, as needed to make the question more
interesting or well-defined, one may assume some other requirements
on the $c_{ijk}$, but at the moment I'll leave that open.)

Now let us define the following function on our probability simplex:
$$
W(\vec{p})=\sum_{i,j,k} c_{ijk}\, p(i)p(j)p(k)\;.
$$
I chose the letter $W$ for the function, in order to emphasize that
it is a ``weird object.''  The question is, does an object of this
variety---or perhaps under specific, interesting choices for the
$c_{ijk}$---obtain any kind of information theoretic meaning?  It is
a purely classical question as far as I can tell; nothing to do with
quantum mechanics. That is my question.  Can one find any information
theoretic motivation or use for a function of the form $W(\vec{p})$?
Has it ever been seen before?
\eq

\section{21-08-07 \ \ {\it Your Bayesian vs.\ Frequentist Talks} \ \ (to R. {\Schack})} \label{Schack124}

I just wrote the abstract below for the quantum foundations summer school here, coming up next week.  I wonder if you can give me some inspiration on how to present the flaws in frequentism?  Could I ask you to send me some sample presentations (or send me a pointer where I can download them) from your repertoire where you particularly bash the frequentists?
\begin{center}
Quantum States as Uncertainty, pure and simple.  But, Uncertainty about What?
\end{center}
\bq\noindent
     David Deutsch implores us to ``take quantum mechanics seriously.''  In these lectures we will take quantum mechanics \emph{deadly} seriously, but not in a way that would please Prof.\ Deutsch.  Here we lay the groundwork for viewing quantum mechanics as a branch of decision theory, specialized to decision-making agents \emph{immersed} in an objective world of some particular characteristic---for need of a name, the quantum world.  That is to say, the view presented here is that quantum mechanics is less a direct picture of the world, and more a method of survival in it.  Its statements about the world are therefore oblique, but nonetheless firm and, from some points of view, more exciting because of the creative ontology they seem to hint at.

     Topics of the lecture will include:  Einstein's pre-EPR argument for the incompleteness of quantum states, contrast between frequency and Bayesian interpretations of probability, Dutch book arguments for the structure of probability theory, the no-cloning and no-broadcasting theorems of quantum mechanics, classical no-cloning and yes-broadcasting examples, the quantum de Finetti representation theorem, the Kochen--Specker theorem, words on a Wolfgang Pauli'an style ontology in the light of Kochen--Specker, fiducial measurements for defining quantum states (for instance, SICs and MUBs), and the shape of quantum state space.
\eq

\section{22-08-07 \ \ {\it Facts, Values and Quanta} \ \ (to D. M. {\Appleby})} \label{Appleby22}

I'm re-reading your ``Facts, Values and Quanta'' in preparation for my summer school lectures advertised below (which I have to give Tuesday and Wednesday).  It's been a while since I've thought about these things!  I'm learning a lot from you.  I'm going to make the paper or the other version required reading for my tutorial group.

\section{22-08-07 \ \ {\it No Fuchs' School} \ \ (to G. L. Comer)} \label{Comer106}

Yeah, I am {\it finally\/} here.  It's my third day at the office, and the measly money from my house sale is somewhere in electronic transit.  (Did I tell you that we ended up having to accept almost \$100K less than the original offer we got on the house just before the flood?  \$94K less, to be exact.)  The move hurt, but I'm moving on!  It's absolutely great here, from the scientific possibilities, to the support staff, to the view, to the food.  And we already love our neighborhood too.

Really, it is a kind of absolute heaven here.  Check out the virtual tour of the facilities, posted here: \ldots\  Particularly look at our lunch and dinner room, the Black Hole Bistro.  We have wine and cheese there every Friday too.  Also, their choice of office picture is a bit drab, but it conveys the idea.  Look at the size of our chalkboards, and the floor to ceiling windows the senior researchers get.  (Only problem is I've got a big glob of bird shit on my window, and there's no way to get to it to wipe it off!)

Went to my first dark-matter/dark-energy seminar yesterday.  Something about trying to predict galaxy cluster distributions.  Didn't understand a thing.  But I did walk away with the feeling that ``there are just two dark clouds on the horizon,'' as {\Poincare} said just before the discovery of quantum mechanics.\footnote{\editornote It was Kelvin, not \Poincare; the statement comes from a lecture entitled ``Nineteenth Century Clouds over the Dynamical Theory of Heat and Light,'' delivered to the Royal Institution in 1900.  Dispelling one cloud required special relativity, and the other, quantum theory.}

\section{23-08-07 \ \ {\it Robin}\ \ \ (to M. Sasaki)} \label{Sasaki6}

I understand that the time for your meeting in Boston is drawing near.  I must apologize for doing this to you, but my early days here at PI are turning out to be so hectic that I am realizing I am going to have to cut some things from my travel schedule.  I hope you will accept my apology.

On the other hand, I did not want to leave your conference high and dry, without any representation of ``maximally quantum'' sets of states!  So I have arranged for an excellent speaker to be in my place, if you will accept that.  He is Dr.\ Robin Blume-Kohout, one of the very founders of the SIC states, and he would talk on that subject, reporting some results from my recent paper with Appleby and Dang, and some results from his own papers on SICs; also he is hoping to have new numerical results to report before then.  Robin has an impressive resume:  He did his undergraduate honors thesis with Ben Schumacher at Kenyon College, he did his Ph.D. with Wojciech Zurek at Los Alamos, then he was a postdoc at Caltech with John Preskill.  Now he is a postdoc with us at Perimeter Institute.

Please let me know if this is acceptable to you.  He has his own travel funding from PI and will be able to pay his own expenses:  He is just looking forward to the opportunity to meet some of your other speakers and introduce the subject of SIC-sets.

\section{24-08-07 \ \ {\it Are You Nuts?}\ \ \ (to G. L. Comer)} \label{Comer107}

\bgc
\bq\noindent\rm
By the way, you need to read more about Alchemy.  One of my big projects for the upcoming year is to familiarize myself much more with alchemical thought.
\eq
Are you nuts?!?
\egc

Changing lead to gold per se is not the interesting part of old alchemy.  It's this:
\begin{itemize}
\item
W.~Heisenberg, ``Wolfgang Pauli's Philosophical Outlook,'' in his book {\sl Across the Frontiers}, translated by P.~Heath, (Harper \& Row, New York, 1974), pp.~30--38.
\begin{quote}
The elaboration of Plato's thought had led, in neo-Platonism and Christianity, to a position where matter was characterized as void of Ideas.  Hence, since the intelligible was identical with the good, matter was identified as evil.  But in the new science the world-soul was finally replaced by the abstract mathematical law of nature.
Against this one-sidedly spiritualizing tendency the alchemistical philosophy, championed here by Fludd, represents a certain counterpoise.  In the alchemistic view ``there dwells in matter a spirit awaiting release. The alchemist in his laboratory is constantly involved in nature's course, in such wise that the real or supposed chemical reactions in the retort are mystically identified with the psychic processes in himself, and are called by the same names.  The release of the substance by the man who transmutes it, which culminates in the production of the philosopher's stone, is seen by the alchemist, in light of the mystical correspondence of macrocosmos and microcosmos, as identical with the saving transformation of the man by the work, which succeeds only `Deo concedente.'\,''  The governing symbol for this magical view of nature is the quaternary number, the so-called ``tetractys'' of the Pythagoreans, hich is put together out of two polarities.  The division is correlated with the dark side of the world (matter, the Devil), and the magical view of nature also embraces this dark region.
\end{quote}
\end{itemize}

The most relevant part: ``The release of the substance by the man who transmutes it \ldots\ is seen by the alchemist \ldots\ as identical with the saving transformation of the man by the work.''  Transforming the metal transforms the soul---that was the real goal.  In place of soul, put quantum state, and you can start to see where I'm coming from.

Yes, I am nuts.  That's why the only place that would have me is PI (and even they aren't so sure).

Actually see the note below, that I wrote you two years ago!  I've been nuts for a long time.  [See 19-06-05 note ``\myref{Comer72}{Philosopher's Stone}'' to G. L. Comer.]

\section{24-08-07 \ \ {\it Leifer} \ \ (to G. Brassard)} \label{Brassard50}

I hope you're doing well.  I'm finally in Canada and loving it!  It's my fourth day in the office at PI, and it's just fantastic here.  Let's hope I put these years to good use.

But the reason I'm writing right now is that I was just talking to Matt Leifer and he got me sad.  He was telling me about his take on the hazards of doing foundational work before obtaining a faculty position---it's true, and it always brings sadness.  Particularly because, in my eyes, the stuff we've all been involved with is just doing good physics.  Doing good physics, period.  And why should a physicist be ashamed of doing that?

\section{24-08-07 \ \ {\it My Email} \ \ (to R. {\Schack})} \label{Schack125}

No, I didn't get it.  So, thanks for sending something, but try again!  Frankly, I'm a little scared about presenting in any detailed way why one should drop frequentism---it's so ingrained in everybody, it strikes one as an almost lost battle to try to sway the masses.  I gain new respect for your efforts over the years, as I try to figure out how on earth I'm going to do this.

\section{25-08-07 \ \ {\it How Pleased!}\ \ \ (to R. Blume-Kohout)} \label{BlumeKohout2}

How pleased I was to stumble across the poster of the students you had tutored and overhear a student debating the merits of a Bayesian approach to probability over a frequentist approach!  It was the greatest feeling.  Even some of the phrases in the poster were just gems.  ``Measurement:  An action that tells you something about something.''  I'm probably reading more into it than I should, but I was particularly pleased by the choice of the word {\it action\/} in the definition.  For the idea of measurement as action has taken on a much bigger role in my own thinking lately.  In case you're interested, have a look at the letters ``{\,}\myref{vanFraassen7}{`Action' instead of `Measurement'}{\,}'' and ``\myref{vanFraassen10}{Questions, Actions, Answers, and Consequences}'' and ``\myref{vanFraassen12}{Canned Answers}'' (all to Bas van Fraassen) in this file:
\myurl{http://perimeterinstitute.ca/personal/cfuchs/nSamizdat-2.pdf}.

Action's where it's at!

\section{26-08-07 \ \ {\it Your Note to {\Ingemar}} \ \ (to D. M. {\Appleby})} \label{Appleby23}

Yes, I had a hard time following the point you really wanted to make to {\Ingemar}.  Also, I wasn't sure how I fit into the context of the discussion.

My reply to this particular part of {\Ingemar}'s note
\bq\noindent
I expect that if I did settle the MUB question in $N = 6$ (say), then
my feeling would be that I had found the correct way to look at an
existing structure. {\it And\/} at a structure that existed also before Hilbert was born.
\eq
might simply have been this:  That may be your feeling; it has a very strong tendency in everyone (including me).  But it is only a feeling.

There is some (small, but I think instructive) analogy with issues surrounding the EPR criterion of reality:
\bq\noindent
   If, without in any way disturbing a system one can [gather the
   information required to] predict with certainty (i.e., with
   probability equal to unity) the value of a physical quantity, then
   there exists an element of physical reality corresponding to this
   physical quantity.
\eq
I already modified the language of EPR a little bit (adding the stuff in square brackets) in order to make it make any sense at all, but let me go a little further and change the language so that the discussion is more in line with the CFS-Bayesian take on certainty:
\bq\noindent
If, without in any way disturbing a system, one can become {\it certain\/} of the value of a physical quantity (i.e., assigning probability equal to unity for the value), then there exists an element of physical reality corresponding to this physical quantity.
\eq

Well, CFS reject that.  It is a category error---or more accurately an illegitimate move to marry distinct categories (epistemic and ontic) at a kissing point.  Having the feeling of certainty does not make an element of reality so.  Having certainty for the outcome of a quantum mechanical measurement does not make the outcome pre-exist the process of measurement, CFS say.  That is the little point of quantum mechanics, I think.  It is a species of the big point of pragmatism.  And it is that which more directly impinges on {\Ingemar}'s feeling.

Now, I feel more secure in quantum mechanics than pragmatism in general, and to that extent am less secure in my assertion about $N=6$ above.  But some flavor of this discussion, or perhaps some careful tempering of it, will, I think, continue to stand with me ultimately.  I thought briefly about cc'ing this note to {\Ingemar}, but I think I'll just keep it between us for now (for no particular reason that I can identify other than that, for once, I'd like to think a little more before speaking more).

Back to work on my lectures.

\section{05-09-07 \ \ {\it QF Seminars} \ \ (to L. Hardy)} \label{Hardy22}

My joke at the beginning of the note was going to be this:
\bq\noindent
``Of course, I'd be very happy to organize the seminar.  You know I always enjoy an opportunity to bias the random walk when I can.''
\eq
\ldots\ Or something like that.  It takes a long time to hone a joke.

\section{07-09-07 \ \ {\it Diploma Work on Gleason's Theorem} \ \ (to H. Granstr\"om)} \label{Granstrom1}

Ingemar Bengtsson told me that you returned from the PI summer school ``full of enthusiasm''.  If true, I'm glad to hear that we served a good purpose.

One thing that I do regret is that I didn't more diligently pursue an understanding of your ideas on how quantum measurements should be viewed in relational terms.  My head was just too full of all the noise around me to think on deeper subjects.  But, in fact, one of my first-year goals here at PI is to get a better handle on what I think about such accounts of measurement and how they compare and contrast with the ``quantum Bayesian'' account that colleagues and I have been trying to construct.  So, I'd like to hear more about what you're thinking if you've got the time during my upcoming visit to Stockholm for {\AA}sa's defense.  I hope you do.  I arrive in Stockholm (the airport at least) 9:30 AM Friday Sept 14 and depart the morning of Sept 19.  I don't have a clue where I'll be during that range of time, but Ingemar should know how to put you in contact with me.  As far as I know, my weekend is free.

At the moment, I'm not inclined to believe that a relational account can capture the real essence of quantum measurement---it seems to me too static and doesn't seem to capture the ``birthy-ness'' of quantum measurement outcomes, which is the deep point I think the KS theorem signifies (i.e., that they are little acts of creation in their own right)---but it's hard for me to really judge until I know more.  Attached is one of my crazier pieces where I try a little to build a picture of what I mean by all this.  [See ``Delirium Quantum,'' \arxiv{0906.1968v1}.]  At this stage, of course, everything is vague and just defines a direction of research, but maybe the document will give you some sense of what I'm thinking:  That the universe is still under construction and the structure of quantum measurement is our biggest clue for that.

\section{07-09-07 \ \ {\it Consciousness Essay} \ \ (to A. Kent)} \label{Kent10}

Here's a transcript of the {\James} essay I was telling you about, ``Does `Consciousness' Exist?''i.

And here's the quote from it that I was trying to repeat:
\bq
To deny plumply that consciousness exists seems so absurd on the face of it -- for undeniably thoughts do exist -- that I fear some readers will follow me no farther. Let me then immediately explain that I mean only to deny that the word stands for an entity, but to insist most emphatically that it does stand for a function. There is, I mean, no aboriginal stuff or quality of being, contrasted with that of which material objects are made, out of which our thoughts of them are made; but there is a function in experience which thoughts perform, and for the performance of which this quality of being is invoked. That function is knowing. Consciousness is supposed necessary to explain the fact that things not only are, but get reported, are known. Whoever blots out the notion of consciousness from his list of first principles must still provide in some way for that function's being carried on.
\eq

Here's a link to the ``Are We Automata?''\ essay:
\bv
\myurl[http://en.wikisource.org/wiki/Are_We_Automata\%3F]{http://en.wikisource.org/wiki/Are\_We\_Automata\%3F}
\ev

\section{07-09-07 \ \ {\it That {\Wheeler} Quote} \ \ (to M. A. Nielsen)} \label{Nielsen7}

My memory blurred yesterday.  I was thinking of some quotes I had used in a later letter.  Anyway, the quote was actually this.

From J.~A. {\Wheeler} (with K.~W. Ford), {\sl  Geons, Black Holes, and Quantum Foam:~A Life in Physics}, (W.~W. Norton, New York, 1998):
\bq
Many students of chemistry and physics, entering upon their study of quantum mechanics, are told that quantum mechanics shows its essence in waves, or clouds, of probability.  A system such as an atom is described by a wave function.  This function satisfies the equation that Erwin {\Schroedinger} published in 1926.  The electron, in this description, is no longer a nugget of matter located at a point.  It is pictured as a wave spread throughout the volume of the atom (or other region of space).

This picture is all right as far as it goes.  It properly emphasizes the central role of probability in quantum mechanics.  The wave function tells where the electron might be, not where it is.  But, to my mind, the {\Schroedinger} wave fails to capture the true essence of quantum mechanics.  That essence, as the delayed-choice experiment shows, is {\it measurement}.  A suitable experiment can, in fact, locate an electron at a particular place within the atom.  A different experiment can tell how fast the electron is moving.  The wave function is not central to what we actually know about an electron or an atom.  It only tells us the likelihood that a particular experiment will yield a particular result.  It is the experiment that provides the actual information.
\eq

Or the single nugget of John's\index{Wheeler, John Archibald} thinking that I wanted to extract:  Unitarity ``fails to capture the true essence of quantum mechanics.''

\section{08-09-07 \ \ {\it The More Complete Quote} \ \ (to A. Kent)} \label{Kent11}

I just read {\James}'s ``Are We Automata?''\ and am now moving on to ``Does `Consciousness' Exist?''  I noted that the first of the two was written in 1879, while the second was written in 1904.  Twenty-five years between the two.  That, I note, is a little important because of these lines I just read, preceding the quote starting ``To deny plumply \ldots'' which I sent you yesterday:
\bq\noindent
For twenty years past I have mistrusted `consciousness' as an entity;
for seven or eight years past I have suggested its non-existence to
my students, and tried to give them its pragmatic equivalent in
realities of experience. It seems to me that the hour is ripe for it
to be openly and universally discarded.
\eq
Thus presumably in the first article, he was still treating ``consciousness as an entity''---whatever that means.  I suppose this was, at least in part, what you were referring to yesterday.

\section{08-09-07 \ \ {\it Finished It} \ \ (to A. Kent)} \label{Kent12}

I just wanted to record one thought, having finished the second of the two {\James} essays.  It is true that he argues in it that consciousness is a function, rather than an entity, but I suspect the resemblance between his doctrine and the functionalism Wallace speaks of ends at the selection of the word `function' itself---i.e., there being no resemblance beyond that.

For instance, Wikipedia says this:
\bq
   An important part of some accounts of functionalism is the idea of
   multiple realizability. Since, according to standard functionalist
theories, mental states are the corresponding functional role, mental
states can be sufficiently explained without taking into account the
   underlying physical medium (e.g.\ the brain, neurons, etc.) that
   realizes such states; one need only take into account the
   higher-level functions in the cognitive system. Since mental states
   are not limited to a particular medium, they can be realized in
   multiple ways, including, theoretically, within non-biological
   systems, such as computers. In other words, a silicon-based machine
   could, in principle, have the same sort of mental life that a human
   being has, provided that its cognitive system realized the proper
   functional roles. Thus, mental states are individuated much like a
valve; a valve can be made of plastic or metal or whatever material,
so long as it performs the proper function (say, controlling the flow
   of liquid through a tube by blocking and unblocking its pathway).
\eq

For James, as I understand him, the whole point of the article was to try to argue that consciousness does not reside on a material substrate.  For material and mental, for him, are two distinct aspects of something that is neither.  His philosophical playground wasn't the kind of physicalism that Wallace, for instance, seems to presuppose in his thought.

So I apologize for my quick judgment the other day, when I said, ``He's probably talking about the same thing that Adrian called functionalism the other day.''

\section{10-09-07 \ \ {\it  Equations of Motion} \ \ (to D. J. Bilodeau)} \label{Bilodeau8}

Thanks for the note and the update on your life.  Good luck in your retirement!  I like the flavor of your question:  Might quantum state spaces be tangent spaces to something?  Holevo is still in Russia as far as I know, though I think he visits Reinhard Werner's group a lot (wherever Werner is in Germany).

\subsection{Doug's Preply}

\bq
I made a pleasant discovery during the 2 months I was off.  Even though the time was not very relaxing [\ldots] I found myself being brought back to mathematics and theoretical physics in general and quantum theory in particular.  It had been so long since I was last genuinely engrossed by these subjects (in that curiosity-driven way which can't be coerced or willed, but just sprouts organically) that I had finally sold most of my technical books via Amazon or half.com.

I'm starting out with a review of analysis (real, complex and functional) as a fresh way of approaching the mathematical foundations of QM, keeping only the corner of my eye for now on the physical phenomena.  I'm still most curious about that intuition I had about the resemblance between quantum state space and a differential tangent space on a manifold -- linearization, physical meaning through projection, in this case statistical rather than deterministic.  I am trying to imagine, for now in a purely mathematical way, what it is that a quantum state space could possibly be tangent to.  If there is any mathematical clarity to be had on that point, then it might show the way to greater clarity for the physical concepts, too.  At worst, it's a harmless way to keep an eccentric old man occupied and out of trouble.  And it's just plain fun, for the first time in a number of years.

I have a copy of Holevo's book from the library here, the one you reviewed a while back.  Do I remember rightly that he is also currently at the Perimeter Institute?  But rather than putting too much effort into attacking it (the book, not the institute) directly for now, I want to focus on letting my curiosity and self-discovery of concepts lead the way.
\eq

\section{10-09-07 \ \ {\it Statistic on the Konstanz Meeting} \ \ (to all BBQWers)}

\noindent Dear BBQWer,\medskip

May I ask how old were you as of Aug 5, 2005?  At a challenge by Michael Nielsen, I'm working out the age statistic for the meeting we had in Konstanz in 2005.

\subsection{Postscript}

\bq
\noindent The average age was just below 40.  A far cry from the original 57 that Michael had speculated.
\eq

\section{10-09-07 \ \ {\it Several Agents} \ \ (to S. Hartmann)} \label{Hartmann15}

Now I've read your note!  Thanks for updating me on all.  I'm glad you like your new home; I had expected you would.  And your position sounds ideal.  Use it as a chance to change the world!

\bsh
Bayesianism always assumes that there is just one agent who has
beliefs which are then updated. But what if there are several
agents who have beliefs and one wants to combine these believes to a
collective belief? Maybe it would be interesting to see how this
works in a quantum setting.
\esh

I agree that would be a very, very interesting topic.  We have done a very little bit (infinitesimal, but nonzero) in that direction with this paper:  \quantph{0206110}.

\section{11-09-07 \ \ {\it Title and Abstract} \ \ (to A. Kent)} \label{Kent13}

Will this fit the bill?\medskip

\noindent {\bf Title:}  13 Quotes from Everettian Papers and Why They Unsettle Me \medskip

\noindent {\bf Abstract:} 101 years ago William James wrote this about the Hegelian movement in philosophy:  ``The absolute mind which they offer us, the mind that makes our universe by thinking it, might, for aught they show us to the contrary, have made any one of a million other universes just as well as this.  You can deduce no single actual particular from the notion of it.  It is compatible with any state of things whatever being true here below.''  With some minor changes of phrase---for instance ``mathematical structure'' in place of ``absolute mind''---one might well imagine morphing this into a remark about Everettian quantum mechanics.  This point, coupled with the observation that the Everett interpretation has been declared complete and consistent for the selfsame number of years that its supporters have been trying to complete it, indicate to me that perhaps the Everett approach is more a quantum-independent mindset than a scientific necessity.  So be it, but then it should be recognized as such.  In this talk, I will try to expand on these suspicions.

\section{11-09-07 \ \ {\it Arrival, Plans} \ \ (to I. Bengtsson)} \label{Bengtsson2}

This is just to let you know that I should be arriving at the Stockholm airport ARN at 9:30 AM, Friday September 14.  \ldots

First question:  What should I do after arriving at the airport?  Where should I proceed to?  And by what means?

Second question:  Have you drawn me up any kind of schedule?  If so, if possible, I'd like to insure a little time to talk to Prof.\ Lindblad.  Not long necessarily, I just wanted to meet him, as I have long respected him.  Also, I'd like to build in a little time to talk to Helena Granstr\"om.  She said some things at the summer school about ``relational'' ideas of quantum measurement that intrigued me, and I would like to have a chance to get her to expand on that.  Beyond that, you can do with me as you will.  Though I would like some time during the weekend to work and think on the talk I must give at the {\sl 50 Years of Everett\/} conference.  That is going to be a very tough talk for me, as I am sure to be attacked most roundly while attempting to give it.  (Abstract below.)  [See 11-09-07 note ``\myref{Kent13}{Title and Abstract}'' to A. Kent.] Thus, perhaps not too heavy of a schedule over the weekend, but also I'm not averse to interactions during it either.

\section{11-09-07 \ \ {\it The World Is Made of Catalysts} \ \ (to L. Hardy)} \label{Hardy22.1}

Here's that Feynman quote I was telling you about:
\bq
   If, in some cataclysm, all of scientific knowledge were to be
   destroyed, and only one sentence passed on to the next generation
   of creatures, what statement would contain the most information in
   the fewest words?  I believe it is the atomic hypothesis \ldots\
   that all things are made of atoms \ldots
\eq
I came across it a few minutes ago as I was organizing my lecture for Tuesday.

\section{18-09-07 \ \ {\it Shakespeare in Sweden} \ \ (to P. G. L. Mana)} \label{Mana8}

Here's that Shakespeare quote I was trying to remember last night:
\bv
These our actors,\\
As I foretold you, were all spirits, and\\
Are melted into air, into thin air:\\
And like the baseless fabric of this vision,\\
The cloud-capp'd tow'rs, the gorgeous palaces,\\
The solemn temples, the great globe itself,\\
Yea, all which it inherit, shall dissolve,\\
And, like this insubstantial pageant faded,\\
Leave not a rack behind. We are such stuff
\\ As dreams are made on \ldots
\ev
At least that's the part of it that John Wheeler always used when talking of quantum mechanics (though the stanza is a little bit longer).

\section{20-09-07 \ \ {\it Further Reading} \ \ (to H. Granstr\"om)} \label{Granstrom2}

Here's the further reading I was going to suggest to you.  Go to pages 411 to 416 of [my samizdat] and look at the notes I wrote to Bas van Fraassen.

I would appreciate it if you would ultimately write me back with some commentary both on 1) the previous ``delirium quantum'' piece I gave you, and 2) these notes to van Fraassen.  Tell me where I need to clarify.  Tell me where you disagree.  Pinpoint those places where my conception of measurement seems to disagree with yours, and also where they seem to agree instead.  It would be valuable for me to get this information.

The two notes below, incomplete in comparison to our discussion the other day though they are, are the ones I had written to Steven van Enk on the Belly Theory.  [See 10-02-07 note ``\myref{vanEnk11}{Inside and Outside}'' and 02-08-07 note ``\myref{vanEnk12}{The One-Belly Theory of the Universe}'' to S. J. van Enk.]  Let me know if you get any good ideas.  In a couple weeks I should have the time to collaborate on this if you wish (unless you find a quick counterexample!!).

\subsection{Helena's Reply, ``Some Random Comments on Delirium Quantum''}

\bq
Enclosed is a document with some of the reflections I've come up with so far. It does not contain much of any expansions on my view (to the extent that I have such a thing as a `view') of the interpretation of QM, but it does however contain a few loose thoughts provoked by the reading.

As I understand it, the Darwinist idea would apply to all worldly phenomena including immaterial ones as the laws of physics. This would mean that the Bayesian rule that quantum mechanics constitutes, is functional in the current context, but that may very well change.

I must admit I am not quite sure about the implications of this idea. Take gravitation as an example. If it one day turns out that the concept of gravitation is no longer functional as a guide about how to be in and interact with reality, would that necessarily be due to a change that had taken place in the minds of human beings, or could it be because of a gradual, evolutionary-type change in the factual interaction between massive bodies? (I regret using the word ``factual'' here, but I'm hoping that you get the distinction that I'm after, at least well enough to decide if you think that it's a meaningful one.)

Also, it would be interesting to know how the view of theories as ``extensions of our biological brains'' put forward on page 13 related to the statement quoted on page 12, that the laws of physics should be applicable to creatures ``sharing none of our sensory modalities''. If one were to take the idea that all that theories formulated by human beings do (an can possibly do) is to describe that way that a human being will experience and interact with her surroundings, the consequence would be that our theories are applicable only in relation to a very specific class of agents. What qualifies an agent to be a member of this class is not clear, though: just referring to some idealized humanity would probably be tending towards religion.

One aspect about the Darwinian idea that I find worth considering (given that I've understood correctly the sense in which you're using the term), is its inherent contextuality. That is, our eyes cannot be said to be a mechanism for seeing independently of context. Indeed, they have gotten their form in dialogue with caribou, mayfly and nuthatch; our tongues are a product of clear spring water and ripe plums---same thing with all our ways of perceiving and thinking about the world. All that we experience as meaningful, even the most abstract construction, we experience using thought that took form during millions of years on vast savannas, and under wet rocks.

As you also point out, this view provides great reason to question the idea of observer non-detachedness, and the idea that striving to find things out about the world, rather than about the nature of our interplay with the world, holds meaning.

Concerning the issue of abandoning the idea of truth as a symbolic correspondence, I fully agree that this is a reasonable direction to take. I would, however, like to see the casual interpretation of truth more expanded upon. I guess that this relates to my earlier question: exactly in what way do you consider the concept of justification to be, as you put it, temporal? From what I understand, though, the basic idea is one very much in accord with the Bayesian view, i.e. that ``laws of physics'' are rather guidelines for building a functional relationship with the world around us.

As for your interpretation of the index of POVM elements, I just printed out your further references. As was probably more than clear during our conversation, something about your statement about the indexes puzzles me, but I'll have to get back to you on that.

One of the ideas in the paper that I find particularly intriguing (and that I have thought about some myself, although not really getting anywhere with it as far as the theory of science is concerned) is what you refer to as radical pluralism. One of the most important qualities of the scientific method is probably its systematic reduction, its radical neglection of all aspects of reality that are not immediately relevant. Relevancy, it should also be said, is largely measured according to some criterion determined by the current world view and thereby to a large extent value laden.

I would say, however, that this categorizing is central just not to science but to western civilized thought in general. And many people would be prone to say that it has been quite successful.  I guess that the problem is just one of ambition: Just as QM need not necessarily be discarded as a theory, but most likely has to be abandoned or at least severely revised as a theory about the way the world is, so does categorizing and neglecting not necessarily have to be abandoned as a scientific strategy, but we have to be aware what are the flaws and consequences of this approach and the limitations it puts on the scientific quest.

Another, more general, remark on the Bayesian view: I agree that several quantum ``paradoxes'' are lifted when the quantum state is regarded as a state of belief rather than an objective state of reality.  I cannot say, however, that I've fully grasped it, more than superficially. That is, I think I understand most of what I've read on the subject (most of it written by you), but some things still trouble me, even though I cannot at the moment quite pin them down. I think it has something to do with the ``updating of information'' taking place during a measurement procedure. From my point of view, that is not a satisfactory description.

As for the asymmetry that I was complaining about during our talk. I think, again, that it comes down to a questions of aspiration, and maybe I am confusing the issues. There are, as I see it, (at least) two, namely:
\begin{itemize}
\item[1)]	How does one formulate a theory that captures the reciprocity and symmetry of interaction? Is this even possible within the current framework of what is considered a physical theory? (I ask this question for example in relation to the above comment about radical pluralism.)
\item[2)]	How does one interpret quantum mechanics?
\end{itemize}

It is in no way obvious that the two are intimately related. That is, even though I would prefer to see a theory such as described in 1) formulated, it might be the case that QM does have a fundamental asymmetry built into it. This would mean that QM satisfactorily captures enough of some aspects of our (MY!) interaction with reality to be able to give accurate predictions, but is also in a fundamental way flawed, which may be part of what is causing all the confusion leading to emails such as this one.

I guess these notes sum up some, if not all, of my thoughts on your paper. Hopefully you will be able to get something out of them. Reading your text gave me several associations and ideas (not all of them relevant to expand on here) that I will try to take further, and I very much appreciate getting the opportunity to read it. On my way home from work I will pick up the two books by James that you referenced, for a start.
\eq

\section{20-09-07 \ \ {\it Easy Questions} \ \ (to P. W. Shor)} \label{Shor4}

I have a few small questions for you, and I hope you have the time to write me back fairly quickly.  I'm preparing a talk for the ``50 Years of Everett'' meeting at PI.  And I wanted to say a few words on the influence and/or lack of influence of the Everett view on various aspects of quantum information.  I intend to be fair and balanced, and faithfully represent what I am told.  With that in mind, I wonder if you would mind answering the following:
\begin{itemize}
\item[1)]  Did you know of the Everett interpretation before starting work on your factoring algorithm?

\item[2)]  If so, was the many-worlds view or the idea of ``parallel computations via parallel worlds'' something that was integral to your thinking for finding the algorithm?  If it was part of the main imagery that steered your mathematics, can you say in what way?  If it wasn't part of your main imagery, can you say what was?

\item[3)]  Would you say that the developments in quantum information and computation are evidence that something is really right about the Everett view?  Or do you think the developments in QI are relatively neutral toward it?
\end{itemize}

I hope you have the time to answer me.  I've got to give the talk Sunday!

\subsection{Peter's Reply}

\bq
\noindent
{\bf Question 1)  Did you know of the Everett interpretation before starting work on your factoring algorithm?}\medskip

Yes.\medskip

\noindent {\bf Question 2)  If so, was the many-worlds view or the idea of ``parallel
computations via parallel worlds'' something that was integral
to your thinking for finding the algorithm?  If it was part
of the main imagery that steered your mathematics, can you
say in what way?  If it wasn't part of your main imagery, can
you say what was?}\medskip

No, the idea was really more to use periodicity, and inspired by Simon's algorithm. \medskip

\noindent
{\bf Question 3)  Would you say that the developments in quantum
information and computation are evidence that something is
really right about the Everett view?  Or do you think the
developments in QI are relatively neutral toward it?}\medskip

You should have heard my after-dinner talk in Japan.  Both the Everett view and the Copenhagen view are misleading in thinking about quantum computation (although misleading in quite different ways).
\eq

\section{20-09-07 \ \ {\it Easy Questions} \ \ (to D. R. Simon)} \label{SimonD1}

I just sent the note below to Peter Shor.  [See 20-09-07 note ``\myref{Shor4}{Easy Questions}'' to P. W. Shor.]  I know I don't know you as well as Peter, but I wonder if I might ask you the same questions.  Just in every place where I ask about Peter's algorithm, let me instead ask you about yours.  Thanks so much if you can answer these.

By the way, do you have any interest in visiting us at the Perimeter Institute in the coming year?  If so, let me know.

\subsection{Dan's Reply}

\bq
No problem--I'll do my best to answer them.  Thanks very much for the invitation, but my interests have drifted about as far away from quantum physics as they can get by now, and having a toddler in the family makes travel less attractive than it used to be, so a visit to your neck of the woods would be pretty hard for me to justify.  Hope you're having fun, though \ldots\medskip

\noindent
{\bf Question 1)  Did you know of the Everett interpretation before starting work on your quantum algorithm?}\medskip

Who's Everett, and what's his interpretation?\medskip

\noindent {\bf Question 2)  If so, was the many-worlds view or the idea of ``parallel
computations via parallel worlds'' something that was integral
to your thinking for finding the algorithm?  If it was part
of the main imagery that steered your mathematics, can you
say in what way?  If it wasn't part of your main imagery, can
you say what was?}\medskip

I was approaching the problem purely from a computer scientist's perspective.  I learned the absolute bare minimum of physics I needed to be able to understand the computer science question, which (as I saw it) was, ``these crazy people are claiming that if you add these very-weird-yet-theoretically-physically-implementable functions to a computer, then you should be able to do amazing things with them.  Prove them right or wrong.''  I actually started out trying to prove that quantum computing was useless, and eventually narrowed down the difficult, unsimulateable part to, ``rotate, compute, rotate''.  That helped guide my search for a computationally interesting quantum algorithm.

From my perspective, though, the ``interpretation'' of the quantum mechanical operations I was given was irrelevant---I was told that a quantum computer could do these things, and it was my job to figure out how computationally useful ``these things'' were.
\medskip

\noindent
{\bf Question 3)  Would you say that the developments in quantum
information and computation are evidence that something is
really right about the Everett view?  Or do you think the
developments in QI are relatively neutral toward it?}\medskip

As a non-physicist, I'm perfectly comfortable living with only the haziest mental model of quantum reality.  Distinctions between different mental models that are experimentally indistinguishable are altogether lost on me.  For that matter, I wonder why physicists, whose mental models have (or at least should have) all the precision of the mathematical formulae underlying them, would care about fuzzily unquantifiable comparisons among fuzzily unquantifiable mental models.

(I'm assuming here that the ``interpretations'' in question here aren't experimentally distinguishable--else they'd be ``theories'', right?)
\eq

\section{20-09-07 \ \ {\it Easy Questions} \ \ (to L. K. Grover)} \label{Grover5}

I just sent the note below to Peter Shor, trying to compile something interesting to say at the Everettfest.  [See 20-09-07 note ``\myref{Shor4}{Easy Questions}'' to P. W. Shor.] I wonder if I might ask you some similar ones---I hope you've got a little time to answer these, as whatever your answers may be, I think they'll bring an aspect to the meeting that no one else will be representing.

Simply, in place of where I ask Peter about the factoring algorithm, would you answer similar questions pertaining to your search algorithm and any of the other algorithms you developed?  If you can do this, it'd be much appreciated!

\subsection{Lov's Reply}

\bq
Sorry, I had missed your message --- only just got it --- hope your talk went well.
\eq

\section{20-09-07 \ \ {\it Easy Questions} \ \ (to D. Gottesman)} \label{Gottesman5}

I just sent the note below to Peter Shor, trying to compile something interesting to say at the Everettfest.  [See 20-09-07 note ``\myref{Shor4}{Easy Questions}'' to P. W. Shor.] I wonder if I might ask you some similar questions---I hope you've got a little time to answer these, as whatever your answers may be I think they'll bring an aspect to the meeting that no one else will be representing.

Simply, in place of where I ask Peter about the factoring algorithm, would you answer similar questions pertaining to your work on the stabilizer formalism (or anything else you want to mention)?  If you can do this, it'd be much appreciated!

\subsection{Daniel's Reply}

\bq
\noindent
{\bf Question 1)  Did you know of the Everett interpretation before starting work on stabilizer codes?}\medskip

Yes, certainly.\medskip

\noindent {\bf Question 2)  If so, was the many-worlds view or the idea of ``parallel
computations via parallel worlds'' something that was integral
to your thinking for finding the algorithm?  If it was part
of the main imagery that steered your mathematics, can you
say in what way?  If it wasn't part of your main imagery, can
you say what was?}\medskip

There wasn't imagery, per se, although there was intuition.  Mostly
it came about by accident.  I was trying to study degenerate codes,
and it seemed to make sense to think about codes whose degeneracies
were characterized by Pauli identities -- e.g., $Z_1$ and $Z_2$ act
the same, so $Z_1 Z_2$ acts as the identity.  I tried writing down a
code with lots of degeneracies, and was surprised to see that it
corrected a variety of different errors without any extra effort
on my part.  I had some intuition that anticommutation was the
right thing to look at -- I'm not completely sure why.  It's
possible the intuition had some genesis with ``incompatible observables''
\`a la Copenhagen, but certainly I was not thinking about that explicitly.

I can't say that any Everettian idea was at all relevant.\medskip

\noindent
{\bf Question 3)  Would you say that the developments in quantum
information and computation are evidence that something is
really right about the Everett view?  Or do you think the
developments in QI are relatively neutral toward it?}\medskip

Really wrong isn't an option?

In any case, as I believe I have told you before, I subscribe to
the pantheistic interpretation, that all interpretations are valid
(excluding those which have experimental differences from standard
QM).  I think work in quantum computation supports that -- sometimes
it is helpful to think about one interpretation to understand a
QI development, at other times a different interpretation is most
helpful.  In that sense, I would say QI has been mostly neutral
towards Everett.

There's one Everettian idea, however, that I think has been very
strongly supported by quantum information, which is that basically
everything is unitary.  I tried to think of supporting arguments
for this view, though, and had trouble.  The ``Church of the larger
Hilbert space'' supports it in the sense that says we can always
think of evolution as unitary, but also opposes it in the sense that
we can do this whether the system has a real-world purification or
not.  Quantum error correction kind of supports it in the sense that
it says we should be able to build arbitrarily large subspaces which,
with appropriate control, behave arbitrarily close to unitary, but
again that's not really the same as saying the underlying dynamics
without control are unitary.  Perhaps the strongest argument is the
study of decoherence mechanisms in real systems, where interaction
with the environment always seems adequate to account for the
decoherence.

On the other hand, in a quantum computer, while we frequently think
of a standard computational basis for convenience, there really is
no preferred basis most of the time, and this creates a large regime
that seems resistant to a ``many-worlds'' interpretation.

I think one big impact of QI work for foundations is that it changes
the list of what are considered signature quantum behaviors.  People
outside of QI frequently seem to think that Planck's constant and the
{\Schroedinger} equation are the be-all and end-all of quantumness, and
that view seems totally incompatible with QI work, since we get lots
of quantum behavior without ever mentioning either.  This is very harsh
on some interpretations (like Bohmian, which depends on a wave equation),
but Everett doesn't seem to depend on those much.
\eq

\section{20-09-07 \ \ {\it Easy Questions} \ \ (to J. Preskill)} \label{Preskill16.1}

I just sent the note below to Peter Shor, trying to compile something interesting to say at the Everettfest at PI this weekend.  [See 20-09-07 note ``\myref{Shor4}{Easy Questions}'' to P. W. Shor.]  I wonder if I might ask you some similar questions---I hope you've got a little time to answer these, as whatever your answers may be, I think they'll bring an aspect to the meeting that no one else will be representing.

Simply, in place of where I ask Peter about the factoring algorithm, would you answer similar questions pertaining to your work on security proofs in quantum crypto (or anything else you think worth mentioning)?  If you can do this, it'd be much appreciated!

\subsection{John's Reply}

\bq
In my zeal to reply, I seem to have inadvertently deleted your message, but not before reading the questions.\medskip

\noindent 1) Did you know about many worlds when \ldots? \medskip

I've known about it since 1973. I was a junior at Princeton and read a book I saw in the university bookstore, edited by DeWitt and Graham, called {\sl The many-worlds interpretation of quantum mechanics}. I think the book was quite new at the time. It included a reprint of Everett's paper. (I was working on a ``junior paper'' about Bell inequalities, which seemed timely because the Freedman-Clauser experiment had been published just months earlier.)\medskip

\noindent 2) Do you think the many-worlds interpretation influenced your work on \ldots?\medskip

No.\medskip

\noindent 3) Does recent work on quantum information strengthen the case for the many-worlds interpretation? Weaken? Is it neutral?\medskip

Neutral. \medskip

\noindent [Yes, I don't usually answer you so promptly, but I'm waiting to board an airplane and I'm kind of bored.]
\eq

\section{20-09-07 \ \ {\it Easy Questions} \ \ (to A. Kent)} \label{Kent14}

I just sent the note below to Peter Shor, trying to compile something interesting to say at the Everettfest at PI this weekend.  [See 20-09-07 note ``\myref{Shor4}{Easy Questions}'' to P. W. Shor.]  I wonder if I might ask you some similar questions---I hope you've got a little time to answer these, as whatever your answers may be, I think they'll bring an aspect to the meeting that no one else will be representing.

Simply, in place of where I ask Peter about the factoring algorithm, would you answer similar questions pertaining to your work on security proofs in quantum crypto (or anything else you think worth mentioning)?  If you can do this, it'd be much appreciated!

\subsection{Adrian's Reply}

\bq
\noindent 1) Did you know of the Everett interpretation before starting work on quantum information theory? \medskip

Sure did. \medskip

\noindent 2) If so, was the many-worlds view or the idea of ``parallel computations via parallel worlds'' something that was integral to your thinking for finding the algorithm?  If it wasn't part of your main imagery, can you say what was?\medskip

Nope. As for mental imagery, I'm not sure there's any one big thing --- lots of confused images of vectors and spheres and squashings and twirlings and things.  \medskip

\noindent 3) Would you say that the developments in quantum information and computation are evidence that something is really right about the Everett view?  Or do you think the developments in QI are relatively neutral toward it? \medskip

Entirely neutral.
\eq

\section{20-09-07 \ \ {\it Easy Questions} \ \ (to A. K. Ekert)} \label{Ekert1}

I just sent the note below to Peter Shor, trying to compile something interesting to say at the Everettfest.  [See 20-09-07 note ``\myref{Shor4}{Easy Questions}'' to P. W. Shor.]  I wonder if I might ask you some similar questions---I hope you've got a little time to answer these, as whatever your answers may be, I think they'll bring an aspect to the meeting that no one else will be representing.

Simply, in place of where I ask Peter about the factoring algorithm, would you answer similar questions pertaining to your entanglement-based quantum key distribution protocol?  If you can do this, it'd be much appreciated!

\subsection{Artur's Reply}

\bq
\noindent {\bf Question 1)  Did you know of the Everett interpretation before starting to work on your key distribution protocol?}\medskip

Yes and I was intrigued by it.\medskip

\noindent {\bf Question 2)  If so, was the many-worlds view or the idea of ``parallel
computations via parallel worlds'' something that was integral
to your thinking for finding the algorithm?  If it was part
of the main imagery that steered your mathematics, can you
say in what way?  If it wasn't part of your main imagery, can
you say what was?}\medskip

Sometime in the mid 1989 I read the EPR paper and my attention was drawn to the sentence  ``\ldots If, without in any way disturbing a system, we can predict with certainty \ldots\ the value of a physical quantity, then there exists an element of physical reality corresponding to this physical quantity.''  This was a definition of perfect eavesdropping.  I guess I was lucky to read it in this particular way.  This was my starting point.  Thinking about locality, reality and security was in a way going in the opposite direction than Everett but it was very productive nonetheless. \medskip

\noindent
{\bf Question 3)  Would you say that the developments in quantum
information and computation are evidence that something is
really right about the Everett view?  Or do you think the
developments in QI are relatively neutral toward it?}\medskip

Yes, QI enhances my predilection for the Everett view.
\eq

\section{20-09-07 \ \ {\it Easy Questions} \ \ (to A. C. Yao)} \label{Yao1}

I just sent the note below to Peter Shor, trying to compile something interesting to say at the Everettfest at the Perimeter Institute this weekend.  [See 20-09-07 note ``\myref{Shor4}{Easy Questions}'' to P. W. Shor.]  I wonder if I might ask you some similar questions---I hope you've got a little time to answer these, as whatever your answers may be, I think they'll bring an aspect to the meeting that no one else will be representing.

Simply, in place of where I ask Peter about the factoring algorithm, would you answer similar questions pertaining to any of your early work on quantum Turing machines and communication complexity?  If you can do this, it'd be much appreciated!

\subsection{Andrew's Reply, ``Easy Questions on Early Quantum Computing''}

Quick answers to your questions:\medskip
\begin{enumerate}
\item
Yes, I was aware of the Everett interpretation since college days.
\item
No, the Everett interpretation was not on my mind when I did my work on quantum Turing machines and quantum communication complexity.
\item
I am inclined to think that QI developments are relatively independent of the interpretations of quantum mechanics.
\end{enumerate}
Hope this found its way to you in time!

\section{20-09-07 \ \ {\it Easy Questions} \ \ (to H. J. Briegel)} \label{Briegel5}

I just sent the note below to Peter Shor, trying to compile something interesting to say at the Everettfest.  [See 20-09-07 note ``\myref{Shor4}{Easy Questions}'' to P. W. Shor.]  I wonder if I might ask you some similar questions---I hope you've got a little time to answer these, as whatever your answers may be, I think they'll bring an aspect to the meeting that no one else will be representing.

Simply, in place of where I ask Peter about the factoring algorithm, would you answer similar questions pertaining to your work on cluster-state computation?  If you can do this, it'd be much appreciated!

I will send Robert a separate note, as I would like your answers to be independent of each other.

\subsection{Hans's Reply}

\bq
I guess it is far too late to answer this e-mail, which I have just stumbled over, searching in my mailbox for something else \ldots\

I was aware of the many world interpretation long before I started working on quantum information and in particular on measurement-based quantum computation.

I never liked the Everett interpretation and it certainly did not play any role in our studies on cluster states and what they could be used for.

Later however, once the one way model had been established, I thought indeed that the one-way quantum computer might be a good playground to challenge different ontological preferences, i.e.\ interpretations of q.m., and in particular the many-world interpretation. I think we even discussed this at some point.

As you know very well, Andrew Steane then wrote a note entitled ``A quantum computer needs only one universe'' which pretty much expressed my views on the subject, too. So I did not find any strong need to elaborate on this.

Did anything new come out of the 50th anniversary party?

I hope you had an interesting meeting, but I also hope that the number of many-world believers has not increased by too many.
\eq

\section{20-09-07 \ \ {\it Easy Questions} \ \ (to P. A. Benioff)} \label{Benioff1}

I just sent the note below to Peter Shor, trying to compile something interesting to say at the Everettfest at the Perimeter Institute this weekend.  [See 20-09-07 note ``\myref{Shor4}{Easy Questions}'' to P. W. Shor.]  I wonder if I might ask you some similar questions---I hope you've got a little time to answer these, as whatever your answers may be, I think they'll bring an aspect to the meeting that no one else will be representing.

Simply, in place of where I ask Peter about the factoring algorithm, would you answer similar questions pertaining to your own work on unitary quantum computation?  If you can do this, it'd be much appreciated!

\subsection{Paul's Reply}

\bq
\noindent Here are my answers to your questions:
\begin{enumerate}
\item
I knew of Everett's work before my work on quantum computation.
\item
His work was not integral to mine and did not play a role.
\item
I do not think developments in QI show anything about the validity or nonvalidity of Everett's work.
\end{enumerate}
Hope your talk goes well.  Also I hope to see you sometime soon.
\eq

\section{20-09-07 \ \ {\it Easy Questions} \ \ (to D. P. DiVincenzo)} \label{DiVincenzo3}

I just sent the note below to Peter Shor, trying to compile something interesting to say at the Everettfest at PI this weekend. I wonder if I might ask you some similar questions---I hope you've got a little time to answer these, as whatever your answers may be, I think they'll bring an aspect to the meeting that no one else will be representing.

Simply, in place of where I ask Peter about the factoring algorithm, would you answer similar questions pertaining to any of your early work in quantum information (say, the gates papers or unextendable product bases)?  If you can do this, it'd be much appreciated!

\subsection{David's Reply}

\bq
\noindent {\bf Question 1)  Did you know of the Everett interpretation before starting to work on your algorithms?}\medskip

Yes.\medskip

\noindent {\bf Question 2)  If so, was the many-worlds view or the idea of ``parallel
computations via parallel worlds'' something that was integral
to your thinking for finding the algorithm?  If it was part
of the main imagery that steered your mathematics, can you
say in what way?  If it wasn't part of your main imagery, can
you say what was?}\medskip

No.

I think that I am relatively ``interpretationless'', the rules of quantum mechanics (viz., Charlie's stone tablets) speak for themselves, to me.
Or perhaps I prefer the {\Schroedinger} 1935-6 interpretation, summed up by, ``poor pussy''.\medskip

\noindent
{\bf Question 3)  Would you say that the developments in quantum
information and computation are evidence that something is
really right about the Everett view?  Or do you think the
developments in QI are relatively neutral toward it?}\medskip

In my view, neutral.
\eq

\section{20-09-07 \ \ {\it Easy Questions} \ \ (to R. Cleve)} \label{Cleve3}

I just sent the note below to Peter Shor, trying to compile something interesting to say at the Everettfest at PI this weekend.  [See 20-09-07 note ``\myref{Shor4}{Easy Questions}'' to P. W. Shor.]  I wonder if I might ask you some similar questions---I hope you've got a little time to answer these, as whatever your answers may be, I think they'll bring an aspect to the meeting that no one else will be representing.

Simply, in place of where I ask Peter about the factoring algorithm, would you answer similar questions pertaining to your work on communication complexity (or anything else you think worth mentioning)?  If you can do this, it'd be much appreciated!

\subsection{Richard's Reply}

\bq
This might be too late to be useful to you. My answers are below:

\noindent {\bf Question 1)  Did you know of the Everett interpretation before starting to work on your algorithms? [communication complexity]}\medskip

I had a vague idea of ``many worlds'' interpretations as corresponding to many universes in superposition.

My viewpoint of the quantum information framework is more along the lines of ``probability theory with minus signs''. (Or ``probability theory on steriods'' \smiley.)

Probability theory already gives a notion like ``superposition'', but without interference occurring.

I know that PT is different in that one can choose to think of a probabilistic state as actually being in a definite state, about which we are ignorant. Quantum states cannot be viewed that way.\medskip

\noindent {\bf Question 2)  If so, was the many-worlds view or the idea of ``parallel
computations via parallel worlds'' something that was integral
to your thinking for finding the algorithm?  If it was part
of the main imagery that steered your mathematics, can you
say in what way?  If it wasn't part of your main imagery, can
you say what was?}\medskip

It wasn't part of my imagery. My imagery was based on Bell inequality violation stuff. \medskip

\noindent
{\bf Question 3)  Would you say that the developments in quantum
information and computation are evidence that something is
really right about the Everett view?  Or do you think the
developments in QI are relatively neutral toward it?}\medskip

Neutral. Whatever works intuitively to help inspire is fine with me.
\eq

\section{20-09-07 \ \ {\it Easy Questions} \ \ (to D. Deutsch)} \label{DeutschD1}

I know you've written about this adequately, but Adrian Kent suggested I ask you too nevertheless.  So, I think I will.

I just sent the note below to Peter Shor, trying to compile something interesting to say at the Everettfest.  [See 20-09-07 note ``\myref{Shor4}{Easy Questions}'' to P. W. Shor.]  I wonder if I might ask you some similar questions---I hope you've got a little time to answer these, as whatever your answers may be, I think they'll bring an aspect to the meeting that no one else will be representing.

Simply, in place of where I ask Peter about the factoring algorithm, I wonder if you would answer similar questions pertaining to your own participation (and any other algorithms you were involved with)?  If you can do this, it'd be much appreciated!

\subsection{David's Reply}

\bq
My replies to the questions, mutatis mutandis, would be:\medskip

\noindent {\bf Question 1)  Did you know of the Everett interpretation before starting to work on your algorithms?}\medskip

Yes.\medskip

\noindent {\bf Question 2)  If so, was the many-worlds view or the idea of ``parallel
computations via parallel worlds'' something that was integral
to your thinking for finding the algorithm?  If it was part
of the main imagery that steered your mathematics, can you
say in what way?  If it wasn't part of your main imagery, can
you say what was?}\medskip

Yes. \medskip

\noindent
{\bf Question 3)  Would you say that the developments in quantum
information and computation are evidence that something is
really right about the Everett view?  Or do you think the
developments in QI are relatively neutral toward it?}\medskip

In a sense. I think the case for Everett was already watertight before quantum computation. Quantum computation provided new, more dramatic, forms of the same arguments, and also provided some better tools for understanding the multiverse.
\eq

\section{20-09-07 \ \ {\it Easy Questions} \ \ (to R. Jozsa)} \label{Jozsa7}

I just sent the note below to Peter, trying to compile something interesting to say at the Everettfest.  [See 20-09-07 note ``\myref{Shor4}{Easy Questions}'' to P. W. Shor.]  I wonder if I might ask you some similar ones---I hope you've got a little time to answer these, as whatever your answers may be, I think they'll bring an aspect to the meeting that no one else will be representing.

Simply, in place of where I ask Peter about the factoring algorithm, I wonder if you would answer similar questions pertaining to your own participation (and any other algorithms you were involved with)?  If you can do this, it'd be much appreciated!

\subsection{Richard's Reply}

\bq
Nice to hear from you after such a long time! And congrats on your move back into rarefied academia! --- I'm sure it much better suits the nature of your main research inclinations.
I'm not sure my answers to your questions will be of much interest but here they are:\medskip

\noindent {\bf Question 1)  Did you know of the Everett interpretation before starting to work on your algorithms?}\medskip

I've known of the Everett interpretation since the mid 1970's and never really adopted/liked it, even from outset. It always was (and still is) a very vague and incomplete framework to me.\medskip

\noindent {\bf Question 2)  If so, was the many-worlds view or the idea of ``parallel
computations via parallel worlds'' something that was integral
to your thinking for finding the algorithm?  If it was part
of the main imagery that steered your mathematics, can you
say in what way?  If it wasn't part of your main imagery, can
you say what was?}\medskip

I'm not aware that the Everett ideas have ever played any significant
role in my thinking on quantum things.
I don't have a clear impression of any particular imagery that I could
name, underlying or guiding my quantum thoughts.
\medskip

\noindent
{\bf Question 3)  Would you say that the developments in quantum
information and computation are evidence that something is
really right about the Everett view?  Or do you think the
developments in QI are relatively neutral toward it?}\medskip

I do not see that any quantum comp/info developments particularly
support the Everett view in any way compared to any other prospective
interpretations.\medskip

Hope you have a good talk on Sunday!
\eq

\section{20-09-07 \ \ {\it Free Will} \ \ (to {\AA}. {\Ericsson})} \label{Ericsson1}

For Renouvier's argument that I was telling you about, see \ldots\ [15-08-08 note titled ``Free Will and Renouvier'' to R. {\Schack}].

Also, here's another little passage I happen to have in my files.  (Read it slowly and three times over; I think it's worth it.)  LaTeX it up if you want slightly easier reading.

\bq
\noindent From W.~James, ``Some Metaphysical Problems Pragmatically Considered,'' in his {\sl Pragmatism, a New Name for Some Old Ways of Thinking:
Popular Lectures on Philosophy}, (Longmans, Green and Co., New York, 1922).\medskip

Let me take up another well-worn controversy, {\it the free-will problem}. Most persons who believe in what is called their free-will do so after the rationalistic fashion. It is a principle, a positive faculty or virtue added to man, by which his dignity is enigmatically augmented. He ought to believe it for this reason. Determinists, who deny it, who say that individual men originate nothing, but merely transmit to the future the whole push of the past cosmos of which they are so small an expression, diminish man. He is less admirable, stripped of this creative principle. I imagine that more than half of you share our instinctive belief in free-will, and that admiration of it as a principle of dignity has much to do with your fidelity.

But free-will has also been discussed pragmatically, and, strangely enough, the same pragmatic interpretation has been put upon it by both disputants. You know how large a part questions of {\it accountability\/} have played in ethical controversy. To hear some persons, one would suppose that all that ethics aims at is a code of merits and demerits. Thus does the old legal and theological leaven, the interest in crime and sin and punishment abide with us. `Who's to blame? whom can we punish? whom will God punish?'---these preoccupations hang like a bad dream over man's religious history.

So both free-will and determinism have been inveighed against and called absurd, because each, in the eyes of its enemies, has seemed to prevent the `imputability' of good or bad deeds to their authors.
Queer antinomy this! Free-will means novelty, the grafting on to the past of something not involved therein. If our acts were predetermined, if we merely transmitted the push of the whole past, the free-willists say, how could we be praised or blamed for anything? We should be `agents' only, not `principals,' and where then would be our precious imputability and responsibility?

But where would it be if we {\it had\/} free-will? rejoin the determinists. If a `free' act be a sheer novelty, that comes not {\it from\/} me, the previous me, but {\it ex nihilo}, and simply tacks itself on to me, how can {\it I}, the previous I, be responsible? How can I have any permanent {\it character\/} that will stand still long enough for praise or blame to be awarded? The chaplet of my days tumbles into a cast of disconnected beads as soon as the thread of inner necessity is drawn out by the preposterous indeterminist doctrine. Messrs.\ Fullerton and McTaggart have recently laid about them doughtily with this argument.

It may be good {\it ad hominem}, but otherwise it is pitiful. For I ask you, quite apart from other reasons, whether any man, woman or child, with a sense for realities, ought not to be ashamed to plead such principles as either dignity or imputability. Instinct and utility between them can safely be trusted to carry on the social business of punishment and praise. If a man does good acts we shall praise him, if he does bad acts we shall punish him,---anyhow, and quite apart from theories as to whether the acts result from what was previous in him or are novelties in a strict sense. To make our human ethics revolve about the question of `merit' is a piteous unreality---God alone can know our merits, if we have any. The real ground for supposing free-will is indeed pragmatic, but it has nothing to do with this contemptible right to punish which has made such a noise in past discussions of the subject.

Free-will pragmatically means {\it novelties in the world}, the right to expect that in its deepest elements as well as in its surface phenomena, the future may not identically repeat and imitate the past. That imitation {\it en masse\/} is there, who can deny? The general `uniformity of nature' is presupposed by every lesser law.
But nature may be only approximately uniform; and persons in whom knowledge of the world's past has bred pessimism (or doubts as to the world's good character, which become certainties if that character be supposed eternally fixed) may naturally welcome free-will as a {\it melioristic\/} doctrine. It holds up improvement as at least possible; whereas determinism assures us that our whole notion of possibility is born of human ignorance, and that necessity and impossibility between them rule the destinies of the world.

Free-will is thus a general cosmological theory of {\it promise}, just like the Absolute, God, Spirit or Design. Taken abstractly, no one of these terms has any inner content, none of them gives us any picture, and no one of them would retain the least pragmatic value in a world whose character was obviously perfect from the start. Elation at mere existence, pure cosmic emotion and delight, would, it seems to me, quench all interest in those speculations, if the world were nothing but a lubberland of happiness already. Our interest in religious metaphysics arises in the fact that our empirical future feels to us unsafe, and needs some higher guarantee. If the past and present were purely good, who could wish that the future might possibly not resemble them? Who could desire free-will? Who would not say, with Huxley, `let me be wound up every day like a watch, to go right fatally, and I ask no better freedom.' `Freedom' in a world already perfect could only mean freedom to {\it be worse}, and who could be so insane as to wish that? To be necessarily what it is, to be impossibly aught else, would put the last touch of perfection upon optimism's universe. Surely the only {\it possibility\/} that one can rationally claim is the possibility that things may be {\it better}.
That possibility, I need hardly say, is one that, as the actual world goes, we have ample grounds for desiderating.

Free-will thus has no meaning unless it be a doctrine of  {\it relief}. As such, it takes its place with other religious doctrines.
Between them, they build up the old wastes and repair the former desolations. Our spirit, shut within this courtyard of sense-experience, is always saying to the intellect upon the tower:
`Watchman, tell us of the night, if it aught of promise bear,' and the intellect gives it then these terms of promise.

Other than this practical significance, the words God, free-will, design, etc., have none. Yet dark tho they be in themselves, or intellectualistically taken, when we bear them into life's thicket with us the darkness {\it there\/} grows light about us. If you stop, in dealing with such words, with their definition, thinking that to be an intellectual finality, where are you? Stupidly staring at a pretentious sham! ``Deus est Ens, a se, extra et supra omne genus, necessarium, unum, infinite perfectum, simplex, immutabile, immensum, aeternum, intelligens,'' etc.,---wherein is such a definition really instructive? It means less than nothing, in its pompous robe of adjectives. Pragmatism alone can read a positive meaning into it, and for that she turns her back upon the intellectualist point of view altogether. `God's in his heaven; all's right with the world!'---{\it That's\/} the real heart of your theology, and for that you need no rationalist definitions.

Why shouldn't all of us, rationalists as well as pragmatists, confess this? Pragmatism, so far from keeping her eyes bent on the immediate practical, foreground, as she is accused of doing, dwells just as much upon the world's remotest perspectives.

See then how all these ultimate questions turn, as it were, upon their hinges; and from looking backwards upon principles, upon an {\it erkenntnisstheoretische Ich}, a God, {\it a Kausalit\"atsprinzip}, a Design, a Free-will, taken in themselves, as something august and exalted above facts,---see, I say, how pragmatism shifts the emphasis and looks forward into facts themselves. The really vital question for us all is, What is this world going to be? What is life eventually to make of itself? The centre of gravity of philosophy must therefore alter its place. The earth of things, long thrown into shadow by the glories of the upper ether, must resume its rights. To shift the emphasis in this way means that philosophic questions will fall to be treated by minds of a less abstractionist type than heretofore, minds more scientific and individualistic in their tone yet not irreligious either. It will be an alteration in `the seat of authority' that reminds one almost of the protestant reformation. And as, to papal minds, protestantism has often seemed a mere mess of anarchy and confusion, such, no doubt, will pragmatism often seem to ultra-rationalist minds in philosophy.  It will seem so much sheer trash, philosophically. But life wags on, all the same, and compasses its ends, in protestant countries. I venture to think that philosophic protestantism will compass a not dissimilar prosperity.
\eq

\section{21-09-07 \ \ {\it Easy Questions, RPWS1} \ \ (to P. W. Shor)} \label{Shor5}

Thanks very much.  What you've already said was very useful.

But could I ask you to expand on this (just for me personally):
\bpws
You should have heard my after-dinner talk in Japan.  Both the Everett
view and the Copenhagen view are misleading in thinking about quantum
computation (although misleading in quite different ways).
\epws
I hate I missed that!  Could you expand briefly on what you said?  Or was the talk recorded and available somewhere?  I am really, really curious.

\subsection{Peter's Reply}

\bq
I believe I said that the Bohm interpretation was like the man that the
man in the balloon asked directions from, that Deutsch's intelligent
quantum computer would have to answer ``I forget,'' and that the Copenhagen
interpretation was responsible for my success (as this is the reason
that nobody discovered the factoring algorithm before me).  I don't
believe that I mentioned the spiders in the basement of the castle in
the forest.

Maybe you should ask Charlie what I said.  I'll expand if I can find my
notes.
\eq

\section{22-09-07 \ \ {\it Easy Questions, RPWS2} \ \ (to P. W. Shor)} \label{Shor6}

What I was hoping particularly was that you'd expand on this statement:
\bpws
Both the Everett view and the Copenhagen view are misleading in
thinking about quantum computation (although misleading in quite different ways).
\epws
Since this is an Everett meeting, do you mind relating at least that one.  How do you see it as misleading when it comes to quantum computation?

\subsection{Peter's Reply}

\bq
Computer scientists who hear about the Everett interpretation construct a mental model that a quantum computer is many worlds in parallel that can all interact, so that you should be able to do polynomial-depth exponential number of processors classical computation.  This is the exponential analog of the complexity class NC, and is much, much more powerful computationally than the real class BQP.

The Copenhagen interpretation, on the other hand, leads people to think about collapse of the wave function as some kind of actual real process, which is also very misleading.
\eq

\section{27-09-07 \ \ {\it Easy Questions, RDG} \ \ (to D. Gottesman)} \label{Gottesman6}

Thanks for the detailed reply.  I didn't get a chance to use it in my talk---I ran out of time---but I had a good chunk of your Answer to \#3 inserted into a transparency.  Sorry I didn't get to it.  There is indeed a certain methodological analogy between purifying states and introducing scalar and vector potentials in E\&M.  And by my own criterion in the talk, it's something I should take note of.  I was going to say that, but then ran out of time.

\section{27-09-07 \ \ {\it The Old Foil} \ \ (to J. Preskill)} \label{Preskill17}

I didn't get a chance to roll in many of the replies to my Everettian questionnaire in my talk (other than Shor, Simon, Deutsch, and Jozsa's replies), but I did nonetheless use you as a foil in my discussion on quantum cosmology in it.  In case you'd be interested to see it, here's the link:  \pirsa{07090068}.  I fell a little flat in my presentation of how the ``external'' system can be extended all the way around the observer---and how the very reason we use quantum mechanics is because we are within the universe, not external to it, and that nothing about that changes when we get to cosmology---and so, I have every bit as much right to write down a wavefunction for the universe as an Everettian (more so really), but at least the points are all there.  I think I need some drama lessons.  Anyway, thanks for being the foil again, 11 years later.

\section{27-09-07 \ \ {\it The Joint-Stock Society} \ \ (to H. R. Brown)} \label{BrownHR3}

Thanks for your remark just before Adrian's talk.  I enjoyed it, and would ultimately like to pursue your point in a (relatively, never too) serious discussion.

Here are two further quotes of {\James} that I had prepared for the talk, but didn't get a chance to actually show.  They have to do with the way {\James} is thinking about ``chance''---not in a Lewisian sort of way, as a synonym for objective numerical probabilities---but as a statement of an ultimate pluralism in things and a rejection of the block-universe conception.  At the moment, I tend to think that does indeed seem to capture the endpoint of my quantum research program---and it is something your joke pretty accurately reflected!

\bq
[Chance] is a purely negative and relative term, giving us no information about that of which it is predicated, except that it happens to be disconnected with something else---not controlled, secured, or necessitated by other things in advance of its own actual presence. As this point is the most subtle one of the whole lecture, and at the same time the point on which all the rest hinges, I beg you to pay particular attention to it. What I say is that it tells us nothing about what a thing may be in itself to call it ``chance.'' It may be a bad thing, it may be a good thing. It may be lucidity, transparency, fitness incarnate, matching the whole system of other things, when it has once befallen, in an unimaginably perfect way.
All you mean by calling it ``chance'' is that this is not guaranteed, that it may also fall out otherwise. For the system of other things has no positive hold on the chance-thing. Its origin is in a certain fashion negative: it escapes, and says, Hands off!\ coming, when it comes, as a free gift, or not at all.

This negativeness, however, and this opacity of the chance-thing when thus considered {\it ab extra}, or from the point of view of previous things or distant things, do not preclude its having any amount of positiveness and luminosity from within, and at its own place and moment. All that its chance-character asserts about it is that there is something in it really of its own, something that is not the unconditional property of the whole. If the whole wants this property, the whole must wait till it can get it, if it be a matter of chance. That the universe may actually be a sort of joint-stock society of this sort, in which the sharers have both limited liabilities and limited powers, is of course a simple and conceivable notion.
\eq

\noindent And

\bq
The more one thinks of the matter, the more one wonders that so empty and gratuitous a hubbub as this outcry against chance should have found so great an echo in the hearts of men. It is a word which tells us absolutely nothing about what chances, or about the modus operandi of the chancing; and the use of it as a war cry shows only a temper of intellectual absolutism, a demand that the world shall be a solid block, subject to one control,---which temper, which demand, the world may not be found to gratify at all. In every outwardly verifiable and practical respect, a world in which the alternatives that now actually distract your choice were decided by pure chance would be by me absolutely undistinguished from the world in which I now live. I am, therefore, entirely willing to call it, so far as your choices go, a world of chance for me. To yourselves, it is true, those very acts of choice, which to me are so blind, opaque, and external, are the opposites of this, for you are within them and effect them. To you they appear as decisions; and decisions, for him who makes them, are altogether peculiar psychic facts. Self-luminous and self-justifying at the living moment at which they occur, they appeal to no outside moment to put its stamp upon them or make them continuous with the rest of nature. Themselves it is rather who seem to make nature continuous; and in their strange and intense function of granting consent to one possibility and withholding it from another, to transform an equivocal and double future into an unalterable and simple past.
\eq

\section{28-09-07 \ \ {\it Easy Questions, RHJB} \ \ (to H. J. Briegel)} \label{Briegel6}

Thanks for the belated reply.  Too late for the talk, of course, but still enlightening.  The funniest chain of replies I had were from Peter Shor and Dan Simon.  Shor said, ``No I wasn't thinking about parallel worlds, I was thinking about periodicity and Simon's algorithm.''  Then Simon said, ``Everett?  Who's Everett?  And what's his interpretation?''  Another interesting dualism came from the separate replies of Deutsch and Jozsa on the Deutsch--Jozsa algorithm---you can imagine how that played out.

\section{28-09-07 \ \ {\it Truth?}\ \ \ (to G. L. Comer)} \label{Comer108}

\bgc
I teach Intro to Physics 111.  It's for physics majors; a one credit hour course just to get freshmen into the flow of the program. Today we talked about the nature of TRUTH! \ A few were distressed when I said I don't believe in truth.
\egc

\bv
I don't believe in Elvis\\
I don't believe in Zimmerman\\
I don't believe in Beatles\\
I just believe in me\\
Yoko and me\\
\ev
Apt.  You made me think again on the pragmatist conception of truth.  ``I just believe in me.''

Not really fair for me to send you a link to my new talk, but here it is:  \pirsa{07090068}.  It, too, touches a little bit upon the pragmatic notion of truth.

\section{01-10-07 \ \ {\it The Joint-Stock Society} \ \ (to A. Wilce)} \label{Wilce16}

Thanks for the encouragement about my talk.  It came off a little flatter than I had hoped, so it was nice at least to hear that you thought it was fun.  Particularly, I think I could have done much better in getting the point across that quantum cosmology presents no problem for the quantum Bayesian, but you live and you learn \ldots\ and I have a proclivity for saying things over and over, so eventually I'll get it right.

You might enjoy two of the further quotes that I ran out of time to present.  They start to make the point of what James had in mind when he used the word ``multiverse''---it certainly wasn't the big block universe the Everettians have in mind. [See 27-09-07 note ``\myref{BrownHR3}{The Joint-Stock Society}'' to H. R. Brown.]

\section{01-10-07 \ \ {\it The Joint-Stock Society, 2} \ \ (to A. Wilce)} \label{Wilce17}

\baw
What do you make of Adrian's ``real world'' branching-histories proposal?
\eaw

Do you remember my transparency where I made fun of ``speculative ontologies (before the very last moment)''.  Adrian's is well before the moment.

\section{01-10-07 \ \ {\it Shakespeare in Sweden, 2} \ \ (to P. G. L. Mana)} \label{Mana9}

Thanks for relieving the mystery of PIPPO.  

\bpglm
Speaking of Wheeler, I read that he once said ``Philosophy is too
important to leave to the philosophers''. Do you know if he was joking
or meant that seriously? I was cordially disappointed when I read that.
\epglm

Yes, he certainly said that, and I think he certainly meant it.  But what do you think he meant by it?  Why did it disappoint you?

I don't know if you will enjoy my new talk, but it is on PIRSA:
\pirsa{07090068}.

\section{01-10-07 \ \ {\it A Quote You May Like} \ \ (to W. C. Myrvold)} \label{Myrvold8}

\bwm
I arrived back in London to find a copy of David {\Mermin}'s new book {\bf Quantum Computer Science} waiting for me.  Browsing through it, the following caught my eye (p.\ 38):
\bq{\rm
Before drawing extravagant practical, or even only metaphysical, conclusions from quantum parallelism, it is essential to remember that when you have a collection of Qbits in a definite but unknown state, \emph{there is no way to find out what that state is}.

If there \emph{were} a way to learn the state of such a set of Qbits, then everyone could join in a rhapsodic chorus. (Typical verses: ``Where were all those calculations done? In parallel universes!'' ``The possibility of quantum computation has established the existence of the multiverse.'' ``Quantum computation achieves its power by dividing the computational task among huge numbers of parallel worlds.'') But there is no way to learn the state. The only way to extract any information from Qbits is to subject them to a measurement.}
\eq
\ewm

I do like it!  Thanks.  I'm sorry I didn't get a chance to talk to you more this meeting.  I'm also sorry my talk came out a little flatter than I had wanted it to (there were several points that I didn't think I emphasized correctly).

Since you're an aficionado of ``chance'' let me forward a note I had written to Harvey Brown in the aftermath of the meeting.  It contains two quotes that I didn't get a chance to read in my talk.  I think they decently capture the contrast between James's notion of chance and a David Lewis style chance, and the contrast between James's multiverse (his usage should get priority in any case, as he invented the term) and the limp, pale ``multiverse'' of the Everettians (a big block universe, whose `multi' aspect is nothing other than the possibility of viewing it from any of its facets)---a dead, lifeless place.

The week that Brian Skyrms is in London, I'd like to come visit you all during some of it.  Keep me abreast.

\section{03-10-07 \ \ {\it Nonlocality Again?}\ \ \ (to J. {\Barrett})} \label{Barrett4}

If you have time, I'd really like to follow up on the discussion we started at lunch \ldots\ in private, so we can hear each other think.  The reason is, I think statements like ``Bell inequalities have nothing to do with quantum mechanics'' and ``their derivation can be posed independently of whether the world operates according to quantum mechanics, and one sees then that they are simply violated by quantum mechanics'' or some such forms (as I think you were saying), are incorrect if one is already taking a quantum-Bayesian-like stance.  That is, I think I can argue that the usual derivation is simply inoperative from that world view.  It takes a completely different world view to get a Bell dilemma going.  This is something I've been thinking about a while but I've never tested it flesh and blood on anyone, nor have I written much about it.  So, you're the perfect candidate and I would enjoy talking about it with you.

Below is a sketch of what I'll say.  (Clearly I never finished my promised notes to van {\Enk}.)  [See 10-02-07 note ``\myref{vanEnk11}{Inside and Outside}'' and 02-08-07 note ``\myref{vanEnk12}{The One-Belly Theory of the Universe}'' to S. J. van Enk.]

\section{03-10-07 \ \ {\it Wave Function of the Universe and Sipe} \ \ (to S. J. van {\Enk})} \label{vanEnk13}

\bsve
Last night we had dinner with Sipe (Toronto) and he was saying how during a talk you had written down a wave function of the universe, something like:
$$
|\psi_{\rm universe}\rangle
$$
I maintained you'd never assign a pure state to the universe, in fact not even to the best ion in Wineland's ion trap. Please tell me I'm right! In other words, I'm sure you must have used some qualification when you wrote down $|\psi_{\rm universe}\rangle$. What was the context here?
\esve

I've been meaning to write you.  We have voted that we would like you to give a quantum foundations seminar here (to talk on your ``toy model'').  They're Tuesdays at 4:00.  Take your pick, most slots are open.  The sooner you can come the better, and of course we would pay for everything.

I did write down a pure state, but that was just meant to be symbolic of a quantum state as a whole.  In practice, no, {\it most\/} people who ``accept quantum mechanics'' (in the Fuchsian sense) would not write down a pure state for the external universe (to themselves)---but, one cannot forget that that statement is ultimately dependent upon a prior.  I can imagine an extreme case where 1) one accepts QM (i.e., as an addition to decision theory), but 2) has an extreme belief of certainty about some particular question one can ask of the external world.  Then that hypothetical agent might well indeed write down a pure state for the external world.

Note the use of ``external world'' when I'm talking about writing down $|\psi_{\rm universe}\rangle$.  The point I was trying to make in the lecture is that that is all one needs for doing quantum cosmology.  And that really is all one wants.  In the Quantum Bayesian view, all the Born rule signifies in any case is a statement of ``coherence'' in a sense closely analogous to de Finetti's.  The agent writes down his beliefs about baryon number, say.  He writes down his expectations for the matter distribution in this era of the universe.  He writes down his expectations for the Hubble constant.  And so on and so on.  Then he distils all those expectations into a quantum state assignment.  All that is doing for him is telling what he should expect for all the other questions that he might later ask.  For instance, what does he expect the mean inhomogeneity in the cosmic microwave background to be?  His belief about that should be coherent with all of his other beliefs.

I tried to make the whole point dramatic in the talk, but I think it kind of came out a flop.  I'll send a link in a minute.  I drew a picture of an agent and a quantum system near him.  Then I made the quantum system bigger, and pointed out that nothing changes.  Then I made it bigger and bigger, all the while saying nothing conceptually changes.  Finally, I had the system completely surrounding him, and made the point that even there nothing changes.  Here's the way I wrote John Preskill the other day:
\bq\noindent
I fell a little flat in my presentation of how the ``external'' system can be extended all the way around the observer---and how the very
reason we use quantum mechanics is because we are within the universe, not external to it, and that nothing about that changes when we get
to cosmology---and so, I have every bit as much right to write down
a wavefunction for the universe as an Everettian (more so really), but at least the points are all there.  I think I need some drama lessons.
\eq
Here's the link to the talk, \pirsa{07090068}.

\section{03-10-07 \ \ {\it Getting the Word Out}\ \ \ (to S. Aaronson)} \label{Aaronson9}

\bsa
Well, from my perspective, it's about information, probabilities, and
observables, and how they relate to each other.
\esa

Congratulations!  With one case of corporate abuse, you've made more of an imprint of that idea on the world than all my years of evangelizing combined!  And exponentially more so.  You're a dream come true.  Thank Darwin's soup for Scott Aaronson!

I sure wish you had stayed at PI.

\section{03-10-07 \ \ {\it Getting the Word Out, 2}\ \ \ (to S. Aaronson)} \label{Aaronson10}

\bsa
I'm so happy you like the infamous quote (though I think it worked better in its original context).  Your talks and papers are a large part of what gave me the courage to talk about quantum mechanics this way.  Indeed, if you scroll down to the ``Further Reading'' section of the plagiarized lecture (\myurl{http://www.scottaaronson.com/democritus/lec9.html}), you'll see ``Pretty much anything Chris Fuchs has written.''
\esa

That was very sweet of you.  I hadn't looked at your lecture before, but I already like it.  Particularly, this phraseology comes through to me:
\bsa
So, what is quantum mechanics? Even though it was discovered by
physicists, it's not a physical theory in the same sense as
electromagnetism or general relativity. In the usual ``hierarchy of
sciences'' -- with biology at the top, then chemistry, then physics,
then math -- quantum mechanics sits at a level between math and
physics that I don't know a good name for. Basically, quantum
mechanics is the operating system that other physical theories run on
as application software (with the exception of general relativity,
which hasn't yet been successfully ported to this particular OS).
There's even a word for taking a physical theory and porting it to
this OS: ``to quantize.''
\esa
I think that's right on the mark, particularly if one particularizes math to ``decision theory''.  I hope that comes through about halfway through this lecture, \pirsa{07090068} (starting around page 28 in the PDF transcript, at least).

``It is not a physical theory in the same sense \ldots .''  It's hard to get that across at foundational conferences!

\section{03-10-07 \ \ {\it Aspects of Lunch} \ \ (to M. A. Nielsen)} \label{Nielsen8}

Thanks for telling me about Scott's\index{Aaronson, Scott} blog yesterday.  I looked it up last night during a fit of insomnia and saw the commercial on YouTube.  It blows my Capt.\ Stabbin story away by miles and miles (and miles and miles)!  Of course, I had a general disgust with the business practices of the ad company (and the CEO's ridiculous denial of plagiarism in the {\sl Sydney Morning Herald}).  But, in contrast, I was also quite tickled by the exact message that they did copy---that quantum mechanics is about information and probabilities.  The creeps couldn't have had a better accident!  To get that idea into pop culture would be so wonderful---it won't happen easily, of course, but this commercial at least gave it a new epsilon.

So, I wrote Scott\index{Aaronson, Scott} with my pleasure (though I didn't tell him about the captain!), and he wrote me back the nice note below.  It took me quite by surprise to learn that even I had an epsilon role in that too.  \ldots\ And that one really is an epsilon.  Still, these are my lucky days, Capt.\ Stabbin and Capt.\ Scott!

I then looked at the actual lecture. It is nice, particularly the philosophy.  A much better quote from it---one that I think strikes particularly deep---is from the paragraph just before the infamous one:
\bq\noindent
So, what is quantum mechanics? Even though it was discovered by physicists, it's {\it not\/} a physical theory in the same sense as electromagnetism or general relativity. In the usual ``hierarchy of sciences'' -- with biology at the top, then chemistry, then physics, then math -- quantum mechanics sits at a level {\it between\/} math and physics that I don't know a good name for. Basically, {\it quantum mechanics is the operating system that other physical theories run on as application software\/} (with the exception of general relativity, which hasn't yet been successfully ported to this particular OS).
\eq
I think that's a very good way of putting it.  I'm sure I'll try to say these kinds of things to Alain Aspect (again), and he'll think I'm just as crazy as he has the last couple of times.

\section{03-10-07 \ \ {\it Feynman and Bell} \ \ (to C. M. {\Caves})} \label{Caves96.1}

Once upon a time you told me a story about the reason Feynman does not cite Bell in his early quantum computing paper---namely, at the spot where he derives a Bell inequality.  As I recall you said it was because Feynman claimed that he had told Bell of the argument sometime before his own publishing of it.  Is that the story you told me?  Can you fill in details?  If you got that second hand, who did you get it from?

I ask because I related it to Alain Aspect at lunch and he gave several reasons to doubt it.  So I just want to get the facts straight hereafter.  It could be that Aspect is accurate and you are too, given Feynman's proclivity for simply making stories up that made him look better than he already was.  (I've compiled several of these---so I know they exist.)

If you could write me back before dinner time when I have to eat with Aspect again, that would be great!

\subsection{Carl's Reply}

\bq
Maybe you're having a late dinner.

This was in a talk Feynman gave at Caltech in the early 80s.  He presented the hidden-variable model for a qubit that colors the Bloch sphere and then showed you couldn't do this with two qubits.  Feynman would never have paid attention to what Bell did, but I believe he did say that he told Bell about this before Bell did his stuff.

There is a chance that this talk was published in Caltech's {\sl Engineering and Science\/} magazine, in which case you might find it relatively easily at that web site.  It is almost certain that some transcript of the talk exists in the Feynman archives at Caltech, so if you're really serious, you could track it down, I think.

Read Gleick's biography.  It's really about how Feynman defined himself in terms of stories that had a grain of truth, but were always embellished to put him in what he perceived as the best possible light.  Ralph Leighton transmitted these stories unaltered in his Feynman books.  Murray got into trouble in the {\sl Physics Today\/} Feynman issue precisely because he made this point.
\eq

\section{04-10-07 \ \ {\it Feynman Question} \ \ (to J. Preskill)} \label{Preskill18}

Yesterday, I had this conversation with Carl Caves (and Alain Aspect).  [See 03-10-07 note ``\myref{Caves96.1}{Feynman and Bell}'' to C. M. {\Caves}.]  I wonder if you can shed any further light on it.  Did you happen to be at the same talk?  Do you have any recollection?  Also, do you have easy access to the magazine or archives Carl mentions; if not could you give me a name of someone I can contact?

Thanks for the help.

\subsection{John's Reply}

\bq
Well, I remember hearing Feynman give a talk in the early 80's on the foundations of quantum theory, called ``Negative probability.'' A version of this was later published in a book, but I don't have a copy handy. I don't recall the talk well, except that I remember that I was not impressed by it. (I heard it in a filled auditorium at MIT, so maybe it was before I left Harvard in 1983.) Maybe Carl heard a similar talk, or maybe he is thinking of a different one.

Anyway, I don't regard Feynman's failure to cite Bell as evidence that Feynman thought he had the idea first. He often did not cite people in those days. I talked to him often about confinement in QCD, and pointed out to him that his ideas about magnetic disorder had been anticipated by 't Hooft, Polyakov, and others (whose work was actually deeper in my opinion), but he still did not cite the earlier work when he wrote about the subject. He just didn't want to bother.

By the way, I have a more vivid recollection of a talk that Alain gave at Caltech in the mid-80s, in which he quoted Feynman's statement: ``I cannot define the real problem, therefore I suspect there's no real problem, but I'm not sure there's no real problem.'' For some reason this seemed very funny, and it got a big laugh. But no one laughed harder than Feynman (I was sitting next to him in the front row.)
\eq

\section{04-10-07 \ \ {\it Renewing Interest in Aristotle} \ \ (to P. Goyal)} \label{Goyal1}

Reviewing my {\Pauli} notes, I do indeed see that I should be a little more interested in Aristotle and his potentia, and try to figure out what he is getting at with the idea.  See {\Pauli}'s letter to Jung (dated 27 Feb '53), starting on page 142 of the collection I gave you (and reprinted below).  Also see the entry from Heisenberg's ``Wolfgang {\Pauli}'s Philosophical Outlook,'' on page 62.

In case you want to follow up on our lunchtime conversation (on how I want I want to view quantum measurement), I might suggest my samizdat {\sl My Struggles with the Block Universe\/}, where I wrote a few letters on the subject to Bas van Fraassen.

\subsection{Letter from Pauli to Jung, 27 February 1953}
\bq
\indent
I cannot anticipate the new {\it coniunctio}, the new {\it hieros
gamos\/} called for by this situation, but I will nevertheless try to
explain more clearly what I meant with the final part of my Kepler
essay: the firm grip on the ``tail''---that is, physics---provides me
with unhoped for aids, which can be utilized with more important
undertakings as well, to ``grasp the head mentally.'' It actually
seems to me that in the {\it complementarity of physics}, with its
resolution of the wave-particle opposites, there is a sort of {\it
role model or example of that other, more comprehensive
coniunctio}.\footnote{I had interesting discussions about these
matters with Mr.\ M.~Fierz, to whom I am most grateful.} For the
smaller {\it coniunctio\/} in the context of physics, completely
unintentionally on the part of its discoverers, has certain
characteristics that can also probably be used to resolve the other
pairs of opposites listed on p.~3. The analogy is on these
lines:\\ \\
\parbox[t]{2.6in}{\small Quantum physics.} \ \ \ \
\parbox[t]{2.6in}{\small Psychology of the individuation
process and the unconscious in general.}
\\ \medskip \\
\parbox[t]{2.6in}{\small Mutually exclusive complementary
experimental setups, to measure position as well as momentum} \ \ \ \
\parbox[t]{2.6in}{\small Scientific thinking --
intuitive feeling.}
\\ \medskip \\
\parbox[t]{2.6in}{\small Impossibility of subdividing the
experimental setup without basically changing the phenomenon.} \ \ \ \
\parbox[t]{2.6in}{\small Wholeness of man consisting of
consciousness and unconsciousness.}
\\ \medskip \\
\parbox[t]{2.6in}{\small Unpredictable intervention with
every {\it observation}.} \ \ \ \
\parbox[t]{2.6in}{\small Change in the conscious and the
unconscious when consciousness is acquired, especially in the process
of the {\it coniunctio}.}
\\ \medskip \\
\parbox[t]{2.6in}{\small The result of the observation is an
irrational actuality of the unique occurrence.} \ \ \ \
\parbox[t]{2.6in}{\small The result of the {\it coniunctio\/}
is the {\it infans solaris}, individuation.}
\\ \medskip \\
\parbox[t]{2.6in}{\small The new theory is the objective,
rational and hence symbolic grasping of the {\it possibilities\/} of
natural occurrences, a sufficiently broad framework to accommodate
the irrational actuality of the unique occurrence.} \ \ \ \
\parbox[t]{2.6in}{\small The objective, rational, and hence
symbolic grasping of the psychology of the individuation process,
broad enough to accommodate the irrational actuality of the unique
individual}
\\ \medskip \\
\parbox[t]{2.6in}{\small One of the means used to back up the
theory is an abstract mathematical sign ($\psi$), and also complex
figures (functions) as a function of space (or of even more
variables) and of time.} \ \ \ \
\parbox[t]{2.6in}{\small The aid and means of backing up the
theory is the concept of the unconscious. It must not be forgotten
that the ``unconscious'' {\it is our symbolic sign for the potential
occurrences in the conscious, not unlike that ($\psi$)}.}
\\ \medskip \\
\parbox[t]{2.6in}{\small The laws of nature to be applied are
statistical laws of probability. An essential component of the
concept of probability is the motif of  ``the One and the Many.''} \ \ \ \
\parbox[t]{2.6in}{\small There is a generalization of the law
of nature through the idea of a self-reproducing ``figure'' in the
psychic or psychophysical occurrences, also called ``archetype.'' The
structure of the occurrences that thus come into being can be
described as ``automorphism.'' Psychologically speaking, it is
``behind'' the time concept.}
\\ \medskip \\
\parbox[t]{2.6in}{\small The atom, consisting of nucleus and
shell.} \ \ \ \
\parbox[t]{2.6in}{\small The human personality, consisting of
``nucleus'' (or Self) and ``Ego.''}
\\ \medskip

I should like to add just a few epistemological remarks to this
provisional schema. By allowing for occurrences and the utilization
of possibilities that cannot be apprehended as predetermined and
existing independent of the observer, the type of interpretation of
Nature characteristic of quantum physics clashes with the old
ontology that could simply say ``Physics is the description of
reality,''\footnote{{\it Einstein's\/} words.} as opposed to
``description of what one simply imagines.''\footnote{{\it
Einstein's\/} words.} ``Being'' and ``nonbeing'' are not unequivocal
characterizations of features that can be checked only by statistical
series of experiments with various experimental setups, which in
certain circumstances are mutually exclusive.

In this way, the confrontation between ``being'' and ``nonbeing''
that was begun in ancient philosophy sees its continuation. In
antiquity, ``nonbeing'' did not simply mean not being present but in
fact always points to a {\it thinking problem}. Nonbeing is that
which cannot be thought about, which cannot be grasped by thinking
reason, which cannot be reduced to notions and concepts and cannot be
defined. It was along these lines, as I see it, that the ancient
philosophers discussed the question of being or
nonbeing.\footnote{You got involved in this old discussion when you
came across the Neoplatonist formula that evil is ``nonbeing,''  is
simply  a  {\it  privatio  boni}. Your characterization of this
statement as ``nonsense''  [{\sl Answer to Job\/}] I attribute more
to the bad habit of modern theologians of using old words whose
meaning they have long ceased to understand rather than to the
original statement itself. For me personally, modern theologians are
totally uninteresting, but on the other hand it seems to me {\it
imperative\/} in such discussions to go back to the original roots of
the words and expressions used.

What the ancients meant when they said ``nonbeing'' was what we would
more accurately describe today as ``irrational'' or ``dark.''

Now ever since Socrates and Plato, Good has been understood and
considered as {\it Rational\/} (the virtues are even teachable!),
unlike Evil, which does not lend itself to any conceptual
definition---a great idea, or so it seems to me. According to this
interpretation, the latter regards Good in the same way that matter
regards the ideal (``being'') mathematical object. With Plato, matter
is actually {\it defined\/} as that which distinguishes the empirical
object from the ideal geometrical object. What they both have in
common is the Comprehensible, the Positive, the Good in the empirical
body; what makes them different---matter---is the Incomprehensible,
later Evil. Hence, matter has only the {\it passive\/} function of
adopting the geometrical ideas hypostasized as ``being'' (it is the
``receptacle'' or ``wet nurse'' of these ideas). Thus, in later
Platonism the {\it privatio boni\/} means: Expressed in general terms
and understood from the point of view of the ``{\it one},''
unchangeable ``being'' idea, {\it like Euclid's geometry}, Evil can
be {\it rationally\/} characterized as the absence of Good, the lack
of ideas.

(It is odd how reading your books always transports me back to
antiquity. It is obviously a personal effect you have on me; before
reading {\sl Aion}, I was not all that interested in antiquity.)} And
it was especially along these lines that the process of becoming and
the changeable, hence also matter, appeared in a certain form of
psychology as nonbeing---a mere {\it privatio\/} of ``Ideas.'' By way
of contrast, Aristotle, evading the issue, created the important
concept of {\it potential being\/} and applied it to {\it hyle}.
Although hyle was actually ``nonbeing'' and simply a {\it privatio\/}
of  ``form'' (which is what he said instead of ``Ideas''), it {\it
was\/} potentially ``being'' and not simply a {\it privatio}. This is
where an important differentiation in scientific thinking came in
Aristotle's further statements on matter (he clung firmly to the
Platonic notion of matter as something passive, receiving) cannot
really be applied in physics, and it seems to me that much of the
confusion in Aristotle stems from the fact that being by far the less
able thinker, he was completely overwhelmed by Plato. He was not able
to fully carry out his intention to grasp the potential, and his
endeavors became bogged down in the early stages. It is on Aristotle
that the peripatetic tradition and, to a large extent, alchemy is
based ({\it vide\/} Fludd). Science today has now, I believe, arrived
at a stage where it can proceed (albeit in a way as yet not at all
clear) along the path laid down by Aristotle. The complementary
characteristics of the electron (and the atom) (wave and particle)
are in fact ``potential being,'' but one of them is always ``actual
nonbeing.'' That is why one can say that science, being no longer
classical, is for the first time a genuine theory of becoming and no
longer Platonic. This accords well with the fact that the man who is
for me the most prominent representative of modern physics, Mr.\ Bohr,
is, in my opinion, the only truly non-Platonic thinker\footnote{The
English philosopher A. N. Whitehead once said that the whole of
European philosophy consisted of footnotes to Plato.}: even in the
early '20s (before the establishment of present-day wave mechanics)
he demonstrated to me the pair of opposites ``Clarity-Truth'' and
taught me that every true philosophy must actually start off with a
{\it paradox}. He was and is (unlike Plato) a {\it
dekranos\footnote{``Double-head''---nickname for disciples of
Heraclitus given by disciples of Parmenides.} kat exochen}, a master
of antinomic thinking.

As a physicist familiar with this course of development and this way
of thinking, the concepts of the gentlemen with the stationary
spheres\footnote{I have Parmenides and Kepler in mind.} are just as
suspect to me as the concepts of ``being'' metaphysical spaces or
``heavens'' (be they Christian or Platonic), and ``the Supreme'' or
``Absolute.''\footnote{This is an allusion to Indian philosophy. Even
those Indian philosophers who, like Prof.\ S.~Radhakrishnan, avoid
applying the word ``illusion'' to the empirical world have no other
way of commenting on the Mysterium of the connection between
``ultimate reality'' and the empirical world, except to call it
``Maya.''

The Absolute always has the tendency to place itself at an
immeasurable distance from man and nature. I happily quote your own
words [{\sl Answer to Job\/}]: ``Only that which affects me do I
acknowledge as real. But what does not affect me may as well not
exist.''} With all of these entities, there is an essential paradox
of human cognition (subject-object relation), which is not expressed,
but sooner or later, when the authors least expect it, it will come
to light!

For these reasons I should like to suggest also applying the
Aristotelean way out of the conflict between ``being'' and
``nonbeing'' to the concept of the unconscious. Many people still say
that the unconscious is ``nonbeing,'' that it is merely a {\it
privatio\/} of consciousness.''\footnote{Cf.\ also {\sl Psychologie
and Religion}, p.\ 153.} (This probably includes all those who
reproach you with ``psychologism.'') The counterposition is that of
placing the unconscious and the archetypes, like ideas in general, in
supracelestial places and in metaphysical spaces. This view strikes
me as equally dubious and contradictory to the law of the Kairos.
This is why I have opted for the third road in my analogy schema in
interpreting the unconscious (as well as the characteristics of the
electron and the atom) as ``potential being.''\footnote{Cf.\ ibid.,
p.\ 186 below: archetypes as formal possibility.} It is a legitimate
description by man for potential occurrences in the conscious and as
such belongs to the genuine symbolic reality of the ``thing in
itself.'' Like all ideas, the unconscious is in {\it both man and
nature}; ideas have {\it no\/} fixed abode, not even a heavenly
one.''\footnote{This point of view was also put forward by Mr.\ Fierz
in the discussion referred to.} To a certain extent, one can say of
{\it all\/} ideas {\it ``cuiuslibet rei centrum, cuius circumferentia
est nullibi''\/} (the center of all things---a center whose periphery
is nowhere), which, according to ancient alchemistic texts, is what
Fludd said of God; see my Kepler article, p.\ 174. As long as
quaternities are kept ``up in heaven'' at a distance from people
(however pleasing and interesting such endeavors, seen as omens, may
be), no fish will be caught, the {\it hieros gamos\/} is absent, and
the psychophysical problem remains unsolved.

The psychophysical problem is the conceptual understanding of the
possibilities of the irrational actuality of the unique (individual)
living creature. We can only come close to dealing with this problem
when we can synthetically resolve the pair of opposites
``materialism-psychism'' in natural philosophy. When I say
``psychism,'' I do not mean ``psychologism'' nor something peculiar
to psychology\footnote{As a psychologist, you have an understandable
aversion to all forms of reality that are not just psychic. And just
as everything that King Midas touched turned to gold, everything you
looked at seemed to me to turn psychic and only psychic. This
aversion to the nonpsychic was probably also one reason why you did
not mention the psychophysical problem in your book {\sl Answer to
Job}. However, in the passage already quoted in {\sl Aion\/} (p.\
372), you put forward a point of view on the ultimate unity of physis
and psyche that coincides with mine. See also note below.} but simply
the opposite of materialism. I could also have said ``idealism,'' but
that would have restricted it in time to the famous currents of
philosophy prevailing in the 19th century after Kant. These currents
(including Schopenhauer), as well as the whole of Indian philosophy,
fall into this category of  ``psychism.''

But as the alchemists correctly surmised, matter goes just as deep as
the spirit, and I doubt whether the goal of any development can be
absolute spiritualization. Sciences made by man---whether or not we
wish or intend it and even if it is natural sciences---will always
contain {\it statements about man}.\footnote{Cf.\ my Kepler essay,
p.\ 163, n. 7.} And that is also precisely what I was trying to
express with the analogy schema in this section.

Thus the aim of science and of life will ultimately remain man, which
is actually the note on which your book {\sl Answer to Job\/} closes:
In him is the ethical problem of Good and Evil, in him is spirit and
matter, and his wholeness is depicted with the symbol of the
quaternaty.

It is today the archetype of the {\it wholeness\/} of man from which
natural science, now in the process of becoming quaternary, derives
its emotional dynamics. In keeping with this, the modern
scientist---unlike those in Plato's day---sees the rational as both
good and evil. For physics has tapped completely new sources of
energy of hitherto unsuspected proportions, which can be exploited
for both good and evil. This has led initially to an intensification
of moral conflicts and of all forms of opposition, both in nations
and in individuals.

This wholeness of man\footnote{This gives rise to the question very
closely connected with the psychophysical problem: Is the archetype
of wholeness restricted to man, or does it also manifest itself in
nature? See your essay ``Der Geist der Psychologie,'' Eranos Jahrbuch
1946, p.\ 483f, where you treat the archetypes as not just psychic.}
seems to be placed in two aspects of reality: the symbolic ``things
in themselves,'' which correspond to ``potential being,'' and
concrete manifestations, which correspond to the actuality of
``being.'' The first aspect is the rational one, the second the
irrational one\footnote{The older ancient philosophers since
Parmenides have correspondingly described concrete phenomena as
``nonbeing.'' By way of contrast, all general concepts and ideas with
unchangeable characteristics (``form'' in Aristotle), especially {\it
geometrical\/} concepts, were ``being.'' There are ancient
astronomical papers that set themselves the task of {\it ``saving''
phenomena\/} $\sigma\omega$ $\zeta\varepsilon\iota\nu$ $\tau\alpha$
$\varphi\alpha\nu\iota o\mu\varepsilon\nu\alpha$. Apparently, they
did not use the word ``explain.'' I am not going into the question of
pure mathematics here.} (I use these adjectives analogously, as you
did in the typology theory for the characterization of the various
functions.) The interplay of the two aspects creates the process of
becoming.

Is it in keeping with the Kairos and the quaternity to call these
fragments of a philosophy ``critical humanism''?
\eq

\section{04-10-07 \ \ {\it Feynman Question} \ \ (to K. S. Thorne)} \label{Thorne1}

Yesterday, I had the conversation below with Carl Caves (and Alain Aspect).  [See 03-10-07 note ``\myref{Caves96.1}{Feynman and Bell}'' to C. M. {\Caves}.] I wonder if you can shed any further light on the issue.  Carl thinks that you were at the same lecture.  Do you have any recollection?  Does your memory bear Carl's remark out, or do you have a dissenting memory?  Now that I've brought the issue up with Aspect, I feel obligated to try and get it straightened out.

Thanks for any help you can give.

\subsection{Kip's Reply}

\bq
My memory is really lousy.  I do remember Feynman's lecture, but
I have no memory of what he may or may not have said about
the relationship of his own thoughts about this to Bell's.

It is true that there might be an audio tape of Feynman's lecture in
the Caltech Archives, but it was not routine to tape lectures in those
days.  If there is a tape, it is because it was Feynman speaking.
\eq

\section{04-10-07 \ \ {\it The Surprise} \ \ (to G. Brassard)} \label{Brassard51}

It was such a pleasant surprise to see you last night---it really shocked me.  And I had a great time just before Aspect's talk [finally!]\ getting a chance to discuss some of the deeper issues about what is ontic and what is epistemic within quantum mechanics.

We didn't get a chance to get anywhere near this far, but I think if I had to put into a slogan what's going to be ultimately found and quantified in this research program of mine, it's this:  That the ontic of quantum systems is that they are CATALYSTS.  That is their conceptual role.  The thing that is intrinsic to quantum systems themselves (and the thing dimensionality is ultimately a quantification of) is that they bring about transformations in things external to them.  Quantum systems are transformers.  There is an element of {\Mermin}'s ``correlation without correlata'' in this idea, but it is a much more active thing, and it is careful to relegate quantum states to the epistemic (which {\Mermin} does not do).  Also, it shares a small piece of similarity to Rovelli's relationalism, but ditto what I just said about {\Mermin}.

Anyway, there are deep things to discuss, and I hope we'll get a much longer chance soon.  I'll work hard to get to {\Montreal} in the spring if we can home in on a good time for both of us.

\section{05-10-07 \ \ {\it Easy Questions, RLKG} \ \ (to L. K. Grover)} \label{Grover6}

\blkg
Sorry, I had missed your message --- only just got it --- hope your talk went well.
\elkg

You might answer anyway if you get the chance.  I got replies from many; it'd be nice to know your answers too \ldots\ for the day when I become a historian.

\section{05-10-07 \ \ {\it My Sick Questions}\ \ \ (to D. M. Appleby \& S. T. Flammia)} \label{Appleby23.1} \label{Flammia0}

See attachment.  I'm sending it to you too Steve, because it gives a little more explanation (though not much more than epsilon) for why I'm asking some of these questions.

Now, it's time for PI's Friday Wine and Cheese.  I love this place!

\bq
\begin{center}
{\large \bf SICk Questions?}\\
A Small Attempt to Deepen Our Understanding of These Bases
\end{center}

\subsection{Preliminaries}

I will alternatively call a set of normalized states $|\psi_i\rangle\in{\cal H}_d$ or the set of projectors associated with them $\Pi_i=|\psi_i\rangle\langle\psi_i|$, with index $i$ running from 1 to $d^2$, ``SIC'' or ``a SIC'' if and only if
\be
\tr\,\Pi_i\Pi_j=|\langle\psi_i|\psi_j\rangle|^2=\frac{1}{d+1}\quad\forall i\ne j\;.
\ee
In all questions that follow I am NOT restricting myself to Weyl--Heisenberg covariant SICs.  (Though I hope that the Weyl--Heisenberg SICs will come out as the answer to one or more of these questions.)

Needless to say, we don't even know if SICs exist.  But the spirit of all the following is, ``{\it Supposing\/} they do exist, can one show properties $X$, $Y$, and $Z$?''

\subsection{A SIC-Schmidt? 1}

Recall the usual Schmidt decomposition theorem.  Given a state $|\psi_{AB}\rangle$ on a bipartite system with Hilbert space ${\cal H}_A\otimes{\cal H}_B$ (each component with dimension $d$), there always exists an orthonormal basis $|i\rangle$ on ${\cal H}_A$ and an orthonormal basis $|i^\prime\rangle$ on ${\cal H}_B$ such that,
\be
|\psi_{AB}\rangle=\sum_{i=1}^d \alpha_i |i\rangle|i^\prime\rangle\;.
\ee
QUESTION:  Under what conditions does a bipartite state $|\psi_{AB}\rangle$ have a decomposition of the form
\be
|\psi_{AB}\rangle=\sum_{i=1}^{d^2} \beta_i |\psi_i\rangle|\psi^\prime_i\rangle\;,
\ee
for some SICs $\{|\psi_i\rangle\}$ and $\{|\psi_i^\prime\rangle\}$?  Like with the usual Schmidt decomposition theorem, is it always possible?  If not, give a characterization of the special cases where it can be done.

\subsection{A SIC-Schmidt? 2}

But maybe a Schmidt-like theorem is not best posed at the state-vector level.  Maybe instead it lives most naturally at the density operator level.

Thus, QUESTION:  Under what conditions will a bipartite density operator $\rho_{AB}$ have a decomposition of the form
\be
\rho_{AB}=\sum_{i=1}^{d^2} \gamma_i\, \Pi_i\otimes\Pi_i^\prime;,
\ee
where $\{\Pi_i=|\psi_i\rangle\langle\psi_i|\}$ and similarly $\{\Pi_i^\prime\}$ are two SICs?  Is it always possible?  If not, give a characterization of the cases where it can be done.

\subsection{Give Me an ONB}

Given an orthonormal basis $|i\rangle$, how can one characterize the closest SICs?  (How can one characterize the farthest?)  Here's one idea that comes to mind.  Given a SIC $\{|\psi_i\rangle\}$, introduce a partition of it into $d$ sets of $d$ vectors.  Relabel them $\{|\psi_{ij}\rangle\}$, $i$ and $j$ now both running between 1 and $d$.  Then tabulate this number
\be
F=\sum_i\left(\sum_j |\langle i|\psi_{ij}\rangle|^2\right).
\ee
The larger this number, the closer the SIC to the original orthonormal basis.  How large can this quantity be?  Could it be that, {\it by a miracle}, the absolute max of $F$ is achieved by a Weyl--Heisenberg SIC, i.e., $|\psi_{ij}\rangle=X^i Z^j |\psi\rangle$, for some fiducial state $|\psi\rangle$.

\subsection{A SIC von Neumann Entropy}

The von Neumann entropy of a density operator $\rho$ can be defined by the following procedure.
\begin{enumerate}
\item
Introduce an orthonormal basis $|i\rangle$, with $\Pi_i=|i\rangle\langle i|$, $i=1,\ldots,d$.
\item
Tabulate $p(i)=\tr\rho\Pi_i$, and then consequently
\item
Tabulate the Shannon entropy $H=-\sum_{i=1}^d p(i)\log p(i)$ of the distribution.
\item
Finally minimize $H$ over all orthonormal bases $|i\rangle$.
\end{enumerate}
The resulting minimal such entropy is the von Neumann entropy $H(\rho)=-\tr\big(\rho\log\rho\big)$, and the orthonormal basis that achieves it is the eigenbasis of $\rho$.

However, we can ask a similar question of SICs.  What is the SIC-entropy of a density operator $\rho$?  It would be defined by the following procedure.
\begin{enumerate}
\item
Introduce a SIC $\{|\psi_i\rangle\}$, with $\Pi_i=|\psi_i\rangle\langle \psi_i|$, $i=1,\ldots,d^2$.
\item
Tabulate $p(i)=\tr\rho\Pi_i$, and then consequently
\item
Tabulate the Shannon entropy $H=-\sum_{i=1}^{d^2} p(i)\log p(i)$ of the distribution.
\item
Finally minimize $H$ over all SICs $\{|\psi_i\rangle\}$, to define a function $S(\rho)$.
\end{enumerate}
Does the SIC-entropy $S(\rho)$ have any interesting functional form?  Could it be a multiple of the von Neumann entropy?  What is the relation between the optimal SICs for this procedure and the eigenbases of $\rho$?

\subsection{Counting Common Elements}

Suppose one has two nonidentical SICs $\{\Pi_i\}$ and $\{\Pi_i^\prime\}$.  What is the maximal number of common elements between the two sets?  If we can't give the number precisely, can we bound it in any interesting way?

\subsection{Really, Just Give Me an ONB}

Going back to the question in ``Give Me an ONB.''  Actually it was motivated by the following question.  Suppose one is given an arbitrary SIC.  Is there any method of deriving a canonically interesting (or interesting class of) orthonormal basis from it $|i\rangle$ from it?  Question ``Give Me an ONB'' was an attempt to tackle this in one way.  Maybe there are others.  The over-riding reason for this question is that I think we are accustomed (at least subconsciously, if not logically) to thinking of Hilbert spaces as coming equipped with a standard orthonormal basis.  Think of the phrase, ``computational basis.'' I want to break that mentality down somewhat, by first equipping the cone of positive operators with a SIC and then taking it from there.   But there's no denying that orthonormal bases are useful.  So I'd like a standard construction of one or more to fall out of a SIC \ldots\ in something of an analogy to the way the Schmidt orthogonalization procedure gives us a way to go from an arbitrary basis to an orthogonal one.

\subsection{A SIC-Gleason Theorem}

Let $d\ge 3$ and $\cal P$ be the set of one-dimensional projectors $\Pi=|\psi\rangle\langle\psi|$.  We will call a function $f:{\cal P}\rightarrow[0,1]$ a SIC-frame function if for any SIC $\{\Pi_i\}$,
\be
\sum_{i=1}^{d^2} f(\Pi_i)=1\;.
\ee
Can one show that for each SIC-frame function, there is a unique density operator $\rho$ such that
\be
f(\Pi)=\frac{1}{d}\tr\,\rho\Pi\;?
\ee
This question was on my mind a long time ago, and ultimately led to ``Gleason-Type Derivations of the Quantum Probability Rule for Generalized Measurements'' [\quantph{0306179}], where we showed that it does not work if $d=2$. However, I have recently regained hope that it may work in $d=3$ or greater.

\eq

\section{05-10-07 \ \ {\it Start of an Answer} \ \ (to P. Goyal)} \label{Goyal2}

\bphg
Since the operational approach to QT is so important to my thinking, I would like to ask you to give some thought to the question of how someone without knowledge of the quantum formalism (QF) might discover/implement a SIC.

But maybe this is simply impossible.  You see, I would be quite keen on the possibility of trying to derive the quantum formalism from a set of postulates where SICs are taken as basic primitive givens, for this has a strong built-in attraction, namely that the outcome probabilities of a single measurement completely determine the state of the system one is measuring.  However, to proceed along this direction in a well-motivated way, the challenge would be to motivate SICs purely operationally (which is of course easy with PVMs since we have paradigmatic examples like SG measurements to abstract from).
\ephg

Thanks again for the question.  It's helped focus my thoughts.  Here's a little start of an answer maybe.

What does it mean to be a quantum system?  Here's one way I might tackle the question.  It means that there is some {\it special\/} action I can take upon a system that has $d^2$ consequences for my experience, and for which I feel I should structure my uncertainties for those consequences in the following way.

{\bf 1)} If I perform the action once, and get outcome $i$, then if I perform the same action again, the uncertainties I find reasonable for a second outcome $j$ are
\begin{eqnarray}
p(j|i) &=& \frac{1}{d} \qquad \qquad \quad\mbox{for}\quad j=i
\\
p(j|i) &=& \frac{1}{d(d+1)}  \qquad\mbox{for}\quad j\neq i \;.
\end{eqnarray}

{\bf 2)} More generally, I (subjectively) accept the following theory of priors $p(i)$ for the consequences of this kind of action.  That valid priors form the convex hull of these two equations:
\begin{eqnarray}
\sum_i p(i)^2 &=& \frac{2}{d(d+1)}
\\
\sum_{i,j,k} c_{ijk}\, p(i)p(j)p(k) &=& \frac{d+7}{(d+1)^3}\;,
\end{eqnarray}
where the coefficients $ c_{ijk}$ have been hard won through previous experience, and, of course, have some very remarkable properties.  (We shouldn't know this yet, but they are the structure coefficients for the Jordan algebra generated by a SIC.)

And that's pretty much it.  If one can give a better motivated argument for those wacky equations---that would be the hard part!!---then one would at least have the start of an operational theory based on SICs.  The main reason I wanted to spell it out like this, however, has to do with our conversation yesterday.  Mathematically at least, ingredient 1) above is really just a minor modification of your postulate of repeatable measurements.  As we discussed yesterday, in the worldview of ``quantum measurements revealing pre-existent properties'' the postulate of repeatable measurements is pretty well motivated.  But then I don't find that worldview tenable because of the Kochen--Specker construction (in conjunction with locality).  So the question that should be thrown back at me is:  What, then, from the alchemical worldview---that's what I'll call my view, for want of a pithier name at the moment---what from the alchemical worldview motivates this looser kind repeatability hypothesis?  {\it I don't know.}  I don't know, but at least I did say this is only the start of an answer.

Now, you may ask, the above certainly can't be all of quantum mechanics?  But it comes pretty close, and this something I'll probably need to explain to you at the board.  For instance, one might say, we'll you've only hypothesized one kind of action you can take on a system.  What are the other ones?  (The answer should be all other POVMs, but how does one get there?)  In fact the answer is not so difficult once one has this much of the framework.  One can answer:  Actions are in one to one correspondence to the refinements one can make to a general (non-boundary) $p(i)$.  And in that way, one recovers all POVMs.  I'll explain that at the board.

\section{07-10-07 \ \ {\it Pragmatism} \ \ (to J. E. {\Sipe})} \label{Sipe14}

\bjes
This seems to me the real strength of a realist approach (i.e., one in which the abstract elements of the theory refer to what actually exists in the world), even if it is something as weak as {\Putnam}'s ``internal realism.''  It gives a narrative in the strong sense (i.e., not a ``just-so story'').
\ejes

I thought what I have been talking about {\it is\/} something as weak as {\Putnam}'s ``internal realism''?  (Though I would call it powerful and liberating, rather than weak.)  It is ``pragmatism,'' and a lot of learned people out there say that internal realism is a species of the same---I think I even learned that from {\Putnam}!

\section{10-10-07 \ \ {\it The Writing Side of My Brain} \ \ (to H. R. Brown)} \label{BrownHR4}

I enjoyed the conversation today.  You've challenged me, and I hope to be able to eventually answer your questions in a convincing way.  \ldots\ At least in a way that's convincing for me, if not for you.  At the moment, I am not even to that first stage yet!

In the meantime, let me give you a link to some letters I wrote to Bas van Fraassen expressing my point of view on the word ``measurement.''  As you get a chance, I hope you will look at the letters to him in \myurl{http://www.perimeterinstitute.ca/personal/cfuchs/nSamizdat-2.pdf}.  The writing side of my brain is, I believe, a little clearer than the speaking side of my brain, and since I'd like to continue this conversation with you, I'd like to make our starting point as clean as possible.

\section{10-10-07 \ \ {\it When No One is Looking} \ \ (to H. R. Brown)} \label{BrownHR5}

Let me also give you a link to this paper of Matt Leifer's ``Conditional Density Operators and the Subjectivity of Quantum Operations'':
\quantph{0611233}. He does a better job of doing what I was trying to argue to you the other day (and a little bit today).

\section{10-10-07 \ \ {\it Here, but Delayed}\ \ \ (to G. L. Comer)} \label{Comer109}

\bgc
I mean you talk about Hilbert spaces and the stuff that lives in these spaces; but don't you also have to discuss how things ``move'', ``evolve'', etc.\ in this space?  Surely an electromagnetic field ``state'' vector, or density operator, must take into account the (at least) $\infty^4$ values that the QM ``thingy'' has to map to?  Doesn't the light in my room, right now, vary from point to point, as well as varying at each point at different times?  Or stated differently, don't I have the freedom to perform measurements at any point of my room at any time?  Don't I have to see Hilbert space in a fiber bundle sense, where spacetime is the base manifold?
\egc

My short answer to your question is that:  I think the {\it formal\/} structure of quantum mechanics has nothing to do with space, time, or spacetime.  And, for instance, quantum entanglement has nothing per se to do with those concepts either.  But I emphasize the words ``formal structure.''  A particular application of quantum theory may be a question to do with spacetime, and that is OK.  But you want a longer answer, with some argumentation justifying my point, and applied to something particular, like a neutron star, say, and for that I'm just running dry at the moment.  Please forgive me!  Eventually I'll counter-rant.

\section{11-10-07 \ \ {\it Modernizing {\James}} \ \ (to J. E. {\Sipe})} \label{Sipe15}

\bjes
Hmmm\ldots. OK, but internal realism does allow for narratives.
Pragmatism, at least in a certain sense, does too.  I don't see how
the Bayesian approach to quantum mechanics that you and co-workers are
developing will result in a narrative for the universe.
\ejes

Well, somehow I see our q-Bayesian program as a beautiful example (and better justified species, in fact) of {\James}ian pragmatism!  Do a search in this document on the word ``James''.  Some of the stuff in the XX instances that pulls up might surprise you.

The narrative is that the world is a creative, plurality.  That is what I want ultimately to come out of the q-Bayesian program.  Two sketches of a manifesto below.  [See 17-06-04 note titled ``\myref{Mabuchi12}{Preamble}'' to H. Mabuchi and 12-02-07 note titled ``\myref{Hardy20}{Lord Zanzibar}'' to J. Christian and L. Hardy.]

\section{11-10-07 \ \ {\it Hi!}\ \ \ (to R. E. Slusher)} \label{Slusher20}

I've been missing you too.  Particularly as I've been plunging into trying to refine the idea that what is ontic of a quantum system is that it is a catalyst.  It is a little core that transforms the world external to itself.  Your grimaces when I say these silly things would be much appreciated---you helped keep me on the straight and narrow!  I would have written you sooner, but a) you never gave me your email address, and b) despite my looking, I could never find it on the web!  I'm glad you finally sent it to me.

Here are three of my recent talks that you might enjoy:
\begin{itemize}
\item \pirsa{07080042}

\item \pirsa{07080043}

\item \pirsa{07090068}
\end{itemize}

I hope Georgia's ultimately living up to your expectations, and that you and your wife are both happy there. Drop me more notes as you get a chance.

\section{12-10-07 \ \ {\it Critchley} \ \ (to myself)} \label{FuchsC17}

\begin{itemize}
\item
Frank Critchley, Paul Marriott, and Mark Salmon, ``On Preferred Point Geometry in Statistics,'' J. Stat.\ Planning \& Inference {\bf 102}, 229--245 (2002).
\end{itemize}

\section{18-10-07 \ \ {\it Copyrights?}\ \ \ (to N. D. {\Mermin})} \label{Mermin133}

Good to hear from you!  I was just thinking about you today.  I'll tell you why, but first:  I'm glad to hear about your recovery from knee surgery.  I never imagined you were going through that.

Here's why I was thinking about you.  Do you think one can get a noncolorable set (in the Kochen--Specker sense) of quantum states ALL of which are maximally entangled states on some bipartite system?

A consistent set of histories (in the Griffiths sense) is {\it just\/} a POVM (the elements being the histories).  (Not an answer to your question, but something I felt like saying.)  So all of the hullabaloo G makes over consistent sets is simply to prefer certain kinds of POVM.

I'm glad to hear you'll come visit us, and I'm glad I've got that in writing!

\section{18-10-07 \ \ {\it Noncolorable Histories} \ \ (to N. D. {\Mermin})} \label{Mermin134}

\bdm
OK, I see what you mean.   But you would never say that the results of measuring a POVM
were there (modulo a framework), whether or not the information was actually acquired.
So I take it as a purely formal remark that, however, misses what is essential for Griffiths.
\edm

Well, I suppose my take on it is that Griffiths is just deluded.  The histories formalism is full of sound and fury, but in the end signifies nothing.  That thing which is ``essential'' for Bob is, in the end, just of no consequence for quantum interpretation.  For me the essential question is whether there are ``noncolorable'' sets of consistent sets of histories.  Sets of sets.  Essentially the KS question, just now applied to histories.  If so, there is no good sense in which one can even say that within a consistent set one history is TRUE (using Bob's way of writing it).  TRUE is only relative to the consistent set.  And what determines the consistent set under attention?  The agent---nothing more and nothing less than that which sets an orthonormal basis in the usual formulation of quantum mechanics.  It is just that the agent now is turning his attention to the class of POVMs that Bob prefers to single out (i.e., sets of ``histories'') rather than the POVMs that Bohr was concerned with (i.e., sets of orthonormal bases).

I tried to say this better in \quantph{0105039}, in the single (long) letter there to Griffiths, starting on page 178.  Looking back at it, it is a horrible job, but maybe I still massaged some issues the right way there.

On your more technical question, yes I think one can generalize consistent sets to sequences of POVMs with a L\"uders collapse rule in between.  {\Carl} {\Caves} did it once upon a time---as I recall---but never published, and then I think someone else published something similar.  I'll try to find that reference for you tomorrow.

So good to have you back!  Keep thinking about the new KS question I asked you.  I am hoping it is useful for readdressing the issue of the objectivity of unitaries.

\section{18-10-07 \ \ {\it Digging Up Bones} \ \ (to N. D. {\Mermin})} \label{Mermin135}

In case it is useful to you, here is Bob's\index{Griffiths, Robert B.} reply to that old letter of mine.  I really had to dig to pull this one up in my system.  It was dated June 13, 1997.

Most relevant to your present query are these words of Bob's:
\bq
   Attempts to work out consistency conditions for POVMs has been
   attempted by Oliver Rudolph in Hamburg.  His earlier work was enormously
   complicated and the results uninteresting.  He claims to have made further
   progress, but I haven't looked at it.  My own opinion is that POVMs (and
   density matrices) represent calculational tools which can certainly be useful
   for certain purposes; however, I don't see any reason to employ them as ``basic
   events'' in a sample space, and hence no point in working out consistency
   conditions for them.
\eq

Have fun in Oregon.  I think I give a colloquium in Eugene in February (either before or after Australia).  I go to China for the first time Nov 1.  When I first went to Japan you told me not to eat the fugu.  Any advice for China?

\section{24-10-07 \ \ {\it Maroney} \ \ (to C. M. {\Caves})} \label{Caves96.2}

I was a little impressed by some of his remarks the other day when Jon Barrett gave us a talk on whether it is possible to give a many-worlds interpretation to all (or a significant subclass, including classical probability theory) of the generalized probabilistic models he and Matt and Howard and Alex have been playing with.  Jon's point being that if such could be done, then it would lend some credence to ``Fuchs's and others'\.'' intuition that the many-worlds interpretation is actually contentless, saying nothing specific to quantum mechanics.  Owen sided with Jon somewhat, pointing out the sense in which one might say there is a ``measurement problem'' in classical statistical mechanics (if, I would say, one takes the wrong point of view).  Still, it was a trenchant remark.

\section{24-10-07 \ \ {\it A SIC Idea} \ \ (to C. M. {\Caves})} \label{Caves96.3}

\bcc
I was discussing SIC-POVMs with Andrew Scott and Nick Menicucci in a desultory fashion, but came up with the following idea.

I think you should organize at PI a two-week workshop on SICs.  The idea would be to bring together everyone who has worked seriously on the topic and to throw them together for two weeks (with no talks) to see if the problem will yield.  If it doesn't, everyone can go home and forget about it, without having to destroy the rest of their careers working on it.
\ecc

I think that is an excellent idea and will very likely follow through.

\section{30-10-07 \ \ {\it Talk Still?}\ \ \ (to J.-{\AA} Larsson)} \label{Larsson4}

\bjal
I would like to spend {\bf some} time before arriving reading and thinking about SICs.
\ejal

OK, I'll change the schedule then.  Of course, the {\it best\/} single paper to look at is \arxiv{0707.2071}.  A whole lot of interesting things happened while Appleby was visiting.  For instance, he classified all Weyl--Heisenberg SICs in~$d=3$. Lane Hughston has also thrown in looking at algebraic geometry aspects in~$d=3$.  He'll give a talk on that today, and in a couple of days, you should be able to find it on PIRSA if you want to see it.  [See \pirsa{07100040}.]  Steve Flammia has had some good number theoretic ideas too (which would  be a little difficult for me to explain).  And me, when I'm not learning from everyone else, I've just been plodding along trying to see if my reduction in equations to~$3d$ (instead of the original $d^2$) doesn't fail in~$d=24$.  There is a fear in me, since a related conjecture explored by Joe Renes fails there.  Any idea how to make the attached {\sl Mathematica\/} code ``Old Notebook'' more efficient?  Basically all I'm trying to do is minimize the function {\tt EC} and then check that the vector that minimizes it makes the matrices {\tt TT} and {\tt SS} look right.  (The matrix {\tt SS}, for instance, should have 1 in the top left entry and $1/(d+1)$ in all the rest.)  The first cell just basically finds me a good random seed, and then in the second cell I use that seed to verify that my vector has all the properties it should have.  Anyway, this method can be pushed to work all the way up to about~$d=18$ if one has patience.  But there's probably a much better way to do it that would squeak out at least 10 more dimensions.  For instance, Renes's notebook which checks the other conjecture pushes out to~$d=23$ without problem.  On the other hand, I've been too stupid to modify mine to work like his (I just cannot figure what's wrong), and Renes himself has gone missing!

Anyway, so glad you're interested in thinking about SICs!  (The ``official'' pronunciation now being Sikhs \ldots\ also to connote their religious aspect.)

\section{05-11-07 \ \ {\it Holevo Hirota} \ \ (to D. Gottesman)} \label{Gottesman7}

I just got this letter from Holevo this morning,
\bq\noindent
If I am correct, next June Hirota will be 60. This is an important age, notably in Japan. I am not sure how better to mark this date, perhaps with a collection of papers in a dedicated  journal?  What is your opinion, perhaps you have better or more concrete suggestions. In any case I would be happy to support any reasonable initiative.
\eq
and it creates in me the following idea.  (I'm trying to be creative, not having it in me to be an editor or any similar thing at the moment.)  What would you think about PI hosting a small meeting (birthday party) in honor of Hirota for his 60th birthday?  He has been an indefatigable defender of quantum information in Japan, and has brought significant funding to the field for the QCMC conferences, and supported the likes of Holevo in Russia's troubled financial times.  Anyway, I'm just thinking something small (like 10 people or less) composed of some of his best scientific friends, like Holevo, Bennett, Yuen, etc.  In fact it would be a good excuse for finally getting Holevo to visit us, and similarly with Charlie.  And maybe we could even unearth Helstrom, etc.

What do you think?  Does PI have an easy mechanism (and a small budget) for making such a thing happen?

\section{07-11-07 \ \ {\it Listening in Beijing} \ \ (to J. W. Nicholson)} \label{Nicholson27}

After hearing of your eating the fugu, I felt a new urgency to listen to your music.  That first one (i.e., two) is a lot of fun!  It sort of put me in the same mood as when I was watching one of those old, cheesy light-hearted Italian flicks, the kind one used to find on TV in the early days of HBO.  I don't know how to articulate the feeling better than that.  The second seems to represent your {\sl Rubber Soul\/} transition.  How do you hold so many styles in your head, man?  I only got one, and it alone gets burdensome at times.

Now I be off to the office.  I'm gonna crack this problem of quantum mechanics, I tell you.  Visiting the Temple of Heaven yesterday and the Chinese apothecaries, and thinking about their practices at both, gave me renewed strength to think about alchemy and the analogy in it of quantum state preparation and measurement.  When I can make the point precise, people will listen.

\ldots\ OK, I probably shouldn't have listened to that {\sl Rubber Soul\/} stuff so early in the morning.

\section{08-11-07 \ \ {\it Hirota's Birthday}\ \ \ (to A. S. Holevo)} \label{Holevo8}

\bash
If I am correct, next June Hirota will be 60. This is an important age, notably in Japan. I am not sure how better to mark this date, perhaps with a collection of papers in a dedicated  journal?  What is your opinion, perhaps you have better or more concrete suggestions. In any case I would be happy to support any reasonable initiative.
\eash

I'm not opposed to your suggestion---I think it is a good idea---but I just don't have the stomach myself to be an editor (or co-editor) of yet another special issue for a while.  On the other hand, if you would like to pursue an issue as editor, I'm quite sure that QIC ({\sl Quantum Information and Computation}, the journal I am associated with) would be more than happy to have you edit a special issue, and I think I could easily help facilitate that for you.

But here's another idea that could be concurrent with that, or separate from it, either way, that I'd be happy to pursue personally.  How about a small scientific ``birthday party'' for Osamu at PI?  The sort of thing I am thinking is a small meeting of maybe 10 or so colleagues, some of Osamu's best friends or greatest influences, ``Osamu Hirota, between Distinguishability and Noncausality''.  For instance, you, Charlie Bennett, Yuen, and some others?  You and I could be co-organizers, but the administrative staff would take care of almost all the mundane affairs of the meeting like travel details for the participants, etc.  And we would have a pleasant banquet with some nice wine in PI's Bistro.  Our main duty would be in choosing the participants and contacting them with invitations, etc.

What do you think of such an idea?  I have already contacted Daniel Gottesman and he was happy with the idea of something like that (I thought it best to get his opinion before pursuing things further).  If it is a kind of thing you too find worth pursuing, I will write an official proposal for PI.  If the conference committee agrees, then we should be able to cover most (if not all) of our participants' expenses.  Let me know what you think, and if you are in agreement, also send me a list of suggested participants so that we might starting thinking about how to structure my proposal.

I am in Beijing, by the way, and my connection into PI's intranet is so slow, that I'll probably have to wait until getting back next week before pursuing anything further.  But it'd be nice to get your commentary before then.

\section{08-11-07 \ \ {\it Uncle Fierz Needs You} \ \ (to H. C. von Baeyer)} \label{Baeyer28}

Greetings from Beijing!  I have been meaning to write you for ages, but one thing would always lead to another, and I would always end up remembering just at a time when I was not at my computer.  Anyway, having a long, contemplative walk around the Temple of Heaven yesterday and visiting a Chinese apothecary apparently made my mind disciplined enough that I'm writing you now.

The subject is, I would very much like you to give us a visit at PI, and when you're not just imbibing the excitement and atmosphere, use the time to start hitting your Pauli--Fierz project in a detailed way.  By being together physically for discussions, I think it could be a great experience for both of us.  More than ever I feel like I've got to get my understanding of quantum measurement into shape---measurement as release of the stone---and the best starting point, I am really convinced, must be the Pauli -- Fierz -- Jung -- von Franz discussions.

I have some visitor funds I need to spend before the end of December, and so I think that is the perfect spur for trying to get together for some long conversations and scholarship.  I know this is short notice, but I hope you can come (and be interested in coming for the project I propose!).  I would be able to pay your travel expenses, put you up in a PI apartment, and provide a per diem for your meals.  Come 1 week, 2 weeks, 3 weeks, or even 4, whatever amount of time you can afford.  The more the better, the more we could get done.

So please do think about it.  If it helps, have a look at a typical dinner menu for the Bistro and its wine list
and browse the cultural venue and maybe take a virtual tour of the facilities.  Also take note of the Friday afternoon wine and cheese socials, and that's not even to mention all the physics, physics, physics that'll be all around you all the time.

\begin{itemize}
\item
W.~Heisenberg, ``Wolfgang Pauli's Philosophical Outlook,'' in his book {\sl Across the Frontiers}, translated by P.~Heath, (Harper \& Row, New York, 1974), pp.~30--38.
\end{itemize}
\bq
The elaboration of Plato's thought had led, in neo-Platonism and Christianity, to a position where matter was characterized as void of Ideas.  Hence, since the intelligible was identical with the good, matter was identified as evil.  But in the new science the world-soul was finally replaced by the abstract mathematical law of nature.
Against this one-sidedly spiritualizing tendency the alchemistical philosophy, championed here by Fludd, represents a certain counterpoise.  In the alchemistic view ``there dwells in matter a spirit awaiting release.'' The alchemist in his laboratory is constantly involved in nature's course, in such wise that the real or supposed chemical reactions in the retort are mystically identified with the psychic processes in himself, and are called by the same names.  The release of the substance by the man who transmutes it, which culminates in the production of the philosopher's stone, is seen by the alchemist, in light of the mystical correspondence of macrocosmos and microcosmos, as identical with the saving transformation of the man by the work, which succeeds only `Deo concedente.'\,''  The governing symbol for this magical view of nature is the quaternary number, the so-called ``tetractys'' of the Pythagoreans, which is put together out of two polarities.  The division is correlated with the dark side of the world (matter, the Devil), and the magical view of nature also embraces this dark region.
\eq

\section{11-11-07 \ \ {\it A Question on Gleason's Theorem for SICs} \ \ (to Z. Ji)} \label{Ji1}

Yes, maybe you do not understand the statement of the problem.  For, as you state it below, the standard Gleason theorem would not be correct either---for that theorem too can be posed solely in terms of rank-1 operators alone.  Then your argument would have gone:
\bq\noindent
     The point is, von Neumann measurements consist of rank-one operators only, the function could
     be $f(A) = \mbox{rank}(A)/d$, and $\sum f(\Pi_k) = 1$ will trivially hold for all $\{\Pi_k\}$ being an orthonormal
     basis. But the rank function is not linear and cannot be written as required.
\eq

The key is that I am only considering functions whose domain of definition is on the rank-1 projectors.  After one gets a characterization of them, perhaps one will want to extend their domain (and it will be doable uniquely if they turn out to be linear), but at the outset the domain of definition is the set of rank-1 projectors.

The statement is:  Let $d\ge 3$.  Suppose $f$ is any given function from the rank-1 projectors to the range $[0,1]$  with the property that $\sum f(\Pi_k) = 1$ whenever the $\{\Pi_k\}$ form a SIC.  Then can one show that there always exists a density operator $\rho$ such that $f(\Pi)= (1/d) \tr(\rho\Pi)$?

About your example, when restricted to the proper domain of definition there is indeed a density operator with the desired property, namely the maximally mixed state.  And then that function $f$ will have a unique linear extension to other operators.

Perhaps it would be helpful for you to look at this old paper of ours, where we considered a precursor question, but with qubits:  \quantph{0306179}.  The section relevant to what we are talking about above is Section IV.

\section{12-11-07 \ \ {\it Autumn in Beijing} \ \ (to D. B. L. Baker)} \label{Baker16}

One day last week, as I was walking through a park watching the leaves fall and the Chinese children play with them---a sentimental scene from every year and every place---I thought I would write you a note with this title.  Imagine Frank singing that.  Of course, I thought I would write the note to you that very day, but as of now I'm making my way from Hong Kong to Anchorage.  That's the way I write now, always in delayed mode.  [See 03-02-06 note ``\myref{Baker14}{Red Spears}'' to D. B. L. Baker.]

I hope you're doing well.  I thought you might enjoy a few pictures I took during my stay in Beijing.  For the first time in my life this trip, I actually took a camera with me.  Pay close attention to what's in the right hand of the Chairman Mao talisman.  After leaving that market, and thinking about it a little, I decided I could kick myself for not trying to buy that as a little decoration for my new house.  How many opportunities I've missed in life!
%

\section{27-11-07 \ \ {\it PI QI Group Meeting Today (Yesterday)}\ \ \ (to M. Mosca \& L. Sheridan)} \label{Mosca1} \label{Sheridan1}

Here a couple of those old van Enk papers I was talking about:
\begin{itemize}
\item
\quantph{0207142}
\item
\quantph{0410083}
\end{itemize}

It would be good to get together and talk about issues to do with ``self-calibration.''  I'd like to see how it fits in my worldview that ``quantum operations are Bayesian too''.  You probably have no clue what I'm talking about, but you can read about it here (in the form of debate with Caves, Mermin, Schack, etc.): [some unknown pages in this samizdat].  The banter there still remains better than any of the more formal stuff I've written.

By the way, what does research on ``the fundamental object of conscious existence'' mean?

\section{27-11-07 \ \ {\it Tables of Contents}\ \ \ (to M. Mosca \& L. Sheridan)} \label{Mosca2} \label{Sheridan2}

\ldots\ just thinking again about yesterday's talk.  For fun I just looked in both Nielsen and Chuang and David Mermin's books on quantum computing.  I did an index search on ``energy conservation'' and ``conservation laws''.  Didn't find anything in either book on either topic.

\section{29-11-07 \ \ {\it Adam's Conversation}\ \ \ (to J.-{\AA} Larsson)} \label{Larsson5}

See page 83, note titled ``I See Why Bit Commitment,'' in:
\begin{center}
\arxiv{quant-ph/0105039}.
\end{center}

\section{29-11-07 \ \ {\it Much Better} \ \ (to W. G. {\Demopoulos})} \label{Demopoulos19}

I just finished reading the final version of your paper and actually feel better about it this time around.  I'm not sure what you changed, but I remember being bothered by more aspects before \ldots\ and today I detected almost nothing that got in the way of my appreciation.  I will forward on the paper to John Sipe, for I think it clarifies some aspects of my ongoing debate with him.  I will also forward it on to {\Carl} {\Caves} because I think page 15's discussion of indeterminacy exposes some dangers in his ways of thinking about the timeless properties of quantum systems.

Where I do take issue with what you write is the very difficult issue of probability 1.  If I understand you correctly, I disagree with the first sentence of the first full paragraph from page 22:
\bq\noindent
The representation of an elementary particle as a function which, when
presented with an experimental configuration, yields an effect, is
interchangeable with its representation as a class of propositions only when the effects are predictable with 0-1 probability.
\eq
For, I would say, not even then.  That is the point of the attached paper, \quantph{0608190}, which you didn't seem to get the point of the last time I sent it to you.  It is also the point of my discussions with Bas van Fraassen on pages 411 to 417 (and in much greater detail elsewhere in the collection) of {\tt nSamizdat-2.pdf}.

\section{30-11-07 \ \ {\it Relationalism vs Pragmatism at PI} \ \ (to B. C. van Fraassen)} \label{vanFraassen16}

I've been meaning to write you for a long time, and the note below from my colleague Nielson on Perimeter Institute's sabbatical program has helped spur me to finally take action.  (Our last conversation was interrupted by the great April Nor'easter floods in New Jersey; my basement took 23,000 gallons of water just as I put my house on the market---you were at the Bubfest that very same weekend.  That on top of the market bubble bursting led to a lot of pain between then and now.  \ldots\ Anyway, that's my excuse for the silence.)

But I'm established in Waterloo now, and I'd like to increase the rate of our understanding of what a relational view of QM is all about.  Thus I had been meaning to invite you for some visits if you've got the time.  For short terms, I could pay all your expenses from my visitors budget.  But something like the sabbatical program below would be much better still!

In case you're not familiar with PI, I'd say, take a look at a typical dinner menu for the Bistro and its wine list at [\ldots], browse the cultural venue at [\ldots], and maybe take a virtual tour of the facilities at [\ldots].  Also take note of the Friday afternoon wine and cheese socials.  Beyond that, of course, are all the people.  With regard to philosophy Harvey Brown will be here for the coming year, and there's everyone in London and Toronto.  So, I think you could have some good fun.

Think about both propositions and let me know.

\section{05-12-07 \ \ {\it The Misak Book} \ \ (to R. W. {\Spekkens})} \label{Spekkens48}

\brws
I saw this book while browsing the Oxford University Press website, and thought that you might be interested.
\bq\noindent{\rm
{\bf New Pragmatists}\smallskip\\
Edited by Cheryl Misak, University of Toronto\smallskip\\
Pragmatism is the view that our philosophical concepts
must be connected to our practices --- philosophy must
stay connected to first-order inquiry, to real examples,
to real-life expertise. The classical pragmatists, Charles
Sanders Peirce, William James, and John Dewey, put
forward views of truth, rationality, and morality that
they took to be connected to, and good for, our
practices of inquiry and deliberation. In this volume,
some of our very best contemporary philosophers
explore this and develop the pragmatist project,
showing that pragmatism is a strong current in
philosophy today.\smallskip\\
Contributors: Jeffrey Stout, Ian Hacking, Arthur
Fine, Cheryl Misak, Huw Price, David Macarthur, David
Bakhurst, Terry Pinkard, Danielle Macbeth.\smallskip\\
March 2007 | 208 pages | Clarendon Press\smallskip\\
978-0-19-927997-5, HARDBACK \pounds25.00/\$45.00}
\eq
\erws

Thanks for pointing the book out:  It definitely needs to be in the collection.  (Look at the beautiful bookshelves, in fact, that I just had built in my house. PICT0380 particularly, captures the part of the library devoted to pragmatism.)  Also, it was nice to learn from the advert that Misak is in Toronto---I hadn't realized that.

\section{07-12-07 \ \ {\it Campbell Soup Cans}\ \ \ (to G. L. Comer)} \label{Comer110}

Well, I'm pretty sure Leonard Susskind just thinks I'm a crank.  But that's OK, I'm quite sure that most everything he spouted that used the word ``information'' was just nonsense.  He had no clue about the concept.  And yet the sort of arrogance that \ldots\  Well, I could go on.

Anyway, the performance went well by all accounts.  We had the audience laughing throughout, and that was a good indicator that things were on the right track.  I even drew a round of applause somewhere in the first 10 minutes when I pinned Seth on his idea that the universe is a computer computing itself.  I asked him if the answer would be 42.

Frankly, I did not have fun.  But I don't think you'd be able to tell it by looking at me.  (The discussion should be on PIRSA within a week,\footnote{\editornote See \pirsa{07120048}.} I guess.)  It was grueling and the idea of 500,000 listeners being out there really did unnerve me at first.  But I was genuinely surprised by all of the praise I got after it and the next day from my colleagues (and one of the postdocs' girlfriend, who is not a physicist).  Two postdocs were so enthusiastic, they told me I ``clearly won'' the debate.  Who knows.   But I guess, really, the main thing that turned me off was the forum---the idea of an adversarial presentation, with all its distractions from making an effective point, etc.  I think if given the chance, I wouldn't so mind making another appearance in this public lecture series, but I'd like it on my own terms:  You know, using my transparencies with stick figures.  You see, Seth was able to say, ``The world is a computer.''  And Leonard was able to say, ``The world is a hologram.''  Two noun-verb-noun sentences.  In my heart, I wanted to get the idea of a noun-verb-adjective sentence across.   That, ``The world is malleable.''  But there was just no real way to approach that in such a forum, and particularly not with the personalities I was up against.

\section{10-12-07 \ \ {\it SICs, FQXi, \& the Way Chris Thinks} \ \ (to D. M. {\Appleby})} \label{Appleby24}

\bma
Are you available to talk?  I am getting really hung up on this 500 word summary.  My mind has got jammed.  I don't like what I have written so far.  But I can't think of anything to put it in its place.  And I badly want to get back to SICs.  Perhaps you could help me to unblock myself.
\ema

\bma
The proposed research will form part of a larger project, to develop a novel way of looking at quantum mechanics.  Classical physics is based on the correspondence theory of truth:  it is supposed that each symbol in the theory is the token for an element of reality (for instance, the symbol E, as it occurs on paper, is the token for an actually existent object, the electric field vector). Wave function collapse makes it very difficult to reconcile that idea with quantum mechanics.  However that has not stopped people trying.  In fact most work in the foundations of quantum mechanics consists in a variety of ingenious attempts to preserve the classical assumption, that the constructions of mathematical physics in general, and the quantum state in particular, should be conceived as being in one-one correspondence with mind-independent realities.  It appears to us, however, that the real message of quantum mechanics is that we need to abandon this rather simple-minded way of thinking about physical theories in favour of something richer, and subtler, and potentially
much more fruitful.   Instead of thinking of the quantum state as a
peculiar kind of physical object, we take it to have a fundamentally probabilistic significance.  Furthermore we take an epistemic, or Bayesian view of probability.  So although the quantum state encapsulates our expectations regarding the world, and to that extent makes a statement about the world, it is not the depiction of something in the world (it cannot be such a depiction because of the way it changes discontinuously consequent on learning a measurement outcome). Preliminary work along these lines has been reported in previous publications by two of the investigators.  However, a great deal more remains to be done. {\rm [\ldots]}
\ema

I saw two movies this week, one titled ``Dr.\ Goldfoot and the Bikini Machine'' and one titled, ``Dr.\ Goldfoot and the Girl Bombs.''  The first (1965) was starred by Vincent Price and Frankie Avalon; the second (1966) was starred by Price and the singer Fabian.  Anyway, the fun thing was that Avalon and Fabian were employed by the ``Secret Intelligence Command'' and were known as SIC-men.

I'm tinkering on your draft to see how it would come out in my own voice. I'll send you the result tomorrow. You're under no obligation to keep any of those changes; I'm just doing it to give further ideas, and perhaps making the hard right turn a little more dramatic.  \ldots\
\ldots\ \ldots\

Below is the result of my tinkering with your draft proposal.  I'm sorry to get this back to you so late, but the quantum gravity meeting actually turned out to be interesting and distracting!  (Have any doubts that SICs are important?  Patrick Hayden today presented a proposed toy-model solution to the so-called black-hole information paradox by him and John Preskill that made use of 2-designs on finite dimensional Hilbert spaces.  He said, ``imagine the black hole enacts a random unitary.'' I said, ``Imagine instead a demon in the hole that measures a SIC and outputs a quantum state corresponding to his result.''  The two expressions are equivalent.)

Anyway, the passages below are what your draft inspired me to write.  They may be indicative of nothing but my own take on the deeper reasons for the research.  I tried to express that as best I could.  The count in this version stands at 556 words.  The main thing I tried to capture a little better was the hard right turn we talked about on the phone the other night.  What I really wanted to say is that our understanding of QM is really doomed without SICs.  And that what we're sikhing is a true bending of our worldview.  I wanted to say to FQXi, ``you're really called for here!''  I don't know that I really succeeded at those goals, but that's the sort of thing that was on my mind, as I modified emphasis away from density operators and made it a little more toward probabilities.

I'll send it to you now without tinkering further.  By the way, I didn't incorporate any of Lane's\index{Hughston, Lane P.} last suggestions, as I didn't know what they were when I started this project.  Anyway, as I told you last night, feel free to completely trash this version.  I have no emotional attachment (mostly it was an exercise in clarifying my own thinking), and ultimately want whatever will turn out with hindsight to have been the best move!  I.e., I want you to win at all costs!

\bq
The proposed research forms part of a larger project to develop a novel way of looking at quantum mechanics.  Classical physics is based on the correspondence theory of truth:  It is supposed that each symbol in the theory is a token for an element of reality.  Wave-function collapse upon the acquisition of information, however, makes it difficult to reconcile this idea with quantum mechanics.  Nonetheless, this has not stopped many researchers from trying---much work in quantum foundations consists in a variety of ingenious attempts to preserve the classical assumption that quantum states should be conceived as being in one-one correspondence with mind-independent realities.  In contrast, it appears to us that the real message of quantum mechanics is that we need to abandon this way of thinking in favour of something richer, subtler, and potentially of great significance for our worldview.  Instead of thinking of quantum states as peculiar kinds of physical objects, we take them to have a fundamentally epistemic significance, particularly along Bayesian lines.  So although quantum states encapsulate our expectations regarding the world, and to that extent make statements about the world itself, they are not the depiction of something in the external world. Preliminary work along these lines has been reported in a number of earlier publications by the investigators.

The deepest development of this idea, however, is significantly impeded by the present formulation of quantum mechanics.  In fact, from our perspective, the usual Hilbert space formulation of quantum mechanics is a good part of the whole interpretational problem, as it insinuates a fundamental distinction between quantum states and the probability assignments they entail.  Thus we seek a mathematically elegant formulation of quantum mechanics in which the main objects are probability assignments, rather than quantum states.  This calls for a deeper understanding and classification of the operator bases allowed within the set of quantum states.  Based on several arguments, we believe the so-called ``symmetric informationally complete'' sets of states (SICs) are the most fundamental such bases and thus should be our starting point.  There is only one catch:  Do SICs exist?  As it turns out, this is an extremely nontrivial mathematical problem, with an origin 35 years ago in coding theory.  Recently though, much progress has been made on the issue, with deep connections to other areas of quantum information theory and algebraic geometry---so much so that with concerted effort, one imagines light at the end of the tunnel.  For instance, SICs are now known (numerically) to exist in Hilbert spaces up to dimension 45, and analytic constructions have been given up to dimension 12.

The first phase of our project is to settle the existence question and to acquire much needed mathematical insight into the geometrical and group-theoretical structures exhibited by SICs. The second phase will involve a number of applications, particularly getting back to the motivational roots of this problem, as well as issues in quantum tomography, cryptography, and a variety of other problems in quantum measurement theory.  The Investigators all have well-established track records for innovative work in quantum theory, and have variously undertaken a number of successful preliminary investigations that are relevant to the proposed research. As a consequence we feel our chance of success is high, and even if only partial success is achieved the results obtained would still amply justify the investment.
\eq

\section{11-12-07 \ \ {\it Last Night -- Seminar -- Teleportation}\ \ \ (to M. Knechtel)} \label{Knechtel1}

Thanks for the interest.

Yes, you do have to have a ``stack of raw building blocks'' at both ends of the teleportation scheme.  For instance, if one wants to teleport the quantum state of an electron, then one needs 1) the original electron situated near Alice, 2) a ``raw'' electron also situated near her, and 3) finally a further raw electron situated near Bob (that is ``entangled'' with Alice's raw electron).  At the end of the protocol, the quantum state that was associated with the original electron will now be associated with the electron in Bob's possession.

What this means quite literally is that any information one had about the original electron is now relevant to the other electron instead.  It's not nearly as exciting as the word ``teleportation'' might have indicated.  ``My information is no longer about this, but instead about this other thing far away.''

Hope that helps.

\subsection{Marguarite's Preply}

\bq
Good round table last night.  Thank you for your input and your humour.   I just wanted to clarify teleportation. It sounded to me like the only way teleportation works is if there is, for example, a stack of building blocks at this end resembling a structure of some sort and a stack of building blocks in a pile at a distance.  The structure does not teleport from one place to another, but the information about how to assemble the building blocks at the other end, or how the original structure is built, is what gets transported and then the building blocks can be formed into a similar looking structure. Is that correct?  My brain needs to process info in examples I am familiar with.
\eq

\section{11-12-07 \ \ {\it Any Comments on This as a Proposal?}\ \ \ (to L. Hardy)} \label{Hardy23}

It's almost 500 words, the question is even at this stage, how do you think it would fly for Marcus?

\bq
The proposed research forms part of a larger project to develop a novel way of looking at quantum mechanics.  Classical physics is based on the correspondence theory of truth:  It is supposed that each symbol in the theory is a token for an element of reality.  Wave-function collapse upon the acquisition of information, however, makes it difficult to reconcile this idea with quantum mechanics.  Nonetheless, this has not stopped many researchers from trying---much work in quantum foundations consists in a variety of ingenious attempts to preserve the classical assumption that quantum states should be conceived as being in one-one correspondence with mind-independent realities.  In contrast, it appears to us that the real message of quantum mechanics is that we need to abandon this way of thinking in favor of something richer, subtler, and potentially of great significance for our worldview.  Instead of thinking of quantum states as peculiar kinds of physical objects, we take them to have a fundamentally epistemic significance, particularly along Bayesian lines.  So although quantum states encapsulate our expectations regarding the world, and to that extent make statements about the world itself, they are not the depiction of something in the external world. Preliminary work along these lines has been reported in a number of earlier publications by the investigators.

The deepest development of this idea, however, is significantly impeded by the present formulation of quantum mechanics.  In fact, from our perspective, the usual Hilbert space formulation of quantum mechanics is a good part of the whole interpretational problem, as it insinuates a fundamental distinction between quantum states and the probability assignments they entail.  Thus we seek a mathematically elegant formulation of quantum mechanics in which the main objects are probability assignments, rather than quantum states.  This calls for a deeper understanding and classification of the operator bases allowed within the set of quantum states.  Based on several arguments, we believe the so-called ``symmetric informationally complete'' sets of states (SICs) are the most fundamental such bases and thus should be our starting point.  There is only one catch:  Do SICs exist?  As it turns out, this is an extremely nontrivial mathematical problem, with an origin 35 years ago in coding theory.  Recently though, much progress has been made on the issue, with deep connections to other areas of quantum information theory and algebraic geometry---so much so that with concerted effort, one imagines light at the end of the tunnel.  For instance, SICs are now known (numerically) to exist in Hilbert spaces up to dimension 45, and analytic constructions have been given up to dimension 12.

The first phase of our project is to settle the existence question and to acquire much needed mathematical insight into the geometrical and group-theoretical structures exhibited by SICs. The second phase will involve a number of applications, particularly getting back to the motivational roots of this problem, as well as issues in quantum tomography, cryptography, and a variety of other problems in quantum measurement theory.  The Investigators all have well-established track records for innovative work in quantum theory, and have variously undertaken a number of successful preliminary investigations that are relevant to the proposed research. As a consequence we feel our chance of success is high, and even if only partial success is achieved the results obtained would still amply justify the investment.
\eq

\section{11-12-07 \ \ {\it Why I Care}\ \ \ (to S. T. Flammia)} \label{Flammia1}

Below is a draft of a ``500-word'' research summary that I helped Marcus write yesterday for his FQXI application.  [See 11-12-07 note ``\myref{Hardy23}{Any Comments on This as a Proposal?}''\ to L. Hardy.]  Of course I know you'll disagree with some of it (the part about light at the end of the tunnel, particularly), but maybe the first paragraph and the first half of the second paragraph will shed some light on why I actually care about this problem.  We are doing important things, even if we spin our wheels 98 percent of the time, and I hope you never lose sight of that.  I feel very lucky that I arrived at PI at the same time as you.  Seriously.

\section{12-12-07 \ \ {\it For Lane Hughston} \ \ (to myself)} \label{FuchsC18}

\bq\noindent
On the windowsill of my home on an island in Maine I keep a rock from the garden of Academe, a rock that heard the words of Plato and Aristotle as they walked and talked. Will there someday arise an equivalent to that garden where a few thoughtful colleagues will see how to put it all together and save us from the shame of not knowing ``how come the quantum''? \\
\hspace*{\fill} --- John Archibald Wheeler
\eq

\section{14-12-07 \ \ {\it No Subject} \ \ (to C. Snyder)} \label{Snyder5}

I know you'll never be able to capture it on paper the way you did in conversation, but I thought your description of your morning coin tossing contained a particularly deep idea.  Could you, would you, please, take a shot at writing it down?  I also liked that distinction---if I remember it correctly---between ``expressing yourself'' (in the form of a coin toss) and ``expressing what you think of yourself'' (in the form of an action with a predictable consequence, like pushing the pint across the table).  Say it all again!

If you're interested, you can find my description of the inveterate gambler and the nurturing-wives and bad-girlfriends penal colonies in this talk:
\pirsa{05070097}.  I think I tell the story of the Jewish mystic in that talk too (couldn't figure out the right way to spell ``moyal'', so I just called him ``mystic'').  Also, a couple of stories you might enjoy (and ones that potentially indicate why I liked your story last night) can be found in the attached pseudo-paper.

It was good seeing you again!

\subsection{Christian's Reply, 11-04-12, ``Data Sets and Who We Think We Are''}

\bq
Long time no chin wag.

I came across a piece of ephemera, a napkin to be exact, with your handwriting on it the other day. It reminded me that I have neglected (for five years) to get back to you about my thoughts on creating a random binary data set (dropping coins on the shitter) and how that informs the self. To be honest, I have nothing too earth-shattering to report (for further reading on ``not too earth-shattering'' see my reports on chicken vs.\ egg and tree in forest). That being said, through the skein of ones and zeros I have managed the kernel of an idea that might be worth some beer if you're around in the next bit.
\eq

\section{21-12-07 \ \ {\it Do You Know This Guy?}\ \ \ (to G. L. Comer)} \label{Comer111}

\myurl[https://en.wikipedia.org/wiki/Rick_Norwood]{http://en.wikipedia.org/wiki/Rick\underline{ }Norwood} ??  Ran across him while reading about Nietsche, Scho\-penhauer, and ``the will''---``the will'' being the only part of those three that I really care about.  Maybe I should title my book, in contrast to Schopenhauer, {\sl The World as Wills and Representation}.  Be sure to take note of the additional S.

\section{22-12-07 \ \ {\it Postdoc Positions}\ \ \ (to H. Price, G. Bacciagaluppi, and others)} \label{Price8}

It looks like I missed a lot of activity the day I was away from home last week.  Luckily, Lucien says I'm absolved of my sins.  Anyway, I promise to be more active during the next level of culling.  That's the real purpose of this message.

But I also wanted to comment on one line of Huw's, because it seems like a fun thing to do.  Namely, this:
\bhp
I agree with Guido that foundations and French philosophy is a very
unpromising mix.
\ehp
For, I would say, ``Ah, but if the French philosophers were Renouvier, Boutroux, and Bergson, that might just be a different story!  \ldots\ Because \ldots\ ahem \ldots\ I actually do think they're a promising mix for quantum interpretation problems.''  But alas, I looked at the guy's CV and found no Renouvier or Boutroux.  So, he fell off my list too.

\section{22-12-07 \ \ {\it One More Comment!}\ \ \ (to W. C. Myrvold)} \label{Myrvold9}

Sometimes it takes me a long time to answer an email.  But until I do, my correspondent's mail sits faithfully in my inbox with me from time to time reading over it again.  Such was this morning.  Anyway, after all this time, I find I've not much to say:  I didn't find a lot in your note that I did (or could) disagree with.  Maybe in fact my only point of departure from you would be with the penultimate sentence, where you write ``assumption -- that spatially separated systems have independent physical states --- is wrong.''  For, I would go further and say ``the assumption that systems have physical states is wrong.''  More accurately, I would say the lesson of the conjunction of EPR, Bell, and Kochen--Specker is that the assumption that systems have physical states is perhaps not wrong, but rather so limp that it's not worth propping up if one desires real progress in physics.

{\it However}, by this rewriting I only mean that the assumption that systems have {\it intrinsic dynamical\/} (i.e., time-changing) states is off track---``intrinsic'' being the most important word.  The main point about this, is that it still leaves the door open for an ultimately realist construal of quantum systems---though one, I think, along more exotic lines than the usual realist has a taste for.  For instance, I can still imagine some kind of relational account of quantum mechanics like {\Spekkens} desires as a possibility.  I myself don't think that's the right direction, but I see it as a remaining possibility.  My own favorite imagery at the moment has to do with a metaphor of  catalysts---I want to see quantum systems as something like catalysts, or when I'm feeling particularly poetic, ``philosopher's stones.''  That timeless property to enact change on that which they interact with is their only intrinsic property.  Catalyst, philosopher's stone, hyle (as that which underlies change), words like that, are an attempt to express the ontic character I toy with as underlying quantum mechanics.

By the way, for the fun of it, let me record some quotes of William {\James} that give his take on indeterminism.  I feel they capture my own take on indeterminism in quantum theory.  (So, when you see that Aristotelian word above, don't think {\Peirce}, but think {\James} \ldots\ at least in relation to me.)

\bq
[Chance] is a purely negative and relative term, giving us no information about that of which it is predicated, except that it happens to be disconnected with something else---not controlled, secured, or necessitated by other things in advance of its own actual presence. As this point is the most subtle one of the whole lecture, and at the same time the point on which all the rest hinges, I beg you to pay particular attention to it. What I say is that it tells us nothing about what a thing may be in itself to call it ``chance.'' It may be a bad thing, it may be a good thing. It may be lucidity, transparency, fitness incarnate, matching the whole system of other things, when it has once befallen, in an unimaginably perfect way.
All you mean by calling it ``chance'' is that this is not guaranteed, that it may also fall out otherwise. For the system of other things has no positive hold on the chance-thing. Its origin is in a certain fashion negative: it escapes, and says, Hands off!\ coming, when it comes, as a free gift, or not at all.

This negativeness, however, and this opacity of the chance-thing when thus considered {\it ab extra}, or from the point of view of previous things or distant things, do not preclude its having any amount of positiveness and luminosity from within, and at its own place and moment. All that its chance-character asserts about it is that there is something in it really of its own, something that is not the unconditional property of the whole. If the whole wants this property, the whole must wait till it can get it, if it be a matter of chance. That the universe may actually be a sort of joint-stock society of this sort, in which the sharers have both limited liabilities and limited powers, is of course a simple and conceivable notion.
\eq
and
\bq
The more one thinks of the matter, the more one wonders that so empty and gratuitous a hubbub as this outcry against chance should have found so great an echo in the hearts of men. It is a word which tells us absolutely nothing about what chances, or about the modus operandi of the chancing; and the use of it as a war cry shows only a temper of intellectual absolutism, a demand that the world shall be a solid block, subject to one control,---which temper, which demand, the world may not be found to gratify at all. In every outwardly verifiable and practical respect, a world in which the alternatives that now actually distract your choice were decided by pure chance would be by me absolutely undistinguished from the world in which I now live. I am, therefore, entirely willing to call it, so far as your choices go, a world of chance for me. To yourselves, it is true, those very acts of choice, which to me are so blind, opaque, and external, are the opposites of this, for you are within them and effect them. To you they appear as decisions; and decisions, for him who makes them, are altogether peculiar psychic facts. Self-luminous and self-justifying at the living moment at which they occur, they appeal to no outside moment to put its stamp upon them or make them continuous with the rest of nature. Themselves it is rather who seem to make nature continuous; and in their strange and intense function of granting consent to one possibility and withholding it from another, to transform an equivocal and double future into an unalterable and simple past.
\eq

\subsection{Wayne's Preply, from 09-02-07}

\bq
Woke up this morning and had a thought: I don't think I adequately expressed {\it why\/} I don't think that your view commits you to using probabilities only for bets whose outcome has an effect on you.

The worries about Wigner's friend and associated paradoxes, which you expressed as saying you don't find events in the density matrix, come from construing the quantum state ontologically.  But for you, a density matrix is a compendium of probabilities regarding all the experiments that might be performed on the system.  Use it to extract the probabilities of outcomes of experiments that are actually performed (whether or not the outcome affects you); the rest --- probabilities of outcomes of experiments that could have been performed, but weren't --- are irrelevant to any decisions you make.

However: that means that the response that you started to make to Brian's question about EPR-Bell correlations --- which, I think, was that as far as you are concerned the correlations don't exist until you compare the results (at which time the experiments are all in your backwards light cone) --- is not available to you.  Which is good, I think, because it permits you to use such correlations as clues to what the world outside the purple line is like.

We associate with EPR-Bell experimental setups certain credences about the outcomes of the measurements.  These are credences that we, as Bayesians, have learned through experience (by conditionalizing on the results of experiments) are more appropriate than certain other credences (such as those yielded by the toy theories that Bell constructs in his original paper) for such setups.

These credences have an interesting feature: they can't be mixtures of probability assignments on which the outcomes at the two ends are probabilistically independent.   That is, these correlations are different from classical correlations.   We can imagine a world in which the experiments had turned out differently --- in which classical correlations sufficed.  But we don't live in such a world.

The degrees of belief that we learn, through experience, to associate with a given physical situation, tell us something about that situation (that is the lesson, I think, of the two islands).  I think that the conclusion we ought to draw from EPR-Bell experiments is that Einstein's separability assumption --- that spatially separated systems have independent physical states --- is wrong.  This is perfectly compatible with an absence of action-at-a-distance, and compatible with special relativity. It is of course not the only possible conclusion, but it's the one that I think is best supported by all available empirical evidence.
\eq

\section{27-12-07 \ \ {\it A Question in Two Keys}\ \ \ (to H. Price)} \label{Price9}

1)  If you have a preference, how would you spell ``Price-ian''?  Pricean? Price-ian?  Prician?  (I only give the last as a possibility, even though I really dislike it, because I recently read about the views of Duns Scotus and saw it called Scotism \ldots\ completely bastardizing the last name.  So, I don't know, maybe it's common in philosophy to go so far.)

2)  Where do you have a more elaborate description or working out of your key metaphor than in the research proposal you sent me?
\bq
The project begins with the hypothesis that science and
philosophy have often been dominated by a shallow and misleading conception of the relation of
the information we use and process to the world we inhabit. Thought, language and scientific theory
are often viewed as passive mirrors, which (when all goes well) offer a kind of image of some facet
of that world. On this view, the factual content of thought is simply what the image reveals about
the world. This project aims to show that this passive, one-way, conception of factual information is
both mistaken in itself, and a major obstacle to a proper understanding of seemingly independent
matters of great importance in philosophy and physics. The project offers an alternative view of
factual information, as an active, practical resource, shaped not only to the contours of our
environment, but also, crucially, to a range of different aspects of our own needs, natures and
situations. In place of the dominant metaphor of information as a {\it mirror\/} or {\it picture}, the project
substitutes the metaphor of a {\it key}: a practical, two-ended device, shaped at one end to the outlines of
some part of the environment, and at the other end to the shape and needs of its users.
\eq
I'd like to read more about it in preparation for what I'll say in Sydney.

Hope you had a nice holiday.  Attached are two pictures of the new Waterloo Library of Pragmatism!  I'll be thinking about your keys in there.

\subsection{Huw's Reply}

\bq
Did I ever send you the draft Introduction to my planned collection of essays? It doesn't have the key metaphor explicitly, I think, but does have some working out of the ideas -- copy attached.

Re \#1, I guess ``Pricean'' makes most sense! There's ``Pricey'', too, but while I'd like to think my views are valuable, I also like to emphasize how economical they are, compared to their metaphysical rivals -- so I'm torn on that one \smiley

Looking forward to seeing you! I've got my head down trying to finish a written version of my Everett talk -- I promised the editors I've had it done by the end of the month.
\eq

\section{27-12-07 \ \ {\it A Little Christmas Pragmatism} \ \ (to W. G. {\Demopoulos}, J. E. {\Sipe}, \& R.~W. {\Spekkens})} \label{Demopoulos20} \label{Sipe16} \label{Spekkens48.1}

For Christmas fun, I've been reading Edward C. Moore's book, {\sl American Pragmatism:\ {\Peirce}, {\James}, and {\Dewey}}.  And I came across a passage on pages 35 and 36 that made me think of all three of you.  Therefore, for a little more Christmas fun, it seemed worthwhile to type it into my computer.  The result is attached.

As I perceive it, there is an affinity between this description of {\Peirce}'s form of realism and Bill's take on a quantum system as a ``function.''  Though Bill seems to want to de-anthropocentrize (or de-agentize) the idea a little more than the pragmatist might.  Anyway, it has been to the extent of this overlap that I have been endorsive of Bill's work (if I can coin a term).  So, I'm not reporting anything particularly new here, I just liked Moore's articulation better than my own.

To put this passage in the context of my discussion with John, I would say:  Part of the message here is that {\Peirce} is analogous to a ``consequence Bayesian'' when it comes to percepts, whereas he is analogous to a ``traditional Bayesian'' when it comes to concepts.  The two doctrines peacefully coexist for him, as I believe they peacefully coexist for me.

\begin{center}
The Orientation Toward Pragmatism
\end{center}
\bq
With the above statement of {\Peirce}'s realism as a background it is not difficult to see what he means by saying that pragmatism ``could scarcely have entered a head that was not already convinced that there are real generals [[`universals' in today's terminology, Chris]]'' (5.503).  Pragmatism is a method for defining general concepts (5.8).  If one is a nominalist and believes that there is no such thing as ``triangularity'' anywhere---that triangularity is only a fiction---then there is no place that he can look to see what triangularity really is.  But if he is a realist and believes that triangularity may be found in any triangle, then he knows how to define it; it may be defined as part of what one will experience when one examines a triangle.

It is at this point that we must get a firm grasp on the elusive doctrine of immediate perception.  Triangularity may be defined as a part of the experience of a triangle because ``the percept is the reality'' (5.568).  ``The experience of a triangle'' and ``a triangle'' are epistemologically identical.  Therefore, if we list all of the possible experiences one might have of a triangle, these experiences {\it are\/} the triangles.

Now how would we go about it if we wanted to list these experiences?  We could just state the bald list of experiences, but the list by itself might not suffice to enable the individual for whose benefit the definition is being made actually to obtain those precise experiences. To insure that he would have the proper experiences, perhaps the best approach would be to prescribe for him a certain action such that, if he accomplished it, he would then be confronted by the required experience.  Such a prescription would be a plan or a guide for action. One who performed the prescribed action would have the requisite experience and would then know---by experience of it---the property being defined.  Of course, such a plan for action would necessarily be complex, but if it is sufficiently detailed so as actually to give a perceptual acquaintance with the property being defined, then it would serve as a definition.  {\Peirce} gives an example of this procedure:
\bq
If you look into a textbook of chemistry for a definition of {\it lithium}, you may be told that it is that element whose atomic weight is 7 very nearly.  But if the author has a more logical mind he will tell you that if you search among minerals that are vitreous, translucent, grey or white, very hard, brittle, and insoluble, for one which imparts a crimson tinge to an unluminous flame, this mineral being triturated with lime or witherite rats-bane, and then fused, can be partly dissolved in muriatic acid; and if this solution be evaporated, and the residue be extracted with sulphuric acid, and duly purified, it can be converted by ordinary methods into a chloride, which being obtained in the solid state, fused, and electrolyzed with half a dozen powerful cells, will yield a globule of a pinkish silvery metal that will float on gasolene; and the material of {\it that\/} is a specimen of lithium.  The peculiarity of this definition---or rather this precept that is more serviceable than a definition---is that it tells you what the word lithium denotes by prescribing what you are to {\it do\/} in order to gain a perceptual acquaintance with the object of the word (2.330).
\eq

One might generalize this approach by holding that a concept may be defined by prescribing that:  If you act in a certain manner, then you will have certain experiences, and the sum of the ideas resulting from these experiences constitutes the meaning of the concept being defined. The development of this thesis leads to {\Peirce}'s version of pragmatism.

Such a theory of meaning can only be accepted if one believes that concepts are real, that is, if he believes that the concepts have a real external counterpart.  If he believes this, and wants to know where to look for this counterpart, then a pragmatic definition will give him a practical guide for actions that will result in an experience of the counterpart.  But if he does not believe that concepts are real, then when he follows out the pragmatic definition he will not believe that what he experiences will be the external counterpart, or the referent of the concept, for he does not believe that the concept has a referent.  In short, to accept pragmatism is to accept metaphysical realism with reference to concepts.

As {\Peirce} says (1.27), the realist-nominalist controversy is a question to which only two answers are possible: yes or no.  If one admits that concepts are general ideas [[i.e., universals again]] and then asks, is there anything in reality that stands in a one-to-one relation to the concept, an affirmative answer is only possible on a realist position; a negative answer relegates concepts to the realm of fictions.  It follows from this that pragmatism could not be accepted by anyone who does not also accept metaphysical realism, and that the former could scarcely have entered the head of anyone who did not already understand the latter (5.503).
\eq

\section{28-12-07 \ \ {\it A Little Christmas Pragmatism, 2} \ \ (to W. G. {\Demopoulos})} \label{Demopoulos21}

\bwd
I found the quote from Peirce very suggestive. I think Moore's commentary is not so good though. Perhaps he has elsewhere explained metaphysical realism, but it strikes me that one can understand what Peirce is saying without using that notion.
\ewd
Perhaps I was a little over-enthusiastic for the passage---it is true that I don't really know what one (and, say, Peirce in particular) means by metaphysical realism.  Attached are three things I dug up in light of your points.  1) Moore's original article, from which the chapter I read yesterday appears to be only a minor modification (save for the stuff about Quine and Carnap). [Edward C. Moore, ``The Scholastic Realism of C. S. Peirce,'' {\sl Philosophy and Phenomenological Research} {\bf 12}, 406--417 (1952).]  In it, he says a little of what he means by ``metaphysical realism'' with respect to Peirce.  2)  An interesting rebuttal by Ralph Bastian.  [Ralph J. Bastian, ``The `Scholastic' Realism of C. S. Peirce,'' {\sl Philosophy and Phenomenological Research} {\bf 14}, 246--249 (1953).] 3) A review of Moore's book (with particular emphasis on the same passage as yesterday) by Richard Rorty.  [Richard M. Rorty, {\sl Ethics} {\bf 72}, 146--147 (1962).]

\bwd
I also think that you're right to see an affinity between Peirce's remarks on lithium and the idea of an effect.
The atomic weight is an eternal property of lithium, but what Peirce emphasizes is not this but
the way it interacts with things that may be taken to test for it. It is relevant but incidental to
this account -- the emphasis on effects -- that we have knowledge of lithium by performing
such tests.  All of which is what I think you mean when you say I would de-agentize Peirce
while still enjoying your endorsiveness.
\ewd
No, the issue isn't ``knowledge'' as a point of conflict between you and me (I, like you, do consider that incidental, though I would couch the issue in terms of `belief' rather than `knowledge' \ldots\ being a good follower of Richard Jeffrey), but rather that there is an agent behind the actions Peirce is speaking of.  That is to say, you want to define ``effect'' without reference to one of my favored terms, ``consequence for the agent.''

In any case, thanks for keeping me honest!  I'll comment on your earlier note to me on ``certainty'' once I return to the real world from holiday mode.

\section{28-12-07 \ \ {\it Qbit} \ \ (to N. D. {\Mermin})} \label{Mermin136}

In case you haven't seen it yet, Scott Aaronson wrote a review of your quantum computing book on his blog: \myurl{http://www.scottaaronson.com/blog/?p=296}.  I found it fun that he raised the standard complaint that I've now heard (probably) ten times over.


\section{31-12-07 \ \ {\it A More Intelligent Model} \ \ (to J. B. Lentz \& S. J. Lentz)} \label{LentzB10} \label{LentzS8}

Here's the YouTube link to the commercial:
\begin{itemize}
\item \myurl{http://www.youtube.com/watch?v=saWCyZupO4U}
\end{itemize}
Here's the {\sl Sydney Morning Herald\/} story:
\begin{itemize}
\item \myurl[http://www.smh.com.au/news/technology/professor-claims-ad-agency-cribs-lecture-notes/2007/10/03/1191091161163.html]{http://www.smh.com.au/news/technology/professor-claims-ad-agency-cribs- \\ lecture-notes/2007/10/03/1191091161163.html}
\end{itemize}
Here's Scott's original blog [post]:
\begin{itemize}
\item \myurl{http://scottaaronson.com/blog/?p=277}
\end{itemize}

\chapter{2008: The First 7 in Quantum Information}

\section{02-01-08 \ \ {\it Conference} \ \ (to A. Y. Khrennikov \& J.-{\AA}. Larsson)} \label{Khrennikov22} \label{Larsson5.1}

I was just looking at the conference venue and noticed that Cabello is in your invited speaker list, along with Jan-{\AA}ke and myself.  Thus the following idea came to mind.  If there's interest, maybe one of the themes of the meeting could be a special focus on the import and implications of the Kochen--Specker theorem, and maybe the latest technical developments with it too.  I also happen to know that Ingemar Bengtsson and Helena Granstr\"om are working on issues to do with the KS theorem themselves.  So that's two more.  If there's any room in the speaker list (and if you're interested in the idea), maybe some of the remaining speaking slots could be tilted in that direction.  If you want any suggestions for other speakers in that regard, I could probably develop a few further ideas for speakers in that specialty.  (You both know I'm always more than eager with ideas!)

\section{03-01-08 \ \ {\it Jerusalem Probability Workshop December 2008} \ \ (to M. Hemmo)} \label{Hemmo0}

I apologize for taking so long to get back to you.  When your first note came I was deeply under the gun on many things and fell behind in my correspondence.  But then when your second one came I had already moved into the Christmas vacation with my family.  Anyway, I am here now, and I want to thank you for the invitation.  It would be an honor to honor Itamar!  So, yes, please do count me in.  I haven't been to Israel in some years, and can already feel a visit may lead to great inspiration.

Please keep me up to date as the planning proceeds.

\section{03-01-08 \ \ {\it 20 Years} \ \ (to G. L. Comer)} \label{Comer112}

Didn't we meet in Fall of '89?

\bgc
So, what's this about there being no spacetime?
\egc
You provoked this retort in me.  ``Yeah, there's only coordinate charts.  Sometimes they can be stitched together, in which case we call it spacetime.  Sometimes they can't, and then we call it quantum mechanics.''  I'm sure that's way off the mark in formulation, but there's a little of me that thinks there's a grain of truth in it.

\section{04-01-08 \ \ {\it Screech!}\ \ \ (to G. Bacciagaluppi \& H. Price)} \label{Bacciagaluppi6} \label{Price10}

Well, I had purchased the tickets!!!  I apologize, I had completely spaced on the Feb 4 meeting!  But at least I've emended things now.  I changed my flight to leave on the same schedule but on Feb 5.

\bgba
Is that a live option?
\egba

Now, you know I like that language.  Very Jamesian of you Guido!  I immediately thought back to beginning of James's essay ``The Will to Believe'':
\begin{quote}
[L]et us call the decision between two hypotheses an {\it option}.  Options may be of several kinds.  They may be --- 1, {\it living\/} or {\it dead}; 2, {\it forced\/} or {\it avoidable}; 3, {\it momentous\/} or {\it trivial}; and for our purposes we may call an option a {\it genuine\/} option when it is of the forced, living, and momentous kind.

1. A living option is one in which both hypotheses are live ones.  If I say to you: ``Be a theosophist or be a Mohammedian,'' it is probably a dead option, because for you neither hypothesis is likely to be alive.  But if I say:  ``Be an agnostic or be a Christian,'' it is otherwise:  trained as you are, each hypothesis makes some appeal, however small, to your belief.
\end{quote}

About a year ago, {\Ruediger} {\Schack}, {\Carl} {\Caves}, and I had a heated debate on whether a probability assignment of unity (i.e., certainty) could be made for a {\it dead\/} option (i.e., one that has already passed and thus couldn't have been otherwise than it actually was).  I.e., whether talk of probability was even sensible in that situation.  For Huw's potential entertainment, I'll paste an excerpt from my take on the issue below.  [See 13-12-06 note titled ``\myref{Schack113}{Offline Discussion}'' to R. {\Schack}.]

Huw, by the way, I finally nabbed a copy of {\sl Facts \& the Function of Truth}.  It looks like the presentation will be a little tough-going for me (my lacking the proper vocabulary at this point), but I'll try to get my head around it some nonetheless.

\section{06-01-08 \ \ {\it 21 Years Ago}\ \ \ (to H. J. Bernstein)} \label{Bernstein9}

Finally, finally after all these years, I ran across a copy of the book you edited with Marcus Raskin, {\sl New Ways of Knowing}.  In it I read your article ``Idols of Modern Science'', and as always, felt much in tune with your way of thinking.  And you were already thinking those things more than 21 years ago!  It was a great article.

Accordingly, I've updated my long-in-the-making document {\sl The Activating Observer\/} to reflect what I most liked in your article.  You can see the result starting on page 9 of the attached (all lovingly typed in with my own two hands).  In connection with those passages, you might also enjoy reading or re-reading Pauli's letter to Markus Fierz, dated 10 August 1954, starting on page 96.  [See 02-01-06 note ``\myref{Baeyer14}{The Oblique {\Pauli}}'' to H. C. von Baeyer.] Pauli's words in that letter are very reminiscent of your own.

Thanks for stimulating my mind once again old friend. \medskip

\noindent Quotes from:
\begin{itemize}\item
H.~J. Bernstein, ``Idols of Modern Science and the Reconstruction of Knowledge,'' in {\sl New Ways of Knowing: The Sciences, Society, and Reconstructive Knowledge}, edited by M.~G. Raskin and H.~J. Bernstein, (Rowman \& Littlefield, Totowa, NJ, 1987), pp.~37--68.
\end{itemize}
\bq
In their quest to extract consistency from nature, scientists relentlessly pursue very particular and specialized knowledge.
\eq
and
\bq
As a physicist, I began this essay with critical self-examination of science as a particular form of organized human behavior.  In this light, one aspect stands preeminent.  The scientific study of the world upholds experience as the final arbiter of knowledge.  Not any experience, but only the specially controlled and preconditioned experience of experiment.  Science entails as much creativity, originality, and as many ``free inventions of the human spirit'' as other intellectual activities (like art, poetry, philosophy, or social studies), but always inventions which can be checked against the experience of nature; not direct, emotional experience like smelling a rose or hiking to exhaustion in the High Sierras or ``psyching out'' the answer by intuition, but rather as stylized, conventionalized and, indeed, bizarre behavior which traps nature and emphasizes its regularity.  Think of the most ordinary scientific operation, such as finding the length of this page with a ruler.  You use an instrument of great complexity, with a straight edge supplied by a brass strip held in the fine groove of a specially stamped and printed piece of wood, calibrated against an arbitrary standard.  For more detailed knowledge of the world, science obviously goes beyond the simple rule, but note how value-laden even that device is:  could any but an industrialized and commercialized society provide so many millions of brass and wooden copies of a central arbitrary definition of length, or the training and attitudes needed to use it correctly?
\eq
and
\bq
Physics shows as much variety and method as any science.  Indeed, the first attack on a new problem is to try all the tricks that previously worked; conversely, every new trick is quickly applied to all the currently interesting problems if it succeeds on any one of them.  There are always new techniques and new approaches being invented.  But some form of mathematics always appears, so we might attempt to deduce a general method from this fact.  At least one eminent theoretical physicist, Eugene Wigner, has marveled at the ``unreasonable effectiveness'' of mathematics very extensively, and that mathematical elegance and newly created mathematics are often guides to new physical theories.  It is as if Nature herself were secretly, somehow, mathematical \ldots\ and man's creation of mathematics mimics nature's creation of phenomena.

But from the present point of view, it is rather bizarre behavior of experiment which projects consistency onto nature.  We are, in effect, making phenomena appear rational for our own intellectual and emotional pleasure.  Current scientific data are the result of manipulations of natural objects with the specifically numerical outcome of measurement. While doing scientific experiments we do not assemble those rare substances and produce those unusual conditions in order to ``groove on the vibrations.''  Men's and women's minds invent facts of science through their intellectualized, and mathematized, perceptions of phenomena.  It is not at all surprising that the highly similar invention of mathematics---with its emotional, nonrational, and unconscious motivation, its own search for consistency as opposed to Truth, and its insistence on producing a rational outcome---has great power and relevance in science.  Out of the vast range our minds possess, mathematics and science seem to have tapped the self-same mode.

Asking why the world turns out to be so mathematical is almost like wondering how vegetables come to be so well-suited to feeding animals. Nature's construction of all the organic forms (we now know) is of a piece; of all the various ways to combine elements of earth, only certain carbon-based molecules, controlled by extraordinarily similar genetic codes, form the compounds of all life.  Compatibility is no mystery.  What sunlight streaming through the biosphere has done for animals and plants, nonrational inspirations flowing into the human mind have done for science and mathematics.  What seems most marvelous in this is the ``unreasonable effectiveness'' of emotions in producing rationality both in mathematics and in physics.
\eq
and
\bq
To Einstein, the good of physics resided precisely in its objective ability to comprehend reality, not to manipulate it.  Only contemplative rationality could serve the morally commendable goal of pure knowledge.  In the light of Forman's analysis, Einstein's choice was to reject the standard science, with its hidden acceptance of political pressure, in favor of older values of objective realism.  He would not have put it this way; he did not identify the Copenhagen (acausal) interpretation with world politics, but simply with an epistemological error in science itself---namely, confusing study of what we know with a study of what actually is.  But most physicists could conveniently relegate such issues to philosophy.  They substituted their own construction, knowledge of the universe, in the place of the real world itself as object of investigation {\it without\/} taking full responsibility for the power of making real that which they would choose to know.  The quantum physicists hoped to banish unverifiable analogies and fictions.  But moral visions are also fictions, images, and analogies.  In bypassing them (as in ignoring the question of {\it what\/} to make real) they excluded this whole realm of thought.  Without addressing any moral implications, they took quantum mechanics and its interpretation as the very definition of disciplinary excellence.

The consequent events were extraordinary.  Quick to apply the ``best'' new theories to nuclear puzzles (once they solved those of atomic electrons), physicist hurtled us willy-nilly into a new age, fraught with powerful new dangers.  In 1928, barely two years after Born invented the statistical interpretation, Gamow applied quantum mechanics to nuclear physics; in fact, his data on alpha decay happened to start from Uranium-238.  In the thirties, experimentalists discovered fission, presaging nuclear energy; the Manhattan Project, triggered by a letter from Einstein to Roosevelt, eventually made World War II our first nuclear war.
\eq
and
\bq
\indent
In the case of quantum mechanics, the defining shift from ``it explains'' of classical physics to ``it works'' inspired a different sort of debate.  For many years Einstein argued that quantum mechanics was physically incomplete, that there were elements of physical reality which it could not represent.  Bohr argued that only the actual results of experiments could be considered real---and these results all obey the probabilities given by quantum mechanics.  Both Bohr and Einstein realized that physics had discovered a limit to its own knowledge of reality.  Neither of them realized that this limit posed questions of moral dimensions:  not whether we were slipping from an old morality of objective (and omniscient) knowledge but new questions, questions about knowledge that works:  ``Works for what?'' ``For whom?'' ``Knowledge to what end?''

All of these questions arise because the quantum revolution entailed not merely abandoning a priori definitions of reality (upon which Einstein seemed to insist) but adopting a new direction of gaze as well---one which is self-reflexive of the fact that it involves a choice.  Jauch compares the phenomena of nature to random messages, waiting to be deciphered by scientists:
\bq\noindent
But since the code is not absolute, there may be several messages in the same raw material of the data, so changing the code will result in a message of equally deep significance in something that was merely noise before and {\it conversely}: In a new code a former message may be devoid of meaning.  Thus a code presupposes a free choice among different, complementary aspects, each of which has equal claim to {\it reality}, if I may use this dubious word.
\eq

That physics has found itself selecting a reality may seem striking.  For our archeological approach, however, the cross-linkage of parallel developments in several fields is even more important.  We might even say the analysis could only be correct if the realization that we create reality were reached simultaneously in many aspects of modern culture.
\eq
and
\bq
Consider again quantum mechanics, where the problems of measurement theory remain unresolved after more than fifty years:  How much does the experimenter control reality?  What attributes of the experimenter are relevant for the exercise of this control?  How different would physics be if {\it more\/} of the experimenter were allowed into the science, if more of the motivation and social connections were considered part of the experiment itself?  These are some of the questions that can now be posed within the discipline of physics.
\eq

\section{07-01-08 \ \ {\it What You've Written} \ \ (to B. C. van Fraassen)} \label{vanFraassen17}

Could you send me all that you've written so far on relational quantum mechanics, \`a la Rovelli etc.?  The title of my talk at the PIAF meeting in Sydney is ``Relational Quantum Mechanics, Jamesian Pragmatism, Pricean Pragmatism, and What I Want from This Collaboration,'' and I want to use the better part of the time to give a survey of the first proper noun in that.

Please send me everything you have!

\subsection{Bas's Reply}

\bq
I really have only the paper I gave at the conference, it was the end of an evolution involving two mss.\ and a day with Carlo and others in France last year, but I have improved it a bit more since the conference, and attach the version that is to appear in Foundations of Physics.

But fyi I am attaching the paper by Groenewold that I think presages Rovelli's --- as I think is visible as soon as we compare the equations --- amazingly early for a discussion of information theory in this context!!
\eq

\section{08-01-08 \ \ {\it What You've Written, 2} \ \ (to B. C. van Fraassen)} \label{vanFraassen18}

Thanks.  I've printed them out now.  You might be interested in Nancy Cartwright's '72 opinion of Groenewold.  I'll attach the pdf. [See N. Cartwright, ``Review of {\sl Quantum Theory and Beyond\/} by Ted Bastin,'' Phil.\ Sci.\ {\bf 39}, 558--560 (1972).]

\section{08-01-08 \ \ {\it Many Worlds Note} \ \ (to B. C. van Fraassen)} \label{vanFraassen19}

Whatever happened of this note from 2006?
\bvf
I read a couple of student papers about recent work on many worlds interpretations, and talking with them and others I kept thinking about just what the appeal could be.  Late one night I wrote up the worry, it's just two pages --- what do you think?
\evf
Did you modify it, publish it, get any devastating feedback on it?

Let me bring this paper by Alex Wilce to your attention.  Attached.  [See A. Wilce, ``Formalism and Interpretation in Quantum Theory.''] The discussion on page 3, I think, captures the crucial issue for me:
\bq
Thus, we have two problems, somewhat in tension with one another. If
we view quantum mechanics as a linear dynamical theory, in which physical
states are wave functions, evolving according to the {\Schroedinger} equation,
then the theory's analytical apparatus is not especially problematic (Hilbert
spaces were, after all, {\it invented\/} to describe just this sort of thing); but its
probabilistic content seems mysterious and {\it ad hoc}. If, on the other hand, we
accept the theory's minimal probabilistic interpretation as unproblematic,
then it is the theory's formal apparatus that seems mysterious and {\it ad hoc}.

For purposes of this paper, I'll refer to the former problem as the {\it problem
of interpretation}, and to the latter problem as the {\it problem of the formalism}.
Both have proved remarkably refractory, withstanding decades of sustained,
and often brilliant, effort by physicists, mathematicians, and philosophers of
science, and, in the process, generating substantial technical literatures. It is
quite remarkable, therefore, that these two obviously related problems have
been pursued somewhat in isolation from one another. Superficially, perhaps,
this is understandable, as each problem begins where the other wishes to end;
nevertheless, when a tunnel is being dug through a mountain, it is usual for
those working from opposite sides to coordinate their efforts.

In this paper, I want to urge that each project has something to contribute
to the other. It has become increasingly clear in recent years
that many of the most puzzling ``quantum'' phenomena---in particular, phenomena associated with entanglement, and including, as I'll show, a version
of the measurement problem---are in fact quite generic features of essentially
all non-classical probabilistic theories, quantum or otherwise. This suggests
that many of the interpretive ideas that have been advanced in connection
with quantum mechanics can be carried over to a much more general setting. This exercise has something to offer to both foundational projects. On
the one hand, an interpretation of quantum mechanics that can't be made
sense of absent certain special structural features of quantum mechanics, is
potentially a source of fruitful ideas with which to approach the problem of
the formalism. On the other hand, if an interpretation can be kept aloft
even in the thin atmosphere of a completely general non-classical probabilistic theory, then perhaps it has little to tell us about the physical content of
quantum theory. To compress this idea into a slogan: a completely satisfactory interpretation of a physical theory should be capable of yielding (or at
least, constraining!)\ its own formalism.
\eq
Namely, despite all the posturing of the Everettians, the pages and pages of elaborated formalism, I simply see {\it nothing\/} uniquely implied by quantum mechanics in the ontology they offer us.  Quantum mechanics makes the view no more compelling than it was 101 years ago, when William James wrote:
\bq
   [I]f you are the lovers of facts I have supposed you to be, you find
   the trail of the serpent of rationalism, of intellectualism, over
   everything that lies on that side of the line.  You escape indeed the
   materialism that goes with the reigning empiricism; but you pay for
   your escape by losing contact with the concrete parts of life.  The
   more absolutistic philosophers dwell on so high a level of
   abstraction that they never even try to come down.  The absolute mind
   which they offer us, the mind that makes our universe by thinking it,
   might, for aught they show us to the contrary, have made any one of a
   million other universes just as well as this.  You can deduce no
   single actual particular from the notion of it.  It is compatible
   with any state of things whatever being true here below.  And the
   theistic God is almost as sterile a principle.  You have to go to the
   world which he has created to get any inkling of his actual
   character:  he is the kind of god that has once for all made that
   kind of a world.  The God of the theistic writers lives on as purely
   abstract heights as does the Absolute.  Absolutism has a certain
   sweep and dash about it, while the usual theism is more insipid, but
   both are equally remote and vacuous.
\eq
I.e., I'm willing to call Many Worlds little more than a modern species of the genus.

\section{09-01-08 \ \ {\it Wednesday, just before Toronto} \ \ (to D. M. {\Appleby})} \label{Appleby25}

I haven't said hi in a while either.  Some people rise to the holidays; myself I was buried by them.  I'm only now working my way back out.  I drop the in-laws off at the Toronto airport this afternoon.

It's great to hear that you are ploughing on in $d=5$.  Myself I spent a little bit of the holidays working on $d=6$ finally.  I got three independent eigenvectors out of your happy unitary and started to work out the equations for the fiducials.  Two levels of elimination finally gave me some linear relations between the coefficients, but I haven't gotten a chance to finish that up yet.  I want to get that before Wednesday, because I have to give the colloquium here, and I want to close with Steve's ``constructible number'' hypothesis if it's true.  One thing's for sure, though, the Grassl ``solution'' (quotes because I only presently trust the 3rd and 6th components of it) is definitely not an eigenvector of happy unitary.  It had been my hope that it would be, and that that would help me eliminate a couple of equations on its own.

The title and abstract for the colloquium are below.  I'm also giving it in Bristol the following week.  One thing I wanted to ask you.  Would you mind if I advertise your tomographic result showing that SICs are sometimes optimal there?  You never did publish that, did you?  If you don't mind my saying something, could you send me a precise statement of the theorem?

I'm glad to hear there's a small number of parameters in the triple products of $d=5$.  There's no chance that the parameters can be adjusted so that {\it all\/} triple products (of distinct projectors) have the same real part, is there?  How I would love that to always be the case!  In $d=3$, looking at your old notebook though, it looks like ``all but nine'' is the best one can do.  So this leads to another question that perhaps can be explored in $d=5$.  Might it possible to make all but 25 have the same value (the real parts that is), and the remaining 25 have all the same of another value?

In general, one can ask can any useful bound be put on the maximal number of Re(triple products) all obtaining the same value?  What does that tell us about the shape of Hilbert space?

\section{09-01-08 \ \ {\it Pragmatism at the Perimeter}\ \ \ (to L. Smolin)} \label{SmolinL9}

Well, the Waterloo Library of Pragmatism is finally open for business!  Look at these beautiful solid-oak bookshelves I had built in my house!  Sadly, I've only now ordered Unger's book for it, but I {\it have\/} ordered it, and I pick it up in Toronto today after dropping my in-laws at the airport.  I'll come to your seminar tonight---but to warn you---I might be a little late.

Going back to the theme of the library, I'd like to discuss a couple of things with you when you get a chance.  What I'd really like to get straight in my mind is how to ultimately shake off any last vestige of a ``block-universe'' conception in (the next) physical theory.  You know that I think pragmatism has its heart in the right place, but it doesn't have the technical brawn; on the other hand, I feel strongly that quantum mechanics shows some of the technique, but it's only a first hint of the direction we need to go.

Thus, I have wondered whether:
\begin{enumerate}
\item
You might have any interest in a little semi-periodic discussion session on the confluence of subjects, with emphasis on the pragmatism side of things?  It might consist of us two, probably John Sipe and Philip Goyal, maybe a couple of others.  Maybe Appleby when he's visiting, etc.  I haven't thought about the format too much yet, but if you have some interest, maybe we could talk about this.
\item
Might it be interesting to have a small meeting at PI this year on the subject?  The title I've played with in my head is ``Pragmatism at the Perimeter.''  I'm not sure who I'd invite, but Ian Hacking and Cheryl Misak from U. Toronto phil.\ dept, Unger, Michel Bitbol, Chris Timpson, {\Ruediger} {\Schack}, Marcus Appleby, and Huw Price come to mind pretty quickly.  And you probably have some ideas too.  Would you be interested in co-organizing something along these lines with me?
\end{enumerate}
Those are the questions.  As I say, as you get a chance, I'd like to talk about these things some more.

\section{10-01-08 \ \ {\it Philosophy Books To Get Rid Of} \ \ (to W. G. {\Demopoulos})} \label{Demopoulos22}

Do you have any interest in having any of these books?
\begin{enumerate}
\item
Harald {\Hoffding}, {\sl A History of Modern Philosophy}, Vol.\ 2, paperback, Dover.
\item
Donald W. Sherburne, {\sl A Key to Whitehead's Process and Reality}, paperback, Indiana University Press, 1966.
\item
John Dewey, {\sl The Philosophy of John Dewey:\ The Structure of Experience}, edited with intro and commentary by John J. McDermott, paperback, G. P. Putnam, 1973.
\item
John Dewey, {\sl The Philosophy of John Dewey:\ The Lived Experience}, edited with intro and commentary by John J. McDermott, paperback, G. P. Putnam, 1973.
\end{enumerate}
I had double copies in my library.  If you want any, I'll reserve the ones until I see you again.  Otherwise they'll be taken to The Old Goat for 50 cents credit or so.  (They're all in good shape, it's just I know I won't get anything for them.)

{\Hoffding}, in case the name doesn't ring a bell, had been Bohr's philosophy mentor in Copenhagen.  Some historians make a big to-do about his influence on Bohr.

\section{11-01-08 \ \ {\it Tomorrow?\ Sunday?}\ \ \ (to S. T. Flammia)} \label{Flammia2}

Are you going to be in at PI tomorrow or Sunday?  I think I've got a good method for bounding the max number of identical triangle areas.  A little cubing of the identity, a little combinatorics, a little Schwarz inequality (or maybe arithmetic-geometric mean inequality), and poof!  Or so it seems.  Anyway, it'd be nice to talk it through with someone.  Unfortunately, I've got a social engagement this evening, and then housework preparing for Emma's party tomorrow, but sometime in the afternoon I'll become free.

\section{15-01-08 \ \ {\it Brain Pulp:\ Quotes on Planck} \ \ (to M. A. Nielsen)} \label{Nielsen9}

\bq
Planck traced the discovery of his vocation to the teaching of an instructor at the gymnasium, Hermann Muller, who awakened an interest, which became a passion, to `investigate the harmony that reigns between the strictness of mathematics and the multitude of natural laws.'  In 1878, at the age of twenty, Planck chose thermodynamics as the subject of his doctoral dissertation, which he wrote in four months. He recalled that his professor at the University of Munich, Philipp von Jolly, had counselled against a career in physics on the ground that the discovery of the principles of thermodynamics had completed the structure of theoretical physics.
That had not dissuaded Planck, who had his compulsion and also an objective far removed from the principal ambition of today's physicists.  He had no wish to make discoveries, he told Jolly, but only to understand and perhaps to deepen the foundations already set.
  --- J. L. Heilbron
\eq

\bq
Many kinds of men devote themselves to Science, and not all for the sake of Science herself.  There are some who come into her temple because it offers them the opportunity to display their particular talents.  To this class of men science is a kind of sport in the practice of which they exult, just as an athlete exults in the exercise of his muscular prowess.  There is another class of men who come into the temple to make an offering of their brain pulp in the hope of securing a profitable return.  These men are scientists only by the chance of some circumstance which offered itself when making a choice of career.  If the attending circumstance had been different they might have become politicians or captains of business.  Should an angel of God descend and drive from the Temple of Science all those who belong to the categories I have mentioned, I fear the temple would be nearly emptied. But a few worshippers would still remain---some from former times and some from ours.  To these latter belongs our Planck.  And that is why we love him.  --- A. Einstein
\eq

\section{15-01-08 \ \ {\it What I Think I Mean by Pluralism and Indeterminism}\ \ \ (to L. Hardy)} \label{Hardy23.5}

It's the idea of a joint-stock society, rather than a universe.  Quotes below.  [See 27-09-07 note titled ``\myref{BrownHR3}{The Joint-Stock Society}'' to H. R. Brown.]  Don't forget to forward to Nielsen.

\section{15-01-08 \ \ {\it Potentest of Premises}\ \ \ (to L. Hardy)} \label{Hardy24}

I never did send you that note on the potentest of my premises, did I?  The note I just sent you on the joint-stock society conception of nature reminded me of that.  [See 15-01-08 note titled ``\myref{Hardy23.5}{What I Think I Mean by Pluralism and Indeterminism}'' to L. Hardy.]  Anyway the short of an answer to you is below.  [See 17-06-04 note ``\myref{Mabuchi12}{Preamble}'' to H. Mabuchi.]  It is that, the key physical element of quantum theory---that is, that in our interactions with the world there is creation---must ultimately be distilled, and then, to it, added a Copernican principle that there is nothing particularly special about us, other than our perspectives.  We have evidence of the creation because we partake in it, but we cannot be the whole story.  I can't think of a more exciting and more positive conception of nature than that it always, everywhere, and without exception, creates.

\section{20-01-08 \ \ {\it Q{\&}Q}\ \ \ (to T. Slee)} \label{Slee1}

Thanks for the compliments, but I'm not fooling myself:  I thought the radio show came out pretty boring honestly.  I hope the Rogers Cable version worked better, with its 40 minutes of more material and untrimmed explanations.  Also the visuals should have helped on some jokes.

\btsl
I think I'd like to try reading some of your papers at some point and see if I can understand
what you are on about if that's possible. I didn't quite see how the ``quantum mechanics
describes the information in a system'' rather than ``quantum mechanics describes a system'' is
more than a linguistic sidestep, but I know there's physics behind what you do and if there is
something accessible I'd be interesting in reading it.
\etsl

Well, I wouldn't use the phrase ``QM describes the information in a system.''  I only make (louder and, I hope, more consistent) the old time-honored observation of Einstein:  That a quantum state cannot be a property inherent in a system---owned by it and it alone---but rather must be something else, of the flavor of one's information.  Else QM would imply action-at-a-distance and a load of other conceptual difficulties (Wigner's friends and so on). The research program I push, and which I think has had decent success, is to build up the structure of quantum mechanics from that idea alone, along with an adequate answer to the deeper question, ``Information about what?''

I'd be flattered if you'd read a paper of mine or two.    Let me give this one as maybe the best starting point:
\bv
\arxiv{quant-ph/0205039}.
\ev
For a study in the sociology of my developing views, you might have some fun with
\bv
\arxiv{quant-ph/0105039}.
\ev
Particularly the correspondence with Mermin and Peres, might be useful.  And finally further fun can be found (if you like Sudoku) in this paper by Rob Spekkens:
\bv
\arxiv{quant-ph/0401052}.
\ev
He doesn't reconstruct quantum mechanics by any means, but he does demonstrate that many of the phenomena touted in quantum information studies are simply generic to the idea of information (a better word would be ``simple uncertainty for whatever the reason'').  As Rob puts it in his abstract, ``The diversity and quality of these analogies is taken as evidence for the view that quantum states are states of incomplete knowledge rather than states of reality.''  The only place where Rob and I differ is that he would ultimately like to revert back to hidden variables in his hoped for reconstruction of QM, whereas I argue vigorously that it's time to move on.  Reality is trying to tell us something really interesting with quantum phenomena---something far more interesting than hidden variables---and we should listen.

\section{21-01-08 \ \ {\it Value Added!}\ \ \ (to H. C. von Baeyer)} \label{Baeyer29}

If you do indeed arrive on March 24, it'll be a great day for alchemical ideas in QM!  It looks like Marcus {\Appleby}---one of my two toppermost intellectual soulmates at the moment---will be arriving the very same day.  Marcus also has a deep interest in Pauli, Jung, Fierz, alchemy, \ldots\ and SIC-POVMs!  So, we'll have a wonderful time, I'm sure.

\section{21-01-08 \ \ {\it Potential Topic of Discussion} \ \ (to R. {\Schack})} \label{Schack126}

I'll let you know Thursday or so when I plan to arrive in Egham.  In the mean time let me send you this little drunken thing I just wrote up (a consequence of my disappointment with Kiki's \$8,000 counters in the kitchen).  If you've got nothing on the agenda for us to talk about when I visit, then maybe this is a worthy subject.  The point is, I think I'm homing in better on the idea that the Born rule is ``an addition to coherence,'' in contrast to the usual idea that ``it is a probability setting rule.''

Looking very much forward to seeing you!

\bq
\begin{center}
{\bf Hidden Assumption in Coherence Arguments? \\
\rm (21 January 2008)}\bigskip
\end{center}

Suppose there is an agent who believes that if he takes a certain action it will result in one of $n$ consequences for him, and he deems the probabilities of those consequences $p(i)$.  The $p(i)$, of course, are all nonnegative and $\sum p(i)=1$, as demanded by coherence.

Alternatively, consider some {\it other\/} action that the agent might perform with consequences $j$ (drawn from a potentially different number $m$ of consequences).  Furthermore suppose the agent has a firm set of beliefs for which consequence $j$ he would find if he had first found $i$ in the other action previously considered.  That is, suppose the agent has a family of probability distributions $p(j|i)$ that he feels are acceptable degrees of belief.  To say it again,
$p(j|i)$ is the probability of consequence $j$ for the second action, supposing the agent finds consequence $i$ for the first.

Question:  Given all that was described above, does coherence make any requirements on the agent's unconditioned probability assignment for $j$?  That is, can we infer what the agent ought to assign for $p(j)$ straight out from the $p(i)$ and the $p(j|i)$?  If we had dropped the language of ``actions'' and ``consequences,'' and substituted ``observed'' and ``found,'' I think most people familiar with Bayesianism would say we could:  If $i$ were some ``event'' in the usual language, with probability $p(i)$, and $p(j|i)$ were the probability of another event $j$, given knowledge of $i$, then
\be
p(j)=\sum_i p(i) p(j|i)
\label{OldFart}
\ee
would be the assignment demanded by coherence for $j$ itself.

But consider $i$ the outcome of some SIC-POVM measurement, and let $j$ denote the outcomes of some {\it other\/} SIC-POVM.  Then one might well be in possession of $p(i)$ and $p(j|i)$, without anything initially further.  If so, what should one assign for $p(j)$?  That depends.  If one really is making the measurement of $i$ intermediately (but somehow not becoming aware of the outcome), then one should assign a $p(j)$ as in Eq.\ (\ref{OldFart}) above.  But suppose one has all the information above and goes straight for the jugular---i.e., one wants to calculate the probability of a $j$ result, knowing full well that one is not going to perform the intermediate measurement of $i$.  Then quantum mechanics still gives a means for calculating the $p(j)$, but it is an interesting modification of the above.  That one will make the intermediate measurement or not actually makes a difference for how one should calculate one's ultimate probabilities!  Also---and this is {\it quite\/} surprising---it makes a difference in a very controlled way.  Taking the $p(i)$ and the $p(j|i)$, one just modifies the formula above to
\be
p(j)=(d+1)\!\left(\sum_i p(i) p(j|i)\right)-\frac{1}{d}\;,
\label{YoungBuck}
\ee
where $d$ is the dimensionality of the Hilbert space.  That is, the probability one {\it should\/} assign is given by the one would imagine as coming about by a normal coherence argument $p'(j)$, but then ``stretching'' it according to $(d+1) p'(j)-\frac{1}{d}$.  It is as if quantum mechanics gives a {\it reward\/} for not actually performing the first measurement $i$.  One can take advantage of the conceptual device of using the $p(j|i)$ for calculating and ultimate $p(j)$, but one is rewarded in comparison to the classical standard if one does not bother to actually take the action to collect a value for $i$. I mean by ``reward'' that one is allowed more certainty generally for one's answer than would be expected classically.

All of this makes me wonder if there is a hidden assumption of ``reality'' in the standard Dutch book argument based on conditional bets that I had not previously appreciated.  I don't have an answer, but that's the sort of thing I'd like to discuss in my brief time in Egham.

In contrast, suppose the measurement of $j$ was of the character of a von Neumann measurement consisting of rank-1 projection operators (while $i$ still denotes the outcome of a SIC-POVM).  Then the modification to standard coherence would be
\be
p(j)=(d+1)\!\left(\sum_i p(i) p(j|i)\right)-1\;,
\label{SimonBoy}
\ee
Again, this is a deviation from the classical formula, but in a controlled way.

Why such simple modifications to coherence!?!?  Why does QM reward us for {\it not\/} performing the intermediate actions we so freely imagine for use in our calculations?

Wilder idea:  Is this the source of the power of quantum computation?  That we have more certainty about the consequences of our actions than a classically honed intuition expects we ought to have?
\eq

\section{21-01-08 \ \ {\it Talk and Diverse} \ \ (to P. G. L. Mana)} \label{Mana10}

As long as you send me your abstract two weeks in advance of your talk, it will be enough time for me.  What are the precise dates you'll be visiting?

\bpglm
I am studying {\Schroedinger}'s equation and its `second quantized' form in this period. There are some nice results which can be easily derived from Holevo's convex framework (we soon need to find an appropriate name for this framework, since `convex', `vector', or `r-p' are misinformative or uninformative names).

First, it can be shown that any physical (`ontic') theory behind the wave function cannot be founded on particle-like entities only, but needs field-like entities as well. More precisely, the physical objects behind the wave function must have a continuum of degrees of freedom, hence cannot be just particles. I know of Beltrametti and Bugajski's extension (which is a special case of a theorem by Holevo from the '80s), but the above is really a mathematical proof.  Which is also interesting, because what is proven is partly `metaphysical'!
\epglm

This sounds similar to Lucien's ``excess ontic baggage'' result.  He never posted the paper on the archive.  If you haven't heard of it before, you might ask him for a copy of the paper.

We will have many things to talk about during your visit.  Unfortunately, I can have no extra mind now, as I am preparing to give a PhD exam in England this week, then I come home for one day before departing for the Sydney meeting, where I will talk on ``Relational QM, Jamesian Pragmatism, and Pricean Pragmatism.''  Particularly, wish me luck for the latter!

\section{23-01-08\ \ {\it In Town} \ \ (to R. Jozsa)} \label{Jozsa8}

I should let you know I'm in town.  I just had one of those kinds of sleeps that one can only have in a foreign hotel in an upside-down time zone after a sleepless flight.  They're the best sleeps of the year!  Now I think I'll slowly wake up, get a shower, and wander out for some food.  I've still got a lot of reading to do in the thesis, so most of the rest of my day ought to be taken up with that.  Did you have any particular plans for me that I should be aware of, though?  Or should I just plan to be a free agent today?

\section{23-01-08\ \ {\it Old Books} \ \ (to R. Jozsa)} \label{Jozsa9}

While I'm thinking about it, I don't want to forget:  Are there any used or antiquarian book stores within walking distance of the hotel?  I'd particularly appreciate ones with good philosophy sections.  Maybe I'll tool around there while you're busy tomorrow morning.  There's a running joke at PI, that Daniel Gottesman won't read anything that's not published electronically, and I won't read anything that's not at least a hundred years old.

\section{24-01-08 \ \ {\it Counter-Bayesians!}\ \ \ (to J. A. Smolin)} \label{SmolinJ9}

You've got to have a look at this article:
\bq\noindent
``When Several Bayesians Agree That There Will Be No Reasoning to a Foregone Conclusion,'' by Joseph B. Kadane, Mark J. Schervish, Teddy Seidenfeld,  {\sl Philosophy of Science}, Vol.\ 63, No.\ 3, Supplement. {\sl Proceedings of the 1996 Biennial Meetings of the Philosophy of Science Association}. Part I: Contributed Papers (Sep., 1996), pp.\ S281--S289.
\eq
\bq\noindent
{\bf Abstract:} When can a Bayesian investigator select an hypothesis $H$ and design an experiment (or a sequence of experiments) to make certain that, given the experimental outcome(s), the posterior probability of $H$ will be lower than its prior probability? We report an elementary result which establishes sufficient conditions under which this reasoning to a foregone conclusion cannot occur. Through an example, we discuss how this result extends to the perspective of an onlooker who agrees with the investigator about the statistical model for the data but who holds a different prior probability for the statistical parameters of that model. We consider, specifically, one-sided and two-sided statistical hypotheses involving i.i.d.\ Normal data with conjugate priors. In a concluding section, using an ``improper'' prior, we illustrate how the preceding results depend upon the assumption that probability is countably additive.
\eq

\section{24-01-08\ \ {\it Bohm-Biederman Correspondence} \ \ (to R. Jozsa)} \label{Jozsa10}

The name of the artist was Charles Biederman.  Here's the Amazon.com link:\medskip\\
\myurl[http://www.amazon.com/Bohm-Biederman-Correspondence-Vol-Creativity-Science/dp/0415162254]{http://www.amazon.com/Bohm-Biederman-Correspondence-Vol-Creativity-Science/dp/ \\ 0415162254}.

\section{25-01-08 \ \ {\it Bristol, 3 AM} \ \ (to D. Gottesman)} \label{Gottesman8}

The other day, after talking to you about how the usual Bayesian ``law of total probability'' gets modified in the context of quantum measurements when one uses a SIC representation of it all, I wrote this little bit of poetry in my notebook:
\bq\noindent
WILD SPECULATION:  This little reward for a physical step not taken is ultimately responsible for the power of quantum computation.
\eq
What it's referring to is this.  Suppose $p(i)$ gives a quantum state in a SIC representation, i.e., it is the probability for getting outcome $i$ for some imagined SIC measurement.  Now consider an arbitrary PVM measurement consisting of rank-1 projection operators, labeled by $j$---this is the measurement that one is {\it actually\/} going to perform.  Finally consider the probability for an outcome $j$ that one would write down if one {\it imagines\/} first measuring the SIC and finding $i$; I'll write that as $r(j|i)$.  I emphasize ``imagines'' because one is not really only going to do the $i$ measurement; we're just going to use these numbers as conceptual pieces in deriving a number $q(j)$ representing the probability for the outcomes of the $j$ measurement that we're actually going to perform.  Putting it all together, the answer turns out to be $$
q(j) = (d+1)\left(\sum_i p(i)r(j|i)\right) - 1 ,
$$
where $d$ is the dimension of the Hilbert space.  That is to say, one can calculate $q(j)$ by an ever slight change of the law of total probability.  Slight change mathematically, but big change conceptually.  For the upshot of the transformation $x \rightarrow (d+1)x-1$ is that it spits out a probability distribution that is a little more pure (a little less mixed) than one would have imagined classically.  That's the reward I was talking about.

Anyway, I'm writing all this down mostly because I feel like writing something down at this funny hour.  But I've also been rereading the Steane paper that I had told you about.  It is: \quantph{0003084}.  I find that I had pretty severely misquoted it when talking to you.  I'd be curious to get your reaction to the real thing, if you feel like reading it.

\section{25-01-08 \ \ {\it Morning Report} \ \ (to J. E. {\Sipe})} \label{Sipe17}

Well, I'm up to page 148 in Morris's book now. [Charles Morris, {\sl The Pragmatic Movement in American Philosophy}.] My opinion seems to be different than what I thought you expressed last week:  I think it's mostly a throw-away book; pretty worthless really.  It is hardly ``the clearest explanation of American pragmatism ever written'' that the attached review claims it to be!!!  [See R. Ginsberg, ``Review of {\sl The Pragmatic Movement in American Philosophy} by Charles Morris,'' {\sl Annals of the American Academy of Political and Social Science\/} {\bf 396}, 184--185 (1971).] Still, at least I did get a little thought out of pages 128--136 on Mead's cosmology, which I hadn't appreciated before as having some similarity to James's ``joint-stock society'' conception of nature.  To the extent that I understand it, I like it.

I've been thinking, maybe a much more relevant thing for us to tackle would be to jointly read, H. S. Thayer's massive history of pragmatism and its roots, {\sl Meaning and Action}.  I have it on my shelf at home and have read bits throughout, though not a complete linear reading yet.  I think it is excellent and much more worthy than the present junk.

\section{28-01-08 \ \ {\it Thinking about Unitarity}\ \ \ (to K. Wiesner)} \label{Wiesner1}

I looked at some of your papers with Crutchfield over the weekend.  I see why you might be interested in the things we talked about at dinner the other night.  We should have you come by PI to give us a talk sometime, maybe next Fall if you've got the time.

Let me point out the two papers of my own that might be most relevant for further thinking about the stuff we discussed the other day.  I'd say, this:
\bv
\arxiv{quant-ph/0205039}
\ev
and this
\bv
\arxiv{0707.2071}.
\ev
After meeting with Schack right after Bristol, I'm even more hopeful that the representation of unitarity I showed you will help unlock some doors in thought.

In all, good meeting you!

\section{29-01-08 \ \ {\it Knowing Me, Knowing Huw}\ \ \ (to H. Price)} \label{Price11}

\ldots\ God, that's a bad ABBA pun, and you must have already heard it hundreds of times in your life \ldots\ but given my excitement, it's the best I can do!  (Ouch, I did it again.)  No need to send me more reading for my flight, I've already got plenty of it, and it's all by Huw Price!

Seriously, I am very excited since reading your ``Naturalism without Mirrors'' on my flight back from Bristol the other day, and I've got about 10 more of your papers printed to read (and reread).  At least at this stage, I couldn't find a single thing in the said article that I didn't like.  I was very, very impressed:  You are an amazingly clear thinker, and I am sorry I had not appreciated your depth of thought before.  I knew that I was attracted to some of your ideas previously, but I think my mind wasn't quite prepared the last time I was looking at your stuff (on naturalism and truth).

Particularly I'm excited by this:  I now think you far outstrip me in clarifying what it is {\it I\/} really want out of quantum mechanics.  It was probably captured best by this passage:
\bq
   Very crudely put --- ignoring, for example, all the obvious grounds for holism ---
   the explanatory project goes something like this. We find our speakers disposed
   to say ``P'' (i.e., `P' appears in the list of statements on the left of the
   model). We now ask, ``Why do they say that?''; and in general (without pretending
   that this distinction is sharp) we look for an explanation that refers both to
   features of the speakers, and to features of their natural environment. Note
   that in our own case, this attitude always looks sideways on, or ironic. We say
   that P, and then wonder why we said so, how we came to be making a claim of
   that kind --- looking for something deeper as an answer, of course, than merely
   ``Because we realised that P''. (This is the kind of irony characteristic of
   practitioners of the human sciences, of course, who cannot help but view
   themselves as examples their own objects of enquiry.)
\eq
For I would say what I have been on about the last few years, is thinking of the interpretational issues of QM as a {\it laboratory\/} for this very kind of thought:
\bq
   Very crudely put, the explanatory project goes something like this. We find our
   agents disposed to accept the structural features of quantum mechanics into
   their catalogue of probability assignments.  We now ask, ``Why do they say
   that?''; and in general (without pretending that this distinction is sharp) we
   look for an explanation that refers both to features of the speakers, and to
   features of their natural environment.
\eq

So, indeed, I'm really looking forward to fleshing out some serious ideas with you in this collaboration.  It's a much more specific project than the one you have in mind, but I think it'll be a very instructive laboratory, and maybe you would get something out of my prods.

Anyway, that paper went down fairly easily for me (probably because it was only an introduction, rather than a technical article), and was enough to hook me.  I've since reread ``Naturalism without Representationalism'' and again come away with the feeling that I am indeed a subject naturalist.  But there I understood your arguments far less well.  You will have to walk me through some.  At the moment, I'm on the paper with Quasi-Realism in the title, which you co-authored with someone.  I hope to have two or three more papers read by the time we see each other in Sydney.

\section{29-01-08\ \ {\it Wittgenstein, Etc.}\ \ \ (to R. Jozsa)} \label{Jozsa11}

\brj
It was certainly a pleasure to catch up last week and I hope you had a good trip back!

I just wanted to send you a link to the must-read book I mentioned about Wittgenstein/Popper (titled ``Wittgenstein's Poker''): [\ldots]
(the American edition appears to have a really ghastly poker picture on the cover! \ldots\ unlike the UK edition). Also the recent Beatles compilation CD I mentioned (titled ``Love'' put together by George Martin), most recommendable.

So, hope you like these, and look fwd to maybe seing you again in Waterloo sometime this year.
\erj

Thanks for that---I'll certainly be getting both.  It was good seeing you too.  I'm on my way to Australia now; very busy week!

I enjoyed talking with your student Ashley too.  But watch out, the guy has Everettian tendencies.  (He said it's his favorite interpretation.)  Remember it's part of your duty as his mentor to disillusion him of that!  Be sure to give him a complete education!

\section{31-01-08 \ \ {\it Tracking Last Night}\ \ \ (to H. Price)} \label{Price12}

It dawned on me that the part of last night's discussion centering on whether there might be two possible threads to follow when focussing your `naturalism without mirrors' project onto quantum mechanics \ldots\ \  It dawned on me that that issue tracks pretty well with Chris Timpson's discussion on pages 26--28 of the attached paper [C. G. {\Timpson}, ``Quantum Bayesianism:\ A Study''], where he discusses potential ontologies that might underly the quantum Bayesian effort of {\Caves}, {\Schack}, and me.  It's easy reading, and just three pages:  Tell me whether you think whether I got it right that it tracks our conversation, or whether I'm still missing something.

\section{31-01-08 \ \ {\it Tracking Last Night, 2}\ \ \ (to H. Price)} \label{Price13}

It doesn't look like you read far enough.  I would think the paragraph starting with, ``It would seem that the cleanest setting for the proposal is \ldots'' on page 28 captures a good bit of what you really have going on in the back of your head:  the universe is just a big block of facts (with no easy local description), and it is our perspectival limitations that give rise to the use of QM and potential explanations of the need for that use in terms of retrocausation.  The retrocausation is an artifact of our perspectival situation; the ontology---the end result of the Copernican project---though is that of a block.

\section{31-01-08 \ \ {\it Potentest of My Premises}\ \ \ (to E. G. Cavalcanti)} \label{Cavalcanti1}

\noindent Hi (I'm sorry I've already forgotten your first name again),\medskip

Below are the relevant quotes about a Copernican Project based on agency.  [See 15-01-08 note ``\myref{Hardy24}{Potentest of Premises}'' to L. Hardy.] There's also a little more context for them in the attached pseudo-wacky-paper-thing.  [See ``Delirium Quantum,'' \arxiv{0906.1968v1}.]  Sorry, it's not much to go on yet.  You'll see some resemblance---I think---between the James quotes below and the remark you made in Vienna that impressed me.

\section{02-02-08 \ \ {\it Practicing French} \ \ (to R. {\Schack})} \label{Schack127}

Here's an article that you and Friedemann might use as a focus for your morning French practice [``Bruno de Finetti: l'Origine de Son Subjectivisme,'' by Simona Morini].  I just stumbled across it as I was preparing for my PIAF talk.  Is there anything of interest in the paper?  See also attached quotes for my own talk.  I'm going to talk on the modified coherence stuff this morning.  I think I put together a pretty nice talk on it.  I hope it stands up in the face of Howard Wiseman \ldots\ and {\Carl} {\Caves}!

\bq
{\it Very\/} crudely put \ldots\ the explanatory project goes something like this. We find our speakers disposed to say ``P''  (i.e., `P' appears in the list of statements on the left of the model). We now ask, ``Why do they say that?''; and in general (without pretending
that this distinction is sharp) we look for an explanation that refers both to features of the speakers, and to features of their natural environment. Note that in our own case, this attitude always looks sideways on, or ironic. We say that P, and then wonder why we said so, how we came to be making a claim of that kind---looking for something deeper as an answer, of course, than merely ``Because we realised that P''. (This is the kind of irony characteristic of practitioners of the human sciences, of course, who cannot help but view themselves as examples their own objects of enquiry.)
\begin{flushright} --- Huw Price, ``Naturalism without Mirrors'' \end{flushright} \bigskip

{\it Very\/} crudely put, the explanatory project goes something like this. We find our agents disposed to accept the structural
features of quantum mechanics into their catalogue of probability
assignments.  We now ask, ``Why do they say that?''; and in general
(without pretending that this distinction is sharp) we look for an
explanation that refers both to features of the speakers, and to
features of their natural environment.   \begin{flushright} --- CAF, stealing from Huw Price \end{flushright}
\eq

\section{03-02-08 \ \ {\it Tsallis and {\Daroczy}}\ \ \ (to O. J. E. Maroney)} \label{Maroney1}

Tsallis's 1988 entropy can be found here:
\bq
\myurl{http://en.wikipedia.org/wiki/Tsallis_entropy}
\eq
A little report on {\Daroczy}'s 1970 entropy can be found on pages 30--34 of here:
\bq
\arxiv{quant-ph/9601020}
\eq
cf.\ Eq.\ (2.130).

\section{03-02-08 \ \ {\it Decoherence}\ \ \ (to M. Schlosshauer)} \label{Schlosshauer1}

It was good meeting you.  I was just looking at some of your papers, and noticed various lines with the sentiment of this one (from your comment on Wiebe and Ballentine), ``This leads the authors to the general conclusion that decoherence is not essential to explanations of the classical behavior of macroscopic systems.  We show that these claims are not warranted \ldots.''  I would be curious to know your reaction to Johannes Kofler's two most recent entries on the archive:
\myurl{http://xxx.lanl.gov/find/quant-ph/1/au:+Kofler/0/1/0/all/0/1}.
Is it really being done without decoherence, or is it, as you were saying in yesterday's conversation, there in a hidden way?

\section{04-02-08 \ \ {\it Tsallis, {\Daroczy}, Gibbs \& Shannon}\ \ \ (to O. J. E. Maroney)} \label{Maroney2}

\bom
Fair 'nuff!

Of course, if I wanted to be pedantic, I could always say that if information theorists insist on calling the Gibbs entropy, Shannon information, there's no reason condensed matter theorists can't call the {\Daroczy} information, Tsallis entropy!! {\rm \smiley}
\eom

Have a look at Eq.\ (3) on page 6 of
\bq
\arxiv{quant-ph/9601025}.
\eq
Notice which name comes first in the line below that.

\section{05-02-08 \ \ {\it Q{\&}Q, 2}\ \ \ (to T. Slee)} \label{Slee2}

In LA at the moment, on the way  back to Waterloo from Australia.

\btsl
I confess the bit that puzzles me comes right in the first three words of that section ``When two systems \ldots'' I
just don't understand what ``two systems'' means.
\etsl

I could see that in the earlier blog entry you gave me.  It's true:  That's where your lack of familiarity with quantum information shows.  One wavefunction does not mean one system.  The notion of system is prior to any talk of wavefunctions.  Two spins are two spins:  They may be assigned a single entangled wavefunction (to describe one's knowledge of them), say, but they remain two spins.  Operationally, a system is defined by the parts of the whole that can be manipulated separately.  Formally, that is captured by writing down a tensor product Hilbert space ${\cal H} \otimes {\cal H}$ when describing the whole.  One ${\cal H}$ belongs to one component, one to the other, and ${\cal H} \otimes {\cal H}$ belongs to the whole.  A unitary operation on the left-hand system, but no action on the right-hand system, would be captured by a bipartite unitary $U \otimes I$ --- $U$ being the local unitary on the left, and $I$ being the identity operator on the right.  An action on the right, but not on the left, would be denoted by $I \otimes U$.  ``Local operations and classical communication (LOCC)'' is the bread and butter of all quantum information theory.  ``Bipartite'' is a common adjective in the field, not of my invention.

\btsl
A single closed system has a state vector, no?
\etsl

No.  A good Copenhagenist wouldn't say that (after a flavor, I am no more than a Copenhagenist myself).  Wavefunctions are not properties systems possess independent of considerations of what an observer knows about them.  Consequently a perfectly isolated system may be described by a pure state by one observer, but a density operator by another observer, and still a different density operator by another.  It simply depends upon what one knows.  This sort of situation arises all the time in quantum crypto.  Alice, the preparer, writes down a pure state for her system; Eve, the eavesdropper, writes down a density operator.  Nothing to do with ensembles:  There's only one copy of the system.  Instead, it has everything to do with knowledge or information.

\btsl
But the main frustration I felt was that, while you repeated in a couple of places that you reject hidden
variable theories (hooray!)\ I couldn't get what you were putting in their place (boo!). \ldots
 didn't get the answer to ``information about what?''
\etsl

All I can say is have patience.  That is the research program.  The whole bit about information is only the first part of the program:  It's the bushwhacking to clear the weeds---to make clear that it's not the quantum state that is the objective property of a system.  In the end, the ``information about what'' will be of the flavor of Bohr's answer, but finally made precise.  And as far as the ontology, what does the system have when there is nobody considering it?  That's neither the information, nor what the information is about.  But something else.  My presently favorite image is that---at a very deep level---quantum systems are catalysts.  That is their ontological role.

By the way, I now have a DVD of the public performance.  It's way way better than the radio program.  If you want to see Tony again, I could lend it to you.

Now, I fly.

\section{08-02-08 \ \ {\it Chipping at the Block}\ \ \ (to J. A. Vaccaro)} \label{Vaccaro1}

It was good meeting you last week.  I hope my personality didn't come off as too audacious the one day we talked \ldots\

But I've been having another crazy idea lately.  I'm thinking about organizing a meeting at PI on issues in pragmatist philosophy (perhaps with Lee Smolin), and I've been thinking about titling it, ``Pragmatism at the Perimeter: Chipping at the Block Universe.''  In that regard, I've become quite taken with your painting ``Broken Block''.  Three (not necessarily intersecting) questions about it:
\begin{itemize}
\item[1)] Would you send me a higher resolution .jpg of it, so that I might view it more closely?

\item[2)] Are you interested in selling it?  I'm starting to fancy it in my home library of pragmatism.  See attached photos (though wall color subject to change).

\item[3)] If said conference materializes, would you be averse to its being used in some way as a logo for the meeting?
\end{itemize}

Just exploring the issue.

\subsection{Joan's Reply}

\bq
It was a pleasure to meet you as well.  I seem to have the impression that you are fearless in your research so I didn't find you too audacious in person!

Regarding the painting, I am rather chuffed that you are interested in it. And I am delighted about the connection with chipping away at the block universe.  At the moment it is sitting in a pile in a cupboard somewhere so you are most welcome to have it, for free of course.  I would be happy to send it to you.
Just let me know of the address.  It is not very big, perhaps 10in x 8in.  It's not framed either.  I would be extremely chuffed if you happened to use it for a logo.  Golly.

The workshop idea sounds interesting.
\eq

\section{08-02-08 \ \ {\it Chipping at the Block Universe}\ \ \ (to L. Smolin)} \label{SmolinL10}

I'm back from Australia now if you want to pick up talking about pragmatic things.  Particularly, I'm thinking harder about that conference idea I mentioned to you.  In fact I think I'd like the running theme of it or its focus to be on arguments con (and for fun, a couple of talks pro) the ``block universe'' conception.  Thus I've tweaked the title a little bit to ``Pragmatism at the Perimeter: Chipping at the Block Universe,'' (and a few variations thereof) and have found myself staring at this painting of Joan Vaccaro's as a potential logo
\begin{center}
\myurl{http://members.optusnet.com.au/j.vaccaro/gallery/}.
\end{center}

PI conference proposals are due Feb 15.  If you have an interest, we should talk very quickly.

I read the first half of Unger's book, btw.  It was hard for me to see precision in a lot of it, but I did pick up a couple of ideas that were novel to me and I liked in the chapter on time.

\section{10-02-08 \ \ {\it Invitation and Potential Anti-Block Universe Conference, 1} \ \ (to C. Misak)} \label{Misak1}

We have never met, but I have two friends, Huw Price and Harvey Brown, who recommend you very highly and suggested that I contact you.  I am a researcher at the Perimeter Institute for Theoretical Physics in Waterloo (see our website at \myurl{http://perimeterinstitute.ca/} if you've never heard about us), who has a great private interest in pragmatism---particularly Jamesian, Deweyian, Rortyan---believing that the interpretational issues of quantum mechanics lead us straight up that path.  Witness the library I just had built for the subject in my new home in Waterloo!  (See attached photo; two books edited by you can be seen in the center section, fourth shelf from the top.)

Anyway, I write to you for two reasons.

1) Harvey and I would like you to give us a visit sometime this Spring, and, if you will agree, we'd very much like to you give us a seminar on Peirce and his ideas on chance, indeterminism, tychism, and such stuff.

2) Lee Smolin (also at PI) and I are thinking hard about organizing a conference here, tentatively titled ``Pragmatism at the Perimeter: Chipping at the Block Universe''.  The reason is that a few us here think our studies in quantum foundations and quantum gravity lead in a natural way to abandoning the block-universe picture.  However, we're somewhat at a loss about how to directly incorporate that insight more deeply into our physics.  Thus it seems like a good time to build a larger community of colleagues we could talk to in this regard, and we're thinking an event like this might help kick off the effort.  PI has a large budget for visitors and conferences.

The question is who might we invite as speakers?  I have a decently strong knowledge of classical pragmatism and thinkers, but I am not so familiar with the crowd of {\it living\/} (non-emeritus) pragmatists (and Bergsonians)---i.e., ones who might best contribute to a meeting a like this.  Do you have any recommendations?  Or, can you point me to someone who would be in a better position to make recommendations?  \ldots\ keeping in mind that the main topic of the meeting will be what's wrong with the block-universe conception and how one might formalize the idea of the opposite.

So, 1) I hope you will give us a visit, and 2) any suggestions you can give regarding the conference will be most appreciated!

\section{10-02-08 \ \ {\it Invitation and Potential Anti-Block Universe Conference, 2} \ \ (to C. Misak)} \label{Misak2}

\bcm
As for people you might invite to the conference (apart from Huw), I'd
consider Arthur Fine and Ian Hacking. Both of course can talk physics.
Arthur is a self-described pragmatist, while Ian is only described as
a pragmatist by me and others.
\ecm

That is already helpful.  I had, by myself, thought of Hacking (whom I met at a ``pragmatism and quantum mechanics'' meeting in Paris last February, a wonderful guy!), but I had not thought of Fine.  Would he be a pragmatist who waves the banner of the universe being on the make?  Or would he be a mushy pragmatist like Huw, who accepts the methodology and the anti-representationalist point on truth, but {\it rejects\/} the ``joint-stock society'' conception of nature (as James put it)?

\bcm
Someone who is less well-known, but is nonetheless excellent is Randy Dipert at SUNY Buffalo. He knows Peirce's mathematics very well.
\ecm

I will look into Dipert right away.

I'm not necessarily looking for people who know anything of the details of quantum mechanics.  Particularly, too much thought on the mysteries of QM might be a bad thing on their resume!  Clouding the view, so to speak.  If the meeting materializes, I would like several of the participants to indeed be immersed in the classical literature.  For instance, an ideal participant might be Ralph Barton Perry, if it weren't too late for that.  And Richard Rorty would have been wonderful.  I have wondered whether we might still be able to attract Hilary Putnam, or whether he's too old now.  It would be great to have an expert on Peirce's metaphysics (that seems easy, given the input you've already given), but also one on James's, one on Dewey's, and one on Mead's.  One on Bergson's?  One on Deleuze's?  \ldots\ Maybe I shouldn't push it too far!  But certainly a good grounding on all the classical pragmatists would be most helpful.

About your talk here, just pick a few possible dates:  our regular quantum foundations seminar is every Tuesday at 4:00, and you could take one of those slots.  The sooner the better!  We'd love to talk to you.

The attached document won't reveal too much of my particular interests, but it may help you to see how I see pragmatism as fitting into the bigger game (for me) of quantum issues.  It is a teaching proposal I wrote several years ago, for a potential joint position between physics and philosophy.

Thanks for the prompt response.  I'm grateful, and will let you know how things develop wrt the conference, and if you have more ideas please do let me know.

\section{10-02-08 \ \ {\it Dues Request, and a Question}\ \ \ (to D. M. Hester)} \label{Hester1}

Thank you for letting me know that my William James Society membership has expired.  I will renew right away.

In the meantime, I'd like to ask you a question.  I am now a researcher at the Perimeter Institute for Theoretical Physics in Waterloo, Canada (see our website at \myurl{http://perimeterinstitute.ca/}), and a colleague, Lee Smolin, and I are thinking hard about organizing a conference here, tentatively titled ``Pragmatism at the Perimeter:\ Chipping at the Block Universe''.  The reason is several of us here think our studies in quantum foundations and quantum gravity lead in a natural way to abandoning the block universe picture.  However, we're somewhat at a loss about how to directly incorporate that insight more deeply into our physics.  Thus it seems like a good time to build a larger community of colleagues we could talk to in this regard, and we're thinking an event like this might help kick off the effort.  PI has a large budget for visitors and conferences.

The question is who might we invite as speakers?  I have a decently strong knowledge of classical pragmatism and thinkers---witness my home library on the subject in the attached photo---but I am not so familiar with the crowd of {\it living\/} (non-emeritus) pragmatists (and Bergsonians)---i.e., ones who might best contribute to a meeting a like this.  Do you have any recommendations?  Or, can you point me to someone who would be in a better position to make recommendations?  \ldots\ keeping in mind that the main topic of the meeting will be what's wrong with the block universe conception and how one might formalize the idea of the opposite.

Thanks for any help you can give!

\section{11-02-08 \ \ {\it Potential Anti-Block Universe Conference}\ \ \ (to D. M. Hester)} \label{Hester2}

Sorry to continue bothering you with this, but you're my only connection to the world of William James.  Let me expand on the question I sent you yesterday:  Below is part of a conversation I had with Cheryl Misak yesterday.  [See 10-02-08 notes ``\myref{Misak1}{Invitation and Potential Anti-Block Universe Conference}, 1 \& \myref{Misak2}{2}'' to C. Misak.]  It says a little more about the cross-section I'm thinking of for the proposed meeting.

What are your thoughts on Tom Burke?  I was favorably impressed by his book {\sl Dewey's New Logic}.  Might he fill the shoes of an expert on Dewey's metaphysics I mention below?  Or are their better choices?

\subsection{Micah's Reply}

\bq
Hacking and Fine are good choices.  Putnam is too, but he is so busy, and his wife does not ambulate as well as she used to.  You can try, but I suspect you will get a polite decline.

Burke is good; Ray Boisvert at Sienna wrote a book on Dewey's metaphysics (though he hasn't written much on it since).  John McDermott would have something to say, but I' don't know that he would be able to make given his schedule.

What about someone like Christopher Hookway?  Russel Goodman at New Mexico might be good as well.

I'm just brainstorming \ldots\ as for James's metaphysics, Bill Gavin at Southern Maine might be a good choice, and Richard Gale (now in Knoxville) is always a kick.
\eq

\section{11-02-08 \ \ {\it Anti-Block Meeting Invitees (working list)}\ \ \ (to L. Smolin)} \label{SmolinL11}

Here's where I've gotten so far in my thinking about who might be invited to this meeting.  I'll go ahead and send this malleable list even at this stage, so that we might have a starting point to discuss things tomorrow.

I'm open to loads of suggestions and dollops of advice.  Please give.  The point of the meeting, as I see it, is to get PI people to stop thinking in block-universe ways.  If we can surpass that barrier, the thought is, our theory-making intuitions will be opened to grander vistas.  But the first step is to get a glimpse of what's over the barrier, and that's what this meeting should be mostly about.  We add a little talk on quantum mechanics (witness Bitbol, Fine, Shimony, and Price) to bring the focus back in---to give some sense that the issue is relevant to existing physics---but the bigger point is peering over the barrier, so that we can get to the next physics.

I hope we can get together tomorrow.

\begin{itemize}
\item
Michel Bitbol ---  director CNRS, \'Ecole Polytechnique, Paris.  Philosopher of physics, organizer of recent international meeting ``Pragmatism and Quantum Mechanics.''
\item
Raymond Boisvert --- Siena College.  Author of {\sl Dewey's Metaphysics}; see attached review.  [See William M. Shea, ``Review of {\sl John Dewey's Theory of Art, Experience and Nature: The Horizons of Feeling\/} by Thomas M. Alexander and {\sl Dewey's Metaphysics} by Raymond D. Boisvert,'' Am.\ J. Edu.\ {\bf 98}, 83--88 (1989).]
\item
Arthur Fine --- University of Washington.  Philosopher of science with a long history of influential research in quantum foundations; author of {\sl The Shaky Game:\
Einstein Realism and Quantum Theory}; pragmatist credentials stated vividly by Richard Rorty, who called him ``my favorite philosopher of science''.
\item
Richard Gale --- University of Pittsburgh.  Expert on James's metaphysics. Vita found here: \myurl[http://www.pitt.edu/~rmgale.vita.htm]{http://www.pitt.edu/$\sim$rmgale/vita.htm}.  See also attached review of one of his books.  [See Gerald E. Myers, ``Review of {\sl The Divided Self of William James\/} by Richard M. Gale,'' Philosophy and Phenomenological Research {\bf 64}, 491--494 (2002).] He looks fantastic actually; comes recommended by a contact in the William James Society.
\item
Russell Goodman --- University of New Mexico.  Author of {\sl Wittgenstein and William James}.  Vita found here: \myurl[http://www.unm.edu/~rgoodman/Vitae.html]{http://www.unm.edu/$\sim$rgoodman/Vitae.html}.
\item
Ian Hacking --- University of Washington.  Renowned philosopher of science, with a deep interest in the history of probability and chance.  Author of 11 books.  In a recent paper, ``On Not Being a Pragmatist:\ Eight Reasons and a Cause,'' he tries to fend off his fellow philosophers who classify him as a pragmatist, but regardless of the reasons for them he has well explored a set of thoughts quite relevant to this meeting.
\item
Louis Menand --- Harvard University.  Author of the Pulitzer prize winning {\sl The Metaphysical Club}---ostensibly a biography of Peirce, James, Dewey, and Oliver Wendell Holmes Jr., but one of the deepest and far ranging explorations of pragmatism that I have read.
\item
Cheryl Misak --- University of Toronto.  Influential scholar of Peircean pragmatism.
\item
Huw Price --- Sydney University.  Perhaps the most influential Australian philosopher of science, and our partner in the PIAF (PI--Australian Foundations) collaboration.  Deeply pragmatist about methodology, philosophy of language, and the meaning of truth, he is a surprise in his ideas about time (on which he has published extensively):  He may be the one pragmatist on earth who believes in the block universe picture!
\item
Hilary Putnam --- Harvard University.  The most famous pragmatist in this list without question!  Deeply influential in all branches of philosophy, and a philosopher who has given a significant amount of thought to quantum mechanics.
\item
Abner Shimony --- Boston University.  His credentials in quantum mechanics are well known.  But also he is a great admirer of Peirce's metaphysics.
\item
Roberto Unger ---
\item
Another expert on William James's metaphysics ---
\item
Expert on George Herbert Meade's metaphysics ---
\item
Expert on Bergson's metaphysics ---
\item
???  Who else ???
\end{itemize}

\section{12-02-08 \ \ {\it Conference, 2} \ \ (to A. Y. Khrennikov)} \label{Khrennikov23}

I apologize!!  I long since forgot about this.  Below are the few lines you requested:
\bq
\noindent We will also encourage a discussion of the significance and meaning of Kochen--Specker style theorems and the issue of contextuality in quantum mechanics, as recently there has been a resurgence of interest in these issues in the quantum information and foundations communities.  To this end, there will be a special session on the topic organized by C. A. Fuchs.  Speakers will include J. Emerson, C. A. Fuchs, I. Bengtsson, J-{\AA}. Larsson, J. A. Smolin, and others.
\eq
I'm so sorry, I was late on this.  I hope you will still add it to you conference webpage.

\section{12-02-08 \ \ {\it Anti-Block Meeting Invitees (working list), 2}\ \ \ (to L. Smolin)} \label{SmolinL12}

\bls
These all sound like good people, but the problem is that this is sounding like a philosophy conference and the people whose minds you want to change are not likely to go to pure philosophy talks.  You can go ahead anyway and make it as interesting as possible, and at least form a network of alliances of people and ideas.  To that end I would broaden it in the direction of other sciences with analogous or homologous issues like biology and economics.  Or you can focus it on the relevant physics issues concerned with time and invite people with the full spectrum of views from pro-blockers to believers in evolution of laws and make it a conference to discuss and argue differences of opinions.
\els

I recognize the truth in what you say.  On the one hand, I worry that my working list is too overtly philosophical, but on the other hand, I worry about fostering once again the usual stalemate that one finds at foundational meetings.  I guess I was imagining something different at this conference:  One that would lay the groundwork for people not too familiar with the subject of a different worldview from the usual one presented in our physics training.

At some point, one just has to jump off the diving board, I think to myself.  I'm reminded of the diary entry William James wrote when he was on the brink of suicide in 1870. [See 05-08-02 note ``\myref{Comer16}{The Spirit that Breathes Life}'' to G. L. Comer.]  At some point, one simply must {\it choose\/} to start physics afresh, to obdurately pursue the anti-block picture.  It, like free will, is a {\it choice\/} one must make before one can have it.

I'm just coming in to the office now (once I get my car out of the snow).  And I'm sure to be in a more sober-talking mood by then.  We can hammer this out!

\section{12-02-08 \ \ {\it Chipping at the Block, 2}\ \ \ (to J. A. Vaccaro)} \label{Vaccaro2}

Well, I'm very flattered that you'd honor me with your painting.  I'll take it!  Simply send it to my PI address below; it'll be better protected from the weather that way (rather than chancing the mail man leaving it on my porch in the snow).  It'll look great near all my volumes on William James (the philosopher who invented the very phrase ``block universe''):
\bq
What does determinism profess? It professes that those parts of the universe already laid down absolutely appoint and decree what the other parts shall be. The future has no ambiguous possibilities bidden in its womb; the part we call the present is compatible with only one totality. Any other future complement than the one fixed from eternity is impossible. The whole is in each and every part, and welds it with the rest into an absolute unity, an iron block, in which there can be no equivocation or shadow of turning. \ldots

Indeterminism, on the contrary, says that the parts have a certain amount of loose play on one another, so that the laying down of one of them does not necessarily determine what the others shall be.  It admits that possibilities may be in excess of actualities, and that things not yet revealed to our knowledge may really in themselves be ambiguous.  Of two alternative futures which we conceive, both may now be really possible; and the one become impossible only at the very moment when the other excludes it by becoming real itself. Indeterminism thus denies the world to be one unbending unit of fact. It says there is a certain ultimate pluralism in it; and, so saying, it corroborates our ordinary unsophisticated view of things.
\eq

He also invented the word ``multiverse'', though he used it in a very different sense than the Everettians do.

I keep my fingers crossed that the conference will materialize.  The difficulty is that I want it to be particularly philosophical, and PI may not be the right venue for that.  Despite Lee's warnings, I'm finding it difficult to give up on the idea.  Thus I'm starting to half toy with the idea of seeing if a philosophy department might throw in with us.  We'll see.

\section{14-02-08 \ \ {\it What To Do About Lewis?}\ \ \ (to R. {\Schack})} \label{Schack128}

Now, I've got a new urgency!  Sorry to worry you.  I just haven't been able to keep up with my obligations, in the face of all this damned construction around me.  It was worse when I got back from Australia than when I left.

To your question:  I'm not sure that apples and apples are being compared.  The argument you gave me a few days ago, is essentially the one of Lewis's paper, if I understand correctly.  And it's the one Wayne Myrvold has spit back at us on several occasions:  That the hypothesis of a chance is a hypothesis like any other.  Consequently, one can have a (subjective) ``credence'' about it and update accordingly in the light of new evidence.

\brs
What saves her from an infinite or finite regress is the fact that she
is a good subjectivist Bayesian. Her belief is her belief and does not
need to be justified. There is thus no need for her to give an
unambiguous description of the experimental setup that she is
repeating. Consequently, my whole anti-Everett argument falls apart.
\ers

There is no need to give an unambiguous description of how to produce the chance if she is treating it like a hypothesis, of which she is gaining confidence by finding the truth values of various ancillary propositions (i.e., the outcomes of those experimental trials).  This is no different from any (valid) conditioning.  By finding the truth value of {\it this\/} proposition, I update my degree of belief for {\it that\/} proposition, but I am under no obligation to give an exact description of the origin of that proposition.  I am under no obligation to further discuss the mechanics of what might physically make it true or false.

This is why I think you are comparing apples to oranges.  Your earlier argument---the one you attempted at the Everett meeting---was an attempt to attack elsewhere.  It was an attempt to attack the logical coherence of the ``truthmaker'' for the chance.  (``Truthmaker'' is actually a technical term in some nonpragmatist philosophies; see attached paper of Huw Price.  See also attached plays-on-words by me; I found much in the Price article that would be of value to us.)  The usual ``truthmaker'' for a Lewisian chance is ``identical chance situation,'' whatever that might mean.  Your attack was on the ``What means identical?''\ part of it, and the argument went like this.  To specify identical situations, one must supply a precise description of those situations.  But it is not good enough to give a classical description for a preparation device.  For, despite Bohr's exhortations, a classical description is not a precise description of a preparation.  One would have to supply the machinist with more than classical drawings if one wants him to build the device one truly has in mind.  And that more precise description would require quantum states.  And so one has an infinite regress in journey to find a precise description of a preparation.  That is to say, the only coherent truthmaker available for a chance is a quantum state, and the only coherent truthermaker for a quantum state is another, more all encompassing quantum state.  And so on, and so on.  So goes your argument.

But that has nothing to do with the Myrvoldian-style distraction expressed above, which is just a species of the original Lewis argument:  That chances are something we can have subjective credences about.  Your argument is not in contestation of that, but rather, ``What would make a chance true in the first place?''  I.e., even supposing Lewis and Myrvold are not flawed, what are these things we imagine we're building up credences for?  Answering that question leads to the infinite regress you point out.

The only thing I see potentially as a flaw in the argument is the Everettian's move to simply stop the regress by declaring a quantum state for the universe.  I think that might be partially why Simon Saunders is so bold as to say only the Everettian view solves all the deep issues in the interpretation of probability.  He is implicitly asserting that probability will ultimately only make sense if chance makes sense.

But that said, you know my own opinion is that there is simply nothing at all in the Everettian view; it really is ``sound and fury signifying nothing''.  They act as if it hangs together, but I don't even see it rise at all from the flat ground.  Thus, when they declare, ``See, there's no problem at all,'' they're not saying anything that's actually consistent.

Anyway, that's my thought.  If you spot a flaw in it, let me know.

\section{18-02-08 \ \ {\it Hirota Meeting Proposal} \ \ (to Perimeter Institute)}

\begin{center}
{\bf Conference Proposal:  Osamu Hirota, between Distinguishability and Noncausality \\
Principle Organizers:  Christopher A. Fuchs and Alexander Holevo}
\end{center}

\bq\noindent
\subsection{Synopsis}

Osamu Hirota has been a key figure in bringing two generations of quantum information theorists together---old QI and new QI---who might not have otherwise been so easily aware of each other.  His tireless conference organization, particularly the major semi-annual QCMC series, fundraising for visitor programs to Japan, and establishment of the QCMC awards for theorists and experimentalists in quantum information and computing, have been no small components in making this community what it is.  Thus, in honor of his 60th birthday (an important birthday in Japan), we would like to hold a small conference at PI, the participants being some of his best friends.   It so happens that those friends are some of the key figures in quantum information, and this is a very good excuse to attract them to PI to report on what they are thinking.

By old QI, we mean the first era of studying the classical-information carrying capacities of quantum channels and the distinguishability of quantum states in the early 1970s.  The key papers during that time being written by Levitin, Holevo, Belavkin, Helstrom, Yuen, Kennedy, Lax, and others.  By new QI, we mean everything after the discovery of quantum teleportation in 1993, the time when it was realized that the successful transmission of quantum states could be an end in itself.  In the first days of the newer era, these were completely separate communities.  But the international Quantum Communication, Measurement, and Computing conferences organized by Hirota saw an end to that.  Perhaps a personal anecdote will help convey the importance of these meetings.  In 1996, a topic of great interest with the newer generation was the classical-information carrying capacity of a set of nonorthogonal quantum states, for which the answer was not known.  In fact, an amazing quantum effect---the superadditivity of capacities---had only recently started to be understood in this connection in that community.  But at the 1996 QCMC meeting, to which Hirota  had drawn Alexander Holevo to after his not being involved in QI issues for many years, it became clear that this ``new'' effect was in fact quite old.  It was reported in a 1978 paper by Holevo, but the paper somehow fell off the map and was simply unknown to those of us in new QI.   Soon after this meeting, progress was rapid, and we can thank it for bringing Holevo and several others back to the field.  Stories like this abound and make Osamu Hirota's contribution to the quantum information community very honorable indeed.

\subsection{Research Areas Benefited}

A broad range of quantum information studies at PI should benefit from the meeting.  The invited speakers---Charles Bennett, Peter Shor, Alexander Holevo, etc.---will virtually ensure that.  But, the focus is likely to lean toward capacity and superadditivity issues in their various guises.  There may also be discussion on the foundations of quantum information itself (particularly in D'Ariano, Ozawa, and Fuchs's talks).  One long-standing open problem that many of the participants have contributed to is the issue of the minimum entropy-making inputs to tensor product quantum channels:  That will most certainly be a hot topic of discussion.  Several researchers at PI and IQC, I am quite sure, will be poised to build on what they learn from the talks and discussions.

\subsection{Relevance and Timeliness of Event}

The timeliness of the event is dictated by Hirota's 60th birthday, which is in June.  The relevance is that their deep friendship with him will bring some of the world leaders in quantum information to PI to share their thoughts.

To emphasize the caliber of the invitees, we list their associations here in more detail than can be given in the space set aside for it on page 6.  Charles Bennett---one of the very fathers of new-QI, having been a co-discoverer of several of the most important effects in the field:  1) quantum teleportation, 2) super-dense coding, 3) the B92 quantum crypto protocol, 4) entanglement distillation, 4) the entanglement assisted capacity, and the list goes on.  John Smolin---also from IBM Research, seminal contributions in all aspects of quantum information theory, especially through the power of his numerical work (it would be hard to count the number of interesting quantum information items first found by his ``minimizer'', for instance the five-qubit code).  Alexander Holevo, played a role much like Charles Bennett, but in old-QI.  Jeffrey Shapiro is director of MIT's historical Research Laboratory of Electronics.  Masanao Ozawa and Horace Yuen were the first to prove that the so-called ``standard quantum limit'' to measuring successive positions of a particle could be exceeded; since both have had long distinguished careers in many aspects from the very theoretical to the quantum optical sides of quantum information.  Masahide Sasaki is the director a major quantum information laboratory in Japan.  Peter W. Shor, founder of the quantum factoring algorithm and the first to discover the existence of quantum error correcting codes.

\subsection{Desired Outcomes}

A report of the state of the art in various quantum channel questions---from the IBM Research team (represented by Bennett and Smolin), the  Pavia team (represented by D'Ariano), the efforts at MIT (represented by Shor and Shapiro), the Tohoku team (represented by Ozawa)---should be of  benefit to all.  A worthwhile goal would be to make good progress on one or more of the additivity issues.
\eq

\section{19-02-08 \ \ {\it These Eyes} \ \ (to D. M. {\Appleby})} \label{Appleby26}

\bma
It is true:  I had forgotten about Hans, and yes I would really like
to see him.  As you say, my purview, like yours, is much much bigger
than SICs.
\ema

And I should keep you up to date on what I've been thinking:  It is what I think is maybe THE ultimate reason for our studying SICs.  I now have a good sense in which a SIC representation of quantum states gives the very best unentangling possible between Bayesian probability theory and what is different in quantum mechanics.  It is clean and simple, and had been right in front of me for over a year.  Below is a weak summary of it.  [See 25-01-08 note ``\myref{Gottesman8}{Bristol, 3 AM}'' to D. Gottesman.]  Sorry that it's not more detailed, and I don't want to take the time to write it more properly now (showing the sense in which this is the best possible disentanglement, etc.).  But in my mind, this (and also the similar representation of unitarity) are now the real reasons I use for justifying the study of SICs.

At the PIAF lecture in Sydney, I put all this in Dutch book terms, but unfortunately there's only an audio file of the lecture here:
\myurl{http://idisk.mac.com/centre.for.time-Public?view=web}.
I think one really needs visuals.  We can talk in detail when you get here.

There's all kind of poetry and prose that little equation evokes in me:
\begin{description}
\item[C:] ``World, I need your help factoring this large number.''
\item[W:] ``Fine.  Push on me the right way, and I'll cough up the answer for you.''
\item[C:] ``But be warned, I'm in a hurry.''
\item[W:] ``Then you too be warned.  Be gentle, don't push on me more than you have to; I'll
    give you a reward in the end.''
\end{description}

\section{19-02-08 \ \ {\it These Eyes, 2} \ \ (to D. M. {\Appleby})} \label{Appleby27}

\bma
Also, am I right to assume that the reduction in entropy (if that is
the measure of ``mixedness'' you are using?) as compared with what you
would have classically is greatest possible for a SIC?
\ema

No, it's that the quantum ``residue''---the last remaining hint of quantum mechanics, thought of as a deviation from the law of total probability---is the smallest.  I.e., when using a SIC in the Bureau, one gets an expression for the $j$ measurement in terms of the $i$ measurement that is as close in form as possible to the law of total probability.  There is still a deviation from it, but that deviation is functionally as small as possible---and that I presume in some shape, form, or fashion, represents the core of quantum mechanics.  It is a hidden expression, I think, of the source of the Kochen--Specker theorem:  unperformed measurements have no results.

\bma
I find this interesting for another reason as it touches on a question
I discussed with Lane, possibly in a note I forgot to copy to you.  We
were speculating whether SICs are really and genuinely what they
superficially seem to be:  namely, the maximally symmetric rank $d^2$
POVM.  The reason I thought they might not be is the triple products,
which seem to be quite nasty for $d>3$.  I was wondering if there might
be a POVM in which, the squared moduli of the overlaps are not all the
same, but in which the triple products are nicer.  Perhaps even a POVM
in which the triple products are all the same?  If it is true (as I
think it might be) that the triple products are more fundamental than the overlaps such a POVM might have a better title to the term ``symmetric'' than SIC.  At any rate it might have an equally good title.
\ema

I find that question interesting too.  But for the present problem, it is the symmetric measurements in our usual sense (of any rank)---rather than the tri-SIC measurements if they exist---that touch closest to the Bayesian representation.  I show why in the colloquium I gave here:  \pirsa{08010004}.  It's only the symmetric measurements (of any rank) that give a quasi-diagonal transformation between the $P$ and $Q$ functions, so to speak.  And then, the rank-1 symmetrics further minimize the constants as much as possible.

So, the most important layer for the questions that titillate me---the ones about this issue of total probability---is that we first satisfy SICness completely.  Thereafter, we try to satisfy tri-SICness as close as possible in one or another senses.

It would have been lovely if we could have had them both at once, SICness and tri-SICness!  But there's certainly something to be learned here by finding that our easiest fantasies are being blocked by something.

\section{19-02-08 \ \ {\it Speaking of ``Blocked by Something''} \ \ (to D. M. {\Appleby})} \label{Appleby28}

Here's a lovely quote I ran across this weekend in William James's 1904 review of F.~C.~S. Schiller's book {\it Humanism}.  (Humanism was Schiller's word for pragmatism.)
\bq
But humanistic empiricism will have many other steps forward to make before it conquers all antagonisms.  Grant, for example, that our human subjectivity determines {\it what\/} we shall say things are; grant that it gives the ``predicates'' to all the ``subjects'' of our conversation.  Still the fact remains that some subjects are there for us to talk about, and others not there; and the farther fact that, in spite of so many different ways in which we may perform the talking, there still is a grain in the subjects which we can't well go against, a cleavage-structure which resists certain of our predicates and makes others slide in more easily.  Does not this stubborn {\it that\/} of some things and not of others; does not this imperfect plasticity of them to our conceptual manipulation, oppose a positive limit to the sphere of influence of humanistic explanations?  Does not the fact that so many of our thoughts are retroactive in their application point to a similar limit?  ``Radium,'' for example; humanistically, both the {\it that\/} and the {\it what\/} of it are creations of yesterday.  But we believe that ultra-humanistically they existed ages before their gifted discoverers were born.  In what shape?  There's the rub!\ for we have no non-humanistic categories to think in.  But the {\it that\/} of things, and their affinity with some of our {\it whats\/} and not with others, and the retroactive force of our conceptions, are so many problems for Humanism over which the battle is sure to rage for a long time to come.
\eq

\section{19-02-08 \ \ {\it Octahedra in Higher D}\ \ \ (to S. T. Flammia \& D. Gottesman)} \label{Flammia3} \label{Gottesman8.1}

Are there any standard names for higher dimensional analogs of regular octahedra?  Analogs in any sense, I suppose.  But particularly, polytopes with $d(d+1)$ vertices whose ambient space is a $d^2$ dimensional linear vector space?  Does that kind of object have a name?

\section{19-02-08 \ \ {\it Octahedra in Higher D, 2}\ \ \ (to S. T. Flammia \& D. Gottesman)} \label{Flammia4} \label{Gottesman8.2}

\bstf
I don't know of any such object.  The only generalization of the octahedron that I know of is called the cross polytope, and it is simply the polytope of the vectors having unit 1-norm.  If the ambient space has dimension $d$, then it has $2d$ vertices, so it isn't what you are looking for.
\estf

Well, it might be, and I'm just asking the question in a confused way.  What I'm really asking is this.  Take a complete set of MUBs and connect the projectors to make a polytope.  Does that kind of polytope have a name?  Bengtsson and Ericsson called it ``the complementarity polytope'' when they discussed it in their paper, but I was wondering whether there is a more common name for the structure, if one strips away considerations about it being embedded in the set of quantum states.

\section{19-02-08 \ \ {\it Lunch at 12:10?}\ \ \ (to S. T. Flammia)} \label{Flammia5}

\bstf
I'm at UNM until Friday, then I'm going on the ski trip Monday.  I'll be in next Tuesday, the 25th.
\estf

OK, I'll see you next Tuesday.  You're probably going to wince when you see this, but here's the conference proposal I ended up submitting for us.  Lucien got a chuckle from my verbal report of it, and told me he'd let me know before it goes to committee if there's anything too outrageous in it.

\bq
\begin{center}
{\Large\bf Conference or Workshop Program Proposal}\bigskip
\end{center}

\noindent {\bf Name of Conference or Workshop:}

Seeking SICs:\ An Intense Workshop on Quantum Frames and Designs at PI \medskip

\noindent {\bf Principal Organizers:}

Steven T. Flammia and Christopher A. Fuchs \medskip

\noindent {\bf Topic and Research Area:}

Quantum Foundations, Quantum Information Theory \medskip

\noindent {\bf Synopsis (Please make this accessible to non-experts. See sample proposal.)}

What is the shape of Hilbert space?  More precisely, what is the geometry of the set of quantum states, thinking here of quantum states as a subset of the Hermitian operators?  What symmetries does this set have?  This question is of deep importance to several issues in quantum information theory, and it is of deep importance to a surging school of thought in quantum foundations, the quantum Bayesian approach.  We propose an intense workshop to settle, or at least make significant progress with regard to a seemingly simple, but extremely recalcitrant, version of this question:  The question of the existence of minimal symmetric informationally-complete (SIC) sets of pure quantum states.

The question is simply this:  Take the set of pure quantum states (one-dimensionsal projection operators) for a $d$-level system; these operators span the $d^2$ dimensional space of Hermitian operators.  Can one create a regular simplex (the higher dimensional analog of a regular tetrahedron) of $d^2$ vertices with elements drawn from this set?  This is seemingly an almost trivial question---one of the most basic questions one can ask of a convex set---but it has a long unsolved history, ranging back, in one guise or another, at least 35 years.  A related question, and a kind of scratchpad for the more pressing problem, is the question of existence of ``complete sets of mutually unbiased bases (MUBs)''---a kind of higher dimensional analog of regular octahedra---in composite dimensions.  In all, these are examples of so-called ``2-designs'' and ``tight frames'' with very specialized properties, hence the subtitle of our workshop proposal.

SICs and MUBs, if and when they exist, are already known to have some remarkable properties, both for quantum information processing and quantum foundations.  For instance, thinking of them as measurements that can be performed in the laboratory, they are known to be optimal for certain ways of quantifying efficiency in quantum tomography.  And SICs in particular may be used to form ``maximally sensitive'' alphabets for eavesdropping detection in quantum cryptography.  For foundational purposes, SICs, if and when they exist, satisfy a sense of being as close to an orthonormal basis in the space of density operators as that structure will allow, and give rise to phase-space representations of quantum states and unitary evolution that are as simple as they are allowed to be in such representations.  Furthermore, both SICs and MUBs may provide deeper insight into and new techniques in quantum computation, as both structures appear to have very interesting connections to the Clifford group.  Thus there is ample (and ever growing) reason for trying to settle this question of existence.

The idea of the meeting is to gather the best people in quantum information theory who have given this problem significant thought, put them all in one place, and see progress made, damned be, by any means!  We'll start off with a fiery rendition of Henry V's St.\ Crispin's day speech:
\bv
      This story shall the good man teach his son;
\\      And Crispin Crispian shall ne'er go by,
\\      From this day to the ending of the world,
\\      But we in it shall be remembered---
\\      We few, we happy few, we band of brothers;
\\      For he to-day that sheds his blood with me
\\      Shall be my brother; be he ne'er so vile,
\\      This day shall gentle his condition;
\\      And gentlemen in England now-a-bed
\\      Shall think themselves accurs'd they were not here,
\\      And hold their manhoods cheap whiles any speaks
\\      That fought with us upon Saint Crispin's day.
\ev
roll up our sleeves, brace for the scars, and get to work.  Each day (save the first) will consist of two morning talks, each on some aspect of the problem, with the afternoons reserved for round-table/chalkboard working sessions in the Alice Room.  The simple idea is to run the troops into the breach opened by these last few years of research, and finally defeat this cantankerous problem!\medskip

\noindent {\bf What research areas at Perimeter Institute will benefit from the event and how? How does this event benefit your own research?}

Both Quantum Information and Quantum Foundations research at PI will benefit for the reasons described in the synopsis.  In the case of Flammia, the problem has been a significant part of his research in the last few years.  In the case of Fuchs, it's now a religious quest---being a significant ingredient in his sought-for interpretation of quantum mechanics. \medskip

\noindent {\bf Please include a justification of the relevance and timeliness of the event (feel free to explain in technical terms).}

The earliest known posing of the SIC problem was in 1973 by Lemmens and Siedel in the context of the maximal number of ``equiangular lines'' that can be supported by a vector space.  In real vector spaces, an upper bound of $d(d+1)/2$ on the answer could be proved, and in complex spaces a similar bound of $d^2$ could be similarly proven.  Early on, however, it was understood that the upper bound generally could not be achieved in the real case; in fact, the actual maximal number turned out to be a very complicated function of $d$.  Because of this, perhaps, it was thought to be similarly so in the complex case, and interest in the problem languished---until 1999 it was only known that the complex-case bound could be achieved in dimensions 2, 3, and 8.  With the study of Renes et al.\ in 2003, however, all that changed:  Numerical work now indicates that SICs exist in dimensions 2 through 47, and analytic constructions (most via the help of computer algebra packages) exist in dimensions 2 through 14.  Thus progress has been very rapid.  At least 46 papers on the {\tt quant-ph} archive cite the original Renes et al.\ paper, and something like 10 of those have been devoted in an substantial way to the existence problem.  Of the related problem with MUBs, over 100 papers on the archive have the words ``mutually unbiased bases'' somewhere in the record.

Here at PI, at least three residents have devoted a significant amount of brain pulp to the SIC existence problem (Blume-Kohout, Flammia, and Fuchs) in the last couple years, with brief forays by some others (Gottesman, Hardy).  Also we can list several visitors of PI that have given a significant amount of time to the problem (months to years in some cases):  D. M. Appleby, L. Hughston, J.-{\AA}. Larsson, H. Barnum, C. M. Caves, M. Roetteler, A. J. Scott, and W. K. Wootters.  The point is the problem has interested many who have walked in these corridors.  It is simply time to settle it if at all possible. \medskip

\noindent {\bf What are the goals and desired outcomes of the conference or workshop?}

To amass as much new information about SICs and MUBs as possible, and possibly to prove existence of SICs in all finite dimensions (or at least an infinite class of dimensions).\medskip

\noindent {\bf For the invited participants and speakers please include their full name and affiliation.} \medskip

\begin{supertabular}{ll}
D. Marcus Appleby & Queen Mary University of London \\
Howard Barnum & Los Alamos National Laboratory \\
Ingemar Bengtsson & Stockholm University \\
Robert Calderbank & Princeton University \\
Carlton M. Caves & University of New Mexico \\
Steven T. Flammia & PI \\
Christopher A. Fuchs & PI \\
Markus Grassl & University of Innsbruck \\
David Gross & Imperial College \\
Lane Hughston & King's College London \\
Andreas Klappenecker & Texas A\&M University \\
Martin Roetteler & NEC Laboratories \\
Andrew J. Scott & Griffith University \\
Peter W. Shor & MIT \\
Neil J. A. Sloane & AT\&T Research \\
William K. Wootters & Williams College \\
\end{supertabular}

\eq

\section{20-02-08 \ \ {\it Your 60th Birthday}\ \ \ (to O. Hirota)} \label{Hirota6}

Alexander Holevo made me aware that your 60th birthday will happen soon.  It is in June, if I am not mistaken?  Thus, in your honor, we would like to organize a small ``birthday party'' for you here at the Perimeter Institute.

In that regard, we have just written a detailed conference proposal and submitted it.  It is titled, ``Osamu Hirota, between Distinguishability and Noncausality.''  If it is accepted by the conference committee at PI, then we would be very much honored by your attending the meeting.  I should know shortly whether we will get the funding.

Under the assumption that it will be approved, let me tell you a little bit about the proposal.  Preliminarily, I have listed the following colleagues as invited speakers:
\bv
Charles Bennett \\
Mauro D'Ariano\\
Christopher Fuchs\\
Osamu Hirota\\
Alexander Holevo\\
Debbie Leung\\
Masanao Ozawa\\
Masahide Sasaki\\
Jeffrey Shapiro\\
Peter Shor\\
John Smolin\\
Horace Yuen
\ev

We probably have room for 1 or 2 more invited speakers if you would like to add anyone to the list.  Perhaps further colleagues or important students from Japan?  Or others that we have forgotten?

If the proposal is approved, we should be able to cover everyone's travel and local expenses, and we would also be able to have a nice banquet for you here at PI's Black Hole Bistro.

An important question for you to be thinking about, is what dates you would like it.  The meeting should be 3 days long, but when?  If you approve, please also send me a range of dates for which it would be possible.  Summer is a very busy time for the facilities here at PI, so I may have to fight for an empty slot in the planning.  Thus, the more possible dates you have the better.

I hope you will accept our proposed birthday gift for you!  You have been a wonderful friend over the years.

\section{25-02-08 \ \ {\it Newsletter Submission}\ \ \ (to E. Goheen)} \label{Goheen1}

\beg
We would like to prepare an article on the recent PIAF workshop and what this new partnership/project means for PI for the March edition of our newsletter, Inside the Perimeter. Lucien suggested I might be able to ask you a few questions about it.
\eeg

Here are my answers finally.  I'm sorry for the delay.

\beg
What is the partnership, who is involved?
\eeg

It's called PIAF (and pronounced as the name of the great French chanteuse Edith Piaf) but stands for the PI-Australian Foundations partnership.  The participants on the Australian side are Sydney University, the University of Queensland, and Griffith University, with principle investigators Huw Price and Stephen Bartlett at Sydney, Gerard Milburn at Queensland, and Howard Wiseman at Griffith.  The partnership will fund three postdoctoral positions based in Australia, all with significant visiting periods at PI.  Thus, between faculty, existing postdocs, new postdocs, and graduate students, there's probably on the order of 15--20 people involved.

\beg
How did this project come to be?
\eeg

It was the brain-child of Howard Burton, really.  Howard recognized that the PI quantum foundations effort cannot live in a vacuum if it is going to achieve its potential.  This is because quantum foundations is unique among the research areas pursued by PI in that, outside our walls, there simply isn't much of an infrastructure for the field.  This contrasts with string theory, for instance, where there are several internationally recognized research centers for it (PI, Princeton, Stanford, Caltech, etc.).  Thus a flow can be established for incoming and outgoing postdocs at those institutions.  But quantum foundations is isolated in comparison---PI might just be the end of the career for many a good quantum foundations researcher.  Thus, if we really want the field to flourish, part of our effort should be of the outreach variety, making physics departments around the world aware of the return one can get from quantum foundations research, and to convince them it's worth investing some of their own resources to this field.  PIAF represents our first effort in that direction.

\beg
How will PI benefit from these types of partnerships?
\eeg

The benefits should  be manifold, and not only for the reasons already mentioned.  PIAF also establishes a vigorous visitor program between our institutions, and we can all imagine the benefits of that.  Sometimes a calculation long processed in the back of one's mind simply needs a jolt of crisp, wintery Ontario air to bring it to life.  Sometimes too a heated mid-Winter (i.e., mid-Summer!)\ discussion at a sidewalk cafe in Sydney might be just the thing to thaw loose a key idea previously frozen away.  Thus we expect many good things to come from this collaboration.  Finally, PIAF intends to establish the premier annual international meeting in our subject.  In general relativity, there are the famous Marcel Grossman meetings; in quantum information, there were the famed Torino meetings; in quantum foundations, there will be PIAF!

\beg
How did the first workshop go? Are there more planned in the future?
\eeg

It was fantastic, far exceeding expectations.  The original intent for this meeting was for it to be a kind of ``kick-off meeting'', establishing what the various researchers would like to get out of this collaboration.  In the end, it somewhat-spontaneously turned into a full-fledged research conference with about 35 participants (17 speakers), with a lively round-table discussion one session.  Our next meeting will be here at PI and will be the first of the international series described above.  This year's theme will be, ``Time and Quantum Foundations.''  But the theme will change yearly, so stay tuned!  Particularly, as it has been noticed that Hawaii lies exactly halfway between Waterloo and Sydney on the geodesic \ldots

I hope these were the sorts of answers you were looking for.

\section{26-02-08 \ \ {\it Great News} \ \ (to A. Wilce)} \label{Wilce18}

That's great news indeed about your tenure!  Well deserved and all that old boy, harrumph.  \ldots\  But, for a serious note:  You are now in a position of security, and that is a call to do something great with your life.  Don't pass up the opportunity.  If you were hammering away at quantum mechanics already, it's time now to pull out the jackhammers.  We've got to understand this damned theory!  We've got to understand it by any means!

So, indeed, very great news from you!  Give my regards to your wife.

I'm soon off for a whirlwind tour of Eugene, Oregon.  One day flying there, and one day flying back, all for the opportunity of spreading the word of SICs for one day in the middle.

\section{28-02-08 \ \ {\it Even a Title} \ \ (to D. Gottesman)} \label{Gottesman9}

And here's the title I'll give the talk, whenever I do give it:  ``Dressing Richard Feynman in Bayesian Clothes''.  It just came to me.  Despite the title, it'll be a chalk board talk with plenty of equations.

\section{29-02-08 \ \ {\it Seattle Note}\ \ \ (to H. B. Dang)} \label{Dang8}

One of our postdoc candidates (Piero Mana) impressed me very much in many ways during his visit here.  Particularly, I learned that he carried a sheet of paper in his pocket marked off with 365 $\times$ 75 squares --- each square was to represent a day in his expected lifespan.  He told me he kept it with him to remind him of the urgency of life and so that he would not forget to prioritize.

\section{29-02-08 \ \ {\it Past the Halfway Mark} \ \ (to R. {\Schack})} \label{Schack129}

\brs
Exchangeability is not a property
of the world, but it depends on the agent's prior judgement.
\ers
I would say ``rather,'' rather than ``but''.

\brs
The concept of chance is therefore simply redundant in a classical
deterministic theory.
\ers
That's not quite what you want to say.  You've already pointed out how the idea of chance in a deterministic theory is simply inconsistent.  ``Unneeded'' perhaps, rather than ``redundant''.

\brs
It turns out that the attempt to define such a preparation procedure
leads to the same regress as in the classical case. The state prepared
by any preparation device depends on the quantum state of the
preparation device itself. To guarantee that the same
system state is prepared in each trial, one has to make assumptions
about the quantum state of the preparation device, which means that
one has simply moved the problem up one level. This leads to a regress as claimed.
\ers
Why are you always so terse?  This is a significant point of the paper.  It seems to me that you could expand it to the level of our verbal discussions.

\brs
From the perspective of a full quantum Bayesian theory the answer
is that objective quantum states, and therefore the objective
wavefunction of the universe, have no useful place in quantum theory.
\ers
And in fact cause great trouble.

\section{29-02-08 \ \ {\it Readers' Rights} \ \ (to R. {\Schack})} \label{Schack130}

\brs
Instead of two fundamental concepts---objective quantum states and
decision-theoretic preferences---and their awkward connection via the
principal principle, we now have a single fundamental concept. This
translates into much greater conceptual and mathematical simplicity.
\ers
I think it would be within the reader's rights to ask:  Now why, really, is that an awkward connection?  It still looks very natural to me.  How have you convinced me that it's awkward?  So much of what you think and feel is left silent in your papers.  Is it really that I'm carried away by your oratorical skills in our personal conversations, and when it all boils down, the content is what you have recorded on these pages?

\brs
In the quantum Bayesian approach, the prior is a single density
operator on ${\cal H}_n$. In the many-worlds approach, this prior is a
probability distribution over all density operators on ${\cal H}_n$, a much
more complex mathematical object
\ers
I like something of the flavor of this, but it is the same mathematical object (the representation theorem gives the isomorphism).  What are you really trying to say when say that it's more complex?

I doubt I was very helpful but those are the most thoughts I've been able to muster this morning.  (I'm in Seattle now; just finished with a long talk with a soldier going back to Iraq.  I would love to see George Bush tarred and feathered.)

Send me the next iteration of the paper; I'll try to think a little more deeply once I'm back in Ontario.

\section{03-03-08 \ \ {\it t.tex} \ \ (to R. {\Schack})} \label{Schack131}

Your use of ``t'' for the names of almost all files is as mysterious to me as the word ``magma'' for the most all inclusive algebraic structure in Bourbaki.  (Any idea why they use that name?)\footnote{\editornote Peter Shor writes,
\begin{quotation}
\noindent It may be a pun. Looking up \emph{magma} in the French [W]ikipedia, another name for \emph{magma} in French is \emph{groupo\"\i{}de de Ore}. Here Ore is a Norwegian mathematician, but \emph{ore} in English is mineral-bearing rock, whereas \emph{magma} (in both English and French) is molten rock. [\ldots] Larousse says the figurative definition of magma is \emph{M\'elange confus, inextricable de choses abstraites} and gives the example usage ``\emph{Ces propositions constituent un magma incoh\'erent.}'' Looking at some instances of its use, I would say the English definition would be something like \emph{confusing, inseparable, and worthless jumble}; this isn't a bad name for the algebraic structure, although possibly the pun [\ldots]\ may have also influenced Bourbaki's choice of name.
\end{quotation}
See \myurl{http://english.stackexchange.com/questions/63210/etymology-of-magma-in-abstract-algebra}.}

I read the appropriate areas; sorry for the delay.  It's better now.  Send it off---not much more to do.  But I do continue to fear that the regress argument will still not be accepted by most.  Why {\it must\/} one assign a quantum state to a preparation device?  At that point it loses its role as a preparation device, a Bohrian would say.  But this paper is not written for the Bohrians, I suppose.

\section{08-03-08 \ \ {\it Chipping at the Block, 3}\ \ \ (to J. A. Vaccaro)} \label{Vaccaro3}

\noindent Dear JV, (just looking at your art page again, where I see you call yourself that),\medskip

I don't know how I can thank you enough.  The piece is just gorgeous, and as I said before I'm very flattered that you'd let me have it.  It'll hold a place of honor in my library.

Things are starting to look good for the conference too.  John Sipe contributed a crucial idea that I think will make it start to look favorable in the eyes of PI, and Lee Smolin now seems to be pretty on board after hearing that.  So, soon, I'll put a formal proposal in with PI.  The idea is to have the philosophers read formal papers on some issue exposing cracks in the block universe conception, and then have physicists reply.  The papers would be prepared well in advance of the meeting, so that the chosen physicists would have to time to develop a relevant response, explaining how something the philosopher said might be relevant (or not) with respect to a given issue, or rather arguing that the philosopher is only spouting some antiquated 19th century idea that has since been made very unlikely by modern physics, etc.  If at all possible, I'll get your painting into the conference poster.

I'll write again when the painting arrives.

\section{10-03-08 \ \ {\it Stabilizer States} \ \ (to M. G. Raymer)} \label{Raymer2}

My, it takes me a long time to reply to emails, doesn't it?!  I'm sorry.  But thanks for the hospitality in Oregon; I really enjoyed the trip.  I was also impressed by your sensitivity to ``pragmatic'' issues.  Attached is a picture of the Waterloo Library of Pragmatism, aka Chris's study---you'll see I'm quite serious!

Thanks also for the note.  It sounds exciting!  Is it true?  Like Caroline Thomson said of me, I really do not have an intuition.  Honestly, I didn't understand why you say ``the answer is independent of the angle.''    Have you tried passing it by the van Enk test?  If it gets that far, then it'll probably be a hop skip and a jump from a formal proof, and I'll try to weigh in on that.  In the meantime, I have learned of a result saying that in finite dimensions, there are no SIC fiducial states with equal amplitudes, nor even such states with periodic amplitudes.  So, if your ansatz does work out in the infinite limit, that will be quite interesting.  For it would mean the infinite limit comes about in a nontrivial way.

\section{10-03-08 \ \ {\it That Paper} \ \ (to M. G. Raymer)} \label{Raymer3}

Also here's the paper I was telling you about on the eight-port detector being able to measure all Weyl--Heisenberg covariant POVMs (making use of an appropriate ``parameter'' state in one port): \arxiv{0708.4094}.

\section{12-03-08 \ \ {\it Reality Check} \ \ (to A. Kent)} \label{Kent15}

\bak
I was really calling just to be quite sure we've reached closure on the possibility of you contributing to the
many worlds book.
\eak

I held off on replying to you because I had hoped there was some chance I'd pull through and have something relevant written by now.  But reality rears its face again.  I think, honestly, there's not a chance I'd have something finished by ``end of March (absolutely final firm deadline)''.  So, thanks for encouraging me, but I will disappoint you again.

In the end, these foundational wars---weighing one stale point of view against another stale point of view---just get old.  What new physics comes from the effort?  What new vista opens before our eyes?  A complete rewrite of quantum mechanics is called for, and none of those guys see it.  Nor will they ever.  So, I think it is better for me to keep going on my own way.

\section{17-03-08 \ \ {\it 0803.1264} \ \ (to C. H. {\Bennett}, G. Brassard, and J. A. Smolin)} \label{Bennett60} \label{SmolinJ9.1} \label{Brassard52}

\bcb
This looks like the kind of thing you (and CBH) were looking for.  Do you understand what their convex framework of theories is, and how they define entanglement within it without automatically limiting the theory to standard quantum theory?  Also, is their convex framework broad enough to encompass theories like Smolin's Pangloss universe, or Intelligent Design?
\ecb

To answer Charlie's question, the convex framework certainly encompasses a Pangloss universe (even if it's not exactly John's version), simply because it contains quantum mechanics.  And the quantum world---we're bound to eventually learn---is surely the best of all possible worlds.

\section{18-03-08 \ \ {\it Application}\ \ \ (to H. B. Dang)} \label{Dang9}

I am presently in the Austin airport, delayed---perhaps indefinitely---by thunderstorms.  Makes for a very miserable day.  But Appleby will arrive at PI Sunday for a week, and then he's coming again for two months in the summer.  So, I have something to look forward to.

\bhbd
I guess what I'm still missing is how probabilistic quantum mechanics
can result in a more intuitive formulation.
\ehbd

Intuitive is not necessarily the watchword.  Reread my \quantph{0205039}.  I just want to get at the core of what QM is telling us.  And that core, by conventional standards, might appear to be the conjunction of two or three {\it seemingly\/} contradictory statements (like, for instance, is the case with special relativity).  The point of a probabilistic representation is that it most directly gets at the (correct) meaning of the quantum state and the correct meaning of the Born rule.  With the underbrush cleared away, then one should be able to more easily see the true forest.

\section{18-03-08 \ \ {\it Needing a Little Jung Myself} \ \ (to D. M. {\Appleby})} \label{Appleby29}

I should record this snippet of a dream while I happen to be thinking of it again.  (Emma and I were just talking about the importance of dreams.)  I remember little of last night's dream except that, in it, I thought I was giving a particularly eloquent exposition of what I was thinking about quantum mechanics.  And included in that exposition was a phrase that went like this:  ``The structures that {\it act\/} on Hilbert space---density operators, unitaries, and such---are the ones that represent the epistemic.  The things that {\it live\/} in Hilbert space are the ones that represent the ontic.''

I have no idea what that means, particularly as I cannot think of anything that ``lives in Hilbert space''.  Clearly I'm making a distinction between $\cal H$ and $B({\cal H})$, and saying the ontic/epistemic cut resides there.  But what ``lives'' in ${\cal H}$?

By the way, our flights were indeed canceled.  We are having to stay in Austin tonight.  I've been amused by your latest notes on $\sqrt{3}$ and such.

\section{19-03-08 \ \ {\it More Spam} \ \ (to D. M. {\Appleby})} \label{Appleby30}

Just reading the Barack Obama's speech from today before knocking off to bed.  Came across these lines, which I like:
\bq\noindent
Understanding this reality requires a reminder of how we arrived at this point. As
William Faulkner once wrote, ``The past isn't dead and buried. In fact, it isn't even past.''
\eq
\ldots\ particularly if you read them out of context and think about QM.

\section{20-03-08 \ \ {\it Jungian Analysis} \ \ (to D. M. {\Appleby})} \label{Appleby31}

\bma
I have managed to find a few more minutes.   Your dream is interesting
because it seems on the face of it to represent almost the opposite of
what you consciously think.  At least it does if one reads the word
``live'' the way I would naturally read it.  As I would understand it
the only things that ``live'' in Hilbert space are vectors.  But in your
waking life you would never say that the state vector has ontic
significance (curiously, though, the density matrix is said to have
epistemic significance:  so to that extent the dream seems to agree with what you consciously think).
\ema

Yeah, that's exactly what troubles me.  What beside vectors ``lives'' in Hilbert space?  I thought to myself, ``Well, does dimensionality `live' in the space?''  If so, that seems like a forced usage of the word.  On the other hand, for instance, one might say that maximal sets of equiangular lines do live in the space.  Is that part of what the dream was trying to bring to the surface?  This much is for sure:  I have long (many years) puzzled over why quantum mechanics seems to be built on a two layer structure:  the states (density operators) are not just in a vector space, but in a vector space $B({\cal H})$.  ($\cal H$ being the bottom layer, and $B({\cal H})$ floating above it.)  What is the significance of that very basic statement?  My interpretation of the dream is that it was aiming at an answer to that question.  And if so, there may be a little progress here.  For, I had never thought of the ontic/epistemic cut as perhaps having anything to do with the $\cal H\,$/$\,B({\cal H})$ cut before.  But now I find myself taking this as a direction worth exploring.

\section{20-03-08 \ \ {\it Group Meeting} \ \ (to D. Gottesman)} \label{Gottesman10}

I should give something of an abstract since my title is so uninformative.  It is below:
\bq
\noindent In this talk, I'll add some details on why I'm so interested in the existence of SIC (symmetric informationally complete) sets of quantum states.  Some {\sl leisure\/} readings relevant to the talk, in case you're interested, are this old paper by Feynman,
\begin{center}
\myurl{http://projecteuclid.org/euclid.bsmsp/1200500252}.
\end{center}
Also Chapter 1 of Volume III in his {\sl Lectures on Physics}, and finally, his original article on quantum computing, Int.\ J. Theor.\ Phys.\ {\bf 21} (1982), page 467.
\eq

\section{24-03-08 \ \ {\it Those Pages} \ \ (to H. C. von Baeyer)} \label{Baeyer30}

Could you send me an electronic version of the translation you made of the Pauli lecture?  I.e., the one you gave me a paper copy of when we first spoke yesterday.  I had only read half of that this morning, but I forgot to put it in my backpack so I could read the rest of it tonight.

See you tomorrow at about 11:00.  I'm really enjoying your being around here; I'm already hoping we can lure you back.

\subsection{Hans's Reply, 24-03-08}

\bq
Chris, ``those pages'' sounds a bit scurrilous, like ``that woman'', but here they are.

I'm having a great time.  My conversations, including dinners with Marcus, are helping me to formulate questions and projects.  I'm thinking very hard!

\bq
\subsubsection{Wolfgang Pauli:    The influence of  archetypal concepts on the formation of Kepler's scientific theories.}

\noindent (In {\sl Wolfgang Pauli: Das Gewissen der Physik}, Charles P. Enz and Karl von Meyenn, eds., Vieweg 1988, p.~509)\medskip

{\it [Pauli reports in a passive voice on a lecture he delivered to the Psychology Club of Zurich in 1947/48.  He poses the question: What is the bridge between a scientist's intuition and the development of scientific concepts and theories?  He uses Kepler as an example. He cites Plato's notion that the feeling of satisfaction evoked by the understanding of nature stems from the ``coming into congruence'' of pre-existing, innate images in the human psyche with the observation of external objects and their behavior.  Kepler calls such images {\it archetypal}.  This notion agrees largely with Jung's introduction of the same word into psychology.  Archetypes could also be called symbols, and furnish the sought-after bridge.

Pauli analyzes Kepler's work, starting with the identification of the three-dimensional sphere as symbol of the trinity of the Christian deity and of the Sun as symbol of God the father.  Thus the insistence on the Sun as the center of the universe turns out to be an expression of religious fervor.   Pauli emphasizes that there is no hint of quaternity in Kepler, and relates this omission to the absence of the concept of time from most of his system.

Kepler's work was attacked vehemently by the respected Oxford physician and Rosicrucian Robert Fludd, who tried to restore the quaternity.  For Pauli this is a symbol of the completeness of experience, which includes emotional components in addition to material ones, and is thus superior to the scientific point of view.  What follows is my translation of the three concluding paragraphs of Pauli's essay.]}\medskip

Finally there is the attempt to connect this problem originating in the 17th century with today's commonly expressed wish for a more unified {\it Weltbild\/} or world picture.  First it is suggested that the significance of the pre-scientific stage of thinking for the development of scientific understanding should be accounted for.  This step can be accomplished by complementing the study of scientific insights on the external world by the study of the inner meaning of these insights.   While the former aims to adapt our understanding to external objects, the latter should illuminate the archetypal images used in the development of scientific concepts.  Complete understanding would seem to be achieved only by combining these two directions of investigation.

Secondly, it is pointed out that modern microphysics has led to the result that today we have natural sciences, but no longer a scientific world picture.  This could be alleviated by progress toward a unified total world picture in which the sciences are only a part.  Indeed, modern quantum physics has begun to return toward the quaternary point of view that opposed the nascent science of the 17th century by treating the role of the observer in a more satisfactory manner than classical physics.  In contrast to the ``detached observer'' of the latter, the former postulates an uncontrollable interaction between observer or apparatus and the observed system in the course of every measurement. In this way the deterministic description of phenomena becomes impossible.  According to modern physics an observation, which makes specific choices and interrupts the game that is proceeding according to predetermined rules, is in essence not an automatic process, and can be compared to a creation in the microcosm, or with a transformation with unpredictable outcome.

The reaction of the discovery on the discoverer [or of observation on the observer]\footnote{The German words are {\it Erkenntnis auf den Erkennenden}. (In trying to translate them, I found out that currently there is no online etymological dictionary in German.)  Ordinary dictionaries yield  {\it Erkenntnis\/} $=$ knowledge, perception, recognition, realization, insight, discovery and cognition.  A hint comes from {\it Erkenntnistheorie} which is simply epistemology or theory of cognition.

In ordinary conversation, {\it erkennen\/} is to recognize something or someone, with an overtone of recovering something you knew beforehand (Plato).  But I think that element is not dominant here.   Since Pauli writes in the preceding about {\it Beobachtung\/} and {\it Beobachter}, which are simply observation and observer, I put those in square brackets.  Intuitively I still go with discovery.}, which leads to transformative religious experiences, and for which alchemy as well as the heliocentric idea furnish useful examples, transcends science.  It can only be captured by symbols which simultaneously express the emotional side of experiences through images and furnish a vivid relationship to all human knowledge and to the actual process of discovery.  Precisely because the possibility of such symbolism has become estranged in our time, it might be of interest to go back to another time which did not yet know about the mechanics we now call classical, but which enables us to provide proof of the existence of symbols with simultaneous religious and scientific functions.\medskip

{\it [Finally I translate the epigraph attached to this piece by the editors of the memorial volume.  It is from a letter written three years after the lecture and exemplifies the unique role of Pauli's correspondence with Markus Fierz.]}\medskip

``I came upon Kepler as a trinitarian and Fludd as a quaternarian, and felt a resonance between their polemic and my own inner conflict.  I have certain traits of both, but now, in the second half of my life, I should switch over to a quaternary attitude.  The problem is that the positive value of the trinitarian attitude may not be sacrificed in this move \ldots\

By the way, I would like to remark that back in Hamburg my journey to the exclusion principle had to do with just this difficult switch from 3 to 4:  namely the necessity of ascribing to the electron, besides its three translations, another, fourth degree of freedom.  To struggle through to the understanding that contrary to the na\"{\i}ve attitude a fourth quantum number is a property of the same electron --- this was actually the principal labor \ldots''\medskip

{\it [Fierz replied modestly that he himself was not as far along in his journey from three to four, that he didn't know whether he was a trinitarian or a quaternarian, and that he might never know.  But by this very admission Fierz reveals himself as the ideal mediator between the divine Pauli and the rest of us.]}
\eq
\eq

\section{25-03-08 \ \ {\it Those Pages, 2} \ \ (to H. C. von Baeyer)} \label{Baeyer31}

\bq
But I want to say one thing to the quantum information community.  I want you to listen to me; I'm going to say this again. [finger wagging]  I did not have intellectual relations with those pages [pause] Mr.\ Pauli's.  I never told anyone to venture into alchemy, not a single time, never.  These allegations are false, and I need to go back to work for the quantum information community \ldots
\eq

Thanks, I got them last night and finally read them. There is some good stuff there.  I'm particularly intrigued by his remarks (of a flavor I have seen before, but this brings it up again) that, ``The reaction of the discovery on the discoverer [or of observation on the observer], which leads to transformative religious experiences, and for which alchemy as well as the heliocentric idea furnish useful examples, transcends science.''  The reaction of the discovery on the discoverer?  The Bayesian would say, yeah, that's just the transformation of one's subjective beliefs from $P(h)$ to $P(h|d)$.  But he means something more or something else than that.

On my end, I became very frustrated looking for the originals of the Bohr--Pauli letters.  They should have been in my files under B, P, or F (for Folse, who gave them to me).  But they're simply missing---very unusual for me.

\section{25-03-08 \ \ {\it Funny Thing I Found on the Web} \ \ (to H. C. von Baeyer)} \label{Baeyer32}

While trying to find Henry Folse's website (so I might recover my missing Bohr--Pauli correspondence), I came across the following article by Robert Pirsig, ``Subjects, Objects, Data and Values'': \myurl{http://www.moq.org/forum/Pirsig/emmpaper.html}. Kind of funny, given that I just bought the 25th anniversary edition of {\sl Zen and the Art of Motorcycle Maintenance}.

\section{25-03-08 \ \ {\it Pauli--Bohr Exchange} \ \ (to H. C. von Baeyer)} \label{Baeyer33}

See, I did have them once!  Below is a letter from Henry Folse, dated 2 April 2001, explaining that they were written in English and had some handwritten marginalia by Pauli.

How on earth could I lose these gems?!?

\subsection{Letter from H. J. Folse, dated 2 April 2001}

\bq
You're quite right that this an interesting exchange between Bohr and Pauli.  I suspect that many have ignored them because of the late date.

The letters are in English and typed, but at least on one Pauli inserted several comments in handwriting.  Since it's about 15 pages or so, I've photocopied and mailed all four of them to you at your Bell Labs address.
The copies are too poor to scan very easily.  The Pauli letters are photocopies of the originals; the Bohr letters are photocopies of Bohr's carbon copies.

Hope you enjoy reading them.
\eq

\section{25-03-08 \ \ {\it That Fierz Paper} \ \ (to H. C. von Baeyer)} \label{Baeyer34}

Here's a link to the Fierz article I was telling you about that had such an influence on me.  It's reprinted in full on pages 405--407 of \myurl[http://www.perimeterinstitute.ca/personal/cfuchs/SamizdatSE.pdf]{http://www.perimeterinstitute.ca/personal/cfuchs/ SamizdatSE.pdf}.

In this case I've pieced together why I can't find the original.  It's because I possessed it before the fire and haven't replaced it.  But where the correspondence went (which I obtained after the fire) still eludes me.

See you later this morning.

\section{25-03-08 \ \ {\it Another Reference} \ \ (to H. C. von Baeyer)} \label{Baeyer35}

The chapter by Inga R. Gammel, ``Mission Impossible?\ Wolfgang Pauli's Idea of a Neutral Language,''
in {\sl Towards Otherland: Languages of Science and Languages Beyond}, edited by R. E. Zimmermann and V. G. Budanov:
\begin{center}
\myurl{http://www.upress.uni-kassel.de/online/frei/978-3-89958-107-2.volltext.frei.pdf}
\end{center}
looks potentially relevant to our discussions.  (I haven't read it yet; just noting it.)

\section{25-03-08 \ \ {\it The Coin Tosser} \ \ (to D. M. {\Appleby})} \label{Appleby32}

Looking over the note [14-12-07 note ``\myref{Snyder5}{No Subject}'' to C. Snyder], I see I missed something crucial in my report to you tonight.  An action with a predictable consequence, he told me, ``expresses what you think of yourself.''  I said I thought it was particularly deep, but really, I guess, I just thought it was particularly interesting.  It certainly does have an affinity with my general orientation that any action initiated by one's bodily motions, and with a predictable consequence, might as well be thought of as an extension of one's body to begin with it.

\section{26-03-08 \ \ {\it (Purely) Measurement-Based Quantum Computation} \ \ (to H. C. von Baeyer)} \label{Baeyer36}

Here a set of good articles on the subject.

Here is Nielsen's elementary exposition of the idea:
\begin{center}
\arxiv{quant-ph/0504097}
\end{center}
Here is a kind of review article by the inventors (Robert Raussendorf and Hans Briegel), but with prettier pictures than Nielsen's:
\begin{center}
\arxiv{quant-ph/0301052}
\end{center}
and a shorter version of the same:
\begin{center}
\arxiv{quant-ph/0207183}
\end{center}
Here is Steane's important conceptual paper that makes use of this kind of computation to put a nail in the many-worlds idea:
\begin{center}
\arxiv{quant-ph/0003084}
\end{center}
Here's a Zeilinger paper on making a 4-qubit cluster state:
\begin{center}
\arxiv{quant-ph/0507086}
\end{center}
And here's one doing an experimental demonstration of Deutsch's algorithm with such states:
\begin{center}
\arxiv{quant-ph/0611186}
\end{center}
There's plenty more out there to read on the subject, but this will get you started.

\section{27-03-08 \ \ {\it Fantastic Quote} \ \ (to H. C. von Baeyer)} \label{Baeyer37}

That was a fantastic quote you sent to Marcus last night of Pauli comparing evil and acausality.  Could you send it to me?  I was surprised by Pauli identifying hyle with nonbeing.  Very different from F. C. S. Schiller's use of the term, and very different from what I find in my book, {\sl The Concept of Matter in Greek and Medievel Philosophy}.  Particularly the chapter, ``Matter as Potency.''

\subsection{Hans's Reply, 28-03-08}

\bq
Chris, I'll reply as soon as I find a Greek etymological dictionary.  Hyle means matter, but Pauli implies that its roots mean something negative.  I'll check it out -- fortunately I attended a German humanistic boarding school where Latin and Greek were compulsory for six years.

I REALLY enjoyed PI, have banished the Demon, and have already started stocking my library with Pauli material.  Thank you for inviting me, and thank you especially for introducing me to Marcus who is clearly a unique person --- not only intellectually but humanly as well.
\eq

\subsection{Hans's Further Reply, ``Pauli Reassessment,'' 03-04-08}

\bq
Pauli died on 15 December 50 years ago.  It occurred to me that one might write a reassessment of his legacy in PT or something.

The elements of such a piece would be:

Downplay Jung.  The late Pauli gave plenty of evidence that while he owed a great debt of gratitude to Jung, he was fed up with Jungians.

Downplay Pauli effect and Pauli anecdotes in favor of his serious legacy.  The principal witness to the idea that Pauli took his effect seriously was Fierz, and he is dead.

Divide the Pauli legacy into two parts:  The private part asks how his dreams and beliefs influenced the creation of new scientific concepts.  Gerald Holton is the expert on that, and his invention, the ``thematic content'' of science, deals with it well. The public part is Marcus's concern: the re-connection of the rational with the irrational in our world picture, the ``coniunctio'' of inner and outer, the healing of the cut between subject and object \ldots

I feel that your program is a step in the right direction.  Consider this good
quote from a letter to Fierz (1953):
\bq\noindent
Einstein has neither the courage nor the
mental agility to admit the essential incompleteness of science within  life -- {\it and for this reason\/} speaks erroneously about the incompleteness of quantum mechanics within physics.
\eq
In other words, accept Copenhagen and go on to larger questions.

What do you think about this idea?  I feel comfortable about writing everything except the last part -- your work.
\eq

\subsection{Hans's Preply to Marcus {\Appleby}, ``Pauli's Darkness,'' 26-03-08}

\bq
Pauli's own footnote in his letter to Richard von Weizsaecker 5 May 1953:
\bq
I also see certain parallels between evil and the acausal.  Both are, after all, the
``dark'', in a manner of speaking, which escapes the rational (``enlightened'') order
of things.   Just like the acausal, evil must also be recognized as real --- in
contrast to neoplatonism, which wanted to declare both evil and matter unreal,
by calling the latter hylae (non-being) and the former, euphemistically,
privatio boni $=$ absence of good.  Just like observation in quantum mechanics,
there are instances in life where ``the calculation doesn't compute'' (e.g.\ Plato in
Sicily). One can neither get rid of these instances, nor avoid them by fleeing into
the other side of the pair of opposites.  One can only approve of them and
accept them into a symbolic reality.
\eq

In the interest of speed I took some liberties in translating, and I don't know
what he is talking about in the second half.  Plato screwed up in Sicily and had to
return home in disgrace. The point of the passage is that Pauli uses in one
passage evil, dark, and acausal.
\eq

\subsection{Hans's Preply to me, ``Postprandial Nugget,'' 26-03-08}

Pauli to Jung 27 May 1953:  ``\ldots It was, if I am not mistaken, in 1931 that I got to know you personally.  At that time I experienced the subconscious like a new dimension \ldots''

\section{27-03-08 \ \ {\it Exposing the Cracks in the Block Universe}\ \ \ (to J. A. Vaccaro)} \label{Vaccaro4}

I got your great painting in the mail yesterday.  Thanks again so much; I'm so tickled with it.  The painting made it briefly to my house yesterday, but at the moment, I have it propped on my office desk at PI.  I started thinking after taking it home that it'd be nice to show it off to more people for a while, before it lives out the rest of {\it my\/} life in my quiet library.

Some questions if you don't mind my asking.  I notice it's signed JV00; do you recall what month you painted it?  Can you articulate your intention for painting it?  What concept you were trying to capture?  (I've already explained my own reason for wanting it.  But I'm sure you were trying to express something completely different in painting it.  And I'm just curious.)

\section{02-04-08 \ \ {\it Need Flammiazation}\ \ \ (to S. T. Flammia)} \label{Flammia6}

When in the hell are you going to be back?

\section{04-04-08 \ \ {\it Hirota's Birthday Party}\ \ \ (to A. S. Holevo \& the invitees)} \label{Holevo9}

\noindent Dear colleague, \medskip

Alexander Holevo and I would like to invite you to a three-day scientific birthday party at the Perimeter Institute for Theoretical Physics, June 25--27, in honor of Osamu Hirota's 60th birthday.  The meeting, ``Osamu Hirota, a True Quantum Communication Channel,'' plans to be a relaxed venue where we'll be able to interact and convey what we each think is presently the most exciting (or most perplexing) in quantum information theory.  Our budget should be able to cover your travel and local expenses, should you need it.

The proposed speaker list includes, along with yourself,
\bv
Charles Bennett     (confirmed) \\
G. Mauro D'Ariano               \\
Steven van Enk                  \\
Christopher Fuchs   (confirmed) \\
Alexander Holevo    (confirmed) \\
Debbie Leung                    \\
Masanao Ozawa                   \\
Masahide Sasaki                 \\
Jeffrey Shapiro                 \\
Peter Shor          (confirmed) \\
John Smolin                     \\
Horace Yuen                     \\
\ev
Please confirm with me whether you will be able to come enjoy Professor Hirota's birthday.

The PI facilities may also be able to accommodate a few of us for a limited time before and after the meeting (up to two weeks) for extended collaboration.  When you reply to this note, please let me know whether you might like to stay longer than the meeting, and we will see what can be done.  PI can be a very good place for thinking and gaining new ideas!\medskip

\noindent Best regards,\medskip

\noindent Chris Fuchs (and Alexander Holevo)

\section{04-04-08 \ \ {\it } \ \ (to T. A. Brun)} \label{Brun9}

On behalf of the GQI executive committee, I am pleased to let you know that your student Bilal Shaw was the winner of the APS March Meeting Best Student-Presentation Award in the theory category.  Could you please pass the word on to him?

The judging panel consisted of Prof.\ John Sipe, chair (Toronto), Dr.\ Howard Barnum (LANL), Prof.\ Dagmar Bruss (Duesseldorf), Prof.\ Barry Sanders (Calgary), Prof.\ Lorenza Viola (Dartmouth), and Prof.\ Harald Weinfurter (Ludwig Maximilians, Munich). The candidates were judged on four equally weighted categories:  content, clarity, style, and q{\&}a.  All the candidates were, in fact, very strong in comparison to many, many more-senior speakers at the meeting, but Mr.\ Shaw's presentation was deemed the best.

The award will be announced in the next issue of {\sl The Quantum Times}.  The award carries a 500 USD cash prize which will be presented to the winner through the American Physical Society, but comes from funds established by the Perimeter Institute for Theoretical Physics in Waterloo, Canada.  Also, PI would like to invite Mr.\ Shaw to Canada to give a longer version of his presentation.  Thus, please put him in contact with me so that we can make arrangements for a visit.

\section{06-04-08 \ \ {\it Librarians}\ \ \ (to O. Hirota)} \label{Hirota7}

I am glad you are happy.

Holevo pointed out to me that between 6--11 July, there is an IEEE Symposium on Information Theory in Toronto.  So, he will stay at PI the two weeks preceding that (roughly 22 June to 5 July), and then he will go to the IEEE meeting.

As I said, I will be happy to show you my library.  You know, I have told you how important I think the philosophy of William James, John Dewey, and F. C. S. Schiller is for our understanding of quantum mechanics (it was quantum mechanics that pushed me toward them).  That philosophy has the name ``pragmatism'' and my library has nearly 400 on books on some aspect of the subject.  In the middle of the shelves is a volume I've recently purchased titled {\sl Contemporary Japanese Philosophical Thought}, published in 1969.  Chapter 3 of it is devoted to pragmatism and related philosophies from 1901--1925.  What is interesting is that I found that there was actually a Japanese pragmatist of some renown.  His name was Tanaka Odo.  Perhaps your son with an interest in philosophy can explain to us the particularities of Tanaka Odo's views!

\section{08-04-08 \ \ {\it PIAF Invitation} \ \ (to the invitees)}

The Perimeter Institute for Theoretical Physics in Waterloo, Canada and three Australian universities (University of Sydney, University of Queensland, and Griffith University) have recently banded together to seed growth in the field of quantum foundations.  In the long run, this PI-Australian Foundations (PIAF) collaboration has as its biggest objective to make it possible for physicists and philosophers to pursue sustainable academic careers in quantum foundations.  More immediately though, within our portfolio of activities, we plan to establish a major international conference for the yearly exposition of the best work in quantum foundations.  This year's meeting ``The Clock and the Quantum: Time in Quantum Foundations'', will be held in Waterloo, 28 September to 2 October, 2008, and we are hoping to give the series a very strong start by having an outstanding set of invited speakers.

In that regard, we would be honored if you would be one of those invited speakers.  All your travel and local expenses would be covered by PIAF.  Would you be able to come?  For your reference, here are the other invitees:   \medskip

\begin{supertabular}{ll}
Samson Abramsky    &   (Oxford U.)\\
Yakir Aharonov     &   (Tel Aviv U.)\\
Julian Barbour     &   (Oxfordshire)\\
Harvey Brown       &   (Oxford U.)                      \\
Phil Dowe          &   (U. Queensland)                  \\
Brian Greene       &   (Columbia U.)                    \\
Gerard 't Hooft    &   (U. Utrecht)                     \\
Matthew Leifer     &   (IQC, Waterloo and PI)           \\
Paul Kwiat         &   (U. Illinois, Urbana-Champagne)  \\
John Norton        &   (U. Pittsburgh)                  \\
Wayne Myrvold      &   (U. Western Ontario)             \\
Roger Penrose      &   (Oxford U.)                      \\
John Preskill      &   (Caltech)                        \\
Carlo Rovelli      &   (U. Mediterranee)                \\
Lee Smolin         &   (PI)                             \\
Roderich Tumulka   &   (Rutgers U.)                     \\
William Unruh      &   (U. British Columbia)            \\
Lev Vaidman        &   (Tel Aviv U.)                    \\
Anton Zeilinger    &   (U. Vienna)            \medskip          \\
\end{supertabular}

Please let us know as soon as possible if you will join us (so that we may set the poster printing in motion).  We believe this promises to become an important conference series, and your participation would be a great coup for us. \medskip

\noindent Sincerely,\smallskip

\noindent Chris Fuchs, Lucien Hardy, and Ward Struyve (the local organizers)\medskip

\underline{Organizing Committee}   \medskip

\begin{supertabular}{ll}
Guido Bacciagaluppi &  (U. Sydney)   \\
Christopher Fuchs   &  (PI)          \\
Lucien Hardy        &  (PI)          \\
Ward Struyve        &  (PI)    \medskip      \\
\end{supertabular}

\underline{Advisory Committee} \medskip

\begin{supertabular}{ll}
Stephen Bartlett    &  (U. Sydney)     \\
Gerard Milburn      &  (U. Queensland) \\
Huw Price           &  (U. Sydney)     \\
Howard Wiseman      &  (Griffith U.)   \\
\end{supertabular}

\section{08-04-08 \ \ {\it PIAF Invitation} \ \ (to M. S. Leifer)} \label{Leifer10}

Matt, I'll also add a word of personal encouragement:  This conference {\it needs\/} you, so I hope you can come.  Like I told Lucien, ``if his work on conditional density operators ain't about time in quantum mechanics, nothin' is.''  And you'd be the only representative of that point of view.

\section{09-04-08 \ \ {\it Yo Yo and JFK on Wiki}\ \ \ (to L. Hardy)} \label{Hardy25}

Thanks for reminding me.  Impressive.  I wish I had gotten the opportunity to meet the guy.

The frame of this also reminds me in turn to look up the James bang quote I told you and Vanessa about.  Here it is:
\bq
It is a common belief that all particular beings have one origin and
source, either in God, or in atoms all equally old. There is no real
novelty, it is believed, in the universe, the new things that appear
having either been eternally prefigured in the absolute, or being
results of the same {\it primordia rerum}, atoms, or monads, getting
into new mixtures. But the question of being is so obscure anyhow,
that whether realities have burst into existence all at once, by a
single `bang,' as it were; or whether they came piecemeal, and have
different ages (so that real novelties may be leaking into our
universe all the time), may here be left an open question, though it
is undoubtedly intellectually economical to suppose that all things
are equally old, and that no novelties leak in.
\eq
It comes from his book, {\sl Some Problems of Philosophy: A Beginning of an
Introduction to Philosophy}, published soon after his death in 1910.

\section{14-04-08 \ \ {\it Your Dad} \ \ (to L. Wheeler Ufford)} \label{Ufford2}

I learned the news of your father's passing away today.  It has been a very sad day for me.  But I have him very literally to thank for my career:  The questions I have worked on these many years were the questions he asked so many years before that.  I was lucky enough to meet him in 1983 and absorb those questions.  He was a very great man.

\section{14-04-08 \ \ {\it John Wheeler's Death} \ \ (to G. L. Comer)} \label{Comer113}

In case you hadn't heard about it yet:
\begin{center}
``John A. Wheeler, Physicist Who Coined the Term `Black Hole,' Is Dead at 96''\medskip\\
\myurl{http://www.nytimes.com/2008/04/14/science/14wheeler.html?scp=1\&sq=Wheeler\&st=nyt}
\end{center}

\section{14-04-08 \ \ {\it Sad Day} \ \ (to D. Overbye)} \label{Overbye7}

Thank you for the nice article on John Wheeler today.  It is a sad day for me.  His thoughts have pushed me for years.

One minor inaccuracy, I believe; I don't know if it is too late for you to fix it.  I believe it was Bryce DeWitt who coined the term ``many worlds'' rather than John.  I believe Wheeler himself always called it ``the relative state formulation of quantum mechanics.''

\section{14-04-08 \ \ {\it Times and the End of Time} \ \ (to W. P. Schleich)} \label{Schleich3}

I became very disorganized, and I apologize for keeping you waiting.  But today, with the sad news of John Wheeler's passing away, I was shaken back to business.  John was a very great man, and how lucky I was to meet him in 1983.  He provided the questions that have framed my career ever since.  Mostly at the moment, I feel like I should work harder:  It is what I owe him for his wonderful gift.

Thanks again for inviting me to Ulm.  The MPQ has kindly offered to pay for my transatlantic flight; so I'll make those guys the major part of my visit.  But I would still like to come to Ulm and give a talk if you will have me \ldots\ and if I can synchronize my family's schedule with the window of time you'll be in town.  It'd be nice to spend a night in Ulm and get you guys revved up about the SIC representation of quantum states.

I should have a precise schedule worked up in a couple of days.  I'll get back in touch then if it looks like I'll be able to come to Ulm July 21 or July 22, say.

I know this day has probably been rough on you.  I could see how close you were to John in Princeton, Feb 2006.

\section{14-04-08 \ \ {\it Epistemic Meeting} \ \ (to H. Westman)}  \label{Westman1}

\bhwe
There has been some interest circulating about getting together and talk
about the epistemic interpretation of QM (\`a la {\Spekkens}) and whether or
not is it possible. It could be that one could construct a no-go theorem
or perhaps a concrete working model.
\ehwe

It struck me that ``the'' is an awfully singular word.  So I thought it would be fun to revisit my own early uses of the word.  Here are a couple of instances I dug from my \quantph{0105039}.  Thus the word did have some currency before {\Spekkens}, and particularly before it was hijacked to the purpose of hidden variable theories.

Still, yes, it would be fun and informative to discuss these things.  It's just a question of me finding the time in these coming weeks.

\bq

\subsection{12 December 1999, to John Preskill, ``Freedom''}

\bjp
Free will usually means the ability of conscious beings to influence their own future behavior. Its existence would seem to imply that different physical laws govern conscious systems and inanimate systems.  I know of no persuasive evidence to support this viewpoint, and so I am inclined to reject it.
\ejp

Can't agree with your second sentence.  When you think of physical law, you should say the chant epistemic, epistemic, epistemic.  Then you can chime in with MLK:  free at last, free at last, thank god, I'm free at last!

Had fun talking to you the other day.

\subsection{20 March 2000, to Paul Benioff, ``Small Addendum''}

I agree with you wholeheartedly that in quantum mechanics the ``randomness'' of measurement outcomes is NOT epistemic.  However that does not preclude my view that all probabilities (including quantum mechanical ones) are epistemic in nature.  They quantify how much we can say about a phenomena based upon what we know.  It so happens in the quantum world that we cannot tighten up our knowledge to the point of removing all ignorance (about the consequences of our interventions), and in that sense the randomness is ontological---it is a property of the world that was here long before we ever showed up on the scene.  But it takes epistemic tools to describe that property.  That's the direction I'm coming from.

\subsection{26 January 2001, to Jeremy Butterfield, ``Worlds in the Everett Interpretation''}

Anyone who knows me knows that I am rather down on attempts to interpret quantum mechanics along Everett-like lines.  I think the most funny and telling statement of this in the present context is that, whereas Mr.\ Wallace speaks of ``Everettians,'' I often speak of ``Everettistas.''  Thus, I am almost surprised that you sent me this paper to referee.

My difficulties come not so much from thinking that an Everett-like interpretation is inherently inconsistent or that parallel worlds tax the imagination too much.  It's more that this line of thought strikes me, at best, as a complete dead end in the physical sense. At worst, I fear it requires us to tack on even more ad hoc structures to quantum theory than we already have.  (Here, I'm thinking of a preferred basis for the Hilbert space and a preferred tensor-producting of it into various factors.)  For these reasons, among umpteen others, I have always been inclined to an epistemic interpretation of the quantum state.  Doing this has helped me
(personally) to focus the issue to asking, ``What is this {\it property\/} of the quantum world---i.e., reality---that keeps us from ever knowing more of it than can be captured by the quantum state?''
To that extent, I consider myself something of a realist who---just as David {\Deutsch}---takes the wavefunction absolutely seriously.
BUT absolutely seriously as a state of knowledge, not a state of nature.  I do well believe we will one day shake a notion of reality from the existing theory (without adding hidden variables, etc.), but that reality won't be the most naive surface term floating to the top (i.e., the quantum state).  When we have it, we'll really have something; there'll be no turning back. Physics won't be at an end, but at a beginning.  For then, and only then, will we be able to recognize how we might extend the theory to something bigger and better than quantum mechanics itself.
\eq

\section{21-04-08 \ \ {\it Born's Rule?}\ \ \ (to R. Laflamme)} \label{Laflamme3}

{\Ruediger} told me last week that you're planning to do some experimental test of the Born rule.  I'd like to learn more about that, particularly as it impinges on our quantum Bayesian program.  If you've got some time this week or next, I wouldn't mind an introduction to the ideas and what's being done.

\section{22-04-08 \ \ {\it Ariana Margaret Blume-Kohout}\ \ \ (to R. Blume-Kohout \& M. E. Blume-Kohout)} \label{BlumeKohout3} \label{BlumeKohoutME1}

Congratulations.  It takes me back years.  It's a truism that your life will change, and it is a truism that it will change for the better.  Myself, I believe that life and creation is the big principle in the universe.  Tell Ariana Margaret that---that her birth played a crucial part in making the universe fly.  She should understand that about herself.

\section{23-04-08 \ \ {\it Dylan Quote on Time} \ \ (to S. Weinstein)} \label{Weinstein2}

\bq
The first thing you notice about New Orleans are the burying grounds---the ceme\-ter\-ies---and they're a cold proposition, one of the best things there are here.  Going by, you try to be as quiet as possible, better to let them sleep.  Greek, Roman, sepulchres---palatial mausoleums made to order, phantomesque, signs and symbols of hidden decay---ghosts of women and men who have sinned and who've died and are now living in tombs.  The past doesn't pass away so quickly here.  You could be dead for a long time.
\eq

\section{24-04-08 \ \ {\it PIAF Abstract}\ \ \ (to L. Hardy)} \label{Hardy26}

\blh
Does time exist between measurements?
\elh

Pauli would be very proud of you for including his question.  Here was his own version of it (from a 1947 letter to Markus Fierz):
\bq
I'm more and more expecting a further revolutionizing of the basic
concepts in physics.  In connection with this particularly the manner
in which the space-time continuum is currently introduced into it
appears to me to be increasingly unsatisfactory.  \ldots\ Something
only really happens when an observation is being made, and in
conjunction with which, as Bohr and Stern have finally convinced
me, entropy necessarily increases.  Between the observations nothing
at all happens, only time has, ``in the interval'', irreversibly
progressed on the mathematical papers.
\eq
More serious comments on all manner of things later.

\section{24-04-08 \ \ {\it Topos Theory Today}\ \ \ (to L. Hardy)} \label{Hardy27}

Here is my own {\it present\/} feeling about the research program of Doering, Isham, and associates.  It seems to be a significant amount of mathematical obfuscation for saying, in the end, something that has been tried, tested, and debated since 1935, but generally in more accessible language: It is that something ascribed probability one should also be ascribed an element of reality.  By implication, some aspect of a pure quantum state then corresponds to an element of reality.  Here's the way Roger Penrose put the essential point:
\bq\noindent
   One of the most powerful reasons for rejecting such a subjective
   viewpoint concerning the reality of $|\psi\rangle$ comes from the
   fact that whatever $|\psi\rangle$ might be, there is always---in
   principle, at least---a {\it primitive measurement\/} whose {\bf YES}
   space consists of the Hilbert-space ray determined by $|\psi\rangle$.
   The point is that the physical state $|\psi\rangle$ (determined by
   the ray of complex multiples of $|\psi\rangle$) is {\it uniquely\/}
   determined by the fact that the outcome {\bf YES}, for this state, is
   {\it certain}.  No other physical state has this property.  For any
   other state, there would merely be some probability, short of
   certainty, that the outcome will be {\bf YES}, and an outcome of {\bf
   NO} might occur. Thus, although there is no measurement which will
   tell us what $|\psi\rangle$ actually {\it is}, the physical state
   $|\psi\rangle$ is uniquely determined by what it asserts must be the
   result of a measurement that {\it might\/} be performed on it.
\eq
What I have been left wondering is what does the category theoretic approach add to that essential point?  Particularly, in what way---and for what compelling reasons---is their approach any less ad hoc than any of the other 500 approaches to quantum logic?  (And I would guess there really are easily 500 other approaches along those lines, without exaggeration.)  And what is the grand plan for where it'll lead (beyond saying we ought to get rid of the continuum)?  Neither has ever come across to me.  Though admittedly I have never worked hard to let it come across to me either:  Instead, because there was no immediate (subjective) mental ``hook'' to it for me, I treated it as somewhat like a patent clerk receiving another detailed plan for a perpetuum mobile.  Complicated, very complicated, interesting move, ingenious, I never thought of it that way \ldots\ \ldots\ \ldots\ but ultimately doomed to failure.  Red stamp.

John Wheeler was too much of an influence on me:  ``The solution is not in some magical mathematics, but in some magical idea.  A magical idea we've yet to see.'' (paraphrase)

So, there that's what I feel about the research program---as I say, {\it presently}.  But, just because I'm guessing category theory will be barren in this particular usage of it, it does not mean that I deem category theory itself barren.  Lee, for instance, has told me more than once how I might need it for my own research program.  Maybe he's right.  And it is always good to have a table of integrals about.

\section{24-04-08 \ \ {\it An Entry That Made Me Giggle Again}\ \ \ (to L. Hardy)} \label{Hardy28}

Just inserted this one.  [See 13-05-04 note ``\myref{Schack83}{10 Lines and MaxEnt}'' to R. {\Schack}.] Made me think about Moses wandering the desert for 40 years!  Just no place we can call our home.

\section{24-04-08 \ \ {\it Which Stern?}\ \ \ (to H. Atmanspacher)} \label{Atmanspacher7}

I found a quote in one of your papers from a 1947 letter from Pauli to Fierz (it is a quote I had known from one of Laurikainen's books, but your version is more complete):
\bq
More and more I expect a further revolution of basic notions in physics, where I am particularly dissatisfied with the way in which the spacetime continuum is introduced at present. (Of course it is ingenious to disband time from ordering causal sequences and --- `as once in May' --- use it as a romping place for probabilities. But if one replaces ingenious by impudent, this is not less true. In fact, something happens only during an observation, where --- as Bohr and Stern finally convinced me --- entropy increases necessarily. Between observations nothing happens at all, only time has reversibly proceeded on our mathematical papers!) This spacetime continuum has now become a Nessus shirt which we cannot take off again! (Instead of `Nessus shirt' you can also say `prejudice', but this would, first, sound too harmless and, second, shift the mistake too much from a mere conception to a judgment.)
\eq
I wonder if you can tell me a couple of things about it?
\begin{itemize}
\item[1)]  Which Stern is Pauli referring to?

\item[2)]  What is a ``Nessus shirt''?
\end{itemize}
The main question is 1), as I'm preparing a better-indexed version of my ``Notes on a Paulian Idea'' and I want to re-post it at the same time as the posting of my new samizdat (another 500 pages, roughly May 10).

By the way, I'll be in Munich July 23--August 9, letting my children (and wife) visit with the grandparents.  I'll take a desk at the MPQ in Garching, but I don't believe I am committed to too much time there.  Would there be any interest in having me give a seminar in Freiburg?  I could talk about these very symmetric quantum measurements we have been spending so much time on at PI, and relate them to quantum foundations problems.

I hope all is well with you.

\subsection{Harald's Reply}

\bq
Thanks for your note, good to hear from you!  Of course, by all means we should organize something while you are in Germany in summer. We can make detailed plans at Paris in June, where you and I will meet anyway.  If you want to have a definitive schedule earlier, please let me know.

Stern is Otto Stern (of the Stern--Gerlach experiment).

The Nessus shirt refers to Greek mythology, see
\bq
 \myurl{http://en.wikipedia.org/wiki/Shirt_of_Nessus}.
\eq
A powerful metaphor for a present that brings harm and mischief.
\eq

\section{25-04-08 \ \ {\it Which Stern?, 2}\ \ \ (to H. Atmanspacher)} \label{Atmanspacher8}

Thanks!  I figured it was Otto Stern, but I seemed to have recalled another Stern that Pauli once referred to, and I wanted to make sure I didn't make a mistake.

Yes, let's plan in June; see you in Paris.  I think it'd be fun to come to Freiburg for a day or a day and a night.

\section{05-05-08 \ \ {\it Appleby Papers to Look At}\ \ \ (to K. Martin)} \label{Martin1}

\begin{itemize}
\item
\quantph{0412001}
\item
\quantph{0611260}
\item
\arxiv{0707.2071}
\item
\quantph{0308114}
\item
\arxiv{0710.3013}
\end{itemize}

\section{05-05-08 \ \ {\it Kant Cola}\ \ \ (to M. Friedman)} \label{Friedman1}

It was good meeting you.  I certainly appreciated your sympathetic ear for Quantum Bayesian ideas.

I thought I'd get your email address from the web so that I'd have it in my address book.  Seeing your interest in Kant (and hearing the things you laughed at last night!), maybe you'll get a chuckle from this diary entry of mine:  page 140 of \myurl[http://www.perimeterinstitute.ca/personal/cfuchs/SamizdatSE.pdf]{http://www.perimeterinstitute.ca/personal/ cfuchs/SamizdatSE.pdf} (that's actually page 162 in machine counting, if you include preface and title pages etc.).  Or you can simply search on the phrase ``Kant Cola''.

\subsection{From 27 February 1996 note ``{\Kant} Cola'' to Greg Comer}

\bq
{\Kiki} and I went into the Outr\`emont area Friday evening in search
of an interesting restaurant \ldots\ and what a find we made!  Let me
tell you a strange little story.  During the summer of 1985, I was
reading a book by C.~F. von {\Weizsacker} titled {\sl The Unity of
Nature}.  Most of the book was about quantum mechanics and {\Kant}ian
philosophy. Apparently it spurred me to have the following dream. I
was in a little hole-in-the-wall joint somewhere in Austin; my old
friend David\label{Baker, David B. L.} was there, also John
{\SimpsonJ} and Marshall {\Burns}. The place really stood out in my
mind because of the Bohemian feel to it:  dark, smoky, mystical
almost. The night wasn't filled with much of interest:  David only
wanted to talk about getting drunk, John only wanted to talk about
finding a girl, and Marshall only wanted to talk about philosophy. In
those days John didn't drink alcohol, so, at some point, when he
asked for a drink, I thought we'd be out of luck.  But upon looking
around, I saw a refrigerator in the middle of the bar near the pool
table.  We walked over to it and took a look. It was filled with all
different sorts of vegetable drinks.  John grabbed one, and I looked
through it for something more interesting. At the very back, I found
one lone can of ``{\Kant} Cola.''  That was written on the label,
along with a small portrait of Immanuel {\Kant}.  I opened the can,
took a drink, \ldots\ and, for a miraculous moment, I understood all
the intricacies of the world---I understood the necessity of quantum
mechanics.  When I came out of my trance, the can was empty and I
knew that I would never see the light again.  Then I awoke.  I was so
taken with this dream that the next day I sketched out the layout of
the joint and made a record of the dream.  That was over ten years
ago.

So back to the restaurant of Friday. The place was called ``City
Pub''; it was such a strange little place:  dark, smoky, mystical
almost.  The food was excellent---far better than it should have
been for the price.  Each option in the place was only \$4.99.  I had
steak, fries, and a vegetable.  {\Kiki} had potato soup, quiche,
fries, and veggie. They had a special on beer, three for the price
of one (so we had six). The music was some sort of strange mesh of
things that I suspect you'd only hear in some little bar in Germany
where everyone wears black.  Anyway, we had quite a time there.
However, just a little while before leaving I started to note how
similar this place was to the place in my dream 10 years ago.  I
told {\Kiki} the whole story. Then I looked around and---strangely
enough---there was a refrigerator in the middle of the room near the
pool table!  I was so taken by this that upon my way to the
restroom, I took a look into it.  What a disappointment: it only
contained beer.  However, the restroom did have a surprise for me.
In the middle of all the graffiti (about Qu\'ebec's hoped for
independence) was something written in bold black letters:
\begin{center}
\parbox{1.4in}{De nobis ipsis silemus\\
\hspace*{\fill} --- {\it E. {\Kant}}}
\end{center}
That made my evening. I wrote down the words so that I wouldn't
forget and went home to look up my old notes on the dream.  Sure
enough, there were similarities in the layout of the two places, and
moreover, I saw that the name of the original place in my dream was
``Hole in the Wall Pub.''  Very strange.  I asked {\Ruediger}
{\Schack} to translate the words for me, and he came up with ``About
ourselves we remain silent.''
\eq

\section{05-05-08 \ \ {\it The Hilary Had No Clothes}\ \ \ (to A. Wilce and H. Barnum)} \label{Wilce19} \label{Barnum23}

Yesterday at the {\Demopoulos} Bash, I saw Hilary Putnam for the first time.  My God, how I was looking forward to it.  The title of the talk was ``Quantum Mechanics and Ontology,'' and I had been wondering all week what flavor of subtle pragmatic consideration it would end up expressing on that subject.  Well, I set myself up for the singular biggest hero-worship crash I ever had in my life.  The guy---this year's version of him anyway---is a friggin' Bohmian and deeply under the influence of Tim Maudlin.  I was clinically depressed the remainder of the afternoon.  I didn't even seek the energy to talk to or meet with him; maybe better to let him pass through my life like a wraith.

He spoke briefly about information-based approaches to quantum foundations, like Clifton-Bub-Halvorson and your own efforts, and maybe my own and Bill {\Demopoulos}'s, I don't know.  He ended his brief survey by declaring them either ``instrumentalism with post-modern sauce'' or ``positivism with post-modern sauce''---I don't remember which as he was using both terms in the talk, but it was one of the two.  Either way, it was equally insulting.

I guess I write you because I remember the little conversation we three had about him in Atlanta at a fast-food restaurant of some sort.\medskip

\noindent Recovering, but sad again for having recorded this,

\section{05-05-08 \ \ {\it Pages, Coffee!, and Stairs}\ \ \ (to R. {\Schack})} \label{Schack132}

I just posted an updated collection.  Your name appears several more times now.  We're up to 623 pages at the moment.

I'm in a London hotel, and I'll drive back to Waterloo as soon as I get a new cup of coffee in me and get a shower.  I just discovered my first cup was wasted:  It was decaffeinated!

I've been at the {\Demopoulos} Bash the last two days.  Yesterday, there was a very fine talk by Allen Stairs titled ``A Loose and Separate Certainty.''  You can guess the subject:  our quantum Bayesian program, and particularly our certainty paper.  I hadn't realized how photogenic we three are!  Anyway, as indicated, the talk was quite good, but what it indicated about our paper was not so:  The guy was genuinely confused by the logic/goals/layout of the paper.  And in most of the cases, I could really tell that we were the cause.  For instance, he spent five minutes arguing that ``Gleason's theorem could not be the greatest triumph of Bayesianism'' hanging on that dreadful sentence {\Carl} insisted on.  Stairs' only sin was that he read the sentence literally, and then pointed out that there was nothing to bar an objectivist about probabilities from the premises of Gleason's theorem.  And so it went with several things.  I had a long lunch with him, and then sat next to him at the dinner.  He strikes me as being very sympathetic to the program (and sympathetic to the idea of preserving locality at all costs); he just wants to get things straight.  I invited him to stay a couple or three weeks at PI and encouraged him to write a ``critical article'' on us.  My guess is, if he delves into this further, it won't end up being so critical, as simply clarifying.

By the way, Hilary Putnam was an absolutely huge disappointment.  See story below.  [See 05-05-08 note titled ``\myref{Wilce19}{The Hilary Had No Clothes}'' to A. Wilce and H. Barnum.]

\section{08-05-08 \ \ {\it Triple Products in Dimension 3} \ \ (to D. M. {\Appleby})} \label{Appleby33}

Now, that is really cool!  I think there is indeed no need to bother with a picture.  I think I can figure out a good way to present this.  It'll be in the context of finding the algebraically simplest possible equations for quantum state space.

As I say, I had intended to present the talk as joint work with you.  I hope you won't mind:  I know that I can be something of a wildcard in a talk and you may not (or quite probably will not) agree with everything I say.  Actually another option is that I can up-front present the technical work as joint work with you, but make a clear statement that you may not agree with anything beyond that:  Time will tell.  That'd probably better as I intend to soon write this conceptual piece with business about modifications of Dutch-book coherence when it comes to SICs, etc., and I had intended to write that on my own.

Among the things I want to emphasize is the contrast between what I see as my own foundational program and the convex sets approach of Lucien and Barnum, {\Barrett}, Leifer, Wilce, etc.:

\begin{enumerate}
\item
They at the outset take measurements as linear maps.  I, on the other hand, want to first establish a Bureau of Standards and think of measurements as refinements of probabilities w.r.t.\ the BoS.  Linearity comes secondarily when  disparate agents come to agreement that they are performing the same measurement.
\item
Best measurement for BoS must be the SICs or Super-SICs.  But they give no natural tensor product structure.  This tracks metaphorically with the de Finettian / {\Appleby}an idea that ``probabilities are single case or nothing'', and I think is a very important point.  This contrasts with the usual convex approaches in that it does not take the composition of systems as a fundamental operation to be axiomatized.  Instead one seeks to characterize the state space of a single system, and {\it when\/} one wants to think of separate systems, one does it by decomposing not by composing.
\item
The essence of the Born rule, when expressed in terms of SICs is conceptually almost the law of total probability.  There is a hidden assumption in the usual Dutch book argument connecting conditional and joint probabilities, that is apparently being surpassed here:  It is that ``unperformed measurements have no outcomes''.  Fine, but then why is there any connection at all between the Born rule and old law?
\end{enumerate}
Those are maybe the major components.

Thanks, by the way, for the extensive notes on the title page.  Interesting the contrasting reactions I've gotten between you, Lucien, Guido, {\Ruediger}, and Kiki.  You and Kiki are all for the ``My Struggles''---enthusiastic about it even.  Guido and {\Ruediger} are offended by it, for some dainty European reason as far as I can tell (but one I don't want to discount, because there are a lot of Europeans in my potential reading audience).  And Lucien's first thought when he heard it was ``Mein Kampf''.  It was Lucien's initial reaction (and Guido throwing his two cents in at the same time) that sowed a seed of doubt in me.  At the very least, I think I'm definitely keeping the ``My''; what I'll do with the subtitles I'm still thinking about.

\section{08-05-08 \ \ {\it Wigner Connectives}\ \ \ (to R. {\Schack})} \label{Schack133}

Thanks for all the advice.  I'll explain in Zurich why I probably won't follow it, but I needed to get a clearer sense of what I might be getting myself into.

Schackcosms fixed as requested.

I'm trusting good discussions in Zurich.  I think we really broke ground the last time we were together.  Your anti-Zurekian diagram is still at the top of my board.  And you should have seen Bill Harper's eyes light up when he understood it.  (BTW, it was Harper and Skyrms who laid the first criticism on van Fraassen when he claimed that reflection was a requirement of rationality.  They pointed out that it only holds when one judges that recognizing the data is ``learning something.''  I.e., not whacking.)

Below is something I dug up as I was compiling the samizdat.  Why I called it a review of general covariance, I'm not completely sure, but it seems to fit now.  Linearity of the probability rule---if you take this point of view that actions are defined as sets of likelihoods---arises as a kind of covariance between different agents.  It comes about as a kind of intersubjective agreement.  I.e., Dutchbook coherence is a one-agent phenomenon, but maybe, just maybe, linearity is a two-agent phenomenon?  I.e., I think I'm quarreling with your discomfort with the idea again.  [See 23-03-06 note titled ``\myref{Comer86}{A Review of General Covariance}'' to G. L. Comer.]

\section{09-05-08 \ \ {\it Keye Bayesian?}\ \ \ (to K. Martin)} \label{Martin2}

My god, you have an interest in Bayesian ideas too?  (I just dug up an old paper of yours on {\tt quant-ph}.)  How on earth did we not meet for so long?  Just do a Google search on ``Fuchs quantum Bayesian'' and you'll get a few hits.

\section{09-05-08 \ \ {\it Change of Plan} \ \ (to S. J. van {\Enk})} \label{vanEnk14}

\bsve
How close were we to writing a paper on counterfactuals?? I do remember having fun making up titles (``Had we found a better title we would have used it'' still looks nice!), but I remember almost nothing about the possible contents \ldots

So far I found no remarks of mine in your samizdat that I regret: probably you edited them so they were not insulting to anyone. Yes, please tell me I did at least insult one person!
\esve

I don't know how close we were, but I think we need to do it.  It is a point that has not come up in my other writings.  Of course you insulted someone, and of course I was careful not to reprint those words \ldots\ One of the entries, by the way, is that very title.  I must have written the note because you said it, and I didn't want to forget it.  [See 04-04-06 note ``\myref{vanEnk6}{Title}'' to S. J. van {\Enk}.]

How about this for the real paper:  ``Had we found a better title we would have used it:  The Use of Counterfactuals in Quantum Bayesianism''?

Did you read about our new director here at PI?  Ever heard of the guy.  BTW, I tried to get 't Hooft to bite on the SIC problem today, but he's a slippery fish.  (I explained how useful they'd be as a starting point for his hidden variable considerations.)

\section{09-05-08 \ \ {\it Change of Plan, 2} \ \ (to S. J. van {\Enk})} \label{vanEnk15}

\bsve
{\rm [CAF wrote:]
\bq\noindent
``Did you read about our new director here at PI?''
\eq
}
Nope \ldots\ does he/she like quantum information??
\esve

The more important question is does he like quantum foundations!  (My guess is cosmologists rarely do.)

\section{09-05-08 \ \ {\it Change of Plan, 3} \ \ (to S. J. van {\Enk})} \label{vanEnk16}

\bsve
While biking back from work I was thinking about counterfactuals: do many-worlders have a different view than Bayesians on this subject??
\esve

I guess it'd be hard to even imagine a counterfactual in a deterministic world---in a world that is one ``unbending unit of fact''.

\section{12-05-08 \ \ {\it Foundations Curriculum}\ \ \ (to S. L. Braunstein)} \label{Braunstein14}

\bslb
If you were to teach a graduate course in foundations of QM what topics would you cover?

I need to decide between this and a course in quantum information pretty soon, so a quick response would be very much appreciated.
\eslb

There's just one delay after another with me.  You're right when you say, ``sounds like you're doing a lot of traveling.''  Here's something I wrote another friend a couple of days ago:
\bq\noindent
   I've adopted a truly insane travel schedule again.  I leave for seven days in
   Zurich tomorrow.  Next month a week in Paris.  2.5 weeks in Munich in July.
   A week in Sweden in August.  Innsbruck in September.  Sydney in November.
   Jerusalem in December.  And I had already been to England, Australia, Oregon,
   and Texas in the previous months of this year.  I probably listened to the
   Allman Brothers Band too much when I was young.
\eq
Let's see if you get the final allusion in that.

Here are the kinds of topics I myself would cover in a quantum foundations course.  (Of course keeping in mind that I have little sympathy for hidden-variables, GRW, or many-worlds:  I.e., I have an agenda.  An agenda which takes as a rallying point something Asher first drove home to me, ``Unperformed Measurements Have No Outcomes.''  Thus quantum foundations, as I see it, is about exploring the truth in that phrase at the same time as {\it not\/} giving up on the more basic agenda that world would still be here if we petty humans happened to drop dead tomorrow.  That physics is an attempt to say something about the world as it is without us.)  Thus, the topics below, roughly in the order I'd weave them together. \medskip

1) As warm up, selections from Mermin's book {\sl Boojums All the Way Through}. and/or his Rev.\ Mod.\ Phys.\ {\bf 65}, 803 -- 815 (1993) paper, ``Hidden variables and the two theorems of John Bell.''\medskip

2) Keeping with that theme some selections from Appleby's paper ``The Bell--Kochen--Specker Theorem,'' \quantph{0308114}.  Maybe also for fun, go over Asher's 33-ray state-independent proof, and Clifton's 8-ray state-dependent proof (old papers, not on {\tt arXiv}).\medskip

That material sets the stage for the slogan, and leaves the tension of ``My god, where do we go from here?''.  Accept action at a distance?  Or accept that quantum states are not elements of reality, but more akin to incomplete information?  (Leaving aside for the moment, the question of information about what.)  At this point I would start to move down the informational track, with a prelude of showing off various of the phenomena of quantum information theory:  no-cloning, no-broadcasting, teleportation, entanglement monogamy, superdense coding, such things.  And then backtrack, by hitting the students over the head with:\medskip

3) Spekkens's toy model, \quantph{0401052}.  That should clench any doubt that quantum states are analogous to incomplete information.\medskip

But information about what?  At that point introduce them to two parallel themes (that are not yet married but to some extent flirt with each other):\medskip

4) The quantum Jamesian pragmatism of Fuchs \quantph{0205039}, the quantum Bayes\-ianism of Caves, Fuchs, and Schack \quantph{0404156}, \quantph{0608190}.\medskip

and\medskip

5) The relational ideas of Rovelli \quantph{9609002}.\medskip

Those are both attempts to speak to the issue of information about what (in a way that is not hidden variables).

You might bolster that with a side tour on Bayesian probability itself:\medskip

6) Appleby's two papers are about the best things out there:  \quantph{0402015} and \quantph{0408058}.\medskip

But if all those ideas on the right track, how do they imply the mathematical shape of quantum mechanics?  Well we haven't figured that out yet, but there have nonetheless been some valiant attempts to derive the shape of the theory recently and they should be reviewed for the insight they might give:\medskip

7) Hardy's Five Axioms \quantph{0101012} (Hardy has a new version not yet posted that he'd probably share with you).\medskip

8) The convex sets work of Barnum, Barrett, Leifer, and Wilce: \arxiv[quant-ph]{0707.0620}, \arxiv[quant-ph]{0803.1264}, \quantph{0611295}, \arxiv[quant-ph]{0712.2265}.\medskip

I'd guess you're at a whole course by now.  But if you still have time, and want to leave on a positive note, maybe explore what all this might mean for the one-way model of quantum computation.  Honestly, I think that's a grand playground for this kind of point of view in quantum foundations:  For there it's obvious that by kicking the cat (state), you really do set the world in motion.

Hope all of this is of some value to your thinking.

\section{12-05-08 \ \ {\it Fascination with Words} \ \ (to W. G. {\Demopoulos})} \label{Demopoulos23}

Can you explain in simple English and give an example or two of what Carnap meant by a ``functor''?  The term came up at PI's wine and cheese Friday as I joked that I'd like to see quantum measurement outcomes not as indicative of properties of the system, or even indicative of a relation between the system and agent, but rather as a functor of the two objects (system $+$ agent).  Then I mumbled some things about the sexual interpretation of QM, and finally admitted that I had no idea what a functor is, but just really liked the sound of the word.

On the other hand, just now I read this in wiki:
\bq\noindent
   Carnap used the term ``functor'' to stand in relation to functions analogously
   as predicates stand in relation to properties. [See Carnap, {\sl The Logical Syntax
   of Language}, p.~13--14, 1937, Routledge \& Kegan Paul.]  For Carnap then, unlike
   modern category theory's use of the term, a functor is a linguistic item.
\eq
which intrigues me.

Greetings from Zurich.  I just arrived a couple of hours ago, and the spirit of Pauli already seems palpable.

\section{12-05-08 \ \ {\it Fascination with Words, 2} \ \ (to W. G. {\Demopoulos})} \label{Demopoulos24}

\bwd
A functor, for Carnap, is just a functional expression---an expression for a function. So this is a case where Wiki is a reliable guide to the truth.
\ewd

You may be a distinguished, honored professor, but you don't read instructions very well!  Now I'm more confused than before.  Trouble is I have no training in philosophy.  Thus, what you've written appears to me to contradict wiki.  Probably because I don't really know what ``functional expression'' is, nor do I really know what ``predicate'' is.  If you could draw up a couple of examples, that'd probably really help me.

\subsection{Bill's Reply}

\bq
Let me try to keep it simple; I can give you the long answer when we get together.

The linguistic expression
$$
\mbox{x is even}
$$
is used to express a property of integers. Upon substitution of a numeral name or other singular term for the variable, the result is either a true or a false sentence.

But the linguistic expression
$$
\mbox{x + y}
$$
is used to express a function from pairs of numbers to numbers; upon substitution for the variables, the result is a singular term which picks out a particular number:
$$
\mbox{x is even/x+y :: predicate/functor.}
$$
And by implication, predicates go with properties, functors with functions.

I'll be curious to know if this settles your perplexity, which I suspect is not at the level of words.

I don't recall what a category theorist means by `functor' but I think it's a {\it special\/} kind of function, not a general term with the same meaning as `function.'
\eq

\section{12-05-08 \ \ {\it Fascination with Words, 3} \ \ (to W. G. {\Demopoulos})} \label{Demopoulos25}

That helps quite a bit.  I'll keep thinking.  I like the word so much, I really do want to use it!  Homework challenge for you:  find a place where a functor arises in your thoughts on effects.

\section{13-05-08 \ \ {\it Incendiary Stuff} \ \ (to C. Misak)} \label{Misak3}

Thanks for the abstract.  It'll be good to start a dialog with you.  But watch out, the charge of postmodernism is incendiary stuff!  It is true, that I think quantum mechanics indicates a certain malleability to reality---that nature can be sculpted to some extent---that is indeed a description I like.  But it's a far cry from that to the more extreme idea that reality is just a conversation.

See Sections 4 and 5 in the attached pseudo-paper.  A reality that gave no resistance to our whims (or conversations) wouldn't be of much use, I would think.

I'll get your abstract into the PI schedule.

Greetings from Zurich, in case I haven't already sent them!

\section{13-05-08 \ \ {\it More on Malleability} \ \ (to C. Misak)} \label{Misak4}

This is secondary material -- certainly I hope you'll read the stuff in the last note I sent you before touching this.  But while I'm bombarding your email box, I thought I'd go all out.

You can find the aspects of William James that I take seriously here in the attached, starting at item 237.  Also, in the Menand item 302, you'll see some of the strains of pragmatism I like best.  Schiller in 411.  The pieces of Rorty I toy with taking seriously start at 395.  But you should never lose sight that I am an experimentalist in thought (part of the nature of being a physicist).  Please resist lumping me in with a set school.

I guess your abstract made me sensitive; the last thing I need in this serious effort to get quantum mechanics straight is a misplaced charge of postmodernism.  It is bad advertising.

\section{13-05-08 \ \ {\it Misak Tuesday} \ \ (to J. E. {\Sipe})} \label{Sipe18}

These are interesting thoughts indeed!  I've never thought about things like this.  What a place for them to hit me, here at the home of Pauli!  (I'm in Zurich at the moment.)

The meeting has not been forgotten; it's just been put into the planning of next year.  Next week I will start discussing it with Misak.  You might be amused by her abstract.  Here it is:
\bq\noindent
Some theoretical physicists, Chris Fuchs among them, take quantum mechanics to go hand in hand with an anti-representationalist accounts of truth and reality such as that offered by the American pragmatists --- William James, Charles Peirce, Richard Rorty, etc. On this view, scientific theories are instruments, rather than mirrors of the real world. In this talk, I'll suggest that if the quantum physicist is to team up with the pragmatist, he'd do best to join not with James and Rorty, who see the world as radically plastic or malleable. He would do best to join with the founder of pragmatism, Peirce, who argued that a regulative assumption of inquiry is that there is a right or determinate answer to the question at hand. It may look as if the anti-representationalist quantum theorist will be unhappy with this suggestion, but I'll argue that this would be a mistake. The trail of the human serpent, as James said, is over everything but, as Peirce saw, this does not toss us into the sea of post-modern arbitrariness, where there is nothing to say about what is true and what is real.
\eq
I love the way she talked about ``some theoretical physicists'' in the plural!

\subsection{John's Preply}

\bq
Sorry I'll miss the talk on pragmatism; it sounds great.   But I'm waiting to hear the announcement for that great workshop you'll be running on pragmatism! \smiley

A quick thought:  I was musing about that workshop and how one of the goals is to confront the the idea of a block universe with a more pragmatic one.  That suggests looking for particular instances where `block universe thinking' and your `reach out and touch someone thinking' give one different perspectives or different strategies for dealing with various issues.  Just for fun, here's a kind of sci-fi issue:  Consider the idea of time travel.  The usual objections or difficulties in thinking about it is that you might be able to go back in time and inadvertently kill your grandfather, and then how could it be that you would have been there to do that, and so on and so on. These difficulties seem to me to follow from at least implicitly thinking in a `block universe' way, with the past and present and future all carved out in the block.  It's that that seems to lead to the kind of contradictory situations that one might be able to get into were time travel possible.   But if one abandons the block universe view completely, and doesn't think of space and time in this way at all, are the problems of time travel so severe?   I'm not asking for a technical solution here, of course, but a discussion involving the concepts.

Grist for a philosopher's mill.  How is the pragmatic analysis of the possibility of time travel different from the block universe analysis?

If you don't think this is too stupid, you might ask Cheryl.
\eq

\section{14-05-08 \ \ {\it Many Thanks for Those Papers} \ \ (to C. Misak)} \label{Misak5}

\bcm
Now that I've read the papers you sent yesterday, I've got a much better set of things to say. You really are a Peircean, not a Jamesian or a Rortian. And that's a good thing!
\ecm

My guess is that I'm a nuance that's taken elements from all three.  And those elements are likely not consistent \ldots\ but they strive to become so in a greater unity motivated by quantum foundational considerations.  In any case, I'm glad you seem to have enjoyed what you read.

See you Tuesday.

\section{15-05-08 \ \ {\it 1:30 AM Note -- Decompressing the Day} \ \ (to D. M. {\Appleby})} \label{Appleby34}

A very short note.  I believe I made some inroads today with the talk.  I think it went quite well.  Ben Toner called it ``a beautiful talk''.  Mauro got very excited as usual, and reprimanded me (as usual) for talking about things not yet published (not because he cares about my career, I think, but because he momentarily wants to get his students working on the stuff).  Reinhard Werner was attentive, and that was flattering enough---I respect Reinhard a lot.  Mathias Christandl really liked the stuff on modified coherence.  Howard Barnum tolerated it kindly, even though he surely understood all this stuff long ago.  \v{C}aslav Brukner was pretty happy, though he pointed out that I did not mention about how he and Anton where the first to make a big deal of the quadratic entropy in quantum mechanics---I really should have; the ellipsis wasn't intentional, but I felt very bad.  Alex Wilce said he much better understood the foundational importance of SICs and my obsession with them after listening today; and wanted to get the summer students in his REU working on the project.  So I was very happy in general.  Gisin was dismissive as usual, but I did not expect less.  I had an edifying conversation with him.  One thing he emphasized was how wrong he thinks my attitude is that one should take the single system as basic in an axiomatization of QM (rather than giving composition rules like Hardy and everyone else).  My point was that tensor products should be derived from decomposition, rather than composition.  Gisin said I would miss entanglement that way.  I explained that I would not, could not.  And I explained how the move was much more in line with your dictum, ``Probabilities are single case or nothing''; it is a Bayesian move demanded almost by consistency itself.  He said, ``Well, Jauch and Piron could get the tensor product from their approach too, but then it took out the mystery of entanglement.  Of entanglement, all they could say was `so what?'.''  And all I could say was, ``Exactly!''  ``You cling to a mystery; whereas I want to get enough understanding to move to the next stage.''

One highlight of the day was meeting Ernst Specker.  We talked on the bus, and then shuffled very slowly across the campus.  He is 88 and an amazing man.  He explained to me how the Kochen--Specker theorem (which he had first published by himself five years before the usually cited joint paper) arose from a theological question.  At the time, he really wanted to know whether God could know what the world would have been like if Hitler were never born.  He had all kinds of insight to share with me in that hour I had with him, and I loved it.  He dismissed the nullification of Meyer as not understanding what real numbers are really about (I didn't completely understand him, but the best I could surmise he thought Meyer was committing a category error).  The thing he wanted to emphasize is how often we are caught up in thinking we have a good grasp of reality.  The way he viewed it, almost all of physics and our usual discourse is just of ``surface phenomena.''  He said we just have no clue of how much more there is to the world than that.  In his lecture, he was very funny, but very serious.  He said we should always strive to see the world from the perspective of the things around us:  For instance, try very hard to imagine the world as it ``appears'' to a flag wrapped around a flagpole.  I think he was very serious on that point, actually.  Wonderful man from what I could tell.

OK, not such a short note, but a grand day.  Tomorrow, I recover from these too many beers, and get back to work on the grant proposal.

Thanks for all the help getting me here.

\section{16-05-08 \ \ {\it 2:35 Happy Blues}\ \ \ (to R. {\Schack})} \label{Schack134}

I know you're sound asleep at the moment (it's 2:45 AM), and that you probably won't read your email until the weekend is over.

But I thought I'd give you a progress report.  I still can't see how to get the triple product constraint, but I have been able to tell this much:  The constraint that the probabilities lie on a sphere is not sufficient to insure that the modified coherence formula be nonnegative.  Here's a simple example of two probability vectors which will make the quantity go negative:
\begin{align}
p &= (0, x, y, y, \ldots, y)
\nonumber\\
q &= (x, 0, y, y, \ldots, y)\nonumber
\end{align}
Fulfilling the proper constraints:
\begin{align}
x + (d^2 - 2)y &= 1\nonumber\\
x^2 + (d^2 - 2)y^2 &= \frac{2}{d(d+1)}\nonumber
\end{align}
one finds that the Born rule quantity
$$
d(d+1)\sum_i p(i)q(i) - 1
$$
already goes negative at $d=3$.

That is very good.  Because I think it now much more likely that we really will need the extra triple-product constraint to keep things from going negative.

I hope you have a good trip back to London in four hours.

\section{17-05-08 \ \ {\it Springtime in the Air} \ \ (to G. Musser)} \label{Musser21}

Yes, definitely desirable and soon.  My writing cap is on again, in fact.  I was stirred by John Wheeler's recent death and had started to compose a note to you the day after.  It would be wonderful if I could combine a story on quantum Bayesianism with a story of the ``law without law'' and ``it from bit'' of John Wheeler.

I'm in Zurich, and traveling back to Canada tomorrow; fighting a deadline to get a grant proposal in by Wednesday (which I absolutely have to get out the door); and sparring with a pragmatist Tuesday who likely will say some bad things about me.  See: [\pirsa{08050041}, possibly?].  (It might be a cup of tea that'll interest you, when it's broadcast on the web starting Wednesday; follow the link on the page I just gave you.)

I'll get back in touch with you Wednesday.  Thanks for the amazing stamina you've shown in trying to get me to write.  It is very flattering.

\section{19-05-08 \ \ {\it Plurals, for Your Reference} \ \ (to C. H. {\Bennett})} \label{Bennett61}

\bcb
Umlauting the vowel and appending an -e is one of the commonest German plural patterns, as in Ansatz $\Rightarrow$ Ans\"atze  or Fuchs $\Rightarrow$ F\"uchse.
\ecb

I've never been pluralized before.  Thanks for letting me know how to do it!

\section{20-05-08 \ \ {\it SICs at PI}\ \ \ (to H. B. Dang)} \label{Dang10}

I thought I'd keep you up to date on my activities.  Attached is a proposal that I just sent off to ONR.  If you dig around, you'll find your name mentioned in the proposal briefly.  If by wild chance I get the money, you would have the first shot at it, if you want it.  That would free you from having to do anything with quantum crypto, and you could devote yourself fully to foundations (ahem, the quantum information theory of SICs).

Schack and I made great, great progress last week in Zurich.  I believe we are on the verge of having something I would be willing to call ``the fundamental equation of quantum theory.''  It's turning into quite a beautiful story, but the work is not complete yet.

\section{21-05-08 \ \ {\it John Wheeler}\ \ \ (to S. K. Stoll)} \label{Stoll3}

Thanks; I wish I could be there to hear it.  When John died, my collaborator {\Ruediger} Schack was visiting, and it was funny how, spontaneously, all week long I would keep coming back to stories of him or some aspect of his thought.  He's a big part of my mind.  I didn't know you knew that.

I hope all is well with you.

\section{21-05-08 \ \ {\it Determinacy, Determinates and Determinable} \ \ (to J. R. Brown)} \label{BrownJR1}

It was good meeting you yesterday.  I was able to dig up the correspondence I had with Demopoulos about the terms above.  Sorry my memory was so frazzled last night.  You can see what he meant by going to text pages 596 and 601, page numbers at the bottom of the pages (or pages 619 and 624, respectively, by the raw pdf page count on the top) of the following file:  \myurl{http://www.perimeterinstitute.ca/personal/cfuchs/nSamizdat-2.pdf}.
At least my memory had served me that W. E. Johnson was the right character. [In the present document, see 03-02-07 note ``\myref{Demopoulos9}{Bill's Thoughts on QL and QI Frameworks}'' and 04-02-07 note ``\myref{Demopoulos11}{The Painful Ambiguity of Language},'' both to W. G. {\Demopoulos}.]

The document I link you to is currently under construction, so I apologize for its present lack of an introduction and its incomplete index.  (And I apologize if the page numbers have changed by the time you take a look at it; in which case, just use the search tool to search on ``determinacy,'' etc.)  I'm somewhat infamous for publishing my foundational emails; this will be a new collection of them, posted officially relatively soon.

\subsection{Jim's Reply}

\bq
Thanks for the passages from Demopoulos.  The idea that we should chop up the world in a very different way (different from objects with properties and the probability of specific instances) is somewhat radical, but may well be right.  Certainly worth considering, at the very least.

It turns out I have a copy of Johnson's Logic, though until now I never read it.  The determinable-determinate distinction comes up when he's trying to distinguish ``Plato is a man'' from ``Red is a colour''.   Grammatically they are the same, but (he rightly claims) they are quite different logically.  In a type hierarchy it is easy to see:  ``Plato'' is an individual object, ``is a man'' and ``is red'' are first-order properties of objects, ``is a colour'', is a second-order property, ie, a property of properties.  For Johnson, a determinable is like a second-order property and a determinate is a first-order property.  Eg, Mary (individual) has the property weight (determinable) and the magnitude of her weight (determinable) is 60 Kg.  I presume the QM analog would be: An electron (individual) has observable momentum (determinable) and the eigenvalue is $p$ (determinate).  This might turn out to be useful terminology.  It is certainly better than the standard ``observable.''

I'm attaching an invitation to Dubrovnik next April when we're doing physics as a topic.  There will be lots of people, including many you know.  The place is stunningly beautiful and the conference lots of fun.  Try hard to make it.

\eq

\section{22-05-08 \ \ {\it Struggles with the Block Universe} \ \ (to C. Misak)} \label{Misak6}

\bcm
Many thanks for the invitation Chris. I had a great time and learned much!
\ecm

No, thank you for coming!  Meeting you has made me wish we could get a serious study group going about these things---too bad you're (I'm) so far away.  (By the way, my knowledge of Chauncey Wright came from Philip Wiener's book, ``Evolution and the Founders of Pragmatism''; have you seen that one?  It might be good material for your present study.)

I'm really angered that I seem to have lost the notes I took at your talk.  There were a few things I wanted to reply to.  I guess I'll just have to listen to your talk again eventually, and take it from there.

But there is at least one issue that stayed in my mind from the lunch conversation that I can potentially reply to.  It is that I got the impression you were dubious of mixing Peircean tychism and Bayesian probability into the same pot.  Don't let it be so!  (If it is so.)  They're perfectly compatible, and not getting confused on the subject, as many fans of Peirce are (like Abner Shimony), is the first step to making progress.

Here's a place to read a little bit about the way to think of it. [\ldots]

\section{22-05-08 \ \ {\it Contextuality} \ \ (to R. W. {\Spekkens})} \label{Spekkens49}

\brws
Speaking of the ``recent resurgence of interest in [contextuality] in quantum information and quantum foundations communities,'' I forgot to tell you about the QIP application of contextuality that Ben Toner and I have finally nailed down: parity-oblivious multiplexing.  See \arxiv{0805.1463}.
\erws

Your new results sound juicy.  I'm surprised:  I didn't realize you were an experimentalist.

Watch Cheryl Misak's talk on PIRSA [see \pirsa{08050041}]:
\bq
\noindent {\bf American Pragmatism and the Construction of the Universe}\medskip

Some theoretical physicists, Chris Fuchs among them, take quantum mechanics to go hand in hand with an anti-representationalist account of truth and reality such as that offered by the American pragmatists --- William James, Charles Peirce, Richard Rorty, etc. On this view, scientific theories are instruments, rather than mirrors of the real world. In this talk, I'll suggest that if the quantum physicist is to team up with the pragmatist, he'd do best to join not with James and Rorty, who see the world as radically plastic or malleable. He would do best to join with the founder of pragmatism, Peirce, who argued that a regulative assumption of inquiry is that there is a right or determinate answer to the question at hand. It may look as if the anti-representationalist quantum theorist will be unhappy with this suggestion, but I'll argue that this would be a mistake. The trail of the human serpent, as James said, is over everything but, as Peirce saw, this does not toss us into the sea of post-modern arbitrariness, where there is nothing to say about what is true and what is real.
\eq
It was quite a good talk.   But she made me sound plumb danged like a philosopher.  (And that surprised me too.)

\section{22-05-08 \ \ {\it Unstoppable Forces, Immovable Objects} \ \ (to H. R. Brown)} \label{BrownHR6}

\bhb
Before she left Cheryl [Misak] told me how much she enjoyed the visit to PI yesterday; you must be pleased with the outcome. And thanks for responding to my criticisms (again) with such good grace yesterday!
\ehb

Your criticisms are a pleasure, precisely because they're much needed.  They are reminders that I don't have as adequate or as pithy of answers for you as I would like to have.

This question of why update probabilities when one's eyes are closed is a good one.  I understand your point well.  If one supposes ontic variables (hidden or otherwise) underwriting the results of quantum measurement outcomes, then one has an immediate story for why one is updating one's probabilities (for measurement outcomes, that is) when one's eyes are closed and time flows by:  One's degrees of belief should change, because one is supposing the ontic variables to be changing by some objective dynamics.

But we quantum Bayesians presumably have no recourse to such a story, as we reject hidden variables Bohm-style, as well the objective wave functions Everett-style.  Our probabilities are probabilities {\it directly\/} about experience---directly about the consequences of an agent's interaction with a system external to himself.  In the vernacular, probabilities directly about measurement outcomes.  So, the point is, you surmise, why would we ever write $p_0(h)$ now, and $p_1(h)$ later, when learn no new data in the meantime?

However, as Marcus Appleby emphatically replied (when he was last visiting and we were discussing your point):  one writes $p_0(h) \rightarrow p_1(h)$, {\it precisely\/} because one does believe SOMETHING is changing.  And I have to agree with that.  The only question is what is changing?  What do we q Bayesians believe is changing with the flow of time?

This is where my answer becomes inadequate, even by my own lights.  An agent in this quantum setting changes his probabilities because he believes something integral between himself and the system is changing.  But to break it down into a change of variables internal and a change of variables external I will not do; that would be a step backward from our point of view.

I would love to have a more precise answer for you, but I do not have it presently.  And when I do have it, I won't only have it for you, but the whole research program will have moved forward significantly by the understanding it'll bring.  So, I welcome the question, but one final point on it:

Not having an answer to one isolated question is not a blemish on the program in my eyes.  It rather means the program is alive, in contrast to the usual way I view the Everettian program as dead:  New PHYSICS will come of it.  It has to; it's forced to.  Physicists are not born into the world with all the answers.  Rather the lucky ones are born with a good nose that'll lead them in directions of progress.

The bit of progress we have made in the last few years is that we now understand much better how to view quantum states in purely probabilistic terms (purely Bayesian or subjectivistic terms, I would even say).  And it hangs together with a formal coherence that surprised even us.  That, we see as an overpowering clue that we're on the right track.  Here's the way, Rob {\Spekkens} put it very eloquently (with regard to his particular approach) in his original toy-theory paper:
\begin{quote}
Because the theory is, by construction, local and non-contextual, it does not reproduce quantum theory.  Nonetheless, a wide variety of quantum phenomena have analogues within the toy theory that admit simple and intuitive explanations. \ldots\ The diversity and quality of these analogies provides compelling evidence for the view that quantum states are states of knowledge rather than states of reality, and that maximal knowledge is incomplete knowledge.  A consideration of the phenomena that the toy theory fails to reproduce, notably, violations of Bell inequalities and the existence of a Kochen--Specker theorem, provides clues for how to proceed with a research program wherein the quantum state being a state of knowledge is the idea upon which one never compromises.
\end{quote}
The point is, we know which idea NOT TO COMPROMISE on anymore, and we let that partially lead the way to new physics.  Your great question is part of that process.

Now to another issue.  It is an important psychological question, I think, as to the roots of why we don't see eye-to-eye.  I look 1) to the sheer number of phenomena in quantum information theory (from no-cloning to teleportation to correlation monogamy to 25 other things) that come about immediately for any kind of incomplete knowledge (i.e., without any regard to quantum issues), and 2) to the formal near equivalence of quantum theory and probability theory simpliciter.  Those two things together lead me down this research path.

For point number 2), SICs make it particularly obvious.  See the attached file, which contains the core of the scientific part from the proposal I just wrote for ONR---it is the best introduction I have at the moment.  In that language, the Born rule is nothing other than a mathematically minor variation of the law of total probability.  And unitary evolution?  Well, it is exactly the same thing---formally, the same expression!  Look at the last two equations on page 5 and the first equation on page 6:
\bq
Let us describe a unitary evolution $\rho\longrightarrow U\rho U^\dagger$ in these terms.  In terms of probabilities, this is just a transfer from an initial probability distribution $p(i)$ to a final one $q(j)$.  What is amazing is the way the formula, to within a sharpening factor and a renormalization, looks exactly like a classical stochastic evolution:
$$
q(j) = (d+1) \sum_i p(i) t(j|i) - \frac{1}{d}\;,
$$
where the $t(j|i)$ are matrix elements of a doubly-stochastic transfer matrix, which itself is a SIC representation of the unitary operator $U$:
$$
t(j|i)=\frac{1}{d}\tr\Big(\Pi_j U\Pi_i U^\dagger\Big)\;.
$$
In other words, aside from the universal sharpening factor $(d+1)$ and the renormalization $1/d$---both of which are some kind of ``quantum magic'' independent of the particular unitary $U$---a unitary evolution in this language is mathematically identical to a noisy classical channel.

Now, the SIC measurement is only there for the representation, but what we are generally interested in are the probabilities of outcomes for some {\it actual\/} measurement, perhaps a measurement of the computational basis at the conclusion of a quantum computation.  For instance, suppose the measurement of real consideration is a von Neumann measurement consisting of $d$ outcomes characterized by an orthornormal basis $|j\rangle$. Then the usual calculational tool for quantum probabilities, the Born rule, turns into a mathematically minor variation of the {\it classical\/} law of total probability!  To see this, let $p(i)$ represent the probabilities of the imagined measurement as before, and let $Q(j)$ represent the probability of outcome $j$ in the measurement that will actually be performed.  Furthermore, given an outcome $i$ for the imagined measurement, let $T(j|i)$ represent the probability of an outcome $j$ in the actual measurement, but now thought of as occurring immediately after the imagined one.  Then,
$$
Q(j) = (d+1) \sum_i p(i)\, T(j|i) - 1\;.
$$
In fact, aside from the differing normalization factor (a 1 instead of a $1/d$), this formula is mathematically identical to the one that represents unitary evolution itself.
\eq

My physicist's sense says that points 1) and 2) cannot be coincidence.  Your philosopher's sense, as far as I can tell, says that it can be NOTHING BUT a coincidence.  Thus you lob back the challenge to me, ``But what of the intricate structure of the {\Schroedinger} equation?''

And I have an answer to that:  The {\Schroedinger} equation is simply our method for setting the conditional probabilities of the last equation of page 5.  In character, in fundamentals, it is a method of no great conceptual distinction from Ed Jaynes' introduction of the maximum entropy principle for setting prior probability assignments in statistical mechanical problems.  That, I already feel is an adequate answer.  But my relations with you are such as to lead me to be believe that you'll think I'm simply not getting the point.

So, I wonder why {\it you\/} don't get the point?  And you wonder why {\it I\/} don't get the point!?!  And with that, I start to speculate that perhaps there is something so deeply ingrained in our individual worldviews that it is blocking translation between the two of us.  That is the psychological point.

In any case, all I can say is there's work to be done, and I have not dismissed your question!

I hope you won't mind, but I plan to forward this note to Marcus Appleby, {\Ruediger} {\Schack}, and Rob {\Spekkens} (by the end of the day), as it should impinge upon their own thinking.

\section{23-05-08 \ \ {\it From Waterloo}\ \ \ (to M. P\'erez-Su\'arez)} \label{PerezSuarez23}

You will have discovered that I'm not the email man I used to be:  It takes me forever to reply to things now.  \ldots

Lately I have been more excited about SICs than ever.  I'm writing a paper at the moment on why they're so very special.  I will be sure to send it your way when it's done.  If you're interested in seeing some of my recent talks, you can go here:  \myurl{http://pirsa.org/}.  Just do a search on my last name.  Another you might find interesting is Cheryl Misak's comparing and contrasting some of the things I've written with C. S. Peirce's writings.  Schack by the way, has had some interesting developments in analyzing quantum random number generators from a Bayesian perspective.  You might write to him for the latest.

\section{27-05-08 \ \ {\it The Best Time}\ \ \ (to R. Renner \& O. C. O. Dahlsten)} \label{Renner2} \label{Dahlsten1}

Sometimes it takes a while for the things on my mind to come out on paper.  I want to tell you guys how much I enjoyed your conference, how much I learned, and how much I was taken with the city of Zurich.  Despite my spending so much time locked in my room writing a grant application (which had to be turned in on the day of my return), the week in Zurich turned out to be the best conference of the preceding year for me.  {\Ruediger} Schack and I broke an impasse in our q-Bayesian program while there, and we now have the glimmer of something we think is going to be very, very fruitful.  And my moments with Specker were priceless---it was a once in a lifetime chance.  So thank you both again.  The conference was simply fantastic.

\section{28-05-08 \ \ {\it Unstoppable Forces, Immovable Objects, 2} \ \ (to W. G. {\Demopoulos})} \label{Demopoulos26}

Let me send you this note I wrote to Harvey last week for two reasons.  [Sentence parsed as, ``Let me send you (this note I wrote to Harvey last week) for two reasons.'']  The first is because the attached document lays out the formalism I showed you on the board yesterday (though, unfortunately, it is buried in the middle of a grant proposal).  Second, it dawned on me overnight, that Harvey, once he's thought about it enough, will surely make the same point with you that he continually makes with me.

The criticism would be this:  If every property intrinsic to a particle is eternal and timeless from your point of view, then why would we ever write down dynamical (time evolving) probabilities for effects?  What is changing if not the particle itself?  That would be his question.  Further explication below. [See 22-05-08 note titled ``\myref{BrownHR6}{Unstoppable Forces, Immovable Objects}'' to H. R. Brown.]

I got to page 5 of your paper last night; I'll try to come back to the rest of it over the weekend.  I can say this much right now:  I wish I could see the force of your argument for determinacy better; it still seems elusive to me.

\section{29-05-08 \ \ {\it Your Vita} \ \ (to C. M. {\Caves})} \label{Caves96.4}

\bcc
I was just looking at your online CV, and noticed that you need the following reference:
\begin{itemize}
{\rm \item C. M. Caves, C. A. Fuchs, and R. Schack, ``Subjective Probability and Quantum Certainty,'' Studies in History and Philosophy of Modern Physics {\bf 38}, 255--274 (2007).}
\end{itemize}
\ecc

OK, I fixed it and reposted it.  I hadn't told you how Allen Stairs gave a detailed talk on that paper at the Demopoulos Festschrift.  The talk was titled ``A Loose and Separate Certainty'', and was good---he nailed us on every misformulated statement we made.  I'm hoping to bring him to PI for a month in the Fall; he is genuinely interested in getting all this sorted out.

\section{29-05-08 \ \ {\it Unstoppable Forces, Immovable Objects and their Points of Contact} \ \ (to H. R. Brown)} \label{BrownHR7}

Would you give me a reference (or, better, a copy) of your paper with Guido, where you take issue with the EPR criterion of reality?

\section{29-05-08 \ \ {\it Determinacy, Determinates and Determinable, 2} \ \ (to W. G. {\Demopoulos})} \label{Demopoulos27}

\bwd
I copied the Specker paper and put it in the mail today. It's very short and very elementary, but it shows that he had the main theorem before 1960.
There are a number of interesting things about it, not least his implicit general understanding of the significance of the theorem as having to do with prediction, things I'd forgotten. Oh and I forgot to say that Allen [Stairs] did the translation.
\ewd

Regarding your remark, the title of Specker's talk at the meeting I attended was ``Infuturabilia and Quantum Logic.''  Looking on the web, you won't find any entries on this strange word, but if you look up ``futurabilia'' you will find plenty.  Particularly you'll find things about Leibniz's discussion of God's knowledge of counterfactuals.  Kind of fun.

\section{29-05-08 \ \ {\it Determinacy, Determinates and Determinable, 3} \ \ (to W. G. {\Demopoulos})} \label{Demopoulos28}

\bwd
Regarding Harvey's question, no doubt the particle is changing; but (I
hope I'm not being too glib here) isn't the point of the effects
framework precisely this: we can ignore how the particle changes, how
it evolves dynamically, and still recover the phenomena---still recover
the effects particles produce in various controlled settings.
\ewd

You're not being glib.  But also I don't think you are helping yourself by talking that way.  To say, ``we can ignore how the particle changes'' is, linguistically at least, pretty much on the verge of invoking hidden variable imagery.  Something much better is called for, I would say, and that is why Harvey's question is a good experimental playground.

\section{29-05-08 \ \ {\it Infuturabilia} \ \ (to E. Specker)} \label{Specker1}

It was very good meeting you on the bus before your talk at the ``Information Primitives and Laws of Nature'' conference a couple of weeks ago.  It was also very nice walking with you across the campus.  I hope you remember me.  Your ``colorability'' results (as in your construction with Simon Kochen) have long interested me, and I well believe they are the key conceptual component of  quantum mechanics.

Thus I wonder if you would entertain an email conversation with me?  To some extent I have become something of a historian for my small community within physics---the field of quantum information---and I enjoy recording stories on the origins of this and that all around.  An example of some I have recorded can be found here:  \myurl[http://www.perimeterinstitute.ca/personal/cfuchs/SamizdatSE.pdf]{http://www.perimeterinstitute.ca/personal/cfuchs /SamizdatSE.pdf}.

In that vein, would you record some of what you told me when we met?  Would you tell me again about what was on your mind when you first started thinking about (what we in my field call) Kochen--Specker constructions?  What were the years?  Were you initially thinking about quantum mechanics, or rather theology?  I believe I understood from you that these things were on your mind well before you had met Si Kochen---but I want to get all the facts straight.

If you will share this history with me it would be much appreciated.  I learn in many ways---some by logical thought, but some by simply compiling history.  (In my own way, I too have been concerned with infuturabilia for a long time; see this collection of thoughts in particular \myurl{http://www.perimeterinstitute.ca/personal/cfuchs/nSamizdat-2.pdf}.)

\subsection{E. Specker's Reply, ``Futurabilia''}

\bq
Thank you for your mail --- I shall try to answer your questions as far as possible, but I need some time, for instance I have no idea when the seminar of Gonseth and Pauli was held. In the mean time you have perhaps a look at my article ``Was is ein Beweis'', published in a book {\sl Grenzen des Wissens}, where there is a reference to the notion of ``double truth'' of Ibn Rushed. I hope that you understand German. By the way, we had in Cornell a friend named Wolfgang Fuchs. We liked to discuss in German --- and not, as an other emigr\`e called it, American baby talk. If you do not understand the article, please let me know and I shall stick to English.
\eq

\section{30-05-08 \ \ {\it Futurabilia} \ \ (to E. Specker)} \label{Specker2}

Thank you much for your note.  I look forward to your more detailed reply.  I apologize, but I don't speak a word of German.  Unfortunately, my ancestors came to the United States (Texas actually) in the 1870s--1880s, so that by the time I stuck my head into the world, there was only English around me.  So, I know that this will make it more difficult on you (and you don't deserve that at 88), but German would be completely lost on me.

I am intrigued by your mention of Gonseth and Pauli, and wonder how they fit into your story.  Did these two people influence your thinking about infuturabilia?  I looked in my own database of Pauli'an things, and found these two items having something to do with Gonseth, but otherwise, I don't know who Gonseth was.

\bq\noindent
F.~Gonseth, ``Remarque sur l'id\'ee de compl\'ementarit\'e,''
Dialectica {\bf 2}, 413--420 (1948).  Wolfgang Pauli writes of this article in his introduction to the special issue of Dialectica:
\bq
Many of the articles mention possible applications of the idea of complementarity outside physics, as for instance to questions connected with biology or psychology.  I shall not discuss these questions in this introductory survey but wish to draw the reader's attention to the interesting attempts of Gonseth to formulate the idea of complementarity so generally that no explicit references is made anymore to physics in [the] proper sense.  This is, of course, only possible by the use of a language to which the physicists are not accustomed, which uses expressions like ``horizons of reality,''
``profound horizon'' and ``apparent horizon,'' ``events of a certain horizon.''  The word ``phenomenon,'' however, is used in this article strictly in the above mentioned sense given to it by Bohr. To the ``profound horizon'' of Gonseth belong the symbolic objects to which conventional attributes can not be assigned in an unambiguous way, while the ``traces'' of Gonseth are identical with the ``phenomena''
in our sense.  I wish again to stress here the circumstance that the free choice of the observer can produce either the one or the other of two ``traces'' and that every phenomenon or ``trace'' is accompanied by an unpredictable and irreversible change in the ``profound horizon.''
\eq

\noindent W.~Pauli, ``Theory and Experiment,'' Dialectica {\bf 6},
141--142 (1952).  Translated in W.~Pauli, {\sl Writings on Physics and
  Philosophy}, edited by C.~P.  Enz and K.~von~Meyenn, and translated
by R.~Schlapp (Springer-Verlag, Berlin, 1994), pp.~125--126.
\bq
F. Gonseth's dualistic standpoint in regard to the ``Dialogue between Experiment and Theory'' appears to me to be a special case of the more general relation of internal (psychical) and external (physical).  In the situation of {\it cognition\/} we are concerned with the relation of the cognizant to the known.  The purely empiricist standpoint, which seeks to reduce every ``explanation'' to a ``description''
(albeit a general and conceptual one) leaves out of account the fact that in every case the setting up of a concept or system of concepts (and hence also a law of nature) is a {\it psychical reality\/} of decisive importance.
\eq
\eq

\section{30-05-08 \ \ {\it Answers and Stories}\ \ \ (to O. J. E. Maroney)} \label{Maroney3}

I'm curious to know whether I answered the question you wanted me to answer yesterday?  Or was I off on a tangent on another subject?

Funny story, I don't know if I ever told you.  Do you know of how much of a big deal Ed Jaynes made of the original rejection of his first MaxEnt paper?  He framed the referee report to show and put it on the wall in his office to show students how inane referees can be.  Well, as it turns out, Landauer's first erasure paper was rejected too.  It put a great smile on my face when I learned that Jaynes was the referee.  If you want, sometime I'll show you the referee report (which I don't think I'm allowed to distribute yet).  [See 26-10-04 note ``\myref{Grandy3}{More `More on Landauer'}\,'' to W. T. Grandy, Jr.]

If you're interested in lunch, come get me on your way up.

\section{02-06-08 \ \ {\it Unstoppable Forces, Immovable Objects, 3} \ \ (to W. G. {\Demopoulos})} \label{Demopoulos29}

\bwd
As for instrumentalism, there is certainly this difference: an instrumentalist will refrain
from saying that a theory is true---it's merely instrumental for getting predictions. But I
think QM gives the true description of the algebraic structure of effects, just as CM
sought to give the true description of the algebraic structure of propositions. So in this
respect I still think of my view as realist. But it's also realist in a more mundane sense
insofar as it holds that the reality of the micro world is an observer independent one; it
therefore coincides with the realism of Einstein, which was rejected by Born and others.
\ewd

First quick comment.  I think you're being too hard on Born.  In an important and probably never cited essay, he made it clear that he believed in the same kind of man-independent reality that Einstein believed in.  Born was not simply positivist.  In the essay (I wish I could remember where I read it), he identified the man-independent stuff with the symmetries quantum theory was built on top of.\footnote{\editornote Born used the term \emph{invariance}: ``The feature which suggests reality is always some kind of invariance of a structure independent of the aspect, the projection.''  [M.\ Born, ``Physical Reality,'' The Philosophical Quarterly {\bf 3,} 11: 139--49 (1953).]}  That is a position, I think, not unlike {\Carl} {\Caves}'s that John Sipe tries to capture in the essay I sent you.

I will try to remember the source.

Other comments later as they bubble into existence.

\section{02-06-08 \ \ {\it My Christmas Present?, 2}\ \ \ (to S. T. Flammia)} \label{Flammia7}

[See 27-12-06 note ``\myref{Comer100}{My Christmas Present?}''\ to G. L. Comer, S. J. van Enk, and J. W. Nicholson.]  Apparently my reputation has somewhat recovered since that time.  Google brings up only two pages of hits now, whereas apparently then it brought up three.

It seems that the CSIQM page is indeed gone now; I guess even porn doesn't help sell some subjects.

\section{02-06-08 \ \ {\it Emailing:\ 453566c} \ \ (to G. L. Comer)} \label{Comer114}

\bgc
What does it mean that the Information is always conserved in Quantum Mechanics?  It's in the short ``Theoretical Physics'' piece in the attached file.  {\rm [``Better Out than In,'' {\sl Nature\/} {\bf 453}, 566 (29 May 2008).]}
\egc

As far as I can tell, it really means nothing---it is just something that people like to say without thinking very deeply.  Information is not a fluid, like phlogiston, with a conservation law.  Information is not a fluid.

\section{03-06-08 \ \ {\it Dow Commercial} \ \ (to myself)} \label{FuchsC19}

``It is a world that responds to our touch.''

\section{04-06-08 \ \ {\it Hammerhead} \ \ (to G. L. Comer)} \label{Comer115}

\bgc
There is a number 7 in one of your [SIC] equations in the proposal.  Now, doesn't that imply something wrong?  I mean, I've never seen a 7 before in a fundamental equation. 8, $\pi$, $e$, are good numbers.  But 7?
\egc

I've got a funny story about the 7.  I first presented the result at an APS meeting a couple years ago.  Charlie Bennett was in the audience and asked, ``Is that a 7?''  I said, ``Yep, it's really a 7.''  Charlie said, ``Well then, it's the first 7 I've ever seen in quantum information.''  And what else would you expect from a truly fundamental equation?!  Indeed it is a 7, and well checked many times by myself and independently by my students.  In fact, just the other day by the latest, Ryan Morris, who first found a 6 instead \ldots\ but then ultimately found a 7.

\section{05-06-08 \ \ {\it Hammerhead, 2} \ \ (to G. L. Comer)} \label{Comer116}

And of course I should have pointed out that 7 is a lucky number.

\section{05-06-08 \ \ {\it The First SIC} \ \ (to G. L. Comer)} \label{Comer117}

\bgc
I just found a news item from an Indian Physics Department:
\begin{center}
	{\rm Sikhs Seek SICs}
\end{center}
\vspace{-6pt} Bad?
\egc

Worse has been done.

\section{05-06-08 \ \ {\it Death of Auchentoshan} \ \ (to J. W. Nicholson)} \label{Nicholson28}

Funny that you wrote me this morning.  I was just going to write you, to tell you of the demise of the lovely bottle of Auchentoshan you brought me in Dublin.  The poor old fellow passed away peacefully about 10:30 yesterday evening.  He'll be missed.

Yes, the new Lazarides money was a bit of a shock last night.  We knew there was going to be a big announcement, but we didn't know about what.  I suspect the money will be used to bolster our expansion plans:  Somehow (architectural nightmare) the office space in the building is going to be doubled, add a new bigger lecture hall, double the size of the Bistro, etc.  The announcement was made just before Bill Phillips's public lecture---I'm really annoyed that I forgot about it, not because of the announcement, but because I bet it was a heck of a lecture: ``Time and Einstein in the 21st Century:\ The coolest stuff in the universe''.

I talked to Lazarides briefly after my own public lecture thing (panel discussion); he had dinner with the group of us, and hung out till midnight.  It was funny, kind of like Jon Waskan once described his meeting with Bill Clinton, how he almost melted when he shook his hand.  In my own case, I had this overwhelming feeling of gratitude---uncontrollably, somehow the first thing out of my mouth when I had him alone was, ``Thank you for this place!''  He said, ``Physicists are a really good investment; you get a really good return on very little money.''

Sciencewise, I'm very happy at the moment.  {\Ruediger} and I have made some serious inroads toward squeezing Hilbert space out of a single inequality.  If we make it all the way, this will be the best scientific work I've ever done.  I just hope the IF isn't too big of one.

If I discover that Philips mentioned your name in his lecture (when I view it on PIRSA), I'll let you know!

\section{09-06-08 \ \ {\it Pauli Understood Entanglement}\ \ \ (to R. Blume-Kohout)} \label{BlumeKohout4}

Points all well taken, of course.  The moral is I should have titled my note ``Pauli had an inkling of entanglement,'' which is certainly no insult for a 1927 thought.

\subsection{Robin's Preply}

\bq
I presume you mean:
\bq
Now it should of course be emphasized that such reductions first of all are not necessary when all the measuring instruments are included in the system. In order to be able to describe the results of observation theoretically at all, one must ask what can be said about one part of the total system on its own. And then one sees as a matter of course the complete solution---that the omission of the instruments of observation in many cases (not always, of course) may formally be replaced by such discontinuous reductions.
\eq
Yah?

I started to write a long analysis, then realized I don't have time to think that deeply!  So I'll content myself to observe that Pauli obviously {\it did\/} have a grasp on some good stuff \ldots\ and then I'll dig at the end bit a little.

He writes (``as a matter of course'') that the solution is to replace ``omission of the instruments of observation'' with ``discontinuous reductions'' \ldots\ while noting that this can only be done some of the time.  Now, granted that the first part of his letter implies at least that he understood that it's {\it possible\/} to describe the supersystem unitarily \ldots\ and that such a description doesn't say anything useful about ``the results of observation'' \ldots\ is there really any content in this last bit?  He seems to be saying that the process of disregarding the apparatus is, sometimes, equivalent to collapse.  This is a rather unhelpful statement because of the ``many cases (not always, of course)'' ambiguity.  Furthermore, it seems to be Copenhagen (``Shut up and collapse'') beyond that \ldots\ and finally it seems to conflict with our idea that you represent ``omission'' of a subsystem by partial tracing.

Perhaps a better way to put that is that we represent omission {\it either\/} by partial tracing {\it or\/} by collapse, but Pauli provides no guidance on which to use.

Anyway, this vague last sentence makes me wonder how to interpret the earlier sentences.  I am not sure what he means by ``such reductions first of all are not necessary when all the measuring instruments are included in the system''.  The best way (IMHO) to interpret that is in a relative-state framework -- i.e., that we can describe the joint state as being in an eigenstate of the ``One system knowing about the other one'' observable, and therefore can confidently answer questions like ``Does the observer know about the system?''\ even if we can't answer questions like ``So, what does she know?''  But I'm not sure if I should interpret Pauli this way!
\eq

\section{25-06-08 \ \ {\it } \ \ (to K. Brading)} \label{Brading2}

By interesting coincidence I discovered that our meeting the other night was our third meeting \ldots\ not our second.  This memory of mine is becoming so horrible!  To see what I mean, download [my samizdat] and search on ``Katherine Brading''.

Well, anyway, it was great meeting you for real this time:  Now you'll last in my memory until there really is none left of it.  I wish it had worked out that we had talked earlier in the week.

In the document linked, there are also some tidbits about {\Poincare} that might interest you.  Particularly, I'm thinking about de Finetti's giving {\Poincare} credit for coming to within a hair of full-blown Bayesianism (and being de Finetti's greatest influence).  And also {\Poincare}'s discussion of Boutroux's philosophy.  [See the 28-06-03 note ``\myref{Mermin87}{Utter Rubbish and Internal Consistency, Part II}'' to R. Schack, C. M. Caves \& N. D. Mermin, and the 10-03-06 note ``\myref{Halvorson15}{Singularities and Evolutionary Laws}'' to H.\ Halvorson.]

Hope you met up in London with your family easily enough.  Take care, and I hope to see you again without a lapse of four years.

\section{25-06-08 \ \ {\it Darn!}\ \ \ (to H. C. von Baeyer)} \label{Baeyer38}

Likewise for me, it was great having you a that meeting:  It was nice seeing a mind of sense floating in that sea of contrariety.  I wish you could come to PI; the time with Marcus is going to be grand.  And as I told him recently, I fear you need a booster shot of Fierz.

Attached is a copy of Bitbol's talk, in case you wanted to review it again.  I so wish you could have been there for Rovelli's talk.  It struck much deeper, it seemed to me, than his papers on ``relational quantum mechanics''.  I took a long walk with him after the meeting, and felt that he is seriously searching \ldots\ even if he over-cautiously fears the agent's introduction into physics.  Anyway, I like that in a person.

I'll keep you up-to-date as I respond to others about the meeting with some relevant points.  But at the moment, I must go be the MC for the HirotaFest!

\section{03-07-08 \ \ {\it Painful Choices of Language}\ \ \ (to A. Stairs)} \label{Stairs1}

\bAllS
Good to hear from you. I'd like very much to work out a way to come to Perimeter, though let me toss in a couple of qualifications.

The first is that I'm very much more a philosopher than a philosopher of physics. By that I mean my technical skills and knowledge are shockingly limited. (I could easily embarrass myself by elaborating\ldots) So {\bf caveat sponsor}.
\eAllS

In all seriousness, it is almost never the case that a philosopher of physics interests me.  The technical skills that you speak of, as far as I can tell, are almost always wasted because of deeper wrong-headednesses that have nothing to do with physics per se.  \ldots\ Thus maybe why I'm attracted to you.

\section{03-07-08 \ \ {\it PIPPO!}\ \ \ (to P. G. L. Mana)} \label{Mana11}

It should not be my business to pry, but I think you should exercise more control with PIPPO.  I am sympathetic to his quantity of self-respect, but there are limits necessary within any civilization---one can only go so far and still expect to maintain an orderly and positive life for all citizens.  No one living being can be so singled out at the expense of others.

Concerning the carrion ``collapse,'' I do not know when the word was introduced, but von Neumann made a distinction between Type I and Type II evolutions for quantum states already in his 1932 book.  Of course, if one changes the word from ``collapse'' to ``conditionalize'' (in the sense of a subjectivistic approach to probability), I myself think we are talking about something completely innocuous.

\section{07-07-08 \ \ {\it Philip Goyal and the Information-Geometric Reconstruction of QM} \ \ (to C. H. {\Bennett})} \label{Bennett62}

\bcb
Does this guy's stuff have anything to do with your goals?
\ecb

Unfortunately, no.  Philip often says the word ``Bayesian'' like I do, but that's where the resemblance ends.  Look harder, and you'll see that he's proposing that quantum theory is a kind of statistical inference theory about a hidden variable.  Roughly, whenever a quantum measurement is performed, two outcomes happen instead of one; it is just that one is hidden from us, and we can't tell which of the two we have.  My own program (for instance the flavor you saw with the SICs the other day) is pretty anti-hidden variable.

And by the way, I do not write things really big to compensate for the weakness of my own ideas.  I write really big because I assume some in my audience are near-sighted \ldots

\section{08-07-08 \ \ {\it Two Places to Read}\ \ \ (to A. S. Holevo)} \label{Holevo10}

It dawned on me to maybe follow up on one of our lunchtime conversations.  With regard to what criterion was used for (imperfect) teleportation in the Kimble experiment, you can go here: \quantph{9910030}.
With regard to our discussion on how the details of an ``instrument'' are ``washed out'' when we consider the state update on the far half of an entangled state, you can look here:  \quantph{0205039}. The relevant part is the discussion around Eqs.\ (98)--(101).

I hope you had a safe flight back to Moscow.

\section{08-07-08 \ \ {\it Wednesday Instead}\ \ \ (to R. {\Schack})} \label{Schack135}

I won't be able to call you this morning after all:  I've got this damned talk to give at IQC, which I'm still preparing for.

Appleby and I had another long discussion on the ueberungleichung yesterday.  The origin of these $\alpha_{ijk}$ is (remains) just damned confusing.  One thing valuable that came though is that maybe ueber is not the best prefix.  We toyed with the idea that ``Ur'' might be better.  Can you tell us the precise meaning of ``Ur''?  Urungleichung, does it work?

\section{09-07-08 \ \ {\it Wednesday Instead, 2}\ \ \ (to R. {\Schack})} \label{Schack136}

\brs
Wednesday OK. Urungleichung is a much better term. Ueberungleichung has been a joke and shouldn't become the official name.

Ur, as a prefix, means original or primitive (it is also a noun meaning aurochs). It is a very powerful prefix. Here are a few examples:
\bv
Ursprung $=$ Ur leap $=$ origin\\
Ursache $=$ Ur thing $=$ cause\\
Urbewohner $=$ Ur inhabitants $=$ first inhabitants, aborigines\\
Urknall $=$ Ur bang $=$ big bang
\ev
\ers
Beautiful!  Urungleichung it is!

\section{09-07-08 \ \ {\it SIC It Will Be} \ \ (to C. H. {\Bennett})} \label{Bennett63}

\bcb
\bq\rm
\noindent [Chris said:] When you were visiting, you seemed to naturally pronounce ``sic'' as
``seek''.  Is that based on a grammatical rule I can look up?  Is
everyone else wrong when they pronounce it ``sick''?
\eq
See attached for my pronunciation of {\bf sic}.  I assumed you were using it in the sense of the Latin textual annotation, not as an abbreviation for symmetric informationally complete, or any of the other meanings for {\bf sic}. According to my dictionary, even the Latin word may be pronounced like sick in English, but I studied Latin in high school, so I pronounce it like {\bf seek}.  By the way, the noun corresponding to {\bf pronounce} is {\bf pronunciation}, not {\bf pronounciation} (sic).  Don't ask me how the second {\bf o} got lost.
\ecb
Thanks!  That settles it.

\section{09-07-08 \ \ {\it Life/Etc.}\ \ \ (to D. R. Terno)} \label{Terno5.1}

Appleby, Schack, and I are working quite hard to ``SICken'' the axioms of quantum mechanics, and I think we're making some very nice progress.  I have {\it hope\/} now that the formal structure of quantum mechanics may be pulled out of essentially a single inequality.  It would be the greatest of all my dreams.  Pulling all of (the formal structure of) quantum mechanics out of a quantification of the idea that ``unperformed measurements have no outcomes''.  It would be nice to be able to say this with some confidence at Asher's summer school.

\section{09-07-08 \ \ {\it HirotaFest Picture}\ \ \ (to O. Hirota, A. S. Holevo, \& the invitees)} \label{Hirota8} \label{Holevo10.1}

Thank you all again for coming to PI last week.  It was a fantastic time.

Charlie Bennett's conference photo has now been posted on the web.  You can find it by going here:
\myurl{http://www.perimeterinstitute.ca/images/conferences/hirota_fullsize.jpg}.

\section{09-07-08 \ \ {\it Tables of Contents}\ \ \ (to S. Gharibian \& L. Sheridan)} \label{Gharibian1} \label{Sheridan3}

Thanks for the interest.

You can find a better, more careful version of my teleportation example here: [See 25-10-05 note ``\myref{vanFraassen5}{GOBs, Bobs, Steering \& Teleportation}'' to H. Halvorson and B. C. van Fraassen.]
In any case, in the talk version, I was careful to say that when Alice checks the parity, she then randomizes the coins (i.e., she shuffles them so that Charlie no longer knows which is which).  Thus from Charlie's perspective, he now knows nothing about the coin in the original box he gave Alice, and would write down a 50-50 distribution for heads/tails in it.  It is in that sense that Charlie's original state is ``destroyed'' for his original coin.

About your second question, my own formalism was just quantum mechanics rewritten.  Thus, there is everything in it that there is in quantum mechanics, including Kochen--Specker phenomena and Bell-inequality violations.  What I said ``must break'' is some feature of Spekkens's toy model.  As presently written (in terms of epistemic states for local hidden variables), it cannot accommodate these kinds of phenomena.  In my own case, it is my belief that my rewrite of the formalism is such that it helps pinpoint the essential difference between the classical and quantum worldviews:  It is that ``unperformed experiments have no outcomes'' (as Asher Peres would say).  Something new, that was not there before, comes out of each and every quantum mechanical measurement.  That is, even at this very low level of things, there is something analogous to birth in the world.  If that is an example of something you would call an ``inherent disconnect'' between the classical and quantum worlds, then, yes, I believe there is such.  It was only that I think SICs help massage the formalism into a form that makes this the clearest yet (by giving the formalism a more direct comparison to classical Dutch book arguments).

But do these damned SICs exist generally?  That's the real moneymaker of a question.

\section{09-07-08 \ \ {\it The History}\ \ \ (to G. Gutoski)} \label{Gutoski1}

Thanks for your comments yesterday.  In case you're still curious about some of the prehistory about these lines of thought, let me refer you to the Brassard section of \quantph{0105039}.

The idea that the existence of QKD might be a key part of the heart of quantum mechanics is an older one, exhibited for instance in the 6 July 1998 letter to Rolf Landauer titled ``Evolution and Physics''.  But the idea that the nonexistence of bit commitment might also be a potential ingredient in QM was born over a bottle of single-malt scotch with Gilles.  I'm still very proud of the allegory I created to go with it.  Let me particularly direct you to that.  It's in the 23 January 2000 letter to Gilles titled ``I See Why No Bit Commitment!''  See that and the note ``Fuchsian Genesis'' it refers to.

I hope I provided some entertainment and some food for thought for your group yesterday.

\section{14-07-08 \ \ {\it Harper, Fuchs, and James}\ \ \ (to R. Blume-Kohout)} \label{BlumeKohout5}

I read the Berlinski essay.  I didn't much like it actually, or maybe more to the point, him.  There was something about the writing that made me feel he's pretty arrogant.  As you suspected, I guess, there were indeed points I'm sympathetic to (about ``things not seen'' and about $M+x$), but then he clouded things up with too much about faith in God (Gods?)\ for my taste.

You can find my own favorite rebuttal of Clifford here:
\myurl[http://www.perimeterinstitute.ca/personal/cfuchs/nSamizdat-2.pdf]{http://www.perimeterinstitute.ca/ personal/cfuchs/nSamizdat-2.pdf}.
Do a search on the phrase ``Pragmatism versus Positivism'' to get to the right spot.  (It'll take you to text page 138 in the present build.  [See 21-11-01 note ``\myref{Schack37}{Pragmatism versus Positivism}'' to R. Schack.])  Also there are a couple of other tidbits on Clifford if you search on his name.

\section{14-07-08 \ \ {\it Wheeler Scholars} \ \ (to W. P. Schleich, B. W. Schumacher \& W. K. Wootters)} \label{Schleich4} \label{Wootters23.1} \label{Schumacher15}

I've had a little correspondence with Michael Brooks at {\sl New Scientist}, and he told me about how he bought a copy of Wheeler's booklet {\sl Frontiers of Time\/} from a mail order place and was pleased to find a little inscription by John in it.  It is dedicated to a ``Ruth'' and dated 3 May 1986.  See attached photo.

The question is, who is the mysterious Ruth?  Do any of you have an idea?

\subsection{Wolfgang's Reply}

\bq
Many thanks for your email concerning the mysterious ``Ruth''. This is indeed an intriguing problem. I did call up one of Wheeler's old secretaries, Emily Benett. She did have some interesting information but not concerning ``Ruth''. She had no idea who she might have been.

However, from the date of that dedication, May 1986, I got some ideas.  First of all, the dedication was written during the time when I was in Austin. I remember that exactly at this time John had his heart operation. From the dedication it sounds as if it was a nurse who took care of him in the hospital in Houston. This theory would also explain why the book was resold. Anybody in physics would have recognized the value of such a dedication and would have not sold the book.

I have tried to get in touch with one of his daughters, Allison, but she did not pick up the phone. Most likely they are in Maine on the island right now.

By the way, the only Ruth that I associate with America is Dr.\ Ruth Westheimer. But I seriously doubt that John would have written such a dedication to her despite the fact that she seems to be very much into ``creating islands of time''. I wonder if she is still alive.
\eq

\subsection{Bill's Reply}

\bq
Adrienne suggested that Ruth might be his secretary at High Island.
He certainly had a longtime secretary there, and I see that Ruth Bentley is thanked along with his Austin secretary Zelda Davis in the preface of the book {\sl Quantum Theory and Measurement}.
\eq

\section{15-07-08 \ \ {\it Paris on the Way to Munich} \ \ (to H. C. von Baeyer)} \label{Baeyer39}

Thank you very much.  I reduced the font and the line spacing, and I printed it [your manuscript on the Parisian parks through the eyes of a physicist]!  Now I have something to read on our flight to Munich Tuesday.  (The family and are staying with Kiki's parents for 18 days \ldots\ though I'm taking a desk at the Max Planck in Garching and also slipping off to Freiburg for three days.)  I will stay conscious of the absence of Lili's drawings, and be surprised again when I see the final product.

This morning, I found a nice new representation of density operators in terms of SICs.  It's embarrassingly simple, but somehow I hadn't noticed it before.  {\Appleby}, {\Schack}, and I are working away to try to derive the structure of quantum state space almost solely from the ``urungleichung''.  I think I may have told you about the hope while at the philosophy meeting last month.  Everyday, the problem yields a little more.  But unfortunately so far, little really means little (it is no hyperbole).  Thank God for the integral!

I keep promising more:  Eventually you'll get it.

\section{16-07-08 \ \ {\it Title/Abstract for FPP-5} \ \ (to A. Y. Khrennikov \& G. Adenier)} \label{Khrennikov24} \label{Adenier1}

In case I didn't already send you a title and abstract for FPP-5, let me send you the one below.  It's the tweak of a tweak of an earlier abstract (which I may or may not have sent you), and I like it better.

\bq
\noindent Charting the Shape of Hilbert Space, or SICkening Axioms of Quantum Mechanics\medskip

\noindent Abstract: As physicists, we have become accustomed to the idea that a theory's content is always most transparent when written in coordinate-free language.  But sometimes the choice of a good coordinate system is very useful for settling deep conceptual issues.  This is particularly so for an information-oriented or Bayesian approach to quantum foundations:  One good coordinate system may (eventually!)\ be worth more than a hundred blue-in-the-face arguments.  This talk will motivate and chronicle the search for one such class of coordinate systems---the so-called Symmetric Informationally Complete measurements---which has caught the attention of a handful of us at PI, a handful of our visitors, and a handful of other colleagues around the world.  Finally, I will turn the tables and discuss how one might hope to get the formal content of quantum mechanics out of the very existence of such a coordinate system.
\eq

\section{18-07-08 \ \ {\it Is Probability 1 Probably Certain?}\ \ \ (to N. D. {\Mermin})} \label{Mermin137}

\bdm
I thought as authorities on $P=1$ you would like the exchange reported by the AP in this morning's Ithaca Journal.
\bv\rm
Congressman Gallegly R-Calif:\smallskip\\
\hspace{.2in}      Had we not [tortured people] would the probability of another\\
\hspace{.2in}      attack not only be a probability but a certainty?\smallskip\\
Former Attorney General Ashcroft\smallskip\\
\hspace{.2in}      It could well have been.
\ev
\edm

Funny to hear from you today:  Marcus {\Appleby} and I were just having a conversation about you last night!  And subject was {\Mermin}, subjectivity, and his dual attractions and distastes of the concept.

But that was a break from other matters.  We've been making great progress this week in understanding the ``shape of Hilbert space.''  By that I mean, what density operator space looks like when represented as a subset of the probability simplex.  There is serious hope that we can reduce everything to the satisfaction of a single inequality between pairs of probability distributions.  This is a program of study {\Ruediger} and I started a couple of months back and it has continued to surprise us evermore \ldots\ by looking ever more right.  And the inequality, where does it come from?  A direct (but quantitative) expression of the idea that ``unperformed measurements have no outcomes.''  Why the particular quantification (where do the constants come from), I don't know:  But it all seems to hang together.

\section{18-07-08 \ \ {\it ``Experience'', with quotes and without}\ \ \ (to H. J. Bernstein)} \label{Bernstein10}

\bhbe
OK so how come you are not a Whiteheadian? Those throbs of experience which constitute Whiteheadian events (at least as Shimon Malin describes
them) are the closest idea I have ever encountered to your famous ``Zing!''---attributes of the interaction of quantum stuff with real world observers addicted to classical answers through the force of ``white man's'' European post Renaissance history, power, hegemony etc etc.

love out of the blue, herb B
\ehbe

You brightened my evening Wednesday.  ``Throbs of experience''---I love the phrase, and hadn't made note of it before.  It's a deep phrase, isn't it?  The quantum measurement mysteries teach us that ``experience'' can't be ignored as a component of things.  But the lessons of the Copernican revolution also teach us that we shouldn't hoard ``experience'' only for ourselves---``ourselves'' being the agents of which the present quantum formalism takes as the localized centers of the universe.

Who knows, maybe I am a Whiteheadian.  But I am so slow.  Whitehead himself (as he explains at the beginning of {\sl Process and Reality\/}) takes his starting point in William James and Henri Bergson.  That's where I myself am at the moment, but I also recognize that it's only the beginning of a road.

I'm glad you came out of the blue.

\section{21-07-08 \ \ {\it Strawson Quote?}\ \ \ (to myself)} \label{FuchsC20}

From Huw Price at breakfast in Sydney:
\bv
``If you prise off language from the world, you prise off the facts, but you don't prise off the objects.'' --- (or something like that), P. F. Strawson.\footnote{The actual quote is, ``If you prise the statements off the world you prise the facts off it too; but the world would be none the poorer.  (You don't also prise off the world what the statements are about --- for
this you would need a different kind of lever.)'' P. F. Strawson, ``Truth,'' Proceedings of the Aristotelian Society (1950).}
\ev

Also look at Ryle --  ``The Concept of Mind'' -- relevant to the ``a implies b'' debate.

\section{28-07-08 \ \ {\it Small Calculation}\ \ \ (to J. I. Cirac)} \label{Cirac1}

I think the attached partially answers your question.  This is at least one sense in which randomly chosen states are always different from SIC states, even asymptotically.

I enjoyed the style of the questions you were asking today.  It was a good dose for me because I do need to think about these things more. \medskip

\bq
\begin{center}
{\bf \large $d^2$ Random States}\bigskip
\end{center}

Recall the function $K$ I defined in the talk (now specialized to pure states):
\be
K=\sum_{i\ne j}|\langle \psi_i|\psi_j\rangle|^4\;.
\ee
One has that the states $|\psi_i\rangle$ form a SIC if and only if
\be
K=\frac{d^2(d-1)}{d+1}\;.
\ee
This was the value I exhibited today; I'll leave the proof aside.

Let us compare this quantity with the value that obtains on average if the $d^2$ states are not SIC, but instead pulled out of a hat according to the unitarily invariant measure.  The key ingredient is the value of the integral (which I looked up in one of my papers, and is of course much older and well known):
\be
\int |\langle\psi|\phi\rangle|^4 d\Omega_\phi =\frac{2}{d(d+1)}\;.
\ee
Then
\begin{eqnarray}
\langle K\rangle &=& \int\int\sum_{i\ne j}|\langle \psi_i|\psi_j\rangle|^4 d\Omega_{\psi_i}d\Omega_{\psi_j}\\
&=&
(d^4-d^2)\int |\langle\psi|\phi\rangle|^4 d\Omega_\phi\\
&=&
2d(d-1)\;.
\end{eqnarray}

The point is, asymptotically min $K$ goes as $d^2$, whereas $\langle K\rangle$ goes as $2d^2$.
\eq

\section{30-07-08 \ \ {\it The Lost Pauli Letter}\ \ \ (to H. Atmanspacher)} \label{Atmanspacher9}

Just so I don't forget:  Please send me the coordinates of that letter from Pauli to Jung that you told me about today.  (The one that Jung published somewhere, but did not appear in the Pauli--Jung correspondence.)

Also, don't forget about the letters to Fierz that you were going to select:  Ones on neutral language and neutral monism.

\section{30-07-08 \ \ {\it Seeking Peirce?}\ \ \ (to R. W. {\Spekkens})} \label{Spekkens50}

\brws
The collected papers of Charles Sanders Peirce, vols.\ I -- VI.  This doesn't happen to be something that you desperately need, does it?
They're selling here second hand for 45 GBP for the lot.
\erws

That shows how much Peirce is valued in the UK!  If they're in good shape, and it is really {\it only\/} 45 GBP for the whole lot (rather than the individuals), then {\it indeed\/} get them!!!  I would repay you in money, kindness, and brotherly love.

I just looked in my records:  Vol 8 alone of the collected papers of Bertrand Russell cost me 20 GBP!  (The only reason I bought that one is because he has correspondence in that one concerning pragmatism.)

You are a saint for bringing this to my attention.

\section{03-08-08 \ \ {\it European Tour} \ \ (to H. C. von Baeyer)} \label{Baeyer40}

Good morning.  I'm writing you as I take a morning train from Freiburg to Munich, having spent two days with Harald Atmanspacher talking of Pauli, neutral monism, and panpsychism.  Actually, we started at his very pleasant home in the beautiful Swiss village of Amden (over Lake Walensee).

Finally I get a chance to write you properly.  As I had promised, your {\sl Swing Under the Eiffel Tower\/} was a companion to me on my flight over.  I very much enjoyed it and got a better sense of some of the things I will see and do when Emma and I are in Paris next Spring.  (I don't know that I've ever shown you pictures of Emma, by the way.  Attached are two:  One at the piano just before our leaving Canada, and one taken her first morning in Munich!)  I came away from my reading with a sense of envy for your years in Paris, and envying your ability to look at things with a physicist's eye.  I think you pulled it off well in your text, mingling these two aspects of your life, neither of which I really have.  Frankly, or I should say honestly, I learned some physics from it.  Too bad I couldn't have seen Lili's illustrations at the same time as my first reading.

If I didn't think it would only cause more trouble in our relationship, I would send the first two paragraphs in your Introduction to my mother-in-law:
\bq
With this book I return to a question that has been rattling around in the lumber room of my brain for thirty years.  Although I have rarely asked it explicitly, the answer informs all of my writing, which the French, with barbed Gallic poignancy, would characterize as {\it vulgarisation de la science}.  The question is this: How does a physicist see the world?  As all of us go about our daily tasks, our perceptions of what we see and feel are molded by our professional knowledge in ways we are hardly aware of.  Artists, for example, see things in terms of colors and shapes, engineers notice structure, biologists pay attention to plants and animal, and musicians to the cacophony of the soundscape they are immersed in.  The interest of mathematicians is drawn to numerical and geometric patterns, and that of historians to the traces of the past that are encoded in the way things are arranged in the present.  It is said that dogs experience their surroundings as a panorama of smells --- in the same way people see primarily that to which they are most sensitive.  Accordingly, we physicists perceive the world in the light of the laws of nature that are indelibly etched in our brains.  What exactly does that imply about our relationship to the universe?  How does it shape our perceptions?  In what way do physicists think differently from other people?  Many of our non-scientific friends would like to understand what makes us tick, but most of them haven't a clue.

I will try to describe some of the images and associations that flit through my mind as I wander through my daily life.  I hasten to admit, though, that generally I am not thinking about physics at all, but about the things everybody else is thinking about too.  Today's grocery list, my daughter's dental appointment, the news from Iraq, and the itch on the back of my neck jostle each other in the mundane meanderings of my thoughts --- normal human concerns.  So my world is infused with meaning not so much by what I {\it think\/} as by what I {\it know\/} about physics.  Newton's laws and Maxwell's equations go with me wherever I go; though they are normally hidden from view, they are powerful in their influence on my world view.  I don't think about physics most of the time, but I am aware of it throughout my waking hours.  Physics provides the spectacles, as it were, through which I see the world.
\eq
They are perfect.  She tells me over and over how I need to learn to relax and enjoy life---which of course only means that I do not walk through life in the same way she does.  I never think of myself as particularly unhappy---certainly at least when I'm traveling away from the construction at my house!---{\it except\/} when I'm being told how unhappy I am.  (I unfortunately really blew up at her Sunday morning when she pointed out to me for the hundredth time in only five days how much of life I'm missing.  How much of life I'm missing?!)  When she sees me as wanting some quiet time to think, or to write, she sees no relaxation in it at all.  When she sees me pull out a pad to sketch an equation, then of course that must mean miserable work.  \ldots\ But I digress.

I'll record some things from your draft that caught my eye while reading.

{\bf Les Tuileries}.  This was my favorite chapter.  It struck my sensibilities, I suppose, because of my meanderings in the last two years into the issue of Wigner's friend.  Einstein taught us that we each ``carry'' our own proper time.  What does quantum mechanics teach us?  In part, I think it teaches us to recognize in a concrete way that we each carry our own ``experience''.  Each ``I'' is the one who has it; no one else does.  The lesson is that there is a public part to the world, but there is also a private part to the world.  You say, ``[R]elative to the park, I am indeed in motion.  But relative to myself, which is to say measured in a reference frame that is attached to my own body, and has its origin at my navel, I am always securely at rest, while the park recedes backward at walking speed.  Both these claims, that I am moving relative to one reference frame and stationary relative to another, are correct \ldots''  In quantum mechanics, something like this goes too, though in the opposite direction for its particulars:  It's in the agent's own frame where the ``motion'' so to speak occurs---for that's where the experience is that quantum mechanics gives us the tools for gambling upon.

Some things to read in this regard, in case you're interested.  First pick up the latest version of {\sl My Struggles with the Block Universe}.
Thereafter, do a search on
\begin{enumerate}
\item
``Wigner's friend''
\item
``Rivers of Nows''
\item
``Wootters' Preply'' (that's actually a note Bill wrote, and probably played some inspiration for my own thoughts)
\item
and finally have a look at the quotes of William James in the frontispiece of the document.
\end{enumerate}

In this chapter of yours, I was reminded of how much I would have liked you to have heard Rovelli's talk in Paris.  I thought it was very deep, and certainly within reach of (both from and to) the cluster of ideas mentioned above.  I think he misses a lot by trying so hard to never refer to agents and states of belief, but on the other hand, he gives hints of having grasped the personal nature of ``experience'' (he calls it the ``relativity of facts'') in a deeper fashion than I had hitherto.

{\bf Parc Monceau}.  I found your challenge in this chapter inspiring:
\bq
     [T]hings can't go on this way.  In 1926, when Erwin {\Schroedinger} was inventing his version of the quantum theory, he wrote to a friend: `Physics does not consist only of atomic research, science does not consist only of physics, and life does not consist only of science.  The aim of atomic research is to fit our empirical knowledge concerning it into our other thinking.'  We have ignored that wisdom, and remained locked in our secret garden.  It's time we invited the people in.  I am confident that in the twenty-first century we will finally heed {\Schroedinger}'s advice, and bring quantum mechanics into the world picture of ordinary people. Creating a more accurate and suggestive atomic logo would be a good start! \ldots

     If I were as rich as the Duke of Chartres, I would sponsor an international competition for the design of a modern atomic logo.  The jury would include writers, painters, sculptors, physicists, chemists, and teachers. Somewhere, I feel, there must be a mind creative enough to come close to translating the well-understood mathematical language of quantum mechanics into visual terms \ldots
\eq

I will be your publicist, as well as taking my own shots at entry!  First though, a few thoughts, as the challenge simmers in my brain.  It's not quite a public slogan yet, or an image that can be drawn on paper, but it is starting to strike me that perhaps the driving idea behind quantum mechanics might just come down to a single inequality.  {\Schack}, {\Appleby}, and I have been calling it the ``urungleichung''.  And it, to my mind, finally starts to give an image to Hilbert space.  It is something like a very symmetric spider web stretched over a sphere.

Let me tell you a little bit about it for the fun of it (and for practice).  The story goes back to three transparencies I used in Paris last month.  Why should I believe Asher Peres's phrase, ``Unperformed measurements have no outcomes''?  And how might I turn it into a quantitative statement?  Until recently, I believed that the real essence of Asher's slogan was in the Kochen--Specker theorem and, better, a variation of Allen Stairs' use of it (which Carl, {\Ruediger}, and I review in our last paper together).  But now I hope/believe it can be traced back to a deeper level, or at least to a deeper quantitative statement in overtly Bayesian terms.

In a separate email, I'll send you scans of three of my transparencies, and a little \LaTeX\ document besides (some working notes from my collaboration with {\Ruediger} and Marcus).  The essence of the story is in the diagram.  It asks us to compare an apple and an orange.  ``What probability we should assign for the outcomes of the lower right measurement device {\it if\/} we're going to take our quantum system directly to it.''  --Versus--  ``What probability we should assign for the outcomes of the lower right measurement device {\it if\/} we first send the quantum system through the measurement device in the sky.''  Apple versus orange.  Well, we know how to calculate the orange in terms of the given data in the diagram.  We make use of the ``law of total probability'', which has a perfectly sound Bayesian derivation in terms of de Finettian coherence for {\it conditional bets}.  But how to calculate the apple?  There are no conditional bets in this scenario.  A Bayesian not steeped in quantum mechanics would say, ``Well then, there's no reason to be probabilistically coherent with any conditional betting.  The probabilities in apple are stand-alone entities; they are priors in the proper sense, tied logically to nothing else.''  That ``apple DOES NOT EQUAL orange'' is a raw statement that unperformed measurements have no outcomes---unperformed SIC measurements at least.  This, any Bayesian who has understood de Finetti's Dutch-book derivation of the connection between conditional and joint probabilities should appreciate.

But quantum mechanics says something that surprises even the well-prepared Bayesian.  Though it certainly re-enforces the raw statement above, in the end it {\it does\/} let us compare the apple to the orange:  For there is a precise relation between the two.  Not equality; but seemingly a simple linear relation.  Who ordered that?!?!?1

Well, I do not know.  But that simple relation seems to contain within it much, if not all, of the formal structure of quantum mechanics.  Particularly, by demanding that apple be between 0 and 1, one is forced to constraints on the probabilities for the outcomes of the SIC-in-the-sky measurement.  We do not yet know whether the resulting space of allowed probabilities is completely equivalent to the standard quantum mechanical set, but it certainly shares many features with it.  And we are guardedly optimistic that it will eventually yield the whole structure.  The Introduction to the {\LaTeX}ed notes defines the problem more precisely than I have done so here and lays out the tools for getting to the image I spoke of previously.

Coming back to your challenge and the diaphanous web.  We start by constructing the extreme points of the convex set it will specify.  The inequality says that they will lie on the surface of a $d^2$ dimensional sphere of a specific radius.  Now, we get the rest of the state-space structure by removing points from the sphere.  Go to any valid point, the urungleichung tells us, and its antipode will not be in the set.  Moreover, a spherical region surrounding the antipode must go too:  It must be excised from the original sphere.  The result of all this pruning, if our hope turns out to be correct, is a diaphanous manifold of dimension $2(d-1)$ living at the intersection of all these excisions.

Maybe not an emotive image yet.  But one that, in some deep way, carries the meaning of quantum mechanics.

But going back to your chapter:  Is the phrase ``put paid'' a typo?  Or rather an English phrase that I do not know?

{\bf Parc Montsouris}.  This passage struck me:
\bq
     Great revolutionaries don't stop at half measures if they can go all the way.  For Newton this meant an almost unimaginable widening of the scope of his new-found law.  Not only Earth, Sun, and planets attract objects in their vicinity, he conjectured, but all objects, no matter how large or small, attract all other objects, no matter how far distant.  It was a proposition of almost reckless boldness, and it changed the way we perceive the world.
\eq
For you are right:  That was indeed an amazingly bold move---and I had not appreciated that before.  Put in that light, one looks at Newton in awe.

{\bf Jardin du Luxembourg}.  Of course, I was flattered to find my name in this chapter.  It teaches me it is good to hobnob with good writers!  But how does one translate ``Chris Fuchs (rhymes with books)'' into French?!  On my trip to Amden from Munich the other day, I passed through a town Buchs, and was quite surprised when the conductor announced it (just before I had looked out the window).  Finally I learned another German word rhyming with my name!

Contentwise, I liked the paragraph on Zeilinger's phrase ``the two freedoms''.  For I think I agree at some level that the secret of quantum mechanics is to be found in the tension between the two freedoms.  I have certainly said things along those lines too at selected times in my life.  (Where, by the way, does Zeilinger write of the ``two freedoms''; I'd like to get the exact flavors of how he says it.)  But also, I think the formulation leaves something out.  Certainly at least in my own previous versions.  For I imagine there is something truly autonomous in the outcomes of quantum measurement.  The agent and the system supply the raw materials so to speak (like parents), but the event thereafter is a life of its own.  This is why I have been moved at times to call this view the ``sexual interpretation'' of quantum theory.  (Speaking of flavors, you might want to do a search on that phrase too in link.)

{\bf Parc des Buttes Chaumont}.  It's a funny thing, these multiverse and literal-landscape worldviews.  At the surface, the arguments behind them must be nonsense---no deeper or more scientific than the story I tell of the priest in the section titled ``\myref{Halvorson5}{Cash Value}'' (17-11-05 note to H. Halvorson).  At the level these guys speak of, we experience one world, but, the argument goes, this world can only make sense (or even exist at all) if it is literally one among many.  Yet so many of our colleagues are ready to accept such a statement out of hand, feeling that they are driven to it by science itself.

{\bf Parc George Brassens}.  There are two typos in the penultimate paragraph:  ``fields are the as essential'' and ``power of the basic idea not been compromised''.

I hope you are enjoying your summer in the parks of Paris!  It wouldn't do to stop your time in them now that your book is finished!

\subsection{Hans's Reply}

\bq
``Fuchs se prononce comme foucs.''  Unfortunately fou means crazy.
\eq

\section{11-08-08 \ \ {\it Too Exhausted}\ \ \ (to K. Martin)} \label{Martin3}

\bkma
By the time I convinced the security guy at PI
to give me a key to my room, I was also tired \ldots

Did you know that in the room I am in they have some
books under the tv for casual reading -- one of them
is a book that you wrote!
\ekma

No, I did not know that.  I had always wondered what happened to the two copies I sent to PI years ago.  I have noted that they are not in the library catalog.  But it might just be someone's personal copy that they didn't care to keep.  There's not a dedication on the first or second page to someone, is there?

\section{11-08-08 \ \ {\it The Two Freedoms} \ \ (to H. C. von Baeyer)} \label{Baeyer41}

That paper?  Actually I'm quite familiar with that one:  I liked it very, very much.  Here's what I have quoted in my Pauli'an Resource file about it:
\bq
\indent
{\bf From:}  A.~Zeilinger, ``On the Interpretation and Philosophical Foundation of Quantum Mechanics,'' in {\sl Vasta\-koh\-tien todellisuus, Juhlakirja Professori K.~V. Lau\-rikaisen 80-vuotisp\"aiv\"an\"a}, edited by U.~Ketvel, A.~Airola, R.~H\"am\"a\-l\"ai\-nen, S.~Laurema, S.~Liukkonen, K.~Rainio and J.~Rastas (Helsinki University Press, Helsinki, 1996), pp.~167--178.\medskip

When investigating various interpretations of quantum mechanics one notices that each interpretation contains an element which escapes a complete and full description. This element is always associated with the stochasticity of the individual event in the quantum measurement process. It appears that the implications of this limit to any description of the world has not been sufficiently appreciated with notable exceptions of, for example, Heisenberg, Pauli and Wheeler. If we assume that a deeper foundation of quantum mechanics is possible, the question arises which features such a philosophical foundation might have. It is suggested that the objective randomness of the individual quantum event is a necessity of a description of the world in view of the significant influence the observer in quantum mechanics has. It is also suggested that the austerity of the Copenhagen interpretation should serve as a guiding principle in a search for deeper understanding.
\eq
But I too don't recall a discussion of the ``two freedoms'' there.  Could it instead have been in the little one page thing he published in {\sl Nature\/} a couple of years ago?

\subsection{Hans's Reply}

\bq
In my frustration I googled ``the two freedoms zeilinger'' and the first entry hit paydirt:
\begin{center}
\myurl{http://www.signandsight.com/features/614.html}.
\end{center}
\eq

\section{11-08-08 \ \ {\it The Two Freedoms Found} \ \ (to H. C. von Baeyer)} \label{Baeyer42}

Cool!  Our emails just crossed each other in mid air!  Synchronicity!  Synchronicity!

I'm not lying your note arrived just as I had hit ``send'' on mine!

\section{13-08-08 \ \ {\it Reflection} \ \ (to V. Hardy)} \label{HardyV1}

At the end of our conversation outside the restaurant last night, I sensed that I had crossed the line of acceptable teasing.  However, even if the line would not have been crossed, I find this morning that I very much regret the things I said.  Birth is to be respected, not feared, and my behavior was no better than a schoolboy's in a playground.  In honesty---without the teasing now---Dina's assessment strikes me as about right.  But more importantly, and the thing I think one should keep in mind through these months up to the moment, is the flipside.  I remember how Charlie Bennett's wife, Theo, put it once:  ``The births of my three children were the three most intense moments of my life.  I was, and have never been since, so alive.''  It is something men never feel, except in the abstract, and in this way, I know that women's lives are more complete than ours can ever hope to be.

Just wanted to say that \ldots

\section{13-08-08 \ \ {\it The Train}\ \ \ (to K. Martin)} \label{Martin4}

Your note did take me back to the South, and brought to mind those Hank Williams words:
\bv
   Hear that lonesome whippoorwill\\
   He sounds too blue to fly\\
   The midnight train is whining low\\
   I'm so lonesome I could cry
\ev
Notes for my (some future day) write-up of the Christian-Snyder morning coin toss story are below. [See 14-12-07 note ``\myref{Snyder5}{No Subject}'' to C. Snyder.]

\section{14-08-08 \ \ {\it Physical Theories as Women} \ \ (to G. L. Comer)} \label{Comer118}

If you haven't seen this before, have a look:  ``Physical Theories as Women,'' by Simon Dedeo, \myurl{http://www.mcsweeneys.net/links/lists/physical.html}.  I love the description of quantum mechanics particularly:
\bq\noindent
Quantum mechanics is the girl you meet at the poetry reading. Everyone thinks she's really interesting and people you don't know are obsessed about her. You go out. It turns out that she's pretty complicated and has some issues. Later, after you've broken up, you wonder if her aura of mystery is actually just confusion.
\eq
No wonder I've never been able to get her off my mind.

\section{14-08-08 \ \ {\it Possible Collaboration?}\ \ \ (to T. Duncan)} \label{Duncan4}

\btd
I've been taking some time this summer to reflect \ldots\ kind of connecting the dots to see the theme that drives and underlies my work, so that the next phase will stay on track and help develop this core theme further. I've started using the phrase ``Looking for meaning in the modern scientific universe'' to describe what I see myself doing. Everything I find myself working on ties somehow to the question, ``Is there a perspective or world view one can hold which satisfies the longing for meaning/purpose to our lives that we feel, which is also accurate as far as we can tell (i.e.\ consistent with all of our current scientific understanding of the world)?'' For some reason I still feel rather squeamish about revealing this theme for my research to other physicists \ldots\ afraid I won't be understood or won't be taken seriously, perhaps? But from some of our conversations, and rereading your collection of papers and notes, I suspect it's a line of thinking you might share or at least be sympathetic to.  Anyway I may have an opportunity to build a more formal collaboration around this theme, so I wanted to share a bit of what I'm thinking and see if you'd like to be involved in the discussions or collaboration in some way.

A few months ago I submitted a proposal for a research program and book on the theme of ``looking for meaning in the modern universe'' to the U of Chicago Arete Wisdom Initiative. In the end my proposal wasn't funded, but I was told by the review panel that it was important work and my proposal made it quite far in the process. They encouraged me to submit the proposal directly to the Templeton Foundation. So I'm in the process of reworking and possibly broadening the proposal to submit to Templeton.  I've attached a very rough draft of the letter of intent.  Please take a look when you have a chance and let me know if you have thoughts, and if it looks like something you'd want to be involved in.
\etd

Thanks.  I'll look at the proposal, and try to write you with any thoughts soon.  There's not a chance you'll be at the {\Vaxjo} meeting at the end of next week, is there? \ldots

Actually this is very interesting.  Plus I have always been impressed by your thoughts in this regard.  Let me think for a few days about how I might fit in.  What were the sorts of things you were envisioning for my role?

\section{15-08-08 \ \ {\it Free Will and Renouvier}\ \ \ (to R. {\Schack})} \label{Schack137}

Attached is the latest version of my quote collection.  The thing to do is just search on the term Renouvier (just to make sure you miss nothing).  But in particular, that will eventually take you to page 101, Logue's discussion.  Then, starting on page 105, you will find Long's discussion.  Funny coincidence that they are back to back.  Then there is some discussion by Menand on pages 111 and 112.  You can find a list of Renouvier's original works on page 158.

Now about Free Will more broadly, you can do a search on that term too.  I myself like James's discussions on pages 67--68 and 78--79.

Hope this helps.\medskip

\noindent {\bf W. Logue, {\sl Charles Renouvier:\ Philosopher of Liberty}, (Louisiana
State University Press, Baton Rouge, 1993):}

\bq
\indent
The discussion of political freedom and human rights had become a
major theme in Western thought only beginning in the seventeenth
century, while the theological and philosophical problem of the
freedom of the will, whether limited to the question of salvation or
more generally understood, had a long history in Western thought, had
indeed been one of the perennial big questions of
philosophy.\footnote{{\sl Deuxieme essai}, xiv--xv. Free will was a
problem in religion before becoming one in philosophy but a problem
most Christians avoided simply by believing in free will and
predestination simultaneously. Greek philosophy, the first to give
this question nonreligious consideration, was inclined to
psychological determinism, though there were, Renouvier believed,
some defenders of freedom---Aristotle, Pyrrhus, Epicurus ({\sl
Esquisse}, I, 227--28, 238, 240--43).} The question of the relations
between this freedom of the will and the more concrete issues of
political freedom and individual rights had rarely been examined, if
only because political and individual freedom were rarely issues
before modern times. But even when they became issues, few saw any
need to connect these practical questions with the abstruse debates
of philosophy.

Unreflective opinion had two choices: Either there was no connection
between free will and public freedom, or there was some direct but
not very clear connection between the two. The former attitude,
common from the beginning of the nineteenth century, enabled thinkers
to be simultaneously defenders of political freedom and exponents of
deterministic and materialistic philosophies that denied free will.
This was a secular counterpart of the traditional Christian attitude
that emphasized free will for reasons of practical morality but could
not reconcile it with the theological determinism of an all-powerful,
all-knowing God.\footnote{Esquisse, I, 249--50; see also I, 248--51.}
Few thought that there was any serious conflict, and still fewer
imagined that the diffusion of deterministic philosophies would serve
to undermine the cause of political and individual liberty.

Renouvier was, I think, the first, at least the first major
philosopher, to see clearly into this danger and to devote himself to
drawing attention to it while proposing a remedy. The development of
his neocriticist philosophy was itself a response to what he had come
to see as the anti-liberty forces dominant in nineteenth-century
thought. Under the influences of discussions with his friend Jules
Lequier, he became convinced of the reality of human free will and
its central importance for the understanding of everything
else.\footnote{Renouvier's account of his ``conversion'' to free will
is in the last part of Vol.\ II of the {\sl Esquisse}. Lionel Dauriac
(``Les Moments de la philosophie de Charles Renouvier,'' {\sl
Bulletin de la soci\'et\'e francaise de philosophie}, IV [1904], 23)
defined the high point of Lequier's influence---the writing of the
{\sl Deuxi\`eme essai}---as one of four ``moments'' in Renouvier's
philosophical development.} This conviction came to Renouvier while
he was still deeply under the influence of his first contact with the
Saint-Simonians. He experienced, not an overnight liberation from
their deterministic viewpoint, but a more gradual readjustment of his
views, which became a complete detachment from them only after 1851.
Perhaps the failure of the socialist movements in 1848, rooted as
they were in the would-be scientific philosophies of the preceding
three decades, finally persuaded him of the dangers of rejecting free
will.\footnote{Mouy ({\sl Id\'ee de progr\`es}, 43) argues that the
disappointments of 1848 played a key role in shaping Renouvier's idea
of liberty. For Renouvier, free will came to be seen as the ultimate
basis of political liberty ({\sl Deuxi\`eme essai}, 551). See {\sl
Histoire}, IV, 431, and especially {\sl Esquisse}, II, 382, and {\sl
Deuxi\`eme essai}, 371{\it n}1.} Alienated from political life during
the Second Empire, he would spend nearly two decades in the
construction and elaboration of his philosophy of liberty,
establishing its foundations and exploring its consequences.

Renouvier was aware that for a long time the question had been of
mainly religious significance: whether man's salvation depended on
free will or on predestination.\footnote{Renouvier saw free will as
one of the basic concepts of both philosophy and Christian doctrine
({\sl Histoire}, IV, 277).} This debate had reached its peak, in both
vehemence and subtlety, in the famous exchange between Erasmus and
Luther in the sixteenth century. The emergence of a secular debate
over free will was a result of the rise of the scientific worldview
in the seventeenth century. The ascendancy of the idea that the world
was governed by invariable laws, taking the role previously occupied
by an all-powerful, all-knowing God, seemed to leave less and less
room for the view that man was somehow an exception to the general
rule. The most heroic task for the modern philosopher was to find a
means of validating science and free will simultaneously, and the
most heroic effort of the eighteenth century was that of Immanuel
Kant. But for many in the next century, it seemed that Kant had saved
free will only at the cost of making it irrelevant.\footnote{What
does it matter to man if he has freedom in the world of the noumena
if his world of phenomena is entirely determined? Renouvier later
felt that Kant held on to free will solely for the sake of morals
while not really believing in it ({\sl Quatri\`eme essai}, 35--36).}
Fichte tried to rescue Kantian philosophy from this unhappy outcome,
but in the general opinion \ldots\ his effort led to the fairyland of
absolute idealism, denying reality to the material
world.\footnote{Renouvier praised Fichte as a defender of freedom and
criticized him as a mystic ({\sl Quatri\`eme essai}, 46).}

Against the rising tide of determinism, Renouvier would try to show
that Kant could be the launching pad for a defense of free will that
would maintain its practical relevance and demonstrate its
compatibility with natural science, properly understood. He did not
claim to be presenting any new arguments in favor of free will; he
felt they were in any case unnecessary.\footnote{Renouvier was
concerned to establish a rationalist and not an empiricist view of
science. He saw free will as perhaps the main issue dividing the
rationalists and empiricists ({\sl Histoire}, IV, 262). He indicated
that there had been no new arguments in favor of free will since Kant
and Rousseau ({\sl Esquisse}, I, 280).} Renouvier's reasons for
coming to the defense of free will were partly shared with Kant and
partly his own. As we have seen in the previous chapter, the shared
part was the most familiar: a concern for the connection between free
will and moral behavior. Free will was for Kant the essential basis
of {\it practical reason}; without it, the whole idea of moral
obligation ceased to have meaning. For Renouvier, this consideration
remained central. Without moral responsibility, man would not be
distinct from the rest of the animal kingdom, and the whole of
civilization would be meaningless. But this was not the sole basis
for his concern with free will, and this additional concern moved
Renouvier beyond Kant and Fichte, bringing him closer to our own
time.\footnote{Renouvier saw Kant's German disciples as having
abandoned liberty for determinism, optimism, and pantheism ({\sl
Histoire}, IV, 467).}

It is not just the moral aspect of civilization that hangs on the
reality of free will, in Renouvier's opinion, but the whole of our
intellectual life. Free will is also the foundation on which
philosophy and the natural sciences rest.\footnote{See {\sl
Deuxi\`eme essai}, 227.} Without free will, our ability to know
anything, whether about man or about nature, is fatally undermined.
Scientists do not need to believe in free will, and as he knew, they
prefer to avoid this sort of question. In practice, they can
legitimately do so because in their narrow spheres of inquiry they
have developed techniques of investigation that work even when the
scientist is unconscious of the fundamental assumptions on which his
method rests. But without free will, the certainty of scientific
truths becomes illusory; a consistent determinism must lead to a
profound skepticism.\footnote{{\sl Histoire}, IV, 399; {\sl
Deuxi\`eme essai}, 327.} Renouvier would never despair of convincing
the scientists that just as our concepts of right and wrong depend on
free will, so do our concepts of true and false. Indeed, without free
will, we could not even talk sensibly about things being true or
false.

If, as he pointed out, I hold such and such a view to be true and I
am determined by forces outside my control to hold this view, the
person who disagrees with me is equally determined by outside forces
in his position. If these mutually contradictory positions are
equally necessary, what grounds can we have for the certainty that
either view is the correct one?\footnote{{\sl Histoire}, IV, 399;
{\sl Deuxi\`eme essai}, 306--307 (according to Hamelin, {\sl
Syst\`eme de Renouvier}, 242). Necessity destroys truth: ``If
everything is necessary, error is necessary just as much as truth is,
and their claims to validity are comparable'' ({\sl Deuxi\`eme
essai}, 327). For a restatement of his argument that freedom is
essential to the certainty of our knowledge, see {\sl Esquisse}, II,
270--74; see also, {\sl Science de la morale}, II, 377.} If our
belief that our ideas are determined is itself determined, so is the
other person's belief in free will determined. Under these conditions
how could it make any sense to speak of one view as ``right'' and the
other as ``wrong''? If, on the other hand, our choices are free, I
may freely choose to believe in free will or in spite of the apparent
contradiction, to believe in universal determinism. Of the four
possible positions revealed by this analysis, the only one that can
serve as a foundation for a rational certainty in the truth of our
beliefs is to freely believe in freedom.\footnote{The four are (1) we
are determined to believe in freedom; (2) we are determined to
believe in determinism; (3) we freely believe in determinism; (4) we
freely believe in freedom. See {\sl Deuxi\`eme essai}, 478; {\sl
Histoire}, IV, 399; Hamelin, {\sl Syst\`eme de Renouvier}, 273--74.}
But as Renouvier insists, this means that we must give up any
pretension to the absolute certainty of our
beliefs.\footnote{``Certitude is not and cannot be an absolute. It
is, as is too often forgotten, a condition and an action of man: not
an action or a condition where he grasps directly that which cannot
be directly grasped---that is to say, facts and laws which are
outside or higher than present experience---but rather where he
places his conscience such as it is and as he supports it. Properly
speaking, there is no certitude; there are only men who are certain''
({\sl Deuxi\`eme essai}, 390). For Renouvier's battle against the
idea of evident truths, see {\sl Histoire}, IV, 75, 261; certitude is
a sort of ``personal contract,'' ``a real contract that a man makes
with himself'' (Lacroix, {\sl Vocation personnelle}, 114).} The truth
of free will cannot be proved so that no rational person can doubt
it. It is a relative truth, like all our other truths, but more
important because it plants a relativism at the very core of our
thought.\footnote{{\sl Deuxi\`eme essai}, 309--10 (according to
Hamelin, {\sl Syst\`eme de Renouvier}, 242). There were, however,
``great probabilities in its [free will's] favor'' ({\sl Deuxi\`eme
essai}, 475). ``It ought to be a universally accepted maxim that {\it
everything that is in the mind is relative to the mind}'' ({\sl
Deuxi\`eme essai}, 390). Philosophy needs to take into account the
existence of disagreement among philosophers ({\it ibid}., 414).
Renouvier's approach to the existence of these disagreements is one
of the distinctive features of his philosophy.}

Scientists, Renouvier thought, should have no difficulty
understanding and accepting this because science is built on an
awareness of the conditional character of our knowledge, an openness
to the discovery of new truths and the abandonment of old
ones.\footnote{On the use of hypotheses in science, see {\sl Premier
essai}, 200.} In fact, he had to admit, many scientists were still
under the sway of older metaphysical conceptions of truth, except in
the conduct of their personal research, and were unaware of any
inconsistency in their position.\footnote{Renouvier credited the
English empiricists, following Hume, with freeing science from the
metaphysical concept of cause ({\sl Histoire}, IV, 273).} Some who
were aware were evidently afraid that to admit that an act of belief
was at the base of scientific knowledge would risk undermining the
claim of science to objectivity and, even worse, open the way to the
proliferation of pseudoscientific beliefs.\footnote{This is the
concern of Parodi ({\sl Du positivisme \`a l'id\'ealisme}, 184--85),
who finds in Renouvier a dangerous fideism. So does Brunschwicg ({\sl
Progr\`es de la conscience}, 625). For Renouvier's praise of
Boutroux's argument that the contingency of the laws of nature is not
a threat to science, see {\sl Histoire}, IV, 673--74.} In reality,
pseudoscientific beliefs were already proliferating under the aegis
of the belief in determinism. Without a critical analysis of the
nature and limits of scientific knowledge, however, our intellectual
life is subject to a constant abuse of the name and prestige of
science.

The abuse of science takes many forms: the application of research
methods to fields where they do not apply, the application of
particular concepts to areas other than those where they originated,
the confusion of ``Science'' with the operations of particular
sciences. One of the main intellectual trends of the nineteenth
century, which Renouvier called {\it scientisme}, usually rendered as
``scientism'' in English, was the product of this abuse. Renouvier's
relativism does not justify believing in whatever we want to
believe.\footnote{Dauriac (``Moments de la philosophie,'' 30--32)
strongly makes this point. It would be interesting to compare
Renouvier's conclusion on this point with the similar view expressed
by Richard Rorty, coming from a rather different direction.} It
insists on submitting our opinions to every possible test of logic,
experiment, and experience. But we have to admit that our logic, the
hypotheses on which our experiments are based, the schemas of thought
by which we interpret our experience, all rest ultimately on acts of
belief and not on absolute certainties.\footnote{{\sl Histoire}, IV,
692.}

If free will is thus essential to both morals and science, just what
does he mean by it? Over the centuries, most of the debate over free
will has failed to advance our understanding because of the lack of
agreement about what is meant by the term. I cannot solve that
problem, but I think we will see that Renouvier's view makes the
issue more comprehensible.

Free will, for Renouvier, is a capacity possessed by human beings,
and only by human beings, that enables them to choose whether to
accept one idea or another, whether to perform one act or a different
one. It is thus a rejection of the doctrine that holds that all
events, mental or physical, are absolutely determined and cannot be
other than what they are.\footnote{See definition of free will in
{\sl Histoire}, IV, 337; on liberty as choice, see {\sl Deuxi\`eme
essai}, 466; on real alternatives, see {\it ibid}., 339. Renouvier is
rejecting a causal necessity, not analytic necessity, as in the
syllogisms of logical operations ({\sl Premier essai}, 232--36).}
Free will is also a rejection of the doctrine of chance, for it is an
active power and not the ``liberty of indifference'' so belabored by
determinists.\footnote{{\sl Deuxi\`eme essai}, 330--34, 336, 337;
Hamelin, {\sl Syst\`eme de Renouvier}, 242--43,249.} Chance is also
hostile to liberty, since it denies man a real power of decision.

The existence of free will requires a measure of indetermination in
the universe but could not exist if nature were essentially
indeterminate.\footnote{``Liberty does not require the complete
indetermination of particular future events, even of those that are
directly connected to it'' ({\sl Deuxi\`eme essai}, 459); see also
{\it ibid}., 357; Hamelin, {\sl Syst\`eme de Renouvier}, 244.} Our
acts of free will are the beginnings of chains of consequences and
would have no meaning if their consequences were not subject to cause
and effect. ``Free acts are not effects without causes; their cause
is man, the ensemble and fullness of his functions. They are not
isolated, but are always closely attached to the preceding condition
of the passions and of knowledge. {\it A posteriori\/} they seem
henceforth indissoluble parts of an order of facts, although a
different order was possible {\it a priori}.''\footnote{{\sl
Deuxi\`eme essai}, 359; see also {\sl Science de la morale}, II,
361--62.} The laws that permit us to say this is followed by that do
not admit of an infinite regression into the past, according to
Renouvier. Therefore, every series of phenomena---and indeed the
existence of any phenomena---must have a beginning that we cannot
explain in terms of antecedents.\footnote{{\sl Premier essai}, 237;
{\sl Science de la morale}, II, 360--61. Most scientists today reject
the idea that infinite regression is an absurdity; {\sl Esquisse},
II, 378--79. We cannot explain beginnings because they are by
definition at the limits of our possible knowledge.}

The act of creation of the universe is thus replicated (in a much
smaller way!)\ in every act of free will. Every act of free will is
the creation of a new series of phenomena, a series that would not
otherwise have existed.\footnote{{\sl Esquisse}, II, 196--97.} These
new chains of cause and effect are not simply the product of the
intersection of existing but independent series, as A.-A. Cournot
argued, for such intersections, though they appear random from the
point of view of any one of the colliding series, would be necessary
from a higher viewpoint.\footnote{A.-A. Cournot, {\sl
Consid\'erations sur la marche des id\'ees et des \'ev\'enements dans
le temps modernes\/} (Paris, 1973), 9--10, Vol.\ IV of Cournot, {\sl
Oeuvres compl\`etes}, ed.\ Andr\'e Robinet.} They must be new
beginnings, arising from a conjuncture in which, given the
antecedents, more than one consequence was possible: ``ambiguous
futures,'' Renouvier called them.\footnote{{\sl Deuxi\`eme essai},
210. ``The real indetermination of various phenomena envisaged in the
future'' ({\sl Premier essai}, 240). ``[A determinist] would renounce
everything called reflection and reason, for these functions do not
work without the consciousness of a {\it representative
self-motivation}, which is itself linked to an awareness of the {\it
real ambiguity of future conditions\/} before it takes action'' ({\sl
Histoire}, IV, 769). See also {\sl Troisi\`eme essai}, xlvii;
Hamelin, {\sl Syst\`eme de Renouvier}, 230.} Free will is the
capacity to opt for one or another of those futures.
\eq
and
\bq
\indent
Renouvier's assault on the optimistic doctrine of inevitable progress
was carried out with both philosophical and historical arguments. The
philosophical arguments were probably the most conclusive for him,
but he understood that it is possible to differ over philosophical
positions. What he found especially difficult to understand was how
anyone could look at human history and still believe in continuous
and automatic progress. The great system builders were also great
oversimplifiers: ``Hegel and Comte, as well as Bossuet and Vico,
treat history the way Eudoxus and Ptolemy treated astronomy, with
their ideal spheres.'' Rather than looking at all of history, most
empirical arguments for progress lean heavily on the history of
Western science or in more recent years the advance of material
comforts or life expectancy. The progress of the sciences in the past
has also favored growth of the illusion that science will in the
future be able to solve all our problems.

In the nineteenth century there were also some who argued the reality
of moral improvement as well as an intellectual advance, and
especially the superiority of the nineteenth century over its
predecessors. Renouvier's measure of progress was the level of
individual freedom, certainly a good measure from which to criticize
the belief in inevitable progress but not an easy one to apply in any
precise way. Unlike many of his contemporaries, Renouvier was not
convinced that individual freedom had made great strides in the
nineteenth century, but what worried him most was the illusion that
the gains made were secure and that further expansion was certain.

The principal reason why there could be (and had been occasionally)
progress was the same reason there could be no inevitable, necessary
progress: free will. Even if, as Renouvier admitted, most of the
decisions men make are psychologically, socially, or otherwise
determined, the presence of even a few free acts prevents any
long-term necessary process of development. Free will also makes
possible not just failure to progress but also regression, and he
found enough historical examples to support his conclusion. Modern
optimism about progress, he observed, took root in the eighteenth
century, but its most celebrated advocate, Condorcet, did not
consider it absolutely inevitable, nor did he hold regression
impossible. The domination of historical determinist views in the
nineteenth century seemed to Renouvier an abusive extension or
corruption of more realistic eighteenth-century views. In philosophy,
the main culprit was Hegel, though most of the main schools were also
guilty.
\eq

\noindent {\bf W.~H. Long, {\sl The Philosophy of Charles Renouvier and Its
Influence on William James}, Ph.~D. thesis, Harvard University, 1927:}

\bq
\indent
In 1834, at the age of nineteen, Renouvier entered l'\'Ecole
Polytechnique in Paris \ldots\ Here at the \'Ecole he met two
fellow-pupils, Felix Ravaisson and Jules Lequier, destined to
contribute significantly to the life of Renouvier. Ravaisson, in his
{\sl Rapport sur la philosophie francaise au XIXe siecle}, was one of
the first to recognize the neo-criticist as one of the great thinkers
of the age. Jules Lequier, an original mind but too sensitive of
imperfection to produce published works, was a direct influence in
developing in Renouvier that psychological and ethical voluntarism
which came to have such a dominant role in his thought, and hence
indirectly also an important influence in the thought of William
James.

Jules Lequier was born on January 30, 1814, at Quintin
(C\^otes-du-Nord), and, having studied at various colleges, entered
1'\'Ecole polytechnique in 1834. Having remained here for two years,
the young man entered military service. In 1838 occurred the death of
his father, inspiring in the son an impassioned sense of the problem
of life and of reality. Resigning from military service, he gave
himself up to an intense study of philosophy, passing through a
critical spiritual and mental struggle, winning at last a faith in
freedom and the divine destiny of man. Strangely enough, Lequier
remained a loyal Roman Catholic through life---``la faiblease de ce
grand esprit,'' in Renouvier's words. He was in spirit, however,
protestant, insisting upon the authority of the primitive tradition
rather than the unique Roman point of view.

It was an unfortunate loss to the world when Lequier, modern Pascal,
burned most of his manuscripts. His sensitive spirit was too keen to
imperfection, and too restless to achieve the ideal formulation of
his thought, to allow him to produce written work. At his death, in
1862, there remained only some few scattered collections of notes,
the major portion of which Renouvier, faithful friend and loyal to
the name of one whom he called his master, published in 1865 under
the title of {\sl La recherche d'une primi{\`e}re v{\'e}rit{\'e},
fragments posthumes de Jules Lequier}. Renouvier's loyalty and
devotion is quite striking, for he acknowledges in practically all of
his works his indebtedness to his friend, and reaffirmed on his
death-bed this debt to the mind which had stimulated in him the
voluntaristic point of view in speculation.

For example, a noteworthy statement by Renouvier in the introduction
of his {\sl Manuel de la Philosophie Ancienne}, written in 1844, is
of value in the history of pragmatic voluntarism.  He writes that
\bq\noindent
Il est surtout une chose que je ne dois poing n\'egliger de dire: de
longs entretiens sur les questions fondamentales de la m\'etaphysique
avec un ami qu'il n'est pas temps encore de nommer, et qui en a fait
depuis plusieurs ann\'ees l'objet habituel, de ses r\'eflexions, ont
imprim\'e quelquefois sur mes id\'ees les traces des siennes. Ces
communications mutuelles ont \'et\'e, malgr\'e les diff\'erences
profondes qui nous s\'eparent, favoris\'ees par un accord frappant
dans quelques-uns des principes de la haute logique. Accoutum\'e,
comme il l'\'etait lui-m\^eme, \`a donner une grande place \`a la
croyance dans les fondements de la science, j'ai mis \`a profit des
analyses sur la foi; sur la libert\'e, sur l'intervention de l'id\'ee
de libert\'e dans celles du savoir et de la certitude, qui sont, pour
lui, le r\'esultat d'\'etudes s\'erieuses et de m\'editations
suivies, et qui joueront un r\^ole important dans ses travaux: la
publication n'en tant pas prochaine, c'\'etait une raison de plus
pour constater ice que les vues generales qui les dominent lui sont
tout \`a fait personnelles.
\eq
Elsewhere Renouvier writes concerning the influence of Lequier upon
his doctrines of freedom, certitude, and belief, that he owed to his
friend and master,
\bq\noindent
tout ce qui concerne d'une mani\`ere essentielle, dans mon livre,
l'\'etab\-lis\-se\-ment de la li\-ber\-t\'e et de ses rapports avec
la certitude, \`a un philosophe, M. Jules Lequier \ldots. Il ne
d\'epend pas de moi de donner \ldots\ satisfaction \`a sa m\'emoire.
Je peux du moins reconna\^{\i}tre, encore qu'il soit difficile de
l'exprimer en termes vraiment suffisants, l'incomparable obligation
que j'ai contract\'ee envers l'homme qui a fait tomber un certain
jour l'\'ecaille de mes yeux, qui m'a montr\'e la faiblesse des
doctrines dont j'\'etais l'adh\'erent, m\^eme involontaire, et m'a
apris ce que c'est que libert\'e, ce que c'est que certitude, et
qu'un agent moral est tenu moralement de {\it se faire\/} des
convictions touchant des v\'erit\'es, dont les penseurs rationalistes
ont la mauvaise habitude de mettre la preuve sur le compte de
l'\'evidence et de la n\'ecessit\'e.
\eq

For Lequier, in Renouvier's words, a man's philosophy centers around
a master thought and an active faith. The  true Odyssey of the human
mind, according to Lequier, must begin with the Cartesian search for
a first verity, for certitude. Certitude requires evidence. What,
then, is evidence?  ``Impossible, \'evident, l\'egitime, que de
rapports mal d\'em\^eles!'' ejaculates this remarkable seeker for
truth.  What is evidence?  That which ``est impossible de douter avec
bonne foi.'' But there is apparent and false evidence as well as
true. Hence the Cartesian methodical doubt is the only true method;
for true evidence and apparence [sic] are not identical. For unknown
errors may creep into the most apparent truths. What, first of all,
shall we say of the pretended separation of belief and science? What
shall determine our choice when the head and the heart, science and
belief, are in conflict?  ``J'aviserai, je verrai: je sacrifierai
l'une \`a l'autre,'' writes this modern Pascal. ``Tu dis vrai, mais je
ne veux pas t'entendre; ne l'ai-je pas dit souvent? Ou plut\^ot je
dirai \`a ma raison: Tu dis vrai et je le voie, mais je ne te crois
pas, et m'aidant de ne te croire pas pour m'emp\^echer de voir que
mon coeur me trompe, je trouve que tu as tort et que c'est mon coeur
qui a raison. Je pr\'ef\`ere la sagesse de mon coeur qui m'\'el\`eve
et me satisfait, \`a ta lumi\`ere qui ne me montre que mon abaissement
et mon d\'esespoir. Quand tu affirmes que ce qu'il affirme est
\'evidemment faux pourquoi te croirais-je, puisqu'il affirme que ce
que tu affirmes est faussement \'evident?

``Mais il serait mieux que ma raison et mon coeur eussent raison
ensemble.''

The search for a first verity, avoiding the snares of prestige and
false evidence, ends in a vicious circle, since first truths are
begged.
\bq\noindent
Les~v{\'e}rit{\'e}s primitives ne peu\-vent s'\-{\'e}\-tab\-lir par
l'{\'e}vidence puisque l'{\'e}vidence est d{\'e}\-duc\-tive.
\eq

To escape the vicious circle and its accompanying doubt requires an
act of will:
\bq\noindent
Franchir ce cercle vicieux c'est pos\-s{\'e}d\-er en quelque fa\c{c}on,
c'est cr{\'e}er, c'est faire que ce qui n'{\'e}tait pas soit; c'est
faire en moi la lumi{\`e}re, mais la faire en effet \ldots\ Agir,
c'est commencer. Je le franchis donc en agissant, ce cercle vicieux,
dans mon effort qui se produit lui-m{\^e}me; cet effort qui l'instant
d'avant n'{\'e}tait pas et qui tout {\`a} coup devenant, par
lui-m{\^e}me {\`a} lui-m{\^e}me sa cause, est, c'est-a-dire s'est
produit, s'est fait de rien. C'est l{\`a} vouloir.
\eq

Thus knowledge begins with free act, and destroys the illusory
necessity.

Lequier's own spiritual history in similar to that of William James,
and it is in fact the former whose doctrine brings the latter out of
his slough of despond. The striking mental regeneration of Lequier is
so interesting in the history of American pragmatism that it is
worthy of quotation. From the whole of the remarkable memoire of this
young men seeking freedom the following item is given:
\bq
\indent
Une seule, une seule id{\'e}e, partout reverber{\'e}e, un seul
soileil aux rayons uniformes: Cela que j'ai fait {\'e}tait
n{\'e}cessaire, Ceci que je pense est n{\'e}cessaire, L'absolue
n{\'e}cessit{\'e} pour quoi que ce soit d'{\^e}tre {\`a} l'instant et
de la mani{\`e}re qu'il est, avec cette cons{\'e}quence formidable:
le bien et la mai confondus {\'e}gaux, fruits n{\'e}s de la m{\'e}me
s{\`e}ve sur la m{\^e}me tige. A cette id{\'e}e, qui r{\'e}volta tout
mon {\^e}tre, je poussai un cri de d{\'e}tresse et d'effroi: la
feuille {\'e}chappa de mes mains, et comme si j'eusse touch{\'e}
l'arbre de la science, je baissai la t{\^e}te en pleurant.

Soudain je la relevai. Ressaisissant la foi en ma libert{\'e} par ma
libert{\'e} m{\^e}me, sans raisonnement, sans h{\'e}sitations, sans
autre gage de l'excellence de ma nature que ce t{\'e}moignage
int{\'e}rieur que se rendait mon {\^a}me cr{\'e}{\'e} {\`a} l'image
de Dieu et capable de lui r{\'e}sister, puisqu'elle devait lui
ob{\'e}ir, je venais de me dire, dans la s{\'e}curit{\'e} d'une
certitude superbe: Cela n'est pas, je suis libre.

Et la chim{\`e}re de la n{\'e}cessit{\'e} s'{\'e}tait {\'e}vanouie,
pareille {\`a} ces fant{\^o}mes formes pendant la nuit d'un jeu de
l'ombre et des lueurs du foyer, qui tiennent immobile de peur sous
leurs yeux flamboyants, l'enfant, r{\'e}veill{\'e} en sursaut, encore
{\`a} demi perdu dans un songe: complice du prestige, il ignore qu'il
l'entretient lui-m{\^e}me par la fixit{\'e} du point de vue, mais
sit{\^o}t qu'il s'en doute, il le dissipe d'un regard au premier
mouvement qu'il ose faire.
\eq
Thus by an act of will Lequier asserted his freedom and renounced the
power of fate and necessity.

True, Lequier insists, it is impossible to demonstrate freedom; but
at the same time it is impossible to demonstrate necessity.  In this
situation the individual has the right to affirm freedom as a
postulate of life, and this affirmed truth receives a new dignity and
value, namely in being freely chosen by the individual in a moral
act. In the absence of proof there arises a definite dilemma and
challenge:
\begin{list}{}{}
\item To affirm Necessity necessarily; or
\item To affirm Necessity freely; or
\item To affirm Freedom necessarily; or
\item To affirm Freedom freely.
\end{list}
In this alternative of freedom and necessity, Lequier bids us choose
``entre l'une et l'autre, avec l'une ou avec l'autre.'' He preferred
to, choose freedom freely; ``j'embrasse le certitude,'' he declares,
``dont je suis l'auteur.'' And in this he finds the first verity
which he sought, namely, ``freedom the positive condition of
knowledge, and the means to knowledge.''

It is thus shown that Lequier was a determining factor in Renouvier's
theory of belief, certitude, and freedom, end hence, as it will be
shown, the immediate source of William James' ``will to believe''
doctrine, and his theory of freedom, is to be traced to this
little-known modern Pascal, Jules Lequier.
\eq

\indent {\bf L.~Menand, {\sl The Metaphysical Club}, (HarperCollins, London,
2001):}

\bq
\indent
In his own way, and despite Holmes's distaste, William James was a
bettabilitarian, too. But he did believe in free will---what would it
mean to bet, after all, if we were not free to choose the stakes? He
was repelled by Wright's reduction of the world to pure
phenomena---he thought Wright made the universe into a ``Nulliverse''
and he regarded the abyss Wright insisted on placing between facts
and values as a fiction. James thought that Wright's decision to
separate science from metaphysics was itself a metaphysical
choice---that Wright's disapproval of talk about values was just an
expression of Wright's own values. Wright was a positivist because
positivism suited his character: moral neutrality was his way of
dealing with the world---and that, in James's view, is what all
beliefs are anyway, ``scientific'' or otherwise.

Wright was a regular visitor to the James family home in Cambridge
long before 1872, and in any case William James did not need the
Metaphysical Club to reach his conclusion about the nature of
beliefs. He had already arrived there by experimentation on what was
always his favorite human subject, himself. When he was living in
Germany in the late 1860s, he had got caught up in the speculative
frenzy about free will and determinism inspired by Buckle's book. As
usual, he found merits on both sides. ``I'm swamped in an empirical
philosophy,'' he wrote to Tom Ward shortly after getting back to
Cambridge in 1869; ``---I feel that we are Nature through and
through, that we are {\it wholly\/} conditioned, that not a wiggle of
our will happens save as the result of physical laws, and yet
notwithstanding we are en rapport with reason \ldots. It is not that
we are all nature {\it but\/} some point which is reason, but that
all is Nature {\it and\/} all is reason too.''

After he took his M.D. from Harvard, in June 1869, James collapsed.
He descended into a deep depression, exacerbated by back pains, eye
trouble, and various other complaints. His diary for the winter of
1869--70 is a record of misery and self-loathing. Then in the spring,
after reading the second installment, published in 1859, of a
three-part work called the {\sl Essais de critique g\'en\'erale\/} by
the French philosopher Charles Renouvier, he had a breakthrough.

Renouvier was a French Protestant from a family active in liberal
politics, but he had quit political life after the rise of the Second
Empire, in 1848, to devote himself to the construction of a
philosophical defense of freedom. Renouvier's argument was that ``the
doctrine of necessity'' is incoherent, since if all beliefs are
determined, we have no way of knowing whether the belief that all
beliefs are determined is correct, and no way of explaining why one
person believes in determinism while another person does not. The
only noncontradictory position, Renouvier held, is to believe that we
freely believe, and therefore to believe in free will. Even so, we
cannot be absolutely certain of the truth of this belief, or of
anything else. ``Certainty is not and cannot be absolute,'' he wrote
in the second {\sl Essai.} ``It is \ldots\ a condition and an action
of human beings \ldots. Properly speaking, there is no certainty;
there are only people who are certain.''

This was, in effect, Wright without the nihilism, and it was entirely
appealing to James. ``I think that yesterday was a crisis in my
life,'' he wrote in his diary on April 30, 1870.

\bq
I finished the first part of Renouvier's 2nd Essays and see no reason
why his definition of free will---the sustaining of a thought {\it
because I choose to\/} when I might have other thoughts---need be the
definition of an illusion. At any rate I will assume for the
present---until next year---that it is no illusion. My first act of
free will shall be to believe in free will \ldots. Hitherto, when I
have felt like taking a free initiative, like daring to act
originally, without carefully waiting for contemplation of the
external world to determine all for me, suicide seemed the most manly
form to put my daring into; now, I will go a step further with my
will, not only act with it, but believe as well; believe in my
individual reality and creative power.
\eq

As bold as this resolution sounds, it did not release James from his
depression. He seems to have been incapacitated by psychosomatic
disorders---in particular, an inability to use his eyes for reading
or writing---for another eighteen months, and he suffered chronically
from depression, eyestrain, and insomnia all his life. Henry's
mention of the formation of the Metaphysical Club in January 1872 is
one of the first signs, after the diary entry about Renouvier,
written a year and a half earlier, that William was socially active
again.

Still, James believed that Renouvier had cured him, and he sent him
thanks. ``I must not lose this opportunity of telling you of the
admiration and gratitude which have been excited in me by the reading
of your {\sl Essais},'' he wrote to Renouvier in the fall of 1872.
``Thanks to you I possess for the first time an intelligible and
reasonable conception of freedom \ldots. I can say that through that
philosophy I am beginning to experience a rebirth of the moral life;
and I assure you, sir, that this is no small thing.'' Renouvier had
taught James two things: first, that philosophy is not a path to
certainty, only a method of coping, and second, that what makes
beliefs true is not logic but results. To James, this meant that
human beings are active agents---that they get a vote---in the evolving
constitution of the universe: when we choose a belief and act on it,
we change the way things are.
\eq

\begin{itemize}
\item
C.~Renouvier, {\sl Science de la Morale}, 2 volumes, (F. Alcan,
Paris, 1869).

\item
C.~Renouvier, {\sl Essais de Critique G\'en\'erale.\ Premier Essai.\
Analyse G\'en\'erale de la Connaissance.\ Bornes de la Connaissance.\
Plus un Appendice sur les Principes G\'en\'eraux de la Logique et des
Math\'ematiques}, (Paris, 1854). Second edition: {\sl Essais de
Critique G\'en\'erale.\ Premier Essai.\  Trait\'e de Logique
G\'en\'erale et de Logique Formelle}, 3 volumes, (A. Colin, Paris,
1875).

\item
C.~Renouvier, {\sl Essais de Critique G\'en\'erale.\ Deuxi\`eme
Essai.\ L'homme: la Raison, la Passion, la Libert\'e, la Certitude,
la Probabilit\'e Morale}, (Paris, 1859). Second edition: {\sl Essais
de Critique G\'en\'erale.\ Deuxi\`eme Essai.\ Trait\'e de Psychologie
Rationelle}, 3 volumes, (A. Colin, Paris, 1875).

\item
C.~Renouvier, {\sl Esquisse d'une Classification Syst\'ematique des
Doctrines Philosophiques}, 2 volumes, (Paris, 1885-86).

\item
C.~Renouvier, {\sl Essais de Critique G\'en\'erale.\ Troisi\`eme
Essai.\ Les Principes de la Nature}, (Paris, 1864). Second edition:
{\sl Les Principes de la Nature}, 2 volumes, (A. Colin, Paris, 1896).

\item
C.~Renouvier, {\sl Essais de Critique G\'en\'erale.\ Quatri\`eme
Essai.\ Introduction \`a la Philosophie Analytique de l'Histoire},
(Paris, 1864). Second edition: {\sl Introduction \`a la Philosophie
Analytique de l'Histoire, Les Id\'ees; les Religions; les
Syste\`emes}, (A. Colin, Paris, 1896).

\item
C.~Renouvier, {\sl Histoire et Solution des Probl\`emes
M\'etaphysiques}, (F. Alcan, Paris, 1901).
\end{itemize}

\section{18-08-08 \ \ {\it SICs and Partial Orders}\ \ \ (to K. Martin)} \label{Martin5}

Thanks for sending your tutorial.  This looks like good stuff.  I very much like this idea of a partial order on the probability simplex, and I would like to think about it more.  Seeing your example about Grover's algorithm the other day convinces me that there's nothing intrinsically bad about building a partial order on the density operators by introducing a fixed BASIS and then relying on a well-known order on the simplex.  Introducing a SIC is doing the same conceptual thing, but the SIC has the added bonus of being informationally complete.  Thus all density operators become comparable under it.

Maybe one interesting technical question that could be considered is what properties of the partial order are preserved when one changes from one SIC to another.  Maybe it's not the case that everything blows to the wind, but at least some features are preserved.

In any case, it was really good meeting you.  And how grateful I am that you're interested in this passion of mine.  I genuinely do think that a lot of stuff (even revolutionary stuff) will come out of the study of SICs.

\section{20-08-08 \ \ {\it Want to See a Glacier?}\ \ \ (to D. Gottesman)} \label{Gottesman11}

Having now gone to Bristol, Sydney, Eugene, Zurich (twice), Paris, and Garching already this year, and leaving again for Sweden next week, I woke up two nights ago with the severe pain of a broken travel bone.  I decided I just can't take it anymore.  Not for a while at least.

Thus I wonder if you'd be interested in taking my place at this meeting:
\bq
\myurl{http://www.uibk.ac.at/th-physik/qics-obergurgl2008/}.
\eq
It would be nice for PI to have some representation there, and I think it is a meeting (and a subject) you would enjoy in any case.

If you're interested, let me know, and when I tell the organizers I'm out, I'll ask them to consider transferring the invitation to you.  (To be honest, it'd be a very good deal for them.)

\section{20-08-08 \ \ {\it Progress}\ \ \ (to E. G. Cavalcanti)} \label{Cavalcanti2}

\bec
By the way, the argument I gave in Vienna, now that I remembered, was along these lines:

Suppose you believe that (i) A quantum state is just an encapsulation of your subjective degrees of belief; (ii) The lesson of Kochen--Specker and Bell's theorems are that there are no objective facts to further specify your degrees of belief even in principle, at least in the cases where you assign a pure state to a situation; and (iii) You are no more central to the universe than any other agent.

Then considering a Wigner's friend scenario, assumptions (i) and (ii) imply that before you look into the box where your friend is measuring a quantum system, you cannot assign a reality to the measurement outcome of your friend. However, assumption (iii) makes you believe that your friend {\bf can} assign a definite outcome, whichever it is. That implies that as far as you are concerned, before you open the box that event of your friend measuring the system never happened. But you believe that relative to your friend it did happen. You conclude that the very existence of `events' is relative to the observer. The world is not made of a collection of events aspersed in space-time, with well-defined causal relations between each pair of events, and between each event and each observer.

To reinforce that view, you can devise two situations: in one you open the box and ask your friend ``Did you see any outcome before I opened the box?''.  Of course she'll say your opening the box had no influence on that, and by assumption (iii) you'll believe her. So you do the same experiment, but now you perform a transformation that takes all the contents of the box, including your friend, back to its initial state, a little time after you presume the measurement has been done. The fact that the evolution was fully reversible is evidence of the fact that your friend was still in a coherent entangled state with the measurement apparatus, and therefore by (ii) you believe there's no objective fact of the matter relative to you as to which outcome she observed, and since you coherently reversed the interaction, there never will be. Yet, you still believe that relative to her own previous position as an observer inside the box, she must have observed something before you reversed the whole thing. But that observation is forever outside any causal relation with you or even your current time-reversed friend. In other words, that event is in a sense not in the same space-time as you are now, if by definition every event in a space-time have {\bf some} causal relation with every other event.

The only way out of this conclusion, it seems to me, is to accept a hidden-variable interpretation in which there {\bf are} objective matters of fact about events, we just don't know what they are, or extreme solipsism.  But I would agree with you if you say that the alternative above sounds much more exciting.
\eec

It took me a long time, but at least I've made a little progress.  I finally read the nice set of notes you sent me.  I will work on a response during my flight(s!)\ to Sweden.  But I think you'll be there so we can discuss things anyway.

My main qualm is with what you call solipsism at the end.  I instead would call it extreme pluralism.  It'll be fun talking again.

\section{22-08-08 \ \ {\it Possible Collaboration?, 2}\ \ \ (to T. Duncan)} \label{Duncan5}

The role you foresee for me is something I can decently fulfill.  As you say, we can bat around ideas---maybe I can become more involved if there are some concrete ideas worth pursuing.  I love the way you quote William James's ``Sentiment of Rationality''.  Maybe the right call is his essay on man's strenuous nature.  I tell my daughter every day that she can change the world---one can suspect that the world is malleable, as the classical pragmatists did.  But that is only reinforced by quantum mechanics---that idea is what drives me.  So there is plenty of room to talk.

\section{22-08-08 \ \ {\it The Two Freedoms Found, 2} \ \ (to H. C. von Baeyer)} \label{Baeyer43}

Thanks for that.  I read the interview a couple or three days ago, and both enjoyed it and grit my teeth too (at the choice of some phrasings).  Getting caught up on email.

\ldots\ Well I started that much of a note to you earlier (about five days ago) and never finished.  But now I'm off to Sweden for one of Andrei's crazy conferences.  But it unites {\Schack}, {\Appleby}, and me together for a while and there are some other interesting people there too (Bengtsson, Cabello, Larsson).

My key grit was with the word ``property''.  I think it is dangerous business to say that it is a particle's properties that are teleported.  See {\sl My Struggles\/} starting at the section ``\myref{SmolinJ2.1}{Qubit and Teleportation Are Words}''.

But there was other good stuff in the interview that was really good stuff too.  I should come back to an exposition of the positive things.

Let's talk soon again about the Pauli--Fierz project.  Have you given it any further thought?  Any progress?  Or has it been forgotten for the meantime.  I'll try to write some words in Sweden (particularly if I see a reply from you).

\section{22-08-08 \ \ {\it Invite!}\ \ \ (to H. Mabuchi)} \label{Mabuchi14}

Thanks for letting me off the hook.

\bhm
I'm thinking we should have another go at trying to pitch some
kind of Perimeter-Stanford quantum foundations thing \ldots
\ehm

Actually I would like that very much.  My grant writing skills are much better than they used to be.  And I have a feeling Rob would be keen too.  Do you have a time frame on which we should think about working something up?

You tell me you can do an experiment that can map out the weirdities of my extended Bloch sphere, and I'll be out to Stanford soon after Christmas.  Actually ``Bloch sphere'' is in a way a misnomer: We now understand it's much more like a spider web sitting on a sphere. A whole lot of spherical caps are missing, and the states really reside at the intersections (the webbing).

Onward to Sweden; I am just about to land at JFK along the way.

\section{25-08-08 \ \ {\it Quotes} \ \ (to A. Plotnitsky)} \label{Plotnitsky20}

Please don't forget to send me the quotes from your talk today.

\subsection{Arkady's Reply}

\bq\noindent
On one hand, the definition of the state of a physical system, as ordinarily understood [i.e.\ in classical physics], claims the elimination of all external disturbances.  But in that case, according to the quantum postulate [of Planck], any observation will be impossible, and, above all, the concepts of space and time lose their immediate sense.  On the other hand, if in order to make observation possible we permit certain interactions with suitable agencies of measurement, not belonging to the system, an unambiguous definition of the state of the system is naturally no longer possible, and there could be no question of causality in the ordinary sense of the word.  The very nature of the quantum theory thus forces us to regard {\it the space-time co-ordination and the claim of causality, the union of which characterizes the classical theories, as complementary but exclusive features of the description}, symbolizing the idealization of observation and definition respectively.
\begin{flushright}
--- Niels Bohr (1927), {\sl PWNB} 1, pp.\ 54--55 (emphasis added)
\end{flushright}
\eq

\section{02-09-08 \ \ {\it Real Lasers} \ \ (to J. W. Nicholson)} \label{Nicholson29}

Learning that one of my high-school acquaintances is now involved in this:
\bq
\myurl{http://www.thequantumalliance.com/}

[[at the time the site had some kind of ``laser product'' for bio-feedback]]
\eq
I couldn't help but forward it on to you.  Now that's real cutting-edge laser work, isn't it?

\section{03-09-08 \ \ {\it Two (Important!)\ Things} \ \ (to C. Rovelli)} \label{Rovelli1}

I have been meaning to write you ever since departing Paris, but the usual malaise in writing came about as soon as I left---too much sociality takes a toll on my correspondence.  The reason I wanted to write you is because after meeting you, listening to your talk, and listening to the lunchtime conversation you had with Si Kochen, I felt a rather deep connection between several aspects of our individual (i.e., your and my) takes on quantum mechanics.  Even the common imagery we both seem to use struck me as a surprise.  What particularly intrigues me is the way we've come at this from different directions, but both end up on a radical thought:  That even facts (particularly quantum outcomes) are relative.  The two directions---and they certainly seem to be distinct---are 1) for you, relationalism more generally, and 2) for me, an understanding of quantum states in terms of Bayesian probabilities.  I would not have initially thought of those roads as traveling to the same point, but there is something to think on here.  Whereas you speak of ``relative facts'', we speak of ``facts for the agent'', but at least in some aspects they are surely the same thing.

I would be flattered if you would read some of my poetry on the idea before we meet and discuss these things again.  Particularly, I want to advertise to you the new compilation I'm putting together as a point of departure.  You can pick it up here:  [See this samizdat, but with updated page numbers.]  I hope that you'll find parts of it entertaining and thought provoking.  Of interest for the present purposes, you might read a bit around where I discuss this phrase ``fact for the agent'' (and see how it compares or contrasts with what you are striving for).  Here are some points of entry:
\begin{itemize}
\item[1)] Page 687, starting at ``\myref{RepliesToReferee4}{Replies to Referee 4}''
\item[2)] Page 650, starting at ``\myref{Schack107}{Our Professor}''
\item[3)] Page 646, starting at ``\myref{Finkelstein9}{My Two Cents}''
\item[4)] Page 714, starting at ``\myref{Baeyer27}{Facts-in-Themselves}''
\item[5)] Page 716, starting at ``\myref{Palge3}{Anti-Algebra, the Reprise}''
\end{itemize}
Also, maybe my correspondence with Bas van Fraassen about what I was perceiving as the difference and similarities between my view and yours:  See pages 550--557.  Also points to {\Ruediger} Schack on the same (and Wigner's friend), pages 559--564.  [See 10-11-05 note ``\myref{vanFraassen9}{Our Own Rovellian Analysis}'' to B. C. van Fraassen, 21-11-05 note ``\myref{Schack87}{Pull!}'' to R. Schack and 21-11-05 note ``\myref{Schack88}{Notes to van Fraassen}'' to R. Schack.]

That's one important thing.  Now, to the next question.  When {\it shall\/} we discuss these things again?  When we parted in Paris, you said pretty firmly (and I don't believe I am making this up), ``I will see you in September.''  Yet, our conference coordinator has not heard from you yet.  Please do write and say you're coming.  And particularly, because I would like to ask a further thing of you:  Would you give the general PI Colloquium that week in fact?  The talk you gave in Paris, from my point of view, would be perfect for it---mostly because it would open up to the wider PI community this difficult idea that we're trying to get at, but also because it was a simply beautiful talk.

Will you be around tomorrow or Friday?  I would very much like to call and talk about this (Lee gave me your number):  A colloquium from you would be great for PI.

\section{03-09-08 \ \ {\it Er Er} \ \ (to D. M. {\Appleby})} \label{Appleby35}

\bma
A conference on multiverses  I could cheerfully miss.  It is strange:  I am not a Bohmian, but the Bohm theory never annoys me.  Quite the contrary.  I don't believe it, but I think it is an interesting hypothesis.  I also enjoy talking to Bohmians---Owen is just the latest in a long list.  But when it comes to multiverses I feel strong aversion.   It strikes me as, not so much a theory, more an intellectual disease.  Perhaps even a moral and spiritual disease:  for if you think that everything happens it seems to me that you cannot help ending up with the belief that nothing really matters.  Everett et al.\ trivialize the world, and perhaps for that reason I don't think I can even enjoy talking to an Everettian.  At least, I can't think of an occasion when I ever have.
\ema

Regarding your first paragraph.  There was an interesting passage in the conversation where Hilary Greaves had David Albert draw a tree structure on the board, and then David Albert started to explain the idea of an ``indexical fact'' to Paul Davies.  Davies, said something like, ``So some branches are lit up because they contain rational agents; but some branches aren't lit at all.''  Albert protested about the idea of some branches being lit up by the presence of an agent---i.e., that that's not what he was getting at.  But the funny thing was that Hilary, who was sitting beside me, went further.  I overhead her say to herself, ``Yeah, that {\it would\/} be crazy.''   I think that fares well with your assessment below.

\section{03-09-08 \ \ {\it Your Old Thesis} \ \ (to G. Valente)} \label{Valente9}

Could you send me a copy of the thesis you wrote after studying in Oxford.  I don't have a copy of the final thesis, and recently I bought the Musil book---I'd like to review what you said about him (or which quote you used of his) in connection with QM.

\subsection{Giovanni's Reply}

\bq
Glad to hear from you! Funny that you write to me because I'm in Paris at the moment\ldots\ and I just got an unexpected e-mail from that Croatian guy who wrote his thesis in Oxford on your work five years ago\ldots

Unfortunately, I don't have a file with my master thesis anymore. It got lost when my old PC collapsed. I may have a disk where I saved the file at home in Italy and I certainly have a paper copy of the work there, but there's nothing I can send you right now. In any case, I recall that I made a reference to Musil only in the introduction, starting out by quoting the passage below which is taken from the section ``If there is a sense of reality, there must also be a sense of possibility'' (it's one of the first in the book). According to my original quantum reading, it conveys the idea that the world is under construction, as a pragmatist would put it; that the realists (e.g.\ Ghirardi, Albert, Maudlin, etc.), who interpret the quantum state as a state of reality, would regard the possibilists (e.g.\ Chris Fuchs), who interpret the quantum state as a state of belief, as dreamers or troublemakers\dots; that all we're given is a bunch of possibilities, namely a set of probabilities, but the latter are of course determined by reality, namely by the outcomes of measurements. I think at the end I even joked that, in connection with the Bayesian interpretation of QM, in quantum theory we're really dealing with ``systems without qualities'', as we can't say anything about them before a measurement.

I'm sure you'll enjoy the book. Perhaps it'll awaken the deep Mitteleuropean spirit that harbours in your soul. I myself was thinking of reading it again soon.

\bq\noindent
From {\sl The Man without Qualities\/} by Robert Musil:\medskip

To pass freely through open doors, it is necessary to respect the fact that they have solid frames. This principle, by which the old professor had lived, is simply a requisite of the sense of reality. But if there is a sense of reality, and no one will doubt that it has its justifications for existing, then there must also be something we can call a sense of possibility.

Whoever has it does not say, for instance: Here this or that has happened, will happen, must happen; but he invents: Here this or that might, could, or ought to happen. If he is told that something is the way it is, he will think: Well, it could probably just as well be otherwise. So the sense of possibility could be defined outright as the ability to conceive of everything there might be just as well, and to attach no more importance to what is than to what is not. The consequences of so creative a disposition can be remarkable, and may, regrettably, often make what people admire seem wrong, and what is taboo permissible, or, also, make both a matter of indifference. Such possibilists are said to inhabit a more delicate medium, a hazy medium of mist, fantasy, daydreams, and the subjunctive mood. Children who show this tendency are dealt with firmly and warned that such persons are cranks, dreamers, weaklings, know-it-alls, or troublemakers.

Such fools are also called idealists by those who wish to praise them. But all this clearly applies only to their weak subspecies, those who cannot comprehend reality or who, in their melancholic condition, avoid it. These are people in whom the lack of a sense of reality is a real deficiency. But the possible includes not only the fantasies of people with weak nerves but also the as yet unawakened intentions of God. A possible experience or truth is not the same as an actual experience or truth minus its ``reality value'' but has---according to its partisans, at least---something quite divine about it, a fire, a soaring, a readiness to build and a conscious utopianism that does not shrink from reality but sees it as a project, something yet to be invented. After all, the earth is not that old, and was apparently never so ready as now to give birth to its full potential.

To try to readily distinguish the realists from the possibilists, just think of a specific sum of money. Whatever possibilities inhere in, say, a thousand dollars are surely there independently of their belonging or not belonging to someone; that the money belongs to a Mr.\ Me or a Mr.\ Thee adds no more to it than it would to a rose or a woman. But a fool will tuck the money away in his sack, say the realists, while a capable man will make it work for him. Even the beauty of a woman is undeniably enhanced or diminished by the man who possesses her. It is reality that awakens possibilities, and nothing would be more perverse than to deny it. Even so, it will always be the same possibilities, in sum or on the average, that go on repeating themselves until a man comes along who does not value the actuality above idea. It is he who first gives the new possibilities their meaning, their direction, and he awakens them.

But such a man is far from being a simple proposition. Since his ideas, to the extent that they are not idle fantasies, are nothing but realities as yet unborn, he, too, naturally has a sense of reality; but it is a sense of possible reality, and arrives at its goal much more slowly than most people's sense of their real possibilities. He wants the forest, as it were, and the others the trees, and forest is hard to define, while trees represent so many cords of wood of a definable quality. Putting it another and perhaps better way, the man with an ordinary sense of reality is like a fish that nibbles at the hook but is unaware of the line, while the man with that sense of realty which can also be called a sense of possibility trawls a line through the water and has no idea whether there's any bait on it. His extraordinary indifference to the life snapping at the bait is matched by the risk he runs of doing utterly eccentric things. An impractical man---which he not only seems to be but really is---will always be unreliable and unpredictable in his dealings with others. He will engage in actions that mean something else to him that to others, but he is at peace with himself about everything as long as he can make it all come together in a fine idea. Today he is still far from being consistent. He is quite capable of regarding a crime that brings harm to another person merely as a lapse to be blamed not on the criminal but on the society that produced the criminal. But it remains doubtful whether he would accept a slap in the face with the same detachment, or take it impersonally as one takes the bite of a dog. The chances are that he would first hit back and then on reflection decide that he shouldn't have. Moreover, if someone were to take away his beloved, it is most unlikely that he would today be quite ready to discount the reality of his loss and find compensation in some surprising new reaction. At present this development still has some way to go and affects the individual person as a weakness as much as a strength.

And since the possession of qualities assumes a certain pleasure in their reality, we can see how a man who cannot summon up a sense of reality even in relation to himself may suddenly, one day, come to see himself as a man without qualities.
\eq
\eq

\section{04-09-08 \ \ {\it My Own Update} \ \ (to D. M. {\Appleby})} \label{Appleby36}

I'll read your longer note in detail tomorrow.  One quick point, I used the full phrase ``rational agent'' in my stories because that was the phrase bandied about in the Greaves/Albert discussion.  But I myself don't know what rational is or should mean.  The key part of it for me, the part that played any role in my mind at all, is ``agent''---by definition, ``that which activates.''

\section{04-09-08 \ \ {\it Need Students} \ \ (to H. C. von Baeyer)} \label{Baeyer44}

I have just learned that I got a US grant with enough funding to support 2 graduate students fully at U Waterloo through 4 years each.  It's of no great use however for good Canadian students, as they will already have their own NSERC funding.  Thus I figure it's most wisely spent on really good non-Canadians who want to be up here in the Mecca of quantum info/foundations.  So, I'm writing to see if you have any good students that might be interested?  The work I'm hoping to emphasize is, of course, ``How come the quantum?'', tackled from a quantum-information/Bayesian perspective.

Any leads you can give me would be most appreciated.

\section{08-09-08 \ \ {\it Progress, 2}\ \ \ (to E. G. Cavalcanti)} \label{Cavalcanti3}

That was helpful.  But, OK, I give up on trying to reply to you:  Too many things that I unfortunately {\it have to\/} write are getting in my way, and in any case, I will see you soon enough.

See you at {\sl The Clock and the Quantum}.

\subsection{Eric's Preply}

\bq
I guess I didn't express myself very well in those end remarks. I should have said that the argument {\it seems to lead\/} to the conclusion that either there are hidden variables or to an extreme solipsism. But when I say I agree with you that the ``alternative above'' sounds more exciting, I mean exactly the alternative which your research program is trying to find, the alternative which we hope will arise out of realising the full significance of the remark from Pauli, that the observer cannot be separated from the things it observes.

Extreme pluralism is probably a better name for that worldview. However, I also have the strong hunch that this pluralism is of a form which is so beautifully symmetric that one cannot avoid the feeling of an unity behind the different perspectives or worldviews. The reason why we seem to be able to glimpse an order, a pattern behind these different perspectives is probably because behind the apparent pluralism stands an unifying simplicity. I think the search for the basis of this unification is the next step after the recognition of the pluralism.
\eq

\section{09-09-08 \ \ {\it Rationalizing Agent} \ \ (to D. M. {\Appleby})} \label{Appleby37}

I've now read both of your notes on the subject.  The short answer is, yes I do agree with you.  I like the way you put this:
\bma
If the word ``rational'' is understood in that way then I don't believe
a rational agent is capable of lighting up anything.  Not even itself.
\ema
The point is, the light---the will---comes first (or at least has a significant priority I think).  Not only do we see something like this in alchemistic strains of thought, but it is a good piece of James and Dewey's psychology.

All this talk of rationality and where it stands in relation to the light, reminds me of something very different.  G. H. Mead's discussion of causality in light of true indeterminism.  I'll paste in a passage below about that.  [See A. E. Murphy quote in the 23-09-03 note ``\myref{Savitt3}{The Trivial Nontrivial}'' to S. Savitt.]  Maybe you'll see why my mind loosely jumps between the subjects.

Anyway, reflecting on the two issues simultaneously led to the title of this note.  ``Rational agent'' doesn't seem right indeed.  But put ``rationalizing agent'' in its place, and it seems to ring of a little truth.

\section{09-09-08 \ \ {\it Panpsychism} \ \ (to D. M. {\Appleby})} \label{Appleby38}

\bma
Which brings me back to the term ``rational agent''.  The word to which I take particular exception here is, not ``rational'', but ``agent'' (I do object to ``rational'', for the reasons I gave before, but I object even more to ``agent'').  To my mind this is an example of the ``impersonal expressed in the passive voice''.  It should be ``rational person'' (or ``rationalizing person'').  Or ``rational human'', in all his or her laughing, crying, breathing, shitting, infinitely layered, ineffable totality.

Reasoning is something that {\bf people} do.  Not boxes, with an operating system.
\ema

I continue to like ``agent'' over ``person'' or ``human'' because I think that is the relevant piece of the equation.  However, in my case, I don't think it's because there is a Cartesian deep inside me that's trying to re-emerge, but rather a Copernican.  That by learning about myself---my position---I learn something about everything else's position.  Here's the way I put it to Sipe once.  The relevant part is the last item.  It is partially, though perhaps not completely, relevant to my discussion with you.

\begin{enumerate}
\item
When I posit a physical system about which I will speak (by assigning it a Hilbert space, etc.), I am doing that in an almost na\"{\i}ve realistic way.  I.e., the $\cal H$ I write down, represents a piece of the world that is out there independently of me.

\item
When I assign a dimension $d$ to that Hilbert space, I am hypothesizing an inherent {\it property\/} of that system.  I might be right, or I might be wrong, but $d$ is something I hypothesize of it, even if only provisionally.  Realism again.

\item
When I draw a quantum state out of the space of operators defined by $\cal H$, however, I am expressing a bundle of my expectations. These are not properties inherent in the system.  They are subjective expectations that I bring into the picture (presumably because they have served me well in the past, or at least done me no harm).  If I were to conceptually delete myself from the picture, these expectations would disappear with me.  To that extent, the view might sound a little like---or at least be confused with---idealism (but that's only if one---as Howard Wiseman often does---forgets elements 1 and 2 above and one of the further elements below).

\item
The subject matter of those expectations, i.e., what they are about, refers both to me and the system I posit.  They are MY expectations for the consequences (for ME) of MY interactions with the system.  That, you might think is a kind of operationalism:  For if I were to conceptually delete myself from the picture, those interactions would disappear too.

\item
On the other hand, the reason we use the formal structure of quantum mechanics to bundle our expectations, to manipulate and update them, to do all that we do with them, is to better cope with the world.  It is a means to help our species to survive and propagate.  That is a kind of instrumentalism.  That part of quantum mechanics is a tool like a hammer; it can be used to fix a lot of things, or simply as an aid to help defend ourselves.

\item
Still one can never forget the ultimately uncontrollable nature of each quantum measurement outcome---and through Kochen--Specker, at least the way I view it, the non-pre-existence of those ``outcomes''. That smacks of realism in the oldest, most time-honored way.  The world surprises us and is not a creation of our whims and fancies. Back to realism \ldots\ but the twist is quantum mechanics, as used by each individual user, only refers to the outcomes HE helps generate. (That smacks of alchemy \ldots\ dangerous to say so, but I call it like it is.)

\item
Nevertheless, having learned a little from Copernicus, it seems we should ultimately try to abstract away from these personal encounters with the world (having learned what we could from the formalism concerned with gambling on them).  If a tiny little system and I create something new in the world when we get together---i.e., we give rise to a birth or new fact---so must it be likely, it seems to me, that any two things give rise to a birth when they get together. That is ``F-theory'' or the ``sexual interpretation of quantum mechanics'' \ldots\ but you'll have to wait for the movie if you question me any further on that \ldots

\end{enumerate}

\section{09-09-08 \ \ {\it Panagentism} \ \ (to D. M. {\Appleby})} \label{Appleby39}

Just playing with titles.  This might have been better for the last note.

\section{10-09-08 \ \ {\it Easier Answers} \ \ (to D. B. L. Baker)} \label{Baker17}

\bdb
Are you planning on staying long?  What's your job like?
\edb

I plan to stay forever if I can.  This is the most fantastic job ever for the likes of someone like me.  I make good money, and the intellectual freedom is wonderful.  It gives me hope that I really will solve the problems of quantum mechanics before my days are up.  And the work environment is like nothing I've ever seen:  good coffee out for us continuously, muffins in the morning, cakes in the afternoon, wine and cheese every Friday afternoon.  You can take a virtual tour of the place here \myurl{http://www.perimeterinstitute.ca/en/About/Facilities/Facilities_Overview/} and you can see our wine list here \myurl[http://www.perimeterinstitute.ca/en/Outreach/Black_Hole_Bistro/Black_Hole_Bistro_Overview]{http://www.perimeterinstitute.ca/en/Outreach/Black\underline{ } Hole\underline{ }Bistro/Black\underline{ }Hole\underline{ }Bistro\underline{ }Overview/}.
So, it's just an amazing place to be at every day.

\section{10-09-08 \ \ {\it Your Old Thesis, 2} \ \ (to G. Valente)} \label{Valente10}

Thanks a million for the quote.  Yes, everything is going OK here, and Appleby, Schack, and I are on the tail of something big we think:  A nice derivation of finite dimensional quantum state-space structure from a single inequality.  Or, at least that's what we hope.

Good luck with all these academic things you're doing!  Geesh, a degree in physics to complement your degrees in philosophy.  Very impressive.

One technical point on what you said:
\bgv
I think at the end I even joked that, in connection with the Bayesian
interpretation of QM, in quantum theory we're really dealing with
``systems without qualities'', as we can't say anything about them before a measurement.
\egv
I would say that you can't say anything particular about them {\it after\/} a measurement either.

\section{10-09-08 \ \ {\it Really Weird and Not-So Weird Stuff}\ \ \ (to M. D. Sanders)} \label{SandersMD3}

\bmds
OK, don't laugh at me, but I am curious if you think the CERN particle accelerator is BS or it might actually reveal something, \myurl[http://public.web.cern.ch/Public/Welcome.html]{http://public.web.cern.ch/Public/Welcome.html}.  I read a few of your latest papers (\,\myurl[http://www.pitp.ca/personal/cfuchs/nSamizdat-2.pdf]{http://www.pitp.ca/personal/cfuchs/nSamizdat-2.pdf}) indicating that you think the next development is just a matter of viewing things differently, (``Likewise, I'll bet the next big step in physics will only require that we see something right here in front of us.'')\ so I wonder if you think that this CERN BS will actually change anything.

I didn't go too deep cuz it is like 800 pages, but I tried.  I also wonder about the skeptics that think the world will end when it fires up.  (Sounds like Star Trek followers, but I don't know.) I don't believe it, but it makes me wonder how crazy people are.

Anyway, if you have time I would like your opinion.  I understand if you think this is stupid based upon where you are \ldots\ you don't have to go too deep, just let me know if you think it will reveal anything or if you think it is a huge waste of money.  Also, I'm curious if you are at the pajama party {\rm \smiley}\  (\,\myurl[https://web.archive.org/web/20080913201536/http://ohmyafly.wordpress.com/2008/09/08/largest-hadron-collider-pajama-party/]{http://ohmyafly.wordpress.com/2008/09/08/largest-hadron-collider-pajama-party/})?

Sorry, but you are the only one I know close to this stuff \ldots
\emds

The really weird stuff is that you were digging around through my notes.  Now, that's an old friend!  I hope you don't think less of me now!  But did you see my TV debut:  \pirsa{07120048}.  Losing a little hair on top, aren't I?  My first comment is lame, but if you have the patience to wait, I get better.  (Yeah, right!?)

Now to the not-so weird stuff:
\begin{itemize}
\item[1)]	Safety of LHC.  Here's something to read, and it strikes me as decently authoritative:
\bv
\myurl[https://en.wikipedia.org/wiki/Safety_of_the_Large_Hadron_Collider]{http://en.wikipedia.org/wiki/Safety\underline{ }of\underline{ }the\underline{ }Large\underline{ }Hadron\underline{ }Collider}.
\ev
I think there's nothing to worry about whatsoever.  For me personally, the most conclusive piece of evidence is this.  At peak the LHC will be producing 14 TeV collisions (7 TeV each beam), but events like that happen in the upper atmosphere all the time, from cosmic rays slamming into us.  The only difference with the LHC is that we'll be able to make the collisions on demand and in the presence of a good detector.  Furthermore, the energies of the LHC are really nothing on that scale:  Cosmic rays have been recorded not with a measly 14 TeV, but with $10^8$ TeV.  See:
\bv
\myurl{http://en.wikipedia.org/wiki/Cosmic_ray}\\
or\\
\myurl{http://en.wikipedia.org/wiki/Ultra-high-energy_cosmic_ray}.
\ev
So, even if these wackos like R\"ossler were by accident right about black-hole production, etc., (and they're not), then we'd already have black holes flowing through the earth all the time.

\item[2)]	Cost of the LHC\@.   Estimates run up to \$10 billion.  I'm pretty sure most of that money is European,
 though I would like to know what the US contribution is.  In my own opinion, it's worth it.  For one thing, a key point of the ``standard model'' of particle physics, the Higgs particle, keeps not being seen.  Why does the rest of the model work if this particular major prediction of it keeps hiding?  That's more than a 25 year mystery (I first heard of it in 1983), and if the LHC doesn't find it, it'll finally put the nail in the coffin.  Secondly, there's the infrastructure for physics as a whole that CERN brings with it.  Throughout the world, it helps keep physics education going, and as you can imagine, I'm all for that.  To put the costs in perspective, note that the Iraq war has been costing us \$8 to \$12 billion A MONTH since 2003.  See for instance:
 \bv
 \myurl{http://en.wikipedia.org/wiki/Financial_cost_of_the_Iraq_War}.
 \ev
 You can bet your bottom dollar I would trade two months of the war to solve a 25 year old mystery in physics.

\item[3)]	Efficacy of LHC.  It remains true that I believe every word of what I said before:  ``I'll bet the next
 big step in physics will only require that we see something right here in front of us.  It'll be something no big multi-billion dollar particle accelerator will be needed for.  We just have to figure out how to take note of it.''  But by big step, I mean a REALLY BIG step---the stuff that has historically happened only every two to three hundred years, and can't be planned for beforehand.  I think here particularly on Galileo's noticing that any two masses (regardless of their sizes and weights) accelerate toward the ground at the same rate.  It is the key insight that ultimately underpinned Einstein's theory of gravitation (containing the essence of his ``principle of equivalence''), which in turn led us to predict a load of things, from black holes to the big bang itself.  Well, that fundamental phenomenon was right there for anyone to see and take note of in any of the previous 2,000 years of civilization.  But no one did.  No one had simply said, ``Wow, two different rocks, two different metals, two anythings, dropped at the same time always hit the ground simultaneously.  I wonder what the significance of that is?''  It's observations like that that make the true revolutions in physics.  And the LHC is not in that category---that's what I was alluding to.  However, I don't belittle little steps in physics either.

There is clearly something really big brewing on the horizon of physics.  Where is all the ``missing mass'' and ``dark energy'' in the universe?  See
\bv
\myurl[https://en.wikipedia.org/Missing_mass]{http://en.wikipedia.org/wiki/Missing\underline{ }mass}.
\ev
Is it really just unseen ``stuff'', or do we have something more basic wrong in our calculations?  Why is the universe's expansion accelerating rather than decelerating?  These are bigger questions (potentially revolutionary), and the LHC's answers to smaller questions may give us guidance there.  At the very least, they may say, ``Give up with your usual ways of thinking.  Look instead for something that's probably right in front of you.''

\item[4)]	No, I wasn't at the pajama party!  I'm too old for that.  The last time I was at CERN, I don't even
 think rap music had been invented yet.
\end{itemize}

I hope that was of some help.

\section{15-09-08 \ \ {\it PanSICism} \ \ (to D. M. {\Appleby})} \label{Appleby40}

What is the new idea for SIC existence?

\section{15-09-08 \ \ {\it  Israel December} \ \ (to A. Wilce)} \label{Wilce20}

\baw
PS -- how did Victor Colussi's week at PI turn out?
\eaw

We loved him, had a lot of fun with him.  He got a chance to see two old geezers in experimental math mode.  While he was here, we were finding an obscene number of new properties for quantum state space when written in SIC language, all by Marcus doing numerical testing as the questions arose.  Of course, we didn't have an idea how to prove anything at the time (we have now), but we really opened up a good vein in the mine while he was here.  I hope that helped excite him about science.

\section{16-08-08 \ \ {\it 81 vs 95}\ \ \ (to C. M. {\Caves})} \label{Caves96.5}

I wish you were coming to the SIC meeting---you in fact were the original source of the idea.  And there's nothing wrong with using it as a learning time.

\section{16-09-08 \ \ {\it What Titles Could Not Be} \ \ (to J. Ismael)} \label{Ismael1}

I don't think we have your title and abstract yet for The Clock and the Quantum.  Could I pressure you to send that in very, very soon?

\subsection{From a 3 May 1998 note titled ``Quantum Probabilities'' to Jenann Ismael}

\bq
I have just read your nice paper ``What Chances Could Not Be'' (Brit.\ J. Phil.\ Sci.\ 1996).  In particular, I am quite enthusiastic about your closing discussion on how quantum mechanical probabilities may be construed as solely epistemic in character.  Have you developed this line of thought further in print?  If so, I would very much like a reference to your follow-up articles.

Carlton Caves and I have discussed what I think is a very similar point of view in:
C. M. Caves and C. A. Fuchs,
``Quantum information: How much information in a state vector?''
in {\sl The Dilemma of Einstein, Podolsky and Rosen -- 60 Years Later
  (An International Symposium in Honour of Nathan Rosen -- Haifa,
  March 1995)}, edited by A. Mann and M. Revzen,
  Annals of The Israel Physical Society, vol.\ 12, pp.\ 226--257, 1996.
Also available at the Los Alamos World Wide Web E-print Archive \myurl[http://xxx.lanl.gov/abs/quant-ph/9601025]{http:// xxx.lanl.gov/abs/quant-ph/9601025}.
(See, in particular, the discussion of Section 5.)

Presently, we're in the process of writing a paper solely devoted to the idea, and I would like our bibliography to be as complete as possible.  (This is how I ran across your paper in the first place.)

Thank you for your help.
\eq

\section{17-09-08 \ \ {\it PanSICism} \ \ (to P. W. Shor)} \label{Shor7}

I hope you'll be able to come to the SIC workshop that Steve Flammia and I are putting together.  I'd really like to see this problem knocked out, and think that an intense workshop with friends (a little talking time, a lot more time at the chalkboard instead) might help shake something loose.

I'm keeping my fingers crossed \ldots

\section{17-09-08 \ \ {\it Seeking {\Spekkens}?}\ \ \ (to R. W. {\Spekkens})} \label{Spekkens51}

\brws
How was the contextuality session in {\Vaxjo}?
\erws

The contextuality session seemed to be enjoyed by everyone participating, but all of the rest of the audience (the ones usually riveted about Bell inequality discussions) slept through most of it.  I was a little disappointed that there weren't more conceptual discussions.  Mostly there was focus on technicalities, etc.

\section{22-09-08 \ \ {\it Threat of Double Secret Probation} \ \ (to A. Kent, A. Steinberg, and W. G. Unruh)} \label{Kent16} \label{Steinberg1} \label{Unruh7}

You three are our last hold-outs for supplying titles for your talks at the Clock and Quantum conference next week.  It really would be nice to get them posted on the webpage for anyone making a last minute decision about coming.  Here's the present schedule:
\begin{center}
\myurl{http://www.perimeterinstitute.ca/conferences/clock-and-quantum}.
\end{center}
Think of me as Dean Wormer.  I really need those titles (and abstracts would be nice too).

\section{24-09-08 \ \ {\it Friends of the Tensor Product}\ \ \ (to S. Abramsky, H. Barnum, B. Coecke, K. Martin, A. Wilce and others)} \label{Abramsky1} \label{Barnum24} \label{Coecke1} \label{Wilce21} \label{Martin6}

At the beginning of the summer, my next-door neighbor Keith Rowe, an emeritus professor at University of Waterloo, died after a battle with cancer.  Keith was a category theorist, with a love for his children and Canadian football.  At the reception for his funeral, I believe I spied a paper titled, ``Tensor Products of Categories,'' but I have not been able to find a trace of it since---maybe I am mistaken.  On the other hand, at the end of the summer, Keith's wife Rosemary, came back from her summer stay in their cottage and showed me the papers listed below.  I thought I might share the titles with you as a small tribute to him.  If any of them look to be potentially useful in your own research (particularly the one that may not have been published), I can get you a copy the next time I see you.

\begin{itemize}
\item
K. A. Rowe, ``Images in Topoi,'' Canad.\ Math.\ Bull.\ {\bf 20} (1977), 471--478.
\item
K. A. Rowe, ``Nuclearity,'' Canad.\ Math.\ Bull.\ {\bf 31} (1988), 227--235.
\item
D. A. Higgs and K. A. Rowe, ``Nuclearity in the Category of Complete Semilattices,'' J. Pure App.\ Alg.\ {\bf 57} (1989), 67--78.
\item
K. A. Rowe, ``All Tensor Products are Coequalizers,'' preprint, don't know if it was published.  My summary: ``Let $E$ be a category equipped with a bifunctor.  In most examples, the bifunctor will be an actual tensor product with special properties such as symmetry or associativity, but we make no such assumptions.  We demonstrate that under very mild conditions, every tensor product of algebras with respect to a certain type of monad on $E$ is a coequalizer in the category of algebras.''
\end{itemize}

\subsection{Samson's Reply}

\bq
I wrote a paper a few years back, on `nuclear and trace ideals in tensor $^*$-categories', with Prakash Panangaden and Rick Blute, for which the Higgs-Rowe paper was a very important reference.
\bv
Nuclear and trace ideals in tensored $^*$-categories\\
S. Abramsky, R. Blute and P. Panangaden\\
In {\sl J. Pure and Applied Algebra}, Vol.\ 143, pages 3--47, 1999.
\ev
\eq

\section{24-09-08 \ \ {\it Two Examples}\ \ \ (to R. Healey)} \label{Healey1}

Attached are two examples making the logically pristine point that frequency data {\it alone\/} (nor any empirical data, for that matter) {\it never\/} determine future probability assignments.  I.e., one must feed in a (prior) probability to obtain a (posterior) probability.  Of course, this is just the subjective Bayesian point.  But it is always nice to be reminded just how simple and uncomplicated that ultimate point is.

In the file {\tt t.pdf} look pages 12--17 (of the total 55).  In the file {\tt s.pdf}, look at pages 25--28 (of the total 47).  The latter has the example I just showed you in the stair well.

A really good discussion of this point---I think---can be found in {\Appleby}'s pedagogical paper:  \quantph{0402015}.

As I've tried to convey to you before though, I {\it do not\/} believe that:  Though probabilities are always subjective, the quantum world leaves no empirical mark.  Rather it leaves its mark elsewhere than in telling us how to {\it set\/} probabilities.  I believe it leaves its mark in giving us additional rules to live by beyond Dutch book coherence (when gambling upon the results of our interactions with the external world).  In particular, there is a way to look at the Born rule that shows it not as a rule for {\it setting\/} probabilities, but rather as a rule for normatively {\it relating\/} probabilities.  The subjective component of the prior never disappears; instead the world makes it empirical mark in encouraging us to tie our assignments together in stricter ways than de Finetti and Ramsey had envisaged.

We'll get down to business tomorrow.

\section{26-09-08 \ \ {\it Keynes and Zabell Quotes}\ \ \ (to R. Healey)} \label{Healey2}

Two quotes I had wanted to send you earlier in the week, but forgot to.

Here is a the Keynes quote on Ramsey:
\bq
The application of these ideas [regarding formal logic] to the logic of probability is very fruitful.  Ramsey argues, as against the view which I had put forward, that probability is concerned not with objective relations between propositions but (in some sense) with degrees of belief, and he succeeds in showing that the calculus of probabilities simply amounts to a set of rules for ensuring that the system of degrees of belief which we hold shall be a {\it consistent\/} system.  Thus the calculus of probabilities belongs to formal logic.  But the basis of our degrees of belief---or the {\it a priori}, as they used to be called---is part of our human outfit, perhaps given us merely by natural selection, analogous to our perceptions and our memories rather than to formal logic.
\eq
And a long quote by Sandy Zabell, from ``Ramsey, Truth, and Probability'':
\bq
The key point is that previous attempts to explain induction had attempted to model the process by a unique description of prior beliefs [[references]], or by a very narrow range of possibilities [[references]].  De Finetti realized that because probability is a logic of consistency, one can never---{\it at a given instance of time}---uniquely dictate the partial beliefs of an individual; at most one can demand consistency.  The essence of inductive behavior, in contrast, lies not in the specific beliefs that an individual entertains at any given point in time, but the manner in which those beliefs evolve over time.  [[In this way it is exactly like classical logic:  One is not judged as irrational for starting with the incorrect truth value for some proposition in one's considerations; one is judged irrational only if one makes an incorrect inference in the proof process.---CAF]] \ Let us pause briefly over this point.

I change my mind slowly; you do so with rapidity; you think I am pigheaded, I think you are rash.  But neither of us is of necessity irrational.  Disagreement is possible even if we share the same information; we may simply be viewing it in a different light.  This is what happens every time the members of a jury disagree on a verdict.  Of course it can be argued that the members of the jury do not share the same body of facts: each brings to the trial the sum total of his life experiences, and on juror tries to persuade another in part by drawing upon those experiences and thus enlarging the background information of their fellow jurors.  It is the credibilist view of probability that if you knew what I knew, and I knew what you knew, then you and I would---or at least should---agree.

Such a metaphysical stance may well be, as I. J. Good says, ``mentally healthy''.  But it is an article of faith of no real practical importance.  None of us can fully grasp the totality of our own past history, experience, and information, let alone anyone else's.  The goal is impossible:  our information cannot be so encapsulated.
\eq

\section{27-09-08 \ \ {\it The Old Jenann}\ \ \ (to R. Healey)} \label{Healey3}

From, J.~Ismael, ``What Chances Could Not Be,'' Brit.\ J. Phil.\ Sci.\ {\bf 47}, 79--91 (1996):
\bq
\indent There is one possibility which I have refrained from mentioning because it amounts to the denial that there is anything properly called chance, insofar as it is distinct from any species of subjective probability.  What I have in mind is the view that chance is just epistemic probability of a particular sort:  consider two systems $A$ and $B$, which differ with respect to the probability pertaining to $A$ of some future event type (say a collapse into a state $\psi$ at $t$).  $A$ either will or will not collapse into $\psi$ at $t$, $B$ either will or will not collapse into $\psi$ at $t$, and there is nothing we can now ascertain about either system (e.g.\ no measurement we can perform) which will decide the case for certain.  The occurrence of $e$ is, however, differently correlated with facts about $A$ and $B$ which we {\it can\/} determine and with respect to which they differ, e.g.\ $e$-type events occur to $l/2$ of systems relevantly like $A$ and only $l/4$ of those like $B$.  This means that we can place $A$ and $B$ in ensembles of systems---as alike to them as we can presently ascertain---in which the frequency of $e$-type events is $l/2$ and $l/4$, respectively, and this is what the probabilities refer to.  The difference between systems with different biographies of probabilities (so long as these assign non-maximal probabilities [i.e.\ $0<pr<l$ ] to the same set of events), is not a difference in the properties of those systems but a difference in what we know about them.

Something about so interpreting probabilities in quantum mechanics has struck physicists as objectionably subjective.  Wrongly so, in my opinion.  The epistemic probability of a proposition $p$ can be (and often is) represented as a relation between $p$ and a given agent, but can just as well be represented as a relation between $p$ and a set of propositions, those representing the body of information possessed by the agent in question.  The $t$-chance of $e$ on $A$ can be defined as the probability of $e$ relative to the propositions which describe $A$'s state at $t$ or its pre-$t$ history, i.e.\ as the personal probability of an agent who has no foreknowledge but is perfectly apprised of $A$'s history and the laws which govern its evolution.  There is nothing scientifically disreputable [about] chances so construed, indeed it is not unlikely that many physicists have been operating with such a conception without distinguishing it carefully from the view that the $t$-chances pertaining to $A$ describe objective properties of $A$ such as mass, spin, and the like. The difference between the two views is just the difference there is in general between the view that some `parameter' $P$ is a redescription of those in another set $\{P_1, \ldots, P_n\}$, and the view that $P$ is ontologically distinct from [them] but happens to covary with $\{P_1, \ldots, P_n\}$.  In either case, fixing the values of $P_1, \ldots, P_n$ fixes the value of $P$ in all physically possible worlds, but the value of $P$ varies independently of those of $P_1, \ldots, P_n$ in the set of {\it metaphysically\/} possible worlds only in the latter case.  The value of $P$ fails to supervene on those of $P_1, \ldots, P_n$ {\it iff\/} $P$ is ontologically distinct from $\{P_1, \ldots, P_n\}$.

Interpreted objectively, quantum probabilities describe properties ontologically independent of the system's other properties and partially characterize its intrinsic state, and it must be
(metaphysically) possible for two systems---otherwise alike---to differ with respect to their chances.  Interpreted epistemically, the intrinsic state of a system at $t$ is imperfectly correlated with aspects of its post-$t$ state (the events to which it assigns non-maximal probabilities).  In this case, the probabilities reflect only uncertainty about the future of the system based on knowledge of its present state (excluding, of course, the chances), and it is impossible for two systems---otherwise intrinsically alike at $t$---to differ with respect to the $t$-chances pertaining to each.
This view seems to me completely adequate to the role of probabilities in physics (though it would take a much longer paper to show this), and---of all those described---the most natural.
\eq

\section{29-09-08 \ \ {\it Tumulka's Reply To You} \ \ (to W. G. Unruh)} \label{Unruh8}

I got a kick out of your exchange with Tumulka yesterday---particularly, the part where you proposed your universal 7-second-in-the-barrier theory.  Roderich's reply that his theory has equations, whereas yours has none, makes all the difference---that that is what makes his a {\it scientific\/} theory---took me back a few years to a report I once wrote.  I'll place it below in case you might enjoy it.  Also a little further commentary below it too.  [See 17-11-05 note ``\myref{Halvorson5}{Cash Value}'' to H. Halvorson.]

\section{01-10-08 \ \ {\it Taking Up for {\Spekkens} (this time)} \ \ (to W. C. Myrvold)} \label{Myrvold10}

\noindent From the abstract:
\bq\noindent
The diversity and quality of these analogies is taken as evidence for the view that quantum states are states of incomplete knowledge rather than states of reality. A consideration of the phenomena that the toy theory fails to reproduce, notably, violations of Bell inequalities and the existence of a Kochen--Specker theorem, provides clues for how to proceed with this research program.
\eq
From a section in near the end:
\bq\noindent
Contextuality and nonlocality. The Kochen--Specker theorem [28, 30] and Bell's theorem [31] state that any hidden variable theory that is local or noncontextual cannot reproduce all the predictions of quantum theory. The toy theory is, by construction, a local and noncontextual hidden variable theory. Thus, it cannot possibly capture all of quantum theory. In the face of these no-go theorems, a proponent of the epistemic view is forced to accept alternative possibilities for the nature of the ontic states to which our knowledge pertains in quantum theory. It is here that the novel conceptual ingredients are required. Note that since nonlocality is an instance of contextuality [57], the latter can be considered as the more fundamental of the two phenomena. Indeed, if quantum theory can be derived from a principle asserting that maximal information is incomplete and some other conceptual ingredient, then contextuality may be our best clue as to what this other conceptual ingredient must be.
\eq
From the concluding section:
\bq\noindent
A principle stating that maximal knowledge is incomplete knowledge is likely to serve as a foundational principle in a simple axiomatization of quantum theory. This is the claim that we argue is made plausible by the strength of the analogy between the toy theory and quantum theory.  Nonetheless, this principle is insufficient for deriving quantum theory.  It is intriguing to speculate that we are lacking just one additional conceptual ingredient, just one extra principle about reality, from which all the phenomena of quantum theory, including contextuality and nonlocality, might be derived. To find a plausible candidate for a second such principle, it may be useful to adopt a similar strategy to the one used here to argue for the first principle: do not attempt to derive all of quantum theory, but rather focus on the more modest goal of reproducing a variety of quantum phenomena, even if only qualitatively and in the context of some incomplete and unphysical theory. In particular, attempt to reproduce those phenomena that the toy theory fails to reproduce. Armed with a conceptual innovation that captures the essence of the missing quantum phenomena, a path to quantum theory might suggest itself.
\eq

\section{05-10-08 \ \ {\it Titles}\ \ \ (to R. {\Schack})} \label{Schack138}

Just in from mowing the lawn.  I want to record these titles before I forget them.  More in a few days.
\begin{itemize}
\item
Quantum Bayesian Quantum Certainty Is Not a Moore Sentence
\item
How a Quantum Bayesian Sees Decoherence and Einselection:\ Van Fraassen Reflection Turned Backwards
\end{itemize}

\section{06-10-08 \ \ {\it Kent Peacock} \ \ (to A. Kent)} \label{Kent17}

Speaking of double-slit toy models, see: \quantph{0209082}, the attachment, and the proposal below.  Any interest in having this guy out?

On the double-slit, I also checked out Jammer today to see if I could see anything on historical toy models.  Not too enlightening, but it did lead me to a tidbit that sounds a little bit like something I might have said.  So, all in all, this exercise has been good for me, even if not you!

\section{06-10-08 \ \ {\it SICkening Quantum Mechanics} \ \ (to P. W. Shor)} \label{Shor8}

I wonder if I can push you for an answer on the SIC meeting we're having here Oct 26--30.  We sure hope you can come, but will understand if you cannot.  So far, the confirmed participants are (besides Flammia and myself):
\bv
D. Marcus Appleby
\\ Ingemar Bengtsson
\\ Howard Barnum
\\ Markus Grassl
\\ David Gross
\\ Andrew Scott
\\ Lane Hughston
\\ Martin Roetteler
\ev
I'm also trying hard to bring John Conway and Simon Kochen here.  Roger Penrose has also recently taken an interest in the problem.  I think there's progress to be had if we can just put our minds together for a little while.

\section{06-10-08 \ \ {\it Penrose} \ \ (to D. M. {\Appleby})} \label{Appleby41}

\bma
If I now believe that the wave-function is non-physical that is in large part {\bf because} I spent so long thinking about Bohm.
(The philosophy teacher at my place of work has a quote pinned to the wall of his classroom:  ``the mark of an educated mind is the ability to entertain propositions without believing them''.  In other words, a readiness to explore the consequences of an hypothesis without being first convinced of its truth.) Taken in that spirit I think the Bohm theory is extremely valuable (and let me also say that my impression is that Bohm himself took his theory in precisely that spirit).
\ema
You and Wayne Myrvold both made similar points to me.  And I accept your arguments.

\section{06-10-08 \ \ {\it Rethinking} \ \ (to M. Hemmo)} \label{Hemmo1}

I want to open up and tell you about a struggle I'm having.  As all the travels I've committed to for this year have unfolded (and ones climbed into the middle unexpectedly), I'm afraid I've become more and more broken.  It's come to the point where I can hardly stomach the idea of another trip abroad.  Basically, one trip abroad each month is killing me---professionally, family-wise, everything.

{\it Yet}, this meeting of yours ``The Probable and the Improbable:\ The Meaning and Role of Probability in the Foundations of Contemporary Physics'' {\it subject-wise\/} is one of the most important ones of the year for me and this Bayesian research program {\Caves}, {\Schack}, {\Appleby} and I have been trying to put together.  The idea of not giving our research some representation in Jerusalem is, thus, killing me almost as badly as the thought of traveling again.

So, I'd like to explore the following with you.  Would you consider trading me out with {\Ruediger} {\Schack} as a representative of our research program?  I have asked him if he would give a broad talk, particularly addressing the criticisms we've received over the years (from yourself and others), and emphasizing the kind of ontology that goes with our quantum Bayesian stance (something not usually emphasized elsewhere).  He has said that he would do that, and that he would be very pleased to go to Jerusalem if it is possible.  I think he would be great for the meeting, and Itamar, and all the participants.  And I keep my fingers crossed that you will consider this proposition.

I await what you have to say.

\section{06-10-08 \ \ {\it SIC States}\ \ \ (to S. Abramsky and Y. J. Ng)} \label{Abramsky2} \label{Ng4}

Thanks for giving me an opportunity to describe my beloved ``SICs'' at lunch the other day.  Here are two things you can read that'll give you a little further introduction to the subject (particularly the aspects of it we discussed at lunch).

\begin{enumerate}
\item
The proof that SICs are as close as one can get to an orthonormal basis on the cone of positive operators:
\arxiv{0707.2071}.
\item
The proof that SICs are maximally sensitive to eavesdropping:
\quantph{0404122}.
To put things more into the context of your axiomatization, Samson, of orthogonal bases, it would be better to have a result proving that SICs form a ``least clonable'' set of states, but at the moment I don't have that.  Though I suspect it is true---for I would guess that any spherical two-design is a least clonable set.  Still, maybe this will be enough to get you thinking on the idea.
\end{enumerate}

Finally, let me attach some of the unpublished stuff I was telling you about---namely, how SICs give a particularly clean way to think of the Born rule as simple modification to the law of total probability.  The attached document comes from a grant proposal I wrote recently, but I hope to get those equations into an actual paper soon.  Anyway, the words there will, I hope, also convey some feeling for the big picture of my foundations program.

Samson, I guess I will see you in New Orleans next week.  Thank you both again for coming to {\sl The Clock and the Quantum}.  I hope you enjoyed it roundly.

\section{06-10-08 \ \ {\it SIC Invitees}\ \ \ (to S. T. Flammia)} \label{Flammia8}

I've updated the Google document.  And also nudged one last time Shor, Conway, and Kochen.  I think if they can't come, we keep it a small intimate meeting and be done with it.

I also got Roger Penrose fairly interested in the problem last week.  He worked out dimension 2, 3, 4 solutions in his head during Harvey Brown's talk.  (Same method, as far as I can tell, as Lane Hughston's \ldots\ though he made his work in $d=4$, and the last I heard Lane hadn't.)  He said the problem related to something he never quite solved in his PhD thesis.  Funny.  I encouraged him to come too, but he said he had too many other obligations.

\section{06-10-08 \ \ {\it SIC Invitees, 2}\ \ \ (to S. T. Flammia)} \label{Flammia9}

\bstf
Wow, Penrose worked out the $d=4$ solution in his head?!?  I'm impressed.  I'd be interested to know what other problems it relates to.
\estf

OK, I updated it again, after hearing from Shor.

Unfortunately, I didn't understand anything he said either about his thesis or $d=4$.  One question, though, did arise that I can get my head around.  Do we know that we can't construct a SIC on two qubits purely of tensor product states?  Of course one can't tensor two SICs and get a new SIC, but that's a different question.  The only demand here is that each SIC element simply be a product state of any variety.

Also, there appears to be a dangerous bug in this Google document thing.  I can look at the documents of another Chris Fuchs out there, one at NDSU (wherever that is).

\section{06-10-08 \ \ {\it A December Visit} \ \ (to K. A. Peacock)} \label{Peacock1}

Yesterday, I read your paper ``Aristotle's Sea Battle and the Kochen--Specker Theorem,'' and enjoyed it very much.  By my reckoning, what you write is still at the stage of an ``idea for an idea'', but it is a good idea for an idea!  It'd be great to see how to give it some firm substance.  I myself tend to believe as well that the ultimate import of KS is that it indicates that the universe isn't a block.  I'll attach a document that says as much.  [See ``Delirium Quantum,'' \arxiv{0906.1968v1}.]  Surely you won't like aspects of it, but perhaps the first and the last sections will tickle you nonetheless.  You might also get a kick out of perusing a document I'm presently putting together titled ``My Struggles with the Block Universe''.
Do a search on the terms ``block'' and ``Kochen'', and it should take you to the right places.

\section{07-10-08 \ \ {\it Rethinking, 2} \ \ (to M. Hemmo)} \label{Hemmo2}

Thank you so much.  And I hope that Itamar will not be too hurt.  I feel absolutely bad, but everything I told you yesterday is true, and I don't feel that I will be able to pull myself across the ocean again in December.  In fact, my wife just reminded me that my second daughter's birthday is Dec 17, and I had forgotten.

So again thank you for going easy on me.  And thank you very much for allowing {\Ruediger} the opportunity to represent our work and be an ear for any difficulties with it that might be sounded at the meeting.

\section{07-10-08 \ \ {\it Kauffman's Email}\ \ \ (to L. Smolin)} \label{SmolinL13}

Don't forget to send me Kaufman's email address.  I'll send him an invitation to give a foundations seminar.

I should tell you a funny story.  A couple months ago, while I was in Austria, I got a note from Rob Spekkens in Cambridge saying that he had run across a nice set of all six volumes of the collected papers of C. S. Peirce, and he offered to buy them for me.  When he told me the price, I was floored:  The whole set (hardcover) was only 45 pounds!  In contrast, I looked in the index I've made of my home library and found that I had paid 20 quid for Vol 8 {\it alone\/} of the collected papers of Bertrand Russell (when I myself was in Cambridge last).  When I wrote back to Rob accepting the offer, I noted that the bargain he found reveals just how much the Brits value American pragmatism!

I think the Kauffman will be a lot of fun, particularly if he gets good and radical.

\section{07-10-08 \ \ {\it SICkening Quantum Mechanics}\ \ \ (to S. T. Flammia)} \label{Flammia10}

Well, we came close.  See Kochen's note below: ``I couldn't come, but John thought he might be able to make it.  Today he told me that he has to stay in Princeton for family reasons.''

If you give no objection, I think I'll close the ranks now.  We'll have a meeting of:
\bv
D. Marcus Appleby\\
Ingemar Bengtsson\\
Howard Barnum\\
Steve Flammia\\
Chris Fuchs\\
Markus Grassl\\
David Gross\\
Andrew Scott\\
Lane Hughston\\
Martin Roetteler
\ev
and any locals who might want to show up.

\section{08-10-08 \ \ {\it Another Title}\ \ \ (to R. {\Schack})} \label{Schack139}

\begin{itemize}
\item
Reviving Feynman's `Only Mystery' of Quantum Mechanics in Bayesian Terms
\end{itemize}

\section{09-10-08 \ \ {\it REU Pestering}\ \ \ (to L. O. Clark)} \label{Clark1}

I liked Mark and Meg's work\footnote{Maegen Demko and Mark Layer.}, but it is true that in the end I was not able to use much of it.  To be honest, the most important thing I got from it was learning about Seth Sullivant's work (Ref 2 in the report):
\bq
\noindent ``Statistical Models are Algebraic Varieties,''\\
\noindent \myurl[http://www.math.harvard.edu/~seths/lecture1.pdf]{http://www.math.harvard.edu/$\sim$seths/lecture1.pdf}.
\eq
That work puts the SIC representation of quantum states into a nice context---the set of quantum states corresponds to a statistical manifold associated with an intriguing quadratic variety.  It was invaluable for me to learn that.  If you think you can turn it into a publication for other purposes, go for it!  It'd be great for them.

\section{10-10-08 \ \ {\it Block U Alternative}\ \ \ (to T. Duncan)} \label{Duncan6}

\btd
What about ``contingent universe'' as a name for the alternative to the block universe?
\etd

I think I like that better than ``open universe,'' which is the only competitive term that I think I've noticed before.  \ldots\  Well actually, John Wheeler used the phrase ``participatory universe'' and surely he was contrasting it to (the ideas underlying) the block universe.
\begin{flushright}
\baselineskip=3pt
\parbox{3.5in}{
\bq
\noindent
[T]he universe is wild---game flavored as a hawk's wing.
\\
\hspace*{\fill} --- Benjamin Paul Blood
\eq
}
\end{flushright}

\subsection{Todd's Reply}

\bq\noindent
Yes, I also like contingent better than open, especially since open has a geometrical/GR meaning.  Malleable universe is also not bad, in the spirit of your Pauli Project notes. The definition of contingent that made me latch onto it was ``possible but not certain to occur,'' from \myurl{http://wordnetweb.princeton.edu/perl/webwn?s=contingent}.
\eq

\section{10-10-08 \ \ {\it Block U Alternative, 2}\ \ \ (to T. Duncan)} \label{Duncan7}

This one goes too far, but inspired by looking in wiki definitions:  ``alchemical universe''.  Thinking in particular of definition 3:  ``Of or pertaining to the creation of something special out of a common material.''

\section{14-10-08 \ \ {\it New Orleans Books} \ \ (to H. Barnum)} \label{Barnum25}

By the way, at least there's excellent book shopping here in the French Quarter.  So, maybe not all of the trip is a waste.  I just picked up these (recording mostly for my own record):
\begin{enumerate}
\item
Jonathan Arac and Barbara Johnson, {\sl Consequences of Theory}, PB, \$5.00.
\item
Justus Buchler, {\sl Metaphysics of Natural Complexes}, PB, \$12.50.
\item
James A. Connor, {\sl Kepler's Witch}, HC, \$9.00.
\item
Christopher Norris, {\sl Derrida}, PB, 8.00.
\item
Harald Ofstad, {\sl An Inquiry into the Freedom of Decision}, HC, \$18.00.
\item
Joachim Schulte, {\sl Wittgenstein}, PB, 6.00.
\end{enumerate}

I also found a two volume set of Lotze's {\sl Microcosmus}, 1885 edition.  Thinking hard about that one, but it's \$100 and they wouldn't bargain down.

\section{15-10-08 \ \ {\it Guilty, but Committed} \ \ (to S. Capelin)} \label{Capelin1}

I must plead guilty.  I got an upgrade on my flights yesterday and found that I essentially slept the whole way here.  Then upon landing, I couldn't resist the idea of reading Wittgenstein in a New Orleans oyster house, while eating fried catfish and hushpuppies.  Finally, I fell completely weak and went shopping at the used bookshops in hopes of finding a new gem to tingle my philosophy (record below).  [See 14-10-08 note ``\myref{Barnum25}{New Orleans Books}'' to H. Barnum.]

{\it But}, I will do my dead level best to hammer you something out tonight.  So please don't give up on me \ldots\ but don't get too angry with me either!  All of these things are important for the grand scheme I have in mind.

\section{16-10-08 \ \ {\it Aaronson's Book Proposal} \ \ (to S. Capelin)} \label{Capelin2}

Every other Christmas or so for the last few years, I have inevitably included one of chef Jamie Oliver's cookbooks in my wife's melange of presents.  I suppose I do this because we enjoy his shows on television, and when gift time comes, it's an easy idea.  I suspect that's the case for a lot of other husbands out there as well, but probably more importantly, it may also hold for thousands of women who reward themselves from time to time with the purchase of another of Oliver's books.  Reading over Scott Aaronson's manuscript reminds me of the happy times I have had reading his blog, and leads me to wonder whether this, at least in small part, might be the business model destined for the present book.  A blog is not the same as a syndicated television show, of course, but it is a wide and ever-renewing vehicle of attention nonetheless.  Certainly, for those of us in the field of quantum information, Scott is as close as we have to a ``Naked Chef.''

I like this book.  I like it a lot.  Reading parts of it have made me feel young again.  And parts of it reminded me of speech patterns I used to speak in---``a polynomial shitload,'' ``a problem called DUH,'' ``The-Mojo-All-Along Theorem''---but have gotten out of the habit of, presumably because of the culture around me that scientific writing should be above that.  I love David Mermin's and Nielsen and Chuang's books on quantum computing, but neither have made me spontaneously laugh out loud like this one.  To the extent that pleasures like this open the thought receptors, I count its method a success.  And there are so many tidbits of wisdom in it---from the mathematical to the philosophical to the social\footnote{Here's an example of the social that tickled my fancy:  ``In my view, probability theory is yet another example where mathematicians immediately go to infinite-dimensional spaces, in order to solve the problem of having a nontrivial problem to solve!''}---even I was surprised by what I found myself thinking about as I absorbed Scott's thought style.

Which leads me to what I view as the greatest intellectual strength of this book---the ebb and flow of topics, and the unique way Scott presents them flowing them into one another.  What one learns from this book that one will not learn from any other book yet in print is that quantum computing is a very {\it big}, very {\it deep\/} subject.  The thoughts surrounding it are not narrow and specialized, but rather have very much to contribute theoretical physics as a whole.  Scott gives a sense in this book that one really might tie this vast landscape together---that one really might one day put all these subjects (from cosmology to free will to quantum foundations to the foundations of mathematics) in service of each other.  Having this in print would be a valuable service to the physics community.

But the presentation of the book clearly needs work, and this is where I have the hardest time pinpointing what I feel.  When I first started reviewing this manuscript, I thought ``Oh no!''  One chapter, two chapters, and the beginning of the third were consumed, and despite my predisposition for not wanting to think it, I thought things were a bit silly.  How would one get into a book like this?  Slowly and imperceptibly initially, however, I did indeed get into it.  Once I started to take the discussions at Scott's pace and Scott's way, and to think about the questions for the non-lazy, I found that my thinking the exposition silly and flippant simply disappeared---the stuff I was reading pretty much flowed into me without resistance.  That very much impressed me.  But I also know that I gave it probably more of a college try than a new reader would---I have seen Scott lecture, and I have read his blog, and felt that there must be something of substance here.  So I plodded on until the tractor beam took hold, but maybe not everyone else would.

How to smooth this rough and risky (risky for sales) edge off, I don't know:  I'm not the author; I'm not the editor.  But I do very much think this book should be out there.

\section{23-10-08 \ \ {\it SIC Conference -- Titles and Abstracts}\ \ \ (to L. P. Hughston)} \label{Hughston1}

We'll miss you!

I talked to Roger Penrose about the SIC-existence problem when he was here a couple weeks back for a quantum foundations conference.  The first exchange on it was cute.  I briefly described the problem to him before the start of a talk.  At the end of the talk, he tapped me on the shoulder and said, ``Well, I can see it works in $n=2$, 3 and 4.''  As far as I could tell, his $n=3$ solution sounded pretty much the same as your own.  When it got to $n=4$, I didn't understand him as well.  But the main cute thing was, it seems he did it all in his head in those 45 minutes.  To the extent that he got a chance to think about it later, I think most of his thought was devoted to seeing if he could recover his $n=3$ solution from his $n=4$ solution by an appropriate projection.  I think things there were less conclusive then.

\section{24-10-08 \ \ {\it P.S.}\ \ \ (to A. M. Steinberg)} \label{Steinberg2}

\bams
By the way, I'm also still curious to hear why my suggestion about
a Bayes-like rule for noisy projectors fails!  (As for the more general
discussion of what probability ``means'' and what valid gambling
strategies are, maybe that will await a pub one day.)
\eams

I will.  In fact, I plan to put it in my talk on Sunday.  I'll send you a link to that after it is posted on {\tt pirsa.org}.

About the second issue, it is not nearly so philosophical as to require a pub.  I'll send you the appropriate place in our writings to read about it.

Thanks for the opportunity of interaction with your group today.  At Atticus, I got one amazing bargain (all five volumes of {\sl Early Defenders of Pragmatism\/} for \$65.00 total) and two other good books (on pragmatism of course) but not such great bargains.

\section{25-10-08 \ \ {\it Fuchs Bio and Blurb for TGQI}\ \ \ (to C. M. {\Caves})} \label{Caves97}

\noindent {\bf Bio:}\medskip

Christopher A. Fuchs is currently a Long-Term Visitor at the Perimeter Institute for Theoretical Physics in Waterloo, Canada and an Adjunct Professor at the University of Waterloo.  For seven years previous to this, he was a research staff member at Bell Labs, Lucent Technologies in Murray Hill, New Jersey, both in the computer science and the physical science divisions.  Fuchs received a PhD in 1996 from the University of New Mexico, under the supervision of Carlton M. {\Caves}.  He then held postdoctoral positions at the University of {\Montreal}, Caltech (where he was a Lee A. DuBridge Prize Postdoctoral Fellow), and Los Alamos National Laboratory.  His awards include the Albert A. Michelson Prize Lectureship at Case Western Reserve University and an E.T.S. Walton Award from Science Foundation Ireland.  He was an associate editor for the journal Quantum Information and Computation from 2000--2008, and the organizer of a dozen international meetings in quantum information and quantum foundations.  He is an author of more than 50 scientific papers and two outlandish books of email correspondence with many founders of the field.  More information can be found by googling the terms ``Fuchs'' and ``quantum'' simultaneously---at the moment, he comes to the top of the heap.\medskip

\noindent {\bf Blurb:}\medskip

Ten years ago, the field of quantum information was hardly a recognized branch of physics, even by the most forward-looking physics departments.  Just ask anyone who was in the job market at the time.  But things have changed, and the very existence of the APS Topical Group on Quantum Information is a harbinger of the idea that quantum information not only has a home within physics, but is a school of technique that will have an impact on physics exploration as a whole.  By our ways, we can help change the world of physics.  In fact, with over 900 members presently, it does not look far-fetched that with an enthusiastic recruitment effort, the GQI could attain APS division status.  This would give our field a clout and visibility that should be of great benefit to our fresh graduates and help further infrastructure and funding stability the field is sorely in need of.  We should also establish premier awards---paying careful attention to make them significantly competitive and desired---for the best PhD work and research in general.  This would bring the best in our field to full view, with all the benefit that entails for everyone.  Furthermore we should work hard to lower the gender barrier of physics, and as it seems more women are already attracted to quantum information than many other branches of physics, we have the opportunity to build on an existing base to make a real impact.  Finally the GQI can do more to disseminate awareness of the potential of quantum informational techniques for physics problems everywhere, and through the organization of strong specialized conferences and their associated advertisement in departments across the States, we can help make our field a resource for physics that's long here to stay.

\section{05-11-08 \ \ {\it As You Wake Up}\ \ \ (to R. {\Schack})} \label{Schack140}

Thank god America has finally had some sense!  The news of our new president is very good indeed---maybe there is a chance now that our world can move forward, rather than back.

\section{05-11-08 \ \ {\it More Atticus}\ \ \ (to myself)} \label{FuchsC21}

\bq\noindent
Moore, Addison W.\medskip\\
{\sl The Collected Writings of Addison W. Moore}. 3 Volumes \medskip\\
Bristol: Thoemmes Pr 2003, 1st Edition, No Jacket, Hardcover, As New Book \medskip\\
Description: After John Dewey, Addison W. Moore was recognized as the chief spokes\-man for the instrumentalist version of pragmatism. Never before available, this complete collection of Moore's work contains dozens of philosophical articles, essays, book reviews, writings by other philosophers, and reviews of his work, together with his book, {\sl Pragmatism and its Critics\/} (1910). Vol 1: {\sl Writings, 1896--1910}. Vol.\ 2: {\sl Pragmatism and Its Critics}. Vol.\ 3: {\sl Writings}, 1911--1927.\medskip\\
Price: USD 45.00 other currencies, order no.\ H2409
\eq

\section{05-11-08 \ \ {\it Slusher Lunch} \ \ (to L. Hardy and R. W. {\Spekkens})} \label{Spekkens51.1} \label{Hardy29}

Would either of you be available for an early lunch tomorrow?  Dick Slusher, my old manager from Bell Labs (and now a director of a quantum information lab at Georgia Tech), is visiting IQC and giving their colloquium tomorrow.  I'm in charge of him from 12:00 to 1:30, and I planned to lunch him at PI and give him a tour.  Dick has always had a kind of inchoate interest in quantum foundations, and I thought it'd be fun for him if we could razzle, dazzle him during lunch with some foundationsy things.  Dick always wants to know, ``how is it relevant for my physics?''  I'd really appreciate it if we could give him something to think on, as this is exactly what we should be about.  Would a 12:30 lunch work for either of you?  I hope you can come.  Please let me know if you can, and I'll dig you up at lunch time.

BTW, seeing Wilczek's vision of ``the grid'' as the ultimate reality was quite revealing to me.  Particularly, his use of ``quantum fluctuations'' made me think he could use a good dose of the quantum foundations mysteries---it might serve him well in these visions of ultimate visions.

\section{06-11-08 \ \ {\it Ghandi} \ \ (to G. L. Comer)} \label{Comer119}

I have no answers for you, of course.  But as you can see from my own emailing methods, I am a fan of the one-hand clapping method.  I write most always with the intention that I will find my own answers in the writing itself.

\section{06-11-08 \ \ {\it The New Things} \ \ (to R. E. Slusher)} \label{Slusher21}

I feel like I didn't answer you very adequately last night when you asked if there's been any progress in SIC world.  What I wanted to express is that though not much new is known about the existence of these things, we have learned plenty about what quantum mechanics looks like when expressed in terms of them.  On the one hand the state space seems to have a really intricate structure, but on the other hand we have hints that all the intricacy may arise from the simple equation I showed you last night.

Unfortunately, I don't have any of this in a completed paper yet.  But, I attach three pieces of evidence to show you that I haven't been sitting on my duff.  1) The scientific piece of a grant proposal I wrote (and obtained).  It is actually a decent overview of the new simple equations I showed you last night (the expression for unitary evolution and the modification of the law of total probability).  2) The beginnings of a paper on the actual shape of quantum state space.  3) A piece on rederiving that state space, taking the modified law of total probability as an axiom.

I hope you'll find some of this a little enjoyable.

\section{06-11-08 \ \ {\it The Man Who Knew Certainty}\ \ \ (to R. {\Schack} \& C. M. {\Caves})} \label{Caves98} \label{Schack141}

\myurl[http://pages.stern.nyu.edu/~abranden]{http://pages.stern.nyu.edu/$\sim$abranden/}

\section{06-11-08 \ \ {\it Wilczek World} \ \ (to L. Hardy)} \label{Hardy30}

\blh
I didn't go to Wilczek's talk.  I probably should have done ``for the cause'' but needed a break.
\elh

Yeah, it was funny:  I was thinking of the things I had heard about how he was somewhat against PI having involvement in foundations, and at the same time seeing him talk about things that were in bad need of some foundational analysis.  The most fundamental stuff he calls, ``the grid.''  As far as I could tell, he means by this that spacetime is filled with a quantum field, continually undergoing ``quantum fluctuations''.  It was the usual particle-physicist speak.  But what does this phrase quantum fluctuation mean?  The only sense I've ever been able to make of it is that it signifies a quantum state in superposition.  The can of worms that's opened up here is precisely the meaning of the wavefunction and the attendant issue of quantum measurement.  Maybe it'd be worth my while to read his latest book first, and then try to start up a dialog with him.

\section{07-11-08 \ \ {\it Thoughts on Consciousness}\ \ \ (to T. Duncan)} \label{Duncan8}

\btd
I've been drafting a paper to organize some thoughts on how to incorporate subjective awareness into our understanding of the natural world. I've attached the latest version in case you have time to take a look \ldots\ critical feedback is most welcome.
\etd

Thanks, I enjoyed that.  I particularly liked this image of trying to recover color from a black and white drawing.

Your considerations reminded me of some of the ones in {\Schroedinger}'s little book, {\sl Nature and the Greeks}.  As I recall, he too held the opinion that any physical model must exclude the idea of subjective experience by its very nature, by its construction.  Thus it is no surprise that subjective experience can never be found back in it.  But then Copenhagen quantum mechanics perplexed him greatly, for he found subjective experience to be an integral part of the Copenhagen model.  How could that be---he thought it must be a confusion in the Copenhageners.  I think the book is worthwhile reading and deeply connected to your paper.

Also, I have to admit, I finally printed out ``An Ordinary World.''  I hope to take it to Australia with me next week.

\section{10-11-08 \ \ {\it Interview Questions -- The Better Me}\ \ \ (to D. T. Pope)} \label{Pope1}

\bdtp
What do you think of the many worlds interpretation?  (For the purposes of this question, you can take the many worlds interpretation to be the idea that the universe splits
into multiple copies whenever a measurement happens.)
\edtp

I think that despite all the attention that has been lavished on it by the popular science magazines (and science fiction magazines), the many worlds interpretation has little to no content.  You say, ``the universe splits into multiple copies whenever a measurement happens,'' but the whole motivation for the interpretation is to get away from ever making mention of ``measurement'' at all---the many-worlds interpretation is one species of reaction to the fear that the word ``measurement'' is too loose and fuzzy and ill-defined to be a significant piece of physical theory.

How does it tackle the problem that the quantum world seems to be making an on-the-fly decision whenever it gives us the outcome of a measurement?  It says instead, the quantum world is making these decisions all the time, everywhere and everywhen.  But since the quantum formalism gives no mechanism for this, and also hints of no bias for which way these decisions will go, the many worlders postulate instead that the world as a whole branches.  It doesn't decide; it just branches.  So that for those events in which there are observers around, in one branch the observer sees this outcome, and in the other branch he sees that one.

The trouble I see is that it is just a story.  A story that could have been made up, and in fact was made up, before quantum mechanics.  There is nothing particular to the quantum formalism about it.  Everything that can happen, does happen---you might find that in the longing poetry of a parent who has recently lost a child.  Why is my world the way it is?  {\it Because\/} I am on this particular branch.  Which branch?  {\it My\/} branch.  But there's always a better branch where everything is happy again.

\bdtp
What do you think of the pilot-wave interpretation?
\edtp

I think it has had 51 years of a chance but has never made a serious impact on physics (since its formulation in 1957).  The de Broglie -- Bohm idea was to supplement the equations of standard quantum mechanics for the wave function alone, with equations for where the particles should actually be at all times \ldots\ all said {\it on the condition\/} that one knows where the particles were initially.  But one never knows such.  So the pilot-wave theory adds a layer of equations that are never in the end used, outside of giving a warm fuzzy feeling to those inclined to take it.  It would be different to me if the addition of those equations gave rise to simplifications in the calculations I need to do as a physicist.  Then I would respect them more.  But as it is, they just seem to be dangling appendages that have yet to offer any help in our getting to the next stage of physics.

\bdtp
What do you think of the Copenhagen interpretation?  (For the purposes of this question, you can take the Copenhagen
interpretation to be the idea that in quantum physics we can only know about the results of measurements.  I.e. that it's meaningless talk about what electrons are doing when
we're not measuring them.)
\edtp

The key idea behind modern variants of the Copenhagen interpretation is that quantum mechanical wave functions represent statistical information, full stop.  That is, the wave function is not a property or value or quality {\it intrinsic\/} to a quantum system.  Instead, when a physicist writes down a wave function, he is writing down his supposed information and nothing more.  Information about what?  His information about the {\it results\/} or {\it outcomes\/} of measurement interactions.  It is not information about unknown hidden variables or pre-existent properties.  Rather, it is his best statistical information on what {\it will come about\/} if he interacts with a quantum system this way or that way.

Opponents of the Copenhagen interpretation often present it in a negative light.  They see it as an arbitrary injunction to ask no questions about what a quantum system is {\it really\/} doing.  But I would say it is simply silent on that, recognizing instead an aspect of reality that is much deeper and wholly unexpected before quantum mechanics came onto the scene.  It recognizes that in interaction, new things actually do come under the sun.  That the big bang is not an isolated far off event, but that little {\it metaphorical\/} versions of it are happening all the time.  And we have direct knowledge of this; for we see it any time a quantum experiment is done.

There is no story of what electrons are {\it really\/} doing simply because there is no completed story of what the universe as a whole is doing anyway.  Parts of it are being created on the fly all the time.  Quantum wave functions being purely information in character is but a stark recognition of this wonderful property of existence.

Now, how does this contrast with my insults to the many-worlds interpretation?  Why would I say these ideas have content when the MWI does not?  The recent wealth of discoveries in quantum information has taught us particularly the value of looking at quantum mechanics through information theoretic eyes.  Particularly, to do quantum information, we've all discovered the necessity of first going out and purchasing all the textbooks on information theory (written by authors who knew nothing of quantum mechanics).  To our great surprise:  When one looked at the properties of information written abstractly, one found the properties of quantum wave functions.  If it looks like a duck, walks like a duck, honks like a duck \ldots\ it's probably a duck.

\bdtp
In the double-slit experiment with electrons, when you put detectors
next to each of the slits, the interference pattern disappears.
Why is this? (This is for a section on measurement disturbance. If possible, it would be fantastic for you to talk about the idea that measurement at
the quantum level is a somewhat disruptive and invasive process. E.g.\ kind of like kicking something.)
\edtp

It is a different experiment.  Why would anyone have ever expected its results to be the same as the original one?  Aha, only if they had expected the experiment itself---the interaction itself---to be an inconsequential component in the phenomenon under study.  But apparently it is not, and that is the lesson of quantum mechanics.

Things like this are encountered all the time in the sphere of psychological study.  My wife asks me as we drive down the road, ``What do you want for dinner?''  I first say, ``I have no idea, I'm not hungry.''  But soon I find myself revising the opinion, feeling the stomach acids start to move in my stomach.  If she had not asked me, I might have remained without appetite for a few more hours of driving.  But that's psychology.  What is interesting is that we now find this phenomenon in our most fundamental physical theory.  But that is no reason to think that the phenomenon must then be secondary and that we have been fooled by thinking we had a fundamental theory after all.  Instead it is a call to recognize, ``It is a world sensitive to our touch.''

\bdtp
How big is the economic impact of quantum physics due to technologies
such as computers, iPODs and other familiar electronic devices?
\edtp

Quantum computers will be very big.  Ten years ago we already knew they would be exciting for what they can do in certain number theoretic problems, like factoring large numbers, and the promise they held for efficiently simulating more exotic physical systems.  But make some quantum computing hardware available to hackers and lord knows what we will find can be done with them.  I think we have no clue and cannot foresee.  All we can do is fulfill this urge to prod matter to do new, exciting things for us.

\bdtp
In the double-slit experiment with electrons, what happens in between
the source and the detector?
What are the electrons doing in between the source and the detector
and how do they pass through the slits?
\edtp

These questions are not the subject matter of quantum mechanics.  See above.  To think that they should have an answer is to think that the universe is like a book already written.

\bdtp
How important is quantum physics to physics overall and our
understanding of the universe?
\edtp

I think we (physicists, mathematicians, computer scientists, philosophers) have taken only the tiniest baby steps toward incorporating the lessons of quantum mechanics into our worldview.  I think on this passage of G. K. Chesterton,
\bq
   There are some people---and I am one of them---who think that the
    most practical and important thing about a man is still his view of
    the universe.  We think that for a landlady considering a lodger it
    is important to know his income, but still more important to know his
    philosophy.  We think that for a general about to fight an enemy it
    is important to know the enemy's numbers, but still more important to
    know the enemy's philosophy.  We think the question is not whether
    the theory of the cosmos affects matters, but whether in the long run
    anything else affects them.
\eq
From this point of view, it is imperative to get the story of quantum mechanics straight.  What will matter in the end is only how it affects our view of the cosmos.

\section{10-11-08 \ \ {\it Dry Run} \ \ (to A. Kent)} \label{Kent18}

Here was a dry run I took for myself before the interview with Damian that you egged me on about.  [See 10-11-08 note ``\myref{Pope1}{Interview Questions -- The Better Me}'' to D. T. Pope.] Of course, I didn't repeat anything written below (my mind doesn't seem to work that way)---so the dry run was really a personal run.  I suspect you'll agree with very little of what I write here, but maybe that's why I send it to you.  I hate being in the limelight like this; I find it physically painful really.  So, I send part of the pain on to you.

\section{11-11-08 \ \ {\it Deleuzean Difference}\ \ \ (to S. Diamond)} \label{Diamond1}

It was good meeting you the other day, when we were all talking with Stu Kauffman.

I wanted to follow up on what I had asked about Deleuze and ``difference''.  Reading the Wikipedia article on Deleuze this morning, I came across this description of his metaphysics:
\bq
Deleuze's main philosophical project in his early works (i.e., those prior to his collaborations with Guattari) can be baldly summarized as a systematic inversion of the traditional metaphysical relationship between identity and difference.   Traditionally, difference is seen as derivative from identity: e.g., to say that ``$X$ is different from $Y$'' assumes some $X$ and $Y$ with at least relatively stable identities.  To the contrary, Deleuze claims that all identities are effects of difference.  Identities are not logically or metaphysically prior to difference, Deleuze argues, ``given that there exist differences of nature between things of the same genus.''  That is, not only are no two things ever the same, the categories we use to identify individuals in the first place derive from differences.  Apparent identities such as ``$X$'' are composed of endless series of differences, where ``$X$'' $=$ ``the difference between $x$ and $x^\prime$'', and ``$x$'' $=$ ``the difference between\ldots'', and so forth.  Difference goes all the way down.  To confront reality honestly, Deleuze claims, we must grasp beings exactly as they are, and concepts of identity (forms, categories, resemblances, unities of apperception, predicates, etc.)\ fail to attain difference in itself.  ``If philosophy has a positive and direct relation to things, it is only insofar as philosophy claims to grasp the thing itself, according to what it is, in its difference from everything it is not, in other words, in its internal difference.''
\eq

I suppose reading something like this once before was the source (in the dark recesses of my mind) that led of my query.  The Deleuzean idea must have caught my eye because of my own thoughts on how best to interpret what is happening in the quantum mechanical measurement process.  Attached is a little artsy piece of mine that lays that out a little more.  [See \arxiv{0906.1968}.] Sections 1 and 3.2 to 3.4 give some sense of what I'm talking about.

In any case, meeting you has brought Deleuze back onto my radar screen.  I'm packing up his book {\sl Bergsonism}, which has been unread on my shelf, to take with me on a short trip to Australia in a couple of days.  Thereafter, I hope to get hold of his book {\sl Difference and Repetition}.

\section{12-11-08 \ \ {\it Quantum Systems Are Cranberries?}\ \ \ (to J. A. Smolin and others)} \label{SmolinJ10}

If I had known the answer would turn out to be this simple, I would have lost interest in the hunt years ago!
\bq\noindent
\myurl{http://www.nytimes.com/2008/11/12/dining/12cran.html?8dpc}
\eq
``The Zing Starts Here''

\section{12-11-08 \ \ {\it Slightly Zingier Explanation} \ \ (to A. Kent)} \label{Kent19}

See Footnote 10, at the bottom of page 9, of
\bv
\arxiv{quant-ph/0205039}.
\ev

\section{12-11-08 \ \ {\it Feynman Comment?}\ \ \ (to N. D. {\Mermin})} \label{Mermin138}

Did you somewhere in print comment on this quote of Feynman:
\bq\noindent
  We choose to examine a phenomenon which is impossible, {\it absolutely\/}
   impossible, to explain in any classical way, and which has in it the heart
   of quantum mechanics.  In reality, it contains the {\it only\/} mystery.
   We cannot make the mystery go away by ``explaining'' how it works.  We will
   just {\it tell\/} you how it works.  In telling you how it works we will
   have told you about the basic peculiarities of all quantum mechanics.
\eq
It is from the first chapter in volume 3 of the Feynman Lectures, and concerns the double slit experiment with individual electrons.  It seems like you had a discussion somewhere of how these double slit considerations cannot be nearly as decisive as Bell tests.  I'd like to track that down and use some of the material in a paper I'm putting together.\footnote{The remark in mind may have been from Braunstein and Caves, rather than Mermin:
\bq
So what is special about Bell inequalities if the two-slit experiment already rules out objective properties?  The immediate problem is that the Feynman argument does not provide a clear-cut criterion for objectivity.  If one tries to formulate such a criterion, as we do below, one confronts the difficulty that the probabilities $p(x|+1)$ and $p(x|-1)$ posited by Feynman are probabilities involving noncommuting observables (electron position at the screen and electron position at the slits); such probabilities are not defined in quantum mechanics, so it is not surprising that a quantum description is inconsistent with an objective description that posits such probabilities.  If one tries to \emph{measure} the probabilities $p(x|+1)$ and $p(x|-1)$, one must modify the experimental apparatus so radically---perhaps by blocking one slit or the other---that the interference pattern is destroyed.  We can never be sure that the inconsistency pointed out by Feynman is not due to our having ``disturbed'' the electron's final position on the screen by our measurement of which slit it went through.

The surprising feature of the standard Bell inequalities is that they avoid this difficulty.  They provide a criterion for objectivity that involves only pairs of commuting observables---thus probabilities that are defined in quantum mechanics---even though the criterion is derived from a ``grand'' joint probability that is not defined in quantum mechanics.  It is perhaps worth considering more fully this contrast between Feynman's argument, which attempts a criterion for objectivity for a single system, and Bell inequalities, which involve two physically separated systems.  Information Bell inequalities provide a natural vehicle for such consideration.
\eq
From S.\ L.\ Braunstein and C.\ M.\ Caves, ``Wringing out better Bell inequalities,'' Ann.\ Phys.\ {\bf 202} (1990), 22--56.}

Hope you're well and still thinking about coming to PI for a visit.

\section{12-11-08 \ \ {\it Feynman Comment?, 2} \ \ (to N. D. {\Mermin})} \label{Mermin139}

I've looked all through {\sl Boojums\/} (of which I again have a copy; got it for \$8.95 in Madison, NJ as I recall), but haven't found anything.

\bdm
What's the paper about?
\edm

It's about this: [See ``Quantum-Bayesian Coherence,'' \arxiv{0906.2187v1}.]\
particularly the diagrams on pages 45 and 50 and the magical equation on page 46.  It then turns the table and asks whether the magical equation (and its relatives on page 51) might be the fundamental idea behind quantum mechanics.

I'm glad I'm not going to that conference; glad I wasn't even invited.  After my eighth (round) trip across an ocean in a year's span (and numerous other side trips in the States), I broke down.  I've canceled trips left and right.  I'm committed to staying home, home, home, only to break my stride when I take Emma to Paris in April for her 10th birthday.

Keep thinking; I really like to use a relevant quote to bolster this passage of mine:
\bq
Richard Feynman wrote these words for the opening chapter on quantum mechanics in his monumental {\sl Feynman Lectures on Physics}.  It was his lead-in for a discussion of the double slit experiment with individual electrons or photons. Imagine if you will, however, someone well-versed in the quantum foundations discussions of the last 25 years---since the Aspect experiment, say---yet surprisingly unaware of when Feynman wrote actually this.  What might he conclude Feynman was speaking of?  Would it be the double slit experiment?  Probably not.  To the modern sensibility, a good guess would be that he was speaking of something to do with quantum entanglement or Bell inequality violations.
\eq

\section{12-11-08 \ \ {\it Autumn at PI} \ \ (to N. D. {\Mermin})} \label{Mermin140}

\bdm
I attach a {\tt .doc} file giving the program of the Tempe workshop.
No website that I'm aware of.
\edm

Liked that title by Derek Abbott (whoever he is): ``What happens when the laws of physics change?''

Give him these quotes by {\Poincare}.  [See 10-03-06 note ``\myref{Halvorson15}{Singularities and Evolutionary Laws}'' to H. Halvorson.]

\section{12-11-08 \ \ {\it Bergsonism} \ \ (to N. D. {\Mermin})} \label{Mermin141}

\bdm
Great stuff.   I hadn't seen it before.

The final paragraph is pertinent to my own remarks:
\bq\noindent\rm
No doubt many readers will be dismayed to note that I seem constantly to substitute for the world a system of simple symbols. This is not due simply to a professional habit of a mathematician; the nature of my subject made this approach absolutely necessary. The Bergsonian world has no laws; what can have laws is simply the more or less distorted image which the scientists make of it.
\eq
\edm
Yeah, it is great stuff.  I love how he foresaw the meaning of the big bang.

\bdm
What does he mean by ``the Bergsonian world''?
\edm
Something like in my credo to Joy Christian and Lucien Hardy. [See 12-02-07 note ``\myref{Hardy20}{Lord Zanzibar}'' to J. Christian and L. Hardy.]

\section{12-11-08 \ \ {\it Boutrouxisme} \ \ (to N. D. {\Mermin})} \label{Mermin142}

Funny, you didn't ask about who Boutroux was.  I remember the great joy I had one evening when I did a google search on the simultaneous names James, Renouvier, and Boutroux.  To my amazement one of the references mentioning all three of these characters was a book by Teddy Roosevelt!

Here's a couple of quotes from the old man himself (Boutroux that is):

\begin{itemize}
\item
E.~Boutroux, {\sl Natural Law in Science and Philosophy}, (David Nutt, London, 1914); translated by F.~Rothwell, from E.~Boutroux, {\it De l'Id\'ee de Loi Naturelle dans la Science et la Philosophie Contemporaine}, (Soci\'et\'e Fran\c{c}ais D'Imprimerie et De Librairie, Paris, 1895). This is an early work concerning the idea that nature's laws themselves might be evolutionary and not immutable.
\bq
\indent
The theory upheld in the present work is that no absolute coincidence exists between the laws of nature as science assumes them to be, and the laws of nature as they really are.  The former may be compared to laws proclaimed by a legislator and imposed {\it a priori\/} upon reality.  The latter are harmonies towards which we ascertain that the actions of different beings really tend.
\eq
and
\bq
That which we call the laws of nature is the sum total of the methods we have discovered for adapting things to the mind, and subjecting them to be moulded by the will.
\eq
\item
E.~Boutroux, {\sl The Contingency of the Laws of Nature}, (Open Court, Chicago, 1920); translated by F.~Rothwell, from E.~Boutroux, {\it De la Contingence des Lois de la Nature}, (Balli\`ere, Paris, 1874).
\end{itemize}

\section{12-11-08 \ \ {\it PIAF, Peres School, Etc.}\ \ \ (to D. R. Terno)} \label{Terno5.2}

\bdt
Another matter: are you happy with the tentative titles that I gave to your lectures?
\edt

If you could change the one titled ``Non-locality in quantum mechanics'' to ``The Significance of Entanglement'', I'd appreciate that.  It'd be more honest to my attitude:  I don't like to use the word nonlocality anywhere in relation to quantum mechanics.  (For instance, the conclusion I draw from Bell inequality violations is NOT that locality goes by the wayside, but rather that ``unperformed measurements have no outcomes''.)

\section{13-11-08 \ \ {\it Hanneke's Thesis}\ \ \ (to R. Blume-Kohout and others)} \label{BlumeKohout6}

Spurred by Hanneke's tragic death, I printed out her 166 page Master's Thesis to have a copy in my office.
(It can be found online at \myurl{http://philsci-archive.pitt.edu/archive/00004224/}.) However, I will be traveling to Australia for a week starting tomorrow.  Thus if anyone would like to borrow the paper copy for a while, just let me know---there's no sense in cutting down more trees.

I think it would be nice if someone would report on the thesis at one of our Thursday foundations group meetings.  And given the subject matter, I think it would be particularly nice if Robin would weigh in for at least part of that, telling us what he thinks of the argument.

\section{13-11-08 \ \ {\it Can't Dubrovnik}\ \ \ (to J. R. Brown)} \label{BrownJR2}

I must apologize, but it is now clear to me that I will not be able to come to the Dubrovnik meeting.  The note below, which I paste in, explains what is happening.  On April 18, the last day of the Dubrovnik meeting, I will be flying from Toronto to Paris.

However you have built such an expectation of this meeting in my mind that I have come to think it is pretty important to have some representation of our quantum-Bayesian/quantum-pragmatist thoughts there \ldots\ lest we be forgotten when the history books are written.  Thus I wonder if I might convince you to invite two of my strongest and most interesting colleagues in my stead?  The two colleagues are {\Ruediger} {\Schack} and Marcus {\Appleby} (both in the UK).  {\Schack} is one of the masterminds of all that we have done in this quantum Bayesian turn, and Appleby is an amazing polymath who has contributed much in the last few years.  Here are a couple of representative papers:
\bv
{\Schack}, \arxiv{quant-ph/0404156} \\
and\\
Appleby, \arxiv{quant-ph/0402015}.
\ev
Both physicists are very philosophically minded, steeped in knowledge of Wittgensteinian ideas.  And I know that {\Schack} has also put significant effort into the classical pragmatists, whereas Appleby has done the same with Hume.

The point is neither would disappoint you.  Finally, let me mention this:  I gather from a side remark in a note from Chris Timpson that he plans to be at the Dubrovnik meeting.  Timpson has recently posted an excellent review and critique of our quantum Bayesian effort:  \arxiv{0804.2047}.
It would be very nice if the simultaneous presence of these three would add some real sparks to your conference, and at the same time expose directions of progress for us.

I will keep my fingers crossed \ldots

\section{13-11-08 \ \ {\it Can't Dubrovnik, but Maybe You Can}\ \ \ (to R. {\Schack} \& D. M. {\Appleby})} \label{Appleby42} \label{Schack142}

In my continuing effort to return to an honest living, I have just canceled two trips in April that I had previously promised to undertake.  One of them is a meeting in Dubrovnik, Croatia that is historically famous in the philosophical circles.  As I understand, the place is truly strikingly beautiful and the food wonderful.  My invitation to that meeting (from Jim Brown) and a poster for it are attached.

The reason I'm telling you all this is that when I canceled, I suggested to Jim that he invite you two.  The reason is I would like to get some representation there of our quantum Bayesian, pragmatist, Pauli'an, and Wittgensteinian thoughts.  Jim had already made me feel that he wanted to develop discussion along these lines, and that this would be a good forum for people to start thinking on these things.   Thus I hope he will take the bait and invite you, AND also that you will go and talk on this subject if given the opportunity.  I happen to know that Timpson will also be there, so I suggested to Jim that he might want to take the opportunity to generate some intellectual sparks via a session showcasing all three of you.

\section{14-11-08 \ \ {\it Ein Begriffe Again} \ \ (to H. C. von Baeyer)} \label{Baeyer45}

Do you remember the discussion we had on the phrase ``ein begriffe'', which I thought had turned up in the letter from Pauli to Bohr?  What note was that, can you send it to me again?  I have finally unearthed the original letters.  It looks like Folse probably did transcribe that marginal note incorrectly.

If I don't hear back from you in the next hour, I'll get back to you in a couple days to continue the discussion.  I'm on my way to Australia this afternoon, so that'll cause a little delay.

\section{16-11-08 \ \ {\it SICs Exist!}\ \ \ (to D. M. {\Appleby} and S. T. Flammia)} \label{Appleby43} \label{Flammia11}

I now know it's unquestionably true!  I met one on my way between Sydney Airport and Darling Harbour this afternoon.  My taxi driver was a Punjabi SI\ldots\ wait a minute, he must have meant Sikh not SIC.  Aw crap.  Back to reality.

\section{26-11-08 \ \ {\it A Draft}\ \ \ (to A. Stairs)} \label{Stairs2}

\bAllS
I've attached a draft of what I want to talk about in December. It starts with things I had been thinking about in the talk I gave at Western, but I think you may find that it's actually quite congenial to a good deal of what you want to say.
\eAllS

That's funny, I already felt pretty damned with this sentence of yours:
\bq\noindent
One might protest that this ignores the difference between probability one and truth, but
there's not much to be gained by following that thought \ldots
\eq

\section{26-11-08 \ \ {\it Appreciation} \ \ (to A. Plotnitsky)} \label{Plotnitsky21}

I fear I haven't yet expressed adequately how much I appreciated your visit.  So, let me take this opportunity before signing off for Thanksgiving.  I got loads out of our discussions, and today's discussion particularly.  I hope you will believe me; I don't always show it in my face.  Conversations with you inevitably teach me how to say things and think things in ways quite different than I am accustomed to, and that is a currency in itself.  Your insights today, I felt were particularly keen, and I'm working to incorporate your ways of thinking into my mathematical developments.  I want to get back to this concept of efficacity and develop it much further, as I have always felt that dimension $d$ to some extent quantifies it.  You helped me a see a clearer relation with $d$ and the SICs today, and for that I'm grateful.

All my best for a safe trip home, and please give my best regards to Paula.

\section{28-11-08 \ \ {\it Why Things Fall}\ \ \ (to N. C. Menicucci)} \label{Menicucci3}

I was misremembering the Wootters result.  What he tried to explain was why things fall, if we take gravitational time dilation as given.  See:
\bq
\myurl{http://www.springerlink.com/content/g7863165m0w47665/fulltext.pdf}
\eq

\section{29-11-08 \ \ {\it A Colleague}\ \ \ (to R. W. {\Spekkens})} \label{Spekkens52}

Well reasoned.  Particularly, I had not thought of this angle on it:
\brws
Nonetheless, he would undoubtedly benefit from spending some time here.  Given that he seems
keen to pursue foundations, and will ultimately be a representative of our field wherever he goes,
it is probably a good idea for us to invite him.
\erws

Strange thing, I go up and I go down on him.  The more I analyze it, though, the more it seems to be a Necker cube phenomenon, rather than something objective about him.  When I think of the science, I'm very down on him.  When I think of the human behind these vague (almost surely non-productive) ideas, a human full of feelings with a wife and child, a different kind of instinct kicks in with me.  Probably the only difference between him and any of a dozen other questionable thinkers who write me is that I've met him in person a few times.  That associates a human struggle with the quack idea, and I find it hard to ignore.

\section{01-12-08 \ \ {\it Prolegomenon} \ \ (to H. C. von Baeyer)} \label{Baeyer46}

Thanks for sending your draft on Pauli's journey inward.  I enjoyed reading it very much, and---you're going to curse me for this---found myself hoping that it is really just a prolegomenon to a more extensive project.  It was very good indeed, and I think the writing will serve the purpose of pulling public interest into the man.  Why did you struggle with it all year?  Can you pinpoint what the difficulty was?

I caught a very small number of typos.  Here they are: [\ldots]

Now a question:  In note 28, I didn't understand why you brought up the 137 story.  Did it connect to something in the text that I missed?

I enjoyed the Fierz quote on page 16 very much.

Another question.  What does ``Gewissen'' in note 3 refer to?

By the way, thanks for forcing me to look up ``retorts''.  My reading of ``retort'' in the Heisenberg quote,
\bq\noindent
   The alchemist in his laboratory is constantly involved in
    nature's course, in such wise that the real or supposed
    chemical reactions in the retort are mystically identified
    with the psychic processes in himself, and are called by
    the same names.
\eq
had always been as ``response'' or ``answer back''---clearly not satisfactory, but that's what I was thinking.  I never thought of chemical vessel!!  (Shows what I know of chemistry.)

Really, let's talk more about your struggle with writing this.  I'd like to be able to peer into your mind a little more.

BTW, Australia is long since past.  I got back almost before I went.  Still, it took its toll as expected, and I have canceled all other travels until my April Parisian outing.  I am going to write and write in the next three months.

\section{01-12-08 \ \ {\it The Slowest Reader on Earth} \ \ (to W. G. {\Demopoulos})} \label{Demopoulos30}

By the way, in preparation of your visit, I finally read the draft you sent me October 28th.  Sorry to have taken so long.  Not many comments at the moment:  Where we agreed before, I think we continue to agree.  And where we disagreed before, I think we continue to disagree.  Particularly, in your treating measurement devices as NONquantum in the sense of being ``unproblematically propositional.''  I.e., ``effects are traces left on systems whose associated propositions are unproblematically determinately true or false.''

One point of wonderment to me, however, was this.  You say,
\bq\noindent
    To my knowledge it has never been maintained that the
    source of the interpretive problem that the Kochen--Specker
    theorem poses is the propositional framework, the framework
    that takes propositions belonging to particles as the
    subject of probability assignments.
\eq
I would say this is exactly what my colleagues Peres, {\Caves}, {\Schack}, Appleby, and I have been on about for a quite some years.  But maybe I really am missing a subtle (or not so subtle) distinction between your system and ours.  We should get this straight, whether I'm missing something.

It'll be great to see you again tomorrow.  I only fear that the discussion may be a little too multi-pronged with so many people wanting to see Allen!

\section{02-12-08 \ \ {\it My Sloppy Reading} \ \ (to H. C. von Baeyer)} \label{Baeyer47}

\bhcvb
Just off the top of my head, did you notice that the page number in the note was 137?
\ehcvb
No, I had not noticed that.  Of course, now that you say it, it is glaringly obvious.  You might have pity on the other readers similarly blind as me, modifying what you write to something like:  ``Note the page number.  Pauli would have been very pleased by it turning out to be 137.  He considered this to be the most important number in physics, and its mystical connotations did not pass him by \ldots

\bhcvb
The idea of entangling the observer with the observation is, of course, nonsense.
\ehcvb
That is right, and somehow I had not noticed that.  But also, now that you bring it up, perhaps the definition could use some tuning.  You write, ``The term referred to the scientific view of the world as seen `from the outside', as it were, by an observer who neither influences nor is influenced by the world.''  I wouldn't go so far as to say that the observer is not influenced by the world.  That would preclude any (classical) Bayesian from updating his expectations upon the acquisition of data.  A significant trouble I always perceived Pauli had with the ``ideal E'' was its one-sidedness.

Here's one of the formulations I've given the point from time to time (this from 2003):
\bq
Everybody has their favorite speculation about what powers quantum information and computing.  Some say it is the superposition principle, some say it is the parallel computation of many worlds, some say it is the mysteries of quantum entanglement, some say it is the exponential growth of computational space due to the tensor product.  For my own part though, my favorite speculation is that it is Newton's Third Law:  For every action, there is an equal and opposite reaction.  Indeed I sometimes wonder if the very essence of quantum mechanics isn't just this principle, only carried through far more consistently than Newton could have envisioned.  That is to say, absolutely NOTHING is exempt from it.

What do I mean by this?  What might have been exempt from the principle in the first place?  To give an answer, let me note an equivalent formulation of old Newton.  For every REACTION, there is an equal and opposite ACTION.  Strange sounding, but there's nothing wrong with it, and more importantly, this formulation allows for the possibility of an immediate connection to information theory.  In particular, we should not forget how information gathering is represented in the Shannon theory.  An agent has gathered information---by the very definition of the process---when something in his environment has caused him to REACT by way of revising a prior expectation $p(h)$ (for some phenomenon) to a posterior expectation $p(h|d)$ (for the same phenomenon).

When information is gathered, it is because we are reacting to the stimulation of something external to us.  The great lesson of quantum mechanics may just be that information gathering is physical.  Even something so seemingly unimportant to the rest of the universe as the reactions that cause the revisions of our expectations are not exempt from Newton's Third Law.  When we react to the world's stimulations upon us, it too must react to our stimulations upon it.

The question is, how might we envision a world with this property---i.e., with such a serious accounting of Newton's law---but in a way that does not make a priori use of the information gathering agent himself?  If the question can be answered at all, the task of finding an answer will be some tall order.  For never before in science have we encountered a situation where the theorizing scientist is so inextricably bound up with what he is trying to theorize about in the first place.

It's almost a paradoxical situation.  On the one hand we'd like to step outside the world and get a clear view of what it looks like without the scientist necessarily in the picture.  But on the other hand, to even pose the question we have to imagine an information gathering agent set in the middle of it all.  You see, neither Shannon nor any of modern information theory has given us a way to talk about the concept of information gain without first introducing the agent-centered concept of an expectation $p(h)$.

So, how to make progress?  What we do know is that we actually are in the middle of the world thinking about it.  Maybe our strategy ought to be to use that very vantage point to get as close as we can to the goal.  That is, though we may not know what the world looks like without the information gathering agent in it, we certainly do know something about what it looks like with him in:  We know, for instance, that he ought to use the formal structure of quantum mechanics when thinking about physical systems.  Beyond that, we know of an imaginary world where Newton's Third Law was never taken so
seriously:  It is the standard world of classical physics and Bayesian probability.

Thus, maybe the thing to do first is to look inward, before looking outward.  About ourselves, at the very least, we can ask how has the formal structure of our {\it behavior\/} changed since moving from what we thought to be a classical Bayesian world to what we now believe to be a quantum world?  In that DIFFERENTIAL---the speculation is---we may just find the cleanest statement yet of what the quantum world is all about.  For it is in that differential, that the world without us surely rears its head.

To do this, we must first express quantum mechanics in a way that it can be directly compared to classical Bayesian theory, where the information-gathering agent was detached from the world.  That is what this lecture is about \ldots\
\eq

Rereading this, I enjoyed how it made a connection to the title of your draft [``Wolfgang Pauli's Journey Inward''].

\section{02-12-08 \ \ {\it Prolegomenon (I'd better not use this word in the subject line, lest I accept it)} \ \ (to H. C. von Baeyer)} \label{Baeyer48}

Well, I hope that is the first step toward accepting it!  Really, you could do so much good, and of all the people I've met nominally interested in these matters, you're the only one with the right combination of skills to be able to pull off something of value if you set your mind to it.

I am relieved to find out the reasons for your blockage.  Both yin and yang look reparable to me.

The point for me is not that all he was thinking (or hoping) will turn out to be right, but that SOME of what he was thinking is inspirational for a solid technical turn we can make in physics.  I have gotten immense technical inspiration from the little bits I know about the Pauli (and Pauli--Fierz) correspondence.  I crave more, and I think knowing more will get us to that technical point oh so much faster than if we have to slog our way there with our own less-than-Pauli'an minds.

I really do applaud this article.  I thought I had completely lost you.  Emma came home with nearly straight A's yesterday (except in ``dance'' which I didn't care about, and French, which I do care about, but in this case can blame on the teacher and the broken Ontario system for teaching French).  I was so proud of her.  So I was already on a bit of a high, but your article reinforced the same feeling.  I am proud that Pauli'an things are still on your mind.

\subsection{Hans's Preply}

\bq
I have now fixed the typos. Thanks!   [Gewissen] is short for a
book entitled {\sl Wolfgang Pauli -- Das Gewissen der Physik}.  I don't have the
bibliographic reference, which somehow got erased, here with me.

Your question about my block is very good, and worth talking about.

I had no trouble writing about Pauli's career in physics, up to page 14.  I did that
early in the summer, and have been polishing it since then.

But then I stopped for two (possibly related) reasons, one literary, and the other
psychological.

The literary reason was that there is just too much material:  The detached
observer, the significance of symbols, archetypes, Fierz, the psycho-physical
``neutral language'', Kepler \ldots\ldots\ldots\ I couldn't figure out how to organize it all.  Just
taking one topic after another made it a laundry list, and no matter what I tried
after p.~14, it all sounded tedious.

Then I came across the letter in which Pauli sends the dreams, and I had an
inspiration.  I removed Fierz from the first 14 pages, popped in some ***s and
told the story about the trip to Princeton.  That provided a turning point, and
then I decided to concentrate on just the letters from that period.  If the topics I
mentioned above came up, fine, if not, not.  The neutral language, for example,
made it only into one sentence as ``common language''.  The ocean voyage fit
perfectly with the metaphor of voyage that I had decided on long before.  Now I
had a structure!

The other reason for my block is more serious. Crudely put, the question is: Do I
believe all this stuff?  Res Jost, a famous theorist and colleague of Pauli, did not.
Aldous Huxley did not.  My brother Carl, a smart psychologist in Saskatoon,
does not. They are all ``rationalists''.  What am I?

The reason I'm taking it seriously is because Pauli is so damn smart.  And so
were Kepler, Newton, and Goethe, who were also mystics.  Well, uncle Fierz
comes to the rescue when he says ``you know, I can really agree with you
{\it provided\/} you express the reconciliation of science and mysticism as a goal for
the future'' (hundreds of years from now, {\Appleby} told me).  And lo!\ when you
read Pauli, he says over and over again that he doesn't know how the two will
be reconciled.  So I'm off the hook!  I don't have to believe all this mystical stuff,
only that some day it may come back into vogue.

So you see, I had an inner and an outer conflict with this story.  Yin and yang.
\eq

\section{02-12-08 \ \ {\it Volumes} \ \ (to H. C. von Baeyer)} \label{Baeyer49}

Wild idea.  If I were to purchase one or more volumes of Pauli correspondence to have by my side, which would be the most important ones with respect to what I want to learn?  I already have a couple of German-English dictionaries.  Who knows what a little will power might accomplish.

\section{03-12-08 \ \ {\it Seeking SICs Story}\ \ \ (to E. Goheen)} \label{Goheen2}

I hope you can use the whole thing.  I spent quite some time in the middle of the night getting all the words just right, so that I might use this as an opportunity to educate the noninitiated in the PI community about the problem.  Partly educational, partly entertaining.  Story below.

I have to take a guest to breakfast.  I'll be back online as of 11:00 if you have any questions.

\bq
\begin{center}
{\bf Seeking SICs:\ An Intense Workshop on Quantum Frames and Designs}\medskip
\end{center}

At the heart of quantum mechanics is Hilbert space.  But what is at the heart of Hilbert space?  Surprisingly, there are still some very basic---indeed, very fundamental---questions about finite dimensional complex vector spaces desperately seeking an answer.  The theme of the workshop ``Seeking SICs'' was one of these questions.

For a Hilbert space of finite dimension $d$, can one always find a set of $d^2$ equiangular lines---that is, a set of lines for which the angle between any two is always the same?  Such structures are called SICs (pronounced ``seeks''), because in another language they can be thought of as ``symmetric informationally complete'' sets of quantum states.  The phrase informationally complete captures the idea that the projection operators onto such a set of lines form a complete basis for the vector space of Hermitian operators, and hence of quantum states.  A SIC, however, wouldn't be just any old basis for the space; it would be a nearly magical one.  For, fixing a SIC in the background gives a means to express quantum states in a way that maps the whole structure of quantum mechanics into as mild a variation of (classical) probability theory as one can imagine:  quantum states themselves become simple probability distributions; unitary evolution becomes, a up to a single constant, classic stochastic evolution; the Born probability rule becomes a simple variation on the classic law of total probability.  The list of similarities grows.  Thus SICs, if they exist, would be a highly valued commodity in the quantum foundations game.  They would give a new, powerful way to rewrite (or dismiss!)\ many an old conceptual problem.

But do they exist?  That's no easy question!  The winner of the best intuition for it of all recent PI visitors is Sir Roger Penrose, who could ``see'' that they exist in $d=2$, 3, and 4.  But for the rest of us it's been a rough slog.  After 9 years of toil since the problem first arose, nearly 20 papers, and some involvement from the best minds in quantum information theory (like Peter Shor, William Wootters, and Marcus Appleby), the most that is known either through hand proof, machine proof, or numerical investigation is that they (appear to) exist for all $d$ from 2 to 47.  Beyond that, nothing definitive is known \ldots\ but most people believe!  And that's the fuel that can organize a conference.

The purpose of ``Seeking SICs'' was to amass as many of the best in the field as we could to have an intense go at it in the companionship and moral support of each other.  The venue consisted of 11 organized talks across five days, with the remaining time devoted to informal idea sessions and chalkboard discussions.  Several from the local quantum information community joined the meeting from time to time, bringing the total number of participants up to about 20 for some of the days.  The meeting, as the name promised, was an intense event.  By all accounts it was a success.  Several of the participants particularly commented on how surprised they were to learn that so much UNpublished progress had been made since the last major works were posted 3--4 years ago.  There was a great flood of information sharing, and perhaps one of the lasting achievements of the conferences was the establishment of a SIC wiki for building a strong online community to work on this problem.

What do we know now that we didn't know at the beginning of the meeting?  Well, we now know by machine proof that SICs do indeed exist in $d=14$ (it was only numerical before), and we know a method that leads to their hand construction in $d=6$.  More than that, we know the community has been reinvigorated and that maybe in the coming months we'll get a proof for the remaining (discrete) infinity of dimensions!
\eq

\section{03-12-08 \ \ {\it I Twitter, Therefore I Am}\ \ \ (to A. Stairs)} \label{Stairs3}

Funny to learn about twittering from you yesterday nearly at the same time as a discussion about Jungian synchronicity \ldots\ and then to find this article in the {\sl Washington Post\/} this morning: \myurl[http://www.washingtonpost.com/wp-dyn/content/article/2008/12/02/AR2008120202935.html?hpid=opinionsbox1]{http://www.washingtonpost.com/wp-dyn/content/article/2008/12/02/AR2008120202935.h\\ tml?hpid=opinionsbox1}.
Then, to read the content of this and find that it has something mildly to do with Zeilinger's view of quantum mechanics.  ``In the Information Age: Knowing equals Being,'' so said.

Surely there's a meaningful coincidence in there somewhere!

From a strange hour of the night \ldots

\section{09-12-08 \ \ {\it Quantum Random Numbers}\ \ \ (to R. {\Schack} \& C. M. {\Caves})} \label{Caves99} \label{Schack143}

I'm writing to make you both aware of a talk Ben Schumacher gave here at PI last week.  It can be found at this site:  \pirsa{08120020}.  The talk is not at all on quantum random number generators, but I took it that the deep idea in it is the same as the one that you two make use of.  {\bf ``quantum correlations are monogamous''} $=$ {\bf ``no inside information''} $=$ {\bf ``unperformed measurements have no outcomes''}.  Anyway, I think the talk was extremely exciting and helps put your work in a more general context.

Check it out.

\section{10-12-08 \ \ {\it All the Qualifiers} \ \ (to R. W. {\Spekkens})} \label{Spekkens53}

(Metaphorically) load your paper tray, print out
\begin{center}
\myurl{http://www.perimeterinstitute.ca/personal/cfuchs/nSamizdat-2.pdf}
\end{center}
and thumb through the document until you get to the pages listed below.

Note all the qualifiers.  By my own admission I don't have a completely consistent story yet, but I don't think you can ignore the overwhelming theme.  See particularly the last three quotes.

Obsession is my trademark, I suppose.

Funny:  One interesting self-psychology thing I learned from this exercise is that I didn't mention ``law of thought'' even once in the last 130 pages of the document.  I wonder what exactly that signifies.

\begin{itemize}
\item
p.\ 14, ``more a \ldots\ than'' in:\smallskip

But that is precisely why I would call quantum mechanics more a ``law of thought'' than a ``law of nature.''  Just as Boole did of probability theory.  And just as Jaynes did with
(classical) statistical mechanics.  (In fact, either of you two---i.e., you or Jaynes---might just as well have written your quote above.)

You might say, and I suspect you will say, the distinction I draw between ``nature'' and ``thought'' is only a semantic one.  But I don't believe that:  I think it is precisely in making that distinction clear (and operational) that we have a chance of closing the quantum foundations debate and moving on.

Physical theory is about two things:  what is and what we know (or what we believe).  It's the process of putting the two things together that gives a prediction in any practical setting.  Quantum theory, interestingly, seems to be a nontrivial jumble of those two things.  I think that is a rather deep statement about the world, and one that we have not yet come to grips with.

\item
p.\ 16, ``with a \ldots\ side-input'':\smallskip

Since the overarching belief here
is that ``quantum mechanics is a law of thought'' (with an almost trivial side-input of honest-to-god physics), it strikes me as being in a category much more like relativity.  In the words of J.~Bub, it is a ``framework theory.''

\item
p.\ 38, ``PLUS'' (in letter to J. Bub):\smallskip

I know I suggested I would write a longer letter soon, but I'm going to wimp out of it again for now.  It would concern the main point of distinction I see between us (and also between myself and Pitowsky).
Namely, A) that I view a large part of quantum mechanics as merely classical probability theory (which on my view may be an a priori ``law of thought'') PLUS an extra assumption narrowing down the characteristics of the phenomena to which we happen to be applying it to at the moment, while B) you are more tempted to view quantum mechanics as a {\it generalization\/} of classical probability theory (and with it information theory).  I know that my view is not fully consistent yet, especially as I have always distrusted mathematical Platonism---which you pointed out to me I am getting oh so close to---but it still feels more right (to me, of course).

\item
p.\ 137, ``substantial part of'':\smallskip

I think our greatest hint of that comes from quantum mechanics. I would say that what we're learning in a precise way from it is that there is something about the stuff of the world that makes it uncaptureable with a purely physical picture.  We find that we cannot even draw a picture of the world without including our beliefs and belief changes as a crucial background in the sketch. (How could we if the world's not completed yet?)

Does that make me an extreme subjectivist?  I don't know.  Whatever it is though that I should be called, I think this willingness to accept a substantial part of quantum mechanics as simply ``law of thought'' will keep me from going down a misguided path.  I.e., the path of trying to ascribe all the easiest terms in the theory a kind of physical reality independent of our presence as active agents.

\item
p.\ 205, ``something that is \ldots\ not law of thought'':\smallskip

Here's where I really think you sell yourself short by advertising your system as an extension or generalization of classical probability theory (with classical probability theory as a special case that's gotten by deleting one of the axioms).  For I would say that your framework of ``states'' as vectors and ``measurements'' as applications of Bayes' rule is {\it classical probability theory}, full stop. Or, I should just say ``{\it probability theory}, full stop''---for, the word ``classical'' seems to imply that it is a subject somehow within empirical science (rather than ``law of thought'' that antecedes science).  In showing me that even quantum ``measurements'' can be viewed legitimately as nothing more than applications of Bayes' rule, you have done me a great service. For you demonstrate to me more clearly than ever that the concept of POVM ought to be put onto the subjective side of the shelf when I tear quantum mechanics into its two components.  But your other intriguing axioms---like the simplicity and composite-system axioms---which you think give the possibility of generalizing upon classical probability, I would say are nothing of the sort. Instead, I would say they express just the opposite.  These axioms seem to me to say something about what we are positing of nature.  They express something that is not subjective and is not ``law of thought.''

\item
p.\ 272, ``to the extent that'':\smallskip

Now, to quantum mechanics.  You find something contradictory about my liking both quantum mechanics and {\Rorty}.  Here is the way I would put it.  Presently at least, I am not inclined to accept quantum mechanics ``to be real, an `intrinsic nature of reality','' except insofar as, or to the extent that, it is a ``law of thought,'' much like simple (Bayesian) probability theory.  Instead, I view quantum mechanics to be the first {\it rigorous\/} hint we have that there might actually be something to {\James}'s vision.

\item
p.\ 343, ``there is a part that remains'':\smallskip

Lee Smolin asks me how I could possibly imagine that the linear structure of quantum mechanics will remain when one moves into such a nonlinear regime as that given by the laws of gravity?  I say I'm not fazed at all:  Most of what he means when he speaks of quantum mechanics as an expression of physics, is for me but a law of thought.  A wave function and its evolution are not properties intrinsic to the system for which they are about.  Rather, if they are properties of anything at all, they are properties of their user's head---for they capture all his judgments about what might occur if he were to interact with the system of interest.

The quantum foundational task as I see it is to baldly accept that A LARGE PART of the theory is simply not about a world without
observers:  It is only about our interface with the world.  But there is a part that remains, and that part must be given a firm identification.  For only once we know how to do that will we know how to move forward when it comes to gravity.  Only then will we see that almost all of the ways that have hitherto been considered for combining quantum mechanics with general relativity were far too
unconstrained:  I am willing to bet that they all essentially boil down to sheer speculation.

\item
p.\ 445, ``branch of decision theory that is contingent upon properties of the world'':\smallskip

Be warned that by the phrase I don't mean something like a Kantian a priori category, i.e., a position like von {\Weizsacker}'s in his book {\sl The Unity of Nature}.  I don't mean something like, ``an understanding using the terms of quantum mechanics is the precondition for possible experience.''  Rather I have started to toy rather strongly with a Darwinian kind of idea:  Using the rules of quantum mechanics for manipulating and updating our expectations (i.e., as a ``law of thought'') is the presently best known means for survival, given that we are immersed in the particular world we are.
That is, I want to view quantum theory as a branch of decision theory that is contingent upon properties of the world we live in \ldots\ and it is something we locked into only in our most recent turn in evolutionary development.

\item
p.\ 488, ``retains a trace of'':\smallskip

When I say that QM is a theory about a very small part of the world, you should literally think of a map of the United States in relation to the rest of the globe.  The map of the US is certainly incomplete in the sense that it is obviously not a map of the whole globe.  But on the other hand it is as complete as it can be (by definition) as a representation of the US.  There are no hidden variables that one can add to the US map that will magically turn into a map of the whole globe after all.  The US map is what it is and need be nothing more.

Does that help any?

I think a good bit of the problem comes from something that was beat into most of us at an early age.  It is this idea:  Whatever else it is, quantum theory should be construed as a theory of the world.  The formalism and the terms within the formalism somehow reflect what is out there in the world.  Thus, if there is more to the world than quantum theory holds out for, the theory must be incomplete.  And we should seek to find what will complete it.

But my tack has been to say that that is a false image or a false expectation.  Quantum theory from my view is not so much a law of nature (as the usual view takes), but rather a law of thought.  In a
slogan:  Quantum mechanics is a law of thought.  It is a way of plagiarizing George Boole who called probability theory a law of thought.  (Look at the first couple of entries in the {\Ruediger} {\Schack} chapter of {\sl Notes on a Paulian Idea}.)  Try to think of it in these terms, and let's see if this helps.

\ldots\ [Long story about probabibility theory, then continues with:] \ldots\
So I say with quantum mechanics.  The story is almost one-to-one the
same:  You just replace probability distributions with quantum states.  \ldots\ But then you reply, ``But there's a difference; quantum theory is a theory of physics, it is not simply a calculus of thought.''  And I say, ``That's where you err.''  Quantum theory retains a trace of something about the real, physical world but predominantly it is a law of thought that agents should use when navigating in the (real, physical) world.  In particular, just like with probability theory, we should not think of quantum theory as incomplete in the usual sense.  If it is incomplete in any way, it is only incomplete in the way that the US map is incomplete with respect to the globe: There's a lot more land and ocean out there.

Teasing out (your words) the trace of the physical world in the formalism---i.e., the part of the theory that compels the rest of it as a useful law of thought---is the only way I see to get a solid handle on what quantum mechanics is trying to tell us about nature itself.

With this let me now go back to the US map for one final analogy.  I said that there is a sense in which the US map is as complete as it can be.  However there is also a sense in which it tells us something about the wider world:  If we tabulate the distances between cities, we can't help but notice that the map is probably best drawn on the surface of a globe.  I.e., the US already reveals a good guess on the curvature of the world as a whole---it hints that the world is not flat.  And that's a great addition to our knowledge!  For it tells a would-be Columbus that he can safely go out and explore new territories.  Exploring those new territories won't make the US map any more complete, but it still means that there is a great adventure in front of him.

\item
p.\ 532, ``THIS \ldots\ rather than THAT'':\smallskip

Concerning, ``But I did think it was a Law of Thought for you.''
\ldots\ you never cease to shock me \ldots.  And you never cease to cause me to strive to try to convey the very simple little idea more effectively! I wonder when I'm gonna finally hit the sweet spot?
Quantum THEORY, a law of thought:  Yes.  Resoundingly yes.  But the quantum WORLD---i.e., that situation, that world, that reality, which conditions us to choose THIS law of thought rather than THAT law of thought (in other words some alternative or imaginary law of thought)---is something else entirely.  It's the stuff that's here whether there are any law-of-thoughters around or not.  That's what I really want to get at; that's what I've always really wanted to get at.

\item
p.\ 533, ``mostly'':\smallskip

This talk tried to set the tone of the meeting by demonstrating that much of the content of finite-dimensional quantum mechanics reduces to two simple modifications of Bayesian
ideology---1) the setting of a theory of prior probabilities with regard to the outcomes of a single special quantum measurement, and
2) a modification of the standard Bayesian conditionalization rule for updating probabilities in the light of new information.  From this perspective, the formal structure of quantum mechanics becomes {\it mostly\/} a ``law of thought'' (in the same sense that George Boole called probability theory a ``law of thought'') rather than a ``law of nature.''  Where nature still rears its head---i.e., makes its contingently given empirical content known---is through the higher-level set of reasons for why decision-making agents in this world should use {\it this\/} law of thought (i.e., quantum
mechanics) rather than {\it that\/} law of thought (i.e., some foil theory other than quantum mechanics).

\item
p.\ 642, ``partially tongue in cheek''

I am in agreement with your last sentence.  Attempts like the former effectively erase the empirical content (or contingency) of quantum mechanics, and I just don't see that.  When I myself invoke the slogan ``Quantum Mechanics is a Law of Thought,'' I only do it partially tongue in cheek, quickly correcting myself.  For if quantum mechanics were {\it only\/} law of thought, it would be like one of Kant's a priori categories of the understanding. On the other hand, I come dangerously close to viewing ``probability theory'' (i.e., the theory of coherent gambling) in such a Kantian kind of way.  And, indeed, it is partially because of this that I would not want to view quantum mechanics as a {\it generalized\/} probability theory.
\end{itemize}

\section{10-12-08 \ \ {\it Myrvold's ``Chance''}\ \ \ (to R. {\Schack})} \label{Schack144}

BTW, I had wanted to give you a heads-up about Wayne Myrvold's talk in Jerusalem.  He proposes a kind of epistemically based notion of ``chance'' that to some extent tracks the sort of thing I say below (from a letter to {\Mermin} two years ago):
\bq
With regard to your phase-diagram remark in particular [I.e.\ ``Don't be so quick to dismiss Leggett as an ideologue, just because he takes quantum states to be objective.  It's hard to spend decades worrying about the phase diagram of helium-3 without believing that quantum states are objective.''], I think it's got to be instructive to try to imagine what a turn-of-the-century physical chemist (trained in the worldview of classical physics) should have thought of the tables of heat capacities and whatnot that he had spent much of his life compiling.  What can those numbers represent in the classical worldview?  Why would two distinct chemists compile the same tables of numbers if those numbers didn't signify something objective and intrinsic to the chemicals themselves?  Well, a Maxwell demon wouldn't compile the same tables.  And there is surely a continuum of positions between us and him.  What those tables represent is something about the relation between us, the coarseness of our senses, our inabilities to manipulate bulk materials, all kinds of things like that, and the materials themselves.  Those numbers are just as much about us (``anyman'') as about ``it.''

Anyway, forgetting about Leggett, how would I convince the classical physical chemist of such a point of view?  It would be damned hard (as history has already shown).  Instead of skirting so close to such an anti-Copernican idea---that knowledge gained in science is as much about ``us'' as about ``it''---he'd probably spend his life trying to poke holes in the conceptually simple Maxwell demon idea and trying to return his beloved thermodynamical quantities to their pristine objectivity (a place where they'll last forever).
\eq

I was shocked!  I had never imagined I would agree with Myrvold about nearly anything.  Anyway, I thought it was a good talk.  I want to see if you have the same reaction, or whether I missed something devastatingly inconsistent between his idea and our program.

When do you leave for Jerusalem?  I presume you'll have email contact there for chit-chat about the draft I'm working on.  You will take your laptop, won't you?!?

\section{11-12-08 \ \ {\it A Sorry Comrade} \ \ (to R. W. {\Spekkens})} \label{Spekkens54}

I'm feeling very bad about this, because I've hardly had any interaction with the candidates and feel responsible to do so, but I'm just in too much pain to come in today. [\ldots]

At the moment I've got my foot in our lobster pot, surrounded by a bath of warm Epsom salt water, and my laptop on my lap, writing on my ``coherence'' paper.  Digging up things on Feynman for the paper, I came across the following quote from David {\Mermin} in his article ``Spooky Actions at a Distance:\  Mysteries of the Quantum Theory''.  I figure it might be of use when we finally put together our ``roots'' article.
\bq\noindent
   Most physicists, I think it is fair to say, are not bothered.  A minority would
   maintain that this is because the majority simply refuse to think about the
   problem, but in view of the persistent failure of any new physics to emerge from
   the puzzle in the half century since Einstein, Podolsky, and Rosen invented it,
   it is hard to fault their strategy.
\eq

\section{12-12-08 \ \ {\it Myrvold's ``Chance,'' 2}\ \ \ (to R. {\Schack})} \label{Schack145}

\brs
Isn't this more or less the point of Jaynes' approach to thermodynamics and all the semi-objective ways of getting entropy to increase: coarse-graining, mixing interaction with an environment, ignoring all details that are irrelevant for pressure, volume etc? The idea would be to define a standard of ignorance, that any honest and nondeluded person would subscribe to.
\ers

Roughly, but with a slight twist.  In place of ``ignoring all details'' something more like ``biologically incapable of.''  ``Ignoring'' implies that you could get at these things if you wanted them.  With hindsight, surely it is a kernel of thought motivated by your hypersensitivity program.

More in a minute.

\section{12-12-08 \ \ {\it Lunch Break}\ \ \ (to R. {\Schack})} \label{Schack146}

I'm sitting in my office eating a curried chicken sandwich.

My only worry in what you say below is the phrase ``classical deterministic theory'' which is usually taken to mean something about time dependence of variables.  The crucial distinction is only in that classical variables are assumed to be ``determinate'' in the sense of possessing a truth value independent of measurement.

\section{12-12-08 \ \ {\it F-Theory} \ \ (to R. W. {\Spekkens})} \label{Spekkens55}

\brws
Did you realize that the string theorists are studying F-theory?
\erws

I'm sure there are a lot of people named (and a lot of expletives starting with) F.

What a loon they would think of me, if they came across my credo below:  [See 12-02-07 note ``\myref{Hardy20}{Lord Zanzibar}'' to J. Christian and L. Hardy.]

\section{14-12-08 \ \ {\it Our Dismissal}\ \ \ (to C. M. {\Caves} \& R. {\Schack})} \label{Caves100} \label{Schack147}

Here's a cute dismissal of us.  See Footnote 7 in
\begin{center}
\myurl{http://philsci-archive.pitt.edu/archive/00002839/01/epist_rev.pdf}.
\end{center}

\section{14-12-08 \ \ {\it FPP5} \ \ (to G. Adenier)} \label{Adenier2}

Attached is my first contribution for the volume; it is a joint paper with {\Ruediger} {\Schack} titled ``Priors in Quantum Bayesian Inference.''

I still owe you my paragraphs for the preface.  I should definitely be able to get those to you tomorrow, but it will be in the afternoon (my time).

I would also like to have a second contribution to the volume; it is another joint paper with {\Schack}, but I think this paper will be of lasting significance, somewhat like my earlier paper ``Quantum Mechanics as Quantum Information (and only a little more)'' was in one of your volumes several years ago.  But because of that, it is a much more complex affair to write, and I have been working very hard at it.  Is there any chance you could hold off until Wednesday for it?  I would very much like to have it in the volume---as it is another important manifesto for me---but I understand I must be pushing you to the limit.

\section{15-12-08 \ \ {\it Are You Out There?}\ \ \ (to J.-{\AA} Larsson)} \label{Larsson6}

How late will you be out there today reading email and such?  I'm about an hour or two away from getting our preface written (OK, I have to admit I just started writing), and I'd like to pass it by you for comment.  Guillaume was hoping I'd submit it today.

Just checking.  Now, I'm writing away.  It shouldn't be hard.

P.S. Great news in SIC world today.  Andrew Scott has finally pushed the numerics further:  SICs are now known to exist (at least to the 38 significant figures) in dimensions up to $d=60$ !  (I placed a space before the exclamation mark so that you wouldn't think I meant 60 factorial.)

\section{15-12-08 \ \ {\it Draft}\ \ \ (to J.-{\AA} Larsson)} \label{Larsson7}

Attached is the draft.  I feel that a sentence or two could be added about your and Cabello's papers.  Could you volunteer that?  Also, I surely missed some important papers in the biblio; please add anything you'd like.

I hope the writing style doesn't offend you.

\bq
\begin{center}
{\bf Foreword: Unperformed Experiments Have No Results} \medskip
\end{center}

This year in {\Vaxjo} we thought we would try an experiment---it felt high time for a new result.  Much of the foundations discussion of previous years has focussed on EPR-style arguments and the meaning and experimental validity of various Bell inequality violations.  Yet, there is another pillar of the quantum foundations puzzle that has hardly received any attention in our great series of meetings:  It is the phenomenon first demonstrated by Kochen and Specker, quantum contextuality.  Recently there has been a rapid growth of activity aimed toward better understanding this aspect of quantum mechanics, which Asher Peres sloganized by the phrase, ``unperformed experiments have no results.''  Below is a sampling of some important papers on the topic for the reader not yet familiar with the subject.

What is the source of this phenomenon?  Does it depend only on high level features of quantum mechanics, or is it deep in the conceptual framework on which the theory rests?  Might it, for instance, arise from the way quantum mechanics amends the classic laws of probability? What are the mathematically simplest ways contextuality can be demonstrated?  How might the known results be made amenable to experimental tests?  These were the sorts of discussions we hoped the session would foster.

There were eight speakers in our special session---D. M. Appleby (Queen Mary U London, United Kingdom), G. Bacciagaluppi (U Sydney, Australia), I. Bengtsson (Stockholm U, Sweden), A. Cabello (U Sevilla, Spain), J. Emerson (U Waterloo, Canada), C. A. Fuchs (Perimeter Institute, Canada), J.-{\AA}. Larsson (Link\"opings U, Sweden), and R. Schack (Royal Holloway U London, United Kingdom)---and the talks were of uniformly high quality and all very enlightening.  In this volume, seven of the talks are represented with written contributions.\footnote{The content of J. Emerson's presentation can be found in the paper: C. Ferrie and J. Emerson, ``Frame Representations of Quantum Mechanics and the Necessity of Negativity in Quasi-Probability Representations,'' J. Phys.\ A {\bf 41} 352001 (2008).  In relevance to this session, it is connected to the Spekkens reference on negativity below.}  Three of the papers are devoted very directly to Kochen--Specker constructions, inequalities derivable from them, and questions of experimental feasibility:  I. Bengtsson, ``A Kochen--Specker Inequality''; A. Cabello, ``Kochen--Specker Meets Experiments''; J.-{\AA}. Larsson, ``The Kochen--Specker Paradox and Great-Circle Descents''.  G. Bacciagaluppi's paper ``Leggett's Theorem without Inequalities'' defines a new no-go theorem that quantum mechanics once again violates: the conjunction of parameter independence and a new condition called ``conditional parameter independence.''  The paper by C. A. Fuchs and R. Schack, ``Priors in Quantum Bayesian Inference,'' is meant to be a follow-up on C. M. Caves, C. A. Fuchs, and R. Schack, ``Subjective Probability and Quantum Certainty,'' (Stud.\ Hist.\ Phil.\ Mod.\ Phys.\ {\bf 38}, 2007, p.\ 255), where A. Stairs' modification of the Kochen--Specker noncolorability result was developed into an argument that even probability-1 assignments do not signify the pre-existence of measurement values.  The paper by D. M. Appleby, ``SIC-POVMS and MUBs: Geometrical Relationships in
Prime Dimension,'' though not directly about issues of contextuality, is part of a larger effort to rewrite quantum mechanics in a language that will be much more serviceable to these kinds of questions.  Finally, the paper by C. A. Fuchs and R. Schack, ``From Quantum Interference to Bayesian Coherence and Back Round Again,'' is an example of that larger effort:  It argues that the essence of Peres's slogan boils down to the fact that the law of total probability is replaced by a modification when considering counterfactual SIC measurements alongside actualized quantum measurements.

We were very happy to have an opportunity to organize this special session in {\Vaxjo}, where there is a long tradition of delightful, intense, and thought-provoking foundational discussions.  As the reader will find, this experiment was one of those well worth performing. \medskip

\noindent C. A. Fuchs \smallskip

\noindent J.-\AA. Larsson \medskip

\begin{enumerate}

\item
A. M. Gleason, ``Measures on the Closed Subspaces of a Hilbert Space,'' J. Math.\ Mech.\ {\bf 6}, 885 (1957).

\item
S. Kochen and E. P. Specker, ``The problem of hidden variables in quantum mechanics,'' J. Math.\ Mech.\ {\bf 17}, 59 (1967).

\item
A. Peres, ``Unperformed Experiments Have No Results,'' Am.\ J. Phys.\ {\bf 46}, 745 (1978).

\item
N. D. Mermin, ``Hidden Variables and the Two Theorems of John Bell,'' Rev.\ Mod.\ Phys.\ {\bf 65}, 803 (1993).

\item
A. Cabello, J.~M. Estebaranz and G. Garcia-Alcaine, ``Bell--Kochen--Specker Theorem: A Proof with 18 Vectors,'' Phys.\ Lett.\ A {\bf 212}, 183 (1996).

\item
J.-{\AA}. Larsson, ``A Kochen--Specker Inequality,'' Europhys.\ Lett.\ {\bf 58}, 799 (2002).

\item
D. M. Appleby, ``The Bell--Kochen--Specker Theorem,'' Stud.\ Hist.\ Philos.\ Mod.\ Phys.\ {\bf 36}, 1 (2005).

\item
R. W. Spekkens, ``Contextuality for Preparations, Transformations, and Unsharp Measurements,'' Phys.\ Rev.\ A {\bf 71}, 052108 (2005).

\item
J. Conway and S. Kochen, ``The Free Will Theorem,'' Found.\ Phys.\ {\bf 36}, 1441 (2006).

\item
R. W. Spekkens, ``Negativity and Contextuality are Equivalent Notions of Nonclassicality,'' Phys.\ Rev.\ Lett.\ {\bf 101}, 020401 (2008).

\item
A. Cabello, ``Experimentally Testable State-Independent Quantum Contextuality,'' Phys.\ Rev.\ Lett.\ {\bf 101}, 210401 (2008).

\end{enumerate}
\eq

\section{15-12-08 \ \ {\it Bad Foot} \ \ (to K. A. Peacock)} \label{Peacock2}

Another interesting person for you to talk to is Rafael Sorkin (on the same floor as you).  He too believes that part of the lesson of quantum mechanics is that the universe is continually ``giving birth.''

\section{17-12-08 \ \ {\it Anything}\ \ \ (to C. M. {\Caves})} \label{Caves100.1}

\bcc
I'm going to be going over your document pretty carefully.  Somehow I think that the evolution you've been pointing to has got to be important.  It's just doubly stochastic update of probabilities, followed by jacking up the final distribution so that it has the same purity (and, hence, the quantum state has the same purity) as the initial state.  This is, of course, true even when nothing at all is happening.  What do you make of that?
\ecc

Since you're thinking about it, go to the presentation I gave in Sydney a couple weeks ago:
\bv
\myurl[http://www.perimeterinstitute.ca/personal/cfuchs/Quantum\%20Bayesian\%20Coherence.pdf]{http://www.perimeterinstitute.ca/personal/cfuchs/Quantum\%20\\\quad Bayesian\%20Coherence.pdf}
\ev
The paper, when complete, should track it decently.  The main points are at the end.

A companion paper with Appleby fleshes out all the intricacies of the cubic equation for pure states in great detail.  The $c_{ijk}$ have a very nice structure.

But for your question, the best I can answer at the moment is:  We mark the flow of time by rotating our SIC in the sky with respect to our SIC on the ground (and demand the urgleichung of course).

Yes, it's very definitely important \ldots\ but I wish I knew more completely why.

\section{18-12-08 \ \ {\it Israel}\ \ \ (to R. {\Schack})} \label{Schack148}

\brs
I think my talk went well. There was a long discussion.

David Albert interrupted me after 5 seconds with the remark that the mottos ``unperformed experiments have no outcomes'' and ``the concept of probability is unaltered in qm'' are hardly controversial. His main objection in the discussion was that the Einstein criterion of reality is like ``bachelors are unmarried'', i.e.\ analytic, and can not be disproved by empirical arguments. Hence the Stairs argument shows that locality has to be abandoned.
\ers
It may be for an objective notion of probability (though I doubt it even there), but it certainly isn't for a subjective notion.  It's amazing that these guys don't even realize the boxes they think in.  And they're the ones who are supposed to be the philosophers.

\section{20-12-08 \ \ {\it Friday}\ \ \ (to R. {\Schack})} \label{Schack149}

\brs
Before you count Wayne as a new ally, be warned that he strongly
believes in objective quantum states.
\ers
Oh, don't worry; I know Wayne all too well.  I just thought he was clear on this point, and I had wondered whether I was missing something.  (My fear of having misheard something comes from the knowing-Wayne-all-too-well-point just mentioned.)

\section{21-12-08 \ \ {\it OK, You Should Be Back Home} \ \ (to N. D. {\Mermin})} \label{Mermin143}

OK, you should be back home from Tempe by now.  You told me to write you after your return.  PI visit, when?

Lucien told me about the meeting.  Makes me double glad I wasn't there.

But I wish {\Carl} had been there, so that you two might have had a chance to talk.  I sense you haven't thought as much about (the reasons for) quantum mechanics as you ought to have lately.  {\Carl} sent me a link to the talk he would have given.  (See, it's all about the counterfactuals.)  Link below.  I'll forward it to you in case you want to peruse it.

Did you see Louisa Gilder's book?  (She must have sent you a copy if she sent me a copy.)  I don't know how well she did with everyone else's dialog yet, but she certainly got me on the mark.  I was very pleased with that.  (And pleasantly surprised.  Since meeting her briefly three years ago, I had forgotten about her.)

\section{22-12-08 \ \ {\it Nonlocal Boxes and Ehrenfest} \ \ (to N. D. {\Mermin})} \label{Mermin144}

Just back in from shoveling snow.

\bdm
Have you ever run across this particular way of presenting history --- in terms of
highly imaginary conversations based on texts, memoirs, reminiscences?
\edm

The only thing approaching it that I can think of is Susan Haack's article ``We Pragmatists.''  She managed to put together a good conversation between herself, Peirce and Rorty---with P and R's contributions being solely from published stuff.

I haven't read more than a few beginning pages of the book yet.  I do know that already at page 7 I didn't like the sentence, ``But no one following Bohr, Heisenberg, Pauli, Dirac, or Born dared grasp, measure, or even name the deepest of all puzzles, entanglement.  Then along came John Bell.''  I feel that Pauli, for instance, already had a decent grasp of entanglement in 1927.  It's just simply that the lesson of it was taken to be something else than the Einstein-{\Schroedinger}-Bell camp thought it should be, and so it was marginalized in their thought.  The lesson wasn't nonlocality for Pauli, but nondetachedness of the observer.  And ``degree of detachedness'' is something he did try to quantify.

Pet peeve of mine recently, with this incessant talk of ``nonlocal boxes'' everywhere I go.  So, I put these lines in my ``coherence'' paper (the one I'm writing at the moment).  Particularly the footnote:
\bq
So, what is it that Bell inequality violations teach us that the double-slit experiment does not?  A common, if quick and dirty, answer is that ``local realism fails.''  Unpacking this phrase, one means more precisely the conjunction of two statements:  1) that actions or experiments in one region of spacetime cannot instantaneously affect matters of fact at far away regions of spacetime, and 2) that measured values {\it pre-exist\/} the act of measurement, which merely ``reads off'' the values, rather than enacting or creating them by the process itself.  The failure of local realism means the failure of one or the other or both of these statements.  This, many would say, is the essential ``mystery'' of quantum mechanics.

But the mystery, as already emphasized, has two sides.  Of the options, it seems the majority of physicists who care about these matters think it is locality (Option 1 above) that goes down the drain with a Bell inequality violation---i.e., that there really are ``spooky actions at a distance.''$^{\rm fn}$  But there is a small minority that thinks the abandonment of Option 2 is the much more sensible conclusion and among these are the {\it quantum Bayesians}. In a slogan near to that of Asher Peres's, ``unperformed measurements have no outcomes.''

$^{\rm fn}\,$Indeed, it flavors almost everything they think of quantum mechanics, including the {\it interpretation\/} of the {\it imaginary\/} games they use to better understand quantum mechanics itself. Take the recent flurry of work on Popescu--Rohrlich boxes.  These are imaginary devices that give rise to greater-than-quantum violations of various Bell inequalities.  Importantly, another common name for these devices is the term ``nonlocal boxes.''  Their exact definition comes via the magnitude of a Bell inequality violation---which entails violation of locality or pre-existence of values or both---but the commonly used name opts only to recognize nonlocality.  They're not called anti-realism boxes, for instance.  The nomenclature is psychologically telling.
\eq
By the way, I really could use that reference to your remarks on double-slits.  Please continue to search your mind.

\bdm
Did you know that (as she claims) the famous Bohr--Einstein
clock-in-the-box exchange was (from Einstein's point of view) not an
attack on the energy-time uncertainty relation but an early version of
EPR, the point of which Bohr simply missed.  I've never seen it
presented that way.  Haven't had a chance to check out the evidence
for that.  It surprised me.  Is that in Arthur Fine's book?  Or Don Howard?
\edm

Well, I do know it morphed into that sometime after the Solvay meeting.  I know that from this passage in Max Jammer's article ``The EPR Problem In Its Historical Development,'' in {\sl Symposium on the Foundations of Modern Physics: 50 Years of the Einstein--Podolsky--Rosen Gedankenexperiment}, edited by P. Lahti and P. Mittelstaedt (World Scientific, Singapore, 1985), pp.\ 129--149, which I have quoted in the original samizdat:
\begin{quote}
[W]hat Einstein had in mind is confirmed by a letter which Ehrenfest wrote to Bohr on July 9, 1931.  As Ehrenfest reports, Einstein uses the photon-box no longer to disprove the uncertainty relation but ``for a totally different purpose.''  For the machine, which Einstein constructs, emits a projectile; well after this projectile has left, a questioner can ask the machinist, by free choice, to predict by examining the machine alone {\it either\/} what value a quantity A {\it or\/} what value an even conjugate quantity B would have if measured on the projectile. ``The interesting point,'' continued Ehrenfest, ``is that the projectile, while flying around isolated on its own, must be able of satisfying totally different non-commutative predictions without knowing as yet which of these predictions will be made\ldots.''
\end{quote}

It would be intriguing if this is what Einstein was trying to get at in the first place.\footnote{[And apparently it was, as I later learned.  See Don Howard's presentation on the subject at \myurl[http://www.nd.edu/~dhoward1/Early\%20History\%20of\%20Entanglement/sld026.html]{http://www.nd.edu/$\sim$ dhoward1/Early\%20History\%20of\%20Entanglement/sld026.html}.]}  On what page of Gilder's book are you referring to?  I'd like to read it.

\section{22-12-08 \ \ {\it Nonlocal Boxes and Ehrenfest, 2} \ \ (to N. D. {\Mermin})} \label{Mermin145}

I think I found the part in Gilder's book you were talking about.  Pages 128--133.  The history is going to be fun to read for me, but the commentary I can already feel is going to be quite painful.  ``They have no states of their own, and, as far as quantum theory is concerned, a measurement performed on one instantly affects its twin.''  More humbug!

\section{22-12-08 \ \ {\it Reading} \ \ (to N. D. {\Mermin})} \label{Mermin146}

Now that I read two more paragraphs down, I see this pisses me off too:
\bq\noindent
Bohr's books and papers---full of careful prohibitions about what cannot be contemplated \ldots---have become holy writ \ldots\  From the point of view of the history of entanglement, they are not worth one clear sentence from Einstein, {\Schroedinger}, de Broglie, or John Bell, \ldots
\eq
Harrumph!!

\chapter{2009: The Most Beautiful Shape}

\section{05-01-09 \ \ {\it After-afterthought} \ \ (to H. C. von Baeyer)} \label{Baeyer50}

\bma
Another example.   You might write down a list of all the things you
know about a person.  It would never occur to you that the description was complete.  Still less would you be inclined to identify the person with your description of them:  to suppose that the person, like the
description, actually consisted of a string of words.   But the
classical physicists were inclined to make an assumption rather similar to that.
\ema

With regard to Marcus's quote below and Hans's remark of liking it, surely in some other context I've sent you this Jamesian beauty before.  But it is so apt here again, surely it is worthwhile sending to you again!

\bq
Let me give the name of `vicious abstractionism' to a way of using concepts which may be thus described: We conceive a concrete situation by singling out some salient or important feature in it, and classing it under that; then, instead of adding to its previous characters all the positive consequences which the new way of conceiving it may bring, we proceed to use our concept privatively; reducing the originally rich phenomenon to the naked suggestions of that name abstractly taken, treating it as a case of `nothing but' that concept, and acting as if all the other characters from out of which the concept is abstracted were expunged. Abstraction, functioning in this way, becomes a means of arrest far more than a means of advance in thought. It mutilates things; it creates difficulties and finds impossibilities; and more than half the trouble that metaphysicians and logicians give themselves over the paradoxes and dialectic puzzles of the universe may, I am convinced, be traced to this relatively simple source. {\it The viciously privative employment of abstract characters and class names\/} is, I am persuaded, one of the great original sins of the rationalistic mind.
\eq

\section{05-01-09 \ \ {\it What I Really Want Out of a Pauli/Fierz-Correspondence Study} \ \ (to H.~C. von Baeyer \& D.~M. {\Appleby})} \label{PauliFierzCorrespondence}

\noindent My two dear friends! \medskip

Of no one else but {\Schack} do I feel such an intellectual kinship---you three are my triumvirate.  Thank you both for the great New Year's reading gift of the last couple of days.  It's funny how these things work out; it arrived at just the right time for me.  For, I was able to put it in juxtaposition with my reading of the latest issue of {\sl William James Studies}.  The two together caused me to think very much about how I want to frame the coming year.  And it struck me as worthwhile to try to record as clearly as I can at the moment what I am {\it personally\/} seeking to get out of a better knowledge of the byways explored by Pauli and Fierz (and potentially Jung).  There is no implication in this that this is what you {\it should\/} be seeking as well---it is only an effort to tie the strands of my life together a little better.

Here is what it is really all about for me---it is the root of the root and the bud of the bud, as e~e cummings put it.  A good vehicle for setting up what I want to say is one of Hans's passages:
\bq
\noindent Pauli summarily dismissed two extreme attitudes -- total separation of science from religion, and complete surrender to mystical experience.  The former approach was advocated in our times by the late paleontologist Stephen Jay Gould, who coined the phrase ``nonoverlapping magisteria'' for the respectful noninterference of the realms of nature and of morality, of what is and what should be.   Discussing a book by a German physicist who also believed in the separateness of science and religion, Pauli had once written to Fierz that he was appalled, and continuing:
\bq
\noindent ``[The book] is a reversion to the 19th century when religion and science lived in separate sections of the human soul -- politely exchanging greetings at a distance, while continually reassuring each other that they had nothing to do with each other -- and when the soul seemed to reside outside the boundaries of science.''
\eq
For Pauli it was obvious that science should be able to deal with the soul, and that the soul in turn can inform science.
\eq
in conjunction with one of Marcus's comments thereafter:
\bq
\noindent It is certainly true that I attach more weight to the opus than I do to quantum mechanics.  That is I am interested in quantum mechanics because I am interested in the opus, rather than the other way round.  For Chris it is, perhaps, a little different.  Though I don't think it is all that different, actually. It is true that Chris doesn't use the language of souls (at least I can't recall him doing so in my hearing).  But he is deeply concerned with, for example, the question of human freedom.  He will have to speak for himself, but I believe that for him too his interest in quantum mechanics is secondary to other considerations.
\eq
For it is true that I rarely speak of ``souls,'' ``religion,'' or ``redemption.''  These terms are mostly dead terms for me---they don't stir my soul, so to speak---or maybe I simply don't understand them well enough yet to see their ultimate usefulness for what I {\it do\/} want to get at.  (Much like I have never understood what the search for ``elegance'' can possibly mean when it comes to forming physical theories, say, as a criterion for string theory:  it is a term that is dead to me.)  It is maybe in this way, or more carefully, {\it in this detail}, that I part company from your tentative feelings on the opus.

{\it Nonetheless}, there is no doubt that I believe there is a place---a very important place---for humanistic concerns within physics proper.  It seems to me it goes to the core of what quantum mechanics is trying to tell us.  You'll find the point made over and over in my ``Activating Observer'' resource-material document, which both of you have versions of.  But I thought the point was made very nicely in the {\it setting of pragmatism more generally\/} in this article that I was reading at the time your emails arrived---``\,`The Many and the One' and the Problem of Two Minds Perceiving the Same Thing'' by Mark Moller in Vol.~3 of {\sl William James Studies}, posted at 
\myurl{http://williamjamesstudies.org/3.1/moller.html}:
\bq
Each of these claims about reality is crucial to James's attempt to offer an alternative to the metaphysical theories of the absolute idealists. The importance of the claim that reality is continuous and in flux is that it offers an alternative to the absolute idealists' view that the universe is ``known by one [infinite] knower in one act, with every feature preserved, and every relation apprehended.'' This means that for the absolute idealists, the universe is forever fixed so as to make real change impossible. We make no difference in such a universe. We neither improve upon it through our efforts nor make it worse. James rejects this view completely. His aim is to argue for a conception of the universe that allows real change to occur in it and where our efforts have a role in causing it. He goes on in the passage from {\sl The Many and the One\/} manuscripts quoted above to make his point:
\bq
\noindent
    This picture of the irremediably pluralistic evolution of things, achieving unity by experimental methods, and getting it in different shapes and degrees and in general only as a last result, is what has made me give to my volume the title of {\sl The Many and the One}.
\eq
According to James, we, as conscious agents in the universe, have an active role in introducing new content and unity into it. Such a view thus aligns his radical empiricism with his meliorism. In earlier essays, eventually published together as {\sl The Will to Believe\/} (1897), and in lectures that he gave to teachers, eventually published as {\sl Talks to Teachers on Psychology\/} (1899), James took the position against the absolute idealists that the ultimate fate of the universe has yet to be decided. He insisted that it is an open question as to whether evil will triumph over good or the other way around. This, in turn, led him to claim that our choices and actions do make a difference in the universe, and, in fact, a crucial one. They help to decide how ``the everlasting battle of the powers of the light with those of darkness'' will turn out. This melioristic attitude only makes sense if the universe is malleable to human action, and, thus, it is one of James's aims in developing his metaphysics to explain how this malleability is possible.
\eq
The very phrases ``our choices and actions do make a difference in the universe'' and ``this melioristic attitude only makes sense if the universe is malleable to human action'' mean outright that there must be room for a humanistic element of some sort within physics itself.

Here's the way I put it to Lucien and Joy Christian when I was in a poetic mood last year:
\bq
\noindent
My view of the universe is that it is many---that it ultimately cannot be unified, for it is alive and changing and creative in a very deep sense.  Moreover, that reality is, to some not-yet well-understood extent, plastic:  It can be molded by our actions.  Thus, though humanity is quite well a Darwinian accident, now that it is here, it is a significant component of the universe that must be reckoned with.
\eq
Finally, it seems worthwhile for me to let James say it himself (from his article ``Pragmatism and Humanism''). It captures with his beautiful sweep the romantic thought that really keeps me going from day to day:
\bq
In many familiar objects every one will recognize the human element.
We conceive a given reality in this way or in that, to suit our
purpose, and the reality passively submits to the conception. You can
take the number 27 as the cube of 3, or as the product of 3 and 9, or
as 26 {\it plus\/} 1, or 100 {\it minus\/} 73, or in countless other
ways, of which one will be just as true as another. You can take a
chess-board as black squares on a white ground, or as white squares
on a black ground, and neither conception is a false one.

You can treat the adjoined figure as a star, as two big triangles
crossing each other, as a hexagon with legs set up on its angles, as
six equal triangles hanging together by their tips, etc. All these
treatments are true treatments---the sensible {\it that\/} upon the
paper resists no one of them. You can say of a line that it runs
east, or you can say that it runs west, and the line {\it per se\/}
accepts both descriptions without rebelling at the inconsistency.

We carve out groups of stars in the heavens, and call them
constellations, and the stars patiently suffer us to do so,---{\it
though\/} if they knew what we were doing, some of them might feel
much surprised at the partners we had given them. We name the same
constellation diversely, as Charles's Wain, the Great Bear, or the
Dipper. None of the names will be false, and one will be as true as
another, for all are applicable.

In all these cases we humanly make an {\it addition\/} to some
sensible reality, and that reality tolerates the addition. All the
additions `agree' with the reality; they fit it, while they build it
out. No one of them is false. Which may be treated as the {\it
more\/} true, depends altogether on the human use of it. If the 27 is
a number of dollars which I find in a drawer where I had left 28, it
is 28 minus 1. If it is the number of inches in a board which I wish
to insert as a shelf into a cupboard 26 inches wide, it is 26 plus 1.
If I wish to ennoble the heavens by the constellations I see there,
`Charles's Wain' would be more true than `Dipper.' My friend
Frederick Myers was humorously indignant that that prodigious
star-group should remind us Americans of nothing but a culinary
utensil.

What shall we call a {\it thing\/} anyhow? It seems quite arbitrary,
for we carve out everything, just as we carve out constellations, to
suit our human purposes. For me, this whole `audience' is one thing,
which grows now restless, now attentive. I have no use at present for
its individual units, so I don't consider them. So of an `army,' of a
`nation.' But in your own eyes, ladies and gentlemen, to call you
`audience' is an accidental way of taking you. The permanently real
things for you are your individual persons. To an anatomist, again,
those persons are but organisms, and the real things are the organs.
Not the organs, so much as their constituent cells, say the
histologists; not the cells, but their molecules, say in turn the
chemists.

We break the flux of sensible reality into things, then, at our will.
We create the subjects of our true as well as of our false
propositions.

We create the predicates also. Many of the predicates of things
express only the relations of the things to us and to our feelings.
Such predicates of course are human additions. Caesar crossed the
Rubicon, and was a menace to Rome's freedom. He is also an American
schoolroom pest, made into one by the reaction of our schoolboys on
his writings. The added predicate is as true of him as the earlier
ones.

You see how naturally one comes to the humanistic principle: you
can't weed out the human contribution. Our nouns and adjectives are
all humanized heirlooms, and in the theories we build them into, the
inner order and arrangement is wholly dictated by human
considerations, intellectual consistency being one of them.
Mathematics and logic themselves are fermenting with human
rearrangements; physics, astronomy and biology follow massive cues of
preference. We plunge forward into the field of fresh experience with
the beliefs our ancestors and we have made already; these determine
what we notice; what we notice determines what we do; what we do
again determines what we experience; so from one thing to another,
altho the stubborn fact remains that there is a sensible flux, what
is {\it true of it\/} seems from first to last to be largely a matter
of our own creation.

We build the flux out inevitably. The great question is: does it,
with our additions, {\it rise or fall in value}? Are the additions
{\it worthy\/} or {\it unworthy}? Suppose a universe composed of
seven stars, and nothing else but three human witnesses and their
critic. One witness names the stars `Great Bear'; one calls them
`Charles's Wain'; one calls them the `Dipper.' Which human addition
has made the best universe of the given stellar material? If
Frederick Myers were the critic, he would have no hesitation in
`turning down' the American witness.

Lotze has in several places made a deep suggestion. We na\"{\i}vely
assume, he says, a relation between reality and our minds which may
be just the opposite of the true one. Reality, we naturally think,
stands ready-made and complete, and our intellects supervene with the
one simple duty of describing it as it is already. But may not our
descriptions, Lotze asks, be themselves important additions to
reality? And may not previous reality itself be there, far less for
the purpose of reappearing unaltered in our knowledge, than for the
very purpose of stimulating our minds to such additions as shall
enhance the universe's total value. {\it `Die Erh\"ohung des vorgefundenen Daseins'\/} is a phrase used by Professor Eucken
somewhere, which reminds one of this suggestion by the great Lotze.

It is identically our pragmatistic conception. In our cognitive as
well as in our active life we are creative. We {\it add}, both to the
subject and to the predicate part of reality. The world stands really
malleable, waiting to receive its final touches at our hands. Like
the kingdom of heaven, it suffers human violence willingly. Man {\it
engenders\/} truths upon it.

No one can deny that such a role would add both to our dignity and to
our responsibility as thinkers. To some of us it proves a most
inspiring notion. Signore Papini, the leader of Italian pragmatism,
grows fairly dithyrambic over the view that it opens of man's
divinely-creative functions.

The import of the difference between pragmatism and rationalism is
now in sight throughout its whole extent. The essential contrast is
that {\it for rationalism reality is ready-made and complete from all
eternity, while for pragmatism it is still in the making, and awaits
part of  its complexion from the future}. On the one side the
universe is absolutely secure, on the other it is still pursuing its
adventures.

We have got into rather deep water with this humanistic view, and it
is no wonder that misunderstanding gathers round it. It is accused of
being a doctrine of caprice. Mr.\ Bradley, for example, says that a
humanist, if he understood his own doctrine, would have to `hold any
end, however perverted, to be rational, if I insist on it personally,
and any idea, however mad, to be the truth if only some one is
resolved that he will have it so.' The humanist view of `reality,' as
something resisting, yet malleable, which controls our thinking as an
energy that must be taken `account' of incessantly (tho not
necessarily merely {\it copied}) is evidently a difficult one to
introduce to novices. The situation reminds me of one that I have
personally gone through. I once wrote an essay on our right to
believe, which I unluckily called the {\it Will\/} to Believe. All
the critics, neglecting the essay, pounced upon the title.
Psychologically it was impossible, morally it was iniquitous. The
`will to deceive,' the `will to make-believe,' were wittily proposed
as substitutes for it.

{\it The alternative between pragmatism and rationalism, in the shape
in which we now have it before us, is no longer a question in the
theory of knowledge, it concerns the structure of the universe
itself.}

On the pragmatist side we have only one edition of the universe,
unfinished, growing in all sorts of places, especially in the places
where thinking beings are at work.

On the rationalist side we have a universe in many editions, one real
one, the infinite folio, or {\it \'edition de luxe}, eternally
complete; and then the various finite editions, full of false
readings, distorted and mutilated each in its own way.
\eq

My anti-Platonist tendencies certainly come out in my feeling of agreement with this quote.  There is no sympathy in it to the idea that one might be {\it attuned\/} with a divine mind privy to ``THE way'' the universe is constructed---as Marcus, I think, rightly points out, seems to have been Einstein's modus operandi when theorizing.  At most, one might be attuned to the content of this other quote of James:
\bq
If you follow the pragmatic method, you cannot look on any [theory] as closing your quest.  You must bring out of each [theory] its practical cash-value, set it at work within the stream of your
experience.  It appears less as a solution, then, than as a program
for more work, and more particularly as an indication of the ways in
which existing realities may be {\it changed}.

{\it Theories thus become instruments, not answers to enigmas, in
which we can rest.}  We don't lie back upon them, we move forward,
and, on occasion, make nature over again by their aid.
\eq
Physical theories, by this view, are conceptual means and tools for making change in the world.\footnote{This is not the dry instrumentalism one incessantly hears of in philosophy of science circles---that an instrumentalist is one who believes a scientific theory is {\it solely\/} an instrument for {\it prediction}.  That is the term of insult they usually throw at Bohr, our quantum Bayesianism, and similar efforts: ``it is merely instrumentalism,'' they say.  (See Asher Peres's funny story at the beginning of \quantph{0310010v1}.)  But prediction only takes a secondary role here, at best.  What's being talked about instead is a far richer concept than those guys---Jeff Bub, Harvey Brown, David Albert, and the like---are even aware of.  For this kind of ``instrumentalism'' (if it should even be called that) the image to associate with a theory is not so much Babbage's analytical engine, but a hammer and a socket set.  [[In a first draft of this footnote I had written ``Babbage's prediction engine''; unfortunately that was not historically correct (as I just learned), and I couldn't bring myself to change the term purely for literary license.]]}  Can one be attuned to such a thing in the Platonic sense that Marcus explores?  Perhaps.  But if so, it's not the medieval Platonism Marcus was talking about---it is instead being attuned to how best be an agent of change.  (Perhaps the exemplar of this modified Platonism would be Barack Obama instead of Einstein \ldots\ so that a pragmatically modified Platonist would look on him and say, ``Now, that man is someone who feels the Old One!'')

In any case, what's important here is that I genuinely do see a humanistic role in the very maintenance of the universe.  And all of this is potentially independent, and certainly broader, than considerations to do with quantum mechanics.  The universe is partially powered by the inhuman, to be sure.  But it is also partially powered by \underline{\it belief}---that I do believe full well.  And to the extent that religion or religious feeling are sources of belief, they do indeed play a role in the very construction of the universe.  However, from this point of view religion is a special case of something much bigger---namely, belief generally.

I'll quote James one last time, so that you get a precise sense of what I mean here, but then I'll get back to the quantum and Pauli/Fierz and tell you way I'm saying all this.  From ``The Sentiment of Rationality'':
\bq
Now, I wish to show what to my knowledge has never been clearly
pointed out, that belief (as measured by action) not only does and
must continually outstrip scientific evidence, but that there is a
certain class of truths of whose reality belief is a factor as well
as a confessor; and that as regards this class of truths faith is not
only licit and pertinent, but essential and indispensable. The truths
cannot become true till our faith has made them so.

Suppose, for example, that I am climbing in the Alps, and have had
the ill-luck to work myself into a position from which the only
escape is by a terrible leap. Being without similar experience, I
have no evidence of my ability to perform it successfully; but hope
and confidence in myself make me sure I shall not miss my aim, and
nerve my feet to execute what without those subjective emotions would
perhaps have been impossible. But suppose that, on the contrary, the
emotions of fear and mistrust preponderate; or suppose that, having
just read [W.~K. Clifford's] {\sl Ethics of Belief}, I feel it would be sinful to act upon an assumption unverified by previous experience---why, then
I shall hesitate so long that at last, exhausted and trembling, and
launching myself in a moment of despair, I miss my foothold and roll
into the abyss. In this case (and it is one of an immense class) the
part of wisdom clearly is to believe what one desires; for the belief
is one of the indispensable preliminary conditions of the realization
of its object. {\it There are then cases where faith creates its own
verification}. Believe, and you shall be right, for you shall save
yourself; doubt, and you shall again be right, for you shall perish.
The only difference is that to believe is greatly to your advantage.

The future movements of the stars or the facts of past history are
determined now once for all, whether I like them or not. They are
given irrespective of my wishes, and in all that concerns truths like
these subjective preference should have no part; it can only obscure
the judgment. But in every fact into which there enters an element of
personal contribution on my part, as soon as this personal
contribution demands a certain degree of subjective energy which, in
its turn, calls for a certain amount of faith in the result---so
that, after all, the future fact is conditioned by my present faith
in it---how trebly asinine would it be for me to deny myself the use
of the subjective method, the method of belief based on desire!

In every proposition whose bearing is universal (and such are all the
propositions of philosophy), the acts of the subject and their
consequences throughout eternity should be included in the formula.
If $M$ represent the entire world minus the reaction of the thinker
upon it, and if $M + x$ represent the absolutely total matter of
philosophic propositions ($x$ standing for the thinker's reaction and
its results)---what would be a universal truth if the term $x$ were
of one complexion, might become egregious error if $x$ altered its
character. Let it not be said that $x$ is too infinitesimal a
component to change the character of the immense whole in which it
lies imbedded. Everything depends on the point of view of the
philosophic proposition in question. If we have to define the
universe from the point of view of sensibility, the critical material
for our judgment lies in the animal kingdom, insignificant as that
is, quantitatively considered. The moral definition of the world may
depend on phenomena more restricted still in range. In short, many a
long phrase may have its sense reversed by the addition of three
letters, {\it n-o-t}; many a monstrous mass have its unstable
equilibrium discharged one way or the other by a feather weight that
falls.
\eq

These, however, are very big things and very big thoughts James is speaking of.  And even though they stir my heart, they remain too vague to transform science as a whole (and indeed the world) in the way that I hope we'll one day transform it.  I love the sound of what I'm hearing, but at the end of the day, I'm still not completely sure what I {\it am\/} hearing.  How do I know I'm not simply fooling myself with pretty words?  The main point I want to make at this juncture is that these ideas which drive me forward---these still-too-vague ideas---are {\it supra}- quantum mechanics.  It is 19$^{\rm th}$ century philosophy, not physics.

What now is the role of quantum mechanics within this system of thought I'm laying out?  It is that it is a miniature version of these general points.  BUT, though it is a miniature version, it is an extremely {\it precise\/} version!  And that's its ace in the hole, as Hans said of Pauli's discussion with Huxley.  Here's where I agree with Marcus,
\bq
\noindent Quantum mechanics is important because it is as close to a refutation of the classical world picture as one could hope to get.  It is very rare that something gets refuted in science as completely and finally as the classical hypothesis, of the world machine, has been refuted.
\eq
Thus, quantum mechanics is a precision laboratory for defining and testing these things {\it we think we see\/} on the horizon.  When we've got it in quantum mechanics, we know we have it.

This finally is where I can express the value for me of learning more about the Pauli--Fierz and the Pauli--Jung correspondence.  There is an amazing amount of development of the metaphysics of a malleable world in the writings of William James and F.~C.~S. Schiller\footnote{F.~C.~S. should not be confused with the famous Schiller, namely Friedrich.  F.~C.~S. stands for Ferdinand Canning Scott---a mostly forgotten pragmatist, who was one of the deepest of the lot if you ask me.}.\footnote{And no disrespect intended, but Pauli and Fierz have got nothing on the sheer volume of even James's and Schiller's correspondence on these subjects, much less their published works.  Pauli was an amateur in comparison \ldots\ as was necessarily the case, being a professional physicist most hours of the day.}  And lo and behold, in which direction does that development lead?  In significant parts at least, it leads in quite the same direction as Pauli's metaphysic.  I don't know that I've ever emphasized this to you two.  When James, for instance, asks the question, ``What are the materials of the universe's composition?,'' the answer he tries to develop is that they are things {\it neutral\/} to the material/mental or physical/psychical distinction.  In fact, he saw the material and mental as {\it complementary}, but exclusive, aspects of the basic stuff.  So, you see the similarity.\footnote{At this point, I'd definitely encourage you to read the whole of the Moller paper mentioned above.  It gives quite a decent summary of the Jamesian metaphysic.  Moreover, in it not only will you see a resemblance between James's ideas and Pauli's ``neutral language'' considerations, but also, in the objections to James's pluralism, you will find some difficulties that we quantum Bayesians (who take an ``alchemical view'' on quantum measurement) must eventually address.}

But James and Schiller knew no quantum mechanics of course.  Particularly, they were not privy to this precision development laboratory for their metaphysic that Pauli and Fierz knew so well.  Therefore, I need, I crave, to know more of the precise things Pauli and Fierz were talking about for just this reason.  The suspicion is that it will inspire us and help us connect the dots between the quantum formalism and the bigger, much richer, and far-from-really-developed idea of a malleable/alchemical world.

Marcus says,
\bq
\noindent Quantum mechanics, as it stands now, is little more than a set of calculational procedures.  The calculations show that the vision of the world-machine is completely without empirical foundation.  However, the mechanical vision of things has not been replaced with any other.
\eq
of which I have a minor qualm, but I think he is right in further saying, ``The classical vision of things \ldots\ was deeply wrong about the relation of mind to matter.'' For I think this neutral stuff that Pauli and James speak of---James called it ``pure experience'' but I am not completely happy with that term---will indeed be the ``shocking'' ingredient that will replace the mechanical vision of things.  So, it is not that we have not started the process of replacement; it is only that it has been slow going.

Hans, I suspect, has wondered why I place so much emphasis on wanting to know the most, within all of Pauli's non-$x$-ing writings and correspondence, about the part to do with the idea of detached and nondetached observers (activating observers).  It is because, as I see it, a thorough study of what it means for an observer or agent to be nondetached from the phenomena he helps bring about is the very starting point for a development of James and Pauli's neutral-stuff ontology.  If we cannot get the activating-observer idea right and understand all its facets, then I feel we have no chance whatsoever of developing a full-blown ``neutral monism''\footnote{Neutral monism is what it is sometimes called in the literature, but a much better name for the idea in the Jamesian context would be ``neutral pluralism.''  Until ten seconds after writing this, I thought I was the first person to make up this term.  But apparently Ruth Anna Putnam beat me to it (damned the power of this internet; it is the second time it has foiled me today):
\bq
[Putnam writes:] Another key element of James's radical empiricism is his rejection of mind/matter dualism as well as its reduction to either materialism or idealism.  In its place, he offers---it is the title of one of his essays---a world of pure experience.  In that world consciousness {\it as an entity\/} does not exist.  But neither is consciousness a function of matter, for matter {\it as an entity\/} also does not exist.  Ultimately there are only pure experiences (and, perhaps, experienceables---that is a difficult interpretative question), experiences which only in retrospect are {\it taken\/} either as part of a stream of thought or as physical objects.  Although one is tempted to call this view a neutral monism, it is, in my opinion, more properly thought of as a neutral pluralism---neutral in not favoring either thought or matter, plural because ``there is no {\it general\/} stuff of which experience at large is made.  There are as many stuffs as there are `natures' in the things experienced \ldots\ and save for time and space (and, if you like, for `being') there appears no universal element of which all things are made.''
\eq
}.  It is the very place to start, for it forces the issue of this new ontology in its very statement.  What is an observer if not psychical?  What is an activator if not physical?  How can we really combine these two aspects of the phenomenon via the help of a Paulian neutral language?

Unfortunately in the technical development of this first step, Marcus and I have hit a great snag by not yet understanding the structure (a.k.a.\ knowing how to prove the existence of) these damned SICs.  The reason I say this is because I see no way forward to a more precise definition of the activating-observer concept than by a kind of ``generalized Newton's third law'' of the style I wrote you about last month.  And I see no way to make progress in that intermediate program without the ability to completely rewrite quantum mechanics (and particularly the Born rule) in terms of SICs.  For completeness, let me reproduce that little essay:
\bq
Everybody has their favorite speculation about what powers quantum
information and computing.  Some say it is the superposition
principle, some say it is the parallel computation of many worlds,
some say it is the mysteries of quantum entanglement, some say it is
the exponential growth of computational space due to the tensor
product.  For my own part though, my favorite speculation is that it
is Newton's Third Law:  For every action, there is an equal and
opposite reaction.  Indeed I sometimes wonder if the very essence of
quantum mechanics isn't just this principle, only carried through far
more consistently than Newton could have envisioned.  That is to say,
absolutely {\it nothing\/} is exempt from it.

What do I mean by this?  What might have been exempt from the
principle in the first place?  To give an answer, let me note an
equivalent formulation of old Newton.  For every {\it reaction}, there is an equal and opposite {\it action}.  Strange sounding, but there's nothing wrong with it, and more importantly, this formulation allows for the possibility of an immediate connection to information theory.  In particular, we should not forget how information gathering is
represented in the Shannon theory.  An agent has gathered
information---by the very definition of the process---when something
in his environment has caused him to {\it react\/} by way of revising a prior expectation $p(h)$ (for some phenomenon) to a posterior expectation $p(h|d)$ (for the same phenomenon).

When information is gathered, it is because we are reacting to the
stimulation of something external to us.  The great lesson of quantum
mechanics may just be that information gathering is physical.  Even
something so seemingly unimportant to the rest of the universe as the
reactions that cause the revisions of our expectations are not exempt
from Newton's Third Law.  When we react to the world's stimulations
upon us, it too must react to our stimulations upon it.

The question is, how might we envision a world with this
property---i.e., with such a serious accounting of Newton's law---but
in a way that does not make a priori use of the information gathering
agent himself?  If the question can be answered at all, the task of
finding an answer will be some tall order.  For never before in
science have we encountered a situation where the theorizing
scientist is so inextricably bound up with what he is trying to
theorize about in the first place.

It's almost a paradoxical situation.  On the one hand we'd like to
step outside the world and get a clear view of what it looks like
without the scientist necessarily in the picture.  But on the other
hand, to even pose the question we have to imagine an information
gathering agent set in the middle of it all.  You see, neither
Shannon nor any of modern information theory has given us a way to
talk about the concept of information gain without first introducing
the agent-centered concept of an expectation $p(h)$.

So, how to make progress?  What we do know is that we actually are in
the middle of the world thinking about it.  Maybe our strategy ought
to be to use that very vantage point to get as close as we can to the
goal.  That is, though we may not know what the world looks like
without the information gathering agent in it, we certainly do know
something about what it looks like with him in:  We know, for
instance, that he ought to use the formal structure of quantum
mechanics when thinking about physical systems.  Beyond that, we know
of an imaginary world where Newton's Third Law was never taken so
seriously:  It is the standard world of classical physics and
Bayesian probability.

Thus, maybe the thing to do first is to look inward, before looking
outward.  About ourselves, at the very least, we can ask how has the
formal structure of our {\it behavior\/} changed since moving from
what we thought to be a classical Bayesian world to what we now
believe to be a quantum world?  In that {\it differential}---the
speculation is---we may just find the cleanest statement yet of what
the quantum world is all about.  For it is in that differential, that
the world without us surely rears its head.

To do this, we must first express quantum mechanics in a way that it
can be directly compared to classical Bayesian theory, where the
information-gathering agent was detached from the world.
\eq

So, you see, there's work to be done all the way down to the SICs, and all the way up to a full-blown ontology of neutral pluralism and a malleable/alchemical reality.  In the case of the SICs, I'm actually more confident:  That problem is going to be solved if I have to recruit all of Hannibal's army.  But in the case of the ontology it's just verbal hints we have presently---there was only so much James, Schiller, Dewey, and others could discern from general premises, without detailed empirical input.\footnote{And only so much patience I have for straining through a set of thoughts framed so far away from direct quantum mechanical considerations.}  Clearly we will have to see our way through in that big task.  But Pauli and Fierz are the only two quantum physicists I know of who have already waded these waters to any depth.  If I could get hold of what little they may have already strained from the rocks below, it would be invaluable for this effort.  And that is what the title of this New Year's note is all about:  What I Really Want Out of a Pauli/Fierz-Correspondence Study.

Thank you both for not losing faith that there is something really important here. \medskip

\noindent Happy New Year!

\section{05-01-09 \ \ {\it Quick Comment} \ \ (to D. M. {\Appleby})} \label{Appleby44}

\bma
\bq\rm
\noindent [Chris said:] An agent has gathered information---by the very definition of the process---when something
in his environment has caused him to REACT by way of revising a prior
expectation $p(h)$ (for some phenomenon) to a posterior expectation
$p(h|d)$ (for the same phenomenon).
\eq
Which is true.  But it needs also to be stressed that what the agent acquires is, not something like this ``je ejszwjm r cly gwjn c yn vqrg eugl srb xqmdbs'' but something like this ``a meaningful statement the agent understands''.

The Shannon theory completely disregards meanings.  This is reasonable enough if you look at it from the narrow perspective of a
communications engineer (it is not the engineer's business to read
the messages he transmits).  But I don't think it is reasonable at all if you look at it from our broader perspective.
\ema

That is true.  However, the long tail of meaning (like the long tail of a rat trying to hide under the cupboard) is not lost altogether.  It is in the setting of the sample space upon which the probabilities are defined.  The sample space is not given by nature alone.  It is the agent himself who sets it.  And that setting reflects the meaning he draws from the various outcomes, and the level of graining with which he decides to take note of nature.

Nonetheless your point on meaning back-reaction is well taken.  If I can dig it up, I have a faint memory of Rorty commenting on Derrida that, for Derrida it is almost that what something is named actually modifies what it is.

Just read your note titled ``Synchronicity???''.  Surely there is no need to rush a reply back to me, or even read what I've sent.  All shakes out in due time!

\section{05-01-09 \ \ {\it F-Theory, 2} \ \ (to R. W. {\Spekkens})} \label{Spekkens56}

I sent the attached essay off to Marcus Appleby and Hans Christian a little while ago, and thought you might enjoy it too.  Especially as the last two pages of it refer to something you once told me you liked in my melange of ideas.  (The rest of it you probably won't like!)  Still, it gives a few more hints of the weird ontology that swirls in the back of my head trying to find substance.  [See 05-01-09 note ``\myref{PauliFierzCorrespondence}{What I Really Want Out of a Pauli/Fierz-Correspondence Study}'' to H. C. von Baeyer and D. M. Appleby.]

\subsection{Rob's Reply}

\bq
You're right -- the last two pages of that note did resonate with me.  In particular, I like this comment:
\bq\noindent
That is, though we may not know what the world looks like without the information gathering agent in it, we certainly do know something about what it looks like with him in: We know, for instance, that he ought to use the formal structure of quantum mechanics when thinking about physical systems. Beyond that, we know of an imaginary world where Newton's Third Law was never taken so seriously: It is the standard world of classical physics and Bayesian probability.
\eq

One of the things that I've been exploring in my toy theories is: what happens if we do take Newton's Third law seriously in the imaginary world of classical physics and we model observers as classical systems that are subject to the law?  The answer is that, assuming no reliable supply of systems in dispersion-free ensembles (to use von Neumann's terminology), observers necessarily cause a disturbance to one variable whenever they gain information about a complementary variable.

Please see the discussion beginning at time index 40:53 and ending at around 45:00 in my talk at \pirsa{08020051}.

Incidentally, there was a piece on religion in today's Guardian which discussed James' views at length.  Given the content of the first half of your note, you may want to have a look:
\myurl[http://www.guardian.co.uk/commentisfree/andrewbrown/2008/nov/07/religion-psychology-william-james]{http://www.guardian.co.uk/commentisfree/andrewbrown/2008/nov/07/re ligion-psychology-william-james}.
\eq

\section{06-01-09 \ \ {\it De Raedt about Bell's Theorem} \ \ (to D. Kobak)} \label{Kobak1}

\bdk
I've recently discovered the papers of Hans De Raedt from Netherlands (there are plenty of them on {\tt arXiv}), whose views on QM seem to be very peculiar. In a series of papers he and his coauthors claim to have constructed local realist ``down-to-earth'' models, mimicking quantum behaviour in different cases. They call it ``event-based computer simulation models'' -- of more or less everything from double-slit single-photon interference to EPR correlations. As far as I understand these models are not supposed to be hypotheses about the real world, but more something like counterexamples to the common misconceptions.

This is certainly most interesting by itself, but what I find completely puzzling is that I failed to find any reply to those papers from other people in quantum foundations community, or any discussions of these ideas whatsoever. Either I'm not looking well enough, or these works are constantly left completely unnoticed. This is maybe my first question: have you heard about these papers, and if yes -- do you have any opinion about that?
\edk

I remember you well from the summer school, and I am flattered that you would think me worthy to try to lead you through these conundrums.  I wish we could have kept you in quantum foundations and that you might have come here to study at PI.  However with your question, I have had enough experience to know that these things are always the same---they are analogous to the experience a patent clerk feels when he is presented with yet another proposed perpetuum mobile.  The poor clerk has the unedifying task of finding where the mistake is hidden.

So, in advance, I'll say I am sorry.  I do not have the stomach for this kind of work.  However, I will do the dirty trick of cc'ing this note to two friends of mine who have found some pleasure in finding the flaws in these kinds of arguments before---Profs.\ Larsson and Gill.  For instance, they both have published experience tackling the Hess--Philipp objections to Bell and also Accardi's very similar sounding claim to De Raedt's of being able to ``simulate'' Bell violations.  Thus maybe they know this work of De Raedt's already and can give you a quick answer.  And if they do not have a quick answer, perhaps they may find it nonetheless a worthy enough windmill to slay.

Good luck to you all and may God save your souls!

\section{06-01-09 \ \ {\it Pauli for the Library?}\ \ \ (to P. Goyal)} \label{Goyal3}

Are you still the foundations representative for the library committee?

My colleague Hans Christian von Baeyer is considering coming from time to time as he is working on a new book on the Pauli--Fierz correspondence on quantum foundations.  Crucial to that program, however, is that he have access to all 9 volumes of the Pauli correspondence (there are 8 volumes presently in existence, and a final one to be published this year).  He has all the volumes at his home department at William \& Mary, of course, but for visits to PI he would have to pack them up each and every time.  And they're not small books.

So the question is, could we purchase them for the library?  They'd certainly be a long-term resource for the institute, as they include Pauli's correspondence with Einstein, Bohr, {\Schroedinger}, Heisenberg, Born, David Bohm, and many, many others.

I'll keep my fingers crossed \ldots

\section{06-01-09 \ \ {\it Pauli for the Library?, 2}\ \ \ (to P. Goyal)} \label{Goyal4}

You might enjoy Hans Christian's book, {\sl Information:\ The New Language of Science}.  I thought it was a great (popular, but serious) account of what people are thinking.  He was a professor of physics (just retired) at William \& Mary, but has also been a writer for many years and has several books (the most recent being on Parisian parks as seen through the eyes of a practicing physicist).  At the moment, aside from working on this new book, he's the book review editor at {\sl American Journal of Physics}.

He's perfect for the job of this Pauli--Fierz (and Pauli--Jung) thing:  a) he's fluent in English and German, b) he's a seasoned writer, c) he's a serious scholar beside, d) he's had a long interest in Jung's thoughts, e) his father was actually Fierz's best friend in college and a life-long friend thereafter, f) Fierz's twin brother, the psychologist, was Hans Christian's godfather, and g) Hans Christian's own brother is a psychologist.  Perfect confluence of events!

Let me attach a little thing for both you and Luca that I wrote yesterday.  [See 05-01-09 note titled ``\myref{PauliFierzCorrespondence}{What I Really Want Out of a Pauli/Fierz-Correspondence Study}'' to H. C. von Baeyer and D. M. {\Appleby}.]  It's a letter to Hans Christian and Marcus Appleby and lays out my metaphysical side fairly decently (with respect to Hans Christian's project).  You might enjoy it for some light early-year reading.

\section{06-01-09 \ \ {\it Pauli Volumes} \ \ (to H. C. von Baeyer)} \label{Baeyer51}

I looked up links between Jung and James yesterday, and was interested to learn that they had actually met in 1909.  Jung was much impressed with James.  They mostly had conversation on parapsychology, and learned that Jung had apparently read {\sl Varieties of Religious Experience\/} and the book {\sl Pragmatism\/} (along with {\sl Principles of Psychology\/} of course).  James in his correspondence with Flournoy only acknowledged meeting ``Yung'' with ``made a very pleasant impression'' and didn't say anything further.  He did however have something snide to say of his meeting with Freud (Freud and Jung were together for the visit); he said he seemed to be a man ``obsessed by fixed ideas''.

\section{06-01-09 \ \ {\it One German Sentence} \ \ (to H. C. von Baeyer)} \label{Baeyer52}

By the way, could I ask you to translate the thing Prof.\ Eucken said in the longer James quote I sent in yesterday's note?

\subsection{Hans's Reply}

\bq
The correct (and this is not negotiable)  capitalization is: ``Die Erh\"ohung des vorgefundenen Daseins.''

A quick translation is ``the enhancement of what is found to exist.''  I have no idea what it means, beyond what precedes it in the passage.

Erh\"ohung is a word one might use to describe the enhancement of  a sensation by means of a chemical stimulant.  It also means intensification.  Vorgefunden means ``as found'' or ``pre-existing''.  Dasein is a typically German fuzzy word meaning ``being'' or ``existence''.  (It can also mean daily life, but not here.)
\eq

\section{06-01-09 \ \ {\it One German Sentence, 2} \ \ (to H. C. von Baeyer)} \label{Baeyer53}

Well, if he could misspell Jung as Yung, I have no doubt he could get German capitalization wrong as well!

Thanks for the translation!

\section{06-01-09 \ \ {\it Causaloid, Meet Psychoid} \ \ (to L. Hardy)} \label{Hardy31}

Yesterday, I put together a little note to try to spur on Hans Christian von Baeyer in the lonely task of distilling what Pauli and Fierz were on about in their correspondence.  (The note is in the attached file, which was actually to both H. C. and Appleby.)  [See 05-01-09 note titled ``\myref{PauliFierzCorrespondence}{What I Really Want Out of a Pauli/Fierz-Correspondence Study}'' to H. C. von Baeyer and D. M. {\Appleby}.]  Hans has already put together an introductory article on that side of Pauli, and I've been encouraging him to tackle a full-blown book project.

Anyway, in my research for the note, I came across this word, which I had never seen before.  So:  Causaloid meet psychoid; psychoid, this is causaloid.
\bq\noindent
{\bf psychoid\/} \ adj. In analytical psychology, soul-like, a term that Carl Gustav Jung (1875--1961) applied to the collective unconscious, which `cannot be directly perceived or ``represented'', in contrast to the perceptible psychic phenomena, and on account of its ``unrepresentable'' nature I have called it ``psychoid''\/' (Collected Works, 8, paragraph 840).  [From Greek psyche the soul $+$ -oid indicating likeness or resemblance, from eidos shape or form]
\eq

In any case, I thought you might enjoy the note for some light reading in these lazy days of the New Year.

\section{06-01-09 \ \ {\it The Metaphysics of SICs}\ \ \ (to K. Martin)} \label{Martin7}

BTW, for your amusement, attached is a letter I sent to Marcus Appleby and Hans Christian von Baeyer yesterday.  [See 05-01-09 note ``\myref{PauliFierzCorrespondence}{What I Really Want Out of a Pauli/Fierz-Correspondence Study}'' to H. C. von Baeyer and D. M. {\Appleby}.]  Hans Christian has just written an article on the more spiritual side of Wolfgang Pauli, and I've been encouraging him to turn his efforts toward writing a full-blown book.  (No ulterior motives here!)  This letter is one of the many mortars I've sent to try to soften him up on the idea; and it reveals a good bit on the weird ontology that I think will ultimately be extracted from quantum mechanics.  By the last three pages, the letter ties SICs back into the picture.  It should be fun reading for you (even if it convinces you I'm insane).  You see, we are working on noble things in this project!

\section{06-01-09 \ \ {\it Remember to get Article on Lotze} \ \ (to myself)} \label{FuchsC22}

\begin{itemize}
\item
Otto F. Kraushaar, ``Lotze's Influence on the Pragmatism and Practical Philosophy of William James,'' J. Hist.\ Ideas {\bf 1}, 439--458 (1940).
\end{itemize}

\section{07-01-09 \ \ {\it De Raedt about Bell's Theorem, 2} \ \ (to D. Kobak)} \label{Kobak2}

\bdk
De Raedt's works can well be completely false, but I wouldn't say that they are just crackpot (regarding what Chris said in the beginning about ``perpetuum mobile'' proposals). The reason I say so is how I learned about those papers: I was told to read them by a prominent solid state physicist, who is interested in quantum foundations as well. He is definitely a very good and very well known theoretical physicist and he was estimating De Raedt very high, as they had chance to work together (on some decoherence issues, not these ``quantum simulators'' of De Raedt).

That's why I became interested in those works, and that's why I was surprised to find almost complete lack of responses from the QM foundations community. I have of course seen the comment by Michael Seevinck and Jan-{\AA}ke Larsson on {\tt arXiv}, but then there's a reply to that comment from De Raedt, and a very detailed one, I would say (\arxiv{0706.2957}\/). As a result, there's an impression that those papers are just left unanswered, and since they don't seem to be ``evidently crackpot'', this situation is somewhat puzzling. I'm honestly sorry if all those topics are completely clear and non-interesting for you, but I can't help wishing that there were a detailed and bottom-line answer to De Raedt's writings, if they are really so false.
\edk

What the lack of response reveals is this:  That almost all the people who run across a paper like this have the same feeling I did when I saw it---i.e., the analogue of, ``yet another perpetuum mobile.''  Then, they don't respond publicly, or even take the time to work through the paper privately, because they don't see that it is worth their while to do so, or that they do not have a lot of understanding to gain from the exercise.

The reason is new ``refutations'' of Bell (and all the other related inequalities and no-go theorems, CHSH, GHZ, Hardy's, etc., etc.)\ crop up almost literally {\it every\/} year.  And this runs for many years back.  De Raedt is by far not an isolated incident.  In the mean time, many very good people (Gill, Larsson, Appleby, {\Mermin}, Shimony, Hardy, van Enk, Kent, Seevinck, more) do in fact, from time to time, make the effort to find the flaws in the new argumentation \ldots\ AND they {\it always\/} find the flaws.  I used to be among that industry when I was younger (20 years ago), though I never published.  I suspect many other people are like me.

So, the issue is mostly one of age and previous experience.  You are still young enough that you have not seen these things come and go for 20 years.  Therefore the attitude you exhibit below is completely healthy and to be commended.  It is not at all that the refutation of each new argument is obvious (at least not for me).  On the contrary, it is often very difficult to ferret out the flaws, and it may take a lot of hard work.  But that doesn't make the exercise less like a proposed perpetuum mobile analysis, it makes it very much of the same flavor.

An education in quantum foundations should consist in exactly what you are doing:  Questioning well received opinion.  But I feel very confident {\it in this particular case\/} that if you
\begin{itemize}
     \item[a)] study very carefully the logic in Bell's argument from
        some good references (I will list some of my personal
        favorites below), and
     \item[b)] very carefully work out for yourself where the De Raedt
        proposal goes astray (perhaps initially being guided by
        the remarks Larsson and Gill have already made),
\end{itemize}
you will come out strengthened from the process, and see more clearly that Bell's reasoning was not arbitrary, but very beautiful instead.

References below.
\begin{itemize}
\item[1)] N. D. {\Mermin}, ``Hidden Variables and the Two Theorems of John Bell''\\
\myurl[http://lin25.thphys.uni-heidelberg.de/~wetzel/qm2005/protected/hidden_var_mermin.pdf]{http://lin25.thphys.uni-heidelberg.de/$\sim$wetzel/qm2005/ \\ protected/hidden\underline{ }var\underline{ }mermin.pdf}
\item[2)] N. D. {\Mermin}, {\sl Boojums All the Way Through\/} (Cambridge U. Press, 1990)\\
\myurl[http://books.google.com/books?hl=en\&lr=\&id=qT7J3Zl_OesC\&oi=fnd\&pg=PR11\&dq=Mermin+Boojums\&ots=R6bDgt_MOy\&sig=CS5E947mMQr7yOPt4IydEElus7c]{http://books.google.com/books?hl=en\&lr=\&id=qT7J3Zl\underline{ }OesC\&oi=fnd\&pg=PR11\&dq=\\ {\Mermin}+Boojums\&ots=R6bDgt\underline{ }MOy\&sig=CS5E947mMQr7yOPt4IydEElus7c}

\item[3)] S. L. Braunstein and C. M. {\Caves}, ``Wringing Out Better Bell Inequalities''\\
\myurl{http://www-users.cs.york.ac.uk/schmuel/papers/90/bcB90.pdf}
\end{itemize}

\section{07-01-09 \ \ {\it Some History of the Block Universe} \ \ (to D.~M. {\Appleby} \& H.~C. von Baeyer)}

I am sorry, this is one paragraph I had not responded to:

\bma
I was thinking particularly of what in the notes I call the ``classical world-machine''.  I didn't use the phrase ``block-universe'' because I had in mind the whole development of classical physics.  I don't believe that Newton was thinking in terms of the block-universe:  that particular version of the nightmare only comes with relativity.  I guess you could say that the block-universe represents the final stage in the evolution of the world-machine.  The senescence of the world-machine?  The corpse of the world-machine?
\ema

I have no problem with your new term; I quite like it.  But I just wanted to point out that historically the ``block universe'' phrase/imagery did in fact precede relativity by quite a substantial time.  I believe the first public appearance of the concept was in William James's 1884 lecture ``The Dilemma of Determinism,'' where he writes:
\bq\noindent
     What does determinism profess? \ldots\ It professes that those
     parts of the universe already laid down absolutely appoint
     and decree what the other parts shall be. The future has no
     ambiguous possibilities hidden in its womb; the part we call
     the present is compatible with only one totality. Any other
     future complement than the one fixed from eternity is
     impossible. The whole is in each and every part, and welds
     it with the rest into an absolute unity, an iron block, in
     which there can be no equivocation or shadow of turning.
\eq
Later in the essay, he writes:
\bq\noindent
     The more one thinks of the matter, the more one wonders that
     so empty and gratuitous a hubbub as this outcry against
     chance should have found so great an echo in the hearts of
     men. It is a word which tells us absolutely nothing about
     what chances, or about the modus operandi of the chancing;
     and the use of it as a war-cry shows only a temper of
     intellectual absolutism, a demand that the world shall be
     a solid block, subject to one control, which temper, which
     demand, the world may not be bound to gratify at all.
\eq
And finally again:
\bq\noindent
     A mind to whom all time is simultaneously present must see
     all things under the form of actuality, or under some form
     to us unknown. If he thinks certain moments as ambiguous in
     their content while future, he must simultaneously know how
     the ambiguity will have been decided when they are past. So
     that none of his mental judgments can possibly be called
     hypothetical, and his world is one from which chance is
     excluded. Is not, however, the timeless mind rather a
     gratuitous fiction? And is not the notion of eternity being
     given at a stroke to omniscience only just another way of
     whacking upon us the block-universe, and of denying that
     possibilities exist? --- just the point to be proved. To
     say that time is an illusory appearance is only a
     roundabout manner of saying there is no real plurality, and
     that the frame of things is an absolute unit. Admit
     plurality, and time may be its form.
\eq

In my further records in my computer, I also have this use of the term from his posthumous book {\sl Some Problems of Philosophy}.  Here, he asks rhetorically, ``If the time-content of the world be not one monistic block of being, \ldots'' \ Now, the manuscript for that book was certainly written after the discovery of special relativity (though certainly before 1910 when James died), but I am pretty sure he had no knowledge of relativity.  Let me quote the passage in detail a) because I can (it's already in my computer), and b) because it makes some connection to our discussion of religion, and that may be useful in the future. What I found in re-reading this passage is that though talk of the soul doesn't seem to stir my soul (as I wrote to Hans and you in the previous note), apparently I am to some extent in tune with a moralistic take on things.  [Good God, my American upbringing!]

\bq
But pluralism, accepting a universe unfinished, with doors and
windows open to possibilities uncontrollable in advance, gives us
less religious certainty than monism, with its absolutely closed-in
world. It is true that monism's religious certainty is not rationally
based, but is only a faith that `sees the All-Good in the All-Real.'
In point of fact, however, monism is usually willing to exert this
optimistic faith: its world is certain to be saved, yes, is saved
already, unconditionally and from eternity, in spite of all the
phenomenal appearances of risk.

A world working out an uncertain destiny, as the phenomenal world
appears to be doing, is an intolerable idea to the rationalistic
mind.

Pluralism, on the other hand, is neither optimistic nor pessimistic,
but melioristic, rather. The world, it thinks, may be saved, on
condition that its parts shall do their best. But shipwreck in
detail, or even on the whole, is among the open possibilities.

There is thus a practical lack of balance about pluralism, which
contrasts with monism's peace of mind. The one is a more moral, the
other a more religious view; and different men usually let this sort
of consideration determine their belief.

So far I have sought only to show the respective implications of the
rival doctrines without dogmatically deciding which is the more
true. It is obvious that pluralism has three great advantages:---

1. It is more `scientific,' in that it insists that when oneness is
predicated, it shall mean definitely ascertainable conjunctive forms.
With these the disjunctions ascertainable among things are exactly on
a par. The two are co-ordinate aspects of reality. To make the
conjunctions more vital and primordial than the separations, monism
has to abandon verifiable experience and proclaim a unity that is
indescribable.

2. It agrees more with the moral and dramatic expressiveness of life.

3. It is not obliged to stand for any particular amount of plurality,
for it triumphs over monism if the smallest morsel of
disconnectedness is once found undeniably to exist. `Ever not quite'
is all it says to monism; while monism is obliged to prove that what
pluralism asserts can in no amount whatever possibly be true---an
infinitely harder task.

The advantages of monism, in turn, are its natural affinity with a
certain kind of religious faith, and the peculiar emotional value of
the conception that the world is a unitary fact.

So far has our use of the pragmatic rule brought us towards
understanding this dilemma. The reader will by this time feel for
himself the essential practical difference which it involves. The
word `absence' seems to indicate it. The monistic principle implies
that nothing that is can in any way whatever be absent from anything
else that is. The pluralistic principle, on the other hand, is quite
compatible with some things being absent from operations in which
other things find themselves singly or collectively engaged. Which
things are absent from which other things, and when,---these of
course are questions which a pluralistic philosophy can settle only
by an exact study of details. The past, the present, and the future
in perception, for example, are absent from one another, while in
imagination they are present or absent as the case may be. If the
time-content of the world be not one monistic block of being, if some
part, at least, of the future, is added to the past without being
virtually one therewith, or implicitly contained therein, then it is
absent really as well as phenomenally and may be called an absolute
novelty in the world's history in so far forth.

Towards this issue, of the reality or unreality of the novelty that
appears, the pragmatic difference between monism and pluralism seems
to converge. That we ourselves may be authors of genuine novelty is
the thesis of the doctrine of free-will. That genuine novelties can
occur means that from the point of view of what is already given,
what comes may have to be treated as a matter of {\it chance}. We are
led thus to ask the question: In what manner does new being come? Is
it through and through the consequence of older being or is it matter
of chance so far as older being goes?---which is the same thing as
asking: Is it original, in the strict sense of the word?
\eq

BUT WAIT!!!  BULLETIN!!!  I love the power of the internet!  I just discovered that James used the block universe idea as early as 1882.  This is a lower bound that I had not known of before!  Moreover this passage is funny to boot; it comes from an article titled, ``On Some Hegelisms.''  Here it is:
\bq
\noindent Why may not the world be a sort of republican banquet of this sort, where all the qualities of being respect one another's personal sacredness, yet sit at the common
table of space and time?

To me this view seems deeply probable.  Things cohere, but the act of cohesion itself implies but few conditions, and leaves the rest of their qualifications indeterminate. \ldots\

[I]f we stipulate only a partial community of partially independent powers, we see
perfectly why no one part controls the whole view, but each detail must come and be actually given, before, in any special sense, it can be said to be determined at all.  This is the moral view, the view that gives to other powers the same freedom it would have itself,---not the ridiculous `freedom to do right,' which in my mouth can only mean the freedom to do as {\it I\/} think right, but the freedom to do as {\it they\/} think right, or wrong either.  After all, what accounts do the nethermost bounds of the universe owe to me?  By what insatiate conceit and lust of intellectual despotism do I arrogate the right to know their secrets, and from my philosophic throne to play the only airs they shall march to, as if I were the Lord's anointed?  Is not my knowing them at all a gift and not a right?  And shall it be given before they are given?  {\it Data!\  gifts!}\ something to be thankful for!  It is a gift that we can approach things at all, and, by means of the time and space of which our minds and they partake, alter our actions so as to meet them.

There are `bounds of ord'nance' set for all things, where they must pause or rue it. `Facts' are the bounds of human knowledge, set for it, not by it.

Now, to a mind like Hegel's such pusillanimous twaddle sounds simply loathsome. Bounds that we can't overpass! Data!\ facts that say, ``Hands off, till we are given''!\ possibilities we can't control!\ a banquet of which we merely share! Heavens, this is intolerable; such a world is no world for a philosopher to have to do with.  He must have all or nothing.  If the world cannot be rational in my sense, in the sense of unconditional surrender, I refuse to grant that it is  rational at all.  It is pure incoherence, a chaos, a nulliverse, to whose haphazard sway I will not truckle.  But, no!\ this is not the world.  The world is philosophy's own,---a single block, of which, if she once get her teeth on any part, the whole shall inevitably become her prey and feed her all-devouring theoretic maw.  Naught shall be but the necessities she creates and impossibilities; freedom shall mean freedom to obey her will; ideal and actual shall be one: she, and I as her champion, will be satisfied on no lower terms.
\eq

Near the end of the article, there's still one more use of the term:
\bq\noindent
In the universe of Hegel the
absolute block whose parts have no loose play, the
pure plethora of necessary being with the oxygen of
possibility all suffocated out of its lungs there can
be neither good nor bad, but one dead level of mere
fate.
\eq

Now I've got the fire in me to try to really figure out the first time the phrase was used either in correspondence or print.  I never imagined I would actually use any of the early volumes of {\sl The Correspondence of William James}, so I only accumulated Vols.\ 7--12 for my library.  But that only delves back to 1890.  It's time to go book shopping again!

\section{08-01-09 \ \ {\it Fuchs--Graaf Inequality} \ \ (to A. Uhlmann)} \label{Uhlmann1}

\bau
Mike Hellmund and I just sent a short paper to the archive, [\arxiv{0812.0906}], extending an inequality of you and Graaf to arbitrary dimensions.
You have used the inequality for an entanglement of assistance estimate. But this we could not do in higher dimensions yet.
\eau

It is so good to hear from you, though I must apologize for taking so long to respond.  The holidays were unusually extensive this year.

That is a nice result of yours.  I will try to tuck it away in the bookshelf of my mind.  Particularly lately I have been wondering about the ``significance'' of low order trace powers.  For instance, suppose $A$ is a Hermitian matrix.  Then one can characterize when $A$ is a rank-1 projection in the following way.  $A$ is a rank-1 projection if and only if $\tr A^2 = \tr A^3 = 1$.  One need not consider higher trace powers (and consequently higher order elementary symmetric functions).  The statement comes about trivially mathematically, but I keep wondering about whether there is any deep information theoretic significance to it (for the purpose of quantum foundations, that is).

Here is a more technical question on my mind.  I wonder if you have any insight.  Let us have a d-dimensional Hilbert space, and let us {\it suppose\/} a SIC exists.  That is, suppose there is a set of $d^2$ rank-1 projections $P_i$ such that $\tr(P_i P_j)=1/(d+1)$.  If you are not familiar with the concept, see the attached paper.  Now here's the question:

\bq\noindent
Let $A$ be a Hermitian matrix such that
\begin{itemize}
\item[1)] $\tr A = 1$,
\item[2)] $\tr A^2 = 1$, and
\item[3)] $0 \le \tr(A P_i) \le 1$, for all $i=1, \ldots, d^2$.
\end{itemize}
Is there an interesting characterization of the further conditions required to ensure that $\tr A^3 = 1$ (and hence we have a rank-1 projection)?
\eq

I hope you are doing well.  How is your health?  What is your age now?  It has been many years since we have seen each other.

\section{08-01-09 \ \ {\it My Book Selections} \ \ (to Cambridge University Press)} \label{CUPBookSelections}

Simon Capelin instructed me to select a few ``Cambridge books to the value of about US\$200'' and contact you in remuneration for a report I wrote for him.  See note below.

After a few hours at your website (it was a lot of fun!), I came to the following four selections:
\begin{itemize}
\item
{\sl William James and the Metaphysics of Experience}\\
David C. Lamberth\\
(Hardback) ISBN: 052158163X\\
99.00 USD
\item
{\sl The Divided Self of William James}\\
Richard M. Gale\\
(Paperback) ISBN: 0521037786\\
39.99 USD
\item
{\sl Wittgenstein and William James}\\
Russell B. Goodman\\
(Paperback) ISBN: 0521038871\\
37.99 USD
\item
{\sl The Dappled World}\\
Nancy Cartwright\\
(Paperback) ISBN: 0521644119\\
34.99 USD
\end{itemize}
The total for these four comes to 211.97 USD.  I hope that this is an appropriate amount.

\section{09-01-09 \ \ {\it That Pragmatism/Science Conference} \ \ (to C. Smeenk)} \label{Smeenk1}

Thanks for bringing that to my attention.  I just wrote the note below to one of the organizers.  I don't know about the organization behind it (the ``Center for Inquiry Transnational''), but their ``VP of Research'' is John Shook.  He has like a gazillion volumes on classical pragmatism that he's edited (great collections of old papers pro and con).  I have literally 20 volumes on my bookshelf, and he has at least one proper book {\sl Dewey's Empirical Theory of Knowledge and Reality}.  (I know that because I have it on my shelf as well; apparently I picked it up in brand new condition for \$6.98, while the list price is \$49.95.)

Is there any chance that you and/or Wayne might go to the conference?

\section{09-01-09 \ \ {\it Wild Chance -- James Correspondence} \ \ (to C. Misak)} \label{Misak7}

I'm wracking my brain over who I know that might be remotely interested in obtaining a good price on some volumes of {\sl The Correspondence of William James}.  I think the chances are slim that you'll be interested, but I'll ask anyway.

My dilemma is this:  I presently own volumes 7--12, but have been thinking I really should back-fill volumes 1--6 (particularly as I've realized I'm missing some crucial years when James was starting to rebel from the block universe idea; I've lower bounded the phrase ``block universe'' to 1880--1882, but want to know if it goes further back).  So, I want to get vols 1--6.  But these volumes generally cost in the \$70--90 US range per volume $+$ shipping, etc.  On the other hand, Amazon.com has a special package for all 12 volumes that works out to \$33US/vol.

On wild chance would you have any interest in buying the top half of the volumes while I buy the bottom half?

Just an idea!  But if you have no interest, do have an idea of anyone else who might?  Or any other ideas in general?

\section{09-01-09 \ \ {\it Wild Chance -- James Correspondence, 2}\ \ \ (to C. Smeenk, W. C. Myrvold, D. Fraser, and S. Dea)} \label{Smeenk2} \label{Dea1} \label{Fraser1} \label{Myrvold10.1}

I just wrote the letter below to Cheryl Misak, but as I figured she
wasn't a taker (already has them all nearby).  So I'm widening the net
now.  Please see note below.  \$33/volume new really is a fantastic price
(except for volumes 1 and 12, for some reason, which crop up here and
there for lower); see \myurl[http://www.abebooks.com/servlet/SearchResults?an=Skrupskelis\&sts=t\&tn=Correspondence+of+William+James\&x=86\&y=13]{http://www.abebooks.com/servlet/SearchResults?an=Skrupskelis\&sts=t\&tn= Correspondence+of+William+James\&x=86\&y=13}.

Is there any chance any of you would be interested in the top half of
James's correspondence, i.e., 1890--1910?  There's a significant number
of letters to and from C. S. Peirce in there.  Supposing you're not
interested, do you know anyone who might be?

\section{09-01-09 \ \ {\it Pragmatism Library}\ \ \ (to S. Dea)} \label{Dea2}

By the way, if you ever need anything for short loan, I'm building up quite a pragmatism library at home.  Just let me know.  Attached is my present book list.

To that, I've also got a few things on the way.  CUP owed me for some reviewing I did for them, so I've just gotten them to send:
\begin{itemize}
\item
{\sl William James and the Metaphysics of Experience} \\
David C. Lamberth\\
(Hardback) ISBN: 052158163X; 99.00 USD

\item
{\sl The Divided Self of William James}\\
Richard M. Gale\\
(Paperback) ISBN: 0521037786; 39.99 USD

\item
{\sl Wittgenstein and William James}\\
Russell B. Goodman\\
(Paperback) ISBN: 0521038871; 37.99 USD

\item
{\sl The Dappled World}\\
Nancy Cartwright\\
(Paperback) ISBN: 0521644119;
34.99 USD
\end{itemize}

And then I went on a purchasing binge with Atticus and have these on the way:
\begin{itemize}
\item
Herman J. Saatkamp, Jr. ed.\\
{\sl Rorty \& Pragmatism: The Philosopher Responds to His Critics}\\
Norman, Oklahoma, U.S.A.: Vanderbilt Univ Pr 1995, 1st Edition Fine Hardcover Fine.\\
Price: USD 14.00

\item
Hodgson, Shadworth H.\\
{\sl The Metaphysic of Experience}, 4 Volumes\\
Bristol: Thoemmes 2000, 1st Edition No Jacket Hardcover. \\
Price: USD 90.00

\item
Wright, Chauncey (1830--1875)\\
{\sl The Evolutionary Philosophy of Chauncey Wright}. ( 3 Volumes) \\
Chicago, Illinois, U.S.A.: Thoemmes Pr 2000, 1st Edition No Jacket Hardcover.\\
Price: USD 90.00
\end{itemize}

Finally, I've got John Shook's annotated bibliography of pragmatism, 1898--1940, coming too.  (It's got over 4,000 references in it.)

\section{09-01-09 \ \ {\it New Propositions} \ \ (to S. Capelin)} \label{Capelin3}

It's a long time since I've gotten in contact with you, particularly with regard to your nice note below.  But the new year is here, and it is time to come up for some air.  I have several things I'd like to discuss with you.  I'll number them individually.

1)  I finally took up your kind offer of ``about US\$200'' in CUP books in payment of the Aaronson review.  I sent the attached note to Laura Clark but have not heard back from her yet.  [See 08-01-09 note ``\myref{CUPBookSelections}{My Book Selections}'' to Cambridge University Press.] Might I ask you to confirm that she got the note?

2)  With regard to ``My Struggles with the Block Universe'' you will see from the running compilation on my web page that it has not developed too much since you last saw it.  When I was most committed to developing the document last year and working at it fanatically, my stride was broken by an emergency grant proposal I had to write, and there is something about getting a fanatical stride broken:  I never quite returned to the project.  It's a bit like a horse tripping in a race, as opposed to tripping in a quiet pasture.

Still, I will return to it.  I have every intention of posting a big---overabundantly complete---version of it on {\tt arXiv.org} before the summer is here.  And after that I would very much like to pursue publishing an abridged version of it with someone, maybe CUP if you're still interested.  In the meantime, however, I'd like to field two more immediate proposals with you \ldots\ that are in fact more likely to be sounder business propositions for CUP anyway.

3)  The first has to do with the fact that I never ended up following through with Springer on their proposal to publish my {\sl Notes on a Paulian Idea}.  A bit before I first met you, they wrote me (23 May 2008):
\bq\noindent
   Should we not receive any response from you in the coming 6 weeks, we will
   consider the project stopped and the contract Null \& Void.
\eq
I never replied, and I think I am happier for it.  Particularly, I am free of them, and the more I think about it and have gotten to know the things you have published, the more I am intrigued by the idea that CUP is the right home for the document anyway.  Only last week did I learn that you published Mermin's {\sl Boojums\/} as well as his more recent books.

So, well before thinking about {\sl My Struggles}, what would you think of publishing {\sl Notes on a Paulian Idea}?  You have seen Mermin's foreword to it, and my previous self-compilation of praise on it (oh those mirrors!)---I sent you that last June---but let me show off two more recent comments, one from a young author, and one from an old one.  I guess what they suggested to me is that the book really might be of broader readership than I had contemplated when Springer first said ``yes'' so long ago.  [I.e., I think I really am ready to publish it now and want it done right.]

The first is praise from Louisa Gilder who wrote this in the acknowledgments of her new book {\sl The Age of Entanglement}:
\bq\noindent
     Thanks also to Steve Weinstein, who gave me {\sl The Shaky Game}; and
     especially to Andrew Whitaker, for writing the beautiful {\sl Einstein,
     Bohr, and the Quantum Dilemma}, and to Chris Fuchs for his wonderful
     samizdat, {\sl Notes on a Paulian Idea}---these two books sat on my desk
     rather than on the shelf.
\eq
And she accompanied the book she sent me with a personal letter saying,
\bq
     Three years ago you gave me your samizdat and it has been my constant
     desk-side companion ever since.  My copy is now dog-eared, bookmarked,
     further indexed, and underlined everywhere and yet I'm still finding
     wonderful unexplored sections (just recently I read your letters to
     Henry Folse for the 1st time, culminating in the great story about
     Ben Schumacher and Star Wars!) \ldots

     P.S.  Have you ever read the tiny book {\sl 84 Charing Cross Road}?  It
     has exactly nothing to do with quantum information theory (it is a
     series of letters between a book lover in N.Y.C. and a bookseller in
     London) but somehow it feels to me like a companion volume to {\sl Notes
     on a Paulian Idea}---that atmosphere of erudite but unsolemn enthusiasm,
     whether for books or P.O.V.M.s, is very satisfying and so vital.
\eq

The second praise came from Hans Christian von Baeyer.  You must have seen his many books popularizing physics, or perhaps know him.  One New Year's Eve, after probably too much wine, he wrote me the very sweet letter in the second attachment.  [See H. C. von Baeyer's preply to my 03-01-07 note ``\myref{Baeyer26}{New Year's Alchemy}''.]

So there:  Between these quotes and the ones I had sent you previously, I think I can do no more salesmanship.  If you want the book, you can have it, and I would say on very short order:  I think little more needs to be done in further preparation of it than my adding a good subject index.  The presently best indexed version can be found at my website in case you would like to re-peruse it.

4)  This leads me to another project that I would like to pursue in short order---something I dreamed up only recently---and I wonder whether CUP might have an interest in it?  Asher Peres was a very dear friend of mine, and recently I gave some lectures in Sydney in his honor at the first Asher Peres International (summer) School of Physics.  My lectures were titled ``Quantum Foundations, Asher Peres Style,'' and the associated transparencies can be found here
\begin{center}
\myurl{http://www.perimeterinstitute.ca/personal/cfuchs/Peres\%20School\%201-110.pdf}
\end{center}
in case you want to peruse them too (though they won't be of much use without lecture recordings).

Anyway, in preparation for those lectures, I reread several of his foundational papers, and I was transported back to my youth in physics.  I had simply forgotten how influential those papers were on me.  And, indeed, how wonderfully written and concise and forcefully argued they all were.  They lay the foundation for a vision of quantum mechanics that is diametrically opposite to the vision of it in Oxford (the vision of David Deutsch, Simon Saunders, Harvey Brown, Jeremy Butterfield, David Wallace, and Artur Ekert), and it struck me:  What better place to have those papers republished than Cambridge!  I think one could easily make a nice volume of his papers, and the result would get beautiful reviews from the likes of David Mermin, Abner Shimony, Ben Schumacher, and others.  (Do you recall Mermin's fantastic review of Peres' book?)  I think I counted that Asher had some 70 foundational papers, and we could pick and choose the very best and most influential on the community.  The running theme in most every one of them is the idea that ``Quantum States Do Not Exist'' (a phrase of mine, but you'll find his precise quote on page 5 of my lecture; unbeknownst to me, his was the precursor to my own)---and he explores every aspect of that idea in these papers.  I made that the theme of my lectures, and I propose to make it the title of the book.  I would write an appropriate introduction explaining the far-influences of Asher's thought (and how it is still developing in pockets of the world) and put Asher, the man, in context with respect to the development of quantum information and the quantum information community.  Attached is the piece on him I wrote for his memorial volumes of {\sl Foundations of Physics}; it may give a sense of my writing style with respect to an introduction for the book.  I would very much like to tribute Asher in this new way---giving him a new chance to influence the world of quantum foundations the way he influenced me.

That's my four topics.  I hope I didn't write you too much to read in one go.

And I hope all is well for you and your family in this new year.

\section{11-01-09 \ \ {\it More on Pauli} \ \ (to D.~M. {\Appleby} \& H.~C. von Baeyer)}

Boy you sure make it hard for me to feel literary!  I thought I was simply being clever with my choice of words---i.e., that talk of the soul doesn't stir my soul!  It wasn't an accident.  Nonetheless, you have many good points; I liked the soul vs.\ mind analysis.

We should start thinking about when the two of you might converge on PI again.  When do you plan to come next, Marcus?  Will you come for your school's Spring break?

It is fantastic what precipitated out of this New Year's discussion of ours.  I've been reading voraciously again.  I think I'm actually going through a second pragmatic awakening, of the force of the one Matthew Donald's prod set alight nearly 10 years ago.  A whole new set of pieces have fallen into place.  This time it's not about the theory of truth, but about pure experience.  A quantum measurement is pure experience!  (In the technical sense.)  It is an example of the neutral stuff Pauli was seeking to describe.  I hope to give you both an extended report in the coming week.

\subsection{Marcus's Preply, 11-01-09}

\bq
I'll start (and, I fear, possibly finish) with some thoughts provoked by these passages
from Hans's notes
\bq\noindent
Gieser's section ends: ``Science as a discipline must in turn realize that science
created by man always includes statements about man. The object of science
will therefore always be man himself and his totality; in him is the ethical
conflict between good and evil, in him is spirit and matter.''
\eq
and
\bq\noindent
That's the thing that eternally amazes me. How can the cynical modern
materialists get around the fact that since time immemorial most of humanity
has been religious?
\eq
and this from Chris's
\bq
For it is true that I rarely speak of ``souls,'' ``religion,'' or ``redemption.'' These
terms are mostly dead terms for me---they don't stir my soul, so to speak---or
maybe I simply don't understand them well enough yet to see their ultimate
usefulness for what I {\it do\/} want to get at. (Much like I have never understood what
the search for ``elegance'' can possibly mean when it comes to forming physical
theories, say, as a criterion for string theory: it is a term that is dead to me.) It is
maybe in this way, or more carefully, {\it in this detail}, that I part company from
your tentative feelings on the opus.

{\it Nonetheless}, there is no doubt that I believe there is a place---a very important
place---for humanistic concerns within physics proper. It seems to me it goes to
the core of what quantum mechanics is trying to tell us.
\eq
(I could have quoted much more from Chris in this connection, but I won't because
cutting and pasting from Acrobat is a pain).

Indeed. As I hope I made clear in my last notes I don't think there is any {\it essential\/}
difference between Chris and myself. It is just that he tends to focus strongly on the
question of freedom while I tend to cast my net a bit wider. For instance, I talk of
souls whereas he doesn't.

{\it Why\/} do I talk of ``souls''? I used to have a conscience about using this word because I
felt I didn't know what it means. And I still don't know what it means: if someone
were to ask me what exactly I hope to convey by the word ``soul'' which I could not
equally well convey by the word ``mind'' I would struggle. Struggle badly, in fact.
Nevertheless, I now do use the word, because, although I don't have a very clear idea
of what I mean by it, I am very sure that I do mean {\it something}. Something important.

The word ``soul'' is a word you simply don't use in respectable scientific discourse.
Which is curious because there was a time when the words ``soul'' and ``mind'' were
regarded as synonymous (if I remember correctly Descartes treats them as synonyms,
for instance). But in the contemporary scientific culture it seems that that is no longer
the case. The word ``mind'' is frequently used in the scientific literature; the word
``soul'' almost never. I suppose one might argue that it isn't used because it is
regarded as archaic (in the way that the word ``verily'' is archaic, for instance).
However, I don't think that can be the explanation because the word is in common
use outside the world of science. Not only would every scientist understand what
Chris means when, in the passage I just quoted, he says ``they don't stir my soul''. I
think that on occasion they would use that phrase themselves. However, they would
not do so in a scientific paper. The word seems to be regarded as being somehow
``unscientific''. The curious thing is that if you asked most scientists what exactly is
wrong with the word ``soul''---what offensive feature it has got which the word
``mind'' hasn't---I think they would probably struggle. Indeed, I think they would find
it as hard to explain why they {\it don't\/} use the word as I find it to explain why I {\it do}.
Nevertheless, they do seem to have a strong feeling that the word is inappropriate
when used in a scientific context. This is an observable fact about the psychology of
the average scientist. I take it to be evidence that the word has as much meaning for
the average scientist as it does for me, even though neither of us can explain quite
what the meaning is.

I think the answer to the question ``why do I use the word?'' may be that I am using it
reactively. In avoiding the word ``soul'' I think the average scientist means to deny
something. And whatever it is that he/she wants to deny, I want to affirm.

But, although I can't give an exact definition, perhaps I can give a rough indication.
Consider once again the passage from Chris's notes which I just quoted. He begins
by saying that the terms ``soul'', ``religion'' and ``redemption'' are dead for him, and
then for additional emphasis he says that they ``don't stir my soul, so to speak''. I
can't resist pointing out that there seems to be a contradiction here: for doesn't the
fact that Chris uses the phrase ``stir my soul'' in this way show that the word ``soul'' is
very far from being dead for him? Be that as it may, however. Whatever Chris meant
or didn't mean by it, that phrase ``they don't stir my soul'' gets right to the heart of the
matter. I think the word ``soul'' is an important word because it enables one to say
things like this:
\bv
It stirs my soul\medskip\\
I felt it in my soul\medskip\\
He put his soul into it\medskip\\
\ev
etc.\ etc. Whatever it is that is conveyed by such phrases cannot be conveyed in mind-language.
Just consider:
\bv
It stirs my mind\medskip\\
I felt it in my mind\medskip\\
He put his mind into it\medskip\\
\ev
So what exactly is the difference between these two groups of phrases? ---Part of it
is, I think, that the word ``mind'' is generally used to refer to the cold-blooded,
ratiocinative aspects of psychic functioning. There has always been a tendency in
modern science to attach overwhelming importance to reason. But in the course of
the 400 years that have elapsed since the genesis of modern science in the 17th century
the concept of reason has itself become narrower and narrower. The end result is that
nowadays the word ``mind'' tends to convey little more than the idea of an extremely
complicated calculating machine. The word ``soul'', by contrast, conveys a concept
which is admittedly more primitive but which is also (it seems to me) richer, and
more complete.

For instance when Chris wanted to expand on the idea of a term being ``dead'' to him
he was forced to use the word ``soul'' rather than the word ``mind'' because for a
calculating machine every term is dead.

I make a point of using the word ``soul'' because I just flat don't believe that a human
being is nothing more than a calculating machine. I should say that I do so with
misgivings. For it is, of course, true that the concept of the soul is primitive. It has a
distinctly medieval feel to it. So there is a danger of my giving the impression that I
am a Luddite, who wants to undo the work of the last five centuries and return to the
Middle Ages. Of course, that isn't really the idea at all. My admiration of the 17th
century creators of modern science is heartfelt. It is just that I share Pauli's belief
that, along with their great achievements, the 17th century thinkers made some serious
mistakes. My aim in all of this is only to do what I can to correct those mistakes.
However, I do worry that my use of medieval language will obscure that fact.
\eq

\section{12-01-09 \ \ {\it Happy Birthday William James} \ \ (to H. C. von Baeyer)} \label{Baeyer54}

Thank you for that fine birthday present for William James.  And thank you for teaching me that James's birthday precedes my daughter Emma's [today] by just one day!

I did find the point about thin language interesting.  When you and Marcus arrive in Waterloo, perhaps you can help thicken my sauce a bit!

\subsection{Hans's Preply, ``Happy Birthday William James,'' 11-01-09}

\bq
Here's a thought I wrote this morning before reading your two contributions.
\begin{center}
\bf On Language
\end{center}
In connection with the role of theology and the opus in our study of quantum mechanics I thought it would be useful to hear what Fierz had to say.  Contrary to Pauli, who was an atheist by conventional Judeo-Christian standards, Fierz was a Christian, who felt free to interpret scripture in strictly metaphorical ways of his own choosing.  Here's what he wrote to Pauli on 11 February 1956:
\bq
\ldots\ The old story of paradise --- which of course is much older than the creation story --- culminates in humans eating from the tree of knowledge and then getting thrown out of the garden of Eden.  In my opinion this story does not mean anything sexual, but the sexual just happens to be part of the problem.  The story primarily means simply that we have now internalized the apple of {\it Erkenntnis\/} (knowledge, discovery, recognition) and that there is no way back.

From this I conclude --- for the story is believable and good --- that the way to salvation is mediated by {\it Erkenntnis}.  But this {\it Erkenntnis\/} is not a revelation.  It is work --- by the sweat of our brow --- thistles and thorns --- an {\it opus}.  Besides thistles the field yields fruit, too, and we may enjoy those.  But the enjoyment of the fruits of our {\it Erkenntnis\/} is not yet an enjoyment of power.

Erkenntnis also implies that opposites are recognized --- Adam ``recognized'' Eve; namely as woman who was the opposite of him, the man.\footnote{[Fierz's footnote.]  Contra Plotin and other observers: {\it Erkenntnis\/} is the the work of {\it Erkenntnis}, which requires a methodology.  Contraries are recognized, i.e.\ one recognizes that they exist, not that they actually don't exist!}

The discovered opposites will probably have to remain separated, at least for now.  That's all right, because, or inasmuch as, they have been truly discovered.  In the discovery they are unified insofar as there is someone who can see them both and who is prepared to carry the burden that is thereby put on him --- i.e.\ the cross.

All this sounds a bit theological; but I don't know a better language.  The philosophical-conceptual language is too thin for me, too lacking in images, and then I have no idea what I'm talking about.

From {\it Erkenntnis\/} I expect salvation (wholeness, well-being).  Yes, I mean even from scientific {\it Erkenntnis\/} or research (for it has a methodology).  This is not a rationalizing point of view.  For the object of {\it Erkenntnis\/} is precisely the irrational.  Internal images and fantasies are just as much objects as the external world.  In order to recognize them we need a methodology with which we can operate on them (experiment!).  In this methodology I include associations, amplification of images; your [Pauli's] comparison with mythological and other materials, etc.
\eq
I thought the bit about language would interest Chris, who's on the other side of the fence.
\eq

\section{13-01-09 \ \ {\it The Remarkable Theorem} \ \ (to A. Uhlmann)} \label{Uhlmann2}

\bau
Yes, this is an unexpected characterization by Jones and Linden
[\quantph{0407117}] and, if I am right, independently by Patrick Hayden.
\eau

[See 08-01-09 note ``\myref{Uhlmann1}{Fuchs--Graaf Inequality}'' to A. Uhlmann for a statement of the characterization.] I am pretty sure that I reported it to Patrick, though my memory may not serve me well.  The only independent discoverer (beside Jones and Linden) I know of is Steve Flammia, at the time a student of Carlton Caves.  When he first showed Carl the result, Carl didn't believe him.  Later I was told about it, and I was shocked:  I thought, surely it must be a folk theorem that Caves and I had somehow missed.  So, I got in the habit of asking everyone I knew about whether they had ever seen the little theorem.  And the interesting thing was that no one had!  Here is a list of people I asked about it:  Holevo, Leib, Ruskai, Calderbank, Lindblad, Gottesman, John Conway, and many more than that.  No one had.  So, it wasn't a folk theorem after all, even though the result is almost trivial.  Thus, in my presentations, I got in the habit of naming it ``the remarkable theorem'' and giving profuse credit to Flammia and Jones and Linden for unearthing this beautiful little mushroom.

\bau
Of course, the set of faithful density operators can be described by
a finite set of strict inequalities, you mentioned the elementary
symmetric functions.
\eau

That's right; I am aware of this.

\bau
I do not know. But the manifold of rank one projections has at every
of its points an individual tangential hyperplane giving a necessary
linear inequality. Therefore, I think, the additional conditions you
are looking for cannot be linear if there are finitely many. I am also
pessimistic about substituting a degree three polynominal equation by
some degree two ones --- but this may be a wrong expectation if there
are in addition some distinguished linear equations or inequalities.
\eau

Yes, I guess I was {\it hoping\/} (no intuition, just hoping) that the inequalities might bring the remaining conditions down to some degree two conditions.  I suppose I'm pessimistic too.  May I ask you, however, to expand on your remark about a ``necessary linear inequality''.  (I.e., your second sentence in this paragraph.)  I don't know what you are saying here.

\bau
If $P_1$, $P_2$, \ldots\ is a SIC, one perhaps needs some insight of what phases occur in the traces of their products.
\eau

Appleby and I will soon post a very detailed study of the ``triple products'':  I mean $\tr( P_i P_j P_k )$.  They have a very remarkable structure.

\bau
Next month I become 79.  Up to now I can be satisfied.  Of course, there are always things to be repaired!
\eau

Very good!  It is always good to become prime once again!

\section{14-01-09 \ \ {\it The Remarkable Theorem, 2} \ \ (to A. Uhlmann)} \label{Uhlmann3}

\bau
There is a curious convex set, given by the intersection of \ $\tr\, \omega^2 \le 1$ and $\tr\, \omega^3 \le 1$. This convex set is definitely larger than the state space if $\mbox{dim} > 2$. Do you know whether somebody has looked at it?
\eau

I had contemplated looking at it at the Cambridge workshop on ``foils to quantum theory'' two summers ago.  But then nothing ever materialized of the hope---other things got in the way.  I also wanted to look at this convex subset of the probability simplex over $d^2$ points:  The one where $\sum p(i)^2 \le 1$ and $\sum p(i)^3 \le 1$.  I don't know that anyone has looked at these things in detail.  If you get any results, I would love to hear about them!

\section{15-01-09 \ \ {\it Waterloo Morning} \ \ (to D. M. {\Appleby} and Several Others)} \label{Appleby45}

\noindent Dear old collaborators (in the warm parts of the world),\medskip

Don't you wish you were up here in beautiful Waterloo with me this morning?  It's $-29$C outside ($-20$F for the Americans)!  Think of the grand discussions we could have on the quantum on our walk to PI!

\section{15-01-09 \ \ {\it Exchange of the Delusional}\ \ \ (to K. Martin)} \label{Martin8}

Thanks for sending that.  I had a great time reading it to my wife last night and trying to get the voices right.  Beautiful writing---it felt it must have really captured the circumstances and atmosphere.  But don't think for a moment there's no quantum there!  I like the way William James put it:
\bq
Taken as it does appear, our universe is to a large extent chaotic.  No one single type of connection runs through all the experiences that compose it.  If we take space-relations, they fail to connect minds into any regular system.  Causes and purposes obtain only among special series of facts.  The self-relation seems extremely limited and does not link two different selves together.  Prima facie, if you should liken the universe of absolute idealism to an aquarium, a crystal globe in which goldfish are swimming, you would have to compare the empiricist universe to something more like one of those dried human heads with which the Dyaks of Borneo deck their lodges.  The skull forms a solid nucleus; but innumerable feathers, leaves, strings, beads, and loose appendices of every description float and dangle from it, and, save that they terminate in it, seem to have nothing to do with one another.  Even so my experiences and yours float and dangle, terminating, it is true, in a nucleus of common perception, but for the most part out of sight and irrelevant and unimaginable to one another.  This imperfect intimacy, this bare relation of withness between some parts of the sum total of experience and other parts, is the fact that ordinary empiricism over-emphasizes against rationalism, the latter always tending to ignore it unduly.  Radical empiricism, on the contrary, is fair to both the unity and the disconnection.  It finds no reason for treating either as illusory.
\eq
In the vision of the world I hope to construct, there'll always be room for the likes of you, me, and even the phone lady.  My experience is a string, yours a bead, and hers a feather.  They are no lesser parts of the world than the results of the LHC.  And it is quantum mechanics that gives us the most compelling case for that indication!

Attached is a little document I put together for Marcus Appleby and Hans Christian von Baeyer at New Year's.  [See 05-01-09 note ``\myref{PauliFierzCorrespondence}{What I Really Want Out of a Pauli/Fierz-Correspondence Study}'' to H. C. von Baeyer and D. M. {\Appleby}.]  In other contexts I've been calling it ``The Metaphysics of SICs.''\footnote{It is a strange thing that those ``other contexts'' would be my note to Keye just nine days earlier, in which I had sent him the very same attachment.  However, I have a faint memory that I did this on purpose for one reason or another.}  You think I'm spending all this time just trying to find better descriptions of quantum channels?  No, I've got much bigger things in mind!

I hope you enjoy this exchange of the delusional!

\section{17-01-09 \ \ {\it It's Always Einstein}\ \ \ (to P. G. L. Mana)} \label{Mana12}

\bpglm
As for correspondences, I suppose you already have the book `Letters on Wave Mechanics:\ {\Schroedinger}, Planck, Einstein, Lorentz'?
\epglm

I read that book many years ago.  I remember a most amazing letter from Einstein to {\Schroedinger}, soon after S had found his time-independent equation.  Einstein misread the draft of the paper and got the equation wrong.  So he wrote to {\Schroedinger} a letter saying what he didn't like about the (incorrectly recalled) equation.  Then he said, ``You will notice that the following equation'' (in which case he wrote down the proper {\Schroedinger} equation as if he were the inventor of it) ``has all the appropriate properties.''  Finally he ended the letter by saying, ``however I can find no interpretation of this function $\psi$ which is its solution''.

Beautiful!  Tell me if I'm right with my memory.\footnote{\editornote See footnote \ref{Einstein1926} on page~\pageref{Einstein1926}.}

\section{19-01-09 \ \ {\it Phishing for Filosophy}\ \ \ (to R. Blume-Kohout)} \label{BlumeKohout7}

I never have quick responses to anything \ldots\ because I never have quick thoughts on anything.  The best I can ever muster is to attach an old, partially relevant email, as I will do here.  But it was good to juxtapose Fish's article today (which I had not read before your note) with this other one in today's NY Times that I had:  \myurl{http://www.nytimes.com/2009/01/19/books/19read.html?hp}.  The attachment concerns my way of saying the humanities are important to the quantum world, as the quantum world is important to the humanities.  (It was a New Year's letter written to Appleby and Hans Christian von Baeyer.)  [See 05-01-09 note ``\myref{PauliFierzCorrespondence}{What I Really Want Out of a Pauli/Fierz-Correspondence Study}.''] I'll leave it to you to tell me whether it means I think physics is a liberal art.

\section{20-01-09 \ \ {\it From Pragmatism to Pure Experience} \ \ (to M. B\"achtold)} \label{Baechtold2}

Thank you for sending your new email address.  I hope you are enjoying your new position and finding it a productive place to think.  I can see that you are somewhere in France, but could not figure any more detail than that from your email address.  Where are you now?

Thank you for enquiring on the paper.  I am in the middle of its construction and hope to have it for you in about a month.  It has been very tough going for me---first with the collapse of Bell Labs, then the flood of our home in New Jersey, and then the move to Canada.  But all is clear intellectually now, and this paper has become more important than ever for me.  In the course of developing it, I have greatly clarified my thinking.  Particularly, I ended up in a place I had not expected to.  What I try to argue now is that in a very serious sense, quantum measurement is an instance of Jamesian ``pure experience''---and that quantum mechanics is our laboratory for finally making that notion precise.  This is quite a turn for me.  My initial attraction to pragmatism had been solely for ``theory of truth'' reasons:  It was a means of making sense of the idea of quantum measurements being generative (i.e., as not simply revealing pre-existent facts), yet, at the same time, allowing that one could be ``certain'' of a measurement outcome via an initial pure-state assignment.  The certainty must be interpreted as subjective Bayesian certainty (i.e., it lives in the realm of ideas), and in any measurement interaction, it can be {\it made\/} true or false.  So, the ``truth'' arising from a quantum state assignment is something that comes after the fact---at the conclusion of measurement---and is not something pre-existent with the state itself.  (For after all, ``quantum states do not exist'' as the Bayesian slogan goes.)  But that was my initial attraction to the pragmatist line of thought.  Where I go now is to understand quantum measurement as an instance of a more general phenomenon---something that is neutral between the physical and the mental (agential would be a better second term).  I try to view these ``pure experiences'' as the active monads of the world, similar to James and similar to John Wheeler with his ``elementary acts of observer-participancy'' being the building blocks of the world.  What I am trying to do in the paper is to show that this is the logical endpoint of my earlier less-radically-empirical pragmatism.  And I hope to set up a framework for making these issues much more precise than hitherto possible.

I hope you will like the final result, and that it will have been worth your wait.

\section{21-01-09 \ \ {\it Old Poetry} \ \ (to G. L. Comer)} \label{Comer120}

\bgc
Congratulations on your new post!  Will you get to meet the head of NSA?
\egc
I'm aiming for the Obama cabinet.  I suspect he needs some advice on quantum foundations.  \ldots\ Though he already shows promise, indeed:  He seems to already have good pragmatist credentials.  Did you notice the line about our ``uncertain destiny'' in yesterday's speech?

\section{21-01-09 \ \ {\it More Obagmatism} \ \ (to G. L. Comer)} \label{Comer121}

This one from Paul Begala's report of the speech on CNN.  (Paul was at the University of Texas just before me, or maybe there was some overlap.  He lost against Hank the Hallucination, a cartoon character, when he ran for student-body president.)
\bq
The president closed by quoting the words from Tom Paine that Gen.\ Washington ordered read aloud at Valley Forge. ``Let it be told to the future world,'' Washington said, ``that in the depth of winter, when nothing but hope and virtue could survive \ldots\ that the city and the country, alarmed at one common danger, came forth to meet [it].''

The quote shows Obama's belief in unflinching courage, unblinking realism and an unrelenting faith that by coming together we Americans can bend history to our will.
\eq
What other philosophy than pragmatism would admit that we ``can bend history to our will''?

\section{21-01-09 \ \ {\it Pauli Retally} \ \ (to R. W. {\Spekkens} and L. Hardy)} \label{Spekkens57} \label{Hardy32}

The title's got to give you a giggle.

After Dawn's high number yesterday for these volumes, I decided to
look into it myself.  The result is attached.  A more accurate number
is around \$1600 CAD total (or \$1270 US), excluding shipping.

I think Rob is right.  Except for {\Appleby}'s and my quest for clues
toward constructing a ``psycho-physically neutral'' language (as Pauli
called it) for quantum phenomena---which I do very much consider
modern research in quantum foundations---this purchase really should
be considered more in the line of a resource for the history of
physics.  I.e., in most anyone's hands but ours, it will be historical reading.
And maybe that is not what PI should be in the business of---at least
not at such a price premium.

Thus, I want to make the following offer and see how you'll consider
it.  If PI will pony up half the expenses, I'll cover the other half with
my ONR grant.  And PI gets a great set of books (however you classify
the subject) that it can house forever.  Or at least until the inevitable
fire that comes to all book collections.

Does that sound reasonable?  If so, then I'd like to get to action so
that some volumes might be here before {\Appleby}'s and von Baeyer's
visit Feb 15.

\section{21-01-09 \ \ {\it Pauli Retally, 2} \ \ (to H. C. von Baeyer and D. M. {\Appleby})} \label{Appleby46} \label{Baeyer55}

Well, I tried (see below), but it seems neither of my senior colleagues in foundations are willing to stand up for a Pauli-correspondence purchase for the PI library.  The view of Rob, at least, was that though it may be a resource for the history of physics that is not our business.  And when he discovered that the volumes are not translated even, that really pushed him over the edge.  Oh you lucky ones at the ``liberal arts'' institutions; your libraries are so much more tolerant.

I even tried to sweeten the pot by promising to launder half the expenses from Marcus and my (still itinerant) ONR grant.  But it was a no go.  And somehow I couldn't find myself laundering the full expenses when I honestly feel it should be housed at PI.

Thus before you arrive, I will try to get as many of the volumes as I can by Guelph--Waterloo library loan.

Maybe I should apply for a humanities grant to ease my conscience.

\subsection{Marcus's Reply}

\bq
Compared with the size of the ONR grant \$1270 isn't that much.  I wouldn't mind at all if you used the grant to cover the whole cost.
Though maybe it ought to be in the possession of someone who can read German (definitely not me!).

I think more hangs on the distinction between ``physics'' and ``history of physics'' than meets the eye.  In other words the fact that it appears obvious to almost  everyone that you can pursue research in ``fundamental physics'' whilst having, at best, only a very sketchy, journalistic understanding of the  ``history of physics'' shows that almost everyone is looking at physics in a profoundly wrong manner.  I think it has to do with objectivism.

The mistake the objectivists make is to misunderstand the relation between (on the one hand) the world and (on the other hand) our thoughts about the world.  In particular objectivists tend to identify the conceptual/linguistic/mathematical structures pertaining to our understanding of the object with the object itself (e.g.\ Everettians identify the state vector of the universe with the universe itself).
It is a really weird mistake when you come to think of it.  As if I were to identify the name ``Pauli'' with the man of whom it was the name and, furthermore, to assume that anyone who challenged that identification was denying the reality of the man himself.  But people don't see that it is weird.  And that shows that the error is very deep-seated.

The idea that ``history of physics'' is irrelevant to physics itself is due, I think, to the fact that people are thinking that ``history of physics'' is to do with us (our ways of understanding), while physics itself they tend to identify with the object (what we are trying to understand). So that makes it look as though ``history of physics'' and ``physics'' are two entirely different things.

It is not so, of course.  Physics is an intellectual structure, of our own creation.  It is true that it has what the philosophers call
``intentionality'':  it points at something different from us.  But just as a pointing finger, though it points at something distinct from the body, is itself part of the body; so physics, though it is about things different from us, is still itself part of us.

The fact that physics is specifically {\it our\/} intellectual structure means that it is, like us, something organic.  And it is fundamental to organic entities that they evolve:  in other words, that they have a history.  It would never occur to a neurophysiologist that you can try to understand the brain whilst completely ignoring the brain's evolutionary history.  The same principle applies (or should apply) to physics which, after all, is a production of the brain. (I was going to say it is a production of the brain at the software level.  But I am not sure that is right.  One of the differences between brains and computers is that there is not the same clear-cut distinction between software and hardware in the case of the brain.  A brain which knows quantum field theory is, probably, physically very different from one which only knows how to make stone axes.)

I am presently trying to understand why some {\sl Mathematica\/} code is giving me the wrong answer.  To do that I am going back over all my working, starting from the beginning.  In other words, I am examining the history of how I got to this point.  It is surely the natural way to proceed.

Similarly with quantum mechanics.  The reason that there is a whole industry devoted to the foundations of quantum mechanics, but no such industry devoted to the foundations of relativity, is that quantum mechanics seems, as judged from the perspective usual among physicists, at least on the face of it, to make no sense  (wave function collapse, etc etc).  In tackling this problem most people start from the assumption that it only {\it seems\/} to make no sense when judged from our usual perspective.  However, it is surely no less reasonable to think that there is something wrong with the usual perspective.  To track down the error it is surely very reasonable to examine the evolution of that perspective.  To look at the history of it, in other words.  And it seems to me that such an examination obviously belongs to physics proper.
\eq

\section{22-01-09 \ \ {\it Bose-Einstein-Bayes}\ \ \ (to N. C. Menicucci)} \label{Menicucci4}

\bnm
What would you say is the key difference between quantum and classical
indistinguishability of particles (e.g., considering bosons/fermions
versus classical particles), if any?
\enm

I wish I had a real answer for you, but I don't.  The only thing I know is that I still stick by the tenets below.  [See 12-03-07 note ``\myref{Comer103}{More Serendipity}'' to G. L. Comer.]

\section{23-01-09 \ \ {\it Feynman Comment?, 3} \ \ (to N. D. {\Mermin})} \label{Mermin147}

Are you still out there?

Did you ever have any insight in the question below?  If you cannot recall any of your own writings, can you recall anyone else's?

I've had an excruciatingly hard time writing this paper, but it is slowly coming together.  The upshot is I want to return ``interference of probabilities'' to the place in quantum foundations where Feynman thought it was.  (Though Bayesianified and quantum informationified, of course.)  It'll help the drama if I have some quotes of opposition to fight it out with.

You never told me you didn't like the title {\sl Notes on a Paulian Idea\/} before.

\section{23-01-09 \ \ {\it First Reports on NPI} \ \ (to S. Capelin)} \label{Capelin4}

Well, Ref B is pretty clearly Scott Aaronson.  There's only one lunatic out there who's ever read the whole thing!\footnote{\editornote No longer true.}

\section{24-01-09 \ \ {\it Pauli Retally, 3} \ \ (to D. M. {\Appleby} and H. C. von Baeyer)} \label{Appleby47} \label{Baeyer56}

[Concerning {\Appleby}'s reply to ``\myref{Appleby46}{Pauli Retally, 2}'' \ldots]  That was indeed an eloquent defense!  Would you mind if I forward it to Rob and Lucien?  I've given up on the book issue, but I think it's something they should think about for the future.

The reason I've been reluctant to purchase the set solely for ourselves is in significant part the language issue.  Putting it in the library and making it available for everybody was the way I had planned to ease my conscience.  I mean, ideally I'd like to go through the volumes page for page and catalogue potentially interesting items, but that's most of the use I'd have for them.  (Anticipating my only having the patience to scan for significant words, since I don't know the language.)  If I felt I was doing more good for the community, I could see doing that.  But without PI (as embodied by Rob and Lucien) recognizing some value in the gift, it kind of galls me to give it to them.

\section{02-02-09 \ \ {\it Your PhD Thesis}\ \ \ (to P. G. L. Mana)} \label{Mana13}

\bpglm
During the last few days I've been thinking about our lunch discussion, about what Chiribella said at the panel discussion, and about how you enthusiastically commented on his words. There's much in the current language we use to describe quantum-mechanical phenomena that throws sands on our eyes and precludes us from thinking better, deeper, unbiassedly, and along new paths. Bell criticised the word `measurement', but there are many, many other words that lurk around in our physical speech and cover our eyes.
Yesterday at the group meeting we discussed, at some point, on whether
we could do without the concept of particle or of wave in quantum
mechanics and quantum field theory. Then, in presenting an experiment
(Mach--Zehnder), sentences like `there is one photon in this mode' came
forth. In presenting experiments with such a language we are already
committing to an idea, and there is therefore little hope for us to be
capable to reinterpret the physics behind them in new ways.
\epglm

This is a point that Ed Jaynes made very beautifully somewhere as well, I recall.

\bpglm
Personally, I can speak of any quantum experiment whatever without
ever mentioning the word `particle' even once, and translate what the
others say in that particle-less language.
\epglm

Similarly for me:  the concepts of both particle and wave hold no use in my own understanding of the world.

\bpglm
Take for example the sentence `The probability of finding the system
in such and such a state', which I hear almost daily.
\epglm

You {\it never\/} hear that language from me.

\bpglm
In classical analytical mechanics people speak all the time about the `evolution of the probability distribution' on phase space.
That's nonsense because a probability is timeless; there is not even
any updating in the Bayesian sense when we speak about that
`evolution'. And even the term `updating' is misleading indeed. Bayes'
formula simply relates probabilities having different contexts.
\epglm

I believe de Finetti put the latter point like this:  It is only the following out of our set opinion.

\section{03-02-09 \ \ {\it Title Thoughts} \ \ (to S. Capelin)} \label{Capelin5}

I'm jumping the gun here, but I'm sober enough to know that if you guys do accept the book, the title at the least will have to be changed.  I just wanted to record these thoughts while they're on my mind, and you seem like the appropriate repository for my finding the note again.
\begin{itemize}
\item
Coming of Age with Quantum Information
\item
What's Under Quantum Information?
\item
Quantum Information, Quantum Reality, Quantum Personality
\end{itemize}
Just playing with words---I know it's got to make connection to a wider audience, but it also has to be true to what the book was intended to be.

Hope you don't mind my bothering you like this.

\section{04-02-09 \ \ {\it Title Thoughts, 2} \ \ (to S. Capelin)} \label{Capelin6}

\bsc
Actually, I like the suggestion that the book should include brief biographical sketches of the people your
emails were sent to.
\esc

I do as well.  It is a good idea and, I think, it would make for a fun project (to try to say something fun about my old friends).

\bsc
Would you be willing to tighten it up a little?
\esc

Potentially.  Since the beginning, I have been acutely aware of the repetition factor.  I tried very hard to guard against it when I first put the thing together.  But the trouble remained that I have variations and variations on the formulations of the ideas I was seeking.  Some of them never converged; sometimes the different formulations continued to hold positive aspects that I did not want to throw away.  But I will think about the idea of tightening---one thing I fear is that it would be a major undertaking.

Also, there is the comment of Referee A.  One of the reasons {\it I believe\/} you can dip into the book almost anywhere and pick up the thread is precisely that so much of the book is repetitious, working out variations on variations.  After all, it is just notes on an idea.

I didn't like my new titles either much.  So I'm glad you're of two minds.

\section{04-02-09 \ \ {\it That Savage}\ \ \ (to R. {\Schack})} \label{Schack150}

Do you remember a place where L. J. Savage said something {\it like\/} ``there is a continuous scale between pure coherence and objective probability  and the issue is, where does one make the cut?  Objectivists go to one extreme, radical personalists go to the other.  One may well think there a further normative rules beyond coherence, but when does one stop the process of adding more structure before the abyss of objective probability?''

At least I recall a discussion like that somewhere.  I'd like to quote a bit of it for our paper, but cannot find it.  Any hints you have would be most useful.

\section{05-02-09 \ \ {\it Savage Morning}\ \ \ (to R. {\Schack})} \label{Schack151}

By the way, I came across this paper, which I found interesting:  L. J. Savage, ``Difficulties in the Theory of Personal Probability,'' {\sl Philosophy of Science\/} {\bf 34}, 305--310 (1967).

The penultimate paragraph harkened back to some of our discussion from the early days of ``you're going to tell me a beamsplitter isn't a beamsplitter!'':
\bq
The idea of facts known is implicit in the use of the preference theory. For one
thing, the person must know what acts are available to him. If, for example, I ask
what odds you will give that the fourth toss of this coin will result in heads if the
first three do, it is normally implicit not only that you know I will keep my part of
the bargain if we bet but also that you will know three heads if you see them. The
statistician is forever talking about what reaction would be appropriate to this or
that set of data, or givens. Yet, the data never are quite given, because there is
always some doubt about what we have actually seen. Of course, in any application,
the doubt can be pushed further along. We can replace the event of three
heads by the less immediate one of three tallies-for-head recorded, and then take
into our analysis the possibility that not every tally is correct. Nonetheless, not
only universals but the most concrete and individual propositions are never really
quite beyond doubt. Indeed, as you know better than I, such seeming statements
of fact must actually be recognized as universals. Is there, then, some avoidable
lack of clarity and rigor in our allusion to known facts? It has been argued that
since indeed there is no absolute certainty, we should understand by ``certainty''
only strong relative certainty. This counsel is provocative but does seem more to
point up, than to answer, the present question.
\eq

I also liked this line from the bottom of page 307 / top page 308:
\bq\noindent
Yet, in the very notion of this choice of a framework there are impressive difficulties.  Is it good, or even possible, to insist, as this preference theory does, on a usage in which acts are without influence on events and events without influence on well-being?
\eq

\section{05-02-09 \ \ {\it The Paulian Idea} \ \ (to S. Capelin)} \label{Capelin7}

I'm OK with what you say about the title.

But here's another potential idea for the future with regard to the title:  Use `Notes on a Paulian Idea' as a subtitle.  {\sl Coming of Age with Quantum Information:\ Notes on a Paulian Idea}.  That would preserve a bit of continuity with the previous edition, but possibly put a more public face on it.

The choice of title came about because reading Pauli particularly turned my conception of quantum mechanics upside down.  Not Bohr, not Heisenberg, but Pauli.  It was through him that I started to realize that one might have it all:  A Copenhagen-style understanding of quantum measurement {\it without\/} giving up on reality itself (as so many are apt to accuse me of).  [See note titled ``\myref{Preskill2}{The Reality of Wives}'' [in this volume] to get a sense of the constant defense I'm on.]  The beginnings of my transition are documented in the first three pages of the Letters to Greg Comer chapter.  Then really, almost all the correspondence throughout built up around the core of Pauli's idea \ldots\ seen through the lens of quantum information and Bayesianism.  The two quotes on page vii are what I consider the essence of the Paulian idea.  Nondetached observers, yes, ones who participate in shaping reality, but nonetheless raw reality too---i.e., the world is not a dream or an expression of my fancy.  That was the idea that hit me over the head, and it is why I paid homage to it with my title.  (Note: Those early letters to Comer were included for completeness; otherwise all correspondence in the book comes from my postdoc period, starting in 1996.)

I would like to deliver the book to you {\it no later\/} than June 1.  I'm trying my hardest to make this Spring a great house-cleaning exercise.

\section{06-02-09 \ \ {\it Swedes, PDFs, and Experiments} \ \ (to R. Laflamme)} \label{Laflamme4}

Speaking of the geometry of Hilbert space, do you think you could do an experiment like the one in Figure 1 of the attached paper?!?!  [See ``Quantum-Bayesian Coherence,'' \arxiv{0906.2187v1}, but Figure 2 instead.]  It would be another way to test the Born Rule, and I got the impression you have some interest in that.  The first interesting system with regard to these considerations, however, would be a quTRIT.  The paper is still under construction.  In fact, it is far from complete (many of the main results are not in it yet, sections unfinished, tentative paragraphs, etc.), but the present draft does at least contain the definitions needed to make sense of what I'm talking about.  Let me know if you have an interest in learning more.  (This is the stuff I'm getting my ONR funding for.)

\section{06-02-09 \ \ {\it 4.1 and PI}\ \ \ (to R. {\Schack})} \label{Schack152}

By the way, this may have been the Savage quote I was thinking of:
\bq
All views of probability are rather intimately connected with one another.  For example, any necessary view can be regarded as an extreme personalistic view in which so many criteria of consistency have been invoked that there is no role left for the person's individual judgment.  Again, objectivistic views can be regarded as personalistic views according to which comparisons of probability can be made only for very special pairs of events, and then only according to such criteria that all (right-minded) people agree in their comparisons.
\eq
It is not nearly as detailed as I remembered, but then again my mind very often fills in a lot of things that were not there originally!

\section{06-02-09 \ \ {\it Something Concrete} \ \ (to {\AA}. {\Ericsson})} \label{Ericsson2}

It dawned on me that it might be fun to give you something concrete to think about in your off hours as you wait to come to Canada.  As you get a chance, have a look at this paper and this presentation:
\bq
\noindent ``Statistical Models are Algebraic Varieties,'' by Seth Sullivant\smallskip\\
\noindent \myurl{http://www.math.harvard.edu/~seths/lecture1.pdf}\smallskip

and

\noindent [Something now lost], by Seth Sullivant\smallskip\\
\noindent \myurl[http://www2.warwick.ac.uk/fac/sci/statistics/staff/research_students/zwiernik/wag_june.pdf]{http://www2.warwick.ac.uk/fac/sci/statistics/staff/research\underline{ }students/ \\ zwiernik/wag\underline{ }june.pdf}
\eq

One thing I'd like to explore with you after you arrive and get settled is whether it is fruitful to think of quantum state space as a ``statistical model''.  And if so, what standard tools we might bring from that area of research to better understand quantum states.

To put this in context, let me attach a paper I'm constructing at the very moment.  It is far from complete (many of the main results are not in it yet, sections unfinished, tentative paragraphs, etc.), but the present draft does at least contain the equations you'll need to make sense of what I'm talking about.  Look at equation (30) or alternatively equations (27) and (28).  These allow us to think of the set of pure quantum states as an algebraic variety on the probability simplex.  By the definition of the links I gave you, quantum state space becomes a ``statistical model''.  The first question that can be asked then is whether it is a meaningful statistical model?  Is it fruitful to think of it that way?

I'm looking forward to your teaching me some geometry!!

\section{07-02-09 \ \ {\it De Raedt Papers} \ \ (to C. Ferrie)} \label{Ferrie0}

\bcf
Certainly a few of the informal conclusions will drive away some of the more war-torn veterans of Bell's theorem.  But, I really do get the sense that this new paper has some merit in its formal mathematical results.  The abstract does not do it justice.  The paper, it seems to me, is really about data analysis and the assumptions which go into forming models of the data.
\ecf

Sorry, Chris, but it's my firm belief that your mind is presently darting around some very nasty places.  Basically, I think you are charmed by a complicated, but ultimately untenable patent application.  I'm alluding to my perpetuum mobiles again.  So, I reissue my challenge.  With regard to the two attached papers [N. D. {\Mermin}, ``Quantum Mysteries for Anyone,'' J. Phil.\ {\bf 78}(7), 397--408 (1981); N. D. {\Mermin}, ``Bring Home the Atomic World:\ Quantum Mysteries for Anybody,'' Am.\ J. Phys.\ {\bf 49}(10), 940--943 (1981)]:  Pinpoint {\it exactly\/} where {\Mermin} makes an unjustified assumption in his analysis.  You may use any insights you want from De Raedt {\it et al}.\ or Jaynes or anyone else, but stick to the issue.  What is wrong with {\Mermin}'s specific argument?

The point here is that (I well believe) there is some very clear thinking going on behind Bell.  And papers like De Raedt's are just so many obfuscations.  Stick with a crystal clear argument like {\Mermin}'s and tell me what's wrong with it.

\section{09-02-09 \ \ {\it Qu}\ \ \ (to O. J. E. Maroney \& H. Westman)} \label{Maroney4} \label{Westman2}

I should comment on Owen's note before walking into the next meeting this afternoon.

\bom
This sounds interesting, but is definitely going to put off the
physics side as presented, which would be a shame as the last one got a good mix. [I.e., the last PIAF meeting.] --- Is it possible to lump ``Perspectivalism'' together with ``The role of agents'' in some way?  That might reduce the overall scariness to the physicists of the list.
\eom

My own opinion is, ``then too bad for the so-called `physicists'.''  It may well just be that the scariness of terms like ``agent'' is what has kept  us in the quantum muddle all these years.  The fear of a single little idea:  that there is a distinction to be drawn between the objective and subjective, and that quantum theory rather than being exclusively about the one or the other, is about both.  So one has to get the distinction clear, or there'll be no progress at all.  It's not quite a 12-step program like Alcoholics Anonymous, but recovery has to start somewhere.

There, I said my piece, now I'll go defend it in chamber.

\section{10-02-09 \ \ {\it Quanta Vista} \ \ (to R. W. {\Spekkens}, L. Hardy, and H. Westman)} \label{Spekkens58} \label{Westman3} \label{Hardy33}

\brws
Hey guys, here's the synopsis:
\bq\noindent
{\rm {\bf Quanta Vista: Frame, gauge, context and agent in quantum theory}\smallskip

Topics:
\bv
$\bullet$ Reference frames\\
$\bullet$ Relationalism\\
$\bullet$ Contextuality\\
$\bullet$ Gauge and symmetry\\
$\bullet$ Subjective elements in the quantum formalism
\ev }
I've slightly modified the topic list.  I've replaced ``role of the agent'' with ``Subjective elements in the quantum formalism'' to avoid the impression that we're talking about consciousness causing collapse.
\eq
\erws

My first quick impressions.

1)  I like ``subjective elements in the quantum formalism'' very much.  It captures the issue just right from my perspective.

2)  But your fear of agent didn't reach up to the title.  In any case, I don't think the title is snappy enough now.  Personally, I think you would do better to drop agent from it but {\it leave\/} context:  ``Quanta Vista: Frame, gauge, and context in quantum theory.''  Sounds physicsy, non-frightening, and captures the matter.  With the bullet on subjective elements, I get my cake and can eat it too \ldots\ but in a less threatening way.

3)  I disagree on throwing out contextuality---more precisely Kochen--Specker---as a subject matter.  These are topics the community ought to have a chance to think of in proximity of each other if progress is going to be had.  If anything sticks out like a sore thumb, it's consistent histories.  I'd say drop that subject this round.  I really don't see how it fits.

4)  I fear that the philosophers are a bit under-represented this time.  Last year we had 7/29.  I think we ought to maybe think harder there.

5)  I personally would like to bump up Timpson---he's the only philosopher in the standard philosophy-of-physics crowd who really gets the subjective/objective divide.  I.e., he's not like most other philosophers of physics who basically just see information as a new kind of fluid.  Also, given van Fraassen's paper or papers on Rovelli (one had a title ``Rovelli's World'') I think it's entirely appropriate to try to recruit him.  Though, I think like Rovelli himself, he's unlikely to come.

6)  Carlton is spelled Carlton.

I'll have more impressions later.  This is becoming a worthwhile exercise.

\section{10-02-09 \ \ {\it Quanta Vista, 2} \ \ (to R. W. {\Spekkens}, L. Hardy, and H. Westman)} \label{Spekkens59} \label{Westman4} \label{Hardy34}

\brws
Speaking as someone who thinks about each of these things, the connection between frame-dependence,
agent-dependence and context-dependence is pretty weak at present.
\erws

I read over the note again and disagree with this even more.  Each topic touches on the idea of something ``numerically additional'' to the quantum system (to use a William Jamesian phrase).   Frame, agent, and context all capture this.  As well as gauge.  It is a unified topic if you look at it this way.  They are all about things non-intrinsic.

\brws
I would be very surprised to see many connections drawn between these threads at the meeting.
\erws

Like a good democracy, the only way to set it at work is to set it at work.  Give the participants a chance to think about {\it all\/} these things in a compact span of time.

\section{10-02-09 \ \ {\it Instance of James}\ \ \ (to R. W. {\Spekkens})} \label{Spekkens60}

\bq\noindent
Do you wish, like so many of my enemies, to force me to make the truth out of the reality itself?  I cannot:  the truth is something known, thought or said about the reality, and consequently numerically additional to it.\smallskip\\
\hspace*{\fill} --- from Chap 15 of {\sl The Meaning of Truth}
\eq

\section{10-02-09 \ \ {\it Your Thesis}\ \ \ (to J. R. Gustafson)} \label{Gustafson1}

Hans Christian von Baeyer has told me of your PhD thesis.  If you have it in electronic form, may I ask you to email me a copy?

Hans Christian is taking on a project to potentially write a book (or at least some articles) on the Pauli--Fierz correspondence, and will be coming here to the Perimeter Institute to get started on the project next week.  And I myself have had a long interest in Pauli (see for instance, my {\sl Notes on a Paulian Idea}, posted at my webpage, which Cambridge U. Press will soon republish).

I think your thesis will be an invaluable resource for our discussions.

\subsection{John's Reply}

\bq
By now I assume you have received the electronic copy of my Ph.D. thesis.  I hope you find it helpful.

Your request got me to pick up my thesis again and to think of unfinished business.  Since you or Hans Christian von Baeyer are in a better position than I am to follow up on one of the unfinished activities, let me pass on to you that there is one source of Pauli materials likely remaining untapped \ldots\ the 1300 or so Pauli dreams collected by Jung.   Jung published only some of them (see for example Sir Herbert Read, Michael Fordham, and Gerhard Adler, ed., {\sl The Collected Works of C.~G. Jung, Vol.\ 12, Psychology and Alchemy\/} (London: Routledge \& Kegan Paul Publihers, 1953).  The rest may still remain in Jung's archives, perhaps in Switzerland.   Since Jung excluded many of Pauli's dreams from publication because they might reveal the name of the author, and because Pauli often mentioned in his dreams ``physics stuff,'' I think the unpublished dreams \ldots\ if they still exist \ldots\ might be excellent resources for deciphering a small part of the mind of Pauli.  I thought I might pass this trivia on to you and ask you to forward it to H.~C. von Baeyer.
\eq

\section{11-02-09 \ \ {\it Invitation to Speak at University of Rochester} \ \ (to B. Weslake)} \label{Weslake1}

I remember you well.  Congratulations on your position in Rochester.

And I'm flattered that you're discussing this Bayesian business.  I want to encourage it!

I could give a talk on your March 25 slot if you'd like.  The topic would be the attached paper.  [See ``Quantum-Bayesian Coherence,'' \arxiv{0906.2187v1}.] {\it Beware}, it is still under construction.  In fact, it is far from complete (many of the main results are not in it yet, sections unfinished, tentative paragraphs, etc.), but it'll give you some sense of the topic.  And if you guys are still in ongoing discussions on our Qubayesian point of view, it may help give some sense of the constructive side of the program (which isn't much revealed by my older papers).

I've marked my calendar, but be sure to remind me again anyway as the time draws nearer.

\section{11-02-09 \ \ {\it The Still Way-Incomplete Paper} \ \ (to H. C. von Baeyer)} \label{Baeyer57}

Well I didn't make much more progress on it since last writing you.  It still has a long way to go---many of the main results are not in it yet, sections unfinished, tentative paragraphs, etc.---but I wanted to get it to you before your flight nonetheless, in case you might want something small to read.

Particularly with regard to our conversations this coming week, I plan to completely scrap the old ending of the paper and start over (that ending was from the preliminary version sent to a conference proceedings).  The new, yet-to-be-written ending will make a transition from Peres to Pauli and have a discussion on the undetached observer.

Anyway, I just wanted to give you some sense of how this all fits together in my mind.

\subsection{Hans's Reply, ``Reaction,'' 13-02-09}

\bq
I'm in Toronto with the three most important volumes, so we'll have plenty of grist.  Let the old mill grind!

I read your paper on and between planes and like it very much.  Here are a couple of reactions.

Objectivity. This is a very heavily laden word and I wish we could get away from it.  Intuitively it is defined mostly by its opposite --- subjectivity.  I notice that you have started using ``personalized'' as an alternative, and I'm not sure that's much better.   A lot of resistance to the Bayesian view stems from the perception that it is solipsistic, and therefore not ``objective.''

Pauli thought that q.m., which he believed to be all about information, is objective in the following sense:  Once the observer has set up an experiment and has turned on the switch, it proceeds on its own without influence from his feelings or desires --- unlike a poem or a work of art.  This doesn't say much about the DESCRIPTION of an experiment, but he felt that he had to insist on the objectivity of physics.

Examples.  I would love to see some examples of simple phenomena, especially since you start with the double slit, and mock other text-book exercises. Those are good rhetorical anchors, because all readers are familiar with them, but I would enjoy seeing you return to them and deal with them explicitly in your new language. (here is a reference to a recent paper of yours that might contain examples of descriptions starting with different priors.)

See you soon!
\eq

\section{12-02-09 \ \ {\it On the Charge of ``Bias''} \ \ (to N. D. {\Mermin})} \label{Mermin148}

Referee 1 wrote this:
\bq\noindent
The paper is proselytizing a particular point of view, rather than drawing together different sources to provide an unbiased review (i.e.\ one which would discuss the pros and cons of this idea, and compare it seriously with other interpretations of QM.)
\eq

Referee 2 wrote this:
\bq\noindent
The authors illustrate their main points with examples and a large number of (text) citations. They do this in such a strong manner that I felt they almost try to inculcate the reader with their viewpoint. In fact, the paper naturally touches on various philosophical aspects (such as the interpretation of quantum states), which are partly lacking a consensus in the scientific community. While the authors, of course, have a right to expose and justify their points of view (and, actually, do this in a very convincing way), I am not sure the style is appropriate for RMP.
\eq

With regard to 1, I thought, ``Well yeah it's proselytizing, of course.  Why should I bother to `compare it seriously with other interpretations' when the whole point of all my papers is to show that a QBist-style interpretation---even without all the details hammered out---is simply natural, obvious, and should hit the reader over the head.''

With regard to 2, I thought, ``To justify one's points `in a very convincing way' is inappropriate for RMP?  What would you really want instead?''

Anyway, I expand on these initial thoughts, by quote my old friend Willy James from his lecture ``The Types of Philosophic Thinking'':
\bq
What distinguishes a philosopher's truth is that it is reasoned. Argument, not supposition, must have put it in his possession. Common men find themselves inheriting their beliefs, they know not how. They jump into them with both feet, and stand there. Philosophers must do more; they must first get reason's license for them; and to the professional philosophic mind the operation of procuring the license is usually a thing of much more pith and moment than any particular beliefs to which the license may give the rights of access. Suppose, for example, that a philosopher believes in what is called free-will. That a common man alongside of him should also share that belief, possessing it by a sort of inborn intuition, does not endear the man to the philosopher at all---he may even be ashamed to be associated with such a man. What interests the philosopher is the particular premises on which the free-will he believes in is established, the sense in which it is taken, the objections it eludes, the difficulties it takes account of, in short the whole form and temper and manner and technical apparatus that goes with the belief in question. A philosopher across the way who should use the same technical apparatus, making the same distinctions, etc., but drawing opposite conclusions and denying free-will entirely, would fascinate the first philosopher far more than would the {\it na{\i}f\/} co-believer. Their common technical interests would unite them more than their opposite conclusions separate them. Each would feel an essential consanguinity in the other, would think of him, write {\it at\/} him, care for his good opinion. The simple-minded believer in free-will would be disregarded by either. Neither as ally nor as opponent would his vote be counted.

In a measure this is doubtless as it should be, but like all professionalism it can go to abusive extremes. The end is after all more than the way, in most things human, and forms and methods may easily frustrate their own purpose. The abuse of technicality is seen in the infrequency with which, in philosophical literature, metaphysical questions are discussed directly and on their own merits. Almost always they are handled as if through a heavy woolen curtain, the veil of previous philosophers' opinions. Alternatives are wrapped in proper names, as if it were indecent for a truth to go naked. The late Professor John Grote of Cambridge has some good remarks about this. `Thought,' he says, `is not a professional matter, not something for so-called philosophers only or for professed thinkers. The best philosopher is the man who can think most {\it simply}. \ldots\ I wish that people would consider that thought-and philosophy is no more than good and methodical thought-is a matter {\it intimate\/} to them, a portion of their real selves \ldots\ that they would {\it value\/} what they think, and be interested in it\ldots. In my own opinion,' he goes on, `there is something depressing in this weight of learning, with nothing that can come into one's mind but one is told, Oh, that is the opinion of such and such a person long ago. \ldots\ I can conceive of nothing more noxious for students than to get into the habit of saying to themselves about their ordinary philosophic thought, Oh, somebody must have thought it all before.' Yet this is the habit most encouraged at our seats of learning. You must tie your opinion to Aristotle's or Spinoza's; you must define it by its distance from Kant's; you must refute your rival's view by identifying it with Protagoras's. Thus does all spontaneity of thought, all freshness of conception, get destroyed. Everything you touch is shopworn. The over-technicality and consequent dreariness of the younger disciples at our American universities is appalling. It comes from too much following of German models and manners. Let me fervently express the hope that in this country you will hark back to the more humane English tradition. American students have to regain direct relations with our subject by painful individual effort in later life. Some of us have done so. Some of the younger ones, I fear, never will, so strong are the professional shop-habits already.

In a subject like philosophy it is really fatal to lose connexion with the open air of human nature, and to think in terms of shop-tradition only. In Germany the forms are so professionalized that anybody who has gained a teaching chair and written a book, however distorted and eccentric, has the legal right to figure forever in the history of the subject like a fly in amber. All later comers have the duty of quoting him and measuring their opinions with his opinion. Such are the rules of the professorial game---they think and write from each other and for each other and at each other exclusively. With this exclusion of the open air all true perspective gets lost, extremes and oddities count as much as sanities, and command the same attention; and if by chance any one writes popularly and about results only, with his mind directly focussed on the subject, it is reckoned {\it oberfl{\"a}chliches zeug\/} and {\it ganz unwissenschaftlich}.
\eq
Rereading that passage, I think it gets to the point.

\section{13-02-09 \ \ {\it It Wasn't Real Complex} \ \ (to R. W. {\Spekkens}, L. Hardy, and R. Blume-Kohout)} \label{BlumeKohout8} \label{Spekkens61} \label{Hardy35}

Here's what I should have said yesterday.  See attached.  It's not hard to prove after all that there is no minimal informationally complete POVM for real Hilbert spaces that will give me my infamous diagram.

Suppose I want a map $\Phi: {\mathcal S}(\mathbb{R}^d)\longrightarrow {\mathcal S}(\mathbb{R}^d)$ of this form
\be
\Phi(X)=\frac{1}{n}\sum_i \Pi_i X\Pi_i\;,
\ee
where ${\mathcal S}(\mathbb{R}^d)$ denotes the $\frac{1}{2}d(d+1)$-dimensional space of symmetric operators, the $\Pi_i=|\psi_i\rangle\langle\psi_i|$ are rank-1 projectors, and the set operators $E_i=\frac{1}{n}\Pi_i$ form a minimal informationally complete POVM.  Then we must have
\be
\frac{1}{n}\sum_{i=1}^{\frac{1}{2}d(d+1)} \Pi_i=I\;,
\ee
and consequently
\be
n=\frac{2}{d+1}\;.
\ee
Furthermore suppose $\Phi$ has this characteristic
\be
\Phi(\rho)=\alpha I+\beta \rho
\ee
for all density operators $\rho$.  To be trace preserving,
\be
\alpha d+ \beta = 1\;.
\ee
These are the properties needed to be able to draw the diagram I always draw and have the Born Rule interpretable as a simple modification of the law of total probability.

The question now is, are there any operators $\Pi_i$ that can do this for me?  Here's a way to see that there generally aren't.  Let $\Phi$ act on one of the $\Pi_k$:
\be
\frac{1}{n}\sum_i \Pi_i\Pi_k\Pi_i=\alpha I+\beta\Pi_k\;.
\ee
Since $\Pi_i\Pi_k\Pi_i=\tr(\Pi_i\Pi_k)\Pi_i$, it follows that we must have
\be
(1-\alpha-n\beta)\Pi_k + \sum_{i\ne k}\Big(\tr(\Pi_i\Pi_k)-\alpha\Big)\Pi_i=0\;.
\ee
But the $\Pi_i$ are linearly independent.  Consequently, it is necessary that
\be
\alpha+n\beta=1
\ee
and
\be
\tr(\Pi_i\Pi_k)=\alpha \qquad \forall i\ne k\;.
\ee
That is, the $E_i$ must form a SIC-POVM.

For real vector spaces, however, SIC-POVMs generally do not exist.

\section{16-02-09 \ \ {\it The Historical Roots of Our SICs} \ \ (to D. M. {\Appleby} and H.~C. von Baeyer)} \label{Appleby48} \label{Baeyer57.1}

It's entirely appropriate that I came across this passage in the Atmanspacher and Primas article tonight:
\bq\noindent
As a consequence of the non-conservation of parity, Pauli discovered a group of transformations relating left- and right-handed neutrinos and antineutrinos---the so-called Pauli group (Pauli 1957).  In 1957, he started an initially enthusiastic collaboration with Heisenberg on his nonlinear spinor equation which should provide a unified description of all elementary particles.  One reason for Pauli's excitement about this project was that this spinor equation was invariant under the Pauli group.
\eq
Pauli 1957:  ``On the conservation of the lepton charge,'' Nuovo Cimento {\bf 6}, 204--215.

\section{17-02-09 \ \ {\it Historical Question} \ \ (to D. Gottesman)} \label{Gottesman12}

I've got Marcus Appleby and Hans Christian von Baeyer here this week working on identifying the various notions at work in the phrase ``detached observer'' in the correspondence between Pauli, Fierz, and Bohr, and of course Marcus is always working on SICs.  So I was delighted last night when I came across the funny little tidbit below, particularly upon learning that Pauli's interest in his failed ``world theory'' with Heisenberg centered on its invariance under the group.  [See 16-02-09 note ``\myref{Appleby48}{The Historical Roots of Our SICs}'' to D. M. {\Appleby} and H.~C. von Baeyer.]

Anyway, completing the tale in my mind, leads in part to the following historical question.  Who invented the phrase ``generalized Pauli group''?  Was it you?  And its first use of the group in quantum information was you?  Or Manny Knill?  Or still someone else.  Presumably it's also called Weyl--Heisenberg group because Weyl and Heisenberg used it.  But I wonder if Heisenberg was post-Pauli in this case.  (I think Weyl goes back to some discussion in his book on group theory and QM.)

For fun, it'd be nice to compile these facts.

\section{17-02-09 \ \ {\it Dedication} \ \ (to A. Y. Khrennikov)} \label{Khrennikov25}

Thanks for all the patience with me in sending in the conference proceedings.  In trying to make this one an important paper, I lost sight of my deadlines!  So, I turned in what I could for the conference proceedings, and then decided to keep plugging away at the paper making a more complete version of it before posting on the archive.  I'm still not finished!!  But I did just write a little dedication to you on the front page on the still-incomplete draft.  [See ``Quantum-Bayesian Coherence,'' \arxiv{0906.2187v1}.]  \ See footnote 1 in the attached:
\bq
\noindent This is a much expanded version of a paper originally appearing as C. A. Fuchs and R. Schack, ``From Quantum Interference to Bayesian Coherence
and Back Round Again,'' in {\sl Foundations of Probability and Physics -- 5, V\"axj\"o, Sweden, 24--27 August 2008}, edited by L.~Accardi, G.~Adenier, C.~A. Fuchs, G.~Jaeger, A.~Khrennikov, J.~{\AA}. Larsson, and S.~Stenholm, AIP Conference Proceedings Vol.~1101, (American Institute of Physics, Melville, NY, 2009), pp.~260--279. Per always, we thank Andrei Khrennikov for organizing this wonderful series of meetings, which has been so instrumental in crystalizing our quantum-Bayesian thoughts.
\eq
It's heartfelt.  These meetings have been very important for my thought.  Hopefully I'll get the paper finalized and posted next week.

\section{18-02-09 \ \ {\it Pop Culture} \ \ (to D. M. {\Appleby} and H.~C. von Baeyer)} \label{Appleby49} \label{Baeyer57.2}

``Reality is at the deepest level the response to the observer.''

Just heard that on the CBC show ``Being Erica'' (a story about a woman whose psychologist keeps sending her back in time to revisit her past) in a small discussion about quantum mechanics.

\section{18-02-09 \ \ {\it Saint of the Retort} \ \ (to H. C. von Baeyer)} \label{Baeyer58}

Thanks for this!

\subsection{Hans's Preply, ``Augmented Index Entry,'' 18-02-09}

\bq
I have found an important reference to the Detached Observer in Pauli's Kepler Study.  Hence the new last item below:

\noindent Here is an entry for the (non-existent) subject index of {\sl Wolfgang Pauli: Writings on Physics and Philosophy}, Enz and Meyenn, ed. (Springer 1994).\medskip

Detached Observer: pp.\ 23, 24, 33, 40, 43, 47, 122, 132, 134, 152, 260

Compiled by HvB\medskip

PS In 1953 Pauli wrote to Panofsky that he is trying to introduce the term ``detached observer'' (in English) to physics.  Maybe we should give him a posthumous hand.
\eq

\section{18-02-09 \ \ {\it Talk} \ \ (to A. Wilce)} \label{Wilce22}

\baw
I suppose it's not too early to be thinking about what talk to give. Let me give you some choices, and ask you to tell me what's of most interest:
\begin{itemize}
\item[(1)] I could talk about my paper on formalism and interpretation -- basically the same talk I gave in Jerusalem (though I think perhaps Jon has given a similar talk in the last year or so);

\item[(2)] I could talk about an axiomatization of QM by way of the characterization of homogeneous self-dual cones (the notes I sent you before Christmas);

\item[(3)] I could give a slightly updated version of the talk I gave in Obergurgl (adding some remarks making contact with Lucien's axioms).
\end{itemize}
Let me know what sounds good.
\eaw

(1)!!  (If it's the one to do with Everett potentially being contentless.)

\subsection{Alex's Reply}

\bq
Below are two different abstracts for more-or-less the same talk, emphasizing different aspects thereof. Let me know which one you like better, and I'll send it along to Karen.

\bq\noindent
Title: \\ Entanglement and measurement in general probabilistic theories.\medskip \\
Abstract: \\ Quantum mechanics is a non-classical probability
theory, but hardly the most general one imaginable: any compact
convex set can serve as the state space for an abstract
probabilistic model (classical models corresponding to
simplices). From this altitude, one sees that many phenomena commonly
regarded as ``characteristically quantum" are in fact
generically ``non-classical". In this talk, I'll show that almost
any non-classical probabilistic theory shares with quantum mechanics
a notion of entanglement and, with this, a version of the so-called
measurement problem. I'll then discuss what's required for an
abstract probabilistic theory to admit a somewhat simplified version
of Everett's response to this problem -- an exercise that turns out
to be instructive in several ways.\medskip
\eq

\bq\noindent
Title: \\ Measurement dynamics in general probabilistic theories.\medskip \\
Abstract: \\ One can view quantum mechanics as a dynamical theory with
a familiar mathematical apparatus, but a mysterious probabilistic interpretation,
seemingly incompatible with the dynamics. Alternatively, one can view QM
as a probabilistic theory with a more or less standard interpretation,
but a mysterious formal apparatus, badly under-determined by that interpretation.
In this talk, I'll try to take both of these complementary points of view at once,
by asking what's required for a general (non-classical) probabilistic theory to support
a minimally reasonable measurement dynamics -- an exercise that turns out to be instructive
in several ways.
\eq
\eq

\section{19-02-09 \ \ {\it Transcript from ``Being Erica''} \ \ (to D. M. {\Appleby} and H.~C. von Baeyer)} \label{Appleby50}

A more accurate transcript from last night's episode \ldots

{\bf Erica:}  Here I am trying to relive the perfect day and I can't.  Because every time something changes, even if it's a little bit, I freak out.

{\bf Dr.\ Tom:}  Yeah.  (Picking up a leaf from the ground.)  Quantum mechanics.  You see, we can never know with any kind of certainty how an atom will behave naturally.  (Pointing to the leaf.)  Because the very instant that we look at an atom we alter it.  The very act of looking is never a passive thing.  It has an effect.  In fact physicists tell us that reality at its deepest level is the response of the observer.  You know, it's kind of like you reliving past events.

\section{20-02-09 \ \ {\it My Own Index} \ \ (to H. C. von Baeyer)} \label{Baeyer59}

Uses of the word ``detached'' in ``The Activating Observer''
\bv
14, Atmanspacher and Primas\\
48, Bohr\\
49, Bohr\\
119, Enz\\
122, Enz\\
138, Feyerabend\\
149, Folse\\
153, Folse\\
159, Folse\\
163, Folse\\
165, Folse\\
167, Folse\\
169, Folse\\
170, Folse\\
190, Fuchs and Peres\\
309, von Meyenn\\
338, Pauli\\
340, Pauli\\
345, Pauli\\
346, Pauli\\
353, Pauli to Jung, 27 May 1953\\
408, Rosenfeld \\
455, von {\Weizsacker}\\
\ev

\section{21-02-09 \ \ {\it PhD Acquired}\ \ \ (to G. {\Plunk})} \label{Plunk16}

\bgp
I just defended my thesis successfully at UCLA and wanted to announce this to the people who were especially influential along the way.  I just wanted to let you know that the path of research that you launched me on in the summer of 2002 has concluded well!  I have taken a post-doctoral job at the University of Maryland and expect that I will stay in theoretical physics for the long haul.
\egp

That's excellent news!  Congratulations.  I hope you like College Park.  I'll be there next Friday and Saturday for an APS meeting.

Funny coincidence you would write me now.  Just the other day, I started working on a section of the paper (that I'm presently constructing) that makes use of one of your results from our summer together.  See Section 5.3, page 23, of the present draft attached.  [See ``Quantum-Bayesian Coherence,'' \arxiv{0906.2187v1}.]  Too bad the paper's not completed yet, so that I could have replied with an even more forceful ``funny coincidence''!

Take care.

\section{22-02-09 \ \ {\it James, Gustafson, and the Urungleichung} \ \ (to D. M. {\Appleby})} \label{Appleby51}

\bma
I have been reading Gustafson, and the following passage struck me.  I am not sure if it bears on this afternoon's discussion, but I feel it might do.  Anyway, I thought it worth recording the point before I forgot it.

On p.~58 Gustafson says this:
\bq\rm
Walter Moore, in describing {\Schroedinger}'s renunciation of Mach's philosophy, could also have been describing Pauli's:
\bq
There are a number of fairly obvious defects in presentational phenomenalism
[positivism]. For instance, it fails to explain the close relationship between
mathematical reasoning and theoretical physics; mathematical operations and
symbols do not denote empirical sensations, and yet one cannot do science
without them. Also, experiments are planned interactions of the scientist with the
environment; how can they be explained as mere collections of sensations? Mach
fails to explain the enormous predictive power of physical theories; how can it be
that [Paul A.~M.] Dirac predicts a positive electron and [Carl David] Anderson
finds it in a cloud chamber?
\eq
Pauli would become fascinated with the deep meaning of mathematical symbols. His ``mind's eye'' was opened through his explorations of complex numbers, Maxwell's equations, and special-relativity transformations.
\eq\it
I think this is a fair description of the way Pauli felt---i.e.\ a fair description of the way in which Pauli was definitely not an empiricist (a Platonist even, I think I have seen it said). But be that as it may.  The reason I am citing it is not that, but its bearing on what you said about James's belief that relations are part of experience.  Could we say that the word ``relations'', as used by James, includes mathematical relations?
\ema
I think James would include mathematical relations as well, insofar or to the extent they are part of experience.

\section{22-02-09 \ \ {\it The Shape of Hilbert Space} \ \ (to G. L. Comer)} \label{Comer122}

Asher Peres used to put it this way:  ``Quantum phenomena do not occur in a Hilbert space; they occur in a laboratory!''

If you've got nothin' to do today, tell me if you like the feel of my latest effort.  Explanatory note below, and unfinished draft of paper attached.  [See ``Quantum-Bayesian Coherence,'' \arxiv{0906.2187v1}.]

\section{22-02-09 \ \ {\it The Shape of Hilbert Space, 2} \ \ (to G. L. Comer)} \label{Comer123}

\bgc
One question: should I prepare for terabytes of musings on a Feynmanian Idea?
\egc

Nope.  Just Jamesian ones.  I had a new epiphany at the beginning of the year, and I've been gearing up to write a lot again.  Wheeler's ``elementary quantum phenomenon'' IS a modern, precisified version of James's ``pure experience''---I get it now after all these years.  So I REALLY MUST get {\sl My Struggles with the Block Universe\/} on the archive this Spring and then start a new chapter.  When the real conclusion of the present paper is written (I'm scrapping the old conclusion), it'll allude to some of that.

BTW, it looks like Cambridge U Press is going to reissue {\sl Notes on a Paulian Idea}.  This time, that name will be the subtitle under {\sl Coming of Age with Quantum Information}.  I never ultimately went with Springer, and I'm glad I didn't.  This way I'll be in the same house with Bell's book, Mermin's boojums, and Nielsen and Chuang's text on quantum computing.  I have to write a biographical sketch of each of my correspondents for the preface:  So, watch out man!

\section{24-02-09 \ \ {\it Good Thinking}\ \ \ (to R. {\Schack})} \label{Schack153}

Do you have a complete reference on I. J. Good's article on the number of kinds of Bayesian?  It is driving me crazy that I have not been able to find it on the web.

\section{24-02-09 \ \ {\it HCvB and CAF} \ \ (to H. C. von Baeyer)} \label{Baeyer60}

I would be honored to write something with you.  I agree with your assessment, only too bad we didn't have it ready to go by mid-December '08!  If it remains OK with you, I would expect my participation to be somewhere neutrally (appropriately) between ``by me with your help'' and ``jointly''.  I.e., you could take the lead, and I would try to be constructive from the side at first, but becoming more involved as more structure starts to arise.  Furthermore, as a point of personal policy for me this time around, you would not find me insisting on idiosyncratic terms like ``imprimatur,'' etc!  I will be very happy to see something of good substance get out into the airwaves, and I can't really see how to do that without your being at the helm.

Perhaps the only sensitive issue I foresee at the moment is the desire to ``tie cautiously into the science/religion debate.''  You can probably already sense I would want to be very cautious with that.  But I also feel that something should be said about the humanistic side of physics---so there is definitely room for negotiation here.

I feel like I will learn a lot with this project.  Certainly I learned very, very much from my effort with Asher on the {\sl Physics Today\/} piece.  Somehow in my mind, this will be a bit like that:  A great opportunity to crystallize thought.

I hope your cold goes away soon.  Marcus and I spent yesterday $x$'ing and testing things on the computer.  We discovered that apparently the conditions arising from the urungleichung also imply $0 \le \tr \rho^3 \le 1$.  (We haven't shown it yet; it's only numerical at this stage.)  This inequality is still not full quantum state space, but it is a little closer.  I.e., another piece of evidence that if we keep squeezing on the urungleichung, eventually the full condition on quantum state space will indeed pop out.

\subsection{Hans's Preply, ``The Detached Observer,'' 24-02-09}

\bq
I am working on the essay for home use that I began at PI.  In addition, though, I want to reply to your remark that I should consider publishing something.

An article about the DO at the level of AJP sounds very appropriate. It would be a fine tribute to Pauli 50 years after his death, it would be good PR for Bayesians, and it would tie cautiously into the science/religion debate.  It would talk about the classical DO, about Pauli's quantum mechanical NDO, about your effort to define and quantify the NDO in the language of information, and finally about Pauli's unfulfilled dreams of pushing the NDO toward a more neutral language.

Such an article could be written by you, by me with your help, or jointly.  Since you are the senior partner in this effort (my grey locks notwithstanding) I would like to ask you to choose among these three options.  I am totally happy with each one!

Unfortunately I brought home from Waterloo not Marcus's cold, but the Canadian arctic cold!
\eq

\subsection{Hans's Pre-Preply, ``Detached Observer,'' 20-02-09}

\bq
Here is the losgel\"oste Beobachter.
\begin{center}
{\bf Pauli's Detached Observer}
\end{center}

\subsubsection{Bibliography}

[A] Karl von Meyenn {\it et al}.\ (eds.)\ {\sl Wolfgang Pauli: Wissenschaftlicher Briefwechsel mit Bohr, Einstein, Heisenberg u.a.} (Springer 1979--2005).

[B] Wolfgang Pauli {\sl Writings on Physics and Philosophy}. (Springer 1994).

[C] Suzanne Gieser {\sl The Innermost Kernel}. (Springer 2005).

\subsubsection{Contents}

\begin{enumerate}
\item
Nomenclature
\item
Commentaries
\item
The classical detached observer
\item
The non-detached observer in quantum mechanics and the dispute with Bohr
\item
The non-detached observer of the future
\end{enumerate}

\subsubsection{Nomenclature}

According to von Meyenn, the phrase {\it detached observer\/} (DO) first appeared in February 1949, in a lecture on complementarity in Zurich.\footnote{[A] vol.\ IV part II, p.\ 149.  The essay is translated into English in [B] where DO occurs on page 47.}   The first mention in the letters seems to be in January 1951.\footnote{[A] vol.\ IV part I p.\ 247.}   The German version is {\it losgel\"ster Beobachter}, which Pauli explicitly translates as {\it detached observer}.\footnote{[A] vol.\ IV part II p.\ 237. \label{ManoMano}}   (He also used {\it abgekapselter Beobachter},\footnote{[A] vol.\ IV part III p.\ 665.} meaning an {\it encapsulated observer}, and once he wrote the English phrase {\it loose and untied observer}\footnote{[A] vol.\ IV part I p.\ 436}.)  Bohr, in German, used {\it aussenstehender Beobachter},\footnote{[A] vol.\ IV part III p.\ 106.} or {\it observer standing outside}.  In French Pauli liked {\it observateur d\'etach\'e}.\footnote{Cf.\ footnote \ref{ManoMano}.}   In the English translation of the essays there are 11 references to DO,\footnote{[B] pp.\ 23, 24, 33, 40, 43, 47, 122, 132, 134, 152, 260.} in the letters many more.  In 1953 Pauli remarks that he has tried several times to introduce the DO into physics.\footnote{Cf.\ footnote \ref{ManoMano}.}  In 1956 he notes that following his dispute with Bohr about the use of the phrase DO neither he nor Bohr ever used it again in print.\footnote{[A] vol.\ IV part III p.\ 784.}

It would be useful to find a compelling antonym for DO.  Pauli used {\it non-detached observer}\footnote{[A] vol.\ IV part III p.\ 147.} once,  and {\it miteinbezogener Beobachter}\footnote{[A] vol.\ IV pp.\ 697, 738, 824.} several times.  The latter adjective can be translated as {\it involved\/} or {\it included (in the description of the phenomenon)}; the term {\it bezogen\/} means {\it related}. Fuchs uses {\it activating}.  Other suggestions include {\it attached, active, bound, participating, linked, joined, entangled, enmeshed}.   In this essay we stick to {\it non-detached observer\/} (NDO).

The only antonym I have found is {\it non-detached observer}\footnote{[A] vol.\ IV part III p.\ 147.} (NDO).

\subsubsection{Commentaries}

Folse\footnote{Henry J. Folse {\sl The Philosophy of Niels Bohr\/} (Amsterdam 1985) p.\ 214.}  comments on the DO, but von Meyenn does not believe that Folse has fully understood Pauli.\footnote{[A] vol.\ IV part III p.\ 107.}  Gieser devotes five pages to the DO.\footnote{[C] p.\ 131.}  Enz comments on the DO in his biographical sketch.\footnote{[B] p.\ 22 ff.}  The editorial apparatus of the letters includes numerous footnotes about the DO.\footnote{[A] vol.\ IV part I p.\ 343; vol.\ IV part II p.\ 149; vol.\ IV part III p.\ 107 etc.}

\subsubsection{The classical detached observer}

Pauli defined the DO in virtually identical words in his 1949 lecture on Complementarity and his 1952 essay on Kepler,\footnote{[B] p.\ 40 and p.\ 260.  The latter version is more grammatical. \label{Mini-a-Mini}} which was published in a book co-authored by Jung: ``\ldots\ [T]here is a basic difference between the observers, or instruments of observation, which must be taken into consideration by modern microphysics, and the detached observer of classical physics.  By the latter I mean one who is not necessarily without effect on the system observed but whose influence can always be eliminated by determinable corrections.''

The DO serves to characterize classical physics.  Einstein claimed: ``There is such a thing a the real state of a physical system, which exists objectively, independently of any observation or measurement, and can in principle be described by the modes of expression used in physics.''\footnote{[B] p.\ 47.}   Pauli calls this point of view the {\it ideal of the detached observer\/} and points out that it is only one special form of physics.  Indeed,  ``this ideal is now revealed as a special case of more general possibilities of explaining nature.''\footnote{[A] vol.\ IV part II p.\ 149.}

Heisenberg calls ``the idea of an objectively real world, whose smallest parts exist objectively in the same way as stones and tree, independently of whether we observe them or not,'' the {\it ontology of materialism}.  Pauli prefers to call the same idea the {\it ideal of the detached observer}, pointing out that the fundamental elements of the real world might be, for example, fields, which don't resemble stones and trees and are not ``small.''\footnote{[A] vol.\ IV part III, p.\ 121.}

Fierz uses the DO in his definition of absolute space:  ``\ldots\ I believe that absolute space is appropriate for classical physics.  The {\it classical objective real world}, which I like to call {\it the absolute world\/} because it is detached from the observer, is located in absolute space.  This space is absolute because it is independent of the physical reality that fills it.''\footnote{[A] vol.\ IV part I p.\ 383.}

\subsubsection{The non-detached observer in quantum mechanics and the dispute with Bohr}

I will examine the NDO with particular reference to two sources --- the essays in which it was first introduced,\footnote{Cf.\ footnote \ref{Mini-a-Mini}} and the final epistolatory exchange with Bohr.\footnote{[A] letters [2015] and [2041] from Pauli, [2035] and [2047] from Bohr.}   In the case of the Kepler essay, Pauli's tireless exchanges with Panofsky concerning the translation into English bear witness to the pains he took to explain himself precisely.  Later, in the Bohr letters, he writes explicitly: ``I shall try to make my point logically clear, by defining my concepts, replacing hereby the disputed phrase by other words.''  Other references in the letters and essays serve as glosses on these two primary sources.  An advantage for anglophones is that the essays have been translated, and the Bohr letters were written English.

In passing from the classical DO to the quantum mechanical NDO a complication arises.  The context of the DO has two parts depicted as ({\it system\/} $|$ {\it observer\/}).   The context of the NDO, on the other hand, is tripartite: ({\it system\/} $|$ {\it instruments\/} $|$ {\it observer\/}).  Later in this essay, the first of the two vertical lines will be referred to as the ``Pauli cut'' and the second one as the ``Bohr cut.''

The first definition of the NDO read:  ``In microphysics \ldots\ every observation is an interference of indeterminable extent, both with the instruments of observation and with the observed system, and interrupts the causal connection between phenomena preceding and subsequent to it \ldots\. In this sense we may say that irrationality presents itself to the modern physicist in the shape of selecting (ausw\"ahlende) observation.''   A footnote refers to the technical elaboration of this process in the ``reduction of wave packets.''

The second definition adds, implicitly, the element of {\it information}: ``In microphysics \ldots\ the natural laws are of such a kind that every bit of knowledge gained from a measurement must be paid for by the loss of other, complementary items of knowledge.  Every observation, therefore, interferes on an indeterminable scale both with the instruments of observation and with system observed and interrupts the causal connection of the phenomena preceding it with those following it.''
\eq

\section{24-02-09 \ \ {\it First Consequences of Our Conversation} \ \ (to H. C. von Baeyer and D. M. {\Appleby})} \label{Appleby52} \label{Baeyer61}

Since our conversations last week, I have not been able to shake the nasty feeling of the complete sterility of what Huw Price, Harvey Brown, Rob {\Spekkens}, Daniel Dennett, and others in that lot think of when they say the word ``reality.''  Marcus's displays of disgust have been quite contagious to me!  So I couldn't quite contain myself today as I was improving and polishing my Introduction for the ``Coherence'' paper.  You'll see what I mean if you look at the last paragraph of Section 1, attached.  Now I have to hope that the concluding section (not written yet) can live up to the expectations I plant there!

\bq
Finally with Section 7 we close the paper by discussing how our work is still far from done:
Hilbert space, from a quantum-Bayesian view, has not yet been derived, only indicated. Nonetheless
the progress made here gives us hope that we are inching our way forward to a formal expression
of the ontology underlying a quantum-Bayesian vision of quantum mechanics: It really does have
to do with the Peres slogan, but tempered with a kind of `realism' that Peres would probably not
have accepted forthrightly. On the other hand, it is not a `realism' we expect to be immediately
accepted by most modern philosophers of science either. (See Footnote 2 and Ref.\ [50, Sec.\ 4.1] for
a sampling of relevant instances of opposition, and Refs.\ [63, 64, 65] for the vision more generally.)
Part of what must go is the ontology of the block universe [66, 67, 68], as well as the ontology of
the `detached observer' [69, 70]. The `realism' of the standard vogue is too narrow a concept to be
used for our purposes. Reality, the stuff of which the world is made, the stuff that was here before
agents and observers, the stuff that prevents us from vanishing in a dream---it strikes us---is more
interesting than that.
\eq

\section{24-02-09 \ \ {\it Good Thinking} \ \ (to W. C. Myrvold)} \label{Myrvold11}

Do you have a complete reference on I. J. Good's article on the number of kinds of Bayesianism?  It is driving me crazy that I have not been able to find it on the web.

\subsection{Wayne's Reply}

\bq
\noindent Attached:

i) Original publication, as a letter to the editor of {\sl The American Statistician}.

ii) Scan of reprint in {\sl Good Thinking}, which (I think) I got from Branden Fitelson's web site, together with an article in which the categories are explained more fully.
\eq

\section{25-02-09 \ \ {\it Undetachedly Denigrating Chance} \ \ (to H. C. von Baeyer)} \label{Baeyer62}

I think the 13 October 1951 letter from Pauli to Fierz is one we'll want to get straight translation-wise.  I was apprised of it from Laurikainen's {\sl Beyond the Atom}, pp.~196--197.

Compare Pauli to the three paragraphs below Eq.\ (1) in the CFS paper \quantph{0608190}:
\bq
Within the context of experimental situations with large sample sizes, where Bayes\-ian
updating leads to similar posteriors for exchangeable priors, the geometric analogy, combined
with the Principal Principle to connect chance with probability, would appear to work quite
well. This gives rise to the idea that the Principal Principle accounts for the concept of
objective chance in physics. However, from a Bayesian perspective, the introduction of
chance is completely unmotivated. More urgently, in those cases where the idea
is not already fraught with obvious difficulties, it serves no role that Bayesian probability
itself cannot handle.

To illustrate one such difficulty, return to the coin-tossing example discussed above, and
assume that there is an objective chance $q$ that a coin-tossing event will produce Heads. As
we have seen in the discussion above, the chance cannot be deduced from physical properties
of the coin alone, because the probability of Heads also depends on initial conditions and
perhaps other factors. An advocate of objective chance is forced to say that the chance is
a property of the entire ``chance situation,'' including the initial conditions and any other
relevant factors. Yet a sufficiently precise specification of these factors would determine the
outcome, leaving no chance at all. The circumstances of successive tosses must be different
to give rise to chance, but if chance aspires to objectivity, the circumstances must also be the
same. Different, but the same---there is no way out of this conundrum as long as objective
and chance are forced to co-exist in a single phrase. Subjective probabilities easily dispense
with this conundrum by maintaining the category distinction. The differences between
successive trials are differences in the objective facts of the initial conditions; the sameness
is an agent's judgment that he cannot discern the differences in initial conditions and thus
assigns the same probability to every trial.

But what of probabilities in quantum mechanics? Given the last paragraph, one might
well think---and many have thought---there is something different going on in the quantum
case. For, in repeating a preparation of a pure state $|\psi\rangle$, aren't all the conditions of preparation
the same by definition? Any subsequent probabilities for measurement outcomes will
then be determined by applying the Born rule to $|\psi\rangle$. They are not subjective probabilities
that come about by an inability to take all circumstances into account. Thus quantum
states (and hence quantum ``chances'') are objective after all, and the Principal Principle
is just the kind of thing needed to connect these quantum chances to an agent's subjective
probabilities---or so a very beguiling account might run.
\eq
(At that stage in his thinking at least) Pauli seems to get round our ``different, but same'' conundrum by saying the difference is the observer:  He ain't detached.  No two ``identical'' experiments are the same, because no two observers are the same.  Another way to say it is, there is an extra index around, and that index is the observer himself.

\subsection{Hans's Reply, ``The Same and Not the Same,'' 26-02-09}

\bq
I don't have Laurikainen's book.  Here's the context.  I will of course be happy to refine, expand, and revise as needed.\medskip

\begin{center}
{\bf The same and not the same}
\end{center}

[Square brackets enclose my comments.]

From the letter of Pauli to Fierz, dated 13 October 1951, number 1289 of Pauli's Scientific Correspondence:

\underline{In part 1} of the letter Pauli, in an irritated tone, corrects Fierz's alleged misunderstandings of general relativity.

\underline{In part 2} he refers back to the natural philosophy of the Italian Renaissance.  (cf.\ Atmanspacher, Primas, \& Wertenschlag 1995 p.\ 242 ff).  Then:

``The {\it anima mundi\/} [world soul], which was also an {\it anima movens\/} [motion causing soul] belonged necessarily to the Neoplatonism of the Renaissance (cf.\ Ficino).  Every planet had its individual soul, but how did these relate to each other:  spiritually, via the {\it anima mundi}, of which they are a part.  (N.B. I see Mr.\ Fludd immediately wrinkling his brow at the mention of ``part'' --- so let's say, for his sake: via the {\it anima mundi}, with which the individual souls are identical, insofar as they belong to the light principle.)

But in the 17th century the {\it anima mundi\/} went out of fashion --- the idea {\it paled}.  (I would happily learn the attitude of your epigone Henry More and his circle toward this idea --- what kind of a Neoplatonism is that, anyhow, {\it without\/} the anima mundi?)  And exactly through the resulting gap, proportion, geometry, mathematics pushed into the ideas about motion, and pushed toward empiricism, toward measurement.   One sees this process clearly not only in Kepler, but also in Galileo.  The latter {\it rejected\/} not only the Aristotelian-Peripatetic tradition, but also {\it Neoplatonism}, including the anima mundi, and referred back the {\it Pythagoreans and to Plato himself}. (``The more ancient is always the new!'')

But with this progress (analytic geometry, Newtonian mechanics), space moved up to the Olympus of the Absolute and the relationship between soul and matter became a special problem which vanished in the twilight of ``parallelism'', just as Venus vanishes in the morning twilight.

But now we seem to be beginning to suffer from the fact that one went too far in the 17th century (cf.\ my Kepler study) and from then come ``revenants'' that haunt me during the night, and occasionally also during the day --- the way Venus returns as evening star.  [Pauli's footnote:  I believe that {\it every\/} addition to consciousness proceeds in such a way that something which was previously conscious disappears into the unconscious, and returns much later.  That's what I want to express with the image of Venus, and also with the maxim: ``The more ancient is always the new!'']  When something becomes invisible, it still endures and remains effective.  General relativity has brought back, in its space-time rippled by matter, the idea attributed to the Peripatetics, of the physical quality of space points (places) --- in the transmuted form of the $g_{ik}$ field (even if g.r. couldn't bring back the entire {\it horror vacui}!)''

\underline{Part 3} ``Now comes the great crisis of the quantum of action: one has to sacrifice the unique case [or event] and its ``meaning'', in order to save an objective and rational description of the phenomena.   When two observers do the same thing it is really even physically no longer the same thing; in general only the {\it statistical averages\/} remain the same.  {\it The physically unique can no longer be detached from the observer\/} --- and therefore slips through the net of physics.  The unique case is {\it occasio\/} [opportunity, the fleeting moment] and not {\it causa\/} [cause, of which there are 32 varieties in philosophy].   I am inclined to see in ``occasio'' --- which includes the observer and his choice of the experimental arrangement --- a ``revenant'' (of course in ``transformed form'') of the {\it anima mundi\/} that was repressed in the 17th century.  La donna \`e mobile [the lady is fickle] --- including the {\it anima mundi\/} and the {\it occasio}.

Something has remained open here, which formerly seemed closed, and my hope is that through this gap {\it new concepts\/} will penetrate in place of ``parallelism'', concepts which should be uniformly physical and psychological simultaneously.  May a ``happier progeny'' achieve this.''

\underline{Part 3} then concludes by saying that the concept of {\it archetype\/} does not have the right attributes for this purpose, because it cannot be applied naturally to atomic physics.  Maybe {\it automorphism\/} will work.

\underline{Part 4} praises Fierz's proposal to divide every quaternity into two pairs of opposites:  one compensatory labeled $+$ and $-$, and one complementary labeled $p$ and $q$.  This scheme yields four elements $+p$, $-p$, $+q$ and $-q$.
\eq

\section{25-02-09 \ \ {\it Image of Occasio} \ \ (to H. C. von Baeyer)} \label{Baeyer63}

See
\begin{center}
\myurl[http://web.archive.org/web/20110704222210/http://www.iisg.nl/occasio/occasio-image.php]{http://web.archive.org/web/20110704222210/http:// www.iisg.nl/occasio/occasio-image.php},
\end{center}
where it is written:
\bq
The image of Occasio is taken from a work by famous Czech philosopher and pedagogue Johan Amos Comenius (Moravia 1592 -- Amsterdam 1670), the {\it Orbis sensualium pictus\/} (1658). The book contains pictures with text in Latin and the vernacular, and was intended for teaching. Occasio is the Opportunity we must seize before it flies away and vanishes: ``{\it Occasioni (quae, fronte capillata sed vertice calva, ad hoc alata facile elabitur) attendit captatque eam}'', or, ``She watches Opportunity (which, having a forelock but being bald at the back of its head and being winged, escapes easily) and seizes it.'' ``She'' is Prudentia, looking at the past with one of her two faces and at the future with the other. Occasio was chosen by Tjebbe van Tijen as an emblem appropriate for a project aiming to archive such volatile materials as Internet documents. An other image of Occasio can be found at Alciato Emblem 122 (Latin) from the 1621 edition of Andrea Alciato's {\sl Book of Emblems}. See also Prudentia, from {\sl Comenius Orbis Sensualium Pictus} at the Universidad Nacional de Educaci\'on a Distancia.
\eq

\section{26-02-09 \ \ {\it Undetachedly Denigrating Chance, 2} \ \ (to H. C. von Baeyer)} \label{Baeyer64}

Thanks.  Here was Laurikainen's translation of the part of ``The same and not the same'':
\bq
Now there comes the major crisis of the quantum of action:  one has to sacrifice the unique individual and the ``sense'' of it in order to save an objective and rational description of the phenomena.  If two observers do the same thing even physically it is, indeed, really no longer the same:  only the {\it statistical averages\/} remain, in general, the same.  {\it The physically unique individual is no longer separable from the observer}---and for this reason it goes through the meshes of the net of physics.  The individual case is {\it occasio\/} and not {\it causa}.  I am inclined to see in this {\it occasio\/} which includes within itself the observer and the selection of the experimental procedure which he has hit upon---a revenue of the {\it anima mundi\/} which was pushed aside in the seventeenth century (naturally ``in an altered form'').  La donna \`e mobile---so are the {\it anima mundi\/} and the {\it occasio}.
\eq
Your word revenant is certainly better than his revenue.

I've been Pauli'ing most of the day when not talking to drivers.  I'm finding I really enjoy Gieser's book, though I'm only 43 pages into it.

\subsection{Hans's Reply}

\bq
``Unique individual'' sounds like a person, which is wrong. The original is ``das Einmalige.''  This word has a sense of being unique in space and time --- ``something that occurs only once''.  In relativity it is called an event, and since Pauli has just mentioned GR this might be an appropriate word.  \eq

\section{27-02-09 \ \ {\it Vivienne and the Universe} \ \ (to L. Hardy and V. Hardy)} \label{Hardy36} \label{HardyV2}

How tickled I was this morning to awake in this lonely Washington hotel to find an announcement of Vivienne's birth.  And what a name!  One that means being alive itself.  My thoughts of course wandered to my favorite vision for an ontology of this world we live in---that the big bang is here and now, everywhere, being the sum total of all acts of creation.  I hope that one day Vivienne will come to understand how deep her role goes in the vast order of things.

Please let me offer her this little passage from William James to commemorate the occasion:
\bq
   Lotze has in several places made a deep suggestion. We naively
   assume, he says, a relation between reality and our minds which
   may be just the opposite of the true one.  Reality, we naturally
   think, stands ready-made and complete, and our intellects
   supervene with the one simple duty of describing it as it is
   already.  But may not our descriptions, Lotze asks, be themselves
   important additions to reality?  And may not previous reality
   itself be there, far less for the purpose of reappearing
   unaltered in our knowledge, than for the very purpose of
   stimulating our minds to such additions as shall enhance the
   universe's total value.  `Die Erhoehung des vorgefundenen
   Daseins' [`the enhancement of what is found to exist.'] is a
   phrase used by Professor Eucken somewhere, which reminds one of
   this suggestion by the great Lotze.
\eq
Congratulations to both of you, and congratulations especially to Vivienne!

\section{02-03-09 \ \ {\it Not Motto, Truth} \ \ (to I. Bengtsson)} \label{Bengtsson2.1}

\begin{flushright}
\baselineskip=3pt
\parbox{3.6in}{
\bq
\noindent
Now my own suspicion is that the Universe is not only queerer than we
suppose, but queerer than we can suppose.\medskip
\\
\hspace*{\fill} --- J.~B.~S. {\Haldane}
\eq
}
\end{flushright}\medskip

\bib
The real question, for me, is: ``Is Haldane right?'' You know the Einstein quote that goes in exactly the opposite direction.
\eib

Yes, Haldane is the right one.  Without doubt.  The comprehensibility is in the laws of thought, not in the ways of the world.

\section{02-03-09 \ \ {\it Getting Facts Straight}\ \ \ (to D. C. Cassidy)} \label{Cassidy1}

It was good meeting you Saturday.  I was grateful for the accident that brought me to your table!  I meant what I said about your book:  I loved it.  I read it about the same time as Moore's biography of {\Schroedinger}, and there was no comparison.

I'd like to make sure I get the facts straight about von Meyenn.  Would you mind saying everything again, but this time in email.  A) Is his biography of Pauli actually finished, or is it something he is writing on?  More importantly, B) please repeat what you said about the delay in the Briefwechsel project.  Is it that Springer lost patience with him?  Or simply that they're not foreseeing a sufficient \$\$ return on the project to go another volume?  It frightens me immensely that this project might slip into oblivion:  I have a vested interest in this side of Pauli (for various lines of research it might inspire in quantum foundations), and I would like to get oriented on how I might help.

In case you're interested, you might peruse my book {\sl Notes on a Paulian Idea\/} to see why I believe these more private matters of Pauli can have an impact on the technical side of physics.  It's the third item down on my webpage.   (It was previously published by a small university press; but I just learned this weekend that Cambridge U. Press will be reissuing it soon with the title {\sl Coming of Age with Quantum Information:  Notes on a Paulian Idea}.)

Thanks for your help.

\section{03-03-09 \ \ {\it Dates}\ \ \ (to R. {\Schack})} \label{Schack154}

\brs
Very provisionally, would this work: after {\Vaxjo}, you come to Egham, we drive to Hay-on-Wye, then we both fly to Perimeter, where I stay for two weeks.
\ers
That arrangement sounds perfect from my perspective.  We'd have a nice continuous time to really get some things thought out.

\brs
If I want to find out how James overcame the Miller--Bode objection, i.e., how he solved the problem of two observers, where
would I look?
\ers
Here's what Lamberth says in his (excellent) book {\sl William James and the Metaphysics of Experience}:
\bq
When James unveiled his radical empiricism in the 1904--5 {\sl Journal of Philosophy\/} ``series,'' both the issue of the compounding of consciousness and the related problem of solipsism (or of direct realism and a shared world) were on his mind.  ``Does Consciousness Exist?''\ goes straight to the issue of the compounding of consciousness, as well as James's direct epistemological view, while ``The Thing and its Relations'' and ``How Two Minds Can Know One Thing'' pursue the philosophical issues that I am construing broadly as having to do with co-consciousness or a shared world.  That James himself remained unsatisfied with the position in those articles, however, is clear from the notebooks he kept between 1905 and 1908 on Dickinson Miller's and B. H. Bode's objections to his articles on radical empiricism.
\eq

So, those three articles are places for you to start, and they can be found on the web.  Unfortunately, I don't yet have the book {\sl Manuscript Essays and Notes\/} which contains the spoken-of notebook.  What I suspect is that this is an issue we're going to have to ultimately work out for ourselves---those old notebooks, when dug up, will only get us so far.  Still I'll have them waiting here when you arrive.

\section{05-03-09 \ \ {\it Graduate Openings} \ \ (to M. A. Graydon)} \label{Graydon1}

All commendable answers.  Sorry to pry like that.  I'm feeling pretty comfortable with my inclination of decision.  I hope you'll join the effort here:  I think we can make some great progress getting quantum mechanics sorted out.  And I'm quite sympathetic to one of the lines in your statement of interest:  What we want in this quantum foundations game is to get at a clear-cut description of the reality underlying quantum mechanics.  The quantum-Bayesian approach is an oblique approach (because it takes as its starting point agents and decisions), but it is a firm and careful approach:  The ultimate residue will be the reality itself, whatever it is.

I think I have two very good other students lined up as well, along with a postdoc.  We could make a great team.  I'm going to work to get everyone in the team some desk space at the Perimeter Institute, for much easier interaction.

Attached is a still further improved version of the paper.  See Eq.\ (44) describing unitary time evolution:  I hope it'll knock your socks off.  [See ``Quantum-Bayesian Coherence,'' \arxiv[quant-ph]{0906.2187}.]

\section{07-03-09 \ \ {\it The Beautiful Shape} \ \ (to D. M. {\Appleby}, cc H. C. von Baeyer)} \label{Appleby53} \label{Baeyer65}

Thanks for the alternate proof of Tommy.  {\Asa} {\Ericsson} and I discussed Tommy a lot yesterday---she's just moved to Waterloo---and so I forwarded her the note.  Hope you don't mind.  I'm trying to get her in the thick middle of this problem.

Let me tell you and Hans a story.

Yesterday as I was parking the car, Katie asked me, ``What's your favorite color?''  ``Blue,'' I said, as if she hadn't already heard the same answer a thousand times before.  But then she asked me something she's never asked, ``What's your favorite shape?''  I was just about to say ``ball'' when I caught myself.  ``What a wonderful question Katie!  My favorite shape is Hilbert space; I've never told you that.''  She said, ``What does it look like?''  I said, ``I can't tell you yet, because we're still trying to figure it out.  That's what I work on every day at PI.''  ``If you don't know what it looks like, how do you know you like it?''  ``Because it's a beautiful shape, the most beautiful shape ever.  That much we can already tell!''  She looked perplexed; I came away delighted.

If Pauli is right, that shape must already be in our archetypes.  On this rainy grey Saturday, I think I'll sit down with a book I ran across in DC the other day. The title is {\sl Robert Fludd\/} and its back cover says, ``All Fludd's important plates are collected here for the first time, annotated and explained \ldots''  Maybe that's another way to look for our lovely, but still hidden shape.

\section{09-03-09 \ \ {\it The Beautiful Shape, 2} \ \ (to D. M. {\Appleby}, cc H. C. von Baeyer)} \label{Appleby54} \label{Baeyer66}

Regarding: \bma I am slightly puzzled, though, because the thoughts in
your note (the beauty of the hidden shape, and the idea that it is
already there in our archetypes) are, to my way of thinking, decidedly
Platonic in character.  \ldots\ I have had the impression that you
didn't wish to go down that path \ema It is a bit like the story of
Niels Bohr's horseshoe.  Upon seeing it hanging over a doorway someone
said, ``But Niels, I thought you didn't believe horseshoes could bring
good luck.''  Bohr replied, ``They say it works even if you don't
believe.''\footnote{\editornote I find it interesting that nobody
  seems to know where this story comes from.  The place where I first
  read it was a jokebook, {\sl Asimov's Treasury of Humor} (1971),
  which happens to be three years older than the earliest appearance
  Wikiquote knows about.  In this book, Isaac Asimov tells a lot of
  jokes and offers advice on how to deliver them.  The Bohr horseshoe
  is joke~\#80.  Asimov's commentary points out a difficulty with
  telling it.  ``To a general audience, even one that is highly
  educated in the humanities, Bohr must be defined---and yet he was
  one of the greatest physicists of all time and died no longer ago
  than 1962.  But defining Bohr isn't that easy; if it isn't done
  carefully, it will sound condescending, and even the suspicion of
  condescension will cool the laugh drastically.''

Note the slight dusting of C.\ P.\ Snow.

Asimov proposes the following solution.  ``If you despair of getting
the joke across by using Bohr, use Einstein.  Everyone has heard of
Einstein and anything can be attributed to him.  Nevertheless, if you
think you can get away with using Bohr, then by all means do so, for
all things being equal, the joke will then sound more literate and
more authentic.  Unlike Einstein, Bohr hasn't been overused.''

I find this, except for the last sentence, strangely appropriate in the
context of quantum-foundations arguments.

Heisenberg's essay ``Science and Religion'' relates the story as an
anecdote told by Bohr around 1927.  However, in that version, the one
with the horseshoe is not Bohr himself, but his neighbor in Tisvilde.}

Yes, you have read me correctly in the past.  Saturday's note was mostly a reflection of my wispy mood of the moment.  But you've caused me to reflect.  Perhaps my flipping through the Fludd book was more akin to this.  With the warmer weather, Kiki has started to dream of what she will do with the porch, whether she'll put rocking chairs on it, a table, what kind of table, where they'll be placed, things like that.  But in her inability to act (the weather is not that warm yet), she's lately taken to looking through catalogs for quite long lengths of time.  She's not doing it, however, to buy anything---just to get ideas, she says.  I think she uses the catalogs mainly to stimulate the right parts of her brain.  And now come to think of it, maybe the monkey in me was doing the same.  Kiki had been looking at catalogs all afternoon, and at some level I fell in line with her:  I reached for the closest thing I had to a catalog.

The book I have is this one: \myurl[http://www.amazon.com/Robert-Fludd-Hermetic-Philosopher-Surveyor/dp/0933999690/ref=pd_bbs_sr_2?ie=UTF8\&s=books\&qid=1236592436\&sr=8-2]{http://www.amazon.com/Robert-Fludd-Hermetic-Philosopher -Surveyor/dp/0933999690/ref=pd\underline{ }bbs\underline{ }sr\underline{ }2?ie=UTF8\&s=books\&qid=1236592436\&sr=8-2}.

There's all sorts of wonderful pictures in there, even ones with decidedly SICish features: a seven pointed rose, a tetragrammaton, or, say, this one \myurl[http://data5.blog.de/media/739/3229739_d0eb33369c_m.jpeg]{http://data5.blog.de/media/739/3229739\underline{ }d0 eb33369c\underline{ }m.jpeg} which evokes in me a whiff of the urungleichung.  But as you've told me with the {\sl I Ching}, one cannot force these things.  And in my looking, I have the overwhelming feeling that I'm probably forcing these interpretations.

\subsection{Marcus's Reply}

\bq
The link to amazon didn't work for some reason.    And when I did a search on Amazon for Robert Fludd it came up with a long list (rather to my surprise:  I hadn't realized he was so well known).  Is the book you are talking about the one by Joscelyn Godwin?

Anyway, Plato.  I wanted to write at length, but I have had an enormous number of distractions today. I have done nothing since lunch-time, and I badly want to get back to the calculation I was doing.  So I hope you will forgive me if I am a little brief.

But yes, the word ``wispy'' would describe my own state of mind when I use the word ``Platonic''.  I don't really know, with any exactitude, what I mean by it.  Just that there is something there which feels right, and which I feel needs development. Badly needs, in fact.  Plato, the man, is not too important really. What is important is that I strongly feel that there is something there (I am not sure what exactly, but something) that matters, and which is being ignored  in the current climate.  In my mind this is all tied up with our discussions about Pauli.  In fact I am pretty sure I have seen Pauli described as a Platonist (though I can't remember by whom).

When we were talking once you said something (again, I can't remember what exactly) which suggested that in your mind Platonism was all tied up with the block universe.  I think you were thinking that the Platonic archetypes are something immutable.  But immutability is certainly not the point for me.  For me the word ``Platonism'' has no such connotation.

Rather ``Platonism'' (as I use the word) has to do with the connection between the inner and the outer.  The idea of people like Popper is that physics progresses in two stages.  First one guesses a hypothesis (Popper uses the word ``conjecture''), and then one tests it.
Philosophers of science focus on the second stage:  testing hypotheses.  But to my mind the really interesting, and by far the most important part is the process by which one comes up with the hypothesis in the first place.  Testing hypotheses is a non-trivial activity.  But it hardly takes genius.  Whereas I think it does take genius to think of the hypothesis.  At least it does if the hypothesis has any depth to it.  If, for example, it is like Faraday's hypothesis of field lines, or Kepler's hypothesis of elliptical orbits.

I should explain that I am using the word ``genius'' in what I believe is the correct sense, which differs significantly from what has become the usual sense.  Nowadays the word ``genius'' means an extremely intelligent person.  But originally it meant something quite
different:  a guiding or tutelary spirit (according to Wikipedia).  I think I have also seen it described as a ``fertile spirit''.  At any
rate a genius, in the original sense of the word, is not a person.
Genius isn't something anyone can {\it be}.  It is something one can {\it have\/} (``have'' in the sense one can {\it have\/} a daughter, or a father, or a friend) (and just as one can't own a daughter, or a father, or a friend, so one cannot own genius) (if anyone---anyone at all, Einstein and Faraday not excluded---has the temerity to think they own their genius---has the arrogance to identify themselves with their genius---they will
surely and deservedly lose it).   I think this is a very important
distinction.  The point is not that Faraday or Einstein were especially intelligent people.  Doubtless they were above averagely smart.  But I would question whether they were so much more smart than lots of other people, whom no one now remembers.  I don't think smartness is the point about someone like Faraday.  Rather it is the fact that Faraday was somehow communicating with something in the depths of his soul.  And, what is more, something maybe not all that personal to Faraday.  Something, in short, for which the word ``genius'', in the old sense, seems appropriate.

And that is what Platonism means to me.  It is the idea that the sources of science are as much internal as external.  Science, as practiced by a man like Faraday, isn't just a matter of taking careful note of data coming in through the external senses.  It is also, and perhaps even more importantly, a matter of listening to something located deep down in his own inner being.

That is also the point about archetypes, as I understand them.  And I feel that you had in mind something similar in the note you sent me about the beautiful shape.
\eq

\section{10-03-09 \ \ {\it Vienna Saturday Afternoon (9 years later)} \ \ (to {\AA}. {\Ericsson})} \label{Ericsson3}

I wanted to say thanks for yesterday, for your commentary on my paper:  It will help me improve it.  So, please don't stop if you've got a mind to keep going.  Attached is the latest version.  It still hasn't incorporated most of yesterday's comments, but I hope to address that soon.  In any case, it already incorporates a ``thanks'' to you!

This morning I was looking through my files for stuff on an old summer student, and I ran across the note below from December 2000.  Look how excited I was about an idea of mine.  In the end, there was a fundamental flaw in the approach I was taking.  But notice the similarity between my phrases below and the things I say in Section 3.1 of the present paper.  Nearly nine years:  I am nothing if not stubborn!

\subsection{From a letter to Greg Comer dated 2 December 2000}

\bq
Did I tell you I'm in Vienna?  Tomorrow I go to Budapest and then I come back to Vienna until Dec 12.  It's been a very busy last few days for me:  I've finally done it!  I have a fully Bayesian derivation of complete positivity for quantum time evolutions (that's the generalization of unitarity for density matrices, taking into account measurement-disturbance nastiness).  This is the best work I've done in six years.  As suspected, quantum time evolutions boil down to {\it nothing\/} but Bayes' rule in disguise.  I've been working furiously to dot the i's and cross the t's and get something written up before I leave here.

So, I may not be writing you email until my return to the States.  Keep thinking about those dictionaries:  that is a very deep train of thought.
\eq

\section{12-03-09 \ \ {\it Cover Art} \ \ (to S. Capelin)} \label{Capelin8}

Attached is one idea.  Maybe a real artist could base something on it.  It is a transparency I have used in my talks for years, and is meant literally to convey what I have been calling ``The Paulian Idea.''  It captures the following points:
\begin{enumerate}
\item
That the quantum state lives in the head of the observer, not in the system being measured.  Quantum states are epistemic.
\item
That measuring devices should not be considered independent of the measuring agent, but rather are more like prosthetic hands than external pieces of the world.
\item
That the quantum system (which is an external piece of the world) is something like an alchemist's ``philosopher stone''---its conceptual role is like a catalyst that transforms the alchemist.
\item
That the result of a measurement is like a miniature act of creation.  These are the sparks depicted as flying between the measuring device (measuring device) and the quantum system.
\end{enumerate}
Anyway, that's one idea, if it is something that can be made tasteful.

Another idea stems from another rhetorical trick I play in my talks.  See pages 2, 7, and 9 of \myurl{http://www.perimeterinstitute.ca/personal/cfuchs/Peres\%20School\%201-110.pdf}.
I often put a picture of a block down and say, ``This is a quantum system.''  But then I superpose a state vector $|\psi\rangle$ on top of it, and say, ``This symbol represents {\it my\/} knowledge of it.''  ``No me, in the room?  Then no state vector in the room.''  And then I lift the vector off the block.  Then, I say, ``But the quantum system remains; it hasn't disappeared, only the state vector.''  That's another way of putting some of what I said above---it is an essential aspect of The Paulian Idea.  That quantum systems exist in a substantive way, even if state vectors do not.  So, another idea for cover art would be of a cubical block with a stylistic state vector rising off of it.

That's the two best ideas I have at the moment.

\section{13-03-09 \ \ {\it Graduate Openings, 2} \ \ (to M. A. Graydon)} \label{Graydon2}

You should make sure you are making the right decision for yourself.  Ross Diener was just out here, and we spent a good bit of time together.  You two should get together and share notes, dreams, plans.  Research is a big step:  It's not at all like class work.  You write your own questions, and you seek your own answers in byways that no one has ever traveled.  In a way, you put your own personality {\it into\/} the world itself.  Emotion, personality, and science become pretty inextricably entwined.  So, take into account the whole package in your thinking.

\section{13-03-09 \ \ {\it The Unfinished Universe} \ \ (to M. A. Graydon)} \label{Graydon3}

In rereading what I just wrote, I found myself thinking of a passage from William James that I've always liked, and I thought I'd send it too.  File attached.  [See passage in 05-01-09 note ``\myref{PauliFierzCorrespondence}{What I Really Want Out of a Pauli/Fierz-Correspondence Study}'' to H. C. von Baeyer \& D. M. {\Appleby}.]  Read it as a metaphor for the group I'd like to assemble.  Picking up on the last line in it---``the universe, unfinished, growing in all sorts of places''---I see our "additions" as helping the universe grow in all sorts of places.

\section{23-03-09 \ \ {\it Bayesian Chance} \ \ (to W. L. Harper)} \label{Harper3}

I want to cite your Bayesian Chance paper with Chow and Murray in the paper I'm writing at the moment.  Has it been published anywhere, or posted anywhere?

I started reading it this morning.  It gives me a great foil for some remarks I want to make at the end of my own paper, where I presently have the placeholder: ``sketch how to build a fully Bayesian theory of objective indeterminism without invoking objective chance.''

\section{26-03-09 \ \ {\it The Varieties of Optimistic Experience} \ \ (to A. Ney \& B. Weslake)} \label{Ney1} \label{Weslake2}

Thanks again for the conversation yesterday and last night.  Particularly, last night, it helped refocus my thoughts on the issue of temperament in science and philosophy.  Here's a quote from William James that I've always had some respect for:
\bq
The history of philosophy is to a great extent that of a certain clash of human temperaments.  Undignified as such a treatment may seem to some of my colleagues, I shall have to take account of this clash and explain a good many of the divergencies of philosophies by it.  Of whatever temperament a professional philosopher is, he tries, when philosophizing, to sink the fact of his temperament. Temperament is no conventionally recognized reason, so he urges impersonal reasons only for his conclusions.  Yet his temperament really gives him a stronger bias than any of his more strictly objective premises.
It loads the evidence for him one way or the other, making a more sentimental or more hard-hearted view of the universe, just as this fact or that principle would.  He {\it trusts\/} his temperament.
Wanting a universe that suits it, he believes in any representation of the universe that does suit it. He feels men of opposite temper to be out of key with the world's character, and in his heart considers them incompetent and `not in it,' in the philosophic business, even though they may far excel him in dialectical ability.

Yet in the forum he can make no claim, on the bare ground of his temperament, to superior discernment or authority.  There arises thus a certain insincerity in our philosophic discussions:  the potentest of all our premises is never mentioned.  I am sure it would contribute to clearness if in these lectures we should break this rule and mention it, and I accordingly feel free to do so.
\eq

I was acutely aware that there was no time in my talk to explain or to even sketch the things that drive me the most in this quantum-Bayesian quest---it is the pursuit of a malleable world.  Let me just point to these papers as a momentary band-aid:
\arxiv{quant-ph/0404156}
and
\arxiv{quant-ph/0608190}

Also, let me share the present attachment, which very much addresses the optimism/pessimism issue discussed last night.  [See ``Delirium Quantum,'' \arxiv{0906.1968v1}.]  It is a conversation I had with Howard Wiseman on the subject:  He too labeled quantum Bayesianism a kind of pessimism, and I did my best to lay out what I find ``wildly optimistic'' in it.  Actually, relooking, the whole document is devoted to this issue.  See, for instance, the paragraph on page 15 that starts ``Doesn't that just make you tingle?''  Anyway, it is the potentest of the premises that drive me personally in physics---for instance, the drive that leads me to reformulate quantum mechanics in this way or that---and it is indeed worth declaring forthrightly.

\section{26-03-09 \ \ {\it Review of the Magic Equation} \ \ (to C. R. Stroud, Jr., J. H. Eberly, A. Ney, Y. Shapir \& B. Weslake)} \label{Ney2} \label{Weslake3} \label{Stroud1} \label{Eberly1}

Thanks for the seeming interested in my talk yesterday; it's much appreciated.  However, this morning as I awake, I find myself feeling disappointed that the key idea from near the end---that of taking the ``magic'' formula as a fundamental axiom of quantum mechanics---didn't have a chance to come across so well.  Thus, let me ease my conscience by sharing a draft of a paper I'm presently writing with Schack on the subject.  [See ``Quantum-Bayesian Coherence,'' \arxiv{0906.2187v1}.] It is attached.  The first five sections of it are stabilized now and should be completely readable.  I hope it will give you a much better picture of where I think quantum mechanics comes from.

If you have any constructive or deconstructive comments on the draft, I would love to try to incorporate them into an improved paper.

It was great discussing with you all.  Thank you for that and the fine wine and the fine hospitality.

\subsection{Excerpt of Joe Eberly's Reply}

\bq
I agree that it's very attractive to have the possibility of an axiom that can encompass several desiderata at once, including the Born Rule, even if only to 38 decimal places.
\eq

\section{26-03-09 \ \ {\it QB Coherence} \ \ (to W. G. {\Demopoulos})} \label{Demopoulos31}

\ldots\ from a hotel in Rochester.  I apologize for taking so long to reply to you.  I kept hoping to finish the draft of my paper so that I could send it along in a reply.  It hasn't happened; I continue to keep fiddling with the manuscript, and I presently judge myself only about 80\% complete.

Still I think it is finally at a stage where it might benefit from your input.  Particularly Section 2, of which I am not completely happy with at the moment.  Sections 2 and 4 could probably use the most Demopoulizing actually.

Any input you might have would be invaluable.

To answer your other questions.  I'd like to stay over when I visit London, if it fits with my other obligations; but I'll tell you definitively as the time draws nearer.

About Smeenk's talk, it was {\it very}, {\it very\/} good.  I enjoyed it tremendously and indirectly it connected with several things that I worry about most in my forming view of quantum mechanics.

\section{27-03-09 \ \ {\it New Slogan}\ \ \ (to R. {\Schack})} \label{Schack155}

QM is only about the clicks you put yourself in a position to see (feel, experience).

\section{27-03-09 \ \ {\it New Slogan, 2}\ \ \ (to R. {\Schack})} \label{Schack156}

\brs
Ahhh, we need some time to discuss these things. The point is that qm is only about your active experience, right?
\ers
Yep!  Sad to think this formulation took so long.  It came to me while driving back from Rochester yesterday.  It looks like we have two strong supporters in Joe Eberly and Carlos Stroud (both involved in quantum optics, both students of Jaynes).  Do you know them?

\section{27-03-09 \ \ {\it Quantum Fiddling and Twiddling} \ \ (to N. D. {\Mermin})} \label{Mermin149}

A while back I mentioned a paper to you that I was just starting to construct; I asked whether you could remember where you had remarked on Feynman's ``only mystery'' of quantum mechanics.  Well, I'm still twiddling on the draft.  But it is now about 80\% complete, and particularly the first 5 sections are pretty stabilized.  Still, I'm not completely happy with the introductory section.  If you've got the time/interest, might I ask you if you have any constructive feedback for the introduction.  No promises that I'd incorporate anything of course, but you've shaken me from my dogmatic slumbers before, and you might again.  And you're certainly the best test kitchen I know.

Hope things are going well for you.

I invented a new slogan yesterday:  QM is only about the clicks you put yourself in a position to see (feel, experience).  You're the second eyes to see it.

\section{28-03-09 \ \ {\it Clicks Happen -- (the bumper sticker)} \ \ (to N. D. {\Mermin})} \label{Mermin150}

Last night I reported your note to Kiki after we'd had a little wine, and I found myself speaking in quite a mocking voice \ldots\ ``David's spennnndinggg threeee months in Cooopenhaggggen.''  That sort of thing.  Hearing myself, I realized I was actually childishly jealous!  You lucky boy!

Funny, we just had a discussion on the Bohr, Mottelson, Ulfbeck paper in our quantum foundations group meeting the week before last.  The consensus was it's damned unclear.  But I've said that before in the singular.  I got much more from the more poetic Ulfbeck--Bohr paper (the one sans Mottelson).  There is an idea in there that I like; that each quantum event is very literally unique.

Yesterday's slogan was meant to imply that stuff happens all the time, everywhere.  But when I use quantum mechanics, I'm talking very strictly about the stuff that impinges on ME (and results partially because of ME).

If you do decide to delve into the paper a few days from now, let me know.  I'll send you whatever is the latest draft at that point.

I'll think about how to make use of the Bohr archive.  Interesting proposition.

\section{30-03-09 \ \ {\it What If?}\ \ \ (to W. G. {\Demopoulos})} \label{Demopoulos32}

\bwd
Here's a simple idea apropos our conversation about conditionalization:  What would be anti-subjectivist about the idea that there is convergence to the same algorithm for assigning subjective probabilities? Of course this has the effect of converging to the same assignment on discovering an outcome, but agreeing to a common algorithm is not the same as agreeing to base our probability assignments on an objective and physical property of things, as I assume an objective chance theorist would hold. The situation is similar to our agreement to rules of inference in logic. (Here I'm not talking about logics of events, etc., where there is something like a physical interpretation of what's going on, but of pure logic.)
\ewd

I don't think there is anything wrong with converging to the ``same algorithm'' (L\"uders' Rule for instance).  Nothing wrong with two agents starting with different state assignments initially, but having a common assignment after viewing a particular piece of data.

\section{01-04-09 \ \ {\it Lunchtime Conversation}\ \ \ (to R. W. {\Spekkens})} \label{Spekkens62}

Our lunchtime conversation yesterday has prodded me write a new essay.  Hopefully I'll have it generated next week after Marcus is gone.  The subject was really all about this slogan I'm trying to polish into something pithy:  ``QM is only about the clicks you put yourself in a position to see (feel, experience).''  It is the very point of a theory from the inside of the world (i.e., if there is no God's eye view, that is what you have).  I want to flesh that out and put it in the context of the Norsen argument (which I view as a recalcitrant attempt to view quantum states onticly\footnote{Or ontically?  How do you spell it?}, via the apparatus of the EPR criterion of reality).  It will be a good exercise.

\section{05-04-09 \ \ {\it E on Time's Illusion} \ \ (to D. M. {\Appleby})} \label{Appleby55}

Here's a quote close to the one you were thinking of.  It's from a letter of Einstein's to the son and sister of Michele Besso, 21 March 1955:
\bq\noindent
Now he [Besso], too, has just preceded me in his departure from this strange world.  This means nothing.  To us believing physicists the separation of past, present and future has only the significance of an illusion, albeit a stubborn one.
\eq

\section{06-04-09 \ \ {\it Where's ONR?}\ \ \ (to K. Martin)} \label{Martin9}

Hey, have you watched ``The DiVincenzo Code'' on YouTube?  It redeems Oxford Physics in a way.  (Nothing can redeem Oxford Philosophy.)  Seriously.  Get some beers and watch it slowly---I did that in Australia a few months ago---it's a lot of fun.

Isn't it like 4 AM there right now?

\section{06-04-09 \ \ {\it A History-of-Knowledge Thing} \ \ (to D. P. DiVincenzo \& C. H. {\Bennett})} \label{DiVincenzo4} \label{Bennett63.1}

A fitting story for John Wheeler.  John himself would slide between two formulations, that our quantum measurements here and now determine what we can say of the past, and that our measurements here and now create the past.  As far as I can tell he never completely decided between the two very different ideas.

I've always been intrigued by an argument of George Herbert Mead (who by the way was born in South Hadley, MA) that in a world with real indeterminism, the past must indeed be open or malleable, something like John's second formulation.  Below is Arthur Murphy's summary of Mead's argument. [See Murphy's quote in 23-09-03 note ``\myref{Savitt3}{The Trivial Nontrivial}'' to S. Savitt.]

Coming back down to earth though, I recall how disappointed I was when I learned that John, though he always gets (and took?)\ credit for it, did not invent the word ``black hole''.  Someone in an audience called it out after one of his talks on gravitational collapse.  This story of Xe-135 might be of a similar type.

\subsection{David's Preply}

\bq
I read with interest some of the stories of Wheeler in the new {\sl Physics Today}.  One that really caught my eye was the story that, in a few minutes of work at Hanford, it was Wheeler who realized that Xe-135 was the principal poisoner of nuclear-pile reactions.  This is one of the profound engineering insights of the 20th century --- it was the lack of this knowledge that may well have blown up Chernobyl.

But then, in browsing a book called {\sl Nuclear Technology}, I see a detailed story attributing this insight to Fermi!  (He is rushed by train to Hanford, does some serious thinking, etc.)  So, I wonder, what is right?  Is this one of these things that can be made to come out either way by those exoplanet residents 65 light years away that are just receiving the first Hanford light?
\eq

\section{06-04-09 \ \ {\it Completing the Story of Incompleteness}\ \ \ (to L. Freidel)} \label{Freidel1}

Here's the lines from that old paper of mine that I was telling you about:
\bq
There are two issues in this \ldots\ that are worth disentangling.
1) Rejecting the rigid connection of all nature---that is to say, admitting that the very notion of {\it separate systems\/} has any meaning at all---one is led to the conclusion that a quantum state cannot be a complete specification of a system.  It must be information, at least in part.  This point should be placed in contrast to the other well-known facet of Einstein's thought: namely, 2) an unwillingness to accept such an ``incompleteness'' as a necessary trait of the physical world.

It is quite important to recognize that the first issue does not entail the second.  Einstein had that firmly in mind, but he wanted more. His reason for going the further step was, I think, well justified {\it at the time\/}:
\bq\noindent
There exists \ldots\ a simple psychological reason for the fact that this most nearly obvious interpretation is being shunned.  For if the statistical quantum theory does not pretend to describe the individual system (and its development in time) completely, it appears unavoidable to look elsewhere for a complete description of the individual system; in doing so it would be clear from the very beginning that the elements of such a description are not contained within the conceptual scheme of the statistical quantum theory.  With this one would admit that, in principle, this scheme could not serve as the basis of theoretical physics.
\eq

But the world has seen much in the mean time.  The last seventeen years have given confirmation after confirmation that the Bell inequality (and several variations of it) are indeed violated by the physical world.  The Kochen--Specker no-go theorems have been meticulously clarified to the point where simple textbook pictures can be drawn of them.  Incompleteness, it seems, is here to stay:  The theory prescribes that no matter how much we know about a quantum system---even when we have {\it maximal\/} information about it---there will always be a statistical residue.  There will always be questions that we can ask of a system for which we cannot predict the outcomes.  In quantum theory, maximal information is simply not complete information.  But neither can it be completed.  As Wolfgang Pauli once wrote to Markus Fierz, ``The well-known `incompleteness' of quantum mechanics (Einstein) is certainly an existent fact somehow-somewhere, but certainly cannot be removed by reverting to classical field physics.''  Nor, I would add, will the mystery of that ``existent fact'' be removed by attempting to give the quantum state anything resembling an ontological status.

The complete disconnectedness of the quantum-state change rule from anything to do with spacetime considerations is telling us something
deep: The quantum state is information. Subjective, incomplete information. Put in the right mindset, this is {\it not\/} so intolerable.  It is a statement about our world. There is something about the world that keeps us from ever getting more information than can be captured through the formal structure of quantum mechanics. Einstein had wanted us to look further---to find out how the incomplete information could be completed---but perhaps the real question is, ``Why can it {\it not\/} be completed?''

Indeed I think this is one of the deepest questions we can ask and still hope to answer.
\eq
The real question is why can it not be completed?  I still like that formulation and still think it has a lot of fruit to bear.

\section{07-04-09 \ \ {\it Quantum Mechanics and Pragmatism} \ \ (to M. S. Leifer)} \label{Leifer11}

There's nearly as many varieties of pragmatism as there are Bayesians:
\begin{itemize}
\item
I. J. Good, ``46656 Varieties of Bayesians,'' in I.~J. Good, {\sl Good Thinking: The Foundations of Probability and Its Applications}, (University of Minnesota Press, Minneapolis, 1983), pp.~20--21.
\end{itemize}

\section{07-04-09 \ \ {\it Your Thoughts} \ \ (to N. Bao)} \label{Bao1}

You wrote, ``My largest concern is with whether or not I am absolutely dead-set sure that I want to do Quantum Information / Quantum Foundations in grad school.'' It is hard for me to see how that reconciles with what you wrote in your statement of purpose for U. Waterloo:  ``At this point, I truly believe that quantum information theory is what I want to do for the rest of my life.''  My guess is you were seduced somewhat by the old-school prestige of Stanford in comparison to U. Waterloo and the self-confidence of the students there.  But those things make little impression on me:  Great physics is had by great passion.  Almost nothing else matters.  On the other hand, there is something John Wheeler once said that did make a great impression on me.  He was asked by someone in an audience, ``What differences do you see between the students at Princeton and the students at University of Texas?''  Wheeler replied, ``Only that the students of Princeton {\it know\/} they're smart.''  The implication was that the local students just needed more nurturing, but therein stopped the difference.  And so it was true:  No student at Princeton invented quantum information theory---that came almost exclusively from the University of Texas, with Schumacher, Wootters, Deutsch, and Zurek there at the time.  It was the place that mattered, not the prestige of the place.  Nor was it the breadth of the place---it was really all about focus and the sense of exploring a true frontier:  That brought out the best in everyone involved.

\section{07-04-09 \ \ {\it Question} \ \ (to A. Plotnitsky)} \label{Plotnitsky22}

Shamefully I'm still working on the SIC paper I told you about.  Attached is the present draft; it's still only about 80\% complete!  [See ``Quantum-Bayesian Coherence,'' \arxiv{0906.2187v1}.]  Feel free to cite it however; it will I hope soon be posted on the {\tt quant-ph} archive.  (OH, there was also a very preliminary version in Andrei's conference proceedings; a citation can be found in the footnote on the first page.)

Though it's only about 80\% complete, the first five sections are relatively stabilized.  Any comments are most welcome!

You can read about Glauber's coherent states here:
\bq
\myurl{http://en.wikipedia.org/wiki/Coherent_state}
\eq
and find the original references therein.  That the coherent states can be used as an informationally complete POVM to give a Bayesian representation of quantum states, can be read about here
\bq
\myurl{http://en.wikipedia.org/wiki/Husimi_Q_representation}.
\eq
Beware though, that that article calls $Q$ a ``quasi-probability'' when in fact it is an honest-to-god probability---nothing quasi about it.

\section{08-04-09 \ \ {\it Media Interview for FQXi about Hugh Everett}\ \ \ (to G. Stemp-Morlock)} \label{StempMorlock1}

Thanks.  I much appreciate that.  Only two changes I'd make to the quote; I'll set them off with brackets:
\bq\noindent
``It was a reaction to the Copenhagen interpretation that went in a particular direction, and it was a healthy move because ultimately one does want to get away from thinking about observers as an integral part of quantum mechanics and it does do that,'' said Fuchs. ``But, what price do you pay [for] a view of the world that is not very particular to [that] world?''
\eq

Might I also suggest you modify this sentence slightly:  ``believes it is a contentless interpretation, that is, it doesn't tell us anything beyond the Copenhagen interpretation that could lead to new developments in quantum foundations.''  In the present formulation, if MWI is contentless, then the only conclusion is that Copenhagen is contentless as well!  I'd never agree with that.  So I would prefer for you to simply strike the words ``beyond the Copenhagen interpretation''.  For instance, you could make it:
\bq\noindent
Christopher Fuchs \ldots\ believes it is a contentless interpretation, that is, it doesn't tell us anything that could lead to new developments in quantum foundations.
\eq
The main point is that one could make up a many-worlds interpretation of {\it any\/} theory.  That is, despite all the posturing of the believers, there is nothing particular to quantum mechanics that actually pushes it upon us.  That is why I say it is contentless.

I wish I could find the quote of John Wheeler where he says something similar to the latter part of your sentence, i.e., ``it doesn't tell us anything that could lead to new developments in quantum foundations.''  If you want me to look, I can do it when I get to the office.

Good luck with your article.

\section{08-04-09 \ \ {\it {\Mermin} the Epistemicist}\ \ \ (to R. W. {\Spekkens})} \label{Spekkens63}

If I read you correctly yesterday, you seemed a little surprised when I said that {\Mermin} ought to be in the epistemic team.  I discovered he says it pretty succinctly here:
\bq\noindent
\arxiv{0808.1582},
\eq
starting particularly at the last paragraph on page 2.

\section{08-04-09 \ \ {\it Plastino y Plastino}\ \ \ (to R. W. {\Spekkens})} \label{Spekkens63.1}

It dawned on me in the shower, ``Wait, I did publish the classical no-cloning argument.''  I think yesterday I said that I hadn't.  I suppose what annoyed me about the P-y-P paper was that I hadn't had the gumption to make it a PRL.  Instead, I probably did my usual of burying it in a footnote of a 40 page paper.

\section{08-04-09 \ \ {\it AJP Resource Letter}\ \ \ (to R. W. {\Spekkens})} \label{Spekkens63.1.1}

Dear Rob,

Please review the note below [from R. H. Stewer, of the {\sl American
    Journal of Physics\,}].  I guess I'm stuck doing this now; I said
I'd do it for them, though I was mostly hoping they'd ``boot me out
the door''.  Would you have any interest in joining me as a coauthor?
I believe your input would help make the selection of materials much
less one sided.  Plus you'd be a morale builder (keeping me on task),
even if I did most of the writing.\footnote{\editornote The AJP
  Resource Letter never came to pass.  The version here is based on a
  21 August 2009 draft by CAF and RWS, with some corrections for the
  present samizdat.}

Maybe it'd be a good time to re-approach the proposed article with Matt for
Phys Today as well.  They could be companion pieces.

\begin{center}
\large{\bf Resource Letter IQM-3: Quantum Foundations in the Light of Quantum Information}
\bigskip\bigskip\\
Christopher A.\ Fuchs and Robert W.\ Spekkens
\end{center}

\bq
\subsection{Introduction}

It is well known that Peter W.\ Shor's quantum factoring algorithm shook two worlds in 1994:  the world of physics and the world of computer science.  But it is not so well known that, like an earthquake, it was precursed by a deep murmur from foundations far below:  In this case, the foundations of quantum theory.

\subsection{Textbooks}
\small
\begin{enumerate}
\item
{\bf Quantum Theory:\ Concepts and Methods}, A. Peres (Kluwer Academic Publishers, Dordrecht, 1995).
\item
{\bf Quantum Computation and Quantum Information}, M. A. Nielsen and I. L. Chuang (Cambridge University Press, 2000).
\item
{\bf Quantum Computer Science: An Introduction}, N. D. Mermin (Cambridge University Press, 2007).
\item
{\bf Quantum Processes, Systems and Information}, B. Schumacher and M. Westmoreland (Cambridge University Press, 2009).
\end{enumerate}
\normalsize

\subsection{Monographs}
\small
\begin{enumerate}
\item
{\bf Coming of Age with Quantum Information:\ Notes on a Paulian Idea,} C. A. Fuchs, (Cambridge University Press, 2010); previously posted as C. A. Fuchs, {\sl Notes on a Paulian Idea:  Foundational, Historical, Anecdotal \& Forward-Looking Thoughts on the Quantum}, \quantph{0105039v1} (2001).
\item
{\bf States, Effects, and Operations: Fundamental Notions of Quantum Theory. Lecture Notes
in Physics, vol.\ 190}, K. Krauss, (Springer-Verlag, Berlin, 1983).
\end{enumerate}
\normalsize

\subsection{Prehistory: Quantum States as Information before Information Theory}

\small
\begin{enumerate}
\item
{\bf ``Autobiographical Notes,''} A. Einstein, in {\sl Albert Einstein:
Philosopher-Scientist,} edited by P. A. Schilpp, (Tudor Publishing Co., New
York, 1949).
\item
{\bf Quantum Theory,} D. Bohm, (Prentice Hall, 1951).
\item
{\bf ``The Development of the Interpretation of the Quantum Theory,''}
W. Heisenberg, in {\sl Niels Bohr and the Development of
  Physics:~Essays Dedicated to Niels Bohr on the Occasion of His
  Seventieth Birthday}, edited by W.~Pauli with the assistance of
L.~Rosenfeld and V.~Weisskopf (McGraw-Hill, New York, 1955),
pp.~12--29.
\item
{\bf ``Continuity, determinism, and reality,''} M.\ Born, in {\sl
  Festskrift til Niels Bohr [Commemorative Volume in Honour of Niels
    Bohr on the occasion of his 70th birthday]} (Royal Danish Academy
of Sciences and Letters, 1955.)
\item
{\bf ``,''} W. Pauli,.
\item
{\bf ``Probability in Quantum Theory,''} E. T. Jaynes, in {\sl Complexity, Entropy, and the Physics of Information,} edited by W. H. Zurek (Addison-Wesley, 1990).
\item
{\bf ``In Defence of `Measurement,'\,''} R. Peierls, Phys.\ World {\bf 4,}
19--20 (1991).
\end{enumerate}
\normalsize

\subsection{Early Wheeler and Wheeler-Entourage Papers}

\small
\begin{enumerate}
\item
{\bf ``\,`Relative state' formulation of quantum mechanics,''} H.
Everett III, Rev.\ Mod.\ Phys.\ {\bf 29,} 454--462 (1957).
\item
{\bf ``Include the Observer in the Wave Function?,''} J. A. Wheeler, in {\sl Quantum Mechanics, a Half Century Later: Papers of a Colloquium on Fifty Years of Quantum Mechanics, Held at the University Louis Pasteur, Strasbourg, May 2--4, 1974}, J. Leite Lopes and M. Paty, eds., (D. Reidel, Dordrecht, 1977), pp.\ 1--18.  This paper represents Wheeler's turning away from an Everettian interpretation of quantum states and marks his first step toward the view that wave functions represent ``information,'' instead of full-blown states of reality.  From the abstract:  ``One is led to recognize that a wave function `encompassing the whole universe' is an idealization, formalistically perhaps a convenient idealization, but an idealization so strained that it can be used only in part in any forecast of correlations that make physical sense.  For making sense it seems essential most of all to `leave the observer out of the wave function'.''
\item
{\bf ``Genesis and Observership,''}  J. A. Wheeler, in {\sl Foundational Problems in the Special Sciences.\ Part Two of the Proceedings of the Fifth International Congress of Logic, Methodology and Philosophy of Science, London, Ontario, Canada -- 1975}, R. E. Butts and J. Hintikka, eds., (D. Reidel, Dordrecht, 1977), pp.\ 3--33.
\item
{\bf ``The Computer and the Universe,''} J. A. Wheeler, Int.\ J. Theor.\ Phys.\ {\bf 21}, 557--572 (1982).  From the abstract:  ``The reasons are briefly recalled why (1) time cannot be a primordial category in the description of nature, but secondary, approximate and derived, and (2) the laws of physics could not have been engraved for all time upon a tablet of granite, but had to come into being by a higgledy-piggledy mechanism. It is difficult to defend the view that existence is built at bottom upon particles, fields of force or space and time. Attention is called to the `elementary quantum phenomenon' as potential building element for all that is. The task of construction of physics from such elements is compared and contrasted with the problem of constructing a computer out of `yes, no' devices.''
\item
{\bf ``Law without Law,''} J. A. Wheeler, in {\sl Quantum Theory and Measurement}, J. A. Wheeler and W. H. Zurek, eds., (Princeton University Press, 1983), pp.\ 182--213.
\item
{\bf ``,''}.
\item
{\bf The Acquisition of Information from Quantum Measurements}, W. K. Wootters, PhD thesis, University of Texas (1980).
\item
{\bf ``On Wheeler's Notion of `Law without Law' in Physics,''} D. Deutsch, Found.\ Phys.\ {\bf 16}, 565--572 (1986).
\item
{\bf ``Computation and Physics:\ Wheeler's Meaning Circuit?,''} R. Landauer, Found.\ Phys.\ {\bf 16}, 551--564 (1986).
\item
{\bf ``,''}.
\item
{\bf ``,''}.
\item
{\bf ``,''}.
\item
{\bf ``,''}.

\end{enumerate}
\normalsize

\subsection{Maxwell's Demon, Landauer's Principle, and Reversible Computation}

\small
\begin{enumerate}
\item
{\bf ``Irreversibility and heat generation in the computing process,''} R.
Landauer, IBM J.\ Res.\ Dev.\ {\bf 5,} 183--191 (1961).
\item
{\bf ``Logical Reversibility of Computation,''} C. H. Bennett, IBM J. Res.\ Dev.\ {\bf 17}, 525--532 (1973).
\item
{\bf ``The thermodynamics of computation---a review,''} C. H. Bennett, Int.\
J.\ Theor.\ Phys.\ {\bf 21,} 905--940 (1982).
\item
{\bf ``The Computer as a Physical System: A Microscopic Quantum Mechanical Hamiltonian Model of Computers as Represented by Turing Machines,''} P. Benioff, J. Stat.\ Phys.\ {\bf 22}, 563--591 (1980).
\item
{\bf Feynman Lectures on Computation.}  R. P. Feynman, J. G. Hey, and
R. W. Allen.  (Addison-Wesley Longman Publishing Co., Inc., 1998.)
\item
{\bf ``,''}.
\item
{\bf ``,''}.

\end{enumerate}
\normalsize

\subsection{No-Cloning and Its Inspirations}

\small
\begin{enumerate}
\item
{\bf ``A Single Quantum Cannot Be Cloned,''} W. K. Wootters and W. H. Zurek, Nature {\bf 299}, 802--803 (1982).
\item
{\bf ``Communication by EPR Devices,''} D. Dieks, Phys.\ Lett.\ {\bf 92A}, 271--272 (1982).
\item
{\bf ``The Probability of the Existence of a Self-Reproducing Unit,''} E. P. Wigner, in {\sl The Logic of Personal Knowledge:  Essays Presented to Michael Polanyi on his Seventieth Birthday}, (Routledge \& Kegan Paul, London, 1961); reprinted in E. P. Wigner, {\sl Symmetries and Reflections:\ Scientific Essays}, (Ox Bow Press, 1979).  This article is of interest for pure psychological reasons.  To put it in perspective, we report something W. K. Wootters said to one of us about his own cloning paper: ``I remember asking someone for his opinion of a draft of that paper, and he said, in a very friendly way, that it would be a tough paper to referee, because on the one hand probably lots of people already knew the result, but on the other hand it may not have ever been written down.''  What is so interesting about the Wigner paper is the way Wigner actually misses the no-cloning theorem and even gets it wrong.  Wigner writes, ``[I]t will be assumed that `the living state' is completely given in the quantum mechanical sense: it has one definite state vector \ldots. \ Before multiplication, the state vector of the system, organism $+$ nutrient, is $\Phi = v \times w$ \ldots  When multiplication has taken place, the state vector will have the form $\Psi=v\times v\times r$, that is, two organisms, \ldots [where] the vector $r$ describes the \ldots\ rejected part of the nutrient \ldots''  He then goes further to say, ``There must be many states [$v^k$] all of which represent a living organism. \ldots\ Then every linear combination of the $v^k$ will also represent a living state.''  But this of course is barred automatically by the no-cloning theorem!  How could a great mind like Wigner's miss this simple point?  No one knows, but we suspect it could have only been a psychological block brought on by thinking in too classical a way.  It was the breaking of that psychological block for the rest of the community that gives the theorem a deep and lasting value.
\item
{\bf ``Quantum Information:\ How Much Information in a State Vector?,''} C. M. Caves and C. A. Fuchs, Ann. Israel Phys.\ Soc.\ {\bf 12}, 226--257
(1996).
\item
{\bf ``Noncommuting Mixed States Cannot Be Broadcast,''} H. Barnum, C. M. Caves, C. A. Fuchs, R. Jozsa, and B. Schumacher, Phys.\ Rev.\ Lett.\ {\bf 76}, 2818--2821 (1996).
\item
{\bf ``A General No-Cloning Theorem,''} G. Lindblad, Lett.\ Math.\ Phys.\ {\bf 47}, 189--196 (1999).
\item
{\bf ``Generalized No-Broadcasting Theorem,''} H. Barnum, J. Barrett, M. Leifer, and A. Wilce, Phys.\ Rev.\ Lett.\ {\bf 99}, 240501 (2007).
\end{enumerate}
\normalsize

\subsection{Quantum Cryptography and Its Inspirations}

\small
\begin{enumerate}
\item
{\bf ``Quantum Cryptography: Public Key Distribution and Coin Tossing,''} C. H. Bennett and G. Brassard, in {\sl International Conference on Computers, Systems and Signal Processing, Bangalore, India, December 10--12, 1984}.  Though the conference proceedings this article appeared in is effectively nonexistent, this article is available online from several sources.
\item
{\bf ``Quantum Cryptography Based on Bell's Theorem,''} A. K. Ekert, Phys.\ Rev.\ Lett.\ {\bf 67}, 661--663 (1991).
\item
{\bf ``Quantum Cryptography Using Any Two Nonorthogonal States,''} C. H. Bennett, Phys.\ Rev.\ Lett.\ {\bf 68}, 3121--3124 (1992).
\item
{\bf ``Quantum Cryptography without Bell's Theorem,''} C. H. Bennett, G. Brassard, and N. D. Mermin, Phys.\ Rev.\ Lett.\ {\bf 68}, 557--559 (1992).
\item
{\bf ``,''}.
\item
{\bf ``,''}.
\item
{\bf ``,''}.

\end{enumerate}
\normalsize

\subsection{Quantum Teleportation and Its Inspirations}

\small
\begin{enumerate}
\item
{\bf ``Optimal Detection of Quantum Information,''} A. Peres and W. K. Wootters, Phys.\ Rev.\ Lett. {\bf 66}, 1119--1121 (1991).
\item
{\bf ``Teleporting an Unknown Quantum State via Dual Classical and Einstein-Podolsky-Rosen Channels,''} C. H. Bennett, G. Brassard, C. Cr\'epeau, R. Jozsa, A. Peres, and W. K. Wootters, Phys.\ Rev.\ Lett.\ {\bf 70}, 1895--1899 (1993).
\item
{\bf ``Quantum Theory Needs No `Interpretation',''} C. A. Fuchs and A. Peres, Phys.\ Tod.\ {\bf 53}(3), 70--71 (2000).  Of course it does, and the title of this paper was meant to be a tongue-in-cheek reference to Rudolf Peierls' statement that, ``The Copenhagen interpretation {\it is\/} quantum theory.''  Part of the paper expands on the idea that, ``The peculiar nature of a quantum state as representing information is
strikingly illustrated by the quantum teleportation process.''
\item
{\bf ``Classical Teleportation of Classical States,''} O. Cohen, \quantph{0310017v1} (2003).
\end{enumerate}
\normalsize

\subsection{Experimental Teleportation}
\small
\begin{enumerate}
\item
{\bf ``Experimental Realization of Teleporting an Unknown Pure Quantum State via Dual Classical and Einstein-Podolsky-Rosen Channels,''} D. Boschi, S. Branca, F. De Martini, L. Hardy, and S. Popescu, Phys.\ Rev.\ Lett.\ {\bf 80}, 1121--1125 (1998).
\item
{\bf ``Experimental Quantum Teleportation,''} D. Bouwmeester, J.-W. Pan, K. Mattle, M. Eibl, H. Weinfurter, and A. Zeilinger, Nature {\bf 390}, 575--579 (1997).
\item
{\bf ``Complete Quantum Teleportation Using Nuclear Magnetic Resonance,''} M. A. Nielsen, E. Knill, and R. Laflamme, Nature {\bf 396}, 52--55 (1998).
\item
{\bf ``Unconditional Quantum Teleportation,''} A. Furusawa, J. L. S{\o}rensen, S. L. Braunstein, C. A. Fuchs, H. J. Kimble, and E. S. Polzik, Science {\bf 282}, 706--709 (1998).
\end{enumerate}
\normalsize

\subsection{Generalized Measurements and Quantum Operations}

\small
\begin{enumerate}
\item
{\bf ``An Operational Approach to Quantum Probability,''} E. B. Davies and J. T. Lewis, Comm.\ Math.\ Phys. {\bf 17}, 239--260 (1960).
\item
{\bf ``General State Changes in Quantum Theory,''} K. Kraus, Ann.\ Phys.\ {\bf 64}, 311--335 (1971).
\item
{\bf ``Information-Theoretical Aspects of Quantum Measurement,''} A. S. Kholevo, Prob.\ Info.\ Trans. {\bf 9}, 110--118 (1973).
\item
{\bf ``Information and Quantum Measurement,''} E. B. Davies, IEEE Trans.\ Info.\ Theory {\bf IT-24}, 596--599 (1978).
\item
{\bf ``,''}.
\item
{\bf ``,''}.
\item
{\bf ``,''}.
\item
{\bf ``,''}.

\end{enumerate}
\normalsize

\subsection{Quantum Channel Capacities and Their Inspirations}

\small
\begin{enumerate}
\item
{\bf ``Quantum Coding,''} B. Schumacher, Phys.\ Rev.\ A {\bf 51}, 2738--2747 (1995).  The submission date on this paper is actually 9 April 1993.
\item
{\bf ``A New Proof of the Quantum Noiseless Coding Theorem,''} R. Jozsa and B. Schumacher, J. Mod.\ Opt.\ {\bf 41}, 2343--2349 (1994).
\item
{\bf ``,''}.
\item
{\bf ``,''}.

\end{enumerate}
\normalsize

\subsection{Entanglement Distillation, Its Roots and Inspirations}

\small
\begin{enumerate}
\item
{\bf ``Concentrating Partial Entanglement by Local Operations,''}, C. H. Bennett, H. J. Bernstein, S. Popescu, and B. Schumacher, Phys.\ Rev.\ A {\bf 53}, 2046--2052 (1996).
\item
{\bf ``Mixed-State Entanglement and Quantum Error Correction,''} C. H. Bennett, D. P. DiVincenzo, J. A. Smolin, and W. K. Wootters, Phys.\ Rev.\ A {\bf 54}, 3824--3851 (1996).
\end{enumerate}
\normalsize

\subsection{Quantum Computation}

\small
\begin{enumerate}
\item
{\bf ``Simulating Physics with Computers,''} R. P. Feynman, Int.\ J. Theor.\ Phys.\ {\bf 21}, 467--488 (1982).
\item
{\bf ``Quantum Theory, the Church-Turing Principle and the Universal Quantum Computer,''} D. Deutsch, Proc.\ Roy.\ Soc.\ London A {\bf 400}, 97--117 (1985).
\item
{\bf ``Quantum Computing via Measurements Only,''} R. Raussendorf and H. J. Briegel, \quantph{0010033}.
\item
{\bf ``Universal Quantum Computation Using Only Projective Measurement, Quantum Memory, and Preparation of the 0 State,''} M. A. Nielsen, \quantph{0108020}.
\item
{\bf ``A Quantum Computer Only Needs One Universe,''} A. M. Steane, Stud.\ Hist.\ Phil.\ Mod.\ Phys.\ {\bf 34}, 469--478 (2003).
\item
{\bf ``Three connections between Everett's interpretation and
  experiment,''} \\D. Deutsch.  In R. Penrose and C. J. Isham (eds.),
{\sl Quantum Concepts in Space and Time} (Oxford University Press, 1986).

\end{enumerate}
\normalsize

\subsection{Toy Theories}

\small
\begin{enumerate}
\item
{\bf ``Evidence for the Epistemic View of Quantum States:\ A Toy Theory,''} R. W. Spekkens, Phys.\ Rev.\ A {\bf 75}, 032110 (2007).
\item
{\bf ``Can Quantum Cryptography Imply Quantum  Mechanics?,''} J. A. Smolin, \quantph{0310067v1} (2003).
\item
{\bf ``,''}.
\item
{\bf ``,''}.
\item
{\bf ``,''}.
\item
{\bf ``,''}.
\item
{\bf ``,''}.

\end{enumerate}
\normalsize

\subsection{Quantum Bayesianism}

\small
\begin{enumerate}
\item
{\bf ``,''}.
\item
{\bf ``,''}.
\item
{\bf ``,''}.
\item
{\bf ``,''}.
\item
{\bf ``,''}.

\end{enumerate}
\normalsize

\eq

\section{09-04-09 \ \ {\it Testing Quantum Certainty}\ \ \ (to R. {\Schack})} \label{Schack157}

How does that sound for a title for yet another paper we must write this year?

Can you spot the flaw in \quantph{0601205} from our perspective? Apparently the only way I'll ever build (true and enduring) respect with my colleague {\Spekkens} is to respond thoroughly to this paper.

By the way, it's an easy (and fun) exercise.  I encourage you to try it.

\section{09-04-09 \ \ {\it QBism, Certainty, and Norsen} \ \ (to R. {\Schack})} \label{Schack158}

Between the two files attached is a draft for a draft of a paper responding to Norsen.  [Handwritten notes originally intended for Rob Spekkens's Quantum Foundations Group Meeting that day.] Let me know what you think.  Using him as a foil is, I think, a chance to do our certainty paper much better.

I'd like the new paper to begin with something like this:
\bq
Travis Norsen has recently given a very clear exposition on the underpinnings of the idea that quantum mechanics all on its own implies a violation of Bell Locality, without any need for a hidden variable assumption.  The paper has been a great service to the community.  However, in making his claim, Norsen made significant and unabashed use of a variant of the EPR criterion of reality.  In contrast, the present authors some years ago wrote a not-so-very-clear paper arguing in effect that the EPR criterion of reality is at odds with a personalist Bayesian account of probability, and that holding strictly to the Einstein notion of locality indeed forces its bankruptcy in the quantum context.  In this paper, we plan to do the argument much better justice by using the clarity of Norsen's paper as a counterpoint with which we must contend.  By sharpening ourselves with respect to Norsen blah blah blah \ldots
\eq

\subsection{Draft for a Draft of a Paper}

\begin{center}
{\bf Chris's Lonely View of the \emph{Formal Structure} of Quantum Mechanics }
\end{center}

\begin{itemize}
\item[1)] Primitive Notions: me, things external to me, my actions on the things, the consequences they return to me.

\item[2)] The formal structure of QM is a theory of how I ought to organize my (Bayesian) probabilities for the consequences of all my potential actions.  Implicit in this is a theory of the structure of actions.

\item[3)] The theory works as follows:
\begin{itemize}
\item[A)] When I posit a thing, I posit a Hilbert space $\mathcal{H}_d$ as the arena for all my considerations.

\item[B)] Actions upon the thing are captured by sets of operators $\{E_i\}$ on $\mathcal{H}_d$ such that $E_i \geq 0$, $\sum_i E_i = I$.
Consequences of the actions are modeled by the individual $E_i$. I.e.,
\begin{center}
ACTION = $\{E_i\}$, CONSEQUENCE = $E_k$.
\end{center}
(Parallel development in Cox, where questions are treated as sets and answers are treated as set elements.)

\item[C)] QM organizes my beliefs by saying I should strive to find a single $\rho \in \mathcal{L}(\mathcal{H}_d)$ such that my degrees of belief will always satisfy \begin{equation}
\hbox{Prob}(E_k|\{E_i\}) = \hbox{tr}\,\rho E_k.
\end{equation}

\item[D)] Unitary time evolution and quantum operations address belief-change issues, and will not be so much the topic of conversation now.

\item[E)] When I posit \emph{two things} external to me, the arena for all my considerations becomes $\mathcal{H}_d \otimes \mathcal{H}_d$.
Actions and consequences now become sets of operators $E_i \in \mathcal{L}(\mathcal{H}_d\otimes\mathcal{H}_d)$.

\item[F)] \emph{Yet,} I can isolate a notion of an action on a single one of the things: These are sets of the form $\{E_i\otimes I\}$.

\item[G)] Resolving the consequence of an action on a single one of the things may cause me to update my expectations, my beliefs, my gambling strategies \emph{for} the consequences of any of \emph{my actions} on the \emph{other} thing.

\qquad But for those latter consequences to come about, I must \emph{elicit} them through an actual action on the second system.

\item[H)] EXAMPLE: The things external to me are two electron spins, my taking $\mathcal{H}_2 \otimes \mathcal{H}_2$ as the arena for my considerations.  Particularly, suppose I organize my degrees of belief according to \begin{equation} \ket{\psi} = \frac{1}{\sqrt{2}} \left(\ket{+1}\ket{-1} -
\ket{-1}\ket{+1}\right)
\end{equation}
(making reference to Norsen's Eq.\ (4)).\footnote{\editornote All of Norsen's equations cited in this note refer to T. Norsen, ``Bell Locality and the Nonlocal Character of Nature,'' \quantph{0601205}.}

\qquad Suppose I perform an action solely on the left particle:
$\{\ket{\vec{n}} \bra{\vec{n}} \otimes I, \ket{-\vec{n}} \bra{-\vec{n}} \otimes I\}$.  My adopted $\ket{\psi}$ above will cause me to give even odds on the two potential consequences of this action.

\qquad But supposing I perform it, I will then expect this: If I further perform an action $\{ I \otimes \ket{\vec{n}} \bra{\vec{n}}, I \otimes \ket{-\vec{n}} \bra{-\vec{n}}\}$ isolated to the right-hand particle, I will expect with great certainty (perfect even) to get the same consequence as I got previously.  I.e., I will accept infinite odds that that will be the case.

\qquad Does that mean I have effectively accepted an \emph{element of reality} corresponding to the predicted consequence on the right-hand side?
I.e., that I believe I will find that consequence because it is already there, or at least determined by some objective condition, independent of the performance of the right-hand action?

\qquad \emph{No.}  It \emph{only} means that I will accept infinite odds on a latter bet.  I will bet my house; I will bet my life.

\qquad Within the context of today's discussion about Norsen's paper, it means that I reject the transition he makes between Eqs. (11) and (12), and the discussion in the two paragraphs following.

\qquad This is the gist of my paper ``Subjective Probability and Quantum Certainty'' with Caves \& Schack.  After all, from our point of view the very assignment of $\ket{\psi}$ arises from (or is the expression of) \emph{a prior.}  And priors, though they may lead to statements of certainty for this and that, \emph{are not infallible,} by definition.

\qquad Let me make one final remark in this regard (i.e., with regard to the contrast between CFS and Norsen on this example).  We reject on general grounds (not specific to QM) the transition between Eqs. (11) and (12), \emph{but} we explore the possibility of whether one can at least contemplate making it in the quantum case.  Perhaps such a transition causes no inconsistency and so, therefore, why not make it on a case by case basis---``physical theories'' for instance being very good candidates.

\qquad Here is where there is a mild convergence in the arguments of the two papers.  Norsen takes Bell Locality and the transition $(11) \to (12)$ as given, and derives a contradiction (via referenced use of Bell inequalities).  He then relinquishes Bell Locality, thinking the transition $(11) \to (12)$ sacrosanct.  CFS on the other hand take a \emph{FORM} of locality and the transition $(11) \to (12)$ and derive a contradiction (via a Stairs style construction, using KS).  So something must give, but in contrast to Norsen, we relinquish the sacrosanctity of the transition $(11) \to (12)$.  We were already suspicious of it to begin with.
\end{itemize}
\end{itemize}

\emph{Back to the general development}:

\begin{itemize}
\item[I)] \emph{LOCALITY}: What does ``locality'' mean within this ``quantum Bayesian'' view (i.e., Chris's lonely view)?  (Hereafter QBism.)

\qquad In positing two distinct things in assigning $\mathcal{H}_d \otimes \mathcal{H}_d$, why throw away the notion in the end?  (This really is what I view as the upshot of the Norsen argument: We thought we had two things, but our hands, our operations, reach across both---might as well call them one thing after all.)

\qquad Here's the notion of locality.  When I perform an action $\{E_i \otimes I\}$ on the left-hand system, it cannot lead to any further consequences upon me, now or in the future, from the right-hand system.  If I want to get consequences from the right-hand side, I have to get them by performing a new action $\{I \otimes F_j\}$.  When the right-hand's future light cone reaches me there will be no residual surprise, no residual consequence of that previous action on the left.  All I can say is if I were to perform a new action on a piece of that light cone, my beliefs concerning its consequences will be conditioned by my initial beliefs $\rho \in \mathcal{L}(\mathcal{H}_d \otimes \mathcal{H}_d)$, the consequence of my first action $E_k \otimes I$, and the model I use for propagating my beliefs on the right-hand side (i.e., which trace-preserving quantum operation I assign to that side of the world and its expanding light cone).

\qquad All this is to say, in Einsteinian terms, the ``being thus'' on the right-hand side is not affected by my actions on the left.  The quantum formalism builds that in for me---and the reason it does is because it only makes oblique, indirect statements about the ``being thus'' to begin with.  In large part it is silent of it.  (I will try to explain these last two sentences better in a couple pages.)

\qquad There is no-signalling in this conception precisely because actions on the left elicit no ontic changes on the right full-stop.

\qquad And with that, let me come out of my general development of QBism to discuss Bell Locality per se.

\pagebreak

\item[II)] \emph{Bell Locality As Contrasts QBism.}

\begin{center}
\includegraphics[width=10cm]{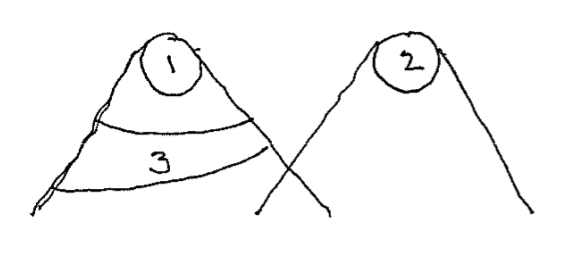}
\end{center}

\qquad ``A theory will be said to be [Bell Local] if the probabilities attached to values of local beables in a space-time region 1 are unaltered by specification of values in a space-like separated region 2, when what happens in the backward light cone of 1 is already sufficiently specified, for example by a full specification of local beables in a spacetime region 3.''

\qquad Is there even a way to make sense of this definition within the QBism view?

\qquad For instance, what probabilities?  And more hurtingly, \emph{WHOSE} probabilities?  Agents---the possessors of Bayesian probabilities---live their actual lives within light cones.

\qquad Imagine an agent in region 1.  He writes down probabilities based on his previous learning, experiences, perhaps genome, generally all kinds of influences in his past light cone.  Now, Bell exhorts us to give the agent ``a specification of the values of local beables in region 2.''
We'll perform a test and see if the agent updates his probabilities for
1 any.  For a Bayesian (specifically a QBayesian as well), that is the only possible meaning that can be given to testing ``if the probabilities attached to values \ldots\ are unaltered.''  For real agents with no god's eye, how could that information ever be conveyed to him in the first place?  It is not from his backward light cone.

\qquad To get this supposed information, the diagram would have to evolve to this.

\begin{center}
\includegraphics[width=10cm]{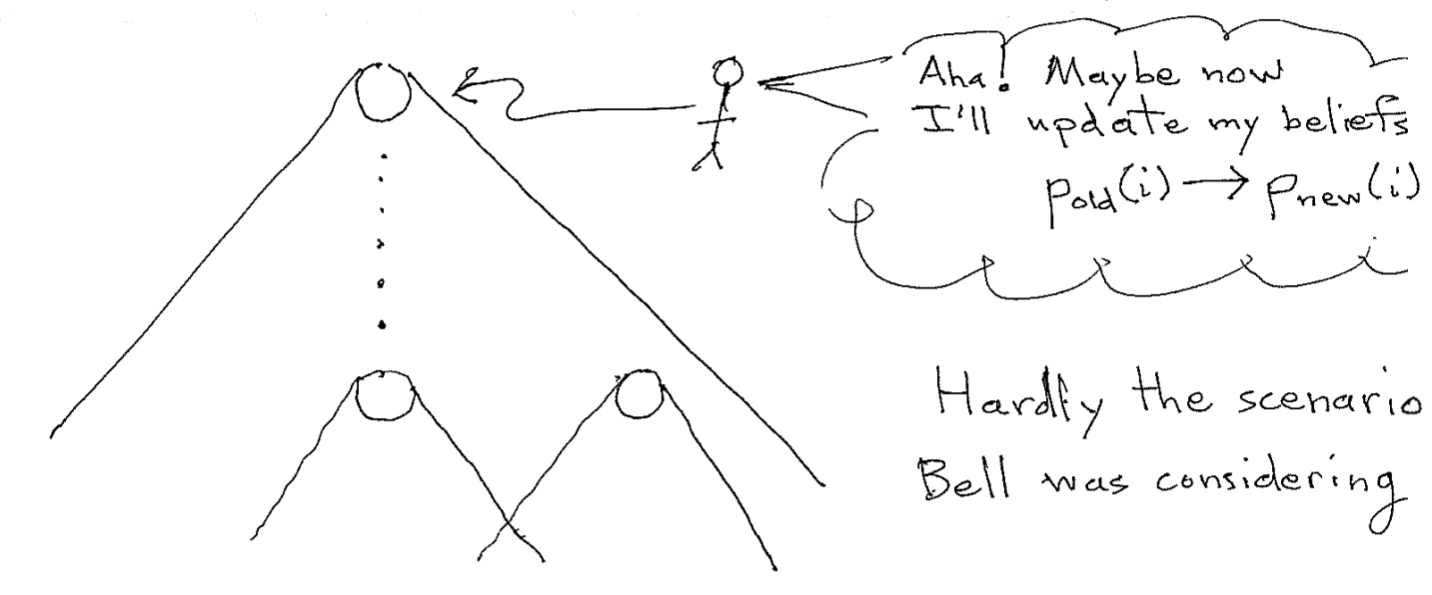}
\end{center}

\qquad Already basic trouble is appearing to come from a \emph{subliminal} message in the language ``probabilities attached to values'' \ldots\ persuading away an image of ``probabilities \emph{about} values but \emph{attached} to the agent.''

\qquad Here's another way to say it all:  What are these ``beables'' Bell is speaking of conditioning upon?  The only thing a QBayesian knows how to condition a probability on is his actual experience (consequences of his actions in this venue, QM).

\item[III)] \emph{Final Remarks on Norsen}

\begin{itemize}
\item[1)] Rob wrote, ``I strongly disagree with your claim that Norsen or Bell's arguments assume an ontic status for $\ket{\psi}$.''

\qquad Everything I wrote above contra Norsen hinged on his use of the transition $(11) \to (12)$.  I, as you ought to know by now, view this as a subtle and usually unquestioned move to re-onticize $\ket{\psi}$.
Once one accepts that the probability-1 statements generated by a quantum state assignment ``are ontic'', one has it for all other probabilities generated from $\ket{\psi}$ as well, via Born's Rule.

\qquad But I'll put that aside, and leave it to the horse's mouth instead.  In the paragraph before Eq.\ (5) he writes, ``Also note that we of course leave open the possibility that $\Lambda_\psi$ is a set with only one element, e.g., the wave function itself.''  In that possibility, $\ket{\psi}$'s onticity is surely reinstated.

Finally,

\item[2)] Travis writes, ``If motivated by the orthodox quantum philosophy, one excludes from the beginning any talk about the ``features of a putative underlying reality,'' then there is literally nothing else \ldots\ to discuss.  The vague anti-realism of the orthodox quantum philosophy thus seems to rule out the very kind of talk that is absolutely required to show that nature violates some locality condition---namely talk of nature!''

\qquad Just for the record, I do not regard the view I put forth at the beginning of these notes as ``ruling out talk of nature.''  Because that is exactly what this whole view is about:  how we should organize our probabilities in response to how nature is.  If nature were otherwise we would organize our probabilities differently.  For instance we might not restrict our state to the convex hull of $\rho^2 = \rho$, but to a cube instead.  Who ordered that?  The reason why---when we find it---will be a statement about nature.  And more than that \ldots\ but I exhaust myself!
\end{itemize}
\end{itemize}

\begin{center}
Look at that: 13 pages!

Surely an unlucky number.
\end{center}

\section{09-04-09 \ \ {\it Request to Review Manuscript} \ \ (to The Physics Teacher)}

I'm sorry, I read the first two sentences of this paper---``In 1964 John Bell proved a theorem allowing the experimental test of whether what Einstein derided as ``spooky actions at a distance'' actually exist.  We will see that they do.''---and I feel a revulsion.  All I can do is reject it.  This is such an old issue, such an old mistake, much like the continued proposal of a perpetuum mobile.

What the violation of a Bell inequality demonstrates is that ``local realism fails.''  Unpacking the term local realism, one means more precisely the conjunction of two statements:  1) that actions or experiments in one region of spacetime cannot instantaneously affect matters of fact at far away regions of spacetime, and 2) that measured values pre-exist the act of measurement, which merely ``reads off'' the values, rather than enacting or creating them by the process itself.  The failure of local realism means the failure of one or the other or both of these statements.  It does not mean what the author says above.  See, for instance:
\begin{itemize}
\item
S. L. Braunstein and C. M. Caves, ``Wringing Out Better Bell Inequalities,'' Ann.\ Phys.\ {\bf 202}, 22 (1990).

\item
N. D. Mermin, ``Hidden Variables and Two Theorems of John Bell,'' Rev.\ Mod.\ Phys.\ {\bf 65}, 803 (1993).
\end{itemize}
for very thorough discussions of this point.  Please save the average physics teacher by sending this article back home.

\section{10-04-09 \ \ {\it Testing Quantum Certainty, 2}\ \ \ (to R. {\Schack})} \label{Schack159}

\brs
The exercise is indeed easy. Unperformed experiments have no results. Even probability 1 predictions are subjective. There is nothing in the world which guarantees the outcome of a probability one measurement. You name it.

This ``Nature IS nonlocal'' theme is everywhere at the moment. Have you seen Albert's paper in {\bf Scientific American} (February I think)?
\ers
I haven't had the courage to look at it, though everyone wants me to.

Long discussion with Rob yesterday.  It seems he's not unsympathetic to ``even probability 1 predictions are subjective'' most generally, but that the general point may become inoperative in physical theories like quantum mechanics.  That is to say, it seems he wants the role of physical theory to be a bridge between belief and fact.  Something of that flavor.

\section{10-04-09 \ \ {\it SICing a Symbol} \ \ (to H. C. von Baeyer and D. M. {\Appleby})} \label{Appleby56} \label{Baeyer67}

Good Good Friday to both of you.  I slowly continue my refreshed study of Pauli, and last night, just at midnight, got to Gieser's chapter on {\it The Reality of the Symbol}.  Previous to this, I had never paid much attention to this concept of a ``symbol.''  However, here are some of Gieser's words that quite struck me, and I haven't been able to get them off my mind all night:
\bq
Jung sees a living symbol as the best possible expression of something divined but not yet fully known -- something which cannot be represented in a more characteristic way than in the form taken by the symbol.  If one says that the cross is a symbol of divine love, then according to Jung one gives a semiotic {\it explanation\/} of the cross, which is something quite different from seeing it as a symbol.  If on the other hand one believes that the cross is beyond all conceivable explanation, but that it is still the most apt expression of an as yet unknown and incomprehensible fact, then one has a symbolic attitude to the cross.  The symbol always consists of a known or rational part and an unknown or irrational part, which is not accessible to reason.  The known part of the symbol is represented by its current form while the unknown part opens up to the non-visual aspect of the archetype.  The state of tension between known and unknown gives the symbol a numinous character, which lends it a power and attraction.  Our fascination with and manipulation of the symbol gradually leads to a discovery of the true characteristics of the object and the symbol increasingly produces real knowledge.  In this way the unknown is made conscious and thus the symbol loses its power and attraction and `dies'.
\eq

It's obvious what I thought upon reading this.  By these lights, the SICs are surely a symbol for Marcus and me.  There's almost a discussion in Gieser's description that Marcus and I had several times over in the last few years.  For somehow we felt, long before we could articulate why---long before the urungleichung or anything like that---that the SICs were somehow the ``most apt expression'' of quantum-state space.  There has been a ``power and attraction'' in these things that neither of us have understood and neither of us have been able to yield to (in our own ways).

I'm quite thankful I read this passage, as I suppose I should be, given that today commemorates the birth of the symbol Gieser uses as her example.

\subsection{Marcus's Reply}

\bq
Yes.  Absolutely.

But maybe I would like to add a couple of qualifications.  If a symbol is defined to be something that
\bq\noindent
consists of a known or rational part and un unknown or irrational
part, which is not accessible to reason
\eq
then I think that the concept of divine love {\it also\/} has a good deal of the symbol about it.  I am sure I don't understand what is meant by it.  So in that sense it is certainly unknown (at least to me).  But at the same time it doesn't mean nothing (at least to me).  Religious language like this leaves me perplexed.  But it doesn't leave me feeling perplexed in the way I would feel perplexed if someone were to say (or rather write) ``oub\#\#4 b12xp''.

I had a conversation with Hulya the other night which may be relevant here.  She has been reading {\sl Zen and the Art of Motorcycle Maintenance}, and she criticized it on the grounds that it isn't deep (though I should perhaps add that she was at pains to stress that she hasn't got very far into it yet and may change her mind).  This led to a discussion as to what counts as deep.  We decided that it was best to begin with an example that is the complete opposite of deep.  Suppose one were to take a USB stick and put onto it a list of the names and addresses of everyone living in Waterloo.  That would strike Hulya and I as totally shallow.  The reason is (we decided) that it contains exactly what one consciously chose to put there, no more, and no less.  If someone lives in Waterloo then their name and address will be on it, but their telephone number won't be.  There is no ambiguity to it at all.  But if something is deep there is always a lot of ambiguity.

Of course mere ambiguity is not sufficient.  Otherwise one could make any book deep simply by leaving it out in the rain, so that the ink runs, and then getting a dog to chew the pages.  It is important also that the ambiguity be suggestive.   Of course that is not enough
either:  for it mustn't be suggestive of just anything but specifically of something that strikes us as --- what exactly?  Here I find myself at a loss.  The best I can do is to say that it must be suggestive of something that strikes us as deep.  Which, it may appear, does not get us very far.

Still I think it does get us somewhere.  If a book is deep that means (Hulya and I decided) that in some sense it reaches out beyond itself.  It actually contains more than the author consciously chose to put in it.  A deep book is one that the author themselves might read and re-read and each time discover something new.  Just as if they were re-reading a book written by someone else.  In short a book is deep if the supposed author isn't exactly the author at all.  At least not completely the author.  As if the author had a hidden co-author.

Hardy, in {\sl A Mathematician's Apology}, says something about this.  He says that a chess puzzle is shallow because, even if it is a very difficult problem, it is open and shut.  Once you have solved it then you have solved it.  There is nothing more to do.  By contrast Euclid's proof that there are infinitely many primes is deep (according to Hardy).  Mathematicians have been coming back to it for thousands of years, and they keep on finding new connections.

I think what I so hate about classical physics (even General Relativity---pace Hans) is that it is built on the idea of what Gieser calls a semiotic explanation.  The universe, classically conceived, consists of a great mass of dead facts.  You can imagine putting these facts on a cosmic USB stick and what would result would be exactly like a list of the names and addresses of everyone in Waterloo.  Ugh!

Quantum mechanics is wonderful because it dispels this nightmare.
Classical physicists see the fact that quantum mechanics doesn't permit complete knowledge as a disaster.  They couldn't be more wrong.  Just because  it forbids a complete description (the kind of description which would allow us to reliably predict the result of any
experiment) it permits the possibility of depth.  It is not a disaster but a liberation.

Which brings me back to Jung and symbols.  I profoundly disagree (indeed profoundly dislike) his idea that
\bq\noindent
Our fascination with and
manipulation of the symbol gradually leads to a discovery of the true
characteristics of the object and the symbol increasingly produces
real knowledge.  In this way the unknown is made conscious and thus
the symbol loses its power and attraction and `dies'.
\eq

One sees from this that Jung was classical at heart.  As if one were to say that our inability to know position and momentum simultaneously is only temporary.  Eventually we will discover Einstein's true theory, and then we can go back to thinking of the universe as an enormous telephone directory.

And, to go back to the point with which I began, I think his concept of symbol is too narrow.  As he sees it the cross is symbolic, but divine love not.  It seems to me that just about every word in the gospel stories is charged with meaning.  Deep meaning.  I don't understand them.  I am certainly not saying that I am a Christian in any conventional sense (though I probably am in some extremely unconventional sense).  But they are shot through with the quality which Jung attributes to a symbol:
\bq\noindent
a numinous character, which lends it a power and attraction.
\eq
I would say everything in them is symbolic.  If one is going to use that word.

But I agree completely about SICs.
\eq

\subsection{Hans's Reply}

\bq
Dear Marcus, thank you for that excellent and useful gloss on the word symbol.  I agree with your critique of the last part of the passage, but I don't know whose words they are.  Probably Gieser's, possibly Jung's, but certainly not Pauli's.  I can't imagine Pauli talking about ``true characteristics of'' or ``real knowledge about'' an object.  Or to give up a symbol as dead, after making it the centerpiece of his enterprise.  In fact the sentence reminds me of the point that Pauli and you have made often: that most people, even good physicists, remain convinced, or at least hopeful, classical thinkers, and that it will therefore take a long time for the scientific community to switch to a radically new worldview.
\eq

\subsection{Chris's Reply to the Replies}

\bq
Oh, you guys make me feel guilty!  I too had disliked the latter part of Gieser's description, but I decided to keep the whole paragraph for completeness sake.  I should have made a comment!  \ldots\ Please don't think less of me!
\eq

\section{11-04-09 \ \ {\it QBism, Certainty, and Norsen, 2} \ \ (to R. {\Schack})} \label{Schack160}

\brs
I have now studied your handwritten notes. I like the parts where you explain why Norsen's argument falls to pieces from a QBism perspective.
I am not too sure about your explanation of what locality means from a QBism perspective, though. We
had these discussions last year. It looks to me as if locality is built into the QBism formalism from the
start, so that it becomes impossible within that formalism to ask, e.g., does nature allow superluminal
signalling?
\ers

Indeed, we don't have a very good story of signaling yet, superluminal or otherwise.  It dawned on me for the first time this morning that until we can say something intelligible about that we're going to always be exposed to the charge of solipsism.  We must understand how to describe ``communication'' within the QBism framework!  I'm a slow learner.

\section{13-04-09 \ \ {\it Easter Pictures and the Beginning of Depth Psychology} \ \ (to D. B. L. Baker)} \label{Baker18}

Fantastic pictures, I enjoyed seeing them very much.  Made me miss you again old man.  Do you still go to church regularly?

Attached is a representative of my own Easter pictures.  Sorry, not much better fare at the moment.  The girls are on the balcony outside of Katie's room.

I'd like to write you a little essay on ``depth'' and ``deep thoughts'' harkening back to our days in high school.  I've been thinking about what we found to be ``deep'' then, when we would discuss this or that, songs of Paul Simon, Rush, The Clash.  This has come about because in reading a book about Carl Jung, I came across a definition of ``symbol'' that struck my fancy, making sense of what I've been doing the last four years.  It's not so very different from our thinking about a song over and over.  See note below.

The particular example being discussed---a thing I call a SIC (pronounced ``seek'')---is a certain mathematical structure we've been trying to prove the existence of, on and off since 1999.  In the last four years or so, it's turned into a real obsession.  You can read about it, if you're interested, in somewhat general terms (which I wrote up for the Navy) in the attached PDF.

I'll also place further down a part of Appleby's reply, which reminded me of our old high school discussions indeed.

\section{14-04-09 \ \ {\it Your Thesis, 2}\ \ \ (to J. R. Gustafson)} \label{Gustafson2}

I think you should post it, for instance on the quant-ph archive.

In case you're interested, I attach my ``Resource Material for a Pauli'an / Wheeler'ish Conception of Nature'' file.  It's been a long time since I've updated it.  For instance, it doesn't have your thesis listed yet, or Gieser's book, etc.  But there's plenty there for you to read and think about, if you want to start thinking in these directions again.  (And if you find any typos, please let me know!)  One of these days, I'm going to finish putting it all together and post it as well.

\section{15-04-09 \ \ {\it Math Question} \ \ (to P. Hayden)} \label{Hayden3}

\bpah
Thanks for coming to dinner on Wednesday (and for the grace under pressure at the lecture). We were unfortunately seated at more than a correlation length apart so didn't get a chance to talk. Did you still want to pose that math question?
\epah

I'm just now replying to this note!  Let me start out with an apology, not for the sloth here, but for forgetting to say goodbye to you after dinner the other evening!  Wine does mysterious things.  And let me apologize too for not doing a better job of storytelling.  With just a tad of thought, I could have pulled off a relevant (and funny) punch line for the story I did tell.  I've been kind of beating myself.  I wish I could blame the wine for this one, but truth be told it was only the aging of my synapses.

With regard to the math problem, it was really just a way to lure you in to thinking about technical aspects of my QBism program.  Attached is a draft (about 80\% complete) of a paper I'm writing at the moment.  [See ``Quantum-Bayesian Coherence,'' \arxiv{0906.2187v1}.]  Sections 1 to 5 are now relatively stable, so they should at least make some sense at this point.  The basic mathematical question is how much of the structure of density operator space is implied simply from having a maximal set with respect to the inequality in (97)
\be
\frac{1}{d(d+1)}\,\le\, \sum_i p(i) s(i) \,\le\, \frac{2}{d(d+1)}\;?
\label{SayHiToYourKnee}
\ee
It would have been better to tell about these things at the board, so you wouldn't have to wade through all this nonsense to get to the basic question \ldots\ but time flies.  If you do wade, though, and have any insight, I'm all ears.  Else we can talk at the ONR meeting in May.

\section{15-04-09 \ \ {\it The Egocentric Paradigm}\ \ \ (to R. W. {\Spekkens})} \label{Spekkens63.2}

The right-hand column of Lindley's discussion (attached) reminds me of our conversation after the group meeting Thursday.

\bq\noindent
The Bayesian, subjectivist, or coherent, paradigm is egocentric. It is a tale of one person contemplating the world
and not wishing to be stupid (technically, incoherent). He realizes that to do this his statements of uncertainty must
be probabilistic. This is important on its own for it rules out a large class of behavior patterns, like sampling theory
statistics, but is it enough? Once I coined the aphorism ``Coherence is all.'' Was I right? It is when we consider
two coherent Bayesians that new features arise. Does their egocentric behavior allow them to talk to one another?
In one respect it does. Suppose that you and I are both coherent and you tell me that your probability for
$A$, were $B$ to be true, is $\alpha$, say. How does this knowledge
affect my probability for $A$ given $B$? It is easy for me to do the coherent calculations in terms of my assessments
that were $A$ and $B$ both true, you would say $\alpha$, and that were $\overline{A}$ and $B$ both true, you would say $\alpha$. For to me $\alpha$
is just data and the likelihood ratio updates my probability in the usual way. A weather forecaster who announces
rain whenever it is subsequently dry, and vice versa, is badly calibrated but very useful. So I can respect your
egocentricity; and you, mine.
\eq

\section{15-04-09 \ \ {\it My Own Credo on Hidden Variables}\ \ \ (to D. M. {\Appleby} \& R. W. {\Spekkens})} \label{Appleby57} \label{Spekkens64}

Just to make sure everyone understands my honest position on hidden variables, I paste in an old note (and an old referee report) below.  [See 17-11-05 note ``\myref{Halvorson5}{Cash Value}'' to H. Halvorson.]  It has similarities to Rob's shape-like remark.  I'm certainly not opposed to postulating entities beyond the clicks.  But the entities had better have ``cash value'' to get my attention.  And I haven't seen a hidden-variable (or ontic-variable theory if you will) yet that has gotten my attention in that way.

\section{15-04-09 \ \ {\it The Well-Calibrated Bayesian} \ \ (to R. {\Schack})} \label{Schack161}

\brs
Internet access speed truly awful. Not sure that I'll be able to access Dawid's paper. I expect that I'll be a lot less negative now. I have a worry that this kind of reasoning might lead to a concept of ``correct'' probability assignments. Well calibrated probabilities are ``better'' than others. But that might not be so bad in the light of our discussions at IHOP. I would pay more for forecasts from a well-calibrated Bayesian.
\ers

Please remind me of those discussions as well.

In the present case (i.e., with regard to the particular paper I sent you), I was more intrigued by the coherence issues to do with the concept.  What he tries to establish in this paper is that any forecaster will expect himself with probability 1 to be well-calibrated.  Then he asks, how can that be, since every forecaster surely knows that he is fallible.  Second to that, though, I wonder if his calibration might just be arbitrary (and also, as some of the commentators expressed, operationally without content).

About Lindley, I very much like the sound of him, except for this Cromwell's Rule business.  It is surely inoperable---same point de Finetti made.

\section{15-04-09 \ \ {\it Marcus on the Symbol 3.5 Years Ago} \ \ (to H. C. von Baeyer)} \label{Baeyer68}

I've been putting together a coffee table book called {\sl The Unpublished {\Appleby}}, and consequently rereading some old notes and paper drafts he's written over the years.  Attached is one example.  Have a look at the paragraph just preceding Section 5:
\bq
The recognition, that quantum mechanics doesn't supply us with an exact
mental replica of the world, may initially be experienced as a disappointment.
In the same way, the little boy I mentioned above may feel disappointed when,
in spite of repeated attempts, he finds that he is simply unable to get his foot
physically into the picture of the shoe. And perhaps his first reaction will be
throw the picture away in disgust, as completely useless. However, he will
eventually come to realize, with time, and a great deal of mental effort, that the
power of a symbol does not depend on its being an exact replica. More than
that: he will come to realize that symbols are powerful precisely because they
do not replicate their objects.

I believe that progress depends on our relearning that nursery lesson.
\eq
So, I guess this symbol talk is nothing new between us!  (You see, I'm a very slow learner.)

\section{15-04-09 \ \ {\it Two Letters to Track Down} \ \ (to H. C. von Baeyer)} \label{Baeyer69}

Here are two letters that it might be interesting to look at in more detail:  Pauli to Jaff\'e, Aug.\ 1954 [1865], PLC IV/2 Pauli to von Franz, 12 Nov.\ 1953 [1672], PLC IV/2.

I was intrigued by there being in the first a variation of the thing Pauli told Fierz about a ``black mass'':
\bq
Suddenly I had a remarkable {\it feeling\/} experience.  The ``observation'' of microphysics appeared to me to be a kind of black mass and I felt remorse.  Remorse with regard to matter, which appeared to me to be a maltreated {\it living thing}.  (Biological implication.) --- The practice of this black ``mass of measuring'' in the external world transforms only {\it its\/} condition, not that of the observer.  [Translation by Gieser, presumably.]
\eq
Second letter apparently says more about the black mass concept.

\subsection{Hans's Reply}

\bq
Chris, [1865] is a very short postcard to Jaff\'e.  It mentions the black mass, as did the long important letter to Fierz [1864] written on the same day.

[1672] to von Franz does not mention the black mass.  (It is nevertheless a useful letter because it spells out like none other the significance of the yin/yang in Pauli's thinking.)

So far I have not seen other mentions of the black mass, but I have not searched.  This is why volume 9 --- online --- would come in very handy!
\eq

\section{15-04-09 \ \ {\it Some Similarity to Your Mumbles on Probability} \ \ (to D. M. {\Appleby})} \label{Appleby58}

See the description of the book below.  It seems to have some similarity to an idea I've seen you groping for.

Happy holiday with Hulya.

\bq\noindent
{\sl Projective Probability}\medskip\\
James Logue\medskip\\
Description:
This book presents a novel theory of probability applicable to general reasoning, science, and the courts. Based on a strongly subjective starting-point, with probabilities viewed simply as the guarded beliefs one can reasonably hold, the theory shows how such beliefs are legitimately ``projected'' outwards as if they existed in the world independent of our judgements.\medskip\\
Product Details:
192 pages; tables; ISBN13: 978-0-19-823959-8 ISBN10: 0-19-823959-9\medskip\\
About the Author: James Logue, Fellow and Tutor in Philosophy, Somerville College, Oxford; and Lecturer in Philosophy, Oxford University
\eq

\section{15-04-09 \ \ {\it QBism, Certainty, and Norsen, 2} \ \ (to R. {\Schack})} \label{Schack162}

By the way, the reason I was re-looking at Dawid originally is because I was thinking of how to respond to Lucien's passages below:
\blh
From a mathematical point of view, any interpretation of probability that allows us to help ourselves to the basic mathematical properties of probabilities will suffice. These properties are that probabilities are positive, sum to one over mutually exclusive outcomes, and the validity of Bayes' rule for converting between joint and conditional probabilities.   One approach that does this might be called the {\bf best prediction available interpretation}.  In this we interpret the probability to be the best prediction available of a theory for the empirical relative frequency (the ratio of times a particular outcome is seen to the total number of times the experiment is repeated when the latter is very large).  The best prediction available interpretation asserts that the purpose of a theory is not to actually make exact predictions (i.e.\ we do not need to say exactly what the empirical relative frequency is for a given denominator) but to rather give the best prediction it can.  This approach remains neutral with respect to what might be regarded as the deeper interpretation of probability but is consistent with such deeper interpretations.  It is consistent with the idea that probability is a degree of belief as promoted by the Bayesians.  It is also consistent with the idea that there is some deeper structure in the world supporting some objective property that might be called a propensity.
\elh
\blh
What about a ``best prediction available'' interpretation for probability. Probability is the best prediction available of a theory for the empirical relative frequency.   We do not expect our theory to give a better prediction than this and we are content that nature only agrees so far with the prediction of theories.  This is a sort of probability squared approach.  Once we let go of the idea that a theory has to predict everything --- even probabilistically --- we are free to take such an approach. This ``best prediction available'' works well for the next definition.  Best prediction available could be regarded as being like the degrees of belief idea but more radical. Alternatively, we might say that there is structure in the world such that the best prediction available is the prob.  This now starts to sound like a propensity approach but where the propensity ``resides'' in the deeper structure rather than just the object itself.  Interesting that the best prediction available approach can be interpreted as both the degrees of belief and the propensity interpretations.  It is a sort of withholding commitment approach.
\elh
I wanted to ask him, how would one know if one made the best prediction available?  It could only be settled with respect to a prior, and now how do you interpret that prior?  Something along those lines.  But it should be fleshed out better.  How would you respond?

When all is said and done though, I still harbor those old fears of being indefensible to someone like Dawid who writes:
\bq
I have been justly chastised by the discussants for spreading alarm about the health of the body Bayesian.  Certainly it has held together more successfully than any other theory of statistical inference, and I am not predicting its imminent demise.  But no human creation is completely perfect, and we should not avert our eyes from its deficiencies.  In its solipsistic satisfaction with the psychological self-consistency of the schizophrenic statistician, it runs the risk of failing to say anything useful about the world outside.
\eq
It probably doesn't help that yet again (after writing up those hand-written notes for Rob), Rob called the position solipsistic.

\section{16-04-09 \ \ {\it QBism, Certainty, and Norsen, 3} \ \ (to N. D. {\Mermin})} \label{Mermin151}

I'm not sure I understood either of your remarks concerning page 2 and page 3.  Could I ask you to try again?  Particularly, I'm lost on this one:
\bdm
Probably what you mean to say is that only a minority thinks that the
(well-known) fact that measured values do not pre-exist the act of measurement, is enough to undermine the view that locality has to abandoned.
\edm
???  Please say it again in a different way.

Not unrelated:  Have you seen \quantph{0601205v2} by Travis Norsen?  Attached are my lecture notes contra that paper.  You will find in them an unequaled clarity with respect to my usual writings.  Anyway, the page 2 and 3 sentences you commented on were molded by the attitude I take in those lecture notes.

How long are you staying in Copenhagen?  And we can't even get you to Waterloo for a minuscule week of conversation!

\section{16-04-09 \ \ {\it QBism, Certainty, and Norsen, 4} \ \ (to N. D. {\Mermin})} \label{Mermin152}

\bdm
We're about to leave for a long weekend in Berlin, so it may be a
while before I can reply.  I'm surprised that you didn't understand what
bothered me.   So I guess it's not that you were being careless (as I had
assumed) but that there is a real difference in point of view here.
\edm

I am never careless in papers.  I am sometimes wrong, but I am never careless.

Indeed!

\section{16-04-09 \ \ {\it QBism, Certainty, and Norsen, 5} \ \ (to N. D. {\Mermin})} \label{Mermin153}

\bdm
I would never have suggested that you were careless in a paper.  But I thought you had said that this was a preliminary draft. Why are you up so early in the morning?
\edm

Chronic insomnia.  It has gotten particularly acute in Canada.

OK, it's a draft.  Mostly I want to understand how you found those few sentences confusing, and I did not understand you at your previous attempt.

Is it that I should change:
\bq\noindent
     But there is a small minority that thinks the abandonment of
     condition 2, that measured values pre-exist the act of
     measurement, is the more warranted conclusion and among these
     are the quantum Bayesians.
\eq
To something more like:
\bq\noindent
     But there is a small minority that thinks the abandonment of
     condition 2, that measured values pre-exist the act of
     measurement, at the same time as the absolute holding fast of
     condition 1, is the more warranted conclusion and among these
     are the quantum Bayesians.
\eq
Is that the sort of thing you thought needed fixing?

\section{16-04-09 \ \ {\it QBism, Certainty, and Norsen, 6} \ \ (to N. D. {\Mermin})} \label{Mermin154}

\bdm
On page 3 you say
\bq\noindent\rm
     But there is a small minority that thinks the abandonment of
condition 2, that measured values pre-exist the act of
measurement, is the more warranted conclusion  [to reach from Bell inequality violations]
\eq
This implies that were it not for violations of Bell inequalities, that small minority (along with the majority) would have believed that
measured values pre-exist the act of measurement.    That's just not
so.  Practically nobody believed that measured values pre-exist the act
of measurement.   They didn't believe it before Bell.   They didn't
believe it before EPR.  EP\&R were condemned as heretics for suggesting that measured values might pre-exist the act of measurement.
It was almost a truism, that values were brought into existence by the act of measurement --- that measurement outcomes were a joint manifestation of the system in combination with the process of measurement.

So I would say that what you should have said is
\bq\noindent\rm
But there is a small minority that thinks that the non-existence of values that pre-exist the act of
measurement is enough to account for Bell inequality violations,  without requiring the abandonment
of condition 1).
\eq

But of course that only makes sense if  you  do not assert earlier on page 2 that measured values pre-existing the act of measurement is one of the two pillars of local realism.  But indeed it is not a pillar.  In Bell's original paper (and in EPR) it is a deduction from the EPR
reality criterion.   In the post-Bell era
there are many assumptions much weaker than the EPR reality criterion.
   For example that joint
distributions exist for the outcomes of all the experiments --- the one you actually did and the ones you might have done but didn't --- without making any commitment to whether such experiments reveal pre-existing values.

Or you can  assume the ``conditionally independent'' form for the joint distributions  implicitly used  by CHSH and explicitly invoked by Jon Jarrett which I'm told goes all the way back to Reichenbach, though I've never found it there.  That's what I was verbally trying to describe in my note to you yesterday.

Hope that helps.  I don't think I'm just being perverse.  I think most people in the Bell-EPR business will react in the same way I have done.

Time to run.  More from Berlin, if we have wireless in the hotel room and I also have insomnia.
\edm

You are right, and now I understand.  I was leaving too much out by not mentioning the EPR considerations.  I will fix this.  Good thing it was only a draft!!  (I now understand I can still wear the label `careless' when it comes to a draft.)  Have a safe trip to Berlin.

\section{16-04-09 \ \ {\it Gieser's Book, Preliminary and Completed Notes} \ \ (to H. C. von Baeyer)} \label{Baeyer70}

Some things that interested me with the book, along with corresponding page numbers.  Don't know whether it'd be of any use to you, but one never knows.  Sadly, I've yet to get back to finishing the book.  I hope to get back to it soon after returning from Paris.  Before Paris, too busy; during Paris, I only want to think about Pure Experience. [Fragment from 09-04-09 note.] \ldots

OK, I got more active than I thought I would before Paris.  Below is my complete set of points of interest with regard to Gieser's book.  Strangely, I enjoyed almost the whole book, even though looking back on it now, I don't think I learned anything at all particularly.  It's probably all just about acquiring a taste, somewhat like I once did for beer many years ago.

I believe there is at least one typo below from my previous partial compilation, but I haven't been able to re-find it.

I'll answer your questions on von Meyenn later today (or tomorrow morning).

\bv
\underline{Suzanne Gieser, \sl The Innermost Kernel: Depth Psychology and Quantum Physics}.\\ \underline{\sl Wolfgang Pauli's Dialogue with C.~G. Jung}\medskip\\
``he aroused the malice of the object'' -- 20, 21\\
QM being incomplete -- 22\\
Schopenhauer \& synchronicity -- 29\\
Religion -- 31, 32\\
Kierkegaard \& free choice -- 32\\
Ortega y Gasset -- 32\\
Schopenhauer -- 37\\
Pauli on God -- 40, 41\\
Mach and James's pure experience -- 45, 47\\
Predilection for indeterminism -- 57, 58\\
For Hardy -- 59\\
Dugald Murdoch on James -- 61\\
Kierkegaard and indeterminism -- 62\\
{\Hoffding} on continuity -- 62, 63\\
{\Hoffding} sounding like Mead -- 63\\
{\Hoffding} and pluralism -- 64\\
Three points of Bohr -- 71\\
Pauli as conscience of physics -- 72\\
Hendry's article -- 74\\
Pauli contradicting Bohr's classical language requirement -- 75\\
Pauli and James -- 78\\
Heisenberg quote -- 80\\
Bohr's realism -- 82, 83, 84\\
Pauli two-face -- 95, 281\\ Creation -- 96\\
Jordan and pure experience -- 97, 99\\
Jordan and radical empiricism -- 100, 101, 159\\
James on {\Hoffding} -- 104\\
James on theories -- 105\\
Bohr on QM and psychological epistemological situation -- 105\\
Complentarity from James?\ -- 107, 108, 109\\
Jung read {\sl Pragmatism} -- 114\\
Jung on reality -- 115, 117, 118\\
James's full fact -- 117\\
Gieser off track (seemingly); James rejected the very existence of the `thing in itself' -- 119\\
Jung against `vicious abstractionism' -- 122\\
``preconceived concepts'' (recall James's remark on Freud) -- 122\\
Value of ambiguous language -- 123\\
A fact is not a simple thing -- 123\\
Reality of phobia -- 123\\
Psychologist as nondetached observer -- 124, 125\\
Gieser trusting Folse's assessment of Bohr -- 120, 121\\
{\Moller} on {\Hoffding}? -- 128\\
Bohr's inconsistency -- 132, 133\\
Stark's article -- 136\\
Pauli on `irrational' -- 138, 139\\
Jung's {\sl Psychology \& Religion\/} speaks of Pauli -- 143\\
Von Franz's memories not to be trusted (at all!)\ -- 151\\
Wasp phobia (Pauli and Emma) -- 151\\
Hermeneutic or hermetic?\ -- 153\\
Second reference to Jung's power with women -- 161\\
Gieser's healthy attitude -- 162\\
Pauli's healthy attitude -- 163\\
Not true; the unibomber game many clues -- 163, 164\\
Heinrich Fierz!\ -- 164\\
Pauli quote on inner \& outer -- 170, 177\\
Pauli as Platonist -- 173, 209, 210\\
Pauli as non-Platonist -- 173\\
Agency in nature -- 174\\
Sexual interpretation of QM -- 180, 181\\
And a SIC for a qubit has four outcomes -- 184, 185\\
Empty space is like a bellows -- 190\\
Fludd on unity of observer and observed -- 193\\
Pauli's unintended consequences (like atomic bomb, etc) -- 195\\
Heisenberg's essay on Pauli -- 195\\
`sacrificing' information -- 196\\
`basic idea' in alchemy (for Gieser) -- 199\\
Lapis having transformative qualities -- 200, 202, 203, 204\\
Mercurius -- 206\\
Neutral language -- 207\\
Einstein to Besso -- 210\\
Pauli's incompleteness of QM -- 210, 211\\
Pauli's chart comparing QM and depth psychology -- 212\\
Archetype as apprehension -- 215\\
The unconscious and acts of creation -- 215\\
Sexuality as substance -- 219\\
Potential being -- 219\\
Neutral monism -- 219, 220, 227\\
God must be born again and again -- 223\\
Spirit -- 227, 228\\
Question of objectivity -- 230\\
God wants to be born -- 231\\
``in need of release'' as a character(istic?)\ missing in Schopenhauer's `will' -- 235, 236\\
Man can alter his `future' -- 235\\
Agnosia -- 236\\
Good quote on alchemy -- 237\\
Numinosum -- 239\\
Material bodies having `latent psyche' -- 243, 244\\
Incarnation as creation -- 248\\
Bohr as an attitude of resignation -- 252\\
Expansion of Heisenberg's description of Pauli on alchemy -- 256, 257\\
Linking matter with evil \& nonbeing -- 259\\
Heraclitus and becoming -- 260\\
Letter to Panofsky -- 260\\
Coniunctio giving something third and unique -- 261\\
QM and becoming -- 261, 262\\
Neutral monism in Spinoza?\ -- 264, 265\\
Observation as activity -- 265\\
Everett as Parmenidean?\ -- 266\\
Jung caught in a lie -- 267\\
Symbolic attitude to SICs -- 268, 269\\
Quantum outcome as miracle or act of creation -- 273\\
Correspondentia -- 274\\
Chinese time -- 279\\
Time as creation of conscious mind -- 279\\
Psychoid -- 281\\
Radioactivity as synchronicity -- 284 fn.\ 870\\
Pernicious influence of statistical method -- 285, 286\\
Pauli on unique events -- 285, 286\\
Schopenhauer on synchronicity -- 288\\
Hans Bender at Freiburg -- 290\\
Archetypes \& prior probability -- 292, 293\\
Classical physics as incomplete -- 294, 295\\
Creatio continua -- 296, 344, 345\\
`I spoke and thus saved my soul' -- 298\\
Pauli on astrology vs I Ching -- 298\\
Pauli evolving away from static worldview -- 299\\
M\"obius strip -- 303\\
Tyche cult -- 307\\
Mathematics as aspect of nature -- 308, 309\\
Weyl on numbers -- 309\\
Automorphism \& Fuchsian function -- 310\\
Numbers as psychoid -- 313\\
Number 3 associated with birth -- 313\\
Yin-Yang and neutral monism -- 323\\
quantum measurement and black mass -- 323, 324, 342, 343\\
Quarternary system and tetractys -- 330\\
Primas article -- 337\\
Newton and alchemy -- 338\\
Inner experiences as autonomous existences -- 338, 339\\
Archetypes structure matter?\ -- 347
\ev

\section{17-04-09 \ \ {\it Pearls of Bayesianism}\ \ \ (to R. W. {\Spekkens})} \label{Spekkens65}

This paper looks quite interesting to me:
\bq\noindent
\myurl{http://ftp.cs.ucla.edu/pub/stat_ser/r284-reprint.pdf}
\eq
I'd like to (read it and) discuss it, if you're interested, after I get back from Paris on the 28th.  If he is right that ``probability theory deals with beliefs about an uncertain, yet static world, while causality deals with [BELIEFS ABOUT] changes that occur in the world itself'' (my crucial addition), then I'm half-Bayesian as well.

\section{17-04-09 \ \ {\it Two Quotes Before I Forget}\ \ \ (to R. W. {\Spekkens})} \label{Spekkens66}

I copied down two quotes yesterday that bear a little on our discussion last week.  Let me record them:

1) From ``Subjective Probability and Its Measurement,'' by J. M. Hampton, P. G. Moore and H. Thomas (J. Roy.\ Stat.\ Soc.\ A {\bf 136}, 21--42 (1973)):
\bq\noindent
Bruno de Finetti believes there is no need to assume that the probability of some event has a uniquely determinable value.  His philosophical view of probability is that it expresses the feeling of an individual and cannot have meaning except in relation to him.
\eq

2) From Dennis V. Lindley's comment on A. P. Dawid's article (J. Am.\ Stat.\ Assoc.\ {\bf 77}, 611--612 (1982)):
\bq
The Bayesian, subjectivist, or coherent, paradigm is egocentric.  It is a tale of one person contemplating the world and not wishing to be stupid (technically, incoherent).  He realizes that to do this his statements of uncertainty must be probabilistic.  This is important on its own for it rules out a large class of behavior patterns, like sampling-theory statistics, but is it enough?  Once I coined the aphorism ``Coherence is all.''  Was I right?  It is when we consider two coherent Bayesians that new features arise.  Does their egocentric behavior allow them to talk to one another? \ldots
\eq

Particularly what caught my attention was the descriptor ``egocentric,'' which I decided I like very much.  There is a world of difference between ``egocentric'' (which is an accurate account of my proposed take on QM) and ``solipsistic'' (which is not).

\section{26-04-09 \ \ {\it Server Config} \ \ (to D. M. {\Appleby})} \label{Appleby59}

\bma
I am glad you and Emma are enjoying your time in Paris.  I know what you mean.  Strolling round big cities, with a long history behind them, does indeed give one insights one can get no other way.  Just the intangible feel of the place, that one somehow picks up without even setting foot in a museum or gallery or cathedral or whatever.  And Paris must be one of the most important foci in the whole of human history.  So many of the things around us (the very existence of the United States) depend on events which took place in those few square miles.
\ema

Hope you get well soon!  I really liked that phrase ``foci of human history''!  Maybe that indeed is how the spirit sinks in.  (I'm turning on to the word ``spirit'', by the way, since our soon-after-new-year's discussions; for instance, I now think Hans should use the labels ``spirit'' and ``matter'' on his M\"obius strip logo.)

\section{28-04-09 \ \ {\it Spirit and Matter} \ \ (to D. M. {\Appleby} and H. C. von Baeyer)} \label{Appleby60} \label{Baeyer71}

Here was the quote I had wanted to send you last night, but I got too lazy to look it up (I was only up briefly from the jetlag).

It comes from the very beginning of William James's article ``Does Consciousness Exist'' (which I started to read for the third time while in Paris).  Previously, I had not noticed any of this, but I suppose our recent discussions set me up for it this time.

\bq
`Thoughts' and `things' are names for two sorts of object, which common sense will always find contrasted and will always practically oppose to each other.  Philosophy, reflecting on the contrast, has varied in the past in her explanations of it, and may be expected to vary in the future.  At first, `spirit and matter,' `soul and body,' stood for a pair of equipollent substances quite on a par in weight and interest.  But one day Kant undermined the soul and brought in the transcendental ego, and ever since then the bipolar relation has been very much off its balance.  The transcendental ego seems nowadays in rationalist quarters to stand for everything, in empiricist quarters for almost nothing.  In the hands of such writers as Schuppe, Rehmke, Natorp, Munsterberg -- at any rate in his earlier writings, Schubert-Soldern and others, the spiritual principle attenuates itself to a thoroughly ghostly condition, being only a name for the fact that the `content' of experience is known.  It loses personal form and activity -- these passing over to the content -- and becomes a bare Bewusstheit or Bewusstsein \"uberhaupt of which in its own right absolutely nothing can be said.
\eq

\section{28-04-09 \ \ {\it Spirit and Matter, 2} \ \ (to D. M. {\Appleby} and H. C. von Baeyer)} \label{Appleby61} \label{Baeyer72}

\bma
How did James answer the question?

Isn't it a very peculiar question?  Suppose someone were to write an article entitled ``Does sleep exist?''\  or ``Does perception exist?''.   Or how about ``Does smelling exist?''.

They wouldn't do it of course.  Or, if they did, no one would publish it.  But why not?  If one thinks the existence of consciousness is doubtful, then surely the existence of perception is doubtful too.  Isn't perceiving something more or less the same as to be conscious of something?

Gassendi's response to the Cartesian ``cogito ergo sum'' was to argue ``I walk therefore I am''.  How if someone were to write an article entitled ``Does walking exist?''
\ema

It should be taken in the spirit of de Finetti's provocative remark, ``PROBABILITY DOES NOT EXIST.''  Just a tad further down than my quote, James writes:
\bq
To deny plumply that `consciousness' exists seems so absurd on the face of it -- for undeniably `thoughts' do exist -- that I fear some readers will follow me no farther. Let me then immediately explain that I mean only to deny that the word stands for an entity, but to insist most emphatically that it does stand for a function. There is, I mean, no aboriginal stuff or quality of being, contrasted with that of which material objects are made, out of which our thoughts of them are made; but there is a function in experience which thoughts perform, and for the performance of which this quality of being is invoked. That function is knowing. `Consciousness' is supposed necessary to explain the fact that things not only are, but get reported, are known. Whoever blots out the notion of consciousness from his list of first principles must still provide in some way for that function's being carried on.
\eq

This is the article where James introduces the idea that reality is made of a neutral stuff that is neither thought nor thing.  But that thoughts and things are aspects of the neutral stuff (when taken in relation to other patches of the stuff).

\section{07-05-09 \ \ {\it Bell's Inequality}\ \ \ (to B. Dreiss)} \label{Dreiss1}

Thanks for your thoughtful note.  The key to understanding that statement from my old paper with Asher Peres is in focusing on the part:  ``any objective theory giving experimental predictions identical to those of quantum theory''.  That means any theory that gives quantum states an objective level of reality, rather than a Bayesian-style significance (as I strongly support), will have this spooky action-at-a-distance character.  Bell was correct on that point.  Asher and I were putting such extensions of quantum theory in contrast to quantum theory itself.

Attached are a couple of pieces that help clarify this.  One is a paper that I am presently constructing (it's about 80\% complete) representing the state of the art with respect to our quantum-Bayesian research program.  The opening section very much addresses your query, and among other things you'll find a very extensive collection of references there to do with the QBism program more generally.  The other two files are copies of some lecture notes I wrote a few weeks back to do with the QBism {\it rejection\/} of nonlocality as the lesson to be learned from Bell inequality violations.  The lesson instead is that quantum ``measurements'' are not measurements at all in the historic sense, but {\it actions\/} one can take on external systems that elicit {\it consequences}.  Bell inequality violations negate not one bit the reality of those external systems.  The external world is there whether you act upon it or not.  Bell inequality violations only negate the pre-existent reality of the consequences to those actions:  No actions on the external world, no consequence.  That is the subject matter of quantum theory, from our take.

Now to Jaynes.  You're right, I should have a specific reply to him.  Many years ago---about 15!---{\Carl} {\Caves} (certainly a disciple of Jaynes in his own right) explained to me what was wrong with Jaynes's argument, and I haven't looked at that aspect of his work since.  Jaynes is certainly one of my heroes, but that doesn't absolve him from making mistakes.  I will read his paper carefully again and form a statement on it (and send it to you).  But that will probably take me a good few weeks:  I've got the present papers to pump out first, and a meeting each week for the next four to travel to!

I was born in Cuero, TX; do you know the town?  Where do you live in Australia?  I've been flying to the continent about once to twice a year lately, and I'll be in Sydney in December.

\section{07-05-09 \ \ {\it Curious George}\ \ \ (to H. Price)} \label{Price14}

Lund?  Your auto-reply says you're on ``research leave''.  Any interesting enclaves of pragmatism there?

By the way, I gave a joint physics/philosophy colloquium in Rochester last month for Brad Weslake.  It was a good time.  But I had a funny encounter with him and Alyssa Ney when I mentioned that the term ``block universe'' originates with William James' 1882 article ``On Some Hegelisms'' (as far as I can tell).  A) They didn't know that, but then B) they assured me that James had the {\it wrong\/} definition of a block universe!

\section{07-05-09 \ \ {\it Obsession with a Footnote}\ \ \ (to H. Price)} \label{Price15}

That's what David Albert would say of my whole philosophy, I'm sure \ldots

I just haven't been able to get that damned footnote exactly right.  Here's the latest attempt:
\bv
\verb+\footnote{See Footnote \ref{Misunderstandings} and Ref.\ +\\
\verb+\cite[Sec.~4.1]{Timpson08} for a sampling of instances of+\\
\verb+opposition to our inchoate ontology, and see Refs.\ + \\
\verb+\cite{Nagel89,Dennett04,Price97} for details on the ``view from + \\
\verb+nowhere'' / ``view from nowhen'' {\it weltanschauung\/} more+\\
\verb+generally.}}+
\ev
I think I'm finally feeling better about this one \ldots\ and wanted to share the moment!

\section{08-05-09 \ \ {\it Your Talk at UWO} \ \ (to W. C. Myrvold)} \label{Myrvold12}

Let me just give you the title and abstract from the new paper.

\begin{center}
Quantum-Bayesian Coherence
\end{center}
\bq
In a quantum-Bayesian take on quantum mechanics, the Born Rule cannot be interpreted as a rule for setting measurement-outcome probabilities from an objective quantum state.  But if not, what is the role of the rule?  In this talk, I will argue that it should be seen as an  empirical addition to Bayesian reasoning itself.  Particularly, I show how to view the Born Rule as a normative rule in addition to usual Dutch-book coherence.  It is a rule that takes into account how one should assign probabilities to the consequences of various intended measurements on a physical system, but explicitly in terms of prior probabilities for and conditional probabilities consequent upon the imagined outcomes of a special counterfactual reference measurement. This interpretation is seen particularly clearly by representing quantum states in terms of probabilities for the outcomes of a fixed, fiducial symmetric informationally complete (SIC) measurement.  We further explore the extent to which the general form of the new normative rule implies the full state-space structure of quantum mechanics.  It seems to get quite far.
\eq
Initially the words ``objective'' and ``counterfactual'' were italicized, but I figured that would get in your way, and so I stripped it out.

Does that sound OK?

\section{09-05-09 \ \ {\it The Excruciating Hangover}\ \ \ (to O. J. E. Maroney)} \label{Maroney5}

Below is the passage from Marcus I was thinking of.  It was in the middle of a very lengthy (and not all that coherent) letter he wrote to Rob.

Attached also is a paper I'm writing at the moment.  [See ``Quantum-Bayesian Coherence,'' \arxiv{0906.2187v1}.]  It still has a good way to go to being completed, but all the advertisements are already in place.  Please see the final paragraph on page 7 (ending on page 8) and the two associated footnotes.  Like my care to put the Martin Gardner quote on realism at the beginning of my recent talks, I'll be writing that section with you in mind \ldots\ \smiley

\subsection{Excerpt from D. M. Appleby's note to R. W. {\Spekkens}, 14 April 2009:}

\bq
I would like to stress that I really meant what I said about finding hidden variables interesting.  I have been arguing in this negative way because I felt you were trying to force me in a direction I do not choose to go, and I felt obliged to give some reasons for resisting.
But now that I have done so let me be a little more positive.

Perhaps you ought to know that Chris and I do not think entirely alike.  To some people, indeed, it seems that we think quite differently.  When I was at PI in the summer I had several lengthy discussions with Owen.  At a fairly early stage in those discussions Owen exclaimed, in tones of incredulity, ``are you aware that Chris goes around telling people that you are a quantum Bayesian?''   It took many many beers, and some truly excruciating hangovers, to convince him that Chris was not misrepresenting me.  And perhaps he never was fully convinced.  To Chris and I it seems that the differences between us are of very minor importance.  But that is not always how it seems to other people.

Anyway one of these minor differences is that, although I do not believe in hidden variables, I do find them interesting.   Another difference is our previous intellectual history.  Though never a true believer, I was for a long time very close to the hidden variables school of thought, to such an extent that people often assumed that I was a Bohmian. (And perhaps there was a sense in which I really was a Bohmian---for Bohm himself was not a Bohmian in the sense this term has come to have.  In particular he didn't believe in hidden variables.)  I was also a convinced anti-Copenhagenist.  When the paper that Chris wrote with Peres ``Quantum Mechanics Needs no Interpretation'' first came out my sympathies were all with the critics.  But then I actually met Chris, and that brought about a revolution in my thinking.

I should explain what exactly this revolution consisted in.  I didn't suddenly decide that the reasons I had previously had  for rejecting the Copenhagen Interpretation were invalid.  On the contrary all the bad things I used to say about the Copenhagen Interpretation, before I met Chris, I continue to say now.  I thought in the past, and I still do think  that Bohr is impossibly obscure, and Heisenberg and Pauli not a lot better.  I used to think, and I still do think that Bell was absolutely bang on the nail when he described the Copenhagen interpretation as ``unprofessionally vague and ambiguous'' (I am quoting from memory, and may not have reproduced his words exactly).  Chris didn't cause me to see clarity where I had formally seen obscurity. But what he did bring about was a fundamental change in my attitude to that obscurity.  Specifically he led  me to think that buried in all the obscurity there were the germs of something true.  Formerly I had seen the Copenhagen interpretation as dirt, fit only for the bin.  But now I see it as soil, in which something good may eventually be got to grow.
\eq

\section{12-05-09 \ \ {\it Read Me}\ \ \ (to R. F. Wachter)} \label{Wachter1}

\brfw
I'm looking for a few bullets about research directions in the area of ``quantum-classical'' computing \& communications.  Sorry for the vagueness.  I could use a few --- just what comes to mind as you read this.
\erfw

\begin{itemize}
\item
(In the application of quantum cryptography) Quantum information knows when it has been read.  (Or leaves a trail that it has been read.)
\item
Why is quantum information a better resource for communication channels than classical information?
\item
What powers quantum information's sensitivity to eavesdropping?
\item
Quantum communication is what you get when you build the world of components that are inherently sensitive to the touch.  If you push, they go further than might have been expected.
\end{itemize}
Since I have no idea what you need, I don't know if you want questions or answers.  Just making things up.
\begin{itemize}
\item
Need formalism that makes quantum and classical information directly comparable, so that inherent differences are revealed in a direct way.
\item
Quantum information is sexy.
\item
{\it Mechanica Quantica Lex Cogitationis Est\/} (this is the quantum Bayesian version of {\it Semper Fi})
\item
Quantum information is the way it is because the world is the way it is, but there is still much confusion on exactly what way the quantum world is.
\item
Quantum information exploits the fantastic creative power of quantum matter.
\item
The field of quantum information is just applied quantum foundations.
\end{itemize}
Give me more guidance, and I'll try to make up something more relevant.
\begin{itemize}
\item
quantum cryptography protocols that are even more sensitive to eavesdropping than present protocols
\item
better quantum error correction through better understanding of the basic structure of quantum channels
\item
understanding geometry of quantum-state space is key to understanding the power and potential of quantum information processing
\end{itemize}
Do you say uncle yet?

\section{12-05-09 \ \ {\it Found It!}\ \ \ (to R. {\Schack})} \label{Schack163}

I finally found that damned William James quote I'd been wanting to use!  It was in his ``unpublished'' manuscripts (and they apparently are not scanned to the web as so many of his other things are).
\bq\noindent
Of every would be describer of the universe
one has a right to ask immediately two general
questions. The first is: ``What are the materials of your universe's composition?'' And the
second: ``In what manner or manners do you represent them to be connected?''
\eq

\section{12-05-09 \ \ {\it Figure} \ \ (to R. {\Schack})} \label{Schack164}

Have you looked at the whole paper now?  If so, then I guess sit tight for a while.

I find myself completely burned out tonight.  The most energy I've been able to muster is to start reading about the Bode--Miller objections, and James's 5 year(!)\ struggle with them.  When it's our turn, we've got to do better!  We at least have a leg up on him when it comes to `certainty'.  Strangely, he only seemed to get halfway there in his conception:
\bq\noindent
When an event that is not yet actual is nevertheless {\it certain to occur\/} in [the] future, it is more than a bare possibility.  We speak of its enjoying even now a {\it virtual or potential existence}.
\eq

Hopefully I'll wake up in the middle of the night and do some work on the paper.  I still want to add a couple of points in Section 5.2.  Add a $d=3$ discussion to Section 5.5.  I want to add another paragraph to 6.1.  And then there's the dreaded Section 8.  I have a vision that the paper should be 54 pages long.

\section{15-05-09 \ \ {\it Emailing:\ 459164a} \ \ (to G. L. Comer)} \label{Comer124}

At first I thought I was going to hate this article [Martin Heisenberg, ``Is Free Will an Illusion,'' {\sl Nature\/} {\bf 459}, 164--165 (14 May 2009)], but then I found that I loved it.  I want to print it out when I get back to the office.  I particularly like the phrase ``their expectations about the consequences of their actions''---because I say that is exactly what a quantum state is.

I read the article (and am writing to you) in the back of a van returning me to Waterloo from the Toronto airport.

Free will in the back of a van \ldots\  I haven't done that since high school.

\section{16-05-09 \ \ {\it Measurement Problem and the Reality of Wives} \ \ (to R. D. Gill, J. Emerson, A. Plotnitsky, and T. M. Nieuwenhuizen)} \label{Emerson0} \label{Plotnitsky22.1} \label{Gill8} \label{Nieuwenhuizen1}

\noindent Dear Richard, Joseph, Arkady, and Theo,\medskip

Thanks for the entertaining banter.  I found seven emails in my box when I returned from today's outing to buy hostas and suddenly felt very popular \ldots\ that is, {\it until\/} I actually read the notes and learned that I'm merely imaginary.  Oh well, I guess I have to live with that.

But it reminded me of a story I wrote down for John Preskill several years ago, titled ``The Reality of Wives.''  I reproduce it below for anyone interested. [See 21-07-01 note ``\myref{Preskill2}{The Reality of Wives}'' to A. J. Landahl \& J. Preskill.] \medskip

\noindent Imaginary wishes,

\section{17-05-09 \ \ {\it Reply to Several Things}\ \ \ (to K. Martin)} \label{Martin10}

\bkma
James told me you had over 50,000 mp3's, is that right? If so, when combined with your email collection and obvious love of
books, one can begin to formulate interesting theories about you.
\ekma
53,694 to be precise, as of today at least.  And 930 books.  Finally, here's a number I've never tabulated.  (It took me about 20 minutes to do it this morning.)  Since May 1997, I apparently have written 31,675 emails.  As you suspected, I like to keep lists.

\section{20-05-09 \ \ {\it Measurement Problem and Dates} \ \ (to R. {\Schack})} \label{Schack165}

I'm in London, ON, where I presented our paper to the philosophy dept yesterday.  Wayne Myrvold had one point that may call for some modification to our Section 4.2.  He complained that we are giving a diachronic reading to the Principal Principle, but that its raw statement (as a requirement on conditional probabilities) is a synchronic one.  I think he's at least partially right.  He points out that one would only adjust one's updated probability to the conditional if one thinks that being apprised of the value of a chance is actual ``learning.''  I.e., it doesn't so shake up one's world picture that one adjusts for this new knowledge by something other than strict conditioning.  That is certainly a technical point that is right in other contexts.  On the other hand, I think the reading we gave the PP is the usual one:  I.e., that {\it implicit\/} in it is the idea that knowing the chance is the pinnacle of knowledge, i.e., implicit in the concept is the diachronic usage of it.  I want to go back to Lewis's original statements to see more precisely how he's using it.  I hope it doesn't cause me too many annoying acrobatics in recomposing the paragraphs to reflect Wayne's point.

Another thing that must be rewritten somewhat is Section 4.1.  When I was in Washington, Patrick Hayden pointed me to a later paper than the one we cite, where nonminimal (but still finite) two-designs are systematically constructed in every finite dimension.  I believe they have an exponential number of outcomes though---I've got to get the reference---though that's not all that important.  Anyway, we just have to admit straight up that it's the minimal two-designs we like.  So, I'm going to shift the argument to emphasize that with larger two-designs, though the urgleichung holds, the priors must then be restricted to a lower dimensional subspace of their defining simplex.  That is surely an unpalatable feature.

\section{28-05-09 \ \ {\it PIAF Postdoc at UQ} \ \ (to G. J. Milburn)} \label{Milburn3}

Thanks for this.  I will distribute it appropriately.

By the way, I could stand to get some references from you.  Attached is a paper I'm constructing at the moment.  [See ``Quantum-Bayesian Coherence,'' \arxiv{0906.2187v1}.]  The references I need are for the middle paragraph of page 26, the one starting with ``The most telling reason \ldots'':
\bq
But the quantum-Bayesian view cannot abide by this.  For, the essential point for a quantum-Bayesian is that there is no such thing as {\it the\/} quantum state.  There are potentially as many states for a given quantum system as there are agents.  And that point is not diminished by accepting the addition to coherence described in this paper.  Indeed, it is just as with standard (nonquantum) probabilities, where their subjectivity is not diminished by normatively satisfying standard Dutch-book coherence.

The most telling reason for this arises directly from quantum statistical practice.  The way one comes to a quantum-state assignment is ineliminably dependent on one's priors.  Quantum states are not god-given, but have to be fought for via measurement, updating, calibration, computation, and any number of related pieces of work.  The only place quantum states are ``given'' outright---that is to say, the model on which much of the notion of an objective quantum state arises from in the first place---is in a textbook homework problem.  For instance, a textbook exercise might read, ``Assume a hydrogen atom in its ground state. Calculate \ldots.''  But outside the textbook it is not difficult to come up with examples where two agents looking at the same data, differing only in their prior beliefs, will asymptotically update to distinct (even orthogonal) pure quantum-state assignments for the same system.  Thus the basis for one's particular quantum-state assignment is always {\it outside\/} the formal apparatus of quantum mechanics.
\eq
I remember your once expressing very eloquently this point with respect to quantum process tomography.  (It may have been at the Konstanz meeting, but it might have been elsewhere as well.)  It would be nice to cite some of the nitty-gritty of coming to a particular quantum state assignment (how one must do tomography and calibration and stuff like that).  Any suggestions?

Are you here at PI this week?  If so, I could also just print this out and give you a copy.

\section{29-05-09 \ \ {\it The Elusive Nature of the Quantum State} \ \ (to G. J. Milburn)} \label{Milburn4}

John Wheeler used to say, if you really want to get a job done, give it to a busy man!  It looks people are using that philosophy in spades with you.

\bgjm
Your paper looks fascinating. I am not sure if I will get time today to give a considered
response to your request, but I should be able to get back to you early next week. Would that
be soon enough?
\egjm
It surely would.  I've still got about five more pages of material to write in it anyway.

BTW, aside from the particular section I asked you the question about, if you want also to get a quick summary of what the paper is about (i.e., without reading the whole thing), just go directly to the figure on page 20.  The point of the paper is that the equation
$$
q(j)=(d+1)\sum_{i=1}^{d^2} p(i) r(j|i) - \frac{1}{d}\sum_{i=1}^{d^2} r(j|i)
$$
holds some chance of being taken as the {\it single\/} postulate of quantum mechanics.

\section{01-06-09 \ \ {\it Old Talk}\ \ \ (to C. M. {\Caves})} \label{Caves100.2}

\bcc
\bq\noindent\rm
[CAF wrote:]
Do you still have a copy around of that talk you gave in Snowbird about 3 or 4 years ago?  Nonlocality vs reality, which should you desire to keep?  If so, could I have a view of it?  I was telling one of the students here about it tonight at dinner.
\eq
I'm a little uncertain what you want to know.  The technical parts of the talk have been published, so you're probably talking about the vague discussion of locality vs.\ efficiency.
\ecc

Oh, it was just the business about how abandoning locality is not such an innocent move as so many involved in quantum foundations would think, but rather opens a can of worms \ldots\ in the end endangering the very idea of physical theories tout court.  Yeah, it was certainly vague; it just struck a chord in me and I wanted to repeat it properly.  Nothing to worry about; I'll eventually write it down in my own way.

\section{02-06-09 \ \ {\it Unprofessionally Vague and Ambiguous}\ \ \ (to D. M. {\Appleby})} \label{Appleby62}

How have things been going the last week?  When you can muster the energy and thought, we should discuss your post-{\Vaxjo} plans.

On an immediate matter:  I've heard you use the phrase in the title of this note several times.  I know it is something John Bell said, but can you give me the precise coordinates of it.  In which essay does it appear.  I have his unspeakable book at home, so I should be able to find it if just given a hint.

Using him, will be part of the completion of these paragraphs of mine:
\bq
\begin{flushright}
\baselineskip=13pt
\parbox{2.8in}{\baselineskip=13pt\small
Of every would be describer of the universe one has a right to ask immediately two general questions.  The first is: ``What are the materials of your universe's composition?'' And the second: ``In what manner or manners do you represent them to be connected?''} \medskip\\ \small --- William James, 1903--1904
\medskip
\end{flushright}

This paper has focussed on adding a new girder to the developing structure of quantum-Bayesianism (`QBism' hereafter).  As such, we have taken much of the previously developed program for the background of the present efforts.  If a reader wants to know the core arguments for why we choose a more `personalist Bayesianism' rather than a so-called `objective Bayesianism', she should refer to Refs.\ [17,20,55].  If she wants to know why a subjective, personalist account of certainty is crucial for breaking the impasse set by the EPR criterion of reality, she should refer to Refs.\ [21,55].  This paper was meant to get a new phase of the program off the ground without dwelling too much on the past.

Still, fearing James' injunction, we know it is our duty to discuss anew one term of this paper that has so far been waived about most uncritically:  It is `measurement.'  How can one really understand the proclamation `Unperformed measurements have no outcomes!'\ in a deep, soul-satisfying way?  Answering this question, we feel, is the first step toward characterizing ``the materials of our universe's composition.''

Believe it or not, we take our cue from John Bell:  Despite our liberal use of the term so far, the word `measurement' should be banished from fundamental discussions of quantum mechanics.
\eq

\section{04-06-09 \ \ {\it Tomorrow.\ Tonight?}\ \ \ (to K. Martin)} \label{Martin11}

I'm sorry to hear this.  [\ldots]

Look, Keye, take care.  Write a poem about this experience, and look for whatever insight it gives you on quantum mechanics (remembering that every single experience should be used to give insight on quantum mechanics).

\section{10-06-09 \ \ {\it Typos and Experience}\ \ \ (to N. D. {\Mermin})} \label{Mermin155}

Sadly, I'm still working on that paper I showed you last.  I put it on the shelf for a while, and now I'm back to it, bound and determined to get it posted by the end of the week.  (There'll be smaller, twin paper after my visit with {\Ruediger}, titled ``Quantum-Bayesian Decoherence'' where we show how Zurek's thinking is {\it exactly\/} backwards, when seen through Bayesian eyes.  [See \arxiv[quant-ph]{1103.5950}.])

I was working on Footnote 29 (where I mention you), when I found your typo.  I might as well attach the latest version.  I reworked the intro after the last comments you sent me.  It was indeed crucial that I make explicit the EPR-criterion-of-reality business prior to Bell's argument, and I've done that now.  Still, I suspect you won't like the new version either---for one thing, I didn't give any respect to all the alternatives to EPR-criterion that you mentioned.

Have a look at Figure 3 and the long caption; you might enjoy it.  I notice that you too have been leaning on the word ``experience'' lately.  At least in the paper that I cite in Fn 29, but also I ran across your {\sl Phys Today\/} piece a couple days ago.  [I wonder too whether you've yet appreciated the beauty of the second equation in the caption to Figure 2.  You've never mentioned it if you have.]

Comments are still welcome, but you're probably tired of me.

More on my new views of ``experience'' below.  [It concerns another one of my papers in the works, that one for a collection on ``pragmatism and quantum mechanics.'']

\section{11-06-09 \ \ {\it The Paulian Idea, Full Frontal} \ \ (to R. {\Schack})} \label{Schack166}

Regarding,
\brs
``From within any part, the future is undetermined'' is exactly what a many worlder would say. It suggests that the evolution of the whole is determined, which is not at all what we want to say. Of course you point that out in the footnote. But there you say that there is no whole. Why do we have to take a view on whether there is a whole beyond saying that q.m.\ isn't about it if it exists? Why do we have to go into this discussion at all? Why not simply point out that from any agent's perspective, there is uncertainty, full stop? I also don't like ``treating a part as an agent''. We are not treating parts of the universe as agents. Agents are primitives at the moment.
\ers
I think it is better to just write a better footnote, incorporating your remarks into that.  The thing is, I do want to ``take a view'' \ldots\ that is the whole point of that and the next paragraph, and the previous two James quotes.

Let me try to finesse it all.  Give me another shot.  Perhaps ``treating a part as'' is not the best choice of words, but this whole last part of the discussion was to build toward James's (and Pauli's) ``pure experience'' \ldots\ where agenthood is not primitive, but rather one of two complementary aspects of something neutral.

Did you read the version where I had the Wheeler quote at the end?

\section{11-06-09 \ \ {\it The Paulian Idea, Full Frontal, 2} \ \ (to R. {\Schack})} \label{Schack167}

Did you see my ``apologetic'' comment on my {\tt quant-ph} posting of ``Delirium Quantum'':
\bq\noindent
Stuffed away in a conference proceedings for years, I finally worked up the nerve to post this because I needed to cite it in a paper with some proper equations. Stay posted.
\eq
Funny thing is I just got a letter from the editor of the {\sl NeuroQuantology Journal\/} saying that they would like to publish the paper!!

Different worlds, different worlds!

\section{11-06-09 \ \ {\it The Paulian Idea, Full Frontal, 3} \ \ (to R. {\Schack})} \label{Schack168}

\brs
I don't understand ``But perhaps this republican-banquet vision of the world that so seems to fit with a QBist understanding\ldots''  It's not even clear to me from this sentence if the republican-banquet vision is ours or not.
\ers

It is certainly my present vision.  I thought I had already declared enough ownership in the sentence: ``It is almost as if one can hear in the very formulation of the Born Rule one of William James's many lectures on chance and indeterminism.''  \ldots\ which itself refers to urgleichung stuff a couple of sentences above.

Think like William James:  absolute clarity might be the enemy of poetic flow.

\section{12-06-09 \ \ {\it In Defense of a Nomenclature that is Psychologically Rather Mature} \ \ (to J. {\Barrett})} \label{Barrett5}

Thanks Jon.  Your note actually pushed me over the edge in some way:  Before I had always said ``anti-realism box'' to myself tongue in cheek.  But now, knowing that it'd be a pleonasm, I'm starting to take it quite seriously.  Actually, it really does capture the idea I was shooting for in Section 8!

BTW, attached is a more extended diatribe on why I don't like to call qm ``nonlocal''.  {\Ruediger\/} and I plan to make that into a proper paper later in the year.

\subsection{Jon's Preply}

\bq\noindent
\bq\noindent
``Indeed, it flavors almost everything they think of quantum mechanics, including the interpretation of the imaginary
games or toy models they use to better understand quantum mechanics itself. Take the recent flurry of work on Popescu--Rohrlich boxes [5]. These are imaginary devices that give rise
to greater-than-quantum violations of various Bell inequalities. Importantly, another common name for these devices is
the term `nonlocal boxes' [6]. Their exact definition comes via the magnitude of a Bell-inequality violation-which
entails the non-pre-existence of values or a violation of locality or both-but the commonly used name opts only to
recognize nonlocality. They're not called anti-realism boxes, for instance. The nomenclature is psychologically
telling.''
\eq

The phrase ``nonlocal box'' is made up of two words: ``nonlocal'' and ``box''.
The second word here is intended to call to mind a ``black box'' --- that is
a device or process that admits an input, and produces an output, but
whose internal workings are either unfathomed or unfathomable. By this
light, the terminology does already recognise the non-preexistence of
values as a possible interpretation, and ``anti-realism box'' would be a
pleonasm.

More seriously: the paper looks very interesting and I'm looking forward
to a better look. Will either of you be at the Reconstructing Quantum
Theory conference at PI in August?
\eq

\section{13-06-09 \ \ {\it Reasons for Rewriting Quantum Mechanics}\ \ \ (to A. S. Holevo)} \label{Holevo11}

I finally wrote down a paper version of the thoughts that were going through my head the last time I saw you.  Let me call your attention to it:
\arxiv{0906.2187}. I hope that, being in paper form now, it will help convince you that what I have in mind is the very antithesis of a hidden-variable program.  (I remember well the pain of your remarks at the end of my talk!)

I hope things are going well with you and that maybe we will have a chance to see each other a little when you are in Toronto.

\section{13-06-09 \ \ {\it At Last, At Last!}\ \ \ (to G. M. D'Ariano)} \label{DAriano8}

I finally fulfilled the promise I made to you long ago:  I finally posted a paper on the foundational significance of SICs.  You can find it here:
\arxiv{0906.2187}.

It'll be great seeing you again this week; I'm just on the drive between Waterloo and Toronto on my way to Sweden.

\section{13-06-09 \ \ {\it Still Seeking SICs} \ \ (to J. Bub \& I. Pitowsky)} \label{Pitowsky5} \label{Bub22}

I have finally written up the stuff I was telling you about one evening at Bill Demopoulos's party.  Since you seemed interested, let me give you a pointer to it:  \arxiv{0906.2187}. I hope maybe it'll give you a little food for thought, and show a little more why I resist thinking of quantum theory as a {\it generalization\/} of Bayesian probability theory.

I hope you are both doing well.  I'm on my way to Sweden and plan to have a thoughtful midsummer there.

\section{13-06-09 \ \ {\it Discussion Points}\ \ \ (to W. K. Wootters)} \label{Wootters24}

I'm so glad you're coming to PI this Fall.  Do you yet know when you will arrive?  It's going to be a fantastic year here, I deem:  Beside you Howard Barnum, Richard Healey, and Steve Bartlett will also come.  For different periods of time each, I suppose, though I do know that Howard is coming for a full year.  Furthermore, we have another added treat:  Marcus Appleby will come for a month-long visit every other month.  So, I'm going to be one happy boy ping-ponging between all you guys.

I finally put down on paper my reasons for thinking SICs are important for quantum foundations, and I wonder if you might have an interest.  If so, you can find the paper here: \arxiv{0906.2187}.  Now that I think about it, I should have cited your ``probability tables'' paper in there, and I can think of exactly where (page 6).  I'm sorry about that.  It was surely the root of the root of the tree that led me up to this point.  I'll fix it in the revised version.  One thing is for sure though, I hope my enthusiasm for SICs in that paper is contagious enough that we'll have many long discussions on related items when you visit.

\section{13-06-09 \ \ {\it Update}\ \ \ (to D. M. {\Appleby})} \label{Appleby63}

Thanks.  I'll print your notes out and read them on my flight later today.  We can discuss in Sweden.

Likewise, would you mind having a read of Section 8 (only section 8, don't worry about the rest) in my: \arxiv{0906.2187}.  I'd like to discuss some of that with you as well.

The other day when I was digging up Wheeler quotes for the above said Section 8, I ran across a little exchange between Wheeler and Richard Rorty.  Wheeler had just given a talk where he explores the idea of taking the ``elementary quantum phenomenon'' as the fundamental building block of the universe.  Rorty's question came afterward.  What struck me was how prescient the question was respect to the title of your talk in {\Vaxjo}.  I thought you might enjoy:
\bq\noindent
RORTY: What puzzles me is the degree to which the surprisingness of
quantum physics is an empirical, and the degree to which it is a
philosophical matter. That is, what I don't know enough to grasp is
whether the philosophical overlay on science, the nineteenth century
philosophical account of the nature of science, was so inadequate in
itself that practically anything would make one doubt it. As it
happened, quantum phenomena did. But it may have been so faked up
that it merely needed a suggestion to collapse in favor of something
like a pragmatist understanding of science. But I take it that
Professor Wheeler feels that it is quite specific quantum results
that simply contradict fundamental empirical theses which had been
presupposed by the older theories. So the relation is much more
direct.
\eq

\section{13-06-09 \ \ {\it From Pragmatism to Pure Experience, 2} \ \ (to M. B\"achtold)} \label{Baechtold3}

I apologize again for keeping you waiting so long.

Yes, I am still working on the paper---intellectualistically, even if the writing has come to a temporary standstill.  I can quite understand if you have well lost patience with me, but I really want to get this paper written in one form or another.  If you need me to drop out, I can do that, but I am trying very hard to pull my quantum Bayesianism into an ontology, and until I get this paper out of my system, I feel it's all going to be clouds to me.  I need to write this to clarify things to myself:  It is a bit like giving birth.  So, I would still like to have a contribution to your volume.

The bad thing is I've gotten all hung up over the Miller--Bode objections.  They are as relevant in this context (the quantum context), I am discovering, as they ever were.  They turned the skeletal manuscript I told you about below into a kind of scarecrow.  {\Ruediger} {\Schack} and I are going to meet June 18 and spend until July 3 intensely trying to come to a conclusion on this, and I very much hope a baby will be born.

Could you possibly give me until the end of August for a submission?  To demonstrate some piece of my good faith that this is topmost on my mind, let me show you another paper much closer to completion (five more pages to go!)\ that may give you some sense of the one I'm trying to produce for you.  The one for you would be an infinitely more philosophical piece (with little to no equations), but it would build off the last paragraph on page 7 / first paragraph of page 8, and particularly all of Section 8 (only 3/4 complete itself).  Please read through those parts, even though the punchline is not included yet.  The title of Section 8 is a direct allusion to my paper for you.  I plan to get the present paper submitted to the archive this week before flying to Sweden Saturday.  I just thought the writing there might give you some flavor of what is to come.

What I am aiming for is a pluralistic ontology something like ``pure experience.''  Quantum measurement should be an instance of it.  But it is so very painful to get the ideas straight.

I will live with whatever your opinion is.

\section{13-06-09 \ \ {\it Measurements Generate Reality} \ \ (to A. Zeilinger)} \label{Zeilinger3}

I finally posted something proper on the foundational import of the representation of quantum states I was speaking of the last time I saw you:  It's all about ``Unperformed experiments have no outcomes.''  I hope you will enjoy it, as I know this is our greatest common interest.  The paper can be found here \arxiv{0906.2187}.

I suspect you guys just had a great conference in Vienna.

\section{13-06-09 \ \ {\it Reply} \ \ (to H. C. von Baeyer)} \label{Baeyer73}

\bhcvb
I have loaded my Kindle with {\bf Pragmatism} for free, and I'm reading it with great pleasure.  ``Theories thus become instruments, not answers to enigmas, in which we can rest.''  So no rest for you!
\ehcvb

Good for you!  I've always liked the continuation of that line:  ``We don't lie back upon them, we move forward, and, on occasion, make nature over again by their aid.''  What a slap in the face to rationalism.  A good theory is not a description of what is there, fixed and unchangeable, but a {\it tool}, not different in kind from a hammer, for actually reshaping nature.

I'm as far as Boston now.  I'm looking forward to plenty of insights this week.  I don't know how it happens, but I always find being in Sweden shakes loose at least one new QBist thought.

\section{13-06-09 \ \ {\it Abstract to QTRF5} \ \ (to I. Bengtsson)} \label{Bengtsson3}

\bib
If possible I would be delighted to have you in ``my'' session; I
have some premonitions about the contents.
\eib

Your premonition is probably correct.  You can find the full story here now: \arxiv{0906.2187}.

I've just started my long journey to Sweden. I hope we get a good chance to discuss things there.  {\Asa} told me that you may now have more sympathy for a ``quantum mechanics within probability theory''  (as opposed to ``quantum mechanics as a generalization of probability theory'') than you once had.  I'd like to hear what you're thinking.

\section{13-06-09 \ \ {\it Single-User Theory} \ \ (to J. Emerson)} \label{Emerson1}

Here's that note I started to construct but never finished.  Also included are the attachments to have gone with it.  I think at this stage I won't go ahead and finish the note after all.  That's because part of the answer was superseded by my recent posting \arxiv{0906.2187} which includes some of the material from the note below anyway.  So, why don't you have a look at Section 1 and Section 8 in the posting, and the notes attached here.  Maybe that'll at least give you enough information so that you can say with good reason that you do indeed disagree with ``what Fuchs thinks about reality''.\medskip

\noindent --- --- --- --- --- --- --- ---\medskip

I'll respond quietly (aka, privately) to a couple of points in one of your emails.  Happy Victoria Day, by the way.  After this note, I'm out to work in the garden again.

\bje
By the way, I was a student of Ballentine's and I agree almost
entirely with his point of view. I'm not quite sure what Fuchs thinks
about reality and I probably would disagree with it.
\eje

If you're not sure what I think, how do you know that you would {\it probably\/} disagree with it?  \smiley

I believe of reality what Martin Gardner believes of it.  The first attachment is a quote from him:
\bq\noindent
   The hypothesis that there is an external world, not dependent on
    human minds, made of {\em something}, is so obviously useful and
    so strongly confirmed by experience down through the ages that we
    can say without exaggerating that it is better confirmed than any
    other empirical hypothesis.
\eq
As for quantum mechanics, I believe it is the best tool we (each of us) presently have for seeing our way through our {\it encounters\/} with that world.  Another way to put this particular point:  I see quantum mechanics as a single user theory.  I can use it, you can use it, she can use it.  Any of us.  And when any of us uses it, we are using it to better prepare for our own (personal) encounters with the world.  We use it to make sure our {\it beliefs\/} about the consequences of those encounters are {\it consistent\/} with each other.  It is a calculus for ``consistifying'' our beliefs.

Saying this much---but only this much---is not inconsistent with a Ballentine--Emerson program of trying to (or hoping to) complete quantum mechanics.  At least as far as I can tell, it is not inconsistent with that program.  For it could be that quantum mechanics is a multi-user theory as well, and that I, Chris, am ignoring an extra, rich layer of the theory.  For this option to be had, those things that I take to be ostensibly personal encounters with the world, would have to be more public things.

\section{13-06-09 \ \ {\it Once a Solipsist, Always a \ldots}\ \ \ (to H. M. Wiseman)} \label{Wiseman19}

Our old conversations have been on my mind again lately.  (That's a good thing.)  Anyway, to honor it, I made an oblique reference to you in my most recent posting \arxiv{0906.2187}. A little challenge:  Let's see if you can find yourself.

\subsection{Howard's Reply}

\bq
A little challenge indeed. How many friends do you have left in Australia? \smiley

I don't think I would have said ``Rubbish! \ldots\ It would mean quantum mechanics
collapses into a kind of solipsism---a theory that there is only the self.'' but rather something like
``Yes indeed, and this crypto-solipsism is, as I think Einstein saw, why quantum mechanics
is not complete, why it does not describe the world in which the many observers
live and make predictions etc, and why we should try to find a theory which does.''

However, exactly what I would have said I could only judge after a careful reading of your
paper rather than the cursory one needed to meet your challenge. I'll try to provide that
sometime soon.

Thanks for the honour.
\eq

\section{14-06-09 \ \ {\it Once a Solipsist, Always a \ldots, 2}\ \ \ (to H. M. Wiseman)} \label{Wiseman20}

\ldots\ from a Heathrow lounge \ldots\

\bhw
How many friends do you have left in Australia? {\rm \smiley}
\ehw
Good point!

One point of language:  I was a little worried that an Australian would ever use the word ``rubbish''.  Rob {\Spekkens} told me that Terry Rudolph would most certainly say ``bullshit''.  But I could swear I've heard you say ``rubbish'' before (though I don't actually remember to what).

\subsection{Howard's Reply}

\bq
Probably. I'm more polite than Terry.
\eq

\section{14-06-09 \ \ {\it Seeking a Home for the SIC'ening Bayesians}\ \ \ (to J. I. Cirac)} \label{Cirac2}

{\Ruediger} {\Schack} and I just posted a rather long paper covering two subjects rather dear to me:  1) what quantum mechanics looks like in terms of symmetric info complete measurements (i.e., what I talked about during my last visit to you), and 2) what this all means for the ``quantum Bayesian'' interpretation of quantum mechanics that Caves, Schack, Peres, Appleby, etc., and I have been putting together for some time.

Here's a link: \arxiv{0906.2187}.

May I ask you to have a look at it and give me an opinion on whether you think Rev.\ Mod.\ Phys.\ is an appropriate home for it?  It's quite extensively referenced, both on things to do with SICs and things to do with the quantum Bayesians (pro and con).  It's definitely partially review article, and no doubt ``in depth'' as your webpage calls for.

If it is a style of article you think worthwhile, we'll submit it to RMP and see what the referees say.  Otherwise, we'll seek another home for it, no problem.

I'm glad to hear you'll be spending a little time at PI.  Do you know when you will likely come in the coming year?

\section{15-06-09 \ \ {\it Your Paper with RS}\ \ \ (to C. M. {\Caves})} \label{Caves100.3}

Thank you for the kind remarks on the paper.  Your words warmed my heart in a way that two southern men probably shouldn't discuss.  I'll leave it at that.

I'll be back after I get some coffee in me.

\bcc
I was able to read your paper (with {\Ruediger}) on the several planes home from Poland.  I thought it was terrific, very much in your voice, with just a leavening of {\Ruediger}'s precision, and thus both entertaining and informative.  I will be trying to convince everyone I encounter who has an interest in foundations to read it.  More to the point, I think you and {\Ruediger} might have pushed things to the point where there is enough to chew on that others will start following your path---and I don't mean just being polite enough to point out what they find right and wrong with QBism, but actually pushing forward the technical program.  I hope that turns out to be the case, because otherwise, the program is going to falter.
\ecc

Yes we've felt that for a long time.  And we have noticed for some time that Rob Spekkens' program is getting maybe more attention than it really deserves (though I don't denigrate it too much) precisely because it can give an industry of busy work.  This convex set, that convex set --- it's not a hell of a lot different than the 3,000 papers devoted to little variations of the Jaynes--Cummings model.  I thought it was really funny at our quantum foundations group meeting a couple of weeks ago, when I gave a five minute lecture titled ``My Favorite Convex Set'' just picking out a small piece of the paper.  All the ears perked up, and three of the students who could not normally care less what I'm on about came up to talk to me afterward about problems.

\bcc
I find your current ideas much more interesting than what you were
doing previously on updating rules,
\ecc

Oh, you just wait and see; everything meshes together in the end.

\bcc
[A]lthough I have to admit that I haven't yet found the current ideas
anything like as compelling as the Dutch-book argument itself.   Indeed, I
would say that I was quite sympathetic to the second-to-last paragraph
of Sec.\ VII, where you express clearly your own opinion of the
incompleteness of the program.
\ecc

Yes, that is a glaring gap.  Whatever the principle is though, it can't be logical.  It is an empirical statement and rests on whatever it is that quantum mechanics is really trying to tell us [about the world].  It's all to do with the world being composed of stuff that can make the equations fly.  [See story ``It's a Wonderful Life'' in 05-12-01 note ``\myref{Schumacher4}{Lucky Seven}'' to B. W. Schumacher.]  But I can't see my way any further than that at the moment.

\bcc
I think you very effectively formulate the idea that qm is a theory of single users interacting with the world (I take it that not saying this clearly was the major defect with our paper on certainty).  I think most physicists won't buy it, but then you shouldn't expect that.  You're writing for the next generation.  I believe a more enduring physical objection will be somewhat different.  The great success of quantum mechanics is not in predicting probabilities of measurement results ``correctly''---more precisely, providing a framework for pragmatically successful probabilistic inference about the world---but in explaining the structure and properties of matter and energy in terms of ground states (or thermal states) in
situations that have nothing to do with measurements.  Surely the ontology of quantum mechanics must include this success,
unless you want to go down the road of Rorty, who in one memorable
part of the book you urged me to read, says that the description of a
giraffe as made up of electrons and nuclei is merely a construct that
we find useful, at least some of the time. Please assure me that you're not going to go there.
\ecc

It has been claimed that there is some affinity between ways I sometimes talk and ``structural realism''.  (Look it up in wiki or the Stanford encyclopedia.)  But philosophers always want to box me in.  Anyway, I see our discussion of Gardner didn't alleviate your fears!

Rorty wasn't perfect---rest his soul---but he had a lot of good things to say.  Particularly, he tried his best to get a handle on what it would take for the world at its deepest roots to be malleable to our actions.  You have to be bold and take hefty risks if you want to explore that direction.

Something about the way you put the giraffe thing doesn't sound right, and I don't recall how he said it.  He certainly strikes me as being over the edge at times.  Here is the way I would put it:  We have to get out of the mindset that things (like giraffes) are ``made of'' atoms.  Heavy emphasis on ``made of.''  It is not like God puts bricks and mortar together and builds everything up in that fashion.  Rather I would say this.  Between any two conceptual pieces of the world, there are various distinctions and various things in common.  The subject of physics as we presently understand it concerns pieces of the world with a certain common denominator (be it described by the words ``atoms'', ``electrons'', or whatnot).  That is, physics focuses on at least one common denominator and ignores the rest.  It is not a question of all descriptions being equally useful, as you (maybe rightly) impart to Rorty above.  But I wouldn't say physics (as presently understood) is primal either.  It is just about common denominators---that's its subject matter, and there's nothing wrong with that.

So, with this program, we are aiming to understand the character of some of the common denominators.  After all these years, I'm back to thinking John Wheeler was on the right track in thinking it is something to do with what he called the ``elementary quantum phenomenon.''  It's more like James and Pauli's ``neutral stuff''---James called it ``pure experience'' but I don't like that name much.  Anyway, it is the spirit that breaths life, and that's what I want to get a handle on.

Thanks again for the encouraging comments; they breathed a little life into me.

\section{16-06-09 \ \ {\it Feeding Quantum Mechanics} \ \ (to B. C. van Fraassen)} \label{vanFraassen20}

You once wrote me this (seven years ago actually!):
\bvf
Now when it comes to theories that give us probabilities, whether
absolute or conditional, I'll agree with scientific realists that
literally read they say that there are objective probabilities in
nature.  But accepting such a theory does not involve believing that.
Rather it involves appointing the theory as an `expert' for guidance
of our subjective probabilities concerning observable events.  The
metaphor of `expert' is cashed out (as by Haim Gaifman) as follows.
Suppose that I appoint Peter as my expert on snuffboxes.  That means
for my subjective probability $P$ and Peter's subjective probability
$q$ the constraint:
$$
   P\Big(A\,|\,q(A) = x \Big) = x
$$
with generalizations of this to intervals, odds, conditional
probabilities, for statements A that are about snuff boxes.

Thus the issue of whether there are objective probabilities in nature
or whether to believe in them is finessed:  there are only the
theory's probabilistic {\bf pronouncements} accepted as input and my
own subjective probabilities.

That is clearly not how you are approaching it overall.  But perhaps
there are connections?  I'd like to know how the QM probabilities are
fed into your subjective probability as a whole -- I wonder if it
will not be similar.  After all, even if a quantum state is read as a
compendium of probabilities, and you say something like ``this
material is in quantum state such and such'', your own subjective
probability function has a domain much larger than facts pertaining
to this material.
\evf

I now have a very direct answer for you.  (Sorry, sometimes it takes me a while!)  If you have a small amount of time, I'd be very flattered if you'd look at my new paper to see what I mean: \arxiv{0906.2187}. (The key idea is in the figure on page 20.)  I hope it gives you some food for thought.

(By the way, it will actually be followed in a month or so with a sibling paper titled ``Quantum-Bayesian Decoherence.''  In that one, we make use of your Reflection Principle to argue that Zurek's decoherence program is just exactly opposite the way a good Bayesian would think about what's going on.  [See \arxiv[quant-ph]{1103.5950}.])

Hope life in California is treating you well.

\section{16-06-09 \ \ {\it Latter-Day Wheeler}\ \ \ (to W. H. Zurek)} \label{Zurek1}

Here's that passage I was thinking of.  I was wrong though, it's not from his December 2000 op-ed in the {\sl New York Times\/}, but rather his 1998 autobiography.
\bq
Many students of chemistry and physics, entering upon their study of quantum mechanics, are told that quantum mechanics shows its essence in waves, or clouds, of probability.  A system such as an atom is described by a wave function.  This function satisfies the equation that Erwin {\Schroedinger} published in 1926.  The electron, in this description, is no longer a nugget of matter located at a point.  It is pictured as a wave spread throughout the volume of the atom (or other region of space).

This picture is all right as far as it goes.  It properly emphasizes the central role of probability in quantum mechanics.  The wave function tells where the electron might be, not where it is.  But, to my mind, the {\Schroedinger} wave fails to capture the true essence of quantum mechanics.  That essence, as the delayed-choice experiment shows, is {\it measurement}.  A suitable experiment can, in fact, locate an electron at a particular place within the atom.  A different experiment can tell how fast the electron is moving.  The wave function is not central to what we actually know about an electron or an atom.  It only tells us the likelihood that a particular experiment will yield a particular result.  It is the experiment that provides the actual information.

Measurement, the act of turning potentiality into actuality, is an act of choice, choice among possible outcomes.  After the measurement, there are roads not taken.  Before the measurement, all roads are possible---one can even say that all roads are being taken at once.
\eq

In the {\sl New York Times\/} pieces, he only says this and doesn't say anything explicit on the wave function:
\bq\noindent
My mentor, the Danish physicist Niels Bohr, made his peace with the quantum. His ``Copenhagen interpretation'' promulgated in 1927 bridged the gap between the strange\-ness of the quantum world and the ordinariness of the world around us. It is the act of measurement, said Bohr, that transforms the indefiniteness of quantum events into the definiteness of everyday experience.
\eq

The last paragraph of the more relevant quote expresses a world of difference between John's view at the time and standard many worlds (of Deutsch, Wallace, and the rest of the Oxford crowd), and one I thought you did good justice expressing in your {\sl Physics Today\/} piece:  Before the measurement ``one can say all roads are being taken at once''; ``after the measurement there are roads not taken''.  Most many-worlders---I'm thinking of Tegmark in particular---would probably dismiss that as a literary flourish, one that's not crucial for how John visualized things.  The typical many-worlder would say, ``Before measurement, all roads at once; after measurement, all roads at once.  And anyway, what is this thing called measurement?  Let's just get rid of it.  To the extent that it's there, it's a messy concept that should be derived and secondary.''  For Wheeler, as far as I can tell---and I have read everything he's ever written---it held the most central place.

Anyway, I was very proud of your piece with Misner and Thorne, and thought it was just wonderful.

\section{19-06-09 \ \ {\it Your Talks}\ \ \ (to D. M. {\Appleby})} \label{Appleby64}

From Hay-on-Wye; what a fantastic town!!  {\Ruediger} and I have been discussing QBism, and I feel it is just a glorious day for the project.  We'll be in the UK for the next three days, then 2 weeks together in Waterloo.  I have this feeling really good things are going to happen.

Thanks for sending your talk.  I read it last night before bed and definitely enjoyed it.

\section{21-06-09 \ \ {\it Rainy Day Train} \ \ (to C. Eriksson)} \label{Eriksson2}

Thanks for the note.  Emma is 10 now, and Katie is 7.  Attached is a recent picture of the two of them just after they got out of bed one morning.  They're good kids; I'm very proud of them---all the time I tell them they'll ``change the world''.\footnote{Because the world is malleable and ``ever not quite''.}

Just flying home from London.  Between Sweden and now, I went book shopping in a strange little town in rural Wales.  They have 35 bookstores, with nearly 3/4 million books between them!  (I bought 27 books in point of fact.)  A great vacation!

\section{21-06-09 \ \ {\it Renege}  \ \ (to Y. J. Ng)} \label{Ng5}

It is no problem.  I am just flying back home from London.  Right after the meeting in Sweden, {\Ruediger} {\Schack} and I took a little vacation to the town Hay-on-Wye in rural Wales.  It's an amazing place:  There are 35 second-hand bookshops there!  And so I'm coming home with a suitcase stuffed with 27 specimens for my library.  Speaking of {\Ruediger}, you might enjoy our new paper:  \arxiv{0906.2187}.  It tells a more complete version of the story I told you during your last visit.

Paul Frampton was at the same meeting that I was in Sweden.  It was good seeing him again after all these years.

\section{22-06-09 \ \ {\it More Thoughts}\ \ \ (to D. M. {\Appleby}, H. C. von Baeyer and R. {\Schack})} \label{Appleby65} \label{Baeyer74} \label{Schack169}

1) Reading your note below,

and

2) thinking of the story where Wheeler commands the equations ``Now fly!''

and

3) thinking of this quote that I copied into my first samizdat:  ``I forbade any simulacrum in the temples because the divinity that breathes life into nature cannot be represented.''

and finally,

4) thinking of this passage from ``my friend'' James:
\bq
I am as confident as I am of anything that, in myself, the stream of thinking (which I recognize emphatically as a phenomenon) is only a careless name for what, when scrutinized, reveals itself to consist chiefly of the stream of my breathing. The `I think' which Kant said must be able to accompany all my objects, is the `I breathe' which actually does accompany them. There are other internal facts besides breathing (intracephalic muscular adjustments, etc., of which I have said a word in my larger {\sl Psychology}), and these increase the assets of `consciousness,' so far as the latter is subject to immediate perception; but breath, which was ever the original of `spirit,' breath moving outwards, between the glottis and the nostrils, is, I am persuaded, the essence out of which philosophers have constructed the entity known to them as consciousness. That entity is fictitious, while thoughts in the concrete are fully real. But thoughts in the concrete are made of the same stuff as things are.
\eq
I found myself asking {\Ruediger} how to say ``primal breath'' in German.  He proposed:  urhauch.  (Now Marcus, try to pronounce that one!)  The neutral stuff fighting it out at the republican banquet, and by way of that making our universe, is urhauch.

Just trying the word on for size.

\subsection{Marcus's Preply}

\bq
I have just been thinking about your friend James.

[\ldots]

Anyway, back to James, and pure experience.  I take your point about what he meant, and I think perhaps I semi-agree with him.  However, I have some significant qualifications.

In the first place I think he chose the wrong name.  Instead of ``pure experience'' it might be better to call it, say, ``ur-stuff''.  Or perhaps ``primal-stuff''.  The trouble with ``experience'' is that the name isn't really neutral, it seems to me.  It is strongly biased on the mind side.  Also, ``experience'' suggests (to me) that what we are talking about is all out on the surface.  But I think that what we are talking about is deep:  behind or below what we see is loads of stuff that we can hardly guess at.  ``Experience'' suggests to me a sheet of water only millimeters thick. But I think what we talking about is an ocean, and it goes way way down.

{\it Both\/} aspects go way way down.  Not only the external matter aspect (the fact that quantum systems are unpredictable).  But also the internal mind aspect.

In this room where I am now writing the principal cause of uncertainty is me (so far as the things I can observe without ancillary apparatus are concerned).  I am pretty confident that the book on the table in front of me will still be there in 5 minutes time, unless I decide to move it.  I am likewise pretty certain what the clock will read in 5 minutes time, unless I decide to adjust it.  But I have no idea what I myself will be thinking/feeling/writing/doing in 5 minutes time.  In a very real and immediate sense I am a mystery to myself.

(I would guess I will still be writing this in 5 minutes time, assuming I don't get up to stretch my legs.  But what will I be writing?  ---It is almost as much a mystery to me who is writing it as it is to you who is reading it.  The ideas are forming themselves as I write, in a way which, though not independent of my will---is in fact an expression of my will---is nevertheless far from being fully known to me.)

I prefer ``ur-stuff'' or ``primal-stuff'' to ``pure experience'' because I think it conveys this sense of something mysterious, not fully known.

Another objection to ``pure experience'' is that it doesn't convey the notion of agency.  At least it doesn't to me.  One just sits there and has experiences, like it or not.   Experiencing, as I understand the word, is passive.

It is of course true that if you go into the neurophysiology of it experiencing things turns out to be far from passive.  It is highly active, in fact.  But I don't think that is in the ordinary language meaning of the word.

Perhaps ``ur-stuff'' doesn't have the connotation of agency either.  But it does at least have the advantage that it doesn't really have any connotations at all, beyond the connotation of something elemental.  Or even pre-elemental.  It is anyone's guess what an ur-thing might do or feel or be.\medskip

\noindent ************* \smallskip

However, though I prefer ``ur-stuff'' or ``primal-stuff''  to ``pure experience'' I am still not totally happy.  The term suggests something structureless.  Something, in other words, like a vector (c.f.\ the slides for my talk) which owes all of its properties to its relations with other vectors, and has no intrinsic properties of its own, considered in isolation.  And I don't like that because I think that as soon as you start to think that way you are going to end up with something which is like classical field theory, in as much as it invites the misunderstanding that what is in question is a complete description.

Perhaps I wouldn't be happy with any name at all (c.f.\ the  Tao Te Ching:  ``it was from the nameless that Heaven and Earth sprang'').  But I guess that doesn't get us very far.

At all events I think it is essential that we avoid the error of thinking that any of our terms denote structureless simples.

In this connection I very much liked your response to Bell's criticism of the way the concept of measurement is used in the Copenhagen Interpretation in section 8 of your paper.  Bell, of course, wants a complete description of the measurement process, framed in the language of beables (as does Rob and many others). He is wrong about that.  Not only is it OK to use terms which aren't defined in the way that Bell would like:  it is absolutely essential that we do so.

We need in fact to recognize that it is not only necessary, but actually essential that we use terms which are similar to the terms of ordinary language in as much as there is no attempt to resolve them into atomic or other elemental components.  No one thinks it necessary to define the meaning of the word ``chair'' by giving a detailed description at the atomic level.  The word ``chair'' is defined relative to its human uses, as something one can sit on.  Similarly with the word ``measurement''.

You do that in your latest paper, and I think you are absolutely right.

(I do have a question however.  My own inclination is to take a measurement to be any process by which I or someone else acquires information.  If I look at the chair opposite me, and see that it is about 6 inches from the window, then that is a measurement I am inclined to say.  Similarly, if an astronomer tells me that such-and-such a white dwarf star has such-and-such mass, then that too is a measurement.  Moreover I am tempted to say that I have measured the star [as opposed to saying that I have measured the astronomer] [because it is the star that I have acquired information about].  My question is:  do you go along with that?  And, if not, can you make it a bit clearer what you do count as a measurement?)\medskip

\noindent ************* \smallskip

A quick point about pragmatism.  I was meaning to tell you in {\Vaxjo} that I am not totally unsympathetic to pragmatism.

The  reason I believe quantum mechanics---the reason that I have thought it worth spending years of my life trying to understand it better---is that quantum mechanics works.  I have always thought (not just now, but back in the days when I was sympathetic to Bohm) that the fact quantum mechanics works means that it must be in some sense true.

So I will agree with James that the concept of ``truth'' is intimately related to the concept of ``what works''.  I also share James's conviction that we need to get away from the idea that knowing the truth equates to the possession of some kind of map.

My only disagreement is that I feel it can't be right to actually identify truth and ``what works''.  I feel the concepts are logically distinct, and that it is important to keep them that way.\medskip

\noindent *************\smallskip

One last remark.  The thought in my last email (about Wheeler-Feynman) grew out of a conversation I had with David Craig after my talk. A little to my surprise he actually liked the talk.  He especially liked the point about the counter-factual element in the classical field concept, and the point that this means a classical field is much more similar to the Bayesian view of the wave-function than is generally realized.  I had the impression that I had got to him there.  However, he then asked what I would say about the space-time manifold.  I think that is a good question.  For of course as soon as you start thinking in space-time terms you do find yourself being driven in a block-universe direction.  At any rate, I find myself being driven in that direction.  The point about Wheeler-Feynman grew out of my attempts to find some way of getting round that difficulty. Trying to find some more appropriate way of thinking about space and time.
\eq

\section{23-06-09 \ \ {\it Fragments}\ \ \ (to D. M. {\Appleby} and H. C. von Baeyer)} \label{Appleby66} \label{Baeyer75}

Let me reply to a couple of your points.

To
\bma
I do have a question however.  My own inclination is to take a
measurement to be any process by which I or someone else acquires
information.  If I look at the chair opposite me, and see that it is
about 6 inches from the window, then that is a measurement I am
inclined to say.  Similarly, if an astronomer tells me that
such-and-such a white dwarf star has such-and-such mass, then that too
is a measurement.  Moreover I am tempted to say that I have measured
the star [as opposed to saying that I have measured the astronomer]
[because it is the star that I have acquired information about].  My
question is:  do you go along with that?  And, if not, can you make it
a bit clearer what you do count as a measurement?
\ema
and
\bhcvb
I'm still re-reading Section 8 and appreciate Marcus's question about
measurement.
\ehcvb
here is the way I respond.

With every movement and every breath, I take actions upon the world surrounding me:  These are actions with potentially unpredictable consequences for my experience, and actions that the rest of the world could not have ``foreseen'' coming either.  With every breath I contribute to the ongoing making of the world.  ``Quantum measurement'' refers to the special cases of this broad idea where one feels comfortable conceptualizing a distinct number of consequences arising from one's actions, imagines one can write down actual numerical expectations (probabilities) for those consequences, and indeed bothers to conceptualize the actions in the first place to some mathematical precision.  In other words, ``quantum measurement,'' as an idea, arises only in very controlled and introspective settings.  But wash away the numerics and the need for mathematical precision, and it is something happening all the time, everywhere around us.

I like the way Misner, Thorne and Zurek put it in their {\sl Physics Today\/} piece on John Wheeler last month.  It carries at least part of what I was trying to say above (though ``relation'' sounds so much more static than what I, and I think John himself, was trying to get at).  Let me copy it:
\begin{quote}
Quantum mechanics in 1976 was the backbone of atomic, nuclear, and condensed-matter physics as well as quantum chemistry.  Yet the essence of ``the quantum'' was a mystery that gripped Wheeler's attention; quantum measurement, he thought, was the crux of the mystery.  Why?  Because quantum measurement is only a euphemism for the relation between observers---us---and the rest of the physical universe.  So quantum measurement, he said, is where ``the quantum gets personal.''
\end{quote}
Getting a more precise handle on this cluster of ideas, I feel, is the next stage of our research.  We are long from getting there, but that is the direction to start turning as we make smaller bits of progress.

\bma
The Wheeler story and the quote from your samizdat are just so totally
appropriate.  I love them.  (Where did you get the quote in your
samizdat from, by the way?---Just to save me the trouble of looking it up).
\ema
I got the quote many years ago (1997) while reading some crazy book by Jean Baudrillard, but I think he himself got it from another source.  Let me see if I can find it on the web.  \ldots\ Well, I didn't.  Maybe Baudrillard made it up, but he put quotes around it as if it came from somewhere else.\footnote{\editornote It did---Voltaire.  See footnote~\ref{Voltrillard} on p.~\pageref{Voltrillard}.}

\bma
I am not so sure about the quote from James.  I do think that
introspection, considered as a route to understanding consciousness,
is totally useless.  And perhaps that is all that James is trying to
say.  But I must say that whenever I look into myself, in an effort to
find out what consciousness is, I don't notice any particular connection with breathing.
I don't notice a connection with anything else, mind you.
\ema
You miss the point of the quote.  It is that ``knowing,'' for us, is a bodily function, neither more nor less sacred than breathing itself.  It is that Descartes well should have said (at least for the track Descartes was pursuing), ``Spiro, ergo sum.''  But in James's exact passage, Kant was his target.  His point was that it is not ``subject'' that always accompanies ``object,'' but that it is ``breath.''

\section{24-06-09 \ \ {\it Those Einstein Quotes}\ \ \ (to W. H. Zurek)} \label{Zurek2}

Sorry for the delay:  Here are those Einstein quotes I said I would send you.  It turned out that I only had two of them in my computer in an acceptable form; the third and most important, that from the Schilpp volume, had only been an abridged version I use in my talks from time to time---so I wanted to add all the missing words before sending it on to you.  Anyway, everything is collected in the attached {\tt .pdf}.

It was good sparring with you the other day.  We should do it more often---it'd certainly lead to more mutual understanding, which would be most useful for me.

\begin{center}
{\bf Einstein's Personal Argument for the Incompleteness of Quantum Mechanics\\
(a selection)}
\end{center}

Einstein, in a letter to Erwin {\Schroedinger} dated 19 June 1935, referring to the EPR paper:

\begin{quote}
For reasons of language this [paper] was written by Podolsky after much discussion.  Still, it did not come out as well as I had originally wanted; rather the essential thing was, so to speak, smothered by the formalism.
\end{quote}
\indent Now, to let Einstein speak for himself.  Particularly note in the quotes below Einstein never invokes what has become known as the EPR criterion of reality:  ``If, without in any way disturbing a system, we can predict with certainty (i.e., with probability equal to unity) the value of a physical quantity, then there exists an element of physical reality corresponding to this physical quantity.''  [See Arthur Fine's discussion in the Afterword to the Second Edition of his book {\sl The Shaky Game}, where he has taken to calling it snidely the Podolsky Criterion.  Fine also makes the point somewhere that these quotes are not atypical; never once after the EPR paper did Einstein use the precise argument of the paper.]

Einstein, in his ``Autobiographical Notes'' section of the 1949 Schilpp volume (with some translation details modified by {\Ruediger} {\Schack}):

\begin{quotation}
If one asks: does a $\psi$-function of the quantum theory represent a real factual situation in the same sense in which this is the case of a material system of points or of an electromagnetic field, one hesitates to reply with a simple ``yes'' or ``no''; why?  What the $\psi$-function (at a definite time) asserts, is this.  What is the probability for finding a definite physical magnitude $q$ (or $p$) in a definitely given interval, if I measure it at time $t$?  The probability is here to be viewed as an empirically determinable, and therefore certainly as a ``real'' quantity which I may determine if I create the same $\psi$-function very often and perform a $q$-measurement each time.  But what about the single measured value of $q$?  Did the respective individual system have this $q$-value even before the measurement?  To this question there is no definite answer within the framework of the theory, since the measurement is a process which implies a finite disturbance of the system from the outside; it would therefore be thinkable that the system obtains a definite numerical values for $q$ (resp.~$p$) the measured numerical value, only through the measurement itself.  For the further discussion I shall assume two physicists, $A$ and $B$, who represent a different conception with reference to the real situation as described by the $\psi$-function.
\begin{itemize}
\item[A.]
The individual system (before the measurement) has a definite value of $q$ (resp.~$p$) for all variables of the system, and more specifically, {\it that\/} value which is determined by a measurement of this variable.  Proceeding from this conception, he will state:  The $\psi$-function is no exhaustive description of the real situation of the system but an incomplete description; it expresses only what we know on the basis of former measurements concerning the system.
\item[B.]
The individual system (before the measurement) has no definite value of $q$ (resp.~$p$).  The measurement value only arises through the act of measurement in cooperation with the probability which is given to it in view of the $\psi$-function.  Proceeding from this conception, he will (or, at least, he may) explain: the $\psi$-function is an exhaustive description of the real state of the system.
\end{itemize}

We now present to these two physicists the following instance:  There is to be a system which at time $t$ of our observation consists of two partial systems $S_1$ and $S_2$, which at this time are spatially separated and (in the sense of classical physics) are without significant reciprocity.  The total system is to be completely described through a known $\psi$-function $\psi_{12}$ in the sense of quantum mechanics.  All quantum theoreticians now agree upon the following:  If I make a complete measurement of $S_1$, I get from the results of the measurement and from $\psi_{12}$ an entirely definite $\psi$-function $\psi_2$ of the system $S_2$.  The character of $\psi_2$ then depends upon {\it what kind\/} of measurement I undertake on $S_1$.

Now it appears to me that one may speak of the real factual situation at $S_2$. Of this real factual situation, we know to begin with, before the measurement of $S_1$, even less than we know of a system described by the $\psi$-function.  But on one supposition we should, in my opinion, absolutely hold fast:  the real factual situation of the system $S_2$ is independent of what is done with the system $S_1$, which is spatially separated from the former. According to the type of measurement which I make of $S_1$, I get,
however, a very different $\psi_2$ for the second partial system $(\psi_2, \psi_2^1,\ldots)$. Now, however, the real situation of $S_2$ must be independent of what happens to $S_1$.  For the same real situation of $S_2$ it is possible therefore to find, according to one's choice, different types of $\psi$-function.  (One can escape from this conclusion only by either assuming that the measurement of $S_1$ (telepathically) changes the real situation of $S_2$ or by denying independent real situations as such to things which are spatially separated from each other.  Both alternatives appear to me entirely unacceptable.)

If now the physicists, $A$ and $B$, accept this consideration as valid, then $B$ will have to give up his position that the $\psi$-function constitutes a complete description of a real factual situation.  For in this case it would be impossible that two different types of $\psi$-functions could be coordinated with the identical factual situation of $S_2$.

The statistical character of the present theory would then have to be a necessary consequence of the incompleteness of the description of the systems in quantum mechanics, and there would no longer exist any ground for the supposition that a future basis of physics must be based on statistics.
\end{quotation}

Einstein in a 1952 letter to Michele Besso:

\begin{quotation}
What relation is there between the ``state'' (``quantum
state'') described by a function $\psi$ and a real deterministic
situation (that we call the ``real state'')?  Does the quantum state
characterize completely (1) or only incompletely (2) a real state?

One cannot respond unambiguously to this question, because each
measurement represents a real uncontrollable intervention in the
system (Heisenberg). The real state is not therefore something that
is immediately accessible to experience, and its appreciation always
rests hypothetical. (Comparable to the notion of force in classical
mechanics, if one doesn't fix {\it a priori\/} the law of motion.)
Therefore suppositions (1) and (2) are, in principle, both possible.
A decision in favor of one of them can be taken only after an
examination and confrontation of the admissibility of their
consequences.

I reject (1) because it obliges us to admit that there is a rigid
connection between parts of the system separated from each other in
space in an arbitrary way (instantaneous action at a distance, which
doesn't diminish when the distance increases).  Here is the
demonstration:

A system $S_{12}$, with a function $\psi_{12}$, which is known, is
composed of two systems $S_1$, and $S_2$, which are very far from
each other at the instant $t$. If one makes a ``complete''
measurement on $S_1$, which can be done in different ways (according
to whether one measures, for example, the momenta or the
coordinates), depending on the result of the measurement and the
function $\psi_{12}$, one can determine by current
quantum-theoretical methods, the function $\psi_2$ of the second
system. {\it This function can assume different forms}, according to
the {\it procedure\/} of measurement applied to $S_1$.

But this is in contradiction with (1) {\it if one excludes action at
a distance}. Therefore the measurement on $S_1$ has no effect on the
real state $S_2$, and therefore assuming (1) no effect on the quantum
state of $S_2$ described by $\psi_2$.

I am thus forced to pass to the supposition (2) according to which
the real state of a system is only described incompletely by the
function $\psi_{12}$.

If one considers the method of the present quantum theory as being in
principle definitive, that amounts to renouncing a complete
description of real states.  One could justify this renunciation if
one assumes that there is no law for real states---i.e., that their
description would be useless.  Otherwise said, that would mean: laws
don't apply to things, but only to what observation teaches us about
them.  (The laws that relate to the temporal succession of this
partial knowledge are however entirely deterministic.)

Now, I can't accept that.  I think that the statistical character of
the present theory is simply conditioned by the choice of an
incomplete description.
\end{quotation}

Einstein in a 1948 letter to Walter Heitler:

\begin{quote}
[T]hat one conceives of the psi-function only as an incomplete description of a real state of affairs, where the incompleteness of the description is forced by the fact that observation of the state is only able to grasp part of the
real factual situation. Then one can at least escape the singular conception that observation (conceived as an act of consciousness) influences the real physical state of things; the change in the psi-function through observation
then does not correspond essentially to the change in a real matter of fact but rather to the alteration in {\it our knowledge\/} of this matter of fact.
\end{quote}

\section{24-06-09 \ \ {\it The Section 8 Interpretation of Agency}\ \ \ (to W. H. Zurek)} \label{Zurek3}

And by the way, speaking of the sparring, I really would appreciate it if you'd read through Section 8 of the paper I gave you (Sec 8 only, don't bother with the rest):  \arxiv{0906.2187}.  I feel it's the best I've done yet for answering your perennial question, ``So what is the observer made of?''  My answer has always been, ``Well he's made of the same things as you.  And that is not greatly distinct in physical kind from what dogs, or amoebas, or even tables and chairs are made of.''  What goes wrong in your question, from my view, is that it goes backwards.  ``Agent'' refers to a position within quantum theory, not to the detailed architecture of some specific types of quantum systems.  Anyway, read my Section 8:  I do myself better there.

\section{24-06-09 \ \ {\it (again?)}\ \ \ (to S. Savitt)} \label{Savitt7}

\bss
[T]he quote from Will Durant in your newest paper is terrific.
I am stealing that for my metaphysics course. Of course, I also am
inclined to think that freedom in an indeterministic universe is also illusory.
\ess

The Jamesian idea (which really goes back to / comes from Renouvier) is that freedom is a more basic concept than either determinism or indeterminism.  Freedom is not something that subsists in one or the other kind of universe, but rather an expression of an ultimate kind of (ontological) pluralism, that can allow to some extent for both.

\section{25-06-09 \ \ {\it Review of the Magic Equation, 2} \ \ (to C. R. Stroud, Jr., J. H. Eberly, A. Ney, Y. Shapir \& B. Weslake)} \label{Ney3} \label{Weslake4} \label{Stroud2} \label{Eberly2}

Getting a note from Andrew Jordan on some APS GQI business reminded me of you all.  I finally posted the paper associated with the talk I gave you guys.  If you're interested, it's here: \arxiv{0906.2187}.  (I'll give a prize to the first person who finds the instances of ``rubbish,'' ``salacious,'' and ``bastardization'' in it.)

\section{25-06-09 \ \ {\it Section 2} \ \ (to A. Y. Khrennikov)} \label{Khrennikov26}

I meant to tell you before leaving Sweden:  Given our discussions on ``contextuality'', you might want to read my discussion in Section 2 of the paper I gave you.  In case you've lost it, you can regain it here: \arxiv{0906.2187}.

\section{25-06-09 \ \ {\it Down with Unitarity} \ \ (to T. Rudolph, cc J. Rau)} \label{Rudolph12} \label{Rau1}

I was talking with Jochen Rau the other day, and he told me about a result of his that was quite similar, if not identical, to your old one that any unitary operation can be simulated by a sequence of closely spaced measurements.  I figure I should put you two in touch with each other.  (You're both addressed on this email.)  If that result still hasn't been pulled out of the drawer, it'd be nice for it to appear out in the external world so more people can discuss it.  (I'm also thinking of John Wheeler's words, which I dug up for Wojciech the other day.  Read below.)

P.S. Terry, by the way, if you haven't seen this one of mine \arxiv{0906.2187}, I'll give you a challenge:  If you find all the instances of ``womb,'' ``rubbish,'' ``salacious,'' and ``bastardization'' in it, I'll reward you with the first annual {\sl Chris Fuchs Creepy Word Prize\/} the next time I see you, to be redeemed at the pub of your choice, either in London or Waterloo.  (No cheating; Adobe search tool invalidates challenge.)

Wheeler quote below.  [See 16-06-09 note ``\myref{Zurek1}{Latter-Day Wheeler}'' to W. H. Zurek.]

\section{26-06-09 \ \ {\it Up, Up, and Away!}\ \ \ (to N. D. {\Mermin})} \label{Mermin156}

{\Ruediger} is here with me for two weeks; we're hard on the trail of what can be meant by ``others' experiences'' from a quantum-Bayesian view.  Our first task is to get the simple idea of ``communication'' straight.  Harder than you might think \ldots\ or at least harder than we might think.

\section{26-06-09 \ \ {\it This and That}\ \ \ (to M. Tait)} \label{Tait1}

I had several things I wanted to tell you by email soon after my last visit to London, but now that too much time has passed, I have forgotten most of the things!  Sorry about that:  But at least I finally write.

Here's a partial list.
\begin{itemize}
\item[1)]  The philosopher I was trying to remember who interprets Bohr in a very Kantian way is John Honner.  The book I read was titled, {\sl The Description of Nature:\ Niels Bohr and the Philosophy of Quantum Physics}.  The main thing I remember is how the guy meticulously discussed something like more than 20 drafts of the Como lecture.

\item[2)]  I finally posted the paper which I was speaking on at your seminar.  You can find it here: \arxiv{0906.2187}.  You might enjoy Section 8 of it, where I turn to more philosophical matters, and lay my heart on the line, so to speak.

\item[3)]  Attached is a hefty piece of paper, that's far, far from complete, but since you have some interest in these matters, I thought it might be of some use to you as a point of entry into some literature.  You might have a look at the Folse entries in particular; he is my personal favorite Bohr interpreter.
\end{itemize}

Maybe we should get you to PI for some more extensive discussions.  Any interest in dropping in for maybe a week or two in the Fall or Spring?

\section{26-06-09 \ \ {\it Laws Workshop} \ \ (to S. Weinstein)} \label{Weinstein3}

You said, women.  What about some of these people:

Jane E. Ruby, who wrote:
\begin{itemize}
\item
J.~E. Ruby, ``The Origins of Scientific `Law','' J.\ Hist.\ Ideas {\bf 47}, 341--359 (1986).
\end{itemize}

Or Lorraine Daston, who co-edited:
\begin{itemize}
\item
{\sl Natural Law and Laws of Nature in Early Modern Europe}.
\end{itemize}

I don't know the people at all, but I do remember enjoying Ruby's article.

Here's another one:  I think Cheryl Misak (from U. Toronto) would be excellent.  She could talk on C. S. Peirce's conception of evolving laws, for instance. And what about Nancy Cartwright?

Moving away from women, what about Ian Hacking?  I think he's always great. What about Shimony?
\begin{itemize}
\item
A.~Shimony, ``Can the Fundamental Laws of Nature Be the Results of Evolution?,'' in {\sl From Physics to Philosophy}, edited by J.~Butterfield and C.~Pagonis (Cambridge U. Press, Cambridge, 2000), pp.~208--223.

\item
Y.~Balashov, ``On the Evolution of Natural Laws,'' Brit.\ J.
Phil.\ Sci.\ {\bf 43}, 343--379 (1992).
\end{itemize}

You see in my mind the ideas of ``information theoretic laws'' and ``law without law'' are fundamentally linked.  That's how these particular names came to mind.

Arthur Fine?

And maybe from the other side of the fence:  How 'bout Daniel Dennett?

Hope that's been of some help.

\section{27-06-09 \ \ {\it Acz\'el}\ \ \ (to P. G. L. Mana)} \label{Mana14}

Thanks; these references are useful.

Also let me take a moment to mend my evil ways:  I was pretty rude, cutting {\Ruediger} off yesterday.  Here is the idea he was trying to express to you, and I got in the way of his expressing it adequately.  As Appleby says ``probability is single case, or nothing'' (see Marcus's paper \quantph{0408058}).  That is, everything that can said of the meaning of probability applies to the single-case alone---consideration of long sequences of trials adds nothing to our foundational understanding of the subject.  Similarly for all meaningful concepts within probability theory, and ``expectation'' is one of those.  It is perfectly meaningful in the single case, and Dutch-book considerations (like in Section 2 of our \arxiv{0906.2187}) can show that.

Consider a single lottery ticket of the form:
\begin{center}
\parbox{2.25in}{\boxedalign{\mbox{If outcome is 1, pay $x_1$ dollars.} \nonumber \\ \mbox{If outcome is 2, pay $x_2$ dollars.} \nonumber \\ \mbox{If outcome is 3, pay $x_3$ dollars.} \nonumber \\ \mbox{Etc.} \nonumber \\ \mbox{If outcome is $n$, pay $x_n$ dollars.} \nonumber \\ \nonumber}}
\end{center}
where the outcomes $i=1,\ldots,n$ are a mutually exclusive exhaustive set, and place that in contrast to a set of other lottery tickets, $n$ of them, each of the form
\begin{center}
\parbox{2.05in}{\boxedalign{\mbox{If outcome is $i$, pay 1 dollar.}\nonumber \\ \nonumber}}
\end{center}

The amounts one will freely buy or sell the latter tickets for define one's ``probabilities'' $p_i$ for the events.  The question now is at what price should one freely buy or sell the first (conglomerate) lottery ticket above for?  The Dutch-book answer is $E$ dollars, where
$$
E = \sum_i p_i x_i\;.
$$
If an agent buys or sells the lottery ticket for anything other than that amount, then a Dutch-bookie can force him to a {\it sure\/} loss of money {\it in a single trial}.  No repetition of trials is needed at all.

And that single case consideration gives meaning to the concept of ``expectation.''

\section{27-06-09 \ \ {\it Abstracts for Reconstructing Quantum Theory} \ \ (to P. Goyal)} \label{Goyal5}

\noindent {\bf Title:}  Quantum-Bayesian Coherence (or, My Favorite Convex Set) \medskip

\noindent {\bf Abstract:}
In a quantum-Bayesian delineation of quantum mechanics, the Born Rule cannot be interpreted as a rule for setting measurement-outcome probabilities from an objective quantum state.  (A quantum system has potentially as many quantum states as there are agents considering it.)  But what then is the role of the rule?  In this paper, we argue that it should be seen as an  empirical addition to Bayesian reasoning itself.  Particularly, we show how to view the Born Rule as a normative rule in addition to usual Dutch-book coherence.  It is a rule that takes into account how one should assign probabilities to the outcomes of various intended measurements on a physical system, but explicitly in terms of prior probabilities for and conditional probabilities consequent upon the imagined outcomes of a special counterfactual reference measurement. This interpretation is seen particularly clearly by representing quantum states in terms of probabilities for the outcomes of a fixed, fiducial symmetric informationally complete (SIC) measurement.  We further explore the extent to which the general form of the new normative rule implies the full state-space structure of quantum mechanics.  It seems to go some way.\medskip

Here are three papers associated with the approach:
\begin{itemize}
\item
\arxiv{0906.2187}
\item
\arxiv{quant-ph/0205039}
\item
\arxiv{quant-ph/0404156}
\end{itemize}

\section{29-06-09 \ \ {\it `Acquiring Information' Is Not the Primary Idea}\ \ \ (to D. M. {\Appleby})} \label{Appleby67}

A short answer to your long letter!

\bma
However, there are some points on which I do agree with Bell.  One is
that I think he was right to insist on the maximum possible degree of
clarity and precision, and right again when he accused Bohr et al.\ of falling short
of that standard.   Hence my question.  If the term ``quantum measurement''
refers to every process by which we acquire information then it is
sufficiently sharp for me.  But if it only refers to a certain subset
of such processes the question arises:  which particular subset,
defined precisely how?

I am not expecting you to answer this question here and now.  Like I
said, I am groping myself, and I appreciate that you are in the same
position.  I am only saying that this is, as it seems to me, a
question which eventually has to be answered.
\ema
I thought I had answered that (even to some precision) with my lines:
\bq\noindent
With every movement and every breath, I take actions upon the world surrounding me:  These are actions with potentially unpredictable consequences for my experience, and actions that the rest of the world could not have ``foreseen'' coming either.  With every breath I contribute to the ongoing making of the world.
\eq

At this point let me send you to some pages of my collection:
Particularly three notes to Bas van Fraassen:
``\,\myref{vanFraassen7}{`Action' instead of `Measurement'},'' on pages 550--551, ``\myref{vanFraassen10}{Questions, Actions, Answers, \& Consequences},'' on pages 553--555, and ``\myref{vanFraassen12}{Canned Answers},'' on pages 555--557.

I can do the story of mutual information (in the second note above) better now, thinking of both random variables in terms of the consequences of actions (i.e., experiences), but the present words should be a decent start.

In short, I am answering you with the ``refers to every process by which we'' you say above; I just protest at the phrase ``acquire information.''

\section{29-06-09 \ \ {\it Quantum Bayesianism $+$ Book}\ \ \ (to M. Schlosshauer)} \label{Schlosshauer2}

In a recent email David {\Mermin} mentioned knowing you, and it caused me to remember how kind you were in {\Vaxjo}.  If I haven't already expressed it, thanks for the interest in QBism.  If you have any further, good pointed questions, feel free to send them.  The exercise is definitely good for me, whatever this research program ultimately morphs into.

By the way, I recall your questions mostly centered on ``certainty.''  Let me forward you a note I wrote to {\Appleby} this morning, where I address some of that (particularly in the third note to van Fraassen).  [See 29-06-09 note ``\,\myref{Appleby67}{`Acquiring Information' Is Not the Primary Idea}'' to D. M. {\Appleby}.]

On another note, {\Ruediger} and I will soon write a much smaller companion piece to the last paper---this one titled ``Quantum-Bayesian Decoherence''---where we try to put decoherence in QBist terms.  When the time comes, I'll forward you a draft if you don't mind.  [See \arxiv[quant-ph]{1103.5950}.]

\section{29-06-09 \ \ {\it Off-the-Shelf Ontology vs.\ the Republican Banquet}\ \ \ (to C. G. {\Timpson})} \label{Timpson13}

I was just talking to Rob {\Spekkens} and he told me he hadn't heard back from you concerning the PIAF meeting in late September.  I sure hope you can come!  I know that you're supposed to be coming to the Reconstructing Quantum Theory workshop a month earlier, but I feel your insights will be needed even more at the PIAF meeting.  Particularly, I would love a proper philosopher to give David Albert the ``what for'' (old Southern language) if he happens to show up and throws his usual dismissal of anything approaching an epistemic view of quantum states.  It would be nice to have a balanced discussion for once.  So, please, please, please do come; I so hope you'll find a way to make it happen.

On another note, I finally posted a summary of the things my little team has  been working on for the last year.  You can find it here if you haven't already seen it: \arxiv{0906.2187}. I hope you will enjoy it (and see some real progress in it).  In my usual way, I've included some footnotes that are bound to cause me to be banned from Oxford.  On the more philosophical side, you might jump to Section 8, which can be read independently of the rest of the paper.

{\Ruediger} and I have quite a palette of things to write up this year, and I'm hopeful that we'll pull it all together.  The next paper will be titled ``Quantum-Bayesian Decoherence'' (you can guess what that will be about), but the one following that will be ``Quantum Bayesian Quantum Certainty Is Not a Moore Sentence'' \ldots\ and you can guess what we'll try to argue there as well!  [See \arxiv[quant-ph]{1103.5950}.]

Anyway, please let me encourage you again to come to PIAF and help set the stage.  Help keep me on the run, and maybe we'll both grow from the exercise!  (It's in the reciprocity of our names!)

\section{29-06-09 \ \ {\it Unpacking a Little Rhetorical Traction}\ \ \ (to H. M. Wiseman \& E. G. Cavalcanti)} \label{Wiseman21} \label{Cavalcanti4}

Sorry to keep you waiting so long for a reply.  It would have come much more quickly if Howard hadn't thrown us for a real loop with his final note:  It has been impressively difficult for us to decide whether Howard exists or not!  We just didn't know what to answer, and all expediency of reply was lost at that point.

But we've somewhat recovered now.  Let me tackle your notes in little spurts.

First and foremost I apologize that our little rhetorical flourish at the beginning of the paper---i.e., our two-paragraph statement on ``local realism''  designed to get some traction on the intro---would cause so much trouble and consternation.  As you surely know, it had very little to do with the paper as a whole (issues to do with locality were not mentioned again).  The trouble is localized, we believe, in your not being able to ``read off'' our intentions when it came to the phrase ``reads off'' (in quotes in the text).  We certainly didn't mean the naive realism that Howard rightly points out would have been the setting up of a straw man.  Instead, we meant very much [double quotes being Howard's own words]:
\bq
`realism' (in the sense of the theorem) $=$ ``measurement results are determined by inputs \ldots\ plus any number of other `hidden' variables specified in the past of all measurements'' {\it without recourse to\/} ``the experimenter (who is assumed for this purpose to be outside the description of the physical system)''
\eq

Howard is right when he says,
\bhw
I don't think Chris and {\Ruediger}'s paper claims that {\bf local causality}
can be saved, only that {\bf locality} can be saved.  Whether {\bf locality}
is a concept worth saving is then a point for discussion.
\ehw
We do indeed think locality can be saved.  And that is because we reject:
\bhw
any agent who is a genuine realist, and who believe[s] he can predict
with certainty the result of a measurement, must believe that there is
a pre-existing value ``out there'' in the world. (This is the
Bayesian-friendly version.)
\ehw

This is the point of our previous paper with Carl, ``Subjective Probability and Quantum Certainty''.  Of course, you are free to define a ``genuine realist'' any way you like, and by rejecting this sentence we then become nongenuine realists.  But realists, we believe we remain.

Attached are a set of notes I wrote up for one of our group meetings here a couple months ago.  They represent my response to Norsen's paper ``Bell Locality and the Nonlocal Character of Nature'' (I keep talking about Eqs.\ (11) and (12) in that paper).  The notes are most certainly tentative and not carefully formulated enough yet, but they represent the skeleton of a paper {\Ruediger} and I are going to write on the subject before the summer is out.  Feel free to look over the notes with this proviso.  Maybe they possibly shed some light on what we were thinking when we wrote our little bit of rhetorical traction in the posted paper.

Finally, with regard to,
\bhw
So in summary it seems to me \ldots\ that perhaps you don't appreciate the
import of Bell's theorem.
\ehw
Bell's theorem is absolutely significant to us:  Rejecting the EPR criterion of reality, while proposing to hold on to Einstein locality, Bell's theorem causes us to reject the particular definition of `realism' in the equation above.  That's pretty earth-shaking to us:  I would never want to think we ``minimize Bell's work''; we just take home a different conclusion than the usual one.

Thus ends my first installment.  More replies later.

\section{29-06-09 \ \ {\it Disturbing the Solipsist}\ \ \ (to H. M. Wiseman \& E. G. Cavalcanti)} \label{Wiseman22} \label{Cavalcanti5}

\bhw
I disagree with your idea that this formulation shows that QM
probabilities have {\bf more structure\/} than classical probabilities. I
think it's just a different structure. I mean if we allow you to use
SIC-POVMs as a tool, then surely we should allow a classical
probabilist the concept of non-disturbing measurements as a tool. Thus
I would argue that in exactly the same way that you have derived
constraints on QM probabilities, we can ``derive'' the ``constraint''
(which is usually taken for granted but which you
disavow) that $p(B) = \sum_A p(B|A)p(A)$ even when $A$ is only a
hypothetical event. You seem to argue this is illogical because $A$ may
disturb $B$. But I could equally claim that your SIC-POVM measurement
could disturb the quantum system {\bf more than is necessary}. If you are
allowed minimally-disturbing SIC-measurements, then in the classical
case you have to accept no disturbance.
\ehw

We do not argue that the law of total probability, i.e., $p(B) = \sum_A p(B|A)p(A)$, is ``illogical''.  Only that it is a judgment that one may or may not make, when comparing the counterfactual upper path (which necessarily generates the $p(B)$ you mention, via Dutch-book coherence) to the `factualized' lower one (which generates the probability $q(B)$ we discuss).  We certainly don't bar any Bayesian from making the judgment that $q(B)=p(B)$; we only point out that one need not have to make it.  In that sense, you are right, quantum theory is just a ``different structure'' (or rather a different addition to raw Bayesianism) \ldots\ {\it but\/} I don't believe we claimed anything stronger than that in the paper.  Try reading that section again and see if you still find it coming out so overly strong.

Thus ends my second installment.  More replies later.

\section{29-06-09 \ \ {\it Forgotten Reason for the Title}\ \ \ (to H. M. Wiseman \& E. G. Cavalcanti)} \label{Wiseman23} \label{Cavalcanti6}

In case you were wondering about the last note's title, it was referring to:
\bhw
Thus in the end, the difference comes down to the fact that quantum measurements
necessarily induce disturbance (of our probabilities), which is indeed an old lesson in QM
that we don't need Bell's theorem for.
\ehw
but I had forgotten to mention that.

\section{29-06-09 \ \ {\it Eric's Note}\ \ \ (to E. G. Cavalcanti \& H. M. Wiseman)} \label{Wiseman24} \label{Cavalcanti7}

Thanks for defending me as a ``metaphysical realist in the sense that he believes in the existence of a world outside of oneself.''  Rarely do I have a friend strong enough to tread those waters!

\bec
Now I think really Chris should say that being a Bayesian does not a
priori entail anything about ontology whatsoever, since that is just a
statement about epistemology.
\eec

This is certainly true, and I would not have wanted to convey otherwise.  This is why I have the attached slide prepared for my talks:  It is meant to express that one can equally validly talk of one's Bayesian degrees of belief about ``states of pre-existent reality'' as one can talk of one's Bayesian degrees of belief about the ``consequences of one's actions.''  Nothing about the {\it raw\/} structure of probability theory changes in the move from one subject matter to the other.  It just so happens---from our point of view---that quantum mechanics represents a use of the latter type.

But that is Bayesianism per se.  It still leaves me room to disagree with this:
\bec
It is just saying that QM is a theory of epistemology, and that this
leaves open the question of ontology.
\eec
And that is because I don't view QM as an expression of raw probability theory.  It is {\it dressed\/} so to speak with an extra normative assumption concerning how to gamble on factuals in terms of counterfactuals (the urgleichung being one such example), and consequently gives in part a theory of priors (i.e., the convex set of states on the simplex).  I imagine the adoption of this dressing as a statement supervening on an ontology yet to be completely fleshed out.  So, QM as a theory of decisions still has an essential empirical component.

\bec
I think Chris seems to want to say something about ontology, and I
think he really wants to say something that sounds like a relational ontology.
(Whatever that is). Maybe it would be easier if he were explicit about it?
\eec
You don't much (if ever) hear me say things about ``relational ontologies'' because right now that doesn't seem like the right direction to develop to me.  This is one of the reasons I was so intrigued by our first conversation together:  You seemed to get that distinction in my thought.  You asked something like, ``Do two quantum Bayesians live in the same spacetime?,'' and I answered something (mystical) like, ``Only when they interact with each other.''  More seriously, I am much taken with the image of a ``republican banquet'' kind of ontology, like that expressed in the two quotes by James and the one by Durant in Section 8.1 of the paper.  Relational ontologies, to the extent that anyone has sketched any detail in them to me, always seem to rely on an image of a static, block-world lacking any good notion of struggle and creation.  Things just are, though now the things that are are relations between nodes rather than facts residing on nodes.  That's not a picture that attracts me much.  The other night {\Ruediger} and I were semi-jokingly saying that the dimension $d$ of a quantum object quantifies how much ``will'' it has.  But, maybe some variant of that will not turn out to be so much of a joke after all.

Thus ends my third installment.  Tomorrow comes the difficult one:  Whether Howard Wiseman indeed exists.

I bid you good night.

\section{29-06-09 \ \ {\it Relationalism vs.\ the Republican Banquet} \ \ (to C. Rovelli)} \label{Rovelli2}

Rob Spekkens walked into my office a few minutes ago with some very disappointing news:  You won't be coming to our PIAF meeting this year, just as you didn't come last year.  I suspect you won't reconsider, but I want you to anyway:  If you want the seeds of relationalism you have planted in quantum interpretation to thrive and grow, there is no more fertile ground than here at PI.  We really, truly need to hear your voice on the subject.  Discussion would proceed so much faster that way.  Please do reconsider!

It's funny too that Rob walked in as I had just sent off the note below to a few colleagues who are reading a paper I posted a couple of weeks ago: \arxiv{0906.2187}.  [See 29-06-09 note ``\myref{Wiseman24}{Eric's Note}'' to E. G. Cavalcanti \& H. M. Wiseman.]  I don't know that you would be interested in the whole thing (the paper), but you might find some amusing things in Sections 8 and 8.1 with regard to the perennial issues.  The note below addresses whether the ontology I'm shooting for is a relational one.  At the moment, I don't see that it would be, as Eric Cavalcanti had suggested.  I'd be curious to get your reaction if you find anything you can pinpoint.  And it'd be better if I could finally lure you here to hear it in person!

\section{30-06-09 \ \ {\it A New Name for Some Old Ways of Thinking}\ \ \ (to M. Schloss\-hauer)} \label{Schlosshauer-newname}

\bmaxs
In your note to van Fraassen entitled ``Canned Answers,'' you mention a file
{\tt certainty.pdf}. Might this be helpful to me too in bringing about further
enlightenment?
\emaxs

I'll do better than that:  Let me point you to pages 330--335 of ``My Struggles''---you'll find there a couple of notes titled ``Utter Rubbish and Internal Consistency'' (\myref{Mermin86}{Parts I} and \myref{Mermin87}{II}).  The long de Finetti quote is in there, but also I bear my soul and pay penance for the greatest mistake of my Bayesian life.  It's a bit embarrassing really \ldots\  but maybe there is something to be learned from it.

\bmaxs
I believe you know everything you need to know about decoherence, but if
there's anything I can help answer or comment on, let me know. Also, if
you ever would like a (free) copy of my book on decoherence, I'm happy to
have Springer send you one.
\emaxs
Yes, that would be most helpful for the upcoming project.  I would be very proud to have a copy!  You can use the address below.

\bmaxs
Finally, I'm quite interested in getting into some William James. If you
had to choose and recommend just one of his books, which one would it be?
\emaxs
That's an easy choice.  It is the book titled {\sl Pragmatism}.  It conveys all the spirit, while leaving aside all the dry technical arguments (of Dewey and Schiller predominantly).  Below is a collection my favorite quotes from the book.

Happy reading, and tell me what you think in the end.

\bq
\noindent W.~James, ``The Present Dilemma in Philosophy,'' in his {\sl
Pragmatism, a New Name for Some Old Ways of Thinking: Popular
Lectures on Philosophy}, (Longmans, Green and Co., New York, 1922).

\bq
\indent
For a hundred and fifty years past the progress of science has
seemed to mean the enlargement of the material universe and the
diminution of man's importance. The result is what one may call the
growth of naturalistic or positivistic feeling. Man is no lawgiver
to nature, he is an absorber. She it is who stands firm; he it is
who must accommodate himself. Let him record truth, inhuman though
it be, and submit to it! The romantic spontaneity and courage are
gone, the vision is materialistic and depressing. Ideals appear as
inert by-products of physiology; what is higher is explained by what
is lower and treated forever as a case of `nothing but'---nothing but
something else of a quite inferior sort. You get, in short, a
materialistic universe, in which only the tough-minded find
themselves congenially at home.
\eq

\noindent W.~James, ``What Pragmatism Means,'' in his {\sl Pragmatism, a New
Name for Some Old Ways of Thinking: Popular Lectures on Philosophy},
(Longmans, Green and Co., New York, 1922).

\bq
\indent
Metaphysics has usually followed a very primitive kind quest. You
know how men have always hankered after unlawful magic, and you know
what a great part in magic {\it words\/} have always played. If you
have his name, the formula of incantation that binds him, you can
control the spirit, genie, afrite, or whatever the power may be.
Solomon knew the names of all the spirits, and having their names, he
held them subject to his will. So the universe has always appeared to
the natural mind as a kind of enigma, of which the key must be sought
in the shape of some illuminating or power-bringing word or name.
That word names the universe's principle, and to possess it is after
a fashion to possess the universe itself. `God,' `Matter,' `Reason,'
`the Absolute,' `Energy,' are so many solving names. You can rest
when you have them. You are at the end of your metaphysical quest.

But if you follow the pragmatic method, you cannot look on any such
word as closing your quest. You must bring out of each word its
practical cash-value, set it at work within the stream of your
experience. It appears less as a solution, then, than as a program
for more work, and more particularly as an indication of the ways in
which existing realities may be {\it changed}.

{\it Theories thus become instruments, not answers to enigmas, in
which we can rest}. We don't lie back upon them, we move forward,
and, on occasion, make nature over again by their aid. Pragmatism
unstiffens all our theories, limbers them up and sets each one at
work.
\eq

\noindent W.~James, ``Some Metaphysical Problems Pragmatically Considered,'' in
his {\sl Pragmatism, a New Name for Some Old Ways of Thinking:
Popular Lectures on Philosophy}, (Longmans, Green and Co., New York,
1922).

\bq
\indent
Let me take up another well-worn controversy, {\it the free-will
problem}. Most persons who believe in what is called their free-will
do so after the rationalistic fashion. It is a principle, a positive
faculty or virtue added to man, by which his dignity is enigmatically
augmented. He ought to believe it for this reason. Determinists, who
deny it, who say that individual men originate nothing, but merely
transmit to the future the whole push of the past cosmos of which
they are so small an expression, diminish man. He is less admirable,
stripped of this creative principle. I imagine that more than half of
you share our instinctive belief in free-will, and that admiration of
it as a principle of dignity has much to do with your fidelity.

But free-will has also been discussed pragmatically, and, strangely
enough, the same pragmatic interpretation has been put upon it by
both disputants. You know how large a part questions of {\it
accountability\/} have played in ethical controversy. To hear some
persons, one would suppose that all that ethics aims at is a code of
merits and demerits. Thus does the old legal and theological leaven,
the interest in crime and sin and punishment abide with us. `Who's to
blame? whom can we punish? whom will God punish?---these
preoccupations hang like a bad dream over man's religious history.

So both free-will and determinism have been inveighed against and
called absurd, because each, in the eyes of its enemies, has seemed
to prevent the `imputability' of good or bad deeds to their authors.
Queer antinomy this! Free-will means novelty, the grafting on to the
past of something not involved therein. If our acts were
predetermined, if we merely transmitted the push of the whole past,
the free-willists say, how could we be praised or blamed for
anything? We should be `agents' only, not `principals,' and where
then would be our precious imputability and responsibility?

But where would it be if we {\it had\/} free-will? rejoin the
determinists. If a `free' act be a sheer novelty, that comes not {\it
from\/} me, the previous me, but {\it ex nihilo}, and simply tacks
itself on to me, how can {\it I}, the previous I, be responsible? How
can I have any permanent {\it character\/} that will stand still long
enough for praise or blame to be awarded? The chaplet of my days
tumbles into a cast of disconnected beads as soon as the thread of
inner necessity is drawn out by the preposterous indeterminist
doctrine. Messrs.\ Fullerton and McTaggart have recently laid about
them doughtily with this argument.

It may be good {\it ad hominem}, but otherwise it is pitiful. For I
ask you, quite apart from other reasons, whether any man, woman or
child, with a sense for realities, ought not to be ashamed to plead
such principles as either dignity or imputability. Instinct and
utility between them can safely be trusted to carry on the social
business of punishment and praise. If a man does good acts we shall
praise him, if he does bad acts we shall punish him,---anyhow, and
quite apart from theories as to whether the acts result from what was
previous in him or are novelties in a strict sense. To make our human
ethics revolve about the question of `merit' is a piteous
unreality---God alone can know our merits, if we have any. The real
ground for supposing free-will is indeed pragmatic, but it has
nothing to do with this contemptible right to punish which has made
such a noise in past discussions of the subject.

Free-will pragmatically means {\it novelties in the world}, the right
to expect that in its deepest elements as well as in its surface
phenomena, the future may not identically repeat and imitate the
past. That imitation {\it en masse\/} is there, who can deny? The
general `uniformity of nature' is presupposed by every lesser law.
But nature may be only approximately uniform; and persons in whom
knowledge of the world's past has bred pessimism (or doubts as to the
world's good character, which become certainties if that character be
supposed eternally fixed) may naturally welcome free-will as a {\it
melioristic\/} doctrine. It holds up improvement as at least
possible; whereas determinism assures us that our whole notion of
possibility is born of human ignorance, and that necessity and
impossibility between them rule the destinies of the world.

Free-will is thus a general cosmological theory of {\it promise},
just like the Absolute, God, Spirit or Design. Taken abstractly, no
one of these terms has any inner content, none of them gives us any
picture, and no one of them would retain the least pragmatic value in
a world whose character was obviously perfect from the start. Elation
at mere existence, pure cosmic emotion and delight, would, it seems
to me, quench all interest in those speculations, if the world were
nothing but a lubberland of happiness already. Our interest in
religious metaphysics arises in the fact that our empirical future
feels to us unsafe, and needs some higher guarantee. If the past and
present were purely good, who could wish that the future might
possibly not resemble them? Who could desire free-will? Who would not
say, with Huxley, `let me be wound up every day like a watch, to go
right fatally, and I ask no better freedom.' `Freedom' in a world
already perfect could only mean freedom to {\it be worse}, and who
could be so insane as to wish that? To be necessarily what it is, to
be impossibly aught else, would put the last touch of perfection upon
optimism's universe. Surely the only {\it possibility\/} that one can
rationally claim is the possibility that things may be {\it better}.
That possibility, I need hardly say, is one that, as the actual world
goes, we have ample grounds for desiderating.

Free-will thus has no meaning unless it be a doctrine of  {\it
relief}. As such, it takes its place with other religious doctrines.
Between them, they build up the old wastes and repair the former
desolations. Our spirit, shut within this courtyard of
sense-experience, is always saying to the intellect upon the tower:
`Watchman, tell us of the night, if it aught of promise bear,' and
the intellect gives it then these terms of promise.

Other than this practical significance, the words God, free-will,
design, etc., have none. Yet dark tho they be in themselves, or
intellectualistically taken, when we bear them into life's thicket
with us the darkness {\it there\/} grows light about us. If you stop,
in dealing with such words, with their definition, thinking that to
be an intellectual finality, where are you? Stupidly staring at a
pretentious sham! ``Deus est Ens, a se, extra et supra omne genus,
necessarium, unum, infinite perfectum, simplex, immutabile, immensum,
aeternum, intelligens,'' etc.,---wherein is such a definition really
instructive? It means less than nothing, in its pompous robe of
adjectives. Pragmatism alone can read a positive meaning into it, and
for that she turns her back upon the intellectualist point of view
altogether. `God's in his heaven; all's right with the world!'---{\it
That's\/} the real heart of your theology, and for that you need no
rationalist definitions.

Why shouldn't all of us, rationalists as well as pragmatists, confess
this? Pragmatism, so far from keeping her eyes bent on the immediate
practical, foreground, as she is accused of doing, dwells just as
much upon the world's remotest perspectives.

See then how all these ultimate questions turn, as it were, upon
their hinges; and from looking backwards upon principles, upon an
{\it erkenntnisstheoretische Ich}, a God, {\it a
Kausalit\"atsprinzip}, a Design, a Free-will, taken in themselves, as
something august and exalted above facts,---see, I say, how
pragmatism shifts the emphasis and looks forward into facts
themselves. The really vital question for us all is, What is this
world going to be? What is life eventually to make of itself? The
centre of gravity of philosophy must therefore alter its place. The
earth of things, long thrown into shadow by the glories of the upper
ether, must resume its rights. To shift the emphasis in this way
means that philosophic questions will fall to be treated by minds of
a less abstractionist type than heretofore, minds more scientific and
individualistic in their tone yet not irreligious either. It will be
an alteration in `the seat of authority' that reminds one almost of
the protestant reformation. And as, to papal minds, protestantism has
often seemed a mere mess of anarchy and confusion, such, no doubt,
will pragmatism often seem to ultra-rationalist minds in philosophy.
It will seem so much sheer trash, philosophically. But life wags on,
all the same, and compasses its ends, in protestant countries. I
venture to think that philosophic protestantism will compass a not
dissimilar prosperity.
\eq

\noindent W.~James, ``The One and the Many,'' in his {\sl Pragmatism, a New
Name for Some Old Ways of Thinking: Popular Lectures on Philosophy},
(Longmans, Green and Co., New York, 1922).

\bq
\indent
It is possible to imagine alternative universes to the one we know,
in which the most various grades and types of union should be
embodied. Thus the lowest grade of universe would be a world of mere
{\it withness}, of  which the parts were only strung together by the
conjunction `and.' Such a universe is even now the collection of our
several inner lives. The spaces and times of your imagination, the
objects and events of your day-dreams are not only more or less
incoherent {\it inter se}, but are wholly out of definite relation
with the similar contents of any one else's mind. Our various
reveries now as we sit here compenetrate each other idly without
influencing or interfering. They coexist, but in no order and in no
receptacle, being the nearest approach to an absolute `many' that we
can conceive. We can not even imagine any reason why they {\it
should\/} be known all together, and we can imagine even less, if
they were known together, how they could be known as one systematic
whole.

But add our sensations and bodily actions, and the union mounts to a
much higher grade. Our {\it audita et visa\/} and our acts fall into
those receptacles of time and space in which each event finds its
date and place. They form `things' and are of `kinds' too, and can be
classed. Yet we can imagine a world of things and of kinds in which
the causal interactions with which we are so familiar should not
exist. Everything there might be inert towards everything else, and
refuse to propagate its influence. Or gross mechanical influences
might pass, but no chemical action. Such worlds would be far less
unified than ours. Again there might be complete physico-chemical
interaction, but no minds; or minds, but altogether private ones,
with no social life; or social life limited to acquaintance, but no
love; or love, but no customs or institutions that should systematize
it. No one of these grades of universe would be absolutely irrational or
disintegrated, inferior tho it might appear when looked at from the
higher grades. For instance, if our minds should ever become
`telepathically' connected, so that we knew immediately, or could
under certain conditions know immediately, each what the other was
thinking, the world we now live in would appear to the thinkers in
that world to have been of an inferior grade.

With the whole of past eternity open for our conjectures to range in,
it may be lawful to wonder whether the various kinds of union now
realized in the universe that we inhabit may not possibly have been
successively evolved after the fashion in which we now see human
systems evolving in consequence of human needs. If such an hypothesis
were legitimate, total oneness would appear at the end of things
rather than at their origin. In other words the notion of the
`Absolute' would have to be replaced by that of the `Ultimate.' The
two notions would have the same content---the maximally unified
content of fact, namely---but their time-relations would be
positively reversed.

After discussing the unity of the universe in this pragmatic way, you
ought to see why I said in my second lecture, borrowing the word from
my friend G. Papini, that pragmatism tends to {\it unstiffen\/} all
our theories. The world's oneness has generally been affirmed
abstractly only, and as if any one who questioned it must be an
idiot. The temper of monists has been so vehement, as almost at times
to be convulsive; and this way of holding a doctrine does not easily
go with reasonable discussion and the drawing of distinctions. The
theory of the Absolute, in particular, has had to be an article of
faith, affirmed dogmatically and exclusively. The One and All, first
in the order of being and of knowing, logically necessary itself, and
uniting all lesser things in the bonds of mutual necessity, how could
it allow of any mitigation of its inner rigidity? The slightest
suspicion of pluralism, the minutest wiggle of independence of any
one of its parts from the control of the totality would ruin it.
Absolute unity brooks no degrees,---as well might you claim absolute
purity for a glass of water because it contains but a single little
cholera-germ. The independence, however infinitesimal, of a part,
however small, would be to the Absolute as fatal as a cholera-germ.

Pluralism on the other hand has no need of this dogmatic rigoristic
temper. Provided you grant {\it some\/} separation among things, some
tremor of independence, some free play of parts on one another, some
real novelty or chance, however minute, she is amply satisfied, and
will allow you any amount, however great, of real union. How much of
union there may be is a question that she thinks can only be decided
empirically. The amount may be enormous, colossal; but absolute
monism is shattered if, along with all the union, there has to be
granted the slightest modicum, the most incipient nascency, or the
most residual trace, of a separation that is not `overcome.'

Pragmatism, pending the final empirical ascertainment of just what
the balance of union and disunion among things may be, must obviously
range herself upon the pluralistic side. Some day, she admits, even
total union, with one knower, one origin, and a universe consolidated
in every conceivable way, may turn out to be the most acceptable of
all hypotheses. Meanwhile the opposite hypothesis, of a world
imperfectly unified still, and perhaps always to remain so, must be
sincerely entertained. This latter hypothesis is pluralism's
doctrine. Since absolute monism forbids its being even considered
seriously, branding it as irrational from the start, it is clear that
pragmatism must turn its back on absolute monism, and follow
pluralism's more empirical path.

This leaves us with the common-sense world, in which we find things
partly joined and partly disjoined.
\eq

\noindent W.~James, ``Pragmatism and Common Sense,'' in his {\sl Pragmatism, a
New Name for Some Old Ways of Thinking: Popular Lectures on
Philosophy}, (Longmans, Green and Co., New York, 1922).

\bq
\indent
But the scientific tendency in critical thought, tho inspired at
first by purely intellectual motives, has opened an entirely
unexpected range of practical utilities to our astonished view.
Galileo gave us accurate clocks and accurate artillery-practice; the
chemists flood us with new medicines and dye-stuffs; Amp\`ere and
Faraday have endowed us with the New York subway and with Marconi
telegrams. The hypothetical things that such men have invented,
defined as they have defined them, are showing an extraordinary
fertility in consequences verifiable by sense. Our logic can deduce
from them a consequence due under certain conditions, we can then
bring about the conditions, and presto, the consequence is there
before our eyes. The scope of the practical control of nature newly
put into our hand by scientific ways of thinking vastly exceeds the
scope of the old control grounded on common sense. Its rate of
increase accelerates so that no one can trace the limit; one may even
fear that the {\it being\/} of man may be crushed by his own powers,
that his fixed nature as an organism may not prove adequate to stand
the strain of the ever increasingly tremendous functions, almost
divine creative functions, which his intellect will more and more
enable him to wield. He may drown in his wealth like a child in a
bath-tub, who has turned on the water and who can not turn it off.
\eq
and
\bq
\indent
The most fateful point of difference between being a rationalist and
being a pragmatist is now fully in sight. Experience is in mutation,
and our psychological ascertainments of truth are in mutation---so
much rationalism will allow; but never that either reality itself or
truth itself is mutable. Reality stands complete and ready-made from
all eternity, rationalism insists, and the agreement of our ideas
with it is that unique unanalyzable virtue in them of which she has
already told us. As that intrinsic excellence, their truth has
nothing to do with our experiences. It adds nothing to the content of
experience. It makes no difference to reality itself; it is
supervenient, inert, static, a reflexion merely. It doesn't {\it
exist}, it {\it holds\/} or {\it obtains}, it belongs to another
dimension from that of either facts or fact-relations, belongs, in
short, to the epistemological dimension---and with that big word
rationalism closes the discussion.

Thus, just as pragmatism faces forward to the future, so does
rationalism here again face backward to a past eternity. True to her
inveterate habit, rationalism reverts to `principles,' and thinks
that when an abstraction once is named, we own an oracular solution.

The tremendous pregnancy in the way of consequences for life of this
radical difference of outlook will only become apparent in my later
lectures.
\eq

\noindent W.~James, ``Pragmatism's Conception of Truth,'' in his {\sl
Pragmatism, a New Name for Some Old Ways of Thinking: Popular
Lectures on Philosophy}, (Longmans, Green and Co., New York, 1922).

\bq
\indent
When Clerk-Maxwell was a child it is written that he had a mania for
having everything explained to him, and that when people put him off
with vague verbal accounts of any phenomenon he would interrupt them
impatiently by saying, `Yes; but I want you to tell me the {\it
particular go\/} of it!' Had his question been about truth, only a
pragmatist could have told him the particular go of it. I believe
that our contemporary pragmatists, especially Messrs.\ Schiller and
Dewey, have given the only tenable account of this subject. It is a
very ticklish subject, sending subtle rootlets into all kinds of
crannies, and hard to treat in the sketchy way that alone befits a
public lecture. But the Schiller-Dewey view of truth has been so
ferociously attacked by rationalistic philosophers, and so abominably
misunderstood, that here, if anywhere, is the point where a clear and
simple statement should be made.

I fully expect to see the pragmatist view of truth run through the
classic stages of a theory's career. First, you know, a new theory is
attacked as absurd; then it is admitted to be true, but obvious and
insignificant; finally it is seen to be so important that its
adversaries claim that they themselves discovered it. Our doctrine of
truth is at present in the first of these three stages, with symptoms
of the second stage having begun in certain quarters. I wish that
this lecture might help it beyond the first stage in the eyes of many
of you.

Truth, as any dictionary will tell you, is a property of certain of
our ideas. It means their `agreement,' as falsity means their
disagreement, with `reality.' Pragmatists and intellectualists both
accept this definition as a matter of course. They begin to quarrel
only after the question is raised as to what may precisely be meant
by the term `agreement,' and what by the term `reality,' when reality
is taken as something for our ideas to agree with.

In answering these questions the pragmatists are more analytic and
pains\-taking, the intellectualists more offhand and irreflective. The
popular notion is that a true idea must copy its reality. Like other
popular views, this one follows the analogy of the most usual
experience. Our true ideas of sensible things do indeed copy them.
Shut your eyes and think of yonder clock on the wall, and you get
just such a true picture or copy of its dial. But your idea of its
`works' (unless you are a clockmaker) is much less of a copy, yet it
passes muster, for it in no way clashes with the reality. Even though
it should shrink to the mere word `works,' that word still serves you
truly; and when you speak of the `timekeeping function' of the clock,
or of its spring's `elasticity,' it is hard to see exactly what your
ideas can copy.

You perceive that there is a problem here. Where our ideas cannot
copy definitely their object, what does agreement with that object
mean? Some idealists seem to say that they are true whenever they are
what God means that we ought to think about that object. Others hold
the copy-view all through, and speak as if our ideas possessed truth
just in proportion as they approach to being copies of the Absolute's
eternal way of thinking.

These views, you see, invite pragmatistic discussion. But the great
assumption of the intellectualists is that truth means essentially an
inert static relation. When you've got your true idea of anything,
there's an end of the matter. You're in possession; you {\it know};
you have fulfilled your thinking destiny. You are where you ought to
be mentally; you have obeyed your categorical imperative; and nothing
more need follow on that climax of your rational destiny.
Epistemologically you are in stable equilibrium.

Pragmatism, on the other hand, asks its usual question. ``Grant an
idea or belief to be true,'' it says, ``what concrete difference will
its being true make in any one's actual life? How will the truth be
realized? What experiences will be different from those which would
obtain if the belief were false? What, in short, is the truth's
cash-value in experiential terms?''

The moment pragmatism asks this question, it sees the answer: {\it
True ideas are those that we can assimilate, validate, corroborate
and verify. False ideas are those that we can not}. That is the
practical difference it makes to us to have true ideas; that,
therefore, is the meaning of truth, for it is all that truth is known
as.

This thesis is what I have to defend. The truth of an idea is not a
stagnant property inherent in it. Truth {\it happens\/} to an idea.
It {\it becomes\/} true, is {\it made\/} true by events. Its verity
{\it is\/} in fact an event, a process: the process namely of its
verifying itself, its veri-{\it fication}. Its validity is the
process of its valid-{\it ation}.
\eq

\noindent W.~James, ``Pragmatism and Humanism,'' in his {\sl Pragmatism, a New
Name for Some Old Ways of Thinking: Popular Lectures on Philosophy},
(Longmans, Green and Co., New York, 1922).

\bq
\indent
What hardens the heart of every one I approach with the view of truth
sketched in my last lecture is that typical idol of the tribe, the
notion of {\it the\/} Truth, conceived as the one answer, determinate
and complete, to the one fixed enigma which the world is believed to
propound. For popular tradition, it is all the better if the answer
be oracular, so as itself to awaken wonder as an enigma of the second
order, veiling rather than revealing what its profundities are
supposed to contain. All the great single-word answers to the world's
riddle, such as God, the One, Reason, Law, Spirit, Matter, Nature,
Polarity, the Dialectic Process, the Idea, the Self, the Oversoul,
draw the admiration that men have lavished on them from this oracular
r\^{o}le. By amateurs in philosophy and professionals alike, the
universe is represented as a queer sort of petrified sphinx whose
appeal to men consists in a monotonous challenge to his divining
powers. {\it The\/} Truth: what a perfect idol of the rationalistic
mind! I read in an old letter---from a gifted friend who died too
young---these words: ``In everything, in science, art, morals and
religion, there {\it must\/} be one system that is right and {\it
every\/} other wrong.'' How characteristic of the enthusiasm of a
certain stage of youth! At twenty-one we rise to such a challenge and
expect to find the system. It never occurs to most of us even later
that the question `what is {\it the\/} truth?'\ is no real question
(being irrelative to all conditions) and that the whole notion of
{\it the\/} truth is an abstraction from the fact of truths in the
plural, a mere useful summarizing phrase like {\it the\/} Latin
Language or {\it the\/} Law.

Common-law judges sometimes talk about the law, and schoolmasters
talk about the latin tongue, in a way to make their hearers think
they mean entities preexistent to the decisions or to the words and
syntax, determining them unequivocally and requiring them to obey.
But the slightest exercise of reflexion makes us see that, instead of
being principles of this kind, both law and latin are results.
Distinctions between the lawful and the unlawful in conduct, or
between the correct and incorrect in speech, have grown up
incidentally among the interactions of men's experiences in detail;
and in no other way do distinctions between the true and the false in
belief ever grow up. Truth grafts itself on previous truth, modifying
it in the process, just as idiom grafts itself on previous idiom, and
law on previous law. Given previous law and a novel case, and the
judge will twist them into fresh law. Previous idiom; new slang or
metaphor or oddity that hits the public taste;---and presto, a new
idiom is made. Previous truth; fresh facts:---and our mind finds a
new truth.

All the while, however, we pretend that the eternal is unrolling,
that the one previous justice, grammar or truth are simply
fulgurating and not being made. But imagine a youth in the courtroom
trying cases with his abstract notion of `the' law, or a censor of
speech let loose among the theatres with his idea of `the'
mother-tongue, or a professor setting up to lecture on the actual
universe with his rationalistic notion of `the Truth' with a big T,
and what progress do they make? Truth, law, and language fairly boil
away from them at the least touch of novel fact. These things {\it
make themselves\/} as we go. Our rights, wrongs, prohibitions,
penalties, words, forms, idioms, beliefs, are so many new creations
that add themselves as fast as history proceeds. Far from being
antecedent principles that animate the process, law, language, truth
are but abstract names for its results.

Laws and languages at any rate are thus seen to be man-made things.
Mr.\ Schiller applies the analogy to beliefs, and proposes the name
of `Humanism' for the doctrine that to an unascertainable extent our
truths are man-made products too. Human motives sharpen all our
questions, human satisfactions lurk in all our answers, all our
formulas have a human twist. This element is so inextricable in the
products that Mr.\ Schiller sometimes seems almost to leave it an
open question whether there be anything else. ``The world,'' he says,
``is essentially $\upsilon\lambda\eta$, it is what we make it. It is
fruitless to define it by what it originally was or by what it is
apart from us; it {\it is\/} what is made of it. Hence \ldots\ the
world is plastic.'' He adds that we can learn the limits of the
plasticity only by trying, and that we ought to start as if it were
wholly plastic, acting methodically on that assumption, and stopping
only when we are decisively rebuked.

This is Mr.\ Schiller's butt-end-foremost statement of the humanist
position, and it has exposed him to severe attack. I mean to defend
the humanist position in this lecture, so I will insinuate a few
remarks at this point.

Mr.\ Schiller admits as emphatically as any one the presence of
resisting factors in every actual experience of truth-making, of
which the new-made special truth must take account, and with which it
has perforce to `agree.' All our truths are beliefs about `Reality';
and in any particular belief the reality acts as something
independent, as a thing {\it found}, not manufactured. Let me here
recall a bit of my last lecture.

{\it `Reality' is in general what truths have to take account of;\/}
and the {\it first\/} part of reality from this point of view is the
flux of our sensations. Sensations are forced upon us, coming we know
not whence. Over their nature, order and quantity we have as good as
no control. {\it They\/} are neither true nor false; they simply {\it
are}. It is only what we say about them, only the names we give them,
our theories of their source and nature and remote relations, that
may be true or not.

The {\it second\/} part of reality, as something that our beliefs
must also obediently take account of is the {\it relations\/} that
obtain between our sensations or between their copies in our minds.
This part falls into two subparts: 1) the relations that are mutable
and accidental, as those of date and place; and 2) those that are
fixed and essential because they are grounded on the inner natures of
their terms. Both sorts of relation are matters of immediate
perception. Both are `facts.' But it is the latter kind of fact that
forms the more important subpart of reality for our theories of
knowledge. Inner relations namely are `eternal,' are perceived
whenever their sensible terms are compared; and of them our
thought---mathematical and logical thought so-called---must eternally
take account.

The {\it third\/} part of reality, additional to these perceptions
(tho largely based upon them), is the {\it previous truths\/} of
which every new inquiry takes account. This third part is a much less
obdurately resisting factor: it often ends by giving way. In speaking
of these three portions of reality as at all times controlling our
beliefs formation, I am only reminding you of what we heard in our
last hour.

Now however fixed these elements of reality may be, we still have a
certain freedom in our dealings with them. Take our sensations. {\it
That\/} they are is undoubtedly beyond our control; but {\it which\/}
we attend to, note, and make emphatic in our conclusions depends on
our own interests; and, according as we lay the emphasis here or
there, quite different formulations of truth result. We read the same
facts differently. `Waterloo,' with the same fixed details, spells a
`victory' for an Englishman; for a Frenchman it spells a `defeat.'
So, for an optimist philosopher the universe spells victory, for a
pessimist, defeat.

What we say about reality thus depends on the perspective into which
we throw it. The {\it that\/} of it is its own; but the {\it what\/}
depends on the {\it which\/}; and the which depends on {\it us}. Both
the sensational and the relational parts of reality are dumb; they
say absolutely nothing about themselves. We it is who have to speak
for them. This dumbness of sensations has led such intellectualists
as T.~H. Green and Edward Caird to shove them almost beyond the pale
of philosophic recognition, but pragmatists refuse to go so far. A
sensation is rather like a client who has given his case to a lawyer
and then has passively to listen in the courtroom to whatever account
of his affairs, pleasant or unpleasant, the lawyer finds it most
expedient to give.

Hence, even in the field of sensation, our minds exert a certain
arbitrary choice. By our inclusions and omissions we trace the
field's extent; by our emphasis we mark its foreground and its
background; by our order we read it in this direction or in that. We
receive in short the block of marble, but we carve the statue
ourselves.

This applies to the `eternal' parts of reality as well: we shuffle
our perceptions of intrinsic relation and arrange them just as
freely. We read them in one serial order or another, class them in
this way or in that, treat one or the other as more fundamental,
until our beliefs about them form those bodies of truth known as
logics, geometrics, or arithmetics, in each and all of which the form
and order in which the whole is cast is flagrantly man-made.

Thus, to say nothing of the new {\it facts\/} which men add to the
matter of reality by the acts of their own lives, they have already
impressed their mental forms on that whole third of reality which I
have called `previous truths.' Every hour brings its new percepts,
its own facts of sensation and relation, to be truly taken account
of; but the whole of our {\it past\/} dealings with such facts is
already funded in the previous truths. It is therefore only the
smallest and recentest fraction of the first two parts of reality
that comes to us without the human touch, and that fraction has
immediately to become humanized in the sense of being squared,
assimilated, or in some way adapted, to the humanized mass already
there., As a matter of fact we can hardly take in an impression at
all, in the absence of a preconception of what impressions there may
possibly be.

When we talk of reality `independent' of human thinking, then, it
seems a thing very hard to find. It reduces to the notion of what is
just entering into experience and yet to be named, or else to some
imagined aboriginal presence in experience, before any belief about
the presence had arisen, before any human conception had been
applied. It is what is absolutely dumb and evanescent, the merely
ideal limit of our minds. We may glimpse it, but we never grasp it;
what we grasp is always some substitute for it which previous human
thinking has peptonized and cooked for our consumption. If so vulgar
an expression were allowed us, we might say that wherever we find it,
it has been already {\it faked}. This is what Mr.\ Schiller has in
mind when he calls independent reality a mere unresisting
$\upsilon\lambda\eta$, which is only to be made over by us.
\eq
and
\pagebreak
\bq
\indent
In many familiar objects every one will recognize the human element.
We conceive a given reality in this way or in that, to suit our
purpose, and the reality passively submits to the conception. You can
take the number 27 as the cube of 3, or as the product of 3 and 9, or
as 26 {\it plus\/} 1, or 100 {\it minus\/} 73, or in countless other
ways, of which one will be just as true as another. You can take a
chess-board as black squares on a white ground, or as white squares
on a black ground, and neither conception is a false one.

You can treat the adjoined figure as a star, as two big triangles
crossing each other, as a hexagon with legs set up on its angles, as
six equal triangles hanging together by their tips, etc. All these
treatments are true treatments---the sensible {\it that\/} upon the
paper resists no one of them. You can say of a line that it runs
east, or you can say that it runs west, and the line {\it per se\/}
accepts both descriptions without rebelling at the inconsistency.

We carve out groups of stars in the heavens, and call them
constellations, and the stars patiently suffer us to do so,---{\it
though\/} if they knew what we were doing, some of them might feel
much surprised at the partners we had given them. We name the same
constellation diversely, as Charles's Wain, the Great Bear, or the
Dipper. None of the names will be false, and one will be as true as
another, for all are applicable.

In all these cases we humanly make an {\it addition\/} to some
sensible reality, and that reality tolerates the addition. All the
additions `agree' with the reality; they fit it, while they build it
out. No one of them is false. Which may be treated as the {\it
more\/} true, depends altogether on the human use of it. If the 27 is
a number of dollars which I find in a drawer where I had left 28, it
is 28 minus 1. If it is the number of inches in a board which I wish
to insert as a shelf into a cupboard 26 inches wide, it is 26 plus 1.
If I wish to ennoble the heavens by the constellations I see there,
`Charles's Wain' would be more true than `Dipper.' My friend
Frederick Myers was humorously indignant that that prodigious
star-group should remind us Americans of nothing but a culinary
utensil.

What shall we call a {\it thing\/} anyhow? It seems quite arbitrary,
for we carve out everything, just as we carve out constellations, to
suit our human purposes. For me, this whole `audience' is one thing,
which grows now restless, now attentive. I have no use at present for
its individual units, so I don't consider them. So of an `army,' of a
`nation.' But in your own eyes, ladies and gentlemen, to call you
`audience' is an accidental way of taking you. The permanently real
things for you are your individual persons. To an anatomist, again,
those persons are but organisms, and the real things are the organs.
Not the organs, so much as their constituent cells, say the
histologists; not the cells, but their molecules, say in turn the
chemists.

We break the flux of sensible reality into things, then, at our will.
We create the subjects of our true as well as of our false
propositions.

We create the predicates also. Many of the predicates of things
express only the relations of the things to us and to our feelings.
Such predicates of course are human additions. Caesar crossed the
Rubicon, and was a menace to Rome's freedom. He is also an American
schoolroom pest, made into one by the reaction of our schoolboys on
his writings. The added predicate is as true of him as the earlier
ones.

You see how naturally one comes to the humanistic principle: you
can't weed out the human contribution. Our nouns and adjectives are
all humanized heirlooms, and in the theories we build them into, the
inner order and arrangement is wholly dictated by human
considerations, intellectual consistency being one of them.
Mathematics and logic themselves are fermenting with human
rearrangements; physics, astronomy and biology follow massive cues of
preference. We plunge forward into the field of fresh experience with
the beliefs our ancestors and we have made already; these determine
what we notice; what we notice determines what we do; what we do
again determines what we experience; so from one thing to another,
altho the stubborn fact remains that there is a sensible flux, what
is {\it true of it\/} seems from first to last to be largely a matter
of our own creation.

We build the flux out inevitably. The great question is: does it,
with our additions, {\it rise or fall in value}? Are the additions
{\it worthy\/} or {\it unworthy}? Suppose a universe composed of
seven stars, and nothing else but three human witnesses and their
critic. One witness names the stars `Great Bear'; one calls them
`Charles's Wain'; one calls them the `Dipper.' Which human addition
has made the best universe of the given stellar material? If
Frederick Myers were the critic, he would have no hesitation in
`turning down' the American witness.

Lotze has in several places made a deep suggestion. We na\"{\i}vely
assume, he says, a relation between reality and our minds which may
be just the opposite of the true one. Reality, we naturally think,
stands ready-made and complete, and our intellects supervene with the
one simple duty of describing it as it is already. But may not our
descriptions, Lotze asks, be themselves important additions to
reality? And may not previous reality itself be there, far less for
the purpose of reappearing unaltered in our knowledge, than for the
very purpose of stimulating our minds to such additions as shall
enhance the universe's total value. {\it `Die Erh\"ohung des
vorgefundenen Daseins'\/}\footnote{H.~C. von Baeyer corrected the capitalization of this from James's original, and translated it as, ``the enhancement of what is found to exist.''} is a phrase used by Professor Eucken somewhere, which reminds one of this suggestion by the great Lotze.

It is identically our pragmatistic conception. In our cognitive as
well as in our active life we are creative. We {\it add}, both to the
subject and to the predicate part of reality. The world stands really
malleable, waiting to receive its final touches at our hands. Like
the kingdom of heaven, it suffers human violence willingly. Man {\it
engenders\/} truths upon it.

No one can deny that such a role would add both to our dignity and to
our responsibility as thinkers. To some of us it proves a most
inspiring notion. Signore Papini, the leader of Italian pragmatism,
grows fairly dithyrambic over the view that it opens of man's
divinely-creative functions.

The import of the difference between pragmatism and rationalism is
now in sight throughout its whole extent. The essential contrast is
that {\it for rationalism reality is ready-made and complete from all
eternity, while for pragmatism it is still in the making, and awaits
part of  its complexion from the future}. On the one side the
universe is absolutely secure, on the other it is still pursuing its
adventures.

We have got into rather deep water with this humanistic view, and it
is no wonder that misunderstanding gathers round it. It is accused of
being a doctrine of caprice. Mr.\ Bradley, for example, says that a
humanist, if he understood his own doctrine, would have to `hold any
end, however perverted, to be rational, if I insist on it personally,
and any idea, however mad, to be the truth if only some one is
resolved that he will have it so.' The humanist view of `reality,' as
something resisting, yet malleable, which controls our thinking as an
energy that must be taken `account' of incessantly (tho not
necessarily merely {\it copied}) is evidently a difficult one to
introduce to novices. The situation reminds me of one that I have
personally gone through. I once wrote an essay on our right to
believe, which I unluckily called the {\it Will\/} to Believe. All
the critics, neglecting the essay, pounced upon the title.
Psychologically it was impossible, morally it was iniquitous. The
`will to deceive,' the `will to make-believe,' were wittily proposed
as substitutes for it.

{\it The alternative between pragmatism and rationalism, in the shape
in which we now have it before us, is no longer a question in the
theory of knowledge, it concerns the structure of the universe
itself.}

On the pragmatist side we have only one edition of the universe,
unfinished, growing in all sorts of places, especially in the places
where thinking beings are at work.

On the rationalist side we have a universe in many editions, one real
one, the infinite folio, or {\it \'edition de luxe}, eternally
complete; and then the various finite editions, full of false
readings, distorted and mutilated each in its own way.
\eq

\noindent W.~James, ``Pragmatism and Religion,'' in his {\sl Pragmatism, a New
Name for Some Old Ways of Thinking: Popular Lectures on Philosophy},
(Longmans, Green and Co., New York, 1922).

\bq
\indent
Let us apply this notion to the salvation of the world. What does it
pragmatically mean to say that this is possible? It means that some
of the conditions of the world's deliverance do actually exist. The
more of them there are existent, the fewer preventing conditions you
can find, the better-grounded is the salvation's possibility, the
more {\it probable\/} does the fact of the deliverance become.

So much for our preliminary look at possibility.

Now it would contradict the very spirit of life to say that our minds
must be indifferent and neutral in questions like that of the world's
salvation. Any one who pretends to be neutral writes himself down
here as a fool and a sham. We all do wish to minimize the insecurity
of the universe; we are and ought to be unhappy when we regard it as
exposed to every enemy and open to every life-destroying draft.
Nevertheless there are unhappy men who think the salvation of the
world impossible. Theirs is the doctrine known as pessimism.

Optimism in turn would be the doctrine that thinks the world's
salvation inevitable.

Midway between the two there stands what may be called the doctrine
of meliorism, tho it has hitherto figured less as a doctrine than as
an attitude in human affairs. Optimism has always been the regnant
{\it doctrine\/} in European philosophy. Pessimism was only recently
introduced by Schopenhauer and counts few systematic defenders as
yet. Meliorism treats salvation as neither necessary nor impossible.
It treats it as a possibility, which becomes more and more of a
probability the more numerous the actual conditions of salvation
become.

It is clear that pragmatism must incline towards meliorism. Some
conditions of the world's salvation are actually extant, and she can
not possibly close her eyes to this fact: and should the residual
conditions come, salvation would become an accomplished reality.
Naturally the terms I use here are exceedingly summary. You may
interpret the word `salvation' in any way you like, and make it as
diffuse and distributive, or as climacteric and integral a phenomenon
as you please.

Take, for example, any one of us in this room with the ideals which
he cherishes and is willing to live and work for. Every such ideal
realized will be one moment in the world's salvation. But these
particular ideals are not bare abstract possibilities. They are
grounded, they are {\it live\/} possibilities, for we are their live
champions and pledges, and if the complementary conditions come and
add themselves, our ideals will become actual things. What now are
the complementary conditions? They are first such a mixture of things
as will in the fullness of time give us a chance, a gap that we can
spring into, and, finally, {\it our act}.

Does our act then {\it create\/} the world's salvation so far as it
makes room for itself, so far as it leaps into the gap? Does it
create, not the whole world's salvation of course, but just so much
of this as itself covers of the world's extent?

Here I take the bull by the horns, and in spite of the whole crew of
rationalists and monists, of whatever brand they be, I ask {\it why
not?} Our acts, our turning-places, where we seem to ourselves to
make ourselves and grow, are the parts of the world to which we are
closest, the parts of which our knowledge is the most intimate and
complete. Why should we not take them at their facevalue? Why may
they not be the actual turning-places and growing-places which they
seem to be, of the world---why not the workshop of being, where we
catch fact in the making, so that nowhere may the world grow in any
other kind of way than this?

Irrational!\ we are told. How can new being come in local spots and
patches which add themselves or stay away at random, independently of
the rest? There must be a reason for our acts, and where in the last
resort can any reason be looked for save in the material pressure or
the logical compulsion of the total nature of the world? There can be
but one real agent of growth, or seeming growth, anywhere, and that
agent is the integral world itself. It may grow all-over, if growth
there be, but that single parts should grow {\it per se\/} is
irrational.

But if one talks of rationality---and of reasons for things, and
insists that they can't just come in spots, what {\it kind\/} of a
reason can there ultimately be why anything should come at all? Talk
of logic and necessity and categories and the absolute and the
contents of the whole philosophical machine-shop as you will, the
only {\it real\/} reason I can think of why anything should ever come
is that {\it some one wishes it to be here}. It is {\it
demanded},---demanded, it may be, to give relief to no matter how
small a fraction of the world's mass. This is {\it living reason},
and compared with it material causes and logical necessities are
spectral things.

In short the only fully rational world would be the world of
wishing-caps, the world of telepathy, where every desire is fulfilled
instanter, without having to consider or placate surrounding or
intermediate powers. This is the Absolute's own world. He calls upon
the phenomenal world to be, and it is, exactly as he calls for it, no
other condition being required. In our world, the wishes of the
individual are only one condition. Other individuals are there with
other wishes and they must be propitiated first. So Being grows under
all sorts of resistances in this world of the many, and, from
compromise to compromise, only gets organized gradually into what may
be called secondarily rational shape. We approach the wishing-cap
type of organization only in a few departments of life. We want water
and we turn a faucet. We want a kodak-picture and we press a button.
We want information and we telephone. We want to travel and we buy a
ticket. In these and similar cases, we hardly need to do more than
the wishing---the world is rationally organized to do the rest.

But this talk of rationality is a parenthesis and a digression. What
we were discussing was the idea of a world growing not integrally but
piecemeal by the contributions of its several parts. Take the
hypothesis seriously and as a live one. Suppose that the world's
author put the case to you before creation, saying: ``I am going to
make a world not certain to be saved, a world the perfection of which
shall be conditional merely, the condition being that each several
agent does its own `level best.' I offer you the chance of taking part
in such a world. Its safety, you see, is unwarranted. It is a real
adventure, with real danger, yet it may win through. It is a social
scheme of co-operative work genuinely to be done. Will you join the
procession? Will you trust yourself and trust the other agents enough
to face the risk?''

Should you in all seriousness, if participation in such a world were
proposed to you, feel bound to reject it as not safe enough? Would
you say that, rather than be part and parcel of so fundamentally
pluralistic and irrational a universe, you preferred to relapse into
the slumber of nonentity from which you had been momentarily aroused
by the tempter's voice?

Of course if you are normally constituted, you would do nothing of
the sort. There is a healthy-minded buoyancy in most of us which such
a universe would exactly fit. We would therefore accept the
offer---``Top! und Schlag auf Schlag!'' It would be just like the
world we practically live in; and loyalty to our old nurse Nature
would forbid us to say no. The world proposed would seem `rational'
to us in the most living way.

Most of us, I say, would therefore welcome the proposition and add
our {\it fiat\/} to the {\it fiat\/} of the creator.
\eq
\eq

\subsection{Max's First Reply}

\bq
I finally received my copy of {\sl Pragmatism\/} and have been reading it with much enjoyment ever since. I found myself nodding in agreement many times as I was reading along. I also enjoyed James' prose and choice of words. I like the idea of a world that might be pluralistic at its core, a world whose future is not yet hammered out, a world that is not subsumed into a monistic, eternal, static One. I like the open-minded, non-dogmatic and practical (yet not mundane) nature of the pragmatist approach. Among many things, reading the book has further cemented my ever-growing antipathy to Everett's many worlds. Not a bad outcome, I think, given that I used to be quite partial to this type of interpretation. Well, once I've finished the book, I hope to be able to communicate some more concrete thoughts and comments.

Also, I just re-read your paper ``Subjective probability and quantum certainty,'' and I feel it has helped me in coming to grips with the idea of a purely subjective nature of certainty. I suppose my initial slight unease with this issue had something to do with the question you address toward the end: ``Isn't not asking for a further explanation a betrayal of the very purpose of science, namely, never to give up the quest for an explanation?'' I think that, for most of us, the idea of ``explanation'' is usually indeed intuitively linked with notions of causality and
(moreover) determinism. And hence the initial discomfort when our feeling of certainty is not further related to some underlying mechanism that ``indeed'' (as we would then say) ensures an objective basis for our certainty. To some extent, much of the spirit of science may be built on the quest for such a mechanism: we observe patterns, regularities, and want to have them set in stone, in form of mechanistic physical laws, for example. But the upshot of QM may indeed be to look no longer for such mind-soothing tracks to underlie the journey of our experiences.

I'm very curious to see how your thoughts on ``decoherence in QBist terms'' will evolve. I too have sometimes found myself contemplating the precise role decoherence could assume in such an interpretive framework, and what the particular input of decoherence would be here.

As with similar questions about the implications and views of decoherence in the various interpretations of QM, it is apparent that, on the one hand, decoherence is simply an application of the QM formalism and hence cannot really ``solve'' any foundational riddles; but also that, on the other hand, decoherence may sharpen or inform or lessen certain worries of foundational relevance. I understand that you believe that decoherence has rather little (if any) input for q-foundations, which is a fair point (though I'd add that its force may also in turn depend on one's interpretive commitments and thinking, and thus on the particular form one chooses to state and rank foundational problems).

Of course, QBism in no way keeps us from applying the full machinery of decoherence analysis. I guess we may say that decoherence teaches us something like this: In order for the agent to improve his predictions of the consequences of future actions on the system, he ought to take into account the openness of any realistic quantum system. He could take actions on the environment and on the composite system-environment complex to bring about facts that in turn help him update is predictions of what he may (or, more often in the case of decoherence, may NOT) expect to experience if he interacts with the system. Or, in absence of access to most environmental degrees of freedom, he might just adjust his prior according to some good guess about what the environment is like and what its interactions with the system are likely to be. In turn, I suppose he could also use his experiences of interactions with the system to update his state assignments to the environment.

These are just a couple of first, painfully vague attempts at wrapping my head around the decoherence $+$ QBist complex. Maybe I've missed the target. At any rate, if you think it may be beneficial/worthwhile for you too, I would certainly enjoy any discussions surrounding QBist decoherence while you and {\Ruediger} work on the draft.
\eq

\subsection{Max's Second Reply}

\bq
I {\it finally\/} finished James' book today. I had read three quarters of it in one go (enthusiastically, in fact!), and then somehow got sidetracked by other things and didn't seriously pick it up again until today. The earlier preliminary assessment I had given to you in my email of July 28th (you got that one, didn't you?)\ still stands, though:
\bq\noindent
``I found myself nodding in agreement many times as I was reading along.
I also enjoyed James' prose and choice of words. I like the idea of a world that might be pluralistic at its core, a world whose future is not yet hammered out, a world that is not subsumed into a monistic, eternal, static One. I like the open-minded, non-dogmatic and practical (yet not
mundane) nature of the pragmatist approach.''
\eq

Most importantly, maybe, I have already applied some of the pragmatic ways of thinking to various issues that have come before me lately, and I've had some nicely head-clearing epiphanies in the process. So, thanks again for pointing me to this book. I shall certainly hear James' voice in the back of my head from now on.

Below are some of my favorite quotes -- I haven't checked explicitly, but it's likely they'll have some overlap with the ones you sent me.
Well, I guess that just proves the universal power and appeal of the Jamesian ideas \ldots

\bq
\subsubsection{The Present Dilemma in Philosophy}

Whatever universe a professor believes in must at any rate be a universe that lends itself to lengthy discourse. A universe definable in two sentences is something for which the professorial intellect has no use.
No faith in anything of that cheap kind!

**

The history of philosophy is to a great extent that of a certain clash of human temperaments.

**

The more absolutistic philosophers dwell on so high a level of abstraction that they never even try to come down. The absolute mind which they offer us, the mind that makes our universe by thinking it, might, for aught they show us to the contrary, have made any one of a million other universes just as well as this. You can deduce no single actual particular from the notion of it. It is compatible with any state of things whatever being true here below.

**

The actual universe is a thing wide open, but rationalism makes systems, and systems must be closed. For men in practical life perfection is something far off and still in process of achievement. This for rationalism is but the illusion of the finite and relative: the absolute ground of things is a perfection eternally complete.

**

The books of all the great philosophers are like so many men. Our sense of an essential personal flavor in each one of them, typical but indescribable, is the finest fruit of our own accomplished philosophic education. What the system pretends to be is a picture of the great universe of God. What it is---and oh so flagrantly!---is the revelation of how intensely odd the personal flavor of some fellow creature is.

**

But almost everyone has his own peculiar sense of a certain total character in the universe, and of the inadequacy fully to match it of the peculiar systems that he knows. They don't just cover HIS world. One will be too dapper, another too pedantic, a third too much of a job-lot of opinions, a fourth too morbid, and a fifth too artificial, or what not. At any rate he and we know offhand that such philosophies are out of plumb and out of key and out of `whack,' and have no business to speak up in the universe's name.

\subsubsection{What Pragmatism Means}

So the universe has always appeared to the natural mind as a kind of enigma, of which the key must be sought in the shape of some illuminating or power-bringing word or name. That word names the universe's PRINCIPLE, and to possess it is, after a fashion, to possess the universe itself. `God,' `Matter,' `Reason,' `the Absolute,'
`Energy,' are so many solving names. You can rest when you have them.
You are at the end of your metaphysical quest.

**

When the first mathematical, logical and natural uniformities, the first LAWS, were discovered, men were so carried away by the clearness, beauty and simplification that resulted, that they believed themselves to have deciphered authentically the eternal thoughts of the Almighty.  (\ldots)

But as the sciences have developed farther, the notion has gained ground that most, perhaps all, of our laws are only approximations. The laws themselves, moreover, have grown so numerous that there is no counting them; and so many rival formulations are proposed in all the branches of science that investigators have become accustomed to the notion that no theory is absolutely a transcript of reality, but that any one of them may from some point of view be useful.

**

Purely objective truth, truth in whose establishment the function of giving human satisfaction in marrying previous parts of experience with newer parts played no role whatever, is nowhere to be found. The reasons why we call things true is the reason why they ARE true, for `to be true' MEANS only to perform this marriage-function.

\subsubsection{Some Metaphysical Problems Pragmatically Considered}

To treat abstract principles as finalities, before which our intellects may come to rest in a state of admiring contemplation, is the great rationalist failing.

**

When we look at what has actually come, the conditions must always appear perfectly designed to ensure it. We can always say, therefore, in any conceivable world, of any conceivable character, that the whole cosmic machinery MAY have been designed to produce it.

\subsubsection{The One and the Many}

I must therefore treat the notion of an All-Knower simply as an hypothesis, exactly on a par logically with the pluralist notion that there is no point of view, no focus of information extant, from which the entire content of the universe is visible at once. ``God's consciousness,'' says Professor Royce, ``forms in its wholeness one luminously transparent conscious moment''---this is the type of noetic unity on which rationalism insists. Empiricism on the other hand is satisfied with the type of noetic unity that is humanly familiar.
Everything gets known by SOME knower along with something else; but the knowers may in the end be irreducibly many, and the greatest knower of them all may yet not know the whole of everything, or even know what he does know at one single stroke:---he may be liable to forget. Whichever type obtained, the world would still be a universe noetically. Its parts would be conjoined by knowledge, but in the one case the knowledge would be absolutely unified, in the other it would be strung along and overlapped.

\subsubsection{Pragmatism and Common Sense}

The scope of the practical control of nature newly put into our hand by scientific ways of thinking vastly exceeds the scope of the old control grounded on common sense. Its rate of increase accelerates so that no one can trace the limit; one may even fear that the BEING of man may be crushed by his own powers, that his fixed nature as an organism may not prove adequate to stand the strain of the ever increasingly tremendous functions, almost divine creative functions, which his intellect will more and more enable him to wield. He may drown in his wealth like a child in a bath-tub, who has turned on the water and who cannot turn it off.

**

But now if the new kinds of scientific `thing,' the corpuscular and etheric world, were essentially more `true,' why should they have excited so much criticism within the body of science itself? Scientific logicians are saying on every hand that these entities and their determinations, however definitely conceived, should not be held for literally real. It is AS IF they existed; but in reality they are like co-ordinates or logarithms, only artificial short-cuts for taking us from one part to another of experience's flux. We can cipher fruitfully with them; they serve us wonderfully; but we must not be their dupes.

**

Ought not the existence of the various types of thinking which we have reviewed, each so splendid for certain purposes, yet all conflicting still, and neither one of them able to support a claim of absolute veracity, to awaken a presumption favorable to the pragmatistic view that all our theories are INSTRUMENTAL, are mental modes of ADAPTATION to reality, rather than revelations or gnostic answers to some divinely instituted world-enigma?

\subsubsection{Pragmatism's Conception of Truth}

But the great assumption of the intellectualists is that truth means essentially an inert static relation. When you've got your true idea of anything, there's an end of the matter. You're in possession; you KNOW; you have fulfilled your thinking destiny. You are where you ought to be mentally; you have obeyed your categorical imperative; and nothing more need follow on that climax of your rational destiny. Epistemologically you are in stable equilibrium.

**

The truth of an idea is not a stagnant property inherent in it. Truth HAPPENS to an idea. It BECOMES true, is MADE true by events. Its verity is in fact an event, a process: the process namely of its verifying itself, its veri-FICATION. Its validity is the process of its valid-ATION.

**

The true thought is useful here because the house which is its object is
useful. The practical value of true ideas is thus primarily derived from
the practical importance of their objects to us. Their objects are,
indeed, not important at all times.

**

Primarily, and on the common-sense level, the truth of a state of mind
means this function of A LEADING THAT IS WORTH WHILE. When a moment in
our experience, of any kind whatever, inspires us with a thought that is
true, that means that sooner or later we dip by that thought's guidance
into the particulars of experience again and make advantageous connexion
with them.

**

Such is the large loose way in which the pragmatist interprets the word
agreement. He treats it altogether practically. He lets it cover any
process of conduction from a present idea to a future terminus, provided
only it run prosperously. It is only thus that `scientific' ideas,
flying as they do beyond common sense, can be said to agree with their
realities. It is, as I have already said, as if reality were made of
ether, atoms or electrons, but we mustn't think so literally. The term
`energy' doesn't even pretend to stand for anything `objective.' It is
only a way of measuring the surface of phenomena so as to string their
changes on a simple formula.

Yet in the choice of these man-made formulas we cannot be capricious
with impunity any more than we can be capricious on the common-sense
practical level. We must find a theory that will WORK; and that means
something extremely difficult; for our theory must mediate between all
previous truths and certain new experiences. It must derange common
sense and previous belief as little as possible, and it must lead to
some sensible terminus or other that can be verified exactly. To `work'
means both these things; and the squeeze is so tight that there is
little loose play for any hypothesis. Our theories are wedged and
controlled as nothing else is. Yet sometimes alternative theoretic
formulas are equally compatible with all the truths we know, and then we
choose between them for subjective reasons. We choose the kind of theory
to which we are already partial; we follow `elegance' or `economy.'
Clerk Maxwell somewhere says it would be ``poor scientific taste'' to
choose the more complicated of two equally well-evidenced conceptions;
and you will all agree with him. Truth in science is what gives us the
maximum possible sum of satisfactions, taste included, but consistency
both with previous truth and with novel fact is always the most
imperious claimant.

**

Truths emerge from facts; but they dip forward into facts again and add
to them; which facts again create or reveal new truth (the word is
indifferent) and so on indefinitely. The `facts' themselves meanwhile
are not TRUE. They simply ARE. Truth is the function of the beliefs that
start and terminate among them.

**

The rationalist's fallacy here is exactly like the sentimentalist's.
Both extract a quality from the muddy particulars of experience, and
find it so pure when extracted that they contrast it with each and all
its muddy instances as an opposite and higher nature. All the while it
is THEIR nature. It is the nature of truths to be validated, verified.
It pays for our ideas to be validated. Our obligation to seek truth is
part of our general obligation to do what pays. The payments true ideas
bring are the sole why of our duty to follow them.

\subsubsection{Pragmatism and Humanism}

I read in an old letter---from a gifted friend who died too young---these
words: ``In everything, in science, art, morals and religion, there MUST
be one system that is right and EVERY other wrong.'' How characteristic
of the enthusiasm of a certain stage of youth!

**

`REALITY' IS IN GENERAL WHAT TRUTHS HAVE TO TAKE ACCOUNT OF; [fn: Mr.
Taylor in his Elements of Metaphysics uses this excellent pragmatic
definition.] and the FIRST part of reality from this point of view is
the flux of our sensations. Sensations are forced upon us, coming we
know not whence. Over their nature, order, and quantity we have as good
as no control. THEY are neither true nor false; they simply ARE. It is
only what we say about them, only the names we give them, our theories
of their source and nature and remote relations, that may be true or not.

**

Lotze has in several places made a deep suggestion. We naively assume,
he says, a relation between reality and our minds which may be just the
opposite of the true one. Reality, we naturally think, stands ready-made
and complete, and our intellects supervene with the one simple duty of
describing it as it is already. But may not our descriptions, Lotze
asks, be themselves important additions to reality? And may not previous
reality itself be there, far less for the purpose of reappearing
unaltered in our knowledge, than for the very purpose of stimulating our
minds to such additions as shall enhance the universe's total value.
``Die Erh\"ohung des vorgefundenen Daseins'' is a phrase used by Professor
Eucken somewhere, which reminds one of this suggestion by the great Lotze.

It is identically our pragmatistic conception. In our cognitive as well
as in our active life we are creative. We ADD, both to the subject and
to the predicate part of reality. The world stands really malleable,
waiting to receive its final touches at our hands. Like the kingdom of
heaven, it suffers human violence willingly. Man ENGENDERS truths upon it.

**

The essential contrast is that for rationalism reality is ready-made and
complete from all eternity, while for pragmatism it is still in the
making, and awaits part of its complexion from the future. On the one
side the universe is absolutely secure, on the other it is still
pursuing its adventures.

**

And this, exactly this, is what the tough-minded of that lecture find
themselves moved to call a piece of perverse abstraction-worship. The
tough-minded are the men whose alpha and omega are FACTS. Behind the
bare phenomenal facts, as my tough-minded old friend Chauncey Wright,
the great Harvard empiricist of my youth, used to say, there is NOTHING.
When a rationalist insists that behind the facts there is the GROUND of
the facts, the POSSIBILITY of the facts, the tougher empiricists accuse
him of taking the mere name and nature of a fact and clapping it behind
the fact as a duplicate entity to make it possible. That such sham
grounds are often invoked is notorious. At a surgical operation I heard
a bystander ask a doctor why the patient breathed so deeply. ``Because
ether is a respiratory stimulant,'' the doctor answered. ``Ah!'' said the
questioner, as if relieved by the explanation. But this is like saying
that cyanide of potassium kills because it is a `poison,' or that it is
so cold to-night because it is `winter,' or that we have five fingers
because we are `pentadactyls.' These are but names for the facts, taken
from the facts, and then treated as previous and explanatory. The
tender-minded notion of an absolute reality is, according to the
radically tough-minded, framed on just this pattern. It is but our
summarizing name for the whole spread-out and strung-along mass of
phenomena, treated as if it were a different entity, both one and previous.

\subsubsection{Pragmatism and Religion}

In short the only fully rational world would be the world of
wishing-caps, the world of telepathy, where every desire is fulfilled
instanter, without having to consider or placate surrounding or
intermediate powers. This is the Absolute's own world. He calls upon the
phenomenal world to be, and it IS, exactly as he calls for it, no other
condition being required. In our world, the wishes of the individual are
only one condition. Other individuals are there with other wishes and
they must be propitiated first. So Being grows under all sorts of
resistances in this world of the many, and, from compromise to
compromise, only gets organized gradually into what may be called
secondarily rational shape. We approach the wishing-cap type of
organization only in a few departments of life. We want water and we
turn a faucet. We want a kodak-picture and we press a button. We want
information and we telephone. We want to travel and we buy a ticket. In
these and similar cases, we hardly need to do more than the wishing---the
world is rationally organized to do the rest.
\eq
\eq

\section{30-06-09 \ \ {\it Disturbing the Solipsist, 2}\ \ \ (to H. M. Wiseman)} \label{Wiseman25}

\bhw
Perhaps not this section, but the introduction states:
\bq\noindent
{\rm For, quantum mechanics-we plan to show in this paper-gives a resource that raw Bayes\-ian
probability theory does not: It gives a rule for forming probabilities for the outcomes of factualizable experiments (experiments that may actually be performed) from the probabilities one assigns for the outcomes of a designated counterfactual experiment (an experiment only imagined,
and though possible to do, never actually performed).}
\eq
Maybe I'm misunderstanding. I thought you were contrasting classical Bayesian reasoning
with quantum Bayesian reasoning, and implying the latter had more structure. This would be
wrong, IMO, for the reasons I gave earlier. But maybe you mean something else by ``raw'' Bayesianism.
\ehw

You are indeed a friend in Australia:  For who else would read the paper this thoroughly!

I don't want the passage to be misleading, but strictly speaking it remains correct and consistent with what I said yesterday.  In a literal reading, the only thing this passage expresses is that the {\it judgment\/} $q(B)=(d+1)p(B) - 1$ (for relating lower path to upper path in the diagram) is a judgment beyond raw Bayesian probability theory (i.e., beyond Dutch-book coherence).  As you called me on in your earlier note, a judgment that $q(B)=p(B)$ would also be an addition to Bayesian probability theory, just a different one.  And that is true.  The passage you quote is simply silent on that.

There is a bit of discussion of this just after Eq.\ (169).  And of course there is a bit of discussion just preceding Eq.\ (4).  Finally some discussion on page 53 near the displayed equations.

I don't know that I should modify anything in the introduction, since as I say the passage is not inaccurate.

\section{01-07-09 \ \ {\it Quick, Before {\Ruediger} Sees \ldots}\ \ \ (to H. M. Wiseman)} \label{Wiseman26}

Yesterday didn't work out the way I planned, and today {\Ruediger} says I can't write so much email (we need to work on our ``Quantum-Bayesian Decoherence'' paper  [see \arxiv{1103.5950}] in his last days in Waterloo).  So, once again, you'll have to wait to learn whether you exist.

\bhw
Are you going to modify your introduction then, if you admit it misrepresents your position?
\ehw
It doesn't so much misrepresent it, as it just doesn't say it in any detail.  (That's the way introductions are supposed to work.)  My feeling was that I had covered myself with respect to {\it most\/} people by using the phrase ``quick and dirty,'' by making the allusion to ``EPR-criterion-of-reality considerations'' along with Bell, and by explicitly putting ``reads off'' in quotes.  We'll see.  If I can find a pithy way to make it more accurate (i.e., without adding more than an extra sentence), I might try.

\bhw
I'll have to read your response to Norsen, but in my correspondence with Norsen I
disagreed with his claim that one can (and Bell did) derive his theorem from locality (in the
strict sense Einstein and Bell used it) alone. I think it is clear that locality plus determinism, or
local causality (which is slightly weaker) is needed. This was one thing I wanted to make clear
in writing a review of RL \& AT. So I suspect we agree on this.
\ehw
If I understand you correctly, then I think we agree to a large extent.

\bhw
I'm not sure what your ``WITHOUT RECOURSE TO'' is supposed to mean.
\ehw
I'll come back to that when I tell you whether you exist.

\bhw
Regarding who is a genuine realist, what I find puzzling is that you say you believe there is a
world out there because you believe it can surprise you. But in cases where you believe the
world can't surprise you, you don't believe it's real. That is, you seem to take predictability as
the {\bf counter-evidence\/} for reality, which is the opposite to the usual scientific argument for
there being a real world.
\ehw

It probably helps explain my attraction to James, Dewey, and Schiller.  But you mangle the point a bit:  Predictability is not {\it counter}-evidence; it's just null.  I hope I formulate the point relatively carefully in my ``Anti-{\Vaxjo} Interpretation of QM'' pseudo-paper.  Michel Bitbol makes the point pretty well in a letter responding to that paper.  I'll paste it in below (point 4 in particular).  [See preply to 10-12-03 note ``\myref{Bitbol1}{First Meeting}'' to M. Bitbol.]

I'll be back when {\Ruediger}'s not watching.

\section{02-07-09 \ \ {\it Your Objections}\ \ \ (to G. M. D'Ariano)} \label{DAriano9}

Thanks for the extensive notes.  They are good food for thought for me, and let us discuss at length when you come to Waterloo.  At the moment, let me just comment on two of your items.

\bgmd
1) You said (publicly) that you gave up the problem of deriving QM from ``operational'' principles.
\egmd

I'm very sorry to cause you distress on this.  I can't exactly recall what I said, but I do remember this much:  What I said was meant to be something of a joke (expressing that I have been failing so far), but it went way wrong.  So much so, that you didn't know that I was joking!  So please be less stressed:  I still believe operational principles are the great starting point for our considerations.

\bgmd
2) You mentioned possible ``empirical'' motivation for QM.
\egmd

Maybe I have not used the best word here, I don't know.  I am discovering that it has connotations with many (like yourself) that I had not intended.  I still very much want a great principle.  But I believe that that great principle will express something about the character of our particular world.  A different world would have had a different principle.  I insert an excerpt from a note to Eric Cavalcanti below, where I reply to some points in his own correspondence with Howard Wiseman.  [See 29-06-09 note ``\myref{Wiseman24}{Eric's Note}'' to E. G. Cavalcanti \& H. M. Wiseman.] Perhaps these notes will help take some of the sting of my choice of words (at least they give the sense of what I meant by ``empirical'') \ldots\ and perhaps I should modify my words to better ones in the future.

I look forward to seeing you soon \ldots\ so I may better clear my name and restore your faith in me!

\section{03-07-09 \ \ {\it Emailing:\ 0907.0416v1} \ \ (to G. L. Comer)} \label{Comer125}

\bgc
From your point of view, is it even necessary to talk about the ``same quantum state''?  Can't I just
forget about the constituents, and say that there is a ``widget'' I'm gonna latch onto and that I'm
gonna write down this wave function for?  Then if I introduce ``fundamental'' constituents, it's only
because I have some prior knowledge about their behavior that I want to take advantage of in
trying to extract info from the widget?
\egc

Yep.  Things don't {\it have\/} wave functions, we make them up.  A line from my most recent paper:
\bq
For, the essential point for a quantum-Bayesian is that there is no such thing as {\it the\/} quantum state.  There are potentially as many states for a given quantum system as there are agents.
\eq
And that applies to anything whatsoever.

\section{06-07-09 \ \ {\it The Verdict}\ \ \ (to H. M. Wiseman)} \label{Wiseman27}

\bhw
The discussion on p.~55 again opens you to accusations of solipsism. I don't care about your saying you're not a solipsist because the world can still surprise you. Tell me what your attitude to other people is! You talk about ``categories of thought''. That seems like a veil to hide behind. Do you believe other people are agents, equal to you, who exist even when you are not observing them, doing things when they are space-like separated from you?

If yes, then you believe in hidden variables, despite your disparagement of them.

If no, then you are a solipsist in my book.

And I don't mean you as a hypothetical agent telling me the answer your quantum theory / philosophy tells you to say. I mean you, Chris Fuchs, the individual I believe in even when I'm not interacting with him. What do you believe about me?
\ehw

I finally come to the reply you have been waiting for \ldots

Of course you exist!  If I didn't believe that, I wouldn't write you emails and expectantly await all the consequences your replies will bring about for me.  You are no different in kind from any other quantum system I interact with.  (I'm sorry if that is an insult.  And by the way, you should note the similarity between the language in the second sentence of this paragraph and that used in Section 8.1---it is no accident.)  Was this not already written so clearly in the paper?

But I continue to reject your statement, ``If yes, then you believe in hidden variables, despite your disparagement of them.''  My understanding of ``hidden variables'' or, better, ``ontic variables''---the sort of thing Bohmians and the like hope to reasonably show as undergirding quantum mechanics---is that these, when found, will give a {\it closed\/} description of the essential content of ``quantum measurement'' (which of course means a broad range of issues/things) without reference to the agent using the theory.  For instance, the contextual hidden-variablists (Bohmians being one species) are happy to put a hidden variable in the device as well as the particle, and think that that takes all the trouble out of the quantum interpretation conundrum.  But a device ain't no agent using the theory as a theory of gambling, and that is why I keep myself at a distance from the hidden-variable conception.  I think we will have understood something deep when we understand why the agent is only so stubbornly removed from the conception of the theory.  Frankly, I think it can't be done without incurring the even greater mysteries the advocates of those approaches simply ignore.  (The endpoint of those researches, I am quite convinced, will be lifeless, block universes of one variety or another---ones that can never be resuscitated, no matter what amount of Dennettian-style sophistry.)  Worse than that, I see the ontic variable program so conceived as a dead end for developing the next stage of physics.

Quantum mechanics (and all its babies, like quantum field theory or even, potentially, though not likely, string theory) cannot be the end of the story for our physical understanding.  I tend to think it is only the beginning.  But I don't see a development going forward that doesn't recognize the agent as the central point (the ``center of narrative gravity'') of any particular use of the theory.

I know these answers leave you deeply unsatisfied.  Your worry was that I don't believe in your existence, but that's the easy part of the conundrum.  The hard part is in assuring my own existence and efficacy, and that is what every physical theory up until quantum mechanics left out of its worldview.  You shouldn't let the small corrective that quantum mechanics provides frighten you into thinking it gives the opposite of what it actually does!

\bhw
See, I'm not really an anti-Bayesian ogre. {\rm \smiley}
\ehw

I never thought you were.  You read our papers more {\it thoroughly\/} than anyone else on the planet, even in those moments when you are unsympathetic; surely that is revealing something.

\section{06-07-09 \ \ {\it Emailing:\ 0907.0416v1, 2} \ \ (to G. L. Comer)} \label{Comer126}

\bgc
Now, for a little Devilishism:

Is there not a uniqueness property for the {\Schroedinger} equation?  That is, for a given set of initial conditions, $\{\Psi(0,x),\dot{\Psi}(0,x)\}$, one gets a unique solution for $\Psi$.  Sure, I can decompose in terms of a particular basis, but that doesn't change $\Psi$, right?  For a specific example, take the classic Hydrogen atom problem.  We specify boundary conditions, a coordinate system, energy eigenstates, and a solution is the result.  Moreover, we get the Balmer, Lyman, etc series for the spectra, whose accuracy is without doubt.

Is there not a ``best'' wave function for this system, independently of the agent?  I mean, if I'm a stupid agent---which is a pretty good assessment---I might place my bets based on the analogous solution for uranium.  Obviously, the point is that the {\Schroedinger} equation is a very good friend for those who want to make bets.
\egc

That is the perennial question, and it goes much deeper than quantum mechanics.  It is a question about probability theory more generally.  The Bayesian says, the apparatus of probability theory gives no means for saying whether one probability assignment is better than another (none that doesn't depend on a more primal probability assignment further back in the stream), just like the apparatus of mathematical logic gives no means for specifying which truth value a proposition ought to have (true or false).  Truth-value assignments and probability assignments come from outside logic and probability {\it theory}.

It's a hard intuition to shake.  Without opening a very long email exchange (which would surely be required), I'm afraid the best I can do is say to read Dennis Lindley's book {\sl Understanding Uncertainty}: \myurl[http://books.google.com/books?id=z0ArJ_CDnssC&dq=Lindley+Understanding+Uncertainty&source=gbs_navlinks_s]{http://books.google.com/books?id=z0ArJ\underline{ }CDnssC{\&}dq=Lindley+Understanding+Un \\ certainty{\&}source=gbs\underline{ }navlinks\underline{ }s}.

The bottom line is, our common quantum-state assignments ``work'' in the way that a 50-50 assignment ``works'' for most coin tosses.  There is nothing absolute in the statement.  Only an expression of our doing our best to get along in a world that's not completely revealed to us.

\section{06-07-09 \ \ {\it Test the Born Rule Another Way} \ \ (to R. Laflamme)} \label{Laflamme5}

Now it is me who is a week behind on email!

Yes, of course, I'm still interested in an experiment.  The final version of the paper I sent you can be found here: \arxiv{0906.2187}.  (It grew by a factor of two since the draft I had previously sent---the proposed experiment is now to be found in Figure 2 on page 20.)  The paper just got an ``invitation'' from {\sl Reviews of Modern Physics}---so that will be its final coordinates.

I'd be very flattered if you'd think about how it might be done with techniques readily available to your team.  Aephraim Steinberg already has plans to do something similar---he tells me he's got two students on it, the equipment ordered, and plans drawn up.  So I'm pleased that something serious will happen there.  But I figure synergy, cooperation, and/or competition from you guys could only be a good thing.  (De Martini, as well, seemed to take quite an interest in the idea when we met in Sweden, but that may mean nothing in the end.)

Anyway, if you'd like to talk, tell me when to come by.

\section{06-07-09 \ \ {\it Refeeding Quantum Mechanics} \ \ (to B. C. van Fraassen)} \label{vanFraassen21}

Thanks for the notes, and indeed thanks for the interest!  Thinking of your being in San Francisco, I should send you a picture of the painted lady my wife and I are making here in Waterloo.  It is a shock to the system of the locals, though we think the color scheme is pretty standard/simple for such things.

I read your notes several times over to make sure I wasn't missing something:  If you don't mind, you'll have to tell me!  I think the main confusion might be localized in my usage of the label ``Bayesian.''  Sorry to have caused you trouble over that.

First, responding to
\bvf
For the orthodox Bayesian, conditional probability on $A$, if $p(A) = 0$
has no sense.
\evf
and
\bvf
I notice first of all that when you introduce conditional probability
with equation (1): $p(A,B) = p(A)p(B|A)$  it is presented in a form
that is satisfied also if $p(A) = 0,$ provided only that $p(B|A)$ has some
numerical value or other. But I guess you did not write it that way
for that purpose
\evf

I think this means I'm a heterodox Bayesian then.  Because I had put (what I call) ``Bayes Rule'' in the given form for just that reason.  And I do think $p(B|A)$ has sense when $p(A) = 0$.  Funny, just the other day, {\Ruediger} {\Schack} wrote me this:
\brs
     I'll be waiting for you at arrivals, where you emerge from customs. If
     I am late, wait a little. We are only 8 minutes from Heathrow, so if I
     am not there (extremely unlikely, but we Bayesians can condition even
     on probability zero events), you can always call us at home.
\ers
If the label is appropriate, maybe I should call myself something a ``Richard-Jeffrey Bayesian''.  For he takes the same stance in his book {\sl Subjective Probability, the Real Thing}.  See the discussion in Section 1.4 (pages 18--20) of \myurl[http://www.princeton.edu/~bayesway/Book*.pdf]{http://www.princeton.edu/$\sim$bayesway/Book*.pdf}.

Does that now fix everything up in your mind?  Perhaps when we repost a new version of the paper, we should be more careful to define our particular flavor of Bayesianism (i.e., the Richard-Jeffrey style, etc.).  What label would you give to the flavor?  Is one already existent in the literature?

On one of your other points,
\bvf
Really important: that coherence argument assumes that the events $A$
are not just mutually exclusive but form an exhaustive list.
\evf
you are absolutely right, and it was a mistake for me to leave that implicit.

I know I've invited you before, but any interest in giving us a visit in Waterloo sometime soon?  (The wine list at our institute isn't half bad.)

\section{07-07-09 \ \ {\it Deriving the Urungleichung from Fundamental Probability Theory}\ \ \ (to R. Adler)} \label{Adler1}

{\Ruediger} and I have had a chance to study your note.  Thanks again for the interest.  I don't believe we've ever heard the compliment of anyone ``obsessing over [our work] for months''!  I assure you, it gives us a very pleasant feeling!

Anyway, to your point, it would be nice if it were so simple \ldots

But you'll note that your $x$ or $Pr(\mbox{state in the sky is } \Pi_i \, |\, \rho)$ is not actually a probability distribution.  If you sum over the index $i$, you do indeed get 1, but the individual $x$ can go negative.  For the extreme case of this, note that $Pr(\Pi_i \mbox{ observed on } \rho)$ can vanish if $\rho$ is a density operator orthogonal to $\Pi_i$; thus $x$ can go all the way down to
 $-1/d$.  In fact, there is no good way to think of the $\Pi_i$ as ontic variables (i.e., elements of reality).

It is true that for any {\it particular\/} measurement on the ground, one can scheme up a hidden variable explanation to go \underline{from} $q(j)$ \underline{to} $s(j)$ in our notation (though not the other way around).  See Eqs.\ (52) and (53) in the paper.  That's a little similar to your idea, though it takes the state $q(j)$ on the ground as more basic than the state in the sky $p(i)$.  But that is only true for any {\it particular\/} measurement.  When one stitches together all the urgleichung diagrams (like Fig 2) for {\it all possible\/} measurements, if full quantum mechanics does indeed arise from this process, then an all-encompassing hidden-variable interpretation (like the one just mentioned, but working for all measurements on the ground) must be excluded.    We've been thinking hard about adding a subsection to the paper explaining this latter point (as we have already had to make it to Bacciagaluppi, Uffink, and {\Spekkens} and think we'd now like to preempt its being asked again!).  If we do write that and repost, I'll certainly let you know when the time comes.

You're a lucky person, living in New York.  When I lived in New Jersey, I would always get such great inspiration by spending a day in the Village (and a morning in The Strand).

\section{07-07-09 \ \ {\it Why Bell Is My Friend}\ \ \ (to B. Dreiss)} \label{Dreiss2}

I apologize for keeping you waiting so long.  Right after Sweden (where I wrote you from last), {\Schack} came to Waterloo for two and a half weeks of collaboration, and all else got dropped.  I'm only now catching up on emails.

The reason my paper confuses you is because I believe there is absolutely nothing wrong with Bell's derivation.  And that remains true even if Bell himself wasn't clear-headedly Bayesian (he died too young).  The derivation of Bell's original inequality and many other similar inequalities (like the Clauser--Horne--Shimony--Holt one or Hardy's, etc.)\ are just right, and Jaynes was wrong on this point.  It is not that the derivation fails as soon as one takes into account a properly Bayesian understanding of probability.

And there are other, crisper ways to see this as well, without any inequalities at all.  My favorite originally comes from Allen Stairs.  I'll tell you about that, but first please reread the EPR section of Jaynes' ``Clearing Up the Mysteries,'' particularly the first three paragraphs.  The ultimate thing that powers Bell inequality violations in quantum mechanics is that EPR's reasoning reported there (and accepted by Jaynes) is just wrong.  One can see that by not just considering two observables $P$ and $Q$ as EPR do, but by considering many overlapping ones, i.e., ones that have some (but not all) common eigenstates in an interesting interlocking fashion as one conceptually travels from one observable to the next.  For {\it each\/} individual observable, under the assumption of locality, one can run through EPR's reasoning to assert that IT ``must have had existence as [a] definite physical quantity before the measurement.''  But when you consider the whole lot, you find a contradiction---the observables are so interlocked that one and the same proposition ultimately has to be evaluated as both true and false, and that cannot be.  So it is the EPR premise that is wrong---it is wrong to assert an observable ``must have had existence as [a] definite physical quantity before the measurement.''  You can read this argument in more detail in Section V of my paper \arxiv{quant-ph/0608190}, and also in the next mailing you can peruse one of my presentations of it.  (The simplest proof of the statement can be had by contemplating two entangled ``four level'' systems rather than two qutrits; have a good study of Cabello's vectors in that presentation, and I think it will hit you over the head.)

Anyway, Bell's treatment is just (an earlier) probabilification of that kind of argument, and it is perfectly airtight:  To be consistent with quantum mechanics, one either has to give up Einstein locality or preexistence of measurement values or some combination of the two.  We QBists think that only preexistence need be given up---which is distinctly a positive for the William James side of my mentality.

Like I said, I'll send you one of my presentations.  But also, let me attach two things to the present note.  One is an old, very clear and nicely written article on the Bell point by David {\Mermin}.  I suggest reading it very carefully; it'll help get these things into your bones.  The other article is also by {\Mermin}.  I attach it because it carefully attacks one of the more technical points Jaynes was trying to make in ``Clearing up the Mysteries.''  Jaynes thought time-evolving hidden parameters would play a role in clearing everything up.  But that path has since been developed extensively by Hess and Philipp and also by De Raedt, and it is still wrong (as {\Mermin} shows in the HP case, others have shown for De Raedt).

In summary, here's where things stand between Jaynes and we QBists.  Jaynes was absolutely right that quantum probabilities can only in the end be subjective, Bayesian probabilities, and that quantum states cannot themselves be elements of reality.  He was also right when he said:
\bq\noindent
     Our present QM formalism is a peculiar mixture describing in part
     laws of Nature, in part incomplete human information about Nature---all
     scrambled up together by Bohr into an omelette that nobody has seen how
     to unscramble. Yet we think the unscrambling is a prerequisite for
     any further advance in basic physical theory.  For, if we cannot separate the subjective and objective aspects of the formalism, we cannot know what we are talking about; it is just that simple.
\eq
and
\bq\noindent
     Of course, the QM formalism also contains fundamentally important and
     correct ontological elements \ldots\ It seems that, to unscramble the
     epistemological probability statements from the ontological elements we
     need to find a different formalism, isomorphic in some sense but based
     on different variables; it was only through some weird mathematical
     accident that it was possible to find a variable $\psi$ which scrambles
     them up in the present way.
\eq
That is what our latest paper is indeed an attempt to do.  Where we differ from Jaynes is in thinking that quantum measurement outcomes cannot represent features of some underlying unmeasured reality.  That, we believe, is what Bell inequality violations (and the simpler Stairs style arguments) demonstrate.  Jaynes held to his view because he thought that taking something like the QBist view would be tantamount to declaring ``the Universe runs on psychokinesis.''  But we feel the Paulian Idea (Section 8 in our paper) is enough to block that extreme.  The gambling agent is a crucial participant in the making or coming to be of a quantum outcome, but he cannot will the outcome he pleases.  That is a crucial distinction, and with it one can have one's cake and eat it too.

I hope that helps.

So strange that you drove through Cuero.  Not much to see there; it's great that you at least saw the one thing that's worth looking at.  For some reason, I was almost about to say, ``You're probably the closest to celebrity the town has ever seen --- someone who has surfed both in Hawaii and Australia!''  But then I remembered that John Wayne used to like to bird hunt at the Hamilton ranch.  And when I worked at the corner gas station (where the McDonald's is now), one day I met a man who claimed to be Marlboro Man in their famous painting.

\section{07-07-09 \ \ {\it PI Kids Quote?}\ \ \ (to N. Waxman)} \label{Waxman1}

\bnw
I thought that was a very neat observation from your daughter---may I put it in the newsletter some month (maybe this month?), in the little ``PI Kids are Saying'' section?

I wrote it down as follows: \ldots
\enw

It was Katie, when she was 5\@.  Sure, you can use it.  But please change the quote to:
\begin{center}
``Wow! It's bigger on the inside than it is on the outside!''\footnote{At the time, we had no knowledge of ``the Doctor'' and his TARDIS---now, the whole family and especially Katie are avid fans of the {\sl Doctor Who\/} television series.  We also now know that Katie's exclamation is where the resemblance between the TARDIS and Perimeter Institute ends:  One represents an adventure in an infinitely surprising universe, whereas the other is a paean to a state of death, a.k.a.\ the ``final theory'' that most of PI's residents yearn for.  From today's standpoint, I wonder whether a better appropriation for Katie's wonderment might be to use it to describe quantum systems themselves: Even the minuscule qubit must have a huge interior.  See 30-10-09 note ``\myref{Barnum26}{My Interiority Complex}'' to H. Barnum and other instances of the word ``interiority'' in the present document and  ``QBism, the Perimeter of Quantum Bayesianism,'' \arxiv{1003.5209v1}.}
\end{center}
Adding the ``even'' slightly takes away from the paradoxical nature of what she said, and in fact, I believe I have the quote exact.

\section{08-07-09 \ \ {\it My Favorite Convex Set} \ \ (to R. {\Schack})} \label{Schack170}

That boy wouldn't know the idea of simplicity if it bit him in the ass.  It is clear as day why a SIC representation is to be preferred, and I have tried to express it to him countless times.  He has never had a positive (or detailed) thing to say with regard to our program (though he professes to love Jaynes), and it starts to grate.  Mostly it gets very old putting up with all these alpha-male wannabes---that is what his note was purely an expression of---and I thought it was time to express that clearly.

\section{09-07-09 \ \ {\it Trying to Make a New Start} \ \ (to C. Ferrie)} \label{Ferrie1}

\bcf
On that note, I'd like at some point to discuss the various approaches
to subjective probability with you and hear the case for the Dutch
book as opposed to the others, e.g.\ Jaynes' robot.
\ecf

Attached is one interesting thing to read with regard to this (and Hacking has many more than that).  [I. Hacking, Brit.\ J. Phil.\ Sci.\ {\bf 16}(64), 334--339 (1966).]  Hacking gives a review of Kyburg and Smokler's collection on subjective probability, and in it he is careful to make a distinction between two kinds of Bayesianism, the ``conservative,'' and the ``radical.''  The first roughly corresponds to Jaynes' thought, and the second to roughly de Finetti and Ramsey (and made somewhat more radical still by {\Caves}, {\Schack}, and me).

I found this little article very useful to read myself, and I have you to thank for that.  Particularly, Hacking's small remarks on J. S. Mill being an ``ancestor'' of the approach.  That made a light-bulb flash in me.  For, though I had realized I was led to de Finetti's version of Bayesianism partially because it gave me a tool with which to preserve {\it locality\/} in quantum mechanics, I had not quite appreciated that I was probably also drawn to it by my predilection for a Jamesian version of indeterminism (the republican-banquet pluralism I write of in the paper).  The latter, as James knew, has its roots in Mill.  (James dedicates his book {\sl Pragmatism\/} to Mill.)  And it is always good to know where one's prejudices are coming from.

\section{09-07-09 \ \ {\it Potentially Useful} \ \ (to C. Ferrie)} \label{Ferrie2}

This one seems to be a potentially useful point of entry as well:  Peter C. Fishburn, ``The Axioms of Subjective Probability,'' Stat.\ Sci.\ {\bf 1}(3), 335--345 (1986).

\section{09-07-09 \ \ {\it My Favorite Convex Set} \ \ (to C. Ferrie)} \label{Ferrie3}

\bcf
You may be interested in our recent paper:
\myurl[http://www.iop.org/EJ/abstract/1367-2630/11/6/063040/]{http://www.iop.org/EJ/abstract/ 1367-2630/11/6/063040/}.
It mentions SICs.
\ecf

I have added a citation to this, and will elevate the point to an enlarged discussion in the section introducing SICs.  {\Ruediger} commented this when we first got your note:
\brs
     From his paper, I suspect that this guy knows a lot about
     quasiprobabilities. From his perspective, our paper shows just one out
     of many possible ways of representing q.m.\ using probabilities. What he
     is asking is why this particular way of doing it and not another.
     Itamar asked a similar question. I am not too sure how best to reply.
\ers

Here is the way I would reply.  The point of all the various representations of quantum mechanics (quasi-probability reps, as well as things like Heisenberg vs.\ {\Schroedinger} picture issues and even path-integral formulations), is that they give a means for isolating or emphasizing one or another aspect of the theory---they help bring a particular aspect into plain view, even if all the representations are logically equivalent.  In our case, we want to bring into plain view (and try to make compelling) the idea that quantum mechanics is an {\it addition\/} to Bayesian probability, not a generalization of it.  With that goal in mind, the SIC representation has always struck me as particularly powerful tool.  With it, one can see the Born Rule as ``really'' a {\it function\/} of a usage of the Law of Total Probability in another context (one different than the actual).  That feature, as far as I can tell, does not leap out in the same way from the more general ``deformed probability representations'' you explore in your papers with Joseph.  That in a nutshell is the reason for my love affair with SICs.

\section{09-07-09 \ \ {\it Articulation} \ \ (to C. Ferrie)} \label{Ferrie4}

\bcf
Here is what I was trying to say.  The sentence ``The only way anyone
has seen how to do it is \ldots'' is very uncharacteristic of your
otherwise beautifully written exposition.  It stuck out and then stuck
in my mind as I read on.  It called out to me ``well \ldots\ this isn't
quite right but it was the best we could come up with \ldots''  It has a
very apologetic tone which forced the skepticism I felt while reading on.
\ecf

I really would appreciate it---and this is not just apologetics now---if you would articulate in greater detail the source of your skepticism.  What in detail, indeed, made you feel ``It seems an ad hoc addition to coherence contrived to bridge the Peresian slogan to your functional relation between $q$ and $p$.''  As I tried to express last night, part of the reason behind my bad reaction was not that you were talking about an isolated page or two in the paper (I have no illusions that the paper is perfect, I certainly know that it could be made much better, or even rewritten from the start).  It was that you were referring to the very point of the paper with your sentence.  (It only dawned on me after your reply yesterday that perhaps you didn't know this, that maybe you thought you were referring to some small thing rather than the paper as a whole.)  So, it's important for me to get it straight.  \ldots\ And it'll be good practice for you in articulation. \smiley

\section{13-07-09 \ \ {\it Articulation, 2} \ \ (to C. Ferrie)} \label{Ferrie5}

Contrived is a very different word than ``unnecessary and distracting.''  ``Contrived,'' at least in my usage, usually conveys bad intent.

Anyway, there are a few reasons I resist the Coxian understanding of probability.  The most important is this:  In the usual way the system is developed (and maybe it is necessarily so), $\mbox{Pr}=1\,\, \Rightarrow \,\, \mbox{TRUTH}$.  But it is utterly important for {\Caves}, {\Schack}, and me that this isn't so; there is an ultimate disconnect between probability statements and truth values for us, and that in particular allows us to preserve locality in quantum mechanics in spite of Bell inequality violations.  And Dutch-book definition of probability allows for just what we need.

Still, let me think further about the notation.

\subsection{Chris's Preply}

\bq
I'm no gambler and perhaps this is why the Dutch book argument has never sat well with me.  I found Jaynes' robot argument much more intuitive.  So let me conjure up his ghost (perhaps a younger, more naive and better looking version).

A bookie asks the agent to commit to $p(B)$ given a conditional $A$-lottery.  The agent gives $p(A)$ and $P(B|A)$ and is forced through coherence to bet on $p(B)$ as calculated through the law of total probability.  But the bookie reveals that there is no $A$-lottery.  You would say that the agent is no longer committed to $p(B)$ as calculated before but now is committed to $q(B)$ (which is $p(B)$ without the $A$-lottery).  Now Jaynes would say this is {\it inconsistent\/} as the 1st law of robots is a robot may not injure a human being \ldots\ wait \ldots\ no, it's consistency (the Jaynesian equivalent of Dutch book coherence I presume).  There is a unique value of $p(B)$ no matter how you calculate it: $p(B)$ is $p(B)$ is $p(B)$ \ldots.

If the agent is told something new then he is now committed to $p(B|\mbox{something new})$ and not $p(B)$.  For the bet, the bookie really asked for $B|C$, where $C=$``The $A$-lottery has been performed'', and then he later reveals that he will take bets on just $B$.  Now ``$q(B) \ne p(B)$'' is just ``$p(B) \ne p(B|C)$''.  This is something which should be obvious and much more clearly formalizes the statement ``measuring $A$ matters even if we don't reason about the outcome''.

This is why I think the conditional lottery is unnecessary and distracting.
\bq\noindent
\verb+\begin{sarcasm}+\\ You didn't get that from ``contrived''?\\ \verb+\end{sarcasm}+
\eq
\eq

\section{13-07-09 \ \ {\it Quantum Randomness}\ \ \ (to K. Martin)} \label{Martin12}

When Schack was visiting me last week, I got him to give me a copy of the paper he and Caves are writing on quantum random number generators (as seen from our quantum Bayesian perspective).  Attached is---he tells me---a very ROUGH DRAFT.  Still, I thought it might be useful to send it to you.  Is that the sort of thing you were interested in?

Things are progressing somewhat on the SIC end.  Schack and I spent most of the last two weeks hashing out our next paper---a companion piece to our ``Quantum Bayesian Coherence.''  This one will be called ``Quantum Bayesian Decoherence'' and will be an attempt to put Zurek's decoherence in its place.   [See \arxiv{1103.5950}.]  Appleby is presently chasing down an idea in dimensions $1\,\mbox{mod}\,6$ and is as happy as fox in a henhouse.  And get this, Aephraim Steinberg's group at University of Toronto are starting up a $d=3$ SIC experiment.

\section{15-07-09 \ \ {\it Some $d=3$ Measurements}\ \ \ (to R. Laflamme)} \label{Laflamme6}

Let $d=3$ and $\omega=e^{\frac{2\pi i}{3}}$.\medskip

Set 1:\medskip
\begin{eqnarray}
{\veec0{-1}1} & {\veec0{-\omega}{\overline{\omega}}} & {\veec0{-\overline{\omega}}\omega}
\nonumber\\
{\veec1 0 {-1}} & {\veec1 0 {\overline{\omega}}} & {\veec1 0 {-\overline{\omega}}}
\nonumber\\
{\veec{-1} 1 0} & {\veec{-1} {\omega} 0} & {\veec{-1} {\overline{\omega}} 0}
\nonumber
\end{eqnarray}\medskip

Set 2:\medskip
\begin{alignat}{5}
{\veec{-2} 1 1} & \quad & {\veec{-2} {\omega} {\overline{\omega}}} & \quad & {\veec{-2} {\overline{\omega}} {\omega}}
\nonumber\\
{\veec1 {-2} 1} & \quad & {\veec1 {-2\omega} {\overline{\omega}}} & \quad & {\veec1 {-2\overline{\omega}} {\omega}}
\nonumber\\
{\veec1 1 {-2}} & \quad & {\veec1 {\omega} {-2\overline{\omega}}} & \quad & {\veec1 {\overline{\omega}} {-2{\omega}}}
\nonumber
\end{alignat}

\section{15-07-09 \ \ {\it Measurement in the Sky without Magic} \ \ (to C. Ferrie)} \label{Ferrie6}

\noindent Dear CFe, \medskip

I know you want me to change my notation.  Eventually I'll tell you why I resist that.  The text surrounding the definitions as well as Footnotes 5 and 6, it seems to me, already capture everything I wanted to say.

Also I'm not quite sure what you want of the paper.  (Your remarks, or at least your phraseology, do continue to put me off a little; perhaps I continue to misunderstand their goal.)  My own goal isn't to be ``surprising,'' only clarifying.  My big dream is to say what the content of quantum mechanics is without ever making a priori mention of ``positive semi-definite operators,'' or ``homogenous self-dual cones,'' or the like, or any issues to do with nonlocal-boxology.  Indeed I want quantum mechanics to shake out to be an absolutely trivial theory in the end, at least from the proper point of view.   In that regard at least, I feel the urgleichung gives me some hope, and I feel we have made some progress by realizing it might be taken as one of the foundational principles.  To the extent that I want there to be any surprise at all, it is that people eventually slap themselves on the head and say, you mean the essence of QM might be so very simple?!

\bq\noindent
\verb+\begin{relevant non-sarcastic mild modification of William James quote}+\\
I fully expect to see the [quantum Bayesian view of quantum mechanics] run through the classic stages of a theory's career.  First, you know, a new theory is attacked as absurd; then it is admitted to be true, but obvious and insignificant; finally it is seen to be so important that its adversaries claim that they themselves discovered it.  Our doctrine \ldots\ is at present in the first of these three stages, with symptoms of the second stage having begun in certain quarters. I wish that this lecture might help it beyond the first stage in the eyes of many of you.\\
\verb+\end{relevant non-sarcastic mild modification of William James quote}+
\eq
If I understand your point, you appear to be using the language of Stage 2.\medskip

\noindent Confused regards,\medskip

\noindent rare-earth Chris (old and quite oxidized)

\section{15-07-09 \ \ {\it Measurement in the Sky without Magic, 2}\ \ \ (to C. Ferrie)} \label{Ferrie7}

\bcf
I'm generally not a very excitable guy.  But this stuff really gets me going.  I guess I've adopted you as my sounding board.  Perhaps that is not acceptable conduct for a student.  I suppose I could understand if it wasn't.  Your inbox is probably bombarded with half-baked ideas.
\ecf
To be honest, yes it is.  Attached is another reaction to the paper.  Here's an exercise for you:  Tell me what's wrong with what the guy says.

But your own note wasn't half baked like that.  I just didn't see that you were telling me something I didn't know \ldots\ so it was hard for me to put it in context as representing a breakthrough in your understanding.  As far as I could tell, you wrote down Eq.\ (54) from the paper with changed notation:
\begin{itemize}
\item
I.e., I wrote    $q(j)            = (d+1) s(j)            - 1$
\item
You wrote      $p(D_j|N) = (d+1) p(D_j|Y) - 1$
\end{itemize}
And I thought, yeah, that's what I wrote.  (Where, though, it is key that $Y$ and $N$ don't refer to just any conditions, but an intermediate conditional lottery.)

Clearly---with hindsight---the changing of notation means something deep to you.  I just haven't gotten what that something deep is.

I am {\it trying\/} to be a better person \ldots

\section{15-07-09 \ \ {\it Bohmian Mechanics} \ \ (to J. Emerson)} \label{Emerson2}

\bje
I am putting together a list of topics / lecturers for another edition of the interpretations/foundations course for this upcoming Winter. I'd like to put a bigger emphasis on de Broglie -- Bohm theory and would appreciate your recommendations regarding potential lecturers.

I would also like to hear your recommendations for many worlds and decoherent histories.
\eje

You might consider Wayne Myrvold for the many-worlds slot.  He strikes me as the rare person who might be able to do this:
\bje
I'm tempted to go with someone who will overview the approaches but who can also be critical of the conceptual and technical shortcomings of those approaches.
\eje
What's rare is someone seriously knowing the details of one of these interpretations (to the last technical bone) without having been a full believer to begin with.  I.e., belief most always comes first, details, if any, are far secondary.

\section{16-07-09 \ \ {\it Some Things to Think About}\ \ \ (to C. Ferrie)} \label{Ferrie8}

Let me give you something specific and mathematical to think about if you're interested.  See pages 25 and 26 of the attached talk [``QBism at QTRF-5''].  I define something called a ``maximal consistent'' subset (MCS) of the probability simplex (over $d^2$ outcomes) there.  It is not difficult to show that all maximal consistent subsets must be convex.

Also, all maximal consistent subsets have vaguely quantum-state-space-like properties.  But how far does that go, I wonder?  Thus let me pose a couple of specific questions to show the flavor of things that we might profit from exploring.
\begin{itemize}
\item[1)]	Let $S$ be any MCS.  Must its extreme points form a connected set?  Or can one find a counterexample?
\item[2)]	Let $S$ be any MCS.  Must its extreme points form a smooth manifold?
\item[3)]	If so, what can be said about the dimensionality of that manifold?  Can one place any bounds on the dimensionality?
\end{itemize}

These are the sorts of things {\Asa} and I are thinking about at the moment, and in the Fall when my students get here, I want to greatly expand the effort.  {\Asa} has some intuition that the answer to 2) should be yes, but I've got no strong feeling, and certainly no idea how to prove anything in this regard.  If you have any input, it'd be wonderful.

\section{20-07-09 \ \ {\it Remembering Hans}\ \ \ (to H. C. von Baeyer)} \label{Baeyer76}

The last couple of days I've been writing my intro for the CUP edition of my samizdat.  Attached is where I've gotten so far.  It formalizes a bit of the story I first told you and Marcus in the Black Hole Bistro one day (I had never told anyone else before):
\bq
I should tell you the strange story of how I came to be a physicist in the first place.  It had nothing to do with seeing or feeling some beautiful order in nature, like the kind of thing I would read about in the physics biographies.  It was not like Einstein's childhood experience with a compass, where he came to the conviction that there had to be ``something behind things, something deeply hidden.''

My career, instead, is to be blamed on the 1970s television stations of San Antonio, Texas, ninety miles down the road from my little hometown Cuero.  In those days, there was a steady stream of science fiction and horror films coming from the antennas every Friday, Saturday, and Sunday.  And every weekday there was an episode of {\sl Star Trek} or {\sl Lost in Space} waiting for me just as I got home from school.  I gained a taste for two things:  the image of the noisy, swirling atom that concluded each commercial break of Channel 5's late-late-night {\sl Project Terror},\footnote{Look it up on YouTube.} and the idea that mankind knew no insurmountable bounds.  I dreamed my friends and I could fly to the stars and back if we wanted.  Television molded my idea of what the world ought to be:  one of danger, constant adventure, and yet, with luck and hard work, one ultimately malleable to what we want to make of it.

That thought led me into junior high school, where I decided life was too short:  If I wanted to see myself flying to the stars with my loved ones (successfully) waiting behind to cheer my return, I had better get to the technical details of making it happen.  So I read popular physics book after popular physics book looking for the ways and means.  To my genuine surprise and deep sadness, no real-world method emerged; I was left with the science fiction I had started with.  I only learned from the wisdom of physics that my dream would stay a dream.

I suppose that is where my story might have ended, but it did not.  Instead, it came to me slowly, and then more and more firmly, ``Physics must be wrong!''  I told myself that, and I was quite serious about it.  I knew I had to become a physicist, not for the love of physics, but for the distrust of it.
\eq
There's much more to come in it, but I think I'll go ahead and send you the draft so far.  As if you need more prodding, even in my autobiography you'll see this Pauli thing runs very deep with me.  It was you who brought up the opportunity of writing something fresh; now I don't want to see it slip away!

\section{22-07-09 \ \ {\it Charlie Stories}\ \ \ (to D. W. Leung)} \label{Leung1}

Thanks for the comments.  I cleaned up the Charlie story still a little more; I think it's reached its final form now.  There's still so many stories to tell about everybody else, but I'll attach the latest draft for your amusement.  It'll be a fun exercise when I finally get to John Smolin's story.

\bdwl
Saying teleportation marks the beginning of quantum information is a
bit like saying that Shor's algorithm marks the beginning of quantum
computation {\rm \smiley}\ (guess what's the omitted punchline \ldots)
\edwl

That's right, I would say Shor's algorithm marked the beginning of quantum computation {\it as a field}.  I.e., that's when outsiders widely started to pay attention to it.  To help stave off confusion a bit, I have now italicized ``as a field'' in the footnote.  Thanks for bringing this up to me.

\section{23-07-09 \ \ {\it Fact Check?}\ \ \ (to B. W. Schumacher)} \label{Schumacher16}

I've been working on an introduction for one of my books, and I just finished up a story on Brassard.  In it I use an anecdote I heard from you, and I wonder if I can ask you to check its accuracy?  I'm quite sure I screwed up some details---my memory is not that good.  In any case, I'll certainly be tweaking the story (trying to get the best drama), but with this much down on paper, you can already see the kind of thing I'm aiming for.  I just mostly need to get the facts straight.  See attached file, starting page 6:
\bq
{\Gilles} {\Brassard}, with Charles {\Bennett}, is a father of quantum cryptography:  The {\Bennett}-{\Brassard} 1984 quantum key distribution protocol started the field.  But, from my perspective, he is dearest to me for all that he has enabled.  Ben {\Schumacher} once told me an anecdote worth repeating in this context.  Ben says that when he first heard of the discovery of (Peter {\Shor}'s) quantum factoring algorithm, he thought it was a joke that {\Gilles} had propagated---he didn't take it seriously.  The reason he didn't is because a year earlier, at one of the ``quantum information parties'' {\Gilles} regularly organized in Montr\'eal,\footnote{Like the one in 1993 that led to the discovery of the quantum teleportation protocol.} {\Gilles} joked around, ``We're going to use quantum computers to factor numbers.'' This may have had no direct influence on {\Shor}, but the tendrils of scientific discovery are sinuous, and one really never knows all the influences that come together in the making of an event. The greatest things happen not because of lone scientists in isolation, but because of communities.
\eq

\subsection{Ben's Reply}

\bq
The anecdote agrees with my own memory.  As I recall it, we were in {\Montreal} and Dan Simon described the work he'd been doing there on quantum period finding.  Gilles made this comment, grinning.  I seem to recall that Shor had read Simon and was motivated thereby; I do not know if Shor had heard Gilles' joke (or if it had occurred to him independently).
\eq

\section{26-07-09 \ \ {\it Specifically Paulian}\ \ \ (to H. C. von Baeyer)} \label{Baeyer77}

I'm sitting on my new front porch, waiting for a thunderstorm to roll in, and thinking Paulian thoughts.  See attached photo; you can just see the chair I'm sitting in.  (The porch isn't painted yet, but you'll get some indication of its ultimate colors from the second photo attached.)

\ldots\ Storm just came, and I had to run inside.  It's a fierce one! \ldots\ \ [Some time passes.]

The storm has mellowed a bit, and I'm now back on the porch.  I'll finally reply to your note of yesterday.  (Sorry that yesterday turned me into a liar.)

\bhcvb
I'm reading at this moment: H. Atmanspacher and H. Primas (eds.)
{\sl Recasting Reality: Wolfgang Pauli's Philosophical Ideas and Contemporary Science}, Springer 2009.
\ehcvb
Your writing this reminds me that I still do not have my copy.  I purchased one at discount, while in Sweden, but Springer had to ship it to me.  I will get on them.

\bhcvb
I find satisfaction in the overlap between some of its entries and my
own essay, which is now on the web in the emag {\sl The Global Spiral},
July/August 2009, search under Authors.  (I'm not sure how to formulate a proper web citation.)
\ehcvb
You give a website?  I have printed it out; it doesn't look so bad as you made it out to be.  I haven't re-read it yet.  Is it substantially different from the version I read in January (which I understood to be the finalized one)?  In any case, I will read/re-read it at some point.  I want to be better in your head, if nothing else to help anticipate the passages you might explore.

\bhcvb
The DO.  The term ``detached observer'' does not merit an entry in
the book's index.  I think it is more important than ever to
promulgate the idea, and have accordingly moved our joint project to
the top of my writing agenda.
\ehcvb

Of course, I am very pleased to hear this!  Interesting to hear that DO issues have very little representation in that book.  I wonder why that's the case?  If I recall correctly, I believe I did read one article there that {\it did\/} have some discussion on the subject.  Could it be that it was just an oversight in the index?

It's probably worthwhile mentioning my self-analysis for why I don't completely remember the article, or even for sure whether it was in that book.  The trouble is most everything these guys write is always the same and so becomes a blur.  Nor, do they ever go an ounce beyond where Pauli went (or, at least, where the part of Pauli they've read has gone).  Did you read the Enz entry in my Intro?  The guy wrote 19 papers on Pauli, 19!---I read every one of them---and in the end didn't learn much more than I might have in three well chosen ones from the full list.

I had a similar experience with Folse on Bohr.  31 papers!  They were all great on so many other details, but I never could get a sense of what Bohr really meant by ``the quantum postulate.''  Folse used the term liberally.  So I thought I would invite him to the ``Shannon Meets Bohr'' session I organized in {\Vaxjo} in 2001 and get the issue settled.  (This was the great meeting we had with {\Caves}, Greenberger, {\Mermin}, Peres, Schumacher, Smolin, etc., in attendance.)  But to no avail.  Folse said nothing new on that subject; he could say nothing beyond repeating the three word phrase.  The thing that shocked me was that he seemed to have no sense that one should treat these things as {\it springboards\/} to deeper thoughts.

For once I'd like to see someone say (with reasonable analysis), ``This is what Pauli would have made of the no-cloning theorem \ldots''  Or, ``Pauli would have understood that the no-cloning theorem didn't at all carry the essence of quantum mechanics.''  Or, ``No, Pauli would have seen the deep statement of his unDO in the Kochen--Specker theorem, and here's why.''  Or, ``Here's where quantum information theory might help us formalize a bit of what Pauli was after.''

We can't go that far in our article, I know.  But one thing I think would be nice is if we could push the reader to think of Pauli's ideas as the beginning of a bigger and better quest.  Somehow give the reader the sense that there's something solid to do.

\bhcvb
Pauli and religion.  There is a lot of research on Pauli's legacy in
such fields as philosophy, metaphysics, psychology, biology,
cognitive science and the philosophy of mathematics, but I don't see
much on religion per se.  I will continue my interest in that.
\ehcvb
And I hope you will never stop keeping me apprised of what you're thinking even here.  I try my best to gather data from all sources.

\section{26-07-09 \ \ {\it A Memory}\ \ \ (to H. C. von Baeyer)} \label{Baeyer78}

By the way, I remember you once writing me this about the CFS use of Kochen--Specker to establish the non-preexistence of measurement values:
\bhcvb
I'm reading {\Caves}, Fuchs, and {\Schack} on quantum certainty, and thinking
that I'm getting the drift, when I come up to an example -- which is my
way of understanding -- and whammo, on page 10 you hit me with 33
states, which are (the cute little ``of course'' rubs salt in the
wound) connected by 96 rotations and one shoulder separation.  I can't
envision that.  Can't you use a simpler manifestation of KS like {\Mermin}'s
very cogent $3\times3$ matrix, or maybe it was $4\times4$?  As popularizer I'm always
on the lookout for simpler versions.
\ehcvb
Did I ever make you aware of Cabello et al.'s version of KS?  I well believe it can get no simpler.  See attached file, page 11.  Using it, the CFS-variation-of-Stairs argument is then powered via the use of two qudits, where $d=4$.

\section{27-07-09 \ \ {\it Under Your Belt} \ \ (to J. W. Nicholson)} \label{Nicholson30}

I'm gearing up to write a story about you.  See attached, page 11.  In preparation for it, may I ask you how many publications you have now?  I just want to make a small mention of the number, before I plunge into bar stories.  And do you have a webpage?  I couldn't find it.\medskip

\noindent Greetings from 4:58 AM,

\section{28-07-09 \ \ {\it Red Naugahyde} \ \ (to J. W. Nicholson)} \label{Nicholson31}

This still isn't a proper reply to your nice notes---I'll work on that soon.  But let me send you the first draft of the Nicholson entry to try on for size.  You'll find your entry on page 12 (close to the new one of Asher Peres).

If there's anything you want changed, let me know and I'll consider.

\bq\noindent
{\bf Jeff Nicholson:} Whether or not it is acceptable to say so in an academic publication such
as this, I'll just record it: Jeff Nicholson was the best beer-drinking buddy I ever had in graduate
school or since, and I was proud to be the best man at his wedding. Jeff is currently a research staff
member at OFS Laboratories in Somerset, New Jersey and writing papers with titles like, ``Continuum Generation Using a Femtosecond Erbium-doped Fiber Laser and a Silica Nonlinear
Fiber.'' We started together as officemates at the University of New Mexico in 1993, and then
found ourselves together again at Bell Labs many years later. Jeff is the taller, stronger, and faster
of the two of us, with over 170 publications and patents. In the days that were, we had many a
good round of philosophy at the ever atmospheric {\sl Jack's Liquor and Lounge\/} in Albuquerque---red
naugahyde seats, and plenty of talk of reality creation. I will well bet that J. W. Nicholson is the
only person on earth to have handled a polarization maintaining, dispersion managed, femtosecond
figure-eight fiber laser {\it and\/} read Richard Rorty.
\eq

\subsection{Jeff's Reply}

\bq
I love it. \smiley

Did I ever actually read Richard Rorty?  I can't recall \ldots.

Also, you forgot my Nobel-cited paper of which I am waayyyyy too inordinately proud. \smiley
\begin{itemize}
\item
Brian R. Washburn, Scott A. Diddams, Nathan R. Newbury, Jeffrey W. Nicholson, Man F. Yan, and Carsten G. Jorgensen, ``Phase-locked, erbium-fiber-laser-based frequency comb in the near infrared,'' Optics Letters {\bf 29}(3), 250--252 (2004).
\end{itemize}
\eq

\section{29-07-09 \ \ {\it Pseudo-QBist State Spaces} \ \ (to {\AA}. {\Ericsson})} \label{Ericsson4}

I thought of a title for your paper.  See above.

Also, thinking about your gingerly reactions yesterday, I thought I'd remind you of some of the words I put in the Mlodinow review:
\bq\noindent
     And the malaise of Professor Gardening (whose real identity is
     protected in the book), though I had not seen in it a professor, I
     had seen it in plenty of graduate students. The burdens of not just
     doing physics, any physics, but {\it important\/} physics, is sometimes
     too much---it is too elusive a phantom to even know how to get
     started.
\eq
The same goes with papers:  Not every paper has to be important or conclusive, and in fact most aren't.  Sometimes the only role of a paper is to get a problem started.  Caltech tolerates elusive phantoms, I don't!

\section{31-07-09 \ \ {\it Preskill Done}\ \ \ (to D. M. {\Appleby})} \label{Appleby68}

I finally got the Preskill story down.  Boy that was hard!  Here's an excerpt:

\bq
When we have discussed the interpretational issues of quantum theory, I have gotten the sense that what John finds most suspicious of the quantum Bayesian approach I promote is that he {\it thinks\/} it treats observers as unphysical systems.  I say, ``Not at all; John, you are a physical system to me.''  The real issue is one of inside versus outside.  Contrary to the textbook exposition of quantum mechanics, a wave function that I write down about him is descriptive not of the outside, but of the inside, namely me:  It captures what I believe will happen to me if I interact with him.  If he knew what I believed, he'd probably do something to surprise me.

I think John's secondary worry is that unless one takes something like the Everettian approach to quantum mechanics, one cannot do quantum cosmology---the idea is that the observer must be excised from a fundamental role in the theory if one is going to address cosmological issues.  (In a phrase, what is desired is a theory that is all outside and no inside.\footnote{On first pass, the Texan in me wanted to write ``all hat and no cattle,'' but I repressed that.})  The Everettians think they do that, but a quantum Bayesian would say, ``only in a way that leaves the observer's clothes behind.''  I tried to impress John once by writing down $|\,\Psi_{\rm \scriptscriptstyle universe}\,\rangle$ and saying, ``Look, I can do that too.''  I don't think it worked.

Most recently, I've tried a new technique in my lectures.  I draw a stick figure on the chalkboard along with a little blob near him, and say, ``There's an observer with a small physical system in front of him:  He can profitably use the formalism of quantum mechanics to estimate what might happen to him if he interacts with it.''  Then I say, ``But there's nothing to keep us from talking about a larger physical system.''  I extend the blob, making it larger.  I do this a couple of times, at first imperceptibly putting a bend in the blob, making it a little kidney-bean shaped with the observer at the center of its curvature.  I make it more pronounced with each iteration and finally go for broke:  The observer is completely surrounded by the physical system he is considering, and I declare in great triumph, ``That's quantum cosmology!''

One of these days maybe I'll work up the nerve to show John.
\eq

\subsection{Marcus's Reply}

\bq
I realize that you have a lot to do so don't even think of trying to answer this now.  But when you have the time I am curious:  what is the criterion for being a physical system?  What would Preskill have to do in order to convince you that he wasn't a physical system?

Also, how does this fit in with Pauli's synthesis of science and religion?  Would Pauli agree that Preskill is a physical system? (just a physical system? [shades of vicious abstractionism]).  And if he wouldn't, would he be wrong?  Do you think?

I guess virtually every religion (absolutely every religion?) would say that a human being isn't just a physical system.  I am not sure what they mean by this (I would ask of virtually every religion the question I just asked of you:  how does one tell whether something is a physical system or not?  What is the criterion?).  But whatever they mean, I get the impression they do say it.  I also get the impression it is an important feature of religion.

I am being a little disingenuous here, you see.   I pretended not to know what is meant by saying a human being isn't just a physical system.  And in one way that is true:  I don't know how to explain what is meant (which is why I asked the question: perhaps you can explain?).  But the statement kind of resonates.  In another way (though not in a way I know how to explain) I do understand what  is meant.

So, although very obscure, I do feel that this idea, that a human being isn't just a physical system, is an important  feature of religion.  Perhaps even an essential feature.  So the question is:  is this a feature of religion which has to be dropped in Pauli's synthesis?  And, if it does have to be dropped, what is left?  What kind of synthesis would it be which leaves out this feature of religion?

The old Soviet Union claimed to be a synthesis.  Notionally Uzbekistan and the Russian Republic were equal partners.  But that was, of course, a cheat.  And I am thinking that a Paulian synthesis in which Preskill is just another physical system (vicious abstractionism again) would be something like that, with science playing the role of the Russian republic and religion playing the role of Uzbekistan.

Also, how does all this fit in with your wish to deconstruct physics?  Given that aim, why do you say that Preskill is specifically a physical system?  As opposed to just a system?

Perhaps we should say that Preskill is a deconstructed physical system.  But in that case why not simply say that he is an unphysical system, and be done with it?

In short:  what is the burden of the qualifier ``physical'' here?
\eq

\subsection{Hans Christian von Baeyer's Comment}

\bq
Marcus, the question seems to me to be related to reductionism --- to ``nothing buttery.''  A cathedral is a pile of bricks, but more. Even a shed is a pile of wood, but more. Steven Weinberg, the quintessential reductionist, admitted that he was not an uncompromising reductionist, but a compromising one. So you can learn a lot about a physical system, but you hit a snag of complexity when you want to understand what Preskill is.

The prettiest thought I learned in Phoenix the other day was this:  Human nature is so exquisitely complex that you can't describe it, image it, limn it if you will, without recourse to a more complex language than that of mathematics.  That language is poetry.  So, to come to grips in any meaningful way at all with questions of morality, say, you need to read King Lear -- neither legalese nor quantum mechanics have anywhere the resources required to do the job.

Preskill is a physical system all right --- unfortunately he's so complex that this statement doesn't get us very far.
\eq

\section{31-07-09 \ \ {\it Shelter Island and Bennett, Preskill, etc.\ Stories} \ \ (to D. Gottesman)} \label{Gottesman13}

I learned the source of confusion over Shelter Island:  The official name for the meeting was ``Shelter Island Conference on the Foundations of Quantum Mechanics''.  On the other hand, a standard description of the conference goes:
\bq\noindent
The general problem the conferees were asked to address was the impasse that elementary particle theory was perceived to have struck over the preceding decade and a half or so. Specific difficulties in quantum electrodynamics (QED) and in upper atmosphere meson phenomena were of particular interest, especially in light of Lamb's and Retherford's recent experimental findings on the fine structure of hydrogen, and Rossi's experiments with cosmic rays. Both Lamb and Rossi were in attendance, and both were asked to report on their findings. Discussion of the problems raised by these experiments, and the papers that resulted from these discussions, produced significant advances in the development of QED.
\eq
And this is reflected in the list of participants:
\bv
	Hans Bethe
\\	David Bohm
\\	Gregory Breit
\\	Karl K. Darrow
\\	Herman Feshbach
\\	Richard Feynman
\\	Hendrik Kramers
\\	Willis Lamb
\\	Duncan MacInnes
\\	Robert Eugene Marshak
\\	John von Neumann
\\	Arnold Nordsieck
\\	J. Robert Oppenheimer
\\	Abraham Pais
\\	Linus Pauling
\\	Isidor Isaac Rabi
\\	Bruno Rossi
\\	Julian Schwinger
\\	Robert Serber
\\	Edward Teller
\\	George Uhlenbeck
\\	John Hasbrouck van Vleck
\\	Victor Frederick Weisskopf
\\	John Archibald Wheeler
\ev
I'd only consider Bohm, Feynman, von Neumann, and Wheeler as ``quantum foundations'' in the usual PI sense.

Speaking of history, you might enjoy some of the stories of people we know in the attached file.  I just finished it; it's hot off the keyboard.  (People stories start up on page 4.)

\section{03-08-09 \ \ {\it  Interpretation of Bayesian Probabilities} \ \ (to D. H. Wolpert)} \label{Wolpert1}

Oh, I believe you:  In practice people are most always ``incoherent'' in the Dutch book sense.  I read Dutch book coherence (and hence the axioms of probability) normatively:  It is what you should strive for.

The quote you refer to is not van Enk himself, but Frank Ramsey.

\subsection{David's Preply}

\bq
Skimming over your fun document, I came across your description of van Enk's view. I don't know what he precisely meant by saying that to measure ``a person's belief \ldots\ propose a bet, and see what are the lowest odds which he will accept''. But it is well-established experimentally that as far as bets are concerned, people have no consistent ``degrees of belief'' in events. In some situations, even if money is on the line, a person will at the same time take a bet that requires $P(a) > P(b)$ and also another bet that requires $P(b) > P(a)$.

Look up ``Allais' paradox'' on Google. Another good one is the Ellsberg paradox, if you want to be cured of the belief that people use Bayesian degrees of belief in the real world (as opposed to in philosophy land).

Here are some other great references:
\begin{itemize}
\item
C. F. Camerer, {\sl Behavioral Game Theory: Experiments in Strategic Interaction}, (Princeton University Press, 2003).

\item
C. Starmer, ``Developments in Non-Expected Utility Theory: The Hunt for a Descriptive Theory of Choice under Risk,'' Journal of Economic Literature {\bf 38}, 332--382 (2000).
\end{itemize}
\eq

\section{03-08-09 \ \ {\it Where I've Been} \ \ (to R. {\Schack})} \label{Schack171}

\brs
Sorry, it isn't Logan, it's James Logue, {\bf Projective Probability}.
\ers
Oh, I see.  Has it had any impact on reflection issues?  What do you think of the main thesis of the book:  ``Based on a strongly subjective starting-point, with probabilities viewed simply as the guarded beliefs one can reasonably hold, the theory shows how such beliefs are legitimately `projected' outwards as if they existed in the world independent of our judgements.''  ??

\section{03-08-09 \ \ {\it Abner Story} \ \ (to A. Shimony)} \label{Shimony15}

I wrote a story about you:  It's in the attached document, an introduction for the Cambridge U. Press re-issue of my original email samizdat.  It'll be titled {\sl Coming of Age with Quantum Information}.  Your story starts on page 18, and I also say some things about you in the Zeilinger story on page 21.  I hope you enjoy.

Let me also alert to a new paper of mine with {\Schack}; it is my pride and joy from the last few months.  The title is ``Quantum-Bayesian Coherence,'' and you can find it here:  \arxiv{0906.2187}.  In it, we show a new way of writing the Born rule, a way that makes it look very, very close in form to the Law of Total Probability.  We use that then to argue that the Born rule should be viewed as an extra (contingent, empirical) addition to Dutch book coherence, to be used when one gambles on the outcomes of one's interactions with a quantum system.  It is a way of having our Bayesian cake and eating it too.  Section 8 addresses broader philosophical issues that may interest you:  How (I see) to remain subjectivist Bayesian about quantum probabilities and still accept objective indeterminism as one of the deep conclusions of quantum theory.

I hope all is going well for you.  I have not seen you in quite a while.

\subsection{Abner's Reply}

\bq
Your email novel (samizdat) is somewhat repetitious in plot, but as a compensation it is rich in characterization.

I was glad to be reminded of quoting Edward Lear's ``They all went to sea in sieve, they did'', which sums up my skeptical attitude towards a subjectivist interpretation of the probabilities of qm. Seeing the quotation reminded me that it was not original with Lear, since one of the witches in the fist act of {\sl Macbeth\/} says, ``But in a sieve I'll thither sail, And like a rat without a tail I'll do, I'll do, and I'll do'' --- wonderfully creepy!
\eq

\section{04-08-09 \ \ {\it Counterfactualizable} \ \ (to R. {\Schack})} \label{Schack172}

\brs
What do you think of this quote from Logue, attributed to L. J. Cohen's book {\bf The Dialogue of Reason}:
\bq\noindent{\rm
\ldots\ people are naturally inclined to think in terms of counterfactualizable rather than non-counterfactualizable probabilities wherever possible, reflecting a general dominance of causal reasoning \ldots\ over reliance on bare statistics.}
\eq
\ers
Nice.  Does this mean the upper path in our diagram is the more basic?

\section{04-08-09 \ \ {\it Personal Probability as \ldots\ Mode of Self-Attribution} \ \ (to R. {\Schack})} \label{Schack173}

\brs
If you find it, maybe you could read those 1.5 pages. It's section 1, entitled ``Personal Probability as Propositional Attitude and as Mode of Self-Attribution''. I know nothing about the context or where in philosophy this kind of idea is discussed, but it sounds a lot like our conviction that one must not think of events or measurement outcomes as existing independently of the agent.
\ers
Like usual, I have great difficulty {\it even\/} reading van Fraassen.  I kind of doubt that that is what he is getting at, but I don't know.  I did definitely note a kinship between us and Hacking, however.  (At least if I read him correctly.)  See section 6 of the attached on ``personal possibility''.

\section{04-08-09 \ \ {\it Charlie and Sappho Sitting in a Tree, \ldots} \ \ (to C. H. {\Bennett})} \label{Bennett64}

Well, I'm not too proud to help you remember.  {\Ruediger} sent me the note below, and I'm not ashamed to forward it on to you (to help you do so).

\subsection{{\Ruediger}'s Preply}

\bq
I had a long talk with Charlie Bennett at Dagstuhl, after his after-dinner speech on forgetting. I'd had quite a bit of that good Dagstuhl red wine, so although the discussion started with the reality of $\psi$, we then moved to logical depth, Homer, Borges, Kodaly, and Charlie's father's compositions in the octatonic scale which he played to me on the castle's grand piano.

His talk was implying there is a sense in which Sappho's lost poems are not real. I said then that if I had to choose between the reality of Sappho's poems and the reality of $\psi$, I would choose the former.  He clearly liked this formulation and said he would try to remember it (although when he tried to remember it the next day he misquoted it by stating the alternative as having a theory that does not give a complete picture of the world).
\eq

\section{04-08-09 \ \ {\it Preskill and Sappho As Well} \ \ (to C. H. {\Bennett})} \label{Bennett65}

To rub a little salt in the wound, see the Preskill entry in the attached.  Particularly, I hope you enjoy Footnote 18:
\bq
In a phrase, what is desired is a theory that is all outside and no inside. [Footnote: On first pass, the Texan in me wanted to write `all hat and no cattle,' but I repressed that.]
\eq

\subsection{Charlie's Reply}

\bq
I like ``all hat and no cattle'' much better, and am glad you didn't repress it (putting it in a footnote does not constitute a successful instance of repression).  I wish you wouldn't try so hard to persuade me of your interpretation of quantum mechanics. I think we can agree that the difference between preferring the reality of Sappho's lost poems vs.\ the reality of the wave function is a difference of interpretation, a matter of aesthetic or religious preference, not susceptible to being decided by rational argument or proof/refutation, since my interpretation logically contains yours and yours logically contains mine.  Moreover, I don't think it would even be desirable for one or another interpretation to win out, because that would impede the intuitions, often leading to substantive progress, that are stimulated, at least in some people, by other interpretations.   Perhaps one could say that the interpretations of quantum mechanics enjoy a complementarity reminiscent of quantum mechanics.

I think {\Ruediger}'s way of casting the question, as to which you would prefer if you couldn't have both, is an especially good way of explaining (to a lay audience likely to be bored by such arcane matters) what these different interpretations of quantum mechanics are all about, and why they are alternative interpretations, not alternative theories.
\eq

\section{04-08-09 \ \ {\it Charlie and Sappho Sitting in a Tree, \ldots, 2} \ \ (to C. H. {\Bennett})} \label{Bennett66}

\bcb
\ldots\ not susceptible to being decided by rational argument or
proof/refutation, since my interpretation logically contains yours and
yours logically contains mine.
\ecb

Aye, there's the rub:  Because I do not believe that that is true.

I'm gonna go put on my hat now \ldots

\section{04-08-09 \ \ {\it The {\Mermin} Challenge}\ \ \ (to C. G. {\Timpson})} \label{Timpson14}

I thought of you the other day when I was rereading through {\Mermin}'s paper, ``Writing Physics.''  Particularly at this point where he writes:
\bq
The puzzlement only arises when you try to combine what quantum mechanics tells you about the possible results of a group of mutually incompatible experiments. When you actually do any one such experiment you lose the ability to do any of the others. Why worry about what might have happened in the experiment you didn't do, if you no longer can do it? That's a problem for philosophers, not physicists.

I'm inclined to agree. The problem is that most philosophers who do worry about quantum mechanics differ from the physicists who refuse to worry only because they worry and the physicists don't. The philosophers have, by and large chosen to embrace nonlocality as a natural phenomenon, rather than homing in on what bad habits of thought and expression make so implausible an inference so hard to resist. Uncharacteristically for philosophers, they ought to be more worried about the nature of language, how it can trap us into formulating questions that have no sensible answers, and whether it is possible to restructure ordinary language in a way that liberates us from those built-in errors that make it so hard to think clearly about quantum physics. They ought, in short, to be worried about writing physics.
\eq
I thought, ``That's perfect for Timpson!''  So, I'll call it {\Mermin}'s Challenge and pass it on to you.  In that regard, let me also give you a little thing I just wrote (and the reason I was referring back to {\Mermin}'s article).  See attached.  The Benioff, {\Mermin}, and Preskill entries have a little bit to do with {\Mermin}'s challenge, and so may be entertaining to you.

Looking forward to seeing you again finally!

\section{10-08-09 \ \ {\it My Attraction to de Finetti's Side of the Force}\ \ \ (to P. G. L. Mana)} \label{Mana15}

Attached are the two pieces of reading I was telling you about.  Putting the two together, I think, starts to give some explanation for my attraction to the de Finettian side of Bayesianism. [See I. Hacking, ``Subjective Probability,'' {\sl The British Journal for the Philosophy of Science\/} {\bf 16}, 334--339 (1966) and the Introduction to C. A. Fuchs, {\sl Coming of Age with Quantum Information}, (Cambridge University Press, 2011).]

\section{11-08-09 \ \ {\it From Moore-ish Sentences to Bohrish Ones}\ \ \ (to C. G. {\Timpson})} \label{Timpson15}

In a better formulation, what is being claimed is this:
\begin{itemize}
\item ``I am certain that if I interact with {\it it}, I will have experience $x$; but there is no intrinsic property possessed by {\it it\/} corresponding to $x$.'' Or,

\item ``I am certain that if I interact with {\it it}, I will have experience $x$; but my certainty is not correspondingly of an intrinsic property possessed by {\it it}.''
\end{itemize}
Put this way, I feel the charge of our stating a Moore-ish sentence is diffused.  Maybe you still disagree?

I hope we get a chance to discuss further today.

By the way, I'm not sure, but I think you may have taken my {\Mermin} Challenge a bit too literally.  My (unstated) focus was only on this part of his quote:
\begin{quote}
for philosophers, they ought to be more worried about the nature of language, how it can trap us into formulating questions that have no sensible answers, and whether it is possible to restructure ordinary language in a way that liberates us from those built-in errors that make it so hard to think clearly about quantum physics.
\end{quote}
Of which, I thought this of you:  ``If anyone is in a position to see that that is the central point, it is {\Timpson}.''  Though, even I'm being a bit too loose here---for as I tried to express in the collection of stories I sent to you, it is not quite right that we need to ``restructure ordinary language,'' but to find better choices of words for describing what is being talked about.

Have you by chance read Section 8 in my \arxiv{0906.2187}?

\section{12-08-09 \ \ {\it More on Moore}\ \ \ (to M. Schlosshauer)} \label{Schlosshauer3}

Thanks for the remarks on the Moore-ish business.  The way I further diffuse your (and {\Timpson}'s) higher level Mooresm in 1) is to have the agent admit in those cases, ``Yeah, that's right, what I'm saying is, I can make it happen.  I fully believe it.''  Yesterday at coffee break, the (probably over the top) imagery I used was that, in cases of certainty, the agent is viewing the quantum system a bit as a surrogate mother.  He is confident that if he plants his seed there, the expected consequence will come about.  (Of course, the world still has the right to surprise him infinitely.) There is a bit more on this on page 667 of the present build of ``My Struggles'' in my last reply to {\Timpson}.  (Maybe I was confused on other points there---I haven't reread it completely---but on this particular point I don't think I've changed my mind any.)  [See 17-11-06 note ``\myref{Timpson11}{Certain Comments}'' to C. G. {\Timpson}.]

Anyway, remember the notion of indeterminism I am aiming for in Section 8 of the last paper (it is along the lines of that spelled out in the James quotes):  It is only that each piece of the world truly has the power to make a contribution to the whole, a contribution that can't be seen from the outside \ldots\ but may well be surmised from the inside.  And it is from the inside that the agent draws his probability assignments.

It'll be fun to read your choice James quotes (probably this weekend).

\subsection{Max's Preply}

\bq
Thanks for your further comments re: Moore-ism. It sounds like the key point of your argument is indeed, as I had suspected, to emphasize that bringing about the ``outcome'' requires the interaction of agent and system. In your note to {\Timpson} you write:
\bq\noindent
	It just means that we should be a little more
	careful to say, ``there is no fact in the object
	guaranteeing the outcome.'' Of course, there is no
	fact in the agent guaranteeing the outcome either;
	otherwise his certainty wouldn't be subjective
	certainty.
\eq
I understood the problem that's supposedly evidenced by {\Timpson}'s Moore-ish sentences as setting up a tension between ``I am certain about what I will experience if I interact with the system'' on the one hand, and ``as a good Qbist, I must consciously proclaim that there's no fact guaranteeing what I will experience'' on the other. Where could the fact reside? Per your quote above, it doesn't resides in either system or agent alone. Is there a fact in agent-plus-system? Plus the rest of the world? Is there ever a fact ANYWHERE that guarantees a particular experience to occur? I guess for the QBist the answer would be no; in your email you say: ``Of course, the world still has the right to surprise him infinitely'' -- and wouldn't the agent, as a die-hard Qbist, have to take this insight to heart?

On the other hand, this just seems to be a completely generic issue. As long as the agent maintains that there are simply no facts, anywhere, before the agent/system interaction that guarantee a certain outcome, aren't we back to {\Timpson}'s original charge? (As I've said before, I don't believe this charge is a serious objection to begin with, but this whole discussion and attempt to rephrase the Moore-ish sentences is obviously built on the premise that it is indeed an objection, to an extent at least.) From Sec.\ 8 of your last paper:
\bq\noindent
	Whenever ``I'' encounter a quantum system, and
	take an action upon it, it catalyzes a
	consequence in my experience that my experience
	could not have foreseen.
\eq
Does my assessment of ``could not have foreseen'' take on any different flavor when I assign probability one (i.e., when I am certain about which consequence I will experience)?

(Well, I hope I'm not beating the proverbial dead horse here. Feel free to tell me if I do, and feel free to give this discussion a rest if you
like.)

By the way, I like the ideas, however vague, spelled out in your note ``Philosopher's Stone.'' Especially this one:
\bq\noindent
	Reciprocally, there should be a transmutation of
	the system external to the agent. But the great
	trouble in quantum interpretation--I now think--is
	that we have been too inclined to jump the gun
	all these years: We have been misidentifying where the
	transmutation indicated by quantum mechanics (i.e.,
	the one which quantum theory actually talks about, the
	``measurement outcome'') takes place. It should be the
	case that there are also transmutations in the external
	world (transmutations in the system) in each quantum
	``measurement'', BUT that is not what quantum theory
	is about. It is only a hint of that more interesting
	transmutation.
\eq
I look forward to seeing these thoughts fleshed out; in the meantime, I'm just going to enjoy letting them play on my mind for a while.
\eq

\section{12-08-09 \ \ {\it History Repeating Itself \ldots}\ \ \ (to C. G. {\Timpson})} \label{Timpson16}

\bcgt
I'm stuck in bed with flu-like symptoms. I've emailed Philip to say I probably have to quarantine myself and stay in my room to avoid infecting everyone else, just in case it's swine flu; so I don't think I'm going to be able to give my talk tomorrow.  Hopefully I'll be recovered to give it b4 the end of the conf, but I thought it might make a difference to your own talk preparation if I'm not going to be able to give mine before you tomorrow. I was just going to summarise my SHPMP paper on q Bayesianism (tho leaving out the Moore's sentence stuff as it takes too long to explain).
\ecgt

I'm really crappy.  I'm sorry to hear that.  Do get better, and I really hope you can ultimately give your talk.

Indeed I was going to bank on it for a set-up on why I would seek an interpretation of the Born rule in terms of ``an addition to coherence.''  But I'll survive!!

Do get better:  I hope this is not a recurring thing {\it every\/} time we get together.

\section{12-08-09 \ \ {\it History Repeating Itself \ldots, 2}\ \ \ (to C. G. {\Timpson})} \label{Timpson17}

Rereading my note, I see I wrote in my first sentence:  ``I'm really crappy.''  That may be true!  But I meant to say, ``That's really crappy.''

I think I'll stick with my present time, so as to get it over with.  But I'll advertise your potential talk, and start off a bit like Bill Wootters did:  ``Suppose for whatever reason you ended up with an interpretation that looks like this.  How could that interpretation {\it imply\/} any of the actual structure of quantum mechanics?''  Then I'll just go down the SICkening path.  With regard to why I cared to end up with an interpretation like that as my starting point, I'll refer to your (hoped for) talk on the strengths and weaknesses of the position.

If you feel well enough before going back to England, I'd like to have you and Marcus {\Appleby} over for dinner (and wine and single malt) one evening.  I'm very proud of my new library and liquor cabinet and like to show it off to the best people.

Get well!

\section{15-08-09 \ \ {\it Roiling Mess}\ \ \ (to C. G. {\Timpson} and R. W. {\Spekkens})} \label{Timpson18} \label{Spekkens67}

When you talked about the ontology of the roiling mess today, you brought my mind back to some lines in William James's essay ``The Dilemma of Determinism.''  Maybe you've never seen them before:
\bq
Let me, then, without circumlocution say just this. The world is enigmatical enough in all conscience, whatever theory we may take up toward it. The indeterminism I defend, the free-will theory of popular sense based on the judgment of regret, represents that world as vulnerable, and liable to be injured by certain of its parts if they act wrong. And it represents their acting wrong as a matter of possibility or accident, neither inevitable nor yet to be infallibly warded off. In all this, it is a theory devoid either of transparency or of stability. It gives us a pluralistic, restless universe, in which no single point of view can ever take in the whole scene; and to a mind possessed of the love of unity at any cost, it will, no doubt, remain forever unacceptable. A friend with such a mind once told me that the thought of my universe made him sick, like the sight of the horrible motion of a mass of maggots in their carrion bed.

But while I freely admit that the pluralism and the restlessness are repugnant and irrational in a certain way, I find that every alternative to them is irrational in a deeper way. The indeterminism with its maggots, if you please to speak so about it, offends only the native absolutism of my intellect,---an absolutism which, after all, perhaps, deserves to be snubbed and kept in check. But the determinism with its necessary carrion, to continue the figure of speech, and with no possible maggots to eat the latter up, violates my sense of moral reality through and through. When, for example, I imagine such carrion as the Brockton murder, I cannot conceive it as an act by which the universe, as a whole, logically and necessarily expresses its nature without shrinking from complicity with such a whole. And I deliberately refuse to keep on terms of loyalty with the universe by saying blankly that the murder, since it does flow from the nature of the whole, is not carrion.
\eq
I thought you might enjoy the imagery of maggots.  I'll cc this to Rob {\Spekkens} for fun.

\section{15-08-09 \ \ {\it Nice Word} \ \ (to R. W. {\Spekkens})} \label{Spekkens68}

Well, I suspect our conversation didn't actually send you to wiki to learn about empiricism, but it did to me.  Here's a nice word I just learned:
\myurl{http://en.wikipedia.org/wiki/Hypokeimenon}.

I suppose I am a bit of a Lockean.

\subsection{Rob's Reply}

\bq\noindent
That is a good word, but too long to remember.  Henceforth, I'll refer to it as your ``Pokemon philosophy''. \ \smiley
\eq

\section{17-08-09 \ \ {\it Footprints}\ \ \ (to M. Schlosshauer)} \label{Schlosshauer4}

\bmaxs
I just came across this sentence in Roberto Calasso's novel {\bf The Marriage of Cadmus and Harmony}:
\begin{center}
{\rm ``For every step, the footprint was already there.''}
\end{center}
A nice way of expressing the spirit of determinism, don't you think?
\emaxs
Yes, that's perfect!

And even ``indeterminism'' of the block-universe type that {\Timpson} defended so strongly yesterday in our extended discussion.  (There is a bit of that point of view in his talk at \pirsa{C09016}.)

\section{17-08-09 \ \ {\it Timezones (a conversation)}\ \ \ (to M. Schlosshauer)} \label{Schlosshauer5}

Yeah, it's a bit after 4:00 AM for me.  I should get back to bed at some point; woke up at 2:30 and couldn't get back to sleep.

By the way, in that PIRSA talk, there are about 10 transparencies that you didn't see in {\Vaxjo}.  I drew them Wednesday in my PI office with your jazz streaming in the background.  That's really very nice stuff.  (I love jazz of all kinds, and have about 15,000 tracks of it in my iTunes folder.)  Very impressive, that you perform and do serious physics as well.  I'd like to see you and your group some day.

\bmaxs
Thanks a lot for your kind words about my music! It's nice to hear that the tunes served as a soundtrack for your drawing session. I'll check out the video of your talk shortly and will try to figure out which transparency goes with what song.

As you may have read on the website, all tracks are 100\% live
improvised. I.e., there's no planning, no sheet music, no pre-arranged
chords or melody, nothing whatsoever, before we go on stage. So it's
all created in the moment in an act of collective improvisation.
\emaxs
I've wondered in the past whether that might be a good metaphor for the kind of reality I'm trying to envision.

\bmaxs
That's very nice!! Indeed it would seem like an apt metaphor.

So once again it seems that the ``temperament'' (as James would say) informs quite consistently the way we each choose our favorites, be it in regards to our conception of reality, our style of music, or whatever else \ldots
\emaxs
Yeah, it's weak but see pages 495--496 in {\sl My Struggles}.  Also I'd like to think that's what I was thinking when I wrote the note to Garisto below \ldots\ but who knows.  [See 13-08-01 note ``\myref{Garisto3}{Out of the Frying Pan \ldots}'' to R. Garisto.]

In any case, you're much better positioned to develop the metaphor than me!  Contemplate it.

Signing off for the evening \ldots

\bmaxs
Thanks! I just had a look at pp.\ 495--496. I guess there were two notes
of relevance? (``The Free City,'' on Christiania and ``Inventory,'' on
jazz and pragmatism.) Sweet little ideas and recollections in each of
them, I thought.

I'm sure you know that there's generally quite an affinity between physicists and music. Heisenberg and his piano playing come to mind, and of course our very own David {\Mermin}. Somehow, though, there also appears to exist a bias toward classical music with most of these people! In some sense, classical music seems to me like the prototypical deterministic edifice, an ``unrolling of the fabric of the past.'' (Don't get me wrong, I enjoy listening to this kind of music very much; it's just the spirit of actually playing it myself that never kindled for me.) To be sure, each musician brings his own ``interpretation'' to a piece, but the notes are all there and laid out already (could this be a metaphor for the usual interpretive programs of QM?!). Jazz, on the other hand, is spontaneous creation, as loosely constrained by rules and theory as one wishes.

I think I really like the jazz-and-QM metaphor \ldots!
\emaxs
Oops!  I didn't realize the proximity of the Free City story.  Oh well.

[See 13-10-04 note ``\myref{Baker11}{The Free City}'' to D. B. L. Baker and 25-10-04 note ``\myref{LentzS3}{Inventory}'' to J. B. Lentz \& S. J. Lentz.]

\section{17-08-09 \ \ {\it {\Timpson}}\ \ \ (to M. Schlosshauer)} \label{Schlosshauer6}

\bmaxs
I watched {\Timpson}'s talk this afternoon. Indeed a very nice and well-done presentation. I noticed that he made no mention of the Moore-isms. Did conversations with you convince him that there was maybe no serious problem after all?
\emaxs

I think it shook his faith sufficiently that he preferred not to get into it in public.  But of course I don't know for sure.  I think that, at the end, he was still inclined to think he was right \ldots\ but couldn't vouch for it so confidently anymore.

I think the key point is, if one adopts a block-universe picture at the outset, the CFS claim will always look suspicious.  For {\Timpson}, there is just no sensible notion of ``facts coming into being'' and I think that might lie at the root of his continued worry.  But that's a longer story than I feel like writing down at the moment.  I wish I could have had a tape recorder for yesterday's 3-hour conversation between him, {\Appleby}, Wootters, and me.  I'll try to get it all down eventually, and then send you a draft at that point.

We just had a nice (but nearly trivial) result:  Maximal consistent sets must be compact.

\section{17-08-09 \ \ {\it Your Thoughts on Timpson}\ \ \ (to M. Tait)} \label{Tait2}

You missed a really good meeting.  Timpson's talk was excellent; I'd really recommend you watch it.  All the talks can be found here:
\pirsa{C09016}.  I think my talk was fairly decent as well---unfortunately though, in this way of viewing things, you won't be able to see to where I'm pointing in the diagrams.

Timpson has another paper as well that should interest you.  This one is directly about QBism.  You can find it here:  \arxiv{0804.2047}.  We had several discussions this week on the Moore-sentence point he brought up there.  I'm pretty sure I've got that all safely wrapped up now.  If you ever get to that point and need my notes; I don't mind sharing them.  Sometime this year, {\Ruediger} and I will write a formal reply.

Sorry I haven't answered about your visit request.  The second week of November is not so good for me; I have to go to North Carolina during that time.  I'm free however Oct 27 -- Nov 7, so I tentatively marked the time in my calendar.  If that time is good for you, I'll send an official request in to the secretaries to check accommodation availability.

Attached are some stories I wrote about various characters in quantum information and quantum foundations.  You might enjoy some of them---I don't know.  A couple were also intended to carry philosophical messages as well (like the ones on {\Mermin} and Preskill).  [See Introduction to C. A. Fuchs, {\sl Coming of Age with Quantum Information}, (Cambridge University Press, 2011).]

\subsection{Morgan's Preply}

\bq
I wanted to say how sorry I am to have missed Chris Timpson's talk on Saturday (I had another commitment). I've read a couple of Timpson's papers on quantum Bayesianism:
\begin{enumerate}
\item
`Information, Immaterialism, Instrumentalism:\ Old and New in Quantum Information' forthcoming in A. Bokulich and G. Jaeger (eds.)\ {\sl Philosophy of Quantum Information and Entanglement\/} (Cambridge University Press 2008)
\item
`Philosophical Aspects of Quantum Information Theory' forthcoming in D. Rickles (ed.)\ {\sl The Ashgate Companion to the New Philosophy of Physics\/} (Ashgate 2008)
\end{enumerate}
and I would have dearly liked to see your reaction to his talk. I don't suppose anyone was filming the talk? If not, I hope we can have a conversation at some point about Timpson's take on your views (if in fact you think he is representing your view at all!).
\eq

\section{17-08-09 \ \ {\it The Activating Observer} \ \ (to J. Rau)} \label{Rau2}

It was good seeing you again.  Sorry to be so very ``thick''---I am a really slow thinker.  I think the line of thought you (or you and Terry) are pursuing is a good one.  After your prodding me yesterday, it does seem to me deeply connected to the point of view I'd like to see developed that unitaries be representable as sequences of measurement.  You'll note for instance, in Section 8 of my recent \arxiv{0906.2187}. I reserve the notion of an ``action'' on a quantum system to correspond to a POVM\@.  I don't, for instance, think of unitaries as actions---which of course goes against the grain of almost everyone else in quantum information.  (Where unitaries are usually viewed as the things you want to ``do'' to a system, and measurement is, well, something mysterious \ldots\ but more analogous to ``looking.'')  What you guys are developing may help fill the gap for me.

I have always thought of the ``engagement'' of the observer in quantum phenomena as best expressed (or demonstrated) by the Kochen--Specker theorems (particularly Stairs's version of it, combining KS with considerations to do with entangled states).  It would be interesting to see if a more direct connection between your considerations and KS could be fleshed out.

Anyway, I endorse.  And at some point I'll come back and watch your talk again.

Let me now back up to the broad philosophical point.  Attached are a couple of files that might entertain you.  Maybe I gave you a paper copy of the ``Resource Materials'' file on your last visit?  (If I haven't, maybe it will help give you further spiritual strength in wavering times.)  But one file I know you haven't seen before---it has some funny stories in it, and some sad stories as well.  [See Introduction in C. A. Fuchs, {\sl Coming of Age with Quantum Information}, (Cambridge University Press, 2011).]  The entry on Nielsen is directly relevant to our conversation above.  (It makes a bit more sense if you read the Carl Caves entry first.)

\section{18-08-09 \ \ {\it Question Regarding Your Preprint} \ \ (to B. R. La Cour)} \label{LaCour1}

\bbrlc
I'm reading your paper, ``Quantum-Bayesian Coherence,'' which you kindly gave preprints of during the {\Vaxjo} conference this summer.  In it you (and {\Ruediger}) speak of a ``measurement in the sky'' involving a SIC POVM.  Now, in a $d$-dimensional Hilbert space, the measurement of any given observable can have at most $d$ outcomes.  How, then, do you construe that a SIC POVM may be used to perform, even hypothetically, a single measurement with $d^2$ outcomes?
\ebrlc

The measurement in the sky is a positive operator valued measure (POVM).  POVMs are not restricted to $d$ outcomes.  A good place to start reading about this is:
\bv
\myurl{http://en.wikipedia.org/wiki/POVM}
\ev
and references therein.

\section{19-08-09 \ \ {\it Question Regarding Your Preprint, 2} \ \ (to B. R. La Cour)} \label{LaCour2}

\bbrlc
I understand that these are POVMs, so you assume that they may be
measured simultaneously.  But they are also (scaled) projection
operators and, as such, may be thought of individually as standard,
von Neumann measurements.

The trouble is that, as a member of a SIC-POVM, they are assigned one probability and, as isolated measurements, they have another, larger probability.  If you'll forgive the rhetorical question, how does one member of a SIC-POVM know that the others are being measured?  Do you consider this a reflection of quantum contextuality?
\ebrlc

You talk about a POVM as if it is a plural---your usage of ``they'' gives it away.  Rather a SIC-POVM is a {\it single\/} measurement with $d^2$ possible outcomes.  It is true that a POVM is written down mathematically as a set of hermitian operators, but those operators are only meant to {\it index\/} the outcomes of the measurement.  (Much as the $d$ eigenprojectors can be used to index the outcomes in a standard, von Neumann measurement.)  In this context, the operators shouldn't be thought of as separate observables themselves.

If you want to think of a POVM in terms of the old von Neumann notion of measurement, you must make use of helper or ancillary systems.  For instance, one way to instantiate a SIC-POVM on a three-level atom is to first let the three-level atom unitarily interact with a nine-level atom (i.e., $3^2$), then take the nine-level off and perform a standard von Neumann measurement on it alone.  That gives rise to a measurement with 9 outcomes and by way of the preceding interaction may be thought of as saying something indirectly about the three-level atom.  (But the ``indirect'' here is a bit of a misnomer, since by this method, one probes the three-level atom more deeply than can be done by any standard measurement.)

So, in all, I would say you are contemplating two kinds of distinct measurement:  One, a POVM with $d^2$ outcomes, $\frac{1}{d}\Pi_i$, $i=1,\ldots,d^2$.  The other, a POVM with two outcomes, $\Pi_k$ and $I-\Pi_k$, say.  How does the system know which is being measured on it?  It takes two distinct kinds of interaction to enact them.  In old terms, the first requires interaction with an ancilla, while second is made directly on the system (no ancilla needed).

Let me this time recommend something further than the wiki article.  Have a look at John Preskill's notes referenced therein.\footnote{\editornote See \myurl{http://www.theory.caltech.edu/people/preskill/ph229/\#lecture}.}  As I recall, they were an excellent introduction to the nuts and bolts of POVMs.

Hope you are braving the Texas heat.  I've been mowing the grass today and it's about killing me at 75 degrees!

\section{19-08-09 \ \ {\it Utterly the Last Time} \ \ (to the QBies)} \label{QBies1}

The more I think about it, the more I think I should make today the day of my concerted effort in the lawn.  Kiki's trying to put a push to finish the porch today; so I'm going to put a push on getting the lawn ready for her parent's visit.

Marcus (sorry {\AA}sa), if you need a break from pure writing, and want to talk pure philosophy for a while---i.e., how the world is not sentence shaped, but made of neutral stuff, etc., I can imagine that happening while I'm moving/stacking rocks and edging, as long as you're within ear distance and willing to sort of follow me around.  If you get bored, feel free to drop by.

\section{20-08-09 \ \ {\it More on the Cover Story} \ \ (to S. Capelin)} \label{Capelin9}

I took the words literally.  It didn't say provide an {\it idea\/} for an image; it said provide an image.  (One that would need to be approved, but a seemingly final image no less.)  It didn't give me a sense that you have a design staff that does the real artwork.

Anyway, now that I understand that you are only seeking ideas, and that there is a real artist behind the scenes, I will try harder.  The cover of Schumacher and Westmoreland's book is indeed very nice, beautiful even.  And the meaning (I'd bet money) was not lost on me.  The wave symbolizes a solution to the {\Schroedinger} equation; the spots at the bottom left symbolize ``its'' (electrons, photons, whatever), the zeros and ones in the upper right symbolize ``bits.''  The whole image is an encoding of John Wheeler's phrase ``it from bit''.  Of course, many in the public won't know that, but it surely must be satisfying to Ben and Mike to have his deep idea encoded on the cover of his book in that way.

The two images I've sent you, amateurish though they were, were similarly infused with meaning for me.  And in the hands of an artist with some imagination---I would well imagine---could be made to amount to something.  You'd get your sales, and I'd go to bed in the evenings with a satisfied soul.

\bsc
If you really can't think of anything else, then we'll ask the
designer to try to make something abstract using the psi image you
sent.  Is there any reason why what you supplied has the symbol three
times?  Or can we use just one?
\esc

I don't know what to tell you at the moment, but I will try to think harder.  I'd hope you'd listen harder as well, or put me in touch with your graphic artist directly so we could more efficiently bandy about ideas.  What that image was trying to get at was a wave function $|\psi\rangle$ rising off of a physical object.  That was the base idea I was trying to convey in some way.  The wave function initially on the object, flies off of it, and moves still further away \ldots\ strengthening in color as it does.  (I'm not wedded to that:  I'd prefer some variation on my stick man if an artist with some imagination could make it more acceptable.  But I already know it does nothing for you, so that idea is probably already dead in the water.)  The symbol was there three times because I didn't know how else to give a sense of it flying off the object (the cube).

Could something be done with a $|\psi\rangle$ and an image of a Bohr atom?  Maybe make a $|\psi\rangle$ fly out of a Bohr atom in some interesting way?  It'd be easier if I could have a conversation with an artist directly.  The image of $|\psi\rangle$ carries for me the very core of quantum mechanics.  (It was a symbol invented in Cambridge, by the way, by Dirac.)

I'm not trying to cause trouble for you.  If you want an image---as I now understand you to very much want---I'll try to work with you.  (My anger stemmed only from having wasted a lot of time via the misunderstanding, trying to play a graphic artist when I am not.)  If you just need hand sketches and ideas, I can shoot them out whenever they come.  But it would be death to my spirit if the endpoint were to be something like the image on the book you mentioned---I really do dislike it, and its meaninglessness is very apparent to me.

\section{20-08-09 \ \ {\it That Formulation}\ \ \ (to L. Hardy)} \label{Hardy37}

Boy did I botch that at lunch.  Below is the true blue thing.  [See 06-07-09 note ``\myref{Wiseman27}{The Verdict}'' to H. M. Wiseman.] The key point is that previous to quantum mechanics agents had reason to doubt their own existence (at least in any deep sense).  That is, it is not that quantum mechanics causes me to doubt your existence, but rather to have better reasons for believing in my own.

\section{21-08-09 \ \ {\it Quantum Locality}\ \ \ (to R. B. Griffiths)} \label{Griffiths1}

\brbg
I have posted the item indicated below on \arxiv{0908.2914}; you will find the title and abstract at the end of this note.

In addition to getting the science right I want to give proper recognition to other people's work, cite the most relevant material, present my own ideas in as clear a manner as possible, and remain courteous in disagreements.  Any comments or suggestions in these or other respects are welcome.
\bq\noindent{\rm
TITLE: Quantum Locality\medskip\\
ABSTRACT: It is argued that while quantum mechanics contains nonlocal or entangled states, the instantaneous or nonlocal influences sometimes thought to be present due to violations of Bell inequalities in fact arise from mistaken attempts to apply classical concepts and introduce probabilities in a manner inconsistent with the Hilbert space structure of standard quantum mechanics. Instead, Einstein locality is a valid quantum principle: objective properties of individual quantum systems do not change when something is done to another noninteracting system. There is no reason to suspect any conflict between quantum theory and special relativity.}
\eq
\erbg

Thanks for the heads up.  I'll read your paper with interest.

I'd like to think that I too have been a rather staunch one of the ``some dissenting voices'' you mention in the first paragraph of your introduction.  Indeed it has been what has driven my epistemic (more particularly Bayesian) view of quantum states, expressed for instance in the opinion piece Peres and I wrote for {\sl Physics Today\/} way back (that you and Todd commented on), more recently in \arxiv{quant-ph/0608190} and most recently in \arxiv{0906.2187}. Timpson also makes the point pretty clearly in his review of us:  \arxiv{0804.2047}.

So, with me you preach to the choir.  But it is good indeed to see things from all angles, and it'll be fun reading your paper.

\section{22-08-09 \ \ {\it A Stupid Question for Saturday}\ \ \ (to N. D. {\Mermin})} \label{Mermin157}

\bdm
If it turned out that there is a dimension above which there are no
SIC POVMs, would this wipe out their foundational significance?
\edm

Yep, it certainly would.  But as my wise student from Vietnam, Hoan Dang, tells me every time we find a new wonderful property for the SICs, ``They have just become more expensive.''  My opinion really is this:  THEY HAVE TO EXIST.  Too many things simply become too pretty (supposing only their existence) for them not to.

\section{24-08-09 \ \ {\it The Penrose Tale}\ \ \ (to N. D. {\Mermin})} \label{Mermin158}

\bdm
Tell Hoan Dang to read Fermat's Last Theorem, by Harold  M. Edwards, pp.~76--79, especially the last paragraph on the top of p.~78 for a cautionary historical parallel.

Another cautionary tale, relevant to your own attitude.  When I was a graduate student there was a story, no doubt apocryphal, about a guy who wrote a Ph.D. Thesis on analytic functions of a quaternion variable.   He proved all kinds of spectacular
properties --- much more beautiful than analytic functions of a complex variable.    When had written it
all up, somebody discovered a proof that the only analytic functions of a quaternion variable are linear.

To be sure, you have many nontrivial examples, so the second cautionary tale is less frightening.
But nevertheless, be wary!  And challenge every mathematician you meet to find a proof
that they exist in all $d$, or a counterexample.
\edm

Indeed I've been doing that for quite some time.  Simon Kochen, John Conway, Elliott Lieb, Peter Shor, \ldots\  The list is quite big.  John Conway had a cute reaction --- he had worked on the real-vector-space version in the late sixties and sometime in the seventies.  There, the corresponding upper bound on the maximal number of equiangular lines is $d(d+1)/2$.  But it is well known that the bound is saturated only in a few sparse dimensions.  When I told John that it appears that the $d^2$ bound is always achieved in the complex case, he was quite delighted.  He thought for just a couple of minutes and declared that he thinks the bound will indeed be achievable in all dimensions.   But he quickly followed with, ``But it will be very hard to prove it.  There are too many of them.''  Unfortunately, I have no idea what he meant by that; we were interrupted immediately afterward, and I could never get back to him.

Anyway, I wanted to tell you a story about Roger Penrose.  You probably know that the known $d=4$ SICs are already quite nontrivial.  See equations 26 to 29 in \arxiv{quant-ph/0310075}.  Here's a conversation I had with Penrose just before (and continuing immediately after) Harvey Brown's talk at last year's PIAF meeting.

\bv
CHRIS:  I have a mathematical problem that I think will interest you, and it's one I could use some guidance with.  Here's the question.  For a complex vector space of dimension $d$, what is the maximal number of equiangular lines you can have in it?

PENROSE:  You mean any two lines have the same angle between them?

C:  Yep, exactly that.

P:  Oh, that'd be a very hard question.  [A bit of a grimace on his face.]

C:  Believe me!  I and several others have been working on it for quite some time now.  The interesting thing is the answer seems to be simply $d^2$.

P:  You mean order $d^2$.

C:  No, I mean precisely $d^2$.

P:  Oh!  [His face lighting up like a child's.]  That's very interesting.  You mean the maximal number appears to be precisely $d^2$?

C:  Yep, out to dimension 47 at least.  [Now we know it goes out to 67 at least.]

[The talk begins; we sit down.  As the talk proceeds, I can see Penrose out of the corner of my eye; he's just sitting there politely, apparently listening to the talk. When the talk is over, he turns to me and says:]

P:  Well, I can see that they exist in dimensions 2, 3, and 4.  What's difficult to see is if one can construct a 3-dimensional set by suitably projecting a 4-dimensional one.
\ev

I was floored.  When he started to tell me more details about the $d=3$ case, I could tell he had used the same construction technique that Lane Hughston had used in his 2 months of work on the problem (Lane classified all solutions in $d=3$ by studying cubic equations and Hessian forms or some such).  In $d=4$, Penrose said the same sort of thing generalized, but I remembered well that Lane had never been able to make it work.  And Penrose did that much without pencil or paper!

\section{26-08-09 \ \ {\it The Poetry of Chance}\ \ \ (to N. Gisin)} \label{Gisin1}

It was fun sparring with you yesterday as ever.  And also as ever, I was struck that there are some points upon which we seem to agree on (or at least seem to feel as the right direction forward).  One of them is the issue of will, if I understand you correctly.  I want to send you two quotes of William James that I think are quite pretty and express the variant of chance and will that I was trying to convey to you at lunch.  Maybe you will enjoy them, or maybe you will disagree.  If it is the latter, I would like to understand your point.

\bq
[Chance] is a purely negative and relative term, giving us no information about that of which it is predicated, except that it happens to be disconnected with something else---not controlled, secured, or necessitated by other things in advance of its own actual presence.  \ldots\ What I say is that it tells us nothing about what a thing may be in itself to call it ``chance.''  \ldots\ All you mean by calling it ``chance'' is that this is not guaranteed, that it may also fall out otherwise.  For the system of other things has no positive hold on the chance-thing. Its origin is in a certain fashion negative: it escapes, and says, Hands off! coming, when it comes, as a free gift, or not at all.

This negativeness, however, and this opacity of the chance-thing when thus considered {\it ab extra}, or from the point of view of previous things or distant things, do not preclude its having any amount of positiveness and luminosity from within, and at its own place and moment.  All that its chance-character asserts about it is that there is something in it really of its own, something that is not the unconditional property of the whole.  If the whole wants this property, the whole must wait till it can get it, if it be a matter of chance.  That the universe may actually be a sort of joint-stock society of this sort, in which the sharers have both limited liabilities and limited powers, is of course a simple and conceivable notion.
\eq

\bq
Why may not the world be a sort of republican banquet of this sort, where all the qualities of being respect one another's personal sacredness, yet sit at the common table of space and time?

To me this view seems deeply probable.  Things cohere, but the act of cohesion itself implies but few conditions, and leaves the rest of their qualifications indeterminate.  As the first three notes of a tune comport many endings, all melodious, but the tune is not named till a particular ending has actually come,---so the parts actually known of the universe may comport many ideally possible complements.  But as the facts are not the complements, so the knowledge of the one is not the knowledge of the other in anything but the few necessary elements of which all must partake in order to be together at all.  Why, if one act of knowledge could from one point take in the total perspective, with all mere possibilities abolished, should there ever have been anything more than that act?  Why duplicate it by the tedious unrolling, inch by inch, of the foredone reality?  No answer seems possible.  On the other hand, if we stipulate only a partial community of partially independent powers, we see perfectly why no one part controls the whole view, but each detail must come and be actually given, before, in any special sense, it can be said to be determined at all.  This is the moral view, the view that gives to other powers the same freedom it would have itself,---not the ridiculous `freedom to do right,' which in my mouth can only mean the freedom to do as {\it I\/} think right, but the freedom to do as {\it they\/} think right, or wrong either.
\eq

For myself, part of this vision of the universe as a ``republican banquet'' is to recognize that probabilities are not rigidly connected to nature, but are of the personalist Bayesian variety even in quantum mechanics.  Nonetheless we do not have to agree on that to feel some overlap in our thoughts.

I have a fuller sketch of how I see the notion of chance above as fitting with my understanding of quantum mechanics, and also why I insistently call myself a ``realist'', in Section 8 of \arxiv{0906.2187}.  You might find that section fun reading, even if you throw it in the bin afterward.

\section{27-08-09 \ \ {\it `Empirical Coherence' and Invitation}\ \ \ (to I. Hacking)} \label{Hacking1}

You may not remember me, but we met at a meeting in the Carr\'e des Sciences in Paris, June 2008.  You seemed somewhat interested in the personalist Bayesian account of quantum probabilities that a few colleagues and I have been working out.  And you also seemed somewhat intrigued by my remarks that we could still have (and quite desire) an ``objective indeterminism'' from quantum mechanics without ever once invoking objective chances (for instance Lewisian) or propensities.  In that regard, let me call your attention to this paper that we've posted recently:
\arxiv{0906.2187}.

It is crucial to us that there is no such thing as a ``right and true'' quantum state.  But on the other hand the Born rule for calculating quantum probabilities is indeed used all the time, and thus calls for an explanation in our terms.  Our solution is not to think of it as defining a propensity, but something more along the lines of probabilistic coherence.  In other words, the Born rule should be viewed as a normative principle for relating probabilities.  The idea being that if one does not make sure his probability assignments (for the outcomes of various potential measurements) are related according to the dictum of the Born rule, Nature is liable to smack his hands.  In contrast to Dutch-book coherence, however, the origin of the normative rule is not of a purely logical nature, but should be seen rather as dependent upon contingent features of our actual world.

Anyway, that's the sort of stuff in the paper, and I hope you will find parts of it entertaining and thought provoking.  Particularly, any feedback you have for me on the Intro, the Dutch-book section, and the concluding Section 8 (on a William Jamesian ``republican banquet'' non-block universes), would be most appreciated.  And I would be flattered to hear back from you.  (Most recently I've been studying your ``Slightly More Realistic Personal Probability'' from 1967, and think parts of it are quite in line with the views we are working out, and crucial to us actually.)

That said, let me change subjects abruptly.  I saw you at Nicolas Gisin's talk two days ago, but I was not able to catch you afterward to convey the following invitation.  It would have saved me some writing!

Steve Weinstein (philosophy, U.\ Waterloo), David Wolpert (NASA), and I are organizing a meeting of physicists, philosophers, and mathematicians at the Perimeter Institute in Waterloo for next year titled, ``Laws of Nature:\ Their Nature and Knowability.''  The target date presently is May 20--22, but we have some flexibility to modify that as needed to attract the best people.  We envision about 20 external participants, plus 6 from PI.  I have been assigned to contact you, Cheryl Misak, Huw Price, and Bas van Fraassen for a sort of pre-invitation before we send out other invitations more generally.  I.e., we really want you to come, and think you'd be a great asset for attracting some of the others that we plan to invite.  Of course, all your expenses would be paid by PI, and I personally am itching to have some refreshed face-to-face conversations with you.

Attached is a list of topics for the meeting written by my colleagues.  I hope it will give you some sense of what the meeting will address.

I very much hope you will say yes, and I'll wait on needles until I hear back from you!

\section{27-08-09 \ \ {\it Progress and Invitation} \ \ (to B. C. van Fraassen)} \label{vanFraassen22}

It is nice to have a reason to write you again, but unfortunately it is not yet to send you a draft of our paper ``Quantum Bayesian Decoherence.''   [See \arxiv{1103.5950}.]  Writing it has been painfully slow, but we do feel that we are really clarifying things with it.  At the moment, we're working on an island within the paper that may become a paper on its own:  We are making a strong endorsement of your reflection principle and want to give that a lot of care in presentation to ward off some of the evils we have seen you endure!!

Instead at the moment, I need to write you about something else.  Steve Weinstein (philosophy, U. Waterloo), David Wolpert (NASA), and I are organizing a meeting of physicists, philosophers, and mathematicians at PI for next year titled, ``Laws of Nature:\ Their Nature and Knowability.''  The target date presently is May 20--22, but we have some flexibility to modify that as needed to attract the best people.  We envision about 20 external participants, plus 6 from PI.  I have been assigned to contact you, Ian Hacking, Cheryl Misak, and Huw Price for a sort of pre-invitation before we send out other invitations more generally.  I.e., we really want you to come, and think you'd be a great asset for attracting some of the others that we plan to invite.  Of course, all your expenses would be paid by PI, and I personally am itching to have some refreshed face-to-face conversations with you.

Attached is a list of topics for the meeting written by my colleagues.  I hope it will give you some sense of what the meeting will address.

I very much hope you will say yes, and I'll wait on needles until I hear back from you!

\section{29-08-09 \ \ {\it Eight-Port Homodyne Detector}\ \ \ (to M. Sasaki)} \label{Sasaki7}

I almost forgot, here is the reference I was telling you about, \arxiv{0708.4094}. For myself, it is far too technical to read, but I believe I understand the conclusion:  By suitably adjusting the ``parameter'' mode, one can measure any Weyl--Heisenberg covariant observable.  In finite dimensions (up to $d=67$ at least), one can always construct a Weyl--Heisenberg SIC-POVM by suitably choosing the right fiducial state.  (See \arxiv{quant-ph/0310075}.) The question then would be, for continuous variables, what sequence of ``parameters'' would converge to something like a SIC-POVM.

As I say, the Kiukas/Lahti reference is too mathematical for my tastes, but it should have references in it that are more understandable.

\section{29-08-09 \ \ {\it QBism House}\ \ \ (to N. Waxman)} \label{Waxman2}

\bnw
All the best (and remind me to send Kiki pictures of my mom's 150 year old house on a cliff here---it was once an epic fixer upper she would truly appreciate!),
\enw

OK, I'll get in touch with you Monday.  And it'll be fun to see the pictures.  I myself am thinking about doing a feature for the newsletter if you'll take it, titled ``QBism House Open for Business''.  I'd like to use a couple pictures of our house in it, particularly one showing our porch with chalkboard installed and my floor to ceiling library.  I'd explain how my group of six (myself, Appleby, Ericsson, and three grad students) will be using the areas as an annex to PI for research.  Say some things about what the research is about, and where we want to go with it.  And welcome anyone else who might have an interest in the subject to join us:  In good weather, the plan is to have a weekly group meeting on the porch.

You should see the house, if you haven't seen it in a while:  It's just about finished on the outside now (except for planting some better grass).  And 95 percent finished on the bottom floor.  I'm quite proud of it.

Let me know when you're available next week, and we can get together on the webpage stuff.

\section{01-09-09 \ \ {\it Schumacher \& Westmoreland Endorsement} \ \ (to S. Capelin)} \label{Capelin10}

Here you go:
\bq\noindent
This is a fantastic book, with one of the authors no less than the very inventor of the word and idea of a qubit.  When I opened the book for the first time, I found I couldn't stop reading through it and working out some of the problems.  I should be ashamed to admit, but after 15 years as a professional quantum information theorist, I was still learning some elementary aspects of quantum mechanics.  In presentation, the book strikingly reminded me of a conversation I had had with Schumacher several years ago.  He told me that just after seeing the new {\sl Star Wars\/} movie, he became quite depressed because he thought it was just a fantasy and he would never himself be a Jedi knight; he would never know the ways of ``the force.''  But one day it dawned on him, ``Wait a minute, I am learning the ways of the quantum!  I'm exploring its great mysteries, its real mysteries.  I am a Jedi after all!''  And so will be the reader who invests his mind in this book.  There's no book out there I would recommend more for learning the mechanics of this quantum world.
\eq
If it is too long for you, you can trim some of it out, but please pass any changes you make by me before going to press.

\section{01-09-09 \ \ {\it Interpretations, Closed Timelike Curves} \ \ (to C. H. {\Bennett})} \label{Bennett67}

\bcb
I would be interested to hear what you think of our (Debbie, John, Graeme, and me) new paper asserting the computational and discriminatory impotence of closed timelike curves.  It involves questions about what is a reasonable way to formalize slippery notions like discrimination and (especially) computation.  Such conundrums are very much up your alley, and I think you'll have interesting things to say about them, though I'm a little worried you'll go all theocratic again like you did in your emails about me and Sappho sitting in a tree.  I'm not her type, anyway.
\ecb

I think your paper is indeed important for foundational issues.  For instance, I think linearity of evolutions is absolutely required for an epistemic interpretation of quantum states.  With nonlinearity, that interpretation goes right out the window.

Will study it with a fine-toothed comb.  (And resist the temptation to infer that I'm devilishly referring to lice.)  With regard to your remark on Sappho, oh I don't know:  Maybe you don't watch the same movies I do.

\section{02-09-09 \ \ {\it Sam Hurt's Eyebeam}\ \ \ (to N. Waxman)} \label{Waxman3}

Here are a couple of links that might jog your memory:
\begin{itemize}
\item
\myurl{http://en.wikipedia.org/wiki/Eyebeam_(comic)}
\item
\myurl{http://eyebeam.com/}
\end{itemize}

\section{02-09-09 \ \ {\it Inconsistencies} \ \ (to R. W. {\Spekkens})} \label{Spekkens69}

Two little points of thought connected to yesterday's conversations.

1)	With regard to your inter-theoretic pragmatism and intra-theoretic realism.  I said I have a good bit of sympathy with that.  But mulling over your particular choice of words, I'm not sure that one of the things you said makes sense.  Intra-theory, you said, you could feel free to invoke a correspondence theory of truth.  But I myself couldn't go that far.  For I'd have to wonder, ``correspondence with what {\it something\/} that's independent of the prior pragmatic choice?''  There's nothing that the theory can give me a hook to by admission.  So, really what I meant when I said I was in some agreement was something more along the lines of the following.

Once a theory like quantum mechanics is accepted (say for purely pragmatic reasons---i.e., for reasons that have directly to do with our practice and our survival in the world)---then one can step back and ask, ``Well, what is implicit in this theory that is not directly to do with my practice?''  That is, in positing the theory, I may be positing many things, not least of all the theorist's postulated limitations in interacting with the world, but I may also be positing things about the world itself (without the theorist).  I wouldn't deny that, and that is our point of overlap.  But those things are a distillate of the original pragmatic choice, and not girded up by a correspondence (that we could never know as actually holding or not).

Actually, my present position is decently close to F. C. S. Schiller's.  See
\bq\noindent
\myurl{http://en.wikipedia.org/wiki/Ferdinand_Canning_Scott_Schiller}
\eq
or better, his paper ``Axioms as Postulates'' directly:
\bq\noindent
\myurl{http://en.wikisource.org/wiki/Personal_Idealism/Axioms_as_Postulates}
\eq
The outline at the beginning of this one, I think, is particularly useful.  In the Wikipedia article, there is this nice explanation of one aspect of the thought.  See below.

2)	Yesterday, you said you preferred not to be political.  What I am about to say is not an outright inconsistency, but I think it is an exhibition of expediency.  It is about our conversation with Natasha on the cartoon.  She wants some reference to many worlds because it will make quick connection to what little the readers may already know of the quantum foundations debate.  But I think that, without finding a way to insult the view at the same time, that's a kind of cheap move, only further promoting an incorrect image of the theory in the public eye.  I.e., on this issue I can see some real (albeit subtle and long-term) damage being done.

Anyway, thanks for provoking my thoughts.  More on Schiller below.  (Many typos fixed from the Wiki article.)

\bq
In ``Axioms as Postulates'' Schiller vindicates the postulation by its success in practice, marking an important shift from {\sl Riddles of the Sphinx}. In {\sl Riddles}, Schiller is concerned with the vague aim of connecting the ``higher'' to the ``lower'' so he can avoid skepticism, but by 1903 he had clarified the connection he sees between these two elements. The ``higher'' abstract elements are connected to the lower because they are our inventions for dealing with the lower; their truth depends on their success as tools. Schiller dates the entry of this element into his thinking in his 1892 essay ``Reality and `Idealism'\,'' (a mere year after his 1891 {\sl Riddles}).

\bq
The plain man's `things,' the physicist's `atoms,' and Mr.\ Ritchie's `Absolute,' are all of them more or less preserving and well-considered schemes to interpret the primary reality of phenomena, and in this sense Mr.\ Ritchie is entitled to call the `sunrise' a theory. But the chaos of presentations, out of which we have (by criteria ultimately practical) isolated the phenomena we subsequently call sunrise, is not a theory, but the fact which has called all theories into being. In addition to generating hypothetical objects to explain phenomena, the interpretation of reality by our thought also bestows a derivative reality on the abstractions with which thought works. If they are the instruments wherewith thought accomplishes such effects upon reality, they must surely be themselves real.
\eq
The shift in Schiller's thinking continues in his next published work, {\sl The Metaphysics of the Time-Process\/} (1895): The abstractions of metaphysics, then, exist as explanations of the concrete facts of life, and not the latter as illustrations of the former [\ldots.] Science [along with humanism] does not refuse to interpret the symbols with which it operates; on the contrary, it is only their applicability to the concrete facts originally abstracted from that is held to justify their use and to establish their `truth.'

Schiller's accusations against the metaphysician in {\sl Riddles\/} now appear in a more pragmatic light. His objection is similar to one we might make against a worker who constructs a flat-head screwdriver to help him build a home, and who then accuses a screw of unreality when he comes upon a Phillips-screw that his flat-head screwdriver won't fit. In his works after {\sl Riddles}, Schiller's attack takes the form of reminding the abstract metaphysician that abstractions are meant as tools for dealing with the ``lower'' world of particulars and physicality, and that after constructing abstractions we cannot simply drop the un-abstracted world out of our account. The un-abstracted world is the entire reason for making abstractions in the first place. We did not abstract to reach the unchanging and eternal truths; we abstract to construct an imperfect and rough tool for dealing with life in our particular and concrete world. It is the working of the higher in ``making predictions about the future behavior of things for the purpose of shaping the future behavior of things for the purpose of shaping our own conduct accordingly'' that justifies the higher.

\bq
To assert this methodological character of eternal truths is not, of course, to deny their validity [\ldots.] To say that we assume the truth of abstraction because we wish to attain certain ends, is to subordinate theoretic `truth' to a teleological implication; to say that, the assumption once made, its truth is `proved' by its practical working [\ldots.] For the question of the `practical' working of a truth will always ultimately be found to resolve itself into the question whether we can live by it.
\eq
A few lines down from this passage Schiller adds the following footnote in a 1903 reprint of the essay: ``All this seems a very fairly definite anticipation of modern pragmatism.'' Indeed, it resembles the pragmatist theory of truth. However, Schiller's pragmatism was still very different from both that of William James and that of Charles Sanders Peirce.
\eq

\section{04-09-09 \ \ {\it Little Lemma, Big Theorem}\ \ \ (to N. S. Jones)} \label{Jones1}

Thanks for the nice note.  Have we ever met in person?  Thanks too for telling me more of the story of ``the remarkable theorem.''  That's what I actually call it in my presentations; in a few of them I've even promoted it to the stature of the no-cloning theorem---pointing out particularly that it's trivial to prove, but that it nonetheless says something very deep about the ``shape'' of quantum-state space.  Also I always tell the story of how many big-time quantum information people simply weren't aware of it---it was a well-hidden little gem.  I know because I have asked around very thoroughly.  Here are the people I can recall who did not know the theorem until I pointed it out to them:  Holevo, Lindblad, Lieb, Nielsen, Bengtsson, Calderbank, Ruskai, Gottesman, Kochen, Uhlmann, Mermin, Brukner, \ldots\ and there were surely others as well.  It's a bit like the story I just wrote up on Wigner and the no-cloning theorem.  See page 2 of the attached (barely started) draft of an upcoming paper.

I'm sorry to hear that you've left quantum information, but I hope you're happy in your new line of work.

\subsection{Nick's Preply}

\bq
Dear Chris -- I was just scoping around and found, in an interesting paper of yours, that you liked my little lemma. I always liked the lemma myself, but Noah and I wrangled about whether it was worth hyping up more. From a mathematical perspective it's trivial to validate; yet (probably because I'm an incompetent) it took me two or three weeks to prove (because, when you only know the left hand side \ldots). I note Flammia apparently scooped me; I presented the idea in 2003 at the Informal Quantum Information Gathering, Max-Plank Institute for Quantum Optics 2003 and had proved it in the winter/spring of '03 I think (it was also in my thesis transfer report of that year). Noah then made me sit on it for a long time (to the frustration of a cocky graduate student) \ldots\ however, there doesn't seem a whole lot of point setting the record straight, as I suspect (like Noah) that, given the simplicity of the result, it must have been proved sometime in the 60's (and then yearly ever since) but we just don't know it. I think I remember being careful in the paper that we didn't explicitly claim that the result was new. Something I never quite seem to get around to is going to town with these equations from a Real Algebraic Geometry perspective.

[My recent favourite cute thing is that the digit sum of a random walk on the integers, shifted so that it never goes below zero, shows $1/f$ noise \ldots\ maybe all those papers on Self Organised Criticality and $1/f$ are just random walks represented wrong.]

You'll be pleased to hear that I left QI disheartened by my ability to make a scientific difference and am now something of a biologist (and also increasingly Bayesian)!

Btw---a smidgeon of flattery---it was some of your work that pulled me into the field of QI.
\eq

\section{07-09-09 \ \ {\it Labor Day Recreations}\ \ \ (to R. W. {\Spekkens}, D. M. {\Appleby}, W. K. Wootters, and R. Blume-Kohout)} \label{Appleby69} \label{BlumeKohout9} \label{Wootters25} \label{Spekkens69.1}

On this Labor Day morning (enjoying my new porch), I thought I'd spend a few minutes capping off our Friday lunchtime conversation.  I never quite gave my own answer to Rob's question of ``What is the purpose of science?''  (Shy in a crowd, I am.)  As usual, I'll answer the question with a couple of quotes written by someone far more eloquent than me.  Here are two from William James that have struck me very deeply.  The first answers the question straight out:  [See James quote at the end of the 02-09-01 note titled ``\myref{Mermin35}{Intersubjective Agreement}'' to N. D. {\Mermin}.]

I love the last two paragraphs especially.  For in my own mind they capture the essence of the story.  Our scientific theories are not descriptions of what the world {\it is}, nor are they positivist {\it prediction\/} devices predominantly---that idea fairly appals me---but they are tools for changing existing realities.  They are like hammers in a very real sense:  tools with which we can build and destroy, and on occasion finesse what we want to make happen.

The following quote probably paints my stripes still more brightly.  I submit it with the understanding that you all tolerate me more or less.

\bq
In many familiar objects every one will recognize the human element.  We conceive a given reality in this way or in that, to suit our purpose, and the reality passively submits to the conception.  You can take the number 27 as the cube of 3, or as the product of 3 and 9, or as 26 {\it plus\/} 1, or 100 {\it minus\/} 73, or in countless other ways, of which one will be just as true as another.  You can take a
chess-board as black squares on a white ground, or as white squares on a black ground, and neither conception is a false one.

You can treat the adjoined figure as a star, as two big triangles crossing each other, as a hexagon with legs set up on its angles, as six equal triangles hanging together by their tips, etc.  All these treatments are true treatments---the sensible {\it that\/} upon the paper resists no one of them.  You can say of a line that it runs east, or you can say that it runs west, and the line {\it per se\/} accepts both descriptions without rebelling at the inconsistency.

We carve out groups of stars in the heavens, and call them constellations, and the stars patiently suffer us to do so,---{\it though\/} if they knew what we were doing, some of them might feel much surprised at the partners we had given them.  We name the same constellation diversely, as Charles's Wain, the Great Bear, or the Dipper.  None of the names will be false, and one will be as true as
another, for all are applicable.

In all these cases we humanly make an {\it addition\/} to some sensible reality, and that reality tolerates the addition.  All the additions `agree' with the reality; they fit it, while they build it out.  No one of them is false.  Which may be treated as the {\it more\/} true, depends altogether on the human use of it.  If the 27 is a number of dollars which I find in a drawer where I had left 28, it is 28 minus 1.  If it is the number of inches in a board which I wish to insert as a shelf into a cupboard 26 inches wide, it is 26 plus 1.  If I wish to ennoble the heavens by the constellations I see there, `Charles's Wain' would be more true than `Dipper.'  My friend Frederick Myers was humorously indignant that that prodigious star-group should remind us Americans of nothing but a culinary utensil.

What shall we call a {\it thing\/} anyhow?  It seems quite arbitrary, for we carve out everything, just as we carve out constellations, to suit our human purposes.  For me, this whole `audience' is one thing, which grows now restless, now attentive.  I have no use at present for its individual units, so I don't consider them.  So of an `army,' of a `nation.' But in your own eyes, ladies and gentlemen, to call you `audience' is an accidental way of taking you.  The permanently real things for you are your individual persons.  To an anatomist, again, those persons are but organisms, and the real things are the organs.  Not the organs, so much as their constituent cells, say the histologists; not the cells, but their molecules, say in turn the chemists.

We break the flux of sensible reality into things, then, at our will.  We create the subjects of our true as well as of our false propositions.

We create the predicates also.  Many of the predicates of things express only the relations of the things to us and to our feelings.  Such predicates of course are human additions.  Caesar crossed the Rubicon, and was a menace to Rome's freedom.  He is also an American schoolroom pest, made into one by the reaction of our schoolboys on his writings.  The added predicate is as true of him as the earlier ones.

You see how naturally one comes to the humanistic principle: you can't weed out the human contribution.  Our nouns and adjectives are all humanized heirlooms, and in the theories we build them into, the inner order and arrangement is wholly dictated by human considerations, intellectual consistency being one of them.  Mathematics and logic themselves are fermenting with human rearrangements; physics, astronomy and biology follow massive cues of preference.  We plunge forward into the field of fresh experience with the beliefs our ancestors and we have made already; these determine what we notice; what we notice determines what we do; what we do again determines what we experience; so from one thing to another, altho the stubborn fact remains that there is a sensible flux, what is {\it true of it\/} seems from first to last to be largely a matter of our own creation.

We build the flux out inevitably.  The great question is: does it, with our additions, {\it rise or fall in value}?  Are the additions {\it worthy\/} or {\it unworthy}?  Suppose a universe composed of seven stars, and nothing else but three human witnesses and their critic.  One witness names the stars `Great Bear'; one calls them `Charles's Wain'; one calls them the `Dipper.'  Which human addition has made the best universe of the given stellar material?  If Frederick Myers were the critic, he would have no hesitation in `turning down' the American witness.

Lotze has in several places made a deep suggestion. We na\"{\i}vely assume, he says, a relation between reality and our minds which may be just the opposite of the true one.  Reality, we naturally think, stands ready-made and complete, and our intellects supervene with the one simple duty of describing it as it is already. But may not our descriptions, Lotze asks, be themselves important additions to
reality?  And may not previous reality itself be there, far less for the purpose of reappearing unaltered in our knowledge, than for the very purpose of stimulating our minds to such additions as shall enhance the universe's total value. {\it `Die Erh\"ohung des vorgefundenen Daseins'\/} is a phrase used by Professor Eucken somewhere, which reminds one of this suggestion by the great Lotze.

It is identically our pragmatistic conception.  In our cognitive as well as in our active life we are creative.  We {\it add}, both to the subject and to the predicate part of reality.  The world stands really malleable, waiting to receive its final touches at our hands.  Like the kingdom of heaven, it suffers human violence willingly.  Man {\it engenders\/} truths upon it.

No one can deny that such a role would add both to our dignity and to our responsibility as thinkers.  To some of us it proves a most inspiring notion.  Signore Papini, the leader of Italian pragmatism, grows fairly dithyrambic over the view that it opens of man's divinely-creative functions.

The import of the difference between pragmatism and rationalism is now in sight throughout its whole extent.  The essential contrast is that {\it for rationalism reality is ready-made and complete from all eternity, while for pragmatism it is still in the making, and awaits part of  its complexion from the future}.  On the one side the universe is absolutely secure, on the other it is still pursuing its
adventures.

We have got into rather deep water with this humanistic view, and it is no wonder that misunderstanding gathers round it.  It is accused of being a doctrine of caprice.  Mr.\ Bradley, for example, says that a humanist, if he understood his own doctrine, would have to `hold any end, however perverted, to be rational, if I insist on it personally, and any idea, however mad, to be the truth if only someone is resolved that he will have it so.'  The humanist view of `reality,' as something resisting, yet malleable, which controls our thinking as an energy that must be taken `account' of incessantly (tho not necessarily merely {\it copied}) is evidently a difficult one to introduce to novices. \ldots

{\it The alternative between pragmatism and rationalism, in the shape in which we now have it before us, is no longer a question in the theory of knowledge, it concerns the structure of the universe itself.}

On the pragmatist side we have only one edition of the universe, unfinished, growing in all sorts of places, especially in the places where thinking beings are at work.

On the rationalist side we have a universe in many editions, one real one, the infinite folio, or {\it \'edition de luxe}, eternally complete; and then the various finite editions, full of false readings, distorted and mutilated each in its own way.
\eq

Happy holiday.

\section{07-09-09 \ \ {\it Awaits Part of Its Complexion from the Future \ldots}\ \ \ (to R. W. {\Spekkens})} \label{Spekkens70}

You should  be able to tell that I've been reading and thinking about your Friday diary entry this morning.  You are an honest intellect, and I very much like that about you.

At the moment, I'm tumbling over the last paragraph in your entry (as I tumbled it over when you gave your brief introduction last Friday).  There is something about the word ``teleological'' that I do {\it not\/} like; but there is something about the title of this note that I very much do like.  Is that a contradiction?  I'm not sure:  So I keep tumbling these things about.

In the meantime I supply you of two more quotes that I think hint of why the word teleology (as often used) troubles me a bit.

The first is from William James:
\bq
Our sense of `freedom' supposes that some things at least are decided here and now, that the passing moment may contain some novelty, be an original starting-point of events, and not merely transmit a push from elsewhere.  We imagine that in some respects at least the future may not be co-implicated with the past, but may be really addable to it, and indeed addable in one shape {\it or\/} another, so that the next turn in events can at any given moment genuinely be ambiguous, i.e., possibly this, but also possibly that.

Monism rules out this whole conception of possibles, so native to our common-sense.  The future and the past are linked, she is obliged to say; there can be no genuine novelty anywhere, for to suppose that the universe has a constitution simply additive, with nothing to link things together save what the words `plus,' `with,' or `and' stand for, is repugnant to our reason.

Pluralism, on the other hand, taking perceptual experience at its face-value, is free from all these difficulties.  It protests against working our ideas in a vacuum made of conceptual abstractions.  Some parts of our world, it admits, cannot exist out of their wholes; but others, it says, can.  To some extent the world {\it seems\/} genuinely additive: it may really be so.  We cannot explain conceptually {\it how\/} genuine novelties can come; but if one did come we could experience {\it that\/} it came.  We do, in fact, experience perceptual novelties all the while. Our perceptual experience overlaps our conceptual reason: the {\it that\/} transcends the {\it why}.  So the common-sense view of life, as something really dramatic, with work done, and things decided here and now, is acceptable to pluralism. `Free will' means nothing but real novelty; so pluralism accepts the notion of free will.

But pluralism, accepting a universe unfinished, with doors and windows open to possibilities uncontrollable in advance, gives us less religious certainty than monism, with its absolutely closed-in world.  It is true that monism's religious certainty is not rationally based, but is only a faith that `sees the All-Good in the All-Real.'  In point of fact, however, monism is usually willing to exert this optimistic faith: its world is certain to be saved, yes, is saved already, unconditionally and from eternity, in spite of all the phenomenal appearances of risk.

A world working out an uncertain destiny, as the phenomenal world appears to be doing, is an intolerable idea to the rationalistic mind.

Pluralism, on the other hand, is neither optimistic nor pessimistic, but melioristic, rather.  The world, it thinks, may be saved, on condition that its parts shall do their best.  But shipwreck in detail, or even on the whole, is among the open possibilities.

There is thus a practical lack of balance about pluralism, which contrasts with monism's peace of mind. The one is a more moral, the other a more religious view; and different men usually let this sort of consideration determine their belief.
\eq

The second comes from Richard Rorty:
\bq
In this essay I shall focus on Whitman's phrase `counts \ldots\ for her justification and success \ldots\ almost entirely upon the future'. As I see it, the link between Whitmanesque Americanism and pragmatist philosophy---both classical and `neo-'---is a willingness to refer all questions of ultimate justification to the future, to the substance of things hoped for. If there is anything distinctive about pragmatism it is that it substitutes the notion of a better human future for the notions of `reality', `reason' and `nature'. One may say of pragmatism what Novalis said of Romanticism, that it is `the apotheosis of the future'.

As I read Dewey, what he somewhat awkwardly called `a new metaphysic of man's relation to nature', was a generalization of the moral of Darwinian biology. The only justification of a mutation, biological or cultural, is its contribution to the existence of a more complex and interesting species somewhere in the future.  Justification is always justification from the point of view of the survivors, the
victors; there is no point of view more exalted than theirs to assume. This is the truth in the ideas that might makes right and that justice is the interest of the stronger. But these ideas are misleading when they are construed metaphysically, as an assertion that the present status quo, or the victorious side in some current war, stand in some privileged relation to the way things really are. So `metaphysic' was an unfortunate word to use in describing this generalized Darwinism which is democracy. For that word is associated with an attempt to replace appearance by reality.

Pragmatists---both classical and `neo-'---do not believe that there is a way things really are. So they want to replace the appearance-reality distinction by that between descriptions of the world and of ourselves which are less useful and those which are more useful. When the question `useful for what?'\ is pressed, they have nothing to say except `useful to create a better future'. When they
are asked, `Better by what criterion?', they have no detailed answer, any more than the first mammals could specify in what respects they were better than the dying dinosaurs. Pragmatists can only say something as vague as: Better in the sense of containing more of what we consider good and less of what we consider bad. When asked, `And what exactly do you consider good?', pragmatists can only say, with Whitman, `variety and freedom', or, with Dewey, `growth'. `Growth itself,' Dewey said, `is the only moral end.'

They are limited to such fuzzy and unhelpful answers because what they hope is not that the future will conform to a plan, will fulfil an immanent teleology, but rather that the future will astonish and exhilarate. Just as fans of the avant garde go to art galleries wanting to be astonished rather than hoping to have any particular expectation fulfilled, so the finite and anthropomorphic deity celebrated by James, and later by A. N. Whitehead and Charles Hartshorne, hopes to be surprised and delighted by the latest product of evolution, both biological and cultural. Asking for pragmatism's blueprint of the future is like asking Whitman to sketch what lies at the end of that illimitable democratic vista. The vista, not the endpoint, matters.

So if Whitman and Dewey have anything interesting in common, it is their principled and deliberate fuzziness. For principled fuzziness is the American way of doing what Heidegger called `getting beyond metaphysics'. As Heidegger uses it, `metaphysics' is the search for something clear and distinct, something fully present. That means something that does not trail off into an indefinite future \ldots
\eq

\section{08-09-09 \ \ {\it Awaits Part of Its Complexion from the Future \ldots, 2}\ \ \ (to R. W. {\Spekkens})} \label{Spekkens71}

Thanks, that's useful.  Dennett's definition of ``teleological explanation'' makes it clear that what I'm talking about with the phrase ``awaiting part of its complexion from the future''---whatever it is---is not that.

The world Dennett sets you to thinking about, the flavor of everything he says, is foreign to the landscape I see in James's writings.  (It's certainly foreign to me.)  ``This exchange reveals one of the troubles with teleology: where does it all stop?''  Part of the very point is that it doesn't.  Nor is there one and only one all-inclusive chain (or `hierarchy' in his terms) that one can imagine embedding the discussion in.  An essential piece of what Dennett balks at is that no finite story can arise from what he views as the other side (i.e., teleological blah blah blah).  To the extent that his other side coincides with my own landscape, {\it yes}, that is right:  It shouldn't be any other way.  But also, there is all this talk of ``explanation'' (in him and in you)---we just hit the rationalist/pragmatist (monist/empiricist) divide again.  It is what sets us apart.  He wants the world reduced; we say it cannot be.  We say that to the extent that one can contemplate any toy reduction, one only does so by leaving essential things out.  It is a divide that has plagued worldviews for a long time.

\section{09-09-09 \ \ {\it Till Tomorrow} \ \ (to R. {\Schack})} \label{Schack174}

I've done just a little to the abstract, intro, and acknowledgments.  I'm going back to bed now.  I'll try to get an hour more in before leaving for Niagara at 9:00 AM.  Then I'll send you (what little) I have, and pick back up on it this evening.

One thing that confuses me (worries me!) is this line:
\brs
Quantum Bayesian state assignments are personalist in the sense that they are functions of the agent as well as the world outside the agent.
\ers
Why the reversal from our now very infamous position?

\section{10-09-09 \ \ {\it Phrases That Came to Mind}\ \ \ (to D. M. {\Appleby})} \label{Appleby70}

``Theory is a focal point for designing counterfactuals.''

``Theory is a focal point that aids in designing counterfactuals.''

\section{10-09-09 \ \ {\it Euler and Classical Mechanics}\ \ \ (to P. G. L. Mana)} \label{Mana16}

Marcus and I had a long discussion about your note yesterday as we were at Niagara (doing work-permit business and viewing the amazing falls).  These are very important points you make.

Do you mind if I share your note with Rob Spekkens if the issue comes up in our on-going debate about the value (or unvalue in my opinion) of assuming a deterministic world?

\subsection{Luca's Preply}

\bq
I hope your research is going fine there! Unfortunately I am going to miss the new perspectives on the quantum state.

I wanted to send you the translation of Euler's passage and some comments on it a while ago, but I got the swine flu (no different from any other flu) and have been to Sweden and Rome afterwards.

Here are the passages from Euler that I found important. Recall that in that paper he is studying the condition for fluid equilibrium:
\bq\noindent
Now, all comes to properly establishing the basic idea on which to base the reasonings to arrive at our goal: the idea of the nature of fluidity in general. For the equilibrium laws for fluids will not have to differ from those of solids but in the fact that the nature of fluids is different from that of solids. So the point is to know the genuine and essential difference that distinguishes the fluids from the solids --- a question well bustled about amongst Philosophers and Physicists: but from all they have said about it we cannot deduce anything suitable to our plan. It is maybe true that the smaller elements of a fluid have no link with one another, and that they are in a perpetual movement; but this truth would be completely sterile for our researches \ldots.
\eq
\bq\noindent
Inasmuch as this essential property of fluids is to provide the principles of Hydrostatics, I find it only in this: that a fluid mass will not be in equilibrium unless it be acted upon all points of its surface by equal forces perpendicular to the surface. \ldots
\eq
\bq\noindent
Conceive an immaterial diaphragm in the interior of the fluid, cutting out a generic portion of it. Since this portion is in equilibrium, all the elements of the diaphragm will sustain forces of the same intensity \ldots.
\eq

This paper is one of the most important in the history of classical mechanics, and it is full of gems. In the first paragraph, for example, Euler has very clear the difference between a general law  (``the laws for fluids will not have to differ from those of solids'') and a constitutive equation (``but in the fact that the nature of fluids is different from that of solids''). I know not a few physicists who do not have that difference clear today.

It is obvious that Euler is here looking for a model with which to describe and possibly deduce new properties of fluid equilibrium. He says clearly that he is not interested about ``the real situation'', because this may be too complicated and irrelevant for the phenomena under investigation. Rather, he wants a simple conceptual device from which to derive, mathematically and as simply as possible, the phenomena of interests; in this case, fluid equilibrium.

Note how his goal is limited to a particular range of phenomena, not {\it all\/} phenomena of this world; and how he adapts the means to the end.

And here `conceptual' is in its original sense of something {\it conceived\/} or {\it conceivable}: something that can be taken in, imaginable. It does not matter if what is imagined contradicts ``the real situation''.

In fact, in this case Euler's device is to {\it imagine\/} any portion of a fluid (in fact, of any body) enclosed by an immaterial diaphragm, and to study the imaginary forces that would act on this diaphragm from the fluid outside it. A very simple thing to do with our imagination! But this obviously goes against the idea, of that time, of a fluid made of elements in perpetual motion and with no interaction. But who cares?

With this device all the equilibrium of fluids is summarised in a simple principle: the forces per unit area on any such diaphragm must be everywhere equal in intensity and perpendicular to it. With a very simple generalisation, from the same principle Euler obtains in the subsequent paper all phenomena of non-turbulent fluid motion. Isn't this genius?

This device was revolutionary and very fertile; indeed, every engineer learns it today: via Cauchy, it became our hodiern concept of stress, used even in general relativity. Thanks to it we can cross bridges, build very high buildings, make robust containers, fly from Canada to Australia, and do many other things (despite the fact that no such diaphragms nor stress forces appear in a molecular model, as every statistical mechanician knows).

It is also clear that the forces acting upon that imaginary diaphragm are hardly measurable. This was clear to Euler, and is clear to the engineer today. The way classical mechanics has always proceeded has been this: make some hypotheses about your model, work out the consequences, and if these match experience and experiment, feel more confident about your hypotheses.

Classical physics continually uses many fundamental hypotheses which {\it in principle\/} cannot be tested. Take the case of the Newton-Euler first law and inertial frames. The law states that, for an observer {\it in an inertial frame},
$$
\sum_i \mathbf{F}_i = m \mathbf{a}\;,
$$
the sum being extended to all forces acting on the body in question. How can we test whether we are in an inertial frame? Well, given all forces, the only way is to see whether the body's acceleration is proportional to their sum. But how can we be sure that we have not neglected some force? Well, given that we are in an inertial frame, we can check whether the sum of the forces considered is proportional to the body's acceleration. But how can we test whether we are in an inertial frame? We're going around in a circle. We cannot test both hypotheses, about the forces and about the frame: we have to {\it assume\/} the one and test the other. If and as long as everything works, we are confident about our untested assumptions.

So I think Euler's paper is one of the many which shows that:
\begin{itemize}
\item Classical physics is concerned about describing the world, but `describing' does not necessarily mean `saying how thing really are'. (In fact, after Wittgenstein we should know that `how the world really is' has no meaning at all.)

\item Classical physics is not based upon all-measurability and all-knowability: in it, hypotheses are everywhere made never to be proven, but to be held as long as their consequences agree with our experience, experiments, and taste.

\item The models and hypotheses of classical physics are based on our imagination: the very concepts we use in it are extremely anthropomorphic, sublimations of our daily experience; sometimes they even contradict our thoughts about how things `really' are.
\end{itemize}
What many people today say about `classical physics', is not what it is and was, but what {\it they think\/}  it {\it should\/} be and have been. Indeed, they very rarely present any single text of our past Masters (apart from the usual two lines by Laplace) to support their claims.
\eq

\section{14-09-09 \ \ {\it Your Introduction}\ \ \ (to C. M. {\Caves})} \label{Caves100.4}

\bcc
I finally had enough bus rides to plow through the intro to your book.  I enjoyed reading the profiles very much.  And I was struck by that last one of Zeilinger.  How appropriate to finish off with the statement that you should be talking and writing for the back rows, not the front rows.  That is indeed precisely where you have to aim your fire now.
\ecc

I'm glad you enjoyed reading them.  Steven van Enk told me he was surprised I didn't use my ``Gmail spam'' story about Preskill.  If you don't know that one, I'll have to tell you some time.  (If I wrote it down, your spam filter would catch it.)

\section{15-09-09 \ \ {\it The Primacy of Measurement}\ \ \ (to D. M. {\Appleby})} \label{Appleby71}

Fun reading on a depressing night.  {\Ruediger} and I have a mistake in our paper and I've spent three days trying to fix it.  (Luckily it doesn't infect my work with you and {\Asa}, but it remains a mistake in the public eye.)  Anyway, we must talk much more of time when all this sickening paper-writing is done.

\subsection{Marcus's Preply to Jochen Rau}

\bq
I much enjoyed our discussions too.  I also enjoyed your very stimulating paper.

I hope you won't take the following remarks as criticism.  They are merely a few  thoughts provoked by reading your paper. I believe that quantum mechanics is challenging us to develop a way of thinking about physical reality which is completely different from the one to which 300 years and more of classical physics have accustomed us (I say 300 years because that is the time elapsed since Newton's {\sl Principia}---and I include the 20th century in that time span because it seems to me that, notwithstanding the mathematical innovations introduced by {\Schroedinger}, Heisenberg et al,  on a conceptual level we have hardly begun to get beyond the psychological cramp which seventeenth century science induced in us).  I have the impression your attitude is similar to mine.  If so you will agree that there is a long way still to go.  And that is the motivation for these comments.  I am thinking ``what next?''

One of the obsessions of classical physics was the desire for what might be called {\it total\/} objectivity:  a picture of the world from which every trace of the knowing subject has been removed.  A picture of ``things as they are in themselves'', independent of them being conceived by us.  I said something about this in my talk at {\Vaxjo}.   Like all simple-minded slogans it seems, on the face of it, an entirely reasonable demand.  It seems (on the face of it) as though any qualification of the demand---any suggestion that things aren't quite as simple as it assumes---represents a surrender of  the foundational principle of modern science, without which it wouldn't deserve to be called science.  I think in many peoples' minds there is almost a moral dimension to this:  that many people in physics regard (for instance) the Bayesian interpretation of the quantum state rather in the way that fundamentalist Christians regard liberal theologians.  But the thing about fundamentalist Christianity is that it is doomed to failure for purely internal reasons.  You cannot base yourself on a perfectly literal interpretation of the Bible because the Bible isn't consistent, either with itself or with the acknowledged facts (fundamentalist Christians often deny Darwin;  not so easy for them, in these days of space  travel, to deny Copernicus).     And I think something similar is true of the fundamentalist demand for {\it total\/} objectivity.  Its own internal inconsistencies mean that is simply not possible  to carry the project through.    This is obvious, really, just from the fact that it is a totally objective {\it picture\/} (or theory, or conceptualization) which is being demanded.   For a picture (or theory, or conceptualization) is a human construction and it must inevitably bear the marks of that.  Complaining that a theory has ``subjective'' elements (elements which reflect the fact that it is specifically {\it our\/} theory) is like complaining that a photograph is printed on paper, using ink---that it isn't literally identical with whatever it is a photograph of.  As I said in my talk in {\Vaxjo} I think quantum mechanics is finally forcing us to face up to this obvious point.  Really it should have been obvious all along.  However, there were features of classical physics which made it comparatively easy for those who wanted to conceal the truth from themselves.  Nowadays it is much harder.  Not that people don't still try.  In fact, most  of the effort in quantum foundations is devoted to the attempt to interpret quantum mechanics along objectivist lines.  However, these attempts fail to carry conviction---which is why there continue to be all these arguments.

Anyway, that is by the way of preamble.  Now for my first question.  I am deeply interested in your idea, that maybe we should think of measurement as the primary thing.  The contrary idea, that unitary evolution is the fundamental concept, and that the aim should be to explain measurement in terms of it, is closely connected with the classical ideal of {\it total\/} objectivity.  Anyone who thinks in those objectivist terms must inevitably find the fact that measurement is treated in quantum mechanics (when interpreted along Copenhagen type lines) as an unexplained primitive, unacceptable.  But if you reject that classical assumption the suggestion that we treat the concept of measurement as a conceptual primitive is no more objectionable than the suggestion that we treat the concept of logical inference as a conceptual primitive.  Sure the idea of measurement makes essential reference to the agent who is doing the measuring.  But so what?  And since I take that view I am naturally very attracted to the idea that we should turn things round completely, and seek to explain unitary evolution in terms of measurement.  However, it raises a question in my mind.  It should, I hope, be clear that in rejecting the ideal of {\it total\/} objectivity I am not committing myself to a rejection of objectivity absolutely, in any shape or form.  There is a (big) universe out there, and I believe it is the job of physics to say something about it.  The fact that  any statement we may make about a distant star (for instance) will contain subjective elements (will contain an essential reference to us, who are making the statement) doesn't mean that we simply can't make any valid statement.  It only means that we need to be much more careful, and subtle, than classical physicists were wont to be.  Objectivists would like to believe that unitary evolution is occurring in distant stars.  I definitely don't believe that.  And I suppose you don't believe it either.  But the question is:  what do we believe?  Do we say that measurements are going on in distant stars?  And if we do say that, what has become of the concept of measurement?  Are they measurements without agents to do the measuring?  Or what?

(Let me hasten to add that I am not expecting you to be able to answer the question.  I certainly couldn't answer it myself.  I am very uncertain what quantum mechanics says about distant stars.  I am just wondering if you have any thoughts on the subject, however tentative and uncertain.)

My other question concerns your reference to ``future attempts at constructing a truly timeless---and possibly measurement-based---theory of quantum gravity''.   I suspect that the classical demand for {\it total\/} objectivity has led to some deep-seated confusions about the nature of time.   You talk about a ``timeless'' theory of quantum gravity.  But it seems to me that there is an important sense in which the classical theory of relativity is already ``timeless''.  That is, it leaves out some of the essential features of what we ordinarily call time.  In the common sense conception of time the present plays a distinguished role.  But in classical relativity there isn't really any concept of the present at all.

Now of course you might say that there is no reason to pay any attention to common sense on this.  Copernicus simply ignored what was once the common sense assumption, that the Earth is {\it obviously\/} at rest.  So why shouldn't Einstein ignore the common sense assumption, that the concept of ``the present moment'' is something more than a subjective illusion? In response to that question, I would agree that we are as free to ignore common sense on this as we are on anything else.  But only if we can do it consistently. And I don't think modern physics does do it consistently.  It seems to me that, on the contrary, although denying the concept of the present at the level of official theory (the 4 dimensional space time manifold), modern physics continually smuggles the concept in unofficially, at the level of actual thought and practice.  Moreover, it does this surreptitiously, without comment, and mostly without anyone noticing.

To justify that statement I need to digress a little and say something about counter-factuality.  I will come back to time in a moment.   At least so far as its practical applications are concerned physics is a tool for answering counter-factual questions.  For instance, an engineer wants to design a bridge.  To do that s/he needs to ask themselves such questions as ``if I put the girder {\it here\/} what will be the stress {\it there}?''  Before arriving at the final design, which is actually built, the engineer runs through a multitude of other designs which never have more than a purely conceptual existence, inside the engineer's mind.  In short, designing a bridge requires us to know the answers to questions about what {\it would\/} happen if one {\it were\/} to do something which, however, one never actually does do.  In other words it requires us to answer counter-factual questions.    And what goes for bridges also goes for just about any other technological application of physics.  To a very large extent  the practical utility of physics depends on its ability to provide answers to such counter-factual ``what would happen if?''\ questions.   I also think that to a considerable extent theoretical understanding depends on them too.  For instance, understanding Newtonian gravity doesn't just mean knowing what the orbits of the planets {\it actually\/} are.  It also means knowing what the orbits {\it would\/} be if they were (counter-factually) perturbed in some way.  Similarly, you don't understand the geometry of a black hole if you don't know how (for example) to calculate how the geometry counter-factually would be affected if the black hole counter-factually did become more massive.  So I would say that counter-factual reasoning is absolutely basic to physics.  To a considerable extent it is what physics is about.  But because such reasoning involves the consideration of situations which are in a certain sense imaginary it is felt that it is not really part of physics proper.   (I use the prejudicial word ``imaginary'' because that is how physicists of a certain bent tend to regard it.   However, I think one might question whether it is really the case that it is only true {\it in imagination\/} that this glass would have broken if I had dropped it [but didn't].)

Counter-factual reasoning is closely connected with the concept of possibility (it was {\it possible\/} for the engineer to span the river:  meaning that s/he counterfactually {\it could\/} have built a satisfactory bridge if s/he had really tried), and that in turn is closely connected with the concept of free will (the engineer was {\it free\/} to build a bridge instead of a tunnel, if s/he had wanted to).  But, notwithstanding the fact that one of the main practical uses of physics is to answer questions as to what is physically possible in various situations, these concepts are regarded with grave suspicion by what is still the dominant tendency in physics.  Witness Bell's embarrassment ({\sl Speakable and Unspeakable}, revised edition, p.~101 ff) when he was accused of postulating the existence of free will, and his strenuous attempts to explain away what is commonly called freedom as really something else.  The reason that Bell didn't want to accept free-will (and the related concept of counter-factuality, which also plays a crucial role in Bell's argument) as a conceptual primitive is, I would suggest, the same as the reason  he didn't want to accept measurement as a conceptual primitive.  My own instincts are the exact reverse of his.  Not only am I ready to embrace measurement as a primitive concept.  I am also ready to embrace counterfactuality and free will as primitive concepts.  And perhaps quite a few other things too.  I think measurement is only the tip of the iceberg.  I think the reason we still don't satisfactorily understand quantum mechanics is that we have hardly got started on the road it naturally suggests.

Which brings me back to time.  I believe the ordinary, naive concept of time includes, as an essential element, the idea that we are free to alter the future, but not free to alter the past.  In other words, the future is the domain in which we can use counter-factual reasoning to practical effect.  Counter-factual questions can be interesting.  For example, engineering students are taught how the designers of the Tacoma Bay bridge counter-factually {\it should\/} have designed it.  Counter-factual questions are also highly relevant in a court of law.  But the future is where counter-factual reasoning really comes into its own:  because that is when we can do something about it.

Now, as I said, the fact that something forms an essential part of the common-sense conception is not in itself a reason for accepting it.  However this way of thinking about time {\it also\/} plays an essential role in the application of modern physics (no one is currently trying to build a quantum computer in the past).  Every modern physicist does {\it in fact\/} think in this way about time.  It is just that they don't want to face up to the fact that they are thinking this way.  Moreover, I strongly suspect that the reason they don't want to is nothing more than a philosophical prejudice, arising from the feeling that all talk of the observer and his or doings has no place in physics.  I believe that does provide a reason for taking it seriously.

Though let me stress that I am not suggesting that we simply go back to common sense ideas about time.  Quite the contrary.  I would speculate that if we were to follow this line of thought seriously, and try to formulate relativity in a way that did proper justice to counter-factuality, and agency, we might arrive at new, and suprising physics.

In short I would suggest that perhaps what we need is a theory of quantum gravity in which, not only measurement, but also certain concepts such as counter-factuality and agency play a central role.  I would also suggest that the theory which resulted wouldn't really be timeless.  It wouldn't get rid of the concept of time.  Rather what it would do is to conceive of time in a very different way from that to which we have become classically accustomed.

Which brings me to my question:  does this make any sense to you?
\eq

\section{16-09-09 \ \ {\it PIAF Quantum Foundations Conference -- Late Reminder} \ \ (to M. Bitbol)} \label{Bitbol3}

\bmb
Thank you for your kind invitation. Unfortunately, I can't come. But I am delighted to see that this topic is now a top priority. The very word ``state'' is doubtful, and generates a lot of misunderstandings.
\emb

Indeed I am in much agreement with your point.  In connection with that, you may enjoy the little story I wrote about David {\Mermin} (and quantum interpretation) in the attached file.  It is partially on linguistical issues to do with quantum interpretation.  See attached file, starting on page 21:

\bq
{\bf David {\Mermin}:} I cannot honestly call David {\Mermin} a father of quantum information, but I can call him the godfather of it---he blessed the field through its formative years and continues to watch with pride as it approaches maturity.  There are many reasons I say this, but one of them is told well in his own words in, ``Copenhagen Computation:\ How I Learned to Stop Worrying and Love {\Bohr},'' \arxiv{quant-ph/0305088}.

Then again, David always tells things well.  N. David {\Mermin} is the Horace {\White} Professor of Physics Emeritus at Cornell University, and for years has been the writer extraordinaire of our field.  I very much knew the latter when I first wrote to him, and believe me, it did take some effort to work up the nerve to critique his papers.  But I did it for the reasons I have already explained.  A further aspect that I had not completely appreciated at the beginning is something David argues so beautifully in his essay, ``Writing Physics'':
\begin{quotation}
\noindent Physicists traditionally replace talk about physics by a mathematical
formalism that gets it right by producing a state of compact
nonverbal comprehension. The most fascinating part of writing physics
is searching for ways to go directly to the necessary modifications
of ordinary language, without passing through the intermediate
nonverbal mathematical structure. This is essential if you want to
have any hope of explaining physics to nonspecialists. And my own
view is that it's essential if you want to understand the subject
yourself.\footnote{Mild paraphrasing from David's original.}
\end{quotation}
I found that when I wrote to David, I wrote better---self assessment, of course---than I did to any of my other friends.  It was a question of wanting to keep up with the Joneses, I suppose, but it had this lovely unintended consequence: I felt I understood things better and better just by touching my keyboard and having a new David-{\Mermin} letter in mind to write.  Better writing, not just any writing, made for better understanding.

I started to realize that the major part of the problem of our understanding quantum mechanics had come from bad choices of English and German words (maybe Danish too?)\ for various things and tasks in the theory.  Once these bad choices got locked into place, they took on lives of their own.  With little exaggeration, I might say that badly calibrated linguistics is the predominant reason for quantum foundations continuing to exist as a field of research.  Measurement?  I agree with John {\Bell}---it is a horrible word and should be banished from quantum mechanics.  But it is not because it is ``unprofessionally vague and ambiguous'' as {\Bell} said of it.  It is because it conveys the {\it wrong\/} image for what is being spoken of.  {\it The\/} quantum state?  That one is just about as bad.  Who would have thought that so much mischief could be made by the use of a definite article?  So many of these things came to mind as I would strive to be clear and entertaining in my writings to David.

I once had a piece of trinitite, that lovely green glass born in the heat of the Trinity test explosion, 16 July 1945.  I was very proud of it and kept it displayed on a shelf in my library until the Cerro Grande Fire took it from me.  It was a material reminder of Stanis{\l}aw {\Ulam}'s words, ``It is still an unending source of surprise for me to see how a few scribbles on a blackboard or on a sheet of paper could change the course of human affairs.''  But from my standpoint today, it is not those equations that changed the world.  They were merely a center of attention for the layers and layers of verbal apparatus and practical action required to give them life.  It is language that powers things, and in this thought David {\Mermin} has been my leader.
\eq

\section{16-09-09 \ \ {\it PIAF Quantum Foundations Conference -- Late Reminder}\ \ \ (to J.-{\AA} Larsson)} \label{Larsson8}

\bjal
Sorry, I've booked a trip to idQuantique over just those dates.

I've forgotten to remind you that I'd very much enjoy meeting you and Marcus at some point. (But the nearest that I can manage is probably November.)
\ejal

Just let me know as your ideas firm up.  Say hello to Nicolas Gisin for me.  See the subtitle on page 3 of the attached file---it is a talk I gave at the meeting where Gisin won the John Bell Prize. [Title: Quantum Bayesian Coherence. Subtitle: The Anti-Nicolas-Gisin Interpretation of Quantum Mechanics.] I don't think he liked it!  But in great part we see the world diametrically oppositely; so I thought why not be honest about it.

\section{16-09-09 \ \ {\it The Primacy of Measurement, 2}\ \ \ (to D. M. {\Appleby})} \label{Appleby72}

\bma
Do we say that measurements are going on in distant stars?  And if
we do say that, what has become of the concept of
measurement?  Are they measurements without agents to do the
measuring?  Or what?
\ema
These are the crucial questions.  My ideas for an answer have been forming since last New Year's Eve when I started to write you and Hans on the subject.  I'll be curious to hear what Jochen answers you on this.

\section{16-09-09 \ \ {\it QF Group Meeting, Thursday 4pm Bob Room}\ \ \ (to R. W. {\Spekkens})} \label{Spekkens72}

\brws
This week's topic of discussion: Is the possibility of doing science contingent on the universe having a certain kind of locality?  Einstein famously claimed something to this effect: ``Without such an assumption of the mutually independent existence (the `being-thus') of spatially distant things, an assumption which originates in everyday thought, physical thought in the sense familiar to us would not be possible. Nor does one see how physical laws could be formulated and tested without such a clean separation.''
\erws

Thanks for the quote.  It reminded me of why I've been pushed to my rather radical Bayesian position about quantum probabilities.  It's because over and over again, I always find myself agreeing with Einstein.

\section{16-09-09 \ \ {\it Till Tomorrow} \ \ (to R. {\Schack})} \label{Schack175}

Suppose we have already enforced the following:  $n b = a - 1$ and $a m = n (b + 1)$.  (The latter condition being the one that arises from fixing the length of a posterior vector of an ISU measurement to be the length of the basis vectors.)

Now look at Eq.\ (163) in the RMP version of the paper.  From the latter constraint it follows that we have $m$ unit vectors with pairwise inner products $-b$.  I'm going to make use of this in a moment.  First though recall the precursor observation I made to you yesterday:

Take $q$ normalized vectors $|f_j\rangle$ such that
\begin{enumerate}
\item
$\sum |f_j\rangle = 0$, and
\item
$\langle f_j | f_k \rangle = c$ for $j\ne k.$
\end{enumerate}
Then $c$ is fixed to $-1/(q-1)$.  The converse is true as well:  If $c=-1/(q-1)$ and 2), then 1) must hold necessarily.

But we have $m$ unit vectors with pairwise inner products of value $-b$.  Think of $b$ as the free variable.  Suppose $b=1/(m-1)$.  Then these vectors could not possibly all be proper probability distributions:  For not all of them could live in the positive ``quadrant'' of the space and still give condition 1 above.  On the other hand, if $b=1/m$, it seems to me nothing obviously or immediately prevents us from augmenting our set of $m$ supposed probability distributions with 1 more vector to get a simplex centered at the origin.

I.e., thinking of $m$ as fixed, only $b=1/m$ and smaller should be viable values.

I think this sketches the solution to our conundrum.  Please think hard about it and see if you see any flaws.  And see if you see a way to make it more rigorous---I'm going back to bed now.  (Shockingly, if the reasoning above is correct, it vindicates most of the old argument, with a proper understanding of the direction of inequality now.  But we should definitely scrap the old argument and use a rigorized version of the one above.)

May I call you at work after I reawaken (and get the kids off to school).  Say about 10:00 AM my time?

\section{16-09-09 \ \ {\it Ciao} \ \ (to G. Valente)} \label{Valente11}

Many congratulations on your new position!  That's wonderful to hear---though Alexei did clue me in some.  I fully endorse your learning some true physics and very much sympathize with your sentiment about those ``just pretending to know physics, as many philosophers do.''  To us on the inside, it is quite clear that many of them are indeed just pretending---they can't hide it.

I must admit that I now have Musil's book (two volumes) on my bookshelves, but I have not read it yet.

Let me point you to a new paper of mine: \arxiv{0906.2187}.  I hope some of the philosophy at the end (Section 8) will bring you back to some of our earliest discussions on the malleability of reality and alchemy.  We've missed the real Giovanni on this side of the fence!

\section{17-09-09 \ \ {\it Bumpy Yard}\ \ \ (to D. M. {\Appleby})} \label{Appleby73}

It's moles man!  It's moles!  Now, how do you get rid of moles?  The bumpy yard has gotten much worse since you were here, and I've finally started to catch on to the pattern.  It's most definitely moles.

Very depressing day of quantum foundations at PI.  We had a big QF group meeting, and once again I realized I was orthogonal to most everyone in the audience.  It's kind of a lonely and frustrating place.

\section{18-09-09 \ \ {\it Happen?}\ \ \ (to the QBies)} \label{QBies2}

Will either of you happen to go to the University today before coming to PI?  If so, could one of you to pick up the following book while you are there?
\bq\noindent
Title: Conceptual problems of quantum gravity\\
Publisher: Boston :  Birkh\"auser, c1991. \\
Location: UW Davis. Book Stacks. Main Floor\\
Call Number: QC178.C63 1991
\eq
I think we should follow up thinking about yesterday's group meeting topic, and there is an important article by Stachel on the subject in that book.

\section{19-09-09 \ \ {\it The Cabibbo Angle} \ \ (to R. {\Schack})} \label{Schack176}

Cabibbo has an angle.  Weinberg has an angle.  Why can't we QBists too have an angle?  Ours satisfies $$\cos\theta=  \frac{ (n-m)}{(m-1)^2 + (n-1)}$$ and we declare, for reasons ``of a strong Bayesian flavor'' that $\cos\theta=1/2$.

But what are those reasons?  We need something good.  I can see at least three paths of attack, but beyond seeing that they're paths, I don't presently know how to proceed any distance down them.  1) is to try to supply a reason for your observation that adding an $(m+1)$'th $|\alpha_k\rangle$ gives a set that sums to the zero vector.  2) Is to try to supply a reason for the value $\cos\theta=1/2$ outright.  And here's a 3).  Let
$|r\rangle=\sum_k |p_k\rangle$.  You can confirm that $\langle r| r\rangle=\frac{m^2}{n}$.  Can we supply a reason that this quantity should be 1?

Note that $|r\rangle$ is not the mixture of the $|p_k\rangle$, but the sum.  If we take the mixture, all useful information disappears because, no matter what $m$, the vector $\frac{1}{m}\sum_k |p_k\rangle$ is the uniform distribution over $n$ outcomes.

2) very much looks like an extra symmetry assumption.  Could we justify it from that angle?  1) really looks like a pretty direct statement about the spider-web character of Hilbert space.  (See:
\myurl{http://www.starmagic.com/MAGIC-LOOPS-Lotus-Flower.html} or do a search in Google Images of ``flexi-sphere''.)  And 3) surely seems like a raw normalization condition of some variety.

Confused and confused.  I wonder if Cabibbo was ever confused \ldots\

\section{20-09-09 \ \ {\it One of the Essences of Things, Maybe?}\ \ \ (to R. {\Schack})} \label{Schack177}

I spent a restless night alternating between dreams and meditating on $\cos\theta=1/2$ and, sometimes, dreaming of meditating on the latter and meditating on dreaming of the latter.  It was a wild mix of neuronal firings.  The small residue left behind is perhaps more a vague feeling than anything else, but I'll share it nonetheless.

Previously we've been treating the {\it form\/} of the urgleichung as nearly the only empirical addition (i.e.\ nonBayesian) component in our reasoning.  Certainly our emphasis was there.  But important for this consideration, shouldn't we also include the scale of $\beta$?  By considering measurements for which can have certainty over the outcomes, we are able to make a connection between $m$, $n$, and $\beta$, but the importance of the scale of $\beta$ remains.  I tend to think that that is the best way to think of it.  What is interesting is that quantum mechanics takes $\beta$ to be neither the smallest it can be $0$, nor the largest $(n-2)/n$.

It's a bit like this maybe:  Could one posit {\Schroedinger}'s equation as having empirical content, without at the same time as positing $h$ to have such content as well?

Anyway, I wonder if we may have been off-base a bit, hoping to get a bit too much from ``assumptions of a strong Bayesian flavor'' as well as pure geometrical considerations.  I.e., maybe indeed it is only empirics that gives $\cos\theta=1/2$.

Or, am I giving up too easily?\medskip

\noindent Continuingly-Miserable Chris

\section{20-09-09 \ \ {\it All Ways of Expression} \ \ (to R. {\Schack})} \label{Schack178}

Maybe another way to put it is this.  Would we really want an urgleichung so self-consistent that it had no adjustable parameters at all signifying {\em which\/} world we live in?  Would we really want it completely fixed, save for the size of the system we were applying it to?  I go up and I go down on this.

\section{20-09-09 \ \ {\it A Way To Salvage Much of the Old} \ \ (to R. {\Schack})} \label{Schack179}

I feel confident that what I am about to say is sound, but---warning!---I've had three glasses of wine, and I might not express myself as ``clearly'' as usual.

Suppose a device with $m$ outcomes that achieves the ideal of certainty.  Then that device defines $m$ states in our proposed state-space.  Let us make the restriction that a basis state can arise as the posterior consequent upon the outcome of such a measurement.  I.e., we are in the usual setting of this last week, where $\beta = (n-m)/[n(m-1)]$.  As we know, without further assumptions, there are no restrictions on $m$ (other than that $2 \le m \le n$).  BUT---and I'm only now starting to appreciate this---{\it we can still ask the question}:  Are there any {\it further\/} potential states with the same symmetry as the first $m$ of them?  There's no claim here that such an extension {\it need\/} correspond to a measurement that achieves the ideal of certainty:  By construction, there cannot be one.  Instead, it is purely a question of symmetry.  We are asking, is our space maximal in the sense of not allowing the addition of one more state with the appropriate relation with the previous ones?  For given $n$ and $m$, we know(?) we can do this if $m$ is already sufficiently large in comparison to $n$.  Thus what we ask is merely (humbly), what is the smallest $m$ for which this symmetry property fails.  To put it in still other words, what we are asking is that our ISU measurement that achieves the ideal of certainty generate a maximal set of states in the sense that adding even one more potential state must push it outside the valid state-space.

Now we use the mathematics from our previous argument (or an {\Asa}-ified version) to show that $m\ge \mbox{blah blah blah}$.  We choose $m$ to be the extremal case as our postulate.

So the assumption is not one of impossibility of construction, but one of the symmetry properties of our space.  Of course, why should one care about symmetry properties?  I am staunchly belligerent to Jochen Rau whenever he invokes symmetry groups in his reconstructions of QM.  But maybe we can really think of it more, again, as a maximality property.  Our supposed measurements achieving the ideal of certainty have posteriors that ``just fit into'' the space.

OK, now four glasses of wine.  Surely there will be problems above, but maybe it still indicates a productive path.

\section{20-09-09 \ \ {\it PIAF Quantum Foundations Conference -- Late Reminder} \ \ (to A. Plotnitsky)} \label{Plotnitsky23}

\barkp
On a different (or ever the same) subject, I have been reading some recent commentaries on the experimental verification of Hardy's paradox, which commentaries I found (unsurprisingly) confusing.  My sense from Hardy's initial article and Mermin's article on it is that it is just a special and subtler case of quantum entanglement, which does not essentially change anything in terms of epistemology, although of course it is not about to stop the usual debates.  I think more and more that nothing in any reasonably near future will stop them, and Hardy's paradox was used by Bohmians to support their view, just as the EPR experiment was.
\earkp
Yes, that's right:  Just more of the same.

\barkp
For instance, it seems to me one would still be easily able to maintain the Bayesian view you guys take, just as one can in the case of the EPR experiment. No? That is, does one need any special adjustments?  Once again, sorry to miss your paper.
\earkp
No, not at all.  The lesson of the Hardy paradox is as it is with the usual Bell inequality violation:  If one assumes Einstein locality, then the outcomes of quantum measurements cannot pre-exist the act of measurement.

Sorry to hear you can't come.  But I know we'll see each other again soon, somewhere, some sunny day.

\section{21-09-09 \ \ {\it Wavering} \ \ (to R. {\Schack})} \label{Schack180}

I'm starting to become disenchanted with the three-glass solution.  I guess the transition is coming about because I started to appreciate tonight that it really seems to rely on the $\alpha_k$ representation for its statement.  If you take the $|w_k\rangle$ representation, there doesn't seem to be an analogue.  So, in what way is it an interesting symmetry assumption that there is no $(m+1)$'st state?

\section{21-09-09 \ \ {\it Fantastic Talk} \ \ (to G. J. Chaitin)} \label{Chaitin1}

Just wanted to let you know, I thought that was a fantastic talk today!  Even though I knew most everything from having studied your papers in great detail in the mid '90s, I found it quite inspiring still.

Among other things, your remarks at the end on experimental mathematics struck a chord with my present situation.  There's a certain kind of mathematical structure we (myself with Carl Caves and {\Ruediger} Schack) need to exist for our ``quantum Bayesian'' program in quantum foundations.  But so far we only know it exists for spaces with dimension 2 to 67.  Beyond that it has so far been too hard computationally.  Nonetheless I plow ahead assuming these things exist in all dimensions, and I become amazed at how pretty the formal structure of quantum mechanics appears in terms of these structures.  So as you say of the new axiom added to ZF set theory:  Everyone believes these things exist, but we shouldn't let their lack of proof stand in our way.  In case you're interested, here's a paper that gives an introduction to the formalism: \arxiv{0906.2187}.

And yes, I fully believe ``God plays dice'' with the universe.  Our QBism program takes that as the essence of quantum mechanics.  See Section 8.1 in the paper cited above.

\section{22-09-09 \ \ {\it The Great Hardy} \ \ (to R. {\Schack})} \label{Schack181}

I just had a conversation with Lucien and he made a very keen observation.  By construction, our $\cos\theta$ must be a rational number.  Suppose we could give some desideratum so that it is a fixed number independent of $n$ and $m$ (exactly as we are doing with the Cabibbo angle postulate that it is 1/2).  Then generally there will be no integer $n$ and $m$ solutions to the equation.  He was guessing that $\cos\theta=0$ and $\cos\theta=1/2$ might be the only rational solutions.

But, Lord God, how would we prove that?

\section{22-09-09 \ \ {\it Our Diophantine Equation} \ \ (to R. {\Schack})} \label{Schack182}

Let $N=n-1$ and $M=m-1$.  Then our demand is that there be integers $N$, $M$, $q$, $r$, with $q\le r$, such that
$$
(q-r) N + r M + q M^2 = 0.
$$
Wouldn't it be nice if that were a famous Diophantine equation that we could just look up?

Just solving the quadratic equation for $M$.  We get that
$$
r^2 + 4 q (r-q) N
$$
must at least be a perfect square.  Which I think means that
$$
4 q (r-q) N
$$
itself must be a perfect square.

Yow, this is too hard for me!!  But I think it is almost surely our salvation!!

\section{22-09-09 \ \ {\it Further Point} \ \ (to R. {\Schack})} \label{Schack183}

I had previously said:
\bq
\noindent Suppose we could give some desideratum so that it is a fixed number
independent of $n$ and $m$ (exactly as we are doing with the Cabibbo
angle postulate that it is 1/2).
\eq
But we wouldn't even need that further desideratum.  The question is simply, when is there a solution.  If the only strictly positive one is $\frac{1}{2}$, then we're done.

I've asked Peter Shor if he could help on the Diophantine equation front.  No reply yet, but it's the sort of email that Peter actually replies to.

What a great turn of fortune.  I'm so glad I had that off-hand conversation with Lucien today.

Cheers, as you wake up to a brighter morning.

\section{22-09-09 \ \ {\it Router Bit}\ \ \ (to N. D. {\Mermin})} \label{Mermin159}

\bdm
I'm happy to bring it along. Have it sent to me at 75 Hickory Road, Ithaca, NY 14850-9606.
\edm
Ah, you mean the place made famous by the 75 Hickory Road Interpretation of Quantum Mechanics?

\section{22-09-09 \ \ {\it Oxford-Austin Cuisine}\ \ \ (to T. Norsen)} \label{Norsen0}

I don't believe we've ever met, but I'm looking forward to some productive conversation with you at the PIAF meeting.  I know Rob Spekkens says great things about you, and I have very much enjoyed the clarity in your Bell papers:  They are great reading.  Today I finally started (thinking about) preparing my talk for the meeting.  I hope you at least enjoy the title [``Texas-Bavarian Home Cooking:\ A Quantum Bayesian Reply to Bell's (and Norsen's) la Nouvelle Cuisine,'' \pirsa{09090087}], even if the content ends up appalling you.

I wonder if I might ask a favor of you, even before our first real hello?  There are two talks on {\tt pirsa.org} that I'd love you to have under your belt before we start talking in person:  One by Chris Timpson, \pirsa{09080010}; and one by me, \pirsa{09080018}.  I think they much lay the groundwork for the particular way in which I will disagree with you in my PIAF talk, and I'm quite fearful that it is not something I can really convey properly in the 20 minutes allotted me.  But if I could come to some productive conversation with you---learning and explaining and identifying subtle points---that's all I really care about.  So, I hope you won't mind giving me this much of a head-start:  I think it'd much help for building the necessary translation dictionary we'll need for comparing our worldviews.

\section{23-09-09 \ \ {\it All QBbibbo Angles} \ \ (to R. {\Schack})} \label{Schack184}

You were probably already aware of this, but let me record it since you didn't mention it.

You can find all valid QBibbo angles by rewriting the relation between $n$ and $m$ as such
$$
n = m + \frac{2c}{1-c}\frac{m(m-1)}{2}.
$$
In order for $n$ to be an integer this means $\frac{2c}{1-c}$ must be an integer.  (Take the $m=2$ case if you have any suspicions.)  This completely classifies all possible QBibbo angles.  Letting $k$ be that integer, we get
$$
n = m + k \frac{m(m-1)}{2}.
$$

The three lowest $k$-value cases give:
\begin{itemize}
\item
$k=0$, the classical law of total probability
\item
$k=1$, something that looks like real-vector-space quantum mechanics
\item
$k=2$, quantum mechanics
\end{itemize}

Discounting $k=0$, it is darned frustrating that there is the $k=1$ case in there below the beautiful quantum solution $k=2$.  Otherwise, we could have used an extremization-style axiom.  I've been thinking all day in the back of my head how we might shut the RVS case out.  Nothing so far, but 99\% of my energy was thrown into {\Asa}'s paper.  Indeed we were wrong in our reasoning in the last version of the big paper, but we were oh so close to having the right answer.  Is that meaningful?  Or at least a hint about the next step.

On the upside, it is kind of cute that we get RVS-QM, when Hardy's axioms shut it out from the beginning, and have not so far found a way to be loosened so as to allow for it.  (I know these things from lunch with Hardy and Wootters, the latter being quite interested in the real case.)

I think I remain with the conviction I outlined Sunday that this may be our most ``physics filled'' postulate.  But given that it is nearly at the end of our scale above, maybe I'll reassess that in light of further, yet to be discovered, facts.

You might want to put some of the above in our paper; I don't know how much.  In any case, we should thank Hardy at the end along with Myrvold.  The conversation yesterday with him inspired much of this.

\section{24-09-09 \ \ {\it Haute Cuisine}\ \ \ (to N. D. {\Mermin})} \label{Mermin160}

What exactly is it that you like so much about Bell's nouvelle cuisine article?  In what way does it spook you?  What is it in there that should really impress me or spook me?  I read it, and I mostly felt nothing.  Can you put it in words?

\subsection{David's Response}

\bq
I don't have it with me now, but I'll try to send you some comments before I leave on Saturday. As I remember, it's the combination of the Jarrett argument, which always struck me as the most compelling way of isolating the assumptions behind the conditionally independent representation of joint distributions that underlies Bell inequalities, with Bell's talents as a writer, which I've always very much admired.

More later, I hope. \ldots

Took a quick look. It's really only two pages: Sections 6.9 and 6.10.
Nothing about EPR arguments, determinism, probability 1 being certainty.
Just a very general notion of what ought to count as an explanation of correlation.
Lambda doesn't have to be EPR elements of reality. It can be attendant circumstances. As I remember my point was that this was the kind of argument that your critical remarks at the beginning of your paper ought to be addressing --- not straw men.

And of course he writes beautifully.
\eq

\section{24-09-09 \ \ {\it That Bungling {\Mermin}}\ \ \ (to N. D. {\Mermin})} \label{Mermin161}

Have you seen has how this guy \arxiv{quant-ph/0607057}\index{Norsen, T.} portrays all your expositions on Bell inequality stuff as bungled misunderstandings?  How does his cuisine taste in your mouth?

As you can see, I've started preparing for my talk \ldots

\section{24-09-09 \ \ {\it That Bungling {\Mermin}, 2}\ \ \ (to N. D. {\Mermin})} \label{Mermin162}

The way he wants to argue (same as Bell but without the charm) is that ``local causality'' alone implies ``local {\Mermin} instruction sets'' in an EPR scenario.  Then Bell demolishes with his inequality local instruction sets.  Consequently local causality all alone is demolished.

I want to say:  no, no, no, local causality {\it alone\/} does not imply {\Mermin}'s instruction sets.  There is a principle that you, Norsen (and I think Bell), take as so sacrosanct you don't even notice that it comes into play.  It is that probability one assignments for a proposition should imply that the proposition has the truth value of TRUE.  This is a significant component of what hides behind the EPR criterion of reality and a part that you accept without question.  But we quantum Bayesians don't accept it at all.  And therefore we don't have to accept that Bell inequality violations deny local causality.  Instead they cap off our quantum Bayesian suspicions about the role of certainty (probability one).

\section{24-09-09 \ \ {\it Timpson's Talk on Sept.\ 25}\ \ \ (to the QBies)} \label{QBies3}

I hope you guys can come to Timpson's talk tomorrow [\pirsa{09090029}].  He's a really good speaker.  Particularly, I thought he did a good job last month in this one \pirsa{09080010} which might be a good lead in for tomorrow's talk.

\bq\noindent
What would a consistent instrumentalism about quantum mechanics be? Or, why Wigner's friendly after all. \smallskip\\
Chris Timpson\\
University of Oxford \medskip

Instrumentalism about the quantum state is the view that this mathematical object does not serve to represent a component of (non-directly observable) reality, but is rather a device solely for making predictions about the results of experiments. One honest way to be such an instrumentalist is a) to take an ensemble view ($=$ frequentism about quantum probabilities), whereby the state represents predictions for measurement results on ensembles of systems, but not individual systems and b) to assign some specific level for the quantum/classical cut. But what happens if one drops (b), or (a), or both, as some have been inclined to? Can one achieve a consistent view then? A major worry is illustrated by the Wigner's friend
scenario: it looks as if it should make a measurable difference where one puts the cut, so how can it be consistent to slide it around (as, e.g., Bohr was wont to)? I'll discuss two main cases: that of Asher Peres' book, which adopts (a) but drops (b); and that of the quantum Bayesians Caves, Fuchs and Shack [\emph{sic}], which drops both. A view of Peres' sort can I, think, be made consistent, though may look a little strained; the quantum Bayesians' can too, though there are some subtleties (which I shall discuss) about how one should handle Wigner's friend.
\eq

\section{25-09-09 \ \ {\it Some Geometry If You Want It}\ \ \ (to N. D. {\Mermin})} \label{Mermin162.1}

\bdm
What is a Qbist?
\edm

\noindent {\bf QBism} -- the foundational program of quantum Bayesianism.
(See first sentence of Section 8 of the big paper.) \medskip

\noindent {\bf QBist} -- a practitioner of QBism.\medskip

\noindent {\bf The QBies} -- myself, along with my group of three grad students and {\AA}sa our postdoc; maybe Appleby, our long-term visitor, as well.\medskip

\noindent {\bf The QBibbo Angle} -- 60 degrees; see Assumption 7 and equation 24 of my short paper with {\Ruediger}.\medskip

\noindent {\bf The QBicle} -- (a common meeting place for the QBies, yet to be determined)

\subsection{David's First Reply}

\bq
So I guess that would make me a QBissel?
\eq

\subsection{David's Second Reply}

\bq
Bissel is German, that's your problem.   It means just a little.
(As in:
     Sprechen Sie Deutsch?
     Nur ein Bissel.)\medskip

\noindent {\bf QBissel} -- One who is intrigued by QBism but not yet convinced.
\eq

\section{25-09-09 \ \ {\it Why Such Radical Moves?}\ \ \ (to R. W. {\Spekkens})} \label{Spekkens73}

It was interesting to watch your reaction again yesterday when I emphasized that in my picture of things, quantum measurement outcomes---the things the particular user of the theory, the agent, is gambling upon---are {\it personal\/} experiences, not happenings completely outside the agent.  You showed a surprise as if you had never heard the idea before.  Thinking about it tonight, I remember you having a similar reaction about a year ago, as if you had never heard of it before at that time either.  Yet the personal aspect of the ``dings'' has been in my thought for at least four years, probably five.  And I have presented it consistently so in my talks ever since.  It is part of the content of the attached diagram which I know you've seen many times, and it was a significant part of my old tale of the two islands (one of nurturing wives, and one of bad girlfriends) in Konstanz and several talks later.  Most recently I thought I gave a very careful statement at the reconstruction meeting, \pirsa{09080018}, and I believe even Timpson did me right in \pirsa{09080010} by emphasizing that the quantum Bayesians have an ``extra layer'' of agential involvement over the classical, radical Bayesian (say Richard Jeffrey, Brian Skyrms, etc.).  You can confirm for yourself.

It's a learning experience for me:  seeing this apparent potential barrier in you, and by extrapolation, to many, many of the ears I have been speaking to over the years.  Come to think of it, there are only a few people who have challenged me on this aspect of the worldview.  (One is Wayne Myrvold who tells me that I don't really believe it, and one is John Sipe who made the definition ``consequence Bayesian'' to try to capture the idea.)  The lack of more and varied systematic challenges probably demonstrates that the idea isn't even getting to the level of awareness in people's minds where they can say, ``ick!''\ if they want to.  There's a conceptual potential barrier that's not even letting the idea get into people's heads.  Somehow I'll use this information.

Anyway, yesterday, one of the things you prodded me about was:  Why take such a radical stance on the character of measurement outcomes?  I said to ``preserve locality'' among other things---but you talk too fast for me (and drop into attack mode too quickly sometimes), for me to give you more adequate responses.  (I don't like that and wish it didn't have to be.)  Locality---in my mind, in my usage---is a codeword for the idea that individual things have independent existences / independent essences of each other.  It is as Einstein put it in part of the quote you sent out last week:  ``\ldots\ it appears to be essential for this arrangement of things introduced in physics that, at a specific time, these things claim an existence independent of one another \ldots''  I like that idea very much, that things ``claim an existence independent of one another,'' so much so that I see Einstein's holy grail of a deterministic field physics---the thing he held to most dearly based on these considerations---as infringing on the very idea.  He imagines a world where specifying a field on a boundary surrounding a spacetime point (a purported individual existence) completely determines the field at that point.  I say, then the independent existence is only illusory:  If you are {\it determined by}, you are not {\it independent of}.  Period.

I see our quantum Bayesianism as the beginning of a correction in that regard; it embodies at its core the idea of individual existences closed off, in part, from the rest of the world.  That is part of the lesson of quantum measurement outcomes as personal experiences.  It gives me the means to identify something in physics that really is closed off from everything else.  And all of this overconfident and little-justified talk of quantum mechanics {\it implying\/} ``nonlocality outright'' (Norsen's writing style does get under my skin) as very much a threat to the worldview of individual existences.  But that is a bit of an aside.

When I set out to write this note, I simply wanted to say that there are more reasons for my adopting the radical move of measurement outcomes as personal experiences.  And most particularly is the issue of a consistent account of Wigner's friend from the QBist point of view.  I think that is part of the subject of Chris Timpson's talk tomorrow and hope you will go to it.  Maybe that would be a good starting point for us to talk about it further some time.

There is a big, tightly knit package in my mind of thoughts on quantum things and a worldview arising from it that I am trying to think my way through.  And it has a landscape so different from the quantum foundations chatter I hear around me all the time that it does make it very difficult to speak on the fly.  I guess I am just asking for patience.

\section{25-09-09 \ \ {\it The Scale of Engagement}\ \ \ (to H. C. von Baeyer)} \label{Baeyer79}

Let me tell you the story of the last couple weeks:  They have been quite miserable.  {\Ruediger} and I were preparing a Ruedigerized version of Section 6 of our last paper, making it a stand-alone paper and hoping to linearize the thinking to make it clearer.  Well, we linearized it indeed!  So much so that we caught a flaw in our earlier derivation.  Much, much gut wrenching followed, I can tell you, as we tried to salvage the result if not the derivation.  And clarity only came very slowly.  In the end, we had to face the inevitable---that our result would not be as powerful as it had appeared previously (giving something that looks identical to the quantum expression), but instead something that needs a further (possibly ad hoc) assumption to get to the quantum case.

Still with every cloud there is a silver lining!  And now I think there is a greater opportunity than ever to frame the whole issue in terms of just how much the observer is engaged in quantum theory.  Please look at the attached paper [\arxiv{0912.4252v1}].  I think you'll this one more of a pleasure to read than previously.  (And I am now 99.9\% confident that there are no mistakes in it other than grammatical ones!)  Here is what leaps out that didn't before.  The parameter $\beta$ in the urgleichung
$$
q(j)= \sum_{i=1}^n \Big(\alpha p(i)-\beta\Big) r(j|i)
$$
signifies the degree to which the observer in quantum mechanics is undetached or engaged---for in the end, it signifies the amount of deviation there is from the law of total probability in this counterfactual games.  When $\beta=0$, we have a detached observer.  The keen and interesting thing is quantum mechanics is not the very next step up, but rather the {\it second\/} step up in the hierarchy of possibilities.  There is a potential theory with observer engagement more than classical but less than quantum.  Then there are wild possibilities above quantum.  See the discussion after assumptions 6 and 7.

I am now convinced that we ended up with a better playground than we had previously for our deeper foundational ideas.  I hope reading this will inspire you, and I hope reinject you with some enthusiasm, so that maybe you'll re-pick back up on our AJP Pauli article.  My guess it has slipped under the rug for a while.

I'd like to hear any thoughts you have (on any subject, much less the above).

\section{25-09-09 \ \ {\it Beta is the Parameter to Rule them All} \ \ (to R. {\Schack})} \label{Schack185}

But thank God there weren't nine parameters.  See my story to Hans Christian below.  [See 25-09-09 note ``\myref{Baeyer79}{The Scale of Engagement}'' to H. C. von Baeyer.] In the order of ideas, I am fairly convinced now that it is $\beta$, a direct expression of the deviation from the Law of Total Probability, that is the key constant that we should be focusing on.  Particularly I have always been reluctant to take $m_0$ (with the particular meaning as the number of distinguishable states) as the ultimate parameter.  It evoked too much of a similarity to classical ideas.

Stream of thought.  Associated with a system is a value of $\beta$.  [Or maybe I could be convinced of $\alpha$ since $\alpha$ increases with dimension.  But I sure like that minus sign in front of the $\beta$.]  Wonder postulate:  There are ISU measurements that achieve the ideal of certainty.  That there can ever be certainty should remain a miracle from our point of view.  $\beta$ then determines the number of outcomes we can have in such a measurement.  Self-consistency forces various rational and integer values.  But $\beta$ in the end rules them all.  The constant $q$ chooses which world we are in.  Apparently one in which the observer is just a little engaged.

Student just came \ldots

\section{25-09-09 \ \ {\it More On Normativity and a Quantitative Measure on Degrees of Detached Observers}\ \ \ (to C. G. {\Timpson})} \label{Timpson19}

I woke up from my nap thinking I wasn't nearly as generous with you at lunch as I should have been.  Your talk was just fantastic, and in point of fact, did in its way help me firm my thoughts on the Wigner friend issue.  I very much liked the detailed exposition on ``level discipline'' (I think you called it) and that really is the key.  Also you caused me to want to explore more thoroughly the tension between my ``internalist''/``personalist'' account of measurement and my respect for a kind of Copernican principle.

If you haven't seen it by now, I should tell you that I've always been in love with this paragraph of John Wheeler's, and with all these QBist explorations, every now and then I start to think I {\it might\/} know what it {\it might\/} mean:
\bq\noindent
How did the universe come into being? Is that some strange, far-off
process beyond hope of analysis? Or is the mechanism that comes into
play one which all the time shows itself? Of all the signs
that testify to ``quantum phenomenon'' as being the elementary act
and building block of existence, none is more striking than its utter
absence of internal structure and its untouchability. For a process
of creation that can and does operate anywhere, that is more basic
than particles or fields or spacetime geometry themselves, a process
that reveals and yet hides itself, what could one have dreamed up out
of pure imagination more magic and more fitting than this?
\eq

You'll see that John has that tension himself---on the one hand the ``untouchability'' and on the other the ``anywhere.''  One of these days I so hope all of this will really fit together.  Could I encourage you to write a proper paper on your talk?  It would be so helpful for all of us and could give us a good focal point for thinking things further and pushing things along.

On another matter:  Attached is one of my latest.  It's an attempt to do much better on Section 6 from the last paper.  There is a discussion on the idea of taking the Born rule normatively that might interest you.  And also the way I now identify a parameter ($q$ at the end of the paper) that is meant to characterize how ``engaged'' (meant to be the opposite of Pauli's ``detached'') an observer is in this theory.   Classical has $q=0$, quantum has $q=2$.  Why $q$ is not 1 identically or strictly $>2$, I do not know.

\section{26-09-09 \ \ {\it Logo}\ \ \ (to H. C. von Baeyer)} \label{Baeyer80}

\bhcvb
In my wanderings I came across a nice confirmation of our M\"obius strip logo.  It occurs in the book {\sl Recasting Reality\/} (Atmanspacher and Primas, eds., Springer 2009) in the article entitled ``Complementarity of Mind and Matter'' by Primas.  He defines: ``Here, a Boolean description refers to a Boolean domain where all propositions are either true or false, characterized by the postulate of the excluded middle.''  And then, since a Boolean theory of matter is impossible, he goes on to say: ``The most important example to get a globally non-Boolean description is to patch local Boolean descriptions together smoothly.''  How did we characterize the physics/psychic dichotomy?  Locally contradictory, globally identical --- Pauli's psychophysical reality combining mind and matter into one world (unus mundus).
\ehcvb

To my surprise I could only find one piece of email on the subject.  [See 26-04-09 note ``\myref{Appleby59}{Server Config}'' to D. M. {\Appleby}.]  It's below.  I recall ``locally different / globally the same.''  Possibly it was ``distinct'' instead of different.  I'm not sure I like ``contradictory.''  Are they contradictory?  The thought I'd like to articulate or give some flesh to is that they're different categories (whatever that might mean).

\section{30-09-09 \ \ {\it That Lovely Quote} \ \ (to \v{C}. Brukner)} \label{Brukner3}

\bq
The value of a [pluriverse], as compared with a universe, lies in this, that where there are cross-currents and warring forces our own strength and will may count and help decide the issue; it is a world where nothing is irrevocably settled, and all action matters. A monistic world is for us a dead world; in such a universe we carry out, willy-nilly, the parts assigned to us by an omnipotent deity or a primeval nebula; and not all our tears can wipe out one word of the eternal script. In a finished universe individuality is a delusion; ``in reality,'' the monist assures us, we are all bits of one mosaic substance. But in an unfinished world we can write some lines of the parts we play, and our choices mould in some measure the future in which we have to live. In such a world we can be free; it is a world of chance, and not of fate; everything is ``not quite''; and what we are or do may alter everything.\medskip

\hspace*{\fill} --- Will Durant, describing William James's philosophy
\eq

\section{30-09-09 \ \ {\it 37 Percent}\ \ \ (to R. W. {\Spekkens})} \label{Spekkens74}

You piqued my paranoia:  So I counted my transparencies from my talk yesterday.  18 of the 49 were completely new.  Two others were major modifications to something old.  Minor changes on about five others.

But no doubt, my core ideas are the same as they ever were.  \ldots\ same as they ever were \ldots\ same as they ever were \ldots

\section{02-10-09 \ \ {\it Experience Primary, Events Secondary?}\ \ \ (to R. Blume-Kohout)} \label{BlumeKohout10}

Events as intersections of experiences?  Does that phrase or the one in the title bear any resemblance to what you were saying to me yesterday?

This is probably not your kind of reading, but let me put it on the table:
\bq\noindent
\myurl{http://psychclassics.yorku.ca/James/experience.htm}.
\eq

\section{03-10-09 \ \ {\it From Dewey to Unger \ldots}\ \ \ (to T. Homer-Dixon)} \label{HomerDixon1}

It was good meeting you---I'm back in my office at PI, immediately after the Unger talk \ldots\ scratching my head, wondering what I heard.   It could be that I'm just missing crucial vocabulary.  In any case, my education in James, Dewey, Schiller, Putnam, Rorty, etc., didn't carry me to the brink of understanding Unger.  Maybe better luck to me next time.

Still, one doesn't denigrate opportunities:  I'm glad to have met another person in Waterloo interested in pragmatism.  In case it might entertain you, let me offer one of my own spoutings on the subject.  See section 8.1 of this paper: \arxiv{0906.2187}.  That section is much independent of the remainder and gives some orientation on how I think the physical world is put together along ``fallibalist'' lines.

\section{06-10-09 \ \ {\it QF/Southwestern Ontario Philosophy of Physics Talk This Friday}\ \ \ (to D. Fraser \& J. Berkovitz)} \label{Berkovitz1} \label{Fraser2}

I'm so sorry I'm going to miss this.  I don't get back to Waterloo from New Orleans until late afternoon.  Good luck Jossi!  The more de
Finetti these guys are exposed to, the better!

\subsection{Doreen's Preply}

\bq
Jossi Berkovitz (U Toronto) will be giving a talk this Friday, October
9th at 11a.m.\ in the Alice Room:
\bq\noindent
The World According to De Finetti\\
Jossi Berkovitz (U Toronto)\medskip\\
Abstract: Bruno de Finetti is one of the founding fathers of the
subjectivist school of probability, where probabilities are
interpreted as rational degrees of belief. His work on the relation
between the theorems of the probability calculus and rationality is
among the corner stones of modern subjective probability theory. De Finetti maintained
that rationality requires that an agent's degrees of belief be
coherent. I argue that de Finetti held that the coherence conditions of degrees
of belief in events depend on their verifiability. On this view, the
familiar constraints of coherence only apply to sets of degrees of
belief that could in principle be jointly verified. Accordingly, the
constraints that coherence imposes on degrees of belief are generally
weaker than the familiar ones. I then consider the implications of
this interpretation of de Finetti for probabilities in quantum
mechanics, focusing on the EPR/Bohm experiment and Bell's theorem.
\eq
\eq

\section{06-10-09 \ \ {\it New Orleans, RMP, and Capacities}\ \ \ (to N. D. {\Mermin})} \label{Mermin163}

Sorry I've kept you waiting so long for a reply.  After that horrible meeting, I needed a couple of days away from anything professional.  Then I jutted off to New Orleans for the next meeting.  And I've been kept busy in all kinds of way since.  Finally a moment to myself:  I'm writing you from the poshest hotel bar I've ever written you from.

Just had dinner with Henry Folse (the Bohr scholar) and had a tour of his amazing house near the Garden District.  It was the poshest house I ever visited before the poshest bar.  An 1868 home with 14 ft ceilings and 10,000 square feet!  The front hall was just amazing---about 12 feet wide and running all the way from the front of the house to the back.  The dining room had two massive mirrors taken from an old New Orleans bordello.  There was just so much character to the house and gardens.  If I had a place like that, my thinking would be of such a higher order!

Anyway, keep the router bit, and thanks for the heads-up about the second referee.  I hope to get you the revised Section 6 by Friday evening.  I'll probably work on it mostly on Friday's flights (I got upgrades to business class).

On my flight over here, I read 90 pages of Nancy Cartwright's {\sl The Dappled World}.  {\Timpson} was right:  I do find much in her of value.  I want to revise my language of what Hilbert-space dimension signifies---I rather like her terminology of ``capacities.''  True enough it is ``real'' as I have emphasized before, but it should be thought of as a capacity like mass or charge.  Its ontological role is of something of the same flavor.  The way I want to emphasize things now is that what quantum mechanics has taught us is that associated with any piece of the physical universe is a previously overlooked capacity.  Newton taught us that associated with any piece of the universe is a capacity to attract other bodies, quantified by the body's mass.  I'm thinking something like that for Hilbert-space dimension.  It's not like I haven't used language like this before, but now it seems to be more entrenched in my big picture of things.  Don't know why, but the idea makes me tingle more than it had previously.

I wish you had been there for Friday's panel discussion.  It was quite annoying but revealing.  Psychologically, that is.  This time Valentini called me a solipsist.  What is it about these guys?!?  Got a note from Norsen Saturday calling me a solipsist again too.  I didn't bother to reply.  If you want to stay on my good side, never call me a solipsist.

I'm glad you got something out of the meeting and (convinced yourself that) you enjoyed yourself.  It meant a lot to me to see you again.

\section{16-10-09 \ \ {\it The More and the Modest} \ \ (to L. Hardy)} \label{Hardy38}

I started writing the note below while in New Orleans, but only finished it today.  It seemed a shame to rewrite the introduction, so I send it as is.

\begin{flushright}
\baselineskip=3pt
\parbox{3.7in}{
\bq
\noindent
It don't mean a thing if it ain't got that Zing!\medskip
\\
\hspace*{\fill} --- Dean Rickles
\eq
}
\end{flushright}\medskip

Greetings from the land of the crawdad.  This note is to say thanks again for your remarks at the panel discussion Friday.  You did something that none of the other ones did---you caused me to think and clarify, rather than piss me off from a display of overt closed-mindedness.  It amazes me the way Norsen and Valentini preach to me without understanding an ounce of the way I'm trying to see things.  (I just got another note from Norsen yesterday accusing me yet again of solipsism.)

Your point on Hilbert-space dimension on the other hand was a productive one, and thinking about how I should reply to you has helped me realize that my language needs a significant overhaul.  With better language, we inevitably get better understanding.  Here goes.

I used to ask, ``What is real of a quantum system?,'' and give ``Zing'' as a placeholder for the answer.  What is Zing?  The only thing I've been able to comfortably identify for it in the quantum formalism is Hilbert-space dimension.  But that certainly can't be enough to get the world going---or at least, I think you would (and did) say something like this.  John Sipe once wrote, ``the quantum world of this interpretation is a fixed, static thing.  It is a frozen, changeless place.''

That is {\it at face value}, there's not enough engine, not enough gears and pinions, in this bare ``reality'' to account for motion and change and indeterminism and novelty and growth and evolution and \ldots\ make the list just as big as you want \ldots\ all the things we see around us.   What is weird is that I agree.  Such a minimalist world doesn't even track with what I myself believe.  So it becomes a question of why I have doggedly resisted saying more.

I believe I have tracked down the troublemaker in my mode of expression, and it is this.  My trouble is {\it not\/} that I believe that there is {\it nothing\/} in the world, or perhaps a minimalist number of things (too minimal), as the solipsism-chargers seem so obsessed with accusing me of.  It is rather that my point of view admits {\it too many\/} things into the world---too many things of an independent and self-sustaining reality, things for which there are no equations; realities of which I am only willing to {\it point\/} to and say effectively, ``Yeah the world includes that too.''

This is because the ontology I imagine is the one of the humanist, rather than the trained or indoctrinated scientist.  Take Democritus as an example:  for him, the universe literally was ``atoms and void.''  That simple phrase was meant to account for all that is.  Similarly, the universe Einstein sought was literally meant to be a differentiable manifold supporting the solution to a very clever (though never found) differential equation---all else, even such as the flow of time, was simply illusion.  (Recall {\Mermin}'s talk at PIAF.)  The unifying theme in these two visions is that chairs and thumbs and bricks and all the like from our common experiences are secondary things---effective (operational) descriptions or illusions---having no real, primary existence of their own.  For a most relevant example, consider David Wallace's universe.  As far as I can tell, it is literally a single quantum state on a Hilbert space.  That vector, timelessly unchanging in a rotating frame, is the universe's whole substance.  The apparent indeterminism of the quantum world is subordinate to the greater monistic determinism and timelessness of the whole.

But here's the funny thing:  It is exactly of Einstein's and Wallace's universes, not my own, that I would say, ``the world of this interpretation is a fixed, static thing.  It is a frozen, changeless place.''  To my mind, these are both completely barren visions for the world.  I think our distaste for each other's proclivities comes from this:  It is a very old philosophical divide, that between the rationalist attitude and the empiricist attitude.  It has only become clear to me recently, but at my core I am an extreme empiricist.

Here is the way my friend William James put the distinction when he was in a relatively mild mood:
\bq
By empiricism I mean the tendency which lays most stress on the part, the element, the individual, treats the whole as a collection, and calls the universal an abstraction. By rationalism I mean the tendency to emphasize the universal, and to make the whole prior to the part, in the order both of logic and of being. The temper of rationalism is dogmatic: it willingly claims necessity for its conclusions.  Empiricism is more modest, and professes to deal in hypotheses only.
\eq
A more full-blooded statement of the issues involved, and just beautiful reading, can be found in a long quote I've plucked out for you from his essay ``The Sentiment of Rationality.''  See subsection at the end of this note.  If you find something worth reading in this note of my own composition, I hope you will take the time to read the longer quote of James.  In the end, it has become very crucial to my thinking.

Let me give a briefer hint of the issue here.  It has to do with the way James put the dangers of rationalism when he was in a fiercer mood:
\bq
Let me give the name of `vicious abstractionism' to a way of using concepts which may be thus described: We conceive a concrete situation by singling out some salient or important feature in it, and classing it under that; then, instead of adding to its previous characters all the positive consequences which the new way of conceiving it may bring, we proceed to use our concept privatively; reducing the originally rich phenomenon to the naked suggestions of that name abstractly taken, treating it as a case of `nothing but' that, concept, and acting as if all the other characters from out of which the concept is abstracted were expunged. Abstraction, functioning in this way, becomes a means of arrest far more than a means of advance in thought. It mutilates things; it creates difficulties and finds impossibilities; and more than half the trouble that metaphysicians and logicians give themselves over the paradoxes and dialectic puzzles of the universe may, I am convinced, be traced to this relatively simple source. {\it The viciously privative employment of abstract characters and class names\/} is, I am persuaded, one of the great original sins of the rationalistic
mind.
\eq

The way this is relevant to me is the following.  Far from thinking the world is an empty place, a place only with me in it.  I think it is full of things, overflowing with things.  {\it All\/} distinct things, from head to toe.  And literally so.  It is not a world made of six flavors of quarks glued together in various combinations.  It is not a world that maps to a single algorithm running on Rob {\Spekkens}'s favorite version of Daniel Dennett's mechanistic cellular automaton.  It is a world of heads and toes and doorknobs and dreams and ambitions and every kind of particular.  (And that is not a typo:  It is a world in which even dreams and ambition have substance.)  It is a world in which Vivienne Hardy is a distinct entity, not ``constructed'' of anything else, but a true-blue crucial piece of the universe as it is today---no less crucial than spacetime itself.

In modern parlance, I am not a reductionist.  And when my world is judged as empty (of all but me), I claim it is because I am being interpreted from the standpoint of an (explicit or implicit) reductionist worldview.  It is true that my envisioned world may be as cockamamie as James' heads of Borneo:
\bq
Taken as it does appear, our universe is to a large extent chaotic.  No one single type of connection runs through all the experiences that compose it.  If we take space-relations, they fail to connect minds into any regular system.  Causes and purposes obtain only among special series of facts.  The self-relation seems extremely limited and does not link two different selves together.  Prima facie, if you should liken the universe of absolute idealism to an aquarium, a crystal globe in which goldfish are swimming, you would have to compare the empiricist universe to something more like one of those dried human heads with which the Dyaks of Borneo deck their lodges.  The skull forms a solid nucleus; but innumerable feathers, leaves, strings, beads, and loose appendices of every description float and dangle from it, and, save that they terminate in it, seem to have nothing to do with one another.  Even so my experiences and yours float and dangle, terminating, it is true, in a nucleus of common perception, but for the most part out of sight and irrelevant and unimaginable to one another.
\eq
But it is not an {\it empty\/} world.

With this as a background, let me now return to Zing, Hilbert-space dimension, and the needed overhaul of my language.  I ask, ``What is real of a quantum system?'' and answer, ``its Hilbert-space dimension,'' but that is a very funny thing to say.  There is {\it so much\/} that is real of a quantum system, why would I ever say that?  I wouldn't say it of Vivienne Hardy, for instance,---I already declared this quite vocally---and she is a perfectly good example of a quantum system.  So, what am I really up to?

It is a bit like this:  Forget quantum mechanics, and think back to the days when the only physics known was basically Newtonian gravity.  Would any Newtonian have ever answered the question, ``What is real of a physical system?,'' with the declaration, ``Only its gravitational mass is real of it.''?  Most definitely not.  But that is because Newton merely/boldly taught us but a singular thing:  That every body in the universe, that every thing that can be carved out from it, had a previously undisclosed capacity---a numerical capacity to (try to) attract every other body in the universe.  On the one hand, I say ``merely'' because looked at in this way, it really is a very {\it modest\/} move---modest at least by the lights of the Steven-Weinberg--Stephen-Hawking-theory-of-everything generation:  It was no theory of everything---never pretended to be.  On the other hand, I say ``boldly'' because I certainly agree with my friend Hans Christian von Baeyer who wrote in one of his books:
\bq
Great revolutionaries don't stop at half measures if they can go all the way.  For Newton this meant an almost unimaginable widening of the scope of his new-found law.  Not only Earth, Sun, and planets attract objects in their vicinity, he conjectured, but all objects, no matter how large or small, attract all other objects, no matter how far distant.  It was a proposition of almost reckless boldness, and it changed the way we perceive the world.
\eq

And therein lies the key, I think, to how I should modify my language.  When I ask myself what have we learned with quantum mechanics, I want to say first and foremost that we have learned how we should more consistently gamble upon the consequences of our interactions with external physical systems.  But why this new calculus for gambling?  Because quantum mechanics is uncovering that every object in the universe has a previously undisclosed {\it capacity}.

I would have liked to have said ``uncovered,'' but at this stage of research I must still settle for ``is uncovering''---it is an unfinished project to understand the significance of quantum mechanics in these terms.  Hilbert-space dimension, like gravitational mass, is representative of some universal capacity.  That's the real idea.  Dimension is a quality a body possesses in a way that, in a QBayesian understanding, it does not ``possess'' a quantum state.  If we conceptually delete an agent gambling upon the consequences of his interactions with a quantum system---the QBist says---we also conceptually delete its quantum state.  But there is no reason to believe the system and the system's capacity are deleted as well.

Capacity for what?  That's where the hard part begins.  Sometimes I say the capacity for birth and creation.  Sometimes I say the capacity to entertain counterfactuals.  Sometimes I say it is a capacity that can be harnessed to aid of computation, as in quantum computation.  The truth is, I don't yet know what I mean in any precise way.  I only know that I have a strong inner tug to thinking that the SIC calculus will help reveal a precise idea.

Anyway, I write all this to put the key idea into perspective and to practice a way of speaking that I had not used to any great extent before.  [Chris {\Timpson} had pointed out that there are several similarities between the way I speak and the way the philosopher Nancy Cartwright speaks of ``capacities.''  And on this trip to New Orleans, I have confirmed it is very much so by starting to read her book, {\sl The Dappled World}.  So, at least I'm not alone in the world.]  You were indeed right to suspect that Hilbert-space dimension doesn't have enough gears and pinions to get things going, but neither did gravitational mass.  Nonetheless, the disclosing of that old capacity---gravitational mass---``changed the way we perceive the world'' (von Baeyer).  And so I think of our new capacity.

Thus I end this instalment by saying thanks again for pointing out a great deficiency in my choice of words and for consequently causing me to think.  I'll leave you with a sentence that I wrote in this file, but could not quite find a way to use in the essay.  It seems a shame to throw it away:
\begin{center}
Quantum mechanics calls out, ``I will not be a representative of your monistic dreams!''
\end{center}
When Norsen, Valentini, and Wiseman call me solipsist (or worse), I think it is because they are working from the middle of a monistic dream.

\subsection{Passage from William James' essay ``The Sentiment of Rationality,'' (Compiled for Lucien)}

\subsection{Part of Immediate Interest}

\bq
The facts of the world in their sensible diversity
are always before us, but our theoretic need is that
they should be conceived in a way that reduces their
manifoldness to simplicity. Our pleasure at finding
that a chaos of facts is the expression of a single
underlying fact is like the relief of the musician at
resolving a confused mass of sound into melodic or
harmonic order. The simplified result is handled
with far less mental effort than the original data; and
a philosophic conception of nature is thus in no
metaphorical sense a labor-saving contrivance. The
passion for parsimony, for economy of means in
thought, is the philosophic passion {\it par excellence};
and any character or aspect of the world's phenomena which gathers up their diversity into monotony
will gratify that passion, and in the philosopher's
mind stand for that essence of things compared with
which all their other determinations may by him be
overlooked.

More universality or extensiveness is, then, one
mark which the philosopher's conceptions must possess. Unless they apply to an enormous number of
cases they will not bring him relief. The knowledge
of things by their causes, which is often given as a
definition of rational knowledge, is useless to him
unless the causes converge to a minimum number,
while still producing the maximum number of effects.
The more multiple then are the instances, the more
flowingly does his mind rove from fact to fact. The
phenomenal transitions are no real transitions; each
item is the same old friend with a slightly altered
dress.

Who does not feel the charm of thinking that the
moon and the apple are, as far as their relation to the
earth goes, identical; of knowing respiration and
combustion to be one; of understanding that the
balloon rises by the same law whereby the stone
sinks; of feeling that the warmth in one's palm when
one rubs one's sleeve is identical with the motion
which the friction checks; of recognizing the difference between beast and fish to be only a higher
degree of that between human father and son; of
believing our strength when we climb the mountain
or fell the tree to be no other than the strength of
the sun's rays which made the corn grow out of
which we got our morning meal?

But alongside of this passion for simplification
there exists a sister passion, which in some minds---though they perhaps form the minority---is its rival.
This is the passion for distinguishing; it is the impulse to be {\it acquainted\/} with the parts rather than to
comprehend the whole. Loyalty to clearness and
integrity of perception, dislike of blurred outlines, of
vague identifications, are its characteristics. It loves
to recognize particulars in their full completeness,
and the more of these it can carry the happier it is.
It prefers any amount of incoherence, abruptness, and
fragmentariness (so long as the literal details of the
separate facts are saved) to an abstract way of conceiving things that, while it simplifies them, dissolves
away at the same time their concrete fulness. Clearness and simplicity thus set up rival claims, and make
a real dilemma for the thinker.

A man's philosophic attitude is determined by the
balance in him of these two cravings. No system
of philosophy can hope to be universally accepted
among men which grossly violates either need, or
entirely subordinates the one to the other. The fate
of Spinoza, with his barren union of all things in one
substance, on the one hand; that of Hume, with
his equally barren ``looseness and separateness'' of
everything, on the other---neither philosopher owning any strict and systematic disciples today, each
being to posterity a warning as well as a stimulus---show us that the only possible philosophy must be
a compromise between an abstract monotony and a
concrete heterogeneity. But the only way to mediate
between diversity and unity is to class the diverse
items as cases of a common essence which you discover in them. Classification of things into extensive ``kinds'' is thus the first step; and classification of their relations and conduct into extensive ``laws''
is the last step, in their philosophic unification. A completed theoretic philosophy can thus never be anything more than
a completed classification of the world's ingredients; and its
results must always be abstract, since the basis of every
classification is the abstract essence embedded in the living
fact---the rest of the living fact being for the time ignored by the
classifier. This means that none of our explanations are complete.
They subsume things under heads wider or more familiar; but the last
heads, whether of things or of their connections, are mere abstract
genera, data which we just find in things and write down.

When, for example, we think that we have rationally explained the
connection of the facts $A$ and $B$ by classing both under their
common attribute $x$, it is obvious that we have really explained
only so much of these items as {\it is x}. To explain the connection
of choke-damp and suffocation by the lack of oxygen is to leave
untouched all the other peculiarities both of choke-damp and of
suffocation---such as convulsions and agony on the one hand, density
and explosibility on the other. In a word, so far as $A$ and $B$
contain $l$, $m$, $n$, and $o$, $p$, $q$, respectively, in addition
to $x$, they are not explained by $x$. Each additional particularity
makes its distinct appeal. A single explanation of a fact only
explains it from a single point of view. The entire fact is not
accounted for until each and all of its characters have been classed
with their likes elsewhere. To apply this now to the case of the
universe, we see that the explanation of the world by molecular
movements explains it only so far as it actually {\it is\/} such
movements. To invoke the ``Unknowable'' explains only so much as is
unknowable, ``Thought'' only so much as is thought, ``God'' only so
much as is God. {\it Which\/} thought? {\it Which\/} God?---are
questions that have to be answered by bringing in again the residual
data from which the general term was abstracted. All those data that
cannot be analytically identified with the attribute invoked as
universal principle, remain as independent kinds or natures,
associated empirically with the said attribute but devoid of rational
kinship with it.

Hence the unsatisfactoriness of all our speculations. On the one
hand, so far as they retain any multiplicity in their terms, they
fail to get us out of the empirical sand-heap world; on the other, so
far as they eliminate multiplicity, the practical man despises their
empty barrenness. The most they can say is that the elements of the
world are such and such, and that each is identical with itself
wherever found; but the question Where is it found? the practical man
is left to answer by his own wit. Which, of all the essences, shall
here and now be held the essence of this concrete thing, the
fundamental philosophy never attempts to decide. We are thus led to
the conclusion that the simple classification of things is, on the
one hand, the best possible theoretic philosophy, but is, on the
other, a most miserable and inadequate substitute for the fulness of
the truth. It is a monstrous abridgment of life, which, like all
abridgments, is got by the absolute loss and casting out of real
matter. This is why so few human beings truly care for philosophy.
The particular determinations which she ignores are the real matter
exciting needs, quite as potent and authoritative as hers. What does
the moral enthusiast care for philosophical ethics? Why does the {\it
\AE sthetik\/} of every German philosopher appear to the artist an
abomination of desolation?
\bq
Grau, teurer Freund, ist alle Theorie \\
\indent Und gr\"un des Lebens goldner Baum.
\eq
The entire man, who feels all needs by turns, will take nothing as an
equivalent for life but the fulness of living itself. Since the
essences of things are as a matter of fact disseminated through the
whole extent of time and space, it is in their spread-outness and
alternation that he will enjoy them. When weary of the concrete clash
and dust and pettiness, he will refresh himself by a bath in the
eternal springs, or fortify himself by a look at the immutable
natures. But he will only be a visitor, not a dweller, in the region;
he will never carry the philosophic yoke upon his shoulders, and when
tired of the gray monotony of her problems and insipid spaciousness
of her results, will always escape gleefully into the teeming and
dramatic richness of the concrete world.
\eq

\subsection{Of Secondary Interest to the Present Discussion, but Worth Reading If You've Gotten This Far and Have Nothing Better To Do}

\bq
So our study turns back here to its beginning. Every way of
classifying a thing is but a way of handling it for some particular
purpose. Conceptions, ``kinds,'' are teleological instruments. No
abstract concept can be a valid substitute for a concrete reality
except with reference to a particular interest in the conceiver. The
interest of theoretic rationality, the relief of identification, is
but one of a thousand human purposes. When others rear their heads,
it must pack up its little bundle and retire till its turn recurs.
The exaggerated dignity and value that philosophers have claimed for
their solutions is thus greatly reduced. The only virtue their
theoretic conception need have is simplicity, and a simple conception
is an equivalent for the world only so far as the world is
simple---the world meanwhile, whatever simplicity it may harbor,
being also a mightily complex affair. Enough simplicity remains,
however, and enough urgency in our craving to reach it, to make the
theoretic function one of the most invincible of human impulses. The
quest of the fewest elements of things is an ideal that some will
follow, as long as there are men to think at all.

But suppose the goal attained. Suppose that at last we have a system
unified in the sense that has been explained. Our world can now be
conceived simply, and our mind enjoys the relief. Our universal
concept has made the concrete chaos rational. But now I ask, Can that
which is the ground of rationality in all else be itself properly
called rational? It would seem at first sight that it might. One is
tempted at any rate to say that, since the craving for rationality is
appeased by the identification of one thing with another, a datum
which left nothing else outstanding might quench that craving
definitively, or be rational {\it in se}. No otherness being left to
annoy us, we should sit down at peace. In other words, as the
theoretic tranquillity of the boor results from his spinning no
further considerations about his chaotic universe, so any datum
whatever (provided it were simple, clear, and ultimate) ought to
banish puzzle from the universe of the philosopher and confer peace,
inasmuch as there would then be for him absolutely no further
considerations to spin.

This in fact is what some persons think. Professor Bain says ---
\bq
\indent
A difficulty is solved, a mystery unriddled, when it can be shown to
resemble something else; to be an example of a fact already known.
Mystery is isolation, exception, or it may be apparent contradiction:
the resolution of the mystery is found in assimilation, identity,
fraternity. When all things are assimilated, so far as assimilation
can go, so far as likeness holds, there is an end to explanation;
there is an end to what the mind can do, or can intelligently desire
\ldots. The path of science as exhibited in modern ages is toward
generality, wider and wider, until we reach the highest, the widest
laws of every department of things; there explanation is finished,
mystery ends, perfect vision is gained.
\eq

But, unfortunately, this first answer will not hold. Our mind is so
wedded to the process of seeing an {\it other\/} beside every item of
its experience, that when the notion of an absolute datum is
presented to it, it goes through, its usual procedure and remains
pointing at the void beyond, as if in that lay further matter for
contemplation. In short, it spins for itself the further positive
consideration of a nonentity enveloping the being of its datum; and
as that leads nowhere, back recoils the thought toward its datum
again. But there is no natural bridge between nonentity and this
particular datum, and the thought stands oscillating to and fro,
wondering ``Why was there anything but nonentity; why just this
universal datum and not another?''\ and finds no end, in wandering
mazes lost. Indeed, Bain's words are so untrue that in reflecting men
it is just when the attempt to fuse the manifold into a single
totality has been most successful, when the conception of the
universe as a unique fact is nearest its perfection, that the craving
for further explanation, the ontological wonder-sickness, arises in
its extremest form. As Schopenhauer says, ``The uneasiness which
keeps the never-resting clock of metaphysics in motion, is the
consciousness that the non-existence of this world is just as
possible as its existence.''

The notion of nonentity may thus be called the parent of the
philosophic craving in its subtilest and profoundest sense. Absolute
existence is absolute mystery, for its relations with the nothing
remain unmediated to our understanding. One philosopher only has
pretended to throw a logical bridge over this chasm. Hegel, by trying
to show that nonentity and concrete being are linked together by a
series of identities of a synthetic kind, binds everything
conceivable into a unity, with no outlying notion to disturb the free
rotary circulation of the mind within its bounds. Since such
unchecked movement gives the feeling of rationality, he must be held,
if he has succeeded, to have eternally and absolutely quenched all
rational demands.

But for those who deem Hegel's heroic effort to have failed, nought
remains but to confess that when all things have been unified to the
supreme degree, the notion of a possible other than the actual may
still haunt our imagination and prey upon our system. The bottom of
being is left logically opaque to us, as something which we simply
come upon and find, and about which (if we wish to act) we should
pause and wonder as little as possible. The philosopher's logical
tranquillity is thus in essence no other than the boor's. They differ
only as to the point at which each refuses to let further
considerations upset the absoluteness of the data he assumes. The
boor does so immediately, and is liable at any moment to the ravages
of many kinds of doubt. The philosopher does not do so till unity has
been reached, and is warranted against the inroads of those
considerations, but only practically, not essentially, secure from
the blighting breath of the ultimate Why? If he cannot exorcize this
question, he must ignore or blink it, and, assuming the data of his
system as something given, and the gift as ultimate, simply proceed
to a life of contemplation or of action based on it.
\eq

\section{16-10-09 \ \ {\it Degree of Engagement}\ \ \ (to H. C. von Baeyer)} \label{Baeyer81}

You might also contemplate this transparency from my last talk:
$$
Q(D_j)=\left(\frac{1}{2} qd+1\right)\sum_{i=1}^{n} P(H_i) P(D_j|H_i) - \frac12 q
$$
with $q=0,1,2,\ldots$ (character of the zing) and $d=2,3,4,\ldots$ (value of a beable, how much zing).

For ``character of the zing'' read ``degree of observer engagement'' or ``degree of observer activation''.  Or maybe the quantity $q$ as ``amount of alchemical potential.''  Mostly playing with words.

\section{16-10-09 \ \ {\it The Sister Passion and the Shape of F-Theory to Come}\ \ \ (to R. W. {\Spekkens})} \label{Spekkens75}

I wrote the note below with Lucien in mind, but I might send it to you as well.  [See 16-10-09 note ``\myref{Hardy38}{The More and the Modest}'' to L. Hardy.]  When composing it, I was struck by how much the first three paragraphs of the associated attachment sounded like what I deem of your philosophy of what makes good science.  (Tell me if I'm right.)  Also, in the actual note (i.e., below, not attached), I further flesh out some thoughts that really, rightly started in a discussion in your office.  So, really you deserve some credit as well.

\section{17-10-09 \ \ {\it The More and the Modest, 2} \ \ (to N. D. {\Mermin})} \label{Mermin164}

Some reading you might enjoy on a Saturday morning---see below.  [See 16-10-09 note ``\myref{Hardy38}{The More and the Modest}'' to L. Hardy.]  I'll get back to fixing section 6 in the Revs Mod Phys paper Tuesday, after my visiting brother leaves.  It's been hard to focus with the New Orleans meeting and then my brother's visit.

You never did say what you thought of that Weyl quote.

The word is out, by the way, that Grassl and Scott have now proven SICs to exist in dimensions 35 and 48.  Two more dimensions in the list now!  Infinity can't be all that far away.

\section{17-10-09 \ \ {\it The More and the Modest, 3} \ \ (to N. D. {\Mermin})} \label{Mermin165}

I'm the youngest of five:  two brothers and two sisters.  The order was sister, brother, brother, sister, me.  Weyl quote below:
\bq\noindent
Finally and above all, it is the essence of the continuum that it cannot be grasped as a rigid existing thing, but only as something which is in the act of an inwardly directed unending process of becoming \ldots\ .  In a given continuum, of course, this process of becoming can have reached only a certain point, i.e.\ the quantitative relations in an intuitively given piece $\cal S$ of the world [regarded as a four-dimensional continuum of events] are merely approximate, determinable only with a certain latitude, not merely in consequence of the limited precision of my sense organs and measuring instruments, but because they are in themselves afflicted with a sort of vagueness \ldots\ .  And only ``at the end of all time,'' so to speak, \ldots\ would the unending process of becoming $\cal S$ be completed, and $\cal S$ sustain in itself that degree of definiteness which mathematical physics postulates as its ideal \ldots\ .  Thus the rigid pressure of natural causality relaxes, and there remains, without prejudice to the validity of natural laws, room for autonomous decisions, causally absolutely independent of one another, whose locus I consider to be the elementary quanta of matter.  These ``decisions'' are what is actually real in the world.
\eq

\section{19-10-09 \ \ {\it Decisions} \ \ (to N. D. {\Mermin})} \label{Mermin166}

\bdm
Calling the constituents of the real present ``decisions'' is a bit
strange, though.
\edm

Expand on that a bit if you can.  I'm quite attracted to what I think I read of his idea.

\section{19-10-09 \ \ {\it Cartwright}\ \ \ (to M. Tait)} \label{Tait3}

\bmt
I should also say that unlike Chris Timpson, I don't think that
imposing a Cartwrightian metaphysics on QBism, or any other
ready-made ontology, is necessarily helpful.
\emt

Well, I'm with you on the ``ready-made'' part.  I shouldn't think anybody's system will fit me perfectly.  On the other hand, I have now read some of Nancy Cartwright's {\sl Dappled World}, and I think Chris Timpson was right to perceive some similarities between Cartwright and me.  You'll see what I mean if you read the note below and the attached.  [See 16-10-09 note ``\myref{Hardy38}{The More and the Modest}'' to L. Hardy.]

Thanks for your proposal; it was good reading.

\section{19-10-09 \ \ {\it Purple Haze}\ \ \ (to M. Tait)} \label{Tait4}

\bmt
I'm still a little hazy on the notion that the Born rule should be
seen as an `empirical addition' to Bayesian reasoning. It is clear
that it is a normative rule that goes beyond Dutch book
coherence; it less clear to me how we ought to interpret this
additional constraint, though I'm still working through the meatier
parts of your paper with {\Schack} `Quantum Bayesian Coherence'.
\emt

Well, I'll make no bones about it:  I'm hazy as well.  There is however a slightly better discussion in the new (attached) paper.  [See \arxiv{0912.4252v1}.] It'd be great if you could contribute to getting these ideas straight.

Also, there was quite a mistake (a technical one) in Section 6 of the bigger paper.  That is fixed in the attached one.  I still need to modify the big one and re-post.  Anyway, fixing the mistake has led to some really good stuff:  A new parameter crops up that I like to think of as quantifying the degree to which the observer is engaged in quantum mechanics.  (By ``engaged observer'' I am aiming for a contrasting term for Pauli's ``detached observer''.)  See other attachment for a quick illustration:
$$
p(D_j)=\left(1+\frac{1}{2}qd\right)\sum_i p(H_i)p(D_j|H_i)-\frac{1}{2}q\;,
$$
where $q=0, 1, 2, \ldots$ signifies the ``character of the zing,'' and $d=2, 3, 4, \ldots$ signifies the ``value of a (local) beable, how much zing.''
[See also 25-09-09 note ``\myref{Baeyer79}{The Scale of Engagement}'' to H. C. von Baeyer.]

\section{20-10-09 \ \ {\it Howard vs Jammer} \ \ (to W. C. Myrvold)} \label{Myrvold13}

Remember our discussion on Don Howard's versus Max Jammer's version of the Einstein-Ehrenfest conversation.  See Don's discussion at:\medskip

\myurl[http://www.nd.edu/~dhoward1/Early\%20History\%20of\%20Entanglement/sld026.html]{http://www.nd.edu/$\sim$dhoward1/Early\%20History\%20of\%20Entanglement/sld026.html}.\medskip

I think you were right:  Don definitely wins.

\section{20-10-09 \ \ {\it My First Dean Rickles Quote} \ \ (to D. Rickles)} \label{Rickles1}

You'll now find yourself immortally samizdatized.  You can find it here:
\bq\noindent
\myurl{http://www.perimeterinstitute.ca/personal/cfuchs/nSamizdat-2.pdf}
\eq
starting at page 785 of text (page 813 wrt the raw pdf page numbering), in a note titled, ``The More and the Modest.''  [See 16-10-09 note ``\myref{Hardy38}{The More and the Modest}'' to L. Hardy.]

Does this capture a little bit of (or have any overlap with) the way you were trying to defend me against Valentini?

\subsection{Dean's Reply}

\bq
I'm deeply honoured.

The `more' versus `modest' distinction hits the nail on the head actually.  I was thinking about (what I think is) an analogous situation in modal logic, where you have your basic propositional calculus together with operators, including a modal operator. You define rules for these things (specifying when you have a well formed formula). Quantum mechanics can be set up initially like this: a formal schema. The job of interpretation is to latch this on to something. In modal logic the formal system can latch on to a range of things that are related only by their functional relations. For example, you might think in terms of possible worlds, of states in a computer program, of temporal instants, etc. These provide the semantics: the interpretation. They can all satisfy the basic formal schema, but they are obviously very different beasts. And indeed, some semantics (e.g.\ states in a computer program) seem more modest than others (e.g.\ possible worlds).

This reminded me of what was going on in the debate you and {\Ruediger} were having with Anthony. You were saying: the formal system isn't latching on to the stuff ``out there, hidden from view'' (or it is only in some very general sense: Hilbert space dimension) and Anthony was saying, No, it Is precisely latching onto stuff out there and hidden from view. One is indeed more Modest and one is More: but both satisfy everything that needs to be satisfied, formally and empirically.  One says QM is limited in what it can say about ``the world''; one doesn't.
\eq

\section{21-10-09 \ \ {\it Since It's My Birthday \ldots}\ \ \ (to H. C. von Baeyer)} \label{Baeyer82}

Since it's my birthday, and I'm feeling myself get older by the minute---very, very finite today---I'll feel free to try to stoke your coals again.

Adding to the note I had sent you last month [see 25-09-09 note ``\myref{Baeyer79}{The Scale of Engagement}'' to H. C. von Baeyer], I'll remind you of that cool Pauli quote where he mentions a ``degree of detachment of the observer.''  It is quite nice.

W.~Pauli, letter to Niels Bohr, dated 11 March 1955, photocopy
obtained from the Niels Bohr Institute via Henry Folse.

\bq
\noindent Dear Bohr,\smallskip

I find your letter of March 2nd very youthfull, which is just the
reason that it is not easy for me to answer. Although we have the
same view ``as regards the fundamental physical problems which fall
within the scope of the present quantum mechanical formalism'' and
although I agree with some parts of your letter, the situation is now
complicated by your use in a publication of a phrase like ``detached
observer'' (without comment!)\ which I used already in some
publications in a very different way.  I believe that this should be
better avoided to prevent a confusion of the readers\footnote{An
explaining remark about it in your {\it new\/} article would be most
welcome!} and I don't cling at all to particular words myself. I also
felt, already before your letter arrived, that my brief
characterisation of the observer in quantum theory as
``non-detached'' is in one important respect misleading. As is well
known to both of us, it is essential in quantum mechanics that the
apparatus can be described by classical concepts. Therefore the
observer is always entirely detached to the {\it results} of his
observations (marks on photographic plates etc.), just as he is in
classical physics. I called him, however in quantum physics
``non-detached'', when he chooses his experimental
arrangements.\footnote{I still believe today that this more
restricted use of my terminology is very good and that it has been
unhappily obscured in your article in a non-logical way!}

I shall try to make my point logically clear, by defining my
concepts, replacing hereby the disputed phrase by other words. As I
was mostly interested in the question, {\it how much informative
reference to the observer an objective description contains}, I am
emphasizing that a communication contains in general {\it
informations on the observing subject}.

Without particularly discussing the separation between a subject and
the informations about subjects (given by themself or by other
persons), which can occur as elements of an ``objective
description'', I introduced a concept ``degree of detachment of the
observer'' in a scientific theory to be judged on the kind and
measure of informative reference to the observer, which this
description contains.  For the objective character of this
description it is of course sufficient, that every individual
observer can be replaced by every other one which fullfills the same
conditions and obeys the same rules.  In this sense I call a
referency to experimental conditions an ``information on the
observer'' (though an impersonal one), and the establishment of an
experimental arrangement fulfilling specified conditions an ``action
of the observer''---of course not of an individual observer but of
``the observer'' in general.

In physics I speak of a detached observer in a general conceptual
description or explanation only then, {\it if it does not contain any
explicit reference to the actions or the knowledge of the observer}.
The ideal, that this should be so, I call now ``the ideal (E)'' in
honor of Einstein.  Historically it has its origin in celestial
mechanics.

There is an important {\it agreement\/} between us that we find
Einstein not consequent in this formulation of the ``ideal E''.
Indeed, there is no a priori reason whatsoever to introduce here a
difference between the {\it motion\/} of the observer on the one
hand, and the realization of specified experimental conditions by
the observer on the other hand.  If Einstein were consequent he had
to ``forbid'' also the word coordinate system in physics (as not
being objective).  That the situation in quantum mechanics has a
deep similarity with the situation in relativity is already shown by
the application of mathematical groups of transformation in the
physical laws in both cases.

In this way I reached the conclusion to distinguish sharply between
the ``ideal of an objective description'' (meaning science) on the
one hand (which I warmly supported just as you do) and the ``ideal
of the detached observer'' on the other hand (which I rejected as
much too narrow).

What really matters for me is not the word ``detached'', but the more
active role of the observer in quantum physics, which is already
implied in your [constatation?] of the ``indivisibility of the
phenomena and the essential irreversibility involved in the very
concept of observation''.  According to quantum physics the observer
has indeed a new relation to the physical events around him in
comparison with the classical observer, who is merely a spectator:
The experimental arrangement freely chosen by the observer lets
appear {\it single\/} events {\it not\/} determined by laws, the
ensembles of which are governed by {\it statistical\/}
laws.\footnote{In this way we obtain just the logical foundations of
an ``{\it objective\/} description'' of the incidents (Ein begriffe)
which the quantum mechanical observer makes within his surroundings
with his experimental arrangements.  Attention: there is {\it no\/}
logical contradiction between a word like ``trouble'' and a
possibility of its objective observation and description.}  It is not
relevant to me, if you say the same thing using {\it different\/}
terminologies (but please use [essentially?] different words than I).
They will only confirm my statements again as all these statements on
the observer are part of an ``objective description''.

I confess, that very different from you, I do find sometimes
scientific inspiration in mysticism\footnote{By the way: the
``Unity'' of everything has always been one of the most prominent
ideas of all mystics.} (if you believe that I am in danger, please
let me know), but this is counterbalanced by an {\it immediate\/}
sense for mathematics.  The result of both seems to be my kind of
physics, whilst I consider epistemology merely as a logical comment
to the application of mathematics in physics.\footnote{We are here
{\it both\/} in our letter in a realm of information on the writing
subject, which do {\it not\/} belong to the ``objective content of
the communications''.}  Thus when I read a sentence as ``how to
eliminate subjective elements in the account of experience'' my
immediate association is ``group theory'' which then determines my
whole reaction to your letter.  Although the first step to
``objectivity'' is sometimes a kind of ``separation'', this task
excites in myself the vivid picture of a superior common order to
which all subjects are subjected, mathematically represented by the
``laws of transformations'' as the key of the ``map'', of which all
subjects are ``elements''.

I hope that it will be possible to find a terminology which will
turn out to be satisfactory for both of us, but it is no hurry with
it.  I propose to resume this discussion only when your new article
will be ready, which I am eagerly awaiting.  It will show me your
terminologies in more general cases of objective descriptions, of
which I am most interested in the application to biology, in
connection with your new expression ``natural evolution''.

From March 16th till about 27th I am away in Germany and Holland and
when I come back I hope either to see you or to hear from you (I
wrote to Basel to get informations on your lecture
there).\footnote{Meanwhile I heard from P. Huber in Basel, [Fierz is
in the United States], that your lecture there is on March 30.  On
this date I am very glad, because I shall be back from my trip by
then. Paa Gensje!}

Hoping that you will in the future (just as I do myself) enjoy the
enrichment coming from the different kind of access to science by
different scientists, expressed in different, but not contradicting
terminologies, I am sending, also in the name of Franca, all good
wishes to yourself, to Margrethe and to the whole family,
\begin{flushright}
\noindent as yours complementary old \quad \smallskip

\noindent W.~Pauli \quad
\end{flushright}
\eq

\section{21-10-09 \ \ {\it All the Things that Got in My Way} \ \ (to R. {\Schack})} \label{Schack186}

Here is the question that has been haunting me since yesterday's drive to Guelph.  Suppose I judge two systems to be of Hilbert-space dimensions $d_1$ and $d_2$, and that they are each localized in space, a distance $r$ from each other.  Other than that, I make no commitments in what I think of them.  I.e., whether they are spin systems, elaborate multi-level quantum dots, a combination of the two things, caffeine molecules, ion traps, etc., etc.  Just ``raw'' Hilbert spaces.  What further things should I judge of them, even with this minuscule, isolated information?

\section{21-10-09 \ \ {\it PI/FQXI Workshop on Laws of Nature (May 2010)} \ \ (to S. Weinstein \& D. H. Wolpert)} \label{Weinstein4} \label{Wolpert2}

Sorry for the long absence.

I'm not as impressed with Unger as Lee is:  I tried reading his {\sl Pragmatism Unbound\/} and he just impressed me as an amateur who like to use big words and confusing constructions more than anything.  In the end I got less than halfway through it.  [You guys do know, or should know, I try to read everything there is on pragmatism.]  I had a similar reaction at his recent talk at CIGI---I didn't understand a single word of it; I didn't even know what it was about \ldots\ though maybe that's just ignorance on my part. Maybe he does better in this video blog \myurl[http://www.cigionline.org/blogs/2009/10/roberto-unger-cigi09-video-blog]{http:// www.cigionline.org/blogs/2009/10/roberto-unger-cigi09-video-blog}. Tell me what you think.  STILL---with regard to my own leanings with regard to these ``law without law'' issues ---I can certify that at least his heart is in the right place.  Thus I don't see any harm in inviting him, particularly if his presence will help Lee be a more effective cheerleader.

Concerning Hacking, I'm writing him again right now.

Concerning Nancy Cartwright, I finally took the time to read her book {\sl The Dappled World}---I have never read any of her, despite Timpson's urging.  And I find her truly interesting, and I would guess that she's got quite a unique point of view with respect to the other participants.  My sense of adventure says, ``She should certainly be invited, and if we have the budget why not?''

\section{22-10-09 \ \ {\it Coming of Age with Quantum Information} \ \ (to S. Capelin)} \label{Capelin11}

\bsc
I was wondering where we stand with your book.  the main thing we're still in need of before we can hand the book over to production is an idea for the front cover.

I tried googling Quantum Computer Images and came up with some interesting pictures.  You could try it.  Given the problems we're having finding something striking and suitable, how would you feel about using something like that?
\esc

I came to a standstill precisely because of the issue of the cover.  (And to be honest, because of several other things that have taken my time as well.)  But having wine with my brother visiting the other night may have led to a potentially viable idea.  What would you think about doing something with the image on my webpage.  (Using it as a core with perhaps some other things around it or in the background or vice versa or something.)  That's a photo my brother took of me when I was about 11 or 12.  He still has the actual photo; so we could get a higher version.  \ldots\ I know you'll be ``brutally honest'' so I'm bracing myself \ldots

\section{22-10-09 \ \ {\it QBism House Draft}\ \ \ (to N. Waxman)} \label{Waxman4}

Attached is the result of my day's labor.  My \LaTeX\ program shows 920 words; that's probably not exactly right, but is probably pretty close.  Be honest in telling me what you think:  It won't hurt my feelings, but you may find me being a stubborn old fart if I have a vested interest in this or that phrase you'd like to see dashed.  If you think I should cut it some I can work at it (cutting part of the details of Kiki's reconstruction, or our research group's projects, or some of the big quote).  But, of course, I like the balance of the essay as it is now---wouldn't have written that way otherwise---and if I can get away with keeping it mostly intact, I'd like to do that.  But I certainly understand your space limitations.

Thanks for encouraging me on this.  It was satisfying to write, and your enthusiasm flatters me.  (I like to be flattered.)

\bq
\begin{center}
\Large QBism House Opens for Business \medskip
\end{center}

Ideas, like children, need homes where they can be loved, nurtured, and raised to independence.  PI researcher Chris Fuchs brought an idea to Waterloo that he feels finally has a proper home.  The home is called QBism House, and the idea, if it can be put in a few words, is that a consistent foundation for quantum mechanics reveals our universe as no universe at all, but a {\it pluriverse\/} instead.  It is the sort of world the philosopher William James described as a ``republican banquet \ldots\ where all the qualities of being respect one another's personal sacredness, yet sit at the common table of space and time.''  It is a world of ``partially independent powers [where] each detail must come and be actually given, before, in any special sense, it can be said to be determined at all.''  Will Durant wrote in his masterful history of philosophy:
\bq
\noindent The value of a [pluriverse] as compared with a universe, lies in this, that where there are cross-currents and warring forces our own strength and will may count and help decide the issue; it is a world where nothing is irrevocably settled, and all action matters.  A monistic world is for us a dead world; in such a universe we carry out, willy-nilly, the parts assigned to us by an omnipotent deity or a primeval nebula; and not all our tears can wipe out one word of the eternal script.  In a finished universe individuality is a delusion; ``in reality,'' the monist assures us, we are all bits of one mosaic substance.  But in an unfinished world we can write some lines of the parts we play, and our choices mould in some measure the future in which we have to live.  In such a world we can be free; it is a world of chance, and not of fate; everything is ``not quite''; and what we are or do may alter everything.
\eq

Take an 1886 home with 10 and 12 foot ceilings, the willpower of Kiki Fuchs to restore it to its historical grandeur, a good carpenter to build a library of solid-oak bookshelves to house nearly 1,000 books on this subject, a covered porch and a chalkboard to discuss these matters at length in the warm summer air, and one has the base for some fantastic, new physics.  The house is called QBism House in honor of its ``quantum Bayesian'' heritage---a certain point of view about quantum mechanics developed by Fuchs, along with Carlton Caves, R\"udiger Schack, Marcus Appleby, and Howard Barnum.  From that view, a quantum state is not a real thing, like a rock or a tree or a quark, but ``a biological function, a means of orientation in life, of enabling and facilitating action, of taking account of reality and dominating it.''  Without doubt, this is a soft-edged idea, but it is one with a sharply defined mathematical core, and it is the latter that the research program of quantum Bayesianism is about.  Still, sharp cores and soft edges complement each other, much like the sharp glassed lines of Perimeter Institute and the curlicued Victorian corbels of QBism House.

Making the home young again, while preserving its memories and wisdom, has been a labor of love for Kiki Fuchs.  The walls had to be opened to replace the rusting pipes and the early-century electrical wiring; the chimneys had to be reconstructed from scratch; the basement had to be closed to the elements it had been subjected for years; crushed sewer pipes dug up and replaced; walls stripped of wallpaper and repainted; floors refinished; kitchen modernized; the dull, crumbling paint on the external walls given full-life Victorian colors; fencing, landscaping, and gardening---the to-do list went on and on for two years.  The home that emerged is now a Waterloo landmark.

Bolstered by a grant of \$512,000 from the United States Office of Naval Research and PI's contribution of much-valued office space and computer support, Fuchs is building a group to plow ahead on the technical aspects of the QBist project:  particularly, developing a good probabilistic representation of quantum states and dynamics by considering the so-called symmetric informationally complete (SIC) quantum measurements.  At the moment, the group is busy exploring whether these structures exist in all finite-dimensional Hilbert spaces, what the geometry of quantum-state space looks like when written in these terms, how quantum mechanics itself might be derived from the supposition of their existence, and characterizing the convex theories (generalizations of quantum mechanics) this representation most naturally fits into.  All told, the QBies---yes, the QBies---consist of PI visiting researcher Marcus Appleby, associate postdoctoral fellow {\AA}sa Ericsson, University of Waterloo PhD students Hoan Dang and Gelo Tabia, and MSc student Matthew Graydon.  The students' desks can be found in room 415 at PI: the QBicle, of course.

This Fall and Winter, the group will meet once a week at PI (at a day and time yet to be settled).  But once the weather is warm, the weekly meeting will be on the porch of QBism House, where the ghosts of pragmatist philosophers past (William James, John Dewey, and F.~C.~S. Schiller) can listen in with satisfaction about how their ideas of an unfinished, malleable world take a more exact form with the help of modern quantum mechanics.  Anyone at PI is welcome to join the meetings, research, and discussion.  The QBies and QBism House are open for business!
\eq

\section{22-10-09 \ \ {\it Happy}\ \ \ (to C. M. {\Caves})} \label{Caves100.5}

Some years are better than others, and some are worse.  This year was a tough one:  I made the mistake of seeing the new movie version of Sendak's book {\sl Where the Wild Things Are}.  It was the first book I had ever checked out of my school library.  The movie version is much darker and more a psychological play than the book was.  It made me realize that all my demons then are all my demons now.  No progress has been made in all these years.

Below is a philosophical piece that might entertain or appal you.  I view it as an important change in my language and emphasis.  [See 16-10-09 note ``\myref{Hardy38}{The More and the Modest}'' to L. Hardy.]

\section{23-10-09 \ \ {\it SICs and Reality}\ \ \ (to C. M. {\Caves})} \label{Caves101}

\bcc
I have long understood that you believe that the world is more than our subjective description of it.  But a danger that you must perceive---and that might make you hesitant to point to anything objective---is that that you can easily slip over into saying that there are things actually out in the world, complex things that are going on, but we are forced by some principle to limit our description to that provided by the quantum-mechanical state, which is ours alone as tool for setting betting odds, not something out in the world.  But then where the heck is all that the complex, objective stuff that's going on?  Is it outside quantum mechanics, which is what your talk of attributing more to the world than a physicist does is likely to lead?  Is it some actual stuff that a hidden-variable guy would be happy with?

I know you're not headed in that direction, but then when where are you headed?  There is, I would guess, little doubt that given the world, we can cook up a quantum state that is a good description of it, i.e., one we are willing to bet on.    These quantum states lie off in some teeny-tiny piece of Hilbert space (really density-operator space), but what can that mean?
At the least you must have a way of relating that corner of Hilbert space to what you want to describe, and that at the least means having a lot of handles on Hilbert space, i.e., what I have called attributes.  I have never understood why you can't at least acknowledge that.  Even so, if all reasonable descriptions lie in that tiny sector of Hilbert space, then most physicists would say that that tiny sector corresponds to something real.
\ecc

Strange question, coming from you at least.  I got the impression that you had really gotten the point of the big paper with {\Ruediger}.  (I'm now depressed a little bit.)  Let me attach another attempt at explanation:  This one is predominantly from {\Ruediger}'s beautifully crisp pen.  [See C. A. Fuchs and R. Schack, ``A Quantum-Bayesian Route to Quantum-State Space,'' \arxiv{0912.4252}.]

The role of a SIC in these diagrams is purely as a counterfactual (an ``action'' that could be taken on the system but isn't).  The idea is that the Born rule {\it supplements\/} the usual rules of probability with a coherence-like (or Dutch-book-style) rule for:  How probabilities should be related between a factual situation and a counterfactual one.

There is no implication here that the SIC elements should be thought of as representing potential ``elements of reality'' in the EPR sense.  (I.e., that there is one TRUE one of the $d^2$ potential ones, maybe not known, but TRUE nonetheless.)  Indeed, if they were to be thought of in that way, then one should relate the factual and the counterfactual via the usual law of total probability:  One might say that usual Dutch-book coherence requires it.

The agent is involved in bringing about measurement outcomes in this picture, just as much as he has always been for me.  The measurement ``on the ground'' is a factual {\it action\/} that the agent takes upon the system; the measurement ``in the sky'' is a counterfactual {\it action}.  No self-supporting hidden variables here:  Outcomes that come about, come about because of the agent's actions.

With regard to another of your points:  One should not think of the SICs as supreme in any way over other informationally complete measurements---instead, only as convenient.  They should be thought of as a convenient ``coordinate system for the problem at hand.''  Here's the way I put it in an abstract once:
\bq\noindent
As physicists, we have become accustomed to the idea that a theory's content is always most transparent when written in coordinate-free language.  But sometimes the choice of a good coordinate system is very useful for settling deep conceptual issues.  Think of how Eddington-Finkelstein coordinates settled the longstanding question of whether the event horizon of a Schwarzschild black hole corresponds to a real spacetime singularity or not.  Similarly we believe for an information-oriented or Bayesian approach to quantum foundations:  That one good coordinate system may (eventually!) be worth more than a hundred blue-in-the-face arguments.
\eq
And here's the way I explained a related point to a student here:
\bq
The point of all the various representations of quantum mechanics (quasi-probability reps, as well as things like Heisenberg vs.\ {\Schroedinger} picture issues and even path-integral formulations), is that they give a means for isolating or emphasizing one or another aspect of the theory---they help bring a particular aspect into plain view, even if all the representations are logically equivalent.  In our case, we want to bring into plain view (and try to make compelling) the idea that quantum mechanics is an {\it addition\/} to Bayesian probability, not a generalization of it.  With that goal in mind, the SIC representation has always struck me as particularly powerful tool.  With it, one can see the Born Rule as ``really'' a {\it function\/} of a usage of the Law of Total Probability in another context (one different than the actual).  That feature, as far as I can tell, does not leap out in the same way from the more general ``deformed probability representations'' you explore in your papers with Joseph.  That in a nutshell is the reason for my love affair with SICs.
\eq
The SICs just emphasize and make clear the point that I want made.  At the end of the day---after all our foundational worries are overcome---they can be thrown away as mere scaffolding.

Does that clarify anything for you?  Did I miss something that you think is a more important point than I've addressed?

I think I lay out the whole program particularly clearly in these two talks: \pirsa{09080018/} and \pirsa{09090087}.
Probably much better than anything I could write to you.  I'd be very flattered if you'd watch them, and even give me feedback (good and bad, though keep in mind the weight watcher's thing \ldots\ don't make fun of me).  [Sometimes Firefox doesn't work so well with these files; if you have any trouble---like not having the pictures of my transparencies, etc---change to Internet Explorer.]  Do they help clarify the worldview (that I partially ascribe to you, as you'll see)?

\section{23-10-09 \ \ {\it QBism House Draft, 2}\ \ \ (to N. Waxman)} \label{Waxman5}

\bnw
Attached is an version of QBism House, with my proposed edits. These aim to focus the piece more on the ideas of QBism House, while still paying homage to the work of physical recreation involved. I also moved the Durant quote to become a stand-alone ``sidebar'' item. One other question I have is ``authorship''---I'd think it quite logical to adjust the 3rd person references to first, and have you attributed as the author of the piece, unless you're uncomfortable with this in any way. Just let me know your preference.
\enw

My first reaction is that it doesn't sound like me anymore.  (First sentences and first paragraphs are always very powerful things for me; when I see changes there I get spooked right away.)  When do you really need the final draft?  I'd rather give you a reasoned reply, rather than an emotional reaction, and it'll probably take me a few hours to get my head in the right place.  But I'll work by your deadlines.

\section{24-10-09 \ \ {\it Dimension as Capacity}\ \ \ (to D. M. {\Appleby})} \label{Appleby74}

How are you feeling today?  I hope better, and I hope you've taken my advice to relax some.

I was feeling sentimental this morning, so I had a look back at the introduction and concluding section of this paper, \arxiv{quant-ph/0404122},
where I first start thinking out loud of dimension as a kind of capacity.  I also cryptically hint at a connection between dimensionality and gravitational concerns.  As long as you're relaxing, you might join me in on these wild musings if you wish.

Get well!

\subsection{Marcus's Reply}

\bq
I am certainly feeling better than I did yesterday, perhaps because I did take your advice to relax.  Though it had got to the point where I didn't have much choice:  I couldn't have worked even if I had wanted to.  But this afternoon I decided I felt sufficiently reinvigorated to come into PI to see if I could do some work.  Which is when I looked at my email and found your two notes.

Let me say first of all that what you said about Hilbert space dimension and the holographic principle really struck a chord in me.  I have the strong sense that there is something there.  But what exactly?  How to take it forward?  ---Here I struggle.  What follows is just a few incoherent thoughts. I am going to ramble, in other words.  Because rambling is all I can do.

I like your story about the guy who asked ``what is energy?''  It is a very good question.  It is curious, however, that most people would instinctively agree with your assessment that this is an easier question than ``what is Hilbert space dimension?''  People think they know what energy is.  At least I assume they do.  It is, at any rate, an observable sociological fact that whole conferences will spend days agonizing over Hilbert space.  But I have never heard of anyone agonizing about the meaning of energy.

I wonder why that is?  It is a priori evident that Hilbert space is a much more obscure concept than energy?

Here I am reminded of my experience as a teacher.  How do you explain what energy is to someone who has never met the concept before?  ---I used to find it very hard.  What I usually used to do was to start off with light bulbs.  Light bulbs are labeled with their wattage, so most kids have already met the idea of that a 100W bulb produces more light than a 60W one.  I used to get them to calculate the relative costs of the different light bulbs:  how many pence per hour.  Then I would go on to computers, televisions, cookers etc, and the relative costs of those (amazing how few people realize that an electric cooker is vastly more expensive to run than a light bulb).  Then I would introduce electric motors, and from that it would be a natural progression to something non-electrical such as car engines.  And from that I would introduce them  to kinetic energy.  Then I would move on to cranes, and introduce potential energy.  Then perhaps I would talk about bows and arrows, and medieval siege engines  so as to introduce elastic potential energy (the medieval allusion being desirable because I don't want the concept to be tied in their minds to modern high-tech shiny boxes).  Then maybe I would talk about Rumford's experiments with boring gun barrels.   And so on.  In short, I would adopt what in Jamesian language might be called a shrunken head approach.  That is, I would introduce them to a richly variegated mass of diverse phenomena, all articulated by this mysterious number, the energy.

I say ``mysterious''.  But is it mysterious?   What makes it seem mysterious?

I suppose because one can't give a simple one sentence definition:  energy is {\it this}.  End of story.  But then one can't give a simple one sentence definition of length.  Is length mysterious?   Well, actually, yes.  It is to me.  But clearly it doesn't seem mysterious (the sociological evidence suggests that it doesn't seem mysterious) to the average quantum foundationalist.  There isn't a discipline called ``length foundations''.  Or a discipline called ``energy foundations''. (Perhaps there should be?)

Like I told you, this is an unashamed ramble.  I am just putting down thoughts as they occur to me, in a directionless way, without any idea of trying to build a coherent argument.

I suppose energy seems mysterious (not to the average quantum foundationalist, but to the crazed guy in the elevator at the University of Texas, or to me faced with the problem of giving an articulate, intellectually respectable  account of it to a class of 16 year olds) because actually the only way I know of introducing it to someone who hasn't met the concept before is by using the kind of  shrunken head approach I outlined above.

You said that I was an empirical mathematician:  that I had a shrunken head attitude to mathematics.   And there is some truth to that.  But I am a recent convert.  I used to be an out-and-out rationalist in such matters.  I used to hate number theory, just because it is all shrunken heads, and no unifying rational theme.  My conversion happened as a direct consequence of my work on the SIC problem, which forced me to think about number theory.  On the other hand I have been teaching physics for 30 years, and for almost all of that time I was forced to say things which went clean against my rationalist conscience.  In particular I was forced to teach energy in a shrunken head spirit, not because I liked it, but because there is no other way of doing it.  At least no other way of doing it that I could discover.

I used to particularly dislike the way I found myself forced to start with electrical energy.  I felt that logically one ought to begin with the simplest case:  projectile motion and the like.  Talking about light bulbs raises the question, where the energy is coming from.  So logically one shouldn't talk to a physics class about light bulbs until one has introduced them to generators, and currents and voltages and ohm's law and so on and so on.  I used to feel it was positively immoral to treat a light bulb as an unexplained primitive.  But I had to do it because it works:  the kids understand if you start off with light bulbs because that connects more directly with the things they already know.

I suppose what I am trying to say is that my rationalist instincts were pulling me in a reductionist direction:  building the complex up out of the simple.  But we are organic beings, and we do not learn reductively.  Instead we learn by a process of assimilation.  Learning is like eating:  a process similar to  the way a tree sucks nutrients up from its roots, and works them into its already existing structure.

Anyway, as I say, I used to have a really bad conscience about the way I taught energy.  So after I had given them my shrunken heads spiel I would ask the class:  ``So what have I told you about energy?''  Can you now say what energy is?  ---And, of course, they never could.  For the very good reason that I couldn't myself.  It is exactly because it is impossible to say what energy is that I was forced to teach it in the way I did.

It didn't usually bother them I hadn't told them what energy is.  It was {\it me}---my rationalist conscience---that was bothered.  Not them.

Kids are funny, in the questions they ask and the questions they don't.  They would often complain that I hadn't explained why two bodies attract with a force proportional to the product of the two masses and inversely proportional to the square of their distance apart.  And if I responded with the rhetorical question, whether the biology teacher had explained any of the innumerable number of individual facts of which biology consists, they would think I was cheating.  I suppose they take it for granted that it is the job of a physics teacher to explain things, but not the job of a biology teacher.

Perhaps it is the influence of all those popular science programs on TV?  All this nonsense about theories of everything has penetrated the consciousness of the average 16 year old, to such an extent that they come to the subject expecting the teacher to explain literally {\it everything}, with no bald assertions of empirical fact at {\it all\/}?

If so it is interesting that they don't seem to have picked up the reductivist disease. Because, it is of course the reductivist bacillus in me that makes me (I use the present tense here because I don't I think I am completely free of the disease even now) unhappy with a shrunken head approach to the concept of energy.   But then now I come to think of it reductionism isn't the kind of thing that usually is stressed in popular science programs.  I wonder why that is?  ---Perhaps  it is just that claims to know the mind of God are sexy, in a way that reductionism isn't.

Anyway, the kids as they come into the educational system, with only the prejudices they have absorbed from the popular culture, do not, in my experience, usually have a bias towards reductionism.  In fact I am not sure they even have the concept of reductionism.  And for that reason a shrunken head approach to energy doesn't bother them.

Of course it doesn't bother them.  Because when you come right down to it I don't believe an explanation can ever be anything but a shrunken head type of explanation.  If Einstein had succeeded in his program he would have been able to say something along the lines of ``there just is this object, the stress-energy tensor, and ``energy'' just is the name we give to the 0-0 component of it''.  And it is something like that the person with reductionist intellectual tendencies  (such as I used to have, and to an extent still do have) always has in mind as the model of how a fundamental physical concept {\it ought\/} to be defined.   But they are deluding themselves of course, and this I think should be obvious to anyone who has had the job of teaching fundamental physical concepts from the ground up (the word ``should'' needs to be stressed here:  because I was doing that job for many many years before this obvious point actually did occur to me).  If you went into a class of 16 year olds and gave them something along the lines of Einstein's ideal definition you would be met with blank stares.  They wouldn't have a clue what you were talking about.  The statement would be meaningless.  And I really do mean {\it literally\/} meaningless.  It would make no impact at all.  It wouldn't even arouse a sense of puzzlement.  It would be like presenting a musical score to someone who didn't know how to read music---who didn't even have the concept that it is possible to read music.

Einstein's ideal definition makes sense to someone like the mature Einstein.  But it wouldn't make any kind of sense to the immature Einstein.  Come to that it probably wouldn't make sense even to the mature Einstein, before Grossman had taught him tensor calculus.

There is only one way for the abstract statements of theoretical physics to acquire meaning, and that is by a long-drawn out process of assimilation, by which the abstract concept acquires numerous connections with the pre-existing body of knowledge and understanding.  In short:  shrunken heads are unavoidable.  People who think otherwise (the majority of our colleagues, I guess) are deluding themselves.

(I say ``pre-existing'' body of knowledge and understanding because I think the tabula rasa idea of the 18th century empiricists is nonsense.  A child grows from a fertilized egg cell.  Not from nothing.  And similarly with a child's understanding.  And, what is more, learning really is a process of {\it growth}.  Not imprinting.  The process by which a human being learns  is of an observably different nature from the process by which one programs a computer).

Anyway, what has all this to do with the holographic principle and Hilbert space dimension?  Possibly nothing.  Like I say I feel confused and all I aimed to do was to spout forth, and see what my fingers typed.  But now that they have finished typing maybe I can see a connection.

For one thing I suddenly felt much more interested in your emphasis on Hilbert space dimension when you made a connection with entropy (I know you have been doing that for years, but this was the first time I really took it in).  It is like the kids and the light bulbs:  here is a connection with something I already know.

For another thing.  There was a suggestion to my mind that you are introducing length alongside Hilbert space dimension as a kind of conceptual primitive (perhaps I have got you wrong?).  Which is fine.  I don't mind conceptual primitives at all.  The only thing is that, as soon as I think of length, I find myself thinking of time, and as soon as I think of length and time together, I think of differentiable 4 manifolds and there I am, in two blinks of an eye, with the whole block universe nightmare.  So length is OK.  But I want something to hang on to, to keep myself from sliding down all the way down to the bottom of that horrible slope.  So I guess I have had at the back of my mind the last few days the thought that, although it is OK to talk about length (of course it is OK), we have somehow got to find a different twist on it.  A new angle, which doesn't descend into differentiable 4 manifolds.  And that has kind of got me to asking:  how do we {\it define\/} length?  What is length?   ---which, of course, is similar to the question ``what is energy?''  So you can regard the foregoing as a meditation on the question:  what sort of definition should we be looking for here?  How do we set about defining a fundamental physical concept?

And, finally, for a third thing.  I think one of my worries about (to quote from your note to Greg Comer) ``just the single, lonely number---the dimension'' has always been precisely the fact that it is lonely.  I think what I would like is something like the shrunken head definition of energy.  A richly textured, multi-faceted phenomenological account.

Maybe.  Or maybe not.

I don't actually agree with James that it is {\it either\/} rationalism {\it or\/} shrunken heads.  Can't one contrive to have both?
\eq

\section{24-10-09 \ \ {\it Dimension as Capacity, 2}\ \ \ (to D. M. {\Appleby})} \label{Appleby75}

\bma
I don't actually agree with James that it is either rationalism or
shrunken heads.  Can't one contrive to have both?
\ema

I think he does contrive to have both.  A simple piece of evidence is in the longer quote I had attached for Lucien.  I think you said you hadn't read it.  Here are a couple of small excerpts:
\bq\noindent
A man's philosophic attitude is determined by the balance in him of these two cravings [rationalism and empiricism].  No system of philosophy can hope to be universally accepted among men which grossly violates either need, or entirely subordinates the one to the other.
\eq
\bq\noindent
Each additional particularity makes its distinct appeal. A single explanation of a fact only explains it from a single point of view. The entire fact is not accounted for until each and all of its characters have been classed with their likes elsewhere. To apply this now to the case of the universe, we see that the explanation of the world by molecular movements explains it only so far as it actually is such movements. To invoke the ``Unknowable'' explains only so much as is unknowable, ``Thought'' only so much as is thought, ``God'' only so much as is God.
\eq
In summary:  There is nothing wrong with invoking some rationalism.  But it explains only so much as is rational.

I'm glad you're starting to feel better.  I was actually starting to get a bit worried since I hadn't heard from you in so long.

\section{24-10-09 \ \ {\it Detached Observer \ldots}\ \ \ (to H. C. von Baeyer)} \label{Baeyer83}

I'm not sure what you're asking.  Can you make your question more specific?

The first part of the Pauli quote refers to what the philosopher Thomas Nagel famously called ``the view from nowhere''.  Google his book of the same title.

With regard to the second part, it's funny but I found a bit of similarity between it and a quote I read in a James biography I just started reading.  (It seems every two years or so, I feel the need to read a new biography of him.)  The quote was this:  ``Individuality is founded in feeling and the recesses of feeling, the darker, blinder strata of character, are the only places in the world in which we catch real fact in the making, and directly perceive how events happen, and how work is actually done.''

Does he explain why he finds the NO-view-from-nowhere view ``more satisfactory''?  That would be interesting to see.

I'm happy to hear you've been dragged back to the subject!!

\subsection{Hans's Preply}

Dear Chris, your persistent encouragement has had the effect of dragging me back to the subject.  The price you have to pay is that occasionally I will ask for advice.  Here is a snippet from a letter to Bohm --- almost the last words before Pauli broke off the correspondence because he thought Bohm was a ``fool'' (his term in a letter to Fierz):
\bq\noindent
Since Descartes it was the ideal of natural philosophy to conceive a system of laws in which an entirely loose and untied observer is looking from outside at a part of the world completely determined by these laws.  For me, however, it is much more satisfactory if the laws of nature themselves exclude in principle the possibility even to conceive the disturbances in the observers (sic) own body and own brain connected with his own observations.
\eq

This was written in English, so if it's difficult to understand the fault lies with Pauli's command of the language, not with the translator.

Without going into the whole background of this statement, what do you make of it?

\subsection{Hans's Reply}

\bq
Bohm says in the P.S. to his letter of 20 Nov.\ 1951:
\bq\noindent
Finally I shall say that even though we cannot at present correct for the disturbance due to the measuring apparatus, we can conceive of how the human being disturbs the apparatus in precise terms.  It is true that such a conception presupposes that the person who conceives of this is left outside the system.  \ldots\ This means that as in classical physics one can conceive of the part of the world under consideration as existing in a state that is substantially independent of what is going on in the mind of the man who is conceiving of it.
\eq
To which Pauli replies on 3 December as I wrote you previously.

I understand Bohm's claim, and Pauli's first sentence.  But in his second sentence, why does he focus on a disturbance {\it in the observer}?  Why doesn't he say that the laws of nature should exclude conceiving of the state of the system without including the effect of the observation on {\it the system\/} as well as {\it on the observer}?
\eq

\section{25-10-09 \ \ {\it Metaphysical Club}\ \ \ (to D. M. {\Appleby})} \label{Appleby76}

\bma
Is it true to say that only one of those guys (Peirce) had any appreciation of physics?  It seems clear to me that neither James nor Holmes did.  And I think that is significant.  If you don't know physics then you are missing something important.
\ema
I think that's probably true.  James's real thing was biology/physiology, before psychology.  James's father though had much wanted William to go into the sciences.  And he at least seemed to have some amount of respect for physics:  He once went to England to meet with Faraday, and he was taught by Joseph Henry in Princeton.  (You probably don't recall, but John Wheeler was the ``Joseph Henry Professor of Physics'' at Princeton.)

\section{25-10-09 \ \ {\it Another Factoid}\ \ \ (to D. M. {\Appleby})} \label{Appleby77}

James met with Mach in Prague for four hours of conversation in 1882.  He also went one of Mach's lectures on gravity.  He wrote to his wife, ``I don't think anyone ever gave me so strong an impression of pure intellectual genius. \ldots\ He apparently has read everything and thought about everything, and has an absolute simplicity of manner and winningness of smile when his face lights up that are charming.''

Sometime later that year or the next, James went to a lecture of Helmholtz on gravity and wrote that he heard Helmholtz, ``give the most idiotic lecture I ever listened to.''

\section{25-10-09 \ \ {\it More Factoids}\ \ \ (to D. M. {\Appleby})} \label{Appleby78}

Of course,  James was in no way a physicist like Peirce truly was.  But more factoids from Richardson's biography of James:\medskip

\noindent p.\ 51,
\bq
On the opening page of a notebook he started this fall of 1862, William copied down comments that show his new interest in the nature of force.  William Grove's {\sl Correlation of Forces\/} maintains \ldots\  William had also been reading Michael Faraday's {\sl Experimental Researches in Chemistry and Physics}. \ldots\ [In his notebook] he quoted Faraday: ``For my own part, many considerations urge my mind towards the idea of a cause of gravity which is not resident in the particles of matter merely, but conjointly in them and in all space.''  James, from now on, had a marked interest in physics, especially in questions of energy and force, an interest he encapsulated some time later as ``matter is motion, motion is force, force is will.''
\eq

\noindent p.\ 165,
\bq
But it was science, especially new science and its procedures and assumptions, that now commanded his deepest, steadiest attention.  He maintained his interest in Darwin and Darwinism; in 1875 he was reading in modern physics as well. \ldots\ In May 1875 James read and reviewed a book called {\sl The Unseen Universe} by physicist and mathematician Peter Guthrie Tait \ldots\ and physicist and meteoriologist Balfour Stewart [a book on Maxwell, ether, etc].
\eq

\noindent [{\bf NOTE:} The spelling of ``Qubism'' in the title of this note differs markedly from my standard QBism.  I am ashamed of this (what surely was a) mistake, but I leave it for historical accuracy.]\medskip

Sorry for my tantrum the other day.  Scientists with ambitions of writing are like little kids.

Attached is a modified version of my story.  I used some of your suggestions, but in a way that made me feel like the whole thing was still coming from my voice.  Overall, I was shooting for a kind of artistic effect, rather than journalistic effect.  I hope that helps explain some of my extraneous phrases, like ``all told,'' etc.  Still, I was able to trim to my satisfaction, and I think to yours.  Your version from late last week had 860 words (after all the instructions, etc., were stripped from it).  The present version has 858, under the same conditions.

I think I would like to keep the article in third person, but signatureless.  For some reason it feels more right to me that way.

Picture coming soon.

Thanks for putting up with me!

\bq
\begin{center}
\Large QBism House Opens for Business\medskip
\end{center}

Ideas, like children, need homes where they can be loved, nurtured, and raised to independence.  PI researcher Chris Fuchs brought an idea to Waterloo that he feels finally has a proper home.  The home is called QBism House, and the idea, if it can be put in a few words, is that a consistent foundation for quantum mechanics reveals our universe as no universe at all, but a {\it pluriverse\/} instead.  It is the sort of world the philosopher William James described as a ``republican banquet \ldots\ where all the qualities of being respect one another's personal sacredness, yet sit at the common table of space and time,''  a world of ``partially independent powers [where] each detail must come and be actually given, before, in any special sense, it can be said to be determined at all.''

Take an 1886 home with 10 and 12 foot ceilings, the willpower of Kiki Fuchs to restore it to its historical grandeur, a good carpenter to build a library of solid-oak bookshelves to house nearly 1,000 books on this subject, a covered porch and a chalkboard to discuss these matters at length in the warm summer air, and one has the base for some fantastic, new physics.  The house is called QBism House in honor of its ``quantum-Bayesian'' heritage---a point of view about quantum mechanics developed by Fuchs, along with Carlton Caves, R\"udiger Schack, Marcus Appleby, and Howard Barnum.

In the QBist view, a quantum state is not a real thing, like a rock or a tree or a quark, but ``a means of orientation in life, of enabling and facilitating action, of taking account of reality and dominating it'' (Tilgher).  Without doubt, this is a soft-edged idea, but it is one with a sharply defined mathematical core, and the latter is QBism's research program.

Yet sharp cores and soft edges complement each other, much like the sharp lines of Perimeter Institute and the curlicued Victorian corbels of QBism House. Making the home young again, while preserving its memories and wisdom, has been a labor of love for Kiki Fuchs.  The walls had to be opened to replace rusting pipes and early-century electrical wiring; chimneys had to be reconstructed from scratch; floors refinished and the kitchen modernized; the dull, crumbling paint on the exterior refreshed to full-of-life Victorian colors---the work went on for more than two years.  The home that emerged is now a Waterloo landmark.

Bolstered by a grant of \$512,000 from the United States Office of Naval Research, Fuchs is building a group to plow ahead on the technical aspects of the QBist project, particularly, developing a probabilistic representation of quantum states through so-called symmetric informationally complete (SIC) quantum measurements.  At the moment, the group is busy exploring whether these structures exist in all finite-dimensional Hilbert spaces, what the geometry of quantum-state space looks like in such terms, how quantum mechanics itself might be derived from the supposition of SICs, and characterizing the convex theories (generalizations of quantum mechanics) this representation most naturally fits into.

All told, the QBies (yes, the QBies) consist of PI visiting researcher Marcus Appleby, associate postdoctoral fellow {\AA}sa Ericsson, University of Waterloo PhD students Hoan Dang and Gelo Tabia, and MSc student Matthew Graydon.  The students' desks can be found in room 415 at PI: the QBicle, of course.  This Fall and Winter, the group will meet once a week at PI, Fridays 1:00--3:00.  But once the weather is warm, the meeting will be on the porch of QBism House, where the ghosts of pragmatist philosophers past (William James, John Dewey, and F.~C.~S. Schiller) can listen in with satisfaction about how their ideas of an unfinished, malleable world take a more exact form with the help of modern quantum mechanics.

{\bf Anyone at PI is welcome to join the meetings, research, and discussion; contact Chris Fuchs at cfuchs@perimeterinstitute.ca.  The QBies and QBism House are open for business!}

\subsubsection{Boxed Sidebar:  The Optimism of Pluralism and Indeterminacy}

Will Durant, from his masterful {\sl The Story of Philosophy\/} (1926):
\bq
\noindent The value of a [pluriverse] as compared with a universe, lies in this, that where there are cross-currents and warring forces our own strength and will may count and help decide the issue; it is a world where nothing is irrevocably settled, and all action matters.  A monistic world is for us a dead world; in such a universe we carry out, willy-nilly, the parts assigned to us by an omnipotent deity or a primeval nebula; and not all our tears can wipe out one word of the eternal script.  In a finished universe individuality is a delusion; ``in reality,'' the monist assures us, we are all bits of one mosaic substance.  But in an unfinished world we can write some lines of the parts we play, and our choices mould in some measure the future in which we have to live.  In such a world we can be free; it is a world of chance, and not of fate; everything is ``not quite''; and what we are or do may alter everything.
\eq

\subsubsection{Potential Boxed Sidebar: A Pretty Equation}

\Large
$$
q(D_j)=(d+1)\sum_i {p(H_i)p(D_j|H_i)} - 1
$$
\medskip
\normalsize

\noindent This equation is something of a logo for the QBists:  It represents the Born Rule when written in SIC terms, rather than in standard quantum-state/measurement-operator terms.  $q(D_j)$ is the usual quantum probability for the outcomes of a von Neumann measurement; $p(H_i)$ is the probability for the outcomes of a standardized SIC measurement; $p(D_j|H_i)$ are the conditional probabilities of one measurement outcome given the other; $d$ represents the Hilbert-space dimension of the system in question.  To Bayesian eyes, this form is particularly beautiful as it shows the Born Rule to be a subtle variation (with major consequences!)\ of the old Law of Total Probability.
\eq

\section{27-10-09 \ \ {\it Partially Sipe Inspired} \ \ (to J. E. {\Sipe})} \label{Sipe19}

\bjes
Thanks for this! Very interesting \ldots.  {\rm [See 16-10-09 note ``\myref{Hardy38}{The More and the Modest}'' to L. Hardy.]}

To perhaps start a discussion on the points you raise here: I have not read anything by Cartwright yet, but how do I distinguish between the kind of ``capacity'' you are arguing for and the kind of Aristotelian idea of ``potentia'' that Heisenberg advocated in his later years, and saw embodied in the wave function. Are you, by chance, sleep-walking towards Copenhagen? {\rm \smiley}
\ejes

Maybe I shouldn't have made the aside in that note to Lucien about Cartwright.  Her name apparently carries lots of baggage (she seems to be the first thing everyone with a little philosophical background, like Healey, Timpson, etc., picks up on).  Just think of it along the lines of gravitational mass as you normally understand it, as I tried to give an image of in the note.

\section{28-10-09 \ \ {\it Gauge Freedom as Room for Art}\ \ \ (to R. Healey)} \label{Healey4}

Sorry, my mind wasn't completely engaged in our group meeting yesterday:  I couldn't seem to tear my mind away from an inequality I wanted to derive.  But I did want to make a remark right at the end of the session:  As usual I was too late.  So here it is just to get it off my chest.

It struck me that the discussion of whether the ``equivalence class'' was the real thing or not was off the mark with respect to actual physical practice (i.e., actual physicists' practice).  When {\it trying\/} very hard to solve one problem or another---i.e., not just pontificating on the nature of reality or the character of the world, but actually using the theory to do something---we are often very happy for gauge and coordinate freedoms.  These freedoms give the practicing physicist some room for art---they help him solve problems that he might not have been able to solve otherwise.  In other words, a theory, among perhaps other things, gives us a catalog of tools for our disposal---tools for actually doing something with the theory.  Thinking of it that way, then, for instance, the spacetime manifold shouldn't be thought of as a topological space upon which, {\it if we wish\/} we can draw some coordinate charts, but rather is the other way around:  It is more literally the collection of charts we might avail ourselves of when trying to work out some problem.

This thought has its origin in the opening lines of an abstract I wrote for a talk some time ago.  I'll record part of it here for completeness.
\bq
As physicists, we have become accustomed to the idea that a theory's content is always most transparent when written in coordinate-free language.  But sometimes the choice of a good coordinate system is very useful for settling deep conceptual issues.  Think of how Eddington-Finkelstein coordinates settled the longstanding question of whether the event horizon of a Schwarzschild black hole corresponds to a real spacetime singularity or not.  Similarly we believe for an information-oriented or Bayesian approach to quantum foundations:  That one good coordinate system may (eventually!) be worth more than a hundred blue-in-the-face arguments.
\eq
John Stachel makes a similar point (about the manifold issue) in one of his historical articles on Einstein, but I don't have it with me at home to quote from it.  If anyone wants to see it, I have it in my office.

\section{30-10-09 \ \ {\it Abelard Etc.}\ \ \ (to J. R. Brown)} \label{BrownJR3}

Abelard doesn't exist anymore, just like Atticus.  We went to BMV near the university, and one other one nearby, and then to Steven Temple Books on Queen.  The latter guy said he was the last used-book store left on Queen.  This is really a shame.

Nonetheless I managed to find 13 books that resonate somewhat with where I want to take things in physics.  Attached is a strange little piece on my house, my research group, and my ambitions for the world, written for the PI newsletter.  In it you'll find that Durant quote I put up at the very end of the meeting:
\bq
The value of a [pluriverse], as compared with a universe, lies in this, that where there are cross-currents and warring forces our own strength and will may count and help decide the issue; it is a world where nothing is irrevocably settled, and all action matters. A monistic world is for us a dead world; in such a universe we carry out, willy-nilly, the parts assigned to us by an omnipotent deity or a primeval nebula; and not all our tears can wipe out one word of the eternal script. In a finished universe individuality is a delusion; ``in reality,'' the monist assures us, we are all bits of one mosaic substance. But in an unfinished world we can write some lines of the parts we play, and our choices mould in some measure the future in which we have to live. In such a world we can be free; it is a world of chance, and not of fate; everything is ``not quite''; and what we are or do may alter everything.
\eq

Thanks for the radio programs; I'll listen.  And thanks for the opportunity for good interaction yesterday.

\section{30-10-09 \ \ {\it My Interiority Complex} \ \ (to H. Barnum)} \label{Barnum26}

Here's that quote of William James on matter that really moved me:
\bq\noindent
To anyone who has ever looked on the face of a dead child or parent the mere fact that matter {\it could\/} have taken for a time that precious form, ought to make matter sacred ever after. It makes no difference what the principle of life may be, material or immaterial, matter at any rate co-operates, lends itself to all life's purposes. That beloved incarnation was among matter's possibilities.
\eq

Attached are three further quotes by the man that connect to our conversation last night:  The theme is that indeterminism is a manifestation of (some kind of notion of) interiority.  I say further things on this in connection to QM in Section 8 of \arxiv{0906.2187}.

Thanks; I had fun last night.

\subsection{Howard's Reply}

\bq
Thanks.  That is indeed a powerful quote.  It captures something of what I think about the mind-ishness, or perhaps spirit, of matter.  I'm not sure, that it is in itself opposed to some kind of compatibilism, or ephiphenomenalism, about mind.  In my more happy-with-compatibilism-and-epiphenomenalism/reductionism moods, I used to say to myself that ``I {\it am\/} that assemblage of matter, so even if it is the evolution of that matter according to deterministic laws that is my doing this, or experiencing that, it is still {\it me\/} doing it''.  I guess I still half-believe that.  But still, on that picture, I find it hard to understand why there should be something it is like to be that assemblage of matter.  And I think that there is probably something deeply wrong about confusing the picture drawn by the equations, with the thing.  The deeper question, though, is whether consciousness and will --- interiority --- are essentially connected with that gap between the picture drawn by the equations and the thing.  I'd like to think it is so, indeed I think it probably is, and that that has something to do with it.

I used to also think that quantum indeterminism, or other probabilistic indeterminism involving probabilities prescribed by a scientific theory, could not help with e.g.\ the problem of free will, because a free choice is not supposed to be random.  It's not an idealized computer-science coin toss.  Although the idea that it's more like an idealized classical-physics coin toss---determined, but from the point of view of the theory that involves the probabilities, we just don't know what determines it---is interesting!

I now understand your quantum Bayesian point of view in its present incarnation as perhaps taking something like the point of view that last sentence suggests---that {\it from the outside point of view\/} it looks like probabilities, but {\it from the inside point of view\/} it looks like choice.  That seems quite explicit in what you said last night about ``every little thing having some ability to make choices'' (apologies if the quote is garbled).

A long time ago, I used to worry that you rejected potentially realist interpretations of the quantum state just on the basis that you wanted room for free will.  The idea that quantum probabilities were making room for free will seemed to me a bit dubious because of the view that choices shouldn't be random (see above), but I now think I may have been misguided about that.  I also thought that the viability, even if unpleasant, of the relative-state interpretation made it clear that QM wasn't {\it forcing\/} us to make room for free will.  Now, as you know, I'm more inclined to think relative states is not viable, and some of the above discussion makes me more willing to think that QM might be telling us that the limits to our attempts to take the Nagelian ``outside'' view are not just manifest in our sense that this is somehow incompatible with the fact that there is---{\it really is\/} an inside view---but are manifest in the scientific theory that actually results from our attempts to take this outside view:  that theory itself has to make room for interiority.  I would really like that to be right.  A more fully developed version of this might end up involving a ``sewing together of all the perspectives'' of the sort I hoped for in my long SHPMP paper.  But there I probably wasn't thinking of such a deep connection to metaphysical questions as your view of perspectivity in QM.   I just wanted some sense of an overall picture that includes all the perspectives in a consistent way, but in which the ``outside'' perspective is just one of all these perspectives somehow coexisting, not a possibly nonsensical perspective similar to the ones of beings within the world but ``from outside the world''.  I am not sure you are as concerned with the kind of overall consistency I would still sort of like, though.
\eq

\section{31-10-09 \ \ {\it Dennerlein, Sacred Matter, Capacities}  \ (to M. Schlosshauer)} \label{Schlosshauer7}

I ran across someone playing the Hammond B3 in my (still largely unexplored) jazz mp3 collection -- Barbara Dennerlein -- and I thought of you.  What I'm hearing sounds like a very different style than you, but it was enough to set the mental wheels in motion on this thoughtful, Fall-leaved morning.  I've been up since 7:00 AM, reading on (yet) another biography of James---I seem to do a new one about every two years, following near upon a birthday.  This one is Richardson's.  Not remarkable so far, but it is helping to remind me of some old things and, here and there, introducing me to a turn of phrase that I might incorporate into my collection.  For instance, here's a quote of James that I missed upon my (now ancient) reading of {\sl Pragmatism}, but that very much strikes me now:
\bq\noindent
To anyone who has ever looked on the face of a dead child or parent the mere fact that matter {\it could\/} have taken for a time that precious form, ought to make matter sacred ever after. It makes no difference what the principle of life may be, material or immaterial, matter at any rate co-operates, lends itself to all life's purposes. That beloved incarnation was among matter's possibilities.
\eq

I hope you are doing well and absorbing all the insights into life that only Copenhagen can give.

Below (with an accompanying attachment) is a recent further amble down the QBist path.  [See 16-10-09 note ``\myref{Hardy38}{The More and the Modest}'' to L. Hardy.]

Just saying hello.

\section{31-10-09 \ \ {\it Thinking of Our Friend \ldots}\ \ \ (to R. W. {\Spekkens})} \label{Spekkens76}

\ldots\ as I read these words in Richardson's book this morning:
\bq
Every university has around it somewhere one person who is more loved and listened to than most, who tutors and prepares his friends for their exams but who never takes his own, who is generally acknowledged as the most brilliant and gifted of all, who has every gift except the ability to use his gifts.  He is the one who never quite finishes, never quite succeeds, never quite writes the great book --- a person like Sherlock Holmes's older brother Mycroft (who solves Sherlock's thorniest problems by sheer force of mind without stirring from his chair), or like a member of Melville's choice circle of the Divine Inert who stands wholly, tragically superior to the working world and the arts the working world requires for success and fame.  In the American Cambridge of the 1860s and 1870s this figure, the village Socrates who lived in the minds of his great students was Chauncey Wright.
\eq

\section{02-11-09 \ \ {\it Furthermore}\ \ \ (to C. Ferrie)} \label{Ferrie9}

\bcf
Sciencewise, I've been thinking about a lot of things trying to find a
thesis topic.  Over the past few days I've been thinking about this
quote from Jeffreys (from ``Epistemology in Modern Physics'', 1941):
\bq\noindent
{\rm But in accounts of quantum theory we are presented on the first page with
at least $\sqrt{-1}$, and usually with matrices or numbers not satisfying the
commutative rules of multiplication. These cannot be necessary. They may
be convenient, but if so it should be shown that rules satisfied by the
probabilities are such as to make them convenient.}
\eq
\ecf

I certainly agree with this.  And as you know, I contend that the SICs will help us best in seeing why $\mbox{GL}(d,\mathbb{C})$ gives the simplest form for the calculus.

Do you have that paper in electronic form?  Can I get a copy from you?  (I couldn't find it on Google Scholar.)

I have not thought about the Markov stuff before.  So, sorry I can't give any guidance here.

By the way, see the flyer I wrote for this month's PI Newsletter.  You're welcome to come any of these meetings you wish.

\section{02-11-09 \ \ {\it One Good Rant Deserves Another}  \ (to M. Schlosshauer)} \label{Schlosshauer8}

Thanks for the further Hammond pointers.  It'll be fun looking them up a little later in the week.

\bmaxs
At a recent workshop in Utrecht I heard a talk by Carl Hoefer, who you may know? He quite vigorously defended a form of objective Bayesianism, which seemed motivated by his view that subjective probabilities are insufficient. Here is how he put it in a recent essay:
\bq\noindent{\rm
Now, I have nothing to say against the idealization of human agents as Bayesian subjective probability-bearing entities; I am sure it is useful for many purposes, and reasonably faithful to reality in a few. But the ontic weaknesses of subjective probabilities, I would say, make them an unsuitable bedrock on which to try to mount a program that aims to capture {\it all\/} probabilities. \ldots\ The exquisite precise and reliable probabilities found in casinos and in physics laboratories deserve to be accounted for in {\it some\/} way or the other.}
\eq
After reading de Finetti and some other texts, I became quite convinced that the subjective viewpoint is the only reasonable one to opt for. I wonder what your general take would be on a program like Hoefer's?
\emaxs

You asked some good questions.  ``How you would try to respond to those rationalist, hard-nosed scientists?''  In the coming months, I plan to work on this, honing out a satisfactory response.  One thing I intend to do for practice is read some things by (and about) Nancy Cartwright, just to get straight where we are the same, and where we differ, and study the shape of the tomatoes that have been thrown at her.  The Carl Hoefer response is easier; I'd just need some time to compose it properly.  The key to the story is this thing {\Ruediger} wrote me after reading Logue's book {\sl Projective Probability}:
\bq\noindent
    Logue makes this interesting point. Often coherence is
    regarded as a minimalist constraint on probabilities.
    But coherence has incredibly far-reaching consequences.
    Coherence alone guarantees that two agents with
    exchangeable priors that are nonzero everywhere will
    converge to a joint belief. Coherence alone leads an
    agent to certainty that he is calibrated in the long
    run. I like this. I like the idea that Dawid's
    calibration result is a strength of subjectivism, not
    a problem for it.
\eq
The key point is that if one is coherent with as many of one's beliefs as one can, that leads to all kinds of constraints on one's isolated assignments.  So many constraints as to start misleading one into thinking the end results must be objective.  But as I say it'd take me some time to do this job right.

Finally, for that awaited rant.  In the week before the PIAF meeting, I tried to do a really honest job of reading many of Travis Norsen's papers.  {\Spekkens}, you see, thinks Norsen is the clearest thinker since Socrates, and I thought it was my duty to try to understand this clarity \ldots\ so that I might find the small chink in the armor, if there be one.  I didn't realize what I was in for:  The end result wasn't enlightenment, just depression.  A really very serious depression.

Check out \pirsa{09090098}, the panel discussion that Norsen and I both participated in.  One of the questions was, what would it take for you two to switch sides.  I laid out a set of criteria for what would turn me Bohmian, and I was honest.  When it came to Norsen's turn, he said, ``Nothing!''  Not surprising at all now that I know him.

\section{03-11-09 \ \ {\it Noncolorable, Maximally Consistent Sets}\ \ \ (to A. Cabello)} \label{Cabello5}

I almost forgot to send you electronic versions of the papers we discussed yesterday.  The two main ones on the {\tt arXiv} are \arxiv{0910.2750} (this one I gave you) and \arxiv{0906.2187} (this is the one with the most meat and motivation in it, for this whole way of looking at quantum mechanics).

Also, I gave you a paper that I have not yet posted.  I will attach it to the present note.

Thanks for spending a little time with me yesterday.  There are insights my QBist program needs that I think you are nearly the only person on earth to be consulted on!  I look forward to any insights you might have.  Part of the question is what is the best question to make progress on!

(Attached also, by the way, is a little blurb written for the PI newsletter.  It reveals in a little more poetic way, where I think ``QBism'' is heading at the ``metaphysical'' level.  See particularly the side bar near the end.)

\section{04-11-09 \ \ {\it QK, QB, QR -- First Read} \ \ (to H. Barnum)} \label{Barnum27}

I gave your ``quantum knowledge'' draft a {\it first\/} read through yesterday.  (I'll be reading it again before saying anything too substantial.)  At the moment, let me just quickly comment (in my usual time-saving way) on two points in the draft.

With regard to the first point, I'm not quite sure which piece of your text I'd like to pluck out for display with regard to what I'm responding to, but maybe this one will do well enough:
\bq\noindent
I worry that they are underselling the extent to which we can have both all the objectivity anyone could reasonably want for at least some quantum state ascriptions---a degree of objectivity which, moreover, it seems to me would be unreasonable not to grant in some cases---while also not reifying the quantum state as an object out in the world, and especially, not as an object localized at the system whose state it is.
\eq
Please go to
\myurl{http://www.perimeterinstitute.ca/personal/cfuchs/nSamizdat-2.pdf} and find the 08-11-06 note to David {\Mermin} titled ``\myref{Mermin124}{November 8th}'' on page 655 (page numbering at the bottom of each page).  With regard to overall pdf page counting, that is page 683 in the file.  The point put there is a mundane one:  {\it Classical\/} thermodynamical heat tables might be described in your terms as ``having all the objectivity anyone could reasonably want'' (publishing houses, for instance, found it profitable to publish them because of their seemingly universal use), but in what way did physics {\it not\/} benefit by Jaynes's constant reminder that all these numbers have their origin in {\it subjective\/} probability assignments?  These same tables would be of no use and no universality for a Maxwellian demon for instance.  And I say it is good to recognize that.

The danger, it seems to me, is not in overemphasizing the subjectivity of probability assignments (and then backing off in those cases when agents, for whatever reason, have similar-enough priors as to potentially come into agreement later or even to be in agreement presently), but rather in {\it forgetting\/} their subjective origin.  The usual physicist's knee-jerk reaction is to do just that, to reify something that shouldn't be \ldots\ to the distraction of all the conundrums that mistake leads them to.  I view the great lesson of Ed Jaynes' work to be this:  Remind always of the principle (that probabilities are subjective), rather than of the convenience (that in many cases agents agree and then sloppily get in the habit of thinking a probability to be of the same logical category as a fact).

Read the piece to {\Mermin}:  It is better.

Now on to this one:
\bq\noindent
I agree that {\it if one does\/} give a quantum description of the measuring device, the operation it performs will be dependent upon one's assignment of a quantum state to it---one's beliefs about it.  But who says a quantum analysis of the apparatus is needed?  From a point of view from which what we take to be quantum depends upon our purposes and the experiments we plan to do or other projects we plan to undertake, to suddenly require that we do a quantum analysis of the apparatus seems to put us in danger of becoming unwitting victims of the cult of the larger Hilbert space.
\eq
Please see page 694 in the file (page 722 by raw pdf accounting), picking up at the line starting with, ``The logic goes like this.''  Please read that paragraph along with the referee's comments above it that I am addressing.  It is true that {\Carl} often seems to skirt quite close to being an ``unwitting victim of the cult of the larger Hilbert space,'' but I do not.  [See ``\myref{RepliesToReferee4}{Replies to Referee 4}'' in the 07-12-06 note ``\myref{Caves93}{(Select) Replies to Referees}'' to R. Schack and C. M. Caves.]

That circuit argument is Carl's favorite one, but it is not mine.  Mine is simply this:  With respect to a fiducial informationally complete measurement, {\it any\/} other measurement can be represented as a collection of conditional probability distributions.  Think, for instance, of my beloved urgleichung.  If all probability assignments are subjective (in the sense of being functions solely of the agent rather than the object), then so are these conditional ones, and then so is the POVM associated with a measuring device.

I'll certainly read your draft again in a day or two (and I'll certainly have more to say).

\section{04-11-09 \ \ {\it QK, QB, QR -- First Read, 2} \ \ (to H. Barnum, D. M. {\Appleby}, and M. Tait)} \label{Barnum28} \label{Appleby79} \label{Tait5}

Rereading this note, and also some others of my old writings, let me also recommend the note ``Prepare Yourself'' on page 711 (page 739 by raw pdf accounting).  It further expands on the subjectivity-because-of-input-state-in-quantum-circuit account of the subjectivity of POVM ascription, and it also says some things on {\Mermin}'s ``disembodied-fact account of quantum states'' which seems to have some similarity to what Howard is striving for.  [See 28-12-06 note ``\myref{Mermin126}{Prepare Yourself}'' to N. D. Mermin.]

\section{06-11-09 \ \ {\it The Interiority of John Preskill} \ \ (to H. Barnum)} \label{Barnum29}

I had forgotten that I had tried to make a statement on the interiority business in my John Preskill story.  ``If he knew what I believed, he'd probably do something to surprise me.'' And, ``In a phrase, what is desired is a theory that is all outside and no inside. [Footnote: On first pass, the Texan in me wanted to write `all hat and no cattle,' but I repressed that.]''  Have a look at the Preskill section, pages 26--28, of the attached:
\bq
When we have discussed the interpretational issues of quantum theory, I have gotten the sense that what John finds most suspicious of the quantum Bayesian approach I promote is that he {\it thinks\/} it treats observers as unphysical systems.  I say, ``Not at all; John, you are a physical system to me.''  The real issue is one of inside versus outside.  Contrary to the textbook exposition of quantum mechanics, a wave function that I write down about him is descriptive not of the outside, but of the inside, namely me:  It captures what I believe will happen to me if I interact with him.  If he knew what I believed, he'd probably do something to surprise me.
\eq

\section{06-11-09 \ \ {\it Renouvier on Kant}\ \ \ (to M. Tait)} \label{Tait6}

See entry 291 (pages 102--106) of ``The Activating Observer'' document I sent you.  See also
\begin{center}
 \myurl{http://en.wikipedia.org/wiki/Charles_Bernard_Renouvier}.
\end{center}
And finally the attached files if you're interested.  [See Shadworth H. Hodgson's ``M. Renouvier's Philosophy--Logic,'' {\sl Mind\/} {\bf 6}(21), 31--61 (1881) and ``M. Renouvier's Philosophy--Psychology,'' {\sl Mind\/} {\bf 6}(22), 173--211 (1881).] The one on Logic is particularly on Kantian things.

\section{06-11-09 \ \ {\it Quotable Schopenhauer}\ \ \ (to M. Tait)} \label{Tait7}

It was this passage of James from an 1870 diary entry that struck me as sounding a bit Schopenhauer-esque when you brought it to my attention:
\bq\noindent
Hitherto, when I have felt like taking a free initiative, like daring to act originally, without carefully waiting for contemplation of the external world to determine all for me, suicide seemed the most manly form to put my daring into; now, I will go a step further with my will, not only act with it, but believe as well; believe in my individual reality and creative power.
\eq

\subsection{Morgan's Preply, ``Quotable Schopenhauer''}

\bq
\noindent Anticipating James, though (perhaps) with a less cynical moral:
\bq\noindent
The discovery of truth is prevented more effectively, not by the false appearance things present and which mislead into error, not directly by weakness of the reasoning powers, but by preconceived opinion, by prejudice.
\eq

\noindent Take heart QBists:
\bq\noindent
All truth passes through three stages. First, it is ridiculed. Second, it is violently opposed. Third, it is accepted as being self-evident.
\eq

\noindent On Hegel:
\bq\noindent
\ldots\ a commonplace, inane, loathsome, repulsive and ignorant charlatan, who with unparalleled effrontery compiled a system of crazy nonsense that was trumpeted abroad as immortal wisdom by his mercenary followers \ldots
\eq

\noindent On Kant's early reception:
\bq\noindent
Great minds are related to the brief span of time during which they live as great buildings are to a little square in which they stand: you cannot see them in all their magnitude because you are standing too close to them.
\eq

\noindent On Suicide:
\bq\noindent
They tell us that suicide is the greatest piece of cowardice \ldots\ that suicide is wrong; when it is quite obvious that there is nothing in the world to which every man has a more unassailable title than to his own life and person.
\eq
\eq

\section{09-11-09 \ \ {\it Vienna Indeterminism}\ \ \ (to A. Zeilinger)} \label{Zeilinger4}

Thank you for the very nice letter---nice in so many aspects.  A week for a visit to Vienna is definitely doable; a month sounds very intriguing.  What is the most beautiful period of Vienna's springtime?  But maybe more practically, when would be the best time for the meeting you want to organize?  Or more practical still, if you decide you have an interest in doing a SIC experiment, perhaps I should fly over earlier rather than later to talk about these things more deeply.  (I could always come back for a longer visit later \ldots\ ?)

I'll write you more about the SICs in a moment, in a separate note.

Your mention of Laurikainen reminds me that I once read a note you and someone else wrote on his passing away.  Do I have that right?  I'm pretty sure it was you.  If so, would it be possible for you send me a copy?  I did a search for the article on the web, but could not find anything.  Hans Christian von Baeyer and I are writing an article for {\sl American Journal of Physics\/} on Pauli's notion of the detached observer (more particularly its opposite), and it'd be nice to have all possible resource materials available.  (I think you know Hans Christian, but you may not have known that Markus Fierz's brother was Hans's godfather, and that Fierz himself was Hans's father's best friend in their college days.)

With regard to Exner, that's not a quote by him.  It is from Cormac McCarthy, a novelist that {\Carl} {\Caves} likes.  I only recently learned that Exner was my academic great-great-great-great grandfather, but I had known about his argument for indeterminism for a very long time.  I learned of it from Paul Hanle's article ``Indeterminacy before Heisenberg: The Case of Franz Exner and Erwin {\Schroedinger}'' (Historical Studies in the Physical Sciences, vol {\bf 10}, 225--269, 1979).  I tried to get a copy of that from the web to send to you, but I couldn't.  Still, I did find some newer tidbits on him that I had not read before, by an author named St\"oltzner.  I'll attach them for you---they look quite interesting.

\section{09-11-09 \ \ {\it Seeking SICs}\ \ \ (to A. Zeilinger)} \label{Zeilinger5}

Let me now address this part of your letter:
\baz
These SIC-POVMs: where can I learn more about them? Why
do you think they are among the deepest structures of
quantum mechanics? You probably mean deepest in a
conceptual/foundational way.
\eaz

I make my most comprehensive statement in this paper ``Quantum-Bayesian Coherence'':
\bq
\arxiv{0906.2187}.
\eq
Here's the way to read it.  First, jump to page 50 to see the color figure, and read the figure caption.  That sets the agenda.  Then, jump to page 20 and look at that figure and caption.  That is a schematic of the experiment I would like to see done:  It could be billed as an experimental test of the Born Rule.  The idea would be:  First, to perform a set of runs with only one or another regular quantum measurement performed on the systems.  But second, perform a set of runs with a SIC measurement first, before the regular quantum measurement at the end.  Then, one just shows that the equation at the bottom of page 20 is satisfied to within experimental tolerance.

Where I think the experiment would be most interesting and dramatic would be for $d=3$ and $d=4$ (single qutrits and two qubits).  Particularly, for $d=3$, one can make easy comparisons to Gleason's theorem, and also my students would have the best chance of providing some visual aids to thinking (graphs and slices will more effectively depict lower-dimensional spaces).  Equation 24 on page 15 shows an exact expression for a $d=3$ SIC---performing that measurement would be the difficult part of the experiment.

Two things to beware of in the paper:
\begin{enumerate}
\item
The introduction is going to be completely rewritten.  The referee thinks it gives Feynman too much credit, that Feynman wasn't anywhere near the key idea of this paper, and that I undersell myself and the result by diverting so much attention upon him.
\item
Section 6 has a very interesting, but isolated, mistake in it.  I doubt you'll read that section, but just in case, I thought I should warn you.  I say ``interesting'' because in fixing the mistake, one finds something really deep:  an extra handle within quantum mechanics that I believe quantifies how much the observer is ``attached'' or ``undetached'' in this theory.  I will soon fix the mistake and repost the paper, but in the meantime let me offer another paper where the mistake is fixed; it is attached.  I also attach an image to show how the equation becomes modified with this new handle---the classical world is the case $q=0$, the quantum world is the case $q=2$.  The case $q=1$ is something else still, something between classical and quantum.  A value $q=3$ or above, means an even more ``engaged'' observer---quantum mechanics is nowhere near the top of the hierarchy!  If you're interested, see a more philosophical explanation below that I had originally sent to Hans Christian.
\end{enumerate}

So there, that's the way to learn more about SICs; the papers I pointed you to are the best I can do for saying why I think they're deep without talking to you directly.

\section{10-11-09 \ \ {\it QF Group Meeting Tuesday at 4pm in the Alice Room}\ \ \ (to R. W. {\Spekkens})} \label{Spekkens77}

\brws
Our next group meeting will be tomorrow in Alice at 4pm.  The topic of discussion will be: what conceptual ingredients of quantum theory are responsible for its superiority for computation (to the extent that we believe that it is superior).  There will be no official discussion leaders. Rather, those who have an opinion should come prepared to say a few words about it.
\erws

Metaphors, it's always metaphors with me.  My weakness, and hopefully (in longer term) my strength.   I'm not sure I will be in the office this afternoon to be able to come to the meeting:  Emma is at home with an ear infection and ongoing fever.  But below is my take on the matter.  See old note to Gottesman, pasted in below.  [See 25-01-08 note ``\myref{Gottesman8}{Bristol, 3 AM}'' to D. Gottesman.] It is a metaphor only, of course, but one that I feel captures the essence.  QM---for the QBist---seems to be a calculus for calculating probabilities for {\it factual\/} situations from {\it counterfactual\/} situations.  And what is most interesting is that the quantum rule generically gives tighter distributions than the corresponding classical rule would.  ``It gives a reward of more certainty for this or that when an intermediate step (in the intermediate reasoning) is not taken physically.''

Take factoring.  The purpose of building a computer is so that, after an appropriate number of steps, one is fairly certain that if one looks at it, one will find the unknown factors.  One cannot gamble (better than a flat distribution) what the factors will be initially, but one {\it can\/} gamble with near certainty that one will obtain them by having a look at the computer's output at the appropriate time.  That's the point of writing an algorithm.  But in this way of putting things, factoring becomes more obviously a probabilistic question.  Analyze the primitive steps, and then make a probability calculation:  Have an actual look when the probability for finding what you want rises to nearly one.  Now transfer to the metaphor of my beloved urgleichung.   First analyze conceptually in classical probabilistic terms---using conditioning, the law of total probability, etc.---but at the end of the day, use the magical quantum modification to that classical reasoning.  Generically, it will give more {\it certainty\/} for some aspect of something.  And more certainty---so the hope goes---should signify the need for fewer ``computational steps'' in between.

Not without connection to this metaphor is Andrew Steane's (less metaphorical) paper, ``A Quantum Computer Only Needs One Universe,'' \arxiv{quant-ph/0003084}.  I definitely recommend discussion of it, if someone has sufficient knowledge of it.

\section{10-11-09 \ \ {\it Broken Ends} \ \ (to H. Barnum)} \label{Barnum30}

Coming back to a Toronto conversation:  Heidegger was Husserl's student, as I had faintly remembered.  I just read it in wiki as I was trying to get a feel for what ``phenomenology'' is.  Still don't know the latter, but I was pleased to run across the former.

\section{10-11-09 \ \ {\it Hippy Shack} \ \ (to D. B. L. Baker)} \label{Baker19}

\bdb
Somehow, only you could turn what would be a nerdfest into something that could possibly be cool.  Since the name ``Quantum Cowboy'' is already taken, I guess you qualify as a Hilbert Space Hippy.  Perhaps it is you, not I, that needs a beret.
\edb

Did I ever recommend Louis Menand's book, {\sl The Metaphysical Club:\ A Story of Ideas in America}, to you?  Look it up at {\tt amazon.com}.  A Pulitzer prizewinner.  It's got lots of good stuff about Emerson, the Civil War, the 54th Massachusetts Regiment (Shaw's), Oliver Wendell Holmes, all intertwined with stuff on Darwin, my hero William James, and the philosophical vision I'm shooting for.  It was the very best book I read in 2003.

Hilbert Space Hippy is a way cool moniker; I'll wear it proudly!

\section{11-11-09 \ \ {\it The Peirce Quote}\ \ \ (to R. Healey)} \label{Healey5}

\begin{quote}
If you look into a textbook of chemistry for a definition of {\it lithium}, you may be told that it is that element whose atomic weight is 7 very nearly.  But if the author has a more logical mind he will tell you that if you search among minerals that are vitreous, translucent, grey or white, very hard, brittle, and insoluble, for one which imparts a crimson tinge to an unluminous flame, this mineral being triturated with lime or witherite rats-bane, and then fused, can be partly dissolved in muriatic acid; and if this solution be evaporated, and the residue be extracted with sulphuric acid, and duly purified, it can be converted by ordinary methods into a chloride, which being obtained in the solid state, fused, and electrolyzed with half a dozen powerful cells, will yield a globule of a pinkish silvery metal that will float on gasolene; and the material of {\it that\/} is a specimen of lithium.  The peculiarity of this definition---or rather this precept that is more serviceable than a definition---is that it tells you what the word lithium denotes by prescribing what you are to {\it do\/} in order to gain a perceptual acquaintance with the object of the word (2.330).
\end{quote}

\section{12-11-09 \ \ {\it Probability is Single Case, or Nothing}\ \ \ (to R. Healey)} \label{Healey6}

\brh
From the abstracts, I expect to be quite sympathetic to {\Appleby}'s line. I certainly don't defend a frequentist view of probability, and agree that probability applies to the single case. But I don't think that makes all probabilities subjective, just as I don't think there are objective chances. I'm still holding out for probabilities that are authoritative for a rational agent in a specific physical situation, in the sense that his degrees of belief should match these probabilities if he has good empirical reasons to accept a theory that prescribes them.
\erh

``Oh, where are the lines of quantum theory?''  (I think I want to use that line in a play someday.)

\section{16-11-09 \ \ {\it A Happy QBism House}\ \ \ (to N. Waxman \& E. Goheen)} \label{Waxman6} \label{Goheen3}

Thanks a million for the work on putting that together.  I never imagined we'd make the front page too---I'm very flattered and grateful.  And what a great coincidence that Bill Wootters's favorite equation is the same equation as mine, just written in different form!  (Or did you set that up Natasha?)

Is there any chance I could get an electronic copy of the newsletter (either Word or PDF formats)?  I'd like to send the issue to my ``watchers'' at the Navy.  Also, could I get one more hard copy?  I'd like to send my mother.  (It's kind of like being on the cover of {\sl The Rolling Stone}.)

\section{18-11-09 \ \ {\it In the Maelstrom of Modernism}\ \ \ (to J. E. {\Sipe})} \label{Sipe20}

Finally finished the book.  I liked it OK, but I was struck at how abruptly it ended:  It really skimped on the last 8 years of James's life, the period when he was actually the hardest working and most productive in his life.  And that period when James turned to full-blown philosophical efforts is, of course, the most intellectually interesting one for me personally---so I felt like I invested a lot in reading the early parts of the book only to be let down at the end.  I honestly got the sense that the author had just gotten tired of writing on this project.

Still, there was plenty for me in there.  For one thing I gained an appreciation for how much I need to read James's three psychology books.  Particularly, I gained an appreciation for how much his ontology (of ``pure experience'') is based on his model of the mind.  [``Pure experience,'' by the way, I think is an unfortunate name---he only means by that something that is neither material nor mental, but both and neither.]  Also, because of Richardson, I finally feel the need to read James's {\sl Varieties of Religious Experience}.  Finally, I gained an appreciation for the influence of James's wife Alice on some of his thoughts---I hadn't realized she played such a large part in his intellectual life, and I want to learn more about that.

So, thanks for recommending the book to me.  Below are the points of interest I recorded while reading it.  A note to you seems like a good depository for me to record them in.

Oh, one last thing:  I picked up a quote from this book that I had certainly read before (in James's {\sl Pragmatism}), but strangely it didn't strike me then (I have no memory of reading it before).  Reading it now though makes my skin tingle and come alive.  It is surely the right idea:
\begin{quote}
To anyone who has ever looked on the face of a dead child or parent the mere fact that matter {\it could\/} have taken for a time that precious form, ought to make matter sacred ever after. It makes no difference what the principle of life may be, material or immaterial, matter at any rate co-operates, lends itself to all life's purposes. That beloved incarnation was among matter's possibilities.
\end{quote}

\bv
Radical empiricism as phenomenology -- xiv \\
Fechner's unimaginably rich universe -- 4 \#12 \\
Elementary psychological fact -- 6 \\
Matter sacred -- 7, 256-257 \\
Dangers of writing tone being misunderstood -- 28 \\
Concave man -- 31 \\
Muddying the waters -- 49 \\
John Wilkes Booth -- 50 \\
Bain on belief -- 79 \\
William vs Father philosophically -- 83, 84 \\
Faust \& pragmatism -- 92 \\
James reads Kant \& Renouvier -- 94 \\
Multiple points of view -- 100 \\
James while still a determinist -- 101 \\
James ``shall I throw this moral business overboard?'' -- 111 \\
James's vastation -- 117 \\
James reading Renouvier \& choosing free will -- 120-122 \\
Accept or protest the universe -- 124 \\
Chauncey Wright similarity to RBK -- 130 \\
Chauncey Wright quote -- 131 \\
Moments getting into a habit of following each other -- 135 \\
Tychasm, anancasm, agapasm -- 136 \\
Philosophical distinction between Peirce \& James -- 136 \\
Taine's ideas -- 143 \\
First letter to Renouvier -- 145 \\
No rebound in students -- 146 \\
Emerson's philosophy -- 151 \\
Anyone with something to teach -- 160 \\
Book by James's ghost -- 160 \\
Science as no synonym for truth -- 163 \\
Renouvier's strategy -- 177 \\
Being and nothingness -- 178 \\
Spencer before Darwin -- 178, 179 \\
Mind gets a vote -- 183 \\
``ignoring data is the most popular mode of obtaining unity'' -- 184 \\
``such decisions seem acts by which we are voting what sort of a universe this shall intimately be'' -- 187 \\
Royce as opposite to James -- 188 \\
``Rationality, Activity, and Faith'' -- 201 \\
Emerson's ``enveloping now'' -- 202 \\
It is the act that matters -- 207 \\
Hodgson \& Renouvier -- 209 \\
``On Some Hegelisms'' -- 214-216 \\
Temperament -- 239, 248 \\
Free will -- 246, 247 \\
Biographies of Henry Sr -- 250 \# 15 \\
Chocorua, 14 \\
outside doors? -- 271, 279 \\
``What the Will Effects'' -- 280, 281 \\
Santayana thesis on Lotze -- 286 \\
Dimensions of James's library -- 290 \\
James's difficulty in writing -- 297, 298 \\
James as originator of phenomenology -- 304 \\
William's son on action -- 327 \\
Reaction -- 342 \\
James on Jung-style stuff -- 348, 349 \\
Science only the minutest glimpse -- 355 \\
``Is Life Worth Living'' great quote -- 356 \\
Pluriverse -- 364 \\
James becoming professor of philosophy -- 369 \\
``On a Certain Blindness'' -- 381 \\
Crab, ``I am myself'' -- 392 \\
When religious experience is complete -- 393 \\
B. P. Blood, ``all days are judgment day'' -- 413 \\
James--Flournoy correspondence -- 573 \#2 \\
Needing too much to do -- 424 \\
Panpsychism -- 424 \\
Bergson on time -- 426, 427 \\
Rockeffeler -- 428, 429 \\
Bergson on matter -- 427 \\
Ultra-gothic -- 440 \\
Chance -- 443, 444 \\
William on Henry's writing -- 424-425, 464 \\
Richardson mistake on ``Many \& One'' -- 446 \\
McDermott on radical empiricism -- 446 \#1 \\
Russell on James -- 449, 450 \\
Fechner -- 447 \\
Rheomode -- 450 \\
Whitehead on James -- 450 \\
Radical empiricism as big scientific revolution -- 452 \\
James impressed by Bradley (once?)\ -- 454 \\
And discussion sounding like Many Worlds Interp -- 455 \\
Good Royce point -- 457 \\
Faith; the faith ladder -- 469 \\
Dreams as full-standing experiences -- 470 \\
Papini in English?\ -- 479 \#3 \\
Actions growing reality -- 487 \\
Pluralism adjectivized -- 491 \\
Dinner with Mark Twain -- 492 \\
Hard to believe what you don't believe -- 494 \\
William on Henry's {\it American Scene\/} -- 496 \\
Fechner again -- 500-502 \\
Good story of death -- 502 \\
What really exists is things in the making -- 503 \\
World is well because of certain forms of death -- 504 \\
Wider consciousness -- 504-505, 509 \\
Essay on B. P. Blood -- 517
\ev

\section{19-11-09 \ \ {\it Psychology and Physics}\ \ \ (to H. C. von Baeyer)} \label{Baeyer84}

Psychology has been much on my mind the last two weeks.  The note below [see 18-11-09 note ``\myref{Sipe20}{In the Maelstrom of Modernism}'' to J. E. {\Sipe}] gives some story of the cause:  It was in reading Robert Richardson's biography of William James, where a lot of attention is paid to his construction of {\sl The Principles of Psychology}.  I also started to realize from the book that there may be a more continuous thread from James to Jung than I would have expected.  For instance, there are already relatively explicit statements of the collective conscious in late James (pre Jung).  There are so many veins I must still explore!  Our key to physics (as Pauli was already guiding) really must be in all this stuff.

\section{20-11-09 \ \ {\it Chris's Picks}\ \ \ (to R. Healey)} \label{Healey7}

Needless to say, I enjoyed yesterday's discussion from beginning to end.

And I liked one of the things you said near the end, which I think you described as a ``new attitude'' about reading some philosophy, to mine it for ideas, and worry about the logic later.  I think that's a healthy thing.  (But maybe I only say that because I'm a physicist, and physicists are opportunists.)

For fun, I was thinking this morning that I might suggest a collection of readings.  (The list started because I wasn't sure you'd remember the name of the James book I suggested, but then it gathered steam on its own.)

Anyway, here goes:
\begin{itemize}
\item
Rorty, {\sl Philosophy and the Mirror of Nature}.
\item
Rorty, {\sl Philosophical Papers Volume 1: Objectivity, Relativism, and Truth}.  This is the one that starts off with this spiel that I have recorded in my computer:
\bq
The six papers that form Part I of this volume offer an
antirepresentationalist account of the relation between natural
science and the rest of culture.  By an antirepresentationalist
account I mean one which does not view knowledge as a matter of
getting reality right, but rather as a matter of acquiring habits of
action for coping with reality.

Philosophers in the English-speaking world seem fated to end the
century discussing the same topic---realism---which they were
discussing in 1900.  In that year, the opposite of realism was still
idealism.  But by now language has replaced mind as that which,
supposedly, stands over and against ``reality.'' So discussion has
shifted from whether material reality is ``mind-dependent'' to
questions about which sorts of true statements, if any, stand in
representational relations to nonlinguistic items.  Discussion of
realism now revolves around whether only the statements of physics
can correspond to ``facts of the matter'' or whether those of
mathematics and ethics might also.  Nowadays the opposite of realism
is called, simply, ``antirealism.''

This term, however, is ambiguous.  It is standardly used to mean the
claim about some particular true statements, that there is no
``matter of fact'' which they represent.  But, more recently, it has
been used to mean the claim that {\it no\/} linguistic items
represent {\it any\/} nonlinguistic items.  In the former sense it
refers to an issue within the community of
representationalists---those philosophers who find it fruitful to
think of mind or language as containing representations of reality.
In the latter sense, it refers to antirepresentationalism---to the
attempt to eschew discussion of realism by denying that the notion of
``representation,'' or that of ``fact of the matter,'' has any useful
role in philosophy. Representationalists typically think that
controversies between idealists and realists were, and controversies
between skeptics and antiskeptics are, fruitful and interesting.
Antirepresentationalists typically think both sets of controversies
pointless.  They diagnose both as the results of being held captive
by a picture, a picture from which we should by now have wriggled
free.
\eq
\item
James, {\sl Some Problems of Philosophy}.  For me a wonderful little book because it pretty well expresses what I think quantum mechanics is indicating at least for the small philosophical scale \ldots\ how far one can carry it I don't know.  It is that everything has an inside---a true inside---that cannot be mathematized from the outside.  This is a bit of Pauli's thought as well.
\item
James, {\sl Pragmatism}.  There's nothing you'd call a rigorous argument in here, but it paints a picture quite beautiful to me, one that certainly changed my life.  There are all kinds of beautiful gems in here, including this quote:
\begin{quote}
To anyone who has ever looked on the face of a dead child or parent the mere fact that matter {\it could\/} have taken for a time that precious form, ought to make matter sacred ever after.  It makes no difference what the principle of life may be, material or immaterial, matter at any rate co-operates, lends itself to all life's purposes.  That beloved incarnation was among matter's possibilities.
\end{quote}
\item
Ralph Barton Perrry, {\sl The Thought and Character of William James, Volume II: Philosophy and Psychology}.  This book is still the best of its type around; it is the book that caused me to ``want to be'' William James.  If you want to know what you should watch out for, so as not to be sucked into the James cult, approach this one cautiously.
\item
Dewey.  I know I should suggest something by Dewey, but it is hard, as I just cannot stay awake while reading this guy.  Still I have the impression that he is ``mostly James'' (in contrast to Peirce, say), but rigorized.  Flipping through Rorty will tell you what of Dewey to read if you want any suggestions---Dewey is Rorty's hero.  Still, let me offer this much:
\item
F. Thomas Burke, et al, {\sl Dewey's Logical Theory: New Studies and Interpretations}.   The introduction gives some nice orientation.
\item
H. S. Thayer, {\sl Meaning and Action: A Critical History of Pragmatism}.  This is a really very good reference; it covers literally everything of the pre-Rorty as far as I can see.
\item
And finally, if you (late in the game, after you've read some of the stuff above) really want to have a look into the wild side---I reveal my biggest dreams here---have a look at some of F. C. S. Schiller (remember, mining, mining):  F. C. S. Schiller, ``Axioms as Postulates''.  (I can make a copy and send it to you.)
\end{itemize}

\section{20-11-09 \ \ {\it Wheeler on Quine}\ \ \ (to R. Healey)} \label{Healey8}

This was the Quine quote that Wheeler liked to use when arguing that ultimately physics should shed the continuum:
\bq\noindent
Just as the introduction of the irrational numbers \ldots\ is a convenient myth [which] simplifies the laws of arithmetic \ldots\ so physical objects are postulated entities which round out and simplify our account of the flux of existence.  \ldots\  The conceptual scheme of physical objects is [likewise] a convenient myth, simpler than the literal truth and yet containing that literal truth as a scattered part.
\eq
It comes from his paper, ``On What There Is''.

And by the way, I told you that the Pragmatism Cybrary classifies Quine as a ``recent pragmatist''; you can find their classification here:
\myurl{http://pragmatism.org/}.

\section{21-11-09 \ \ {\it SICs, Reason, and Faith} \ \ (to the QBies)} \label{QBies4}

Continuing on with our discussion from Wine \& Cheese yesterday:  Particularly, on how {\it belief\/} itself can {\it sometimes\/} be the very cause of a reality.  That, without a belief first, without a {\it faith\/} that something {\it ought\/} to be,---in some cases!---it simply would not be.  I'm thinking of that part of our conversation.   Anyway, on these things, I was led back to William James's speech ``Reason and Faith.''  I attach the full thing, and excerpt below the part of it that is important to our little group.

I can do nothing but admit that in coming to the conclusion that the foundation of quantum mechanics should be built on SICs, I climbed a ``faith-ladder'' in the sense that James defines it below:
\bq\noindent
SICs are {\it fit\/} to be true.  It would be well if they {\it were\/} true.  They {\it might\/} be true; they {\it may\/} be true; they {\it ought\/} to be true.  They {\it must\/} be true!  They {\it shall\/} be true!  For me at least!!
\eq

And so with it, faith brought a little bit of reality into being.  It is the ONR funding and our little group, and all the results positive and negative we shall find over the course of the next few years.

\bq
Faith uses a logic altogether different from Reason's logic. Reason claims certainty and finality for her conclusions. Faith is satisfied if hers seem probable and practically wise.

Faith's form of argument is something like this: Considering a view of the world: ``It is {\it fit\/} to be true,'' she feels; ``it would be well if it {\it were\/} true; it {\it might\/} be true; it {\it may\/} be true; it {\it ought\/} to be true,'' she says; ``it {\it must\/} be true,'' she continues; ``it {\it shall\/} be true,'' she concludes, ``{\it for me}; that is, I will treat it as if it {\it were\/} true so far as my advocacy and actions are concerned.''

Obviously this is no intellectual chain of inferences, like the {\it Sorites\/} of the logic-books. You may call it the ``faith-ladder,'' if you like; but, whatever you call it, it is the sort of slope on which we all habitually live. In no complex matter can our conclusions be more than {\it probable}. We use our feelings, our good-will, in judging where the greater probability lies; and when our judgment is made, we practically turn our back on the lesser probabilities as if they were not there. Probability, as you know, is mathematically expressed by a fraction. But seldom can we {\it act\/} fractionally---half-action is no action (what is the use of only half-killing your enemy?---better not touch him at all); so for purposes of action we equate the most probable view to 1 (or certainty) and other views we treat as naught.
\eq

\section{22-11-09 \ \ {\it More Reason and Faith} \ \ (to H. C. von Baeyer)} \label{Baeyer85}

\bhcvb
I will gladly read  ``Reason and Faith'' in the spirit you propose, with one reservation.  James, in his context, speaks about religious faith, but I would like to distinguish between the concepts of faith and religion.  In my Pauli speech I referred to a book by James P. Carse entitled {\bf The religious case against belief} (2008).  My brother, a psychologist in Saskatoon, was so taken with it when I told him about it that he gave a sermon to his Unitarian church (appropriate since James was speaking to Unitarians too.) The abstract of the sermon put it succinctly:
\bq\noindent
\rm Belief systems are wilfully closed and certain, and are the enemy of religions, which are so open to debate, uncertainty, and reinterpretation that it takes thousands of years for these discussions to run their course.  In his 2008 book, theologian James P. Carse explains how the essence of religion is poetry or mystery or a higher form of ignorance rather than a didactic set of claims or beliefs.  This book turned my understanding of religion on its head.
\eq
The relevance of all this to your thinking is further complicated by the difference between the words ``faith'' and ``belief'', which I am not prepared to discuss here.  All I want to do is to make sure that religion is properly decoupled from the conversation about SICs.

\ehcvb

Thanks for your brother's abstract; I think I'd like to take a look at this book during the Christmas holidays.  I don't think you have anything to worry about with regard to James pushing `belief systems' in the sense you mention.

Particularly, the more I thought about it today, I think he might have been served well to use the word ``decision'' rather than ``belief.''  This can be seen in the last essay I sent, but it can be seen even more clearly in the attached essay.  (It is from his book {\sl Some Problems of Philosophy}, but I picked it up from a badly scanned copy on the internet; then I spent some time cleaning it up in \TeX.  It struck me as the sort of thing I'd like to have a good copy of in my computer.)  This particular essay is less about religion, but more about the idea that the world can be changed by our actions.  The faith-ladder once again makes an appearance in it.\medskip

\bq
\begin{center}
\large {\bf Faith and the Right to Believe}\smallskip \\ (Appendix from William James's book {\sl Some Problems of Philosophy})\bigskip
\end{center}

`Intellectualism' is the belief that our mind comes upon a world complete in itself, and has the duty of ascertaining its contents; but has no power of re-determining its character, for that is already given.

Among intellectualists two parties may be distinguished. Rationalizing intellectualists lay stress on deductive and `dialectic' arguments, making large use of abstract concepts and pure logic (Hegel, Bradley, Taylor, Royce). Empiricist intellectualists are more `scientific,' and think that the character of the world must be sought in our sensible experiences, and found in hypotheses based exclusively thereon (Clifford, Pearson).

Both sides insist that in our conclusions personal preferences should play no part, and that no argument from what {\it ought to be\/} to what {\it is}, is valid.  `Faith,' being the greeting of our whole nature to a kind of world conceived as well adapted to that
nature, is forbidden, until purely intellectual {\it evidence\/} that such {\it is\/} the actual world has come in.
Even if evidence should eventually prove a faith true, the truth, says Clifford, would have been `stolen,' if assumed and acted on too soon.

Refusal to believe anything concerning which `evidence' has not yet come in, would thus be the
rule of intellectualism. Obviously it postulates certain conditions, which for aught we can see need
not necessarily apply to all the dealings of our minds with the Universe to which they belong.

1. It postulates that {\it to escape error\/} is our paramount duty. Faith {\it may\/} grasp truth; but also it
may not. By resisting it always, we are sure of escaping error; and if by the same act we renounce
our chance at truth, that loss is the lesser evil, and should be incurred.

2. It postulates that in every respect the universe is finished in advance of our dealings with it;

That the knowledge of what it thus is, is best gained by a passively receptive mind, with no
native sense of probability, or good-will towards any special result;

That `evidence' not only needs no good-will for its reception; but is able, if patiently waited for, to
neutralize ill-will;

Finally, that our beliefs and our acts based thereupon, although they are parts of the world, and
although the world without them is unfinished, are yet such mere externalities as not to alter in any
way the significance of the rest of the world when they are added to it.

In our dealings with many details of fact these postulates work well. Such details exist in advance
of our opinion; truth concerning them is often of no
pressing importance; and by believing nothing, we
escape error while we wait. But even here we often
cannot wait but must act, somehow; so we act on
the most {\it probable\/} hypothesis, trusting that the
event may prove us wise. Moreover, not to act on
one belief, is often equivalent to acting as if the
opposite belief were true, so inaction would not
always be as `passive' as the intellectualists assume. It is one attitude of will.

Again, Philosophy and Religion have to interpret
the total character of the world, and it is by no
means clear that here the intellectualist postulates
obtain. It may be true all the while (even though
the evidence be still imperfect) that, as Paulsen
says, `the natural order is at bottom a moral order.'
It may be true that work is still doing in the world-
process, and that in that work we are called to bear
our share. The character of the world's results may
in part depend upon our acts. Our acts may depend
on our religion,---on our not-resisting our faith-tendencies, or on our sustaining them in spite of
`evidence' being incomplete. These faith-tendencies in turn are but expressions of our good-will
towards certain forms of result.

Such faith-tendencies are extremely active psychological forces, constantly outstripping evidence.
The following steps may be called the `faith-ladder':
\begin{enumerate}
\item
There is nothing absurd in a certain view of the
world being true, nothing self-contradictory;
\item
It {\it might\/} have been true under certain conditions;
\item
It {\it may\/} be true, even now;
\item
It is {\it fit\/} to be true;
\item
It {\it ought\/} to be true;
\item
It {\it must\/} be true;
\item
It {\it shall\/} be true, at any rate true for {\it me}.
\end{enumerate}

Obviously this is no intellectual chain of inferences, like the {\it sorites\/} of the logic-books. Yet it is
a slope of good-will on which in the larger questions of life men habitually live.

Intellectualism's proclamation that our good-will,
our `will to believe,' is a pure disturber of truth, is
itself an act of faith of the most arbitrary kind. It
implies the will to insist on a universe of intellectualist constitution, and the willingness to stand in
the way of a pluralistic universe's success, such
success requiring the good-will and active faith, theoretical as well as practical, of all concerned, to
make it `come true.'

Intellectualism thus contradicts itself. It is a sufficient objection to it, that if a `pluralistically'
organized, or `co-operative' universe or the `melioristic' universe above, were really here, the veto
of intellectualism on letting our good-will ever have any vote would debar us from ever admitting that
universe to be true.

Faith thus remains as one of the inalienable birth-rights of our mind. Of course it must remain a
practical, and not a dogmatic attitude. It must go with toleration of other faiths, with the search for
the most probable, and with the full consciousness of responsibilities and risks.

It may be regarded as a formative factor in the
universe, if we be integral parts thereof, and co-determinants, by our behavior, of what its total
character may be.\medskip

\noindent {\bf How We Act on Probabilities} \medskip

In most emergencies we have to act on probability, and incur the risk of error.

`Probability' and `possibility' are terms applied to things of the conditions of whose coming
we are (to some degree at least) ignorant.

If we are entirely ignorant of the conditions that
make a thing come, we call it a `bare' possibility.
If we know that some of the conditions already
exist, it is for us in so far forth a `grounded' possibility. It is in that case {\it probable\/} just in proportion as the said conditions are numerous, and few hindering conditions are in sight.

When the conditions are so numerous and confused that we can hardly follow them, we treat a
thing as probable in proportion to the {\it frequency\/} with which things of that {\it kind\/} occur. Such frequency being a fraction, the probability is expressed by a fraction. Thus, if one death in 10,000 is by
suicide, the antecedent probability of my death
being a suicide is 1-10,000th. If one house in 5000
burns down annually, the probability that my house
will burn is l-5000th, etc.

Statistics show that in most kinds of thing the
frequency is pretty regular. Insurance companies
bank on this regularity, undertaking to pay (say)
5000 dollars to each man whose house burns, provided he and the other house-owners each pay
enough to give the company that sum, plus something more for profits and expenses.

The company, hedging on the large number of
cases it deals with, and working by the long run,
need run no risk of loss by the single fires.

The individual householder deals with his own single case exclusively. The probability of his house
burning is only 1-5000, but if that lot befall he
will lose everything. He has no `long run' to go by,
if his house takes fire, and he can't hedge as the
company does, by taxing his more fortunate neighbors. But in this particular kind of risk, the company helps him out. It translates his one chance in
5000 of a big loss, into a certain loss 5000 times
smaller, and the bargain is a fair one on both sides.
It is clearly better for the man to lose {\it certainly}, but
{\it fractionally}, than to trust to his 4999 chances of no
loss, and then have the improbable chance befall.

But for most of our emergencies there is no insurance company at hand, and fractional solutions are
impossible. Seldom can we {\it act\/} fractionally. If the
probability that a friend is waiting for you in Boston is 1-2, how should you act on that probability?
By going as far as the bridge? Better stay at home!
Or if the probability is 1-2 that your partner is a
villain, how should you act on that probability?
By treating him as a villain one day, and confiding
your money and your secrets to him the next?
That would be the worst of all solutions. In all such
cases we must act wholly for one or the other horn of
the dilemma. We must go in for the more probable
alternative as if the other one did not exist, and
suffer the full penalty if the event belie our faith.

Now the metaphysical and religious alternatives
are largely of this kind. We have but this one life
in which to take up our attitude towards them, no
insurance company is there to cover us, and if we
are wrong, our error, even though it be not as great
as the old hell-fire theology pretended, may yet be
momentous. In such questions as that of the {\it character\/} of the world, of life being moral in its essential
meaning, of our playing a vital part therein, etc.,
it would seem as if a certain {\it wholeness\/} in our faith
were necessary. To calculate the probabilities and
act fractionally, and treat life one day as a farce,
and another day as a very serious business, would
be to make the worst possible mess of it. Inaction
also often counts as action. In many issues the
inertia of one member will impede the success of
the whole as much as his opposition will. To refuse,
{\it e.\ g.}, to testify against villainy, is practically to
help it to prevail.\footnote{Cf.\ Wm.\ James: {\sl The Will to Believe}, etc., pp.\ 1--31, and 90--110.}
\medskip

\noindent {\bf The Pluralistic or Melioristic Universe}\medskip

Finally, if the `melioristic' universe were {\it really\/}
here, it would require the active good-will of all of
us, in the way of belief as well as of our other activities, to bring it to a prosperous issue.

The melioristic universe is conceived after a
social analogy, as a pluralism of independent powers. It will succeed just in proportion as more of
these work for its success. If none work, it will fail.
If each does his best, it will not fail. Its destiny
thus hangs on an {\it if}, or on a lot of {\it ifs}---which
amounts to saying (in the technical language of
logic) that, the world being as yet unfinished, its
total character can be expressed only by {\it hypothetical\/} and not by {\it categorical\/} propositions.

(Empiricism, believing in possibilities, is willing to formulate its universe in hypothetical propositions. Rationalism, believing only in impossibilities and necessities, insists on the contrary on their
being categorical.)

As individual members of a pluralistic universe,
we must recognize that, even though we do {\it our\/} best,
the other factors also will have a voice in the result.
If they refuse to conspire, our good-will and labor
may be thrown away. No insurance company can
here cover us or save us from the risks we run in
being part of such a world.

We {\it must\/} take one of four attitudes in regard to
the other powers: either

1. Follow intellectualist advice: wait for evidence; and while waiting, do nothing; or

2. {\it Mistrust\/} the other powers and, sure that the
universe will fail, {\it let\/} it fail; or

3. {\it Trust\/} them; and at any rate do {\it our\/} best, in
spite of the {\it if}; or, finally,

4. {\it Flounder}, spending one day in one attitude,
another day in another.

This 4th way is no systematic solution. The 2d
way spells faith in failure. The 1st way may in
practice be indistinguishable from the 2d way.
The 3d way seems the only wise way.

`{\it If\/} we do {\it our\/} best, and the other powers do {\it their\/} best, the world will be perfected'---this proposition expresses no actual fact, but only the complexion of a fact thought of as eventually possible.
As it stands, no conclusion can be positively deduced from it. {\it A conclusion would require another
premise of fact, which only we can supply. The original proposition per se has no pragmatic value whatsoever, apart from its power to challenge our will to produce the premise of fact required.} Then indeed
the perfected world emerges as a logical conclusion.

We can {\it create\/} the conclusion, then. We can and
we may, as it were, jump with both feet off the
ground into or towards a world of which we trust
the other parts to meet our jump---and {\it only so\/}
can the {\it making\/} of a perfected world of the pluralistic pattern ever take place. Only through our precursive trust in it can it come into being.

There is no inconsistency anywhere in this, and
no `vicious circle' unless a circle of poles holding
themselves upright by leaning on one another, or a circle of dancers revolving by holding each other's
hands, be `vicious.'

The faith circle is so congruous with human
nature that the only explanation of the veto that
intellectualists pass upon it must be sought in the
offensive character to {\it them\/} of the faiths of certain
concrete persons.

Such possibilities of offense have, however, to be
put up with on empiricist principles. The long run
of experience may weed out the more foolish faiths.
Those who held them will then have failed: but
without the wiser faiths of the others the world
could never be perfected.

(Compare G. Lowes Dickinson: ``Religion, a Criticism and a Forecast,'' N. Y. 1905, Introduction; and chaps.\ iii, iv.)
\eq

\section{23-11-09 \ \ {\it CoE Proposal}\ \ \ (to H. Price)} \label{Price16}

Sorry to keep you waiting.  That's a genuinely nice document \ldots\ even if I did cringe at the thought of being associated with something that says this:  ``A striking example is the attempt to understand quantum `nonlocality' (the `spooky action at distance' in quantum theory, as Einstein famously called it). Nonlocality was characterised mathematically by John Bell in the 1960s, and later confirmed experimentally by Aspect and others.''  (The implication is pretty clearly that `spooky action at a distance' is `confirmed'.  And you know of course, that that's about the last thing I think Bell inequality violations confirm.)  \ldots\ Still, I was seduced back by the small homage to these, I would say, closer-to-correct words, ``the real breakthrough will come when we start to realise the connections between reality, knowledge and our actions.''  \ldots\ And finally overpowered by that beautiful closing argument, ``The quest to understand reality is one of humanity's fundamental projects. Each generation inherits new challenges on the shoulders of the last. In our generation, some of the deepest puzzles require the joint skills of physics and philosophy. Our goal is to lead the world in directing the best available intellectual resources to these challenges.''

I'd be proud to be associated with this project.  And you can list me for a 20\% time commitment as you wish.

One final thing, I was amused by this line in the proposal, ``What is the status of probability in Quantum Bayesianism?''  I would have thought that is the one thing that is a settled issue by now!  Attached is a draft I'm writing at the very moment, where effectively I say so.  Rue the day your funding agents ever see my own writings about this question!  They might just take that portion of the money back.

\section{23-11-09 \ \ {\it CoE Proposal, 2}\ \ \ (to H. Price)} \label{Price17}

By the way, I like that way of writing ``Quantum Bayesianism'' as you do in the proposal, with both ``Quantum'' and ``Bayesianism'' capitalized.  I just adopted the convention myself!

\section{23-11-09 \ \ {\it My Visit, James, Lotze, and the Malleable World}\ \ \ (to A. Zeilinger)} \label{Zeilinger6}

Then let's settle it.  I will come for the week of Jan 17--23, with possibly some extra time around the front or back edge of that range.

Funny, just this morning (because of a further grant proposal our Australian partners in the PI-AF quantum foundations collaboration are working up), I read the {\sl New Scientist\/} thing on you from last month.  They write,
\bq\noindent
     Zeilinger specialises in quantum experiments that demonstrate the
     apparent influence of observers in the shaping of reality. ``Maybe the
     real breakthrough will come when we start to realise the connections
     between reality, knowledge and our actions,'' he says.
\eq
Nice.  With that in mind, I give you a little gift:  One of my favorite discussions from William James's little book {\sl Pragmatism}.  I hope you enjoy the attached short file, and I am very much looking forward to January.

\section{23-11-09 \ \ {\it GRW and Bayes} \ \ (to G. J. Milburn)} \label{Milburn5}

\bgjm
I gave a `reading' of your paper ``Subjective probability and quantum certainty'' to our PIAF group on Wednesday. Some way through the discussion the question of what to make of spontaneous localisations in GRW came up.  Have you, {\Carl} or {\Ruediger} thought about this from a quantum bayesian point of view?
\egjm

I would guess that neither {\Carl} nor {\Ruediger} nor I have thought much about GRW in these terms.  In my case, not only would I guess, but I know that I haven't!  It's a symptom of the thing I described in the draft I sent you and Huw, etc., a couple of hours ago.  GRW is usually thought of in terms of reified states doing the collapsing, i.e.\ the states {\it are\/} the dynamical variable.  The main point of the quantum Bayesian movement is to explicitly remove the quantum state from the observer-INdependent world, leaving no trace behind.  If GRW can be posed in such a way that the thing jumping is not a quantum state, then maybe there's a chance of analyzing it in our terms.  But I don't know of such a way.

I hope that helps answer things, and that our old paper didn't give you or the other guys too much of a headache.

\section{23-11-09 \ \ {\it While Translating}\ \ \ (to J. Berkovitz)} \label{Berkovitz2}

Thanks much for this remark in your paper:
\bq\noindent
``Probability, too, if regarded as something endowed with some kind of objective existence, is no less a
misleading misconception, an illusory attempt to exteriorize or materialize our true [i.e.\ actual]
probabilistic beliefs.'' (de Finetti 1974a, p.\ x) \ldots\
Here the translation from Italian seems incorrect. The word
`vero' could be translated as `actual' or `true', and it is clear
that here it should be translated as `actual'.
\eq
It cleared up a good mystery for me.  I had never liked the seeming use of true there.  But it prompts another question for me.  In the fuller quote below, should it really be an ``or Fairies and Witches'' or rather an `and'.  The `or' has always struck me as out of place.
\bq
The abandonment of superstitious beliefs about the existence
of Phlogiston, the Cosmic Ether, Absolute Space
and Time, \ldots, or Fairies and Witches, was an essential step
along the road to scientific thinking. Probability,
too, if regarded as something endowed with some kind of
objective existence, is no less a misleading conception, an
illusory attempt to exteriorize or materialize our true probabilistic
beliefs.
\eq

I haven't forgotten that I want to comment extensively on your draft.  I sure hope to do it in the coming couple of weeks.

\subsection{Jossi's Reply}

\bq
As for the translation, the word `o' in Italian is usually translated as `or' but could sometime be translated as `and'.
\eq

\section{24-11-09 \ \ {\it New Apologies:\  Me, Me, Me}\ \ \ (to N. Waxman)} \label{Waxman7}

Another week has passed and I still haven't put the {\sl Physics in Canada\/} paper in your hand.  I feel completely ashamed.  On the other hand, I can honestly admit to having put an embarrassing amount of time into writing.  (You'd probably not believe me if I told you.)  Trouble is I've been proceeding with my two paper concept being very careful with my word choices:
\bq\noindent
{\it To wit, from my previous note to you and Rob Myers}:  How strict is the 2500 word-count limit going to be for the {\sl Physics in Canada\/} thing?  I ask because I just reached that number and still have a ways to go with my draft.  I think what I am going to do, now that I'm on a roll and really liking what I've written so far, is just plunge ahead, writing what feels right internally, and then let Natasha go at it with an ample scalpel.  (I justify this in my mind by thinking, at the end I'll have two articles for the price of one.  I'll find some place to send the longer one off to, even if only {\tt arXiv.org}.)
\eq
The manuscript is, at the moment, just at 4000 words.  Hopefully, if I get it done right, a significant portion of that will be simply detachable, and then we can smooth it out thereafter.  But there's no denying, I've been writing the big version for me, me, me---trying to say things in a way that will have lasting value and persuasiveness.

Anyway, I felt I should write you this morning because, actually, I must take two days off this week and I feel very guilty about that:  today and Thursday.  Today, because I have to go to U. Toronto to work with the experimental team that's doing an experiment on one section of my {\it Physics in Canada\/} article (something called a SIC).  Then Thursday because of American Thanksgiving, which is a big family tradition for us.  But Wednesday and Friday, I'll give the paper my all---I promise.

To make a long grovel short:  Can you---{\it please}---wait until Monday for the finished product???

\section{25-11-09 \ \ {\it Modest Theories (with a capital M)} \ \ (to L. K. Shalm)} \label{Shalm1}

OK, Krister, your turn.  (I'll cc Aephraim in case he's interested in these wispier things too.)

Thanks for your continued interest in this ``QM as a single-user theory business''---I really do appreciate your efforts to get it straight, whether you accept OR reject in the end.  The endpoint doesn't matter as much as the effort; the effort alone is flattering.  (Plenty of so-called professional philosophers of science, for instance, reject the view at the outset without ever making an ounce of effort to understand what is new in it.)

First, on your particular question from yesterday.  For it, please refer to a long note I constructed for Lucien Hardy after the last PIAF meeting.  It can be found \myurl[http://www.perimeterinstitute.ca/personal/cfuchs/nSamizdat-2.pdf]{http://www.perimeterinstitute. ca/personal/cfuchs/nSamizdat-2.pdf} starting at page 813 by raw pdf accounting (with regard to \LaTeX\  page numbering, instead, that is page 785).  The note is titled, ``The More and the Modest''.  [See 16-10-09 note ``\myref{Hardy38}{The More and the Modest}'' to L. Hardy.] The idea argued for there is the sense in which Quantum Theory is a very modest theory.  Not modest in the sense of saying it is merely about ``black boxes''---as you were describing yesterday---but {\it Modest\/} in the way of Newton's theory of universal gravitation.   I say this to contrast it with Weinberg, Witten, or Hawking's dream of a ``theory of everything''.  Newton's theory is not that, but on the other hand it is still a sweeping statement about nature.  So too I think for quantum theory---that though it is a single-user theory, it is nonetheless a sweeping and universal statement about the character of the things in the world.  Here's the best way I know how to put the point:  What I am aiming for is that quantum theory is a theory of ONE ASPECT of everything.

Read the note to Hardy (including the supplementary material in it); it'll do a better job than what I can write here in a single paragraph.

Further, let me attach a paper that I'm writing at the very moment.  [See ``QBism, the Perimeter of Quantum Bayesianism,'' \arxiv{1003.5209v1}.] Sorry, it's only halfway finished, but the beginning material may do a better job of motivating the single-user aspect of the program than previously.  I have been trying to write it in an entertaining way; so hopefully it will draw you in.

If this stuff shakes anything loose in your thinking, please let me know.  Or, if there is an aspect of it all that still strikes you of solipsism (or sophism!), please try to articulate it, and I'll give it my best shot of a thought-out reply.  From every challenge I take on, I learn a little more.

\section{30-11-09 \ \ {\it QELS}\ \ \ (to Z. E. D. Medendorp \& A. M. Steinberg)} \label{Medendorp1} \label{Steinberg3}

\begin{itemize}
\item
This is Bayes' Rule or Bayes' Theorem:\\
\myurl{http://en.wikipedia.org/wiki/Bayes\%27_theorem}\\
See the first equation in.

\item
On the other hand, this is the Law of Total Probability:\\
\myurl{http://en.wikipedia.org/wiki/Law_of_total_probability}\\
See the third equation in.
\end{itemize}
My rewrite of the Born Rule (using SICs) is an analogue of the latter rather than the former.

\section{30-11-09 \ \ {\it  NOT Recommended {\em AMS Notices} Article} \ \ (to M. B. Ruskai)} \label{Ruskai2}

I'm sorry to take so long to get back to you.  I read the Chernoff article, and the end of it is indeed horrible:  ``In addition, experiments have been done which suggest that influence from one system to the other propagates enormously faster than light.  These experiments point toward instantaneous transfer of information.''  There was just no call for any of that kind of discussion in an article about Gleason.  On the other hand, is it worth a battle?  This kind of crap happens all the time, everywhere you look---Nicolas Gisin is guilty of it, Alain Aspect is guilty of it:  The list of otherwise good physicists just goes on and on.  I try to fight the fight when I have to---for instance, see the present article under construction (and required of me from PI for a special issue)---but then sometimes I just get tired of building the background, trying to get the uninitiated to see the trouble after they've already had an untrustworthy introduction.  [See ``Quantum Bayesianism at the Perimeter,'' \arxiv{1003.5182}.]

I don't know, I partially fear taking on another project that'll probably leave me unsatisfied and more frustrated with the community in the end, at this time when I'm already far behind on all the other things I need to do.

\section{01-12-09 \ \ {\it Hawaii in April?}\ \ \ (to H. Price)} \label{Price18}

It dawned on me that if we're going to have that meeting in April, I probably ought to get on planning it before December is out.  [\ldots]

As I said when I volunteered for this, I'd like to make the theme of the workshop:  Contextuality, Perspectivalism, Pragmatism, or some combination/variation of that.  In that regard, what do you think about inviting any non-PIAFers?  I ask mostly because Richard Healey visited us for a couple months, and you should see the change in this guy (to the far-more interesting):  He is now calling his research project for the next few years something like ``developing a consistent pragmatism''.  See this fantastic talk he gave, \pirsa{09110136}, at the end of his visit.  Anyway, I think I'd like to invite at least him into the mix if we've got the budget for it.

Once we set some dates, I'll probably get Holly onto taking care of the hotel booking and logistics, etc.  But first I want to hear back from you, then I'll start testing the waters  more broadly within PIAF.

By the way, did you know you can be found on YouTube?  I found you as I was looking up pragmatism talks!

\section{01-12-09 \ \ {\it Maybe the Best Rejection I'll Ever Get} \ \ (to G. L. Comer)} \label{Comer127}

Since you sent me an early Christmas present (a record in a way of your achievements this year), let me send you one of my own.

I just found out that my big paper of the year got rejected from {\sl Rev.\ Mod.\ Phys}.  Three reports---three all glowing about the content, but three rejecting the style of presentation and two declaring it thusly completely inappropriate for RMP.\footnote{\editornote The joke's on them. See C.\ Fuchs and R.\ Schack, ``Quantum-Bayesian Coherence,'' Rev.\ Mod.\ Phys.\ {\bf 85}, 1693--1715 (2013).}

Still by any standard, I've got to look on this as my greatest rejection ever.  Look at what Referee \#3 wrote in particular:
\bq\noindent
   This is as strikingly novel a way of looking at the
   quantum theory as Feynman's sum of amplitudes over
   histories was when it appeared in the late 1940s.
   Whether it will prove as fruitful and durable remains
   to be seen, but it ought to be known to a broad audience,
   and I can think of no better venue for accomplishing this
   than {\sl Reviews of Modern Physics}.
\eq
Oh well, crap.  My smart mouth got me in trouble again.

\section{04-12-09 \ \ {\it Proc.\ Roy.\ Soc.\ A} \ \ (to G. L. Comer)} \label{Comer128}

\bgc
Nils and I managed to break into the general world of fluids, meaning that the article is not entirely geared for neutron star applications.  Our first {\bf Proc Roy Soc} article.  I think one of the messages is how one can treat entropy as a ``fluid'' with its own form of inertia.  The thing is that we recover traditional approaches based on the idea of heat, so the equations are ``correct'' in that sense.  For example, superfluid helium four.  Always in the back of my head is entropy as information, and what it means that entropy can be treated as a fluid.

In this work I realized that the flow of matter equations in kinetic theory has the same attribute as entropy.  What I mean is that the equations only describe a statistical flow of particles; so if someone wants to complain that entropy is not ``real'' (i.e., as a practical matter, can be treated as a fluid with inertia) then I would respond that the matter flow has the same character.  At gut level, what one is really determining with matter flow equations is how the probability distribution for the particles evolves and changes from one point to the next.  So, matter, entropy, heat, their equations have more to do with accounting based on some statistic.
\egc

Sounds very nice.  Congratulations!

I thought of you when I wrote the little line on neutron stars in my upcoming {\sl Physics in Canada\/} (kinda' their version of {\sl Phys Today}):
\bq\noindent
In these days of unceasing media coverage for the H1N1 virus, we are reminded more than usual that a healthy body can be stricken with a fatal disease that to all outward appearances is nearly identical to a common yearly annoyance: simple seasonal flu.  Different though the subject matter be, there are lessons here for quantum mechanics.  In the history of physics, there has never been a healthier body than quantum theory; no theory has ever been more all-encompassing or more powerful.  Its calculations are relevant at every scale of physical experience, from subnuclear particles, to table-top lasers, to the cores of neutron stars and even the first three minutes of the universe as a whole.  Yet, since its founding days, there has been a fear in many physicists that the yearly annoyance may one day turn fatal:  It is the perennial unease that something at the bottom of the theory does not make sense.
\eq
I liked your answer about ``matter.''  Indeedle-dee-dee, I think.

\section{07-12-09 \ \ {\it Eric's Note, 2}\ \ \ (to E. G. Cavalcanti)} \label{Cavalcanti8}

\bec
I found this message gathering some dust in my mailbox.
\eec

Taking your cue, I think I'm going to have to let your own note gather some dust in my mailbox.  I just don't have time for a proper reply at the moment.  I'll try to come back to you more properly after the New Year!

\bec
I am still very confused about this, even after reading your paper and listening to your talk. If this ``dressing'' of raw probability theory has an empirical component, doesn't it have a fundamentally distinct origin as the rest of probability theory? The extra structure that is added by quantum mechanics seems to have, in this view, a contingent nature, dependent on the specific ontology or the laws of nature of our world, and it is not a pure ``law of thought'' as plain-vanilla Dutch-book coherence.

I mean, I could, in a very similar way, propose a theory of ``gravitational decision theory'', which claims that, on top of Dutch-book coherence, we should add a normative principle about gambling on the outcomes of experiments involving the relative movement of massive objects. But really, the best way of thinking about this, of course, is that Dutch-book coherence is all there is to probability theory, and anything added on top is just an instance of a good-old law of physics. In this case it would be a category mistake to confound the two. What is essentially different in your view that makes it not the same type of category mistake?

Ok, I will partially answer my own rhetorical question: underlying the probabilities in the example above is a deterministic, noncontextual hidden-variable theory. And now comes a non-rhetorical question: is there something about probabilities which do not allow noncontextual hidden-variables that makes your kind of position immune against the category-mistake criticism above?

On another note, another important feature of quantum probabilities is that they refer to {\bf fundamentally} counterfactual (or complementary, in Bohr's words) measurements, as opposed to the classical ones which could always in principle be performed simultaneously (and complementarity is a necessary---perhaps sufficient?---condition for the failure of noncontextual hidden variables. That is, in a world where all measurements can be performed at once there is always a noncontextual hidden variable model). Do you think perhaps you could say something along the lines that the extra normative assumption you propose can be justifiably called a normative rule of decision theory (as opposed to a law of nature) in a world where complementarity exists?
\eec

In the meantime, I know there is a lot of repetition with what you have already seen, but let me recommend these two presentations:
\begin{itemize}
\item
\pirsa{09090087}
\item
\pirsa{09080018}
\end{itemize}
I explicitly address some of your questions about the interaction between normative rules and physical law there.

Hope things are going well for you.

I certainly liked reading this line of yours:
\bec
And perhaps even Leibniz's monadology can accommodate something like
the ``republican banquet'' you (and I!)\ have in mind.
\eec

And finally let me give you a hint of an answer to this:
\bec
\bq\noindent\rm
[CAF wrote:]
The other night {\Ruediger} and I were semi-jokingly saying that the
dimension $d$ of a quantum object quantifies how much ``will'' it has.
But, maybe some variant of that will not turn out to be so much of a
joke after all.
\eq
That sounds interesting. Can you expand on that?
\eec

See attached excerpt from ``My Struggles \ldots''  [See 16-10-09 note ``\myref{Hardy38}{The More and the Modest}'' to L. Hardy.]

\section{10-12-09 \ \ {\it Lawless World / Malleable World / Pluriverse}\ \ \ (to N. Cartwright)} \label{Cartwright1}

I was glad to see today that Steve Weinstein could send you an invitation to our meeting on laws.  (I am only a minor player in organizing this meeting.)  I thought I might tell you independently how much I really hope you can come.  Sad to say, I've only recently discovered your work, but that said, now that I know it, I am greatly enthusiastic about it.

I offer evidence that our thoughts on laws, etc.\ are in some sympathy with a few pieces of my own writing (see attached):
\begin{itemize}
\item[1)]	The introduction to my Cambridge U Press book {\sl Coming of Age with Quantum Information}---you might find some of my early life (especially my distrust of ``laws of physics'' at the age of 12 or so) reported in the first few pages amusing.  You might also enjoy some of my stories in there about some people you surely know---for instance, Abner Shimony and Jeff Bub.

\item[2)]	An excerpt from my samizdat {\sl My Struggles with the Block Universe\/} where I describe my extreme empiricism and how I am starting to see the Hilbert-space dimension of quantum systems as a ``capacity'' in a sense (I think) quite similar to your own. [See 16-10-09 note ``\myref{Hardy38}{The More and the Modest}'' to L. Hardy.]

\item[3)]	The beginning of a paper I'm writing for {\sl Physics in Canada\/} describing my ``Quantum Bayes\-ian'' program for interpreting quantum mechanics.  Though, I don't have the section on ``Hilbert dimension as capacity'' written yet, it will give you some introduction to the way I am starting to think of quantum mechanics in ``normative terms,'' and maybe you can use your imagination on how to fill in the last sections.
\end{itemize}
Anyway, I hope these things whet your appetite a bit---i.e., to see that there is a (practicing) physicist who very much believes that there is no law that cannot be broken, that the world is malleable, and that the concept of a capacity is just what the doctor ordered for understanding quantum mechanics.

Of all our participants, I feel I could learn the most from a few well-placed conversations with you.  I very much hope you can come.

\section{14-12-09 \ \ {\it Invitation to Lecture} \ \ (to J. Emerson)} \label{Emerson3}

If I understand correctly, my lectures are March 9 and 11.  Assuming that's correct, I've marked my calendar.\medskip

\noindent Title: Quantum Bayesianism at the Perimeter \medskip

\noindent Abstract: Theses lectures give a summary of the Quantum Bayesian point of view of quantum mechanics, originally developed by D. M. Appleby, H. Barnum, C. M. {\Caves}, A. Peres, R. {\Schack}, myself, and others.  It is a view that arose from and depends crucially upon the tools of quantum information theory.  Work at Perimeter Institute continues the development and is focussed on the hard technical problem of a finding a good representation of quantum mechanics purely in terms of probabilities, without amplitudes or Hilbert-space operators.  The best candidate representation involves a mysterious entity called a symmetric informationally complete quantum measurement.  Contemplation of it gives a way of thinking of the Born Rule as an addition to probability theory, operative when one contemplates gambling on the consequences of one's interactions with a new universal capacity:  Hilbert-space dimension.  (Newton gave us gravitational mass as a universal capacity; with hindsight one can say that the founders of quantum mechanics gave us another universal capacity.)  The lectures end by showing that the egocentric elements in this point of view represent no impediment to pursuing quantum cosmology and outlining some directions for future work.

(I stripped both title and abstract for a {\sl Physics in Canada\/} article I'm just finishing this week.)

\section{18-12-09 \ \ {\it Sorters and Jokes}\ \ \ (to D. Bacon)} \label{Bacon6}

Hey, you're a man who likes a good stiff joke.  Tell me whether a couple of my anti-Everett snide remarks hit the right tone in this paper I'm writing at the moment.  See attached.  Number 1 can be found in the last paragraph just before the section ``Quantum States Do Not Exist'' begins:
\bq\noindent
The Everettian world, on the other hand, purports to have no observers,
but then it has no probabilities either. What are we
then to do with the Born Rule for calculating quantum
probabilities? Throw it away and say it never mattered? It
is true that much effort has been made by the Everettians
to rederive the rule, but to many in the outside world, it
looks like the success of these derivations depends upon
where they are assessed:  for instance, be it Oxford [8] or
Cambridge [9] instead.
\eq
Number 2 is in the mid paragraph of the right column of page 3; scan for the word ``lizard'':
\bq\noindent
The point is, far from being an appendage cheaply tacked
on to the theory, the idea of quantum states as information has a simple unifying power that goes a significant way
toward explaining why the theory has exactly the mathematical structure it does. By contrast, who could take
the many-worlds idea and derive the structure of quantum
theory back out of it? This would be a bit like trying to
regrow a lizard from the tip of its chopped-off tail: The
Everettian conception never purported to be more than a
reaction to the formalism in the first place.
\eq

\section{29-12-09 \ \ {\it \ldots\ And Happy New Year} \ \ (to H. C. von Baeyer)} \label{Baeyer86}

\noindent \ldots\ or, better, Happy New Decade

Thank you for your nice Christmas Day note.  I'm sorry to say it has taken me this long to reply to you.  The Saturday before Christmas I started coming down with (I guess) a sinus infection, general congestion, deep cough, chest congestion, muscle aches, etc.  At first I thought it was the flu (perhaps brought on by the flu shot I got the day before), but since I never developed fever, my guess now is that it was just probably a badly placed bacterium.  Anyway, I have been very slow to get over it, and the thoughtful/philosophical/reading-filled break I was hoping to have from PI is now starting to disappear.  (Don't get me wrong, the holiday with the family was still quite good and satisfying family-wise; I just haven't had the thought-filled time I had planned for.)

I had dreams of writing you and Marcus all kinds of stuff on the ``neutral stuff'' that I'd like to start thinking of ``quantum events'' (called ``measurement outcomes'' when one starts to think of one term as an agent) as exemplary of.  It was a theme of thinking I started last year at this time and has been in the background of my thoughts since \ldots\ trying, I can tell, to solidify into some kind of articulable entity.  Though it is not there yet, at least a little progress has been made.  If I could just put what little bit I've got onto paper, I know that would accelerate the process.

As a poor substitute, let me attach another draft in progress.  It at least helps lead up to the point I want to get to in personal conversation \ldots\ in ways perhaps that I have not articulated before.  Even though incomplete, I hope you enjoy.  [A few warnings:  1) Figure 2 is not quite right; that should be $Q(D_j)$ in the box---it's an old picture and will be redrawn.  2) Most of the exposition is stabilized up until the last two paragraphs of the ``capacity'' section.  Thereafter, all bets are off, including the references I cite.  3) If you print this, the hand-drawn illustrations, until better cleaned up, will come out much better on a color printer.]

Maybe too, I'll leave you with an intellectual challenge if you're looking for a little diversion.  I'll give it the code name, ``What Did He Read and When Did He Start Saying It?''  Pauli, as you know, used the terms ``psychophysically neutral,'' ``neutral language'' and ``neutral'' itself quite often.  But where did he pick the term up from?  Is it a self-invention?  Or did he pick it up from Jung?  Or in reading William James or Bertrand Russell or still someone else?  Did Pauli read Russell's book {\sl The Analysis of Mind\/} (1921) where the word is splattered everywhere?

James himself has these two passages in ``Essays on Radical Empiricism'':
\bq\noindent
     First of all, this will be asked: ``If experience has not `conscious' existence,
     if it be not partly made of `consciousness,' of what then is it made? Matter
     we know, and thought we know, and conscious content we know, but
     neutral and simple `pure experience' is something we know not at all. Say
     {\it what\/} it consists of --- for it must consist of something --- or be willing to
     give it up!
\eq
and
\bq\noindent
     In `Does Consciousness Exist?' I have tried to show that when we call an
     experience `conscious,' that does not mean that it is suffused throughout
     with a peculiar modality of being (`psychic' being) as stained glass may be
     suffused with light, but rather that it stands in certain determinate relations
     to other portions of experience extraneous to itself. These form one
     peculiar `context' for it; while, taken in another context of experiences, we
     class it as a fact in the physical world. This `pen,' for example, is, in the
     first instance, a bald {\it that}, a datum, fact, phenomenon, content, or
     whatever other neutral or ambiguous name you may prefer to apply. I
     called it in that article a `pure experience.' To get classed either as a
     physical pen or as some one's percept of a pen, it must assume a {\it function},
     and that can only happen in a more complicated world.
\eq
Did Pauli, by chance, read this book?  It's not a very well written article, but one easy source for you to find some other literature that Pauli might have read is the Stanford Encyclopedia article on neutral monism: \myurl{http://plato.stanford.edu/entries/neutral-monism/}.

Anyway, there are the few thoughtful thoughts I can muster this morning.  For the rest of the day my alter ego must come out and work on Marcus's latest masterwork.  I'm giving it my final run-through, and with some luck we'll get it posted before the week is out---it should have been posted two weeks ago, except for my dragging things on like this.  I'm just trying to catch typos at this stage, but I've got nearly 30 pages to go, and I'm finding it particularly taxing in this condition.  In case you need a diversion for your brain's other hemisphere, I'll attach it as well.

January 15 to 25 I'll be in Vienna giving the Zeilinger group some lectures on SICs.  It would sure be wonderful if you could be in town at the same time!  Now that's a place to walk and talk and get some insight on quantum mechanics!

All the best to you and Barbara, and really a happy New Year.

\subsection{Hans's Reply}

\bq
I'm looking forward to reading your {\sl Physics in Canada\/} piece at leisure.

In the meantime, I have done a quick check of my Pauli library to find what seems to be the first reference to ``neutral language.''  Provisionally I believe it occurs in an unpublished 1948 essay reproduced on p.\ 179 of {\sl Atom and Archetype: The Pauli/Jung Letters}. Do you have it to hand? ``Neutral'' appears on p.\ 182.

Interestingly Pauli goes on to answer your question (What did he read?)\ on the bottom of p.\ 188.  As is often the case, he throws us a curve!  Pauli does not always look to higher authority, as most of us do, and as you suggest in your list of possible inspirations.  Often he looks to {\it lower\/}  authority, by which I mean that he reads something with which he strongly disagrees, but which he then corrects and extends.  In this case it is by an author whom he batters into the ground before picking his brain.  At the end he has the priceless comment: ``Regardless of any justified criticism, however, we can again ascertain from the spontaneous manifestations of this author's unconscious that\ldots''  I love the pat on his own back implied by the word ``justified'', and the idea that the poor guy wasn't even {\it conscious\/} of what he taught Pauli.
\eq

\section{29-12-09 \ \ {\it New Year, New Decade, New Physics!}\ \ \ (to M. Schlosshauer)} \label{Schlosshauer9}

These coincidences keep happening to us, and it is a strange thing.  The very day you wrote your note below, I was literally thinking about you as I was walking home from PI.  I was thinking it would be nice to send you some of the latest foundational thoughts on quantum mechanics, jazz, nonreductionism, and metaphysical pluralism, all stirred into a soup---the time seemed ripe again.  But as I walked home, I started to feel worse and worse and found myself in the beginning of a holiday-long sinus/cold/cough/congestion/flu sort of thing, and I'm only starting to recover decently now.  BUT that evening when I opened my email at home, I saw your greeting and, indeed, noted the new round of synchronicity!

I hope you have a great coming year, and indeed a great new decade, filled with all the same things you wished for me.

It is a sorry substitute for saying something more personal, and particularly using this time to think more deeply about the context of jazz, but let me attach something that I hope to return to writing in the next few days after I really get recovered.  Most of the document is pretty well stabilized, at least up to the last two paragraphs of the ``capacity'' section.  Thereafter, I guess all bets are off.  [And warning: Figure 2 is not quite right.  That should be $Q(D_j)$ in the box---it's an old picture and will be redrawn eventually.]  Comments for improvement certainly welcome, though if my health keeps improving I hope to have it posted next week.

Sorry to say, {\Ruediger} still hasn't come through on the promised decoherence draft.   [See \arxiv{1103.5950}.]  I'm dropping in on him for a day Jan 25, on my way between Vienna and Texas, and I'm hoping to tighten the screw to a very uncomfortable level then.

And by the way, we need to come back to the idea of a PI visit this Spring or Summer if you'd still like to do it.  If you could flag potential dates, that'd be great.  At the moment, it looks like I should be able to pay all your expenses.

Anyway, once again:  Happy New Year and happy New Physics!

\section{30-12-09 \ \ {\it Gentle Rib} \ \ (to H. C. von Baeyer)} \label{Baeyer87}

\bhcvb
Like all Americans you are color-blind to the distinction between ie and ei.  The lady's name is GIESER as in Fierz (which you always get right).
\ehcvb
That's funny:  I tried to be so careful about that!  (I am aware of my shortcomings.)  I remember even looking at the cover of the book before writing my sentence ``to make sure'' I got it right.  So much for surety.

Anyway, look at Gieser's book, page 45.  There she says:  ``According to Mach direct experience is {\it psychophysically neutral}.''  Now a question on my mind is whether Mach himself ever used that kind of terminology.

BTW, I often check my spelling of Fierz too.

\section{30-12-09 \ \ {\it Happy New Year Again} \ \ (to H. C. von Baeyer)} \label{Baeyer88}

\bhcvb
Provisionally I believe it occurs in an unpublished 1948 essay reproduced on p.\ 179 of {\sl Atom and Archetype: The Pauli/Jung Letters}. Do you have it to hand? ``Neutral'' appears on p.\ 182.

Interestingly Pauli goes on to answer your question (What did he read?)\ on the bottom of p.\ 188.
\ehcvb

Thanks, I had a look at that.  From Gieser's book I find that he wrote Fierz on the subject at almost the same time, 21 August 1948.  See PLC III [971] on page 561.  It might be good to look at that.

But even earlier than that I note that Pauli writes Jung in a 28 October 1946 letter, ``It appears that Fludd's voice, which was ignored at the time, is imbued with new meaning, since for the moderns the objectifying of space had only limited validity.  The {\it neutral language of the `Blond'\/} in the dream (he did not employ such terms as `physical' or `psychic' but just talks of people who `know what rotation is' and those who do not know) seems to be reanimating that intermediary layer where the infans solaris used to be.  The modern unconscious speaks here of a `radioactive nucleus'.''

Somehow I suspect though that it really must go earlier than that, but I've got no evidence.  If you read the Preface and first 8 pages of the Introduction to E. C. Banks' book {\sl Ernst Mach's World Elements: A Study in Natural Philosophy\/} (on Google Books) you'll see his take on Mach was that he was actually espousing a neutral monism with his world elements.  And Pauli would certainly, I would think, have read Mach very carefully.  But see Pauli's letter to Jung dated 31 March 1953---there he definitely describes Mach's elements as psychical rather than neutral.  [By the way, doesn't Banks' book look fascinating?]

I might write Gieser eventually.  She seems to have noted what books precisely were in Pauli's library.  It would be interesting to know exactly which James books (if any), which Russell books (if any), etc., Pauli gives evidence of having read.

On another subject, it looks like we're going to get that {\Appleby} masterwork (now finalized) on the {\tt arXiv} this evening!  With that, my 2009 work is done and I myself can look toward the new decade!

\section{31-12-09 \ \ {\it Last Posting of the Decade} \ \ (to C. H. {\Bennett})} \label{Bennett68}

I've got to say I was jealous this morning when I saw that you had the last posting of the decade.  That's because a few weeks ago I was dreaming of myself trying to nab that coveted spot.  But I ended up pushing myself so close to the edge of the deadline that I didn't even finish the paper!  (With another paper, I did at least get position \#4 of the new year.)

Anyway, with the end of the day here, and the thing still not finished, I thought I might send you the present draft of the less fortunate one.  It might give you a smile (and some significant horror) to see what might have, with more diligence, been in the place of the beautiful reverse Shannon theorem.  (You might also particularly enjoy Figure 3.)

I hope you and Theo and all the kids and grandkids are doing well and perched for a happy and productive new decade.  Greetings to you all from Kiki, the kids, and me.  [The kids know of you from the (around our house, famous) Charlie and the Christmas-Tree Cookies story.]

\section{31-12-09 \ \ {\it QB at the Perimeter of Understanding} \ \ (to H. C. von Baeyer)} \label{Baeyer89}

\bhcvb
I have spent a very pleasant last afternoon of 2009 reading your draft.  It is, of course, perfect for me in my role as Sagredo to your Salviati and your critics' Simplicio. Since the decade will be over in a few hours, I do have a wish.  I think your argument has, indeed, stabilized, but it should not be allowed to fossilize.  I read your accounts of Bayesianism and SICs with growing understanding.  Nevertheless. I still don't have a truly internalized grasp of Figure 2.  Seeing it yet again does not help me.  So I have the hope and wish that in the new decade you will invent a new version that looks at the same issue from a slightly different point of view.  (Feynman praised the ability to recast the same problem into different  formulations.)  Maybe your visit to Vienna will result in a simplified version of what Anton will actually DO to verify equation 6 -- and what he would expect to find in a classical world.  Or maybe there is a useful analogy to the double slit experiment.

So that's the challenge:  Your words are new and fresh, your formulas are immutable, can you come up with new pictures?
\ehcvb

I will do my best, and you will probably find me consulting with you on the issue.  (Maybe I need to be like a salesman in a dress shop.)

In the meantime, let me attach the figure of the Steinberg group qutrit SIC experiment (in an abstract submitted to the QELS meeting).  It doesn't depict the meaning of the old diagram, but maybe it will give you some sense of one way of to do the experiment.

I send this to you, and then I go drink my first round of bubbles for the evening!

\subsection{Hans's Reply}

\bq
I trust the bubblies were uplifting!

Thank you for the Steinberg paper.  Maybe it shook something loose in my brain.  The attempt is to make {\it actual\/} SIC measurements, and this brings them down to Earth for me.  My trouble was precisely with the point you labeled ``unheard of in physical science'', namely the reference to counterfactuals.  These have always been vaguely mysterious to me.  First you called them measurements in the Bureau of Standards.  Then they moved to the sky.  For me this language was perhaps a little too mystical.  Now the SIC measurements have moved down to Steinberg's or Zeilinger's lab, so you might call them ``Toronto measurements'' or, narcissistically, ``Perimeter measurements.''

I guess I had clothed them in magnificent raiments, when they should have been as naked as the emperor.  They simply provide a unique universal characterization of a quantum system, usurping the role of the wave function.  Is the word counterfactual a little too suggestive here?  I mean, when I step on my bathroom scale, I am tacitly saying that ``if I were to weigh myself, counterfactually, against the kilogram in Paris, this is what I would find.''  This statement is then transferable to any scale in the world. (Of course that's what you had in mind in the first place.)  But I never use the word counterfactual in my bathroom.

I will try to find language, such as ``standard description'' or ``naked description'' or something, that I'm more comfortable with.

I'm sorry to be so obtuse and thank you for providing me with an AHA moment to start the decade with.
\eq

\chapter{2010: Jazz and the Bourbon Court}

\section{03-01-10 \ \ {\it New Decade, Three Days In}\ \ \ (to M. Schlosshauer)} \label{Schlosshauer10}

I've had a wonderful afternoon with the Hammond B3.  My three girls went out skiing, and I built a fire and put Larry, Larry, Barbara, Max, and all the Water Babies on in the background, just letting the music loop and loop.  A very relaxing and thoughtful afternoon---exactly the sort of thing I had been hoping for all vacation \ldots\ but just as PI is about to open up again!  Too bad \ldots\ but at least I got one day of it.

\bmaxs
Thank you for the attached paper. \ldots\ I was happy to see that you're including a little section on the Deutschian quantum-cosmology argument in your paper. I completely agree with you that the argument is ``nonsense.'' Deutsch's quote breathes the same spirit as those shopworn claims that it's only the Everett picture that ``takes QM seriously.'' Here's another statement by Deutsch -- you probably know it already -- that seems to fall into a similar trap:
\begin{quote}{\rm
But in fact, there is only one known interpretation of quantum theory. \ldots\ Perhaps one reason why the dichotomy between `formalism' and `interpretation' has been accepted so uncritically is that the debate has been conducted almost exclusively among theorists. Thus it has revolved around the question, `what exactly does quantum theory imply about reality?' Putting it that way can make it seem natural to try to separate the `scientific' (mathematical, predictive) core of the theory from its explanatory structure, and to keep the former fixed while adjusting the latter to one's philosophical prejudices. But no good can come of such an exercise. The formalism of quantum theory did not come out of nowhere. It is the solution of a scientific problem and, as always in science, the problem was not primarily what mathematical formula best predicts the outcomes of experiments. It was what mathematical structures correspond best to reality. If we alter the `interpretation' of the theory without regard to the second question, we can conjure up virtually any world we like. But it will not be the real world. The real world is the multiverse, and it does contain many universes.}
\end{quote}
\emaxs

And thanks for the long note, Deutsch thought, typo, and all the rest.  Most especially for the news that you and your wife will be having a baby!  (You look so young to me that I had never expected you to be married \ldots\ much less nearing fatherhood.)  As David {\Mermin} (I think!)\ told me just before our first was born nearly eleven years ago:  Nothing in your life will be harder or nothing more rewarding.  Your children will shape you just as much as you shape them.  And if you're like me, they will inspire sentences and paragraphs in your papers, and all kinds of natural philosophy in your mind.

I think I doomed myself by writing you last week, ``I hope to have it [my draft] posted next week.''  That means that, of course, the draft changed nearly none at all from the first version I sent you!  And there's no way it will now be posted this week, as---first thing---I've got to stop working on the extended version and get to work on the smaller extract (long overdue and must be gotten in to the editor this week).  But I did read through the whole thing again, tweaking things a little when I could.  One substantial point though:  There was something quite misleading in the quantum de Finetti discussion.   Since you indicated you might be reading those early parts, let me send you the present draft even though I still haven't completed the damned thing (and, as I say, probably will have to leave developing for much of the week):  I just want to divert as much potential damage as possible in your thoughts on de Finetti.

There are a few more lines at the end of the capacity and cosmology sections, but not too much.  Particularly, though, I found myself playing with Shakespeare as I was driving home from the library today, and I just had to record it.  And now apparently I just have to send it to you as well!  (I get tickled with myself when I feel I have made a good phrase.)  Somehow I will make those lines fit properly into the paper when it's all done:
\bv
These our actors, as I foretold you, were all spirits\\
And are melted into air, into thin air \ldots\\
We are such stuff as\\
Quantum measurements are made on.
\ev

At the moment, all of July seems to be open for me.  (Jan Vienna, Feb Nagoya, Mar Portland, Apr Chapel Hill NC, May nothing, Jun {\Vaxjo}, Jul nothing, Aug Leipzig possibly, \ldots)  So, let's tentatively pencil in July.  If you have any weeks to suggest, let me know, and at least that will provide some points of resistance if anything tries to intrude.  (We don't have to set them in stone yet, but points of resistance are good for helping promote well-meant plans to reality.)

I did very much enjoy (finally!) learning about the two Larry's today---I'm sorry I put it off for so long.  That is really good stuff.  Larry Young's version of Monk's Theme, for instance, really took me.  Please do give me more recommendations as you think of them.  Tell me about some of the Jimmys I should be looking out for.

Congratulations again on the baby.  That's really nice news.  What's your wife's name?  Where's she from, etc.

\section{03-01-10 \ \ {\it Slightly Better Rhyme}\ \ \ (to M. Schlosshauer)} \label{Schlosshauer11}

Well, the girls came home from their day of skiing, full of stories of grand adventure.  That went on for some hours as their mother and I poured more and more wine, eating cheese, laughing, wincing, and smiling at their stories, enjoying our everlasting pride in them.  (When I said ``three girls'' earlier, by the way, I meant their mother and the two younger ones---total of three.)  But then they left me all alone tending this fire that simply won't die!  (I have tried to speed it along, but it is as procrastinating as I am in writing papers.)  Anyway, I had nothing to do in this stuporous state but write a little more on the {\sl Physics in Canada\/} draft---and you are the guinea pig.  Attached is the result of a drunken fit:  I rearranged a rhyme!  See the Shakespearian modification!

Danger becoming one of my correspondents, it is!

\section{04-01-10 \ \ {\it Birth Notes}\ \ \ (to M. Schlosshauer)} \label{Schlosshauer12}

More later on various subjects in your note, when I need a second break!  For the moment, I'll just say I do know Bill Evans very well (love him), and let me comment on this:
\bmaxs
Thanks for your congratulatory wishes on the baby! \ldots\ The assessment of ``nothing in your life will be harder or nothing more rewarding'' is wonderful. What I like about the whole prenatal period is this exhilarating mixture of certainty and uncertainty. One knows that there will be a child added to your life in a mere few weeks. Yet there's so little one can figure out in terms of what the whole experience will actually be and feel like, until it all happens. (Excuse the crude and somewhat geeky analogy, but it's almost like with quantum measurement:  You know there will be an outcome, but you don't know what it will be.
\emaxs

It's not geeky at all.  You remember my story about Christiania?  Below are some more respectable variations on the moniker that story was about.  [See notes:  28-07-00, ``The Role of Registered Phenomena,'' to David Baker in {\sl Coming of Age with Quantum Information}; 03-09-01, ``\myref{Nielsen1}{Subject/Object},'' to Michael Nielsen; 20-09-01, ``\myref{Wootters2}{Praise, Folly, Enthusiasm},'' to Bill Wootters; 19-07-03, ``\myref{Mermin99}{Definitions from Britannica},'' to David {\Mermin}; 18-05-04, ``\myref{Woodward2}{Agents, Interventions, and Surgical Removal},'' to Jim Woodward; 17-06-04, ``\myref{Mabuchi12}{Preamble},'' to Hideo Mabuchi; 27-02-09, ``\myref{Hardy36}{Vivienne and the Universe},'' to Lucien Hardy.]  My opinions have certainly changed on some things since the earliest notes, but the common thread, that quantum measurement and birth are identical in a very deep sense, seems to have weathered the storms.

Thanks for giving me the impetus for this little stroll down memory lane.  I once read a book by Ludwik Fleck, {\sl Genesis and Development of a Scientific Fact}.  It was mostly a study of how the diagnosis and treatment of syphilis developed over the years.  What was so interesting to me in it was how the science that developed did so because, long before there was any science at all to the matter, and long before anyone had a right to think so, syphilis came to be viewed as a ``blood condition.''  And this of course suggested the idea of blood tests for syphilis, and treatments that had nothing to do with the outward symptoms, etc.  You're wondering why I'm telling you this.  Well, in this stroll down the memory lane of birth, creation, and measurement, I started thinking how I had thought I was writing something fresh in this new paper of mine!  Calling Hilbert space dimension a capacity, etc., etc.  {\it But no}, I was calling it a capacity already in 2004, long before I had a right to think so, and particularly long, long before Eq.\ (9) in the present paper was even dreamed of!

No lesson here yet, but it is a curiosity to me.

\section{05-01-10 \ \ {\it Three Attachments \ldots\ No, Four}\ \ \ (to M. Schlosshauer)} \label{Schlosshauer13}

Thanks for your musings on the psychology of quantum foundations.  It is a vexing issue, and one that I come to in my own head from time to time.  {\Appleby} and I also discuss it quite often, usually in frustration over something Rob {\Spekkens}, or Harvey Brown, or Jeremy Butterfield, or Matthew Donaldson, has said---it seems those four are very often the foci of our conversations on these matters.  They are our friendly foils.  {\Appleby} gets particularly riveted---you should see him in action---going on about how they desire a dead universe or how they can't possibly {\it really\/} believe what they say.  He claims that if one {\it really\/} believed in a deterministic world (despite one's going about life in a normal way), it would be tantamount to insanity; so he thinks deep down they're all speaking lies.

Sometimes, I tend to think the rationalists derive a kind of religious comfort from determinism.  It takes all the responsibility off their shoulders at some very deep level.

I think you hit the nail on the head when you talk about the influence of the PhD advisors---certainly in the vast majority of the cases.  Of course, I feel immune on this issue myself, as I tried to make clear in my ``How to Stuff a Wild Samizdat,'' but I do really think it true for the vast majority of the cases.  There was one Bohmian postdoc we had here at PI (actually you saw him speak at a PIAF meeting in Sydney when I first met you) who very clearly could have fallen into any line of physics if his PhD advisor had just been a different one.  And when he was having trouble finding his next position, I told Kiki several times (she was concerned because she knew the wife) that, ``If he had just had a better luck of the draw field-wise \ldots''  Anyway, I definitely worry about this with my own students:  I almost wish they would disagree with me more and try to articulate reasons for holding their ground.  It worries me that they sometimes seem to accept QBism a little too easily.

\bmaxs
This desire might be so ingrained in us, not only
intuitively but also in terms of the general
scientific mindset, that anything that departs from
pursuing a deterministic program is akin to giving
up on of the very ideals of science itself. (I bet
you've heard people say that about your program!)
\emaxs
Indeed I have!  Many times over.  Attached is a report to Asher Peres of the first time I can recall of it---it's from {\tt Notes on a Paulian Idea}.  It was the first time David Albert beat me up (though it wasn't the worst licking he ever gave me).  If you look at the ``disclaimers'' page in NPI, you'll find these words:
\bq\noindent
     Various deletions of text have been made to the
     original letters. The purpose of the vast majority
     of these is to spare the reader of the ``merely
     personal'' in my life. A smaller fraction are for
     the sake of protecting the innocent, protecting the
     correspondents, and protecting the illusion that I
     am good-natured.
\eq
Well, I intentionally left one personal insult in that document (and one by accident).  The intentional one was about Albert.  (I'll tell you about his worst thrashing of me the next time we see each other---it's a story I don't feel like writing at the moment.)    [\ldots]  And, yes, I rearranged the Shakespearian thing again---this time I finally got it right, and will make it a piece of the ending of the paper.

\bmaxs
Basically, the upshot is this. Genomics tries to
make us believe that life is basically nothing but
DNA.
\emaxs
And funny you bring that up; I used DNA among my examples today in the ``nothing but'' section of the paper.  In the PiC draft, do a search on the word ``Carbon''---you might enjoy some of things in my list that DNA has characteristics in common with.

\bmaxs
How about yourself? I know --- from your Samizdats --- that
your wife's (nick-?)\ name is Kiki and that you have two
daughters. But that's about it.
\emaxs
Yep, Kiki is a nickname.  When she was a baby, her older brother couldn't pronounce Kristen, and it stuck.  The happy story of how I met her is attached (written in a larger context of a rather sad story).  [See 16-02-02 note titled ``\myref{ShouldNotPass}{Some Things Should Not Pass}.''] Her present career is refurbishing this rotten old house, but she was an elementary-school teacher in the past (until Emma, aka 11 years ago).  Final attachment(!)\ is a story on the interplay/intersection of our two careers.

OK, I should stop boring you!  (And certainly don't read any of that crap if you don't feel like it.  It's just my long habit of regurgitating stories---sometimes I can't stop.  I'll not be bothered at all if you've got better things to do.)

\subsection{Max's Preply, ``Randomness, Temperament, Genomics''}

\bq
Thanks so much for these very thoughtful and thought-provoking musings about birth. They're very inspirational, in the best sense of the word.
(I.e., they're the proverbial opposite of the ``inspirational'' associated with the tacky tele-evangelist. Not that I'd ever confuse you with a member of that creed.)

Many thoughts came to my mind when I was reading these notes of yours.
Two of them really stuck out. They're both, in a way, personal contemplations about why other people hold certain opinions, or, in particular, arrive at particular conceptions of the universe (or quantum mechanics, for that matter). I hope you don't mind if I just throw them at you. They probably amount to hardly anything new.

The first one is -- and it's something that's occupied me for a while -- the issue of determinism. During the era of classical physics, determinism may have seemed like an inevitable consequence of what science was telling us. And people -- philosophers, scientists, and lay persons -- tried to come to terms with this development. They did so either by trying to find an escape route from the confines of the block universe, or by accepting determinism while still trying to seek out a niche where our intuition about an open future that's in continual creation could continue to exist in some way.

But then quantum mechanics came along. And suddenly we had the strongest indicator (not proof, of course) of some form of intrinsic
(``irreducible'') randomness -- indeterminism -- that one could possibly dream up. (Really, could you imagine a {\it better\/} way? So simple, so abstract, so mind-boggling, and so far-reaching?)

And what seem many, if not most, people working in foundations these days engaged in? To me, it seems to be programs geared toward restoring determinism, may it be in form of Bohmian trajectories, a global Everettian ``quantum universe,'' or whatever. I'm not inclined to judge these approaches, I simply and plainly find them puzzling.  And so I wonder (a lot) why so many people pursue such programs.

I have no good answer. Right now, I can mainly think of one reason, namely, that the ideal of science itself all along has all been one of determinism. Not just one of causality, but one of linking the occurrence of event A to the definite consequence of event B. This desire might be so ingrained in us, not only intuitively but also in terms of the general scientific mindset, that anything that departs from pursuing a deterministic program is akin to giving up on of the very ideals of science itself. (I bet you've heard people say that about your
program!) I know I'm probably not saying or observing anything new here; I'm just trying to make some sense of the sociology of foundational scholarship.

The other thought is actually something I was contemplating again earlier today, and your note struck the same chord. (Another one of those coincidences, it seems.) More precisely, it was your quote of James (a quote that I highlighted, {\it and\/} later marked in the margins, when I was reading James' book):
\begin{quote}
The history of philosophy is to a great extent that of a certain clash of human temperaments.  Undignified as such a treatment may seem to some of my colleagues, I shall have to take account of this clash and explain a good many of the divergencies of philosophies by it.  Of whatever temperament a professional philosopher is, he tries, when philosophizing, to sink the fact of his temperament. Temperament is no conventionally recognized reason, so he urges impersonal reasons only for his conclusions.  Yet his temperament really gives him a stronger bias than any of his more strictly objective premises.  It loads the evidence for him one way or the other, making a more sentimental or more hard-hearted view of the universe, just as this fact or that principle would.  He {\it trusts\/} his temperament.  Wanting a universe that suits it, he believes in any representation of the universe that does suit it. He feels men of opposite temper to be out of key with the world's character, and in his heart considers them incompetent and `not in it,' in the philosophic business, even though they may far excel him in dialectical ability.

Yet in the forum he can make no claim, on the bare ground of his temperament, to superior discernment or authority.  There arises thus a certain insincerity in our philosophic discussions:  the potentest of all our premises is never mentioned.  I am sure it would contribute to clearness if in these lectures we should break this rule and mention it, and I accordingly feel free to do so.
\end{quote}

If you look back at the set of questions I sent you for the interview book that I'm putting together, there was one question that asked precisely about this issue of predisposition, of temperament, of philosophical prejudice, with respect to one's choice of foundational program. This influence may start very plainly, very simply, at the level of one's PhD supervisor. As you may have observed yourself, there are many people in foundations who work on a particular interpretation that ``happens'' to also be the favorite interpretation of their adviser. But of course it goes much deeper than that. A Bohmian may not be a Bohmian simply because he likes so much the form of the guiding equation and the double-slit imagery of trajectories, but also because it fits a more deeply rooted fundamental affinity toward a classical particle-based world view. Again, I'm not stating anything new here, but I think that some of the emotional debates in foundations would have some chance of leveling off to some extent if everyone just clearly stated their philosophical convictions and these convictions' influence on their (i.e., the people's) attitudes.

This is a very long-winded way of saying that I'm very fond of the birth imagery in quantum mechanics. If I see a lifeboat that has promise of staying afloat, and may actually take me to a new island, why hold on to the old sinking ship of determinism?

OK, I think I'm entering the ``rambling'' territory, so I better stop before this ship is drifting even more off course.

PS. Against all better intuition, here's another off-topic addition to the theme of our conversation. Last night I finished reading a little book on Darwin, genomics, and the ``book of life.'' (The book is in German, but it's not really relevant anyway.) Basically, the upshot is this. Genomics tries to make us believe that life is basically nothing but DNA. What we are, what makes us human, is encoded in this sprawling sequence of ACGTs. Figure out the sequence, and what proteins it codes, and you've got it all. The author of the book argues, quite persuasively I thought (mostly by appealing to recent scientific evidence, but also via more philosophical insights), that this is a gross oversimplification, and that we ought to be ever-cautious of such simplifications. In a very vague sense, this reminded me of the debate in quantum mechanics: how we try to read the formalism as reality, as describing everything there is, thus taking reductionism to its extreme. (Witness the Deutsch quote I sent you.) In this terribly vague and surely oversimplified way, quantum states strike me as akin to the ACGT of molecular biology. Both are {\it our\/} ways of dealing with the world, of deciphering it and coming to terms with it and trying to manipulate it. But they are not the world (or the human, or life) itself.
\eq

\section{05-01-10 \ \ {\it That Thoughts Matter}\ \ \ (to M. Schlosshauer)} \label{Schlosshauer14}

I worked hard to find something in Marcus's email corpus that would instantiate/corroborate what I told you of his reactions to determinism-seeking psychologies.  I don't feel I've been all that successful (I'm sure there's something better out there), but I give up for the time being.  Anyway, attached are two things:  1) A draft of an article Hans Christian von Baeyer was writing at the time, and 2) Marcus's reaction to it.  It's the latter that I wanted to exhibit to you, but I included the former for completeness.

Marcus's writing is in PauliNotes.doc.  For entertainment, you might pick about at page 6 where he starts talking about believing positivism as having the potential to lead to suicide.  And then again at the bottom of page 8, where he asks ``why people find it so hard to accept that thoughts actually matter.''  Oh, here's a choice one:  Look on page 10 where he writes, ``Materialists affect to believe---believe themselves to believe---that we are nothing more than complicated machines.''  Thereafter, there's some interesting stuff up to page 13 at least.  Actually, there's further good stuff throughout.  Just food for thought.

Now, maybe, I go to bed!!

\section{05-01-10 \ \ {\it Importance of Humanism and Dogs}\ \ \ (to M. Schlosshauer)} \label{Schlosshauer15}

Wow!  Maybe even more than for the music, let me thank you for this!  Particularly for the parts between the 1st and 2nd diamonds:

\bmaxs
I was very pleased to see that you gave an explicit run-down of the Wigner's-friend
scenario (page 6). I think most readers will agree that the worry over an assignment
of different quantum states by Wigner and his friend is dissolved in the personalist
framework.

Still, your argument may not---at least not at this point of your paper---dissolve {\it all\/}
the concerns people get tangled up in when thinking about this thought experiment.
What they might continue to worry about may not have so much to do with quantum-state
assignments but with ``what is really out there.'' On the one hand, Wigner's
friend is believed to have seen one outcome. On the other hand, Wigner could, in
principle, perform an experiment demonstrating interference effects between different
outcomes. So the casual reader will ask: ``How can it be that the friend has observed
one outcome, but Wigner can do an experiment that would seem to require that no
single outcome has occurred?'' And furthermore, ``Is this mystery, after all, reducible
to the subject of quantum-state assignments? Or do I need to go further and embrace
a different worldview, like the one you outline in the `Capacity' section? Or do these
two questions come down to the same thing?'' You do point to this new mindset
already on page 6:
\bq\noindent
The point of view here is that a quantum measurement is nothing other
than a well-placed kick upon a physical system---a kick that leads to
unpredictable consequences for the very agent who made it.
\eq
So things {\it should\/} be clear, worries dissolved, even for the casual reader, even at this
point of the paper---but maybe they're not, and I wonder if the argument on page 6
could therefore be made even more forcefully? \`A la, ``This outcome is {\it mine\/} only. If
{\it you\/} want an outcome too, you have to come over and elicit one for yourself.''

Just a thought.
\emaxs
I had already had a nagging fear that I wasn't doing that part quite right, and you've pushed me over the edge.  It's true I need to do that better.  This sentence:  ``This outcome is mine only. If you want an outcome too, you have to come over and elicit one for yourself.''  Is that one of mine from somewhere?  Or is it your own?  Either way, I love it and definitely want to use it!  You are also quite right in the parts between diamonds 2 and 3:
\bmaxs
In the context of discussing Wigner's friend, you also mention unitary evolution and
describe it as something that adjusts ``the beliefs with the
flow of time.'' I think it would be useful to elaborate a little bit on this issue.

After all, there will be people who, just as they've opened their minds and hearts
and have readied themselves to embrace a personalistic notion of quantum states,
are suddenly stopped dead in their tracks by asking themselves why, in the world,
their personal beliefs should change (or be updated) according to the {\Schroedinger}
equation. They will point to the usual text-book understanding of unitary evolution
as being given by the Hamiltonian, which in turn is usually thought of as something
objective defined by the physical situation under consideration. And thus they will
perceive a clash between subjective quantum states and the supposedly ``objective''
evolution of these states.

So maybe it would be helpful for the reader to hear exactly what the status of unitary
evolution in the QBist framework is, and how and why consistency of the subjective
viewpoint requires one to understand evolution operators (and, generally, quantum
operations) as subjective entities, too.
\emaxs
I should put more about unitary evolution there (and/or also in the SIC section).  At this point in the ``extended'' version (whatever it means for it to be ``extended'' over something when it is the original draft, and wherever it is going to go), there is not much point in being overly economical with words.  [That, by the way, was my original reason for not broaching time evolution in any detailed way.]  I'll fix it!  Your first comment, I will probably pass on and take my chances:
\bmaxs
Page 2:
\bq\noindent{\rm
The trouble with all these interpretations as quick fixes for Bell's hard-edged
remark is that they look to be just that, really quick fixes. They
look to be interpretive strategies hardly compelled by the particular details
of the quantum formalism, giving only more or less arbitrary appendages
to it.}
\eq
I certainly agree with this assessment. However, I wonder what an Everettian would
say in response to your charge that his interpretation depends on ``more or less arbitrary
appendages'' to the formalism. I imagine him exclaiming, ``My interpretation
is the most economic of them all. In fact, I don't need anything else besides the bare
formalism. So what are these appendages you're talking about?'' You mention the
need of making sense of probabilities, but I suppose the Everettian might dispute to
what extent this need constitutes an ``appendage,'' rather than just an exploration
of the consequences and features of the formalism itself. Similarly for the ``many
worlds''---to an Everettian, they're not an ontological appendage, but a consequence of
the formalism.
\emaxs
Trying to beat off a committed Everettian is like opening a can of worms or Pandora's box---it's not something that can be won.  (Too many conversations, too many experiences \ldots\ this is exactly the sort of thing we end up talking about when Marcus and I get riveted over Harvey Brown.  The gulf is deeper than quantum interpretation, it's whole worldview definitely.)  I'll just remain with my well-placed insult about Cambridge vs Oxford.

The sentence may not be quite right as it is presently, but it was meant to be ``bald'' (naked, bare, etc.).

Anyway, I really, really, really appreciate these comments.  They were very constructive for me.  I'll try not to let you down.

You know, your understanding of this business is certainly getting to the point where you could express aspects of it in your own voice and probably do a much better job than me.  I was serious before, but I'm more serious now:  You might try your hand at telling your own version of the story, or how it relates/contrasts to Everettism, etc., being critical and pointing out all the holes in QBist thinking wherever need be.  Understanding criticism is far, far better than un-understanding criticism, and there is no doubt we QBists need continued stretching:  Life is, and should be, a struggle.

BTW:  Do you ever have any interaction with Ulfbeck there?  I have always had some sympathy with this sentence from one of his papers with Aage Bohr:
\begin{quote}
The click with its onset is seen to be an event entirely beyond law. \ldots\ [I]t is a unique event that never repeats \ldots\ The uniqueness of the click, as an integral part of genuine fortuitousness, refers to the click in its entirety, with all the complexity required for a break-through onto the spacetime scene.  \ldots\ [T]he very occurrence of laws governing the clicks is contingent on a lowered resolution.
\end{quote}
From my one meeting with him, I suspect he is a recalcitrant sort, but there are definitely some interesting elements in his thought.

\section{05-01-10 \ \ {\it If You Take Quantum Mechanics Seriously \ldots}\ \ \ (to M. Schlosshauer)} \label{Schlosshauer16}

\bmaxs
I imagine him exclaiming, ``My interpretation is the most
economic of them all. In fact, I don't need anything else
besides the bare formalism. So what are these appendages
you're talking about?''
\emaxs

By the way, one of the buzz phrases that bugs me the most from the Deutsch--Saunders--Wallace--Greaves set is, ``if you take quantum mechanics seriously'' (always, to my ears, delivered with immense smugness).  Those words, even when nothing further is said, so get under my skin.   Deutsch {\it et al}.\ act as if it is a completely unambiguous, value-free statement, containing no prejudice whatsoever, and I think, ``Hey, I take quantum mechanics seriously too!  {\bf I} just take it seriously in the way {\bf it} is pleading to be taken.''

Attached are three slides from my (awful) talk ``13 Quotes from Everettian Papers and Why They Unsettle Me'' on {\tt pirsa.org}.  It really is an awful talk---quotes far too long, printed too small; it was duller than anything else I've ever done, and a lot of people fell asleep---but it did have a couple of zingers in it that I remain proud of \ldots\ and I guess one of them is the title!  One thing of interest with regard to yesterday's discussion is that I just found out by reviewing the first five minutes of it that I presented it as ``a case study in psychology''---I had forgotten that.

Read the slides in the order:  ``Seriously,'' ``Saunders,'' ``Mirror-World.''  The last slide could have been done much better and made more effective with some work, but it says something of the point.

\subsection{Seriously}

\noindent Taking Quantum Mechanics Seriously, 1 \medskip

\bq
\noindent At the conference, the strongest ``Everettian'' was physicist
David Deutsch of the Center for Quantum Computation at University of
Oxford. Deutsch asserted that people have been wasting time for
decades debating whether or not the Copenhagen Interpretation is
meaningful because, ``Only Everett's theory consists of taking
quantum mechanics seriously.''\\
\hspace*{\fill}--- Peter Byrne, ``Many Worlds in Oxford,'' at {\tt
blog.sciam.com}\bigskip
\eq

\bq
\noindent It's sad enough when cranks churn out this tawdry old excuse
for refusing to contemplate the implications of science [{\it i.e.,
the Everett interpretation\/}], but when highly competent physicists
--- quantum physicists --- dust it off and proudly repeat it, it's a
crying shame.\medskip\\
\hspace*{\fill}--- David Deutsch, posted on {\tt
Fabric-of-Reality@egroups.com},\\
\hspace*{\fill} 16 July 2000, commenting on a paper by Fuchs and Peres\bigskip
\eq

\bq
\noindent Let quantum be quantum.
\hspace*{\fill}--- Wojciech Zurek, his talk at this meeting
\eq

\subsection{Saunders}

\noindent {Taking Quantum Mechanics Seriously} \\ (Simon Saunders version,
this meeting)

\begin{itemize}
\item
The theory is to apply universally.

\item
Without any special mention of `the observer' or `measurement'.

\item
And without any special interpretative assumptions or additional
[???] \ldots\
\end{itemize}

\subsection{Mirror-World}

\noindent Taking Quantum Mechanics Seriously \\ (Quantum Bayesian
Version)

\begin{itemize}
\item
The decision theory should be a normative ideal for all physical
agents, universally.

\item
Without any special mention of a speculative ontology behind the
agent's experiences (at least before the first necessary moment).

\item
And without any additional ad hoc structures introduced for the fuzzy
feeling of an {\it ontologized\/} dynamics (i.e., that which is
usually called ``unitarity'').
\end{itemize}

\section{05-01-10 \ \ {\it Out with the Old, In with the New!}\ \ \ (to C. H. {\Bennett})} \label{Bennett69}

\ldots\ oh I can't resist.  At first I was disappointed that I didn't get the last {\tt quant-ph} number of the last decade, but learning this morning that I got the first one of the new decade, I got quite tickled.  Maybe it means there's some hope for my quantum foundations program yet!  \ldots\ dig, dig, dig.

\section{05-01-10 \ \ {\it New Decade, Five Days In}\ \ \ (to D. M. {\Appleby})} \label{Appleby80}

I saw this morning that the toil of your paper-writing troubles turned out to be the first paper of the decade on the {\tt quant-ph} archive (not number 4 as I thought it would be).  And with that, I finally got happy about my screw-up, my long dragging of feet:  Maybe it is actually a beautiful sign that we nabbed that spot. (The last paper of the last decade stars Charlie Bennett, a devoted Everettian, as its first author.) Maybe this means that our way of looking at things, or at least the one we are trying to develop, will find a prosperous home in this coming decade.  As I'm learning from you and Hans and Pauli and Jung:  Symbols are important.

Myself, I have been slaving to write something lasting:  It started as a super-long extension of that PiC paper you saw me working on last month, but now I feel it is much more than that---that it will capture the hopes and dreams of where this program of ours can be taken and will serve as a focal point for people to develop this worldview-to-be.  I hope to unveil the paper to you and {\Ruediger}, etc., by next week:  We shall see.

But why I really write this note is not to tell you all that.  I wanted to thank you:  You have profoundly influenced me the last three years, and I thought you should know that.  What struck me particularly as I worked through your Lie-algebra draft last week is your utter fearlessness.  It is remarkable, really.  The calculations were so elementary in this case, but it was {\it you\/} and you alone who had the bravery to push them through.  You were the one fearless enough to step into the river of experience and see where it took you.   And it dawned on me, you're always like that---that is what I have come to know of you:   It permeates your whole being.  There really is some hope that we're going to do big things this coming decade, but the hope is here because you are here.

Happy New Year, Happy New Decade.

\section{06-01-10 \ \ {\it You Dog!}\ \ \ (to C. H. {\Bennett})} \label{Bennett70}

\bcb
Quantum Theory Needs No `Interpretation' Besides Ours.
\ecb

I just found this on the web while I was trying to find something nasty that David Deutsch had said about me way back (so I could make a gibe at him to Max Schlosshauer).  I find Charles Bennett instead!\footnote{\editornote The figure in question is now reproduced in C.\ A.\ Fuchs and B.\ C.\ Stacey, ``Some Negative Remarks on Operational Approaches to Quantum Theory,'' \arxiv{1401.7254}.}  With friends like you, who needs \ldots\ \  How does that saying go?

\section{06-01-10 \ \ {\it Deutsch's Proof}\ \ \ (to M. Schlosshauer)} \label{Schlosshauer17}

\bmaxs
By the way, what also irks me about Deutsch is that he
claims flat-out that interference experiments PROVE the
existence of parallel worlds -- conveniently leaving
out all the assumptions required for such a claim, not
to mention the philosophical predispositions etc.
Somehow I just don't find that, well, academically honest.
\emaxs
Yes, indeedy, I was absolutely appalled when I got the impression of that in his book.  Attached is the passage I'm thinking of (from my 13 quotes talk), and I'm pretty sure somewhere in the book he calls it a proof:
\pagebreak
\bq\noindent
{\bf Deutsch ``explaining'' one-particle-at-a-time
interference:}\medskip

So, if the photons do not split into fragments, and are not being
deflected by other photons, what does deflect them?  When a single
photon at a time is passing through the apparatus, what can be coming
through the other slits to interfere with it? \ldots

I shall now start calling the interfering entities `photons'.  That
is what they are, though for the moment it does appear that photons
come in two sorts, which I shall temporarily call {\it tangible\/}
photons and {\it shadow\/} photons.  Tangible photons are the ones we
can see, or detect with instruments, whereas shadow photons are
intangible (invisible) -- detectable only indirectly through their
interference effects on tangible photons. \ldots

Thus we have inferred the existence of a seething, prodigiously
complicated, hidden world of shadow photons.  They travel at the
speed of light, bounce off mirrors, are refracted by lenses, and are
stopped by opaque barriers or filters of the wrong colour.  Yet they
do not trigger even the most sensitive detectors.  The only thing in
the universe that a shadow photon can be observed to affect is the
tangible photon it accompanies.  That is the phenomenon of
interference. \ldots

\ldots\ Thus we have reached the conclusion of the chain of reasoning
that begins with strangely shaped shadows and ends with parallel
universes.  Each step takes the form of noting that the behaviour of
objects that we observe can be explained only if there are unobserved
objects present, and if those unobserved objects have certain
properties.
\eq
Is that what you're referring to?\footnote{\editornote One is irresistibly reminded of a scene in Carl Sagan's \emph{Cosmos} where Sagan talks about what people thought Venus was like after
they knew it was shrouded in clouds, but before they knew the ``hot enough
to melt metal'' part.  Clouds meant water, and clouds all the time over the
whole planet meant a lot of water, so Venus had to be mostly ocean.  And
if there were any land, it would be covered with marshes and swamps.  And
in the swamps would live ferns, which would be eaten by primitive
swamp-animals.  So, ``Observation: you couldn't see a thing. Conclusion:
dinosaurs!''}

Does it shock you that I made an attempt to read his book?  Well, the truth is my arm was twisted by circumstance.  Deutsch was scheduled to speak at the QCMC conference in Cambridge, MA a few years ago, and I had never forgotten something that Michael Nielsen alerted me to in 2000.  It was a chat at this place (just found it after that aside on Charlie Bennett):
\myurl{http://groups.yahoo.com/group/Fabric-of-Reality/message/17}.
Deutsch said, ``They really need to read FoR, don't they?''  So I read much of the damned thing (as much as I could stomach) with the intention of shaking his hand at QCMC, and introducing myself with, ``Hi.  I'm Chris Fuchs, and I read the Fabric of Reality.''  But he foiled me!  He accepted his award at the meeting and gave his talk by video conference.

\section{06-01-10 \ \ {\it 13 Quotes}\ \ \ (to M. Schlosshauer)} \label{Schlosshauer18}

By the way, that title was a play on Arthur Lovejoy's old essay, ``The Thirteen Pragmatisms.''

\section{06-01-10 \ \ {\it 13 Quotes, 2}\ \ \ (to M. Schlosshauer)} \label{Schlosshauer19}

\bmaxs
Aha! I hadn't spotted that. Thanks for enlightening me!
\emaxs

Well, I would have been very surprised if you would have spotted it.  I suspect I'm nearly the only physicist who's ever heard of it, and there might well be less than a handful of the (so-called) ``philosophers of physics'' who've ever heard of it themselves.  (I remember Bill {\Demopoulos} saying that he had never heard of it, or the book of the same title, even though he had great respect for Lovejoy and had read many of his books.)  Particularly, I was tickled, amused, and appalled when Chris {\Timpson} apparently felt a need to include a footnote like this, defining pragmatism!, in his paper	
\arxiv[quant-ph]{0804.2047v1}:
\bq\noindent
Pragmatism is the position traditionally associated
    with the nineteenth and early twentieth century American
    philosophers Pierce\footnote{And note the mispelling of Peirce.}, James and Dewey; its defining
    characteristic being the rejection of correspondence
    notions of truth in which truths are supposed to mirror
    an independently existing reality after which we happen
    to seek \ldots
\eq
Mind you this paper appears in {\it Stud.\ Hist.\ Phil.\ Mod.\ Phys.}---a philosophy of science journal!

\section{06-01-10 \ \ {\it New Decade, Five Days In, 2}\ \ \ (to D. M. {\Appleby})} \label{Appleby81}

May I forward this inspiring note (excluding my personal bit) to Max Schlosshauer, who I've been in a lot of contact with recently.  Max is one who has made the transition away from Everettism (after even writing a book on decoherence and all that) and is now leaning very much in our direction.  And he too is very much concerned with ``what will come next,'' now that he finally feels something WILL come next.

I see you're very late, but I guess I should expect no less of you.

\subsection{Marcus's Preply}

\bq
I felt it was significant that we were the first paper of the decade.  Though actually I don't think we need rely on Jung and Pauli and symbolism here.  I think it's automatic.  At least, I think it is automatic that Everett (Hawking, Turok, etc etc) will go out of fashion some time soon (``soon'' as measured on a generational timescale: in this context 20 years would count as ``soon'').   The only question is what will replace them?  That, of course, is a question of will.  It depends on us, and it is far from automatic.  It is in the sphere of human freedom.  In this case your freedom, and mine.

When I am faced with the dominant orthodoxy as it presents itself during (for example) lunch-time conversations in the Black Hole Bistro  I always think of the court of the Bourbons in the 1760's or 1770's.  A visitor to the French court in the second half of the eighteenth century would doubtless have been very impressed by the surface magnificence.  Superficially it might have looked as though the monarchy would last for ever.  But in fact it collapsed more or less overnight, almost without bloodshed.  There was lots of bloodshed afterwards, of course, but that was the revolutionaries fighting each other.  In 1789 the power of the monarchy simply evaporated, with no significant fighting at all.

I think something similar applies to the situation which faces us.  In conversations at conferences, the Black Hole Bistro etc etc it always seems as though you and I are in a tiny minority.  Which, I guess, we are.  But if one looks a little more closely I think one can see that the dominant orthodoxy in early 21st century physics, like the 18th century Bourbons, is supported by virtually nothing. When the time comes it will be blown away like smoke.

When I was talking to {\Asa}'s office mate (I have forgotten his name:  the post doc who came up to you, me and Howard at wine and cheese, just before you left) he strongly emphasized (a) his atheism and (b) his strong belief in Everett.  Afterwards I got to wondering what exactly was the difference between his belief in Everett and the religious belief he rejects.  He cannot claim that his belief in many worlds is better supported by the empirical evidence than a belief in God.  Everettianism, being expressed in terms of equations, may superficially look more scientific than the Bible, or the Koran.  But when it comes to the crucial test, of empirical support, it fails completely.  There is no more empirical reason for believing in many worlds than there is for believing that God created the world in seven days.  Moreover, when you ask an Everettian what their belief does rest on they will typically appeal to simplicity and beauty.  I think someone might give similar grounds for believing in religion.  Fundamentalist religion is certainly simple.  And, although religion does not appeal to the aesthetic sensibilities of the average PI resident, it surely does appeal, very strongly, to the sensibilities of the devout (and even the non-devout: many aetheists like to listen to Gregorian chants, or Verdi's requiem mass).  In these respects Everettianism is very like a religion.  However, there is one important difference.  Religious belief speaks to something deeply rooted in the human soul, and it is very tenacious.  Everettianism is an altogether more superficial psychic phenomenon.  I can't see people giving up a belief in the approximate validity of Newtonian mechanics within its domain of application, because of the weight of empirical evidence, and because of its practical utility.  Nor can I see them giving up a belief in religion any time soon, because of its deep-seated psychic attractiveness.  But  Everettianism lacks both the practical usefulness of science, and the psychic appeal of religion.  It is nothing more than a transient intellectual fashion.  Sooner or later the fashion will change, as fashions always do.

So I don't think we need worry about defeating the current orthodoxy.  That will take care of itself.  As in 1789 the important question is, not the revolution itself, but what comes afterwards.

I am looking forward to reading your new paper.
\eq

\section{07-01-10 \ \ {\it Writing Is So Very Hard!}\ \ \ (to H. C. von Baeyer)} \label{Baeyer90}

I can't get papers finished; I can't even get emails finished!  How on earth could you make writing such a large part of your career without having committed suicide by now?

Below is a note I started writing you some time back.  (It was originally titled, ``New Decade, Three Days In,'' if that gives you a hint.)  I'd like to finish writing that tonight or tomorrow, or at least pick up where I left off.  At the moment, though, I don't really want to break my train of thought of trying to reduce the big paper to the little one.  (Read the note below to see what I'm talking about.) And I thought:  If Hans is out there with nothing better to do (yeah, right), maybe I could ask your advice (with all your experience of editing).  Of course, feel free to say ``Go away pest!''\ and in the harshest terms!

Anyway, attached is the big paper and the latest partial reduction of it.

At the moment, I've managed to cut it down from 16 pages (12,217 words) to 7 pages (4,886 words).  And my understanding is, I really should reduce still $\frac{3}{4}$ of page more text-wise.  The trouble is, at this stage, I can't see where to trim it anymore!!  Already the smaller document feels lifeless, soulless to me.  But maybe as an outside (relatively uninitiated) reader you can more easily see some large chunk that can be lopped off without affecting the drama, conceit, and essential content of the message.  Maybe at this stage, at some point in the draft, I'm still fighting battles from philosophers past---battles that don't need to be aired to the general public in this venue.

Just a thought:  If you see something you can suggest to lop off wholesale, please let me know what you think.

More on counterfactuals later.

\bq\noindent
Dear Hans,

Thanks for the notes.  I'm certainly glad to give you an AHA moment when I can.  This morning, I've got to get back to work on the articles---with the ``extract'' version taking top priority.  It absolutely must be finished, or I will be in deep trouble with PI.  That article will only cover the stuff {\it before\/} the capacity section, leaving the parts less friendly to the usual physics prejudices (i.e., the thoughts of those who say, ``We're going after a theory of everything!  That's what physics is all about!'') for another venue.  Contrast that with the parts in the longer version, where I write some lines like this:
\bq\noindent
     Everything experienced, everything experienceable, has no less an
       ontological status than anything else.  You tell me of your experience, and
       I will say it is real, even a distinguished part of reality.  A child awakens in
       the middle of the night frightened that there is a monster under her bed,
       soon to reach around and grab her arm---that experience has no less a
       hold on onticity than a Higgs-boson detection event would if it were to
       occur at a fully operational LHC.  This is because the world of the empiricist
       is not a sparse world like the world of Democritus ({\it nothing but\/} atom
       and void) or Einstein ({\it nothing but\/} unchanging spacetime manifold
       equipped with this or that field), but a world overflowingly full---full
       beyond anything grammatical (rule-bound) expression can articulate.
\eq
Yep, that's not going into an issue of {\sl Physics in Canada}, true as it is, and true as every practicing physicist should recognize it to be.

Anyway, I attach the latest version of the big one for the heck of it.  The last three sections are still in transition, far from finished, but I guess mostly I send it to you because of the earlier part that you did read:  There was something quite misleading in the discussion of the quantum de Finetti theorem, and I have that fixed up now.  You shouldn't be misled any longer than you have to be!

Speaking of which \ldots
\bhcvb
They simply provide a unique universal characterization of a quantum
system, usurping the role of the wave function.  Is the word
counterfactual a little too suggestive here?  I mean, when I step on my
bathroom scale, I am tacitly saying that ``if I were to weigh myself,
counterfactually, against the kilogram in Paris, this is what I would
find.''  This statement is then transferable to any scale in the world. (Of
course that's what you had in mind in the first place.)  But I never use the
word counterfactual in my bathroom.
\ehcvb
I think this may indicate a little confusion, and I want to see if I can clear it up.  You're absolutely right that one would not normally speak in the manner of, ``If I were to weigh myself, counterfactually, against the kilogram in Paris, this is what I would find''---the point of oddity being the word ``counterfactual.''  But that is not how I am using it when I am using it.
\eq

\section{07-01-10 \ \ {\it The Question Is:  What comes next?}\ \ \ (to M. Schlosshauer)} \label{Schlosshauer20}

Since you and I have been in so much contact recently, I asked Marcus if I could forward you the note he sent me last night.  [See 06-01-10 note ``\myref{Appleby81}{New Decade, Five Days In, 2}'' to D. M. {\Appleby}.]  I thought it hit the mark, and in an eloquent way:  What comes next?  That's the only really important question.

I'm feeling doubly good today.  I feel I had a very deep insight this morning just as I was getting into the shower.  It is the inevitable conclusion to draw from this QBist trek I've been taking, but I only feel that I really saw it with clarity this morning.  In a single sentence, it goes like this:  Just as quantum theory is not in conflict with probability theory, only an {\it addition\/} to it, quantum mechanics is not in conflict with the normal, empirically perceived world of common experience (sometimes called the classical world); it is only an {\it addition\/} to it.  It is not that the classical world undergirds quantum mechanics, as I read Bohr as saying; it is not that the classical world should be {\it derivable\/} from quantum mechanics (a thought that you are very familiar and probably still quite sympathetic with).  It is that one is an {\it addition\/} to the other, just as quantum theory is an addition to Bayesian probability.  It is an overpowering feeling I have in my chest, a feeling like I've only had a few times in my career---my whole body is most definitely reacting to this thought.

Anyway, I pray (metaphorically) that I can articulate this properly in some short number of paragraphs.  We shall see---I will try my best and whatever the result, it will be incorporated into the manuscript you have been reading over.  I feel I have to say it now and not wait to write another paper.

In lesser progress, I have the distillated version down to 6 pages now.  Somehow I've got to get rid of still one more, even though it already feels like a dead, lifeless document in comparison to its parent.  The consolation is that it will be in a journal that no one reads.  But the antithesis to this is that many of people who will read this particular issue are my PI colleagues (those who wouldn't normally read me, cosmologists, string theorists, etc.)---and I would much rather them read the long version!

\section{08-01-10 \ \ {\it Oh Editorship} \ \ (to H. C. von Baeyer)} \label{Baeyer91}

Thanks again for helping me out.   I followed through on nearly all of your suggestions I believe, save maybe 2 or 3.  Some of your suggestions too, weren't just about size, but flow---I'm particularly grateful for those things, especially when they were with regard to sentences that I could just never get right.  For instance, that very first sentence of the paper---I must have rewritten it a 100 times, with all kinds of little variations.

Attached is the version of the paper I finally turned in.  It didn't cover nearly as many subjects as I had originally intended for it.  (For instance, since I knew PI people would be reading it, I wanted to make sure I said something about quantum cosmology \ldots\ but alas.)  Still, for what it does cover, I don't think it does a bad job.

Really, thank you very much for the help.  The experiences I've had with your editing have been quite new to me---they're the first to ever give me respect for the trade:  My writing is visibly better after your input.  I would be hard pressed to think of a time I've ever thought that about anyone else's input (even in isolated instances).

Funny too, that this has happened at a time when I had read this article in the {\sl New York Times\/}:
\myurl[http://www.nytimes.com/2010/01/03/opinion/03galassi.html?scp=1&sq=electronic\%20books&st=cse]{http://www.nytimes.com/2010/01/03/opinion/03galassi.html?scp=1{\&}sq=electronic{\%}20\\ books{\&}st=cse}
on the value of a good editor.

Tomorrow, I'll try to write you an extended version of the clarifying stuff on counterfactuals if you still need it.  For the moment, I'll just write the definition from {\tt dictionary.com\/} and put it into context: {\it Counterfactual: a conditional statement the first clause of which expresses something contrary to fact, as ``If I had known.''}

``If I had thrown the system up to the sky and let it cascade through that device back down to the one on the ground, I would have gambled this way (usual Law of Total Probability).''  But that is contrary to fact:  I do not actually throw it up to the sky, I throw it directly to the one on the ground.  Quantum theory says,  ``Don't waste your old calculation.  Just stretch it a little, and subtract a bit off afterward (Eq.\ 5 in the present draft).''  The claim is, that's the meaning of the Born Rule---it tells us how we ought to rewrite our counterfactual probabilities.

That said, could I ask you to try to articulate what you do not like about Figure 2, or what confuses you about it?  [Recall, it's still a bit inaccurate---I haven't redrawn it yet to fix the new notation.  The quantity in the lower left box should be $Q(D_j)$ rather than $P(D_j)$.]

Have a look at the new conclusion as well.  It gives the hint that I'm going to get downright alchemical before the bigger version is all over with.  Eventually, you'll see what I mean.

Lucky Barbara, to be in Paris for as long as she can.  I've been trying to convince Emma she should go to college there so I could come visit her often!

\section{09-01-10 \ \ {\it The Duke of Chartres} \ \ (to H. C. von Baeyer)} \label{Baeyer92}

One last point to address then,
\bhcvb
So, forgive me, but I don't like your ``sky''
nomenclature any more than the NBS.  But unlike {\Mermin}
with his qbits I'm not going to fight you, as long as I
and others understand you.
\ehcvb
Indeed I wouldn't want a fight.  But I wouldn't mind suggestions either.  Following your advice, I should not fossilize.  There is nothing about my previous attempts at expression (sky, NBS, etc.) written in stone --- they were just the best that came to mind at the point.

My student Matthew Graydon once gave a presentation to our group, where the SIC was placed near the bottom of a wishing well.  You throw the quantum system down the well and, after interacting with the SIC, it is blown back up to ground level by a fan.  We all had a good giggle.

You want a more drastic change than that.  I'm open:  I just need a better understanding of how the present imagery is ineffective and some suggestions for what should be put in its place.  What part of what is going on is not being captured adequately?

At times, I've toyed with making a modification of my usual activating observer image (Fig 1 in the paper):  One frame showing him giving a 1-2 punch with his measurement-device hands, and one frame showing him giving a real whammy with just one hand, a knockout punch.  But that's probably nothing like you're thinking.

In your Swing book, I read this:
\bq\noindent
    If I were as rich as the Duke of Chartres, I would
    sponsor an international competition for the design
    of a modern atomic logo.  The jury would include
    writers, painters, sculptors, physicists, chemists,
    and teachers. Somewhere, I feel, there must be a
    mind creative enough to come close to translating
    the well-understood mathematical language of
    quantum mechanics into visual terms \ldots
\eq
and ever since then, whenever I see a Bohr atom (as I have multiple times recently on some cable channel), I think back to it and wonder how I could do better.

You've got the bully pulpit secured; use it!

\section{09-01-10 \ \ {\it Challenge} \ \ (to H. C. von Baeyer)} \label{Baeyer93}

\bhcvb
That's quite a challenge but I'll give it a shot.  Before I do, though, I need to go through a check list of properties of SIC measurements, strictly in words.  Is it correct that:
\ehcvb

\bhcvb
The quantum system does not include the agent or his instruments.
\ehcvb
True.

\bhcvb
The quantum system comes with a description of the apparatus that
has prepared it
\ehcvb
False.  At this moment, I'm thinking of a particularly big, particularly interesting quantum system:  Hans Christian von Baeyer.  I mentally delimit him as a system (a decently defined piece of the world), yet have not at all made any attempt to write down all the things I believe might come about if I were to interact with him.  (I.e., give him a well-placed kick of one sort or another.)  Thus I hold the thought of him in my mind separately from any quantum state I might ultimately write down for him.

In my last few papers, you probably won't find the word ``prepared'' anywhere in them.  This will be connected with one of the further answers below.

\bhcvb
It may be simple or complicated
\ehcvb
True.

\bhcvb
A spin 1/2, after passing through a bunch of magnets, is a simple
system with dimension 2
\ehcvb
Without having seen you for several months, I posit that you are about 160 pounds.  As a hypothesis, let's say you're exactly 160 pounds.  I didn't need to put you on a scale to make that hypothesis---if I wanted to check it, I might put you on a scale, etc., but that is secondary in the logical order of things.  Your weight is something you possess independently of any of my manipulations of you.  Similarly with an ejectum of a hot piece of tungsten---I posit that, with regard to the part of it I might tweak with a Stern--Gerlach ``hand'' (Fig 1), it's about 2-d worth.  If I wanted to check that, I could put a SG device near the tungsten and see whether I get two stripes, or three, or more, but that is secondary in the logical order of things.

Think this way:  ``system'' is akin to a ``rock'', ``HS dimension'' is akin to the ``rock's mass'', ``quantum state'' is akin to ``the chance I think this rock will break that window, if I throw it as hard as I can.''

\bhcvb
An electron, with all its properties, is an example of a
complicated quantum system
\ehcvb
More complicated at least.  Better imagery is simply, ``has more stuff.''  Has more dimension.

\bhcvb
A SIC measurement is often difficult, but can ACTUALLY be done
anywhere
\ehcvb
True.  Anywhere, anytime.

\bhcvb
A system can be prepared in the same way over and over again
\ehcvb
Nothing wrong with that so far as it goes.  But this word ``preparation'' often carries with it a way of thought that is not-so-Bayesian.  {\it Sometimes\/} quantum states do come about because of ``preparations''---precise laboratory procedures---and for those cases, I might be willing to write down the same state over and over after each new instance of the procedure.

More on this below.

\bhcvb\label{hottie}
If a complete set of SIC measurements is done on this system, and
statistics have been collected, an agent has as much information as
can possibly be gathered.
\ehcvb
Sounds very unBayesian.  Think first of simple probability assignments:  Bayesians leave unanalyzed where the priors come from---the answer to the question of why one possesses this or that prior is outside the apparatus of probability theory.

The same we say for a quantum state:  It is only (and you have to take ``only'' very seriously here) a state of belief.  Writing a quantum state down is tantamount to expressing what one believes of the $d^2$ possible outcomes to a single SIC measurement.  My beliefs with regard to what will happen just once.  As I can have a belief about whether the world will end or not tomorrow, I can hold a belief about a singular SIC measurement.  And if I do hold such a belief, THAT is a quantum state assignment.

I don't use the word ``information'' much anymore to say what quantum states are, except to make the transition to speaking of quantum states as beliefs.  So, also, for instance, I would not speak of a pure-state assignment to your spin-1/2 as ``having as much information about the system as I can have''.  A pure state is a belief state of a certain variety, but I might hold that belief state for religious reasons rather than ``having gathered statistics''.  Bayesians leave the origins of priors unanalyzed; the calculus kicks in once one has been set.  Similarly of quantum mechanics.

\bhcvb
``Thinking of a quantum state as literally an agent's probability
assignment for the outcomes of a potential SIC measurement'' implies
making many measurements for each i on the system as prepared.  Why do
you use the singular of the word measurement?
\ehcvb
See discussion above, for von Baeyerism \ref{hottie}.  To make the point extreme, consider the pulsar PSR B1919+21.  Astronomers have had their eyes on it since 1967.  Wiki says it has a period of 1.3373 seconds and a pulse width of 0.04 second.  And I'm sure lots of other things are known about its spectra.  My friend Greg Comer, who works on the fluid and superfluid mechanics of neutron stars, could take that data, make various assumptions, and impose a ``maximum entropy'' calculation \`a la Jaynes, to come up with a quantum state for the star's core.  There's no preparation here, no really significant statistics in the usual probabilistic sense, and no {\it unique\/} way to turn the observational data into a quantum state assignment.  But with the tools of MaxEnt, one can give {\it some\/} quantum state assignment, and that's what people usually do.  The QBist says, ``Ah, what's going on here is that the theorist is expressing his full catalogue of beliefs about that system.''  And that is tantamount to posing the question of what the theorist believes will come about if a super-duper-big SIC were brought up to the neutron star and allowed to kick it.  There would be one outcome, and only one, and the quantum state the extent to which one believes it will be this or that or that.

Maybe this is why I habitually put the SIC in the sky---it makes it inaccessible.  I don't want anyone to think of it as a measurement that need be performed in order to have a quantum state assignment in the first place.  An initial quantum state assignment is a {\it prior\/} from the point of view of quantum theory.  A prior about what?  About my SIC up in the sky.

\bhcvb
For a spin 1/2 a SIC requires finding the probabilities for
pointing up or down in the four directions defined by a tetrahedron
\ehcvb
Yes.

\bhcvb
Equation 5 relates the probability of finding a spin
1/2 pointing up in the direction defined by the geographical
coordinates of Waterloo, say, to the measurements along path~2.
\ehcvb
Yes, if I understand correctly.

\bhcvb
For a spin 1/2 system only one kind of Von Neumann measurement (a
Stern--Gerlach) is possible, but for a more complicated system all
sorts of conventional measurements are possible -- but only one set of
SIC measurements (except for a little bit of choice).
\ehcvb
If by ``one kind'' you mean the class of all Stern--Gerlach orientations.  SICs start there:  I.e., if you give me one SIC, I can mathematically produce a continuous infinity of other SICs by acting on the original with an arbitrary unitary operation.  $\Pi_i \longrightarrow U \Pi_i U^\dagger$.  But it's even more interesting than that.  There are even more SICs than that.  By that I mean, not all SICs are unitarily equivalent.  But for the Bureau of Standards, you only need one.  Pick any one you want and don't change it.  I.e., pick your coordinate system and stick with it.

Thanks for the book review; I'll read it tomorrow morning in my comfy library chair and my coffee.  Now I'm off to build a fire and a chance to beat Kiki at Scrabble.  (It happens once every blue moon.)

\section{10-01-10 \ \ {\it SIC}\ \ \ (to G. Zauner)} \label{Zauner1}

Thank you for your kind note.  It is quite nice meeting you, and seeing that you still have some interest in this baby of yours.  Perhaps you would like to visit our group at Perimeter Institute sometime and become even more involved?  We could easily bring you in for 1 week, 2 weeks, or even a month and pay your expenses if you have the time.

I am most curious about your last remark on the quantum Bayesian program.  Would you expand a little?  What sort of thing are you thinking?  The reason I am most interested in SICs is because 1) they are 2-designs, and thus give me a way of thinking of the Born Rule as a kind of ``quantum law of total probability'', but 2) give rise to a probability space with the same dimensionality as the space of density operators.  Any 2-design with a larger number of outcomes would have the valid quantum probability distributions confined to a proper subspace within the probability simplex.  It's for this reason that I have thought of SICs as ``closer to the truth'' and emphasized using them for the purpose of the QBism program.  But I wonder what sort of criterion you have in mind?

Attached is a kind of semi-popular account of the QBism program just finished for the magazine {\it Physics in Canada}. [See ``Quantum Bayesianism at the Perimeter,'' \arxiv{1003.5182}.]

\section{11-01-10 \ \ {\it Analogy?}\ \ \ (to H. C. von Baeyer)} \label{Baeyer94}

\bhcvb
Is the following analogy too misleading?

SIC measurements are like Fourier analysis. An experimentalist launches a square wave pulse of light at a window. He can either predict directly what he will see, or {\bf theoretically} decompose his square wave into sine waves and then launch the sine waves at the window.  The sine waves are universal, in that they work for any input signal whatsoever.  In order for this to work he needs an equation that relates the two paths to each other.  That this relationship is far from trivial is demonstrated by the Gibbs phenomenon (discovered by Wilbraham).

I realize that this analog pays no attention to Bayesianism, but it seems to me to reproduce certain features of your scheme.  Not the least of which is that it justifies an experimental check of Equation 5, just as the Gibbs phenomenon only gained traction when it was investigated experimentally.
\ehcvb

It's not misleading, it's exact.  Eq.\ (5)
$$
Q(D_j) =
(d+1)\sum_{i=1}^{d^2} P(H_i) P(D_j|H_i) - 1
$$
comes about because of Eq.\ (3)
$$
\rho = \sum_{i=1}^{d^2}\left( (d+1)\,P(H_i) - \frac1d \right)\Pi_i\,,
$$
and Eq.\ (3) simply expresses that any quantum state (pulse) can be decomposed into a linear combination of SIC elements (sine waves).  The sine waves are a basis for the vector space of all continuous functions; the SIC elements are a basis for the vector space of $d \times d$ matrices.

And like launching your pulse at the window:  If an agent knows how he would gamble on the ground outcomes {\it given\/} a ``preparation'' $\Pi_i$, and he also knows how he will gamble on the $\Pi_i$'s themselves (up in the sky), then he will also know how to gamble on the ground outcomes directly.  The analog of reintegrating the Fourier analysis is Eq.\ (5).

So that much of what say is exact.  Maybe the Gibbs phenomenon part of what you say is a bit trickier.  The Gibbs phenomenon comes about because one is trying to decompose a function outside of the vector space (a discontinuous function) with a basis for the continuous functions.  I'm not sure that I know an analog of that for our SICs.

\bhcvb
Not the least of which is that it justifies an experimental check of
Equation 5, just as the Gibbs phenomenon only gained traction
when it was investigated experimentally.
\ehcvb
Most recently in my correspondence (describing Steinberg's or Zeilinger's interest), I've gotten in the habit of calling it a ``demonstration'' rather than ``experiment.''   I think the word ``experiment'' should really only be used, say, as with the LHC, when one genuinely does not know what the outcome will be (whether the Higgs will be seen or not).  In the case of ``checking'' Eq.\ (5), there is no doubt it will be confirmed.

Still, my nitpicking aside, you are completely right about the traction thing.  My feeling has been that if Anton were to do a demo, it would bring a lot of attention to the SICs.  That's really my most significant motivation for him to play with them.

By the way, Marcus, Steve, and I got a very nice letter from Gerhard Zauner yesterday saying he had read our most recent paper, some of the quantum Bayesian papers, and watched the talks from our ``Seeking SICs'' conference from a couple years back.  Zauner, you probably don't know, was the {\it real\/} inventor of SICs!  His 1999 thesis
\bq
 \myurl{http://gerhardzauner.at/documents/gz-quantumdesigns.pdf}
\eq
was on them, though I didn't know of it until many years later.  And in fact, until yesterday, I always thought he had fallen off the planet (he works for a financial firm or something).  I always credit {\Caves} too with the invention of the idea (summer of 1999), but we were really taking baby steps in comparison to what Zauner had already done.  Zauner had already had the connection to the Weyl--Heisenberg group conjectured and proven their existence up to dimension 5 when we only knew about dimension 2 and 3 and didn't about the Weyl--Heisenberg connection at all (til Robin Blume-Kohout discovered it in 2002 or 2003).  Zauner had also already had Marcus's (2004) order-3 symmetry conjecture way back then.  (Not that any of us have ever read the thesis; we just know these things from Markus Grassl.)

Just as I write all this, it finally registers in me that that PhD was done in Vienna!!  This is perfect for my talk next week!  I'll definitely use it.

I've always wondered what Zauner might have been up to when he was studying these things---why was he interested in the question.  And just now, I see from some of his references (Gudder, Accardi) that he had some knowledge or interest in quantum foundations.  I'm glad I looked up the thesis to give you a link!  If you've got some time, would you mind snooping through it to see what he gives as a motivation for what he's doing.  Clearly almost all of the thing is about mathematics, but because of the references there must be something about foundations.  (For some reason, I can't seem to do a word search in the document, otherwise I'd try to give you some page numbers.)

In our case, the original motivation came from trying to prove the quantum de Finetti theorem.  Before we had a proof of it, at some point {\Carl} thought we would need these structures in our proof---that's what motivated him to think of the idea, and he knew they existed in $d=2$ because Feynman had mentioned them in a paper, but he didn't know whether they existed in any other dimension.  Luckily though we did not actually need them---I figured out how to make the de Finetti proof work with {\it any\/} minimal informationally complete POVM with rank-1 elements.   So, the symmetry was an extra condition that was not needed.  But then the SIC question took on a life of its own---for no good reason really.  They were just so damned pretty.  And their elusiveness made them that much more desirable.  I got the feeling (for no good reason again) that they had to be very deep structures, revealing something very important about quantum mechanics.  And that thought ever so slowly carried me all the way to the urgleichung.

But maybe Zauner was already there, long, long ago.

Further note attached below (written to Max Schlosshauer) on the mysterious byways that lead to scientific discovery.  [See 04-01-10 note ``\myref{Schlosshauer12}{Birth Notes}'' to Max Schlosshauer.]

\section{11-01-10 \ \ {\it Born, Heisenberg, and the QBies} \ \ (to H. C. von Baeyer)} \label{Baeyer95}

Have a look at this nice article:
\begin{center}
\myurl[http://philsci-archive.pitt.edu/archive/00004759/01/SHPMP_paper_07_10_09.pdf]{http://philsci-archive.pitt.edu/archive/00004759/01/SHPMP\underline{ }paper\underline{ }07\underline{ }10\underline{ }09.pdf}
\end{center}
by Guido Bacciagaluppi and a student.  I was knocked over with a Wow! when I read it a little while ago.  Particularly, see the discussion right before and right after Eqs.\ (2) and (3) in it.  It was Born and Heisenberg's terminology that took me: ``[I]t should be noted that this `interference' does not represent a contradiction with the rules of the probability calculus.''

Compare that to my discussion around Eq.\ (7) starting with ``But beware:''\ in the present draft of my longer paper (attached).

Maybe it's the ghosts of Heisenberg and Born that have been haunting me all along, and not Pauli after all!

\section{11-01-10 \ \ {\it Psi-ontologists}\ \ \ (to R. W. {\Spekkens})} \label{Spekkens78}

In the PiC paper I sent you earlier, there is a line that goes:
\bq\noindent
What considerations like this tell the objectifiers of quantum states is that, far from being an appendage cheaply tacked on to the theory, the idea of quantum states as information has a unifying power that goes a significant way toward explaining why the theory has the mathematical structure it does.
\eq
And the extended version of the paper, the corresponding line goes:
\bq\noindent
What considerations like this tell the objectifiers of quantum states---i.e., those who to attempt to too-quickly remove the observer from quantum mechanics by giving quantum states an unfounded ontic status (status as a state of reality)---was well put by {\Spekkens}:  [A quote of you follows, then I pick up with the ``far from being \ldots.'']
\eq

These strike me as the perfect place to make a joke.  I'm toying with the idea of coopting your term, if you'll let me, in place of ``objectifiers''.  If I were to use the term, and put a footnote right at it giving you credit for inventing the term, would you mind?  I'd probably write it as $\psi$-ontologists, and the footnote would read something like \verb+\+footnote\{Not to be confused with the Scientologists.  This term was coined by R. W. {\Spekkens} blah, blah, blah.\}.

If you want to reserve the term for one of your own papers, I won't be hurt.  But it is such a cool term, and so fits my pre-existing sentences, the desire to use it is just getting the better of me.

\section{11-01-10 \ \ {\it \ldots\ and the Smell of Postmodernity}\ \ \ (to M. Schlosshauer)} \label{Schlosshauer21}

\noindent [Replying to Max's note ``The Spell of Modernity'' \ldots] \medskip

Sorry to take so long to get back to you!  I've listened to Brad Mehldau, Jimmy Smith, Dr.\ Lonnie Smith, and Jazz Side of the Moon now.  All really great stuff!

Attached is the final version of the essay for {\it Physics in Canada}.  Nothing new in there for you to read, but I send it to you for completeness since I thank you in the back.

I'm going to hold off on personally answering your questions on classicality for the moment, and just try to do it really well in the draft.  If I don't address everything, you can take another shot at me at that point.

\section{12-01-10 \ \ {\it \ldots\ and the Smell of Postmodernity, 2}\ \ \ (to M. Schlosshauer)} \label{Schlosshauer22}

\bmaxs
I retrospectively thought my from-the-gut reply to your ideas on
``addition'' and ``classicality'' wasn't very pointed, and I apologize if
it came across as either not too bright or as just plain insufficient
in responding to your enthusiasm.
\emaxs

There is absolutely nothing to apologize for!  I was very flattered to see your immediate response that day.  I've just been swamped taking care of PI things, getting ready for Vienna, spending time with my students, trying to figure out how to get an internet link at my mom's house while I'm visiting her (after Vienna), etc.  I was just saving myself time writing, and I knew I'd be addressing all your points in a later draft of the paper.  They were exactly the questions that needed asking---it's not that they were not too bright, but TOO bright!  And I knew it would take me some work.  So I took the easy way out until I could recover my forces.

I'll probably be more in touch once I get to Vienna and get away from family duties etc.  Already got one experiment on the old SICs going (see attached).  If I can get Zeilinger interested in showing a demo in his own way that would be really cool---probably a fat chance, but it's worth a shot.

Send that balanced Everettian paper when you've got it done.  It's about my turn to read something of yours!

\section{12-01-10 \ \ {\it May 2010 Laws of Nature}\ \ \ (to R. Couban)} \label{Couban1}

Thanks for your intriguing note.  I had never thought of efforts to understand quantum mechanics as being clouded or thwarted by capitalism before!  It does make me smile, but on the other hand I am intrigued as well.  What could he mean?  I snooped on the web a bit and saw that your professor has some interesting credentials---for instance, a senior thesis with H. S. Thayer, who wrote the masterful {\sl Meaning and Action}.  It's a history of pragmatism that sits on my bookshelf.

I also think there is a lot to learn from DeLeuze's metaphysics of ``difference.''  I'm very much intrigued by his idea that difference goes all the way down.  (I might also recommend that you read the beginning parts of Nancy Cartwright's {\sl The Dappled World\/} and John Dupr\'e's {\sl The Disorder of Things}.)

About the meeting, I'm sorry, it's an invitation only thing, and we have it in an already tight room.  The idea was to put everyone in each other's face, and I think we succeeded at that.  However you can watch all the talks on {\tt pirsa.org}---they'll be taped with good quality, showing up on the website the very next day after they're given live.  I hope that will be a good substitute for you.

\subsection{Rachel's Preply}

\bq
I was just curious about the upcoming conference, where I would be interested in keeping my ears open and my mouth shut. I don't belong to any websites, blogs or societies.

I work at a research institute, and last year a new post-doc was seated in the cubicle next to mine and conversed with me about some theoretical problems of statistics. These conversations reminded me of George Caffentzis' arguments on the knowability of natural laws (``physics is not only `about' Nature and applied `just' to technology, its essential function is to provide models of capitalist work''), which Prof.\ Caffentzis explained to me are influenced by his 1978 thesis that ``we will not be able to have a coherent conception of quantum physics until humanity transcends capitalism.'' (!)

Meanwhile at my home in Oakville I have been cleaning the basement and sent home movie film and other ephemera from the 1920's to the Tuxedo Park Historical Society in New York. (One of my predecessors had a place in Tuxedo Park.) I read about the Loomis Laboratory at Tuxedo Park, and it reminded me of the Perimeter Institute.
I think Caffentzis' arguments may be answered by Gilles Deleuze's series of  discussions of paradox in his book {\sl The Logic of Sense}, but I lack the cognitive ability to formalize this opinion. I am going to order the 1978 thesis ({\sl Does Quantum Mechanics Necessitate a Theoretical Revolution in Logic?})\ through InterLibrary Loan. Maybe reading this will provide some traction for my idea of the Deleuzian (non-mystical) paradox.
\eq

\section{16-01-10 \ \ {\it The Tetragrammaton} \ \ (to H. C. von Baeyer and D. M. {\Appleby})} \label{Appleby82} \label{Baeyer96}

Greetings, dear alchemical friends, from Vienna!

Earlier today I sent Anton that quote from Heisenberg's essay ``Wolfgang Pauli's Philosophical Outlook'' that I like so much:
\bq\noindent
In the alchemistic view ``there dwells in matter a spirit awaiting release. The alchemist in his laboratory is constantly involved in nature's course, in such wise that the real or supposed chemical reactions in the retort are mystically identified with the psychic processes in himself, and are called by the same names.  The release of the substance by the man who transmutes it, which culminates in the production of the philosopher's stone, is seen by the alchemist, in light of the mystical correspondence of macrocosmos and microcosmos, as identical with the saving transformation of the man by the work, which succeeds only `Deo concedente'.''  The governing symbol for this magical view of nature is the quaternary number, the so-called ``tetractys'' of the Pythagoreans, which is put together out of two polarities.
\eq
and since then I've been walking around the area of Franz Joseph's imperial palace thinking all kinds of crazy thoughts.  Just stopping into a coffee shop to write you.

Lovely imagery:  The SIC of a qubit (``two polarities'') has four outcomes (``tetractys'').  How much closer to the unspeakable name of God could we come with this!

With a smile, and a hunger for schnitzel \ldots

\section{16-01-10 \ \ {\it Paulian Alchemy, Sunday Reading}\ \ \ (to A. Zeilinger)} \label{Zeilinger7}

You asked me last night what I have been working on.  I said, ``SICs,'' but something like these lines from Werner Heisenberg's essay ``Wolfgang Pauli's Philosophical Outlook'' might have been closer to the mark!

\bq
In the alchemistic philosophy, he had been captivated by the attempt to speak of material and psychical processes in the same language.  Pauli came to think that in the abstract territory traversed by modern atomic physics and modern psychology such a language could once more be attempted \ldots

In the alchemistic view ``there dwells in matter a spirit awaiting release. The alchemist in his laboratory is constantly involved in nature's course, in such wise that the real or supposed chemical reactions in the retort are mystically identified with the psychic processes in himself, and are called by the same names.  The release of the substance by the man who transmutes it, which culminates in the production of the philosopher's stone, is seen by the alchemist, in light of the mystical correspondence of macrocosmos and microcosmos, as identical with the saving transformation of the man by the work, which succeeds only `Deo concedente'.''  The governing symbol for this magical view of nature is the quaternary number, the so-called ``tetractys'' of the Pythagoreans, which is put together out of two polarities.
\eq
The connection between what I said last night and what I say today is that (a little tongue in cheek, but only a little):  Performing a SIC would ``release the spirit that dwells in matter.''  (It is interesting too that a SIC for a qubit has four outcomes---a SIC is its tetractys!)

Anyway, if you want some light reading for Sunday (and it actually is), see the attached draft.  It is an essay I am hoping to finish in Vienna---to give it the right spirit so to speak.  All that's left really is finding a respectable way of saying this crazy alchemical stuff in ``Hilbert-Space Dimension as a Universal Capacity'' section!

A very pleasant day today walking around the city and thinking.

\section{17-01-10 \ \ {\it Oh Translator} \ \ (to H. C. von Baeyer)} \label{Baeyer97}

This passage of Pauli is now taking on such an important role in my newest version of that damned paper that I'm wondering if it is translated correctly:
\bq
The objectivity of physics is however fully ensured in quantum mechanics in the following sense.  Although in principle, according to the theory, it is in general only the statistics of series of experiments that is determined by laws, the observer is unable, even in the unpredictable single case, to influence the result of his observation---as for example the response of a counter at a particular instant of time.  Further, personal qualities of the observer do not come into the theory in any way---the observation can be made by objective registering apparatus, the results of which are objectively available for anyone's inspection. Just as in the theory of relativity a group of mathematical transformations connects all possible coordinate systems, so in quantum mechanics a group of mathematical transformations connects the possible experimental arrangements.
\eq
There's something awkward about the English in a couple of places of it.  For instance, ``the observer is unable, even in the unpredictable single case.''

The translation comes from pp.\ 117--123 in {\sl Writings on Physics and Philosophy}.  You think there's any way we can dig up the original text and check?  Any idea how we can get our hands on that old journal?  So far I haven't had any luck with Google Scholar, but maybe I'm not being too smart.

The longer passage of interest to me (but not all of it is used in the present paper) is below.

\bq
Einstein's opposition to it [the so-called ``Copenhagen interpretation''] is reflected in the papers which he published, at first in collaboration with Rosen and Podolsky, and later alone, as a critique of the concept of reality in quantum mechanics. We often discussed these questions together, and I invariably profited very greatly even when I could not agree with Einstein's views. ``Physics is after all the description of reality,'' he said to me, continuing, with a sarcastic glance in my direction, ``or should I perhaps say physics is the description of what one merely imagines?''  This question clearly shows Einstein's concern that the objective character of physics might be lost through a theory of the type of quantum mechanics, in that as a consequence of its wider conception of objectivity of an explanation of nature the difference between physical reality and dream or hallucination become blurred.

The objectivity of physics is however fully ensured in quantum mechanics in the following sense.  Although in principle, according to the theory, it is in general only the statistics of series of experiments that is determined by laws, the observer is unable, even in the unpredictable single case, to influence the result of his observation---as for example the response of a counter at a particular instant of time.  Further, personal qualities of the observer do not come into the theory in any way---the observation can be made by objective registering apparatus, the results of which are objectively available for anyone's inspection. Just as in the theory of relativity a group of mathematical transformations connects all possible coordinate systems, so in quantum mechanics a group of mathematical transformations connects the possible experimental arrangements.

Einstein however advocated a narrower form of the reality concept, which assumes a complete separation of an objectively existing physical state from any mode of its observation. Agreement was unfortunately never reached.
\eq

\subsection{Hans's First Reply, 17-01-10}

\bq
OK, here's the story.  On 7 January 1958, in the middle of his fight with Heisenberg about their doomed theory, Pauli gave an address at the ETH on the occasion of the unveiling of an Einstein bust. The talk was entitled ``Albert Einstein in der Entwicklung der Physik.''  Five days later it was published in the Neue Z\"urcher Zeitung, a national daily newspaper.  It was collected in {\sl Aufs\"atze und Vortr\"age \"uber Physik und Erkenntnistheorie} (Viehweg 1961 and 1984) {\it which I do not own}.  This book was translated by Robert Schlapp into the volume you cited.  Schlapp, whose name means ``slack'', was a decent translator, but not a great one. I think checking is a good idea.

Pauli's German book is quite prominent, so there are many ways to get a copy.  I will ask my distant cousin Klaus Hepp, emeritus at the ETH (and Planck medal winner!)  to send me a scan of the article.  In Vienna there are copies galore.  I will also try interlibrary loan.  U. of Toronto might well own a copy!

If you find the passage in the institute library or somewhere -- maybe even Anton's private library -- please send me a copy.  I will proceed at this end.
\eq

\subsection{Hans's Second Reply, 19-01-10}

\bq
With the German and the English texts in front of me I can try to answer questions.  In general the translation is good, but you are talking about nuances of meaning that I can't disentangle by changing a few words here and there.  Instead, you have to tell me what's bothering you.

According to the paragraph preceding, Einstein thought that QM sacrifices objectivity.  To which Pauli answers: Objectivity is vouchsafed in the following sense:
\begin{enumerate}
\item
Theory can determine only the statistics of a series of experiments.
\item
However, the observer cannot influence the unpredictable outcome of a specific, single measurement.  (This sounds like ``a kick that leads to unpredictable consequences \ldots'')
\item
Furthermore, the theory contains no references to the observer's personal characteristics or properties.   The measurement can even be made by an apparatus.
\end{enumerate}

How are the Viennese reacting to your views?
\eq

\section{17-01-10 \ \ {\it Holocaust Memorial} \ \ (to H. C. von Baeyer)} \label{Baeyer98}

BTW, I went to see the memorial today.  Stirring, as all such memorials are.  It was an empty square on a grey day, and the name kept ringing in my ear:  Judenplatz.

I liked your imagery.  I'm full tilt on quantum mechanics while here and getting close to writing the section in the paper where I say that each piece of the world has something within it that the rest of the world cannot get.  (Your words start to work a little for me:  It is not hidden variable, but something like soul.)  Each piece of the world is a little like your box inside out---it contains a universe within.

\section{17-01-10 \ \ {\it Oh Translator, 2} \ \ (to H. C. von Baeyer)} \label{Baeyer99}

You can see why I want to get the passage right:  I disagree with it!  See words at top left corner of page 7:
\bq
{\it Whose information?} ``Mine!'' {\it Information about what?}
``The consequences (for {\it me}) of {\it my\/} actions upon the
physical system!'' It's all ``I-I-me-me mine,'' as the Beatles
sang.

The answer to the first question surely comes as no surprise
by now, but why on earth the answer for the second?
``It's like watching a Quantum Bayesian shoot himself
in the foot.'' Why something so egocentric, anthropocentric, psychology-laden, myopic, and positivistic as {\it the
consequences (for me) of my actions upon the system}?
Why not simply say something nice and neutral like ``the
outcomes of measurements,'' or fall in line with Wolfgang
Pauli and say,
\bq
The objectivity of physics is \ldots\ fully ensured in quantum
mechanics in the following sense. Although in principle,
according to the theory, it is in general only the statistics
of series of experiments that is determined by laws, the
observer is unable, even in the unpredictable single case,
to influence the result of his observation---as for example
the response of a counter at a particular instant of time.
Further, personal qualities of the observer do not come
into the theory in any way---the observation can be made
by objective registering apparatus, the results of which
are objectively available for anyone's inspection.
\eq

To the uninitiated, our answer to {\it Information about what?}\ surely appears to be a rash abandonment of realism. But it
is the opposite. The answer we give is the very injunction
that keeps the potentially conflicting statements of Wigner
and his friend in check (Pauli's wouldn't have done that),
and, more importantly, gives each agent a hook to the
external world in spite of QBism's egocentric approach.

You see, for the QBist, the real world, the one each agent
is embedded in---with its objects and events---is taken for
granted. What is not taken for granted is each agent's
access to the parts of it he has not touched.
\eq

Gotta give my talk tomorrow, but, of course, haven't prepared anything specific on it yet (having such a bank of old talks).  First thing tomorrow I've got to get some new shoes---with the coming of REAL winter in Waterloo, I had forgotten that I had a hole in my more casual shoes.  Then I'll get ready for my talk!

But sleep?  Maybe a little later.

\section{19-01-10 \ \ {\it Translation} \ \ (to H. C. von Baeyer)} \label{Baeyer100}

\bhcvb
How are the Viennese reacting to your views?
\ehcvb
Probably a better reaction than {\Caves} would give me!

More details tomorrow when I'm sober again.  Just back from my time with Gerhard Zauner.  Apparently---and I believe it!---he had proven the existence of MUBs in prime-power dimensions before (and not knowing of) Wootters.  And SICs?  He first got the idea of the concept around 1991 (mentioned in his master's thesis)!  Though he (mistakenly, stubbornly) believed them to {\it only\/} exist in prime-power dimensions---he simply thought they could not exist otherwise.  In 1996 he got the idea to check in composite dimensions, and (nearly trivially with a quick check) they existed there too, so he decided to follow through with a PhD on it (despite his initial advisor's discouragement).  He kept going on about how two-designs (SICs and MUBs and a set of similar objects) were his objects of study because they had the potential to express quantum mechanics in a non-algebraic way.  I didn't understand what he meant, but I definitely felt he was on the tip of something.

There is no doubt, he knows things that Marcus does not know, but he only sporadically thinks about all these things.  For instance, in 2004 when in a hospital for 6 weeks with a head injury; he had amnesia for a bit, but the first thing that came back to him was a two-design question that he had first come to him in the early 1990s.  I am much, much intrigued.  Tomorrow, when I write, I'll tell you about Rankin-Bass and the ``Island of Misfit Toys'':  I swear, Marcus, Zauner, and I all come from that place.  Zauner never published (not even one paper!)\ because when he solicited the level of interest of anyone he could think of that should have been interested, they yawned.  It is a tragedy for the development for quantum mechanics:  People could have been thinking about SICs since 1991 if someone (probably nearly anyone) had just given some encouragement to the guy!

More tomorrow, as I say.  Too much beer!  I've had a great time here this visit.  It is good to literally get in the lab (fascinating actually):  I felt so alive this morning when I saw their Kochen--Specker experiment.  I joked that I always talk of reality creation, but these guys, with that equipped lab table, are actually doing it.  Anton said, ``Well, you yourself are doing it all the time, you just don't know it.''  I agreed, and offered my daughters as examples.

I'll try to make my questions on the Pauli article pointed when I come back.

\section{19-01-10 \ \ {\it The Reason I Like Your Axiom 1} \ \ (to \v{C}. Brukner)} \label{Brukner4}

\ldots\ particularly the second part, ``All systems of the same information carrying capacity are equivalent'' \ldots\ has to do with the ideas I'm trying to formulate on ``dimension'' as a capacity analogous to (or in the spirit of) gravitational mass.  Somewhere in your axiom 1 is an analogy to the E\"otv\"os experiment.

See the two elevator stories in \arxiv{quant-ph/0404122} (i.e., see the introduction and conclusion of this paper, skipping everything else).  Platinum and magnalium were the two elements in the E\"otv\"os experiment.

The yet-to-be-written capacities section will have some of that in it and some of the attached rant, originally written to Lucien.  [See 16-10-09 note ``\myref{Hardy38}{The More and the Modest}'' to L.\ Hardy.] I'll definitely be citing your new paper too.

\section{20-01-10 \ \ {\it Bohr was Bayesian?}\ \ \ (to C. Ferrie)} \label{Ferrie10}

Good to hear from you.  I'm in Vienna trying to get some SIC demonstrations going with the Zeilinger group, and trying to write this damned paper.  Since you seem to need some good Bayesian-support-group amens, I'll go ahead and attach the paper as it stands.  The last three sections still have a have a good ways to go, but much of the rest of it is stabilized.

I suspect there'll be aspects you like, and aspects you hate \ldots\ but I can't do much about that.  The key, for your respect though, I think, is that I am a realist---there is no doubt about it---I am just not a hidden-variable realist.  (And the same statement is true of Bohr.)

Anyway, such as it is, see attached.  And hold tight through those quantum foundations classes!  I'll see you in a couple weeks.

\subsection{Chris's Preply}

\bq
The usual attacks were made on Bohr at the latest lecture of the interpretations course.  Those with enough intellectual responsibility to read the Bohr quotes without joining in on the laughter saw that what Bohr actually said and the ``Copenhagen Interpretation'' philosophy which is attributed to him are in fact polar opposites.  I sat there thinking Bohr was really a Bayesian but didn't know it.  Then I thought that you must have come to this realization as well, provided you had read any Bohr, which I assumed you did.  So I started searching your PDFs for ``Bohr'' and found that exact phrase!

It is now obvious to me that you are far more well read in Bohr than I am.  Ironically, the only things of Bohr I have read are cherry-picked quotes intended to support the claims that Bohr was either schizophrenic or an evangelist for the ``Copenhagen Interpretation'', which I now doubt ever even existed.  Fortunately, I committed to neither claim then and re-reading them now (after my Bayesian enlightenment) his words are more clear than unclear.

I felt sorry for Bohr that you were not there to defend his legacy.  But at the time I could only sit in silence as I tried to think of a coherent defense of Bohr coupled with a rebuttal to the inevitable charges of solipsism or antirealism.
\eq

\section{20-01-10 \ \ {\it Bohr was Bayesian?, 2}\ \ \ (to C. Ferrie)} \label{Ferrie11}

And yes, the letters may not completely indicate it (all that crap is too sporadic), but I have tried to read everything Bohr wrote on quantum interpretation pretty darned carefully.  (Also I've read plenty of things by the ``Bohr scholars'' Folse, Faye, Plotnitsky, etc.)  He is not my hero in the game (Pauli is), but there is plenty in Bohr's gropings to think hard about and try to make clear ``in the new modern way'' (from a country song, Tom T. Hall).  Attached is a picture of one of my visits to the old man himself.  Below is a story about his desk.  [See 03-02-04 note ``\myref{Comer49}{The Land of Bohr}'' to G. L. Comer.]

I've got to say, I had the greatest experience the other day.  At some point in a big discussion in Anton Zeilinger's office, I got up to write something on the chalkboard.  The chalkboard was pretty rough, not nearly as nice as the one Kiki installed on our porch (it was more like the plywood she had once painted with chalkboard paint).  \ldots\ AND \ldots\ I was just about to say something, making some smart remark that would come off as clever and sound like a joke---that's the thing I try to do all the time.  Bad habit.  But just in time I saw a brass plaque at the bottom of the board \ldots\ and caught myself!  It said it was Ludwig Boltzmann's chalkboard!  I almost fell over.  I was so excited; my heart skipped a beat, and I acted like a little boy.  Thereafter I tried to find every excuse I could to write some more on it.  Anton caught on, and that turned out to be my big joke.

Look, hold on tight.  You've survived my bastardly wrath---you're a strong guy.  Your Bayesianism is on the right track.  And the very fact that you've had an epiphany (one of any sort, any sort whatsoever), already sets you apart from the rest of the foundations pack.  Few foundations folk are ever really stirred by anything that takes them by surprise.  Just learn to articulate yourself, defend yourself, and the battles themselves will lead to clarity.  Crown yourself emperor, and you may get a nation.

\section{20-01-10 \ \ {\it Translation, 2} \ \ (to H. C. von Baeyer)} \label{Baeyer101}

I guess it's only the phrase ``even in the unpredictable single case.''  It's the ``even'' that confuses me.  Is it really part of his construction?  Is it a term of contrast?  For instance, who might have thought he could ``influence'' the ``unpredictable single case''?  ``Nope, even in that case, I can't do it.''  That's the sort of thing that's confusing me and making me wonder whether it is an artifact of the translation or whether he really meant it that way?

\subsection{Hans's Reply}

\bq
OK -- here we go. I have added some emphases.

\bq\noindent
Pauli: {\bf Obwohl} nach der Theorie im Prinzip im allgemeinen nur die Statistik von Versuchsreihen gesetzm\"assig bestimmt ist, kann der Beobachter {\bf auch} im nicht voraussagbaren Einzelfall das Resultat der Beobachtung  --  wie zum Beispiel das Ansprechen eines Z\"ahlers in einem bestimmten Zeitmoment  -- nicht beeinflussen.
\eq

\bq\noindent
Schlapp:  {\bf Although} in principle, according to the theory, it is in general only the statistics of series of experiments that is
determined by laws, the observer is unable, {\bf even} in the unpredictable single case, to influence the result of his observation---as for example
the response of a counter at a particular instant of time.
\eq

The word {\bf auch} could be translated as even, or as also.  But the initial {\bf although} demands the stronger contrast provided by {\bf even}.

It seems to me that the fault is Pauli's, not in the word ``auch'' but in the phrase ``kann\ldots\ nicht''  which is translated as ``is unable.''  Maybe he should have said ``does not'' rather than ``is unable.''  What I have in mind is the difference between classical and quantum. In a series of coin tosses the beautiful law emerges without the help of the observer. But the observer certainly {\it does\/} influence a single throw.  If we knew enough about it, we could predict the throw.  In QM, on the other hand, the observer does not influence the throw even in principle.

This uncertainty in principle was not Pauli's emphasis here. He wanted to establish the objectivity of physics -- almost by introducing a detached observer who is at the mercy of the objective world out there. He can't even influence a single observation.  And of course the bell curve is not his creation either.

Once your head is clear I would be grateful for your keen analysis.
\eq

\section{20-01-10 \ \ {\it Translation} \ \ (to H. C. von Baeyer)} \label{Baeyer102}

OK, I guess I've got nothing to say:  So much for my ``keen analysis.''  I will let it stew in the back of my head however---maybe your point about the classical coin toss is a good one.

Tonight is my last dinner with Anton, and tomorrow the taxi arrives at the hotel at 5:45 AM---I'm off for a day and an evening with {\Ruediger} {\Schack} in London.  Then it's to the hicksville of Texas, where I hope the cellular modem I had shipped there will work with my computer \ldots\ otherwise I'm going to be in a world of cybermisery.

\section{21-01-10 \ \ {\it Coffee and Spirits} \ \ (to \v{C}. Brukner)} \label{Brukner5}

I think I'm going to go work in that caf\'e of the building where Exner, Hassenohrl, and {\Schroedinger} worked.  I'll show up at the institute early afternoon.  Writing is a very painful thing for me to do.  I'm hoping Exner's spirit might give me a bit of guidance.

\section{21-01-10 \ \ {\it God Likes Qutrits} \ \ (to C. Schaeff \& the Vienna Qutrit Guys)} \label{Schaeff1}

I just set a tab in my address book titled ``The Vienna Qutrit Guys''.  If there's anyone out there who doesn't want to be in this list, let me know.  Also if I've missed anyone who should be here, let me know that too.

I've been trying to articulate why the qutrit is a really important state space, and the best I have at the moment is this:  Gleason's theorem shows that if you can characterize the set of probability functions allowed by the structure of quantum measurement in $d=3$, then you can characterize it in all dimensions.  So all the weight of the probabilistic interpretation of quantum mechanics bears down on the qutrit.  From the experimental perspective, if an experimentalist wants to say ``our experiments have mapped out the essence of quantum theory'' then he has to be ready for the rejoinder, ``Well then, do you have the capability of performing every kind of measurement in dimension 3, including POVMs?''  ``Unless you can do that, you have not fully explored the features of that state space.''  I want to say something like that, and find a way to make it pithy and convincing.

\section{21-01-10 \ \ {\it On Realism}\ \ \ (to C. Ferrie)} \label{Ferrie12}

Here are some notes I took on Henry Folse's papers on Bohr.  I'd be curious to hear what you think of his reading of Bohr.

You're absolutely right on the six billion varieties of realism.  When I said about myself that ``I am a realist'' I meant something somewhat close to the attached Martin Gardner quote---that the world is made of something that doesn't a priori require the existence of human minds.  That's all I meant.  But as you point out, you surely wouldn't be able to figure that out by looking at wiki.

\bcf
It seems that what is overlooked in all these debates about
``interpretations'' is the assumptions that quantum theory is either ``The
Truth'' or the next logical step to it.  But what if that assumption is
wrong?
\ecf

Well, even with all my love for trying to get the meaning of quantum theory straight, you would not hear me say that quantum theory is the Truth or the next logical step to it.  Do you know the old Dr. Hook song, ``On the Cover of the Rolling Stone''?  If not, look at:  \myurl[http://www.youtube.com/watch?v=-Ux3-a9RE1Q]{http://www.youtube.com/ watch?v=-Ux3-a9RE1Q}.
I just had a copy of the {\sl Oxford Dictionary of American Quotations\/} sent to my mother so that we can flip through it when I visit her for her 81st birthday.  One of the best merit badges I've ever attained in my life---they actually quote me!  Here it is, I think (haven't actually seen the precise page yet):
\bq\noindent
``There is no one way the world is because the world is still in creation, still being hammered out.''
\eq
If there is not one way the world is because the world is still in creation, there is no everlasting scientific theory either.  Quantum mechanics will one day meet its demise in one way or other.  The point for me for interpreting it (in a way that it ought to be) is to get a safer suggestion for how to take the next step.

\section{22-01-10 \ \ {\it Is the Big Bang Here?}\ \ \ (to A. Zeilinger)} \label{Zeilinger8}

See attached for the Wheeler quotes I promised you.  Most of the quotes are from his presentation (which in their own way discuss the question above), but the actual question came from Richard Elvee at the end of the talk.  Wheeler answered,  ``A lovely way to put it---`Is the big bang here?' I can imagine that we will someday have to answer your question with a `yes'.''

\bq
It is difficult to escape asking a challenging question. Is the
entirety of existence, rather than being built on particles or fields
of force or multidimensional geometry, built upon billions upon
billions of elementary quantum phenomena, those elementary acts of
``observer-participancy,'' those most ethereal of all the entities
that have been forced upon us by the progress of science?

At first sight no question could seem more ridiculous. How fantastic
the disproportion seems between the microscopic scale, of the typical
quantum phenomenon and the gigantic reach of the universe!
Disproportion, however, we have learned, does not give us the right
to dismiss. Else how would we have discovered that the heat of the
carload of molten pig iron goes back for its explanation to the
random motions of billions of microscopic atoms and the shape of the
elephant to the message on a microscopic strand of DNA? Is the term
``big bang'' merely a shorthand way to describe the cumulative
consequence of billions upon billions of elementary acts of
observer-participancy reaching back into the past?

Stepping stone though this question may be to a new outlook, it is
beset before and beyond by traps. One trap is a misjudgment of the
role of ``consciousness.'' The other is an exaggerated estimate of
the category of ``time.''
\eq
and
\bq
\indent
Guided by these warnings and encouraged by the view that there is
nothing at all of physics more elementary than ``elementary quantum
phenomena,'' are we destined some coming century to see all of
existence derived out of this utterly primitive unit? And on the way;
are we not surely destined to find some single simple idea that will
lend itself to statement in a single sentence, so compelling that we
will all say to one another, ``Oh, how simple!'' and ``How stupid we
all were!'' and ``How could it have been otherwise?''

Have I been making the world sound like a very mysterious place? It
is! But amid all mysteries, we remember those great words of Leibniz,
``Although the whole of this life were said to be nothing but a dream
and the physical world nothing but a phantasm, I should call this
dream or phantasm real enough, if, using reason well, we were never
deceived by it.''

There are many sources of energy. There is solar energy and the
energy of fossil fuels, coal and oil. There is the energy of uranium,
but a source of energy greater than any of these is the energy of the
human heart, the energy that makes us reach for an understanding of
the universe and our place in it, our responsibilities, our
opportunities, our hopes. No questions are more pertinent to this
comprehension than ``How did the world come into being?'' and ``How
is the world constructed?'' I know of no clue more likely to allow us
someday to grasp an understanding of these questions than the
quantum.
\eq
and
\bq
\noindent ELVEE: Dr.\ Wheeler, who was there to observe the universe when it
started? Were we there? Or does it only start with our observation?
Is the big bang here?\medskip

\noindent WHEELER: A lovely way to put it---``Is the big bang here?'' I can
imagine that we will someday have to answer your question with a
``yes.'' If there is any conclusion that follows more strongly than
another about the nature of time from the study of the quantum nature
of space and time, it is the circumstance that the very idea of
``before'' and ``after'' is in some sense transcended.

There are two aspects of this idea. First, Einstein's theory of space
and time tells us that in order to predict all of space and time for
time to come, we have to know what the conditions of space are now
and how fast they're changing. Only then do we have enough
information to predict all the future. The uncertainty principle of
quantum theory tells us that if we know the condition of space now,
we cannot know how fast it's changing. Or if we know how fast it's
changing now, we cannot know what the geometry is now. Nature is so
built with this complementary feature that we cannot have the
information we need to give a deterministic account of space geometry
evolving with time.

That deterministic account of space evolving in time is what we mean
by spacetime. Everything that we say in everyday language, about time
is directly built on that concept. And with determinism out, the very
ideas of before and after are also out. For practical everyday
matters, this indeterminism, this indefinability of spacetime is of
no concern. The uncertainties only show up effectively at distances
of the order of $10^{-33}$ cm. Nobody at present has equipment fine
enough to reach down to a distance so small.

What does all this have to do with the big bang? At the very
beginning of time we know that---according to Einstein's
account---the universe was indefinitely small. Things were
indefinitely compact. When we talk about time when the universe
itself is so fantastically small, we deal with a state of affairs
where the very words ``before'' and ``after'' lose all meaning. This
circumstance puts one heavy restriction on the usefulness of the word
``time.'' There is another.

When we do our observations in the here and the now on photons,
quanta of light, hunks of energy coming from distant astrophysical
sources, we ourselves have an irretrievable part in bringing about
that which appears to be happening. We can put it this way: that
reality is, in a certain sense, made up of a few iron posts of
definite observation between which we fill in, by an elaborate work
of imagination and theory, all the rest of the construction that we
call reality. In other words, we are wrong to think of the past as
having a definite existence ``out there.'' The past only exists
insofar as it is present in the records of today. And what those
records are is determined by what questions we ask. There is no other
history than that. This is the sense in which we ourselves are
involved in defining the conditions of individual elementary quantum
phenomena way back at the beginning of the big bang.

Each elementary quantum phenomenon is an elementary act of ``fact
creation.'' That is incontestable. But is that the only mechanism
needed to create all that is? Is what took place at the big bang the
consequence of billions upon billions of these elementary processes,
these elementary ``acts of observer-participancy,'' these quantum
phenomena? Have we had the mechanism of creation before our eyes all
this time without recognizing the truth? That is the larger question
implicit in your comment. Of all the deep questions of our time, I do
not know one that is deeper, more exciting, more clearly pregnant
with a great advance in our understanding.
\eq

\section{25-01-10 \ \ {\it Silly Name Play}\ \ \ (to A. Zeilinger)} \label{Zeilinger9}

Just playing around with words on my flight to Texas.  Here are some potential names for your Copenhagen conference.
\begin{itemize}
\item	Breathing New Life into Bohr
\item	Breathing New Life into the Copenhagen Interpretation
\item	Quantum Information's Impact on the Copenhagen Interpretation
\item	New Perspectives on the Copenhagen Interpretation
\item	Lifting of the Fog from the North:  New Perspectives on the Copenhagen Interpretation
\item	Reading Bohr, Heisenberg, and Pauli in the Language of Quantum Information
\item	Bohr's Relevance Today
\item	The Copenhagen Interpretation of Quantum Mechanics as a Modern Interpretation of Quantum Mechanics
\item	Teleporting Bohr to Modern Times:  Quantum Information and the Copenhagen Interpretation
\item	Beam Me Up, Professor Bohr!
\item	Seeking Bohr's Guidance Afresh:  How the Copenhagen Interpretation Enlightens Quantum Information
\item	Seeking the Soul of Quantum Information Theory \ldots\ One Bohr, Heisenberg, and Pauli at a Time
\item	Searching Copenhagen's Soul:  Quantum Information and the Future of Physics
\item	I Like the Copenhagen Interpretation, Why Don't You?
\item	Bohr 2, Einstein 1 (the poster could show Einstein and Bohr in football clothing kicking qubits)
\item	Protecting Bohr with Fault-Tolerant Quantum Error Correction
\item	If Bohr Knew Then What We Know Now (What Would He Say Differently)
\item	Bohring the Tears Out of Einstein
\end{itemize}

The ``lifting the fog from the north'' remark refers to a quote of John Wheeler:  ``It may be, as one French physicist put it, `the fog from the north,' but the Copenhagen interpretation remains the best interpretation of the quantum that we have.''  (I think the French physicist was de Broglie.)

\section{28-01-10 \ \ {\it Morning Time, Cuero} \ \ (to D. B. L. Baker)} \label{Baker20}

Getting your note yesterday still strikes me as such a strange coincidence.  Everything you said was exactly on my mind---both, about not going home and never really leaving.

Attached is a paper I've been working on (and still have a good ways to go before finishing).  [See ``QBism, the Perimeter of Quantum Bayesianism,'' \arxiv{1003.5209v1}.]  It is an example of your point on ``never really leaving''---there are Texanisms and Cueroisms sprinkled throughout.  Little things from our youth.  For instance, when I wrote Footnote 7, I kept thinking of [Cindy X]'s dad calling Terry ``you common son of a bitch'' the night he caught them together.  And I thought of a beer joint jukebox when I wrote the ``hell hath no fury'' line on page 5.  There are others, as well.  Funny too you write me about the Bible just as I was putting in Figure 2.

And the Ike story.  Before coming to Cuero, I had a bunch of books shipped here from Amazon etc.  Things that I couldn't get in Canada without waiting a ridiculous time, or having to pay exorbitant shipping costs.  One of them was the {\sl Oxford Dictionary of American Quotations}, which I discovered only a couple of weeks ago to really have a quote of mine---I found my name in the index of the Amazon preview of it, as I was buffing up my resume, but that was all I could see.  (The dictionary contacted me years ago, asking permission, but as time went on I had forgotten about it, until I was searching for merit badges to beef up my resume to try to get a promotion.)  It's really very flattering being there with Jefferson and Emerson and Mark Twain, even if you know it was someone's arbitrary decision.  I'm rather proud of it really.  Anyway, for a brief few minutes I thought about writing Linda Henderson with the news.  But I backed off, thinking something exactly like your Ike Eichholz story.

Tomorrow morning early, I go back to Waterloo to face the cold and the snow.  But then in two weeks, I go to Japan for my 10th time.  (Nagoya this time; never been there actually.)  It still smarts to hear my old brother-in-law from years back, saying, ``Why would you want to go over there?''

Funny the synchronicity between us.  I do think of you often.  As I walked the streets of Vienna and the quiet square of Judenplatz, where there is a little memorial for the 65,000 Austrian Jews killed by the Nazis, I was thinking of the time you brought the Blue Danube into our tent at Boy Scouts.  It was one of my first introductions to any of that side of the world.  You brought a perspective that I had never seen.  You were the only culture I ever knew in Cuero, and there's no doubt the little bit you pulled me into, I fought kicking and screaming.  I sometimes wish I had it all to do over again---to try to keep the good with the good from that little town, but reject the parts that gave me such a slow start.

I learned yesterday from my mom that Michael's dad died a few weeks back.  I should send Michael a note this morning.

\section{02-02-10 \ \ {\it Permission?}\ \ \ (to C. E. Granade)} \label{Granade1}

I'm working on an article that reviews my research program, and
there's a spot in it that one of your neologisms [psi-ontologists] would be perfect for.
See footnote 2 (and the paragraph to which it refers) on page 3 of the
attached draft:
\bq\noindent
Not to be confused with the Scientologists. This useful descriptor
was coined by Chris Granade, a PSI student at Perimeter Institute,
and brought to the author's attention by R. W. {\Spekkens}, who
immediately understood its psychological value.
\eq
The article is not actually going to {\sl Physics in
Canada}, despite the formatting---I just haven't changed it yet.  I'm
not yet quite sure where I will send the thing.  In any case though,
do I have permission to use your beautiful word?  If no, then I'll
strip it out.  If yes, then is there anything you want me to change in the way that I give you credit?

\section{03-02-10 \ \ {\it Three Cheers for Measurement} \ \ (to J. Rau)} \label{Rau3}

Thanks for the revised version of your paper.  [See J. Rau, ``Measurement-Based Quantum Foundations,'' \arxiv{0909.1036}.] ``We are such stuff as quantum measurement is made on.''  That is a line that will end one of my own forthcoming compositions.  [See ``QBism, the Perimeter of Quantum Bayesianism,'' \arxiv{1003.5209v1}.]  I'll attach it as it stands---it still has a good ways to go before being finished, but perhaps you can start to see the outline of things to come.  At the end, it will agree with you that the notion of measurement is the very primitive.

\section{03-02-10 \ \ {\it Giving QM and Pragmatism a Chance to Flirt with Each Other} \ \ (to R. {\Schack}, D. M. {\Appleby}, H. C. von Baeyer)} \label{Appleby83} \label{Baeyer103} \label{Schack187}

The Templeton Foundation has announced two funding opportunities that seem to fit our proclivities.  See:
\begin{center}
\myurl[http://www.templeton.org/what_we_fund/2010_funding_priorities/]{http://www.templeton.org/what\underline{ }we\underline{ }fund/2010\underline{ }funding\underline{ }priorities/}
\end{center}
I will be applying for the quantum one, and it looks like the free-will one very much matches up with {\Ruediger}'s discussion with me at Windsor Castle.  Maybe everyone might want to take a shot at it?

I have no firm ideas for myself yet, but I think the direction I want to pursue is to find a way to get quantum foundational thinkers to mix and mingle with pragmatist philosophers.  The thought I have is that it is the physicists who could use some education, and I want to think of ways to foster that.  If anyone has ever seen a model of similar ilk work before or has any ideas along such lines, I'd enjoy hearing.

\section{09-02-10 \ \ {\it Village Voice} \ \ (to D. B. L. Baker)} \label{Baker21}

\bdb
Thanks for attaching a sample of your latest writing.  I definitely heard your voice as I read it.  I have to admit that the only science journalism I ever read is stuff you forward to me so it is kind of hard to compare it to anything else.  How much of what is published is written in such a colloquial, conversationalist style?  Even the hard science/technical writing that you send me sounds the same.  Maybe it's just that I know you that I hear it so clearly.  What do your Ivy League and foreign colleagues say?

I don't mean to swell your already large sized head but I would compare the experience somewhat to reading Jack Kerouac.  Perhaps Ada was wrong and it is not Thoreau you have a feeling for.  Even when I read Ginsberg or Burroughs it's Kerouac I hear.  He just seems/sounds more real so theirs must be borrowed.  Maybe it's just a Cuero thing, having to explain things to people who may not be interested or even understand; people who never wanted to leave town or think about anything that doesn't happen there.  Hal Ketchum's ``Small Town Saturday Night'' definitely on the jukebox there.  ``Lucy, you know the world must be flat 'cause people leave town, they never come back.''

PS. You can use ``Quantum Kerouac'' if you want, though I do enjoy being cited.
\edb

Thanks for the nice compliment---the idea of being on the road with Kerouac is nice.  But there are lightyears between him and me.  The very big difference is that his writing was natural:  He typed his first book on a paper towel roll, as I recall!  And I don't think it took him more than a few weeks. I pain over every sentence, changing some probably a hundred times before I get it right.  It's completely fake with me.  Kerouac was the real deal.

But thanks!  I'll send you the final version of the paper when it's finished.  I got sidetracked with lots of things once I got back to Waterloo.  My collaborator Appleby is here from England, and we've been at the chalkboard pretty much nonstop.  Those damned consternating SICs.

\section{10-02-10 \ \ {\it Boyle} \ \ (to D. Kobak)} \label{Kobak3}

Thank you very much for that---it was thought provoking.  Soon I will put a paper on the {\tt arXiv} titled ``The Perimeter of Quantum Bayesianism'' where I expand in more detail on the character and origin of quantum indeterminism.  It is a bit as you say, but in place of ``divine will'' I would write ``divine wills''.  You will see what I mean when I finally release the paper.

\subsection{Dmitry's Preply}

\bq
As I feel that I'd better introduce myself: we met at PI during one of the quantum foundations schools several years ago, and then last year had a brief communication about the infamous anti-Bell works of De Raedt. This time I have a much more pleasant topic.

I have recently stumbled upon a short text about Robert Boyle -- an extremely interesting account of his theological thoughts driving his scientific research. Here it is: \myurl{http://shkrobius.livejournal.com/189428.html} (don't be discouraged by the domain: it's a blog of a chemist and one of the most educated people I know). In short, he is saying that Boyle -- who basically invented modern scientific method and was largely responsible for starting science as an enterprise as we know it -- that Boyle did all that trying to disprove the idea of natural physical laws. He didn't believe in laws, he believed in Divine will, and so for example wanted to disprove, and did disprove the ``law'' that Nature ``abhors vacuum''. I will stop retelling this story and encourage you instead to read that post.

The reason I think it might be interesting for you, is that it strikes me how similar Boyle's views would be to the Bayesian view of QM, where quantum laws are just conceits of mind and there seems to be no law whatsoever behind an outcome of quantum measurement -- nothing but the pure Divine will. Would Boyle be a Bayesian?\ldots

I also wrote this in a bit more detail in the comments section of that post. Actually, before writing to you I made a search for ``Boyle'' in all your samizdats and found zero results. So I thought you might be interested. If you indeed are, I would be happy to know what you think about it.
\eq

\section{10-02-10 \ \ {\it Three Cheers for Measurement, 2} \ \ (to J. Rau)} \label{Rau4}

\bjr
Given that you are such a philosophically minded person (after all, I owe to you my first encounter with American pragmatism), I was wondering whether you had ever come across the Vienna Circle? I first learnt about this group of philosophers (Schlick, Carnap, Neurath, Frank, \ldots) when I visited \v{Ca}slav Brukner and Anton Zeilinger in Vienna last October, and I discovered that their ``logical empiricism'' is very much in tune with the Copenhagen interpretation of quantum theory. In fact, there was a fruitful exchange between members of the Vienna Circle and Niels Bohr. I attach a little paper discussing this exchange -- maybe you'll enjoy it.
\ejr

Thank you for the Faye article.  I have read him before and find him very interesting.  Particularly, I read his book {\sl Niels Bohr, His Heritage and Legacy\/} in very careful detail about 15 years ago.  Particularly it was very important for me in learning about {\Hoffding} (which connects to my present infatuation with pragmatism, James and {\Hoffding} were correspondents).  It would be so nice to read this all again, knowing (thinking) what I know (think) now.  Faye and Henry Folse, by the way, are good friends.  I visited Henry at his magnificent home in New Orleans in October, and he told me a story of the time Faye lived in his basement there.  Anyway, though they are good friends, they have very different readings of Bohr.  And for my money, I think there is more truth in Folse's reading---which of course can only mean operationally that Folse's reading agrees with my own reading of Bohr.  (I may not reflect it in my writings, but I have put significant scholarly effort into Bohr himself.)  Still, the healthiest reading involves both these countering opinions.  I will study the article you sent me with care.

\section{11-02-10 \ \ {\it Itamar} \ \ (to M. Hemmo)} \label{Hemmo3}

This hits me like a hammer.  It is a sad day.  He was a great man, from whom I learned so much.

\subsection{Meir's Preply}

\bq
\noindent Dear friends,\medskip

I am very sorry to report that Itamar died last night.
There are no words to express this great loss for all of us.

The funeral will take place tomorrow, Friday 12 February, at 10:00 in Jerusalem.\medskip

\noindent Meir
\eq

\section{11-02-10 \ \ {\it Itamar}\ \ \ (to W. G. {\Demopoulos})} \label{Demopoulos33}

I had already heard about Itamar from Meir and sent my condolences.  It is very tough.  Itamar was a great man.  I learned all the deep, secret details of Gleason's theorem from him.

My mind this morning keeps coming back to a quote of William James:
\begin{quote}
To anyone who has ever looked on the face of a dead child or parent the mere fact that matter {\it could\/} have taken for a time that precious form, ought to make matter sacred ever after. It makes no difference what the principle of life may be, material or immaterial, matter at any rate co-operates, lends itself to all life's purposes. That beloved incarnation was among matter's possibilities.
\end{quote}
I want say that Itamar's work in quantum theory was a foundation for the same conclusion, having come at it from a more rational side.  Matter is sacred.

I will remember him all my life.

\section{15-02-10 \ \ {\it  Notes} \ \ (to M. B. Ruskai)} \label{Ruskai3}

Yes, thank you, that would be nice.  There is a quote of Primas's that I keep in my computer that I think is very deep:

\bq\noindent
It is a tacit assumption of all engineering and experimental sciences that nature can be manipulated and that the initial conditions required by experiments can be created by interventions using means external to the object under investigation. That is, {\it we take it for granted that the experimenter has a certain freedom of action which is not accounted for by first principles of physics}.  Man's free will implies the ability to carry out actions, it constitutes his essence as an actor. Without this freedom of choice, experiments would be impossible. {\it The framework of experimental science requires this freedom of action as a constitutive though tacit presupposition.}
\eq

I'm off to Japan early tomorrow morning.

\subsection{Mary Beth's Preply}

\bq
Don't know if you're familiar with the work of Hans Primas on
quantum theory.  Going through some papers, I found a set of lecture notes from
a course he gave at the ETH about 1970 on
\bv
Problem der Interpretation der Quantnemechanik Grosser
     Molekular Systeme
\ev
and I wonder if you might be interested.   I also have a few
preprints from that time.

If you'd like them, I'll see if Bei Zeng can bring them back to
Waterloo.
\eq

\section{17-02-10 \ \ {\it Jet Lagged Titular Thought} \ \ (to R. {\Schack}, D. M. {\Appleby}, H. C. von Baeyer)} \label{Appleby84} \label{Baeyer104} \label{Schack188}

``Institute for Quantum and Freedom''

Better suggestions?

\section{17-02-10 \ \ {\it Strong Radio Silence}\ \ \ (to M. Schlosshauer)} \label{Schlosshauer23}

You weren't kidding about ``radio silence'':  I'm the king of it!  Not only am I not in Vienna anymore, but last night I arrived in Japan.  In Nagoya.  And to my surprise this morning, I learned that the meeting starts today, not tomorrow as I had thought!  So my head is really in the clouds somewhere.

Let us shoot for early June with respect to my getting the interview paper in.

It must be getting very close to your baby's due date now.  I hope everything is going OK in that regard.

On my flight over, I read your paper with Camilleri on Bohr's doctrine of classical concepts.  It was quite good, and I learned a lot.  Particularly, I did not know about the controversy between Bohr and Heisenberg on the movability of the cut/schnitt.  I want to delve into that more carefully as I get a chance.  I was quite surprised by this line of Heisenberg's too:  ``[A] dividing line must be drawn between, on the one hand, the apparatus which we use as an aid in putting the question and thus, in a way, treat as part of ourselves, and on the other hand, the physical systems we wish to investigate.''  I thought it was Pauli only who went so far as to take the measuring device as a prosthetic hand.  That point is crucial for QBism in fact; so clearly I should learn more about its roots.

Attached finally is a bit more careful version of what I brought up with you after the New Year, my little epiphany.  I am not completely pleased with the exposition---it remains too sketchy---but it is starting to take shape, and it gives me something to try to do better in a proper article later.  The argument starts kicking in at page 5 with the discussion of Wigner's friend, comes back a little again at the end of the Seeking SICs section, and then climbs to a fevered pitch starting at page 16 in the Universal Capacity section.

You'll see that everything has grown a bit since the last version I sent you.  I took into account almost all your comments, for instance.  And indeed, your comments helped set me up for my New Year's epiphany---so I owe you a lot.  One thing I haven't done yet that you suggested is to have a discussion on the subjective-Bayesian nature of unitaries---that's actually crucial now, so it's coming.  In the present version, I am a bit harsh on the ``quantum-to-classical transitionists.''  I hope you will be provisionally tolerant of me and not get too angry.  I can work harder to set the tone right and still make my point.  Those paragraphs were written on the flight over here; so they're pretty raw.  They are like Homer Simpson thinking out loud.  Still, I thought I'd go ahead and show them to you now and take into account your reactions (and levels of anger) as they are for real, rather than try to anticipate them.

Scanning over your note again \ldots\  On your question about Anton:  Yes, it seems he is actually going to get one of his teams to do a qutrit SIC demo.  So tugging on his shirt was a good thing.  In fact, he's now quite interested in MUBs and SICs and will be holding a meeting in Vienna on the subject in July.

My apology again on the long silence.  One should not go quiet on friends.  But some not-good things happened right after Vienna (an old friend died), and I just ended up shutting up for a while.

\section{17-02-10 \ \ {\it Oh, Repository} \ \ (to R. {\Schack})} \label{Schack189}

Isn't your new ``potential schism'' the very core of our considerations already?  The Law of Total Probability is operative when we expect to actually have an experience $E_i$.  The Born Rule (urgleichung) tells us how to modify that old estimate into something relevant when we don't expect an intermediate $E_i$ experience.  It is a postulate of quantum mechanics, over and above coherence considerations.

I'm probably still not getting your point.

Random thought.  Just recording it even it's not relevant to the present discussion.  Decoherence is really just a cheap way of extending ``the self'' without recognizing it as a such.  It tries to say, ``that experience is really there waiting for me to have it'' when in fact, all it is doing is lengthening the arm of the agent (with a complicated story in between).  In this regard, William James's notion of self is important:  ``In its widest possible sense, however, a man's self is the sum total of all that he can call his, not only his body and his psychic powers, but his clothes and his house, his wife and children, his ancestors and friends, his reputation and works, his lands and horses, and yacht and bank account.''

\subsection{{\Ruediger}'s Preply}

\bq
I hope you made it safely to Japan!

I read the first 10 pages of the paper carefully and have some comments (see below), but when reading the Wigner's friend bit and then the measurement in the sky story, I got hung up on a problem that may turn out to be of central importance.

If I intend to perform the experiment in the sky, I expect to get outcome $E_i$ with probability $P(E_i)$, and I expect that in the case I actually experience $E_i$ my new probability for $D_j$ will be my current $P(D_j|E_i)$. Reflection then gives me the law of total probability.

But what if I only {\it imagine\/} performing the experiment? I.e., what if I believe there is decoherence (I can see you shudder!)\ and I want to use reflection to set my probabilities according to the law of total probability as above. I would have to say my beliefs are {\it as if\/} I expected to actually experience $E_i$. This sounds like vile as-if-ism to me. If I don't expect to actually experience $E_i$, how to I justify to myself setting $P(D_j)$ according to the law of total probability in one case, and according to the Urgleichung in the other case?

Does this have the potential for a schism?
\eq

\section{18-02-10 \ \ {\it Eliot}\ \ \ (to M. Schlosshauer)} \label{Schlosshauer24}


This synchronicity thing between us seems to have no end.  I was on a plane reading your paper, thinking about how you think, thinking about how you will react, and all kinds of things like that.  Then, along the way (as I stop in Dallas), {\Ruediger} finally sends his QB-decoherence draft.   [See \arxiv{1103.5950}.]  Then, I learn that you and your wife were in Copenhagen participating in the biggest event of your lives.

Many, many congratulations.  You will have a strong son; he was born in the land of Niels Bohr.\medskip


\section{19-02-10 \ \ {\it Potentially Perfect Potential Aaron Problem}\ \ \ (to A. Fenyes \& D. M. Appleby)} \label{Fenyes1} \label{Appleby84.1}

The PPPAP. [Notation and equations refer to D. M. Appleby, S. T. Flammia, and C. A. Fuchs, ``The Lie Algebraic Significance of Symmetric Informationally Complete Measurements,'' \arxiv[quant-ph]{1001.0004}.]

Marcus has known for some time that I keep hoping that the $R_r$ will imply everything we need to know about the $J_r$.  If the raw shape of quantum state space is more fundamental than anything else, then the $R_r$ have to be more fundamental.  I still hold out this hope.

This morning as I was putting together my talk for Japan, I was struck that we haven't asked the obvious question.  What is the analogue of Theorem 7 for the Jordan algebra induced by the anti-commutator?

In other words, equip the vector space of Hermitian matrices with a symmetric product (the anti-commutator), and let $\Pi_t$ be a basis for the algebra.  (No assumptions about $\Pi_t$ being a SIC or anything.)  Now define a matrix $R_r$ by Eq.\ (16) in the Lie algebra paper, and suppose the $R_r$ have an analogue of the $Q$-$Q^T$ property.  Namely that the $R_r$ can be written as in Eq.\ (103), where the $|{\!}|e_r\rangle\!\rangle$ are defined as EXACTLY as in Eq.\ (116) and the $Q_r$ are rank $d-1$ projectors satisfying Eq.\ (105).  Does it follow in analogue to Theorem 7 that a SIC must exist?

If it does, this will be immensely exciting.  How are the $\Pi_t$ related to this SIC, in analogue to Eq.\ (164)?  And if such a theorem does not follow, then what does?  What is the most one can say about the basis $\Pi_t$?

Aaron, I think this could be a PSI project for you (or the beginning of one) if you are interested.  If you get a positive result, it may not have so many applications to the SIC existence problem as the Lie algebra version might (I say might), but for my money it'll be a deeper result about quantum mechanics itself.

Part of the point is, if there is a result like this, then one has a nice way to start with a Jordan algebra and then define a natural Lie algebra associated with it.

Let me know what you guys think?  Marcus, what are your feelings/thoughts?

\section{19-02-10 \ \ {\it Postscript}\ \ \ (to D. M. {\Appleby})} \label{Appleby85}

\bma
Hope you are enjoying Japan.
\ema

No enjoyment yet.  I've just been writing, working on ONR things, trying to plan (unsuccessfully) my PI colloquium, and trying to get Belavkin and Ozawa to respect us.  In the wee hours of the morning when I have my hot bath, I've been trying to think about gravity.  The bold proposition is that simply the presence of a quantum system of any kind in an otherwise empty region of space must bend light around it.  Just like those beautiful pictures we saw at the colloquium on gravitational lensing.  No talk of coupling to the EM field (in the usual sense), no talk of the quantum state one ascribes to it, simply the presence of a system.  There has to be an equivalence principle lurking here.

I am very intrigued by what you seem to have found.  You are right about the flavor seeming to hint that all SICs are WH SICs or simple transformations of them.

Write as you wish, but there is no necessity.  If you prove SIC existence, I'll buy you the next bottle of single malt!  And a very, very rare one at that.

\section{21-02-10 \ \ {\it Pierce Quote}\ \ \ (to V. P. Belavkin)} \label{Belavkin1}

J. R. Pierce (IEEE Trans.\ Inf.\ Theory, vol {\bf IT-19}, 3--8 (1973)):
\bq\noindent
I think that I have never met a physicist who understood information theory.  I wish that physicists would stop talking about reformulating information theory and would give us a general expression for the capacity of a channel with quantum effects taken into account rather than a number of special cases.
\eq

\section{21-02-10 \ \ {\it Ways of Expression}\ \ \ (to D. M. {\Appleby})} \label{Appleby86}

Here's the way I expressed it tonight to Gen Kimura:  If SICs exist, then there's a beautiful way of thinking of quantum state space that had not been thought of before.  It is a union of a bunch of regular simplices, each over $d^2$ extreme points.  ``Quantum state space is a union of regular simplices.''\footnote{See Footnote \ref{OhCrap} for a technical, but remediable, mistake in the original statement of this. \label{OhCrap2}}

\section{22-02-10 \ \ {\it Schelling, Quantum, Creation}\ \ \ (to I. Ojima)} \label{Ojima1}

I am very happy about our chance conversation today---now I have another refreshing thinker on nature's creativity to read and think about.  Schelling strikes me as very interesting.

Attached is the essay I've been writing at this very meeting.  It is still unfinished, but it already demonstrates why I was so interested in my conversation with you today.

Here are a few sentences from my paper that may whet your appetite:
\begin{itemize}
\item
Page 6, figure caption)\\
Measurement devices are depicted as prosthetic hands to make it clear that they should be considered an integral part of the agent. The sparks between the measurement-device hand and the quantum system represent the idea that the consequence of each quantum measurement is a unique creation within the previously existing universe.

\item
Page 8)\\
QBism says when an agent reaches out and touches a quantum system---when he performs a quantum measurement---that process gives rise to birth in a nearly literal sense. With the action of the agent upon the system, the no-go theorems of Bell and Kochen--Specker assert that something new comes into the world that wasn't there previously:  It is the ``outcome,'' the unpredictable consequence for the very agent who took the action. John Archibald Wheeler said it this way, and we follow suit, ``Each elementary quantum phenomenon is an elementary act of `fact creation.' That is incontestable.''

\item
Page 14)\\
The Quantum Bayesian, however, with a different understanding of probability and a commitment to the idea that quantum measurement outcomes are personal, draws a different conclusion from the theorem. And it is profound: Bell's theorem for the Quantum Bayesian is the deepest reaching statement ever drawn from quantum theory. It is that quantum measurements are moments of creation.

\item
Page 20)\\
To put it still differently, and now in the metaphor of music, a jazz musician might declare that a tune once heard thereafter plays its most crucial role as a substrate
for something new. It is the fleeting solid ground upon which something new can be born.
\end{itemize}
Anyway, those are the sorts of things you'll find argued for in the essay.

Below are my already favorite Schelling quotes:
\bq\noindent
Has creation a final goal? And if so, why was it not reached at once? Why was the consummation not realized from the beginning? To these questions there is but one answer: Because God is Life, and not merely Being. \smallskip \\ \hspace*{\fill} --- Schelling, {\sl Philosophical Inquiries into the Nature of Human Freedom}, 1809
\eq
and
\bq\noindent
Only he who has tasted freedom can feel the desire to make over everything in its image, to spread it throughout the whole universe. \smallskip \\
\hspace*{\fill} --- Schelling, {\sl Philosophical Inquiries into the Nature of Human Freedom}, 1809
\eq

\section{22-02-10 \ \ {\it Hilbert-Space Dimension in Second-Quantized Theories}\ \ \ (to N. C. Menicucci \& E. G. Cavalcanti)} \label{Cavalcanti9} \label{Menicucci5}

\bnm
Eric Cavalcanti raised a question that I've thought about, as well, in relation to your claim in ``Quantum Mechanics as Quantum Information (and only a little more)'' that the dimension of a Hilbert space is real: what happens if the dimension is not well defined, as is the case in QED?  One can have a superposition of photon-number states, which results in a superposition of number of qubits available and thus a superposition of dimensions of the overall state space (e.g., H/V qubit encoding).  Is the answer trivial (e.g., the overall dimension is infinite, which is a real property of the world), or is there something more subtle at work here?  Have you (or anyone else that you know of) written about this?
\enm

No I haven't written about this before, but I am just about going to!  There'll be a discussion in the attached paper which I am in the process of writing at this very moment.  I'll go ahead and send you this much, and sketch very briefly an answer to your present question.  The paper helps give the setting, and the precise sense in which I'm thinking of quantum dimension now.

Anyway, to answer your question:
\begin{enumerate}
\item If quantum fields really did exist, the answer would be the one you call the trivial one:  $\mbox{dimension} = \infty$.

\item Even if they don't exist, one might think of this case as like the moral equivalent of an infinite heat bath:  a limiting case or approximation.

\item But in fact, I'm starting to suspect the nonexistence of infinite dim Hilbert spaces.  This is because I am starting to take this holographic principle stuff more seriously.  But I think these guys are misidentifying the principle as giving an entropy bound (how could you do that with a subjective quantity?).  What I suspect it really gives is a dimension bound.

\item In any case, if such is the case, it keeps my ``dimension as valence'' thinking from collapsing.  So, something like that ought to be the case.
\end{enumerate}
These are the things I'll eventually write more thoroughly (and more poetically!)\ in the completed paper.

I hope you find some food for thought in the paper, even as it stands.

Best wishes from Nagoya, Japan (in the middle of a talk on ``nonlocality'' \ldots\ as if such crap were true).

\section{23-02-10 \ \ {\it Kitaro and James}\ \ \ (to I. Ojima)} \label{Ojima2}

I have now read this article on Nishida Kitaro's conception of ``pure experience'' in contrast to William James's conception of it.  [See D. Dilworth, ``The Initial Formations of `Pure Experience' in Nishida Kitaro and William James,'' {\sl Monumenta Nipponica} {\bf 24}(1/2), 93--111 (1969).] Indeed---a bit as you warned---Kitaro does not look like such an interesting philosopher to me.

The article ``Nature in American Philosophy,'' by Stefano Poggi, on Google Books looks to be very interesting.  I have just started reading it, but it looks like it does a good job of tracing Schelling's influence on Fechner, and Fechner's influence in turn on James.

\section{26-02-10 \ \ {\it Kitaro and James, 2}\ \ \ (to I. Ojima)} \label{Ojima3}

Thank you so much for the papers.  I have so far read the first [I. Ojima, ``Nature vs.\ Science,'' Acta Inst.\ Phil.\ Aesth.\ {\bf 10}, 55 (1992)] and 1/4 of the second on my flight back to North America.  These are very deep papers, and we are very much on similar wavelengths.  Technically, I enjoyed your argument on the nonexistence of (ontic) repeatability in the first paper.  You gave repeatability a logical inconsistency (rather than an empirical establishment) that I had not appreciated before.  To get repeatability you call for ``approximation'', and in my own naming scheme, I call it ``judgment''.  But it is much the same thing---certainly our hearts are in the same place.

I look forward to finishing your longer paper as I have a chance.

I have gotten further in the writing of the paper I sent you last week.  When it is finally finished, I will surely send it to you.

\section{01-03-10 \ \ {\it My Lectures for You} \ \ (to J. Emerson \& R. Laflamme)} \label{Emerson4} \label{Laflamme7}

I plan to work through the night putting the finishing touches on the attached (still slightly incomplete) essay.  [See \arxiv{1003.5209}.]  Here's what I'd like to do for my lectures for your course.  Let me check that you're OK with this.

Rather than give the students a post-lecture homework assignment of any sort, I would like to give them a pre-lecture suggestion or assignment:  That they read the attached essay before I show up next week (tomorrow morning's version more precisely).  Then through the week or over the weekend, they can email me any questions or discussion points they would like to be brought up in my lectures.  In place of rigid lectures then, I'd try to spend much or most of the time replying to their questions, dodging tomatoes, etc.  Similarly with any issues that arise spontaneously in the class.  I think they'd get a lot more out of this than ``yet another lecture on yet another point of view.''  I.e., it puts me on the spot, and they (and you) can try to pin me down as much as they (and you) want!  The essay was written with an intention of being entertaining and provocative; so it should be quite easy reading for them.

If it's OK with you, I'd like to spend less than one minute tomorrow saying this is how I want to run things for my lecture slots.  And I'd put a pile of these papers at the front so that they could pick a copy up.  (How many students are there?)  Then I'd quickly disappear so as not to interrupt your present lecturer.

Please, let me know, and I'll come prepared to tomorrow's lecture with a pile in hand.  And I hope to give you some fun next week.

\section{01-03-10 \ \ {\it Curiosity}\ \ \ (to C. Ferrie)} \label{Ferrie13}

Thanks for this.  It is a deeply thoughtful essay, and I enjoyed it very much.  You have no need to worry that you cannot ``articulate so well''---it simply is not true.  Your message came across loud and clear.  Are you a different person than when I met you?  Or was I just stubbornly blind/deaf?

Just because I caught them, I'll point to some of your typos: \ldots

Now, not a typo, just a disagreement:
\bcf
``Humans possess freedom of will.  This is a privilege which the Earth, for example, does not seem to benefit from.''
\ecf
Want to bet?

\section{02-03-10 \ \ {\it Our Unfinished Universe} \ \ (to A. Zeilinger \& \v{C}. {\Brukner})} \label{Brukner6} \label{Zeilinger10}

Well, I have finally finished (within epsilon at least), that essay I was writing on in Vienna---the one in which I try to capture my whole (present) worldview.  I hope you will both enjoy and not find the ending sections (which were not yet written in Vienna) too outlandish!  If you have any comments on anything you think I should change or add or delete!, please do tell me.  I'm hoping to post the thing at the end of the week.

\v{C}aslav, in the footnote on Kofler, you'll see that our conversation the last days of the visit very much influenced me.

Brothers in arms for a creative universe!

\section{02-03-10 \ \ {\it Wheelerism Full-Throttle!}\ \ \ (to B. W. Schumacher)} \label{Schumacher17}

I'm nearly finished with an essay I've been pouring my soul into, and I thought I'd send you a draft before posting since it's chock-full of Wheelerish stuff.  I attributed Wheeler's ``great smoky dragon'' to a paper by he and Warner Miller; do you happen to know whether that is the right one?  Anyway, if you see anything that should be corrected or amended, please don't hesitate to say.  I'll probably shoot for posting this Friday.

Hope you and Carol and the whole family are well.

\subsection{Ben's Reply}

\bq
I won't be able to read this for a couple of days. Looks fun! The Great Smoky Dragon cartoon appears as Plate 27 (p.\ 399) of the 1988 {\sl Between Quantum and Cosmos\/} book. The notes at the end (p.\ 619--20) give a reference that appears to agree with your Ref.~70.
\eq

\section{02-03-10 \ \ {\it Nearly Final Product}\ \ \ (to H. C. von Baeyer)} \label{Baeyer105}

Attached is the nearly final product; within epsilon it is finished!  You will find it has changed substantially since the partial draft you read.  Quantum measurement, for instance, is no longer a kick, but a birth!  I hope you enjoy and not find the whole thing too outlandish.

From YOU (but only you), editorial comments always welcome!  Don't be shy if you see something that really could be improved.

\subsection{Hans's Replies}

\bq
Dear Chris -- witnessing a birth is vastly more pleasant than giving or receiving a kick!  I am reading carefully, and am only on page ten.  Soon I will send you a list of small suggestions. \ldots

As I read that the world is not made of [Essence X] I recall a relevant quote by the American poet Muriel Rukeyser: ``The world is not made of atoms; it is made of stories.''
\eq

\section{02-03-10 \ \ {\it Fatherhood/Paperhood}\ \ \ (to M. Schlosshauer)} \label{Schlosshauer25}

How are you and your wife making it?  Hopefully it's been relatively easy; my recollection is that babies are pretty easy at the beginning.

Attached is an actual (potentially) FINAL version of the essay.  I bet you thought you'd never see that coming.  I beefed up the quantum-to-classical transition discussion a bit---I hope it helps matters a small amount.  Only one thing in your ``to do'' list that I did not do:  I just couldn't bring myself to writing on the subjective meaning of unitaries.  I lost steam, and couldn't find an easy place for it to fit into the present structure.  If you do think the essay is really lacking because of this, please do tell me, and I'll rethink.  A day or two of rest often brings a second wind.  I value your input; I hope you know that.  And I'll value your criticism as well:  If I can't get it right with you, I won't be able to get it right with anyone.

I've got a big talk coming up next week that has me completely frightened:  It's part of trying to establish myself a permanent position here.  Who would think a 45 year old man could be scared like a little boy?  But it happens.

I met one of your colleagues from Copenhagen in Japan, Teiko Heinosaari.  He seems to be a good guy.  He's even working on SICs now!

Please give my regards to your wife.  And say hello to Niels from close up; he's been on my mind again lately.

\section{02-03-10 \ \ {\it Another Attempt To Get You To Call Me a Realist!}\ \ \ (to H. M. Wiseman)} \label{Wiseman28}

Since I implicate you in this essay, I should probably send it to you before posting!  [See ``QBism, the Perimeter of Quantum Bayesianism,'' \arxiv{1003.5209v1}.]  I drew one of the figures for you.  Seriously, thanks for the great challenges you've given me over the years.  Any comments you have for improvement---so long as they're relatively easy to implement---will be welcome.

\section{02-03-10 \ \ {\it QBism}\ \ \ (to C. M. {\Caves})} \label{Caves101.1}

Here's a little essay that I'm just on the verge of finishing (awaiting various readers' comments in case they lead to some relatively easy improvements).  [See ``QBism, the Perimeter of Quantum Bayesianism,'' \arxiv{1003.5209v1}.]  If I understand you correctly, you think of nonreductionism as antithetical to science, but I don't see it that way, and I hope I did a good job of dispelling at least some of your scepticism in this.  I tried at least.  In any case, I do think a good bit of what I'm talking about {\it is\/} science.

I do hope you enjoy some of the passages in the essay, even if you find you disagree with the majority of the thing.  If you have any comments, I will accept them gracefully---it's a pledge.

I hope you're doing well.  And maybe I'll see you at the March meeting?

\section{03-03-10 \ \ {\it My Worldview} \ \ (to A. Kent)} \label{Kent20}

Since I implicate you in one of my sly insults, I probably ought to send you this essay before posting it:
\bq\noindent
Or take the Everettians. Their world purports to have no
observers, but then it has no probabilities either. What
are we then to do with the Born Rule for calculating
quantum probabilities, the very core of the theory from
the experimental perspective? Throw it away and say
it never mattered? It is true that much effort has been
made by the Everettians to rederive the rule, but to many
in the outside world, it looks like the success of these
derivations depends upon where they are assessed---for
instance, whether in Oxford [9] or Cambridge [10].
\eq
I hope you enjoy (some of) it.  It is my latest attempt to form a worldview.

I hope you come to PI again soon; it's always a blander place when you're not around.

\section{03-03-10 \ \ {\it A Jaume Stamp?}\ \ \ (to J. Gomis)} \label{Gomis1}

I finally finished that essay I was telling you about.  [See ``QBism, the Perimeter of Quantum Bayesianism,'' \arxiv{1003.5209v1}.]  It's attached.  I didn't do as good a job connecting Hilbert-space dimension to mass/energy as I had hoped to, but I do believe there is something much deeper to the analogy between ``fungibility of quantum information'' and the equivalence principle than meets the eye.  Anyway, the essay expresses the direction of thought I'll be exploring.

\section{03-03-10 \ \ {\it Novelty, Creation, QBism}\ \ \ (to L. Smolin)} \label{SmolinL14}

I haven't seen you in a long time, and I've just written something that I hope will intrigue you.  [See ``QBism, the Perimeter of Quantum Bayesianism,'' \arxiv{1003.5209v1}.]  It's an essay on my worldview, at least as it stands at the moment.  See attached.  Section V onward, particularly, is about how my QBism research program mounts onto the idea of a pluriverse in constant creation.  I don't know to what extent my ``interiority'' has a connection to Verlinde's ``our starting point was that space has one emergent holographic direction,'' but I am tempted to think something like this is going on \ldots\ but from a completely different end of the philosophical spectrum.  I feel it is a big research program ahead.

I hope to post the essay next week maybe.  If you have any feedback on how the draft might be improved or made clearer, I'd much appreciate hearing your thoughts.

\section{04-03-10 \ \ {\it My Worldview, 2} \ \ (to A. Kent)} \label{Kent21}

\bak
Thanks.  I haven't had a chance to read yet, beyond
searching for the flagged implication.   I have to be
honest, I'm rather disappointed by it: it seems a cheap and lazy way
of avoiding serious discussion, and quite inappropriate for an
academic article.
\eak

Guilty.  And I misstated when I said ``sly''---I forgot the extra {\it i\/} and {\sl l}.  Silly.  But the cheapness of it was meant to be a direct comparison to John Bell's statement.  The beginning of the paragraph announced it so.  There are some things, we each of us, just do not take seriously.  For instance a thoroughgoing epistemic interpretation of quantum states, as Bell didn't (and maybe you, don't).  Anyway, much of the remainder of the article is meant to show how one can recover from the sting of Bell's original cheap shot.

The only thing I worry about now is that you won't look at the rest of the article, where I try to explain the mindset that would make Bell's remark a ``cheap shot'' rather than a ``forceful way of conveying the essential point''.

\section{04-03-10 \ \ {\it Edits}\ \ \ (to H. C. von Baeyer)} \label{Baeyer106}

\bhcvb
Figure 2 is unnecessary, and the inclusion of the ass among the Ten
Commandments is wrong.
\ehcvb
See:  \myurl[http://en.wikipedia.org/wiki/Ten_Commandments]{http://en.wikipedia.org/wiki/Ten\underline{ }Commandments}:
\bq\noindent
The firstborn of a donkey you shall redeem with a lamb, or if you will not redeem it you shall break its neck. All the firstborn of your sons you shall redeem.
\eq
And from another page:
\bq\noindent
But the firstling of an ass thou shalt redeem with a lamb: and if thou redeem him not, then shalt thou break his neck. All the firstborn of thy sons thou shalt redeem. And none shall appear before me empty.
\eq
But you are right, strictly speaking, it is not one of the 10 Commandments.  I'll fix the drawing and caption before posting.  It was an attempt at humor; it definitely works in talks, but maybe it would not in the paper \ldots\ and it might even be offensive to some.  Still, I don't think the discussion is unnecessary:  You are the rare physicist who has ever heard of a normative rule.  And generally, in both physics and philosophy of science, the Born Rule is thought to be of a character as Maxwell's and Einstein's equations:  A statement of what is in the world.  A quantum state IS, and the probability it carries with it IS.  I want to hammer it home that it should be thought of as a recommendation, nothing more.

I keep drawing myself away from my talk preparations to look at your notes!

\section{04-03-10 \ \ {\it Fuchsianism / Ones For Which}\ \ \ (to R. Blume-Kohout)} \label{BlumeKohout11}

I finally put it all on paper, everything in my head.  It's empty now.  Anyway, you always say that whatever I believe of quantum mechanics is a mystery to you:  I hope it won't remain so if you read this paper.

And don't forget:
\begin{enumerate}
\item
Scaffolding paragraph.
\item
``Hilbert-Space Dimension as a Universal Capacity'' section
\item
And I might as well point out these lines too:
\end{enumerate}
\bq
You see, for the QBist, the real world, the one both agents are embedded in---with its objects and events---is taken for granted.  What is not taken for granted is each agent's access to the parts of it he has not touched.  Wigner holds two thoughts in his head: 1) that his friend interacted with a quantum system, eliciting some consequence of the interaction for himself, and 2) after the specified time, for any of Wigner's own further interactions with his friend or system or both, he ought to gamble upon their consequences according to $U\big(\rho\otimes|\psi\rangle\langle\psi|\big)U^\dagger$.  One statement refers to the friend's potential experiences, and one refers to Wigner's own.  So long as it is kept clear that $U\big(\rho\otimes|\psi\rangle\langle\psi|\big)U^\dagger$ refers to the latter---how Wigner should gamble upon the things that might happen to him---making no statement whatsoever about the former, there is no conflict.  The world is filled with all the same things it was before quantum theory came along, like each of our experiences, that rock and that tree, and all the other things under the sun; it is just that quantum theory provides a calculus for gambling on each agent's own experiences---it doesn't give anything else than that.  It certainly doesn't give one agent the ability to conceptually pierce the other agent's personal experience.  It is true that with enough effort Wigner could enact Eq.~(1), causing him to predict that his friend will have amnesia to any future questions on his old measurement results.  But we always knew Wigner could do that---a mallet to the head would have been good enough.
\eq

\section{04-03-10 \ \ {\it For Fun}\ \ \ (to C. M. {\Caves})} \label{Caves101.2}

By the way, the baseball example comes from you.  I should cite you, but didn't remember where you had put it in print.  You did, didn't you?  Tell me where, and I'll mark it.

\bcc
I think I might have talked to you about what an outfielder does in trying to catch a fly ball.  Certainly not integrating Newton's equations, more like a dog catching a Frisbee, by comparing the observed trajectory against movies stored in the brain.  I do think they're only observing position directly and comparing the observed trajectory against stored data.  Anyone who thinks we directly measure velocity has never been riding a bike up a hill and finding that drivers wait at intersections just as though you were going as fast as a car.
\ecc
Your example is a good one.  But actually Appleby told me last week that it seems the brain has a position processing center and velocity processing center, and that the two centers can be damaged independently \ldots\ allowing windows into some very strange human behaviour.  I'll ask him if he can tell me a reference.

\begin{itemize}
\item
S, Zeki, ``Cerebral Akinetopsia (Visual Motion Blindness),'' Brain {\bf 114}, 811--824 (1991).
\item
M. Rizzo, M. Nawrot, and J. Zihl, ``Motion and Shape Perception in Cerebral Akinetopsia,'' Brain {\bf 188}, 1105--1127 (1995).
\end{itemize}

\section{05-03-10 \ \ {\it Laws of Nature Conference} \ \ (to G. Musser)} \label{Musser22}

I finally finished that essay that I had shown you part of.  It's attached.  [See ``QBism, the Perimeter of Quantum Bayesianism,'' \arxiv{1003.5209v1}.]  Do you think you can cut it with meat cleaver and scalpel into something for Sci Am?  Would you be interested?  Section V is definitely the antithesis of that David Albert article you had.

\section{05-03-10 \ \ {\it Torture} \ \ (to A. Kent)} \label{Kent22}

\ldots\ like a grain of sand in an oyster shell, you have been with me for the last 24 hours.  It makes no pearl with me, but I do come to appreciate the annoyance the little fellow feels.

I do insist on insulting Everett, but your scolding (beside making me feel ashamed) did make me realize that I wasn't getting to the essential point.  I try to do that in the present version.  I will probably tweak it a hundred times more before posting, but the final result will be something more like this than yesterday's:
\bq\noindent
Or take the Everettians. Their world purports
to have no observers, but then it has no probabilities either. What are we then to do with the Born Rule for
calculating quantum probabilities? Throw it away and say it never mattered? It is true that quite an effort
has been made by the Everettians to rederive the rule
from decision theory. Some sympathizers think it works
[9], some don't [10]. But outside the {\it sprachspiele\/} who
could ever believe? No amount of sophistry can make
``decision'' anything other than a hollow concept in a deterministic universe.
\eq

\section{05-03-10 \ \ {\it Higher Road}\ \ \ (to R. W. {\Spekkens})} \label{Spekkens79}

By the way, I did choose a (slightly) higher road with the insult.  Now it targets the idea more than the people.  You were right.  I also rejigged the footnote on Granade and you to something a little less lame:
\bq
Not to be confused with Scientologists. This neologism was
coined by Chris Granade, a Perimeter Scholars International student at Perimeter Institute, and brought to the author's
attention by R. W. {\Spekkens}, who pounced on it for its beautiful subtlety.
\eq
If you're interested, latest version attached.

\section{05-03-10 \ \ {\it A Wild Shock!}\ \ \ (to O. C. O. Dahlsten)} \label{Dahlsten2}

Thanks.  On Daki\'c and Brukner, I personally don't think ``information capacity'' is the right idea---it is still too close to treating a quantum system as a bucket into which one can pour a certain amount of stuff---but I felt I should cite them in some way.  Maybe I'll leave the reference, but take the quote out.  I liked Hardy's (old) way of putting it better.  It is just that no distinction of quantum theoretical properties can be made between two systems of dimension $N$.

I look forward to seeing your draft.  In my own way, I might want to see systems dropped as well---that's partly what the mumbo-jumbo about ``pure experience'' was about at the end.  But things are not clear in my mind yet.  Extra stimulation would be good!

\subsection{Oscar's Preply}

\bq
Thanks a lot, I enjoyed reading it!

I also take PSI-epistemic view, that PSI represents our knowledge about something out there. I even agree that the Hilbert space dimension seems to be something real out there.

A small remark on the Daki\'c and Brukner paper which you mention in a footnote, which I otherwise like, is that `information capacity' may be a strange choice of word. Many other hypothetical systems, including gbits have the same information capacity as a qubit, i.e., one can only communicate one classical bit with them (in the sense that Bob can distinguish at most two of Alice's preparations after Alice has sent him the system). We (Roger, Renato and I) are writing something about these things at the moment, i.e., about different notions of how much information a system can be used to convey.

In my way of thinking of things I am inclined to think that the notion of system itself should actually be dropped. Am writing this up at the moment and will send you a draft in case you have time/interest to look at it.
\eq

\section{08-03-10 \ \ {\it Another Attempt To Get You To Call Me a Realist!, 2}\ \ \ (to H. M. Wiseman)} \label{Wiseman29}

\bhw
I've started to reading it, and I like the tenor of it so far. You seem to have taken a deep intellectual breath
before writing this time.

Did I ever share with you my idea of Escher's ``Print Gallery'' as a metaphor for the measurement problem?
The boy is the observer, the picture is what he is observing, the town is the universe. The blur in the middle is the flaw
which means we will never be able to solve the problem.
\ehw

I'll keep my fingers crossed that that lasts!  I don't believe you'd ever told me the Escher metaphor.\footnote{\editornote See \myurl{http://escherdroste.math.leidenuniv.nl/}, and also B.\ de Smit and H.\ W.\ Lenstra Jr., ``The Mathematical Structure of Escher's Print Gallery,'' Notices of the AMS \textbf{50}(4), pp.~446--51 (2003).}  My own favorite metaphor comes from a painting as well:  Joan Vaccaro's ``Broken Block'' which she sent me, and I have hanging in my study at QBism house.  It beautifully expresses all the cracks in the block universe idea.  You can still see it at Joan's webpage, and attached is a little blurb on the house.

\section{08-03-10 \ \ {\it Lee, from the Inside}\ \ \ (to L. Smolin)} \label{SmolinL15}

Actually a Google search on the phrase ``quantum cosmology from the inside'' gives {\it Precisely\/} one hit \ldots\ and it was one of your papers!

\section{10-03-10 \ \ {\it Meliorism}\ \ \ (to L. Smolin)} \label{SmolinL16}

I read your essay and enjoyed it very much.  We are both meliorists!  See Footnote 31, page 19 in the essay I gave you for a bit on the term.  (I'll attach a copy anew in case you don't have the essay in front of you.)  [See ``QBism, the Perimeter of Quantum Bayesianism,'' \arxiv{1003.5209v1}.]  I also like this idea of ``merging nature and technology''---it is long deep in my thinking.  It is, in a way, why I always draw a quantum measurement device as a prosthetic hand.

I also enjoyed the section on democracy.  It reminded me of something I wrote to Carl Caves once, when I sent him Louis Menand's book {\sl The Metaphysical Club\/} and Richard Rorty's book {\sl Philosophy and Social Hope\/} as a holiday gift (2003).  Digging up the note, I ended it with:
\bq\noindent
       Anyway, I learned a lot about quantum mechanics from these books.
       And I learned a lot about true Americanism (i.e., the kind of Americanism
       we need to strive to get back to \ldots\ and actually mould into a more stable,
       long-lasting form).  Done right, I think the two subjects are probably the
       same thing.
\eq
It very much fits with the ``republican banquet'' ontology I see quantum mechanics hinting at.

Anyway, just first thoughts.  I'll read it again when things are less hectic.

\section{10-03-10 \ \ {\it The Great Lotze}\ \ \ (to L. Smolin)} \label{SmolinL17}

Thinking a bit more on your avowal of human agency as a real element in the world, I thought you might enjoy this passage from William James if you've never seen it.  [See long William James quote near the beginning of the 05-01-09 note ``\myref{PauliFierzCorrespondence}{What I Really Want Out of a Pauli/Fierz-Correspondence Study}'' to H. C. von Baeyer \& D. M. {\Appleby}.]

\section{10-03-10 \ \ {\it QBism on the Make} \ \ (to R. D. Sorkin)} \label{Sorkin1}

I didn't feel I answered your questions very well today---neither to my satisfaction nor yours.  I'd like to get a chance again if you've got some time to chat tomorrow.  Attached is the essay I just wrote on the big picture of the QBism program.  If you have time to give me some feedback on it, I'd sure appreciate it before I post it:  I'm sure there are still plenty of places it can be tweaked for clarity, as well as to make sure the right message is coming across.

If you want a short introduction to the technical part about the SIC rewrite of QM, maybe I could recommend this one:
\arxiv{0912.4252}.

\section{10-03-10 \ \ {\it Lighter Fare}\ \ \ (to L. Smolin)} \label{SmolinL18}

I feel I didn't answer your questions very well today, at least not to the best of my ability.  I'd like to get your questions straight.  Would you have any time to chat further tomorrow?  At the moment, I'm free all day except 2:00 to 4:00.

Attached is a lighter version of some of the ``nonreductionism pieces'' in the essay I gave you; it also encompasses some of the things I tried to say in answering your questions.  [See 16-10-09 note ``\myref{Hardy38}{The More and the Modest}'' to L. Hardy.]  The point being that QBism sees its rewrite of the Born rule as an identification of a new ``active power'' inherent in all matter.  And that the form of the equations are simple enough that we have strong hope that they alone imply the formal structure of quantum mechanics.  That should be first thing they do before we get to the next step beyond quantum mechanics, which I definitely see coming.  I want to give a shot at conveying that to you.

Also, if you want a short introduction to the technical part about the SIC rewrite of QM, maybe I could recommend this paper to you: \arxiv{0912.4252}.

\section{11-03-10 \ \ {\it Lighter Fare, 2}\ \ \ (to L. Smolin)} \label{SmolinL19}

\bls
I thoroughly enjoyed your talk and even felt like I understood most of it.
\els

Thanks; I was afraid you were put off by the talk.

\bls
I can put a query about what you mean by non-reductionism this way; to me strong reductionism is the claim that the properties needed to understand a composite system can always be translated completely into the language used to describe the properties of the parts. Non-reductionism denies this and allows the whole to have properties not expressible in the language used to describe the properties of the parts.

Your proposal, if I understand it, is that the parts have no
properties at all, ie you claim there is no realist microscopic
description at all. This seems a denial of more than reductionism.

Am I misunderstanding?
\els

Yes.  Or it could be that I used some choice of words that accidentally conveyed this.  I certainly would not have meant to!  I think the two paragraphs below from my essay directly address your question.  (And for fun, I'll leave in a footnote that I had opted to leave out of the actual paper.)  Maybe a way of putting it is that the view here is that reality runs deep:  It is to be found in a new and unique way at every level.

There may be some small similarity to something you said in your open-universe essay (or at least the wording of it).  Let me dig up the line:  ``a lot of progress has recently been made on approaches to quantum cosmology in which time is fundamental rather than emergent.''  Reality runs deep in the sense that there is something nontranslatable from every level to every other level.  And similarly from every ``time'' to every other ``time''.  (Though I probably shouldn't be using the language of ``level'' at all:  The view here is that there is no ordering or even overly broad partial ordering (though there may be within islands).)

Maybe another troublemaker is my slogan ``Quantum States Do Not Exist.''  Maybe it is this that is giving you the impression that I am saying ``the parts have no properties at all.''  All the slogan is really meant to convey is that quantum states are not microscopic properties, but rather macroscopic.  Namely, a quantum state is a property of the agent who holds it.  It is a property of the head of the agent, for it reflects how he will gamble on the consequences of the actions he contemplates taking upon a system.

Two paragraphs below.  I hope this helps.

\bq
Physics---in the right mindset---is not about identifying the bricks with which nature is made, but about identifying what is {\it common to\/} the largest range of phenomena it can get its hands on.  The idea is not difficult once one gets used to thinking in these terms.  Carbon?  The old answer would go that it is {\it nothing but\/} a building block that combines with other elements according to the following rules, blah, blah, blah.  The new answer is that carbon is a {\it characteristic\/} common to diamonds, pencil leads, deoxyribonucleic acid, burnt pancakes, the space between stars, the emissions of Ford pick-up trucks, and so on---the list is as unending as the world is itself.  For, carbon is also a characteristic common to this diamond and this diamond and this diamond and this.  But a flawless diamond and a purified zirconium crystal, no matter how carefully crafted, have no such characteristic in common:  Carbon is not a {\it universal\/} characteristic of all phenomena.  The aim of physics is to find characteristics that apply to as much of the world in its varied fullness as possible.  However, those common characteristics are hardly what the world is made of---the world instead is made of this and this and this.  The world is constructed of every particular there is and every way of carving up every particular there is.
\eq

\bq
An unparalleled example of how physics operates in such a world can be found by looking to Newton's law of universal gravitation.  What did Newton really find?  Would he be considered a great physicist in this day when every news magazine presents the most cherished goal of physics to be a Theory of Everything?  For the law of universal gravitation is hardly that!  Instead, it {\it merely\/} says that every body in the universe tries to accelerate every other body toward itself at a rate proportional to its own mass and inversely proportional to the squared distance between them.  Beyond that, the law says nothing else particular of objects, and it would have been a rare thinker in Newton's time, if any at all, who would have imagined that all the complexities of the world could be derived from that limited law.  Yet there is no doubt that Newton was one of the greatest physicists of all time.  He did not give a theory of everything, but a Theory of One Aspect of Everything.  And only the tiniest fraction of physicists of any variety, much less the TOE-seeking variety, have ever worn a badge of that more modest kind.  It is as H.~C. von Baeyer wrote in one of his books [{\sl Petite Le\c{c}ons de Physique dans les Jardins de Paris}],
\begin{quote}
Great revolutionaries don't stop at half measures if they can go all the way.  For Newton this meant an almost unimaginable widening of the scope of his new-found law.  Not only Earth, Sun, and planets attract objects in their vicinity, he conjectured, but all objects, no matter how large or small, attract all other objects, no matter how far distant.  It was a proposition of almost reckless boldness, and it changed the way we perceive the world.
\end{quote}
Finding a theory of ``merely'' one aspect of everything is hardly something to be ashamed of:  It is the loftiest achievement physics can have in a living, breathing nonreductionist world.\footnote{Theories of everything, by contrast, belong to dead worlds, ones whose lives have already been completed.  But perhaps I should not go so far as to say.}
\eq

\section{11-03-10 \ \ {\it Survival} \ \ (to R. {\Schack})} \label{Schack190}

OK, I survived giving this: \pirsa{10030036}.  Now today, I give the follow on to this lecture (grilling), \pirsa{10030005}, for Laflamme and Emerson's class, and further write my report for the Navy (due Friday 4:00 PM).  Saturday, I finally turn my attention to your reflection/decoherence draft and our Templeton proposal.  I'm sorry for all these delays; it's just that I am so damned linear.

\section{11-03-10 \ \ {\it New Footnote}\ \ \ (to C. E. Granade)} \label{Granade2}

Thanks for the question yesterday.  You'll find a better answer than I gave you in the discussion starting on page 23, and particularly footnote 44, of this nearly final draft of my essay.
[See ``QBism, the Perimeter of Quantum Bayesianism,'' \arxiv{1003.5209v1}.] But that's not the ``new'' footnote I was referring to in the title.  I've rejigged the footnote about you.
\bq\noindent
Not to be confused with Scientologists. This neologism was
coined by Chris Granade, a Perimeter Scholars International student at Perimeter Institute, and brought to the author's
attention by R. W. Spekkens, who pounced on it for its beautiful subtlety.
\eq
The whole thing is getting very near posting now.

\section{12-03-10 \ \ {\it Counterexample} \ \ (to J. Emerson \& R. Laflamme)} \label{Emerson5} \label{Laflamme8}

Just to follow up on one of my remarks yesterday.  Take some finite dimensional Hilbert space of dimension $d>4$, and imagine performing (sequentially) two {\it projection\/} valued measurements on it.  Suppose the first consists of $n$ non-rank-1 projection operators $P_i$.  For the second, let it be a PVM with $d$ rank-1 projections $Q_j$.  Further suppose that none of the $Q_j$ commute with any of the $P_i$.

We can think of this as a single measurement with $nd$ outcomes $(i,j)$, and POVM elements $P_i Q_j P_i$.  That is to say, a POVM with $nd$ rank-1 elements:  Let's call them $G_{ij}$.  To further simplify notation, let's use lexicographic order to associate ordered pairs $(i,j)$ with a single integer $k=1,\ldots, nd$.  In other words, let us write $G_k$ for the POVM elements.

Now, suppose you hadn't read any of the above, and I walk up to you out of the blue and say, ``Joseph, I have a POVM with a these rank-1 operators $G_k$ as its elements.  I'd like you to rig up a measurement of it for me.''  My guess is that your knee-jerk reaction would be to say, ``Let me use the Neumark extension theorem to write this as a unitary interaction with an ancilla plus a projection valued measurement on an $nd$-dimensional Hilbert space.''

And I would reply, ``Yes, that's one way to think of it, but it is not how I described it to you above.  Above, I mapped it out as a sequential measurement.''

Similarly, a common technique used by the Zeilinger group and others, is to use extra timing information in their measurements.  I would be resistant to call that the introduction of extra Hilbert space.  The Steinberg SIC experiment is a bit of an example of that as well.  I'll attach Medendorp's QELS abstract.  There, a perfect SIC is enacted by a photon running around and around a loop and generally evolving unitarily, except encountering very weak partial beam splitters along the way.  Zeilinger will also be doing a qutrit SIC experiment, and in the present plans, he will make use of the Neumark theorem and do it that way, i.e., as a PVM on a 9-dimensional system, but that is not the only way to do it.

I will take you up on those beers when I'm back from the APS meeting.

\section{12-03-10 \ \ {\it To Contemplate}\ \ \ (to R. Laflamme \& J. Emerson)} \label{Laflamme9} \label{Emerson5.1}

Attached is the vector I showed yesterday from a $d=6$ SIC.  The remaining 35 vectors are generated by acting $X^j Z^k$ on it, for all values of $j$ and $k$.  ($X$ and $Z$ are the usual shift and phase operators.)

And I'll remind you of the interest in dimension 6:  No one has ever been able to construct a complete set of mutually unbiased bases in that dimension, and it is widely believed (though not proved) that no one ever will.  This says the SIC is the only optimal measurement around in that dimension, and thus shows a divergence of the two concepts, SICs and MUBs.

The Singapore group has done a SIC in $d=2$.  Steinberg has done (something resembling) a SIC in $d=3$, and Zeilinger will soon be doing a more honest-to-god SIC in $d=3$ as well.  It'd be nice to see IQC do the next interesting dimension, $d=6$.  And I suspect your techniques are much more tailored to the problem anyway.

I'll be back from the APS meeting on the $21^{\mbox{st}}$.  Ray, I'll come by your office the following Wednesday.

\section{12-03-10 \ \ {\it Portland!}\ \ \ (to the QBies)} \label{QBies5}

I came downstairs grumbling about having to go to Portland, thinking, ``Why in the hell did I commit myself to going so early in the meeting?''  {\it But then\/} I Googled ``used books Portland'' and to my absolute surprise, shock, and pleasure found that the downtown area is chock full of bookstores!   And Powell's City of Books (1 million in stock, and the building fills a whole city block!)\ is just four blocks from my hotel!!

I'm taking an empty duffle along with my suitcase---one would not believe it but there are still many books on the anti-block (universe) that I do not have.  I suggest you all do the same!

\ldots\ Now this trip is worth making.

\section{13-03-10 \ \ {\it Conceptual Barrier!}\ \ \ (to the QBies)} \label{QBies6}

Ever since I first exclaimed to some audience, ``SICs are so \underline{\phantom{beautiful}} (beautiful?\ right?\ I can't remember the exact phrase) for quantum mechanics that, even if they DO NOT EXIST, they OUGHT TO exist!'' \ldots\ ever since then, I have been fascinated with what my subconscious was trying to express.  And it only got fueled one day when Bob Coecke said, ``Yeah, there's some truth to that.  It's a game that's worth playing.  If you can't prove they exist, you just add them as an axiom.''  You both have heard my allusions to William James's faith ladder in this context (I reprise the story below), and the power of the SICs as a symbol like the Christian cross.

But today it finally all came together!  And no less than in a philosophy of science seminar!  I have many times said that I never get much of value from a philosopher of science, but this one was different.  It was at Mark Wilson's seminar at UW.  His abstract is below, but it gives no great hint of the depth of the talk.  It seemed to be based on his book {\sl Wandering Significance:\ An Essay on Conceptual Behavior}, which I am certainly going to buy.

His main lesson was this:  There has never been any stabilized, finished thing called Maxwell's theory of electromagnetism.  Instead, the meaning of all the terms in it have been subtly changing since the beginning (and even before).  Meanings have been modified for ``parts'' of the ``theory'' in light of this and that for other ``parts.''  Moving away from Maxwell's theory he had a nice example in Dedekind cuts.  Finally he landed with the lesson of the definition of a Riemannian manifold.  It was an analogy that I really took to heart.  The point he emphasized was that one uses a structure that is not quite right as the basis for getting the structure one really wants.  He had a beautiful picture of a curvy manifold with tangent spaces tacked on everywhere:  The manifold is the limiting thing that the collection of tangent spaces is not.

[Next day.  I started the note above before Kiki and I went to Jazz at the Bistro last night.]

Lesson:  There is nothing wrong in using a structure that is not quite right to build up one that is more satisfactory for this or that purpose.  So here is what I mean when I say we will {\it make\/} SICs exist:  If SICs exist, the Born Rule
$$
Q(D_j)\, =\, \tr\, (\,\hat\rho\, \hat D_j\,)
$$
is allowed to take the form of the urgleichung
$$
Q(D_j)\, =\,
(d+1)\sum_{i=1}^{d^2} P(H_i) P(D_j|H_i) - 1\;.
$$
We know that the Born Rule has that privilege in at least 67 dimensions; we know that the quantum-state space has the privilege of being a ``maximal consistent set'' in at least 67 dimensions.  In 67 dimensions, the Born Rule represents the simplest modification one can imagine to the Law of Total Probability for taking into account the counterfactuality of the sky path vs the ground path.  If living counterfactuality is the message quantum theory is trying to tell us, it has to find a simple expression somewhere in the theory.  In at least 67 dimensions, quantum mechanics is isomorphic to ``SIC mechanics,'' but if SICs don't exist generally within the usual quantum formalism, so much the worse for quantum mechanics:  It means that quantum mechanics is not quite right.  Close, but not quite right.  There was never a commandment written in stone that the stories we tell of quantum phenomena must be written in the language of linear operators.

It is hard for me to express how liberating I found this thought yesterday.  We have identified an essentially simple core to quantum mechanics in the urgleichung---and it is this that must be saved at all costs.  For physics should be built on the conceptually simple.  It is only our habit of thinking in terms of orthogonal bases that makes the SICs look difficult.  If God had given us ``SIC mechanics'' to begin with, we would be finding it similarly difficult to establish ``orthonormal bases'' for representing the theory, and our nearby colleagues would be wondering why we would even want to.   For though we would know that we could  do it in dimensions 2 through 67 at least, we would not be able to hide the fact that the expression of these bases always look exceedingly complicated in terms of their original ``SICs.''  It would just be our faith that for every extreme point in the set there should be a measurement for which it gives certainty that drives us forward and makes us slave to find such a structure.

One technical question for everyone to think about:  Let us not try to build up a maximal consistent set one point at a time, but $d^2$ points at a time.  That is, let us try to build it up one regular simplex at a time (what in quantum mechanical terms would amount to demanding that every pure state be an element in at least one SIC).  Let us build a maximal consistent set by taking a maximal union of simplexes, subject only to the urungleichung.  Is there now, modulo rotations, one unique set?  If so, then under the assumption of SIC existence, that structure would be isomorphic to quantum state space.\footnote{We now know that, even if SICs do exist, quantum state space {\it is not\/} generally a simple union of regular simplexes. Rather, on the supposition of SIC existence, it is the {\it convex hull\/} of a union of regular simplexes.  This is because there are quantum states for all $d\ge 4$ which are not expressible as a convex combination of SIC elements for any SIC. \label{OhCrap}}  And if SICs do not always exist, what physical phenomena would recommend that we not pass from the one theory to the other?  Maybe we wouldn't be able to find a reason at all.

James's faith ladder and Wilson's abstract below.

\begin{quotation}
Faith uses a logic altogether different from Reason's logic. Reason claims certainty and finality for her conclusions. Faith is satisfied if hers seem probable and practically wise.

Faith's form of argument is something like this: Considering a view of the world: ``It is {\it fit\/} to be true,'' she feels; ``it would be well if it {\it were\/} true; it {\it might\/} be true; it {\it may\/} be true; it {\it ought\/} to be true,'' she says; ``it {\it must\/} be true,'' she continues; ``it {\it shall\/} be true,'' she concludes, ``{\it for me}; that is, I will treat it as if it {\it were\/} true so far as my advocacy and actions are concerned.''

Obviously this is no intellectual chain of inferences, like the {\it Sorites\/} of the logic-books. You may call it the ``faith-ladder,'' if you like; but, whatever you call it, it is the sort of slope on which we all habitually live. In no complex matter can our conclusions be more than {\it probable}. We use our feelings, our good-will, in judging where the greater probability lies; and when our judgment is made, we practically turn our back on the lesser probabilities as if they were not there. Probability, as you know, is mathematically expressed by a fraction. But seldom can we {\it act\/} fractionally---half-action is no action (what is the use of only half-killing your enemy?---better not touch him at all); so for purposes of action we equate the most probable view to 1 (or certainty) and other views we treat as naught.
\end{quotation}
\vspace{-12pt}
\underline{\phantom{separate}}
\bq
\noindent ``How `Wave Front' Found its Truth-Values''\\
Mark Wilson, Department of Philosophy, University of Pittsburgh\medskip

Scientific theories rarely serve as islands entire of themselves.  Such hidden connections sometimes force novel readings upon venerable doctrines, dramatically altering our understanding of their basic ``subject matter'' in the process. Conventional Maxwellian electromagnetism underwent an upheaval of this type in the mid twentieth century.  The case (which will be recounted in non-technical terms) offers interesting insights for both philosophy of language and philosophy of science, for it supplies a portrait of semantic stabilization unfolding in a different manner than conventional ``natural kind'' models suggest.
\eq

\section{13-03-10 \ \ {\it All the More Reason}\ \ \ (to R. {\Schack}, D. M. {\Appleby}, {\AA}. {\Ericsson})} \label{Appleby87} \label{Schack191} \label{Ericsson5}

Let me add one more reason for wanting the technical question in the last note to work out.  It has to do with the axiom {\Ruediger} and I called ``the availability of certainty'' in the QB Coherence paper.  I've never really liked it, and the footnote attached to our statement of the axiom explains why:
\bq
     In several axiomatic developments of quantum theory---see for instance
     [Goyal08] and [Hardy01]---the idea of repeated measurements
     giving rise to certainty (and the associated idea of ``distinguishable states'')
     is viewed as fundamental to the whole effort.  The usual justification is that
     the existence of such kinds of measurement is nearly a self-evident
     necessity.  However, from the quantum-Bayesian point of view where
     {\it all\/} measurements are generative of their outcomes---i.e., outcomes
     never pre-exist the act of measurement---and certainty is always subjective
     certainty [Caves07], the consistency of adopting a state of certainty as
     one's state of belief, even in what is judged to be a repeated experiment, is
     not self-evident at all.  In fact, from this point of view, why one ever has
     certainty is the greater of the mysteries.
\eq
Couple that with something I said in the last note:
\bq
     If God had given us ``SIC mechanics'' to begin with, we would be finding it
     similarly difficult to establish ``orthonormal bases'' for representing the
     theory, and our nearby colleagues would be wondering why we would
     even want to.   For though we would know that we could  do it in
     dimensions 2 through 67 at least, we would not be able to hide the fact
     that the expression of these bases always look exceedingly complicated in
     terms of their original ``SIC''.  It would just be our faith that for every
     extreme point in the set there should be a measurement for which it gives
     certainty that drives us forward and makes us slave to find such a structure.
\eq
That is, the reason we'd be driven is that we would be working to {\it establish\/} the availability of certainty as a {\it theorem}.  And that is indeed the way it should be from a QBist perspective.  That any certainty at all could come out of quantum theory {\it should\/} be a theorem.  If something is the ``greater of the mysteries'' it really shouldn't have a comfortable spot as one of the very axioms.

So now I really find myself hoping the ``union of simplexes'' idea carries a lot of power.\footnote{See Footnote \ref{OhCrap} for a technical, but remediable, mistake in the original statement of this. \label{OhCrap3}}  It would be a very clean foundation for quantum theory.

\section{13-03-10 \ \ {\it The Tetra Pak}\ \ \ (to H. C. von Baeyer and D. M. {\Appleby})} \label{Appleby88} \label{Baeyer107}

Yesterday I wrote these lines down:
\bq\noindent
Ever since I first exclaimed to some audience, ``SICs are so \underline{\phantom{beautiful}} (beautiful?\ right?\ I can't remember the exact phrase) for quantum mechanics that, even if they DO NOT EXIST, they OUGHT TO exist!'' \ldots\ ever since then, I have been fascinated with what my subconscious was trying to express.
\eq

Today, it dawned on me that maybe there really is some kind of subconscious tug on me in this regard.  As I was sitting on a bench in the shopping mall where Marcus gets his groceries here, I remembered my brother coming home one evening in 1968 or '69, when I was 4 or 5, with a small tetra pak of coffee creamer from a diner or Dairy Queen or something.  (Look up \myurl[http://en.wikipedia.org/wiki/Tetra_Pak]{http://en.wikipedia.org/wiki/Tetra\underline{ }Pak} if you don't know what I mean.)  He put it in the refrigerator, but I walked behind him and took it out.  I thought it was the most fascinating shape I had ever seen.  In fact, it was in the refrigerator for some time, and whenever I would open the door, I would have a look at the little thing.  Today was the first time my memory has ever gone back to that.

The Tetra Pak \ldots\ Pauli's fascination with the Tetractys \ldots\ and the hope that quantum state space is {\it nothing but\/} a grand union of simplices (subject only to the restriction of the urungleichung).\footnote{See Footnote \ref{OhCrap} for a technical, but remediable, mistake in the original statement of this. \label{OhCrap4}}  Today is one of those days when I can't seem to stop myself from thinking about the SICs and the Tetragrammaton in the same thought.

\section{13-03-10 \ \ {\it Comments}\ \ \ (to H. C. von Baeyer)} \label{Baeyer108}

\bhcvb
p.\ 22 footnote 37.  Magnalium is showing off.  I had no idea what it was, and it certainly doesn't matter.  Didn't he use other, more common materials as well, like maybe brass?
\ehcvb

Cool!  Just looked through; there is indeed new stuff.  Thanks a million, million!

Magnalium isn't showing off---it is what he used, and I had never heard of the stuff either.  In the first paper where I make a comparison between dimension and mass, I wrote in the acknowledgements:
\bq\noindent
     This paper is dedicated to Alexander Holevo on the occasion of his 60th birthday. I
     thank Osamu Hirota, Chris King, Masahide Sasaki, and Peter Shor for useful
     discussions, and especially Greg Comer for keeping me close to gravity. I thank Ken
     Duffy for teaching me about magnalium.
\eq

\subsection{Hans's Reply}

\bq
Many years ago I started a column on absolute zero (an impenetrable barrier in physics) this way: ``The other day, as I was having morning coffee in my solarium, seventeen yellow waxwings crashed into the window and dropped dead into the flower bed below.''   This sounded ridiculous, so I changed it to seven, and then to three.   That's how it was published.

But, Chris, seventeen was, indeed, the truth.  Sometimes, in writing, you have to lowball in order to appeal.
\eq

\section{14-03-10 \ \ {\it Sadness and Opportunity}\ \ \ (to G. Bacciagaluppi)} \label{Bacciagaluppi7}

I'm so very sorry, but this time I'm going to have to miss the meeting for sure.  It's a shame, I really would like to show my support for you and reach out to the UK audience once again.  The meetings in Hull and Belfast were very important for me; they shaped my research life really.  But July is just too full a month for me.  I have to go to Brisbane for a 19 to 23 meeting, and Vienna for a 27 to 31 meeting.  And in June there's the {\Vaxjo} meeting as well, 13 to 18, of which I'm already committed to speaking.

On another subject, I can't tell you how much I enjoyed your recent article with someone (can't remember the name) on Heisenberg's schnitt argument.  It was very thought provoking for me.  I wish I had it in front of me while writing you (I'm on my way to Portland to show off my students at the APS March meeting):  There was a quote by Heisenberg and I think Born on the meaning of probability surviving the transition from the classical venue to the quantum venue that might have come from my own mouth.  Attached is a new essay where I say as much (and hopefully a lot more), that I plan to post Wednesday (before Thursday's talk in Portland).  Time is short, but if you have any feedback I may be able to take it into account.  At the very least---come to think of it---I may be able to cite your new article at some appropriate spot.

\section{14-03-10 \ \ {\it Beginnings of Subconscious}\ \ \ (to the QBies)} \label{QBies7}

\bv
Ev'rybody's talkin' 'bout\\
Bagism, Shagism, Dragism, Madism, Ragism, Tagism\\
This-ism, that-ism, ism ism QBism\\
All we are saying is give peace a chance \ldots
\ev

Sing it with me boys!  Over the weekend, I've gotten quite taken with the technical question below.  [See 13-03-10 note ``\myref{QBies6}{Conceptual Barrier!}''\/ to the QBies.] Now a question for you guys to contemplate.  Do you think there's any way we might check it numerically for $d=3$?  Or check any aspect of it?  Could quantum state space be built up one simplex at a time?

See you in Portland.

\section{14-03-10 \ \ {\it Shunning Certainty and Recovering It} \ \ (to L. Hardy)} \label{Hardy39}

I thought you might enjoy the notes below that I launched off yesterday to the guys in the crew.  [See the 13-03-10 notes ``\myref{QBies6}{Conceptual Barrier!}''\ and ``\myref{Appleby87}{All the More Reason}'' to the QBies.]  I think they capture another distinction of direction where the path I'm following seems to take me away from your original axiomatic scheme.  Ultimately I'd like to get away from taking the notion of distinguishable states as a primitive for the theory, and rederive that there are some as a consequence of other structures.  Below starts to flesh out how that might go, and why I would want to go that way.

It seems to me a rather pretty fact that quantum-state space in SIC representation is exactly a union of regular simplexes.\footnote{See Footnote \ref{OhCrap} for a technical, but remediable, mistake in the original statement of this. \label{OhCrap5}}  And a maximal union as well, the addition of even one more point would make my urungleichung violated.  Only recently has it struck me however that that might be a significant postulate in itself.

\section{16-03-10 \ \ {\it Landed in Portland}\ \ \ (to M. Schlosshauer)} \label{Schlosshauer26}

You flatter me through and through.  And you make it hard for me to function in the REAL world, where no one likes the sound of anything I say.  It's much easier to read Max's notes over and over and then take a nice nap.

Anyway, thank you, thank you so much for your feedback.  And then thank Eli and Kari for me, for letting you go for a while to spend some time reading and composing notes.  I fully realize that this is a very busy time of your life.

[I wrote the lines above Sunday evening; it is now Tuesday morning.]

But OK, beyond all the nicety, you pushed me:  I couldn't stop myself without an effort to do your requests all the way through.  Come to think of it, it was the opposite of reading your note and then taking a nice nap!  You are subtle---a man who will go places.  Let me rush this off to you so that it gets to you before you are asleep.  I really want to post the thing tonight or tomorrow morning come hell or high water.  Aside from making a couple cross references, and placing the figures more properly it should be done!  See footnote 22 on page 13.

On another occasion (soon) we'll talk more about {\it measuring\/} dimension.  As for the paper, I'll just leave it with the one reference to Scarani for the time being.

More later.

\section{16-03-10 \ \ {\it Progress Report} \ \ (to {\AA}. {\Ericsson})} \label{Ericsson6}

Six books so far, but the two most notable are:
\begin{itemize}
\item[{1)}]	A picture book of alchemical symbols,

\item[{2)}]	A book with a serious Jamesian style argument for human immortality
\end{itemize}
So far, I've only gotten through the first third (plus epsilon) of Powell's philosophy section.  Following that I've got about four other bookstores to go to in town!

Haven't seen Hoan, Gelo, and Matthew yet.  I'm staying at about the fanciest hotel I've ever stayed at (paid for by the APS)---beautiful lobby I'm writing you from, rich wood everywhere.  The other guys on the other hand are probably staying at a dump.  It makes me feel guilty.

Not one note from anyone on my weekend epiphany---I thought I had made a great conceptual leap.  It has kind of disappointed me.  But still I have plodded on thinking about unions of regular simplices.

\section{16-03-10 \ \ {\it Sadness and Opportunity, 2}\ \ \ (to G. Bacciagaluppi)} \label{Bacciagaluppi8}

\bgba
Thanks for the appreciation of the Heisenberg article (my co-author is Elise Crull, currently finishing her PhD with Don Howard at Notre Dame). Is the quote the one about the `quite usual probabilities'? It is actually not clear to me what picture of probabilities Born and Heisenberg have in mind (if they have a single one in mind!), and I suspect it is something like a propensity view (so I am actually wondering whether a quote would be out of context!).
\egba

The issue is not Bayesian vs.\ Frequentist probability (Heisenberg and Born, at least in that quote, seem to be treating probability frequentistically, not propensitistically), but that whatever notion is chosen is stable under the transition from classical to quantum.  Certainly not everyone believes that---they (and Feynman) did.  Both emphasize a rule change for {\it calculating\/} probabilities when comparing a ``factual'' to a ``counterfactual'' (though they didn't say it that way).  They both emphasized that the {\it notion\/} of probability itself does not change.  What I particularly enjoyed about Born and Heisenberg is that they say explicitly this entails no break from the usual probability calculus:  The quantum mechanical rule is rather an addition to it!

Attached is a modified draft where I reference you.  See footnote 19 on page 12.  Please share with Elise if you wish.

The final version of the paper should be posted Wednesday.

\section{17-03-10 \ \ {\it We Are the Small Axe \ldots}\ \ \ (to M. Schlosshauer)} \label{Schlosshauer27}

I didn't get back to you today as I had hoped, and now I'm a bit drunk from the TGQI ``executive'' dinner (as if I could be the executive of anything).  Blame John Preskill (even though he doesn't drink, funny all the causal connections).  So I'll keep it short.  It's no payment for all the help you've given me, but I thought I'd make you an ``honorary'' member of the group by spewing to you like I spew to all the rest of them.  (I'm listening to Bob Marley as I write you this; do you know the song in the title?  I am no executive, but I might be a small axe.)  Below are two notes capturing a good experience I had at the end of last week.  [See the 13-03-10 notes to the QBies titled ``\myref{QBies6}{Conceptual Barrier!}''\ and ``\myref{Appleby87}{All the More Reason}.''] I don't seriously believe that QM doesn't always contain the SICs, but even if it {\it does\/} contain them properly, I now understand that the urgleichung is truly the more basic structure:  QM, at best, is secondary.  Either way, the equations already tell us that we must {\it make\/} these damned SICs exist.  I hope you enjoy James's faith ladder; I suspect that you more than anyone else in the group will feel it intuitively.

Just sharing with you, and trying to thank you in the deepest way I can for the ends you've pushed me to.  The paper is way better because of you.  I will finally post it tomorrow morning!  It will be so good to get it off my chest.

I hope all is well with you and Kari and Eli.

\section{17-03-10 \ \ {\it Bob Marley Words} \ \ (to {\AA}. {\Ericsson})} \label{Ericsson7}

Just heard a lovely phrase in a Bob Marley song, ``mighty God is a living man.''  (I've known it for a long time, but hearing it again brought it to mind.)  Anyway, isn't that what I've been saying myself?

I'll give you a better report tomorrow, but it looks like 17 books so far.  More tomorrow!

\section{18-03-10 \ \ {\it Very Final Version}\ \ \ (to R. W. {\Spekkens})} \label{Spekkens80}

Here is the very final version of the paper, the one that'll be posted Friday.  I looked at footnote 37 on Wootters' kid:
\bq
That is, every piece of the universe had better be hard-wired for the contingency that an agent might experience it somewhere, somehow, no matter how long and drawn out the ultimate chain might be to such a potential experience.  Does this mean even ``elementary'' physical events just after the Big Bang must make use of {\it concepts\/} that, to the reductionist mind, ought to be 15 billion years removed down the evolutionary chain?  You bet it does.  But a nonreductionist metaphysic need make no apology for this---such things are in the very idea.  John Wheeler's great smoky dragon comes into the world biting its own tail.%

W.~K. Wootters tells a lovely story of an encounter he had several years ago of with his young son Nate.  Nate said, ``I wish I could make this flower move with my mind.''  Wootters reached out and pushed the flower, saying, ``You can. You do it like this.''  From the perspective here, this is an example of an interaction between two nonreductionist realms.  Each realm influences the other as its turn comes.  There is a kind of reciprocality in this, an action-reaction principle, that most reductionist visions of the world would find obscene.
\eq
One aspect of what it says is the very idea you were trying to defend against Tumulka:  That ``no operational distinction'' should correspond to ``no ontic distinction.''  Tumulka would say, ``What business does ontology have in knowing what you, a high level agent, can or cannot distinguish?''  ``Ontology was written down before agents were ever thought of; their limitations are of no consequence for ontology.'' I imagine him saying.  ``You have the conceptual flow going the wrong way.''

There is a bit of William James in you yet.

\section{18-03-10 \ \ {\it We Are the Small Axe \ldots, 2}\ \ \ (to R. W. {\Spekkens})} \label{Spekkens81}

Here are some of the other things I was telling you tonight, about how I should have answered Lee.  [See 13-03-10 note ``\myref{QBies6}{Conceptual Barrier!}''\ to the QBies.]

Hierarchies of theories are all well and good, but when it comes time to modify quantum mechanics, I suspect it's going to be something much more subtle than just jumping to the next level in one's premade class.   One of the things that struck me when I came across the thought below is that I actually found it believable---for the first time I felt the tug of why one would even think of modifying QM. Though as I wrote Max:
\bq\noindent
     I don't seriously believe that QM doesn't always contain the SICs, but even
     if it {\it does\/} contain them properly, I now understand that the urgleichung
     is truly the more basic structure:  QM, at best, is secondary.  Either way, the
     equations already tell us that we must {\bf make} these damned SICs exist.
\eq

\section{19-03-10 \ \ {\it Dimension Witnesses}\ \ \ (to V. Scarani)} \label{Scarani1}

Thanks.  I was just about to post the thing today, but then was thwarted by some upgrade they've made to their system.  Now, with my travels, it probably won't be until next week that I'll be able to figure the new {\tt arXiv\/} system out.  So I attach for you the near-posted version.  [See ``QBism, the Perimeter of Quantum Bayesianism,'' \arxiv{1003.5209v1}.] The language near the end of the paper is very close to some things I note in Nicolas Gisin lately (quantum events as moments of creation, etc.), but I suspect he would never accept of any of my starting points or the reasoning that led me to these similar conclusions.

I enjoyed meeting you in Japan, and hope to see you again in the decent future.

\subsection{Valerio's Preply}

\bq
Here are a few articles beyond ours on dimension witnesses:
\begin{itemize}
\item
\arxiv{0812.1572} (where it is proved that even with binary outcomes you can have dimension witnesses for any $d$, just by changing the number of measurement settings);
\item
\arxiv{0901.2542} (relating dimension witnesses to evolution)
\end{itemize}
\eq

\section{19-03-10 \ \ {\it {\tt arXiv} New submission $\rightarrow$ 1003.3223 in {\tt physics.hist-ph} from {\tt jgus@tds.net}}\ \ \ (to J. R. Gustafson)} \label{Gustafson3}

That's great news!  Attached is something I'm just on the verge of posting myself.  (I tried to post it today, but they thwarted me with a new system that's incompatible with my old version of \LaTeX!  So, now I'm the beginner again!)

\subsection{John's Reply}

\bq
I am enjoying your new ``bigqbism'' article that you sent to me, making me think of some Pauli quotes I had inserted into my dissertation \ldots\ roughly pages 137 and 138 of the PDF file that is now posted to arXiv (excerpt appears below).  Pauli's remarks imply to me he may have been thinking of issues we now relate to quantum entanglement.  Now I'm motivated to dig further into his later writings.  Here's that excerpt.\\
\underline{\phantom{Gustafson}}
\bq\noindent
In any case, Pauli saw in the concept of synchronicity a direct link to Arthur Schopenhauer's philosophy, which he made clear in a letter to Jung of June 28, 1949:
\bq\noindent
The idea of {\it meaningful coincidence}--i.e., simultaneous events not cau\-sally connected--was expressed very clearly by Schopenhauer (1788 -- 1860) in his essay \ldots.  There he postulates an ``ultimate union of necessity and chance,'' which appears to us as a ``force,'' ``which links together all things, even those that are causally unconnected, and does it in such a way that they come together just at the right moment.''
\eq
To my knowledge, Pauli never connected his exclusion principle to sychronicity.   He did talk about his exclusion principle and ``action-at-a-distance,'' however, in a course on {\it Wellenmechanik\/} at the Federal Institute of Technology (ETH) in Zurich in 1956-1957, according to notes taken by his students:
\bq
\noindent From a superficial consideration of the exclusion principle, it might be thought that a sort of action-at-a-distance is being postulated, as a result of which even two widely separated particles are aware of one another (``sign a contract'').  However, this is not so, because the exclusion principle is valid only as long as the wave packets of the two particles overlap.
\eq
Still, Pauli later called for a unification of physics and psychology, which would have  entailed both his exclusion principle and the concept of synchronicity.
\eq
\eq

\section{19-03-10 \ \ {\it Lunchtime Guilt}\ \ \ (to R. W. {\Spekkens})} \label{Spekkens82}

You won't get this note until I can drop it off at the airport, but I compose it at the famous Rocco's, caddy-corner to Powell's.  I just want to say I've been walking around the city feeling ashamed and guilty since my outburst the morning.  Ostensibly, it was initially meant to be a kind of biting joke \ldots\ until I could see the hurt on your face.  But the more I've examined the moment in my meanderings through the streets, the more I've become convinced that its real source was a rather deep bitterness in me.  Looking back on it, I think I can even identify that I had an image of Travis Norsen in my mind as I lashed out at you.  I could be fooling myself, but it feels very palpable at the moment.  The sad fact is, like Pavlov's dogs, I carry associations in my mind, and the association you inadvertently provoked my reaction to was between those who give up on Einstein locality and those who call me solipsist.  The latter has indeed been a thorn in my side for many years.  But you---of all people!---in no way deserved that reaction from me.  Even though we at times disagree about the future direction of physics, I {\it do\/} recognize that you think harder and more deeply about quantum mechanics than anyone else I know.  And I don't say that lightly; I mean it.  So, I hope you will accept my apology.  It's useful for me to recognize my demons, but there is no reason for my friends to be the targets of my primal screams.

I got quite a lot out of our dinnertime conservation the other night.  Some stuff of lasting thought.  I want to continue that from time to time without your fearing that I'll crack.

Please don't forget to send me your spellcraft talk.  Emma will be thrilled to see it.

\section{19-03-10 \ \ {\it Book Report} \ \ (to {\AA}. {\Ericsson})} \label{Ericsson8}

\begin{enumerate}
\item Harold Morick, editor, {\sl Challenges to Empiricism}, PB, 6.95 USD.

\item Myles Brand, editor, {\sl The Nature of Human Action}, PB, 9.95 USD.

\item F. W. J. Schelling, {\sl Philosophical Inquiries into the Nature of Human Freedom}, PB, 9.95 USD.

\item G. E. Moore, {\sl Philosophical Studies}, HC, 12.95 USD.

\item John Dupr\'e, {\sl Humans and Other Animals}, HC, 15.95 USD.

\item Gilles Deleuze, {\sl Difference \& Repetition}, translated by Paul Patton, PB, 24.95 USD.

\item Bina Gupta, {\sl The Empirical and the Transcendental: A Fusion of Ideas}, HC, 23.97 USD.

\item Maurice Freedman, {\sl To Deny Our Nothingness: Contemporary Images of Man}, PB, 6.95 USD.

\item Frederick C. Beiser, {\sl German Idealism: The Struggle Against Subjectivism, 1781-1801}, PB, 29.95 USD.

\item Barry Allen, {\sl Truth in Philosophy}, PB, 9.95 USD.

\item Alexander Roob, {\sl Alchemy \& Mysticism: The Hermetic Cabinet}, HC, 7.99 USD.

\item Willaim James, {\sl Essays in Radical Empiricism \& A Pluralistic Universe}, HC, 15.95 USD.

\item Eugene Fontinell, {\sl Self, God, and Immortality: A Jamesian Investigation}, HC, 19.95 USD.

\item Irving J. Good, {\sl Good Thinking: The Foundations of Probability and Its Applications}, PB, 10.95 USD.

\item Abner Shimony, {\sl Search for a Naturalistic World Views, Volume I: Scientific Method and Epistemology}, PB, 26.95 USD.

\item Dale H. Porter, {\sl The Emergence of the Past:  A Theory of Historical Explanation}, PB, 8.50 USD.

\item John K. Sheriff, {\sl Charles Peirce's Guess at the Riddle: Ground for Human Significance}, PB, 10.00 USD.

\item Roger Poole, {\sl Towards Deep Subjectivity}, PB, 3.95 USD.
\item Ishtiyaque Haji, {\sl Incompatibilism's Allure: Physical Argument for Incompatibilism}, PB, 26.95 USD.

\item Arhur O. Lovejoy, {\sl The Great Chain of Being: A Study of the History of an Idea}, PB, 12.50 USD.

\item Richard M. Gale, {\sl John Dewey's Quest for Unity: The Journey of a Promethean Mystic}, HC, 32.95 USD.

\item Edwin B. Holt, Walter T. Marvin, William Pepperrell Montague, Ralph Barton Perry, Walter R. Pitkin, and Edward Gleason Spaulding, {\sl The New Realism: Cooperative Studies in Philosophy}, HC, 65.00 USD.

\item R. W. B. Lewis, {\sl The Jameses: A Family Narrative}, HC, 6.95 USD.
\end{enumerate}

\section{19-03-10 \ \ {\it Counterexample, 2} \ \ (to J. Emerson)} \label{Emerson6}

Sorry it's taken me so long to get back to you.  I've been in Portland all week and this meeting has dulled all my senses.  (Though I am presently writing you from a nice Argentinean restaurant.)  To give a short reply, let me just say we should discuss this paper by Oreshkov and Brun in the context of your framework:  \arxiv{quant-ph/0503017}.  I think it gives a formal account of what I was claiming about Aephraim's SIC-POVM technique.  Particularly, it shows a way of trading off continuous interaction for ancilla.

My general philosophical beef is this.  If one must recognize measurement as a ``primitive'' in the QM framework, then I see no reason to prefer one kind of primitive (PVM) over another kind of primitive (POVM).  One can think of the PVMs as a subset of the POVMs, or one can think of the POVMs as a subset of the set of all possible two stage thingies (unitary coupling to ancilla plus PVM).  Since POVMs are formally powerful, that tips the scale for me.  Furthermore for a given POVM there is no unique Naimark-extension way of thinking of it:  The POVM corresponds to an equivalence class.  And where there is such ``gauge freedom'' I say, one has the wrong object---one should mod out the freedom, and find the mathematical object that remains.

In any case, in my truest philosophy, a POVM is {\it nothing other than\/} a collection of conditional probability distributions.  See footnote 22 on page 13 of this (still unposted!)\ paper attached.  [See \arxiv{1003.5209}.]  If one takes that point of view, the unitary $+$ ancilla $+$ PVM take on POVMs is rather unnatural way of thinking.  That's my truest philosophy, independent of the Oreshkov--Brun stuff.  I just threw it out by way of ammunition (on the suspicion that you won't find my identification of a POVM with a set of $P(D_j|H_i)$ as compelling as I do).

I'm getting primed for those beers you offered.

\section{22-03-10 \ \ {\it Pong} \ \ (to R. {\Schack})} \label{Schack192}

Sorry, the March Meeting turned out to be more taxing than I had imagined.  I'm back home now and getting back on track.  Today I plan to get my email box thoroughly cleaned out and get to work on two things:  1) Templeton proposal, and 2) Research statement for PI.  I have no intention of letting Templeton slip through my fingers.

Quick answers to some old items:
\brs
Did we ever make progress on communication?
\ers
No we didn't, or at least no great amount.  And I don't think much of what we did talk about is in the correspondence.  I am now much happier with my presentation of Wigner's friend.  But it leaves communication out other than to speak of it as another quantum measurement.  This leaves much, much to be desired.  There is no doubt we have to come back to this.

\brs
I feel strongly that Wigner's friend, communication, counterfactuals
and reflection should be coming together in one coherent story \ldots
\ers
Yes, I couldn't agree more.  In most ways, we're just at the beginning of this project.

\brs
You still haven't given me any feedback at all on the reflection draft
paper. It would help me enormously to discuss it with you.
\ers
I understand this.  I know so much that I must get on it and help you out.  I did read the full draft in Portland, and I must get that conveyed to you.  But my guess is, with honesty, we're going to have to knock Templeton out of the way first.

\brs
I have been reading, pondering, and rereading your two notes. There is lots in there which I like a lot. Most of all I like the idea of coming up with a theory that is different from quantum mechanics.  I am not sure the urgleichung is sufficiently compelling to warrant such a departure, but there are other reasons to want it. \ldots\ If the new theory derived in a world in which SICs exist only as approximations, or only in small dimensions, had features that make a block universe look unnatural, this would be marvelous.
\ers
My guess is that a departure will not be required---my guess is that SICs actually DO exist always.  That aside, what I was hoping to express is that my mind is now prepared for what should be done on the off chance that they do not.  In my bones, I do feel the urgleichung is sufficiently compelling, but there is no doubt that must be articulated further:  For what we already have, in a way that has never been seen before (certainly not from Gleason's theorem, or the raw injunction of a linear rule as in Hardy or Barnum, {\Barrett} and Wilce, etc.), shows that the Born Rule {\it is\/} a rule connecting probabilities to probabilities---between counterfactual and factual.  It is the ``engaged observer'' in crystal clear form, and one that pulls the essential factor to the very front of the equation.  We need a more metaphysical story to shore it up with, but that's why the equation already feels completely right to me as the foundation of things.

\brs
A more important one is that some aspects of quantum mechanics make it
very hard not to be driven to a many-worlds view. The church of the
larger Hilbert space can be regarded as simply a mathematical
convenience, but any defender of Everett can claim that it is
extremely suggestive.
\ers
You should watch this talk or read the associated {\tt arXiv} paper:  \pirsa{10020070}. It is a bit frightening how much power ``purification'' can have as an axiom.  (As I write this, it starts to strike me that there might possibly be a connection between it and our ``posteriors equal priors'' postulate.  Very slippery thought, but something we might think about.)

\brs
That is my vision at the moment. To come up with a theory that makes the agent dependence explicit in a far stronger way than quantum mechanics does it. To derive quantum mechanics from the requirement that the theory rules out a block universe.

I am certainly very sympathetic to deriving the existence of certainty in some way. Maybe even finding that certainty only exists in the approximate theory called quantum mechanics.

And I derive a small amount of satisfaction from the fact that you
have finally come round to my way of phrasing maximality and abandoned
building up maximal consistent sets point by point.
\ers
This is important!  I had not realized you proposed this.  Clearly my memory is not infallible.  What was the context of your earlier proposal?  What were the sorts of things we were thinking about at the time?

\brs
Are you somewhere out there? Do you want to give me a ring?
What time zone are you in?
\ers
If it's OK with you, let me try to stay away from the phone for a couple of days.  I'm working from my study in the house today, and trying to avoid contact with other humans (even the best of the other humans).  My social spring got stretched beyond its elastic limit at the March meeting.

\section{22-03-10 \ \ {\it A Line I Shall Use in Templeton} \ \ (to R. {\Schack})} \label{Schack193}

Below is a note I had gotten from one of the students, Ian Hincks, in the Emerson-Laflamme quantum foundations class after my first lecture.  His last two sentences are a lovely testimonial to what you and I are up to:
\bq\noindent
``I like your theory because it returns to me as much freedom as I feel that I have. Such freedom is lost or partially lost in other interpretations.''
\eq
I'm going to try to find some way to use this as a testimonial in our Templeton proposal.

\section{22-03-10 \ \ {\it Haeccity Through and Through} \ \ (to G. Smith)} \label{Smith1}

Wiki says this of it:
\bq\noindent
{\bf Haecceity} (from the Latin {\it haecceitas}, which translates as ``thisness'') is a term from medieval philosophy first coined by Duns Scotus which denotes the discrete qualities, properties or characteristics of a thing which make it a particular thing. Haecceity is a person or object's ``{\it thisness}''.
\eq

Thisness would be easier to say.  Anyway, it imbues my philosophy of things through and through, and looking back on it, maybe that's why I liked John's lockboxes.  Attached is the new paper that I've had trouble posting with the new {\tt arXiv.org} system.  [See ``QBism, the Perimeter of Quantum Bayesianism,'' \arxiv{1003.5209v1}.]  The haeccity business becomes quite apparent in Section VI.  You might enjoy the big eyeball as well.

\section{24-03-10 \ \ {\it Your Discussion with Tumulka}\ \ \ (to R. W. {\Spekkens})} \label{Spekkens83}

Coming back to the point I wrote you about earlier, about why the psi-ontologists generally do not find your notion of noncontextuality so compelling (as you reported of Tumulka).   Previous note pasted below.  [See 18-03-10 note ``\myref{Spekkens80}{Very Final Version}'' to R. W. {\Spekkens}.] I was just cutting and pasting the footnote mentioned there for something else and discovered that I had originally started to write a follow-on footnote as well, but apparently decided not to use it in the paper.  Since it's relevant, let me at least use it in this note:
\bq
It is perhaps worthwhile contrasting this to the ``hidden-variable vision'' of the quantum world, say of the Bohmian pilot-wave theory mentioned previously.  The implication of a hidden variable is that it is not directly experienceable, and to the extent that it does ultimately trickle into experience, it is only so that it props up the quantum mechanical equations.
\eq

\section{24-03-10 \ \ {\it Notes on the Open Future}\ \ \ (to L. Smolin)} \label{SmolinL20}

I had a chance to read your draft on ``The Philosophy of the Open Future'' again.  Attached are some notes on it.

\begin{center}
\large Notes on Lee Smolin's 15 February 2010 draft of\\
{\bf The Philosophy of the Open Future}\medskip\\
\end{center}

With the APS March Meeting behind me, I've gotten a chance to reread your draft on the philosophy of the open future.  It's got some big ideas in it, and I like it very much for the message it's starting to formulate.

I've got nothing very deep to say in response at the moment, but I'll react to a few points in your draft.

First a very general thing.  Of the philosophers I have read, the thoughts in your essay sound to me the most like John Dewey's.  I say this not to mean that ``it's all been said before,'' but that you two might make good conversation partners.  Particularly, you might find some congenial inspiration in his writings.  With regard to the subject matter of your essay, I think his sweep and interests are similar to your own.  Dewey's thoughts, of course, will not be informed by today's physics or today's economical issues, but he has nonetheless a well-developed system that might be a good backdrop for your thinking.  Picking a nearly random book off my shelf ({\sl Dewey's Logical Theory: New Studies and Interpretations}, edited by Burke {\it et al}.), I find this key point about what Dewey was thinking:
\bq
Dewey begins from a Darwinian premise of interaction. On this view, experience is ``an affair of the intercourse of a living being with its physical and social environment'' and thus an affair ``primarily of doing.''  Experience then is not a receiving and filtering sensory data from an external world.  It is rather an {\it exchange}, a {\it transaction}, between an organism and the physical and social factors within its environment:  ``When we experience something, we act upon it, we do something with it; then we suffer or undergo the consequences.  We do something to the thing and then it does something to us in return.''

Dewey's placement of experience in the interactions and transactions between an organism and its environment is further augmented with the recognition of a stabilization propensity characteristic of living beings.  Experience is episodic, punctuated by occasions of disturbance and resolution, of imbalance and regained composure.  Thus experience is not only transformational.  It also has force and direction, impelled by an innate drive of the living being to maintain its own {\it well}-being.  In short, experience is an activity in and by which an organism maintains integration with its environment.
\eq
And here's another way the point was made in an article on Sidney Hook's intellectual debt to John Dewey:
\bq
In Dewey's naturalistic metaphysics, a contextual and empirical account of the origin and function of human knowledge recognized man's quest for adaptation to and control of his environment. This emphasis upon human adaptation and control characterized the recognition within pragmatic philosophy of its debt to Darwinism in the development of its theory of mind's place in nature and of the nature of human knowledge.

The thesis that human beings were capable of transforming the natural environment and the seemingly ``given'' facts of nature was a feature of Deweyan naturalistic metaphysics that was relevant for the pragmatic theory of knowledge. Hook followed Dewey in stressing the connection between a naturalistic metaphysics and the identification of human knowledge as an instrument in transforming the real nature of Nature. Significantly, this feature of naturalistic metaphysics heightened the pragmatic emphasis upon the capacity of human agency to function as an activist transformer of the natural environment. To a significant degree, what was natural, whether in the sphere of the universe or the sphere of human affairs, could be considered in the idiom of pragmatism not so much as a ``given,'' but as a possibility for a ``taken.'' The theme of human transformation of the natural environment carried Hook from the initial interest in the metaphysics of pragmatism to the possibilities of pragmatism as a philosophy which could apply scientific method for social change.
\eq

When I read your sentences,
\bls
The problem requires not a choice between nature and technology but a reorientation of their relationship to each other.
\els
\bls
As long as we have the capability to do it, we must prevent or moderate major changes in the climate, for the same reason we must look out for and destroy asteroids that might collide with earth. Once we have resolved this emergency we will be committed to a continuing regulation of the climate to keep it in a range where human beings can thrive. This means melding our technologies with the natural cycles and systems that already regulate the climate.
\els
\bls
The economy and the climate will not be two things, they will be aspects of a single system.  Thus, to survive the climate crisis we have to conceive of and bring into existence a novel kind of system, which will be a symbiosis of the biological processes which determine the climate with our technological civilization.
\els
I hear an echo of the Dewey above.  For, like you, Dewey saw these transactions with nature occurring at all scales, from simple organisms to large societies, nations, and economies.  It is particularly the last quote of you that I was thinking of when I said that your essay contains some big ideas---the economy and the climate as two aspects of a single thing.  I liked your way of phrasing the issue.

Dewey called his philosophy ``instrumentalism,'' but by that he meant something {\it not at all like\/} what most modern philosophers of science (Harvey Brown, Richard Healey, Jeff Bub, and the like) mean when they say instrumentalism.  The common, modern definition goes something like this:  ``instrumentalism is the view that a concept or theory should be evaluated by how effectively it explains and predicts phenomena, as opposed to how accurately it describes objective reality'' (wiki), or ``instrumentalism can be formulated as the thesis that scientific theories---the theories of the so-called `pure' sciences---are nothing but computational rules (or inference rules)'' (Popper).  For Dewey, a scientific theory is a good instrument to the extent that it can be used to mold the world to our needs---some of that might be about prediction, but much of it is not.  It is all about the actions we can reasonably take upon the world in our present stage of development.  Here is the way William James put it, but this strain is certainly in Dewey too:
\bq
[I]f you follow the pragmatic method, you cannot look on any such word as closing your quest.  You must bring out of each word its
practical cash-value, set it at work within the stream of your experience.  It appears less as a solution, then, than as a program
for more work, and more particularly as an indication of the ways in which existing realities may be {\it changed}.

{\it Theories thus become instruments, not answers to enigmas, in which we can rest.}  We don't lie back upon them, we move forward, and, on occasion, make nature over again by their aid.
\eq
The purpose of a good theory is not to draw an accurate image of a pre-existent static world, but to be like a hammer, saw, or screwdriver:  It is a thought structure that helps us to participate in the inevitable flow and change of the world as best we can.  And as we evolve/devolve/change such laws evolve/devolve/change as well (and reciprocally so).

Certainly I believe this of quantum mechanics---that it is a hammer we use in light of the world possessing a certain active, life-breathing power---for I think this is much of its very message.  (I probably believe the hammer metaphor of other theories as well, like classical mechanics and general relativity, but there it is not so ``in your face'' as it is with quantum mechanics.)  Maybe it's worthwhile putting it like this.  In your essay and my own, one common theme comes out:  Human agency matters.  ``We can on occasion make nature over.''  Though I myself like to emphasize that agency matters from everything so large as the ecosystem and climate, to everything so small as a quantum measurement.

If you would like to borrow my copy Dewey's {\sl Reconstruction in Philosophy\/} (a good introduction to his thought), or most anything else for that matter that he's written, let me know.

Now on to more specific matters.

\bls
This hierarchy applies also to the natural/artificial divide, in that the natural is valued
over the artificial because it is closer to absolute perfection, and therefor closer to
timelessness.
\els
Typo.

\bls
But if the search for knowledge is essentially the task of remembering truths which have been ever-present, then the possible solutions to any problem
already exist in fixed menus of possibility, and novelty is an illusion. A world without time is a world with fixed categories and fixed possibilities that cannot be transcended.
\els
I like that phrase ``fixed menus of possibility'' and can see myself using it. Is the phrase your invention?  Or Unger's or Kauffman's or still someone else's?

\bls
To fully realize the conception of a time bound world, we need to invent a notion of truth which removes the equation of truth with timelessness and replaces it with a notion of truth that is no less objective, but which has room for surprise, invention and novelty.
\els
Somewhere---and I have not been able to pinpoint where this morning---William James makes the point that the pragmatic theory of truth he was involved in developing was only a secondary aim for him.  It might be written somewhere in {\sl The Meaning of Truth\/} or {\sl Essays in Radical Empiricism}.  Instead James's main aim was something very much like what you say here:  One needs more looseness than a correspondence theory of truth allows if one is going to have a malleable, creative world.  Truth has to have a different meaning in a world that is ``ever not quite.''  If I can dig up a reference, I'll send it to you.

[Time elapsed.]  Aha, I found it!  It's in the conjunction of two prefaces---that for {\sl The Will to Believe\/} and that for {\sl The Meaning of Truth}. Let me quote the relevant pieces:
\bq
Were I obliged to give a short name to the attitude in question, I should call it that of {\it radical empiricism}, in spite of the fact that such brief nicknames are nowhere more misleading than in philosophy. I say `empiricism,' because it is contented to regard its most assured conclusions concerning matters of fact as hypotheses liable to modification in the course of future experience; and I say `radical,' because it treats the doctrine of monism itself as an hypothesis, and, unlike so much of the half-way empiricism that is current under the name of positivism or agnosticism or scientific naturalism, it does not dogmatically affirm monism as something with which all experience has got to square. The difference between monism and pluralism is perhaps the most pregnant of all the differences in philosophy. {\it Prim\^{a} facie\/} the world is a pluralism; as we find it, its unity seems to be that of any collection; and our higher thinking consists chiefly of an effort to redeem it from that first crude form. Postulating more unity than the first experiences yield, we also discover more. But absolute unity, in spite of brilliant dashes in its direction, still remains undiscovered, still remains a {\it Grenzbegriff}. ``Ever not quite'' must be the rationalistic philosopher's last confession concerning it. After all that reason can do has been done, there still remains the opacity of the finite facts as merely given, with most of their peculiarities mutually unmediated and unexplained. To the very last, there are the various `points of view' which the philosopher must distinguish in discussing the world; and what is inwardly clear from one point remains a bare externality and datum to the other. The negative, the alogical, is never wholly banished. Something---``call it fate, chance, freedom, spontaneity, the devil, what you will''---is still wrong and other and outside and unincluded, from {\it your\/} point of view, even though you be the greatest of philosophers. Something is always mere fact and {\it givenness}; and there may be in the whole universe no one point of view extant from which this would not be found to be the case. ``Reason,'' as a gifted writer says, ``is but one item in the mystery; and behind the proudest consciousness that ever reigned, reason and wonder blushed face to face. The inevitable stales, while doubt and hope are sisters. Not unfortunately the universe is wild,---game-flavored as a hawk's wing. Nature is miracle all; the same returns not save to bring the different. The slow round of the engraver's lathe gains but the breadth of a hair, but the difference is distributed back over the whole curve, never an instant true,---ever not quite.''

This is pluralism, somewhat rhapsodically expressed.  He who takes for his hypothesis the notion that it is the permanent form of the world is what I call a radical empiricist. For him the crudity of experience remains an eternal element thereof. There is no possible point of view from which the world can appear an absolutely single fact. Real possibilities, real indeterminations, real beginnings, real ends, real evil, real crises, catastrophes, and escapes, a real God, and a real moral life, just as common-sense conceives these things, may remain in empiricism as conceptions which that philosophy gives up the attempt either to `overcome' or to reinterpret in monistic form.
\eq
And
\bq
Most of the pragmatist and anti-pragmatist warfare is over what the word ``truth'' shall be held to signify, and not over any of the facts embodied in truth-situations; for both pragmatists and anti-pragmatists believe in existent objects, just as they believe in our ideas of them. The difference is that when pragmatists speak of truth, they mean exclusively something about the ideas, namely their workableness; whereas when anti-pragmatists speak of truth they seem most often to mean something about the objects. Since the pragmatist, if he agrees that an idea is ``really'' true, also agrees to whatever it says about its object; and since most anti-pragmatists have already come round to agreeing that, if the object exists, the idea that it does so is workable; there would seem to be so little left to fight about that I might well be asked why instead of reprinting my share in so much verbal wrangling, I do not show my sense of ``values'' by burning it all up.

I understand the question and I will give my answer. I am interested in another doctrine in philosophy to which I give the name of radical empiricism, and it seems to me that the establishment of the pragmatist theory of truth is a step of first-rate importance in making radical empiricism prevail.
\eq

\bls
A new philosophy is needed which anticipates the merging of the natural and the artificial by achieving a consilience of the natural and social sciences, in which human agency has a rightful place in nature. This is not a relativism in which anything we want to be true can be. To survive the challenge of climate change it matters a great deal what is true.
\els
And here I have to say you led my thoughts back to F.~C.~S. Schiller's paper, ``Axioms as Postulates.''  It is just a wonderful paper.  Schiller emphasized that {\it methodologically\/} it is important to start by supposing the world to be plastic and malleable to the fullest extent imaginable (i.e., that we {\it can\/} have anything we want), and only in exhaustion---when the world does not yield it after our impassioned tries---to accept otherwise.

Let me quote him at length; I think you'll find it fun:
\bq
[M]y fourth and most important point is that the world is {\it plastic}, and may be moulded by our wishes, if only we are determined to give effect to them, and not too conceited to learn from experience, {\it i.e.}\ by trying, {\it by what means\/} we may do so.

That this plasticity exists will hardly be denied, but doubts may be raised as to how far it extends. Surely, it may be objected, it is mere sarcasm to talk of the plasticity of the world; in point of fact we can never go far in any direction without coming upon rigid limits and insuperable obstacles. The answer surely is that the extent of the world's plasticity is not known {\it a priori}, but must be found out by trying. Now in trying we can never start with a recognition of rigid limits and insuperable obstacles. For if we believed them such, it would be {\it no use\/} trying. Hence we must assume that we can obtain what we want, if only we try skilfully and perseveringly enough. A failure only proves that the obstacles would not yield to the method employed: it cannot extinguish the hope that by {\it trying again\/} by other methods they could finally be overcome.

Thus it is a {\it methodological necessity\/} to assume that the world is {\it wholly plastic}, {\it i.e.}\ to {\it act as though\/} we believed this, and will yield us what we want, if we persevere in wanting it.

To what extent our assumption is true in the fullest sense, {\it i.e.}\ to what extent it will work in practice, time and trial will show. But our faith is confirmed whenever, by acting on it, we obtain anything we want; it is checked, but not uprooted, whenever an experiment fails.

As a first attempt to explain how our struggle to mould our experience into conformity with our desires is compatible with the `objectivity' of that experience, the above may perhaps suffice, though I do not flatter myself that it will at once implant conviction. Indeed I expect rather to be asked indignantly---`Is there not an objective nature which our experiments do not make, but only discover? Is it not absurd to talk as if our attempts could alter the facts? And is not reverent submission to this pre-existing order the proper attitude of the searcher after truth?'

The objection is so obvious that the folly of ignoring it could only be exceeded by that of exaggerating its importance. It is because of the gross way in which this is commonly done that I have thought it salutary to emphasise the opposite aspect of the truth. We have heard enough, and more than enough, about the duty of humility and submission; it is time that we were told that energy and enterprise also are indispensable, and that as soon as the submission advocated is taken to mean more than rational methods of investigation, it becomes a hindrance to the growth of knowledge. Hence it is no longer important to rehearse the old platitudes about sitting at the feet of nature and servilely accepting the kicks she finds it so much cheaper to bestow than halfpence. It is far more important to emphasise the other side of the matter, viz.\ that unless we ask, we get nothing. We must ask often and importunately, and be slow to take a refusal. It is only by asking that we discover whether or not an answer is attainable, and if they cannot alter the `facts,' our demands can at least make them appear in so different a light, that they are no longer {\it practically\/} the same.

For in truth these independent `facts,' which we have merely to acknowledge, are a mere figure of speech. The growth of experience is continually transfiguring our `facts' for us, and it is only by an {\it ex post facto\/} fiction that we declare them to have been `all along' what they have {\it come to mean for us}. To the vision of the rudimentary eye the world is {\it not\/} coloured; it becomes so only to the eye which has developed colour `sensitiveness': just so the `fact' of each phase of experience is relative to our knowledge, and that knowledge depends on our efforts and desires to know. Or, if we cling to the notion of an absolutely objective fact of which the imperfect stages of knowledge only catch distorted glimpses, we must at least admit that only a final and perfect rounding-off of knowledge would be adequate to the cognition of such fact. The facts therefore which {\it we\/}  as yet encounter are not of this character: it may turn out that they are not what they seem and can be transfigured if we try. Hence the antithesis of subjective and objective is a false one: in the process of experience `subject' and `object' are only the poles, and the `subject' is the `positive' pole from which proceeds the impetus to the growth of knowledge. For the modifications in the world, which we desire, can only be brought about by our assuming them to be possible, and {\it therefore\/}  trying to effect them.
\eq

\bls
What is needed is instead a relationalism, according to which the future is restricted by, but not determined from the present, so that genuine novelty and invention, while rare, are real and possible.
\els
Why relationalism?  I don't see a connection between (what I know of) that word and much of the rest of your essay.

\bls
The notion that truth is eternal and lies outside of the world we experience is not only a religious idea. It is in science as well, as is exemplified by the Platonic view of mathematics.
\els
You remind me here of a point that Richard Rorty once made forcefully.  Unfortunately, I seem to have lost my copy and I haven't been able to get hold of the full paper online (titled ``Pragmatism as Anti-authoritarianism'').  But here is a part of the abstract for your enjoyment:
\bq
There is a useful analogy to be drawn between the pragmatists' criticism of the idea that truth is a matter of correspondence to the intrinsic nature of reality and the Enlightenment's criticism of the idea that morality is a matter of correspondence to the will of a Divine Being. The pragmatists' anti-representationalist account of belief is, among other things, a protest against the idea that human beings must humble themselves before something non-human, whether the Will of God or the Intrinsic Nature of Reality. Seeing anti-representationalism as a version of anti-authoritarianism permits one to appreciate an analogy which was central to John Dewey's thought: the analogy between ceasing to believe in Sin and ceasing to accept the distinction between Reality and Appearance.
\eq

\bls
This view is not popular among philosophers due to a powerful and commonsense objection to it, which is that to the extent that mathematical objects
live in a separate realm outside of time and space, there is no way for human beings to gain access to or knowledge about them.
\els
Typo?  ``gain access to or have knowledge'' maybe?

\bls
This was the view of Einstein and many 20th Century philosophers (although, one of them. the logical positivist Carnap, recounts
conversations in which Einstein regretted the loss of a place for the present moment).
\els
Typo: period after them.  Also, it is not completely clear that the word ``regretted'' should be used here.  Here's the Carnap quote,
\bq
\noindent
Einstein said that the problem of the Now worried him seriously. He explained that the experience of the Now means something special for man, something essentially different from the past and the future, but that this important difference does not and cannot occur within physics. That this experience cannot be grasped by science seemed to him a matter of painful but inevitable resignation.
\eq
On the other hand compare that to what he wrote to Michele Besso's widow,
\bq
\noindent
In quitting this strange world he has once again preceded me by just a little. That doesn't mean anything. For we convinced physicists the distinction between past, present, and future is only an illusion, however persistent.
\eq
Just a question:  Does ``painful resignation'' imply ``regret''?

\bls
This field is as much speculative metaphysics as science. A lot of its literature is taken up with issues that can have no bearing on experiment,
such as how probabilities are to be defined in an infinite and timeless realm \ldots
\els
Funny, this is just about the same thing I would say of the Everettians.

\bls
There is a powerful if unconscious attraction to theories in which time plays no role. It gives theorists the
impression of standing outside the world, in a timeless realm of pure truth, against
which the time and contingency of the real world pale.
\els
I love the way Arthur Lovejoy puts a similar point in his book {\sl The Great Chain of Being}:
\bq
Some examples of metaphysical pathos in the stricter sense ought \ldots\ to be given. A
potent variety is the eternalistic pathos---the aesthetic pleasure which the bare abstract idea of immutability gives us.
The greater philosophical poets know well how to evoke it. In English poetry it is illustrated by those familiar lines in Shelley's
{\it Adonais\/} of which we have all at some time felt the magic:
\begin{verse}
The One remains, the many-change and pass,\\
Heaven's light forever shines, earth's shadows fly \ldots.
\end{verse}
It is not self-evident that remaining forever unchanged should be regarded as an excellence; yet through the associations and
the half-formed images which the mere conception of changelessness arouses---for one thing, the feeling of rest which its
{\it innere Nachahmung\/} induces in us in our tired moods---a philosophy which tells us that at the heart of things there is a
reality wherein is no variableness nor shadow that is cast by turning, is sure to find its response in our emotional natures, at
all events in certain phases of individual or group experience.  Shelley's lines exemplify also another sort of metaphysical
pathos, often conjoined with the last---the monistic or pantheistic pathos. That it should afford so many people a peculiar
satisfaction to say that All is One is, as William James once remarked, a rather puzzling thing. What is there more
beautiful or more venerable about the numeral {\it one\/} than about any other number? But psychologically the force of the monistic
pathos is in some degree intelligible when one considers the nature of the implicit responses which talk about oneness produces.
It affords, for example, a welcome sense of freedom, arising from a triumph over, or an absolution from, the
troublesome cleavages and disjunctions of things. To recognize that things which we have hitherto kept apart in our
minds are somehow the same thing---that, of itself, is normally
an agreeable experience for human beings. \ldots  So, again, when a monistic philosophy declares, or suggests, that one is
oneself a part of the universal Oneness, a whole complex of obscure emotional responses is released. The deliquescence of
the sense---the often so fatiguing sense---of separate personality,
for example, which comes in various ways (as in the
so-called mob-spirit), is also capable of excitation, and of
really powerful excitation, too, by a mere metaphysical
theorem. Mr.\ Santayana's sonnet beginning ``I would I
might forget that I am I'' almost perfectly expresses the mood
in which conscious individuality, as such, becomes a burden.
Just such escape for our imaginations from the sense of being a
limited, particular self the monistic philosophies sometimes
give us.
\eq
Of course, when he goes on to opine about my own pathos (the voluntaristic one), I'm not so enamored with his sharp tongue!

\bls
We live in a world in which it is impossible in principle to anticipate and list most of the
contingencies that will arise in the future. Neither the political context, nor the
inventions, nor the fashions, nor the weather nor climate are specifiable in advance.
\els
Nor the outcome of a single minuscule quantum measurement.  All these instances, if you ask me, are cut from a single cloth.  Where quantum mechanics has pedagogical value in comparison to the others, however, is with its shock value.  For it is in fundamental physics that one {\it previously\/} would not have expected such a thing.  Most importantly, it gives a sense that the issue is central to our world.

\bls
Moreover, the higher level laws can and often do influence the lower levels of
description. There are molecules which only exist on Earth because they are
components of biological systems and hence byproducts of evolution. Suppose we
wanted to know how many hemoglobin molecules are present on Earth. Hemoglobin is
just a big molecule, completely describable by the laws of quantum mechanics. But
there is no way the question of how many exist at a given time could be arrived at just
by solving the laws of quantum mechanics. To explain why there are any hemoglobin
molecules on Earth, and to estimate their number, one must reason in terms of higher
order laws which apply to animals and are not completely reducible to quantum
mechanics.
\els
In my present stage of thought, I myself would never say ``hemoglobin is
a big molecule {\bf completely\/} describable by the laws of quantum mechanics.''  As I wouldn't say it of humans, nor of diamonds (see Section VI in my essay), I wouldn't say it of hemoglobin.  It is the ``completely'' thing that contradicts my own nonreductionism.  For me, quantum mechanics is {\it additive\/} to whatever is particular in an object/concept.  Be that as it may, there are definitely similarities in our thinking here.  Here's a little passage and footnote from my essay:
\bq
The metaphysics of empiricism can be put like this.  Everything experienced, everything experienceable has no less an ontological status than anything else.

Footnote:  That is, every piece of the universe had better be hard-wired for the contingency that an agent might experience it somewhere, somehow, no matter how long and drawn out the ultimate chain might be to such a potential experience.  Does this mean even ``elementary'' physical events just after the Big Bang must make use of {\it concepts\/} that, to the reductionist mind, ought to be 15 billion years removed down the evolutionary chain?  You bet it does.  But a nonreductionist metaphysic need make no apology for this---such things are in the very idea.  John Wheeler's great smoky dragon comes into the world biting its own tail.

W.~K. Wootters tells a lovely story of an encounter he had several years ago of with his young son Nate.  Nate said, ``I wish I could make this flower move with my mind.''  Wootters reached out and pushed the flower, saying, ``You can. You do it like this.''  From the perspective here, this is an example of an interaction between two nonreductionist realms.  Each realm influences the other as its turn comes.  There is a kind of reciprocality in this, an action-reaction principle, that most reductionist visions of the world would find obscene.
\eq

\bls
The best that theory may be able to do is describe what complexity theorist Stuart Kauffman calls the adjacent possible, that is discern some of the very next forms that may emerge. Or there may be singularities where not even this is possible.
\els
I've always been intrigued by this passage from Henri Poincar\'e's essay, ``The Evolution of Laws,'' in his {\sl Derni\`eres Pens\'ees}:
\bq
\indent
Mr.~Boutroux, in his writings on the contingency of the laws of
Nature, queried, whether natural laws are not susceptible to change
and if the world evolves continuously, whether the laws themselves
which govern this evolution are alone exempt from all variation.
\ldots\ I should like to consider a few of the aspects which the
problem can assume. \ldots\

In summary, we can know nothing of the past unless we admit that the
laws have not changed; if we do admit this, the question of the
evolution of the laws is meaningless; if we do not admit this
condition, the question is impossible of solution, just as with all
questions which relate to the past. \ldots

But, it may be asked, is it not possible that the application of the
process just described may lead to a contradiction, or, if we wish,
that our differential equations admit of no solution?  Since the
hypothesis of the immutability of the laws, posited at the beginning
of our argument would lead to an absurd consequence, we would have
demonstrated {\it per absurdum\/} that laws have changed, while at the
same time we would be forever unable to know in what sense.

Since this process is reversible, what we have just said applies to
the future as well, and there would seem to be cases in which we would
be able to state that before a particular date the world would have to
come to an end or change its laws; if, for example, our calculations
indicate that on that date one of the quantities which we have to
consider is due to become infinite or to assume a value which is
physically impossible.  To perish or to change its laws is just about
the same thing; a world which would no longer have the same laws as
ours would no longer be our world but another one.
\eq
Your remark about singularities brought that to mind.

\bls
The first person to ask this question cogently seems to have been the founder of the
American pragmatist school of philosophy, Charles Sanders Pierce. He wrote ``The only
possible way of accounting for the laws of nature and of uniformity in general is to
presume them results of evolution''.
\els
Typo: should be Peirce.  With regard to Peirce being the first, I suppose it depends upon what you mean be ``cogently.''  Peirce wrote those lines in ``The Architecture of Theories'' in 1891.  On the other hand \'Emile Boutroux published his book {\sl The Contingency of the Laws of Nature\/} in 1874. And then there was Bergson, whose books started appearing in 1889.  James in his essay, ``On the Notion of Reality as Changing,'' writes:
\bq
Volumes i, ii, and iii of the {\sl Monist\/} (1890--1893) contain a
number of articles by Mr.\ Charles S. Peirce, articles the
originality of which has apparently prevented their making an
immediate impression, but which, if I mistake not, will prove a
gold-mine of ideas for thinkers of the coming generation. Mr.\
Peirce's views, tho reached so differently, are altogether congruous
with Bergson's. Both philosophers believe that the appearance of
novelty in things is genuine. To an observer standing outside of its
generating causes, novelty can appear only as so much `chance'; to
one who stands inside it is the expression of `free creative
activity.' Peirce's `tychism' is thus practically synonymous with
Bergson's `devenir r\'eel.' The common objection to admitting
novelties is that by jumping abruptly in, {\it ex nihilo}, they
shatter the world's rational continuity. Peirce meets this objection
by combining his tychism with an express doctrine of `synechism' or
continuity, the two doctrines merging into the higher synthesis on
which he bestows the name of `agapasticism' ({\it loc.\ cit.}, iii,
188), which means exactly the same thing as Bergson's `\'evolution
cr\'eatrice.' Novelty, as empirically found, does n't arrive by jumps
and jolts, it leaks in insensibly, for adjacents in experience are
always interfused, the smallest real datum being both a coming and a
going, and even numerical distinctness being realized effectively
only after a concrete interval has passed. \ldots

I can give no further account of Mr.\ Peirce's ideas in this note,
but I earnestly advise all students of Bergson to compare them with
those of the french philosopher.
\eq
Thus, I myself, would probably credit the three as a triumvirate.

\bls
This provides further motivation for believing in the reality of time and the openness of
the future. If the selection of the laws of nature that determine what elementary
particles exist and how they interact is a result of evolution analogous to natural
selection, then there must be a time that this evolution plays out in. And it must be a
time that exists prior to the laws, for the laws must have scope to evolve in time.
\els
Yes, but I wonder whether there might not {\it in principle\/} be as many ``reality of times'' as there are agents (more broadly, objects)---time being single valued only to the extent that groups of agents interact and ``synchronize.''  My QBist program already captures a bit of an image like this.  Wheeler had his ``many-fingered time,'' I need a good name for mine.

\bls
The big bang may even be an event that brought into being a novel form of organization in
which particles move in space; if this is so then the laws we probe with particle
accelerators like the LHC emerged then.
\els
I'm not sure I understand how to parse this sentence.  I.e., there's something about it that confuses me; I'm not sure what it means.

\bls
But we also have evolved the capacity to think quickly and act decisively at the snap of a twig.
\els
Nice; immediately gave me imagery of early man huddled in the night.

\bls
Fundamentalist communities and open communities can be distinguished by their understanding of time and the possibilities for knowledge of the future. Fundamentalists believe that the future already exists, and are prone to believe that time is altogether an illusion. They believe in mythological stories according to which the ever changing world we perceive is an illusion that hides a true timeless reality.
\els
This took my mind back to the fact that Travis Norsen, beside calling me a solipsist at every turn just because I say ``quantum states do not exist,'' also happens to be a follower of Ayn Rand.

\bls
Open communities teach that the future is not yet real, what is real instead is time and the processes by which the future
unfolds and emerges out of the present. For a fundamentalist, human agency is an
illusion, for a member of an open society human agency is the necessary means of
constructing a future that will not otherwise exist.
\els

I might as well end things with another long quote---Richard Rorty and William James this time:
\bq
\noindent By way of conclusion, I want to offer a gloss on the passage from William James which supplied the title for this conference. James writes
\bq
What really exists is not things made but things in the making. Once made, they are dead, and an infinite number of alternative conceptual decompositions can be used in defining them. But put yourself in the making by a stroke of intuitive sympathy with the thing and, the whole range of possible decompositions coming at once into your possession, you are no longer troubled by the question of which of them is the more absolutely true. Reality falls in passing into conceptual analysis; it mounts in living its own undivided life-it buds and burgeons, changes and creates. Once adopt the movement of this life in any given instance and you know Bergson calls the {\it devenir r\'eel}, by which the thing evolves and grows. Philosophy should seek this kind of living understanding of the movement of reality-not follow science in vainly patching together fragments of its dead results.
\eq
There are various things wrong with this passage \ldots. But if you read this passage as a meditation on the relation between history and philosophy, it takes on a non-metaphysical meaning, and says something important: namely that the human future will always, with a little luck and a lot of imagination, be so different from the past that it is pointless to look for a set of concepts that will cover both. Bergson was right to think that the metaphysical attempt to see things under the aspect of eternity was a failure \ldots.

Reading James in this way lets one draw the moral: do not think that making the past ideas coherent with one another will ever enable you to find a substitute for imagination. Do not think that philosophy will ever succeed in its attempt to trump poetry and the arts. Do not look to philosophers for anything different than the sort of inspiration that you get from poets, painters, musicians, and architects. For their ability to find coherence will never be more than a perspicuous archival arrangement of the imaginative products of the past. They will never provide authoritative guidance for the imagination of the present.

John Rajchman describes James as preoccupied with ``the problem of novelty''---the problem of how to deal with ``things in the making'', how to see and respond to the emergence of things for which we have no preset manner of seeing or responding.  I do not think that there is a solution to this problem, and therefore, as a good verificationist, I do not see it as a problem. The thing to do with novelty is just to be grateful for it, and to create the socio-political conditions which will ensure that there will be a lot more of it. There is a political problem about how to encourage novelty without weakening communal solidarity and social order, but this is a problem to be solved ambulando, experimentally, and democratically. All that philosophy can do to help out with this political problem is to keep reminding us of what is likely to happen if the past is allowed to dictate terms to the future.
\eq
For a good bit of the point of your essay is in ``reminding us of what is likely to happen if the past is allowed to dictate terms to the future.''

Thanks for sharing your essay and giving me the opportunity to think a little.  I hope it's well-received wherever you send it.

\section{24-03-10 \ \ {\it Revealing}\ \ \ (to D. M. {\Appleby}, R. {\Schack}, and H. C. von Baeyer)} \label{Appleby89} \label{Baeyer109} \label{Schack194}

I've just had a little exchange with Adrian Kent on my lines,
\begin{quote}
What are we then to do with the Born Rule for calculating quantum probabilities?  Throw it away and say it never mattered?  It is true that quite an effort has been made by the Everettians to rederive the rule from {\it decision theory}.  Some sympathizers think it works [Wallace09], some don't [Kent09].  But outside the {\it sprachspiel\/} who could ever believe?  No amount of sophistry can make ``decision'' anything other than a hollow concept in a predetermined world.
\end{quote}
from the new paper.

My guess is that you'll really enjoy the contortions in his reply, so I'll share it with you:
\begin{quote}
It seems to me that (a) we can sensibly discuss the preferences of agents in a multiverse --- i.e.\ what successor states they might prefer if they had freedom to choose amongst them --- and so that it isn't ridiculous to formulate a decision theory in this context.  But also that (b) the project of trying to recover Copenhagen predictions from such a theory ultimately fails. Maybe we disagree on (a)?
\end{quote}
Just look at that lovely counterfactual!  ``what successor states they {\it might\/} prefer if they {\it had freedom to choose\/} amongst them''!!!  No, we wouldn't dare say outright that they {\it have\/} freedom; we only talk about what they would do if they {\it had\/} freedom!  What a sad statement on the common philosophies of our time!

\section{25-03-10 \ \ {\it Very, Very Final Version}\ \ \ (to the QBies \& C. Ferrie)} \label{QBies8} \label{Ferrie13.1}

You may be interested in Howard Barnum's posting this morning: ``Quantum Knowledge, Quantum Belief, Quantum Reality: Notes of a QBist Fellow Traveler,'' \arxiv{1003.4555}.

Also, attached is the very, very final version of my own screed.  [See ``QBism, the Perimeter of Quantum Bayesianism,'' \arxiv{1003.5209v1}.]  I'm posting the thing today finally.  Since the version you've all read, the very important Footnote 22 has been added.

\section{25-03-10 \ \ {\it Possible?}\ \ \ (to {\AA}. {\Ericsson})} \label{Ericsson9}

\bae
If you should assume everything to be possible because you have no proof of the opposite, wouldn't that lead you to assume just anything ``to be the case'' (or whatever one should say without using the word ``true'') for which there is no proof of the opposite?  But there cannot really be any such rigorous proof.  And we cannot live like that.  We form beliefs from our experiences, which leads us (at least me!)\ not to believe in {\em [just]} anything and also not to believe everything is possible for us.
\eae

If we did not believe that we could prolong human life with better technologies, we as a society would not invest in medical research.  But we do.  And what is the limit of the prolongation?  How far can we go?  Can we make people last 100 years on average?  150?  Why should there be any end in sight?  We console our personal selves to the idea of eventual death because the alternative is not within reach now---not within our expected lifetimes.  And that is as we should do.  But if death is a {\it necessary\/} end, why worry about the difference between 80 and 100 years?  {\it Methodologically\/} (as Schiller says), we are behaving as if eventually it can be completely overcome.  All our industry is geared toward that, one step at a time.  First we wipe out HIV, then we wipe out heart disease, then we wipe out cancer (all cancers!).  There is no end to the process.  And we should not believe that there is one.

\section{25-03-10 \ \ {\it Torture and Relief} \ \ (to A. Kent)} \label{Kent23}

Thanks for your note.

(1)  OK, I back off on the label ``sympathizer.''  The insult now reads:
\bq\noindent
   Or take the Everettians.  Their world purports to have no
   observers, but then it has no probabilities either.  What
   are we then to do with the Born Rule for calculating
   quantum probabilities?  Throw it away and say it never
   mattered?  It is true that quite an effort has been made
   by the Everettians to rederive the rule from {\it decision
   theory}.  Of those who take the point seriously, some
   think it works \verb+\cite{Wallace09}+, some don't \verb+\cite{Kent09}+.
   But outside the {\it sprachspiel\/} who could ever
   believe?  No amount of sophistry can make ``decision''
   anything other than a hollow concept in a predetermined
   world.
\eq

You were right to call me on that, for all I had really meant was, ``of those who take the point seriously.''  My plan is to finally post the darned thing today.

(2)  Because I don't really.  For me it is not that the Everettian interpretation is ``bizarre'' (as many people say) or carries too much ``ontological baggage'' (as others say), but that it is simply empty of content.  It is a great effort full of sound and fury, in the end signifying nothing.  When I heard David Wallace once admit (at the MWI meeting you organized here at PI) that one {\it could\/} give an Everettian interpretation of classical Liouville mechanics, that capped it all off for me.  His defence was only, ``but there is no great need in that case.''  And so I say of QM anyway.  But the essential point is, one can play the MWI game with {\it any\/} probabilistic theory.  And when done, it is a ``more or less arbitrary appendage'' to the theory.

(3a)  I have to admit getting a really good laugh from this part of your sentence:
\bak
   [W]e can sensibly discuss the preferences of agents in a
   multiverse -- i.e. what successor states they might prefer
   if they had freedom to choose amongst them \ldots
\eak
For the Everettians, each one of them, are great monists at heart; in a different century they would have been Hegelians. The real thing behind their motivation is that paper thin. They contemplate a world with no true counterfactuals from within, but then to make sense of this world they invoke a counterfactual at the outset, ``what successor states they might prefer if they had freedom to choose amongst them.''  When I read your sentence, a little sarcastic voice in me came out fighting, ``No, we wouldn't {\it dare\/} say outright that they {\it have\/} freedom; we only talk about what they would do if they {\it had\/} freedom!''

\bak
Obviously, though, I'd be intellectually unhappy with the suggestion
that there isn't anything substantive to discuss.
\eak

My psychoanalysis is that this is why my first version of the Everettian insult stung you, when I had intended you to be  ``on the good side of it''.  I am sorry about that---I don't mind insulting people at times if I think it might help shake a nut or bolt loose, but I did not want to insult you.  On the other hand, I can't hide what you suspect---that I do believe there is not too much substantive to discuss in the MWI view {\it at this stage in our thinking}.  (There once was, even for me, before I had wholly given up on psi-ontology.  But now I think it is time to move on.)

One technical point related to this about my present attitude on the Born Rule.  It is that it is not something to be {\it derived\/} (in Wallace's way, or Gleason's way, or any other way).  Rather my present feeling is that it is the fundamental statement of what quantum mechanics is about---it is a rule that takes probabilities in, and gives new probabilities back out.  (See the attached ``press release'' on my house; what I mean in the last sentence is made formal by the ``pretty equation'' at the bottom of it.  {\Ruediger} and I call it the urgleichung.)  Two further notes below on this ``attitude'' in case you're interested.

I'd love to come to Cambridge again (if you still want me after the note above).  Thanks for the invitation.  Maybe sometime this Fall.

\section{25-03-10 \ \ {\it What Bayesians Expect of Each Other} \ \ (to G. Chiribella)} \label{Chiribella1}

This is the paper I was talking about.  [See M. J. Bayarri and M. H. DeGroot, ``What Bayesians Expect of Each Other,'' J. Am.\ Stat.\ Assoc.\ {\bf 86}, 924--932 (1991).] It'd be nice to get a quantum version of this theorem.  If you'd like to discuss and think about this, let me know.    It'd be great if we could get something solid.

\section{26-03-10 \ \ {\it Templeton OFI} \ \ (to R. {\Schack})} \label{Schack195}

There is very little on paper so far, but I have been thinking much in my head.  Fear not:  It will come together!

Any preference on these potential titles?
\begin{itemize}
\item That the World Can Be Otherwise: The Essence of Quantum Theory

\item That the World Could Be Otherwise: The Essence of Quantum Theory

\item That the World Has the Power to Be Otherwise:  The Essence of Quantum Theory

\item The World's Power to Be Otherwise:  The Essence of Quantum Theory

\item The Essence of Quantum Theory:  That the World Can Be Otherwise

\item Quantum Theory's Essence:  That the World Can Be Otherwise
\end{itemize}

\section{26-03-10 \ \ {\it Title} \ \ (to R. {\Schack})} \label{Schack196}

``That the World Can Be Shaped:  Quantum Mechanics, Counterfactuality, Free Will''

\section{26-03-10 \ \ {\it Thinking, Scheming} \ \ (to S. Weinstein \& D. Fraser)} \label{Weinstein5} \label{Fraser3}

{\Ruediger} {\Schack} and I are working on a Templeton Foundation proposal, and I find myself trying to think big.  At the moment, I am playing with, ``That the World Can Be Shaped: Quantum Mechanics, Counterfactuality, Free Will,'' for its title.  Anyway, the reason I write you is that I am also playing with the idea of using part of the money to hire a QBism-pragmatism-Cartwrightism sympathetic philosopher as a two-year postdoc.  (Beware!)  But one of the things I'm bumping into is the space limitations at PI and the ``second-class status'' of our ``associate postdocs'' (those that are housed at PI but have funding coming from external sources).  Which leads me to the following idea:  How receptive would the UW philosophy department be to a postdoc freebie of this variety?  There is the space issue already mentioned, but also I'm thinking it might be healthier for the person to be around there more than here anyway.  He/she and I could have weekly meetings, lunch, etc., but otherwise it'd probably be good for the person to mingle more closely with proper philosophers.

A couple of questions if this is a possibility.  What is the standard salary of a philosophy postdoc at UW?  And how much overhead would your department require?  (Templeton will give up to 15\%.  But of course the smaller the percentage I can get away with the happier I'd be.)

Just some thoughts.

By the way, if you don't know what I mean by QBism see the attached new paper and ``press release'' on my house.  [See ``QBism, the Perimeter of Quantum Bayesianism,'' \arxiv{1003.5209v1}.] Particularly the connection between QBism and pragmatism / ``Cartwrightism'' is described in Section VI, starting page 19.

\section{26-03-10 \ \ {\it QBist Double Slit} \ \ (to M. A. Graydon)} \label{Graydon4}

\bmag
Question: Have you written anything on a QBist interpretation of Young's experiment? In the text-book explanations we find references to free-particle wavefunctions in infinite dimensional Hilbert spaces \ldots\ so does this mean that one cannot translate the math into SIC language?
\emag

The Mach--Zehnder interferometer captures the essence of that, and it is a finite dimensional system.  Gelo is looking at this in detail at the moment---you might get his take on that.

With regard to infinite dimensional systems more generally, I really don't know how the answer is going to go:  It could be that SICs do after all have an infinite dimensional limit, but it could be that they don't.  Gutwise, I kind of hope that they don't have an infinite limit.  So much the worse for infinite dimensional Hilbert spaces, I say.  Physics has been wrenching with the difficulties of field theory and (the absurdity of) renormalization for so long, that maybe it is time for worldview re-adjustment.  Infinite dimension might be like infinite inertial mass---a handy approximation to make at times, but ultimately not something that is ever really true.

In general I advocate what Jaynes calls the ``the cautious approach policy'':
\bq\noindent
{\bf Our ``Cautious Approach'' Policy} \medskip

The derivation of the rules of probability theory from simple desiderata of rationality and consistency in Chapter 2 applied to discrete, finite sets of propositions. Finite sets are therefore our safe harbor, where Cox's theorems apply and nobody has ever been able to produce an inconsistency from application of the sum and product rules. Likewise, in elementary arithmetic finite sets are the safe harbor in which nobody has been able to produce an inconsistency from applying the rules of addition and multiplication.

But as soon as we try to extend probability theory to infinite sets, we are faced with the need to
exercise the same kind of mathematical caution that one needs in proceeding from finite arithmetic
expressions to infinite series. The ``parlor game'' at the beginning of Chapter 15 illustrates how
easy it is to commit errors by supposing that the operations of elementary arithmetic and analysis,
that are always safe on finite sets, may be carried out also on infinite sets.

In probability theory, it appears that the only safe procedure known at present is to derive
our results first by strict application of the rules of probability theory on finite sets of propositions;
then after the finite set result is before us, observe how it behaves as the number of propositions
increases indefinitely. There are, essentially, three possibilities:
\begin{itemize}
\item[(1)] It tends smoothly to a finite limit, some terms just becoming smaller and dropping
out, leaving behind a simpler analytical expression.
\item[(2)] It blows up, i.e., becomes infinite in the limit.
\item[(3)] It remains bounded, but oscillates or fluctuates forever, never tending to any definite
limit.
\end{itemize}
In case (1) we say that the limit is ``well behaved'' and accept the limit as the correct solution
on the infinite set. In cases (2) and (3) the limit is ill-behaved and cannot be considered a valid
solution to the problem. Then we refuse to pass to the limit at all.

This is the ``Look before you leap'' policy: in principle, we pass to a limit only after verifying
that the limit is well-behaved. Of course, in practice this does not mean that we conduct such a
test anew on every problem; most situations arise repeatedly, and rules of conduct for the standard
situations can be set down once and for all. But in case of doubt, we have no choice but to carry
out this test.

In cases where the limit is well-behaved, it may be possible to get the correct answer by
operating directly on the infinite set, but one cannot count on it. If the limit is not well-behaved,
then any attempt to solve the problem directly on the infinite set would have led to nonsense, {\it the
cause of which cannot be seen if one looks only at the limit, and not the limiting process}.  The
paradoxes noted in Chapter 15 illustrate some of the horrors that have resulted from carelessness
in this regard.
\eq
Recommended reading!

\section{28-03-10 \ \ {\it  Game Theory and Quantum Mechanics} \ \ (to D. H. Wolpert)} \label{Wolpert3}

\bdhw
There's been a lot of stuff on applying game theory to situations where the game players have quantum-mechanical apparatus to play with.  ``Quantum games'' it's called.
IMO, it's physicists looking for something to do.

There's also been work where people somehow try to modify the Nash equilibrium concept to be ``quantum mechanical''. The point of which eludes me.

Both of these are simply importing physics into game theory. Yawn.

Now there's another set of ideas, which nobody has managed to run to ground, which would be truly profound, and (I think) related to what you're talking about. This is importing game theory into physics, rather than vice-versa. Basically, can one cast various physics scenarios in terms of particles that are ``players in a game'' in some sense.
\edhw

I like the sound of such an idea.  A bit like exporting ``agency'' to the ``inanimate'' world.

\section{29-03-10 \ \ {\it Symmetry of Quantum State Space}\ \ \ (to A. Fenyes)} \label{Fenyes2}

A SIC measurement allows a one-to-one correspondence between $d \times d$ density operators (positive semi-definite matrices with trace 1) and probability distributions over $d^2$ points.  The result is an affine transformation from one set to the other.  Affine transformations preserve various convexity properties (like mixtures, extreme points, and maybe some other things), but they don't necessarily preserve geometric symmetry.  For instance, we can get from an ellipsoid to a sphere by an affine transformation.  And here is a nice symmetry property of the sphere that a proper ellipsoid doesn't share:  The sphere (the ball really) can be thought of as a union of {\it regular\/} simplexes each of full dimension.

If SICs exist, then the space of density operators has that property as well---quantum state space can be thought of as a union of regular simplexes.\footnote{See Footnote \ref{OhCrap} for a technical, but remediable, mistake in the original statement of this. \label{OhCrap6}}  In either representation, I guess:  I.e., represented as complex matrices or as probability distributions.

I've always thought of the SIC-induced affine mapping as the best way to ``straighten out'' out the space of quantum states.  For instance, if for a qubit we do not use a SIC, but some other informationally complete POVM to induce a mapping from quantum states to probability distributions, we will not get a sphere, but an ellipsoid.  That's a kind of trivial change, of course, but I don't know the full implications of the change in higher dimensions.  Maybe there are better reasons to resist ``ellipsifying'' the sphere there.  (Given my remarks above, if SICs exist, quantum-state space is already ``straightened out'' in a way.  I.e., the density operator formalism already captures this nice symmetry---one doesn't have to map on to the probability simplex to see it.  If on the other hand, SICs don't exist---then I think one must always ``ellipsify'' the Bloch ball when transforming from density operators to probability distributions via an informationally complete POVM with rank-1 elements.)

Anyway, I didn't know how sensitive $\mbox{sym}(x,S)$ was to affine transformations when I wrote my first note to you---I hadn't really looked at the paper to any depth.  But now, looking over the paper, I guess Proposition 3 caps off this question. [See A. Belloni and R. M. Freund, ``On the Symmetry Function of a Convex Set,'' Math.\ Programming {\bf 111}, 57--93 (2008).]

Still though, one can ask whether better knowing the structure of the function $\mbox{sym}(x,S)$ for the set of quantum states might give any clues on SIC existence or nonexistence, or even just reveal some nice features of the state-space that we hadn't known before.

I like your question of comparing the scaling of $S$ to that of a simplex as dimension grows.  Also, what is the symmetry value of a sphere of the same dimension?  (I'm bad:  When I say sphere, I most always mean ball.)  I guess it's 1.

\section{29-03-10 \ \ {\it Premature QBism!}\ \ \ (to N. D. {\Mermin})} \label{Mermin167}

To reply:
\bdm
1.  Of the three examples at the end of page 1, the second
(spontaneous collapse) is more than an interpretive strategy, since it
alters the quantitative predictions of the theory.  Of course it's
motivated by an implicit interpretive strategy.
\edm
Agreed, it is ``more than,'' but I had meant to cover that with this:  ``The trouble with all these interpretations as quick fixes for Bell's hard-edged remark is that they look to be just that, {\it really quick fixes}.  They look to be interpretive strategies hardly compelled by the particular details of the quantum formalism, giving only more or less arbitrary appendages to it.''  Looking back at the draft, I see I had cut out the appendage part in the small version of the paper.  I hope you'll give me a pass on this one:  I'd rather not complexify the issue too much in a ``quick introduction.''

\bdm
2.  Remark on pilot-wave theories on upper left of page 2.
Do you really mean ``pretending'' (and not, for example,
``presenting'')?   It's a peculiar use of ``pretend'' and I
don't understand what it means here.  The term veneer is already
disparaging enough.  And what is it about counting angels on the head
of a pin that is supposed to resemble pilot-wave theories.   (When Pauli used the image in his
letter to Born he was illustrating worrying about things you can't
know anything about.)
\edm
You're right.  I was worried there was something wrong with that sentence, but could never quite put my finger on it.  Is this better (basically what you suggest):
\bq\noindent
     If there were no equations to give the illusion of
     science, this would have been called counting angels
     on the head of a pin.
\eq
And yes, it is precisely the Pauli point.  There is a fun exchange between Bill Unruh and Roderich Tumulka during one of the PIAF meeting panel discussions a couple years back.  Tumulka said something like, ``Bohmian mechanics allows one to calculate answers to some questions that all the other interpretations can't even formulate.  For instance, the amount of time a particle is on the inside of a potential barrier.''  Unruh said, ``That's not true!  My interpretation allows it.  Every particle stays in every potential barrier precisely seven seconds.''  Tumulka said, ``But this is nonsense.  We're not just making some number up; we have a scientific theory.''  Unruh said, ``Your theory is every bit as scientific as mine.''

\bdm
3.  Remark about Bell on upper right of page 2.   You
attribute to Bell a fear that is not in the quotation you cite on page
1 but in your own broader remarks in the paragraph that follows it.
I'm see no evidence that Bell feared that the conceptual weakness of
the theory might lead to a breakdown in its empirical validity.
\edm
Not a fear of a breakdown in QM's empirical validity, but that physics is way, way on the wrong track because of its seeming need to invoke the observer.  I've always interpreted Bell as thinking along the lines of something Einstein said:
\bq\noindent
     It may appear as if all such considerations were just
     superfluous learned hairsplitting, which have nothing
     to do with physics proper.  However, it depends
     precisely upon such considerations in which direction
     one believes one must look for the future conceptual
     basis of physics.
\eq

\bdm
And I'm not sure what you mean when you say that the only thing to
fear is that particular fear itself, regardless of whom you attribute
it to.  Are you saying that we should take QM as correct and try to
understand it better?
\edm
Yes.  Or more accurately that the observer is essential in formulating quantum mechanics, and we should try to understand that better.  FDR:  ``The only thing we have to fear is fear itself.''

\bdm
Finally, I don't see anything ``shrill'' in Bell's rhetoric, and I don't
see why your simple resolution of his worry about the collapse of the
wave-function of the universe warrants calling his concern ``the least
of his worries.''
\edm
In that spot, I had originally written:
\bq\noindent
     The real substance of Bell's fear is just that---fear
     itself. To succumb to it is to block the way to really
     understanding the theory.  Moreover, the harsher shrills
     of Bell's rhetoric are the least of the worries:
\eq
Upon which, Hans Christian von Baeyer who read the draft wrote:
\bq\noindent
     I don't think shrill is a noun.  For ``hasher shrills'' I
     would put ``shriller notes.''
\eq
I just basically followed through with his suggestion.  But you're right, ``shrill'' is too strong.  How about simply,
\bq\noindent
     the harsher notes of Bell's rhetoric are the least of
     the worries:
\eq

\bdm
This leads the reader to expect an even bigger problem --- not the
solution to the problem.
\edm
Because that is basically where Asher would have left it \ldots\ never trying to make his view of quantum mechanics completely consistent.

\bdm
4.  Still upper right, page 2.  I'm not sure how you get from
antibodies in some of the elderly to the antibody itself being
elderly.  And I'm not sure you want to call it ineffective, when it's
the direction you want to move in.
\edm
Right again:  Strange for me to call the antibody itself elderly.  How about simply,
\bq\noindent
     But this much of the solution is only a somewhat
     ineffective antibody.  Its presence is mostly a
     call for more research.  Luckily the days for this
     are ripe, and it has much to do with the development
     of the field of quantum information
\eq

\bdm
Nothing else for now.
\edm
The bigger paper is a better paper, but of course inherits many of the bad word choices you mention above.

The only trouble with your being right with your remarks, is that now I'll have to repost the damned thing!  And frankly I'm frightened of that.

I hope you'll still read the big paper, even though you've read the little one---it contains a load of mini-epiphanies for me.  For instance, Footnote 44, where I turn ``the quantum-to-classical transition'' around.  I'd like to hear whether you think the paper makes any progress toward something else you said:
\bdm
I suppose it's ineffective in that it hasn't convinced everybody, and
is not yet sharply enough developed.
\edm

\section{30-03-10 \ \ {\it Quantum Bayesianism} \ \ (to J. Wright)} \label{Wright1}

This is a very late reply; I'm sorry about that.  Let me just respond to two points.

\bjw
My concern is that on one hand the Bayesian view takes a very
forward stance, saying so much as ``quantum states are not real'' and
then you said in your lecture that you'd be willing to accept other
interpretations on pragmatic grounds --- in the sense of
electromagnetic fields.
\ejw
and
\bjw
It was an example used in jest but then moments later you talked
about the problem of quantum channels and how you wouldn't stand
against an interpretation if thinking in that manner provided a
solution to such a problem.
\ejw

The root of these blurbs is not an inconsistency in my thinking or a religious reliance on pragmatism, but a politician's way of negotiating---a way of demonstrating the possibility of compromise.  I give a criterion for what {\it would\/} move me toward their views.  And I'm honest about that.  But it is a bit hollow nonetheless:  For I make this ``bargain'' in very strong confidence that they won't be able to deliver the goods.  50 years of history already shows a world of difference between electromagnetic potentials and the Bohmian trajectories.  Potentials, though unobservable except through the forces ``they give rise to,'' are used every day in the aid of hundreds of problems in electromagnetism classes.  Bohmian trajectories, even if assumed unobservable except through the measurement clicks ``they give rise to,'' are {\it nearly never\/} used.  The only exceptions I know of are in the demo problems invented to make them look like a healthy option (like the question of how long in a potential barrier).  No actual practical problem is ever helped by their aid.

I hope that makes my strategy a little more understandable, even if a bit slimy.

\section{30-03-10 \ \ {\it That Old Shoe} \ \ (to S. Lloyd)} \label{Lloyd1}

I nearly forgot:  I believe I thanked you for the first time in a paper.  Have a look at the end of \arxiv{1003.5209}; the wording might amuse you.  The paper is my latest attempt to verbalize a worldview that's been growing in me. Apparently that '92 lecture of yours left a lasting impression on me.  (Though in the text of Section VI, I transformed your shoe into a diamond to fit the rest of the story.  It deserved to be a diamond anyway!)

\section{30-03-10 \ \ {\it I Knew He'd Just Call Me Solipsist Again, and he did}\ \ \ (to N. D. {\Mermin})} \label{Mermin168}

I made a probability one assignment, and the world complied to what I believed with all my heart.  Does that synchronicity call out for further explanation?  I made the probability assignment because all my experience thus far had pointed me in just that direction.  I compacted all that I felt and saw around me, all my previous dealings with people and {\it particular\/} people, my choice of words while writing, and everything else in my memory into a singular supremely strong probability assignment.  And the world complied:  The action I took in posting my paper to {\tt arXiv.org} led another part of the world to a reaction that I was quite sure I could see in advance.

I suppose it'd be no solace to Travis to know that I really do believe in his autonomy---that he is more than the sum total of my provocations.  See below.

With a smile of satisfaction,

\subsection{Original note from Travis Norsen, ``Bell, locality, etc,'' dated 30 March 2010}

\bq
I've just read the short QBism paper you put on arxiv the other day,
and skimmed the longer one, and thought I'd write and just say hi and
register a couple of very quick comments.  First, I appreciated the
compliment in the footnote of the longer paper, even if it was, well,
somewhat qualified!  =)  More seriously, I think there is a kind of
deeper disagreement between us (or between you and Bell) on this stuff
than you acknowledge.  Let me try to explain what I think the issue is.

I think that, from the very beginning, you are just on the premise of
understanding probability statements as referring to subjective degrees
of belief (or something in that neighborhood).  And, on that basis,
your reaction to (for example) the passage from one of my papers that
you quote in that footnote, is completely understandable.  If you
understand probability statements to be, in principle, about degrees
of belief, and then you see $P(a) = 1$, then you are absolutely right:
that just means that somebody believes ``a'' with all their heart.  It
doesn't necessarily mean or warrant some further claim about there
being some real element of reality, ``a'', out there independent of
human intervention, awareness, etc.

So on that part we agree.  But where we disagree is about whether
that's what the probabilities in this particular context actually
mean.  As a perfectly uncontroversial factual/historical matter, that
(i.e., subjective degrees of belief) is {\it not\/} what Bell meant by the
probabilities he writes down in his various papers.  He is just coming
from a very different place.  He's taking for granted what I've
elsewhere called ``metaphysical realism'' (i.e., that there is some kind
of physical world out there independent of us, which it is the goal of
candidate physical theories to describe).  Then the specific concern
which brings probabilities into the discussion is the desire to avoid
{\it assuming\/} that determinism is true.  That is, he very carefully and
deliberately sets things up so that we are open, in principle and from
the very beginning (and here I'm mostly talking about his formulation
of ``local causality'' from, e.g., ``la nouvelle cuisine''), to candidate
physical theories (i.e., candidate descriptions of external physical
reality) which involve irreducible stochasticity.  So he gives, for
example, a mathematical definition of ``local causality'' that is a
statement about certain probabilities -- and then literally from there
(plus the additional assumption of ``no conspiracies'', aka ``freedom'')
you can run the EPR argument for the existence of certain things,
which then run afoul of experiment via Bell's inequality.

Now the key point here is just that the probabilities that are in the
mix here -- that is, the ones in Bell's definition of ``local
causality'', because that's the premise from which everything else
follows -- are {\it not\/} subjective degrees of belief.  They are rather
the ``irreducible dynamical probabilities'' of some candidate theory.
In particular, they are the probabilities assigned by the theory to
certain physical happenings, not on the basis of anyone's subjective
and perhaps-incomplete knowledge of various goings-on, but rather on
the basis of a complete description of physical goings-on in some
relevant spacetime region.  (And note here that both the ``complete''
and ``physical goings-on'' parts of the previous sentence mean:
according to, i.e., relative to, some particular candidate physical
theory.  So there simply is no place here where anything relating to
knowledge or subjective belief or {\it anything like that\/} enters {\it at
all}.)  And so, given {\it this\/} understanding of the probability
statements that are in the mix here, you can surely see why it would
be quite inappropriate to say ``Just because $P(a) = 1$ doesn't mean
there's some fact of reality out there corresponding to `a'.''  Yes, it
does mean that, because (with this alternative understanding of
probabilities in place) ``$P(a) = 1$'' literally just means that,
according to some particular candidate physical theory, the physical
fact corresponding to ``a'' definitely happens.  And that means:  really
happens, out there in reality (according to this theory).  And that is
all that is needed for the whole argument to go through.

Now, don't get me wrong.  I'm not saying ``Well, Bell meant something
different by probabilities, and so you just have to accept his view.''
I appreciate that you could -- and I'm sure do -- disagree with Bell.
That is, I suspect you think that Bell's whole way of understanding
probabilities is just wrong, and that it just doesn't make any sense
to interpret probability statements as propensities (if that is the
right terminology for Bell's view) rather than subjective degrees of
belief.  Fine.  But then you should I think more openly acknowledge
that you have this more basic disagreement -- instead, I mean, of
writing as if you and Bell and everybody obviously just agree that
probabilities have to mean subjective degrees of belief, and then
pointing to the end of some argument (like the one you quote from me)
and saying ``obviously this conclusion doesn't follow''.  Because doing
that makes your argument into a kind of petty straw man sort of thing,
right?

Let me also say that -- {\it if\/} you really accept what I called
``metaphysical realism'' above, as I believe you {\it claim\/} to do -- I
don't see how you can possibly argue that it is meaningless/nonsense
to talk about probabilities in Bell's ``propensity'' sense.  If you
think there's a physical world out there, and you think that in
principle its fundamental laws might be stochastic rather than
deterministic, then you just have to accept the possibility of
candidate physical theories like the ones Bell contemplates.   And so
I have the very strong suspicion -- or perhaps I should say this
strengthens my very strong suspicion -- that (your own protestations
to the contrary notwithstanding) you actually reject ``metaphysical
realism''.  Because, frankly, it is only if you reject that, that I
think it then makes any kind of logical sense to insist that {\it all\/}
probability statements {\it must\/} just refer to subjective degrees of
belief.  And that, of course, would just make you a \underline{\bf solipsist}.\footnote{My emphasis!}  Which
I know you vehemently deny.  But in your heart of hearts, come on --
you must understand why you keep hearing this accusation.

Let me also just say that I haven't forgotten that you owe ``me''
(really, the world) some kind of more explicit and careful QBist
analysis of (in particular) Bell's formulation of ``local causality''
from ``la nouvelle cuisine''.  This is what you thought you were going
to talk about at PIAF, right, but then you ended up going in a
different direction?  Anyway, I'm being tongue in cheek in saying ``you
owe me'' -- the point really is just that if you want to have a serious
and possibly-productive discussion about Bell, locality, QBism, etc.,
I think this is going to have to be a central part of the discussion.

OK, enough for now.
\eq

\section{30-03-10 \ \ {\it I Knew He'd Just Call Me Solipsist Again, and he did, 2}\ \ \ (to M. Schlosshauer and R. {\Schack})} \label{Schack197} \label{Schlosshauer28}

You two might enjoy this one as well, given your own previous run-ins with Travis Norsen.  [See 30-03-10 note ``\myref{Mermin168}{I Knew He'd Just Call Me Solipsist Again, and he did}'' to N. D. {\Mermin}.] I've got a good racket going:  I post a paper; I get called a solipsist.  If I could only figure out how to bottle this stuff and sell it for profit!

\section{30-03-10 \ \ {\it Soliciting Your Thoughts on Intro to Our Steering Paper} \ \ (to H. Barnum)} \label{Barnum31}

\bhb
I meant to ask you this earlier\ldots\ but when writing the intro to our steering paper, I said
``The Bohrian notion of complementarity, for example, can be understood in terms of one type
of knowledge about a system precluding another.''  I felt a bit uncomfortable about this\ldots\ I
thought one or both of you might object based on the idea that what one acquires in a measurement is not really ``knowledge about a system'' but ``knowledge about the outcome of my just-completed interaction with a system''.  There is a similar issue with the next sentence, about information-disturbance relations.

I wonder if that's right\ldots\ and if either of you can think of a better way of pithily summarizing complementarity that still relates it to information/information processing in the way we want to in our intro.
\ehb

Let me think on this for a bit.  I'm not sure at all about how I would approach trying to say something about ``complementarity''.  As for ``steering'' being such a prominent component, its real essence (from a QBie perspective) must be about the possibilities for conditioning.  I keep thinking there must be a deep connection between it and the axiom that {\Ruediger} and I toy with:  That in our urgleichung set-up, any posterior for the sky could have been taken as a prior for it.  It's this that sets up a duality between states and measurements in our scheme.  But I can't quite put my finger on how this and ``steering'' (god I hate that name) should amount to the same thing.

\section{30-03-10 \ \ {\it Qbism Questions} \ \ (to C. Ududec)} \label{Ududec2}

Wow!  I just read this all again.  I feel I can't even begin to answer you.  Would you hold it against me for a still further while?  Question 1, for instance, is a research project in itself.  Maybe we should just ``talk'' more.  How about we get together sometime next week for an oral version of this?

\subsection{Cozmin's Preply}

\bq
\noindent [Concerning a draft I handed out at the Laflamme--Emerson quantum foundations course and which became ``QBism, the Perimeter of Quantum Bayesianism,'' \arxiv{1003.5209}.] \bigskip

Sorry for being so late with this, but I didn't get a chance to look at your essay until this Friday.  I have a few long winded (and probably still confused) questions and comments.  I'd like to think about the following more, so maybe this should be a kind of first draft of what I'd like to ask and comment on.

1) In section three, you argue that since probability does not exist, quantum states do not exist.  Presumably this also means that CP maps do not exist.  I'm inferring this since if teleportation (which I think you used in an older paper) can be used to argue that quantum states do not exist, then gate teleportation can be used to argue that CP maps do not exist.  So what about tensor product decompositions of a Hilbert space, do they exist?  There is a version of the Choi--Jamio{\l}kowsky isomorphism (introduced by Matt Leifer in 2006) which shows that specifying a bipartite quantum state $\rho_{AB}$, and two POVMs, $M_A$, $N_B$ (which commute), is equivalent to specifying an initial state on $A$ (given by an $M_A$ preparation), and a CPTP map from $A$ to $B$.  So if one agent says that he is working with a bi-partite system and an entangled state, another agent can say he is working with one system which is mapped to another by some CP dynamics.  This is analogous to how specifying a joint probability for two random variables $P(X,Y)$ is equivalent to specifying a pair of $\big(P(X), P(Y|X)\big)$ where $P(X)$ is now understood as a distribution for $X$ at some initial time, and $P(Y|X)$ is a channel, taking $P(X)$ to $P(Y)$ at a later time.

Also, would it be worth your discussing at least that CPTP maps do not exist in class?  Is this implicit in your Wigner's friend argument?

2) Could you comment a little on someone like Leslie Balletine's views on the issues from section III, and his best argument for his views and how you understand them?  This may be a strange request, but I think it's interesting to see how people with opposing views evaluate and argue against each others' best arguments.  I think of it as a kind of second order evaluation of two positions, with the first order being just looking at both and evaluating them personally.

3) On page 7 you discuss the issue of ``Whose information?  Information about what?'', as well as Pauli's statement that ``the results of [an observation] are objectively available for anyone's inspection.''  Would you still say that the fact of an observation/intervention are objectively available for anyone's inspection?  I'm not sure if this is right, but I had the thought that maybe the strong reactions you get to your answer to the information about what question is partly based on confusion over the difference between a statement about the consequences (for me) of my actions on a system, and the consequences of my actions on the system, on someone else's beliefs, or alternatively, the difference between the result of an observation/intervention, and the fact of that intervention.  Talking about `the fact of' an observation already seems to be presuming the answer here, but I'm not sure how to phrase this better right now.

4) In the convex sets, or general probabilistic theories framework which Luca, How\-ard, Lucien, myself, and others are constantly going on about, we often talk about equivalence classes of preparation devices, and often say (maybe imprecisely) that a  `state'  is something like a concise representation of --- everything that is statistically relevant about --- this equivalence class, i.e., it's a collection of probabilities for the outcomes of future measurements, conditional on the system being prepared by a device from this equivalence class of procedures.  From this point of view, statements about unknown quantum states become statements about unknown quantum preparation devices or more generally unknown quantum interactions between the system of interest and another piece of the world.  So an agent using some quantum state, is equivalent to (?)\ it making a statement about the past history of the system of interest.  Could you comment on this, and how it fits with your view of quantum states, the quantum de Finetti theorem, and the discussion of the `right' quantum state?

5) Regarding your comments about the elementary notion of what it means to be two objects rather than one, you say that if one gives up the autonomy of one system from the other, it surely amounts to saying that there were never two systems there after all, and ask why the quantum formalism would engender us to formulate our description in terms of a tensor product of two four dimensional Hilbert spaces, rather than simply a raw 16 dimensional Hilbert space.  The conceptual distinction that the tensor product brings in is purely at the operational level, while the question of the autonomy of systems seems to have a much more ontological flavor.  The tensor product is a structure we use within quantum theory (which you've argued is a manual that any agent can use to make wiser decisions) given certain other assumptions about some physical situation and what statistics we expect to see (i.e. No-Signaling statistics).  My question (1) above also seems very relevant here, but I'm not sure how to phrase things properly in relation to that \ldots.

For example, I can imagine a world where at some basic ontological level, there are objects in the world, which after interacting with each other in a precise way no longer claim an existence independent from each other.  However at the operational level, the level with agents and preparation and measurement boxes, it appears as if --- the statistics of various measurements show that --- they do have an existence independent of each other.  So the agents in this world, would use something like a tensor product decomposition for their description of the statistics of their experiments.

6) A warning about the following: this is partly my attempt to understand the various issues surrounding things like scientific explanation, causation, Bell's theorem, and last but not least, Bell's theorem Qbism style.

I don't think it would be controversial to say that probability theory conceived of as a calculus of consistency in the face of uncertainty (or language for dealing with uncertainty coherently?)\ does not have the structure or syntax to express the concept of causation.  For example, the syntax of probability theory does not allow us to express the statement that  ``symptoms do not cause disease''.  There is nothing in the joint distribution of symptoms and diseases to tell us that curing the former would or would not cure the latter.

Why am I bringing this up? Well, I would argue that research in most areas of science such as physics, and biology, as well as in engineering, law, economics, and immunology, is not statistical but is causal in nature.  Many questions in these fields are not just about discovering correlations between different variables, but are questions about the process by which the data was generated.  They are questions of an explanation of the data, or of what caused or generated the data.  We can look at the world, and the things in it as black boxes whose internal workings we can't directly observe. What we want to do is open the black boxes and expose their inner mechanisms.

Why should we care about this notion of explanation through mechanisms?  Many laws or statements in physics do not express causal relations, but just regularities.  Some examples are A) the daily regularity of the tides, and their correlation with the positions of the sun and moon in the sky, B) the ideal gas law, which relates pressure, volume, and temperature for a given sample of gas, and C) Kepler's laws of planetary motion which describe relationships between various parameters of the orbits of the planets.

There are also many laws or theories which do express causal relations.  For example, the increase in pressure of a gas when we change the volume with a piston is explained causally by postulating particles which collide with the walls of the container more frequently when the volume is decreased.  So we have a noncausal regularity which we explain on the basis of some postulated underlying causal mechanism.  Similarly, Newton's laws of motion and gravitation explain such regularities as Kepler's laws, as well as the tides.  The theories (not just in physics) which postulate some causal processes and explain some observed regularities tend to be more precise or powerful (and let us sleep better at night).  They also tend to come with more baggage, like postulated particles or fields, and their dynamics.

Now, it seems to me that the way Bell's theorem is usually presented is as a statement about causality, or about the structure (the functional relationships between various postulated variables) of physical theories which explain certain correlations in certain well specified experiments.  I say explain, rather than account for, since in a sense quantum theory already accounts for the correlations, i.e., given a belief about how a bipartite system was prepared, and some beliefs about the structure of two further interactions with the systems, the correlations between the outcomes of the interactions are predicted by the theory.  Quantum Theory seems more like the examples A, B, C than say General Relativity or Statistical Mechanics.  So the way I see the ontological model's program, and more specifically, Bell's theorem, is as an attempt to find a mechanism for the observed correlations, i.e., to find a General Relativity for the Kepler's laws of Quantum Theory.

Given the above, I'm worried that analyzing Bell's theorem purely from a ``probability theory conceived of as a calculus of consistency in the face of uncertainty'' doesn't do justice to it.  So I'm curious what you think of the above considerations of explanation and causation.

You also argue that a probability-one assignment lays no necessary claim on what the world is.  Most scientific theories have many probability-one assignments, and they do seem to make claims about the way the world is.  I'm wondering how you would rephrase those theories (like classical Electromagnetism, modern genetics, and chemistry experiments where catalysis is important) and the claims they make, to be more consistent with the ``probability theory conceived of as a calculus of consistency in the face of uncertainty'' point of view.  Also, how would you look at a statement like $Pr(X=i | B, T)$, i.e., ``Given background knowledge $B$, and postulated theory $T$ (with all the extra variables, and functional relationships between variables which it postulates), the value of observable $X$ will be $i$, with probability one''?  It seems to me like the quote in your footnote 27 has the structure of a statement like the above, rather than a statement like ``this probability one assignment for variable $X$ (given background knowledge $B$), necessarily implies the world is such and such \ldots''.

Do you think that Bell's theorem (or maybe some slightly weakened version of it) could be formulated without something explicitly or implicitly equivalent to the EPR criterion of reality?  In other words, is taking something like that criterion a necessary condition for Bell like no-go theorems?

Finally, do you mind sending me an electronic copy of the Qbism essay?
\eq

\section{31-03-10 \ \ {\it An Important Message from the QBomancer} \ \ (to the QBies)} \label{QBies9}

(OK, I need a better name for myself or my role in this group.  ``Cubomancy'' is defined as ``divination by throwing dice.'' \ldots\ Given that that's what I've been doing all my life, it seemed a little appropriate.  QBomancy---divination by performing quantum measurements.)

Anyway on to more serious matters: [\ldots]

\section{31-03-10 \ \ {\it Bell, Locality, Etc.}\ \ \ (to T. Norsen)} \label{Norsen1}

Thank you for your note.  I have studied it carefully, and plan to study it still more carefully on second and third passes.  But your remark, ``I've just read the short QBism paper \ldots\ and skimmed the longer one,'' coupled with your second-to-last paragraph where the S-word is mentioned, made me smile and reminded me of an encounter I once had with Leslie Ballentine at a meeting in Sweden.  Leslie had sent me his previous conference proceedings a bit before my travel to the new meeting, and we met in the coffee line the first morning.  I had skimmed his paper, and to outward appearances, it looked to have no innovations over his previous discussions; so I left it at that.  But as you can guess, at coffee the conversation very quickly came to the issue of quantum probability.  Leslie said, ``Did you read my new paper?''  I said, ``No, actually.''  He crisply turned 90 degrees and said as he walked away, ``Then we have nothing to discuss.''

Much of Section VI was written with your inevitable challenge at the back of my mind.  Your image arose fleetingly as I composed Section V, but it was the composition of Section VI that more than once brought me back to thinking of you.  (Maybe it was the big eye in Figure 5 that did it.)

When you have read the whole paper, like I have read your whole note, we can discuss.

\section{31-03-10 \ \ {\it Erwin Schrolipsism}\ \ \ (to N. D. {\Mermin})} \label{Mermin169}

\bdm
As one of the earliest people to accuse you of solipsism (do a search for the word in our correspondence, if you're well enough organized to do that --- it was a very long time ago) and as one who understands how wrong-headed that was of me, I can only sympathize.
\edm

Well you called Erwin {\Schroedinger} a solipsist long before you called me one!  That note dates from 25 January 2000:
\bq\noindent
P.P.S.  Have you seen a little book by {\Schroedinger} called ``My View of
the World''?  It contains two extended essays on the nature of
objective reality, one in 1925 (before his Equation!) and one in 1960.
They're surprisingly similar.  They turn upside down the notion that
objective reality is a valid inference from the fact of
intersubjective agreement.  He argues instead (I think) that
intersubjective agreement is a manifestation of the unity of all
consciousness, which is also what creates [the illusion of] objective
reality.  A kind of global mass-solipsism.  But very beautifully put.
Makes me think I should read Spinoza, which I never have done.  So the
IIQM is being tugged in one direction by QIP and quite another by
goddamed mysticism.  It's really delicious that {\Schroedinger} is the
hero of those who believe Bohr abandoned hard-headed rationality.
\eq

\section{31-03-10 \ \ {\it I Knew He'd Just Call Me Solipsist Again, and he did, 2}\ \ \ (to N. D. {\Mermin})} \label{Mermin170}

\bdm
On the other hand I'm sympathetic to Travis's complaint that you owe the world the critique you promised but never delivered on Bell's analysis of what counts as an explanation of correlations in distant places, that I commended to you in his essay on the new cooking (and in Bertlmann's Socks).  I remember pressing {\Ruediger} on it when I visited him in London several years ago, and the best he could come up with was to shrug his shoulders and say something like,  ``Why should correlations have to have explanations?''  I still regard the absence of a better rejoinder than that as the weakest point in our battle against the purveyors of nonlocality.
\edm

The key issue is this:  WHAT QUANTUM MECHANICS, AS A SINGLE-USER THEORY, IS TALKING ABOUT ARE NOT IN {\it DISTANT\/} PLACES.  They ARE wherever the agent is whenever he performs the action (you'll see what I mean when you read Sections III and V).  Of course, I am using William James's notion of `the self':
\bq\noindent
     ``In its widest possible sense, however, a man's
     self is the sum total of all that he can call his,
     not only his body and his psychic powers, but his
     clothes and his house, his wife and children, his
     ancestors and friends, his reputation and works,
     his lands and horses, and yacht and bank account.''
\eq
But I'd never get you to write nice things about me in recommendation letters if I told the world the real source of all my ideas!

\section{31-03-10 \ \ {\it Bell, Locality, Etc., 2}\ \ \ (to T. Norsen)} \label{Norsen2}

\btn
Your point is fair and well taken.  For what it's worth, I actually have skimmed the whole paper -- that is to say, I was already aware that there was significant explicit discussion of the S-word in these later sections.  But you're absolutely right that there's no point
discussing it until I've read those sections fully and carefully.
I'll be trying to get around to that in the next few days, and so I'll just let you know what I think afterwards.

That said, though, and to try to explain further why I wrote what I did yesterday even before having read the whole paper carefully, let me say this:
the subjective probability I assign to my coming to believe that your sections 5 and 6 address all of my concerns, is very low.
\etn
Don't think that I myself didn't already assign a subjective probability of nearly one to this proposition long before reading it this morning!

\subsection{Travis's Reply}

\bq
I figured you did.  More precisely, I assigned a subjective probability near one to the proposition that you did.

By the way, despite really loathing your solipsism and even more the fact that you're in denial about it, I love and respect your openness and sense of humor.

OK, maybe some serious discussion in a couple days \ldots
\eq

\section{31-03-10 \ \ {\it Truth and Happening}\ \ \ (to N. M. Boyd)} \label{Boyd1}

I'm looking forward to your seminar on ``truth'' that was just announced.  Just to throw in a vote:  I hope you will have the time to say a few words (pro or con) on the notion of truth in American pragmatism (particularly William James's and John Dewey's) and that this important notion from a century past will not get lost in the shuffle.  Here's one of my favorite sentences of all time by William James:
\bq\noindent
The truth of an idea is not a stagnant property inherent in it. Truth
{\it happens\/} to an idea.  It {\it becomes\/} true, is {\it made\/}
true by events.  Its verity {\it is\/} in fact an event, a process:
the process namely of its verifying itself, its veri-{\it fication}.
Its validity is the process of its valid-{\it ation}.
\eq

Attached is a page of something I wrote to Lee Smolin a few days ago in response to reading one of his drafts.  [See 24-03-10 note ``\myref{SmolinL20}{Notes on the Open Future}'' to L. Smolin.] It says a bit of why this notion of truth is important to me.  The relevant discussion is between Smolinisms 9 and 10.  It says that this notion of truth is important for making possible a certain kind of ontology.  The reason that in turn is important to me is that it happens to be the kind ontology I myself am aiming for.  If you have any interest in my motivations, you might skim the last three sections of my recent essay:  [See ``QBism, the Perimeter of Quantum Bayesianism,'' \arxiv{1003.5209v1}.]

As I say, looking forward to hearing some discussion on the important issue of ``truth'' at PI.

\section{31-03-10 \ \ {\it Working Glossary}\ \ \ (to the QBies, R. {\Schack} \& N. D. Mermin)} \label{QBies10} \label{Schack197.1} \label{Mermin170.1}

I thought it'd be useful to update our group glossary.  Please bring any omissions of words we have used before to my attention.  Creativity is welcome as well; though I reserve final say on quality control.  Particularly, I'd like a good word for myself or my role in this group.  Matthew suggested QBoss or QBishop, but that feels a bit too aggrandizing for me to use as a description of myself---I need something a little more Chris-safe.  (Of course, you can call me anything you want in private!  Even QBastar\ldots.)

Also it'd be a nice to find a grammatically correct nickname for Joan Vaccaro's painting {\sl Broken Block\/} that I think illustrates so well the inevitable cracks in the block universe conception of things.\medskip

\noindent ------------------------------ \medskip

\noindent {\bf QBism} -- the quantum foundational program of quantum Bayesianism; see \arxiv{1003.5209}.\medskip

\noindent {\bf QBist} -- a practitioner of QBism.\medskip

\noindent {\bf The QBies} -- C. A. Fuchs, along with his group of three grad students (Hoan Dang, Matthew Graydon, Gelo Tabia), PI associate postdoc Asa Ericsson, and PI visiting researcher Marcus Appleby.\medskip

\noindent {\bf QBism House} -- 49 George Street, Waterloo, Ontario; see attached ``press release''.\medskip

\noindent {\bf The QBibbo Angle} -- 60 degrees; see Assumption 7 and Equation 24 of \arxiv{0912.4252}.\medskip

\noindent {\bf The QBicle} -- the office of the student QBies, PI room 415.\medskip

\noindent {\bf QBic Equation} -- the most important equation underlying the shape of quantum state space; see Equation 19 of \arxiv{1001.0004} or Gelo's beautiful rendition of it in $d=3$:
$$
\sum_i p(i)^3 \; - \; 3 \sum_{(ijk)\in Q} p(i) p(j) p(k) \;=\; 0
$$
where $Q$ consists of all lines in the affine plane,
$$
\begin{array}{ccc}
  1 & 2 & 3 \\
  4 & 5 & 6 \\
  7 & 8 & 9
\end{array}\;
\qquad\qquad\mbox{I.e.,}\qquad\qquad
Q \;\; =\;\;
\begin{array}{ccc}
(123) & (456) & (789) \\
(147) & (258) & (369) \\
(159) & (267) & (348) \\
(168) & (249) & (357)
\end{array}
$$

\noindent {\bf QBissel} -- one who is intrigued by QBism but not yet convinced (compare German word ``bissel'').\medskip

\noindent {\bf QBicity} -- the quality or state of being properly oriented with respect to quantum interpretational issues (compare ``cubicity'' and ``state of nirvana'').\medskip

\noindent {\bf QBomancy} -- divination through quantum measurements (compare ``cubomancy'').\medskip

\noindent {\bf QBomancer} -- one who divines with the help of quantum measurement (compare ``Raussendorf--Briegel model of quantum computation'')\medskip

\noindent {\bf QBalla} -- ??

\section{01-04-10 \ \ {\it Too Short an Answer} \ \ (to T. Norsen)} \label{Norsen3}

This is no answer to your nice note --- I won't be able to do that for a few days.  (I am not so quick as you.)  But I want to try to make one point very tersely, in case it might open up some understanding.  It is related to your Ballentine bullet.

With regard to the issue of quantum states (and probabilities) being ``epistemic'' (broadly construed), there are two points that must be addressed:

\begin{enumerate}
\item
Are they?  I.e., are they ``epistemic'' (broadly construed as ``in your head rather than out there'')?  My stance, and Ballentine's (possibly, since he also will say things like ``a hydrogen atom ground state is no one's degree of belief'') and Einstein's and Peres's and Peierls's and {\Spekkens}'s (all to varying degrees of consistency), is that they are.
\item
But epistemic about what?  There is more than one possibility here, and that seems to be one of the things getting in the way of your seeing where I am coming from.  The epistemic states could represent uncertainties about hidden/ontic variables completely external to the agent---that is the common story of Einstein's take on it, and it seems to me what Ballentine must be talking about implicitly.  (Though as I say you never know about Ballentine.)  It is definitely Rob {\Spekkens}'s desired interpretation of the epistemicity of quantum states.  But it is not mine.  My whole paper is about exploring a different view:  That they are an agent's epistemic states for his own potential experiences {\em under the premise} that those experiences will arise as a consequence of his actions on (``interactions with'' is too loaded of a term for me to want to use it here) the world external to him.
\end{enumerate}

The imprint of the character of the external world (and the world as a whole, including the world before man)---this is an idea that approaches your definition of ``metaphysical realism''---does find its way into the formalism (in my view), but when it does it does so as an additional normative rule for the agent's gambling attitudes on those potential experiences.  Where ``normative rule'' does not mean that one must adopt this or this or this {\it particular\/} probability assignment, but that one should try to make all of one's  gambling attitudes (for all kinds of things) mesh together according to a certain scheme.  In the case of quantum phenomena (when made explicit with SIC language), the Born Rule is seen as a normative rule for meshing together beliefs between ``a one-step factual measurement'' and ``a two-step counterfactual one.''

I'll leave it at that for now, and just ask that you mentally readdress your questions, issues, and reading of the paper with distinction number 2 at the forefront of your mind.  It is the predominant thing that sets QBism apart from the ``epistemic-state, ontic-variable program.''  And it is underneath everything I say about the locality issue:  For the quantum state, on this view, is only calibrated to what the agent believes he himself will experience (which in common language one might say as ``happens in his vicinity'').

I'll come back to your many other issues at the end of Easter.

I'll cc this note to Rob {\Spekkens} since I impute a definite statement onto him, and he may disagree.  (I'd cc it to Einstein, Peierls, and Peres too, but they're dead.)

\section{01-04-10 \ \ {\it Strange Anton Dream} \ \ (to A. Zeilinger \& H. C. von Baeyer)} \label{Zeilinger11} \label{Baeyer110}

Since H. C. von Baeyer is convincing me more of the importance of dreams, I'll record this one from the night before last.  You (Anton) were visiting Waterloo, and I had had some house work done a week or so before your visit.  There was a fresh concrete sidewalk laid and some concrete steps leading up to my house.  The concrete was dry and hard---it was already a week old.  When you arrived at my house, there was some melting snow on the ground; the new sidewalk was a bit wet.  We started discussing matters quantum mechanical on my porch, or maybe at my chalkboard at the wall.  I told you that there is some pattern you must understand---it was a very important thing.  The pattern was apparently on a piece of paper or a leaf, or ``both'' if you can figure out what that means.  (I have conflicting images of both.)  But the wind had blown it away.  I walked into my yard looking for it in the snow, but finally found it wet and crumbling on the new sidewalk.  I knew that it could not be recovered without destroying it if I tried to peel it from the sidewalk.  I became very sad at the loss, and I tried to peel it anyway.  It did indeed tear into a hundred pieces, or was a bit like oatmeal as I tried to pull it up.  But to my astonishment the pattern had been transferred into the concrete---the concrete now had etchings and lines and bumps.  And it was a bit surreal even; I seemed to be able to see it more clearly now than I could before.  I said, ``Anton, Anton, you must see this!''  You walked over and we studied it very carefully, more excited about the transference than about the original pattern that I had thought was so important.  Then I looked up and down the sidewalk and steps and found all sorts of other things recorded in it.  For instance, on one of the steps, there was a very deep footprint (I think yours) showing every detail of the underside of your shoe.  I thought, ``But this is finished concrete; how can this be?''  And what struck me again, was the surreality of it:  The edges were extremely crisp, almost as if the shoe had been pressed into a kind of plastic.  I thought, this is so much like the {\sl Magic Eye\/} books, where when the 3D image finally becomes visible it is always strangely crisper and clearer than it should be.

And that is all I remember.  (See \myurl[http://en.wikipedia.org/wiki/Magic_Eye]{http://en.wikipedia.org/wiki/Magic\underline{ }Eye} if you don't know what I mean by the {\sl Magic Eye\/} books.)  Of course, I have no real idea what it means (if anything), but I wonder if it is not connected to an excited note I wrote to my colleagues a couple weeks about about the SIC problem.  I'll append part of that below for completeness.

On a not unrelated matter:  Did you ever make contact with Karl von Meyenn about the funding issue?  I continue to worry that Pauli's work on dreams and alchemy will be lost forever to the broader audience if von Meyenn does not complete his project.

\section{01-04-09 \ \ {\it Beyond QT?}\ \ \ (to M. A. Graydon)} \label{Graydon5}

\bmag
I've been thinking harder these days about more conceptual issues (must be the hot weather), and I wonder if you might agree with the following:

The reality underlying quantum theory consists of systems, and those systems are defined exclusively by a single, objective, ontic property denoted by $D$.
Of course, the QBist need not posit that the whole of physical reality may be described in these terms, for the QBist is primarily concerned with an interpretation of the quantum theory. In QBism, there are no mathematical symbols in the quantum formalism that point to any other ontic property other than $D$. In particular, the quantum state represents one subjective agent's degrees of belief regarding the outcomes of future measurement interactions with an objectively existent external physical system. An `agent' is one in the world who may identify systems, contemplate their dynamics, and participate in measurement interactions. A `measurement interaction' involves an agent and a system. Operationally speaking, the agent passes the system through some measuring device and learns that one of $n$ outcomes has occurred. But then it seems that, at least for the QBist --- {\bf measurement interactions cannot change the (quantum) ontic nature of the system}, for the system's only property is $D$. Would you agree? Must we remove the `(quantum)' and look for a deeper theory?
\emag

Pretty darned good summary.  I was with you all the way until the last two sentences.  I think my best reply is to suggest re-reading Section VI in my recent long paper ``QBism, the Perimeter \ldots'' on {\tt arXiv.org} now.  I think you'll see in that how your last two sentences contradict what I'm thinking.  I'm prepared to think that quantum measurements (like any human action) change the world in a very deep way.  I'll send you some supplementary material in a minute, but do reread Section 6 (and 7 and 8 as well).

\section{01-04-10 \ \ {\it Explanatory Note} \ \ (to M. A. Graydon)} \label{Graydon6}

``More to Read $x$'' where $x=1,2,3,4$.  They're all about how ``measurement'' changes the world.  Happy Easter!

I was just reading your note again before filing it.  [See 01-04-09 note ``\myref{Graydon5}{Beyond QT?}''\ to M. A. Graydon.] Part of me wants to say ``broader'' is a better term.  But I do get where you're coming from in the last question.  Absolutely, there must be a ``next step'' in physics.  My own favorite dream is that whatever ``quantum measurement'' is, the world is made of it---it was a dream instilled in me by John Wheeler long ago.  QBism, as it presently exists, has for me always been preparatory work for getting at that deeper question.  (There, I used ``deeper'' myself.)  John's sense was that quantum measurement has more to do with all kinds of things than many imagine (for instance, ``the big bang'').

So, getting back to $D$.  I view it as a quantitative measure of ``one aspect'' of everything.  But what is that aspect?  That's the broad question we're gearing up to with QBism.  The aspect I think has to do with is ``whatever quantum measurement is.''

\section{02-04-10 \ \ {\it More to Read 5} \ \ (to M. A. Graydon)} \label{Graydon7}

[See 01-04-10 note ``\myref{Zeilinger11}{Strange Anton Dream}'' to A. Zeilinger and H. C. von Baeyer.] Here is my interpretation at the moment.  That the urgleichung is the ``imprint'' in the concrete (world), and that QM and SICs were the original ``pattern'' that I was trying to tell Anton about.  The imprint is surreally crisper than the original.

\section{02-04-10 \ \ {\it The Quantum Bayesian Glossary -- Some Updates} \ \ (to N. D. {\Mermin} and R. {\Schack})} \label{Mermin171} \label{Schack198}

Boys, it's just like the old days!  Remember when we put so much effort into constructing:\medskip
\begin{center}
{\Large \bf Mechanica Quantica Lex Cogitationis Est}\smallskip
\end{center}
Makes me feel young again!

\section{02-04-10 \ \ {\it The Friday Philosopher}\ \ \ (to {\AA}. {\Ericsson})} \label{Ericsson10}

I'm sorry:  It's turned out to be such a pleasant day outside, that I think I'll just stay home.  In fact, just a little while ago, I wrote Max Schlosshauer the lines below.

It's funny, earlier today I was thinking I would ask you if would be interested in going to see {\sl Clash of the Titans\/} with me at the 1:30.  (I figured you wouldn't be so interested---none of my girls were---but I thought it'd be worth a shot.)  Then I went outside for breakfast, and it was just so pleasant, it rearranged my whole mentality.

If you want a reading suggestion, for something to print out and take under a tree, here's what I'm dipping into from time to time today (I've already told you about it before):
\begin{center}
\myurl{http://en.wikisource.org/wiki/Personal_Idealism/Axioms_as_Postulates}.
\end{center}

\section{02-04-10 \ \ {\it Congratulations!}\ \ \ (to J. Emerson)} \label{Emerson7}

\bje
Thought you might be happy to hear, Quantum Bayesianism won the vote in class today for ``favorite interpretation''!
Well done!  PS: Many worlds got zero votes\ldots\ So the tides may be turning!
\eje

That is great news!  What were the other interpretations on the table?  And what was the margin?  Definitely let's get some beer Wednesday.

\section{02-04-10 \ \ {\it Congratulations!, 2}\ \ \ (to the QBies)} \label{QBies11}

Joseph Emerson just wrote me this:
\bje
Thought you might be happy to hear, Quantum Bayesianism won the vote in class today for ``favorite interpretation''!
Well done!  PS: Many worlds got zero votes\ldots\ So the tides may be turning!
\eje
Those of you in the class, tell me more about this!  Was there much of a margin or was it slim?  What were the other interpretations on the table?

\subsection{Hoan's Reply}

\bq
QBism got 8 most-favorite votes and 3 least-favorite votes, resulting in +5 (we could have got +6 had Gelo had not to leave for Europe).

If I remember right, ensemble and operationalism both got +4. Many-worlds got $-9$, Bohmian mechanics got $-3$.

There were totally 8 or 9 interpretations on the table, but some of those we never studied carefully so most of the people didn't vote on those.
\eq

\section{02-04-10 \ \ {\it Congratulations!, 3}\ \ \ (to the QBies)} \label{QBies12}

Who voted ``least favourite'' for QBism.  Do you know the names of anyone?  I won't throw eggs at their houses:  I'm just curious.  It'd be interesting to learn their reasons.

\subsection{Hoan's Reply}

\bq
I think they were all undergrads, but I didn't recognize them. One thing I remember is that when they raised their hands to vote QBism as the least favorite, one other undergrad (who voted most favorite for QBism) shouted out ``come on, it's not solipsism!''
\eq

\section{03-04-10 \ \ {\it The Quantum Bayesian Glossary -- Some Updates, 2}\ \ \ (to M. Schlosshauer)} \label{Schlosshauer29}

\bmaxs
Q'ran (noun) -- the Holiest of Holy Scriptures telling the story of
QBism from the ground up. (This is of course nothing but a thinly
veiled attempt at letting the idea of a book on QBism resurface.
Honestly, I've been thinking about writing something, sometime. But it
remains a ridiculously vague plan for now.)
\emaxs
Now, that's an idea I like!  You do the writing, and I'll do the praising!  I say, find a way to make it a less vague plan!  One of my own troubles is that I don't know where the ground is yet.  Also, I don't have the broad perspective that you do:  I've been a QBist-in-the-making since the beginning, with all kinds of petty prejudices against other directions since the beginning.  With honesty, I know that I've never given most of the other foundational programs a fair shake.  What the program really needs is someone who can give it a reasonably careful comparison to the other extant programs.  Someone too who can more detachedly discuss QBism's weaknesses.

And I'm getting good at ``praising.''  I wrote something for the back of Schumacher and Westmoreland's new book.  And I just wrote this a few days ago for the Nielson and Chuang's 10th anniversary edition:
\bq\noindent
Nearly every child who has read Harry Potter believes that if you just
say the right thing or do the right thing, you can coerce matter to do
something fantastic.  But what adult would believe it?  Until quantum
computation and quantum information came along in the early 1990s,
nearly none.  The quantum computer is the Philosopher's Stone of our
century, and Nielsen and Chuang is our basic book of incantations.
Ten years have passed since its publication, and it is as basic to the
field as it ever was.  Matter will do wonderful things if asked to,
but we must first understand its language.  No book written since
(there was no before) does the job of teaching the language of quantum
theory's possibilities like Nielsen and Chuang's.
\eq
Don't know whether the editor will let me keep it in this form (or make me make it more technical), but anyway, just imagine what'd I'd write for you!

As far as the glossary term Q'ran goes, though, I know you've recognized that it breaks the B rule.  QBie's already taken (that's what I call my group members.)  And QBest was sweet, but David {\Mermin} already beat you out:
\bq\noindent
QBlue -- a fanatical, unshakeable QBist; codename for C. A. Fuchs.
\eq
I'll think about QBitter.  How about this variation:
\bq\noindent
QBitter -- despair experienced by a QBist every time he is called a solipsist (again!); not to be confused with QBiteer (see next term).
\eq

\section{03-04-10 \ \ {\it QBar!}\ \ \ (to M. Schlosshauer)} \label{Schlosshauer30}

\bmaxs
Here's another one:
\bq\noindent
{\rm QBar (or QBa/QBah) -- the cocktail lounge at QBism house, serving exclusively mix drinks from Havana and the rest of the forbidden island.}
\eq
Oh, bad, bad, bad. I stop now.
\emaxs

OK, that was bad!  But it did inspire me to think that we need a \LaTeX\ symbol \verb+\qbar+,
analogous to \verb+\hbar+.  It is QBism's version of Bohr's ``finite quantum of action.''  I'm thinking here of the value \verb+\qbar=2+ for Eq.\ 12 (and discussion following) in the new paper.  {\Appleby} when he writes a lower-case \verb+q+ vector on the chalk board, always puts a line through the tail.  That's the sort of thing I'm imagining.  The value \verb+\qbar=2+ corresponds to 60 degrees for the QBibbo angle.

Thanks!  I wanted to respond to your science of philosophy note, but I finally got into my hotel room and am getting a bit sleepy.  It might wait until after a nap, i.e., you'll have some morning reading.  (The family is in Buffalo today and Niagara tomorrow, and while the girls have been out shopping, I've been working in the hotel lobby.  We're having the Easter egg hunt in the hotel room tomorrow morning.)

\section{03-04-10 \ \ {\it Philosophys}\ \ \ (to M. Schlosshauer)} \label{Schlosshauer30.1}

\bmaxs
Namely, have {\bf you\/} ever considered crossing over to a philosophy
department?
\emaxs

Too many times.  But the philosophers never wanted me.

John Preskill and I once worked to get Caltech to make a joint physics/philosophy position for me through a very broad ``information department'' initiative.  Attached are my proposals for that.  [See 12-02-04 note ``\myref{Preskill11.1}{The House Philosopher}'' to J. Preskill and H. Mabuchi.]

Now a nap.  More later on how at times, I have actually gotten something from the philosophers.  I exaggerated a bit in that sentence you quoted.  (And I should be careful before samizdatizing it.)

\section{03-04-10 \ \ {\it The Science of Philosophy -- 2}\ \ \ (to M. Schlosshauer)} \label{Schlosshauer30.2}

Now to this:
\bmaxs
It was a nice coincidence that you again appended your note
``Conceptual Barrier!''\ to your email. After reading it the last time, I
was struck by these words of yours: ``I have many times said that I
have never once gotten anything of any value from a philosopher of
science \ldots''
\emaxs

I'm glad you did mention Arthur Fine's name, because it reminded me that there are many ways to get ``something of value'' from people.  And from Arthur, I got quite a lot from his historical work on Einstein and {\Schroedinger}.  A lot of {\sl The Shaky Game\/} was quite good for me.

Also his ``Do Correlations Need to Be Explained?'' (or a similar title) gave lots of food for thought for me, particularly with his discussion of ``the hidden hand.''  Did I ever recommend Bruno de Finetti's ``Probabilismo'' to you?  He has a lovely discussion on this:  Suppose you make a probability-nearly-one assignment and then upon looking at the event you find that it did indeed happen.  Do you then have the right to be surprised?  Do you then have the right to declare that the fact of the outcome actually occurring was a great mystery, needing more explanation than it already had?  Well, ``correlation'' is just a two-variable version of that, and Arthur came so close to the point.  If he hadn't gotten confused with frequentism, he would have had it.  So, looking back, I should recognize that he likely planted a seed in my mind that would help me better appreciate de Finetti when that later time came.

Also, I remember reading Richard Rorty say that Arthur Fine was his favorite philosopher of science.  And that means something to me.  I have gotten so much from reading Richard Rorty.  If there are some confluences in their thoughts, then more familiarity with Fine might do me some good.

Another example is Bas van Fraassen.  There's been very little I've gotten from him, and for instance, when I sat in his seminar with Halvorson on the philosophy of quantum information, I thought both of them so amazingly confused and far from clear thinking that I wondered how they could be bestowed with the credential of ``philosopher'' in the first place.  {\it But\/} Bas's ``reflection principle'' within probability theory is a {\it golden nugget\/}.  It has turned out to be the key to a QBist's understanding of why everyone {\it thinks\/} (i.e., gets confused) that decoherence {\it precedes\/} the registration of a measurement outcome (and thus seeks out detailed measurement models to make that turn out to be the case).  Decoherence as a \underline{\it physical process}\/ does not precede registration:  It is only the agent ``reflecting'' in van Fraassen's sense that causes him to gamble \underline{\it as if}\/ it had been so.  And if {\it {\Ruediger}} and I ever write this up, we will owe Bas an immense intellectual debt.   [See \arxiv[quant-ph]{1103.5950}.]

Finally, there is no doubt that Jeff Bub, Bill Demopoulos, Richard Healey, Wayne Myrvold, Huw Price, Allen Stairs, and Chris Timpson have provided a kind of tension in my conversations with them that I have found extremely useful for organizing my thoughts.

So now I qualify myself, I should never say ``never once gotten anything of any value from a philosopher of science.''

\section{06-04-10 \ \ {\it In Reverse}\ \ \ (to M. Schlosshauer \& H. C. von Baeyer)} \label{Baeyer111} \label{Schlosshauer31}

I'll send this to Hans Christian too.

We're back home today, and I've really got to kick in the work the next nine days.  But I want to first record an experience that quite knocked me for a loop yesterday.  It has to do with your remark [concerning my 01-04-10 note to A. Zeilinger and H. C. von Baeyer titled, ``\myref{Zeilinger11}{Strange Anton Dream}'']:
\bmaxs
That's an impressively surreal dream you've had there! I'm also
flabbergasted at the amount of detail you could remember.
\emaxs
Sometimes I get a little too close to dreams.  And vice versa.  I'll try to make the story short and simple; I suspect nothing I write will convey the impact those few minutes had on me.

It was yesterday, early afternoon and we had been walking around the cheesy part of Niagara for a couple of hours---that is, the amusement-park looking area, with wax museums, haunted houses, a Harley-Davidson memorabilia shop, etc.  We started by taking a ride on the big Ferris wheel, letting the girls buy a couple of postcards, and ended with a walk through a park closer to the Falls.  When it was time to go, we trudged back up the hill toward where our mini-van was parked.  I remember well holding Emma's hand for about the last block of the walk, and just as we neared the car I gave it a firm squeeze before letting go.  Kiki said with a bit of triumph, ``All right, we're back.''  The kids went straight to the car door on the sidewalk side, and I went to the street side of the car to get into the driver's seat.  Nothing out of the ordinary you would say.

There was a bit of traffic, so I waited for a second to open the door.  I did note to myself, ``Funny that Kiki didn't unlock the door with the clicker.  I guess she had left it in the car.''  I inserted the key, used the button to unlock everyone else's doors, and looked up.  There was no one with me!  No one behind the car, no one near it, and all of this happened within the small flash of a second.  Everyone was gone.  I walked back onto the sidewalk, and looked into four nearby store fronts---no sight of them.  I can't describe the feeling of how disorienting this was.  I even came to the car to look in and see if I hadn't missed them somehow.  It was an absolutely strange experience.  I thought to myself, ``Have I been living in a dream for months?  Was this happy family life and all of today just a dream?  Could it be that Kiki left me long ago, and I've been living on my own and out of my head?''  I was in shock.  I looked down at my clothes and thought, ``They look clean, not like the clothes of a transient.''  I looked in all the stores again, this time more thoroughly, and nothing.  My mind kept coming back to the image of the firm squeeze I had given Emma just before we parted---I remembered it vividly.  I had this really palpable fear that I had been living a dream and just become conscious.  I started to look for some evidence of the reality of my life.  I went into the shade under an awning and pulled out my digital camera.  To my great relief, there were images of the morning.  Still I waited a lifetime---or so it felt---and they never came.  The fear kept nagging me.

Finally I went back into a store and started to pull out my digital camera as I asked a lady at the front, ``Have you seen a woman in a blue coat with two young girls in the last few minutes?''  She said, ``No, but you might check downstairs; sometimes I don't pay attention to everyone coming and going.''  Sure enough, just as I turned to the stairs, Kiki and the kids emerged from the lower floor.  I nearly cried with relief, and it took me much of the time driving home to completely recover.

With hindsight, what I find most interesting about this is the question of why my mind was so susceptible to the theory that I might have just become conscious from a dream.  There is a little bit of a cheap answer in that we had just watched the movie {\sl 50 First Dates\/} with Drew Barrymore and Adam Sandler two nights before.  It is the story of a man who romances a woman that had suffered a head injury:  Each night when she falls asleep, her short term memory disappears.  Each day he must romance her again, and eventually, with the help of a growing video montage, becomes her memory.  They marry, and in the last scene of the movie, he introduces her to their 10 year-old daughter.  It was a cute movie and must have been in the recesses of my mind.    Also there is another Hollywood precedent in my mind, a movie I saw 21 years ago:  {\sl Dad\/} starring Jack Lemmon, Ted Danson, and Olympia Dukakis.  Lemmon had two completely separate lives that he had been sustaining for years---one in a dream, and one in the real world, and only when he became quite ill did he realize that there were these two separate lives and he was one person.  The memory of that movie flashed through my mind yesterday as well.

But neither of these thoughts feel like they go deep enough to explain my susceptibility.  It's absolutely freaky that I ``freaked out'' to the extent described above.  So, I record the story here with you two because you have been involved in my dream thoughts lately (in all kinds of ways), and I continue to think about what this might mean in the wider context.

\section{06-04-10 \ \ {\it QBism Glossary}\ \ \ (to the QBies and Other Friends)} \label{QBies13}

The long awaited QBism Glossary is near completion.  If you have any suggestions for improvement, or any terms so good that they'd knock some of the existing ones off the page, now's your time to speak up:  I will finalize and post on my door soon.

I'm not all that happy with the look of the present QBar.  Better might be to have a $q$ with an extended tail and a bar through it, but I couldn't figure out how to do that with \LaTeX.   If anyone has any ideas on how to improve the symbol, please say so.

\subsection{Luca Mana's Reply}

\bq\noindent
There could also be QBismillah, from bismillah \ldots\ Good for quantum fundamentalists.
\eq

\section{06-04-10 \ \ {\it QBism Glossary}\ \ \ (to M. Schlosshauer)} \label{Schlosshauer32}

\bmaxs
Love it!

And thanks for the credit -- which I don't even deserve, since it looks like none of my suggestions made the cut anyway. (Which is absolutely fine I hasten to add: I wasn't too electrified myself about my submissions.)
\emaxs
That is not true:  You directly inspired QBar.  One of the most important concepts in the whole list.  Attached is {\Asa}'s variant:  When blown up like this, it's beautiful:
\begin{center}
{\Huge $q\hspace{-1.10ex}{\rule[-0.155ex]{0.45em}{0.09ex}}$}
\end{center}
Unfortunately it doesn't work so well in small font.

\section{07-04-10 \ \ {\it Paper and Glossary} \ \ (to D. B. L. Baker)} \label{Baker22}

I never did send you a final version of the paper.  You might enjoy the obscenities in the closing paragraphs of Section VIII.  [See the paper ``QBism, the Perimeter of Quantum Bayesianism,'' \arxiv{1003.5209}.]  (It starts with ``mount'' and goes downhill from there.)  More than that though, you might enjoy the further glossary of terms, fresh off the keyboard this morning.

\section{07-04-10 \ \ {\it SICs, Counterfactuals, and Other Stuff} \ \ (to O. Cohen)} \label{Cohen2}

\boc
Secondly, I wanted to mention that I have been following your recent work on Quantum Bayesianism with interest. I think that the SIC structure is very elegant and compelling. I am still digesting the longer of your two recent postings on {\tt quant-ph}, and may get back to you in due course with some comments, if that's OK with you. Your formulation in terms of counterfactual vs actual measurements is of particular interest to me right now as \ldots

Thirdly, I have been drafting up some recent musings of mine, which were triggered after re-visiting some of Bell's papers. I have been mainly concerned with looking at Bell/CHSH from a different perspective, which focuses on the actual vs counterfactual relationship when viewed from a local perspective. I think there may be a connection with your formulation, though I haven't had time to explore this yet. I am attaching a draft of my paper, which I am thinking of posting on {\tt quant-ph}, though I am a bit tentative as it's possible that what I have done is trivial and/or nonsense. If you are able to give any comments or feedback, that would be great and I would really value your opinion. But if you are too busy to spend time looking at the work of amateurs like me, that's also fine and I fully understand!
\eoc

Well, I certainly like this line in your paper, ``By exploring possibilities such as this, as a means to unravel Bell's theorem, we are moving the focus away from locality and on to counterfactuals.''  I've only skimmed your paper just a little, but that philosophical or programmatic position seems to be on the right track to me.

Hopefully I'll get a chance to look at your paper in seriousness in a couple of weeks, after I get some grant proposal things cleared off my desk.  Also, I hope you don't mind, but I'm going to encourage Howard Barnum, our seminar chair, to invite you to give us a talk on the subject.

\section{07-04-10 \ \ {\it Templeton}\ \ \ (to H. Price)} \label{Price19}

\bhp
Dean and I are putting together a proposal for the Templeton Foundation, to try to get money for a series of three conferences (and a postdoc). One of them is intended to be a ``motivations for thinking that physics needs a flow of time'' meeting. You'll see the general line, in the attached printout of our draft from their online application page. Can we list your name, as one of our intended participants/collaborators?
\ehp

Absolutely.  Thanks:  It's nice to be called, even if implicitly, one of ``the world's leading philosophers and physicists of time.''   You might consider Lee Smolin as well, since he now writes things like,
\bq\noindent
This provides further motivation for believing in the reality of time and the openness of the future. If the selection of the laws of nature that determine what elementary particles exist and how they interact is a result of evolution analogous to natural selection, then there must be a time that this evolution plays out in. And it must be a time that exists prior to the laws, for the laws must have scope to evolve in time.
\eq

It would seem the meeting you wish me to play a role in has a little bit to do with this gem of a book I found in Portland:  {\sl Incompatibilism's Allure}, \myurl[http://metapsychology.mentalhelp.net/poc/view_doc.php?type=book&id=4752]{http://metapsychology.mentalhelp.net/ poc/view\underline{ }doc.php?type=book\&id=4752}.  Just what is this allure of Chris and others to something {\it you\/} cannot even formulate a meaning for?  \smiley

Anyway, maybe we'll get a chance to play north and south pole within the Templeton Foundation.  {\Ruediger} {\Schack} are at the moment writing a proposal, ``That the World Can Be Shaped: Quantum Mechanics, Counterfactuality, and Free Will'' and you know me well enough to guess where that's going!

\section{07-04-10 \ \ {\it Yet Another Workshop} \ \ (to \v{C}. Brukner)} \label{Brukner7}

Unfortunately I think it is going to be almost impossible for me to attend this, as I must be at QCMC in Brisbane, and its dates are July 19--23.

This is really sad, because you know your meeting's topic is much closer to my heart!  To your title question, you would do me honor to read one paragraph from my paper in one of the discussions.

``What exists in the quantum world?''  Tell them, ``Everything!''\ and read:
\bq
Physics---in the right mindset---is not about identifying the bricks with which nature is made, but about identifying what is {\it common to\/} the largest range of phenomena it can get its hands on.  The idea is not difficult once one gets used to thinking in these terms.  Carbon?  The old answer would go that it is {\it nothing but\/} a building block that combines with other elements according to the following rules, blah, blah, blah.  The new answer is that carbon is a {\it characteristic\/} common to diamonds, pencil leads, deoxyribonucleic acid, burnt pancakes, the space between stars, the emissions of Ford pick-up trucks, and so on---the list is as unending as the world is itself.  For, carbon is also a characteristic common to this diamond and this diamond and this diamond and this.  But a flawless diamond and a purified zirconium crystal, no matter how carefully crafted, have no such characteristic in common:  Carbon is not a {\it universal\/} characteristic of all phenomena.  The aim of physics is to find characteristics that apply to as much of the world in its varied fullness as possible.  However, those common characteristics are hardly what the world is made of---the world instead is made of this and this and this.  The world is constructed of every particular there is and every way of carving up every particular there is.
\eq

If you were to say that, it'd be like my being at the meeting anyway!  For, in spilling those I would already exhaust all my thoughts up to this date!  (Though I miss a chance to learn from all of you.)

\section{07-04-10 \ \ {\it Is the Compromise Worth It?}\ \ \ (to N. D. {\Mermin})} \label{Mermin172}

Waking up in the middle of the night, I'm torn worse than earlier.  {\Asa} wrote me this,
\bq
{\Mermin}'s comment makes sense for someone with a broad knowledge. But I believe your response is correct. When it comes to understanding these words I don't think I am much worse than many others at PI. The glossary looks good now. With footnotes it will likely look cluttered. To make the `Compare ``\underline{\hspace{.33in}}''\/' not stick out as much, I suggest you use `cf.\ \underline{\hspace{.33in}}', perhaps in parentheses or with ; before instead of starting a new sentence (but this is perhaps incorrect, I don't know rules for semi-colon). The `` can be taken away if the words are italicized instead.
\eq
My gut feeling is that she's more right for the common folk and that your opinion, though more pure (and godlike for it, I emphasize), will lose me too much of an audience.

Your QBit was different.  The QB here is to emphasize ``Quantum Bayesian'' and the two letters come as a package.  Moreover, you were fighting the long usage of qubit.  In contrast, before QBism there was no QBism!

Of course, you are right about ``consistency.''  If I had not been driven by insane consistency, I would not have arrived at QBism as it presently stands.  I would rather have ended at a sort of ugly halfway house, like Peres and Ballentine did, each in their own way.  Thus, if I were to revert, I should give QBlue a ``compare'' as well.

Well, I will try to go back to sleep in spite of these deep troubles, and hope that a solution comes in the morning.

\section{07-04-10 \ \ {\it Up Front} \ \ (to {\AA}.\ {\Ericsson}, N. D. {\Mermin}, and R. {\Schack})} \label{Mermin173} \label{Schack199} \label{Ericsson10.1}

\noindent My three most trusted consultants, \medskip

I think the attached is my own favorite version:  I decided to put the cognate words right up front.  In my mind that somehow diffuses David's criticism \ldots\ but I don't know, he might find it even more annoying now.  I'll listen to all your opinions carefully.  Clearly this is something I want to get right!

\begin{center}
{\huge \bf Glossary of Quantum Bayesianism} \medskip \\
\normalsize (with contributions from the QBies, Kiki Fuchs, N. David {\Mermin}, {\Ruediger} {\Schack}, and Maximilian Schlosshauer)
\end{center}

\noindent {\bf QBism} -- the quantum foundational program of quantum Bayesianism; see ``QBism, the Perimeter of Quantum Bayesianism,'' \arxiv{1003.5209v1}.\bigskip

\noindent {\bf QBist} -- adjectival form of QBism and a practitioner thereof.\bigskip

\noindent {\bf QBlue} -- cf.\ {\it true blue}; a fanatical, unshakeable QBist; codename for C. A. Fuchs.\bigskip

\noindent {\bf QBies} -- everyone supported by ONR grant N00014-09-1-0247:  QBlue, along with graduate students Hoan Dang, Matthew Graydon, and Gelo Tabia, PI associate postdoc {\Asa} {\Ericsson}, and PI visiting researcher Marcus {\Appleby}.\bigskip

\noindent {\bf QBism House} -- 49 George Street, Waterloo, Ontario, ``where the ghosts of pragmatist philosophers past listen to how their ideas of an unfinished, malleable world take a more exact form with the help of quantum theory.''\bigskip

\noindent {\bf QBicle} -- cf.\ {\it cubicle}; the office of the student QBies, PI room 415.\bigskip

\noindent {\bf QBuki} --  cf.\ {\it kabuki}; the theater of the QBies' weekly group meeting outside PI office 355, Fridays 1:00--3:00 PM.\bigskip

\noindent {\bf QBicity} -- cf.\ {\it cubicity\/}; the state of being properly oriented on quantum interpretational issues.  ``Along with nirvana, one should strive for QBicity.'' \bigskip

\noindent {\bf QBar} -- the QBist's {\it elementary quantum of action}, $\qbar=2$. When $q=\qbar$, the generalized urgleichung
$$
Q(D_j)=\left(\frac{1}{2} qd+1\right)\sum_i P(H_i) P(D_j|H_i) - \frac12 q
$$
becomes the usual Born Rule if SICs exist.\bigskip

\noindent {\bf QBibbo Angle} -- cf.\ {\it Cabibbo angle}; maximum angle $\theta_{\scriptscriptstyle\rm QB}$ between quantum states in a QBist representation. See Assump.~7 and Eq.~(24) of ``A Quantum-Bayesian Route to Quantum-State Space,'' \arxiv{0912.4252v1},
$$
\cos\theta_{\scriptscriptstyle\rm QB} = \frac{q}{q+2}\;.
$$
When $q=\qbar$, $\theta_{\scriptscriptstyle\rm QB}=60$ degrees.\bigskip

\noindent {\bf QBissel} -- cf.\ Bavarian {\it bissel\/} and {\it N.~D. {\Mermin}}; one who is intrigued by QBism but not yet convinced.\bigskip

\noindent {\bf QBisl} -- cf.\ Yiddish {\it bisl}; one who is intrigued by QBism but is too busy studying the Talmud to give it the time it requires.\bigskip

\noindent {\bf QBitz} -- cf.\ {\it kibitz}; a common social interaction between a QBist and a QBissel---QBist as the QBitzer of course.\bigskip

\noindent {\bf QBic Equation} -- most important equation underlying the shape of quantum state space; see Eq.~(19) in ``The Lie Algebraic Significance of Symmetric Informationally Complete Measurements,'' \arxiv{1001.0004v1},
$$
\frac{d}{2}\sum_{ijk}R_{ijk}\, p(i)p(j)p(k)=\frac{d+7}{(d+1)^3}\;.\medskip
$$

\noindent {\bf QBomancy} -- cf.\  {\it cubomancy\/}; divination through quantum measurements, as with the Raussendorf \& Briegel model of quantum computation. \bigskip

\noindent {\bf QBosh} -- cf.\ {\it kibosh}; rare moment when a QBist, by force of argument, gets the holder of an alternative quantum interpretation to flinch.  {\sl Inside the Perimeter\/} headline: ``QBist Puts the QBosh on Visiting Everettian!''  \bigskip

\noindent {\bf QBlah} -- cf.\ Islamic {\it kiblah};  the direction QBists face when contemplating the beauty of SICs---inward! \bigskip

\noindent {\bf QBalah} -- cf.\ {\it kabalah}; mystical practice of demanding SICs to exist in all dimensions, or, if mathematics won't yield, that quantum mechanics be scrapped in favor of ``urgleichung mechanics.'' See ``Quantum-Bayesian Coherence,'' \arxiv{0906.2187v1}.

\section{07-04-10 \ \ {\it Finished Product} \ \ (to J. Rau)} \label{Rau5}

\bjr
Many thanks! I really enjoy the wit and clarity with which you expose your scientific and philosophical point of view. I was wondering, actually (and asked this question to {\Ruediger} who didn't know the answer) whether your acronym QBism was intentionally close to cubism, and whether you would like to become the Picasso of quantum mechanics? {\rm \smiley}
\ejr

Thanks for the papers.  On your questions, I have quite wondered how I might use an association with cubism to further my agenda, but so far I have fallen short.  On the other hand, you might enjoy a few of the other words in the QBism Glossary.  See attached.

\section{07-04-10 \ \ {\it Mid Conversation Frustration} \ \ (to T. Norsen)} \label{Norsen4}

\btn
Anyway, running with the photocopy metaphor, I see our dispute this
way.  We both look at something and agree:
this is definitely a photocopy.  I say ``there's got to be an original
somewhere, let's start working on finding it''.
You say, ``ah, but maybe this photocopy is only a copy of another
photocopy, in which case we don't have to talk any longer about
originals and all the genuinely foundational issues (though not some
interesting technical ones) evaporate.''
\etn
I know that you like to think that you are a careful reader, but all the evidence I have ever gathered from interacting with you shows that the thing you read best is your own mind.

If we had agreed, why would I have written these (unnoticed?) lines in the paper?
\bq
A fairer-minded assessment is that the accusation springs from our
opponents ``hearing'' much of what we do say, but interpreting it in
terms drawn from a particular conception of what physical theories
{\it always ought to be}:
Attempts to directly represent (map, picture, copy, correspond to,
correlate with) the {\it universe} \ldots

Quantum Bayesianism sidesteps the poisoned dart, as the previous
sections have tried to convey, by asserting that quantum theory is
just not a physical theory in the sense the accusers want it to be.
Rather it is an addition to personal, Bayesian, normative probability
theory.  Its normative rules for connecting probabilities (personal
judgments) were developed in light of the {\it character of the
world}, but there is no sense in which the quantum state itself
represents (pictures, copies, corresponds to, correlates with) a part
or a whole of the external world, much less a world that {\it just
is}.  In fact the very character of the theory seems to point to the
inadequacy of the representationalist program when attempted on the
particular world we live in.
\eq
Part of me sometimes wants to say, ``Oh, why don't you stop talking so much, and just go out and find an equation that someone will believe?''

But I hope my frustration will eventually wane, and I'll get back to ``fairer-minded assessment'' mode.  Maybe indeed much does depend on what you mean by the phrase ``physical world out there independent of us.''  As I read you, if I don't mean by the phrase {\it exactly\/} what you mean, then the
view {\it must\/} collapse into a belief in no external world at all.  It's probably a useful strategy for starting a church, but it's a waste of your
time (trying to straighten me out) if your aim really is to find a bigger, better equation that will pass the tests of other physicists' notice.  (If
you're not worried about it passing the test of other practitioners finding it useful in some way, then you're the solipsist, not me.)

What is the thing you truly want to accomplish in life?  And are your energies really turned to it?

\section{07-04-10 \ \ {\it And After a Deep Breath} \ \ (to T. Norsen)} \label{Norsen5}

More nicely,

That the external world is at times resistant to my actions upon it, and participating in the consequences of my actions upon it, is enough of a notion of ``independence'' for me.  It is a view that gives both conceptual players (agent and world) an autonomy that doesn't take away from the other's.  Indeed, I suspect that this is not what you mean by ``physical world out there independent of us.''  Rather what I can tell of your view is that the world underneath us acts upon us insofar as we are part of its great cog, but that there is no meaningful reciprocity (for we are parts, and it is the whole).  This is the kind of thing I call ``monism'' in the paper, in distinction to ``pluralism.''  Solipsism, as far as I can tell, is a monism, just a monism at the other end of speaking:  for, the agent is now the whole, rather than a part.

If you want some philosophical literature to rip into, I would suggest this paper by F. C. S. Schiller:
\myurl[http://en.wikisource.org/wiki/Personal_Idealism/Axioms_as_Postulates]{http://en.wikisource.org/wiki/Personal\underline{ }Idealism/Axioms\underline{ }as\underline{ }Postulates}

Calmer wishes,

\section{07-04-10 \ \ {\it Second Deep Breath} \ \ (to T. Norsen)} \label{Norsen6}

This is an example of what I find so frustrating in trying to discuss things with you.  My paper says this:
\bq
The metaphysics of empiricism can be put like this.  Everything experienced, everything experienceable has no less an ontological status than anything else.  You tell me of your experience, and I will say it is real, even a distinguished part of reality.  A child awakens in the middle of the night frightened that there is a monster under her bed, one soon to reach up and steal her arm---that {\it we-would-call-imaginary\/} experience has no less a hold on onticity than a Higgs-boson detection event would if it were to occur at the fully operational LHC.  They are of equal status from this point of view---they are equal elements in the filling out and making of reality.  This is because the world of the empiricist is not a sparse world like the world of Democritus ({\it nothing but\/} atom and void) or Einstein ({\it nothing but\/} unchanging spacetime manifold equipped with this or that field), but a world overflowingly full of variety---a world whose details are beyond anything grammatical (rule-bound) expression can articulate.
\eq
And then it goes on, I think rather clearly and carefully, to say that ``dimension'' is a numerical {\it quality\/} common to all pieces of the physical world, first prefacing it with
\bq
Physics---in the right mindset---is not about identifying the bricks with which nature is made, but about identifying what is {\it common to\/} the largest range of phenomena it can get its hands on.  The idea is not difficult once one gets used to thinking in these terms.  Carbon?  The old answer would go that it is {\it nothing but\/} a building block that combines with other elements according to the following rules, blah, blah, blah.  The new answer is that carbon is a {\it characteristic\/} common to diamonds, pencil leads, deoxyribonucleic acid, burnt pancakes, the space between stars, the emissions of Ford pick-up trucks, and so on---the list is as unending as the world is itself.  For, carbon is also a characteristic common to this diamond and this diamond and this diamond and this.  But a flawless diamond and a purified zirconium crystal, no matter how carefully crafted, have no such characteristic in common:  Carbon is not a {\it universal\/} characteristic of all phenomena.  The aim of physics is to find characteristics that apply to as much of the world in its varied fullness as possible.  However, those common characteristics are hardly what the world is made of---the world instead is made of this and this and this.  The world is constructed of every particular there is and every way of carving up every particular there is.
\eq
And yet you say,
\btn
you {\bf claim} to believe in reality (and say some things always at the end about how maybe at the end of the day we'll learn about the dimension of the hilbert space of some system, which you take to be as far as I can tell the one-and-only external fact that exists or that we might get access to in the near future, etc.), \ldots
\etn

Where in THIS paper did it say that dimension is the one-and-only external fact?  Two objects have {\it this\/} quality; five objects have {\it another\/} one.  But dimension is something that they all seemingly have---it is a quality, and I would say the only one that quantum theory identifies.  But I was not born with this answer in my head; I have been working the idea out slowly, and this paper represents the present status of formulation.  What you write above may refer to an inarticulate way in my previous expression from the 2002 paper, but I thought the issue at hand was what is in this paper.

\section{07-04-10 \ \ {\it Me, Me, Me Again!}\ \ \ (to N. D. {\Mermin})} \label{Mermin174}

You always delight me!

\bdm
14R.  I was surprised to read that ``quantum
measurement outcomes'' are personal.   I had
thought the probabilities of those outcomes
were personal.  I will peruse the earlier sections
to see if you really meant this.  It sounds (ahem)
like solipsism.
\edm

It goes back {\it at least\/} to the attached note to YOU (and {\Ruediger}) in 2003 [titled ``\myref{Mermin101}{Me, Me, Me}'' and dated 12-08-03], and the idea has been in every one of my papers since then!

Here is the way I put it to Norsen in yet the latest attempt to get at the two-level personalism of QBism.  [See 01-04-10 note ``\myref{Norsen3}{Too Short an Answer}'' to T. Norsen.] (He in his oh-so-careful reading didn't even notice the personal bit, and couldn't figure out the distinction between Ballentine and me.  At least you noticed it!)  {\it Even\/} Einstein recognized that the probability statements of QM might refer to ``personal outcomes''---that was the reason for that very long quote!

\section{07-04-10 \ \ {\it LNC Section 6.8} \ \ (to N. D. {\Mermin})} \label{Mermin175}

\bdm
There are no probabilities of 1 in the argument in LNC {\em [J. S. Bell's article ``La Nouvelle Cuisine'']}.  (Indeed
arguments of that type were originally developed explicitly to avoid
using probability-one assignments, which were held to be unphysical
idealizations of the probabilities one could assign to the outcomes of
actual experiments.)  No agent is certain about anything.
\edm

\bq
That ordinary quantum mechanics is not locally causal was pointed out by Einstein, Podolsky and Rosen, in 1935.  Their argument was simplified by Bohm in 1951.  \ldots\  Each of the counters considered separately has on each repetition of the experiment a 50\% chance of saying ``yes''.  But when one counter says ``yes'' so also always does the other, and when one counter says ``no'' the other also says ``no'', according to quantum mechanics.  The theory requires a perfect correlation of ``yeses'' or ``nos'' on the two sides.  So specification of the result on one side permits a 100\% confident prediction of the previously totally uncertain result on the other side.  Now in ordinary quantum mechanics there just is nothing but the wavefunction for calculating probabilities.  There is then no question of making the result on one side redundant on the other by more fully specifying events in some space-time region 3.  We have a violation of local
causality.

Most physicists were (and are) rather unimpressed by this.  That is because most physicists do not really accept, deep down, that the wavefunction is the whole story.  They tend to think that the analogy of the glove left at home is a good one.  If I find that I have brought only one glove, and that it is right-handed, then I predict confidently that the one still at home will be seen to be left handed.  But suppose we had been told, on good authority, that gloves are neither right- or left-handed when not looked at.  Then that, by looking at one, we could predetermine the result of looking at the other, at some remote place, would be remarkable.  Finding that this is so in practice, we would very soon invent the idea that gloves are already one thing or the other even when not looked at.
\eq

QBISM BLOCKS THAT MOVE RIGHT THERE.  QBists would not go with EPR and Bell and declare ``ordinary quantum mechanics is not locally causal'' just because of that.  And it is not by a quick assumption of action-at-a-distance, but because for the QBist, probability-1 carries no force in truth assignments.

Yet, the QBist, might have gone {\it partially\/} with EPR and said,
\begin{quote}
They decided that the wavefunction, making no distinction whatever between one possibility and another, could not be the whole story.
\end{quote}

I.e., the QBist might have said, ``Yet it gets me into no trouble to provisionally accept that my probability-1 assignment does imply that I actually believe in a pre-existent truth value.''  And that is where Bell's reasoning shows him a contradiction.  So long as the QBist holds on to locality (and he can because of the blocked move above), he {\it cannot\/} entertain that his ``probability-1 assignment for a measurement outcome implies that he actually believes in a pre-existent truth value for it.''

The issues are exactly the same in the algebraic version of it all that I gave in the paper.

Bell and Norsen (and most everyone else but the three QBists) rely on the same crutch when they say that raw, ordinary quantum theory is ``not locally causal''.

Geesh!

\section{07-04-10 \ \ {\it Me, Me, Me Again!, 2} \ \ (to N. D. {\Mermin})} \label{Mermin176}

\bdm
What, for example, does it mean to make a bet that the reading will be
0 when the Dutch bookie I'm placing the bet with, watching the same measurement gate, may disagree?
Do you need this additional layer of looseness to get rid of ``the measurement problem''?
\edm

{\Ruediger} is always careful to point out that Dutch-book coherence really is always an internal requirement (i.e., not in last analysis a two-person adversarial thing).  It is really only that probabilities be consistent with logic.  I'll cc him on this in case he wants to send you something he's written on this take of it.

\section{07-04-10 \ \ {\it Births and Preclusivity} \ \ (to R. D. Sorkin)} \label{Sorkin2}

By the way, I much appreciated your talk yesterday; I got a lot out of it and it helped clarify things to me.  (You're on my mind because of a current email debate I'm having with Travis Norsen.)  There is certainly a broad way in which our two research programs overlap:  it is that quantum mechanics is ultimately about events being born, and that what we presently call ``measurement'' should be considered instances of such, though that is not the exclusive provenance of such events.  Or at least this is what I thought I understood from you.

Where the programs seem to part is in 1) the manner in which that broad statement above is implemented, and 2) the exact usage/meaning of probability within QM.

For more personal (or petty) reasons I was quite happy to see your {\it axiom\/} of preclusivity laid out so explicitly and so honestly.  For, by recognizing it as an axiom, you show that you implicate it (at least in the case of your ``quantum measures'') as {\it no\/} logical necessity.  You take preclusivity as an axiom connecting the measures to truth values.  Nothing wrong with that, but it is an axiom---in principle, one can take it or leave it.  Of course, you tried to motivate this by an analogy to old-time probability theory---but my guess is that you would admit that preclusivity is really an axiom there too.  I don't know; am I wrong?  Anyway, it was useful for me to see these things.  For in my program of QBism, one of the key points is that preclusivity is not a logical necessity in personalist Bayesian probability theory either (though I haven't yet used the word ``preclusivity'').  In fact, the whole approach is based on trying to get some interpretative traction by going the opposite way from you on this issue!  For instance, it is the point of Footnote 29, page 18, of my recent posting, \arxiv{1003.5209}, where I take Norsen and Bell's nonlocality on.

I've just printed your new paper out and will be studying it.

\section{08-04-10 \ \ {\it Me, Me, Me Again!, 3} \ \ (to N. D. {\Mermin})} \label{Mermin177}

\bdm
Do you need this additional layer of looseness to get rid of ``the
measurement problem''?
\edm

In a word, yes.  That's what the section on Wigner's friend is about.  If quantum theory is a ``user's manual'' for one's personal experiences, then there is no measurement problem.  Each of us uses it, each of us get the consequences of our actions upon the world, but we never let our own uses of the manual be dictatorial over others' experiences.

\section{08-04-10 \ \ {\it Me, Me, Me Again!, 4} \ \ (to N. D. {\Mermin})} \label{Mermin178}

\bdm
What, for example, does it mean to make a bet that the reading will be
0 when the Dutch bookie I'm placing the bet with, watching the same
measurement gate, may disagree?
\edm

And it's not that he would disagree, but that ``to see what I have seen'' he would have to interact with me.  An old and metaphorical idea:  No two poets see the same world.  But seriously.  It is the point of Footnote 35 and the text to which it is attached.

Sorry I didn't catch this yesterday.

\section{08-04-10 \ \ {\it Your Weltanschauung} \ \ (to N. D. {\Mermin})} \label{Mermin179}

\bdm
So I don't think you should have been surprised at my
reaction to 14L in the context of your paper, though I
really have to take more time to inspect the document
before deciding that I should have been reading you
more carefully.  But my present view is that my
confusion on this point is at least as much your fault
as mine.
\edm

All fair enough.  But I think I was referring to a conversation that I had had with you when I talked about shooting myself in the foot in the passage below (I was never completely sure, but someone definitely said it):
\bq
The answer to the first question surely comes as no surprise by now, but why on earth the answer for the second?  ``It's like watching a Quantum Bayesian shoot himself in the foot,'' a friend once said. Why something so egocentric, anthropocentric, psychology-laden, myopic, and positivistic (we've heard any number of expletives) as {\it the consequences (for me) of my actions upon the system}?  Why not simply say something neutral like ``the outcomes of measurements''?  Or, fall in line with Wolfgang Pauli and say:
\begin{quote}
The objectivity of physics is \ldots\ fully ensured in quantum mechanics in the following sense.  Although in principle, according to the theory, it is in general only the statistics of series of experiments that is determined by laws, the observer is unable, even in the unpredictable single case, to influence the result of his observation---as for example the response of a counter at a particular instant of time.  Further, personal qualities of the observer do not come into the theory in any way---the observation can be made by objective registering apparatus, the results of which are objectively available for anyone's inspection.
\end{quote}
To the uninitiated, our answer for {\it Information about what?}\ surely appears to be a cowardly, unnecessary retreat from realism.  But it is the opposite.  The answer we give is the very injunction that keeps the potentially conflicting statements of Wigner and his friend in check,\footnote{Pauli's statement certainly wouldn't have done that.  Results objectively available for anyone's inspection?  This is the whole issue with ``Wigner's friend'' in the first place.  If both agents could just ``look'' at the counter simultaneously with negligible effect {\it in principle}, we would not be having this discussion.} at the same time as giving each agent a hook to the external world in spite of QBism's egocentric quantum states.
\eq

I hope you didn't take my exclamation marks yesterday too seriously.  I resonated deeply to what you said with, ``I have trouble enough keeping a grip on my own Weltanschauung, so it's not surprising that I can't hang on to yours.''  It demands too much of a QBissel.  And {\Carl} worked like hell to suppress drawing direct attention to it---to the extent that he could do so without speaking an untruth---in the paper on ``quantum certainty'' that you refereed for us \ldots\ which could have better prepared the ground for you.  That is why that paper marked our last-ever joint conceptual-writing project together.  (I leave open the possibility that we could write something technical together again.  But something where words are particularly important, no.)

I'm glad your daughter is coming by for a visit.

\section{08-04-10 \ \ {\it Your Weltanschauung $=$ My Weltanschauung$-$Epsilon?}\ \  (to N. D. {\Mermin})} \label{Mermin180}

You mean 14R (should have told you that earlier).  ``Outcome'' seems to be too loaded of a term, for you and probably for others as well.  Would ``experience'' in the caption to Figure 1 have alleviated some of your pain?  I use it in the figure itself, but I had not in the caption:  ``consequence $=$ experience'' the figure says.  Similarly for Figure 5.  Read the caption to Fig 5 with ``experience'' in place of ``outcome.''

As a long exchange with you once taught me to stop using the word ``knowledge'' for quantum states (that's where ``belief'' came from), maybe you are teaching me that I should be more honest and forthright with the word ``experience.''  For it is definitely where I am heading---that the world before man, after man, in between and everywhere, is a patchwork of ``pure experience'' (as James called it).  See 27L and 27R for this said rather explicitly.

In some ways, this strikes me as a lot in the spirit of your old ``correlation without correlata'' --- that the world becomes, like Noah filled the arc, bipartite by bipartite (instead of the more unary ``event by event'' for instance).  But to get at the ultimate idea, one must jettison the notion of quantum state in its formulation.  That is what always clouded me away from your earlier program---it made explicit use of quantum states.  But if one strips away the earlier technical formulation, there might be some of your spirit in this present stage of my thinking.  Though still another difference might be my emphasis on ``becomes.''  It is the same thing I meant when I told Rob {\Spekkens} that his hoped for view was not ``sexual enough'' in the attached pages [see 24-09-03 note ``\myref{Spekkens21}{The Helping Hand}'' to R. {\Spekkens}.]  With him, the idea of true becoming, is a non sequitur (he embraces the block universe conception of things).  With you, I haven't been so sure, but I always felt your ``correlation without correlata'' at least in those days leaned in that direction.

Does this sound credible to you?  Are we making progress?

\section{08-04-10 \ \ {\it Too Silly?}\ \ \ (to N. D. {\Mermin})} \label{Mermin181}

\bdm
Yes, they're too silly.   CVs should be concise.  People spend a lot of time reading them.   You
don't want to waste their time with ephemera, however entertaining they might be.   But breaking up
the writings into categories is a good idea.   It's surprising how many people don't.
\edm

This is a good word to know actually.  ``Ephemeron'' and ``Ephemera''.  The world ``becomes'' ephemeron by ephemeron.  Doesn't capture the bipartite by bipartite part of it though.

Thanks for the advice.

\section{08-04-10 \ \ {\it Starting It Over} \ \ (to N. D. {\Mermin})} \label{Mermin182}

Your last hatching system was confusing too me; line breaks got deleted or something.  So, I'll just respond to your responses directly.

\bdm
So even you, whose writing I admire, use the redundant
``of''!  Then the battle is lost.
\edm
Not really:  I had thought there was something strange in what I wrote---I recognized that, but couldn't figure out what it was.

\bdm
Are you willing to say up front that different
agents can experience different readings of a single
measurement gate at a single moment.
\edm
To paraphrase William James,
\bq\noindent
     In its widest possible sense, however, an agent
     is the sum total of all that he can call his,
     not only his body and his psychic powers, but his
     clothes and his house, his wife and children, his
     ancestors and friends, his reputation and works,
     his lands and horses, and yacht and bank account.
\eq
See footnote 35 on 20R.  More seriously, if ``two'' agents are part of a single common experience, then they are ``one'' with respect to the experience.  Quantum theory as a ``single agent's user's manual'' only has two slots:  one for the agent and one for the system.  The theory as it stands, gives no tool for even posing what it means for ``two different agents to read a single measurement gate at a single time.''  In a way, that's been the conundrum since the very beginning of the theory---QBism is just an attempt to embrace that as positively as possible.

\bdm
If so, then you ought to spell out explicitly why
QBism requires this.
\edm
I thought I had done that with the Wigner's friend discussion on 5R through 8L.  What's lacking there?

\bdm
Yes indeed.  And also in the spirit of that Physics
Today column on reifying abstractions --- and the talk
I gave at PI last October.
\edm
I definitely want to watch that again after Templeton.  I remember liking it, and also the column, because you did use the word ``experience.''

\bdm
but now I lose you.  The correlata are the unary
experiences, which belong to me, but not to physics.
What's bipartite here? Is the pair experience-outcome?
\edm
Maybe so (i.e., you lose me), because I'm wanting to put the experiences out into the world as its atoms.  That is to say, to put it into physics.  See the Dupr\'e quote on 19L and the thing on Bohr in Footnote 34 on 20L.  Bipartite refers to the way we usually think of what is happening in quantum measurement---one part is the agent, one part is the system.  A clearer view might get rid of the cut.  William James would say, the cut is {\it additive\/} to the pure experience; it is something placed on top of it.  I'm still working out how far I can go with that.  But all of this is far down the road for me.  The first step is to understand what the QBist's ``elementary quantum of action'' QBar is telling us in the usual (bipartite) picture.  At the very least, when I put an experience into spacetime, it becomes bipartite.

\section{08-04-10 \ \ {\it UNC Visit}  \ \ (to Y. J. Ng)} \label{Ng6}

I'm just about to remove some silly lines that I had provisionally put in my CV, but I thought you might enjoy seeing them first.  They're the two entries just above the ``Awarded and Endowed Lectures'' section on page 2:
\bv
A true-to-form conversation with Terry Rudolph on the meaning of quantum mechanics is immortalized
in Louisa Guilder's history of quantum entanglement, {\sl The Age of Entanglement:\ When Quantum
Physics was Reborn}, (Alfred A. Knopf, New York, 2008), pp.\ 332--336.\medskip\\
Still Sillier Things: Academic Lineage --- F. S. Exner, F. Hasen\"ohrl, K. Herzfeld, J. A. Wheeler, K. S.
Thorne, C. M. Caves, C. A. Fuchs. Erd\H{o}s number --- 3 (but who doesn't have one). Einstein number ---
3 (Einstein--Rosen--Peres--Fuchs). Wolfgang Pauli number --- 4 (Pauli--Einstein--Rosen--Peres--Fuchs).
\ev
Looking at it, they're too silly.  But maybe they'll give you some material for my intro!  Exner was {\Schroedinger}'s advisor, and actually is someone who embraced indeterminism even before quantum mechanics.  And Herzfeld (according Mehra and Rechenberg's history) was the first person to coin the term ``quantum mechanics.''

\section{08-04-10 \ \ {\it What I Mean}\ \ \ (to P. G. L. Mana)} \label{Mana17}

Yes, I like the Truesdell quote very much.

\subsection{Luca's Preply}

\bq
Physics is a kind of art, like painting (C. A. Truesdell):
\bq
A theory is a mathematical model for an aspect of nature.  One good theory extracts and exaggerates some facets of the truth. Another good theory may idealize other facets. A theory cannot duplicate nature, for if it did so in all respects, it would be isomorphic to nature itself and hence useless, a mere repetition of all the complexity which nature presents to us, that very complexity we frame theories to penetrate and set aside.
\eq
and
\bq
If a theory were not simpler than the phenomena it was designed to model, it would serve no purpose. Like a portrait, it can represent only a part of the subject it pictures. This part it exaggerates, if only because it leaves out the rest. Its simplicity is its virtue, provided the aspect it portrays be that which we wish to study. If, on the other hand, our concern is an aspect of nature which a particular theory leaves out of account, then that theory is for us not wrong but simply irrelevant. For example, if we would analyse the stagnation of traffic in the streets, to take into account the behavior of the elementary particles that make up the engine, the body, the tires, and the driver of each automobile, however ``fundamental'' the physicists like to call those particles, would be useless even if it were not insuperably difficult. The quantum theory of individual particles is not wrong in studies of the deformation of large samples of air; it is simply a model for something else, something irrelevant to matter in gross.
\eq
\eq

\section{09-04-10 \ \ {\it Starting It Over, 2} \ \ (to N. D. {\Mermin})} \label{Mermin183}

\bdm
But now you're telling me that the two agents can't both assign
probabilities to the readings of one and the same measurement gate
when it interacts with that system?
\edm
That's right:  The agent IS the measurement gate.  The prosthetic hands are crucial to Figure 1.

\bdm
That not only are the probabilities personal to the agent but so is
the event to which the agent assigns those probabilities?
\edm
Well, it is shared between him and the system.  So, in a sense it's personal to the system as well.

\bdm
If they can then we're back to my original question --- can one
experience 0 while the other experiences 1?
\edm
Right, I'm trying to nullify your original question.

\bdm
If they can't, what does it mean for them both to place bets on that
reading?
\edm
I don't think it means anything within the quantum formalism.  If they want to discuss both seeing something from the same system, they have to ``talk'' to each other and that is another quantum measurement---one shared between the two of them:  From one point of view, agent and system, and from another point of view, system and agent.

I don't want to infuriate you, but this is the way I'm seeing it.  And I recognize that these questions are good for me.

\section{09-04-10 \ \ {\it Moving Along} \ \ (to N. D. {\Mermin})} \label{Mermin184}

\bdm
since for Bohr (and for me, and I had thought for you) communication
between agents was at an entirely different level.
\edm

No, for me, every action an agent takes on the external world (which can be putting an SG device in front of an electron or talking to another agent) falls within the framework of a ``quantum measurement''.  Please see Footnote 46, the paragraph to which it is attached, and the one before that on page 24.  Whacking a baseball is a quantum measurement from this point of view.

Here's where I hope you'll say that ``strikingly original'' phrase again.  It is that quantum mechanics is ``additive'' to the classical world.

Give that little bit of reading a shot, and I'll try to fill the further head-scratching it might engender.

Glad to hear you're not infuriated (and I hope not appalled either).

\section{09-04-10 \ \ {\it And Coming Back a Bit} \ \ (to N. D. {\Mermin})} \label{Mermin185}

\bdm
But now you're telling me that the two agents can't both assign
probabilities to the readings of one and the same measurement gate
when it interacts with that system?

That not only are the probabilities personal to the agent but so is
the event to which the agent assigns those probabilities?
\edm

Thinking upon this, you may not realize, but it is why I have always been fascinated with Pauli rather than Bohr.  I never felt Bohr's position was consistent, and I tried very hard to believe that it was (and read nearly every philosophical thing he wrote).  It is in exactly whether the measurement device is a prosthetic hand, rather than a piece of equipment in the laboratory separate from the agent, that the two men disagreed.  This is why the old book is {\sl Notes on a Paulian Idea}---this very issue is the deep part of THE Paulian Idea.  (OK, so Pauli himself is not completely consistent, and I take him to task in the present paper, but he came closer to consistency than Bohr, I always felt.)  In a way, it means you never quite took me seriously on this ``detail.''  (Quotes because who would take this seriously as ``just a detail.'')

Actually, I'm sorry I didn't realize that we weren't communicating on this for all this time!  Wow, I'm feeling idiosyncratic at the moment!  It makes me wonder how many people who think they follow me (to some extent at least, like Max) don't realize how radical I (and Pauli) get.

I deeply appreciate having you as a reader!

\section{09-04-10 \ \ {\it QB Decoherence} \ \ (to M. Schlosshauer)} \label{Schlosshauer33}

I talked with {\Ruediger}, and he was OK with my showing you the draft as it stands.  As I wrote to him, ``Max's feedback may well spur us on to finish the thing and make it an easier paper to write in the long run.''

Let me try to sketch a bit of the idea.  When von Neumann was not using ``Type 1 process / Type 2 process'' language for collapse and unitaries, but rather discussing ``measurement models'' he described it as the two-step process you described nicely in your paper with Camilleri.  First there is entanglement between the system and device (and with the addition of Zeh and Zurek, entanglement with the environment), and then---mysteriously---selection.  (Of course the Everettians try to drop that last step, but they're not my target here.)  You can probably guess that I would say, ``von Neumann's setting the issue of measurement in these terms was the great original sin of the quantum foundational debate,''  and I would.  The question is, what would a QBist put in its place and what connection, if any, is there to the previous notion of decoherence occurring before the magical selection step?

For a QBist, the only {\it physical\/} process in a quantum measurement is the selection step---i.e., the part the decoherence program tries to ignore---the whack, the action/reaction, the data that leads to a new state of belief about the system.  Decoherence?  There's no room for it, no role for it in this picture \ldots\ or is there?  See attached photo (of my chalkboard before it was painted).  The ghostly arrows represent the vN-Z-Z way of looking at measurement:  We go from an initial quantum state and a specification of a ``measurement interaction'' (using something closer to my own terms, I have it written the specification as a set of completely positive maps $\Phi_i$) to a decohered quantum state to a selected final quantum state (one of the $\rho_i$ with probability $p(i)$).  The arrows are ghostly because for us, that's all imaginary.  What happens in reality is the dark upper arrow labeled ``whack.''  It would appear that the decohered state $\rho_w$ is never the state of belief of anybody.  And in a way, that is true.  For though the usual story is that the outcome of the measurement is ``out there'' simply unknown to the agent, and so he should adopt this as his new state of belief, we QBists reject that:  There is no outcome for the agent except in his  own experience.  Like the thing John Wheeler used to say about the umpire, ``They ain't nothin' till I calls 'em.''

Nonetheless, one can imagine a situation in which an agent would adopt $\rho_w$ for his gambling attitudes.  It has to do with van Fraassen's reflection principle.  Suppose I know that I'm going to whack the system {\it tomorrow morning\/} and that consequent to it I will adopt one of the $\rho_i$ for my future beliefs.  But you, Dutch bookie, offer me a lottery ticket {\it today\/} for a second whack that I'll perform {\it tomorrow afternoon}.  What should I regard as a fair price for the ticket?  The reflection principle requires I adopt $\rho_w$, or I can be Dutch booked.

And that from the QBist side of the world is the story of decoherence.  Decoherence doesn't come conceptually before a ``selection,'' but rather is predicated on the possibilities for the next quantum state.  Decoherence comes conceptually {\it after\/} the recognition of the future possibilities (i.e., what used to be called ``selection'').  This is why I've been going around for a year saying that Zurek has it exactly backwards.  [See \arxiv[quant-ph]{1103.5950}.]

\section{10-04-10 \ \ {\it Introduction, Pure Experience, Quantum Bayesianism}\ \ \ (to D. C. Lamberth)} \label{Lamberth1}

First off, let me say that I am a great fan of your book on James and the metaphysics of pure experience, as well as your contribution to the {\sl Cambridge Companion to William James}.  Both writings influenced me deeply, and I think it is time I met you (electronically at least).

I am a physicist at the Perimeter Institute for Theoretical Physics in Waterloo, Canada with a deep interest in pragmatism and radical empiricism.  You can read about Perimeter here if you care:  \myurl{http://www.perimeterinstitute.ca/}.  As well, I attach my CV, so that you will have some evidence that I'm not a lunatic coming out of the woodwork.

I also attach a recent paper on the research program into the foundations of quantum theory that I head (Quantum Bayesianism, or QBism), and a ``press release'' on my research group and the pragmatism library I have built up for everyone.  I don't expect you to be aware much of quantum theory, but my guess is that the last three sections of the paper should still be quite readable to you nonetheless (because of their broader nature).  Most importantly, they give some hint of why I am so interested in James's notion of pure experience, and why I cite you.  My feeling is that this phenomenon in my field called ``the quantum measurement problem'' is as close to an expression of pure experience as we have ever encountered in physics.  Moreover, that thinking in these terms will ultimately delete the word ``problem'' from the phrase ``measurement problem'' and open up great new vistas of physical exploration.

That is the general reason I contact you---to ultimately set up a dialogue on this subject.  I think much progress can be made, but it will require a dialogue between physicists and the right kind of philosophers (pragmatists).

The more specific reason that I contact you now, rather than say three months from now, is that I am working on a pre-proposal for the Templeton Foundation for their call on research in quantum theory (due Thursday, April 15).  The title of the proposal (if it gets invited) will be, ``That the World Can Be Shaped:  Quantum Bayesianism, Counterfactuals, Free Will,'' and the research it describes will have a large component in making the things I mentioned above more concrete.  Below is the ``Executive Summary'' I have written for it, and it may give you some hint of what I am aiming for.  With my reputation, I believe the pre-proposal has a strong chance of being invited as a proposal and then ultimately fly as an honest-to-god grant.

You will note that I mention ``two international workshops'' in the summary.  I am thinking of something fairly intimate, with approximately 15 people for each.  One will be targeted for the Alps near Zurich in the land of Wolfgang Pauli and his particular contributions on this subject (along with his interactions with Carl Jung).  But for the other workshop, I would like to devote it to ``Quantum Theory and Pure Experience'' and make it a dialogue between physicists, philosophers, and theologians.  Since you are my favorite contemporary thinker on pure experience, it seems natural to apprise you of this, and test whether you might have any interest in being involved.  We have excellent conference facilities at Perimeter Institute, but I would likely try to hold it in a more inspiring setting.

Involvement only means spiritual involvement, organizational involvement, intellectual involvement.  I would provide all of the funds for this workshop if it materializes.

A student wrote me recently (and it is certainly a line I will use in the proposal),  ``I like your theory because it returns to me as much freedom as I feel that I have. Such freedom is lost or partially lost in other interpretations [of quantum mechanics].''  I hope the same point will help cause you to take notice of what I am writing you here.

\bq
\noindent \underline{Executive Summary}\medskip

Speaking on scientific theories, the philosopher William James once said, ``If you follow the pragmatic method, you cannot look on [a completed theory] as closing your quest. You must set it at work within the stream of your experience. It appears less as a solution than as a program for more work, and more particularly as an indication of the ways in which existing realities may be changed. Theories thus become instruments, not answers to enigmas. We don't lie back upon them, we move forward, and, on occasion, make nature over again by their aid.'' On this conception, a theory is not a statement about what the world is, but is a tool, like a hammer, to aid in making the world what we want it to be. The world may resist, but to some extent it is plastic.

The research program of Quantum Bayesianism (or QBism) is a take on the quantum interpretation problem that reveals with mathematical precision, not poetry, that such plasticity is quantum theory's greatest lesson. With every quantum measurement set by an experimenter's free will, the world is shaped just a little; and so of every action of every agent everywhere.

To further develop QBism and explore its broad implications, we request funds for two postdocs, two international workshops, and regular travel for collaborators.
\eq

\section{11-04-10 \ \ {\it QBism Comes to Life!}\ \ \ (to C. Ferrie)} \label{Ferrie14}

\bcf
I wanted to play around with the website below and the first dialogue that came to mind was the one in your paper: \myurl{http://www.xtranormal.com/watch/6389801/}.  What do you think?
\ecf

QBism already was alive!  It's the most living interpretation of quantum mechanics there ever was!

But I'm very proud that that was the first dialogue that came to your mind!  You flatter me.

What do you think about this one from 2000?

\bq
In the first frame {\God} starts to speak to {\Adam} at a time just
before Genesis, ``{\Adam}, I am going to build you a world.  Do you
have any suggestions?'' \medskip

{\bf {\Adam}}:  Mostly I don't want to be alone.  I want to have
friends \ldots\ and enemies to spice things up \ldots\ and generally
just plenty of people to talk to.\medskip

{\bf {\God}}:  Done.  I'll give you a world populated with loads of
other people.  But you ask for a bit of an engineering feat when you
ask to be able to talk to them.  If you want to communicate, the
world can't be too rigid; it has to be a sort of malleable thing. It
has to have enough looseness so that you can write the messages of
your choice into its properties.  It will make the world a little
more unpredictable than it might have been for me---I may not be able
to warn you about impending dangers like droughts and hurricanes
anymore---but I can do that if you want.\medskip

{\bf {\Adam}}:  Also {\God}, I would like there to be at least one
special someone---someone I can share all my innermost thoughts with,
the ones I'd like to keep secret from the rest of the world.\medskip

{\bf {\God}}:  Now you ask for a tall order!  You want to be able to
communicate with one person, and make sure that no one else is
listening?  How could I possibly do that without having you two
bifurcate into a world of your own, one with no contact whatsoever
with the original?  How about we cut a compromise?  Since I'm already
making the world malleable so that you can write your messages into
it, I'll also make it sensitive to unwanted eavesdropping.  I'll give
you a means for checking whether someone is listening in on your
conversations:  whenever information is gathered from your
communication carriers, there'll be a reciprocal loss in what you
could have said about them otherwise.  There'll be a disturbance.
Good enough?  You should be able to do something clever enough with
that to get by.\medskip

{\bf {\Adam}}:  Good enough!\medskip

{\bf {\God}}:  Then now I'll put you in a deep sleep, and when you
awake you'll have your world.\medskip

{\bf {\Adam}}:  Wait, wait!  I overlooked something!  I don't want an
unmanageable world, one that I'll never be able to get a scientific
theory of.  If whenever I gather information about some piece of the
world, my colleagues lose some of their information about it, how
will we ever come to agreement about what we see?  Maybe we'll never
be able to see eye to eye on anything.  What is science if it's not
seeing eye to eye after a sufficient amount of effort?  Have I doomed
myself to a world that is little more than chaos as far as my
description of it goes?\medskip

{\bf {\God}}:  No, actually you haven't.  I can do this for you:
I'll turn the information-disturbance tradeoff knob just to the point
where you'll still be able to do science.  What could be better?  You
have both privacy and science.\medskip

So {\Adam} fell into a deep sleep, and {\God} set about making a
world consistent with his desires.  And, poof(!), there was QUANTUM
MECHANICS.
\eq
Or this version of it from 2010:
\bq
Strictly speaking, meliorism is the doctrine ``that humans can, through their interference with processes that would otherwise be natural, produce an outcome which is an improvement over the aforementioned natural one.''  But we would be reluctant to take a stand on what ``improvement'' really means.  So said, all we mean in the present essay by meliorism is that the world before the agent is malleable to some extent---that his actions really can change it.  Adam said to God, ``I want the ability to write messages onto the world.''  God replied, ``You ask much of me.  If you want to write upon the world, it cannot be so rigid a thing as I had originally intended.  The world would have to have some malleability, with enough looseness for you to write upon its properties.  It will make your world more unpredictable than it would have been---I may not be able to warn you about impending dangers like droughts and hurricanes as effectively as I could have---but I can make it such if you want.'' And with that Adam brought all host of uncertainties to his life, but he gained a world where his deeds and actions mattered.
\eq

\section{11-04-10 \ \ {\it Templeton, Schlosshauer}\ \ \ (to A. M. Steinberg)} \label{Steinberg4}

\bams
Obviously, we'd be very happy to have any of your turncoats around.  Does
sound like philosophy or IHPST are the natural homes for him, but the more
he interacted with us in Physics, the happier we'd be.  I believe I could pretty
easily secure him a (shared) postdoctoral office if all else fell through, but that
might not be his first choice.
\eams

Thanks on Max.  With regard to Templeton, not to worry:  {\it Everyone\/} I know is applying for this.  There's no real competition between us unless you too are calling to make connections between quantum theory, William James's notion of ``pure experience,'' and Wolfgang Pauli's interest in alchemy and Jungian psychology.  You aren't \ldots\ are you?  Though I could imagine an experiment or two that'd tidy up some of those connections \ldots

\section{12-04-10 \ \ {\it Truth and Happening, 2}\ \ \ (to N. M. Boyd)} \label{Boyd2}

Thanks, I enjoyed that!

I'm glad you mentioned Richard Rorty.  I think there's just a wealth of good stuff in his writings (though I don't follow him on many things).  I have a long quote of him, that at least indicates what I find most useful to cherry-pick from him, in this pseudo paper:  ``Delirium Quantum,'' \arxiv{0906.1968}.

\section{12-04-10 \ \ {\it QB Glossary}\ \ \ (to the QBies)} \label{QBies14}

By the way, here's the finalized version of the QB glossary. \medskip

\noindent QBlue signing off!

\section{12-04-10 \ \ {\it Notes on a First and Second Reading}\ \ \ (to D. M. {\Appleby} and H. Barnum)} \label{Barnum32} \label{Appleby90}

Last night, I read Marcus's note ``quantum belief and quantum reality'' for the first time and Howard's paper for the second.  Call me pigheaded, but my reaction was like that of a Japanese friend who wrote recently, ``I do not understand your long sentence.''

In the part that I take issue with (i.e., {\it not\/} the part about meatiness, where I do believe there is much work to be done, see Footnote 46 in my recent posting), I only felt that I saw two laments---one of five thousand words, and one of ten thousand---that boiled down to saying a single sentence, ``In some cases, there {\it should\/} be a notion of a {\it right\/} probability assignment independent of the agent's personal mesh of beliefs.''  Why?  ``Well there should be.''  Why?  ``Well there should be!''

I continue to think such a notion is only trouble, a cloud that keeps us from seeing over the hill.  Why would I adopt it when I can do everything your gut tells you that you need to do without it?  1) I walk into a lab, make a judgment about this or that piece of equipment; the experimentalist tells me that I'm ``wrong.''  Fair enough:  But, he does that based on his personal mesh of beliefs.  2)  I use {\it my\/} mesh of beliefs to extract a probability-1 statement about some potential experience.  I wait for the experience, and it does not happen.  I say to myself, ``That was stupid; I was wrong!''  Fair enough:  But ``wrong'' comes {\it after\/} the fact, and only then can I see that I should readjust my mesh of beliefs.

Particularly, the ``wrong'' does not come before the actual event.  And, for me, that point is crucial for breaking the block.  The block universe.  What I see in your laments is a feeling that the probabilities should be right or wrong beforehand, and right or wrong based on something impersonal.

I accuse that idea of being a cloud.  When it comes to quantum mechanics particularly, it is a cloud that keeps us from having a solution to Wigner's friend already in hand.  And it leaves us with an action-at-a-distance that I find unacceptable (block issue again).  Howard admits as much with his last sentence:
\bq\noindent
We seem on solid ground if we wish to maintain that the fact about how
Alice should bet is not a fact about how things are at Bob's site.
It's not reasonable to say that quantum physics recommends that Bob
should immediately change his betting behavior to the one it
recommends for Alice. (Though I suppose it does claim that he'd be
better off if he did.)
\eq
And I tried to make this point at length on page 5, right column, and footnote 7, of my last big paper.  Furthermore, the issue has nothing to do with no-signalling or incompatibility with Lorentz invariance for me:  It is all about the issue of the block.  See page 18L.

Finally I'll say Marcus sizes me up wrong in the part of the discussion about my old phrase ``(and only a little more)''.  I agree with William James that ``the trail of the human serpent is over everything.''  Every statement of the form ``I hypothesize the world to be made (in part) of stuff X'' without doubt exhibits the trail of the human serpent.  But still such statements have a function, and that function can be better or worse served by different choices of X.  My {\it only\/} claim is that if one invokes a statement that boils down to ``I hypothesize that the world is made (in part) of probability'' (as you two seem to me to be doing) one does oneself no favor.

With regard to the trail of the human serpent, I believe that I agree with this paper of F. C. S. Schiller to a great extent:
\myurl[http://en.wikisource.org/wiki/Personal_Idealism/Axioms_as_Postulates]{http://en.wikisource.org/wiki/Personal\underline{ }Idealism/Axioms\underline{ }as\underline{ } Postulates}. Reading it might help diffuse the repeated accusal of my believing otherwise.

\section{13-04-10 \ \ {\it After Hans, 1}\ \ \ (to H. C. von Baeyer)} \label{Baeyer112}

Opening section now reads:
\bq
Of scientific theories, the philosopher William James once said, ``You cannot look on [a completed theory] as closing your quest. You must set it at work within the stream of your experience. It appears less as a solution than as a program for more work, and more particularly as an indication of the ways in which existing realities may be changed. Theories thus become instruments, not answers to enigmas. We move forward, and, on occasion, make nature over again by their aid.'' On this conception, a theory is not a statement about what the world is, but a tool, like a hammer, to aid in making the world what we want it to be. The world may resist, but to some extent it is malleable.

The research program of Quantum Bayesianism (QBism) is an approach to quantum interpretation that reveals, with mathematical precision not poetry, that quantum theory's greatest lesson is the world's plasticity. With every quantum measurement set by an experimenter's free will, the world is shaped just a little; and so of every action of every agent everywhere.

To develop QBism's mathematics, amalgamate its philosophical implications, put it before a wider jury, and write a book on the subject, we request funds for two postdocs, two international workshops, and regular visits from collaborators.
\eq

\section{13-04-10 \ \ {\it A Worry}\ \ \ (to H. C. von Baeyer)} \label{Baeyer113}

\bhcvb
The opening quote by James is wonderful, BUT the key phrase, for which you selected it, is in brackets, giving the impression that YOU supplied it to make it fit your purpose.  Maybe James was talking about something different from a theory!  Is there a way to use more of his own phrasing here, fixing it up with brackets only minimally?
\ehcvb

I guess strictly speaking, he was talking about ``metaphysics'' just before those sentences, but a couple of paragraphs down, he goes on to say:
\bq
One of the most successfully cultivated branches of philosophy in our time is what is called inductive logic, the study of the conditions under which our sciences have evolved. Writers on this subject have begun to show a singular unanimity as to what the laws of nature and elements of fact mean, when formulated by mathematicians, physicists and chemists. When the first mathematical, logical, and natural uniformities, the first laws, were discovered, men were so carried away by the clearness, beauty and simplification that resulted, that they believed themselves to have deciphered authentically the eternal thoughts of the Almighty. His mind also thundered and reverberated in syllogisms. He also thought in conic sections, squares and roots and ratios, and geometrized like Euclid. He made Kepler's laws for the planets to follow; he made velocity increase proportionally to the time in falling bodies; he made the law of the sines for light to obey when refracted; he established the classes, orders, families and genera of plants and animals, and fixed the distances between them. He thought the archetypes of all things, and devised their variations; and when we rediscover any one of these his wondrous institutions, we seize his mind in its very literal intention.

But as the sciences have developed farther, the notion has gained ground that most, perhaps all, of our laws are only approximations. The laws themselves, moreover, have grown so numerous that there is no counting them; and so many rival formulations are proposed in all the branches of science that investigators have become accustomed to the notion that no theory is absolutely a transcript of reality, but that any one of them may from some point of view be useful. Their great use is to summarize old facts and to lead to new ones. They are only a manmade language, a conceptual shorthand, as some one calls them, in which we write our reports of nature; and languages, as is well known, tolerate much choice of expression and many dialects.

Thus human arbitrariness has driven divine necessity from scientific logic. If I mention the names of Sigwart, Mach, Ostwald, Pearson, Milhaud, {\Poincare}, Duhem, Ruyssen, those of you who are students will easily identify the tendency I speak of, and will think of additional names.
\eq

That he could have been talking strictly of ``scientific theories'' in the passage from my Executive Summary follows from this line in the paragraphs above:
\bq
\noindent Their great use is to summarize old facts and to lead to new ones.
\eq
Thanks for worrying.  But I'm not quite sure what to do.

\section{13-04-10 \ \ {\it Possibly Final Version of Section 2}\ \ \ (to H. C. von Baeyer)} \label{Baeyer114}

I turned you into a writer and added one more Jamesian phrase at the very end.  0 characters remaining:
\bq
QBism is an interpretation and technical development of quantum theory due predominantly to C. M. {\Caves}, C. A. Fuchs and R. {\Schack}, with early contributions from N. D. {\Mermin} and A. Peres and historical roots in an eclectic mix of the thoughts from W. Pauli, J. A. Wheeler, and B. de Finetti.

\label{ForIntroduction2} Three characteristics set QBism apart from other existing interpretations of quantum mechanics. First is its crucial reliance on the mathematical tools of quantum information theory to reshape the look and feel of quantum theory's formal structure. Second is its stance that two levels of radical ``personalism'' are required to break the interpretational conundrums plaguing the theory. Third is its recognition that with the solution of the theory's conundrums, quantum theory does not reach an end, but is the start of a great journey.

The two levels of personalism refer to how the ``probabilities'' and ``measurement events'' of quantum theory are to be interpreted. With regard to quantum probabilities, QBism asserts that they be interpreted as personal, Bayesian degrees of belief. This is the idea that probability is not something out in the world that can be right or wrong, but a personal accounting of what one expects. The implications of this are deep, for one can see with the help of quantum information theory that it means that quantum states, too, are not things out in the world. Quantum states rather represent personal accounting, and two agents speaking of the same quantum system may have distinct state assignments for it. The second level of personalism appears with the meaning of a quantum-measurement outcome. On this QBism holds with Pauli that a measurement apparatus must be understood as an extension of the agent herself, not something foreign and separate. A quantum measurement device is like a prosthetic hand, and the outcome of a measurement is an unpredictable, undetermined ``experience'' shared between the agent and external system. Quantum theory, thus, is no mirror image of what the world is, but rather a ``user's manual'' that any agent can adopt for better navigation in a world suffused with creation: The agent uses it for her little part and participation in this creation.

The project we propose is to further develop this point of view, both mathematically and philosophically, by way of our own efforts as well as getting a wider community involved in its consideration. On the mathematical side, much work remains to develop a formalism that most crisply expresses QBism's worldview. Fuchs already has a group of six at Perimeter Institute tackling this head on--a key issue is understanding the structure of symmetric informationally complete (SIC) positive-operator measures. For with them, quantum theory takes the cleanest possible expression in terms of probabilities. A key collaborator D. M. {\Appleby} lives in England--travel and accommodation expenses would allow him to become more intimately involved with the group. Co-investigator {\Schack} would advise a postdoc in developing the notions of ``quantum randomness'' and ``private information'' from a purely Bayesian point of view. Its mathematics, we believe, particularly will take us to the next stage of understanding the ``interiority'' (Pauli) of quantum phenomena. Fuchs would advise proposed postdoc M. Schlosshauer in writing a full-length book on QBism, much like his well-recognized book on decoherence. On the philosophical side, we would organize two 10-day workshops. The first, with the help of Pauli scholar H. Atmanspacher and writer H. C von Baeyer, would be devoted to meshing Pauli's prosthetic-hand notion of measurement with modern developments. The second, in conjunction with the William James Society, would explore J. A. Wheeler's idea that ``the big bang is here all around us'' with each quantum measurement, which in the hands of QBism starts to look like a modern and promising version of James's ontology of ``pure experience'' where creation comes ``in spots and patches.''
\eq

\section{14-04-10 \ \ {\it Final Cut?}\ \ \ (to H. C. von Baeyer)} \label{Baeyer115}

Thanks a million!  (But quit telling me to break a leg!  The imagery is too vivid.)  On point number 1, you're right:  It's funny how many times I've had a little battle with my conscience before showing you something, and then when I do, you come out as ``the good'' voice that had already been in me but not listened to.  Anyway, it's stripped away!  On 2, I modified it to this:
\bq
QBism is an example of the unexpected places to which striving for consistency can lead. When Quantum Bayesianism started 15 years ago, all it attempted initially was to view quantum states as knowledge to dispel some foundational conundrums. But the urge for consistency kept pushing it further, all the way to the two-level personalism we have today. The full implications of this are starting to sink in: From an original reductionist disposition, we arrived at a NONreductionist view of one of the most fundamental theories of physics! See \arxiv{1003.5209}. We now see the many efforts to derive the ``quantum-to-classical transition'' proceeding backward. Instead, quantum theory should be seen as ``additive'' to common experience--the union of the two make something greater. This turns the traditional debate on its head. The proposed JTF project is part of a greater goal to develop the view until other physicists feel it in their bones and use it themselves to reshape physics.
\eq

Now, if I can get {\Ruediger}'s CV turned into .doc format without too much distortion, I'll be done!

\section{14-04-10 \ \ {\it Willing Free Will}\ \ \ (to D. Rideout)} \label{Rideout1}

If I understood you correctly on our walk to the bus stop, you think ``free will'' is a presupposition for science.  In any case, I certainly do!   See for instance the left column, page 18, of \arxiv{1003.5209}.

In case you're interested, attached is the thing I proposed to the Templeton Foundation today (actually it was a web form, but I weakly {\TeX}ified it---that's what I'm sending to you).  Honestly, I was rather impressed with John Templeton's Donor Intent; see:\medskip

\myurl{http://www.templeton.org/what_we_fund/donor_intent/philanthropic_vision/}.\medskip

Let's talk about this again some time.  It's my favorite subject on earth.

\section{16-04-10 \ \ {\it Dangerous Jaynesian Tendencies}\ \ \ (to C. Ferrie)} \label{Ferrie15}

I sent {\Ruediger} your term paper yesterday, and he replied thusly:
\brs
I just read his term paper. There is some very nice
stuff in it. He seems to have dangerous Jaynesian
tendencies, though.
\ers
Beware those Jaynesian tendencies!!

\section{16-04-10 \ \ {\it QBies Activate!}\ \ \ (to the QBies and R. {\Schack})} \label{QBies15} \label{Schack199.1}

Today, 1:00 group meeting.  It sounds like the big discussion will be on linear dependencies.

And never forget these lines from Henry V'th:
\bv
      This day is call'd the feast of Crispian.                       \\
      He that outlives this day, and comes safe home,                 \\
      Will stand a tip-toe when this day is nam'd,                    \\
      And rouse him at the name of Crispian.                          \\
      He that shall live this day, and see old age,                   \\
      Will yearly on the vigil feast his neighbours,                  \\
      And say `To-morrow is Saint Crispian.'                          \\
      Then will he strip his sleeve and show his scars,               \\
      And say `These wounds I had on Crispian's day.'                 \\
      Old men forget; yet all shall be forgot,                        \\
      But he'll remember, with advantages,                            \\
      What feats he did that day. Then shall our names,               \\
      Familiar in his mouth as household words---                     \\
      Harry the King, Bedford and Exeter,                             \\
      Warwick and Talbot, Salisbury and Gloucester---                 \\
      Be in their flowing cups freshly rememb'red.                    \\
      This story shall the good man teach his son;                    \\
      And Crispin Crispian shall ne'er go by,                         \\
      From this day to the ending of the world,                       \\
      But we in it shall be remembered---                             \\
      We few, we happy few, we band of brothers;                      \\
      For he to-day that sheds his blood with me                      \\
      Shall be my brother; be he ne'er so vile,                       \\
      This day shall gentle his condition;                            \\
      And gentlemen in England now-a-bed                              \\
      Shall think themselves accurs'd they were not here,             \\
      And hold their manhoods cheap whiles any speaks                 \\
      That fought with us upon Saint Crispin's day.
\ev

We will break this SIC problem one day!  With {\Ruediger} in the list, we are all seven here!\medskip

\section{17-04-10 \ \ {\it QBism's Foster Home}\ \ \ (to G. L. Comer)} \label{Comer129}

I was just about to listen to your new recording, but I think I've got a ``better yet.''  I'll listen to it tomorrow evening in The Carolina Inn.  I'll be in Chapel Hill for the first time in 10 years, to give the physics colloquium and a seminar Monday and Tuesday.  I'll listen to QBism from the old foster home (where the ideas were first fostered in me from reading a paper by Braunstein and {\Caves}).

\section{17-04-10 \ \ {\it World Elements} \ \ (to M. Schlosshauer)} \label{Schlosshauer34}

For a couple of days I had been planning to write you on the ``pure experience'' idea, but your last note, replying to ``Reasons Why'' stole the thunder.  You're on the right track, and what you wrote is about all I was going to write you anyway.

But this is all so sketchy at the moment.  You wrote:
\bmaxs
And I suppose this is where your link to James' ``pure experience''
comes in---provided I understand correctly what may be meant by
this term.
\emaxs

I think the way to treat James here is only as giving the hint of a hint.  And the major part of the synthesis will have to be done by us ourselves, informed by the actual structure of quantum mechanics.  This idea of ``pure experience'' (broadly construed) being the stuff of the world runs wide and deep in the particular lot of people I have taken the time to study over the years.  It can be found in the writings of James, Pauli, and Wheeler.   A good bit of the hidden Pauli, for instance in his personal letters to Fierz, Jung, and von Franz, was devoted to understanding this ``neutral stuff'' that quantum theory is the first indication of and dreaming of a ``neutral language'' to describe it all in.  Also, there's some of Bohr even in this regard---revisit the Folse quote I used in the last paper.

But still, much of this only starts to click for me now:  It took me nearly 26 years to get this far---i.e., the part where I start to get the hint of the hint and to see that it actually is in context (fitting for quantum mechanics).  The clock started when I met John Wheeler and first read things like the quote you quote.

Anyway, I feel like I need to make a big push on reading everyone who has contributed to the neutral-stuff idea.  And the group is bigger than I had previously imagined.  For instance, have a look at the preface and introduction to Erik Banks' book {\sl Ernst Mach's World Elements\/} on Google Books.  Also, I've recently learned that Bertrand Russell embraced good bits of the Jamesian idea in his two books {\sl Analysis of Mind\/} and {\sl Analysis of Matter}.  And too Ralph Barton Perry's ``New Realism.''  Also Shadworth Hodgson.  I feel like I need to get really coordinated on this, and go at it systematically for a bit, before I try to combine it with all this talk of dimension as a kind of valence.

That's where it stands at the moment.  Tomorrow morning I fly to Chapel Hill, North Carolina to give the UNC colloquium and a seminar.  Greetings to Kari and Eli as well.

\subsection{Max's Preply}

\bq
Well, it had already dawned on me over the past few days that my query could have something to do with the Jamesian conception of ``pure experience.'' There are certainly some hints of this in Section VIII of your paper; I did pick up on those earlier, but I feel the gears are only starting to mesh now. For example, there's the Wheeler quote that opens your section:
\bq\noindent
	Is the entirety of existence, rather than being built on
	particles or fields of force or multidimensional geometry,
	built upon billions upon billions of elementary quantum
	phenomena, those elementary acts of `observer-
	participancy'\ldots?
\eq

So I suppose that part of the answer to my query would be to refer to exactly this speculation of Wheeler's: that, in fact, there are no other ``physical processes'' besides creation-by-quantum-measurement, and that the primary ``ontology'' are these acts of creation. Then there would be no ``particles scattering other particles.'' It would be merely a figurative way of saying that two parts of the world have entered into a relationship by which they give rise to something new, namely, the ``experience'' that's shared between the two systems.

And so on.

The obvious next question would of course be: what precisely constitutes these ``experiences''? How do they affect the future evolution of the world?

And I suppose this is where your link to James' ``pure experience'' comes in---provided I understand correctly what may be meant by this term. For instance, the {\sl Stanford Encyclopedia of Philosophy\/} says:
\bq
James's fundamental idea is that mind and matter are both aspects of, or structures formed from, a more fundamental stuff---pure experience---that (despite being called ``experience'') is neither mental nor physical. Pure experience, James explains, is ``the immediate flux of life which furnishes the material to our later reflection with its conceptual categories \ldots\ a that which is not yet any definite what, tho' ready to be all sorts of whats \ldots'' (ERE, 46). That ``whats'' pure experience may be are minds and bodies, people and material objects, but this depends not on a fundamental ontological difference among these ``pure experiences,'' but on the relations into which they enter. Certain sequences of pure experiences constitute physical objects, and others constitute persons; but one pure experience (say the perception of a chair) may be part both of the sequence constituting the chair and of the sequence constituting a person. Indeed, one pure experience might be part of two distinct minds, as James explains in a chapter entitled ``How Two Minds Can Know One Thing.''
\eq
So, in this way of thinking, one would really close the circle back to the Wheeler quote!

(I like it!)
\eq

\section{19-04-10 \ \ {\it QBism}\ \ \ (to G. L. Comer)} \label{Comer130}

Boy did that take me back to some youthful experiences in old John Hamilton's dance hall in Cuero TX, when a band would be just freestyle jamming before or after their paid performance and nearly everyone was gone (but old David and me).  I have a particular memory in mind, though I can't express it adequately.

No doubt you caught an aspect of QBism---a world with creation all around, near and far.  There was something about your licks that had an almost spatial feel, as if I could feel things all around me.  That was really cool; thanks!

Things around here are significantly similar to the way they were 20 years ago (though all traces of Jim York are gone).  I went into The Cave last night and drank a PBR, and shopped at The Bookshop tonight.  Tonight's stash below.

Even Eugen Merzbacher was the same!  He's 89 now, and was without doubt my most attentive listener!  I have lunch with him tomorrow.

\begin{enumerate}
\item Murray N. Rothbard, {\sl Ludwig von Mises:\ Scholar, Creator, Hero}, PB, 6.50 USD.

\item Edward H. Madden, {\sl Chauncey Wright and the Foundations of Pragmatism}, HC, 8.50 USD.

\item George Herbert Mead, {\sl Movements of Thought in the Nineteenth Century:\ Works of George Herbert Mead, Vol.\ 2},  edited with introduction by Merritt H. Moore, PB, 6.50 USD.

\item Robert Avens, {\sl Imagination is Reality: Western Nirvana in Jung, Hillman, Barfield and Cassirer}, PB, 7.99 USD.

\item James Cambell, {\sl Understanding John Dewey:\ Nature and Cooperative Intelligence}, PB, 8.50 USD.

\item Donald D. Palmer, {\sl Structuralism and Poststructuralism for Beginners}, PB , 5.99 USD.

\item Jim Powell, {\sl Derrida for Beginners}, PB, 5.99 USD.

\item David Baggett and Shawn E. Klein, editors, {\sl Harry Potter and Philosophy:\ If Aristotle Ran Hogwarts}, PB, 8.99 USD.

\item Maurice Friedman, editor, {\sl The Worlds of Existentialism:\ A Critical Reader}, HC, 6.99 USD.

\item Philip P. Wiener, {\sl Evolution and the Founders of Pragmatism}, with foreword by John Dewey, PB, 3.99 USD (slated as gift for Lee Smolin).
\end{enumerate}

\section{19-04-10 \ \ {\it Pauli, Fechner, Schelling}\ \ \ (to H. C. von Baeyer)} \label{Baeyer116}

Good morning.  I'm a bit closer to your neck of the woods this morning---I'm in Chapel Hill, NC.  (Though there was no gravy for the biscuits this morning:  Maybe there are a bit too many souths for my taste.)

I'm reading a very nice book by Michael Heidelberger on Gustav Fechner, {\sl Nature from Within}.  Fechner was the father of psychophysics and a philosopher who argued that nature is permeated with souls.  I'm halfway thinking I might write a review on this book for you if you're interested in it.

For the moment though, I'm just curious:  We know that Pauli read Schopenhauer and Mach, but is there any evidence that he read Schelling, Oken, or Fechner?  (Mach was greatly influenced by Fechner.)  Anyway, all three of these guys thought nature was alive and in continuous creation, which you can guess intrigues me.

\subsection{Hans's Reply}

\bq
Welcome to the Confederacy.  Fechner is the very first reference in Karl von Meyenn's contribution to the Almanspacher/Primas book.  (The chapter is entitled ``Wolfgang Pauli's philosophical ideas viewed from the perspective of his correspondence.'')  So, to answer your questions I will hunt for clues in the yellow bible.

The idea of atoms with souls has a history reaching back to Lucretius.  Bernard Pullman, in {\sl The Atom in the History of Human Thought}, cites the lovely conception of Denis Diderot (1713--1784) of animate atoms. Diderot hopes that he will be buried next to his beloved Sophie, because his atoms and hers may still retain vestiges of feeling, and may try to mingle with hers.  Atoms in love!
\eq

\section{19-04-10 \ \ {\it Fuchs\underline{ }Penta-Chart\underline{ }2010(2).ppt}\ \ \ (to R. F. Wachter)} \label{Wachter2}

Sorry for the delay.  It looks pretty good.  Here are my answers to your queries. [\ldots]

3.  {\it Every\/} quantum measurement is a little act of creation.  Disturbance is too weak of a concept; it goes further than that.  In a wispier philosophical mood, I wrote recently:  ``The research program of Quantum Bayesianism (or QBism) is an approach to quantum interpretation that reveals, with mathematical precision not poetry, that quantum theory's greatest lesson is the world's plasticity. With every quantum measurement set by an experimenter's free will, the world is shaped just a little; and so with every action of every agent everywhere.''  Do you dare say that in front of your admirals?  You might enjoy the attached ``Glossary of QBism,'' particularly with respect to SICs, ``QBalah.'' [See 07-04-10 note ``\myref{Mermin173}{Up Front}'' to {\AA}.\ {\Ericsson}, N. D. {\Mermin}, and R. {\Schack}.]

\section{20-04-10 \ \ {\it Fuchs\underline{ }Penta-Chart\underline{ }2010(2).ppt, 2}\ \ \ (to R. F. Wachter)} \label{Wachter3}

\brfw
Enjoy this well-known alternative dictionary,
\begin{center}
\myurl{http://www.gutenberg.org/etext/972}.
\end{center}
\erfw

You compare my QBism glossary to the Devil's Dictionary?

\section{20-04-10 \ \ {\it Adam and God Again, Again}\ \ \ (to \v{C}. Brukner)} \label{Brukner8}

I just booked my flights to Jeff Bub's meeting.  So I'll actually be there.  For a bit I wavered because I had waited so long in booking my hotel that the price had jumped up to \$272/night.  But {\Spekkens} is letting me room with him; so that took the sting out of it.  Now I get to witness the fireworks between you and Tumulka first-hand!

On this trip, I've been reading a nice book on the life and thought of the physicist/philosopher Gustav Fechner.  (I'm in North Carolina at the moment.)  I'm finding many things thought-provoking in it.  Here is a little passage on the thought of theologian Christian Hermann Weisse:
\bq
   All that is real originates in acts of freedom,
   originates in ``voluntary action in general,'' either
   acts of God or acts of his creation.  These acts are
   voluntary because they totally lack subjection to
   necessity.  In creating the world, God willingly limits
   his own power and thereby continuously establishes
   the spontaneity and freedom of the beings he created.
   Spontaneity and freedom of action occur within time
   and thus make God a historical being, unfolding
   himself over time.  These two traits also restrict
   his capacity to entirely foresee the future.\footnote{\editornote From M.\ Heidelberger, {\sl Nature from Within: Gustav Theodor Fechner and His Psychophysical Worldview} (Princeton University Press, 2004).}
\eq
It strikes me as hauntingly similar to my Adam/God story.  I didn't realize that I'd be branching into reading theology soon!

\section{20-04-10 \ \ {\it My Itinerary for DC}\ \ \ (to R. W. {\Spekkens})} \label{Spekkens84}

I've been talking with Eugen Merzbacher today:  Wonderful firsthand stories of Einstein, Oppie, Lothar Nordheim (now there's a name for you to search your memory), Wheeler, Rosen, Podolsky, Glauber, Bryce DeWitt, Everett, Wigner, Dirac.  Very nice.

\section{21-04-10 \ \ {\it Very Nice}  \ \ (to Y. J. Ng)} \label{Ng7}

I just finished reading the von Baeyer article.  Your idea with Henk was very nice.  I am just about to board the plane.  I bought six more books last night.  Everything fit in one suitcase.  Thanks again for all the hospitality---I think that is the best I have ever been treated just for giving a couple of seminars!

(And lobby your department chair to instate a modern quantum information course; it is sad when one of the {\sl few\/} people at the colloquium to already be aware of generalized quantum measurements was an 89 year old!)

\section{21-04-10 \ \ {\it World Elements, 2} \ \ (to M. Schlosshauer)} \label{Schlosshauer35}

Now you flatter me too much.  Lost wanderers shouldn't get such easy rides; it's not healthy.  I just think there is something solid to work on here.

How is your Everettian analysis going?  Are you still working on that somewhere on the side?  I do hope you'll finish that.  This morning I'm going to read Zeh's article, ``Feynman's Quantum Theory.''  Before the Los Alamos fire, I had a copy of the 1957 conference proceedings he discusses, and I was aware of the dialogue he cites (nearly 20 years ago!).  I happened upon it at the University of North Carolina in a room that was opened briefly so that an old, mostly out-of-commission copier could be used while the main one was broken.  There was an old file cabinet in the room labeled ``preprints,'' so I rummaged through it.  Lots of stuff from when Bryce DeWitt was still there, and at the very back was a folder with two copies of this proceedings.  One was labeled, ``Last Copy. Do Not Take.''  But the other one said nothing, and I took it.  (I later told Jim York about this, and he took the last copy, ignoring the warning label.)

I had some nice discussions with Eugen Merzbacher, who was the first to {\it really\/} teach me quantum mechanics.  He's 89 now, and was actually the most attentive person in my colloquium.  He moves slowly, but his mind is as sharp as it was when I first met him, without exaggeration.  It was nice to hear so many first-hand stories of Einstein, Oppenheimer, Lothar Nordheim (now there's a name for you to search for -- it was at his house that Wheeler's much advertised version of twenty questions really happened), Wheeler, Teller, Rosen, Podolsky, Glauber, DeWitt, Everett, Wigner, Dirac, Schlosshauer!, and others.  Merzbacher gave me a first edition of von Neumann's book that he had had bound long ago.  It's in beautiful shape.

\section{21-04-10 \ \ {\it Sci Am}\ \ \ (to H. C. von Baeyer)} \label{Baeyer117}

We never did write that Pauli article.  (My guess at this stage is that, unless I sketch a first draft, it's not going to happen.  But I might be wrong?)  Anyway, beside it or on top of it or instead of it, here's an idea:  See correspondence with George Musser from {\sl Scientific American\/} below, where he writes:
\bq
I'm not sure this paper lends itself to straightforward abridgment for
a Sci Am article, though, because the style is that of a manifesto --
an effort to rally one's colleagues, address the misgivings they have,
and set them on the right path. A general reader (and I include myself
in that category) is coming to this from a very different point of view.
We'd need a more straightforward account: here are the mysteries of
quantum mechanics; here are the usual responses; here is why those
responses fail; here is a better way. I suspect that such a reader
will actually be more receptive to the notion that quantum states
represent a form of gambling odds than a physicist or philosopher
steeped in the subject is; we have less baggage to shed. {\it But\/} we {\it do\/}
share the desire to know ultimately what the information is about and,
even if this question cannot be answered, will want to know how the
Bayesian approach advances that goal.
\eq
Trouble is, I don't think I know how to write something that is {\it not\/} a manifesto---it is a character flaw.  That's where your writing discipline could come in.  What would you think about us throwing in together on something like this?  Does the idea evoke any emotions in you?

\section{21-04-10 \ \ {\it Book Review Notes}\ \ \ (to H. C. von Baeyer)} \label{Baeyer118}

Notes from a mid-flight reading below:

At the nine-volume remark, should mention or cite von Meyenn \ldots\ at the very least so that readers who do not know the word ``Briefwechsel'' will know what to search on.

I've never liked it when people say Pauli and Jung ``co-authored'' a book.  It gives the sense that they wrote the two papers together.  It is a book of two papers by two distinct authors.  They didn't even write an introduction together, did they?

Planck length:  Miller didn't say that in the context of quantum gravity?  Was there a discussion of gravity about?

Your page 100 remark, about Pauli getting to the Born Rule before
Born.  I have a faint (actually not so very faint) memory that Miller
may be close to right on this.  Trouble is, at the moment, I can't
remember my source.  It was a scholarly article by Fine or someone
else (probably not Fine).  Anyway, the point they made was that Born
received his Nobel prize so very late because everyone in Copenhagen
had already understood probabilities for measurement results as
squared amplitudes.  It was a vestige of the Bohr--Kramers--Slater
theory, and even some Einsteinian work earlier than that.  So Born's
Nobel-prize referees (surely Bohr was in that lot) were not so very
keen to give him a Nobel prize on that in any great hurry.  I have a
rather strong impression of this.\footnote{\editornote The article in
  question may have been A. Pais (1982), ``Max Born's Statistical
  Interpretation of Quantum Mechanics,'' Science 218 (4578) 1193--98.
\begin{quotation}
\noindent {\bf Paisism 1:} I will return shortly to the significant
fact that Born originally associated probability with $\Phi_{mn}$
rather than with $|\Phi_{mn}|^2$.  As I learned from recent private
discussions, Dirac had the same idea at that time.  So did Wigner, who
told me that some sort of probability interpretation was then on the
minds of several people and that he, too, had thought of identifying
$\Phi_{mn}$ or $|\Phi_{mn}|$ with a probability.  When Born's paper
came out and $|\Phi_{mn}|^2$ turned out to be the relevant quantity,
``I was taken aback but soon realized that Born was right,'' Wigner
said.
\end{quotation}
Pais then has some interesting speculations about which Einsteinian work
was influential on Born.  Einstein's 1916 work on radiation theory, which
may have been on other people's minds and (along with
Bohr--Kramers--Slater) had them thinking of probabilities as squared
amplitudes, wasn't what was on Born's mind, at least not initially,
suggests Pais.  Instead, Born was thinking of some unpublished Einstein
work of the early 1920s, which explains why Born originally fingered
$\Phi_{mn}$ as a probability but then turned to $|\Phi_{mn}|^2$.  The part
which maybe pertains to the ``why was Born's Nobel so late?''\ question is in
the postscript.
\begin{quotation}
\noindent {\bf Paisism 2:} It is a bit odd---and caused Born some
chagrin---that his papers on the probability concept were not always
adequately acknowledged in the early days.  Heisenberg's own version
of the probability interpretation, written in Copenhagen in November
1926, does not mention Born.  One finds no reference to Born's work in
the two editions of Mott and Massey's book on atomic collisions, nor
in Kramers' book on quantum mechanics.  In his authoritative
\emph{Handbuch der Physik} article of 1933, Pauli refers to this
contribution by Born only in passing, in a footnote. [Ha! ---Ed.]
J\"orgen Kalckar from Copenhagen wrote to me about his recollections
of discussions with Bohr on this issue.  ``Bohr said that as soon as
Schroedinger had demonstrated the equivalence between his wave
mechanics and Heisenberg's matrix mechanics, the `interpretation' of
the wave function was obvious\ldots\ For this reason, Born's paper was
received without surprise in Copenhagen. `We had never dreamt that it
could be otherwise,' Bohr said.''  A similar comment was made by Mott:
``Perhaps the probability interpretation was the most important of all
[of Born's contributions to quantum mechanics], but given
Schroedinger, de Broglie, and the experimental results, this must have
been very quickly apparent to everyone, and in fact when I worked in
Copenhagen in 1928 it was already called the `Copenhagen
interpretation'---I do not think I ever realized that Born was the
first to put it forward'' [from Mott's introduction to Born's \emph{My
    Life and My Views} (1968)].  In response to a query, Casimir, who
started his university studies in 1926, wrote to me: ``I learned the
Schroedinger equation simultaneously with the interpretation.  It is
curious that I do not recall that Born was especially referred to.  He
was of course mentioned as co-creator of matrix mechanics.''  The same
comments apply to my own university education, which started a decade
later.
\end{quotation}
This is the closest approximation I've seen to the sentiments
attributed to Fine or somebody else (probably not Fine).}  I believe
there is a letter from Einstein to Born in their letter collection,
where Einstein laments the lateness with which the prize
came.\footnote{\editornote The following can be found beginning on
  p.~228 of \emph{The Born--Einstein Letters} (1971), where it is
  letter~117.
\begin{quotation}
\noindent \emph{Dear Born}

I was very pleased to hear that you have been awarded the Nobel Prize,
although strangely belatedly, for your fundamental contributions to
the present quantum theory.  In particular, of course, it was your
subsequent statistical interpretation of the description which has
decisively clarified our thinking.  It seems to me there is no doubt
about this at all, in spite of our inconclusive correspondence on the
subject.
\end{quotation}
See \myurl{http://archive.org/details/TheBornEinsteinLetters}.}  But
Einstein was not part of the Copenhagen crowd.  It'd be interesting to
look up what Pauli wrote to Born about the award, and what Pauli wrote
to his other friends at about that time.

\section{23-04-10 \ \ {\it WRT Your Talk on Quantum-Bayesianism} \ \ (to I. N. Hincks)} \label{Hincks1}

I never did get back to you on your queries.  I hope you've found some of the answers to your questions in a re-reading of the paper.  But I did want to tell you how beautiful I found the closing sentences in your note.  In fact, I have made use of them, and I hope you don't mind.  Please see the attached proposal to the Templeton Foundation; I quote you in Section 7.  If I ever use them again in a more public document, I will certainly cite you by name.  Thank you for inspiring me.

\subsection{Ian's Preply}

\bq
Your talk today turned some things up side down for me.

I find the underlying principle of the thing rock solid, at least for now. I have questions of clarification below. (If you have already written down the answers somewhere, please refer me there.)

\begin{enumerate}
\item
On the first page of the essay you handed out was a quote from Bell, about the qualifications involved in being an observer (does it need a PhD, etc.). Does your theory have a set of such qualifications? I imagine your response might be (it's fun to put words into other people's mouths) that such a list is not necessary, it is after all, a single-user theory, and that, perhaps, the trick is to use empirical observation to infer that (most) other humans are qualified to be agents. And perhaps an ordinary rock is just as capable an observer, but having no useful way of communicating with a rock, we choose to draw no conclusion. But let's remove the focus from other (potential) observers, or agents, onto ourselves. What I would like to get at is this: your overhead slide with the stick-man with a moustache and fan hands was drawn by you as, well, a stick figure. Was the reason for this iconic, or literal? I can perform measurements with my dial hands. If I get into an accident I can perform (albeit, more slowly) observations with just my left hand. But with no hands, what am I? Well, I get robotic arms implanted to continue with my measurements. My point is that your theory seems, at least to me, that your diagram must have been iconic, simply because it is hard to ``nail the soul to the body'' in a consistent way. (I had some of Walt Whitman's words, from {\sl I Sing the Body Electric}, running through my head during your talk today: ``And if the body were not the soul, what is the soul?'')
\item
This is how I understood one of your points: it is unfair to necessitate the deconstruction of the shoe into its constituents (being ``elementary particles''), and call this deconstruction the shoe. Rather, a shoe is a shoe, and an elementary particle is an elementary particle, and when we are lucky, physics will give us relationships between the two. This has a tinge of Platonism to me. It seems like ``shoeness'' must exist in a realm, and it must be the physical realm since this is where our fan hands measure it. Would you agree? And if you substitute ``Love'' for ``shoe'', where is Love's realm? The physical realm? These are earnest questions I have, they are not posed as a criticism.
\end{enumerate}

I like your theory because it returns to me as much freedom as I feel that I have. Such freedom is lost or partially lost in other interpretations.
\eq

\section{24-04-10 \ \ {\it Implications} \ \ (to N. D. {\Mermin})} \label{Mermin186}

How's Berlin?  I just wrote the note below to Huw Price, and because of it, I reread the thing I had sent off to Templeton last week.  It dawned on me that maybe you should know how I have implicated you in this business!  You'll find your name mentioned at the top of Section 3.  Thanks for all the good tension over the years!

\section{24-04-10 \ \ {\it Weyl's Book and John Wheeler}\ \ \ (to the QBies)} \label{QBies16}

\begin{itemize}
\item
1928. {\sl Gruppentheorie und Quantenmechanik}. transl.\ by H. P. Robertson, {\it The Theory of Groups and Quantum Mechanics}, 1931, rept.\ 1950 Dover.
\end{itemize}

Wheeler, on the other hand, was born in 1911 and got his PhD in 1933 (having skipped a bachelor's degree).  He wrote:
\bq
I first knew Weyl before I first knew him. Picture a youth of nineteen seated in a Vermont hillside pasture, at his family's summer place, with grazing cows around, studying Weyl's great book, {\sl Theory of Groups and Quantum Mechanics}, sentence by sentence, in the original German edition, day after day, week after week. That was one student's introduction to quantum theory. And what an introduction it was! His style is that of a smiling figure on horseback, cutting a clean way through, on a beautiful path, with a swift bright sword. Some years ago I was asked, like others, I am sure, to present to the Library of the American Philosophical Society the four books that had most influenced me. {\sl Theory of Groups and Quantum Mechanics\/} was not last on my list. That book has, each time I read it, some great new message.\footnote{\editornote From J.\ A.\ Wheeler, ``Hermann Weyl and the Unity of Knowledge,'' American Scientist \textbf{74,} pp.~366--75 (1986).}
\eq
I never realized that Weyl's book was written before von Neumann's.  Indeed, it was even written before Dirac's!  (von Neumann's was 1932, and Dirac's was 1930.)

And if I'm not mistaken, our beloved Weyl--Heisenberg group goes back to that book of Weyl's.  Does anyone know where to find it in there?  Despite Andrew Gleason's spirit residing in my copy, he hasn't given me enough hints to find the right section on perusal.

\section{24-04-10 \ \ {\it Saturday Morning Alchemy}\ \ \ (to the QBies)} \label{QBies17}

It's one of those mornings when I find myself thinking that what we're doing is {\it really}, {\it really\/} important.  Because it goes to the very heart of what matter is.  Lately, I've become fixated on the imagery of ``the philosopher's stone'' as the root of the right way to think of quantum systems.  My daughter Emma was reading to me from a book {\sl The Sorcerer's Companion\/} about the real-life sources of the things in Harry Potter, and it put me in an extraordinarily sentimental/contemplative mood.  ``Medieval alchemists like Nicholas Flamel now spoke of producing a new substance---an extraordinarily potent catalyst that when added to common metals would trigger their transmutation \ldots''

I usually only share this earthy base of my thinking with Marcus, but I'm feeling rather open mouthed this morning.  I say ``earthy'' because of this thing Marcus wrote a couple weeks back:
\bq\noindent
   One of the difficulties with this programme is that a
   building has to rest on something.  Actual modernistic
   buildings rest on the same wormy, fungus and bacteria-
   impregnated earth that the medievals used to build
   their mud-huts on.  With the modernistic structures we
   build in our minds it is even worse:  they don't
   simply rest on the same soil, they actually rest on
   the mud hut itself, entire and unreconstructed.
   Despise the mud-hut at the bottom of the whole thing,
   as contemporary physicists almost all do, and you are
   despising the foundations on which everything else
   depends.
\eq

The earthy is definitely welling up in me lately.  For instance, here's a blurb I wrote for the back of the new edition of Nielsen and Chuang's book:
\bq
Nearly every child who has read Harry Potter believes that if you just say the right thing or do the right thing, you can coerce matter to do something fantastic.  But what adult would believe it?  Until quantum computation and quantum information came along in the early 1990s, nearly none.  The quantum computer is the Philosopher's Stone of our century, and Nielsen and Chuang is our basic book of incantations.  Ten years have passed since its publication, and it is as basic to the field as it ever was.  Matter will do wonderful things if asked to, but we must first understand its language.  No book written since (there was no before) does the job of teaching the language of quantum theory's possibilities like Nielsen and Chuang's.
\eq
I guess I just want to say something nearly the same to you QBies:  Believe in the philosopher's stone!

Attached is an extract of the pre-proposal I sent off to the Templeton Foundation last week.

Top of the mornin' to you,

\section{24-04-10 \ \ {\it Weyl's Book and John Wheeler, 2}\ \ \ (to the QBies)} \label{QBies18}

The section in Weyl's book seems to be Section IV.D.14, starting on page 272.  I'm not quite sure what it is saying, but it sounds awfully deep.  It seems he views the generalized Pauli operators as the very defining feature of a quantum system, and bases everything on the consideration of them.  I think it means SICs really do form the right ``phase space'' for QM.

\section{25-04-10 \ \ {\it The Importance of Gadflycity} \ \ (to N. D. {\Mermin})} \label{Mermin187}

Glad Berlin is living up to your dreams.  That's one German city I've never been to.

\bdm
I'm honored to be mentioned with Asher as a kind of
godfather of Qbism, but I'm not sure what I did to
deserve the accolade.  I think of myself as more of
a gadfly.  At best.
\edm

I thought long and hard about putting you in that line, but decided that if I were to say something about Asher, I must also say something about you.  Do not underestimate the contributions of a good gadfly---that was exactly your role; you hit the nail on the head.   And I'm proud to say it; if it weren't for the tension you (more than anyone else) have given me, QBism would be just another lazy, half-way interpretation of quantum theory.   Indeed another line in the proposal was an implicit statement about your gadflycity:  ``QBism is an example of the unexpected places to which striving for consistency can lead. When Quantum Bayesianism started 15 years ago, all it attempted initially was to view quantum states as knowledge to dispel some foundational conundrums. But the urge for consistency kept pushing it further, all the way to the two-level personalism we have today.''  The ``urge for consistency'' was really just ``saving face'' in light of your queries.

\bdm
Does Templeton require a religious twist? Is it
necessary that Qbism end up pointing the finger
at God?
\edm

They do require an applicant to declare the relation between his proposal and ``Sir John's Donor Intent.''  See Section 13 in the thing I sent you.  But I enjoyed reading his donor intent, and found it quite liberal and humble.  My reading is that it says, in essence, never throw away a possible source of enlightenment---no matter the source.  My reading of it didn't sound a lot different (though less thorough and eloquent) from my readings of William James.  Anyway, everything I said in this proposal is genuine me, and if that is religious, it is that I'm not going to let the world completely tell me what to do:  The world must listen to man (in the end), as much as man must listen to the world.  If someone construes  that saying agency is real, has something to do with pointing a finger at God, then QBism is already pointing there.  That's the thing that spooks Norsen and {\Spekkens} (and {\Caves}?)\ and the like---they think that if science doesn't reduce mankind's actions to a kind of nothingness, then all hope is lost.  But of course, I just think QBism points to agency, NONreductionism, and plasticity:  It points to the idea that the world is malleable.  I wouldn't call that God.

\bdm
How did you hook up with Max Schlosshauer?  I met him
for the first time in Copenhagen last year, and liked
him a lot.
\edm

That's a story worth recording on a slow Sunday.  My first contact with him was pretty inconsequential; I had just asked him to comment on a Kofler/Brukner paper that I think does the quantum-to-classical transition better than anything the decoherence program had ever dreamed up.  Max's reply was predictable (defensive of the status quo) and not enlightening.

It was really my second contact with him that got things going in a positive direction.  He wrote me in his capacity as an editor, asking if I would consider writing a book on QBism.  In that note, he wrote furthermore:
\bq
First, I'd like you to know that for quite some time
I have really enjoyed following the development of
the quantum Bayesian program and the intellectual
discourse surrounding it. I have found your writings
both illuminating and entertaining, and I feel that
they have instilled in me an ever-increasing
appreciation of the Bayesian viewpoint and of the
motivations and philosophical attitudes underlying
the approach. You see, I have been doing a lot of
work in the decoherence field, and as you know
several of the original forces in this area are
openly or semi-openly Everettian, sometimes to the
point of where one may get the impression that an
Everett view is somehow {\it implied\/} by the insights
brought about by decoherence. Even I had not been
immune to this influence for quite some time, though
I finally feel some recovery coming on; the fog is
lifting.
\eq
I thought, ``Yeah right, good salesman; you just want me to write a book for this series.''  And I pretty much wrote it off as a sales tactic.

But then I saw him at the {\Vaxjo} meeting last year, and gave him a copy of the QBcoherence paper that you refereed.  I thought I was just being devilish.  (Zurek was there too, and Max was hanging out with him mostly.)  To my surprise, throughout the week, I spied Max actually reading the paper during the talks.  On my return home, I stopped in London for a night with {\Ruediger} and reported all this to him, saying how it surprised me and that maybe ``he's not a fake'' after all.  Then, while in {\Ruediger}'s living room, I looked up Max's webpage and found to even more surprise this line: ``More recently, I have become interested in information-theoretic axiomatizations of quantum mechanics, quantum Bayesianism, and generalized quantum-like theories.''  I said to {\Ruediger}, ``Fakes don't put lines like that in their webpages.''

Thus we were off to a new start.  Sometime later, Max wrote asking for what of William James I recommend he read.  Lots of correspondence ensued.  I think maybe what really capped off this transition in him was when I said that my vision of the world was not so different than what happens in jazz improv.  That seemed to go to his bone.  (You know that he's a jazz musician, don't you?)  Since then, he's been a good force behind QBism, helping keep me honest, pushing me to expand points and write more clearly, etc. Then at some point he brought up the idea that he write a book on QBism, and I ran with it.

That's the story roughly.

\bdm
I just turned 75.
\edm
Congratulations.  You are the most un-reified 75-year-old I have ever met.  I'll be happy when you're ready to take up that gadfly role again!  The program needs you.

All the best from drizzly Waterloo.

\section{27-04-10 \ \ {\it Midnight Reading}\ \ \ (to H. C. von Baeyer)} \label{Baeyer119}

Attached are a couple of articles you might be interested in.  Particularly the Bernstein article.  I like the way you have your review written now, and I wouldn't want you to change it, but midnight curiosity still drives me to find some evidence for what I claimed of Pauli vs Born.  (I am not there yet, but I might be closing in \ldots\ or giving up \ldots\ we shall see what another hour gives.)

Anyway, reading these articles was quite nice for all sorts of reasons for me.  Most particularly, I have recently taken an interest in the historical development of our beloved ``displacement operators'' or ``generalized Pauli operators,'' by which we generate SICs.  I had the most pleasant shock the other day:  Weyl in his 1928 book declares, roughly, ``A quantum system is one for which there are two quantities $A$ and $B$ that satisfy, $AB = e^{i\phi}BA$.''  I.e., He {\it defines\/} a quantum system by the existence of our generalized Pauli operators!  (And consequently with hindsight some time from now, if all works out for us, by the existence of SICs.)  See Section IV.D.14, starting on page 272, Dover edition, and notes below.  I don't have the book with me at the moment, but I {\it believe\/} he takes this as the essence of Heisenberg's postulate (he calls it) when it comes to finite dimensions.  Thus it was a real surprise to learn that this commutation relation (in ``Heisenberg form'') is actually Born and Jordan's and inscribed on Born's gravestone!

Of course, it has to be that after all this work of ours we will end up back at quantum mechanics, but it is interesting to see how some of the earliest math in one of the earliest formulations of the theory (Weyl's) came so close to what we are working with today.

By the way, the remark below about Gleason is referring to the fact that, thanks to a chance encounter of one of my students in Powell's Bookstore in Portland (and his reticence to buy it), I now own Andrew Gleason's copy of Weyl's book.  He told me the story of the stamp in the book, and I got on the web and bought it right away!

\section{27-04-10 \ \ {\it Questions for Aaron's Consideration}\ \ \ (to A. Fenyes)} \label{Fenyes3}

Here's a document I hastily threw together this morning.  Could we meet to discuss some tomorrow?  I will be out of town Thursday and Friday.

\bq
Here are some things I'd be quite interested in knowing the answer to.  If anything strikes you, let's work on it.  All three questions are geared toward pieces of rederiving quantum-state space from urungleichung considerations.  If these question don't lead to directions for progress, and something else strikes you, feel free to modify at will. [\ldots]

\subsection{3. Building Up Quantum-State Space One Simplex at a Time?}

Let me call a set $\mathcal S$ of probability distributions {\it consistent\/} if no two points violate
\be
\frac{1}{d(d+1)}\le \sum_r p_r q_r \le \frac{2}{d(d+1)}\;.
\label{Hamstring}
\ee
(All probability distributions $p_r$ assumed to be over $d^2$ points.)  Let me call such a set {\it maximal\/} if no further probability distribution can be added to $\mathcal S$ without violating (\ref{Hamstring}), but otherwise the set must be as full as can be.  One can prove that all maximal consistent sets must be convex and closed.  (See paper with Appleby and Ericsson.)  Moreover, quantum-state space is an example of a maximal consistent set.

Tell me as much as you can about maximal consistent sets with the following extra bit of structure:  Every extreme point $\vec{p}\in\mathcal S$ must be embeddable in some set of $d^2$ extreme points $\vec{p}_\alpha$ that form a regular simplex.\footnote{This appears to be a more careful statement than what I was guilty of so many other times. See Footnote \ref{OhCrap}, as well as Footnotes \ref{OhCrap2}, \ref{OhCrap3}, \ref{OhCrap4}, \ref{OhCrap5}, and \ref{OhCrap6}.}  I.e., for every extreme $\vec{p}$ there must exist some set $\vec{p}_\alpha$ of extreme points with $(\vec{c}-\vec{p}_\alpha)\cdot(\vec{c}-\vec{p}_\beta)=\mbox{constant}$ (with $\alpha\ne\beta$ and $\vec{c}$ being the center point of the probability simplex), and $\vec{p}=\vec{p}_\alpha$ for one of the values.  That is to say, what happens if we build up a maximal consistent set $\mathcal S$ one regular simplex at time?  What further properties does it share with quantum-state space?  Might it just be isomorphic to quantum-state space?!?
\eq

\section{28-04-10 \ \ {\it Stueckelberg??}\ \ \ (to L. Hardy)} \label{Hardy40}

\blh
Do you know what Stueckelberg did in his paper on real Hilbert space quantum theory?  I haven't read it and it doesn't seem to be available on line.  In particular, did he come up with the notion of local tomography?  Bill and I are hoping to finish our paper in the next few days and it would be good to cite Stueckelberg properly.
\elh

No, Stueckelberg didn't have that notion.  He just showed that one could recover ``single system'' quantum mechanics (or ``full system'' rather) using a real Hilbert space, plus superselection rule.  Particularly demanding that all observables commute with a certain operator $J$ such that $J^2=-I$.

I think the first person to use a local tomography axiom independently of Bill was Araki.  It should be at the end of this paper if recall correctly:  H. Araki, ``On a Characterization of the State Space of Quantum Mechanics,'' Comm.\ Math.\ Phys.\ {\bf 75}, 1 (1980).  He noted the usual problem in fields other than the complex.

Stueckelberg, by the way, went quite insane at some point.  I read a story a while back about how he showed up at a conference once thinking (and acting) that he was a horse. He was eventually institutionalized.  Maybe it means you shouldn't think about real Hilbert spaces too much!

\section{29-04-10 \ \ {\it  Spectra of $G=\sum_{r=1}^{d}\Pi_{r}$} \ \ (to M. A. Graydon)} \label{Graydon8}

Thanks Matthew!  Interesting, especially $d=5$.  I've been reading on your Emersonian essay again, with my after-dinner beer.  (Demopoulos's essay had more urgency, since I have to meet with him tomorrow.  So he got the dinner reading.)  I've just gotten to the Wigner's friend part, which I haven't started yet.  That'll have to wait for another quiet spell.  Eventually, I'll write you up some notes on it.  You do a really good job, but I'll nitpick some.

I'm steeling myself for Spekkens's arrival.  We're rooming together for this conference; he should arrive in a half hour.  Imagine the shear in the spacetime continuum of this room tonight:  The left-hand side a block universe, the right-hand side an anti-block!  Something strange is bound to happen.

\section{04-05-10 \ \ {\it Gout and Regout}\ \ \ (to W. G. {\Demopoulos})} \label{Demopoulos34}

It's slowly getting better.  I soaked it in Epsom salts water a couple times yesterday.  Sunday was absolute misery getting through the airports---from Washington Reagan, I had a connection in Chicago O'Hare.  Getting from H12 to G8 was a triumph of human will!  But by yesterday afternoon, I was able to stand on it well enough to pull dandelions (of which there has been an explosion here that I can't afford to turn my back on).

I'm glad to hear you're back home safely, and enjoyed my time with you this weekend.  I had said that I would be in London for the Saturday talks of Harperfest, but I'm starting to have a few misgivings:  There is so much that needs to be done in this yard, and I hate to see another weekend evaporate.  (I've been gone the last two for conferences.)

Yesterday, I found Bertrand Russell's {\sl Collected Papers}, Vol 7, which I had ordered ages ago.  So, I stopped reading my book on Fechner for the evening to indulge in a little Russellania.  The reason I bought this volume is because it contains the only book-sized manuscript that he never published, and it records his transition from being against to being for William James's (and Mach's) ``neutral monism'' (as Russell called it).

The thing in {\Schroedinger}'s {\sl Nature and the Greeks\/} I was telling you about is what he calls the ``principle of objectivation'':
\bq
By this [i.e., by ``the principle of objectivation''] I mean what is also frequently called the `hypothesis of the real world' around us. I maintain that it amounts to a certain simplification, which we adopt in order to master the infinitely intricate problem of nature.  Without being aware of it and without being rigorously systematic about it, we exclude the Subject of Cognizance from the domain of nature that we endeavor to understand. We step with our own person back into the part of an onlooker who does not belong to the world, which by this very procedure becomes an objective world.
\eq
He also goes on about it in a lengthier way in {\sl What is Life?}  As I recall, it is all really good reading.  Though of course, with my present pragmatism and radical empiricism, I don't much subscribe to the principle, at least when it comes to quantum issues:  QBism's view of quantum theory as a normative single-agent ``user's manual'' leaves the theory necessarily in great part about the agent himself.

I will dig up the Forman article for you, about {\Schroedinger}'s and Weyl's and Exner's and others early leanings toward indeterminism, once the library restores my privileges to retrieve things online.  (I discovered that I was overdue with some books.)

\section{04-05-10 \ \ {\it Gout and Regout, 2}\ \ \ (to W. G. {\Demopoulos})} \label{Demopoulos35}

\bwd
Russell's {\bf Theory of Knowledge} ms is somewhat misleadingly titled since it's really more about his theory of propositions and propositional understanding than anything else. I believe the best development of the views of his that you're interested in is in {\bf The Analysis of Matter} (oddly, better than {\bf The Analysis of Mind} on these topics, a book that I regard as something of an abortion that he should have discreetly buried). But there are parts of {\bf Theory of Knowledge} that were published (in
{\bf The Monist}, under the title ``The Principle of Acquaintance'' that deal explicitly with James and Mach and which are likely to interest
you, if I'm remembering correctly.
\ewd

Indeed, that is the very part of the book that interests me.  Also I now have both {\sl Analysis of Mind\/} and {\sl Analysis of Matter\/} on my bookshelf.  All of this is part of my summer reading program.  (And I will keep your criticism of the former in mind as I read it.)

\section{05-05-10 \ \ {\it 1003.5209}\ \ \ (to C. H. {\Bennett})} \label{Bennett71}

When you think of me, why is it always prompted by a bad philosophical paper?  What do you think about when you see the good philosophical papers?  I suspect you never write Scott Aaronson, for instance, saying, ``Scott, What do you think of `QBism, the Perimeter of Quantum Bayesianism' by the physicist Fuchs?''  \smiley

Scott was at Jeff Bub's annual meeting last week and gave an excellent talk on the ``mud fight'' (he called it) between himself and Watrous, on the one side of the ring, and you, Debbie, John, and Graeme on the other side.  Best talk of the whole meeting, for sure.  I had some nice discussions with Scott before and after.  He hadn't realized that various assumptions of linearity are pushed upon you nearly automatically by an epistemic interpretation of quantum states.  For instance, if you could clone with quantum theory, you would {\it never\/} call a quantum state a state of knowledge.  Which is the same as saying:  If you think there's enough evidence to call a quantum state a state of knowledge, you wouldn't expect that you could clone (even if you didn't yet have a proof in hand).  The ``linearity trap'' is a very happy home for the quantum-state epistemicist.

Attached are two pictures of {\Spekkens} and me at that meeting, sent to me by Michael Seevinck.  (The old man with his mouth agape is Charles Misner.)

Yeah---I know you could guess---I don't think much of the Lyre paper:  It's the standard 80-year-old fare; no creativity.  Quantum measurement problem, humbug.  ``Quantum theory is in conflict with [Lyre's] CMP,'' humbug again.  My own feeling is that ``measurement outcomes'' and the world of common experience are just given (i.e., undeniable), but that quantum theory is ``additive'' to that.  It is means of achieving uncommon experience in addition to the usual common stuff.  I'd be curious to hear your reaction of pages 22 (right-hand column) through 25 (left column, end of first paragraph) of \arxiv{1003.5209}.

Well, it's a work in progress:  There's no pretense that it's a final story.  I am absolutely enamored, however, with Eq.\ (8) on page 12.  If these SICs always exist they give a very pretty way of thinking of the Born Rule for calculating probabilities.  I joked with Scott (about my friend Hardy and others), ``If your axioms for quantum theory only took you a few months to construct, then it's too cheap a solution.''

Things are going well here.  My little group is discovering all kinds of things about these SIC measurements.  I.e., even if we can't see how far down the beach they go, each one seems to be pretty in its own right, and every now and then one comes across two that have something in common.  For instance, we just discovered something very nice in dimensions divisible by~3!

\section{06-05-10 \ \ {\it The Urgleichung as a Normative Rule}\ \ \ (to R. {\Schack})} \label{Schack200}

Your note from yesterday confronts the central issue, and makes it clear that we have much to discuss and work out in Zurich (what better place than where the whole thing was born!)\ and {\Vaxjo}.  My feeling is that when we adequately answer the issues you bring up, we will at the same time understand the precise reason for {\it this\/} modification to the LTP rather than some other---i.e., why the urgleichung takes the precise form it does.  I may comment a little more later after I get some things cleared off my desk.

\subsection{{\Ruediger}'s Preply}

\bq
For the last few weeks, I have spent a lot of time thinking about the status of the Urgleichung. Regarding it as an addition was a stroke of genius. What I am struggling with is where exactly between the normative and the empirical it is located.

It is clearly not normative in the Dutch book coherence sense, i.e., it is not a completely internal rule. It says something about the world. It gives an agent who uses it an advantage. A piece of the world has a new property, dimension. Suppose I measure a system with dimension $d$. Then I can analyze the measurement in terms of a particular counterfactual measurement, one that is appropriate for dimension $d$. The Urgleichung tells me how my betting odds for the counterfactual measurement should relate to those for the actual measurement.

What happens to me if I use different betting odds, i.e., odds that violate the urgleichung? One possible answer, which is problematical and not really what I want, is that if I violate the urgleichung, I will not be well calibrated. One way in which this might work is sketched below. Do you have any thoughts on this? This is tied up with the very definition of dimension, of course. If the definition of dimension is that it enables an agent to set probabilities according to the urgleichung, then one simply could not have a system of dimension $d$ and not use the urgleichung. But that is begging the question.

One way to bring in calibration is through a game in which the agent announces probabilities for the ground both for the case that the measurement in the sky is counterfactual and the case that the measurement in the sky is factualized. An opponent then chooses if the measurement in the sky is performed. The statement would be that if this game is repeated many times, this creates two subsets of outcomes. The statement would be that the agent cannot be well calibrated for both subsets unless he uses the urgleichung.

Several things are wrong with this construction. If a measurement is an extension of the agent, how could an opponent choose which measurement is performed? Also, for this construction to work one would have to ascertain that the measurement in the sky is an actual SIC and not something else, which elevates ``being a SIC'' to a property of a piece of the world. Bad.
\eq

\section{06-05-10 \ \ {\it {\Vaxjo} Abstract} \ \ (to M. Schlosshauer)} \label{Schlosshauer36}

Attached is my draft of an abstract for your session.  If it fits the bill of what you wanted, let me know and I'll hit ``submit'' on the registration form.  If you don't like it, or you catch any typos, let me know.

Any further word one way or the other on whether WHZ will be there?  Morally, I really ought to confront him real-time with this.
\bq
\noindent \begin{center}
Quantum-Bayesian Decoherence\\
Christopher A. Fuchs and {\Ruediger} {\Schack}
\end{center}

The usual story of quantum measurement since von Neumann has been that it occurs in two steps:  First the quantum system becomes entangled with the measuring device; then, mysteriously, there is a selection of one of the entangled state's components in order to single out an actual measurement result. But the entangled wave function alone (with its freedom to be expressed in any basis) cannot say how it should be decomposed so that the selection can even be effected. The theory of quantum decoherence, developed by Zeh, Zurek, and others, attempts to shim up this deficiency in the von Neumann story by supplementing it with a further story of interaction between the device and an environment: The idea is that the specific form of the interaction with the environment specifies how the system$+$device state ought to be decomposed. But then what of the mysterious selection? Decoherence theorists usually leave that question aside, implicitly endorsing one variety or another of an Everettian interpretation of quantum mechanics.

In contrast, the Quantum Bayesian view of quantum theory developed by {\Caves}, Fuchs, {\Schack}, {\Appleby}, Barnum, and others, leaves most of the von Neumann chain aside:  Instead of unitary evolution, it takes the idea of common experience as first and foremost. In this view, quantum ``measurement'' is nothing other than an agent acting upon the world and experiencing the consequences of his actions---the very same thing he was doing before quantum theory was even thought of. A quantum measurement is most generally an agent's  ``whack'' upon the external world and its unpredictable consequence for him.  To say it still differently, the Quantum Bayesian story is the preface and conclusion of the von Neumann one, without the dramatic artifice of unitary evolution, entanglement, and decoherence in between: selection is the consequence of a whack and nothing more.

Thus, it would seem there is no foundational place for decoherence in the Quantum Bayesian program. And that is true. Nonetheless, one can identify a two-time gambling situation (gamble now on the outcome of a measurement in the future, given that an intermediate measurement will be performed in the nearer future) for which consistent gambling behavior mimics a belief in decoherence. The consistent gamble makes use of a quantum version of van Fraassen's reflection principle, and explains to some extent the seduction (but misleadingness) of the decoherence program. This work builds on [1], [2], and [3], where the Born Rule is viewed as an empirical addition to probability theory and quantum theory is seen as ``additive'' to common experience, rather than ``underlying it'' in a reductionistic sense.

\begin{enumerate}
\item C. A. Fuchs and R. {\Schack}, ``Quantum-Bayesian Coherence,'' \arxiv{0906.2187}.

\item C. A. Fuchs and R. {\Schack}, ``A Quantum-Bayesian Route to Quantum-State Space,'' \arxiv{0912.4252}.

\item C. A. Fuchs, ``QBism, the Perimeter of Quantum Bayesianism,'' \arxiv{1003.5209}.
\end{enumerate}
\eq

[Now, see also C. A. Fuchs and R. Schack, ``Bayesian Conditioning, the Reflection Principle, and Quantum Decoherence,'' \arxiv{1103.5950}.]

\section{06-05-10 \ \ {\it {\Vaxjo} Abstract, 2} \ \ (to M. Schlosshauer)} \label{Schlosshauer37}

\bmaxs
Judging from the abstract and what you've told me before, you surely have a point: the final step of ``selection'' is the notorious achilles'
heel of the decoherence program. But at the same---I'd say---the program (understood properly!) never claimed to remedy this worry.  Instead, it assumed that one would tack on in the end whatever interpretation of QM one wanted, in order to effect selection. In other words, decoherence (again, properly understood) assumed that one works in a picture where selection happens at some stage of the vN chain.
\emaxs
Understood.  But the way I said it was this:
\bq\noindent
But then what of the mysterious ``selection''? Decoherence theorists
usually leave that question aside, implicitly endorsing one variety or
another of an Everettian interpretation of quantum mechanics.
\eq
And you have to admit that that is {\it usually\/} the case.  Furthermore, the Oxford Everettians think that decoherence is precisely the remedy---that selection is never needed and decoherence is precisely the missing link.

In any case, I'll make this claim:  It may be true that, logically speaking, the decoherence program is noncommittal on the interpretive program, but my suspicion is that, for all those who take it seriously as a foundational ingredient, the die is loaded.  That von Neumann Type 2 (unitarity) is considered the more {\it natural\/} thing for physics to be talking about than von Neumann Type 1.  The assumption is, unitarity is no mystery---that's the starting point.

\bmaxs
Now, to my mind, the new aspect that the QBist approach is bringing to the table is to say:
all unitaries are subjective. So the whole entanglement business of decoherence appears in a much less physically relevant light. I think most (if not all) other interpretations would ascribe some sort of objective character to unitaries (and thus the process of entanglement \`a la vN), and thus there'd be a real place for decoherence in this picture.
\emaxs

\section{06-05-10 \ \ {\it Movie Title?}\ \ \ (to S. Savitt)} \label{Savitt8}

It was good talking to you again.  May I ask you to send me the movie title you mentioned that brought up some imagery of the Jamesian theory of truth?  I'd like to watch that.

I read the Gerard Manley Hopkins poem:  It is indeed powerful!  I wish I could read it with the same verve you demonstrated the other night!

\subsection{Steven's Reply}

\bq
The movie is {\sl Playtime}, which fans on the IMDB say should be watched only in a theatre in the 70mm version, which one never has a chance to do. Every film by Tati is wonderful in its own way.
\eq

\section{10-05-10 \ \ {\it The Urgleichung as a Normative Rule, 2}\ \ \ (to R. {\Schack})} \label{Schack201}

[Here's the completion of the note I had started last week.]

You do have several good points in this note.  [See R. {\Schack}'s Preply to the 06-05-10 note ``\myref{Schack200}{The Urgleichung as a Normative Rule}.'']  And I do think that Zurich is going to be a grand time to think about it deeply.  Here are some shorter, preliminary replies at the moment.

\brs
What I am struggling with is where exactly between the normative and
the empirical it is located.
\ers
My answer at the moment is that I don't want to see it as ``between'' at all.  The urgleichung is a normative statement, full stop.  It is that a normative statement (a recommendation for behavior) has been latched onto for empirical reasons.  The usual vision of a physical law is that it is a) intended to be descriptive (in a correspondence sense) and b) empirically motivated.  What we QBists do is simply replace a) by ``intended to be normative.''

\brs
One way to bring in calibration is through a game in which the agent
announces probabilities for the ground both for the case that the
measurement in the sky is counterfactual and the case that the
measurement in the sky is factualized. An opponent then chooses if the
measurement in the sky is performed.
\ers
Yes, like you, I don't like this.  It is similar to Itamar's way of posing coherence in
\bq \arxiv{quant-ph/0208121} and \arxiv{quant-ph/0510095}.\eq
As you always emphasize, coherence should be posed as a purely internal criterion.  That's a first consideration, but then there is also this further crucial point that you make (specific to quantum theory): ``If a measurement is an extension of the agent, how could an opponent choose which measurement is performed?''  And finally, on calibration, I'd say we can't waiver:  ``Probability is single case or nothing!''

My feeling is that if we get an argument for the specific form of the urgleichung, we'll at the same time start to understand what the agent is supposing of his world so that he would view any other connection between the probabilities as ill-advised.  And for this, we're going to have to dig deep into our souls again.  I wrote this to Max yesterday:
\bq
   {\Ruediger} and I are now set for {\Vaxjo}.  We're in Zurich
   the week before, and then we transfer to Sweden, via
   Copenhagen, on Sunday, June 13.  At the moment, we're
   scheduled to stay in {\Vaxjo} 'till the end, because I'd
   like to talk with Anton if he does actually show up.
   But on the guess that there's a significant chance that
   he'll be a no-show with only a few days notice (as he
   often is at conferences), we're contemplating uprooting
   to Copenhagen for a day or more.  Two weeks of sublime
   thought!  I'm really looking forward to it.  The main
   project on the table is to find a convincing motivation
   for the urgleichung.  Consistency, like always, should
   carry the day!  It's just a question of following its
   instructions.
\eq
It must be that there is something in our conception of quantum measurement that pushes us to the urgleichung (or at least the linear form of it).

I wish I had more to say, but at the moment, I don't.

\section{10-05-10 \ \ {\it Reference and Weyl}\ \ \ (to W. G. {\Demopoulos})} \label{Demopoulos36}

\bwd
I'll be copying the Weyl today and should be able to put it in the mail very soon. Do you have the page number for your quote from Feynman about the atomic hypothesis?
\ewd

I wish I had marked that page number down!  For, this morning, I haven't been able to find it.  However, I am quite certain I found it in that book---I believe I pulled it off my shelf and tracked it down when I was writing that section of the paper.  Anyway, a web search this morning shows me that it can (also?)\ definitely be found in {\sl The Feynman Lectures on Physics}, Volume 1, beginning Section 1-2.  Unfortunately, I cannot see the page number on that.  \ldots\ Wait!  I can do better, that IS the page number, 1-2 (that's the convention of the book).  The Section is titled ``Matter is Made of Atoms.''

\ldots\ Interesting.  Just as I was putting my copy of {\sl The Character of Physical Law\/} back on the shelf, I noticed that it was sitting exactly beside Gerald Feinberg's book, {\sl What is the World Made Of?}  It's been about 30 years since I've read that one, but my guess is, his answer wasn't, ``The world is made of this and this and this.  The world is constructed of every particular there is and every way of carving up every particular there is.''

\section{10-05-10 \ \ {\it 1003.5209, 2}\ \ \ (to C. H. {\Bennett})} \label{Bennett72}

Beautiful article on pre-mineralarian mineralarianism!  On Rob, I think he was just temporarily knocked for a loop.  I had just commented that I swallowed six pills at once.  He seemed surprised, ``Six at once!?''  I said, ``No gag reflex; I'd be great in the porn industry.''  Then I was so tickled with myself, I couldn't stop laughing.  Rob had difficulty recovering himself as well.

I'm jealous of your visit to Seven Pines.  You and I were there together previously in 2004.

The 200th anniversary celebration sounds perfect.  Kiki and I would have loved to have come (i.e.\ we would definitely have done it), if I weren't already scheduled for Europe on the same day.  I'll be in Zurich on the 12th just finishing Renato Renner's meeting, and then traveling to Sweden on the 13th for Max Schlosshauer's special session on ``quantum decoherence'' at the annual {\Vaxjo} meeting.  So, from Zurich to Zurek for me.  See attached trouble-making abstract. [See 06-05-10 note ``\myref{Schlosshauer36}{{\Vaxjo} Abstract}'' to M. Schlosshauer.] I get to confront Wojciech directly with this.

\subsection{Charlie's Preply}

\bq
I think of you often, and not mainly in connection with bad philosophical papers.  I suspected it was bad, just needed you to tell me.  I am at a conference on decoherence in Minnesota with, it turns out, a lot of talk about quantum foundations, the Seven Pines Symposium, sponsored by a wealthy gentleman physicist whose hobby is refurbishing 1910 era cars, especially Mercedes-Benz.  Others include Bill Unruh, Amit Hagar, Phil Stamp, Hans Briegel, David Wallace, and Alastair Rae.  In the pictures you sent, why is Rob holding his book upside down?  Is that part of the SIC-POVM philosophy?  In reading the suggested part of your latest tract, I especially liked the new uses of diamonds and the ability to calculate the additional attraction of a hungry child toward an apple.

I have been waging a quixotic battle of my own, trying to persuade the curator of the Museum of Hoaxes to correct his site's entry on Mineralarianism, where he says minearlarians eat rocks. In fact they eat synthetic chemicals made from rocks (and air and water), but most people are surprised to hear this is possible. I have found an amazing number of otherwise well educated people who think that human nutritional requirements are so complex that a person could not possibly survive on an all-chemical diet---surely there must be some undiscovered substances in natural foods without which humans would sicken and die, just as they do when deprived of known vitamins or essential amino acids.   In the course of this I uncovered a piece of research done by NASA exploring just this kind of diet.  See attached article.

Theo and I are celebrating the 200th anniversary of our house in June---see attached invitation.  It's maybe a bit far for you and Kiki to come, but I think it's the sort of event you would enjoy.
\eq

\section{12-05-10 \ \ {\it Correct Woody Allen Quote}\ \ \ (to the QBies)} \label{QBies19}

You might want to know the correct quote as well:
\begin{center}
``I hate reality, but, you know, where else can you get a good steak dinner?''
\end{center}

\section{12-05-10 \ \ {\it Today's Colloquium}\ \ \ (to the QBies)} \label{QBies20}

Today's colloquium speaker Verlinde strikes me as potentially interesting:  He strikes me as having a good ``idea for an idea'' (but it probably shouldn't be thought of as more rigorous than that).  Anyway, the talk might give you some food for thought in relation to QBism, or ``additionism'' as Marcus calls the broader program.

\section{12-05-10 \ \ {\it Definitely Over the Limit} \ \ (to M. Schlosshauer)} \label{Schlosshauer38}

\bmaxs
I discovered [this album] just the other day: it's pianist Jacky Terrasson with vibes player Stefon Harris locked in spontaneous interplay. Another great example for how standards are just some very vague foil for creating something entirely new.
\emaxs
And so too of space itself.  See this talk we just had at PI today, \pirsa{10050022} (it's not posted yet, but should be by tomorrow morning my time \ldots\ or probably by the time you read this note).  Space, like the jazz standard, is the foil in the background---or at least that's a thought that intrigues me.  You can also see this paper by Verlinde, \arxiv{1001.0785}.  It's all vague and hand-wavy at the moment (dimensional arguments really), and from the wrong philosophical point of view, but I can't help the feeling that there is something really deep here.  It says that space is the stuff on the ``outside,'' and I quite like that.  It is connected (I think or hope) to what I meant in the last paper when I said that QBism ``tinkers with spacetime.''

But on to the music!  There is deep stuff here as well.  I loved this line in the album you suggested:  ``Our two instruments became one instrument with four hands.''  That, I think, is the right way to think of Alice and Bob in most quantum information protocols.  When in collaborative communication, two (ostensible) agents are really one with respect to the concerns of personalist, Bayesian probability theory.  I must have sent you William James's notion of the self before:
\bq\noindent
     In its widest possible sense, however, a man's
     self is the sum total of all that he can call his,
     not only his body and his psychic powers, but his
     clothes and his house, his wife and children, his
     ancestors and friends, his reputation and works,
     his lands and horses, and yacht and bank account.
\eq

Thanks a million again for the suggestion.  It pushes me to new thoughts, and for that I am always grateful.

\section{13-05-10 \ \ {\it Contact from the Stellenbosch Institute for Advanced Study}\ \ \ (to H. B. Geyer)} \label{Geyer1}

It looks like it's been seven years since we've ``seen'' each other!  Great to hear from you.  It sounds like a lot has been happening down there.

Thanks for the flattering remarks on my latest paper.  I am intrigued by your proposition, but let me try to understand it better.  Are you wondering whether I would be interested in an existing project?  Or are you asking whether I might consider being a ``project leader'' (phrase you used)?

It is true that I look for every opportunity possible to promote progress on understanding these magical quantum measurements, the symmetric informationally complete (SIC) positive operator valued measures.  (See all the references in this paper by Scott and Grassl, if you want to get a sense of the breadth of the problem: \arxiv{0910.5784}.
You might also get a laugh out of the ``press release'' on my group's involvement with these things; attached.  And here's a meeting I organized on the problem a couple years back:
\pirsa{C08025}.)  Would a program like that be something of interest to STIAS?

It would indeed be nice to give you a shorter visit and see what it is like there.  Thank you for the invitation.  But with all the travel I'm already committed to this summer, I couldn't possibly do it before September or October.  Would something like that work for you?

\subsection{Hendrik's Preply}

\bq
A voice from the past -- you may recall that you helped guide Robert Schumann 10 years ago through a thesis on Quantum Information Theory which I supervised; I think we sporadically exchanged a few e-mails after that.

Much has happened here since; I/we managed to establish a National Institute for Theoretical Physics in South Africa, funded by the SA Dept of Science and Technology. Neil Turok will be able to tell you something about it; amongst other things Neil was instrumental in getting Stephen Hawking here in 2008 at the inauguration (see \myurl{http://www.nithep.ac.za/}).

My own career has recently somewhat diverged from theoretical physics; I am presently the director of the Stellenbosch Institute for Asdvanced Study (STIAS) -- see \myurl{http://www.stias.ac.za/}.  (Lenny Susskind has been a long time supporter, by the way, and spent time here in 2009 -- see \myurl{http://www.stias.ac.za/news.html}). STIAS has an ambitious programme covering all disciplines, with a focus on inter-/transdisciplinary projects. Finding projects within, or primarily rooted in, the natural sciences proves to be quite a challenge.

I recently read your papers on Qbism, Perimeter etc., finding them stimulating, became aware of your work at Perimeter and wondered whether you would be interested to explore a quantum information related project at STIAS. Such projects typically entail that a project leader and a group of researchers identified by/in consultation with him/her are invited as STIAS Fellows who then spend time here simultaneously -- from one to several months (not unlike the Kavli Institute setup, but on a smaller scale of course).

Perhaps you would consider first coming to STIAS on your own to get a feel for the place. I can invite you as my guest for a week or two (or longer, if your programme would allow it). This will also provide an opportunity to interact with people at NITheP, apart from whoever is here as STIAS Fellows at the time.

Please give it a thought, and let me know.
\eq

\section{14-05-10 \ \ {\it Definitely Over the Limit, 2} \ \ (to M. Schlosshauer)} \label{Schlosshauer39}

\bmaxs
As for considering Alice and Bob as one agent: do you then look at this agent as possessing two independent prosthetic hands with which ``he'' can perform two different---and potentially synchronous---actions upon the world, at different points in space?
\emaxs

Your question on Alice and Bob is good one, a deep one, an important one.  I promise you an answer in {\Vaxjo}!  Monday, I go back to DC, this time to report on the QBies to ONR.  So I'm scrambling once again!

\section{14-05-10 \ \ {\it Qutrit SIC POVM experiment} \ \ (to L. A. Rozema)} \label{Rozema1}

\blar
Is there anything that makes the 3d sic povm more interesting than the 2d sic?
\elar

There's a quote by Richard Feynman that goes:  ``A poet once said, `The whole universe is in a glass of wine.'  We will probably never know in what sense he said that, for poets do not write to be understood.  But it is true that if we look in glass of wine closely enough we see the entire universe.''

Similarly one can say of the qutrit:  The whole of the probabilistic structure of quantum mechanics is already in it.  What I mean by this less poetically is that Gleason's theorem, which underlies the Born Rule for quantum probabilities, makes crucial use of the qutrit in its proof.  Particularly, Gleason shows that if the theorem is true in dimension three, then it is true in all dimensions (even infinite dimensions), and as it turns out, the hard part of  the proof is precisely the dimension-three part.  So everything stands or falls on the qutrit case.

Now, we Quantum Bayesians want to take the ``quantum law of total probability'' (i.e., the Born rule written in SIC language) as a new foundation for quantum theory.  So, it is just nice to see it firmly established (experimentally) in dim$=$3, and by way of this, one can make a direct comparison (in word and deed) to Gleason's theorem when advertising the new point of view.

That is to say, Gleason (effectively) shows the QLTP as a {\it theorem\/} in dim 3.  (I say ``effectively'' because he didn't know of SICs at the time, and thus could not rewrite the Born Rule into that language.)  We on the other hand want to take the QLTP as an axiom.  So, it really ought to have some experimental grounding.  And showing that it has experimental grounding in dim 2 is of no great interest, because Gleason's theorem doesn't even work there:  I.e., the QLTP has no competition from Gleason in dimension 2.  All the direct competition is in dim 3.

Finally, let me attach one of Gelo's very pretty equations.  See Eq.\ (2) in the attached file.  The ``QBic equation'' along with the ``Bloch-sphere-type'' equation uniquely specifies which probability distributions for SIC outcomes correspond to pure quantum states.\footnote{\editornote In $d = 3$, the QBic equation can be reduced through a clever choice of SIC to
\begin{displaymath}
\sum_i p(i)^3 - 3 \sum_{(ijk) \in S(9)} p(i) p(j) p(k) = 0.
\end{displaymath}
Here, $S(9)$ denotes the \emph{Steiner triple system} of order 9, a set of 12 elements which can be found by cyclically tracing all the horizontal, vertical and diagonal lines in the array
\begin{displaymath}
\begin{array}{ccc}
1 & 2 & 3\\
4 & 5 & 6\\
7 & 8 & 9
\end{array};
\qquad\hbox{ that is to say, }\qquad
S(9) =
\begin{array}{ccc}
(123) & (456) & (789) \\
(147) & (258) & (369) \\
(159) & (267) & (348) \\
(168) & (249) & (357)
\end{array}.
\end{displaymath}
See G.\ N.\ Tabia, ``Experimental scheme for qubit and qutrit SIC-POVMs using multiport devices,'' Phys.\ Rev.\ A\ \textbf{86}(6), 062107 (2012), \arxiv[quant-ph]{1207.6035}.  For more games one can play with this construction, see B.\ C.\ Stacey, ``SIC-POVMs and Compatibility among Quantum States,'' \arxiv[quant-ph]{1404.3774}.}  Only in dimension 3, so far as we know, does the QBic equation take such a simple and beautiful form.  If simplicity unlocks secrets, it's going to be found there.  Still another reason for looking at qutrits.  (And once again, another reason that qubits are too trivial.  In the qubit case, only the Bloch sphere equation is active---for the QBic equation reduces to it in that case.  Quantum life only becomes interesting in dimension 3.)

Hope that helps some.

\section{17-05-10 \ \ {\it  Chris and the Navy} \ \ (to D. H. Wolpert)} \label{Wolpert4}

\bdhw
Navy? Related to things like quantum computing/encryption?
\edhw

No, quantum foundations.  They fund my QBism (Quantum Bayesian) effort.  Specifically, the technical part of it---to develop a {\it clean\/} formalism for quantum theory that dispenses with quantum states and instead uses probabilities only.  The key part of the apparatus is the SIC (symmetric informationally complete) positive operator valued measures---an object we believe (but have not yet proven) to exist in all finite Hilbert-space dimensions.

See attached glossary. [See 07-04-10 note ``\myref{Schack199}{Up Front}'' to {\AA}.\ {\Ericsson} {\it et al}.]

\section{20-05-10 \ \ {\it Civil War Times} \ \ (to D. B. L. Baker)} \label{Baker23}

Yesterday, on my drive between Shepherdstown, West VA and Dulles Airport, I took a small side jaunt to the Antietam Battleground.  The view was breathtaking; I wish you could have been there.  Such a beautiful place, juxtaposed with the idea of so much killing.  I cried a bit before getting back in the car.

\section{20-05-10 \ \ {\it My Weyl Quote} \ \ (to H. Price)} \label{Price20}

I'll show you mine; now you show me yours \ldots

Unfortunately, I don't know the original source (where Weyl wrote it).  I do know however that I got this from Paul Forman's article, ``Weimar Culture, Causality, and Quantum Theory, 1918--1927:  Adaptation by German Physicists and Mathematicians to a Hostile Intellectual Environment.''  Presumably (hard to remember, it being 11 years since I copied this down) the phrase ``[regarded as a four-dimensional continuum of events]'' was Forman's addition (i.e., not mine, but I can't guarantee it at the moment).\footnote{\editornote The bracketed phrase does appear in Forman's quotation of Weyl.}

\bq\noindent
Finally and above all, it is the essence of the continuum that it cannot be grasped as a rigid existing thing, but only as something which is in the act of an inwardly directed unending process of becoming \ldots\ .  In a given continuum, of course, this process of becoming can have reached only a certain point, i.e.\ the quantitative relations in an intuitively given piece $\cal S$ of the world [regarded as a four-dimensional continuum of events] are merely approximate, determinable only with a certain latitude, not merely in consequence of the limited precision of my sense organs and measuring instruments, but because they are in themselves afflicted with a sort of vagueness \ldots\ .  And only ``at the end of all time,'' so to speak, \ldots\ would the unending process of becoming $\cal S$ be completed, and $\cal S$ sustain in itself that degree of definiteness which mathematical physics postulates as its ideal \ldots\ .  Thus the rigid pressure of natural causality relaxes, and there remains, without prejudice to the validity of natural laws, room for autonomous decisions, causally absolutely independent of one another, whose locus I consider to be the elementary quanta of matter.  These ``decisions'' are what is actually real in the world.
\eq

\section{21-05-10 \ \ {\it About Changing the World}\ \ \ (to L. Freidel)} \label{Freidel2}

Returning to last night's discussion \ldots\  The part about ``changing the world'' (literally) with the aid of a scientific theory.  I expand on some of what I said in Sections 6, 7, and 8 of this paper:  \arxiv{1003.5209}.

But more directly, you might enjoy reading the first few paragraphs of the attached proposal I wrote for the Templeton Foundation.  It says it all a bit more flowingly, and I was pretty upfront about it with them.  I.e., I actually believe this stuff!

\section{21-05-10 \ \ {\it Workshop Talks}\ \ \ (to O. C. O. Dahlsten)} \label{Dahlsten3}

Let's give it the following provisional title:\\
``Some Properties of QBist State Spaces, whichever Ones {\Ruediger} Schack Does Not Cover''

How does that work for you?

\section{21-05-10 \ \ {\it A Third Way}\ \ \ (to M. Gleiser)} \label{Gleiser1}

I was reading your piece in the {\sl Harvard Divinity Bulletin\/} from 2005.  I just want to quickly respond to these lines of yours near the beginning:
\bq
After reading hundreds of creation myths I realized they all fall within a
simple classification scheme, based on how each answered the question
``Did the world come to be at a specific moment in the past?'' That is,
``Was there a moment of creation?'' The answer can only be ``yes'' or
``no.'' A ``yes'' means the universe has a finite age, just as we do; it
appeared some time in the past and is still around today. A ``no'' can
mean two things: either the universe has existed forever, an eternal,
uncreated cosmos, or it is created and destroyed in a cyclic succession
that repeats itself throughout boundless time.
\eq
But there is a third way to envision things \ldots\ at least, and probably more than that, but one I care about.  It is that creation comes in ``spots and patches'' as William James puts it:
\bq
Our acts, our turning-places, where we seem to ourselves to
make ourselves and grow, are the parts of the world to which we are
closest, the parts of which our knowledge is the most intimate and
complete. Why should we not take them at their facevalue? Why may
they not be the actual turning-places and growing-places which they
seem to be, of the world---why not the workshop of being, where we
catch fact in the making, so that nowhere may the world grow in any
other kind of way than this?

Irrational! we are told. How can new being come in local spots and
patches which add themselves or stay away at random, independently of
the rest? There must be a reason for our acts, and where in the last
resort can any reason be looked for save in the material pressure or
the logical compulsion of the total nature of the world? There can be
but one real agent of growth, or seeming growth, anywhere, and that
agent is the integral world itself. It may grow all-over, if growth
there be, but that single parts should grow {\it per se\/} is
irrational.

But if one talks of rationality---and of reasons for things, and
insists that they can't just come in spots, what {\it kind\/} of a
reason can there ultimately be why anything should come at all?
\eq
I like it because it tracks with what I think is going on with quantum measurement.  But we can discuss.

I can't wait until this meeting is over, so that I get a chance to read some of these writings of yours that I just discovered!!

\section{22-05-10 \ \ {\it The Coffeemaker Revisited}\ \ \ (to L. Hardy)} \label{Hardy41}

I just got the coffee brewing this morning and imagined the following.  A devotee of Sarah Palin says, ``God created the world just as it is.  All the complexity, all the detail you see, to its last detail, was in God's great plan.''  And that of course, includes the bad Sunbeam coffeemaker design I deal with everyday (the carafe always drips onto the counter).  A devotee of Daniel Dennett says, ``The way to understand the world is as Conway's Game of Life.  All the complexity, all the detail you see, to its last detail, was in the game's initial condition.''  And that of course, includes the bad Sunbeam coffeemaker design I deal with everyday (the carafe always drips onto the counter).

\section{26-05-10 \ \ {\it Invitation to Speak at Science and Nonduality 2010}\ \ \ (to S. E. Sobottka)} \label{Sobottka1}

I have been thinking hard about whether I might take you up on your invitation---the conference distinctly intrigues me---but in the end I must say ``no.''  It is not that I am not sympathetic to {\it some\/} aspects of what I read at the conference website.  In my own opinion, quantum theory might well imply, or at least hint at, a kind of pantheism.  But when so, it does as a kind of pluralism (something along the lines of ideas in William James's book {\sl A Pluralistic Universe}).  So, it would be a kind of spiritualistic universe, but not one of ``oneness''---instead a kind of ``manyness,'' a pluriverse.  However, that's just a line of thought and research---one that I'm sure will take me several years and several developments in the formalism to see through clearly.  Not something written in stone---but a research direction---and ideally it might be good for the general public to know something about this.

The reason I say ``no'' to the invitation is what I perceive as the hodgepodge nature of the meeting and the speakers there.  The ideas I read are really all over the map:  It feels to me like a million stories by a million lonely souls.  Take F. David Peat, Stuart Hameroff, Amit Goswami, and me.  The {\it only\/} thing we'd really have in common is the {\it word\/} ``quantum''---there's no common substance beyond that.  Goswami's website contains nearly the opposite of everything I'd say.  There's nothing wrong with debate of course:  But a debate at the service of a pre-established cause (say ``nonduality'') troubles me.  Moreover, I worry about the level, or to be more honest, lack, of serious critical thinking in many that I see in the speaker list.

In all, I think the conference is just not the right forum for me.  At Perimeter Institute, I carry a coffee cup that says, ``Life is not about finding yourself, but about creating yourself.''  All the evidence in front of me indicates a conference largely consisting of people who've spent years ``finding themselves'' rather than ones who've jumped into the stream of life and contributed to making the world ``go.''  In the end, I'd likely be pretty uncomfortable there.

But thanks for your consideration, and best wishes.

\section{27-05-10 \ \ {\it Stream of Life}\ \ \ (to J. B. Lentz)} \label{LentzB11}

Just yesterday, I turned down an invitation to speak at one of the strangest conferences I've ever been invited to.  If you have a look at the other speakers, you'll understand why.  See:
\myurl{http://www.scienceandnonduality.com/speakers.shtml}. (The main page of the conference is \myurl{http://www.scienceandnonduality.com/}.)  As it turns out, I ended my note to the organizers with these words:
\bq\noindent
   In all, I think the conference is just not the right  forum for me.  At
   Perimeter Institute, I carry a coffee cup that says, ``Life is not about
   finding yourself, but about creating yourself.''  All the evidence in
   front of me indicates a conference largely consisting of people who've
   spent years ``finding themselves'' rather than ones who've jumped
   into the stream of life and contributed to making the world ``go.''  In
   the end, I'd likely be pretty uncomfortable there.
\eq

Before meeting Kiki and meeting you, I'm not sure I could have written some of those sentences.  The way you live has been an inspiration to me since the beginning:  You more than most others I know have certainly ``jumped into the stream of life and contributed to making the world go.''

Happy 70th birthday Brad!  I think now is when life gets really good.

\section{29-05-10 \ \ {\it God Wouldn't Scam, Would He?}\ \ \ (to H. B. Dang \& the QBies)} \label{Dang10.1}

\bhbd
I've bought many things on eBay, and this is a principle I've been always followed to avoid scams: if it sounds too good to be true it probably is. {\rm [\ldots]}

As Ingemar-Kate pointed out, in $d=6$, among the 984 normal vectors, we can find sets of 4 vectors whose squared inner products are all 1/3. Moreover, in each such set, those four 6-component vectors span a 2-D subspace, so they indeed form a 2-D SIC. This is very interesting, because one doesn't need to start with a 6-D SIC fiducial. Start from any vector in the Zauner subspace (which we know how to calculate), we will still get the same linear dependency structure and still get 2-D SICs out of it.

The big question is, of course, will this kind of miracle happen in $d=9$? {\rm [\ldots]}

In $d=9$ there are 79767 normal vectors (corresponding to that many hyperplanes), and there are over 3 billion inner products to be checked. Instead of taking a long walk like Ingemar did, I slept, played pool, talked to Chris, etc.\ while waiting for 20 hours for my desktop and the QBism server to finish the calculation. Indeed, there are pairs of normal vectors that have the right inner product. Those pairs come from 54 normal vectors.

Next small question: among those 54 vectors, can one find set of 9 vectors whose squared inner product between any 2 of them is 1/4? {\rm [\ldots]}

Last small question: for those set of 9 vectors, do they live in a 3-D subspace?

The answer is yes! So they do form a 3-D SIC!
\ehbd

God?  It is {\it our\/} efforts and actions that complete nature!  You have added to nature, and it has tolerated it.  (See one of my favorite William James quotes attached.)  [See quote in 05-01-09 note ``\myref{PauliFierzCorrespondence}{What I Really Want Out of a Pauli/Fierz-Correspondence Study}'' to H.~C. von Baeyer \& D.~M. {\Appleby}.]

Mighty good work!  This gives me a very happy feeling this morning.

\section{31-05-10 \ \ {\it References} \ \ (to N. D. {\Mermin})} \label{Mermin188}

\bdm
I'm refereeing a paper that claims that virtually nobody but Fine and de Munyck dissent
from the view that quantum mechanics is nonlocal.   Could you send me a
few references to published papers by yourself {\it et al}.\ that take issue with nonlocality.
\edm

I'm sorry, I forgot to reply to this one!  My ``inbox'' is an absolute mess.  Is this still relevant to you?  If so, I'll put a list together a little later in the day.  Also, Zeilinger and Brukner and Zukowski emphasize that there is nothing nonlocal about QM.  Griffiths and Gell-Mann as well.  And I'm sure more names will come to mind if I think harder.

\section{31-05-10 \ \ {\it References, 2} \ \ (to N. D. {\Mermin})} \label{Mermin189}

\bdm
Any other names that come to mind (me, for example) should be added to my list.
\edm

With regard to yourself, I'll let you make your own judgements. With regard to me, these three are fairly explicit:
\begin{enumerate}
\item
C. A. Fuchs and A. Peres, ``Quantum Theory Needs No `Interpretation','' Physics Today {\bf 53}(3), 70--71 (2000).
\item
C. M. {\Caves}, C. A. Fuchs, and R. {\Schack}, ``Subjective Probability and Quantum Certainty,'' Studies in History and Philosophy of Modern Physics {\bf 38}, 255--274 (2007).
\item
C.~A. Fuchs, ``QBism, the Perimeter of Quantum Bayesianism,'' (2010), \arxiv{1003.5209v1}.
\end{enumerate}

With regard to Brukner and Zukowski, see: \arxiv{0909.2611}.  They tell me they are writing a masterwork on the subject, coming flat out against nonlocality, but I don't see it on the {\tt arXiv} yet.  You might contact them directly.  Here's an older one by Marek: \arxiv{quant-ph/0605034}.  As I recall, here's one where Duvenhage came out fighting for locality: \arxiv{quant-ph/0203070}.  There were a gazillion things by Peres.  Here's one of the later ones: \arxiv{quant-ph/0310010}.  Is that enough to get you started?

It sounds like Griffiths exaggerated, but I concur with the gist of what he was trying to get after.  The vast majority of quantum info people, at least, seem to buy spooky action at a distance out of hand.

\section{31-05-10 \ \ {\it Another on the Side of Locality} \ \ (to N. D. {\Mermin})} \label{Mermin190}

This one from Itamar Pitowsky: \arxiv{quant-ph/0510095}.

\section{31-05-10 \ \ {\it And Bill {\Demopoulos}} \ \ (to N. D. {\Mermin})} \label{Mermin191}

Also, Bill {\Demopoulos} has a paper coming out that is sympathetic
with the side of locality.  It's not on the web yet, but I'll cc this
note to him and maybe he'll send you the present
draft.\footnote{\editornote One can, in fact, find defences of
  locality from across the spectrum of quantum-foundational views.
  For example, Berthold-Georg Englert comes out swinging for locality
  in \arxiv{1308.5290}, a position statement which sounds
  rather Peierls--Peresian.  Pierre Hohenberg, well-known for work in
  nonequilibrium statistical physics, advocates locality in the
  context of consistent-history-ology.  See \arxiv{0909.2359}, where
  he says that the
  ``pervasive use of the term `nonlocal' to describe quantum
  mechanics'' is ``maximally misleading''.  And if one \emph{must}
  have an Everettista on the list, one can add the late Sidney
  Coleman.  In his ``Quantum Mechanics in Your Face'' lecture (Harvard
  has the video online, at \myurl{http://media.physics.harvard.edu/video/?id=SidneyColeman\_QMIYF})
  he says that a physical theory which is consistent with what quantum
  mechanics predicts for the GHZ experiment must have either quantum
  mechanics or nonlocality, but not both.  If you're willing to
  embrace nonlocality, says Coleman, you can stick with the de
  Broglie--Bohm pilot-wave theory and keep your hidden variables.}

\section{31-05-10 \ \ {\it {\Caves} Talk} \ \ (to N. D. {\Mermin})} \label{Mermin192}

You might enjoy this talk of Carl's ``What are the laws of physics?\ Resisting reification'' at
\myurl[http://info.phys.unm.edu/~caves/talks/talks.html]{http://info.phys.unm.edu/$\sim$caves/talks/talks.html}.

\section{31-05-10 \ \ {\it The Gleason--Weyl Thing} \ \ (to D. B. L. Baker)} \label{Baker24}

Speaking of bad mental places, sorry I've kept you so long with a reply.  This Spring has turned to insanity for me.  Some men can't keep their pants on; my problem is I can't stay off a damned airplane.  I'll try my best to write you a longer note from Zurich next week; the town is the kind of place that makes me think, and I'll surely be lonely in the soul once again.

I certainly still want to steal away with you, but the demons have me booked up all the way into the Fall.  Maybe when (more carefully, if) my position becomes permanent here, I can use that as a good excuse to blow some money on a celebration with you.

When I was in Portland in March, I couldn't keep myself out of Powell's Bookstore.  It was the most amazing place I had ever seen.  And you want to talk about hippie chicks!  First off the town is full of them, but then the concentration at Powell's reaches truly dangerous levels.  Nonetheless, I nearly single-mindedly kept my attention on the books:  I came home with over \$400 worth!

But my best purchase of all from them had to wait until my return home---it came from the internet.  When I got home, I asked my three students, ``Did you go to Powell's.''  ``Only to the technical book annex a few blocks away from the main store,'' they responded.  I was like, ``You losers!  You missed the greatest bookstore ever.  All the history, art, architecture, philosophy, biography \ldots''  Then I said, ``Well, did you buy anything?''  They all responded, ``Nah.''  But Matthew, the youngest, said, ``Well, I did see an old book on group theory, with a stamp on the first page that said, `Andrew M. Gleason'.''  I said, ``Which book on group theory?''  He said, ``An old book by Weyl.''  I said, ``{\sl Theory of Groups and Quantum Mechanics}?''  He said, ``Yeah, that was it; an old hardcover version of it.''  I said, ``You came across Gleason's copy of Weyl's book and you didn't buy it?!?!''  ``How much did they want for it?''  He said, ``\$17.95.''  My jaw dropped.  I said, ``What on earth were you thinking?''  He said, ``Well, I didn't know if that was really Gleason's stamp or not; somebody else might have just put it there to sell the book more easily.''  I said, ``Now who would do that?!?!  Who on earth besides us even knows who Gleason is?''  Well, I got on the internet and found that Powell's had one hardcover copy of the book for \$17.95 and bought it immediately.  Now, the mingling of Gleason and Weyl's spirits hanging out in my study have been giving me all kinds of inspirations.  Report of one such below.  (And still further below, John Wheeler's story of how he learned quantum mechanics from Weyl's book when he was 19.)  [See 27-04-10 note ``\myref{Baeyer119}{Midnight Reading}'' to H. C. von Baeyer and 24-04-10 note ``\myref{QBies16}{Weyl's Book and John Wheeler}'' to the QBies.]

In another note, I'll tell you about who Gleason was and why anyone would care, and I'll record a funny story about the one time I met him.

\section{01-06-10 \ \ {\it SIC POVMs and the Law of Total Probability} \ \ (to B. R. La Cour)} \label{LaCour3}

I got your note and have read it.  Thanks for your interest in these matters.  Near the closing, you write,
\bbrlc
The use of SIC POVM and projection measurements leads to a result
which, at first, appears to be at variance with the law of total
probability. A more careful examination reveals that the apparent
discrepancy is due to \ldots
\ebrlc
I think, implying that {\it I\/} think [sic] that there is some discrepancy between the quantum rule and the law of total probability.  But I do not think that.  Rather I view the Born Rule as a {\it supplement\/} to raw probability theory that allows probability assignments in {\it one\/} experiment to be expressed in terms of the probability assignments in another (incompatible) experiment (where there need not have been any a priori connection).  Please see Section IV of my paper ``QBism, the Perimeter of Quantum Bayesianism,'' \arxiv{1003.5209v1}. Particularly, the part below Eq.\ (8) where it says,
\bq\noindent
But beware: One should not interpret Eq.\ (8) as
invalidating probability theory itself in any
way: For the old Law of Total Probability has no
jurisdiction in the setting of our diagram, which
compares a `factual' experiment (Path 1) to a
`counterfactual' one (Path 2).
\eq

If you'll be in {\Vaxjo}, we can discuss then.  I'm just getting ready for it, by way of Zurich the week before.

\section{01-06-10 \ \ {\it URGENT QCMC Invited Speakers ABSTRACT} \ \ (to QCMC 2010)}

Attached is the abstract for my talk, both in LaTeX and as a PDF file.

\bq
\begin{center}
{\bf Charting the Shape of Hilbert Space}\medskip
\end{center}

The space of quantum states for a $d$-level system is usually thought of as a smooth, featureless place---simply, a linear vector space ${\mathcal H}_d$ over the complex field.  But in fact, the space of quantum states corresponds to the convex set of trace-1 positive semi-definite operators {\it on top of\/} ${\mathcal H}_d$:  This is a body that is anything but smooth and featureless.  And in its shape may lie the key to a deeper conceptual understanding of quantum mechanics---at least this is the point of view of the Quantum Bayesian approach to quantum theory developed by Carlton Caves, {\Ruediger} Schack, the author, and others [Fuchs10].

The reason for this is that Quantum Bayesianism, or QBism, is based on the idea that quantum theory is best understood as an {\it addition\/} to probability theory, not as a theory separate from standard probability theory or one including standard probability theory as a special case.  It is an addition invoked in a situation where one wants to analyze a given physical experiment in terms of another, never-actualized (or counterfactual) experiment.  As such, the first step to finding a deeper understanding quantum theory is to find a manageable representation of quantum states purely in terms of probabilities, without amplitudes or Hilbert-space operators.  In this talk, we review the efforts the Perimeter Institute QBism group has made to find such a good representation.

The best candidate so far involves a mysterious entity called a ``symmetric informationally complete positive-operator-valued measure,'' or SIC (pronounced ``seek'') for short.  This is a set of $d^2$ operators $H_i=\frac1d \Pi_i$ on ${\mathcal H}_d$, where the $\Pi_i=|\psi_i\rangle\langle\psi_i|$ are rank-one projection operators such that
\begin{equation}
\big|\langle\psi_i|\psi_j\rangle\big|^2=\frac{1}{d+1}\quad \mbox{whenever} \quad i\ne j\;.
\label{Mojo}
\end{equation}
We say mysterious because, despite 10 years of growing effort since the definition was first introduced [Zauner99,Caves99] (there are now nearly 50 papers on the subject), no one has been able to show that SICs exist in general finite dimensions.  All that is known firmly is that they exist in dimensions 2 through 67 [Scott09].  Dimensions $2\,$--$\,15$, 19, 24, 35, and 48 are known through direct or computer-automated analytic proof; the remaining solutions are known through numerical simulation, satisfying Eq.~(\ref{Mojo}) to within a precision of $10^{-38}$.

What is most intriguing about a SIC is that the probabilities $P(H_i)$ for its outcomes uniquely determine the system's quantum state $\rho$, and they do so through an amazingly simple formula,
\begin{equation}
\rho = \sum_{i=1}^{d^2}\left( (d+1)\,P(H_i) - \frac1d \right)\Pi_i\;.
\label{Ralph}
\end{equation}
Making us of this formula, one finds a novel way to think of the Born Rule for quantum probabilities. For instance, consider a von Neumann measurement with outcomes $D_j=|j\rangle\langle j|$ (the vectors $|j\rangle$ forming an orthonormal basis), and let $P(D_j|H_i)$ be a conditional probability for finding outcome $D_j$ if the system had been prepared in state $\Pi_i$.  Then the probability for $D_j$ given by the standard Born Rule
\begin{equation}
Q(D_j)=\tr (\rho D_j)
\end{equation}
becomes
\begin{equation}
Q(D_j) =
(d+1)\sum_{i=1}^{d^2} P(H_i) P(D_j|H_i) - 1\;.
\label{ScoobyDoo}
\end{equation}
Compare this to the expression one would expect from classical probability theory (i.e., the Law of Total Probability),
\begin{equation}
P(D_j) = \sum_{i=1}^{d^2} P(H_i) P(D_j|H_i)\;.
\end{equation}
What a tiny modification to the classical law!  In fact, the Born Rule seems to be nothing but a kind of Quantum Law of Total Probability.

Recent work at Perimeter Institute has been devoted much to the SIC existence problem, rewriting the problem in Lie algebraic terms [Appleby10],  and trying to see how much of the shape of quantum-state space is implied by the very consistency of Eq.\ (\ref{ScoobyDoo}) [Fuchs09a,Fuchs09b].  Surprisingly, one can glean quite a bit about the structure of quantum states from the requirement that $Q(D_j)$ always be a proper probability distribution.  The talk will end with a list of open problems and avenues for further research.

This work was supported in part by the U.~S. Office of Naval Research (Grant No.\ N00014-09-1-0247).

\begin{enumerate}
\item
{C. A. Fuchs, ``QBism, the Perimeter of Quantum Bayesianism,'' \arxiv{1003.5209}, (2010).}

\item
{G. Zauner, {\sl Quantum Designs -- Foundations of a Non-Commutative Theory of Designs} (in German), PhD thesis, University of Vienna, 1999.}

\item
{C.~M.~Caves, ``Symmetric Informationally Complete POVMs,'' University of New Mexico internal report (1999).}

\item
{A. J. Scott and M. Grassl, ``SIC-POVMs:\ A New Computer Study,'' [later published as J.\ Math.\ Phys.\ {\bf 51}, 042203 (2010)] \arxiv{0910.5784}, (2009).}

\item
{D.~M. Appleby, S.~T. Flammia, and C.~A. Fuchs, ``The Lie Algebraic Significance of Symmetric Informationally Complete Measurements,'' \arxiv{1001.0004}, (2010).}

\item
{C.~A. Fuchs and R.~Schack, ``Quantum-Bayesian Coherence,''  \arxiv{0906.2187}, (2009).}

\item
{C. A. Fuchs and R. Schack, ``A Quantum-Bayesian Route to Quantum-State Space,'' \arxiv{0912.4252}, (2009).}
\end{enumerate}
\eq

\section{01-06-10 \ \ {\it Antietam} \ \ (to D. B. L. Baker)} \label{Baker25}

Thanks for the lesson.  That was great!

It's funny, I know so little about the Civil War myself, but I could still envision a vague sense of the battle.  Probably old memories from nearly forgotten movies replaying in my head.  But it was real enough at the moment to pull the emotions forth.  Things were probably already set in motion by the massive cemetery I drove past in Sharpsburg before getting to the battleground.  The weather, dreary and drizzly, was probably conducive as well.

I didn't have much time because I felt I was running late for the airport, but I also stopped at an oak grove in Sharpsburg where Robert E. Lee had camped (for three days I think).  It was a beautiful setting---big, massive trees.  Very peaceful.  It goes to show you the depth of things hidden in the calm airs around us.

\section{01-06-10 \ \ {\it Flattery Will Get You Everywhere} \ \ (to J. Ismael)} \label{Ismael2}

You flatter me (but I bet you say such things to all the boys).  In any case, it worked!  I printed out your papers and will be carrying them to Europe with me.  I hope to get back to you soon with my thoughts on them.

I too enjoyed our conversation.  I feel like I'm learning a lot on Weyl, Ramsey, and the embedded perspective.

\subsection{Excerpt from the Draft ``Volition, Time, and Becoming'' Jenann Had Sent}

\bq
What makes the world itself appear to be in the process of becoming as it is experienced (rather than simply the otherwise natural idea that we simply have a changing appearance of a fixed reality) is that we observe the results of our own actions. The world partakes in the unsettled, unresolved, remaining to be decided character of volition because what I freely decide to do changes the future course of experience.  If we just were just observers of history, there would be no reason that becoming wouldn't be a feature of our experience of time, rather than a feature of time itself. It's the discovery that what happens depends on my will, that makes time itself partake in the contingency of my present and future willings.  I can no more regard time itself as a static dimension of which I have a varying appearance than I can regard my own volition that way. Of course there is a transcendent perspective from which both my decision processes and their downstream consequences are part of the fixed, eternal manifold of events.  But from any given temporally embedded perspective with a life, as it is being lived, there are any number of futures, any one of which might yet be chosen.

In sum, then, our conception of time differs crucially from our conception of space in that we don't think of ourselves as simply `viewing' different parts of time in temporal sequence as we think of ourselves as viewing different parts of space. We think of the future as existing only in potential until experienced because the future depends on the movements of my will, and the movements of my will are up for grabs, from my own perspective, up until the very moment of choice.
\eq

\section{03-06-10 \ \ {\it Traunkirchen Again}\ \ \ (to the QBies)} \label{QBies21}

On the other workshop, I'm fine with that for anyone who wants to.  I myself had to turn down their invitation since I'm going to be in Australia at another meeting.  It doesn't look quite so exciting as the second meeting, but it'd be interesting to see how people react to Richard Healey's talk, since he's turning partial QBist.  It would give you more time in the Alps to think, contemplate, calculate, get inspired, eat schnitzel, whatever.  (The urungleichung, after all, was born in the Alps.)

\section{03-06-10 \ \ {\it Padmanabhan}\ \ \ (to L. Smolin)} \label{SmolinL21}

I found the talk by your friend yesterday extremely exciting.  (And much better/plausible than Verlinde's.)  It is hard not to feel that there is something really deep in pursuing this direction.

When he talked about observer-dependent entropy, I wanted to say, ``Entropy is {\it always\/} observer dependent in a Bayesian understanding of statistical mechanics.''  I.e., the kind of Ed Jaynes thermodynamics/stat-mech that {\Caves} and I have been promoting all these years.  But I was too shy:  It was too trivial a point in comparison to the other deep ideas I was hearing.

I feel though that there is something wrong about the ``micro-state counting'' part of Padmanabhan's picture.  What came through loud and clear to me instead yesterday is that an ``accelerating observer'' is an ``agent taking an action upon space'' (like in the picture on my office door---it is just that the cube is replaced with empty space).  To accelerate is to initiate a quantum measurement (i.e., taking an action on the external world and suffering its consequence).

Anyway, huge foods of thought here!

\section{03-06-10 \ \ {\it Padmanabhan, 2}\ \ \ (to L. Smolin)} \label{SmolinL22}

Thanks; that was fun!  Who could ever say that the standard quantum foundational issues aren't absolutely crucial for issues to do with gravity.  The Rindler observer {\it is\/} Wigner's friend!

\section{03-06-10 \ \ {\it Gravity and Wigner}\ \ \ (to L. Hardy)} \label{Hardy42}

By the way, yesterday you missed one of the best talks I've ever heard coming through the quantum gravity group here.  You should watch it as you get a chance:  \pirsa{10060000}.

The reason I think I'm so excited is that I saw more clearly than previously that one should think of an accelerating observer as one who is taking an {\it action\/} (quantum measurement) on empty space (the quantum system).  It starts to fit QBist terms for me now, and that always makes me feel happy.

\section{03-06-10 \ \ {\it Elementary Question}\ \ \ (to W. G. Unruh)} \label{Unruh9}

I have a probably stupid question, but maybe you have an elementary explanation.  As I understand it, a uniformly accelerated observer will see black-body radiation in a case where an inertial observer will see none.  People say this arises from the existence of a horizon, blah, blah, blah.  But what of the case of an observer standing on the surface of the earth?  In the idealized case of a free-falling observer seeing {\it no\/} blackbody radiation, would the stationary observer on the surface see some?  There's no horizons here, are there?  Yet, what I understand of the equivalence principle, it would seem that the fellow stuck on the surface of the earth should see the same thing as the fellow in the elevator.\medskip

\noindent Confused in Waterloo,

\section{03-06-10 \ \ {\it A (really, really) Recommended Talk}\ \ \ (to G. L. Comer)} \label{Comer131}

Which email address should I use for you now?  I'm confused; I didn't see your university one come up in my address book.

Anyway, there is a link to what I thought was a fantastic talk below.  And the reasons why I thought it fantastic too.  [See 03-06-10 note ``\myref{Hardy42}{Gravity and Wigner}'' to L. Hardy.]  Still one more conceptual link:  I saw Bill Unruh give a talk on (your old) dumb holes last month in DC.  Afterward I made a comment to him that when two things look too much like each other, one might should stop calling one the analogy of the other.  For instance, one should stop thinking of the numerical identification between inertial and gravitational mass as accidental, but rather the consequence of a deep principle.  I said, ``Maybe you should be looking for your own equivalence principle.''  He joked that he asks his students, ``What do elephants and steam ships have in common?''  and then replies, ``They both carry trunks.  But that doesn't make them the same thing, does it?  Well, it's amazing how often it works in physics to act as if they are.''

But I think I'm not joking.  I think the principle here is that acceleration (through any medium) enacts a quantum measurement upon it.  It is an action upon it.

\section{05-06-10 \ \ {\it A Question on Informational Gravitation}\ \ \ (to H. Poirier)} \label{Poirier3}

I definitely remember you.  And I definitely feel a resonance between Verlinde's latest turn and my QBism program.  See, for instance, this paper of mine:  \arxiv{1003.5209}.  See the end of Section VI on page 25; I cite Verlinde's paper there.  Particularly, I am intrigued with the way he treats spacetime as having emerged ``outside of the boundary'', while inside the boundary there is not yet such a thing as spacetime.

It is all very early in this stuff still, though.  What he seems to have is really more of ``an idea for an idea'' (as John Wheeler would say) than any truly solid idea, much less an actual ``derivation'' of Newton's laws, etc., as he claims in the paper.

Have a look at the papers by this guy, Padmanabhan.  For instance,
\begin{itemize}
\item
\arxiv{0911.5004}
\item
\arxiv{0911.1403}
\item
\arxiv{0910.0839}
\end{itemize}
They support a similar point of view, but I think are much more solid.

I fly to Zurich tomorrow morning, and so may be delayed in my emailing for a day or two, but if you have any further questions, let me know.

\section{05-06-10 \ \ {\it The Consequences of Tommy}\ \ \ (to the QBies)} \label{QBies22}

Saturday morning philosophizing.  Thinking about how the urgleichung fits in with Unruh radiation and the principle of equivalence.  Listening to The Who's album {\sl Tommy}.  And at this very moment, it's going by the words:
\bv
Right behind you I see the millions.\\
On you I see the glory.\\
From you I get opinions.\\
From you I get the story.
\ev

I like that line, ``Right behind you I see the millions.''  For I do:  It's one of those days when I feel like we're doing really very big things.  Small, small steps, all on the path toward something really big.  And from you guys, ``I get the story.''

Let me recommend this interview with John Wheeler for your own Saturday philosophizing:  \myurl{http://www.bigear.org/vol1no4/wheeler.htm}.  Among other things, I think it characterizes well, the role you guys play in my life, or at least it's what I'd like to strive for:
\vspace{-12pt}\bq\noindent
\begin{description}
\item[Interviewer:]  Do you have some thoughts about educating students?

\item[Wheeler:]  Shouldn't you rephrase your question? After all, I'm sure that it is really the students who educate me! We all know that the real reason universities have students is to educate the professors. But, in order to be educated by the students, one has to put good questions to them. You try out your questions on the students. If there are questions that the students get interested in, then they start to tell you new things and keep you asking more new questions. Pretty soon you have learned a great deal.
\end{description}
\eq

Thought of the day in the attachment:
\bq\noindent
It is difficult to escape asking a challenging question. Is the
entirety of existence, rather than being built on particles or fields
of force or multidimensional geometry, built upon billions upon
billions of elementary quantum phenomena, those elementary acts of
``observer-participancy,'' those most ethereal of all the entities
that have been forced upon us by the progress of science?\medskip
\\
\hspace*{\fill} --- John Archibald Wheeler
\eq

\section{07-06-10 \ \ {\it Time Flies} \ \ (to J. D. Norton)} \label{Norton2}

Greetings from Zurich, where I happened to bring a few John Norton papers with me (most on historical matters in GR).  The physicist in me thinks that before using the words ``quantum'' and ``gravity'' in the same sentence, we should take all that we've learned about quantum mechanics and go back to 1907.

Anyway, not quite related to that project, but because I was intrigued, I read your ``Time Really Passes'' and I enjoyed it very much.  I'm glad you've now joined those who believe time flies.  For your enjoyment a little story about John Wheeler below drawn from my email collection (it was originally a letter to Greg Comer).

\subsection{From a 17 December 1997 note to Greg Comer, ``It's a Wonderful Life''}

\bq
Good holidays to you.  This morning, as I was driving to work, it
dawned on me that roughly this day 10 years ago, I was conferred my
degrees at the University of Texas.  Time does fly.

It made me think of a little anecdote about John Wheeler that I
heard from John Preskill a few days ago.  In 1972 he had
Wheeler for his freshman classical mechanics course at
Princeton.  One day Wheeler had each student write all the
equations of physics s/he knew on a single sheet of paper.  He
gathered the papers up and placed them all side-by-side on the stage
at the front of the classroom.  Finally, he looked out at the
students and said, ``These pages likely contain all the fundamental
equations we know of physics.  They encapsulate all that's known of
the world.''  Then he looked at the papers and said, ``Now fly!''
Nothing happened.  He looked out at the audience, then at the
papers, raised his hands high, and commanded, ``Fly!'' Everyone was
silent, thinking this guy had gone off his rocker.  Wheeler said,
``You see, these equations can't fly.  But our universe flies.
We're still missing the single, simple ingredient that makes it all
fly.''
\eq

\section{07-06-10 \ \ {\it Time Flies, 2} \ \ (to J. D. Norton)} \label{Norton3}

\bjdn
Nice to hear from you and thanks for the Wheeler story. There have been just a few extraordinarily people like Wheeler, full of ideas and able to enthuse students.

I expect to be excommunicated from the Church of Philosophers of Physics!
\ejdn

Well, in my eyes, that represents genuine progress.  Here's the way William James once put it with regard to German philosophy:
\bq
In a subject like philosophy it is really fatal to lose connexion with the open air of human nature, and to think in terms of shop-tradition only.  In Germany the forms are so professionalized that anybody who has gained a teaching chair and written a book, however distorted and eccentric, has the legal right to figure forever in the history of the subject like a fly in amber.  All later comers have the duty of quoting him and measuring their opinions with his opinion.  Such are the rules of the professorial game --- they think and write from each other and for each other and at each other exclusively.  With this exclusion of the open air all true perspective gets lost, extremes and oddities count as much as sanities, and command the same attention \ldots
\eq
And I must say, I find myself thinking that of the Philosophy of Physics a lot.  Extremes and oddities sit side by side with sanities, and it is all tolerated in the forum.  Your essay is an example that makes it clear---why should you fear that you'll be excommunicated but that the subject has gone so very wrong.  Don't get me wrong, I think I am personally deeply philosophical, and it is a very worthy business, but most of my influences have not come from the clique of philosophy that calls itself Philosophy of Physics.

Thinking of the ending remarks in your essay, it strikes me that the time is right in your life to rethink James's {\sl Principles of Psychology}, particularly the bit on the perception of time, if you haven't read it recently.  If nothing else, I'll guarantee that the writing will captivate you.

All the best (and I wish you were here for dinner time discussions as I make a bit of progress in your old papers).

\section{09-06-10 \ \ {\it Appleby's Mud Hut}\ \ \ (to R. W. {\Spekkens})} \label{Spekkens85}

With regard to the conversation we just had, here are some excerpts from a ridiculously long email from Marcus to Howard and me:
\bma
The usual attitude, which not only drives the Bohmians and Everettians, but is also dominant among physicists generally, is to regard ordinary language/common-sense as a kind of medieval mud-hut.   A way of thinking which was useful in the past, but whose day is now gone.  Survivals from the past often induce reverence.  One seeks to preserve them.  And indeed that is the attitude of archaeologists to actual medieval huts found in the course of excavation.  However, there are limits.  No archaeologist would want to live in a medieval mud-hut, all dank and dark and smelly (except perhaps for a week or two, just to see what it is like).     Similarly with common-sense, in the minds of most physicists.  It may have a place in the museum.   But so far as serious thinking is concerned the only thing to do is tear it down and replace it with some modernistic structure, all glass  and polished steel.

One of the difficulties with this programme is that a building has to rest on something.  Actual modernistic buildings rest on the same wormy, fungus and bacteria-impregnated earth that the medievals used to build their mud-huts on.  With the modernistic structures we build in our minds it is even worse:  they don't simply rest on the same soil, they actually rest on the mud hut itself, entire and unreconstructed.  Despise the mud-hut at the bottom of the whole thing, as contemporary physicists almost all do, and you are despising the foundations on which everything else depends.

[\ldots]

The dominant tendency among physicists is to identify physics with the developed mathematical theory.  With the equations, in other words.   But if you were only allowed to write equations, and nothing else at all, then I doubt that it would be possible even to get started on the problem of explaining (in response to a question from an undergraduate, for example) just what is so puzzling about quantum mechanics.  To get puzzled by quantum mechanics in the first place you have to draw on ideas which reach all the way down to that poor old primitive mud-hut.  Dank and dark and smelly it may be.  But there is no getting away from it.  Trying to get away from it:  this would be like a tree trying to pull itself up by the roots.  The attempt is unlikely to succeed.  But if it did it could only end in disaster.

Moreover, I don't think it is only quantum foundations which has this dependency on common-sense and ordinary language.  I think the same is true of every other area of physical thinking, without exception.  Walk along the corridors of any physics department and, though you will certainly see people writing on chalkboards, you will also see them talking and writing, using ordinary words.  Moreover  if you take the trouble to inspect what has been written on the chalkboards, though you will doubtless see equations, you are likely also to see simple-minded drawings, assimilable at the level of mud-hut thinking.  Again, any half-way decent talk includes an attempt to provide background, and intuitive motivation:  and those are things which simply can't be done just by writing equations.  It takes words.  Ordinary words.
\ema

\section{12-06-10 \ \ {\it Pauli, Four, and Me}\ \ \ (to R. Renner)} \label{Renner3}

Thanks again for a wonderful, thought-provoking conference.  Schack and I are soon to head out to explore the city and the hills nearby.

I wanted to show you one thing that Heisenberg wrote on Pauli in his essay ``Wolfgang Pauli's Philosophical Outlook'':
\bq\noindent
In the alchemistic view ``there dwells in matter a spirit awaiting release. The alchemist in his laboratory is constantly involved in nature's course, in such wise that the real or supposed chemical reactions in the retort are mystically identified with the psychic processes in himself, and are called by the same names.  The release of the substance by the man who transmutes it, which culminates in the production of the philosopher's stone, is seen by the alchemist, in light of the mystical correspondence of macrocosmos and microcosmos, as identical with the saving transformation of the man by the work, which succeeds only `Deo concedente.' ''  The governing symbol for this magical view of nature is the quaternary number, the so-called ``tetractys'' of the Pythagoreans, which is put together out of two polarities.
\eq
If you compare that to what I wrote for the back cover of the $10^{\mbox{\footnotesize th}}$ anniversary edition of Nielsen and Chuang:
\bq\noindent
Nearly every child who has read Harry Potter believes that if you just say the right thing or do the right thing, you can coerce matter to do something fantastic.  But what adult would believe it?  Until quantum computation and quantum information came along in the early 1990s, nearly none.  The quantum computer is the Philosopher's Stone of our century, and Nielsen and Chuang is our basic book of incantations.  Ten years have passed since its publication, and it is as basic to the field as it ever was.  Matter will do wonderful things if asked to, but we must first understand its language.  No book written since (there was no before) does the job of teaching the language of quantum theory's possibilities like Nielsen and Chuang's.
\eq
you might see why I have a fascination with Pauli!  (I have a big portrait of him on my wall at PI.)

Anyway, thanks again.  Schack and I have done some good science in this city.  It seems to bring out the best in people.

\section{14-06-10 \ \ {\it Displacements All the Way Down} \ \ (to A. Plotnitsky)} \label{Plotnitsky24}

Here is this paper on Weyl that has interested me greatly.  [Erhard Scholz, ``Weyl Entering the `New' Quantum Mechanics
Discourse,'' conference contribution to {\sl History of Quantum Physics 1}, Berlin, July 2--6, 2007.]  The main reason is that my beloved SICs seem to be the dual structure to Weyl's ``phase space.''  And that phase space goes all the way back to at least 27 September 1925!

I'd like to look up the dates (or narrow the range) of when Heisenberg and, particularly, Born found the unitial commutation relations.

Best wishes from just behind you \ldots

\section{15-06-10 \ \ {\it Human Observations}\ \ \ (to R. {\Schack})} \label{Schack202}

BTW, did you notice how the guy from the Austrian Academy of Sciences seemed to insist on invading the space between you and me?  What was up with that?  When I would rotate toward you, he would reinsert himself between us, only looking toward me, keeping the back of his head more toward you.

Very strange.

\section{15-06-10 \ \ {\it How To \underline{Make} SICs Exist}\ \ \ (to J.-{\AA} Larsson)} \label{Larsson9}

Attached is my spiel on willing SICs into existence.  It starts at the heading titled ``Conceptual Barrier!'' [See 13-03-10 note ``\myref{QBies6}{Conceptual Barrier!}''\ to the QBies.] (You'll see that I'm not crazy \ldots\ just very close to being so.)

\section{15-06-10 \ \ {\it Gravity as Thermodynamics}\ \ \ (to F. De Martini)} \label{DeMartini1}

Following on from our conversation on the bus this morning:  The guy I was telling you about is Padmanabhan.  Here a couple of links to his papers:
\begin{itemize}
\item
\arxiv{0911.5004}
\item
\arxiv{0910.0839}.
\end{itemize}
And here is a link to a video of the talk he gave us at PI:
\begin{itemize}
\item \pirsa{10060000}.
\end{itemize}
It'll be good to talk more when you're at PI.

\section{21-06-10 \ \ {\it QBist Interference} \ \ (to J. I. Rosado)} \label{Rosado3}

\bjir
I read with great interest your paper ``QBism, the Perimeter of Quantum Bayes\-ianism''. I think the ideas exposed in this paper not only are useful to find a QBist explanation of the Born rule but also to explain the fundamental concept of quantum interference.
\ejir

Thanks for your continued interest during all these years.  I'm glad you are working out some of these ideas yourself.  I make a bit of a more thorough analogy to interference in \arxiv{0906.2187}.  You might have a look at that paper if you have some interest.

\section{24-06-10 \ \ {\it Idea This Morning}\ \ \ (to R. Renner)} \label{Renner4}

This morning I was thinking about a proposal I'll be writing to the Templeton Foundation in a month or so, and I had an idea that might involve you if you have any interest.  Attached is the pre-proposal I had written for them earlier, and they have given me the go ahead for writing a full proposal.  I think my chances for obtaining the funding are quite good, given their mission and my subject matter \ldots\ [yeah, right]\footnote{Not in original note; addition for the purposes of this samizdat.}.

If you read through this thing, you'll note a certain emphasis on Wolfgang Pauli's thoughts being brought into the modern context.  And particularly, in Sections 3 and 6 you'll find mentions of a workshop on Paulian ideas.  The tentative title for the meeting is, ``A Malleable World?\ Wolfgang Pauli's Dreams for Physics in the Light of Quantum Information.''  It doesn't say so in the proposal yet, but I've been thinking I'd like to have that meeting in Switzerland, perhaps in a place like the small village where Harald Atmanspacher lives, Amden.  (Beautiful setting.)  The point is to have an inspiring setting, and I'll make the argument we'll more easily get some Pauli scholarly work going in Switzerland than anywhere else.  I.e., that this is the natural place for such a meeting.

The idea I had this morning is this:  Might the Pauli Center have any interest in being involved in sponsoring such a workshop?  Or at least lending its name to the proposal, even if no dollars?   As I say, I think my chances of getting this funding are already fairly good, but it never hurts to get as much legitimacy as possible before pursuing these things.  And who knows, maybe the Pauli Center would even get something out of my efforts in return.

Just an idea.  I'd welcome your thoughts pro and con and even sideways.

\section{26-06-10 \ \ {\it We're in the Top Ten!}\ \ \ (to the QBies)} \label{QBies23}

Have a look at this old blog entry of Scott Aaronson's:
\bv
\myurl{http://scottaaronson.com/blog/?p=112}
\ev
Particularly, look at the rules in Number 10.  It was fun to learn that already in 2006 someone voted our research subject one of the ``ten most annoying questions in quantum computing.''

\section{26-06-10 \ \ {\it Things to Read} \ \ (to P. Wells)} \label{Wells1}

Here are a couple of the blogs I was telling you about, by Perimeter-style thinkers, even if some are not presently at Perimeter:
\begin{itemize}
\item
Scott Aaronson's ``Shtetl Optimized'':
\bv
\myurl{http://www.scottaaronson.com/blog/}.
\ev
Scott had been a postdoc with us some years ago; now he's a professor at MIT.
(Funny too:  I was pleased to learn in flipping through it a few minutes ago, as I retrieved the address for you, that my group's main research problem was voted one of the ``ten most annoying questions in quantum computing.''  Quite an honor!)

\item
Michael Nielsen's blog:
\bv
\myurl{http://michaelnielsen.org/blog/}.
\ev
Michael wrote the major textbook in quantum computing, was a professor at U. Queensland, and finally on faculty at Perimeter until he decided to drop out of physics entirely.

\item
John Baez's ``This Week's Finds in Mathematical Physics'' can be found here:
\bv
\myurl{http://math.ucr.edu/home/baez/}.
\ev
John is a cousin of Joan Baez the singer, as I was telling you, and apparently is just about to make a career change as Nielsen did.  (Unrelated, but did you know that Olivia Newton-John is a granddaughter of the great physicist Max Born, one of the founders of quantum mechanics?  I learned recently that he had his beautiful equation
$$
p q - q p = \frac{h}{2\pi i}
$$
carved into his gravestone.)

\item
Sabine Hossenfelder's BackReaction:
\bv
\myurl{http://backreaction.blogspot.com/}.
\ev
Sabine had been a postdoc with us; she works in quantum gravity.

\item
Finally, let me mention Howard Barnum's ``Wine, Physics, and Song'':
\bv
\myurl{http://winephysicssong.com/}.
\ev
Howard is a visiting researcher with us, on leave from Los Alamos National Lab, and working in the quantum foundations group.  He hasn't written on his blog in a while, but some of the entries in it on wine at the Black Hole Bistro are real gems.  What amazes me in reading some of these entries, is that I can well remember being with him at the moments the entries describe \ldots\ and to all outward appearances we were having technical discussions about details of quantum mechanics.  But in the back of his mind, he was clearly writing up his tasting notes!
\end{itemize}

I myself am not a blogger, but I do fancy myself something of a writer.  In prep for our discussions next week, I might point you to Sections 1 and 6 of this paper:
\begin{quotation}
\arxiv{1003.5209}
\end{quotation}
and Section 1 of this one:
\begin{quotation}
\arxiv{quant-ph/0205039}.
\end{quotation}
I think they're written in a way that a general reader can get the point from, and they tell the story of why I think quantum foundations is important.

Also, let me attach the intro to my Cambridge University Press book:  It tells the story of why I'm in physics in the first place (not for the search of truth and beauty, but for science fiction).  You might find it a bit entertaining.

\section{30-06-10 \ \ {\it Two Pictures from Copenhagen}\ \ \ (to the QBies)} \label{QBies24}

The Niels Bohr Institute, one innocent day in June, 2010---what was happening on the outside, and what was happening on the inside.  You'll find the eight new isms I think QBism is pointing us to:
\begin{enumerate}
\item interiorism

\item agentialism

\item meliorism

\item jazzism

\item nonreductionism

\item empiricism / ``the world is not sentence shaped''-ism

\item additivism

\item experientialism

\end{enumerate}

\section{02-07-10 \ \ {\it Version 6, Reply 1}\ \ \ (to H. C. von Baeyer)} \label{Baeyer120}

OK, organize my thoughts, I need to do.  (Spoken like Yoda in my head.)  It's not easy!

I liked your lede (except for the little-bit distracting parenthetical detail on Einstein vs Bohr).  And I wondered what came next!  To make the transition from the (ancient) wave vs particle to the (modern) probability vs click is probably a good way to go for Sci Am readers.

A long time ago when I first thought about writing an article for them, I thought about starting off with something like this:
\bq
In the holiday movie classic {\sl It's a Wonderful Life}, the protagonist George Bailey proclaims in a moment of anguish, ``I suppose it'd have been better if I'd never been born at all.''  It was the very idea his guardian angel Clarence needed to save his soul.  ``You got your wish.  You've never been born.''  The story develops with George seeing how disturbingly different the world would have been without his presence.  As Clarence told it, ``You've been given a great gift, George---a chance to see what the world would be like without you.''  George came to realize how integral his life and his actions were for the very shape of the world around him. There was great wisdom in that movie:  For in the 85 years since the advent of quantum theory, we have started to learn that this may be just be its greatest lesson.  That our actions and their consequences are so integral to what the theory is about, to try to imagine what the theory is saying about a world without us simply cannot be done.
\eq

This certainly was not the vision of science before quantum theory, and is very much the source of why quantum theory often seems to mysterious even today.  Science, and certainly physics, was once thought to be {\it exactly\/} about what the world is like without {\it us\/} playing any active role in its course.  A whole cottage industry has developed \ldots

Anyway, that's old stuff.  As I wrote you at the start of this project, ``Trouble is, I don't think I know how to write something that is {\it not\/} a manifesto---it is a character flaw.''  And that remains true.  I think you're probably much more right on your track.  After getting back from dropping the kids off at the Toronto airport tomorrow morning, I plan to put much more quality time into your old outline and give you some proper feedback, and also to try to figure out what my role can be here---i.e., how I can help you, while letting your seasoned science-writing instinct reign.

Kids off to Texas tomorrow!  And as I say, I proper letter back to you tomorrow.

\section{04-07-10 \ \ {\it Idea This Morning, 2}\ \ \ (to R. Renner)} \label{Renner5}

Thank you for your thoughtful message. Maybe it is best not to push the issue too far with the Pauli Center.

Still!, given what you have written, you make my mind turn:  I start to wonder whether it might be useful to consider eventually writing a Pauli-Center-specific proposal.  Maybe for a follow-on conference for the year after the Templeton one (if that proposal flies).  I can easily imagine a theme that a) would be politically safer than the Templeton stuff, but b) still in line with what I think is most worth developing in quantum foundations.  For instance, a meeting with a title like ``Quantum Contextuality:\  From Wolfgang Pauli to Kochen--Specker and Beyond.''  Maybe you'd have some interest even?  It could be partially historical, partially philosophical, and partially technical.  I write all the time that I think quantum mechanics is just the beginning of something much bigger; the workshop could be on that---giving theorists with good technical skills some better perspective (historical and philosophical) on why they should be pursuing the mathematics that they are pursuing.

Just the ``idea this evening''!  Let it roll around in your head a bit and see if we might do anything with it in the next year.  I, of course, would do most of the legwork if you have any interest.  And if you don't have any interest, that is OK too:  As always, I'm just trying to see what I can help make happen.

\section{06-07-10 \ \ {\it Comments}\ \ \ (to C. Ferrie)} \label{Ferrie16}

\bcf
Are you still around the IQC?  If not, this is just an e-mail to remind myself to follow up on your question in my talk.
\ecf

Sorry, I'm gone already.  Here's the kind of thing I was thinking---please allow me to use a SIC representation of quantum states to get the problem started.  Suppose I have a measurement whose clicks are labeled $j$.  Then the quantum probability $q(j)$ given by the Born rule can be written in terms of the SIC basis according to:
$$
q(j)=\sum_i \left[(d+1)p(i) - \frac{1}{d}\right] r(j|i)
$$
where $p(i)$ is the probability distribution for SIC outcomes and $r(j|i)$ is the probability for a $j$ outcome given that the preparation was the $i$ SIC vector.

In this context, I would call the quantum state corresponding to $p(i)$ ``classical'' in the case that
$$
(d+1)p(i) - \frac{1}{d} \ge 0.
$$
for all $i$.  The reason I would call it that is because then $q(j)$ is calculated according to the classic Law of Total Probability for any measurement at all.

This defines a convex set of the $p(i)$.  In this case, a regular simplex.

Anyway, I was thinking the ``classical states'' must always correspond to an affine transformation of that basic set.

\section{07-07-10 \ \ {\it Leipzig}\ \ \ (to {\AA}. {\Ericsson})} \label{Ericsson10.2}

I'm thinking harder of cleaning my schedule of the Leipzig meeting, and you did a fantastic job with your talk in {\Vaxjo}.  I.e., I trust you now.  I thought about giving a talk titled, ``Probably the Best Statistical Manifold for Quantum Theory,'' as a play on the Carlsberg beer advertisement, but if you could give a talk expressing the same idea, even if not the same title!, this might be something to consider.  Here is a link to the conference venue:
\bv
\myurl{http://www.mis.mpg.de/calendar/conferences/2010/infgeo/} \smallskip \\
and \smallskip \\
\myurl{http://www.mis.mpg.de/calendar/conferences/2010/infgeo/program.html}.
\ev
Please let me know your thoughts.  Is it something you have any interest in going to?  Would you ``strive to'' do an effective sales job on our program while there (not only during your talk, but lunches and dinners and coffee breaks \ldots\ a bit like I would do)?

\section{11-07-10 \ \ {\it Marley Lyrics of the Morning}\ \ \ (to {\AA}. {\Ericsson})} \label{Ericsson11}

\noindent Lyrics to ``Put It On'' by Bob Marley:
\bv
Feel them spirit
\\Feel them spirit
\\Feel them spirit
\\Lord, I thank you
\\Lord, I thank you\bigskip
\\Feel alright now
\\Feel alright now
\\Feel alright now
\\Lord, I thank you
\\Lord, I thank you\bigskip
\\I'm gonna put it on, I put it on already
\\I'm gonna put it on, and it was steady
\\I'm gonna put it on, put it on again
\\Good Lord, help me
\\Good Lord, help me\bigskip
\\I'm not boastin'
\\I'm not boastin'
\\I'm not boastin'
\\Feel like toastin'
\\Feel like toastin'\bigskip
\\I rule my destiny, yeah
\\I rule my destiny
\\I rule my destiny
\\Lord, I thank you, yeah
\\Lord, I thank you\bigskip
\\No more cryin'
\\No more cryin'
\\No more cryin'
\\Good Lord, hear me
\\Good Lord \ldots
\ev
I really like that line about ruling my destiny.

\section{12-07-10 \ \ {\it Dentist Office} \ \ (to P. Wells)} \label{Wells2}

It was part of this:
\bq
For my own part, I imagine the world as a seething
orgy of creation \ldots\ There is no one way the world is
because the world is still in creation, still being
hammered out. It is still in birth and always will be \ldots
\eq
I think the dictionary only used the middle sentence of that.

\section{12-07-10 \ \ {\it Interpretation}\ \ \ (to {\AA}. {\Ericsson})} \label{Ericsson12}

After you left, I had a read through some of the discussion on that Marley song.

One person wrote:
\bq\noindent
{\it his voice sounds ``otherworldly'' here}
\eq
Another one wrote:
\bq\noindent
{\it I know exactly what you mean. I have felt a spiritual presence while listening to his music. It's as if when he sang he became a vessel for something beyond even his own comprehension. This is probably my fav song of his. It's so bare.}
\eq
And earlier, still another had written:
\bq\noindent
{\it 1 year ago 2 I had just left a graveyard after visiting an old beau's headstone with his mom. I had this song on in the car and felt my seat belt constrict for no apparent reason. The song completely spoke to me and that moment and I realized so much about my relationship with God and my old boyfriend. ``Put on the spirit of Christ.'' The reason I love Bob and reggae! ``Good Lord, thank you!''}
\eq

\section{12-07-10 \ \ {\it Interpretation, 2}\ \ \ (to {\AA}. {\Ericsson})} \label{Ericsson13}

My own interpretation is something like the last one.  By ``feeling them spirit'' and ``putting on (something like) Christ'' (i.e., like putting on clothing), he could rule his destiny.  And like the other commentator, one of my own attractions to the song is how bare it is.  I think it is basically a prayer.

\section{12-07-10 \ \ {\it Interpretation, 3}\ \ \ (to {\AA}. {\Ericsson})} \label{Ericsson14}

``put it on" $=$ take, receive, try out.  ``I'm going to wear it.''  ``I'm going to make it part of me.''

\section{12-07-10 \ \ {\it Interpretation, 4}\ \ \ (to {\AA}. {\Ericsson})} \label{Ericsson15}

Here, actually listen to this version: \myurl{http://www.youtube.com/watch?v=McJu3cRcjeo}.
(It is slower than some other versions.)  Definitely has the sense of a chant or a prayer.  I think he was asking for strength and received it.

\section{12-07-10 \ \ {\it QBism, Installments 1, 2, and 3}\ \ \ (to H. C. von Baeyer)} \label{Baeyer121}

I like your introduction even better now.

I'm not sure how I should tackle giving you feedback.  Perhaps recording my ``stream of thought'' might be the most efficient way to go forward at the moment.  At present, I'm finding it extremely difficult to {\it organize\/} any thoughts.

1) As you know, I love it when I learn something just from the very act of writing.  And I love it too when I learn something from my co-author's very act of writing!  These lines were quite thought provoking for me:  ``The resolution of the incongruity between waviness and graininess was effected by a compromise.  Quantum mechanics avoids the description of an electron as a particle or a wave.  Instead, it introduces the new, somewhat nebulous notion of the quantum state.''  I think that is a really good way to put it, and I don't think I've ever quite thought in terms of a compromise before.

2) Whenever/wherever possible, might we use ``a quantum state'' rather than ``the quantum state''?  Remember these lines from my introduction of {\sl Coming of Age}:
\bq\noindent
     Better writing, not just any writing, made for better
     understanding. \ldots\ I started to realize that the major
     part of the problem of our understanding quantum mechanics
     had come from bad choices of English and German words
     (maybe Danish too?) for various things and tasks in the
     theory.  Once these bad choices got locked into place,
     they took on lives of their own.  With little
     exaggeration, I might say that badly calibrated
     linguistics is the predominant reason for quantum
     foundations continuing to exist as a field of research.
     Measurement?  I agree with John Bell---it is a horrible
     word and should be banished from quantum mechanics.  But
     it is not because it is ``unprofessionally vague and
     ambiguous'' as Bell said of it.  It is because it conveys
     the {\sl wrong\/} image for what is being spoken of.  {\it The\/}
     quantum state?  That one is just about as bad.  Who would
     have thought that so much mischief could be made by the
     use of a definite article?
\eq

3) I know I've got to fight my manifesto-izing instincts, but I wonder if we couldn't come off more strongly in the closing paragraph of the introduction?  I think it would be nice to give the reader a sense that something thunderous is happening in our understanding of the quantum before he plunges into the rest of the article.  Somehow I'd like to send a big rush of enthusiasm.  Also, I think it'd be good to take a swipe there at the science-fiction solutions to the quantum conundrum the regular readers of that magazine have surely seen.  I'm thinking particularly of parallel universes and action-at-distance.  Remember, Musser had written this to me just before I thought of proposing the project to you:
\bq\noindent
     We'd need a more straightforward account: here are the
     mysteries of quantum mechanics; HERE ARE THE USUAL
     RESPONSES; here is why those responses fail; here is a
     better way. I suspect that such a reader will actually be
     more receptive to the notion that quantum states represent
     a form of gambling odds than a physicist or philosopher
     steeped in the subject is; we have less baggage to shed.
     But we do share the desire to know ultimately what the
     information is about and, even if this question cannot be
     answered, will want to know how the Bayesian approach
     advances that goal.
\eq
I all-capped the part I was thinking of particularly.  I wouldn't want to dwell on those other ``responses,'' but I think I would like to say something to the effect, ``those sensationalistic responses are all well and good for a Hollywood movie, but in the end they're not very imaginative (like a Hollywood movie) and don't come close to expressing how really cool/awesome/nifty/exciting/unusual the quantum world really is.''  The world is wired in such a way that our actions matter; and that is certainly not the world of Banville's left-hand group.

4) I'm OK with 1/2 of Banville's distinction, but I'm not sure his right-hand group is the right way to put where I sit personally. \ldots\ [MORE TO COME]

I'll come back to that point in minute.  In the meanwhile, let me send this much to you:  I'm not sure when you go to bed, and I want to make sure you have {\it something\/} from me today before you close your eyes.

Picking back up \ldots

4) I'm OK with 1/2 of Banville's distinction, but I'm not sure his right-hand group is the proper way to identify where I (or QBism) sit(s) in the spectrum.  To pinpoint it, maybe I can say this:  I'm not so sure I like to think that a physical theory imposes {\it order\/} on an unruly world, so much as it imposes our will (to the extent that it can presently be imposed).  Remember the description I gave of William James's view in the Templeton pre-proposal:
\bq
     Of scientific theories, the philosopher William James
     once wrote, ``You cannot look on any such [theory] as
     closing your quest. You must set it at work within the
     stream of your experience. It appears less as a solution
     than as a program for more work, and more particularly as
     an indication of the ways in which existing realities may
     be changed. Theories thus become instruments, not answers
     to enigmas. We move forward, and, on occasion, make nature
     over again by their aid.'' On this conception, a theory is
     not a statement about what the world is, but a tool, like
     a hammer, to aid in making the world what we want it to
     be. The world may resist, but to some extent it is
     malleable.
\eq
I'm not saying not to work with Banville's distinction or use it as a hook, I'm just pointing the difference of flavor in my own thought.

5) I did however like Banville's phrase (and your use of it) of a ``boiling chaos''.  It's a bit similar to Chris {\Timpson}'s description of my view:
\bq\noindent
     The four-dimensional pattern is too unruly: it does not
     admit of any parsing into laws, or even weaker forms of
     generalisation, not even statistical ones.

     This, then, is the micro-level we have dubbed unspeakable;
     to which we are denied direct descriptive access. The
     picture is of a roiling mess. Fuchs adds:
\bq\noindent
        For my own part, I imagine the world as a seething
        orgy of creation\ldots\ There is no one way the world is
        because the world is still in creation, still being
        hammered out. It is still in birth and always will
        be\ldots\ (To Sudbery-Barnum 18.8.03)
\eq
\eq
Part of that quote of me, by the way, made it into the {\sl Oxford Dictionary of American Quotations}.  (Surely I've told you this before, but my memory gets worse and worse.)

6) ``Mathematically it is described \ldots''  In line with what I said previously, ``the function'' $\longrightarrow$ ``a function''?  ``can usually not be visualized'' $\longrightarrow$ ``can almost never be visualized''

7) ``Max Born'' \ Worth parenthetically mentioning (grandfather of the pop singer Olivia Newton-John)?  Probably not, but I thought I'd throw it in for consideration.

8) ``psi should be interpreted in terms of a probability'' $\longrightarrow$ ``psi is a tool that can be used to calculate probabilities''?  The reason I bring this up is because (part of) QBism's contribution has been to emphasize that it is ``absolutely nothing more'' than probabilities; it has no existence beyond them.

9) ``negative or even imaginary'' $\longrightarrow$ ``negative or even an imaginary number''?

Since I just got a response from you, I'll chop this much off now and send it.  Back later with stuff on section 2 and beyond.

Picking up again with the random stream of thought (this time more random and vague than previously).

10) ``Atom of Information,'' let me just play around that phrase, or use it as a focus, for a while.  A journalist from {\it Science et Vie\/} this morning wrote me, ``what is for you the meaning of information (in a physics sense) --- a physical substance?''  To which I plan to reply, ``God no!''  One of the key things I want to stress with him, and I guess with you, is that when Shannon quantified information, he quantified it in terms of ``surprise.''  To receive a ``bit of information'' is to say one will be completely surprised by the answer to the relevant question.  (Or, in the quantum case, the consequence of the relevant action.)  If I write down a 0, as opposed to a 1, on a piece of paper, and then announce to you that I wrote down a 0, when I finally give you the piece of paper you will not receive any bits at all:  The paper gives no information because it gives no surprise.  Somehow I'd like to see that crucial point worked into the story:  Information (in a Bayesian understanding) is not a new fluid that we pour into receptacles.

Still, ``atom of information.''  It is a phrase I rather like.  Particularly because it helped me think of qubit as an ``atom of surprise.''  No matter how I take an action on it, it always has the ability to surprise me with the response I receive.  And that is a crucial piece of QBism.

And maybe too emphasizing that ``bit'' quantifies ``surprise'' rather than ``substance'' or ``fluid content'' can help make the transition to what Bayesian probability is all about, ``disciplined uncertainty accounting.''

11) Actually I wouldn't mind having a small box, not particularly on SICs, but on the urgleichung.  A little visual on the sky vs ground measurements, and the simple rule that connects them.  It is a statement that one can do without quantum states completely.  Instead the Born rule is really about ``taking probabilities as inputs and returning probabilities as outputs.''  Does it go too far for {\sl Sci Am\/} to actually show one equation?  Kind of in the way that I did in the attached?

But the way, the SIC was brought out of the sky and placed a little closer to the ground a couple weeks ago.  See:
\arxiv{1006.4905}.

12) Going back to point 10 above.  After laying down what Bayesian probability is about (uncertainty, surprise, not objective features ``already there''), maybe then it is a good time to come to the ``puzzles of qm'' (item 4 in your old outline)?  One could contrast their standard accounts with the QBist stories of the same.

13) Maybe I'll leave it at this for now, not giving an unlucky point 13.

After I hear some of your responses, I'll think harder about a more ``look ahead'' outline.

Time to start heading for home so that I can start heading for the airport to pick up the kids---those wonderful disciplinary resources for keeping old dad working rather than slacking off.

\section{13-07-10 \ \ {\it QBism, Installment 4}\ \ \ (to H. C. von Baeyer)} \label{Baeyer122}

\bhcvb
Do you think we should go through the list of current
``interpretations'' of q.m.\ in the text, the way you did in ``Perimeter'', or
should that be a box?  I agree that the list is necessary, but am on the
fence about the issue of a box.  Whatever the answer, the list has to
come early in the piece.
\ehcvb

14) I tend to think it ought to come in the text directly.  This is part of what I was trying to suggest in Point 3 (Installment 1).  I think that, with some good planning, one should be able to knock out a couple of them relatively quickly.  Or perhaps, another way to go is to put a sentence or two in the text (I would prefer taking a little swipe at them for some spice), but then have a larger box saying a little bit about each mentioning a short-coming:  A) Many worlds, the beloved one of science fiction writers, could have been imagined under any physics, and indeed was.  Aka, it is contentless, despite all the bluster about it.  And even those who think it has content still can't agree whether it can generate probability.  B) Bohm.  Must have faster-than-light ``signals'' but yet, miraculously, they cannot be used to actually signal.  C) GRW-Flash.  The ``flash'' events can always be adjusted to be just out of experimental reach.  Furthermore, as {\Spekkens} emphasizes, the flashes are ad hoc additions that give no unifying power.  I.e., supposing them doesn't help us understand the structure of quantum theory in any way.  D) Consistent Histories (Griffiths version).  Despite its bluster, it doesn't go an inch beyond Copenhagen.  E) Decoherence (Zurek version).  Many worlds in disguise.  Disguised mostly by all the bling it adds to the costume.

\section{15-07-10 \ \ {\it Wednesday Soir\'ee} \ \ (to P. Wells)} \label{Wells3}

Similar to Latham, I want to make an addendum to last night's discussion.

Actually a corrigendum.  When I went onto the porch for my coffee this morning, I realized to my horror that there was a typo in the equation I wrote for you!

I wrote
$$
q(j)=\sum_i p(i)\! \left[ (d+1) r(j|i) - \sum_k r(j|k) \right]
$$
when I should have written
$$ q(j)=\sum_i p(i)\! \left[ (d+1) r(j|i) - \frac{1}{d} \sum_k r(j|k) \right]\;.
$$

Sorry!

\section{15-07-10 \ \ {\it Wheeler's 20 Questions} \ \ (to P. Wells)} \label{Wells4}

Here's an extract (less than five pages) from a big, fat document (194 pages) of mine.  The first page records Wheeler's 20-questions-in-reverse story, the other pages (pages 183--186) expand on it a little more.  It looked like you were interested in the story last night, so I thought you might enjoy reading it in the old man's own words.

\bq
The Universe can't be Laplacean.  It may be
higgledy-piggledy.  But have hope.  Surely someday we will see the
necessity of the quantum in its construction.  Would you like a
little story along this line?

Of course!  About what?

About the game of twenty questions.  You recall how it goes---one of
the after-dinner party sent out of the living room, the others
agreeing on a word, the one fated to be a questioner returning and
starting his questions.  ``Is it a living object?''  ``No.''  ``Is it
here on earth?''  ``Yes.''  So the questions go from respondent to
respondent around the room until at length the word emerges: victory
if in twenty tries or less; otherwise, defeat.

Then comes the moment when we are fourth to be sent from the room.
We are locked out unbelievably long.  On finally being readmitted,
we find a smile on everyone's face, sign of a joke or a plot.  We
innocently start our questions.  At first the answers come quickly.
Then each question begins to take longer in the answering---strange,
when the answer itself is only a simple ``yes'' or ``no.''  At
length, feeling hot on the trail, we ask, ``Is the word `cloud'?''
``Yes,'' comes the reply, and everyone bursts out laughing.  When we
were out of the room, they explain, they had agreed not to agree in
advance on any word at all.  Each one around the circle could
respond ``yes'' or ``no'' as he pleased to whatever question we put
to him.  But however he replied he had to have a word in mind
compatible with his own reply---and with all the replies that went
before.  No wonder some of those decisions between ``yes'' and ``no''
proved so hard!

And the point of your story?

Compare the game in its two versions with physics in its two
formulations, classical and quantum.  First, we thought the word
already existed ``out there'' as physics once thought that the
position and momentum of the electron existed ``out there,''
independent of any act of observation.  Second, in actuality the
information about the word was brought into being step by step
through the questions we raised, as the information about the
electron is brought into being, step by step, by the experiments
that the observer chooses to make. Third, if we had chosen to ask
different questions we would have ended up with a different
word---as the experimenter would have ended up with a different
story for the doings of the electron if he had measured different
quantities or the same quantities in a different order.  Fourth,
whatever power we had in bringing the particular word ``cloud'' into
being was partial only.  A major part of the selection---unknowing
selection---lay in the ``yes'' or ``no'' replies of the colleagues
around the room.  Similarly, the experimenter has some substantial
influence on what will happen to the electron by the choice of
experiments he will do on it; but he knows there is much
impredictability about what any given one of his measurements will
disclose.  Fifth, there was a ``rule of the game'' that required of
every participator that his choice of yes or no should be compatible
with {\it some\/} word. Similarly, there is a consistency about the
observations made in physics.  One person must be able to tell
another in plain language what he finds and the second person must
be able to verify the observation. \medskip
\\ \hspace*{\fill} --- {\it John Archibald Wheeler} $\qquad$
\\ \hspace*{\fill} Frontiers of Time, 1979 $\qquad$
\eq

\section{16-07-10 \ \ {\it IGAIA III -- Missing Titles and Abstracts} \ \ (to IGAIA III)}

I am sorry to keep you waiting.  Here it is.

\bq\noindent
Title:  Rewriting Quantum State Space as a (Nice) Statistical Manifold\medskip\\
Abstract:  Recent work in the field of quantum information theory has pointed to the significance of a very special kind of quantum measurement---the so-called symmetric informationally complete quantum measurements (or SIC).  This structure is only known to exist for finite-dimensional quantum state spaces with dimension $d = 2$ to $67$, but they are widely believed to always exist for all finite dimensions.  If they do exist, then quantum state space (i.e., the convex set of trace one positive semi-definite matrices) can be rewritten as a $d^2$ dimensional (very symmetric) statistical manifold within the $d^2$ dimensional probability simplex.  Furthermore, in this language, the Born Rule for calculating probabilities for all other quantum measurements transforms into a very simple variant of the classic Law of Total Probability.  In this talk, I will present all the above in greater depth and list several open questions that the information geometry community may be in a position to answer.  Some details can be found at:
\arxiv{0906.2187} and \arxiv{0910.2750}.
\eq

\section{16-07-10 \ \ {\it Where Women Roar and Men Thunder} \ \ (to D. B. L. Baker)} \label{Baker26}

It's been a long time since I've written you from a long flight.  I've got nothing particular to say, but {\it it is a long flight\/} that I am on, and I suppose I hope that this will inspire me to say something.  I'm on my way to Los Angeles from Toronto, the first leg of a trip to Brisbane, Australia.  I arrive in LA at 9:10 PM, and then depart for Brisbane at 11:30---two hours of misery in a way.  My body will be thinking it's between the hours of 12:10 and 2:30, but yet I dare not sleep.  At least once I'm on the plane, 14 hours later, I'm in Brisbane.

I believe this is my sixth trip to Australia, though only my second to Brisbane.  The other ones have been to Sydney.  I'm going to the QCMC conference (Quantum Communication, Measurement, and Computation), where I pick up my \$3K award and make an after-dinner speech.  I hate making after-dinner speeches:  For some reason they put me on edge like a scientific talk never does.  Maybe it has something to do with something Paul Simon said about song writing.  He said, in writing a song, he strives to start with a statement of truth; after that everything flows smoothly.  Maybe that's why I'm disoriented with the after-dinner thing:  I don't know where the truth lies in those contexts.

We were talking about having a walk-about together.  I still think that's a good idea.  Just got to figure out how to pull it off.  I tabulated the other day that I've flown nearly 49,000 miles already since January 1 (46K of it on American Airlines where it earns me free flights more efficiently).  Immediately after this trip, I take the family to Munich (separating to go first to a conference in Austria and then a conference in Leipzig).  Then in October, I go to Cape Town, South Africa (Stellenbosch, a wine region, actually).  And finally in December I go back to Brisbane.  So, I plan to have a butt-load of frequent-flyer miles at the end of this year.  We just have to do something once I get some of this work cleared out of the way.

My kids took their first great Texas adventure last week.  Emma (11) and Katie (8) flew for the first time by themselves.  My sister Cathy picked them up in Houston, and then got them back to the airport at the end of the week.  Next year, they fly to Europe for the first time without Kiki and me.  It's all part of the accelerated growth plan.  I was 30 already before I said anything new and useful about quantum mechanics, or natural philosophy more generally.  My thinking is that they ought to beat me by at least 10 years.

After reading all the {\sl Harry Potter\/} books three times, and the {\sl Percy Jackson\/} series, I've finally bamboozled Emma into reading the {\sl Lord of the Rings\/} trilogy.  She's fought it tooth and nail the last couple of years.  But I offered to pay her 10 cents a sentence for every email she writes me about what she's read in the books while I'm away in Australia, and that seemed to tip the scales.  She's been saving for an iPod Touch.  Ah, priorities!!

Hey, check out this blurb I wrote for the back cover of Nielsen and Chuang's textbook on quantum computing (10th anniversary edition).  It made me feel literary when I put it together:
\begin{quote}
Nearly every child who has read Harry Potter believes that if you just say the right thing or do the right thing, you can coerce matter to do something fantastic.  But what adult would believe it?  Until quantum computation and quantum information came along in the early 1990s, nearly none.  The quantum computer is the Philosopher's Stone of our century, and Nielsen and Chuang is our basic book of incantations.  Ten years have passed since its publication, and it is as basic to the field as it ever was.  Matter will do wonderful things if asked to, but we must first understand its language.  No book written since (there was no before) does the job of teaching the language of quantum theory's possibilities like Nielsen and Chuang's.
\end{quote}
If Emma hadn't become so fascinated with {\sl Harry Potter}, I'd have never been able to write that.  There's so much to learn from our kids.  How are your three doing?  What sorts of growth and what sorts of mischief have they been up to?

PS.  I must say, I still haven't had a vegemite sandwich.

\section{18-07-10 \ \ {\it Where Women Glow and Men Plunder} \ \ (to D. B. L. Baker)} \label{Baker27}

It seems my memory failed me yesterday.

Oh well \ldots\ head full of zombie,

\section{20-07-10 \ \ {\it An Honest Attempt}\ \ \ (to C. M. {\Caves})} \label{Caves102}

\bcc
Well, of course, this is exactly why I didn't want to be answered by an abstract, because it doesn't even start to answer my question.  I'm fully aware that a good reason for doing things---jeez, a really good reason that I often use myself---is that one is exploring fundamental issues that might---who knows?---turn out to be important.  But I was looking for something more than that in this case.
\ecc

OK, an attempt at an actual answer.

It has to do with the trail of thought I've been following since 2001 or so.  The only way I know how to express it is in analogy to Newton's law of universal gravitation.  Newton's law is that every two masses in the universe attract each other with a force proportional to a product of two numbers intrinsic to each:  Their gravitational masses.  Newton identified a property common to all matter.  But beyond that, it was silent on what is real of the world.  It was not a nuts and bolts ontology in the sense that one usually seems to want of quantum mechanics---i.e., that the wave function is real, or the Hamiltonian is real, or the Bohmian trajectory is real, or the multiverse with its universal wave function is the sole reality, etc., etc.  All Newton's theory does is say, ``Look here.  Here is a property common to all matter, in all its manifest and yet to be discovered forms.  Every piece of the world has gravitational mass, and the significance of that number is blah, blah, blah.''

Now, I think that is the best way to understand quantum mechanics.  I.e., to scale back our reductionist dreams for it and say, ``It is not at all a theory tailor made for supplying an ontology in the traditional sense.  Instead it is like Newton's law in identifying a single, important aspect of all pieces of the world.''  That aspect, I think, is what we conventionally call Hilbert space dimension, $d$.  Newton's law was a great achievement for physics even though it didn't pretend to be a theory of everything, and the same I say of quantum mechanics.  If it can solidly identify one---just one---aspect of this world common to all things, even if it says nothing beyond this, that is enough for me.

So, my whole thought is geared toward pinpointing what the essence of this quantity --- dimension --- really is.

That is why I am attracted so much to the urgleichung.  This is because it displays Hilbert space dimension front and center {\it as\/} ``deviation from a naive (or unjustified) application of the law of total probability.''  When we deal with a certain piece of matter, quantum theory advises that we should deviate---by a particular amount, $d$---from the law of total probability.  The more $d$ a system is hypothesized to have, the more we should deviate from a naive application of the LTP in the diagram I always draw.  That is something that the usual form of the Born Rule just buries away; it isn't seen at all in the usual complex-Hilbert-space way of writing the rule.

Now, for my reason for pursuing a deep detailed study of the urUNgleichung.  These ``maximal consistent sets'' seem to have a rich enough structure to hint of the possibility that the full structure of quantum-state space might just come from the preservation of the consistency of the urgleichung.  In other words:  It is that the principle that brings dimensionality out front and center as a deviation from the LTP is the same principle that sets the full structure of quantum state space.

If that {\it turns out\/} to be true, then I see it as a great unifying step for QBism.  Our thought hangs together better than it used to, and such a theorem would clear out just that much more underbrush.  And with that, the hope is we'll be able to see the full forest much better for it.

To a person who wants to build a better gravity wave detector {\it today}, I would guess all of this will make no difference to his life.  But to the person who wants to go the next step in physical theory (say, to talk about gravity and quantum in the same sentence, but not necessarily this), my gut says it will make all the difference in the world.  My gut says, despite all the erosions in confidence the live-for-today physicist tries to dole out upon me, that this is the right way to go.  At least it is the right way for me.  My gut says the fruits of this effort will have lasting value that will long outlive me.

I honestly believe what John said, ``Until we see the quantum principle with this simplicity [in one clear, simple sentence] we can well believe that we do not know the first thing about the universe, about ourselves, and about our place in the universe.''  To the extent of my limited ability, that is what I want to be about.  And I feel our emphasized rewriting of the Born Rule is a step in that direction.

\section{20-07-10 \ \ {\it An Honest Attempt, 2}\ \ \ (to C. M. {\Caves})} \label{Caves103}

\bcc
Thanks for the answer.  I was hoping for something that would, without appealing to the esoterica of quantum gravity, tell me why you're working on this particular program.
\ecc
But I wasn't at all appealing to quantum gravity for any kind of justification.  I was appealing to the idea of the value of identifying one property common to all things.  I.e., nonreductionism.  Gravitational mass just happened to be the example.  OK, no more emails: I will desist.

\section{21-07-10 \ \ {\it Those James Quotes} \ \ (to G. J. Milburn)} \label{Milburn6}

Here they are.  The first concerns how I think you should view your new center, as an addition to your ``self'':
\bq\noindent
     In its widest possible sense, however, a man's
     self is the sum total of all that he can call his,
     not only his body and his psychic powers, but his
     clothes and his house, his wife and children, his
     ancestors and friends, his reputation and works,
     his lands and horses, and yacht and bank account.
\eq
With this new arm, you may do things you had never previously imagined possible.

The second quote concerns German philosophy in the early 1900s, but my experience tells me it is completely apropos of the present-day ``philosophy of physics'' (considered a professional field).
\bq
     In a subject like philosophy it is really fatal
     to lose connexion with the open air of human nature,
     and to think in terms of shop-tradition only.  In
     Germany the forms are so professionalized that
     anybody who has gained a teaching chair and written
     a book, however distorted and eccentric, has the
     legal right to figure forever in the history of the
     subject like a fly in amber.  All later comers have
     the duty of quoting him and measuring their opinions
     with his opinion.  Such are the rules of the
     professorial game --- they think and write from each
     other and for each other and at each other
     exclusively.  With this exclusion of the open air
     all true perspective gets lost, extremes and oddities
     count as much as sanities, and command the same
     attention \ldots
\eq

The quote on the self comes from his {\sl Principles of Psychology}.  I've never read the whole book, but Scott Aaronson wrote me once:  ``I'm working through it now.  I've decided to recommend it to people as `the most up-to-date, state-of-the-art book about consciousness' (without telling them the publication date).''

The quote on German philosophy is from his book, {\sl A Pluralistic Universe}.

The first book that really made an impression on me was simply titled, {\sl Pragmatism}.

\section{23-07-10 \ \ {\it Subjectivism in Classical Statistical Mechanics}\ \ \ (to P. G. L. Mana)} \label{Mana18}

A key point (I think in agreement with what you say), is to be found in the second and third paragraphs after Merminition 150 of ``Extract for Luca.''  Also something to think about is Keynes' remark on Ramsey in Extract 2, that the ``basis for our degrees of belief \ldots\ is \ldots\ perhaps given merely by natural selection.''  [See 08-11-06 note ``\myref{Mermin124}{November 8th}'' to N. D. Mermin, Mermition \ref{MerminitionForLuca} now, and the subsection titled ``The Projection Postulate as Bayesian Updating'' in the 24-06-06 note ``\myref{Bub21}{Notes on `What are Quantum Probabilities'}\,'' to J. Bub.]

\subsection{Luca's Preply}

\bq
This question became more interesting to me after I finally proved to myself that the maximum-entropy principle is just as `subjective' as the rest of plausibility logic \arxiv{0911.2197}.

It seems as if there is much to say, and very little to say, about it. The most honest books on statistical mechanics make clear that the various distributions used are just postulated --- they are used because they work. Any `derivation' is only an a posteriori complacency.

Indeed, one could say that we have arrived at them through a sort of historical Bayesian updating: we have tried some until we arrived at those that work, i.e., that are not further updated by new data. This is corroborated by the fact that the usual distributions do not work in some cases, for granular materials e.g., and new distributions are used there --- trial and error again.
\eq

\section{23-07-10 \ \ {\it What GK Say}\ \ \ (to C. M. {\Caves})} \label{Caves104}

From G. K. Chesterton's book, {\sl Heretics}:
\bq
There are some people---and I am one of them---who think that the
most practical and important thing about a man is still his view of
the universe.  We think that for a landlady considering a lodger it
is important to know his income, but still more important to know his philosophy.  We think that for a general about to fight an enemy it
is important to know the enemy's numbers, but still more important to know the enemy's philosophy.  We think the question is not whether
the theory of the cosmos affects matters, but whether in the long run
anything else affects them.
\eq

\section{26-07-10 \ \ {\it Just Noticed!}\ \ \ (to A. Kent)} \label{Kent24}

I just noticed the name change:
\begin{center}
Centre for Quantum Information and Foundations (formerly Centre
for Quantum Computation)
\end{center}
That's great!  Congratulations.

\section{03-08-10 \ \ {\it Noncolorable Sets}\ \ \ (to A. Cabello)} \label{Cabello6}

\bac
Would it be interesting to have a finite non-(KS)colorable set of:

(a) trines

(b) regular tetrahedrons?
\eac

Regular tetrahedrons for sure.  (Assuming you are talking about for a
qubit.)  More generally, it would be quite nice to know if in arbitrary finite dimension $d$, there are always noncolorable SIC sets.\medskip

\noindent Cheers from Munich (working at my in-laws' house),

\section{04-08-10 \ \ {\it Help}\ \ \ (to C. M. {\Caves})} \label{Caves105}

\bcc
I gave a public lecture here the other night on why quantum mechanics is strange and tried to show that it means that things don't have objective properties.  Afraid I offended my officemate, Cyril Branciard, who is pointing me to Gisin's work contending that nonlocality is the only
assumption in proving Bell inequalities.   I don't have time to read and
come up with good answers, so give me some help.  I will read it, but I'm sure you've thought about it much more deeply than I will.
\ecc
Yeah indeed people say that \ldots\ all the time.  Read for instance this ``very clear'' exposition of Travis Norsen: \arxiv{quant-ph/0601205}.

My own opinion is that each and every time this claim is made, the claimer is either explicitly or implicitly (and in a roundabout way) assuming the EPR criterion of reality.  And an essential piece of that is that {\it certainty\/} implies {\it truth\/} (in the ontic or correspondence sense).

I'll send you some of my conversations with {\Spekkens}, {\Mermin}, etc., on the subject if I can dig it up.

I think my own best exposition on the subject is Section V (starting page 14) of \arxiv{1003.5209} [``QBism, the Perimeter of Quantum Bayesianism''].  I think that section can be read independently of the rest of the paper.

\section{07-08-10 \ \ {\it  Scientific Report} \ \ (to M. A. Graydon)} \label{Graydon9}

\bmag
I want to thank you again for making the trip to Traunkirchen possible. I had the pleasure of meeting and discussing physics with some really interesting people while I was there --- making the long trip well worth its while. After you left, I gave a short informal talk to a small group on quaternion qt and the general Jordan-Banach algebraic structures that have been on my mind lately. Now back home, I am realizing just how beautiful our surroundings were. It's left me with a sense of optimism for the future, both personally and work-wise.
\emag

They were fantastic surroundings.  Mother earth is big, and she brings her size and grandeur up front for your inspection in the Alps.  I hope you'll use every opportunity like this to see the world from a different angle.  There'll be many opportunities in your future.  The mind is an amazing connected web of thoughts, plans, feelings and facts, and one shouldn't think for a minute that an emotion in the Alps may not play a role in a mathematical theorem somewhere down the road in Waterloo or anywhere else.

\section{10-08-10 \ \ {\it Whaddya Think?}\ \ \ (to T. Jacobson)} \label{Jacobson1}

Every new paper like this gives me occasion to recall my wisdom teeth being pulled.  No matter how many times it's recalled, it never becomes more pleasant.

\subsection{Ted's Preply}

\bq
\arxiv{1008.1066}
\eq

\section{12-08-10 \ \ {\it Slide Permissions Question} \ \ (to J. Gleick)} \label{Gleick1}

Your publisher sent the note below to me, assuming that you had already written me.  I'm not quite sure which slide she is talking about.  Maybe you can shed light.  Also, it'd be nice to know what your book is about.

I read your {\sl Chaos\/} book many years ago and enjoyed it much.  And your Feynman book has sat on my shelf for a couple years calling me, calling me.

\subsection{James's Reply}

\bq
Oh, gosh. I've been meaning to call or write. Thanks for emailing me directly.

The book is a sort of history of information and information theory. It has Claude Shannon as its core, and goes back to the invention of writing and forward to quantum information science, and meanwhile explores many issues that are more cultural than scientific.

The slide of yours that I want to use, by way of illustration, is attached. Contrary to what Jill says, I don't think a higher resolution version is needed. In the text, I quote something you said in ``Quantum Mechanics as Quantum Information (and Only a Little More)''---the passage about holy tumult over quantum foundations, and your comment, ``Quantum mechanics has always been about information; it is just that the physics community has forgotten this.''

Of course I'll send you the book when it's released. For that matter, would you want to see it in galleys? I'd certainly welcome any comments or corrections you might have. I'd ask only that you keep it entirely to yourself for now.

Thank you also for the kinds words about {\sl Chaos}, and I'm glad the Feynman book has at least made it as far as your shelf.
\eq

\section{16-08-10 \ \ {\it Whaddya Think?, 2}\ \ \ (to T. Jacobson)} \label{Jacobson2}

Sorry for the delay:  I wrote you my ``smart ass'' note just as I was preparing to come home from Europe, and then I needed a few days of ``recovery from the vacation'' before looking at email again.

I'm going to refer you to Carlton Caves for this one, since he and {\Ruediger} Schack have a nice paper on the subject:
\bq
\arxiv{quant-ph/0409144}
\eq
(I'd refer you to {\Ruediger} too, but I know he's hiking in France for the next two weeks.)

\subsection{Ted's Preply}

Let's take it out of the cosmology context.

Do you object a priori to the crazy idea that there {\it are\/} an infinite number of copies of the system, and that the composite {\it is\/} always an exact eigenstate of the relative frequency operator?

If not, that leaves the question whether and in what sense it implies the Born rule for {\it us\/} (one of those copies), when {\it we\/} do the experiment a {\it finite\/} number of times, say 100. A brief perusal of their paper gave me the impression that they were analyzing this latter situation by positing that we sample the infinite number of copies with equal ``statistical'' probability, so we measure, statistically, the relative frequency operator.  But I'm not sure this is what they were saying.
And it is suspect, since our 100 repetitions of the experiment either breaks the symmetry between us and the rest of the subsystems, or all subsystems do it, i.e., the whole infinite composite does the experiment 100 times. Then one can ask about the relative frequency operator of just our outcomes, and we know the system cannot be in an eigenstate of this, because that would preclude the 1/sqrt(100)=10\% fluctuations that we know QM gives \ldots\ which suggests that we can't deduce the Born rule after all \ldots

\section{16-08-10 \ \ {\it Slide Permissions Question, 2} \ \ (to J. Gleick)} \label{Gleick2}

Thanks for the information.  (Hee hee.)  Sorry to contact you quickly and then leave you high and dry, but I wrote you just as I was traveling back home from Europe \ldots\ then there was the weekend.

Your new book sounds interesting.  Thanks very much for your kind offer to send me a copy!  Attached is a little ``obituary'' I had written on Shannon for the participants of a meeting I was organizing when he died.  I wonder now whether you were the author of whatever {\sl New York Times\/} article I linked to then?  [See 27-02-01 note ``\myref{ClaudeShannonDeath}{Claude Shannon's Death}'' to the ``Shannon meets Bohr'' invitees.]

And thanks for letting me know which slide you need.  I think I can supply you nonetheless with a cleaner scan than the one you have.  Just give me a couple of days to dig it up.

Your publisher asked me about ``licensing fees'' for the slide, but how about we just do this?  If you'd cite a couple of my papers on the subject (along with the old one you quote), as well as the upcoming Cambridge University Press reprint of my book {\sl Notes on a Paulian Idea\/} (in this incarnation it is called {\sl Coming of Age with Quantum Information}), I'd be most appreciative.  Attached is a copy of the Introduction to the new edition of the book; you might enjoy some of the character sketches of quantum information figures in it (like Landauer, Bennett, Peres, etc).  And the more up-to-date paper I'm thinking for the citing is this one: \arxiv{1003.5209}.
I would think that Section IV sounds particularly relevant to the thing from the old paper that you're already quoting.  For instance, this paragraph would seem to indicate so:
\bq\noindent
     In this regard, no question of QBism tests nature's
     tolerance more probingly than this. If quantum theory
     is so closely allied with probability theory, if it
     can even be seen as an addition to it, then why is it
     not written in a language that starts with probability,
     rather than a language that ends with it? Why does
     quantum theory invoke the mathematical apparatus of
     complex amplitudes, Hilbert spaces, and linear
     operators?
\eq

Anyway, sound like a deal?

\subsection{James's Reply}

\bq
No need to make a deal. I don't really think you want to charge a licensing fee for the slide, whatever my publisher says, and in any case I cite papers I have consulted and found useful, which includes your wonderful {\sl Notes on a Paulian Idea\/} among several others. Would you please send me the correct bibliographic information for the CUP book, so that I can update the reference?

Thank you also for the QBism paper; I can see that it's going to cause me to update the book. Luckily it's not too late.
\eq

\section{17-08-10 \ \ {\it Hitler's Fireplace} \ \ (to D. B. L. Baker)} \label{Baker28}

Thanks for the BBC story.  I've been meaning to write you for some time.  Particularly, I wanted to write you from Germany, but then things got in my way---I'm certainly not the correspondent I used to be.

What I had wanted to tell you about was my trip to Berchtesgaden, and the emotions it brought with it.  Chances are, it's too late to express them properly now.  Probably one of those things where you had to be there.  It hit me as I was explaining to my girls that Hitler's Eagle Nest (or Kehlsteinhaus) was more than just a place to have a very fancy tea party or to simply impress the visiting diplomats.  It was intended to put fear into the hearts of those visiting diplomats as well.  It was a subtle way of saying, ``Look what the power of this nation can do.  Take note.  What we can build for ourselves, we can destroy of yours.''  You really get a sense of this as you drive up the mountain, going through about five tunnels.  Then you get to the level where Hitler's Berghof used to be, and you walk into a tunnel deep into the mountain crafted by Italian stone cutters, finally entering Hitler's original (massive) bronze plated elevator to get the rest of the way to Kehlsteinhaus.  So, I was really feeling this sense of the Third Reich's strength and power \ldots\ when I ran across a picture of Eisenhower and Montgomery viewing a room I had just been standing in.  I think it was the room with Hitler's fireplace, now quite damaged from all the chips of marble soldiers had broken from it.  There was Eisenhower, standing in the damage, a symbol of America.  And I turned to the girls and said, ``All that power, and America's in the end was greater.''  ``They put all these stones together on this high, high mountain to show their power and start a war.  And we ended the war by putting 14 pounds of plutonium in just the right place.  And just 14 pounds!  What subtlety.  The thousands of beautifully cut stones meant nothing in comparison; power is measured by what you can make happen.  14 pounds in just the right place.''

For a few minutes, I was very proud again to be an American and to think of those old victories of mind and industry.

If you want to remind yourself of Hitler's Eagle Nest, this seems to be a pretty good site:  
\myurl{http://www.songsofwar.info/hitlermountain/Kehlsteinhausnew.html}.

Attached are a couple of pictures from the trip.  The first is of the whole family, taken at the level of Hitler's Berghof and Goering's home (i.e., not all the way at the top).  The second is of Katie and me at a biergarten back in Munich.  (Don't worry: That's an apfelschorle in Katie's hand, not a beer!)

\section{18-08-10 \ \ {\it Your Paper} \ \ (to J. I. Rosado)} \label{Rosado4}

Congratulations!  You did it all without me.  {\sl Foundations of Physics\/} sent me the paper to referee, and I didn't accept to review it:  I thought it would be better for everyone if you had impartial reviewers.  And it succeeded even without my help.  This is a very good sign for the science!  [See J. I. Rosado, ``Representation of quantum states as points in a probability simplex associated to a SIC-POVM,'' Found.\ Phys.\ {\bf 41}, 1200--1213 (2011).]

\section{21-08-10 \ \ {\it The Chartreuse Microbus}\ \ \ (to H. C. von Baeyer)} \label{Baeyer123}

\bhcvb
One of the important ingredients of our paper is going to have to be an analogy for the SIC Born rule.  Your measurement in the sky or at NIST won't do it, because analogies must not refer back to the concept being explained -- quantum measurements in this case. Analogies must be to images that are instantly familiar.  Even Fourier series fails on that count, tho it's closer.  I'll suggest alternatives later. However, for now, I seem to recall that you wrote a paper with a student about the Fourier analogy but I can't find it in the {\tt arXiv}.
Does it exist?
\ehcvb

I think you're thinking of the conversation below.  [See 11-01-10 note ``\myref{Baeyer94}{Analogy?}''\ to H. C. von Baeyer.] The equation numbers must refer to the version of the paper I had sent you just before.  I'll re-attach that version for reference.

Just like Fourier expansion is a very special functional expansion (into sine waves), so is a SIC expansion (into a nice set of positive semi-definite operators).

For argument's sake though, I don't think the bureau of standards conception is circular at all.  It is more like the Lord of the Rings:
\bv\it
Three Rings for the Elven-kings under the sky,\\
Seven for the Dwarf-lords in their halls of stone,\\
Nine for Mortal Men doomed to die,\\
One for the Dark Lord on his dark throne\\
In the Land of Mordor where the Shadows lie.\\
One Ring to rule them all, One Ring to find them,\\
One Ring to bring them all and in the darkness bind them\\
In the Land of Mordor where the Shadows lie.\\
He paused, and then said in a deep voice,\\
``This is the Master-Ring, the One Ring to rule them all.''
\ev
I'll expand and explain once I can get my hands more firmly on the keyboard.  (I'm presently having cheese and sausage and sliced turkey for lunch---finger food all.)

\section{21-08-10 \ \ {\it One Ring to Bind Them, One Ring to Rule Them All}\ \ \ (to H. C. von Baeyer)} \label{Baeyer124}

OK, my fingers are clean.

Here's what I meant by saying that the SIC is a bit like the Lord of the Rings.  Like all quantum measurements, it is an action that an agent can take on a system with some number of unpredictable consequences for him---so it is not special in that regard, or in any way explanatory of the concept of quantum measurement.  However, it does have a special defining feature that sets it apart from all other measurements.  It is an action (and the only one) for which the agent can judge the following:  That if he contemplates performing it twice on his system (getting outcomes $i$ and $j$ respectively), no matter what his Bayesian probabilities for the first outcome $i$, he will assign an almost flat conditional probability $p(j|i)$ for the second outcome $j$.  That is, he will assign a $p(j|i)$ of the form $(a, a, \ldots, a, b, a, a, \ldots, a)$, where the value $b$ inhabits the $i$'th slot of the probability vector, no matter what his prior $p(i)$ was.

And that one simple defining feature---in the end---gives the SIC all its expressive power for the Born Rule for all other measurements.  What a nice primitive for a world believed to be indeterministic to its core.

\ldots\ Ah, now that I've written that I don't feel I've shed the light that I wanted.  I'm not getting to the key point in the right way.  I may try again later in the day.

\section{22-08-10 \ \ {\it Things from the Middle of the Night}\ \ \ (to H. C. von Baeyer)} \label{Baeyer125}

Let me think on the new draft.  Two broad thoughts at the moment.
\begin{enumerate}
\item
Something like the red paragraph at the end of Section 2 could work, but it should be re-written to be more in Hans style than Chris style.  At present it kind of stands out a bit too obviously (and even when the color was changed from red to black).
\item
Section 3 is more sympathetic with frequentism presently than I can personally sign on to.  W. K. Clifford once wrote, ``It is wrong always, everywhere, and for anyone, to believe anything upon insufficient evidence''---which is not something I subscribe to (give me James's ``will to believe'' any day).  But I do find Clifford's wording appropriate for frequentism: ``It is wrong always, everywhere, and for anyone.''  And that opinion will cause havoc for using the Nagel quote as a pivot as well.  In the game of probability, one cannot give an inch without losing a mile.  I think these two papers by our friend {\Appleby} make the point better than nearly any other source that I've read:
\begin{center}
\arxiv{quant-ph/0408058}\\
\arxiv{quant-ph/0402015}.
\end{center}
``Probability is single case, or nothing.''
\end{enumerate}
I shall meditate a bit.

\section{23-08-10 \ \ {\it Probability}\ \ \ (to H. C. von Baeyer)} \label{Baeyer126}

2:15 AM, at the moment; I had started to become restless at 12:39.  I.e., it's another one of those nights.

Thank you so much for this note.  I don't think I can adequately express the relief in seeing it.  When I finally roused myself from bed five minutes ago, I thought that I would be writing you a small essay that started like this:

A frequentist says probability is long run frequency (always).  A Bayesian says probability is judgment {\it always}. \ldots\ A young man has been tossing a favorite coin for a long time, 10,000 trials and has noted the frequency of heads and tails is almost exactly 50/50.  Now it comes time to gamble on the next toss---now it comes time to put his observations to use.  The young man confidently writes down 50/50 for his probability assignment \ldots\ \ BUT WAIT!  Just before he's had to act on that assignment he notes the strangest feature of his sequence of coin tosses:  By some miracle, it is a binary representation of the digits of $\pi$!  He uses {\sl Mathematica\/} to check that there's not one error.  How could that be?  Is there some hidden mechanism behind the toss?  ``Hogwash, it can't be,'' he says, ``I've been doing the tossing myself.''  Still a bookie walks from around the corner, and offers a lottery ticket on the young man's next toss.  ``Receive \$1,000,000 if heads'' for the price of \$65,000.  Will he buy the ticket?  Or not?  That will depend upon his judgement.  He thinks to himself, ``It could have been extraordinary luck.  If probability is frequency, then this very string of heads and tails was among the possible.  But it has no hold over the precise detail of the next coin toss.  I will use my study of the frequency to temper myself.  It would be foolish to buy such a ticket.  That is my salary for a whole year; I would be devastated if I lose that much money.''  But if instead he judges the finding of $\pi$ significant, miraculous even---for instance that God is with him in some way in his coin tossing (and I would think myself very likely to judge something like that in such an extreme case)---then he would accept the ticket.  All his experience in the past---not just some coarse grained variable like frequency---tells him that it is the right thing to do.  He lives by the dictum, ``Never throw away information,'' and proceeds to buy the ticket.  Which judgement is the right one?  The young man will have time to contemplate that when the moment to buy or not to buy the ticket has passed and the outcome of the next coin toss is in hand.

Moral:  A (past) frequency can be used to determine one's (future) probability assignment if one {\it judges\/} it to be relevant to that assignment.  The judgement, however, is always there in the background; it cannot be gotten rid of.  The Bayesian is often considered shrill for pointing this out over and over, but losing sight of it is the great original sin of frequentism.  Past frequency is one thing; future gambling (i.e., that which is measured by one's probability assignment) is another.

Anyway, I thought I was going to have to write you an essay like that, but a much more careful one, one that tried to anticipate your every countermove.  ``But what about this?  But what about that?''  It would have taken a lot of thought.  And then perhaps a long email debate would have ensued---it has happened to me before. \ldots\ And I must say I was dreading it!  BUT you saved the day!!!  AND I bet I can fall back asleep now!!!

\subsection{Hans's Preply}

\bq
OK, I see, I'll pour the weak tea down the drain.  Somehow I thought that we could define the two prevailing approaches to probability before launching into QBism, but I see now that we must enthrone Bayesian probability before we get to q.m.

Don't bother to save parts of section 3.  I'll start it over, taking comfort only in the implication of {\Appleby} that it IS possible to derive probabilities from finite ensembles, the way everybody does, provided only that one admits to sloppy thinking:
\bq\noindent
Frequentists are impressed by the fact that we infer probabilities from frequencies
observed in finite ensembles. What they overlook is the fact that we do not infer
probabilities from just any ensemble, but only from certain very carefully selected
ensembles in which the probabilities are, we suppose, constant (or, at any rate,
varying in a specified manner). This means that statistical reasoning makes an
essential appeal to the concept of a single-case probability: for you cannot say that
the probability is the same on every trial if you do not accept that the probability
is defined on every trial.
\eq
This will take a little longer.
\eq

\section{23-08-10 \ \ {\it Probabilismo!}\ \ \ (to H. C. von Baeyer)} \label{Baeyer127}

I don't know how much prodding to cap off your apparent ascendancy to personalist Bayesianism, but in case you need just a bit more, let me give another reading recommendation.  This one is de Finetti's soul-searching article Probabilismo.  More than any other thing I've read in the foundations of probability, this article affected me the most deeply.  (It took four readings, I should add, before it had this effect on me, but the final result was nonetheless real.)

I hope I didn't cause you too much grief yesterday---as {\Ruediger} and I know too well, schisms in our thoughts are always painful even if necessary.

\section{23-08-10 \ \ {\it B. Russell You Might Like} \ \ (to D. M. {\Appleby})} \label{Appleby91}

I'm having a horribly boring day going through the final proofs of my Cam U Press book, but I came across a quote of Bertrand Russell that I had sent to Bill Wootters in 1999 that you might quite like:
\bq\noindent
Physics is mathematical not because we know so much about the physical world, but because we know so little:  it is only the mathematical properties we can discover.
\eq
Looking forward to your return to Waterloo \ldots\

\section{24-08-10 \ \ {\it Your Eloquent Point}\ \ \ (to R. W. {\Spekkens})} \label{Spekkens86}

I remember you once making the point that to really hope to contribute to quantum foundations, one should know as much of modern physics as possible.  And I thought you made the point rather eloquently.  Have you ever written that kind of thing down anywhere?  I want to use it to help give my student Matthew some guidance on choosing his courses.

\section{25-08-10 \ \ {\it Particle Physics Course} \ \ (to M. A. Graydon)} \label{Graydon10}

Sorry to keep you waiting.  I remembered that my colleague Spekkens had something rather elegant to say on the issue, and I wanted to get that quote from him and send it in my reply to you.  But Spekkens has no clue what I'm talking about.  So, let me just wing it.

Rob says that one can't really hope to make a deep contribution to quantum foundations unless one has a good working knowledge of all the other things going on in  modern physics.   (He said it more elegantly.)  I think there's some truth in that.  And from that perspective, a high energy course would be just the right thing.

\section{25-08-10 \ \ {\it Time for Templeton}\ \ \ (to D. C. Lamberth)} \label{Lamberth2}

I hope you remember our brief encounter from April.  I'll place my original letter of introduction below, as well as your reply to it.  You might also go to the Perimeter Institute website \myurl{http://www.perimeterinstitute.ca/} if you want to get a picture of me in your head; there happens to be a banner on the homepage at the moment picturing me for a recent award I won.

This time I'm writing you because the Templeton Foundation pre-proposal that I sent you got past their first round of refereeing, and I have been invited to write a full proposal.  (I'll attach the pre-proposal again for your convenience.) Thus I'm gearing up to do that---I just started today, and it is due September 15.

In that regard, I'd like to explore again the idea that you join me as a co-organizer in planning a workshop on pure experience and quantum measurement.  I presently envision a workshop of about 15--20 people, for about 10 days, in a nice setting.  Its participation would consist of a carefully chosen combination of quantum foundational physicists, pragmatist philosophers (possibly a smattering of other neutral monist philosophers), and theologians, with plenty of time for long discussions---a way for the communities to get to know each other and get to know each other's considerations.  I am imagining the meeting sometime in 2012, and ideally I think I would like it to be held somewhere in New England, where the spirit of James is particularly strong:  Harvard, near Chocorua, don't know.

If you are amenable, I would like to list you among the ``personnel'' in this proposal, and start up a dialog, even as I write it in the next twenty days, over some of the details of the shape of such a workshop.  The only obligation on you with regard to the grant would be in having an active participation in the organization of the workshop.  I would try to get enough money from the Templeton Foundation so as to completely cover all the expenses for the meeting.  Also, I would of course give you an opportunity to edit any piece of the proposal that has to do with you personally.

I understand it is quite bold of me to contact you like this without either of us knowing each other:  But life is short, and I want to make things happen.  It is time for physicists and pragmatist philosophers to start talking to each other, and I finally see an opportunity that can give that some possibility of happening.

I hope to hear back from you soon, whatever your opinion.

\section{27-08-10 \ \ {\it Excerpts}\ \ \ (to L. Freidel)} \label{Freidel3}

Here's a link to the New York Times article on how the structure of one's native language influences what his attention is drawn to:
\bq\noindent
MAGAZINE   | August 29, 2010 \\
Does Your Language Shape How You Think? \\
\myurl{http://www.nytimes.com/2010/08/29/magazine/29language-t.html?emc=eta1}\\
By GUY DEUTSCHER \\
The idea that your mother tongue shapes your experience of the world may be true after all.
\eq
On the other hand, here is an excerpt from the introduction to my book:
\bq
I started to realize that the major part of the problem of our understanding quantum mechanics had come from bad choices of English and German words (maybe Danish too?) for various things and tasks in the theory.  Once these bad choices got locked into place, they took on lives of their own.  With little exaggeration, I might say that badly calibrated linguistics is the predominant reason for quantum foundations continuing to exist as a field of research.  Measurement?  I agree with John Bell---it is a horrible word and should be banished from quantum mechanics.  But it is not because it is ``unprofessionally vague and ambiguous'' as Bell said of it.  It is because it conveys the {\it wrong\/} image for what is being spoken of.  {\it The\/} quantum state?  That one is just about as bad.  Who would have thought that so much mischief could be made by the use of a definite article?  So many of these things came to mind as I would strive to be clear and entertaining in my writings to David \ldots
\eq
As I came to realize while talking to you, I think that's why I was so interested in the language article this morning.

Finally another excerpt related to our discussions---this one from one of Rob Spekkens's papers:
\bq
We shall argue for the superiority of the epistemic view over the ontic view by demonstrating how a great number of quantum phenomena that are mysterious from the ontic viewpoint, appear natural from the epistemic viewpoint.
These phenomena include interference, noncommutativity, entanglement, no cloning, teleportation, and many others. Note that the distinction we are emphasizing is whether the phenomena can be understood conceptually, not whether they can be understood as mathematical consequences of the formalism, since the latter type of understanding is possible regardless of one's interpretation of the formalism. The greater the number of phenomena that appear mysterious from an ontic perspective but natural from an epistemic perspective, the more convincing the latter viewpoint becomes. For this reason, the article devotes much space to elaborating on such phenomena.

Of course, a proponent of the ontic view might argue that the phenomena in question are not mysterious if one abandons certain preconceived notions about physical reality.
The challenge we offer to such a person is to present a few simple physical principles by the light of which all of these phenomena become conceptually intuitive (and not merely mathematical consequences of the formalism) within a framework wherein the quantum state is an ontic state. Our impression is that this challenge cannot be met. By contrast, a single information-theoretic principle, which imposes a constraint on the amount of knowledge one can have about any system, is sufficient to derive all of these phenomena in the context of a simple toy theory, as we shall demonstrate.
\eq
It is that point that really drives me:  By adopting a way of thinking of quantum mechanics so that it seems to come from a single, very far reaching principle---rather than simply coming out of the blue---I feel we'll more quickly get to the next stage of physics.

\section{30-08-10 \ \ {\it Some Bayesian Crap}\ \ \ (to P. G. L. Mana)} \label{Mana19}

\bpglm
Bayesian crap apart, what still surprises me is how students don't ask ``but what determines the individual outcomes?'', as any good physicist should do. How many of such questions do you get? Are we just teaching them to become good parrots?  Attached is the image of a `vacuum noise' current measurement, from Leonhardt's book. How can people be happy to know what's the distribution of the peaks without asking why and how it's going up and down that way?

It still baffles me.

How could Bohr, Heisenberg, and other philosophical clowns have such a big influence?
\epglm

Well, the philosophical clowns influenced me deeply.  I had no illusions that they ``had all the answers'' or even ``most of the answers,'' but---to me at least---they did have interesting directions of thought.  Here is a recent formulation of my own:
\bq
Most of the time one sees Bayesian probabilities characterized (even by very prominent Bayesians like Edwin T. Jaynes) as measures of ignorance or imperfect knowledge.  But that description carries with it a metaphysical commitment that is not at all necessary for the personalist Bayesian, where probability theory is an extension of logic.  Imperfect knowledge?  It sounds like something that, at least in imagination, could be perfected, making all probabilities zero or one---one uses probabilities only because one does not know the true, pre-existing state of affairs.  Language like this, the reader will notice, is never used in this paper.  All that matters for a personalist Bayesian is that there is {\it uncertainty\/} for whatever reason.  There might be uncertainty because there is ignorance of a true state of affairs, but there might be uncertainty because the world itself does not yet know what it will give---i.e., there is an objective indeterminism.  As will be argued in later sections, QBism finds its happiest spot in an unflinching combination of ``subjective probability'' with ``objective indeterminism.''
\eq
It comes from \arxiv{1003.5209}.

Appleby as well has some interesting points on the clowns in connection with Bayesianism in the introduction to \arxiv{quant-ph/0402015}.

\section{31-08-10 \ \ {\it From the Vault}\ \ \ (to A. Kent)} \label{Kent25}

\bak
But scientifically I'm a little puzzled.    I'd thought probably most people at
this point would guess that SICs exist in all dimensions.   (Am I wrong?)
\eak

No, you're right.  It was just one of those things that had been nagging my soul.  At the end of a talk I would often get the question, ``But what if they don't exist in all dimensions?  What of QBism then?''  I found that I would always get into apologetic mode, squirming around a bit, and saying something like, ``Well, it'd really make no philosophic difference to QBism, but the ultimate formalism would be uglier.  The world might have been better, but it wasn't.''  Maybe the audience couldn't see it, but I'd always shrink a bit inside---sweaty palms and all that.  Now instead, I feel I'd rise to the occasion.  When someone asks, ``But what if they don't exist?'' I'll shoot back, ``Even better!!  Then I'd have a theory---not quantum mechanics---of my very own!  I'd be complete in a way Richard Feynman never felt he was!''  (From a story in {\sl Surely You're Joking\/} or somewhere like that.)  You get the point.\footnote{\editornote The story is in {\sl Surely You're Joking, Mr.\ Feynman!} (1985), and refers to his work on the weak interaction in 1957.  ``It was the first time, and the only time, in my career that I knew a law of nature that nobody else knew.  The other things I had done before were to  take somebody else's theory and improve the method of calculating, or take an equation, such as the Schr\"odinger Equation, to explain a phenomenon, such as helium.''  According to Gleick's biography, ``Gell-Mann's rage could be heard through the halls of Lauritsen Laboratory, and he told other physicists that he was going to sue'' (p.~411).  Later editions of {\sl SYJ} included a parenthetical disclaimer:  ``Of course it wasn't true, but finding out later that at least Murray Gell-Mann---and also Sudarshan and Marshak---had worked out the same theory didn't spoil my fun.''  Much later, Gell-Mann told Gleick, ``He was not at all a thief of ideas---even very generous in some ways.  It's just that he was not always capable of regarding other people as really existing.''  Cf.\ the 03-10-07 note ``\myref{Caves96.1}{Feynman and Bell}'' to C.\ M.\ Caves and the 04-10-07 note ``\myref{Preskill18}{Feynman Question}'' to J.\ Preskill.}

\section{01-09-10 \ \ {\it Opening the Mind of Max}\ \ \ (to M. Schlosshauer)} \label{Schlosshauer40}

BTW, ever since you called my house a palazzo, I've kind of fancied to the term and use it all the time.  This morning I coined Palazzo del QBismo and am thinking about modifying the QBism glossary accordingly.

\section{02-09-10 \ \ {\it No Laughing Matter}\ \ \ (to R. W. {\Spekkens})} \label{Spekkens87}

OK, I'm procrastinating from my Templeton theological efforts \ldots\  {\it but}, two points for me tonight.  (I'll make a rule---first person to five gets a free beer.)  I looked up the twins story.  (1) It wasn't laughs, it was smiles.  And (2) Sacks interjected an 8-digit prime.  Corroborating passages below.

\bq
The second time they were seated in a corner together, with a mysterious, secret smile on their faces, a smile I had never seen before, enjoying the strange pleasure and peace they now seemed to have.  I crept up quietly, so as not to disturb them.  They seemed to be locked in a singular, purely numerical, converse.  John would say a number---a six-figure number.  Michael would catch the number, nod, smile and seem to savour it.  Then he, in turn, would say another six-figure number, and now it was John who received, and appreciated it richly.  They looked, at first, like two connoisseurs wine-tasting, sharing rare tastes, rare appreciations.
\eq
and
\bq
I returned to the ward the next day, carrying the precious book of primes with me.  I again found them closeted in their numerical communion, but this time, without saying anything, I quietly joined them.  They were taken aback at first, but when I made no interruption, they resumed their `game' of six-figure primes.  After a few minutes I decided to join in, and ventured a number, an eight-figure prime.  They both turned towards me, then suddenly became still, with a look of intense concentration and perhaps wonder on their faces.  There was a long pause---the longest I had ever known them to make, it must have lasted a half-minute or more---and then suddenly, simultaneously, they both broke into smiles.\footnote{\editornote From Oliver Sacks' {\sl The Man Who Mistook His Wife for a Hat} (Summit Books, 1985).}
\eq

\section{03-09-10 \ \ {\it Three Rabbis in a Bar}\ \ \ (to H. C. von Baeyer)} \label{Baeyer128}

By the way, I have read the document they're wanting comment upon in that question, ``The Philanthropic Vision of Sir John Templeton,'' very thoroughly and twice.  I do realize his disciples will know him inside and out, but my reading of that document is that QBism really is on his side (and vice versa) in many, many of the matters expressed there.

But as you say, best not to speak directly of the tetragrammaton.

Still, quoting Sir John himself:
\bq\noindent
Possibly, we can become servants of creation or even helpers in divine creativity.  Possibly, we are a new beginning, the first creatures in the history of life on earth to participate consciously in the ongoing creative process.
\eq

\section{03-09-10 \ \ {\it The Strip Tease}\ \ \ (to H. C. von Baeyer)} \label{Baeyer129}

Well, Kiki and the kids finally got home from Buffalo.  So I guess I have to go now.  If I follow your suggestions in the quickest fix way, just stripping out the presumption about Sir J, and rearrange the remark on radical pluralism (see below), that leaves me with 115 more characters to fill in.  But now, what to say!?!?

I'll have to think on it, but Kiki's here now.  If there's anything that you'd like to see said, suggest away!

\bq
A student who heard the applicant give a guest lecture on QBism in a graduate course on quantum foundations sent an email saying, ``I like your theory because it returns to me as much freedom as I feel that I have. Such freedom is lost or partially lost in other interpretations [of QM].'' This in fact is the heart of the matter for QBism's development of quantum theory and the heart of the matter, as we see it, with respect to Sir John's Donor Intent.

By solid science, QBism has arrived at a stage of seeing human freedom as no less illusory and no less crucial to quantum theory than an electron's freedom as it ``passes through'' a Stern--Gerlach device. We feel that we have a window on quantum theory that has never been peered through before.  Through the tools of quantum information theory and a sense that Wolfgang Pauli and John Wheeler were on the right track to understand the meaning and significance of ``quantum measurement'' (our two answers to the two Big Questions of this Funding Priority), we have come to a vision of quantum theory that finds its most natural expression in the humanistic terms William James used to convey his radically pluralistic ontology.

As we hope to make clear in the more detailed parts of this proposal, what is within grasp through a methodical rewrite of quantum probabilities into Bayesian probabilities is that, at the same time as they bring relief to the quantum foundational conundrums, they indicate a world not ``one'' in its composition, but ``many''--a partial community of partially independent powers, a pluriverse.

Will Durant put it most beautifully, ``The value of a [pluriverse] as compared with a universe, lies in this, that where there are cross-currents and warring forces our own strength and will may count and help decide the issue; it is a world where nothing is irrevocably settled, and all action matters.''
\eq

\section{03-09-10 \ \ {\it The Strip Tease, 2}\ \ \ (to H. C. von Baeyer)} \label{Baeyer130}

\bhcvb
I'm sorry, but I still don't really understand what you mean by freedom in your opening passage.  Freedom from what?  Could your 115 characters help here?
\ehcvb
Freedom of will, of course.  The boy was saying QBism rejects the block universe picture of things, and he liked that.  No other interpretation of quantum mechanics gives him that, he says.  Would it help if I simply put the first and second paragraphs back together (as they were in the preproposal and didn't seem to set of any alarm bells in you then)?

I'm trying to pin down what exactly is confusing you.

Strange morning:  For the first time ever, our big golden retriever decided he was frightened of walking down the staircase and wouldn't go down.  He wouldn't let himself be dragged either without bringing the house down with him.  Even the bribery of food wouldn't work.  He was absolutely frightened.  Now, where did that come from?

\section{03-09-10 \ \ {\it Putnam Story and Others}\ \ \ (to D. C. Lamberth)} \label{Lamberth3}

Hilary Putnam was your PhD advisor, wasn't he?  I've been looking for a little story I wrote up on him once \ldots\ expressing my near horror over his turn toward the Bohmian interpretation of quantum mechanics in 2008 (in a talk at the Bill {\Demopoulos} Fest), but so far I've had no luck.  When I find it---if it's nice and joking enough!---I'll send it to you.  As I recall, he called information theoretic approaches to quantum theory like my own, ``instrumentalism with a postmodern sauce.''

Anyway, I like writing stories on people.  Attached are some I wrote for the introduction of my Cambridge University Press book.  A couple of the stories are about philosophers, I don't know if you know any of these guys---Jeffrey Bub, Allen Stairs, Richard Healey (another student of Hilary's), and Abner Shimony.  You might know Abner through proximity.  I tried to display his sweetness in a story starting on page 30, and he plays a bit part in the less serious Zeilinger story on page 34.

Hope you enjoy---the Zeilinger story is particularly relevant for my interactions with you:
\bq\noindent
Personal anecdotes aside, {\Anton} {\Zeilinger} is certainly a philosophical uncle to the very core of this book, the Paulian idea, and maybe he deserves the title for that alone.  Here's the way he put it in a recent interview:
\begin{quotation}
\noindent I call that the two freedoms: first the freedom of the experimenter in choosing the measuring equipment---that depends on my freedom of will; and then the freedom of nature in giving me the answer it pleases. The one freedom conditions the other, so to speak. This is a very fine property. It's too bad the philosophers don't spend more time thinking about it.
\end{quotation}
Simple and clean the notion is:  Too bad indeed the philosophers don't spend more time thinking about it!
\eq

\section{04-09-10 \ \ {\it House Warming Bubbles}\ \ \ (to H. C. von Baeyer)} \label{Baeyer131}

Of course I wouldn't want you to enter an empty house!  This time I hope I have successfully navigated your concerns.

The references in the Donor Intent section refer to these passages in one of Sir JT's books:
\bq
Although we seem to be the most sophisticated species at present on our planet, perhaps we should not think of our place as at the end of cosmogenesis. Should we resist the pride that might tempt us to think that we are the final goal of creation? Possibly, we can become servants of creation or even helpers in divine creativity. Possibly, we are a new beginning, the first creatures in the history of life on earth to participate consciously in the ongoing creative process. [{\sl Possibilities}, p.\ 41]
\eq
and
\bq
The laws of the spirit refer to patterns of voluntary human behavior, not to the involuntary behavior of physical objects. A person is free to choose to act in accord with these spiritual laws or to try to defy them. This being the case, the patterns which these laws express are not uniformly exhibited by humans at all times. Rather, they represent the ideal patterns to which humans may aspire.

Conformity to the laws of the spirit is a free choice of all responsible humans. So perhaps, to avoid misunderstandings, we should call them spiritual principles. [Possibilities, p.\ 154]
\eq
In the latter case, think of interpreting the Born Rule as a normative rule, and my figure with the tablet of the ten commandments (which I know you still hate).  Anyway, they're solid, defensible references for me.  In the terminology of Sir JT, QBism says that the Born Rule is a law of spirit or spiritual principle!

\section{06-09-10 \ \ {\it On Dice, Divinity, and Telephone Companies}\ \ \ (to D. M. {\Appleby})} \label{Appleby92}

I walked over to tell Hulya that Rogers (or Bell or whoever you used for your telephone line) said that they'd be by to install it sometime between 11:00 and 2:00 today.  She said that she'd be there.  Kiki told the telephone people that if for any reason they could not get into your apartment, they should give us a call here at home.

By the way, I reread your paper ``Concerning Dice and Divinity'' over the weekend.  I liked your remark near the end, ``By contrast, the epistemic interpretation obliges us to concede to the cat a degree of metaphysical privacy.''  Looking back, that remark probably had quite some influence on me, expressed today, for instance, in my long rants in Sections 3 and 6 of the long recent QBism paper, \arxiv{1003.5209}.

\section{08-09-10 \ \ {\it Research Proposal}\ \ \ (to H. B. Dang)} \label{Dang11}

\bhbd
Maybe this would cheer you up if I tell you that I start to understand the ``ladder of faith'' now.

When I first looked at the Vanier Scholarship it was like ``I want this, but this seems hopeless''.
Then when I read about the selection criteria more carefully: ``I do think I have some chance''.
After I talked to you about it: ``I might actually get it''.
After you submitted your letter: ``I ought to get it''.
I'm not yet at ``I must get it'', but I will fight for that!
\ehbd

Indeed, James's ideas in use always cheers me up.  I am impressed that you paid attention to the faith ladder.

\section{09-09-10 \ \ {\it Your Manuscript LT12790 Medendorp}\ \ \ (to A. M. Steinberg \& Z. E. D. Medendorp)} \label{Medendorp2} \label{Steinberg5}

I will return to a civilized state (as opposed to the very furry Neanderthal-ish cave-dwelling persona I've taken on in the last few days), when my full proposal to the John Templeton Foundation has been turned in September 15.  (It's on SICs and theology, of course.)  In the meantime, maybe someone else will have some useful input on Aephraim's question about how to respond to the instrument matrix thing.  Sounds like it's too late to reconstruct it now.  It seems the question is now how to find the most positive and politic way to say, in essence, ``Well we didn't do it, and we ain't gonna either.''

\section{07-09-10 \ \ {\it Agony}\ \ \ (to D. M. {\Appleby})} \label{Appleby93}

In agony over this letter writing (long story that I'll have to make meaningful for you eventually), I can't sleep \ldots\ and I can't write either.  I turned to the {\sl New York Times\/} for diversion.  Anyway, here is something I just read that struck me as worth recording for myself:
\bq\noindent
Science too has its share of mysteries (or rather: things that must simply be accepted without further explanation). But one aim of science is to minimize such things, to reduce the number of primitive concepts or primitive explanations. The religious attitude is very different. It does not seek to minimize mystery. Mysteries are accepted as a consequence of what, for the religious, makes the world meaningful.
\eq
In a sense, I think, QBism incorporates religion into science at just that point:  Every quantum system is a little core of mystery.

\subsection{Marcus's Reply}

\bq
You rang just as I started to reply to this.  I absolutely agree.  I think this question of ``mystery'' and ``explanation'' is extremely important.   People in the {\Spekkens}--Hawking--Weinberg mould invariably regard mystery rather in the way that catholics regard a black mass, as an evil entire and unadulterated.  And explanation as a corresponding good.  That is a good in the absolute sense.  For myself I certainly wouldn't want to reverse the emotional loading.   Obviously, explanation is OK in its place, and there are mysteries which are crying out to be resolved (for instance the mystery of why SHW believe what they do).   However, it really does need to be kept in its place.  The word ``to explain'' means, literally, to lay something out flat.  The idea being that when one lays it out flat one can see what is going on.  Which can often be an interesting and useful thing to do.  But of course SHW aren't thinking of just explaining this thing or that thing, in some limited context, as the occasion may arise.  They want to explain everything, in some absolute way.  EVERYTHING in capital letters.   And that I think is a perversion.  Laying the entire world out flat:  if that were possible wouldn't it mean that the world was very boring?  Indeed, the word ``flat'' is often used more or less a synonym for ``boring''.

Explanation is certainly a good.  But only in the way that food is a good.  How would it be if someone got so obsessed with eating that they came to see stuffing their face as the meaning of existence?  I think the way SHW regard explanation is similarly pathological.

Incidentally, I think the definition of mystery as something that ``must simply be accepted without further explanation'' misses the point.  The religious attitude of mind is not only willing, it positively desires to contemplate a mystery.  Roll it around the mind rather in the way that one rolls wine around the mouth in order to appreciate it.  Verbal articulation has a place in this process.  Just as Howard will articulate the meaning that a wine has for him, so there are any number of religious writers who will articulate the meaning of a religious mystery.  What makes it mysterious is not the fact that it just has to be accepted without further words or thought. Rather it is the fact that it is recognized that there is more to it than words can fully capture.  That is obviously true of wine (reading Howard's wine tasting notes is one thing; drinking the wine quite another).  Why can we not approach quantum mechanics in a similar spirit?  ---not as a ``mystery'' in the pejorative sense that SHW attach to this word, but rather in the way that anyone, them included, appreciates a glass of wine?  (it may, incidentally, be apposite to remark that wine is an essential part of one of the central Christian mysteries).

We should talk about this some more, once you have finished your letter and Templeton writing.  Also those articles you sent me on language and artificial intelligence.  Perhaps over a glass of wine?
\eq

\section{08-09-10 \ \ {\it Mixed Signals}\ \ \ (to H. C. von Baeyer)} \label{Baeyer132}

Uh oh.  I thought the guy went out of his way to say {\it no\/} manifesto.  So when I show him a piece of another manifesto (i.e., a piece of the JTF proposal), I didn't quite expect a reaction like this:
\bgm
Damn, this is great:
\bq\noindent\rm
The research program of Quantum Bayesianism (or QBism) is an approach
to quantum theory that hopes to show with mathematical precision that its
greatest lesson is the world's plasticity. With every quantum measurement set
by an experimenter's free will, the world is shaped just a little as it takes part in
a moment of creation. So too it is with every action of every agent everywhere,
not just experimentalists in laboratories. Quantum measurement represents
those moments of creation that are sought out or noticed.
If this vision of quantum theory stands scrutiny, it will mean that modern
physics itself already speaks of a humanistic world---a world of hope \& struggle
\& possibility \& change. That would be the project's enduring impact.
\eq
It almost makes me angry that you haven't written the Sci Am article yet!  Because it's going to be a great article.
\egm
Maybe he's just bowled over by the ideas.

What are you going to be up to in Spain?  How long will you be gone?  I don't think I've been to Spain for 12 years, but I still have an overpowering memory of garlic soup, red beef, and Rioja wines.

Bit.  Stone.  Qubit.  Philosopher's Stone.  We shall come back to all that after Spain.  Have you ever watched the Dustin Hoffman movie, {\sl Mr.\ Magorium's Wonder Emporium}?  My family thinks it's a really wonderful movie, despite all the critics' pans of it.  Recommended viewing for all QBies, in fact.  Note the similarity between the Congreve Cube and the QBist qubit.

\begin{description}
\item[Magorium:]  Come with me.  This, my lovely, is for you.

\item[Mahoney:]  Thank you.  What is it?

\item[Magorium:]  It's the Congreve Cube.

\item[Mahoney:]  It looks like a big block of wood.

\item[Magorium:]  It is a big block of wood.  But now, it's your big block of wood.

\item[Mahoney:]  Thank you.  I was just saying last night I don't have enough big blocks of wood.

\item[Magorium:]  Unlikely adventures require unlikely tools.

\item[Mahoney:]  Are we going on an adventure?

\item[Magorium:]  Well, my dear, we're already on one.  All I will say is this: With faith\ldots\ love\ldots\ this block\ldots\ and a counting mutant, you may find yourself somewhere you've never imagined.
\end{description}

\section{08-09-10 \ \ {\it Two Dialogues on QBism}\ \ \ (to the QBies)} \label{QBies25}

Both from the movie {\sl Mr.\ Magorium's Wonder Emporium} with Dustin Hoffman as Magorium and Natalie Portman as Mahoney.\medskip

\noindent Part 1)
\vspace{-18pt}
\bq\noindent
\begin{description}
\item[Magorium:]  Come with me.  This, my lovely, is for you.

\item[Mahoney:]  Thank you.  What is it?

\item[Magorium:]  It's the Congreve Cube.

\item[Mahoney:]  It looks like a big block of wood.

\item[Magorium:]  It is a big block of wood.  But now, it's your big block of wood.

\item[Mahoney:]  Thank you.  I was just saying last night I don't have enough big blocks of wood.

\item[Magorium:]  Unlikely adventures require unlikely tools.

\item[Mahoney:]  Are we going on an adventure?

\item[Magorium:]  Well, my dear, we're already on one.  All I will say is this: With faith\ldots\ love\ldots\ this block\ldots\ and a counting mutant, you may find yourself somewhere you've never imagined.
\end{description}
\eq
\vspace{-6pt}
\noindent Part 2)
\vspace{-18pt}
\bq\noindent
\begin{description}

\item[Mahoney:] Then why are you leaving?

\item[Magorium:] It's my time to go.

\item[Mahoney:] That's it?

\item[Magorium:] What else could there be?

\item[Mahoney:] What are we gonna do without you?

\item[Magorium:] Run the store.

\item[Mahoney:] Sir, I don't know how.

\item[Magorium:] That's why I gave you the Congreve Cube.

\item[Mahoney:] But it just sits there.

\item[Magorium:] What have you done with it?

\item[Mahoney:] I don't know what to do with it.  It's a block of wood.

\item[Magorium:] Can you think of nothing?

\item[Mahoney:] Well, I'm sure I could think of a million things to do with it.

\item[Magorium:] There are a million things one might do with a block of wood, but, Mahoney, what do you think might happen if someone just once \ldots\ believed in it?

\item[Mahoney:]  Sir, I don't understand.
\end{description}
\eq

\section{10-09-10 \ \ {\it Ouch!}\ \ \ (to S. Hartmann)} \label{Hartmann16}

I was just thinking it'd be nice to have the book on Nancy Cartwright's Phil Sci that you helped edit on my bookshelf \ldots\ and then I saw the price!  Ouch; maybe not.

Lately, I've taken an interest in her (never noticed her before) because of my newest thoughts on Hilbert space dimension as a ``capacity''.  In that regard, Sections 6, 7, and 8 of this paper \arxiv{1003.5209} might amuse you.

By the way, is it true that Cartwright doesn't take email (as it advertises on her home page)?  In my files, I do actually have an email from her (turning down a conference invitation).  I guess I ask because I am thinking about writing her, but I'd like to understand beforehand how to interpret the silence I'll most likely get in response!!

How are things going in your new chair?  You must be getting settled in comfortably by now.

\section{12-09-10 \ \ {\it Interpreting the Universe after a Social Analogy}\ \ \ (to D. C. Lamberth)} \label{Lamberth4}

I guess because you're on my mind, I reread your article in the {\sl Cambridge Companion\/} last night.  It was as nice as I remembered it to be.  I hope you realize from all that I've sent you (especially my last paper, ``QBism, the Perimeter \ldots'') that your title (that Jamesian phrase)--- Interpreting the Universe after a Social Analogy--- is exactly what I'm after and what I think our QBistic analysis of ``Wigner's Friend'' in quantum theory (in Section III) leads straight away to.  There's a world of interesting physics in front of us, and the time is ripe---I've just got to muster the resources and sustain the community's attention long enough for this line of thought to sink in!

There's one sentence in your paper that quite captures a distinction between my own approach to quantum foundations and an opposing theme that gets more and more airplay---relationalism (though so far, no one has put much substance behind it).  See, for instance: \myurl[https://en.wikipedia.org/wiki/Relational_quantum_mechanics]{http://en.wikipedia .org/wiki/Relational\underline{ }quantum\underline{ }mechanics}.
You pointed out that ``James favors a philosophy in which all our dynamic relations in the world are cast (metaphysically) as reciprocal rather than merely relational.''  Certainly for me.  The relational interpretations of quantum theory are all monistic to the core, and all too block-universey for my taste.

I liked your remarks on ``compenetration'' on page 246 and ``novelty'' on pages 256--257 as well.

\subsection{David's Reply}

\bq
Thanks for this.  I am amused, because this piece has received the most mixed reviews of things I've written, and most readers don't really know what to make of it.  The comments you make are right on the mark with what I was after, and what I find remarkable and useful in James's understanding of experience.  He had to resort to the language of panpsychism, which is confusing at base because most panpsychists understood consciousness precisely in the way James criticized in his ``Does Consciousness Exist.''  So most have missed the point.  This connection, and whether it can be ramified in terms of physics, is what interests me about your proposal here and from the outset.

I expect to be sending you the two documents needed by tomorrow afternoon---I have time mapped out to finish them then
\eq

\section{12-09-10 \ \ {\it New Questions}\ \ \ (to D. C. Lamberth)} \label{Lamberth5}

I've already asked you where you might think a good location for the conference might be (and I hope you'll answer!), but I'm gearing up to write the ``conference section'' of the proposal, and I'm starting to realize that I'll need more information.

Could you supply me with a list of 5 to 10 or more (the more the better for sure at this stage) people that we can list as potential invitees.  Range your mind over:
\begin{enumerate}
\item	Philosophers of pure experience
\item	Philosophers of neutral monism
\item	Pragmatists
\item	Disciples of the new realists (are there any left?)\ like R. B. Perry, W. T. Marvin and that lot
\item	Panpsychists
\item	Interesting Deweyians who have something to say on ``experience''
\item	Anyone who has anything interesting to say on Mach's world elements (I for instance have been quite taken with Erik Banks' work, maybe there are others)
\item Any Bergsonians?
\item Pragmatists you have ongoing debates with?  (Like the trouble-in-river-city guy!?)
\item	Any disciples of Russell's neutral monism?
\end{enumerate}

I know plenty of standard philosophers of science who know a little about pragmatism (and even say they are pragmatists or turning pragmatist) like Huw Price, Arthur Fine, Richard Healey, and some others---and they, I imagine, will provide some good connective material for the meeting (go-betweens on quantum and philosophy), but they're not quite optimal for the things I really want to get at.

\section{12-09-10 \ \ {\it I Don't Know That I Had Ever Seen This One Before}\ \ \ (to R. {\Schack})} \label{Schack203}

\noindent Richard Jeffrey, ``De Finetti's Probabilism,'' {\sl Synthese\/} {\bf 60}(1), 73--90 (1984).

\section{12-09-10 \ \ {\it Nor This One}\ \ \ (to R. {\Schack})} \label{Schack204}

\noindent Richard Jeffrey, ``Carnap's Voluntarism,'' in {\sl Logic, Epistemology, and Philosophy of Science IX}, edited by D. Prawitz, B. Skyrms, and D. Westerst{\aa}hl, (Elsevier, Dordrecht, 1994).

\section{12-09-10 \ \ {\it Books:\ Free To Good Homes}\ \ \ (to T. Slee)} \label{Slee3}

\btsl
We have finally accepted that there are many books on our shelves that we are not going to read again.

For the next few days these books will be on our driveway at 178 Herbert Street, free to good homes. Please feel free to browse and to take as many as you want.
\etsl

I just discovered them and gave some Jungian readings (I promise) a good home!

\section{13-09-10 \ \ {\it Pauli's Dreams}\ \ \ (to T. Slee)} \label{Slee4}

I just noticed that I should probably thank Lynne more particularly for the Jung books:  I noticed the name Shelagh Supeene is signatured in the front.  (A relative? Or another name for Lynne?)

Anyway, on page 116 of the Dreams book, the following line can be found:  ``The material consists of over a thousand dreams and visual impressions coming from a young man of excellent scientific education.''

That young man was Wolfgang Pauli (revealed years later).  And I learned from Atmanspacher---one of the personnel in the proposal I'm writing---that another 1,000 pages of Pauli's dreams are soon to be published.  (This time, not 1,000 dreams, but 1,000 pages!  Or at least that's what he said.)  Another one of the personnel in the proposal is my friend Hans Christian von Baeyer. (Maybe you've seen some of his semi-popular books on physics? The one on `information' is excellent.)  I attach one of his articles on Pauli, the man, ``Wolfgang Pauli's Journey Inward,'' in case you're interested.

Thanks to Lynne again!

\section{14-09-10 \ \ {\it The Latest Cleanest Latest!}\ \ \ (to H. C. von Baeyer)} \label{Baeyer133}

\bhcvb
The project description reads well.  My only negative reaction is to the {\Timpson} bit, which sounds a little immature.  After all, when Feynman gave a lecture to physicists, he would go out of his way to describe Newton's second law.  So to accuse philosophers of science (no less) of not knowing their subject, on the basis of one man's careful definition of terms, seems wrongheaded.
\ehcvb

Well \ldots\ as an initial compromise to throw on the table, I took out the ``no less!''\ \ldots\ which, after your remark, I started to think was a bit uppity my own self.

\ldots\ BUT \ldots\ BUT \ldots\ At the same time I discovered that he had misspelled Peirce (as Pierce).  Now what does that tell you about these philosophers of science?

Here's the present version of the passage (I took the liberty of fixing the spelling on Peirce):
\bq
Crucial to our whole effort is fielding the ideas behind QBism more thoroughly in a larger community of expertise.  No research effort with a hope to grow and flower can be an island unto itself.  But as should be clear by now, QBism is in quite a predicament when it comes to placing itself in a community of appropriate peers.  Part of its roots are in quantum information, but how many quantum information theorists have ever read the correspondence between Wolfgang Pauli, Markus Fierz and Carl Jung, where so many of Pauli's ideas on ``background physics'' were worked out?  How many pragmatist philosophers and experts on William James know the distinctions between the frequentist and Bayesian conceptions of probability? How many know the details of quantum mechanics?  Vice versa, how many quantum information theorists are there out there who think the philosophy of pragmatism simply means, ``Shut up and calculate!''?  Maybe to the greatest surprise, it is telling that C. G. {\Timpson} in his review article on QBism in {\sl Studies in History and Philosophy of Modern Physics}---a philosophy of science journal---felt compelled to include a footnote stating, ``Pragmatism is the position traditionally associated with the \ldots\ American philosophers Peirce, James and Dewey; its defining characteristic being \ldots''  Apparently even the philosophers of science who lay judgment on quantum theory are not particularly familiar with pragmatism!
\eq

\section{14-09-10 \ \ {\it From Coleman to Cuero}\ \ \ (to D. C. Lamberth)} \label{Lamberth6}

Many thanks for this!  [\ldots]

It was interesting learning that you were born in Coleman.  I know I've traveled near it, if not through it, on the small-road crazy route I used to take between Albuquerque and my own hometown, Cuero.  Apparently we were born just 262 miles from each other.  And probably temporally close as well; I was born October '64.

The new name I'm toying for the conference is
\begin{center}
From Pure Experience to Elementary Quantum Phenomena: \\ William James, John Wheeler, Quantum Interpretation
\end{center}
and I've tentatively put it ``near Chocorua.''

\section{14-09-10 \ \ {\it The James Meeting}\ \ \ (to R. {\Schack})} \label{Schack205}

Lamberth pointed out this place as a possibility for the James meeting:
\myurl[http://www.asticou.com]{http://www.asticou. com/} say, in June.  (He said tourists don't start coming until July.)

Looking in Richardson's biography of James, I see that I can give as a narrative that it was just after the completion of this time in Maine, at the Northeast Bay, that James brought to a close and bound his notes on radical empiricism.

Or at least that's the general target area, and maybe the prices are typical for something we'd find up there.
\bq
For summer jaunts James went twice in 1906 to Maine.  The first time, in July, he went to Northeast Harbor and Mount Desert, and the second time, in August, to Penobscot Bay, Rockland, North Haven, Isle au Haut, and Swan's Island.  But it was pragmatism that was on his mind.  Between trips he wrote exultingly to Schiller that `things are drifting tremendously in our direction.' Never hamstrung by modesty, James wrote, `It reminds me of the Protestant reformation!'

There was one casualty of James's rush toward pragmatism: somewhere along the way he benched radical empiricism. \ldots

\ldots\ Perhaps it was his turning away from what he scornfully called `technical' work, perhaps it was the force of `The Miller--Bode Objections,' perhaps it had something to do with his fervid new interest in pragmatism.  Whatever the reason, radical empiricism moved to the background as a program and a professional pursuit for William James \ldots
\eq
Actually that latter set of places are much closer to where John Wheeler's home was:  High Island, Maine.  It might be an interesting theme if we could find a good hotel closer to that region.

\section{15-09-10 \ \ {\it Your Offer Taken!}\ \ \ (to R. {\Schack})} \label{Schack206}

\brs
I attach my attempt at the budget narrative, at the moment excluding the workshops. Regarding the project timeline and the cost effectiveness sections, they should refer to the outcomes and outputs sections. Do we have such sections?

Regarding outputs, it's relatively easy. 2 books, say 4 papers each on Jamesian things, crypto and randomness things, SICish things, and 2 workshops.

Outcomes, I am almost tempted to say that it is all towards the one big outcome, a new conception of the world or something like this. The cost effectiveness factor would be something like \$720,000 per revolutionary world view, which is a bargain indeed.
\ers

I loved that!!!!  I laughed and laughed.  The family already thinks I'm going crazy because I stood in the yard yelling and wagging my finger at a squirrel the day before last.  They almost had a look of horror when I kept laughing and laughing at your ``\$720,000 per revolutionary world view, which is a bargain indeed.''

I dare say I like the idea too!  It meshes well with something Hans said, late, just before I went up to bed.  Placed below, and cc'd to Hans, so each of you can see each other's thoughts on telling Templeton ``we're important.''

I'm on my first cup of coffee and getting into the groove \ldots

\section{15-09-10 \ \ {\it Outcome 1}\ \ \ (to R. {\Schack})} \label{Schack207}

What will be different and who will be affected:

Mathematical Research:
\bq
QBism is one of the first efforts in quantum interpretation that we are aware of that has said, ``A proper interpretation of quantum mechanics should almost close its eyes to the existing mathematical formalism of the theory, and instead IMPLY what the formalism OUGHT to be.'' In that regard we have already had some success--our rewriting of the Born Rule is an example of an interpretation-driven change in the very formalism itself--but of course, we want more. In the end, we expect of a Quantum Bayesian interpretation of quantum mechanics that it should actually imply the full mathematical formalism of the theory. An interpretation that did that alone would already be revolutionary in the quantum foundations wars.

Who will be affected? The whole quantum foundations community, for it will hold up an example of how things ought to be done:  Interpretation first, quantum formalism next. (Not like, say, the Many Worlds Interpretation views the order of affairs.)
\eq

What are the indicators of the difference and the measurable change you expect:
\bq
Recently, we got a referee report for a submission to {\sl Reviews of Modern Physics\/} that had these lines in it, ``This is as strikingly novel a way of looking at the quantum theory as Feynman's sum of amplitudes over histories was when it appeared in the late 1940s. Whether it will prove as fruitful and durable remains to be seen, but it ought to be known to a broad audience \ldots.'' When we get referee reports like that, we certainly get the sense that we are making a useful contribution to the physics world as a whole.

We want more referee reports like that! And we will strive our best to get them.

But an obvious indicator of how much a scientific result is being noticed is how many times the result is cited per year. We will be watching that closely, but also noticing simply how much ``buzz'' the view gets as develops ever more.
\eq

\section{15-09-10 \ \ {\it Outcome 2 (half of it)}\ \ \ (to R. {\Schack})} \label{Schack208}

Conceptual Research:
\bq
We feel that in the last few years we (QBists) have turned a corner in our understanding by realizing that the deepest import of QBism is that it implies a significantly nonreductionist reading of quantum theory. One of the two great, fundamental theories of physics by its own structure, puts a stop to the reductionist dream that is usually identified with ``science itself.'' This, we feel, instates a humanistic face to physics that has the potential not only to affect physical practice, but also the worldview of even lay scientific enthusiasts (readers of {\sl Scientific American\/} and such).

Also with the success of this reading of quantum theory, we expect to see the philosophy of physics community for the first time take note of old-style American pragmatism, and perhaps even squeeze their own golden nuggets from it.
\eq
Now the question is, what to write for the ``indicators'' half of this!!

\section{15-09-10 \ \ {\it Outcome 2's INDICATOR}\ \ \ (to R. {\Schack})} \label{Schack209}

Indicator:
\begin{itemize}
\item Count the number of philosophers who stop talking so seriously about Many Worlds (more generally, views that take quantum states as objective properties) and start talking about Pauli's ideas, Bayesianism and pragmatism instead. This is not a joke. It is the method we have used for years to measure the power of our view. We can list (and point fingers to appropriate publications of) philosophers Bub, {\Demopoulos}, Healey, and Stairs, as examples exhibiting themselves as cross-overs in the Bohr--Einstein debate, adopting one or another element close in spirit to the QBist view (even if not remotely the whole QBist package).

We certainly want more of that and will have our eyes open for it and pass on any anecdotal information to JTF.

\item
More generally the usual markers: invitations to conferences, keynote speeches, citation counts, and reports in the lay press. Also invitations to write articles ourselves for the lay public (as in our recent invitation from {\sl Scientific American}).
\end{itemize}

\section{15-09-10 \ \ {\it Outcome 3 and Indicators}\ \ \ (to R. {\Schack})} \label{Schack210}

If you can see anything to add, please suggest.\medskip

Workshops:
\bq
The key goal of the two workshops is to get parts of the disparate communities of quantum information physicists, Bayesians, pragmatist philosophers, and Pauli scholars all talking to each other and becoming aware that there is overlap in their interests and concerns. With this, we hope to start to build a community capable of understanding QBist ideas, and even capable of building upon them (at least in spirit, even if not completely in letter). QBism needs a wider community to interact with--to educate and to learn from itself--and the time appears ripe for making this happen.

The conference proceedings arising from this activity will be invaluable resources for the worldwide Bayesian, Paulian, quantum, and pragmatist communities to get to know each other.
\eq

Indicators:
\bq\noindent
\vspace{-.2in}
\begin{itemize}
\item As in the example provided, pre- and post-questionnaires, testing the various community's knowledge of each other is a very good idea, actually.

\item Collect anecdotal evidence of friendships and professional ties made at the meetings.

\item Collect data six months and a year down the road from the meetings on the network of active email ties between participants.

\item Track sales and professional reviews of the conference proceedings.

\item Collect data on conference participants starting to take part in one of the other community's conferences.
\end{itemize}
\eq

\section{15-09-10 \ \ {\it Unruh's Talk}\ \ \ (to the QBies)} \label{QBies26}

You can imagine I'm still doing Templeton things---it is literally the hardest grant proposal I've ever written.  But I did want to encourage you all to go to Unruh's talk today in my place!  The great WG is a man who has made a very deep contribution to physics (that walking through space heats it up), and it's actually rare that we get physicists coming through PI of that stature in nature's eyes.\footnote{\editornote See \pirsa{10090068}.}

\section{16-09-10 \ \ {\it The Enduring Impact}\ \ \ (to R. {\Schack})} \label{Schack211}

\bq
Realistically, we hope the efforts of this project, when all is said and done, will be identified with a small span of history when a tide was turned.  It would be nice to change the world fully and deeply in two years, but such things can never be planned for (and perhaps dare not be hoped for). A turning of the tide of this 85-year-old debate on quantum foundations would already be a revolution in itself. QBism points in a direction that says quantum theory's great lesson is not that IT is incomplete (that there is another theory waiting to be uncovered from underneath it), but that the world itself is unfinished---it is still in the furnace of creation. If our efforts contribute to getting our fellow physicists to take this possibility seriously, even if only enough that its residue shapes their work and their questions ever so slightly below the threshold of awareness, then the results of this project will be very pleasing indeed. Physical thought is a collective effect; one physicist affects another in turn with every paper and every experiment. If QBism does point in the right direction, as we believe it does, and if a tide is turned, things will eventually take care of themselves.
\eq

\section{16-09-10 \ \ {\it The Final Submission}\ \ \ (to H. C. von Baeyer)} \label{Baeyer134}

FYI, here's the final thing, warts and all.  You'll certainly note, if you care to amuse yourself, that my answers become more and more uncouth as the deadline drew nearer (particularly in the budget justification section, where some of their questions became truly ridiculous).  Story below to {\Ruediger}; I got the thing in just 2 minutes before its deadline!

Wow, what work!  And what work for nothing but a chance!  Still, there were lessons learned from the exercise, and maybe that's what in the end counts:

1) Let me attach {\Appleby}'s letter of support.  In it he makes a good point about quantum theory being initially based on a mechanical analogy, and that a wrong turn was taken for interpretation just right there.  I liked it and had never quite thought about it that way.  It'd be nice if we could squeeze something about that into the {\sl Sci Am\/} piece in some way.

2)	Like with a good joke, I amuse myself sometimes.  Here's something I wrote for one of their more worthless fields of questioning:
\bq
QBism is one of the first efforts in quantum interpretation that we are aware of that has said, ``A proper interpretation of quantum mechanics should almost close its eyes to the existing mathematical formalism of the theory, and instead IMPLY what the formalism OUGHT to be.''  In that regard we have already had some success---our rewriting of the Born Rule is an example of an interpretation-driven change in the very formalism itself---but of course, we want more. In the end, we expect of a Quantum Bayesian interpretation of quantum mechanics that it should actually imply the full mathematical formalism of the theory. An interpretation that did that alone would already be revolutionary in the quantum foundations wars.
\eq
That really is a key point, and I don't think I've said it that way before.  So I should thank the JTF.  Or maybe I {\it have\/} said it that way.  Anyway, the key thing is this.  What sets QBism apart from Many-Worlds, say, is that
\begin{enumerate}
\item	{\it they\/} take the formalism as given (they've got no clue why Hilbert space vectors) and try to make up a story for it w.r.t.\ what it could mean in common life, but
\item {\it we\/} take a story, and try to show how it implies the mathematical theory.
\end{enumerate}
There's no doubt our job is harder than theirs:  It's easy to make up stories to fit something one doesn't understand.  Or another (slightly sophomoric) way to put it:  We're doing what physics is all about---noting things, expressing them, and then trying to find a mathematical description that captures their essence---while they're obsessed with a {\it homework\/} problem.  (Imagine the voice of Darla in {\sl Our Gang}.)  ``Teacher said `Hilbert space'.  Now we've got to make a story that uses the word five times.''

Anyway, like with {\Appleby}, it'd be nice to try to really bring this point home in the {\sl Sci Am\/} article:  I.e., The story has to come first, and then the equations second, or it is not {\it deep\/} physics, it's just a homework problem.  (Taylor and Wheeler?)

I do hope you're still engaged with QBism and not getting fed up with all my shenanigans.  Thanks again (for the near hundredth time) for all the help you've given me.  You're teaching me discipline in writing; even {\Mermin} didn't do that; and I do appreciate (even when I stray otherwise).  You're my mentor.

Good night, I say, \ldots\ as you're probably nearly about to wake up \ldots

\section{16-09-10 \ \ {\it New Religion} \ \ (to D. B. L. Baker)} \label{Baker29}

It's nearly 2:00 AM for me, and I've been celebrating getting this thing (attached) in literally just two minutes before the bloody deadline!  (I.e., a wee dram of Scotch and such things.)  But, not lying; two minutes.  Have a read.  Is it just Methodism all over again?  Or is it something else?  I'd almost like to think it's a new religion.  And I've got credentials too \ldots\ I'm asking for money!  (Not even based in Houston, TX, but I'm asking for money.  How else would you define ``religion''?)  But to Sir John Templeton, I told him it's science.

Religion or science?  Not such a clean line is it?  Who would expect otherwise from two Cuero degenerates?

\section{17-09-10 \ \ {\it Free Will Saved James's Life}\ \ \ (to the QBies)} \label{QBies27}

\begin{center}
\bf William James's Diary Entry, 30 April 1870
\end{center}
\bq
I think that yesterday was a crisis in my life. I finished the first
part of Renouvier's second ``Essais'' and see no reason why his
definition of Free Will---``the sustaining of a thought because I
choose to when I might have other thoughts''---need be the definition
of an illusion. At any rate, I will assume for the present---until
next year---that it is no illusion. My first act of free will shall
be to believe in free will. For the remainder of the year, I will
abstain from the mere speculation and contemplative {\it Gr\"ublei\/}
in which my nature takes most delight, and voluntarily cultivate the
feeling of moral freedom, by reading books favorable to it, as well
as by acting. After the first of January, my callow skin being
somewhat fledged, I may perhaps return to metaphysical study and
skepticism without danger to my powers of action. For the present
then remember: care little for speculation; much for the form of my
action; recollect that only when habits of order are formed can we
advance to really interesting fields of action---and consequently
accumulate grain on grain of willful choice like a very miser; never
forgetting how one link dropped undoes an indefinite number. {\it
Principiis obsta}---Today has furnished the exceptionally passionate
initiative which Bain posits as needful for the acquisition of
habits. I will see to the sequel. Not in maxims, not in {\it
Anschauungen}, but in accumulated acts of thought lies salvation.
{\it Passer outre}. Hitherto, when I have felt like taking a free
initiative, like daring to act originally, without carefully waiting
for contemplation of the external world to determine all for me,
suicide seemed the most manly form to put my daring into; now, I will
go a step further with my will, not only act with it, but believe as
well; believe in my individual reality and creative power. My belief,
to be sure, can't be optimistic---but I will posit life (the real,
the good) in the self-governing resistance of the ego to the world.
Life shall [be built in] doing and suffering and creating.
\eq

\section{17-09-10 \ \ {\it Motivated By Our Pluriverse Discussion}\ \ \ (to the QBies)} \label{QBies28}

\bv
\myurl[https://en.wikipedia.org/wiki/Universal_set]{http://en.wikipedia.org/wiki/Universal\underline{ }set}
\ev

\section{20-09-10 \ \ {\it Back from the Mountaintop} \ \ (to P. Wells)} \label{Wells5}

Thanks for the article!  Sorry to take so long to say ``thanks,'' but when I got your note Thursday I was exhausted, and my writing fingers were broken.  Not an easy trip back down from the mountaintop:  I had just spent several days in a frenzy trying to get a proposal written for the John Templeton Foundation---toughest grant proposal I had ever written.

I loved your James 1878 quote (and your play on it at the end of the article); where did you get it from?  Was it ``The Sentiment of Rationality''?  I liked too your ``read like a forecast of quantum theory.''  That's the heart of the matter; that's why there's a shrine to him.  I was impressed by your sensitivity (at least in my case) to what was going on around you.

Great article.

\section{20-09-10 \ \ {\it Small Vision} \ \ (to H. Barnum \& R. {\Schack})} \label{Barnum33} \label{Schack212}

In broadest outline, here's the way I think maybe I can envision how this could become a report.
\begin{itemize}
\item[{1)}] It could start with a section saying, here is how the convex-operational framework is understood.  States, effects, linear maps, and all that.  Cones, blah, blah, blah.  Maybe most of it can be cut and pasted from one of Howard's papers.

\item[{2)}] On the other hand, here's the QBist framework.  Not states and effects, but modified law of total probability.  No mention of cones directly at all!  The QBist hope is that consistency of the modified law alone will imply a significant nontrivial restriction on the parts of the simplex that should be used (i.e., state space), and similarly with the conditional probabilities through the principle of reciprocity.

\item[{3)}] Now show that the QBist framework can be rewritten in convex-operational terms and comes automatically with these statements about the cone of effects.

\item[{4)}] Speculation on next steps.
\end{itemize}
Enough for a conference proceedings and we'll have a resource for others who are scratching their heads over whether QBism and Coneheadonism have any connections.

\section{21-09-10 \ \ {\it Whitehead on James} \ \ (to D. M. {\Appleby})} \label{Appleby94}

I just haven't been able to concentrate so much today, so I took the easy way out reading ``people magazines'' for the most part.

Here is that quote from the letter Whitehead wrote Hartshorne in January
1936:
\bq
     European philosophy has gone dry, and cannot make any
     worthwhile use of the results of nineteenth century
     scholarship.  It is in chains to the sanctified
     presuppositions derived from later Greek thought \ldots.
     My belief is that the effective founders of the
     renascence in American philosophy are Charles Peirce
     and William James.  Of these men, W. J. is the
     analogue to Plato, and C. P. To Aristotle, though the
     time-order does not correspond, and the analogy must
     not be pressed too far. Have you read Ralph Perry's
     book (2 vols.)\ on James? It is a wonderful disclosure
     of the living repercussions of late 19th century
     thought on a sensitive genius. It is reminiscent of
     the Platonic Dialogues. W. J.'s pragmatic descendants
     have been doing their best to trivialize his meanings
     in the notions of Radical Empiricism, Pragmatism,
     Rationalization. But I admit W. J. was weak on
     Rationalization. Also he expressed himself by the
     dangerous method of overstatement
\eq
And here is another passage I came across; it comes from Whitehead's book {\sl Modes of Thought}.  I'll place it below.  Personality drivel mostly, but still I wanted to record it, and you seemed like the most natural repository.

\bq
In Western Literature there are four great thinkers, whose services to civilized thought rest largely upon their achievements in philosophical assemblage; though each of them made important contributions to the structure of philosophic system. These men are Plato, Aristotle, Leibniz, and William James.

Plato grasped the importance of mathematical system; but his chief fame rests upon the wealth of profound suggestions scattered throughout his dialogues, suggestions half smothered by the archaic misconceptions of the age in which he lived. Aristotle systematized as he assembled. He inherited from Plato, imposing his own systematic structures.

Leibniz inherited two thousand years of thought. He really did inherit more of the varied thoughts of his predecessors than any man before or since. His interests ranged from mathematics to divinity, and from divinity to political philosophy, and from political philosophy to physical science. These interests were backed by profound learning. There is a book to be written, and its title should be, `The Mind of Leibniz'.

Finally, there is William James, essentially a modern man. His mind was adequately based upon the learning of the past. But the essence of his greatness was his marvelous sensitivity to the ideas of the present. He knew the world in which he lived, by travel, by personal relations with its leading men, by the variety of his own studies. He systematized; but above all he assembled. His intellectual life was one protest against the dismissal of experience in the interest of system. He had discovered intuitively the great truth with which modern logic is now wrestling.
\eq

\section{22-09-10 \ \ {\it The Quote on Sacred Matter!!}\ \ \ (to P. Hayden)} \label{Hayden4}

\bq\noindent
To anyone who has ever looked on the face of a dead child or parent the mere fact that matter {\it could\/} have taken for a time that precious form, ought to make matter sacred ever after. It makes no difference what the principle of life may be, material or immaterial, matter at any rate co-operates, lends itself to all life's purposes. That beloved incarnation was among matter's possibilities.
\eq

\section{22-09-10 \ \ {\it The Tube Alloys Project}\ \ \ (to B. W. Schumacher \& M. D. Westmoreland)} \label{Schumacher18} \label{Westmoreland1}

I greatly enjoyed our conversation on ``pure indeterminism'' after your talk.  (Ben will understand the allusion in the title; he can explain.)

Attached are two notes from my collection somewhat to do with the issue of ``probability $+$ x'' that we were talking about.  The first one is a (rather nice) note to Abner Shimony introducing the second one.  The second one is a rather rabid one to a Princeton professor who annoyed me.  (He called me a pansy, basically.)  Anyway, it happens to be the one with the key point (starting right after McDonaldism 6). [See 18-10-06 note ``\myref{Shimony11}{Real Possibility}'' to A. Shimony and 30-01-06 note ``\myref{McDonald6}{Island of Misfit Toys}'' to K. McDonald.  It is McDonaldism \ref{Hoompah!} in the present compilation.]

I liked Ben's image of the alloy:  I think it's much more of the right flavor than the simple ``$+$'' I had been using.

\section{22-09-10 \ \ {\it Tube Alloys 3!}\ \ \ (to B. W. Schumacher \& M. D. Westmoreland)} \label{Schumacher19} \label{Westmoreland2}

Actually I do a much better job on the point in the attached page, drawn from a recent paper.  The good stuff is in the left-hand column, particularly Footnote 14:
\bq
Most of the time one sees Bayesian probabilities characterized (even by very prominent Bayesians like Edwin T. Jaynes) as measures of ignorance or imperfect knowledge. But that description carries with it a metaphysical commitment that is not at all necessary for the personalist Bayesian, where probability theory is an extension of logic. Imperfect knowledge? It sounds like something that, at least in imagination, could be perfected, making all probabilities zero or one---one uses probabilities only because one does not know the true, pre-existing state of affairs. Language like this, the reader will notice, is never used in this paper. All that matters for a personalist Bayesian is that there is uncertainty for whatever reason. There might be uncertainty because there is ignorance of a true state of affairs, but there might be uncertainty because the world itself does not yet know what it will give---i.e., there is an objective indeterminism. As will be argued in later sections, QBism finds its happiest spot in an unflinching combination of ``subjective probability'' with ``objective indeterminism.''
\eq

\section{27-09-10 \ \ {\it Pauli's Eye Quote}\ \ \ (to H. Atmanspacher \& H. C. von Baeyer)} \label{Baeyer134.1} \label{Atmanspacher10}

Greetings from Stockholm!

Can either of you quickly pin down where I can find an English
translation of the nice thing that Pauli wrote Heisenberg, just a bit
before Heisenberg found the uncertainty relation?  It is something
about being able to look through the $x$-eye or the $p$-eye, but not being
able to open both eyes at once.

I want to use that as a paradigm example that not all great
contributions to physics are explicit and based on publication.  A lot
goes on behind the scenes.

\subsection{Harald's Reply}

\bq
It's in P to H Oct 19, 1926 (letter 143, p.~347 in P's corr vol.1), actually in a discussion about collision processes (my translation):
\bq\noindent
It's always the same thing: due to diffraction, there are no arbitrarily thin rays in the wave optics of the $\psi$-field, and it is illegitimate to assign ordinary `c-numbers' to `p-numbers' and `q-numbers' simultaneously.
One may then view the world with the `p-eye' and one may view it with the `q-eye', but if you want to open both eyes at the same time this drives you crazy.
\eq
Note that P does NOT say it is IMPOSSIBLE to open both eyes. He says (in German: ``dann wird man irre'') that it drives you crazy, or maybe a bit weaker, that it is mind-boggling. [Compare the perception of ambiguous figures like the Necker cube.] The letter was sent to H at Copenhagen, and generated a lot of enthusiasm among H, Bohr, Dirac and Hund who were all there at the time (as H reports in his reply of Oct 28).
H's paper ``Ueber den anschaulichen Inhalt der quantentheoretischen Kinematik und Mechanik'' in ZPhys was received there March 23, 1927.

Another example would be P's postcard of 4 Dec 1930 to the Tuebingen conference on radioactivity where he predicts the neutrino (my
translation):
\bq\noindent
\ldots\ in connection with the `wrong' statistics of the N and Li6 nuclei as well as the continuous $\beta$-spectrum, I hit upon a desperate remedy for saving the `alternation law' of statistics and the energy law. This is the possibility that there may be, in the nuclei, electrically neutral particles, which I shall call neutrons [later:\ neutrinos], which have spin 1/2, obey the exclusion principle and, moreover, differ from light quanta by not traveling with the velocity of light. Their mass should be of the same order of magnitude as the electron mass and not greater than 0.01 proton masses. The continuous $\beta$-spectrum would then be understandable under the assumption that in $\beta$-decay a neutron will be emitted together with an electron, such that the sum of their energies is constant. \ldots\ (letter 259, p.~39--40, in P corr vol.2)
\eq
\eq

\section{29-09-10 \ \ {\it Quantum Foundations in the Light of Quantum Information III} \ \ (to G. Brassard)} \label{Brassard53}

Absolutely!  Of course!  Yes!  Wonderful!  Thanks!  Please!  Cool!  Count me in!

\section{29-09-10 \ \ {\it Perfect Contrast}\ \ \ (to H. C. von Baeyer)} \label{Baeyer135}

I am in the middle of G\"oran Lindblad's talk, where he had the most lovely (and arrogant) quote by Steven Weinberg.  Here's an extension of it that I just dug off the web:
\bq\noindent
Bohr had presided over the formulation of a ``Copenhagen interpretation'' of quantum mechanics, in which it is only possible to calculate the probabilities of the various possible outcomes of experiments.  Einstein rejected the notion that the laws of physics could deal with probabilities, famously decreeing that God does not play dice with the cosmos.  But history gave its verdict against Einstein---quantum mechanics went on from success to success, leaving Einstein on the sidelines.

All this familiar story is true, but it leaves out an irony.  Bohr's version of quantum mechanics was deeply flawed, but not for the reason Einstein thought.  The Copenhagen interpretation describes what happens when an observer makes a measurement, but the observer and the act of measurement are themselves treated classically.  This is surely wrong:  Physicists and their apparatus must be governed by the same quantum mechanical rules that govern everything else in the universe.\footnote{See S.\ Weinberg, ``Einstein's Mistakes,'' Phys.\ Today\ \textbf{58}(11), pp.~31--35 (2005).}
\eq

The reason I say it is the perfect contrast to us is because QBism rejects exactly what Weinberg takes as unquestionable: ``Physicists and their apparatus must be {\it governed\/} by the same quantum mechanical rules that govern everything else in the universe.''  QBism says quantum theory doesn't ``govern'' anything in the physical world, in the same way that raw probability theory doesn't {\it govern\/} anything in the physical world.  Quantum theory is something physicists {\it use}---that was implicit in Bohr's view of QT, as any careful reading of him would show.  This is why I say Weinberg is one hell  of an arrogant guy when he calls Bohr's view ``deeply flawed''.  He didn't understand a word of it.  Humbug!

\section{01-10-10 \ \ {\it Great Minds Think Alike} \ \ (to S. Hossenfelder)} \label{Hossenfelder1}

Marek reminded me last night of Wigner's remark on von Neumann.  I had indeed known that once (long ago), but forgot about it.  Sorry I didn't bring it up yesterday.  Anyway, it looks like you're thinking like von Neumann.\footnote{\editornote For consequences of this correspondence, see S.\ Hossenfelder, ``Testing super-deterministic hidden variables theories,'' Found.\ Phys.\ {\bf 41,} 1521--31 (2011), \arxiv[quant-ph]{1105.4326}; and the 01-10-13 blog entry \myurl{http://backreaction.blogspot.com/2013/10/testing-conspiracy-theories.html}.}  See Footnote~1 in the attached paper.  [E. Wigner, ``On Hidden Variables and Quantum Mechanical Probabilities,'' Am.\ J. Phys.\ {\bf 38}, 1005--1009 (1970).]

\section{03-10-10 \ \ {\it Scaling, of a QBist Flavor} \ \ (to M. A. Graydon)} \label{Graydon11}

Just a small break from the logistical matters.  I think your pushing me on the Hardy axioms has been quite needed.  In the following form, I think I might find the composite system law more palatable from a QBist philosophy.

The way I'd like to transform the question is this:  For a given number of outcomes in the sky (for a given measure of how much matter we have), how much of a deviation from the law of total probability should we expect in an urgleichung?  How does the urgleichung scale with the amount of matter?

We could say as an axiom, $\beta$ should be multiplicative in those cases when $n$ is multiplicative.  And my guess is that that would force upon us the same functional equation Hardy explored.

OK, but now why that particular choice of parameters to assume mutual multiplicativity for?  Why not $n$ and $\alpha$, for instance?  Or could one show that any other choice of pairs for multiplicativity would lead to a contradiction with some other axiom?

I'm reminded of a speculation I've told of Bohr.  Somewhere I've read that when Bohr got the idea of quantizing the hydrogen atom, he locked himself away for like three weeks to do the calculations.  But you know that it's no more than a night or two of homework to figure out the energy level spacing from Bohr's assumption of simple integer quantization of angular momentum.  How could it take Bohr three weeks?  I've speculated that at the beginning, he simply didn't know which variable to quantize in terms of simple integer values.  And thus he set out on a random walk to find out which scheme would be simple and believable and would give the needed energy level spacing.\footnote{\editornote Bohr wrote to his brother Harald on 19 June 1912, saying, ``Perhaps I have found out a little about the structure of atoms,'' and confiding that he was taking ``a couple of days off from the laboratory'' to finish his calculations.  On 13 July, he wrote to Harald again: ``Things are going rather well, for I believe I have found out a few things; but, to be sure, I have not been so quick to work them out as I was stupid to think.''  See \myurl{http://www-history.mcs.st-andrews.ac.uk/Biographies/Bohr\_Niels.html}.}

Maybe that's the way to approach this problem.  ``Distinguishable states'' and ``degrees of freedom'' \ldots\ humbug!  (It's nearly hidden-variable talk. Or at least evocative of the imagery.)  But ``how much stuff'' (capacity) and ``how much deviation from a naive use of law of total probability''?  Now we're cookin'!

\section{05-10-10 \ \ {\it The Old Bayesian Motivations}\ \ \ (to P. J. Lahti)} \label{Lahti1}

Thanks very much for your curiosity about my research program.  I was very flattered and very much enjoyed my conversations with you.

Let me point you to three papers in case you have any continuing curiosity.

The first really tries to make a connection between the Bayesian motivations and the direction of formalism I seek:  \arxiv{1003.5209}.  The second lays out some (provisional) axioms:  \arxiv{0912.4252}. And the third tells something about how to get some of the structure of quantum-state space back out of all of this:  \arxiv{0910.2750}.

Much of it, in the end, will have to do with the story you tell over and over:  For the SICs are just a particularly crisp way of simultaneously measuring position and momentum (in Weyl's discrete phase space).

I hope you will view our efforts as complementary!

\section{05-10-10 \ \ {\it Quote of the Day}\ \ \ (to the QBies)} \label{QBies29}

I've rigged Google to give me a ``quote of the day'' every day now.   Here was today's pick:
\bv
Equations are the devil's sentences.\medskip\\
 $\qquad$ --- Stephen Colbert
\ev

\section{05-10-10 \ \ {\it Your Newest Book} \ \ (to J. W. Moffat)} \label{Moffat1}

I read about your new book in the newspaper after getting home from Sweden Sunday (thus I'm not sure which day's newspaper I was looking at).  Anyway, it sounds quite interesting, and you can be sure I'll be going to the bookstore once I hear that it is out.

\subsection{John's Reply}

\bq
Yes, there was an article in last Saturday's {\sl Waterloo Record\/} about me and my new book called: {\sl Einstein Wrote Back}. It will be published by Thomas Allen \& Sons, Toronto, about the middle of this month. My first book was: {\sl Reinventing Gravity}, published 2008 by HarperCollins NY and Thomas Allen, Toronto.

Thanks for your interest in my book.
\eq

\section{05-10-10 \ \ {\it Galoshes} \ \ (to M. A. Graydon)} \label{Graydon12}

I won't tell you what I put in my letter, but I will tell you this much:  You have some damned big shoes to fill now.  I expect to see some really important physics from you.

The more I think about it, the more I'm extremely pleased with the thought direction you've pushed us to.  The game is to get in a position of having Hardy's functional equation {\it without\/} speaking of composite systems and local measurements.  I like having this image in my mind.  I throw one rock into a bag and place it on a scale; the scale goes down a  bit.  I throw two rocks into a bag and put it on the scale, and the scale goes down a bit more.  Now I ask myself, if confronted with a bear, do I want to whack it in the head with the first bag, or the second?  The second of course---it has more oomph!  It's got nothing to do with any local tomographic properties, etc. It's all about how the oomph scales.  It's all about how the oomph scales with the stuff.

So, here's a question.  Suppose {\it before\/} getting to Assumption 6 we make the assumption that $n$ is a function of $\beta$.  And that when
$$
\beta = \beta_1 \beta_2
$$
that
$$
n(\beta_1 \beta_2) = n(\beta_1)\, n(\beta_2)\;.
$$
The sort of thing I'm wondering is, will this allow us to clean up Assumption 6 any?  Will it allow us to put something much better in its place?

\section{07-10-10 \ \ {\it Wolfgang Pauli and Alchemy} \ \ (to W. R. Newman)} \label{Newman1}

I enjoyed meeting you last night.  I feel a bit embarrassed that I was not so coherent in our conversation, but wine does that to me.

I just placed an order with Amazon for your {\sl Promethean Ambitions\/} book.  I look forward to reading it.

Lollygagging away a little time today, I decided to re-read some things in my accumulated notes on the intersection of Pauli and alchemy.  I'll share those with you, in case you have any interest.  They're in the attached file.  Here are the relevant parts:
\begin{itemize}
\item[1)]	The first quote from Werner Heisenberg's article, ``Wolfgang Pauli's Philosophical Outlook'' [217].
\item[2)]	The long quote from Pauli's ``unpublished'' draft ``Modern Examples of `Background Physics'\/'' [336].
\item[3)]	``The Influence of Archetypal Ideas on the Scientific Theories of Kepler'' [339]  (This one is quite good reading.)
\item[4)]	``Ideas of the Unconscious from the Standpoint of Natural Science and Epistemology'' [343].
\item[5)]	Letters from Pauli to Jung [354].
\end{itemize}

\section{08-10-10 \ \ {\it History} \ \ (to A. Wilce)} \label{Wilce23}

\baw
I still owe you (and the world) a screed. Meanwhile, a question: do you know of a good book on the history of probability
theory and/or statistics that talks at any length about early attitudes towards the Normal distribution? Any pointers
towards material of this kind would be most welcome.
\eaw

Sorry for the delay.  Probably the best possible things are Ian Hacking's two books, {\sl The Taming of Chance\/} and {\it The Emergence of Probability}.  Louis Menand's book {\sl The Metaphysical Club\/} (one of the best books ever, in any case) has a nice chapter or two on Quetelet and his reception.

\section{12-10-10 \ \ {\it The APS March Meeting} \ \ (to G. Chiribella)} \label{Chiribella2}

Thanks so much taking me up on the invitation to speak.  The information on the session is below, and the meeting will be March 21--25 in Dallas, Texas.  Anyway, in this, you join some famous company:  Lucien Hardy was the invited speaker one year, Rob Spekkens another, and I think Bill Wootters another. [\ldots]

I never could find the right Feynman quote, by the way.  It's one about how it takes a generation or two to get used to a new physical theory.  The only thing that's really broken in quantum interpretation is just our intuitions, or something like that.  If you know which quote I'm talking about, I'd be most grateful if you could send it to me.\footnote{R. P. Feynman, ``Simulating Physics with Computers'', Int.\ J. Theor.\ Phys.\ {\bf 21}, 471 (1982):  ``Might I say immediately, so that you know where I really intend to go, that we always have had (secret, secret, close the doors!)\ we always have had a great deal of difficulty in understanding the world view that quantum mechanics represents. At least I do, because I'm an old enough man that I haven't got to the point that this stuff is obvious to me. Okay, I still get nervous with it. And therefore, some of the younger students \ldots\ you know how it always is, every new idea, it takes a generation or two until it becomes obvious that there's no real problem. It has not yet become obvious to me that there's no real problem. I cannot define the real problem, therefore I suspect there's no real problem, but I'm not sure there's no real problem.''}

\bq\noindent
23.13.4:  Quantum Information for Quantum Foundations\\
Organizer:  Christopher Fuchs (Perimeter Institute for Theoretical Physics)\medskip

\noindent Description: \medskip

Richard Feynman once famously said, ``\ldots''  This may or may not be true, but if one supposes it to be getting at the right idea, it behooves us to ask, ``Then what would it take to readjust our intuitions so that the discord is alleviated?''  Starting with John A. Wheeler, several in the last two decades have suggested that re-exploring the foundations of quantum theory in the language of quantum information may be the first step in a long needed therapeutic process.  This focus session will be devoted to assessing the impact of quantum information theory in diffusing the perennial quantum mechanical mysteries, as well as providing novel ways to rewrite the theory in terms of informational structures and functions.
\eq

\section{13-10-10 \ \ {\it Congratulations}\ \ \ (to C. G. {\Timpson})} \label{Timpson20}

\bcgt
Sorry you've not heard anything from me for {\bf ages}. It's been an eventful year: Jane just gave birth to our baby daughter Catherine Jane on 5 August --- a most wonderful thing! Such a joy to be a father; though it's interesting trying to teach logic on very little sleep \ldots\ Oddly, it seems ok trying to teach QM \ldots\ not sure what that says (about QM, or me \ldots)
\ecgt

Many congratulations on your daughter; that is wonderful news!  Children are the greatest classroom ever for learning about the world.  Please give my best wishes to Jane, and of course to Catherine too!

I liked your quip on teaching logic.  I think it means logic is forced and strained in our natural humanity, whereas quantum theory is actually a gut feeling we all already have!  (It being simply a refinement of everyday life --- a story of the consequences of agents taking actions upon the world.)  It'd be fun to revive one of F. C. S. Schiller's polemics on logic for a short class reading, like the introduction (as I recall) from his {\sl Formal Logic, a Scientific and Social Problem}.  It'll make everyone laugh out loud.  (No matter how much inspiration I draw from him, the guy was really and truly a nutcase \ldots\ and probably mostly should be kept in the closet.)

\section{13-10-10 \ \ {\it Up in the Air} \ \ (to D. B. L. Baker)} \label{Baker30}

\bdb
I have to admit I haven't been able to completely read your ``New Religion'' paper.  I'm still working on it.  Have you seen this entire quote by Samuel Butler?
\bq\noindent{\rm
A belief in human progress is a matter of faith.  Progress does not necessarily imply a monotonically increased advance, but rather an advance that will eventually occur within the limits of mankind's collective morality and knowledge of its respective environment.  It is common to hear both philosophers and non-philosophers complain that philosophy makes no progress.  Whether such a complaint is justified depends, or course, on one's understanding of the nature of philosophy, and on one's criteria of ``progress''.  All progress is based upon a universal innate desire on the part of every organism to live beyond its income.}
\eq
\edb

Thanks for the recommendation.  Sounds great.  Don't worry about the ``new religion'' thing---it wasn't a paper at all, just a grant proposal \ldots\ all geared toward extracting money.  (Though I definitely was honest in it with regard to my physical and philosophical beliefs.  \ldots\ Now, with regard to the ``outputs'' and ``outcomes'' that they required me to divine, that might be a different story!)

I didn't know the Samuel Butler quote, and in fact I didn't know about Samuel Butler at all before your bringing him to my attention.  So thanks!  I just looked him up on Wiki, and will learn more as time goes on.  I've got to read more of this guy.  The quote certainly resonates with my ``meliorist'' tendencies.  I'll match your Butler with a James or two.  I like the little piece that starts in the right-hand column on the first page of ``Meliorism 1'' at the words, ``Free-will is thus a general cosmological theory of promise \ldots'' [See 20-09-07 note ``\myref{Ericsson1}{Free Will}'' to {\AA}. {\Ericsson}.]  I also like entry 247 in ``Meliorism 2''. [See ``Pragmatism and Religion'' section in 30-06-09 note ``\myref{Schlosshauer-newname}{A New Name for Some Old Ways of Thinking}'' to M. Schlosshauer.]

In fact, I just opened up a nice new collector's item this morning that arrived at the door yesterday---a 1933 printing of Henry Call Sprinkle's Yale PhD thesis, {\sl Concerning the Philosophical Defensibility of a Limited Indeterminism: An Enquiry Based upon a Critical Study of the Indeterministic Theories of James, Renouvier, Boutroux, Eddington, Bergson and Whitehead}.

So, the evidence you present to me (and the corroborating evidence I send to you) say that maybe this ``new religion'' is not so new after all!  It's all old stuff really; it's just that science as usually practiced has forgotten it.

\section{14-10-10 \ \ {\it Temporarily Imposed Titles}\ \ \ (to {\AA}. Ericsson, B.-G. Englert, S. T. Flammia, \& W. K. Wootters)} \label{Ericsson15.1} \label{Englert1} \label{Flammia11.1} \label{Wootters25.1}

\noindent Dear Sessioneers! \medskip

Since I haven't heard back from most of you, I've gone ahead and made up some tentative titles for you and placed them into the APS system for the purpose of the planning meeting which I leave for tomorrow afternoon:  1) in Steve's case, the title of his last paper on the subject, 2) in {\AA}sa's case, a minor variation of the title she sent me, 3) in Berge's case, the title of his second-to-last paper, and 4) in Bill's case, the title of his last talk on the subject.  I.e.,
\begin{itemize}
\item
The Lie Algebraic Significance of Symmetric Informationally Complete Measurements \\ -- Steven T. Flammia, California Institute of Technology

\item
Quantum States as Probabilities from Symmetric Informationally Complete Measurements \\ -- {\AA}sa Ericsson, Institut Mittag-Leffler

\item
On Mutually Unbiased Bases \\ -- Berthold-Georg Englert, Centre for Quantum Tech., National University of Singapore

\item
States with the Same Probability Distribution for Each Basis in a Complete Set of MUBs \\ -- William K. Wootters, Williams College
\end{itemize}

I hope you don't mind me taking this liberty.  But please don't forget that these are just tentative titles and can be changed up until November 19.
For the remaining speaker, I'm aiming for an {\it experimentalist\/} doing MUB and SIC experiments.  (I figure if we have an experimentalist, we might just have someone besides ourselves attending the session!)  Keep your fingers crossed!

\section{14-10-10 \ \ {\it (Even More Urgent) APS March Meeting Invitation}\ \ \ (to A. M. Steinberg)} \label{Steinberg6}

I'm the APS March-Meeting program chair for the Topical Group on Quantum Information this year, and I'm presently in the hurried process of planning out three invited sessions for it.  On two of them, I've been completely selfless (one theoretical, one experimental), but on one I've been quite self-indulgent (as I have noticed has been the tradition in previous years \ldots\ so I don't feel so bad).  Anyway, as you might nearly guess, the title of that session will be ``Symmetric Discrete Structures for Finite Dimensional Quantum Systems''.  (A little bit more detail can be found below in the sample invitation I had sent to Berge Englert.)

Here's a listing of four of its five speakers:
\begin{itemize}
\item
The Lie Algebraic Significance of Symmetric Informationally Complete Measurements \\ -- Steven T. Flammia, California Institute of Technology
\item
Quantum States as Probabilities from Symmetric Informationally Complete Measurements \\ -- {\AA}sa Ericsson, Institut Mittag-Leffler
\item
On Mutually Unbiased Bases \\ -- Berthold-Georg Englert, Centre for Quantum Tech., National University of Singapore
\item
States with the Same Probability Distribution for Each Basis in a Complete Set of MUBs \\ -- William K. Wootters, Williams College
\end{itemize}
The trouble is, I've been having a hard time thinking of a perfectly appropriate fifth speaker.  Most of the guys that come to mind are either already burdened with coming to my Tulane U.\ ``Clifford Lectures'' the week before, or too mathematical to really be good salesmen for the subject at the March meeting, or both.

But today I had an idea!  (I know, shoot me.)  An {\it experimentalist\/} in the list would probably be the very best thing we could do for the fame and fortune of the subject!  (I.e., that way somebody beside the speakers might actually attend the session.)  Would you come and talk on your SIC and MUB experiments?

Please, please, please!  (And please let me know as soon as possible.  I fly to DC for the planning meeting tomorrow afternoon, and really should have all proposals in the online system before I arrive.)

\section{16-10-10 \ \ {\it (Doubling the Enjoyment}\ \ \ (to A. M. Steinberg)} \label{Steinberg7}

Read from bottom to top.  I got tickled at your email ``signature''
\bv
``Shut up and measure.'' \ --- \ Jeff Lundeen (after Mermin)
\ev
and sent it to David Mermin.

\subsection{Aephraim's Reply}

\bq
To be honest, I would've attributed it to Feynman too, but Jeff used it in his
talk with the attribution to Mermin, and the web seemed to agree with him.

I guess the point is that it seems to be what Feynman believed, but in David's
words.  I have to go back to the original essay \ldots\ is that what he was saying?

I should mention that I have a soft spot in my heart for Mermin (whom I've only
met about twice, and don't really know personally).  I doubt I ever told you, but
when in '94 I was first applying for postdoc positions, I wrote a ``chain letter'' job
application, along the lines of ``Karl M\"uller received a copy of this application and
forwarded it to 300 colleagues; he received the Nobel Prize the following week.
Stefan Schmidt ignored the letter, and his entire division of IBM-Zurich was closed
two days later.''  I sent it to people I knew, and to some people I though might
possibly have a job for me but probably wouldn't, and to people I expected would
have senses of humor --- notably Dan Kleppner \& David Mermin, based on their
Reference Frame columns.  My advisor later ran into Kleppner at some meeting,
and Dan said, ``Oh, is that your student who sent me this ridiculous email?  You
should tell him to be careful if he wants people to take him seriously.''  Prof.\ Mozart,
on the other hand, responded very kindly, and even suggested that I apply for a
job with Alex Gaeta, who is indeed an excellent researcher I hadn't previously
known about.

I actually really wish I could find that email!  Since I never knowingly throw anything
(even a backup of the disk from an old computer which probably contains the backup
of the disk from an older computer which probably contains all the mailboxes from
my 1994 unix account) away, it's hard to believe that I can't find it!

Perhaps it's a WriteNow file unreadable on some disk anyway, since I'm not sure
how many of the recipients actually had email addresses at the time!!
\eq

\section{17-10-10 \ \ {\it Modal Quantum Theory}\ \ \ (to  M. D. Westmoreland \& B. W. Schumacher)} \label{Schumacher19.1} \label{Westmoreland3}

This sounds interesting.  I'm especially intrigued by your remark, ``In fact even if you allow some possible results (in the modal sense) to receive a probability of zero (so that zero probability is not identified with impossible), \ldots''

The reason, of course, is that the distinction between probability zero and impossibility, and between probability one and truth, is crucial to the QBistic view of quantum theory that Carl Caves, {\Ruediger} {\Schack}, and I are trying to build up.  So the distinction would be crucial (in my mind) for your modal models as well.  Here's some of the original reading in case you're interested: \arxiv{quant-ph/0608190}, and I tried to do a better job of some of the arguments here \arxiv{1003.5209}, particularly in Section V.

I'll look forward to really understanding your papers once I fight off some of this week's upcoming fires.

\subsection{Mike's Preply}

\bq
Ben and I have posted a paper on the material Ben covered in his colloquium talk on modal quantum theory. Here is the {\tt quant-ph} link: \arxiv{1010.2929}.

We should have a paper on the open systems analogues up in two to three weeks.

We are currently in the midst of writing a paper on the quantum analogues to the Kochen--Specker and the Conway--Kochen (Free Will) results. We hope to have that paper posted within a week.

We have also discovered that, at least in some cases, these modal systems cannot be embedded in any probabilistic model.  That is, in the mobit model, there is a state and measurements such that no probability distribution on the measurement results can assign a positive probability to the results labeled as possible in the modal system. In fact even if you allow some possible results (in the modal sense) to receive a probability of zero (so that zero probability is not identified with impossible), the only possible probability assignment gives a Popescu--Rohrlich system.

It was great to visit with you last week. I hope it will not be another seven years before we see each other again.
\eq

\section{17-10-10 \ \ {\it QBlue's Fall Quantum Information Prize}\ \ \ (to the QBies)} \label{QBies30}

\noindent QBies, \medskip

I had a wacky idea yesterday in my exhaustion from the March Meeting planning session that I might start instating prizes for little non-SIC puzzles (for God's sake!) that I'd like to see solved.  (An effort to broaden us all, if you will.)

So here's my first challenge:  Figure out an application for free will.  More specifically, consider the lovely nine bases for a $d=4$ system found by Cabello et al., and described between Eqs.\ (14) and (17) of my paper \arxiv{1003.5209}.  Challenge: find some quantum information protocol built around these nine bases, and for which these nine bases give the optimal answer.  For instance, one can certainly build a quantum crypto protocol of the style of BB84 out of these.  But is there any advantage to these 9 bases for that task, rather than some other 9 bases?  What is significant about these 9 bases is that they cannot be colored in the Kochen--Specker way.  But what is the practical cash value of that observation?   Is there any cryptographic use for the way these bases lock together?  Is there something unique and interesting about the Gram matrix of the associated 18 vectors?  Little questions.  Add to the list if you can think of anything.

Anyway, I'll award the finder of the best observation, protocol, or theorem (by my judgement) with some old, nice book on the subject of free will and a good heap of Kiki's cooking.\medskip

\noindent For the prize and award committee,\medskip

\noindent QBlue

\section{19-10-10 \ \ {\it Intersections} \ \ (to L. Hardy)} \label{Hardy42.1}

Just taking a two minute break (before my nap!).  Anyway, when all this writing and politicking is over with, it'd be nice to talk to you about some physics.  Most recently (after Matthew's pushing me on the subject), I've been taking your remark from a few months back more seriously---i.e., that maybe we should take an intersection of some of your axioms and some of mine.  Below is the note that started the thinking off.  [See 05-10-10 note ``\myref{Graydon12}{Galoshes}'' to M. A. Graydon.]

We've now got a rigorous argument that our modified law of total probability (MLTP) demands that $K_{12} \ge K_1 K_2$.  I.e., that that much of your Axiom 4 is not a postulate for us, but required by consistency.  Now we're (well, not me really at the moment) trying to see if the MLTP similarly demands that the parameter $\beta$ (in eq 10 of \arxiv{0912.4252}) be multiplicative (or submultiplicative or supermultiplicative) as well.

As I say below, the question from the QBism point of view is how the oomph scales with the stuff.

Keeping my fingers crossed \ldots

\section{21-10-10 \ \ {\it Genetic Engineering of Nature Itself} \ \ (to M. A. Graydon)} \label{Graydon13}

Sections 4 and 5 are the relevant parts.  [See ``The Anti-{\Vaxjo} Interpretation of Quantum Mechanics,'' \arxiv{quant-ph/0204146}.]

\section{22-10-10 \ \ {\it Crazy Talk} \ \ (to C. Smeenk)} \label{Smeenk3}

Thanks for the invitation to the do last night, if you were responsible for it.  Immediately after asking my question to Kitcher, I started regretting how I posed it.  I should have just asked more crisply, ``What would John Dewey have said about Wikipedia?''

Attached is the article from Maclean's Magazine that I was telling you about last night.  The reporter sure didn't do me any favours, juxtaposing my thinking with our director's views on the matter!

\section{22-10-10 \ \ {\it Clarence Irving Lewis} \ \ (to P. Wells)} \label{Wells6}

I was just forwarding your PI article to a philosopher friend at Western, and I noticed for the first time that you said I had C. S. Lewis on my bookshelf.  Not at all!  It's C. I. Lewis.  Let me introduce you to him:
\bq\noindent
\myurl{http://en.wikipedia.org/wiki/Clarence_Irving_Lewis}.
\eq

\section{22-10-10 \ \ {\it Qualities of Matter} \ \ (to C. Smeenk)} \label{Smeenk4}

By the way, if it's in readable form already, I wouldn't mind getting a copy of your ``qualities of matter'' paper with Biener.  Funnily enough, I've recently adopted a bit of terminology like that for ``Hilbert space dimension''---seeing it as a quality or capacity of matter \ldots\ not sure which term is best and what all the existing connotations are.  Perhaps reading you will help.  There's a bit of my present thinking on the matter in Section VI, starting page 19, of this paper:  \arxiv{1003.5209}.  And I even give my own---almost surely distorted---view of Newton there.  It'd be nice to learn about the real man.

\section{25-10-10 \ \ {\it Paul Krugman's the Man!}\ \ \ (to R. W. {\Spekkens})} \label{Spekkens88}

From today's {\sl New York Times}:
\bq\noindent
What we do know is that the inadequacy of the stimulus has been a political catastrophe. Yes, things are better than they would have been without the American Recovery and Reinvestment Act: the unemployment rate would probably be close to 12 percent right now if the administration hadn't passed its plan. But voters respond to facts, not counterfactuals, and the perception is that the administration's policies have failed.
\eq

\section{25-10-10 \ \ {\it Tentative Title}\ \ \ (to H. B. Dang)} \label{Dang12}

\bhbd
The title is actually the hardest line, which I was planning to leave as the last thing to write, with the hope that along the way I will come across a title that best captures this paper.
\ehbd

You should always write the title long before the writing the paper!  I never do otherwise myself.

\section{26-10-10 \ \ {\it A Paper and a Fragment}\ \ \ (to R. {\Schack})} \label{Schack213}

Sorry for all the silence.  I've had a grueling month.  Presently I'm hurriedly working on my tenure application and I've finally come to the ``Future Research'' section \ldots\ and thus have started thinking of what I will say in the subsection, ``Decoherence and the Classical-to-Quantum Transition (sic)''.  Anyway, to make a long story short, I discovered the attached paper on the web.  It looks useful for our purposes (in case you haven't seen it before):  Mitchell S. Green and Christopher R. Hitchcock, ``Reflections on Reflection: Van Fraassen on Belief,'' {\sl Synthese} {\bf 98}(2), 297--324 (1994).

Everything is kosher, by the way, with Meir Hemmo.  He's waiting on us relatively patiently.  I hope to push your draft to the highest possible priority next week while I'm in South Africa getting things set up for a big QBism meeting at STIAS.

\section{26-10-10 \ \ {\it That Old Free Will Again} \ \ (to D. M. {\Appleby})} \label{Appleby95}

Yes, I'm still working on the sordid self-aggrandizement exercise---it's up to 45 pages now.  But I have a small hope that I will finish it today.

Just writing on Bell--Kochen--Specker stuff \ldots\ and came across this reference:
\begin{center}
\arxiv{1002.1410}.
\end{center}
It looks like there's some interesting commentary on your work in there.  Were you aware of it?

\section{30-10-10 \ \ {\it Hilbert Space Dimension as an Occult Quality \ldots}\ \ \ (to C. Smeenk)} \label{Smeenk5}

I thought you might enjoy a couple of sections near the end of the completed draft (or at least I think I just finished it!)\ where I speculate on HS dimension as an occult quality!  It's a joke of course, but I am a bit serious too.  At least to the point that it seems useful for me study some of the historical literature, like your paper, and some other nice looking things I've found on the web.

Once I get a chance to read and think about all this esoteric stuff a bit more deeply, I'm sure to be back in touch!

\section{01-11-10 \ \ {\it And Conway Wasn't Conway}\ \ \ (to A. Karlsson)} \label{Karlsson1}

You probably learned complex analysis from J. B. Conway's book.  The Conway here is John Horton Conway, the very, very powerful mathematician:
\bv
\myurl{http://en.wikipedia.org/wiki/John_Horton_Conway}
\ev

I have a story about Conway and SICs.  I told John that it looks like the Welch bound can be achieved in the complex case.  He was a bit shocked to learn that---for you know, he knows the real case very well, making very important contributions to it.  Anyway, he was very pleased with this new knowledge, thought for a minute, and then declared, ``I do believe the bound can be achieved in the complex case, but it would be very difficult to prove it.''

\section{02-11-10 \ \ {\it Probable Delay}\ \ \ (to R. {\Schack})} \label{Schack214}

From New York JFK.  Just to warn you:  They're presently showing my flight as delayed by 40 minutes.  My scepticism says that that'll turn into a much larger delay before it's all over with.  Anyway, stay on the lookout so that you don't waste too much time in the airport.

I've been reading some very nice stuff on alchemy this evening.  I've learned that Wheeler's ``now fly!''\ metaphor is every bit as old as Aristotle himself.  From a 1997 note of mine:
\begin{quote}
In 1972 [John Preskill] had Wheeler for his freshman classical mechanics course at Princeton.  One day Wheeler had each student write all the equations of physics s/he knew on a single sheet of paper.  He gathered the papers up and placed them all side-by-side on the stage at the front of the classroom.  Finally, he looked out at the students and said, ``These pages likely contain all the fundamental equations we know of physics.  They encapsulate all that's known of the world.''  Then he looked at the papers and said, ``Now fly!'' Nothing happened.  He looked out at the audience, then at the papers, raised his hands high, and commanded, ``Fly!'' Everyone was silent, thinking this guy had gone off his rocker. Wheeler said, ``You see, these equations can't fly.  But our universe flies. We're still missing the single, simple ingredient that makes it all fly.''
\end{quote}
Compare:
\begin{quote}
Plato's distrust of the mimetic arts reflects a widespread ambivalence toward imitation in antiquity.  This mistrust is rooted in the idea that the painter or sculptor, by producing a replica of something natural, is engaging in a sort of counterfeit.  The same attitude existed with regard to the {\it technai\/} more broadly.  Although they might be clever simulacra of nature, they could not themselves be natural.  A clear formulation of this distinction between the products of nature and the products of artifice appear in Aristotle's {\sl Physics}, where the Stagirite distinguishes natural products from artificial ones on the basis of the fact that the natural have an innate principle of movement (or change), whereas the artificial have no inherent trend toward change.  For this reason, Aristotle says, ``men propagate men, but bedsteads do not propagate bedsteads.'' The artificial product is static, having received no intrinsic principle of development.
\end{quote}

On other matters:  The key issue of Weyl's phase space must be that for the measurements associated with it (i.e., $X$ and $Z$) every SIC vector will give a  give a probability distribution for those two observables with identical {\Renyi} 2-entropy.  Every SIC vector (WH SIC at least) will generate an entropy of~$\log\!\left(\frac{2}{d+1}\right)$ for either of those measurements.  It strikes me that that might well be an axiom we'd want to take.  For instance, I have this pet idea (since 4:00 AM this morning) that this might give us the right bound on the number of zero components in a probability vectors.  Our bounds previously have been horrible (only saturated in~$d=3$).  But this looks like a much tougher constraint to me.

But I cannot get too distracted this week:  My purpose is to write, write, write of QBism from the bottom of the world (not derive, derive, derive).  And when we all come back next year, we will turn QBism on its head there!  That's got to be the purpose of the place.

\section{05-11-10 \ \ {\it Sabbatical in South Africa}\ \ \ (to R. {\Schack})} \label{Schack215}

I'm now starting to understand what they want:  Ideally about 5 people (interdisciplinary) to stay for 6 to 8 weeks and work in their facility.  Travel and accommodation covered, as well as a small stipend for food.  But no salary support.  It is stunningly beautiful here; very nice facilities with nice gardens, vineyard, etc.  (Much more my style than Perimeter.)  And a very pleasant community with lots of restaurants and caf\'es.

\section{05-11-10 \ \ {\it Heisenberg's Microscope and Battery}\ \ \ (to H. B. Geyer)} \label{Geyer2}

It looks like we were both right.  See this excerpt from Cassidy's book:
\begin{center}
\myurl{http://www.aps.org/publications/apsnews/199801/heisenberg.cfm}
\end{center}
The relevant lines are:
\bq
As the 21-year-old Heisenberg appeared before the four professors on July 23, 1923, he easily handled Sommerfeld's questions and those in mathematics, but he began to stumble on astronomy and fell flat on his face on experimental physics. In his laboratory work Heisenberg had to use a Fabry--Perot interferometer, a device for observing the interference of light waves, on which Wien had lectured extensively. But Heisenberg had no idea how to derive the resolving power of the interferometer nor, to Wien's surprise, could he derive the resolving power of such common instruments as the telescope and the microscope. When an angry Wien asked how a storage battery works, the candidate was still lost. Wien saw no reason to pass the young man, no matter how brilliant he was in other fields.
\eq

I do, however, think Cassidy's article is wrong near the end of the webpage.  My understanding is that Bohr found his complementarity the same week Heisenberg found his uncertainty principle, and the two of them were in different places at the time.

\section{06-11-10 \ \ {\it QBies for the BQs} \ \ (to N. D. {\Mermin})} \label{Mermin193}

\bdm
You've had much more experience with philosophers than I've had.

Here's a general comment from my draft replies to Max's 17 questions.  (He gave me another 15 day extension. I gather you are also delinquent.)
\bq\rm
When I got into this business I had hoped that philosophers would bring to the conversation their historical expertise in the Big Questions.  What is the nature of human knowledge?  How do people construct a model of the world external to themselves?  How does our mental organization limit our ability to picture phenomena? How does our need to communicate with each other constrain the kinds of science we can develop?  Those kinds of questions.

But, to my disappointment, it seems to me the professional philosophers prefer to behave like amateur physicists.  They don't try to view the formalism as part of a Bigger Picture.  On the contrary, they seem to prefer to interpret it more literally and less imaginatively than many professional physicists.  Because they are less proficient than physicists in using the tools of physics, they tend not to do as good a job on these narrower matters.  Sometimes they strike me as naive and unsophisticated.

So I would say that up to now professional philosophers have not played a significant role in advancing our understanding of quantum foundations. I would not (and could not) discourage philosophers from working in quantum foundations.   But I would urge them to keep their eyes on the Big Questions.
\eq
So go for somebody who worries about BQs --- somebody who brings BQism to QBism.
\edm

Oh, I wish you hadn't sent me that!  I was only a little way into answering that question, but I was going to make a point much like you.  Now, I'll have to think hard so that I don't feel like I'm sounding too much like you.

I was going to base the answer on two recent rants.  One to John Norton and one to Max himself.  I'll put the rant to Norton below. [See 07-06-10 note ``\myref{Norton2}{Time Flies}'' to J. D. Norton.]

Here were the words I had constructed so far for my answer to Q14, ``What is the role of philosophy in advancing our understanding of the foundations of quantum mechanics?''
\bq
If you catch me on a bad day, I'd say ``no role.''  But I'd be lying if I left it at that.  What I mean more particularly is that a large fraction of the philosophers of science who hang out at quantum foundations meetings have never seemed to me to bring much to the table that is forward looking or creative. Except for their willingness to engage in foundational questions in a way most physicists will not, they almost represent an impediment.  There is no doubt that my views would not be what they are today if I had not had a sustained interaction with this community, but their role has always been a negative and resistive one; what I have gotten out of the deal is that it has been a kind of whet stone for sharpening my own thoughts.  I'd rather say that I've learned something {\it directly\/} from them---that my eyes were opened by this one's or that one's considerations---but it hasn't been much so.

The trouble is they advertise that what they are up to is all about the logic of what the physicists present them with, but it has been my experience that it is most often logic used in the service of {\it their own\/} prejudices.  The manipulations of logic work just as well on false values as they do on truth values \ldots\
\eq
Don't send me any more of your answers!\footnote{\editornote Mermin's answers can be read in his ``Annotated Interview with a QBist in the Making,'' \arxiv{1301.6551}.}

Off for drive out into the wine regions.

\section{07-11-10 \ \ {\it (Urgent) APS Invitation !!}\ \ \ (to B. W. Schumacher)} \label{Schumacher20}

I write you (from South Africa!) as a {\it desperately running late\/} organizer for the Topical Group on Quantum Information's portion of the APS March Meeting.  It's going to be in Dallas, March 21--25.

It's not quite a 10 year anniversary of the old PhysComp '92 meeting we were at, but I did find a way to make it an important anniversary for quantum information.  Among the sessions I'm organizing, there'll be one titled ``20 Years of Quantum Information in Physical Review Letters'' (I used Ekert's QCrypto paper as the excuse) and through it, I'm hoping to draw broad attention to the TGQI by having (somewhat) historical talks about the golden age of quantum information.  So far, for that session, I've secured Bennett, Shor, DiVincenzo (who will be at Aachen U by then), and Ekert.

I'm writing \ldots\ {\it urgently!} \ldots\ to ask if you'd be willing to join the crowd?  It'd be a 30 minute talk with 6 minutes of questions.  It would be wonderful to have a talk about the origin of the qubit, quantum compression, and quantum channel capacities by you.

In other sessions, so far, I've secured Bill Wootters, Anton Zeilinger, Jeff Kimble, Chris Monroe, Dick Slusher, Richard Hughes (giving a talk titled ``27 Years of Quantum Cryptography!'').  And present for the executive meeting will be John Preskill and Dave Bacon.  I'm sure I'm forgetting others.  The main thing I'm trying to impress you with is that you should have some fun people to talk to if you would join us!  Oh, and a talk by Till Rosenband on one of the coolest experiments on earth: \myurl[http://www.nist.gov/pml/div688/clocks_092810.cfm]{http://www.nist.gov/pml/div688/clocks\underline{ }092810.cfm}.  (They can measure GR effects on clock synchronization over just 33 cm!!)

Also there'll be lectures by at least one (and maybe both) of the graphene Nobel prize winners.  And since it's a 100th year anniversary of superconductivity, there'll be a special Nobel session with Giaver, Ketterle, Leggett, Mueller, and Wilczek speaking.   Finally, David Mermin has said that he might come just to hang out.  It'll be a big, buzzing zoo with likely 7,500 attendees (I'm told).

Anyway, I hope you can tell me you'll come!  (And I hope you can tell me really {\it quickly\/} whatever the answer be.  I've really got to get an announcement out to the general membership tomorrow or the next day prodding them to submit abstracts before the Nov 19 deadline, and I want to have just as much firepower as I can by then!)

It is a general policy of APS to not extend travel grants or fee waivers to invited speakers, trying to keep the ``exceptional cases'' to where the speaker is a nonphysicist or needing to make a transoceanic flight.  So far, I've only used this power for Peter Shor because of his maths affiliation.  But I don't recall your having a travel grant like all my other guys, so if you need airfare or fee waiver or both, I think I will be able to argue the case (with Ivan Deutsch and Christine Lenihan).

Mostly I want you there!  Please let me know just as soon as possible.  And at the same time, tell me your financial needs \ldots\ and your title! \medskip

\noindent From Stellenbosch, \medskip

\noindent Chris \ldots\ (oh, and there'll be a focus session ``Quantum Information in the Service of Quantum Foundations'' where you and/or Mike your modal stuff for a 12$+$3 talk)

\section{07-11-10 \ \ {\it Ellis on Time}\ \ \ (to H. B. Geyer)} \label{Geyer3}

Here is a short summary of Ellis's view on time that we were discussing yesterday:
\begin{center}
\myurl{http://www.fqxi.org/community/forum/topic/361}
\end{center}

\section{07-11-10 \ \ {\it Another Meaning of Truth.\ Or The Usual One?}\ \ \ (to D. M. {\Appleby})} \label{Appleby96}

These lines struck me deeply tonight:
\bv
Everybody loves the sound of a train in the distance\\
Everybody thinks it's true\\
\ev
They come from Paul Simon's song ``Train in the Distance'':
\bv
She was beautiful as southern skies
\\The night he met her
\\She was married to someone
\\He was doggedly determined that he would get her
\\He was old, he was young
\\From time to time he'd tip his heart
\\But each time she withdrew
\\Everybody loves the sound of a train in the distance
\\Everybody thinks it's true
\\Well eventually the boy and the girl get married
\\Sure enough they have a son
\\And though they both were occupied
\\With the child she carried
\\Disagreements had begun
\\And in a while they fell apart
\\It wasn't hard to do
\\Everybody loves the sound of a train in the distance
\\Everybody thinks it's true
\\Two disappointed believers
\\Two people playing the game
\\Negotiations and love songs
\\Are often mistaken for one and the same
\\Now the man and the woman
\\Remain in contact
\\Let us say it's for the child
\\With disagreements about the meaning
\\Of a marriage contract
\\Conversations hard and wild
\\But from time to time
\\He makes her laugh
\\She cooks a meal or two
\\Everybody loves the sound of a train in the distance
\\Everybody thinks it's true
\\Everybody loves the sound of a train in the distance
\\Everybody thinks it's true
\\What is the point of this story
\\What information pertains
\\The thought that life could be better
\\Is woven indelibly
\\Into our hearts
\\And our brains
\ev
Meliorism is woven into our hearts and our brains.

Greetings from beautiful/ugly, rich/poor, full/empty \ldots\ sad/hopeful \ldots\ South Africa.  This country feels bigger than anything I've ever encountered before.

\section{08-11-10 \ \ {\it Bill's Functionalism}\ \ \ (to W. G. {\Demopoulos})} \label{Demopoulos37}

Two African meals have passed, and I have finished reading your new version of the paper.  It was a pleasure this, as every time before.

I think I caught one typographical error, page 10, 6th line from the top:  ``to the question whether'' $\rightarrow$ ``to the question of whether''?

As you can probably guess, I'm still not very sympathetic with (what I view as) your ontological dualism.  ``Effects are the traces of particle interactions on systems which admit a theoretical description in terms of propositions which belong to them.''  I don't like the idea of the world consisting of two fundamentally different kinds of entity (in reductionistic character at least).  I'm much happier to think ``propositions never belong to systems'' quantum or classical.  If, as opposed to me, you'll have nothing of agents, then I'd much rather see you develop a view where it's effects all the way down.  A particle's effect on a device is that its own effects on anything else can now be something that they would not have been otherwise.  A world that is ``ever not quite'' (Benjamin Paul Blood).

My favorite part of the paper continues to be how ``a particle's characterization has the logical form of a function.''  A broad suggestion for some next-step explorations/research.  It would be nice if you would compare and contrast what you're talking about here with with some of the (so-far only partially developed) relational views on quantum mechanics.  For instance, Rovelli's
\arxiv{quant-ph/9609002}.
What are the differences between Rovelli's ``relationalism'' and Bill's ``functionalism''?  (It seems your dualism is one thing.)  But is there anything in common between them?

Another thing I'd like to hear is a more lingering discussion on, ``What are these eternal properties you speak of?''  Say, an electron:  What are its eternal properties?  And is there any connection between the eternal properties and the set of effects a particle can produce.

I liked Itamar's remark, ``The lesson of EPR is not about locality but about how their criterion of reality is insufficient.''  However, I don't agree with one of his lines a few previous to that---that ``EPR's criterion is at best necessary.''  That gives it too much credit for me.  But that's all tied up with my personalist Bayesianism even for probability 1.

When you feel like writing again, let me know how you're doing.

\section{08-11-10 \ \ {\it There and Back Again}\ \ \ (to H. B. Geyer)} \label{Geyer4}

I didn't catch the name of the NITheP director, and I couldn't find it on their webpage.  You might forward this to him.  It looks like the Google Swedish tool might be a bit better than the German one after all!  Or maybe it's just that the Swedes have a certain sympathy to Jim's (old) way of looking at the world.

Here is a passage from a 1968 paper of Jim Hartle's:
\bq\noindent
A quantum-mechanical state, being a summary of the observers' information about an individual physical system, changes both by dynamical laws and whenever the observer acquires new information about the system through the process of measurement.  The existence of two laws for the evolution of the state vector becomes problematical only if it is believed that the state vector is an objective property of the system.  Then, the state vector must be required to change only by dynamical law, and the problem must be faced of justifying the second mode of evolution from the first.  If the state of a system is defined as [our information of it], it is not surprising that after a measurement the state must be changed to be in accord with [any] new information.  The ``reduction of the wave packet'' does take place in the consciousness of the observer, not because of any unique physical process which takes place there, but only because the state is a construct of the observer and not an objective property of the physical system.
\eq

Here it is after Google's attempt to translate it into German:
\bq\noindent
Ein quantenmechanischer Zustand, wobei eine Zusammenfassung der Beobachter Informationen {\"u}ber eine einzelne physische System, Ver{\"a}nderungen sowohl durch dynamische Gesetze und wenn der Beobachter erwirbt neue Informationen {\"u}ber das System durch den Prozess der Messung. Die Existenz von zwei Gesetze f{\"u}r die Entwicklung des Zustandsvektors wird problematisch, wenn man glaubt, dass der Zustandsvektor eine objektive Eigenschaft des Systems ist. Dann muss der Zustandsvektor verpflichtet, nur {\"a}ndern, indem dynamische Gesetz sein, und dieses Problem muss zur Rechtfertigung der zweite Modus der Evolution von der ersten konfrontiert werden. Wenn der Zustand eines Systems wie [unsere Informationen davon] definiert ist, ist es nicht verwunderlich, dass nach einer Messung den Zustand ge{\"a}ndert werden muss, um im Einklang mit [alle] neue Informationen sein. Die ``Reduktion des Wellenpakets'' findet statt im Bewusstsein des Betrachters, nicht wegen irgendeiner einzigartigen physikalischen Vorgang, der dort, aber nur, weil der Staat ist ein Konstrukt des Beobachters und nicht eine objektive Eigenschaft des physikalischen Systems.
\eq
And here it is after Google's attempt to re-translate it back into English:
\bq\noindent
A quantum mechanical state, with a summary of information obtained on a single physical system, with dynamic changes in both law and if the observer acquires new information about the system through the process of measurement. The existence of two laws for the development of the state vector is problematic if one believes that the state vector is an objective property of the system. Then the state vector required to change only by law to be dynamic, and this problem needs to justify the second mode of evolution are confronted from the first. If the state of a system like [our information like this] is defined, it is not surprising that change after a measurement the state needs to be in accordance with [any] new information. The ``reduction of the wave packet'' will take place in the consciousness of the viewer, not because of any unique physical process that there, but only because the state is a construct of the observer and not an objective property of the physical system.
\eq
Not so good indeed!

Now, I go through the same process, but with Swedish rather than German.  First to Swedish:
\bq\noindent
En kvantmekaniska tillst{\aa}nd, {\"a}r en sammanfattning av observat{\"o}rernas information om en enskild fysisk systemet, {\"a}ndringar av b{\aa}de dynamiska lagar och n{\"a}r Observer f{\"o}rv{\"a}rvar ny information om systemet genom processen av m{\"a}tning. F{\"o}rekomsten av tv{\aa} lagar f{\"o}r utvecklingen av statens vektor blir problematisk endast om det kan antas att staten vektorn {\"a}r ett m{\aa}l egendom av systemet. Sedan m{\aa}ste staten vektor m{\aa}ste {\"a}ndra endast genom dynamiska lag, och problemet m{\aa}ste m{\"o}tas av att motivera det andra s{\"a}ttet f{\"o}r utvecklingen fr{\aa}n den f{\"o}rsta. Om tillst{\aa}ndet hos ett system definieras som [v{\aa}r information om det], {\"a}r det inte f{\"o}rv{\aa}nande att efter en m{\"a}tning staten m{\aa}ste {\"a}ndras f{\"o}r att vara i {\"o}verensst{\"a}mmelse med [n{\aa}gra] nya uppgifter. Den ``minskning av den v{\aa}g paket'' tar plats i medvetandet hos den observat{\"o}r, inte p{\aa} grund av unika fysikalisk process som {\"a}ger rum d{\"a}r, men bara f{\"o}r att staten {\"a}r en konstruktion av observat{\"o}ren och inte en objektiv egenskap hos det fysiska systemet.
\eq
And now back again:
\bq\noindent
A quantum state, is a summary of the observers' information about a single physical system, changes to both the dynamic laws, and when the Observer acquires new information about the system through the process of measurement. The existence of two laws for the development of the state vector becomes problematic if it can be assumed that the state vector is an objective property of the system. Then, the state vector must change only through dynamic team, and the problem must be met to justify the other type of development from the first. If the state of a system is defined as [our information on it], it is not surprising that after a measurement the state must be modified to be consistent with [any] new data. The ``reduction of the wave packet'' takes place in the mind of the observer, not because of the unique physical process that takes place there, but just because the state is a construct of the observer and not an objective property of the physical system.
\eq
I'm not quite sure, but I think Sweden does a little better!

\section{09-11-10 \ \ {\it The Universe on a Social Analogy} \ \ (to P. Cilliers)} \label{Cilliers1}

I am looking forward to dinner with you all tomorrow evening.  At tonight's dinner, I read your paper with Heylighen and Gershenson, ``Complexity and Philosophy.''  Now I understand much better where you're coming from.

You should know that I am quite sympathetic with many things you wrote in your paper.  In fact, my foundational effort in quantum theory (which I call QBism) is based on a {\it denial\/} of one of the sentences in your paper \ldots\ but it is a denial that I think you will be pleased with once you come to understand it:
\bq\noindent
While the notion of uncertainty or indeterminacy is an essential aspect
of the newly emerging world view centering around complexity
(Gershenson \& Heylighen, 2005; Cilliers, 1998), it is in itself not
complex, and the physical theories that introduced it are still in
essence reductionist.
\eq

The denial is that actually for quantum theory to make any foundational sense, it must ultimately be read in nonreductionist terms---in fact, it is what QBism is all about!  Moreover, it gives some of the main statements within quantum theory a normative reading.  You can read about this view in this paper of mine (particularly in Section VI, starting on page 19):  \arxiv{1003.5209v1}.  It is written in an entertaining manner and meant to be easy reading, so I hope you'll give it a go.

If I were to organize a project at STIAS, it would be on the themes in this paper, and thus, I think would have quite a natural overlap with your concerns.

I have always been taken with this quote of William James,
\bq
Why may not the world be a sort of republican banquet of this sort, where all the qualities of being respect one another's personal sacredness, yet sit at the common table of space and time?

To me this view seems deeply probable.  Things cohere, but the act of cohesion itself implies but few conditions, and leaves the rest of their qualifications indeterminate.  As the first three notes of a tune comport many endings, all melodious, but the tune is not named till a particular ending has actually come,---so the parts actually known of the universe may comport many ideally possible complements. But as the facts are not the complements, so the knowledge of the one is not the knowledge of the other in anything but the few necessary elements of which all must partake in order to be together at all. Why, if one act of knowledge could from one point take in the total perspective, with all mere possibilities abolished, should there ever have been anything more than that act? Why duplicate it by the tedious unrolling, inch by inch, of the foredone reality? No answer seems possible. On the other hand, if we stipulate only a partial community of partially independent powers, we see perfectly why no one part controls the whole view, but each detail must come and be actually given, before, in any special sense, it can be said to be determined at all.  This is the moral view, the view that gives to other powers the same freedom it would have itself.
\eq
It sounds similar to some of the ideas in your paper.

\subsection{Paul's Reply}

\bq
I look forward to finally get some time to talk, I have been running around too much. There seems to be a very good basis for a good discussion.

The sentence you ``deny'' was actually aimed primarily at certain forms of complexity theory, mainly those relying heavily on chaos theory. I have been quite  consistent (I hope) in avoiding all reference to quantum theory, mainly because of my own lack of understanding, but also because of a deep suspicion of chicanery when quantum ideas are used by philosophers and other social scientists. This is also something I would like to talk about.

I will probably not have time to read your paper before we meet for dinner, but let us see.

Thanks for taking the trouble to get the discussion going.
\eq

\section{13-11-10 \ \ {\it Quantum Foundations at the APS March Meeting} \ \ (to J. N. Butterfield)} \label{Butterfield7}

Word has gotten out that you maintain a mailing list of foundations folks in the UK.  May I ask that you distribute the announcement below (along with the showy, crucial attachment of invited speakers) to your list?  Nicely, we've already gotten a few European submissions; so it is not out of the question that a European would want to come.  Any submission of quality work would certainly help the cause of making foundations a respected field within the broader APS audience.

And a test of your philosophical prowess:  What phrase below was snatched from a philosopher, and who was it?  Talking about what?

\subsection{Jeremy's Reply}

\bq
Thanks Chris for this! Im afraid I don.t have such a list now, but saw your message go out next day through Bob Coecke.s mail list, which would have covered at least most of my addressees.
May I say dear friend: it is: William James on the stream of consciousness?!
\eq

\section{13-11-10 \ \ {\it Quantum Foundations at the APS March Meeting}\ \ \ (to B. Coecke)} \label{Coecke2}

I agree with Rob, your idea is an excellent idea and long overdue.  Having that list would have been very convenient for this year.

In the meantime, I am running out of time to try to get submissions for the present meeting.  So if you could forward the announcement below to your existing list, that would be much appreciated.  Surprisingly, I have actually gotten a few European and Australian submissions.  So, I guess it is not out of the question that Europeans will come.

Please note that the note below and the attachment are slightly different than the previous ones I sent you.  Some typos have been corrected, etc.  So if you could please use the present versions \ldots

This is much, much appreciated.  And it'd be great to meet someone from your group (or you!)\ at the March Meeting!

\bq\noindent
\verb+============================================+\medskip

\noindent {\bf Deadline:  19 November!}\medskip

\noindent Dear quantum-foundational colleagues,\medskip

When Danny Greenberger and Anton Zeilinger first solicited the American Physical Society to form what later became the Topical Group on Quantum Information (GQI), they also intended it to be a representative body for quantum foundations research.  Our mission statement makes this explicit:
\bq\noindent
       The Group is committed to serving as the home within the American
       Physical Society for researchers in the foundations of quantum mechanics.
       The Topical Group will promote a continuation of the active and beneficial
       exchange of ideas between quantum foundations and quantum
       information science.
\eq
I write to you as the chair-elect of the GQI executive committee and 2011 March Meeting program chair to ask you:  {\bf PLEASE DO} support quantum foundations in the APS by {\bf SUBMITTING A TALK OR POSTER} on your work for the upcoming March Meeting in Dallas, Texas, 21--25 March 2011.

To stay viable, quantum foundations research must stay visible.  It must also contribute to the larger project of physics more generally.  Bringing our work to the attention of the APS and showing its respectability is crucial to that project.  It helps physics, and it may help create foundations-oriented faculty positions for the younger of us out there.

This year we will have a focus session titled ``Quantum Information for Quantum Foundations'' with Giulio Chiribella (Perimeter Institute for Theoretical Physics) as our invited speaker.  His talk will be titled, ``Toward a Conceptual Foundation of Quantum Information Processing.''  Also, Anton Zeilinger (University of Vienna) will be giving a symposium talk, ``Quantum Information and the Foundations of Quantum Mechanics:\ A Story of Mutual Benefit.''

It is very important that we have many other quality submissions to our focus session.  I hope that you too will recognize the importance of this meeting and do what you can.

More details of the full program can be found below and in the attached schedule that I have previously sent to the full GQI membership.  With all the Nobel prize speakers (6+) and so many of the fathers of quantum information, this promises to be one of the most memorable March Meetings in quite some time.

To register for the meeting and submit an abstract, please go to
\bq
\myurl{http://www.aps.org/meetings/march/}
\eq
and note that the {\bf DEADLINE} for abstract submission is {\bf 19 November}, just 6 days away.

Please join us in this ``blooming, buzzing confusion'' of 7,500 physicists and show them what good quantum foundations work is all about!\medskip

\noindent Best wishes,\medskip

\noindent Chris Fuchs

\noindent (more details on program below and attached)\medskip

\noindent \verb+============================================+\medskip

\noindent Dear GQI Membership,\medskip

I write to you as the chair-elect of the GQI executive committee and the program chair of our portion of the 2011 APS March Meeting.  This coming year the meeting will be in Dallas, Texas, 21--25 March 2011.

We believe we have put together an exciting venue of invited talks and focus sessions.  Please have a look at the attachment and you will see.  There will be some astounding experiments reported, and you will also have a chance to meet several of the founders of our field.  2011 is a hallmark year for quantum information as a field within physics  Also we are pleased to announce that one of our talks will be given by one of the two LeRoy Apker Award winners for ``outstanding achievements in physics by undergraduate students.''

I should further mention that the meeting will host a talk from one of this year's Nobel-Prize winners for the discovery of graphene, Konstantin Novoselov.  (Andre Geim may also speak, but has not yet confirmed.)  Moreover, there will be a recognition of the 100th anniversary of the discovery of superconductivity with a session of historical talks devoted to the subject, as well as a  Nobel-laureate session on it.  Speakers will include Ivar Giaever, Wolfgang Ketterle, Sir Anthony Leggett, K. Alexander Mueller, and Frank Wilczek, and there is word that there may be more.

In all, it should be a more-than-usual memorable meeting, with some quite wonderful GQI invited and focus sessions.  The executive committee and I hope the venue will be exciting enough to tip the scales for you if you have been indecisive about attending.

Particularly, we encourage you to submit a talk or poster on your latest  research.  The better showing GQI makes at this meeting, the greater the chance we have of increasing general APS awareness of our field, the better the chance the topical group may recruit enough members to attain APS Division status, and, {\bf MOST IMPORTANTLY}, the better the chance we have of convincing American physics departments that it is worthwhile to create faculty and research positions for all of us.  Your participation is really, truly vital.  Quantum information needs you!

\bv
      This story shall the good man teach his son;
\\      And Crispin Crispian shall ne'er go by,
\\      From this day to the ending of the world,
\\      But we in it shall be remembered---
\\      We few, we happy few, we band of brothers;
\\      For he to-day that sheds his blood with me
\\      Shall be my brother; be he ne'er so vile,
\\      This day shall gentle his condition;
\\      And gentlemen in England now-a-bed
\\      Shall think themselves accurs'd they were not here,
\\      And hold their manhoods cheap whiles any speaks
\\      That fought with us upon Saint Crispin's day.
\ev

Please note that the deadline for abstract submission is {\bf NOVEMBER 19} (less than 6 days away!).  Please submit an abstract yourself; please get your students to submit an abstract too!  Please get your associates to submit an abstract as well!! The place to go to submit and register for the meeting is here:
\bq
\myurl{http://www.aps.org/meetings/march/}
\eq

The GQI executive committee and I hope to see you in Dallas.  It'll be a whoppin' good time!\medskip

\noindent Chris Fuchs\\
Chair-elect of APS Topical Group on Quantum Information\\
GQI Program Chair for 2011 APS March Meeting
\eq

\section{15-11-10 \ \ {\it Quantum Foundations at the APS March Meeting}\ \ \ (to S. L. Braunstein)} \label{Braunstein15}

I hope you have a good student or postdoc who needs some speaking practice or exposure that you can send.   Better yet, come yourself.   Your work on Heisenberg groups would fit right in (perfect for the foundations session, and would complement the discrete structures symposium).  Beside everyone in the lists below and above, I know that Gerard Milburn, Danny Greenberger, several Zeilinger associates, \ldots, will be there.

It really promises to be a very good meeting this year.

\section{18-11-10 \ \ {\it Quantum Foundations in Dallas!}\ \ \ (to I. T. Durham)} \label{Durham1}

OK, I give up.  I'm off to have some dinner.  I tried to get this thing posted on FQXi myself, and I just couldn't get it to stick.  I don't know what I'm doing wrong.  But this is the last you'll hear from me tonight.  Please if you don't mind, use the exact wording I've given below (modulo adding your own title and adjusting to 36 if you wish).  I was given hell from a visiting professor earlier today because of my previous wording (he was offended)---I hope this time I have made it clear that these titles ``can be considered'' foundations and that they are not all in one session, but ``culled from all sessions''.

\noindent \verb+============================================+\medskip

\noindent Dear quantum foundations folk,\medskip

As I write to you, 3400 abstracts have already been submitted for the APS March Meeting, with 140 of those earmarked for the Topical Group on Quantum Information.  Very importantly for quantum foundations, {\bf 35 of those abstracts} (culled from all sessions) can be considered with good justification to be {\bf quantum foundations submissions}!!  In other words, at the moment, we've got 1\% of the whole meeting thinking about the foundations of physics!

Have a look at some of the titles and speakers below; there are going to be some very good talks at this meeting.  It will be a grand opportunity for everyone in our community to mix and mingle and learn from each other.

{\bf Please don't forget that the abstract submission deadline is tomorrow, November 19, at 5:00 PM EST.}

I really encourage everyone who wants to see quantum foundations thrive and be memorable to please submit a talk to this meeting.  Encourage your colleagues and students too.  Let's build a critical mass.  Your voice will count.

The place to go is:
\myurl{http://www.aps.org/meetings/abstract/instructions.cfm}.
You must have an APS membership before submitting (\$128 regular, \$64 for recently completed PhDs, and \$0 for students first joining), but you can still submit an abstract even if you don't have your membership number yet--the instructions at the link explain how to do it.  (It is not necessary, but please do spend the extra \$8 to join the Topical Group on Quantum Information, the official home within the APS for quantum foundations research.)\medskip

\noindent Sincerely,\medskip

\noindent Chris Fuchs

\section{20-11-10 \ \ {\it How Foundations Did} \ \ (to D. M. Greenberger \& A. Zeilinger)} \label{Zeilinger12} \label{Greenberger4}

I thought I'd give you a summary of how foundations ended up doing.  There were 7,700 submissions for the March Meeting as a whole.  The topical group for quantum information got 352 submissions.  That is up from 256 last year (and 227 the year previous to that).  The impressive thing for us is that we got 66 talks that should be of foundational interest.  This means we will commandeer 4 sessions for the submitted foundational talks alone!  I'll put all the titles below; and you can notice that there's not even any crazy ones in the list!  The best we'd ever done on foundations before was two sessions (one mostly solid stuff, and one mostly crazy).

I've got to say, I'm very proud of what we accomplished this year.  Thanks Anton for getting so many of your guys there.  Our community can stand and rise! \bigskip

\noindent {\bf Long Talks:}
\begin{enumerate}
\item
A Brief Prehistory of Qubits
\\ {\it Benjamin Schumacher}

\item
Quantum Information and the Foundations of Quantum Mechanics: A Story of Mutual Benefit
\\ {\it Anton Zeilinger}

\item
Toward a Conceptual Foundation of Quantum Information Processing
\\ {\it Giulio Chribella}

\item
Pairwise complementary observables and their mutually unbiased bases (MUBs)
\\ {\it Berthold-Georg Englert}

\item
Quantum States as Probabilities from Symmetric Informationally Complete
Measurements (SICs)
\\ {\it {\Asa} {\Ericsson}}

\item
The Lie Algebraic Significance of Symmetric Informationally Complete Measurements
\\ {\it Steven T. Flammia}

\item
Experimental access to higher-dimensional discrete quantum systems, towards realizing SIC-POVM and MUB measurements, using integrated optics
\\ {\it Christophe Schaef}

\item
Isotropic States in Discrete Phase Space
\\ {\it William K. Wootters}
\end{enumerate}

\pagebreak

\noindent {\bf Short Talks:}

\begin{enumerate}
\item
Physics as Information
\\ {\it Giacomo Mauro D'Ariano}

\item
Quantum theory cannot be extended
\\ {\it Roger Colbeck, Renato Renner}

\item
The quantal algebra and abstract equations of motion
\\ {\it Samir Lipovaca}

\item
Scaling of quantum Zeno dynamics in thermodynamic systems
\\ {\it Wing Chi Yu, Li-Gang Wang, Shi-Jian Gu}

\item
Mathematical Constraint on Realistic Theories
\\ {\it James Franson}

\item
Uncertainty Relation for Smooth Entropies
\\ {\it Marco Tomamichel, Renato Renner}

\item
Quaternions and the Quantum
\\ {\it Matthew Graydon}

\item
A Linear Dependency Structure Arising from Weyl--Heisenberg Symmetry
\\ {\it Hoan Bui Dang, Marcus Appleby, Ingemar Bengtsson, Kate Blanchfield, {\Asa} {\Ericsson},\\ Christopher Fuchs, Matthew Graydon, Gelo Tabia}

\item
Proofs of the Kochen--Specker theorem based on the 600-cell
\\ {\it P. K. Aravind, Mordecai Waegell, Norman Megill, Mladen Pavicic}

\item
Proofs of the Kochen--Specker theorem based on two qubits
\\ {\it Mordecai Waegell, P. K. Aravind}

\item
Quantum Theory for a Total System with One Internal Measuring Apparatus
\\ {\it Wen-ge Wang}

\item
The thermodynamic meaning of negative entropy
\\ {\it Lidia del Rio, Renato Renner, Johan Aaberg, Oscar Dahlsten, Vlatko Vedral}

\item
Pseudo-unitary freedom in the operator-sum representation
\\ {\it Yong Cheng Ou, Mark S. Byrd}

\item
Quantum Computational Geodesic Derivative
\\ {\it Howard Brandt}

\item
Hardy's paradox and a violation of a state-independent Bell inequality in time
\\ {\it Alessandro Fedrizzi, Marcelo P. Almeida, Matthew A. Broome, Andrew G. White, Marco Barbieri}

\item
Topos formulation of History Quantum Theory
\\ {\it Cecilia Flori}

\item
Quantum Darwinism in an Everyday Environment: Huge Redundancy in Scattered Photons
\\ {\it Charles Riedel, Wojciech Zurek}

\item
Redundant imprinting of information in non-ideal environments: Quantum Darwinism via a noisy channel
\\ {\it Michael Zwolak, Haitao Quan, Wojciech Zurek}

\item
Foundational aspects of energy-time entanglement
\\ {\it Jan-{\AA}ke Larsson}

\item
A Bigger Quantum Region in Multi-Party Bell Experiments
\\ {\it Matty Hoban, Dan Browne}

\item
Qutrits under a microscope
\\ {\it Gelo Noel Tabia}

\item
Quantum systems as embarrassed colleagues: what do tax evasion and state tomography have in common?
\\ {\it Chris Ferrie, Robin Blume-Kohout}

\item
Modal Quantum Theory
\\ {\it Michael Westmoreland, Benjamin Schumacher}

\item
On the Experimental Violation of {\Mermin}'s High-Spin Bell Inequalities in the Schwinger Representation
\\ {\it Ruffin Evans, Olivier Pfister}

\item
Measurement backaction and the quantum Zeno effect in a superconducting qubit
\\ {\it Daniel H. Slichter, R. Vijay, Irfan Siddiqi}

\item
A derivation of quantum theory from physical requirements
\\ {\it Markus Mueller, Lluis Masanes}

\item
Quantum simulation of time-dependent Hamiltonians and the convenient illusion of Hilbert space
\\ {\it Rolando Somma, David Poulin, Angie Qarry, Frank Verstraete}

\item
Time-asymmetry and causal structure
\\ {\it Bob Coecke, Raymond Lal}

\item
Large violation of Bell's inequalities using both counting and homodyne measurements
\\ {\it Valerio Scarani, Daniel Cavalcanti, Nicolas Brunner, Paul Skrzypczyk, Alejo Salles}

\item
Simulating Concordant Computations
\\ {\it Bryan Eastin}

\item
A generalization of Noether's theorem and the information-theoretic approach to the study of symmetric dynamics
\\ {\it Iman Marvian, Robert {\Spekkens}}

\item
MUB Entanglement Patterns by Transformations in Phase Space
\\ {\it Jay Lawrence}

\item
Regrouping phenomena of SIC POVMs covariant with respect to the Heisenberg--Weyl group
\\ {\it Huangjun Zhu}

\item
Quantum networks reveal quantum nonlocality
\\ {\it Daniel Cavalcanti, Mafalda Almeida, Valerio Scarani, Antonio Acin}

\item
Interpreting quantum discord through quantum state merging
\\ {\it Vaibhav Madhok, Animesh Datta}

\item
Long-range spin-coupled interactions: a Gedankenexperiment on the nature of spin
\\ {\it Ian Durham}

\item
Affine Maps of the Polarization Vector for Quantum Systems of Arbitrary Dimension
\\ {\it Mark Byrd, C. Allen Bishop, Yong-Cheng Ou}

\item
Inadequacy of von Neumann entropy for characterising extractable work
\\ {\it Oscar Dahlsten, Renato Renner, Elisabeth Rieper, Vlatko Vedral}

\item
Causality, Bell's theorem, and Ontic Definiteness
\\ {\it Joe Henson}

\item
Entanglement in Mutually Unbiased Bases
\\ {\it Marcin Wiesniak, Tomasz Paterek, Anton Zeilinger}

\item
Experimental Violation of Two-Party Leggett-Garg Inequalities with Semi-weak Measurements
\\ {\it Justin Dressel, Curtis Broadbent, John Howell, Andrew Jordan}

\item
Testing spontaneous localization with ultra-massive cluster interferometry
\\ {\it Stefan Nimmrichter, Klaus Hornberger, Markus Arndt}

\item
Experimental non-classicality of an indivisible system
\\ {\it Radek Lapkiewicz, Peizhe Li, Christoph Schaeff, Nathan Langford, Sven Ramelow, Marcin Wiesniak, Anton Zeilinger}

\item
Violation of local realism with freedom of choice
\\ {\it Johannes Kofler, Thomas Scheidl, Rupert Ursin, Sven Ramelow, Xiao-Song Ma, Thomas Herbst, Lothar Ratschbacher, Alessandro Fedrizzi, Nathan Langford, Thomas Jennewein, Anton Zeilinger}

\item
Matter wave interferometry with large and complex molecules
\\ {\it Stefan Gerlich, Sandra Eibenberger, Mathias Tomandl, Jens T\"{u}xen, Marcel Mayor, Markus Arndt}

\item
Surface based detection schemes for molecular interferometry experiments -- implications and possible applications
\\ {\it Thomas Juffmann, Adriana Milic, Michael Muellneritsch, Markus Arndt}

\item
Integrable matrices with a given number of commuting partners and their exact solution
\\ {\it Haile Owusu, Emil Yuzbashyan}

\item
Closed Systems that Measure Particles
\\ {\it Michael Steiner, Ronald Rendell}

\item
Causal Tapestries
\\ {\it William Sulis}

\item
Measures of non classical correlations
\\ {\it Matthias Lang, Anil Shaji, Carlton {\Caves}}

\item
Information geometric approach to foundations of quantum theory
\\ {\it Ryszard Kostecki}

\item
A non-local quantum eraser
\\ {\it X. Ma, J. Kofler, A. Qarry, N. Tetik, T. Scheidl, R. Ursin, S. Ramelow, L. Ratschbacher, T. Herbst, A. Fedrizzi, T. Jennewein, A. Zeilinger}

\item
Homogeneous Self-Dual Cones and the Structure of Quantum Theory
\\ {\it Alexander Wilce}

\item
Quantum Measurement, Correlation, and Contextuality
\\ {\it Masanao Ozawa}

\item
Operational interpretations of quantum discord
\\ {\it Marco Piani, Daniel Cavalcanti, Leandro Aolita, Sergio Boixo, Kavan Modi, Andreas Winter}

\item
Multipartite Entanglement: Classification, Quantification, Manipulation, Evolution and Applications
\\ {\it Gilad Gour}

\item
Decoherence Free Neutron Interferometry
\\ {\it Dmitry A. Pushin, David G. Cory, Michael G. Huber, Mohamed Abutaleb, Muhammad Arif, Charles W. Clark}

\item
Time-Reversal Symmetry and Temporal Coherent Back-Scattering in a Driven Two-Level System
\\ {\it Simon Gustavsson, Mark Rudner, Jonas Bylander, Leonid Levitov, Will Oliver}
\end{enumerate}

\section{25-11-10 \ \ {\it Pauli's Dreams $+$ Dinner $+$ Additionism}\ \ \ (to F. De Martini)} \label{DeMartini2}

And let us join forces.  From our brief conversation after my return from South Africa, I got the sense that there is a good bit in common between our views of the ``nature of classicality.''  It is a view that Appleby and I call ``additionism''---that quantum behavior is something that comes ``in addition to'' to classical.  As you write on the subject and have your lab perform relevant experiments, please keep me informed.

\section{26-11-10 \ \ {\it Qbism} \ \ (to W. C. Myrvold)} \label{Myrvold14}

\bwm
Did you know that ``Qbism'' is an iPhone game?  See
\begin{center}
\myurl{http://www.blowfishstudios.com/}
\end{center}
\ewm
Very cool!  But too bad they misspelled QBism.

\section{29-11-10 \ \ {\it {\Mermin}'s Cautionary Tale}\ \ \ (to D. M. {\Appleby})} \label{Appleby97}

Below is David's cautionary tale.
\bdm
You've forgotten that my interest {\rm [in SICs]} goes back at least to our independent,
but strongly correlated interactions with Gabe Plunk.   Why is it so
damned hard to show that such sets exist, or to find a counterexample?
(A cautionary tale is unique factorization in the cyclotomic integers---remember the videotape of my lecture at Bell?---which holds for every degree up to 22, before breaking down at 23.  But that's a much more subtle issue.   Why is this apparently trivial one so stubborn?)
\edm

(Independent, poetic) question of the day:  What is the shape of a philosopher's stone?  Prompt\-ed by this passage that I just wrote:
\bq
There is an anecdote I sometimes use for starting my talks on quantum foundations that is particularly relevant to the proposal at hand.  It is about a conversation I had with my (then) seven-year-old daughter.  For some time over the last year or two she had been asking me what is my favorite color.  But one day she took me quite by surprise by asking, ``Dad, what is your favorite shape?'' At first, I started to answer glibly, ``A ball.'' But I caught myself and became quite delighted with the possibilities.  ``It's Hilbert space,'' I said, ``My favorite shape is Hilbert space.''  ``What does it look like?''\ she asked.  The only honest reply I could give was {\it I don't know!}  ``Then how do you know it's your favorite shape?'' ``Because we know enough about it to know that it is the most beautiful shape ever imagined.  We can already see that much even if we don't know all its details.  When I go to work at Perimeter Institute every day that's what I work on---trying to understand this beautiful shape.''  Two days later, my daughter announced that when she grew up she would work where I work!

What the anecdote really refers to is the convex body that makes up the set of quantum states (pure and mixed) for a finite-dimensional quantum system.  It is easy enough to say what this body is in conventional, algebraic terms:  It is the set of trace-one positive semi-definite operators over a complex inner-product space of finite dimension $d$.  But what is the object geometrically?  There a great mystery lies and coming to grips with its answer is of the utmost importance for the foundations of quantum theory.

In the broadest terms, this is what Dr.\ X's proposal to you is all about:  Understanding this wonderful shape that empowers quantum information processing and quantum computing like a philosopher's stone.
\eq

\section{02-12-10 \ \ {\it Your Notes}\ \ \ (to D. M. {\Appleby})} \label{Appleby98}

I have read your notes now, and they were quite interesting to me.  I like this stress of yours between a probability assignment on the one hand, and a ``decision'' on the other.

Also, I liked these lines very much:
\bq
And what is that objective something? It would be fair to say that finding out what that objective something truly is was one of the central projects of classical physics. Trying to answer that question eventually led to the destruction of classical physics. And we still have the problem now, in the shape of interpreting quantum mechanics.
\eq
I am glad you have decided to write a paper on all this.  Now that I've read your notes, I am even more convinced that you'll make a real contribution with this.  Your emphasis that ``when one judges two samples to have two distinct half lives, one is ipso facto committed to the belief that there is an objective difference between them'' is more and more pressing on me.  It is right, and it needs fleshing out.  At the moment, it remains beyond my ability, but somebody has to start the process of unwinding this issue somewhere.

If the topic comes up, as you wish, I will reveal your identity.

Happy birthday, yesterday.  I wish that I had known.  I will try to make it up to you when I am back in town.

\section{02-12-10 \ \ {\it QI Meets GR in a Real Way}\ \ \ (to N. Waxman)} \label{Waxman8}

I'm on my way to Washington for a planning meeting for the APS March meeting (I'm program chair for the 400 talks on quantum information), and I was thinking about one of my invited speakers, Till Rosenband from NIST.  It dawns on me that doing a story on his experiment for {\sl Inside the Perimeter\/} would be a good way to highlight the potential for fruitful interchange between quantum information and general relativity (and gravitational experiments)---i.e., PI so well representing both fields.

Till has built the world's most accurate clock, one so accurate that he can see two such clocks come out of synchronization because of GR effects over only 33 centimeters of height difference!  And the cool thing is that the clock mechanism is based on quantum computing techniques.  NIST calls them ``quantum logic clocks''.  You can read a bit about them here:
\begin{itemize}
\item \myurl{http://www.nist.gov/pml/div688/clocks_092810.cfm}

\item \myurl{http://www.nist.gov/pml/div688/logicclock_020410.cfm}

\item \myurl{http://www.nist.gov/pml/div688/logic_clock.cfm}
\end{itemize}

Just an idea that I thought I'd record.

\section{02-12-10 \ \ {\it The Size of Quantum Information}\ \ \ (to N. Waxman)} \label{Waxman9}

How do you like that title?  Thinking further.  Once everything gets settled on planning this meeting, I might write something for you on the whole breadth of quantum information at this meeting.  The number of talks in quantum information might actually far exceed the 400 I've already told you about.  (The GQI that I chair had 360 submissions, and then there were at least 22 invited talks from other divisions that are on the subject of quantum information as well.  I have no idea how many submitted talks from other divisions will be on quantum information until I see the titles, but I was quite confident there would be far in excess of 20 if there were already 22 invited speakers.)

Then if you wanted to write a little accompanying piece on the Rosenband/Wineland experiment (the one I mentioned earlier)---the two pieces might go together well.

\section{02-12-10 \ \ {\it Probability Distributions}\ \ \ (to S. Gharibian)} \label{Gharibian2}

Good question.  And your intuition is right:  For a general POVM with $|M|$ outcomes and an arbitrary probability distribution $P$ over $|M|$ outcomes, one cannot find a quantum state that will give rise to $P$.

Take for example the SIC POVMs that I always talk about.  This is a POVM on a $d$-dimensional Hilbert space with $d^2$ outcomes.  It turns out that no quantum state can ever give too sharply peaked a probability distribution for its outcomes.  For instance, no probability can ever be larger than $1/d$.  More generally, the probability distributions arising from quantum states are always confined within a very interesting convex subset of the full simplex of probability distributions.  See the story I wrote below about this shape (from a review that I was writing on someone's proposal). [See 29-11-10 note ``\myref{Appleby97}{{\Mermin}'s Cautionary Tale}'' to D. M. {\Appleby}.]

So, I think this answers at least your question 2.  For instance take the valid probability distribution $P=[1,0,0,0,\ldots,0]$ over $d^2$ outcomes.  The closest distribution one can get to that with a SIC POVM is $Q=[1/d, 1/d(d+1), 1/d(d+1), 1/d(d+1), \ldots, 1/d(d+1)]$. You could for instance measure the distance between these two vectors with the standard Euclidean distance.

You can read a bit more about the geometry of this convex set in this paper of mine:
\bq
\arxiv{0910.2750}.
\eq
Of course, it is particular to the SIC POVM but up to an affine transformation it is generic behavior.

Toy theories where your question can be answered in the positive are called ``classical''.  You can read plenty about that in some of Howard Barnum's papers.

If you want to talk more about this, drop by my office some time.  I think it's one of the most important questions in all of quantum foundations.

\section{04-12-10 \ \ {\it Wave Function Volume} \ \ (to A. Ney)} \label{Ney4}

The paper you mention is dated, and I have evolved much since then.  Would you consider this one instead:
\arxiv{1003.5209}.
How does its length look for your needs?  How much would it have to be trimmed?

So, that is a potential ``yes'', depending upon what you say.  Thanks for thinking of me.

PS.  It is very funny:  When I first looked at your email title, I thought that you had some question about the ``volume of Hilbert space''.  See story below that I had just written up for a referee report.  [See 29-11-10 ``\myref{Appleby97}{{\Mermin}'s Cautionary Tale}'' to D. M. {\Appleby}.]

\section{05-12-10 \ \ {\it Quantum Information at the March Meeting} \ \ (to the American Physical Society)}

You asked for a heads-up of some newsworthy items at the March meeting.  I've worked hard to put together a good program on the 20th anniversary of quantum information and quantum computing in APS journals.  Attached is my list of speakers.  Particularly, the 20th anniversary session is filled with several founders of the field.  For instance, Bennett is one of the authors of the original quantum teleportation paper; Ekert is one of the discoverers of quantum cryptography; and Schumacher is the very inventor of the qubit (word and idea).  Wootters, another session, was a discoverer of the no-cloning theorem, and also one of the authors of the quantum teleportation paper.  Two of the guys in the experimental session, Jeff Kimble and Anton Zeilinger, performed the first two quantum teleportation experiments back in 1997/98.

Let me also flag this guy in the experimental session:  Till Rosenband from NIST.  Till, most recently, has built the world's most accurate clock, one so accurate that he can see two such clocks come out of synchronization because of general relativistic effects over only 33 centimeters of height difference!  And the cool thing is that the clock mechanism is based on quantum computing techniques.  NIST calls them ``quantum logic clocks''.  You can read a bit about them here:
\begin{itemize}
\item
\myurl{http://www.nist.gov/pml/div688/clocks_092810.cfm}
\item
\myurl{http://www.nist.gov/pml/div688/logicclock_020410.cfm}
\item
\myurl{http://www.nist.gov/pml/div688/logic_clock.cfm}
\end{itemize}

Anyway, needless to say, I would love to see you work up a bit of a feature on quantum information and computing with the press.  I do think it would be of general interest.

\section{08-12-10 \ \ {\it Receipt and Bill}\ \ \ (to O. J. E. Maroney)} \label{Maroney6}

OK, I was a bit late, but both letters are off now.  In the case of Pitt, I sent the signed version by snail mail, but I also sent an electronic copy to John Norton and asked him to put it into your file.

I wrote the letters in a way to be intellectually honest \ldots\ but you shouldn't worry in any way.  They both came off very strong I believe.  I wish you all the luck.

If you haven't already surmised however, you should know that nothing from me ever comes for free.  Thus, for my labors, let me bill you this:  A reading of two of Marcus Appleby's papers, \arxiv{quant-ph/0402015} and \arxiv{quant-ph/0408058}.  As an un- or non-initiate in personalistic probability, I'd like to know what you find convincing in them (if anything) and what you don't (if anything).

Thanks for the comments last week on Bennett's TIDY classical computers and their potential quantum analogs.  I think it would be good to make that explicit in your research proposal---i.e., that you are talking about those computers particularly.  I am cautious, however, in my belief of you on this one.  To my nose, it smells a bit too much of the stories one used to hear of quantum error correction before Shor, i.e., that the no-cloning theorem would forbid it.  But the key point was the clean subspaces, not the redundancy of $|\psi\rangle \rightarrow |\psi\rangle|\psi\rangle|\psi\rangle$ (which can't be made generally).  Anyway, you might be right; I'm just saying I'm cautious of believing you.  Also, when thinking of the circuit models of quantum computation that everyone uses, my mind goes back to the computer geeks I used to know in college who would laud the beauties of the C programming language.  ``Dude, it's so structured, it won't let you be inefficient!''  It strikes me that the circuit models might already be like that---they won't let you be inefficient (in comparison to Bennett's original FORTRAN).  If the quantum circuit models are inefficient in bit usage, then in what way?

Anyway, this note counts as your receipt---your letters are off; I hope you do well.  And my bill is that you give Appleby some serious thought.

\section{08-12-10 \ \ {\it Diracula!}\ \ \ (to D. Lynch)} \label{Lynch1}

I hope you understood:  Diracula was sheer genius!  I never heard that one before.

Here's the quote from David Deutsch at the bulletin board associated with his book {\sl Fabric of Reality}, dated 16 July 2000.  It refers to Asher Peres and me:

\vspace{-12pt}
\bq\noindent
\bq\noindent
A. Carvalho wrote, {\it ``\,`Contrary to those desires, quantum theory does not describe reality \ldots' I wonder what it can possibly describe. Is there anything else beyond reality?''}
\eq

No, but that's not what they think. They think it describes our
observations, but that we are not entitled to regard this as telling
us anything about a reality beyond our observations. Why? Just for
the Bohring old reason that they don't like the look of the reality
that it would describe, if it did describe reality. Why? --- I have
many speculations, but basically I don't know. I don't understand
why.

It's sad enough when cranks churn out this tawdry old excuse for
refusing to contemplate the implications of science, but when highly
competent physicists --- quantum physicists --- dust it off and
proudly repeat it, it's a crying shame.

They really need to read FoR, don't they?
\eq

By the way, David Deutsch is sometimes described as being vampirish in his habits (being reclusive, only receiving visitors at his home at night, after midnight, etc).  And he was a winner of the Dirac prize in 1998!  So maybe you're on to something.\medskip

\noindent Mischievous wishes,

\section{09-12-10 \ \ {\it The Shape of Quantum Theory}\ \ \ (to N. Waxman)} \label{Waxman10}

Attached are a few paragraphs from a referee report I wrote for the Guggenheim Foundation last week.  Definitely, consider it *only* a first pass on the question you asked me.  I'm sure for the annual report we'll need to sober it all up significantly.  But it is a true story (about Katie) as it presently stands, and conveys in an essential way what's really driving me on the day-to-day basis at PI.

We can fill out, subtract, rearrange, etc., later.  I just wanted to record this now since you asked tonight.

\bq
There is an anecdote I sometimes use for starting my talks on quantum foundations that is particularly relevant to the proposal at hand.  It is about a conversation I had with my (then) seven-year-old daughter.  For some time over the last year or two she had been asking me what is my favorite color.  But one day she took me quite by surprise by asking, ``Dad, what is your favorite shape?'' At first, I started to answer glibly, ``A ball.'' But I caught myself and became quite delighted with the possibilities.  ``It's Hilbert space,'' I said, ``My favorite shape is Hilbert space.''  ``What does it look like?'' she asked.  The only honest reply I could give was, {\it ``I don't know!''}  ``Then how do you know it's your favorite shape?'' ``Because we know enough about it to know that it is the most beautiful shape ever imagined.  We can already see that much even if we don't know all its details.  When I go to work at Perimeter Institute every day that's what I work on---trying to understand this beautiful shape.''  Two days later, my daughter announced that she when she grew up she wanted to work at Perimeter Institute too!

What the anecdote really refers to is the convex body that makes up the set of quantum states (pure and mixed) for a finite-dimensional quantum system.  It is easy enough to say what this body is in conventional, algebraic terms:  It is the set of trace-one positive semi-definite operators over a complex inner-product space of finite dimension $d$.  But what is the object geometrically?  There a great mystery lies and coming to grips with its answer is of the utmost importance for the foundations of quantum theory.

In the broadest terms, this is what Dr. [X]'s proposal to you is all about:  Understanding this wonderful shape that empowers quantum information processing and quantum computing like a philosopher's stone.
\eq

\section{09-12-10 \ \ {\it Time Stories}\ \ \ (to N. Waxman)} \label{Waxman11}

By the way, your son seems to be quite some young physicist!  Below is a little story I wrote up nearly 13 years ago (it's drawn from one of my samizdats).  You can pass it on to your son if you wish and tell him what I think the missing ingredient is:  It's time.  [See story ``It's a Wonderful Life'' in 05-12-01 note ``\myref{Schumacher4}{Lucky Seven}'' to B. W. Schumacher.]

\section{09-12-10 \ \ {\it My Clifford Lectures}\ \ \ (to R. Penrose)} \label{Penrose1}

When you were last visiting Perimeter Institute, you showed some interest in the problem of the ``maximum number of equiangular lines in a complex vector space of dimension $d$''.  Also my postdoc {\AA}sa Ericsson told me that you had expressed a potential interest in participating in a meeting on the subject she is organizing in Banff for 2012.

Anyway, the reason I write you now is that (by some miracle) I was chosen to be the Clifford Lecturer at Tulane University for 2011
(see \myurl{http://www.math.tulane.edu/activities/clifford/})
and a small workshop will be built up around my own lectures, March 14--17.  There should be some quite interesting people there, and maybe some old friends of yours.  Beside whoever else is at Tulane, Lane Hughston, Samson Abramsky, $+$6 more, for instance will be there.  And particularly, my colleague Marcus Appleby, I think you would enjoy meeting:
\bv
\myurl{http://arxiv.org/find/quant-ph/1/au:+Appleby/0/1/0/all/0/1}.
\ev

Most recently we have found a lovely connection between this equiangular line problem and cyclotomic number fields, at least in dimensions 2, 3, 4, 5, and 7.  So we certainly get the sense that some real progress can be made on the problem in the near future, and that this once very specialized problem has tendrils into all sorts of mathematics.

Anyway, in all, would you have some interest in participating?  You could speak on any subject you wish.  But mostly, personally, I would like to get any input and feedback on this interesting problem in algebraic geometry you could give in our personal discussions outside the talks.

If you have an interest, I'll get an official invitation sent to you.  All expenses paid, of course.

\section{10-12-10 \ \ {\it Texan Roots}\ \ \ (to M. E. L. Oakes)} \label{Oakes1}

You probably don't remember me.  I was in two of your courses in 1984/1985, I believe.

Last night I was up with a bout of insomnia, and somehow in all the web surfing, I ended up wandering (first) to the {\sl Daily Texan\/} and (then) to the UT Physics Dept.  It was quite a stroll down memory lane.

I suppose it was a natural consequence of my working on an interview article the last few days for a book someone is compiling.  One of the questions caused me to be semi-autobiographical again, and I suppose Texas has been on my mind.  In fact, this is the third round of autobiography I've had to do this year:  The first was for the introduction to my book {\sl Coming of Age with Quantum Information\/} (Cambridge U. Press, to be released in about a week), and the second was for my tenure application here at Perimeter Institute.

I'll attach both of these in case you have some interest.  They tell the story of how one Texas boy ended up doing physics, and even making a mark in the crazy (usually markless) world of quantum foundations.

I guess I'm saying I haven't done too badly, and seeing your picture last night reminded me of what a great teacher you were.  Without exaggeration, your lectures were the best I encountered in my four years of physics at UT.  They kept me going at a time when I had doubts of why I was even there (the opening pages of the {\sl Coming of Age\/} intro explain why I say this).

There's a line on the second page of the Perimeter research statement that says, ``To the best students who came asking for a research project, graduate and undergraduate alike, John [Wheeler] would say, `Derive quantum theory from an information theoretic principle.'  In my case, I was given a more workaday project in the Regge calculus \ldots''  The student I had particularly in mind when I was writing that was Drew Debelack; he was also in your classes at the same time.  I looked for him on the web but didn't find much---it looks like maybe he dropped out of physics.

Anyway, thanks again for all the time you put into the students (like me) way back when and the ones more recently.  Seeing your title of ``Distinguished Teaching Professor Emeritus'' last night, it was clear that it is the perfect title for you.

\section{12-12-10 \ \ {\it Dallas} \ \ (to A. Wilce)} \label{Wilce24}

\baw
I'm trying to sort out my spring travel plans. It would be helpful to know when, approximately,
we'll be hearing about the acceptance of talks submitted to the APS foundations meeting in March.
Do you have a sense of when we might hear about this?
\eaw

You, of course, are accepted!  And your talk will be in the featured foundations session, along with the invited speaker Chiribella, and submitted talks by D'Ariano, Masanes, Greenberger, and 9 others.  I think the session was called ``Axiomatics and Toy Models''.  In total, we have four foundations sessions this year!  A total of 58 talks in those sessions.  Then there's the discrete structures invited session with Wootters, Englert, Flammia, Ericsson, and Schaeff (an experimentalist) that might interest you.  And the seven Nobel prizers' talks.

Attached is the latest list of all QI relevant invited talks.  In total, there were over 360 GQI submissions.  And I would guess at least 20 to 40 more quantum-information talks that were not classified so.

I can't tell you exactly when yet, but the session you'd speak in will most likely be on the Monday or Tuesday.  And for your reference, I'll be staying in the Adolphus Hotel (with its ghost).

\section{14-12-10 \ \ {\it The Archeology of SICs}\ \ \ (to A. Karlsson \& L. Piispanen)} \label{Karlsson2} \label{Piispanen1}

Two things to read on the archaeology of SICs:
\begin{itemize}
\item
W. A. Fedak and J. J. Prentis, ``The 1925 Born and Jordan paper `On quantum mechanics','' Am.\ J. Phys.\ {\bf 77}, 128--139 (2009).
\item
Erhard Scholz, ``Weyl Entering the `New' Quantum Mechanics Discourse,'' conference contribution to {\sl History of Quantum Physics 1}, Berlin, July 2--6, 2007.\\
\myurl[http://quantum-history.mpiwg-berlin.mpg.de/eLibrary/hq1_talks/mathematics/35_scholz]{http://quantum-history.mpiwg-berlin.mpg.de/eLibrary/hq1\underline{ }talks/mathematics/35 \\ \underline{ }scholz}
\end{itemize}

\section{18-12-10 \ \ {\it Greetings and Thanks!}\ \ \ (to F. De Martini)} \label{DeMartini3}

Thank you for the kind words.  We much enjoyed having you over.

I'm writing to you from a caf\'e in a small ski-resort town.  We came here to celebrate Katie's 9th birthday, and the girls are out on the slopes.  They have all the fun.  For myself, I am using the time to write an ``interview'' paper for a volume that Max Schlosshauer is putting together.  I'll surely send you a copy when it is complete.  One of the questions in it is, ``If you could choose one experiment, regardless of its current technical feasibility, to help answer a foundational question, which one would it be?''  I wonder what your answer would be.  At the moment, I have no idea how I'm going to answer it.

Please give my best regards to Fiorenza, and a very merry Christmas to both of you.

\section{19-12-10 \ \ {\it The Qubit Story}\ \ \ (to A. Stairs)} \label{Stairs4}

\bAllS
The usual lore seems to be that Ben Schumacher coined the term `qubit' in 1995. But the wonderfully nifty NGram tool from google suggests otherwise.
The vertical axis is percent of all words in books in English that Google indexed between 1975 and 2008. (You can also look for several words; separate by commas.)

So if not Ben S., who coined it?
\eAllS

It was definitely Ben.  Story of its origin below.  [See 28-04-04 note
  titled ``\myref{Wootters19}{Quantum History}'' to B. W. Schumacher and W. K. Wootters.]
But I believe it is also an allowed spelling for cubit.  That might
explain why Google shows the word in use before
1992/93.\footnote{\editornote Google Ngram results can be glitchy
  sometimes, thanks to OCR errors, books in their corpus tagged with
  the wrong dates and so forth.  Search for modern terms like
  ``Internet'' to reveal this.  If you do an Ngram search, you can
  click the links below the graph to see Google Books results from
  those years.  ``Qubit'' turns out to be a financial database
  management system from 1979, an Australian manufacturer of marine
  navigation systems and an advertiser's term for a shorter version of
  an infomercial.}

I believe Daniel Gottesman coined the word qudit.  And it would be fun to learn who was the first to write qunit.  I believe I coined the term rebit, for real-Hilbert-space qubits.  I wonder what statistics Google gives on that one!

\section{22-12-10 \ \ {\it Did I Do You Good?}\ \ \ (to R. W. {\Spekkens})} \label{Spekkens89}

\brws
Incidentally, the example of nonlocality without entanglement that has the
closest analogue in the toy theory is the one using three qubits due to Tal
Mor.
\erws

Yes, I knew that it had to be:  Your toy model has no qutrits.  I troubled over this a bit, and your remark causes me to feel (again) that maybe I'm being a bit too disingenuous in my presentation.  The reasons I opted for the qutrit example are 1) that it's closer to my heart---it was the example for which we made the name, 2) it was the only example we had a genuine proof for (but maybe that has changed in the intervening years, I didn't look), and 3) the qutrit example was genuinely more ugly, and I wanted to make a point of that.

But then I also wanted particularly to make the point, ``he can pull a little conceptual model from his pocket and gain quick insight into any number of technical questions in quantum theory, just by having started with the right conception of quantum states!  That is physical insight \ldots''

Part of me says, ``Ah, let them eat cake.''  The other part of me says I should add a sentence to soften the transition.

\brws
My only minor quibble is when you say: ``The toy theory is not quantum theory itself (nor does it
pretend to be a start for deriving the real theory).''  I guess I did think about it as a kind of start to deriving the real thing in the
sense that a proper axiomatization is likely to include an epistemic
constraint (and maybe the knowledge is even about a pre-existing reality!).
\erws

The thing that always struck me was how much power you got from a {\it simple\/} epistemic constraint.  That really did hit me over the head and is probably a significant part of the source of my own faith that, in the right language (SIC language for instance), the ultimate epistemic constraint (which in turn, for me, must arise from some new Bayesian rule for consistency between probability assignments) will itself be simply stated.  Of course, you know that I have my problems with the parenthetical part of your sentence.  Anyway I will think about how I might modify my sentence so that a little bit of ``the toy model being that much of a start'' is on display.

I'm listening to a great album of Ozzy Osbourne doing covers of all kinds of songs.  At the moment, the Beatles' ``In My Life''.  Kind of bizarre.

\section{22-12-10 \ \ {\it More Honesty}\ \ \ (to R. W. {\Spekkens})} \label{Spekkens90}

The attached draft should do your quibble some justice (I hope).  Also, it slips in a clause about the triple-qubit example so that my soul might rest more peacefully tonight.

\section{26-12-10 \ \ {\it Egg Nog}\ \ \ (to L. Hardy)} \label{Hardy43}

Thanks for the sympathetic note.  In the end, I'll get over it --- chalking it up to a learning experience.  When I get my head around what happened, I'll probably end up thinking the new understanding is something we ought to incorporate into our next physical theories!

\ldots\ semi-seriously.  Already I've become more keenly sensitive to something I've been repeating for quite a while, that ``the world is not sentence shaped.''  Take van Fraassen's example, ``Why did Adam eat the apple?''  What does it ask?  Depending upon how you read it and where you put the emphasis, it can be asking any number of things.  Two days ago, I said some nasty things about [Professor X], but if I were to present a transcript of the conversation we had to a court of law, they would be completely puzzled over why I felt slighted.  I've come to realize that most of the ``conversation'' wasn't conveyed through the ``transcript words'' but through the intonation, emphasis, and tone of speech.  Even [Professor X]'s eye movements and chuckle were part of it.  This is really interesting and makes the point very concrete for me, where it was mostly academic before.

\chapter{2011: The Perimeter of the Perimeter}

\section{06-01-11 \ \ {\it Quadits Inside SIC Simplexes} \ \ (to M. A. Graydon)} \label{Graydon14}

\bmag
I have been wondering about the shape of quaternionic quantum state spaces inside complex SIC probability simplexes. In case you are interested, I have attached some notes from the past few days on that subject. The calculations are routine, but the results could serve as food for thought. I must admit that I am finding the idea of dimensionality as an ontic property harder to swallow these days, and I have included some brief thoughts on that in the notes (I also quote one sentence of yours that I am at odds with at the moment.) Today, at least, I feel like there is a lot more to do before I would be happy saying that dimension is agent-independent.
\emag

Take a coin, and imagine flipping it.  We generally write down a (subjective) probability distribution over two outcomes to capture our degrees of belief of which way the flip will go.  But of course it is a subjective judgement that it can only go two ways.  Steven van Enk would say it could always land on its side; so he would always write down a probability distribution over three outcomes.

If one takes $(p_0, p_1)$ as a subjective assignment, the number 2 is objective with respect to it.  Something we imagine or hypothesize about the coin. If one takes the $(p_0, p_1, p_2)$ as a subjective assignment, then the number 3 is objective with respect to it---something we imagine or hypothesize about the coin, that it can fall three ways.  I.e., it will fall one of three ways regardless of what we believe of which of the three ways it will fall.

So objectivity/subjectivity comes in layers.  We call something objective, and then make probability assignments in the subjective layer above it.  But of course, the first ``calling something objective'' was a subjective judgment in itself.

What if I had said it like this:
\bq\noindent
Dimension is something [I hypothesize] a body [to hold] all by itself, regardless of what an agent thinks of it.
\eq
Much like I hypothesize that a coin can fall out two ways, while van Enk hypothesizes that it can fall out three ways.

\section{06-01-11 \ \ {\it Invitation to xQIT Conference May 3--4} \ \ (to S. Lloyd)} \label{Lloyd2}

Yes, I can come.  I haven't been to Cambridge in a few years, and it would be nice to see it again. You can give my talk the title:  ``Still Seeking Those Damned SICs''.

Please keep me up to date on where I should book a room, etc.

\section{07-01-11 \ \ {\it Paper for Itamar's Volume}\ \ \ (to W. G. {\Demopoulos})} \label{Demopoulos38}

\bwd
A quantum question: Would it be fair to say that instead of effects in my sense (whatever that may be) you prefer ``effects'' in the sense of changes in subjective probability assignments by your ``agents'' in response to their ``probing'' the world? Thus an effect is not any kind of property of anything, but a probability assignment or judgement.
\ewd

Congratulations on the reprieve!  I'm glad to hear it.  Very best of luck on all that's to follow.  Your friends' thoughts are with you, and my heart is definitely with you.

On your quantum question:  Yep, that's it.  I have a very old discussion of it somewhere in \arxiv{quant-ph/0205039}.  But the key point is this:  Take a POVM consisting of operators $E_k$.  I.e., these are positive semi-definite operators that sum to the identity operator.  Thus they might seem like mysterious abstract entities (perhaps properties of the system, or properties of the `measuring device', or maybe something still weirder, though surely agent-independent).  But, via a fiducial SIC POVM in the sky (as I talk about in my papers), one can map these operators to a set of conditional probabilities $p(i|k)$ in a one-to-one fashion.  That is, these operators contain nothing over and above the prescription of a probability assignment.  There is nothing to them mathematically but that---nothing is lost by thinking of them in these terms, as probability assignments.  Similarly, the philosophical point should be this:  That there is nothing conceptually more than this either.

So, what should one call the ``outcome'' of a quantum measurement?  Well, for the agent it is an experience ``$k$'' \ldots\ but what meaning does that have for him?  How could he articulate it?   What does it mean to him?  How will he change his behavior with regard to it?  The answer is:  It is $p(i|k)$.  It is a probability distribution conditioned on $k$.  The only operational handle the agent has on $k$ is through the assignment he makes, $p(i|k)$.

But all probabilities (for the personalist Bayesian) are subjectively given (i.e., functions of the agent only, his history and experiences, not the object).  And thus, I guess, my resistance to part of your way of thinking about effects.  (Of course, as you should know by now, I am in serious agreement with you in other ways \ldots\ in fact, probably with you more than with any other living philosopher.)

Attached is a little thing I've just finished for Schlosshauer's interview book.  I hope I'm not too harsh on your friends in it.  You'll see that I'm my usual belligerent self.  Still I hope it will give you a smile from time to time (I can always hope).

\section{08-01-11 \ \ {\it Strange Einstein Remark}\ \ \ (to R. W. {\Spekkens})} \label{Spekkens91}

Having finally watched the video of Ernst Specker that {\Adan} gave me, I wanted to learn a little bit more about this guy Gonseth that Specker mentions (and whose name I had known through a couple of remarks by Wolfgang Pauli).  Anyway, a Google search landed me on one of Mehra and Rechenberg's volumes of {\sl The Historical Development of Quantum Theory\/} (I can't quite figure out which one because of an inconsistency in Google Books, and I'm too lazy to figure it out).\footnote{\editornote J.\ Mehra and H.\ Rechenberg, {\sl The Historical Development of Quantum Theory,} volume 6, part~2 (Springer, 2001), p.\ 1198.}

Anyway, it's talking about the 1948 issue of Dialectica that Pauli edited and says:
\bq\noindent
Einstein, on the other hand, used the opportunity to repeat again his objections that had been familiar since the Einstein--Podolsky--Rosen paper (1935), recalling that `if in quantum mechanics we consider the $\psi$-function as (in principle) a complete description of a real physical situation we thereby imply the hypothesis of action-at-a-distance, a hypothesis which is actually hardly acceptable,' and `if, on the other hand, we consider the $\psi$-function as an incomplete description of a real situation, then it is hard to believe that, for this incomplete description, strict laws of temporal dependence hold' (Einstein, 1948, p.\ 323).
\eq

Of course, their commentary about ``familiar since EPR'' is not exactly right, but I'm struck by the second Einstein quote.  Surely Einstein was familiar with Liouvillean mechanics.  I wonder whether he was really trying to get at something else, or just made a mistake?  Have you ever run across Einstein saying something like this before?  Do you know the wider context?

Here's the link in case you have an interest in looking yourself:
\begin{center}
\noindent\myurl{http://books.google.com/books?id=9l61Dy9FBfYC\&pg=PA1198\&v=onepage\&q\&f=false}
\end{center}

Another thing that was kind of interesting was a little discussion on
Max Born a few pages later, talking about a 1955 paper of
his\footnote{\editornote M. Born, ``Continuity, determinism, and
  reality,'' in \emph{Festskrift til Niels Bohr [Commemorative Volume
      in Honour of Niels Bohr on the occasion of his 70th birthday].}
  Royal Danish Academy of Sciences and Letters, 1955. This paper is also of interest for the prehistory of chaos theory.  Born writes, for example,
\begin{quotation}
\noindent In astronomy, much work has been done to prove the stability of the planetary system.  For our purpose, the results of these investigations are irrelevant.  What matters is that there exist simple mechanical systems of a type familiar in atomic physics (kinetic theory of gases) for which all orbits are instable.  These systems display therefore only what I should call \emph{weak determinism}; the future state can be predicted only if the initial state is defined absolutely sharply, in the sense of the mathematical concept of a point in a continuum; the slightest initial deviation produces an ever increasing vagueness of the final state.
\end{quotation}
He does not, however, go so far as finding exponential divergence of trajectories characterized by Lyapunov exponents.} where (apparently) he first emphasized the similarity of quantum
mechanics to Liouvillean mechanics (and sounded very Bayesian as well,
I might add): \bq It is misleading to compare quantum mechanics with
the deterministically formulated classical mechanics; instead, one
should first reformulate the classical theory, even for a single
particle, in an indeterministic, statistical manner.  After that some
of the distinctions between the two theories disappear, [while] others
emerge with great clarity.  Amongst the first is the feature of
quantum mechanics, that each measurement interrupts the automatic flow
of events and introduces new initial conditions (so called ``reduction
of probability''); this is true just as well for a statistically
formulated classical theory.  \eq

\section{08-01-11 \ \ {\it Rampant Anti-Realism}\ \ \ (to R. W. {\Spekkens})} \label{Spekkens92}

I wonder what this guy would think of me?
\begin{center}
\myurl{http://www.sfu.ca/content/dam/sfu/philosophy/docs/bradley/anti_realism.pdf}.
\end{center}
Anyway some quotes from Max Born's correspondence with him in there.

\section{08-01-11 \ \ {\it Strange Einstein Remark, 2}\ \ \ (to R. W. {\Spekkens})} \label{Spekkens93}

\brws
I love the Born quote.  I hadn't realized he had ever made statements that so clearly espoused an epistemic view of the quantum state.
The view attributed to Einstein in the statement: ``it is hard to believe that, for this incomplete description, strict laws of temporal dependence hold'' is a bit puzzling.  Of course, it's hard to take something that isn't a direct quote too seriously.  Many of his contemporaries couldn't even understand what Einstein was getting at, so I wouldn't be surprised if they failed to summarize it adequately.  If Einstein did say something like this, it doesn't follow that he was endorsing indeterminism at the level of epistemic states for isolated systems.  Perhaps he just meant that discontinuous changes in the epistemic state as a result of measurement on a correlated system are natural.
\erws
No, I think that was a direct quote of Einstein.  (Well, a translation from the German.)\footnote{\editornote It comes from the summary (``Zusammenfassung'') of Einstein's 1948 Dialectica paper, ``Quanten-Mechanik und Wirklichkeit.''  The German version reads, ``Fasst man die $\Psi$-Funktion in der Quantenmechanik als eine (im Prinzip) \emph{vollst\"andige} Beschreibung eines realen Sachverhaltes auf, so ist die Hypothese einer schwer annehm-baren Fernwirkung impliziert. Fasst man die $\Psi$-Funktion aber als eine \emph{unvollst\"andige} Beschreibung eines realen Sachverhaltes auf, so ist es schwer zu glauben, dass f\"ur eine unvollst\"andige Beschreibung strenge Gesetze f\"ur die zeitliche Abh\"angigkeit gelten. --A.E.''  Mehra and Rechenberg quote the English translation of this summary which the journal provides just after the German original.}

\section{08-01-11 \ \ {\it Strange Einstein Remark, 3}\ \ \ (to R. W. {\Spekkens})} \label{Spekkens94}

\brws
Okay, then it is somewhat puzzling.  I take it you think that he may here be endorsing indeterminism of some variety?
\erws
No, not endorsing indeterminism.

The more I think about it, it is maybe this:  That he found a puzzling incongruity in being stuck with epistemic states for the systems themselves, but presumably ``complete'' information about their evolutions.  I.e., that we could still interpret unitary time evolution (Hamiltonians) in ontic-style terms.

I sure would like to read his real article.  How's your German?

\section{08-01-11 \ \ {\it Strange Einstein Remark, 4}\ \ \ (to R. W. {\Spekkens})} \label{Spekkens95}

\brws
Ah, the question of the status of unitaries \ldots\  I see.

Here is essentially the extent of my German:
Ein grosse bier, bitte.
You can get by pretty well in Germany knowing only that \ldots
\erws

Actually, it hadn't dawned on me until now that that might be the issue.  I was just genuinely perplexed that Einstein would say something like that.  At least if he knew about Liouvillean mechanics, that is.  But maybe he didn't.  I don't know.  I guess it also seems strange to me that Max Born didn't seem to understand the similarities between QM and LM until the mid-fifties.

\section{08-01-11 \ \ {\it Strange Einstein Remark, 5}\ \ \ (to R. W. {\Spekkens})} \label{Spekkens96}

\brws
Ah, the question of the status of unitaries \ldots\  I see.
\erws

That does seem to be an important paper of Einstein's that I'd like to get hold of (and get translated).  It also apparently contains this quote:
\bq
The following idea characterises the relative independence of objects far apart in space A and B: external influence on A has no direct influence on B; this is known as the Principle of Local Action, which is used consistently only in field theory. If this axiom were to be completely abolished, the idea of the existence of quasienclosed systems, and thereby the postulation of laws which can be checked empirically in the accepted sense, would become impossible.
\eq

\section{08-01-11 \ \ {\it Strange Einstein Remark, 6}\ \ \ (to R. W. {\Spekkens})} \label{Spekkens97}

And you even cite the paper in your paper with Harrigan.

\section{08-01-11 \ \ {\it Mehra and Rechenberg}\ \ \ (to R. W. {\Spekkens})} \label{Spekkens98}

I had heard before that one should be careful with the Mehra and Rechenberg books---that they are not always that accurate.  Well, I have just had my first case in point.

In the piece I pointed out to you this morning, they first report Pauli's commentary on the Einstein paper, in which Pauli says the following:
\bq
According to my opinion one cannot draw from the particular cases of correlated systems any new conclusions which are not already contained in the previously mentioned requirement in quantum mechanics of giving up the general predictability, of the results of individual observations on a single atomic system in a given state.  In view of both the empirical facts and the existence of the logically consistent quantum mechanical formalism it seems to me that only this renouncement enables us still to use in physics the concept ``closed system'' and the usual perception of space and time, which are so closely connected with each other.  It is in this sense that I consider the quantum mechanical description to be complete.
\eq
I've now read the original Pauli article and he was clearly talking about the issue of locality here.

On the other hand, Mehra and Rechenberg then go on to say:
\bq\noindent
A more explicit description of `closed systems' was provided by Heisenberg (1948) in the same issue of the {\sl Dialectica}.
\eq

And I've now read the Heisenberg article too.  (It was reprinted and translated in his book {\sl Across the Frontiers}, on my shelf.)  That article is titled ``The Notion of a `Closed Theory' in Modern Science'' and had nothing whatsoever to say about ``closed [atomic] systems'' in the Pauli and Einstein articles.  It was instead about how theories don't actually reduce to one another---how for instance classical mechanics and quantum mechanics live side-by-side, both distinct, both valid.  (Which is a position somewhat like QBism's present one.  Makes me wonder whether I ever actually say anything new.  I must have read this article 25 years ago.)

So!  Be careful with those guys.

\section{09-01-11 \ \ {\it Sunday Letter}\ \ \ (to W. G. {\Demopoulos})} \label{Demopoulos39}

Thanks for the new draft  I like the increased discussion on the timeless properties; I have much agreement with that, you know.  You have a little typo at the end of your footnote 6; it ends with a ``,''.

I didn't see that you strengthened your defense against my charge of a kind of dualism.  Maybe I'm still missing something.  When you write, ``Such systems are epistemically accessible to an extent that systems which are characterized only in terms of their eternal properties and their effects are not,'' I would think that you're admitting that there are (at least) two distinct kinds of systems in the world:  Those that have a certain type of epistemic accessibility and those that do not, and that that epistemic accessibility is not characterized by things to do with the observer's situation, but rather the with system's itself.

Another typo on page 17, by the way, ``does not precludes''.

And you don't weaken my accusation any either by writing on page 18:  ``This kind of conceptual dependence, does not preclude the application of quantum mechanics to systems that record effects. Although it is largely a matter of convenience which systems are, and which are not, taken to record effects, it is not wholly a matter of convenience.''  The phrase `not wholly a matter of convenience' again points me to an ontic distinction between two types of system.  Again, a reason for my saying, ``dualism''.  Similarly with respect to your sentence, ``This leaves entirely open the empirical question of why it is that some systems appear to be amenable to descriptions that are expressible in the framework of classical mechanics.''  I.e., you don't explain why, but there is an empirical distinction between two kinds of system.

I'm not saying there's anything necessarily philosophically wrong with a dualism; mostly it is that it just doesn't match my taste, and doesn't feel like the right direction for moving physics forward.  It adds a bigger burden on the physicist than I'd rather him have:  For now, for each lump of matter, he'll have to---in some mysterious way---come to a conclusion of whether it supports dynamical properties or not.  How does he do that?  Where you leave us is where Bohr left us---as far as I can tell, in just telling us ``it must be so, but I'm not going to tell you where/how to make that distinction'' \ldots\ i.e., that there must be two kinds of system for us to build our evidentiary base for the quantumly treatable ones at all.

Imagine my going up to Rainer Blatt and saying, ``Rainer, I just found this nice rock on the beach.  You think you might be able to use it as a component in that quantum computer you want to build?''  Asher Peres would say, ``Of course he can use it as a component; it is just a matter of money.  With enough money, any old rock can be polished into a quantum computer.''  But if it ain't so, then it must be a burden on physics to say when it can and when it cannot be done.  My guess is that Rainer will never be able to codify and make explicit such a criterion of distinction; he'll never be able to tell me which tests he must perform to certify my rock ineligible for quantum computation.

On the question of Rovelli, I don't know what to say.  I leave it to you to come to your own conclusions of whether there's any (strained) similarity between your ``functionalism'' and his ``relationalism''.  I just like to flounder with all kinds of ideas; every now and then one emerges as providing some insight.

On Israel, I go between Feb 13 and 19.

Oh, how's your French?  I was reading through the 1948 issue of Dialectica edited by Pauli yesterday, and came across this description of his of the Gonseth paper in it:
\bq
Many of the articles mention possible applications of the idea of complementarity outside physics, as for instance to questions connected with biology or psychology. I shall not discuss these questions in this introductory survey but wish to draw the reader's attention to the interesting attempts of Gonseth to formulate the idea of complementarity so generally that no explicit reference is made anymore to physics in proper sense. This is, of course, only possible by use of a language to which the physicists are not accustomed, which uses expressions like ``horizons of reality'', ``profound horizon'' and ``apparent horizon'', ``events of a certain horizon''. The word ``phenomenon'', however, is used in this article strictly in the above mentioned sense given to it by Bohr. To the ``profound horizon'' of Gonseth belong the symbolic objects to which conventional attributes can not be assigned in an unambiguous way, while the ``traces'' of Gonseth are identical with the ``phenomena'' in our sense.  I wish again to stress here the circumstance that the free choice of the observer can produce either the one or the other of two ``traces'' and that every phenomenon or ``trace'' is accompanied by an unpredictable and irreversible change in the ``profound horizon''.
\eq
I was struck by the use of the word ``trace'' since I know that you also use it.  Maybe that's where the similarity stops.  Still, I attach the Gonseth paper in case you care to find out.

One last thing with regard to my nasty accusations of dualism, as you get a chance, I'd like to know what you think of Heisenberg's article in that same issue of Dialectica.  It was reprinted and translated as ``The Notion of a `Closed Theory' in Modern Science'' in his book {\sl Across the Frontiers}.

\section{09-01-11 \ \ {\it The Gonseth Connection}\ \ \ (to W. G. {\Demopoulos})} \label{Demopoulos40}

Specker credits a seminar of Gonseth's (attended by Pauli) for his interest in quantum mechanics.  I have a little video of him telling a bit of the story to someone last year or so that I was listening to yesterday morning.  Below is the conversation it spurred on other matters.  If your German is better than your French you might be interested in the attached Einstein paper.  You might be interested in the part about Einstein's general methodological considerations for requiring locality (a bit reported below).

\section{09-01-11 \ \ {\it The Gonseth Connection, 2}\ \ \ (to W. G. {\Demopoulos})} \label{Demopoulos41}

\bwd
Isn't this paper translated and re-printed in the Born--Einstein Letters?
\ewd
Is it????  Certainly he's given variations of the locality argument all over the place---in papers and correspondence.  What intrigued me most was the sentence I was curious about on time evolutions---I had never noticed him saying something like this before.

\section{09-01-11 \ \ {\it Mine! Yours?}\ \ \ (to N. D. {\Mermin})} \label{Mermin194}

``I'll show you mine, if you'll show me yours!''  Attached is the interview I finally sent off to Max last Wednesday.  I think I wrote you in November, ``It's just 17 questions for god's sake.''  Ha!  That'll teach me.  At least it sounds like Zeilinger was every bit as late as I was.  So, if you throw darts in your mind, don't just throw them at me. [\ldots]

But enough of these bad things.  Now that I've got the nerve to read it, I hope you'll send me your own interview for Max's book.  It'd be a pleasure to read it, and it'd make my Sunday a brighter, happier one.  My three girls are out skiing now, and as you can see, I'm catching up on long waiting correspondence etc.  Actually, I should be writing the cover story for the GQI newsletter, The Quantum Times, bragging about the March Meeting \ldots\ but that's only three days overdue!  So my cycles repeat \ldots

\section{09-01-11 \ \ {\it Mine! Yours?, 2} \ \ (to N. D. {\Mermin})} \label{Mermin195}

Reading through your answers slowly, as Kiki and the kids tell me of this week's ski adventure.  (There is a new one every week.)  What is this, ``{\Adan} Cabello's demonstration that, whatever the sense in which correlations have physical reality, it cannot be that their values are EPR `elements of reality'.''  What paper does that refer to?\footnote{\editornote A.\ Cabello, ``Quantum correlations are not local elements of reality,'' \emph{Physical Review A} {\bf 59,} 3 (1999), pp.~113--15, \arxiv{quant-ph/9812088}.}


\bdm
I look forward to the day when some clear-headed gifted writer has
spelled it out so lucidly that everybody is completely convinced that
there is no such problem.
\edm
There go my credentials \ldots


\bdm
This adds a word to Asher's famous title: ``Unperformed experiments
have no conceivable results.'' That addition makes his point just a
little harder to swallow.

But swallowing becomes easier again if I expand Asher's title further
to ``Many different sets of unperformed experiments have no conceivable
sets of results with the result for each local test being the same for
every member of the set of results in which that particular local test
appears.'' (The expanded title, however, is itself harder to swallow.)
\edm
I loved that!!!


\bdm
My intuitions about the nature of quantum mechanics are not coherent
enough to add up to anything I would dignify with the term
``interpretation''. Admittedly, shortly after turning 60 I did write a
few papers setting out what I called the ``Ithaca Interpretation'' (see
also my answer to Q6). But I was young then, innocent, and overly
willing to sacrifice an accurate phrase for an entertaining one.
\edm
Loved this one too!!!  Compare to my, ``Switch sides to what?''


\bdm
It remains entirely possible that some wise, imaginative, and readable
person may in the future lure me away from the position I am trying to
sketch in my answers to these questions.
\edm
There go my credentials again \ldots


\bdm
What I don't like about consistent histories is the
reluctance---verging on refusal---of many of its practitioners to
acknowledge the utterly radical nature of what they are proposing. The
relativity of time was a pretty big pill to swallow, but the
relativity of reality itself is to the relativity of time as an
elephant is to a gnat.
\edm
Three cheers!


\bdm
I would nominate for the most important recent development the
application of quantum mechanics to the processing of information,
starting with the invention of quantum cryptography by Bennett and
Brassard in 1984, and continuing with the development of quantum
computation. As runner-up, I would cite the study of pre- and
post-selected ensembles by Aharanov and his collaborators, and
(perhaps---I still lack a good feeling for it) the ensuing notion of
weak measurement. In third place I would put the consistent histories point of view, as
put forth by Bob Griffiths.
\edm
\ldots\ There go my credentials for the final time of the evening!!!

Good night!!!

\section{09-01-11 \ \ {\it One Endorsement}\ \ \ (to M. Schlosshauer)} \label{Schlosshauer41}

The usual:  I should thank you for everything you've done for me.  In the present case, forcing me to finish the interview for you.  There was a time when I genuinely wanted you to let me off the hook, but you wouldn't do it.  If {\Appleby} is genuine below, at least he got something from my answers---and maybe that ought to be enough for me and my spirits.  But if {\Appleby} is right to the point of there being one really clever, really technically-capable student inspired to develop QBism far beyond what we old guys have been capable of, then I should thank your persistence ever that much more.

\subsection{Marcus's Preply}

\bq
I think your interview is great.   Really, really good.  You sound depressed about it.  But I think this is an example of an author not necessarily being the best judge of his own work.  I think it could have a big impact.  I don't even know who the other contributors are, let alone what  (exactly) they will say.  But I would be willing to bet a lot of money that someone uncommitted---a student say---reading this volume will be deeply impressed by your contribution, and will promptly forget all the others.  And it is the uncommitted, above all the students, whom we most want to catch.  I think this must be the first time that qbism has been set out in this way, side by side with all the other presently (but not for much longer) more fashionable approaches.  At any rate, if it has been done before, it can't have been done often.  There is only one possible winner from such a comparison.   It is going to be just as it was with the course at the University of Waterloo, where the students were asked to vote on their favorite interpretation.  [See 02-04-10 note ``\myref{Emerson7}{Congratulations!}'' to J. Emerson.]  It is not simply the force of the argument (important though that is).  It is a deep down intuitive thing.   You are just so much more alive than everyone else.  People (people with a bit of youth in their soul, that is) feel that:  can sense at a level below words the difference between the fresh, new-baked bread that you are giving them, and the stale old crusts which is all that anyone else has got to offer.

Actually, I found it thought-provoking myself.  This may sound surprising, as I am obviously familiar with the general argument.  Even so, it didn't read like yet another reiteration of the same old thing.  The trouble with familiarity is that one develops intellectual habits.  One's mind starts to run along rails.  But reading your interview I found my mind kept jumping the tracks---following a train of ideas different from the one always suggested before.   I am not sure why this was exactly.  I think it was partly because the writing really is fresh---it isn't simply a cut and paste job.  It is also because I happen to have been reading Schweber's book on the history of QED.  Not that Schweber's book is all that good.  But there are a few things in it which somehow hooked up with your interview and caused me to look at things from a different angle.

It is what Schweber says about Feynman which is relevant.   It started with what you say about telling a story.  Reading that I thought that telling a story is exactly what Feynman wanted to do.  Nowadays the  robots  all think of Feynman diagrams as a mere calculational device.  And perhaps that is how Feynman himself came to think of them in the end.   But it is not how he thought of them in the beginning.  In the beginning he was trying to tell a story.  I  think that is the main reason everyone loves him---why he stands out from all his contemporaries.  He was the only one trying to do that (or nearly the only one---I guess Wheeler should also be included  as it was he, not Feynman, who originally came up with the picture of electrons zig-zagging backwards and forwards in time).  And this  got me to wondering about the differences between the story Feynman told, and the one we want to tell.  Which in turn led to me to thoughts about space-time.  On the face of it Feynman was thinking in very blockist terms.  Positrons moving back in time, advanced and retarded potentials---it is hard to see how such ideas could be meaningful if one was not thinking of the future being somehow given, along with the past, all in one go, as one great block.  On the other hand, it is obvious that Feynman didn't take the idea entirely seriously.  He didn't think it was the literal truth about things.  Only that it was, for certain limited purposes, a useful way to think.  So I don't think that Feynman was really a blockist.  I digress, however.  Although Feynman didn't think of his pictures as the literal truth about things, they clearly do, so far as they go, take the space-time manifold for granted.  And that got me to thinking:  what does a qbist say about space-time?  Well, we have asked ourselves that question  before of course.  But on this occasion I suddenly had the sense of new vistas opening up.

I am still feeling very confused, and I am certainly not up to writing out a coherent argument.  Really I want to talk about this, not write about it.  Perhaps we can do that when I get back to Waterloo?   But I think I would like to jot down a few thoughts while they are fresh.  Not a connected piece of Jamesian prose.  Just a few markers, which may serve as the jumping off point for future discussion.

The basic principle of qbism (so far):  QUANTUM STATES DO NOT EXIST.   I think we should supplement that with the principle:  SPACE-TIME DOES NOT EXIST.  In particular THE PAST DOES NOT EXIST.  THE PRESENT DOES NOT EXIST.  THE FUTURE DOES NOT EXIST.

Perhaps this sounds rather wild to you?  But I think we are strongly pushed to it.   Obviously, if one believes in free will one can't believe the future exists.  If one believes in relativity one can't believe the present exists (there is no preferred space-like slice).  And once one has gone that far isn't it strongly suggested that one should go the rest of the way, and also deny that the past exists?

I think we have discussed this idea before.  But for me, this time, it is different.  Before I somehow felt blocked from pursuing this train of thought.  But now I have a sense of inner conviction.

What was blocking me?  Two things, I think.

The first was Minkowski's suggestion that relativity forces us to think of space-time as a single entity.\footnote{\editornote For Mermin's post-conversion-to-QBism thoughts on this, see ``QBism as CBism: Solving the Problem of `the Now',\,'' \arxiv[quant-ph]{1312.7825}.  For an argument that \emph{general} relativity can be made less blocky than special relativity, see G.\ F.\ R.\ Ellis and R.\ Goswami, ``Spacetime and the passage of time,'' chapter 13 in the {\sl Springer Handbook of Spacetime} edited by A.\ Ashtekar and V.\ Petkov (Springer, 2014), \arxiv[gr-qc]{1208.2611}.}  I have somehow been brain-washed by that into thinking that relativity naturally suggests a blockist conception.  But I don't think it does.  For sure relativity obliges us to abandon the Newtonian conception, of space as a physically existent thing.  But once you have taken that step there are two possible directions in which to go.  One possibility is to take the Minkowski route, and to think of space-time as a physically existent thing.  The other is to reject the idea that either space or time are physically existent things.

The second was the phrase ``frame of reference''.  The word ``frame'' suggests something physical.  Indeed Einstein himself strongly encouraged that, with his talk of measuring rods and clocks.  But now a different characterization occurs to me.  Hartle (it was Hartle wasn't it?)\ characterized the quantum state as a ``catalogue of probabilities''.   I think space-time is similar to that.  Except it is something still more tenuous---something even less like a real existent thing.    Space-time isn't the catalogue itself, only the cataloguing system.  Something like the Dewey system used in libraries.    Not the actual cards.  Only the abstract principle used to organize the cards.

For instance, we have the actual book {\it Galois Theory\/} by Harold M. Edwards.   Then we have the card which has printed on it ``Harold M. Edwards, {\sl Galois Theory\/} [various additional bibliographical information] MATHS E30 EDW''.  (I am looking at the copy I took out from University College London, who it seems don't use the Dewey system---but never mind that).  Now consider the actual event, of Obama being elected president.  This actual event is analogous to the actual book.  Corresponding to the catalogue card we have the proposition ``Obama was elected president in 2008''.  The proposition describing the event is as different from the event itself as the  catalogue card is from the book itself.  And finally, corresponding to the catalogue mark MATHS E30 EDW we have the temporal marker 2008.

Temporal markers are really useful.  The temporal marker 2020 will cause me to think:  well, this hasn't happened yet and it is incumbent on me, either to try to ensure that it really does happen (if I think it desirable), or to ensure that it doesn't really happen (if I think it undesirable).  By contrast the temporal marker 2008 will cause me to think ``well, like it or loathe it, there is nothing I can do about it''.    Like probabilities, temporal markers play a crucial role in organizing our thoughts, and deciding on our actions.  But, also like probabilities, that doesn't mean they are actually existent parts of the world.  Similarly with the concepts of past and future.  Being in the past isn't a property of the event.  Just a feature of the way I think about the event.

The single most important  message of quantum mechanics (as I see it) is that we need to introduce some distance between our concepts and the things they are concepts of.  We are used to the idea that our usual idea that everything has a position doesn't apply to electrons.  I think maybe we need to go much farther than that.  It isn't only microscopic things like electrons which slip through our conceptual net.  It is also macroscopic things, such as Obama (or me, or you).  Obama in his own nature is neither  past nor future.

In a way this is not so different from the standard Minkowskian view, according to which the concepts of past and future have no absolute significance.  On that view one can talk of event  A being past relative to event B, but it makes no sense to talk of an event being past absolutely and intrinsically, in its own nature.  However, Minkowski then introduces the concept of space-time as an absolutely existent entity.  And the associated, truly horrible idea that everything has already happened.

I was intending to go on.  There is certainly quite a lot more in my mind.   But I think I won't because, although it is in my mind, it is very confused.  All I set out to do here is to put down enough on paper to start a conversation when we next meet (I will be back in Waterloo a week today).  And I think I have achieved that already.

But there is one other thing I would like to get down before I forget.  It concerns the Pauli--Jung idea of synchronicity.  I was never very impressed by this.  But now I am suddenly wondering if there might not be something in it.  Consider Feynman's definition of science as ``a method for, and a body of information obtained by, trying to answer only questions which can be put in the form:  If I do this, what will happen?'' (Schweber, pp.~462--3).    I rather like that definition (I like it because it is so strongly agent-relative:  what will happen if I  do this?).  I also think it captures the essence of the concept of causation.  Causation is all to do with control.  For instance, turning the ignition key  causes the car to start (causal thinking is mechanical thinking, as is acknowledged by such phrases as ``mechanical materialism'', or ``the clockwork universe'').  I think this relates to non-locality.  In your discussion of non-locality you focus on ``unperformed experiments have no results''.  Suppose, however, that one has performed the experiments.  One has this great long sequence of measurement outcomes which display certain striking correlations.  This often leads people to ask ``what is the explanation?  How could those correlations have come about?''  And they feel compelled to postulate the existence of  some superluminal causal influence.  It seems to me, however, that it isn't a causal influence by definition, because there is no possibility of control.  Nevertheless one is still left with the existence of those undoubted correlations.  What would be a good name for them?  Well, how about ``synchronicity''?  And is it possible that it was something like this that Pauli actually had in mind?
\eq

\section{10-01-11 \ \ {\it Freedom $=$ ``Each with a Fire of Its Own''} \ \ (to W. K. Wootters)} \label{Wootters26}

I never came back to answering your wonderful note below.  I know I still owe you and need to work my way through the exercise of answering your query 3---it would be very good for me to do that; I know I'll learn from it.  But in the meantime, I wonder if the attached thing I just wrote up for Max Schlosshauer's interview book might lay some groundwork and context for how I should eventually answer.  I think I get close to your question in my answer to Max's question 5.  Unfortunately, that too relies on some of the groundwork in the pages previous to it.  But if you can stomach reading the first 10 pages or so, as you get a chance, please let me if you think this might be working us toward an answer to your question.

It must be the beginning of the semester for you, and I know you're awfully busy.  Best of luck for the new year.

\subsection{Bill's Preply, ``Nate Stories''}

\bq
Greetings from Kigali!  Things here are going well, though I confess we haven't yet tried banana beer.

Thanks for sending me your paper.  I read it immediately but it has taken a while for it to sink in, and I know for a fact that it still hasn't fully sunken in!  You have some really good arguments here, and you do a particularly good job of fending off likely misconceptions.

I have a few comments for now, maybe more later.

1.  It's cool that you were able to use the story about me and Nate and the flower.

2.  A trivial comment regarding the end of Section VI:  I'm not sure I can think of a freshman physics problem that treats the earth's gravitational mass as infinite.  Sometimes we treat the earth's {\it inertial mass\/} as infinite.  That is, we say that the earth doesn't move when a force acts on it.  We do this even though we also say, in the same problem, that the earth exerts a finite gravitational force on, say, an orbiting satellite.

3.  I'm struggling with this important sentence: ``Sometimes one will have no strong beliefs for what will result from the creation (as with the measurement of $H$), and sometimes one will have very strong beliefs (as with the subsequent measurement of $H^{\rm T}\,$), but a free creation of nature it remains.''  I completely agree with the claim that every quantum measurement creates a new fact that was not there before.  It's the word ``free'' that I'm struggling with.  Let me begin with this question: Is there any possibility of quantifying nature's freedom?  Of course an agent can associate an entropy with the outcome of a measurement he is about to perform, and thereby get a personal estimate of nature's freedom for that measurement.  For example, the agent might assign entropy zero to a certain measurement, as in the scenario you imagine.  In that case the agent will say he believes that nature is not free to choose the outcome of that particular measurement.  But you, who are acting not as an agent but rather as a theorizer of agents, say that nature is free even in that circumstance.  Is there any way for you, acting in this capacity, to quantify the freedom of nature?  Or is nature always simply free---there is no ``more free'' or ``less free''?  I think the QBist answer must be that there is no way to objectively quantify nature's freedom in choosing a particular outcome of a measurement.  To do so would suggest a propensity interpretation of probability.  So nature is simply free to choose whatever measurement outcome it wants.

But even in QBism, surely nature is not totally free.  There is order in the world; it is not total chaos.  How does one express the constraints on nature's freedom?  It can't be directly through probabilities, since probabilities belong to the agent, not to nature.  So what does one say?
\eq

\section{10-01-11 \ \ {\it Q3, Part 1} \ \ (to N. D. {\Mermin})} \label{Mermin196}

I quite enjoyed reading your interview last night, despite all the noise around me.  But I would guess you woke up to already find that this morning!

Here's the first of my promised answers:
\bdm
In Q3, Bob Griffiths and Niu {\bf did} discover their spectacular
simplification of the Shor algorithm (replacement of the
2-Qbit gates with 1-Qbit gates in the quantum Fourier
transform) by viewing it from the Consistent Histories
point of view, but that's the only example I know of. I
note that it did not convert you.
\edm
This is true, and you have brought this up with me before.  For instance at the PIAF meeting panel discussion a couple of years ago.

But let me tell you another Griffiths story.  I am quite sure I have written this one up before, but I have not been able to find it in my email system, so I guess there is nothing to do but write it again.

In 1996, Bob and I were office mates at the ITP in Santa Barbara, and we ended up collaborating on a problem that Asher and I had started up.  Something on optimal eavesdropping in BB84.  You can read about it in these two papers: \arxiv{quant-ph/9701039} and \arxiv{quant-ph/9702015}.
You'll note that ``part 1'' of the paper has Fuchs, Gisin, Griffiths, Niu (a student of Bob's), and Peres as authors, while ``part 2'' has just Griffiths and Niu.  The reason the paper bifurcated into two parts was because after all the first calculations, Niu had found a quantum circuit for enacting the eavesdropping strategy, and Bob said that the only way to {\it understand\/} its operation and {\it motivate\/} it was through the consistent history picture.  Well you can imagine that Peres and Gisin would have none of that and didn't want any such thing mentioned in the paper.  Bob was adamant that it needed to be in the explanation.  So we decided on an amicable split.

Anyway, a couple years later Niu came by Caltech and gave me a visit.  As it so happens, I was in correspondence with Bob at the time about CH, and I remembered back to our old collaboration.  So I asked Chi-Sheng about how exactly consistent histories made an impact on his thinking when he discovered the circuit.  He revealed to me that he hadn't used consistent histories at all when finding it.  He ``just played around'' until he got the right unitary.  I got a great laugh out of that one!

Let me tell you another story.  I once did something rather sneaky for a talk I had to give at a ``50 years of Everett'' meeting.  I sent out a questionnaire to various people asking if Everett imagery helped them make their discovery (whatever it was) in quantum computing.  For instance, I asked Shor about his algorithm, Simon about his own, and the pi{\`e}ce de r{\'e}sistance was asking Deutsch and Jozsa individually about the Deutsch--Jozsa algorithm (I had a suspicion about the answers I would get).  Their replies are attached.  Be sure to read them in the order:  Deutsch, Jozsa, Shor, Simon.  It's fun.

[See 20-09-07 note ``\myref{DeutschD1}{Easy Questions}'' to D. Deutsch, 20-09-07 note ``\myref{Jozsa7}{Easy Questions}'' to R. Jozsa, 20-09-07 note ``\myref{Shor4}{Easy Questions}'' to P. W. Shor, and 20-09-07 note ``\myref{SimonD1}{Easy Questions}'' to D. R. Simon.]

\section{10-01-11 \ \ {\it Q3, Part 2} \ \ (to N. D. {\Mermin})} \label{Mermin197}

Continuing,
\bdm
Also in Q3, it wasn't clear to me (as it is clear in the
case of Griffiths) how the epistemic leads directly to
the answer. Is it just that {\Spekkens} has analogous simple
examples?
\edm

Yes, you're on to me.  What really happened was that {\Spekkens} noticed the three-qubit example we had in the paper and thought, ``I bet I can find the same phenomenon in the toy model.''  Sure enough, it was so.  I thought for a while about being less ingenuous in my presentation, but our three-qubit states didn't have the same spark of {\it ugliness\/} about them that the original 9 two-qutrit states did.  I wanted the ugly factor to be there.

A better story might have been the time I thought the SIC states were the only sets of states in Hilbert space to have a certain property.  Rob said, ``Really?''  Then he showed me a set of states in his toy model that had the same property.  A much closer to the process I advertised in the interview:  He just thought for a moment, and then exhibited the set.  Sure enough, once I checked on my own in regular quantum mechanics, I found that I had been wrong.

I should ask Rob if he has an analogous example to our 9 two-qutrit states in one of his later toy models.

\section{10-01-11 \ \ {\it Born--Einstein}\ \ \ (to R. W. {\Spekkens})} \label{Spekkens99}

Bill Demopoulos thinks that the Einstein paper we were talking about a couple of days ago was translated and reprinted in the Born--Einstein letters.\footnote{\editornote See {\sl The Born--Einstein Letters,} edited by M.\ Born and translated by I.\ Born (Macmillan, 1971), pp.\ 168--73.  The book can be read online at \myurl{https://archive.org/details/TheBornEinsteinLetters}.  Unfortunately, the translation does not include the summary, from which Mehra and Rechenberg take the statement under discussion here!}  You don't have that book, do you?  I was kind of shocked that I don't have it on my shelf.  I believe I've read it twice in the past.

\section{11-01-11 \ \ {\it Mr.\ Physics, professional sidekick}\ \ \ (to R. W. {\Spekkens})} \label{Spekkens100}

I realized I didn't explain my story so well yesterday.  As Katie was announcing SuperMom, SuperOma, and SuperOpa, I was in the process of pouring their wine.  I suppose that's how I became Wine Guy.  But as I say, the next morning she relayed to me that she had thought it over a bit, and that I should be called Mr.\ Physics.

This morning at breakfast, she announced that there was nothing wrong in being Mr.\ Physics --- it really is a important role.  ``Mr.\ Physics is SuperMom's sidekick!''

Maybe it's the best I could have hoped for anyway.

\section{11-01-11 \ \ {\it Jaynes}\ \ \ (to R. W. {\Spekkens})} \label{Spekkens101}

\begin{flushright}
\baselineskip=3pt
\parbox{3.9in}{
\bq
\noindent
But our present quantum mechanical formalism is not purely epistemological; it is a peculiar mixture describing in part realities of Nature, in part incomplete human information about Nature --- all scrambled up by Heisenberg and Bohr into an omelette that nobody has seen how to unscramble.  Yet we think that the unscrambling is a prerequisite for any further advance in basic physical theory.  For, if we cannot separate the subjective and objective aspects of the formalism, we cannot know what we are talking about; it is just that simple.
\\
\hspace*{\fill} --- E. T. Jaynes
\eq
}
\end{flushright}
\brws
Incidentally, I presented this slide to the PSI students yesterday.  It reminded me of a comment that Graeme Mitchison once made about the same picture of E. T. Jaynes (which was hanging in my office in Cambridge).  ``That's a haircut that you can set your watch to!''
\erws
And {\Carl} {\Caves} tells me that his politics was just like his hair.

\section{11-01-11 \ \ {\it For January Issue, Part 1}\ \ \ (to I. T. Durham)} \label{Durham2}

Here's the first piece, along with the relevant attachment.  How does it look to you?

\bq
\noindent {\bf March Meeting, Take Note of Quantum Information!}\medskip

Put on your cowboy hats, it's time to go to Dallas!  This year we will have an impressive line-up of quantum information talks for the APS March Meeting in Dallas, Texas, and you won't want to miss it \ldots\ or at least you shouldn't!

This will be the biggest year yet for quantum information and computing at an APS March Meeting.  There will be over 400 talks on the subject, covering nearly every aspect of the field you can imagine---superconducting qubits, semiconducting qubits, optical qubits, ion traps, entanglement, coherent control, decoherence, quantum error correction, and much, much more.  (For gosh sakes, there's even 60 talks on the foundations of quantum mechanics!)  On the back of this page, you can find a full list of GQI invited speakers and topics, as well as a list of speakers and topics from other divisions that are relevant to GQI concerns.  The full schedule of GQI-sponsored talks can be found here:
\myurl{http://meetings.aps.org/Meeting/MAR11/sessionindex2?SponsorID=GQI}.

Two of the invited sessions particularly stand out.  ``20 Years of Quantum Information in Physical Review Letters'' on the Wednesday will feature five retrospective talks from five of the field's founding fathers.  Want to see the man who invented the word qubit (and lived to watch it become an entry in the Scrabble Players' Dictionary)?  He'll be there!  We're hoping this session will be a big draw for the whole APS and serve to entertain and educate them about our growing field.  Before that though, on the Tuesday, there'll be a session titled ``Quantum Information:\ Featured Experiments''.  Who in the general (non-GQI) APS would have thought that the world's presently most accurate clocks are directly reliant on quantum-computing techniques?  We hope it'll be a lesson.  Quantum information is not only crucial for the future of computing; it is forcefully relevant for weights and measures today!

Beyond the talks, GQI will also sponsor a tutorial ``Quantum Simulation and Computing with Atoms'' (given by Ivan Deutsch), two ``Graduate Student Lunches with the Experts'' (hosted by Benjamin Schumacher and Anton Zeilinger), and a ``pizza and beverage-of-your-choice reception'' before the business meeting.  (Yes, we'll have beer; it's Texas.)

At the risk of sounding like a Slap Chop infomercial, ``But wait, there's more!''  The March Meeting in general this year is just going to be an exceptionally good one.  For instance, there'll also be a celebration of the 100th year anniversary of superconductivity, with a Nobel laureate session associated with it.  Speakers include Ivar Giaever, Wolfgang Ketterle, Sir Anthony Leggett, K. Alexander Mueller, and Frank Wilczek.  Furthermore, one of this year's Nobel prize winners for the discovery of graphene, Konstantin Novoselov, will give a special talk.

There are all kinds of reasons to come.  We hope that nearly every member of the GQI will be there.  We have a real chance to show our relevance and importance to the APS this year.  Going to the March Meeting is, of course, about learning and communicating physics, but it is also about building community and friendships and establishing a base for career paths in our field.  Young people need jobs, and one way to see to it that there will be some jobs out there is to get the world physics community to take note of quantum information.

So, please tell your students, tell your teachers and friends, that important things will happen in Dallas this year.   Here's the website to go to for all the details on the meeting:  \myurl{http://www.aps.org/meetings/march/}.   Don't forget, January 21 is the deadline for early registration---thereafter registration fees go up.  Get everyone you can to come.   Texas (like Hilbert space) is a big place and always accommodating!
\eq

\section{11-01-11 \ \ {\it For January Issue, Part 2}\ \ \ (to I. T. Durham)} \label{Durham3}

And how does this sound?

And probably no need to use it, but I've asked Charlie Bennett permission to use this photo in case we might want to.  (But no reply from him yet; so maybe best not to plan around it.)  The photo could be called, ``The World of Quantum Information, When it Could Fit in One Photo Frame''.  This was taken just after Shor's factoring algorithm came out.

I really apologize again for delaying you so long.  It'll surely be great to get this issue out so that some might act before the early registration deadline is up (Jan 21).

\pagebreak

\bq
\noindent {\bf Letter from the Incoming Chair}\medskip

It's hard to compete with last year's letter from last year's chair, Dave ``Pledge Drive'' Bacon---I won't even try.  But pledge drive is what it's all about!  So, I am glad he set the tone.

My guess is that this is going to be a key year for the GQI.  Presently our membership is just a bit below 1,100.  If we can get it up to 1,450 members and sustain that for two years, we can petition to become an APS division.  That might look like a lot, but I don't think it is really---that is why I say this could be a key year.  With our upcoming much-larger-than-previous turn-out at the APS March Meeting (our submissions grew by 40\% from last year!), I think we will be in a very good position to make it happen.  The March Meeting is our crystal.  We just need to give the APS the best showing of quantum information it's ever seen, and that is bound to build excitement and a desire in many to be part of the topical group.  I'll tell two friends, and they'll each tell two friends, and the same for each of you, and we can make this thing happen.

``APS Division of Quantum Information,'' doesn't that sound so sweet to the ear?  And doesn't it sound so respectable?  We would obtain for the first time actual representation within the APS---we'd be respected brothers and sisters to the Division of Condensed Matter Physics, the Division of AMO Physics, and all the others.  American physics would respond in part to our membership's desires and needs.

I will tell two (embarrassing) stories from my history in this field, from earlier times when I was trying to get a faculty position.  At the conclusion of one job colloquium, one of the professors asked me, ``This is nice for weekend physics, but what do you during the week?''  Of another interview, it was leaked to me that in the faculty discussion after my departure, one of the professors implored, ``Well, if he likes quantum mechanics so much, why doesn't he do anything with it!?''  But I was doing quantum information then, as I am doing quantum information now---the subject matter has not changed particularly.

So many of you must have experienced (or will soon experience) something similar in your own careers.  But we want that number to be less and less.  To the extent that such behavior in hiring committees has abated over the years, it has done so only because of the increased awareness and respect the rest of the physics community has come to for this subject we hold  so dear to our hearts.  We have made great progress in the 20 years since Physical Review Letters started publishing papers in quantum information---2011 is an anniversary for us---but there is so much further to go.

I look to 2011 to be a bend in our curve of growth within the APS.  So, my job for the coming year will be to ``tell two friends'' every chance I get about the beauty and promise of quantum information.  I hope all the GQI will do their best to do the same.

We have a world to change.  Let's do it!
\eq

\section{12-01-11 \ \ {\it Quick and Dirty (a conversation)}\ \ \ (to M. Schlosshauer)} \label{Schlosshauer42}

\bmaxs
What are you doing up so late?
\emaxs
Reading Hardy's interview at the very moment.

\bmaxs
It's a good one, isn't it?
\emaxs
Haven't gotten far enough yet to know.  I did jump to the little bit on QBism 'cause he had flagged it---the guy still gets it wrong.  Chairs are no more accountable by quantum theory---on QBism's account---than the particular outcome of a Stern--Gerlach measurement.  QBism's framework is empiricism; Hardy is still looking at it through reductionist glasses.

\bmaxs
Yeah, I know what you mean. I too was surprised by Hardy's insistence on ontology as a necessary part of quantum theory.
\emaxs
Not quite right.  ``Ontology is a necessary part of physics,'' he would say.  He's not sure that quantum theory itself has the tools to give a correct ontology.  This is because he thinks QT is an effective theory.  (Or should that be affective theory?  Told you I always get confused.)

\section{14-01-11 \ \ {\it Great Incendiary Fun}\ \ \ (to I. T. Durham)} \label{Durham4}

I'm sorry!  First, things didn't go nearly as smoothly as I had expected for our transition from Waterloo to here.  (Blue Mountain is in Ontario, on the coast of Lake Huron.)  Then, for some reason, I repeatedly would remember that I needed to take care of this for you, but then would forget again.  Here's the new by-line:
\bq\noindent
Christopher A. Fuchs is a Senior Researcher at the Perimeter Institute for Theoretical Physics in Waterloo, Canada and an Adjunct Professor at the University of Waterloo.  He is a winner of the 2010 International Quantum Communication Award and Chair of the American Physical Society Topical Group on Quantum Information.  His Erd\H{o}s number is 3, but so is his Einstein number.   Mostly, he is very proud that his academic great-great-great-great grandfather Franz Exner (through the lineage Carlton {\Caves}, Kip Thorne, John Wheeler, and up) believed, long before quantum mechanics was around, that our ultimate physics would be indeterministic.  Chris's Cambridge University Press book {\sl Coming of Age with Quantum Information\/} has just appeared at the bookstores, and he's been told, makes for some ``great incendiary fun.''
\eq

On the picture, I still haven't heard back from Charlie.  So, let's just drop it.  I'm quite sorry I've already delayed you this much.  If the notes are going to be effective at all for getting the populace to the polls before the Jan 21 early-reg deadline, the sooner we can get the issue appearing the better.

You're a saint for putting up with contributors like me!

Hope all the snow shoveling you had to do didn't break your back.

\section{18-01-11 \ \ {\it What Is the Scale at which Quantum Theory Applies?}\ \ \ (to R. W. {\Spekkens})} \label{Spekkens102}

\brws
So I spoke to a public audience about quantum theory on the weekend.  It was a ``caf\'e scientifique'' in a bar.  Each of the four panelists tried to answer, in 10 minutes, the question of what quantum theory was and why the audience should care.  After the break, the moderator asked us if we could remind the audience that quantum theory is a theory of stuff at small scales.  That idea always sat badly with me because quantum theory applies to all systems, even macroscopic ones.  So I got to thinking about how to better describe the regime in which quantum effects are important.  Here's what I came up with (and told the audience):  If quantum theory were about particles and forces and motion then one might have expected a length scale to be significant, but given that it is actually about information and knowledge, the scale that is important is a scale of {\it certainty}.  If one is talking about a system about which one has very imprecise information, for instance the position and momentum of a car, then one is nowhere near the regime in which the Heisenberg uncertainty principle is important and classical mechanics suffices.  If one is talking about an atom that is being measured extremely precisely, then one can begin to see the effect of the Heisenberg limit and quantum theory is important.  The reason that quantum effects were first seen in microscopic systems is because these are the systems wherein it is easiest to achieve a great deal of certainty.  It is difficult to do so for macroscopic systems because these generally interact more readily with their environment, and this generally leads to a decrease in our degree of certainty about the system because we typically have a great deal of uncertainty about the environment.

So I have a new slogan: The scale at which quantum theory applies is the scale of small uncertainty.  What do you think?
\erws

Here, here!  I like honesty to the layman, and I can see why the moderator's suggestion wouldn't sit well with you.  I wish I had been there to hear the whole story.  Who and what were the other points of view?  (I had to decline their invitation when they asked me.)

I liked your answer all the way to the bit about ``decoherence'' (though I resumed liking it again at your slogan).  Of macroscopic systems, I would say we never had any certainties to begin with to lose.  For instance, for a cat, no one has ever come close to making a pure-state assignment for it for which decoherence would then be relevant.  In my mind, decoherence as an ``explanation'' of classicality is defunct.  I lay my money on something closer to what Kofler and Brukner propose: \arxiv{quant-ph/0609079} and then in more detail in Kofler's thesis.

Of regular ``decoherence,'' (as used for quantum foundational discussion \`a la Zeh, Zurek, etc.)\ my new view is that it has essentially nothing to do with interactions with an environment.  Instead it has everything to do with how an agent should gamble under the supposition of future certainties.  Attached is a presently poorly written draft of a paper where {\Ruediger} and I finally try to write down this point.  Or at least this is our first stab at it.  The quantum section will be strengthened as I rewrite it (much comes from an earlier conference abstract of mine), but the essential point is there:  Quantum decoherence (in the context of measurement theory) is none other than van Fraassen-ian reflection.  [See ``Bayesian Conditioning, the Reflection Principle, and Quantum Decoherence,'' \arxiv{1103.5950}.]

Maybe as a modification to your slogan that fits a bit better my own proclivities:  Quantum theory applies to the scale of precision where contextuality and counterfactuals start to matter.

\section{19-01-11 \ \ {\it Born's Rule}\ \ \ (to J.-{\AA} Larsson)} \label{Larsson10}

\bjal
\myurl{http://xkcd.com/849/}.
\ejal

Thanks, that was funny.  Attached is my own latest cartoon.  [See \arxiv{1207.2141}.] I can't post it for a few months, but that doesn't stop me from sending it to friends like you.

At the moment, I'm working on a more {\Ruediger}-style paper on ``decoherence'' from the point of view of QBism.  I.e., what I disappointed everyone with at Khrennikov's meeting last year.  I hope we do a better job this time. Here's the abstract for it:
\bq
The probabilities a Bayesian agent assigns to a set of events typically change with time, for instance when the agent updates them in the light of new data.  In this paper we address the question of how an agent's probabilities at different times are constrained by  Dutch book coherence. We review and attempt to clarify the argument that, although an agent is not forced by coherence to use the usual Bayesian conditioning rule to update his probabilities, coherence does require the agent's probabilities to satisfy van Fraassen's [1984] {\it reflection principle\/} (or the similar constraints due to Goldstein [1983] and Shafer [1983]).  Bringing the argument to the context of quantum measurement theory, we show that ``quantum decoherence'' can be understood in purely personalist terms---quantum decoherence (as supposed in a von Neumann chain) is not a physical process at all, but an application of the reflection principle. From this point of view, the decoherence theory of Zeh, Zurek, and others as a story of quantum measurement has the plot turned exactly backward.
\eq

\section{20-01-11 \ \ {\it Combining Probabilities}\ \ \ (to R. W. {\Spekkens} \& M. S. Leifer)} \label{Leifer12} \label{Spekkens103}

This may be a useful paper for you [C. Genest and J. D. Zidek, ``Combining Probability Distributions:\ A Critique and an Annotated Bibliography,'' {\sl Statistical Science\/} {\bf 1}, 114--148 (1986)], in case you don't know about it:
\bq\noindent
\myurl{http://www.jstor.org/pss/2245510}.
\eq

\section{21-01-11 \ \ {\it Section 5}\ \ \ (to R. {\Schack})} \label{Schack216}

And I am confused on the priority of Section 5.  Is that an argument purely due to you?  Or had someone else given it before, perhaps in different form?

Here's the present version of the abstract.  Have I made it too weak with respect to what's actually been done?
\bq
The probabilities a Bayesian agent assigns to a set of events typically change with time, for instance when the agent updates them in the light of new data. In this paper we address the question of how an agent's probabilities at different times are constrained by  Dutch book coherence. We review and attempt to clarify the argument that, although an agent is not forced by coherence to use the usual Bayesian conditioning rule to update his probabilities, coherence does require the agent's probabilities to satisfy van Fraassen's [1984] {\it reflection principle\/} (or a related constraint due to Goldstein [1983]). Bringing the argument to the context of quantum measurement theory, we show that ``quantum decoherence'' can be understood in purely personalist terms---quantum decoherence (as supposed in a von Neumann chain) is not a physical process at all, but an application of the reflection principle.  From this point of view, the decoherence theory of Zeh, Zurek, and others as a story of quantum measurement has the plot turned exactly backward.
\eq

\section{22-01-11 \ \ {\it Woefully Inadequate}\ \ \ (to the QBies)} \label{QBies31}

I felt woefully inadequate yesterday in our discussion of ``facts'' and ``experience''.  I feel the same this morning.  I feel so far from verbalizing whatever it is that's trying to form in my mind.  Attached are my two best shots at it so far (from two recent papers).  Maybe putting it in the middle of poetry, like in my pseudo-Shakespearean  quote, is the most I can hope for at the moment:
\bv
               Our revels are now ended.  These our actors,\\
               As I foretold you, were all spirits and\\
               Are melted into air, into thin air \ldots\medskip\\
               We are such stuff as\\
               \hspace{.33in}        quantum measurement is made on.
\ev

The young Wittgenstein started off the {\sl Tractatus\/} with, ``The world is everything that is the case.  What is the case, the fact, is the existence of atomic facts.''  Somehow, someway, I want to replace his ``facts'' with something closer to ``birth'' in its imagery.  Facts, as usually thought, are kind of dead things.  (I said something like that yesterday.)  In any case, I do very much believe that whatever it is that I'm looking for should be modeled on ``quantum measurement'' (whatever that is) in its first pass.

\subsection{Marcus's Reply}

\bq
I agree totally with this
\bq
 Facts, as usually thought, are kind of dead things.
\eq
and I agree also that there is something wrong with that.    Except ``wrong'' is too weak.  So let me try again.   Black as night.   Dark as the grave.   Life denying.   Soul destroying.  A figment of Satan.

But, like you, I am not sure where to go from here.
\eq

\section{22-01-11 \ \ {\it Creatia!}\ \ \ (to the QBies)} \label{QBies32}

Figment of Satan!  Beautiful.  Attached is just a bit more to read.  It comes from somewhere in the middle of the proposal I wrote for the Templeton Foundation.  In it, I called this new stuff of the world ``creatia.''

\bq
As important as getting straight the ``quantum-theory-as-a-user's-manual conception'' has been for Quantum Bayesianism, it is striking that taking that step leads right away to the very next one.  The journey is not concluded by settling the foundational conundrums; instead, it is just the start of a greater, more important trek.

What we have learned from the user's-manual conception of quantum theory is that every quantum measurement (by definition, a transaction between an agent and a system) represents a moment of creation.  Something new comes into the world that was not there before.  A knee-jerk reaction that one often hears as a response to these ideas is that they're a veiled form of solipsism---the idea that the world consists only of {\it the self\/} (presumably the Project Leader?).  We have argued strenuously, however, in \verb+\cite[Section VI]{Fuchs10b}+ that the source of the knee-jerk reaction is mostly a lack of appreciation that physical theories can have any reading other than reductionistic ones.

The key is not to view quantum theory in reductive terms and try to derive the world we commonly see around us out of the theory's mathematical structure, but rather to view the advice of quantum theory as {\it additive\/} to the world of (pre-quantum) experience. Here is the way it was put in \verb+\cite{Fuchs10b}+,
\bq
The expectation of the quantum-to-classical transitionists is that quantum theory is at the bottom of things \verb+\cite{Schlosshauer07}+, and ``the classical world of our experience'' is something to be derived out of it.  QBism says ``No.  Experience is neither classical nor quan\-tum.  Experience is experience with a richness that classical physics of any variety could not remotely grasp.''  Quantum mechanics is something put on top of raw, unreflected experience.  It is additive to it, suggesting wholly new types of experience, while never invalidating the old.  To the question, ``Why has no one ever {\it seen\/} entanglement in diamond before?,'' the QBist replies:  It is simply because before recent technologies and controlled conditions, as well as lots of refined analysis, no one had ever mustered a mesh of beliefs relevant to such a range of interactions with diamonds.  No one had ever been in a position to adopt the extra normative constraints required by the Born Rule.  For QBism, it is not the emergence of classicality that needs to be explained, but the emergence of our new ways of manipulating, controlling, and interacting with matter that do.
\eq
The backdrop for this program of thought is in recognizing that the happiest philosophical home for QBism is a kind of empiricism---the idea that everything experienced, everything experienceable has no less an ontological status than anything else. From this view, an experienced dream and a Higgs-boson detection event at the LHC are equal elements in the filling out and making of reality.  Instead of a starkly empty world, as a quantum reductionist might have it (``the world is {\it nothing but\/} the universal wavefunction undergoing unitary evolution''), the world of QBism is overflowingly full---it is a world whose details are beyond anything grammatical or rule-bound expression can articulate.  In other words, it avoids the great ``original sin'' of ``vicious abstractionism'' (as William James called it),
\bq\noindent
We conceive a concrete situation by singling out some salient or important feature in it, and classing it under that; then, instead of adding to its previous characters all the positive consequences which the new way of conceiving it may bring, we proceed to use our concept privatively; reducing the originally rich phenomenon to the naked suggestions of that name abstractly taken, treating it as a case of `nothing but' that, concept, and acting as if all the other characters from out of which the concept is abstracted were expunged.
\eq
Previous to QBism's thinking, the seed of that sin was spread over nearly every flavor of quantum interpretation known.

So what are the ``salient and important features'' QBism singles out to be {\it added\/} to the ``previous character'' of common experience?  We have at least one answer to this so far, and it is identified right at the front of
$$
Q(D_j)=(d+1)\sum_{i=1}^{d^2} P(H_i) P(D_j|H_i) - 1\;.
$$
It is the single number $d$, called dimension, associated with each quantum system.  What is so nice about this equation, as opposed to the usual way of expressing the Born Rule, is that dimension is pulled out and displayed prominently:  It represents the {\it deviation\/} from the classical Law of Total Probability one should use in one's considerations for a given system.  In \verb+\cite[Section VI,VII]{Fuchs10b}+, we have argued that dimension should be thought of as a new {\it capacity\/} inherent in all matter, much like ``gravitational mass'' can be thought of as a capacity inherent in all matter \verb+\cite{Cartwright99}+.  It is this new capacity that allows one to do unexpected and wonderful things with quantum systems.

But our thoughts really are nascent in this, and so much work needs to be done to distill it into productive physics.  With JTF's support, having Dr.\ Schlosshauer stationed at Perimeter Institute and Dr.\ Appleby available for more conceptual work, we will be able to have a devoted conceptual-research group set to get straight to the heart of this matter. (For instance, by taking advantage of the 1,000-book ``pragmatism library'' already built up by the Project Leader for the purpose of exploring this line of thought; see supplementary material.) Extended visits by Co-Project Leader {\Schack} will also be decisive in this matter.

Chief among our considerations is making a push to pin down exactly which flavor of empiricism surrounds the technical apparatus of QBism.  At the moment, we think it is a combination of Wolfgang Pauli's thoughts (where the individual stuff of the world is ``neutral,'' neither matter nor psyche) and William James's ``radical empiricism'' of ``pure experience.''  \label{ForIntroduction1} QBism says that every quantum measurement is a moment of creation, and the formal apparatus of quantum theory is an aid for each agent's thinking about those ``creatia'' she is involved with.  But surely a Copernican principle applies just as much to QBism as to any other science.  QBism's solution starts by saying the last point just that much more clearly: ``Quantum measurement represents those moments of creation an agent happens to seek out or notice.'' It does not at all mean that there aren't moments of creation going all around, unnoticed, unparticipated in by the particular agent, all the time.  The larger world of QBism is something aligned with James's vision of a pluriverse where ``being comes in local spots and patches which add themselves or stay away at random, independently of the rest.''

The biggest goal of the research group, of course, would be to say something decisive on John Wheeler's Big Question, ``Is the Big Bang here?'':
\bq\noindent
Each elementary quantum phenomenon is an elementary act of ``fact
creation.'' That is incontestable. But is that the only mechanism
needed to create all that is? Is what took place at the big bang the
consequence of billions upon billions of these elementary processes,
these elementary ``acts of observer-participancy,'' these quantum
phenomena? Have we had the mechanism of creation before our eyes all
this time without recognizing the truth?
\eq

Concretely, in the period of the funding, what will happen is that the total force of our collective thought will make its way into the chapters of Schlosshauer's book---particularly, the parts devoted to the larger worldview of QBism---as well as some papers that Appleby has long been gearing up to write.  Finally, Fuchs and {\Schack} will write a detailed, more specialized monograph on the sum total ``Worldview of QBism''---the monograph be as thorough as possible, expressing all pros, cons, and obstacles identified by the research group.
\eq

\section{23-01-11 \ \ {\it Creatia!, 2}\ \ \ (to the QBies)} \label{QBies33}

A quick example of the trouble that might be caused by a quick choice of phrase.  Marcus writes, ``the radical empiricism of Mach'' in this note.
Whereas in lots of recent papers, I cite James's book {\sl Essays in Radical Empiricism}.  In the case of James, ``radical empiricism'' was his own name for his particular doctrine.  In the case of Marcus, though, I think ``radical'' was chosen (I would guess) as a way to express that Mach went further in a certain way than Locke and Hume.

You can get a quick sense that these might be different ideas by looking at the wiki page:
\myurl[http://en.wikipedia.org/wiki/Radical_empiricism]{http://en.wikipedia.org/wiki/Radical\underline{ }empiricism}.
I don't believe Mach's own philosophy depended upon ``relations themselves''
being directly experienceable.

Anyway, here's how James defines his ``radical empiricism'':
\bq
I give the name of radical empiricism to my Weltanschauung. Empiricism is known as the opposite of rationalism. Rationalism tends to emphasize universals and to make wholes prior to parts in the order of logic as well as in that of being. Empiricism, on the contrary, lays the explanatory stress upon the part, the element, the individual, and treats the whole as a collection and the universal as an abstraction. My description of things, accordingly, starts with the parts and makes of the whole a being of the second order. It is essentially a mosaic philosophy, a philosophy of plural facts, like that of Hume and his descendants, who refer these facts neither to Substances in which they inhere nor to an Absolute Mind that creates them as its objects. But it differs from the Humian type of empiricism in one particular which makes me add the epithet radical.

To be radical, an empiricism must neither admit into its constructions any element that is not directly experienced, nor exclude from them any element that is directly experienced. For such a philosophy, the relations that connect experiences must themselves be experienced relations, and any kind of relation experienced must be accounted as real as anything else in the system. Elements may indeed be redistributed, the original placing of things getting corrected, but a real place must be found for every kind of thing experienced, whether term or relation, in the final philosophic arrangement.
\eq

Anyway, I'm not wedded to the terms ``empiricism'' or ``experience'' myself---I might change to something else radically different in the near future.  They just seemed expedient at the time of my writing.  I didn't know better words and I had been influenced by James, Schiller, and Dewey.

\section{23-01-11 \ \ {\it I Am Available to Talk}\ \ \ (to D. M. {\Appleby})} \label{Appleby99}

Timing was bad.  Your note came a bit after we took off for the movies; we saw the new Narnia movie.  (I much liked it.)  We just got back now.
Maybe we can wait on philosophy till tomorrow.

I read the wiki page on empiricism this morning, and learned that there is a bit of an echo of J. S. Mill in the footnote 37 I sent you yesterday.
Mill called matter ``the permanent possibility of sensation.''

\section{24-01-11 \ \ {\it Putnam}\ \ \ (to W. G. {\Demopoulos})} \label{Demopoulos42}

\bwd
Apropos Hilary's citation of the quote attributed to Bohr by Aage Petersen (text to
Hilary's Fn.\ 19), I would put things differently:
\bq\rm
There is no classical world. There is only an abstract classical mechanical description.
\eq
Hence, I don't see myself as just recycling Bohr, although I do ``echo'' him.
\ewd

I liked your play on the Petersen quote!  Nice.

On your lines,
\bwd
We agreed earlier that on your view, it is the effects the would has on Bayesian agents that forms the subject matter of quantum mechanics. I regard the dualism implicit in this view as entirely tenable---it's the dualism between our intentions and expectations about the future on the one hand, and reality on the other. But I can't follow you in your incorporation of this dualism into the thesis that QM is about the consequences of our actions.
\ewd
There must be a typo in the first sentence; I would like to know what your exact words would have been.  But I think I get the gist.  I found myself thinking, ``What a strangely asymmetric world.  I feel it, but it never feels me.''

I have to admit that I've been frightened ever since you told me that Putnam may be in my audience in Jerusalem.  I guess I'm still trying to get over the ``postmodern sauce'' he flung at Jeff Bub and me at your meeting a couple years ago.  You think there's any chance I could get him to read my \arxiv{1003.5209} before my talk, before he comes out with his ladle full?

I will think harder and deeper on your remarks.  I know you probably don't think much of my ``experience mumbo jumbo'' but your play on Petersen rings a bit with something I wrote in that paper:
\bq
The expectation of the quantum-to-classical transitionists is that quantum theory is at the bottom of things, and ``the classical world of our experience'' is something to be derived out of it.  QBism says ``No.  Experience is neither classical nor quantum.  Experience is experience with a richness that classical physics of any variety could not remotely grasp.''  Quantum mechanics is something put on top of raw, unreflected experience.  It is additive to it, suggesting wholly new types of experience, while never invalidating the old.  To the question, ``Why has no one ever {\it seen\/} superposition or entanglement in diamond before?,'' the QBist replies:  It is simply because before recent technologies and very controlled conditions, as well as lots of refined analysis and thinking, no one had ever mustered a mesh of beliefs relevant to such a range of interactions (factual and counterfactual) with diamonds.  No one had ever been in a position to adopt the extra normative constraints required by the Born Rule.  For QBism, it is not the emergence of classicality that needs to be explained, but the emergence of our new ways of manipulating, controlling, and interacting with matter that do.

In this sense, QBism declares the quantum-to-classical research program unnecessary (and actually obstructive) in a way not so dissimilar to the way Bohr's 1913 model of the hydrogen atom declared another research program unnecessary (and actually obstructive).  Bohr's great achievement above all the other physicists of his day was in being the first to say, ``Enough!  I shall not give a mechanistic explanation for these spectra we see.  Here is a way to think of them with no mechanism.''
\eq

BTW, the rocks on the beach were diamond.  Diamond has recently been learned to be a very promising material for quantum computation.

\section{24-01-11 \ \ {\it Putnam, 2}\ \ \ (to W. G. {\Demopoulos})} \label{Demopoulos43}

\bwd
My talk of there being admissible classical descriptions is just another
way of saying that our experience is classical.
\ewd

Yes, you're right:  There's probably not as much resemblance as I was hoping for.  All I was really meaning to say is that I would be willing to mimic part of what you were saying.  Maybe my own (not so pithy) modification to Peterson would be:
\bq\noindent
There is no quantum world. There is only an abstract quantum mechanical description for some of the things in it (at certain times and places in our relations with it).  But there is no classical world either.  There is only an abstract classical mechanical description of some of the things in it (at certain times and places in our relations).  The former is additive to the latter---in the sense of our operational uses of the world's objects.  The world in itself is richer than either schema---classical or quantum---can accommodate.
\eq

On Putnam, you are probably right.  I will hold my tongue until (and unless) I am forced to.

\section{27-01-11 \ \ {\it Schlosshauer and Mach} \ \ (to A. Zeilinger)} \label{Zeilinger13}

Thank you for the kind letter---your continued patience moves me.  I apologize for not having written you sooner:  I had wanted the latest update from Max Schlosshauer before writing you again, and he had fallen silent.  It turns out he was on a small island somewhere editing his book, and he had had no internet connection for several days. [\ldots]

I enjoyed reading your short interview.  It was helpful for my orientation to see a new portrait of you.  (It was similarly useful with respect to understanding {\Mermin} and Hardy, whose contributions I have read as well.)  I particularly liked the way you put a point near the end:
\bq\noindent
This is not to say that everything is just information or knowledge. What I
mean is that we need a new fundamental concept unifying the notions of
information and reality.
\eq
There is some amount of similarity between that and the far end of my own QBism program.  Or at least I think there is.  I say so because of (for instance) this passage from a recent paper of mine:
\begin{quote}
[W]e have learned enough from Copernicus to know that egocentrism, whenever it can be shaken away from a weltanschauung, it ought to be.  Whenever ``I'' encounter a quantum system, and take an action upon it, it catalyzes a consequence in my experience that my experience could not have foreseen.  Similarly, by a Copernican principle, I should assume the same for ``you'':  Whenever you encounter a quantum system, taking an action upon it, it catalyzes a consequence in your experience.  By one category of thought, we are agents, but by another category of thought we are physical systems.  And when we take actions upon each other, the category distinctions are symmetrical.  Like with the Rubin vase, the best the eye can do is flit back and forth between the two formulations. In the common circles of the philosophy of science there is a strong popularity in the idea that agentialism can always be reduced to some complicated property arrived at from physicalism.  But perhaps this republican-banquet vision of the world that so seems to fit a QBist understanding of quantum mechanics is telling us that the appropriate ontology we should seek would treat these dual categories as just that, dual aspects of a higher, more neutral concept.  That is, these concepts ``action'' and ``unforseen consequence in experience,'' both so crucial for clarifying the very meaning of quantum measurement, might just be applicable after a fashion to arbitrary components of the world---i.e., venues in which probability talk has no place.   Understanding or rejecting this idea is the long road ahead of us.  The development of formalism in this paper we see as at least one promising entrance into that road.
\end{quote}
It is probably a bit hard to understand out of the context for which it was originally written.

Anyway:  What it is trying to get at is, that on the one hand, the quantum mechanical measurer is an agent (in the sense of probability and decision theory) and, on the other hand, it is a physical system like any other.  You use the words ``information'' and ``reality'' \ldots\ and I use the words ``agent'' (bearer of probabilities) and ``system''.  I think we are both thinking ``complementary aspects of \ldots\ something.''

In both cases, I think a philosopher of science would say that he thinks we are seeking a ``neutral monism.''  I don't know that that is the best name for it, but we both are certainly seeking something that breaks the cut between Descartes' {\it res cogitans\/} and {\it res extensa}.

We are not the first to want to want something like this, and it explains in part why I find such inspiration in the writings of William James (who aimed for this explicitly with his concept of ``pure experience'') and the private letters of Wolfgang Pauli (who tended to just call it ``psychophysically neutral'' stuff).

More recently I've taken a great interest in Ernst Mach's own version of it, with his ``sensations'' or ``world elements''.  I have to admit that until recently, I had never understood the depths of his thoughts in this regard.  I had only known of the logical-positivist leaning extensions of his earlier thoughts by others.  So much is hidden to us who cannot read the German language!!  The actual man's thinking is so much more interesting to me than the other schools that arose (in Vienna) immediately after him.

At the moment, I am in Texas to spend a bit of time with my mother for her 82nd birthday, and I brought a book with me that I've been reading voraciously:  Erik C. Banks' {\sl Ernst Mach's World Elements: A Study in Natural Philosophy}.  It is a very useful resource because the whole thing is devoted to giving a ``conclusive demonstration'' that the answer to this question:
\bq\noindent
     What one really wants from Mach is some clear statement that he
     believed in objects or elements that were not human sensations \ldots :
     elements that made up a mind-independent world of nature even when
    no human being is sensing it, similar to what Russell later called sensibilia \ldots
\eq
is ``Yes.''  Mach's elements were psychophysically neutral elements (in the way that Pauli and James wanted such a thing), but not explicitly tied to human existence.

It is a very good book, and I would recommend it wholeheartedly.  It is a certainly a good source of ideas for me, and I suspect it will be for your own variations on the theme as well.

\section{01-02-11 \ \ {\it How To Remove Yourself from Someone's Xmas Card List} \ \ (to M. S. Leifer)} \label{Leifer13}

\bml
I have written up some notes on my objection to your treatment of Wigner's friend in the QBism papers.  It came out somewhat longer than I had intended, because I was trying to get all my thoughts on the subject straight.  Somewhere along the way, it seems to have morphed into a paper.
It sort of depends on the papers I am writing with Rob, so I definitely want to get those finished before I do anything further with it.  That gives you plenty of time to make comments and maybe convince me that I have got it wrong and should junk the whole thing before I decide whether to post this anywhere.
\eml

On the contrary:  That you take me seriously enough to write a paper on my ideas (pro {\it or\/} contra, it doesn't matter) is a powerful reason to keep you in my correspondence that much more!  (We should all beware of the unintended consequences of our actions.)

Thanks for this.  I am sorry it took my so long to get back to you.  I was in the backwoods of Texas last week, visiting my 82-year-old mother, and that took most of my energies.

This paper will be very valuable to me, as I work out how to respond to you.  The Wigner's Friend issue is the central point for me now as I try to move a step beyond quantum mechanics to the kind of neutral-monism ontology (neutral pluralism really) that I promised at the end of my last few papers (``QB Coherence'' and ``QBism, the Perimeter of \ldots'').  Similarly in my defunct Templeton Foundation proposal:
\bq
Chief among our considerations is making a push to pin down exactly which flavor of empiricism surrounds the technical apparatus of QBism.  At the moment, we think it is a combination of Wolfgang Pauli's thoughts (where the individual stuff of the world is ``neutral,'' neither matter nor
psyche) and William James's ``radical empiricism'' of ``pure experience.''
QBism says that every quantum measurement is a moment of creation, and the formal apparatus of quantum theory is an aid for each agent's thinking about those ``creatia'' she is involved with.  But surely a Copernican principle applies just as much to QBism as to any other science.  QBism's solution starts by saying the last point just that much more clearly:
``Quantum measurement represents those moments of creation an agent happens to seek out or notice.'' It does not at all mean that there aren't moments of creation going on all around, unnoticed, unparticipated in by the particular agent, all the time.  The larger world of QBism is something aligned with James's vision of a pluriverse where ``being comes in local spots and patches which add themselves or stay away at random, independently of the rest.''
\eq

The key issue in any response to you, my guess is, is going to be in the meaning of Claim 1---the explicit ``two levels of personalism'' QBism now adopts probably does not mesh well with the current philosophy-of-science usage of the term ``objectively real'' (with strong connotations toward ``publicly accessible'').  Maybe just as importantly will be this thing you call an ``ignorance interpretation''.  I don't know yet, but from the sounds of it (i.e., your choice of wording), it would indicate that you're missing a key point in my thinking \ldots\ or \ldots\ that I'm being grossly inconsistent.  You'll forgive me for thinking it'll be the former rather than the latter---but I could be wrong, and time will tell.

I hope to get back to you from Israel the week after next, where I will be giving the Pitowsky memorial lecture.  They particularly want an explication of the similarity and distinctions of my Bayesianism and his, and all this should be deeply on my mind.

Best wishes, and honestly, thanks for this!

\section{01-02-11 \ \ {\it Mechanism, Sensation, Texarkana}\ \ \ (to C. Smeenk)} \label{Smeenk6}

On other matters, I've been studying Erik Banks' book {\sl Ernst Mach's World Elements} (UWO Series) for the last week and have been greatly enjoying it.  I honestly hadn't appreciated the depths of Mach's thought (even in the 1860s and 1870s already).  It's anything but the logical positivism I had always mislabeled him with.  You read some of his passages, and if you didn't know better, you might think he was in the quantum gravity group at PI.  Anyway, it's been enlightening, and it teaches me once again that some of the best sources of ideas in physics *for me* are in long forgotten 120-year-old books.  I've got to learn much more about Mach now.

Attached is something I just wrote that's bound to get any philosopher of science's dander up.  I hope you get a laugh from it every now and then, even as you grimace.

\section{04-02-11 \ \ {\it Mach on James on Wine, Coffee, and God}\ \ \ (to D. M. {\Appleby})} \label{Appleby100}

This comes from a 1911 letter of Ernst Mach to Anton Thomsen.  James had just died in 1910.
\bq
My personal memories of William James are very pleasant; he visited me while still in Prague in 80 or 81.  [In fact it was 1882. CAF]  I remember no one with whom, despite the divergence of viewpoints, I could discuss so well and fruitfully.  He opposed me almost everywhere and yet I benefited almost everywhere by his objections.  Already at that time he avoided any drop of wine or coffee so that I believed him more of a nervous hypochondriac than a really sick man.  The center of his work certainly lies in his excellent {\sl Psychology}.  I cannot quite come to terms with his Pragmatism.  ``We cannot give up the concept of God because it promises too much.''  That is a rather dangerous argument.
\eq

Reading this again, it reminds me that here is another noteworthy distinction between Mach and James.  Mach was an atheist and wanted to weed out as many ``transcendent'' terms from science as possible (presumably God was one of them).  On the other hand, James's ``radical empiricism'' strives to be as inclusive as possible with all experiences:  ``To be radical, an empiricism must neither admit into its constructions any element that is not directly experienced, nor exclude from them any element that is directly experienced.''  So, now think of what he sought to establish in {\sl The Varieties of Religious Experience}---that the experience of God is a widespread and very real thing for so many people.  A radical empiricism, therefore, cannot exclude it.

\section{07-02-11 \ \ {\it The 2011 QBist Challenge} \ \ (to the QBies)} \label{QBies34}

I attach the QBist Glossary to remind you of the running theme of our work.

The other day at group meeting, I read you all a passage that I had written for a 2004 lecture at Caltech:
\bq
A lecturer faces a dilemma when teaching a course at a farsighted
summer school like this one.  This is because, when it comes to
research, there is often a fine line between what one thinks and what
is demonstrable fact.  More than that, conveying to the students what
one thinks---in other words, one's hopes, one's desires, the
potentest of one's premises---can be just as empowering to the
students' research lives (even if the ideas are not quite right) as
the bare tabulation of any amount of demonstrable fact.  So I want to
use one percent of this lecture to tell you what I think---the
potentest of all my premises---and use the remaining ninety-nine to
tell you about the mathematical structure from which that premise
arises.

I think the greatest lesson quantum theory holds for us is that when
two pieces of the world come together, they give birth.  [Bring two
fists together and then open them to imply an explosion.]  They give
birth to FACTS in a way not so unlike the romantic notion of
parenthood:  that a child is more than the sum total of her parents,
an entity unto herself with untold potential for reshaping the world.
Add a new piece to a puzzle---not to its beginning or end or edges,
but somewhere deep in its middle---and all the extant pieces must be
rejiggled or recut to make a new, but different, whole.  That is the
great lesson.

But quantum mechanics is only a glimpse into this profound feature of
nature; it is only a part of the story.  For its focus is exclusively
upon a very special case of this phenomenon:  The case where one
piece of the world is a highly-developed decision-making agent---an
experimentalist---and the other piece is some fraction of the world
that captures his attention or interest.

When an experimentalist reaches out and touches a quantum
system---the process usually called quantum `measurement'---that
process gives rise to a birth.  It gives rise to a little act of
creation.  And it is how those births or acts of creation impact the
agent's {\it expectations\/} for other such births that is the
subject matter of quantum theory.  That is to say, quantum theory is
a calculus for aiding us in our decisions and adjusting our
expectations in a QUANTUM WORLD\@.  Ultimately, as physicists, it is
the quantum world for which we would like to say as much as we can,
but that is not our starting point.  Quantum theory rests at a level
higher than that.

To put it starkly, quantum theory is just the start of our adventure.
The quantum world is still ahead of us.  So let us learn about
quantum theory.
\eq

As Marcus pointed out, I'm certainly using the word ``facts'' (in all caps) in an awfully nonstandard way there.  For instance, one does not usually think of facts as having ``untold potential for reshaping the world.''

So, I've spent the last few minutes trying to find a better word that fits our group theme.  The one that strikes me most at the moment is
\begin{itemize}
\item
{\bf QBoom} -- cf.\ {\it kaboom}; a fundamental event in analogy to the outcome of a quantum measurement.  ``In the QBist weltanschauung, the universe does not come into being with a Big Bang, but instead arises `in spots and patches' through millions upon millions of little QBooms.''
\end{itemize}
If any of you think of a more elegant exercise in QBistry, I'm open to suggestions.

\section{07-02-11 \ \ {\it More QB Silliness} \ \ (to the QBies)} \label{QBies35}

\begin{itemize}
\item
{\bf QBoom} -- cf.\ {\it kaboom}; the sought-for deanthropocentrized distillate of a quantum measurement that QBism imagines as powering the world.  William James called an early version of something along these lines ``pure experience''.  John Wheeler asked, ``Is the entirety of existence, rather than being built on particles or fields of force or multidimensional geometry, built upon billions upon billions of elementary quantum phenomena, those elementary acts of `observer-participancy,' those most ethereal of all the entities that have been forced upon us by the progress of science?''  and, ``Is what took place at the big bang the consequence of billions upon billions of these elementary processes, these elementary `acts of observer-participancy,'?''  In other words, in place of the Big Bang, the QBist wonders whether it might not be billions upon billions of little QBooms instead?

\item
{\bf QBlooey} -- cf.\ {\it kablooey}; a QBist slur on the usual conception of the Big Bang---i.e., a conception of the universe in which nature already blew its wad at the very beginning.
\end{itemize}

\section{07-02-11 \ \ {\it Renouvier} \ \ (to S. Weinstein)} \label{Weinstein6}

Thanks for the music; I'll have a listen of it when I get home.  And, no, I haven't read Hume.  I should be ashamed of myself---with my abiding interest in free will, it would serve me to.

Attached are some of my notes from a book on Renouvier.  [See 13-08-03 note ``\myref{Schack72}{Renouvier}'' to R. {\Schack}.] For Renouvier, free will was something that had to be fought for.  It didn't come for free; unless you really wanted it, you would not have it.  I also like (the flavor of) his argument that---surprisingly---the concepts of true and false are crucially dependent on freedom.

\section{08-02-11 \ \ {\it From Screed to Screech}\ \ \ (to A. Wilce)} \label{Wilce24.1}

Thanks for the new paper!  I've sent it on to my Jordan-algebra-savvy student Matthew Graydon, to get him to explain it to me.  Also dug up your reference 8 since it looked like it might intrigue him.  Noticed that you had Brukner's name misspelled as Bruckner, and you're missing some diacritical marks as well (on him and his co-author).

Attached is my own latest.  [See ``Interview with a Quantum Bayesian,'' \arxiv{1207.2141v1}.] It won't be posted for two or three months until after the book comes out.  And if you read some of the nasty things I say about our philosophical friends, maybe that's a good thing!

I've got a more technical paper coming down the pipe soon, on diachronic Dutch book arguments and their relation to (what is misconstrued as a physical, rather than epistemic) quantum decoherence.  I'll let you know when it's finally posted.  [See \arxiv{1103.5950}.]

I'm off to Jerusalem Friday to give the first Itamar Pitowsky memorial lecture next week.  A very flattering thing to do.  But I'm also scared stiff:  I've been told Hilary Putnam will be in the audience.

It'll be good to see you again in March.  I'll make sure I reserve some time for you.

\section{09-02-11 \ \ {\it Questions and Answers} \ \ (to R. Colbeck \& G. Chiribella)} \label{Colbeck1} \label{Chiribella3}

I enjoyed both your questions yesterday.  They were completely on the mark, and I appreciated them.

For Roger's question on ``what if they don't exist,'' I attach the write-up of a little epiphany I had while Howard Barnum and I were at some philosophy of science talk at the university.  [See 13-03-10 note ``\myref{QBies6}{Conceptual Barrier!}''\ to the QBies.]  I do well believe that someone will eventually show that SICs always exist, but if they don't I'm prepared for it, and the attached write-up makes the point with some passion.  You might find it entertaining reading.  I also include for your amusement the write-up of a dream I sent to Anton Zeilinger and its interpretation in terms of the epiphany.  [See 01-04-10 note ``\myref{Zeilinger11}{Strange Anton Dream}'' to A. Zeilinger and H. C. von Baeyer.]

The ``allusion to William James's faith ladder'' that I mention in the note refers to this little ditty I had sent to my group and Howard in an earlier email:
\bq\noindent
       SICs are {\it fit\/} to be true.  It would be well if they {\it were\/} true.  They {\it might\/} be true;
       they {\it may\/} be true; they {\it ought\/} to be true.  They {\it must\/} be true!  They {\it shall\/} be
       true!  For me at least!!
\eq

Anyway, what is powering my faith ladder is the simplicity of the displayed equation in the note.  It's consistency $+$ Bayes' rule $+$ an assumption of maximality \ldots\ these three things alone \ldots\ imply an already very rich convex set structure that gets quite some distance toward quantum theory.  If you have some interest, you can read about some of the properties of such sets here:  \arxiv{0910.2750}.

Now on to Giulio's question.  I think it would be quite healthy for me to try to tackle some of the things he suggested --- to try to understand what quantum-like protocols we can perform with this class of theories.  It's a crucial question really.  If you have any suggestions or any ideas or any methods for getting at some answers, I certainly invite you to get involved.

In that regard, attached are some notes that Howard Barnum wrote up after one of our conversations where he goes some way toward translating the QBist program into the more commonly-used convex-operational language.  The equations don't look so pretty in that framework (to me), but it can be done.  One of these days, these notes ought to be extended and  written into a proper paper.  If either of you have an contributions to make, I would surely welcome that as well.

Finally, let me attach an interview I just wrote up that won't be posted on the web for a while (until the book it appears in has been published).  [See ``Interview with a Quantum Bayesian,'' \arxiv{1207.2141v1}.]  In question Q10, starting on page 16, I give my opinion on axiomatic reconstruction efforts.  All my thinking is really in that kind of ``literary'' context.

Thanks again guys.

\section{11-02-11 \ \ {\it On My Way}\ \ \ (to W. G. {\Demopoulos})} \label{Demopoulos44}

I just wanted to say hi before crossing the ocean.  I'm writing from JFK airport, on my way to Jerusalem for Itamar's memorial.  I accompanied myself with several things on neutral monism to read along the way and there, and I just ran across a reference to a paper by you and Friedman, apparently relevant to a discussion of Russell's particular version of it all.

It would be fun to talk to you again sometime soon.  I hope you're doing as well as possible at this point.

\section{12-02-11 \ \ {\it Incompatibility's Allure}\ \ \ (to R. {\Schack})} \label{Schack217}

Ishtiyaque Haji, {\sl Incompatibilism's Allure: Principle Arguments for Incompatibilism}, PB, 26.95 USD.

\section{13-02-11 \ \ {\it Incompatibility's Allure, 2}\ \ \ (to R. {\Schack})} \label{Schack218}

Here's a nice review of the book, in case you haven't read it yet.  This hotel is fantastic!
\begin{center}
\myurl{http://metapsychology.mentalhelp.net/poc/view_doc.php?type=book&id=4752}
\end{center}

\section{13-02-11 \ \ {\it Primary Stuff} \ \ (to A. Zeilinger)} \label{Zeilinger14}

Greetings from a jet-lagged night in Jerusalem.  I am here to deliver a memorial lecture in the honor of Itamar Pitowsky.

I am happy to hear that you have offered a position to Schlosshauer.  Thank you so much; I was very worried about him.  Let's hope this works out to everyone's advantage.

Some small remarks on the deeper stuff in your note.

\baz
I think of {\Schroedinger}, who in his ``Meine Weltanschauung'' regards
himself clearly as a monist. He is an even more radical monist than any of
us, because he believes that we are all part of one single all-encompassing
consciousness!
\eaz

I have read this side of {\Schroedinger} in the past and am very aware of it.  (There is a discussion about it between {\Mermin} and me in my book Coming of Age \ldots\ which, by the way, you should advertise around your group!)  Just to make sure we have a common language for future discussions, though, I want to emphasize the difference between a ``monist'' --- someone who only thinks there is only {\it one thing\/} to the world --- and those people called ``neutral monists''.  The latter should not necessarily be thought of as a species of the former.  ``Neutral monism'' is usually meant as a doctrine that the world consists of {\it one kind of stuff\/} (to put it into contrast, for instance, with the Cartesian mind/matter distinction).  For instance, Mach is sometimes called a neutral monist, but he did not take the world to consist of just one thing (as, for instance, the modern-day Everettians do with their universal wavefunction), but the whole mass of ``sensations.''

My hero William James, went particularly far in the direction opposite to monism (that the world consists of only one thing), and because of that, I almost think the label of ``neutral monism'' is particularly bad in his case.  Thus I tend to call his position ``neutral pluralism'' (along with Ruth Anna Putnam).  This comes about because of these two remarks of his.

First:
\begin{quote}
My thesis is that if we start with the supposition that there is only one primal stuff or material in the world, a stuff of which everything is composed, and if we call that stuff `pure experience,' then knowing can easily be explained as a particular sort of relation towards one another into which portions of pure experience may enter.
\end{quote}
But then a bit later:
\begin{quote}
Although for fluency's sake I myself spoke early in this article of a stuff of pure experience, I have now to say that there is no general stuff of which experience at large is made. There are as many stuffs as there are `natures' in the things experienced. If you ask what any one bit of pure experience is made of, the answer is always the same: ``It is made of that, of just what appears, of space, of intensity, of flatness, brownness, heaviness, or what not.'' \ldots\ Experience is only a collective name for all these sensible natures, and save for time and space (and, if you like, for `being') there appears no universal element of which all things are made.
\end{quote}

A very readable article on neutral monism can be found in the {\sl Stanford Encyclopedia of Philosophy\/}:  \myurl{http://plato.stanford.edu/entries/neutral-monism/}.
I think there is one person mentioned in there that might particularly intrigue you because he builds his neutral monism on the concept of information.  It is a philosopher named Kenneth Sayre (still alive possibly, but 82 years old).  They have a quote of him, for instance, where he says:
\begin{quote}
If the project \ldots\ is successful, it will have been shown not only that the concept of information provides a primitive for the analysis of both the physical and the mental, but also that states of information \ldots\ existed previously to states of mind. Since information in this sense is prior to mentality, but also implicated in all mental states, it follows that information is prior also in the ontological sense \ldots\ \  Success of the present project thus will show that an ontology of informational states is adequate for an explanation of the phenomena of mind, as distinct from an ontology of physical events. [And Sayre adds:] It is a reasonable conjecture that an ontology of information is similarly basic to the physical sciences \ldots\
\end{quote}

\baz
maybe there is a point where we might have some slightly different view.
I personally would see (a) our roles of being agents, and, if I understand
you correctly, (b) physical systems not as something complementary or
mutually exclusive. Rather, I think that we have to find a new vantage point
from which we can unify both into one concept.  \ldots\  So to me, in other
words, when I want to unify the notions of information and reality, I do not
view them as complementary aspects, but as two aspects or features or
manifestations of the same entity. An association which comes to mind is
the fact that we have learned that space and time are the same in a deep
sense rather than complementary.
\eaz

Seeing your remark makes me think that I was probably too sloppy with my choice of the word ``complementary,'' for it certainly does usually carry the connotation of ``mutual exclusivity'' for anyone who has read Bohr.  But I don't think I meant to imply that aspect of Bohr's use of the word.  (Of course, I don't really know what I mean yet---I am most definitely groping in the dark for the right concepts.)  Thus, I think what I do want to say is something closer to an analogue (but with agent and object) of your ``I want to unify the notions of information and reality, I do not view them as complementary aspects, but as two aspects or features or manifestations of the same entity.''

The way James pulls it off, for instance, is to make thought and matter relational aspects of his neutral ``pure experience''.  A given ``pure experience'' when considered in relation to one set of other ``pure experiences'' is called a thought; and when considered in relation to another set of other ``pure experiences'' is called matter.  Moreover, he imagines every gradation in between (I think), and still other aspects in all kinds of dimensions that are neither thought nor matter.

But of course all of this is just the beginning of my thinking on these things (as expressed in my last three or four papers).  I've got a long way to go before I'm going to be able to put some substance to the way I ended my paper \arxiv{1003.5209} where I wrote:
\bq\noindent
       Quantum states, QBism declares, are not the stuff of the world, but
       quantum  measurement might be.  Might a one-day future Shakespeare
       write with honesty,
\bv
               Our revels are now ended.  These our actors,\\
               As I foretold you, were all spirits and\\
               Are melted into air, into thin air \ldots\medskip\\
               We are such stuff as\\
               \hspace{.33in}        quantum measurement is made on.
\ev
\eq

\section{13-02-11 \ \ {\it In Town} \ \ (to M. Hemmo)} \label{Hemmo4}

No need to call.  Thank you for the suggestions.

I've had some sleep, and a bit of a walk outside.  And I think I'm so taken with the tranquility here, that I'd just like to get some work done for the rest of today and tomorrow (in the compound, with only small breaks out into the neighborhood).  If that's OK with you?  I'm rereading Itamar's papers, preparing my talk, and finishing up my paper for you.  What a place to do these things!

Yemima says that I will go to the Dead Sea Thursday.  I will see you Tuesday; is that correct?  Maybe by then my head will be clear enough that I'll starting more broadly about other things to do.

\section{13-02-11 \ \ {\it In Israel?}\ \ \ (to N. Argaman)} \label{Argaman1}

It is true that Norsen's writings get under my skin, so the idea of having to talk about one of his papers for an hour comes with some trepidation.  (See, for instance, our exchange in this panel discussion:  \pirsa{09090098}, as well as my talk here \pirsa{09090087} whose title aims squarely at him.)

I'll make a deal with you.  If you'll read Section 5, ``The Essence of Bell's Theorem, QBism Style,'' of \arxiv{1003.5209} before coming here, we can talk.  (Section 5 alone is adequate; I'm not asking that you read the whole thing.)  There, I try to say as clearly as I can why taking a view of quantum theory that implies the existence of undiminishing, instantaneous causal connections between distant objects carries a far greater danger epistemically than the opposing point of view ever could (i.e., the one of QBism, that quantum measurement acts create their outcomes, rather than reveal pre-existing properties).  Without the idea of an autonomy at some level for all physical systems, how could physics as an analytic science ever get off the ground?  That was Einstein's worry; that is mine.  Giving up the idea of autonomy for physical systems is an immense can of worms, and Norsen closes his eyes to it.  If one gives up on the idea of the autonomy of physical systems, how can one be so self-assured that even experimenters can be autonomous from the systems they are testing?  Anyway, that's what I argue in the paper (and in the talk mentioned above).

If we have a deal, I'll ask Yemima how she has me scheduled for Tuesday, and I'll tell you what time is available for us.

\section{14-02-11 \ \ {\it Corrections?}\ \ \ (to S. T. Flammia)} \label{Flammia12}

I'm up with jetlag, preparing my talk for Wednesday.  Giving one in honor of Itamar Pitowsky, who died a year or two ago.  Funny how little things affect you.  I remembered well the closing lines of the first paper I had read of his.  Seemed like an appropriate way to start the talk.

\bq
\noindent More broadly, Theorem 6 is part of the attempt
to understand the mathematical foundations of
quantum mechanics. In particular, it helps to
make the distinction between its physical content and mathematical artifact clear.\\
\hspace*{\fill} --- Itamar Pitowsky\\
\hspace*{\fill} Closing words of J. Math.\ Phys.\ {\bf 39}, 218--228 (1998)
\eq

\section{16-02-11 \ \ {\it Thursday?}\ \ \ (to I. Belfer)} \label{Belfer1}

Do you know how we're supposed to meet up tomorrow and at what time?

The Dead Sea would be nice, but it's not essential if things get in our way.  I'll follow your advice.  Mostly it'd be nice to talk about the history of quantum information.

\subsection{Roly's Preply}

\bq
My name is Israel (Roly) Belfer. I am a PhD candidate (BIU) working under Prof.\ Silvan Schweber. The subject of my research is the impact of Information Theory and informational terminology on physics.

I was excited to hear that you are coming to speak at the Pitowsky memorial event on the 16th (a friend of mine is the recipient of the prize).
If you have the time, I would very much like to speak with you about issues of information in physics, such as the evolution of QIT (your new book --- which I cannot get here yet); the different approaches to information in science; philosophy-of-physics concerns about information.

If so, I would love the opportunity to buy you a cup of coffee and discuss these topics.
\eq

\section{16-02-11 \ \ {\it Since October} \ \ (to D. B. L. Baker)} \label{Baker31}

I looked in my files and see that I haven't written you since last October.  Pretty pathetic.  Where have I been since then?  South Africa, DC twice, and even Texas once.  And where am I now?  I'm writing from a little caf\'e in the Old City part of Jerusalem, behind the wall, I think (hope) near the Jaffa gate.  Tonight I put on a jacket and tie and give a memorial lecture for a physicist/philosopher/friend who died last year.  But that's five hours from now; at the moment, I'm having a ``premium quality'' Israeli beer \ldots\ if you can believe such a thing (at least that's what it says on its label).  Between typing words I take a bite of chicken schnitzel.  Basic stuff.

Yesterday, after some business, I spent three hours in the in the Israel Museum.  The highlight was a little building called the ``shrine of the book'' where I got to see the Dead Sea scrolls.  I read every word of every display and tried to let my mind float across the centuries.  Funny thing, I thought, ``When I get home, I've got to get the kids to watch {\sl The Ten Commandments\/} again.''

It's really an interesting city; I regret not coming here before.  (I've passed up two opportunities before.)  Tomorrow, I'm taking a drive to the Dead Sea.  More than the salt in the water, I'm intrigued by it being the lowest land point on earth.  (I think 400 meters below sea level.)  I came to Israel 11 years ago, and at that time stayed in Haifa, with visits to Bethlehem, the Sea of Galilee, and the Golan Heights.  Jerusalem is a very different area---much prettier and more meaningful to me.  So far, the only thing a little spooky is watching a soldier or two walk by from time to time carrying a sub-machine gun.  It was well spookier the day I sat outside a caf\'e in Stellenbosch watching a black guy getting beaten with dowels or bamboo or something.  Old write-up of the incident to Kiki below.

\section{17-02-11 \ \ {\it Symmetries of the Magic Measuement}\ \ \ (to J. D. Bekenstein)} \label{Bekenstein1}

Was it perhaps you who asked the question about symmetries at my memorial lecture for Itamar Pitowsky last night?  It's just a hunch that it might have been you; if I mentally erase the mustache I see in your web pictures, it coincides a bit with what I remember of the questioner.

If none of the paragraph above makes sense to you, then I've got the wrong person!  On the other hand, if it was you, and you have some interest, perhaps we could chat for a small time about these mathematical questions while I'm still in Israel Friday.  It was not emphasized last night, but all of this work is a legacy of John Wheeler's imploring for an information-theoretic understanding of quantum theory.  (See attached recent interview.)  [See ``Interview with a Quantum Bayesian,'' \arxiv{1207.2141v1}.]

\section{17-02-11 \ \ {\it Symmetries of the Magic Measuement, 2}\ \ \ (to J. D. Bekenstein)} \label{Bekenstein2}

Thanks for your kind offer of time.  I'm very tempted to take you up on it, even if you aren't the guy who asked the question last night --- it'd be nice to tell you about this math problem I find so consternating, but at the same time holds so much promise as a new formalism for quantum theory.  Besides, I've always wanted to meet you.

However, the more I think about it, the more I think I probably ought to take one last advantage of the historical sites here, to see what kind of insights they might plant in my head.

Would you like to come to the Perimeter Institute sometime to discuss information in physics?  I could bring you as my guest and take care of all expenses, etc.

Attached is an article for a Canadian Physics-Today kind of magazine that explains a bit of the open math problem in its last section. [See ``Quantum Bayesianism at the Perimeter,'' \arxiv{1003.5182}.]

I hope you'll say ``yes'' to the invitation.  As far as I know, you've never visited us at PI.

\section{18-02-11 \ \ {\it On the Difficulties of Communicating across Professional and Philosophical Lines}\ \ \ (to I. Belfer)} \label{Belfer2}

I enjoyed yesterday very much and it was a pleasure to get to know you.  You strike me as having a very sound head on your shoulders, and you gave me some good food for thought.

Here are the things on communicating with Leggett and the like that I mentioned yesterday.  See [this samizdat] at the letters titled ``\myref{Phish1}{Slammed by the Closet Door}'' and ``\myref{Phish2}{November 8th}'' starting at pages \pageref{Phish1} and \pageref{Phish2}, respectively.  Tony Leggett is a very kind and gentle and great man.  A great scholar and a penetrating thinker all around.  And I would never want to hurt his feelings.  But that is not to say he will ever understand the message of the quantum --- there my gamble would be against him.

The Rotman Institute that I was telling you about yesterday can be found here:
\bv
\myurl{http://www.rotman.uwo.ca}.
\ev
I think it would be fantastic to have you in the neighborhood for a couple of years.  Write a good PhD and get yourself noticed!

\section{18-02-11 \ \ {\it The Chances of a Neutral Monism} \ \ (to H. Price)} \label{Price21}

I saw your cryptic job offer a while ago (why was it announced before it was announced?).  Anyway, many, many congratulations!  Here's an issue:  If you become the Bertrand Russell Professor of Philosophy, will you take neutral monism (even neutral pluralism) as seriously as he did?\medskip

\section{18-02-11 \ \ {\it First QBist Church of the Holy Sepulchre}\ \ \ (to H. C. von Baeyer)} \label{Baeyer136}

As I recall, you were never completely happy with either putting the reference SIC up in the sky or putting it in a vault at the Bureau of Standards.  What would you think about putting it in a Holy Sepulchre?  It's a counterfactual measurement in any case and was never truly meant to be approached?  The only thing I don't like about the image of the sepulchre is that it conveys the idea of something dead, whereas the SIC is really something more like the Holy Spirit.

Greetings from {\it somewhere\/} in the Old City of Jerusalem --- I have no idea where.  I have the hope that it is close enough to the original Church that I'll eventually stumble across it.  I've been in town since Sunday to give a memorial lecture for Itamar Pitowsky, who died a bit over a year ago.  The talk turned out pretty nice, I think, or at least I got a lot of praise for it.  I included many anecdotes of Itamar for his family and his nontechnical colleagues, and I tried to give enough technical details to please his spirit up in the sky (and the 15 people in the audience who might understand it).

This has been a place of thought for me, letting my mind rove back to the writing of the Dead Sea scrolls, trying to feel the fear of Masada, watching as Jesus carried the cross down these small streets \ldots\ \ {\it Actually, at the very moment I write this, a fight has erupted at the fountain near me.} \ldots\ I will return eventually from another caf\'e.

\ldots\ OK, it's a few hours later now.  I did find the Church of the Holy Sepulchre, and many other things beside.  At the Western Wall, I told an African couple a story about the time I ran into Arnold Schwarzenegger at the Pasadena flea market.  When a vendor ran after him with a pencil and pad to get an autograph, he said, ``Not now, later, I'll be back.''

And so, I too --- I am back.  I've been wanting to write you for a couple of weeks, but of course things got delayed.  At least this is the best place to write you from anyway.  I told you that I've been following Marcus's therapy.  That's been somewhat successful.  We had good philosophical discussions every day at lunch, and I've launched into a new study of ``neutral monism'' (I hate that term though).  I spent some time reading and contemplating what I was learning from Erik Banks's book {\sl Ernst Mach's World Elements}.  Before this, I really knew very little about Mach --- I was always under the impression that he was a proto- logical positivist, but he really wasn't much like that at all.  A much more interesting man.

Of course, it leaves me with a million questions on Mach's ultimate influence on Pauli.  For instance, I have long known about Pauli's longing for a psycho-physically neutral language for describing physics.  But how much of this is due to his interactions with Jung, and how much more is due to the influence of Mach and Mach's writings upon him?  I had not sufficiently appreciated before that such ``neutral stuff'' was what Mach was seeking/positing with his ``sensations'' or ``world elements''.  Nor had I appreciated how he ultimately wanted to de-anthropocentrize the idea (or at least Banks holds forth enough quotes along these lines to make one think it's true).

Lately I have not been able to get the fake Shakespearean lines I put in the last paper off my mind:
\bq\noindent
       Quantum states, QBism declares, are not the stuff of the world, but
       quantum  measurement might be.  Might a one-day future Shakespeare
       write with honesty,
\bv
               Our revels are now ended.  These our actors,\\
               As I foretold you, were all spirits and\\
               Are melted into air, into thin air \ldots\medskip\\
               We are such stuff as\\
               \hspace{.33in}        quantum measurement is made on.
\ev
\eq

The next step in my education in this regard is to study the work of the ``American New Realists,'' particularly Ralph Barton Perry.  Perry was a student, follower, and biographer of William James (the very best one in fact), who tried to move past his master, particularly, by working to carefully de-anthropocentrize James's ``pure experience''.  I've started the bible of the movement this week, {\sl The New Realism:\ Cooperative Studies in Philosophy}, a collective work by the leading six new realists.  And it'll be my main reading on the flights tomorrow \ldots\ all three of them!  (From a Clash song circa 1982.  ``These are your rights \ldots\  Know your rights \ldots\  All three of them!'')

Where do we go from here?  I don't know at all, really.  I wrote confidently in Schlosshauer's interview:
\bq\noindent
          The first [thing to do] is to find a crisp, convincing way to pose quantum
          theory in such a way that it gets rid of these trouble-making quantum
          states in the first place.  What I mean by this is, if quantum theory is
          actually about how to structure one's degrees of belief, it should become
          conceptually the clearest when written in its own native terms.
\eq
What I mean by that, you know, is to re-write the Born Rule as a statement about probabilities based on a SIC representation --- i.e., the urgleichung.  But does that in any way move us closer to Pauli's psycho-physically neutral language?  This has been on my mind all week.

Suppose I were to get all I wanted out of the urgleichung:  I.e., that its consistency (in the sense of never generating negative probabilities), the identification of priors with posteriors and vice-versa (as we have been doing in the last papers), and a maximality condition, taken with one further mild assumption (yet to be identified), completely specifies quantum-state space.  Suppose I get that holy grail of quantum-state space from the urgleichung predominantly.  Of course I'll be immensely pleased by the simplicity of the product.  It will mean that I have isolated the essence of the formal structure of quantum theory to a single counterfactual situation.  I will have pulled the system's dimensionality to a place of prominence in our understanding --- it is the extent of the deviation from the law of total probability in our counterfactual scheme.

All of this is fine.  But how does it help us get closer to identifying the next step?  That's what I'm not seeing at the moment.  Maybe it's just too distant into the future, and we've got so much to do before then that it's not worth thinking about this {\it yet}.  But what better place to try to see farther than one normally can than in a short visit to the Holy Land?  (How much damage can be done in a short visit?)

Behind quantum mechanics is the idea that ``we could do otherwise'' in any experimental situation.  {\it If\/} all the above is true, through Kochen--Specker type arguments on the final product, we know that in any experimental situation the world can, in return, do otherwise.  I.e., there is no hidden-variable model for our urgleichung considerations.  ``We can do otherwise.''  ``The world can do otherwise.''  Is that the extent of the psychophysical neutral ground Pauli dreamed for?  There's got to be something much bigger, much deeper than this!

And it must play off the particular form of the urgleichung, if it's going to play off anything.  But how so?

How so?  How so??  How so?!?!  I don't want to wait another 10 years for it to come to me.  But I am as blank at the end of this note as I was at the beginning.

Yesterday, I went to Masada with a nice grad student, Roly Belfer, who's doing his PhD with Sam Schweber.  His thesis will be on (roughly) ``information as a style of thinking within physics.''  I very much enjoyed his company.  A cute thing was when he was praising your book on information, I revealed, ``Yes I know the book; we are close friends.''  You should have seen his eyes bug out in only the way the young ones do when they find out something interesting.  It was as if he had met someone who had met a demi-god.

We brought up the possibility of your visiting PI in our last conversation.  What do you think?  It could be really good for giving us re-orientation.  (Marcus needs it as much as I do!)  Marcus will be in Waterloo March 17 to April 9.  And I will be back from the APS March meeting starting on the 27th.  I shouldn't be traveling again until April 28th.  It'd be nice to get the triumvirate together again; we haven't held congress in a while and, no doubt, our philosophies are lacking for it.

Three prayers from the Holy Land!

\section{19-02-11 \ \ {\it Alas!}\ \ \ (to A. Wilce)} \label{Wilce25}

\baw
I hope your talk in Jerusalem went well -- I'll be interested to know what
sort of reactions you provoked.

On a less happy note, in trying to fix what seemed a minor error in the
manuscript I recently sent you, I found the entire thing unraveling. In
fact, it's a tissue of errors, all stemming from one horrifyingly
elementary mistake (the parameter ``s'' in Lemma 3 is generally negative).

I may yet be able to salvage something from this wreck, but for now,
please accept my sincere apologies for troubling you with such a mess! And please pass along both this message,
and my apologies, to Matthew.
\eaw

I'm sorry to hear about this.  I'll keep my fingers crossed that you can salvage a proof, and I'll make sure to buy you an extra beer in Dallas.

The talk went quite well.  The way I wrote H. C. von Baeyer about it:
\bq\noindent
       I included many anecdotes of Itamar for his family
       and his nontechnical colleagues, and I tried to give
       enough technical details to please his spirit up in
       the sky (and the 15 people in the audience who might
       understand it).
\eq

It turned out that Hilary Putnam was not in the audience.  I looked out across the room not knowing that, as I started to tell of the party where I last saw Itamar.  I asked, ``Is Hilary here?,'' mentioning that he too had been at the party.  When I found out that he wasn't, I decided to report the conversation I had had with him that night.  Hilary said, ``Could you hold this for a second,'' handing me his cell phone.  He dug quickly into his wallet for some piece of paper, and then said ``thank you'' as he took the cell phone back.  I nodded.

In depth conversation with the man who sweepingly called {\it all\/} information-oriented reconstructions of quantum theory ``instrumentalism with a post-modern sauce.''  I'll probably never recover from the trauma of that weekend:  I really lost most of the intellectual respect I had for him.  In any case, you can now tell your kids and grandkids, ``Chris, bless his soul, respected me more than Hilary Putnam \ldots\ even after my 4.5 axioms unraveled!''

\section{19-02-11 \ \ {\it Long Time, Far Away, Small World, Big Dreams}\ \ \ (to G. L. Comer)} \label{Comer132}

It's been ages and ages since I've written you---and ages and ages since you've written me.  My excuse is that life falls apart on a daily basis!  (It really does.)  What is yours?  Probably some things are universal.  Where did all the years of calm and solace go to?

Anyway, I had an extra tug to write you properly this week.  I have finally worked my way to that place where you found inspiration already 20 years ago, the place from which you had taught me to write email:  I'm in Jerusalem.  I'm starting this note in a little place where I found that I could buy a pint of Guinness, but I doubt I'll finish it here (the note, that is).

I'm in town---my first time---because Itamar Pitowsky died a bit over a year ago and his family invited me to give the first memorial lecture in his honor.  They're hoping it'll be a yearly thing hereafter.  I was very touched that I had meant that much to Itamar, and I hope that I did everyone a good job.

What a beautiful, fascinating city.  I don't believe I've ever seen so much stone in my life.  I've certainly never seen so many caves in the distance, as I saw on my way to Masada today.  And then more stones and more stones once I was there.  Everywhere!  It's crazy, but I sing a different song in my head for every major city I go to.  I do it every time I arrive in a place  and can't seem to turn it off until I head back home.  In Chicago, I always sing, ``Chicago, Chicago, that toddlin' town.''  In Tokyo, I sing, ``Nature shows again and again the folly of man.  Godzilla!''  Here I've been singing ``Flintstones, meet the Flintstones.''  I haven't meant any disrespect by it, and even (one side of) my own mind seems bothered by my singing it, but it's nothing to do with the culture, the people, or anything like that.  It's all about the stone!!  In a way, even their most modern buildings look like careful piles of stone---they all carry the spirit of Masada, and probably something much deeper still.

On the small-world side of this note, I have to report that I met and had lunch with your old friend Daniel Rohrlich Tuesday.  It was amazing; I remembered his face as he was walking toward me and knew who I would be talking to.  He introduced himself by saying that we have a mutual friend.  I said something like, ``Yes, we met a couple years back and we talked about Greg then.''  He didn't remember meeting me before!  I thought to myself, ``I guess I'm not much of a guy to remember.''  Well, we sat down for the talk and instead of listening, I started digging through my email to find when we actually had met --- you know how I keep records.  To my great shock, it wasn't a ``couple of years ago'' that we met, but in 1997!!!  13 years ago!  And for me it's all compressed into all the other blurs of my life, so much so that I had no clue of all the years that had gone by.

Sad really.  At least I'm glad I still hold on to some things and thoughts, like my dreams for the quantum.  Honestly, I wonder how long that'll last!?  I know you're going to think me a lunatic, but I wrote the note below to Jacob Bekenstein a bit earlier tonight, at the restaurant where I had dinner before arriving at this joint.  What a friggin' fool would give up a chance to meet Jacob Bekenstein?

It's part of the set of symptoms I wrote you about in the first paragraph.  There's a certain sense in which I've been running overload for about three years now.  For instance, I literally travelled over 100,000 miles last year (excluding my uncounted trips to DC to manage my Navy money).  That's four times around the earth, if you can believe.  And what great insight did I get in return?  Maybe something here or there---I can't remember (that's the point)---but certainly nothing to the depths of the old days.

Why do these things?  Because I'm a corporation now, I suppose.  My life is a friggin' corporation.  ME has been lost in a thousand ways.

But I digress!  This is no doubt a city where insight awaits in every
alley, every corner, every turn, every caf\'e, every honk in the
street, every piece of cut STONE, ready to jump out at me.  I have to
return!  I feel it in the air here.  My strength has always been in
combining every personal incident of my life, profound or trivial,
into my physics.  I want to get back here as soon as possible.  It's a
place different than all the rest, and I envy the Greg who got an
early start of it so long ago.

Write me back as you get a chance.

Attached is a piece I've just written for a book of interviews Max Schlosshauer is putting together.  Some parts of it might amuse you, particularly the place where I try to take the wind out of the Oxford professors (at Q7).  I wonder if it'll remind you of our old days together?

\section{20-02-11 \ \ {\it Monism Is the Devil}\ \ \ (to H. C. von Baeyer)} \label{Baeyer137}

Good morning, this time from somewhere in my kitchen.  The world is a kind of blooming, buzzing confusion this morning; everything is kind of shiny out of the sides of my eyes.  I didn't get back to Waterloo until 2:30 this morning --- 31 hours of travel total!  It was a tortured route to begin with (just to get American Airlines miles), but then all the delays \ldots\

I'm sorry to have caused you confusion.  That damned term---``neutral monism''---is trouble, and your letter, along with an earlier one from Anton, convinces me to banish it forever and never use it again!  Hereafter ``neutral monism'' shall be a dead term for me.  I should have explained to you like I explained to Anton after I saw his own confusion.

Here follows an excerpt from the note to Anton.  \ldots\  Actually, let me just copy in the whole note below.  [See 13-02-11 note ``\myref{Zeilinger14}{Primary Stuff}'' to A. Zeilinger.]  The hatched parts, of course, correspond to quotes from his earlier note to me.  In a nutshell the trouble is that ``monism'' is the doctrine that the world is ``one thing,'' whereas ``neutral monism'' is associated the idea that the world is composed of ``a stuff of one kind of character'' (even though it may be plural to the bones).  See the quotes from James below to see how that fits in with his pluralism.  Ruth Anna Putnam calls it ``neutral pluralism,'' but I don't like that term so much either.  It remains a bit lifeless in comparison to the heavy duty it is meant to be called to.  But if ``neutral pluralism'' is lifeless already, ``neutral monism'' is surely dead, dead as doornail.

Trying to find the appropriate term has word for whatever is going on here has been a real thorn in my side, even before this nail in the grave of yours.  That was the cause of my joking note with the word QBoom.  On a more serious note, Marcus has toyed with urspracht \ldots\ but I don't think I like that one either (or at least he hasn't convinced me yet).

Let me know if all this relieves the confusion I induced.  If my family will let me go, I'd like to write you a bit more on the subject.  But now they've kicked me off the table so they could set it for breakfast.

What's your physical location now?

\section{20-02-11 \ \ {\it Urspracht}\ \ \ (to H. C. von Baeyer and D. M. {\Appleby})} \label{Appleby101} \label{Baeyer138}

From our emails this morning.
\begin{description}
\item[Chris said:]  ``Trying to find the appropriate term for whatever is going on here has been a real thorn in my side, even before this nail in the grave of yours.  That was the cause of my joking note with the word QBoom.  On a more serious note, Marcus has toyed with urspracht \ldots\  but I don't think I like that one either (or at least he hasn't convinced me yet).''
\item[Hans said:]  ``The word urspracht is gibberish.  What do you and Marcus mean?''
\end{description}
Well, if it is gibberish, that might be a partial selling point for it (as you'll see in Marcus's last paragraph, which is a point I definitely agree with, about ``keeping the bastards guessing''):
\bq
But I think on reflection that the best option might be to avoid presenting any kind of fixed target for the philosophical labeling guns.  The trouble with the two above suggestions is that they might lead to us being labeled as ``mystics''.  And I no more want Hilary Putnam saying that we are just a pair of mystics, than I want him saying we are just a pair of empiricists.  I want to keep the b***** guessing.  Whilst at the same time being totally clear, and totally explicit.
\eq
But anyway, below is Marcus's original note on the sort of thing he was after, at least at that time.  (Which, there is probably no need to say, may not be exactly the same thing that I'm trying to get after.  But what I am looking for may have a lot of elements of what he is after too---I'm not sure---it's all so much confusion at the moment.)

\section{21-02-11 \ \ {\it Einstein on Religion; Pragmatism on Spinoza}\ \ \ (to {\AA}. {\Ericsson})} \label{Ericsson16}

I greatly enjoyed reading through these quotes.  My guess is you won't as I did; but still I thought I'd make you aware of them:
\myurl{http://www.stephenjaygould.org/ctrl/quotes_einstein.html}.

A lot of Einstein's arguments would be my own.  On the other hand, Spinoza's worldview in its details is certainly foreign to me.  I like the way Ralph Barton Perry put it as the antithesis of pragmatism:
\bq
The perfect antithesis to pragmatism is Spinoza, and it is the perpetuation of Spinozism in objective and absolute idealism that is the real object of the pragmatist attack.  Absolutism is other-worldly, contrary to appearances; pragmatism [is] mundane, empirical.  Absolutism is mathematical and dialectical in method, establishing ultimate truths with demonstrable certainty; pragmatism is suspicious of all short-cut arguments, and holds philosophy to be no exception to the rule that all hypotheses are answerable to experience.  Absolutism is monistic, deterministic, quietistic; pragmatism is pluralistic, indeterministic, melioristic.  That which absolutism holds to be most significant, namely, the logical unity of the world, is for pragmatism a negligible abstraction.  That which for absolutism is mere appearance --- the world of space and time, the interaction of man and nature, and of man and man, is for pragmatism the quintessence of reality.  The one is the philosophy of eternity, the other the philosophy of time.
\eq

\section{22-02-11 \ \ {\it Flesh and Blood}\ \ \ (to R. W. {\Spekkens})} \label{Spekkens104}

Here is that line I was proud of:
\bq
What is the source [of truth-value assignments]?  When it comes to formal logic, one is tempted to think of it as the facts of the world.  The facts of the world set truth values. But it is not the world that is using the calculus of formal logic for any real-world problem (like the ones encountered by practicing physicists).  The ``source'' is rather a finite subscriber to the service, one with limited abilities and resources; the source is always one of us---flesh and blood and fallible through and through---the kind of thing IBM Corporation is taking its first baby steps toward with its {\it Jeopardy!}-playing supercomputer Watson.  The source of truth values in any application of logic are our {\it guesses}.  Thus, it would be better to be completely honest with ourselves:  Applications of formal logic get their truth values from an {\it agent}, pencil and paper in hand, playing with logic tables not so differently than crossword puzzles.  The facts of the world only later let the agent know whether his guesses were acceptable or unacceptable judgments.
\eq

I agree that I should read more Dennett (I have read some) \ldots\ I know there's insight to be gained there.  But I read a little piece in Perry yesterday about how Spinoza is the ``perfect antithesis'' to pragmatism (I'll copy it below), and I think it quite captures my pre-conceived notion of Dennett, i.e., that he is my antithesis.  [See 21-02-11 note ``\myref{Ericsson16}{Einstein on Religion; Pragmatism on Spinoza}'' to {\AA}. {\Ericsson}.]  I'm aware that this is what is providing the friction to my reading.

\section{22-02-11 \ \ {\it Solvay Conferences}\ \ \ (to I. Belfer)} \label{Belfer3}

Thanks for the Galison paper; it looks like it is going to be a lot of fun.

The Jammer paper I was referring to was: Max Jammer, ``The EPR Problem In Its Historical Development,'' in {\sl Symposium on the Foundations of Modern Physics:\ 50 Years of the Einstein--Podolsky--Rosen Gedankenexperiment}, edited by P. Lahti and P. Mittelstaedt (World Scientific, Singapore, 1985), pp.\ 129--149.

Jammer writes,
\bq\noindent
\ldots\ what Einstein had in mind is confirmed by a letter which
Ehrenfest wrote to Bohr on July 9, 1931.  As Ehrenfest
reports, Einstein uses the photon-box no longer to disprove the
uncertainty relation but ``for a totally different purpose.''  For
the machine, which Einstein constructs, emits a projectile; well
after this projectile has left, a questioner can ask the machinist,
by free choice, to predict by examining the machine alone either
what value a quantity A or what value an even
conjugate quantity B would have if measured on the projectile. ``The
interesting point,'' continued Ehrenfest, ``is that the
projectile, while flying around isolated on its own, must be able of
satisfying totally different non-commutative predictions without
knowing as yet which of these predictions will be made \ldots''
\eq

The contrary story comes from Don Howard's translation of the Ehrenfest letter:
\begin{center}
\myurl[http://www.nd.edu/~dhoward1/Early\%20History\%20of\%20Entanglement/sld026.html]{http://www.nd.edu/$\sim$dhoward1/Early\%20History\%20of\%20Entanglement/sld026.html}.
\end{center}
Don has a more extended discussion of this point in one of his papers but I can't find it in my files at the moment.  I think you should be able to find it on his website.

\section{23-02-11 \ \ {\it The Very Improbable Tamworth Iron Works}\ \ \ (to H. C. von Baeyer)} \label{Baeyer139}

A most improbable thing happened to me yesterday, and I've been racing with adrenaline in my veins ever since.  It's a long story, which I will tell below \ldots

\ldots\ But first an introduction from Linda Simon's book {\sl Genuine Reality\/}:
\bq
In September [1886], James set out on a real estate search that ended at the base of Mount Chocorua, in a town called, infelicitously according to James, Tamworth Iron Works.  The ``small farm'' that interested him consisted of seventy-five acres, most of it woods, the rest hayfields.  The shingled, green-trimmed farmhouse, built in the year James was born, needed considerable work, but it was large enough---fourteen rooms, eleven with doors that opened to the outside---included a big barn, and looked out at Chocorua Lake and the mountain.  ``It is only 4 hours from Boston by rail and 1 hour's drive from the station,'' William told Henry.  ``Few neighbors, but good ones; hotel a mile off.  If this is a dream, let me, at least indulge it a week or so longer!''
\eq

The very improbable thing that happened to me yesterday was that I set out to make arrangements for the MIT meeting May 3rd and 4th, and I got distracted by the hotel address.  ``Is that the street William James lived on?''  So I did a Google search on ``william james'', ``home'', and ``cambridge''.  ``Nope, he lived on Irving Street; I should have known that!''  But one of the other things that came up was a YouTube video of the house.  Curiosity got me, and I watched it.  Kind of creepy really; just someone standing outside with his video camera.  But, you know, curiosity sometimes doesn't know what's best for it.  I thought, ``I wonder if the same guy took a video of the Chocorua house?'' and so changed ``cambridge'' to  ``chocorua'' in the Google search.  He didn't, it turned out, but I did find a video by a real estate company instead!

What a place it turned out to be.  4,387 square feet, 44 acres, barn, swimming pool, and woods, woods, woods.  Have a look yourself: [\ldots]  If you click on the little down arrow at the address, a box will open with details on the house.  And if you go to this website, [\ldots] you can get a satellite view of the property, right next to state forest and national conservation ares.  Zoom out enough, and you'll find the nearby lake.  And if you go to this wiki link \myurl[http://en.wikipedia.org/wiki/Chocorua,_New_Hampshire]{http://en.wikipedia.org/wiki/Chocorua,\underline{ }New\underline{ }Hampshire} you'll see it described as ``among the most beautiful lakes in the White Mountains.''  Half hour drive from good skiing (lots of skiing in fact), and an hour and a half drive from the seashore, similarly to Portland, etc.

Their asking price had been \$895,000, but it's off the market.  Curiosity kicked in again, and I called the real estate company.  Why off market?  The agent I talked to said that she would talk to her colleague who had listed it and get back to me.   She said one possibility was the owners simply changed their minds about selling.  Another was that when Winter encroached and it had not sold yet, maybe they took it off the market till Spring when the weather would be better.

Anyway, all of that was enough to spur all kinds of crazy thoughts.  My imagination has been running wild all night.  Ever since Kiki and I uncorked a bottle of wine in 2001 and sat down dreaming over Peter Robertson's book of the early years of the Niels Bohr Institute (including floor plans of the place), we've had a fantasy of a little institute where big things might happen.  And wouldn't you know it, the details of the fantasy were precisely this:  An old farmhouse somewhere in New England, with some forested land to build a few bungalows and a small conference center on.  Kiki would be the caretaker and social director; I would manage the science and philosophy and, somehow, someway, bring the money in.

I remember Dr.\ Ruth Westheimer in the 1980s saying, ``Fantasy is OK, even encouraged, as long as it's not actualized.''  The trouble now is that this {\it fact\/} hits so close to the {\it fantasy}!  I had never imagined that that much of the James land was still intact.  44 acres in an idyllic setting!  And surely of all places on earth, this one is the very spiritual home of QBism.

I've just been overwhelmed emotionally.  How, how, how could I make it happen?  Or at least buy the damned thing for myself and the family, until somewhere down the road a real institute might materialize?  Or even just keep the thing off the market for a few years \ldots\ so that some ``investor'' doesn't get the idea of making it a two million dollar home before I can get to it?

Easy enough, right?  I just sell this house, and buy that one.  Yeah, right!  The trouble with a house is that it ought to be close to one's source of income!  And what income would I have without the Perimeter Institute.

Unfortunately, even my craziest ideas aren't crazy enough.  For instance, I had a little fantasy that I would pay my \$200K of it for my in-laws (to sweeten the pot); they sell their house outside Munich and move back to New Hampshire, with the agreement that we eventually inherit it from them, instead of their German house.  (My father-in-law had started as an assistant professor of history at UNH in Durham, before joining the business world; they still have lots of old friends in Durham.)  But two things count against the idea:  1) I expect them to live a long, long time, and 2) even though my mother-in-law has talked of moving back to the States from time to time, they haven't precisely because my father-in-law loves Germany too much.

Of course, it could be that the house will never come back on market anyway.  It is curious the date the house went on market:  27 August 2010.  The very day after the 100th anniversary of William James's death.  (He had died in the house, by the way.)  Two weeks before that, there had been a symposium in Chocorua, with house tours, etc.  Thus it has struck me that it could  be that the owners put it on market because of the enthusiastic reactions they were getting from the participants, and they thought they might feel out how much money they could get.  Maybe they learned that they couldn't get nearly so much in that area; maybe they learned that they could get a lot more if they play their cards right and market it appropriately.

Anyway, dreaming, dreaming.  ``If this is a dream, let me, at least indulge it a week or so longer!''

There, I had to get that off my chest!  The question now is do you have any ideas?  Do you know of any philanthropists who might be convinced to help finance an institute on pragmatism and the quantum?  Do you know of any physics or philosophy departments near Chocorua that it would be fruitful to inquire into?  (Perimeter tenure coming?  Who would care about Perimeter who could live and think in William James's home?!)  Do you have any wilder ideas still?  A thought on how to make a QBist commune viable?

But dreaming, dreaming \ldots\ \  Now with too much adrenaline and {\it too much coffee}!!

Oh, long note on ``neutral stuff in the pluriverse'' coming soonish; it's been percolating in my head since the weekend.  (We need a better name than ``neutral stuff''!)

\section{23-02-11 \ \ {\it Whitehead on James, 2}\ \ \ (to A. Hamma)} \label{Hamma1}

It was nice to learn of your philosophical inclinations the other day.  I don't have {\sl Process and Reality\/} in front of me to find the lines where Whitehead said that his effort was to extend and make precise James and Bergson.  But below are some quotes I sent to Appleby on Whitehead's regard for James.  [See 21-09-10 note titled ``\myref{Appleby94}{Whitehead on James}'' to D. M. {\Appleby}.]

Attached also is a little interview I wrote up recently.  In the answers to Q2 and Q10, I made the point that you seemed to like.

Let's indeed talk sometime.  I'd love to learn more about Whitehead's thought.

\section{24-02-11 \ \ {\it Whitehead on James, 3}\ \ \ (to A. Hamma)} \label{Hamma2}

No, I didn't know that.  Bryce had said it when I was an undergraduate taking a class on Lagrangean mechanics from him. It was just something that caught my ear then.  If you find the precise Whitehead quote, please send it my way.  I enjoy collecting such things.

\subsection{Alioscia's Preply}

\bq
Thank you for sharing the interview. I am reading it with great interest.

By the way, do you know that what Bryce DeWitt said about mathematics is itself a quote (maybe involuntary) from Whitehead?  Whitehead said the same thing in the introduction to algebra, if I recall. It is interesting that also Hannah Arendt shares a similar opinion. It would be great to find some time to talk about this and other philosophical topics!
\eq

\subsection{Alioscia's Reply}

\bq
This is the quote I promised. It is in {\sl Introduction to Mathematics}, around page 60 (depending on the editions). Whitehead is talking about the importance of symbolism in mathematics. He is showing an example in algebra where identities are very easy to state symbolically while they are difficult to explain in words. Then he says:
\bq
This example shows that, by the aid of symbolism, we can make transitions in reasoning
almost mechanically by the eye, which otherwise would call into play the higher faculties of the brain.

It is a profoundly erroneous truism, repeated
by all copy-books and by eminent people when
they are making speeches, that we should
cultivate the habit of thinking of what we are
doing. The precise opposite is the case.
Civilization advances by extending the number of important operations which we can
perform without thinking about them. Operations of thought are like cavalry charges in
a battle --- they are strictly limited in number, they require fresh horses, and must only
be made at decisive moments.
\eq
\eq

\section{24-02-11 \ \ {\it Whitehead on James, 4}\ \ \ (to A. Hamma)} \label{Hamma3}

And here are the lines from the preface of {\sl Process and Reality}.  My false memory made it much more detailed than it really was.

\bq
Among the contemporary schools of thought, my obligations to the English and American Realists are obvious.  In this connection, I should like especially to mention professor T. P. Nunn, of the University of London. His anticipations, in the {\it Proceedings of the Aristotelian Society}, of some of the doctrines of recent Realism, do not appear to be sufficiently well known.

I am also greatly indebted to Bergson, William James, and John Dewey.  One of my preoccupations has been to rescue their type of thought from the charge of anti-intellectualism, which rightly or wrongly has been associated with it. \ldots
\eq

It's good to know one's intellectual grandfathers.  As a coincidence, I had only recently started studying the ``American New Realists'' (to which Whitehead is undoubtedly referring to), particularly (at the moment) the writings of Ralph Barton Perry.

\section{24-02-11 \ \ {\it The Manic and the Depressive}\ \ \ (to H. C. von Baeyer)} \label{Baeyer140}

If Gleick could get away with {\sl The Information\/} for the title of his book, I figure I can get away with using inappropriate definite articles in my own titles.  (This is not stroking:  I don't like the title.  It could be that when I read the book, I will find it to be very natural and suggestive for the work, but at the moment, to my eye, it sticks out like a sore toe.)

Anyway, to THE subject.  Oh, a soberer mood started to set in last night while Kiki and I looked over the house plans and details, and I started---as I knew I would eventually---to admit defeat to myself.  With honesty, I am no position to see this kind of thing through at the moment.

Any institute would have to be three, four, five, six years down the road.  It would take so much effort to build the idea, the support, the funding, the logistical infrastructure.  The real concern for me (the one that put me into the manic panic) is that the property be secured before some feckless investor or vacation-homer buy it and do more damage.

You are right that the essential idea is James's aura.  That land and house and barn that he purchased for \$750 is the geographical center of his aura.  He died in the house; it was where he wanted to die.  If ghosts live, as James had so hoped to gain evidence for with his dozens of interviews with mediums and his presidency of the American Society for Psychical Research, then his ghost lives in that house.  The left horn of my own dilemma is that I feel I am possibly the only person alive who could channel that ghost to the good of physics.  It may be lunatic, but it is a responsibility I feel.  But the sharper right horn remains:  I am in no position to buy that property on the fly like this.

Your question about the architecture is not lost on me, and certainly not Kiki.  We have pored over the photos and the listing, looking for clues of this and that.  There are certainly places where we can see that the ceiling has been dropped (kitchen being the most obvious example, but also the red-walled foyer).  On the other hand, it looks that quite some of the original does remain.  Only direct eyesight, and Kiki's expertise, with a physical visit will give me a truly accurate picture though.

You might have a look yourself.  The realtor posted the old listing for me: [\ldots]  The little camera at the top of the webpage (for viewing photos) appears to be redundant---there are no more photos than are on the first page---but if you click on the paper clip, you'll find house plans and a survey of the farm.  Here's a sketch of the outside from 1902 for some comparison:  
\myurl{http://www.uky.edu/~eushe2/Pajares/JamesChocorua.html}.
And this photo gives some sense of how the stone walls and hay field have stayed the same since James and Royce gave a similar pose in 1903:
\begin{center}
\myurl{http://www.flickr.com/photos/wjsymposium/5269017904/in/set-72157625617564916/}.
\end{center}

On the other hand, the James house is clearly not preserved as accurately as this other (though much younger) house in the area: [\ldots]
The wood trim really speaks for itself.  (And---though it pierces my heart---Kiki has made it clear to me that she finds this house much more interesting than James's own.)

One thing that becomes clear by comparing the two listings (and some others), is that the owners of the James home know that they are selling his ghost more than the house.  The +\$270,000 / $-1274$ sq ft differences say a lot.  (But there is a $-33$ acre difference too, and maybe I should give that more credit.)

I also know by direct experience that your remark, ``starting with a building often ends in disappointment,'' is accurate.  But then I remind myself of your other remark that the issue is an aura, not a building.  Maybe I am particularly sensitive to this (or particularly illusioned by this) because it all comes just after my visit to Jerusalem's Old City.  In the end, it is not as much the building as the ghost.  Think of how holy the Western Wall is,  the mere remnant of a temple though it be.

Woe is me.  Kiki and I will certainly go see the thing if it is still up for grabs by the time the March Meeting comes and goes.  But it'll almost surely be a painful act of self-flagellation.

The only Rutherfordity I know is, ``To any young man who utters the word `universe' in my laboratory, I ask him to pack his bags and leave.''  (Or something along those lines.)\footnote{\editornote John Wheeler's version is, ``Rutherford, it is well known, did not trust theoretical men. `When a young man in my laboratory uses the word \emph{universe,}' he once thundered, `I tell him it is time for him to leave.' `But how does it come,' he was asked on another occasion, `that you trust Bohr?' `Oh,' was the response, `but he's a football player!'\,''  See J.\ A.\ Wheeler, ``Niels Bohr, the Man,'' Physics Today {\bf 38}, 66 (1985).}

On your suggestion about the James Society, one would think that they would have already been in action, given their symposium in Chocorua last year.  Delving further into the documentation the realtor made available, I note that the owners actually put the house on the market three days before they gave the guided tour of it.  Surely the appropriate people were all abuzz with the news, and the sellers wanted it that way.

\bhcvb
The place seems to be halfway on a straight line between my
brother-in-law in Kennebunk, ME, and my cousin in Weybridge, VT.
A lovely place.
\ehcvb
You never thought about moving closer to your family, did you?

\section{25-02-11 \ \ {\it Home and Home}\ \ \ (to W. G. {\Demopoulos})} \label{Demopoulos45}

I'm glad to hear that you're back home!!  It's much earlier than I expected actually.  I wish you the best possible recovery.

Let me give you some light reading for your iPad; maybe some of it will make you chuckle.  (My guess is you're spending time in bed mostly?)  The first attachment below answers your question about my talk in Jerusalem.  The next two (of many more than that that I won't bore you with) detail a bit of my crazy quest to find some way to buy William James's Chocorua, New Hampshire home.  If you have any ideas, feel free to suggest!  [See 23-02-11 and 24-02-11 notes ``\myref{Baeyer139}{The Very Improbable Tamworth Iron Works}'' and ``\myref{Baeyer140}{The Manic and the Depressive}'' to H. C. von Baeyer.]

Somewhere in my reading in Israel I made a note on ``effects'' (either from Schiller or Perry or, perhaps, the new realist consortium) that I wanted to send to you.  I will try to dig it up tomorrow.  (Unfortunately my wife is forcing me to ski this morning \ldots)

\section{25-02-11 \ \ {\it The Biggest and Best}\ \ \ (to I. Belfer)} \label{Belfer4}

\birb
Can you think of indispensable highlights of the quantum-information
era (like your statement about the inauguration of QIT as a real field
after 1993)?
\eirb

Off the top of my head:
\begin{itemize}
\item[1973:]  Alexander Holevo proves his famous bound that limits the information that can be retrieved from a quantum system

\item[1979:]  Holevo points out that the classical capacity of a quantum channel is nonadditive; Holevo's discovery was essentially unrecognized, until it was rediscovered by Asher Peres and Bill Wootters in 1991, when it became the problem that motivated quantum teleportation later in 1993

\item[1982:]  Richard Feynman's paper ``Simulating Physics with Computers''  (in my opinion still one of the speculations closest to being correct on what powers quantum computation)

\item[1982:]   Wootters and Zurek, and independently Dieks, point out the quantum no cloning theorem

\item[1984:]  Bennett and Brassard's first quantum key distribution protocol

\item[1985:]  David Deutsch's paper ``Quantum theory, the Church-Turing principle and the universal quantum computer'' (first example of a non-quantum-simulation problem that gets a speed up from quantum equipment)

\item[1992:]  Ben Schumacher invents the idea of a qubit on a drive to the Columbus, OH airport, in conversation with Bill Wootters; finally appears first in a 1995 paper (with a 1993 submission date), but had long been a common term in the community by then

\item[1993:]  Quantum Teleportation

\item[1994:]  Peter Shor's factoring algorithm (many would mark the beginning of the field here; but I stick with 1993 as being deeper conceptually)

\item[1995:]  Peter Shor and independently Andrew Steane discover the ideas of quantum error correction and fault tolerance (enabling the idea that quantum computers can actually be built)

\item[1995:]  Cirac and Zoller's paper ``Quantum Computations with Cold Trapped Ions'' (the first serious implementation for quantum computation ever proposed)

\item[2000:] Robert Raussendorf and Hans Briegel's paper ``Quantum Computing via Measurements Only''

\item[2001:]  Fuchs's paper ``Quantum Foundations in the Light of Quantum Information'' points out the true path to progress in the quantum foundations debate!  \smiley
\end{itemize}

At least that's what comes to mind as I have my first cups of coffee this morning.

\section{26-02-11 \ \ {\it Feelings of Guilt} \ \ (to the QBies)} \label{QBies35.1}

\noindent PPS\@.  I refined the definitions of QBoom and QBlooey (with no off color jokes this time either):

\begin{itemize}
\item
{\bf QBoom} -- cf.\ {\it kaboom}; the sought-for deanthropocentrized distillate of quantum measurement that QBism imagines powering the world.  William James called it ``pure experience,'' where ``new being comes in local spots and patches which add themselves or stay away at random, independently of the rest.''  John Wheeler asked, ``Is the entirety of existence, rather than built on particles or fields or multidimensional geometry, built on billions upon billions of elementary quantum phenomena, those elementary acts of `observer-participancy'? \ldots\ Is what took place at the big bang the consequence of billions upon billions of these elementary `acts of observer-participancy'?''  In place of a Big Bang, the QBist wonders whether it might not be myriads and myriads of little QBooms!

\item
{\bf QBlooey} -- cf.\ {\it kablooey}; a QBist slur on the usual conception of the Big Bang, where the universe had its one and only creative moment at the very beginning.
\end{itemize}

\section{27-02-11 \ \ {\it Home and Home, 2}\ \ \ (to W. G. {\Demopoulos})} \label{Demopoulos46}

By the way, not to worry, the version of the story I told at my talk, was a lighthearted, gentle version targeted to make people laugh.  Everything said with a smile.  I didn't say all the business about the postmodern sauce and his newfound Bohmianism, etc.

You are correct that the instrumentalism remark was confined to the conference.  Sorry to conflate two stories into one in the email I forwarded to you.  I wasn't bothered by the charge of instrumentalism; it was the ``postmodern sauce'' that got under my skin.  It's a general symptom of philosophers that they ignore that there are NEW equations and theorems behind this interpretative effort.  And to hear the charge of it being a ``postmodern sauce'' just went too far.

But now you catch me off guard yourself:
\bwd
I'd be very surprised if you can get anything out of neutral monism but would love to talk with you about it; in my view it's not that different from what underlies ``flash ontologies.'' What is distinctive about the BDP point of view is its eschewal of ontology, its embrace of a moderate instrumentalism which, however, is {\bf not} anti-realist.
\ewd
Flash ontologies?!?!?!?  Not at all!!!!  You should expect something much more subtle from me than that by this point in our relationship!  I can see I've got my work cut out for me to try to explain myself (more precisely explain my desires).  Alright, I take the challenge.  But you'll have to wait some time for me to compose myself.

Stay away from germs for the next 83 days!

\subsection{Bill's Reply}

\bq
Looking forward to hearing more from you on neutral monism.

I don't think NM's connection with Flash Ontologies is obvious, and I may be very mistaken about this. But the development of NM in {\sl Analysis of Matter} by Russell is very suggestive of a connection. I appreciate that you don't care for Russell and that you're more interested in James's version. But R's development tries hard to spell out James's metaphors. Let me know what you come up with.

As for Hilary's remark about ``PoMo sauce,'' I'm not sure what he had in mind. Perhaps the notion that instrumentalism can be derived from a more general relativism? I think this is not entirely alien to your view of it and its connection with personalism, but I say this only to insert a further sense of urgency in you to write further on this stuff (pardon the pun)!
\eq

\section{27-02-11 \ \ {\it Pragmatism in Sweden}\ \ \ (to {\AA}. {\Ericsson})} \label{Ericsson17}

\ldots\ a gift for Sunday morning!

[See: Sami Pihlstr\"om, ``Nordic Pragmatism,'' Euro.\ J. Pragma.\ and Amer.\ Phil.\ {\bf 2}(1); online at \myurl{http://lnx.journalofpragmatism.eu/}.]

\section{27-02-11 \ \ {\it SAAP Membership}\ \ \ (to W. T. Myers)} \label{Myers1}

How I wish I could attend your annual meeting, now that I've discovered the society.  But March 13, I head to Tulane U to give their Clifford lectures this year, and then immediately after that it's to Dallas for the big American Physical Society meeting---I organized the part for the Topical Group on Quantum Information this year (being the chair), and we're having over 400 talks!  The point is, though in principle I could get to the SAAP meeting, it hits at an awfully bad time for me.  I'll try my best to be at next year's meeting.

Seeing the link to your webpage below, I got snoopy, and I'm glad I did.  It's good to meet another University of Texas alumnus.  UT Austin is without doubt my spiritual home.  (I tell some of the story in my answer to Question 1 in the interview I sent you yesterday.)

I'm also quite intrigued by your research.  Where I am headed with my own is to develop a kind of ``process physics'' behind quantum measurement theory, though I haven't been so much in the habit of calling it that.  (I do so for you.)  You can see some of the yet-very-vague thoughts I am having at the end of this recent paper of mine, \arxiv{1003.5209}, and also at the end of the interview.  They will certainly be beefed up one day or another.

I would be much appreciative if you could send me a copy of your PhD dissertation.  From the abstract you post, it seems to be just what the doctor ordered.  An electronic version of it would be just fine.

\section{27-02-11 \ \ {\it Lenovo Machine}\ \ \ (to C. M. {\Caves})} \label{Caves106}

Random things now.

1)	I found these words in some old correspondence from you a couple days ago, and I really liked the way you put it:
\bq\noindent
[QBism takes] a three-pronged approach to subjectivity and objectivity: (i) quantum states and probabilities are wholly subjective; (ii) system attributes are wholly objective; and (iii) measurement outcomes are where the rubber meets the road, i.e., where subjective and objective meet to produce something that is not under the control of the agent, but is also not out there in the world.
\eq
Very eloquent.

2)	I really wish you were more a fan of William James and old-school pragmatism.  I discovered last week that his farm in Chocorua, NH will be on the market in the Spring (it was on the market in the Fall, and taken off for the Winter).  It's the house in which James died.  The 4387 sq ft house is relatively historically intact (though a lot of updates) and barn still there, but with swimming pool now in addition.  44 acres of land, 39 of it thick forest, and beautiful hay fields and rock walls still there.  Quiet secluded place.  Perfect for a QBistic Institute!!  (Imagine carving 5 little bungalows into the forest, a lecture room, etc.)  20 minutes from good skiing.  1.5 hours to Portland, ME and Portsmouth, NH.  \$895,000 asking \ldots\ but probably overpriced by \$100K.  I've been wracking my brain trying to figure out a way I could secure the place till said institute could become a reality.  Know any philanthropists with deep pockets for pragmatism or QBism or both?  Seriously; I'm not joking.

Glad to hear you're back in Oz.  You always seem happiest there.

\section{28-02-11 \ \ {\it Home and Home, 3}\ \ \ (to W. G. {\Demopoulos})} \label{Demopoulos47}

\bwd
I think this is not entirely alien to your view of it and its connection
with personalism, but I say this only to insert a further sense of
urgency in you to write further on this stuff (pardon the pun)!
\ewd
Sadly, I didn't catch the pun!  What is it?

\bwd
Russell, and I thought James as well, call the neutral constituents of the world which are neither mental nor physical, ``stuff.''
\ewd

\section{28-02-11 \ \ {\it xQIT Meeting}\ \ \ (to MIT)}

Title and abstract below.
\bq\noindent
Title:   Charting the Shape of Quantum State Space\medskip\\
Abstract:  Physicists have become accustomed to the idea that a theory's content is always most transparent when written in coordinate-free language.  Sometimes though the choice of a good coordinate system can be quite useful for settling deep conceptual issues.  This is particularly so for an information-oriented or Bayesian approach to quantum foundations:  One good coordinate system may be worth more than a hundred blue-in-the-face arguments.  This talk will motivate and chronicle the search for one such class of coordinate systems for finite dimensional operator spaces, the so-called Symmetric Informationally Complete (SIC, pronounced ``seek'') measurements.  The desired class will take little more than five minutes to define, but the quest to construct these objects will carry us down a 35 year journey, with the most frenzied activity only recently.  Beyond this, we will turn the tables and discuss how one might hope to get the formal content of quantum mechanics out of the very existence of such a coordinate system.  It has to do with seeing the Born Rule as a ``tiny'' modification to the Bayesian Law of Total Probability.
\eq

\section{28-02-11 \ \ {\it Postmodern Smoke}\ \ \ (to C. M. {\Caves})} \label{Caves107}

\bcc
As for pragmatism, here's my problem.  Any approach that doesn't recognize that there is a hierarchy of concepts and truths---and not just a set of utilities---has problems, in my view.  Clearly, the truths of the human experience are special to us; they don't even apply to other life on earth.  The truths of life that has evolved on earth are special to earth; they wouldn't apply generally to life on other planets.  The truths of physics might apply to the whole Universe.  And I use truths here advisedly, willing to discuss the right word and concept for what I'm talking about.

Last July at QCMC I got the clear impression that you don't have any ability to assert the fact or truth of climate change, and if that's where pragmatism takes you, well, I ain't going there.  It could just be that you're too wrapped up in your professional and personal concerns to have much of a position on climate change, but I got the impression that even if you did have a position, you wouldn't be willing to assert it in the face of the deniers.  And that just doesn't cut it, in my view.
\ecc

I wish I had never given you that Rorty book.  I actually tried to divert your thought away from it in advance, by writing ``old-school pragmatism'' (to take your mind back to the metaphysical club days) but that was far too small a fin in the water if it was meaningful to you at all.  Sorry, I didn't mean to rile you.

Below is a little bit I wrote on what I think is the issue at hand here (it comes from Max's interview).

Of course, I believe global warming is a clear and present danger.  Just ask Kiki about my own rants.  And I am appalled that much of the rest of the world does not acknowledge it (certainly not America's politicians).  The world is very likely at a turning point, and its fate will be decided by whether we do something about it.

The issue will be decided at the point ``where the rubber meets the road'' as you nicely put it.  (I see no difference between the quantum and all the rest of life.)  We act because of our beliefs, and nature grants passage of the belief {\it or not\/} by its power.

If I understand you correctly, the gulf between us might be this.  I think you would like to say of some beliefs, they are correct, right here and right now, before any action has been made based upon them (and regardless of whether any action will ever be based on them at all).  The truth of an idea is a timeless, tenseless thing on such a conception.  Whereas I wish to reserve ``right'' and ``wrong'' to the sense of ``only time will tell.''  That is, ``right'' and ``wrong'' are always bestowed (to a belief, proposition, sentence, policy, quantum state assignment) after the {\it fact}.  Like that great scene in one of the {\sl Lord of the Rings\/} movies, where Gandalf says commandingly to the balrog, ``You shall not pass!''  The universe made a decision at just that point, I would say, not before.

I know that you are right about the James house.  It's just a silly pipe dream while I continue to wait for Godot (aka tenure).

I am trying to make sure I've got you right.  I hope you'll have patience with me, even if this conversation doesn't continue any further.

\bq
\noindent {\bf Question 6: Quantum probabilities: subjective or objective?}\medskip

``Subjective'' is such a frightening word.  All our lives we are taught that science strives for objectivity.  Science is not a game of opinions, we are told.  That diamond is harder than calcite is no one's {\it opinion}!  Mr.\ Mohs identified such a fact once, and it has been on the books ever since.

In much the same way, quantum theory has been on the books since 1925, and it doesn't appear that it will be leaving any time soon.  That isn't lessened in any way by being honest of quantum theory's {\it subject matter}:  That, on the QBist view, it is purely a calculus for checking the consistency of one's personal probabilities.  If by subjective probabilities one means probabilities that find their only source in the agent who has assigned them, then, yes, quantum probabilities are subjective probabilities.  They represent an agent's attempt to quantify his beliefs to the extent he can articulate them.

Why should this role for quantum theory---that it is a calculus in the service of improving subjective degrees of belief---be a frightening one?  I don't know, but a revulsion or fear does seem to be the reaction of many if not most upon hearing it.  It is as if it is a demotion or a slap in the face of this once grand and majestic theory.  Of course QBism thinks just the opposite:  For the QBist, the lesson that the structure of quantum theory calls out to be interpreted in {\it only\/} this way is that the world is an unimaginably rich one in comparison to the reductionist dream.  It says that the world has excitement, risk, and adventure at its very core.

Perhaps the source of the fear is like I was taught of ``that marijuana'' in my little Texas town:  Use it once, and it will be the first step in an unstoppable slide to harder drugs.  If quantum probabilities are once accepted as subjective, somewhere down the line Mr.\ Mohs' scale will have to disappear in a great puff of postmodern smoke.  There will be no way to enforce a distinction between fact and fiction, and the world will be anything our silly imaginations make up for it!

The first symptom is already there in a much more limited question:  If quantum probabilities are subjective, why would an agent not make them up to be anything he wants?  Why not pull them from thin air?  The defense to this little question is the same as the defense against the ``inevitable'' postmodern horrors.  My colleague Marcus {\Appleby} put his finger on the issue sharply when he once said, ``You know, it is {\it really hard\/} to believe something you don't actually believe!''  Why would one assign arbitrary probabilities---ones that have nothing to do with one's previous thoughts and experiences---if the whole point of the calculus is to make the best decisions one can?  The issue is as simple as that.
\eq

\section{28-02-11 \ \ {\it On the Lighter Side}\ \ \ (to H. B. Geyer)} \label{Geyer5}

And now for something on the lighter side.  I wasn't lying when I said that I was impressed with your fund-raising skills.

Just last week, I learned that my philosophical hero William James's farm is up for sale, and I have been going crazy thinking how I might secure it with the ultimate intent of turning it into a small New England version of something like STIAS (though with the focus being on the intersection between pragmatist philosophy and physics, QBism, and further such things).

For instance, I wrote this description to my old advisor Carlton Caves:
\bq\noindent
I really wish you were more a fan of William James and old-school pragmatism.  I discovered last week that his farm in Chocorua, NH will be on the market in the Spring (it was on the market in the Fall, and taken off for the Winter).  It's the house in which James died.  The 4387 sq ft house is relatively historically intact (though a lot of updates) and barn still there, but with swimming pool now in addition.  44 acres of land, 39 of it thick forest, and beautiful hay fields and rock walls still there.  Quiet secluded place.  Perfect for a QBistic Institute!!  (Imagine carving 5 little bungalows into the forest, a lecture room, etc.)  20 minutes from good skiing.  1.5 hours to Portland, ME and Portsmouth, NH.  \$895,000 asking \ldots\ but probably overpriced by \$100K.  I've been wracking my brain trying to figure out a way I could secure the place till said institute could become a reality.  Know any philanthropists with deep pockets for pragmatism or QBism or both? Seriously; I'm not joking.
\eq
And here was part of a further explanation to another friend (H. C. von Baeyer):
\bq\noindent
Any institute would have to be three, four, five, six years down the road.  It would take so much effort to build the idea, the support, the funding, the logistical infrastructure.  The real concern for me (the one that put me into the manic panic) is that the property be secured before some feckless investor or vacation-homer buy it and do more damage.
\eq
And I've sent out more than handful of significantly more detailed pleas as well.  (None to any avail \ldots\ of course.)

So maybe let me ask your advice.  If you were confronted with a similar situation, how would you tackle it?  What would be your first ideas for building up a funding effort (in a short timescale)?  Any advice you could give would be most appreciated.

\section{01-03-11 \ \ {\it Addiction Control}\ \ \ (to H. C. von Baeyer)} \label{Baeyer141}

Just dropping in to give you an update.  In case you're wondering about the title, I suppose it struck me because I've been watching and reading so much in the news on Charlie Sheen the last couple of days---I've never seen anyone go so over the top, and I've found it strangely fascinating.  Of course, I also flinch in those moments when I think about the impending train wreck:  I fear something really bad is going to happen in his private hours.

But on to my own addiction!!  I haven't shaken it completely, but I feel I have regained control back to a relatively healthy level.  It started by removing the house listing from the open tabs in my browser.  That was a start.  Then I stuffed the house plans and property map into my backpack, out of sight and off my desk.  Now, every day, I think about Chocorua a little less.  A feeling of health is starting to return.

On other matters, you never answered on whether you might come out to play the next time Marcus is in town.  Having visitor money sometimes burns a hole in one's pocket.  Mostly though, I'm trying to look ahead a little and get some inspiration \ldots\ in a kind of ``neutral stuff'' study group.  I want to know more about Pauli's thoughts there and his influences, etc.  Not quite sure what questions I ought to ask, but if we get the ball rolling maybe it'll go somewhere.  ``Neutral stuff'' is surely a better addiction.

\section{01-03-11 \ \ {\it Entities and Effects}\ \ \ (to W. G. {\Demopoulos})} \label{Demopoulos48}

I was just looking at the wiki article on entity realism (because of some little debate on climate change that I'm having with Carlton {\Caves} at the moment), and this line caught my eye:  ``[E]ntity realism claims that the theoretical entities that feature in scientific theories, e.g.\ `electrons', should be regarded as real if and only if they refer to phenomena that can be routinely used to create effects in domains that can be investigated independently.''  It caused me to think of your ``effects'' of course, and I wondered whether you had ever thought about how your own position compares and contrasts to entity realism.

\section{01-03-11 \ \ {\it Whitehead, Dewey, and a Potent Premise}\ \ \ (to W. T. Myers)} \label{Myers2}

Thanks for the article.  Reading it made for a very pleasant lunch in my office.  I do wish you could send me the complete thesis; I would read it with just as much relish.  It'd make a nice addition to my big Palazzo del QBismo library of pragmatism (nearly 700 volumes on the philosophical side of the room).  If you find that you can still print it out, I would love it if you mailed me a copy.  I've got a lot to learn about Whitehead's relation to the other things that feel right to me ({\it pieces\/} of James, Dewey, Schiller, and {\it parts\/} of the new realists), and I'd like to make the process as painless as possible.

Reading your quote of Whitehead on the ``category of the ultimate'' reminded me faintly of something I said at the start of a lecture once.  Quote pasted in below.  [See 07-02-11 note ``\myref{QBies34}{The 2011 QBist Challenge}'' to the QBies.]  My colleague, Appleby, and I have been debating a replacement for the word ``FACT'' used there; the common usage of the word is out of place with respect to the rest of the passage, which makes it clear that I am talking about some kind of ``active entity'' (i.e., not something ``dead'' as one usually thinks of facts).

Sure, you can share the interview; I don't mind.  (The reason for writing is so that people might read.)

\subsection{Bill's Preply}

\bq
I've attached an chapter I wrote for the {\sl Hanbook of Whiteheadian Process Thought\/} on the intellectual relationship between Whitehead and Dewey. This piece distills my thinking on Dewey's metaphysics in a much neater fashion that my dissertation. Also, I have never distilled my dissertation into a single .doc file, so it's messy to send. I wrote the thing in Wordstar, and I'm still not sure it's all been converted properly. I need to tend to that at some point. But, the meat of the interpretation is in this article.

I checked out the interview piece you sent me, and that is fascinating. Would you object to my sharing it will a few of my colleagues? I know a couple of folks who would find that of great interest, folks who know more about the physics than I do.

I hope that some day you'll find your way to our meeting. It's a pluralistic, very congenial group of scholars. The meeting truly represents the highlight of my academic year.

Off to teach Leibniz!
\eq

\section{01-03-11 \ \ {\it Second Rant} \ \ (to M. A. Graydon \& L. Hardy)} \label{Graydon15} \label{Hardy44}

I've been watching the Charlie Sheen case very closely this week, and I think I've learned a pointer or two from him.  Thus I launch into my second rant.

The quote below is from William James and refers to the German philosophers of his time.  But in my mind, I often shift its relevance to be about much of modern-day ``philosophy of science'':
\bq
In a subject like philosophy it is really fatal to lose connexion with the open air of human nature, and to think in terms of shop-tradition only.  In Germany the forms are so professionalized that anybody who has gained a teaching chair and written a book, however distorted and eccentric, has the legal right to figure forever in the history of the subject like a fly in amber.  All later comers have the duty of quoting him and measuring their opinions with his opinion.  Such are the rules of the professorial game --- they think and write from each other and for each other and at each other exclusively.  With this exclusion of the open air all true perspective gets lost, extremes and oddities count as much as sanities, and command the same attention \ldots
\eq

``Extremes and oddities count as much as sanities, and command the same attention \ldots''  Over and over I feel like I see that when I watch philosophers of science discuss physics.

\section{02-03-11 \ \ {\it New Scientist? \ldots\ New Sensationalist, Maybe!}\ \ \ (to H. C. von Baeyer)} \label{Baeyer142}

First you wrote:
\bhcvb
After Valerie Jamieson, physics feature editor for {\bf The New Scientist}, had made very positive noises about an article by me about QBism, I heard nothing from her for a while. Finally I phoned her, and found out that she is  quite knowledgable
(reads {\tt arXiv} and has seen some of my work and yours), but basically
of the opinion that quantum mechanics has done well all these years,
and unless you have a fully worked out new formulation, there's not
much interest.
\ehcvb
Then, because of our new emailing system, I noticed that New Scientist has been sending notices of the latest issues to my gmail account \ldots\ I guess for some time.  (Never noticed that before \ldots\ I usually only use that account when I'm on the road and can't get into my PI account.  But with the new system, there's not so much noise on my gmail account.  Difficult to explain \ldots)\ \ Anyway, the point is, I read this:
\bq\noindent
{\bf This week's top stories: The search for quantum loopholes}\medskip\\
Can the universe really be as weird as quantum theory suggests? Einstein suggested that loopholes might let us escape its more baffling implications and now ingenious experiments are finding out if they actually exist \ldots\
\eq

Therein is the key!  You've got to say that QBism makes the world look really, really weird!  You've got to make all those other doofus interpretations look tame in comparison.

You mean man's free will has the potential to shape reality itself?  How weird, how fantastic, how UNscientific is that!  You mean our actions actually matter?  Things can be moved to places that they might not have been otherwise?  The world is made of a stuff that's neither matter nor mind?!?  \ldots\ Ahh, these people are worthless \ldots\  Not much different than every news outlet in the land updating us on Charlie Sheen's sick mind with each of his latest ``tweets.''\footnote{\editornote In recent years, {\sl New Scientist} has been caught shilling for some remarkably fractured ceramics.  For an overview of physics-related examples up until 2007, see \url{http://www.sunclipse.org/?p=50}.  Biologists gave up in January 2009, when the magazine ran a cover which proclaimed, ``Darwin was Wrong''.  A more accurate description of the research nominally being popularized would have been, ``Horizontal gene transfer in microorganisms, a subject that Darwin knew nothing about, complicates the simple picture of a `tree of life', at least regarding species which Darwin had no idea existed.  Say, why are you bringing up Darwin again?  The {\sl Origin of Species} predates even the Mendelian idea of the gene, let alone everything we have learned since then.''  The editorial ``justifying'' the cover also promulgated some pseudohistory of physics, claiming that Kelvin said, ``There is nothing new to discover in physics.''  Supposedly from a 1900 address to the  British Association for the Advancement of Science, this quotation has yet to be substantiated and, given Kelvin's actual sentiments, is probably apocryphal.  [See Kelvin's ``Nineteenth Century Clouds over the Dynamical Theory of Heat and Light'' (1901).]  It is also a poor reflection of physicists' beliefs at the time; see \url{http://www.sunclipse.org/?p=947}.  For entertainment, see \url{https://twitter.com/NS_headlines}.}

\section{03-03-11 \ \ {\it Normalizing Fiducial Stability Groups} \ \ (to M. A. Graydon)} \label{Graydon16}

I was very impressed by all that you did and all that you understood, of course.  I never expected so much; I just wanted to alert you all to the thing before I ripped into the guy.

A much more subtle and interesting case will be Matt Leifer's paper, ``Retrodiction In Quantum Bayesianism, Or Why Wigner's Friend Is Fuchs' Enemy,'' which I will handle gently and with great care.  In case you're interested it's attached.  I am amazed that Matt gets so much about the program (well, the subjective Bayesian part), but doesn't seem to absorb at all the American pragmatist side of it, i.e., the personal character of the measurement outcomes.  Or the way Carl Caves got close to getting it right with:
\bq\noindent
``[QBism takes] a three-pronged approach to subjectivity and objectivity: (i) quantum states and probabilities are wholly subjective; (ii) system attributes are wholly objective; and (iii) measurement outcomes are where the rubber meets the road, i.e., where subjective and objective meet to produce something that is not under the control of the agent, but is also not out there in the world.''
\eq

Independently of that though, Matt's formalism is worth much thinking about.

\section{03-03-11 \ \ {\it My Consternating Collaborator}\ \ \ (to R. {\Schack})} \label{Schack219}

Section 5 starts out thus:
\bq\noindent
The reflection principle is a constraint on an agent's beliefs about her future probability assignments.
\eq
Why, why, why do you write like that?!?!?!  I would not be able to hold you to it in a court of law, but I am personally confident that the average reader will read it exactly in the way that caused all the confusion over van Fraassen to begin with.  I.e., that the reflection principle demands that the agent set her present beliefs by her supposed future probabilities.  Our task, as I recall, was to try to blunt that debate a little.

Something else would surely do better justice.  Maybe for instance,  ``The reflection principle is a constraint on the relation between an agent's present beliefs of her future beliefs and those future beliefs themselves.''  I'll think some more; there's still much room for improvement.

Oh, I get frustrated so easily by iniquitous choices of words:  Words that cause the very damage that we are trying mend.  Your draft has been more riddled with these kinds of things than usual.

Oh, I'm frustrated, and I've got to finish this damned thing.  Your Dutch book argument in Section 4 still has me totally confused.  It looks to me like the paragraph starting ``So let us first assume\ldots'' is a dangling appendage, not needed at all for the actual Dutch book argument that kicks in at the following paragraph.  I am probably misunderstanding something very deeply, but {\it then I blame it on your exposition\/} in that section.

Can we talk or Skype tomorrow?  I think I desperately need this.

Take my grumpiness with a grain of salt.  Attached is the latest draft.

\section{04-03-11 \ \ {\it Wherefrom Such Confidence}\ \ \ (to {\AA}. {\Ericsson} \& M. A. Graydon)} \label{Ericsson17.1} \label{Graydon16.1}

I've got no good answer except to say, ``From a long, hard fight.''  I have a collection of letters on my website titled {\sl The Post-{\Vaxjo} Phase Transition:  Knowledge $\rightarrow$ Information $\rightarrow$ Belief $\rightarrow$ Pragmatic Commitment\/} that documents the gradual change of my view on quantum states.  (But it only picks up from the days when I was already calling them knowledge; ten years before that I was thinking they were objective propensities of some sort.)  Here's the file:
\myurl{http://www.perimeterinstitute.ca/personal/cfuchs/PhaseTransition.pdf}.

But I don't really know what it shows but a soul in torment.  Or an oyster trying to remove the grains of sand from its shell.  No one argued more strongly that quantum states were personalist Bayesian than me---more and more determinedly, in fact, as I checked my own consistency and found my failings therein.  (In fact, as far as I know, no one had ever argued for such an extreme view at all.)  No one led me; it was my own push and shove that got me there. You'll see it in every letter of exasperation I wrote to Caves, {\Schack}, and Mermin as they fought back and I tried more and more to put together a convincing picture.

The conception was my baby, and maybe that's why I love it and have confidence in it like no one else.

\section{04-03-11 \ \ {\it Among the World's Religions}\ \ \ (to {\AA}. {\Ericsson})} \label{Ericsson18}

Thanks for your thoughtful note.  I've read it and reread it and two times beyond that (and I'll probably read it again), as I've been building and tending this fire for Kiki's return tonight.  She's out at some kind of spousal dinner, and the children are away at a girl scout camp.

Particularly, I keep coming back to these lines:
\bae
The thing is that I find my beliefs as based on something real outside
myself, they are not crucially dependent on my philosophizing and it
includes an explanation to how my beliefs can be strong. To me it is
consistent in a way I can't see how anything I think about physics ever
could be. So then I find it amazing you can believe, if not as much as I,
still close to it. It's not a criticism, it is a conundrum to me.
\eae
[OK, I fixed your spelling.]

I'm not sure you'll understand it, but my mind kept coming back to the image of Napoleon's coronation as Emperor.  He somehow convinced the pope to travel from Rome to crown him in Notre Dame Cathedral rather than at the Vatican.  (Something unheard of before.)  At the very last moment, in front of hundreds of witnesses, as the pope reached for the crown to place it on Napoleon's head, Napoleon grabbed it and placed it on his head himself.  ``I crown myself Emperor!''

It's not unconnected to the last note I sent you.  ``I build my own religion.''  Contemplating the holy mystery of quantum theory is my personal sacrament.  It seeps into every aspect of my life like a religion does.  See for instance, the first two lines of an old proposal I wrote for Caltech once upon a time (attached).  Better still, see the story ``\myref{ShouldNotPass}{Some Things Should Not Pass}'' from 16 February 2002 on page 150 of the web document I noted to you earlier.  All these things are tied together in my mind.  Quantum theory is the closest I've yet come to an articulated  expression of what I strive to say in the line, ``For a small time we have the chance to move around and determine our courses as we please---to leave a trail behind us.''  See also Footnote 33 (page 20) in \arxiv{1003.5209}.

I do hope I am not hurting your feelings with my jesting with you.  In my mind it is playful, though I fear that I do not always convey it that way to you.  I think long and hard, contemplating our disparate positions on religion.  You are a catalyst to me, even if I don't reciprocate to you.  I'm just saying I value these discussions.

\section{06-03-11 \ \ {\it WTB, Reason and Emotion}\ \ \ (to H. Yadsan-Appleby)} \label{Yadsan1}

I am sorry to take so long to reply to you.  I wanted to get home to look at some quotes before responding, and then once I got home, I became ill (as you have probably heard).

I am glad that you are enjoying O'Connell's book [{\sl William James on the Courage to Believe}] and finding some new thoughts in James's philosophy.

At the moment, I can't help you much with the meanings of ``forced'', ``living'', ``momentous'', etc., as it has been a long time since I've read the appropriate essays, but I think you will find the definitions clear enough if you read the actual essay ``The Will to Believe''.

In fact here is what I suggest as a study plan.  I would say the next thing you should read is precisely ``The Will to Believe''.  And then follow that with the magnificent, particularly deep (but harder to read) essay, ``The Sentiment of Rationality.''

On one of your points:
\bhya
I never accepted or even understood the split between reason and faith (or reasoning and feeling emotions).
\ehya
I wanted to bring your attention to a nice paragraph from Robert D. Richardson's biography of James:

\bq
About a month after he sent off his ``Remarks on Spencer's Definition of Mind as Correspondence,'' James sent off, in January 1878, an article called ``The Sentiment of Rationality.''  He later described it as ``the first chapter of a psychological work on the motives which lead men to philosophize,'' and he noted ruefully that it might better have been called ``The Psychology of Philosophizing.''

We may, at this distance, prefer the original title, if only for its fresh and unorthodox, not to say brash, announcement that rationality is at bottom a feeling.  Not a matter of logic or math, not reasoning or ratios, not induction, deduction, or syllogism, not something higher than and detached from the senses, not the opposite of a feeling or emotion---rationality is itself a feeling or emotion.  He might even have called the essay ``The Feeling of Rationality.'' He begins by asking how we recognize the rationality of a conception, and he answers, ``By certain subjective marks, that is, a strong feeling of ease, peace, rest.'' He amplifies, saying, ``This feeling of the sufficiency of the present moment, of its absoluteness---the absence of all need to explain it, account for it or justify it---is what I call the Sentiment of Rationality.''
\eq

This sounds a lot to me like what you were saying above, and I suspect you would have quite an agreement with it.  So I think when you are prepared for it, and your mind has gotten a bit accustomed to James's peculiar writing style, you would get much out of reading ``The Sentiment of Rationality''.  It is a wide-ranging essay on much, more than this subject.

\section{07-03-11 \ \ {\it Einstein, Indeed!}\ \ \ (to R. W. {\Spekkens})} \label{Spekkens105}

\bq
There can \underline{\it be\/} no difference that \underline{\it makes\/} no difference.\medskip
\\
\hspace*{\fill} --- William James, ``What Pragmatism Means'' (1907) \hspace*{.3in}
\eq
[underlines mine for still extra emphasis, italics his own]

\subsection{Rob's Reply}

\bq
Ah, but Einstein took that principle (in the form of the equivalence of acceleration and gravitational force) and derived general relativity!  Surely no amount of prose can provide a better {\it justification\/} of the principle than that.
\eq

\section{07-03-11 \ \ {\it The Battle of the Eyeballs} \ \ (to J. B. Hartle)} \label{Hartle2}

I was amused by the no-eyeball sign in your talk today.  See attached for my dueling point of view. 

I enjoyed your talk, and I'm glad I was able to get there:  Seeing it teaches me that there's a bit of common groundwork between us for this little piece of formalism I want to show you tomorrow or Wednesday.

\bq
{\bf Quantum Cosmology from the Inside.}~The agent can consider measurements on ever larger systems. There is nothing in quantum mechanics to bar the systems considered from being larger and larger, to the point of eventually surrounding the agent.  Pushed far enough, this is quantum cosmology!  Why all this insistence on thinking that ``an agent must be outside the system he measures'' in the cosmological context should mean ``outside the physical universe itself''?  It means outside the system of interest, and that is the large-scale universe. Nor is there any issue of self-reference at hand.  One would be hard pressed to find a cosmologist who wants to include his beliefs about how the beats of his heart correlate with the sidereal cycles in his quantum-state assignment for the external universe.  The symbol $|\,\Psi_{\rm universe}\,\rangle$ refers to the green boxes alone.  [Figure 6 in ``QBism, the Perimeter of Quantum Bayesianism,'' \arxiv{1003.5209v1}.]
\eq

\begin{center}
\includegraphics[width=6cm]{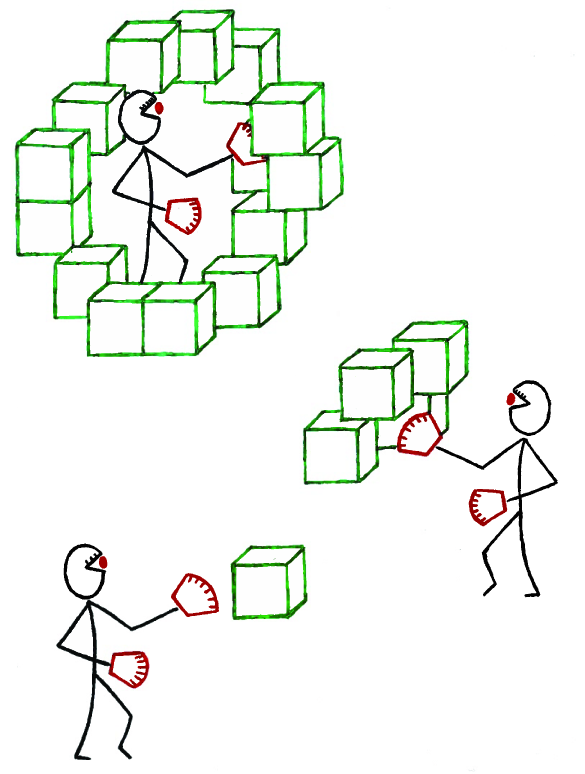}
\end{center}

\section{08-03-11 \ \ {\it Quantum Systems 101} \ \ (to the QBies)} \label{QBies36}

\begin{center}
``If the universe is a quantum mechanical system, then it has a quantum state.'' -- J. B. Hartle
\end{center}
See the first two minutes of Jim Hartle's talk today, \pirsa{11030103}.

Here's an essay question for you QBies:  What does it mean to be a quantum mechanical system?

\section{08-03-11 \ \ {\it Teaching the Teacher?}\ \ \ (to M. A. Graydon)} \label{Graydon17}

\bmag
I have attached the absolute latest version of the paper. A while back I tried to defend my opening sentence. Jim Hartle said something yesterday which might help my cause:
\begin{center}\noindent
{\rm A physical theory provides reliable probabilities for betting on the regularities of the universe. $\qquad$ --- James Hartle (\pirsa{11020124})}
\end{center}
\emag

Jim's statement is too ``objectivist'' about probabilities for my taste.  Reliable?  By whose measure?  With respect to whose priors?  (Don't fool yourself into thinking you can define such a notion without a prior.)  Regularities?  Is that a sneaky way of endowing nature with propensities?

My present thinking is that probability theory is a calculus for flagging ``inconsistency''---nothing more.  By my view, quantum theory does not {\it provide\/} probabilities any more than the raw axioms of probability theory do.  Probabilities come out of a hat that no theory itself has yet entered (i.e., the subjective agent).

I expand on this view in Section III of \arxiv{1003.5209} and my answer to Question 4 in the attached.  [See ``Interview with a Quantum Bayesian,'' \arxiv{1207.2141v1}.]

Thanks for the offer of a tutorial.  I am going to try to get in early tomorrow morning to hopefully give Hartle a session on the urgleichung.  Drop me a note when you're in the building, and I'll let you know if I'm free.  Unfortunately, I have to chair the colloquium tomorrow; so that time will be taken up.

PS.  {\Ruediger} tells me that Logue's book {\sl Projective Probability\/} makes a good case for the idea (in opposition to E. T. Jaynes and the objective Bayesians) that nothing more than coherence is needed for making probability assignments useful tools.  I can lend you the book if you wish.

\section{10-03-11 \ \ {\it Late Great Johnny Ace}\ \ \ (to M. A. Graydon)} \label{Graydon18}

\bv
On a cold December evening\\
I was walking through the Christmas tide\\
When a stranger came up and asked me\\
If I'd heard John Lennon had died\\
And the two of us went to this bar\\
And we stayed to close the place\\
And every song we played\\
Was for the late great Johnny Ace, yeah, yeah, yeah \medskip\\
$\quad\qquad$ --- Paul Simon, ``The Late Great Johnny Ace''
\ev

Your choice of photos for your Beamer example took me down memory lane.  I remember well the day John Lennon died.  I heard it on the radio driving home from school.  I had literally just gotten into the Beatles a few days previous, after reading the {\sl Playboy\/} interview of John and Yoko (sifting through my brother-in-law's stash of the magazine).  It was devastating; I recorded as much Beatles music as I could directly off the radio.  No internet, no YouTube back then; no record store in my hometown.

I later named my first golden retriever Albert Winston Fuchs.  Albert was for Albert Einstein; Winston was for John Lennon.

I remember being home from school sick the day Elvis Presley died.  A ticker scrolled across the bottom of the TV screen saying ``Elvis Aaron Presley, the king of rock and roll, is dead at age of 42.''

I don't remember Hunter Thompson dying, but my records show that I was somewhere between a talk in College Park, Maryland and a talk in Wroc{\l}aw, Poland.

\section{10-03-11 \ \ {\it Your Reference} \ \ (to J. B. Hartle)} \label{Hartle3}

Thanks for the little audience yesterday.  I enjoyed it.  Also, I'm glad to learn that you're now thinking of probabilities in (kind of) Bayesian ways.  As it happens, I'm just starting to slap together a contribution for a conference proceedings, where I've got a footnote which reads, ``This is what particularly distinguishes QBism from several other Bayesian approaches to quantum probability.  See Refs.\ \verb+\cite{...}+ for details of the other approaches.''  Footnote 2 in the attached.

If you have further papers than the one I cited [J. B. Hartle, ``Quantum Mechanics with Extended Probabilities,'' Phys.\ Rev. A {\bf 78}, 012108 (2008)], or I didn't cite the most appropriate one and there is something better, please let me know.

\section{13-03-11 \ \ {\it Hurried Remarks}\ \ \ (to M. A. Graydon)} \label{Graydon19}

Greetings from Chicago airport.  I just had a pleasant time reading your paper (in the Toronto airport and then on the flight).  {\AA}sa and I are in the Admiral's Club, and she's off having a bowl of soup (that she paid for), and I of course am having my complimentary beer (because I'm cheap).

Your paper is extremely well written \ldots\ much like I had expected.  Beautiful clarity and organization.  (And I'm not making this up.)

My only worry at all is that there is somehow still not enough ``oomph'' in the first page, abstract, and conclusion.  Why is the question you're tackling here important?  I feel the reader ought to be left with a memorable sentence in his mind long after the details of your construction have slipped away from him.  Indeed, could you find one sentence that would really help draw potential readers in?  I don't know, maybe something like:  ``Ultimately one would like to use the developing technologies of quantum information science to test the validity of complex over quaternionic quantum mechanics.  Before that, however, one must have a clear conception of the relations and contrasts between the theories with respect to the full apparatus of quantum information theory --- particularly, with respect to generalized quantum measurements and quantum operations, not {\it merely\/} projective measurements and unitary operations as has been worked out previously \verb+\cite{Hermie,Joe,Peter}+.  This paper fills that gap in the literature.''  It's not very elegant or finely tuned (or one sentence long either!), but I just wanted to throw something down before {\AA}sa and I have to leave.

\section{13-03-11 \ \ {\it Quaternions for Jesse Jackson}\ \ \ (to M. A. Graydon)} \label{Graydon20}

Now I'm writing you from the airplane!  First time I've been able to get online on a flight in years.  (Several years ago, it was briefly legal, but then the FAA squashed it for quite some time.)  Now, for \$9.95 I'm in internet bliss.

Funny thing:  I was in seat 3E and {\AA}sa was in 6A.  I came back to the guy who was sitting beside her and said, ``I'll trade you an aisle seat for an aisle seat so that I can sit beside my colleague.  If it'll sweeten the pot for you, Jesse Jackson is sitting in the seat beside me.''  You see, our {\AA}sa is much more important than any politician!

But wouldn't you know, she later said something that made me realize:  If I had only been sitting beside him, I'm might have hit him up for help in buying the William James house.  Darn, foiled again!

But then again, I'm on the internet!  And thus, I can read things like this:
\myurl[http://query.nytimes.com/gst/fullpage.html?res=9B0DE7DC113CF93AA15752C1A961948260&sec=&spon=&pagewanted=5]{http://query. nytimes.com/gst/fullpage.html?res=9B0DE7DC113CF93AA15752C1A961948260{\&}sec={\&}spon={\&}p\\ agewanted=5}.
I guess I did the right thing after all.

So let me get back to your notes!

\section{15-03-11 \ \ {\it Itamar's Volume}\ \ \ (to R. {\Schack})} \label{Schack220}

I guess Meir's ``urgent'' is now putting me on edge.  I hope you've made progress.

I am now up at 3:30 AM working on talk number 2.  I'm not sure talk number 1 did much for the general populace, but it did spur a long discussion with Samson Abramsky and Adam Brandenburger.  (Samson was the Clifford lecturer three years ago.)  Interesting seeing the tensions in Samson.  He does take us seriously, but he's got the usual spook, ``The Bayesian view is attractive, surely the best notion of probability in many situations [implicitly, he was meaning `not in physics'], but surely the probabilities in quantum mechanics come about through `nature's regularities'.''

Today, the de Finetti theorem, fidelity, and no-broadcasting will be the main fare.

\section{19-03-11 \ \ {\it Congratulations.\ Question}\ \ \ (to R. B. Griffiths)} \label{Griffiths2}

I'm sorry for the delay.  I just delivered this year's Clifford lectures at Tulane U in New Orleans, and now I'm having a brief respite with my 82-year-old mom in South Texas on my way to Dallas (for the APS March meeting).

Thanks for the congratulations; it means a lot to me to hear from you.

\brbg
I am trying to write up an item where I would like to put a reference
to your opinion that quantum information is [only] about the outcomes
of future measurements, assuming that I am not mixing you up with
someone else.  Is there a good accessible reference?
\erbg

No, you're probably not mixing me up with anyone else, and {\it in a sense\/} that is right.  {\it However}, my students did recently point me to a passage in one of your 2009 papers where you write:
\bq
       It seems almost inevitable that if quantum mechanics
       is incompatible with ``local realism'' (a term which in
       many minds is essentially synonymous with Bell
       inequalities) but is still somehow local, then it is
       realism that must be discarded, and one finds serious
       arguments to this effect [38]. But if the world of
       atoms is unreal, what shall we say of the macroscopic
       objects which most physicists think are composed of
       such atoms? Are the things we see around us (not to
       mention we ourselves) unreal?
\eq
That does {\it not\/} characterize my position now, and never has, though I'm getting better and better at articulating what it is that I am trying to get after.  Let me thus send you to this paper: \arxiv{1003.5209}. It is probably better read as a whole, but if you want to go straight to the metaphysics (i.e., what is real), go to Sections~VI and VII on page~19.  It is not at all that there is no microscopic (or macroscopic) reality---it never has been---it is that there is too much of it.  As one philosopher once put it, it is that ``The world is not sentence shaped.''  Give the reading a chance; at the least I hope you'll find it entertaining.  Also, if you back up to Section~V, you'll find my own take on the significance of the Bell inequalities.

Finally, let me attach an interview that is not posted yet.  [See ``Interview with a Quantum Bayesian,'' \arxiv{1207.2141v1}.]  I hope it tackles some of these same issues adequately.  In fact these two documents together, I believe, make the clearest and most complete statements yet of what it is that I see as my research program.

I wish I were going to see you at the March meeting, but as far as I can tell you won't be.  It'd be nice to discuss these things with you when they might be fresh on your mind.  Would you be interested in visiting the Perimeter Institute sometime this year?  I could arrange it.

\section{20-03-11 \ \ {\it Notes on Norsen}\ \ \ (to N. Argaman)} \label{Argaman2}

Let me start by quoting a passage from William James's essay ``The Will to Believe'':
\bq
Let us give the name of {\it hypothesis\/} to anything that may be proposed to our belief; and just as the electricians speak of live and dead wires, let us speak of any hypothesis as either {\it live\/} or {\it dead}. A live hypothesis is one which appeals as a real possibility to him to whom it is proposed. If I ask you to believe in the Mahdi, the notion makes no electric connection with your nature,---it refuses to scintillate with any credibility at all. As an hypothesis it is completely dead. To an Arab, however (even if he be not one of the Mahdi's followers), the hypothesis is among the mind's possibilities: it is alive. This shows that deadness and liveness in an hypothesis are not intrinsic properties, but relations to the individual thinker. They are measured by his willingness to act. The maximum of liveness in an hypothesis means willingness to act irrevocably. Practically, that means belief; but there is some believing tendency wherever there is willingness to act at all.
\eq

The truth is, at the moment, I've just got no energy for the nonlocality debate.  Nonlocality is just not a ``live hypothesis'' for me.  And when I've got so many other urgent matters to take care of (like the Clifford lectures I gave last week or the 400 talks I've organized for the APS March meeting this week), it seems a special burden to even think the slightest amount about it.

I'm sorry; I just can't muster the steam.  The best I can say is that Section V of \arxiv{1003.5209} already speaks my piece on the subject.  At the moment, I've got nothing else to say.

But, look, good luck in your journey.

\section{21-03-11 \ \ {\it In Place of Regularities}\ \ \ (to S. Abramsky)} \label{Abramsky3}

I've spent the early morning hours today reviewing a lot of things from last week \ldots\ not only my iniquities, but also the discussion we had at the reception where I shunned your phrase ``regularities of nature''.  Anyway, in that regard, I hope you'll sometime have the chance to read my paper, \arxiv{1003.5209}, particularly Section~VI, starting on page 19.  I'm not quite sure how to put what it's striving at in a pithy way, but I do know that it is an attempt to put something else---``things in common''---in place of your ``regularities''.

\section{27-03-11 \ \ {\it NY Anytime Soon?}\ \ \ (to A. Plotnitsky)} \label{Plotnitsky25}

I've been spending the morning re-organizing my bookshelf to make room for 18 new purchases.  (Photo of one of the two walls in case, I haven't shown you before.)  And I came across my two copies of {\sl Complementarity}!  (The signed paperback you gave me and the hard cover in perfect condition that I found sometime later.)  It sent me on a trail of thinking about our discussions, but it also made me wonder whether you'll be in New York anytime soon, visiting Paula or your sister.  Maybe you're even there now?  I ask because there is a \$300 set I'd like to purchase from The Strand, but I'm wondering what kind of condition it's really in.  And I'd only want to trust a real book lover's opinion.

Oh, if you look in the middle column near the arm of the chair, you'll see the two copies of {\sl Complementarity}.

Will you be in {\Vaxjo} this year?  It'd be nice to see you again and talk more extensively with you than I have the last couple of years.  Recently, I've taken a renewed interest in Bohr and his complementarity.  It has to do with my SIC work actually.  The key is that one can always think of quantum state space as arising from an algebra with {\it only\/} two generators (Weyl--Heisenberg group).  Appleby particularly has me starting to wonder whether Bohr saw this very fact \ldots\ in his own (to me usually inscrutable) way.  I'd love to pick your brain on this!

Hope you're doing well.  All the best \ldots.  Wait, let me attach a new composition that you might enjoy.  It's an interview for Max Schlosshauer's new book.  David Mermin wrote one too; but maybe you should ask him for it directly if you'd like to read it.

\section{28-03-11 \ \ {\it Stream O'} \ \ (to M. Schlosshauer)} \label{Schlosshauer43}

\bmaxs
I was just reading a long essay about Joyce's {\bf Ulysses}, and I learned that it was William James who coined the term ``stream of consciousness.''
I had no idea! This James is really one hell of a personality. As they'd say in sports, ``Go James!''
\emaxs

\begin{itemize}
\item big bang
\item multiverse
\item stream of consciousness
\item turtles all the way down
\item block universe
\end{itemize}
Yep, he's the man.  Have I told you how I've become obsessed with purchasing his old farm (beautiful big house and 44 acres of forest) and turning it into an institute for law-without-law studies?  All I need to do is gather \$895K quickly and then figure out a source of sustaining funds.  Yeah right!

\section{28-03-11 \ \ {\it Case for Being Considered for a Faculty Appointment at PI} \ \ (to Correspondent Y)} \label{CorrespondentY1}

As it turns out, just yesterday I updated my CV and the Research Statement I had sent you previously.  Some things have changed substantially:  [\ldots]

New documents attached.


\begin{center}
\Large{\bf Statement of Research}
\end{center}

\bq
\subsection{1. Introduction}

In 1985, I was fortunate enough to spend a year in the group of John Archibald Wheeler at the University of Texas.  Little did I know then, but I was witnessing the first heaves of birth of what we now call quantum information.  The names of the graduate students and postdocs in that group---David Deutsch (quantum Turing machines, quantum speed-up), Benjamin Schumacher (quantum compression, qubits), William Wootters (no-cloning, quantum teleportation), and Wojciech Zurek (no-cloning, quantum decoherence)---now ring in the ears of young quantum information theorists like the names  of Bedford, Exeter, Warwick, and Talbot in King Henry's famous St.\ Crispin's Day Speech.

By the late 1970s and onward, John had it in his head that the central {\it physical\/} idea of quantum mechanics was ``measurement''---that ``measurement'' itself was the raw stuff of the world (even the world without man)---and the central {\it mathematical\/} idea of the formalism was ``information'' \cite{Wheeler83}.  In his mind, there was no {\it measurement problem\/} for quantum theory, but an opportunity for physical understanding that might reach to the very core of physics.  Black holes, the big bang, the heat radiating from a block of molten pig iron were all equally important considerations for him, and at bottom---he would say---what principle powers one, powers them all. At his boldest, he would ask things like \cite{Wheeler82}, ``Is what took place at the big bang the consequence of billions upon billions of these elementary processes, these elementary `acts of observer-participancy'? Have we had the mechanism of creation before our eyes all this time?'' That was the kind of environment---as disjoint as it might seem from today's more applied interests---that gave birth to part of the field of quantum information (and all the myriad expectations we have of that field's products).

To the best students who came asking for a research project, graduate and undergraduate alike, John would say, ``Derive quantum theory from an information theoretic principle'' \cite{Wootters80,Deutsch86,Schumacher90}.  In my case, I was given a more workaday project in the Regge calculus, enumerating vertices, edges, and faces.  But I was not deterred!  I think it is fair to say that my whole career in physics has been a ``poor man's version''---i.e., a careful delineation and augmentation---of that initial vision of John Wheeler's.  It has taken thinking about the foundations of probability theory in ways that would have likely been foreign to him (Bayesian) and familiarity with quantum effects that he had no knowledge of when he beseeched those young students (the present-day fruits of quantum information theory), but the big goal now appears to be within grasp in ever more convincing terms.

There is a whole field of research here, ``quantum foundations in the light of quantum information'' \cite{Fuchs01}, and like of John Wheeler's earliest hopes for it, it is not a field in the service of dotting I's, crossing T's, and {\it finishing\/} the book of quantum mechanics, but a field in the service of setting the stage for new, great discoveries in physics. Particularly, with the more specialized turn to Quantum Bayesianism (or QBism) \cite{Fuchs09a,Fuchs10b} propounded by the author, it feels that the equations are locking into place at the same pace as the conceptual development.  These are very exciting times. This Statement of Research is about the field and about my place in it.

Briefly put, I feel my four most significant contributions to physics are:
\begin{enumerate}
\itemsep -1pt
\item
The quantum no-broadcasting theorem.
\item
The quantum de Finetti theorem.
\item
Developing Quantum Bayesianism into a robust interpretation of quantum theory.
\item
Realizing that the symmetric informationally-complete quantum-state representations (based on SICs, see Section 3) are not just  novelties, but the basis for a new formulation of quantum mechanics, potentially in a lineage with matrix mechanics (Heisenberg representation), wave mechanics (Schr\"odinger representation), and the path integral.
\end{enumerate}

The plan for the remainder of this document is as follows.  In Section 2, I describe a sampling of my papers and try to give some sense of their impact on the community.  When mentioned, all citation numbers are drawn from Google Scholar.\footnote{Note that Google Scholar sometimes returns more than one record on a given publication---the first one is of the usual style with a link to the paper, whereas the second (and sometimes a third, fourth and more)\ is prefaced with a small marker ``{\tiny [CITATION]}'' and no link to the paper.  When such is the case, I report the total count of citations from all records.  If upon a first look at Google Scholar a discrepancy is noted with what I write here, a more careful look further down the list should reveal this to be the cause.  For example, the accounting for my paper \cite{Fuchs01} (58 cites) comes about by adding up 3 records: 50+6+2.} My total citation count is just over 5,700 at the moment with a Hirsch index of $h=32$, but I will not exhibit all the papers here.  Also to reduce clutter, I do not list citation counts that number less than 50.  In Section 3, I describe the main elements of my research program of the last two years.  In Section 4, I explain some of the next logical steps in the program and my plans for tackling them.  Finally in Section 5, I point out a few things unique to my research style that, by their nature, cannot be gleaned from my publication record.

\subsection{2. Research History}
\label{TheOldIEEEsTRATEGY}

The greater fraction of my research has been written up in 68 scientific papers since the mid-1990s. In this section, I classify the work into two major periods, BC and AD\@.

BC (``before coming out of the closet'') corresponds to a period of time when I carried all the enthusiasm described in the Introduction, but I dared not declare professionally that I considered myself doing quantum foundations. It is by now, however, well documented that my eyes were always turned toward foundational matters as I selected my research topics in that period:  See my book {\it Coming of Age with Quantum Information\/} (Cambridge U. Press, 2010) \cite{Fuchs10}, for instance, or the 508-page original draft of it posted at \quantph{0105039v1}.

AD (``after declaration'') corresponds to the period of time when I started to meet the right-hand side of John Wheeler's vision head on---going straight for the idea that the whole mathematical {\it formalism\/} of quantum theory is, in the end, about information.\footnote{In Sections 3 and 4, I will come back to the left-hand side of John Wheeler's vision.}  By that period, however, I had started to notice an inherent flaw in how the term ``information'' was being used in the quantum information community, as well as the philosophers of science who were hanging out pretty thick on the field by then.  As they talked (and even as I talked, before I became self-conscious), ``information'' was just some new magical {\it fluid\/} added to physics---people talked as if it flowed incompressibly, it was conserved, etc.  And that was exactly what physics DID NOT NEED for addressing any of the quantum foundational conundrums.  No progress could be made from it: The only thing it did was give hope that where the old substances did no good, the new fluid would finally grease all the joints.

It was here that I realized that a truly (or radical) Bayesian conception of probability and information was what was called for: If information was going to do any good for the world of physics, it had to be unequivocally demarcated from being a physical property itself.  The research in the AD period was all geared toward making that rigorous and showing that it had a very real reflection in the quantum formalism \ldots\ even to the point of affecting our description of laser light \cite{vanEnk02a,vanEnk02b}.\footnote{These papers make fundamental use of the Quantum Bayesian ``quantum de Finetti theorem,'' which I will talk about later.}  It was at this stage that Quantum Bayesianism was born, and it was a gut-wrenching experience.  Beside the publications described below, the intellectual toil of that time can be seen in my collected correspondence on the issues: Presently 832 pages of it are compiled for another upcoming book, {\sl My Struggles with the Block Universe} \cite{Fuchs10c}, which will be posted on {\tt arXiv.org} on the 10th anniversary of its predecessor \cite{Fuchs10}.

\subsubsection{2.1 BC -- Quantum Information with an Eye to Quantum Foundations}

What is a quantum state? Is it a physical property of every spacetime point like an electromagnetic field?  Or is it something that only the universe as whole has a unique one of, as in the Everett interpretation?  Under the influence of my undergraduate exposure to the issue, I suppose even very early on I was inclined to believe that quantum states were of the character of information. For it seemed to me, one got the biggest, quickest ``bang for the buck'' in the quantum foundations wars if one did that.  I very much liked the way Jim Hartle put it in a paper that I knew well before graduate school \cite{Hartle68}:
\bq\small
\noindent A quantum-mechanical state, being a summary of the observers' information about an
individual physical system, changes both by dynamical laws and
whenever the observer acquires new information about the system
through the process of measurement.  The existence of two laws for
the evolution of the state vector
\dots\
becomes problematical only if it is believed that the state vector is
an objective property of the system.
Then, the state vector must be
required to change only by dynamical law, and the problem must be
faced of justifying the second mode of evolution from the first.
If the state of a system is defined as [our information of it], it is not surprising that after a measurement the
state must be changed to be in accord with [any] new information.
The ``reduction of the wave packet'' does take place in the
consciousness of the observer, not because of any unique physical
process which takes place there, but only because the state is a
construct of the observer and not an objective property of the
physical system.
\eq
Still there were all kinds of tricky matters that got in the way of a clear understanding for me.  If it is just information, how can we make sense of interference? If the Bell inequality violations throw out the possibility of local hidden variables, then what on earth can the information actually be about?  And, by the way, whose information?  All these questions made my head dizzy, and I knew I shouldn't be doing quantum foundations anyway.

So, I packed up my bags and moved to General Relativity, starting graduate school under the direction of Jim York at the University of North Carolina.  That lasted for a while.  But then I read two papers by Sam Braunstein and Carlton Caves on {\it information-theoretic\/} Bell inequalities \cite{Braunstein88,Braunstein90}, and somehow the concept blew me away:  There was someone out there that was neither shy to talk about Bell inequalities, nor shy to dream of an information theoretic conception of quantum theory.  And both things were even done in one paper!  The old enthusiasm flooded back into me, and I found a way to go to a conference in South Carolina so that I could meet Carl Caves.  I asked him if he would be my PhD advisor, and I enrolled at the University of New Mexico soon after he arrived there.

Caves suggested a sober, metered approach, and I agreed.  It was time to roll up my sleeves and learn about information {\it in\/} quantum theory.  With Peter Shor's discovery of the quantum factoring algorithm in 1994, the postdoctoral job market didn't look so formidable for someone quietly doing that much quantum foundations. And in my case, it was what I absolutely needed to do anyway.

Thus developed a general strategy.  I would ask over and over, what sets quantum uses of information apart from the ones in, say, a standard issue of the {\sl IEEE Transactions on Information Theory}, and I would do anything or join any effort to shed some light on the matter.  My quiet agenda was that I thought one would get the surest handle on identifying the raw stuff of the world by such a method, by studying the difference.  Rolf Landauer's motto that ``information is physical'' was not my motto.  It was that ``information {\it carriers\/} are physical,'' and they indirectly reveal their hidden talents by the way they carry it.

Here is a sampling of things this strategy led me to think about or to join in working out with a set of collaborators.

\subsubsection{2.1.1 Distinguishability and Accessible Information}

Bill Wootters in his PhD work \cite{Wootters81} had shown that the usual notion of distance in projective Hilbert space (i.e., the space of pure quantum states) can be thought of as a measure of the statistical distinguishability in an optimal experiment for discriminating states. Particularly, he started with a statistically-motivated reason for considering a distance measure between probability distributions $\bf p_0$ and $\bf p_1$ which made use of the square-roots of the probability components rather than the probabilities themselves,
\be
d({\bf p_0},{\bf p_1})=\cos^{-1}\!\left(\sum_i \sqrt{p_0(i)}\sqrt{p_1(i)}\right)\;.
\ee
He was then able to show that if these probabilities came about by considering projective measurements on quantum states $|\psi_0\rangle$ and $|\psi_1\rangle$, this quantity would be maximized by a measurement that gives precisely $d\big({\bf p_0},{\bf p_1})=\cos^{-1}|\,\langle\psi_0|\psi_1\rangle|\;$.  The hope was that this would make it more understandable why quantum mechanics uses probability amplitudes in its analyses, rather than the probabilities themselves.

In my mind the interpretation was always debatable, but trying to generalize it to all quantum states $\rho_0$ and $\rho_1$ (not just the pure ones) and all quantum measurements (not just the projective ones) was still a good first problem to sink my teeth into.  It was uncharted territory in quantum information, and it might tell us something interesting geometrically about the full quantum state space.  To my great surprise, the answer to this generalized question turned out to be calculable, and in fact identical to something in a completely different context:
\be
d(\rho_0,\rho_1)\,\equiv\,\min_{\mbox{\small measurements}}\, d\big({\bf p_0},{\bf p_1})\,=\,\cos^{-1}\!\left({\rm tr}\sqrt{\rho_0^{1/2}\rho_1\rho_0^{1/2}}\,\right)\;,
\label{Kiki}
\ee
The quantity on the right-hand side was none other than Uhlmann's ``fidelity'' measure for quantum channel theory \cite{Jozsa94}.

What was striking philosophically was the two disparate foundations for this strange quantity.  The previous time it had cropped up made essential use of the notion of ``purification.''  This is the idea that a mixed quantum state can always be thought of as the restriction to a sub-system of a pure state on a larger, more inclusive system.  The use of this strategy in quantum information is sometimes called ``appealing to the Church of the Larger Hilbert Space'' (i.e., an idea loosely related the Everett Many-Worlds Interpretation of quantum mechanics)\footnote{See \url{http://www.quantiki.org/wiki/The_Church_of_the_larger_Hilbert_space}. To contrast this with the Quantum Bayesian view, see Matt Leifer's  ``church of the smaller Hilbert space'' at \url{http://matt leifer.info/wordpress/wp-content/uploads/2008/11/commandments.pdf}. }, whereas in my derivation, there was only the system itself and all notions surrounding it were defined in operational or experimental terms.  I took a lesson from my result:  The Church of the Larger Hilbert Space may not be so necessary to quantum information as its advocates often advertise it to be.\footnote{A particularly extreme example of this is David Deutsch \cite{Deutsch97}, who claims that the {\it only\/} way to understand the computational speed-up of quantum computers over classical ones is to adopt a Many-Worlds Interpretation of quantum mechanics.}

In any case, this was the launch of a long-term interest in distinguishability measures for quantum states---from generalizations of the one above, to norm distances like
\be
d_{tr}(\rho_0,\rho_1)={\rm tr}\,|\rho_1- \rho_0|\;,
\label{Albert}
\ee
and relative and mutual information measures like
\be
d_S(\rho_0,\rho_1)=S\!\left(\,\frac{1}{2}\rho_0+\frac{1}{2}\rho_1\right)-\frac{1}{2}S(\rho_0)-\frac{1}{2}S(\rho_1)
\label{Wizzy}
\ee
(for $S(\rho)=-{\rm tr}\rho\ln\rho$ the von Neumann entropy), along with all their interrelations---and it became the subject of the larger part of my PhD thesis \cite{Fuchs96a} (144 cites).  Two significant papers of mine from that era are \cite{Fuchs94a} (62 cites)  and \cite{Fuchs99a} (178 cites).

\subsubsection{2.1.2 Cloning and No-Broadcasting}
\label{No-Bro}

It is often underappreciated that taking a stand on the interpretation of quantum states carries with it a fairly distinctive force on one's research.  For, the stand one takes implicitly directs the analogies (and disanalogies) one will seek for comparing classical physics to quantum physics.  At least this is the case for an information theoretic conception of quantum states.

In 1996, I was quite taken with a point that both Asher Peres and Michael Berry emphasized in their discussions of quantum chaos and more broadly:  That in making a comparison between quantum mechanics and classical Hamiltonian mechanics, the proper correspondence is not between quantum states and points in phase space, but between quantum states and {\it Liouville distributions\/} on the phase space.  The key insight is that the points of phase space are meant to represent states of reality, whereas the Liouville distributions are rather explicitly meant to represent one's ignorance of a state of reality.  In an information theoretic conception of quantum states, the quantum states too should not be states of reality, but ignorance of something (maybe not of a state of reality, but nonetheless ignorance of {\it something\/}).

The attitude I developed was that if the analogy worked once, then it should work again and again, at least within a certain range of phenomena.  Furthermore, maybe one could even take the insight the other way around, from quantum to classical.  Thus it came to me one evening that despite all the hoopla over the no-cloning theorem,\footnote{Examples have abound since its discovery, with people sometimes saying it is the key principle behind all of quantum mechanics.  To mention a case where no less than a Perimeter Institute DRC is guilty of it, see \cite{Susskind07}.} there was nothing particularly quantum mechanical about it.  Classical Liouville evolution preserves phase space volume---this any graduate student versed in Goldstein's classical-mechanics book knows---but a less emphasized consequence of it is that Hamiltonian evolution must preserve the overlap between Liouville distributions as well (this is what Peres and Berry had been emphasizing in their quantum chaos work).  But then the same argument that drives the proof of the no-cloning theorem for quantum states would also drive it for classical Liouville distributions---for the no-cloning theorem is a nearly immediate consequence of the fact that unitary evolutions preserve Hilbert-space inner products.  So, nonorthogonal quantum states cannot be cloned, but neither can nonorthogonal classical Liouville distributions \cite{Caves96} (60 cites).

Excited in my act of destruction, I told the result to Ben Schumacher, and he said, ``What one really needs is a task analogous to cloning, but that can be enacted on a set of states if and only if it is a commuting set. That would draw a distinction between classical and quantum {\it even\/} for a Bayesian.'' Richard Jozsa had a key idea that came while he was writing a set of notes we would later publish on mixed-state quantum compression \cite{Barnum01}, ``If two density operators commute, a mutual eigenbasis can always be cloned.''  That is, if $\rho_\alpha\in {\cal L}({\cal H})$ is a commuting set of density operators, then we can find a mutual eigenbasis $|i\rangle$ so that $\rho_\alpha=\sum_i \lambda_\alpha(i)\, |i\rangle\langle i|$, and moreover there is a physical process that can transform one copy of an unknown state $\rho_\alpha$, along with a piece of ``blank tape'' initially in a standard state $\sigma=|1\rangle\langle 1|$, according to,
\be
\left(\sum_i \lambda_\alpha(i)\, |i\rangle\langle i|\right)\otimes \sigma \; \longrightarrow \; \sum_i \lambda_\alpha(i)\, |i\rangle\langle i|\otimes|i\rangle\langle i|\;.
\label{Doofus}
\ee

These observations led to generalizing the cloning question in the following way.  Supposing that a system will be secretly prepared in one state drawn from a given set $\rho_\alpha\in {\cal L}({\cal H}_A)$, under what conditions will it be possible that there exists an auto\-matic device for enacting a transformation $\Phi: {\cal L}({\cal H}_A)\otimes{\cal L}({\cal H}_B)\longrightarrow{\cal L}({\cal H}_A)\otimes{\cal L}({\cal H}_B)$ (i.e., a trace pre\-serv\-ing completely positive linear map) such that for some standard state $\sigma\in{\cal L}({\cal H}_B)$,
\be
\Phi(\rho_\alpha\otimes\sigma)=R_\alpha
\ee
with
\be
{\rm tr}_A (R_\alpha) = \rho_\alpha \qquad\quad \mbox{and} \qquad\quad {\rm tr}_B (R_\alpha) = \rho_\alpha \;\,?
\label{HangYourPants}
\ee
Eq.~(\ref{HangYourPants}) made no demand that $R_\alpha$ be of the form of the right-hand side of Eq.~(\ref{Doofus}); it allowed for completely general entangled states, for instance. Charles Bennett suggested we call this kind of process ``broadcasting'' to mark its distinction from cloning, and the name stuck.

Based on little more than ``hope,'' we hoped that the answer would be that such an automatic device would exist if and only if all the $\rho_\alpha$ commute. The intuition was simply that when all the $\rho_\alpha$ commute, they could be thought of as representing a kind of classical ignorance about some underlying state of reality, $|i\rangle$.  Then just as with a Liouville distribution, the $\rho_\alpha$ themselves could not be cloned, but the underlying states of reality could---the automatic device would just make a nondemolition measurement on the orthogonal set $|i\rangle$ and reproduce the one it found.  This would produce a state of classical correlation.  On the other hand, when the $\rho_\alpha$ do not all simultaneously commute, there could be no such common underlying reality beneath them (even in imagination), and there would be nothing for the automatic device to ``grab onto'' for the broadcasting process.

It was time to role up the sleeves and simply prove the thing.  Well, it was not so easy, and it took us quite some weeks of concerted effort.  Carlton Caves, Howard Barnum, and I had soon lit upon the idea of running an analogy to pure-state cloning by trying to find some quantity of ``overlap'' that could only nondecrease under proper physical evolutions, but must nonincrease for broadcasting to be possible. Presumably, the case when the $\rho_\alpha$ all commuted would be the only one in which the quantity could stay uniformly flat.  But what quantity would be a good one for tracing these needed changes in ``overlap'' for mixed states?  Caves put his money on a quantity analogous to Eq.~(\ref{Wizzy}); Barnum put his on a proper distance measure analogous to Eq.~(\ref{Albert}).  I doggedly stuck with Eq.~(\ref{Kiki}).  Every day we three would meet and compare notes, and the hope would wax and wane.  ``Maybe, after all, broadcasting is too weak a notion and cannot define an `if and only if commutativity' condition?''  That is the kind of fear that crept in, but luckily in the end, the initial intuition sustained itself long enough.

To make a long story short, studying the dynamics of Eq.~(\ref{Kiki}) triumphed, and a proof was eventually in hand \cite{Barnum96a} (322 cites).  Looking back on it, I am still proud of this as one of my best pieces of purely technical work.  G\"oran Lindblad, for instance, in a later extension of the result \cite{Lindblad99}, called the original proof ``ingenious,'' and I was very proud of that.

I linger on the history of this paper in such detail for a second purpose,  however.  It is because its genesis is a particularly clear example of the research style I developed then and have carried with me ever since.  By that time, I had come under the influence of Charles Bennett and the IBM group, and I thought their philosophy that ``a paper should be a party'' was sound.  Cooperation in science, I felt, sped it along far faster than competition.  So, when a big problem comes up, everyone throws in what they can, and everyone gets alphabetical credit in the author list.  The IBM group, with collaborators accumulated around the world, was notorious for huge author lists on their papers, but everyone was having fun, and the science flowed like wine.

That was the way I wanted to see this paper.  Schumacher's contribution had been the remark above, but it was the remark that got the whole effort going in the first place.  Jozsa's contribution too was limited to Eq.~(\ref{Doofus}).  The weeks and weeks of labor trying to show a converse were left to Barnum, Caves, and me.  When it was time to write it all up, my advisor suggested that a three-author scheme was not out of the question, and furthermore he saw me as the first author.  I held fast, and my opinion won the battle.  Since then, I have held to the alphabetical/party mentality in all my publications, excepting only my contributions to experimental work.  Further points on the issue of research style will be expanded in Section 5.

But returning to science: The legacy of the no-broadcasting theorem in my mind was that it taught me that there seems to be an intimate connection between the structure of quantum states and what can be done with them, on the one hand, and that if they are information, what that information could possibly be about, on the other.  The driving intuition behind the no-broadcasting theorem (at least for Caves and me) had been that when states do not commute, then one cannot even pretend there is an unknown state of reality underneath them for which they describe our ignorance of.  Still, which comes first conceptually?  The structure of the states of information, or the tougher question of what the information is about?  Which informs which the most directly?

For a while, I banked that it was better to learn about the structure of states, and the no-broadcasting theorem certainly felt like it had taught me something. Maybe the key point behind quantum mechanics wasn't the shape of distinguishability {\it per se\/} (as was the emphasis in Wootters' original investigations \cite{Wootters80,Wootters81}), but the way distinguishability {\it changes\/} when one attempts attempts to gather information from a quantum system?  Thus I ranged around for more cloning and more broadcasting problems, and more generally any tools that would help me sharpen the issue.  Getting a sense of the the fundamental limits to ``partial cloning'' (and later ``partial broadcasting'') was an example of such an exercise.  One notable paper from that period is \cite{Bruss98} (346 cites).

\subsubsection{2.1.3 Information-Disturbance Tradeoff}

Suppose one stores the value of a single bit in a classical system (the system is either prepared in a state 0 or a state 1) and then passes it off to a courier for delivery elsewhere.  If the courier is very careful, he can in principle have a look at the message en route without leaving a trace of his curiosity.  But this is not so when the bit is encoded in nonorthogonal quantum states $|\psi_0\rangle$ and $|\psi_1\rangle$.  If the courier  tries to extract even an $\epsilon$ of information, he will leave a disturbance in his wake.  That kind of phraseology was as old as quantum mechanics itself.  But in the absence of a hidden variable theory, what could it possibly mean?  Just what was disturbed?  One wondered whether talk of ``disturbance'' might have been a pile of hollow words since the beginning.\footnote{For an extended discussion of this point, see my \cite{FuchsJacobs01} (65 cites).}

Asher Peres and I decided that the place to look for the wake of the information gathering was in $|\psi\rangle$ itself.  If the courier gathers any $\epsilon$ of information about the  identity $|\psi_0\rangle$ vs $|\psi_1\rangle$, it should leave its mark {\it with respect to the sender\/} by transforming these two states into some mixtures, $\rho_0$ and $\rho_1$.  Particularly, it would have to be the case that
\be
d(\rho_0,\rho_1)\le d\Big(|\psi_0\rangle\langle \psi_0|,|\psi_1\rangle\langle \psi_1|\Big)\;.
\ee
We wanted to refine this idea and ask what is the best possible tradeoff between information gathering and reciprocal disturbance.  The grand hope was that we would get a beautiful equation---one that would rival the Heisenberg uncertainty relation $\Delta x \Delta p\ge \hbar/2$ in simplicity---for we both thought that the information-disturbance tradeoff was near the core of what quantum mechanics was all about.  Instead, we got a {\it royal mess\/} \cite{Fuchs96b} (168 cites), and later refinements of it \cite{Fuchs98} (83 cites) did little better:
\be
P_{\rm success}(d)=\frac{1}{2} + \frac{1}{2}\sqrt{1-x^2}\sqrt{
2\sqrt{\frac{d(1-d)}{d_0(1-d_0)}\,}\;-\;
\frac{d(1-d)}{d_0(1-d_0)}\;}\;\;.
\label{Hooha}
\ee
What is expressed here is that if $|\langle\psi_0|\psi_1\rangle|=x$, then in order to get enough information to guess with a probability $P_{\rm success}\ge1/2$ which state is actually there, one must induce (at a minimum) an appropriate disturbance $d\in[0,d_0]$.  The more probability of success one wants, the more disturbance one will have to induce.

Still, there were fruitful foundational insights from seeing how to formulate the problem most generally, and all of this work connected up with a very natural playground of practical application:  namely, the needs of quantum cryptography.  Another paper along the same lines, though more relevant for the BB84 quantum key distribution protocol was \cite{Fuchs97a} (337 cites).

Another paper strangely associated with my fascination for information-disturbance tradeoff issues had to do with a phenomenon we dubbed ``nonlocality without entanglement'' \cite{Bennett99} (357 cites).  This was a multifaceted paper with nearly as many authors as facets, but most of it centered around an example of two qutrits (i.e., two three-level systems) that might have been prepared in one of nine possible states:
\bea
&|1\rangle\otimes|1\rangle& \nonumber\\
|0\rangle\otimes|0+1\rangle && |0\rangle\otimes|0-1\rangle \nonumber\\
|2\rangle\otimes|1+2\rangle && |2\rangle\otimes|1-2\rangle \\
|1+2\rangle\otimes|0\rangle && |1-2\rangle\otimes|0\rangle \nonumber\\
|0+1\rangle\otimes|2\rangle && |0-1\rangle\otimes|2\rangle \nonumber
\eea
where $|0\rangle$, $|1\rangle$, $|2\rangle$ represent an orthonormal basis for each system, and $|0+1\rangle$ stands for the state $\sqrt{\frac{1}{2}}\Big(|0\rangle+|1\rangle\Big)$, etc.  What is fascinating about this set of states (dreamed up by Peter Shor and David DiVincenzo) is that they form an orthonormal basis for the bipartite system {\it yet\/} they cannot be distinguished reliably by separated observers restricted to performing measurements on their own systems alone. And this remains true even if the observers are allowed any sequence of (arbitrarily gentle) local measurements, aided by any amount classical communication between themselves.  For instance, one could imagine very elaborate ping-ponging strategies where the left-hand observer performs a very gentle measurement, syphoning off an $\epsilon$ of information about the preparation, then communicating that to the right-hand observer who optimizes his own gentle measurement in light of what was found on the other side, and then back and forth and back and forth, etc.  Nothing like that would help.  My main involvement in the paper was to actually prove that, with David DiVincenzo.

The technique was to recognize that for localized observers, this was an information-disturbance tradeoff problem.  Orthogonal though the states might be in God's eyes, the possibilities for the individual systems were not, and that was the only currency of any importance for localized observers. Like every other example reviewed so far, this I felt taught me something deep about the foundations of quantum mechanics and sped me toward the Quantum Bayesian view that I will describe later in the ``current research'' section.  In my mind, it was another demotion of the concept of entanglement as being of any great foundational importance.  ``The mystery is there whether you got entanglement or not.''

\subsubsection{2.1.4 Channel Capacity}

It is hard not to think about information all the time without also thinking some about channel capacities.  On this subject, I want to mention two papers I have always felt never quite got the credit they deserved.  The first one \cite{Fuchs97b} (63 cites) was once again spawned by foundational concerns, but nonetheless served a practical question.

Suppose one has access to a quantum channel $\Phi$ that takes pure states $|\psi\rangle$ as its inputs and returns mixed states $\rho$ at its outputs.  A question one can ask is how best to encode classical bits 0 and 1 into quantum states $|\psi_0\rangle$ and $|\psi_1\rangle$ so that the maximum number of bits per transmission can be reliably retrieved by quantum measurements at the receiver's end.  Because of some deep theorems of Holevo \cite{Holevo98} and Schumacher and Westmoreland \cite{Schumacher97}, one could pose this question at once both in its most general form and in a rather tractable way.  It was just a question of maximizing a certain function over the states $|\psi_0\rangle$ and $|\psi_1\rangle$ and a further parameter $x\in[0,1]$.  What struck me, however, was the nonlinear character of the function.  It seemed hard to imagine that it would be maximized by orthogonal $|\psi_0\rangle$ and $|\psi_1\rangle$.  But if not, to get an optimal signal one would have to throw away information at the outset, in the very act of the coding $0\mapsto |\psi_0\rangle$ and $1\mapsto |\psi_1\rangle$.  That is, the recommendation of the theory would be to throw information away before even sending a signal into the channel. Moreover, because of the no-broadcasting theorem described above, it would mean that even though the task at hand was to send a classical bit from one end of the channel to the other, the best coding would choose states for which no underlying classical reality could be envisioned.  So, in a sense, not only would one be throwing away distinguishability with the encoding, but even throwing the bit out of existence in the interim between the sender and receiver.

Still that was all stumbly, mumbly intuition.  What really shored it up was that I had gone through the exercise before, but when there was no coding theorem so general as the Holevo--Schumacher--Westmoreland one.  So long as the receiver had no quantum memory, and therefore had to make a measurement on each quantum signal separately, one could indeed show that the classical intuition held (for qubit channels at least):  Orthogonal input states were always optimal for maximizing classical information capacity \cite{Fuchs97b}.  That problem had corresponded to maximizing a function every bit as nonlinear as the one described above, but the classical intuition held nonetheless.

However in the generalized setting with limitless quantum memory, such was not the case!  Nonorthogonal quantum states did in fact maximize classical information capacity, I was able to show with a counterexample. This genuinely took me by surprise, and it still awaits a deeper understanding, even within my own Quantum Bayesian point of view on quantum theory.  Still, though the effect was called ``very surprising'' \cite{Schumacher01} and the work ``seminal'' \cite{Daems09}, extended in the work of King and Ruskai \cite{King02} and Shor \cite{Shor97}, and cited in well-read review articles \cite{Bennett00} and \cite{Bennett02}, it never garnered much attention.

A more mysterious case arises with the second paper I wish to mention \cite{Bennett97} (70 cites).  One early evening in 1996 in the basement bar of the Villa Gualino in Turin, Italy, Charles Bennett and I realized that we did not know why separate transmissions through a quantum channel (for the purposes of sending classical information) needed to be constructed from a single alphabet at all.  If a quantum channel is represented by a trace-preserving completely positive map $\Phi$, then two actions of the channel on separate transmissions will be represented by $\Phi\otimes\Phi$.  The use of the tensor product symbol for two transmissions made it clear that one could imagine entangling the separate transmissions before sending them on their way.  For instance, if one were to transmit messages by sending a set of possible entangled states $|ent_i\rangle$, it would generally lead to entangled outputs $\Phi\otimes\Phi\big(|ent_i\rangle\langle ent_i|\big)$ , and these might be more distinguishable than if one had restricted oneself to product-state inputs.

Indeed it was the case, as we showed in \cite{Bennett97}.  But we also posed a tougher question at the end of that paper, and as it turned out, a much tougher one.  Might one be able to transmit a greater number of reliable bits per transmission if one dropped the requirement of building the signals from a fixed alphabet of states?  This, like in the case of the last paper I described, was a question of capacity.  Could one achieve a greater capacity by encoding in terms of entangled states rather than unentangled ones?  Before our conversation in the Villa Gualino, it seems the question had never been posed.  And that was very strange, actually.  For, the key ingredient in the Holevo--Schumacher--Westmoreland capacity theorems was the consideration of collective measurements on whole codewords at the receiver's end, making necessary use of Hermitian operators with entangled eigenvectors.  That issue had been recognized as early as 1979 by Holevo himself \cite{Kholevo79}.  But entanglement at the sender's end was new territory.  I remember well asking Holevo in 1996, as he was studying a draft of our paper, whether he had ever considered entangled encodings; he said, ``No.''  I asked, ``Why not?''  He said, ``One thing at a time. It never occurred to me.''  He then went on to say some things about how Bohr's way of thinking about quantum mechanics had maybe blinded him somehow.

The question of whether entangled encodings could beat the single-letterized capacities of the Holevo--Schumacher--Westmoreland theorem became known as the question of the {\it additivity\/} of the classical capacity of a quantum channel, and trying to settle it became a cottage industry in quantum information theory for nearly 12 years.  Particularly, once Peter Shor \cite{Shor04} showed the equivalence of this question to several other (later posed) additivity questions, things took off, and there were {\it dozens\/} of papers on the subject. The issue was finally laid to rest in 2008 with Matt Hastings' discovery of a counterexample to another one of the questions in Shor's equivalence class \cite{Hastings09}.  Consequently, the classical capacity of a quantum channel can after all be increased by the aid of entangled encodings, as we had originally speculated.

But, why so few citations to this work?  It was without doubt influential on the community and cited as inventing the problem in some of the key papers along the 12 year trek, for instance in the Bennett and Shor review article \cite{Bennett02}, as well as in a precursor to Shor's equivalence result \cite{Matsumoto04}.   My best guess is that the question of additivity of classical capacity simply became part of the folklore of quantum information.  Most people probably thought the question was posed already in the original Holevo and Schumacher--Westmoreland papers \cite{Holevo98,Schumacher97}.

\subsubsection{2.1.5 Quantum Teleportation and Measures of Quantumness}
\label{HJeffrey}

Being around a quantum optics group at Caltech, I became a little annoyed with hearing a coherent state of light
\be
|\alpha\rangle=e^{-\frac{|\alpha|^2}{2}}\sum_{n=0}^\infty \frac{\alpha^n}{\sqrt{n!}}\,|n\rangle
\ee
always being called a ``classical state.''  After all, no two coherent states are orthogonal to each other.  And for instance, if $|\langle\alpha_1|\alpha_2\rangle|=x$, then there will be an information-disturbance tradeoff curve Eq.~(\ref{Hooha}) for them, just as for any other two quantum states.  So it struck me that the idea of speaking of a ``classical state'' without explicit consideration of the set of states it is embedded in is a malformed idea.

Take the example of the no-broadcasting theorem again.  A lesson learned from it is that if a set of states do not commute, then one cannot even pretend they capture the potential information one might have about an underlying (clonable) reality.  But, could one quantify the degree to which a set of states where ``truly quantum'' in this or a related sense?  This spurred me to want to develop a notion of ``how quantum'' a set of quantum states are with respect to each other.

There was a range of possible measures, even when the embedding set consisted of just two quantum states \cite{Fuchs98b,Luo09,Luo10}.  I found something intriguing, however, about a measure that attempted to quantify how difficult it would be to substitute a classical fact\footnote{ In my later quantum foundational persona, I would call it an ontic state, along the lines of Spekkens' terminology \cite{Spekkens07} \ldots} for a quantum state\footnote{\ldots\ to contrast it with my predilection for thinking of quantum states as epistemic states.}. I called it the problem of ``trying to squeeze quantum information through a classical channel'' and quantified the quantumness \cite{Fuchs03} of a set of quantum states by the best case average fidelity an eavesdropper could achieve between input and output if she tried to do just that.

Pure quantum foundations considerations again, but when Jeff Kimble asked my advice for a criterion for successful quantum teleportation, I was able to pull my favored quantumness measure off the shelf and offer it for his use \cite{Braunstein00} (132 cites).  Lo and behold, Kimble was teleporting nothing but randomly chosen coherent states, precisely those states called ``classical'' in the usual view---there could not have been a better reason for my wanting to use my quantumness measure for something so grubby as an actual experiment!  The use was simply this:  Without a sufficient amount of entanglement between Alice and Bob, the whole teleportation feat might well have been pulled off by Alice measuring the quantum state, talking to Bob over a telephone (i.e., a classical channel), and then having Bob reproduce a fresh quantum state at his end.  If one could show that the actual laboratory treatment achieved a fidelity between inputs and outputs on average that exceeded the bound placed by the quantumness of the set of states, then one was home free.

I modeled the selection of a random coherent state via selection from a Gaussian distribution centered on the vacuum state, with variance $\frac{1}{2\lambda}$.  Then I was able to show that the quantumness of such an ensemble gave a best case fidelity of
\be
F(\lambda)=\frac{1+\lambda}{2+\lambda}\;.
\ee
In the limit that $\lambda \rightarrow 0$, the variance becomes infinite, and $F\rightarrow \frac{1}{2}$.  This was the number Kimble needed to know, and the Furusawa et al.\ experiment was able to achieve an average case fidelity of $F=0.58\pm0.02$.  We wrote the result up \cite{Furusawa98} (1768 cites), and the paper was listed by the editors of {\sl Science\/} as one of its top ten ``breakthroughs of the year 1998.''  (The {\it real\/} breakthrough of the year with which we shared the list, however, was without doubt the discovery of the acceleration of the universe's expansion.)

To this day, I still find myself intrigued by this notion of quantumness.  Here is the way I put my motivation for continued study of it in  \cite{Fuchs04}:
\bq\small
Associated with each quantum system is a Hilbert space. In the case of finite dimensional ones, it is commonly said that the dimension corresponds to the number of distinguishable states a system can ``have.'' But what are these distinguishable states? Are they potential properties a system can possess in and of itself, much like a cat's possessing the binary value of whether it is alive or dead? If the Bell--Kochen--Specker theorem has taught us anything, it has taught us that these distinguishable states should not be thought of in that way.

In this paper, I present some results that take their motivation \ldots\ in a different point of view about the meaning of a system's dimensionality. From this view, dimensionality may be the raw, irreducible concept---the single property of a quantum system---from which other consequences are derived \ldots. The best I can put my finger on it is that dimensionality should have something to do with a quantum system's ``sensitivity to the touch,'' its ability to be modified with respect to the external world due to the interventions of that world upon its natural course. Thus, for instance, in quantum computing each little push or computational step has the chance of counting for more than in the classical world.
\eq
In that paper, I aimed at finding the maximum quantumness that could be supported on a Hilbert space of dimension $d$.  Particularly, I wanted to see how quantumness scaled as the dimension grew.

First I proved that the maximum quantumness for a given dimension could be achieved with the continuous infinity of states that make up the Hilbert space as a whole.  But then I showed that one could achieve the same quantumness with a finite set of states $|\psi_i\rangle$ with only $d^2$ elements.  The only thing was that the states had to have the right symmetry with respect to each other.  Namely,
\be
|\langle\psi_i|\psi_j\rangle|^2=\frac{1}{d+1} \qquad \mbox{when} \qquad i\ne j \;.
\label{BigBoy}
\ee
To this day, we still do not know whether $d^2$ states with such a symmetry can always be found in every $d$, but they have been taking an ever more important role in my foundational considerations since.  This will be described in Sections 3 and 4, but at the moment it is time to turn back the clock to the story of the more focussed foundational program that led up to this question.

\subsubsection{2.2 AD -- Quantum Foundations in the Light of Quantum Information}

\noindent\textbf{2.2.1 Quantum Theory Needs No `Interpretation'}
\label{Bitterness}

The turning point from BC to AD came from the process of trying to flesh out an ``opinion piece'' on quantum interpretations for {\sl Physics Today\/} with Asher Peres.  Previous to that, {\sl Physics Today\/} had been publishing (what seemed like a long) series of articles on how quantum mechanics was inconsistent at its core.  Asher and I had long been correspondents on foundational matters, and much of my opinion on quantum theory had been developed from reading his papers.  So it seemed natural to throw in together on such a project.  We both thought it was nonsense that quantum mechanics might be called inconsistent; it was just that its scope was being misunderstood by the naysayers.  There were foundational obscurities in the theory for sure---this we both agreed---but these were not marks of inconsistency, instead {\it markers\/} for where to focus our attention and learn the most from it.

Little did I know how life changing it would be to write such a piece and how particular I would become in my choice of words for expressing quantum theory.  If you want to know what you really believe about quantum theory, try to write an article on its interpretation with someone you {\it believe\/} holds the same beliefs as you!  It will make you recognize every nuance in everything you say, and it will lay open every difference in your own thought from your collaborator's in a way that no relaxed reading of his papers can.

The process of writing that two page opinion piece \cite{Fuchs00} (144 cites) was harder than any scientific paper I had ever undertaken.  The words from one of Bryce DeWitt's lectures (with whom I had taken an undergraduate Lagrangian mechanics course) finally hit home:  ``We learn mathematics so that we don't have to {\it think\/} when we do physics.''  That little paper was all thought, and a good amount of negotiation and compromise as well.

We titled it {\sl Quantum Theory Needs No `Interpretation'\/} because Asher was taken with something Rudolf Peierls once said: that, ``The
Copenhagen interpretation {\it is\/} quantum mechanics!'' Of course the whole paper was about our own direction of interpretation.  We both learned so much in writing it.  For instance, for myself it was the first time I had ever articulated in epistemic terms what quantum teleportation was all about.  Even more surprising was that Asher, one of the authors of the original teleportation paper \cite{Bennett93}, had not forced himself to properly understand the phenomenon in those terms (i.e., his very own interpretation).  In 2002, there was a discussion between all the authors of the teleportation paper, as well as David Mermin, Ben Schumacher, John Smolin, and myself on what definition should be adopted by the {\sl American Heritage Dictionary\/} for ``quantum teleportation.''  It was very flattering when Asher weighed in on the discussion with this statement in an email titled ``quantum teleportation defined:\ nothing happens at Bob's end'' \cite{Peres02p}:
\bq
\small\tt
\noindent Dear friends,

When the quantum teleportation process was conceived, we had no clear\\
understanding of what was going on, and this may still be true today.  I\\
learnt a lot from Chris when we wrote our ``Opinion'' essay in Physics\\
Today, March 2000, pp.\ 70,71. Here is the passage about quantum\\
teleportation, mostly due to Chris: \ldots
\eq

Still, I have already emphasized the negotiation and compromise aspects of our {\sl Physics Today\/} paper.  What became clear to me in the writing (as I would try to find words that would placate Asher at the same time as giving me no sense of being dishonest) was that indeed there was no inconsistency in quantum theory, but instead it was we quantum-state epistemicists who were not completely consistent in our descriptions of things.  Much more work was called for, and it was time to break from my ``quiet agenda'' of the previous years of ``learning about information {\it in\/} quantum theory''---it was time to more carefully understand the reverse, how the quantum fits {\it within\/} information theory.

Key to this shift was taking very seriously the personalist Bayesian conception of probability \cite{Bernardo94,Lindley06,DeFinetti90}\@.  How could Peres and I say ``nothing happens at Bob's end'' (upon Alice's measurement in the quantum teleportation protocol) unless the quantum state had no {\sl factivity\/}\footnote{This term was introduced in this context by Chris Timpson. See, for instance, his excellent review article on Quantum Bayesianism \cite{Timpson08}, as well as \cite{Timpson06,Timpson07}.} about it at all.  That is, though we had long been accustomed to calling a quantum state a ``state of information'' or a ``state of knowledge,'' that no longer seemed to be quite the right terminology.  A state of knowledge, under most understandings, is made right or wrong by external circumstances.  Plato, for instance, defined knowledge to be ``justified, true belief.''  If one accepts a definition of knowledge along those lines, then if $|\psi\rangle$ is a state of knowledge, it can only be so by being ``true'' with respect to some external circumstance.  In other words, one might as well say a quantum state corresponds to a fact after all.  But this would mean that instantaneously after Alice's measurement (and before any classical communication between Alice and Bob), the quantum system in Bob's possession had better be prepared to respond to any further measurement upon it in a way appropriate to the teleported $|\psi\rangle$ (i.e., modulo Bob's yet-to-be-performed final unitary operation). It would not matter whether the fact was located upon Bob's system or was instead some kind of delocalized fact possessed by the universe as a whole---either way, Einstein's ``spooky action at a distance'' would be alive and well.

It must be admitted that I present this train of thought to be significantly more linear than it really was at the time. The trouble is I don't know how to present it in any quickly digestible form that is true to the painful (somewhat random) every-day-over-and-over consistency check it really was.  It was damned hard to give up on the idea that a $|\psi\rangle$ should be factive, and all my instincts fought against it initially---it was only with time that my strength rallied, and that itself only happened in great part because of the even greater resistance my collaborators Carlton Caves and R\"udiger Schack and correspondent David Mermin heaped upon me.  I had to defend those nascent thoughts, and I became stronger in the process.  The best I can do is refer the reader to my samizdat {\sl My Struggles with the Block Universe\/} \cite[pages 44--275]{Fuchs10c} if he or she wants to see the logic play out in real time.

The upshot was that I ever more vigorously embraced a view in which {\it all\/} probabilities calculated from quantum states should be taken as personalist Bayesian probabilities.  By this turn, a quantum state would not be described as a state of knowledge, but more accurately as a compendium of personalist Bayesian degrees of {\it belief}.  Sometimes I would simply say that a quantum state itself was a ``state of belief.'' (You can imagine that this might not be something that would go down well in an audience at Caltech, say.  It is a good thing a scheduling conflict arose when I was asked to give the physics colloquium there!\footnote{``Quantum Foundations in the Light of Quantum Information,'' initially scheduled for 17 February 2005.})

In any case, because of this process, I leaned more and more heavily on my collaborators as well as on myself to make sure that we had dotted every I and crossed every T in our conception of quantum theory.  The effort earned us a label in the quantum foundations wars---we became known as the ``Quantum Bayesians'' \cite{Caves02} (117 cites)---as we published our efforts
\cite{Fuchs01,Barnum00,Schack01,Caves02b,Caves02c,Schack02,Fuchs02,Fuchs03b,Fuchs04b,Fuchs04c,Caves05,Appleby05,Fuchs05,Appleby05a,Appleby05b,Leifer06,Caves07,%
Fuchs09} (734 cites cumulative)
to secure an airtight consistency.\footnote{Physicists were generally kind and tolerant of us---every now and then they would even use an equation that we had developed with this effort---but one thing is for sure of philosophers of science:  They don't much like a new interpretation of quantum mechanics coming onto the scene \cite{Mohrhoff01,Hagar03,Dennis04,Marchildon04,Ferrero04,Hagar05,Hagar06,Wallace06,Mohrhoff07,Wallace07,Palge08,Jaeger09}.}

\subsubsection{2.2.2 Quantum de Finetti Theorem}
\label{JuiceNewton}

A case in point in such T crossing is the technical apparatus we had to develop to make sense (from our point of view, of course) of one of the most common phrases in all of quantum information theory---the ``unknown quantum state.''  This was the motivation for formulating and proving a {\it quantum de Finetti representation theorem\/} \cite{Caves02b} (134 cites).

The term ``unknown state'' is ubiquitous in quantum information:  Unknown quantum states are teleported, protected with quantum error correcting codes, used to check for quantum eavesdropping, and arise in innumerable other applications.  For a Quantum Bayesian, though, the phrase can only be an oxymoron:  If quantum states are compendia of beliefs, and not states of nature, then the state is known to someone, at the very least the agent who holds it.  But if so, then what are experimentalists doing when they say they are performing quantum-state tomography in the laboratory?  The very goal of the procedure is to characterize the unknown quantum state a piece of laboratory equipment is repetitively preparing. There is certainly no little agent sitting on the inside of the device devilishly sending out quantum systems representative of his beliefs, and smiling as the experimenter on the outside slowly homes in on those private thoughts through his experiments.

The quantum de Finetti theorem is a result that allows the story of quantum-state tomography to be told purely in terms of a single agent---namely, the experimentalist in the laboratory.  In a nutshell, the theorem is this.  Suppose the experimentalist walks into the laboratory with the very minimal belief that, of the systems his device is spitting out (no matter how many), he could interchange any two of them and it would not change the statistics he expects for any measurements he might perform.  Then the theorem says that ``coherence with this belief alone'' requires him to make a quantum-state assignment $\rho^{(n)}$ (for any $n$ of those systems) that can be represented in the form:
\begin{equation}
\rho^{(n)}=\int P(\rho)\, \rho^{\otimes n}\, d\rho\;,
\label{MushuPork}
\end{equation}
where $P(\rho)\, d\rho$ is some probability measure on the space of single-system density operators and $\rho^{\otimes n}=\rho\otimes\cdots\otimes\rho$ represents an $n$-fold tensor product of identical quantum states.

To put it in words, this theorem licenses the experimenter to act {\it as if\/} each individual system has some state $\rho$ unknown to him, with a probability density $P(\rho)$ representing his ignorance of which state is the true one.  But it is only {\it as if}---the only active quantum state in the picture is the one the experimenter actually possesses in his mind, namely $\rho^{(n)}$.  When the experimenter performs tomography, all he is doing is gathering data system-by-system and updating, via Bayes rule \cite{Schack01}, the state $\rho^{(n)}$ to some new state $\rho^{(k)}$ on a smaller number of remaining systems. Particularly, one can prove that this form of quantum-state assignment leads the agent to expect that with more data, he will approach ever more closely a posterior state of the form $\rho^{(k)}=\rho^{\otimes k}$.  This is why one gets into the habit of speaking of tomography as revealing ``the unknown quantum state.''

One of the very nice things about the quantum de Finetti theorem is that even though it was my first technical exercise in quantum foundations {\it per se\/} in the AD period---{\it i.e., after coming out of the closet on the subject}---it nonetheless fared like a stand-alone result for quantum information theory.  Among other things, it turned out to be useful for proving the security of some quantum key distribution schemes \cite{Lo05,Renner07,Renner08}, it became an important component in the analysis of entanglement detection \cite{Doherty05,Enk07}, and even served in an analysis of the quantum state of propagating laser light by Steven van Enk and myself \cite{vanEnk02a,vanEnk02b} (88 cites).

\subsubsection{2.2.3 Quantum Mechanics as Quantum Information (and only a little more)}
\label{JockeyBoy}

Thinking of quantum foundations as a {\it field\/} within physics, to many minds, almost seems to present a certain amount of paradox.  How can any part of physics be a sustainable, viable field if its whole purpose is to bring a small number of questions to their resounding ends?  To make the  contrast plain, consider condensed matter physics.  Within the American Physical Society, the Division of Condensed Matter Physics has more than 6,000 members, and the workforce is still not large enough to do all that needs to be done!  It is a field whose subject matter has no limitations; there will always be more to do.  But could one say the same of quantum foundations?

This sort of question started troubling me more and more in the time leading up to the AD transition.  On the one hand, I certainly thought that John Wheeler was on to something when he speculated that understanding ``measurement'' itself (or the ``elementary quantum phenomenon,'' as he called it) was the key for taking physics to a new era \cite{Wheeler88a}.  On the other hand, even after all the time my mind had been turned toward these matters, I still didn't quite know what should count as progress in quantum foundations.  I remember walking and talking with Lucien Hardy at a conference in Hull, England in 1997, asking, ``How do you know when you've made progress in quantum foundations?''  Hardy said, ``That's a good question.  We never ask that.''  I was never quite sure, but I was afraid it was not a joke.

I searched inwardly, and after watching the frustrating theater of a foundations meeting in Greece, my thoughts finally crystalized into a kind of manifesto \cite{Fuchs01,Fuchs02,Fuchs03b} (287 cites, cumulative over three editions).  I did my best to express where I thought quantum foundations needed to go, how we would know when we got there, and why we should go in the first place.  Some of the opening lines of that paper were these\footnote{Silently edited so as not to show the distracting ellipses from all the omissions.}:
\bq
\small
Go to any quantum foundations meeting, and it is like being in a holy city in great tumult. You will find all the religions with all their priests pitted in holy war---the Bohmians, the Consistent Historians, the Spontaneous Collapseans, the outright Everettics, and many more beyond that. They all declare to see the light. Each tells us that if we will accept their solution as our savior, then we too will see the light.

But there has to be something wrong with this!  If any of these
priests had shown the light, there simply would not be the
year-after-year conference.  The verdict seems clear:  If
we really care about quantum foundations, then it behooves us as a community
to ask why these meetings are happening and find a way to put a stop
to them.

My view of the problem is this. Despite the accusations one
sees one religion making against the other, I see little to no
difference in {\it any\/} of their canons.  They all look equally
detached from the world of quantum practice to me. For, though each
seems to want a firm reality within the theory---i.e., a single God
they can point to and declare, ``There, that term is what is real in
the universe even when there are no physicists about''---none have
worked very hard to get out of the realm of pure
mathematics to find it.

What I mean by this deliberately provocative statement is that in
spite of the differences in what the churches label to be ``real'' in quantum
theory, they nonetheless all proceed from the
same {\it abstract\/} starting point---the standard textbook
accounts of the {\it axioms\/} of quantum theory.

``But what nonsense is this,'' you must be asking.  ``Where else
could they start?''  The main issue is this. Where present-day
quantum-foundation studies have stagnated in the stream of history
is not so unlike where the physics of length contraction and time
dilation stood before Einstein's 1905 paper on special relativity.

The Lorentz transformations have the name they do for good reason: Lorentz had published
some of them as early as 1895. Indeed one could say that most of the
empirical predictions of special relativity were in place well
before Einstein came onto the scene. But that was of little
consolation to the pre-Einsteinian physics community striving to make sense of electromagnetic phenomena and the luminiferous
ether. Precisely because the {\it only\/} justification for the
Lorentz transformations appeared to be their {\it empirical
adequacy}, they remained a mystery to be conquered. More
particularly, this was a mystery that heaping further {\it ad hoc\/}
(mathematical) structure onto could not possibly solve.

What was being begged for was an
understanding of the {\it origin\/} of that abstract, mathematical
structure---some simple, crisp {\it physical\/} statements with
respect to which the necessity of the mathematics would be
indisputable. Einstein supplied that and became one of the greatest
physicists of all time.  He reduced the mysterious structure of the
Lorentz transformations to two simple statements expressible in
common language:
\begin{verse}
1) the speed of light in empty space is independent of the speed of
its source, \\
2) physics should appear the same in all inertial reference frames.
\end{verse}
The deep significance of this for the quantum problem should stand up
and speak overpoweringly to anyone who admires these principles.

Einstein's move effectively stopped all further debate on the origins
of the Lorentz transformations.  Most
importantly, with the supreme simplicity of Einstein's principles,
physics became ready for ``the next step.'' Is it possible to
imagine that any mind could have made the leap
to general relativity directly from the original, abstract structure
of the Lorentz transformations?  A structure that was only
empirically adequate?  I would say no. Indeed, one can dream of the
wonders we will find in pursuing the same strategy of simplification
for the quantum foundations.

The task is not to make sense of the quantum axioms by heaping more
structure, more definitions, more science-fiction imagery\footnote{Yes, I was talking about the Many Worlds Interpretation at the time.} on top of
them, but to throw them away wholesale and start afresh.  We should
be relentless in asking ourselves:  From what deep {\it physical\/}
principles might we {\it derive\/} this exquisite mathematical
structure?  Those principles should be crisp; they should be
compelling. They should stir the soul. When I was in junior high
school, I sat down with Martin Gardner's {\sl Relativity for
the Million} and came away with an understanding of
the subject that sustains me today:  The concepts were strange, but
they were clear enough that I could get a grasp on them knowing
little more mathematics than simple arithmetic. One should expect no
less for a proper foundation to quantum theory. Until we can explain
quantum theory's {\it essence\/} to a
high-school student and have them walk away with a deep, lasting
memory, we will have not understood a thing about the quantum
foundations.

So, throw the existing axioms of quantum mechanics away and start
afresh! But how to proceed? I myself see no alternative but to
contemplate deep and hard the tasks, the techniques, and the
implications of quantum information theory. The reason is simple,
and I think inescapable.  Quantum mechanics has always been about
information.  It is just that the physics community has somehow
forgotten this.

This, I see as the line of attack we should pursue with relentless
consistency:  The quantum system represents something real and
independent of us; the quantum state represents a collection of
subjective degrees of belief about {\it something\/} to do with that
system. The structure called
quantum mechanics is about the interplay of these two things---the
subjective and the objective.  The task before us is to separate the
wheat from the chaff.  If the quantum state represents subjective
information, then how much of its mathematical support structure
might be of that same character?  Some of it, maybe most of it, but
surely not all of it.

Our foremost task should be to go to each and every axiom of quantum
theory and give it an information theoretic justification {\it if we can}.
Only when we are finished picking off all the terms that can be interpreted as subjective information will we
be in a position to make real progress in quantum foundations.  The
distillate left behind---minuscule though it may be with respect
to the full theory---will be our first glimpse of what quantum
mechanics is trying to tell us about nature itself.

Let me try to give a better way to think about this by using
Einstein again. What might have been his greatest achievement in
building general relativity? I would say it was in his recognizing
that the ``gravitational field'' one feels in an accelerating
elevator is a coordinate effect. That is, the ``field'' in that case
is something induced purely with respect to the description of an
observer. In this light, the program of trying to develop general
relativity boiled down to recognizing all the things within
gravitational and motional phenomena that should be viewed as
consequences of our coordinate choices.  It was in identifying all
the things that are ``numerically additional'' to
the observer-free situation---i.e., those things that come about
purely by bringing the observer back into the picture.

This was a breakthrough.  For in weeding out all the things that
can be interpreted as coordinate effects, the fruit left behind
becomes clear to sight: It is the Riemannian manifold we call
spacetime---a mathematical object, the study of which one can hope
will tell us something about nature itself, not merely about the
observer in nature.

The dream I see for quantum mechanics is this. Weed out all the
terms that have to do with information,
knowledge, and belief, and what is left behind will play the role of
Einstein's manifold. That is our goal.
\eq

That was my goal then; I still want it now. In a recent grant proposal \cite{Fuchs10d}, I put a related thought this way:
\bq
QBism is one of the first efforts in quantum interpretation that we are aware of that has said, ``A proper interpretation of quantum mechanics should almost close its eyes to the existing mathematical formalism of the theory, and instead {\it imply\/} what the formalism {\it ought\/} to be.''  In that regard we have already had some success---our rewriting of the Born rule is an example of an interpretation-driven change in the very formalism itself---but of course, we want more. In the end, we expect of a Quantum Bayesian interpretation of quantum mechanics that it should actually imply the full mathematical formalism of the theory. An interpretation that did that alone would already be revolutionary in the quantum foundations wars.
\eq

The manifesto in its most cited edition \cite{Fuchs02} (189 cites) was titled, ``Quantum Mechanics as Quantum Information (and only a little more),'' and it aimed to get the wheels turning on just that---showing that a good bit of the formalism {\it really, really\/} looked, felt, and smelt of information, and taking the first steps toward identifying one part of the theory that did not (that was the ``only a little more'').  It presented a potpourri of mathematical results drawn from things I had learned in my exposure to quantum information theory to argue that both quantum states and quantum operations (generalized time-evolution maps) were epistemic in character.  Further it aimed to show that the formalism of positive operator-valued measures (POVMs, generalized quantum measurements) took a very natural place as the primitive notion of measurement in quantum theory---a place much more natural for it than the usual textbook justification \cite{Peres95,Nielsen00} in terms of the Church of the Larger Hilbert Space.

One favored result in the paper is that I demonstrated a new way to think of the L\"uders collapse rule for generalized quantum measurements---particularly, finding a way to display {\it both\/} its similarity and its contrast to Bayes' rule in probability theory.  The usual way to state L\"uders' rule is that if one starts with a quantum state $\rho$ and performs a measurement whose outcome operators are $\{E_k\}$, then after receiving an outcome $i$ the transition to an updated quantum state $\rho_i$ (up to normalization) should be,
\be
\rho\quad\longrightarrow \quad \rho_i=\sqrt{E_i}\, \rho\, \sqrt{E_i}\;.
\ee
In contrast, I showed that one could think of the rule instead as composed of two conceptual pieces:
\bea
\rho \quad\longrightarrow \quad \rho_i^\prime=\sqrt{\rho}\, E_i\, \sqrt{\rho} \quad\longrightarrow \quad \rho_i= U_i\,\rho_i^\prime\,U_i^\dagger\;,
\eea
where $U_i$ is some unitary operator that depends both on the outcome and the initial state $\rho$.  What was so pleasing about this was that it could be shown that the first part of the process corresponded {\it exactly\/} to standard Bayesian updating.  Thus, it isolated the mystery of quantum collapse in the ``unitary readjustment'' that one gives to the Bayesian state.  One might say, it captures the measurement device's ``back-action'' upon the quantum system.  Measurement is not only information gathering, but also in some sense invasive to the physical system measured.

Just a touch of new formalism, but look what it lays bare!  Consider two scenarios.  In Scenario 1, Alice has a single system in front of her which she initially describes according to $\rho$ and then performs the measurement $\{E_k\}$.  The state, as we have already discussed, collapses to $\rho_i$.  In Scenario 2, Alice and Bob share a bipartite system that Alice describes according to some $|\psi_\rho\rangle$ for which it is the reduced state on Bob's side alone that is initially $\rho$.  Supposing she measures $\{E_k\}$ on her own system, using L\"uders' rule, what new state would she ascribe to {\it Bob's system\/} after getting outcome $i$?  It won't be $\rho_i$; let's call it $\sigma_i$ instead---it would take a small calculation to find it (first using L\"uders' rule on the bipartite state and then doing a partial trace operation).

Now let us ask the question, of these two collapses which is the more mysterious?  The common sort of answer that one hears in quantum foundational circles is that it is the one in Scenario 2 that is the great mystery. For it is elicited by the ``spooky action at a distance'' that is the very staple of the field. ``There's nothing very mysterious about a back-action at Alice's own site.''  But from the quantum Bayesian perspective, the collapse in Scenario 2 is required simply because Alice is learning something about Bob's system.  Nothing actually changes physically on Bob's side.

Well, this dichotomy of views has been around since 1935.  Nicely enough, though, in the new formalism it is the equations themselves that lean toward the Bayesian story.  For in Scenario 2, if one does the small calculation just mentioned, one will find that Alice using L\"uders' rule on the full bipartite system will lead her to a purely Bayesian updating for Bob's!  That is, there is no further unitary readjustment for Bob's side.  The bipartite system, instead of exposing action-at-a-distance, makes it clear that Alice's measurement on her system gives no back-action whatsoever to Bob's.  ``Who would have expected otherwise?''\ the Quantum Bayesian smirks.  From this point of view it is Scenario 1 that is the greater mystery!

Interestingly enough, that whole point had come to me because of a proof technique I had developed for deriving a certain quantum entropy inequality in \cite{FuchsJacobs01} (65 cites).  There is something interesting here sociologically, and once again it sets QBism apart from certain other efforts in quantum foundations.  There is seemingly a healthy interplay between doing physics for applications and doing physics for foundations---even proof techniques from the former seem to inform the latter.  I have not mentioned it yet, but this was also the case for the quantum de Finetti theorem of the last subsection.  Not only did a foundationally motivated theorem go forward to impact quantum information more broadly, but it also fed back to indicate a direction for still more foundational progress.  Let me explain.

The way the quantum de Finetti theorem was proved was through a reduction of its statement to an analogous theorem in classical probability theory (the classical de Finetti theorem).  What was key to this reduction was the ability to replace quantum states with simple classical probability distributions---not {\it compendia\/} of probabilities (one for every possible measurement) as we were accustomed to thinking of with regard to quantum states, but by a {\it single\/} probability distribution, a unique one for each state.  This may seem impossible to the uninitiated, but it was a story again made possible by the tools of quantum information.  It required the introduction of an single, fiducial informationally complete POVM of certain character:  A set of operators $\{E_i\}$ fulfilling the usual requirements of a POVM (i.e., all the $E_i$ had to be positive semi-definite and $\sum_i E_i=I$), but also all the $E_i$ had to be rank-one, there needed to be $d^2$ of them, and all of them had to be linearly independent.  Such sets did indeed exist, and a reduction could be enacted through them.  A quantum state $\rho$ could now be uniquely identified through the Born rule probabilities for the outcomes of a single measurement:
\be
p(i)={\rm tr}\rho E_i\;.
\ee

It was all just a means to an end in the quantum de Finetti paper \cite{Caves02b}, but in \cite{Fuchs01,Fuchs02,Fuchs03b} I started taking it  seriously as the right way to think about quantum states.  After all, what is done in a usual representation?  One basically picks an orthonormal basis $|i\rangle$ out of a hat (in quantum information, one anoints it with the name ``the computational basis'' \cite{Mermin07}) and then represents all states $\rho$ thereafter via its matrix elements $\langle i|\rho| j\rangle$.  Why not instead pull an informationally complete POVM out of a hat and represent $\rho$ by its probabilities directly?  To my mind, $\langle i|\rho| j\rangle$ just looked like some abstract beast falling from the sky (perhaps from an Everettian cloud), but $p(i)$ looked like information, only information.  That was the way it should be.

But what did the ``only a little more'' refer to?  The partial answer sketched there was the parameter $d$ telling us how many outcomes we need for a fiducial, informationally complete POVM for a system.  It was the dimensionality of the system, denoting how much ``stuff'' is there.  And ``stuff'' isn't ``information'' or ``belief.''  I will have more to say on this after laying the groundwork of my current research.

\subsection{3. Current Research}
\label{CurtisSliwa}

Looking back on it, the papers \cite{Fuchs01,Fuchs02,Fuchs03b} and \cite{Caves07} were a kind of early announcement for a party.  For, particularly in writing \cite{Caves07}, it started to crystalize in me that my attention must turn to showing that the Born rule itself is at the foundation of quantum theory.  That is, instead of thinking of the Born rule as a theorem to be {\it derived}---as it is thought of in nearly every other interpretation of quantum mechanics \cite{Hartle68,Barnum00,Caves05,Bohm53, Gleason57, Wallace09, Kent09,Coleman94,Zurek03,Zurek09,Schlosshauer04,Dieks07,Saunders04,Aguirre10}---it should rather be taken as a fundamental {\it postulate\/} long in advance of any other mathematical structure in the theory.  The reason for this goes directly to the core of Quantum Bayesianism, or QBism as I started calling it at Perimeter Institute.  It was in a concluding remark of \cite{Caves07}:
\bq\small
We have
emphasized that one of the arguments often repeated to justify that
quantum-mechanical probabilities are objective, rather than subjective, is
that they are ``determined by physical law.'' But, what can this mean?
Inevitably, what is being invoked is an idea that quantum states
$|\psi\rangle$ are independent theoretical constructs from the
probabilities they give rise to through the Born rule,
$$
p(i)=\langle\psi| E_i | \psi\rangle\;.
$$
From the Bayesian perspective, however, these expressions are not
independent at all, and what we have argued in this paper is that
quantum states are every bit as subjective as any other Bayesian
probability.  What then is the role of the Born rule?  Can it be
dispensed with completely?

It seems no, it cannot.  But from our perspective, its
significance is quite different than in other developments of quantum
foundations.  For us, the Born rule is not a rule for {\it setting\/}
probabilities, but rather a rule for {\it transforming\/} or {\it
interconnecting\/} them.

For instance, take a complete set of $d+1$ observables $O^k$,
$k=1,\ldots,d+1$, for a Hilbert space of dimension $d$.  Subjectively setting probabilities for the $d$
outcomes of each such measurement uniquely determines a quantum state
$|\psi\rangle$ (via inverting the Born rule).  Thus, as concerns
probabilities for the outcomes of any other quantum measurements,
there can be no more freedom.  All further probabilities are obtained
through linear transformations of the originals.  In this way, the
role of the Born rule can be seen as something of the flavor of
Dutch-book coherence, but with an empirical content added to the top
of raw probability theory:  An agent interacting with the quantum
world would be wise to adjust his probabilities for the outcomes of
various measurements to those of quantum form if he wants to avoid
devastating consequences.  The role of physical law (i.e., the
assumption that the world is quantum mechanical) shows its head in
how probabilities are related, not how they are set.
\eq
This is what I meant by an early announcement for a party.  I use the imagery because I can't imagine anything more fun or more that I would like to invite my friends into than answering the main question coming from this observation.  If quantum theory is so closely allied with Bayesian probability theory, if it can even be seen as an addition to it, then why is it not written in a language that starts with probability, rather than a language that ends with it?  Or to say it differently, why does quantum theory invoke the mathematical apparatus of complex amplitudes, Hilbert spaces, and linear operators {\it first}, rather than {\it last\/}?

I wanted the party, but like with any real party at my house, I knew that a vast cleaning would have to be done before the fun could begin.  This launched me into the deepest study of the quantum formalism I have ever undertaken and brings us to present-day research at Perimeter Institute.

There is no doubt that it is an old idea to represent quantum states purely in terms of probabilities.  This is exactly what the Husimi $Q$-function is about \cite{Husimi40}, and it is crucially related to the work \cite{Glauber63} Roy Glauber got the 2005 Nobel prize for.  However, there might seem to be a certain disincentive to the enterprise.  As Wootters put it in a 1986 paper \cite{Wootters86}:
\begin{quotation}\small
\noindent It is obviously {\it possible\/} to devise a formulation of quantum mechanics
without probability amplitudes. One is never forced to use any quantities in
one's theory other than the raw results of measurements. However, there is
no reason to expect such a formulation to be anything other than
extremely ugly. After all, probability amplitudes were invented for a reason.
They are not as directly observable as probabilities, but they make the
theory simple.
\end{quotation}
What this means is that if we are going to have any hope of taking probability as a foundation for quantum theory (rather than an ornament on top of it), we had better first find the cleanest, crispest way to write the formalism directly in terms of probabilities, with no quantum states whatsoever as intermediaries.  Only in that way will we know what to aim for.

Such a formalism now seems to be within sight, and it turns out be deeply connected to the question posed back in Section 2.1.5, near Eq.~(\ref{BigBoy}), with regard to achieving the ``quantumness of a Hilbert space.''  The key is none other than the sets of states mentioned there, but now thought of as defining a quantum measurement.  This is the structure Carlton Caves dubbed the ``symmetric informationally complete positive-operator-valued measure,'' or SIC-POVM in an email to me \cite {Caves99}---now often called simply SIC for short (but this time pronounced ``seek'').  It consists of $d^2$ rank-one projection operators $\Pi_i=|\psi_i\rangle\langle\psi_i|$ on a $d$-dimensional Hilbert space such that
\begin{equation}
{\rm tr}( \Pi_i\Pi_j)=\frac{1}{d+1}\qquad \mbox{whenever} \qquad i\ne j\;.
\label{Mojo2}
\end{equation}
Because of their extreme symmetry, such sets of operators have several fine-tuned properties. One is that they are linearly independent and span the space of Hermitian operators.  Another is that, after rescaling, they form a resolution of the identity. Thus the operators $H_i=\frac1d \Pi_i$ form not only a measurement, but a minimal informationally complete measurement.

Historically, these operators actually already arose when Caves and I were trying to prove the quantum de Finetti theorem, described in Section 2.2.2.\footnote{Unbeknownst to us, the question of such operators' existence was also discovered by Gerhard Zauner, who was wrote a PhD thesis on them at the University of Vienna \cite{Zauner99} the very same year.}  Briefly Caves had thought we would need such an extra symmetry to prove the theorem; luckily for us---as I'll explain more fully in a moment---it turned out that such symmetry was overkill, and we didn't need it after all.

What is most astounding about this class of measurements is that it gives rise to an {\it extremely\/} simple relation between the Born-rule probabilities $P(H_i)={\rm tr}\big(\rho H_i\big)$ for its outcomes and the expansion coefficients for $\rho$ in terms of the $\Pi_i$:
\begin{equation}
\rho = \sum_{i=1}^{d^2}\left( (d+1)\,P(H_i) - \frac1d \right)\Pi_i\;.
\label{Ralph2}
\end{equation}
There are no other rank-one measurement operators that span the space of quantum states without redundancy (i.e., only having $d^2$ elements) at the same time as giving such a simple formula. This formula is exactly what is needed to meet Wootters' ``no reason to expect such a formulation to be any\-thing other than extremely ugly.''  Particularly, thinking of a quantum state as literally an experimenter's probability assignment for the outcomes of a potential SIC measurement leads to a remarkably Bayesian way of expressing the Born rule for the probabilities associated with any {\it other\/} quantum measurement.

\begin{figure}[t]
\begin{center}
\includegraphics[height=3.2in]{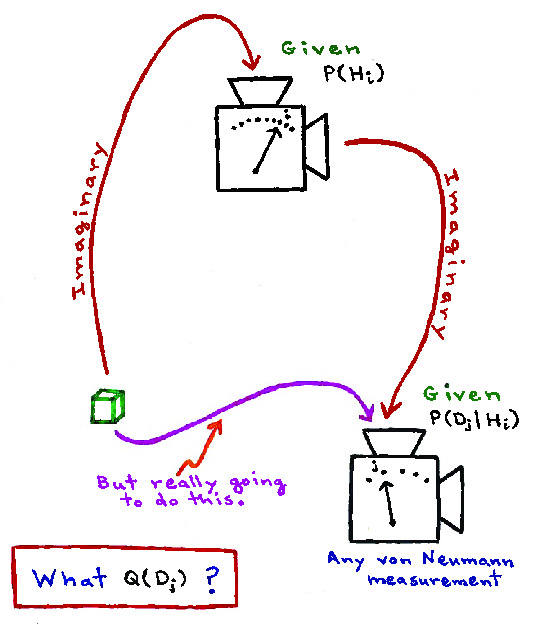}
\end{center}
\end{figure}

Consider the diagram in the figure above.  It depicts a SIC measurement ``in the sky,'' with outcomes $H_i$, and {\it any\/} standard von Neumann measurement $D_j=|j\rangle\langle j|$ ``on the ground'' (the $|j\rangle$ being some orthonormal basis).  We can conceive of two possibilities---better, two ``paths''---for the quantum system to get to the measurement on the ground:  ``Path 1'' is that it proceeds directly to the measurement on the ground.  ``Path 2'' is that it proceeds first to the measurement in the sky and only subsequently to the one on the ground.

Suppose now, we are given the experimenter's personal probabilities $P(H_i)$ for the outcomes in the sky and his conditional probabilities $P(D_j|H_i)$ for the outcomes on the ground subsequent to the sky.  I.e., we are given the probabilities the experimenter would assign on the supposition that the quantum system follows Path 2.  Then the usual rules of probability alone are enough to tell what probabilities $P(D_j)$ the experimenter should assign for the outcomes of the measurement on the ground---it is given by the Law of Total Probability applied to these numbers:
\begin{equation}
P(D_j)=\sum_i P(H_i) P(D_j|H_i)\;.
\label{Magnus}
\end{equation}

A question one can ask is, is this enough information to recover the probability assignment $Q(D_j)$ the experimenter would assign for the outcomes on Path 1 via a normal application of the Born rule, i.e., that
\begin{equation}
Q(D_j)={\rm tr} (\rho D_j)
\label{EmmaPlayStarWars}
\end{equation}
for some $\rho$?  Clearly, $P(D_j)$ itself cannot be the answer.  For, one has
$
Q(D_j)\ne P(D_j)\;
$
simply because Path 2 is {\it not\/} a coherent process with respect to Path 1---there is a measurement that takes place in Path 2 that does not take place in Path 1.

What is remarkable about the SIC representation is that it implies that, even though $Q(D_j)$ is not equal to $P(D_j)$, it is still a function of it.  Particularly,
\begin{eqnarray}
Q(D_j) &=& (d+1) P(D_j) - 1\nonumber
\\
&=&
(d+1)\sum_{i=1}^{d^2} P(H_i) P(D_j|H_i) - 1\;.
\label{ScoobyDoo2}
\end{eqnarray}
From this point of view, the Born rule is a kind of quantum {\it version\/} of the Law of Total Probability \cite{Fuchs09a} (invited article for Reviews of Modern Physics).  No complex amplitudes, no operators---only ``probabilities in'' and ``probabilities out''---and all the while with a remarkable simplicity.\footnote{So that the formula would really stun the eyes, I restricted the main exposition to the case where the $D_j$ represents a von Neumann measurement.  But it is nearly as simple in the case that $D_j$ represents a completely general POVM.  In that case one has,
$$
Q(D_j)=\sum_i\left[(d+1)P(H_i)-\frac{1}{d}\right]\! P(D_j|H_i)\;.
$$
}

One should not, however, interpret Eq.~(\ref{ScoobyDoo2}) as invalidating probability theory itself in any way:  For the regular Law of Total Probability has no jurisdiction in the setting of our diagram, which compares a ``factual'' experiment (Path 1) to a ``counterfactual'' one (Path 2).  Indeed as any Bayesian would emphasize, if there is a distinguishing mark in one's considerations---say, the fact of two distinct experiments, not one---then one ought to take that into account in one's probability assignments (at least initially so).  Thus there is a hidden, or at least suppressed, condition in our notation:  Really we should have been writing the more cumbersome, but honest, expressions $P(H_i|{\mathcal E}_2)$, $P(D_j|H_i,{\mathcal E}_2)$, $P(D_j|{\mathcal E}_2)$, and $Q(D_j|{\mathcal E}_1)$ all along.  With this explicit, it is no surprise that,
\begin{equation}
Q(D_j|{\mathcal E}_1)\ne\sum_iP(H_i|{\mathcal E}_2)P(D_j|H_i,{\mathcal E}_2)\;.
\end{equation}
The message is that quantum theory supplies some\-thing---a new form of ``Bayesian coherence,'' though empirically based (as quantum theory itself is)---that raw probability theory does not.  The Born rule in these lights is an {\it addition\/} to Bayesian probability, not in the sense of a supplier of some kind of more-objective probabilities, but in the sense of giving extra rules to guide the agent's behavior when he interacts with the physical world.

It is a rule for reasoning about the consequences of one's proposed experimental actions in terms of the potential consequences of an explicitly counterfactual experimental action.  It is like nothing else physical theory has contemplated before.  Peres was right to emphasize that ``unperformed experiments have no results'' \cite{Peres78}.  What has not so much been emphasized, and formula (\ref{ScoobyDoo2}) uniquely quantifies, is that the formal apparatus of quantum theory nonetheless fills the gap, allowing one to calculate probabilities for the ``to be performed'' experiment in terms of probabilities from the ``will not be performed'' experiment.

Seemingly at the heart of quantum mechanics from the QBist view is a statement about the impact of {\it counterfactuality\/}.  The impact parameter is metered by a single, significant number associated with each physical system---its Hilbert-space dimension $d$.  The larger the $d$ associated with a system, the more $Q(D_j)$ must deviate from $P(D_j)$.  Of course this point must have been implicit in the usual form of the Born rule, Eq.~(\ref{EmmaPlayStarWars}), all the while.  What is important from the QBist perspective, however, is how the new form puts the significant parameter front and center, displaying it in a way that one ought to nearly trip over.

To get a sense of how this new representation of the Born rule might be taken as a very axiom for quantum theory, I need to give an explanation of the conditions under which Eq.~(\ref{ScoobyDoo2}) can be rightfully used.  Suppose you are given a matrix $A$ and told that it might correspond to a proper quantum state $\rho$. How would you check if it does?  Easy enough:  You check if it is Hermitian, whether its trace is one, and whether its eigenvalues are all nonnegative. But suppose you are given a probability distribution $P(H_i)$; how would you check if it corresponds to a proper quantum state $\rho$?  The point is, we are not allowed to use Eq.~(\ref{ScoobyDoo2}) with just any $P(H_i)$ we please.  Any probability distribution $P(H_i)$ that is plugged into formula (\ref{Ralph2}) will give rise to a Hermitian operator with trace one, but it might not give a positive semidefinite operator.  There are further restrictions.

Any allowed $P(H_i)$ must be in the convex hull of all those probability distributions that make $\rho$ a pure state (a rank one projection operator). Thus if we specify the extreme distributions, in principle at least, we have the answer.  In \cite{Appleby07} we showed that two conditions were necessary and sufficient for specifying our set.  Namely,
\be
\sum_i p(i)^2 \; =\; \frac{2}{d(d+1)}
\label{Jack}
\ee
and
\be
\sum_{i,j,k} c_{ijk}\, p(i)p(j)p(k) \; =\; \frac{d+7}{(d+1)^3}\;,
\label{Jill}
\ee
where the latter equation involves a completely symmetric three-tensor $c_{ijk}$ given by
\be
c_{ijk} \,=\, \mbox{Re}\, {\rm tr}\Big( \Pi_i\Pi_j\Pi_k\Big)
\label{TripleSec}
\ee
which has the most amazing structure.  For instance, defining a set of $d^2$ matrices $C_k$ with matrix entries $(C_k)_{ij}=c_{ijk}$, for each value of $k$, $C_k$ turns out to have the form
\be
C_k=\|m_k\drangle\dlangle m_k\|+\frac{d}{2(d+1)}\, Q_k\;,
\label{Magma}
\ee
where the $k$-th vector $\|m_k\drangle$ is defined by
\be
\|m_k\drangle = \left( \frac{1}{d+1}, \ldots, 1, \ldots, \frac{1}{d+1}\right)^{\!\rm T}\;,
\label{MySchninky}
\ee
and the $Q_k$ are $(2d-2)$-dimensional projection operators that play a significant role in the Lie algebra ${\rm gl}(d,{\bf C})$ \cite{Appleby10b}.

The key point to absorb from this succession of equations is that the set of allowed probability distributions has a rather rich structure.  Even for instance, in the $d=3$ case \cite{Fuchs11a} where one can still\footnote{In the $d=2$ case, Eq.~(\ref{Jill}) reduces to Eq.~(\ref{Jack}), which recovers the simple qubit Bloch sphere we are all used to.  In the cases $d\ge4$, one gets the feeling it is no longer useful to attempt to explicitly write out the condition.} unwind all the terms in Eq.~(\ref{Jill}), it is no simpler than this:
\be
\sum_{i=1}^9 p(i)^3 \; - \; 3 \sum_{(ijk)\in {\bf Q}} p(i) p(j) p(k) \;=\; 0
\ee
where ${\bf Q}$ consists of all lines in an order-3 affine plane,
\be
\begin{array}{ccc}
  1 & 2 & 3 \\
  4 & 5 & 6 \\
  7 & 8 & 9
\end{array}\;
\qquad\qquad\mbox{I.e.,}\qquad\qquad
Q \;\; =\;\;
\begin{array}{cccc}
(123) & (456) & (789)\\
(147) & (258) & (369)\\
(159) & (267) & (348)\\
(168) & (249) & (357)
\end{array}\;.
\ee
Thus, despite the newfound ``Bayesian beauty'' of Eq.~(\ref{ScoobyDoo2}), the most efficient {\it test\/} for its use might well remain to plug a potential probability distribution into Eq.~(\ref{Ralph2}) and diagonalize to see if all the eigenvalues are nonnegative.  But then again, maybe not; it is too early to tell.

What we are really after at this stage, however, is {\it why\/} quantum theory takes the structure that it does.  Taking Eq.~(\ref{ScoobyDoo}) as an axiom gives immediate insight into why there are any restrictions at all on the allowed $P(H_i)$.  For, just look at Eq.~(\ref{ScoobyDoo2}): If $P(D_j)$ is too small, $Q(D_j)$ will go negative; and if $P(D_j)$ is too large, $Q(D_j)$ will become larger than 1.  So, the $P(D_j)$ must be restricted in order for the $Q(D_j)$ to form a proper probability distribution.  But that in turn forces the set of valid $P(H_i)$ to be restricted as well.  And so the argument goes.

In fact, just adding one more simple, Bayesian motivated axiom---that any $P(H_i)$ can come about as the posterior probability for the sky by a suitably chosen measurement on the ground\footnote{In full-blown quantum theory, this corresponds to the observation that any quantum state $\rho$ can be prepared by performing a suitably chosen measurement.}---gives a devilishly interesting condition on the allowed distributions:  If $\cal S$ is the allowed set, for any two probability distributions $P(H_i),R(H_i)\in\cal S$, it must be the case that
\be
\frac{1}{d(d+1)}\le \sum_i P(H_i)R(H_i)\le\frac{2}{d(d+1)}\;.
\label{Nublette}
\ee
Moreover, $\cal S$ must contain all points of the form Eq.~(\ref{MySchninky}), after normalizing properly.

Let us call a subset of the probability simplex with that structure, {\it consistent}.  Let us further say that a consistent set is {\it maximal\/} if attempting to add any single further point to it will cause the new set to violate Eq.~(\ref{Nublette}).  It can be shown that the $P(H_i)$ corresponding to the set of quantum states form a maximal, consistent subset of the simplex.

Are there maximal, consistent subsets other than the space of quantum states?  Yes, there are.  But even with this very simplistic structure one already forces a significant number of properties held by quantum-state space \cite{Fuchs09a,Appleby10,Fuchs10e}.  For instance, all maximal, consistent sets must be closed and convex.  They cannot be polytopes. They are bounded by the same sphere that quantum-state space is, Eq.~(\ref{Jack}).  The $P(H_i)$ are bounded above by $\frac1d$ just as with the quantum case; not too many $P(H_i)$ can be zero, just as with the quantum case; full permutation symmetry of the values of $P(H_i)$ is disallowed, just as with the quantum case.  And here is very nice one:  Suppose one has a family of distributions $P_k(H_i)$, $k=1, \ldots, n$ that saturate the upper and lower bounds of Eq.~(\ref{Nublette}) when one takes their inner products.  Then the cardinality of the family must be bounded in the same way as quantum state space, $n\le d$.

This tells us that if the Born rule is not the whole story of quantum mechanics, it is at least a good part of it. And the exercise of deriving what we have so far definitely gives us hope that we are on the right track.

But it is here that I must {\it own up\/} to another aspect of my research.  I have been in the habit in this Section of calling Eq.~(\ref{ScoobyDoo2}) a new representation of the Born rule.  But of course, that is premised on a set of $d^2$ projection operators $\Pi_i$ satisfying Eq.~(\ref{Mojo2}) to actually exist.  If such operators do not exist, then the Born rule cannot be written that way.

There are two options here:  If SICs do always exist, Eq.~(\ref{ScoobyDoo2}) is always valid, and we are learning to understand quantum theory in new terms from it.  If SICs do not always exist, we are playing with a new theory---one that matches quantum theory perfectly for small systems, but diverges when they get sufficiently large.  My guess is that it will be the former that prevails, but one cannot say at this stage.  What has been so consternating about the SICs is that the answer is still not known.  Despite the 11 years of growing effort since the definition was first introduced (there are now nearly 50 papers on the subject), no one has been able to show that they exist in a general finite dimensions (or even a specific infinite class thereof).  What is known firmly is that they exist in dimensions 2 through 67 \cite{Scott09}.  Dimensions $2\,$--$\,15$, 19, 24, 35, and 48 are known through direct or computer-automated analytic proof; the remaining solutions are known through numerical simulation, satisfying Eq.~(\ref{Mojo2}) to within a precision of $10^{-38}$.\footnote{It is known however that the analogs of SICs in real vector spaces, that is, sets of $d(d+1)/2$ equiangular lines, do {\it not\/} exist in general dimension.  They do in some, but not in all---for example, they exist in $d=2,3,7$, and 23. Apparently for gravity-wave theorists, it is a good thing that they do in $d=3$. As discovered by Latham Boyle \cite{Boyle10}, such a configuration is needed for building an optimal gravity-wave detector!}

Thus a significant part of my present research is devoted not only to going forward with the assumption of SIC existence and trying to derive quantum theory from the reformulations it gives rise to, but going backward as well.  That is, by studying the properties of these supposed objects {\it within\/} normal Hilbert space quantum theory and trying to get insight on the existence question by standard means.  Perhaps the word ``backward'' is more apt than it might first appear. For, chief among the conjectures in the SIC literature is that a SIC can always be generated by the actions of a set of operators introduced by Hermann Weyl at the very beginning of quantum theory.  In common notation, let us define a unitary operator $Z$ whose eigenvalues are powers of the $d$-th primitive root of unity $\omega=e^{\frac{2\pi i}{d}}$:
\be
Z|k\rangle=\omega^k |k\rangle\;, \qquad k=0,1,\ldots, d-1\;.
\ee
Let $X$ be its Fourier transform, so that its action with respect to the same basis is:
\be
X|k\rangle=|k+1\rangle
\ee
where addition is modulo $d$.  These two operators satisfy
\be
ZX=\omega X\!Z
\label{Weyl-E-Coyote}
\ee
which is Weyl's form of the canonical commutation relations.  The conjecture is that one can always find a quantum state $|\psi_{00}\rangle$ such that
the $d^2$ projection operators $\Pi_{mn}=|\psi_{mn}\rangle\langle \psi_{mn}|$ given by
\be
|\psi_{mn}\rangle=X^m Z^n |\psi_{00}\rangle\;,
\label{CovariantGrump}
\ee
form a SIC.  The evidence for this is as before---it is known to be satisfiable in dimensions 2 to 67---and it is to the proof of this refined conjecture that my group and I have applied our greatest effort.

Thus, the ``backward.''  It is perhaps not so well known that the Weyl quasi-commutation relation Eq.~(\ref{Weyl-E-Coyote}) has been with quantum mechanics from a time nearly simultaneous with its very beginning.  In July of 1925, Max Born had the idea that he could solve one of the equations in Heisenberg's seminal 1925 paper if he made an {\it ansatz\/} that the position observable $\bf \hat q$ and momentum observable $\bf \hat p$ satisfied the commutation relation \cite{Bernstein05,Fedak09}
\be
{\bf \hat p}{\bf \hat q}-{\bf \hat q}{\bf \hat p}=\frac{h}{2\pi i}{\bf \hat 1}\;.
\label{Tombstone}
\ee
Pascual Jordan later proved that Born's ansatz was the only one that could work, and the paper they wrote up together was received by {\sl Zeitschrift f\"ur Physik\/} on 27 September 1925.  It was titled ``On Quantum Mechanics.'' On the same day, Max Born received a letter from Hermann Weyl \cite{Scholz08} saying that a previous discussion they had earlier in the month inspired him to generalize Born's relation (\ref{Tombstone}) to Eq.~(\ref{Weyl-E-Coyote}).  Part of what pleased Weyl was that his generalization was not dependent upon infinite dimensional spaces---Weyl's relation would always have a solution in the complex matrices, finite and infinite dimensional.  And by that circumstance he declared to have a way of ``defining a general quantum system''---to each would be associated a ``phase space.''  In the discrete case, the points of the phase space would be associated with the operators
\be
(m,n) \quad \longrightarrow \quad X^m Z^n\;.
\ee
See Weyl's \cite{Weyl27}, as well as Julian Schwinger's later development and extension of the idea in \cite{Schwinger70}.
If a covariant SIC as in Eq.~(\ref{CovariantGrump}) always exists, then it would mean that not only could a phase space be associated with a quantum system abstractly, but that the points of the phase space very directly correspond to the outcomes of a potential measurement.  This again gives credence that these considerations are on the right track.

It is hard to express in these few pages the effort, thought, and enthusiasm I have put into this project.  Deep in my bones I feel it is the right way to go for unlocking quantum theory's greatest secrets---for peculiarly Quantum Bayesian reasons, as well as all the tendrils it seems to have into diverse areas of mathematics, from Lie algebra theory \cite{Appleby10b} to the theory of theta functions, elliptic functions, and modular forms \cite{Appleby10c}.  The topic is quite vast, certainly far beyond me individually, and I pride myself on being one of the research topic's most assiduous enablers as well as one of its outspoken cheerleaders.

I will end this section with a raw list of ways I have worked to enable this topic beyond my own hours of pencil and paper:
\begin{enumerate}
\item
I worked to obtain a U. S. Office of Naval Research grant of \$512,000 US for study\-ing ``SIC Representations for Quantum States and
Quantum Channels'' \cite{FuchsONR}. This money is used to buy out Dr.\ Marcus Appleby from his teaching duties for three years, and supports my two University of Waterloo PhD students Hoan Dang and Gelo Tabia, and MSc student Matthew Graydon.
\item
I supported Dr.\ {\AA}sa Ericsson for an Associate Postdoc at Perimeter Institute so that she could work on this subject.  Most of her funding comes from a Swedish council, but I also use some of my ONR money to fund her travel.
\item
I have recently written a proposal for \$669,000 to the John Templeton Foundation.  The proposal is on SIC and QBism research, its applications within quantum cryptography and quantum random number generation, and additionally asks for funding to enable Maximilian Schlosshauer to write a major book\footnote{Schlosshauer's previous (and critically acclaimed) book {\sl Decoherence and the Quantum-to-Classical Transition\/} \cite{Schlosshauer07} comes from a time previous to his embracing QBism.} on the subject \cite{Fuchs10d}.  The grant will be decided 22 December 2010.
\item
I organized with Steve Flammia a Perimeter Institute workshop titled, ``Seeking SICs:\ An Intense Workshop on Quantum Frames and Designs,'' 26--31 October 2008.  See \pirsa{C08025}.
\item
I have worked hard to bring SICs to the attention of our experimental colleagues.  To this end, I have collaborated with Aephraim Steinberg's group at the University of Toronto to experimentally demonstrate a qutrit SIC, \cite{Medendorp10} (submitted to Physical Review Letters).  Also, presently I am collaborating with Anton Zeilinger's group at the University of Vienna to do their own demonstration of qubit and qutrit SICs with integrated optics; Christoph Schaef is leading that effort in the laboratory and expects to have data before March of 2011.  Finally Bernie Yurke, Michael Vasilyev, and I hold two patents on qubit SIC implementations \cite{FuchsPATENT1,FuchsPATENT2}---these are not experiments, but maybe they serve as another kind of indication that SICs are closer to reality than one might initially think.
\item
When invited to be the Tulane University Mathematics Department 2011 Clifford Lecturer, I set the meeting's theme to be ``SIC Representations of Quantum States.''  The workshop will be held 14--17 March 2011.
\item
I have organized an invited session for the 2011 American Physical Society March Meeting in Dallas, Texas, titled ``Symmetric Discrete Structures for Finite Dimensional Quantum Systems.''
\item
Last year I supervised Perimeter Scholars International student Aaron Fenyes in a research problem on this subject, ``Maximal Consistent Sets and Non-Commutative Probability.''  This year, I will be supervising two more PSI students on the subject, Laura Piispanen and Antti Karlsson.
\item
I will be organizing a month-long research project at the Stellenbosch Institute for Advanced Study in South Africa in 2011 on the subject of QBism and SICs.
\end{enumerate}

\subsection{4. Future Research}

\subsubsection{4.1 The Shape of Quantum-State Space}

In Section 3, I already laid out quite a robust line of research.  It was called ``current research,'' but by its nature it is ``future research'' as well.  There are just so many questions beyond the ones already posed there that need to be asked.  Why, for instance, in this modification of the Law of Total Probability, Eq.~(\ref{ScoobyDoo2}), is the deviation factor $d+1$ rather than $d$ itself or $d^2$ or anything else?  Why are the number of outcomes for the measurement in the sky $d^2$ instead of $d^3$ or $3^d$ or anything else?   We have made a bit of progress in our understanding of this by stepping back to a less specific form for Eq.~(\ref{ScoobyDoo2}) and postulating
\be
Q(D_j)=\sum_{i=1}^n\Big(\alpha P(H_i)-\beta\Big) P(D_j|H_i)\;,
\label{ScoobyDum}
\ee
with floating $\alpha$, $\beta$, and $n$.  Then, invoking a couple of other postulates similar to quantum mechanics (for instance, that there exist ``maximal'' measurements whose outcomes can be predicted with certainty), one can see that there cannot be arbitrary couplings between these constants---actually there is only one integer-valued parameter $q$ connecting all three of them \cite{Fuchs10e}. As an example,
\be
n=\frac{1}{2}qd(d+1)-d\;,
\ee
with $q=0,1,2,\ldots, \infty$.  This gives a sense that QBist quantum theory is embedded in an infinite family of other theories (though a very different family than, say, in Hardy's axiomatic system \cite{Hardy01}).  When $q=0$ we have the conventional understanding of classical probability with the usual Law of Total Probability (i.e., there is no counterfactuality to the measurement in the sky).  And interestingly, when $q=1$ and $q=4$ the structures appear to be quite analogous to real-vector-space \cite{Stueckelberg60} and quaternionic-vector-space \cite{AdlerBook} quantum mechanics.  Analogs to the rest of the family, however, we have never seen before.  Ref.~\cite{Graydon} represents our first foray into trying to understand this better.

But one should ask even more probingly, why the simple affine modification to the usual Law of Total Probability, Eq.~(\ref{ScoobyDum}), in the first place?  Why not something else?  Indeed we ask this question every day and will continue to beat on it and turn it around until an answer finally comes!

The kinds of questions outlined here, though, are pretty clearly generalized from the last section.  They are the next logical steps in the program, if you will.  But we aim for much, much more with QBism and well expect it to make much deeper contributions to ``the great project of physics,'' as Hardy calls it.  The next two sections will say something on two specific items that I think are within our long-range grasp.

\subsubsection{4.2 Decoherence and the Classical-to-Quantum Transition (sic)}

Recently I wrote the following abstract for a meeting in Sweden; I playfully titled it ``Quantum-Bayesian Decoherence'' to contrast it with a paper R\"udiger Schack and I wrote the previous year, ``Quantum-Bayesian Coherence'' \cite{Fuchs09a}. In truth, decoherence plays no positive role in the Quantum Bayesian conception of things, but we have sketched a research program whose aim is to explain some of the ``allure of decoherence'' nonetheless:
\bq
\small
The usual story of quantum measurement since von Neumann has been that it occurs in two steps:  First the quantum system becomes entangled with the measuring device; then, mysteriously, there is a selection of one of the entangled state's components in order to single out an actual measurement result. But the entangled wave function alone (with its freedom to be expressed in any basis) cannot say how it should be decomposed so that the selection can be effected. The theory of quantum decoherence, developed by Zeh, Zurek, and others, attempts to shim up this deficiency in the von Neumann story by supplementing it with a further story of interaction between the device and an environment: The idea is that the specific form of the interaction with the environment specifies how the system$+$device state ought to be decomposed. But then what of the mysterious selection? Decoherence theorists usually leave that question aside, implicitly endorsing one variety or another of an Everettian interpretation of quantum mechanics.

In contrast, the Quantum Bayesian view of quantum theory leaves most of the von Neumann chain aside:  Instead of unitary evolution, it takes the idea of common experience as first and foremost. In this view, quantum ``measurement'' is nothing other than an agent acting upon the world and experiencing the consequences of his actions---the very same thing he was doing before quantum theory was even thought of. A quantum measurement is most generally an agent's  ``whack'' upon the external world and its unpredictable consequence for him.  To say it still differently, the Quantum Bayesian story is the preface and conclusion of the von Neumann one, without the dramatic artifice of unitary evolution, entanglement, and decoherence in between: selection is the consequence of a whack and nothing more.

Thus, it would seem there is no foundational place for decoherence in the Quantum Bayesian program. And that is true. Nonetheless, one can identify a two-time gambling situation (gamble now on the outcome of a measurement in the future, given that an intermediate measurement will be performed in the nearer future) for which consistent gambling behavior mimics a belief in decoherence. The consistent gamble makes use of a quantum version of van Fraassen's reflection principle, and explains to some extent the seduction (but misleadingness) of the decoherence program. This work builds on \cite{Fuchs09a}, \cite{Fuchs10e}, and \cite{Fuchs10b}, where the Born rule is viewed as an empirical addition to probability theory and quantum theory is seen as ``additive'' to common experience, rather than ``underlying it'' in a reductionistic sense.
\eq
The root of a more detailed argument can be found in this soon-to-be-posted paper of ours \cite{Root}.

This, however, is just one of a large cluster of issues that arise from the way QBism turns the usual quantum-foundations debate around.  The essential distinction is that QBism must rest upon a nonreductionist, empiricist conception of the quantum world.  Another passage from another paper may help to put this into perspective \cite{Fuchs10b}:
\bq
\small
The expectation of the quantum-to-classical transitionists is that quantum theory is at the bottom of things, and ``the classical world of our experience'' is something to be derived out of it.  QBism says ``No.  Experience is neither classical nor quantum.  Experience is experience with a richness that classical physics of any variety could not remotely grasp.''  Quantum mechanics is something put on top of raw, unreflected experience.  It is additive to it, suggesting wholly new types of experience, while never invalidating the old.  To [a question like], ``Why has no one ever {\it seen\/} superposition or entanglement in diamond before?''\ the QBist replies:  It is simply because before recent technologies and very controlled conditions, as well as lots of refined analysis and thinking, no one had ever mustered a mesh of beliefs relevant to such a range of interactions (factual and counterfactual) with diamonds.  No one had ever been in a position to adopt the extra normative constraints required by the Born rule.  For QBism, it is not the emergence of classicality that needs to be explained, but the emergence of our new ways of manipulating, controlling, and interacting with matter that do.

In this sense, QBism declares the quantum-to-classical research program unnecessary (and actually obstructive) in a way not so dissimilar to the way Bohr's 1913 model of the hydrogen atom declared another research program unnecessary (and actually obstructive). Bohr's great achievement above all the other physicists of his day was in being the first to say, ``Enough!  I shall not give a mechanistic explanation for these spectra we see.  Here is a way to think of them with no mechanism.''  The important question is how matter can be coaxed to do new things.

All is not lost, however, for the scores of decoherentists this policy would unforgivingly unemploy.  For it only suggests that they redirect their work to the opposite direction.  The thing that needs insight is not the quantum-to-classical transition, but the classical-to-quantum!  The burning question for the QBist is how to model in Hilbert-space terms the common sorts of measurements we perform just by opening our eyes, cupping our ears, and extending our fingers.

Take a professional baseball player watching a ball fly toward him:  He puts his whole life into when and how he should swing his bat.  But what does this mean in terms of the immense Hilbert space a quantum theoretical description would associate with the ball?  Surely the player has an intuitive sense of both the instantaneous position and instantaneous momentum of the baseball before he lays his swing into it---that's what ``keeping his eye on the ball'' means.  Indeed it is from this intuition that Newton was able to lay down his laws of classical mechanics.   Yet, what can it mean to say this given quantum theory's prohibition of simultaneously measuring complementary observables?  It means that whatever the baseball player is measuring, it ain't that---it ain't position {\it and\/} momentum as usually written in operator terms.  Instead, a quantum model of what he is doing would be some interesting, far-from-extremal {\it single\/} POVM---perhaps even one that takes into account some information that does not properly live within the formal structure of quantum theory (the larger arena that Howard Barnum calls ``meaty quantum physics'' \cite{Barnum10}). For instance, that an eigenvector $|i\rangle$ of some Hermitian operator, though identically orthogonal to fellow eigenvectors $|j\rangle$ and $|k\rangle$ in the Hilbert-space sense, might be {\it closer\/} in meaning to $|k\rangle$ than to $|j\rangle$ for some issue at hand.

So the question becomes how to take a given common-day measurement procedure and {\it add to it\/} a consistent quantum description?  The original procedure was stand alone---it can live without a quantum description of it---but if one wants to move it to a new level or new direction, having added a consistent quantum description will be most helpful to those ends.  Work along these lines is nascent, but already some excellent examples exist.  See \cite{Kofler08}.  Of course, unconsciously it is what has been happening since the founding days of quantum mechanics.
\eq

What this amounts to is a call for a new {\it quantization\/} program---not unlike, say, the Weyl and geometric quantization programs, in spirit at least---but where the expression to be quantized is not a classical dynamical variable or field equation.  Rather the thing to be ``quantized'' (in the above sense) is process of {\it observation\/} itself.

This is a big, big question, and QBism's ultimate success really depends on being able to put some technical flesh and bones onto an answer for it.

\subsubsection{4.3 Hilbert-Space Dimension as a Universal Capacity}

If the last question is a necessary one for completeness, the present question is the one that truly pushes QBism toward the future. Here we come back to the issue of what precisely is the ``only a little more'' of Section 2.2.3.  What is being asked for here is not a new dynamical variable or collapse mechanism, say, but {\it a new quality of all matter}.

To give this some sense, I again need to dip into the larger set of issues that form the backdrop of QBism's ontology: nonreductionism and radical empiricism.  Please forgive my quoting \cite{Fuchs10b} again, but it gets to the heart of the matter in the most direct way:
\bq
\small
[Accepting empiricism and nonreductionism] is no statement that physics should give up, or that physics has no real role in coming to grips with the world.  It is only a statement that physics should better understand its function. \ldots

Physics---in the right mindset---is not about identifying the bricks with which nature is made, but about identifying what is {\it common to\/} the largest range of phenomena it can get its hands on.  The idea is not difficult once one gets used to thinking in these terms.  Carbon?  The old answer would go that it is {\it nothing but\/} a building block that combines with other elements according to the following rules, blah, blah, blah.  The new answer is that carbon is a {\it characteristic\/} common to diamonds, pencil leads, deoxyribonucleic acid, burnt pancakes, the space between stars, the emissions of Ford pick-up trucks, and so on---the list is as unending as the world is itself.  For, carbon is also a characteristic common to this diamond and this diamond and this diamond and this.  But a flawless diamond and a purified zirconium crystal, no matter how carefully crafted, have no such characteristic in common:  Carbon is not a {\it universal\/} characteristic of all phenomena.  The aim of physics is to find characteristics that apply to as much of the world in its varied fullness as possible.  However, those common characteristics are hardly what the world is made of---the world instead is made of this and this and this.  The world is constructed of every particular there is and every way of carving up every particular there is.

An unparalleled example of how physics operates in such a world can be found by looking to Newton's law of universal gravitation.  What did Newton really find?  Would he be considered a great physicist in this day when every news magazine presents the most cherished goal of physics to be a Theory of Everything?  For the law of universal gravitation is hardly that!  Instead, it {\it merely\/} says that every body in the universe tries to accelerate every other body toward itself at a rate proportional to its own mass and inversely proportional to the squared distance between them.  Beyond that, the law says nothing else particular of objects, and it would have been a rare thinker in Newton's time, if any at all, who would have imagined that all the complexities of the world could be derived from that limited law.  Yet there is no doubt that Newton was one of the greatest physicists of all time.  He did not give a theory of everything, but a Theory of One Aspect of Everything.  And only the tiniest fraction of physicists have ever worn a badge of that more modest kind.  It is as H.~C. von Baeyer wrote in one of his books,
\begin{quote}\small
Great revolutionaries don't stop at half measures \ldots.  For Newton this meant an almost unimaginable widening of the scope of his newfound law.  Not only Earth, Sun, and planets attract objects in their vicinity, he conjectured, but all objects, no matter how large or small, attract all other objects, no matter how far distant.  It was a proposition of almost reckless boldness, and it changed the way we perceive the world.
\end{quote}
Finding a theory of ``merely'' one aspect of everything is hardly something to be ashamed of:  It is the loftiest achievement physics can have in a non\-re\-duc\-tionist world.

Which leads us back to Hilbert space.  Quantum theory---that user's manual for decision-making agents immersed in a world of {\it some\/} yet to be fully identified character---makes a statement about the world to the extent that it identifies a quality common to all the world's pieces.  QBism says the quantum state is not one of those qualities.  But of Hilbert spaces themselves, particularly their distinguishing characteristic one from the other, {\it dimension}, QBism carries no such grudge.  Dimension is something one posits for a body or a piece of the world, much like one posits a mass for it in the Newtonian theory.  Dimension is something a body holds all by itself, regardless of what an agent thinks of it.

That this is so can be seen already from reasons internal to the theory.  Just think of all the arguments rounded up for making the case that quantum states should be interpreted as of the character of Bayesian degrees of belief.  None of these work for Hilbert-space dimension.  \ldots

The claim here is that quantum mechanics, when it came into existence, implicitly recognized a previously unnoticed capacity inherent in all matter---call it {\it quantum dimension}.  In one manifestation, it is the fuel upon which quantum computation runs \cite{Fuchs04,BlumeKohout02}.  In another it is the raw irritability of a quantum system to being eaves\-dropped upon \cite{Fuchs03,Cerf02}.  In Eq.~(\ref{ScoobyDoo2}) it was a measure of deviation from the Law of Total Probability induced by counterfactual thinking.

When quantum mechanics was discovered, something was {\it added\/} to matter in our conception of it.  Think of the apple that inspired Newton to his law.  With its discovery the color, taste, and texture of the apple didn't disappear; the law of universal gravitation didn't reduce the apple privatively to {\it just\/} gravitational mass. Instead, the apple was at least everything it was before, but afterward even more---for instance, it became known to have something in common with the moon.  A modern-day Cavendish would be able to literally measure the fur\-ther attraction an apple imparts to a child already hungry to pick it from the tree.  So similarly with Hil\-bert-space dimension.  Those diamonds we have used to illustrate the idea of nonreductionism, in very careful conditions, could be used as components in a quantum computer \cite{Prawer08}.  Diamonds have among their many properties something not envisioned before quantum mechanics---that they could be a source of relatively accessible Hilbert space dimension and as such have this much in common with any number of other proposed implementations of quantum computing.  Diamonds not only have something in common with the moon, but now with the ion-trap quantum-computer prototypes around the world.
\eq

A large part of my upcoming research intends to make this precise.  Indeed, it might be said that the whole playground of quantum information and computing is already providing the data for this project.  What we are doing with each protocol, algorithm, and channel issue that arises in the field is getting a feel for how this new capacity ticks. We get a feel for ``the go of it'' \cite{MaxwellBook}, as Maxwell would have said.  But some order must be given to the plethora of phenomena.

The best place to start I believe is to chip away at phenomenon after phenomenon of quantum information, but now in the technical language QBism has been striving to develop---that is, by thinking of the Born rule as an empirical addition to probability theory for cases of counterfactual actions, as well as using the toolkit of SIC technology that comes with it.  It is not really so different from the IEEE methodology I committed myself to in graduate school---see Section 2---but have walked a bit away from in these rather focussed years of QBism's direct development.

To give an example:  I have started to wonder, speculatively, whether our new understanding of the Born rule might be what is actually going on, but in a disguised (and less clarified) form, in the art of quantum computation.  Consider again Eq.~(\ref{ScoobyDoo2}) and the associated diagram in Section 3.  What it expresses is a mismatch between the ``classical reasoning'' that might be used for calculating a final probability and its direct quantum counterpart.  I think it is not a stretch of the imagination to think that there is a connection between this and Andrew Steane's way of looking at the power of quantum computation \cite{Steane03}.

\subsubsection{4.4 Quantum Gravity in the Light of Nonreductionism}

Some years ago, I wrote a whimsical story about elevators to begin, and then end, a paper \cite{Fuchs04} on my measure of quantumness (see Section 2.1.5).  ``Memorable experiences sometimes happen in elevators,'' I said. ``I have had two \ldots''  As unusual as it might be, it is worth repeating more of the story in this context.

Here is the part of the story from the beginning of the paper:
\bq
\small
[T]en years before that second experience, I met an odd
fellow in an elevator at the physics department of the University of
Texas.  All alone, with the doors shut, he looked at me with a crazed
look in his eyes and asked, ``What is energy?'' Feeling
uncomfortable, I turned my own eyes to the ground and was happy that
the doors soon opened. I slipped away, but regaining composure just
before the doors shut again, I replied, ``I don't know, that which
gravitates?'' A cheap answer!  But at least I had something to say.
Looking back over the years I thank my lucky stars he didn't ask a
tough question. He could have asked, ``What is Hilbert space?''
\eq
And here is a part from the end:
\bq
\small
In several pieces of recent literature much to-do has been made of
the fungibility of quantum information~\cite{Caves03}.  Hilbert
spaces of the same dimension are said to be fungible: What can be
done in one can be done in the others.  Thus, for instance, it would
not matter if a quantum cryptographer decided to build a quantum key
distribution scheme based on $d$-dimensional subspaces gotten from
pellets of platinum, or $d$-dimensional subspaces gotten from pellets
of magnalium.  The ultimate security he can achieve---at least in
principle---will be the same in either case.

What is the meaning of this?  One might say that this is the very
reason quantum information is quantum {\it information\/}!  If a
protocol like quantum key distribution depended upon the kinds of
matter used for its implementation, one would hardly be justified in
thinking of it as a pure protocol solely of (quantum)
information-theoretic origin.  That is a useful and fruitful point of
view.

However, another point of view is that this may be a call to
reexamine physics, much like the miraculous equivalence between
gravitational and inertial mass (as revealed by the E\"otv\"os
experiment~\cite{Eotvos22}) was once a call to reexamine the origin
of gravitation.  From the perspective of gravity, the
``implementation'' of the mass is inconsequential---platinum and
magnalium fall with the same acceleration. And that, in the hands of
Einstein, led ultimately to the realization that gravity is a
manifestation of spacetime curvature.

What is Hilbert space?  Who knows!  But in quantum information we are
learning how to make use of it as a raw resource, basking in the good
fortune that it {\it is\/} fungible---that the implementation of a
Hilbert space dimension $d$ is inconsequential. There could be some
very deep physics in that, but it might take another elevator story.
\eq

Both here and in the previous subsection, I have played up a certain conceptual similarity between Newtonian gravitational mass and Hilbert space dimension (within QBism's purview, at least).  What really intrigues me is the question of whether there are situations where the two concepts might be identified with each other.  Or, more exactly, one being a pure function of the other.

Think for instance about a Schwarzschild black hole. For an observer at infinity, its rest mass is $M$ and, according to Hawking, its entropy is $S=4\pi M^2$.  But personalist Bayesians are always a bit suspicious of statements about what an entropy {\it is\/} or {\it must be}.  An entropy is a property of a probability distribution or a quantum state, and from the view of QBism, neither thing is an objective feature of nature.  So what is the right way to think of this entropy-mass connection?  I am in no position to say at the moment!  But I would like to be, and I intend a significant effort in this direction.  I am, for instance, titillated by the idea that perhaps what has been identified as an entropy (and mistreated as objective) has really all along been the logarithm of a Hilbert-space dimension.  I am similarly titillated that perhaps the content of Bekenstein's entropy bound conjecture is a way of associating (finite) Hilbert-space dimension with regions of space.

In any case, I am quite excited by the recent turn toward trying to understand gravity more generally as an entropic phenomenon as, for instance, exhibited in the work of Padmanabhan \cite{Padmanabhan10} and Erik Verlinde \cite{Verlinde10}.  That work, however, being rife with words and imagery like, ``What microscopic degrees of freedom contribute to this entropy?,'' demonstrates that it is coming about in a far-from-QBist atmosphere and understanding of quantum theory.  (See discussion near the end of Section 2.1.5 on Hilbert-space dimensionality.)  I surely feel that QBism's tools will one day, sooner rather than later, be needed here.

\subsubsection{5. My Wider Role}

My publication record will not hide that I generally do not seek out high-impact-factor journals for my papers.  It is a complicated emotion.  It is one of those things I know all the ``success manuals'' tell me I ought to do.  But on the other hand, my annoyance or dread of anonymous referees often wins out, and I think ``Why bother?''  Nonetheless, my academic record has been decently successful in a nonstandard kind of way, and because of this I feel that my idiosyncracies bring some things to the table of science that a more publishable-or-perishable colleague's might not.  This is something I am very proud of.\footnote{A case in point is the quote of mine that appears in the {\sl Oxford Dictionary of American Quotations} \cite{FuchsOx}. To be in the company of Thomas Jefferson, Mark Twain, and Robert Frost.  I think to myself, ``Now that's doing physics!''}

A look at my CV will show that I go to a lot of meetings (giving 170+ invited talks since 1994), organize a fair number of meetings myself (18 since 2000), and beyond all else write lots and lots of email (the two books \cite{Fuchs10,Fuchs10c} represent the tip of an iceberg).  Nearly everything I want to see of physics and every direction I turn in it has been worked out in public as I have confronted other foundational-minded physicists.  It is a bit fantastical, but I might compare my agenda to the firings of a neuron in a larger brain.

There is a passage I once ran across in a letter from Wolfgang Pauli to Werner Heisenberg that has always intrigued me.  On 19 October 1926, Pauli wrote \cite{AtmanspacherTranslation},
\bq\small\noindent
It's always the same thing: due to diffraction, there are no arbitrarily thin rays in the wave optics of the $\psi$-field, and it is illegitimate to assign ordinary `$c$-numbers' to `$p$-numbers' and `$q$-numbers' simultaneously.  One may then view the world with the `$p$-eye' and one may view it with the `$q$-eye', but if you want to open both eyes at the same time this drives you crazy.
\eq
The letter generated quite a bit of enthusiasm in Copenhagen with Bohr, Dirac, and others, as Heisenberg reported in his reply to Pauli nine days later.  By 23 March 1927, Heisenberg had submitted his famous uncertainty-relation paper to {\sl Zeitschrift f\"ur Physik}.  It has always seemed to me that Pauli may well have planted the seed for what turned into Heisenberg's more thorough work.

Far be it from me to compare myself to Pauli as a physicist, but there is one feature of Pauli's {\it social role\/} in physics that I have no qualm in trying my best to emulate.  Pauli has been called ``the conscience of physics'' \cite{Dyson02}. I would be pleased enough if my firing neuron became part of the ``unconscious'' of physics.

One thing my raw publication record cannot capture is the work I do behind the scenes.  Two examples of this---things better left unmentioned under normal circumstances, details of life not visible in author lists and publication numbers---were given already in Sections 2.1.2 and 2.2.1 in my discussions of the no-broadcasting theorem and the dictionary definition of quantum teleportation.  It seems worthwhile to relate a couple more such stories for the purpose of this document.

I am like a gadfly; if something is on my mind I cannot leave people alone about it.  On a couple of occasions, my pestering (to my surprise) helped spur some cottage industries in quantum foundations that were not at all my intention!  One example comes from the Fall--Winter of 1998--1999, when I was going around asking every mathematical sophisticate I knew whether Gleason's theorem for quantum theory might or might not hold in Hilbert spaces over the {\it rational\/} field.  For myself, I was just wanting to get straight what Gleason's theorem was saying about quantum theory structurally---did the field really play a big role in the result (I knew already that it didn't matter if the field is real or complex).  One person I asked was Itamar Pitowsky (email) \cite{KS-Lump1}, another was David Meyer (at a restaurant) \cite{KS-Lump2}.  David said, ``Oh, I know the answer to that.  I had always wondered if there might be a use for that old Godsil and Zaks result.'' (See \cite{Godsil88} and the proof of Lemma 1 in \cite{Meyer99} stating, ``This is an immediate consequence of a result of Godsil and Zaks \ldots'')  My question, despite its near immediate answer when posed to the right expert, opened up a can of worms---for rather than focusing the community's attention on the structure of Gleason's theorem, it mostly re-ignited the old and, to my mind stale, hidden-variables debate (by a quick look, at least 12 papers with 260+ citations pointing back to them).  See \cite{LumpLump} for a small sample of the discussions and \cite{Appleby05} for what I think is a genuinely interesting result to come of it all.

Another---now broader---example, but again one that didn't go in quite the direction I had hoped for with my gadflicity, is some fraction of the recent work on ``operational axioms'' for quantum theory.  Here, there is no doubt that I am {\it no great contributor\/} to the effort.  But I wonder if I might take some credit in a kind of avuncular way for setting the tone and building the atmosphere that ultimately supported the effort?  The real father of the modern turn toward operational axioms is Lucien Hardy through his very influential 2000--2001 paper ``Quantum Theory from Five Reasonable Axioms'' \cite{Hardy01}.  But it seems the time was ripe for a culture change in the foundations community just about then, and Hardy wasn't the only one to turn from a predominantly-Bohmian foundational stance to something more operationally tinged.  Another good example is the influential paper by Clifton, Bub, and Halvorson \cite{Clifton03}.

\begin{figure}
\begin{center}
\fbox{
\includegraphics[height=6.0in]{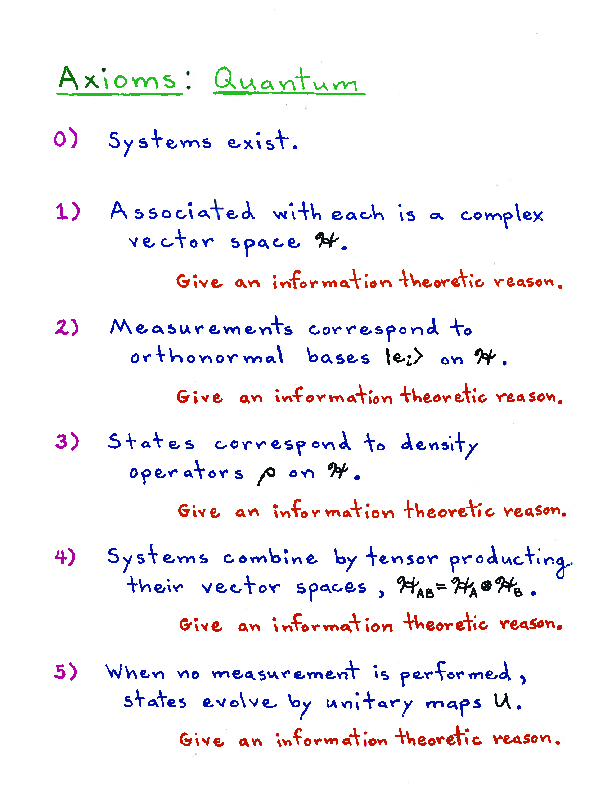}
}
\end{center}
\end{figure}

For some time previous to those early systems,\footnote{Now, operational axiomatic systems seem to come a dime a dozen, with scant motivation for the variations on variations they seem so occupied with.  One starts to wonder, ``What do they see wrong with the other guy's axioms?''} I had been going around giving talks that used a slide like the one on the next page as their rallying cry. The point I was trying to make was that {\it methodologically\/} the thing one should try to do is identify all the components of quantum theory that had something to do with information and see what remained behind. I had not intended a call to axiomatize, axiomatize, axiomatize.  This method, instead, I viewed to be in grand service of the quest to find a small number of {\it physical principles\/} upon which to base quantum mechanics---statements of ontology if you will.  And statements of ontology should not be operational axioms, nor should they be information theoretic ones.  They should be direct statements about the (hypothesized) character of the (hypothesized) stuff of the world.  At least it does seem my efforts were partially successful---see e.g.\ Hardy's acknowledgement in \cite{Hardy01}, Appleby's introduction to \cite{Appleby05a}, and Jeff Bub's introspective account of his thinking in \cite[Chap.\ 9]{Fuchs10}---for, much in the operational axiomatizations has indeed led to a deepened understanding of quantum theory. But operationalism for-its-own-sake was not the end I had in mind, and I was a little disappointed that in all those axioms \cite{AxiomLump}, I never saw my long dreamed for simple, crisp physical principle leap out and sweep me away.

But, who's complaining?!  In the end, I'll be happy if my ``impact factor'' is not based on a journal's title, but on how much science changes in the long run by my own efforts or by the strong, young minds I might inspire.  I was very flattered when I learned what Scott Aaronson had written for the back cover of my book \cite{Fuchs10}:
\bq\small
\noindent I'm delighted that what Chris Fuchs calls his ``samizdat''---seven years' worth of fascinating correspondence about the foundations of quantum mechanics---has finally been published in book form.  I read the samizdat as a beginning graduate student, and it changed my career.  Some of the specific questions Chris ponders here---why, for example, does quantum mechanics involve complex numbers, rather than real numbers or quarternions?---have haunted me ever since.  But more than that, reading Chris's letters made me feel, for the first time, like the great conversation of Bohr, Einstein, and other luminaries in the 1920s was still going on today; that their bewilderment over why God made the universe this way and not that hadn't completely given way to technicalities and jaded commentaries on commentaries.
\eq
If I have done it once, I want to do it a million more times.

\eq

\section{29-03-11 \ \ {\it NY Anytime Soon?, 2}\ \ \ (to A. Plotnitsky)} \label{Plotnitsky26}

Thanks so much for being willing to look into this.  The item is a four volume set titled, {\sl Cambridge School of Pragmatism\/} edited by John R. Shook and Andre de Tienne.  The price is \$300 and the only description is ``very good,'' with no further details.

Please tell me what you think of it, as you have a chance.  If it sounds like it's roughly of the condition of any anything else on my bookshelf (no pencil markings, rips in the cover, etc.), then I will probably order it from them.  I hope this is not too much trouble on you.

Thanks so much for sending me your draft.  I absolutely love the title ``Dark Materials to Create More Worlds''!!  It's just chock full of good stuff.  And I so love that Milton quote.  I'm drawn to it.  ``His dark materials to create more worlds.''  I'm fascinated with that line particularly, even if I ultimately want to misuse it for my purposes:  I want to think of ``quantum systems'' as the ``dark materials,'' rather than the ``chance'' that spews from them in a quantum measurement setting.  In my mind it has a kind of connection to a Pauli quote in the attached excerpt from my new samizdat (the one starting at the end of page 525 about quantum measurement being like a ``black mass''). [See 19-06-05 note ``\myref{Comer72}{Philosopher's Stone}'' to G. L. Comer.]  When things cool down for me a bit, I'd like to continue this discussion on ``dark materials''.

\section{30-03-11 \ \ {\it A Thought I Would Call Religious} \ \ (to the QBies)} \label{QBies37}

\bq
I must say one word about the extraordinarily intimate and important character which the phenomenon of effort assumes in our own eyes as individual men.  Of course we measure ourselves by many standards.  Our strength and our intelligence, our wealth and even our good luck, are things which warm our heart and make us feel ourselves a match for life.  But deeper than all such things, and able to suffice unto itself without them, is the sense of the amount of effort which we can put forth.  Those are, after all, but effects, products, and reflections of the outer world within.  But the effort seems to belong to an altogether different realm, as if it were the substantive thing which we {\it are}, and those were but externals which we {\it carry}.  If the `searching of our heart and reins' be the purpose of this human drama, then what is sought seems to be what effort we can make.  He who can make none is but a shadow; he who can make much is a hero.  The huge world that girdles us about puts all sorts of questions to us, and tests us in all sorts of ways.  Some of the tests we meet by actions that are easy, and some of the questions we answer in articulately formulated words.  But the deepest question that is ever asked admits of no reply but the dumb turning of the will and tightening of our heartstrings as we say, {\it ``Yes, I will even have it so!''}  When a dreadful object is presented, or when life as a whole turns up its dark abysses to our view, then the worthless ones among us lose their hold on the situation altogether, and either escape from its difficulties by averting their attention, or if they cannot do that, collapse into yielding masses of plaintiveness and fear.  The effort required for facing and consenting to such objects is beyond their power to make.  But the heroic mind does differently.  To it, too, the objects are sinister and dreadful, unwelcome, incompatible with wished-for things.  But it can face them if necessary, without for that losing its hold upon the rest of life.  The world thus finds in the heroic man its worthy match and mate; and the effort which he is able to put forth to hold himself erect and keep his heart unshaken is the direct measure of his worth and function in the game of human life.  He can {\it stand\/} this Universe.  He can meet it and keep up his faith in it in presence of those same features which lay his weaker brethren low.  He can still find a zest in it, not by `ostrich-like forgetfulness,' but by pure inward willingness to face it with those deterrent objects there.  And hereby he makes himself one of the masters and the lords of life.  He must be counted with henceforth; he forms a part of human destiny. \ldots\

Thus not only our morality but our religion, so far as the latter is deliberate, depend on the effort which we can make.  {\it ``Will you or won't you have it so?''}\ is the most probing question we are ever asked; we are asked it every hour of the day, and about the largest as well as the smallest, the most theoretical as well as the most practical, things.  We answer by {\it consents or non-consents\/} and not by words.  What wonder that these dumb responses should seem our deepest organs of communication with the nature of things!  What wonder if the effort demanded by them be the measure of our worth as men!  What wonder if the amount which we accord of it were the one strictly underived and original contribution which we make to the world!\medskip

\noindent William James \smallskip \\
{\sl Psychology, A Briefer Course}
\eq

\section{31-03-11 \ \ {\it A Good Day's Impressions} \ \ (to D. Fraser)} \label{Fraser4}

I want to thank you again for yesterday's opportunity; I much enjoyed the experience.  And actually one of the strongest impressions I got was in getting to know you better:  You had a grasp of things, and my own idiosyncrasies, that impressed me much.  Genuinely, if much had been the same, but [Philosophy Professor X, not at U. Waterloo] (for instance) had led the discussion, I don't think I would have encountered such understanding.  That is not to say {\it agreement\/}, which you may not have had (i.e., I well expect a face of neutrality from a lecturer in a situation such as this), but what sets you apart in my mind is that you did not give the impression of hearing only non sequiturs spewing from me (as indeed so many philosophers portray of me).  That built a sense of camaraderie, and that's much better for getting to the heart of these matters.  I can learn something from that, and I appreciate it.

\subsection{Doreen's Preply}

\bq
I'm teaching an undergrad/grad seminar on philosophy of QM this term and I'm wondering whether you'd be willing to make a guest appearance near the end of term.  I've scheduled (most of) two weeks on information-theoretic approaches to QM and it would be great to have you for the second of the two weeks.  I will assign the students readings to complete in advance and the class will be devoted to 15 minute presentations by some of the students and discussion.  The readings are listed as TBA in the syllabus; I'm thinking that I will assign most of Chris Timpson's survey chapter in the {\sl Ashgate Companion to Contemporary Physics\/} and then a paper of yours and Chris' paper ``QB: A Study.''  But I'm open to suggestions!  [\ldots]

I've attached the schedule of readings for the course.  I have 12 undergrads evenly split between honors philosophy majors and (mostly) math/physics majors, 3 philosophy grad students and a handful of auditors.  I'm using Albert's textbook---which has the virtue of being accessible to the philosophy audience---and supplementing it with additional readings.  Teaching this course to the diverse audience is a bit of a challenge, but we've already had some good exchanges between the philosophers and the physicists.
\eq

\section{04-04-11 \ \ {\it Les Papillons}\ \ \ (to H. C. von Baeyer)} \label{Baeyer143}

It is 10:20 AM in Waterloo, and the review meeting will start at 11:00.  I'm sitting at home, hiding, and my stomach is in butterflies, almost to the point of nausea.  I thought it might help to write you; so here I am.

This morning, I have been using the coffee mug Emma bought for her mother at the Eiffel tower, when we were there two years ago.  Somehow, the mug has gotten transferred to my nearly sole use of it, and it's generally considered one of ``dad's mugs'' now.  It's got two scenes of the tower, and Katie was looking at it pretty thoroughly this morning at breakfast.  I started to tell her stories of how she will be so impressed when she walks around a corner and spies it for the first time.

How I wish I were in the city with you today.  I could so use a good serious discussion on QBism, Pauli, and alchemy again.  And to do it in the right atmosphere would cap it all off.  Or this would nice:  Sitting in some nice Parisian caf\'e reading Boutroux and Renouvier!  Ahh, the possibilities you have in your life.

Let me come back to the issue of criticism of QBism.  Attached is a file I put together for the review.  One faculty member had asked if there were any published criticisms of QBism.  So it mostly contains the ones you've already seen, but there are some other ones in there as well:
\begin{enumerate}
\item
D. Styer, S. Sobottka, W. Holladay, T. A. Brun and R. B. Griffiths, P.~Harris, letters to editor in ``Quantum Theory---Interpretation, Formulation, Inspiration,'' Phys.\ Tod.\ {\bf 53}(9), 11 (2000).

\item
A. Hagar, ``A Philosopher Looks at Quantum Information Theory,'' Phil.\ Sci.\ {\bf 70}, 752 (2003).

\item
P. Grangier, ``Contextual Objectivity and Quantum Holism,'' \arxiv{quant-ph/0301001v2} (2003).

\item
E. Dennis and T. Norsen, ``Quantum Theory: Interpretation Cannot be Avoided,''\\ \arxiv{quant-ph/0408178v1} (2004).

\item
L. Marchildon, ``Why Should We Interpret Quantum Mechanics?,'' Found.\ Phys.\ {\bf 34}, 1453 (2004).

\item
M. Ferrero, D. Salgado, and J.~L. S\'anchez-G\'omez, ``Is the Epistemic View of Quantum Mechanics Incomplete?,'' Found.\ Phys.\ {\bf 34}, 1993 (2004).

\item
A. Hagar, ``Quantum Information---What Price Realism?,'' Int.\ J. Quant.\ Info.\ {\bf 3}, 165 (2005).

\item
A. Hagar and M. Hemmo, ``Explaining the Unobserved---Why Quantum Mechanics Ain't Only About Information,'' Found.\ Phys.\ {\bf 36}, 1295 (2006).

\item
U. Mohrhoff, ``Objective Probability and Quantum Fuzziness,'' \arxiv{quant-ph/0703035v1} (2007).

\item
D. Wallace, ``The Quantum Measurement Problem:\ State of Play,'' \arxiv[quant-ph]{0712.0149v1} (2007).

\item
V. {\Palge} and T. Konrad, ``A Remark on Fuchs' Bayesian Interpretation of Quantum Mechanics,'' Stud.\ Hist.\ Phil.\ Sci.\ Part B {\bf 39}, 273 (2008).

\item
C. J. {\Timpson}, ``Quantum Bayesianism:\ A Study'', Stud.\ Hist.\ Phil.\ Mod.\ Phys.\ {\bf 39}, 579 (2008).

\item
A. Stairs, ``A Loose and Separate Certainty: {\Caves}, Fuchs and {\Schack} on Quantum Probability One,'' to appear in Stud.\ Hist.\ Phil.\ Mod.\ Phys.\

\item
M. S. Leifer, ``Retrodiction In Quantum Bayesianism, Or Why Wigner's Friend Is Fuchs' Enemy''

\item
S. Friederich, ``How to spell out the epistemic conception of quantum states,''\\ \arxiv{1101.1975v1} (2011).

\item
J. Bub, ``Quantum probabilities as degrees of belief,'' Stud.\ Hist.\ Phil.\ Mod.\ Phys.\ {\bf 38}, 232 (2007).

\item
G. Jaeger, {\sl Entanglement, Information, and the Interpretation of Quantum Mechanics}, (Sprin\-ger, 2009).
\end{enumerate}
Probably the best recommendations in there are Bub's paper and Stairs's paper.  Their email addresses are {\tt jbub@umd.edu} and {\tt stairs@umd.edu} in case you want to get the papers directly from them.  (I know that Allen's paper was refereed, and so have may have changed somewhat since I last saw it.)  I'll attach the latest version of Matt Leifer's paper, though he said a day or two ago that he is expecting to substantially revise it.  The key thing that Matt misses is the two-levels of personalism, but I expect that the article might be enlightening otherwise.  (I haven't actually read it yet.)

Maybe in the next note, I'll send you the bigger presentation I made up for Lucien.  Who knows, maybe some of that will be useful for your own QBism summary.  It's not too very braggarty (compared to some of my other documents) \ldots\ mostly factual.

Thanks again for the nice note the other evening.  There are times when little compliments help a lot.

\section{07-04-11 \ \ {\it The Betterized {\Mermin}}\ \ \ (to R. W. {\Spekkens} \& M. S. Leifer)} \label{Leifer14} \label{Spekkens106}

I dug this up yesterday, but then I forgot to send it to you.  It is the article where David {\Mermin} thinks he does a better job than he did on the three-man BFM paper.

I've got to say, I'm much taken with your program \ldots\ even though there are several philosophical divides from my own program \ldots\ the deepest being that quantum theory is a {\it generalization\/} of probability theory.  You know that I would now say it is standard personalist probability theory (which includes for me, if I need to say so explicitly, van Fraassen's reflection principle)---not a generalization at all---with some structure about counterfactuals added to it.  I.e., it is a more particular structure than abstract probability theory.  But that doesn't get in the way of my appreciating this great formalism you're developing and all the things you're able to tackle with it.

\section{07-04-11 \ \ {\it Huangjun Zhu Visiting at the Moment}\ \ \ (to R. W. {\Spekkens} \& M. S. Leifer)} \label{Leifer15} \label{Spekkens107}

Huangjun Zhu gave his talk today, and it was quite good.  (Giulio was there, and you may consult him for an independent opinion.)  But the talk can be viewed on PIRSA already:  \pirsa{11040105}.

Only they left out my dramatic introduction for Huangjun, telling the story of how Hermann Weyl became excited after a visit from Max Born in September 1925, where Born revealed that imposing the canonical commutation relations for position and momentum might be the key axiom for a new quantum mechanics.  Weyl's excitement led him to discovering the Weyl commutation relations, so that he could reframe the Born--Jordan assumption in terms of bounded operators, but more importantly, so that he could define a quantum theory for finite systems.  Weyl's letter announcing the discovery was dated 27 September 1925; the very same day the Born--Jordan paper---the very first to write down the full theory of quantum mechanics---was submitted for publication.  After telling the tale, I returned to the subject of Huangjun to say that he would present a result that shows the deep connection between SIC POVMs in prime dimension and that very starting point of quantum theory.

\section{08-04-11 \ \ {\it Thinking of You}\ \ \ (to W. G. {\Demopoulos})} \label{Demopoulos49}

\bwd
I recently ran across a reference to Santayana's {\bf Character and Opinion}. In general, I find him a bit to literary for my taste, but this is not to say he isn't insightful. The book (based on lectures given in California) contains a chapter on James that I thought hit some nails on their heads pretty well. Have a look---there's a new edition, but I'd have to look up the reference again---which I'm happy to do if you can't find it---here it is, edited by James Seaton.
\ewd

That's one Santayana book I don't have; I'll be on the lookout for it.

I read your {\Poincare} introduction this morning with my coffee.  It was nice. I really liked the passage you plucked out:
\bq\ldots
     \ldots\ more than a dozen processes, entirely independent and that I
     would not be able to enumerate without tiring you, lead us to the
     same result. If there were more or less molecules per gram, the
     brightness of the blue sky would be entirely different; incandescent
     bodies would radiate more or radiate less, etc.; one has to admit, we
     see atoms.
\eq
That's the style of reasoning that leads us to epistemic states.  That's why I've always liked this passage from {\Spekkens}'s original toy model paper.
\bq
   We shall argue for the superiority of the epistemic view over the
   ontic view by demonstrating how a great number of quantum phenomena
   that are mysterious from the ontic viewpoint, appear natural from
   the epistemic viewpoint.  These phenomena include [about a million
   things]. \ldots\  The greater the number of phenomena that appear
   mysterious from an ontic perspective but natural from an epistemic
   perspective, the more convincing the latter viewpoint becomes. \ldots

   Of course, a proponent of the ontic view might argue that the
   phenomena in question are not mysterious \emph{if} one abandons
   certain preconceived notions about physical reality.  The challenge
   we offer to such a person is to present a few simple \emph{physical}
   principles by the light of which all of these phenomena become
   conceptually intuitive (and not merely mathematical consequences of
   the formalism) within a framework wherein the quantum state is an
   ontic state.  Our impression is that this challenge cannot be met.
   By contrast, a single information-theoretic principle, which
   imposes a constraint on the amount of knowledge one can have about
   any system, is sufficient to derive all of these phenomena in the
   context of a simple toy theory, as we shall demonstrate.
\eq

I'll go back to the actual {\Poincare} paper later.

\section{18-04-11 \ \ {\it If Words Have Meaning} \ \ (to N. D. {\Mermin})} \label{Mermin198}

Attached is a copy of the one article out there that cites your nonworld nonview paper.  Have you seen it before?  It's quite something.

I was just re-reading this passage in your paper:
\bq
This is the quintessential quantum situation: How can you tell whether your experiment is giving you information about the object of study (particle), the apparatus (box), or some combination of the two?

Figure 2 shows a way to tell. Suppose you can test particles before they get to the box, and on the basis of the results of the test you can predict in advance which bin each particle ends up in. Clearly that rules out the possibility that the box disposes randomly of the particle. Indeed, if you can specify in which bin the particle ends up before it even reaches the box, then, if words have any meaning at all, we can say that two kinds of particles enter the box: those that always get put into the left bin, and those that always get put into the right bin. We can call them particles of type L and particles of type R. There are indeed two kinds of particle, and the box is sorting them out.
\eq
So, I see our debate on ``certainty'' started all the way back in 1992.

\section{18-04-11 \ \ {\it If Words Have Meaning, 2} \ \ (to N. D. {\Mermin})} \label{Mermin199}

\bdm
Apropos of the content, all I'm doing there is repeating the EPR
reality criterion (which is, after all, nothing but common sense.)
The debate started in 1935.
\edm

Aha!  So you indirectly admit that the trouble with Bell is the assumption of the EPR criterion after all!  For, you write in the text:
\bq
Remarkably---amazingly---Bohr was right to do so. But this did not become clear for nearly thirty years. Since the overwhelming majority of physicists believed that Bohr had refuted Einstein in 1935, the remarkable 1964 paper by John Bell showing that Bohr was right to keep his head in the sand, had negligible impact on practicing physicists. Within the last decade, however, there has been a revival of interest in the foundations of quantum mechanics, primarily because experimental techniques are now available that reveal a far greater variety of exotic quantum phenomena than were available to the founders of the theory sixty-five years ago. In 1988 Greenberger, Horne, and Zeilinger came up with a spectacular refinement of the EPR argument and its refutation by Bell, that I shall now describe.
\eq
and
\bq
It was thirty years before John Bell pointed out that there were other aspects of the EPR situation (involving the behavior of the particles at additional types of boxes through which they might be
sent) which, though not contradicting in any way the data on which EPR based their conclusion, were nevertheless inconsistent with each particle possessing both 1-ness and 2-ness. I shall not pursue Bell's argument, because I believe the argument found a few years ago by GHZ is even more dramatic and compelling.
\eq
Shame, shame.

\section{18-04-11 \ \ {\it The One Underived and Original Contribution We Make to the World} \ \ (to N. D. {\Mermin} and C. M. {\Caves})} \label{Mermin200} \label{Caves108}

Attached is the promised quote on the will by William James.  I felt like copying it in again.

\bq
[I] must say one word about the extraordinarily intimate and important character which the phenomenon of effort assumes in our own eyes as individual men. Of course we measure ourselves by many standards. Our strength and our intelligence, our wealth and even our good luck, are things which warm our heart and make us feel ourselves a match for life. But deeper than all such things, and able to suffice unto itself without them, is the sense of the amount of effort which we can put forth. Those are, after all, but effects, products, and reflections of the outer world within. But the effort seems to belong to an altogether different realm, as if it were the substantive thing which we {\it are}, and those were but externals which we {\it carry}. If the ``searching of our heart and reins'' be the purpose of this human drama, then what is sought seems to be what effort we can make. He who can make none is but a shadow; he who can make much is a hero.

The huge world that girdles us about puts all sorts of questions to us, and tests us in all sorts of ways. Some of the tests we meet by actions that are easy, and some of the questions we answer in articulately formulated words.  But the deepest question that is ever asked admits of no reply but the dumb turning of the will and tightening of our heartstrings as we say, {\it ``Yes, I will even have it so!''} When a dreadful object is presented, or when life as a whole turns up its dark abysses to our view, then the worthless ones among us lose their hold on the situation altogether, and either escape from its difficulties by averting their attention, or if they cannot do that, collapse into yielding masses of plaintiveness and fear. The effort required for facing and consenting to such objects is beyond their power to make. But the heroic mind does differently. To it, too, the objects are sinister and dreadful, unwelcome, incompatible with wished-for things. But it can face them if necessary, without for that losing its hold upon the rest of life. The world thus finds in the heroic man its worthy match and mate; and the effort which he is able to put forth to hold himself erect and keep his heart unshaken is the direct measure of his worth and function in the game of human life. He can {\it stand\/} this Universe. He can meet it and keep up his faith in it in presence of those same features which lay his weaker brethren low. He can still find a zest in it, not by ``ostrich-like forgetfulness,'' but by pure inward willingness to face the world with those deterrent objects there. And hereby he becomes one of the masters and the lords of life. He must be counted with henceforth; he forms a part of human destiny. \ldots

Thus not only our morality but our religion, so far as the latter is deliberate, depend on the effort which we can make. {\it ``Will you or won't you have it so?''}\ is the most probing question we are ever asked; we are asked it every hour of the day, and about the largest as well as the smallest, the most theoretical as well as the most practical, things. We answer by {\it consents\/} or {\it non-consents\/} and not by words. What wonder that these dumb responses should seem our deepest organs of communication with the nature of things! What wonder if the effort demanded by them be the measure of our worth as men! What wonder if the amount which we accord of it be the one strictly underived and original contribution which we make to the world!
\eq

\section{21-04-11 \ \ {\it Quick Question}\ \ \ (to H. C. von Baeyer)} \label{Baeyer144}

\bhcvb
Reading {\Timpson} has been very useful.  I do have a question though.  One of his principal complaints is the ``explanatory deficit'' of QBism -- that you cannot match conventional quantum theory in explaining the thermal conductivity of lithium, for example.  In particular, he claims: ``When we are disallowing the possibility of any descriptive theory below this kind of level of generality it seems that there is little we can say about why macroscopic objects have the properties they do\ldots.''

Are you indeed disallowing this possibility?  I thought that the Born rule and the Schr.\ equation are the twin pillars of conventional q.m., and that you have succeeded in a finding a new role for the former, {\bf but have not yet found the replacement for the latter}.  ({\Timpson}, in his last sentence, does allow for the possibility that you may resolve this problem.)

I am not expecting the Schr.\ equation to be translated literally into the language of Bayesian probabilities.  It may look quite different in character -- just as your Born rule does now.  But is the Schr.\ equation {\bf disallowed}?

If {\Timpson} is right, and it {\bf is} disallowed, where do the actual numbers that go into the Born rule come from?
\ehcvb

Well, they don't come from the {\Schroedinger} equation itself:  They come from the largest practical Bayesian web of beliefs.  Do a word search on the term ``mesh'' in \arxiv{1003.5209} and that will take you to some of the appropriate discussions.  And do the same in the attached document.  In fact, the discussion there might be better anyway:
\bq\noindent
Now, our path back to quantum theory is complete because I want to say this:  A quantum state just \underline{\it is}\/ a probability assignment.  The particular character of the quantum world places new, physically-influenced consistency requirements on our mesh of beliefs (like the second equation in my answer to Q2), but in the end, even quantum probabilities must port into probability theory more generally.  A quantum state assignment is only one element in a much larger Bayesian mesh of beliefs each agent inevitably uses for his calculations.  It is a numerical {\it commitment\/} to how he will gamble and make his decisions when he plans to interact with a quantum system.
\eq

But back to the meaning of the {\Schroedinger} equation.  In the end, I want to say it too is a nothing else than a consistency requirement on various beliefs, just like the urgleichung.  In fact, ultimately I want to say it just {\it is\/} the urgleichung in a different guise.  See footnote 22 on page 13 of the {\tt arXiv} paper.  Rob {\Spekkens} and Matt Leifer have some outstanding formal work coming out along these lines; there might be some discussions in there that will help you (forgetting about the new formalism).  I will see if they are willing to share the draft at this stage (it may be too rough still).

Also, maybe I haven't sent you this little summary article for the QCMC proceedings, attached.  It has a little discussion of unitarity as well.  \ldots\  Oh, maybe it doesn't (I just discover).  But I'll send it to you anyway.

On this business of ``explanatory power,'' my best recommendation for preparing to answer is to read Louis Menand's book {\sl The Metaphysical Club}.  It's the very best book I read in 2003 (and I read a lot of books that year).  Here's a passage from it on the subject:
\bq
The world is filled with unique things. In order to deal with the
world, though, we have to make generalizations. On what should we
base our generalizations? One answer, and it seems the obvious
answer, is that we should base them on the characteristics things
have in common. No individual horse is completely identical to any
other horse; no poem is identical to any other poem. But all things
we call horses, and all things we call poems, share certain
properties, and if we make those properties the basis for
generalizations, we have one way of ``doing things'' with horses or
poems---of distinguishing a horse from a zebra, for example, or of
judging whether a particular poem is a good poem or a bad poem. These
common properties can be visible features or they can be invisible
qualities; in either case, we create an idea of a ``horse'' or a
``poem,'' or of ``horseness'' or ``poetry,'' by retaining the
characteristics found in all horses or poems and ignoring
characteristics that make one horse or poem different from another.
We even out, or bracket, the variations among individuals for the
sake of constructing a general type.

Darwin's fundamental insight as a biologist was that among groups of
sexually reproducing organisms, the variations are much more
important than the similarities. ``Natural selection,'' his name for
the mechanism of evolutionary development that he codiscovered with
Alfred Russel Wallace, is the process by which individual
characteristics that are more favorable to reproductive success are
``chosen,'' because they are passed on from one generation to the
next, over characteristics that are less favorable. Darwin regretted
that the word ``selection'' suggested an intention: natural selection
is a blind process, because the conditions to which the organism must
adapt in order to survive are never the same. In periods of drought,
when seeds are hard to find, finches that happen to have long narrow
beaks, good for foraging, will be favored over finches with broad
powerful beaks: more of their offspring will survive and reproduce.
In periods of abundance, when seeds are large and their shells are
hard, the broad-beaked finches will hold the adaptive advantage.
``Finchness'' is a variable, not a constant.

Darwin thought that variations do not arise because organisms need
them (which is essentially what Lamarck had argued). He thought that
variations occur by chance, and that chance determines their adaptive
utility. In all seasons it happens that some finches are born with
marginally longer and narrower beaks than others, just as children of
the same parents are not all exactly the same height. In certain
environmental conditions, a narrower beak may have positive or
negative survival value, but in other conditions---for example, when
seeds are plentiful and finches are few---it may make no difference.
The ``selection'' of favorable characteristics is therefore neither
designed nor progressive. No intelligence, divine or otherwise,
determines in advance the relative value of individual variations,
and there is no ideal type of ``finch,'' or essence of ``finchness,''
toward which adaptive changes are leading.

Natural selection is a law that explains {\it why\/} changes occur in
nature---because, as Darwin and Wallace both realized after reading,
independently, Thomas Malthus's {\sl Essay on the Principle of
Population\/} (1798), if all members of a group of sexually
reproducing organisms were equally well adapted, the population of
the group would quickly outgrow the resources available to sustain
it. Since some members of the group must die, the individuals whose
slight differences give them an adaptive edge are more likely to
survive. Evolution is simply the incidental by-product of material
struggle, not its goal. Organisms don't struggle because they must
evolve; they evolve because they must struggle. Natural selection
also explains {\it how\/} changes occur in nature---by the relative
reproductive success of the marginally better adapted. But natural
selection does not dictate {\it what\/} those changes shall be. It is
a process without mind.

A way of thinking that regards individual differences as inessential
departures from a general type is therefore not well suited for
dealing with the natural world. A general type is fixed, determinate,
and uniform; the world Darwin described is characterized by chance,
change, and difference---all the attributes general types are
designed to leave out. In emphasizing the particularity of individual
organisms, Darwin did not conclude that species do not exist. He only
concluded that species are what they appear to be: ideas, which are
provisionally useful for naming groups of interacting individuals.
``I look at the term species,'' he wrote, ``as one arbitrarily given
for the sake of convenience to a set of individuals closely
resembling each other \ldots. [I]t does not essentially differ from
the term variety, which is given to less distinct and more
fluctuating forms. The term variety, again, in comparison with mere
individual differences, is also applied arbitrarily, and for mere
convenience sake.'' Difference goes all the way down.
\eq
And maybe this one is good to see too, though it may not completely reflect to my own beliefs on the subject.  In any case there is so much food for thought in the book.  And wonderfully written.
\bq
Darwinism was a scandal to many Laplaceans. In the Laplacean
worldview, randomness is only appearance; in the Darwinian, it is
closer to a fact of nature---in some respects, it is {\it the\/} fact
of nature. Herschel, the man who had helped introduce Quetelet to
British readers, wrote in 1850 that if all the literature of Europe
were to perish and only Laplace's {\sl Syst\`eme du monde\/} and {\sl
Essai sur les probabilit\'es\/} remained, ``they would suffice to
convey to the latest posterity an impression of the intellectual
greatness of the age which could produce them, surpassing that
afforded by all the monuments antiquity has left us.'' But when {\sl
On the Origin of Species\/} appeared, in 1859, he ridiculed Darwin's
theory as ``the law of higgledy-pigglety.''\footnote{Charles Darwin to
Charles Lyell, December 10, 1859: ``I have heard by round about channel that Herschel says my Book `is the
law of higgledy-pigglety'\,''.}

What does it mean to say we ``know'' something in a world in which
things happen higgledy-pigglety? Virtually all of Charles Peirce's
work---an enormous body of writing on logic, semiotics, mathematics,
astronomy, metrology, physics, psychology, and philosophy, large
portions of it unpublished or unfinished---was devoted to this
question. His answer had many parts, and fitting them all
together---in a form consistent with his belief in the existence of a
personal God---became the burden of his life. But one part of his
answer was that in a universe in which events are uncertain and
perception is fallible, knowing cannot be a matter of an individual
mind ``mirroring'' reality. Each mind reflects differently---even the
same mind reflects differently at different moments---and in any case
reality doesn't stand still long enough to be accurately mirrored.
Peirce's conclusion was that knowledge must therefore be social. It
was his most important contribution to American thought, and when he
recalled, late in life, how he came to formulate it, he described
it---fittingly---as the product of a group. This was the conversation
society he formed with William James, Oliver Wendell Holmes, Jr., and
a few others in Cambridge in 1872, the group known as the
Metaphysical Club.
\eq

We start our little pilgrimage to Chocorua this afternoon.  But I will contemplate more on a better answer to you \ldots\ in between contemplating the words I'll use for various job enquiries.

\section{21-04-11 \ \ {\it Quick Question, 2}\ \ \ (to H. C. von Baeyer)} \label{Baeyer145}

I probably didn't send you the right passage from {\sl The Metaphysical Club}.  I know there is something more closely related to your question; it's a question of finding it.  I will try my best to give you a proper answer this weekend.

\section{21-04-11 \ \ {\it Quick Question, 3}\ \ \ (to H. C. von Baeyer)} \label{Baeyer146}

\bhcvb
In particular, he claims: ``When we are disallowing the possibility of any
descriptive theory below this kind of level of generality it seems that
there is little we can say about why macroscopic objects have the
properties they do\ldots.''
\ehcvb

Maybe another way to make a stab at answering you is to suggest you read James's essay ``The Sentiment of Rationality''.  Here's one place to get it online: \myurl[http://www.archive.org/details/sentimentration02jamegoog]{http://www.archive.org/details/ sentimentration02jamegoog}.

\section{28-04-11 \ \ {\it A Standing Ovation for Reductionism}\ \ \ (to M. A. Graydon)} \label{Graydon21}

\bmag
Here is a transcription of part of the discussion during Christopher Stubbs's colloquium last week. I remember coming home with these words burning in my mind, and I wanted to preserve them as a reminder.
\bq\noindent{\rm
{\bf Christopher Stubbs:}  I mean, it's funny. I came of age in the 1980s. I went to grad school when the standard model of particle physics was built, right. And running coupling constants, and grand unification, and that had a very definite impact on my outlook as a physicist. Now, whether that was a good or a bad thing, I don't yet know. I crave in my deepest, darkest soul that there is some fundamental theory of everything that is uniquely determined by some symmetry principle that we don't yet understand, and things are the way they are because they had no choice. I want it to be that way in the most fervent [way] you can imagine. \smallskip

{\it ***ovation for reductionism***} \smallskip

\noindent But, just because I want it to be that way doesn't mean it's so. I am personally doing everything I can to continue to push the reductionist paradigm, because that's how I was brought up. \medskip

\noindent {\bf Neil Turok:}  You said that this program [reductionism] has worked for centuries and why do we give it up now for something which doesn't actually give any prediction at all. It's more like: forget about science, let's all go and do art and music \ldots\ I mean that's what I see the anthropic tendency, or spiritualism, dare I say it.}
\eq
\emag

So I guess it didn't help my case to write a section titled ``Quantum Gravity in the Light of Nonreductionism'' in my PI research proposal?

Fires, floods, and bad philosophy---who would have thought the world could carry such destructive tendencies?

\section{29-04-11 \ \ {\it Questions} \ \ (to I. Bengtsson)} \label{Bengtsson4}

\bib
In the not-too-distant future I should give a talk where I make the point that it is not true --- as Bogdan Mielnik, who is to be celebrated, used to complain --- that all Hilbert spaces are alike. Because the dimension matters a lot. I think you sometimes say things that goes in the direction that it is the dimension that really reflects the underlying reality? If you recognize that sentence, can you direct me to the place where you say it most forcefully?
\eib

With regard to some spirited writing of mine on dimension, perhaps see Sections VI, VII, and VIII of \arxiv{1003.5209}.  Another good place to look is to look at the story titled ``\myref{Hardy38}{The More and the Modest}'' starting on pdf file page 883 (alternatively text page 874) on
\bq
\myurl{http://www.perimeterinstitute.ca/personal/cfuchs/nSamizdat-2.pdf}.
\eq
Beyond that, do searches on ``capacity'' and ``alchemy'' and ``degree of detachedness'' and you'll find any number of crazy things.

\section{30-04-11 \ \ {\it Abstract for Conceptual Foundations Conference} \ \ (to G. Chiribella)} \label{Chiribella4}

OK, I bend.  Below is my title and abstract \ldots\ and some evidence that there is no such clean line between science and emotions as the reductionism-lovers might want in a world picture.

\bq\noindent
Title:  Some Negative Remarks on Operational Approaches to Quantum Theory\medskip\\
Abstract:  Over the last 10 years there has been an explosion of ``operational reconstructions'' of quantum theory.  This is great stuff:  For, through it, we come to see the myriad ways in which the quantum formalism can be chopped into primitives and, through clever toil, brought back together to form a smooth whole.  An image of an IQ-Block puzzle comes to mind, \myurl{http://www.prismenfernglas.de/iqblock_e.htm} [``THE IQ-BLOCK, or a game for big and small children you could get addicted to''].  There is no doubt that this is invaluable work, particularly for our understanding of the intricate connections between so many quantum information protocols.  But to me, it seems to miss the mark for an ultimate understanding of quantum theory; I am left hungry.  I still want to know what strange property of matter forces this formalism upon our information accounting.  To play on something Einstein once wrote to Max Born, ``The quantum reconstructions are certainly imposing.  But an inner voice tells me that they are not yet the real thing.  The reconstructions say a lot, but do not really bring us any closer to the secret of the `old one'.''  In this talk, I hope to expand on these points and convey some sense of why I am fascinated with the problem of the symmetric informationally complete POVMs to an extent greater than axiomatic reconstructions.\footnote{\editornote An edited transcript of this talk is now available: C.\ A.\ Fuchs and B.\ C.\ Stacey, ``Some Negative Remarks on Operational Approaches to Quantum Theory,'' \arxiv[quant-ph]{1401.7254}.}
\eq

\section{02-05-11 \ \ {\it A JTF William James Center?}\ \ \ (to H. S. Choi)} \label{Choi1}

I have been informed that you will be visiting Perimeter Institute May 9 and will be scheduled to meet with me for part of that time.  I look forward to meeting you and telling you all about what Appleby and I plan to do with our new JTF grant.  I wish Appleby could be around to meet you as well---you would be impressed by his breadth of knowledge---but he will be in the UK at that time.

The reason I write you today, however, has to do with something completely independent of Perimeter Institute concerns.  My wife and I have been thinking for some years about how we might run a small research center (a place to mingle quantum physics, philosophy and theology and excite a synergy between them).  For the first time a clear-cut opportunity has arisen that might put our vision within reach.

This is because I happen to know that William James's home and property in Chocorua, New Hampshire will soon go on the market.  It is a home built in 1842, the very year James was born, and the home in which he died in 1910.  It is where he wrote the major part of his {\sl Principles of Psychology\/} and had debates with Josiah Royce and Charles Sanders Peirce and any number of other philosophers.  When James first discovered the property in 1886, he wrote to his brother Henry, ``It is only 4 hours from Boston by rail and 1 hour's drive from the station.  Few neighbors, but good ones; hotel a mile off.  If this is a dream, let me, at least indulge it a week or so longer!''

Well, I can tell you, the home remains every bit as dreamlike to this day, though now it is only a 2 hour 20 minute drive from Boston Logan airport.  My wife and I have made one trip to the property already and have inspected it in detail.  (My wife is an expert on old homes; please see the attached article on QBism House.)  I will make a second reconnaissance trip this Thursday to gather more information and photos of the surrounding forest and lakes.

The home is 4,400 sq feet in size and immaculate in condition; very well maintained with a brand new roof.  The house has an easy, elegant charm and flow to it conducive to all that is needed of a research station:  spaces for private work, areas for conversation, solitude for contemplation.   Some distance away from the house, in the yard, there is a rather large artist studio that James built for one of his sons---it could easily be converted into a conference room.

The property comes with 44 acres of land, mostly forest, and abuts a NH state forest.  Chocorua Lake is a two minute walk away.  The town of North Conway is a 15 minute drive (smaller villages in between) and has a variety of restaurants and cafes that puts our downtown Waterloo (where Perimeter Institute is situated) to shame.  The region is filled with hiking trails, ski slopes, lakes, opportunities for cycling, fishing, and much more.

The owners already had the home on the market last Fall, but took it off the market for the Winter.  They plan to put it back on the market again very soon, for Springtime viewing.  When they last listed it, they were asking \$895,000.

The vision I presently have for the research center is that it would provide workspace and interaction space for three postgraduate researchers (two year terms each)---one a physicist, one a philosopher, one a theologian---as well as the director (myself, whose own research spans the first two subjects).  These four researchers would all live off-site.  The bedrooms of the home would be used to house up to three visiting researchers at any one time, with stays ranging from 1 week to 3 months.  From time to time, conferences could also be hosted by the center.  A key component of the idea is to excite a synergy between the three subjects in a setting nearly perfect for their mixture.  What the James home in particular brings to the table is an automatic branding for the product.

As I wrote in the QBism House article, ``Ideas, like children, need homes where they can be loved, nurtured, and raised to independence.''

Anyway, this is all in the way of introduction.  When you visit Waterloo, I wonder whether I might be able to schedule some time with you outside of the Perimeter Institute setting to discuss all this?  The late afternoon or early evening May 8 or May 9 would be ideal from my point of view:  We could sit on the porch of QBism House if the weather cooperates.  It is an atmosphere congenial for discussions of this nature (see attached {\sl Maclean's Magazine\/} article discussing Perimeter Institute, QBism House, and William James, all in one go).

I could have a reasonably polished presentation prepared for you by then \ldots\ with photos, maps, a discussion of the other institutes that I would like to model this one upon (drawing elements from the Traunkirchen Academy, Schloss Dagstuhl, and the Stellenbosch Institute for Advanced Study, all of which I know intimately), a proposed advisory board, funding models, the deep connection between William James's views and Sir John Templeton's, etc., etc.

I realize that I am not approaching the JTF through the standard funding portal (nor during the standard funding timeline) by sending this note, but there is some urgency in me to try to remove the house from the market on fear of another (inevitable) buyer.  If I could get some guidance from you and the John Templeton Foundation, it would be invaluable to me.

To give you a bit more sense of where I'm coming from with the connections I see between William James's philosophy, John Wheeler's post-Everettian cosmology (that ``the Big Bang is here, all around us'' in moments of constant creation), and my Quantum Bayesian interpretation of quantum mechanics, please see the attached recent interview---particularly my answers to Questions 8, 13, 14, and 17. [See ``Interview with a Quantum Bayesian,'' \arxiv{1207.2141v1}.]

Finally I attach a current CV.  It speaks to some of my credentials for seeing a project like this through (for instance detailing my chairmanship of the 1,100-member APS topical group on quantum information, the many conferences I've organized, my recognition in the community through my citation record, etc., etc.).  [\ldots] \medskip

\noindent I apologize for this very long note, so please let me summarize:
\begin{itemize}
\item[1)]	The old William James house in Chocorua, NH will soon be on market.
\item[2)]	I have a vision to turn it into a research center for quantum physicists, philosophers, and theologians to live and learn from each other and make great discoveries on its grounds.
\item[3)]	The key point of the James house is that it comes with a ``brand'' that will be recognized by philosophers, theologians, psychologists, and many more beyond that the world around.
\item[4)]	Might we meet and discuss this in more detail outside your Perimeter Institute schedule during your visit to Waterloo?  I would appreciate any guidance the John Templeton Foundation could give me.
\end{itemize}

Thank you much for your time,

\section{11-05-11 \ \ {\it Thinking of You}\ \ \ (to H. J. Bernstein)} \label{Bernstein11}

I hope everything went well yesterday.  You were in my thoughts yesterday, as your ``realit{\it t\/}y'' has been in my thoughts for years.  Few understand how malleable the world is as you.  Keep making it happen.








\pagebreak


\pagebreak

\thispagestyle{empty}

\vspace*{6in}

\begin{flushright}
\parbox{2.9in}{
My work is purely autobiographical.  It is about myself and my surroundings.  It is an attempt at a record.  I work from the people that interest me, and that I care about and think about, in rooms that I live in and know.  I use the people to invent my pictures with, and I can work more freely when they are there. \\ \hspace*{\fill} --- {\it
Lucian Freud}}\hspace*{0.5in}
\end{flushright}

\end{document}